# A complete catalogue of High-Quality Topological Materials

*M. G. Vergniory*[1,2,3*]

[1]*Donostia International Physics Center,*

*P. Manuel de Lardizabal 4, 20018 Donostia-San Sebastian, Spain*

[2]*IKERBASQUE, Basque Foundation for Science*

*Maria Diaz de Haro 3, 48013 Bilbao, Spain and*

[3]*Applied Physics Department II, University of the Basque Country UPV/EHU,*

*Apartado 644, 48080 Bilbao, Spain*

*L. Elcoro*[1*]

[1]*Department of Condensed Matter Physics,*

*University of the Basque Country UPV/EHU, Apartado 644, 48080 Bilbao, Spain*

*Claudia Felser*[1]

[1]*Max Planck Institute for Chemical Physics of Solids, 01187 Dresden, Germany*

*Nicolas Regnault*[1]

[1]*Laboratoire de Physique de l'Ecole normale supérieure*

*PSL University, CNRS, Sorbonne Université, Université Paris Diderot,*

*Sorbonne Paris Cité, 24 rue Lhomond, 75005 Paris, France*

*B. Andrei Bernevig*[1,2,3†]

[1]*Department of Physics, Princeton University, Princeton, New Jersey 08544, USA*

[2]*Physics Department, Freie Universitat Berlin,*

*Arnimallee 14, 14195 Berlin, Germany and*

[3]*Max Planck Institute of Microstructure Physics, 06120 Halle, Germany*

*Zhijun Wang*[1,2‡]

[1]*Department of Physics, Princeton Univertiy, Princeton, New Jersey 08544, USA*

[2]*Beijing National Laboratory for Condensed Matter Physics,*

*and Institute of Physics, Chinese Academy of Sciences, Beijing 100190, China.*

[*] These authors contributed equally to this work.
[†] bernevig@princeton.edu
[‡] zjwang@princeton.edu

**Using a newly developed formalism called topological quantum chemistry, we perform a high-throughput search of high-quality materials in the Inorganic Crystal Structure Database. We develop several codes to compute all characters of all symmetries of the materials and to identify new topological phases. We find, out of 26,938 stoichiometric materials, 3,307 topological insulators, 4,078 topological semimetals and no fragile phases. For these 7,385 materials we provide the electronic band structure, including some electronic properties (band gap and number of electrons), symmetry indicators, and other topological information. Our results show that more than 27 per cent of all materials in nature are topological. We have added an open-source code to the Bilbao Crystallographic Server; this code checks the topology of any material and allows other researchers to reproduce our results.**

Topological materials have particular topological arrangements in the geometry of their electronic band structures, resulting in robust surface states and unconventional electromagnetic activity. But how many topological materials exist? What are they? How abundant are they? The remarkable field of topological insulators and semimetals has combined deep theoretical insights with almost immediate material predictions and their experimental discovery. Since the field began 13 years ago[1–3], the prevalent impression in the physics and chemistry communities has been that topological materials represent an infinitesimal proportion of the total number of materials existing in nature. Even though it has led to a comparatively large number of successes, the method of predicting new topological classes and corresponding materials has so far been based mostly on educated guesses. In this way, time-reversal[2,3], mirror-symmetric and nonsymmorphic topological insulators[4,5]; Dirac[6,7], Weyl[8–11] and nodal-chain[12] semimetals; new fermions[13] and many other states in realistic materials have been predicted. However, the deficiencies of the method are clear: only about 20 topological semimetals have been predicted and experimentally discovered so far. The problem underlying the efficient discovery of materials has been a missing link between the chemistry of different compounds and their topological properties.

In topological quantum chemistry (TQC)[14], a link has been established between the topology of a nonmagnetic material and its crystal symmetry, position and content of the orbitals. The basis of the approach is elementary band representations (EBRs)[15,16], which represent all of the 10,398 (with/without time reversal: 4,757/5,641) building blocks of bands that come from atomic orbitals, and the Brillouin zone compatibility relations[17,18], which provide rules for all of the ways in which electron bands can connect in the Brillouin zone. Both of these features are tabulated on the Bilbao Crystallographic Server (BCS; http://www.cryst.ehu.es), in program BANDREP[19] for EBRs, and program DCOMPREL[20] for the compatibility relations. If a set of valence bands in a material is not a sum of EBRs, then the material is topological; if a set of bands does not satisfy the compatibility relations, then they can not be separated from others and the material is a semimetal. Insulators satisfy the compatibility relations in the Brillouin zone; and if their valence bands cannot be written as a sum of EBRs, then they are topological. An insulator can differ from a sum of EBRs in several ways. The simplest way is that the characters of representations at a certain collection of high-symmetry points called maximal *k*-points cannot be written as a sum of characters of the EBRs in that group (to access the maximal *k*-vectors in the BCS website[19], choose a space group and ask for the elementary band rep (with or without time reversal); the first column gives the maximal *k*-vectors). These are the 'strong' topological phases, which usually exhibit surface or hinge states[22–26]. There are also finer aspects of topology that TQC uniquely captures fully, namely the 'fragile' topological phases, whereby a set of bands can be written as a difference plus a sum (hereafter a 'difference' refers to 'a difference plus a sum') but not purely as a sum of EBRs[15,27–29], with further refinement of topology as in ref. [30]. TQC can so far depict all of the topological phases that exist in nature, at very fine levels of topological refinement. In the original paper, Bradlyn et al.[14] used TQC to predict a set of about 150 new topological materials by targeting groups with a lot of 'disconnected' (or topological) EBRs. We will show how to use TQC to predict new materials in ref. [31]. Using the EBR data of TQC[14], refs. [32,33] obtain a set of strong topological indices that exhaust the strong topological phases[14,34]. These strong indices are equivalent to the EBR equivalence classes defined here.

In this paper, we present a large list of (strong) topological (nonmagnetic) 'high-quality' materials in nature, along with a wide array of information about their structure, bands and topological properties. This extensive work is composed of several research modules, presented in Fig. 1, each of which is essential for finding materials with strong and fragile topology. We outline the steps and results here, and give further details in the Supplementary Information.

#### Computational approach for finding topological materials

In step A, we wrote a code called VASP2Trace to calculate the characters of all symmetries at all the high-symmetry points of ab initio VASP (Vienna ab initio simulation package)[35] wavefunctions. This information did not previously exist for nonsymmorphic symmetries. Details are given in Supplementary Information appendix A, and the code is made available on the BCS[36]. In step B, we scanned the Inorganic Crystal Structure Database (ICSD; http://www2.fiz-

karlsruhe.de/icsd_home.html) of 184,270 materials, and removed bad or poor-data compounds by applying the ICSD's 'high-quality' filter (for materials whose atomic positions and structure are measured very accurately). This filter picked up 89,240 'high-quality' materials, as follows: hexagonal, 16,974; monoclinic-triclinic, 23,355; orthorhombic, 22,430; tetragonal, 13,631; and cubic, 12,850. We then applied the 'stoichiometric', 'few atoms (<30)', 'not alloy' and 'not-usual magnetic atoms' filters to find the total number of stoichiometric compounds: 26,938. We then separated these compounds into 'problematic' *f*-electron compounds (4,286) and 22,652 high-quality stoichiometric, nonproblematic *f*-electron ones. We detail the procedure and the filters in Supplementary Information appendix B.

In step C, we used VASP and the code developed in step A to obtain the characters of all bands at all the relevant high-symmetry points (the maximal *k*-vectors) for all of the compounds found in step B. The maximal *k*-vectors are the set of high-symmetry points for which the determination of band characters implies their knowledge everywhere in the Brillouin zone. Details about the set of maximal *k*-vectors (which can be found by using the BCS[19], as described above), VASP computational parameters, grid mesh and other benchmarks are available in Supplementary Information appendix C. In step D, we wrote a code to check whether a set of characters at the relevant high-symmetry points satisfies the compatibility conditions[17,18] in the Brillouin zone. Furthermore, we wrote another code to check whether—based purely on the characters at high-symmetry points—a set of bands could be expressed as a linear combination of the characters of EBRs, with positive and, crucially, negative coefficients (for finding fragile topology). Self-contained details about these codes, the compatibility relations and our approach are provided in Supplementary Information appendix D.

In step E, we used our codes and the VASP ab initio results to check, for each material, whether the characters obtained in step C satisfy the compatibility relations. If they did not, they are enforced semimetals (ESs, the prototypical example being sodium bismuth[7], $Na_3Bi$). We further separated the enforced semimetals depending on whether they have a high-symmetry point degeneracy at the Fermi level (enforced semimetals with Fermi degeneracy (ESFDs), the prototypical example being bulk mercury telluride, HgTe) or not. If the characters obtained for a given material from step C satisfy the compatibility relations, it means that an insulating filling is possible. We then checked whether the band characters could be written as a sum, or a difference, of the EBRs in the space group. If they could not be written as either a sum or a difference, they are necessarily strong topological. If they could not be written as a sum but could be written as a difference, they are fragile topological[15,27,28]. We then further distinguished the band topology on the basis of whether the band characters could be written as sums of parts of split EBRs[14]. If they could be written as a sum of split (disconnected) EBRs, the materials are denoted SEBRs (the prototypical example being the large spin–orbit coupling (SOC) material graphene). If the band characters could not be written as a sum of split EBRs but were still not a sum or a difference of EBRs, they are denoted more generically 'no linear combination' (NLC) of EBRs (the prototypical compound being calcium arsenide[37], $CaAs_3$).

This partitioning of topological materials into ES, ESFD, SEBR or NLC is based on physical consequences. The ESFDs can become fully insulating when a symmetry-breaking perturbation is applied to the semimetal, and we analyse in step J whether such an insulator is topological or not. In HgTe, upon symmetry breaking, the system becomes a topological insulator. The SEBR insulators are physically similar to graphene: the bands directly below and above the Fermi level, when taken together, form one EBR. In this case, the gap within an EBR has to be topological. The NLC-type materials are physically similar to $CaAs_3$: the bands below and above the Fermi level together form two EBRs, with neither the valence nor the conduction branch being a part of a split EBR. Hence, NLC compounds can be tuned, by slightly varying some parameters, between insulating trivial and insulating nontrivial topological phases; SEBRs can be tuned either to topologically nontrivial insulators or enforced semimetals. In Supplementary Information appendices D and E, we present the details of our materials classification into the 'physical classes' ES, ESFD, SEBR and NLC. In appendix E, Table I, we present the number of materials, per space group, for each of the classes we define. In Table 1 of this main text, we provide the number of 'high-quality' topological insulators or semimetals per space group.

**Physical and topological characterization**

After finding all the topological materials in our high-quality sample, we moved on to their physical and topological characterization. In step F, we indexed all of the topological materials found in our search by a TQC number (TQC # = SG.$y$.SN#, where SG refers to the space-group number; $y$ = 1, 2, 3 or 4 refers to SEBR, NLC, ES or ESFD, respectively; and SN refers to the serial number of the material) for future reference in the entire paper (see Supplementary Information, Table XIV). For each material we provide the following information, in tables. One set of tables (Supplementary Tables XIV-XVI) contains the material's: F1, formula; F2, direct gap; F3, indirect gap; and F4, general class characterization into ES, ESFD, SEBR or NLC. We also provide a set of tables with (F5) all of the representations of the bands below the Fermi level at all of the maximal *k*-points, for each compound; we do this so that the public can perform independent checks. This information is presented in Supplementary Information section F.

In step G, for each of the topological insulating compounds from step E, we provide the symmetry(ies) that stabilizes the compound. Columns 6 and 7 of Table XIV (Supplementary Information) list: (1), the minimal topological space subgroup of the compound's space group for which the material is still topological; and (2), the maximal trivial space subgroup of the compound's space group for which the material becomes topologically trivial. Lowering the symmetry of the compound to the space subgroup in (2) renders it trivial; the difference in symmetries between the minimal topological subgroup and the maximal trivial subgroup protects the topological nature of the material.

In step H, for each of the topological insulators from step E, in each space group, we separate the compounds into topological classes: two compounds whose band structures are related by addition/subtraction of characters of EBRs are equivalent in strong topology. The EBR equivalence classes are identical to the topological indices in refs. [32,34]. An equivalent set of indices can be obtained by performing a Smith decomposition on the EBR matrix. We pair up the topological com- pounds into these classes and provide the information in columns 6 and 7 of Supplementary Table XIV (also in Supplementary Tables XV and XVI, with and without time-reversal symmetry respectively). In appendices G and H we propose that the topological indices should be understood physically and mathematically as equivalence classes of compounds modulo EBRs, which represents a physical picture of the Smith decomposition of the EBR matrix.

For all enforced semimetals, we provide a different set of relevant information in appendix I of the Supplementary Information. For the enforced semimetals that we found in step E, we provide the largest subgroup of their space group for which the bands below the Fermi level satisfy the compatibility relations and hence allow an insulating filling. For this insulating filling, we check whether the material is topological by performing the test mentioned in step E for insulators. We then provide the topological classifications of these new insulators obtained by applying symmetry-breaking perturbations to the enforced semimetals.

In step J, for the ESFDs found in step E, we provide the largest subgroup of their space group for which the material satisfies the compatibility relations and allows an insulating filling (Supplementary Information, Table XIX). Given that in ESFDs the degeneracy is at the Fermi level, we have several options for how it splits in the subgroup. The former irreducible representation (irrep) becomes reducible into irreps of the subgroup: one has to choose which irrep moves above or below the Fermi level. The material might also have Fermi degeneracy at other points in the Brillouin zone (for example, related by a symmetry in the group, but not in the lower subgroup), resulting in a mesh of different insulators. For each possibility of an insulating filling, we check whether the material is topological by performing the test in step E for insulators. Appendices I and J of the Supplementary Information give all ways in which semimetals can become topological insulators upon lowering the symmetry. This has practical consequences: this information provides experimentalists with the dials to turn in order to turn semimetals into topological insulators. The celebrated bulk HgTe is an example of this phenomenon: it is an ESFD (with fourfold degeneracy at the Fermi level); however, when its symmetry is broken, a gap opens at the Fermi level to produce a three-dimensional topological insulator.

**Computing the electronic band structures**

Having enumerated the topological properties of insulating and semimetallic band structures, for all the compounds from step E we obtain and plot the electronic band structure in the Brillouin zone (step K). We provide benchmark parameters for the computation time on the Cobra supercomputer. This gargantuan task allows for determination of the 'cleanest' topological insulators and semimetals, and is beneficial to experimentalists looking for materials to grow. The band structures are presented in appendix K.

In step L, for an infinitesimal proportion of some novel topological compounds (HgS, $GeTa_3$, $K_2Mg_5Sn_3$, $Na_2OSb_2Ti_2$, LaI and PbSe, whose nontrivial topology is not protected by inversion symmetry), we perform an in-depth analysis that includes Wilson loops and other holonomy arguments in order to reconfirm the topological nature of the materials. For the first time (to our knowledge), we find roto-inversion topological insulator materials in almost all topological classes. We remind the reader of the generic Wilson-loop procedure for obtaining topological invariants in Supplementary Information appendix L. In Fig. 2 we present a very small set of materials; details can be found in appendix L. In the Supplementary Information we also show examples of materials whose topology is described by all symmetry indicators in Table VI and give their band structures in Figs 8–11.

In appendix M of the Supplementary Information, for some 4,286 of the compounds that were discarded by our initial filters—such as materials containing 'problematic' *f*-electrons—we present very short tables/summaries, but caution against trusting the ab initio calculations in these cases. For step N, we have added a page on the BCS

(http://www.cryst.ehu.es/cryst/checktopologicalmat) that allows the user to check whether a material is topological. We provide a short description of all these functionalities in appendix N.

Our EBR method is uniquely suitable to check for the existence of fragile topological phases, which are undetectable by strong indices. We show in appendix O of the Supplementary Information that none of the 26,938 materials checked exhibit fragile topology. Next, in step P, we compute several well known indices of well known insulators such as topological insulators and higher-order topological insulators, and present the identified compounds in column 10 of Table XIV (Supplementary Information). Alternatively, a topological index is a distinct EBR-equivalent class: two bands are considered in the same class of strong topology if they differ only by a sum or difference of EBRs. This is also computed in step P and is identical to the more well known[32,33,38] topological index shown in appendix Q. In step Q, we compute all of the topological indices[32,33,38] of all bands (satisfying the compatibility relations) and show perfect matching with our EBR equivalence class classification (see appendix Q, with a concise version of this shown here in Table 2).

In appendix R of the Supplementary Information, we provide (for most materials) chemistry-based comments on the difficulty of growing the compound, its air stability, and possible Mott or magnetic tendencies. In appendix S we provide some 'low-quality' ICSD materials that are topological, and a list of checks performed on both our materials and our topological indices and classification in order to guarantee the accuracy of our data. We have crosschecked some of our compounds with existing literature. Finally, in appendix T we have compiled a large list of about 200 clean/good band structures (for example, a full gap across the Brillouin zone) from our 7,385 topological materials.

**Discussion**

We have found that roughly 27% of the materials in the world are topological. Roughly 12% are topological insulators. With these findings, we enter an important era of topological material design. One direction will be to compute and discover the physical properties of our large set of materials. Slab calculations should be performed for all of our topological insulators, to reveal their surface states. Compounds whose topological character is protected by symmetries that are locally preserved by specific surfaces probably exhibit specific states on those surfaces. Compounds whose topological character is protected by symmetries that are not locally preserved by any specific surface should exhibit hinge states. Rod calculations should be performed for these compounds (for example, bismuth[26]). Wilson-loop calculations should be performed for all of the compounds herein. The *d*/*f*-electron compounds that we have predicted should be analysed by more accurate dynamical mean-field theory codes, which properly take into account interactions; these codes can potentially give us many new interacting materials with strong topology. Another step will be to analyse the full ICSD database, and not only the 'high-quality' compounds. Given that we have shown more than one-quarter of all materials to be topological, we are guaranteed to find many others in the ICSD database.

### Note added

When finalizing this paper, we became aware of three recent preprints that similarly diagnose topological materials[39–41], covering respectively one, five or eight out of the 230 space groups in their searches.

**Data Availability** All the data is available in the Supplementary Information.

**Online content** Any methods, additional references, Nature Research reporting summaries, source data, statements of data availability and associated accession codes are available at https://doi.org/10.1038/s41586-019-0954-4.

**Acknowledgments** We thank B. Bradlyn, J. Cano and M. Aroyo for countless discussions and collaborations, and for their help with the development of the BANDREP section of the BCS, without which none of the present work would have been possible. We thank H. Gross, S. Parkin, U. Schmidt, M. Rampp and the computational resources of the Max Planck Institute at Halle and Garching, as well as the staff at the Atlas supercomputer of the Donostia International Physics Center. We are grateful to H. Lederer and I. Weidl for allowing us access to the Cobra Supercomputer at the Max Planck Gesellschaft (MPG) computing centre. We also thank H. Borrmann of the Max Planck Institute in Dresden for help with the ICSD database. L.E. was supported by the Government of the Basque Country (project IT779-13), the Spanish Ministry of Economy and Competitiveness (MINECO), and the European Fund for Economic and Regional Development (FEDER; project MAT2015-66441-P). M.G.V. was supported by national project IS2016-75862-P of the Spanish MINECO. B.A.B. and Z.W. acknowledge support for the analytical work from the Department of Energy (de-sc0016239), a Simons Investigator Award, the Packard Foundation, and the Schmidt Fund for Innovative Research. Z.W. was also supported by the CAS Pioneer Hundred Talents Program. The computational part of the Princeton work was performed under National Science Foundation (NSF) Early-concept Grants for Exploratory Research (EAGER): DMR 1643312 NOA-AWD1004957, ONR-N00014-14-1-0330, ARO MURI W911NF-12-1-0461, NSF-MRSECDMR-1420541.

**Reviewer information** *Nature* thanks J. Checkelsky, M. Franz and the other anonymous reviewer(s) for their contribution to the peer review of this work.

**Author Contributions** All authors contributed to performing the calculations and the analysis of the results.

**Competing interests** The authors declare no competing interests.

Additional information

**Supplementary Information** is available for this paper at https://doi.org/10.1038/s41586-019-0954-4.

**Reprints and permissions information** is available at http://www.nature.com/reprints.

**Correspondence and requests for materials** should be addressed to B.A.B. or Z.W.

**Publisher's note:** Springer Nature remains neutral with regard to jurisdictional claims in published maps and institutional affiliations.

**FIG. 1**: **Workflow diagram.** The figure shows the workflow diagram of the many different codes and calculations implemented in the current material search. The open-source code "*VASP2Trace*" and end-user button "*CheckTopologicalMat*" are available online: www.cryst.ehu.es/cryst/checktopologicalmat, and can check the topology of any material. ES, enforced semimetal; TI, topological insulator.

**FIG. 2**: **Band structures of some novel topological compounds**. They are $GeTa_3$ (a), HgS (b), $K_2Mg_5Sn_3$ (c), $Na_2OSb_2Ti_2$ (d), LaI (e), and PbSe (f), whose nontrivial topology is not protected by inversion symmetry. $GeTa_3$ and HgS are $S_4$-protected TIs, which don't have inversion symmetry. The others have inversion symmetry, but their inversion indicators (i.e., z2w and z4 as defined in Ref.[33]) are zeros (see TABLE IV in SI). In fact, the nontrivial topology is protected by rotational symmetries. Their nontrivial topology is further confirmed by the winding of the Wilson loops (see more details in SI (L)).

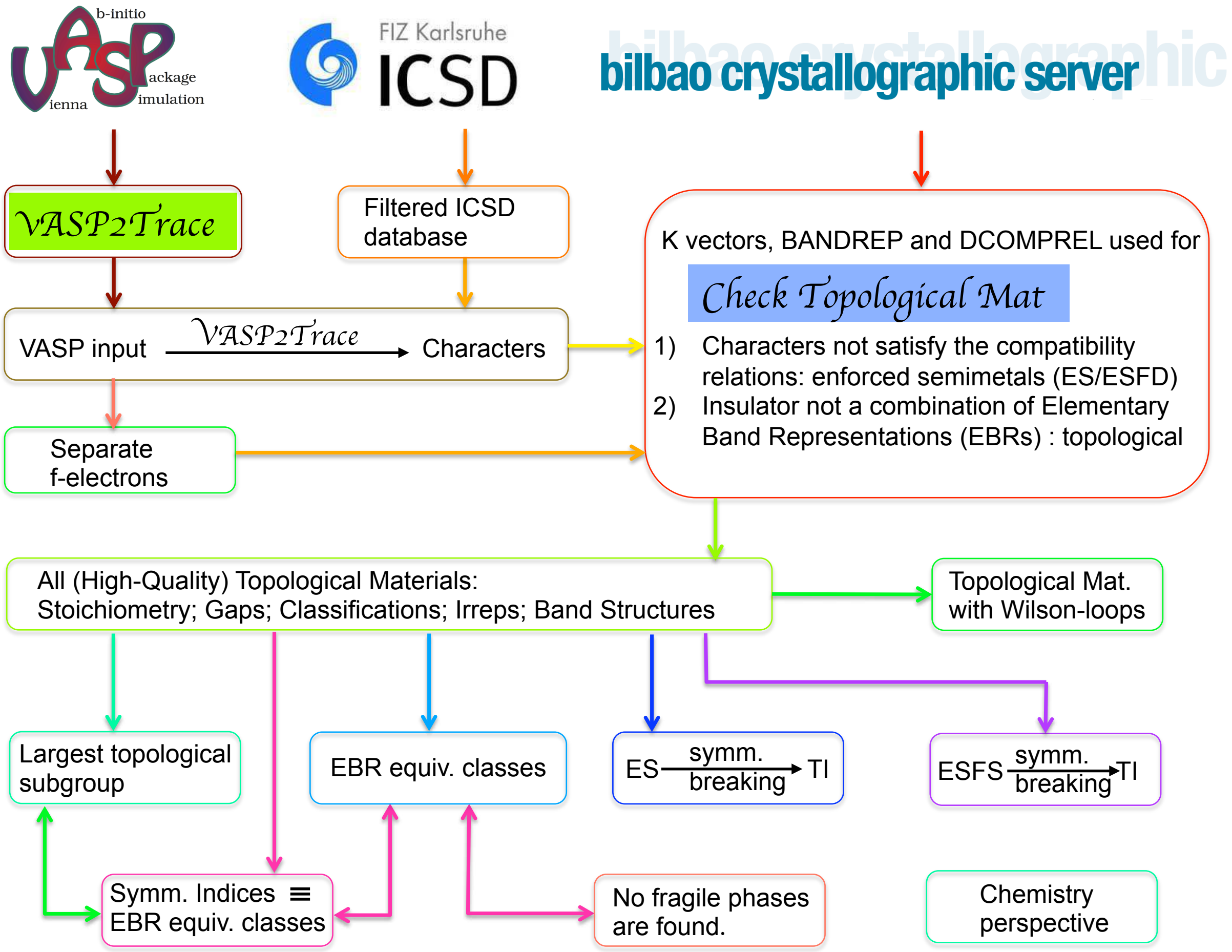

**FIG. 1**: **Workflow diagram**

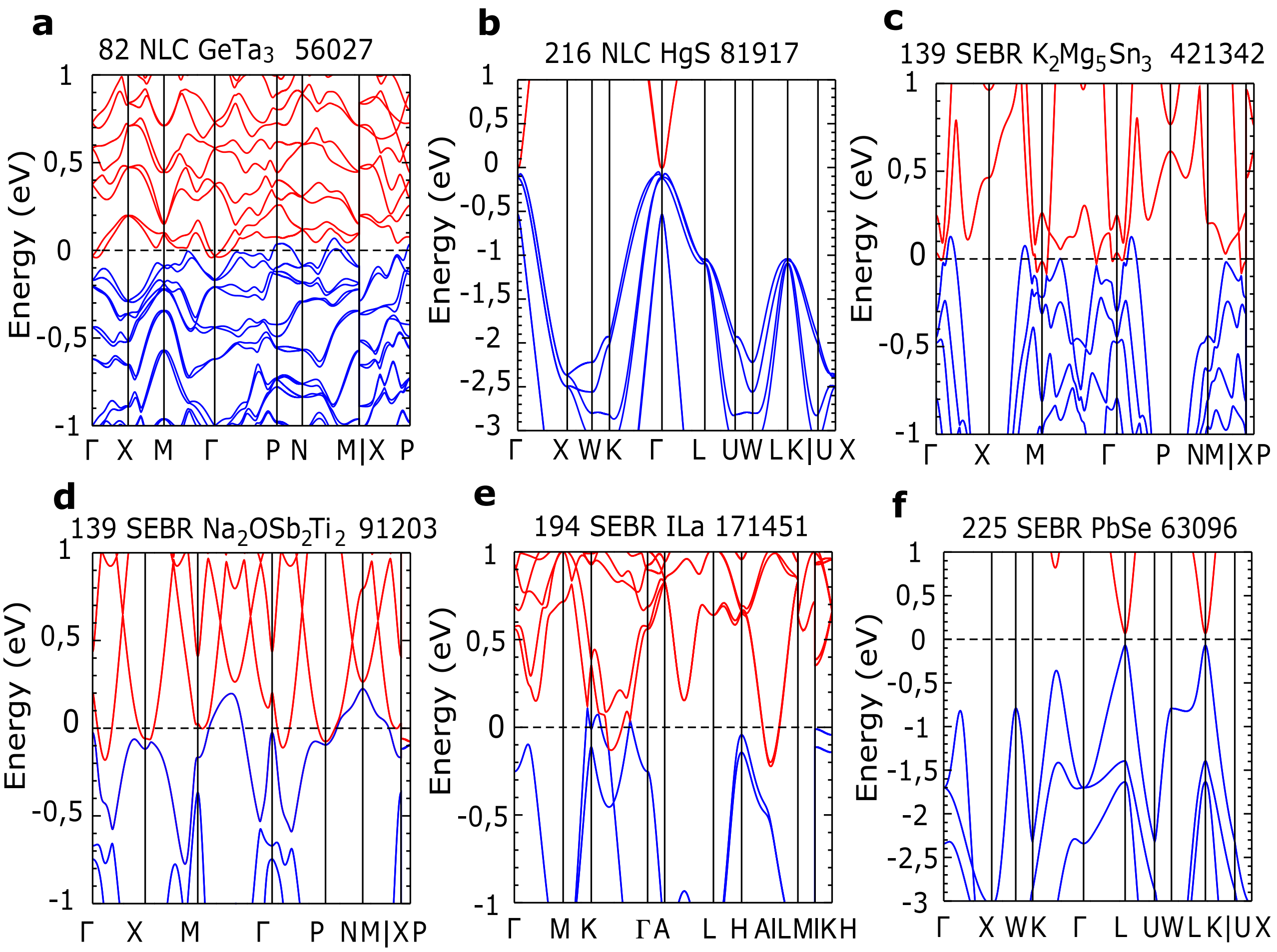

**FIG. 2**: **Band structures of some novel topological compounds**

**TABLE 1. Numbers of topological materials per SG.** The numbers of Topological Insulators (TIs) and enforced metals or semimetals (SM), at the Fermi level, per SG. TIs consist of both SEBR and NLC, while SMs consist of both ESs and ESFDs (See the number for each classes per SG in Table I of SI).

| SG | TIs | SM | SG | TIs | SM | SG | TIs | SM | SG | TIs | SM | SG | TIs | SM | SG | TIs | SM | SG | TIs | SM |
|---|---|---|---|---|---|---|---|---|---|---|---|---|---|---|---|---|---|---|---|---|
| 2 | 101 | 0 | 47 | 13 | 0 | 74 | 25 | 19 | 115 | 5 | 0 | 142 | 0 | 2 | 173 | 0 | 11 | 204 | 24 | 55 |
| 4 | 0 | 8 | 51 | 17 | 30 | 78 | 0 | 1 | 120 | 0 | 1 | 143 | 0 | 1 | 174 | 0 | 1 | 205 | 6 | 11 |
| 7 | 0 | 3 | 52 | 0 | 4 | 79 | 0 | 1 | 121 | 9 | 0 | 145 | 0 | 1 | 176 | 2 | 29 | 212 | 0 | 4 |
| 9 | 0 | 8 | 53 | 2 | 0 | 82 | 8 | 0 | 122 | 15 | 6 | 146 | 0 | 5 | 180 | 0 | 4 | 213 | 0 | 7 |
| 10 | 1 | 0 | 55 | 63 | 18 | 83 | 1 | 3 | 123 | 55 | 24 | 147 | 2 | 1 | 181 | 0 | 1 | 216 | 8 | 29 |
| 11 | 98 | 101 | 57 | 9 | 8 | 84 | 1 | 2 | 124 | 0 | 7 | 148 | 54 | 11 | 182 | 0 | 8 | 217 | 7 | 12 |
| 12 | 390 | 0 | 58 | 12 | 17 | 85 | 1 | 8 | 125 | 5 | 15 | 150 | 0 | 1 | 185 | 0 | 12 | 220 | 4 | 16 |
| 13 | 13 | 29 | 59 | 33 | 46 | 86 | 0 | 2 | 126 | 0 | 2 | 152 | 0 | 1 | 186 | 0 | 115 | 221 | 14 | 70 |
| 14 | 71 | 87 | 60 | 7 | 2 | 87 | 11 | 10 | 127 | 37 | 112 | 155 | 0 | 1 | 187 | 0 | 14 | 223 | 0 | 13 |
| 15 | 95 | 169 | 61 | 2 | 3 | 88 | 0 | 8 | 128 | 1 | 3 | 156 | 0 | 5 | 188 | 0 | 2 | 224 | 0 | 2 |
| 19 | 0 | 15 | 62 | 257 | 308 | 92 | 0 | 9 | 129 | 177 | 433 | 159 | 0 | 5 | 189 | 44 | 47 | 225 | 30 | 71 |
| 20 | 0 | 1 | 63 | 223 | 230 | 97 | 0 | 1 | 130 | 0 | 10 | 160 | 0 | 10 | 190 | 2 | 23 | 226 | 0 | 19 |
| 26 | 0 | 3 | 64 | 26 | 30 | 99 | 0 | 3 | 131 | 2 | 1 | 161 | 0 | 34 | 191 | 20 | 44 | 227 | 88 | 172 |
| 29 | 0 | 1 | 65 | 57 | 0 | 102 | 0 | 2 | 135 | 0 | 11 | 162 | 5 | 7 | 193 | 8 | 160 | 229 | 1 | 23 |
| 31 | 0 | 8 | 67 | 1 | 2 | 103 | 0 | 3 | 136 | 14 | 73 | 163 | 2 | 6 | 194 | 57 | 132 | Total | 3307 | 4078 |
| 33 | 0 | 21 | 69 | 18 | 0 | 106 | 0 | 1 | 137 | 3 | 2 | 164 | 102 | 48 | 197 | 0 | 2 | | | |
| 36 | 0 | 15 | 70 | 23 | 5 | 107 | 0 | 4 | 139 | 251 | 99 | 165 | 2 | 3 | 198 | 0 | 34 | | | |
| 40 | 0 | 3 | 71 | 105 | 0 | 109 | 0 | 4 | 140 | 28 | 144 | 166 | 410 | 209 | 199 | 0 | 4 | | | |
| 43 | 0 | 4 | 72 | 9 | 31 | 113 | 4 | 27 | 141 | 24 | 77 | 167 | 92 | 240 | 202 | 0 | 2 | | | |

**TABLE 2. Numbers of materials for different topological indices.** Column 1: Topological Index of [33] or EBR Equivalence Class - see SI (H) and SI (Q). Column 2: Total number of the topological compounds (per space group) for that index.

| Indices | Number of compound per space group |
|---|---|
| z2w | 87(2), 61(11), 298(12), 9(13), 42(14), 37(15), 10(47), 13(51), 1(53), 38(55), 5(57), 21(59), 122(63), 21(64), 47(65), 1(67), 15(69), 73(71), 6(72), 21(74), 1(83), 1(84), 7(87), 27(123), 5(125), 26(127), 117(129), 1(131), 122(139), 13(140), 1(147), 32(148), 2(162), 63(164), 215(166), 10(191), 13(204), 9(221) |
| z4 | 79(2), 1(10), 83(11), 315(12), 10(13), 62(14), 82(15), 10(47), 15(51), 2(53), 53(55), 7(57), 12(58), 26(59), 7(60), 2(61), 257(62), 189(63), 25(64), 47(65), 1(67), 18(69), 23(70), 86(71), 9(72), 20(74), 1(83), 1(85), 8(87), 42(123), 1(125), 30(127), 1(128), 162(129), 2(131), 14(136), 3(137), 140(139), 21(140), 24(141), 2(147), 50(148), 5(162), 2(163), 86(164), 2(165), 378(166), 92(167), 1(176), 13(191), 7(193), 49(194), 15(204), 6(205), 12(221), 16(225), 88(227), 1(229) |
| z2 | 8(82), 1(83), 1(85), 8(87), 4(113), 5(115), 9(121), 15(122), 31(123), 1(125), 25(127), 84(129), 2(131), 10(136), 2(137), 89(139), 14(140), 20(141), 8(216), 7(217), 4(220), 7(221), 10(225), 76(227), 1(229) |
| z4m,pi | 1(83), 36(123), 31(127), 9(221) |
| z8 | 1(83), 11(87), 53(123), 31(127), 1(128), 233(139), 27(140), 12(221), 30(225), 1(229) |
| z3m,0 | 42(189), 2(190) |
| z3m,pi | 11(189) |
| z6m,0 | 2(176), 19(191), 8(193), 54(194) |
| z6m,pi | 16(191) |
| z12 | 17(191) |
| z12’ | 2(176), 8(193), 57(194) |

## CONTENTS

## CONTENT OF TABLES AND FIGURES

## Appendix A: Traces (Eigenvalues) of Operators In VASP and A New Code: VASP2Trace

A general symmetry operator is defined as $\mathcal{O} = \{R|\vec{t}\}$ where $R$ is a symmetry operation (mirror or roto-inversion), while $\vec{t}$ is a translation vector, which, in non-symmorphic groups can take fractional values of the lattice constant. The all-electron Blöch wave-functions ($\psi_{n\mathbf{k}}(\vec{r}) \equiv \langle r|\psi_{n\mathbf{k}}\rangle$) for a **k**-point are obtained from the DFT package (*i.e.*, VASP). An operator acting on the real space function $f(\vec{r})$ is defined as $\mathcal{O}f(\vec{r}) = f(\mathcal{O}^{-1}\vec{r}) = f(R^{-1}\vec{r} - \vec{t})$. As thus, the eigenvalue (trace) of the operator $\mathcal{O}$ of a given (set of) band(s) is computed as:

$$\chi_{\mathcal{O}}^{\mathbf{k}} = \sum_n \langle\psi_{n\mathbf{k}}|\mathcal{O}|\psi_{n\mathbf{k}}\rangle = \sum_n \int d\vec{r}\psi^*_{n\mathbf{k}}(\vec{r})\mathcal{O}\psi_{n\mathbf{k}}(\vec{r}) \tag{A1}$$

This trace is computed at high symmetry points, where the rotational part of the operator satisfies the relation $R\mathbf{k} = \mathbf{k} + \vec{G}$ where $\vec{G}$ could be any reciprocal vector in the BZ.

The all-electron wave-function is obtained by acting a linear operator $\mathcal{T}$ on the pseudo wave-function: $|\psi_{n\mathbf{k}}\rangle = \mathcal{T}\left|\tilde{\psi}_{n\mathbf{k}}\right\rangle$. The linear operator can be written explicitly as: $\mathcal{T} = \mathbf{1} + \sum_i \left(|\phi_i\rangle - \left|\tilde{\phi}_i\right\rangle\right)\langle p_i|$, where $|\phi_i\rangle$ ($\left|\tilde{\phi}_i\right\rangle$) is a set of all-electron (pseudo) partial waves around each atom and $|p_i\rangle$ is a set of corresponding projector functions on each atom within an augmentation region ($r < R_0$), where $R_0$ is the core part for each atom. In the VASP code, the pseudo wavefunction is expanded in the plane waves:

$$\tilde{\psi}_{n\mathbf{k}}(\vec{r}) \equiv \left\langle r|\tilde{\psi}_{n\mathbf{k}}\right\rangle = \sum_{\vec{G}} C_{n,\mathbf{k}+\vec{G}} e^{i(\mathbf{k}+\vec{G})\cdot\vec{r}} \tag{A2}$$

where the number of the $\vec{G}$ vectors is determined by the condition $\frac{\hbar^2}{2m_e}(\mathbf{k}+\vec{G})^2 < E_{\text{cutoff}}$ with a cutoff $E_{\text{cutoff}}$. Note that $|\phi_i\rangle$ is exactly the same as $\left|\tilde{\phi}_i\right\rangle$ outside the augmentation region ($r > R_0$), and $|\psi_{n\mathbf{k}}\rangle$ is also the same as $\left|\tilde{\psi}_{n\mathbf{k}}\right\rangle$ outside the spheres ($r > R_0$).

Because $\psi_{n\mathbf{k}}$ can be chosen to be an eigenstate of $\mathcal{O}$ in the $k$-little group and $\mathcal{T}$ is a linear operator, $\langle\psi_{n\mathbf{k}}|\mathcal{O}|\psi_{n\mathbf{k}}\rangle = \frac{\langle\psi_{n\mathbf{k}}|\mathcal{O}|\psi_{n\mathbf{k}}\rangle}{\langle\psi_{n\mathbf{k}}|\psi_{n\mathbf{k}}\rangle} = \frac{\langle\mathcal{T}\tilde{\psi}_{n\mathbf{k}}|\mathcal{O}|\mathcal{T}\tilde{\psi}_{n\mathbf{k}}\rangle}{\langle\mathcal{T}\tilde{\psi}_{n\mathbf{k}}|\mathcal{T}\tilde{\psi}_{n\mathbf{k}}\rangle} = \frac{\langle\tilde{\psi}_{n\mathbf{k}}|\mathcal{O}|\tilde{\psi}_{n\mathbf{k}}\rangle}{\langle\tilde{\psi}_{n\mathbf{k}}|\tilde{\psi}_{n\mathbf{k}}\rangle}$, where $\langle\psi_{n\mathbf{k}}|\psi_{n\mathbf{k}}\rangle = 1$ and $\left\langle\tilde{\psi}_{n\mathbf{k}}|\tilde{\psi}_{n\mathbf{k}}\right\rangle \neq 1$. By using this substitution, the Eq. (A1) can be rewritten as :

$$\chi_{\mathcal{O}}^{\mathbf{k}} = \sum_n \frac{\left\langle\tilde{\psi}_{n\mathbf{k}}|\mathcal{O}|\tilde{\psi}_{n\mathbf{k}}\right\rangle}{\left\langle\tilde{\psi}_{n\mathbf{k}}|\tilde{\psi}_{n\mathbf{k}}\right\rangle} = \sum_n \frac{\int d\vec{r}\tilde{\psi}^*_{n\mathbf{k}}(\vec{r})\mathcal{O}\tilde{\psi}_{n\mathbf{k}}(\vec{r})}{\int d\vec{r}\tilde{\psi}^*_{n\mathbf{k}}(\vec{r})\tilde{\psi}_{n\mathbf{k}}(\vec{r})} \tag{A3}$$

The Eq. (A3) can substantially simplify the calculations for the trace of the operations and saves a lot of time compared to the step Eq(A1). The coefficients of pseudo wavefunctions of Eq. (A2) are output in the "WAVECAR" file from the VASP code.

We use VASP to output the pseudo wave-functions of *ab-initio* calculations and wrote a code, called *VASP2Trace* to implement Eq. (A3) to calculate the traces for every symmetry operator. Due to another code we wrote, the symmetry operator can be expressed in any basis. Taking the symmetry operations given, the program IDENTIFY GROUP [2] identifies the symmetry group and the transformation matrix to the standard setting used in the Bilbao Crystallographic server. Of course, the coordinates of the $k$ vectors must be correctly given. These codes are now available for download at (www.cryst.ehu.es/cryst/checktopologicalmat).

We then wrote another code, which we called "Check Topological Mat" and is now available on the BCS, to compare the computed traces from *ab-initio* to the character tables in the BCS, in order to assign the traces of all the symmetry operators to different irreps at a set of $k$ points in the BZ defined in Appendix[C]. This means that for every compound we are able to obtain the set of characters of any of the bands at any point in the BZ under any symmetry operation that leaves that point invariant. A set of ($10 \sim 400$) bands are taken into account for every compound, depending on valence electrons of each element.

## Appendix B: Pre-Filtering of the ICSD Database

The total number of entries in the ICSD data base is 199466. We pre-processed the ICSD database in text file in order to obtain reliable results. An original set of filters had been applied before we started processing the text file of the materials. While usually giving correct results, these filters also removed some small set of well-known topological materials which we then added by hand. They are listed in Table[XII] in Appendix[S]. The upshot, however, is that we have a set of materials whose structure and chemical composition that is certain to be correct.

As a first filter applied to the ICSD data base, we only selected entries that were marked as "high quality" in the database. The high quality data contains structure determination including refinement (in case of powder data including Rietveld refinement), temperature factors, pressure in the range of 0.09-0.11 MPa, temperature within the range of 285-300 K and standard deviation given for cell parameters. These are basically compounds whose crystallographic structure has been well-determined. This filter also removes repetitions of compounds unless the difference were relevant (different temperature or pressure). This filter picked up 89240 compounds.

Second, amongst this selection, we then discarded compounds if any of their atoms had an occupancy different from one (alloys or non-stoichiometric compounds). The alloys require relaxation calculations and their results are likely to be far less reliable than the ones for stoichiometric compounds. Third, we have not consider compounds with more that 30 atoms in their primitive unit cell: the large unit cell again makes *ab-initio* calculations unreliable. Fourth, since we use a single-particle formalism, and in order to perform a reliable search, we pre-select from the ICSD database only materials with weak interactions. We discard some problematic $f$-electron atoms (such as Pu, Lu, Tm, Er, Pr, Nd, Pm, Eu, Gd, Tb, Dy) as well as magnetic compounds and atoms but we keep some less-problematic $f$-electron atoms (such as Ce,Cs,La,Y,Yb,Sm and U). We then took special care of a specific set of $f$-electron materials (containing Ce,Cs,La,Y,Yb,Sm and U) by performing longer, converged, *ab-initio* calculations on them. This gives us a set of 26938 "good" compounds on which we perform *ab-initio* calculations; There are 22652 non-$f$-electron compounds. The further set of 4286 $f$-electron materials was also analyzed, but in less detail. Out of all these we found 3307 topological insulators (TIs) and 4078 topological semimetals.

### Appendix C: Maximal $k$ Vectors and VASP Computational Parameters: Obtaining the Symmetry Characters of Bands

The list of sufficient and necessary $k$ points where the characters need to be computed in order to obtain a complete picture of the band characters in the whole BZ are called maximal $\mathbf{k}$ vectors, can differ from space group to space group but remain the same within a symmetry group, and can be found at [3] or at the header of our irrep tables for materials in every symmetry group in Appendix[F].

Calculations of the wave-functions and energies at maximal $\mathbf{k}$-vectors have been performed within the DFT[4, 5] as implemented in the Vienna Ab-initio Simulation Package (VASP)[6, 7]. The interaction between ion cores and valence electrons was treated by the projector augmented-wave method[8], the generalized gradient approximation (GGA) for the exchange-correlation potential with the Perdew-Burke-Ernkzerhof for solids parametrization [9] and the spin-orbit coupling (SOC) was taken into account by the second variation method[10]. A Monkhorst-Pack $k$-point grid of (11×11×11) for reciprocal space integration and 550 eV energy cutoff of the plane-wave expansion have been used to get a residual error on the energy of $10^{-5}$ meV, resulting in a fully converged electronic structure including SOC. For benchmarking, the time used for this step in the case of a topological insulator $Bi_2Te_3$ (5 atoms in the primitive unit cell) was 16 min 3s. This computation was performed for all the 22652 materials in the filtered ICSD database, as well as for 4286 $f$-electron compounds. After we establish which materials are topological, we compute the full band structure calculation for the topological materials - a step which on average takes 240s per compound in 20 skylake cores at 2.4GHz (for compounds with less than 10 atoms).

### Appendix D: Compatibility Relations and EBR Linear Dependence Codes

We now describe one of the most important parts of our material search: the codes we build, whose input are the characters of the bands below the Fermi level - obtained in Appendix[C] - and whose output are, in effect, the topological properties of a material. There are two such codes: one that checks whether the compatibility relations between characters at maximal $\mathbf{k}$ points are satisfied in the BZ. If the answer is no, then the material is an enforced semimetal, which we further characterize in the following appendices. If the answer is yes, then the material admits an insulating filling (can be made an insulator by smoothly changing the bands around the Fermi level without changing the filling). We then built a second code which checks whether the insulator characters at maximal $\mathbf{k}$-vectors can be written in terms of linear combinations of EBRs (in which case the system is topologically trivial) or not (in which case the system is topological). If topologically nontrivial, we then further characterize the material in the following appendices. This appendix gives the theory behind the two codes we developed, as well as a description of the codes. The presentation is self-contained.

The energy bands, or electronic bands, in a solid are continuous functions defined in the reciprocal (or momentum) space whose dependence on the momentum $\mathbf{k}$ in the BZ is restricted by the space group of the structure. The space group contains both rotational operators as well as translation parts. On the one hand, translational invariance by a lattice constant forces the electronic bands to be periodic functions in momentum space. On the other hand, the rotational operations impose restrictions on the degeneracies of the energy in bands at specific points in the BZ. At each **k**-point in the BZ, the electronic states transform under irreducible representations of the little group of **k**. The set of states form a continuous domain in momentum space, and one can define the concept of "connectivity", i.e. how many bands *have* to be connected (cannot be separated to give rise to an insulating filling) throughout the BZ. In general, different energy bands at general positions of the momentum could merge into degenerate bands at points of high symmetry in the BZ. The connections (and degeneracies) of bands throughout the BZ, are protected by the symmetry (space group) of the system. Bands continuing out of a high-symmetry point in the BZ of a system are, generically, only compatible with a *certain* set of bands continuing out of another high symmetry point in the BZ [11, 12]. This is clear: as two high-symmetry points are connected by a high-symmetry line or plane, one band starting from one of the high symmetry points can connect with another band starting from the other high symmetry point *iff* they carry the same irreps under the symmetry little group of the high-symmetry plane or line [11–13]. The maximal **k** vectors are the high-symmetry points necessary to fully describe the connections of bands in the BZ. By finding all the possible connections in the BZ [11, 13], we could, potentially, find all the number of possible bands in momentum space that can exist.

We dwell on this point slightly further. Any number of bands are *partially* identified by their decomposition into irreps of the little group of **k** upon subduction, for all the **k**-vectors in the BZ. (As any band is a representation of the space-group of a lattice, which is infinite, its decomposition into irreps of the little group of the momentum does not fully determine the band, as it would for representations of a finite group. One also needs the use of Berry phases, or parallel transport, to determine the band. We leave aside this complication for later.) However, it is not necessary to consider the multiplicities of the irreps in the whole BZ due to the *compatibility relations* between the irreps of two connected subsets of **k**-vectors. In fact, what makes the problem of defining the decomposition of bands into irreps of the little group of the continuous variable **k** in the BZ feasible is that the irrep decomposition in the whole BZ is fully fixed by looking at the irrep decomposition of only a finite, very small number of **k** points, called maximal **k** vectors [12–15]. Consider a **k** point (line) in the momentum space that belongs to a line ($l$) $\mathbf{k_l}$ and let's assume that the little group of **k**, $G^{\mathbf{k}}$ is a supergroup of the little group of $\mathbf{k_l}$, $G^{\mathbf{k_l}} < G^{\mathbf{k}}$. For example, **k** can be a high-symmetry point and $\mathbf{k_l}$ can be the line of momenta going through the high symmetry point **k**. Then, the matrices of an irreducible representation $\rho^i$ of the little group $G^{\mathbf{k}}$ associated with the symmetry operations that belong to $G_l^{\mathbf{k}}$ form a representation, not necessarily irreducible, of $G_l^{\mathbf{k}}$. This representation, in general, is the direct sum of irreducible representations $\rho_l^j$ of $G_l^{\mathbf{k}}$,

$$\rho^i \downarrow G_l^{\mathbf{k}} = \bigoplus_{i=1}^{s} m_{ij}\rho_l^j \tag{D1}$$

being $s$ the number of irreducible representations of $G_l^{\mathbf{k}}$ and $m_{ij}$ the (integer) multiplicity of $\rho_l^j$ in the decomposition of $\rho^i$. The program DCOMPREL [16] on the Bilbao Crystallographic Server (BCS), gives the multiplicities for each pair of connected **k**-vectors in the momentum space (point-line, line-plane, etc...) for any double space group [14]. This process of finding the representations of a subgroup included in the representation of a bigger group is called subduction. Applying this further, under subduction, the multiplicity of an irrep of a little group of a plane ($p$) in the momentum space $\mathbf{k_p}$ is determined by the multiplicities of the irreps of any line $\mathbf{k_l}$ that belongs to the plane, and these last multiplicities are determined by the multiplicities of the irreps of any point of higher symmetry in the line. It is enough, then, to consider only the multiplicities of the irreps of the little group of the **k** points of maximal symmetry in the space group to fully characterize a Band Representation (for more details see [12, 14]).

Our first aim is to write a code that checks whether different sets of irreps, at different high-symmetry points, are compatible in the BZ and can create an isolated band - in which case an insulator is possible - or not - in which case an insulator is not possible and the system is an enforced semimetal. The terminology enforced implies that the symmetries create a band connectivity which cannot support an insulating filling at the number of electrons given - the prototypical example is a band with time-reversal invariance (which forces Kramers pairs) but an odd-number of electrons per unit cell. In order to check if the irreps can form an isolated band, it is necessary to identify the intermediate paths between each pair of **k**-points of maximal symmetry to analyze the connectivities of the electronic bands (connectivity graphs). If $\mathbf{k_1}$ and $\mathbf{k_2}$ are two points of maximal symmetry in the momentum space and $\mathbf{k_l}$ is a line or a plane that connects the two points, a set of compatibility relations are defined between $\mathbf{k_1}$ and $\mathbf{k_l}$ on one hand and between $\mathbf{k_2}$ and $\mathbf{k_l}$ on the other hand. As the electronic bands are continuous functions of **k**, the compatibility relations fix the possible ways two connect two sets of irreducible representations at $\mathbf{k_1}$ and $\mathbf{k_2}$. Consequently, the two sets of irreps $\mathbf{k_1}$ and $\mathbf{k_2}$ can form a disconnected subset of electronic bands if and only if the sum of multiplicities of each irrep of the little group of $\mathbf{k_l}$ in the decomposition of the irreps at $\mathbf{k_1}$ and at $\mathbf{k_2}$ is the same [12–14]. Note that a trivially necessary condition to be fulfilled by a set of connected bands is that the number of bands that cross $\mathbf{k_1}$ and $\mathbf{k_2}$ (the sum of the dimensions of the irreps) must be the same. The intermediate paths of a given space group can be obtained by inspection in the intermediate input page of DCOMPREL [16].

Another important concept in band theory is that of an Elementary Band Representation[17–19]. These are sets of bands induced from localized orbitals (representations of the site-symmetry group) of the maximal Wyckoff positions in a space group [13, 17–19]. In a recent series of works ([12–15]) the authors calculated all the 10398 Elementary Band Representations (EBRs) of all the 230 double space groups. The authors of [13] introduced a generic definition of topology. Any set of bands induced from localized orbitals (which transform under some irreducible representation of the site-symmetry group of the *center* of the orbitals) can be expressed as a linear combination of EBRs. They are topologically trivial. However, if such linear combination does not exist, the material must be topologically non-trivial. Then, given a set of bands in an energy range (for example, but not necessarily, below the Fermi level), a simple and extremely fast way to determine the topology of the system is to check if the bands within that energy range in the realistic material can be written as a linear combination of the tabulated EBRs (available on the BCS) in that group. EBRs, being bands, are partially defined by their decomposition into irreps at maximal **k** vectors. The program BANDREP [20] on the BCS gives complete information of all the subduction of EBRs at maximal **k** points for all the space groups. Once the space group has been chosen and assuming that the system has time-reversal symmetry, clicking on "Elementary TR", the program gives, for every EBR, the multiplicities of the irreps at the **k** points of maximal symmetry. These points are shown in the first column of the table (the decomposition of all other EBRs, with or without time-reversal, for single or double groups, is also given).

The combination of DCOMPREL and BANDREP allows us to write a code to perform a systematic search of compounds that can be topological insulators or topological semimetals. We have implemented the program *Check-TopologicalMat* [21] that checks if a given material is topological (insulator or semimetal) or not. The program makes use of DCOMPREL, BANDREP and other programs of the BCS, and needs an input file with the following information (see the example given in [21]):

- Electron filling: stoichiometric number of electrons below the Fermi level.
- A symmetry operation of the (double) space group for each element of the point group in a primitive basis (or in any basis, the program identifies the symmetry).
- Coordinates of a complete set on **k**-vectors of maximal symmetry (can be obtained easily from the BCS.)
- For each **k**-vector, it should be indicated which operations in the given list belong to the little group of **k**. (can be obtained easily from the BCS.)
- For each band at each **k**-vector, the value of the energy and the traces of the symmetry operations of the little group in the given irrep (or reps, if two or more irreps accidentally merge in our energy window).

In the first step, using the list of symmetry operations given, the program IDENTIFY GROUP [2] identifies the space group and the transformation matrix to the standard setting. The BCS hosts a database (REPRESENTATIONS DSG [22]) of irreducible representations containing: the symmetry operation (one for each operation of the point group, in a standard setting), the matrix (thus, the trace) of the representation, and its label. In order to apply the representation reduction formula Eq[D1] we need to transform the given symmetry operations that VASP outputs (in general, in a non-standard setting) to the standard setting. The transformation matrix also allows us to identify the given **k**-vectors of maximal symmetry in the standard setting of the space group. Then we can project the sets of eigenvalues obtained by VASP into the set of irreducible representations at that point and calculate the multiplicity of each irrep.

Once a specific order of the merged sets of irreps at all the maximal **k**-vecs has been chosen, a set of bands can be described as a list of integer numbers that represent the multiplicities of the corresponding irreps. The same lists can be constructed for the EBRs of the space group. Therefore, to check if a set of band representations can be expressed as linear combinations of EBRs or not reduces to the analysis of a system of linear equations with integer coefficients (diophantine equations). We later differentiate between positive and negative coefficients.

Following the above procedure in a systematic way, we can find four different cases in our data:

1. Since we know the electron number, we count up the number of bands at every **k** maximal point, up to the electron number. We can reach a case when the number of bands is not the same at every maximal **k**-vector, i.e., the sum of the dimensions of the irreps is different from the given number of bands at least at one maximal **k**-vector. In this case, the bands below the Fermi level cannot form a subset of interconnected bands separated from the bands above the Fermi level (do not, trivially, satisfy the compatibility relations). This happens when the closest irrep to the Fermi level is degenerate. In fact, the Fermi level crosses the point of degeneracy. In this case the material is an Enforced Semimetal with Fermi Degeneracy (ESFD). HgTe, for example, is in this case.

2. The number of bands below the Fermi level is the same at all the maximal **k**-vecs, but they do not fulfill the compatibility relations. As a consequence, the irreps below the Fermi level at some points must be connected with irreps above the Fermi level at some other points. It is a gapless set of bands and, then, an Enforced Metal or an Enforced Semimetal. We abbreviate these situations ES.

3. The set of irreps at some k-point does not fulfill time-reversal (TR) symmetry, and when TR is artificially imposed, the number of bands at the Fermi level is not the same at every **k** vector (when counted, from below, up to the number of electrons in the system). In non-magnetic materials with TR symmetry, the irreducible representations are derived from the irreps of the group without TR, satisfying the reality of the irrep (see for instance the books of Bradley and Cracknell [23] or Dresselhaus [24]). In some cases (irreps of type (b) and (c) in the given references) the TR symmetry forces two irreps (duplication of two irreps of the same label in the first case and two different irreps in the second case) to be degenerate. In the numerical calculation of the energy bands of some compounds, a degenerate irrep of this type can be *split* into the two irreps that form the pair having assigned a very slightly different energy, probably due to lack of precision in the calculations. Therefore, the set of bands below the Fermi level given in the file does not fulfill the TR symmetry. However, if the two parts of the TR-invariant irrep are merged (we have merged those bands whose energy differ in less than a 0.001 factor to keep TR symmetry), the material could also be considered an ESFD.

4. The number of irreps below the Fermi level is the same at all the **k**-vectors and they fulfill the compatibility relations. In this case, we are at an insulating filling of the system (i.e. the system can be made gapped by changing the quantitative, not qualitative details of the bands), and we treat the bands as if they would form an insulator. In reality, bands might cross a straight Fermi level, but there is always a direct gap where a "winding" Fermi level can be placed. We can distinguish four different cases of this:

   (a) The characters of the set of bands is a linear combinations with positive coefficients of the characters of the EBRs in a space group; the linear equations have solutions with non-negative coefficients. In this case, the material is not guaranteed to be topological, and more calculations, such as Wilson loop calculations, would be necessary in order to establish the topological nature of the material. In most cases it is a trivial insulator. For example, in SG 1, *only* Wilson loop calculations can distinguish if a material is topological or not. As a bonus, the program *Check Topological Mat.* [21] not only checks the possible linear dependence, but also gives the possible decompositions of the set of bands into EBRs. In some cases this number is very large (hundreds of thousands of different combinations) and only a limited number of the possible decompositions is given.

   (b) The characters of the set of bands is a linear combinations of EBRs plus one or more disconnected branches into which a single EBR can be disconnected but is not a linear combination of EBR's with positive coefficients (in which case, it would go to item 4a). In some space groups, some EBRs can be separated into subsets of irreps that fulfill the compatibility relations. These are decomposable EBRs (see [11–15]). The information about the possible decomposition of an EBR and the different ways into which a decomposable EBR can decompose is included in the program BANREP for all the double groups. The program *Check Topological Mat.* gives the possible decompositions into EBRs and branches of EBRs. In this case, the material is a topological insulator with strong or fragile topology. The prototypical case is graphene

   (c) The characters of the set of bands is *not* a linear combinations of EBRs, and is also not a linear combinations of EBRs plus one or more disconnected branches into which a single EBR can be disconnected. This is for example the case of two EBR's which mix, become gapless, and re-open a (topological) gap. Neither the valence nor the conduction band is part of a disconnected EBR, but cannot be expressed as a linear combination of EBRs. This is a somewhat artificial distinction from the previous case, but it makes physical sense as it distinguishes the case of graphene - which is always topological when insulating, from the case of a material like $CaAs_3$, which, in theory, can be made either topological or trivial, by tuning the inversion gap. It is a topological insulator.

   (d) Finally, in cases 4b and 4c, the program then tries to solve for the characters of the bands as linear combinations of EBRs including as acceptable solutions those with negative coefficients. Those materials have a finer topological structure [15, 25, 26], and have been dubbed fragile topological insulators, although their fragility has not been properly investigated and they should be better called "finer" topological insulators. If no solution of the EBR linear dependence including negative coefficients is found, the material is a topological insulator with strong topology. *Importantly, in this case the material can also be a Weyl semimetal with Weyl points at generic points in the BZ! These systems also satisfy compatibility (as Weyls are at generic points in the BZ, where the little group is Identity, are not sums of EBRs but cannot be insulators. An example is given in Appendix[Q]. Further research is needed to separate these cases)*

In most cases, our Mathematica program calculates all the possible solutions in seconds, except when the number of solution is very large. In a first step, it tries to get all the possible linear combinations (cases 4a and 4b), but once a time limit is reached, it calculates only a limited number of solutions. If there is no solution, the program comes to this conclusion in few seconds.

## Appendix E: Enforced Semimetals (ES), Enforced Semimetals with Fermi Degeneracy (ESFD), Split EBRs (SEBR) Topological Insulators, No Linear Combination (NLC) Topological Insulators Separation of Classes of Materials.

We provide in Table[I] meta-data about the different physical classes of materials we find. In the table below we provide, for each symmetry group: First column: space-group number. Second Column: the total number of topological materials in each space group in the ICSD database, after all our filters. Third Column: total number of Split EBR (SEBR) topological insulators in the respective space-group. Fourth Column: total number of No Linear Combination (NLC) topological insulators in the respective space-group. Fifth Column: total number of enforced semimetals (ES) in the respective space-group. Sixth column: total number of enforced semimetals with Fermi Degeneracy (ESFD) in the respective space-group.

Some of the groups contain no topological insulators: this does *not* mean that topological insulators *cannot or do not* exist in that space group. Lack of topological insulators in a certain space-group can be explained by several factors. First, our method cannot distinguish topological insulators that do now have a symmetry imprint: for example, in SG1, which contains no symmetry operators except translational invariance, we see that there are no topological metals, semimetals or insulators listed. This is because, clearly, our symmetry-based method cannot distinguish topological materials in a group that has no symmetries. Certainly Chern Insulators or Time-Reversal Invariant topological insulators can exist in this group (they actually do not even need translations invariance to exist), but they would have to be diagnosed by invariants such as the Wilson loop winding. This happens in more groups than just SG1: there are other groups whose topology cannot be distinguished by the irreps of bands at high symmetry points. In these groups, all bands that satisfy the momentum compatibility relations can be expressed as sums of EBRs. All those groups have zero materials in our search, as they should. These groups coincide with the groups of with no symmetry indicators. Second, for groups with symmetries, another reason - besides the previous ones - for the possible lack of materials is that our initial filters applied to the ICSD database have removed many low-quality data materials, in order to keep only the most reliable data. Unfortunately, this has the potential of removing some known topological materials, such as KHgSb, but is a small price to pay for augmenting the reliability of our predictions.

Lastly, we comment on the NLC vs SEBR separation of insulators. This is a physical, ad-hoc separation, of insulators which basically tells us if an insulator resembles, in character, $CaAs_3$ or graphene with large Spin-Orbit Coupling respectively. In graphene, the bands around the Fermi level form *one* EBR [13]. This EBR can then split/gap and then the bands above or below the gap form part of a split EBR. This material would then be classified within the SEBR group. This situations tells the reader that, if gapped, graphene with $p_z$ orbitals would have to be topological. We have classified *all* the split EBR's on the BCS, and we can immediately check if a given band structure can be a linear combination of SEBRs. The SEBR classification of materials is then much like the one in graphene: these materials "have to be" topological. By contrast, in NLC, the bands around the Fermi level form *two* EBRs. Starting with the atomic limit of one EBR below and one EBR above the Fermi levels, these two EBRs can mix, close the gap, and decompose again into valence and conduction bands neither of which form an EBR. However, they also do not form split parts of a single EBRS. This is the case of $CaAs_3$. These are insulators that, in principle, could be trivial, but because of quantitative details such as gap closing and re-opening, are nontrivial. The NLC stands for No Linear Combination (of EBRs), but could more appropriately be denoted as No Linear Combination of split parts of EBRs.

| SG | Tot. Num. | SEBR | NLC | ES | ESFD |
|---|---|---|---|---|---|
| 1 | 0 | 0 | 0 | 0 | 0 |
| 2 | 101 | 0 | 101 | 0 | 0 |
| 3 | 0 | 0 | 0 | 0 | 0 |
| 4 | 8 | 0 | 0 | 8 | 0 |
| 5 | 0 | 0 | 0 | 0 | 0 |
| 6 | 0 | 0 | 0 | 0 | 0 |
| 7 | 3 | 0 | 0 | 3 | 0 |
| 8 | 0 | 0 | 0 | 0 | 0 |
| 9 | 8 | 0 | 0 | 8 | 0 |
| 10 | 1 | 0 | 1 | 0 | 0 |
| 11 | 199 | 0 | 98 | 0 | 101 |
| 12 | 390 | 198 | 192 | 0 | 0 |
| 13 | 42 | 0 | 13 | 0 | 29 |
| 14 | 158 | 0 | 71 | 23 | 64 |
| 15 | 264 | 0 | 95 | 0 | 169 |
| 16 | 0 | 0 | 0 | 0 | 0 |
| 17 | 0 | 0 | 0 | 0 | 0 |
| 18 | 0 | 0 | 0 | 0 | 0 |
| 19 | 15 | 0 | 0 | 15 | 0 |
| 20 | 1 | 0 | 0 | 1 | 0 |
| 21 | 0 | 0 | 0 | 0 | 0 |
| 22 | 0 | 0 | 0 | 0 | 0 |
| 23 | 0 | 0 | 0 | 0 | 0 |
| 24 | 0 | 0 | 0 | 0 | 0 |
| 25 | 0 | 0 | 0 | 0 | 0 |
| 26 | 3 | 0 | 0 | 0 | 3 |
| 27 | 0 | 0 | 0 | 0 | 0 |
| 28 | 0 | 0 | 0 | 0 | 0 |
| 29 | 1 | 0 | 0 | 1 | 0 |
| 30 | 0 | 0 | 0 | 0 | 0 |
| 31 | 8 | 0 | 0 | 0 | 8 |
| 32 | 0 | 0 | 0 | 0 | 0 |
| 33 | 21 | 0 | 0 | 21 | 0 |
| 34 | 0 | 0 | 0 | 0 | 0 |
| 35 | 0 | 0 | 0 | 0 | 0 |
| 36 | 15 | 0 | 0 | 0 | 15 |
| 37 | 0 | 0 | 0 | 0 | 0 |
| 38 | 0 | 0 | 0 | 0 | 0 |
| 39 | 0 | 0 | 0 | 0 | 0 |
| 40 | 3 | 0 | 0 | 3 | 0 |
| 41 | 0 | 0 | 0 | 0 | 0 |
| 42 | 0 | 0 | 0 | 0 | 0 |
| 43 | 4 | 0 | 0 | 0 | 4 |
| 44 | 0 | 0 | 0 | 0 | 0 |
| 45 | 0 | 0 | 0 | 0 | 0 |
| 46 | 0 | 0 | 0 | 0 | 0 |
| 47 | 13 | 0 | 13 | 0 | 0 |
| 48 | 0 | 0 | 0 | 0 | 0 |
| 49 | 0 | 0 | 0 | 0 | 0 |
| 50 | 0 | 0 | 0 | 0 | 0 |
| 51 | 47 | 0 | 17 | 0 | 30 |
| 52 | 4 | 0 | 0 | 4 | 0 |
| 53 | 2 | 2 | 0 | 0 | 0 |
| 54 | 0 | 0 | 0 | 0 | 0 |
| 55 | 81 | 0 | 63 | 0 | 18 |
| 56 | 0 | 0 | 0 | 0 | 0 |
| 57 | 17 | 0 | 9 | 8 | 0 |
| 58 | 29 | 0 | 12 | 2 | 15 |
| 59 | 79 | 19 | 14 | 0 | 46 |
| 60 | 9 | 0 | 7 | 2 | 0 |
| 61 | 5 | 0 | 2 | 3 | 0 |
| 62 | 565 | 0 | 257 | 308 | 0 |
| 63 | 453 | 103 | 120 | 0 | 230 |

| SG | Tot. Num. | SEBR | NLC | ES | ESFD |
|---|---|---|---|---|---|
| 64 | 56 | 26 | 0 | 0 | 30 |
| 65 | 57 | 26 | 31 | 0 | 0 |
| 66 | 0 | 0 | 0 | 0 | 0 |
| 67 | 3 | 0 | 1 | 0 | 2 |
| 68 | 0 | 0 | 0 | 0 | 0 |
| 69 | 18 | 18 | 0 | 0 | 0 |
| 70 | 28 | 23 | 0 | 0 | 5 |
| 71 | 105 | 105 | 0 | 0 | 0 |
| 72 | 40 | 2 | 7 | 0 | 31 |
| 73 | 0 | 0 | 0 | 0 | 0 |
| 74 | 44 | 0 | 25 | 0 | 19 |
| 75 | 0 | 0 | 0 | 0 | 0 |
| 76 | 0 | 0 | 0 | 0 | 0 |
| 77 | 0 | 0 | 0 | 0 | 0 |
| 78 | 1 | 0 | 0 | 1 | 0 |
| 79 | 1 | 0 | 0 | 1 | 0 |
| 80 | 0 | 0 | 0 | 0 | 0 |
| 81 | 0 | 0 | 0 | 0 | 0 |
| 82 | 8 | 0 | 8 | 0 | 0 |
| 83 | 4 | 0 | 1 | 3 | 0 |
| 84 | 3 | 0 | 1 | 1 | 1 |
| 85 | 9 | 0 | 1 | 0 | 8 |
| 86 | 2 | 0 | 0 | 0 | 2 |
| 87 | 21 | 11 | 0 | 10 | 0 |
| 88 | 8 | 0 | 0 | 1 | 7 |
| 89 | 0 | 0 | 0 | 0 | 0 |
| 90 | 0 | 0 | 0 | 0 | 0 |
| 91 | 0 | 0 | 0 | 0 | 0 |
| 92 | 9 | 0 | 0 | 5 | 4 |
| 93 | 0 | 0 | 0 | 0 | 0 |
| 94 | 0 | 0 | 0 | 0 | 0 |
| 95 | 0 | 0 | 0 | 0 | 0 |
| 96 | 0 | 0 | 0 | 0 | 0 |
| 97 | 1 | 0 | 0 | 1 | 0 |
| 98 | 0 | 0 | 0 | 0 | 0 |
| 99 | 3 | 0 | 0 | 3 | 0 |
| 100 | 0 | 0 | 0 | 0 | 0 |
| 101 | 0 | 0 | 0 | 0 | 0 |
| 102 | 2 | 0 | 0 | 1 | 1 |
| 103 | 3 | 0 | 0 | 0 | 3 |
| 104 | 0 | 0 | 0 | 0 | 0 |
| 105 | 0 | 0 | 0 | 0 | 0 |
| 106 | 1 | 0 | 0 | 1 | 0 |
| 107 | 4 | 0 | 0 | 4 | 0 |
| 108 | 0 | 0 | 0 | 0 | 0 |
| 109 | 4 | 0 | 0 | 0 | 4 |
| 110 | 0 | 0 | 0 | 0 | 0 |
| 111 | 0 | 0 | 0 | 0 | 0 |
| 112 | 0 | 0 | 0 | 0 | 0 |
| 113 | 31 | 0 | 4 | 0 | 27 |
| 114 | 0 | 0 | 0 | 0 | 0 |
| 115 | 5 | 5 | 0 | 0 | 0 |
| 116 | 0 | 0 | 0 | 0 | 0 |
| 117 | 0 | 0 | 0 | 0 | 0 |
| 118 | 0 | 0 | 0 | 0 | 0 |
| 119 | 0 | 0 | 0 | 0 | 0 |
| 120 | 1 | 0 | 0 | 1 | 0 |
| 121 | 9 | 9 | 0 | 0 | 0 |
| 122 | 21 | 15 | 0 | 0 | 6 |
| 123 | 79 | 24 | 31 | 24 | 0 |
| 124 | 7 | 0 | 0 | 0 | 7 |
| 125 | 20 | 4 | 1 | 7 | 8 |
| 126 | 2 | 0 | 0 | 1 | 1 |

| SG | Tot. Num. | SEBR | NLC | ES | ESFD |
|---|---|---|---|---|---|
| 127 | 149 | 0 | 37 | 60 | 52 |
| 128 | 4 | 1 | 0 | 2 | 1 |
| 129 | 610 | 93 | 84 | 103 | 330 |
| 130 | 10 | 0 | 0 | 0 | 10 |
| 131 | 3 | 0 | 2 | 0 | 1 |
| 132 | 0 | 0 | 0 | 0 | 0 |
| 133 | 0 | 0 | 0 | 0 | 0 |
| 134 | 0 | 0 | 0 | 0 | 0 |
| 135 | 11 | 0 | 0 | 8 | 3 |
| 136 | 87 | 14 | 0 | 32 | 41 |
| 137 | 5 | 3 | 0 | 1 | 1 |
| 138 | 0 | 0 | 0 | 0 | 0 |
| 139 | 350 | 251 | 0 | 99 | 0 |
| 140 | 172 | 14 | 14 | 34 | 110 |
| 141 | 101 | 24 | 0 | 19 | 58 |
| 142 | 2 | 0 | 0 | 2 | 0 |
| 143 | 1 | 0 | 0 | 1 | 0 |
| 144 | 0 | 0 | 0 | 0 | 0 |
| 145 | 1 | 0 | 0 | 1 | 0 |
| 146 | 5 | 0 | 0 | 5 | 0 |
| 147 | 3 | 2 | 0 | 1 | 0 |
| 148 | 65 | 54 | 0 | 11 | 0 |
| 149 | 0 | 0 | 0 | 0 | 0 |
| 150 | 1 | 0 | 0 | 0 | 1 |
| 151 | 0 | 0 | 0 | 0 | 0 |
| 152 | 1 | 0 | 0 | 0 | 1 |
| 153 | 0 | 0 | 0 | 0 | 0 |
| 154 | 0 | 0 | 0 | 0 | 0 |
| 155 | 1 | 0 | 0 | 1 | 0 |
| 156 | 5 | 0 | 0 | 5 | 0 |
| 157 | 0 | 0 | 0 | 0 | 0 |
| 158 | 0 | 0 | 0 | 0 | 0 |
| 159 | 5 | 0 | 0 | 2 | 3 |
| 160 | 10 | 0 | 0 | 10 | 0 |
| 161 | 34 | 0 | 0 | 4 | 30 |
| 162 | 12 | 5 | 0 | 7 | 0 |
| 163 | 8 | 2 | 0 | 4 | 2 |
| 164 | 150 | 102 | 0 | 48 | 0 |
| 165 | 5 | 2 | 0 | 1 | 2 |
| 166 | 619 | 410 | 0 | 209 | 0 |
| 167 | 332 | 92 | 0 | 118 | 122 |
| 168 | 0 | 0 | 0 | 0 | 0 |
| 169 | 0 | 0 | 0 | 0 | 0 |
| 170 | 0 | 0 | 0 | 0 | 0 |
| 171 | 0 | 0 | 0 | 0 | 0 |
| 172 | 0 | 0 | 0 | 0 | 0 |
| 173 | 11 | 0 | 0 | 10 | 1 |
| 174 | 1 | 0 | 0 | 1 | 0 |
| 175 | 0 | 0 | 0 | 0 | 0 |
| 176 | 31 | 2 | 0 | 10 | 19 |
| 177 | 0 | 0 | 0 | 0 | 0 |
| 178 | 0 | 0 | 0 | 0 | 0 |
| 179 | 0 | 0 | 0 | 0 | 0 |
| 180 | 4 | 0 | 0 | 3 | 1 |
| 181 | 1 | 0 | 0 | 1 | 0 |
| 182 | 8 | 0 | 0 | 8 | 0 |
| 183 | 0 | 0 | 0 | 0 | 0 |
| 184 | 0 | 0 | 0 | 0 | 0 |
| 185 | 12 | 0 | 0 | 8 | 4 |
| 186 | 115 | 0 | 0 | 53 | 62 |
| 187 | 14 | 0 | 0 | 14 | 0 |
| 188 | 2 | 0 | 0 | 1 | 1 |
| 189 | 91 | 44 | 0 | 47 | 0 |
| 190 | 25 | 0 | 2 | 8 | 15 |
| 191 | 64 | 20 | 0 | 44 | 0 |
| 192 | 0 | 0 | 0 | 0 | 0 |
| 193 | 168 | 8 | 0 | 72 | 88 |
| 194 | 189 | 57 | 0 | 74 | 58 |
| 195 | 0 | 0 | 0 | 0 | 0 |
| 196 | 0 | 0 | 0 | 0 | 0 |
| 197 | 2 | 0 | 0 | 0 | 2 |
| 198 | 34 | 0 | 0 | 0 | 34 |
| 199 | 4 | 0 | 0 | 1 | 3 |
| 200 | 0 | 0 | 0 | 0 | 0 |
| 201 | 0 | 0 | 0 | 0 | 0 |
| 202 | 2 | 0 | 0 | 0 | 2 |
| 203 | 0 | 0 | 0 | 0 | 0 |
| 204 | 79 | 24 | 0 | 0 | 55 |
| 205 | 17 | 0 | 6 | 1 | 10 |
| 206 | 0 | 0 | 0 | 0 | 0 |
| 207 | 0 | 0 | 0 | 0 | 0 |
| 208 | 0 | 0 | 0 | 0 | 0 |
| 209 | 0 | 0 | 0 | 0 | 0 |
| 210 | 0 | 0 | 0 | 0 | 0 |
| 211 | 0 | 0 | 0 | 0 | 0 |
| 212 | 4 | 0 | 0 | 0 | 4 |
| 213 | 7 | 0 | 0 | 0 | 7 |
| 214 | 0 | 0 | 0 | 0 | 0 |
| 215 | 0 | 0 | 0 | 0 | 0 |
| 216 | 37 | 0 | 8 | 6 | 23 |
| 217 | 19 | 7 | 0 | 0 | 12 |
| 218 | 0 | 0 | 0 | 0 | 0 |
| 219 | 0 | 0 | 0 | 0 | 0 |
| 220 | 20 | 0 | 4 | 0 | 16 |
| 221 | 84 | 14 | 0 | 21 | 49 |
| 222 | 0 | 0 | 0 | 0 | 0 |
| 223 | 13 | 0 | 0 | 2 | 11 |
| 224 | 2 | 0 | 0 | 2 | 0 |
| 225 | 101 | 30 | 0 | 21 | 50 |
| 226 | 19 | 0 | 0 | 6 | 13 |
| 227 | 260 | 88 | 0 | 41 | 131 |
| 228 | 0 | 0 | 0 | 0 | 0 |
| 229 | 24 | 1 | 0 | 3 | 20 |
| 230 | 0 | 0 | 0 | 0 | 0 |
| Total | 7385 | 1954 | 1353 | 1721 | 2357 |

TABLE I: Number of materials per space group. Please refer to the text in Appendix E and Appendix D for an explanation of different abbreviations, as well as for an understanding of why some symmetry groups have no compounds found by our method.

## Appendix F: All the Topological Materials In the World: Compound Formula, Direct Gap, Indirect Gap, Representations of Bands Below The Fermi Level Tables

In the tables linked to this appendix, we provide a large set of data for all the topological materials found. For an easier indexing of materials, we assign them a Topological Quantum Chemistry number (TQC #), which will be used in other parts of the paper (as well as in the future material database on the Bilbao Crystallographic Server), specifically in Appendix[G], [H],[I],[J], [K],[L],[Q], to reference back to the material. The TQC number makes it easier to identify the material.

The format of the TQC number is $x.y.z.$ $x$ denotes the number of the symmetry group of the material. The $y$ takes values is $1, 2, 3, 4$ for the SEBR, NLC, ES and ESFD respectively. The $z$ is a number indexing the material. The search function in the PDF document can then be used to search for all the instances/information of a material based on their TQC number.

The data is given in Table XIV, of which, column 1 shows the TQC number; column 2 shows the ICSD number; column 3 shows the total number of valence electrons; columns 4 and 5 show calculated direct and indirect gap, respectively. Columns 6 and 7 of Table XIV provide the minimal symmetry group that keeps the topological nature of the *gapped* compound with and without Time Reversal symmetry and topological classes (Appendix[H]). Columns 8 and 9 of Table XIV provide the largest subgroup (with and without time-reversal) of the space group (Max. SubSG) for which the *gapless* material (either ES or ESFD) satisfies the compatibility relations and hence allows an insulating filling. Column 10 of Table XIV shows the computed symmetry indicators. In last column, we present the stoichiometric formula.

Second, we provide another, less important set of Tables [ XXI, XXII, XXIII, XXIV, XXV, XXVI, XXVII, XXVIII, XXIX, XXX, XXXI, XXXII, XXXIII, XXXIV, XXXV, XXXVI, XXXVII, XXXVIII, XXXIX, XL, XLI, XLII, XLIII, XLIV, XLV, XLVI, XLVII, XLVIII, XLIX, L, LI, LII, LIII, LIV, LV, LVI, LVII, LVIII, LIX, LX, LXI, LXII, LXIII, LXIV, LXV, LXVI, LXVII, LXVIII, LXIX, LXX, LXXI, LXXII, LXXIII, LXXIV, LXXV, LXXVI, LXXVII, LXXVIII, LXXIX, LXXX, LXXXI, LXXXII, LXXXIII, LXXXIV, LXXXV, LXXXVI, LXXXVII, LXXXVIII, LXXXIX ] listing **(6d.)** all the representations of the bands below the Fermi level at all the maximal **k**-points. These representations are given using the notation conventions of the Bilbao Crystallographic Server, at each maximal **k**-point. This is done so that the public can perform independent checks on the materials and compare with our calculations. These tables take up 792 pages, and are given at the very end of the Supplementary Information, after all the other tables.

## Appendix G: Symmetries Protecting the Topological Nature of the Compounds

In this Appendix we provide, from a physical perspective, information about the symmetries necessary to stabilize the topological class of a material. This information can also be obtained from the topological index data in Appendix[Q]. In Appendix[H] we provided all topological insulators for non-magnetic groups, with time-reversal. While we know these materials are topological at this point, we do not yet know what time of topology do they have. We now elucidate this point: we ask, *what non-magnetic symmetries are necessary to protect the topological nature of the material.* The answer to this question provides us with the symmetries one can we break without destroying the topological nature of the compound. In spirit, this question is similar to the questions: if we break time-reversal in a $Z_2$ topological insulator, or mirror symmetry in a mirror Chern number insulator, do we destroy (some of) its topological nature? The answer is yes. We want to find the generic rules for answering this question. Similar information is available through the topological index data in Appendix[H], but we here provide a physical description of the procedure through the means of EBRs.

Stoichiometric materials in nature, under external fixed conditions, have structures classified by different symmetry groups. All the reported structures in the databases have assigned a space group as the set of symmetry operations that keep the atomic positions invariant. However, the change of the external conditions can induce a distortion in the structure, lowering its symmetry. The space group of the "new" (distorted) material is a *subgroup* of the space group of the undistorted structure.

In general, there could be several types of symmetry breaking mechanisms, leading to different "new" materials within different space groups. In the present context, we are interested in the identification of the symmetry operations that protect the topological state of a material (rotational or roto-inversion operations). Then we can limit the calculation of the subgroups of a given group to those subgroups that do not involve a change of the lattice ($\mathbf{k} = \mathbf{0}$ or *translationengleiche* subgroups). Given a space group $S_0$ we first look for the maximal space subgroups $S_1, S_2 \dots$ of $S_0$ (in general, there are more than one), i.e., those subgroups that are not subgroups of a subgroup of $S_0$. Then, we calculate the maximal subgroups of $S_1$, $S_2$, etc... until we reach to the space group P1 (No. 1). Then, we can establish a hierarchy of subgroups forming a group-subgroup diagram. The program SUBGROUPS in the BCS [27] does this calculation, using the option *Alternatively give the modulation wave-vectors* and choosing $\mathbf{k} = \mathbf{0}$. We later explain why we do not allow modulations that change or fold the **k** points. In general, the resulting space groups are not given in the standard setting, but the program calculates a transformation matrix for every subgroup to its standard setting. Using this transformation matrix we can establish the relation between the maximal **k**-vectors in the BZ of $S_0$ and the corresponding maximal **k**-vectors in each subgroup. In the next step, we can calculate the subduced irreducible representations at each maximal **k**-vector of the subgroup from the irreducible representations of the corresponding **k**-vector of $S_0$, and then express the characters of the bands, initially given in space group $S_0$, in terms of the irreps of the subgroup.

These calculations allow us to raise the following question: we know the symmetry characters of the bands in our material in space group $S_0$. Let us assume that we have obtained the confirmation that some bands in a material are topological SEBR or NLC. We now ask: if we add a perturbation that breaks the symmetry to the maximal subgroup (e.g.) $S_1$, will the material remain topological? Since we know the EBR's of $S_1$, all we have to do to answer this question is to repeat the same check that we have done in the space group $S_0$ but now in the space group $S_1$. If we obtain the result that, for all the maximal subgroups, the band representations can be expressed as linear combination of EBRs, then the $S_0$ itself is the group that protects the topological character of the compound; specifically, the symmetries that exist in $S_0$ but not in $S_1$ protect the topological nature of the material. However, if we find a maximal (or several maximal) *topological* subgroups, we know that the symmetries we broke do not destroy all of the topological nature of the material. In each subgroup, the removed symmetry operations from $S_0$ do not remove the full nontrivial topology of the system. We then proceed further down in the subgroup tree repeating the calculation in every subgroup until we obtain a set of minimal topological subgroups whose subgroups are non-topological. The difference in symmetries between the group $S_0$ and its largest subgroup where the system has become trivial (all the symmetry operations together) protect the topological character of the compound. If we break these sets of symmetries down a tree, the system becomes topologically trivial (modulo the 10 “strong” symmetry classes, as we explain next). As subgroups, we consider systems with and without time-reversal.

Our results in this section answer the question up to tensor product of Chern or time-reversal topological insulator $Z_2$ phases which cannot be determined by our method and have to be determined by more complex and time-consuming Wilson loop calculations [28–30]. In order to exemplify this, we pick an example: consider an insulator in space-group $P\bar{1}$ (No. 2), with time-reversal symmetry. We now consider one of the simplest topological insulators, the $Z_2$ time-reversal protected topological insulator. With inversion symmetry, this insulator has to have a specific set of inversion eigenvalues at the high-symmetry (inversion and time-reversal invariant) points in the BZ. At one of those **k** points (which, without loss of generality, we pick to be $\Gamma$), the inversion symmetries for the Kramers pair are $(--)$ while at all other points, they are $(++)$ (inversion symmetry eigenvalues are equal for each partner of a Kramers pair). We clearly know that this insulator *cannot* be a linear combinations of any EBR's: it is trivial to show that *any* EBR in $S_0$ has $4n$ with $n = 0, 1, 2$ eigenvalues $(--)$ Kramer pairs in the whole BZ, and hence one $(--)$ eigenvalue Kramers pair cannot be a sum or differences of EBRs. We now try to break symmetries. $S_0$ then has two maximal subgroups, $S_1 = P1$+TR and $S_2 = P\bar{1}$. We analyze the fate of the strong topological insulator under both these subgroups, and compare the results with our theory.

First, lets break the inversion symmetry, but keep time-reversal: $S_0 \to S_1$. We know that the time-reversal invariant topological insulator does *not* need spatial symmetries to exist (not even translation), and hence a soft breaking of the inversion symmetry which does not close the bulk gap cannot kill the time-reversal invariant topological insulator. However, this topology would *not* be detectable by the first iteration of our method in group $S_1$. In terms of eigenvalues at high symmetry points, in the space group $P1$ everything is trivial, as there is no symmetry. The eigenvalues of our time reversal invariant topological insulator would certainly have bands which can be expressed as a linear combination of EBRs in $P_1$, as the group has no symmetry. However, a Wilson loop calculation would immediately show that we have a TI. Hence, when in our subgroup we find a “trivial” state, we must remember that it is trivial up to addition of *time-reversal* $Z_2$ topological insulators.

Now lets break time-reversal symmetry: the group becomes $S_2 = P\bar{1}$. We know, that the system has stopped becoming a topological insulators with time-reversal symmetry, because TR has been broken. However, since there are two bands with negative inversion eigenvalues at $\Gamma$ (all other eigenvalues at all other points being $+$), and since we know that EBR's (even without time-reversal) can have only $4n$ with $n = 0, 1, 2$ negative eigenvalues in the whole BZ, it is clear that our material is still strong topological. It actually the identical response to that of the topological insulator with time-reversal invariance $\theta = \pi$ term in the Lagrangian of the low-energy theory[31, 32]. In fact, in this case, our supposed material is known in the literature under the name time-reversal symmetry broken Higher Order Topological Insulator (HOTI) which has hinge states on any hinge between two surfaces related by inversion. (We want to emphasize that the Names given in literature to different classes such as time-reversal TI's, mirror symmetric TI's, Higher-Order Topological insulators, Crystalline TI's, Hourglass Insulator [30], Dirac Insulator are rather ad-hoc and they can all be indexed by the topological classes in Appendix[G], [H], [P],[Q] and physically by the symmetry operations that stabilize the state. )

A more involved analysis, based on the EBR's in [13], reveals, from the characters of the bands at maximal **k**

vectors, whether the system is part of the 10 symmetry classes of materials that do not need spatial symmetry. The topological index method of [1, 33, 34] can link, in most but not all symmetry groups, the $Z_2$ topological phases with the characters of the bands at high-symmetry points, and can tell us if a system is $Z_2$ time-reversal topological insulator or not just from its symmetry eigenvalues. Being part of the 10 classes that do not need spacial symmetry to exist [35, 36], this system will remain topological even after the breaking of spatial symmetries.

Lastly, we comment on our choice to limit the calculation of the subgroups of a given group to those subgroups that do not involve a change of the lattice - $\mathbf{k} = \mathbf{0}$ or *translationengleiche* subgroups (using the function *Alternatively give the modulation wave-vectors* and choosing $\mathbf{k} = \mathbf{0}$). We choose this because distinct topological classes can become identical upon a change in translational symmetry or of lattice. For example, consider what has previously been described as a "weak" $Z_2$ topological insulator with time-reversal and inversion symmetry. This is an insulator whose minimal model, which we adopt here, contains *two* occupied bands, with Kramers degeneracy at the high-symmetry points, whose eigenvalues are $(--)$ at *two* high-symmetry (time-reversal invariant) points in the BZ, and $(++)$ everywhere else. Without loss of generality, we denote the two points with $(--)$ inversion eigenvalues as $\Gamma = (0, 0, 0)$ and $Z = (0, 0, \pi)$ on a cubic lattice. This is a state of matter which is different from a Higher Order topological insulator (HOTI): to start with, a HOTI requires a minimal model with at least *four* occupied bands; these four bands then have two Kramers pairs that have both $(--)$ eigenvalues at the *same* momentum in the BZ and all others having $(++)$. Without loss of generality, pick the momentum where the two Kramers pairs of $(--)$ inversion eigenvalues are to be $\Gamma = (0, 0, 0)$. The weak TI and the HOTI differ from each other. It is trivial to see that there is no way to add or subtract EBRs from one and obtain the characters of the other [37].

However, if we now add to the "weak" TI a commensurate Charge density wave fluctuation at momentum $\vec{Q} = (0, 0, \pi)$, the bands of the "weak" TI fold, and we now have 4 bands below the Fermi level. The inversion and time-reversal symmetries are not broken. Moreover, the inversion eigenvalues of $Z$ fold on top of the ones at $\Gamma$ and the system now has two Kramers pairs each with $(--)$ inversion eigenvalues at the new $\Gamma$ point [37]. We can then ask: have we now transformed our weak TI into HOTI? The answer is yes, up to addition or subtraction of EBRs [37]. The $k_z = \pi$ plane on the new, folded lattice has $(++)$ and $(--)$ eigenvalues at all its 4 high-symmetry points, which means that it differs from the $(++)$, and $(++)$ of the HOTI by an EBR induced by $s$-orbitals at the $1a = (0, 0, 0)$ position (in units of the Bravais lattice vectors) and by $s$ orbitals at the $1b = (0, 0, \frac{1}{2})$ position (in units of the Bravais lattice vectors) [37]. Hence, if we want to be able to classify the two phases as different, we have to disallow phase transitions that fold $k$ points into one another.

In columns 6 and 7 of Table XIV, we provide the minimal symmetry group that keeps the topological nature of the compound with and without Time Reversal symmetry, respectively. In Appendix[Q] we provide the link between the usual time-reversal $Z_2$ topological insulators and the topological indices of the material, for some space groups [1]

## Appendix H: Compounds by Topological Classes

While in Appendix G we asked which symmetries protect a topological class, in this Appendix, we separate the different topological materials by (strong) topological classes. We define the topological classes by the following definition:

***Definition 1:*** *If the characters of symmetry operations at the maximal* $\mathbf{k}$ *vectors of two materials differ from each other by sums or differences of characters of EBRs, then these two materials correspond to the same topological class. Otherwise, they are in different strong topological classes. This definition is implicit in the work of Bradlyn et al. [13]*

Our definition of strong topological classes says that two materials are in the same strong topological class if they can be related to each other by adding or removing bands which have the same symmetry representations as atomic limit insulators. We implemented a program that parses the materials in the different topological classes. This can be done by picking one material, checking the definition above versus all other materials, then adding all the materials that satisfy *Definition 1* into one topological class (notice that by definition, the topological classes are transitive: if materials 1 and 2 belong to the same topological class and materials 2 and 3 belong to the same topological class then materials 1 and 3 will belong to the same topological class). The topological class is then removed from the list of topological materials in that space group and the procedure is repeated, allowing for a very fast identification of topological classes.

For each of the topological insulating compounds of Appendix[E], in each space-group, we separate the compounds into topological classes: two compounds whose band structures are related by addition/subtraction of characters of EBR's are considered equivalent (in strong topology). These are akin to topological classes, and the information is provided in a table. We classify all the topological compounds into different classes. Each class should have all the same symmetry indicators.

The equivalence classes in Appendix[G] and Appendix[H]. elevate the equivalence classes of compounds to topological indices, and we propose that the topological indices should be understood physically and mathematically as equivalence classes of compounds modulo EBRs, whose protection is given by the symmetry difference between the space-group and the largest space-subgroup where the materials becomes trivial. The number of equivalence classes are topological indices. In Appendix[Q] we show directly, by computing the topological indices [1, 33, 34] of all our materials, that these EBR equivalence classes, clearly implicit in the work of Bradlyn *et al.* [13], for every group, correspond to the topological indices of [1, 33, 34]. The materials in a given EBR equivalence class have the same set of topological indices. Materials in different EBR equivalence classes must differ by at least one topological index.

Tables [XV] and [XVI] (or columns 6 and 7 of Table XIV) provide, for each symmetry group, the topological classes of each of the compounds in Table XIV, with and without Time-Reversal. Per *each* group, the compounds listen in the same class exhibit the same topology and can be obtained from each other by adding and subtracting atomic limits in that SG with TR (Table [XV] or column 6 of Table XIV) and without TR (Table [XVI] or column 7 of Table XIV). Compounds in the same topological class in Tables [XV] and [XVI] have the same minimal topological group in columns 6 and 7 of Table [XIV]. Compounds in the same topological class in Table [XV] will have the same topological index which we also provide in column 10 of Table [XIV] of Appendix[Q].

## Appendix I: Enforced Semimetals To Topological Insulators: Fate of Enforced Semimetals (ES) Upon Symmetry Breaking Perturbations

In this Appendix, we provide another set of useful (theoretically and experimentally) topological data, this time for semimetals. For every enforced semimetal of Appendix[E] we find the possible insulating phases that it can transition to upon symmetry breaking perturbations. We provide the information of whether these insulating phases are topological or not, their symmetry protection, and place them into topological classes. This information is important for experimentalists asking what kind of perturbations can we add to a semimetal to turn it into a topological insulator. The archetypical example of this phenomenon is HgTe, which we discuss in the next Appendix[J], because it has degeneracy at the Fermi level and is classified as an ESFD; however the general idea is similar. For ES materials, it is likely that, even if adding a certain symmetry-breaking perturbation results in a topological insulator, the material will be a compensated metal. It will have an insulating filling, but with bands crossing back and forth the Fermi level due to band dispersion - and not a clean insulator. However, the symmetry breaking of ESFD materials, if it decomposes the high-degeneracy at the Fermi level into new irreps of the lower symmetry subgroup, will most likely lead to a cleaner insulator. This is the case of HgTe.

For all the enforced semimetals of Appendix[E] we provide, in columns 8 and 9 of Tables XIV, the largest subgroup (with and without time-reversal) of the space group (Max. SubSG) for which the material satisfies the compatibility relations and hence allows an insulating filling. For this insulating filling, we check if the material is topological by performing the test for insulators in Appendix[D]: we check if the material is a linear combinations of EBRs in the maximal subgroup and give, in a column a yes/no answer. When more than one maximal subgroup exists, we check for all maximal subgroups, and give a "yes/no(Max. SubSG)" answer.

If the ES with symmetry breaking term is identified as a topological insulator, we further provide the symmetries that stabilize the topology of the compound and the topological class in which the material sits, i.e. Appendices[G] and [H] as well as the topological indices in Appendix[Q], column 10 of Table[XIV] for the TI's that resulted from ES.

## Appendix J: Enforced Semimetals With Fermi Degeneracy To Topological Insulators: Fate of Enforced Semimetals With Fermi Degeneracy Upon Symmetry Breaking Perturbations

In this appendix we perform a similar analysis as that in Appendix[I] but for the case of ESFD's. While similar as an idea, our procedure in the ESFD case differs in two important ways from the ES case. First, the insulator arising after the symmetry breaking term is applied to the ESFD is likely to have a much cleaner gap than in the similar case of the ES. This is because the system becomes insulating by lifting the degeneracy at the Fermi level in the ESFD case, whereas this is not the case for the ES systems. The second difference is that, because the high degeneracy - that needs to split in order for the system to become insulating - rests at the Fermi level, we need to pick a choice of the energy of the split irreps (if the split irreps are not identical). Alternatively, we need to pick which irrep lifts above the Fermi level after degeneracy breaking and which one sinks below. If we have degeneracy at the Fermi level at more than one maximal $k$-vector, then we could have several choices of lifting the degeneracy at each of the different $\mathbf{k}$ points, and all the different possibilities need to be analyzed separately.

The archetypal example of the ESFD to topological insulator transition upon lowering symmetry is the case of the famous bulk material HgTe. In symmetry group SG 216, this material is a semimetal. We will present a heuristic analysis where we augment the group of HgTe by inversion symmetry. The bands at the Fermi level are "inverted", with the $s$-orbitals below the $p$-orbital bands at $\Gamma$ (this is true even without spin-orbit coupling), due to an effect called lone-pair effect in chemistry (Darwin term in physics). The Fermi level sits in the middle of a 4-fold degeneracy at the Γ point, and the system is an ESFD. Upon straining the sample, a gap can be opened at the Γ point as the 4-fold degeneracy is split into $2 \oplus 2$ and the system admits an insulating filling (and is in fact an insulator if free of vacancies and doping). Irrespective of how the gap is opened, the presence of an $s$-level in the valence bands at Γ implies that the insulator is inverted and hence topological . The symmetry based analysis of HgTe reinforces the heuristic analysis above

For each of the ESFDs of Appendix[E] we provide, in a table[XIX], the largest subgroup (with time-reversal) of their space group for which the material satisfies the compatibility relations and hence allows an insulating filling. Since the degeneracy is at the Fermi level, we now have several options of how this degeneracy splits in the subgroup; the material might also have Fermi degeneracy at other points in the and the result here is a mesh of different insulating possibilities. For this insulating filling, we check if the material is topological by performing the test mentioned (for insulators) in Appendix[D]. In addition, the symmetry indicators are computed and the results are shown in Table[XIX].

Hence Appendices [I] and [J] give all the possibilities for the semimetals to become topological insulators upon lowering the symmetry.

## Appendix K: Band Structures for all the Non-Magnetic Topological Compounds In the World

In this appendix we obtain and provide the electronic band structure in the BZ for all the compounds in Appendix[E]. The materials are indexed by their TCQ number defined in Appendix[F], so that fast cross-referencing with the tables of Appendices[F], [G], [H], [I], [J] is possible.

Since this appendix represents the result of a very involved and tedious task, it occupies a gargantuan amount of time and space in the manuscript. We also will provide all the plots on the BCS. This gargantuan task allows for determination of the “cleanest” topological insulators and semimetals, and is a treasure trove for experimentalists looking for materials to grow and perform experiments on.

We also provide benchmark parameters for the time took to calculate all the bands on the Halle and Hydra supercomputers using in both cases 15 processors and 5Gb of memory it took for the case of a well know topological insulator such as $Bi_2Te_3$ 5 min for the first step of obtaining only the High-Symmetry points, and 15 minutes for the band structure calculation.

The band structure plots take up more than 500 pages of space at the end of the paper, after the tables giving the characters of all bands below the Fermi level. Each plot gives the stoichiometric formula of the material, its TQC number for fast referencing with other tables in the paper, its space group, the bands along high-symmetry lines in the linking the maximal **k**-vectors (usual band-structure plot), and whether the system is a SEBR, NLC, ES and ESFD. To check the band structures of compounds in a space group, click FIG in the next table of contents [U].

## Appendix L: Wilson Loop Analysis of Several Selected Compounds

In this Appendix, we pick an infinitesimal sample of several of the best (as defined by cleanest gap and band structure) topological compounds presented, and we perform an in-depth symmetry analysis including Wilson loops and other holonomy arguments to re-confirm their topological nature. We have obtained, for the first time, regular time-reversal symmetric TI's without inversion symmetry whose indices can be found by a roto-inversion, insulators with strong topology but which are not regular TI's or HOTI's, new topological insulator materials and many other classes of materials.

We now shortly present the Wilson loop methods used. Invented by some of the authors [28, 38], [29, 31], these methods are now standard workhorses for discovering the topology of materials. We also point out, that, in SG1, or in any of the other groups in which the sets of bands that satisfy the compatibility relations can always be decomposed in EBRs, Wilson loops are the *only* methods that can determine the topology of materials. In several other groups, such as the group of the Hourglass Fermion compound KHgSb (SG 194) [30], where simple indices such as the Fu-Kane index did not exist before the development of the EBR theory (where now an hourglass index exists due to [1]), the Wilson loops provided the only way to identify the hourglass phases. In Appendix[P] we provide a new, complete method for defining topological phases by “EBR equivalence classes”, implicit in the work of Bradlyn *et al.* [13], which is identical to the topological index method of [1].

The Wilson line along a contour $\mathcal{C}$ in the BZ, starting at $\mathbf{k}$ and ending at $\mathbf{k}'$ is defined as

$$W_{\mathbf{k}' \xleftarrow[\mathcal{C}]{contour} \mathbf{k}} = P e^{i \int_{\mathcal{C}} \vec{A}(\mathbf{k}) \cdot \mathbf{d}\tilde{\mathbf{k}}} \tag{L1}$$

| Space group number | 2 | 81 | 82 | 83 | 87 | 143 | 174 |
|---|---|---|---|---|---|---|---|
| Space group symbol | P-1 | P-4 | I-4 | P4/m | I4/m | P3 | P-6 |
| Generators | -1 | -4 | -4 | 4,m | 4,m | 3 | -6 |

TABLE II. The list of the space groups, which are shown in the "SubSG" column of the Table XV, and their generators. "n" denotes the $2\pi/n$ rotational symmetry. "-n" denotes the combined symmetry of inversion symmetry and "n". In these notions, "m" is "-2".

where $\vec{A}^{mn}(\mathbf{k}) = \langle \mathbf{u^m}(\mathbf{k})| \partial_{\tilde{\mathbf{k}}} |\mathbf{u^n}(\mathbf{k})\rangle$ is the Non-abelian Berry potential, and $P$ represents the path-ordering operator over the contour $\mathcal{C}$. The indices $m, n$ run over the number of bands whose Wilson line we compute - which usually are the occupied bands, but do not need be. $|u^n(\mathbf{k})\rangle$ is the electrons $n$'th eigenstate at momentum $\mathbf{k}$. Although Eq[L1] is provides the interpretation of the Wilson line as parallel transport, its form is not the one that is usually implemented numerically. In [28, 29, 31, 38] it was shown how to implement the Wilson lines in a numerically efficient fashion. We adopt this approach here. In general, the Wilson line above is gauge covariant, and its eigenvalues depend on the starting point of the loop.

If the contour $\mathcal{C}$ is closed, the Wilson line in Eq[L1] becomes a Wilson loop, its eigenvalues are now gauge independent, and do not depend on the starting point (base point) of the loop. The eigenvalues and their behavior as the contour $\mathcal{C}$ is varied in the BZ have been used as unambiguous proof for topological behavior. In a 2 dimensional plane, the contour $\mathcal{C}$ can then be chosen to be a large loop along $k_y$ in the BZ, at constant $k_x$: $\mathcal{C} = (k_x, 0) \to (k_x, 2\pi)$. Then, if we define the unitary Wilson loop at each $k_x$ as the exponential of a Hermitian Wilson Hamiltonian

$$W_{(k_x,2\pi)\xleftarrow[\mathcal{C}]{contour}(k_x,0)} = \exp(iH_W(k_x)) \tag{L2}$$

then the Hamiltonian $H_W(k_x)$ is supposed to be topologically equivalent to that of the *edge* Hamiltonian with open boundary conditions in the $y$ direction (the direction in which the Wilson loop was taken). For a quantum Hall state, the eigenvalues wind an integer number $C$ (equal to the Hall conductance, in units of $e^2/h$) as $k_x$ is varied around its $1D$ BZ, and for a quantum-spin Hall state, there is a Kramers pair winding in different directions [28, 38]. More complex Wilson loops, where $\mathcal{C}$ is a more complex contour, sometimes bent [29] have been shown to reveal more subtle topology. Wilson loops have also been used to determine the different topologies in non-symmorphic topological insulators [30, 39], as well as in determining the topology of Fermi surfaces around Weyl and Dirac fermions [40, 41] and nodal lines.

In the examples below, we use the Wilson loops to characterize a infinitesimal set of our new materials. We provide informations on the contour of the Wilson loop and their eigenvalue flow. While, of course, confirming the topological nature of our compounds, these Wilson loops also give information about what kind of edge states we would see in a slab calculation.

### 1. Example of topological insulators protected by inversion symmetry

By the symmetry analysis presented in Appendix[G], we can output the minimum space groups that still leave nontrivial materials nontrivial. From this list, we immediately conclude the important physical information of which symmetries protect nontrivial topology of the materials. For those materials with minimum nontrivial space group 2 (P-1) in Table[II], we can conclude that they host nonzero invariants based on the inversion eigenvalues. The same conclusion can be obtained from Appendix[Q] where we compute the topological indices of these materials Since the EBRs always have 4 pairs of Kramers pairs in the BZ, and anything related by adding or subtracting EBRs is considered to be equivalent, we find that with inversion symmetry the index can be a combination of $Z_4$ and $Z_2$. Single band inversions signify strong topological insulators, while double band inversions signify the simplest version of HOTIs.

Scanning the database, we find a lot of inversion symmetry-protected insulators. One can easily check the $Z_4$ invariant, in Table[XVIII]. Beside the Fu-Kane ($Z_2 = 1$) topological insulators, we obtain many ($Z_4 = 2$) HOTI. These indices are a way of relating the current results with well known indices in the literature, but provide less physical insight than the EBRs. These two types of insulators are different as they cannot be related by addition or subtraction of EBR's of the group $P\bar{1}$ - the group with only inversion symmetry, as per our results in Appendix[G]. We stress again: for a symmetry group, two compounds are in two distinct topological classes of that group if they cannot be related by adding or subtracting EBRs. The EBR's equivalence classes should be seen as the indices of topological insulators.

As an example, we pick $BaMn_6S_8$ and $Ba_3Cd_2Sb_4$ - new, previously unknown, topological insulators. Their EBR

analysis, characters and many other data was presented in Appendix[F]. We present the link with the topological indices in Table[III].

| TRIM | Γ | X | Y | Z | S | T | U | R | $(\nu_0;\nu_1\nu_2\nu_3)$ | $Z_4$ |
|---|---|---|---|---|---|---|---|---|---|---|
| $BaMo_6S_8$ | 17 | 17 | 17 | 17 | 18 | 18 | 18 | 17 | (1;000) | 3 |
| $Ba_3Cd_2Sb_4$ | 5 | 6 | 6 | 5 | 5 | 5 | 5 | 5 | (0;110) | 2 |

TABLE III. The total number of time-reversal pairs of parity -1 bands at eight TRIM points. $Z_2$ and $Z_4$ invariants are computed. The momenta are Γ is (0,0,0); X is ($\frac{1}{2}$,0,0); Y is (0,$\frac{1}{2}$,0); Z is (0,0,$\frac{1}{2}$); S is ($\frac{1}{2}$,$\frac{1}{2}$,0); T is (0,$\frac{1}{2}$,$\frac{1}{2}$); U is ($\frac{1}{2}$,0,$\frac{1}{2}$); and R is ($\frac{1}{2}$,$\frac{1}{2}$,$\frac{1}{2}$). (0/$\frac{1}{2}$,0/$\frac{1}{2}$,0/$\frac{1}{2}$) are given in units of the primitive reciprocal vectors.

For those materials with minimum nontrivial space group not containing inversion (i.e. $n \neq 2$ in Table(II)), we can conclude that they must have zero $Z_2$ and $Z_4$ invariants based on the inversion eigenvalues, when their original space group has inversion symmetry. However, this does *not* mean that they cannot be a regular Fu-Kane topological insulator - which needs no special symmetries to exist, see comments in Appendix[G].

We would now like to show interesting topological insulator examples where their nontrivial topology is not protected (characterized) by inversion symmetry.

### 2. Examples of topological insulators not protected by inversion

The symmetry generator of symmetry groups 81(82) $P\bar{4}(I\bar{4})$ is -4, which is the combined symmetry of inversion and the 4-fold rotational symmetry (see Table[II]). We now give the example of $GeTa_3$: a topological insulator in SG 82 without inversion symmetry, whose nontrivial $Z_2$ topological insulator nature is fixed by $S4$ eigenvalues. From Appendix[F] we find that it belongs to the kind "NLC" - i.e. it is a compound that can have both insulating trivial and insulating nontrivial topological phases (as opposed to the SEBR compounds which can have only insulating non-trivial or semi-metallic phases). We computed the Wilson loop spectrum of this material along two parallel time-reversal-invariant planes (*i.e.* $k_3 = 0, \pi$ ). $k_1, k_2, k_3$ are given in units of the primitive reciprocal vectors over $2\pi$. The Wilson loops are computed in the $k_2$ from 0 to $2\pi$ direction. We plotted the Wilson bands as a function of $k_1$ in both the $k_3 = 0$ and the $k_3 = \pi$ planes. We present them in 1(b) and (c). We find that one plane has the Wilson loop winding of a QSH state, while the other plane is trivial. We conclude that the material is a strong TI, identified as nontrivial by our theory.

HgS is a "NLC" TI in SG 216, a group that also does not have inversion symmetry. Our program in Appendix[G] outputs the minimum nontrivial space group of this material as SG 82. SG 216 has a cubic face-centered structure. The three primitive reciprocal vectors are (-2π,2π,2π), (2π,-2π,2π), (2π,2π,-2π) in units of $1/a$. The Wilson loops are computed in the $y$ direction, going from $(k_x, 0, 0(\pi))$ to $(k_x, 4\pi, 0(\pi))$ in Fig[2]. From those two Wilson loop spectra, we conclude that HgS belongs to a strong TI.

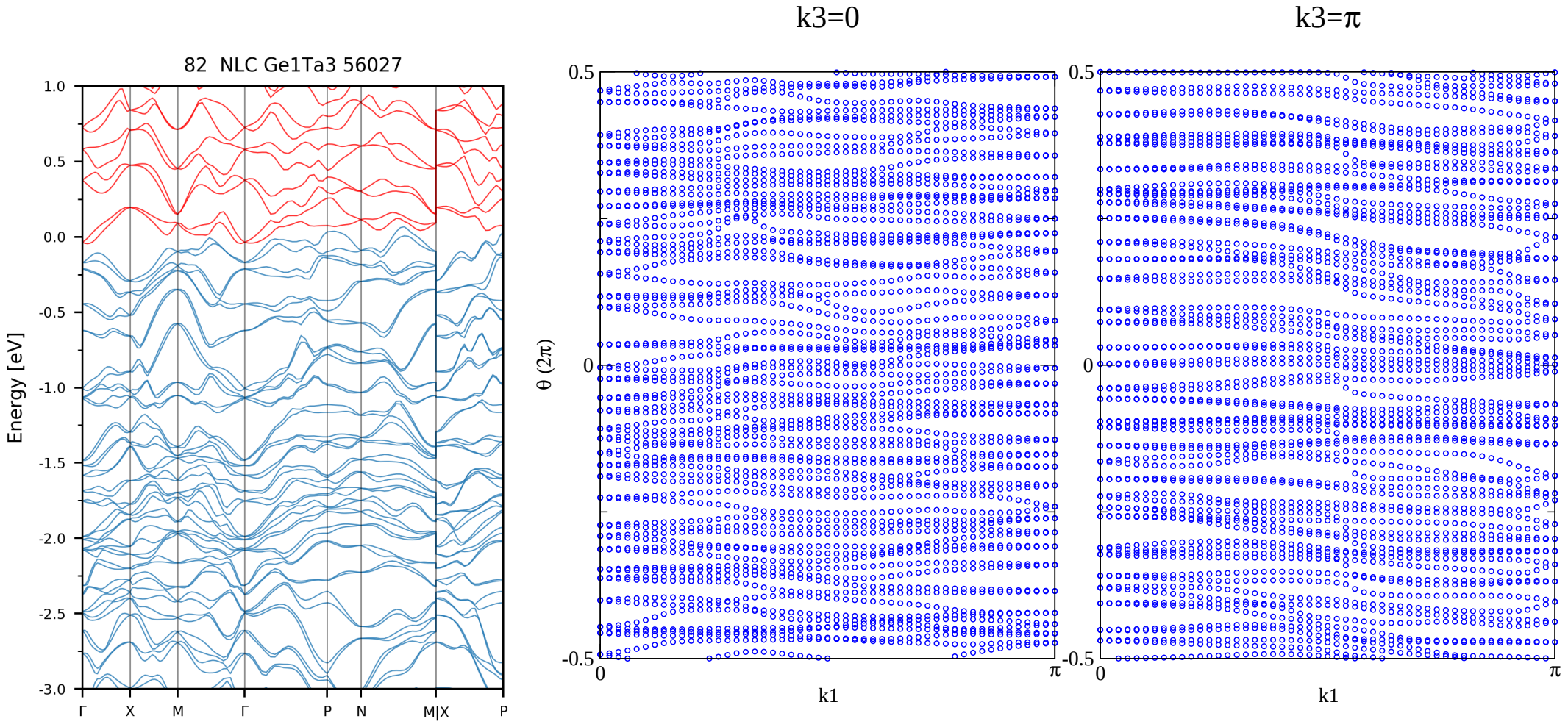

FIG. 1. (Color online) The band structures of $GeTa_3$. Its space groups does not have inversion symmetry. The nontrivial non-trivial topology are confirmed by the Wilson loop spectrum of two parallel high-symmetry planes in momentum space.

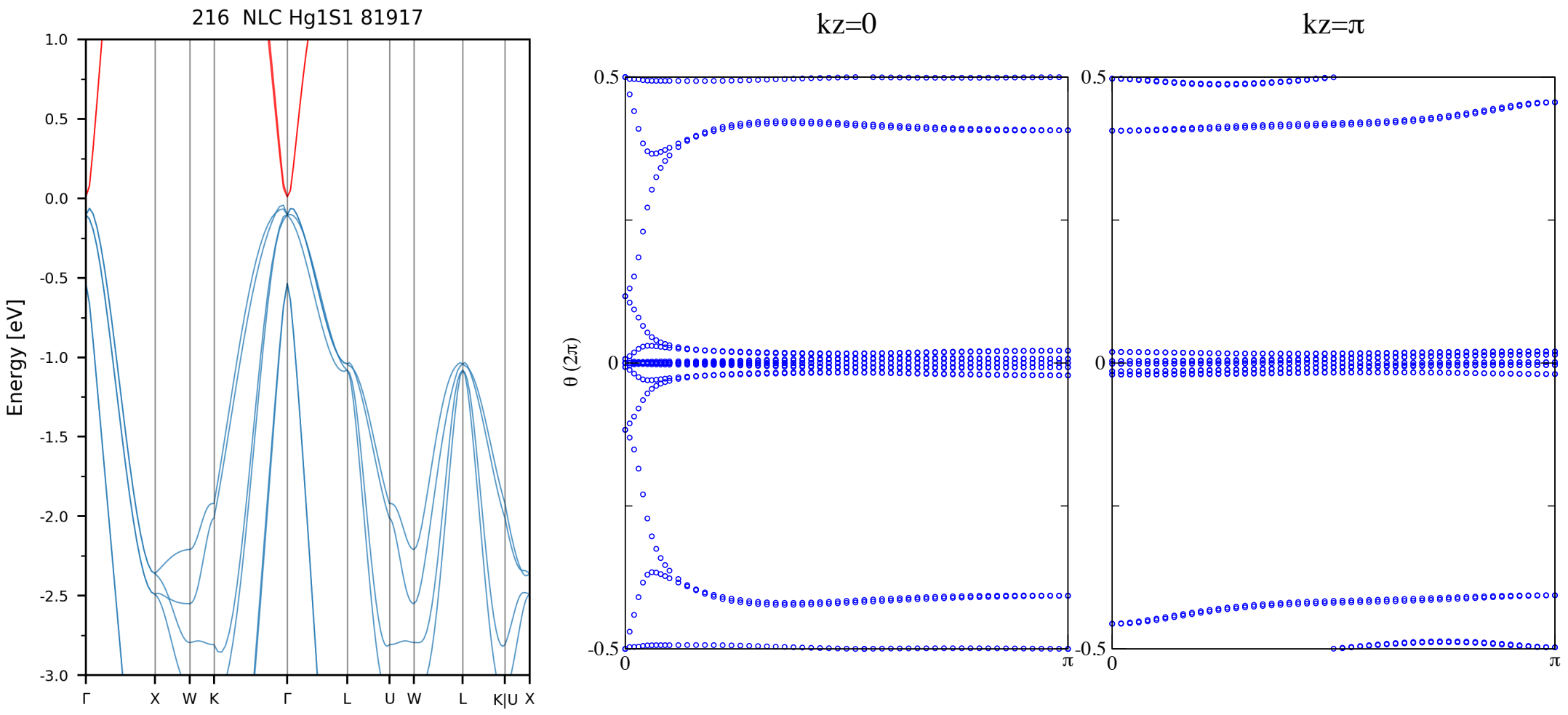

FIG. 2. (Color online) The band structure HgS. Its space group does not have inversion symmetry. The nontrivial non-trivial topology are confirmed by the Wilson loop spectrum of two parallel planes high-symmetry in momentum space.

$K_2Mg_5Sn_3$ and $Na_2OSb_2Ti_2$ are "SEBR" TIs in SG 139, which has a tetragonal body-centered structure (with lattice parameters $a$ and $c$). This means that these materials, as long as no-major band re-arrangement happens around the Fermi surface, can only have two phases: topological insulator or enforced semimetal. They cannot be trivial phases. Even though they have inversion symmetry, their nontrivial topology is not characterized by inversion symmetry. We confirm that, if we focus only on inversion eigenvalues, the characters of the compounds can be written as a linear superposition of EBR's, or, alternatively, that the $Z_2$ and $Z_4$ indices are *zero*. This is shown in Table IV. We provide their nonzero index in Appendix[Q].

However, from the Wilson loop analysis, we discover the topology that matches our theoretical analysis in Appendix[F]. The three primitive reciprocal lattice vectors are (0,2π,2π), (2π,0,2π), (2π,2π,0) in units of $1/a$, $1/a$, $1/c$. The minimum nontrivial SG of these compounds is SG 87 (I4/m) and it has a topological index $z_8$=4 (see Appendix Q and Table[XIV] ). Then we compute the Wilson loops in three non-equivalent planes: $k_z = 0$, $k_x = 0$, and $k_x = k_y$ planes. For $K_2Mg_5Sn_3$, the nontrivial mirror plane is the $k_x = 0$ plane, which is a $M_x$-invariant plane. The Wilson loops are computed in the $k_z$ direction, going from $(0, k_y, 0)$ and $(0, k_y, 4\pi)$. The results are shown in Fig. 3(b), where the red(blue) circles indicate the $M_x = +i(-i)$ sector. It shows the mirror Chern number is 2. For $Na_2OSb_2Ti_2$, the nontrivial mirror plane is the $k_z = 0$ plane, which is a $M_z$-invariant plane. The Wilson loops are computed in the $k_y$ direction, going from $(k_x, 0, 0)$ and $(k_x, 4\pi, 0)$. The results are shown in Fig. 3(d), where the red(blue) circles indicate the $M_z = +i(-i)$ sector. The result shows that the mirror Chern number is 4.

| TRIM | Γ | X | Y | Z | S | T | U | R | $(\nu_0;\nu_1\nu_2\nu_3)$ | $Z_4$ |
|---|---|---|---|---|---|---|---|---|---|---|
| $K_2Mg_5Sn_3$ | 4 | 6 | 6 | 6 | 6 | 6 | 6 | 4 | (0;000) | 0 |
| $Na_2OSb_2Ti_2$ | 5 | 3 | 3 | 7 | 7 | 3 | 3 | 5 | (0;000) | 0 |
| LaI | 0 | 0 | 0 | 1 | 0 | 1 | 1 | 1 | (0;000) | 0 |
| PbSe | 3 | 3 | 3 | 3 | 3 | 3 | 3 | 3 | (0;000) | 0 |

TABLE IV. Inversion topological indices of three of the materials. The total number of time-reversal pairs of parity -1 bands at eight TRIM points. $Z_2$ and $Z_4$ invariants are computed.

LaI is a "SEBR" TI in SG 194, which has a hexagonal structure (with lattice parameters $a$ and $c$). Even though they have inversion symmetry, their nontrivial topology is not characterized by inversion symmetry. We confirm that, if we focus only on inversion eigenvalues, the characters of the compounds can be written as a linear superposition of EBR's, or, alternatively, that the $Z_2$ and $Z_4$ indices are *zero*. This is shown in Table IV. The Wilson loop analysis confirms the nontrivial topological nature discovered in Appendix[F]. $(k_1, k_2, k_3)$ are given in units of the primitive reciprocal vectors over $2\pi$. The minimum nontrivial SG of them is SG 174 (I-6). Then we compute the Wilson loops in some planes. The results are presented in Fig. 4. Both planes show the trivial spectrum. Since we know that this compound is topological, finding a Wilson loop (perhaps on a bent manifold) whose eigenvalues flow, is still outstanding.

PbSe is a "SEBR" TI in SG 225, which has a face-centered structure (with a lattice parameter $a$). Even though they have inversion symmetry, their nontrivial topology is not characterized by inversion symmetry. We confirm that, if we focus only on inversion eigenvalues, the characters of the compounds can be written as a linear superposition of EBRs, or, alternatively, that the $Z_2$ and $Z_4$ indices are *zero*. This is shown in Table IV. The Wilson loop analysis confirms the nontrivial topological nature discovered in Appendix[F]. Then we compute the Wilson loops in the $k_x = k_y$ plane. The results are presented in Fig. 5. The Since we know that this compound is topological, finding a Wilson loop (perhaps on a bent manifold) whose eigenvalues flow, is still outstanding.

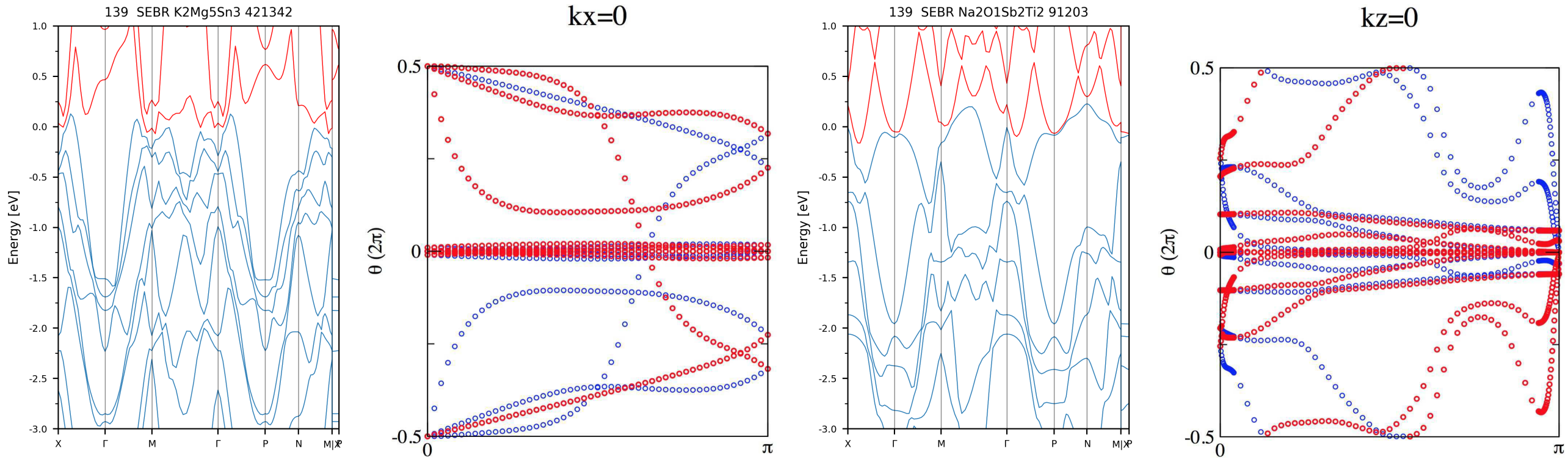

FIG. 3. (Color online) Band structures and the Wilson loop spectrum for the compounds $K_2Mg_5Sn_3$ and $Na_2OSb_2Ti_2$. Since those planes are mirror-symmetric, the Wilson bands are further separated by different color. The mirror eigenvalue $+i$ $(-i)$ is colored in blue (red). The Wilson loops show that the mirror Chern number is 2 in the $k_x = 0$ plane of $K_2Mg_5Sn_3$, and that is 4 in the $k_z = 0$ plane of $Na_2OSb_2Ti_2$.

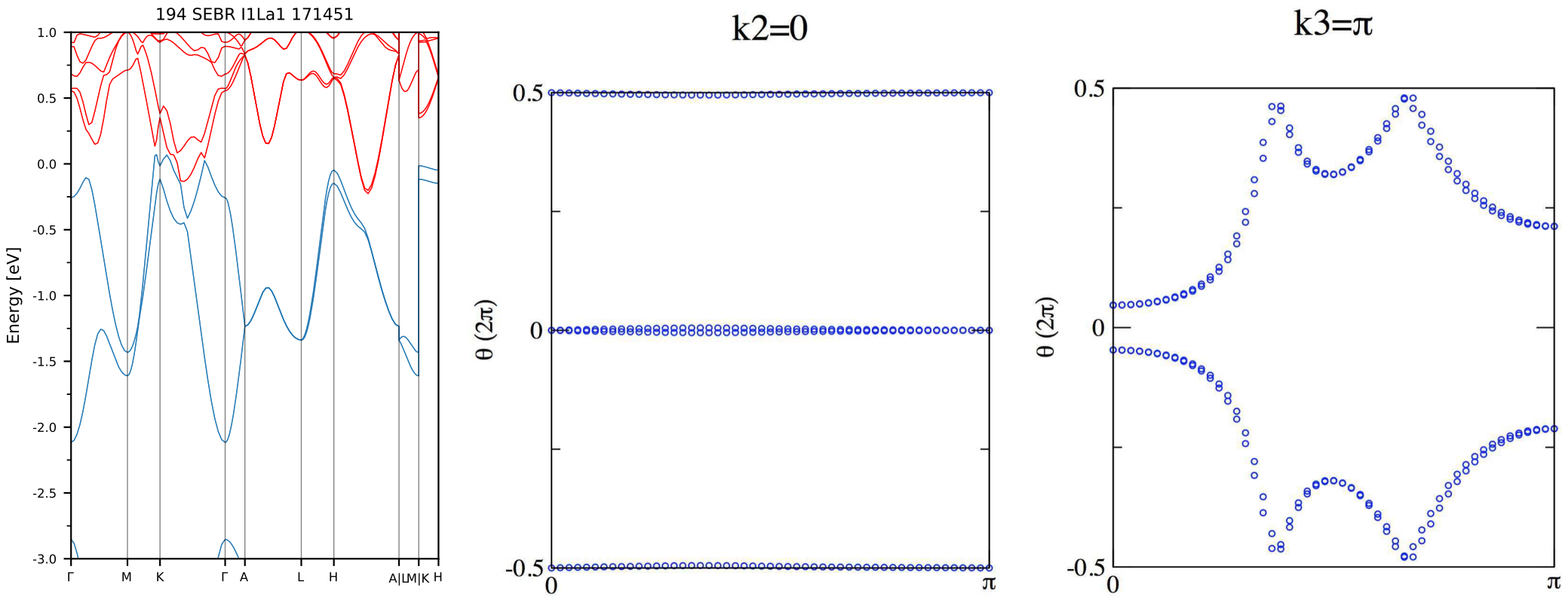

FIG. 4. (Color online) Band structures and the Wilson loop spectrum for LaI. Although we know for certain (both from the EBR analysis in Table[XIV] as well as from the topological indices in Table[XIV]) that this is an insulator with strong topology, finding the winding Wilson loop for this insulator is still an outstanding problem.

### 3. The Wilson loop spectra for MoTe2 and KHgSb

Our "high-quality" ICSD filter has filtered out a small number of known compounds which are known to be topological. We have added them back by hand. Here we present two of the well-known ones.

$\beta$-$MoTe_2$ is in space group 11. It is not expressible as a sum of EBRs of the subgroup SG(2) and hence it has a nontrivial inversion index. It is not a topological insulator - it has $Z_4 = 2$, which implies it could be a higher-order topological insulator. Without SOC, it contains a pair of monopole-charged nodal lines. The Wilson loop spectra are shown in Fig. 6, confirming the non-triviality obtained by the EBR calculation.

KHgSb is the hourglass fermion compound [30, 42], in space group 194. It is not expressible in terms of EBRs of SG194. It is expressible a sum of EBR's of the symmetry group SG(2), and hence inversion symmetry alone is not responsible for its nontrivial topology. Indeed, computing, with inversion symmetry, the topological indices, we find zero $z2w$ and $z4$ indices. However, it has other nontrivial indices: $z_{6m,0} = 2$ , $z'_{12} = 8$. We present its Wilson loop flow in Fig[7], confirming the non-triviality obtained by the EBR calculation.

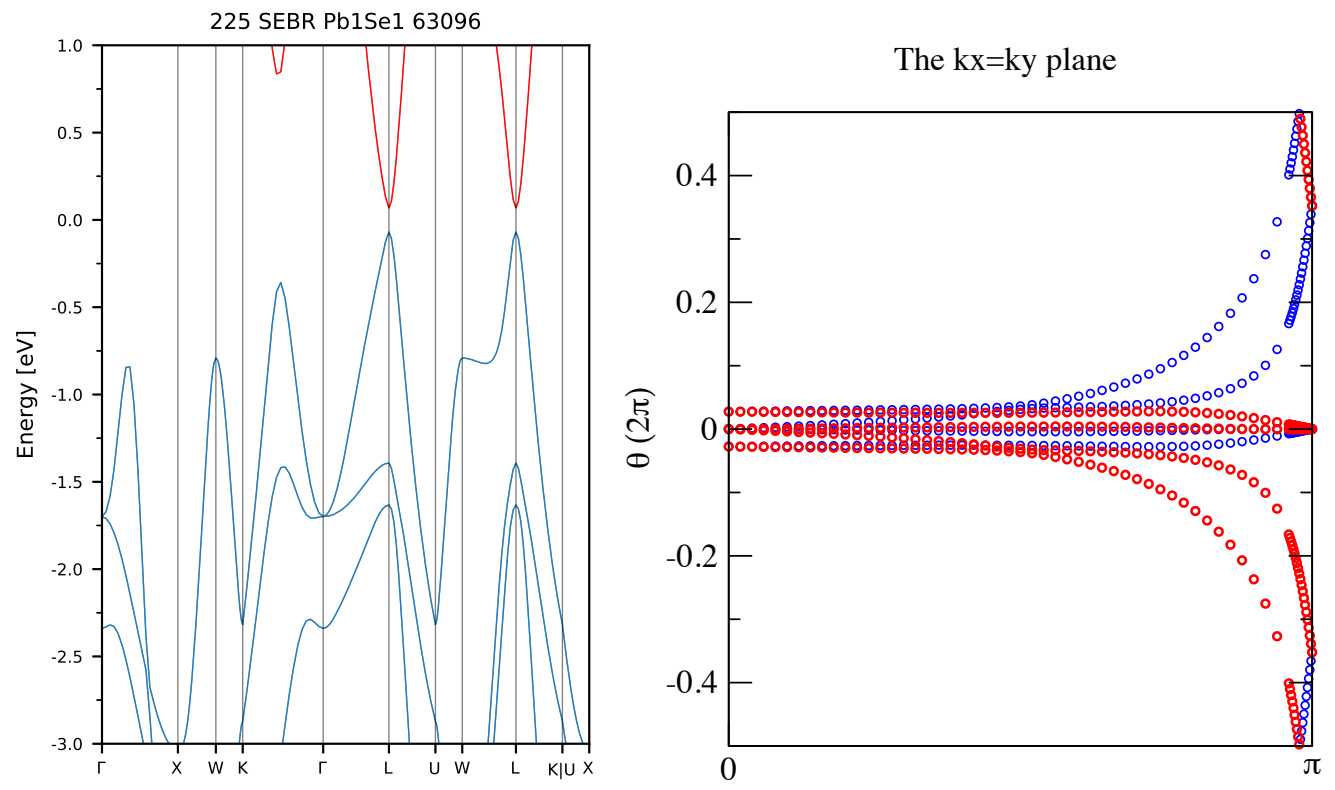

FIG. 5. (Color online) Band structures and the Wilson loop spectrum for PbSe. Since the $k_x = k_y$ plane is mirror-symmetric, the Wilson bands are further separated by different color. The mirror eigenvalue $+i$ $(-i)$ is colored in blue (red). The Wilson loops show that the mirror Chern number is 2 in the plane.

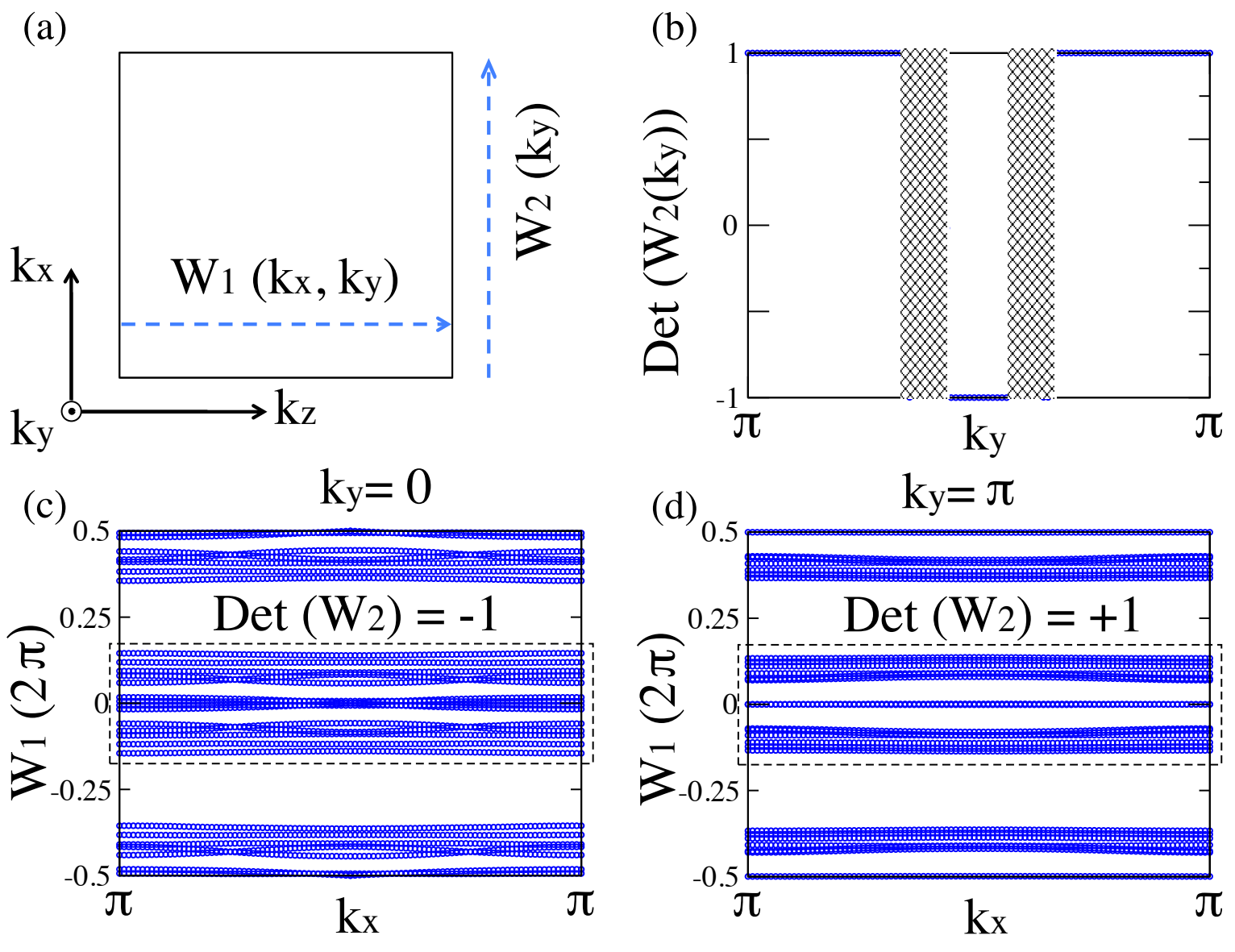

FIG. 6. (Color online) (a) Bulk Wilson loops calculated for $\beta$-$MoTe_2$ from first principles in the absence of SOC. The Wilson loop spectra shown in (c) and (d) are computed in the $k_z$-direction, they are will separated by the $\theta = \pm 0.25$ $(2\pi)$. (b)Calculating the determinant of the *nested* Wilson matrix $W_2(k_y)$ of the Wilson bands within this gap, we find that it is quantized at $\pm 1$ by the combined anti-unitary symmetry $(\mathcal{I} \times \tilde{\mathcal{T}})^2 = +1$, and jumps as it passes over a bulk monopole nodal-line (MNL), indicating that the planes above and below the MNL are topologically distinct.

### 4. Symmetry indicators for examples with no inversion indicators

It's well known that, with inversion symmetry, the time-reversal-invariant topological insulators can be easily diagnosed by either $\mathbb{Z}_{2w,j=1,2,3}$ or $\mathbb{Z}_4$ (Note Fu-Kane strong $\mathbb{Z}_2$ index is included in $\mathbb{Z}_4$). By checking the Table[XIV], we picked some examples, which have *zero* $\mathbb{Z}_{2w,j=1,2,3}$ and $\mathbb{Z}_4$ (defined by inversion symmetry), but have other *non-zero* symmetry indicators (defined in ref[1]). These examples are listed in Table[V], which are consistent with our previous Wilson-loop calculations. In addition, we listed some examples for all different symmetry indicators in Table[VI]. Their band structure are shown in Fig.8.

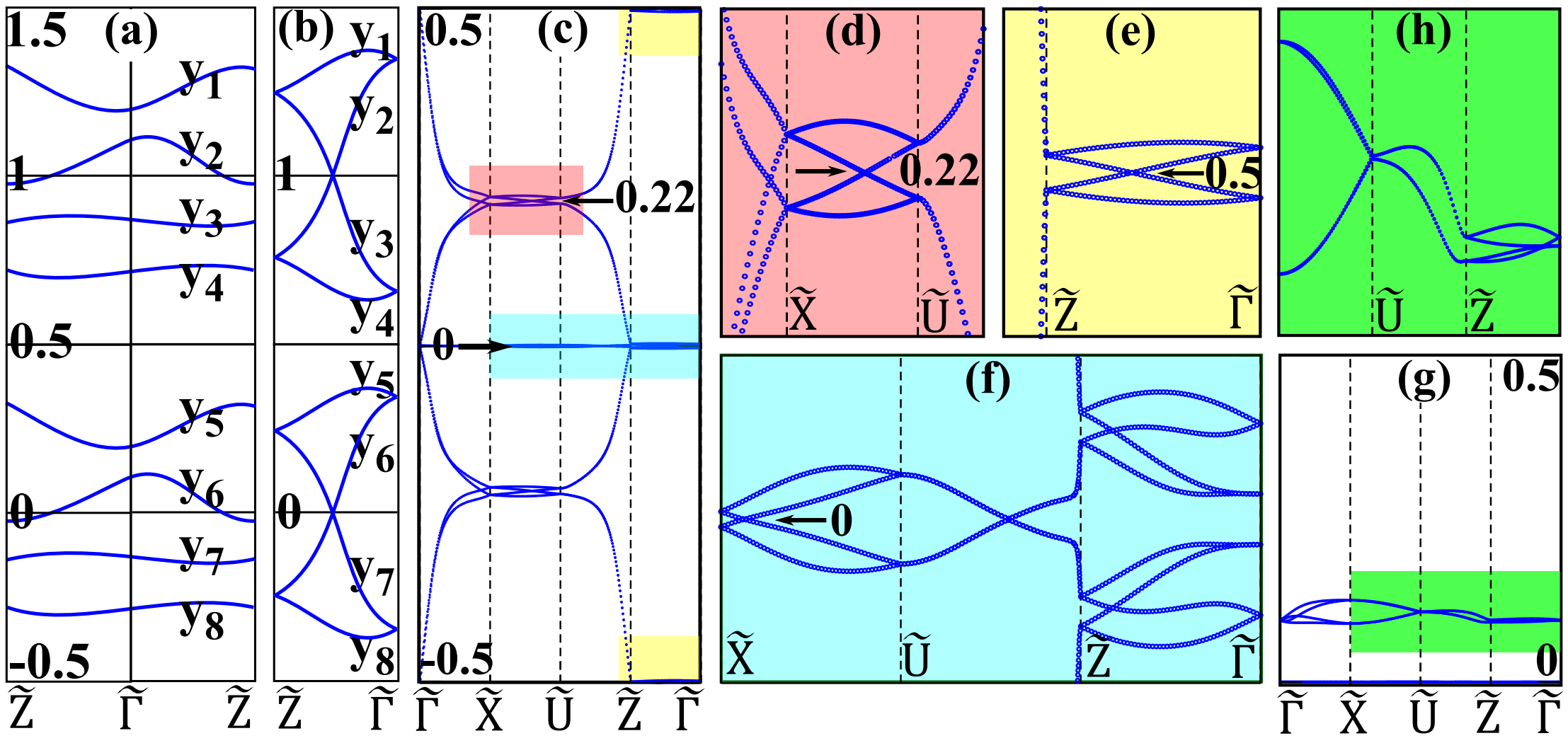

FIG. 7. (Color online) The computed $k_y$-directed Wilson loop as a function of $(k_x, k_z)$. We can see that there is a Hourglass flow along the loop $\Gamma X U Z \Gamma$.

| | | | |
|---|---|---|---|
| Ge1Ta3 | 82.2.6 | 56027 | z2=1 |
| Hg1S1 | 216.2.6 | 81917 | z2=1 |
| K2Mg3Sn3 | 139.1.60 | 421342 | z8=4 |
| Na2O1Sb2Ti2 | 139.1.40 | 91203 | z8=4 |
| I1La1 | 194.1.26 | 171451 | z6m0=2; z12prime=8 |
| Ba1Ge3Mg4 | 123.1.21 | 165631 | z8=4 |
| K3N1 | 193.1.6 | 99999 | z6m0=2; z12prime=8 |
| PbSe | 225.1.4 | 63096 | z8=4 |
| PbS | 225.1.10 | 62190 | z8=4 |

TABLE V. Computed symmetry indicators of topological insulators, which are not protected by inversion symmetry. All the symmetry indicators are defined in ref[1]. Note z2 is defined by the symmetry "-4"($S_4$).

## Appendix M: F-electron Compounds, Initially Filtered Out: Short Summary of Their Topology

In this Appendix we provide information about the "problematic" $f$-electron compounds. These materials were initially discarded from the input set of high-quality materials because they have been intricate band structures and correlation effects that cannot be accurately captured by DFT calculations. We however, after more precise calculations, include them here, albeit with less information than the high-quality compounds. We tested in the upward of 4286 $f$-electron compounds deemed to have band structure suitable for DFT-calculations. There are rare earth compounds difficult to converged in a massive run when we consider $f$ electrons in the valence bands, such as Yb. They required more detailed band structure calculation like removing the $f$ electrons and add them to the core. Out of these, we found 2306 topological compounds, a rate consistent with the percentage for non $f$-electron compounds. For some of the compounds discarded by our initial filters - such as materials containing $f$-electrons, we present very short tables/summaries in Table[XVII], but caution against fully trusting the *ab-initio* calculations in these cases.

## Appendix N: Public Service: Bilbao Crystallographic Server Codes To Check Nontrivial Topology Of Compounds

In this appendix, we provide a short description of the button functionality that we will soon add to the Bilbao Crystallographic Server, available at [21], and which allows for checking if a material is topological. At the moment this button requires, as input, the list of characters that are the combined output of VASP plus our code for computing characters at high-symmetry points, in a specific format given by the examples at [21]. The program then checks compatibility relations and EBR linear independence to output if the system is a topological insulator or a semimetal. While the user has to perform the VASP calculation by him/her/self, the code we provide on the web page takes the VASP output and computes the representation characters. The program generates a file that can be uploaded to the server to then obtain the final result (whether the material is topological or not). This includes the identification of the irreducible representations at each maximal **k**-vector and, when these irreps can be described as a linear combination of EBRs, it lists the different linear combinations. If the material has been identified as topological insulator, it is also indicated the minimal subgroup(s) for which the band structure is still topological. When the material is an ES, i.e., the bands below the Fermi level do not fulfill the compatibility relations, the program gives the maximal subgroup(s) that satisfy the compatibility relations and hence accepts an insulating filling. It then indicates if the resulting set of bands have a topological configuration or not. In the first case, it further also also given the minimal subgroup(s) (of the maximal subgroup that renders the ES to be a TI) under which the TI remains topological (the symmetries, for the same lattice, that protect the topological class).

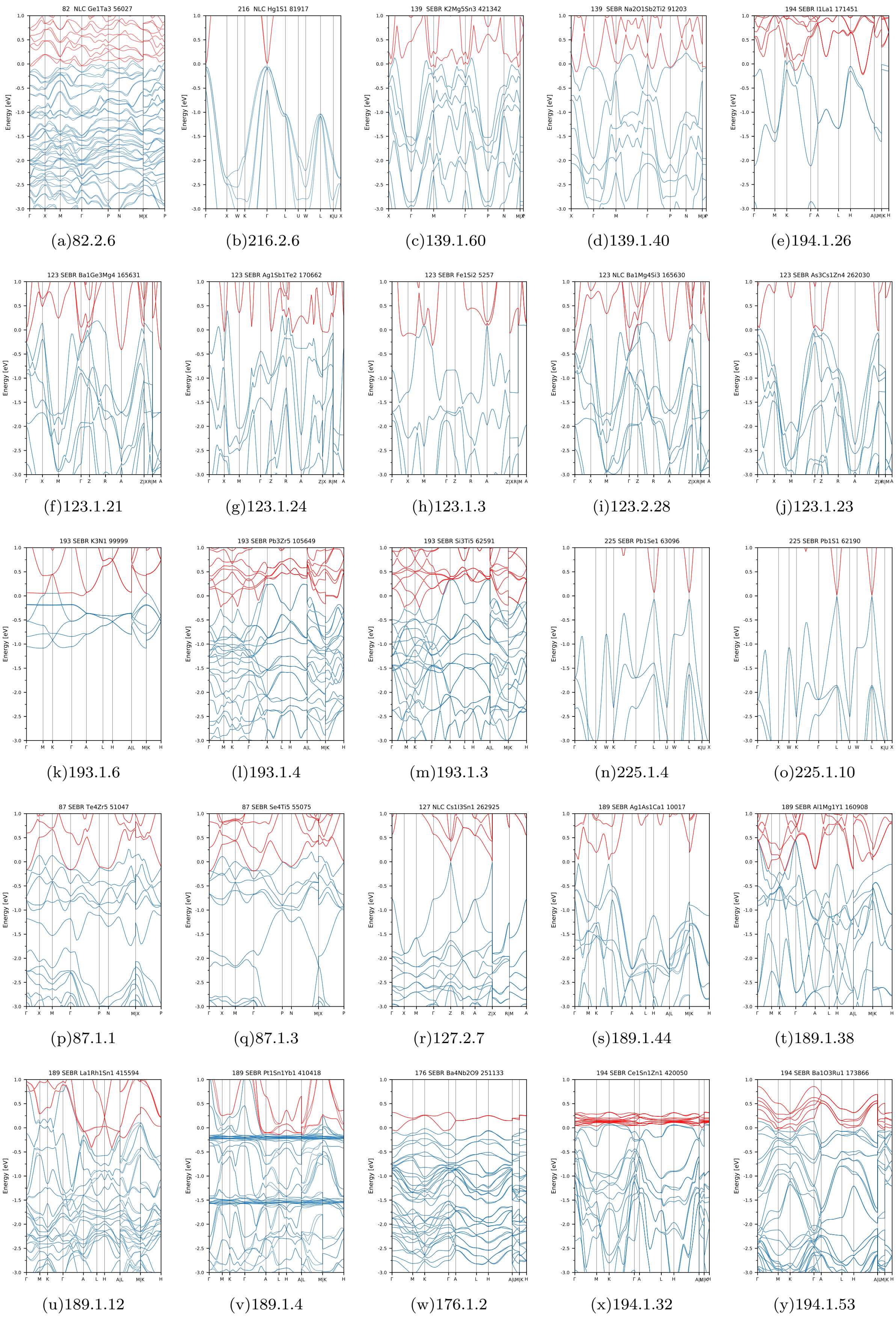

FIG. 8. (Color online) Band structures for the examples of different topolgocial symmetry indices in Table[V] and Table[VI].

| | | | | | | | |
|---|---|---|---|---|---|---|---|
| z2=1 | 82.2.6 | 56027 | Ge1Ta3 | z6m,pi=1 | - | - | - |
| z2=1 | 216.2.6 | 81917 | Hg1S1 | z6m,pi=2 | 191.1.9 | 246238 | Ba1Ga2 |
| z4m,pi=1 | 123.1.21 | 165631 | Ba1Ge3Mg4 | z6m,pi=3 | 191.1.14 | 56723 | B2Ti1 |
| z4m,pi=2 | 123.1.24 | 170662 | Ag1Sb1Te2 | z6m,pi=4 | 191.1.7 | 418614 | Ca1Zn5 |
| z4m,pi=3 | 123.1.23 | 262030 | As3Cs1Zn4 | z6m,pi=5 | 191.1.4 | 246236 | Ga2Sr1 |
| z8=1 | 87.1.1 | 51047 | Te4Zr5 | z12=1,2,3,5,6 | - | - | - |
| z8=2 | 123.1.3 | 5257 | Fe1Si2 | z12=4 | 191.1.15 | 184968 | B2Fe3U1 |
| z8=3 | 123.2.28 | 165630 | Ba1Mg4Si3 | z12=7 | 191.1.7 | 418614 | Ca1Zn5 |
| z8=4 | 225.1.4 | 63096 | PbSe | z12=8 | 191.1.14 | 56723 | B2Ti1 |
| z8=5 | 87.1.3 | 55075 | Se4Ti5 | z12=9 | 191.1.9 | 246238 | Ba1Ga2 |
| z8=6 | 123.1.23 | 262030 | As3Cs1Zn4 | z12=10 | 191.1.16 | 237660 | B1Eu1Pt4 |
| z8=7 | 127.2.7 | 262925 | Cs1I3Sn1 | z12=11 | 191.1.18 | 417869 | B1Co4Gd1 |
| z3m,0=1 | 189.1.44 | 10017 | Ca1Ag1As1 | z12prime=1 | 193.1.3 | 62591 | Si3Ti5 |
| z3m,0=2 | 189.1.38 | 160908 | Al1Mg1Y1 | z12prime=2 | 194.1.27 | 163344 | Gd1Ge1Zn1 |
| z3m,pi=1 | 189.1.12 | 415594 | La1Rh1Sn1 | z12prime=3,4,5 | - | - | - |
| z3m,pi=2 | 189.1.4 | 410418 | Pt1Sn1Yb1 | z12prime=6 | 194.1.53 | 173866 | Ba1O3Ru1 |
| z6m,0=1 | 193.1.4 | 105649 | Pb3Zr5 | z12prime=7 | 194.1.1 | 421015 | Zn1 |
| z6m,0=2 | 193.1.6 | 99999 | K3N1 | z12prime=8 | 193.1.6 | 99999 | K3N1 |
| z6m,0=3 | 194.1.32 | 420050 | Ce1Sn1Zn1 | z12prime=9 | 194.1.32 | 420050 | Ce1Sn1Zn1 |
| z6m,0=4 | - | - | - | z12prime=10 | - | - | - |
| z6m,0=5 | 176.1.2 | 251133 | Ba4Nb2O9 | z12prime=11 | 176.1.2 | 251133 | Ba4Nb2O9 |

TABLE VI. Examples for different *nonzero* symmetry indicators.

## Appendix O: Unique Method to Detect Fragile Topology

In this Appendix, we present several comments on the fragile topological phases in the materials in nature. Fragile topological insulators [15, 25, 26, 43] have been named as insulators which exhibit nontrivial topology for a particular number of bands, but which lose the topology when trivial bands are added to these numbers of bands. Since this is not always the case [44, 45], we would prefer to call them examples of "finer" topology.

One predominant set of fragile topological insulators is those for which the characters of the bands at high-symmetry points cannot be written as a linear superposition with positive coefficients of the characters of EBR's but *can* be written as a linear superposition with positive *and* negative coefficients of the characters of EBR's or, more heuristically $Top = EBR_1 - EBR_2$ where $EBR_1$ and $EBR_2$ are any sums of EBR's in the symmetry group; no EBRs in the sum $EBR_1$ can match the EBRs in the sum $EBR_2$. The intuition, not always satisfied [44, 45] is that if one adds the appropriate number of trivial atomic limit ERB's of negative coefficients to the topological bands, they would become a sum of EBRs, i.e. $Top + EBR_2 = EBR_1$.

We have investigated whether any of our topological materials can be written as a linear combination of EBRs with positive *and* negative coefficients. We found that *none* of the 26938 materials we have analyzed is a fragile phase. We have found this by brute force but also by the following, more subtle argument. For *each* compound we have characterized as not being equal to a *sum* of EBRs, we found that a strong topological index in Appendix[Q]. As EBR's have topological indices zero, this immediately tells us that there are no fragile phases in the 26938 compounds we looked at. We want to emphasize that our method is unique in finding this. Strong-topological index methods that compute strong indices would never been able to come to this conclusion as the strong indices vanish for a fragile phase.

### Appendix P: New HOTIs, and A New Proposal for the Indices of Materials with Strong Topology In Each SG

Due to the recent activity in higher-order topological insulators (HOTI) in this Appendix we compute several previously known indices of well-known insulators - such as HOTI's with inversion symmetry. In Ref[46–49], it was shown that a HOTI can be understood as a doubly inverted system. For example, last year [46] first presented a physical picture of a HOTI as two inverted $Z_2$ topological insulators, each in two distinct rotational subspaces and noted that the rotation is not necessary to protect the state; this introduced the concept of double band inversion. The same authors predicted the first HOTI, in Bismuth [47] Then Ref [48, 49] wrote down a physical index which is that of a double band inversion. We present the compounds we found for these simplest HOTIs in a table XVIII.

We want to stress that while the $Z_2$ invariants are well-known to describe topological classes, it is actually far less clear how to write down a simple index expression for more complicated topological insulators. In general, the invariants that characterize the material have complicated expressions in terms of products or sums of eigenvalues (see for ex KHgSb [30], which has a rather complicated "index" to compute [1], but whose topology can be obtained in a physically transparent fashion by Wilson loop winding [30]). In general, without symmetry, Berry phases (or Wilson loops) around (sometimes) intricate surfaces in momentum space are needed to characterize the material.

Alternatively, through our work in Appendix[H], we propose that the distinct EBR-equivalent classes as an index for the materials in one group. Two materials which differ by addition and subtraction of EBRs belong to the same topological class.

Hence if

$$\textbf{Material A} + \sum_1 EBR's = \textbf{Material B} + \sum_2 EBR's, \tag{P1}$$

then Material A and Material B are in the same strong topological class.

If

$$\textbf{Material A} + \sum_1 EBR's \neq \textbf{Material B} + \sum_2 EBR's, \tag{P2}$$

then Material A and Material B are in different strong topological classes. The number of equivalence classes - different types of Materials not related to one another by Eq[P1] (see Appendix[H]) gives the topological classification of the symmetry group. For each symmetry group, the topological classes of each of the compounds in Table XIV, with and without Time-Reversal are provided in Tables [XV] and [XVI] Compounds in the same topological class inTables [XV] and [XVI] have the same minimal topological group in column 6 and 7 of Table [XIV]. Compounds in the same topological class in Table [XV] will have the same topological index which we also provide in column 10 of Table[XIV] of Appendix [Q]. For a particular set of EBR equivalence classes, we can ask what are the symmetries that protect the topological class (see Appendix[G]).

### Appendix Q: All Topological Indices For *All* Materials In Every Group, Equivalence With the EBR Classification

In this Appendix we provide complete information of the topological indices of *all* of the materials in our search. In Ref [1] [33] [34] the authors used the ERB theory and data of [13] and data on the BCS server to compute what seems to be a complete set of topological indices for each material. This set of topological indices should be equivalent to our EBR equivalence classes. We have computed the topological indices for *all* of our materials and we present them in column 10 of Table[XIV]. We find perfect matching with our EBR equivalence class (two materials in different EBR equivalence classes in the same group have different topological indices). This provides the user with another, more conventional, set of information about the compounds we presented. Furthermore, In column 10 of Table[XIV] and [XIX], we provide the topological indices for the topological insulators that result from breaking the symmetries of topological semimetals in Appendices[I] and [J]

We again want to caution that having a certain topological index nonzero does *not* immediately mean (with the exception of well known cases) that the material is an insulator. It could be a Weyl semimetal with Weyl points at generic $\mathbf{k}$'s in the BZ. There exist a set of rules for that to happen (for example, it cannot happen with both inversion and time-reversal), and for systems with zero SOC, Ref[33] characterized the topological indices that reveal the presence of Weyl points at generic symmetry points - the simplest of which, pointed out in [31], is reviewed below. However, with SOC, such a work has not yet been performed.

We now review the example pointed out in [31, 33], where a system can satisfy compatibility, not be a linear combination of EBRs, but still be a semimetal. Even though this example is for the spinless case, similar examples can exist for spinful systems. Consider a single occupied band in symmetry group $P-1$ with inversion symmetry and broken time-reversal. As there are no high-symmetry lines and planes in this group, *any* inversion eigenvalues at the maximal **k**-vectors in the BZ satisfy the compatibility relations (the little group of a generic **k** point is identity). Hence in the absence of any new criterion, bands with the same number of occupied energy levels at inversion-invariant **k** vectors would be considered at "insulating" filling. Consider a band with inversion eigenvalues $(-+++++++)$ at the 8 inversion symmetric points. Pick the $-$ eigenvalue to be at $\Gamma$ (all other configurations would give the same conclusion). This band is clearly *not* an EBR: all EBRs in groups with inversion symmetry have $4n$, $n \in (0,1,2)$ negative inversion eigenvalues. So it would be classified as a TI. In reality, it is a Weyl semimetal, with Weyl nodes at generic points in the BZ - which hence cannot be seen purely for the linear dependence EBR argument made so far. The $k_z = 0$ plane of the system has an odd Chern number (the product of inversion eigenvalues is the parity of the Chern number [31]), while the $k_z = \pi$ plane has to have an even Chern number (perhaps 0). Hence as a function of the parameter $k_z$ the system has to be a gapless semimetal, but since at generic $k_z$ there is no symmetry, the Weyl nodes can be at generic points in the BZ.

## Appendix R: Chemistry Remarks About the Identified Topological Materials; Several "Low Quality" Compounds

In Table[XX] we provide some chemist advice about every compound. We associate to the compound a letter. A: compound is perfect and can be made easily; B: compound looks good and feasible but maybe no stable in air; C: compound can be AFM Ferro or MOTT. D: compound has to be check - DFT problem! Non transition metal radical such as N3-, or (O2)2- or (N2O4) or La+1 or La+2. E: compound is radio active and mostly magnetic - spin polarized calculation necessary.

In addition, for future DFT and DMFT studies of our compounds, all 3d transition metal atoms which have left over d electrons have to be treated spin polarized - the non0spin polarized results have to be treated with caution and could be wrong. The d-electron materials one needs to take special care of are V+3, V+4, Ti+3, Cr3+, Co2+, Ni2+, Cu2+, Fe3+ Fe2+ Ce3+. All this compounds are AFM especially if they are oxidized with O, N or Halogenides Cl, F, Br, etc. All rare earth atoms such as Ho and Ce must be treated as spin polarized.

We have also analyzed a small set (around 600 compounds) of the data in the "low-quality" compounds of the ICSD database. This data can be obtained by selecting the "low quality" button on the ICSD database website. We present the results in Table[XC] ; we find a similar proportion of materials (around 27%) exhibit topological properties in the "low-quality" set of data as in the high-quality one.

## Appendix S: Consistency checks for Topological Data

In this section we present a series of consistency checks that we have performed on our data in order to guarantee its correctness.

### 1. Impossible ESFD in Electron Number/Symmetry Group combinations

Many symmetry groups do not allow enforced semimetals with Fermi degeneracy for a given number of electrons. To give an example, in SG 2 (inversion) with time-reversal, every band has degeneracy 2. As such, materials with even electron numbers *cannot* have the Fermi level stay in the middle of a degenerate level at a high symmetry point and any material in this SG cannot be an ESFD. More complicated, SG 55 has only 4-fold representations at high-symmetry points R,S,T,U,X,Y (and 2-fold at other high-symmetry points). As a result, any material with electron number divisible by 4 cannot be an ESFD. We have calculated all the pairs Electron Number/SG for which ESFDs are impossible. We present them in Table[VII], which we call "Impossible ESFD".

We have checked that none of our ESFD materials have the Number of electron/SG combinations in the "Impossible ESFD" Table[VII].

| Electron Number Multiple of | Symmetry Groups (SG) |
|---|---|
| 2 | 1- 10,12, 16, 17, 20-25, 28, 35, 38-42, 44-47, 65, 69, 71, 75-79, 81- 83, 87, 89, 91, 93, 95, 97, 99, 107, 111, 115, 119-121, 123, 139, 143-149, 151, 153, 155, 156, 160, 162, 164, 166, 174, 175, 187, 189, 191 |
| 4 | 11, 13-15, 18, 19, 26, 27, 29- 34, 36, 37, 43, 48- 64, 66-68, 70, 72- 74, 84-86, 88, 90, 94, 100-106, 108, 110, 112-114, 116-118, 124-129, 131-134, 136-138, 140, 141, 163, 167, 176, 193, 194, |
| 8 | 130, 135 |

TABLE VII. Impossible Electron Number/SG ESFD. Right Column: Symmetry groups for which ESFDs are impossible if the electron number is a multiple of the Left Column, with Time-Reversal.

| Electron Number Multiple of | Symmetry Groups (SG) |
|---|---|
| 2 | 1- 10,12, 16, 17, 20-25, 28, 35, 38-42, 44-47, 65, 69, 71, 75- 83, 87, 89, 91, 93, 95, 97-99, 107, 111, 115, 119-121, 123, 139, 143- 162, 164, 166, 168-175, 177-183, 187, 189-191, 195-197, 199, 200, 202, 204, 207-211, 214-217, 219-221, 225, 229 |
| 4 | 11, 13-15, 18, 19, 26, 27, 29- 34, 36, 37, 43, 48- 64, 66-68, 70, 72-74, 84-86, 88, 90, 92, 94, 96, 100-106, 108, 109, 110, 112-114, 116-118, 122, 124-129, 131-134, 136-138, 140-142, 163, 165, 167, 176, 184-186, 192-194, 198, 201, 203, 205, 206, 212, 213, 218, 222, 223, 224, 226-228, 230 |
| 8 | 130, 135 |

TABLE VIII. Must Be Electron Number/SG ESFD. Right Column: Symmetry groups for which a given material must be ESFDs if the electron number *not* a multiple of the Left Column, with Time-Reversal.

#### 2. Must Be ESFD in Electron Number/Symmetry Group combinations

In many symmetry groups (but not in all), the electron number can immediately classify a material as ESFD. For example, SG 55 has only 4-fold representations at high-symmetry points R,S,T,U,X,Y. Hence any material in this symmetry group with $4N+2,\quad N\in Z$ electrons *has* to have a degeneracy at the Fermi level at a high symmetry point, and is, therefore, an ESFD. We have calculated all the pairs of Electron Number/SG for which materials must be an ESFD. We present them in Table[VIII], which we call "Must Be ESFD".

We have checked that none of our ES, NLC, or SEBR materials have the Number of electron/SG combinations in the "Must Be ESFD" Table[VII].

#### 3. Check on Insulating Fillings In Non-Symmorphic Groups

For non-symmorphic groups, solving the compatibility relations leads to bands being connected in multiples of certain integers. For example, in the SG 194, bands come connected in multiples of 4 to give rise to Hourglass-type connectivity of the band structures [30]. This implies that one can have insulators only at fillings which are multiples of 4. In Table[IX] we have tabulated all the fillings for non-symmorphic groups

We have checked that all the TIs in these groups have a number of electrons is a multiple of the integer in the first column of the file for the corresponding SG in Table[IX] .

#### 4. Strong Topological Index- EBR Equivalence Class Calculation

In Appendices[Q] and [P], we have showed that the EBR equivalence classes, defined in Appendix[H] provide a set of topological indices for our materials. While the EBR equivalence classes provide a finer classification than the strong topological indices of [1], they include these indices. Having also computed explicitly the indices of [1] for all our materials, we now have a very strong check of the correctness of our results. For all TI's (SEBR, NLC) we have checked that *all* the members of the same topological class (as defined in Appendix[H]) have exactly the same values of the strong topological indices of [1], which we computed in Appendix[Q]. We have also checked that all the compounds that have exactly the same strong topological indices of [1] belong to the same topological class defined by EBR equivalence in Appendix[H]. All checks passed. We further performed this test for *all* TI's that result out of symmetry breaking of an ES or an ESFD. Together, these form a huge set of checks that attest for the correctness and self-consistency of the data.

| Electron Number Multiple of | Symmetry Groups (SG) |
|---|---|
| 4 | 4, 7, 9, 11, 13-15, 17, 18, 20, 24, 26- 28, 30-32, 34, 36, 37, 39, 40, 41, 43, 45, 46, 48-51, 53, 55, 58, 59, 63, 64-68, 70, 72, 74, 77, 80, 84-86, 88, 90, 93, 94, 98, 100-105, 108, 109, 112-114, 116-118, 120, 122, 124-129, 131, 132, 134, 136, 137, 140, 141, 158, 159, 161, 163, 165, 167, 173, 176, 182, 184-186, 188, 190, 192-194, 199, 201, 203, 208, 210, 214, 218-220, 222-224, 226, 227 |
| 6 | 144, 145, 151-154, 171, 172, 180, 181 |
| 8 | 19, 29, 33, 52, 54, 56, 57, 60, 61, 62, 73, 76, 78, 91, 92, 95, 96, 106, 110, 130, 133, 135, 138, 142, 198, 205, 206, 212, 213, 228, 230 |
| 12 | 169, 170, 178, 179 |

TABLE IX. Insulating filling. Right Column: Symmetry groups for which a given material can be an insulator if the electron number is a multiple of the Left Column, with Time-Reversal. For any fillings different than these the system is an ES or an ESFD.

### 5. Checks of Strong Topological Indices of Ref[1]

In TableX we list all the symmetry groups with strong topological phases identified by strong topological indices symmetry indicators. For every compound, we have computed these indices based on the formulas of [1]. One can see that the toWe remark that several conventions needed to be re-derived in order to compute these indices.

A simple check we performed is the following. When any of the following two strong topological indices show up together: $z_4$,$z_2$, $z_8$, $z_{12}$, $z'_{12}$, they have the same parity (they are either both even or both odd). We present a simple proof of this for the case of groups with both inversion ($z_4$ index) and $S_4$ symmetry ($z_2$ index).

We have checked that our computed indices, in all TIs (SEBR and NLC) and ES/ESFD $\rightarrow$ TI, the $z_4$,$z_2$, $z_8$, $z_{12}$, $z'_{12}$ (when they appear together) have the same parity.

#### a. *Proof that The $S_4$ invariant is the same as the Inversion invariant* $\mod 2$

We consider groups that have both $S_4$ and Inversion. Hence, they have separately Inversion $I$ and $C_4 = S_4 I$. Without loss of generality, we take the $C_4$ axis to be $C_{4z}$.

For groups with inversion we have the invariant [1]

$$Z_4 = \sum_{\vec{k}\in TRIM} \frac{1}{2} n_k^- - \frac{1}{2} n_k^+ \mod 4 = (\sum_{\vec{k}\in TRIM} n_k^-) - 4N_{bands} \mod 4 = \sum_{\vec{k}\in TRIM} n_k^- \mod 4 \tag{S1}$$

where $n_k^{+/-}$ are the number of bands of inversion eigenvalue $+/-$ at a certain TRIM. Since $\sum_k n_k^+ = 8N_{bands} - \sum_k n_k^-$; the factor of 8 comes from the 8 TRIM points in the $k$ summation. $N_{bands}$ is the number of bands in the system. Since we have $C_4$ symmetry, and $[C_4, I] = 0$, the number of Kramers pairs of negative inversion eigenvalue satisfies the equations

$$n_{0\pi 0}^- = n_{\pi,0,0}^-; \quad n_{0\pi\pi}^- = n_{\pi,0,\pi}^- \tag{S2}$$

| Indicator | Symmetry Groups (SG) |
|---|---|
| $z_2$weak; $z_4$: (All SG with Inversion) | 2, 10-15, 47-74, 83-88, 123-142, 147,148, 162-167, 175, 176, 191-194, 200-206,221-230 |
| $z_2$: All $S_4$ SG | 81-88, 111-142, 215-230 |
| $z_{4m,\pi}$ | 83,123, 127; 221 |
| $z_8$ | 83; 87; 123; 124; 127; 128; 139; 140; 221; 225; 226; 229 |
| $z_{3m,0}$ | 174, 187, 188, 189, 190 |
| $z_{3m,\pi}$ | 174, 187, 189 |
| $z_{6m,0}$ | 175,191, 192; 176, 193, 194 |
| $z_{6m,\pi}$ | 175, 191, 192 |
| $z_{12}$ | 175; 191; 192 |
| $z'_{12}$ | 176; 193; 194 |

TABLE X. Strong topological indices (symmetry indicators) in symmetry groups; We remark that not all indices are independent. For example $z'_{12}$ is completely defined by $z_{6m,0}$ and $z_4$

We want to look at the parity of the $Z_4$ index, which, taking into account the above, becomes

$$Z_4 \mod 2 = n^-_{000} + n^-_{\pi\pi 0} + n^-_{00\pi} + n^-_{\pi\pi\pi} \tag{S3}$$

The $S_4$ index, which we call $Z_2$ is:

$$Z_2 = \sum_{k\in S_4} \frac{1}{2} n_k^{3/2} - \frac{1}{2} n_k^{1/2} \mod 2 = (\sum_{k\in S_4} n_k^{3/2} - 2N_{bands}) \mod 2 = \sum_{k\in S_4} n_k^{3/2} \mod 2 \tag{S4}$$

Where the sum if over the $S_4$ invariant momenta: $000, \pi\pi 0, 00\pi, \pi\pi\pi$. $n_k^{3/2}$ is the number of Kramers doublets of $S_4$ eigenvalue $-\sqrt{2}$. We now relate this eigenvalue to $C_4$ and Inversion eigenvalue. For the $m$'th Kramers doublet at $k$, let $n_k^{m3/2} \in (0,1)$ be whether the Kramers band $m$ of $S_4$ eigenvalue $-sqrt2$. Notice that it can either be 0 (in which case the doublet is the 1/2 doublet) or 1. Call $n_{C_4,k}^{m3/2} \in (0,1)$ whether the Kramers band $m$ of $C_4$ eigenvalue $-\sqrt{2}$ and $n_k^{m-} \in (0,1)$ whether the inversion eigenvalue of the Kramers $m$'th band is $-$. Then, because $S_4 = C_4 I$ we have

$$n_k^{m3/2} = n_{C_4,k}^{m3/2}(1 - n_k^{m-}) + (1 - n_{C_4,k}^{m3/2}) n_k^{m-} = n_k^{m-} + n_{C_4,k}^{m3/2} - 2 n_{C_4,k}^{m3/2} n_k^{m-} \tag{S5}$$

Plugging this into the $Z_2$ index:

$$Z_2 = \sum_{m,k\in S_4} n_k^{m3/2} \mod 2 = \sum_{m,k\in S_4} n_k^{m-} + n_{C_4,k}^{m3/2} \mod 2 \tag{S6}$$

Since $C_{4z}$ is a symmetry axis, and since we have considered an insulator, satisfying the compatibility relations, the sum of the representations of $C_4$ must not change as we change $k_z$ from 0 to $\pi$. Hence:

$$\sum_m n_{C_4,000}^{m3/2} = \sum_m n_{C_4,00\pi}^{m3/2}, \quad \sum_m n_{C_4,\pi\pi 0}^{m3/2} = \sum_m n_{C_4,\pi\pi\pi}^{m3/2} \tag{S7}$$

Plugging this into the $Z_2$ index we have:

$$Z_2 \sum_{m,k\in S_4} n_k^{m-} \mod 2 = n^-_{000} + n^-_{\pi\pi 0} + n^-_{00\pi} + n^-_{\pi\pi\pi} = Z_4 \mod 2 \tag{S8}$$

### 6. A cross checking of existing literatures

A non-systematic cross checking of existing literatures with the database is presented in Table XI.

TABLE XI: This is a non-systematic cross checking of existing literature with the database. "TI" refers to a strong Fu-Kane topological insulator. White rows do not appear in present database. Green rows appear and seem to agree with existing literatures. Red rows appear and seem to disagree with existing literatures. But, in face, they agree. Gray rows appear and are removed by our filters.

| Compound | Reference | Known to be | Appears in our paper | Classified in our paper as: (TQC) | Comments |
|---|---|---|---|---|---|
| beta-Ag2Te | Zhang et al., PRL 106, 156808 (2011); Bansil-lin-Das, RMP; Sulaev, AIP Adv, 2013 | TI | y | NLC (14.2.69) | consistent |
| Bi14Rh3I9 | Rasche et al., Nat. Mat. 10,422 (2013) | TI (weak TI) | n | - | The topological phase in the ref. is of SG 191, P6/mmm, which is not in the "high-quality" database. |
| Bi2Se3 | H.J. Zhang et al., Nature Physics 5, 438-442 (2009); Hasan-Kane, RMP 82,3045 (2010) | TI | y | SEBR (166.1.331) | consistent |
| Bi2Te2Se | arXiv:1007.5111v1 | TI | y | SEBR (166.1.335) | consistent |
| Bi4Br4 | Zhou et al., Nano lett. 14, 4767 (2014); Liu et al., PRL 116, 066801 (2016) | TI (weak TI) | yes | SEBR (12.1.158) | consistent |
| Ca3NBi | Bansil-Lin-Das, RMP; Goh, 2016 | trivial insulator | n | - | Ca3NBi reported in the ref. is in SG 221, Pm-3m, which is not in the "high-quality" database. |
| Ca3P2 | Xie, APL mater., 2015 | nodal-line SM (theory) | n | - | Ca3P2 is in SG 193, P63/mcm, whose icsd number is 429048. The occupancy of Ca atoms is 0.9 (less than 1). Therefore, we discard it by applying our general filtering rules. |
| Ca3PbO | Bansil-Lin-Das, RMP; Karlyado, 2015 | TCI, massive Dirac bulk | y | SEBR (221.1.9) | Our topological gapped phases (SEBR & NLC) include all the symmetry-protected (e.g., time-reversal and crystalline symmetries) phases, such as, Fu-Kane TI, TCI, Hourglass fermions, HOTI, etc. |
| CaAg(P,As) | Yamakage, JPSJ, 2015; Xu, PRB, 2018 | nodal-line SM | y | SEBR (189.1.43-44) | They are nodal-line SM without SOC. Upon including SOC, they are Fu-Kane TI. We always include SOC in our calculations. Threfore, they should be of gapped phase. |
| Cd3As2 | Z. Wang et al, PRB 88, 125427 (2013); Liu et al., Nat. Mater. 13, 677(2014) | Dirac SM | n | - | The Dirac SM Cd3As2 is in SG 110, I41cd, whose icsd number is 42882 with 80 ($>30$) atoms in a unit cell. We discard it by applying our general filtering rules. |
| CeOs4As12 | Bansil-Lin-Da, RMP; Yan, PRB, 2012 | band touching at G | n | - | CeOs4As12 is in SG 204, Im-3, which is not in the "high-quality" database. |
| GeBi2Te4 | Okamoto, PRB, 2012 | TI | y | SERB (166.1.336) | consistent |
| KHgSb | Z. Wang et al., Nature 532, 189-194 (2016) | TCI | n | - | It's first reported as a topological phase by our NATURE paper. In both ref. [Bansil-Lin-Da, RMP] and [Zhang, PRL, 2011], KHgSb was reported as a topologically trivial insulator. Its icsd number is 56201, but not in the "high-quality" database. Therefore, it's not included. |

Continued on next page

**TABLE XI– continued**

| Compound | Reference | Known to be | Appears in our paper | Classified in our paper as: (TQC) | Comments |
|---|---|---|---|---|---|
| LaAgSb2 | Bansil-Lin-Da, RMP; Shi, PRB, 2016 | massless Dirac (ARPES) | n | - | It is in SG 129, P4/nmm, which is not in the “high-quality” database. |
| LaAlGe | Xu et al., Sci. Adv. 3, e1603266 (2017) | Weyl SM | n | - | The reported LaAlGe is in SG 109, I41md, which is not in the “high-quality” database. |
| LaBi | Nayak et al., Nat. Commun. 8, 12942 (2017) | TI | n | - | The reported LaBi is in SG 225; Fm-3m, which is not in the “high-quality” database. |
| Li2AgBi | Bansil-Lin-Da, RMP; Lin, arxiv, 2010 | J=3/2 band touching at G | n | - | It’s in SG 216; F-43m, which is not in the “high-quality” database. |
| Li2AgSb | Bansil-Lin-Da, RMP; Lin, arxiv, 2010 | band touching | n | - | It’s in SG 216; F-43m, which is not in the “high-quality” database. |
| LiAuSe | Bansil-Lin-Da, RMP; Zhang, PRL 2011 | TI | n | - | There is no record about LiAuSe in SG 194 in the ICSD database. |
| MoTe2 | Sun, PRB, 2015; Wang, PRL, 2016; Tamai, PRX, 2016 | Weyl SM | y | NCL (11.2.35) | The repoted SM phase of MoTe2 is in SG 31, Pmn21, which is not in the “high-quality” database. The phase of MoTe2 in our paper is in SG 11, which is a HOTI [Wang et al., arXiv1806.11116, (2018)]. |
| Na3Bi | Wang et al., PRB 85, 195320 (2012); Liu et al., Science, 2014; Xiong et al., Science 350, 413 (2015) | Dirac SM | n | - | The first report on Na3Bi is the paper [Wang et al. PRB 85, 195320 (2012)]. The reported Na3Bi is in SG 194 (#ICSD 26881), which is not in the “high-quality” database. Therefore, it doesn’t show up in Table XIV. |
| PbTaSe2 | Bian, NC, 2016 | nodal-line SM | n | - | The reported PbTaSe2 is in SG 187; P-6m2, which is not in the “high-quality” database. |
| PtSn4 | Wu, Nat. Phy. 2016 | Dirac SM | n | - | The reported PtSn is in SG 68, Ccca. The reason why it’s reported as a Dirac SM in the ref. is becaues the “Dirac-like” dispersing bands are observed near both Z and X points. But it is lack of symmetries analysis. We have carefully checked electronic band structures and concluded that there is no symmetry-protected Dirac point in the bulk bands. In addition, we find it’s trivial because the valence bands are decomposed into EBRs. Therefore, it doesn’t show up in Table XIV. |
| SnTe | Hsieh et al., NC 3, 982 (2012); Bansil-Lin-Da, RMP | TCI | n | - | The TCI phase of SnTe is in SG 225 (#icsd 188457), Fm-3m, which is not in the “high-quality” database. |
| Sr2Pb3 | Wieder, Science 2018 | TCI | y | NLC (127.2.1) | It’s first reported in our Science paper [Wieder, Science 2018]. The reported Sr2Pb3 is in SG 127, P4/mbm. TCI is included in our topological gapped phases. |
| Sr3NBi | Bansil-Lin-Da, RMP; Goh, 2016 | band touching at G | n | - | The reported Sr3NBi is SG 221 Pm-3m, which is not in the “high-quality” database. |

Continued on next page

**TABLE XI– continued**

| Compound | Reference | Known to be | Appears in our paper | Classified in our paper as: (TQC) | Comments |
|---|---|---|---|---|---|
| SrMnBi2 | Bansil-Lin-Da, RMP; Park, PRL, 2011 | Dirac SM | n | - | The reported SrMnBi2 is in SG 139, I4/mmm. The Dirac point in the system is obtained withour SOC. Upon including SOC, it becomes a trivial gapped phase. Therefore, it doesn't show up in Table XIV. |
| TaAs | Weng et al., PRX 5, 011029 (2015); Xu et al., Science 349, 613 (2015) | Weyl SM | n | - | Our method is based on symmetries. Since the Weyl doesn't require any symmetry-protection (except translational symmetry), our method fails when it comes to general Weyl points. |
| TlBiSe2 | Chen et al., PRL 105, 266401 (2010); Singh et al., PRB 86, 115208 (2012) | TI | n | - | The reported TlBiSe2 is SG 166 (#icsd 197811), which is not in the "high-quality" database. |
| WC | Ma, Nat. Phys. 2018 | three component SM | y | ES (187.3.6) | The reported WC is 187, which is in the "high-quality" database. Since the three-component points is protected by symmetry, our method has to work for it. Therefore, we do find it in the class "ES" in Table XIV. |
| WTe2 | Soluyanov et al., Nature 527, 495-498 (2015) | Weyl SM | n | - | Our method is based on symmetries. Since the Weyl doesn't require any symmetry-protection (except translational symmetry), our method fails when it comes to general Weyl points. |
| ZrSiS | Schoop et al., Nat. Commun. 7, 1169 (2016) | nodal-line SM | n | - | The reported ZrSiS is in SG 129 (#icsd 15569), which is not in the "high-quality" database. |

The End

| Compounds | SG | #ICSD | Class | SIs | Comments |
|---|---|---|---|---|---|
| Ca3P2 | 193 | 429048 | SEBR | z4=3<br>z6m0=5<br>z12prime=11 | [Xie, APL mater., 2015] shows it's a nodal-line SM without SOC. But, with SOC, the hypothetical Ca5P3 is ESFD, and Ca3P2 (with 0.9 filling for Ca atoms) is an SEBR. |
| Cd3As2 | 110 | 42882 | ES | - | consistent |
| KHgSb | 194 | 56201 | SEBR | z6m0=2<br>z12prime=8 | consistent |
| Na3Bi | 194 | 26881 | ES | - | consistent |
| SnTe | 225 | 188457 | SEBR | z8=4 | consistent |
| TlBiSe2 | 166 | 197811 | trivial | - | [Chen *et al.* PRL 105, 266401 (2010)] is an experimental paper. [Singh *et al.* PRB 86, 115208 (2012)] shows the calculations with optimized lattice and internal parameters. The band topology can be sensitive to the lattice and internal parameters. |
| ZrSiS | 129 | 15569 | SEBR | z2w,1=0<br>z2w,3=1 | [Schoop *et al.* Nat. Commun. 7, 1169 (2016)] shows it's a nodal-line SM without SOC. But with SOC, it's a Weak TI. |

TABLE XII. The results for some known topological materials that our filters removed.

The compounds in the gray rows of Table[XI] were removed by our filters in a large search. Since they have been well studied and proved to be topological, we have performed the TQC calculations for them by hand. The results are listed in Table[XII]. Most of them show good agreement with literatures. Some disagreement happens due to the neglecting of SOC, or relaxation of parameters in the literature. Our database is correct.

### Appendix T: Summary of "good" band structures and Comparison with literatures

In this section we scan the 7385 band structures by eye to pick some "good" and "very good" topological compounds, which have relatively less bands around the Fermi level. They are listed in Table XIII. Their band structures are shown in Fig.We also listed some well-known topological materials and compare them with our results. Our database shows good agreement with previous references. For some materials, the disagreement is due to either the neglecting of SOC or the lack of symmetry analysis. Our method cannot immediately apply to Weyl semimetals, since Weyl points do not need any symmetry protection.

TABLE XIII: We picked some "good" and "very good" compounds in this table. "vg" refers to a very good compound.

| TQC | ICSD | Stoich. Formula | if known | Notes and ref. |
|---|---|---|---|---|
| 2.2.1 | 202564 | S1Ta6 | no | |
| 2.2.2 | 41017 | P3Tc2 | no | |
| 2.2.9 | 59657 | O4Pb1Pt2 | no | |
| 2.2.16 | 237587 | Mo3S4 | no | |
| 2.2.26 | 60937 | H1As4Cu3-Na5O16 | no | vg |
| 2.2.48 | 39439 | Cu1Mo1O4 | no | |
| 2.2.64 | 92165 | Cl2Cu3O6Se2 | no | |
| 2.2.68 | 408774 | Cl14Mo4O4-Tl2 | no | vg |
| 2.2.69 | 74479 | Ca1P3 | yes | vg; PRB 95, 045136 (2017) |
| 2.2.84 | 85490 | Ba1Mo6S8 | no | vg |
| 2.2.91 | 86400 | As3Ca1 | yes | vg; PRB 95, 045136 (2017) |
| 2.2.98 | 95832 | Ag2F5 | no | vg |
| 2.2.99 | 421341 | Ag1O4S1 | no | |
| 11.2.70 | 414338 | Cu1Nb1Te2 | no | |
| 11.2.71 | 69990 | Cu1La2O6Sn1 | no | |
| 11.4.16 | 71524 | I3Ir1Y3 | no | |

Continued on next page

**TABLE XIII– continued**

| TQC | ICSD | Stoich. Formula | if known | Notes and ref. |
|---|---|---|---|---|
| 11.4.19 | 248158 | H6Cu1N1-Na3O13P2 | no | vg |
| 11.4.31 | 252429 | H3Cl1Cu1-O3Zn1 | no | |
| 11.4.40 | 200123 | H2Cl2Cu1-O4Pb3 | no | |
| 11.4.72 | 422033 | Cu1N4O10 | no | vg |
| 12.1.7 | 42077 | P2V1 | no | vg |
| 12.1.24 | 77752 | Nb1Sb2 | no | vg |
| 12.1.46 | 239971 | Li2Mn1O3 | no | |
| 12.1.62 | 637671 | Ge2Re1 | no | |
| 12.1.102 | 251718 | Cr1Se2 | no | |
| 12.1.103 | 75420 | Cr1S2 | no | |
| 12.1.134 | 618223 | C3Lu2Mo2 | no | |
| 12.2.131 | 252737 | C2Ca1 | no | |
| 12.1.155 | 251720 | Bi1Se1 | yes | vg; APL 110, 162102 (2017) |
| 12.1.156 | 251717 | Bi1Se1 | yes | vg; APL 110, 162102 (2017) |
| 12.1.157 | 1559 | Bi1I1 | no | vg |
| 12.1.158 | 1560 | Bi1Br1 | yes | vg; PRL 116,066801 (2016) |

Continued on next page

**TABLE XIII– continued**

| TQC | ICSD | Stoich. Formula | if known | Notes and ref. |
|---|---|---|---|---|
| 12.1.161 | 173685 | Ba3Cd2Sb4 | no | vg |
| 12.1.176 | 424760 | As4Ba3Zn2 | no | vg |
| 12.1.177 | 424761 | As4Ba3Cd2 | no | vg |
| 12.1.178 | 43185 | As3W2 | no | |
| 12.1.180 | 43184 | As3Mo2 | no | vg |
| 12.1.181 | 41015 | As3Mo2 | no | vg |
| 12.1.183 | 611574 | As2V1 | no | |
| 12.1.184 | 44069 | As2Ta1 | yes | vg; Sci. Rep. 6,27294 (2016) |
| 12.1.186 | 81218 | As2Nb1 | no | vg |
| 12.1.194 | 409319 | Al4Na4P12Sr8 | no | vg |
| 12.2.79 | 422453 | Cu2Sn2Sr1 | no | |
| 12.2.80 | 182102 | Cu2Sn2Sr1 | no | |
| 12.2.81 | 79536 | Cu2Na2S4Zr1 | no | vg |
| 12.2.123 | 54994 | Ca1Ga4 | no | |
| 12.2.132 | 252734 | C2Ca1 | no | |
| 12.2.138 | 240479 | Br4Cs2I2Pd1 | no | vg |
| 12.2.157 | 423458 | Ba5Cd2O2Sb4 | no | vg |
| 12.2.162 | 420834 | Ba2Cd2Sb3 | no | vg |
| 12.2.166 | 49000 | Ba1Sb3 | no | |
| 12.2.173 | 614701 | B2Lu1Rh3 | no | |
| 12.2.174 | 614578 | B2Ir3Y1 | no | |
| 12.2.177 | 614561 | B2Ir3Sc1 | no | |
| 12.2.180 | 8058 | Au2P3 | no | vg |
| 12.2.183 | 41831 | As3Sr1 | no | |
| 12.2.185 | 86402 | As3Ba1 | no | |
| 12.2.188 | 417968 | Al2Ge4Sr3 | no | vg |
| 14.2.19 | 98698 | H1Cu1F1O1 | no | |
| 14.2.20 | 91088 | H1Cl1Cu1O1 | no | |
| 14.2.21 | 38384 | H1Cl1Cu1O1 | no | |
| 14.2.38 | 1672 | Cu1O4Te1 | no | |
| 14.2.40 | 202451 | Cu1O3Te1 | no | |
| 14.2.46 | 63553 | Co1Sb2 | no | vg |
| 14.2.50 | 4411 | Cl3Mo1O1 | no | |
| 14.2.51 | 16150 | Cl3Mo1O1 | no | |
| 14.2.57 | 86049 | C7Y4 | no | |
| 14.2.59 | 172071 | Br2Cu2O6Sr2Te1 | no | vg |
| 14.2.67 | 74948 | As2Ba1Cu2O8 | no | vg |
| 14.2.69 | 73402 | Ag2Te1 | yes | vg; PRL 106, 156808 (2011) |
| 14.2.70 | 73230 | Ag2Te1 | yes | vg; PRL 106, 156808 (2011) |
| 15.2.58 | 23479 | Cu2O7V2 | no | |
| 15.2.60 | 67316 | Cu2O7P2 | no | |
| 15.2.61 | 28279 | Cu2O7P2 | no | |
| 15.2.62 | 157107 | Cu2O7P2 | no | |
| 15.2.73 | 95789 | Ca2Ir2Si1 | no | |
| 15.2.91 | 86064 | Ba1Cu1N1 | no | vg |
| 15.4.34 | 202385 | Fe1Rb1S2 | no | vg |
| 15.4.35 | 202384 | Fe1Rb1S2 | no | vg |
| 15.4.57 | 40780 | Fe1K1Se2 | no | |
| 15.4.94 | 603 | Cu1O5Se2 | no | |
| 15.4.95 | 245059 | Cu1O5Se2 | no | |
| 15.4.96 | 69094 | Cu1O1 | no | |
| 15.4.131 | 429177 | Ca2Ga1Pt2 | no | vg |
| 15.4.166 | 428674 | Al1Ca2Pt2 | no | vg |
| 19.3.11 | 35451 | Cu1Na1O4P1 | no | vg |
| 52.3.4 | 410665 | Ag2Bi1O3 | no | |
| 58.2.8 | 186851 | B4Cr1 | no | |
| 62.2.97 | 260793 | Ge1Pt1Zr1 | no | |

Continued on next page

**TABLE XIII– continued**

| TQC | ICSD | Stoich. Formula | if known | Notes and ref. |
|---|---|---|---|---|
| 62.2.137 | 2635 | F1Se1Y1 | no | |
| 62.2.213 | 416561 | Ba2Pb1 | yes | PRB 84, 165127 (2011) |
| 62.3.97 | 41804 | H1Cu1I1O4 | no | |
| 63.4.172 | 159037 | Ca1Ir1O3 | no | |
| 63.4.173 | 159036 | Ca1Ir1O3 | no | |
| 63.4.188 | 415220 | C1Cu1N2 | no | |
| 64.1.5 | 23836 | P1 | no | vg |
| 64.1.6 | 150873 | P1 | no | vg |
| 64.4.4 | 78297 | Cu1La2O4 | no | |
| 64.4.5 | 78296 | Cu1La2O4 | no | |
| 64.4.12 | 78289 | Cu1La2O4 | no | |
| 97.3.1 | 35190 | I1Se8Ta2 | no | vg |
| 121.1.4 | 412239 | Cu3S4Sb1 | yes | vg; NJP 13, 085017 (2011) |
| 121.1.5 | 2857 | Cu3S4Sb1 | yes | NJP 13, 085017 (2011) |
| 121.1.6 | 95119 | Cu2Hg1Se4Sn1 | yes | NJP 13, 085017 (2011) |
| 121.1.7 | 152761 | Cu2Ge1Hg1Se4 | yes | NJP 13, 085017 (2011) |
| 127.2.7 | 262925 | Cs1I3Sn1 | no | |
| 136.3.6 | 34234 | O2Pb1 | no | vg |
| 136.3.7 | 23292 | O2Pb1 | no | vg |
| 136.3.31 | 50005 | Bi2Mg1O6 | no | vg |
| 139.1.36 | 91207 | Na2O1Sb2Ti2 | no | |
| 148.1.3 | 37134 | O12Te1Tl6 | no | |
| 148.1.7 | 171584 | Ni1O3Ti1 | no | vg |
| 148.1.8 | 59375 | Mo6Te8 | no | vg |
| 148.1.15 | 89004 | Mo12Rb2Se14 | no | |
| 148.1.46 | 96087 | Ba1Ni2O8V2 | no | vg |
| 152.4.1 | 65692 | Te1 | no | vg |
| 160.3.8 | 42667 | In2S4Zn1 | no | |
| 163.3.1 | 78748 | F6Ga1Li1Pd1 | no | |
| 163.3.3 | 73132 | Al1F6Li1Pd1 | no | |
| 165.1.1 | 154809 | H3Y1 | no | vg |
| 166.1.19 | 291530 | Sb3Te40005 | no | vg |
| 166.1.22 | 651638 | Sb2Te3 | yes | vg; Nature Physics 5, 438-442 (2009) |
| 166.1.23 | 651635 | Sb2Te3 | yes | Nature Physics 5, 438-442 (2009) |
| 166.1.24 | 651629 | Sb2Te3 | yes | Nature Physics 5, 438-442 (2009) |
| 166.1.50 | 87416 | Sb2Se1Te2 | yes | vg; Journal of Applied Physics 116, 193708 (2014) |
| 166.1.51 | 2085 | Sb2Se1Te2 | yes | Journal of Applied Physics 116, 193708 (2014) |
| 166.1.52 | 651646 | Sb1Te2Tl1 | yes | vg; JETP Letters 91, 594-598 (2010) |
| 166.1.55 | 651498 | Sb1 | yes | vg; PRB 80, 085307 (2009) |
| 166.1.268 | 24410 | Cl1La1 | no | |
| 166.1.277 | 409057 | Ca1Si2 | yes | vg; Phys. Rev. B 93, 201114 (2016) |
| 166.1.294 | 96188 | C1Y2 | no | vg |
| 166.1.312 | 20319 | Bi4Se3 | no | vg |

Continued on next page

**TABLE XIII– continued**

| TQC | ICSD | Stoich. Formula | if known | Notes and ref. |
|---|---|---|---|---|
| 166.1.313 | 74348 | Bi2Te3 | yes | vg; Nature Physics 5, 438-442 (2009) |
| 166.1.332 | 20385 | Bi2Se3 | yes | vg; Nature Physics 5, 438-442 (2009) |
| 166.1.335 | 43512 | Bi2Se1Te2 | yes | vg; Phys. Rev. B 85, 235406 (2012) |
| 166.1.336 | 658633 | Bi2Ge1Te4 | yes | vg; PRB, 2012 |
| 166.1.340 | 64704 | Bi1 | yes | vg; Nat. Phys. 14, 918-924 (2018) |
| 166.1.391 | 16517 | As1 | no | vg |
| 166.1.396 | 420092 | Al10Ba7 | no | |
| 166.1.397 | 424311 | Ag4Sb2Sr1 | no | |
| 166.3.132 | 80261 | Cu1I1 | yes | vg; PNAS. 1803599115 (2018) |
| 166.3.182 | 30526 | Bi4Te3 | no | vg |
| 166.3.198 | 262036 | As3Na1Zn4 | no | vg |
| 167.3.86 | 61067 | C1Ni1O3 | no | |
| 167.4.13 | 82660 | Ir1O6Sr3Zn1 | no | vg |
| 167.4.55 | 81902 | Ca4Ir1O6 | no | |
| 176.4.8 | 30752 | K1Mo3S3 | no | |
| 189.1.30 | 413810 | Ba1Bi1Na1 | no | |
| 189.1.43 | 10016 | Ag1Ca1P1 | yes | vg; JPSJ 85, 013708 (2015) |
| 189.1.44 | 10017 | Ag1As1Ca1 | yes | vg; JPSJ 85, 013708 (2015) |
| 190.4.7 | 107474 | Hf1Rh1Sn1 | no | |
| 198.4.1 | 79233 | Rh1Si1 | no | vg |
| 198.4.22 | 260982 | Co1Si1 | no | vg |
| 204.4.55 | 9188 | As3Co1 | no | vg |
| 216.2.6 | 81917 | Hg1S1 | no | vg |
| 216.4.4 | 162198 | In1Sb1 | no | vg |
| 216.4.9 | 290038 | Hg1Se1 | no | vg |
| 216.4.12 | 412294 | Cu1Mg1Sb1 | no | |
| 216.4.20 | 248682 | Au1Bi1Ca1 | no | |
| 221.1.1 | 201036 | O1Sn1Sr3 | no | vg |
| 221.1.2 | 181127 | Ni3Sn1 | no | |
| 221.1.8 | 100787 | Ca3O1Sn1 | no | |
| 221.1.9 | 100789 | Ca3O1Pb1 | yes | J. Phys. Soc. Jpn. 80, 083704 (2011) |
| 221.1.13 | 50313 | B6Sr1 | yes | Phys. Chem. Chem. Phys. 19, 7245-7251 (2017) |
| 221.1.14 | 51495 | B6Ba1 | yes | Phys. Chem. Chem. Phys. 19, 7245-7251 (2017) |
| 221.4.9 | 421118 | N1Na3 | no | vg |
| 223.3.2 | 186819 | Ca1O4Pd3 | no | vg |
| 224.3.1 | 281041 | Ag2O1 | no | vg |
| 225.1.3 | 63097 | Pb1Se1 | no | vg |
| 225.1.6 | 62195 | Pb1Se1 | no | vg |
| 225.1.9 | 62191 | Pb1S1 | no | vg |

The End

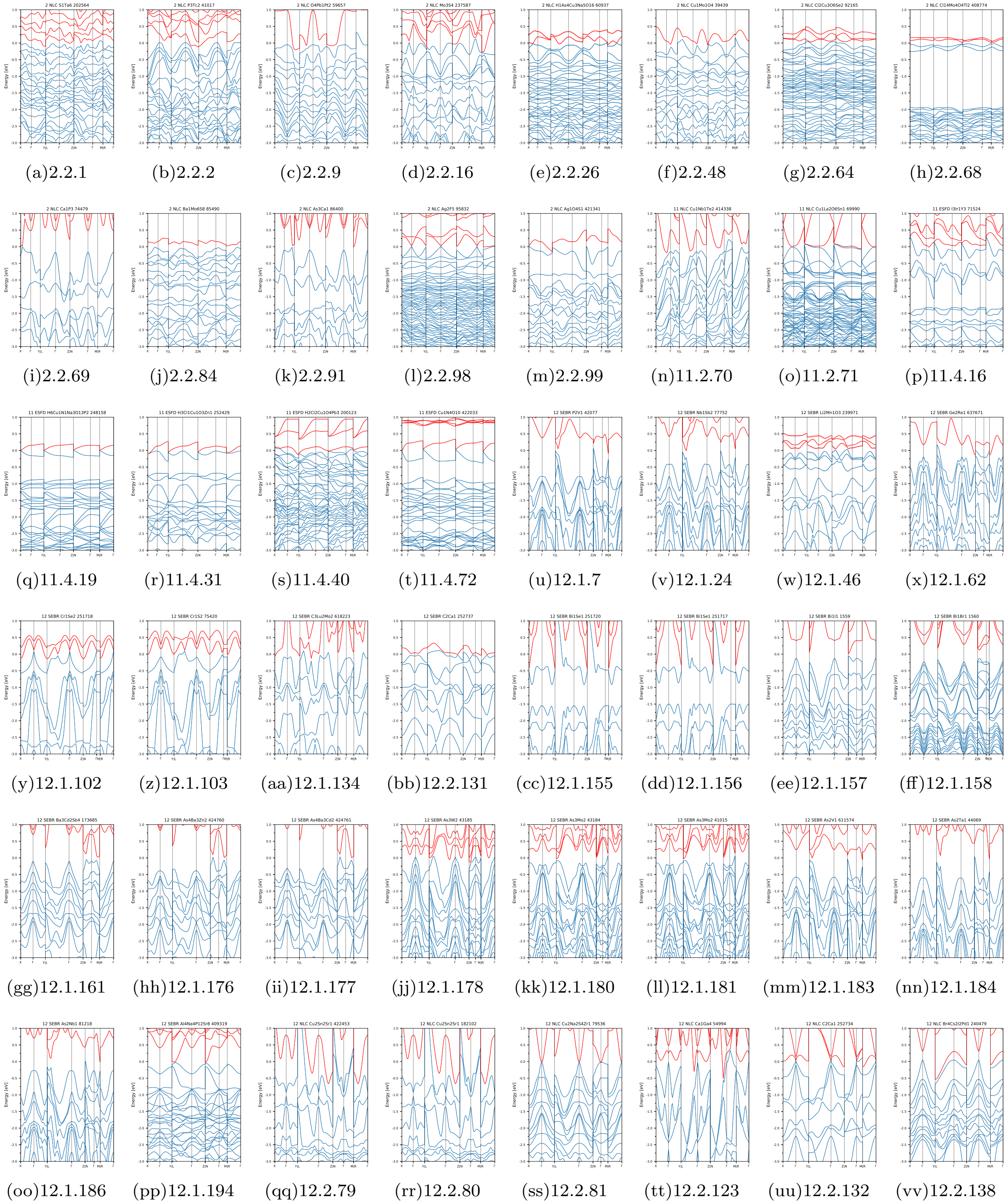

(a)2.2.1 (b)2.2.2 (c)2.2.9 (d)2.2.16 (e)2.2.26 (f)2.2.48 (g)2.2.64 (h)2.2.68
(i)2.2.69 (j)2.2.84 (k)2.2.91 (l)2.2.98 (m)2.2.99 (n)11.2.70 (o)11.2.71 (p)11.4.16
(q)11.4.19 (r)11.4.31 (s)11.4.40 (t)11.4.72 (u)12.1.7 (v)12.1.24 (w)12.1.46 (x)12.1.62
(y)12.1.102 (z)12.1.103 (aa)12.1.134 (bb)12.2.131 (cc)12.1.155 (dd)12.1.156 (ee)12.1.157 (ff)12.1.158
(gg)12.1.161 (hh)12.1.176 (ii)12.1.177 (jj)12.1.178 (kk)12.1.180 (ll)12.1.181 (mm)12.1.183 (nn)12.1.184
(oo)12.1.186 (pp)12.1.194 (qq)12.2.79 (rr)12.2.80 (ss)12.2.81 (tt)12.2.123 (uu)12.2.132 (vv)12.2.138

FIG. 9. (Color online) Band structures for the good and very good compounds in Table[XIII].

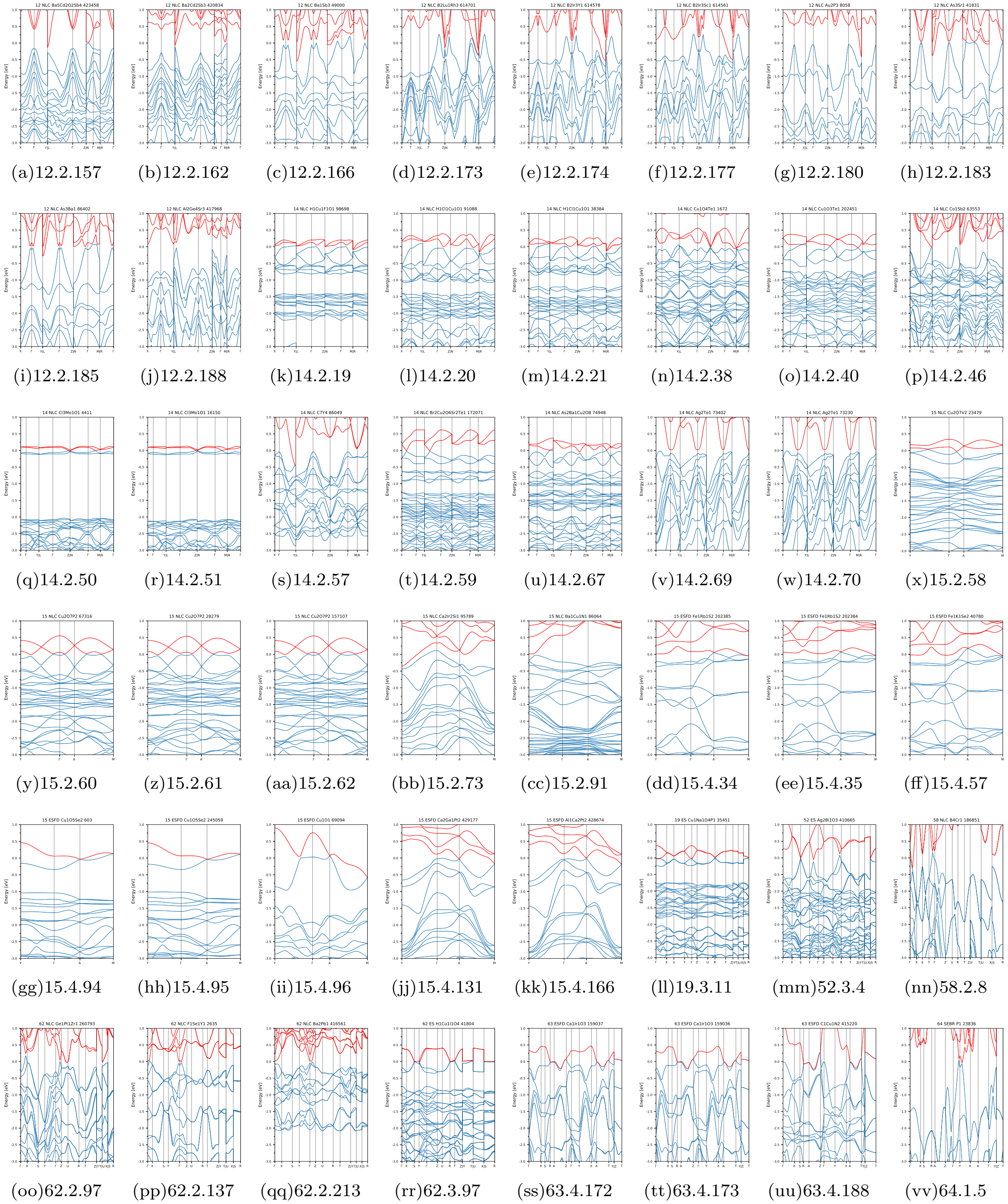

(a)12.2.157 (b)12.2.162 (c)12.2.166 (d)12.2.173 (e)12.2.174 (f)12.2.177 (g)12.2.180 (h)12.2.183

(i)12.2.185 (j)12.2.188 (k)14.2.19 (l)14.2.20 (m)14.2.21 (n)14.2.38 (o)14.2.40 (p)14.2.46

(q)14.2.50 (r)14.2.51 (s)14.2.57 (t)14.2.59 (u)14.2.67 (v)14.2.69 (w)14.2.70 (x)15.2.58

(y)15.2.60 (z)15.2.61 (aa)15.2.62 (bb)15.2.73 (cc)15.2.91 (dd)15.4.34 (ee)15.4.35 (ff)15.4.57

(gg)15.4.94 (hh)15.4.95 (ii)15.4.96 (jj)15.4.131 (kk)15.4.166 (ll)19.3.11 (mm)52.3.4 (nn)58.2.8

(oo)62.2.97 (pp)62.2.137 (qq)62.2.213 (rr)62.3.97 (ss)63.4.172 (tt)63.4.173 (uu)63.4.188 (vv)64.1.5

FIG. 10. (Color online) Band structures for the good and very good compounds in Table[XIII].

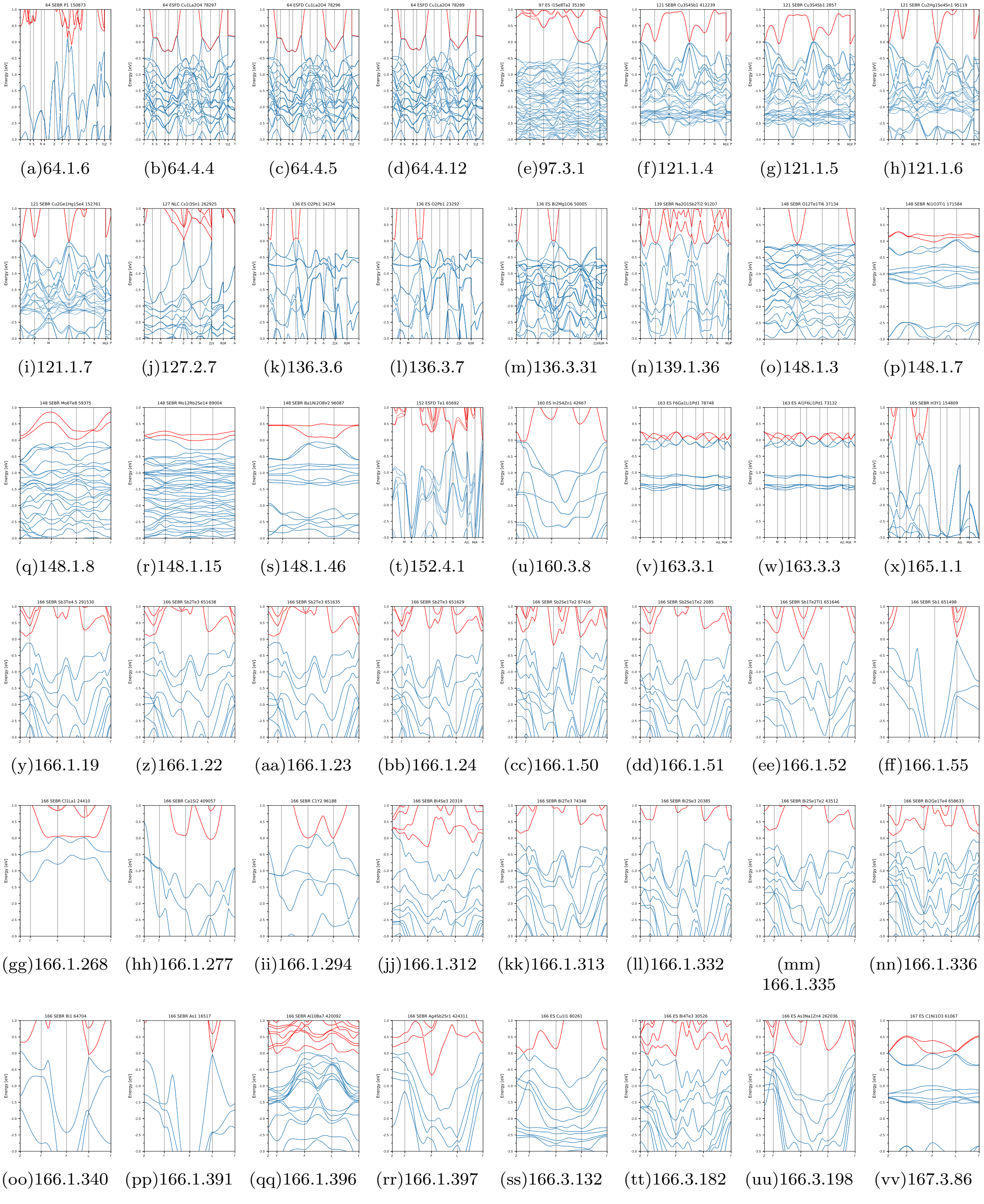

FIG. 11. (Color online) Band structures for the good and very good compounds in Table[XIII].

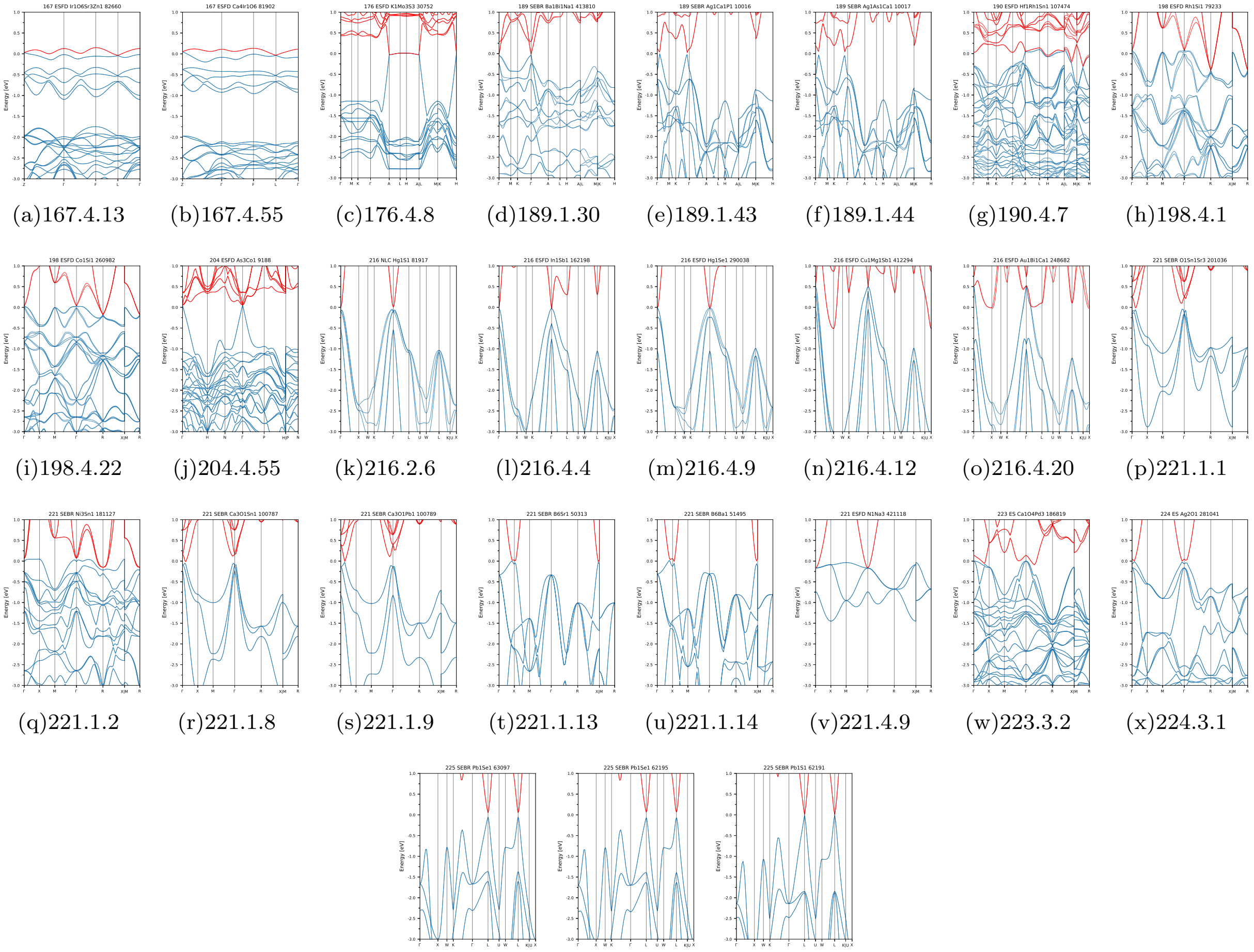

(a)167.4.13 (b)167.4.55 (c)176.4.8 (d)189.1.30 (e)189.1.43 (f)189.1.44 (g)190.4.7 (h)198.4.1

(i)198.4.22 (j)204.4.55 (k)216.2.6 (l)216.4.4 (m)216.4.9 (n)216.4.12 (o)216.4.20 (p)221.1.1

(q)221.1.2 (r)221.1.8 (s)221.1.9 (t)221.1.13 (u)221.1.14 (v)221.4.9 (w)223.3.2 (x)224.3.1

(y)225.1.3 (z)225.1.6 (aa)225.1.9

FIG. 12. (Color online) Band structures for the good and very good compounds in Table[XIII].

TABLE XIV: A full list of all our 7385 “High-Quality” topological compounds. We give, in this order: the TQC number used for fast referencing with other tables, the ICSD database number, the number of electrons (filling), the smallest Direct Gap (DG) and indirect gap (IndG) calculated by our *ab-initio* methods. Columns 6 and 7 provide the minimal symmetry group that keeps the topological nature of the *gapped* compound (either SEBR or NLC) with and without Time Reversal symmetry and topological classes. Columns 8 and 9 provide the largest subgroup (with and without time-reversal) of the space group (Max. SubSG) for which the *gapless* material (either ES or ESFD) satisfies the compatibility relations and hence allows an insulating filling. Column 10 shows the computed symmetry indicators. In last column, we present the stoichiometric formula. See the list of tables.

| TQC | ICSD | NoE | DG | IndG | SubSG (w/ T) | cl | SubSG (w/o T) | cl | Max. Sub. SG | yes or no | Max. Sub. SG | yes or no | SIs (w/ T) | Stoich. Formula |
|---|---|---|---|---|---|---|---|---|---|---|---|---|---|---|
| 2.2.1 | 202564 | 72 | 0.0693 | -0.1697 | 2 | 19 | 0 | 2 | - | - | - | - | z2w,3=1<br>z4=2 | S1Ta6 |
| 2.2.2 | 41017 | 116 | 0.0423 | -0.1282 | 2 | 6 | 2 | 1 | - | - | - | - | z4=1 | P3Tc2 |
| 2.2.3 | 66584 | 158 | 0.0053 | -0.0927 | 2 | 2 | 2 | 1 | - | - | - | - | z2w,1=1<br>z2w,2=1<br>z2w,3=1<br>z4=1 | O9V5 |
| 2.2.4 | 80363 | 154 | 0.0065 | -0.0467 | 2 | 12 | 0 | 2 | - | - | - | - | z2w,2=1<br>z2w,3=1<br>z4=2 | O9P3Ru1 |
| 2.2.5 | 2212 | 124 | 0.0300 | 0.0068 | 2 | 12 | 0 | 2 | - | - | - | - | z2w,2=1<br>z2w,3=1<br>z4=2 | O7V4 |
| 2.2.6 | 19012 | 124 | 0.0029 | -0.0691 | 2 | 17 | 0 | 2 | - | - | - | - | z2w,3=1 | O7V4 |
| 2.2.7 | 61430 | 96 | 0.0150 | -0.0713 | 2 | 19 | 0 | 2 | - | - | - | - | z2w,3=1<br>z4=2 | O6Pb3S1 |
| 2.2.8 | 82204 | 96 | 0.1069 | -0.5142 | 2 | 10 | 0 | 2 | - | - | - | - | z2w,2=1<br>z2w,3=1 | O4Pb1Pt2 |
| 2.2.9 | 59657 | 96 | 0.0770 | -0.5263 | 2 | 10 | 0 | 2 | - | - | - | - | z2w,2=1<br>z2w,3=1 | O4Pb1Pt2 |
| 2.2.10 | 281617 | 98 | 0.0127 | -0.1365 | 2 | 11 | 2 | 1 | - | - | - | - | z2w,2=1<br>z2w,3=1<br>z4=3 | Na1O6Si2Ti1 |
| 2.2.11 | 281616 | 98 | 0.0039 | -0.1497 | 2 | 11 | 2 | 1 | - | - | - | - | z2w,2=1<br>z2w,3=1<br>z4=3 | Na1O6Si2Ti1 |
| 2.2.12 | 281615 | 98 | 0.0053 | -0.1420 | 2 | 11 | 2 | 1 | - | - | - | - | z2w,2=1<br>z2w,3=1<br>z4=3 | Na1O6Si2Ti1 |
| 2.2.13 | 281614 | 98 | 0.0022 | -0.1479 | 2 | 11 | 2 | 1 | - | - | - | - | z2w,2=1<br>z2w,3=1<br>z4=3 | Na1O6Si2Ti1 |
| 2.2.14 | 281613 | 98 | 0.0030 | -0.1464 | 2 | 11 | 2 | 1 | - | - | - | - | z2w,2=1<br>z2w,3=1<br>z4=3 | Na1O6Si2Ti1 |
| 2.2.15 | 65941 | 94 | 0.0374 | -0.0708 | 2 | 3 | 2 | 1 | - | - | - | - | z4=3 | Mo6S8Sr1 |
| 2.2.16 | 237587 | 84 | 0.0306 | -0.3991 | 2 | 16 | 2 | 1 | - | - | - | - | z2w,3=1<br>z4=3 | Mo3S4 |
| 2.2.17 | 61050 | 130 | 0.0024 | -0.0578 | 2 | 3 | 2 | 1 | - | - | - | - | z4=3 | Mn3Na2O7 |
| 2.2.18 | 59328 | 130 | 0.0000 | -0.0235 | 2 | 3 | 2 | 1 | - | - | - | - | z4=3 | Mn1O7Te1V2 |
| 2.2.19 | 431529 | 152 | 0.0130 | -0.1056 | 2 | 9 | 0 | 2 | - | - | - | - | z2w,1=1<br>z4=2 | Lu1Na1P2S6 |
| 2.2.20 | 419880 | 122 | 0.0914 | -0.2013 | 2 | 25 | 2 | 1 | - | - | - | - | z2w,1=1<br>z2w,2=1<br>z4=3 | K1O6Os2 |

Continued on next page

**TABLE XIV– continued**

| TQC | ICSD | NoE | DG | IndG | SubSG (w/ T) | cl | SubSG (w/o T) | cl | Max. Sub. SG | yes or no | Max. Sub. SG | yes or no | SIs (w/ T) | Stoich. Formula |
|---|---|---|---|---|---|---|---|---|---|---|---|---|---|---|
| 2.2.21 | 39887 | 98 | 0.0153 | -0.1237 | 2 | 10 | 0 | 2 | - | - | - | - | z2w,2=1 z2w,3=1 | H2Cu1O5Se1 |
| 2.2.22 | 166101 | 98 | 0.0611 | -0.0736 | 2 | 12 | 0 | 2 | - | - | - | - | z2w,2=1 z2w,3=1 z4=2 | H2Cu1O5S1 |
| 2.2.23 | 167608 | 90 | 0.0000 | -0.0417 | 2 | 9 | 0 | 2 | - | - | - | - | z2w,1=1 z4=2 | H1Fe1Li1O5P1 |
| 2.2.24 | 62742 | 154 | 0.0021 | -0.4908 | 2 | 25 | 2 | 1 | - | - | - | - | z2w,1=1 z2w,2=1 z4=3 | H1Cu2K3O14P4 |
| 2.2.25 | 416823 | 114 | 0.0196 | -0.2406 | 2 | 19 | 0 | 2 | - | - | - | - | z2w,3=1 z4=2 | H1Co1O5P1Zn1 |
| 2.2.26 | 60937 | 154 | 0.0030 | -0.0621 | 2 | 14 | 2 | 1 | - | - | - | - | z2w,2=1 z2w,3=1 z4=1 | H1As4Cu3Na5O16 |
| 2.2.27 | 8063 | 118 | 0.0137 | -0.0147 | 2 | 7 | 2 | 1 | - | - | - | - | z2w,1=1 z2w,2=1 z2w,3=1 z4=3 | H1As1O5Zn2 |
| 2.2.28 | 37436 | 118 | 0.0179 | -0.0049 | 2 | 7 | 2 | 1 | - | - | - | - | z2w,1=1 z2w,2=1 z2w,3=1 z4=3 | H1As1O5Zn2 |
| 2.2.29 | 34341 | 118 | 0.0124 | -0.0146 | 2 | 7 | 2 | 1 | - | - | - | - | z2w,1=1 z2w,2=1 z2w,3=1 z4=3 | H1As1O5Zn2 |
| 2.2.30 | 8065 | 174 | 0.0138 | -0.0249 | 2 | 14 | 2 | 1 | - | - | - | - | z2w,2=1 z2w,3=1 z4=1 | H1As1Ba1Cr2O10 |
| 2.2.31 | 73197 | 78 | 0.0050 | -0.1332 | 2 | 22 | 0 | 2 | - | - | - | - | z2w,2=1 | H18Al2Cl6N6 |
| 2.2.32 | 417328 | 118 | 0.0187 | -0.0267 | 2 | 19 | 0 | 2 | - | - | - | - | z2w,3=1 z4=2 | H12Cs4O6P2S6 |
| 2.2.33 | 35417 | 130 | 0.0293 | -0.0575 | 2 | 14 | 2 | 1 | - | - | - | - | z2w,2=1 z2w,3=1 z4=1 | H12Cl8Hg3Mg1O6 |
| 2.2.34 | 194210 | 68 | 0.0184 | -0.0416 | 2 | 3 | 2 | 1 | - | - | - | - | z4=3 | H12Cl4N4Pt2 |
| 2.2.35 | 240179 | 84 | 0.1492 | -0.4364 | 2 | 4 | 0 | 2 | - | - | - | - | z2w,1=1 | Ga5Rh3 |
| 2.2.36 | 79719 | 130 | 0.0061 | -0.0752 | 2 | 1 | 2 | 1 | - | - | - | - | z2w,1=1 z4=1 | Fe2O14P4Sr1 |
| 2.2.37 | 72390 | 190 | 0.0112 | -0.0117 | 2 | 5 | 0 | 2 | - | - | - | - | z2w,1=1 z2w,2=1 z2w,3=1 z4=2 | Fe1N2Sr2 |
| 2.2.38 | 171431 | 128 | 0.1300 | -0.1728 | 2 | 2 | 2 | 1 | - | - | - | - | z2w,1=1 z2w,2=1 z2w,3=1 z4=1 | Cu4Mo6Se8 |
| 2.2.39 | 190444 | 198 | 0.0154 | -0.0903 | 2 | 9 | 0 | 2 | - | - | - | - | z2w,1=1 z4=2 | Cu3Mo2O9 |
| 2.2.40 | 84558 | 82 | 0.0318 | -0.2953 | 2 | 14 | 2 | 1 | - | - | - | - | z2w,2=1 z2w,3=1 z4=1 | Cu1O4W1 |
| 2.2.41 | 69759 | 82 | 0.0310 | -0.3126 | 2 | 11 | 2 | 1 | - | - | - | - | z2w,2=1 z2w,3=1 z4=3 | Cu1O4W1 |
| 2.2.42 | 4189 | 82 | 0.0066 | -0.3248 | 2 | 21 | 0 | 2 | - | - | - | - | z2w,1=1 z2w,2=1 | Cu1O4W1 |

Continued on next page

**TABLE XIV– continued**

| TQC | ICSD | NoE | DG | IndG | SubSG (w/ T) | cl | SubSG (w/o T) | cl | Max. Sub. SG | yes or no | Max. Sub. SG | yes or no | SIs (w/ T) | Stoich. Formula |
|---|---|---|---|---|---|---|---|---|---|---|---|---|---|---|
| 2.2.43 | 24339 | 82 | 0.0191 | -0.3141 | 2 | 11 | 2 | 1 | - | - | - | - | z2w,2=1 z2w,3=1 z4=3 | Cu1O4W1 |
| 2.2.44 | 182753 | 82 | 0.0284 | -0.3172 | 2 | 11 | 2 | 1 | - | - | - | - | z2w,2=1 z2w,3=1 z4=3 | Cu1O4W1 |
| 2.2.45 | 182752 | 82 | 0.0250 | -0.3198 | 2 | 11 | 2 | 1 | - | - | - | - | z2w,2=1 z2w,3=1 z4=3 | Cu1O4W1 |
| 2.2.46 | 182750 | 82 | 0.0257 | -0.3639 | 2 | 11 | 2 | 1 | - | - | - | - | z2w,2=1 z2w,3=1 z4=3 | Cu1O4W1 |
| 2.2.47 | 16009 | 82 | 0.0310 | -0.3162 | 2 | 11 | 2 | 1 | - | - | - | - | z2w,2=1 z2w,3=1 z4=3 | Cu1O4W1 |
| 2.2.48 | 39439 | 82 | 0.0457 | -0.2290 | 2 | 21 | 0 | 2 | - | - | - | - | z2w,1=1 z2w,2=1 | Cu1Mo1O4 |
| 2.2.49 | 62367 | 174 | 0.0049 | -0.0384 | 2 | 8 | 2 | 1 | - | - | - | - | z2w,1=1 z2w,3=1 z4=3 | Cu1I3Na1O9 |
| 2.2.50 | 75006 | 178 | 0.0026 | -0.0064 | 2 | 6 | 2 | 1 | - | - | - | - | z4=1 | Cu1Gd1O8W2 |
| 2.2.51 | 64660 | 126 | 0.0046 | -0.0579 | 2 | 24 | 2 | 1 | - | - | - | - | z2w,2=1 z4=3 | Cu1F6Pt1 |
| 2.2.52 | 69655 | 132 | 0.0037 | -0.4478 | 2 | 20 | 0 | 2 | - | - | - | - | z4=2 | Cu1F3Na1 |
| 2.2.53 | 200533 | 98 | 0.0064 | -0.1771 | 2 | 14 | 2 | 1 | - | - | - | - | z2w,2=1 z2w,3=1 z4=1 | Cu1F1Nb1O3 |
| 2.2.54 | 62000 | 84 | 0.0395 | -0.4391 | 2 | 19 | 0 | 2 | - | - | - | - | z2w,3=1 z4=2 | Cs1Hg1 |
| 2.2.55 | 150974 | 84 | 0.0399 | -0.4392 | 2 | 17 | 0 | 2 | - | - | - | - | z2w,3=1 | Cs1Hg1 |
| 2.2.56 | 62676 | 68 | 0.0052 | -0.1401 | 2 | 15 | 2 | 1 | - | - | - | - | z2w,2=1 z4=1 | Cr1Na4O4 |
| 2.2.57 | 73191 | 174 | 0.0028 | -0.0041 | 2 | 3 | 2 | 1 | - | - | - | - | z4=3 | Co2O4Rb5 |
| 2.2.58 | 73190 | 174 | 0.0087 | -0.0002 | 2 | 3 | 2 | 1 | - | - | - | - | z4=3 | Co2O4Rb5 |
| 2.2.59 | 167202 | 94 | 0.0064 | -0.1412 | 2 | 9 | 0 | 2 | - | - | - | - | z2w,1=1 z4=2 | Co1F1Li1O4S1 |
| 2.2.60 | 410945 | 150 | 0.0196 | 0.0024 | 2 | 10 | 0 | 2 | - | - | - | - | z2w,2=1 z2w,3=1 | Cl9Te1W1 |
| 2.2.61 | 410101 | 134 | 0.0058 | -0.0390 | 2 | 18 | 2 | 1 | - | - | - | - | z2w,1=1 z4=3 | Cl7Mo1O1Se1 |
| 2.2.62 | 84620 | 82 | 0.0131 | -0.0852 | 2 | 17 | 0 | 2 | - | - | - | - | z2w,3=1 | Cl5Mo1 |
| 2.2.63 | 431558 | 156 | 0.0133 | -0.1473 | 2 | 3 | 2 | 1 | - | - | - | - | z4=3 | Cl3Cu3O3Te1 |
| 2.2.64 | 92165 | 190 | 0.0512 | 0.0095 | 2 | 1 | 2 | 1 | - | - | - | - | z2w,1=1 z4=1 | Cl2Cu3O6Se2 |
| 2.2.65 | 427665 | 120 | 0.0022 | -0.0200 | 2 | 2 | 2 | 1 | - | - | - | - | z2w,1=1 z2w,2=1 z2w,3=1 z4=1 | Cl1O4Tm1W1 |
| 2.2.66 | 50441 | 148 | 0.0034 | -0.0391 | 2 | 18 | 2 | 1 | - | - | - | - | z2w,1=1 z4=3 | Cl18Mo2P2 |
| 2.2.67 | 410649 | 148 | 0.0050 | -0.0357 | 2 | 1 | 2 | 1 | - | - | - | - | z2w,1=1 z4=1 | Cl18Mo2P2 |
| 2.2.68 | 408774 | 152 | 0.0268 | 0.0268 | 2 | 16 | 2 | 1 | - | - | - | - | z2w,3=1 z4=3 | Cl14Mo4O4Tl2 |
| 2.2.69 | 74479 | 34 | 0.0199 | 0.0199 | 2 | 15 | 2 | 1 | - | - | - | - | z2w,2=1 z4=1 | Ca1P3 |

Continued on next page

**TABLE XIV– continued**

| TQC | ICSD | NoE | DG | IndG | SubSG (w/ T) | cl | SubSG (w/o T) | cl | Max. Sub. SG | yes or no | Max. Sub. SG | yes or no | SIs (w/ T) | Stoich. Formula |
|---|---|---|---|---|---|---|---|---|---|---|---|---|---|---|
| 2.2.70 | 42698 | 100 | 0.0014 | -0.0312 | 2 | 14 | 2 | 1 | - | - | - | - | z2w,2=1<br>z2w,3=1<br>z4=1 | C4H16Cl2N8Se4Te1 |
| 2.2.71 | 170721 | 54 | 0.0579 | -0.2697 | 2 | 21 | 0 | 2 | - | - | - | - | z2w,1=1<br>z2w,2=1 | C3H6As1N1 |
| 2.2.72 | 62435 | 120 | 0.0219 | -0.0537 | 2 | 13 | 0 | 2 | - | - | - | - | z2w,1=1<br>z2w,3=1 | C2H8Cl2N4Se2Te1 |
| 2.2.73 | 62436 | 120 | 0.0081 | -0.0680 | 2 | 12 | 0 | 2 | - | - | - | - | z2w,2=1<br>z2w,3=1<br>z4=2 | C2H8Br2N4Se2Te1 |
| 2.2.74 | 186554 | 94 | 0.0067 | -0.2268 | 2 | 27 | 0 | 2 | - | - | - | - | z2w,2=1<br>z4=2 | C2H3Na1O5 |
| 2.2.75 | 196028 | 58 | 0.0096 | -0.3097 | 2 | 22 | 0 | 2 | - | - | - | - | z2w,2=1 | C1Mn1O3 |
| 2.2.76 | 196027 | 58 | 0.0131 | -0.3308 | 2 | 17 | 0 | 2 | - | - | - | - | z2w,3=1 | C1Mn1O3 |
| 2.2.77 | 196026 | 58 | 0.0268 | -0.3169 | 2 | 22 | 0 | 2 | - | - | - | - | z2w,2=1 | C1Mn1O3 |
| 2.2.78 | 196025 | 58 | 0.0350 | -0.3573 | 2 | 22 | 0 | 2 | - | - | - | - | z2w,2=1 | C1Mn1O3 |
| 2.2.79 | 196024 | 58 | 0.0076 | -0.3251 | 2 | 7 | 2 | 1 | - | - | - | - | z2w,1=1<br>z2w,2=1<br>z2w,3=1<br>z4=3 | C1Mn1O3 |
| 2.2.80 | 423462 | 82 | 0.0102 | -0.1147 | 2 | 11 | 2 | 1 | - | - | - | - | z2w,2=1<br>z2w,3=1<br>z4=3 | Br5W1 |
| 2.2.81 | 39784 | 50 | 0.0137 | -0.2091 | 2 | 10 | 0 | 2 | - | - | - | - | z2w,2=1<br>z2w,3=1 | Br3Ti1 |
| 2.2.82 | 51779 | 110 | 0.0101 | -0.7963 | 2 | 17 | 0 | 2 | - | - | - | - | z2w,3=1 | Bi3O7Sb1 |
| 2.2.83 | 97768 | 132 | 0.0000 | -0.1394 | 2 | 26 | 2 | 1 | - | - | - | - | z2w,1=1<br>z2w,3=1<br>z4=1 | Bi1Ni1O3 |
| 2.2.84 | 85490 | 94 | 0.0883 | 0.0273 | 2 | 3 | 2 | 1 | - | - | - | - | z4=3 | Ba1Mo6S8 |
| 2.2.85 | 65775 | 94 | 0.0570 | -0.0735 | 2 | 3 | 2 | 1 | - | - | - | - | z4=3 | Ba1Mo6S8 |
| 2.2.86 | 237763 | 84 | 0.0045 | -0.0345 | 2 | 17 | 0 | 2 | - | - | - | - | z2w,3=1 | Ba1Fe2O8P2 |
| 2.2.87 | 192060 | 146 | 0.0111 | -0.4262 | 2 | 9 | 0 | 2 | - | - | - | - | z2w,1=1<br>z4=2 | Ba1Cu1O7P2 |
| 2.2.88 | 96562 | 108 | 0.0111 | -0.1074 | 2 | 8 | 2 | 1 | - | - | - | - | z2w,1=1<br>z2w,3=1<br>z4=3 | B2Co2O5 |
| 2.2.89 | 24284 | 108 | 0.0184 | -0.1128 | 2 | 7 | 2 | 1 | - | - | - | - | z2w,1=1<br>z2w,2=1<br>z2w,3=1<br>z4=3 | B2Co2O5 |
| 2.2.90 | 66662 | 116 | 0.0305 | -0.1803 | 2 | 6 | 2 | 1 | - | - | - | - | z4=1 | As3Tc2 |
| 2.2.91 | 86400 | 50 | 0.4913 | 0.0927 | 2 | 1 | 2 | 1 | - | - | - | - | z2w,1=1<br>z4=1 | As3Ca1 |
| 2.2.92 | 193 | 34 | 0.0535 | 0.0423 | 2 | 1 | 2 | 1 | - | - | - | - | z2w,1=1<br>z4=1 | As3Ca1 |
| 2.2.93 | 404850 | 216 | 0.0254 | -0.0867 | 2 | 5 | 0 | 2 | - | - | - | - | z2w,1=1<br>z2w,2=1<br>z2w,3=1<br>z4=2 | As2Cu4O9 |
| 2.2.94 | 31787 | 150 | 0.0053 | -0.0756 | 2 | 4 | 0 | 2 | - | - | - | - | z2w,1=1 | As1F6N2S3 |
| 2.2.95 | 92916 | 100 | 0.0112 | -0.1366 | 2 | 10 | 0 | 2 | - | - | - | - | z2w,2=1<br>z2w,3=1 | As1Bi1Ni1O5 |
| 2.2.96 | 428633 | 98 | 0.0077 | -0.1982 | 2 | 2 | 2 | 1 | - | - | - | - | z2w,1=1<br>z2w,2=1<br>z2w,3=1<br>z4=1 | As1Bi1Co1O5 |

Continued on next page

**TABLE XIV– continued**

| TQC | ICSD | NoE | DG | IndG | SubSG (w/ T) | cl | SubSG (w/o T) | cl | Max. Sub. SG | yes or no | Max. Sub. SG | yes or no | SIs (w/ T) | Stoich. Formula |
|---|---|---|---|---|---|---|---|---|---|---|---|---|---|---|
| 2.2.97 | 422342 | 84 | 0.0060 | -0.0983 | 2 | 24 | 2 | 1 | - | - | - | - | z2w,2=1<br>z4=3 | Al2Fe3Si3 |
| 2.2.98 | 95832 | 228 | 0.0255 | -0.0084 | 2 | 3 | 2 | 1 | - | - | - | - | z4=3 | Ag2F5 |
| 2.2.99 | 421341 | 82 | 0.0300 | -0.1590 | 2 | 2 | 2 | 1 | - | - | - | - | z2w,1=1<br>z2w,2=1<br>z2w,3=1<br>z4=1 | Ag1O4S1 |
| 2.2.100 | 51507 | 126 | 0.0124 | -0.0253 | 2 | 1 | 2 | 1 | - | - | - | - | z2w,1=1<br>z4=1 | Ag1F6Pd1 |
| 2.2.101 | 174407 | 172 | 0.0022 | -0.4471 | 2 | 11 | 2 | 1 | - | - | - | - | z2w,2=1<br>z2w,3=1<br>z4=3 | Ag1Cu1F3 |
| 4.3.1 | 83647 | 118 | - | - | - | - | - | - | 1 | yes | 1 | yes | - | Li1Mo1O7P2 |
| 4.3.2 | 53666 | 78 | - | - | - | - | - | - | 1 | yes | 1 | yes | - | Ge2La1Pt2 |
| 4.3.3 | 95751 | 122 | - | - | - | - | - | - | 1 | yes | 1 | yes | - | Fe1Li1O7P2 |
| 4.3.4 | 69197 | 122 | - | - | - | - | - | - | 1 | yes | 1 | yes | - | Fe1Li1O7P2 |
| 4.3.5 | 166611 | 122 | - | - | - | - | - | - | 1 | yes | 1 | yes | - | Fe1Li1O7P2 |
| 4.3.6 | 20157 | 106 | - | - | - | - | - | - | 1 | yes | 1 | yes | - | F6Fe1Na3 |
| 4.3.7 | 4327 | 122 | - | - | - | - | - | - | 1 | yes | 1 | yes | - | Cu1I2O6 |
| 4.3.8 | 240965 | 118 | - | - | - | - | - | - | 1 | yes | 1 | yes | - | Cr1Li1O7P2 |
| 7.3.1 | 281242 | 134 | - | - | - | - | - | - | 1 | yes | 1 | yes | - | I6K2Re1 |
| 7.3.2 | 414303 | 90 | - | - | - | - | - | - | 1 | yes | 1 | yes | - | Ga9Ir2 |
| 7.3.3 | 200680 | 78 | - | - | - | - | - | - | 1 | yes | 1 | yes | - | Co1Ge1Na2O4 |
| 9.3.1 | 69758 | 34 | - | - | - | - | - | - | 1 | yes | 1 | yes | - | Cu1O1 |
| 9.3.2 | 69757 | 34 | - | - | - | - | - | - | 1 | yes | 1 | yes | - | Cu1O1 |
| 9.3.3 | 85632 | 62 | - | - | - | - | - | - | 1 | yes | 1 | yes | - | Co1Na4O3 |
| 9.3.4 | 10501 | 62 | - | - | - | - | - | - | 1 | yes | 1 | yes | - | Co1Na4O3 |
| 9.3.5 | 158488 | 82 | - | - | - | - | - | - | 1 | yes | 1 | yes | - | C4H12Ag1N1Se5 |
| 9.3.6 | 20827 | 58 | - | - | - | - | - | - | 1 | yes | 1 | yes | - | C2Ce1Co1 |
| 9.3.7 | 192517 | 62 | - | - | - | - | - | - | 1 | yes | 1 | yes | - | Bi1Fe1O3 |
| 9.3.8 | 192515 | 62 | - | - | - | - | - | - | 1 | yes | 1 | yes | - | Bi1Fe1O3 |
| 10.2.1 | 417035 | 96 | 0.0073 | -0.0618 | 2 | 30 | 2 | 1 | - | - | - | - | z4=3 | Co2Gd4Mg3 |
| 11.2.1 | 42076 | 44 | 0.0046 | -0.7626 | 2 | 14 | 2 | 1 | - | - | - | - | z2w,1=1<br>z4=1 | Te3Zr1 |
| 11.2.2 | 424108 | 84 | 0.0285 | -0.6782 | 2 | 1 | 2 | 1 | - | - | - | - | z4=3 | Sn2Sr1Zn2 |
| 11.2.3 | 42982 | 56 | 0.1095 | -0.2043 | 2 | 6 | 2 | 1 | - | - | - | - | z2w,1=1<br>z4=3 | Se3Ta2 |
| 11.2.4 | 43213 | 92 | 0.0636 | -0.0633 | 2 | 12 | 0 | 2 | - | - | - | - | z2w,1=1<br>z2w,3=1<br>z4=2 | Se3Ta1 |
| 11.2.5 | 52307 | 40 | 0.1106 | -0.1787 | 2 | 4 | 0 | 2 | - | - | - | - | z2w,1=1<br>z4=2 | Sb2Sr1 |
| 11.2.6 | 171417 | 64 | 0.0168 | -0.2022 | 2 | 7 | 0 | 2 | - | - | - | - | z2w,3=1<br>z4=2 | S3U1 |
| 11.2.7 | 430 | 256 | 0.0030 | -0.1019 | 2 | 10 | 0 | 2 | - | - | - | - | z2w,3=1 | S3Tm2 |
| 11.2.8 | 260980 | 384 | 0.0009 | -0.0868 | 2 | 3 | 0 | 2 | - | - | - | - | z4=2 | S3Tm2 |
| 11.2.9 | 52066 | 112 | 0.0276 | -2.0807 | 2 | 2 | 2 | 1 | - | - | - | - | z2w,1=1<br>z2w,3=1<br>z4=3 | Rh4Si5 |
| 11.2.10 | 24355 | 112 | 0.0248 | -2.0745 | 2 | 11 | 2 | 1 | - | - | - | - | z2w,1=1<br>z2w,3=1<br>z4=1 | Rh4Si5 |
| 11.2.11 | 31692 | 256 | 0.0073 | -0.0493 | 2 | 10 | 0 | 2 | - | - | - | - | z2w,3=1 | Pu1 |
| 11.2.12 | 43283 | 160 | 0.0346 | -2.5862 | 2 | 3 | 0 | 2 | - | - | - | - | z4=2 | Pt6Si5 |
| 11.2.13 | 105804 | 140 | 0.0105 | -0.6631 | 2 | 4 | 0 | 2 | - | - | - | - | z2w,1=1<br>z4=2 | Pt3Ta1 |
| 11.2.14 | 99271 | 84 | 0.0177 | -0.8081 | 2 | 1 | 2 | 1 | - | - | - | - | z4=3 | Ni3Te2 |

Continued on next page

**TABLE XIV– continued**

| TQC | ICSD | NoE | DG | IndG | SubSG (w/ T) | cl | SubSG (w/o T) | cl | Max. Sub. SG | yes or no | Max. Sub. SG | yes or no | SIs (w/ T) | Stoich. Formula |
|---|---|---|---|---|---|---|---|---|---|---|---|---|---|---|
| 11.2.15 | 54304 | 140 | 0.0283 | -0.8707 | 2 | 7 | 0 | 2 | - | - | - | - | z2w,3=1<br>z4=2 | Ni3Ta1 |
| 11.2.16 | 245081 | 28 | 0.0161 | -0.5971 | 2 | 9 | 2 | 1 | - | - | - | - | z4=1 | Ni1Ti1 |
| 11.2.17 | 240195 | 28 | 0.0173 | -0.9053 | 2 | 9 | 2 | 1 | - | - | - | - | z4=1 | Ni1Ti1 |
| 11.2.18 | 166372 | 28 | 0.0038 | -0.7505 | 2 | 9 | 2 | 1 | - | - | - | - | z4=1 | Ni1Ti1 |
| 11.2.19 | 166367 | 28 | 0.0130 | -0.6105 | 2 | 9 | 2 | 1 | - | - | - | - | z4=1 | Ni1Ti1 |
| 11.2.20 | 164155 | 28 | 0.0190 | -0.5423 | 2 | 9 | 2 | 1 | - | - | - | - | z4=1 | Ni1Ti1 |
| 11.2.21 | 164154 | 28 | 0.0218 | -0.4976 | 2 | 9 | 2 | 1 | - | - | - | - | z4=1 | Ni1Ti1 |
| 11.2.22 | 164153 | 28 | 0.0216 | -0.4411 | 2 | 9 | 2 | 1 | - | - | - | - | z4=1 | Ni1Ti1 |
| 11.2.23 | 164152 | 28 | 0.0239 | -0.4421 | 2 | 9 | 2 | 1 | - | - | - | - | z4=1 | Ni1Ti1 |
| 11.2.24 | 164151 | 28 | 0.0220 | -0.3975 | 2 | 9 | 2 | 1 | - | - | - | - | z4=1 | Ni1Ti1 |
| 11.2.25 | 157983 | 28 | 0.0165 | -0.5367 | 2 | 9 | 2 | 1 | - | - | - | - | z4=1 | Ni1Ti1 |
| 11.2.26 | 105415 | 28 | 0.0161 | -0.5971 | 2 | 9 | 2 | 1 | - | - | - | - | z4=1 | Ni1Ti1 |
| 11.2.27 | 105414 | 28 | 0.0113 | -0.7643 | 2 | 5 | 2 | 1 | - | - | - | - | z2w,3=1<br>z4=1 | Ni1Ti1 |
| 11.2.28 | 66316 | 60 | 0.0061 | -0.0373 | 2 | 4 | 0 | 2 | - | - | - | - | z2w,1=1<br>z4=2 | Nd1Ru1Si2 |
| 11.2.29 | 42981 | 88 | 0.0574 | -0.1654 | 2 | 14 | 2 | 1 | - | - | - | - | z2w,1=1<br>z4=1 | Nb2Se3 |
| 11.2.30 | 54283 | 172 | 0.0248 | -0.7698 | 2 | 4 | 0 | 2 | - | - | - | - | z2w,1=1<br>z4=2 | Nb1Pt3 |
| 11.2.31 | 93368 | 132 | 0.0745 | -0.9344 | 2 | 7 | 0 | 2 | - | - | - | - | z2w,3=1<br>z4=2 | Na1O8P1W2 |
| 11.2.32 | 73453 | 60 | 0.0211 | -1.4556 | 2 | 13 | 2 | 1 | - | - | - | - | z2w,3=1<br>z4=3 | Mo2S3 |
| 11.2.33 | 62486 | 60 | 0.0334 | -1.4440 | 2 | 5 | 2 | 1 | - | - | - | - | z2w,3=1<br>z4=1 | Mo2S3 |
| 11.2.34 | 16261 | 60 | 0.0022 | -1.3203 | 2 | 13 | 2 | 1 | - | - | - | - | z2w,3=1<br>z4=3 | Mo2.06S3 |
| 11.2.35 | 14349 | 72 | 0.0610 | -0.2128 | 2 | 3 | 0 | 2 | - | - | - | - | z4=2 | Mo1Te2 |
| 11.2.36 | 59708 | 56 | 0.0168 | -0.1579 | 2 | 4 | 0 | 2 | - | - | - | - | z2w,1=1<br>z4=2 | Li4S4Si1 |
| 11.2.37 | 68366 | 92 | 0.0486 | -2.4150 | 2 | 8 | 0 | 2 | - | - | - | - | z2w,1=1 | In2S7Sn3 |
| 11.2.38 | 26525 | 124 | 0.0123 | -0.0399 | 2 | 14 | 2 | 1 | - | - | - | - | z2w,1=1<br>z4=1 | H8Ba1O7Pt1 |
| 11.2.39 | 437 | 72 | 0.0120 | 0.0030 | 2 | 2 | 2 | 1 | - | - | - | - | z2w,1=1<br>z2w,3=1<br>z4=3 | H6Br3N1O1Sn1 |
| 11.2.40 | 24075 | 204 | 0.0052 | -0.0652 | 2 | 8 | 0 | 2 | - | - | - | - | z2w,1=1 | H5In1O7S1 |
| 11.2.41 | 15199 | 112 | 0.0159 | -0.0817 | 2 | 2 | 2 | 1 | - | - | - | - | z2w,1=1<br>z2w,3=1<br>z4=3 | H5Fe1O7S1 |
| 11.2.42 | 262728 | 192 | 0.0071 | -0.0486 | 2 | 8 | 0 | 2 | - | - | - | - | z2w,1=1 | H4Co2O10P2Pb1 |
| 11.2.43 | 60659 | 60 | 0.0445 | -0.0147 | 2 | 14 | 2 | 1 | - | - | - | - | z2w,1=1<br>z4=1 | H4Ba1O3 |
| 11.2.44 | 151476 | 544 | 0.0009 | -0.0306 | 2 | 13 | 2 | 1 | - | - | - | - | z2w,3=1<br>z4=3 | H4Ag2K4O2S9Sn3 |
| 11.2.45 | 62901 | 224 | 0.0032 | -0.0057 | 2 | 12 | 0 | 2 | - | - | - | - | z2w,1=1<br>z2w,3=1<br>z4=2 | H3Dy3Ge1O8 |
| 11.2.46 | 70101 | 100 | 0.0210 | -0.1531 | 2 | 10 | 0 | 2 | - | - | - | - | z2w,3=1 | H3Cl1Cu2O3 |
| 11.2.47 | 153707 | 100 | 0.0399 | -0.1982 | 2 | 7 | 0 | 2 | - | - | - | - | z2w,3=1<br>z4=2 | H3Cl1Cu2O3 |
| 11.2.48 | 166218 | 76 | 0.0253 | -0.2213 | 2 | 1 | 2 | 1 | - | - | - | - | z4=3 | H2O4P1Rb1 |
| 11.2.49 | 27406 | 28 | 0.0388 | -0.2031 | 2 | 3 | 0 | 2 | - | - | - | - | z4=2 | H2N1Rb1 |
| 11.2.50 | 169093 | 76 | 0.0649 | -0.1979 | 2 | 7 | 0 | 2 | - | - | - | - | z2w,3=1<br>z4=2 | H2K1O4P1 |

Continued on next page

**TABLE XIV– continued**

| TQC | ICSD | NoE | DG | IndG | SubSG (w/ T) | cl | SubSG (w/o T) | cl | Max. Sub. SG | yes or no | Max. Sub. SG | yes or no | SIs (w/ T) | Stoich. Formula |
|---|---|---|---|---|---|---|---|---|---|---|---|---|---|---|
| 11.2.51 | 6163 | 284 | 0.0010 | -0.0368 | 2 | 8 | 0 | 2 | - | - | - | - | z2w,1=1 | H2Ga11Na1O18 |
| 11.2.52 | 36300 | 76 | 0.0658 | -0.1179 | 2 | 7 | 0 | 2 | - | - | - | - | z2w,3=1<br>z4=2 | H2Cs1O4P1 |
| 11.2.53 | 20449 | 60 | 0.0034 | -0.2293 | 2 | 7 | 0 | 2 | - | - | - | - | z2w,3=1<br>z4=2 | H2Cl1O2Y1 |
| 11.2.54 | 260836 | 88 | 0.0107 | -0.0429 | 2 | 10 | 0 | 2 | - | - | - | - | z2w,3=1 | H2Cl1O2Tm1 |
| 11.2.55 | 260832 | 80 | 0.0000 | -0.0260 | 2 | 1 | 2 | 1 | - | - | - | - | z4=3 | H2Cl1O2Tb1 |
| 11.2.56 | 260834 | 84 | 0.0049 | -0.0068 | 2 | 10 | 0 | 2 | - | - | - | - | z2w,3=1 | H2Cl1Ho1O2 |
| 11.2.57 | 6164 | 64 | 0.0113 | -0.0106 | 2 | 1 | 2 | 1 | - | - | - | - | z4=3 | H1O2Tb1 |
| 11.2.58 | 162390 | 52 | 0.0097 | -0.0270 | 2 | 1 | 2 | 1 | - | - | - | - | z4=3 | H1Nd1O2 |
| 11.2.59 | 424942 | 80 | 0.0051 | -0.2190 | 2 | 10 | 0 | 2 | - | - | - | - | z2w,3=1 | H1F4Mn1Na2O3P1 |
| 11.2.60 | 201682 | 116 | 0.0038 | -0.1151 | 2 | 4 | 0 | 2 | - | - | - | - | z2w,1=1<br>z4=2 | H14Cl2N5O2Tc1 |
| 11.2.61 | 54147 | 380 | 0.0399 | -0.1329 | 2 | 3 | 0 | 2 | - | - | - | - | z4=2 | H14Cd1Cu4O18S2 |
| 11.2.62 | 156004 | 228 | 0.0015 | -0.0706 | 2 | 1 | 2 | 1 | - | - | - | - | z4=3 | H12Cl8Hg3N2O2 |
| 11.2.63 | 63219 | 188 | 0.0060 | -0.0221 | 2 | 4 | 0 | 2 | - | - | - | - | z2w,1=1<br>z4=2 | H10Mo3N2O11 |
| 11.2.64 | 418027 | 108 | 0.0380 | -0.2592 | 2 | 9 | 2 | 1 | - | - | - | - | z4=1 | Ge1Hf1Pd1 |
| 11.2.65 | 180364 | 200 | 0.0544 | -0.6250 | 2 | 6 | 2 | 1 | - | - | - | - | z2w,1=1<br>z4=3 | Ga5Pt1 |
| 11.2.66 | 89343 | 148 | 0.0030 | -0.1181 | 2 | 15 | 0 | 2 | - | - | - | - | z2w,1=1<br>z2w,3=1 | F3K1Mn1 |
| 11.2.67 | 94536 | 188 | 0.0138 | -0.0811 | 2 | 2 | 2 | 1 | - | - | - | - | z2w,1=1<br>z2w,3=1<br>z4=3 | Eu1K3O8P2 |
| 11.2.68 | 190487 | 76 | 0.0097 | -0.2004 | 2 | 7 | 0 | 2 | - | - | - | - | z2w,3=1<br>z4=2 | D2O4P1Rb1 |
| 11.2.69 | 67816 | 296 | 0.0242 | -0.0646 | 2 | 10 | 0 | 2 | - | - | - | - | z2w,3=1 | Cu4K1Mg1O13V3 |
| 11.2.70 | 414338 | 72 | 0.0957 | -0.3998 | 2 | 1 | 2 | 1 | - | - | - | - | z4=3 | Cu1Nb1Te2 |
| 11.2.71 | 69990 | 292 | 0.0208 | -0.1099 | 2 | 4 | 0 | 2 | - | - | - | - | z2w,1=1<br>z4=2 | Cu1La2O6Sn1 |
| 11.2.72 | 108896 | 148 | 0.0043 | -0.1559 | 2 | 5 | 2 | 1 | - | - | - | - | z2w,3=1<br>z4=1 | Co1O3Sr1 |
| 11.2.73 | 411140 | 132 | 0.0042 | -0.0121 | 2 | 9 | 2 | 1 | - | - | - | - | z4=1 | Cl7Cs1Ti2 |
| 11.2.74 | 415798 | 112 | 0.0319 | -0.2098 | 2 | 2 | 2 | 1 | - | - | - | - | z2w,1=1<br>z2w,3=1<br>z4=3 | Cl2Co2O3Te1 |
| 11.2.75 | 65346 | 92 | 0.0146 | -0.2124 | 2 | 1 | 2 | 1 | - | - | - | - | z4=3 | Cl1Fe1Li1Mo1O4 |
| 11.2.76 | 54553 | 152 | 0.0114 | -0.4373 | 2 | 1 | 2 | 1 | - | - | - | - | z4=3 | Ce1In2Pt2 |
| 11.2.77 | 862 | 24 | 0.1605 | -0.3797 | 2 | 11 | 2 | 1 | - | - | - | - | z2w,1=1<br>z2w,3=1<br>z4=1 | Ca1Sb2 |
| 11.2.78 | 62428 | 300 | 0.0003 | -0.0122 | 2 | 10 | 0 | 2 | - | - | - | - | z2w,3=1 | C4H2As1F17N2 |
| 11.2.79 | 110059 | 84 | 0.0026 | -0.0302 | 2 | 2 | 2 | 1 | - | - | - | - | z2w,1=1<br>z2w,3=1<br>z4=3 | C4H12Cd1Cl3N1 |
| 11.2.80 | 110260 | 48 | 0.0356 | -0.0174 | 2 | 7 | 0 | 2 | - | - | - | - | z2w,3=1<br>z4=2 | C3H10Cl1N1 |
| 11.2.81 | 20741 | 188 | 0.0503 | -0.1031 | 2 | 9 | 2 | 1 | - | - | - | - | z4=1 | C2H8N4O7Se1U1 |
| 11.2.82 | 416300 | 80 | 0.0708 | -1.8202 | 2 | 3 | 0 | 2 | - | - | - | - | z4=2 | Bi2Pd2Sr1 |
| 11.2.83 | 42679 | 20 | 0.0094 | -1.2823 | 2 | 2 | 2 | 1 | - | - | - | - | z2w,1=1<br>z2w,3=1<br>z4=3 | Bi1 |
| 11.2.84 | 81819 | 96 | 0.0285 | -1.1198 | 2 | 8 | 0 | 2 | - | - | - | - | z2w,1=1 | Ba3Ge2N2 |
| 11.2.85 | 409517 | 40 | 0.0580 | -0.3049 | 2 | 8 | 0 | 2 | - | - | - | - | z2w,1=1 | Ba1Sb2 |
| 11.2.86 | 194985 | 440 | 0.0160 | 0.0096 | 2 | 7 | 0 | 2 | - | - | - | - | z2w,3=1<br>z4=2 | Ba1Nd2O10Ti3 |

Continued on next page

**TABLE XIV– continued**

| TQC | ICSD | NoE | DG | IndG | SubSG (w/ T) | cl | SubSG (w/o T) | cl | Max. Sub. SG | yes or no | Max. Sub. SG | yes or no | SIs (w/ T) | Stoich. Formula |
|---|---|---|---|---|---|---|---|---|---|---|---|---|---|---|
| 11.2.87 | 429621 | 432 | 0.0011 | -0.0028 | 2 | 6 | 2 | 1 | - | - | - | - | z2w,1=1<br>z4=3 | Ba1Ho4O17Si5 |
| 11.2.88 | 167615 | 252 | 0.0013 | -0.0025 | 2 | 5 | 2 | 1 | - | - | - | - | z2w,3=1<br>z4=1 | Ba1Er2O10Si3 |
| 11.2.89 | 416299 | 80 | 0.0749 | -1.5647 | 2 | 3 | 0 | 2 | - | - | - | - | z4=2 | Ba1Bi2Pd2 |
| 11.2.90 | 413519 | 120 | 0.0429 | -0.2337 | 2 | 4 | 0 | 2 | - | - | - | - | z2w,1=1<br>z4=2 | Ba1.5Sb3 |
| 11.2.91 | 261042 | 116 | 0.0043 | -1.2047 | 2 | 3 | 0 | 2 | - | - | - | - | z4=2 | Au2In4Yb1 |
| 11.2.92 | 410884 | 152 | 0.0101 | -0.0236 | 2 | 1 | 2 | 1 | - | - | - | - | z4=3 | Au2Eu2Sn5 |
| 11.2.93 | 416298 | 84 | 0.0246 | -1.2577 | 2 | 1 | 2 | 1 | - | - | - | - | z4=3 | Au2Ba1Sb2 |
| 11.2.94 | 902 | 108 | 0.0268 | -0.2321 | 2 | 2 | 2 | 1 | - | - | - | - | z2w,1=1<br>z2w,3=1<br>z4=3 | As1Cl1Co2O4 |
| 11.2.95 | 43456 | 68 | 0.0031 | -0.4564 | 2 | 12 | 0 | 2 | - | - | - | - | z2w,1=1<br>z2w,3=1<br>z4=2 | Am1 |
| 11.2.96 | 104626 | 156 | 0.0045 | -0.0888 | 2 | 1 | 2 | 1 | - | - | - | - | z4=3 | Al1Ce3 |
| 11.2.97 | 57494 | 56 | 0.1020 | -1.7956 | 2 | 1 | 2 | 1 | - | - | - | - | z4=3 | Al1Au1 |
| 11.2.98 | 104649 | 56 | 0.0659 | -1.9409 | 2 | 1 | 2 | 1 | - | - | - | - | z4=3 | Al1Au1 |
| 11.4.1 | 166509 | 138 | - | - | - | - | - | - | See details in Table XIX | | | | | S3Ta1 |
| 11.4.2 | 15251 | 138 | - | - | - | - | - | - | See details in Table XIX | | | | | S3Ta1 |
| 11.4.3 | 99170 | 58 | - | - | - | - | - | - | See details in Table XIX | | | | | Pr1Ru1Si2 |
| 11.4.4 | 409179 | 98 | - | - | - | - | - | - | See details in Table XIX | | | | | P5Yb1 |
| 11.4.5 | 1239 | 186 | - | - | - | - | - | - | See details in Table XIX | | | | | O2Rb9 |
| 11.4.6 | 424526 | 66 | - | - | - | - | - | - | See details in Table XIX | | | | | Np1Se3 |
| 11.4.7 | 82605 | 86 | - | - | - | - | - | - | See details in Table XIX | | | | | Ni2Ta1Te3 |
| 11.4.8 | 84826 | 46 | - | - | - | - | - | - | See details in Table XIX | | | | | Ni1P1Zr2 |
| 11.4.9 | 71498 | 186 | - | - | - | - | - | - | See details in Table XIX | | | | | Nb1Se3 |
| 11.4.10 | 30002 | 186 | - | - | - | - | - | - | See details in Table XIX | | | | | Nb1Se3 |
| 11.4.11 | 97682 | 58 | - | - | - | - | - | - | See details in Table XIX | | | | | Mo2S2Sb1 |
| 11.4.12 | 104785 | 38 | - | - | - | - | - | - | See details in Table XIX | | | | | Li7Sn3 |
| 11.4.13 | 67845 | 134 | - | - | - | - | - | - | See details in Table XIX | | | | | Li4O8V3 |
| 11.4.14 | 71960 | 54 | - | - | - | - | - | - | See details in Table XIX | | | | | La1Ru1Si2 |
| 11.4.15 | 402125 | 78 | - | - | - | - | - | - | See details in Table XIX | | | | | In5S7Tl1 |
| 11.4.16 | 71524 | 126 | - | - | - | - | - | - | See details in Table XIX | | | | | I3Ir1Y3 |
| 11.4.17 | 84828 | 46 | - | - | - | - | - | - | See details in Table XIX | | | | | Hf2Ni1P1 |
| 11.4.18 | 405616 | 46 | - | - | - | - | - | - | See details in Table XIX | | | | | Hf2Ni1P1 |
| 11.4.19 | 248158 | 226 | - | - | - | - | - | - | See details in Table XIX | | | | | H6Cu1N1Na3O13P2 |
| 11.4.20 | 48018 | 94 | - | - | - | - | - | - | See details in Table XIX | | | | | H6Cl2Cu4O6 |
| 11.4.21 | 16780 | 94 | - | - | - | - | - | - | See details in Table XIX | | | | | H6Br2Cu4O6 |
| 11.4.22 | 250377 | 94 | - | - | - | - | - | - | See details in Table XIX | | | | | H5Ga1N1O5P1 |
| 11.4.23 | 56961 | 262 | - | - | - | - | - | - | See details in Table XIX | | | | | H5Al2As2Ca2F4Na1O11 |
| 11.4.24 | 239168 | 390 | - | - | - | - | - | - | See details in Table XIX | | | | | H4Cu5Na2O19S3Te1 |
| 11.4.25 | 239167 | 422 | - | - | - | - | - | - | See details in Table XIX | | | | | H4Cu5K2O19S3Te1 |
| 11.4.26 | 183584 | 182 | - | - | - | - | - | - | See details in Table XIX | | | | | H4Cu3Mo1O8 |
| 11.4.27 | 262400 | 178 | - | - | - | - | - | - | See details in Table XIX | | | | | H3O8V1Y3 |
| 11.4.28 | 36417 | 198 | - | - | - | - | - | - | See details in Table XIX | | | | | H3O14Sb5 |
| 11.4.29 | 38152 | 126 | - | - | - | - | - | - | See details in Table XIX | | | | | H3Cu2N1O6 |
| 11.4.30 | 27668 | 126 | - | - | - | - | - | - | See details in Table XIX | | | | | H3Cu2N1O6 |
| 11.4.31 | 252429 | 102 | - | - | - | - | - | - | See details in Table XIX | | | | | H3Cl1Cu1O3Zn1 |
| 11.4.32 | 413481 | 218 | - | - | - | - | - | - | See details in Table XIX | | | | | H3Ca1Cs1O12Se3 |
| 11.4.33 | 27493 | 94 | - | - | - | - | - | - | See details in Table XIX | | | | | H3Br1Cu2O3 |
| 11.4.34 | 172060 | 98 | - | - | - | - | - | - | See details in Table XIX | | | | | H2Na3O6Os1 |
| 11.4.35 | 159479 | 182 | - | - | - | - | - | - | See details in Table XIX | | | | | H2K1O9S2Ti1 |
| 11.4.36 | 68173 | 114 | - | - | - | - | - | - | See details in Table XIX | | | | | H2Cu1O6Pb1S1 |
| 11.4.37 | 25714 | 114 | - | - | - | - | - | - | See details in Table XIX | | | | | H2Cu1O6Pb1S1 |
| 11.4.38 | 15915 | 114 | - | - | - | - | - | - | See details in Table XIX | | | | | H2Cu1O6Pb1S1 |

Continued on next page

**TABLE XIV– continued**

| TQC | ICSD | NoE | DG | IndG | SubSG cl (w/ T) | SubSG cl (w/o T) | Max. Sub. SG yes or no | Max. Sub. SG yes or no | SIs (w/ T) | Stoich. Formula |
|---|---|---|---|---|---|---|---|---|---|---|
| 11.4.39 | 15435 | 114 | - | - | - - | - - | See details in Table XIX | | | H2Cu1O6Pb1S1 |
| 11.4.40 | 200123 | 122 | - | - | - - | - - | See details in Table XIX | | | H2Cl2Cu1O4Pb3 |
| 11.4.41 | 260837 | 90 | - | - | - - | - - | See details in Table XIX | | | H2Cl1O2Yb1 |
| 11.4.42 | 260831 | 78 | - | - | - - | - - | See details in Table XIX | | | H2Cl1Gd1O2 |
| 11.4.43 | 260833 | 82 | - | - | - - | - - | See details in Table XIX | | | H2Cl1Dy1O2 |
| 11.4.44 | 26341 | 74 | - | - | - - | - - | See details in Table XIX | | | H1O2Yb1 |
| 11.4.45 | 60675 | 46 | - | - | - - | - - | See details in Table XIX | | | H1La1O2 |
| 11.4.46 | 82535 | 138 | - | - | - - | - - | See details in Table XIX | | | H1K1O7Se2V1 |
| 11.4.47 | 35408 | 62 | - | - | - - | - - | See details in Table XIX | | | H1F3Sr1 |
| 11.4.48 | 36582 | 102 | - | - | - - | - - | See details in Table XIX | | | H1Cu1O5Se1Tl1 |
| 11.4.49 | 36581 | 102 | - | - | - - | - - | See details in Table XIX | | | H1Cu1O5S1Tl1 |
| 11.4.50 | 4333 | 78 | - | - | - - | - - | See details in Table XIX | | | H1Ca1Na1O4Si1 |
| 11.4.51 | 35409 | 62 | - | - | - - | - - | See details in Table XIX | | | H1Ba1F3 |
| 11.4.52 | 420467 | 182 | - | - | - - | - - | See details in Table XIX | | | H1B2F10Nd1 |
| 11.4.53 | 180447 | 62 | - | - | - - | - - | See details in Table XIX | | | H1B1Ca1O3 |
| 11.4.54 | 67314 | 90 | - | - | - - | - - | See details in Table XIX | | | H1As1Cl2O3Pb2 |
| 11.4.55 | 188917 | 90 | - | - | - - | - - | See details in Table XIX | | | H1As1Cl2O3Pb2 |
| 11.4.56 | 423907 | 238 | - | - | - - | - - | See details in Table XIX | | | H1Al4B3O14U1 |
| 11.4.57 | 68016 | 250 | - | - | - - | - - | See details in Table XIX | | | H18Br7N6Ru2Zn1 |
| 11.4.58 | 63297 | 250 | - | - | - - | - - | See details in Table XIX | | | H18Br7N6Ru2Zn1 |
| 11.4.59 | 100691 | 330 | - | - | - - | - - | See details in Table XIX | | | H11Cs4O18P3Te1 |
| 11.4.60 | 76293 | 194 | - | - | - - | - - | See details in Table XIX | | | Ge6Pt4Y3 |
| 11.4.61 | 114 | 86 | - | - | - - | - - | See details in Table XIX | | | Gd1P5 |
| 11.4.62 | 260894 | 206 | - | - | - - | - - | See details in Table XIX | | | Gd1K3O8P2 |
| 11.4.63 | 66513 | 62 | - | - | - - | - - | See details in Table XIX | | | Fe1Ta1Te3 |
| 11.4.64 | 166586 | 126 | - | - | - - | - - | See details in Table XIX | | | F5K1N1O1Os1 |
| 11.4.65 | 424958 | 94 | - | - | - - | - - | See details in Table XIX | | | F3Fe1Na2O4S1 |
| 11.4.66 | 10081 | 54 | - | - | - - | - - | See details in Table XIX | | | Eu1Sb2 |
| 11.4.67 | 280200 | 110 | - | - | - - | - - | See details in Table XIX | | | Eu1K1P1Se4 |
| 11.4.68 | 90775 | 166 | - | - | - - | - - | See details in Table XIX | | | Er1Na1O8S2 |
| 11.4.69 | 33963 | 30 | - | - | - - | - - | See details in Table XIX | | | D1Rb1S1 |
| 11.4.70 | 426627 | 214 | - | - | - - | - - | See details in Table XIX | | | Cu3Ge1O10P2 |
| 11.4.71 | 167596 | 30 | - | - | - - | - - | See details in Table XIX | | | Cu1Zr1 |
| 11.4.72 | 422033 | 182 | - | - | - - | - - | See details in Table XIX | | | Cu1N4O10 |
| 11.4.73 | 2232 | 122 | - | - | - - | - - | See details in Table XIX | | | Cu1I2O6 |
| 11.4.74 | 60759 | 98 | - | - | - - | - - | See details in Table XIX | | | Cu1F2O3W1 |
| 11.4.75 | 182754 | 294 | - | - | - - | - - | See details in Table XIX | | | Cs2Cu3F12Zr1 |
| 11.4.76 | 166587 | 126 | - | - | - - | - - | See details in Table XIX | | | Cs1F5N1O1Os1 |
| 11.4.77 | 409488 | 98 | - | - | - - | - - | See details in Table XIX | | | Cr5P3Zr1 |
| 11.4.78 | 35044 | 98 | - | - | - - | - - | See details in Table XIX | | | Cr2Nb2Se10 |
| 11.4.79 | 1029 | 110 | - | - | - - | - - | See details in Table XIX | | | Co1I4Rb2 |
| 11.4.80 | 99178 | 74 | - | - | - - | - - | See details in Table XIX | | | Ce1P5 |
| 11.4.81 | 409181 | 74 | - | - | - - | - - | See details in Table XIX | | | Ce1P5 |
| 11.4.82 | 413847 | 294 | - | - | - - | - - | See details in Table XIX | | | Ca4Ir8P7 |
| 11.4.83 | 183583 | 174 | - | - | - - | - - | See details in Table XIX | | | Ca1O3Rh1 |
| 11.4.84 | 58880 | 130 | - | - | - - | - - | See details in Table XIX | | | Ca1Cu1 |
| 11.4.85 | 248211 | 282 | - | - | - - | - - | See details in Table XIX | | | C6H7N1O14U2 |
| 11.4.86 | 1138 | 234 | - | - | - - | - - | See details in Table XIX | | | C4H10Cs1O13Ti1 |
| 11.4.87 | 237796 | 130 | - | - | - - | - - | See details in Table XIX | | | C1Cu1Na3O7P1 |
| 11.4.88 | 237797 | 126 | - | - | - - | - - | See details in Table XIX | | | C1Co1Na3O7P1 |
| 11.4.89 | 20270 | 50 | - | - | - - | - - | See details in Table XIX | | | C1Br1Hg1N1S1 |
| 11.4.90 | 151019 | 162 | - | - | - - | - - | See details in Table XIX | | | C10H21I3N4 |
| 11.4.91 | 410948 | 150 | - | - | - - | - - | See details in Table XIX | | | Br9Te1W1 |
| 11.4.92 | 39257 | 118 | - | - | - - | - - | See details in Table XIX | | | Bi9I2 |
| 11.4.93 | 162834 | 62 | - | - | - - | - - | See details in Table XIX | | | Bi1Fe1O3 |
| 11.4.94 | 79480 | 186 | - | - | - - | - - | See details in Table XIX | | | Ba2Cl1F7Mn1Ni1 |
| 11.4.95 | 82063 | 146 | - | - | - - | - - | See details in Table XIX | | | Ba1O8V3 |

Continued on next page

**TABLE XIV– continued**

| TQC | ICSD | NoE | DG | IndG | SubSG (w/ T) | cl | SubSG (w/o T) | cl | Max. Sub. SG | yes or no | Max. Sub. SG | yes or no | SIs (w/ T) | Stoich. Formula |
|---|---|---|---|---|---|---|---|---|---|---|---|---|---|---|
| 11.4.96 | 65932 | 74 | - | - | - | - | - | - | See details in Table XIX | | | | | B3Er1Mo1 |
| 11.4.97 | 65934 | 186 | - | - | - | - | - | - | See details in Table XIX | | | | | B2Ba2Cl1O6Yb1 |
| 11.4.98 | 262073 | 174 | - | - | - | - | - | - | See details in Table XIX | | | | | B2Ba2Cl1Gd1O6 |
| 11.4.99 | 413846 | 294 | - | - | - | - | - | - | See details in Table XIX | | | | | As7Ca4Ir8 |
| 11.4.100 | 427443 | 114 | - | - | - | - | - | - | See details in Table XIX | | | | | As3Ca1Fe5 |
| 11.4.101 | 420736 | 110 | - | - | - | - | - | - | See details in Table XIX | | | | | As1Eu1K1S4 |
| 12.1.1 | 187638 | 22 | 0.0424 | -1.3011 | 2 | 3 | 0 | 1 | - | - | - | - | z2w,3=1<br>z4=2 | Y1 |
| 12.1.2 | 43412 | 36 | 0.0251 | -0.2109 | 2 | 7 | 0 | 1 | - | - | - | - | z2w,1=1<br>z2w,2=1 | Te4Ti3 |
| 12.1.3 | 41047 | 56 | 0.0850 | -0.5151 | 2 | 5 | 0 | 1 | - | - | - | - | z2w,1=1<br>z2w,2=1<br>z2w,3=1<br>z4=2 | Ta2Te3 |
| 12.1.4 | 37122 | 76 | 0.0073 | -0.3026 | 2 | 4 | 0 | 1 | - | - | - | - | z2w,3=1 | Se8Tl1V5 |
| 12.1.5 | 601347 | 36 | 0.0208 | -0.4038 | 2 | 3 | 0 | 1 | - | - | - | - | z2w,3=1<br>z4=2 | Se4Ti3 |
| 12.1.6 | 48176 | 78 | 0.0033 | -0.6858 | 2 | 6 | 0 | 1 | - | - | - | - | z4=2 | Rh1Si2Y2 |
| 12.1.7 | 42077 | 30 | 0.1617 | -0.1291 | 2 | 1 | 0 | 1 | - | - | - | - | z2w,1=1<br>z2w,2=1<br>z2w,3=1 | P2V1 |
| 12.1.8 | 56694 | 50 | 0.1025 | -0.9392 | 2 | 2 | 0 | 1 | - | - | - | - | z2w,1=1<br>z2w,2=1<br>z4=2 | O5Ti5 |
| 12.1.9 | 34417 | 68 | 0.0091 | -0.3959 | 2 | 5 | 0 | 1 | - | - | - | - | z2w,1=1<br>z2w,2=1<br>z2w,3=1<br>z4=2 | O2V1 |
| 12.1.10 | 34416 | 68 | 0.0080 | -0.3528 | 2 | 5 | 0 | 1 | - | - | - | - | z2w,1=1<br>z2w,2=1<br>z2w,3=1<br>z4=2 | O2V1 |
| 12.1.11 | 50409 | 108 | 0.0127 | -0.4184 | 2 | 2 | 0 | 1 | - | - | - | - | z2w,1=1<br>z2w,2=1<br>z4=2 | O13V6 |
| 12.1.12 | 15028 | 108 | 0.0147 | -0.4252 | 2 | 2 | 0 | 1 | - | - | - | - | z2w,1=1<br>z2w,2=1<br>z4=2 | O13V6 |
| 12.1.13 | 646 - | 42 | 0.0323 | -0.4163 | 2 | 1 | 0 | 1 | - | - | - | - | z2w,1=1<br>z2w,2=1<br>z2w,3=1 | Ni1Se4Ti2 |
| 12.1.14 | 646544 | 42 | 0.0166 | -0.4109 | 2 | 1 | 0 | 1 | - | - | - | - | z2w,1=1<br>z2w,2=1<br>z2w,3=1 | Ni1Se4Ti2 |
| 12.1.15 | 646533 | 42 | 0.0443 | -1.4532 | 2 | 4 | 0 | 1 | - | - | - | - | z2w,3=1 | Ni1Se4Si2 |
| 12.1.16 | 48004 | 62 | 0.0204 | -0.6562 | 2 | 6 | 0 | 1 | - | - | - | - | z4=2 | Ni1Sc3Si3 |
| 12.1.17 | 646389 | 74 | 0.0402 | -0.3577 | 2 | 4 | 0 | 1 | - | - | - | - | z2w,3=1 | Ni1S8Ti4 |
| 12.1.18 | 646388 | 42 | 0.0086 | -0.7492 | 2 | 1 | 0 | 1 | - | - | - | - | z2w,1=1<br>z2w,2=1<br>z2w,3=1 | Ni1S4Ti2 |
| 12.1.19 | 646321 | 52 | 0.0112 | -0.6213 | 2 | 5 | 0 | 1 | - | - | - | - | z2w,1=1<br>z2w,2=1<br>z2w,3=1<br>z4=2 | Ni1Rh2Se4 |

Continued on next page

**TABLE XIV– continued**

| TQC | ICSD | NoE | DG | IndG | SubSG (w/ T) | cl | SubSG (w/o T) | cl | Max. Sub. SG | yes or no | Max. Sub. SG | yes or no | SIs (w/ T) | Stoich. Formula |
|---|---|---|---|---|---|---|---|---|---|---|---|---|---|---|
| 12.1.20 | 646318 | 52 | 0.0150 | -0.6225 | 2 | 5 | 0 | 1 | - | - | - | - | z2w,1=1<br>z2w,2=1<br>z2w,3=1<br>z4=2 | Ni1Rh2Se4 |
| 12.1.21 | 15214 | 52 | 0.0123 | -0.6211 | 2 | 5 | 0 | 1 | - | - | - | - | z2w,1=1<br>z2w,2=1<br>z2w,3=1<br>z4=2 | Ni1Rh2Se4 |
| 12.1.22 | 41238 | 222 | 0.0064 | -0.6193 | 2 | 2 | 0 | 1 | - | - | - | - | z2w,1=1<br>z2w,2=1<br>z4=2 | Nb7P4 |
| 12.1.23 | 26183 | 128 | 0.0272 | -0.3145 | 2 | 1 | 0 | 1 | - | - | - | - | z2w,1=1<br>z2w,2=1<br>z2w,3=1 | Nb2Se1 |
| 12.1.24 | 77752 | 46 | 0.2063 | 0.0083 | 2 | 5 | 0 | 1 | - | - | - | - | z2w,1=1<br>z2w,2=1<br>z2w,3=1<br>z4=2 | Nb1Sb2 |
| 12.1.25 | 645350 | 46 | 0.2041 | 0.0102 | 2 | 5 | 0 | 1 | - | - | - | - | z2w,1=1<br>z2w,2=1<br>z2w,3=1<br>z4=2 | Nb1Sb2 |
| 12.1.26 | 57430 | 46 | 0.2063 | 0.0083 | 2 | 5 | 0 | 1 | - | - | - | - | z2w,1=1<br>z2w,2=1<br>z2w,3=1<br>z4=2 | Nb1Sb2 |
| 12.1.27 | 190180 | 46 | 0.2076 | 0.0101 | 2 | 1 | 0 | 1 | - | - | - | - | z2w,1=1<br>z2w,2=1<br>z2w,3=1 | Nb1Sb2 |
| 12.1.28 | 18144 | 46 | 0.1577 | -0.0537 | 2 | 1 | 0 | 1 | - | - | - | - | z2w,1=1<br>z2w,2=1<br>z2w,3=1 | Nb1Sb2 |
| 12.1.29 | 645169 | 46 | 0.2509 | -0.0022 | 2 | 5 | 0 | 1 | - | - | - | - | z2w,1=1<br>z2w,2=1<br>z2w,3=1<br>z4=2 | Nb1P2 |
| 12.1.30 | 29383 | 16 | 0.0504 | -0.5196 | 2 | 2 | 0 | 1 | - | - | - | - | z2w,1=1<br>z2w,2=1<br>z4=2 | N3Na1 |
| 12.1.31 | 29382 | 16 | 0.0225 | -0.3423 | 2 | 2 | 0 | 1 | - | - | - | - | z2w,1=1<br>z2w,2=1<br>z4=2 | N3Na1 |
| 12.1.32 | 29381 | 16 | 0.0000 | -0.3615 | 2 | 2 | 0 | 1 | - | - | - | - | z2w,1=1<br>z2w,2=1<br>z4=2 | N3Na1 |
| 12.1.33 | 29380 | 16 | 0.0344 | -0.3697 | 2 | 2 | 0 | 1 | - | - | - | - | z2w,1=1<br>z2w,2=1<br>z4=2 | N3Na1 |
| 12.1.34 | 29379 | 16 | 0.0350 | -0.3593 | 2 | 2 | 0 | 1 | - | - | - | - | z2w,1=1<br>z2w,2=1<br>z4=2 | N3Na1 |
| 12.1.35 | 29378 | 16 | 0.0383 | -0.3 - | 2 | 2 | 0 | 1 | - | - | - | - | z2w,1=1<br>z2w,2=1<br>z4=2 | N3Na1 |
| 12.1.36 | 29377 | 16 | 0.0321 | -0.3634 | 2 | 2 | 0 | 1 | - | - | - | - | z2w,1=1<br>z2w,2=1<br>z4=2 | N3Na1 |

Continued on next page

**TABLE XIV– continued**

| TQC | ICSD | NoE | DG | IndG | SubSG (w/ T) | cl | SubSG (w/o T) | cl | Max. Sub. SG | yes or no | Max. Sub. SG | yes or no | SIs (w/ T) | Stoich. Formula |
|---|---|---|---|---|---|---|---|---|---|---|---|---|---|---|
| 12.1.37 | 29376 | 16 | 0.0365 | -0.4043 | 2 | 2 | 0 | 1 | - | - | - | - | z2w,1=1 z2w,2=1 z4=2 | N3Na1 |
| 12.1.38 | 29375 | 16 | 0.0403 | -0.3450 | 2 | 2 | 0 | 1 | - | - | - | - | z2w,1=1 z2w,2=1 z4=2 | N3Na1 |
| 12.1.39 | 29374 | 16 | 0.0282 | -0.3527 | 2 | 2 | 0 | 1 | - | - | - | - | z2w,1=1 z2w,2=1 z4=2 | N3Na1 |
| 12.1.40 | 29373 | 16 | 0.0151 | -0.3531 | 2 | 2 | 0 | 1 | - | - | - | - | z2w,1=1 z2w,2=1 z4=2 | N3Na1 |
| 12.1.41 | 29372 | 16 | 0.0062 | -0.3473 | 2 | 2 | 0 | 1 | - | - | - | - | z2w,1=1 z2w,2=1 z4=2 | N3Na1 |
| 12.1.42 | 29371 | 16 | 0.0203 | -0.3267 | 2 | 2 | 0 | 1 | - | - | - | - | z2w,1=1 z2w,2=1 z4=2 | N3Na1 |
| 12.1.43 | 29370 | 16 | 0.0152 | -0.3429 | 2 | 2 | 0 | 1 | - | - | - | - | z2w,1=1 z2w,2=1 z4=2 | N3Na1 |
| 12.1.44 | 151588 | 122 | 0.0046 | -0.2374 | 2 | 6 | 0 | 1 | - | - | - | - | z4=2 | Mn7Na1O12 |
| 12.1.45 | 78328 | 148 | 0.0095 | -0.0384 | 2 | 2 | 0 | 1 | - | - | - | - | z2w,1=1 z2w,2=1 z4=2 | Mn1Mo1O4 |
| 12.1.46 | 239971 | 54 | 0.0313 | 0.0106 | 2 | 3 | 0 | 1 | - | - | - | - | z2w,3=1 z4=2 | Li2Mn1O3 |
| 12.1.47 | 165326 | 38 | 0.0228 | -2.3743 | 2 | 1 | 0 | 1 | - | - | - | - | z2w,1=1 z2w,2=1 z2w,3=1 | Li1Mo1O2 |
| 12.1.48 | 49021 | 58 | 0.0195 | -0.3284 | 2 | 2 | 0 | 1 | - | - | - | - | z2w,1=1 z2w,2=1 z4=2 | K1Li1Mn1O2 |
| 12.1.49 | 240135 | 44 | 0.0466 | -0.9180 | 2 | 6 | 0 | 1 | - | - | - | - | z4=2 | In4Sr1 |
| 12.1.50 | 240133 | 44 | 0.0073 | -0.8990 | 2 | 6 | 0 | 1 | - | - | - | - | z4=2 | In4Sr1 |
| 12.1.51 | 407248 | 192 | 0.0369 | -0.8219 | 2 | 1 | 0 | 1 | - | - | - | - | z2w,1=1 z2w,2=1 z2w,3=1 | I3La5Si5 |
| 12.1.52 | 407247 | 162 | 0.0221 | -0.7584 | 2 | 2 | 0 | 1 | - | - | - | - | z2w,1=1 z2w,2=1 z4=2 | I3La4Si4 |
| 12.1.53 | 83965 | 56 | 0.0348 | -0.7402 | 2 | 6 | 0 | 1 | - | - | - | - | z4=2 | Hf2Te1 |
| 12.1.54 | 20755 | 72 | 0.0448 | -0.1752 | 2 | 5 | 0 | 1 | - | - | - | - | z2w,1=1 z2w,2=1 z2w,3=1 z4=2 | H4Cl2Ni1O2 |
| 12.1.55 | 23361 | 50 | 0.0346 | -0.0498 | 2 | 3 | 0 | 1 | - | - | - | - | z2w,3=1 z4=2 | H3N1Na2S3 |
| 12.1.56 | 62646 | 106 | 0.0317 | -0.0113 | 2 | 4 | 0 | 1 | - | - | - | - | z2w,3=1 | H3Cu2F2K1O8P2 |
| 12.1.57 | 422475 | 94 | 0.0040 | -0.0892 | 2 | 3 | 0 | 1 | - | - | - | - | z2w,3=1 z4=2 | H3Co2Mo2Na1O10 |
| 12.1.58 | 24759 | 126 | 0.0305 | -0.0192 | 2 | 4 | 0 | 1 | - | - | - | - | z2w,3=1 | H2I2O12Sb8 |
| 12.1.59 | 24760 | 126 | 0.0181 | -0.0834 | 2 | 4 | 0 | 1 | - | - | - | - | z2w,3=1 | H2Br2O12Sb8 |
| 12.1.60 | 16023 | 166 | 0.0028 | -0.1384 | 2 | 4 | 0 | 1 | - | - | - | - | z2w,3=1 | H12N2O16Zn5 |
| 12.1.61 | 262220 | 82 | 0.0127 | -0.2285 | 2 | 5 | 0 | 1 | - | - | - | - | z2w,1=1 z2w,2=1 z2w,3=1 z4=2 | Ge6La4Mg7 |

Continued on next page

**TABLE XIV– continued**

| TQC | ICSD | NoE | DG | IndG | SubSG (w/ T) | cl | SubSG (w/o T) | cl | Max. Sub. SG | yes or no | Max. Sub. SG | yes or no | SIs (w/ T) | Stoich. Formula |
|---|---|---|---|---|---|---|---|---|---|---|---|---|---|---|
| 12.1.62 | 637671 | 30 | 0.2123 | -0.3226 | 2 | 1 | 0 | 1 | - | - | - | - | z2w,1=1<br>z2w,2=1<br>z2w,3=1 | Ge2Re1 |
| 12.1.63 | 183378 | 82 | 0.0279 | -0.4508 | 2 | 6 | 0 | 1 | - | - | - | - | z4=2 | Gd1O6Ru1Sr2 |
| 12.1.64 | 430563 | 82 | 0.0025 | -0.4439 | 2 | 1 | 0 | 1 | - | - | - | - | z2w,1=1<br>z2w,2=1<br>z2w,3=1 | Fe5O7 |
| 12.1.65 | 633477 | 48 | 0.0238 | -0.2620 | 2 | 5 | 0 | 1 | - | - | - | - | z2w,1=1<br>z2w,2=1<br>z2w,3=1<br>z4=2 | Fe3Se4 |
| 12.1.66 | 24560 | 48 | 0.0335 | -0.1631 | 2 | 5 | 0 | 1 | - | - | - | - | z2w,1=1<br>z2w,2=1<br>z2w,3=1<br>z4=2 | Fe3Se4 |
| 12.1.67 | 150568 | 48 | 0.0238 | -0.3072 | 2 | 4 | 0 | 1 | - | - | - | - | z2w,3=1 | Fe3Se4 |
| 12.1.68 | 240050 | 164 | 0.0255 | -1.6313 | 2 | 1 | 0 | 1 | - | - | - | - | z2w,1=1<br>z2w,2=1<br>z2w,3=1 | Fe1Zn13 |
| 12.1.69 | 163222 | 164 | 0.0300 | -1.6223 | 2 | 1 | 0 | 1 | - | - | - | - | z2w,1=1<br>z2w,2=1<br>z2w,3=1 | Fe1Zn13 |
| 12.1.70 | 633326 | 72 | 0.0101 | -0.3378 | 2 | 4 | 0 | 1 | - | - | - | - | z2w,3=1 | Fe1S8Ti4 |
| 12.1.71 | 53534 | 72 | 0.0238 | -0.3361 | 2 | 4 | 0 | 1 | - | - | - | - | z2w,3=1 | Fe1S8Ti4 |
| 12.1.72 | 42563 | 40 | 0.0422 | -0.4108 | 2 | 4 | 0 | 1 | - | - | - | - | z2w,3=1 | Fe1S4Ti2 |
| 12.1.73 | 103658 | 160 | 0.0034 | -0.1127 | 2 | 4 | 0 | 1 | - | - | - | - | z2w,3=1 | Fe17Th2 |
| 12.1.74 | 633332 | 40 | 0.0792 | -0.3533 | 2 | 1 | 0 | 1 | - | - | - | - | z2w,1=1<br>z2w,2=1<br>z2w,3=1 | Fe0.5S2Ti1 |
| 12.1.75 | 192506 | 122 | 0.0031 | -0.0132 | 2 | 5 | 0 | 1 | - | - | - | - | z2w,1=1<br>z2w,2=1<br>z2w,3=1<br>z4=2 | F5Fe1K2 |
| 12.1.76 | 106627 | 58 | 0.0092 | -0.4469 | 2 | 3 | 0 | 1 | - | - | - | - | z2w,3=1<br>z4=2 | Eu1In4 |
| 12.1.77 | 151123 | 94 | 0.0024 | -0.0084 | 2 | 1 | 0 | 1 | - | - | - | - | z2w,1=1<br>z2w,2=1<br>z2w,3=1 | Er2O7Si2 |
| 12.1.78 | 419435 | 186 | 0.0026 | -0.0352 | 2 | 3 | 0 | 1 | - | - | - | - | z2w,3=1<br>z4=2 | Er2O3 |
| 12.1.79 | 180224 | 48 | 0.0095 | -0.0074 | 2 | 6 | 0 | 1 | - | - | - | - | z4=2 | Dy1Si1 |
| 12.1.80 | 180223 | 48 | 0.0090 | -0.0080 | 2 | 6 | 0 | 1 | - | - | - | - | z4=2 | Dy1Si1 |
| 12.1.81 | 180222 | 48 | 0.0063 | -0.0067 | 2 | 6 | 0 | 1 | - | - | - | - | z4=2 | Dy1Si1 |
| 12.1.82 | 180221 | 48 | 0.0058 | -0.0066 | 2 | 6 | 0 | 1 | - | - | - | - | z4=2 | Dy1Si1 |
| 12.1.83 | 418227 | 120 | 0.0064 | -0.0446 | 2 | 3 | 0 | 1 | - | - | - | - | z2w,3=1<br>z4=2 | Dy1P2Pt8 |
| 12.1.84 | 421466 | 138 | 0.0143 | -0.0121 | 2 | 7 | 0 | 1 | - | - | - | - | z2w,1=1<br>z2w,2=1 | Cu4Na1O12Te2Tl3 |
| 12.1.85 | 71624 | 106 | 0.0227 | -0.0381 | 2 | 2 | 0 | 1 | - | - | - | - | z2w,1=1<br>z2w,2=1<br>z4=2 | Cs2Ir1Na2O4 |
| 12.1.86 | 26977 | 90 | 0.0000 | -0.1085 | 2 | 1 | 0 | 1 | - | - | - | - | z2w,1=1<br>z2w,2=1<br>z2w,3=1 | Cr7Se8 |
| 12.1.87 | 626888 | 42 | 0.0217 | -0.0114 | 2 | 6 | 0 | 1 | - | - | - | - | z4=2 | Cr3Te4 |
| 12.1.88 | 626884 | 42 | 0.0237 | -0.0558 | 2 | 6 | 0 | 1 | - | - | - | - | z4=2 | Cr3Te4 |
| 12.1.89 | 626874 | 42 | 0.0260 | -0.0146 | 2 | 6 | 0 | 1 | - | - | - | - | z4=2 | Cr3Te4 |
| 12.1.90 | 626716 | 42 | 0.0000 | -0.4013 | 2 | 4 | 0 | 1 | - | - | - | - | z2w,3=1 | Cr3Se4 |

Continued on next page

**TABLE XIV– continued**

| TQC | ICSD | NoE | DG | IndG | SubSG cl (w/ T) | SubSG cl (w/o T) | Max. Sub. SG yes or no | Max. Sub. SG yes or no | SIs (w/ T) | Stoich. Formula |
|---|---|---|---|---|---|---|---|---|---|---|
| 12.1.91 | 626712 | 42 | 0.0000 | -0.3962 | 2 4 | 0 1 | - - | - - | z2w,3=1 | Cr3Se4 |
| 12.1.92 | 26976 | 42 | 0.0067 | -0.3862 | 2 3 | 0 1 | - - | - - | z2w,3=1<br>z4=2 | Cr3Se4 |
| 12.1.93 | 81886 | 42 | 0.0028 | -0.4820 | 2 4 | 0 1 | - - | - - | z2w,3=1 | Cr3S4 |
| 12.1.94 | 626598 | 42 | 0.0077 | -0.4548 | 2 5 | 0 1 | - - | - - | z2w,1=1<br>z2w,2=1<br>z2w,3=1<br>z4=2 | Cr3S4 |
| 12.1.95 | 192944 | 42 | 0.0096 | -0.4336 | 2 4 | 0 1 | - - | - - | z2w,3=1 | Cr3S4 |
| 12.1.96 | 16722 | 42 | 0.0085 | -0.4449 | 2 5 | 0 1 | - - | - - | z2w,1=1<br>z2w,2=1<br>z2w,3=1<br>z4=2 | Cr3S4 |
| 12.1.97 | 74296 | 70 | 0.0091 | -0.1981 | 2 1 | 0 1 | - - | - - | z2w,1=1<br>z2w,2=1<br>z2w,3=1 | Cr2S8Tl1V3 |
| 12.1.98 | 626640 | 40 | 0.0123 | -0.2531 | 2 7 | 0 1 | - - | - - | z2w,1=1<br>z2w,2=1 | Cr2S4Ti1 |
| 12.1.99 | 625968 | 44 | 0.0115 | -0.1601 | 2 4 | 0 1 | - - | - - | z2w,3=1 | Cr2Fe1Se4 |
| 12.1.100 | 155121 | 44 | 0.0150 | -0.1607 | 2 4 | 0 1 | - - | - - | z2w,3=1 | Cr2Fe1Se4 |
| 12.1.101 | 625931 | 44 | 0.0129 | -0.3591 | 2 5 | 0 1 | - - | - - | z2w,1=1<br>z2w,2=1<br>z2w,3=1<br>z4=2 | Cr2Fe1S4 |
| 12.1.102 | 251718 | 18 | 0.0306 | -0.2013 | 2 3 | 0 1 | - - | - - | z2w,3=1<br>z4=2 | Cr1Se2 |
| 12.1.103 | 75420 | 18 | 0.0551 | -0.0637 | 2 3 | 0 1 | - - | - - | z2w,3=1<br>z4=2 | Cr1S2 |
| 12.1.104 | 626579 | 48 | 0.0433 | -0.5814 | 2 5 | 0 1 | - - | - - | z2w,1=1<br>z2w,2=1<br>z2w,3=1<br>z4=2 | Cr1Rh2Te4 |
| 12.1.105 | 626577 | 48 | 0.0269 | -0.4656 | 2 5 | 0 1 | - - | - - | z2w,1=1<br>z2w,2=1<br>z2w,3=1<br>z4=2 | Cr1Rh2Se4 |
| 12.1.106 | 626541 | 56 | 0.0079 | -0.3481 | 2 5 | 0 1 | - - | - - | z2w,1=1<br>z2w,2=1<br>z2w,3=1<br>z4=2 | Cr1Pr2Se4 |
| 12.1.107 | 821 | 70 | 0.0182 | -0.0748 | 2 3 | 0 1 | - - | - - | z2w,3=1<br>z4=2 | Cr1P1S4 |
| 12.1.108 | 56835 | 70 | 0.0178 | -0.0736 | 2 3 | 0 1 | - - | - - | z2w,3=1<br>z4=2 | Cr1P1S4 |
| 12.1.109 | 173875 | 74 | 0.0370 | -0.7352 | 2 5 | 0 1 | - - | - - | z2w,1=1<br>z2w,2=1<br>z2w,3=1<br>z4=2 | Cr1O4Re1 |
| 12.1.110 | 624485 | 52 | 0.0403 | -0.7229 | 2 5 | 0 1 | - - | - - | z2w,1=1<br>z2w,2=1<br>z2w,3=1<br>z4=2 | Co2Ni1Se4 |
| 12.1.111 | 66006 | 70 | 0.0091 | -0.0233 | 2 3 | 0 1 | - - | - - | z2w,3=1<br>z4=2 | Co2Na7O6Rb1 |
| 12.1.112 | 2371 | 50 | 0.0059 | -0.8334 | 2 5 | 0 1 | - - | - - | z2w,1=1<br>z2w,2=1<br>z2w,3=1<br>z4=2 | Co1Si2Zr2 |

Continued on next page

**TABLE XIV– continued**

| TQC | ICSD | NoE | DG | IndG | SubSG (w/ T) | cl | SubSG (w/o T) | cl | Max. Sub. SG | yes or no | Max. Sub. SG | yes or no | SIs (w/ T) | Stoich. Formula |
|---|---|---|---|---|---|---|---|---|---|---|---|---|---|---|
| 12.1.113 | 624952 | 46 | 0.0241 | -0.7125 | 2 | 7 | 0 | 1 | - | - | - | - | z2w,1=1 z2w,2=1 | Co1Sc2Si2 |
| 12.1.114 | 623423 | 52 | 0.0106 | -1.2455 | 2 | 6 | 0 | 1 | - | - | - | - | z4=2 | Co1Ge1 |
| 12.1.115 | 43677 | 52 | 0.0107 | -1.2461 | 2 | 6 | 0 | 1 | - | - | - | - | z4=2 | Co1Ge1 |
| 12.1.116 | 184254 | 52 | 0.0222 | -1.2138 | 2 | 6 | 0 | 1 | - | - | - | - | z4=2 | Co1Ge1 |
| 12.1.117 | 183325 | 106 | 0.0087 | -0.0625 | 2 | 4 | 0 | 1 | - | - | - | - | z2w,3=1 | Co1Gd2Si2 |
| 12.1.118 | 184277 | 106 | 0.0038 | -0.0383 | 2 | 4 | 0 | 1 | - | - | - | - | z2w,3=1 | Co1Gd2Ge2 |
| 12.1.119 | 183326 | 106 | 0.0034 | -0.0376 | 2 | 4 | 0 | 1 | - | - | - | - | z2w,3=1 | Co1Gd2Ge2 |
| 12.1.120 | 427661 | 110 | 0.0064 | -0.0096 | 2 | 3 | 0 | 1 | - | - | - | - | z2w,3=1 z4=2 | Cl1Gd1O4W1 |
| 12.1.121 | 35292 | 110 | 0.0056 | -0.0101 | 2 | 3 | 0 | 1 | - | - | - | - | z2w,3=1 z4=2 | Cl1Gd1O4W1 |
| 12.1.122 | 420055 | 110 | 0.0054 | -0.0128 | 2 | 3 | 0 | 1 | - | - | - | - | z2w,3=1 z4=2 | Cl1Gd1Mo1O4 |
| 12.1.123 | 73096 | 98 | 0.0035 | -0.0183 | 2 | 1 | 0 | 1 | - | - | - | - | z2w,1=1 z2w,2=1 z2w,3=1 | Cl1Dy1O3Ti1 |
| 12.1.124 | 424719 | 86 | 0.0168 | -0.0667 | 2 | 1 | 0 | 1 | - | - | - | - | z2w,1=1 z2w,2=1 z2w,3=1 | Ce4Ge6Mg7 |
| 12.1.125 | 100827 | 126 | 0.0043 | -0.0700 | 2 | 3 | 0 | 1 | - | - | - | - | z2w,3=1 z4=2 | Ca2Fe7O11 |
| 12.1.126 | 20762 | 34 | 0.1209 | -0.9189 | 2 | 2 | 0 | 1 | - | - | - | - | z2w,1=1 z2w,2=1 z4=2 | C6H5Cl2Tl1 |
| 12.1.127 | 63503 | 98 | 0.0210 | -0.8193 | 2 | 7 | 0 | 1 | - | - | - | - | z2w,1=1 z2w,2=1 | C5Nb6 |
| 12.1.128 | 39297 | 98 | 0.0326 | -0.7970 | 2 | 4 | 0 | 1 | - | - | - | - | z2w,3=1 | C5Nb6 |
| 12.1.129 | 20695 | 98 | 0.0328 | -0.8494 | 2 | 4 | 0 | 1 | - | - | - | - | z2w,3=1 | C5Nb6 |
| 12.1.130 | 167353 | 98 | 0.0210 | -0.8193 | 2 | 7 | 0 | 1 | - | - | - | - | z2w,1=1 z2w,2=1 | C5Nb6 |
| 12.1.131 | 619030 | 52 | 0.0228 | -0.3039 | 2 | 5 | 0 | 1 | - | - | - | - | z2w,1=1 z2w,2=1 z2w,3=1 z4=2 | C3U2W2 |
| 12.1.132 | 618707 | 54 | 0.0386 | -0.2204 | 2 | 5 | 0 | 1 | - | - | - | - | z2w,1=1 z2w,2=1 z2w,3=1 z4=2 | C3Re2U2 |
| 12.1.133 | 618357 | 52 | 0.0131 | -0.3067 | 2 | 5 | 0 | 1 | - | - | - | - | z2w,1=1 z2w,2=1 z2w,3=1 z4=2 | C3Mo2U2 |
| 12.1.134 | 618223 | 74 | 0.0787 | -0.2283 | 2 | 5 | 0 | 1 | - | - | - | - | z2w,1=1 z2w,2=1 z2w,3=1 z4=2 | C3Lu2Mo2 |
| 12.1.135 | 88513 | 66 | 0.0034 | -0.0421 | 2 | 4 | 0 | 1 | - | - | - | - | z2w,3=1 | C3Ho2Mo2 |
| 12.1.136 | 88512 | 66 | 0.0031 | -0.0437 | 2 | 4 | 0 | 1 | - | - | - | - | z2w,3=1 | C3Ho2Mo2 |
| 12.1.137 | 88511 | 66 | 0.0029 | -0.0441 | 2 | 4 | 0 | 1 | - | - | - | - | z2w,3=1 | C3Ho2Mo2 |
| 12.1.138 | 602482 | 72 | 0.0329 | -0.2881 | 2 | 5 | 0 | 1 | - | - | - | - | z2w,1=1 z2w,2=1 z2w,3=1 z4=2 | C3Cr2Yb2 |
| 12.1.139 | 617537 | 62 | 0.0000 | -0.0242 | 2 | 5 | 0 | 1 | - | - | - | - | z2w,1=1 z2w,2=1 z2w,3=1 z4=2 | C3Cr2Tb2 |

Continued on next page

**TABLE XIV– continued**

| TQC | ICSD | NoE | DG | IndG | SubSG (w/ T) | cl | SubSG (w/o T) | cl | Max. Sub. SG | yes or no | Max. Sub. SG | yes or no | SIs (w/ T) | Stoich. Formula |
|---|---|---|---|---|---|---|---|---|---|---|---|---|---|---|
| 12.1.140 | 617514 | 74 | 0.0554 | -0.2928 | 2 | 5 | 0 | 1 | - | - | - | - | z2w,1=1 z2w,2=1 z2w,3=1 z4=2 | C3Cr2Lu2 |
| 12.1.141 | 617492 | 64 | 0.0045 | -0.0216 | 2 | 5 | 0 | 1 | - | - | - | - | z2w,1=1 z2w,2=1 z2w,3=1 z4=2 | C3Cr2Dy2 |
| 12.1.142 | 43018 | 64 | 0.0039 | -0.0235 | 2 | 5 | 0 | 1 | - | - | - | - | z2w,1=1 z2w,2=1 z2w,3=1 z4=2 | C3Cr2Dy2 |
| 12.1.143 | 417827 | 48 | 0.0529 | -0.2023 | 2 | 4 | 0 | 1 | - | - | - | - | z2w,3=1 | C3Ce2Mo2 |
| 12.1.144 | 618692 | 56 | 0.0576 | -0.0431 | 2 | 2 | 0 | 1 | - | - | - | - | z2w,1=1 z2w,2=1 z4=2 | C3.5Pu2W2 |
| 12.1.145 | 262897 | 126 | 0.0041 | -0.4143 | 2 | 3 | 0 | 1 | - | - | - | - | z2w,3=1 z4=2 | C1H3La1O6P2 |
| 12.1.146 | 79394 | 74 | 0.0085 | -0.1055 | 2 | 1 | 0 | 1 | - | - | - | - | z2w,1=1 z2w,2=1 z2w,3=1 | C1Fe2Si2Tm2 |
| 12.1.147 | 195010 | 64 | 0.0129 | -0.0910 | 2 | 1 | 0 | 1 | - | - | - | - | z2w,1=1 z2w,2=1 z2w,3=1 | C1Fe2Gd2Si2 |
| 12.1.148 | 67467 | 68 | 0.0069 | -0.0223 | 2 | 5 | 0 | 1 | - | - | - | - | z2w,1=1 z2w,2=1 z2w,3=1 z4=2 | C1Dy2Fe2Si2 |
| 12.1.149 | 391256 | 62 | 0.0058 | -0.0011 | 2 | 2 | 0 | 1 | - | - | - | - | z2w,1=1 z2w,2=1 z4=2 | C1Cl2Eu2N2 |
| 12.1.150 | 603998 | 52 | 0.1343 | -0.1958 | 2 | 5 | 0 | 1 | - | - | - | - | z2w,1=1 z2w,2=1 z2w,3=1 z4=2 | C1.66Mo1U1 |
| 12.1.151 | 416907 | 156 | 0.0144 | -0.5107 | 2 | 1 | 0 | 1 | - | - | - | - | z2w,1=1 z2w,2=1 z2w,3=1 | C10Cl4La8 |
| 12.1.152 | 617015 | 90 | 0.0472 | -0.7940 | 2 | 4 | 0 | 1 | - | - | - | - | z2w,3=1 | Bi2Rh3Si2 |
| 12.1.153 | 420725 | 98 | 0.0075 | -0.4765 | 2 | 4 | 0 | 1 | - | - | - | - | z2w,3=1 | Bi2Rh3S2 |
| 12.1.154 | 261457 | 104 | 0.0600 | -1.4902 | 2 | 2 | 0 | 1 | - | - | - | - | z2w,1=1 z2w,2=1 z4=2 | Bi2Pd3Se2 |
| 12.1.155 | 251720 | 22 | 0.1106 | -1.7263 | 2 | 4 | 0 | 1 | - | - | - | - | z2w,3=1 | Bi1Se1 |
| 12.1.156 | 251717 | 22 | 0.1967 | -1.4612 | 2 | 4 | 0 | 1 | - | - | - | - | z2w,3=1 | Bi1Se1 |
| 12.1.157 | 1559 | 48 | 0.0839 | 0.0839 | 2 | 3 | 0 | 1 | - | - | - | - | z2w,3=1 z4=2 | Bi1I1 |
| 12.1.158 | 1560 | 96 | 0.2559 | 0.2554 | 2 | 6 | 0 | 1 | - | - | - | - | z4=2 | Bi1Br1 |
| 12.1.159 | 65614 | 144 | 0.0070 | -0.0894 | 2 | 1 | 0 | 1 | - | - | - | - | z2w,1=1 z2w,2=1 z2w,3=1 | Ba3Cu2O10Pt1Y2 |
| 12.1.160 | 69569 | 166 | 0.0088 | -0.0060 | 2 | 5 | 0 | 1 | - | - | - | - | z2w,1=1 z2w,2=1 z2w,3=1 z4=2 | Ba3Cu2Er2O10Pt1 |
| 12.1.161 | 173685 | 74 | 0.1460 | 0.0734 | 2 | 2 | 0 | 1 | - | - | - | - | z2w,1=1 z2w,2=1 z4=2 | Ba3Cd2Sb4 |

Continued on next page

**TABLE XIV– continued**

| TQC | ICSD | NoE | DG | IndG | SubSG (w/ T) | cl | SubSG (w/o T) | cl | Max. Sub. SG | yes or no | Max. Sub. SG | yes or no | SIs (w/ T) | Stoich. Formula |
|---|---|---|---|---|---|---|---|---|---|---|---|---|---|---|
| 12.1.162 | 423268 | 86 | 0.0365 | -1.1855 | 2 | 5 | 0 | 1 | - | - | - | - | z2w,1=1<br>z2w,2=1<br>z2w,3=1<br>z4=2 | Ba2Ge4Ni5 |
| 12.1.163 | 238228 | 96 | 0.0270 | -0.0655 | 2 | 1 | 0 | 1 | - | - | - | - | z2w,1=1<br>z2w,2=1<br>z2w,3=1 | Ba1Te4U1 |
| 12.1.164 | 195017 | 80 | 0.0569 | -1.8656 | 2 | 6 | 0 | 1 | - | - | - | - | z4=2 | Ba1Pt2Sb2 |
| 12.1.165 | 154011 | 142 | 0.0066 | -0.2289 | 2 | 4 | 0 | 1 | - | - | - | - | z2w,3=1 | Ba1Mn2O6Y1 |
| 12.1.166 | 173848 | 222 | 0.0054 | -0.0965 | 2 | 2 | 0 | 1 | - | - | - | - | z2w,1=1<br>z2w,2=1<br>z4=2 | Ba1Ir1O3 |
| 12.1.167 | 151699 | 110 | 0.0033 | -0.0062 | 2 | 4 | 0 | 1 | - | - | - | - | z2w,3=1 | Ba1Er2F8 |
| 12.1.168 | 61499 | 66 | 0.1769 | -0.6823 | 2 | 2 | 0 | 1 | - | - | - | - | z2w,1=1<br>z2w,2=1<br>z4=2 | Ba1Bi1O3 |
| 12.1.169 | 63510 | 90 | 0.0035 | -0.0265 | 2 | 3 | 0 | 1 | - | - | - | - | z2w,3=1<br>z4=2 | B4Nd2Ni5 |
| 12.1.170 | 170618 | 84 | 0.0107 | -0.2924 | 2 | 1 | 0 | 1 | - | - | - | - | z2w,1=1<br>z2w,2=1<br>z2w,3=1 | B4La2Ni5 |
| 12.1.171 | 405291 | 182 | 0.0832 | -0.8591 | 2 | 1 | 0 | 1 | - | - | - | - | z2w,1=1<br>z2w,2=1<br>z2w,3=1 | B4Br5La4 |
| 12.1.172 | 615340 | 44 | 0.0178 | -0.3137 | 2 | 6 | 0 | 1 | - | - | - | - | z4=2 | B2Rh3Y1 |
| 12.1.173 | 261932 | 96 | 0.0118 | -0.7541 | 2 | 1 | 0 | 1 | - | - | - | - | z2w,1=1<br>z2w,2=1<br>z2w,3=1 | B2Ni3Zn1 |
| 12.1.174 | 107555 | 254 | 0.0031 | -0.8073 | 2 | 1 | 0 | 1 | - | - | - | - | z2w,1=1<br>z2w,2=1<br>z2w,3=1 | Au8Ga1K4 |
| 12.1.175 | 425401 | 102 | 0.0043 | -0.0528 | 2 | 2 | 0 | 1 | - | - | - | - | z2w,1=1<br>z2w,2=1<br>z4=2 | Au1Ce2O1P2 |
| 12.1.176 | 424760 | 74 | 0.0776 | -0.0069 | 2 | 2 | 0 | 1 | - | - | - | - | z2w,1=1<br>z2w,2=1<br>z4=2 | As4Ba3Zn2 |
| 12.1.177 | 424761 | 74 | 0.0749 | -0.0209 | 2 | 2 | 0 | 1 | - | - | - | - | z2w,1=1<br>z2w,2=1<br>z4=2 | As4Ba3Cd2 |
| 12.1.178 | 43185 | 54 | 0.0287 | -0.1838 | 2 | 5 | 0 | 1 | - | - | - | - | z2w,1=1<br>z2w,2=1<br>z2w,3=1<br>z4=2 | As3W2 |
| 12.1.179 | 15049 | 70 | 0.0130 | -0.4091 | 2 | 6 | 0 | 1 | - | - | - | - | z4=2 | As3V4 |
| 12.1.180 | 43184 | 54 | 0.0709 | -0.0737 | 2 | 5 | 0 | 1 | - | - | - | - | z2w,1=1<br>z2w,2=1<br>z2w,3=1<br>z4=2 | As3Mo2 |
| 12.1.181 | 41015 | 54 | 0.0711 | -0.0645 | 2 | 5 | 0 | 1 | - | - | - | - | z2w,1=1<br>z2w,2=1<br>z2w,3=1<br>z4=2 | As3Mo2 |
| 12.1.182 | 610428 | 64 | 0.0224 | -0.1913 | 2 | 3 | 0 | 1 | - | - | - | - | z2w,3=1<br>z4=2 | As3Eu1 |

Continued on next page

**TABLE XIV– continued**

| TQC | ICSD | NoE | DG | IndG | SubSG (w/ T) | cl | SubSG (w/o T) | cl | Max. Sub. SG | yes or no | Max. Sub. SG | yes or no | SIs (w/ T) | Stoich. Formula |
|---|---|---|---|---|---|---|---|---|---|---|---|---|---|---|
| 12.1.183 | 611574 | 30 | 0.1186 | -0.0128 | 2 | 5 | 0 | 1 | - | - | - | - | z2w,1=1<br>z2w,2=1<br>z2w,3=1<br>z4=2 | As2V1 |
| 12.1.184 | 44069 | 30 | 0.3292 | -0.0648 | 2 | 5 | 0 | 1 | - | - | - | - | z2w,1=1<br>z2w,2=1<br>z2w,3=1<br>z4=2 | As2Ta1 |
| 12.1.185 | 69002 | 72 | 0.0363 | -0.3607 | 2 | 4 | 0 | 1 | - | - | - | - | z2w,3=1 | As2Ni2O7 |
| 12.1.186 | 81218 | 46 | 0.2713 | -0.0262 | 2 | 5 | 0 | 1 | - | - | - | - | z2w,1=1<br>z2w,2=1<br>z2w,3=1<br>z4=2 | As2Nb1 |
| 12.1.187 | 69003 | 66 | 0.0212 | -0.0406 | 2 | 2 | 0 | 1 | - | - | - | - | z2w,1=1<br>z2w,2=1<br>z4=2 | As2Mn2O7 |
| 12.1.188 | 261143 | 66 | 0.0204 | -0.0335 | 2 | 2 | 0 | 1 | - | - | - | - | z2w,1=1<br>z2w,2=1<br>z4=2 | As2Mn2O7 |
| 12.1.189 | 79441 | 174 | 0.0024 | -0.0676 | 2 | 5 | 0 | 1 | - | - | - | - | z2w,1=1<br>z2w,2=1<br>z2w,3=1<br>z4=2 | As2Gd1Pd3 |
| 12.1.190 | 81775 | 156 | 0.0195 | -0.2244 | 2 | 7 | 0 | 1 | - | - | - | - | z2w,1=1<br>z2w,2=1 | As2Ce1Pd3 |
| 12.1.191 | 75454 | 50 | 0.0317 | -0.4053 | 2 | 1 | 0 | 1 | - | - | - | - | z2w,1=1<br>z2w,2=1<br>z2w,3=1 | As2Ba2Mn2O1 |
| 12.1.192 | 68906 | 64 | 0.0087 | -0.0277 | 2 | 4 | 0 | 1 | - | - | - | - | z2w,3=1 | As1F6N1O2 |
| 12.1.193 | 418388 | 186 | 0.0182 | -0.7355 | 2 | 4 | 0 | 1 | - | - | - | - | z2w,3=1 | Al6O1Si8Sr13 |
| 12.1.194 | 409319 | 156 | 0.0146 | 0.0146 | 2 | 4 | 0 | 1 | - | - | - | - | z2w,3=1 | Al4Na4P12Sr8 |
| 12.1.195 | 27569 | 150 | 0.0132 | -0.0618 | 2 | 1 | 0 | 1 | - | - | - | - | z2w,1=1<br>z2w,2=1<br>z2w,3=1 | Al4Cl51Zr12 |
| 12.1.196 | 40332 | 70 | 0.0139 | -3.0404 | 2 | 2 | 0 | 1 | - | - | - | - | z2w,1=1<br>z2w,2=1<br>z4=2 | Al1Cu1 |
| 12.1.197 | 165239 | 68 | 0.0058 | -0.0969 | 2 | 1 | 0 | 1 | - | - | - | - | z2w,1=1<br>z2w,2=1<br>z2w,3=1 | Al14Ca13 |
| 12.1.198 | 429000 | 116 | 0.0000 | -0.0639 | 2 | 3 | 0 | 1 | - | - | - | - | z2w,3=1<br>z4=2 | Ag1Ba1S3Tb1 |
| 12.2.1 | 187496 | 56 | 0.0246 | -0.7004 | 2 | 13 | 2 | 2 | - | - | - | - | z2w,1=1<br>z2w,2=1<br>z2w,3=1<br>z4=3 | Sb2Te3 |
| 12.2.2 | 420133 | 44 | 0.0142 | -0.5980 | 2 | 8 | 2 | 2 | - | - | - | - | z4=1 | Ru1Sc2Si2 |
| 12.2.3 | 61353 | 124 | 0.0512 | -0.3104 | 2 | 9 | 2 | 2 | - | - | - | - | z4=3 | Pt1Se7Ta2 |
| 12.2.4 | 248500 | 26 | 0.0051 | -0.2640 | 2 | 12 | 2 | 2 | - | - | - | - | z2w,3=1<br>z4=1 | Pr1 |
| 12.2.5 | 61005 | 56 | 0.0157 | -0.4290 | 2 | 11 | 2 | 2 | - | - | - | - | z2w,1=1<br>z2w,2=1<br>z4=3 | Pd1Se6Ta2 |
| 12.2.6 | 40281 | 96 | 0.0108 | -0.5588 | 2 | 12 | 2 | 2 | - | - | - | - | z2w,3=1<br>z4=1 | P4Zr7 |
| 12.2.7 | 41016 | 58 | 0.0021 | -1.7609 | 2 | 11 | 2 | 2 | - | - | - | - | z2w,1=1<br>z2w,2=1<br>z4=3 | P3Tc2 |

Continued on next page

**TABLE XIV– continued**

| TQC | ICSD | NoE | DG | IndG | SubSG (w/ T) | cl | SubSG (w/o T) | cl | Max. Sub. SG | yes or no | Max. Sub. SG | yes or no | SIs (w/ T) | Stoich. Formula |
|---|---|---|---|---|---|---|---|---|---|---|---|---|---|---|
| 12.2.8 | 74780 | 98 | 0.0327 | -0.0649 | 2 | 14 | 2 | 2 | - | - | - | - | z2w,1=1 z2w,2=1 z2w,3=1 z4=1 | O7Si2Yb2 |
| 12.2.9 | 65363 | 98 | 0.0226 | -0.0861 | 2 | 14 | 2 | 2 | - | - | - | - | z2w,1=1 z2w,2=1 z2w,3=1 z4=1 | O7Si2Yb2 |
| 12.2.10 | 62356 | 98 | 0.0226 | -0.0862 | 2 | 14 | 2 | 2 | - | - | - | - | z2w,1=1 z2w,2=1 z2w,3=1 z4=1 | O7Si2Yb2 |
| 12.2.11 | 248719 | 96 | 0.0174 | -0.0047 | 2 | 9 | 2 | 2 | - | - | - | - | z4=3 | O7Si2Tm2 |
| 12.2.12 | 75194 | 84 | 0.0045 | -0.3771 | 2 | 15 | 2 | 2 | - | - | - | - | z2w,3=1 z4=3 | O5Ti3 |
| 12.2.13 | 75193 | 84 | 0.0047 | -0.2881 | 2 | 13 | 2 | 2 | - | - | - | - | z2w,1=1 z2w,2=1 z2w,3=1 z4=3 | O5Ti3 |
| 12.2.14 | 38223 | 84 | 0.0032 | -0.0565 | 2 | 9 | 2 | 2 | - | - | - | - | z4=3 | O3Rb2Te1 |
| 12.2.15 | 16779 | 108 | 0.0064 | -0.9823 | 2 | 11 | 2 | 2 | - | - | - | - | z2w,1=1 z2w,2=1 z4=3 | O26V12 |
| 12.2.16 | 10455 | 122 | 0.0118 | -0.1377 | 2 | 10 | 2 | 2 | - | - | - | - | z2w,1=1 z2w,2=1 z4=1 | O13Sr2Ti6 |
| 12.2.17 | 646757 | 46 | 0.1050 | -1.8254 | 2 | 15 | 2 | 2 | - | - | - | - | z2w,3=1 z4=3 | Ni3Sn4 |
| 12.2.18 | 105364 | 46 | 0.0081 | -1.8072 | 2 | 15 | 2 | 2 | - | - | - | - | z2w,3=1 z4=3 | Ni3Sn4 |
| 12.2.19 | 105363 | 46 | 0.1135 | -1.7720 | 2 | 15 | 2 | 2 | - | - | - | - | z2w,3=1 z4=3 | Ni3Sn4 |
| 12.2.20 | 30433 | 72 | 0.0264 | -0.3356 | 2 | 8 | 2 | 2 | - | - | - | - | z4=1 | Ni2O7P2 |
| 12.2.21 | 646556 | 44 | 0.0418 | -0.3360 | 2 | 10 | 2 | 2 | - | - | - | - | z2w,1=1 z2w,2=1 z4=1 | Ni1Se4V2 |
| 12.2.22 | 646555 | 44 | 0.0364 | -0.3382 | 2 | 10 | 2 | 2 | - | - | - | - | z2w,1=1 z2w,2=1 z4=1 | Ni1Se4V2 |
| 12.2.23 | 23970 | 44 | 0.0389 | -0.3403 | 2 | 10 | 2 | 2 | - | - | - | - | z2w,1=1 z2w,2=1 z4=1 | Ni1Se4V2 |
| 12.2.24 | 35141 | 78 | 0.0126 | -0.1608 | 2 | 14 | 2 | 2 | - | - | - | - | z2w,1=1 z2w,2=1 z2w,3=1 z4=1 | Ni1S8V4 |
| 12.2.25 | 646403 | 44 | 0.0112 | -0.3307 | 2 | 10 | 2 | 2 | - | - | - | - | z2w,1=1 z2w,2=1 z4=1 | Ni1S4V2 |
| 12.2.26 | 35140 | 44 | 0.0365 | -0.2953 | 2 | 10 | 2 | 2 | - | - | - | - | z2w,1=1 z2w,2=1 z4=1 | Ni1S4V2 |
| 12.2.27 | 646394 | 42 | 0.0309 | -0.5877 | 2 | 14 | 2 | 2 | - | - | - | - | z2w,1=1 z2w,2=1 z2w,3=1 z4=1 | Ni1S4Ti2 |
| 12.2.28 | 193869 | 42 | 0.0086 | -0.4157 | 2 | 12 | 2 | 2 | - | - | - | - | z2w,3=1 z4=1 | Ni0.5Se2Ti1 |

Continued on next page

**TABLE XIV– continued**

| TQC | ICSD | NoE | DG | IndG | SubSG (w/ T) | cl | SubSG (w/o T) | cl | Max. Sub. SG | yes or no | Max. Sub. SG | yes or no | SIs (w/ T) | Stoich. Formula |
|---|---|---|---|---|---|---|---|---|---|---|---|---|---|---|
| 12.2.29 | 73315 | 234 | 0.0217 | -0.0396 | 2 | 15 | 2 | 2 | - | - | - | - | z2w,3=1<br>z4=3 | Nb8Pt1Se20 |
| 12.2.30 | 280016 | 88 | 0.0118 | -0.2790 | 2 | 9 | 2 | 2 | - | - | - | - | z4=3 | Nb2Te3 |
| 12.2.31 | 192215 | 132 | 0.0813 | -0.1051 | 2 | 12 | 2 | 2 | - | - | - | - | z2w,3=1<br>z4=1 | Nb2Pd1S5 |
| 12.2.32 | 81059 | 160 | 0.0079 | -0.2145 | 2 | 15 | 2 | 2 | - | - | - | - | z2w,3=1<br>z4=3 | Mo1Ni1O4 |
| 12.2.33 | 20296 | 66 | 0.0155 | -0.0326 | 2 | 13 | 2 | 2 | - | - | - | - | z2w,1=1<br>z2w,2=1<br>z2w,3=1<br>z4=3 | Mn2O7P2 |
| 12.2.34 | 15615 | 148 | 0.0021 | -0.0577 | 2 | 11 | 2 | 2 | - | - | - | - | z2w,1=1<br>z2w,2=1<br>z4=3 | Mn1Mo1O4 |
| 12.2.35 | 50001 | 154 | 0.0044 | -0.4321 | 2 | 14 | 2 | 2 | - | - | - | - | z2w,1=1<br>z2w,2=1<br>z2w,3=1<br>z4=1 | Mg2Si10Sr11 |
| 12.2.36 | 180011 | 66 | 0.0501 | -0.3059 | 2 | 12 | 2 | 2 | - | - | - | - | z2w,3=1<br>z4=1 | Li0.5O2Ti1 |
| 12.2.37 | 246807 | 62 | 0.0051 | -0.5206 | 2 | 15 | 2 | 2 | - | - | - | - | z2w,3=1<br>z4=3 | La1Si5 |
| 12.2.38 | 428765 | 84 | 0.0086 | -0.7543 | 2 | 9 | 2 | 2 | - | - | - | - | z4=3 | Hg2Sn2Sr1 |
| 12.2.39 | 157861 | 36 | 0.1010 | -2.5220 | 2 | 9 | 2 | 2 | - | - | - | - | z4=3 | Hg1 |
| 12.2.40 | 405208 | 96 | 0.0055 | -0.8968 | 2 | 9 | 2 | 2 | - | - | - | - | z4=3 | Hf7P4 |
| 12.2.41 | 162649 | 106 | 0.0018 | -0.0717 | 2 | 9 | 2 | 2 | - | - | - | - | z4=3 | H8Cu2Mg1O12V2 |
| 12.2.42 | 33293 | 42 | 0.0611 | -0.4437 | 2 | 15 | 2 | 2 | - | - | - | - | z2w,3=1<br>z4=3 | H6Au1Br3N2 |
| 12.2.43 | 2559 | 60 | 0.0052 | -0.1266 | 2 | 14 | 2 | 2 | - | - | - | - | z2w,1=1<br>z2w,2=1<br>z2w,3=1<br>z4=1 | H4F6Li2O2Sn1 |
| 12.2.44 | 201563 | 116 | 0.0085 | -0.0335 | 2 | 10 | 2 | 2 | - | - | - | - | z2w,1=1<br>z2w,2=1<br>z4=1 | H4Cl8N1O2Rb3Ru2 |
| 12.2.45 | 16130 | 116 | 0.0083 | -0.0808 | 2 | 9 | 2 | 2 | - | - | - | - | z4=3 | H4Cl8K3N1O2Ru2 |
| 12.2.46 | 200012 | 84 | 0.0294 | -0.1108 | 2 | 12 | 2 | 2 | - | - | - | - | z2w,3=1<br>z4=1 | H4Al1Cr2K1O10 |
| 12.2.47 | 35156 | 26 | 0.0072 | -0.3392 | 2 | 10 | 2 | 2 | - | - | - | - | z2w,1=1<br>z2w,2=1<br>z4=1 | H3Li1O2 |
| 12.2.48 | 30223 | 26 | 0.0201 | -0.3294 | 2 | 10 | 2 | 2 | - | - | - | - | z2w,1=1<br>z2w,2=1<br>z4=1 | H3Li1O2 |
| 12.2.49 | 27761 | 28 | 0.0882 | -0.3455 | 2 | 14 | 2 | 2 | - | - | - | - | z2w,1=1<br>z2w,2=1<br>z2w,3=1<br>z4=1 | H3Li1O2 |
| 12.2.50 | 159510 | 26 | 0.0196 | -0.3337 | 2 | 10 | 2 | 2 | - | - | - | - | z2w,1=1<br>z2w,2=1<br>z4=1 | H3Li1O2 |
| 12.2.51 | 24430 | 106 | 0.0057 | -0.0909 | 2 | 8 | 2 | 2 | - | - | - | - | z4=1 | H3Al2O7P1 |
| 12.2.52 | 421319 | 64 | 0.0408 | -0.0018 | 2 | 14 | 2 | 2 | - | - | - | - | z2w,1=1<br>z2w,2=1<br>z2w,3=1<br>z4=1 | H2Bi1Cl3O1 |

Continued on next page

**TABLE XIV– continued**

| TQC | ICSD | NoE | DG | IndG | SubSG (w/ T) | cl | SubSG (w/o T) | cl | Max. Sub. SG | yes or no | Max. Sub. SG | yes or no | SIs (w/ T) | Stoich. Formula |
|---|---|---|---|---|---|---|---|---|---|---|---|---|---|---|
| 12.2.53 | 26919 | 60 | 0.0173 | -0.1419 | 2 | 11 | 2 | 2 | - | - | - | - | z2w,1=1 z2w,2=1 z4=3 | H12Cl2Ni1O6 |
| 12.2.54 | 49911 | 70 | 0.0065 | -0.0064 | 2 | 9 | 2 | 2 | - | - | - | - | z4=3 | H12Cl2N2O6Pt1 |
| 12.2.55 | 200906 | 182 | 0.0216 | -0.0109 | 2 | 15 | 2 | 2 | - | - | - | - | z2w,3=1 z4=3 | H10F6Mn1O5Zr1 |
| 12.2.56 | 76350 | 116 | 0.0029 | -0.0152 | 2 | 13 | 2 | 2 | - | - | - | - | z2w,1=1 z2w,2=1 z2w,3=1 z4=3 | Ge2Ho2Os1 |
| 12.2.57 | 183379 | 82 | 0.0141 | -0.4836 | 2 | 9 | 2 | 2 | - | - | - | - | z4=3 | Gd1O6Ru1Sr2 |
| 12.2.58 | 635673 | 36 | 0.0025 | -0.2916 | 2 | 9 | 2 | 2 | - | - | - | - | z4=3 | Ga4Yb1 |
| 12.2.59 | 165189 | 48 | 0.0063 | -0.1098 | 2 | 10 | 2 | 2 | - | - | - | - | z2w,1=1 z2w,2=1 z4=1 | Ga3Sc5 |
| 12.2.60 | 15044 | 48 | 0.0306 | -0.1714 | 2 | 8 | 2 | 2 | - | - | - | - | z4=1 | Fe3Se4 |
| 12.2.61 | 15043 | 48 | 0.0325 | -0.2411 | 2 | 8 | 2 | 2 | - | - | - | - | z4=1 | Fe3Se4 |
| 12.2.62 | 152533 | 178 | 0.0016 | -0.0736 | 2 | 12 | 2 | 2 | - | - | - | - | z2w,3=1 z4=1 | Fe1Si3Tm3 |
| 12.2.63 | 84205 | 44 | 0.0192 | -0.4576 | 2 | 8 | 2 | 2 | - | - | - | - | z4=1 | Fe1Sc2Si2 |
| 12.2.64 | 76348 | 44 | 0.0192 | -0.4576 | 2 | 8 | 2 | 2 | - | - | - | - | z4=1 | Fe1Sc2Si2 |
| 12.2.65 | 633455 | 44 | 0.0159 | -0.5210 | 2 | 8 | 2 | 2 | - | - | - | - | z4=1 | Fe1Sc2Si2 |
| 12.2.66 | 20661 | 44 | 0.0151 | -0.5488 | 2 | 8 | 2 | 2 | - | - | - | - | z4=1 | Fe1Sc2Si2 |
| 12.2.67 | 633361 | 42 | 0.0218 | -0.4519 | 2 | 10 | 2 | 2 | - | - | - | - | z2w,1=1 z2w,2=1 z4=1 | Fe1S4V2 |
| 12.2.68 | 603354 | 50 | 0.0273 | -0.3192 | 2 | 13 | 2 | 2 | - | - | - | - | z2w,1=1 z2w,2=1 z2w,3=1 z4=3 | Fe1Rh2Se4 |
| 12.2.69 | 632972 | 52 | 0.0356 | -0.5949 | 2 | 13 | 2 | 2 | - | - | - | - | z2w,1=1 z2w,2=1 z2w,3=1 z4=3 | Fe1Ni2Se4 |
| 12.2.70 | 90460 | 72 | 0.0391 | -0.1695 | 2 | 10 | 2 | 2 | - | - | - | - | z2w,1=1 z2w,2=1 z4=1 | Fe1Ge1Yb1 |
| 12.2.71 | 631983 | 48 | 0.0339 | -0.5306 | 2 | 10 | 2 | 2 | - | - | - | - | z2w,1=1 z2w,2=1 z4=1 | Fe1Ge1 |
| 12.2.72 | 53458 | 48 | 0.0448 | -0.5180 | 2 | 10 | 2 | 2 | - | - | - | - | z2w,1=1 z2w,2=1 z4=1 | Fe1Ge1 |
| 12.2.73 | 53457 | 48 | 0.0420 | -0.5191 | 2 | 10 | 2 | 2 | - | - | - | - | z2w,1=1 z2w,2=1 z4=1 | Fe1Ge1 |
| 12.2.74 | 53456 | 48 | 0.0331 | -0.5273 | 2 | 10 | 2 | 2 | - | - | - | - | z2w,1=1 z2w,2=1 z4=1 | Fe1Ge1 |
| 12.2.75 | 53455 | 48 | 0.0305 | -0.5307 | 2 | 10 | 2 | 2 | - | - | - | - | z2w,1=1 z2w,2=1 z4=1 | Fe1Ge1 |
| 12.2.76 | 43055 | 48 | 0.0392 | -0.5192 | 2 | 10 | 2 | 2 | - | - | - | - | z2w,1=1 z2w,2=1 z4=1 | Fe1Ge1 |
| 12.2.77 | 416629 | 94 | 0.0124 | -0.0847 | 2 | 9 | 2 | 2 | - | - | - | - | z4=3 | Eu2Li1Si3 |
| 12.2.78 | 425141 | 166 | 0.0028 | -0.0161 | 2 | 9 | 2 | 2 | - | - | - | - | z4=3 | Dy3O3Sb1 |
| 12.2.79 | 422453 | 40 | 0.1689 | -2.0590 | 2 | 9 | 2 | 2 | - | - | - | - | z4=3 | Cu2Sn2Sr1 |

Continued on next page

**TABLE XIV– continued**

| TQC | ICSD | NoE | DG | IndG | SubSG (w/ T) | cl | SubSG (w/o T) | cl | Max. Sub. SG | yes or no | Max. Sub. SG | yes or no | SIs (w/ T) | Stoich. Formula |
|---|---|---|---|---|---|---|---|---|---|---|---|---|---|---|
| 12.2.80 | 182102 | 40 | 0.1689 | -2.0590 | 2 | 9 | 2 | 2 | - | - | - | - | z4=3 | Cu2Sn2Sr1 |
| 12.2.81 | 79536 | 52 | 0.0146 | -0.0067 | 2 | 9 | 2 | 2 | - | - | - | - | z4=3 | Cu2Na2S4Zr1 |
| 12.2.82 | 426133 | 166 | 0.0059 | -0.3593 | 2 | 8 | 2 | 2 | - | - | - | - | z4=1 | Cu1Ge6Yb2 |
| 12.2.83 | 380416 | 68 | 0.0437 | -0.4343 | 2 | 11 | 2 | 2 | - | - | - | - | z2w,1=1 z2w,2=1 z4=3 | Cs4Ge2Se6 |
| 12.2.84 | 626873 | 42 | 0.0166 | -0.0553 | 2 | 9 | 2 | 2 | - | - | - | - | z4=3 | Cr3Te4 |
| 12.2.85 | 601348 | 42 | 0.0062 | -0.3917 | 2 | 8 | 2 | 2 | - | - | - | - | z4=1 | Cr3Se4 |
| 12.2.86 | 73143 | 114 | 0.0000 | -0.0892 | 2 | 9 | 2 | 2 | - | - | - | - | z4=3 | Cr3Rb1S5 |
| 12.2.87 | 1824 | 90 | 0.0125 | -0.2555 | 2 | 10 | 2 | 2 | - | - | - | - | z2w,1=1 z2w,2=1 z4=1 | Cr3Ga4 |
| 12.2.88 | 626896 | 40 | 0.0072 | -0.1018 | 2 | 9 | 2 | 2 | - | - | - | - | z4=3 | Cr2Te4Ti1 |
| 12.2.89 | 16884 | 46 | 0.0110 | -0.3728 | 2 | 14 | 2 | 2 | - | - | - | - | z2w,1=1 z2w,2=1 z2w,3=1 z4=1 | Cr2Ni1S4 |
| 12.2.90 | 625971 | 44 | 0.0147 | -0.1400 | 2 | 10 | 2 | 2 | - | - | - | - | z2w,1=1 z2w,2=1 z4=1 | Cr2Fe1Se4 |
| 12.2.91 | 626655 | 40 | 0.0240 | -0.4839 | 2 | 10 | 2 | 2 | - | - | - | - | z2w,1=1 z2w,2=1 z4=1 | Cr1S4V2 |
| 12.2.92 | 626543 | 56 | 0.0234 | -0.1844 | 2 | 11 | 2 | 2 | - | - | - | - | z2w,1=1 z2w,2=1 z4=3 | Cr1Pr2Se4 |
| 12.2.93 | 626536 | 44 | 0.0276 | -0.1124 | 2 | 14 | 2 | 2 | - | - | - | - | z2w,1=1 z2w,2=1 z2w,3=1 z4=1 | Cr1Pr2S4 |
| 12.2.94 | 16087 | 28 | 0.0041 | -0.1328 | 2 | 10 | 2 | 2 | - | - | - | - | z2w,1=1 z2w,2=1 z4=1 | Cr1O5Pb2 |
| 12.2.95 | 79865 | 140 | 0.0108 | -0.1583 | 2 | 11 | 2 | 2 | - | - | - | - | z2w,1=1 z2w,2=1 z4=3 | Cr1O4V1 |
| 12.2.96 | 202377 | 76 | 0.0033 | -0.0835 | 2 | 15 | 2 | 2 | - | - | - | - | z2w,3=1 z4=3 | Cr1Nd1O1S2 |
| 12.2.97 | 151873 | 46 | 0.0021 | -0.3530 | 2 | 8 | 2 | 2 | - | - | - | - | z4=1 | Cr1Fe2Se4 |
| 12.2.98 | 74542 | 70 | 0.0264 | -0.1413 | 2 | 8 | 2 | 2 | - | - | - | - | z4=1 | Co2O7P2 |
| 12.2.99 | 41026 | 152 | 0.0028 | -0.0040 | 2 | 12 | 2 | 2 | - | - | - | - | z2w,3=1 z4=1 | Co2Nb4Pd1Se12 |
| 12.2.100 | 408031 | 118 | 0.0142 | -1.0862 | 2 | 15 | 2 | 2 | - | - | - | - | z2w,3=1 z4=3 | Cl2La3Si3 |
| 12.2.101 | 420061 | 122 | 0.0041 | -0.0403 | 2 | 14 | 2 | 2 | - | - | - | - | z2w,1=1 z2w,2=1 z2w,3=1 z4=1 | Cl1Mo1O4Yb1 |
| 12.2.102 | 420060 | 120 | 0.0035 | -0.0067 | 2 | 14 | 2 | 2 | - | - | - | - | z2w,1=1 z2w,2=1 z2w,3=1 z4=1 | Cl1Mo1O4Tm1 |
| 12.2.103 | 420059 | 112 | 0.0032 | -0.0096 | 2 | 9 | 2 | 2 | - | - | - | - | z4=3 | Cl1Mo1O4Tb1 |
| 12.2.104 | 427663 | 116 | 0.0023 | -0.0120 | 2 | 15 | 2 | 2 | - | - | - | - | z2w,3=1 z4=3 | Cl1Ho1O4W1 |
| 12.2.105 | 427664 | 118 | 0.0047 | -0.0049 | 2 | 15 | 2 | 2 | - | - | - | - | z2w,3=1 z4=3 | Cl1Er1O4W1 |

Continued on next page

**TABLE XIV– continued**

| TQC | ICSD | NoE | DG | IndG | SubSG (w/ T) | cl | SubSG (w/o T) | cl | Max. Sub. SG | yes or no | Max. Sub. SG | yes or no | SIs (w/ T) | Stoich. Formula |
|---|---|---|---|---|---|---|---|---|---|---|---|---|---|---|
| 12.2.106 | 420053 | 118 | 0.0040 | -0.0037 | 2 | 15 | 2 | 2 | - | - | - | - | z2w,3=1<br>z4=3 | Cl1Er1Mo1O4 |
| 12.2.107 | 424718 | 124 | 0.0105 | -0.2326 | 2 | 11 | 2 | 2 | - | - | - | - | z2w,1=1<br>z2w,2=1<br>z4=3 | Ce5Ge8Mg8 |
| 12.2.108 | 417665 | 96 | 0.0033 | -0.2116 | 2 | 8 | 2 | 2 | - | - | - | - | z4=1 | Ce1Ru1Sn1 |
| 12.2.109 | 192243 | 144 | 0.0086 | -0.1425 | 2 | 8 | 2 | 2 | - | - | - | - | z4=1 | Ce1Ru1Sn1 |
| 12.2.110 | 192242 | 96 | 0.0077 | -0.2156 | 2 | 8 | 2 | 2 | - | - | - | - | z4=1 | Ce1Ru1Sn1 |
| 12.2.111 | 189952 | 144 | 0.0080 | -0.1378 | 2 | 8 | 2 | 2 | - | - | - | - | z4=1 | Ce1Ru1Sn1 |
| 12.2.112 | 189412 | 144 | 0.0040 | -0.0969 | 2 | 8 | 2 | 2 | - | - | - | - | z4=1 | Ce1Ru1Sn1 |
| 12.2.113 | 189411 | 96 | 0.0052 | -0.2184 | 2 | 8 | 2 | 2 | - | - | - | - | z4=1 | Ce1Ru1Sn1 |
| 12.2.114 | 2284 | 24 | 0.0453 | -0.4202 | 2 | 8 | 2 | 2 | - | - | - | - | z4=1 | Ce1 |
| 12.2.115 | 620729 | 48 | 0.0047 | -0.0941 | 2 | 8 | 2 | 2 | - | - | - | - | z4=1 | Ce1Co1Ga1 |
| 12.2.116 | 620726 | 48 | 0.0047 | -0.0941 | 2 | 8 | 2 | 2 | - | - | - | - | z4=1 | Ce1Co1Ga1 |
| 12.2.117 | 429884 | 48 | 0.0163 | -0.0554 | 2 | 8 | 2 | 2 | - | - | - | - | z4=1 | Ce1Co1Ga1 |
| 12.2.118 | 187960 | 48 | 0.0168 | -0.0599 | 2 | 8 | 2 | 2 | - | - | - | - | z4=1 | Ce1Co1Ga1 |
| 12.2.119 | 171243 | 62 | 0.0219 | -1.0973 | 2 | 10 | 2 | 2 | - | - | - | - | z2w,1=1<br>z2w,2=1<br>z4=1 | Ca6Cu2Sn7 |
| 12.2.120 | 85505 | 88 | 0.0054 | -0.4458 | 2 | 12 | 2 | 2 | - | - | - | - | z2w,3=1<br>z4=1 | Ca1K2N4O8 |
| 12.2.121 | 619274 | 14 | 0.0286 | -0.8700 | 2 | 9 | 2 | 2 | - | - | - | - | z4=3 | Ca1Ga4 |
| 12.2.122 | 58893 | 14 | 0.0308 | -0.8716 | 2 | 9 | 2 | 2 | - | - | - | - | z4=3 | Ca1Ga4 |
| 12.2.123 | 54994 | 14 | 0.0308 | -0.8716 | 2 | 9 | 2 | 2 | - | - | - | - | z4=3 | Ca1Ga4 |
| 12.2.124 | 417717 | 32 | 0.1916 | -2.3157 | 2 | 9 | 2 | 2 | - | - | - | - | z4=3 | Ca1Cu2Sn2 |
| 12.2.125 | 236391 | 68 | 0.0113 | -0.2807 | 2 | 10 | 2 | 2 | - | - | - | - | z2w,1=1<br>z2w,2=1<br>z4=1 | C4Co1Sc3 |
| 12.2.126 | 618350 | 70 | 0.0158 | -0.0971 | 2 | 9 | 2 | 2 | - | - | - | - | z4=3 | C3Mo2Tm2 |
| 12.2.127 | 88516 | 68 | 0.0072 | -0.0447 | 2 | 13 | 2 | 2 | - | - | - | - | z2w,1=1<br>z2w,2=1<br>z2w,3=1<br>z4=3 | C3Er2Mo2 |
| 12.2.128 | 617669 | 68 | 0.0060 | -0.0740 | 2 | 9 | 2 | 2 | - | - | - | - | z4=3 | C3Er2Mo2 |
| 12.2.129 | 617541 | 70 | 0.0105 | -0.0507 | 2 | 9 | 2 | 2 | - | - | - | - | z4=3 | C3Cr2Tm2 |
| 12.2.130 | 617494 | 68 | 0.0000 | -0.0444 | 2 | 9 | 2 | 2 | - | - | - | - | z4=3 | C3Cr2Er2 |
| 12.2.131 | 252737 | 20 | 0.0276 | -0.4062 | 2 | 10 | 2 | 2 | - | - | - | - | z2w,1=1<br>z2w,2=1<br>z4=1 | C2Ca1 |
| 12.2.132 | 252734 | 20 | 0.0165 | -0.2677 | 2 | 15 | 2 | 2 | - | - | - | - | z2w,3=1<br>z4=3 | C2Ca1 |
| 12.2.133 | 252716 | 20 | 0.0025 | -1.5192 | 2 | 15 | 2 | 2 | - | - | - | - | z2w,3=1<br>z4=3 | C2Ca1 |
| 12.2.134 | 617825 | 50 | 0.0356 | -0.3779 | 2 | 8 | 2 | 2 | - | - | - | - | z4=1 | C1Fe2Si2Y2 |
| 12.2.135 | 617663 | 72 | 0.0021 | -0.0522 | 2 | 12 | 2 | 2 | - | - | - | - | z2w,3=1<br>z4=1 | C1Er2Fe2Si2 |
| 12.2.136 | 61141 | 144 | 0.0106 | -0.1658 | 2 | 10 | 2 | 2 | - | - | - | - | z2w,1=1<br>z2w,2=1<br>z4=1 | C10Fe22Mo12 |
| 12.2.137 | 65114 | 166 | 0.0081 | -0.0334 | 2 | 10 | 2 | 2 | - | - | - | - | z2w,1=1<br>z2w,2=1<br>z4=1 | C10Au2Cl8S8 |
| 12.2.138 | 240479 | 70 | 0.1039 | -0.5748 | 2 | 10 | 2 | 2 | - | - | - | - | z2w,1=1<br>z2w,2=1<br>z4=1 | Br4Cs2I2Pd1 |

Continued on next page

**TABLE XIV– continued**

| TQC | ICSD | NoE | DG | IndG | SubSG (w/ T) | cl | SubSG (w/o T) | cl | Max. Sub. SG | yes or no | Max. Sub. SG | yes or no | SIs (w/ T) | Stoich. Formula |
|---|---|---|---|---|---|---|---|---|---|---|---|---|---|---|
| 12.2.139 | 172782 | 62 | 0.0097 | -0.1013 | 2 | 13 | 2 | 2 | - | - | - | - | z2w,1=1 z2w,2=1 z2w,3=1 z4=3 | Br1Mn1S2Sb1 |
| 12.2.140 | 427411 | 126 | 0.0104 | -0.0248 | 2 | 14 | 2 | 2 | - | - | - | - | z2w,1=1 z2w,2=1 z2w,3=1 z4=1 | Bi4Pb7Se13 |
| 12.2.141 | 58840 | 120 | 0.0302 | -0.9739 | 2 | 9 | 2 | 2 | - | - | - | - | z4=3 | Bi2Pd5 |
| 12.2.142 | 616957 | 104 | 0.0407 | -1.5899 | 2 | 13 | 2 | 2 | - | - | - | - | z2w,1=1 z2w,2=1 z2w,3=1 z4=3 | Bi2Pd3Se2 |
| 12.2.143 | 421873 | 104 | 0.0527 | -1.5045 | 2 | 14 | 2 | 2 | - | - | - | - | z2w,1=1 z2w,2=1 z2w,3=1 z4=1 | Bi2Pd3Se2 |
| 12.2.144 | 261459 | 104 | 0.0527 | -1.4980 | 2 | 13 | 2 | 2 | - | - | - | - | z2w,1=1 z2w,2=1 z2w,3=1 z4=3 | Bi2Pd3Se2 |
| 12.2.145 | 261458 | 104 | 0.0521 | -1.5006 | 2 | 13 | 2 | 2 | - | - | - | - | z2w,1=1 z2w,2=1 z2w,3=1 z4=3 | Bi2Pd3Se2 |
| 12.2.146 | 261456 | 104 | 0.0532 | -1.4946 | 2 | 14 | 2 | 2 | - | - | - | - | z2w,1=1 z2w,2=1 z2w,3=1 z4=1 | Bi2Pd3Se2 |
| 12.2.147 | 616954 | 104 | 0.0433 | -1.5266 | 2 | 15 | 2 | 2 | - | - | - | - | z2w,3=1 z4=3 | Bi2Pd3S2 |
| 12.2.148 | 616946 | 40 | 0.1574 | -1.5323 | 2 | 9 | 2 | 2 | - | - | - | - | z4=3 | Bi2Pd1 |
| 12.2.149 | 42565 | 40 | 0.1580 | -1.5535 | 2 | 9 | 2 | 2 | - | - | - | - | z4=3 | Bi2Pd1 |
| 12.2.150 | 1024 | 104 | 0.0142 | -1.4318 | 2 | 13 | 2 | 2 | - | - | - | - | z2w,1=1 z2w,2=1 z2w,3=1 z4=3 | Bi2Ni3Se2 |
| 12.2.151 | 195678 | 92 | 0.0348 | -0.2635 | 2 | 10 | 2 | 2 | - | - | - | - | z2w,1=1 z2w,2=1 z4=1 | Bi2Hg1S4 |
| 12.2.152 | 415138 | 62 | 0.0066 | -0.0959 | 2 | 12 | 2 | 2 | - | - | - | - | z2w,3=1 z4=1 | Bi1I1Mn1Se2 |
| 12.2.153 | 429459 | 154 | 0.0021 | -0.0019 | 2 | 9 | 2 | 2 | - | - | - | - | z4=3 | Bi1Gd3O3 |
| 12.2.154 | 164224 | 186 | 0.0127 | -0.1573 | 2 | 11 | 2 | 2 | - | - | - | - | z2w,1=1 z2w,2=1 z4=3 | Bi1Fe1O3 |
| 12.2.155 | 164223 | 186 | 0.0150 | -0.1536 | 2 | 11 | 2 | 2 | - | - | - | - | z2w,1=1 z2w,2=1 z4=3 | Bi1Fe1O3 |
| 12.2.156 | 246145 | 108 | 0.0596 | -0.9130 | 2 | 15 | 2 | 2 | - | - | - | - | z2w,3=1 z4=3 | Bi16I4 |
| 12.2.157 | 423458 | 106 | 0.0187 | -0.1435 | 2 | 13 | 2 | 2 | - | - | - | - | z2w,1=1 z2w,2=1 z2w,3=1 z4=3 | Ba5Cd2O2Sb4 |
| 12.2.158 | 240016 | 66 | 0.0209 | -0.5380 | 2 | 9 | 2 | 2 | - | - | - | - | z4=3 | Ba3Li4Sn8 |
| 12.2.159 | 62390 | 144 | 0.0093 | -0.0933 | 2 | 9 | 2 | 2 | - | - | - | - | z4=3 | Ba3Cu2O10Pt1Y2 |

Continued on next page

**TABLE XIV– continued**

| TQC | ICSD | NoE | DG | IndG | SubSG (w/ T) | cl | SubSG (w/o T) | cl | Max. Sub. SG | yes or no | Max. Sub. SG | yes or no | SIs (w/ T) | Stoich. Formula |
|---|---|---|---|---|---|---|---|---|---|---|---|---|---|---|
| 12.2.160 | 62391 | 164 | 0.0063 | -0.0029 | 2 | 12 | 2 | 2 | - | - | - | - | z2w,3=1<br>z4=1 | Ba3Cu2Ho2O10Pt1 |
| 12.2.161 | 408938 | 84 | 0.0167 | -0.1819 | 2 | 13 | 2 | 2 | - | - | - | - | z2w,1=1<br>z2w,2=1<br>z2w,3=1<br>z4=3 | Ba2Mg3Si4 |
| 12.2.162 | 420834 | 118 | 0.0274 | -0.0429 | 2 | 10 | 2 | 2 | - | - | - | - | z2w,1=1<br>z2w,2=1<br>z4=1 | Ba2Cd2Sb3 |
| 12.2.163 | 80901 | 80 | 0.0058 | -0.0185 | 2 | 11 | 2 | 2 | - | - | - | - | z2w,1=1<br>z2w,2=1<br>z4=3 | Ba2Bi1O6Tb1 |
| 12.2.164 | 261206 | 44 | 0.0313 | -0.6261 | 2 | 9 | 2 | 2 | - | - | - | - | z4=3 | Ba1Tl4 |
| 12.2.165 | 238227 | 92 | 0.0351 | -1.8122 | 2 | 9 | 2 | 2 | - | - | - | - | z4=3 | Ba1Te4Th1 |
| 12.2.166 | 49000 | 50 | 0.0874 | -0.7628 | 2 | 15 | 2 | 2 | - | - | - | - | z2w,3=1<br>z4=3 | Ba1Sb3 |
| 12.2.167 | 16634 | 180 | 0.0297 | -1.9928 | 2 | 9 | 2 | 2 | - | - | - | - | z4=3 | Ba1O5Ti2 |
| 12.2.168 | 83618 | 280 | 0.0448 | -0.8400 | 2 | 9 | 2 | 2 | - | - | - | - | z4=3 | Ba1Cu10P4 |
| 12.2.169 | 418886 | 242 | 0.0111 | 0.0111 | 2 | 9 | 2 | 2 | - | - | - | - | z4=3 | Ba11Cd6Sb12 |
| 12.2.170 | 63501 | 84 | 0.0307 | -0.3783 | 2 | 11 | 2 | 2 | - | - | - | - | z2w,1=1<br>z2w,2=1<br>z4=3 | B4La2Ni5 |
| 12.2.171 | 170608 | 84 | 0.0251 | -0.3792 | 2 | 11 | 2 | 2 | - | - | - | - | z2w,1=1<br>z2w,2=1<br>z4=3 | B4La2Ni5 |
| 12.2.172 | 615319 | 52 | 0.0026 | -0.2925 | 2 | 15 | 2 | 2 | - | - | - | - | z2w,3=1<br>z4=3 | B2Rh3Tb1 |
| 12.2.173 | 614701 | 58 | 0.0284 | -0.1418 | 2 | 12 | 2 | 2 | - | - | - | - | z2w,3=1<br>z4=1 | B2Lu1Rh3 |
| 12.2.174 | 614578 | 44 | 0.1554 | -0.7429 | 2 | 15 | 2 | 2 | - | - | - | - | z2w,3=1<br>z4=3 | B2Ir3Y1 |
| 12.2.175 | 614571 | 56 | 0.0108 | -0.1150 | 2 | 15 | 2 | 2 | - | - | - | - | z2w,3=1<br>z4=3 | B2Ir3Tm1 |
| 12.2.176 | 614566 | 52 | 0.0080 | -0.1275 | 2 | 15 | 2 | 2 | - | - | - | - | z2w,3=1<br>z4=3 | B2Ir3Tb1 |
| 12.2.177 | 614561 | 36 | 0.0117 | -0.7803 | 2 | 12 | 2 | 2 | - | - | - | - | z2w,3=1<br>z4=1 | B2Ir3Sc1 |
| 12.2.178 | 614531 | 58 | 0.0423 | -0.7996 | 2 | 15 | 2 | 2 | - | - | - | - | z2w,3=1<br>z4=3 | B2Ir3Lu1 |
| 12.2.179 | 614472 | 54 | 0.0032 | -0.1023 | 2 | 15 | 2 | 2 | - | - | - | - | z2w,3=1<br>z4=3 | B2Ho1Ir3 |
| 12.2.180 | 8058 | 74 | 0.1323 | -0.3831 | 2 | 15 | 2 | 2 | - | - | - | - | z2w,3=1<br>z4=3 | Au2P3 |
| 12.2.181 | 429212 | 90 | 0.1466 | -1.6783 | 2 | 9 | 2 | 2 | - | - | - | - | z4=3 | Au2Ni4Si7 |
| 12.2.182 | 425402 | 106 | 0.0033 | -0.0265 | 2 | 13 | 2 | 2 | - | - | - | - | z2w,1=1<br>z2w,2=1<br>z2w,3=1<br>z4=3 | Au1O1P2Pr2 |
| 12.2.183 | 41831 | 50 | 0.0130 | -0.4336 | 2 | 14 | 2 | 2 | - | - | - | - | z2w,1=1<br>z2w,2=1<br>z2w,3=1<br>z4=1 | As3Sr1 |
| 12.2.184 | 80375 | 86 | 0.0170 | -0.2083 | 2 | 13 | 2 | 2 | - | - | - | - | z2w,1=1<br>z2w,2=1<br>z2w,3=1<br>z4=3 | As3Mn4 |
| 12.2.185 | 86402 | 50 | 0.2063 | -0.1980 | 2 | 12 | 2 | 2 | - | - | - | - | z2w,3=1<br>z4=1 | As3Ba1 |

Continued on next page

**TABLE XIV– continued**

| TQC | ICSD | NoE | DG | IndG | SubSG (w/ T) | cl | SubSG (w/o T) | cl | Max. Sub. SG | yes or no | Max. Sub. SG | yes or no | SIs (w/ T) | Stoich. Formula |
|---|---|---|---|---|---|---|---|---|---|---|---|---|---|---|
| 12.2.186 | 69001 | 70 | 0.0312 | -0.1825 | 2 | 8 | 2 | 2 | - | - | - | - | z4=1 | As2Co2O7 |
| 12.2.187 | 239291 | 110 | 0.0253 | -0.7636 | 2 | 9 | 2 | 2 | - | - | - | - | z4=3 | As1Pd5 |
| 12.2.188 | 417968 | 52 | 0.1556 | 0.1113 | 2 | 11 | 2 | 2 | - | - | - | - | z2w,1=1<br>z2w,2=1<br>z4=3 | Al2Ge4Sr3 |
| 12.2.189 | 417659 | 168 | 0.0382 | -0.1296 | 2 | 10 | 2 | 2 | - | - | - | - | z2w,1=1<br>z2w,2=1<br>z4=1 | Al2Ba5Ge7 |
| 12.2.190 | 20632 | 48 | 0.0000 | -0.0490 | 2 | 8 | 2 | 2 | - | - | - | - | z4=1 | Al1Ce1Co1 |
| 12.2.191 | 416360 | 198 | 0.0213 | -0.7029 | 2 | 9 | 2 | 2 | - | - | - | - | z4=3 | Ag5Cl5Hg4P8 |
| 12.2.192 | 27743 | 186 | 0.0043 | -0.0157 | 2 | 8 | 2 | 2 | - | - | - | - | z4=1 | Ag1Er1O8W2 |
| 13.2.1 | 415459 | 104 | 0.0050 | -0.0357 | 2 | 5 | 2 | 2 | - | - | - | - | z4=3 | O4Ta1Tm1 |
| 13.2.2 | 415434 | 96 | 0.0080 | 0.0047 | 2 | 1 | 0 | 1 | - | - | - | - | z2w,1=1<br>z4=2 | O4Ta1Tb1 |
| 13.2.3 | 174299 | 72 | 0.0492 | -0.7317 | 2 | 7 | 0 | 1 | - | - | - | - | z2w,2=1 | O4Sn3 |
| 13.2.4 | 16685 | 80 | 0.0222 | -0.0406 | 2 | 5 | 2 | 2 | - | - | - | - | z4=3 | Ni1O4W1 |
| 13.2.5 | 15852 | 80 | 0.0141 | -0.4066 | 2 | 4 | 2 | 2 | - | - | - | - | z2w,2=1<br>z4=1 | Ni1O4W1 |
| 13.2.6 | 27701 | 96 | 0.0041 | -0.1616 | 2 | 15 | 2 | 2 | - | - | - | - | z2w,2=1<br>z4=3 | Mo8O23 |
| 13.2.7 | 81060 | 80 | 0.0227 | -0.3131 | 2 | 6 | 0 | 1 | - | - | - | - | z2w,1=1<br>z2w,2=1 | Mo1Ni1O4 |
| 13.2.8 | 170685 | 76 | 0.0093 | -0.0211 | 2 | 3 | 2 | 2 | - | - | - | - | z2w,1=1<br>z2w,2=1<br>z4=3 | Mn1O4Re1 |
| 13.2.9 | 415436 | 100 | 0.0049 | -0.0039 | 2 | 1 | 0 | 1 | - | - | - | - | z2w,1=1<br>z4=2 | Ho1O4Ta1 |
| 13.2.10 | 155660 | 100 | 0.0026 | -0.0175 | 2 | 2 | 0 | 1 | - | - | - | - | z2w,1=1 | Ho1O4Ta1 |
| 13.2.11 | 261836 | 164 | 0.0029 | -0.0041 | 2 | 5 | 2 | 2 | - | - | - | - | z4=3 | Ho1Li1O8W2 |
| 13.2.12 | 36623 | 184 | 0.0025 | -0.0015 | 2 | 5 | 2 | 2 | - | - | - | - | z4=3 | Cu1Ho1O8W2 |
| 13.2.13 | 67168 | 132 | 0.0040 | -0.0080 | 2 | 1 | 0 | 1 | - | - | - | - | z2w,1=1<br>z4=2 | Cl7Cs1Ti2 |
| 13.4.1 | 415460 | 106 | - | - | - | - | - | - | See details in Table XIX | | | | | O4Ta1Yb1 |
| 13.4.2 | 261838 | 170 | - | - | - | - | - | - | See details in Table XIX | | | | | Li1O8W2Yb1 |
| 13.4.3 | 16163 | 170 | - | - | - | - | - | - | See details in Table XIX | | | | | Li1O8W2Yb1 |
| 13.4.4 | 154808 | 114 | - | - | - | - | - | - | See details in Table XIX | | | | | H6Cu1O5Sr1 |
| 13.4.5 | 20264 | 94 | - | - | - | - | - | - | See details in Table XIX | | | | | Gd1O4Ta1 |
| 13.4.6 | 155659 | 94 | - | - | - | - | - | - | See details in Table XIX | | | | | Gd1O4Ta1 |
| 13.4.7 | 161847 | 138 | - | - | - | - | - | - | See details in Table XIX | | | | | Fe1Na1O8W2 |
| 13.4.8 | 415437 | 102 | - | - | - | - | - | - | See details in Table XIX | | | | | Er1O4Ta1 |
| 13.4.9 | 20262 | 102 | - | - | - | - | - | - | See details in Table XIX | | | | | Er1O4Ta1 |
| 13.4.10 | 155661 | 102 | - | - | - | - | - | - | See details in Table XIX | | | | | Er1O4Ta1 |
| 13.4.11 | 261837 | 166 | - | - | - | - | - | - | See details in Table XIX | | | | | Er1Li1O8W2 |
| 13.4.12 | 415435 | 98 | - | - | - | - | - | - | See details in Table XIX | | | | | Dy1O4Ta1 |
| 13.4.13 | 20263 | 98 | - | - | - | - | - | - | See details in Table XIX | | | | | Dy1O4Ta1 |
| 13.4.14 | 182751 | 82 | - | - | - | - | - | - | See details in Table XIX | | | | | Cu1O4W1 |
| 13.4.15 | 73748 | 186 | - | - | - | - | - | - | See details in Table XIX | | | | | Cu1Er1O8W2 |
| 13.4.16 | 192430 | 174 | - | - | - | - | - | - | See details in Table XIX | | | | | Cs1Gd1Mo2O8 |
| 13.4.17 | 15851 | 78 | - | - | - | - | - | - | See details in Table XIX | | | | | Co1O4W1 |
| 13.4.18 | 253697 | 110 | - | - | - | - | - | - | See details in Table XIX | | | | | Co1Na4O6Te1 |
| 13.4.19 | 81061 | 78 | - | - | - | - | - | - | See details in Table XIX | | | | | Co1Mo1O4 |
| 13.4.20 | 281235 | 78 | - | - | - | - | - | - | See details in Table XIX | | | | | Co1Mo1O4 |
| 13.4.21 | 82364 | 102 | - | - | - | - | - | - | See details in Table XIX | | | | | Cl4Er1Na1 |
| 13.4.22 | 251426 | 90 | - | - | - | - | - | - | See details in Table XIX | | | | | C6Li1N4 |
| 13.4.23 | 251425 | 90 | - | - | - | - | - | - | See details in Table XIX | | | | | C6Li1N4 |
| 13.4.24 | 251423 | 90 | - | - | - | - | - | - | See details in Table XIX | | | | | C6Li1N4 |
| 13.4.25 | 251421 | 90 | - | - | - | - | - | - | See details in Table XIX | | | | | C6Li1N4 |

Continued on next page

**TABLE XIV– continued**

| TQC | ICSD | NoE | DG | IndG | SubSG (w/ T) | cl | SubSG (w/o T) | cl | Max. Sub. SG | yes or no | Max. Sub. SG | yes or no | SIs (w/ T) | Stoich. Formula |
|---|---|---|---|---|---|---|---|---|---|---|---|---|---|---|
| 13.4.26 | 251420 | 90 | - | - | - | - | - | - | See details in Table XIX | | | | | C6Li1N4 |
| 13.4.27 | 251419 | 90 | - | - | - | - | - | - | See details in Table XIX | | | | | C6Li1N4 |
| 13.4.28 | 251418 | 90 | - | - | - | - | - | - | See details in Table XIX | | | | | C6Li1N4 |
| 13.4.29 | 56734 | 158 | - | - | - | - | - | - | See details in Table XIX | | | | | Al2Cl8Eu1 |
| 14.2.1 | 96154 | 112 | 0.0034 | -0.0185 | 2 | 3 | 2 | 1 | - | - | - | - | z2w,1=1<br>z4=1 | Si4Ti3U2 |
| 14.2.2 | 57480 | 144 | 0.0395 | -0.3870 | 2 | 7 | 0 | 2 | - | - | - | - | z4=2 | Re2Si1 |
| 14.2.3 | 9402 | 140 | 0.0079 | -0.0335 | 2 | 2 | 0 | 2 | - | - | - | - | z2w,1=1 | O4Te1V1 |
| 14.2.4 | 8217 | 72 | 0.0610 | -0.9430 | 2 | 4 | 2 | 1 | - | - | - | - | z2w,1=1<br>z4=3 | O2W1 |
| 14.2.5 | 80829 | 72 | 0.0624 | -0.8984 | 2 | 4 | 2 | 1 | - | - | - | - | z2w,1=1<br>z4=3 | O2W1 |
| 14.2.6 | 67381 | 264 | 0.1655 | 0.1054 | 2 | 2 | 0 | 2 | - | - | - | - | z2w,1=1 | O1S2Yb2 |
| 14.2.7 | 281705 | 176 | 0.0304 | -0.1012 | 2 | 6 | 2 | 1 | - | - | - | - | z4=1 | Ni3O8P2 |
| 14.2.8 | 100194 | 144 | 0.0053 | -0.1536 | 2 | 4 | 2 | 1 | - | - | - | - | z2w,1=1<br>z4=3 | Ni2O7P2 |
| 14.2.9 | 246365 | 180 | 0.0156 | -0.0294 | 2 | 4 | 2 | 1 | - | - | - | - | z2w,1=1<br>z4=3 | Ni1O10P2V2 |
| 14.2.10 | 80000 | 124 | 0.0123 | -0.0817 | 2 | 1 | 2 | 1 | - | - | - | - | z4=3 | Mo3Si4U2 |
| 14.2.11 | 291320 | 172 | 0.0095 | -0.0008 | 2 | 6 | 2 | 1 | - | - | - | - | z4=1 | Mo1O6Sr2Yb1 |
| 14.2.12 | 80830 | 72 | 0.0198 | -0.9829 | 2 | 4 | 2 | 1 | - | - | - | - | z2w,1=1<br>z4=3 | Mo1O2 |
| 14.2.13 | 23722 | 72 | 0.0194 | -1.0268 | 2 | 4 | 2 | 1 | - | - | - | - | z2w,1=1<br>z4=3 | Mo1O2 |
| 14.2.14 | 174231 | 76 | 0.0540 | -1.2101 | 2 | 3 | 2 | 1 | - | - | - | - | z2w,1=1<br>z4=1 | Ir1P2 |
| 14.2.15 | 171490 | 168 | 0.0122 | 0.0105 | 2 | 1 | 2 | 1 | - | - | - | - | z4=3 | Ir1O6Sr2Tb1 |
| 14.2.16 | 153133 | 168 | 0.0154 | 0.0052 | 2 | 1 | 2 | 1 | - | - | - | - | z4=3 | Ir1O6Sr2Tb1 |
| 14.2.17 | 47158 | 100 | 0.0046 | -0.0893 | 2 | 6 | 2 | 1 | - | - | - | - | z4=1 | H2Li1N1O3 |
| 14.2.18 | 84949 | 80 | 0.0282 | -0.2612 | 2 | 4 | 2 | 1 | - | - | - | - | z2w,1=1<br>z4=3 | H1Mn1O2 |
| 14.2.19 | 98698 | 100 | 0.0493 | -0.1311 | 2 | 5 | 0 | 2 | - | - | - | - | z2w,1=1<br>z4=2 | H1Cu1F1O1 |
| 14.2.20 | 91088 | 100 | 0.0340 | -0.2710 | 2 | 2 | 0 | 2 | - | - | - | - | z2w,1=1 | H1Cl1Cu1O1 |
| 14.2.21 | 38384 | 100 | 0.0216 | -0.2141 | 2 | 1 | 2 | 1 | - | - | - | - | z4=3 | H1Cl1Cu1O1 |
| 14.2.22 | 80002 | 124 | 0.0159 | -0.0651 | 2 | 1 | 2 | 1 | - | - | - | - | z4=3 | Ge4Mo3U2 |
| 14.2.23 | 416421 | 156 | 0.0306 | -1.2634 | 2 | 5 | 0 | 2 | - | - | - | - | z2w,1=1<br>z4=2 | Ge2La1Pt2 |
| 14.2.24 | 421335 | 120 | 0.0070 | 0.0032 | 2 | 1 | 2 | 1 | - | - | - | - | z4=3 | Gd1S2 |
| 14.2.25 | 62281 | 96 | 0.0115 | -0.1611 | 2 | 6 | 2 | 1 | - | - | - | - | z4=1 | Ga1La1Ni1 |
| 14.2.26 | 189438 | 136 | 0.0122 | -0.2431 | 2 | 4 | 2 | 1 | - | - | - | - | z2w,1=1<br>z4=3 | Fe2O3 |
| 14.2.27 | 154372 | 164 | 0.0075 | -0.1302 | 2 | 2 | 0 | 2 | - | - | - | - | z2w,1=1 | Fe2K4O5 |
| 14.2.28 | 419955 | 216 | 0.0058 | -0.0103 | 2 | 1 | 2 | 1 | - | - | - | - | z4=3 | F1O3Se1Tm1 |
| 14.2.29 | 412767 | 160 | 0.0041 | -0.0071 | 2 | 3 | 2 | 1 | - | - | - | - | z2w,1=1<br>z4=1 | Eu2P2Se6 |
| 14.2.30 | 412763 | 160 | 0.0054 | 0.0003 | 2 | 3 | 2 | 1 | - | - | - | - | z2w,1=1<br>z4=1 | Eu2P2S6 |
| 14.2.31 | 416422 | 180 | 0.0023 | -0.3930 | 2 | 4 | 2 | 1 | - | - | - | - | z2w,1=1<br>z4=3 | Eu1Ge2Pt2 |
| 14.2.32 | 401556 | 124 | 0.0099 | -0.0971 | 2 | 3 | 2 | 1 | - | - | - | - | z2w,1=1<br>z4=1 | Eu1Ge1Pd1 |
| 14.2.33 | 67379 | 248 | 0.0000 | -0.0200 | 2 | 1 | 2 | 1 | - | - | - | - | z4=3 | Er2O1S2 |
| 14.2.34 | 69772 | 172 | 0.0242 | 0.0067 | 2 | 7 | 0 | 2 | - | - | - | - | z4=2 | Er1O6Ru1Sr2 |
| 14.2.35 | 69771 | 172 | 0.0285 | 0.0064 | 2 | 7 | 0 | 2 | - | - | - | - | z4=2 | Er1O6Ru1Sr2 |
| 14.2.36 | 162113 | 192 | 0.0027 | -0.0170 | 2 | 1 | 2 | 1 | - | - | - | - | z4=3 | Dy1Ni1O3 |
| 14.2.37 | 60653 | 208 | 0.0594 | 0.0521 | 2 | 1 | 2 | 1 | - | - | - | - | z4=3 | Cu2O4Se1 |

Continued on next page

**TABLE XIV– continued**

| TQC | ICSD | NoE | DG | IndG | SubSG (w/ T) | cl | SubSG (w/o T) | cl | Max. Sub. SG | yes or no | Max. Sub. SG | yes or no | SIs (w/ T) | Stoich. Formula |
|---|---|---|---|---|---|---|---|---|---|---|---|---|---|---|
| 14.2.38 | 1672 | 164 | 0.0075 | -0.1188 | 2 | 7 | 0 | 2 | - | - | - | - | z4=2 | Cu1O4Te1 |
| 14.2.39 | 1671 | 164 | 0.0075 | -0.1188 | 2 | 7 | 0 | 2 | - | - | - | - | z4=2 | Cu1O4Te1 |
| 14.2.40 | 202451 | 140 | 0.0059 | -0.0905 | 2 | 1 | 2 | 1 | - | - | - | - | z4=3 | Cu1O3Te1 |
| 14.2.41 | 260194 | 152 | 0.0181 | -0.0203 | 2 | 1 | 2 | 1 | - | - | - | - | z4=3 | Cu1Np1Se2 |
| 14.2.42 | 73398 | 164 | 0.0146 | -0.0597 | 2 | 3 | 2 | 1 | - | - | - | - | z2w,1=1<br>z4=1 | Cu1Na1O4P1 |
| 14.2.43 | 411020 | 132 | 0.0099 | -0.0750 | 2 | 5 | 0 | 2 | - | - | - | - | z2w,1=1<br>z4=2 | Cu1Ge1O3 |
| 14.2.44 | 59291 | 140 | 0.0032 | -0.1327 | 2 | 6 | 2 | 1 | - | - | - | - | z4=1 | Co2O7P2 |
| 14.2.45 | 73738 | 104 | 0.0107 | -0.1306 | 2 | 4 | 2 | 1 | - | - | - | - | z2w,1=1<br>z4=3 | Co1Ta1Te2 |
| 14.2.46 | 63553 | 76 | 0.0260 | 0.0158 | 2 | 4 | 2 | 1 | - | - | - | - | z2w,1=1<br>z4=3 | Co1Sb2 |
| 14.2.47 | 73739 | 136 | 0.0211 | -0.2350 | 2 | 4 | 2 | 1 | - | - | - | - | z2w,1=1<br>z4=3 | Co1Nb1Te2 |
| 14.2.48 | 250364 | 132 | 0.0000 | -0.0312 | 2 | 2 | 0 | 2 | - | - | - | - | z2w,1=1 | Cl4V1 |
| 14.2.49 | 422546 | 192 | 0.0028 | -0.0548 | 2 | 5 | 0 | 2 | - | - | - | - | z2w,1=1<br>z4=2 | Cl4Hg4I2S1 |
| 14.2.50 | 4411 | 132 | 0.0049 | 0.0021 | 2 | 2 | 0 | 2 | - | - | - | - | z2w,1=1 | Cl3Mo1O1 |
| 14.2.51 | 16150 | 132 | 0.0078 | 0.0045 | 2 | 2 | 0 | 2 | - | - | - | - | z2w,1=1 | Cl3Mo1O1 |
| 14.2.52 | 27020 | 96 | 0.0113 | -0.1058 | 2 | 3 | 2 | 1 | - | - | - | - | z2w,1=1<br>z4=1 | Ce1Se2 |
| 14.2.53 | 419346 | 96 | 0.0099 | -0.0698 | 2 | 4 | 2 | 1 | - | - | - | - | z2w,1=1<br>z4=3 | Ce1S2 |
| 14.2.54 | 238150 | 112 | 0.0183 | -0.2182 | 2 | 3 | 2 | 1 | - | - | - | - | z2w,1=1<br>z4=1 | Ca2Fe1O6Os1 |
| 14.2.55 | 238148 | 112 | 0.0040 | -0.2270 | 2 | 3 | 2 | 1 | - | - | - | - | z2w,1=1<br>z4=1 | Ca2Fe1O6Os1 |
| 14.2.56 | 69790 | 64 | 0.0367 | -0.5242 | 2 | 6 | 2 | 1 | - | - | - | - | z4=1 | Ca1Pd1Si1 |
| 14.2.57 | 86049 | 144 | 0.0307 | -0.3943 | 2 | 3 | 2 | 1 | - | - | - | - | z2w,1=1<br>z4=1 | C7Y4 |
| 14.2.58 | 54136 | 120 | 0.0026 | -0.0093 | 2 | 1 | 2 | 1 | - | - | - | - | z4=3 | Br3Cr1In1 |
| 14.2.59 | 172071 | 196 | 0.3928 | 0.3583 | 2 | 5 | 0 | 2 | - | - | - | - | z2w,1=1<br>z4=2 | Br2Cu2O6Sr2Te1 |
| 14.2.60 | 75019 | 180 | 0.0106 | -0.2030 | 2 | 1 | 2 | 1 | - | - | - | - | z4=3 | Ba2P3Pd1 |
| 14.2.61 | 50158 | 144 | 0.0166 | -0.5511 | 2 | 3 | 2 | 1 | - | - | - | - | z2w,1=1<br>z4=1 | Ba1Rh2Si2 |
| 14.2.62 | 50160 | 144 | 0.0271 | -0.3557 | 2 | 3 | 2 | 1 | - | - | - | - | z2w,1=1<br>z4=1 | Ba1Ga2Pt2 |
| 14.2.63 | 88450 | 172 | 0.0146 | -0.0198 | 2 | 4 | 2 | 1 | - | - | - | - | z2w,1=1<br>z4=3 | B1Fe2O4 |
| 14.2.64 | 86862 | 172 | 0.0146 | -0.0198 | 2 | 4 | 2 | 1 | - | - | - | - | z2w,1=1<br>z4=3 | B1Fe2O4 |
| 14.2.65 | 237469 | 204 | 0.1007 | -0.5839 | 2 | 2 | 0 | 2 | - | - | - | - | z2w,1=1 | Au4Ga2Na1 |
| 14.2.66 | 58406 | 108 | 0.0389 | -0.7472 | 2 | 3 | 2 | 1 | - | - | - | - | z2w,1=1<br>z4=1 | Au4Ca5 |
| 14.2.67 | 74948 | 180 | 0.0514 | 0.0514 | 2 | 1 | 2 | 1 | - | - | - | - | z4=3 | As2Ba1Cu2O8 |
| 14.2.68 | 73978 | 148 | 0.0044 | -0.0556 | 2 | 2 | 0 | 2 | - | - | - | - | z2w,1=1 | As1Fe1O4 |
| 14.2.69 | 73402 | 112 | 0.1067 | 0.0354 | 2 | 1 | 2 | 1 | - | - | - | - | z4=3 | Ag2Te1 |
| 14.2.70 | 73230 | 112 | 0.1192 | 0.0374 | 2 | 1 | 2 | 1 | - | - | - | - | z4=3 | Ag2Te1 |
| 14.2.71 | 165661 | 180 | 0.0089 | -0.1479 | 2 | 1 | 2 | 1 | - | - | - | - | z4=3 | Ag1Fe2S3 |
| 14.3.1 | 87926 | 148 | - | - | - | - | - | - | 7 | yes | 7 | yes | - | Pt5Se4 |
| | | | | | | | | | 2 | no(2) | 2 | yes | z2w,2=1<br>z2w,3=1 | |
| 14.3.2 | 26902 | 260 | - | - | - | - | - | - | 7 | yes | 7 | yes | - | P1Pd6 |

Continued on next page

**TABLE XIV– continued**

| TQC | ICSD | NoE | DG | IndG | SubSG cl (w/ T) | | SubSG cl (w/o T) | | Max. Sub. SG | yes or no | Max. Sub. SG | yes or no | SIs (w/ T) | Stoich. Formula |
|---|---|---|---|---|---|---|---|---|---|---|---|---|---|---|
| | | | | | | | | | 2 | no(2) | 2 | no(2) | z2w,1=1 z2w,2=1 z2w,3=1 z4=1 | |
| 14.3.3 | 238697 | 180 | - | - | - | - | - | - | 2 | yes | 7 | no(7) | - | O4P1Pu1 |
| | | | | | | | | | - | - | 2 | yes | - | |
| 14.3.4 | 47164 | 108 | - | - | - | - | - | - | 4 | yes | 7 | no(7) | - | O3Rb1 |
| | | | | | | | | | 2 | no(2) | 4 | yes | z2w,2=1 z2w,3=1 z4=3 | |
| | | | | | | | | | - | - | 2 | no(2) | - | |
| 14.3.5 | 61038 | 100 | - | - | - | - | - | - | 7 | yes | 7 | yes | - | K2Mn2O3 |
| | | | | | | | | | 2 | no(2) | 2 | no(2) | z2w,2=1 z2w,3=1 z4=3 | |
| 14.3.6 | 79753 | 188 | - | - | - | - | - | - | 2 | yes | 7 | no(7) | - | Gd1O4P1 |
| | | | | | | | | | - | - | 2 | yes | - | |
| 14.3.7 | 15501 | 148 | - | - | - | - | - | - | 7 | yes | 7 | yes | - | F4Fe1Na1 |
| | | | | | | | | | 2 | yes | 2 | yes | - | |
| 14.3.8 | 75557 | 148 | - | - | - | - | - | - | 4 | yes | 7 | no(7) | - | F1O1Yb1 |
| | | | | | | | | | 2 | no(2) | 4 | yes | z2w,1=1 z2w,2=1 z2w,3=1 z4=2 | |
| | | | | | | | | | - | - | 2 | yes | - | |
| 14.3.9 | 241141 | 156 | - | - | - | - | - | - | 7 | yes | 7 | yes | - | F1Fe1O3Te1 |
| | | | | | | | | | 2 | no(2) | 2 | yes | z2w,1=1 z2w,2=1 z2w,3=1 | |
| 14.3.10 | 238078 | 156 | - | - | - | - | - | - | 7 | yes | 7 | yes | - | F1Fe1O3Se1 |
| | | | | | | | | | 2 | no(2) | 2 | yes | z2w,1=1 z2w,2=1 z2w,3=1 z4=2 | |
| 14.3.11 | 62900 | 124 | - | - | - | - | - | - | 4 | yes | 7 | no(7) | - | Eu1Ge1Ni1 |
| | | | | | | | | | 2 | no(2) | 4 | yes | z2w,2=1 z2w,3=1 z4=2 | |
| | | | | | | | | | - | - | 2 | yes | - | |
| 14.3.12 | 95380 | 140 | - | - | - | - | - | - | 4 | yes | 7 | no(7) | - | Er1Li1O2 |
| | | | | | | | | | 2 | no(2) | 4 | yes | z2w,2=1 z2w,3=1 z4=3 | |
| | | | | | | | | | - | - | 2 | no(2) | - | |
| 14.3.13 | 66811 | 180 | - | - | - | - | - | - | 4 | yes | 7 | no(7) | - | Cs1Fe1I4 |
| | | | | | | | | | 2 | no(2) | 4 | yes | z2w,1=1 z2w,2=1 z2w,3=1 z4=3 | |
| | | | | | | | | | - | - | 2 | no(2) | - | |
| 14.3.14 | 402070 | 180 | - | - | - | - | - | - | 4 | yes | 7 | no(7) | - | Cs1Fe1I4 |
| | | | | | | | | | 2 | no(2) | 4 | yes | z2w,1=1 z2w,2=1 z2w,3=1 z4=3 | |
| | | | | | | | | | - | - | 2 | no(2) | - | |
| 14.3.15 | 37108 | 140 | - | - | - | - | - | - | 7 | yes | 7 | yes | - | Cr1F4Na1 |

Continued on next page

**TABLE XIV– continued**

| TQC | ICSD | NoE | DG | IndG | SubSG (w/ T) | cl | SubSG (w/o T) | cl | Max. Sub. SG | yes or no | Max. Sub. SG | yes or no | SIs (w/ T) | Stoich. Formula |
|---|---|---|---|---|---|---|---|---|---|---|---|---|---|---|
| | | | | | | | | | 2 | no(2) | 2 | no(2) | z2w,1=1<br>z2w,2=1<br>z2w,3=1<br>z4=3 | |
| 14.3.16 | 99835 | 140 | - | - | - | - | - | - | 4 | yes | 7 | no(7) | - | Ce1O3V1 |
| | | | | | | | | | 2 | no(2) | 4 | yes | z2w,1=1<br>z2w,2=1<br>z2w,3=1 | |
| | | | | | | | | | - | - | 2 | yes | - | |
| 14.3.17 | 162747 | 140 | - | - | - | - | - | - | 7 | yes | 7 | yes | - | Ce1O3V1 |
| | | | | | | | | | 2 | no(2) | 2 | yes | z2w,2=1<br>z2w,3=1 | |
| 14.3.18 | 56772 | 100 | - | - | - | - | - | - | 7 | yes | 7 | yes | - | Ce1Li1O2 |
| | | | | | | | | | 2 | no(2) | 2 | no(2) | z2w,1=1<br>z2w,2=1<br>z2w,3=1<br>z4=1 | |
| 14.3.19 | 189806 | 40 | - | - | - | - | - | - | 7 | yes | 7 | yes | - | Bi1 |
| | | | | | | | | | 2 | no(2) | 2 | no(2) | z2w,2=1<br>z2w,3=1<br>z4=1 | |
| 14.3.20 | 418156 | 160 | - | - | - | - | - | - | 4 | yes | 7 | no(7) | - | As2Pt2Sr1 |
| | | | | | | | | | 2 | no(2) | 4 | yes | z2w,1=1<br>z2w,2=1<br>z2w,3=1<br>z4=2 | |
| | | | | | | | | | - | - | 2 | yes | - | |
| 14.3.21 | 249707 | 196 | - | - | - | - | - | - | 7 | yes | 7 | yes | - | As1Eu1Rb1S3 |
| | | | | | | | | | 2 | no(2) | 2 | no(2) | z2w,2=1<br>z2w,3=1<br>z4=1 | |
| 14.3.22 | 249705 | 164 | - | - | - | - | - | - | 7 | yes | 7 | yes | - | As1Eu1Li1S3 |
| | | | | | | | | | 2 | no(2) | 2 | yes | z2w,1=1<br>z2w,2=1<br>z2w,3=1 | |
| 14.3.23 | 249706 | 196 | - | - | - | - | - | - | 7 | yes | 7 | yes | - | As1Eu1K1S3 |
| | | | | | | | | | 2 | no(2) | 2 | no(2) | z2w,2=1<br>z2w,3=1<br>z4=1 | |
| 14.4.1 | 193236 | 170 | - | - | - | - | - | - | See details in Table XIX | | | | | O6Sr2Ta1Yb1 |
| 14.4.2 | 49500 | 150 | - | - | - | - | - | - | See details in Table XIX | | | | | O6Ru1Sr2Y1 |
| 14.4.3 | 238814 | 134 | - | - | - | - | - | - | See details in Table XIX | | | | | O6Ru1Sr2Y1 |
| 14.4.4 | 192766 | 150 | - | - | - | - | - | - | See details in Table XIX | | | | | O6Ru1Sr2Y1 |
| 14.4.5 | 93361 | 166 | - | - | - | - | - | - | See details in Table XIX | | | | | O6Ru1Sr2Tb1 |
| 14.4.6 | 290794 | 166 | - | - | - | - | - | - | See details in Table XIX | | | | | O6Ru1Sr2Tb1 |
| 14.4.7 | 153132 | 166 | - | - | - | - | - | - | See details in Table XIX | | | | | O6Ru1Sr2Tb1 |
| 14.4.8 | 245796 | 106 | - | - | - | - | - | - | See details in Table XIX | | | | | O6Re2Sc1 |
| 14.4.9 | 238315 | 150 | - | - | - | - | - | - | See details in Table XIX | | | | | O6Os1Sr2Y1 |
| 14.4.10 | 238314 | 150 | - | - | - | - | - | - | See details in Table XIX | | | | | O6Os1Sr2Y1 |
| 14.4.11 | 238313 | 150 | - | - | - | - | - | - | See details in Table XIX | | | | | O6Os1Sr2Y1 |
| 14.4.12 | 238312 | 150 | - | - | - | - | - | - | See details in Table XIX | | | | | O6Os1Sr2Y1 |
| 14.4.13 | 238311 | 150 | - | - | - | - | - | - | See details in Table XIX | | | | | O6Os1Sr2Y1 |
| 14.4.14 | 193235 | 186 | - | - | - | - | - | - | See details in Table XIX | | | | | Nb1O6Sr2Yb1 |
| 14.4.15 | 280907 | 142 | - | - | - | - | - | - | See details in Table XIX | | | | | Mn1P2Rb2S6 |
| 14.4.16 | 69801 | 110 | - | - | - | - | - | - | See details in Table XIX | | | | | Mn1O6Se2 |
| 14.4.17 | 202867 | 178 | - | - | - | - | - | - | See details in Table XIX | | | | | Lu1O6Ru1Sr2 |
| 14.4.18 | 202866 | 178 | - | - | - | - | - | - | See details in Table XIX | | | | | Lu1O6Ru1Sr2 |
| 14.4.19 | 88523 | 158 | - | - | - | - | - | - | See details in Table XIX | | | | | La2O6Rh1Zn1 |

Continued on next page

**TABLE XIV– continued**

| TQC | ICSD | NoE | DG | IndG | SubSG cl (w/ T) | | SubSG cl (w/o T) | | Max. Sub. SG | yes or no | Max. Sub. SG | yes or no | SIs (w/ T) | Stoich. Formula |
|---|---|---|---|---|---|---|---|---|---|---|---|---|---|---|
| 14.4.20 | 88522 | 138 | - | - | - | - | - | - | See details in Table XIX | | | | | La2Mg1O6Rh1 |
| 14.4.21 | 97487 | 134 | - | - | - | - | - | - | See details in Table XIX | | | | | La2Li1O6Ru1 |
| 14.4.22 | 81882 | 158 | - | - | - | - | - | - | See details in Table XIX | | | | | Ir1La2O6Zn1 |
| 14.4.23 | 79492 | 138 | - | - | - | - | - | - | See details in Table XIX | | | | | Ir1La2Mg1O6 |
| 14.4.24 | 238318 | 134 | - | - | - | - | - | - | See details in Table XIX | | | | | In1O6Os1Sr2 |
| 14.4.25 | 238317 | 134 | - | - | - | - | - | - | See details in Table XIX | | | | | In1O6Os1Sr2 |
| 14.4.26 | 65510 | 134 | - | - | - | - | - | - | See details in Table XIX | | | | | I6K2Tc1 |
| 14.4.27 | 67555 | 114 | - | - | - | - | - | - | See details in Table XIX | | | | | H8Br2Cu1O4 |
| 14.4.28 | 280918 | 98 | - | - | - | - | - | - | See details in Table XIX | | | | | H4Cu1O4P2 |
| 14.4.29 | 280917 | 98 | - | - | - | - | - | - | See details in Table XIX | | | | | H4Cu1O4P2 |
| 14.4.30 | 98699 | 150 | - | - | - | - | - | - | See details in Table XIX | | | | | H2Cu3F4O2 |
| 14.4.31 | 290251 | 98 | - | - | - | - | - | - | See details in Table XIX | | | | | F6Li2V1 |
| 14.4.32 | 416295 | 174 | - | - | - | - | - | - | See details in Table XIX | | | | | F6Gd1K3 |
| 14.4.33 | 420874 | 162 | - | - | - | - | - | - | See details in Table XIX | | | | | Eu2Ir1Mg1O6 |
| 14.4.34 | 80576 | 114 | - | - | - | - | - | - | See details in Table XIX | | | | | Cu1O6Sb2 |
| 14.4.35 | 36071 | 98 | - | - | - | - | - | - | See details in Table XIX | | | | | Cu1O4U1 |
| 14.4.36 | 9790 | 50 | - | - | - | - | - | - | See details in Table XIX | | | | | Cu1F2 |
| 14.4.37 | 9789 | 50 | - | - | - | - | - | - | See details in Table XIX | | | | | Cu1F2 |
| 14.4.38 | 9788 | 50 | - | - | - | - | - | - | See details in Table XIX | | | | | Cu1F2 |
| 14.4.39 | 71833 | 50 | - | - | - | - | - | - | See details in Table XIX | | | | | Cu1F2 |
| 14.4.40 | 6024 | 50 | - | - | - | - | - | - | See details in Table XIX | | | | | Cu1F2 |
| 14.4.41 | 173093 | 178 | - | - | - | - | - | - | See details in Table XIX | | | | | Cu0.5O5P1Ti1 |
| 14.4.42 | 195658 | 142 | - | - | - | - | - | - | See details in Table XIX | | | | | Co1O6Sr2Te1 |
| 14.4.43 | 195651 | 142 | - | - | - | - | - | - | See details in Table XIX | | | | | Co1O6Sr2Te1 |
| 14.4.44 | 195650 | 142 | - | - | - | - | - | - | See details in Table XIX | | | | | Co1O6Sr2Te1 |
| 14.4.45 | 153002 | 142 | - | - | - | - | - | - | See details in Table XIX | | | | | Co1O6Sr2Te1 |
| 14.4.46 | 169193 | 118 | - | - | - | - | - | - | See details in Table XIX | | | | | Co1O6Pb2Te1 |
| 14.4.47 | 411351 | 158 | - | - | - | - | - | - | See details in Table XIX | | | | | Cl4F6Ir1 |
| 14.4.48 | 150697 | 110 | - | - | - | - | - | - | See details in Table XIX | | | | | Ca2Fe1O6Re1 |
| 14.4.49 | 98732 | 110 | - | - | - | - | - | - | See details in Table XIX | | | | | Ca2Co1O6W1 |
| 14.4.50 | 417591 | 182 | - | - | - | - | - | - | See details in Table XIX | | | | | C2H2Cu3O8 |
| 14.4.51 | 2934 | 182 | - | - | - | - | - | - | See details in Table XIX | | | | | C2H2Cu3O8 |
| 14.4.52 | 182324 | 182 | - | - | - | - | - | - | See details in Table XIX | | | | | C2H2Cu3O8 |
| 14.4.53 | 158577 | 182 | - | - | - | - | - | - | See details in Table XIX | | | | | C2H2Cu3O8 |
| 14.4.54 | 60858 | 114 | - | - | - | - | - | - | See details in Table XIX | | | | | C2Cu1Na2O6 |
| 14.4.55 | 426579 | 114 | - | - | - | - | - | - | See details in Table XIX | | | | | C2Cu1Na2O6 |
| 14.4.56 | 10355 | 114 | - | - | - | - | - | - | See details in Table XIX | | | | | C2Cu1Na2O6 |
| 14.4.57 | 155317 | 122 | - | - | - | - | - | - | See details in Table XIX | | | | | B2Cu1O6Pb2 |
| 14.4.58 | 89620 | 178 | - | - | - | - | - | - | See details in Table XIX | | | | | Au3F8 |
| 14.4.59 | 58154 | 90 | - | - | - | - | - | - | See details in Table XIX | | | | | Al9Rh2 |
| 14.4.60 | 414304 | 90 | - | - | - | - | - | - | See details in Table XIX | | | | | Al9Rh2 |
| 14.4.61 | 414302 | 90 | - | - | - | - | - | - | See details in Table XIX | | | | | Al9Ir2 |
| 14.4.62 | 59225 | 114 | - | - | - | - | - | - | See details in Table XIX | | | | | Ag3O4 |
| 14.4.63 | 202218 | 114 | - | - | - | - | - | - | See details in Table XIX | | | | | Ag3O4 |
| 14.4.64 | 421461 | 114 | - | - | - | - | - | - | See details in Table XIX | | | | | Ag1F4K2 |
| 15.2.1 | 187497 | 56 | 0.0575 | -1.8309 | 2 | 7 | 0 | 2 | - | - | - | - | z4=2 | Sb2Te3 |
| 15.2.2 | 185954 | 56 | 0.0645 | -1.9384 | 2 | 5 | 2 | 1 | - | - | - | - | z2w,1=1<br>z2w,2=1<br>z4=1 | Sb2Te3 |
| 15.2.3 | 16041 | 144 | 0.1124 | -0.0369 | 2 | 4 | 2 | 1 | - | - | - | - | z4=3 | S1Ta6 |
| 15.2.4 | 24327 | 120 | 0.0384 | -0.1613 | 2 | 3 | 0 | 2 | - | - | - | - | z2w,1=1<br>z2w,2=1 | P2Pt5 |
| 15.2.5 | 261031 | 144 | 0.0055 | -0.0159 | 2 | 3 | 0 | 2 | - | - | - | - | z2w,1=1<br>z2w,2=1 | O6Pb2Sb1Tm1 |
| 15.2.6 | 35148 | 84 | 0.0027 | -0.1130 | 2 | 2 | 2 | 1 | - | - | - | - | z4=1 | O5Ti3 |
| 15.2.7 | 194464 | 84 | 0.0135 | -0.1126 | 2 | 2 | 2 | 1 | - | - | - | - | z4=1 | O5Ti3 |
| 15.2.8 | 109187 | 96 | 0.0082 | -0.0117 | 2 | 4 | 2 | 1 | - | - | - | - | z4=3 | O4Ta1Tb1 |

Continued on next page

**TABLE XIV– continued**

| TQC | ICSD | NoE | DG | IndG | SubSG (w/ T) | cl | SubSG (w/o T) | cl | Max. Sub. SG | yes or no | Max. Sub. SG | yes or no | SIs (w/ T) | Stoich. Formula |
|---|---|---|---|---|---|---|---|---|---|---|---|---|---|---|
| 15.2.9 | 97442 | 56 | 0.0063 | -0.0524 | 2 | 2 | 2 | 1 | - | - | - | - | z4=1 | O3V2 |
| 15.2.10 | 95763 | 56 | 0.0245 | -0.0756 | 2 | 2 | 2 | 1 | - | - | - | - | z4=1 | O3V2 |
| 15.2.11 | 95762 | 56 | 0.0204 | -0.0707 | 2 | 2 | 2 | 1 | - | - | - | - | z4=1 | O3V2 |
| 15.2.12 | 6286 | 56 | 0.0043 | -0.0612 | 2 | 7 | 0 | 2 | - | - | - | - | z4=2 | O3V2 |
| 15.2.13 | 50703 | 172 | 0.0193 | -0.2415 | 2 | 3 | 0 | 2 | - | - | - | - | z2w,1=1<br>z2w,2=1 | Ni2O8Te3 |
| 15.2.14 | 61148 | 100 | 0.0365 | 0.0365 | 2 | 4 | 2 | 1 | - | - | - | - | z4=3 | Ni1Se5Ta2 |
| 15.2.15 | 109181 | 120 | 0.0066 | -0.0128 | 2 | 4 | 2 | 1 | - | - | - | - | z4=3 | Nb1O4Tm1 |
| 15.2.16 | 109178 | 112 | 0.0087 | -0.0097 | 2 | 4 | 2 | 1 | - | - | - | - | z4=3 | Nb1O4Tb1 |
| 15.2.17 | 78342 | 100 | 0.0046 | 0.0046 | 2 | 4 | 2 | 1 | - | - | - | - | z4=3 | Na1O6Si2V1 |
| 15.2.18 | 163133 | 100 | 0.0327 | 0.0327 | 2 | 4 | 2 | 1 | - | - | - | - | z4=3 | Na1O6Si2V1 |
| 15.2.19 | 163132 | 100 | 0.0290 | 0.0290 | 2 | 4 | 2 | 1 | - | - | - | - | z4=3 | Na1O6Si2V1 |
| 15.2.20 | 163131 | 100 | 0.0550 | 0.0292 | 2 | 4 | 2 | 1 | - | - | - | - | z4=3 | Na1O6Si2V1 |
| 15.2.21 | 163130 | 100 | 0.0460 | 0.0085 | 2 | 4 | 2 | 1 | - | - | - | - | z4=3 | Na1O6Si2V1 |
| 15.2.22 | 163129 | 100 | 0.0438 | 0.0082 | 2 | 4 | 2 | 1 | - | - | - | - | z4=3 | Na1O6Si2V1 |
| 15.2.23 | 163128 | 100 | 0.0446 | 0.0199 | 2 | 4 | 2 | 1 | - | - | - | - | z4=3 | Na1O6Si2V1 |
| 15.2.24 | 163127 | 100 | 0.0098 | 0.0098 | 2 | 4 | 2 | 1 | - | - | - | - | z4=3 | Na1O6Si2V1 |
| 15.2.25 | 163126 | 100 | 0.0126 | 0.0126 | 2 | 4 | 2 | 1 | - | - | - | - | z4=3 | Na1O6Si2V1 |
| 15.2.26 | 163125 | 100 | 0.0155 | 0.0155 | 2 | 4 | 2 | 1 | - | - | - | - | z4=3 | Na1O6Si2V1 |
| 15.2.27 | 105057 | 52 | 0.0541 | -0.4896 | 2 | 2 | 2 | 1 | - | - | - | - | z4=1 | Mo1P4 |
| 15.2.28 | 49802 | 156 | 0.0141 | -0.1382 | 2 | 2 | 2 | 1 | - | - | - | - | z4=1 | Mn2P12Ti1 |
| 15.2.29 | 78721 | 112 | 0.0215 | -0.1219 | 2 | 4 | 2 | 1 | - | - | - | - | z4=3 | Li2O3Ru1 |
| 15.2.30 | 180367 | 112 | 0.0752 | -0.0310 | 2 | 7 | 0 | 2 | - | - | - | - | z4=2 | Li2O3Ru1 |
| 15.2.31 | 55164 | 100 | 0.0481 | 0.0379 | 2 | 4 | 2 | 1 | - | - | - | - | z4=3 | Li1O6Si2V1 |
| 15.2.32 | 152074 | 100 | 0.0471 | 0.0344 | 2 | 4 | 2 | 1 | - | - | - | - | z4=3 | Li1O6Si2V1 |
| 15.2.33 | 416039 | 184 | 0.0195 | -0.0108 | 2 | 6 | 0 | 2 | - | - | - | - | z2w,1=1<br>z2w,2=1<br>z4=2 | K3Na1O8Ru2 |
| 15.2.34 | 192833 | 100 | 0.0055 | -0.0143 | 2 | 1 | 2 | 1 | - | - | - | - | z2w,1=1<br>z2w,2=1<br>z4=3 | Ho1O4V1 |
| 15.2.35 | 79499 | 100 | 0.0039 | -0.0133 | 2 | 4 | 2 | 1 | - | - | - | - | z4=3 | Ho1O4Ta1 |
| 15.2.36 | 153363 | 100 | 0.0040 | -0.0183 | 2 | 4 | 2 | 1 | - | - | - | - | z4=3 | Ho1O4Ta1 |
| 15.2.37 | 15271 | 92 | 0.0158 | -0.0643 | 2 | 1 | 2 | 1 | - | - | - | - | z2w,1=1<br>z2w,2=1<br>z4=3 | H8O6Pt1 |
| 15.2.38 | 14329 | 140 | 0.0121 | -0.0251 | 2 | 6 | 0 | 2 | - | - | - | - | z2w,1=1<br>z2w,2=1<br>z4=2 | H4Ba1O8S2 |
| 15.2.39 | 20138 | 136 | 0.0094 | -0.0157 | 2 | 1 | 2 | 1 | - | - | - | - | z2w,1=1<br>z2w,2=1<br>z4=3 | H4Al2Ca1F4O4 |
| 15.2.40 | 14176 | 180 | 0.0174 | -0.0135 | 2 | 3 | 0 | 2 | - | - | - | - | z2w,1=1<br>z2w,2=1 | H1Cr1Hg1O4.5 |
| 15.2.41 | 183508 | 136 | 0.0080 | -0.0439 | 2 | 6 | 0 | 2 | - | - | - | - | z2w,1=1<br>z2w,2=1<br>z4=2 | H12Cl3Mg1O6Rb1 |
| 15.2.42 | 192134 | 136 | 0.0066 | -0.0351 | 2 | 4 | 2 | 1 | - | - | - | - | z4=3 | H12Br3Mg1O6Rb1 |
| 15.2.43 | 237747 | 104 | 0.0217 | -0.0074 | 2 | 6 | 0 | 2 | - | - | - | - | z2w,1=1<br>z2w,2=1<br>z4=2 | Ge2Mn1Na1O6 |
| 15.2.44 | 237746 | 104 | 0.0295 | -0.0018 | 2 | 6 | 0 | 2 | - | - | - | - | z2w,1=1<br>z2w,2=1<br>z4=2 | Ge2Mn1Na1O6 |
| 15.2.45 | 430120 | 220 | 0.0062 | -0.0734 | 2 | 2 | 2 | 1 | - | - | - | - | z4=1 | Gd1Ni8Si3 |

Continued on next page

**TABLE XIV– continued**

| TQC | ICSD | NoE | DG | IndG | SubSG (w/ T) | cl | SubSG (w/o T) | cl | Max. Sub. SG | yes or no | Max. Sub. SG | yes or no | SIs (w/ T) | Stoich. Formula |
|---|---|---|---|---|---|---|---|---|---|---|---|---|---|---|
| 15.2.46 | 151766 | 208 | 0.0160 | -0.0527 | 2 | 6 | 0 | 2 | - | - | - | - | z2w,1=1<br>z2w,2=1<br>z4=2 | Fe7S8 |
| 15.2.47 | 24416 | 100 | 0.0075 | -0.0870 | 2 | 5 | 2 | 1 | - | - | - | - | z2w,1=1<br>z2w,2=1<br>z4=1 | Fe2O5Ti1 |
| 15.2.48 | 154291 | 164 | 0.0084 | -0.0427 | 2 | 4 | 2 | 1 | - | - | - | - | z4=3 | Fe2O5Rb4 |
| 15.2.49 | 71059 | 104 | 0.0174 | -0.0279 | 2 | 2 | 2 | 1 | - | - | - | - | z4=1 | Fe2Li1N3Sr2 |
| 15.2.50 | 41899 | 132 | 0.0122 | 0.0106 | 2 | 4 | 2 | 1 | - | - | - | - | z4=3 | Fe1S5U2 |
| 15.2.51 | 15909 | 32 | 0.0217 | -0.5268 | 2 | 5 | 2 | 1 | - | - | - | - | z2w,1=1<br>z2w,2=1<br>z4=1 | F2Sn1 |
| 15.2.52 | 2739 | 140 | 0.0045 | -0.0174 | 2 | 5 | 2 | 1 | - | - | - | - | z2w,1=1<br>z2w,2=1<br>z4=1 | Er1Na1O2 |
| 15.2.53 | 94884 | 92 | 0.0132 | -0.2551 | 2 | 2 | 2 | 1 | - | - | - | - | z4=1 | D7Al1Sr2 |
| 15.2.54 | 185090 | 172 | 0.0194 | -0.8676 | 2 | 7 | 0 | 2 | - | - | - | - | z4=2 | Cu6Sn5 |
| 15.2.55 | 185089 | 172 | 0.0195 | -0.8789 | 2 | 7 | 0 | 2 | - | - | - | - | z4=2 | Cu6Sn5 |
| 15.2.56 | 150124 | 172 | 0.0346 | -0.8728 | 2 | 7 | 0 | 2 | - | - | - | - | z4=2 | Cu6Sn5 |
| 15.2.57 | 106530 | 172 | 0.0371 | -0.8770 | 2 | 7 | 0 | 2 | - | - | - | - | z4=2 | Cu6Sn5 |
| 15.2.58 | 23479 | 148 | 0.0107 | -0.0559 | 2 | 3 | 0 | 2 | - | - | - | - | z2w,1=1<br>z2w,2=1 | Cu2O7V2 |
| 15.2.59 | 158375 | 148 | 0.0146 | -0.0509 | 2 | 3 | 0 | 2 | - | - | - | - | z2w,1=1<br>z2w,2=1 | Cu2O7V2 |
| 15.2.60 | 67316 | 148 | 0.0054 | -0.0373 | 2 | 6 | 0 | 2 | - | - | - | - | z2w,1=1<br>z2w,2=1<br>z4=2 | Cu2O7P2 |
| 15.2.61 | 28279 | 148 | 0.0280 | -0.0179 | 2 | 6 | 0 | 2 | - | - | - | - | z2w,1=1<br>z2w,2=1<br>z4=2 | Cu2O7P2 |
| 15.2.62 | 157107 | 148 | 0.0051 | -0.0354 | 2 | 6 | 0 | 2 | - | - | - | - | z2w,1=1<br>z2w,2=1<br>z4=2 | Cu2O7P2 |
| 15.2.63 | 14369 | 148 | 0.0056 | -0.0307 | 2 | 6 | 0 | 2 | - | - | - | - | z2w,1=1<br>z2w,2=1<br>z4=2 | Cu2O7P2 |
| 15.2.64 | 83490 | 128 | 0.0224 | -0.0299 | 2 | 5 | 2 | 1 | - | - | - | - | z2w,1=1<br>z2w,2=1<br>z4=1 | Cu2Gd1O4 |
| 15.2.65 | 83494 | 136 | 0.0040 | -0.0161 | 2 | 4 | 2 | 1 | - | - | - | - | z4=3 | Cu2Er1O4 |
| 15.2.66 | 83491 | 132 | 0.0050 | -0.0253 | 2 | 7 | 0 | 2 | - | - | - | - | z4=2 | Cu2Dy1O4 |
| 15.2.67 | 2790 | 52 | 0.0368 | -0.2572 | 2 | 2 | 2 | 1 | - | - | - | - | z4=1 | Cr1P4 |
| 15.2.68 | 26814 | 124 | 0.0149 | -0.1460 | 2 | 2 | 2 | 1 | - | - | - | - | z4=1 | Co1Ge1O3 |
| 15.2.69 | 417870 | 144 | 0.0000 | 0.0002 | 2 | 2 | 2 | 1 | - | - | - | - | z4=1 | Cl8Ga2Ni1 |
| 15.2.70 | 56838 | 164 | 0.0615 | -0.0965 | 2 | 1 | 2 | 1 | - | - | - | - | z2w,1=1<br>z2w,2=1<br>z4=3 | Cl3Cu1Rb1 |
| 15.2.71 | 189154 | 176 | 0.0212 | -0.0285 | 2 | 4 | 2 | 1 | - | - | - | - | z4=3 | Cl2Cu2O6Pb1Se2 |
| 15.2.72 | 247292 | 192 | 0.0020 | -0.0165 | 2 | 1 | 2 | 1 | - | - | - | - | z2w,1=1<br>z2w,2=1<br>z4=3 | Ce4Rh4Si3 |
| 15.2.73 | 95789 | 52 | 0.0794 | -0.4935 | 2 | 6 | 0 | 2 | - | - | - | - | z2w,1=1<br>z2w,2=1<br>z4=2 | Ca2Ir2Si1 |
| 15.2.74 | 82216 | 112 | 0.0326 | -0.1847 | 2 | 2 | 2 | 1 | - | - | - | - | z4=1 | Ca1Ni1O6Si2 |
| 15.2.75 | 69381 | 128 | 0.0057 | -0.2175 | 2 | 2 | 2 | 1 | - | - | - | - | z4=1 | Ca1Ni1O6Si2 |
| 15.2.76 | 202244 | 112 | 0.0122 | -0.2057 | 2 | 2 | 2 | 1 | - | - | - | - | z4=1 | Ca1Ni1O6Si2 |

Continued on next page

**TABLE XIV– continued**

| TQC | ICSD | NoE | DG | IndG | SubSG (w/ T) | cl | SubSG (w/o T) | cl | Max. Sub. SG | yes or no | Max. Sub. SG | yes or no | SIs (w/ T) | Stoich. Formula |
|---|---|---|---|---|---|---|---|---|---|---|---|---|---|---|
| 15.2.77 | 158299 | 180 | 0.0139 | -0.0646 | 2 | 5 | 2 | 1 | - | - | - | - | z2w,1=1<br>z2w,2=1<br>z4=1 | C2H8Cl3Cu1N1 |
| 15.2.78 | 423887 | 96 | 0.0149 | -0.0599 | 2 | 2 | 2 | 1 | - | - | - | - | z4=1 | C2Fe5 |
| 15.2.79 | 423886 | 96 | 0.0190 | -0.0683 | 2 | 2 | 2 | 1 | - | - | - | - | z4=1 | C2Fe5 |
| 15.2.80 | 423885 | 96 | 0.0187 | -0.0769 | 2 | 2 | 2 | 1 | - | - | - | - | z4=1 | C2Fe5 |
| 15.2.81 | 183902 | 120 | 0.0025 | -0.0813 | 2 | 2 | 2 | 1 | - | - | - | - | z4=1 | Bi1Mn1O3 |
| 15.2.82 | 174405 | 116 | 0.0067 | -0.0496 | 2 | 4 | 2 | 1 | - | - | - | - | z4=3 | Bi1Cr1O3 |
| 15.2.83 | 174404 | 116 | 0.0059 | -0.0509 | 2 | 3 | 0 | 2 | - | - | - | - | z2w,1=1<br>z2w,2=1 | Bi1Cr1O3 |
| 15.2.84 | 160458 | 116 | 0.0071 | -0.0367 | 2 | 3 | 0 | 2 | - | - | - | - | z2w,1=1<br>z2w,2=1 | Bi1Cr1O3 |
| 15.2.85 | 160457 | 116 | 0.0061 | -0.0377 | 2 | 3 | 0 | 2 | - | - | - | - | z2w,1=1<br>z2w,2=1 | Bi1Cr1O3 |
| 15.2.86 | 160456 | 116 | 0.0055 | -0.0386 | 2 | 3 | 0 | 2 | - | - | - | - | z2w,1=1<br>z2w,2=1 | Bi1Cr1O3 |
| 15.2.87 | 160455 | 116 | 0.0057 | -0.0408 | 2 | 3 | 0 | 2 | - | - | - | - | z2w,1=1<br>z2w,2=1 | Bi1Cr1O3 |
| 15.2.88 | 245254 | 224 | 0.0442 | -0.0222 | 2 | 4 | 2 | 1 | - | - | - | - | z4=3 | Ba3Ir2O9Sr1 |
| 15.2.89 | 245253 | 224 | 0.0074 | -0.0497 | 2 | 3 | 0 | 2 | - | - | - | - | z2w,1=1<br>z2w,2=1 | Ba3Ca1Ir2O9 |
| 15.2.90 | 71060 | 104 | 0.0162 | 0.0046 | 2 | 2 | 2 | 1 | - | - | - | - | z4=1 | Ba2Fe2Li1N3 |
| 15.2.91 | 86064 | 156 | 0.1985 | 0.1985 | 2 | 1 | 2 | 1 | - | - | - | - | z2w,1=1<br>z2w,2=1<br>z4=3 | Ba1Cu1N1 |
| 15.2.92 | 418385 | 80 | 0.0226 | -0.2758 | 2 | 4 | 2 | 1 | - | - | - | - | z4=3 | B2Ni1O4 |
| 15.2.93 | 58403 | 64 | 0.0380 | -0.3816 | 2 | 3 | 0 | 2 | - | - | - | - | z2w,1=1<br>z2w,2=1 | Au2Ca5 |
| 15.2.94 | 417872 | 144 | 0.0020 | 0.0020 | 2 | 2 | 2 | 1 | - | - | - | - | z4=1 | Al2Cl8Ni1 |
| 15.2.95 | 251835 | 188 | 0.1140 | 0.0132 | 2 | 1 | 2 | 1 | - | - | - | - | z2w,1=1<br>z2w,2=1<br>z4=3 | Al1B2La2Os2 |
| 15.4.1 | 40760 | 138 | - | - | - | - | - | - | See details in Table XIX | | | | | Rh3Si5Y2 |
| 15.4.2 | 74837 | 90 | - | - | - | - | - | - | See details in Table XIX | | | | | Rb1Te6 |
| 15.4.3 | 39762 | 114 | - | - | - | - | - | - | See details in Table XIX | | | | | O7P2V1 |
| 15.4.4 | 160944 | 114 | - | - | - | - | - | - | See details in Table XIX | | | | | O7P2V1 |
| 15.4.5 | 160943 | 114 | - | - | - | - | - | - | See details in Table XIX | | | | | O7P2V1 |
| 15.4.6 | 160942 | 114 | - | - | - | - | - | - | See details in Table XIX | | | | | O7P2V1 |
| 15.4.7 | 51023 | 110 | - | - | - | - | - | - | See details in Table XIX | | | | | O6Re2Sb1 |
| 15.4.8 | 32587 | 90 | - | - | - | - | - | - | See details in Table XIX | | | | | O5V3 |
| 15.4.9 | 27800 | 90 | - | - | - | - | - | - | See details in Table XIX | | | | | O5Sb2V1 |
| 15.4.10 | 2292 | 90 | - | - | - | - | - | - | See details in Table XIX | | | | | O5Sb2V1 |
| 15.4.11 | 95210 | 106 | - | - | - | - | - | - | See details in Table XIX | | | | | O4Ta1Yb1 |
| 15.4.12 | 430121 | 222 | - | - | - | - | - | - | See details in Table XIX | | | | | Ni8Si3Tb1 |
| 15.4.13 | 281609 | 98 | - | - | - | - | - | - | See details in Table XIX | | | | | Na1O6Si2Ti1 |
| 15.4.14 | 281607 | 98 | - | - | - | - | - | - | See details in Table XIX | | | | | Na1O6Si2Ti1 |
| 15.4.15 | 281606 | 98 | - | - | - | - | - | - | See details in Table XIX | | | | | Na1O6Si2Ti1 |
| 15.4.16 | 281604 | 98 | - | - | - | - | - | - | See details in Table XIX | | | | | Na1O6Si2Ti1 |
| 15.4.17 | 248826 | 70 | - | - | - | - | - | - | See details in Table XIX | | | | | Na0.5O2V1 |
| 15.4.18 | 425216 | 126 | - | - | - | - | - | - | See details in Table XIX | | | | | Mn3O4Rb2 |
| 15.4.19 | 280877 | 186 | - | - | - | - | - | - | See details in Table XIX | | | | | K1O8W2Yb1 |
| 15.4.20 | 87184 | 42 | - | - | - | - | - | - | See details in Table XIX | | | | | K1O2 |
| 15.4.21 | 104493 | 286 | - | - | - | - | - | - | See details in Table XIX | | | | | Ir2Lu5 |
| 15.4.22 | 187813 | 222 | - | - | - | - | - | - | See details in Table XIX | | | | | Ir1O3Sr1 |
| 15.4.23 | 391433 | 86 | - | - | - | - | - | - | See details in Table XIX | | | | | In1Pt2Sr2 |
| 15.4.24 | 391432 | 86 | - | - | - | - | - | - | See details in Table XIX | | | | | In1Pd2Sr2 |
| 15.4.25 | 412275 | 98 | - | - | - | - | - | - | See details in Table XIX | | | | | I4La1Sr1 |

Continued on next page

**TABLE XIV– continued**

| TQC | ICSD | NoE | DG | IndG | SubSG cl (w/ T) | SubSG cl (w/o T) | Max. Sub. SG yes or no | Max. Sub. SG yes or no | SIs (w/ T) | Stoich. Formula |
|---|---|---|---|---|---|---|---|---|---|---|
| 15.4.26 | 16738 | 110 | - | - | - - | - - | See details in Table XIX | | | H5Ca1O6P1 |
| 15.4.27 | 16727 | 50 | - | - | - - | - - | See details in Table XIX | | | H4O6V2 |
| 15.4.28 | 412876 | 114 | - | - | - - | - - | See details in Table XIX | | | H4Cl6N1W1 |
| 15.4.29 | 195740 | 70 | - | - | - - | - - | See details in Table XIX | | | H3K2N1Se2 |
| 15.4.30 | 16628 | 142 | - | - | - - | - - | See details in Table XIX | | | H1K2N1O6S2 |
| 15.4.31 | 15499 | 142 | - | - | - - | - - | See details in Table XIX | | | H1Cs1F2O6S2 |
| 15.4.32 | 109186 | 94 | - | - | - - | - - | See details in Table XIX | | | Gd1O4Ta1 |
| 15.4.33 | 40781 | 58 | - | - | - - | - - | See details in Table XIX | | | Fe1Rb1Se2 |
| 15.4.34 | 202385 | 58 | - | - | - - | - - | See details in Table XIX | | | Fe1Rb1S2 |
| 15.4.35 | 202384 | 58 | - | - | - - | - - | See details in Table XIX | | | Fe1Rb1S2 |
| 15.4.36 | 202381 | 58 | - | - | - - | - - | See details in Table XIX | | | Fe1Rb1S2 |
| 15.4.37 | 9671 | 106 | - | - | - - | - - | See details in Table XIX | | | Fe1Na1O6Si2 |
| 15.4.38 | 180313 | 106 | - | - | - - | - - | See details in Table XIX | | | Fe1Na1O6Si2 |
| 15.4.39 | 180312 | 106 | - | - | - - | - - | See details in Table XIX | | | Fe1Na1O6Si2 |
| 15.4.40 | 180311 | 106 | - | - | - - | - - | See details in Table XIX | | | Fe1Na1O6Si2 |
| 15.4.41 | 161982 | 106 | - | - | - - | - - | See details in Table XIX | | | Fe1Na1O6Si2 |
| 15.4.42 | 161980 | 106 | - | - | - - | - - | See details in Table XIX | | | Fe1Na1O6Si2 |
| 15.4.43 | 161979 | 106 | - | - | - - | - - | See details in Table XIX | | | Fe1Na1O6Si2 |
| 15.4.44 | 161978 | 106 | - | - | - - | - - | See details in Table XIX | | | Fe1Na1O6Si2 |
| 15.4.45 | 161977 | 106 | - | - | - - | - - | See details in Table XIX | | | Fe1Na1O6Si2 |
| 15.4.46 | 161976 | 106 | - | - | - - | - - | See details in Table XIX | | | Fe1Na1O6Si2 |
| 15.4.47 | 161975 | 106 | - | - | - - | - - | See details in Table XIX | | | Fe1Na1O6Si2 |
| 15.4.48 | 158082 | 106 | - | - | - - | - - | See details in Table XIX | | | Fe1Na1O6Si2 |
| 15.4.49 | 157733 | 106 | - | - | - - | - - | See details in Table XIX | | | Fe1Na1O6Si2 |
| 15.4.50 | 156561 | 106 | - | - | - - | - - | See details in Table XIX | | | Fe1Na1O6Si2 |
| 15.4.51 | 156560 | 106 | - | - | - - | - - | See details in Table XIX | | | Fe1Na1O6Si2 |
| 15.4.52 | 10219 | 106 | - | - | - - | - - | See details in Table XIX | | | Fe1Na1O6Si2 |
| 15.4.53 | 262314 | 138 | - | - | - - | - - | See details in Table XIX | | | Fe1Li1O8W2 |
| 15.4.54 | 9669 | 106 | - | - | - - | - - | See details in Table XIX | | | Fe1Li1O6Si2 |
| 15.4.55 | 158582 | 106 | - | - | - - | - - | See details in Table XIX | | | Fe1Li1O6Si2 |
| 15.4.56 | 158581 | 106 | - | - | - - | - - | See details in Table XIX | | | Fe1Li1O6Si2 |
| 15.4.57 | 40780 | 58 | - | - | - - | - - | See details in Table XIX | | | Fe1K1Se2 |
| 15.4.58 | 68383 | 58 | - | - | - - | - - | See details in Table XIX | | | Fe1K1S2 |
| 15.4.59 | 202383 | 58 | - | - | - - | - - | See details in Table XIX | | | Fe1K1S2 |
| 15.4.60 | 202382 | 58 | - | - | - - | - - | See details in Table XIX | | | Fe1K1S2 |
| 15.4.61 | 202380 | 58 | - | - | - - | - - | See details in Table XIX | | | Fe1K1S2 |
| 15.4.62 | 167454 | 58 | - | - | - - | - - | See details in Table XIX | | | Fe1K1S2 |
| 15.4.63 | 167453 | 58 | - | - | - - | - - | See details in Table XIX | | | Fe1K1S2 |
| 15.4.64 | 15557 | 58 | - | - | - - | - - | See details in Table XIX | | | Fe1K1S2 |
| 15.4.65 | 180310 | 106 | - | - | - - | - - | See details in Table XIX | | | Fe1Ge2Na1O6 |
| 15.4.66 | 180308 | 106 | - | - | - - | - - | See details in Table XIX | | | Fe1Ge2Na1O6 |
| 15.4.67 | 180307 | 106 | - | - | - - | - - | See details in Table XIX | | | Fe1Ge2Na1O6 |
| 15.4.68 | 180306 | 106 | - | - | - - | - - | See details in Table XIX | | | Fe1Ge2Na1O6 |
| 15.4.69 | 169480 | 106 | - | - | - - | - - | See details in Table XIX | | | Fe1Ge2Li1O6 |
| 15.4.70 | 169479 | 106 | - | - | - - | - - | See details in Table XIX | | | Fe1Ge2Li1O6 |
| 15.4.71 | 162951 | 106 | - | - | - - | - - | See details in Table XIX | | | Fe1Ge2Li1O6 |
| 15.4.72 | 162950 | 106 | - | - | - - | - - | See details in Table XIX | | | Fe1Ge2Li1O6 |
| 15.4.73 | 162949 | 106 | - | - | - - | - - | See details in Table XIX | | | Fe1Ge2Li1O6 |
| 15.4.74 | 162948 | 106 | - | - | - - | - - | See details in Table XIX | | | Fe1Ge2Li1O6 |
| 15.4.75 | 251694 | 118 | - | - | - - | - - | See details in Table XIX | | | F3Nb2O2 |
| 15.4.76 | 262447 | 154 | - | - | - - | - - | See details in Table XIX | | | Eu3Ga2P4 |
| 15.4.77 | 106624 | 114 | - | - | - - | - - | See details in Table XIX | | | Eu3Ga2 |
| 15.4.78 | 423578 | 114 | - | - | - - | - - | See details in Table XIX | | | Eu2In1Pt2 |
| 15.4.79 | 423577 | 114 | - | - | - - | - - | See details in Table XIX | | | Eu2In1Pd2 |
| 15.4.80 | 153364 | 102 | - | - | - - | - - | See details in Table XIX | | | Er1O4Ta1 |
| 15.4.81 | 109189 | 102 | - | - | - - | - - | See details in Table XIX | | | Er1O4Ta1 |
| 15.4.82 | 109304 | 118 | - | - | - - | - - | See details in Table XIX | | | Er1Nb1O4 |

Continued on next page

**TABLE XIV– continued**

| TQC | ICSD | NoE | DG | IndG | SubSG cl (w/ T) | SubSG cl (w/o T) | Max. Sub. SG yes or no | Max. Sub. SG yes or no | SIs (w/ T) | Stoich. Formula |
|---|---|---|---|---|---|---|---|---|---|---|
| 15.4.83 | 109180 | 118 | - | - | - - | - - | See details in Table XIX | | | Er1Nb1O4 |
| 15.4.84 | 157832 | 182 | - | - | - - | - - | See details in Table XIX | | | Er1K1O8W2 |
| 15.4.85 | 109188 | 98 | - | - | - - | - - | See details in Table XIX | | | Dy1O4Ta1 |
| 15.4.86 | 109305 | 110 | - | - | - - | - - | See details in Table XIX | | | Dy1Nb1O4 |
| 15.4.87 | 109179 | 114 | - | - | - - | - - | See details in Table XIX | | | Dy1Nb1O4 |
| 15.4.88 | 202587 | 166 | - | - | - - | - - | See details in Table XIX | | | Cu3F7Na1 |
| 15.4.89 | 83493 | 114 | - | - | - - | - - | See details in Table XIX | | | Cu2O4Y1 |
| 15.4.90 | 83486 | 114 | - | - | - - | - - | See details in Table XIX | | | Cu2La1O4 |
| 15.4.91 | 83489 | 126 | - | - | - - | - - | See details in Table XIX | | | Cu2Eu1O4 |
| 15.4.92 | 72721 | 174 | - | - | - - | - - | See details in Table XIX | | | Cu1O6Pt1Sr3 |
| 15.4.93 | 66347 | 174 | - | - | - - | - - | See details in Table XIX | | | Cu1O6Pt1Sr3 |
| 15.4.94 | 603 | 106 | - | - | - - | - - | See details in Table XIX | | | Cu1O5Se2 |
| 15.4.95 | 245059 | 106 | - | - | - - | - - | See details in Table XIX | | | Cu1O5Se2 |
| 15.4.96 | 69094 | 34 | - | - | - - | - - | See details in Table XIX | | | Cu1O1 |
| 15.4.97 | 67850 | 34 | - | - | - - | - - | See details in Table XIX | | | Cu1O1 |
| 15.4.98 | 291393 | 34 | - | - | - - | - - | See details in Table XIX | | | Cu1O1 |
| 15.4.99 | 291390 | 34 | - | - | - - | - - | See details in Table XIX | | | Cu1O1 |
| 15.4.100 | 291387 | 34 | - | - | - - | - - | See details in Table XIX | | | Cu1O1 |
| 15.4.101 | 160630 | 34 | - | - | - - | - - | See details in Table XIX | | | Cu1O1 |
| 15.4.102 | 16025 | 34 | - | - | - - | - - | See details in Table XIX | | | Cu1O1 |
| 15.4.103 | 80418 | 130 | - | - | - - | - - | See details in Table XIX | | | Cu1Na2O7P2 |
| 15.4.104 | 79868 | 130 | - | - | - - | - - | See details in Table XIX | | | Cu1Na2O7P2 |
| 15.4.105 | 72485 | 130 | - | - | - - | - - | See details in Table XIX | | | Cu1Li2O7P2 |
| 15.4.106 | 160321 | 130 | - | - | - - | - - | See details in Table XIX | | | Cu1Li2O7P2 |
| 15.4.107 | 262273 | 98 | - | - | - - | - - | See details in Table XIX | | | Cu1F1Na1O4S1 |
| 15.4.108 | 97262 | 102 | - | - | - - | - - | See details in Table XIX | | | Cr1Na1O6Si2 |
| 15.4.109 | 97258 | 102 | - | - | - - | - - | See details in Table XIX | | | Cr1Na1O6Si2 |
| 15.4.110 | 97257 | 102 | - | - | - - | - - | See details in Table XIX | | | Cr1Na1O6Si2 |
| 15.4.111 | 9670 | 102 | - | - | - - | - - | See details in Table XIX | | | Cr1Na1O6Si2 |
| 15.4.112 | 192422 | 102 | - | - | - - | - - | See details in Table XIX | | | Cr1Na1O6Si2 |
| 15.4.113 | 192420 | 102 | - | - | - - | - - | See details in Table XIX | | | Cr1Na1O6Si2 |
| 15.4.114 | 192419 | 102 | - | - | - - | - - | See details in Table XIX | | | Cr1Na1O6Si2 |
| 15.4.115 | 192418 | 102 | - | - | - - | - - | See details in Table XIX | | | Cr1Na1O6Si2 |
| 15.4.116 | 10240 | 102 | - | - | - - | - - | See details in Table XIX | | | Cr1Na1O6Si2 |
| 15.4.117 | 10239 | 102 | - | - | - - | - - | See details in Table XIX | | | Cr1Na1O6Si2 |
| 15.4.118 | 152072 | 102 | - | - | - - | - - | See details in Table XIX | | | Cr1Li1O6Si2 |
| 15.4.119 | 152071 | 102 | - | - | - - | - - | See details in Table XIX | | | Cr1Li1O6Si2 |
| 15.4.120 | 260079 | 102 | - | - | - - | - - | See details in Table XIX | | | Cr1Ge2Na1O6 |
| 15.4.121 | 194813 | 102 | - | - | - - | - - | See details in Table XIX | | | Cr1Ge2Li1O6 |
| 15.4.122 | 194812 | 102 | - | - | - - | - - | See details in Table XIX | | | Cr1Ge2Li1O6 |
| 15.4.123 | 174432 | 102 | - | - | - - | - - | See details in Table XIX | | | Cr1Ge2Li1O6 |
| 15.4.124 | 414127 | 122 | - | - | - - | - - | See details in Table XIX | | | Co2Na7O6 |
| 15.4.125 | 157364 | 90 | - | - | - - | - - | See details in Table XIX | | | Co2Ga7Mg3 |
| 15.4.126 | 290052 | 94 | - | - | - - | - - | See details in Table XIX | | | Co1F1Na1O4S1 |
| 15.4.127 | 262275 | 94 | - | - | - - | - - | See details in Table XIX | | | Co1F1Na1O4S1 |
| 15.4.128 | 1242 | 114 | - | - | - - | - - | See details in Table XIX | | | Cl6Cs1W1 |
| 15.4.129 | 418203 | 54 | - | - | - - | - - | See details in Table XIX | | | Ca2In1Pt2 |
| 15.4.130 | 418202 | 54 | - | - | - - | - - | See details in Table XIX | | | Ca2In1Pd2 |
| 15.4.131 | 429177 | 54 | - | - | - - | - - | See details in Table XIX | | | Ca2Ga1Pt2 |
| 15.4.132 | 429176 | 54 | - | - | - - | - - | See details in Table XIX | | | Ca2Ga1Pd2 |
| 15.4.133 | 170824 | 114 | - | - | - - | - - | See details in Table XIX | | | Ca1Cu1Ge2O6 |
| 15.4.134 | 170823 | 114 | - | - | - - | - - | See details in Table XIX | | | Ca1Cu1Ge2O6 |
| 15.4.135 | 170822 | 114 | - | - | - - | - - | See details in Table XIX | | | Ca1Cu1Ge2O6 |
| 15.4.136 | 170821 | 114 | - | - | - - | - - | See details in Table XIX | | | Ca1Cu1Ge2O6 |
| 15.4.137 | 82212 | 126 | - | - | - - | - - | See details in Table XIX | | | Ca1Co1O6Si2 |
| 15.4.138 | 246209 | 110 | - | - | - - | - - | See details in Table XIX | | | Ca1Co1O6Si2 |
| 15.4.139 | 415406 | 102 | - | - | - - | - - | See details in Table XIX | | | C4Cu1O4 |

Continued on next page

**TABLE XIV– continued**

| TQC | ICSD | NoE | DG | IndG | SubSG (w/ T) | cl | SubSG (w/o T) | cl | Max. Sub. SG | yes or no | Max. Sub. SG | yes or no | SIs (w/ T) | Stoich. Formula |
|---|---|---|---|---|---|---|---|---|---|---|---|---|---|---|
| 15.4.140 | 174608 | 166 | - | - | - | - | - | - | See details in Table XIX | | | | | C4Cu1Hg1N4S4 |
| 15.4.141 | 240902 | 110 | - | - | - | - | - | - | See details in Table XIX | | | | | C2H4Mn1O6 |
| 15.4.142 | 407226 | 142 | - | - | - | - | - | - | See details in Table XIX | | | | | C2Gd1K1O6 |
| 15.4.143 | 380495 | 50 | - | - | - | - | - | - | See details in Table XIX | | | | | C2Eu1 |
| 15.4.144 | 419171 | 74 | - | - | - | - | - | - | See details in Table XIX | | | | | C2Cl1La2 |
| 15.4.145 | 109643 | 34 | - | - | - | - | - | - | See details in Table XIX | | | | | C1H1Na1O2 |
| 15.4.146 | 240017 | 74 | - | - | - | - | - | - | See details in Table XIX | | | | | Bi2Cs3 |
| 15.4.147 | 181425 | 110 | - | - | - | - | - | - | See details in Table XIX | | | | | Bi1O6Re2 |
| 15.4.148 | 246363 | 206 | - | - | - | - | - | - | See details in Table XIX | | | | | Ba3Ir2Na1O9 |
| 15.4.149 | 246362 | 206 | - | - | - | - | - | - | See details in Table XIX | | | | | Ba3Ir2Na1O9 |
| 15.4.150 | 246361 | 206 | - | - | - | - | - | - | See details in Table XIX | | | | | Ba3Ir2Na1O9 |
| 15.4.151 | 72448 | 210 | - | - | - | - | - | - | See details in Table XIX | | | | | Ba3Bi1O9Ru2 |
| 15.4.152 | 174292 | 214 | - | - | - | - | - | - | See details in Table XIX | | | | | Ba3Bi1Ir2O9 |
| 15.4.153 | 20905 | 162 | - | - | - | - | - | - | See details in Table XIX | | | | | Ba2Cu1O7Si2 |
| 15.4.154 | 173850 | 222 | - | - | - | - | - | - | See details in Table XIX | | | | | Ba1Ir1O3 |
| 15.4.155 | 202388 | 138 | - | - | - | - | - | - | See details in Table XIX | | | | | Ba1Cu1O6Se2 |
| 15.4.156 | 407832 | 126 | - | - | - | - | - | - | See details in Table XIX | | | | | Ba1Cl5Gd1 |
| 15.4.157 | 24308 | 98 | - | - | - | - | - | - | See details in Table XIX | | | | | B3Ni4 |
| 15.4.158 | 187403 | 98 | - | - | - | - | - | - | See details in Table XIX | | | | | B3Ni4 |
| 15.4.159 | 150561 | 98 | - | - | - | - | - | - | See details in Table XIX | | | | | B3Ni4 |
| 15.4.160 | 39487 | 94 | - | - | - | - | - | - | See details in Table XIX | | | | | B2Ho1Ni2 |
| 15.4.161 | 2125 | 158 | - | - | - | - | - | - | See details in Table XIX | | | | | Au6Pr1 |
| 15.4.162 | 423784 | 154 | - | - | - | - | - | - | See details in Table XIX | | | | | As4Eu3Ga2 |
| 15.4.163 | 415951 | 134 | - | - | - | - | - | - | See details in Table XIX | | | | | Al2Cl8V1 |
| 15.4.164 | 95801 | 94 | - | - | - | - | - | - | See details in Table XIX | | | | | Al1Gd2Ge2 |
| 15.4.165 | 412545 | 94 | - | - | - | - | - | - | See details in Table XIX | | | | | Al1Co2Pr2 |
| 15.4.166 | 428674 | 54 | - | - | - | - | - | - | See details in Table XIX | | | | | Al1Ca2Pt2 |
| 15.4.167 | 428673 | 54 | - | - | - | - | - | - | See details in Table XIX | | | | | Al1Ca2Pd2 |
| 15.4.168 | 416931 | 194 | - | - | - | - | - | - | See details in Table XIX | | | | | Ag4Cu1O6Te1 |
| 15.4.169 | 158336 | 102 | - | - | - | - | - | - | See details in Table XIX | | | | | Ag2Cu2O3 |
| 19.3.1 | 36301 | 164 | - | - | - | - | - | - | 4 | yes | 4 | yes | - | H7N1O6 |
| | | | | | | | | | 4 | yes | 4 | yes | - | |
| 19.3.2 | 27116 | 164 | - | - | - | - | - | - | 4 | yes | 4 | yes | - | H7N1O6 |
| | | | | | | | | | 4 | yes | 4 | yes | - | |
| 19.3.3 | 24674 | 164 | - | - | - | - | - | - | 4 | yes | 4 | yes | - | H7N1O6 |
| | | | | | | | | | 4 | yes | 4 | yes | - | |
| 19.3.4 | 38127 | 44 | - | - | - | - | - | - | 4 | yes | 4 | yes | - | H3N1O1 |
| | | | | | | | | | 4 | yes | 4 | yes | - | |
| 19.3.5 | 88000 | 164 | - | - | - | - | - | - | 4 | yes | 4 | yes | - | H1Na1O4Si1Zn1 |
| | | | | | | | | | 4 | yes | 4 | yes | - | |
| 19.3.6 | 26621 | 100 | - | - | - | - | - | - | 4 | yes | 4 | yes | - | H1I1O3 |
| | | | | | | | | | 4 | yes | 4 | yes | - | |
| 19.3.7 | 9321 | 100 | - | - | - | - | - | - | 4 | yes | 4 | yes | - | H1F1Hg1O1 |
| | | | | | | | | | 4 | yes | 4 | yes | - | |
| 19.3.8 | 27754 | 100 | - | - | - | - | - | - | 4 | yes | 4 | yes | - | H1Cd1F1O1 |
| | | | | | | | | | 4 | yes | 4 | yes | - | |
| 19.3.9 | 187197 | 44 | - | - | - | - | - | - | 4 | yes | 4 | yes | - | Ge1Mn1 |
| | | | | | | | | | 4 | yes | 4 | yes | - | |
| 19.3.10 | 187196 | 44 | - | - | - | - | - | - | 4 | yes | 4 | yes | - | Ge1Mn1 |
| | | | | | | | | | 4 | yes | 4 | yes | - | |
| 19.3.11 | 35451 | 164 | - | - | - | - | - | - | 4 | yes | 4 | yes | - | Cu1Na1O4P1 |
| | | | | | | | | | 4 | yes | 4 | yes | - | |
| 19.3.12 | 156291 | 180 | - | - | - | - | - | - | 4 | yes | 4 | yes | - | Cu1Er1S2 |
| | | | | | | | | | 4 | yes | 4 | yes | - | |
| 19.3.13 | 425075 | 196 | - | - | - | - | - | - | 4 | yes | 4 | yes | - | Ce2Ga3Ru2 |
| | | | | | | | | | 4 | yes | 4 | yes | - | |
| 19.3.14 | 951 | 180 | - | - | - | - | - | - | 4 | yes | 4 | yes | - | Ag1Er1Se2 |

Continued on next page

**TABLE XIV– continued**

| TQC | ICSD | NoE | DG | IndG | SubSG cl (w/ T) | SubSG cl (w/o T) | Max. Sub. SG | yes or no | Max. Sub. SG | yes or no | SIs (w/ T) | Stoich. Formula |
|---|---|---|---|---|---|---|---|---|---|---|---|---|
| | | | | | | | 4 | yes | 4 | yes | - | |
| 19.3.15 | 15302 | 180 | - | - | - - | - - | 4 | yes | 4 | yes | - | Ag1Er1Se2 |
| | | | | | | | 4 | yes | 4 | yes | - | |
| 20.3.1 | 63230 | 66 | - | - | - - | - - | 5 | yes | 5 | yes | - | Ba1S3V1 |
| | | | | | | | 5 | yes | 5 | yes | - | |
| 26.4.1 | 81039 | 166 | - | - | - - | - - | See details in Table XIX | | | | | Cu1Fe1O6Pb2Sr2 |
| 26.4.2 | 75401 | 98 | - | - | - - | - - | See details in Table XIX | | | | | C1Cu1F1K1O3 |
| 26.4.3 | 65082 | 194 | - | - | - - | - - | See details in Table XIX | | | | | Bi1Cu4Fe1Pb1S6 |
| 29.3.1 | 21071 | 180 | - | - | - - | - - | 7 | yes | 7 | no(7) | - | F4Fe1Rb1 |
| | | | | | | | 4 | yes | 7 | yes | - | |
| | | | | | | | - | - | 4 | yes | - | |
| 31.4.1 | 88640 | 82 | - | - | - - | - - | See details in Table XIX | | | | | Li1O5V2 |
| 31.4.2 | 250189 | 114 | - | - | - - | - - | See details in Table XIX | | | | | H6Cu1N1O5P1 |
| 31.4.3 | 81286 | 106 | - | - | - - | - - | See details in Table XIX | | | | | H2K1Mn1O5P1 |
| 31.4.4 | 71178 | 106 | - | - | - - | - - | See details in Table XIX | | | | | H2K1Mn1O5P1 |
| 31.4.5 | 71177 | 106 | - | - | - - | - - | See details in Table XIX | | | | | H2K1Mn1O5P1 |
| 31.4.6 | 87325 | 42 | - | - | - - | - - | See details in Table XIX | | | | | H1Fe1O2 |
| 31.4.7 | 420002 | 42 | - | - | - - | - - | See details in Table XIX | | | | | H1Fe1O2 |
| 31.4.8 | 2257 | 38 | - | - | - - | - - | See details in Table XIX | | | | | H1Cr1O2 |
| 33.3.1 | 23670 | 84 | - | - | - - | - - | 7 | yes | 7 | yes | - | Mn1N2Si1 |
| | | | | | | | 7 | yes | 7 | yes | - | |
| 33.3.2 | 23527 | 84 | - | - | - - | - - | 7 | yes | 7 | yes | - | Mn1N2Si1 |
| | | | | | | | 7 | yes | 7 | yes | - | |
| 33.3.3 | 30619 | 116 | - | - | - - | - - | 7 | yes | 7 | no(7) | - | H3N1O4 |
| | | | | | | | 4 | yes | 7 | yes | - | |
| | | | | | | | - | - | 4 | yes | - | |
| 33.3.4 | 202718 | 132 | - | - | - - | - - | 7 | yes | 7 | no(7) | - | H1Cs1O3Se1 |
| | | | | | | | 4 | yes | 7 | yes | - | |
| | | | | | | | - | - | 4 | yes | - | |
| 33.3.5 | 23528 | 84 | - | - | - - | - - | 7 | yes | 7 | yes | - | Ge1Mn1N2 |
| | | | | | | | 7 | yes | 7 | yes | - | |
| 33.3.6 | 15057 | 52 | - | - | - - | - - | 7 | yes | 7 | no(7) | - | Fe1P1 |
| | | | | | | | 4 | yes | 7 | yes | - | |
| | | | | | | | - | - | 4 | yes | - | |
| 33.3.7 | 23822 | 148 | - | - | - - | - - | 7 | yes | 7 | no(7) | - | Fe1O3Y1 |
| | | | | | | | 4 | yes | 7 | yes | - | |
| | | | | | | | - | - | 4 | yes | - | |
| 33.3.8 | 68984 | 84 | - | - | - - | - - | 7 | yes | 7 | no(7) | - | Fe1Na1O2 |
| | | | | | | | 4 | yes | 7 | yes | - | |
| | | | | | | | - | - | 4 | yes | - | |
| 33.3.9 | 33772 | 84 | - | - | - - | - - | 7 | yes | 7 | no(7) | - | Fe1Na1O2 |
| | | | | | | | 4 | yes | 7 | yes | - | |
| | | | | | | | - | - | 4 | yes | - | |
| 33.3.10 | 33771 | 84 | - | - | - - | - - | 7 | yes | 7 | no(7) | - | Fe1Na1O2 |
| | | | | | | | 4 | yes | 7 | yes | - | |
| | | | | | | | - | - | 4 | yes | - | |
| 33.3.11 | 33770 | 84 | - | - | - - | - - | 7 | yes | 7 | no(7) | - | Fe1Na1O2 |
| | | | | | | | 4 | yes | 7 | yes | - | |
| | | | | | | | - | - | 4 | yes | - | |
| 33.3.12 | 33767 | 84 | - | - | - - | - - | 7 | yes | 7 | no(7) | - | Fe1Na1O2 |
| | | | | | | | 4 | yes | 7 | yes | - | |
| | | | | | | | - | - | 4 | yes | - | |
| 33.3.13 | 33766 | 84 | - | - | - - | - - | 7 | yes | 7 | no(7) | - | Fe1Na1O2 |
| | | | | | | | 4 | yes | 7 | yes | - | |
| | | | | | | | - | - | 4 | yes | - | |
| 33.3.14 | 33765 | 84 | - | - | - - | - - | 7 | yes | 7 | no(7) | - | Fe1Na1O2 |
| | | | | | | | 4 | yes | 7 | yes | - | |
| | | | | | | | - | - | 4 | yes | - | |

Continued on next page

**TABLE XIV– continued**

| TQC | ICSD | NoE | DG | IndG | SubSG (w/ T) | cl | SubSG (w/o T) | cl | Max. Sub. SG | yes or no | Max. Sub. SG | yes or no | SIs (w/ T) | Stoich. Formula |
|---|---|---|---|---|---|---|---|---|---|---|---|---|---|---|
| 33.3.15 | 33764 | 84 | - | - | - | - | - | - | 7 | yes | 7 | no(7) | - | Fe1Na1O2 |
| | | | | | | | | | 4 | yes | 7 | yes | - | |
| | | | | | | | | | - | - | 4 | yes | - | |
| 33.3.16 | 33763 | 84 | - | - | - | - | - | - | 7 | yes | 7 | no(7) | - | Fe1Na1O2 |
| | | | | | | | | | 4 | yes | 7 | yes | - | |
| | | | | | | | | | - | - | 4 | yes | - | |
| 33.3.17 | 27804 | 84 | - | - | - | - | - | - | 7 | yes | 7 | no(7) | - | Fe1Na1O2 |
| | | | | | | | | | 4 | yes | 7 | yes | - | |
| | | | | | | | | | - | - | 4 | yes | - | |
| 33.3.18 | 186309 | 84 | - | - | - | - | - | - | 7 | yes | 7 | no(7) | - | Fe1Na1O2 |
| | | | | | | | | | 4 | yes | 7 | yes | - | |
| | | | | | | | | | - | - | 4 | yes | - | |
| 33.3.19 | 158330 | 84 | - | - | - | - | - | - | 7 | yes | 7 | no(7) | - | Fe1Na1O2 |
| | | | | | | | | | 4 | yes | 7 | yes | - | |
| | | | | | | | | | - | - | 4 | yes | - | |
| 33.3.20 | 240332 | 196 | - | - | - | - | - | - | 7 | yes | 7 | yes | - | F5O1Os1 |
| | | | | | | | | | 7 | yes | 7 | yes | - | |
| 33.3.21 | 89904 | 196 | - | - | - | - | - | - | 7 | yes | 7 | yes | - | Cu1O4P1Rb1 |
| | | | | | | | | | 7 | yes | 7 | yes | - | |
| 36.4.1 | 105935 | 138 | - | - | - | - | - | - | See details in Table XIX | | | | | Rh3Sn5Y2 |
| 36.4.2 | 1182 | 110 | - | - | - | - | - | - | See details in Table XIX | | | | | O3Rb2Tb1 |
| 36.4.3 | 188064 | 90 | - | - | - | - | - | - | See details in Table XIX | | | | | Ni1P1 |
| 36.4.4 | 83613 | 102 | - | - | - | - | - | - | See details in Table XIX | | | | | H5Cl3K1N1O1Pt1 |
| 36.4.5 | 68459 | 50 | - | - | - | - | - | - | See details in Table XIX | | | | | H2Cu1O2 |
| 36.4.6 | 1545 | 42 | - | - | - | - | - | - | See details in Table XIX | | | | | H1Fe1O2 |
| 36.4.7 | 159032 | 58 | - | - | - | - | - | - | See details in Table XIX | | | | | Ca1Ir1O3 |
| 36.4.8 | 184382 | 142 | - | - | - | - | - | - | See details in Table XIX | | | | | Bi3Mn2O7 |
| 36.4.9 | 182795 | 142 | - | - | - | - | - | - | See details in Table XIX | | | | | Bi3Mn2O7 |
| 36.4.10 | 63229 | 66 | - | - | - | - | - | - | See details in Table XIX | | | | | Ba1S3V1 |
| 36.4.11 | 52692 | 66 | - | - | - | - | - | - | See details in Table XIX | | | | | Ba1S3V1 |
| 36.4.12 | 154182 | 66 | - | - | - | - | - | - | See details in Table XIX | | | | | Ba1S3V1 |
| 36.4.13 | 73252 | 110 | - | - | - | - | - | - | See details in Table XIX | | | | | As4Mn5 |
| 36.4.14 | 106267 | 86 | - | - | - | - | - | - | See details in Table XIX | | | | | As1Pb2Pd3 |
| 36.4.15 | 66120 | 90 | - | - | - | - | - | - | See details in Table XIX | | | | | As1Ni1 |
| 40.3.1 | 280199 | 94 | - | - | - | - | - | - | 6 | yes | 6 | yes | - | Eu1Li1P1Se4 |
| | | | | | | | | | 5 | yes | 5 | yes | - | |
| 40.3.2 | 419600 | 134 | - | - | - | - | - | - | 6 | yes | 6 | yes | - | Cu2Eu1S4Sn1 |
| | | | | | | | | | 5 | yes | 5 | yes | - | |
| 40.3.3 | 262583 | 94 | - | - | - | - | - | - | 6 | yes | 6 | yes | - | As1Eu1Na1S4 |
| | | | | | | | | | 5 | yes | 5 | yes | - | |
| 43.4.1 | 20612 | 106 | - | - | - | - | - | - | See details in Table XIX | | | | | Ge5Y3 |
| 43.4.2 | 200779 | 146 | - | - | - | - | - | - | See details in Table XIX | | | | | C2Cu1K2O6 |
| 43.4.3 | 200695 | 146 | - | - | - | - | - | - | See details in Table XIX | | | | | C2Cu1K2O6 |
| 43.4.4 | 98667 | 190 | - | - | - | - | - | - | See details in Table XIX | | | | | Ba5P9 |
| 47.2.1 | 86427 | 64 | 0.0078 | -0.6638 | 2 | 6 | 2 | 7 | - | - | - | - | z4=3 | Cu1O5Sr2Tl1 |
| 47.2.2 | 390001 | 206 | 0.0061 | -1.1607 | 2 | 6 | 2 | 7 | - | - | - | - | z4=3 | Ba4Cu6O13Y2 |
| 47.2.3 | 77745 | 106 | 0.0134 | -2.1218 | 2 | 11 | 2 | 9 | - | - | - | - | z2w,1=1<br>z2w,3=1<br>z4=1 | Ba2Cu3O7Y1 |
| 47.2.4 | 63335 | 106 | 0.0047 | -2.3705 | 2 | 1 | 10 | 1 | - | - | - | - | z2w,2=1<br>z4=2 | Ba2Cu3O7Y1 |
| 47.2.5 | 262809 | 106 | 0.0013 | -2.1786 | 2 | 7 | 10,10,10 | 8 | - | - | - | - | z2w,1=1<br>z2w,2=1<br>z2w,3=1 | Ba2Cu3O6.93Y1 |
| 47.2.6 | 73896 | 206 | 0.0392 | -1.1574 | 2 | 6 | 2 | 7 | - | - | - | - | z4=3 | Ba2Cu3O6.5Y1 |
| 47.2.7 | 68044 | 116 | 0.0000 | -0.0778 | 2 | 5 | 10,10 | 6 | - | - | - | - | z2w,1=1<br>z2w,3=1 | Ba2Cu3Ho1O7 |

Continued on next page

**TABLE XIV– continued**

| TQC | ICSD | NoE | DG | IndG | SubSG (w/ T) | cl | SubSG (w/o T) | cl | Max. Sub. SG | yes or no | Max. Sub. SG | yes or no | SIs (w/ T) | Stoich. Formula |
|---|---|---|---|---|---|---|---|---|---|---|---|---|---|---|
| 47.2.8 | 94046 | 160 | 0.0030 | -0.1449 | 2 | 9 | 2 | 5 | - | - | - | - | z2w,2=1<br>z4=1 | Ba2Co4O11Tb2 |
| 47.2.9 | 94045 | 160 | 0.0000 | -0.1218 | 2 | 9 | 2 | 5 | - | - | - | - | z2w,2=1<br>z4=1 | Ba2Co4O11Tb2 |
| 47.2.10 | 96380 | 150 | 0.0217 | -0.0546 | 2 | 4 | 10,10 | 4 | - | - | - | - | z2w,1=1<br>z2w,2=1<br>z4=2 | Ba1Co2Nd1O5.5 |
| 47.2.11 | 96377 | 150 | 0.0124 | -0.0663 | 2 | 3 | 10,10 | 3 | - | - | - | - | z2w,2=1<br>z2w,3=1 | Ba1Co2Nd1O5.5 |
| 47.2.12 | 187120 | 150 | 0.0250 | -0.0666 | 2 | 2 | 10,10 | 2 | - | - | - | - | z2w,2=1<br>z2w,3=1<br>z4=2 | Ba1Co2Nd1O5.5 |
| 47.2.13 | 71640 | 114 | 0.0172 | -1.6667 | 2 | 1 | 10 | 1 | - | - | - | - | z2w,2=1<br>z4=2 | B3Ir4Zn1 |
| 51.2.1 | 248052 | 208 | 0.0021 | -0.0944 | 2 | 3 | 2 | 2 | - | - | - | - | z4=3 | La2Se9U2 |
| 51.2.2 | 106800 | 68 | 0.0123 | -0.8939 | 2 | 5 | 0 | 4 | - | - | - | - | z2w,3=1<br>z4=2 | In1La1Ni2 |
| 51.2.3 | 370040 | 80 | 0.0000 | -0.0488 | 2 | 13 | 10 | 3 | - | - | - | - | z2w,2=1<br>z4=2 | Co2In1Tb1 |
| 51.2.4 | 421074 | 84 | 0.0027 | -1.1411 | 2 | 1 | 2 | 1 | - | - | - | - | z2w,2=1<br>z4=3 | Ca2Ge2Ni3 |
| 51.2.5 | 194877 | 96 | 0.0073 | -1.2477 | 2 | 4 | 10 | 3 | - | - | - | - | z2w,2=1<br>z2w,3=1<br>z4=2 | C3Al1Y3 |
| 51.2.6 | 69767 | 48 | 0.0181 | -0.3916 | 2 | 9 | 2 | 1 | - | - | - | - | z2w,2=1<br>z2w,3=1<br>z4=1 | C1B2U1 |
| 51.2.7 | 247569 | 148 | 0.0000 | -0.1004 | 2 | 3 | 2 | 2 | - | - | - | - | z4=3 | Ba1Fe2Gd1O5 |
| 51.2.8 | 96391 | 148 | 0.0240 | -0.0246 | 2 | 6 | 10 | 3 | - | - | - | - | z2w,2=1<br>z2w,3=1 | Ba1Fe2Gd1O4.999 |
| 51.2.9 | 99611 | 152 | 0.0000 | -0.0175 | 2 | 8 | 0 | 4 | - | - | - | - | z4=2 | Ba1Dy1Fe2O4.996 |
| 51.2.10 | 107495 | 264 | 0.0353 | -1.3086 | 2 | 7 | 2 | 2 | - | - | - | - | z2w,3=1<br>z4=3 | B3Rh7Zn5 |
| 51.2.11 | 107494 | 264 | 0.0417 | -1.4020 | 2 | 6 | 10 | 3 | - | - | - | - | z2w,2=1<br>z2w,3=1 | B3Ir7Zn5 |
| 51.2.12 | 429116 | 96 | 0.0129 | -1.1860 | 2 | 2 | 2 | 2 | - | - | - | - | z2w,3=1<br>z4=1 | B1Ru1Ta1 |
| 51.2.13 | 426468 | 144 | 0.0022 | -1.0591 | 2 | 2 | 2 | 2 | - | - | - | - | z2w,3=1<br>z4=1 | B1Nb1Ru1 |
| 51.2.14 | 194435 | 144 | 0.0467 | -0.6755 | 2 | 4 | 10 | 3 | - | - | - | - | z2w,2=1<br>z2w,3=1<br>z4=2 | B1Nb1Ru1 |
| 51.2.15 | 249924 | 108 | 0.0918 | -0.5610 | 2 | 3 | 2 | 2 | - | - | - | - | z4=3 | Au3Rb2Tl1 |
| 51.2.16 | 16325 | 60 | 0.0554 | -0.4885 | 2 | 2 | 2 | 2 | - | - | - | - | z2w,3=1<br>z4=1 | Au1I1Te2 |
| 51.2.17 | 9986 | 64 | 0.0062 | -0.5759 | 2 | 1 | 2 | 1 | - | - | - | - | z2w,2=1<br>z4=3 | Al4Co1La1 |
| 51.4.1 | 417836 | 62 | - | - | - | - | - | - | See details in Table XIX | | | | | In4Ir1Sr1 |
| 51.4.2 | 280921 | 98 | - | - | - | - | - | - | See details in Table XIX | | | | | H4Cu1O4P2 |
| 51.4.3 | 89669 | 66 | - | - | - | - | - | - | See details in Table XIX | | | | | Cu1O3Si1 |
| 51.4.4 | 89407 | 66 | - | - | - | - | - | - | See details in Table XIX | | | | | Cu1Ge1O3 |
| 51.4.5 | 89406 | 66 | - | - | - | - | - | - | See details in Table XIX | | | | | Cu1Ge1O3 |
| 51.4.6 | 89405 | 66 | - | - | - | - | - | - | See details in Table XIX | | | | | Cu1Ge1O3 |
| 51.4.7 | 89404 | 66 | - | - | - | - | - | - | See details in Table XIX | | | | | Cu1Ge1O3 |
| 51.4.8 | 89403 | 66 | - | - | - | - | - | - | See details in Table XIX | | | | | Cu1Ge1O3 |
| 51.4.9 | 89402 | 66 | - | - | - | - | - | - | See details in Table XIX | | | | | Cu1Ge1O3 |
| 51.4.10 | 89401 | 66 | - | - | - | - | - | - | See details in Table XIX | | | | | Cu1Ge1O3 |

Continued on next page

**TABLE XIV– continued**

| TQC | ICSD | NoE | DG | IndG | SubSG (w/ T) | cl | SubSG (w/o T) | cl | Max. Sub. SG | yes or no | Max. Sub. SG | yes or no | SIs (w/ T) | Stoich. Formula |
|---|---|---|---|---|---|---|---|---|---|---|---|---|---|---|
| 51.4.11 | 89400 | 66 | - | - | - | - | - | - | See details in Table XIX | | | | | Cu1Ge1O3 |
| 51.4.12 | 56700 | 66 | - | - | - | - | - | - | See details in Table XIX | | | | | Cu1Ge1O3 |
| 51.4.13 | 56699 | 66 | - | - | - | - | - | - | See details in Table XIX | | | | | Cu1Ge1O3 |
| 51.4.14 | 50331 | 66 | - | - | - | - | - | - | See details in Table XIX | | | | | Cu1Ge1O3 |
| 51.4.15 | 50330 | 66 | - | - | - | - | - | - | See details in Table XIX | | | | | Cu1Ge1O3 |
| 51.4.16 | 50329 | 66 | - | - | - | - | - | - | See details in Table XIX | | | | | Cu1Ge1O3 |
| 51.4.17 | 411019 | 66 | - | - | - | - | - | - | See details in Table XIX | | | | | Cu1Ge1O3 |
| 51.4.18 | 411018 | 66 | - | - | - | - | - | - | See details in Table XIX | | | | | Cu1Ge1O3 |
| 51.4.19 | 411017 | 66 | - | - | - | - | - | - | See details in Table XIX | | | | | Cu1Ge1O3 |
| 51.4.20 | 25754 | 66 | - | - | - | - | - | - | See details in Table XIX | | | | | Cu1Ge1O3 |
| 51.4.21 | 62330 | 70 | - | - | - | - | - | - | See details in Table XIX | | | | | Cu1Fe2Ge2 |
| 51.4.22 | 172189 | 114 | - | - | - | - | - | - | See details in Table XIX | | | | | Cs1Te5U1Zr1 |
| 51.4.23 | 79885 | 114 | - | - | - | - | - | - | See details in Table XIX | | | | | Cs1Te5Ti1U1 |
| 51.4.24 | 62856 | 74 | - | - | - | - | - | - | See details in Table XIX | | | | | Co2Cu1Ge2 |
| 51.4.25 | 410891 | 46 | - | - | - | - | - | - | See details in Table XIX | | | | | Ca1In4Rh1 |
| 51.4.26 | 410892 | 46 | - | - | - | - | - | - | See details in Table XIX | | | | | Ca1In4Ir1 |
| 51.4.27 | 173616 | 62 | - | - | - | - | - | - | See details in Table XIX | | | | | Ba1In4Ir1 |
| 51.4.28 | 281204 | 134 | - | - | - | - | - | - | See details in Table XIX | | | | | Ba1Fe2O5Y1 |
| 51.4.29 | 281203 | 134 | - | - | - | - | - | - | See details in Table XIX | | | | | Ba1Fe2O5Y1 |
| 51.4.30 | 154678 | 70 | - | - | - | - | - | - | See details in Table XIX | | | | | Al4Co1Nd1 |
| 52.3.1 | 402064 | 156 | - | - | - | - | - | - | 34 | yes | 34 | yes | - | Br4Fe1Tl1 |
| | | | | | | | | | 30 | yes | 30 | yes | - | |
| | | | | | | | | | 13 | no(2) | 13 | no(2) | z4=3 | |
| | | | | | | | | | 13 | no(2) | 13 | no(2) | - | |
| 52.3.2 | 164987 | 140 | - | - | - | - | - | - | 34 | yes | 34 | yes | - | Bi3Sr2 |
| | | | | | | | | | 30 | yes | 30 | yes | - | |
| | | | | | | | | | 13 | no(2) | 13 | no(2) | z2w,1=1<br>z2w,2=1<br>z4=3 | |
| | | | | | | | | | 13 | no(2) | 13 | no(2) | - | |
| 52.3.3 | 106329 | 140 | - | - | - | - | - | - | 34 | yes | 34 | yes | - | Bi3Sr2 |
| | | | | | | | | | 30 | yes | 30 | yes | - | |
| | | | | | | | | | 13 | no(2) | 13 | no(2) | z2w,1=1<br>z2w,2=1<br>z4=3 | |
| | | | | | | | | | 13 | no(2) | 13 | no(2) | - | |
| 52.3.4 | 410665 | 180 | - | - | - | - | - | - | 34 | yes | 34 | yes | - | Ag2Bi1O3 |
| | | | | | | | | | 30 | yes | 30 | yes | - | |
| | | | | | | | | | 13 | no(2) | 13 | no(2) | z2w,1=1<br>z2w,2=1<br>z4=1 | |
| | | | | | | | | | 13 | no(2) | 13 | no(2) | - | |
| 53.1.1 | 71063 | 108 | 0.0447 | -0.2018 | 2 | 2 | 2 | 1 | - | - | - | - | z2w,2=1<br>z4=1 | Ni2Ta2Te4 |
| 53.1.2 | 71520 | 132 | 0.0083 | -0.3340 | 2 | 1 | 2 | 1 | - | - | - | - | z4=3 | Fe1Nb1Te2 |
| 55.2.1 | 404193 | 96 | 0.0035 | -0.7262 | 2 | 5 | 2 | 1 | - | - | - | - | z2w,3=1<br>z4=1 | Se2Ti9 |
| 55.2.2 | 249665 | 120 | 0.0027 | -0.0599 | 2 | 2 | 2 | 3 | - | - | - | - | z4=1 | Pr2S5Sn1 |
| 55.2.3 | 97282 | 72 | 0.0048 | -0.2628 | 2 | 5 | 2 | 1 | - | - | - | - | z2w,3=1<br>z4=1 | O4Pb3 |
| 55.2.4 | 249666 | 124 | 0.0041 | -0.0099 | 2 | 3 | 2 | 3 | - | - | - | - | z4=3 | Nd2S5Sn1 |
| 55.2.5 | 90183 | 128 | 0.0049 | -0.4960 | 2 | 4 | 10 | 2 | - | - | - | - | z2w,3=1 | Mn2O5Sr2 |
| 55.2.6 | 417818 | 128 | 0.0049 | -0.4849 | 2 | 7 | 10 | 2 | - | - | - | - | z2w,3=1<br>z4=2 | Mn2O5Sr2 |
| 55.2.7 | 417817 | 128 | 0.0056 | -0.4926 | 2 | 7 | 10 | 2 | - | - | - | - | z2w,3=1<br>z4=2 | Mn2O5Sr2 |

Continued on next page

**TABLE XIV– continued**

| TQC | ICSD | NoE | DG | IndG | SubSG (w/ T) | cl | SubSG (w/o T) | cl | Max. Sub. SG | yes or no | Max. Sub. SG | yes or no | SIs (w/ T) | Stoich. Formula |
|---|---|---|---|---|---|---|---|---|---|---|---|---|---|---|
| 55.2.8 | 417816 | 128 | 0.0041 | -0.4992 | 2 | 7 | 10 | 2 | - | - | - | - | z2w,3=1<br>z4=2 | Mn2O5Sr2 |
| 55.2.9 | 417815 | 128 | 0.0060 | -0.4989 | 2 | 7 | 10 | 2 | - | - | - | - | z2w,3=1<br>z4=2 | Mn2O5Sr2 |
| 55.2.10 | 417814 | 128 | 0.0046 | -0.4909 | 2 | 7 | 10 | 2 | - | - | - | - | z2w,3=1<br>z4=2 | Mn2O5Sr2 |
| 55.2.11 | 417813 | 128 | 0.0037 | -0.5083 | 2 | 7 | 10 | 2 | - | - | - | - | z2w,3=1<br>z4=2 | Mn2O5Sr2 |
| 55.2.12 | 417812 | 128 | 0.0046 | -0.5043 | 2 | 7 | 10 | 2 | - | - | - | - | z2w,3=1<br>z4=2 | Mn2O5Sr2 |
| 55.2.13 | 417811 | 128 | 0.0067 | -0.5062 | 2 | 7 | 10 | 2 | - | - | - | - | z2w,3=1<br>z4=2 | Mn2O5Sr2 |
| 55.2.14 | 417810 | 128 | 0.0036 | -0.5140 | 2 | 7 | 10 | 2 | - | - | - | - | z2w,3=1<br>z4=2 | Mn2O5Sr2 |
| 55.2.15 | 417809 | 128 | 0.0000 | -0.5203 | 2 | 7 | 10 | 2 | - | - | - | - | z2w,3=1<br>z4=2 | Mn2O5Sr2 |
| 55.2.16 | 417808 | 128 | 0.0026 | -0.5317 | 2 | 7 | 10 | 2 | - | - | - | - | z2w,3=1<br>z4=2 | Mn2O5Sr2 |
| 55.2.17 | 417807 | 128 | 0.0044 | -0.5251 | 2 | 7 | 10 | 2 | - | - | - | - | z2w,3=1<br>z4=2 | Mn2O5Sr2 |
| 55.2.18 | 417806 | 128 | 0.0023 | -0.5379 | 2 | 7 | 10 | 2 | - | - | - | - | z2w,3=1<br>z4=2 | Mn2O5Sr2 |
| 55.2.19 | 391127 | 100 | 0.0046 | -0.7060 | 2 | 3 | 2 | 3 | - | - | - | - | z4=3 | In4Rh2Zr5 |
| 55.2.20 | 193223 | 132 | 0.0048 | -0.6751 | 2 | 2 | 2 | 3 | - | - | - | - | z4=1 | In1La2Si2 |
| 55.2.21 | 412706 | 100 | 0.0175 | -0.9115 | 2 | 4 | 10 | 2 | - | - | - | - | z2w,3=1 | Hf5In4Rh2 |
| 55.2.22 | 22046 | 84 | 0.0082 | -0.3817 | 2 | 7 | 10 | 2 | - | - | - | - | z2w,3=1<br>z4=2 | Ge1Mn2O4 |
| 55.2.23 | 23226 | 64 | 0.0045 | -1.9998 | 2 | 7 | 10 | 2 | - | - | - | - | z2w,3=1<br>z4=2 | Ga2Mg1 |
| 55.2.24 | 426620 | 164 | 0.0000 | -0.5910 | 2 | 5 | 2 | 1 | - | - | - | - | z2w,3=1<br>z4=1 | F3O1Rb1V1 |
| 55.2.25 | 249403 | 164 | 0.0000 | -0.5910 | 2 | 5 | 2 | 1 | - | - | - | - | z2w,3=1<br>z4=1 | F3O1Rb1V1 |
| 55.2.26 | 82389 | 148 | 0.0270 | -1.3896 | 2 | 2 | 2 | 3 | - | - | - | - | z4=1 | Cu2La2O5 |
| 55.2.27 | 82388 | 148 | 0.0221 | -1.3729 | 2 | 2 | 2 | 3 | - | - | - | - | z4=1 | Cu2La2O5 |
| 55.2.28 | 73399 | 148 | 0.0096 | -1.4041 | 2 | 2 | 2 | 3 | - | - | - | - | z4=1 | Cu2La2O5 |
| 55.2.29 | 82019 | 144 | 0.0721 | -0.5235 | 2 | 3 | 2 | 3 | - | - | - | - | z4=3 | Cu1O2.5Sr1 |
| 55.2.30 | 171414 | 184 | 0.0056 | -0.3481 | 2 | 7 | 10 | 2 | - | - | - | - | z2w,3=1<br>z4=2 | Ce5In4Pt2 |
| 55.2.31 | 2578 | 180 | 0.0000 | -0.0279 | 2 | 2 | 2 | 3 | - | - | - | - | z4=1 | Ce4O4S3 |
| 55.2.32 | 102075 | 312 | 0.0251 | -0.0905 | 2 | 6 | 0 | 4 | - | - | - | - | z4=2 | Cd7Th6 |
| 55.2.33 | 36466 | 172 | 0.0430 | 0.0430 | 2 | 6 | 0 | 4 | - | - | - | - | z4=2 | Ca5Ga2Sb6 |
| 55.2.34 | 35130 | 96 | 0.0000 | -0.6067 | 2 | 4 | 10 | 2 | - | - | - | - | z2w,3=1 | Ca2Mn2O5 |
| 55.2.35 | 35219 | 128 | 0.0067 | -0.6953 | 2 | 2 | 2 | 3 | - | - | - | - | z4=1 | Ca1Mn1O2.5 |
| 55.2.36 | 84869 | 128 | 0.0107 | -1.2818 | 2 | 7 | 10 | 2 | - | - | - | - | z2w,3=1<br>z4=2 | C5Y4 |
| 55.2.37 | 23834 | 68 | 0.0352 | -1.8660 | 2 | 2 | 2 | 3 | - | - | - | - | z4=1 | C2B2Sc1 |
| 55.2.38 | 423924 | 140 | 0.0089 | -1.2447 | 2 | 2 | 2 | 3 | - | - | - | - | z4=1 | C1B2Lu1 |
| 55.2.39 | 412563 | 220 | 0.1178 | -1.3647 | 2 | 4 | 10 | 2 | - | - | - | - | z2w,3=1 | Bi8Ca3Pd4 |
| 55.2.40 | 415576 | 172 | 0.1380 | 0.0278 | 2 | 2 | 2 | 3 | - | - | - | - | z4=1 | Bi6In2Sr5 |
| 55.2.41 | 262861 | 192 | 0.0271 | -0.0592 | 2 | 4 | 10 | 2 | - | - | - | - | z2w,3=1 | Bi2Fe4O9 |
| 55.2.42 | 186442 | 192 | 0.0084 | -0.0948 | 2 | 2 | 2 | 3 | - | - | - | - | z4=1 | Bi2Fe4O9 |
| 55.2.43 | 186441 | 192 | 0.0106 | -0.0676 | 2 | 2 | 2 | 3 | - | - | - | - | z4=1 | Bi2Fe4O9 |
| 55.2.44 | 186440 | 192 | 0.0182 | -0.0344 | 2 | 4 | 10 | 2 | - | - | - | - | z2w,3=1 | Bi2Fe4O9 |
| 55.2.45 | 419361 | 180 | 0.0043 | -0.4510 | 2 | 6 | 0 | 4 | - | - | - | - | z4=2 | B4Rh1Yb1 |
| 55.2.46 | 419358 | 128 | 0.0126 | -0.9147 | 2 | 2 | 2 | 3 | - | - | - | - | z4=1 | B4Rh1Y1 |

Continued on next page

**TABLE XIV– continued**

| TQC | ICSD | NoE | DG | IndG | SubSG (w/ T) | cl | SubSG (w/o T) | cl | Max. Sub. SG | yes or no | Max. Sub. SG | yes or no | SIs (w/ T) | Stoich. Formula |
|---|---|---|---|---|---|---|---|---|---|---|---|---|---|---|
| 55.2.47 | 419360 | 176 | 0.0065 | -0.1751 | 2 | 1 | 2 | 1 | - | - | - | - | z2w,3=1<br>z4=3 | B4Rh1Tm1 |
| 55.2.48 | 409820 | 184 | 0.0113 | -0.8905 | 2 | 3 | 2 | 3 | - | - | - | - | z4=3 | B4Ni1Yb1 |
| 55.2.49 | 425457 | 100 | 0.0089 | -0.7444 | 2 | 6 | 0 | 4 | - | - | - | - | z4=2 | B4Ni1Sc1 |
| 55.2.50 | 61039 | 112 | 0.0181 | -0.7764 | 2 | 5 | 2 | 1 | - | - | - | - | z2w,3=1<br>z4=1 | B4Mg2Ru5 |
| 55.2.51 | 190184 | 176 | 0.0081 | -0.4647 | 2 | 1 | 2 | 1 | - | - | - | - | z2w,3=1<br>z4=3 | B4Fe1Yb1 |
| 55.2.52 | 409626 | 156 | 0.0000 | -0.0222 | 2 | 6 | 0 | 4 | - | - | - | - | z4=2 | B4Er1V1 |
| 55.2.53 | 81542 | 120 | 0.0080 | -0.1818 | 2 | 4 | 10 | 2 | - | - | - | - | z2w,3=1 | B4Cr1Th1 |
| 55.2.54 | 250251 | 180 | 0.0000 | -0.4151 | 2 | 6 | 0 | 4 | - | - | - | - | z4=2 | B4Co1Yb1 |
| 55.2.55 | 417873 | 156 | 0.0022 | -0.0554 | 2 | 1 | 2 | 1 | - | - | - | - | z2w,3=1<br>z4=3 | B4Co1Gd1 |
| 55.2.56 | 78823 | 128 | 0.0197 | -0.8914 | 2 | 5 | 2 | 1 | - | - | - | - | z2w,3=1<br>z4=1 | B10Ni1Th2 |
| 55.2.57 | 78821 | 124 | 0.1844 | -0.5839 | 2 | 4 | 10 | 2 | - | - | - | - | z2w,3=1 | B10Fe1Th2 |
| 55.2.58 | 56973 | 112 | 0.0171 | -0.5015 | 2 | 3 | 2 | 3 | - | - | - | - | z4=3 | As1Ga2Rh5 |
| 55.2.59 | 36364 | 172 | 0.2146 | -0.3488 | 2 | 2 | 2 | 3 | - | - | - | - | z4=1 | Al2Bi6Ca5 |
| 55.2.60 | 260796 | 156 | 0.0087 | -0.9068 | 2 | 4 | 10 | 2 | - | - | - | - | z2w,3=1 | Al1B4Yb1 |
| 55.2.61 | 167532 | 156 | 0.0087 | -0.9064 | 2 | 4 | 10 | 2 | - | - | - | - | z2w,3=1 | Al1B4Yb1 |
| 55.2.62 | 167534 | 160 | 0.0161 | -1.5677 | 2 | 1 | 2 | 1 | - | - | - | - | z2w,3=1<br>z4=3 | Al1B4Lu1 |
| 55.2.63 | 260799 | 160 | 0.2313 | -0.2511 | 2 | 1 | 2 | 1 | - | - | - | - | z2w,3=1<br>z4=3 | Al1.33B5.33Lu1.33 |
| 55.4.1 | 90366 | 114 | - | - | - | - | - | - | See details in Table XIX | | | | | O4Pr1Sr2 |
| 55.4.2 | 237190 | 114 | - | - | - | - | - | - | See details in Table XIX | | | | | O4Pr1Sr2 |
| 55.4.3 | 414506 | 90 | - | - | - | - | - | - | See details in Table XIX | | | | | In4Rh2Sc5 |
| 55.4.4 | 165133 | 174 | - | - | - | - | - | - | See details in Table XIX | | | | | In4Pd2Y5 |
| 55.4.5 | 414505 | 94 | - | - | - | - | - | - | See details in Table XIX | | | | | In4Ni2Sc5 |
| 55.4.6 | 165139 | 314 | - | - | - | - | - | - | See details in Table XIX | | | | | In4Lu5Pd2 |
| 55.4.7 | 106807 | 314 | - | - | - | - | - | - | See details in Table XIX | | | | | In4Lu5Ni2 |
| 55.4.8 | 417906 | 118 | - | - | - | - | - | - | See details in Table XIX | | | | | In3Pd5 |
| 55.4.9 | 10004 | 142 | - | - | - | - | - | - | See details in Table XIX | | | | | Cu4Na3S4 |
| 55.4.10 | 201979 | 118 | - | - | - | - | - | - | See details in Table XIX | | | | | Cl6N1Sc4 |
| 55.4.11 | 201978 | 118 | - | - | - | - | - | - | See details in Table XIX | | | | | Cl6N1Sc4 |
| 55.4.12 | 9136 | 74 | - | - | - | - | - | - | See details in Table XIX | | | | | Cl4Mn1Na2 |
| 55.4.13 | 414344 | 142 | - | - | - | - | - | - | See details in Table XIX | | | | | C2Ge2La5 |
| 55.4.14 | 417442 | 70 | - | - | - | - | - | - | See details in Table XIX | | | | | B2Li2Rh3 |
| 55.4.15 | 201977 | 114 | - | - | - | - | - | - | See details in Table XIX | | | | | B1Cl6Sc4 |
| 55.4.16 | 78822 | 126 | - | - | - | - | - | - | See details in Table XIX | | | | | B10Co1Th2 |
| 55.4.17 | 58186 | 90 | - | - | - | - | - | - | See details in Table XIX | | | | | Al7Th2 |
| 55.4.18 | 55579 | 118 | - | - | - | - | - | - | See details in Table XIX | | | | | Al3Pt5 |
| 57.2.1 | 20933 | 112 | 0.0000 | -0.0368 | 2 | 5 | 2 | 1 | - | - | - | - | z2w,1=1<br>z4=1 | Fe2Hf1Si2 |
| 57.2.2 | 85640 | 128 | 0.0095 | -0.3841 | 2 | 4 | 0 | 2 | - | - | - | - | z4=2 | Cr1La1Sb3 |
| 57.2.3 | 85639 | 128 | 0.0049 | -0.3508 | 2 | 4 | 0 | 2 | - | - | - | - | z4=2 | Cr1La1Sb3 |
| 57.2.4 | 83907 | 128 | 0.0112 | -0.3307 | 2 | 3 | 0 | 2 | - | - | - | - | z2w,1=1 | Cr1La1Sb3 |
| 57.2.5 | 247291 | 248 | 0.0196 | -0.0213 | 2 | 3 | 0 | 2 | - | - | - | - | z2w,1=1 | Ce3Rh2Si2 |
| 57.2.6 | 76352 | 248 | 0.0119 | -0.0267 | 2 | 2 | 0 | 2 | - | - | - | - | z2w,1=1<br>z4=2 | Ce3Ge2Rh2 |
| 57.2.7 | 190838 | 176 | 0.0000 | -0.2116 | 2 | 1 | 2 | 1 | - | - | - | - | z4=3 | Ce1Pd2Sn3 |
| 57.2.8 | 93749 | 192 | 0.0060 | -0.1160 | 2 | 1 | 2 | 1 | - | - | - | - | z4=3 | Ca1Mn2O4 |
| 57.2.9 | 150592 | 184 | 0.0000 | -0.0065 | 2 | 5 | 2 | 1 | - | - | - | - | z2w,1=1<br>z4=1 | Al1Dy1 |
| 57.3.1 | 30725 | 236 | - | - | - | - | - | - | 26<br>18<br>13 | yes<br>yes<br>no(2) | 26<br>18<br>13 | yes<br>yes<br>no(2) | -<br>-<br>z4=1 | Rh2Si2Y3 |

Continued on next page

**TABLE XIV– continued**

| TQC | ICSD | NoE | DG | IndG | SubSG (w/ T) | cl | SubSG (w/o T) | cl | Max. Sub. SG | yes or no | Max. Sub. SG | yes or no | SIs (w/ T) | Stoich. Formula |
|---|---|---|---|---|---|---|---|---|---|---|---|---|---|---|
| | | | | | | | | | 11 | no(2) | 11 | no(2) | z4=1 | |
| | | | | | | | | | - | - | 7 | no(7) | - | |
| 57.3.2 | 188903 | 180 | - | - | - | - | - | - | 26 | yes | 26 | yes | - | Mn3O4 |
| | | | | | | | | | 18 | yes | 18 | yes | - | |
| | | | | | | | | | 13 | no(2) | 13 | no(2) | z2w,2=1<br>z4=3 | |
| | | | | | | | | | 11 | no(2) | 11 | no(2) | z2w,1=1<br>z4=3 | |
| | | | | | | | | | - | - | 7 | no(7) | - | |
| 57.3.3 | 102123 | 180 | - | - | - | - | - | - | 26 | yes | 26 | yes | - | In1Th1 |
| | | | | | | | | | 18 | yes | 18 | yes | - | |
| | | | | | | | | | 13 | no(2) | 13 | no(2) | z4=1 | |
| | | | | | | | | | 11 | no(2) | 11 | no(2) | z4=1 | |
| | | | | | | | | | - | - | 7 | no(7) | - | |
| 57.3.4 | 20939 | 236 | - | - | - | - | - | - | 26 | yes | 26 | yes | - | Ga2La3Ni2 |
| | | | | | | | | | 18 | yes | 18 | yes | - | |
| | | | | | | | | | 13 | no(2) | 13 | yes | z2w,2=1 | |
| | | | | | | | | | 11 | no(2) | 11 | yes | z2w,1=1 | |
| | | | | | | | | | - | - | 7 | no(7) | - | |
| 57.3.5 | 402545 | 108 | - | - | - | - | - | - | 26 | yes | 26 | yes | - | Fe2Sc1Si2 |
| | | | | | | | | | 18 | yes | 18 | yes | - | |
| | | | | | | | | | 13 | yes | 13 | yes | - | |
| | | | | | | | | | 11 | yes | 11 | yes | - | |
| | | | | | | | | | - | - | 7 | no(7) | - | |
| 57.3.6 | 83725 | 180 | - | - | - | - | - | - | 26 | yes | 26 | yes | - | F4Fe1Rb1 |
| | | | | | | | | | 18 | yes | 18 | yes | - | |
| | | | | | | | | | 13 | no(2) | 13 | yes | z2w,2=1<br>z4=2 | |
| | | | | | | | | | 11 | no(2) | 11 | yes | z2w,1=1<br>z4=2 | |
| | | | | | | | | | - | - | 7 | no(7) | - | |
| 57.3.7 | 158115 | 180 | - | - | - | - | - | - | 26 | yes | 26 | yes | - | Cr1Sb3Yb1 |
| | | | | | | | | | 18 | yes | 18 | yes | - | |
| | | | | | | | | | 13 | no(2) | 13 | yes | z4=2 | |
| | | | | | | | | | 11 | no(2) | 11 | yes | z4=2 | |
| | | | | | | | | | - | - | 7 | no(7) | - | |
| 57.3.8 | 419661 | 164 | - | - | - | - | - | - | 26 | yes | 26 | yes | - | Cr1F5 |
| | | | | | | | | | 18 | yes | 18 | yes | - | |
| | | | | | | | | | 13 | yes | 13 | yes | - | |
| | | | | | | | | | 11 | yes | 11 | yes | - | |
| | | | | | | | | | - | - | 7 | no(7) | - | |
| 58.2.1 | 42724 | 40 | 0.0983 | -2.0111 | 2 | 1 | 2 | 1 | - | - | - | - | z4=1 | Ni1Sb2 |
| 58.2.2 | 414575 | 132 | 0.0038 | -1.1601 | 2 | 2 | 2 | 1 | - | - | - | - | z4=3 | Ge6Li2Sr4 |
| 58.2.3 | 409767 | 188 | 0.0073 | -0.0337 | 2 | 2 | 2 | 1 | - | - | - | - | z4=3 | Eu2Ge3Li1 |
| 58.2.4 | 42720 | 32 | 0.0000 | -0.5914 | 2 | 3 | 0 | 2 | - | - | - | - | z4=2 | Cr1Sb2 |
| 58.2.5 | 42601 | 32 | 0.0000 | -0.5837 | 2 | 3 | 0 | 2 | - | - | - | - | z4=2 | Cr1Sb2 |
| 58.2.6 | 81572 | 84 | 0.0301 | -0.6654 | 2 | 1 | 2 | 1 | - | - | - | - | z4=1 | C3Al2Th2 |
| 58.2.7 | 425106 | 36 | 0.0087 | -0.4871 | 2 | 1 | 2 | 1 | - | - | - | - | z4=1 | B4Cr1 |
| 58.2.8 | 186851 | 36 | 0.0143 | -0.4830 | 2 | 1 | 2 | 1 | - | - | - | - | z4=1 | B4Cr1 |
| 58.2.9 | 31387 | 144 | 0.0142 | -0.1575 | 2 | 1 | 2 | 1 | - | - | - | - | z4=1 | B2Ni3O6 |
| 58.2.10 | 2016 | 144 | 0.0162 | -0.1518 | 2 | 1 | 2 | 1 | - | - | - | - | z4=1 | B2Ni3O6 |
| 58.2.11 | 251066 | 260 | 0.0858 | -1.4048 | 2 | 2 | 2 | 1 | - | - | - | - | z4=3 | Au8Sn3Sr3 |
| 58.2.12 | 42605 | 40 | 0.0797 | -2.2426 | 2 | 1 | 2 | 1 | - | - | - | - | z4=1 | As2Ni1 |
| 58.3.1 | 402015 | 192 | - | - | - | - | - | - | 31 | yes | 31 | yes | - | Re4Si2Th1 |
| | | | | | | | | | 18 | yes | 18 | yes | - | |
| | | | | | | | | | 14 | no(2) | 14 | yes | z2w,1=1 | |
| | | | | | | | | | 10 | no(2) | 10 | no(10) | z2w,2=1<br>z2w,3=1 | |

Continued on next page

**TABLE XIV– continued**

| TQC | ICSD | NoE | DG | IndG | SubSG (w/ T) | cl | SubSG (w/o T) | cl | Max. Sub. SG | yes or no | Max. Sub. SG | yes or no | SIs (w/ T) | Stoich. Formula |
|---|---|---|---|---|---|---|---|---|---|---|---|---|---|---|
| | | | | | | | | | - | - | 7 | no(7) | - | |
| 58.3.2 | 42721 | 32 | - | - | - | - | - | - | 31 | yes | 31 | yes | - | Cr1Sb2 |
| | | | | | | | | | 18 | yes | 18 | yes | - | |
| | | | | | | | | | 14 | no(2) | 14 | no(2) | z2w,1=1<br>z4=3 | |
| | | | | | | | | | 10 | no(2) | 10 | no(2) | z2w,2=1<br>z2w,3=1<br>z4=3 | |
| | | | | | | | | | - | - | 7 | no(7) | - | |
| 58.4.1 | 48007 | 42 | - | - | - | - | - | - | See details in Table XIX | | | | | Li1O2Ru1 |
| 58.4.2 | 186047 | 130 | - | - | - | - | - | - | See details in Table XIX | | | | | Fe4Ge2Lu1 |
| 58.4.3 | 186046 | 130 | - | - | - | - | - | - | See details in Table XIX | | | | | Fe4Ge2Lu1 |
| 58.4.4 | 170634 | 154 | - | - | - | - | - | - | See details in Table XIX | | | | | Eu3In2P4 |
| 58.4.5 | 2598 | 206 | - | - | - | - | - | - | See details in Table XIX | | | | | Co3O8U2 |
| 58.4.6 | 86115 | 42 | - | - | - | - | - | - | See details in Table XIX | | | | | Co1Te2 |
| 58.4.7 | 92076 | 106 | - | - | - | - | - | - | See details in Table XIX | | | | | C4Mn1N6 |
| 58.4.8 | 85618 | 110 | - | - | - | - | - | - | See details in Table XIX | | | | | C4Co1N6 |
| 58.4.9 | 422726 | 78 | - | - | - | - | - | - | See details in Table XIX | | | | | C2H2Co1N4 |
| 58.4.10 | 236414 | 78 | - | - | - | - | - | - | See details in Table XIX | | | | | C2H2Co1N4 |
| 58.4.11 | 236413 | 78 | - | - | - | - | - | - | See details in Table XIX | | | | | C2H2Co1N4 |
| 58.4.12 | 194807 | 38 | - | - | - | - | - | - | See details in Table XIX | | | | | B4Mn1 |
| 58.4.13 | 236773 | 138 | - | - | - | - | - | - | See details in Table XIX | | | | | B2Co3O6 |
| 58.4.14 | 10487 | 46 | - | - | - | - | - | - | See details in Table XIX | | | | | B1Pd2 |
| 58.4.15 | 57497 | 250 | - | - | - | - | - | - | See details in Table XIX | | | | | Al1Au2 |
| 59.1.1 | 408080 | 200 | 0.0105 | -1.0451 | 2 | 2 | 0 | 1 | - | - | - | - | z2w,3=1 | Pd1Si1Y1 |
| 59.1.2 | 154823 | 40 | 0.0044 | -0.1641 | 2 | 2 | 0 | 1 | - | - | - | - | z2w,3=1 | Li1Mn1O2 |
| 59.1.3 | 425285 | 68 | 0.0067 | -0.0343 | 2 | 1 | 0 | 1 | - | - | - | - | z2w,3=1<br>z4=2 | Ho1I1S1 |
| 59.1.4 | 154397 | 180 | 0.0596 | -1.2690 | 2 | 1 | 0 | 1 | - | - | - | - | z2w,3=1<br>z4=2 | Ge5N2Sr6 |
| 59.1.5 | 410357 | 216 | 0.0000 | -0.0285 | 2 | 1 | 0 | 1 | - | - | - | - | z2w,3=1<br>z4=2 | Ge1Pr1Pt1 |
| 59.1.6 | 195027 | 88 | 0.0311 | -0.5010 | 2 | 2 | 0 | 1 | - | - | - | - | z2w,3=1 | Fe2O1Se2Sr1 |
| 59.1.7 | 195026 | 88 | 0.0264 | -0.5527 | 2 | 2 | 0 | 1 | - | - | - | - | z2w,3=1 | Fe2O1S2Sr1 |
| 59.1.8 | 195025 | 88 | 0.0170 | -0.5414 | 2 | 2 | 0 | 1 | - | - | - | - | z2w,3=1 | Fe2O1S2Sr1 |
| 59.1.9 | 88203 | 128 | 0.0000 | -0.1228 | 2 | 2 | 0 | 1 | - | - | - | - | z2w,3=1 | Er2Ge5 |
| 59.1.10 | 160543 | 120 | 0.0000 | -0.0057 | 2 | 1 | 0 | 1 | - | - | - | - | z2w,3=1<br>z4=2 | Dy2Sn5 |
| 59.1.11 | 72376 | 140 | 0.0031 | -0.1609 | 2 | 1 | 0 | 1 | - | - | - | - | z2w,3=1<br>z4=2 | Cr6P4U1 |
| 59.1.12 | 30701 | 216 | 0.0560 | -0.1468 | 2 | 3 | 0 | 1 | - | - | - | - | z4=2 | Co8La1P5 |
| 59.1.13 | 410358 | 208 | 0.0000 | -0.0462 | 2 | 3 | 0 | 1 | - | - | - | - | z4=2 | Ce1Ge1Pt1 |
| 59.1.14 | 154398 | 180 | 0.0281 | -0.7161 | 2 | 1 | 0 | 1 | - | - | - | - | z2w,3=1<br>z4=2 | Ba6Ge5N2 |
| 59.1.15 | 421431 | 68 | 0.0104 | -0.9795 | 2 | 1 | 0 | 1 | - | - | - | - | z2w,3=1<br>z4=2 | Ba1Ge1Ni2 |
| 59.1.16 | 428170 | 88 | 0.0187 | -0.5428 | 2 | 2 | 0 | 1 | - | - | - | - | z2w,3=1 | Ba1Fe2O1S2 |
| 59.1.17 | 421524 | 28 | 0.0788 | -1.6440 | 2 | 1 | 0 | 1 | - | - | - | - | z2w,3=1<br>z4=2 | B2Ru1 |
| 59.1.18 | 31871 | 28 | 0.0782 | -1.6483 | 2 | 1 | 0 | 1 | - | - | - | - | z2w,3=1<br>z4=2 | B2Ru1 |
| 59.1.19 | 421523 | 28 | 0.2040 | -2.2130 | 2 | 1 | 0 | 1 | - | - | - | - | z2w,3=1<br>z4=2 | B2Os1 |
| 59.2.1 | 187622 | 28 | 0.1322 | -2.3727 | 2 | 4 | 2 | 2 | - | - | - | - | z4=3 | Ni1Si1 |
| 59.2.2 | 95640 | 40 | 0.0037 | -0.1280 | 2 | 4 | 2 | 2 | - | - | - | - | z4=3 | Li1Mn1O2 |
| 59.2.3 | 408082 | 256 | 0.0000 | -0.0161 | 2 | 7 | 2 | 2 | - | - | - | - | z2w,3=1<br>z4=3 | Gd1Pd1Si1 |

Continued on next page

**TABLE XIV– continued**

| TQC | ICSD | NoE | DG | IndG | SubSG (w/ T) | cl | SubSG (w/o T) | cl | Max. Sub. SG | yes or no | Max. Sub. SG | yes or no | SIs (w/ T) | Stoich. Formula |
|---|---|---|---|---|---|---|---|---|---|---|---|---|---|---|
| 59.2.4 | 81303 | 152 | 0.0075 | -0.8562 | 2 | 6 | 2 | 2 | - | - | - | - | z2w,3=1 z4=1 | Fe1Hf8Te6 |
| 59.2.5 | 408081 | 288 | 0.0000 | -0.0549 | 2 | 6 | 2 | 2 | - | - | - | - | z2w,3=1 z4=1 | Er1Pd1Si1 |
| 59.2.6 | 99042 | 100 | 0.0194 | -1.2587 | 2 | 4 | 2 | 2 | - | - | - | - | z4=3 | Cu2O3Sr1 |
| 59.2.7 | 155492 | 172 | 0.0078 | -0.1211 | 2 | 5 | 2 | 2 | - | - | - | - | z4=1 | Co1Na0.5O2 |
| 59.2.8 | 413690 | 108 | 0.0022 | -0.0217 | 2 | 5 | 2 | 2 | - | - | - | - | z4=1 | Br3Mo1 |
| 59.2.9 | 249645 | 108 | 0.0000 | -1.7404 | 2 | 4 | 2 | 2 | - | - | - | - | z4=3 | Au3K1Sn3 |
| 59.2.10 | 245679 | 104 | 0.0174 | -2.1312 | 2 | 4 | 2 | 2 | - | - | - | - | z4=3 | Au3In3Sr1 |
| 59.2.11 | 107445 | 108 | 0.0251 | -0.2113 | 2 | 4 | 2 | 2 | - | - | - | - | z4=3 | Au1Rb3Sn4 |
| 59.2.12 | 107447 | 108 | 0.0103 | -0.4158 | 2 | 4 | 2 | 2 | - | - | - | - | z4=3 | Au1Pb4Rb3 |
| 59.2.13 | 107448 | 108 | 0.0070 | -0.4042 | 2 | 4 | 2 | 2 | - | - | - | - | z4=3 | Au1Cs3Pb4 |
| 59.2.14 | 107525 | 200 | 0.0027 | -0.0095 | 2 | 6 | 2 | 2 | - | - | - | - | z2w,3=1 z4=1 | Al1Eu1 |
| 59.4.1 | 241187 | 134 | - | - | - | - | - | - | See details in Table XIX | | | | | O7Sr1V3 |
| 59.4.2 | 241186 | 134 | - | - | - | - | - | - | See details in Table XIX | | | | | O7Sr1V3 |
| 59.4.3 | 241185 | 134 | - | - | - | - | - | - | See details in Table XIX | | | | | O7Sr1V3 |
| 59.4.4 | 241184 | 134 | - | - | - | - | - | - | See details in Table XIX | | | | | O7Sr1V3 |
| 59.4.5 | 241183 | 134 | - | - | - | - | - | - | See details in Table XIX | | | | | O7Sr1V3 |
| 59.4.6 | 241182 | 134 | - | - | - | - | - | - | See details in Table XIX | | | | | O7Sr1V3 |
| 59.4.7 | 241181 | 134 | - | - | - | - | - | - | See details in Table XIX | | | | | O7Sr1V3 |
| 59.4.8 | 241180 | 134 | - | - | - | - | - | - | See details in Table XIX | | | | | O7Sr1V3 |
| 59.4.9 | 241179 | 134 | - | - | - | - | - | - | See details in Table XIX | | | | | O7Sr1V3 |
| 59.4.10 | 241178 | 134 | - | - | - | - | - | - | See details in Table XIX | | | | | O7Sr1V3 |
| 59.4.11 | 241177 | 134 | - | - | - | - | - | - | See details in Table XIX | | | | | O7Sr1V3 |
| 59.4.12 | 241176 | 134 | - | - | - | - | - | - | See details in Table XIX | | | | | O7Sr1V3 |
| 59.4.13 | 88354 | 82 | - | - | - | - | - | - | See details in Table XIX | | | | | Na1O5V2 |
| 59.4.14 | 59345 | 82 | - | - | - | - | - | - | See details in Table XIX | | | | | Na1O5V2 |
| 59.4.15 | 409210 | 82 | - | - | - | - | - | - | See details in Table XIX | | | | | Na1O5V2 |
| 59.4.16 | 251182 | 82 | - | - | - | - | - | - | See details in Table XIX | | | | | Na1O5V2 |
| 59.4.17 | 20825 | 174 | - | - | - | - | - | - | See details in Table XIX | | | | | Mo2Ni6P3 |
| 59.4.18 | 88639 | 82 | - | - | - | - | - | - | See details in Table XIX | | | | | Li1O5V2 |
| 59.4.19 | 107524 | 118 | - | - | - | - | - | - | See details in Table XIX | | | | | In6La1Ni3 |
| 59.4.20 | 416462 | 62 | - | - | - | - | - | - | See details in Table XIX | | | | | Gd1I1S1 |
| 59.4.21 | 50194 | 70 | - | - | - | - | - | - | See details in Table XIX | | | | | Er1I1Se1 |
| 59.4.22 | 79107 | 66 | - | - | - | - | - | - | See details in Table XIX | | | | | Dy1I1S1 |
| 59.4.23 | 93966 | 34 | - | - | - | - | - | - | See details in Table XIX | | | | | Cu1Te1 |
| 59.4.24 | 33996 | 98 | - | - | - | - | - | - | See details in Table XIX | | | | | Co1Cu2O3 |
| 59.4.25 | 39314 | 34 | - | - | - | - | - | - | See details in Table XIX | | | | | Cl1O1Ti1 |
| 59.4.26 | 155833 | 34 | - | - | - | - | - | - | See details in Table XIX | | | | | Cl1O1Ti1 |
| 59.4.27 | 261526 | 34 | - | - | - | - | - | - | See details in Table XIX | | | | | Cl1Li1N1Ti1 |
| 59.4.28 | 40963 | 42 | - | - | - | - | - | - | See details in Table XIX | | | | | Cl1Fe1O1 |
| 59.4.29 | 16013 | 42 | - | - | - | - | - | - | See details in Table XIX | | | | | Cl1Fe1O1 |
| 59.4.30 | 21009 | 70 | - | - | - | - | - | - | See details in Table XIX | | | | | Cl1Er1S1 |
| 59.4.31 | 161723 | 222 | - | - | - | - | - | - | See details in Table XIX | | | | | Cl1Cu3O8Se2Y1 |
| 59.4.32 | 4086 | 38 | - | - | - | - | - | - | See details in Table XIX | | | | | Cl1Cr1O1 |
| 59.4.33 | 80005 | 242 | - | - | - | - | - | - | See details in Table XIX | | | | | Ce3Pd6Sb5 |
| 59.4.34 | 79106 | 66 | - | - | - | - | - | - | See details in Table XIX | | | | | Br1Dy1S1 |
| 59.4.35 | 69661 | 38 | - | - | - | - | - | - | See details in Table XIX | | | | | Br1Cr1S1 |
| 59.4.36 | 69660 | 38 | - | - | - | - | - | - | See details in Table XIX | | | | | Br1Cr1S1 |
| 59.4.37 | 69659 | 38 | - | - | - | - | - | - | See details in Table XIX | | | | | Br1Cr1S1 |
| 59.4.38 | 238106 | 182 | - | - | - | - | - | - | See details in Table XIX | | | | | Bi5K6Zn1 |
| 59.4.39 | 69385 | 210 | - | - | - | - | - | - | See details in Table XIX | | | | | Bi1Cl1Cu3O8Se2 |
| 59.4.40 | 54190 | 210 | - | - | - | - | - | - | See details in Table XIX | | | | | Bi1Cl1Cu3O8Se2 |
| 59.4.41 | 280759 | 210 | - | - | - | - | - | - | See details in Table XIX | | | | | Bi1Br1Cu3O8Se2 |
| 59.4.42 | 54011 | 194 | - | - | - | - | - | - | See details in Table XIX | | | | | B3Ge2Ni8 |
| 59.4.43 | 245680 | 118 | - | - | - | - | - | - | See details in Table XIX | | | | | Au3Eu1In3 |

Continued on next page

**TABLE XIV– continued**

| TQC | ICSD | NoE | DG | IndG | SubSG (w/ T) | cl | SubSG (w/o T) | cl | Max. Sub. SG | yes or no | Max. Sub. SG | yes or no | SIs (w/ T) | Stoich. Formula |
|---|---|---|---|---|---|---|---|---|---|---|---|---|---|---|
| 59.4.44 | 92440 | 270 | - | - | - | - | - | - | See details in Table XIX | | | | | As1Pd1Zr1 |
| 59.4.45 | 2721 | 74 | - | - | - | - | - | - | See details in Table XIX | | | | | Ag3Sn1 |
| 59.4.46 | 154084 | 74 | - | - | - | - | - | - | See details in Table XIX | | | | | Ag3Sn1 |
| 60.2.1 | 415270 | 64 | 0.0278 | -0.0222 | 2 | 1 | 2 | 1 | - | - | - | - | z4=3 | O2Pb1 |
| 60.2.2 | 415269 | 64 | 0.0268 | -0.0219 | 2 | 1 | 2 | 1 | - | - | - | - | z4=3 | O2Pb1 |
| 60.2.3 | 415268 | 64 | 0.0378 | -0.0212 | 2 | 1 | 2 | 1 | - | - | - | - | z4=3 | O2Pb1 |
| 60.2.4 | 246148 | 64 | 0.0354 | -0.8745 | 2 | 1 | 2 | 1 | - | - | - | - | z4=3 | C1Mo2 |
| 60.2.5 | 246147 | 64 | 0.0287 | -0.8813 | 2 | 1 | 2 | 1 | - | - | - | - | z4=3 | C1Mo2 |
| 60.2.6 | 246146 | 64 | 0.0235 | -0.8925 | 2 | 1 | 2 | 1 | - | - | - | - | z4=3 | C1Mo2 |
| 60.2.7 | 430298 | 208 | 0.0108 | -0.3112 | 2 | 1 | 2 | 1 | - | - | - | - | z4=3 | Au2Pb1 |
| 60.3.1 | 40944 | 164 | - | - | - | - | - | - | 13 | no(2) | 13 | yes | z2w,1=1 z2w,2=1 z4=2 | Cl3Cu1Rb1 |
| | | | | | | | | | 4 | yes | 7 | no(7) | - | |
| | | | | | | | | | - | - | 7 | no(7) | - | |
| | | | | | | | | | - | - | 4 | yes | - | |
| 60.3.2 | 15522 | 164 | - | - | - | - | - | - | 13 | no(2) | 13 | yes | z2w,1=1 z2w,2=1 z4=2 | Cl3Cu1Rb1 |
| | | | | | | | | | 4 | yes | 7 | no(7) | - | |
| | | | | | | | | | - | - | 7 | no(7) | - | |
| | | | | | | | | | - | - | 4 | yes | - | |
| 61.2.1 | 162404 | 160 | 0.0414 | -0.3541 | 2 | 2 | 2 | 1 | - | - | - | - | z4=3 | Pd1Sn1Te1 |
| 61.2.2 | 415968 | 152 | 0.0057 | -0.4975 | 2 | 1 | 2 | 1 | - | - | - | - | z4=1 | Au1Sn2 |
| 61.3.1 | 9922 | 100 | - | - | - | - | - | - | 4 | yes | 7 | no(7) | - | Ag1F2 |
| | | | | | | | | | 4 | yes | 7 | no(7) | - | |
| | | | | | | | | | 2 | no(2) | 7 | no(7) | z2w,1=1 z2w,2=1 z2w,3=1 z4=2 | |
| | | | | | | | | | - | - | 4 | yes | - | |
| | | | | | | | | | - | - | 4 | yes | - | |
| | | | | | | | | | - | - | 2 | yes | - | |
| 61.3.2 | 66014 | 100 | - | - | - | - | - | - | 4 | yes | 7 | no(7) | - | Ag1F2 |
| | | | | | | | | | 4 | yes | 7 | no(7) | - | |
| | | | | | | | | | 2 | no(2) | 7 | no(7) | z2w,1=1 z2w,2=1 z2w,3=1 z4=2 | |
| | | | | | | | | | - | - | 4 | yes | - | |
| | | | | | | | | | - | - | 4 | yes | - | |
| | | | | | | | | | - | - | 2 | yes | - | |
| 61.3.3 | 6277 | 100 | - | - | - | - | - | - | 4 | yes | 7 | no(7) | - | Ag1F2 |
| | | | | | | | | | 4 | yes | 7 | no(7) | - | |
| | | | | | | | | | 2 | no(2) | 7 | no(7) | z2w,1=1 z2w,2=1 z2w,3=1 z4=2 | |
| | | | | | | | | | - | - | 4 | yes | - | |
| | | | | | | | | | - | - | 4 | yes | - | |
| | | | | | | | | | - | - | 2 | yes | - | |
| 62.2.1 | 418619 | 280 | 0.0049 | -0.8991 | 2 | 2 | 2 | 1 | - | - | - | - | z4=1 | Sr1Zn5 |
| 62.2.2 | 54980 | 88 | 0.0156 | -0.7189 | 2 | 1 | 2 | 1 | - | - | - | - | z4=3 | Sr1Zn1 |
| 62.2.3 | 82295 | 104 | 0.0038 | -0.0651 | 2 | 1 | 2 | 1 | - | - | - | - | z4=3 | Se2U1 |
| 62.2.4 | 10001 | 128 | 0.0433 | -0.0574 | 2 | 1 | 2 | 1 | - | - | - | - | z4=3 | Sb2Sr1Zn1 |
| 62.2.5 | 402457 | 256 | 0.0000 | -0.1117 | 2 | 3 | 0 | 2 | - | - | - | - | z4=2 | S3Tm2 |
| 62.2.6 | 42430 | 104 | 0.0034 | -0.0593 | 2 | 1 | 2 | 1 | - | - | - | - | z4=3 | S1Te1U1 |

Continued on next page

**TABLE XIV– continued**

| TQC | ICSD | NoE | DG | IndG | SubSG (w/ T) | cl | SubSG (w/o T) | cl | Max. Sub. SG | yes or no | Max. Sub. SG | yes or no | SIs (w/ T) | Stoich. Formula |
|---|---|---|---|---|---|---|---|---|---|---|---|---|---|---|
| 62.2.7 | 30398 | 176 | 0.1240 | -1.0206 | 2 | 2 | 2 | 1 | - | - | - | - | z4=1 | Ru4Si3 |
| 62.2.8 | 421663 | 240 | 0.2352 | -1.5888 | 2 | 3 | 0 | 2 | - | - | - | - | z4=2 | Ru2Sn2Zn3 |
| 62.2.9 | 421255 | 120 | 0.0205 | -0.2164 | 2 | 2 | 2 | 1 | - | - | - | - | z4=1 | Rh1Sc3Si3 |
| 62.2.10 | 15248 | 80 | 0.0184 | -0.5032 | 2 | 1 | 2 | 1 | - | - | - | - | z4=3 | Rh1Sc1Si2 |
| 62.2.11 | 159305 | 184 | 0.0270 | -1.2036 | 2 | 1 | 2 | 1 | - | - | - | - | z4=3 | Pt1Yb1Zn1 |
| 62.2.12 | 412012 | 96 | 0.0653 | -1.2738 | 2 | 2 | 2 | 1 | - | - | - | - | z4=1 | Pt1Sn1Sr1 |
| 62.2.13 | 260792 | 72 | 0.0317 | -0.2539 | 2 | 1 | 2 | 1 | - | - | - | - | z4=3 | Pt1Si1Zr1 |
| 62.2.14 | 260794 | 72 | 0.0439 | -0.1355 | 2 | 1 | 2 | 1 | - | - | - | - | z4=3 | Pt1Si1Ti1 |
| 62.2.15 | 79238 | 56 | 0.0076 | -1.3790 | 2 | 3 | 0 | 2 | - | - | - | - | z4=2 | Pt1Si1 |
| 62.2.16 | 2623 | 56 | 0.0676 | -1.3453 | 2 | 3 | 0 | 2 | - | - | - | - | z4=2 | Pt1Si1 |
| 62.2.17 | 105785 | 64 | 0.0193 | -0.1973 | 2 | 3 | 0 | 2 | - | - | - | - | z4=2 | Pt1Sc2 |
| 62.2.18 | 158364 | 96 | 0.0190 | -2.1382 | 2 | 3 | 0 | 2 | - | - | - | - | z4=2 | Pd2Sn1 |
| 62.2.19 | 159304 | 184 | 0.0196 | -1.0236 | 2 | 1 | 2 | 1 | - | - | - | - | z4=3 | Pd1Yb1Zn1 |
| 62.2.20 | 408408 | 152 | 0.0045 | -1.1056 | 2 | 2 | 2 | 1 | - | - | - | - | z4=1 | Pd1Sn1Yb1 |
| 62.2.21 | 151167 | 152 | 0.0046 | -1.1060 | 2 | 2 | 2 | 1 | - | - | - | - | z4=1 | Pd1Sn1Yb1 |
| 62.2.22 | 109415 | 56 | 0.0150 | -1.8651 | 2 | 3 | 0 | 2 | - | - | - | - | z4=2 | Pd1Sn1 |
| 62.2.23 | 79237 | 56 | 0.0065 | -2.1488 | 2 | 3 | 0 | 2 | - | - | - | - | z4=2 | Pd1Si1 |
| 62.2.24 | 15016 | 56 | 0.0034 | -2.1341 | 2 | 3 | 0 | 2 | - | - | - | - | z4=2 | Pd1Si1 |
| 62.2.25 | 1918 | 104 | 0.0021 | -0.0930 | 2 | 1 | 2 | 1 | - | - | - | - | z4=3 | Pd1Pu1 |
| 62.2.26 | 400334 | 56 | 0.0077 | -0.4391 | 2 | 1 | 2 | 1 | - | - | - | - | z4=3 | P2Zr1 |
| 62.2.27 | 417911 | 72 | 0.0059 | -0.2277 | 2 | 1 | 2 | 1 | - | - | - | - | z4=3 | P1Pt1Sc1 |
| 62.2.28 | 100624 | 160 | 0.0000 | -0.1028 | 2 | 1 | 2 | 1 | - | - | - | - | z4=3 | O4S1Sr1 |
| 62.2.29 | 415748 | 200 | 0.0050 | 0.0050 | 2 | 1 | 2 | 1 | - | - | - | - | z4=3 | O4Rb2Ru1 |
| 62.2.30 | 100625 | 136 | 0.0035 | -0.1142 | 2 | 3 | 0 | 2 | - | - | - | - | z4=2 | O4Pb1S1 |
| 62.2.31 | 185833 | 144 | 0.0028 | -0.0629 | 2 | 1 | 2 | 1 | - | - | - | - | z4=3 | O3Pr1V1 |
| 62.2.32 | 172354 | 160 | 0.0073 | -0.0359 | 2 | 1 | 2 | 1 | - | - | - | - | z4=3 | O3Pr1Rh1 |
| 62.2.33 | 172348 | 160 | 0.0073 | -0.0359 | 2 | 1 | 2 | 1 | - | - | - | - | z4=3 | O3Pr1Rh1 |
| 62.2.34 | 413386 | 328 | 0.0220 | 0.0052 | 2 | 3 | 0 | 2 | - | - | - | - | z4=2 | O1Si1Yb3 |
| 62.2.35 | 74785 | 168 | 0.0281 | -0.6337 | 2 | 3 | 0 | 2 | - | - | - | - | z4=2 | Ni2P2Th1 |
| 62.2.36 | 64952 | 192 | 0.0144 | -0.0897 | 2 | 2 | 2 | 1 | - | - | - | - | z4=1 | Ni2O4Si1 |
| 62.2.37 | 35675 | 192 | 0.0110 | -0.0960 | 2 | 2 | 2 | 1 | - | - | - | - | z4=1 | Ni2O4Si1 |
| 62.2.38 | 202376 | 192 | 0.0122 | -0.0934 | 2 | 2 | 2 | 1 | - | - | - | - | z4=1 | Ni2O4Si1 |
| 62.2.39 | 100645 | 192 | 0.0272 | -0.0782 | 2 | 2 | 2 | 1 | - | - | - | - | z4=1 | Ni2O4Si1 |
| 62.2.40 | 100644 | 192 | 0.0217 | -0.0850 | 2 | 2 | 2 | 1 | - | - | - | - | z4=1 | Ni2O4Si1 |
| 62.2.41 | 100643 | 192 | 0.0173 | -0.0895 | 2 | 2 | 2 | 1 | - | - | - | - | z4=1 | Ni2O4Si1 |
| 62.2.42 | 100642 | 192 | 0.0073 | -0.0987 | 2 | 2 | 2 | 1 | - | - | - | - | z4=1 | Ni2O4Si1 |
| 62.2.43 | 105403 | 176 | 0.0124 | -0.3884 | 2 | 2 | 2 | 1 | - | - | - | - | z4=1 | Ni1Th1 |
| 62.2.44 | 54360 | 152 | 0.0000 | -1.0160 | 2 | 2 | 2 | 1 | - | - | - | - | z4=1 | Ni1Sn1Yb1 |
| 62.2.45 | 18188 | 72 | 0.0154 | -0.1906 | 2 | 2 | 2 | 1 | - | - | - | - | z4=1 | Ni1Si1Ti1 |
| 62.2.46 | 95449 | 56 | 0.0046 | -1.8856 | 2 | 3 | 0 | 2 | - | - | - | - | z4=2 | Ni1Si1 |
| 62.2.47 | 95448 | 56 | 0.0342 | -1.8159 | 2 | 3 | 0 | 2 | - | - | - | - | z4=2 | Ni1Si1 |
| 62.2.48 | 190105 | 56 | 0.0342 | -1.8159 | 2 | 3 | 0 | 2 | - | - | - | - | z4=2 | Ni1Si1 |
| 62.2.49 | 248408 | 168 | 0.0002 | -0.0174 | 2 | 1 | 2 | 1 | - | - | - | - | z4=3 | Ni1S3U1 |
| 62.2.50 | 419759 | 104 | 0.0143 | -0.9586 | 2 | 1 | 2 | 1 | - | - | - | - | z4=3 | Ni1Pd1Te1 |
| 62.2.51 | 50990 | 72 | 0.0218 | -0.1264 | 2 | 1 | 2 | 1 | - | - | - | - | z4=3 | Ni1P1Sc1 |
| 62.2.52 | 189299 | 208 | 0.0119 | -0.0178 | 2 | 1 | 2 | 1 | - | - | - | - | z4=3 | Ni1O3Yb1 |
| 62.2.53 | 72290 | 184 | 0.0050 | -0.0082 | 2 | 1 | 2 | 1 | - | - | - | - | z4=3 | Nd2S3 |
| 62.2.54 | 415953 | 112 | 0.0000 | -0.0325 | 2 | 1 | 2 | 1 | - | - | - | - | z4=3 | Nd1Pt1Sn1 |
| 62.2.55 | 82008 | 144 | 0.0037 | -0.0230 | 2 | 1 | 2 | 1 | - | - | - | - | z4=3 | Nd1O3Ti1 |
| 62.2.56 | 67723 | 168 | 0.0000 | -0.1627 | 2 | 1 | 2 | 1 | - | - | - | - | z4=3 | Nd1Ni1O3 |
| 62.2.57 | 67722 | 168 | 0.0026 | -0.0687 | 2 | 1 | 2 | 1 | - | - | - | - | z4=3 | Nd1Ni1O3 |
| 62.2.58 | 75008 | 88 | 0.0058 | -0.5315 | 2 | 1 | 2 | 1 | - | - | - | - | z4=3 | Nb1P1Zr1 |
| 62.2.59 | 49728 | 112 | 0.0156 | -0.2430 | 2 | 1 | 2 | 1 | - | - | - | - | z4=3 | Nb1Ni1P1 |
| 62.2.60 | 41149 | 80 | 0.0072 | -0.1414 | 2 | 3 | 0 | 2 | - | - | - | - | z4=2 | Mn1Rh1Si1 |
| 62.2.61 | 43399 | 48 | 0.0051 | -0.2204 | 2 | 1 | 2 | 1 | - | - | - | - | z4=3 | Mn1P1 |
| 62.2.62 | 43246 | 48 | 0.0052 | -0.2205 | 2 | 1 | 2 | 1 | - | - | - | - | z4=3 | Mn1P1 |
| 62.2.63 | 185418 | 120 | 0.0039 | -0.1365 | 2 | 2 | 2 | 1 | - | - | - | - | z4=1 | Mn1O3V1 |

Continued on next page

**TABLE XIV– continued**

| TQC | ICSD | NoE | DG | IndG | SubSG (w/ T) | cl | SubSG (w/o T) | cl | Max. Sub. SG | yes or no | Max. Sub. SG | yes or no | SIs (w/ T) | Stoich. Formula |
|---|---|---|---|---|---|---|---|---|---|---|---|---|---|---|
| 62.2.64 | 97084 | 152 | 0.0109 | -0.0114 | 2 | 1 | 2 | 1 | - | - | - | - | z4=3 | Mn1O3Pr1 |
| 62.2.65 | 97083 | 152 | 0.0111 | -0.0096 | 2 | 1 | 2 | 1 | - | - | - | - | z4=3 | Mn1O3Pr1 |
| 62.2.66 | 97082 | 152 | 0.0063 | -0.0109 | 2 | 1 | 2 | 1 | - | - | - | - | z4=3 | Mn1O3Pr1 |
| 62.2.67 | 84922 | 152 | 0.0058 | -0.0245 | 2 | 1 | 2 | 1 | - | - | - | - | z4=3 | Mn1O3Pr1 |
| 62.2.68 | 426614 | 88 | 0.0738 | -1.4661 | 2 | 1 | 2 | 1 | - | - | - | - | z4=3 | Mg1Pt1Sr1 |
| 62.2.69 | 97765 | 160 | 0.0288 | -0.1069 | 2 | 2 | 2 | 1 | - | - | - | - | z4=1 | Li1Ni1O4P1 |
| 62.2.70 | 72929 | 160 | 0.0295 | -0.0735 | 2 | 2 | 2 | 1 | - | - | - | - | z4=1 | Li1Ni1O4P1 |
| 62.2.71 | 402760 | 160 | 0.0293 | -0.0939 | 2 | 2 | 2 | 1 | - | - | - | - | z4=1 | Li1Ni1O4P1 |
| 62.2.72 | 417912 | 104 | 0.0307 | -0.1070 | 2 | 1 | 2 | 1 | - | - | - | - | z4=3 | La1P1Pt1 |
| 62.2.73 | 415749 | 200 | 0.0169 | -0.0112 | 2 | 2 | 2 | 1 | - | - | - | - | z4=1 | K2O4Ru1 |
| 62.2.74 | 427252 | 96 | 0.0314 | -1.3402 | 2 | 2 | 2 | 1 | - | - | - | - | z4=1 | Ir1Sn1Y1 |
| 62.2.75 | 427253 | 96 | 0.0075 | -0.6912 | 2 | 2 | 2 | 1 | - | - | - | - | z4=1 | Ir1Si1Y1 |
| 62.2.76 | 411884 | 72 | 0.0362 | -0.3803 | 2 | 3 | 0 | 2 | - | - | - | - | z4=2 | Ir1Si1Ta1 |
| 62.2.77 | 421256 | 120 | 0.0560 | -0.0705 | 2 | 2 | 2 | 1 | - | - | - | - | z4=1 | Ir1Sc3Si3 |
| 62.2.78 | 420416 | 64 | 0.0511 | -0.4056 | 2 | 2 | 2 | 1 | - | - | - | - | z4=1 | Ir1Sc1Si1 |
| 62.2.79 | 422623 | 152 | 0.0042 | -0.9923 | 2 | 2 | 2 | 1 | - | - | - | - | z4=1 | Ir1Sb1Yb1 |
| 62.2.80 | 411882 | 104 | 0.0312 | -0.3546 | 2 | 3 | 0 | 2 | - | - | - | - | z4=2 | Ir1Nb1Si1 |
| 62.2.81 | 246548 | 80 | 0.0167 | -0.1775 | 2 | 3 | 0 | 2 | - | - | - | - | z4=2 | Ir1Mn1Si1 |
| 62.2.82 | 246 - | 80 | 0.0079 | -0.1698 | 2 | 3 | 0 | 2 | - | - | - | - | z4=2 | Ir1Mn1Si1 |
| 62.2.83 | 413857 | 152 | 0.0258 | -0.6568 | 2 | 2 | 2 | 1 | - | - | - | - | z4=1 | Ir1Lu1Si1 |
| 62.2.84 | 240201 | 72 | 0.0426 | -0.5857 | 2 | 1 | 2 | 1 | - | - | - | - | z4=3 | In3Ir1 |
| 62.2.85 | 412913 | 144 | 0.0077 | -0.9940 | 2 | 1 | 2 | 1 | - | - | - | - | z4=3 | In1Rh1Yb1 |
| 62.2.86 | 72291 | 240 | 0.0044 | -0.0255 | 2 | 3 | 0 | 2 | - | - | - | - | z4=2 | Ho2S3 |
| 62.2.87 | 99546 | 168 | 0.0031 | -0.0081 | 2 | 1 | 2 | 1 | - | - | - | - | z4=3 | Ho1O3Sc1 |
| 62.2.88 | 165196 | 184 | 0.0004 | -0.0346 | 2 | 2 | 2 | 1 | - | - | - | - | z4=1 | Ho1Mn1O3 |
| 62.2.89 | 82727 | 88 | 0.0000 | -0.5179 | 2 | 2 | 2 | 1 | - | - | - | - | z4=1 | Hf3P2 |
| 62.2.90 | 260839 | 176 | 0.0050 | -0.0391 | 2 | 3 | 0 | 2 | - | - | - | - | z4=2 | H2Cl1O2Tm1 |
| 62.2.91 | 42912 | 128 | 0.0058 | -2.4793 | 2 | 1 | 2 | 1 | - | - | - | - | z4=3 | Ge3Pt2 |
| 62.2.92 | 16822 | 56 | 0.0247 | -0.3474 | 2 | 1 | 2 | 1 | - | - | - | - | z4=3 | Ge2Mo1 |
| 62.2.93 | 429881 | 48 | 0.0070 | -1.7937 | 2 | 1 | 2 | 1 | - | - | - | - | z4=3 | Ge2Li3Na1 |
| 62.2.94 | 280691 | 128 | 0.0051 | -0.0650 | 2 | 1 | 2 | 1 | - | - | - | - | z4=3 | Ge2In1Li1Sr2 |
| 62.2.95 | 44860 | 88 | 0.0197 | -0.2314 | 2 | 3 | 0 | 2 | - | - | - | - | z4=2 | Ge1Rh2 |
| 62.2.96 | 421092 | 72 | 0.0283 | -0.2482 | 2 | 1 | 2 | 1 | - | - | - | - | z4=3 | Ge1Rh1Ta1 |
| 62.2.97 | 260793 | 72 | 0.0263 | -0.2003 | 2 | 1 | 2 | 1 | - | - | - | - | z4=3 | Ge1Pt1Zr1 |
| 62.2.98 | 419135 | 152 | 0.0055 | -1.1997 | 2 | 2 | 2 | 1 | - | - | - | - | z4=1 | Ge1Pt1Yb1 |
| 62.2.99 | 2624 | 56 | 0.1075 | -1.3524 | 2 | 3 | 0 | 2 | - | - | - | - | z4=2 | Ge1Pt1 |
| 62.2.100 | 409926 | 72 | 0.0230 | -0.1328 | 2 | 1 | 2 | 1 | - | - | - | - | z4=3 | Ge1Pd1Ti1 |
| 62.2.101 | 413387 | 328 | 0.0097 | -0.0161 | 2 | 1 | 2 | 1 | - | - | - | - | z4=3 | Ge1O1Yb3 |
| 62.2.102 | 237302 | 64 | 0.0138 | -1.8689 | 2 | 2 | 2 | 1 | - | - | - | - | z4=1 | Ge1Mg1Ni1 |
| 62.2.103 | 428468 | 64 | 0.0073 | -0.3361 | 2 | 2 | 2 | 1 | - | - | - | - | z4=1 | Ge1Ir1Sc1 |
| 62.2.104 | 411883 | 104 | 0.0228 | -0.2921 | 2 | 3 | 0 | 2 | - | - | - | - | z4=2 | Ge1Ir1Nb1 |
| 62.2.105 | 15534 | 288 | 0.0000 | -0.0182 | 2 | 2 | 2 | 1 | - | - | - | - | z4=1 | Gd3Ni1Si2 |
| 62.2.106 | 280201 | 216 | 0.0000 | -0.0023 | 2 | 1 | 2 | 1 | - | - | - | - | z4=3 | Gd2O1Se2 |
| 62.2.107 | 1375 | 232 | 0.0000 | -0.0086 | 2 | 3 | 0 | 2 | - | - | - | - | z4=2 | Gd1P1S1 |
| 62.2.108 | 419793 | 120 | 0.0037 | -0.0180 | 2 | 2 | 2 | 1 | - | - | - | - | z4=1 | Gd1Ge1Ru1 |
| 62.2.109 | 103904 | 224 | 0.0759 | -0.5714 | 2 | 1 | 2 | 1 | - | - | - | - | z4=3 | Ga2Pd5 |
| 62.2.110 | 420711 | 168 | 0.0021 | -0.0061 | 2 | 1 | 2 | 1 | - | - | - | - | z4=3 | Ga1Ho1O3 |
| 62.2.111 | 55537 | 104 | 0.0250 | -0.1170 | 2 | 2 | 2 | 1 | - | - | - | - | z4=1 | Fe1Si1U1 |
| 62.2.112 | 169317 | 56 | 0.0417 | -0.7604 | 2 | 1 | 2 | 1 | - | - | - | - | z4=3 | Fe1Se1 |
| 62.2.113 | 169316 | 56 | 0.0322 | -0.7996 | 2 | 2 | 2 | 1 | - | - | - | - | z4=1 | Fe1Se1 |
| 62.2.114 | 169314 | 56 | 0.0028 | -0.7035 | 2 | 1 | 2 | 1 | - | - | - | - | z4=3 | Fe1Se1 |
| 62.2.115 | 169312 | 56 | 0.0176 | -0.7873 | 2 | 2 | 2 | 1 | - | - | - | - | z4=1 | Fe1Se1 |
| 62.2.116 | 169310 | 56 | 0.0089 | -0.7958 | 2 | 2 | 2 | 1 | - | - | - | - | z4=1 | Fe1Se1 |
| 62.2.117 | 169308 | 56 | 0.0165 | -0.8542 | 2 | 2 | 2 | 1 | - | - | - | - | z4=1 | Fe1Se1 |
| 62.2.118 | 169305 | 56 | 0.0176 | -0.7093 | 2 | 1 | 2 | 1 | - | - | - | - | z4=3 | Fe1Se1 |
| 62.2.119 | 169282 | 56 | 0.0369 | -0.5630 | 2 | 1 | 2 | 1 | - | - | - | - | z4=3 | Fe1Se1 |
| 62.2.120 | 169281 | 56 | 0.0683 | -0.4722 | 2 | 1 | 2 | 1 | - | - | - | - | z4=3 | Fe1Se1 |

Continued on next page

**TABLE XIV– continued**

| TQC | ICSD | NoE | DG | IndG | SubSG cl (w/ T) | SubSG cl (w/o T) | Max. Sub. SG yes or no | Max. Sub. SG yes or no | SIs (w/ T) | Stoich. Formula |
|---|---|---|---|---|---|---|---|---|---|---|
| 62.2.121 | 169280 | 56 | 0.0794 | -0.4257 | 2 1 | 2 1 | - - | - - | z4=3 | Fe1Se1 |
| 62.2.122 | 169279 | 56 | 0.0122 | -0.4876 | 2 1 | 2 1 | - - | - - | z4=3 | Fe1Se1 |
| 62.2.123 | 169278 | 56 | 0.0148 | -0.4855 | 2 1 | 2 1 | - - | - - | z4=3 | Fe1Se1 |
| 62.2.124 | 169277 | 56 | 0.0040 | -0.4589 | 2 1 | 2 1 | - - | - - | z4=3 | Fe1Se1 |
| 62.2.125 | 169276 | 56 | 0.0377 | -0.4636 | 2 2 | 2 1 | - - | - - | z4=1 | Fe1Se1 |
| 62.2.126 | 169275 | 56 | 0.0517 | -0.4979 | 2 2 | 2 1 | - - | - - | z4=1 | Fe1Se1 |
| 62.2.127 | 169274 | 56 | 0.0137 | -0.4310 | 2 1 | 2 1 | - - | - - | z4=3 | Fe1Se1 |
| 62.2.128 | 169273 | 56 | 0.0347 | -0.4609 | 2 1 | 2 1 | - - | - - | z4=3 | Fe1Se1 |
| 62.2.129 | 35010 | 56 | 0.0021 | -0.4695 | 2 2 | 2 1 | - - | - - | z4=1 | Fe1S1 |
| 62.2.130 | 35009 | 56 | 0.0040 | -0.4678 | 2 2 | 2 1 | - - | - - | z4=1 | Fe1S1 |
| 62.2.131 | 35008 | 56 | 0.0247 | -0.2753 | 2 1 | 2 1 | - - | - - | z4=3 | Fe1S1 |
| 62.2.132 | 49729 | 72 | 0.0328 | -0.3037 | 2 3 | 0 2 | - - | - - | z4=2 | Fe1P1Ta1 |
| 62.2.133 | 153236 | 160 | 0.0070 | -0.1242 | 2 1 | 2 1 | - - | - - | z4=3 | Fe1Nd1O3 |
| 62.2.134 | 32756 | 200 | 0.0168 | -0.0057 | 2 1 | 2 1 | - - | - - | z4=3 | Fe1K2O4 |
| 62.2.135 | 2876 | 200 | 0.0148 | -0.0085 | 2 1 | 2 1 | - - | - - | z4=3 | Fe1K2O4 |
| 62.2.136 | 240891 | 136 | 0.0086 | -0.0682 | 2 1 | 2 1 | - - | - - | z4=3 | F2K1N1O1 |
| 62.2.137 | 2635 | 96 | 0.0240 | -0.0260 | 2 1 | 2 1 | - - | - - | z4=3 | F1Se1Y1 |
| 62.2.138 | 195382 | 136 | 0.0000 | -0.0143 | 2 3 | 0 2 | - - | - - | z4=2 | F1Ho1Se1 |
| 62.2.139 | 405049 | 120 | 0.0000 | -0.1311 | 2 2 | 2 1 | - - | - - | z4=1 | Eu1In1Pt1 |
| 62.2.140 | 404044 | 120 | 0.0022 | -0.0904 | 2 2 | 2 1 | - - | - - | z4=1 | Eu1In1Pd1 |
| 62.2.141 | 423115 | 176 | 0.0000 | -0.0205 | 2 1 | 2 1 | - - | - - | z4=3 | Er1Pt1Zn1 |
| 62.2.142 | 27282 | 192 | 0.0000 | -0.0093 | 2 1 | 2 1 | - - | - - | z4=3 | Er1Fe1O3 |
| 62.2.143 | 419 | 232 | 0.0000 | -0.0185 | 2 1 | 2 1 | - - | - - | z4=3 | Dy2Se3 |
| 62.2.144 | 80768 | 232 | 0.0000 | -0.0211 | 2 1 | 2 1 | - - | - - | z4=3 | Dy2S3 |
| 62.2.145 | 66532 | 232 | 0.0000 | -0.0200 | 2 1 | 2 1 | - - | - - | z4=3 | Dy2S3 |
| 62.2.146 | 66531 | 232 | 0.0000 | -0.0211 | 2 1 | 2 1 | - - | - - | z4=3 | Dy2S3 |
| 62.2.147 | 36603 | 232 | 0.0000 | -0.0145 | 2 1 | 2 1 | - - | - - | z4=3 | Dy2O1S2 |
| 62.2.148 | 180220 | 96 | 0.0000 | 0.0005 | 2 3 | 0 2 | - - | - - | z4=2 | Dy1Si1 |
| 62.2.149 | 280091 | 184 | 0.0030 | -0.0018 | 2 3 | 0 2 | - - | - - | z4=2 | Dy1Fe1O3 |
| 62.2.150 | 27280 | 184 | 0.0029 | -0.0015 | 2 3 | 0 2 | - - | - - | z4=2 | Dy1Fe1O3 |
| 62.2.151 | 86391 | 72 | 0.2357 | -0.3966 | 2 1 | 2 1 | - - | - - | z4=3 | Cu1Sc1Si1 |
| 62.2.152 | 172850 | 184 | 0.0025 | -0.2201 | 2 1 | 2 1 | - - | - - | z4=3 | Cu1S2Tm1 |
| 62.2.153 | 172848 | 176 | 0.0000 | -0.0346 | 2 3 | 0 2 | - - | - - | z4=2 | Cu1Ho1S2 |
| 62.2.154 | 33799 | 200 | 0.0416 | 0.0286 | 2 1 | 2 1 | - - | - - | z4=3 | Cs2O4Ru1 |
| 62.2.155 | 33861 | 200 | 0.0327 | -0.0325 | 2 1 | 2 1 | - - | - - | z4=3 | Cs2Fe1O4 |
| 62.2.156 | 192388 | 176 | 0.0099 | -0.2844 | 2 1 | 2 1 | - - | - - | z4=3 | Cr2Fe1O4 |
| 62.2.157 | 80332 | 152 | 0.0000 | -0.0400 | 2 3 | 0 2 | - - | - - | z4=2 | Cr1Nd1O3 |
| 62.2.158 | 79349 | 152 | 0.0000 | -0.0408 | 2 3 | 0 2 | - - | - - | z4=2 | Cr1Nd1O3 |
| 62.2.159 | 156326 | 152 | 0.0000 | -0.0514 | 2 3 | 0 2 | - - | - - | z4=2 | Cr1Nd1O3 |
| 62.2.160 | 156322 | 152 | 0.0000 | -0.0403 | 2 3 | 0 2 | - - | - - | z4=2 | Cr1Nd1O3 |
| 62.2.161 | 156320 | 152 | 0.0000 | -0.0593 | 2 3 | 0 2 | - - | - - | z4=2 | Cr1Nd1O3 |
| 62.2.162 | 251083 | 184 | 0.0000 | -0.0263 | 2 1 | 2 1 | - - | - - | z4=3 | Cr1Er1O3 |
| 62.2.163 | 183759 | 184 | 0.0000 | -0.0302 | 2 1 | 2 1 | - - | - - | z4=3 | Cr1Er1O3 |
| 62.2.164 | 65751 | 184 | 0.0116 | -0.0978 | 2 2 | 2 1 | - - | - - | z4=1 | Co2O4Si1 |
| 62.2.165 | 6248 | 184 | 0.0128 | -0.0987 | 2 2 | 2 1 | - - | - - | z4=1 | Co2O4Si1 |
| 62.2.166 | 50072 | 184 | 0.0304 | -0.0910 | 2 2 | 2 1 | - - | - - | z4=1 | Co2O4Si1 |
| 62.2.167 | 260092 | 184 | 0.0129 | -0.0989 | 2 2 | 2 1 | - - | - - | z4=1 | Co2O4Si1 |
| 62.2.168 | 260091 | 184 | 0.0109 | -0.0982 | 2 2 | 2 1 | - - | - - | z4=1 | Co2O4Si1 |
| 62.2.169 | 202255 | 184 | 0.0117 | -0.0979 | 2 2 | 2 1 | - - | - - | z4=1 | Co2O4Si1 |
| 62.2.170 | 200705 | 184 | 0.0076 | -0.0952 | 2 2 | 2 1 | - - | - - | z4=1 | Co2O4Si1 |
| 62.2.171 | 192245 | 168 | 0.0041 | -0.0861 | 2 2 | 2 1 | - - | - - | z4=1 | Co1Y3 |
| 62.2.172 | 192244 | 168 | 0.0293 | -0.0375 | 2 2 | 2 1 | - - | - - | z4=1 | Co1Y3 |
| 62.2.173 | 409847 | 72 | 0.0083 | -0.1915 | 2 3 | 0 2 | - - | - - | z4=2 | Co1Si1V1 |
| 62.2.174 | 49726 | 72 | 0.0247 | -0.1529 | 2 1 | 2 1 | - - | - - | z4=3 | Co1P1Zr1 |
| 62.2.175 | 43401 | 56 | 0.0255 | -0.5441 | 2 3 | 0 2 | - - | - - | z4=2 | Co1P1 |
| 62.2.176 | 43249 | 56 | 0.0190 | -0.5227 | 2 3 | 0 2 | - - | - - | z4=2 | Co1P1 |
| 62.2.177 | 172046 | 184 | 0.0037 | -0.0086 | 2 1 | 2 1 | - - | - - | z4=3 | Co1O3Tb1 |

Continued on next page

**TABLE XIV– continued**

| TQC | ICSD | NoE | DG | IndG | SubSG (w/ T) | cl | SubSG (w/o T) | cl | Max. Sub. SG | yes or no | Max. Sub. SG | yes or no | SIs (w/ T) | Stoich. Formula |
|---|---|---|---|---|---|---|---|---|---|---|---|---|---|---|
| 62.2.178 | 88738 | 160 | 0.0000 | -0.0578 | 2 | 1 | 2 | 1 | - | - | - | - | z4=3 | Co1O3Pr1 |
| 62.2.179 | 94437 | 192 | 0.0023 | -0.0036 | 2 | 1 | 2 | 1 | - | - | - | - | z4=3 | Co1O2Rb3 |
| 62.2.180 | 2421 | 80 | 0.0089 | -0.2291 | 2 | 3 | 0 | 2 | - | - | - | - | z4=2 | Co1Mo1P1 |
| 62.2.181 | 150531 | 168 | 0.0053 | -0.2213 | 2 | 3 | 0 | 2 | - | - | - | - | z4=2 | Co1La3 |
| 62.2.182 | 73212 | 192 | 0.0123 | 0.0016 | 2 | 3 | 0 | 2 | - | - | - | - | z4=2 | Co1K3O2 |
| 62.2.183 | 172048 | 192 | 0.0027 | -0.0138 | 2 | 1 | 2 | 1 | - | - | - | - | z4=3 | Co1Ho1O3 |
| 62.2.184 | 102428 | 120 | 0.0000 | -0.0221 | 2 | 1 | 2 | 1 | - | - | - | - | z4=3 | Co1Ga1Gd1 |
| 62.2.185 | 48206 | 112 | 0.0000 | -0.0133 | 2 | 1 | 2 | 1 | - | - | - | - | z4=3 | Cl2Nd1 |
| 62.2.186 | 157332 | 232 | 0.0178 | 0.0053 | 2 | 3 | 0 | 2 | - | - | - | - | z4=2 | Ce3In2Ru2 |
| 62.2.187 | 428424 | 168 | 0.0000 | -0.0286 | 2 | 2 | 2 | 1 | - | - | - | - | z4=1 | Ce2Te3 |
| 62.2.188 | 106433 | 64 | 0.0197 | -0.0984 | 2 | 2 | 2 | 1 | - | - | - | - | z4=1 | Ce1Si1 |
| 62.2.189 | 418546 | 136 | 0.0000 | -0.0992 | 2 | 2 | 2 | 1 | - | - | - | - | z4=1 | Ce1Pt1Zn1 |
| 62.2.190 | 420207 | 136 | 0.0078 | -0.0847 | 2 | 2 | 2 | 1 | - | - | - | - | z4=1 | Ce1Pd1Zn1 |
| 62.2.191 | 409972 | 248 | 0.0262 | -0.3031 | 2 | 1 | 2 | 1 | - | - | - | - | z4=3 | Ce1P2Pt4 |
| 62.2.192 | - 33 | 176 | 0.0120 | -0.6917 | 2 | 1 | 2 | 1 | - | - | - | - | z4=3 | Ce1In4Pd2 |
| 62.2.193 | 99240 | 192 | 0.0000 | -0.2694 | 2 | 3 | 0 | 2 | - | - | - | - | z4=2 | Ce1In2Pd3 |
| 62.2.194 | 150105 | 312 | 0.0158 | -0.1182 | 2 | 3 | 0 | 2 | - | - | - | - | z4=2 | Ce1Cu6 |
| 62.2.195 | 237728 | 128 | 0.0378 | -1.6734 | 2 | 1 | 2 | 1 | - | - | - | - | z4=3 | Cd1Pt1Sr1 |
| 62.2.196 | 12159 | 72 | 0.0115 | 0.0022 | 2 | 1 | 2 | 1 | - | - | - | - | z4=3 | Cd1Na1Sb1 |
| 62.2.197 | 72642 | 64 | 0.1028 | -1.7858 | 2 | 2 | 2 | 1 | - | - | - | - | z4=1 | Ca1Pt1Si1 |
| 62.2.198 | 426616 | 56 | 0.0524 | -1.6261 | 2 | 1 | 2 | 1 | - | - | - | - | z4=3 | Ca1Mg1Pd1 |
| 62.2.199 | 410841 | 56 | 0.0045 | -1.3431 | 2 | 1 | 2 | 1 | - | - | - | - | z4=3 | Ca1In1Rh1 |
| 62.2.200 | 106343 | 64 | 0.0533 | -1.8495 | 2 | 2 | 2 | 1 | - | - | - | - | z4=1 | Ca1Ge1Pt1 |
| 62.2.201 | 291004 | 200 | 0.0221 | 0.0221 | 2 | 2 | 2 | 1 | - | - | - | - | z4=1 | Ca1Fe2O4 |
| 62.2.202 | 426610 | 96 | 0.1031 | -1.6825 | 2 | 1 | 2 | 1 | - | - | - | - | z4=3 | Ca1Cd1Pt1 |
| 62.2.203 | 426611 | 96 | 0.0386 | -1.7386 | 2 | 1 | 2 | 1 | - | - | - | - | z4=3 | Ca1Cd1Pd1 |
| 62.2.204 | 68116 | 112 | 0.0037 | -0.1798 | 2 | 1 | 2 | 1 | - | - | - | - | z4=3 | C2U1W1 |
| 62.2.205 | 16549 | 112 | 0.0054 | -0.2033 | 2 | 1 | 2 | 1 | - | - | - | - | z4=3 | C2Mo1U1 |
| 62.2.206 | 151477 | 104 | 0.0093 | -0.6846 | 2 | 3 | 0 | 2 | - | - | - | - | z4=2 | C2Cr3 |
| 62.2.207 | 15086 | 104 | 0.0052 | -0.6882 | 2 | 3 | 0 | 2 | - | - | - | - | z4=2 | C2Cr3 |
| 62.2.208 | 247128 | 136 | 0.1101 | 0.1101 | 2 | 1 | 2 | 1 | - | - | - | - | z4=3 | C1H2Cl1Cu1N2 |
| 62.2.209 | 236383 | 96 | 0.0108 | -0.9559 | 2 | 1 | 2 | 1 | - | - | - | - | z4=3 | Bi3Co1 |
| 62.2.210 | 193272 | 96 | 0.0082 | -0.9574 | 2 | 1 | 2 | 1 | - | - | - | - | z4=3 | Bi3Co1 |
| 62.2.211 | 428110 | 120 | 0.0092 | -0.1925 | 2 | 1 | 2 | 1 | - | - | - | - | z4=3 | Bi1Li1Yb1 |
| 62.2.212 | 52698 | 96 | 0.0171 | 0.0025 | 2 | 2 | 2 | 1 | - | - | - | - | z4=1 | Ba2Si1 |
| 62.2.213 | 416561 | 96 | 0.0435 | 0.0336 | 2 | 2 | 2 | 1 | - | - | - | - | z4=1 | Ba2Pb1 |
| 62.2.214 | 90336 | 304 | 0.0107 | -0.0316 | 2 | 3 | 0 | 2 | - | - | - | - | z4=2 | Ba1Ho2Te4 |
| 62.2.215 | 424857 | 328 | 0.0719 | -0.5958 | 2 | 1 | 2 | 1 | - | - | - | - | z4=3 | Ba1Hg6 |
| 62.2.216 | 237304 | 96 | 0.0022 | -0.4329 | 2 | 2 | 2 | 1 | - | - | - | - | z4=1 | Ba1Ge1Ni1 |
| 62.2.217 | 424315 | 176 | 0.0000 | -0.2566 | 2 | 1 | 2 | 1 | - | - | - | - | z4=3 | Ba1Fe2Se3 |
| 62.2.218 | 424314 | 176 | 0.0032 | -0.2441 | 2 | 1 | 2 | 1 | - | - | - | - | z4=3 | Ba1Fe2Se3 |
| 62.2.219 | 290595 | 176 | 0.0029 | -0.2855 | 2 | 1 | 2 | 1 | - | - | - | - | z4=3 | Ba1Fe2Se3 |
| 62.2.220 | 185512 | 176 | 0.0069 | -0.2498 | 2 | 1 | 2 | 1 | - | - | - | - | z4=3 | Ba1Fe2Se3 |
| 62.2.221 | 41893 | 112 | 0.0122 | -0.4933 | 2 | 3 | 0 | 2 | - | - | - | - | z4=2 | B2Co1Nb1 |
| 62.2.222 | 421745 | 136 | 0.0080 | -0.0219 | 2 | 1 | 2 | 1 | - | - | - | - | z4=3 | B1O3Pr1 |
| 62.2.223 | 42894 | 72 | 0.0308 | -0.3989 | 2 | 1 | 2 | 1 | - | - | - | - | z4=3 | B1Co1Mo1 |
| 62.2.224 | 50842 | 96 | 0.0000 | -0.0487 | 2 | 1 | 2 | 1 | - | - | - | - | z4=3 | B1Ca2F1N2 |
| 62.2.225 | 263057 | 192 | 0.0270 | -0.5450 | 2 | 1 | 2 | 1 | - | - | - | - | z4=3 | Au3Ga2Rb1 |
| 62.2.226 | 263058 | 192 | 0.0187 | -0.5546 | 2 | 1 | 2 | 1 | - | - | - | - | z4=3 | Au3Cs1Ga2 |
| 62.2.227 | 418239 | 264 | 0.0370 | -1.1479 | 2 | 1 | 2 | 1 | - | - | - | - | z4=3 | Au3Ca3In1 |
| 62.2.228 | 249563 | 152 | 0.0122 | -1.6755 | 2 | 3 | 0 | 2 | - | - | - | - | z4=2 | Au2In2Sr1 |
| 62.2.229 | 249562 | 152 | 0.0000 | -1.3637 | 2 | 2 | 2 | 1 | - | - | - | - | z4=1 | Au2Ba1In2 |
| 62.2.230 | 391422 | 96 | 0.0623 | -1.0412 | 2 | 2 | 2 | 1 | - | - | - | - | z4=1 | Au1In1Sr1 |
| 62.2.231 | 420674 | 160 | 0.0033 | -0.1067 | 2 | 2 | 2 | 1 | - | - | - | - | z4=1 | Au1Eu1Zn1 |
| 62.2.232 | 411544 | 160 | 0.0023 | -0.1822 | 2 | 1 | 2 | 1 | - | - | - | - | z4=3 | Au1Cd1Eu1 |
| 62.2.233 | 408579 | 64 | 0.0471 | -1.3405 | 2 | 2 | 2 | 1 | - | - | - | - | z4=1 | Au1Ca1In1 |
| 62.2.234 | 106273 | 64 | 0.0502 | -1.5422 | 2 | 2 | 2 | 1 | - | - | - | - | z4=1 | Au1Ca1Ga1 |

Continued on next page

| TQC | ICSD | NoE | DG | IndG | SubSG (w/ T) | cl | SubSG (w/o T) | cl | Max. Sub. SG | yes or no | Max. Sub. SG | yes or no | SIs (w/ T) | Stoich. Formula |
|---|---|---|---|---|---|---|---|---|---|---|---|---|---|---|
| 62.2.235 | 168665 | 56 | 0.0040 | -0.3378 | 2 | 1 | 2 | 1 | - | - | - | - | z4=3 | As2Zr1 |
| 62.2.236 | 42446 | 40 | 0.0058 | -0.1819 | 2 | 3 | 0 | 2 | - | - | - | - | z4=2 | As1V1 |
| 62.2.237 | 51201 | 160 | 0.0150 | -0.0004 | 2 | 2 | 2 | 1 | - | - | - | - | z4=1 | As1Li1Ni1O4 |
| 62.2.238 | 245184 | 160 | 0.0173 | -0.0141 | 2 | 2 | 2 | 1 | - | - | - | - | z4=1 | As1Li1Ni1O4 |
| 62.2.239 | 412296 | 144 | 0.0628 | -0.0898 | 2 | 1 | 2 | 1 | - | - | - | - | z4=3 | As1Cu1Mg1 |
| 62.2.240 | 406953 | 72 | 0.0250 | -0.1164 | 2 | 1 | 2 | 1 | - | - | - | - | z4=3 | As1Co1Hf1 |
| 62.2.241 | 48030 | 56 | 0.0191 | -0.5765 | 2 | 3 | 0 | 2 | - | - | - | - | z4=2 | As1Co1 |
| 62.2.242 | 48029 | 56 | 0.0236 | -0.5724 | 2 | 3 | 0 | 2 | - | - | - | - | z4=2 | As1Co1 |
| 62.2.243 | 48028 | 56 | 0.0318 | -0.5675 | 2 | 3 | 0 | 2 | - | - | - | - | z4=2 | As1Co1 |
| 62.2.244 | 48026 | 56 | 0.0377 | -0.5607 | 2 | 3 | 0 | 2 | - | - | - | - | z4=2 | As1Co1 |
| 62.2.245 | 9571 | 72 | 0.0325 | 0.0050 | 2 | 1 | 2 | 1 | - | - | - | - | z4=3 | As1Cd1Na1 |
| 62.2.246 | 249802 | 184 | 0.0099 | -0.1023 | 2 | 1 | 2 | 1 | - | - | - | - | z4=3 | As1.01Ce1Se0.99 |
| 62.2.247 | 150173 | 104 | 0.0038 | -0.0478 | 2 | 3 | 0 | 2 | - | - | - | - | z4=2 | Al1Pr1Pt1 |
| 62.2.248 | 160051 | 184 | 0.0000 | -0.3028 | 2 | 2 | 2 | 1 | - | - | - | - | z4=1 | Al1Ce1Ru1 |
| 62.2.249 | 55623 | 192 | 0.0046 | -0.2721 | 2 | 3 | 0 | 2 | - | - | - | - | z4=2 | Al1Ce1Rh1 |
| 62.2.250 | 160052 | 192 | 0.0044 | -0.2712 | 2 | 3 | 0 | 2 | - | - | - | - | z4=2 | Al1Ce1Rh1 |
| 62.2.251 | 370027 | 152 | 0.0081 | -0.7855 | 2 | 2 | 2 | 1 | - | - | - | - | z4=1 | Al1Au1Yb1 |
| 62.2.252 | 370015 | 64 | 0.0075 | -1.3743 | 2 | 2 | 2 | 1 | - | - | - | - | z4=1 | Al1Au1Ca1 |
| 62.2.253 | 83983 | 160 | 0.0050 | -0.3184 | 2 | 1 | 2 | 1 | - | - | - | - | z4=3 | Ag1Sb1Yb1 |
| 62.2.254 | 420615 | 160 | 0.0000 | -0.0107 | 2 | 3 | 0 | 2 | - | - | - | - | z4=2 | Ag1As2Tb1 |
| 62.2.255 | 98736 | 136 | 0.0000 | -0.0550 | 2 | 1 | 2 | 1 | - | - | - | - | z4=3 | Ag1As2Pr1 |
| 62.2.256 | 420611 | 136 | 0.0000 | -0.0562 | 2 | 1 | 2 | 1 | - | - | - | - | z4=3 | Ag1As2Pr1 |
| 62.2.257 | 412816 | 136 | 0.0000 | -0.0550 | 2 | 1 | 2 | 1 | - | - | - | - | z4=3 | Ag1As2Pr1 |
| 62.3.1 | 104201 | 204 | - | - | - | - | - | - | 26 | yes | 26 | yes | - | Tl1Yb2 |
| | | | | | | | | | 11 | no(2) | 11 | no(2) | z2w,1=1 z2w,3=1 z4=3 | |
| | | | | | | | | | - | - | 7 | no(7) | - | |
| 62.3.2 | 54364 | 92 | - | - | - | - | - | - | 26 | yes | 26 | yes | - | Si1Tb1 |
| | | | | | | | | | 11 | no(2) | 11 | no(2) | z2w,1=1 z2w,3=1 z4=1 | |
| | | | | | | | | | - | - | 7 | no(7) | - | |
| 62.3.3 | 54363 | 92 | - | - | - | - | - | - | 26 | yes | 26 | yes | - | Si1Tb1 |
| | | | | | | | | | 11 | no(2) | 11 | no(2) | z2w,1=1 z2w,3=1 z4=1 | |
| | | | | | | | | | - | - | 7 | no(7) | - | |
| 62.3.4 | 54361 | 92 | - | - | - | - | - | - | 26 | yes | 26 | yes | - | Si1Tb1 |
| | | | | | | | | | 11 | no(2) | 11 | no(2) | z2w,1=1 z2w,3=1 z4=1 | |
| | | | | | | | | | - | - | 7 | no(7) | - | |
| 62.3.5 | 421254 | 116 | - | - | - | - | - | - | 26 | yes | 26 | yes | - | Ru1Sc3Si3 |
| | | | | | | | | | 11 | no(2) | 11 | yes | z2w,1=1 z2w,3=1 z4=2 | |
| | | | | | | | | | - | - | 7 | no(7) | - | |
| 62.3.6 | 420417 | 60 | - | - | - | - | - | - | 26 | yes | 26 | yes | - | Ru1Sc1Si1 |
| | | | | | | | | | 11 | no(2) | 11 | no(2) | z2w,1=1 z2w,3=1 z4=1 | |
| | | | | | | | | | - | - | 7 | no(7) | - | |
| 62.3.7 | 24354 | 172 | - | - | - | - | - | - | 26 | yes | 26 | yes | - | Rh3Si4 |
| | | | | | | | | | 11 | no(2) | 11 | yes | z2w,1=1 z2w,3=1 z4=2 | |
| | | | | | | | | | - | - | 7 | no(7) | - | |

Continued on next page

**TABLE XIV– continued**

| TQC | ICSD | NoE | DG | IndG | SubSG cl (w/ T) | SubSG cl (w/o T) | Max. Sub. SG | yes or no | Max. Sub. SG | yes or no | SIs (w/ T) | Stoich. Formula |
|---|---|---|---|---|---|---|---|---|---|---|---|---|
| 62.3.8 | 42960 | 92 | - | - | - - | - - | 26 | yes | 26 | yes | - | Rh2Sb1 |
| | | | | | | | 11 | no(2) | 11 | yes | z2w,1=1<br>z2w,3=1 | |
| | | | | | | | - | - | 7 | no(7) | - | |
| 62.3.9 | 51842 | 132 | - | - | - - | - - | 26 | yes | 26 | yes | - | Rh1Sb1Tb1 |
| | | | | | | | 11 | no(2) | 11 | no(2) | z2w,1=1<br>z2w,3=1<br>z4=1 | |
| | | | | | | | - | - | 7 | no(7) | - | |
| 62.3.10 | 202438 | 164 | - | - | - - | - - | 26 | yes | 26 | yes | - | Rh1S3U1 |
| | | | | | | | 11 | no(2) | 11 | no(2) | z2w,1=1<br>z2w,3=1<br>z4=1 | |
| | | | | | | | - | - | 7 | no(7) | - | |
| 62.3.11 | 8016 | 172 | - | - | - - | - - | 26 | yes | 26 | yes | - | Pt1Y3 |
| | | | | | | | 11 | no(2) | 11 | yes | z2w,1=1<br>z2w,3=1 | |
| | | | | | | | - | - | 7 | no(7) | - | |
| 62.3.12 | 417531 | 100 | - | - | - - | - - | 26 | yes | 26 | yes | - | Pt1Sn1Y1 |
| | | | | | | | 11 | no(2) | 11 | no(2) | z2w,1=1<br>z2w,3=1<br>z4=1 | |
| | | | | | | | - | - | 7 | no(7) | - | |
| 62.3.13 | 421257 | 124 | - | - | - - | - - | 26 | yes | 26 | yes | - | Pt1Sc3Si3 |
| | | | | | | | 11 | no(2) | 11 | yes | z2w,1=1<br>z2w,3=1 | |
| | | | | | | | - | - | 7 | no(7) | - | |
| 62.3.14 | 79597 | 68 | - | - | - - | - - | 26 | yes | 26 | yes | - | Pt1Sc1Si1 |
| | | | | | | | 11 | no(2) | 11 | yes | z2w,1=1<br>z2w,3=1 | |
| | | | | | | | - | - | 7 | no(7) | - | |
| 62.3.15 | 35087 | 140 | - | - | - - | - - | 26 | yes | 26 | yes | - | Pd2Si1Y1 |
| | | | | | | | 11 | no(2) | 11 | yes | z2w,1=1<br>z2w,3=1 | |
| | | | | | | | - | - | 7 | no(7) | - | |
| 62.3.16 | 183334 | 132 | - | - | - - | - - | 26 | yes | 26 | yes | - | Pd1Y1Zn1 |
| | | | | | | | 11 | no(2) | 11 | yes | z2w,1=1<br>z2w,3=1<br>z4=2 | |
| | | | | | | | - | - | 7 | no(7) | - | |
| 62.3.17 | 424012 | 100 | - | - | - - | - - | 26 | yes | 26 | yes | - | Pd1Sn1Y1 |
| | | | | | | | 11 | no(2) | 11 | yes | z2w,1=1<br>z2w,3=1<br>z4=2 | |
| | | | | | | | - | - | 7 | no(7) | - | |
| 62.3.18 | 420418 | 68 | - | - | - - | - - | 26 | yes | 26 | yes | - | Pd1Sc1Si1 |
| | | | | | | | 11 | no(2) | 11 | yes | z2w,1=1<br>z2w,3=1 | |
| | | | | | | | - | - | 7 | no(7) | - | |
| 62.3.19 | 92441 | 76 | - | - | - - | - - | 26 | yes | 26 | yes | - | Pd1Sb1Zr1 |
| | | | | | | | 11 | no(2) | 11 | no(2) | z2w,1=1<br>z2w,3=1<br>z4=3 | |
| | | | | | | | - | - | 7 | no(7) | - | |
| 62.3.20 | 391182 | 156 | - | - | - - | - - | 26 | yes | 26 | yes | - | Pd1Sb1Yb1 |
| | | | | | | | 11 | no(2) | 11 | yes | z2w,1=1<br>z2w,3=1<br>z4=2 | |

Continued on next page

**TABLE XIV– continued**

| TQC | ICSD | NoE | DG | IndG | SubSG cl (w/ T) | SubSG cl (w/o T) | Max. Sub. SG | yes or no | Max. Sub. SG | yes or no | SIs (w/ T) | Stoich. Formula |
|---|---|---|---|---|---|---|---|---|---|---|---|---|
| | | | | | | | - | - | 7 | no(7) | - | |
| 62.3.21 | 42056 | 44 | - | - | - - | - - | 26 | yes | 26 | yes | - | P1W1 |
| | | | | | | | 11 | no(2) | 11 | no(2) | z2w,1=1<br>z2w,3=1<br>z4=1 | |
| | | | | | | | - | - | 7 | no(7) | - | |
| 62.3.22 | 77856 | 60 | - | - | - - | - - | 26 | yes | 26 | yes | - | P1V2 |
| | | | | | | | 11 | no(2) | 11 | yes | z2w,1=1<br>z2w,3=1 | |
| | | | | | | | - | - | 7 | no(7) | - | |
| 62.3.23 | 30727 | 68 | - | - | - - | - - | 26 | yes | 26 | yes | - | P1Ru1Zr1 |
| | | | | | | | 11 | no(2) | 11 | no(2) | z2w,1=1<br>z2w,3=1<br>z4=3 | |
| | | | | | | | - | - | 7 | no(7) | - | |
| 62.3.24 | 48123 | 76 | - | - | - - | - - | 26 | yes | 26 | yes | - | P1Re2 |
| | | | | | | | 11 | no(2) | 11 | yes | z2w,1=1<br>z2w,3=1 | |
| | | | | | | | - | - | 7 | no(7) | - | |
| 62.3.25 | 186202 | 76 | - | - | - - | - - | 26 | yes | 26 | yes | - | P1Re2 |
| | | | | | | | 11 | no(2) | 11 | yes | z2w,1=1<br>z2w,3=1<br>z4=2 | |
| | | | | | | | - | - | 7 | no(7) | - | |
| 62.3.26 | 186200 | 76 | - | - | - - | - - | 26 | yes | 26 | yes | - | P1Re2 |
| | | | | | | | 11 | no(2) | 11 | yes | z2w,1=1<br>z2w,3=1<br>z4=2 | |
| | | | | | | | - | - | 7 | no(7) | - | |
| 62.3.27 | 84610 | 132 | - | - | - - | - - | 26 | yes | 26 | yes | - | O3Ti1Y1 |
| | | | | | | | 11 | no(2) | 11 | yes | z2w,1=1<br>z2w,3=1 | |
| | | | | | | | - | - | 7 | no(7) | - | |
| 62.3.28 | 84609 | 132 | - | - | - - | - - | 26 | yes | 26 | yes | - | O3Ti1Y1 |
| | | | | | | | 11 | no(2) | 11 | yes | z2w,1=1<br>z2w,3=1 | |
| | | | | | | | - | - | 7 | no(7) | - | |
| 62.3.29 | 84608 | 132 | - | - | - - | - - | 26 | yes | 26 | yes | - | O3Ti1Y1 |
| | | | | | | | 11 | no(2) | 11 | yes | z2w,1=1<br>z2w,3=1 | |
| | | | | | | | - | - | 7 | no(7) | - | |
| 62.3.30 | 8150 | 132 | - | - | - - | - - | 26 | yes | 26 | yes | - | O3Ti1Y1 |
| | | | | | | | 11 | no(2) | 11 | yes | z2w,1=1<br>z2w,3=1<br>z4=2 | |
| | | | | | | | - | - | 7 | no(7) | - | |
| 62.3.31 | 183450 | 140 | - | - | - - | - - | 26 | yes | 26 | yes | - | O3Sr1Tc1 |
| | | | | | | | 11 | no(2) | 11 | yes | z2w,1=1<br>z2w,3=1<br>z4=2 | |
| | | | | | | | - | - | 7 | no(7) | - | |
| 62.3.32 | 66346 | 148 | - | - | - - | - - | 26 | yes | 26 | yes | - | Ni2Ta1Te2 |
| | | | | | | | 11 | no(2) | 11 | no(2) | z2w,1=1<br>z2w,3=1<br>z4=1 | |
| | | | | | | | - | - | 7 | no(7) | - | |
| 62.3.33 | 86174 | 148 | - | - | - - | - - | 26 | yes | 26 | yes | - | Ni2Se1Ta1Te1 |

Continued on next page

**TABLE XIV– continued**

| TQC | ICSD | NoE | DG | IndG | SubSG cl (w/ T) | SubSG cl (w/o T) | Max. Sub. SG | yes or no | Max. Sub. SG | yes or no | SIs (w/ T) | Stoich. Formula |
|---|---|---|---|---|---|---|---|---|---|---|---|---|
| | | | | | | | 11 | no(2) | 11 | yes | z2w,1=1<br>z2w,3=1 | |
| | | | | | | | - | - | 7 | no(7) | - | |
| 62.3.34 | 105456 | 84 | - | - | - - | - - | 26 | yes | 26 | yes | - | Ni1Y1 |
| | | | | | | | 11 | no(2) | 11 | no(2) | z2w,1=1<br>z2w,3=1<br>z4=3 | |
| | | | | | | | - | - | 7 | no(7) | - | |
| 62.3.35 | 105379 | 100 | - | - | - - | - - | 26 | yes | 26 | yes | - | Ni1Sn1Y1 |
| | | | | | | | 11 | no(2) | 11 | yes | z2w,1=1<br>z2w,3=1 | |
| | | | | | | | - | - | 7 | no(7) | - | |
| 62.3.36 | 427285 | 132 | - | - | - - | - - | 26 | yes | 26 | yes | - | Ni1Sn1Tb1 |
| | | | | | | | 11 | no(2) | 11 | no(2) | z2w,1=1<br>z2w,3=1<br>z4=3 | |
| | | | | | | | - | - | 7 | no(7) | - | |
| 62.3.37 | 79598 | 100 | - | - | - - | - - | 26 | yes | 26 | yes | - | Ni1Si1Y1 |
| | | | | | | | 11 | no(2) | 11 | yes | z2w,1=1<br>z2w,3=1<br>z4=2 | |
| | | | | | | | - | - | 7 | no(7) | - | |
| 62.3.38 | 418977 | 68 | - | - | - - | - - | 26 | yes | 26 | yes | - | Ni1Sc1Sn1 |
| | | | | | | | 11 | no(2) | 11 | yes | z2w,1=1<br>z2w,3=1<br>z4=2 | |
| | | | | | | | - | - | 7 | no(7) | - | |
| 62.3.39 | 105338 | 68 | - | - | - - | - - | 26 | yes | 26 | yes | - | Ni1Sc1Sn1 |
| | | | | | | | 11 | no(2) | 11 | yes | z2w,1=1<br>z2w,3=1 | |
| | | | | | | | - | - | 7 | no(7) | - | |
| 62.3.40 | 41800 | 68 | - | - | - - | - - | 26 | yes | 26 | yes | - | Ni1Sc1Si1 |
| | | | | | | | 11 | no(2) | 11 | yes | z2w,1=1<br>z2w,3=1 | |
| | | | | | | | - | - | 7 | no(7) | - | |
| 62.3.41 | 408195 | 76 | - | - | - - | - - | 26 | yes | 26 | yes | - | Ni1Sb1Zr1 |
| | | | | | | | 11 | no(2) | 11 | yes | z2w,1=1<br>z2w,3=1<br>z4=2 | |
| | | | | | | | - | - | 7 | no(7) | - | |
| 62.3.42 | 427254 | 100 | - | - | - - | - - | 26 | yes | 26 | yes | - | Ni1Pb1Y1 |
| | | | | | | | 11 | no(2) | 11 | yes | z2w,1=1<br>z2w,3=1<br>z4=2 | |
| | | | | | | | - | - | 7 | no(7) | - | |
| 62.3.43 | 189296 | 204 | - | - | - - | - - | 26 | yes | 26 | yes | - | Ni1O3Tm1 |
| | | | | | | | 11 | no(2) | 11 | no(2) | z2w,1=1<br>z2w,3=1<br>z4=3 | |
| | | | | | | | - | - | 7 | no(7) | - | |
| 62.3.44 | 93507 | 124 | - | - | - - | - - | 26 | yes | 26 | yes | - | Ni1O3Tl1 |
| | | | | | | | 11 | no(2) | 11 | yes | z2w,1=1<br>z2w,3=1 | |
| | | | | | | | - | - | 7 | no(7) | - | |
| 62.3.45 | 69787 | 164 | - | - | - - | - - | 26 | yes | 26 | yes | - | Ni1O3Pr1 |
| | | | | | | | 11 | no(2) | 11 | no(2) | z2w,1=1<br>z2w,3=1<br>z4=3 | |

Continued on next page

**TABLE XIV– continued**

| TQC | ICSD | NoE | DG | IndG | SubSG (w/ T) | cl | SubSG (w/o T) | cl | Max. Sub. SG | yes or no | Max. Sub. SG | yes or no | SIs (w/ T) | Stoich. Formula |
|---|---|---|---|---|---|---|---|---|---|---|---|---|---|---|
| | | | | | | | | | - | - | 7 | no(7) | - | |
| 62.3.46 | 23281 | 204 | - | - | - | - | - | - | 26 | yes | 26 | yes | - | Nb2P5 |
| | | | | | | | | | 11 | no(2) | 11 | no(2) | z2w,1=1<br>z2w,3=1<br>z4=1 | |
| | | | | | | | | | - | - | 7 | no(7) | - | |
| 62.3.47 | 68103 | 212 | - | - | - | - | - | - | 26 | yes | 26 | yes | - | Nb1Pd1Te5 |
| | | | | | | | | | 11 | no(2) | 11 | no(2) | z2w,1=1<br>z2w,3=1<br>z4=1 | |
| | | | | | | | | | - | - | 7 | no(7) | - | |
| 62.3.48 | 416177 | 108 | - | - | - | - | - | - | 26 | yes | 26 | yes | - | Nb1Pd1Si1 |
| | | | | | | | | | 11 | no(2) | 11 | no(2) | z2w,1=1<br>z2w,3=1<br>z4=3 | |
| | | | | | | | | | - | - | 7 | no(7) | - | |
| 62.3.49 | 88723 | 164 | - | - | - | - | - | - | 26 | yes | 26 | yes | - | Nb1O3Sr1 |
| | | | | | | | | | 11 | no(2) | 11 | yes | z2w,1=1<br>z2w,3=1 | |
| | | | | | | | | | - | - | 7 | no(7) | - | |
| 62.3.50 | 20633 | 132 | - | - | - | - | - | - | 26 | yes | 26 | yes | - | Nb1Ni1P2 |
| | | | | | | | | | 11 | no(2) | 11 | yes | z2w,1=1<br>z2w,3=1<br>z4=2 | |
| | | | | | | | | | - | - | 7 | no(7) | - | |
| 62.3.51 | 100824 | 132 | - | - | - | - | - | - | 26 | yes | 26 | yes | - | Nb1Ni1P2 |
| | | | | | | | | | 11 | no(2) | 11 | yes | z2w,1=1<br>z2w,3=1<br>z4=2 | |
| | | | | | | | | | - | - | 7 | no(7) | - | |
| 62.3.52 | 202906 | 132 | - | - | - | - | - | - | 26 | yes | 26 | yes | - | Na1O4Ti2 |
| | | | | | | | | | 11 | no(2) | 11 | yes | z2w,1=1<br>z2w,3=1<br>z4=2 | |
| | | | | | | | | | - | - | 7 | no(7) | - | |
| 62.3.53 | 172608 | 164 | - | - | - | - | - | - | 26 | yes | 26 | yes | - | Na1O4Ru2 |
| | | | | | | | | | 11 | no(2) | 11 | no(2) | z2w,1=1<br>z2w,3=1<br>z4=3 | |
| | | | | | | | | | - | - | 7 | no(7) | - | |
| 62.3.54 | 170598 | 172 | - | - | - | - | - | - | 26 | yes | 26 | yes | - | Na1O4Rh2 |
| | | | | | | | | | 11 | no(2) | 11 | yes | z2w,1=1<br>z2w,3=1<br>z4=2 | |
| | | | | | | | | | - | - | 7 | no(7) | - | |
| 62.3.55 | 99549 | 132 | - | - | - | - | - | - | 26 | yes | 26 | yes | - | Na1O3U1 |
| | | | | | | | | | 11 | no(2) | 11 | yes | z2w,1=1<br>z2w,3=1 | |
| | | | | | | | | | - | - | 7 | no(7) | - | |
| 62.3.56 | 202889 | 132 | - | - | - | - | - | - | 26 | yes | 26 | yes | - | Na1O3U1 |
| | | | | | | | | | 11 | no(2) | 11 | yes | z2w,1=1<br>z2w,3=1 | |
| | | | | | | | | | - | - | 7 | no(7) | - | |
| 62.3.57 | 108856 | 132 | - | - | - | - | - | - | 26 | yes | 26 | yes | - | Na1O3U1 |
| | | | | | | | | | 11 | no(2) | 11 | yes | z2w,1=1<br>z2w,3=1 | |
| | | | | | | | | | - | - | 7 | no(7) | - | |
| 62.3.58 | 55622 | 116 | - | - | - | - | - | - | 26 | yes | 26 | yes | - | Mn3Sn2 |

Continued on next page

**TABLE XIV– continued**

| TQC | ICSD | NoE | DG | IndG | SubSG cl (w/ T) | SubSG cl (w/o T) | Max. Sub. SG | yes or no | Max. Sub. SG | yes or no | SIs (w/ T) | Stoich. Formula |
|---|---|---|---|---|---|---|---|---|---|---|---|---|
| | | | | | | | 11 | no(2) | 11 | no(2) | z2w,1=1<br>z2w,3=1<br>z4=1 | |
| | | | | | | | - | - | 7 | no(7) | - | |
| 62.3.59 | 150139 | 116 | - | - | - - | - - | 26 | yes | 26 | yes | - | Mn3Sn2 |
| | | | | | | | 11 | no(2) | 11 | no(2) | z2w,1=1<br>z2w,3=1<br>z4=1 | |
| | | | | | | | - | - | 7 | no(7) | - | |
| 62.3.60 | 95000 | 140 | - | - | - - | - - | 26 | yes | 26 | yes | - | Mn1Si1Yb1 |
| | | | | | | | 11 | no(2) | 11 | yes | z2w,1=1<br>z2w,3=1<br>z4=2 | |
| | | | | | | | - | - | 7 | no(7) | - | |
| 62.3.61 | 85947 | 68 | - | - | - - | - - | 26 | yes | 26 | yes | - | Mn1P1Ta1 |
| | | | | | | | 11 | no(2) | 11 | yes | z2w,1=1<br>z2w,3=1 | |
| | | | | | | | - | - | 7 | no(7) | - | |
| 62.3.62 | 158732 | 116 | - | - | - - | - - | 26 | yes | 26 | yes | - | Mn1O3Ti1 |
| | | | | | | | 11 | no(2) | 11 | yes | z2w,1=1<br>z2w,3=1 | |
| | | | | | | | - | - | 7 | no(7) | - | |
| 62.3.63 | 78331 | 76 | - | - | - - | - - | 26 | yes | 26 | yes | - | Mn1O2 |
| | | | | | | | 11 | no(2) | 11 | yes | z2w,1=1<br>z2w,3=1 | |
| | | | | | | | - | - | 7 | no(7) | - | |
| 62.3.64 | 171866 | 76 | - | - | - - | - - | 26 | yes | 26 | yes | - | Mn1O2 |
| | | | | | | | 11 | no(2) | 11 | no(2) | z2w,1=1<br>z2w,3=1<br>z4=3 | |
| | | | | | | | - | - | 7 | no(7) | - | |
| 62.3.65 | 68280 | 100 | - | - | - - | - - | 26 | yes | 26 | yes | - | Mn1Nb1P1 |
| | | | | | | | 11 | no(2) | 11 | yes | z2w,1=1<br>z2w,3=1 | |
| | | | | | | | - | - | 7 | no(7) | - | |
| 62.3.66 | 74784 | 108 | - | - | - - | - - | 26 | yes | 26 | yes | - | Mg1Ni2P1 |
| | | | | | | | 11 | no(2) | 11 | yes | z2w,1=1<br>z2w,3=1<br>z4=2 | |
| | | | | | | | - | - | 7 | no(7) | - | |
| 62.3.67 | 189097 | 148 | - | - | - - | - - | 26 | yes | 26 | yes | - | Li1Mn1O4P1 |
| | | | | | | | 11 | no(2) | 11 | no(2) | z2w,1=1<br>z2w,3=1<br>z4=1 | |
| | | | | | | | - | - | 7 | no(7) | - | |
| 62.3.68 | 427930 | 148 | - | - | - - | - - | 26 | yes | 26 | yes | - | La2Ni1Sb1 |
| | | | | | | | 11 | no(2) | 11 | yes | z2w,1=1<br>z2w,3=1 | |
| | | | | | | | - | - | 7 | no(7) | - | |
| 62.3.69 | 423625 | 60 | - | - | - - | - - | 26 | yes | 26 | yes | - | La1Si1 |
| | | | | | | | 11 | no(2) | 11 | yes | z2w,1=1<br>z2w,3=1<br>z4=2 | |
| | | | | | | | - | - | 7 | no(7) | - | |
| 62.3.70 | 35441 | 212 | - | - | - - | - - | 26 | yes | 26 | yes | - | La1S3Yb1 |
| | | | | | | | 11 | no(2) | 11 | yes | z2w,1=1<br>z2w,3=1 | |
| | | | | | | | - | - | 7 | no(7) | - | |

Continued on next page

**TABLE XIV– continued**

| TQC | ICSD | NoE | DG | IndG | SubSG cl (w/ T) | | SubSG cl (w/o T) | | Max. Sub. SG | yes or no | Max. Sub. SG | yes or no | SIs (w/ T) | Stoich. Formula |
|---|---|---|---|---|---|---|---|---|---|---|---|---|---|---|
| 62.3.71 | 99274 | 100 | - | - | - | - | - | - | 26 | yes | 26 | yes | - | La1Pt1Sn1 |
| | | | | | | | | | 11 | no(2) | 11 | yes | z2w,1=1<br>z2w,3=1<br>z4=2 | |
| | | | | | | | | | - | - | 7 | no(7) | - | |
| 62.3.72 | 416896 | 100 | - | - | - | - | - | - | 26 | yes | 26 | yes | - | La1Pt1Sn1 |
| | | | | | | | | | 11 | no(2) | 11 | yes | z2w,1=1<br>z2w,3=1<br>z4=2 | |
| | | | | | | | | | - | - | 7 | no(7) | - | |
| 62.3.73 | 183335 | 132 | - | - | - | - | - | - | 26 | yes | 26 | yes | - | La1Pd1Zn1 |
| | | | | | | | | | 11 | no(2) | 11 | yes | z2w,1=1<br>z2w,3=1 | |
| | | | | | | | | | - | - | 7 | no(7) | - | |
| 62.3.74 | 422484 | 100 | - | - | - | - | - | - | 26 | yes | 26 | yes | - | La1Pd1Sn1 |
| | | | | | | | | | 11 | no(2) | 11 | yes | z2w,1=1<br>z2w,3=1<br>z4=2 | |
| | | | | | | | | | - | - | 7 | no(7) | - | |
| 62.3.75 | 189419 | 212 | - | - | - | - | - | - | 26 | yes | 26 | yes | - | La1O3Yb1 |
| | | | | | | | | | 11 | no(2) | 11 | yes | z2w,1=1<br>z2w,3=1 | |
| | | | | | | | | | - | - | 7 | no(7) | - | |
| 62.3.76 | 98414 | 132 | - | - | - | - | - | - | 26 | yes | 26 | yes | - | La1O3Ti1 |
| | | | | | | | | | 11 | no(2) | 11 | yes | z2w,1=1<br>z2w,3=1 | |
| | | | | | | | | | - | - | 7 | no(7) | - | |
| 62.3.77 | 8146 | 132 | - | - | - | - | - | - | 26 | yes | 26 | yes | - | La1O3Ti1 |
| | | | | | | | | | 11 | no(2) | 11 | yes | z2w,1=1<br>z2w,3=1 | |
| | | | | | | | | | - | - | 7 | no(7) | - | |
| 62.3.78 | 63575 | 132 | - | - | - | - | - | - | 26 | yes | 26 | yes | - | La1O3Ti1 |
| | | | | | | | | | 11 | no(2) | 11 | yes | z2w,1=1<br>z2w,3=1<br>z4=2 | |
| | | | | | | | | | - | - | 7 | no(7) | - | |
| 62.3.79 | 63574 | 132 | - | - | - | - | - | - | 26 | yes | 26 | yes | - | La1O3Ti1 |
| | | | | | | | | | 11 | no(2) | 11 | yes | z2w,1=1<br>z2w,3=1 | |
| | | | | | | | | | - | - | 7 | no(7) | - | |
| 62.3.80 | 63573 | 132 | - | - | - | - | - | - | 26 | yes | 26 | yes | - | La1O3Ti1 |
| | | | | | | | | | 11 | no(2) | 11 | yes | z2w,1=1<br>z2w,3=1<br>z4=2 | |
| | | | | | | | | | - | - | 7 | no(7) | - | |
| 62.3.81 | 63572 | 132 | - | - | - | - | - | - | 26 | yes | 26 | yes | - | La1O3Ti1 |
| | | | | | | | | | 11 | no(2) | 11 | yes | z2w,1=1<br>z2w,3=1<br>z4=2 | |
| | | | | | | | | | - | - | 7 | no(7) | - | |
| 62.3.82 | 63571 | 132 | - | - | - | - | - | - | 26 | yes | 26 | yes | - | La1O3Ti1 |
| | | | | | | | | | 11 | no(2) | 11 | yes | z2w,1=1<br>z2w,3=1<br>z4=2 | |
| | | | | | | | | | - | - | 7 | no(7) | - | |
| 62.3.83 | 93099 | 156 | - | - | - | - | - | - | 26 | yes | 26 | yes | - | La1O3Pd1 |
| | | | | | | | | | 11 | no(2) | 11 | yes | z2w,1=1<br>z2w,3=1 | |

Continued on next page

**TABLE XIV– continued**

| TQC | ICSD | NoE | DG | IndG | SubSG cl (w/ T) | SubSG cl (w/o T) | Max. Sub. SG | yes or no | Max. Sub. SG | yes or no | SIs (w/ T) | Stoich. Formula |
|---|---|---|---|---|---|---|---|---|---|---|---|---|
| | | | | | | | - | - | 7 | no(7) | - | |
| 62.3.84 | 42731 | 172 | - | - | - - | - - | 26 | yes | 26 | yes | - | Ir3Si4 |
| | | | | | | | 11 | no(2) | 11 | yes | z2w,1=1<br>z2w,3=1<br>z4=2 | |
| | | | | | | | - | - | 7 | no(7) | - | |
| 62.3.85 | 411736 | 148 | - | - | - - | - - | 26 | yes | 26 | yes | - | Ir1Si1Yb1 |
| | | | | | | | 11 | no(2) | 11 | yes | z2w,1=1<br>z2w,3=1<br>z4=2 | |
| | | | | | | | - | - | 7 | no(7) | - | |
| 62.3.86 | 79236 | 52 | - | - | - - | - - | 26 | yes | 26 | yes | - | Ir1Si1 |
| | | | | | | | 11 | no(2) | 11 | yes | z2w,1=1<br>z2w,3=1<br>z4=2 | |
| | | | | | | | - | - | 7 | no(7) | - | |
| 62.3.87 | 410704 | 92 | - | - | - - | - - | 26 | yes | 26 | yes | - | In1Pt1Sr1 |
| | | | | | | | 11 | no(2) | 11 | yes | z2w,1=1<br>z2w,3=1<br>z4=2 | |
| | | | | | | | - | - | 7 | no(7) | - | |
| 62.3.88 | 417907 | 92 | - | - | - - | - - | 26 | yes | 26 | yes | - | In1Pd2 |
| | | | | | | | 11 | no(2) | 11 | yes | z2w,1=1<br>z2w,3=1 | |
| | | | | | | | - | - | 7 | no(7) | - | |
| 62.3.89 | 412912 | 148 | - | - | - - | - - | 26 | yes | 26 | yes | - | In1Lu1Rh1 |
| | | | | | | | 11 | no(2) | 11 | yes | z2w,1=1<br>z2w,3=1 | |
| | | | | | | | - | - | 7 | no(7) | - | |
| 62.3.90 | 26420 | 124 | - | - | - - | - - | 26 | yes | 26 | yes | - | I3Mn1Tl1 |
| | | | | | | | 11 | no(2) | 11 | no(2) | z2w,1=1<br>z2w,3=1<br>z4=1 | |
| | | | | | | | - | - | 7 | no(7) | - | |
| 62.3.91 | 1 - 01 | 100 | - | - | - - | - - | 26 | yes | 26 | yes | - | Ho1Si1 |
| | | | | | | | 11 | no(2) | 11 | no(2) | z2w,1=1<br>z2w,3=1<br>z4=1 | |
| | | | | | | | - | - | 7 | no(7) | - | |
| 62.3.92 | 24876 | 196 | - | - | - - | - - | 26 | yes | 26 | yes | - | Ho1Ni1O3 |
| | | | | | | | 11 | no(2) | 11 | no(2) | z2w,1=1<br>z2w,3=1<br>z4=3 | |
| | | | | | | | - | - | 7 | no(7) | - | |
| 62.3.93 | 427001 | 324 | - | - | - - | - - | 26 | yes | 26 | yes | - | Hg6K1 |
| | | | | | | | 11 | no(2) | 11 | yes | z2w,1=1<br>z2w,3=1 | |
| | | | | | | | - | - | 7 | no(7) | - | |
| 62.3.94 | 68790 | 68 | - | - | - - | - - | 26 | yes | 26 | yes | - | Hf1Rh1Si1 |
| | | | | | | | 11 | no(2) | 11 | no(2) | z2w,1=1<br>z2w,3=1<br>z4=1 | |
| | | | | | | | - | - | 7 | no(7) | - | |
| 62.3.95 | 237382 | 124 | - | - | - - | - - | 26 | yes | 26 | yes | - | H2Cl2Co1O1 |
| | | | | | | | 11 | no(2) | 11 | yes | z2w,1=1<br>z2w,3=1<br>z4=2 | |
| | | | | | | | - | - | 7 | no(7) | - | |

Continued on next page

**TABLE XIV– continued**

| TQC | ICSD | NoE | DG | IndG | SubSG cl (w/ T) | SubSG cl (w/o T) | Max. Sub. SG | yes or no | Max. Sub. SG | yes or no | SIs (w/ T) | Stoich. Formula |
|---|---|---|---|---|---|---|---|---|---|---|---|---|
| 62.3.96 | 245057 | 84 | - | - | - - | - - | 26 | yes | 26 | yes | - | H1Fe1O2 |
| | | | | | | | 11 | no(2) | 11 | no(2) | z2w,1=1<br>z2w,3=1<br>z4=3 | |
| | | | | | | | - | - | 7 | no(7) | - | |
| 62.3.97 | 41804 | 172 | - | - | - - | - - | 26 | yes | 26 | yes | - | H1Cu1I1O4 |
| | | | | | | | 11 | no(2) | 11 | yes | z2w,1=1<br>z2w,3=1<br>z4=2 | |
| | | | | | | | - | - | 7 | no(7) | - | |
| 62.3.98 | 429637 | 220 | - | - | - - | - - | 26 | yes | 26 | yes | - | Ge2Rh4Y1 |
| | | | | | | | 11 | no(2) | 11 | no(2) | z2w,1=1<br>z2w,3=1<br>z4=1 | |
| | | | | | | | - | - | 7 | no(7) | - | |
| 62.3.99 | 409768 | 132 | - | - | - - | - - | 26 | yes | 26 | yes | - | Ge2Li1Yb1 |
| | | | | | | | 11 | no(2) | 11 | yes | z2w,1=1<br>z2w,3=1<br>z4=2 | |
| | | | | | | | - | - | 7 | no(7) | - | |
| 62.3.100 | 162582 | 76 | - | - | - - | - - | 26 | yes | 26 | yes | - | Ge2Li1Sr1 |
| | | | | | | | 11 | no(2) | 11 | no(2) | z2w,1=1<br>z2w,3=1<br>z4=1 | |
| | | | | | | | - | - | 7 | no(7) | - | |
| 62.3.101 | 20520 | 204 | - | - | - - | - - | 26 | yes | 26 | yes | - | Ge2La3Ni1 |
| | | | | | | | 11 | no(2) | 11 | yes | z2w,1=1<br>z2w,3=1<br>z4=2 | |
| | | | | | | | - | - | 7 | no(7) | - | |
| 62.3.102 | 77447 | 148 | - | - | - - | - - | 26 | yes | 26 | yes | - | Ge1Rh1Yb1 |
| | | | | | | | 11 | no(2) | 11 | yes | z2w,1=1<br>z2w,3=1<br>z4=2 | |
| | | | | | | | - | - | 7 | no(7) | - | |
| 62.3.103 | 76355 | 148 | - | - | - - | - - | 26 | yes | 26 | yes | - | Ge1Rh1Yb1 |
| | | | | | | | 11 | no(2) | 11 | yes | z2w,1=1<br>z2w,3=1<br>z4=2 | |
| | | | | | | | - | - | 7 | no(7) | - | |
| 62.3.104 | 44861 | 52 | - | - | - - | - - | 26 | yes | 26 | yes | - | Ge1Rh1 |
| | | | | | | | 11 | no(2) | 11 | yes | z2w,1=1<br>z2w,3=1<br>z4=2 | |
| | | | | | | | - | - | 7 | no(7) | - | |
| 62.3.105 | 428474 | 68 | - | - | - - | - - | 26 | yes | 26 | yes | - | Ge1Pt1Sc1 |
| | | | | | | | 11 | no(2) | 11 | no(2) | z2w,1=1<br>z2w,3=1<br>z4=1 | |
| | | | | | | | - | - | 7 | no(7) | - | |
| 62.3.106 | 42379 | 68 | - | - | - - | - - | 26 | yes | 26 | yes | - | Ge1Pr1 |
| | | | | | | | 11 | no(2) | 11 | yes | z2w,1=1<br>z2w,3=1<br>z4=2 | |
| | | | | | | | - | - | 7 | no(7) | - | |
| 62.3.107 | 42378 | 68 | - | - | - - | - - | 26 | yes | 26 | yes | - | Ge1Pr1 |
| | | | | | | | 11 | no(2) | 11 | yes | z2w,1=1<br>z2w,3=1 | |

Continued on next page

**TABLE XIV– continued**

| TQC | ICSD | NoE | DG | IndG | SubSG cl (w/ T) | SubSG cl (w/o T) | Max. Sub. SG | yes or no | Max. Sub. SG | yes or no | SIs (w/ T) | Stoich. Formula |
|---|---|---|---|---|---|---|---|---|---|---|---|---|
| | | | | | | | - | - | 7 | no(7) | - | |
| 62.3.108 | 427255 | 100 | - | - | - - | - - | 26 | yes | 26 | yes | - | Ge1Ni1Y1 |
| | | | | | | | 11 | no(2) | 11 | yes | z2w,1=1<br>z2w,3=1<br>z4=2 | |
| | | | | | | | - | - | 7 | no(7) | - | |
| 62.3.109 | 428471 | 68 | - | - | - - | - - | 26 | yes | 26 | yes | - | Ge1Ni1Sc1 |
| | | | | | | | 11 | no(2) | 11 | no(2) | z2w,1=1<br>z2w,3=1<br>z4=3 | |
| | | | | | | | - | - | 7 | no(7) | - | |
| 62.3.110 | 186940 | 108 | - | - | - - | - - | 26 | yes | 26 | yes | - | Ge1Nb1Pt1 |
| | | | | | | | 11 | no(2) | 11 | no(2) | z2w,1=1<br>z2w,3=1<br>z4=1 | |
| | | | | | | | - | - | 7 | no(7) | - | |
| 62.3.111 | 425590 | 100 | - | - | - - | - - | 26 | yes | 26 | yes | - | Ge1Mn1U1 |
| | | | | | | | 11 | no(2) | 11 | yes | z2w,1=1<br>z2w,3=1 | |
| | | | | | | | - | - | 7 | no(7) | - | |
| 62.3.112 | 420232 | 156 | - | - | - - | - - | 26 | yes | 26 | yes | - | Ge1Lu1Pt1 |
| | | | | | | | 11 | no(2) | 11 | no(2) | z2w,1=1<br>z2w,3=1<br>z4=1 | |
| | | | | | | | - | - | 7 | no(7) | - | |
| 62.3.113 | 88247 | 156 | - | - | - - | - - | 26 | yes | 26 | yes | - | Ge1Lu1Ni1 |
| | | | | | | | 11 | no(2) | 11 | yes | z2w,1=1<br>z2w,3=1 | |
| | | | | | | | - | - | 7 | no(7) | - | |
| 62.3.114 | 189328 | 140 | - | - | - - | - - | 26 | yes | 26 | yes | - | Ge1La1Pd2 |
| | | | | | | | 11 | no(2) | 11 | yes | z2w,1=1<br>z2w,3=1<br>z4=2 | |
| | | | | | | | - | - | 7 | no(7) | - | |
| 62.3.115 | 423624 | 60 | - | - | - - | - - | 26 | yes | 26 | yes | - | Ge1La1 |
| | | | | | | | 11 | no(2) | 11 | yes | z2w,1=1<br>z2w,3=1<br>z4=2 | |
| | | | | | | | - | - | 7 | no(7) | - | |
| 62.3.116 | 414433 | 148 | - | - | - - | - - | 26 | yes | 26 | yes | - | Ge1Ir1Yb1 |
| | | | | | | | 11 | no(2) | 11 | yes | z2w,1=1<br>z2w,3=1<br>z4=2 | |
| | | | | | | | - | - | 7 | no(7) | - | |
| 62.3.117 | 57070 | 132 | - | - | - - | - - | 26 | yes | 26 | yes | - | Ge1Ho1Ru1 |
| | | | | | | | 11 | no(2) | 11 | no(2) | z2w,1=1<br>z2w,3=1<br>z4=3 | |
| | | | | | | | - | - | 7 | no(7) | - | |
| 62.3.118 | 419895 | 156 | - | - | - - | - - | 26 | yes | 26 | yes | - | Gd1Rh1Zn1 |
| | | | | | | | 11 | no(2) | 11 | yes | z2w,1=1<br>z2w,3=1 | |
| | | | | | | | - | - | 7 | no(7) | - | |
| 62.3.119 | 40391 | 164 | - | - | - - | - - | 26 | yes | 26 | yes | - | Gd1O3V1 |
| | | | | | | | 11 | no(2) | 11 | yes | z2w,1=1<br>z2w,3=1<br>z4=2 | |
| | | | | | | | - | - | 7 | no(7) | - | |

Continued on next page

**TABLE XIV– continued**

| TQC | ICSD | NoE | DG | IndG | SubSG cl (w/ T) | SubSG cl (w/o T) | Max. Sub. SG | yes or no | Max. Sub. SG | yes or no | SIs (w/ T) | Stoich. Formula |
|---|---|---|---|---|---|---|---|---|---|---|---|---|
| 62.3.120 | 95493 | 172 | - | - | - - | - - | 26 | yes | 26 | yes | - | Gd1Mn1O3 |
| | | | | | | | 11 | no(2) | 11 | yes | z2w,1=1 z2w,3=1 | |
| | | | | | | | - | - | 7 | no(7) | - | |
| 62.3.121 | 246649 | 148 | - | - | - - | - - | 26 | yes | 26 | yes | - | Ga1Pt1Yb1 |
| | | | | | | | 11 | no(2) | 11 | no(2) | z2w,1=1 z2w,3=1 z4=1 | |
| | | | | | | | - | - | 7 | no(7) | - | |
| 62.3.122 | 409939 | 92 | - | - | - - | - - | 26 | yes | 26 | yes | - | Ga1Pd2 |
| | | | | | | | 11 | no(2) | 11 | yes | z2w,1=1 z2w,3=1 z4=2 | |
| | | | | | | | - | - | 7 | no(7) | - | |
| 62.3.123 | 291062 | 148 | - | - | - - | - - | 26 | yes | 26 | yes | - | Ga1Ni1Yb1 |
| | | | | | | | 11 | no(2) | 11 | no(2) | z2w,1=1 z2w,3=1 z4=3 | |
| | | | | | | | - | - | 7 | no(7) | - | |
| 62.3.124 | 84203 | 60 | - | - | - - | - - | 26 | yes | 26 | yes | - | Fe1Sc1Si1 |
| | | | | | | | 11 | no(2) | 11 | yes | z2w,1=1 z2w,3=1 z4=2 | |
| | | | | | | | - | - | 7 | no(7) | - | |
| 62.3.125 | 76346 | 60 | - | - | - - | - - | 26 | yes | 26 | yes | - | Fe1Sc1Si1 |
| | | | | | | | 11 | no(2) | 11 | yes | z2w,1=1 z2w,3=1 z4=2 | |
| | | | | | | | - | - | 7 | no(7) | - | |
| 62.3.126 | 43204 | 68 | - | - | - - | - - | 26 | yes | 26 | yes | - | Fe1P1Zr1 |
| | | | | | | | 11 | no(2) | 11 | no(2) | z2w,1=1 z2w,3=1 z4=3 | |
| | | | | | | | - | - | 7 | no(7) | - | |
| 62.3.127 | 43400 | 52 | - | - | - - | - - | 26 | yes | 26 | yes | - | Fe1P1 |
| | | | | | | | 11 | no(2) | 11 | yes | z2w,1=1 z2w,3=1 | |
| | | | | | | | - | - | 7 | no(7) | - | |
| 62.3.128 | 43248 | 52 | - | - | - - | - - | 26 | yes | 26 | yes | - | Fe1P1 |
| | | | | | | | 11 | no(2) | 11 | yes | z2w,1=1 z2w,3=1 z4=2 | |
| | | | | | | | - | - | 7 | no(7) | - | |
| 62.3.129 | 92199 | 148 | - | - | - - | - - | 26 | yes | 26 | yes | - | Fe1O4P1 |
| | | | | | | | 11 | no(2) | 11 | no(2) | z2w,1=1 z2w,3=1 z4=1 | |
| | | | | | | | - | - | 7 | no(7) | - | |
| 62.3.130 | 290337 | 148 | - | - | - - | - - | 26 | yes | 26 | yes | - | Fe1O4P1 |
| | | | | | | | 11 | no(2) | 11 | no(2) | z2w,1=1 z2w,3=1 z4=1 | |
| | | | | | | | - | - | 7 | no(7) | - | |
| 62.3.131 | 80866 | 148 | - | - | - - | - - | 26 | yes | 26 | yes | - | Fe1O3Y1 |
| | | | | | | | 11 | no(2) | 11 | yes | z2w,1=1 z2w,3=1 | |
| | | | | | | | - | - | 7 | no(7) | - | |
| 62.3.132 | 43260 | 148 | - | - | - - | - - | 26 | yes | 26 | yes | - | Fe1O3Y1 |

Continued on next page

**TABLE XIV– continued**

| TQC | ICSD | NoE | DG | IndG | SubSG cl (w/ T) | SubSG cl (w/o T) | Max. Sub. SG | yes or no | Max. Sub. SG | yes or no | SIs (w/ T) | Stoich. Formula |
|---|---|---|---|---|---|---|---|---|---|---|---|---|
| | | | | | | | 11 | no(2) | 11 | yes | z2w,1=1<br>z2w,3=1 | |
| | | | | | | | - | - | 7 | no(7) | - | |
| 62.3.133 | 93835 | 116 | - | - | - - | - - | 26 | yes | 26 | yes | - | Fe1O3Tl1 |
| | | | | | | | 11 | no(2) | 11 | yes | z2w,1=1<br>z2w,3=1 | |
| | | | | | | | - | - | 7 | no(7) | - | |
| 62.3.134 | 171654 | 204 | - | - | - - | - - | 26 | yes | 26 | yes | - | F6Ir1 |
| | | | | | | | 11 | no(2) | 11 | no(2) | z2w,1=1<br>z2w,3=1<br>z4=1 | |
| | | | | | | | - | - | 7 | no(7) | - | |
| 62.3.135 | 16872 | 196 | - | - | - - | - - | 26 | yes | 26 | yes | - | F5O1Os1 |
| | | | | | | | 11 | no(2) | 11 | no(2) | z2w,1=1<br>z2w,3=1<br>z4=3 | |
| | | | | | | | - | - | 7 | no(7) | - | |
| 62.3.136 | 67746 | 116 | - | - | - - | - - | 26 | yes | 26 | yes | - | F3Mn1Na1 |
| | | | | | | | 11 | no(2) | 11 | yes | z2w,1=1<br>z2w,3=1 | |
| | | | | | | | - | - | 7 | no(7) | - | |
| 62.3.137 | 67745 | 116 | - | - | - - | - - | 26 | yes | 26 | yes | - | F3Mn1Na1 |
| | | | | | | | 11 | no(2) | 11 | yes | z2w,1=1<br>z2w,3=1 | |
| | | | | | | | - | - | 7 | no(7) | - | |
| 62.3.138 | 405050 | 132 | - | - | - - | - - | 26 | yes | 26 | yes | - | Eu1Sn1Zn1 |
| | | | | | | | 11 | no(2) | 11 | yes | z2w,1=1<br>z2w,3=1 | |
| | | | | | | | - | - | 7 | no(7) | - | |
| 62.3.139 | 66694 | 156 | - | - | - - | - - | 26 | yes | 26 | yes | - | Eu1Se3Zr1 |
| | | | | | | | 11 | no(2) | 11 | yes | z2w,1=1<br>z2w,3=1 | |
| | | | | | | | - | - | 7 | no(7) | - | |
| 62.3.140 | 420684 | 156 | - | - | - - | - - | 26 | yes | 26 | yes | - | Eu1Pt1Zn1 |
| | | | | | | | 11 | no(2) | 11 | no(2) | z2w,1=1<br>z2w,3=1<br>z4=3 | |
| | | | | | | | - | - | 7 | no(7) | - | |
| 62.3.141 | 404400 | 124 | - | - | - - | - - | 26 | yes | 26 | yes | - | Eu1Pt1Sn1 |
| | | | | | | | 11 | no(2) | 11 | no(2) | z2w,1=1<br>z2w,3=1<br>z4=3 | |
| | | | | | | | - | - | 7 | no(7) | - | |
| 62.3.142 | 107571 | 124 | - | - | - - | - - | 26 | yes | 26 | yes | - | Eu1Pt1Sn1 |
| | | | | | | | 11 | no(2) | 11 | no(2) | z2w,1=1<br>z2w,3=1<br>z4=3 | |
| | | | | | | | - | - | 7 | no(7) | - | |
| 62.3.143 | 420683 | 156 | - | - | - - | - - | 26 | yes | 26 | yes | - | Eu1Pd1Zn1 |
| | | | | | | | 11 | no(2) | 11 | yes | z2w,1=1<br>z2w,3=1 | |
| | | | | | | | - | - | 7 | no(7) | - | |
| 62.3.144 | 404401 | 124 | - | - | - - | - - | 26 | yes | 26 | yes | - | Eu1Pd1Sn1 |
| | | | | | | | 11 | no(2) | 11 | yes | z2w,1=1<br>z2w,3=1<br>z4=2 | |
| | | | | | | | - | - | 7 | no(7) | - | |
| 62.3.145 | 192313 | 156 | - | - | - - | - - | 26 | yes | 26 | yes | - | Eu1O3Zr1 |

Continued on next page

**TABLE XIV– continued**

| TQC | ICSD | NoE | DG | IndG | SubSG cl (w/ T) | SubSG cl (w/o T) | Max. Sub. SG | yes or no | Max. Sub. SG | yes or no | SIs (w/ T) | Stoich. Formula |
|---|---|---|---|---|---|---|---|---|---|---|---|---|
| | | | | | | | 11 | no(2) | 11 | yes | z2w,1=1<br>z2w,3=1 | |
| | | | | | | | - | - | 7 | no(7) | - | |
| 62.3.146 | 160941 | 156 | - | - | - - | - - | 26 | yes | 26 | yes | - | Eu1O3Zr1 |
| | | | | | | | 11 | no(2) | 11 | yes | z2w,1=1<br>z2w,3=1<br>z4=2 | |
| | | | | | | | - | - | 7 | no(7) | - | |
| 62.3.147 | 56611 | 180 | - | - | - - | - - | 26 | yes | 26 | yes | - | Eu1Ni1O3 |
| | | | | | | | 11 | no(2) | 11 | no(2) | z2w,1=1<br>z2w,3=1<br>z4=3 | |
| | | | | | | | - | - | 7 | no(7) | - | |
| 62.3.148 | 237630 | 92 | - | - | - - | - - | 26 | yes | 26 | yes | - | Eu1Mg1Sn1 |
| | | | | | | | 11 | no(2) | 11 | yes | z2w,1=1<br>z2w,3=1 | |
| | | | | | | | - | - | 7 | no(7) | - | |
| 62.3.149 | 412725 | 116 | - | - | - - | - - | 26 | yes | 26 | yes | - | Eu1Mg1Pd1 |
| | | | | | | | 11 | no(2) | 11 | no(2) | z2w,1=1<br>z2w,3=1<br>z4=3 | |
| | | | | | | | - | - | 7 | no(7) | - | |
| 62.3.150 | 428107 | 92 | - | - | - - | - - | 26 | yes | 26 | yes | - | Eu1Li1Sb1 |
| | | | | | | | 11 | no(2) | 11 | no(2) | z2w,1=1<br>z2w,3=1<br>z4=3 | |
| | | | | | | | - | - | 7 | no(7) | - | |
| 62.3.151 | 410699 | 116 | - | - | - - | - - | 26 | yes | 26 | yes | - | Eu1In1Rh1 |
| | | | | | | | 11 | no(2) | 11 | yes | z2w,1=1<br>z2w,3=1<br>z4=2 | |
| | | | | | | | - | - | 7 | no(7) | - | |
| 62.3.152 | 427571 | 92 | - | - | - - | - - | 26 | yes | 26 | yes | - | Eu1Ge1Mg1 |
| | | | | | | | 11 | no(2) | 11 | yes | z2w,1=1<br>z2w,3=1 | |
| | | | | | | | - | - | 7 | no(7) | - | |
| 62.3.153 | 182776 | 132 | - | - | - - | - - | 26 | yes | 26 | yes | - | Eu1Ga2Sb2 |
| | | | | | | | 11 | no(2) | 11 | no(2) | z2w,1=1<br>z2w,3=1<br>z4=3 | |
| | | | | | | | - | - | 7 | no(7) | - | |
| 62.3.154 | 27277 | 172 | - | - | - - | - - | 26 | yes | 26 | yes | - | Eu1Fe1O3 |
| | | | | | | | 11 | no(2) | 11 | yes | z2w,1=1<br>z2w,3=1 | |
| | | | | | | | - | - | 7 | no(7) | - | |
| 62.3.155 | 40393 | 180 | - | - | - - | - - | 26 | yes | 26 | yes | - | Er1O3V1 |
| | | | | | | | 11 | no(2) | 11 | yes | z2w,1=1<br>z2w,3=1<br>z4=2 | |
| | | | | | | | - | - | 7 | no(7) | - | |
| 62.3.156 | 185838 | 180 | - | - | - - | - - | 26 | yes | 26 | yes | - | Er1O3V1 |
| | | | | | | | 11 | no(2) | 11 | yes | z2w,1=1<br>z2w,3=1<br>z4=2 | |
| | | | | | | | - | - | 7 | no(7) | - | |
| 62.3.157 | 97791 | 132 | - | - | - - | - - | 26 | yes | 26 | yes | - | Er1Ge1Mn1 |

Continued on next page

**TABLE XIV– continued**

| TQC | ICSD | NoE | DG | IndG | SubSG cl (w/ T) | SubSG cl (w/o T) | Max. Sub. SG | yes or no | Max. Sub. SG | yes or no | SIs (w/ T) | Stoich. Formula |
|---|---|---|---|---|---|---|---|---|---|---|---|---|
| | | | | | | | 11 | no(2) | 11 | no(2) | z2w,1=1<br>z2w,3=1<br>z4=3 | |
| | | | | | | | - | - | 7 | no(7) | - | |
| 62.3.158 | 81411 | 172 | - | - | - - | - - | 26 | yes | 26 | yes | - | Er1F3 |
| | | | | | | | 11 | no(2) | 11 | no(2) | z2w,1=1<br>z2w,3=1<br>z4=1 | |
| | | | | | | | - | - | 7 | no(7) | - | |
| 62.3.159 | 419897 | 164 | - | - | - - | - - | 26 | yes | 26 | yes | - | Dy1Rh1Zn1 |
| | | | | | | | 11 | no(2) | 11 | no(2) | z2w,1=1<br>z2w,3=1<br>z4=1 | |
| | | | | | | | - | - | 7 | no(7) | - | |
| 62.3.160 | 91711 | 180 | - | - | - - | - - | 26 | yes | 26 | yes | - | Dy1Mn1O3 |
| | | | | | | | 11 | no(2) | 11 | no(2) | z2w,1=1<br>z2w,3=1<br>z4=1 | |
| | | | | | | | - | - | 7 | no(7) | - | |
| 62.3.161 | 157396 | 180 | - | - | - - | - - | 26 | yes | 26 | yes | - | Dy1Mn1O3 |
| | | | | | | | 11 | no(2) | 11 | yes | z2w,1=1<br>z2w,3=1 | |
| | | | | | | | - | - | 7 | no(7) | - | |
| 62.3.162 | 422562 | 132 | - | - | - - | - - | 26 | yes | 26 | yes | - | Dy1Li1O2 |
| | | | | | | | 11 | no(2) | 11 | yes | z2w,1=1<br>z2w,3=1 | |
| | | | | | | | - | - | 7 | no(7) | - | |
| 62.3.163 | 107269 | 284 | - | - | - - | - - | 26 | yes | 26 | yes | - | Cu5Sn1Th1 |
| | | | | | | | 11 | no(2) | 11 | yes | z2w,1=1<br>z2w,3=1 | |
| | | | | | | | - | - | 7 | no(7) | - | |
| 62.3.164 | 71683 | 140 | - | - | - - | - - | 26 | yes | 26 | yes | - | Cu2Li1O2 |
| | | | | | | | 11 | no(2) | 11 | no(2) | z2w,1=1<br>z2w,3=1<br>z4=1 | |
| | | | | | | | - | - | 7 | no(7) | - | |
| 62.3.165 | 71221 | 140 | - | - | - - | - - | 26 | yes | 26 | yes | - | Cu2Li1O2 |
| | | | | | | | 11 | no(2) | 11 | no(2) | z2w,1=1<br>z2w,3=1<br>z4=3 | |
| | | | | | | | - | - | 7 | no(7) | - | |
| 62.3.166 | 402700 | 140 | - | - | - - | - - | 26 | yes | 26 | yes | - | Cu1Yb1 |
| | | | | | | | 11 | no(2) | 11 | no(2) | z2w,1=1<br>z2w,3=1<br>z4=1 | |
| | | | | | | | - | - | 7 | no(7) | - | |
| 62.3.167 | 190123 | 76 | - | - | - - | - - | 26 | yes | 26 | yes | - | Cu1Sn1Zr1 |
| | | | | | | | 11 | no(2) | 11 | no(2) | z2w,1=1<br>z2w,3=1<br>z4=3 | |
| | | | | | | | - | - | 7 | no(7) | - | |
| 62.3.168 | 97747 | 140 | - | - | - - | - - | 26 | yes | 26 | yes | - | Cu1O3Se1 |
| | | | | | | | 11 | no(2) | 11 | yes | z2w,1=1<br>z2w,3=1 | |
| | | | | | | | - | - | 7 | no(7) | - | |
| 62.3.169 | 51847 | 140 | - | - | - - | - - | 26 | yes | 26 | yes | - | Cu1O3Se1 |
| | | | | | | | 11 | no(2) | 11 | yes | z2w,1=1<br>z2w,3=1 | |

Continued on next page

**TABLE XIV– continued**

| TQC | ICSD | NoE | DG | IndG | SubSG cl (w/ T) | SubSG cl (w/o T) | Max. Sub. SG | yes or no | Max. Sub. SG | yes or no | SIs (w/ T) | Stoich. Formula |
|---|---|---|---|---|---|---|---|---|---|---|---|---|
| | | | | | | | - | - | 7 | no(7) | - | |
| 62.3.170 | 498 | 140 | - | - | - - | - - | 26 | yes | 26 | yes | - | Cu1O3Se1 |
| | | | | | | | 11 | no(2) | 11 | yes | z2w,1=1<br>z2w,3=1 | |
| | | | | | | | - | - | 7 | no(7) | - | |
| 62.3.171 | 67531 | 180 | - | - | - - | - - | 26 | yes | 26 | yes | - | Cu1Fe2S3 |
| | | | | | | | 11 | no(2) | 11 | yes | z2w,1=1<br>z2w,3=1<br>z4=2 | |
| | | | | | | | - | - | 7 | no(7) | - | |
| 62.3.172 | 67530 | 180 | - | - | - - | - - | 26 | yes | 26 | yes | - | Cu1Fe2S3 |
| | | | | | | | 11 | no(2) | 11 | yes | z2w,1=1<br>z2w,3=1 | |
| | | | | | | | - | - | 7 | no(7) | - | |
| 62.3.173 | 67529 | 180 | - | - | - - | - - | 26 | yes | 26 | yes | - | Cu1Fe2S3 |
| | | | | | | | 11 | no(2) | 11 | yes | z2w,1=1<br>z2w,3=1<br>z4=2 | |
| | | | | | | | - | - | 7 | no(7) | - | |
| 62.3.174 | 42104 | 180 | - | - | - - | - - | 26 | yes | 26 | yes | - | Cu1Fe2S3 |
| | | | | | | | 11 | no(2) | 11 | yes | z2w,1=1<br>z2w,3=1<br>z4=2 | |
| | | | | | | | - | - | 7 | no(7) | - | |
| 62.3.175 | 14255 | 180 | - | - | - - | - - | 26 | yes | 26 | yes | - | Cu1Fe2S3 |
| | | | | | | | 11 | no(2) | 11 | yes | z2w,1=1<br>z2w,3=1<br>z4=2 | |
| | | | | | | | - | - | 7 | no(7) | - | |
| 62.3.176 | 59157 | 252 | - | - | - - | - - | 26 | yes | 26 | yes | - | Cu1Eu2S3 |
| | | | | | | | 11 | no(2) | 11 | no(2) | z2w,1=1<br>z2w,3=1<br>z4=1 | |
| | | | | | | | - | - | 7 | no(7) | - | |
| 62.3.177 | 150365 | 252 | - | - | - - | - - | 26 | yes | 26 | yes | - | Cu1Eu2S3 |
| | | | | | | | 11 | no(2) | 11 | no(2) | z2w,1=1<br>z2w,3=1<br>z4=1 | |
| | | | | | | | - | - | 7 | no(7) | - | |
| 62.3.178 | 84866 | 116 | - | - | - - | - - | 26 | yes | 26 | yes | - | Cr1Sb1Se3 |
| | | | | | | | 11 | no(2) | 11 | yes | z2w,1=1<br>z2w,3=1 | |
| | | | | | | | - | - | 7 | no(7) | - | |
| 62.3.179 | 43247 | 44 | - | - | - - | - - | 26 | yes | 26 | yes | - | Cr1P1 |
| | | | | | | | 11 | no(2) | 11 | no(2) | z2w,1=1<br>z2w,3=1<br>z4=1 | |
| | | | | | | | - | - | 7 | no(7) | - | |
| 62.3.180 | 160073 | 140 | - | - | - - | - - | 26 | yes | 26 | yes | - | Cr1La1O3 |
| | | | | | | | 11 | no(2) | 11 | yes | z2w,1=1<br>z2w,3=1 | |
| | | | | | | | - | - | 7 | no(7) | - | |
| 62.3.181 | 251089 | 180 | - | - | - - | - - | 26 | yes | 26 | yes | - | Cr1Ho1O3 |
| | | | | | | | 11 | no(2) | 11 | no(2) | z2w,1=1<br>z2w,3=1<br>z4=3 | |
| | | | | | | | - | - | 7 | no(7) | - | |
| 62.3.182 | 94380 | 92 | - | - | - - | - - | 26 | yes | 26 | yes | - | Co2P1 |

Continued on next page

**TABLE XIV– continued**

| TQC | ICSD | NoE | DG | IndG | SubSG cl (w/ T) | SubSG cl (w/o T) | Max. Sub. SG | yes or no | Max. Sub. SG | yes or no | SIs (w/ T) | Stoich. Formula |
|---|---|---|---|---|---|---|---|---|---|---|---|---|
| | | | | | | | 11 | no(2) | 11 | no(2) | z2w,1=1 z2w,3=1 z4=3 | |
| | | | | | | | - | - | 7 | no(7) | - | |
| 62.3.183 | 94379 | 92 | - | - | - - | - - | 26 | yes | 26 | yes | - | Co2P1 |
| | | | | | | | 11 | no(2) | 11 | no(2) | z2w,1=1 z2w,3=1 z4=1 | |
| | | | | | | | - | - | 7 | no(7) | - | |
| 62.3.184 | 74161 | 156 | - | - | - - | - - | 26 | yes | 26 | yes | - | Co1O4S1 |
| | | | | | | | 11 | no(2) | 11 | yes | z2w,1=1 z2w,3=1 z4=2 | |
| | | | | | | | - | - | 7 | no(7) | - | |
| 62.3.185 | 49727 | 108 | - | - | - - | - - | 26 | yes | 26 | yes | - | Co1Nb1P1 |
| | | | | | | | 11 | no(2) | 11 | no(2) | z2w,1=1 z2w,3=1 z4=3 | |
| | | | | | | | - | - | 7 | no(7) | - | |
| 62.3.186 | 87422 | 156 | - | - | - - | - - | 26 | yes | 26 | yes | - | Co1Li1O4P1 |
| | | | | | | | 11 | no(2) | 11 | no(2) | z2w,1=1 z2w,3=1 z4=3 | |
| | | | | | | | - | - | 7 | no(7) | - | |
| 62.3.187 | 400625 | 156 | - | - | - - | - - | 26 | yes | 26 | yes | - | Co1Li1O4P1 |
| | | | | | | | 11 | no(2) | 11 | no(2) | z2w,1=1 z2w,3=1 z4=3 | |
| | | | | | | | - | - | 7 | no(7) | - | |
| 62.3.188 | 291401 | 156 | - | - | - - | - - | 26 | yes | 26 | yes | - | Co1Li1O4P1 |
| | | | | | | | 11 | no(2) | 11 | no(2) | z2w,1=1 z2w,3=1 z4=3 | |
| | | | | | | | - | - | 7 | no(7) | - | |
| 62.3.189 | 247498 | 156 | - | - | - - | - - | 26 | yes | 26 | yes | - | Co1Li1O4P1 |
| | | | | | | | 11 | no(2) | 11 | no(2) | z2w,1=1 z2w,3=1 z4=3 | |
| | | | | | | | - | - | 7 | no(7) | - | |
| 62.3.190 | 247497 | 156 | - | - | - - | - - | 26 | yes | 26 | yes | - | Co1Li1O4P1 |
| | | | | | | | 11 | no(2) | 11 | no(2) | z2w,1=1 z2w,3=1 z4=3 | |
| | | | | | | | - | - | 7 | no(7) | - | |
| 62.3.191 | 247496 | 156 | - | - | - - | - - | 26 | yes | 26 | yes | - | Co1Li1O4P1 |
| | | | | | | | 11 | no(2) | 11 | no(2) | z2w,1=1 z2w,3=1 z4=1 | |
| | | | | | | | - | - | 7 | no(7) | - | |
| 62.3.192 | 173773 | 156 | - | - | - - | - - | 26 | yes | 26 | yes | - | Co1Li1O4P1 |
| | | | | | | | 11 | no(2) | 11 | no(2) | z2w,1=1 z2w,3=1 z4=1 | |
| | | | | | | | - | - | 7 | no(7) | - | |
| 62.3.193 | 55538 | 108 | - | - | - - | - - | 26 | yes | 26 | yes | - | Co1Ge1U1 |
| | | | | | | | 11 | no(2) | 11 | no(2) | z2w,1=1 z2w,3=1 z4=1 | |

Continued on next page

**TABLE XIV– continued**

| TQC | ICSD | NoE | DG | IndG | SubSG cl (w/ T) | SubSG cl (w/o T) | Max. Sub. SG | yes or no | Max. Sub. SG | yes or no | SIs (w/ T) | Stoich. Formula |
|---|---|---|---|---|---|---|---|---|---|---|---|---|
| | | | | | | | - | - | 7 | no(7) | - | |
| 62.3.194 | 102443 | 100 | - | - | - - | - - | 26 | yes | 26 | yes | - | Co1Ga1Pr1 |
| | | | | | | | 11 | no(2) | 11 | yes | z2w,1=1 z2w,3=1 | |
| | | | | | | | - | - | 7 | no(7) | - | |
| 62.3.195 | 72321 | 124 | - | - | - - | - - | 26 | yes | 26 | yes | - | Co1F3Na1 |
| | | | | | | | 11 | no(2) | 11 | yes | z2w,1=1 z2w,3=1 z4=2 | |
| | | | | | | | - | - | 7 | no(7) | - | |
| 62.3.196 | 87902 | 220 | - | - | - - | - - | 26 | yes | 26 | yes | - | Co1Cs2I4 |
| | | | | | | | 11 | no(2) | 11 | no(2) | z2w,1=1 z2w,3=1 z4=1 | |
| | | | | | | | - | - | 7 | no(7) | - | |
| 62.3.197 | 87901 | 220 | - | - | - - | - - | 26 | yes | 26 | yes | - | Co1Cs2I4 |
| | | | | | | | 11 | no(2) | 11 | no(2) | z2w,1=1 z2w,3=1 z4=3 | |
| | | | | | | | - | - | 7 | no(7) | - | |
| 62.3.198 | 87898 | 220 | - | - | - - | - - | 26 | yes | 26 | yes | - | Co1Cs2I4 |
| | | | | | | | 11 | no(2) | 11 | no(2) | z2w,1=1 z2w,3=1 z4=1 | |
| | | | | | | | - | - | 7 | no(7) | - | |
| 62.3.199 | 87897 | 220 | - | - | - - | - - | 26 | yes | 26 | yes | - | Co1Cs2I4 |
| | | | | | | | 11 | no(2) | 11 | no(2) | z2w,1=1 z2w,3=1 z4=1 | |
| | | | | | | | - | - | 7 | no(7) | - | |
| 62.3.200 | 9468 | 228 | - | - | - - | - - | 26 | yes | 26 | yes | - | Cl4Cs2Cu1 |
| | | | | | | | 11 | no(2) | 11 | no(2) | z2w,1=1 z2w,3=1 z4=1 | |
| | | | | | | | - | - | 7 | no(7) | - | |
| 62.3.201 | 91188 | 228 | - | - | - - | - - | 26 | yes | 26 | yes | - | Cl4Cs2Cu1 |
| | | | | | | | 11 | no(2) | 11 | no(2) | z2w,1=1 z2w,3=1 z4=3 | |
| | | | | | | | - | - | 7 | no(7) | - | |
| 62.3.202 | 91187 | 228 | - | - | - - | - - | 26 | yes | 26 | yes | - | Cl4Cs2Cu1 |
| | | | | | | | 11 | no(2) | 11 | no(2) | z2w,1=1 z2w,3=1 z4=3 | |
| | | | | | | | - | - | 7 | no(7) | - | |
| 62.3.203 | 91186 | 228 | - | - | - - | - - | 26 | yes | 26 | yes | - | Cl4Cs2Cu1 |
| | | | | | | | 11 | no(2) | 11 | no(2) | z2w,1=1 z2w,3=1 z4=3 | |
| | | | | | | | - | - | 7 | no(7) | - | |
| 62.3.204 | 15699 | 228 | - | - | - - | - - | 26 | yes | 26 | yes | - | Cl4Cs2Cu1 |
| | | | | | | | 11 | no(2) | 11 | no(2) | z2w,1=1 z2w,3=1 z4=1 | |
| | | | | | | | - | - | 7 | no(7) | - | |
| 62.3.205 | 72066 | 148 | - | - | - - | - - | 26 | yes | 26 | yes | - | Cl3K1Mn1 |
| | | | | | | | 11 | no(2) | 11 | yes | z2w,1=1 z2w,3=1 | |

Continued on next page

**TABLE XIV– continued**

| TQC | ICSD | NoE | DG | IndG | SubSG cl (w/ T) | | SubSG cl (w/o T) | | Max. Sub. SG | yes or no | Max. Sub. SG | yes or no | SIs (w/ T) | Stoich. Formula |
|---|---|---|---|---|---|---|---|---|---|---|---|---|---|---|
| | | | | | | | | | - | - | 7 | no(7) | - | |
| 62.3.206 | 151925 | 124 | - | - | - | - | - | - | 26 | yes | 26 | yes | - | Cl1Mn1S2Sb1 |
| | | | | | | | | | 11 | no(2) | 11 | no(2) | z2w,1=1<br>z2w,3=1<br>z4=1 | |
| | | | | | | | | | - | - | 7 | no(7) | - | |
| 62.3.207 | 418539 | 212 | - | - | - | - | - | - | 26 | yes | 26 | yes | - | Cl1Er1O3Se1 |
| | | | | | | | | | 11 | no(2) | 11 | no(2) | z2w,1=1<br>z2w,3=1<br>z4=3 | |
| | | | | | | | | | - | - | 7 | no(7) | - | |
| 62.3.208 | 412861 | 212 | - | - | - | - | - | - | 26 | yes | 26 | yes | - | Cl1Er1O3Se1 |
| | | | | | | | | | 11 | no(2) | 11 | no(2) | z2w,1=1<br>z2w,3=1<br>z4=1 | |
| | | | | | | | | | - | - | 7 | no(7) | - | |
| 62.3.209 | 412755 | 212 | - | - | - | - | - | - | 26 | yes | 26 | yes | - | Cl1Er1O3Se1 |
| | | | | | | | | | 11 | no(2) | 11 | no(2) | z2w,1=1<br>z2w,3=1<br>z4=3 | |
| | | | | | | | | | - | - | 7 | no(7) | - | |
| 62.3.210 | 418537 | 204 | - | - | - | - | - | - | 26 | yes | 26 | yes | - | Cl1Dy1O3Se1 |
| | | | | | | | | | 11 | no(2) | 11 | no(2) | z2w,1=1<br>z2w,3=1<br>z4=3 | |
| | | | | | | | | | - | - | 7 | no(7) | - | |
| 62.3.211 | 400014 | 132 | - | - | - | - | - | - | 26 | yes | 26 | yes | - | Ce1S3Sc1 |
| | | | | | | | | | 11 | no(2) | 11 | yes | z2w,1=1<br>z2w,3=1 | |
| | | | | | | | | | - | - | 7 | no(7) | - | |
| 62.3.212 | 23294 | 132 | - | - | - | - | - | - | 26 | yes | 26 | yes | - | Ce1S3Sc1 |
| | | | | | | | | | 11 | no(2) | 11 | yes | z2w,1=1<br>z2w,3=1 | |
| | | | | | | | | | - | - | 7 | no(7) | - | |
| 62.3.213 | 99834 | 140 | - | - | - | - | - | - | 26 | yes | 26 | yes | - | Ce1O3V1 |
| | | | | | | | | | 11 | no(2) | 11 | yes | z2w,1=1<br>z2w,3=1 | |
| | | | | | | | | | - | - | 7 | no(7) | - | |
| 62.3.214 | 185832 | 140 | - | - | - | - | - | - | 26 | yes | 26 | yes | - | Ce1O3V1 |
| | | | | | | | | | 11 | no(2) | 11 | yes | z2w,1=1<br>z2w,3=1 | |
| | | | | | | | | | - | - | 7 | no(7) | - | |
| 62.3.215 | 165721 | 140 | - | - | - | - | - | - | 26 | yes | 26 | yes | - | Ce1O3V1 |
| | | | | | | | | | 11 | no(2) | 11 | yes | z2w,1=1<br>z2w,3=1<br>z4=2 | |
| | | | | | | | | | - | - | 7 | no(7) | - | |
| 62.3.216 | 162748 | 140 | - | - | - | - | - | - | 26 | yes | 26 | yes | - | Ce1O3V1 |
| | | | | | | | | | 11 | no(2) | 11 | yes | z2w,1=1<br>z2w,3=1 | |
| | | | | | | | | | - | - | 7 | no(7) | - | |
| 62.3.217 | 195596 | 132 | - | - | - | - | - | - | 26 | yes | 26 | yes | - | Ce1O3Sc1 |
| | | | | | | | | | 11 | no(2) | 11 | yes | z2w,1=1<br>z2w,3=1 | |
| | | | | | | | | | - | - | 7 | no(7) | - | |
| 62.3.218 | 262201 | 84 | - | - | - | - | - | - | 26 | yes | 26 | yes | - | Ce1Ge2Li1 |
| | | | | | | | | | 11 | no(2) | 11 | yes | z2w,1=1<br>z2w,3=1 | |

Continued on next page

**TABLE XIV– continued**

| TQC | ICSD | NoE | DG | IndG | SubSG cl (w/ T) | SubSG cl (w/o T) | Max. Sub. SG | yes or no | Max. Sub. SG | yes or no | SIs (w/ T) | Stoich. Formula |
|---|---|---|---|---|---|---|---|---|---|---|---|---|
| | | | | | | | - | - | 7 | no(7) | - | |
| 62.3.219 | 414509 | 100 | - | - | - - | - - | 26 | yes | 26 | yes | - | Ce1Ge1Ir1 |
| | | | | | | | 11 | no(2) | 11 | no(2) | z2w,1=1<br>z2w,3=1<br>z4=1 | |
| | | | | | | | - | - | 7 | no(7) | - | |
| 62.3.220 | 400393 | 140 | - | - | - - | - - | 26 | yes | 26 | yes | - | Ce1Ga1Pd2 |
| | | | | | | | 11 | no(2) | 11 | no(2) | z2w,1=1<br>z2w,3=1<br>z4=1 | |
| | | | | | | | - | - | 7 | no(7) | - | |
| 62.3.221 | 150269 | 100 | - | - | - - | - - | 26 | yes | 26 | yes | - | Ce1Ga1Ni1 |
| | | | | | | | 11 | no(2) | 11 | yes | z2w,1=1<br>z2w,3=1<br>z4=2 | |
| | | | | | | | - | - | 7 | no(7) | - | |
| 62.3.222 | 426281 | 100 | - | - | - - | - - | 26 | yes | 26 | yes | - | Ce1Cl1Te1 |
| | | | | | | | 11 | no(2) | 11 | no(2) | z2w,1=1<br>z2w,3=1<br>z4=3 | |
| | | | | | | | - | - | 7 | no(7) | - | |
| 62.3.223 | 180009 | 148 | - | - | - - | - - | 26 | yes | 26 | yes | - | Cd1O3Tc1 |
| | | | | | | | 11 | no(2) | 11 | yes | z2w,1=1<br>z2w,3=1 | |
| | | | | | | | - | - | 7 | no(7) | - | |
| 62.3.224 | 428850 | 156 | - | - | - - | - - | 26 | yes | 26 | yes | - | Cd1Eu1Pt1 |
| | | | | | | | 11 | no(2) | 11 | no(2) | z2w,1=1<br>z2w,3=1<br>z4=1 | |
| | | | | | | | - | - | 7 | no(7) | - | |
| 62.3.225 | 156887 | 76 | - | - | - - | - - | 26 | yes | 26 | yes | - | Ca5Ir1 |
| | | | | | | | 11 | no(2) | 11 | no(2) | z2w,1=1<br>z2w,3=1<br>z4=3 | |
| | | | | | | | - | - | 7 | no(7) | - | |
| 62.3.226 | 413803 | 28 | - | - | - - | - - | 26 | yes | 26 | yes | - | Ca2In1 |
| | | | | | | | 11 | no(2) | 11 | yes | z2w,1=1<br>z2w,3=1 | |
| | | | | | | | - | - | 7 | no(7) | - | |
| 62.3.227 | 60830 | 68 | - | - | - - | - - | 26 | yes | 26 | yes | - | Ca1Pt1Sb1 |
| | | | | | | | 11 | no(2) | 11 | yes | z2w,1=1<br>z2w,3=1 | |
| | | | | | | | - | - | 7 | no(7) | - | |
| 62.3.228 | 88978 | 100 | - | - | - - | - - | 26 | yes | 26 | yes | - | Ca1O3V1 |
| | | | | | | | 11 | no(2) | 11 | yes | z2w,1=1<br>z2w,3=1 | |
| | | | | | | | - | - | 7 | no(7) | - | |
| 62.3.229 | 81056 | 100 | - | - | - - | - - | 26 | yes | 26 | yes | - | Ca1O3V1 |
| | | | | | | | 11 | no(2) | 11 | yes | z2w,1=1<br>z2w,3=1 | |
| | | | | | | | - | - | 7 | no(7) | - | |
| 62.3.230 | 261342 | 108 | - | - | - - | - - | 26 | yes | 26 | yes | - | Ca1O3Tc1 |
| | | | | | | | 11 | no(2) | 11 | yes | z2w,1=1<br>z2w,3=1<br>z4=2 | |
| | | | | | | | - | - | 7 | no(7) | - | |
| 62.3.231 | 261341 | 108 | - | - | - - | - - | 26 | yes | 26 | yes | - | Ca1O3Tc1 |

Continued on next page

**TABLE XIV– continued**

| TQC | ICSD | NoE | DG | IndG | SubSG cl (w/ T) | SubSG cl (w/o T) | Max. Sub. SG | yes or no | Max. Sub. SG | yes or no | SIs (w/ T) | Stoich. Formula |
|---|---|---|---|---|---|---|---|---|---|---|---|---|
| | | | | | | | 11 | no(2) | 11 | yes | z2w,1=1<br>z2w,3=1<br>z4=2 | |
| | | | | | | | - | - | 7 | no(7) | - | |
| 62.3.232 | 164775 | 148 | - | - | - - | - - | 26 | yes | 26 | yes | - | Ca1O3Rh1 |
| | | | | | | | 11 | no(2) | 11 | yes | z2w,1=1<br>z2w,3=1 | |
| | | | | | | | - | - | 7 | no(7) | - | |
| 62.3.233 | 85569 | 132 | - | - | - - | - - | 26 | yes | 26 | yes | - | Ca1Nb1O3 |
| | | | | | | | 11 | no(2) | 11 | yes | z2w,1=1<br>z2w,3=1 | |
| | | | | | | | - | - | 7 | no(7) | - | |
| 62.3.234 | 35218 | 108 | - | - | - - | - - | 26 | yes | 26 | yes | - | Ca1Mn1O3 |
| | | | | | | | 11 | no(2) | 11 | yes | z2w,1=1<br>z2w,3=1 | |
| | | | | | | | - | - | 7 | no(7) | - | |
| 62.3.235 | 165699 | 108 | - | - | - - | - - | 26 | yes | 26 | yes | - | Ca1Mn1O3 |
| | | | | | | | 11 | no(2) | 11 | yes | z2w,1=1<br>z2w,3=1<br>z4=2 | |
| | | | | | | | - | - | 7 | no(7) | - | |
| 62.3.236 | 161472 | 108 | - | - | - - | - - | 26 | yes | 26 | yes | - | Ca1Mn1O3 |
| | | | | | | | 11 | no(2) | 11 | yes | z2w,1=1<br>z2w,3=1<br>z4=2 | |
| | | | | | | | - | - | 7 | no(7) | - | |
| 62.3.237 | 161466 | 108 | - | - | - - | - - | 26 | yes | 26 | yes | - | Ca1Mn1O3 |
| | | | | | | | 11 | no(2) | 11 | yes | z2w,1=1<br>z2w,3=1<br>z4=2 | |
| | | | | | | | - | - | 7 | no(7) | - | |
| 62.3.238 | 161465 | 108 | - | - | - - | - - | 26 | yes | 26 | yes | - | Ca1Mn1O3 |
| | | | | | | | 11 | no(2) | 11 | yes | z2w,1=1<br>z2w,3=1 | |
| | | | | | | | - | - | 7 | no(7) | - | |
| 62.3.239 | 157060 | 108 | - | - | - - | - - | 26 | yes | 26 | yes | - | Ca1Mn1O3 |
| | | | | | | | 11 | no(2) | 11 | yes | z2w,1=1<br>z2w,3=1 | |
| | | | | | | | - | - | 7 | no(7) | - | |
| 62.3.240 | 182245 | 116 | - | - | - - | - - | 26 | yes | 26 | yes | - | Ca1Ir1O3 |
| | | | | | | | 11 | no(2) | 11 | yes | z2w,1=1<br>z2w,3=1 | |
| | | | | | | | - | - | 7 | no(7) | - | |
| 62.3.241 | 106348 | 60 | - | - | - - | - - | 26 | yes | 26 | yes | - | Ca1In1Pt1 |
| | | | | | | | 11 | no(2) | 11 | yes | z2w,1=1<br>z2w,3=1<br>z4=2 | |
| | | | | | | | - | - | 7 | no(7) | - | |
| 62.3.242 | 106340 | 60 | - | - | - - | - - | 26 | yes | 26 | yes | - | Ca1Ga1Pt1 |
| | | | | | | | 11 | no(2) | 11 | yes | z2w,1=1<br>z2w,3=1 | |
| | | | | | | | - | - | 7 | no(7) | - | |
| 62.3.243 | 23742 | 108 | - | - | - - | - - | 26 | yes | 26 | yes | - | C2U1V1 |
| | | | | | | | 11 | no(2) | 11 | no(2) | z2w,1=1<br>z2w,3=1<br>z4=3 | |
| | | | | | | | - | - | 7 | no(7) | - | |
| 62.3.244 | 69098 | 148 | - | - | - - | - - | 26 | yes | 26 | yes | - | C2Re1Y2 |

Continued on next page

**TABLE XIV– continued**

| TQC | ICSD | NoE | DG | IndG | SubSG cl (w/ T) | SubSG cl (w/o T) | Max. Sub. SG | yes or no | Max. Sub. SG | yes or no | SIs (w/ T) | Stoich. Formula |
|---|---|---|---|---|---|---|---|---|---|---|---|---|
| | | | | | | | 11 | no(2) | 11 | no(2) | z2w,1=1<br>z2w,3=1<br>z4=1 | |
| | | | | | | | - | - | 7 | no(7) | - | |
| 62.3.245 | 69099 | 236 | - | - | - - | - - | 26 | yes | 26 | yes | - | C2Er2Re1 |
| | | | | | | | 11 | no(2) | 11 | no(2) | z2w,1=1<br>z2w,3=1<br>z4=3 | |
| | | | | | | | - | - | 7 | no(7) | - | |
| 62.3.246 | 412665 | 124 | - | - | - - | - - | 26 | yes | 26 | yes | - | C1Eu1N2 |
| | | | | | | | 11 | no(2) | 11 | no(2) | z2w,1=1<br>z2w,3=1<br>z4=3 | |
| | | | | | | | - | - | 7 | no(7) | - | |
| 62.3.247 | 26024 | 228 | - | - | - - | - - | 26 | yes | 26 | yes | - | Br4Cs2Cu1 |
| | | | | | | | 11 | no(2) | 11 | no(2) | z2w,1=1<br>z2w,3=1<br>z4=1 | |
| | | | | | | | - | - | 7 | no(7) | - | |
| 62.3.248 | 10185 | 228 | - | - | - - | - - | 26 | yes | 26 | yes | - | Br4Cs2Cu1 |
| | | | | | | | 11 | no(2) | 11 | no(2) | z2w,1=1<br>z2w,3=1<br>z4=1 | |
| | | | | | | | - | - | 7 | no(7) | - | |
| 62.3.249 | 7 - 0 | 124 | - | - | - - | - - | 26 | yes | 26 | yes | - | Br3In1Mn1 |
| | | | | | | | 11 | no(2) | 11 | yes | z2w,1=1<br>z2w,3=1<br>z4=2 | |
| | | | | | | | - | - | 7 | no(7) | - | |
| 62.3.250 | 58821 | 100 | - | - | - - | - - | 26 | yes | 26 | yes | - | Bi3Ni1 |
| | | | | | | | 11 | no(2) | 11 | yes | z2w,1=1<br>z2w,3=1<br>z4=2 | |
| | | | | | | | - | - | 7 | no(7) | - | |
| 62.3.251 | 391336 | 100 | - | - | - - | - - | 26 | yes | 26 | yes | - | Bi3Ni1 |
| | | | | | | | 11 | no(2) | 11 | yes | z2w,1=1<br>z2w,3=1 | |
| | | | | | | | - | - | 7 | no(7) | - | |
| 62.3.252 | 162895 | 124 | - | - | - - | - - | 26 | yes | 26 | yes | - | Bi1Fe1O3 |
| | | | | | | | 11 | no(2) | 11 | yes | z2w,1=1<br>z2w,3=1 | |
| | | | | | | | - | - | 7 | no(7) | - | |
| 62.3.253 | 174403 | 116 | - | - | - - | - - | 26 | yes | 26 | yes | - | Bi1Cr1O3 |
| | | | | | | | 11 | no(2) | 11 | yes | z2w,1=1<br>z2w,3=1 | |
| | | | | | | | - | - | 7 | no(7) | - | |
| 62.3.254 | 160454 | 116 | - | - | - - | - - | 26 | yes | 26 | yes | - | Bi1Cr1O3 |
| | | | | | | | 11 | no(2) | 11 | yes | z2w,1=1<br>z2w,3=1<br>z4=2 | |
| | | | | | | | - | - | 7 | no(7) | - | |
| 62.3.255 | 172156 | 124 | - | - | - - | - - | 26 | yes | 26 | yes | - | Bi1Cl1Mn1S2 |
| | | | | | | | 11 | no(2) | 11 | no(2) | z2w,1=1<br>z2w,3=1<br>z4=1 | |
| | | | | | | | - | - | 7 | no(7) | - | |
| 62.3.256 | 26231 | 180 | - | - | - - | - - | 26 | yes | 26 | yes | - | Ba2Mn1S3 |

Continued on next page

**TABLE XIV– continued**

| TQC | ICSD | NoE | DG | IndG | SubSG cl (w/ T) | SubSG cl (w/o T) | Max. Sub. SG | yes or no | Max. Sub. SG | yes or no | SIs (w/ T) | Stoich. Formula |
|---|---|---|---|---|---|---|---|---|---|---|---|---|
| | | | | | | | 11 | no(2) | 11 | no(2) | z2w,1=1 z2w,3=1 z4=3 | |
| | | | | | | | - | - | 7 | no(7) | - | |
| 62.3.257 | 261714 | 92 | - | - | - - | - - | 26 | yes | 26 | yes | - | Ba2In1 |
| | | | | | | | 11 | no(2) | 11 | no(2) | z2w,1=1 z2w,3=1 z4=1 | |
| | | | | | | | - | - | 7 | no(7) | - | |
| 62.3.258 | 160295 | 188 | - | - | - - | - - | 26 | yes | 26 | yes | - | Ba2Co1S3 |
| | | | | | | | 11 | no(2) | 11 | yes | z2w,1=1 z2w,3=1 z4=2 | |
| | | | | | | | - | - | 7 | no(7) | - | |
| 62.3.259 | 160294 | 188 | - | - | - - | - - | 26 | yes | 26 | yes | - | Ba2Co1S3 |
| | | | | | | | 11 | no(2) | 11 | yes | z2w,1=1 z2w,3=1 z4=2 | |
| | | | | | | | - | - | 7 | no(7) | - | |
| 62.3.260 | 100435 | 84 | - | - | - - | - - | 26 | yes | 26 | yes | - | B2Ni1Ta1 |
| | | | | | | | 11 | no(2) | 11 | yes | z2w,1=1 z2w,3=1 z4=2 | |
| | | | | | | | - | - | 7 | no(7) | - | |
| 62.3.261 | 100774 | 156 | - | - | - - | - - | 26 | yes | 26 | yes | - | B2Lu1Ru1 |
| | | | | | | | 11 | no(2) | 11 | no(2) | z2w,1=1 z2w,3=1 z4=1 | |
| | | | | | | | - | - | 7 | no(7) | - | |
| 62.3.262 | 247263 | 212 | - | - | - - | - - | 26 | yes | 26 | yes | - | B1Pd4Sr1 |
| | | | | | | | 11 | no(2) | 11 | no(2) | z2w,1=1 z2w,3=1 z4=3 | |
| | | | | | | | - | - | 7 | no(7) | - | |
| 62.3.263 | 391331 | 44 | - | - | - - | - - | 26 | yes | 26 | yes | - | B1Fe1 |
| | | | | | | | 11 | no(2) | 11 | yes | z2w,1=1 z2w,3=1 z4=2 | |
| | | | | | | | - | - | 7 | no(7) | - | |
| 62.3.264 | 391329 | 44 | - | - | - - | - - | 26 | yes | 26 | yes | - | B1Fe1 |
| | | | | | | | 11 | no(2) | 11 | yes | z2w,1=1 z2w,3=1 z4=2 | |
| | | | | | | | - | - | 7 | no(7) | - | |
| 62.3.265 | 247264 | 212 | - | - | - - | - - | 26 | yes | 26 | yes | - | B1Ba1Pd4 |
| | | | | | | | 11 | no(2) | 11 | no(2) | z2w,1=1 z2w,3=1 z4=1 | |
| | | | | | | | - | - | 7 | no(7) | - | |
| 62.3.266 | 249529 | 180 | - | - | - - | - - | 26 | yes | 26 | yes | - | Au2In4La1 |
| | | | | | | | 11 | no(2) | 11 | yes | z2w,1=1 z2w,3=1 | |
| | | | | | | | - | - | 7 | no(7) | - | |
| 62.3.267 | 237467 | 140 | - | - | - - | - - | 26 | yes | 26 | yes | - | Au2Ga4Na1 |
| | | | | | | | 11 | no(2) | 11 | yes | z2w,1=1 z2w,3=1 | |
| | | | | | | | - | - | 7 | no(7) | - | |
| 62.3.268 | 159306 | 188 | - | - | - - | - - | 26 | yes | 26 | yes | - | Au1Yb1Zn1 |

Continued on next page

**TABLE XIV– continued**

| TQC | ICSD | NoE | DG | IndG | SubSG (w/ T) | cl | SubSG (w/o T) | cl | Max. Sub. SG | yes or no | Max. Sub. SG | yes or no | SIs (w/ T) | Stoich. Formula |
|---|---|---|---|---|---|---|---|---|---|---|---|---|---|---|
| | | | | | | | | | 11 | no(2) | 11 | no(2) | z2w,1=1<br>z2w,3=1<br>z4=1 | |
| | | | | | | | | | - | - | 7 | no(7) | - | |
| 62.3.269 | 262044 | 132 | - | - | - | - | - | - | 26 | yes | 26 | yes | - | Au1Y2 |
| | | | | | | | | | 11 | no(2) | 11 | yes | z2w,1=1<br>z2w,3=1<br>z4=2 | |
| | | | | | | | | | - | - | 7 | no(7) | - | |
| 62.3.270 | 411303 | 148 | - | - | - | - | - | - | 26 | yes | 26 | yes | - | Au1Mg1Yb1 |
| | | | | | | | | | 11 | no(2) | 11 | yes | z2w,1=1<br>z2w,3=1 | |
| | | | | | | | | | - | - | 7 | no(7) | - | |
| 62.3.271 | 404043 | 124 | - | - | - | - | - | - | 26 | yes | 26 | yes | - | Au1Eu1In1 |
| | | | | | | | | | 11 | no(2) | 11 | no(2) | z2w,1=1<br>z2w,3=1<br>z4=3 | |
| | | | | | | | | | - | - | 7 | no(7) | - | |
| 62.3.272 | 418712 | 140 | - | - | - | - | - | - | 26 | yes | 26 | yes | - | Au1Ce1Zn1 |
| | | | | | | | | | 11 | no(2) | 11 | yes | z2w,1=1<br>z2w,3=1<br>z4=2 | |
| | | | | | | | | | - | - | 7 | no(7) | - | |
| 62.3.273 | 411545 | 188 | - | - | - | - | - | - | 26 | yes | 26 | yes | - | Au1Cd1Yb1 |
| | | | | | | | | | 11 | no(2) | 11 | yes | z2w,1=1<br>z2w,3=1<br>z4=2 | |
| | | | | | | | | | - | - | 7 | no(7) | - | |
| 62.3.274 | 426612 | 132 | - | - | - | - | - | - | 26 | yes | 26 | yes | - | Au1Cd1Sr1 |
| | | | | | | | | | 11 | no(2) | 11 | yes | z2w,1=1<br>z2w,3=1<br>z4=2 | |
| | | | | | | | | | - | - | 7 | no(7) | - | |
| 62.3.275 | 58401 | 68 | - | - | - | - | - | - | 26 | yes | 26 | yes | - | Au1Ca3 |
| | | | | | | | | | 11 | no(2) | 11 | yes | z2w,1=1<br>z2w,3=1<br>z4=2 | |
| | | | | | | | | | - | - | 7 | no(7) | - | |
| 62.3.276 | 420574 | 100 | - | - | - | - | - | - | 26 | yes | 26 | yes | - | Au1Ca1Cd1 |
| | | | | | | | | | 11 | no(2) | 11 | yes | z2w,1=1<br>z2w,3=1 | |
| | | | | | | | | | - | - | 7 | no(7) | - | |
| 62.3.277 | 419559 | 84 | - | - | - | - | - | - | 26 | yes | 26 | yes | - | Au1Ba1 |
| | | | | | | | | | 11 | no(2) | 11 | no(2) | z2w,1=1<br>z2w,3=1<br>z4=3 | |
| | | | | | | | | | - | - | 7 | no(7) | - | |
| 62.3.278 | 38412 | 132 | - | - | - | - | - | - | 26 | yes | 26 | yes | - | As2Nb1Ni1 |
| | | | | | | | | | 11 | no(2) | 11 | yes | z2w,1=1<br>z2w,3=1<br>z4=2 | |
| | | | | | | | | | - | - | 7 | no(7) | - | |
| 62.3.279 | 48036 | 52 | - | - | - | - | - | - | 26 | yes | 26 | yes | - | As1Fe1 |
| | | | | | | | | | 11 | no(2) | 11 | yes | z2w,1=1<br>z2w,3=1<br>z4=2 | |
| | | | | | | | | | - | - | 7 | no(7) | - | |
| 62.3.280 | 48035 | 52 | - | - | - | - | - | - | 26 | yes | 26 | yes | - | As1Fe1 |

Continued on next page

**TABLE XIV– continued**

| TQC | ICSD | NoE | DG | IndG | SubSG cl (w/ T) | SubSG cl (w/o T) | Max. Sub. SG | yes or no | Max. Sub. SG | yes or no | SIs (w/ T) | Stoich. Formula |
|---|---|---|---|---|---|---|---|---|---|---|---|---|
| | | | | | | | 11 | no(2) | 11 | yes | z2w,1=1<br>z2w,3=1<br>z4=2 | |
| | | | | | | | - | - | 7 | no(7) | - | |
| 62.3.281 | 48034 | 52 | - | - | - - | - - | 26 | yes | 26 | yes | - | As1Fe1 |
| | | | | | | | 11 | no(2) | 11 | yes | z2w,1=1<br>z2w,3=1<br>z4=2 | |
| | | | | | | | - | - | 7 | no(7) | - | |
| 62.3.282 | 48033 | 52 | - | - | - - | - - | 26 | yes | 26 | yes | - | As1Fe1 |
| | | | | | | | 11 | no(2) | 11 | yes | z2w,1=1<br>z2w,3=1<br>z4=2 | |
| | | | | | | | - | - | 7 | no(7) | - | |
| 62.3.283 | 48031 | 52 | - | - | - - | - - | 26 | yes | 26 | yes | - | As1Fe1 |
| | | | | | | | 11 | no(2) | 11 | yes | z2w,1=1<br>z2w,3=1<br>z4=2 | |
| | | | | | | | - | - | 7 | no(7) | - | |
| 62.3.284 | 428105 | 92 | - | - | - - | - - | 26 | yes | 26 | yes | - | As1Eu1Li1 |
| | | | | | | | 11 | no(2) | 11 | no(2) | z2w,1=1<br>z2w,3=1<br>z4=3 | |
| | | | | | | | - | - | 7 | no(7) | - | |
| 62.3.285 | 72413 | 92 | - | - | - - | - - | 26 | yes | 26 | yes | - | As1Cu1Mn1 |
| | | | | | | | 11 | no(2) | 11 | no(2) | z2w,1=1<br>z2w,3=1<br>z4=3 | |
| | | | | | | | - | - | 7 | no(7) | - | |
| 62.3.286 | 195015 | 44 | - | - | - - | - - | 26 | yes | 26 | yes | - | As1Cr1 |
| | | | | | | | 11 | no(2) | 11 | no(2) | z2w,1=1<br>z2w,3=1<br>z4=1 | |
| | | | | | | | - | - | 7 | no(7) | - | |
| 62.3.287 | 195014 | 44 | - | - | - - | - - | 26 | yes | 26 | yes | - | As1Cr1 |
| | | | | | | | 11 | no(2) | 11 | no(2) | z2w,1=1<br>z2w,3=1<br>z4=1 | |
| | | | | | | | - | - | 7 | no(7) | - | |
| 62.3.288 | 195013 | 44 | - | - | - - | - - | 26 | yes | 26 | yes | - | As1Cr1 |
| | | | | | | | 11 | no(2) | 11 | no(2) | z2w,1=1<br>z2w,3=1<br>z4=3 | |
| | | | | | | | - | - | 7 | no(7) | - | |
| 62.3.289 | 245183 | 156 | - | - | - - | - - | 26 | yes | 26 | yes | - | As1Co1Li1O4 |
| | | | | | | | 11 | no(2) | 11 | no(2) | z2w,1=1<br>z2w,3=1<br>z4=3 | |
| | | | | | | | - | - | 7 | no(7) | - | |
| 62.3.290 | 155305 | 156 | - | - | - - | - - | 26 | yes | 26 | yes | - | As1Co1Li1O4 |
| | | | | | | | 11 | no(2) | 11 | no(2) | z2w,1=1<br>z2w,3=1<br>z4=3 | |
| | | | | | | | - | - | 7 | no(7) | - | |
| 62.3.291 | 261013 | 108 | - | - | - - | - - | 26 | yes | 26 | yes | - | Al2Eu1Ge1 |
| | | | | | | | 11 | no(2) | 11 | no(2) | z2w,1=1<br>z2w,3=1<br>z4=1 | |

Continued on next page

**TABLE XIV– continued**

| TQC | ICSD | NoE | DG | IndG | SubSG cl (w/ T) | | SubSG cl (w/o T) | | Max. Sub. SG | yes or no | Max. Sub. SG | yes or no | SIs (w/ T) | Stoich. Formula |
|---|---|---|---|---|---|---|---|---|---|---|---|---|---|---|
| | | | | | | | | | - | - | 7 | no(7) | - | |
| 62.3.292 | 245328 | 212 | - | - | - | - | - | - | 26 | yes | 26 | yes | - | Al1Pd5 |
| | | | | | | | | | 11 | no(2) | 11 | yes | z2w,1=1<br>z2w,3=1 | |
| | | | | | | | | | - | - | 7 | no(7) | - | |
| 62.3.293 | 168212 | 180 | - | - | - | - | - | - | 26 | yes | 26 | yes | - | Al1O3Yb1 |
| | | | | | | | | | 11 | no(2) | 11 | no(2) | z2w,1=1<br>z2w,3=1<br>z4=3 | |
| | | | | | | | | | - | - | 7 | no(7) | - | |
| 62.3.294 | 78969 | 148 | - | - | - | - | - | - | 26 | yes | 26 | yes | - | Al1Ge3Y2 |
| | | | | | | | | | 11 | no(2) | 11 | yes | z2w,1=1<br>z2w,3=1<br>z4=2 | |
| | | | | | | | | | - | - | 7 | no(7) | - | |
| 62.3.295 | 150355 | 156 | - | - | - | - | - | - | 26 | yes | 26 | yes | - | Al1Gd1O3 |
| | | | | | | | | | 11 | no(2) | 11 | no(2) | z2w,1=1<br>z2w,3=1<br>z4=3 | |
| | | | | | | | | | - | - | 7 | no(7) | - | |
| 62.3.296 | 150354 | 156 | - | - | - | - | - | - | 26 | yes | 26 | yes | - | Al1Gd1O3 |
| | | | | | | | | | 11 | no(2) | 11 | no(2) | z2w,1=1<br>z2w,3=1<br>z4=3 | |
| | | | | | | | | | - | - | 7 | no(7) | - | |
| 62.3.297 | 150353 | 156 | - | - | - | - | - | - | 26 | yes | 26 | yes | - | Al1Gd1O3 |
| | | | | | | | | | 11 | no(2) | 11 | no(2) | z2w,1=1<br>z2w,3=1<br>z4=3 | |
| | | | | | | | | | - | - | 7 | no(7) | - | |
| 62.3.298 | 150352 | 156 | - | - | - | - | - | - | 26 | yes | 26 | yes | - | Al1Gd1O3 |
| | | | | | | | | | 11 | no(2) | 11 | no(2) | z2w,1=1<br>z2w,3=1<br>z4=3 | |
| | | | | | | | | | - | - | 7 | no(7) | - | |
| 62.3.299 | 150351 | 156 | - | - | - | - | - | - | 26 | yes | 26 | yes | - | Al1Gd1O3 |
| | | | | | | | | | 11 | no(2) | 11 | no(2) | z2w,1=1<br>z2w,3=1<br>z4=3 | |
| | | | | | | | | | - | - | 7 | no(7) | - | |
| 62.3.300 | 150350 | 156 | - | - | - | - | - | - | 26 | yes | 26 | yes | - | Al1Gd1O3 |
| | | | | | | | | | 11 | no(2) | 11 | no(2) | z2w,1=1<br>z2w,3=1<br>z4=3 | |
| | | | | | | | | | - | - | 7 | no(7) | - | |
| 62.3.301 | 150349 | 156 | - | - | - | - | - | - | 26 | yes | 26 | yes | - | Al1Gd1O3 |
| | | | | | | | | | 11 | no(2) | 11 | no(2) | z2w,1=1<br>z2w,3=1<br>z4=3 | |
| | | | | | | | | | - | - | 7 | no(7) | - | |
| 62.3.302 | 150348 | 156 | - | - | - | - | - | - | 26 | yes | 26 | yes | - | Al1Gd1O3 |
| | | | | | | | | | 11 | no(2) | 11 | no(2) | z2w,1=1<br>z2w,3=1<br>z4=3 | |
| | | | | | | | | | - | - | 7 | no(7) | - | |
| 62.3.303 | 150172 | 100 | - | - | - | - | - | - | 26 | yes | 26 | yes | - | Al1Ce1Pt1 |

Continued on next page

**TABLE XIV– continued**

| TQC | ICSD | NoE | DG | IndG | SubSG (w/ T) | cl | SubSG (w/o T) | cl | Max. Sub. SG | yes or no | Max. Sub. SG | yes or no | SIs (w/ T) | Stoich. Formula |
|---|---|---|---|---|---|---|---|---|---|---|---|---|---|---|
| | | | | | | | | | 11 | no(2) | 11 | no(2) | z2w,1=1<br>z2w,3=1<br>z4=3 | |
| | | | | | | | | | - | - | 7 | no(7) | - | |
| 62.3.304 | 104635 | 100 | - | - | - | - | - | - | 26 | yes | 26 | yes | - | Al1Ce1Pt1 |
| | | | | | | | | | 11 | no(2) | 11 | no(2) | z2w,1=1<br>z2w,3=1<br>z4=1 | |
| | | | | | | | | | - | - | 7 | no(7) | - | |
| 62.3.305 | 261896 | 60 | - | - | - | - | - | - | 26 | yes | 26 | yes | - | Al1Ca1Pt1 |
| | | | | | | | | | 11 | no(2) | 11 | yes | z2w,1=1<br>z2w,3=1 | |
| | | | | | | | | | - | - | 7 | no(7) | - | |
| 62.3.306 | 57496 | 100 | - | - | - | - | - | - | 26 | yes | 26 | yes | - | Al1Au2 |
| | | | | | | | | | 11 | no(2) | 11 | yes | z2w,1=1<br>z2w,3=1<br>z4=2 | |
| | | | | | | | | | - | - | 7 | no(7) | - | |
| 62.3.307 | 57342 | 84 | - | - | - | - | - | - | 26 | yes | 26 | yes | - | Ag1Ba1 |
| | | | | | | | | | 11 | no(2) | 11 | no(2) | z2w,1=1<br>z2w,3=1<br>z4=1 | |
| | | | | | | | | | - | - | 7 | no(7) | - | |
| 62.3.308 | 260623 | 84 | - | - | - | - | - | - | 26 | yes | 26 | yes | - | Ag1Ba1 |
| | | | | | | | | | 11 | no(2) | 11 | no(2) | z2w,1=1<br>z2w,3=1<br>z4=1 | |
| | | | | | | | | | - | - | 7 | no(7) | - | |
| 63.1.1 | 68416 | 64 | 0.0091 | -0.1923 | 2 | 2 | 0 | 1 | - | - | - | - | z2w,1=1<br>z2w,2=1 | Te3U1 |
| 63.1.2 | 106074 | 28 | 0.0558 | -0.9575 | 2 | 1 | 0 | 1 | - | - | - | - | z2w,1=1<br>z2w,2=1<br>z4=2 | Sn1Sr1 |
| 63.1.3 | 416971 | 68 | 0.0000 | -2.1494 | 2 | 2 | 0 | 1 | - | - | - | - | z2w,1=1<br>z2w,2=1 | Si6Sr1 |
| 63.1.4 | 240779 | 68 | 0.0000 | -2.1494 | 2 | 2 | 0 | 1 | - | - | - | - | z2w,1=1<br>z2w,2=1 | Si6Sr1 |
| 63.1.5 | 78997 | 28 | 0.0163 | -0.7966 | 2 | 1 | 0 | 1 | - | - | - | - | z2w,1=1<br>z2w,2=1<br>z4=2 | Si1Sr1 |
| 63.1.6 | 87915 | 84 | 0.0321 | -1.2779 | 2 | 3 | 0 | 1 | - | - | - | - | z4=2 | Pd3Te2 |
| 63.1.7 | 77902 | 84 | 0.0329 | -1.2745 | 2 | 3 | 0 | 1 | - | - | - | - | z4=2 | Pd3Te2 |
| 63.1.8 | 411798 | 84 | 0.0000 | -1.2607 | 2 | 3 | 0 | 1 | - | - | - | - | z4=2 | Pd1Sn2Yb1 |
| 63.1.9 | 99253 | 212 | 0.0000 | -0.0099 | 2 | 1 | 0 | 1 | - | - | - | - | z2w,1=1<br>z2w,2=1<br>z4=2 | O7Re1Tb3 |
| 63.1.10 | 96410 | 176 | 0.0070 | -0.0207 | 2 | 1 | 0 | 1 | - | - | - | - | z2w,1=1<br>z2w,2=1<br>z4=2 | O7Pr3Re1 |
| 63.1.11 | 96149 | 76 | 0.0000 | -0.1242 | 2 | 3 | 0 | 1 | - | - | - | - | z4=2 | Np1Si1 |
| 63.1.12 | 86102 | 84 | 0.0309 | -0.2289 | 2 | 2 | 0 | 1 | - | - | - | - | z2w,1=1<br>z2w,2=1 | Mg1O5V2 |
| 63.1.13 | 50981 | 84 | 0.0038 | -0.2337 | 2 | 2 | 0 | 1 | - | - | - | - | z2w,1=1<br>z2w,2=1 | Mg1O5V2 |
| 63.1.14 | 50979 | 84 | 0.0046 | -0.2458 | 2 | 2 | 0 | 1 | - | - | - | - | z2w,1=1<br>z2w,2=1 | Mg1O5V2 |
| 63.1.15 | 250752 | 172 | 0.0000 | -0.8875 | 2 | 2 | 0 | 1 | - | - | - | - | z2w,1=1<br>z2w,2=1 | La2Sn1Zn5 |

Continued on next page

**TABLE XIV– continued**

| TQC | ICSD | NoE | DG | IndG | SubSG (w/ T) | cl | SubSG (w/o T) | cl | Max. Sub. SG | yes or no | Max. Sub. SG | yes or no | SIs (w/ T) | Stoich. Formula |
|---|---|---|---|---|---|---|---|---|---|---|---|---|---|---|
| 63.1.16 | 250756 | 172 | 0.0026 | -0.8387 | 2 | 2 | 0 | 1 | - | - | - | - | z2w,1=1<br>z2w,2=1 | La2Pb1Zn5 |
| 63.1.17 | 423611 | 72 | 0.0173 | -0.3091 | 2 | 2 | 0 | 1 | - | - | - | - | z2w,1=1<br>z2w,2=1 | La2Li2Si3 |
| 63.1.18 | 410732 | 112 | 0.0083 | -1.1636 | 2 | 2 | 0 | 1 | - | - | - | - | z2w,1=1<br>z2w,2=1 | La1Rh1Sn2 |
| 63.1.19 | 38311 | 152 | 0.0073 | -0.6067 | 2 | 2 | 0 | 1 | - | - | - | - | z2w,1=1<br>z2w,2=1 | La1Ni5P3 |
| 63.1.20 | 165363 | 96 | 0.0028 | -0.8098 | 2 | 1 | 0 | 1 | - | - | - | - | z2w,1=1<br>z2w,2=1<br>z4=2 | In5Y3 |
| 63.1.21 | 165362 | 96 | 0.0437 | -0.3456 | 2 | 3 | 0 | 1 | - | - | - | - | z4=2 | In5La3 |
| 63.1.22 | 418178 | 64 | 0.0129 | -2.2148 | 2 | 3 | 0 | 1 | - | - | - | - | z4=2 | In4Ni1Sr1 |
| 63.1.23 | 424100 | 80 | 0.0066 | -1.5168 | 2 | 3 | 0 | 1 | - | - | - | - | z4=2 | In2Pt1Yb1 |
| 63.1.24 | 410437 | 80 | 0.0063 | -1.3449 | 2 | 3 | 0 | 1 | - | - | - | - | z4=2 | In2Pd1Yb1 |
| 63.1.25 | 414077 | 80 | 0.0057 | -1.2960 | 2 | 3 | 0 | 1 | - | - | - | - | z4=2 | In2Ni1Yb1 |
| 63.1.26 | 412431 | 52 | 0.0100 | -1.0434 | 2 | 3 | 0 | 1 | - | - | - | - | z4=2 | In2La1Rh1 |
| 63.1.27 | 414479 | 52 | 0.0144 | -1.1080 | 2 | 3 | 0 | 1 | - | - | - | - | z4=2 | In2Ir1La1 |
| 63.1.28 | 200067 | 88 | 0.0066 | -1.1148 | 2 | 3 | 0 | 1 | - | - | - | - | z4=2 | Hf3Ni2Si3 |
| 63.1.29 | 26225 | 36 | 0.0416 | -1.8676 | 2 | 1 | 0 | 1 | - | - | - | - | z2w,1=1<br>z2w,2=1<br>z4=2 | H5Li4Rh1 |
| 63.1.30 | 262431 | 72 | 0.0378 | -0.3414 | 2 | 2 | 0 | 1 | - | - | - | - | z2w,1=1<br>z2w,2=1 | Ge3La2Li2 |
| 63.1.31 | 42375 | 36 | 0.0043 | -0.0165 | 2 | 3 | 0 | 1 | - | - | - | - | z4=2 | Ge1Nd1 |
| 63.1.32 | 42374 | 36 | 0.0030 | -0.0186 | 2 | 3 | 0 | 1 | - | - | - | - | z4=2 | Ge1Nd1 |
| 63.1.33 | 170875 | 208 | 0.0000 | -0.0137 | 2 | 1 | 0 | 1 | - | - | - | - | z2w,1=1<br>z2w,2=1<br>z4=2 | Gd3O7Os1 |
| 63.1.34 | 423067 | 80 | 0.0026 | -0.0902 | 2 | 1 | 0 | 1 | - | - | - | - | z2w,1=1<br>z2w,2=1<br>z4=2 | Ga4Gd1Ni1 |
| 63.1.35 | 189439 | 68 | 0.0065 | -0.5332 | 2 | 2 | 0 | 1 | - | - | - | - | z2w,1=1<br>z2w,2=1 | Fe2O3 |
| 63.1.36 | 261037 | 80 | 0.0080 | -0.0289 | 2 | 1 | 0 | 1 | - | - | - | - | z2w,1=1<br>z2w,2=1<br>z4=2 | Fe1Se3U1 |
| 63.1.37 | 245105 | 200 | 0.0000 | -0.0418 | 2 | 1 | 0 | 1 | - | - | - | - | z2w,1=1<br>z2w,2=1<br>z4=2 | Eu3O7Re1 |
| 63.1.38 | 400536 | 172 | 0.0000 | -0.1597 | 2 | 1 | 0 | 1 | - | - | - | - | z2w,1=1<br>z2w,2=1<br>z4=2 | Eu2O7Ta2 |
| 63.1.39 | 40812 | 164 | 0.0102 | -0.2 - | 2 | 3 | 0 | 1 | - | - | - | - | z4=2 | Eu1Ni5P3 |
| 63.1.40 | 237971 | 76 | 0.0108 | -0.4966 | 2 | 3 | 0 | 1 | - | - | - | - | z4=2 | Eu1In4Ir1 |
| 63.1.41 | 80501 | 164 | 0.0000 | -0.0702 | 2 | 1 | 0 | 1 | - | - | - | - | z2w,1=1<br>z2w,2=1<br>z4=2 | Er3Ge4 |
| 63.1.42 | 171141 | 80 | 0.0000 | -0.1320 | 2 | 2 | 0 | 1 | - | - | - | - | z2w,1=1<br>z2w,2=1 | Er1Ni1Si2 |
| 63.1.43 | 423066 | 88 | 0.0025 | -0.1855 | 2 | 3 | 0 | 1 | - | - | - | - | z4=2 | Er1Ga4Ni1 |
| 63.1.44 | 152349 | 56 | 0.0000 | -0.0078 | 2 | 3 | 0 | 1 | - | - | - | - | z4=2 | Dy1Sn2 |
| 63.1.45 | 180214 | 48 | 0.0000 | -0.0188 | 2 | 3 | 0 | 1 | - | - | - | - | z4=2 | Dy1Si1 |
| 63.1.46 | 154699 | 48 | 0.0000 | -0.0166 | 2 | 3 | 0 | 1 | - | - | - | - | z4=2 | Dy1Si1 |
| 63.1.47 | 57225 | 64 | 0.0000 | -0.0171 | 2 | 1 | 0 | 1 | - | - | - | - | z2w,1=1<br>z2w,2=1<br>z4=2 | Dy1Ge3 |

Continued on next page

**TABLE XIV– continued**

| TQC | ICSD | NoE | DG | IndG | SubSG (w/ T) | cl | SubSG (w/o T) | cl | Max. Sub. SG | yes or no | Max. Sub. SG | yes or no | SIs (w/ T) | Stoich. Formula |
|---|---|---|---|---|---|---|---|---|---|---|---|---|---|---|
| 63.1.48 | 59858 | 228 | 0.0000 | -0.0783 | 2 | 2 | 0 | 1 | - | - | - | - | z2w,1=1<br>z2w,2=1 | Cu3Ho2Rb1S5 |
| 63.1.49 | 174371 | 228 | 0.0000 | -0.0904 | 2 | 2 | 0 | 1 | - | - | - | - | z2w,1=1<br>z2w,2=1 | Cu3Ho2K1Se5 |
| 63.1.50 | 423716 | 220 | 0.0024 | -0.2415 | 2 | 3 | 0 | 1 | - | - | - | - | z4=2 | Cu3Er2S5Tl1 |
| 63.1.51 | 174372 | 232 | 0.0000 | -0.2740 | 2 | 3 | 0 | 1 | - | - | - | - | z4=2 | Cu3Er2K1Se5 |
| 63.1.52 | 95028 | 172 | 0.0081 | -0.0327 | 2 | 2 | 0 | 1 | - | - | - | - | z2w,1=1<br>z2w,2=1 | Cu1Ho2K1Se4 |
| 63.1.53 | 97561 | 172 | 0.0060 | -0.0372 | 2 | 2 | 0 | 1 | - | - | - | - | z2w,1=1<br>z2w,2=1 | Cu1Ho2K1S4 |
| 63.1.54 | 106440 | 80 | 0.0000 | -0.2720 | 2 | 3 | 0 | 1 | - | - | - | - | z4=2 | Cu1Eu1In4 |
| 63.1.55 | 429089 | 112 | 0.0033 | -0.1509 | 2 | 1 | 0 | 1 | - | - | - | - | z2w,1=1<br>z2w,2=1<br>z4=2 | Cs1Sc1Se5U1 |
| 63.1.56 | 413682 | 240 | 0.0022 | -0.4974 | 2 | 3 | 0 | 1 | - | - | - | - | z4=2 | Cs1Cu3Se5Yb2 |
| 63.1.57 | 413677 | 228 | 0.0000 | -0.0893 | 2 | 2 | 0 | 1 | - | - | - | - | z2w,1=1<br>z2w,2=1 | Cs1Cu3Ho2Se5 |
| 63.1.58 | 412074 | 232 | 0.0000 | -0.2340 | 2 | 3 | 0 | 1 | - | - | - | - | z4=2 | Cs1Cu3Er2S5 |
| 63.1.59 | 413674 | 224 | 0.0000 | -0.0467 | 2 | 2 | 0 | 1 | - | - | - | - | z2w,1=1<br>z2w,2=1 | Cs1Cu3Dy2Se5 |
| 63.1.60 | 236871 | 76 | 0.0123 | -0.0776 | 2 | 1 | 0 | 1 | - | - | - | - | z2w,1=1<br>z2w,2=1<br>z4=2 | Cr2Mg1O4 |
| 63.1.61 | 20952 | 192 | 0.0084 | -0.0040 | 2 | 2 | 0 | 1 | - | - | - | - | z2w,1=1<br>z2w,2=1 | Co3Ga9Ho2 |
| 63.1.62 | 427218 | 124 | 0.0000 | -0.0415 | 2 | 2 | 0 | 1 | - | - | - | - | z2w,1=1<br>z2w,2=1 | Ce3Ge3Mn2 |
| 63.1.63 | 250753 | 176 | 0.0177 | -0.0401 | 2 | 3 | 0 | 1 | - | - | - | - | z4=2 | Ce2Sn1Zn5 |
| 63.1.64 | 250757 | 176 | 0.0361 | -0.0384 | 2 | 3 | 0 | 1 | - | - | - | - | z4=2 | Ce2Pb1Zn5 |
| 63.1.65 | 63036 | 76 | 0.0056 | -0.3140 | 2 | 1 | 0 | 1 | - | - | - | - | z2w,1=1<br>z2w,2=1<br>z4=2 | Ce2Ge3Li2 |
| 63.1.66 | 262430 | 76 | 0.0065 | -0.3104 | 2 | 1 | 0 | 1 | - | - | - | - | z2w,1=1<br>z2w,2=1<br>z4=2 | Ce2Ge3Li2 |
| 63.1.67 | 99133 | 68 | 0.0043 | -0.2244 | 2 | 2 | 0 | 1 | - | - | - | - | z2w,1=1<br>z2w,2=1 | Ce1Rh2Si1 |
| 63.1.68 | 421810 | 212 | 0.3095 | 0.1207 | 2 | 1 | 0 | 1 | - | - | - | - | z2w,1=1<br>z2w,2=1<br>z4=2 | Cd2F1Sb5Sr5 |
| 63.1.69 | 424107 | 136 | 0.0265 | -0.8705 | 2 | 3 | 0 | 1 | - | - | - | - | z4=2 | Ca2Sn6Zn3 |
| 63.1.70 | 96228 | 24 | 0.1123 | -0.6738 | 2 | 1 | 0 | 1 | - | - | - | - | z2w,1=1<br>z2w,2=1<br>z4=2 | Ca2In1N1 |
| 63.1.71 | 165193 | 12 | 0.0635 | -1.0315 | 2 | 1 | 0 | 1 | - | - | - | - | z2w,1=1<br>z2w,2=1<br>z4=2 | Ca1Sn1 |
| 63.1.72 | 410774 | 40 | 0.0155 | -1.9921 | 2 | 2 | 0 | 1 | - | - | - | - | z2w,1=1<br>z2w,2=1 | Ca1Pd1Sn2 |
| 63.1.73 | 408880 | 36 | 0.0126 | -2. - 8 | 2 | 3 | 0 | 1 | - | - | - | - | z4=2 | Ca1In2Pt1 |
| 63.1.74 | 408881 | 36 | 0.0122 | -2.3681 | 2 | 2 | 0 | 1 | - | - | - | - | z2w,1=1<br>z2w,2=1 | Ca1In2Pd1 |
| 63.1.75 | 16356 | 184 | 0.0215 | -0.0238 | 2 | 1 | 0 | 1 | - | - | - | - | z2w,1=1<br>z2w,2=1<br>z4=2 | Ca1Fe5O7 |
| 63.1.76 | 75133 | 72 | 0.0189 | -0.2617 | 2 | 2 | 0 | 1 | - | - | - | - | z2w,1=1<br>z2w,2=1 | C1Fe2Si1Th1 |

Continued on next page

**TABLE XIV– continued**

| TQC | ICSD | NoE | DG | IndG | SubSG (w/ T) | cl | SubSG (w/o T) | cl | Max. Sub. SG | yes or no | Max. Sub. SG | yes or no | SIs (w/ T) | Stoich. Formula |
|---|---|---|---|---|---|---|---|---|---|---|---|---|---|---|
| 63.1.77 | 421809 | 212 | 0.2102 | 0.1219 | 2 | 1 | 0 | 1 | - | - | - | - | z2w,1=1 z2w,2=1 z4=2 | Ba5Cd2F1Sb5 |
| 63.1.78 | 411979 | 100 | 0.0617 | -0.2639 | 2 | 1 | 0 | 1 | - | - | - | - | z2w,1=1 z2w,2=1 z4=2 | Ba3Pb5 |
| 63.1.79 | 165184 | 100 | 0.0615 | -0.2639 | 2 | 1 | 0 | 1 | - | - | - | - | z2w,1=1 z2w,2=1 z4=2 | Ba3Pb5 |
| 63.1.80 | 710014 | 68 | 0.0072 | -1.9904 | 2 | 2 | 0 | 1 | - | - | - | - | z2w,1=1 z2w,2=1 | Ba1Si6 |
| 63.1.81 | 245295 | 68 | 0.0068 | -1.8966 | 2 | 2 | 0 | 1 | - | - | - | - | z2w,1=1 z2w,2=1 | Ba1Si6 |
| 63.1.82 | 78998 | 28 | 0.0449 | -0.4571 | 2 | 1 | 0 | 1 | - | - | - | - | z2w,1=1 z2w,2=1 z4=2 | Ba1Si1 |
| 63.1.83 | 427286 | 116 | 0.0590 | -1.0812 | 2 | 3 | 0 | 1 | - | - | - | - | z4=2 | Ba1Ru3Sn6 |
| 63.1.84 | 58664 | 28 | 0.0985 | -0.5328 | 2 | 1 | 0 | 1 | - | - | - | - | z2w,1=1 z2w,2=1 z4=2 | Ba1Pb1 |
| 63.1.85 | 251583 | 68 | 0.0224 | -1.9929 | 2 | 2 | 0 | 1 | - | - | - | - | z2w,1=1 z2w,2=1 | Ba1Ge6 |
| 63.1.86 | 290596 | 88 | 0.0098 | -0.3502 | 2 | 1 | 0 | 1 | - | - | - | - | z2w,1=1 z2w,2=1 z4=2 | Ba1Fe2Se3 |
| 63.1.87 | 290594 | 88 | 0.0266 | -0.4481 | 2 | 1 | 0 | 1 | - | - | - | - | z2w,1=1 z2w,2=1 z4=2 | Ba1Fe2Se3 |
| 63.1.88 | 16307 | 88 | 0.0258 | -0.2295 | 2 | 1 | 0 | 1 | - | - | - | - | z2w,1=1 z2w,2=1 z4=2 | Ba1Fe2S3 |
| 63.1.89 | 64596 | 124 | 0.0041 | -0.4906 | 2 | 2 | 0 | 1 | - | - | - | - | z2w,1=1 z2w,2=1 | B7Fe1Y3 |
| 63.1.90 | 91243 | 76 | 0.0412 | -1.0399 | 2 | 2 | 0 | 1 | - | - | - | - | z2w,1=1 z2w,2=1 | B4Mg1Os3 |
| 63.1.91 | 420401 | 124 | 0.0039 | -0.0080 | 2 | 1 | 0 | 1 | - | - | - | - | z2w,1=1 z2w,2=1 z4=2 | B4Fe1O7 |
| 63.1.92 | 418162 | 108 | 0.0110 | -0.7040 | 2 | 2 | 0 | 1 | - | - | - | - | z2w,1=1 z2w,2=1 | B2Cr2Ir1 |
| 63.1.93 | 237221 | 80 | 0.0000 | -0.2567 | 2 | 1 | 0 | 1 | - | - | - | - | z2w,1=1 z2w,2=1 z4=2 | Au1Eu1In4 |
| 63.1.94 | 33916 | 164 | 0.0084 | -0.2545 | 2 | 3 | 0 | 1 | - | - | - | - | z4=2 | As3Eu1Ni5 |
| 63.1.95 | 104653 | 152 | 0.0259 | -1.1831 | 2 | 3 | 0 | 1 | - | - | - | - | z4=2 | Al9Co3Y2 |
| 63.1.96 | 58157 | 52 | 0.0148 | -0.5634 | 2 | 2 | 0 | 1 | - | - | - | - | z2w,1=1 z2w,2=1 | Al6Ru1 |
| 63.1.97 | 370014 | 80 | 0.0101 | -1.2563 | 2 | 3 | 0 | 1 | - | - | - | - | z4=2 | Al2Ni1Yb1 |
| 63.1.98 | 423614 | 64 | 0.0135 | -0.7683 | 2 | 1 | 0 | 1 | - | - | - | - | z2w,1=1 z2w,2=1 z4=2 | Al2Ge1La2 |
| 63.1.99 | 99138 | 36 | 0.0323 | -1.3476 | 2 | 1 | 0 | 1 | - | - | - | - | z2w,1=1 z2w,2=1 z4=2 | Al1Si1Y1 |
| 63.1.100 | 415373 | 64 | 0.0279 | -1.4183 | 2 | 1 | 0 | 1 | - | - | - | - | z2w,1=1 z2w,2=1 z4=2 | Al1Lu1Si1 |

Continued on next page

**TABLE XIV– continued**

| TQC | ICSD | NoE | DG | IndG | SubSG (w/ T) | cl | SubSG (w/o T) | cl | Max. Sub. SG | yes or no | Max. Sub. SG | yes or no | SIs (w/ T) | Stoich. Formula |
|---|---|---|---|---|---|---|---|---|---|---|---|---|---|---|
| 63.1.101 | 76295 | 36 | 0.0700 | -1.4861 | 2 | 1 | 0 | 1 | - | - | - | - | z2w,1=1 z2w,2=1 z4=2 | Al1Ge1Y1 |
| 63.1.102 | 426322 | 92 | 0.0091 | -1.1897 | 2 | 1 | 0 | 1 | - | - | - | - | z2w,1=1 z2w,2=1 z4=2 | Al1Fe1.5Si2 |
| 63.1.103 | 93684 | 220 | 0.0021 | -0.0359 | 2 | 2 | 0 | 1 | - | - | - | - | z2w,1=1 z2w,2=1 | Ag3Cs1Se5Tb2 |
| 63.2.1 | 43419 | 28 | 0.0084 | -0.9997 | 2 | 7 | 2 | 2 | - | - | - | - | z4=1 | U1 |
| 63.2.2 | 43339 | 28 | 0.0082 | -0.9972 | 2 | 7 | 2 | 2 | - | - | - | - | z4=1 | U1 |
| 63.2.3 | 411981 | 100 | 0.0099 | -0.5635 | 2 | 5 | 2 | 2 | - | - | - | - | z2w,1=1 z2w,2=1 z4=1 | Sn5Sr3 |
| 63.2.4 | 165183 | 100 | 0.0259 | -0.5857 | 2 | 6 | 2 | 2 | - | - | - | - | z4=3 | Sn5Sr3 |
| 63.2.5 | - 56 | 52 | 0.0307 | -1.3057 | 2 | 5 | 2 | 2 | - | - | - | - | z2w,1=1 z2w,2=1 z4=1 | Sn4Sr1 |
| 63.2.6 | 281669 | 52 | 0.0285 | -1.3127 | 2 | 5 | 2 | 2 | - | - | - | - | z2w,1=1 z2w,2=1 z4=1 | Sn4Sr1 |
| 63.2.7 | 73619 | 100 | 0.0020 | -0.0608 | 2 | 7 | 2 | 2 | - | - | - | - | z4=1 | Si3Tb2 |
| 63.2.8 | 95179 | 24 | 0.0070 | -1.0317 | 2 | 5 | 2 | 2 | - | - | - | - | z2w,1=1 z2w,2=1 z4=1 | Si2Zr1 |
| 63.2.9 | 20284 | 16 | 0.0065 | -0.7417 | 2 | 6 | 2 | 2 | - | - | - | - | z4=3 | Si1Zr1 |
| 63.2.10 | 22235 | 68 | 0.0133 | -0.6269 | 2 | 5 | 2 | 2 | - | - | - | - | z2w,1=1 z2w,2=1 z4=1 | Sb2Yb1 |
| 63.2.11 | 163454 | 52 | 0.0124 | -1.3089 | 2 | 7 | 2 | 2 | - | - | - | - | z4=1 | Rh1Zr1 |
| 63.2.12 | 55546 | 28 | 0.0210 | -0.6865 | 2 | 7 | 2 | 2 | - | - | - | - | z4=1 | Pd1Zr1 |
| 63.2.13 | 165577 | 52 | 0.0126 | -2.1441 | 2 | 6 | 2 | 2 | - | - | - | - | z4=3 | Pd1Sr1Tl2 |
| 63.2.14 | 78010 | 172 | 0.0000 | -0.0182 | 2 | 6 | 2 | 2 | - | - | - | - | z4=3 | O7Pr3Sb1 |
| 63.2.15 | 96322 | 28 | 0.0043 | -0.8881 | 2 | 7 | 2 | 2 | - | - | - | - | z4=1 | Ni1Zr1 |
| 63.2.16 | 10 - 8 | 28 | 0.0142 | -0.9104 | 2 | 7 | 2 | 2 | - | - | - | - | z4=1 | Ni1Zr1 |
| 63.2.17 | 71964 | 56 | 0.0085 | -1.2694 | 2 | 7 | 2 | 2 | - | - | - | - | z4=1 | Ni1O1Zr3 |
| 63.2.18 | 63084 | 184 | 0.0048 | -0.0162 | 2 | 6 | 2 | 2 | - | - | - | - | z4=3 | Nd3O7Ru1 |
| 63.2.19 | 190762 | 184 | 0.0026 | -0.0213 | 2 | 6 | 2 | 2 | - | - | - | - | z4=3 | Nd3O7Ru1 |
| 63.2.20 | 71692 | 64 | 0.0000 | -0.0414 | 2 | 5 | 2 | 2 | - | - | - | - | z2w,1=1 z2w,2=1 z4=1 | Nd1Ni1Si2 |
| 63.2.21 | 71691 | 64 | 0.0000 | -0.0410 | 2 | 5 | 2 | 2 | - | - | - | - | z2w,1=1 z2w,2=1 z4=1 | Nd1Ni1Si2 |
| 63.2.22 | 425493 | 48 | 0.0143 | -1.6823 | 2 | 6 | 2 | 2 | - | - | - | - | z4=3 | Mg2Pd1Sr1 |
| 63.2.23 | 163453 | 52 | 0.0247 | -1.5262 | 2 | 7 | 2 | 2 | - | - | - | - | z4=1 | Ir1Zr1 |
| 63.2.24 | 412908 | 92 | 0.0030 | -1.4690 | 2 | 6 | 2 | 2 | - | - | - | - | z4=3 | In4Pt1Yb1 |
| 63.2.25 | 418180 | 64 | 0.0023 | -2.3285 | 2 | 5 | 2 | 2 | - | - | - | - | z2w,1=1 z2w,2=1 z4=1 | In4Pt1Sr1 |
| 63.2.26 | 411398 | 92 | 0.0028 | -1.4204 | 2 | 6 | 2 | 2 | - | - | - | - | z4=3 | In4Pd1Yb1 |
| 63.2.27 | 418179 | 64 | 0.0188 | -2.3671 | 2 | 5 | 2 | 2 | - | - | - | - | z2w,1=1 z2w,2=1 z4=1 | In4Pd1Sr1 |
| 63.2.28 | 415937 | 92 | 0.0224 | -0.7239 | 2 | 5 | 2 | 2 | - | - | - | - | z2w,1=1 z2w,2=1 z4=1 | In4Pb1Sr3 |
| 63.2.29 | 410400 | 52 | 0.0140 | -2.4596 | 2 | 6 | 2 | 2 | - | - | - | - | z4=3 | In2Pt1Sr1 |

Continued on next page

**TABLE XIV– continued**

| TQC | ICSD | NoE | DG | IndG | SubSG (w/ T) | cl | SubSG (w/o T) | cl | Max. Sub. SG | yes or no | Max. Sub. SG | yes or no | SIs (w/ T) | Stoich. Formula |
|---|---|---|---|---|---|---|---|---|---|---|---|---|---|---|
| 63.2.30 | 410402 | 52 | 0.0030 | -2.2738 | 2 | 6 | 2 | 2 | - | - | - | - | z4=3 | In2Pd1Sr1 |
| 63.2.31 | 190810 | 76 | 0.0648 | -1.1873 | 2 | 6 | 2 | 2 | - | - | - | - | z4=3 | Hf2Pt3 |
| 63.2.32 | 15051 | 68 | 0.0472 | 0.0472 | 2 | 4 | 2 | 2 | - | - | - | - | z2w,1=1 z2w,2=1 z4=3 | Hf1Te5 |
| 63.2.33 | 171235 | 172 | 0.0034 | -0.0175 | 2 | 4 | 2 | 2 | - | - | - | - | z2w,1=1 z2w,2=1 z4=3 | Ge6Pr2Zn3 |
| 63.2.34 | 171234 | 176 | 0.0000 | -0.0245 | 2 | 4 | 2 | 2 | - | - | - | - | z2w,1=1 z2w,2=1 z4=3 | Ge6Nd2Zn3 |
| 63.2.35 | 171232 | 164 | 0.0076 | -0.3327 | 2 | 4 | 2 | 2 | - | - | - | - | z2w,1=1 z2w,2=1 z4=3 | Ge6La2Zn3 |
| 63.2.36 | 56047 | 40 | 0.0422 | -0.5362 | 2 | 6 | 2 | 2 | - | - | - | - | z4=3 | Ge2Th1 |
| 63.2.37 | 240343 | 44 | 0.0143 | -1.6036 | 2 | 6 | 2 | 2 | - | - | - | - | z4=3 | Ge2Ni1Sr1 |
| 63.2.38 | 240728 | 36 | 0.0063 | -0.3347 | 2 | 5 | 2 | 2 | - | - | - | - | z2w,1=1 z2w,2=1 z4=1 | Ge1Na2Zn1 |
| 63.2.39 | 163917 | 140 | 0.0000 | -0.0181 | 2 | 7 | 2 | 2 | - | - | - | - | z4=1 | Gd3Ge4 |
| 63.2.40 | 104087 | 56 | 0.0025 | -0.1505 | 2 | 4 | 2 | 2 | - | - | - | - | z2w,1=1 z2w,2=1 z4=3 | Gd1Ni1 |
| 63.2.41 | 423070 | 92 | 0.0023 | -1.4031 | 2 | 6 | 2 | 2 | - | - | - | - | z4=3 | Ga4Ni1Yb1 |
| 63.2.42 | 154676 | 92 | 0.0031 | -1.4499 | 2 | 6 | 2 | 2 | - | - | - | - | z4=3 | Ga4Ni1Yb1 |
| 63.2.43 | 262740 | 80 | 0.0063 | -1.3100 | 2 | 4 | 2 | 2 | - | - | - | - | z2w,1=1 z2w,2=1 z4=3 | Ga2Ni1Yb1 |
| 63.2.44 | 262739 | 80 | 0.0069 | -1.3142 | 2 | 4 | 2 | 2 | - | - | - | - | z2w,1=1 z2w,2=1 z4=3 | Ga2Ni1Yb1 |
| 63.2.45 | 262738 | 80 | 0.0074 | -1.3247 | 2 | 4 | 2 | 2 | - | - | - | - | z2w,1=1 z2w,2=1 z4=3 | Ga2Ni1Yb1 |
| 63.2.46 | 262737 | 80 | 0.0065 | -1.3033 | 2 | 4 | 2 | 2 | - | - | - | - | z2w,1=1 z2w,2=1 z4=3 | Ga2Ni1Yb1 |
| 63.2.47 | 423623 | 28 | 0.0110 | -0.7527 | 2 | 7 | 2 | 2 | - | - | - | - | z4=1 | Ga1La1 |
| 63.2.48 | 2795 | 60 | 0.0089 | -3.4554 | 2 | 6 | 2 | 2 | - | - | - | - | z4=3 | Ga1 |
| 63.2.49 | 290249 | 124 | 0.0053 | -0.2321 | 2 | 7 | 2 | 2 | - | - | - | - | z4=1 | Fe4O5 |
| 63.2.50 | 98 | 80 | 0.0096 | -0.0198 | 2 | 6 | 2 | 2 | - | - | - | - | z4=3 | Fe1S3U1 |
| 63.2.51 | 261040 | 80 | 0.0112 | -0.0180 | 2 | 6 | 2 | 2 | - | - | - | - | z4=3 | Fe1S3U1 |
| 63.2.52 | 261039 | 80 | 0.0092 | -0.0203 | 2 | 6 | 2 | 2 | - | - | - | - | z4=3 | Fe1S3U1 |
| 63.2.53 | 261038 | 80 | 0.0102 | -0.0198 | 2 | 6 | 2 | 2 | - | - | - | - | z4=3 | Fe1S3U1 |
| 63.2.54 | 410700 | 68 | 0.0052 | -0.2054 | 2 | 5 | 2 | 2 | - | - | - | - | z2w,1=1 z2w,2=1 z4=1 | Eu1Ir1Sn2 |
| 63.2.55 | 411268 | 64 | 0.0122 | -0.4008 | 2 | 6 | 2 | 2 | - | - | - | - | z4=3 | Eu1In2Rh1 |
| 63.2.56 | 404599 | 68 | 0.0040 | -0.5386 | 2 | 5 | 2 | 2 | - | - | - | - | z2w,1=1 z2w,2=1 z4=1 | Eu1Ge2Ir1 |
| 63.2.57 | 425774 | 188 | 0.0000 | -0.0429 | 2 | 6 | 2 | 2 | - | - | - | - | z4=3 | Er3Ge3Ru2 |
| 63.2.58 | 55524 | 68 | 0.0023 | -0.2531 | 2 | 6 | 2 | 2 | - | - | - | - | z4=3 | Er1Ge3 |
| 63.2.59 | 55523 | 68 | 0.0000 | -0.2533 | 2 | 6 | 2 | 2 | - | - | - | - | z4=3 | Er1Ge3 |
| 63.2.60 | 59896 | 160 | 0.0000 | -0.0230 | 2 | 5 | 2 | 2 | - | - | - | - | z2w,1=1 z2w,2=1 z4=1 | Cu1Gd2Rb1S4 |

Continued on next page

**TABLE XIV– continued**

| TQC | ICSD | NoE | DG | IndG | SubSG (w/ T) | cl | SubSG (w/o T) | cl | Max. Sub. SG | yes or no | Max. Sub. SG | yes or no | SIs (w/ T) | Stoich. Formula |
|---|---|---|---|---|---|---|---|---|---|---|---|---|---|---|
| 63.2.61 | 416795 | 72 | 0.0029 | -0.1249 | 2 | 5 | 2 | 2 | - | - | - | - | z2w,1=1<br>z2w,2=1<br>z4=1 | Cu1Eu1Sn2 |
| 63.2.62 | 413681 | 236 | 0.0000 | -0.3223 | 2 | 7 | 2 | 2 | - | - | - | - | z4=1 | Cs1Cu3Se5Tm2 |
| 63.2.63 | 96226 | 220 | 0.0000 | -0.0345 | 2 | 6 | 2 | 2 | - | - | - | - | z4=3 | Cs1Cu3Se5Tb2 |
| 63.2.64 | 413680 | 220 | 0.0000 | -0.0342 | 2 | 6 | 2 | 2 | - | - | - | - | z4=3 | Cs1Cu3Se5Tb2 |
| 63.2.65 | 193548 | 76 | 0.0040 | -0.1145 | 2 | 5 | 2 | 2 | - | - | - | - | z2w,1=1<br>z2w,2=1<br>z4=1 | Cr2Mg1O4 |
| 63.2.66 | 192389 | 88 | 0.0081 | -0.1928 | 2 | 5 | 2 | 2 | - | - | - | - | z2w,1=1<br>z2w,2=1<br>z4=1 | Cr2Fe1O4 |
| 63.2.67 | 20951 | 152 | 0.0410 | -0.9891 | 2 | 6 | 2 | 2 | - | - | - | - | z4=3 | Co3Ga9Y2 |
| 63.2.68 | 102378 | 88 | 0.0053 | -0.2614 | 2 | 5 | 2 | 2 | - | - | - | - | z2w,1=1<br>z2w,2=1<br>z4=1 | Co2Eu1Ga3 |
| 63.2.69 | 246169 | 60 | 0.0036 | -0.4230 | 2 | 6 | 2 | 2 | - | - | - | - | z4=3 | Ce1Pt1Si2 |
| 63.2.70 | 42711 | 60 | 0.0097 | -0.1811 | 2 | 5 | 2 | 2 | - | - | - | - | z2w,1=1<br>z2w,2=1<br>z4=1 | Ce1Ni1Si2 |
| 63.2.71 | 154440 | 44 | 0.0103 | -0.5497 | 2 | 7 | 2 | 2 | - | - | - | - | z4=1 | Ce1Ni1 |
| 63.2.72 | 157363 | 68 | 0.0047 | -0.2101 | 2 | 6 | 2 | 2 | - | - | - | - | z4=3 | Ce1In4Pd1 |
| 63.2.73 | 150132 | 56 | 0.0000 | -0.1200 | 2 | 5 | 2 | 2 | - | - | - | - | z2w,1=1<br>z2w,2=1<br>z4=1 | Ce1In2Pd1 |
| 63.2.74 | 106394 | 60 | 0.0000 | -0.1423 | 2 | 5 | 2 | 2 | - | - | - | - | z2w,1=1<br>z2w,2=1<br>z4=1 | Ce1Ge2Ni1 |
| 63.2.75 | 425494 | 88 | 0.0707 | -1.5593 | 2 | 6 | 2 | 2 | - | - | - | - | z4=3 | Cd2Pd1Sr1 |
| 63.2.76 | 425064 | 96 | 0.0000 | -0.0290 | 2 | 5 | 2 | 2 | - | - | - | - | z2w,1=1<br>z2w,2=1<br>z4=1 | Cd2Nd1Pd1 |
| 63.2.77 | 421432 | 120 | 0.0058 | -1.1127 | 2 | 6 | 2 | 2 | - | - | - | - | z4=3 | Ca4Ge3Ni4 |
| 63.2.78 | 58948 | 36 | 0.0051 | -0.5985 | 2 | 4 | 2 | 2 | - | - | - | - | z2w,1=1<br>z2w,2=1<br>z4=3 | Ca3Zn1 |
| 63.2.79 | 78996 | 12 | 0.0289 | -0.9524 | 2 | 7 | 2 | 2 | - | - | - | - | z4=1 | Ca1Si1 |
| 63.2.80 | 190659 | 12 | 0.0233 | -0.9451 | 2 | 7 | 2 | 2 | - | - | - | - | z4=1 | Ca1Si1 |
| 63.2.81 | 410890 | 48 | 0.0210 | -2.5118 | 2 | 5 | 2 | 2 | - | - | - | - | z2w,1=1<br>z2w,2=1<br>z4=1 | Ca1In4Pd1 |
| 63.2.82 | 106347 | 48 | 0.0087 | -2.3380 | 2 | 5 | 2 | 2 | - | - | - | - | z2w,1=1<br>z2w,2=1<br>z4=1 | Ca1In4Ni1 |
| 63.2.83 | 240342 | 44 | 0.0205 | -1.5087 | 2 | 6 | 2 | 2 | - | - | - | - | z4=3 | Ca1Ge2Ni1 |
| 63.2.84 | 710075 | 36 | 0.0090 | -2.4884 | 2 | 6 | 2 | 2 | - | - | - | - | z4=3 | Ca1Ga2Ni1 |
| 63.2.85 | 16355 | 140 | 0.0000 | -0.0758 | 2 | 6 | 2 | 2 | - | - | - | - | z4=3 | Ca1Fe4O6 |
| 63.2.86 | 260482 | 88 | 0.0000 | -0.0764 | 2 | 7 | 2 | 2 | - | - | - | - | z4=1 | C1P1Ru2Tb1 |
| 63.2.87 | 159578 | 84 | 0.0135 | -0.1703 | 2 | 5 | 2 | 2 | - | - | - | - | z2w,1=1<br>z2w,2=1<br>z4=1 | C1Gd1Ru2Si1 |
| 63.2.88 | 50633 | 88 | 0.0000 | -0.0511 | 2 | 6 | 2 | 2 | - | - | - | - | z4=3 | C1Dy1Ru2Si1 |
| 63.2.89 | 40773 | 88 | 0.0021 | -0.0520 | 2 | 6 | 2 | 2 | - | - | - | - | z4=3 | C1Dy1Fe2Si1 |
| 63.2.90 | 411622 | 104 | 0.0038 | -1.7147 | 2 | 5 | 2 | 2 | - | - | - | - | z2w,1=1<br>z2w,2=1<br>z4=1 | C1B3Nb3 |
| 63.2.91 | 107610 | 64 | 0.0130 | -1.7877 | 2 | 6 | 2 | 2 | - | - | - | - | z4=3 | Be1Pd3 |

Continued on next page

**TABLE XIV– continued**

| TQC | ICSD | NoE | DG | IndG | SubSG (w/ T) | cl | SubSG (w/o T) | cl | Max. Sub. SG | yes or no | Max. Sub. SG | yes or no | SIs (w/ T) | Stoich. Formula |
|---|---|---|---|---|---|---|---|---|---|---|---|---|---|---|
| 63.2.92 | 411980 | 100 | 0.0108 | -0.3463 | 2 | 5 | 2 | 2 | - | - | - | - | z2w,1=1 z2w,2=1 z4=1 | Ba3Sn5 |
| 63.2.93 | 167166 | 100 | 0.0026 | -0.0781 | 2 | 5 | 2 | 2 | - | - | - | - | z2w,1=1 z2w,2=1 z4=1 | Ba3Sn5 |
| 63.2.94 | 167163 | 100 | 0.0042 | -0.4455 | 2 | 5 | 2 | 2 | - | - | - | - | z2w,1=1 z2w,2=1 z4=1 | Ba3Ge5 |
| 63.2.95 | 429723 | 148 | 0.0602 | -0.0943 | 2 | 4 | 2 | 2 | - | - | - | - | z2w,1=1 z2w,2=1 z4=3 | Ba2Ge4Ir4Mg1 |
| 63.2.96 | 418610 | 140 | 0.0060 | -1.0756 | 2 | 5 | 2 | 2 | - | - | - | - | z2w,1=1 z2w,2=1 z4=1 | Ba1Zn5 |
| 63.2.97 | 405111 | 60 | 0.0123 | -1.2153 | 2 | 6 | 2 | 2 | - | - | - | - | z4=3 | Ba1Pd1Sb2 |
| 63.2.98 | 14110 | 76 | 0.0192 | -1.6201 | 2 | 7 | 2 | 2 | - | - | - | - | z4=1 | Ba1Ni2Si2 |
| 63.2.99 | 411228 | 52 | 0.0444 | -1.6981 | 2 | 6 | 2 | 2 | - | - | - | - | z4=3 | Ba1In2Pt1 |
| 63.2.100 | 414228 | 52 | 0.0173 | -1.0665 | 2 | 6 | 2 | 2 | - | - | - | - | z4=3 | Ba1In1 |
| 63.2.101 | 429115 | 88 | 0.0316 | -1.3099 | 2 | 6 | 2 | 2 | - | - | - | - | z4=3 | Ba1Cd2Pt1 |
| 63.2.102 | 39663 | 120 | 0.0049 | -0.1639 | 2 | 7 | 2 | 2 | - | - | - | - | z4=1 | B7W1Y3 |
| 63.2.103 | 62282 | 192 | 0.0460 | -0.4472 | 2 | 7 | 2 | 2 | - | - | - | - | z4=1 | B7Co4Nb3 |
| 63.2.104 | 43662 | 48 | 0.0434 | -2.0528 | 2 | 6 | 2 | 2 | - | - | - | - | z4=3 | B1Re3 |
| 63.2.105 | 107505 | 36 | 0.0365 | -3.0906 | 2 | 6 | 2 | 2 | - | - | - | - | z4=3 | Au1In2Na1 |
| 63.2.106 | 411390 | 68 | 0.0000 | -0.1098 | 2 | 5 | 2 | 2 | - | - | - | - | z2w,1=1 z2w,2=1 z4=1 | Au1Eu1In2 |
| 63.2.107 | 107515 | 108 | 0.0075 | -0.0350 | 2 | 5 | 2 | 2 | - | - | - | - | z2w,1=1 z2w,2=1 z4=1 | Au1Co1Na4O5 |
| 63.2.108 | 290262 | 196 | 0.0248 | 0.0049 | 2 | 6 | 2 | 2 | - | - | - | - | z4=3 | As4Cd5Rb2 |
| 63.2.109 | 15032 | 268 | 0.0713 | -0.5622 | 2 | 4 | 2 | 2 | - | - | - | - | z2w,1=1 z2w,2=1 z4=3 | As3Nb4 |
| 63.2.110 | 405110 | 60 | 0.0277 | -1.2671 | 2 | 6 | 2 | 2 | - | - | - | - | z4=3 | As2Ba1Pd1 |
| 63.2.111 | 153384 | 48 | 0.0291 | -0.6819 | 2 | 5 | 2 | 2 | - | - | - | - | z2w,1=1 z2w,2=1 z4=1 | Al2Ba1Si2 |
| 63.2.112 | 415372 | 60 | 0.0038 | -0.1806 | 2 | 4 | 2 | 2 | - | - | - | - | z2w,1=1 z2w,2=1 z4=3 | Al1Si1Tm1 |
| 63.2.113 | 54470 | 56 | 0.0346 | -0.6152 | 2 | 4 | 2 | 2 | - | - | - | - | z2w,1=1 z2w,2=1 z4=3 | Al1La1 |
| 63.2.114 | 251809 | 24 | 0.0923 | -1.0412 | 2 | 5 | 2 | 2 | - | - | - | - | z2w,1=1 z2w,2=1 z4=1 | Al1B1W1 |
| 63.2.115 | 251808 | 24 | 0.0248 | -0.7482 | 2 | 5 | 2 | 2 | - | - | - | - | z2w,1=1 z2w,2=1 z4=1 | Al1B1Mo1 |
| 63.2.116 | 182114 | 120 | 0.0212 | 0.0011 | 2 | 6 | 2 | 2 | - | - | - | - | z4=3 | Al10Ru2U1 |
| 63.2.117 | 99179 | 120 | 0.0077 | -0.0090 | 2 | 6 | 2 | 2 | - | - | - | - | z4=3 | Al10Fe2U1 |
| 63.2.118 | 152354 | 116 | 0.2040 | -0.2537 | 2 | 5 | 2 | 2 | - | - | - | - | z2w,1=1 z2w,2=1 z4=1 | Al10Fe2Th1 |
| 63.2.119 | 59912 | 116 | 0.0479 | -0.1136 | 2 | 5 | 2 | 2 | - | - | - | - | z2w,1=1 z2w,2=1 z4=1 | Al10Ce1Ru2 |

Continued on next page

**TABLE XIV– continued**

| TQC | ICSD | NoE | DG | IndG | SubSG (w/ T) | cl | SubSG (w/o T) | cl | Max. Sub. SG | yes or no | Max. Sub. SG | yes or no | SIs (w/ T) | Stoich. Formula |
|---|---|---|---|---|---|---|---|---|---|---|---|---|---|---|
| 63.2.120 | 238167 | 232 | 0.0208 | -0.1711 | 2 | 4 | 2 | 2 | - | - | - | - | z2w,1=1<br>z2w,2=1<br>z4=3 | Ag1.2Cs0.4Er0.8Te2 |
| 63.4.1 | 422877 | 90 | - | - | - | - | - | - | See details in Table XIX | | | | | Sr3Tl5 |
| 63.4.2 | 380325 | 90 | - | - | - | - | - | - | See details in Table XIX | | | | | Sr3Tl5 |
| 63.4.3 | 416771 | 38 | - | - | - | - | - | - | See details in Table XIX | | | | | Sn2Y1 |
| 63.4.4 | 2239 | 70 | - | - | - | - | - | - | See details in Table XIX | | | | | S3Sc1U1 |
| 63.4.5 | 63105 | 114 | - | - | - | - | - | - | See details in Table XIX | | | | | Rh5Zr3 |
| 63.4.6 | 62932 | 114 | - | - | - | - | - | - | See details in Table XIX | | | | | Rh5Zr3 |
| 63.4.7 | 14382 | 78 | - | - | - | - | - | - | See details in Table XIX | | | | | Rh3Te2 |
| 63.4.8 | 420355 | 54 | - | - | - | - | - | - | See details in Table XIX | | | | | Pt8Sc4Si4 |
| 63.4.9 | 71292 | 90 | - | - | - | - | - | - | See details in Table XIX | | | | | Pt1Ta1Te5 |
| 63.4.10 | 424600 | 50 | - | - | - | - | - | - | See details in Table XIX | | | | | Pt1Sc1Zn1 |
| 63.4.11 | 412465 | 62 | - | - | - | - | - | - | See details in Table XIX | | | | | Pr1Se1Te2 |
| 63.4.12 | 35086 | 130 | - | - | - | - | - | - | See details in Table XIX | | | | | Pd2Si3Y3 |
| 63.4.13 | 424601 | 50 | - | - | - | - | - | - | See details in Table XIX | | | | | Pd1Sc1Zn1 |
| 63.4.14 | 36630 | 230 | - | - | - | - | - | - | See details in Table XIX | | | | | P5Pd5Sr4 |
| 63.4.15 | 192737 | 214 | - | - | - | - | - | - | See details in Table XIX | | | | | O7Ru1Tb3 |
| 63.4.16 | 187410 | 214 | - | - | - | - | - | - | See details in Table XIX | | | | | O7Os1Tb3 |
| 63.4.17 | 36520 | 66 | - | - | - | - | - | - | See details in Table XIX | | | | | O4P1Ti1 |
| 63.4.18 | 83361 | 130 | - | - | - | - | - | - | See details in Table XIX | | | | | O14Si4Sr2V2 |
| 63.4.19 | 33914 | 150 | - | - | - | - | - | - | See details in Table XIX | | | | | Ni5P3Sr1 |
| 63.4.20 | 65760 | 82 | - | - | - | - | - | - | See details in Table XIX | | | | | Ni2Sc3Si3 |
| 63.4.21 | 73314 | 90 | - | - | - | - | - | - | See details in Table XIX | | | | | Ni1Ta1Te5 |
| 63.4.22 | 71694 | 62 | - | - | - | - | - | - | See details in Table XIX | | | | | Ni1Pr1Si2 |
| 63.4.23 | 99823 | 46 | - | - | - | - | - | - | See details in Table XIX | | | | | Ni1Pr1 |
| 63.4.24 | 73316 | 106 | - | - | - | - | - | - | See details in Table XIX | | | | | Nb1Ni1Te5 |
| 63.4.25 | 105166 | 50 | - | - | - | - | - | - | See details in Table XIX | | | | | Na9Sn4 |
| 63.4.26 | 237248 | 50 | - | - | - | - | - | - | See details in Table XIX | | | | | Na1Si6 |
| 63.4.27 | 84911 | 82 | - | - | - | - | - | - | See details in Table XIX | | | | | N5Na1Ta3 |
| 63.4.28 | 99582 | 74 | - | - | - | - | - | - | See details in Table XIX | | | | | Mn1Se3Th1 |
| 63.4.29 | 166874 | 62 | - | - | - | - | - | - | See details in Table XIX | | | | | Mg1Ni1Tb1 |
| 63.4.30 | 426769 | 66 | - | - | - | - | - | - | See details in Table XIX | | | | | Lu1Sn2 |
| 63.4.31 | 249621 | 114 | - | - | - | - | - | - | See details in Table XIX | | | | | Lu1Nd1Se3 |
| 63.4.32 | 50982 | 82 | - | - | - | - | - | - | See details in Table XIX | | | | | Li1O5V2 |
| 63.4.33 | 50980 | 82 | - | - | - | - | - | - | See details in Table XIX | | | | | Li1O5V2 |
| 63.4.34 | 201974 | 82 | - | - | - | - | - | - | See details in Table XIX | | | | | Li1O5V2 |
| 63.4.35 | 281581 | 74 | - | - | - | - | - | - | See details in Table XIX | | | | | Li1Mn1O4V1 |
| 63.4.36 | 165185 | 106 | - | - | - | - | - | - | See details in Table XIX | | | | | La3Sn5 |
| 63.4.37 | 423418 | 98 | - | - | - | - | - | - | See details in Table XIX | | | | | La3Sn4 |
| 63.4.38 | 416718 | 166 | - | - | - | - | - | - | See details in Table XIX | | | | | La3O7Ru1 |
| 63.4.39 | 416818 | 150 | - | - | - | - | - | - | See details in Table XIX | | | | | La3Ni3Zn1 |
| 63.4.40 | 91142 | 190 | - | - | - | - | - | - | See details in Table XIX | | | | | La3Ni2O7 |
| 63.4.41 | 91141 | 190 | - | - | - | - | - | - | See details in Table XIX | | | | | La3Ni2O7 |
| 63.4.42 | 262792 | 118 | - | - | - | - | - | - | See details in Table XIX | | | | | La1Zn4 |
| 63.4.43 | 423421 | 30 | - | - | - | - | - | - | See details in Table XIX | | | | | La1Sn1 |
| 63.4.44 | 413173 | 58 | - | - | - | - | - | - | See details in Table XIX | | | | | La1Se1Te2 |
| 63.4.45 | 413171 | 58 | - | - | - | - | - | - | See details in Table XIX | | | | | La1Se1Te2 |
| 63.4.46 | 158253 | 50 | - | - | - | - | - | - | See details in Table XIX | | | | | La1Mg2Pd1 |
| 63.4.47 | 96152 | 50 | - | - | - | - | - | - | See details in Table XIX | | | | | La1Mg2Ni1 |
| 63.4.48 | 64625 | 102 | - | - | - | - | - | - | See details in Table XIX | | | | | K4P3 |
| 63.4.49 | 245610 | 142 | - | - | - | - | - | - | See details in Table XIX | | | | | K3O4Pd2 |
| 63.4.50 | 14157 | 142 | - | - | - | - | - | - | See details in Table XIX | | | | | K3Ni2O4 |
| 63.4.51 | 165188 | 98 | - | - | - | - | - | - | See details in Table XIX | | | | | In7K2Sr1 |
| 63.4.52 | 411746 | 66 | - | - | - | - | - | - | See details in Table XIX | | | | | In4La1Ni1 |
| 63.4.53 | 152048 | 78 | - | - | - | - | - | - | See details in Table XIX | | | | | In2Rh1Yb1 |
| 63.4.54 | 413459 | 54 | - | - | - | - | - | - | See details in Table XIX | | | | | In2Ni1Y1 |

Continued on next page

**TABLE XIV– continued**

| TQC | ICSD | NoE | DG | IndG | SubSG cl (w/ T) | SubSG cl (w/o T) | Max. Sub. SG yes or no | Max. Sub. SG yes or no | SIs (w/ T) | Stoich. Formula |
|---|---|---|---|---|---|---|---|---|---|---|
| 63.4.55 | 413448 | 54 | - | - | - - | - - | See details in Table XIX | | | In2La1Pd1 |
| 63.4.56 | 410401 | 50 | - | - | - - | - - | See details in Table XIX | | | In2Ir1Sr1 |
| 63.4.57 | 201170 | 70 | - | - | - - | - - | See details in Table XIX | | | I3U1 |
| 63.4.58 | 72343 | 106 | - | - | - - | - - | See details in Table XIX | | | I3K1Tm1 |
| 63.4.59 | 1 - 00 | 50 | - | - | - - | - - | See details in Table XIX | | | Ho1Si1 |
| 63.4.60 | 83917 | 78 | - | - | - - | - - | See details in Table XIX | | | Ho1Ni1Si2 |
| 63.4.61 | 83916 | 78 | - | - | - - | - - | See details in Table XIX | | | Ho1Ni1Si2 |
| 63.4.62 | 107591 | 54 | - | - | - - | - - | See details in Table XIX | | | Hf1Sb1 |
| 63.4.63 | 180189 | 82 | - | - | - - | - - | See details in Table XIX | | | H6Cl2Cu1N2 |
| 63.4.64 | 23318 | 138 | - | - | - - | - - | See details in Table XIX | | | H2Cl5Cs2Fe1O1 |
| 63.4.65 | 409766 | 134 | - | - | - - | - - | See details in Table XIX | | | H2Cl5Cr1Cs2O1 |
| 63.4.66 | 88105 | 98 | - | - | - - | - - | See details in Table XIX | | | Ge4Y3 |
| 63.4.67 | 88075 | 46 | - | - | - - | - - | See details in Table XIX | | | Ge3Y1 |
| 63.4.68 | 169059 | 70 | - | - | - - | - - | See details in Table XIX | | | Ge3Tm1 |
| 63.4.69 | 425769 | 122 | - | - | - - | - - | See details in Table XIX | | | Ge3Ru2Y3 |
| 63.4.70 | 425775 | 194 | - | - | - - | - - | See details in Table XIX | | | Ge3Ru2Tm3 |
| 63.4.71 | 88246 | 214 | - | - | - - | - - | See details in Table XIX | | | Ge3Lu3Ni2 |
| 63.4.72 | 412709 | 210 | - | - | - - | - - | See details in Table XIX | | | Ge3Ir2Lu3 |
| 63.4.73 | 57071 | 182 | - | - | - - | - - | See details in Table XIX | | | Ge3Ho3Ru2 |
| 63.4.74 | 425773 | 182 | - | - | - - | - - | See details in Table XIX | | | Ge3Ho3Ru2 |
| 63.4.75 | 82875 | 186 | - | - | - - | - - | See details in Table XIX | | | Ge3Ho3Ir2 |
| 63.4.76 | 159585 | 62 | - | - | - - | - - | See details in Table XIX | | | Ge2Tm1 |
| 63.4.77 | 189136 | 62 | - | - | - - | - - | See details in Table XIX | | | Ge2Ni1Pr1 |
| 63.4.78 | 85600 | 58 | - | - | - - | - - | See details in Table XIX | | | Ge2La1Ni1 |
| 63.4.79 | 42377 | 34 | - | - | - - | - - | See details in Table XIX | | | Ge1Pr1 |
| 63.4.80 | 42376 | 34 | - | - | - - | - - | See details in Table XIX | | | Ge1Pr1 |
| 63.4.81 | 53647 | 50 | - | - | - - | - - | See details in Table XIX | | | Ge1Ho1 |
| 63.4.82 | 188506 | 202 | - | - | - - | - - | See details in Table XIX | | | Gd3O7Sb1 |
| 63.4.83 | 190729 | 218 | - | - | - - | - - | See details in Table XIX | | | Gd3Nb1O7 |
| 63.4.84 | 411392 | 66 | - | - | - - | - - | See details in Table XIX | | | Gd1In2Rh1 |
| 63.4.85 | 423074 | 66 | - | - | - - | - - | See details in Table XIX | | | Ga4Ni1Y1 |
| 63.4.86 | 423069 | 90 | - | - | - - | - - | See details in Table XIX | | | Ga4Ni1Tm1 |
| 63.4.87 | 423068 | 82 | - | - | - - | - - | See details in Table XIX | | | Ga4Ni1Tb1 |
| 63.4.88 | 423079 | 86 | - | - | - - | - - | See details in Table XIX | | | Ga4Ho1Ni1 |
| 63.4.89 | 192026 | 58 | - | - | - - | - - | See details in Table XIX | | | Ga3Pd1Sr1 |
| 63.4.90 | 103895 | 54 | - | - | - - | - - | See details in Table XIX | | | Ga2Ni1Y1 |
| 63.4.91 | 413557 | 90 | - | - | - - | - - | See details in Table XIX | | | Ga2Ir1 |
| 63.4.92 | 100051 | 10 | - | - | - - | - - | See details in Table XIX | | | Ga1Li2 |
| 63.4.93 | 98966 | 166 | - | - | - - | - - | See details in Table XIX | | | Fe2Te2Y5 |
| 63.4.94 | 166049 | 74 | - | - | - - | - - | See details in Table XIX | | | Fe2S3Tl1 |
| 63.4.95 | 81548 | 86 | - | - | - - | - - | See details in Table XIX | | | Fe2Rb1Te3 |
| 63.4.96 | 81 - | 86 | - | - | - - | - - | See details in Table XIX | | | Fe2Rb1Se3 |
| 63.4.97 | 99505 | 86 | - | - | - - | - - | See details in Table XIX | | | Fe2Rb1S3 |
| 63.4.98 | 424672 | 86 | - | - | - - | - - | See details in Table XIX | | | Fe2K1Se3 |
| 63.4.99 | 424671 | 86 | - | - | - - | - - | See details in Table XIX | | | Fe2K1Se3 |
| 63.4.100 | 99507 | 86 | - | - | - - | - - | See details in Table XIX | | | Fe2K1S3 |
| 63.4.101 | 100180 | 86 | - | - | - - | - - | See details in Table XIX | | | Fe2K1S3 |
| 63.4.102 | 155065 | 74 | - | - | - - | - - | See details in Table XIX | | | Fe1O4P1 |
| 63.4.103 | 60615 | 90 | - | - | - - | - - | See details in Table XIX | | | F4Fe1K1 |
| 63.4.104 | 93668 | 202 | - | - | - - | - - | See details in Table XIX | | | Eu3O7Ru1 |
| 63.4.105 | 416436 | 142 | - | - | - - | - - | See details in Table XIX | | | Eu3Ge5 |
| 63.4.106 | 245230 | 142 | - | - | - - | - - | See details in Table XIX | | | Eu3Ge5 |
| 63.4.107 | 245229 | 142 | - | - | - - | - - | See details in Table XIX | | | Eu3Ge5 |
| 63.4.108 | 416579 | 82 | - | - | - - | - - | See details in Table XIX | | | Eu1Si6 |
| 63.4.109 | 416578 | 82 | - | - | - - | - - | See details in Table XIX | | | Eu1Si6 |
| 63.4.110 | 416577 | 82 | - | - | - - | - - | See details in Table XIX | | | Eu1Si6 |
| 63.4.111 | 416576 | 82 | - | - | - - | - - | See details in Table XIX | | | Eu1Si6 |

Continued on next page

**TABLE XIV– continued**

| TQC | ICSD | NoE | DG | IndG | SubSG cl (w/ T) | SubSG cl (w/o T) | Max. Sub. SG | yes or no | Max. Sub. SG | yes or no | SIs (w/ T) | Stoich. Formula |
|---|---|---|---|---|---|---|---|---|---|---|---|---|
| 63.4.112 | 193724 | 78 | - | - | - - | - - | See details in Table XIX | | | | | Eu1In4Pt1 |
| 63.4.113 | 411391 | 78 | - | - | - - | - - | See details in Table XIX | | | | | Eu1In4Pd1 |
| 63.4.114 | 404251 | 78 | - | - | - - | - - | See details in Table XIX | | | | | Eu1In4Ni1 |
| 63.4.115 | 410438 | 66 | - | - | - - | - - | See details in Table XIX | | | | | Eu1In2Pd1 |
| 63.4.116 | 402571 | 66 | - | - | - - | - - | See details in Table XIX | | | | | Eu1In2Ni1 |
| 63.4.117 | 42420 | 70 | - | - | - - | - - | See details in Table XIX | | | | | Eu1Ge3Mg3 |
| 63.4.118 | 71204 | 206 | - | - | - - | - - | See details in Table XIX | | | | | Cu5K2Te5 |
| 63.4.119 | 54129 | 46 | - | - | - - | - - | See details in Table XIX | | | | | Cu4O8 |
| 63.4.120 | 249533 | 190 | - | - | - - | - - | See details in Table XIX | | | | | Cu3K2S5U1 |
| 63.4.121 | 380487 | 146 | - | - | - - | - - | See details in Table XIX | | | | | Cu2K3O4 |
| 63.4.122 | 82330 | 102 | - | - | - - | - - | See details in Table XIX | | | | | Cu1Se1 |
| 63.4.123 | 83051 | 66 | - | - | - - | - - | See details in Table XIX | | | | | Cu1O2Sr1 |
| 63.4.124 | 77293 | 66 | - | - | - - | - - | See details in Table XIX | | | | | Cu1O2Sr1 |
| 63.4.125 | 77291 | 66 | - | - | - - | - - | See details in Table XIX | | | | | Cu1O2Sr1 |
| 63.4.126 | 202992 | 66 | - | - | - - | - - | See details in Table XIX | | | | | Cu1O2Sr1 |
| 63.4.127 | 150913 | 46 | - | - | - - | - - | See details in Table XIX | | | | | Cu1O2 |
| 63.4.128 | 249967 | 106 | - | - | - - | - - | See details in Table XIX | | | | | Cu1Np1Rb1S3 |
| 63.4.129 | 249966 | 106 | - | - | - - | - - | See details in Table XIX | | | | | Cu1K1Np1S3 |
| 63.4.130 | 84200 | 70 | - | - | - - | - - | See details in Table XIX | | | | | Cu1Ge4Zr2 |
| 63.4.131 | 54101 | 194 | - | - | - - | - - | See details in Table XIX | | | | | Cs2Cu5Se4 |
| 63.4.132 | 281446 | 114 | - | - | - - | - - | See details in Table XIX | | | | | Cs1Gd1Hg1Se3 |
| 63.4.133 | 81549 | 86 | - | - | - - | - - | See details in Table XIX | | | | | Cs1Fe2Se3 |
| 63.4.134 | 187535 | 86 | - | - | - - | - - | See details in Table XIX | | | | | Cs1Fe2Se3 |
| 63.4.135 | 99504 | 86 | - | - | - - | - - | See details in Table XIX | | | | | Cs1Fe2S3 |
| 63.4.136 | 280844 | 122 | - | - | - - | - - | See details in Table XIX | | | | | Cs1Er1Se3Zn1 |
| 63.4.137 | 280842 | 118 | - | - | - - | - - | See details in Table XIX | | | | | Cs1Dy1Se3Zn1 |
| 63.4.138 | 249968 | 106 | - | - | - - | - - | See details in Table XIX | | | | | Cs1Cu1Np1S3 |
| 63.4.139 | 62159 | 70 | - | - | - - | - - | See details in Table XIX | | | | | Cr1O4P1 |
| 63.4.140 | 60825 | 82 | - | - | - - | - - | See details in Table XIX | | | | | Cr1Cu1O4 |
| 63.4.141 | 183216 | 82 | - | - | - - | - - | See details in Table XIX | | | | | Cr1Cu1O4 |
| 63.4.142 | 102490 | 206 | - | - | - - | - - | See details in Table XIX | | | | | Co5Ga1Y5 |
| 63.4.143 | 10044 | 126 | - | - | - - | - - | See details in Table XIX | | | | | Co3Ga1Y3 |
| 63.4.144 | 41745 | 78 | - | - | - - | - - | See details in Table XIX | | | | | Co2Sc3Si3 |
| 63.4.145 | 106479 | 78 | - | - | - - | - - | See details in Table XIX | | | | | Co2Ge3Sc3 |
| 63.4.146 | 33736 | 78 | - | - | - - | - - | See details in Table XIX | | | | | Co1O4S1 |
| 63.4.147 | 55670 | 66 | - | - | - - | - - | See details in Table XIX | | | | | Co1Ga4Th1 |
| 63.4.148 | 417507 | 114 | - | - | - - | - - | See details in Table XIX | | | | | Co1Cs1Ho1Se3 |
| 63.4.149 | 169508 | 58 | - | - | - - | - - | See details in Table XIX | | | | | Ce1Rh1Si2 |
| 63.4.150 | 106411 | 54 | - | - | - - | - - | See details in Table XIX | | | | | Ce1Mn1Si2 |
| 63.4.151 | 106410 | 54 | - | - | - - | - - | See details in Table XIX | | | | | Ce1Mn1Si2 |
| 63.4.152 | 102212 | 42 | - | - | - - | - - | See details in Table XIX | | | | | Ce1Li1Sn2 |
| 63.4.153 | 414474 | 54 | - | - | - - | - - | See details in Table XIX | | | | | Ce1In2Ir1 |
| 63.4.154 | 427867 | 62 | - | - | - - | - - | See details in Table XIX | | | | | Ce1Ga3Pd1 |
| 63.4.155 | 16501 | 58 | - | - | - - | - - | See details in Table XIX | | | | | Ce1Co1Si2 |
| 63.4.156 | 417029 | 90 | - | - | - - | - - | See details in Table XIX | | | | | Cd2La1Pd1 |
| 63.4.157 | 99139 | 114 | - | - | - - | - - | See details in Table XIX | | | | | Cd2Cu1Er1 |
| 63.4.158 | 281437 | 114 | - | - | - - | - - | See details in Table XIX | | | | | Cd1Cs1Gd1Se3 |
| 63.4.159 | 173313 | 122 | - | - | - - | - - | See details in Table XIX | | | | | Cd1Cs1Er1Te3 |
| 63.4.160 | 281434 | 102 | - | - | - - | - - | See details in Table XIX | | | | | Cd1Ce1Cs1Se3 |
| 63.4.161 | 79095 | 230 | - | - | - - | - - | See details in Table XIX | | | | | Ca4P5Pd5 |
| 63.4.162 | 380324 | 42 | - | - | - - | - - | See details in Table XIX | | | | | Ca3Tl5 |
| 63.4.163 | 58894 | 42 | - | - | - - | - - | See details in Table XIX | | | | | Ca3Ga5 |
| 63.4.164 | 410775 | 38 | - | - | - - | - - | See details in Table XIX | | | | | Ca1Rh1Sn2 |
| 63.4.165 | 410773 | 38 | - | - | - - | - - | See details in Table XIX | | | | | Ca1Ir1Sn2 |
| 63.4.166 | 420479 | 58 | - | - | - - | - - | See details in Table XIX | | | | | Ca1Ir1O3 |
| 63.4.167 | 420478 | 58 | - | - | - - | - - | See details in Table XIX | | | | | Ca1Ir1O3 |
| 63.4.168 | 182246 | 58 | - | - | - - | - - | See details in Table XIX | | | | | Ca1Ir1O3 |

Continued on next page

**TABLE XIV– continued**

| TQC | ICSD | NoE | DG | IndG | SubSG cl (w/ T) | SubSG cl (w/o T) | Max. Sub. SG yes or no | Max. Sub. SG yes or no | SIs (w/ T) | Stoich. Formula |
|---|---|---|---|---|---|---|---|---|---|---|
| 63.4.169 | 180015 | 58 | - | - | - - | - - | See details in Table XIX | | | Ca1Ir1O3 |
| 63.4.170 | 160816 | 58 | - | - | - - | - - | See details in Table XIX | | | Ca1Ir1O3 |
| 63.4.171 | 159038 | 58 | - | - | - - | - - | See details in Table XIX | | | Ca1Ir1O3 |
| 63.4.172 | 159037 | 58 | - | - | - - | - - | See details in Table XIX | | | Ca1Ir1O3 |
| 63.4.173 | 159036 | 58 | - | - | - - | - - | See details in Table XIX | | | Ca1Ir1O3 |
| 63.4.174 | 159035 | 58 | - | - | - - | - - | See details in Table XIX | | | Ca1Ir1O3 |
| 63.4.175 | 159034 | 58 | - | - | - - | - - | See details in Table XIX | | | Ca1Ir1O3 |
| 63.4.176 | 159033 | 58 | - | - | - - | - - | See details in Table XIX | | | Ca1Ir1O3 |
| 63.4.177 | 159029 | 58 | - | - | - - | - - | See details in Table XIX | | | Ca1Ir1O3 |
| 63.4.178 | 159028 | 58 | - | - | - - | - - | See details in Table XIX | | | Ca1Ir1O3 |
| 63.4.179 | 159027 | 58 | - | - | - - | - - | See details in Table XIX | | | Ca1Ir1O3 |
| 63.4.180 | 58898 | 62 | - | - | - - | - - | See details in Table XIX | | | Ca1Ga3Ni2 |
| 63.4.181 | 419100 | 10 | - | - | - - | - - | See details in Table XIX | | | Ca1Ga1 |
| 63.4.182 | 260641 | 10 | - | - | - - | - - | See details in Table XIX | | | Ca1Ga1 |
| 63.4.183 | 106332 | 38 | - | - | - - | - - | See details in Table XIX | | | Ca1Cu1In2 |
| 63.4.184 | 411623 | 138 | - | - | - - | - - | See details in Table XIX | | | C2B3Nb4 |
| 63.4.185 | 50632 | 82 | - | - | - - | - - | See details in Table XIX | | | C1Re2Si1Tb1 |
| 63.4.186 | 50634 | 90 | - | - | - - | - - | See details in Table XIX | | | C1Ho1Ru2Si1 |
| 63.4.187 | 260483 | 86 | - | - | - - | - - | See details in Table XIX | | | C1Gd1Os2P1 |
| 63.4.188 | 415220 | 50 | - | - | - - | - - | See details in Table XIX | | | C1Cu1N2 |
| 63.4.189 | 161461 | 50 | - | - | - - | - - | See details in Table XIX | | | C1Cu1N2 |
| 63.4.190 | 161460 | 50 | - | - | - - | - - | See details in Table XIX | | | C1Cu1N2 |
| 63.4.191 | 260480 | 74 | - | - | - - | - - | See details in Table XIX | | | C1Ce1P1Ru2 |
| 63.4.192 | 73828 | 42 | - | - | - - | - - | See details in Table XIX | | | C1B1U1 |
| 63.4.193 | 108984 | 78 | - | - | - - | - - | See details in Table XIX | | | Br3Cm1 |
| 63.4.194 | 1107 | 66 | - | - | - - | - - | See details in Table XIX | | | Br1O2U1 |
| 63.4.195 | 409756 | 54 | - | - | - - | - - | See details in Table XIX | | | Bi1Zr1 |
| 63.4.196 | 50826 | 222 | - | - | - - | - - | See details in Table XIX | | | Ba3Cu1O9Ru2 |
| 63.4.197 | 50825 | 222 | - | - | - - | - - | See details in Table XIX | | | Ba3Cu1O9Ru2 |
| 63.4.198 | 63228 | 66 | - | - | - - | - - | See details in Table XIX | | | Ba1S3V1 |
| 63.4.199 | 58669 | 122 | - | - | - - | - - | See details in Table XIX | | | Ba1Pb6Rh3 |
| 63.4.200 | 195765 | 58 | - | - | - - | - - | See details in Table XIX | | | Ba1La1Si2 |
| 63.4.201 | 411232 | 50 | - | - | - - | - - | See details in Table XIX | | | Ba1In2Rh1 |
| 63.4.202 | 414434 | 50 | - | - | - - | - - | See details in Table XIX | | | Ba1In2Ir1 |
| 63.4.203 | 58648 | 58 | - | - | - - | - - | See details in Table XIX | | | Ba1Cu1Sn2 |
| 63.4.204 | 429001 | 114 | - | - | - cl | - cl | See details in Table XIX | | | Ba1Cu1Gd1Te3 |
| 63.4.205 | 280698 | 114 | - | - | - - | - - | See details in Table XIX | | | Ba1Cu1Gd1Se3 |
| 63.4.206 | 78176 | 122 | - | - | - - | - - | See details in Table XIX | | | Ba1Cu1Er1S3 |
| 63.4.207 | 428663 | 62 | - | - | - - | - - | See details in Table XIX | | | Ba1Co1O1S1 |
| 63.4.208 | 428516 | 62 | - | - | - - | - - | See details in Table XIX | | | Ba1Co1O1S1 |
| 63.4.209 | 64595 | 122 | - | - | - - | - - | See details in Table XIX | | | B7Re1Y3 |
| 63.4.210 | 62630 | 186 | - | - | - - | - - | See details in Table XIX | | | B7Cr1Er3 |
| 63.4.211 | 91244 | 78 | - | - | - - | - - | See details in Table XIX | | | B4Os3Sc1 |
| 63.4.212 | 391409 | 122 | - | - | - - | - - | See details in Table XIX | | | B4Mn1O7 |
| 63.4.213 | 391408 | 130 | - | - | - - | - - | See details in Table XIX | | | B4Cu1O7 |
| 63.4.214 | 420402 | 126 | - | - | - - | - - | See details in Table XIX | | | B4Co1O7 |
| 63.4.215 | 79258 | 38 | - | - | - - | - - | See details in Table XIX | | | B3V2 |
| 63.4.216 | 71643 | 270 | - | - | - - | - - | See details in Table XIX | | | B1Ir2Zn2 |
| 63.4.217 | 44249 | 18 | - | - | - - | - - | See details in Table XIX | | | B1Cr1 |
| 63.4.218 | 410436 | 82 | - | - | - cl | - - | See details in Table XIX | | | Au1In2Yb1 |
| 63.4.219 | 408882 | 38 | - | - | - - | - - | See details in Table XIX | | | Au1Ca1In2 |
| 63.4.220 | 88718 | 114 | - | - | - - | - - | See details in Table XIX | | | Au1Ba1Gd1Se3 |
| 63.4.221 | 33915 | 150 | - | - | - - | - - | See details in Table XIX | | | As3Ni5Sr1 |
| 63.4.222 | 57973 | 50 | - | - | - - | - - | See details in Table XIX | | | Al6Mn1 |
| 63.4.223 | 154855 | 82 | - | - | - - | - - | See details in Table XIX | | | Al4Ni1Tb1 |
| 63.4.224 | 370012 | 38 | - | - | - - | - - | See details in Table XIX | | | Al2Ni1Sc1 |
| 63.4.225 | 415062 | 38 | - | - | - - | - - | See details in Table XIX | | | Al2Cu1Mg1 |

Continued on next page

**TABLE XIV– continued**

| TQC | ICSD | NoE | DG | IndG | SubSG (w/ T) | cl | SubSG (w/o T) | cl | Max. Sub. SG | yes or no | Max. Sub. SG | yes or no | SIs (w/ T) | Stoich. Formula |
|---|---|---|---|---|---|---|---|---|---|---|---|---|---|---|
| 63.4.226 | 370003 | 250 | - | - | - | - | - | - | See details in Table XIX | | | | | Ag7Yb2 |
| 63.4.227 | 55510 | 162 | - | - | - | - | - | - | See details in Table XIX | | | | | Ag7Ca2 |
| 63.4.228 | 247421 | 62 | - | - | - | - | - | - | See details in Table XIX | | | | | Ag1Ce1Si2 |
| 63.4.229 | 57353 | 26 | - | - | - | - | - | - | See details in Table XIX | | | | | Ag1Ca1 |
| 63.4.230 | 54979 | 26 | - | - | - | - | - | - | See details in Table XIX | | | | | Ag1Ca1 |
| 64.1.1 | 411192 | 80 | 0.0120 | -0.6723 | 2 | 1 | 2 | 1 | - | - | - | - | z2w,1=1<br>z2w,2=1<br>z4=3 | Sn3Ti2 |
| 64.1.2 | 410902 | 80 | 0.0109 | -0.6780 | 2 | 1 | 2 | 1 | - | - | - | - | z2w,1=1<br>z2w,2=1<br>z4=3 | Sn3Ti2 |
| 64.1.3 | 413279 | 88 | 0.0072 | -0.9898 | 2 | 7 | 2 | 1 | - | - | - | - | z4=1 | Pd8Sn24 |
| 64.1.4 | 417180 | 20 | 0.0406 | -0.0725 | 2 | 2 | 2 | 1 | - | - | - | - | z2w,1=1<br>z2w,2=1<br>z4=1 | P1 |
| 64.1.5 | 23836 | 20 | 0.0276 | -0.0850 | 2 | 2 | 2 | 1 | - | - | - | - | z2w,1=1<br>z2w,2=1<br>z4=1 | P1 |
| 64.1.6 | 150873 | 20 | 0.0132 | -0.2061 | 2 | 2 | 2 | 1 | - | - | - | - | z2w,1=1<br>z2w,2=1<br>z4=1 | P1 |
| 64.1.7 | 77831 | 68 | 0.0000 | -1.9263 | 2 | 4 | 0 | 2 | - | - | - | - | z2w,1=1<br>z2w,2=1<br>z4=2 | La1S1 |
| 64.1.8 | 79878 | 132 | 0.0000 | -0.1226 | 2 | 2 | 2 | 1 | - | - | - | - | z2w,1=1<br>z2w,2=1<br>z4=1 | K1Na2Ni1O2 |
| 64.1.9 | 40442 | 108 | 0.0070 | -0.7727 | 2 | 6 | 2 | 1 | - | - | - | - | z4=3 | Hf2Ni3Si4 |
| 64.1.10 | 90341 | 72 | 0.0041 | -1.3147 | 2 | 4 | 0 | 2 | - | - | - | - | z2w,1=1<br>z2w,2=1<br>z4=2 | Ge2Ni1 |
| 64.1.11 | 150569 | 84 | 0.0261 | -0.0025 | 2 | 2 | 2 | 1 | - | - | - | - | z2w,1=1<br>z2w,2=1<br>z4=1 | Ga1Sn2V2 |
| 64.1.12 | 106752 | 84 | 0.0266 | -0.0026 | 2 | 2 | 2 | 1 | - | - | - | - | z2w,1=1<br>z2w,2=1<br>z4=1 | Ga1Sn2V2 |
| 64.1.13 | 71196 | 136 | 0.0000 | -0.2915 | 2 | 5 | 0 | 2 | - | - | - | - | z4=2 | Co2Nb2Te4 |
| 64.1.14 | 71519 | 136 | 0.0027 | -0.3026 | 2 | 5 | 0 | 2 | - | - | - | - | z4=2 | Co1Nb1Te2 |
| 64.1.15 | 424937 | 120 | 0.0052 | -0.1826 | 2 | 4 | 0 | 2 | - | - | - | - | z2w,1=1<br>z2w,2=1<br>z4=2 | Co1Cs1O2 |
| 64.1.16 | 40372 | 72 | 0.0195 | -0.0703 | 2 | 4 | 0 | 2 | - | - | - | - | z2w,1=1<br>z2w,2=1<br>z4=2 | Ca5Ga2N4 |
| 64.1.17 | 103352 | 192 | 0.0049 | -0.0691 | 2 | 4 | 0 | 2 | - | - | - | - | z2w,1=1<br>z2w,2=1<br>z4=2 | Ca2Mn1O3.5 |
| 64.1.18 | 186936 | 172 | 0.0044 | -0.5485 | 2 | 3 | 0 | 2 | - | - | - | - | z2w,1=1<br>z2w,2=1 | Ba2Ge4Pd5 |
| 64.1.19 | 58636 | 152 | 0.0232 | -0.7967 | 2 | 2 | 2 | 1 | - | - | - | - | z2w,1=1<br>z2w,2=1<br>z4=1 | Ba2Bi4Cd3 |
| 64.1.20 | 416447 | 152 | 0.0303 | -0.7585 | 2 | 2 | 2 | 1 | - | - | - | - | z2w,1=1<br>z2w,2=1<br>z4=1 | Ba2Bi4Cd3 |

Continued on next page

**TABLE XIV– continued**

| TQC | ICSD | NoE | DG | IndG | SubSG (w/ T) | cl | SubSG (w/o T) | cl | Max. Sub. SG | yes or no | Max. Sub. SG | yes or no | SIs (w/ T) | Stoich. Formula |
|---|---|---|---|---|---|---|---|---|---|---|---|---|---|---|
| 64.1.21 | 416446 | 152 | 0.0311 | -0.7563 | 2 | 2 | 2 | 1 | - | - | - | - | z2w,1=1<br>z2w,2=1<br>z4=1 | Ba2Bi4Cd3 |
| 64.1.22 | 416445 | 152 | 0.0214 | -0.8003 | 2 | 2 | 2 | 1 | - | - | - | - | z2w,1=1<br>z2w,2=1<br>z4=1 | Ba2Bi4Cd3 |
| 64.1.23 | 416444 | 152 | 0.0257 | -0.8017 | 2 | 2 | 2 | 1 | - | - | - | - | z2w,1=1<br>z2w,2=1<br>z4=1 | Ba2Bi4Cd3 |
| 64.1.24 | 412877 | 180 | 0.0024 | -0.0623 | 2 | 5 | 0 | 2 | - | - | - | - | z4=2 | Ba1Fe1K1O3 |
| 64.1.25 | 63708 | 176 | 0.0157 | -0.2519 | 2 | 2 | 2 | 1 | - | - | - | - | z2w,1=1<br>z2w,2=1<br>z4=1 | As2Ni3O8 |
| 64.1.26 | 68347 | 180 | 0.0022 | -0.2660 | 2 | 1 | 2 | 1 | - | - | - | - | z2w,1=1<br>z2w,2=1<br>z4=3 | Ag1Fe2S3 |
| 64.4.1 | 78300 | 114 | - | - | - | - | - | - | See details in Table XIX | | | | | Cu1La2O4 |
| 64.4.2 | 78299 | 114 | - | - | - | - | - | - | See details in Table XIX | | | | | Cu1La2O4 |
| 64.4.3 | 78298 | 114 | - | - | - | - | - | - | See details in Table XIX | | | | | Cu1La2O4 |
| 64.4.4 | 78297 | 114 | - | - | - | - | - | - | See details in Table XIX | | | | | Cu1La2O4 |
| 64.4.5 | 78296 | 114 | - | - | - | - | - | - | See details in Table XIX | | | | | Cu1La2O4 |
| 64.4.6 | 78295 | 114 | - | - | - | - | - | - | See details in Table XIX | | | | | Cu1La2O4 |
| 64.4.7 | 78294 | 114 | - | - | - | - | - | - | See details in Table XIX | | | | | Cu1La2O4 |
| 64.4.8 | 78293 | 114 | - | - | - | - | - | - | See details in Table XIX | | | | | Cu1La2O4 |
| 64.4.9 | 78292 | 114 | - | - | - | - | - | - | See details in Table XIX | | | | | Cu1La2O4 |
| 64.4.10 | 78291 | 114 | - | - | - | - | - | - | See details in Table XIX | | | | | Cu1La2O4 |
| 64.4.11 | 78290 | 114 | - | - | - | - | - | - | See details in Table XIX | | | | | Cu1La2O4 |
| 64.4.12 | 78289 | 114 | - | - | - | - | - | - | See details in Table XIX | | | | | Cu1La2O4 |
| 64.4.13 | 78288 | 114 | - | - | - | - | - | - | See details in Table XIX | | | | | Cu1La2O4 |
| 64.4.14 | 78287 | 114 | - | - | - | - | - | - | See details in Table XIX | | | | | Cu1La2O4 |
| 64.4.15 | 78237 | 114 | - | - | - | - | - | - | See details in Table XIX | | | | | Cu1La2O4 |
| 64.4.16 | 78194 | 114 | - | - | - | - | - | - | See details in Table XIX | | | | | Cu1La2O4 |
| 64.4.17 | 75891 | 114 | - | - | - | - | - | - | See details in Table XIX | | | | | Cu1La2O4 |
| 64.4.18 | 74150 | 114 | - | - | - | - | - | - | See details in Table XIX | | | | | Cu1La2O4 |
| 64.4.19 | 74145 | 114 | - | - | - | - | - | - | See details in Table XIX | | | | | Cu1La2O4 |
| 64.4.20 | 69312 | 114 | - | - | - | - | - | - | See details in Table XIX | | | | | Cu1La2O4 |
| 64.4.21 | 69311 | 114 | - | - | - | - | - | - | See details in Table XIX | | | | | Cu1La2O4 |
| 64.4.22 | 50265 | 114 | - | - | - | - | - | - | See details in Table XIX | | | | | Cu1La2O4 |
| 64.4.23 | 261661 | 114 | - | - | - | - | - | - | See details in Table XIX | | | | | Cu1La2O4 |
| 64.4.24 | 173895 | 114 | - | - | - | - | - | - | See details in Table XIX | | | | | Cu1La2O4 |
| 64.4.25 | 173893 | 114 | - | - | - | - | - | - | See details in Table XIX | | | | | Cu1La2O4 |
| 64.4.26 | 173891 | 114 | - | - | - | - | - | - | See details in Table XIX | | | | | Cu1La2O4 |
| 64.4.27 | 202724 | 146 | - | - | - | - | - | - | See details in Table XIX | | | | | Ba2Cu1O6Tl2 |
| 64.4.28 | 202722 | 146 | - | - | - | - | - | - | See details in Table XIX | | | | | Ba2Cu1O6Tl2 |
| 64.4.29 | 76359 | 94 | - | - | - | - | - | - | See details in Table XIX | | | | | Al3Ge4La2 |
| 64.4.30 | 50977 | 98 | - | - | - | - | - | - | See details in Table XIX | | | | | Al3Ce2Ge4 |
| 65.1.1 | 87827 | 66 | 0.0198 | -2.0903 | 2 | 2 | 10 | 2 | - | - | - | - | z2w,3=1<br>z4=2 | Ni1Si3Y1 |
| 65.1.2 | 78760 | 42 | 0.0000 | -0.1760 | 2 | 3 | 10 | 3 | - | - | - | - | z2w,3=1 | Mn2S4Sn1 |
| 65.1.3 | 240761 | 64 | 0.0000 | -0.0792 | 2 | 4 | 0 | 4 | - | - | - | - | z2w,1=1<br>z2w,2=1 | Mg3Ni2Tb2 |
| 65.1.4 | 261871 | 84 | 0.0291 | -0.4168 | 2 | 5 | 10 | 2 | - | - | - | - | z2w,1=1<br>z2w,2=1<br>z2w,3=1<br>z4=2 | La2Sn5 |
| 65.1.5 | 167637 | 188 | 0.0072 | -1.0470 | 2 | 3 | 10 | 3 | - | - | - | - | z2w,3=1 | In9Ni4Y11 |

Continued on next page

**TABLE XIV– continued**

| TQC | ICSD | NoE | DG | IndG | SubSG (w/ T) | cl | SubSG (w/o T) | cl | Max. Sub. SG | yes or no | Max. Sub. SG | yes or no | SIs (w/ T) | Stoich. Formula |
|---|---|---|---|---|---|---|---|---|---|---|---|---|---|---|
| 65.1.6 | 167624 | 188 | 0.0050 | -0.7152 | 2 | 4 | 0 | 4 | - | - | - | - | z2w,1=1<br>z2w,2=1 | In9La11Ni4 |
| 65.1.7 | 415513 | 156 | 0.0021 | -0.4891 | 2 | 2 | 10 | 2 | - | - | - | - | z2w,3=1<br>z4=2 | In16Pt7Tb2 |
| 65.1.8 | 54612 | 66 | 0.0000 | -0.0329 | 2 | 6 | 65 | 1 | - | - | - | - | z4=2 | Ho2Ni2Pb1 |
| 65.1.9 | 60646 | 66 | 0.0040 | -1.4170 | 2 | 1 | 65 | 1 | - | - | - | - | z2w,1=1<br>z2w,2=1<br>z4=2 | Ge3Ni1Y1 |
| 65.1.10 | 421933 | 62 | 0.0108 | -0.1377 | 2 | 5 | 10 | 2 | - | - | - | - | z2w,1=1<br>z2w,2=1<br>z2w,3=1<br>z4=2 | Gd2Mg3Ni2 |
| 65.1.11 | 20337 | 86 | 0.0031 | -0.1898 | 2 | 3 | 10 | 3 | - | - | - | - | z2w,3=1 | Fe2Si7U3 |
| 65.1.12 | 261512 | 76 | 0.0050 | -0.7134 | 2 | 1 | 65 | 1 | - | - | - | - | z2w,1=1<br>z2w,2=1<br>z4=2 | Fe2O5Sr3 |
| 65.1.13 | 238922 | 86 | 0.0054 | -0.0935 | 2 | 2 | 10 | 2 | - | - | - | - | z2w,3=1<br>z4=2 | Fe2Ge7U3 |
| 65.1.14 | 174439 | 94 | 0.0000 | -0.0652 | 2 | 4 | 0 | 4 | - | - | - | - | z2w,1=1<br>z2w,2=1 | Er3Sn7 |
| 65.1.15 | 165783 | 74 | 0.0000 | -0.0188 | 2 | 4 | 0 | 4 | - | - | - | - | z2w,1=1<br>z2w,2=1 | Dy2Ni1Sn6 |
| 65.1.16 | 418215 | 152 | 0.0020 | -0.1093 | 2 | 1 | 65 | 1 | - | - | - | - | z2w,1=1<br>z2w,2=1<br>z4=2 | Cu1Mg4Tb1 |
| 65.1.17 | 424317 | 132 | 0.0000 | -0.2642 | 2 | 4 | 0 | 4 | - | - | - | - | z2w,1=1<br>z2w,2=1 | Ce7Ge10Li8 |
| 65.1.18 | 63667 | 64 | 0.0065 | -0.3264 | 2 | 3 | 10 | 3 | - | - | - | - | z2w,3=1 | Ce3Sn7 |
| 65.1.19 | 261436 | 62 | 0.0057 | -0.0273 | 2 | 3 | 10 | 3 | - | - | - | - | z2w,3=1 | Cd1Cu2Nd2 |
| 65.1.20 | 419476 | 56 | 0.0023 | -0.0383 | 2 | 5 | 10 | 2 | - | - | - | - | z2w,1=1<br>z2w,2=1<br>z2w,3=1<br>z4=2 | C3B2Tb2 |
| 65.1.21 | 413212 | 154 | 0.0000 | -0.0405 | 2 | 2 | 10 | 2 | - | - | - | - | z2w,3=1<br>z4=2 | Br2Ce4Ga5 |
| 65.1.22 | 78621 | 136 | 0.0101 | -1.1720 | 2 | 1 | 65 | 1 | - | - | - | - | z2w,1=1<br>z2w,2=1<br>z4=2 | Ba2Cu4O8Yb1 |
| 65.1.23 | 154010 | 158 | 0.0038 | -0.2120 | 2 | 1 | 65 | 1 | - | - | - | - | z2w,1=1<br>z2w,2=1<br>z4=2 | Ba1Mn2O6Tb1 |
| 65.1.24 | 106291 | 104 | 0.0735 | -0.7983 | 2 | 1 | 65 | 1 | - | - | - | - | z2w,1=1<br>z2w,2=1<br>z4=2 | Au7Rb3 |
| 65.1.25 | 106290 | 104 | 0.0870 | -0.7942 | 2 | 1 | 65 | 1 | - | - | - | - | z2w,1=1<br>z2w,2=1<br>z4=2 | Au7Rb3 |
| 65.1.26 | 418529 | 64 | 0.0093 | -1.2937 | 2 | 1 | 65 | 1 | - | - | - | - | z2w,1=1<br>z2w,2=1<br>z4=2 | Au4Ce1Si2 |
| 65.2.1 | 15927 | 50 | 0.0052 | -0.1235 | 2 | 7 | 2 | 6 | - | - | - | - | z2w,1=1<br>z2w,2=1<br>z4=1 | Mg1O3V1 |
| 65.2.2 | 104809 | 84 | 0.0125 | -2.0742 | 2 | 10 | 2 | 7 | - | - | - | - | z4=3 | Lu2Ni1Sn6 |
| 65.2.3 | 106923 | 102 | 0.0043 | -1.7023 | 2 | 7 | 2 | 6 | - | - | - | - | z2w,1=1<br>z2w,2=1<br>z4=1 | Lu1Ni1Sn4 |
| 65.2.4 | 83885 | 44 | 0.0056 | -0.1670 | 2 | 9 | 2 | 6 | - | - | - | - | z4=1 | Ge2U1 |

Continued on next page

**TABLE XIV– continued**

| TQC | ICSD | NoE | DG | IndG | SubSG (w/ T) | cl | SubSG (w/o T) | cl | Max. Sub. SG | yes or no | Max. Sub. SG | yes or no | SIs (w/ T) | Stoich. Formula |
|---|---|---|---|---|---|---|---|---|---|---|---|---|---|---|
| 65.2.5 | 82702 | 44 | 0.0120 | -0.1797 | 2 | 9 | 2 | 6 | - | - | - | - | z4=1 | Ge2U1 |
| 65.2.6 | 236995 | 44 | 0.0117 | -0.1817 | 2 | 9 | 2 | 6 | - | - | - | - | z4=1 | Ge2U1 |
| 65.2.7 | 56046 | 40 | 0.0154 | -0.4407 | 2 | 10 | 2 | 7 | - | - | - | - | z4=3 | Ge2Th1 |
| 65.2.8 | 193219 | 32 | 0.0567 | -2.0720 | 2 | 12 | 2 | 8 | - | - | - | - | z2w,1=1 z2w,2=1 z2w,3=1 z4=1 | Ge2Pb1Sr2 |
| 65.2.9 | 40487 | 240 | 0.0036 | -0.2239 | 2 | 7 | 2 | 6 | - | - | - | - | z2w,1=1 z2w,2=1 z4=1 | Ge2Lu1Mn1 |
| 65.2.10 | 104134 | 82 | 0.0039 | -0.0274 | 2 | 11 | 2 | 8 | - | - | - | - | z2w,3=1 z4=1 | Gd3Sn7 |
| 65.2.11 | 91064 | 138 | 0.0026 | -0.1386 | 2 | 13 | 2 | 5 | - | - | - | - | z2w,3=1 z4=3 | Fe4O11Sr4 |
| 65.2.12 | 171189 | 88 | 0.0000 | -0.0944 | 2 | 7 | 2 | 6 | - | - | - | - | z2w,1=1 z2w,2=1 z4=1 | Er1Ni1Si3 |
| 65.2.13 | 154835 | 120 | 0.0000 | -0.0880 | 2 | 7 | 2 | 6 | - | - | - | - | z2w,1=1 z2w,2=1 z4=1 | Dy1Fe4Ge2 |
| 65.2.14 | 419475 | 120 | 0.0000 | -1.4774 | 2 | 9 | 2 | 6 | - | - | - | - | z4=1 | Cu1Mg4Y1 |
| 65.2.15 | 418093 | 88 | 0.0023 | -0.1868 | 2 | 15 | 2 | 5 | - | - | - | - | z2w,1=1 z2w,2=1 z2w,3=1 z4=3 | Co2Ge7U3 |
| 65.2.16 | 72872 | 40 | 0.0454 | -1.4968 | 2 | 13 | 2 | 5 | - | - | - | - | z2w,3=1 z4=3 | Co1O4Re1 |
| 65.2.17 | 16497 | 88 | 0.0045 | -0.4362 | 2 | 12 | 2 | 8 | - | - | - | - | z2w,1=1 z2w,2=1 z2w,3=1 z4=1 | Ce3Ni2Si8 |
| 65.2.18 | 174528 | 52 | 0.0085 | -1.2218 | 2 | 7 | 2 | 6 | - | - | - | - | z2w,1=1 z2w,2=1 z4=1 | Ce2Si7 |
| 65.2.19 | 39221 | 234 | 0.0000 | -0.0211 | 2 | 15 | 2 | 5 | - | - | - | - | z2w,1=1 z2w,2=1 z2w,3=1 z4=3 | C4Er4Ni13 |
| 65.2.20 | 185392 | 148 | 0.0058 | -1.2283 | 2 | 10 | 2 | 7 | - | - | - | - | z4=3 | Br2Cd2Hg2S2 |
| 65.2.21 | 391060 | 184 | 0.0540 | 0.0186 | 2 | 11 | 2 | 8 | - | - | - | - | z2w,3=1 z4=1 | Ba3Ge4 |
| 65.2.22 | 75687 | 134 | 0.0000 | -0.3412 | 2 | 11 | 2 | 8 | - | - | - | - | z2w,3=1 z4=1 | Ba2Cu4Er1O8 |
| 65.2.23 | 74127 | 134 | 0.0027 | -0.3385 | 2 | 11 | 2 | 8 | - | - | - | - | z2w,3=1 z4=1 | Ba2Cu4Er1O8 |
| 65.2.24 | 67642 | 134 | 0.0000 | -0.3317 | 2 | 11 | 2 | 8 | - | - | - | - | z2w,3=1 z4=1 | Ba2Cu4Er1O8 |
| 65.2.25 | 39394 | 96 | 0.0025 | -0.5431 | 2 | 10 | 2 | 7 | - | - | - | - | z4=3 | B6Ni3Yb2 |
| 65.2.26 | 68002 | 98 | 0.0000 | -1.0880 | 2 | 7 | 2 | 6 | - | - | - | - | z2w,1=1 z2w,2=1 z4=1 | B6Lu2Ni3 |
| 65.2.27 | 417990 | 144 | 0.0182 | -1.6268 | 2 | 9 | 2 | 6 | - | - | - | - | z4=1 | B4Rh8Zn5 |
| 65.2.28 | 413191 | 146 | 0.0053 | -0.5824 | 2 | 8 | 2 | 7 | - | - | - | - | z2w,1=1 z2w,2=1 z4=3 | Al5Br2La4 |
| 65.2.29 | 260798 | 78 | 0.0183 | -0.9395 | 2 | 7 | 2 | 6 | - | - | - | - | z2w,1=1 z2w,2=1 z4=1 | Al1B4Yb1 |

Continued on next page

**TABLE XIV– continued**

| TQC | ICSD | NoE | DG | IndG | SubSG (w/ T) | cl | SubSG (w/o T) | cl | Max. Sub. SG | yes or no | Max. Sub. SG | yes or no | SIs (w/ T) | Stoich. Formula |
|---|---|---|---|---|---|---|---|---|---|---|---|---|---|---|
| 65.2.30 | 167533 | 78 | 0.0182 | -0.9394 | 2 | 7 | 2 | 6 | - | - | - | - | z2w,1=1<br>z2w,2=1<br>z4=1 | Al1B4Yb1 |
| 65.2.31 | 167535 | 80 | 0.0444 | -1.4645 | 2 | 13 | 2 | 5 | - | - | - | - | z2w,3=1<br>z4=3 | Al1B4Lu1 |
| 67.2.1 | 167666 | 52 | 0.0259 | -1.0900 | 2 | 1 | 2 | 1 | - | - | - | - | z2w,3=1<br>z4=3 | Al2Cu1Ir1 |
| 67.4.1 | 411905 | 54 | - | - | - | - | - | - | See details in Table XIX | | | | | Ba2Na1O1 |
| 67.4.2 | 59306 | 142 | - | - | - | - | - | - | See details in Table XIX | | | | | Ba1F6Tb1 |
| 69.1.1 | 262559 | 196 | 0.0060 | -0.0545 | 2 | 4 | 69 | 1 | - | - | - | - | z2w,1=1<br>z2w,3=1<br>z4=2 | Pt3Rb2Se6U1 |
| 69.1.2 | 262561 | 196 | 0.0157 | -0.0 - | 2 | 4 | 69 | 1 | - | - | - | - | z2w,1=1<br>z2w,3=1<br>z4=2 | Pd3Rb2Se6U1 |
| 69.1.3 | 262562 | 196 | 0.0101 | -0.0666 | 2 | 4 | 69 | 1 | - | - | - | - | z2w,1=1<br>z2w,3=1<br>z4=2 | Pd3Rb2S6U1 |
| 69.1.4 | 262563 | 196 | 0.0055 | -0.0862 | 2 | 4 | 69 | 1 | - | - | - | - | z2w,1=1<br>z2w,3=1<br>z4=2 | K2Pd3S6U1 |
| 69.1.5 | 414574 | 66 | 0.0072 | -0.5026 | 2 | 1 | 69 | 1 | - | - | - | - | z2w,1=1<br>z2w,2=1<br>z4=2 | Ge6Li2Sr4 |
| 69.1.6 | 68676 | 60 | 0.0113 | -1.7218 | 2 | 3 | 2 | 2 | - | - | - | - | z4=3 | Cu2O3Sr2 |
| 69.1.7 | 262555 | 196 | 0.0063 | -0.0376 | 2 | 4 | 69 | 1 | - | - | - | - | z2w,1=1<br>z2w,3=1<br>z4=2 | Cs2Pt3Se6U1 |
| 69.1.8 | 262556 | 196 | 0.0000 | -0.0627 | 2 | 11 | 2 | 2 | - | - | - | - | z2w,1=1<br>z2w,3=1<br>z4=3 | Cs2Pt3S6U1 |
| 69.1.9 | 262557 | 196 | 0.0082 | -0.0221 | 2 | 4 | 69 | 1 | - | - | - | - | z2w,1=1<br>z2w,3=1<br>z4=2 | Cs2Pd3Se6U1 |
| 69.1.10 | 261242 | 196 | 0.0082 | -0.0221 | 2 | 4 | 69 | 1 | - | - | - | - | z2w,1=1<br>z2w,3=1<br>z4=2 | Cs2Pd3Se6U1 |
| 69.1.11 | 262558 | 196 | 0.0047 | -0.0526 | 2 | 4 | 69 | 1 | - | - | - | - | z2w,1=1<br>z2w,3=1<br>z4=2 | Cs2Pd3S6U1 |
| 69.1.12 | 73160 | 174 | 0.0000 | -0.0222 | 2 | 3 | 2 | 2 | - | - | - | - | z4=3 | Ba2Na1Ni3O6 |
| 69.1.13 | 59588 | 174 | 0.0000 | -0.0257 | 2 | 3 | 2 | 2 | - | - | - | - | z4=3 | Ba2Na1Ni3O6 |
| 69.1.14 | 72328 | 180 | 0.0034 | -0.1262 | 2 | 2 | 2 | 2 | - | - | - | - | z2w,2=1<br>z2w,3=1<br>z4=3 | Ba2Cu3Na1O6 |
| 69.1.15 | 59589 | 180 | 0.0031 | -0.1788 | 2 | 2 | 2 | 2 | - | - | - | - | z2w,2=1<br>z2w,3=1<br>z4=3 | Ba2Cu3Na1O6 |
| 69.1.16 | 401239 | 180 | 0.0070 | -0.1495 | 2 | 2 | 2 | 2 | - | - | - | - | z2w,2=1<br>z2w,3=1<br>z4=3 | Ba2Cu3Li1O6 |
| 69.1.17 | 174521 | 114 | 0.0044 | -0.3066 | 2 | 11 | 2 | 2 | - | - | - | - | z2w,1=1<br>z2w,3=1<br>z4=3 | Au1Eu2Ge3 |
| 69.1.18 | 410522 | 54 | 0.0000 | -1.1478 | 2 | 1 | 69 | 1 | - | - | - | - | z2w,1=1<br>z2w,2=1<br>z4=2 | Ag2Ca4Si6 |
| 70.1.1 | 1089 | 24 | 0.0070 | -1.8215 | 2 | 2 | 2 | 1 | - | - | - | - | z4=3 | Si2Ti1 |

Continued on next page

**TABLE XIV– continued**

| TQC | ICSD | NoE | DG | IndG | SubSG (w/ T) | cl | SubSG (w/o T) | cl | Max. Sub. SG | yes or no | Max. Sub. SG | yes or no | SIs (w/ T) | Stoich. Formula |
|---|---|---|---|---|---|---|---|---|---|---|---|---|---|---|
| 70.1.2 | 97545 | 52 | 0.0100 | -0.9835 | 2 | 2 | 2 | 1 | - | - | - | - | z4=3 | Sb1Sn1Ti1 |
| 70.1.3 | 88369 | 96 | 0.0000 | -0.0257 | 2 | 2 | 2 | 1 | - | - | - | - | z4=3 | O4Tb1V1 |
| 70.1.4 | 106552 | 84 | 0.0055 | -0.9551 | 2 | 1 | 2 | 1 | - | - | - | - | z4=1 | Nb1Sn2 |
| 70.1.5 | 429511 | 68 | 0.0478 | -0.0421 | 2 | 1 | 2 | 1 | - | - | - | - | z4=1 | Ga2Rh2Sr1 |
| 70.1.6 | 89074 | 80 | 0.0034 | -0.1583 | 2 | 2 | 2 | 1 | - | - | - | - | z4=3 | Cr2O4Si1 |
| 70.1.7 | 89073 | 80 | 0.0051 | -0.1562 | 2 | 2 | 2 | 1 | - | - | - | - | z4=3 | Cr2O4Si1 |
| 70.1.8 | 89072 | 80 | 0.0061 | -0.1503 | 2 | 2 | 2 | 1 | - | - | - | - | z4=3 | Cr2O4Si1 |
| 70.1.9 | 89071 | 80 | 0.0000 | -0.1471 | 2 | 2 | 2 | 1 | - | - | - | - | z4=3 | Cr2O4Si1 |
| 70.1.10 | 89070 | 80 | 0.0071 | -0.1109 | 2 | 2 | 2 | 1 | - | - | - | - | z4=3 | Cr2O4Si1 |
| 70.1.11 | 89069 | 80 | 0.0065 | -0.1173 | 2 | 2 | 2 | 1 | - | - | - | - | z4=3 | Cr2O4Si1 |
| 70.1.12 | 75639 | 80 | 0.0068 | -0.1212 | 2 | 2 | 2 | 1 | - | - | - | - | z4=3 | Cr2O4Si1 |
| 70.1.13 | 96610 | 84 | 0.0301 | -0.2170 | 2 | 2 | 2 | 1 | - | - | - | - | z4=3 | Cl2Cu2O1 |
| 70.1.14 | 77143 | 68 | 0.0327 | -0.2875 | 2 | 1 | 2 | 1 | - | - | - | - | z4=1 | Ba1Ge2Ru2 |
| 70.1.15 | 40620 | 84 | 0.0037 | -0.0213 | 2 | 1 | 2 | 1 | - | - | - | - | z4=1 | B2Gd1Ir2 |
| 70.1.16 | 40799 | 68 | 0.0303 | -0.3067 | 2 | 1 | 2 | 1 | - | - | - | - | z4=1 | B2Ce1Ru2 |
| 70.1.17 | 40618 | 72 | 0.0160 | -0.1929 | 2 | 1 | 2 | 1 | - | - | - | - | z4=1 | B2Ce1Ir2 |
| 70.1.18 | 49737 | 52 | 0.0242 | -0.2282 | 2 | 1 | 2 | 1 | - | - | - | - | z4=1 | B2Ca1Rh2 |
| 70.1.19 | 49738 | 52 | 0.1164 | -0.4991 | 2 | 1 | 2 | 1 | - | - | - | - | z4=1 | B2Ca1Ir2 |
| 70.1.20 | 75028 | 52 | 0.0570 | -0.4290 | 2 | 1 | 2 | 1 | - | - | - | - | z4=1 | B1Ir1Li1 |
| 70.1.21 | 75027 | 52 | 0.0673 | -0.4253 | 2 | 1 | 2 | 1 | - | - | - | - | z4=1 | B1Ir1Li1 |
| 70.1.22 | 418163 | 60 | 0.0106 | -0.3762 | 2 | 2 | 2 | 1 | - | - | - | - | z4=3 | B1Cr2 |
| 70.1.23 | 410521 | 228 | 0.0000 | -0.0946 | 2 | 1 | 2 | 1 | - | - | - | - | z4=1 | Ag2Eu4Si6 |
| 70.4.1 | 73610 | 178 | - | - | - | - | - | - | See details in Table XIX | | | | | O10U1V3 |
| 70.4.2 | 84406 | 226 | - | - | - | - | - | - | See details in Table XIX | | | | | Nb3O10U1 |
| 70.4.3 | 94466 | 134 | - | - | - | - | - | - | See details in Table XIX | | | | | Fe2K3O4 |
| 70.4.4 | 40621 | 70 | - | - | - | - | - | - | See details in Table XIX | | | | | B2Ir2Y1 |
| 70.4.5 | 40619 | 82 | - | - | - | - | - | - | See details in Table XIX | | | | | B2Eu1Ir2 |
| 71.1.1 | 82655 | 52 | 0.0216 | 0.0192 | 2 | 7 | 2 | 3 | - | - | - | - | z2w,1=1<br>z2w,2=1<br>z2w,3=1<br>z4=1 | Te2U1 |
| 71.1.2 | 82654 | 52 | 0.0216 | 0.0192 | 2 | 7 | 2 | 3 | - | - | - | - | z2w,1=1<br>z2w,2=1<br>z2w,3=1<br>z4=1 | Te2U1 |
| 71.1.3 | 82653 | 52 | 0.0216 | 0.0192 | 2 | 7 | 2 | 3 | - | - | - | - | z2w,1=1<br>z2w,2=1<br>z2w,3=1<br>z4=1 | Te2U1 |
| 71.1.4 | 82652 | 52 | 0.0216 | 0.0192 | 2 | 7 | 2 | 3 | - | - | - | - | z2w,1=1<br>z2w,2=1<br>z2w,3=1<br>z4=1 | Te2U1 |
| 71.1.5 | 82651 | 52 | 0.0215 | 0.0191 | 2 | 7 | 2 | 3 | - | - | - | - | z2w,1=1<br>z2w,2=1<br>z2w,3=1<br>z4=1 | Te2U1 |
| 71.1.6 | 82650 | 52 | 0.0215 | 0.0190 | 2 | 7 | 2 | 3 | - | - | - | - | z2w,1=1<br>z2w,2=1<br>z2w,3=1<br>z4=1 | Te2U1 |
| 71.1.7 | 82649 | 52 | 0.0215 | 0.0191 | 2 | 7 | 2 | 3 | - | - | - | - | z2w,1=1<br>z2w,2=1<br>z2w,3=1<br>z4=1 | Te2U1 |

Continued on next page

**TABLE XIV– continued**

| TQC | ICSD | NoE | DG | IndG | SubSG (w/ T) | cl | SubSG (w/o T) | cl | Max. Sub. SG | yes or no | Max. Sub. SG | yes or no | SIs (w/ T) | Stoich. Formula |
|---|---|---|---|---|---|---|---|---|---|---|---|---|---|---|
| 71.1.8 | 82648 | 52 | 0.0215 | 0.0191 | 2 | 7 | 2 | 3 | - | - | - | - | z2w,1=1<br>z2w,2=1<br>z2w,3=1<br>z4=1 | Te2U1 |
| 71.1.9 | 82647 | 52 | 0.0217 | 0.0191 | 2 | 7 | 2 | 3 | - | - | - | - | z2w,1=1<br>z2w,2=1<br>z2w,3=1<br>z4=1 | Te2U1 |
| 71.1.10 | 82646 | 52 | 0.0216 | 0.0191 | 2 | 7 | 2 | 3 | - | - | - | - | z2w,1=1<br>z2w,2=1<br>z2w,3=1<br>z4=1 | Te2U1 |
| 71.1.11 | 82645 | 52 | 0.0216 | 0.0191 | 2 | 7 | 2 | 3 | - | - | - | - | z2w,1=1<br>z2w,2=1<br>z2w,3=1<br>z4=1 | Te2U1 |
| 71.1.12 | 82644 | 52 | 0.0215 | 0.0190 | 2 | 7 | 2 | 3 | - | - | - | - | z2w,1=1<br>z2w,2=1<br>z2w,3=1<br>z4=1 | Te2U1 |
| 71.1.13 | 82643 | 52 | 0.0216 | 0.0191 | 2 | 7 | 2 | 3 | - | - | - | - | z2w,1=1<br>z2w,2=1<br>z2w,3=1<br>z4=1 | Te2U1 |
| 71.1.14 | 82642 | 52 | 0.0216 | 0.0191 | 2 | 7 | 2 | 3 | - | - | - | - | z2w,1=1<br>z2w,2=1<br>z2w,3=1<br>z4=1 | Te2U1 |
| 71.1.15 | 63170 | 52 | 0.0218 | 0.0194 | 2 | 7 | 2 | 3 | - | - | - | - | z2w,1=1<br>z2w,2=1<br>z2w,3=1<br>z4=1 | Te2U1 |
| 71.1.16 | 40694 | 52 | 0.0193 | 0.0173 | 2 | 7 | 2 | 3 | - | - | - | - | z2w,1=1<br>z2w,2=1<br>z2w,3=1<br>z4=1 | Te2U1 |
| 71.1.17 | 416820 | 70 | 0.0025 | -0.0559 | 2 | 5 | 71 | 4 | - | - | - | - | z2w,1=1<br>z2w,2=1<br>z2w,3=1<br>z4=2 | Ni2Tb2Zn1 |
| 71.1.18 | 85406 | 120 | 0.0078 | -1.4499 | 2 | 4 | 2 | 1 | - | - | - | - | z2w,1=1<br>z2w,2=1<br>z2w,3=1<br>z4=3 | Ni2P2Sr1 |
| 71.1.19 | 25307 | 54 | 0.0350 | -0.2270 | 2 | 5 | 71 | 4 | - | - | - | - | z2w,1=1<br>z2w,2=1<br>z2w,3=1<br>z4=2 | Li4Sb4Sr3 |
| 71.1.20 | 416817 | 54 | 0.0042 | -0.9428 | 2 | 3 | 2 | 3 | - | - | - | - | z4=1 | La2Ni2Zn1 |
| 71.1.21 | 56747 | 48 | 0.0200 | -0.0362 | 2 | 1 | 2 | 1 | - | - | - | - | z4=3 | H1Li1Nd2O3 |
| 71.1.22 | 56746 | 48 | 0.0206 | -0.0362 | 2 | 1 | 2 | 1 | - | - | - | - | z4=3 | H1Li1Nd2O3 |
| 71.1.23 | 56745 | 44 | 0.0104 | -0.1030 | 2 | 3 | 2 | 3 | - | - | - | - | z4=1 | H1Ce2Li1O3 |
| 71.1.24 | 710017 | 34 | 0.0158 | -0.6477 | 2 | 3 | 2 | 3 | - | - | - | - | z4=1 | Ge5Nd1 |
| 71.1.25 | 246814 | 34 | 0.0154 | -0.6413 | 2 | 3 | 2 | 3 | - | - | - | - | z4=1 | Ge5Nd1 |
| 71.1.26 | 411182 | 128 | 0.0100 | -0.4321 | 2 | 6 | 71 | 4 | - | - | - | - | z4=2 | Ge4Pd4Yb3 |
| 71.1.27 | 262435 | 62 | 0.0000 | -0.0124 | 2 | 1 | 2 | 1 | - | - | - | - | z4=3 | Ge4Li4Nd3 |

Continued on next page

**TABLE XIV– continued**

| TQC | ICSD | NoE | DG | IndG | SubSG (w/ T) | cl | SubSG (w/o T) | cl | Max. Sub. SG | yes or no | Max. Sub. SG | yes or no | SIs (w/ T) | Stoich. Formula |
|---|---|---|---|---|---|---|---|---|---|---|---|---|---|---|
| 71.1.28 | 170342 | 168 | 0.0037 | -0.9242 | 2 | 2 | 0 | 2 | - | - | - | - | z2w,1=1 z2w,2=1 z2w,3=1 | Ge2Ni1Yb1 |
| 71.1.29 | 170343 | 172 | 0.0071 | -1.4238 | 2 | 2 | 0 | 2 | - | - | - | - | z2w,1=1 z2w,2=1 z2w,3=1 | Ge2Lu1Ni1 |
| 71.1.30 | 76509 | 112 | 0.0468 | -1.6977 | 2 | 6 | 71 | 4 | - | - | - | - | z4=2 | Ge2Ir1Y1 |
| 71.1.31 | 88227 | 156 | 0.0033 | -0.1290 | 2 | 2 | 0 | 2 | - | - | - | - | z2w,1=1 z2w,2=1 z2w,3=1 | Ge2Ho1Pt1 |
| 71.1.32 | 300260 | 156 | 0.0050 | -0.0215 | 2 | 1 | 2 | 1 | - | - | - | - | z4=3 | Ge2Ho1Pt1 |
| 71.1.33 | 174530 | 38 | 0.0140 | -0.3055 | 2 | 3 | 2 | 3 | - | - | - | - | z4=1 | Gd1Si5 |
| 71.1.34 | 710016 | 38 | 0.0199 | -0.3096 | 2 | 3 | 2 | 3 | - | - | - | - | z4=1 | Gd1Ge5 |
| 71.1.35 | 154471 | 80 | 0.0000 | -0.2129 | 2 | 4 | 2 | 1 | - | - | - | - | z2w,1=1 z2w,2=1 z2w,3=1 z4=3 | Fe6Ga6U1 |
| 71.1.36 | 187334 | 90 | 0.0164 | -0.0601 | 2 | 3 | 2 | 3 | - | - | - | - | z4=1 | Fe4Ge4U3 |
| 71.1.37 | 83190 | 102 | 0.0000 | -0.0084 | 2 | 2 | 0 | 2 | - | - | - | - | z2w,1=1 z2w,2=1 z2w,3=1 | Cu8Ge8Nd6 |
| 71.1.38 | 102956 | 114 | 0.0000 | -0.0127 | 2 | 2 | 0 | 2 | - | - | - | - | z2w,1=1 z2w,2=1 z2w,3=1 | Cu8Gd6Sn8 |
| 71.1.39 | 413485 | 158 | 0.0063 | -0.8400 | 2 | 4 | 2 | 1 | - | - | - | - | z2w,1=1 z2w,2=1 z2w,3=1 z4=3 | Cu6Sn5Yb3 |
| 71.1.40 | 426132 | 132 | 0.0117 | -0.9149 | 2 | 4 | 2 | 1 | - | - | - | - | z2w,1=1 z2w,2=1 z2w,3=1 z4=3 | Cu4Ge4Yb3 |
| 71.1.41 | 82565 | 102 | 0.0000 | -0.1055 | 2 | 7 | 2 | 3 | - | - | - | - | z2w,1=1 z2w,2=1 z2w,3=1 z4=1 | Cu4Ge4U3 |
| 71.1.42 | 81142 | 102 | 0.0000 | -0.1045 | 2 | 7 | 2 | 3 | - | - | - | - | z2w,1=1 z2w,2=1 z2w,3=1 z4=1 | Cu4Ge4U3 |
| 71.1.43 | 42717 | 114 | 0.0022 | -0.0235 | 2 | 2 | 0 | 2 | - | - | - | - | z2w,1=1 z2w,2=1 z2w,3=1 | Cu4Gd3Ge4 |
| 71.1.44 | 55650 | 126 | 0.0029 | -0.0501 | 2 | 6 | 71 | 4 | - | - | - | - | z4=2 | Cu4Er3Si4 |
| 71.1.45 | 201112 | 58 | 0.0229 | 0.0038 | 2 | 5 | 71 | 4 | - | - | - | - | z2w,1=1 z2w,2=1 z2w,3=1 z4=2 | Cs1Fe1S2 |
| 71.1.46 | 247288 | 88 | 0.0111 | -0.5988 | 2 | 2 | 0 | 2 | - | - | - | - | z2w,1=1 z2w,2=1 z2w,3=1 | Ce3Rh4Si4 |
| 71.1.47 | 174076 | 92 | 0.0155 | -0.7241 | 2 | 4 | 2 | 1 | - | - | - | - | z2w,1=1 z2w,2=1 z2w,3=1 z4=3 | Ce3Pd4Si4 |

Continued on next page

**TABLE XIV– continued**

| TQC | ICSD | NoE | DG | IndG | SubSG (w/ T) | cl | SubSG (w/o T) | cl | Max. Sub. SG | yes or no | Max. Sub. SG | yes or no | SIs (w/ T) | Stoich. Formula |
|---|---|---|---|---|---|---|---|---|---|---|---|---|---|---|
| 71.1.48 | 262434 | 56 | 0.0024 | -0.2090 | 2 | 5 | 71 | 4 | - | - | - | - | z2w,1=1 z2w,2=1 z2w,3=1 z4=2 | Ce3Ge4Li4 |
| 71.1.49 | 151160 | 96 | 0.0063 | -0.1361 | 2 | 1 | 2 | 1 | - | - | - | - | z4=3 | Ce3Cu4Sn4 |
| 71.1.50 | 416816 | 56 | 0.0043 | -0.1314 | 2 | 5 | 71 | 4 | - | - | - | - | z2w,1=1 z2w,2=1 z2w,3=1 z4=2 | Ce2Ni2Zn1 |
| 71.1.51 | 174529 | 32 | 0.0255 | -0.9682 | 2 | 3 | 2 | 3 | - | - | - | - | z4=1 | Ce1Si5 |
| 71.1.52 | 246812 | 32 | 0.0323 | -0.7294 | 2 | 3 | 2 | 3 | - | - | - | - | z4=1 | Ce1Ge5 |
| 71.1.53 | 429175 | 26 | 0.0191 | -1.5738 | 2 | 5 | 71 | 4 | - | - | - | - | z2w,1=1 z2w,2=1 z2w,3=1 z4=2 | Ca2Mg1Pt2 |
| 71.1.54 | 429174 | 26 | 0.0088 | -1.9848 | 2 | 5 | 71 | 4 | - | - | - | - | z2w,1=1 z2w,2=1 z2w,3=1 z4=2 | Ca2Mg1Pd2 |
| 71.1.55 | 42 - 1 | 36 | 0.0191 | -1.7988 | 2 | 5 | 71 | 4 | - | - | - | - | z2w,1=1 z2w,2=1 z2w,3=1 z4=2 | Ca2Cd1Pd2 |
| 71.1.56 | 40950 | 34 | 0.0206 | -1.1475 | 2 | 2 | 0 | 2 | - | - | - | - | z2w,1=1 z2w,2=1 z2w,3=1 | C4Co1Sc3 |
| 71.1.57 | 236393 | 34 | 0.0043 | -1.1346 | 2 | 7 | 2 | 3 | - | - | - | - | z2w,1=1 z2w,2=1 z2w,3=1 z4=1 | C4Co1Sc3 |
| 71.1.58 | 173167 | 34 | 0.0055 | -1.1394 | 2 | 7 | 2 | 3 | - | - | - | - | z2w,1=1 z2w,2=1 z2w,3=1 z4=1 | C4Co1Sc3 |
| 71.1.59 | 94741 | 68 | 0.0423 | -0.7469 | 2 | 5 | 71 | 4 | - | - | - | - | z2w,1=1 z2w,2=1 z2w,3=1 z4=2 | C2Bi2Ca1O8 |
| 71.1.60 | 238028 | 52 | 0.0091 | -0.0807 | 2 | 4 | 2 | 1 | - | - | - | - | z2w,1=1 z2w,2=1 z2w,3=1 z4=3 | C1Co2Pr2Si1 |
| 71.1.61 | 418403 | 56 | 0.0030 | -0.0206 | 2 | 4 | 2 | 1 | - | - | - | - | z2w,1=1 z2w,2=1 z2w,3=1 z4=3 | C1B4Dy2 |
| 71.1.62 | 428114 | 54 | 0.0372 | -0.2638 | 2 | 5 | 71 | 4 | - | - | - | - | z2w,1=1 z2w,2=1 z2w,3=1 z4=2 | Ba3Li4Sb4 |
| 71.1.63 | 290306 | 162 | 0.0174 | -0.6649 | 2 | 2 | 0 | 2 | - | - | - | - | z2w,1=1 z2w,2=1 z2w,3=1 | Ba3Cd1Hg10 |
| 71.1.64 | 428115 | 54 | 0.0455 | -0.6898 | 2 | 5 | 71 | 4 | - | - | - | - | z2w,1=1 z2w,2=1 z2w,3=1 z4=2 | Ba3Bi4Li4 |
| 71.1.65 | 72631 | 96 | 0.0027 | -0.0563 | 2 | 1 | 2 | 1 | - | - | - | - | z4=3 | Ba1Ni1O5Tm2 |

Continued on next page

**TABLE XIV– continued**

| TQC | ICSD | NoE | DG | IndG | SubSG (w/ T) | cl | SubSG (w/o T) | cl | Max. Sub. SG | yes or no | Max. Sub. SG | yes or no | SIs (w/ T) | Stoich. Formula |
|---|---|---|---|---|---|---|---|---|---|---|---|---|---|---|
| 71.1.66 | 71415 | 96 | 0.0029 | -0.0567 | 2 | 1 | 2 | 1 | - | - | - | - | z4=3 | Ba1Ni1O5Tm2 |
| 71.1.67 | 67831 | 96 | 0.0026 | -0.0563 | 2 | 1 | 2 | 1 | - | - | - | - | z4=3 | Ba1Ni1O5Tm2 |
| 71.1.68 | 62761 | 96 | 0.0032 | -0.0580 | 2 | 1 | 2 | 1 | - | - | - | - | z4=3 | Ba1Ni1O5Tm2 |
| 71.1.69 | 62426 | 96 | 0.0032 | -0.0580 | 2 | 1 | 2 | 1 | - | - | - | - | z4=3 | Ba1Ni1O5Tm2 |
| 71.1.70 | 66078 | 88 | 0.0074 | -0.0196 | 2 | 7 | 2 | 3 | - | - | - | - | z2w,1=1<br>z2w,2=1<br>z2w,3=1<br>z4=1 | Ba1Ni1O5Tb2 |
| 71.1.71 | 75035 | 76 | 0.0088 | -0.0320 | 2 | 5 | 71 | 4 | - | - | - | - | z2w,1=1<br>z2w,2=1<br>z2w,3=1<br>z4=2 | Ba1Ni1O5Pr2 |
| 71.1.72 | 72626 | 78 | 0.0358 | -0.0251 | 2 | 1 | 2 | 1 | - | - | - | - | z4=3 | Ba1Nd2Ni1O5 |
| 71.1.73 | 62627 | 78 | 0.0355 | -0.0248 | 2 | 1 | 2 | 1 | - | - | - | - | z4=3 | Ba1Nd2Ni1O5 |
| 71.1.74 | 68425 | 86 | 0.0049 | -0.0186 | 2 | 5 | 71 | 4 | - | - | - | - | z2w,1=1<br>z2w,2=1<br>z2w,3=1<br>z4=2 | Ba1Gd2Ni1O5 |
| 71.1.75 | 62758 | 86 | 0.0037 | -0.0191 | 2 | 5 | 71 | 4 | - | - | - | - | z2w,1=1<br>z2w,2=1<br>z2w,3=1<br>z4=2 | Ba1Gd2Ni1O5 |
| 71.1.76 | 62423 | 86 | 0.0037 | -0.0191 | 2 | 5 | 71 | 4 | - | - | - | - | z2w,1=1<br>z2w,2=1<br>z2w,3=1<br>z4=2 | Ba1Gd2Ni1O5 |
| 71.1.77 | 72630 | 94 | 0.0000 | -0.0323 | 2 | 2 | 0 | 2 | - | - | - | - | z2w,1=1<br>z2w,2=1<br>z2w,3=1 | Ba1Er2Ni1O5 |
| 71.1.78 | 69324 | 94 | 0.0000 | -0.0335 | 2 | 2 | 0 | 2 | - | - | - | - | z2w,1=1<br>z2w,2=1<br>z2w,3=1 | Ba1Er2Ni1O5 |
| 71.1.79 | 69323 | 94 | 0.0000 | -0.0330 | 2 | 2 | 0 | 2 | - | - | - | - | z2w,1=1<br>z2w,2=1<br>z2w,3=1 | Ba1Er2Ni1O5 |
| 71.1.80 | 69315 | 94 | 0.0028 | -0.0342 | 2 | 2 | 0 | 2 | - | - | - | - | z2w,1=1<br>z2w,2=1<br>z2w,3=1 | Ba1Er2Ni1O5 |
| 71.1.81 | 62760 | 94 | 0.0027 | -0.0322 | 2 | 2 | 0 | 2 | - | - | - | - | z2w,1=1<br>z2w,2=1<br>z2w,3=1 | Ba1Er2Ni1O5 |
| 71.1.82 | 62425 | 94 | 0.0027 | -0.0322 | 2 | 2 | 0 | 2 | - | - | - | - | z2w,1=1<br>z2w,2=1<br>z2w,3=1 | Ba1Er2Ni1O5 |
| 71.1.83 | 85046 | 90 | 0.0047 | -0.0395 | 2 | 2 | 0 | 2 | - | - | - | - | z2w,1=1<br>z2w,2=1<br>z2w,3=1 | Ba1Dy2Ni1O5 |
| 71.1.84 | 72627 | 90 | 0.0047 | -0.0389 | 2 | 2 | 0 | 2 | - | - | - | - | z2w,1=1<br>z2w,2=1<br>z2w,3=1 | Ba1Dy2Ni1O5 |
| 71.1.85 | 167153 | 48 | 0.0192 | -0.4876 | 2 | 5 | 71 | 4 | - | - | - | - | z2w,1=1<br>z2w,2=1<br>z2w,3=1<br>z4=2 | B6Fe2U1 |
| 71.1.86 | 81543 | 42 | 0.0267 | -0.7177 | 2 | 1 | 2 | 1 | - | - | - | - | z4=3 | B6Cr2Th1 |
| 71.1.87 | 81544 | 42 | 0.0334 | -0.5776 | 2 | 1 | 2 | 1 | - | - | - | - | z4=3 | B6Ce1Cr2 |
| 71.1.88 | 40791 | 30 | 0.0039 | -1.1261 | 2 | 6 | 71 | 4 | - | - | - | - | z4=2 | B4Cr3 |
| 71.1.89 | 423380 | 18 | 0.0213 | -0.4776 | 2 | 6 | 71 | 4 | - | - | - | - | z4=2 | B4Cr1 |

Continued on next page

**TABLE XIV– continued**

| TQC | ICSD | NoE | DG | IndG | SubSG (w/ T) | cl | SubSG (w/o T) | cl | Max. Sub. SG | yes or no | Max. Sub. SG | yes or no | SIs (w/ T) | Stoich. Formula |
|---|---|---|---|---|---|---|---|---|---|---|---|---|---|---|
| 71.1.90 | 407576 | 68 | 0.0022 | -0.0162 | 2 | 4 | 2 | 1 | - | - | - | - | z2w,1=1 z2w,2=1 z2w,3=1 z4=3 | B2N4Nd3 |
| 71.1.91 | 68610 | 62 | 0.0020 | -0.1550 | 2 | 4 | 2 | 1 | - | - | - | - | z2w,1=1 z2w,2=1 z2w,3=1 z4=3 | B2Ce3N4 |
| 71.1.92 | 281420 | 30 | 0.1234 | -0.1767 | 2 | 1 | 2 | 1 | - | - | - | - | z4=3 | As1Te1Ti1 |
| 71.1.93 | 58046 | 48 | 0.0194 | -0.5055 | 2 | 4 | 2 | 1 | - | - | - | - | z2w,1=1 z2w,2=1 z2w,3=1 z4=3 | Al5Ni2Pr1 |
| 71.1.94 | 418387 | 142 | 0.0027 | -0.7754 | 2 | 5 | 71 | 4 | - | - | - | - | z2w,1=1 z2w,2=1 z2w,3=1 z4=2 | Al4O1Si6Sr10 |
| 71.1.95 | 9564 | 44 | 0.0098 | -1.2442 | 2 | 4 | 2 | 1 | - | - | - | - | z2w,1=1 z2w,2=1 z2w,3=1 z4=3 | Al2Sn2Sr3 |
| 71.1.96 | 100127 | 44 | 0.0086 | -1.2344 | 2 | 4 | 2 | 1 | - | - | - | - | z2w,1=1 z2w,2=1 z2w,3=1 z4=3 | Al2Si2Sr3 |
| 71.1.97 | 52617 | 44 | 0.0124 | -1.2742 | 2 | 4 | 2 | 1 | - | - | - | - | z2w,1=1 z2w,2=1 z2w,3=1 z4=3 | Al2Ge2Sr3 |
| 71.1.98 | 100126 | 20 | 0.0118 | -1.3085 | 2 | 4 | 2 | 1 | - | - | - | - | z2w,1=1 z2w,2=1 z2w,3=1 z4=3 | Al2Ca3Si2 |
| 71.1.99 | 52616 | 20 | 0.0045 | -1.3330 | 2 | 5 | 71 | 4 | - | - | - | - | z2w,1=1 z2w,2=1 z2w,3=1 z4=2 | Al2Ca3Ge2 |
| 71.1.100 | 9565 | 44 | 0.0228 | -0.9183 | 2 | 4 | 2 | 1 | - | - | - | - | z2w,1=1 z2w,2=1 z2w,3=1 z4=3 | Al2Ba3Sn2 |
| 71.1.101 | 100128 | 44 | 0.0248 | -0.8294 | 2 | 3 | 2 | 3 | - | - | - | - | z4=1 | Al2Ba3Si2 |
| 71.1.102 | 52612 | 44 | 0.0156 | -0.8529 | 2 | 3 | 2 | 3 | - | - | - | - | z4=1 | Al2Ba3Ge2 |
| 71.1.103 | 163921 | 162 | 0.0196 | -0.9682 | 2 | 2 | 0 | 2 | - | - | - | - | z2w,1=1 z2w,2=1 z2w,3=1 | Al16Ce2Pt9 |
| 71.1.104 | 157372 | 114 | 0.0024 | -0.0099 | 2 | 2 | 0 | 2 | - | - | - | - | z2w,1=1 z2w,2=1 z2w,3=1 | Ag4Gd3Sn4 |
| 71.1.105 | 417491 | 96 | 0.0072 | -0.1434 | 2 | 1 | 2 | 1 | - | - | - | - | z4=3 | Ag4Ce3Ge4 |
| 72.1.1 | 84206 | 144 | 0.0129 | -0.4749 | 2 | 1 | 0 | 1 | - | - | - | - | z2w,1=1 z2w,2=1 z2w,3=1 z4=2 | Ni3Si5Y2 |
| 72.1.2 | 428478 | 168 | 0.0000 | -0.1362 | 2 | 1 | 0 | 1 | - | - | - | - | z2w,1=1 z2w,2=1 z2w,3=1 z4=2 | Eu2Pt3Si5 |

Continued on next page

**TABLE XIV– continued**

| TQC | ICSD | NoE | DG | IndG | SubSG (w/ T) | cl | SubSG (w/o T) | cl | Max. Sub. SG | yes or no | Max. Sub. SG | yes or no | SIs (w/ T) | Stoich. Formula |
|---|---|---|---|---|---|---|---|---|---|---|---|---|---|---|
| 72.2.1 | 170063 | 76 | 0.0072 | -0.1853 | 2 | 4 | 2 | 2 | - | - | - | - | z4=1 | Pb5Sc6 |
| 72.2.2 | 416512 | 84 | 0.0347 | -0.7938 | 2 | 3 | 2 | 2 | - | - | - | - | z2w,1=1<br>z2w,2=1<br>z2w,3=1<br>z4=3 | Ni1Pd2Te2 |
| 72.2.3 | 416515 | 80 | 0.0123 | -0.3193 | 2 | 3 | 2 | 2 | - | - | - | - | z2w,1=1<br>z2w,2=1<br>z2w,3=1<br>z4=3 | Fe1Pd2Se2 |
| 72.2.4 | 68792 | 176 | 0.0029 | -0.0059 | 2 | 5 | 2 | 2 | - | - | - | - | z2w,1=1<br>z2w,2=1<br>z2w,3=1<br>z4=1 | Er2Ru3Si5 |
| 72.2.5 | 54107 | 100 | 0.0069 | -0.4324 | 2 | 2 | 2 | 2 | - | - | - | - | z4=3 | Cr4Sc2Si5 |
| 72.2.6 | 40758 | 140 | 0.0000 | -0.0249 | 2 | 5 | 2 | 2 | - | - | - | - | z2w,1=1<br>z2w,2=1<br>z2w,3=1<br>z4=1 | C4Fe1Tm2 |
| 72.2.7 | 28232 | 196 | 0.0086 | -0.1002 | 2 | 2 | 2 | 2 | - | - | - | - | z4=3 | Br2Hg5O4 |
| 72.4.1 | 65460 | 74 | - | - | - | - | - | - | See details in Table XIX | | | | | Mn1Rb2Te2 |
| 72.4.2 | 65457 | 74 | - | - | - | - | - | - | See details in Table XIX | | | | | Mn1Rb2Se2 |
| 72.4.3 | 65454 | 74 | - | - | - | - | - | - | See details in Table XIX | | | | | Mn1Rb2S2 |
| 72.4.4 | 40427 | 138 | - | - | - | - | - | - | See details in Table XIX | | | | | La2Rh3Si5 |
| 72.4.5 | 65459 | 74 | - | - | - | - | - | - | See details in Table XIX | | | | | K2Mn1Te2 |
| 72.4.6 | 65456 | 74 | - | - | - | - | - | - | See details in Table XIX | | | | | K2Mn1Se2 |
| 72.4.7 | 65453 | 74 | - | - | - | - | - | - | See details in Table XIX | | | | | K2Mn1S2 |
| 72.4.8 | 86502 | 138 | - | - | - | - | - | - | See details in Table XIX | | | | | Ge5La2Rh3 |
| 72.4.9 | 86503 | 138 | - | - | - | - | - | - | See details in Table XIX | | | | | Ge5Ir3Y2 |
| 72.4.10 | 86499 | 138 | - | - | - | - | - | - | See details in Table XIX | | | | | Ge5Ir3La2 |
| 72.4.11 | 40462 | 86 | - | - | - | - | - | - | See details in Table XIX | | | | | Ge2Mn5 |
| 72.4.12 | 68523 | 42 | - | - | - | - | - | - | See details in Table XIX | | | | | Fe1Li3N2 |
| 72.4.13 | 78933 | 126 | - | - | - | - | - | - | See details in Table XIX | | | | | Cs2Mn3Te4 |
| 72.4.14 | 59276 | 126 | - | - | - | - | - | - | See details in Table XIX | | | | | Cs2Mn3Te4 |
| 72.4.15 | 78932 | 126 | - | - | - | - | - | - | See details in Table XIX | | | | | Cs2Mn3Se4 |
| 72.4.16 | 65461 | 74 | - | - | - | - | - | - | See details in Table XIX | | | | | Cs2Mn1Te2 |
| 72.4.17 | 65458 | 74 | - | - | - | - | - | - | See details in Table XIX | | | | | Cs2Mn1Se2 |
| 72.4.18 | 65455 | 74 | - | - | - | - | - | - | See details in Table XIX | | | | | Cs2Mn1S2 |
| 72.4.19 | 65254 | 138 | - | - | - | - | - | - | See details in Table XIX | | | | | Co3Cs2S4 |
| 72.4.20 | 67391 | 78 | - | - | - | - | - | - | See details in Table XIX | | | | | Co1Rb2Se2 |
| 72.4.21 | 67388 | 78 | - | - | - | - | - | - | See details in Table XIX | | | | | Co1Rb2S2 |
| 72.4.22 | 416514 | 82 | - | - | - | - | - | - | See details in Table XIX | | | | | Co1Pd2Te2 |
| 72.4.23 | 416513 | 82 | - | - | - | - | - | - | See details in Table XIX | | | | | Co1Pd2Se2 |
| 72.4.24 | 67386 | 46 | - | - | - | - | - | - | See details in Table XIX | | | | | Co1Na2S2 |
| 72.4.25 | 67390 | 78 | - | - | - | - | - | - | See details in Table XIX | | | | | Co1K2Se2 |
| 72.4.26 | 67387 | 78 | - | - | - | - | - | - | See details in Table XIX | | | | | Co1K2S2 |
| 72.4.27 | 67392 | 78 | - | - | - | - | - | - | See details in Table XIX | | | | | Co1Cs2Se2 |
| 72.4.28 | 67389 | 78 | - | - | - | - | - | - | See details in Table XIX | | | | | Co1Cs2S2 |
| 72.4.29 | 73169 | 134 | - | - | - | - | - | - | See details in Table XIX | | | | | C4Er2Mn1 |
| 72.4.30 | 69494 | 146 | - | - | - | - | - | - | See details in Table XIX | | | | | Ba2Cu3P4 |
| 72.4.31 | 26253 | 130 | - | - | - | - | - | - | See details in Table XIX | | | | | As2Cu5 |
| 74.2.1 | 106234 | 96 | 0.0098 | -1.3303 | 2 | 4 | 0 | 2 | - | - | - | - | z2w,1=1<br>z2w,2=1<br>z2w,3=1<br>z4=2 | Yb1Zn2 |
| 74.2.2 | 429139 | 88 | 0.0082 | -1.3868 | 2 | 2 | 2 | 1 | - | - | - | - | z4=3 | Ga2La1Rh3 |

Continued on next page

**TABLE XIV– continued**

| TQC | ICSD | NoE | DG | IndG | SubSG (w/ T) | cl | SubSG (w/o T) | cl | Max. Sub. SG | yes or no | Max. Sub. SG | yes or no | SIs (w/ T) | Stoich. Formula |
|---|---|---|---|---|---|---|---|---|---|---|---|---|---|---|
| 74.2.3 | 238760 | 88 | 0.0119 | -0.3388 | 2 | 5 | 2 | 1 | - | - | - | - | z2w,1=1<br>z2w,2=1<br>z2w,3=1<br>z4=1 | Fe2O4Si1 |
| 74.2.4 | 238759 | 88 | 0.0134 | -0.3309 | 2 | 5 | 2 | 1 | - | - | - | - | z2w,1=1<br>z2w,2=1<br>z2w,3=1<br>z4=1 | Fe2O4Si1 |
| 74.2.5 | 238758 | 88 | 0.0124 | -0.3404 | 2 | 5 | 2 | 1 | - | - | - | - | z2w,1=1<br>z2w,2=1<br>z2w,3=1<br>z4=1 | Fe2O4Si1 |
| 74.2.6 | 93229 | 164 | 0.1471 | 0.0472 | 2 | 1 | 2 | 1 | - | - | - | - | z2w,1=1<br>z2w,2=1<br>z2w,3=1<br>z4=3 | Fe2Mo4O7 |
| 74.2.7 | 74288 | 128 | 0.0047 | -0.0625 | 2 | 3 | 0 | 2 | - | - | - | - | z2w,1=1<br>z2w,2=1<br>z2w,3=1 | Cs1F5Ni1Pd1 |
| 74.2.8 | 10156 | 184 | 0.0040 | -0.1433 | 2 | 6 | 0 | 2 | - | - | - | - | z4=2 | Co2O4Si1 |
| 74.2.9 | 246171 | 180 | 0.0936 | -0.0185 | 2 | 5 | 2 | 1 | - | - | - | - | z2w,1=1<br>z2w,2=1<br>z2w,3=1<br>z4=1 | Ce3Pt5Si1 |
| 74.2.10 | 93244 | 180 | 0.0118 | -0.0604 | 2 | 5 | 2 | 1 | - | - | - | - | z2w,1=1<br>z2w,2=1<br>z2w,3=1<br>z4=1 | Ce3Pd5Si1 |
| 74.2.11 | 260381 | 68 | 0.0150 | -1.3657 | 2 | 3 | 0 | 2 | - | - | - | - | z2w,1=1<br>z2w,2=1<br>z2w,3=1 | Cd2Sr1 |
| 74.2.12 | 97094 | 124 | 0.0415 | -0.1280 | 2 | 1 | 2 | 1 | - | - | - | - | z2w,1=1<br>z2w,2=1<br>z2w,3=1<br>z4=3 | Ca2O7Os2 |
| 74.2.13 | 97093 | 124 | 0.0324 | -0.1261 | 2 | 1 | 2 | 1 | - | - | - | - | z2w,1=1<br>z2w,2=1<br>z2w,3=1<br>z4=3 | Ca2O7Os2 |
| 74.2.14 | 97092 | 124 | 0.0349 | -0.1197 | 2 | 1 | 2 | 1 | - | - | - | - | z2w,1=1<br>z2w,2=1<br>z2w,3=1<br>z4=3 | Ca2O7Os2 |
| 74.2.15 | 97091 | 124 | 0.0365 | -0.1166 | 2 | 1 | 2 | 1 | - | - | - | - | z2w,1=1<br>z2w,2=1<br>z2w,3=1<br>z4=3 | Ca2O7Os2 |
| 74.2.16 | 418971 | 52 | 0.0211 | -1.7539 | 2 | 3 | 0 | 2 | - | - | - | - | z2w,1=1<br>z2w,2=1<br>z2w,3=1 | Ca1Zn2 |
| 74.2.17 | 420576 | 52 | 0.0189 | -1.7296 | 2 | 3 | 0 | 2 | - | - | - | - | z2w,1=1<br>z2w,2=1<br>z2w,3=1 | Ca1Cd2 |
| 74.2.18 | 414229 | 32 | 0.0455 | -0.8942 | 2 | 1 | 2 | 1 | - | - | - | - | z2w,1=1<br>z2w,2=1<br>z2w,3=1<br>z4=3 | Ba1In2 |

Continued on next page

**TABLE XIV– continued**

| TQC | ICSD | NoE | DG | IndG | SubSG (w/ T) | cl | SubSG (w/o T) | cl | Max. Sub. SG | yes or no | Max. Sub. SG | yes or no | SIs (w/ T) | Stoich. Formula |
|---|---|---|---|---|---|---|---|---|---|---|---|---|---|---|
| 74.2.19 | 251586 | 60 | 0.0117 | -2.1050 | 2 | 4 | 0 | 2 | - | - | - | - | z2w,1=1<br>z2w,2=1<br>z2w,3=1<br>z4=2 | Ba1Ge5 |
| 74.2.20 | 194880 | 60 | 0.0117 | -2.1050 | 2 | 4 | 0 | 2 | - | - | - | - | z2w,1=1<br>z2w,2=1<br>z2w,3=1<br>z4=2 | Ba1Ge5 |
| 74.2.21 | 260668 | 68 | 0.0202 | -1.1759 | 2 | 3 | 0 | 2 | - | - | - | - | z2w,1=1<br>z2w,2=1<br>z2w,3=1 | Ba1Cd2 |
| 74.2.22 | 55542 | 48 | 0.0132 | -2.1348 | 2 | 2 | 2 | 1 | - | - | - | - | z4=3 | Au2Ca1 |
| 74.2.23 | 72289 | 128 | 0.0036 | -0.0262 | 2 | 4 | 0 | 2 | - | - | - | - | z2w,1=1<br>z2w,2=1<br>z2w,3=1<br>z4=2 | Al1F7Na2Ni1 |
| 74.2.24 | 417225 | 48 | 0.0096 | -2.1315 | 2 | 2 | 2 | 1 | - | - | - | - | z4=3 | Ag2Ca1 |
| 74.2.25 | 417227 | 64 | 0.0517 | -0.6411 | 2 | 1 | 2 | 1 | - | - | - | - | z2w,1=1<br>z2w,2=1<br>z2w,3=1<br>z4=3 | Ag1Ca1Zn1 |
| 74.4.1 | 236759 | 70 | - | - | - | - | - | - | See details in Table XIX | | | | | O3Sr1Tc1 |
| 74.4.2 | 183451 | 70 | - | - | - | - | - | - | See details in Table XIX | | | | | O3Sr1Tc1 |
| 74.4.3 | 260033 | 174 | - | - | - | - | - | - | See details in Table XIX | | | | | La3Pd5Si1 |
| 74.4.4 | 200066 | 66 | - | - | - | - | - | - | See details in Table XIX | | | | | La1Re2Si2 |
| 74.4.5 | 408698 | 66 | - | - | - | - | - | - | See details in Table XIX | | | | | Hg2Rb1 |
| 74.4.6 | 195614 | 78 | - | - | - | - | - | - | See details in Table XIX | | | | | Ge2Na1Pd3 |
| 74.4.7 | 104006 | 58 | - | - | - | - | - | - | See details in Table XIX | | | | | Ga2Tm1 |
| 74.4.8 | 104005 | 58 | - | - | - | - | - | - | See details in Table XIX | | | | | Ga2Tm1 |
| 74.4.9 | 202872 | 138 | - | - | - | - | - | - | See details in Table XIX | | | | | F7Fe1Na2Ni1 |
| 74.4.10 | 202871 | 138 | - | - | - | - | - | - | See details in Table XIX | | | | | F7Fe1Na2Ni1 |
| 74.4.11 | 409889 | 46 | - | - | - | - | - | - | See details in Table XIX | | | | | Eu1Ga2 |
| 74.4.12 | 36535 | 74 | - | - | - | - | - | - | See details in Table XIX | | | | | Cu2O2Tl1 |
| 74.4.13 | 408695 | 66 | - | - | - | - | - | - | See details in Table XIX | | | | | Cs1Hg2 |
| 74.4.14 | 79221 | 94 | - | - | - | - | - | - | See details in Table XIX | | | | | Ce1Ir3Si2 |
| 74.4.15 | 40664 | 74 | - | - | - | - | - | - | See details in Table XIX | | | | | Br4Li2Mn1 |
| 74.4.16 | 249923 | 170 | - | - | - | - | - | - | See details in Table XIX | | | | | Au5K3Tl1 |
| 74.4.17 | 249922 | 170 | - | - | - | - | - | - | See details in Table XIX | | | | | Au5In1K3 |
| 74.4.18 | 413972 | 130 | - | - | - | - | - | - | See details in Table XIX | | | | | As1Cu1F7 |
| 74.4.19 | 245266 | 66 | - | - | - | - | - | - | See details in Table XIX | | | | | Al1Ce1O3 |
| 78.3.1 | 67327 | 116 | - | - | - | - | - | - | 4 | yes | 4 | yes | - | C2Ce1Rh1 |
| 79.3.1 | 76499 | 74 | - | - | - | - | - | - | 5 | yes | 5 | yes | - | Sb2Tl1Zn2 |
| 82.2.1 | 81539 | 72 | 0.0044 | -0.2174 | 82 | 1 | 82 | 1 | - | - | - | - | z2=1 | Te1Zr3 |
| 82.2.2 | 51048 | 72 | 0.0093 | -0.2416 | 82 | 1 | 82 | 1 | - | - | - | - | z2=1 | Te1Zr3 |
| 82.2.3 | 195057 | 68 | 0.0105 | -0.1383 | 82 | 1 | 82 | 1 | - | - | - | - | z2=1 | Sb1Zr3 |
| 82.2.4 | 98373 | 140 | 0.0175 | -0.6512 | 82 | 1 | 82 | 1 | - | - | - | - | z2=1 | Ni3P1 |
| 82.2.5 | 23612 | 32 | 0.0223 | -0.7581 | 82 | 1 | 82 | 1 | - | - | - | - | z2=1 | In1P1S4 |
| 82.2.6 | 56027 | 76 | 0.0280 | -0.1093 | 82 | 1 | 82 | 1 | - | - | - | - | z2=1 | Ge1Ta3 |
| 82.2.7 | 43365 | 116 | 0.0020 | -0.0491 | 82 | 1 | 82 | 1 | - | - | - | - | z2=1 | Fe3P1 |
| 82.2.8 | 23560 | 92 | 0.0031 | -0.1376 | 82 | 1 | 82 | 1 | - | - | - | - | z2=1 | Cr3P1 |
| 83.2.1 | 106354 | 106 | 0.0027 | -1.0866 | 2,81 | 1 | 2,81 | 1 | - | - | - | - | z2w,1=1<br>z2w,2=1<br>z4=3<br>z4mpi=2<br>z2=1<br>z8=7 | Ca7Ni4Sn13 |
| 83.3.1 | 42959 | 108 | - | - | - | - | - | - | 81 | no(81) | 81 | no(81) | z2=1 | C4U1W4 |

Continued on next page

**TABLE XIV– continued**

| TQC | ICSD | NoE | DG | IndG | SubSG (w/ T) | cl | SubSG (w/o T) | cl | Max. Sub. SG | yes or no | Max. Sub. SG | yes or no | SIs (w/ T) | Stoich. Formula |
|---|---|---|---|---|---|---|---|---|---|---|---|---|---|---|
| | | | | | | | | | 10 | no(2) | 10 | no(10) | z2w,1=1<br>z2w,2=1<br>z2w,3=1 | |
| 83.3.2 | 39222 | 108 | - | - | - | - | - | - | 81 | yes | 81 | yes | - | C4U1W4 |
| | | | | | | | | | 10 | no(2) | 10 | yes | z4=2 | |
| 83.3.3 | 999 | 100 | - | - | - | - | - | - | 81 | yes | 81 | yes | - | As2V3 |
| | | | | | | | | | 10 | yes | 10 | yes | - | |
| 84.2.1 | 74390 | 120 | 0.0026 | -0.0377 | 2 | 1 | 0 | 1 | - | - | - | - | z2w,1=1<br>z2w,2=1 | P2S6U1 |
| 84.3.1 | 155153 | 168 | - | - | - | - | - | - | 81 | no(81) | 81 | no(81) | z2=1 | Fe1O4Sb2 |
| | | | | | | | | | 10 | no(2) | 10 | yes | z2w,1=1<br>z2w,3=1 | |
| 84.4.1 | 59987 | 146 | - | - | - | - | - | - | See details in Table XIX | | | | | Cd1F6Tb1 |
| 85.2.1 | 33884 | 136 | 0.0036 | -0.1458 | 2,81 | 6 | 2,81 | 1 | - | - | - | - | z4=3<br>z2=1 | As2Cl4F6 |
| 85.4.1 | 195053 | 230 | - | - | - | - | - | - | See details in Table XIX | | | | | Ni1Sb9Zr5 |
| 85.4.2 | 27315 | 82 | - | - | - | - | - | - | See details in Table XIX | | | | | Mo1O5V1 |
| 85.4.3 | 36095 | 82 | - | - | - | - | - | - | See details in Table XIX | | | | | Mo1O5P1 |
| 85.4.4 | 24894 | 82 | - | - | - | - | - | - | See details in Table XIX | | | | | Mo1O5P1 |
| 85.4.5 | 413979 | 130 | - | - | - | - | - | - | See details in Table XIX | | | | | Hf5Sb9 |
| 85.4.6 | 56963 | 98 | - | - | - | - | - | - | See details in Table XIX | | | | | H4B1Cl1Cu1O4 |
| 85.4.7 | 80646 | 98 | - | - | - | - | - | - | See details in Table XIX | | | | | Ag1B1F5 |
| 85.4.8 | 80645 | 98 | - | - | - | - | - | - | See details in Table XIX | | | | | Ag1B1F5 |
| 86.4.1 | 8078 | 110 | - | - | - | - | - | - | See details in Table XIX | | | | | B4La1Ru4 |
| 86.4.2 | 100717 | 118 | - | - | - | - | - | - | See details in Table XIX | | | | | B4Ir4La1 |
| 87.1.1 | 51047 | 44 | 0.0076 | -0.4239 | 2,82 | 4 | 2,82 | 1 | - | - | - | - | z2w,1=1<br>z2w,2=1<br>z2w,3=1<br>z4=1<br>z2=1<br>z8=1 | Te4Zr5 |
| 87.1.2 | 409337 | 44 | 0.0079 | -0.4251 | 2,82 | 4 | 2,82 | 1 | - | - | - | - | z2w,1=1<br>z2w,2=1<br>z2w,3=1<br>z4=1<br>z2=1<br>z8=1 | Te4Zr5 |
| 87.1.3 | 55075 | 44 | 0.0044 | -0.3844 | 2,82 | 1 | 2,82 | 1 | - | - | - | - | z2w,1=1<br>z2w,2=1<br>z2w,3=1<br>z4=1<br>z2=1<br>z8=5 | Se4Ti5 |
| 87.1.4 | 157026 | 72 | 0.0047 | -1.0926 | 87 | 5 | 0 | 3 | - | - | - | - | z8=4 | Mo1Ni1O6Sr2 |
| 87.1.5 | 157024 | 72 | 0.0230 | -1.0989 | 87 | 5 | 0 | 3 | - | - | - | - | z8=4 | Mo1Ni1O6Sr2 |
| 87.1.6 | 57076 | 140 | 0.0110 | -0.4102 | 2,82 | 4 | 2,82 | 1 | - | - | - | - | z2w,1=1<br>z2w,2=1<br>z2w,3=1<br>z4=1<br>z2=1<br>z8=1 | Ge4Li4Yb5 |
| 87.1.7 | 173488 | 72 | 0.0333 | -0.4312 | 87 | 5 | 0 | 3 | - | - | - | - | z8=4 | Co1O6Re1Sr2 |
| 87.1.8 | 60822 | 54 | 0.0100 | -0.3536 | 2,82 | 3 | 2,82 | 2 | - | - | - | - | z2w,1=1<br>z2w,2=1<br>z2w,3=1<br>z4=3<br>z2=1<br>z8=3 | C4Cr4U1 |

Continued on next page

**TABLE XIV– continued**

| TQC | ICSD | NoE | DG | IndG | SubSG (w/ T) | cl | SubSG (w/o T) | cl | Max. Sub. SG | yes or no | Max. Sub. SG | yes or no | SIs (w/ T) | Stoich. Formula |
|---|---|---|---|---|---|---|---|---|---|---|---|---|---|---|
| 87.1.9 | 40319 | 54 | 0.0000 | -0.3947 | 2,82 | 3 | 2,82 | 2 | - | - | - | - | z2w,1=1 | C4Cr4U1 |
| | | | | | | | | | | | | | z2w,2=1 | |
| | | | | | | | | | | | | | z2w,3=1 | |
| | | | | | | | | | | | | | z4=3 | |
| | | | | | | | | | | | | | z2=1 | |
| | | | | | | | | | | | | | z8=3 | |
| 87.1.10 | 43186 | 50 | 0.0225 | -0.6309 | 2,82 | 2 | 2,82 | 2 | - | - | - | - | z4=3 | As4Mo5 |
| | | | | | | | | | | | | | z2=1 | |
| | | | | | | | | | | | | | z8=7 | |
| 87.1.11 | 66017 | 118 | 0.0129 | -1.3729 | 2,82 | 1 | 2,82 | 1 | - | - | - | - | z2w,1=1 | As4Ba1Cu8 |
| | | | | | | | | | | | | | z2w,2=1 | |
| | | | | | | | | | | | | | z2w,3=1 | |
| | | | | | | | | | | | | | z4=1 | |
| | | | | | | | | | | | | | z2=1 | |
| | | | | | | | | | | | | | z8=5 | |
| 87.3.1 | 51214 | 68 | - | - | - | - | - | - | 82 | no(82) | 82 | no(82) | z2=1 | O2V1 |
| | | | | | | | | | 12 | yes | 12 | yes | - | |
| 87.3.2 | 194696 | 66 | - | - | - | - | - | - | 82 | no(82) | 82 | no(82) | z2=1 | Mg1O6Os1Sr2 |
| | | | | | | | | | 12 | yes | 12 | yes | - | |
| 87.3.3 | 190088 | 206 | - | - | - | - | - | - | 82 | no(82) | 82 | no(82) | z2=1 | In4Pd17Se4 |
| | | | | | | | | | 12 | yes | 12 | yes | - | |
| 87.3.4 | 421410 | 102 | - | - | - | - | - | - | 82 | yes | 82 | yes | - | Ba9In4 |
| | | | | | | | | | 12 | no(2) | 12 | no(2) | z4=3 | |
| 87.3.5 | 16309 | 214 | - | - | - | - | - | - | 82 | no(82) | 82 | no(82) | z2=1 | Ba6Fe8S15 |
| | | | | | | | | | 12 | no(2) | 12 | yes | z4=2 | |
| 87.3.6 | 290783 | 76 | - | - | - | - | - | - | 82 | yes | 82 | yes | - | Ba2Mo1Nd1O6 |
| | | | | | | | | | 12 | no(2) | 12 | no(2) | z4=3 | |
| 87.3.7 | 158358 | 82 | - | - | - | - | - | - | 82 | no(82) | 82 | no(82) | z2=1 | Ba2Ho1O6Ta1 |
| | | | | | | | | | 12 | yes | 12 | yes | - | |
| 87.3.8 | 238713 | 50 | - | - | - | - | - | - | 82 | no(82) | 82 | no(82) | z2=1 | Ba1Fe2Se4 |
| | | | | | | | | | 12 | no(2) | 12 | yes | z2w,1=1 | |
| | | | | | | | | | | | | | z2w,2=1 | |
| | | | | | | | | | | | | | z4=2 | |
| 87.3.9 | 23081 | 50 | - | - | - | - | - | - | 82 | no(82) | 82 | no(82) | z2=1 | Ba1Fe2S4 |
| | | | | | | | | | 12 | no(2) | 12 | yes | z2w,1=1 | |
| | | | | | | | | | | | | | z2w,2=1 | |
| | | | | | | | | | | | | | z4=2 | |
| 87.3.10 | 66016 | 118 | - | - | - | - | - | - | 82 | yes | 82 | yes | - | Ba1Cu8P4 |
| | | | | | | | | | 12 | no(2) | 12 | no(2) | z2w,1=1 | |
| | | | | | | | | | | | | | z2w,2=1 | |
| | | | | | | | | | | | | | z4=1 | |
| 88.3.1 | 16189 | 68 | - | - | - | - | - | - | 82 | yes | 82 | yes | - | O4Pb1W1 |
| | | | | | | | | | 15 | no(2) | 15 | no(2) | z4=1 | |
| 88.4.1 | 78846 | 110 | - | - | - | - | - | - | See details in Table XIX | | | | | O4Pd2Y1 |
| 88.4.2 | 79340 | 110 | - | - | - | - | - | - | See details in Table XIX | | | | | La1O4Pd2 |
| 88.4.3 | 9914 | 106 | - | - | - | - | - | - | See details in Table XIX | | | | | F4Li1Yb1 |
| 88.4.4 | 429261 | 94 | - | - | - | - | - | - | See details in Table XIX | | | | | Eu1O4W1 |
| 88.4.5 | 173264 | 102 | - | - | - | - | - | - | See details in Table XIX | | | | | Cr1Ho1O4 |
| 88.4.6 | 169873 | 102 | - | - | - | - | - | - | See details in Table XIX | | | | | Cr1Ho1O4 |
| 88.4.7 | 38326 | 94 | - | - | - | - | - | - | See details in Table XIX | | | | | Cl4Gd1Li1 |
| 92.3.1 | 33775 | 84 | - | - | - | - | - | - | 5 | yes | 5 | yes | - | Fe1Na1O2 |
| | | | | | | | | | 4 | yes | 4 | yes | - | |
| 92.3.2 | 33774 | 84 | - | - | - | - | - | - | 5 | yes | 5 | yes | - | Fe1Na1O2 |
| | | | | | | | | | 4 | yes | 4 | yes | - | |
| 92.3.3 | 33773 | 84 | - | - | - | - | - | - | 5 | yes | 5 | yes | - | Fe1Na1O2 |
| | | | | | | | | | 4 | yes | 4 | yes | - | |
| 92.3.4 | 33769 | 84 | - | - | - | - | - | - | 5 | yes | 5 | yes | - | Fe1Na1O2 |
| | | | | | | | | | 4 | yes | 4 | yes | - | |

Continued on next page

**TABLE XIV– continued**

| TQC | ICSD | NoE | DG | IndG | SubSG (w/ T) | cl | SubSG (w/o T) | cl | Max. Sub. SG | yes or no | Max. Sub. SG | yes or no | SIs (w/ T) | Stoich. Formula |
|---|---|---|---|---|---|---|---|---|---|---|---|---|---|---|
| 92.3.5 | 33768 | 84 | - | - | - | - | - | - | 5 | yes | 5 | yes | - | Fe1Na1O2 |
| | | | | | | | | | 4 | yes | 4 | yes | - | |
| 92.4.1 | 424580 | 196 | - | - | - | - | - | - | See details in Table XIX | | | | | Ni1O2Rb3 |
| 92.4.2 | 262578 | 196 | - | - | - | - | - | - | See details in Table XIX | | | | | K3Ni1O2 |
| 92.4.3 | 262577 | 196 | - | - | - | - | - | - | See details in Table XIX | | | | | K3Ni1O2 |
| 92.4.4 | 262576 | 196 | - | - | - | - | - | - | See details in Table XIX | | | | | K3Ni1O2 |
| 97.3.1 | 35190 | 130 | - | - | - | - | - | - | 23 | yes | 23 | yes | - | I1Se8Ta2 |
| | | | | | | | | | 22 | yes | 22 | yes | - | |
| 99.3.1 | 247466 | 32 | - | - | - | - | - | - | 35 | yes | 35 | yes | - | Bi1Co1O3 |
| | | | | | | | | | 25 | yes | 25 | yes | - | |
| 99.3.2 | 87466 | 70 | - | - | - | - | - | - | 35 | yes | 35 | yes | - | Ba1Cu1Fe1O5Y1 |
| | | | | | | | | | 25 | yes | 25 | yes | - | |
| 99.3.3 | 71355 | 70 | - | - | - | - | - | - | 35 | yes | 35 | yes | - | Ba1Cu1Fe1O5Y1 |
| | | | | | | | | | 25 | yes | 25 | yes | - | |
| 102.3.1 | 43217 | 420 | - | - | - | - | - | - | 35 | yes | 35 | yes | - | U1 |
| | | | | | | | | | 34 | yes | 34 | yes | - | |
| 102.4.1 | 202705 | 118 | - | - | - | - | - | - | See details in Table XIX | | | | | F6Fe2Li1 |
| 103.4.1 | 62050 | 58 | - | - | - | - | - | - | See details in Table XIX | | | | | Ta1Te4 |
| 103.4.2 | 65129 | 74 | - | - | - | - | - | - | See details in Table XIX | | | | | Nb1Te4 |
| 103.4.3 | 65128 | 74 | - | - | - | - | - | - | See details in Table XIX | | | | | Nb1Te4 |
| 106.3.1 | 190731 | 180 | - | - | - | - | - | - | 37 | yes | 37 | yes | - | Cu1O4Sb2 |
| | | | | | | | | | 32 | yes | 32 | yes | - | |
| 107.3.1 | 161866 | 32 | - | - | - | - | - | - | 44 | yes | 44 | yes | - | Ce1Ru1Si3 |
| | | | | | | | | | 42 | yes | 42 | yes | - | |
| 107.3.2 | 58662 | 32 | - | - | - | - | - | - | 44 | yes | 44 | yes | - | Ba1Ni1Sn3 |
| | | | | | | | | | 42 | yes | 42 | yes | - | |
| 107.3.3 | 174261 | 32 | - | - | - | - | - | - | 44 | yes | 44 | yes | - | Ba1Ge3Pd1 |
| | | | | | | | | | 42 | yes | 42 | yes | - | |
| 107.3.4 | 290391 | 34 | - | - | - | - | - | - | 44 | yes | 44 | yes | - | Al3Cu1Nd1 |
| | | | | | | | | | 42 | yes | 42 | yes | - | |
| 109.4.1 | 414516 | 54 | - | - | - | - | - | - | See details in Table XIX | | | | | Ir1P1Pr1 |
| 109.4.2 | 414514 | 50 | - | - | - | - | - | - | See details in Table XIX | | | | | Ir1La1P1 |
| 109.4.3 | 95191 | 50 | - | - | - | - | - | - | See details in Table XIX | | | | | As1La1Rh1 |
| 109.4.4 | 423921 | 90 | - | - | - | - | - | - | See details in Table XIX | | | | | Ag1Er1S2 |
| 113.2.1 | 246245 | 44 | 0.0025 | -0.3637 | 81 | 1 | 81 | 1 | - | - | - | - | z2=1 | Ca1Fe1O2 |
| 113.2.2 | 39120 | 172 | 0.0000 | -0.0142 | 81 | 1 | 81 | 1 | - | - | - | - | z2=1 | Be2Gd2Ge1O7 |
| 113.2.3 | 160121 | 84 | 0.0029 | -0.0264 | 81 | 1 | 81 | 1 | - | - | - | - | z2=1 | Ag1Tb1Te2 |
| 113.2.4 | 1 - 93 | 84 | 0.0027 | -0.0266 | 81 | 1 | 81 | 1 | - | - | - | - | z2=1 | Ag1Tb1Te2 |
| 113.4.1 | 54208 | 150 | - | - | - | - | - | - | See details in Table XIX | | | | | Ta30 |
| 113.4.2 | 280872 | 150 | - | - | - | - | - | - | See details in Table XIX | | | | | Ta30 |
| 113.4.3 | 250214 | 150 | - | - | - | - | - | - | See details in Table XIX | | | | | Ta30 |
| 113.4.4 | 54206 | 110 | - | - | - | - | - | - | See details in Table XIX | | | | | Ta22 |
| 113.4.5 | 96594 | 150 | - | - | - | - | - | - | See details in Table XIX | | | | | Ta1 |
| 113.4.6 | 89533 | 226 | - | - | - | - | - | - | See details in Table XIX | | | | | P6Rh7Sr2 |
| 113.4.7 | 408045 | 42 | - | - | - | - | - | - | See details in Table XIX | | | | | Na1Sn5 |
| 113.4.8 | 291413 | 110 | - | - | - | - | - | - | See details in Table XIX | | | | | Mn1Mo1O6Te1 |
| 113.4.9 | 239 | 90 | - | - | - | - | - | - | See details in Table XIX | | | | | Cu3Se2 |
| 113.4.10 | 191282 | 162 | - | - | - | - | - | - | See details in Table XIX | | | | | Cu1K2O7P2 |
| 113.4.11 | 262399 | 158 | - | - | - | - | - | - | See details in Table XIX | | | | | Co1Ge2O7Sr2 |
| 113.4.12 | 262398 | 158 | - | - | - | - | - | - | See details in Table XIX | | | | | Co1Ge2O7Sr2 |
| 113.4.13 | 94139 | 126 | - | - | - | - | - | - | See details in Table XIX | | | | | Ca2Co1O7Si2 |
| 113.4.14 | 94137 | 126 | - | - | - | - | - | - | See details in Table XIX | | | | | Ca2Co1O7Si2 |
| 113.4.15 | 94136 | 126 | - | - | - | - | - | - | See details in Table XIX | | | | | Ca2Co1O7Si2 |
| 113.4.16 | 94132 | 126 | - | - | - | - | - | - | See details in Table XIX | | | | | Ca2Co1O7Si2 |
| 113.4.17 | 72543 | 126 | - | - | - | - | - | - | See details in Table XIX | | | | | Ca2Co1O7Si2 |
| 113.4.18 | 186949 | 126 | - | - | - | - | - | - | See details in Table XIX | | | | | Ca2Co1O7Si2 |
| 113.4.19 | 97762 | 162 | - | - | - | - | - | - | See details in Table XIX | | | | | Ba2Cu1O7Si2 |

Continued on next page

**TABLE XIV– continued**

| TQC | ICSD | NoE | DG | IndG | SubSG (w/ T) | cl | SubSG (w/o T) | cl | Max. Sub. SG | yes or no | Max. Sub. SG | yes or no | SIs (w/ T) | Stoich. Formula |
|---|---|---|---|---|---|---|---|---|---|---|---|---|---|---|
| 113.4.20 | 281293 | 158 | - | - | - | - | - | - | See details in Table XIX | | | | | Ba2Co1O7Si2 |
| 113.4.21 | 423351 | 158 | - | - | - | - | - | - | See details in Table XIX | | | | | Ba2Co1Ge2O7 |
| 113.4.22 | 423350 | 158 | - | - | - | - | - | - | See details in Table XIX | | | | | Ba2Co1Ge2O7 |
| 113.4.23 | 290484 | 158 | - | - | - | - | - | - | See details in Table XIX | | | | | Ba2Co1Ge2O7 |
| 113.4.24 | 290483 | 158 | - | - | - | - | - | - | See details in Table XIX | | | | | Ba2Co1Ge2O7 |
| 113.4.25 | 1 - 91 | 90 | - | - | - | - | - | - | See details in Table XIX | | | | | Ag1Er1Te2 |
| 113.4.26 | 160122 | 86 | - | - | - | - | - | - | See details in Table XIX | | | | | Ag1Dy1Te2 |
| 113.4.27 | 1 - 94 | 86 | - | - | - | - | - | - | See details in Table XIX | | | | | Ag1Dy1Te2 |
| 115.1.1 | 93242 | 34 | 0.0020 | -1.9619 | 81 | 1 | 81 | 1 | - | - | - | - | z2=1 | Cu1Sb3Zr2 |
| 115.1.2 | 426313 | 34 | 0.0030 | -1.9422 | 81 | 1 | 81 | 1 | - | - | - | - | z2=1 | Cu1Sb3Zr2 |
| 115.1.3 | 195058 | 34 | 0.0039 | -1.9436 | 81 | 1 | 81 | 1 | - | - | - | - | z2=1 | Cu1Sb3Zr2 |
| 115.1.4 | 93241 | 34 | 0.0048 | -1.6195 | 81 | 1 | 81 | 1 | - | - | - | - | z2=1 | Cu1Sb3Ti2 |
| 115.1.5 | 93243 | 34 | 0.0035 | -1.9387 | 81 | 1 | 81 | 1 | - | - | - | - | z2=1 | Cu1Hf2Sb3 |
| 120.3.1 | 58522 | 122 | - | - | - | - | - | - | 22 | yes | 22 | yes | - | Au4K1Sn2 |
| 121.1.1 | 237706 | 92 | 0.0000 | -0.2946 | 82 | 1 | 82 | 1 | - | - | - | - | z2=1 | Mo3P1 |
| 121.1.2 | 104607 | 84 | 0.0079 | -0.3332 | 82 | 1 | 82 | 1 | - | - | - | - | z2=1 | Ir1Zr3 |
| 121.1.3 | 400652 | 62 | 0.0081 | -0.3837 | 82 | 1 | 82 | 1 | - | - | - | - | z2=1 | Cu3Sb1Se4 |
| 121.1.4 | 412239 | 62 | 0.0419 | 0.0210 | 82 | 1 | 82 | 1 | - | - | - | - | z2=1 | Cu3S4Sb1 |
| 121.1.5 | 2857 | 62 | 0.0410 | -0.0591 | 82 | 1 | 82 | 1 | - | - | - | - | z2=1 | Cu3S4Sb1 |
| 121.1.6 | 95119 | 62 | 0.0052 | 0.0052 | 82 | 1 | 82 | 1 | - | - | - | - | z2=1 | Cu2Hg1Se4Sn1 |
| 121.1.7 | 152761 | 62 | 0.0046 | 0.0046 | 82 | 1 | 82 | 1 | - | - | - | - | z2=1 | Cu2Ge1Hg1Se4 |
| 121.1.8 | 85126 | 58 | 0.0066 | -0.3662 | 82 | 1 | 82 | 1 | - | - | - | - | z2=1 | Cu2Fe1Se4Sn1 |
| 121.1.9 | 165094 | 62 | 0.0026 | -0.2298 | 82 | 1 | 82 | 1 | - | - | - | - | z2=1 | Cd1Cu2Ge1Te4 |
| 122.1.1 | 248539 | 144 | 0.0024 | -0.0022 | 82 | 1 | 82 | 1 | - | - | - | - | z2=1 | O6Rb4 |
| 122.1.2 | 659 | 196 | 0.0000 | -0.0534 | 82 | 1 | 82 | 1 | - | - | - | - | z2=1 | F5U1 |
| 122.1.3 | 16427 | 196 | 0.0032 | -0.0466 | 82 | 1 | 82 | 1 | - | - | - | - | z2=1 | F5U1 |
| 122.1.4 | 16426 | 196 | 0.0000 | -0.0687 | 82 | 1 | 82 | 1 | - | - | - | - | z2=1 | F5U1 |
| 122.1.5 | 16425 | 196 | 0.0000 | -0.0490 | 82 | 1 | 82 | 1 | - | - | - | - | z2=1 | F5U1 |
| 122.1.6 | 16424 | 196 | 0.0000 | -0.0560 | 82 | 1 | 82 | 1 | - | - | - | - | z2=1 | F5U1 |
| 122.1.7 | 84027 | 52 | 0.0179 | 0.0076 | 82 | 1 | 82 | 1 | - | - | - | - | z2=1 | Cu1In1Se2 |
| 122.1.8 | 73351 | 52 | 0.0129 | 0.0064 | 82 | 1 | 82 | 1 | - | - | - | - | z2=1 | Cu1In1Se2 |
| 122.1.9 | 154409 | 52 | 0.0302 | 0.0103 | 82 | 1 | 82 | 1 | - | - | - | - | z2=1 | Cu1In1Se2 |
| 122.1.10 | 154408 | 52 | 0.0295 | 0.0101 | 82 | 1 | 82 | 1 | - | - | - | - | z2=1 | Cu1In1Se2 |
| 122.1.11 | 154407 | 52 | 0.0289 | 0.0100 | 82 | 1 | 82 | 1 | - | - | - | - | z2=1 | Cu1In1Se2 |
| 122.1.12 | 154406 | 52 | 0.0288 | 0.0100 | 82 | 1 | 82 | 1 | - | - | - | - | z2=1 | Cu1In1Se2 |
| 122.1.13 | 154405 | 52 | 0.0194 | 0.0078 | 82 | 1 | 82 | 1 | - | - | - | - | z2=1 | Cu1In1Se2 |
| 122.1.14 | 68325 | 52 | 0.0091 | 0.0091 | 82 | 1 | 82 | 1 | - | - | - | - | z2=1 | As2Sn1Zn1 |
| 122.1.15 | 42098 | 52 | 0.1077 | 0.0298 | 82 | 1 | 82 | 1 | - | - | - | - | z2=1 | As2Cd1Ge1 |
| 122.4.1 | 80095 | 62 | - | - | - | - | - | - | See details in Table XIX | | | | | Cu1Fe1S2 |
| 122.4.2 | 80094 | 62 | - | - | - | - | - | - | See details in Table XIX | | | | | Cu1Fe1S2 |
| 122.4.3 | 261883 | 62 | - | - | - | - | - | - | See details in Table XIX | | | | | Cu1Fe1S2 |
| 122.4.4 | 261882 | 62 | - | - | - | - | - | - | See details in Table XIX | | | | | Cu1Fe1S2 |
| 122.4.5 | 2518 | 62 | - | - | - | - | - | - | See details in Table XIX | | | | | Cu1Fe1S2 |
| 122.4.6 | 156643 | 62 | - | - | - | - | - | - | See details in Table XIX | | | | | Ag1Fe1S2 |
| 123.1.1 | 193701 | 38 | 0.0341 | -0.3003 | 2 | 2 | 10 | 2 | - | - | - | - | z2w,3=1<br>z4=2<br>z4mpi=3<br>z8=6 | In5Rh1U1 |
| 123.1.2 | 103639 | 38 | 0.0584 | -0.0978 | 2 | 2 | 10 | 2 | - | - | - | - | z2w,3=1<br>z4=2<br>z4mpi=3<br>z8=6 | Ga5Ir1U1 |
| 123.1.3 | 5257 | 16 | 0.0449 | -0.3222 | 2 | 8 | 10 | 9 | - | - | - | - | z2w,3=1<br>z4=2<br>z4mpi=1<br>z8=2 | Fe1Si2 |

Continued on next page

**TABLE XIV– continued**

| TQC | ICSD | NoE | DG | IndG | SubSG (w/ T) | cl | SubSG (w/o T) | cl | Max. Sub. SG | yes or no | Max. Sub. SG | yes or no | SIs (w/ T) | Stoich. Formula |
|---|---|---|---|---|---|---|---|---|---|---|---|---|---|---|
| 123.1.4 | 106640 | 78 | 0.0000 | -0.5633 | 2 | 5 | 10,10 | 5 | - | - | - | - | z2w,1=1 z2w,2=1 z2w,3=1 z4=2 z4mpi=3 z8=2 | Fe1Ga8Tm2 |
| 123.1.5 | 102455 | 38 | 0.0334 | -0.1748 | 2 | 2 | 10 | 2 | - | - | - | - | z2w,3=1 z4=2 z4mpi=3 z8=6 | Co1Ga5U1 |
| 123.1.6 | 192621 | 58 | 0.0126 | -1.0311 | 2 | 56 | 10,83 | 8 | - | - | - | - | z2w,1=1 z2w,2=1 z4mpi=2 z8=4 | Ce2In8Pt1 |
| 123.1.7 | 150226 | 36 | 0.0120 | -1.3049 | 83 | 7 | 83 | 7 | - | - | - | - | z4mpi=2 z8=4 | Ce1In5Rh1 |
| 123.1.8 | 110780 | 36 | 0.0103 | -1.4444 | 83 | 7 | 83 | 7 | - | - | - | - | z4mpi=2 z8=4 | Ce1In5Rh1 |
| 123.1.9 | 110779 | 36 | 0.0126 | -1.3869 | 83 | 7 | 83 | 7 | - | - | - | - | z4mpi=2 z8=4 | Ce1In5Rh1 |
| 123.1.10 | 110778 | 36 | 0.0149 | -1.3982 | 83 | 7 | 83 | 7 | - | - | - | - | z4mpi=2 z8=4 | Ce1In5Rh1 |
| 123.1.11 | 110777 | 36 | 0.0152 | -1.3411 | 83 | 7 | 83 | 7 | - | - | - | - | z4mpi=2 z8=4 | Ce1In5Rh1 |
| 123.1.12 | 150225 | 36 | 0.0141 | -1.0862 | 83 | 7 | 83 | 7 | - | - | - | - | z4mpi=2 z8=4 | Ce1In5Ir1 |
| 123.1.13 | 92435 | 64 | 0.0099 | -0.0173 | 83 | 7 | 83 | 7 | - | - | - | - | z4mpi=2 z8=4 | Bi2I1O4Tm1 |
| 123.1.14 | 92408 | 64 | 0.0142 | -0.0210 | 83 | 7 | 83 | 7 | - | - | - | - | z4mpi=2 z8=4 | Bi2Cl1O4Tm1 |
| 123.1.15 | 92421 | 64 | 0.0134 | -0.0215 | 83 | 7 | 83 | 7 | - | - | - | - | z4mpi=2 z8=4 | Bi2Br1O4Tm1 |
| 123.1.16 | 68047 | 110 | 0.0123 | -0.0314 | 2 | 6 | 10 | 6 | - | - | - | - | z2w,3=1 z4mpi=1 | Ba2Cu3Ho1O6 |
| 123.1.17 | 74163 | 98 | 0.0233 | -3.1821 | 2 | 1 | 10,65,83 | 1 | - | - | - | - | z2w,1=1 z2w,2=1 z4=2 z4mpi=2 z8=6 | Ba2Cu2O7Tl1Y1 |
| 123.1.18 | 430061 | 34 | 0.0063 | -0.5830 | 2 | 5 | 10,10 | 5 | - | - | - | - | z2w,1=1 z2w,2=1 z2w,3=1 z4=2 z4mpi=3 z8=2 | Ba1O1Sb2Ti2 |
| 123.1.19 | 430060 | 34 | 0.0061 | -0.5550 | 2 | 5 | 10,10 | 5 | - | - | - | - | z2w,1=1 z2w,2=1 z2w,3=1 z4=2 z4mpi=3 z8=2 | Ba1O1Sb2Ti2 |
| 123.1.20 | 237789 | 34 | 0.0084 | -0.5458 | 2 | 4 | 10,10 | 4 | - | - | - | - | z2w,1=1 z2w,2=1 z2w,3=1 z4mpi=1 | Ba1O1Sb2Ti2 |
| 123.1.21 | 165631 | 30 | 0.0321 | -0.5853 | 2 | 3 | 10 | 3 | - | - | - | - | z2w,3=1 z4mpi=1 z8=4 | Ba1Ge3Mg4 |

Continued on next page

**TABLE XIV– continued**

| TQC | ICSD | NoE | DG | IndG | SubSG (w/ T) | cl | SubSG (w/o T) | cl | Max. Sub. SG | yes or no | Max. Sub. SG | yes or no | SIs (w/ T) | Stoich. Formula |
|---|---|---|---|---|---|---|---|---|---|---|---|---|---|---|
| 123.1.22 | 262038 | 72 | 0.0470 | 0.0040 | 2 | 2 | 10 | 2 | - | - | - | - | z2w,3=1 z4=2 z4mpi=3 z8=6 | As3Rb1Zn4 |
| 123.1.23 | 262030 | 72 | 0.0286 | -0.0411 | 2 | 2 | 10 | 2 | - | - | - | - | z2w,3=1 z4=2 z4mpi=3 z8=6 | As3Cs1Zn4 |
| 123.1.24 | 170662 | 28 | 0.0597 | -0.4653 | 2 | 1 | 10,65,83 | 1 | - | - | - | - | z2w,1=1 z2w,2=1 z4=2 z4mpi=2 z8=6 | Ag1Sb1Te2 |
| 123.2.1 | 98585 | 36 | 0.0262 | -0.0579 | 2,81 | 31 | 2,81 | 23 | - | - | - | - | z2w,1=1 z2w,2=1 z2w,3=1 z4=3 z4mpi=1 z2=1 z8=3 | Nd1Ni1O2 |
| 123.2.2 | 428960 | 16 | 0.0038 | -0.7613 | 2,81 | 11 | 2,81 | 12 | - | - | - | - | z2w,3=1 z4=1 z4mpi=3 z2=1 z8=5 | Mg1Ru1Sc2 |
| 123.2.3 | 157999 | 66 | 0.0114 | -0.1956 | 2,81 | 17 | 2,81 | 22 | - | - | - | - | z2w,1=1 z2w,2=1 z4=3 z4mpi=2 z2=1 z8=7 | Fe5Si3U1 |
| 123.2.4 | 427123 | 58 | 0.0047 | -0.0489 | 2,81 | 43 | 2,81 | 21 | - | - | - | - | z2w,1=1 z2w,2=1 z4=3 z4mpi=2 z2=1 z8=3 | Er2Mg3Ru1 |
| 123.2.5 | 74164 | 98 | 0.0443 | -1.7650 | 2,81 | 16 | 2,81 | 18 | - | - | - | - | z2w,1=1 z2w,2=1 z2w,3=1 z4=3 z4mpi=1 z2=1 z8=7 | Cu2O7Sr2Tl1Y1 |
| 123.2.6 | 85244 | 116 | 0.0030 | -0.5890 | 2,81 | 38 | 2,81 | 20 | - | - | - | - | z2w,1=1 z2w,2=1 z2w,3=1 z4=1 z4mpi=1 z2=1 z8=5 | Cu2Gd1O8Ru1Sr2 |
| 123.2.7 | 240161 | 40 | 0.0105 | -0.2540 | 2,81 | 15 | 2,81 | 19 | - | - | - | - | z4=1 z2=1 z8=1 | Ce1Ga6Pd1 |

Continued on next page

**TABLE XIV– continued**

| TQC | ICSD | NoE | DG | IndG | SubSG (w/ T) | cl | SubSG (w/o T) | cl | Max. Sub. SG | yes or no | Max. Sub. SG | yes or no | SIs (w/ T) | Stoich. Formula |
|---|---|---|---|---|---|---|---|---|---|---|---|---|---|---|
| 123.2.8 | 102108 | 36 | 0.0111 | -0.8729 | 2,81 | 16 | 2,81 | 18 | - | - | - | - | z2w,1=1<br>z2w,2=1<br>z2w,3=1<br>z4=3<br>z4mpi=1<br>z2=1<br>z8=7 | Ce1Co1In5 |
| 123.2.9 | 92425 | 54 | 0.0220 | -0.0835 | 2,81 | 24 | 2,81 | 17 | - | - | - | - | z4=3<br>z2=1<br>z8=7 | Bi2I1O4Pr1 |
| 123.2.10 | 92403 | 60 | 0.0061 | -0.0712 | 2,81 | 24 | 2,81 | 17 | - | - | - | - | z4=3<br>z2=1<br>z8=7 | Bi2Cl1O4Tb1 |
| 123.2.11 | 92398 | 54 | 0.0168 | -0.0383 | 2,81 | 24 | 2,81 | 17 | - | - | - | - | z4=3<br>z2=1<br>z8=7 | Bi2Cl1O4Pr1 |
| 123.2.12 | 92401 | 58 | 0.0105 | 0.0019 | 2,81 | 24 | 2,81 | 17 | - | - | - | - | z4=3<br>z2=1<br>z8=7 | Bi2Cl1Eu1O4 |
| 123.2.13 | 92416 | 60 | 0.0057 | -0.0664 | 2,81 | 24 | 2,81 | 17 | - | - | - | - | z4=3<br>z2=1<br>z8=7 | Bi2Br1O4Tb1 |
| 123.2.14 | 92411 | 54 | 0.0100 | -0.0363 | 2,81 | 24 | 2,81 | 17 | - | - | - | - | z4=3<br>z2=1<br>z8=7 | Bi2Br1O4Pr1 |
| 123.2.15 | 424316 | 134 | 0.0052 | -1.2470 | 2,81 | 47 | 2,81 | 16 | - | - | - | - | z2w,1=1<br>z2w,2=1<br>z2w,3=1<br>z4=1<br>z4mpi=1<br>z2=1<br>z8=1 | Ba3O5Te1Yb2 |
| 123.2.16 | 84739 | 100 | 0.0189 | -2.2294 | 2,81 | 14 | 2,81 | 15 | - | - | - | - | z4=1<br>z2=1<br>z8=5 | Ba2Cu3O6Y1 |
| 123.2.17 | 72254 | 100 | 0.0183 | -2.2367 | 2,81 | 14 | 2,81 | 15 | - | - | - | - | z4=1<br>z2=1<br>z8=5 | Ba2Cu3O6Y1 |
| 123.2.18 | 68290 | 100 | 0.0195 | -2.2139 | 2,81 | 14 | 2,81 | 15 | - | - | - | - | z4=1<br>z2=1<br>z8=5 | Ba2Cu3O6Y1 |
| 123.2.19 | 68046 | 100 | 0.0177 | -2.2217 | 2,81 | 14 | 2,81 | 15 | - | - | - | - | z4=1<br>z2=1<br>z8=5 | Ba2Cu3O6Y1 |
| 123.2.20 | 68008 | 100 | 0.0226 | -2.2209 | 2,81 | 14 | 2,81 | 15 | - | - | - | - | z4=1<br>z2=1<br>z8=5 | Ba2Cu3O6Y1 |
| 123.2.21 | 67016 | 100 | 0.0223 | -2.2262 | 2,81 | 14 | 2,81 | 15 | - | - | - | - | z4=1<br>z2=1<br>z8=5 | Ba2Cu3O6Y1 |
| 123.2.22 | 66903 | 100 | 0.0184 | -2.2271 | 2,81 | 14 | 2,81 | 15 | - | - | - | - | z4=1<br>z2=1<br>z8=5 | Ba2Cu3O6Y1 |
| 123.2.23 | 63269 | 100 | 0.0226 | -2.2209 | 2,81 | 14 | 2,81 | 15 | - | - | - | - | z4=1<br>z2=1<br>z8=5 | Ba2Cu3O6Y1 |

Continued on next page

**TABLE XIV– continued**

| TQC | ICSD | NoE | DG | IndG | SubSG (w/ T) | cl | SubSG (w/o T) | cl | Max. Sub. SG | yes or no | Max. Sub. SG | yes or no | SIs (w/ T) | Stoich. Formula |
|---|---|---|---|---|---|---|---|---|---|---|---|---|---|---|
| 123.2.24 | 203116 | 100 | 0.0239 | -2.2418 | 2,81 | 14 | 2,81 | 15 | - | - | - | - | z4=1<br>z2=1<br>z8=5 | Ba2Cu3O6Y1 |
| 123.2.25 | 98114 | 114 | 0.0209 | -2.2469 | 2,81 | 14 | 2,81 | 15 | - | - | - | - | z4=1<br>z2=1<br>z8=5 | Ba2Cu3Lu1O6 |
| 123.2.26 | 98113 | 114 | 0.0230 | -2.2513 | 2,81 | 14 | 2,81 | 15 | - | - | - | - | z4=1<br>z2=1<br>z8=5 | Ba2Cu3Lu1O6 |
| 123.2.27 | 64639 | 106 | 0.0203 | -2.0963 | 2,81 | 13 | 2,81 | 14 | - | - | - | - | z2w,3=1<br>z4=3<br>z4mpi=1<br>z2=1<br>z8=7 | Ba2Cu2La1O8Ta1 |
| 123.2.28 | 165630 | 30 | 0.0250 | -0.5892 | 2,81 | 12 | 2,81 | 13 | - | - | - | - | z4=3<br>z2=1<br>z8=3 | Ba1Mg4Si3 |
| 123.2.29 | 169074 | 34 | 0.0116 | -0.6127 | 2,81 | 11 | 2,81 | 12 | - | - | - | - | z2w,3=1<br>z4=1<br>z4mpi=3<br>z2=1<br>z8=5 | As2Ba1O1Ti2 |
| 123.2.30 | 174342 | 36 | 0.0343 | -0.6574 | 2,81 | 10 | 2,81 | 11 | - | - | - | - | z2w,3=1<br>z4=3<br>z4mpi=3<br>z2=1<br>z8=3 | Al2B1La1Ru2 |
| 123.2.31 | 251834 | 36 | 0.0142 | -0.4042 | 2,81 | 9 | 2,81 | 10 | - | - | - | - | z2w,1=1<br>z2w,2=1<br>z2w,3=1<br>z4=3<br>z4mpi=3<br>z2=1<br>z8=3 | Al2B1La1Os2 |
| 123.3.1 | 79355 | 196 | - | - | - | - | - | - | 115 | yes | 115 | yes | - | Nb4O6Sr1 |
| | | | | | | | | | 111 | yes | 111 | yes | - | |
| | | | | | | | | | 65 | no(2) | 65 | no(2) | z2w,3=1<br>z4=3 | |
| | | | | | | | | | 47 | no(2) | 47 | no(2) | z2w,3=1<br>z4=3 | |
| 123.3.2 | 412909 | 48 | - | - | - | - | - | - | 115 | no(81) | 115 | no(81) | z2=1 | In5Rh1Yb1 |
| | | | | | | | | | 111 | no(81) | 111 | no(81) | z2=1 | |
| | | | | | | | | | 65 | no(2) | 65 | no(2) | z2w,3=1<br>z4=1 | |
| | | | | | | | | | 47 | no(2) | 47 | no(2) | z2w,3=1<br>z4=1 | |
| 123.3.3 | 413163 | 48 | - | - | - | - | - | - | 115 | yes | 115 | yes | - | In5Ir1Yb1 |
| | | | | | | | | | 111 | yes | 111 | yes | - | |
| | | | | | | | | | 65 | no(2) | 65 | no(2) | z2w,3=1<br>z4=1 | |
| | | | | | | | | | 47 | no(2) | 47 | no(2) | z2w,3=1<br>z4=1 | |
| 123.3.4 | 290932 | 92 | - | - | - | - | - | - | 115 | yes | 115 | yes | - | In2Pd8Se1 |
| | | | | | | | | | 111 | yes | 111 | yes | - | |
| | | | | | | | | | 65 | no(2) | 65 | no(2) | z2w,1=1<br>z2w,2=1<br>z2w,3=1<br>z4=1 | |

Continued on next page

**TABLE XIV– continued**

| TQC | ICSD | NoE | DG | IndG | SubSG cl (w/ T) | SubSG cl (w/o T) | Max. Sub. SG | yes or no | Max. Sub. SG | yes or no | SIs (w/ T) | Stoich. Formula |
|---|---|---|---|---|---|---|---|---|---|---|---|---|
| | | | | | | | 47 | no(2) | 47 | no(2) | z2w,1=1<br>z2w,2=1<br>z2w,3=1<br>z4=1 | |
| 123.3.5 | 236215 | 68 | - | - | - - | - - | 115 | no(81) | 115 | no(81) | z2=1 | Hg1Pd5Se1 |
| | | | | | | | 111 | no(81) | 111 | no(81) | z2=1 | |
| | | | | | | | 65 | no(2) | 65 | no(2) | z4=3 | |
| | | | | | | | 47 | no(2) | 47 | no(2) | z4=3 | |
| 123.3.6 | 187377 | 68 | - | - | - - | - - | 115 | no(81) | 115 | no(81) | z2=1 | Hg1Pd5Se1 |
| | | | | | | | 111 | no(81) | 111 | no(81) | z2=1 | |
| | | | | | | | 65 | no(2) | 65 | no(2) | z4=3 | |
| | | | | | | | 47 | no(2) | 47 | no(2) | z4=3 | |
| 123.3.7 | 418606 | 30 | - | - | - - | - - | 115 | no(81) | 115 | no(81) | z2=1 | Fe1O2Sr1 |
| | | | | | | | 111 | no(81) | 111 | no(81) | z2=1 | |
| | | | | | | | 65 | yes | 65 | yes | - | |
| | | | | | | | 47 | yes | 47 | yes | - | |
| 123.3.8 | 418605 | 30 | - | - | - - | - - | 115 | no(81) | 115 | no(81) | z2=1 | Fe1O2Sr1 |
| | | | | | | | 111 | no(81) | 111 | no(81) | z2=1 | |
| | | | | | | | 65 | yes | 65 | yes | - | |
| | | | | | | | 47 | yes | 47 | yes | - | |
| 123.3.9 | 418603 | 30 | - | - | - - | - - | 115 | no(81) | 115 | no(81) | z2=1 | Fe1O2Sr1 |
| | | | | | | | 111 | no(81) | 111 | no(81) | z2=1 | |
| | | | | | | | 65 | yes | 65 | yes | - | |
| | | | | | | | 47 | yes | 47 | yes | - | |
| 123.3.10 | 173434 | 30 | - | - | - - | - - | 115 | no(81) | 115 | no(81) | z2=1 | Fe1O2Sr1 |
| | | | | | | | 111 | no(81) | 111 | no(81) | z2=1 | |
| | | | | | | | 65 | yes | 65 | yes | - | |
| | | | | | | | 47 | yes | 47 | yes | - | |
| 123.3.11 | 88880 | 98 | - | - | - - | - - | 115 | no(81) | 115 | no(81) | z2=1 | F1K1Nb4O5 |
| | | | | | | | 111 | no(81) | 111 | no(81) | z2=1 | |
| | | | | | | | 65 | yes | 65 | yes | - | |
| | | | | | | | 47 | yes | 47 | yes | - | |
| 123.3.12 | 412910 | 48 | - | - | - - | - - | 115 | no(81) | 115 | no(81) | z2=1 | Co1In5Yb1 |
| | | | | | | | 111 | no(81) | 111 | no(81) | z2=1 | |
| | | | | | | | 65 | no(2) | 65 | no(2) | z2w,3=1<br>z4=1 | |
| | | | | | | | 47 | no(2) | 47 | no(2) | z2w,3=1<br>z4=1 | |
| 123.3.13 | 162956 | 48 | - | - | - - | - - | 115 | yes | 115 | yes | - | Co1Ga5Yb1 |
| | | | | | | | 111 | yes | 111 | yes | - | |
| | | | | | | | 65 | no(2) | 65 | no(2) | z2w,3=1<br>z4=1 | |
| | | | | | | | 47 | no(2) | 47 | no(2) | z2w,3=1<br>z4=1 | |
| 123.3.14 | 153908 | 46 | - | - | - - | - - | 115 | no(81) | 115 | no(81) | z2=1 | Co1Er1In5 |
| | | | | | | | 111 | no(81) | 111 | no(81) | z2=1 | |
| | | | | | | | 65 | no(2) | 65 | no(10) | z2w,3=1<br>z4=2 | |
| | | | | | | | 47 | no(2) | 47 | no(10) | z2w,3=1<br>z4=2 | |
| 123.3.15 | 247864 | 58 | - | - | - - | - - | 115 | yes | 115 | yes | - | Ce2In8Pd1 |
| | | | | | | | 111 | yes | 111 | yes | - | |
| | | | | | | | 65 | no(2) | 65 | no(2) | z2w,1=1<br>z2w,2=1<br>z4=3 | |
| | | | | | | | 47 | no(2) | 47 | no(2) | z2w,1=1<br>z2w,2=1<br>z4=3 | |

Continued on next page

**TABLE XIV– continued**

| TQC | ICSD | NoE | DG | IndG | SubSG (w/ T) | cl | SubSG (w/o T) | cl | Max. Sub. SG | yes or no | Max. Sub. SG | yes or no | SIs (w/ T) | Stoich. Formula |
|---|---|---|---|---|---|---|---|---|---|---|---|---|---|---|
| 123.3.16 | 173438 | 22 | - | - | - | - | - | - | 115 | yes | 115 | yes | - | Ca1Fe1O2 |
| | | | | | | | | | 111 | yes | 111 | yes | - | |
| | | | | | | | | | 65 | yes | 65 | yes | - | |
| | | | | | | | | | 47 | yes | 47 | yes | - | |
| 123.3.17 | 189334 | 36 | - | - | - | - | - | - | 115 | yes | 115 | yes | - | C1Ce1Mo2Si2 |
| | | | | | | | | | 111 | yes | 111 | yes | - | |
| | | | | | | | | | 65 | no(2) | 65 | no(2) | z2w,1=1<br>z2w,2=1<br>z4=3 | |
| | | | | | | | | | 47 | no(2) | 47 | no(2) | z2w,1=1<br>z2w,2=1<br>z4=3 | |
| 123.3.18 | 75342 | 142 | - | - | - | - | - | - | 115 | yes | 115 | yes | - | Ba2Cu2Nd2O11Ti2 |
| | | | | | | | | | 111 | yes | 111 | yes | - | |
| | | | | | | | | | 65 | no(2) | 65 | no(2) | z4=1 | |
| | | | | | | | | | 47 | no(2) | 47 | no(2) | z4=1 | |
| 123.3.19 | 150705 | 74 | - | - | - | - | - | - | 115 | yes | 115 | yes | - | Ba1Mn2Nd1O6 |
| | | | | | | | | | 111 | yes | 111 | yes | - | |
| | | | | | | | | | 65 | no(2) | 65 | no(10) | z2w,3=1<br>z4=2 | |
| | | | | | | | | | 47 | no(2) | 47 | no(10) | z2w,3=1<br>z4=2 | |
| 123.3.20 | 422759 | 76 | - | - | - | - | - | - | 115 | no(81) | 115 | no(81) | z2=1 | Ba1Co2Gd1O5 |
| | | | | | | | | | 111 | no(81) | 111 | no(81) | z2=1 | |
| | | | | | | | | | 65 | no(2) | 65 | no(10) | z2w,3=1<br>z4=2 | |
| | | | | | | | | | 47 | no(2) | 47 | no(10) | z2w,3=1<br>z4=2 | |
| 123.3.21 | 430063 | 34 | - | - | - | - | - | - | 115 | yes | 115 | yes | - | Ba1Bi2O1Ti2 |
| | | | | | | | | | 111 | yes | 111 | yes | - | |
| | | | | | | | | | 65 | no(2) | 65 | no(2) | z2w,1=1<br>z2w,2=1<br>z4=1 | |
| | | | | | | | | | 47 | no(2) | 47 | no(2) | z2w,1=1<br>z2w,2=1<br>z4=1 | |
| 123.3.22 | 79256 | 86 | - | - | - | - | - | - | 115 | yes | 115 | yes | - | As2Ba1Cu6 |
| | | | | | | | | | 111 | yes | 111 | yes | - | |
| | | | | | | | | | 65 | no(2) | 65 | no(2) | z2w,3=1<br>z4=1 | |
| | | | | | | | | | 47 | no(2) | 47 | no(2) | z2w,3=1<br>z4=1 | |
| 123.3.23 | 193115 | 58 | - | - | - | - | - | - | 115 | no(81) | 115 | no(81) | z2=1 | As1In1Pt5 |
| | | | | | | | | | 111 | no(81) | 111 | no(81) | z2=1 | |
| | | | | | | | | | 65 | yes | 65 | yes | - | |
| | | | | | | | | | 47 | yes | 47 | yes | - | |
| 123.3.24 | 193117 | 58 | - | - | - | - | - | - | 115 | yes | 115 | yes | - | As1In1Pd5 |
| | | | | | | | | | 111 | yes | 111 | yes | - | |
| | | | | | | | | | 65 | no(2) | 65 | no(2) | z4=3 | |
| | | | | | | | | | 47 | no(2) | 47 | no(2) | z4=3 | |
| 124.4.1 | 62049 | 58 | - | - | - | - | - | - | See details in Table XIX | | | | | Ta1Te4 |
| 124.4.2 | 75321 | 74 | - | - | - | - | - | - | See details in Table XIX | | | | | Nb1Te4 |
| 124.4.3 | 75319 | 74 | - | - | - | - | - | - | See details in Table XIX | | | | | Nb1Te4 |
| 124.4.4 | 65130 | 74 | - | - | - | - | - | - | See details in Table XIX | | | | | Nb1Te4 |
| 124.4.5 | 61113 | 74 | - | - | - | - | - | - | See details in Table XIX | | | | | Nb1Te4 |
| 124.4.6 | 60599 | 74 | - | - | - | - | - | - | See details in Table XIX | | | | | Nb1Te4 |
| 124.4.7 | 60598 | 74 | - | - | - | - | - | - | See details in Table XIX | | | | | Nb1Te4 |
| 125.1.1 | 412794 | 92 | 0.0065 | -0.0030 | 2 | 1 | 0 | 1 | - | - | - | - | z2w,3=1 | Pr1Rb1Te4 |

Continued on next page

**TABLE XIV– continued**

| TQC | ICSD | NoE | DG | IndG | SubSG (w/ T) | cl | SubSG (w/o T) | cl | Max. Sub. SG | yes or no | Max. Sub. SG | yes or no | SIs (w/ T) | Stoich. Formula |
|---|---|---|---|---|---|---|---|---|---|---|---|---|---|---|
| 125.1.2 | 391204 | 92 | 0.0064 | -0.0053 | 2 | 1 | 0 | 1 | - | - | - | - | z2w,3=1 | K1Pr1Te4 |
| 125.1.3 | 152562 | 60 | 0.0103 | -0.5311 | 2 | 1 | 0 | 1 | - | - | - | - | z2w,3=1 | Ce1Ga6 |
| 125.1.4 | 102167 | 60 | 0.0023 | -0.6502 | 2 | 1 | 0 | 1 | - | - | - | - | z2w,3=1 | Ce1Ga6 |
| 125.2.1 | 400461 | 80 | 0.0207 | -0.3928 | 2,81 | 2 | 2,81 | 2 | - | - | - | - | z2w,3=1<br>z4=3<br>z2=1 | H8I1N4P1 |
| 125.3.1 | 250901 | 84 | - | - | - | - | - | - | 117 | yes | 117 | yes | - | Ga6Yb1 |
| | | | | | | | | | 111 | yes | 111 | yes | - | |
| | | | | | | | | | 67 | no(2) | 67 | no(2) | z2w,3=1<br>z4=3 | |
| | | | | | | | | | 50 | no(2) | 50 | no(2) | z2w,3=1<br>z4=3 | |
| 125.3.2 | 183717 | 136 | - | - | - | - | - | - | 117 | no(81) | 117 | no(81) | z2=1 | Ga12La2Pd1 |
| | | | | | | | | | 111 | no(81) | 111 | no(81) | z2=1 | |
| | | | | | | | | | 67 | no(2) | 67 | yes | z4=2 | |
| | | | | | | | | | 50 | no(2) | 50 | yes | z4=2 | |
| 125.3.3 | 171486 | 136 | - | - | - | - | - | - | 117 | no(81) | 117 | no(81) | z2=1 | Ga12La2Pd1 |
| | | | | | | | | | 111 | no(81) | 111 | no(81) | z2=1 | |
| | | | | | | | | | 67 | no(2) | 67 | yes | z4=2 | |
| | | | | | | | | | 50 | no(2) | 50 | yes | z4=2 | |
| 125.3.4 | 161765 | 136 | - | - | - | - | - | - | 117 | no(81) | 117 | no(81) | z2=1 | Ga12La2Ni1 |
| | | | | | | | | | 111 | no(81) | 111 | no(81) | z2=1 | |
| | | | | | | | | | 67 | no(2) | 67 | yes | z4=2 | |
| | | | | | | | | | 50 | no(2) | 50 | yes | z4=2 | |
| 125.3.5 | 171487 | 140 | - | - | - | - | - | - | 117 | no(81) | 117 | no(81) | z2=1 | Ce2Ga12Pd1 |
| | | | | | | | | | 111 | no(81) | 111 | no(81) | z2=1 | |
| | | | | | | | | | 67 | no(2) | 67 | yes | z2w,3=1<br>z4=2 | |
| | | | | | | | | | 50 | no(2) | 50 | yes | z2w,3=1<br>z4=2 | |
| 125.3.6 | 161766 | 140 | - | - | - | - | - | - | 117 | no(81) | 117 | no(81) | z2=1 | Ce2Ga12Ni1 |
| | | | | | | | | | 111 | no(81) | 111 | no(81) | z2=1 | |
| | | | | | | | | | 67 | no(2) | 67 | yes | z2w,3=1<br>z4=2 | |
| | | | | | | | | | 50 | no(2) | 50 | yes | z2w,3=1<br>z4=2 | |
| 125.3.7 | 50738 | 164 | - | - | - | - | - | - | 117 | yes | 117 | yes | - | Ag5Rb1Se3 |
| | | | | | | | | | 111 | yes | 111 | yes | - | |
| | | | | | | | | | 67 | no(2) | 67 | no(2) | z4=3 | |
| | | | | | | | | | 50 | no(2) | 50 | no(2) | z4=3 | |
| 125.4.1 | 412793 | 94 | - | - | - | - | - | - | See details in Table XIX | | | | | Nd1Rb1Te4 |
| 125.4.2 | 412792 | 94 | - | - | - | - | - | - | See details in Table XIX | | | | | K1Nd1Te4 |
| 125.4.3 | 391206 | 102 | - | - | - | - | - | - | See details in Table XIX | | | | | Gd1Rb1Te4 |
| 125.4.4 | 391205 | 102 | - | - | - | - | - | - | See details in Table XIX | | | | | Gd1K1Te4 |
| 125.4.5 | 59298 | 102 | - | - | - | - | - | - | See details in Table XIX | | | | | Cs1Gd1Te4 |
| 125.4.6 | 161768 | 142 | - | - | - | - | - | - | See details in Table XIX | | | | | Ce2Cu1Ga12 |
| 125.4.7 | 161767 | 142 | - | - | - | - | - | - | See details in Table XIX | | | | | Ce2Cu1Ga12 |
| 125.4.8 | 67656 | 90 | - | - | - | - | - | - | See details in Table XIX | | | | | Ce1K1Se4 |
| 126.3.1 | 421051 | 152 | - | - | - | - | - | - | 118 | no(81) | 118 | no(81) | z2=1 | Mo2O8Pr1Tl1 |
| | | | | | | | | | 112 | no(81) | 112 | no(81) | z2=1 | |
| | | | | | | | | | 68 | yes | 68 | yes | - | |
| | | | | | | | | | 48 | yes | 48 | yes | - | |
| 126.4.1 | 161960 | 154 | - | - | - | - | - | - | See details in Table XIX | | | | | Mo2Nd1O8Tl1 |
| 127.2.1 | 105627 | 64 | 0.0400 | -1.0061 | 2 | 2 | 65,83 | 15 | - | - | - | - | z4=2<br>z4mpi=2<br>z8=2 | Pb3Sr2 |

Continued on next page

**TABLE XIV– continued**

| TQC | ICSD | NoE | DG | IndG | SubSG (w/ T) | cl | SubSG (w/o T) | cl | Max. Sub. SG | yes or no | Max. Sub. SG | yes or no | SIs (w/ T) | Stoich. Formula |
|---|---|---|---|---|---|---|---|---|---|---|---|---|---|---|
| 127.2.2 | 427282 | 124 | 0.0000 | -0.0478 | 2,81 | 28 | 2,81 | 13 | - | - | - | - | z2w,3=1<br>z4=1<br>z4mpi=1<br>z2=1<br>z8=1 | Ni2Sn1Tb2 |
| 127.2.3 | 391441 | 116 | 0.0000 | -0.5489 | 2,81 | 4 | 2,81 | 1 | - | - | - | - | z2w,3=1<br>z4=3<br>z4mpi=3<br>z2=1<br>z8=7 | Mg1Si2Yb2 |
| 127.2.4 | 26200 | 44 | 0.0079 | -3.2359 | 2 | 12 | 65,83 | 14 | - | - | - | - | z4=2<br>z8=6 | Li2Sn5 |
| 127.2.5 | 415117 | 100 | 0.0000 | -0.0115 | 2,81 | 4 | 2,81 | 1 | - | - | - | - | z2w,3=1<br>z4=3<br>z4mpi=3<br>z2=1<br>z8=7 | Dy2Mg1Si2 |
| 127.2.6 | 423452 | 100 | 0.0000 | -0.0084 | 2,81 | 4 | 2,81 | 1 | - | - | - | - | z2w,3=1<br>z4=3<br>z4mpi=3<br>z2=1<br>z8=7 | Dy2Ge2Mg1 |
| 127.2.7 | 262925 | 68 | 0.0291 | 0.0291 | 2,81 | 4 | 2,81 | 1 | - | - | - | - | z2w,3=1<br>z4=3<br>z4mpi=3<br>z2=1<br>z8=7 | Cs1I3Sn1 |
| 127.2.8 | 413849 | 68 | 0.0000 | -0.1379 | 2,81 | 11 | 2,81 | 4 | - | - | - | - | z2w,3=1<br>z4=1<br>z4mpi=3<br>z2=1<br>z8=5 | Ce2Ge2Mg1 |
| 127.2.9 | 157389 | 160 | 0.0177 | -0.5562 | 2,81 | 28 | 2,81 | 13 | - | - | - | - | z2w,3=1<br>z4=1<br>z4mpi=1<br>z2=1<br>z8=1 | Cd1Pd2Yb2 |
| 127.2.10 | 156063 | 160 | 0.0177 | -0.5616 | 2,81 | 28 | 2,81 | 13 | - | - | - | - | z2w,3=1<br>z4=1<br>z4mpi=1<br>z2=1<br>z8=1 | Cd1Pd2Yb2 |
| 127.2.11 | 157385 | 116 | 0.0068 | -0.2320 | 2,81 | 23 | 2,81 | 6 | - | - | - | - | z4=3<br>z2=1<br>z8=7 | Cd1Pd2Pr2 |
| 127.2.12 | 414596 | 116 | 0.0000 | -0.2021 | 2,81 | 23 | 2,81 | 6 | - | - | - | - | z4=3<br>z2=1<br>z8=7 | Cd1Ni2Pr2 |
| 127.2.13 | 107389 | 104 | 0.0242 | -0.6747 | 2,81 | 4 | 2,81 | 1 | - | - | - | - | z2w,3=1<br>z4=3<br>z4mpi=3<br>z2=1<br>z8=7 | Cd1La2Rh2 |
| 127.2.14 | 424179 | 136 | 0.0000 | -0.8374 | 2,81 | 4 | 2,81 | 1 | - | - | - | - | z2w,3=1<br>z4=3<br>z4mpi=3<br>z2=1<br>z8=7 | Cd1Ge2Yb2 |

Continued on next page

**TABLE XIV– continued**

| TQC | ICSD | NoE | DG | IndG | SubSG (w/ T) | cl | SubSG (w/o T) | cl | Max. Sub. SG | yes or no | Max. Sub. SG | yes or no | SIs (w/ T) | Stoich. Formula |
|---|---|---|---|---|---|---|---|---|---|---|---|---|---|---|
| 127.2.15 | 424169 | 84 | 0.0176 | -1.3148 | 2,81 | 10 | 2,81 | 12 | - | - | - | - | z4=1<br>z2=1<br>z8=1 | Cd1Ge2Y2 |
| 127.2.16 | 424178 | 132 | 0.0022 | -0.4177 | 2,81 | 26 | 2,81 | 11 | - | - | - | - | z2w,3=1<br>z4=3<br>z4mpi=1<br>z2=1<br>z8=3 | Cd1Ge2Tm2 |
| 127.2.17 | 424177 | 128 | 0.0039 | -0.0335 | 2,81 | 4 | 2,81 | 1 | - | - | - | - | z2w,3=1<br>z4=3<br>z4mpi=3<br>z2=1<br>z8=7 | Cd1Er2Ge2 |
| 127.2.18 | 424175 | 120 | 0.0000 | -0.0096 | 2,81 | 4 | 2,81 | 1 | - | - | - | - | z2w,3=1<br>z4=3<br>z4mpi=3<br>z2=1<br>z8=7 | Cd1Dy2Ge2 |
| 127.2.19 | 193226 | 32 | 0.0079 | -1.4056 | 2,81 | 21 | 2,81 | 2 | - | - | - | - | z4=1<br>z4mpi=2<br>z2=1<br>z8=5 | Ca2Ge2Sn1 |
| 127.2.20 | 79239 | 184 | 0.0313 | -0.2949 | 2,81 | 9 | 2,81 | 10 | - | - | - | - | z4=3<br>z4mpi=2<br>z2=1<br>z8=3 | C5Ru6Th2 |
| 127.2.21 | 39204 | 184 | 0.0617 | -0.3026 | 2,81 | 8 | 2,81 | 10 | - | - | - | - | z4=3<br>z4mpi=2<br>z2=1<br>z8=7 | C5Ru6Th2 |
| 127.2.22 | 427156 | 64 | 0.0102 | -0.0209 | 2 | 17 | 10 | 9 | - | - | - | - | z2w,3=1<br>z4=2<br>z4mpi=1<br>z8=6 | C2B2Gd1 |
| 127.2.23 | 94042 | 72 | 0.0075 | -0.0468 | 2 | 17 | 10 | 9 | - | - | - | - | z2w,3=1<br>z4=2<br>z4mpi=1<br>z8=6 | C2B2Er1 |
| 127.2.24 | 94040 | 68 | 0.0000 | -0.0390 | 2 | 13 | 10 | 8 | - | - | - | - | z2w,3=1<br>z4=2<br>z4mpi=3<br>z8=2 | C2B2Dy1 |
| 127.2.25 | 94036 | 52 | 0.0061 | -0.0899 | 2 | 7 | 10 | 5 | - | - | - | - | z2w,3=1<br>z4mpi=1 | C2B2Ce1 |
| 127.2.26 | 88561 | 52 | 0.0062 | -0.0908 | 2 | 7 | 10 | 5 | - | - | - | - | z2w,3=1<br>z4mpi=1 | C2B2Ce1 |
| 127.2.27 | 88560 | 52 | 0.0051 | -0.0951 | 2 | 7 | 10 | 5 | - | - | - | - | z2w,3=1<br>z4mpi=1 | C2B2Ce1 |
| 127.2.28 | 280180 | 52 | 0.0050 | -0.0943 | 2 | 7 | 10 | 5 | - | - | - | - | z2w,3=1<br>z4mpi=1 | C2B2Ce1 |
| 127.2.29 | 202173 | 176 | 0.0020 | -0.0175 | 2 | 20 | 10 | 7 | - | - | - | - | z2w,3=1<br>z4mpi=3 | Ba1O5Pd1Tb2 |
| 127.2.30 | 404497 | 184 | 0.0000 | -0.0157 | 2,81 | 6 | 2,81 | 6 | - | - | - | - | z4=3<br>z2=1<br>z8=3 | Ba1Ho2O5Pd1 |
| 127.2.31 | 404496 | 180 | 0.0052 | -0.0027 | 2 | 7 | 10 | 5 | - | - | - | - | z2w,3=1<br>z4mpi=1 | Ba1Dy2O5Pd1 |

Continued on next page

**TABLE XIV– continued**

| TQC | ICSD | NoE | DG | IndG | SubSG (w/ T) | cl | SubSG (w/o T) | cl | Max. Sub. SG | yes or no | Max. Sub. SG | yes or no | SIs (w/ T) | Stoich. Formula |
|---|---|---|---|---|---|---|---|---|---|---|---|---|---|---|
| 127.2.32 | 67222 | 92 | 0.0139 | -0.8238 | 2,81 | 5 | 2,81 | 4 | - | - | - | - | z2w,3=1<br>z4=1<br>z4mpi=3<br>z2=1<br>z8=1 | B4Y1 |
| 127.2.33 | 26777 | 92 | 0.0123 | -0.5016 | 2,81 | 4 | 2,81 | 1 | - | - | - | - | z2w,3=1<br>z4=3<br>z4mpi=3<br>z2=1<br>z8=7 | B4Y1 |
| 127.2.34 | 2360 | 92 | 0.0036 | -0.5832 | 2,81 | 4 | 2,81 | 1 | - | - | - | - | z2w,3=1<br>z4=3<br>z4mpi=3<br>z2=1<br>z8=7 | B4La1 |
| 127.2.35 | 182659 | 132 | 0.0000 | -0.0085 | 83 | 3 | 0 | 3 | - | - | - | - | z8=4 | B4Ho1 |
| 127.2.36 | 51434 | 120 | 0.0103 | -0.6800 | 2,81 | 19 | 2,81 | 2 | - | - | - | - | z4=1<br>z4mpi=2<br>z2=1<br>z8=1 | B2Rh5Sc3 |
| 127.2.37 | 107439 | 132 | 0.0161 | -1.2034 | 2,81 | 1 | 2,81 | 1 | - | - | - | - | z2w,3=1<br>z4=3<br>z4mpi=3<br>z2=1<br>z8=3 | Al1Pt3 |
| 127.3.1 | 73695 | 100 | - | - | - | - | - | - | 117 | no(81) | 117 | no(81) | z2=1 | Si2U3 |
| | | | | | | | | | 113 | no(81) | 113 | no(81) | z2=1 | |
| | | | | | | | | | 65 | no(2) | 65 | no(10) | z2w,3=1 | |
| | | | | | | | | | 55 | no(2) | 55 | no(10) | z2w,3=1 | |
| 127.3.2 | 54347 | 100 | - | - | - | - | - | - | 117 | yes | 117 | yes | - | Rh2Sn1U2 |
| | | | | | | | | | 113 | yes | 113 | yes | - | |
| | | | | | | | | | 65 | no(2) | 65 | no(2) | z2w,3=1<br>z4=1 | |
| | | | | | | | | | 55 | no(2) | 55 | no(2) | z2w,3=1<br>z4=1 | |
| 127.3.3 | 99189 | 92 | - | - | - | - | - | - | 117 | no(81) | 117 | no(81) | z2=1 | Pb1Pd2Y2 |
| | | | | | | | | | 113 | no(81) | 113 | no(81) | z2=1 | |
| | | | | | | | | | 65 | no(2) | 65 | no(65) | z4=2 | |
| | | | | | | | | | 55 | no(2) | 55 | yes | z4=2 | |
| 127.3.4 | 99199 | 140 | - | - | - | - | - | - | 117 | no(81) | 117 | no(81) | z2=1 | Pb1Pd2Tm2 |
| | | | | | | | | | 113 | no(81) | 113 | no(81) | z2=1 | |
| | | | | | | | | | 65 | yes | 65 | yes | - | |
| | | | | | | | | | 55 | yes | 55 | yes | - | |
| 127.3.5 | 99192 | 100 | - | - | - | - | - | - | 117 | yes | 117 | yes | - | Pb1Pd2Pr2 |
| | | | | | | | | | 113 | yes | 113 | yes | - | |
| | | | | | | | | | 65 | no(2) | 65 | no(2) | z2w,3=1<br>z4=1 | |
| | | | | | | | | | 55 | no(2) | 55 | no(2) | z2w,3=1<br>z4=1 | |
| 127.3.6 | 54348 | 60 | - | - | - | - | - | - | 117 | yes | 117 | yes | - | Ni2Sc2Sn1 |
| | | | | | | | | | 113 | yes | 113 | yes | - | |
| | | | | | | | | | 65 | no(2) | 65 | no(65) | z4=2 | |
| | | | | | | | | | 55 | no(2) | 55 | yes | z4=2 | |
| 127.3.7 | 81707 | 296 | - | - | - | - | - | - | 117 | yes | 117 | yes | - | Ni2P4Zr9 |
| | | | | | | | | | 113 | yes | 113 | yes | - | |
| | | | | | | | | | 65 | no(2) | 65 | no(2) | z4=3 | |
| | | | | | | | | | 55 | no(2) | 55 | no(2) | z4=3 | |
| 127.3.8 | 92839 | 128 | - | - | - | - | - | - | 117 | no(81) | 117 | no(81) | z2=1 | Mo4O6Sn1 |

Continued on next page

**TABLE XIV– continued**

| TQC | ICSD | NoE | DG | IndG | SubSG cl (w/ T) | SubSG cl (w/o T) | Max. Sub. SG | yes or no | Max. Sub. SG | yes or no | SIs (w/ T) | Stoich. Formula |
|---|---|---|---|---|---|---|---|---|---|---|---|---|
| | | | | | | | 113 | no(81) | 113 | no(81) | z2=1 | |
| | | | | | | | 65 | no(2) | 65 | no(2) | z4=3 | |
| | | | | | | | 55 | no(2) | 55 | no(2) | z4=3 | |
| 127.3.9 | 412728 | 120 | - | - | - - | - - | 117 | yes | 117 | yes | - | Mg1Pd2Tb2 |
| | | | | | | | 113 | yes | 113 | yes | - | |
| | | | | | | | 65 | no(2) | 65 | no(2) | z4=3 | |
| | | | | | | | 55 | no(2) | 55 | no(2) | z4=3 | |
| 127.3.10 | 411080 | 96 | - | - | - - | - - | 117 | no(81) | 117 | no(81) | z2=1 | Mg1Ni2Pr2 |
| | | | | | | | 113 | no(81) | 113 | no(81) | z2=1 | |
| | | | | | | | 65 | no(2) | 65 | no(65) | z4=2 | |
| | | | | | | | 55 | no(2) | 55 | yes | z4=2 | |
| 127.3.11 | 107328 | 96 | - | - | - - | - - | 117 | no(81) | 117 | no(81) | z2=1 | Mg1Ni2Pr2 |
| | | | | | | | 113 | no(81) | 113 | no(81) | z2=1 | |
| | | | | | | | 65 | no(2) | 65 | no(65) | z4=2 | |
| | | | | | | | 55 | no(2) | 55 | yes | z4=2 | |
| 127.3.12 | 411079 | 100 | - | - | - - | - - | 117 | yes | 117 | yes | - | Mg1Nd2Ni2 |
| | | | | | | | 113 | yes | 113 | yes | - | |
| | | | | | | | 65 | no(2) | 65 | no(2) | z2w,3=1 z4=1 | |
| | | | | | | | 55 | no(2) | 55 | no(2) | z2w,3=1 z4=1 | |
| 127.3.13 | 107329 | 100 | - | - | - - | - - | 117 | yes | 117 | yes | - | Mg1Nd2Ni2 |
| | | | | | | | 113 | yes | 113 | yes | - | |
| | | | | | | | 65 | no(2) | 65 | no(2) | z2w,3=1 z4=1 | |
| | | | | | | | 55 | no(2) | 55 | no(2) | z2w,3=1 z4=1 | |
| 127.3.14 | 99200 | 148 | - | - | - - | - - | 117 | no(81) | 117 | no(81) | z2=1 | Lu2Pb1Pd2 |
| | | | | | | | 113 | no(81) | 113 | no(81) | z2=1 | |
| | | | | | | | 65 | no(2) | 65 | no(65) | z4=2 | |
| | | | | | | | 55 | no(2) | 55 | yes | z4=2 | |
| 127.3.15 | 99190 | 92 | - | - | - - | - - | 117 | no(81) | 117 | no(81) | z2=1 | La2Pb1Pd2 |
| | | | | | | | 113 | no(81) | 113 | no(81) | z2=1 | |
| | | | | | | | 65 | no(2) | 65 | no(65) | z4=2 | |
| | | | | | | | 55 | no(2) | 55 | yes | z4=2 | |
| 127.3.16 | 411081 | 88 | - | - | - - | - - | 117 | no(81) | 117 | no(81) | z2=1 | La2Mg1Ni2 |
| | | | | | | | 113 | no(81) | 113 | no(81) | z2=1 | |
| | | | | | | | 65 | no(2) | 65 | no(2) | z4=1 | |
| | | | | | | | 55 | no(2) | 55 | no(2) | z4=1 | |
| 127.3.17 | 107327 | 88 | - | - | - - | - - | 117 | no(81) | 117 | no(81) | z2=1 | La2Mg1Ni2 |
| | | | | | | | 113 | no(81) | 113 | no(81) | z2=1 | |
| | | | | | | | 65 | no(2) | 65 | no(2) | z4=1 | |
| | | | | | | | 55 | no(2) | 55 | no(2) | z4=1 | |
| 127.3.18 | 192643 | 220 | - | - | - - | - - | 117 | no(81) | 117 | no(81) | z2=1 | In2Nd1Ni9 |
| | | | | | | | 113 | no(81) | 113 | no(81) | z2=1 | |
| | | | | | | | 65 | yes | 65 | yes | - | |
| | | | | | | | 55 | yes | 55 | yes | - | |
| 127.3.19 | 189071 | 220 | - | - | - - | - - | 117 | no(81) | 117 | no(81) | z2=1 | In2Nd1Ni9 |
| | | | | | | | 113 | no(81) | 113 | no(81) | z2=1 | |
| | | | | | | | 65 | yes | 65 | yes | - | |
| | | | | | | | 55 | yes | 55 | yes | - | |
| 127.3.20 | 99197 | 132 | - | - | - - | - - | 117 | no(81) | 117 | no(81) | z2=1 | Ho2Pb1Pd2 |
| | | | | | | | 113 | no(81) | 113 | no(81) | z2=1 | |
| | | | | | | | 65 | no(2) | 65 | no(10) | z2w,3=1 | |
| | | | | | | | 55 | no(2) | 55 | no(10) | z2w,3=1 | |
| 127.3.21 | 107371 | 108 | - | - | - - | - - | 117 | no(81) | 117 | no(81) | z2=1 | Hg2Sr3 |
| | | | | | | | 113 | no(81) | 113 | no(81) | z2=1 | |
| | | | | | | | 65 | yes | 65 | yes | - | |

Continued on next page

**TABLE XIV– continued**

| TQC | ICSD | NoE | DG | IndG | SubSG cl (w/ T) | SubSG cl (w/o T) | Max. Sub. SG | yes or no | Max. Sub. SG | yes or no | SIs (w/ T) | Stoich. Formula |
|---|---|---|---|---|---|---|---|---|---|---|---|---|
| | | | | | | | 55 | yes | 55 | yes | - | |
| 127.3.22 | 107096 | 64 | - | - | - - | - - | 117 | yes | 117 | yes | - | Hf2Ni2Sn1 |
| | | | | | | | 113 | yes | 113 | yes | - | |
| | | | | | | | 65 | no(2) | 65 | no(10) | z2w,3=1<br>z4=2 | |
| | | | | | | | 55 | no(2) | 55 | no(10) | z2w,3=1<br>z4=2 | |
| 127.3.23 | 193225 | 120 | - | - | - - | - - | 117 | yes | 117 | yes | - | Ge2Sn1Yb2 |
| | | | | | | | 113 | yes | 113 | yes | - | |
| | | | | | | | 65 | no(2) | 65 | no(2) | z4=1 | |
| | | | | | | | 55 | no(2) | 55 | no(2) | z4=1 | |
| 127.3.24 | 423455 | 120 | - | - | - - | - - | 117 | yes | 117 | yes | - | Ge2Lu2Mg1 |
| | | | | | | | 113 | yes | 113 | yes | - | |
| | | | | | | | 65 | no(2) | 65 | no(2) | z4=1 | |
| | | | | | | | 55 | no(2) | 55 | no(2) | z4=1 | |
| 127.3.25 | 423453 | 104 | - | - | - - | - - | 117 | no(81) | 117 | no(81) | z2=1 | Ge2Ho2Mg1 |
| | | | | | | | 113 | no(81) | 113 | no(81) | z2=1 | |
| | | | | | | | 65 | no(2) | 65 | no(10) | z2w,3=1<br>z4=2 | |
| | | | | | | | 55 | no(2) | 55 | no(10) | z2w,3=1<br>z4=2 | |
| 127.3.26 | 99194 | 120 | - | - | - - | - - | 117 | no(81) | 117 | no(81) | z2=1 | Gd2Pb1Pd2 |
| | | | | | | | 113 | no(81) | 113 | no(81) | z2=1 | |
| | | | | | | | 65 | no(2) | 65 | no(10) | z2w,3=1 | |
| | | | | | | | 55 | no(2) | 55 | no(10) | z2w,3=1 | |
| 127.3.27 | 93403 | 92 | - | - | - - | - - | 117 | yes | 117 | yes | - | Gd2Ge2Mg1 |
| | | | | | | | 113 | yes | 113 | yes | - | |
| | | | | | | | 65 | no(2) | 65 | no(2) | z2w,3=1<br>z4=1 | |
| | | | | | | | 55 | no(2) | 55 | no(2) | z2w,3=1<br>z4=1 | |
| 127.3.28 | 103988 | 84 | - | - | - - | - - | 117 | no(81) | 117 | no(81) | z2=1 | Ga2Th3 |
| | | | | | | | 113 | no(81) | 113 | no(81) | z2=1 | |
| | | | | | | | 65 | no(2) | 65 | no(10) | z2w,3=1<br>z4=2 | |
| | | | | | | | 55 | no(2) | 55 | no(10) | z2w,3=1<br>z4=2 | |
| 127.3.29 | 260213 | 28 | - | - | - - | - - | 117 | no(81) | 117 | no(81) | z2=1 | Ga2Mg1Sc2 |
| | | | | | | | 113 | no(81) | 113 | no(81) | z2=1 | |
| | | | | | | | 65 | no(2) | 65 | no(10) | z2w,3=1 | |
| | | | | | | | 55 | no(2) | 55 | no(10) | z2w,3=1 | |
| 127.3.30 | 192650 | 236 | - | - | - - | - - | 117 | yes | 117 | yes | - | Er1In2Ni9 |
| | | | | | | | 113 | yes | 113 | yes | - | |
| | | | | | | | 65 | no(2) | 65 | no(2) | z2w,3=1<br>z4=3 | |
| | | | | | | | 55 | no(2) | 55 | no(2) | z2w,3=1<br>z4=3 | |
| 127.3.31 | 99196 | 128 | - | - | - - | - - | 117 | yes | 117 | yes | - | Dy2Pb1Pd2 |
| | | | | | | | 113 | yes | 113 | yes | - | |
| | | | | | | | 65 | no(2) | 65 | no(2) | z2w,3=1<br>z4=3 | |
| | | | | | | | 55 | no(2) | 55 | no(2) | z2w,3=1<br>z4=3 | |
| 127.3.32 | 427283 | 128 | - | - | - - | - - | 117 | yes | 117 | yes | - | Dy2Ni2Sn1 |
| | | | | | | | 113 | yes | 113 | yes | - | |
| | | | | | | | 65 | no(2) | 65 | no(2) | z2w,3=1<br>z4=3 | |

Continued on next page

**TABLE XIV– continued**

| TQC | ICSD | NoE | DG | IndG | SubSG cl (w/ T) | SubSG cl (w/o T) | Max. Sub. SG | yes or no | Max. Sub. SG | yes or no | SIs (w/ T) | Stoich. Formula |
|---|---|---|---|---|---|---|---|---|---|---|---|---|
| | | | | | | | 55 | no(2) | 55 | no(2) | z2w,3=1<br>z4=3 | |
| 127.3.33 | 419472 | 92 | - | - | - - | - - | 117 | no(81) | 117 | no(81) | z2=1 | Cu2Mg1Y2 |
| | | | | | | | 113 | no(81) | 113 | no(81) | z2=1 | |
| | | | | | | | 65 | yes | 65 | yes | - | |
| | | | | | | | 55 | yes | 55 | yes | - | |
| 127.3.34 | 411711 | 92 | - | - | - - | - - | 117 | no(81) | 117 | no(81) | z2=1 | Cu2Mg1Y2 |
| | | | | | | | 113 | no(81) | 113 | no(81) | z2=1 | |
| | | | | | | | 65 | yes | 65 | yes | - | |
| | | | | | | | 55 | yes | 55 | yes | - | |
| 127.3.35 | 411709 | 92 | - | - | - - | - - | 117 | no(81) | 117 | no(81) | z2=1 | Cu2La2Mg1 |
| | | | | | | | 113 | no(81) | 113 | no(81) | z2=1 | |
| | | | | | | | 65 | yes | 65 | yes | - | |
| | | | | | | | 55 | yes | 55 | yes | - | |
| 127.3.36 | 99191 | 96 | - | - | - - | - - | 117 | no(81) | 117 | no(81) | z2=1 | Ce2Pb1Pd2 |
| | | | | | | | 113 | no(81) | 113 | no(81) | z2=1 | |
| | | | | | | | 65 | no(2) | 65 | no(10) | z2w,3=1 | |
| | | | | | | | 55 | no(2) | 55 | no(10) | z2w,3=1 | |
| 127.3.37 | 411012 | 92 | - | - | - - | - - | 117 | yes | 117 | yes | - | Ce2Mg1Ni2 |
| | | | | | | | 113 | yes | 113 | yes | - | |
| | | | | | | | 65 | yes | 65 | yes | - | |
| | | | | | | | 55 | yes | 55 | yes | - | |
| 127.3.38 | 411013 | 96 | - | - | - - | - - | 117 | no(81) | 117 | no(81) | z2=1 | Ce2Cu2Mg1 |
| | | | | | | | 113 | no(81) | 113 | no(81) | z2=1 | |
| | | | | | | | 65 | no(2) | 65 | no(65) | z4=2 | |
| | | | | | | | 55 | no(2) | 55 | yes | z4=2 | |
| 127.3.39 | 107322 | 96 | - | - | - - | - - | 117 | no(81) | 117 | no(81) | z2=1 | Ce2Cu2Mg1 |
| | | | | | | | 113 | no(81) | 113 | no(81) | z2=1 | |
| | | | | | | | 65 | no(2) | 65 | no(65) | z4=2 | |
| | | | | | | | 55 | no(2) | 55 | yes | z4=2 | |
| 127.3.40 | 425067 | 272 | - | - | - - | - - | 117 | yes | 117 | yes | - | Cd2Cu9Pr1 |
| | | | | | | | 113 | yes | 113 | yes | - | |
| | | | | | | | 65 | no(2) | 65 | no(2) | z2w,3=1<br>z4=3 | |
| | | | | | | | 55 | no(2) | 55 | no(2) | z2w,3=1<br>z4=3 | |
| 127.3.41 | 424171 | 96 | - | - | - - | - - | 117 | no(81) | 117 | no(81) | z2=1 | Cd1Ge2Nd2 |
| | | | | | | | 113 | no(81) | 113 | no(81) | z2=1 | |
| | | | | | | | 65 | no(2) | 65 | no(10) | z2w,3=1 | |
| | | | | | | | 55 | no(2) | 55 | no(10) | z2w,3=1 | |
| 127.3.42 | 157387 | 136 | - | - | - - | - - | 117 | no(81) | 117 | no(81) | z2=1 | Cd1Gd2Pd2 |
| | | | | | | | 113 | no(81) | 113 | no(81) | z2=1 | |
| | | | | | | | 65 | no(2) | 65 | no(2) | z2w,3=1<br>z4=1 | |
| | | | | | | | 55 | no(2) | 55 | no(2) | z2w,3=1<br>z4=1 | |
| 127.3.43 | 424173 | 112 | - | - | - - | - - | 117 | no(81) | 117 | no(81) | z2=1 | Cd1Gd2Ge2 |
| | | | | | | | 113 | no(81) | 113 | no(81) | z2=1 | |
| | | | | | | | 65 | no(2) | 65 | no(10) | z2w,3=1 | |
| | | | | | | | 55 | no(2) | 55 | no(10) | z2w,3=1 | |
| 127.3.44 | 157388 | 144 | - | - | - - | - - | 117 | no(81) | 117 | no(81) | z2=1 | Cd1Dy2Pd2 |
| | | | | | | | 113 | no(81) | 113 | no(81) | z2=1 | |
| | | | | | | | 65 | no(2) | 65 | no(10) | z2w,3=1<br>z4=2 | |
| | | | | | | | 55 | no(2) | 55 | no(10) | z2w,3=1<br>z4=2 | |
| 127.3.45 | 419851 | 140 | - | - | - - | - - | 117 | no(81) | 117 | no(81) | z2=1 | Cd1Cu2Gd2 |
| | | | | | | | 113 | no(81) | 113 | no(81) | z2=1 | |

Continued on next page

**TABLE XIV– continued**

| TQC | ICSD | NoE | DG | IndG | SubSG cl (w/ T) | SubSG cl (w/o T) | Max. Sub. SG | yes or no | Max. Sub. SG | yes or no | SIs (w/ T) | Stoich. Formula |
|---|---|---|---|---|---|---|---|---|---|---|---|---|
| | | | | | | | 65 | no(2) | 65 | no(2) | z4=3 | |
| | | | | | | | 55 | no(2) | 55 | no(2) | z4=3 | |
| 127.3.46 | 411009 | 112 | - | - | - - | - - | 117 | yes | 117 | yes | - | Cd1Ce2Pt2 |
| | | | | | | | 113 | yes | 113 | yes | - | |
| | | | | | | | 65 | no(2) | 65 | no(2) | z2w,3=1<br>z4=1 | |
| | | | | | | | 55 | no(2) | 55 | no(2) | z2w,3=1<br>z4=1 | |
| 127.3.47 | 411010 | 112 | - | - | - - | - - | 117 | yes | 117 | yes | - | Cd1Ce2Pd2 |
| | | | | | | | 113 | yes | 113 | yes | - | |
| | | | | | | | 65 | no(2) | 65 | no(2) | z2w,3=1<br>z4=1 | |
| | | | | | | | 55 | no(2) | 55 | no(2) | z2w,3=1<br>z4=1 | |
| 127.3.48 | 195139 | 160 | - | - | - - | - - | 117 | yes | 117 | yes | - | C4Al3Y5 |
| | | | | | | | 113 | yes | 113 | yes | - | |
| | | | | | | | 65 | no(2) | 65 | no(2) | z2w,3=1<br>z4=1 | |
| | | | | | | | 55 | no(2) | 55 | no(2) | z2w,3=1<br>z4=1 | |
| 127.3.49 | 67376 | 168 | - | - | - - | - - | 117 | no(81) | 117 | no(81) | z2=1 | C3La2Ni5 |
| | | | | | | | 113 | no(81) | 113 | no(81) | z2=1 | |
| | | | | | | | 65 | no(2) | 65 | no(2) | z2w,3=1<br>z4=3 | |
| | | | | | | | 55 | no(2) | 55 | no(2) | z2w,3=1<br>z4=3 | |
| 127.3.50 | 62277 | 168 | - | - | - - | - - | 117 | no(81) | 117 | no(81) | z2=1 | C3La2Ni5 |
| | | | | | | | 113 | no(81) | 113 | no(81) | z2=1 | |
| | | | | | | | 65 | no(2) | 65 | no(2) | z2w,3=1<br>z4=3 | |
| | | | | | | | 55 | no(2) | 55 | no(2) | z2w,3=1<br>z4=3 | |
| 127.3.51 | 300285 | 100 | - | - | - - | - - | 117 | yes | 117 | yes | - | Br3Cs1Dy1 |
| | | | | | | | 113 | yes | 113 | yes | - | |
| | | | | | | | 65 | no(2) | 65 | no(2) | z4=3 | |
| | | | | | | | 55 | no(2) | 55 | no(2) | z4=3 | |
| 127.3.52 | 202172 | 152 | - | - | - - | - - | 117 | no(81) | 117 | no(81) | z2=1 | Ba1O5Pd1Pr2 |
| | | | | | | | 113 | no(81) | 113 | no(81) | z2=1 | |
| | | | | | | | 65 | no(2) | 65 | no(10) | z2w,3=1 | |
| | | | | | | | 55 | no(2) | 55 | no(10) | z2w,3=1 | |
| 127.3.53 | 62609 | 156 | - | - | - - | - - | 117 | no(81) | 117 | no(81) | z2=1 | Ba1Nd2O5Pd1 |
| | | | | | | | 113 | no(81) | 113 | no(81) | z2=1 | |
| | | | | | | | 65 | yes | 65 | yes | - | |
| | | | | | | | 55 | yes | 55 | yes | - | |
| 127.3.54 | 81545 | 96 | - | - | - - | - - | 117 | yes | 117 | yes | - | B4Th1 |
| | | | | | | | 113 | yes | 113 | yes | - | |
| | | | | | | | 65 | no(2) | 65 | no(2) | z4=1 | |
| | | | | | | | 55 | no(2) | 55 | no(2) | z4=1 | |
| 127.3.55 | 68649 | 124 | - | - | - - | - - | 117 | yes | 117 | yes | - | B4Tb1 |
| | | | | | | | 113 | yes | 113 | yes | - | |
| | | | | | | | 65 | no(2) | 65 | no(10) | z2w,3=1 | |
| | | | | | | | 55 | no(2) | 55 | no(10) | z2w,3=1 | |
| 127.3.56 | 68650 | 136 | - | - | - - | - - | 117 | no(81) | 117 | no(81) | z2=1 | B4Er1 |
| | | | | | | | 113 | no(81) | 113 | no(81) | z2=1 | |
| | | | | | | | 65 | no(2) | 65 | no(2) | z4=3 | |
| | | | | | | | 55 | no(2) | 55 | no(2) | z4=3 | |
| 127.3.57 | 44214 | 136 | - | - | - - | - - | 117 | no(81) | 117 | no(81) | z2=1 | B4Er1 |
| | | | | | | | 113 | no(81) | 113 | no(81) | z2=1 | |

Continued on next page

**TABLE XIV– continued**

| TQC | ICSD | NoE | DG | IndG | SubSG (w/ T) | cl | SubSG (w/o T) | cl | Max. Sub. SG | yes or no | Max. Sub. SG | yes or no | SIs (w/ T) | Stoich. Formula |
|---|---|---|---|---|---|---|---|---|---|---|---|---|---|---|
| | | | | | | | | | 65 | no(2) | 65 | no(2) | z4=3 | |
| | | | | | | | | | 55 | no(2) | 55 | no(2) | z4=3 | |
| 127.3.58 | 417745 | 96 | - | - | - | - | - | - | 117 | yes | 117 | yes | - | B4Ce1 |
| | | | | | | | | | 113 | yes | 113 | yes | - | |
| | | | | | | | | | 65 | no(2) | 65 | no(2) | z4=1 | |
| | | | | | | | | | 55 | no(2) | 55 | no(2) | z4=1 | |
| 127.3.59 | 51438 | 120 | - | - | - | - | - | - | 117 | no(81) | 117 | no(81) | z2=1 | B2Ga1Rh5Sc2 |
| | | | | | | | | | 113 | no(81) | 113 | no(81) | z2=1 | |
| | | | | | | | | | 65 | no(2) | 65 | no(65) | z4=2 | |
| | | | | | | | | | 55 | no(2) | 55 | yes | z4=2 | |
| 127.3.60 | 409531 | 60 | - | - | - | - | - | - | 117 | yes | 117 | yes | - | Au2Ca2Pb1 |
| | | | | | | | | | 113 | yes | 113 | yes | - | |
| | | | | | | | | | 65 | no(2) | 65 | no(2) | z4=1 | |
| | | | | | | | | | 55 | no(2) | 55 | no(2) | z4=1 | |
| 127.4.1 | 40962 | 122 | - | - | - | - | - | - | See details in Table XIX | | | | | Mo4Na1O6 |
| 127.4.2 | 420085 | 54 | - | - | - | - | - | - | See details in Table XIX | | | | | Li1Rh2Si2 |
| 127.4.3 | 159116 | 218 | - | - | - | - | - | - | See details in Table XIX | | | | | La7Ni2Zn1 |
| 127.4.4 | 68533 | 138 | - | - | - | - | - | - | See details in Table XIX | | | | | K1Mo4O6 |
| 127.4.5 | 401730 | 46 | - | - | - | - | - | - | See details in Table XIX | | | | | In5Ti2 |
| 127.4.6 | 192651 | 214 | - | - | - | - | - | - | See details in Table XIX | | | | | In2Ni9Y1 |
| 127.4.7 | 189069 | 218 | - | - | - | - | - | - | See details in Table XIX | | | | | In2Ni9Pr1 |
| 127.4.8 | 192640 | 214 | - | - | - | - | - | - | See details in Table XIX | | | | | In2La1Ni9 |
| 127.4.9 | 54484 | 122 | - | - | - | - | - | - | See details in Table XIX | | | | | In1Pd2Tb2 |
| 127.4.10 | 54480 | 98 | - | - | - | - | - | - | See details in Table XIX | | | | | In1Pd2Pr2 |
| 127.4.11 | 107372 | 62 | - | - | - | - | - | - | See details in Table XIX | | | | | In1Ni2Zr2 |
| 127.4.12 | 107333 | 58 | - | - | - | - | - | - | See details in Table XIX | | | | | In1Ni2Sc2 |
| 127.4.13 | 54490 | 146 | - | - | - | - | - | - | See details in Table XIX | | | | | In1Lu2Pd2 |
| 127.4.14 | 404531 | 46 | - | - | - | - | - | - | See details in Table XIX | | | | | Hf2In5 |
| 127.4.15 | 402570 | 118 | - | - | - | - | - | - | See details in Table XIX | | | | | Ge2In1Yb2 |
| 127.4.16 | 172494 | 118 | - | - | - | - | - | - | See details in Table XIX | | | | | Ge2In1Yb2 |
| 127.4.17 | 172491 | 98 | - | - | - | - | - | - | See details in Table XIX | | | | | Ge2In1Tb2 |
| 127.4.18 | 414072 | 74 | - | - | - | - | - | - | See details in Table XIX | | | | | Ge2In1Pr2 |
| 127.4.19 | 414071 | 78 | - | - | - | - | - | - | See details in Table XIX | | | | | Ge2In1Nd2 |
| 127.4.20 | 87511 | 66 | - | - | - | - | - | - | See details in Table XIX | | | | | Ge2In1La2 |
| 127.4.21 | 172493 | 106 | - | - | - | - | - | - | See details in Table XIX | | | | | Ge2Ho2In1 |
| 127.4.22 | 249632 | 58 | - | - | - | - | - | - | See details in Table XIX | | | | | Ga5Mn2 |
| 127.4.23 | 85499 | 90 | - | - | - | - | - | - | See details in Table XIX | | | | | F4Fe1Rb1 |
| 127.4.24 | 54487 | 134 | - | - | - | - | - | - | See details in Table XIX | | | | | Er2In1Pd2 |
| 127.4.25 | 107343 | 126 | - | - | - | - | - | - | See details in Table XIX | | | | | Dy2In1Pd2 |
| 127.4.26 | 411708 | 94 | - | - | - | - | - | - | See details in Table XIX | | | | | Cu2In1La2 |
| 127.4.27 | 21111 | 82 | - | - | - | - | - | - | See details in Table XIX | | | | | Cu1F3K1 |
| 127.4.28 | 21110 | 82 | - | - | - | - | - | - | See details in Table XIX | | | | | Cu1F3K1 |
| 127.4.29 | 404847 | 58 | - | - | - | - | - | - | See details in Table XIX | | | | | Co2In1Zr2 |
| 127.4.30 | 107331 | 58 | - | - | - | - | - | - | See details in Table XIX | | | | | Co2In1Zr2 |
| 127.4.31 | 410293 | 94 | - | - | - | - | - | - | See details in Table XIX | | | | | Ce2In1Pt2 |
| 127.4.32 | 150615 | 94 | - | - | - | - | - | - | See details in Table XIX | | | | | Ce2In1Ni2 |
| 127.4.33 | 415108 | 70 | - | - | - | - | - | - | See details in Table XIX | | | | | Ce2Ge2In1 |
| 127.4.34 | 424134 | 250 | - | - | - | - | - | - | See details in Table XIX | | | | | Ca1Cd2Cu9 |
| 127.4.35 | 427155 | 50 | - | - | - | - | - | - | See details in Table XIX | | | | | C2B2Y1 |
| 127.4.36 | 94043 | 74 | - | - | - | - | - | - | See details in Table XIX | | | | | C2B2Tm1 |
| 127.4.37 | 94039 | 66 | - | - | - | - | - | - | See details in Table XIX | | | | | C2B2Tb1 |
| 127.4.38 | 94037 | 54 | - | - | - | - | - | - | See details in Table XIX | | | | | C2B2Pr1 |
| 127.4.39 | 94044 | 78 | - | - | - | - | - | - | See details in Table XIX | | | | | C2B2Lu1 |
| 127.4.40 | 94035 | 50 | - | - | - | - | - | - | See details in Table XIX | | | | | C2B2La1 |
| 127.4.41 | 63497 | 158 | - | - | - | - | - | - | See details in Table XIX | | | | | Ba1Cu1Nd2O5 |
| 127.4.42 | 88317 | 42 | - | - | - | - | - | - | See details in Table XIX | | | | | B2V3 |
| 127.4.43 | 427424 | 122 | - | - | - | - | - | - | See details in Table XIX | | | | | B2Ru5Ta3 |

Continued on next page

**TABLE XIV– continued**

| TQC | ICSD | NoE | DG | IndG | SubSG (w/ T) | cl | SubSG (w/o T) | cl | Max. Sub. SG | yes or no | Max. Sub. SG | yes or no | SIs (w/ T) | Stoich. Formula |
|---|---|---|---|---|---|---|---|---|---|---|---|---|---|---|
| 127.4.44 | 423469 | 170 | - | - | - | - | - | - | See details in Table XIX | | | | | B2Nb3Ru5 |
| 127.4.45 | 69487 | 118 | - | - | - | - | - | - | See details in Table XIX | | | | | B2Ir5Mg2Si1 |
| 127.4.46 | 44316 | 126 | - | - | - | - | - | - | See details in Table XIX | | | | | B2Hf3Ir5 |
| 127.4.47 | 51432 | 118 | - | - | - | - | - | - | See details in Table XIX | | | | | B2Ge1Mg2Rh5 |
| 127.4.48 | 51437 | 130 | - | - | - | - | - | - | See details in Table XIX | | | | | B2Fe1Rh5Sc2 |
| 127.4.49 | 51435 | 126 | - | - | - | - | - | - | See details in Table XIX | | | | | B2Cr1Rh5Sc2 |
| 127.4.50 | 262043 | 110 | - | - | - | - | - | - | See details in Table XIX | | | | | Au2Y3 |
| 127.4.51 | 92455 | 118 | - | - | - | - | - | - | See details in Table XIX | | | | | Al1Si2Yb2 |
| 127.4.52 | 172061 | 102 | - | - | - | - | - | - | See details in Table XIX | | | | | Al1Dy2Ge2 |
| 128.1.1 | 425655 | 160 | 0.0041 | -0.7028 | 2 | 1 | 83 | 1 | - | - | - | - | z4=2<br>z8=2 | B2Nb2Os1 |
| 128.3.1 | 38143 | 120 | - | - | - | - | - | - | 118 | no(81) | 118 | no(81) | z2=1 | Cr1I6Tl4 |
| | | | | | | | | | 114 | no(81) | 114 | no(81) | z2=1 | |
| | | | | | | | | | 66 | yes | 66 | yes | - | |
| | | | | | | | | | 58 | yes | 58 | yes | - | |
| 128.3.2 | 429117 | 96 | - | - | - | - | - | - | 118 | no(81) | 118 | no(81) | z2=1 | B2Os1Ta2 |
| | | | | | | | | | 114 | no(81) | 114 | no(81) | z2=1 | |
| | | | | | | | | | 66 | no(2) | 66 | yes | z4=2 | |
| | | | | | | | | | 58 | no(2) | 58 | yes | z4=2 | |
| 128.4.1 | 245749 | 146 | - | - | - | - | - | - | See details in Table XIX | | | | | Cl6K2Nb1 |
| 129.1.1 | 80190 | 28 | 0.0593 | -0.1695 | 2 | 1 | 0 | 1 | - | - | - | - | z2w,3=1<br>z4=2 | Sn1Te1Zr1 |
| 129.1.2 | 26273 | 48 | 0.0060 | -0.0476 | 2 | 1 | 0 | 1 | - | - | - | - | z2w,3=1<br>z4=2 | Sn1Te1U1 |
| 129.1.3 | 74522 | 28 | 0.0669 | -0.3439 | 2 | 2 | 0 | 1 | - | - | - | - | z2w,3=1 | Si1Te1Zr1 |
| 129.1.4 | 25734 | 28 | 0.0577 | -0.2951 | 2 | 1 | 0 | 1 | - | - | - | - | z2w,3=1<br>z4=2 | Si1Te1Zr1 |
| 129.1.5 | 15572 | 28 | 0.0667 | -0.3134 | 2 | 1 | 0 | 1 | - | - | - | - | z2w,3=1<br>z4=2 | Si1Te1Zr1 |
| 129.1.6 | 25733 | 28 | 0.0176 | -0.3211 | 2 | 1 | 0 | 1 | - | - | - | - | z2w,3=1<br>z4=2 | Se1Si1Zr1 |
| 129.1.7 | 15571 | 28 | 0.0183 | -0.2794 | 2 | 1 | 0 | 1 | - | - | - | - | z2w,3=1<br>z4=2 | Se1Si1Zr1 |
| 129.1.8 | 25732 | 28 | 0.0258 | -0.2948 | 2 | 1 | 0 | 1 | - | - | - | - | z2w,3=1<br>z4=2 | S1Si1Zr1 |
| 129.1.9 | 15570 | 28 | 0.0265 | -0.2912 | 2 | 1 | 0 | 1 | - | - | - | - | z2w,3=1<br>z4=2 | S1Si1Zr1 |
| 129.1.10 | 15569 | 28 | 0.0248 | -0.2990 | 2 | 1 | 0 | 1 | - | - | - | - | z2w,3=1<br>z4=2 | S1Si1Zr1 |
| 129.1.11 | 26271 | 48 | 0.0224 | -0.3765 | 2 | 1 | 0 | 1 | - | - | - | - | z2w,3=1<br>z4=2 | S1Si1U1 |
| 129.1.12 | 50351 | 64 | 0.0138 | -0.1175 | 2 | 3 | 0 | 1 | - | - | - | - | z4=2 | Ru1Sb2U1 |
| 129.1.13 | 50350 | 64 | 0.0132 | -0.1142 | 2 | 3 | 0 | 1 | - | - | - | - | z4=2 | Ru1Sb2U1 |
| 129.1.14 | 61194 | 80 | 0.0119 | -2.3762 | 2 | 3 | 0 | 1 | - | - | - | - | z4=2 | Pd2Sb2Sr1 |
| 129.1.15 | 50349 | 68 | 0.0157 | -0.1091 | 2 | 3 | 0 | 1 | - | - | - | - | z4=2 | Pd1Sb2U1 |
| 129.1.16 | 50348 | 68 | 0.0087 | -0.1136 | 2 | 3 | 0 | 1 | - | - | - | - | z4=2 | Pd1Sb2U1 |
| 129.1.17 | 77853 | 48 | 0.0022 | -0.1731 | 2 | 1 | 0 | 1 | - | - | - | - | z2w,3=1<br>z4=2 | P2U1 |
| 129.1.18 | 25731 | 28 | 0.0591 | -0.1523 | 2 | 1 | 0 | 1 | - | - | - | - | z2w,3=1<br>z4=2 | O1Si1Zr1 |
| 129.1.19 | 85856 | 52 | 0.0054 | -0.0448 | 2 | 1 | 0 | 1 | - | - | - | - | z2w,3=1<br>z4=2 | Nd1Ru1Si1 |
| 129.1.20 | 41250 | 52 | 0.0061 | -0.0441 | 2 | 1 | 0 | 1 | - | - | - | - | z2w,3=1<br>z4=2 | Nd1Ru1Si1 |
| 129.1.21 | 76235 | 44 | 0.0073 | -0.6173 | 2 | 1 | 0 | 1 | - | - | - | - | z2w,3=1<br>z4=2 | Mn1Si1Y1 |
| 129.1.22 | 73039 | 84 | 0.0350 | -0.5510 | 2 | 2 | 0 | 1 | - | - | - | - | z2w,3=1 | Ir2Pu1Si2 |

Continued on next page

**TABLE XIV– continued**

| TQC | ICSD | NoE | DG | IndG | SubSG (w/ T) | cl | SubSG (w/o T) | cl | Max. Sub. SG | yes or no | Max. Sub. SG | yes or no | SIs (w/ T) | Stoich. Formula |
|---|---|---|---|---|---|---|---|---|---|---|---|---|---|---|
| 129.1.23 | 15420 | 72 | 0.0025 | -0.0856 | 2 | 1 | 0 | 1 | - | - | - | - | z2w,3=1<br>z4=2 | I1O1Tm1 |
| 129.1.24 | 25737 | 28 | 0.0559 | -0.4980 | 2 | 1 | 0 | 1 | - | - | - | - | z2w,3=1<br>z4=2 | Hf1Si1Te1 |
| 129.1.25 | 25736 | 28 | 0.0409 | -0.5615 | 2 | 1 | 0 | 1 | - | - | - | - | z2w,3=1<br>z4=2 | Hf1Se1Si1 |
| 129.1.26 | 25735 | 28 | 0.1108 | -0.4743 | 2 | 1 | 0 | 1 | - | - | - | - | z2w,3=1<br>z4=2 | Hf1S1Si1 |
| 129.1.27 | 25740 | 28 | 0.0720 | -0.3194 | 2 | 1 | 0 | 1 | - | - | - | - | z2w,3=1<br>z4=2 | Ge1Te1Zr1 |
| 129.1.28 | 25739 | 28 | 0.0486 | -0.6300 | 2 | 1 | 0 | 1 | - | - | - | - | z2w,3=1<br>z4=2 | Ge1Se1Zr1 |
| 129.1.29 | 15574 | 28 | 0.1079 | -0.4603 | 2 | 1 | 0 | 1 | - | - | - | - | z2w,3=1<br>z4=2 | Ge1Se1Zr1 |
| 129.1.30 | 25738 | 28 | 0.0419 | -0.5955 | 2 | 1 | 0 | 1 | - | - | - | - | z2w,3=1<br>z4=2 | Ge1S1Zr1 |
| 129.1.31 | 15573 | 28 | 0.0631 | -0.5805 | 2 | 1 | 0 | 1 | - | - | - | - | z2w,3=1<br>z4=2 | Ge1S1Zr1 |
| 129.1.32 | 87315 | 48 | 0.0160 | -0.0613 | 2 | 1 | 0 | 1 | - | - | - | - | z2w,3=1<br>z4=2 | Ge1S1U1 |
| 129.1.33 | 26272 | 48 | 0.0154 | -0.2681 | 2 | 1 | 0 | 1 | - | - | - | - | z2w,3=1<br>z4=2 | Ge1S1U1 |
| 129.1.34 | 85863 | 52 | 0.0053 | -0.0860 | 2 | 1 | 0 | 1 | - | - | - | - | z2w,3=1<br>z4=2 | Ge1Nd1Ru1 |
| 129.1.35 | 25743 | 28 | 0.0173 | -0.3431 | 2 | 1 | 0 | 1 | - | - | - | - | z2w,3=1<br>z4=2 | Ge1Hf1Te1 |
| 129.1.36 | 25741 | 28 | 0.0801 | -0.6896 | 2 | 1 | 0 | 1 | - | - | - | - | z2w,3=1<br>z4=2 | Ge1Hf1Se1 |
| 129.1.37 | 25742 | 28 | 0.1136 | -0.6053 | 2 | 1 | 0 | 1 | - | - | - | - | z2w,3=1<br>z4=2 | Ge1Hf1S1 |
| 129.1.38 | 106758 | 92 | 0.0000 | -0.4470 | 2 | 1 | 0 | 1 | - | - | - | - | z2w,3=1<br>z4=2 | Gd1Pt2Si2 |
| 129.1.39 | 180602 | 28 | 0.0214 | -0.4332 | 2 | 1 | 0 | 1 | - | - | - | - | z2w,3=1<br>z4=2 | Fe1Te1 |
| 129.1.40 | 169271 | 28 | 0.0204 | -0.6103 | 2 | 1 | 0 | 1 | - | - | - | - | z2w,3=1<br>z4=2 | Fe1Se1 |
| 129.1.41 | 169269 | 28 | 0.0438 | -0.5843 | 2 | 1 | 0 | 1 | - | - | - | - | z2w,3=1<br>z4=2 | Fe1Se1 |
| 129.1.42 | 169267 | 28 | 0.0388 | -0.5658 | 2 | 1 | 0 | 1 | - | - | - | - | z2w,3=1<br>z4=2 | Fe1Se1 |
| 129.1.43 | 169265 | 28 | 0.0288 | -0.5410 | 2 | 1 | 0 | 1 | - | - | - | - | z2w,3=1<br>z4=2 | Fe1Se1 |
| 129.1.44 | 169263 | 28 | 0.0262 | -0.5199 | 2 | 1 | 0 | 1 | - | - | - | - | z2w,3=1<br>z4=2 | Fe1Se1 |
| 129.1.45 | 169261 | 28 | 0.0283 | -0.5149 | 2 | 1 | 0 | 1 | - | - | - | - | z2w,3=1<br>z4=2 | Fe1Se1 |
| 129.1.46 | 169259 | 28 | 0.0331 | -0.4935 | 2 | 1 | 0 | 1 | - | - | - | - | z2w,3=1<br>z4=2 | Fe1Se1 |
| 129.1.47 | 169257 | 28 | 0.0300 | -0.4870 | 2 | 1 | 0 | 1 | - | - | - | - | z2w,3=1<br>z4=2 | Fe1Se1 |
| 129.1.48 | 169255 | 28 | 0.0238 | -0.4704 | 2 | 1 | 0 | 1 | - | - | - | - | z2w,3=1<br>z4=2 | Fe1Se1 |
| 129.1.49 | 169253 | 28 | 0.0178 | -0.4566 | 2 | 1 | 0 | 1 | - | - | - | - | z2w,3=1<br>z4=2 | Fe1Se1 |
| 129.1.50 | 94475 | 68 | 0.0060 | -0.0249 | 2 | 2 | 0 | 1 | - | - | - | - | z2w,3=1 | F1Ho1S1 |
| 129.1.51 | 42969 | 68 | 0.0046 | -0.1732 | 2 | 3 | 0 | 1 | - | - | - | - | z4=2 | Er1Se2 |
| 129.1.52 | 59594 | 48 | 0.0204 | -0.5110 | 2 | 1 | 0 | 1 | - | - | - | - | z2w,3=1<br>z4=2 | Cu1P1Si1Zr1 |

Continued on next page

**TABLE XIV– continued**

| TQC | ICSD | NoE | DG | IndG | SubSG (w/ T) | cl | SubSG (w/o T) | cl | Max. Sub. SG | yes or no | Max. Sub. SG | yes or no | SIs (w/ T) | Stoich. Formula |
|---|---|---|---|---|---|---|---|---|---|---|---|---|---|---|
| 129.1.53 | 91199 | 124 | 0.0132 | -0.6273 | 2 | 2 | 0 | 1 | - | - | - | - | z2w,3=1 | Cu1Mn1O3S1Sr2 |
| 129.1.54 | 27230 | 164 | 0.0075 | -0.1410 | 2 | 1 | 0 | 1 | - | - | - | - | z2w,3=1 z4=2 | Cs1F4Ti1 |
| 129.1.55 | 62696 | 180 | 0.0000 | -0.1052 | 2 | 1 | 0 | 1 | - | - | - | - | z2w,3=1 z4=2 | Cs1F4Fe1 |
| 129.1.56 | 6043 | 180 | 0.0000 | -0.0924 | 2 | 1 | 0 | 1 | - | - | - | - | z2w,3=1 z4=2 | Cs1F4Fe1 |
| 129.1.57 | 16496 | 48 | 0.0092 | -1.2675 | 2 | 2 | 0 | 1 | - | - | - | - | z2w,3=1 | Cr1N1Nb1 |
| 129.1.58 | 78578 | 84 | 0.0000 | -0.4381 | 2 | 1 | 0 | 1 | - | - | - | - | z2w,3=1 z4=2 | Co2P2U1 |
| 129.1.59 | 15770 | 60 | 0.0164 | -0.2553 | 2 | 1 | 0 | 1 | - | - | - | - | z2w,3=1 z4=2 | Co1K1O2 |
| 129.1.60 | 97019 | 104 | 0.0000 | -0.6775 | 2 | 1 | 0 | 1 | - | - | - | - | z2w,3=1 z4=2 | Cl1Mn1O3Sr2 |
| 129.1.61 | 97018 | 104 | 0.0022 | -0.6809 | 2 | 1 | 0 | 1 | - | - | - | - | z2w,3=1 z4=2 | Cl1Mn1O3Sr2 |
| 129.1.62 | 96044 | 104 | 0.0000 | -0.6208 | 2 | 1 | 0 | 1 | - | - | - | - | z2w,3=1 z4=2 | Cl1Mn1O3Sr2 |
| 129.1.63 | 94745 | 104 | 0.0000 | -0.6208 | 2 | 1 | 0 | 1 | - | - | - | - | z2w,3=1 z4=2 | Cl1Mn1O3Sr2 |
| 129.1.64 | 76171 | 68 | 0.0031 | -0.0530 | 2 | 1 | 0 | 1 | - | - | - | - | z2w,3=1 z4=2 | Cl1Ho1O1 |
| 129.1.65 | 91750 | 108 | 0.0357 | -0.2097 | 2 | 1 | 0 | 1 | - | - | - | - | z2w,3=1 z4=2 | Cl1Co1O3Sr2 |
| 129.1.66 | 90122 | 108 | 0.0128 | -0.2583 | 2 | 1 | 0 | 1 | - | - | - | - | z2w,3=1 z4=2 | Cl1Co1O3Sr2 |
| 129.1.67 | 84529 | 64 | 0.0000 | -0.3103 | 2 | 3 | 0 | 1 | - | - | - | - | z4=2 | Ce1Ni1Sb2 |
| 129.1.68 | 411678 | 76 | 0.0027 | -1.2565 | 2 | 1 | 0 | 1 | - | - | - | - | z2w,3=1 z4=2 | Ce1Ir2Si2 |
| 129.1.69 | 61371 | 80 | 0.0099 | -0.6430 | 2 | 1 | 0 | 1 | - | - | - | - | z2w,3=1 z4=2 | Ce1Ge2Pt2 |
| 129.1.70 | 106393 | 76 | 0.0020 | -1.0681 | 2 | 2 | 0 | 1 | - | - | - | - | z2w,3=1 | Ce1Ge2Ir2 |
| 129.1.71 | 191053 | 76 | 0.0050 | -0.6673 | 2 | 3 | 0 | 1 | - | - | - | - | z4=2 | Ce1Ga2Pd2 |
| 129.1.72 | 92319 | 180 | 0.0000 | -0.0353 | 2 | 1 | 0 | 1 | - | - | - | - | z2w,3=1 z4=2 | Br1Nd1O3Te1 |
| 129.1.73 | 92318 | 144 | 0.0000 | -0.2278 | 2 | 2 | 0 | 1 | - | - | - | - | z2w,3=1 | Bi1I1O3Te1 |
| 129.1.74 | 246945 | 64 | 0.0451 | -0.1110 | 2 | 2 | 0 | 1 | - | - | - | - | z2w,3=1 | Ba1Ni1S2 |
| 129.1.75 | 15289 | 64 | 0.0298 | -0.1170 | 2 | 2 | 0 | 1 | - | - | - | - | z2w,3=1 | Ba1Ni1S2 |
| 129.1.76 | 41915 | 88 | 0.0178 | -0.5449 | 2 | 2 | 0 | 1 | - | - | - | - | z2w,3=1 | Ba1Mg2Sn2 |
| 129.1.77 | 25338 | 44 | 0.0051 | -1.0140 | 2 | 2 | 0 | 1 | - | - | - | - | z2w,3=1 | Ba1Mg2Pb2 |
| 129.1.78 | 370004 | 76 | 0.0865 | -1.9549 | 2 | 3 | 0 | 1 | - | - | - | - | z4=2 | Au2Ga2Sr1 |
| 129.1.79 | 238706 | 68 | 0.0148 | -0.0331 | 2 | 2 | 0 | 1 | - | - | - | - | z2w,3=1 | Au1Bi2Pr1 |
| 129.1.80 | 238707 | 68 | 0.0133 | -0.0350 | 2 | 2 | 0 | 1 | - | - | - | - | z2w,3=1 | Au1Bi2Nd1 |
| 129.1.81 | 66765 | 48 | 0.0062 | -0.1017 | 2 | 1 | 0 | 1 | - | - | - | - | z2w,3=1 z4=2 | As2U1 |
| 129.1.82 | 44081 | 48 | 0.0101 | -0.0968 | 2 | 1 | 0 | 1 | - | - | - | - | z2w,3=1 z4=2 | As2U1 |
| 129.1.83 | 107523 | 48 | 0.0049 | -0.0990 | 2 | 1 | 0 | 1 | - | - | - | - | z2w,3=1 z4=2 | As2U1 |
| 129.1.84 | 56976 | 68 | 0.0033 | -0.1415 | 2 | 3 | 0 | 1 | - | - | - | - | z4=2 | As2Pd1U1 |
| 129.1.85 | 412554 | 88 | 0.0096 | -0.2402 | 2 | 3 | 0 | 1 | - | - | - | - | z4=2 | As2Cu1Tm1 |
| 129.1.86 | 412551 | 80 | 0.0000 | -0.0187 | 2 | 3 | 0 | 1 | - | - | - | - | z4=2 | As2Cu1Tb1 |
| 129.1.87 | 246061 | 64 | 0.0153 | -2.2429 | 2 | 2 | 0 | 1 | - | - | - | - | z2w,3=1 | As1La1Ni1O1 |
| 129.1.88 | 248695 | 112 | 0.0048 | -0.3746 | 2 | 2 | 0 | 1 | - | - | - | - | z2w,3=1 | As1Fe1O3Sr2V1 |
| 129.1.89 | 422004 | 76 | 0.0000 | -0.0655 | 2 | 1 | 0 | 1 | - | - | - | - | z2w,3=1 z4=2 | As1Fe1O1Tb1 |

Continued on next page

**TABLE XIV– continued**

| TQC | ICSD | NoE | DG | IndG | SubSG (w/ T) | cl | SubSG (w/o T) | cl | Max. Sub. SG | yes or no | Max. Sub. SG | yes or no | SIs (w/ T) | Stoich. Formula |
|---|---|---|---|---|---|---|---|---|---|---|---|---|---|---|
| 129.1.90 | 42165 | 48 | 0.0358 | -0.4780 | 2 | 1 | 0 | 1 | - | - | - | - | z2w,3=1<br>z4=2 | As1Cu1Si1Zr1 |
| 129.1.91 | 423103 | 48 | 0.0118 | -0.4283 | 2 | 1 | 0 | 1 | - | - | - | - | z2w,3=1<br>z4=2 | As1Cu1Ge1Zr1 |
| 129.1.92 | 42750 | 28 | 0.0100 | -0.3943 | 2 | 1 | 0 | 1 | - | - | - | - | z2w,3=1<br>z4=2 | Al1Ge1Mn1 |
| 129.1.93 | 190195 | 68 | 0.0000 | -0.0320 | 2 | 1 | 0 | 1 | - | - | - | - | z2w,3=1<br>z4=2 | Ag1Pr1Sb2 |
| 129.2.1 | 52412 | 12 | 0.0000 | -1.3493 | 2,81 | 4 | 2,81 | 2 | - | - | - | - | z4=3<br>z2=1 | Sc1 |
| 129.2.2 | 54346 | 76 | 0.0094 | -1.1339 | 2,81 | 4 | 2,81 | 2 | - | - | - | - | z4=3<br>z2=1 | Pt2Si2U1 |
| 129.2.3 | 38176 | 52 | 0.0186 | -0.1520 | 2,81 | 4 | 2,81 | 2 | - | - | - | - | z4=3<br>z2=1 | O1Te1U1 |
| 129.2.4 | 22204 | 52 | 0.0261 | -0.1183 | 2,81 | 4 | 2,81 | 2 | - | - | - | - | z4=3<br>z2=1 | O1Te1U1 |
| 129.2.5 | 419464 | 64 | 0.0214 | -0.0842 | 2,81 | 7 | 2,81 | 2 | - | - | - | - | z2w,3=1<br>z4=3<br>z2=1 | O1Os1P1Pr1 |
| 129.2.6 | 74786 | 84 | 0.0078 | -0.5616 | 2,81 | 4 | 2,81 | 2 | - | - | - | - | z4=3<br>z2=1 | Ni2P2Th1 |
| 129.2.7 | 54276 | 60 | 0.0030 | -0.0428 | 2,81 | 5 | 2,81 | 2 | - | - | - | - | z4=1<br>z2=1 | Mn1Si1Tb1 |
| 129.2.8 | 75054 | 48 | 0.0080 | -0.0727 | 2,81 | 7 | 2,81 | 2 | - | - | - | - | z2w,3=1<br>z4=3<br>z2=1 | Mn1Pr1Si1 |
| 129.2.9 | 75053 | 48 | 0.0000 | -0.0725 | 2,81 | 7 | 2,81 | 2 | - | - | - | - | z2w,3=1<br>z4=3<br>z2=1 | Mn1Pr1Si1 |
| 129.2.10 | 75052 | 48 | 0.0055 | -0.0743 | 2,81 | 7 | 2,81 | 2 | - | - | - | - | z2w,3=1<br>z4=3<br>z2=1 | Mn1Pr1Si1 |
| 129.2.11 | 50243 | 64 | 0.0000 | -0.1023 | 2,81 | 4 | 2,81 | 2 | - | - | - | - | z4=3<br>z2=1 | Mn1Nd1O1Sb1 |
| 129.2.12 | 50241 | 64 | 0.0050 | -0.1033 | 2,81 | 4 | 2,81 | 2 | - | - | - | - | z4=3<br>z2=1 | Mn1Nd1O1P1 |
| 129.2.13 | 423892 | 32 | 0.0397 | -0.1933 | 2,81 | 4 | 2,81 | 2 | - | - | - | - | z4=3<br>z2=1 | La1N1 |
| 129.2.14 | 62716 | 80 | 0.0114 | -0.1037 | 2,81 | 6 | 2,81 | 2 | - | - | - | - | z2w,3=1<br>z4=1<br>z2=1 | Ir2Si2U1 |
| 129.2.15 | 200099 | 56 | 0.0000 | -0.0743 | 2,81 | 4 | 2,81 | 2 | - | - | - | - | z4=3<br>z2=1 | I1Np1O1 |
| 129.2.16 | 87913 | 48 | 0.0243 | -1.6964 | 2,81 | 4 | 2,81 | 2 | - | - | - | - | z4=3<br>z2=1 | Ge1Ni1Pd1 |
| 129.2.17 | 42369 | 48 | 0.0243 | -1.6964 | 2,81 | 4 | 2,81 | 2 | - | - | - | - | z4=3<br>z2=1 | Ge1Ni1Pd1 |
| 129.2.18 | 85859 | 48 | 0.0000 | -0.0725 | 2,81 | 7 | 2,81 | 2 | - | - | - | - | z2w,3=1<br>z4=3<br>z2=1 | Ge1Mn1Pr1 |
| 129.2.19 | 80012 | 48 | 0.0055 | -0.0687 | 2,81 | 7 | 2,81 | 2 | - | - | - | - | z2w,3=1<br>z4=3<br>z2=1 | Ge1Mn1Pr1 |
| 129.2.20 | 80011 | 48 | 0.0021 | -0.0757 | 2,81 | 7 | 2,81 | 2 | - | - | - | - | z2w,3=1<br>z4=3<br>z2=1 | Ge1Mn1Pr1 |

Continued on next page

**TABLE XIV– continued**

| TQC | ICSD | NoE | DG | IndG | SubSG (w/ T) | cl | SubSG (w/o T) | cl | Max. Sub. SG | yes or no | Max. Sub. SG | yes or no | SIs (w/ T) | Stoich. Formula |
|---|---|---|---|---|---|---|---|---|---|---|---|---|---|---|
| 129.2.21 | 183310 | 76 | 0.0000 | -0.0650 | 2,81 | 6 | 2,81 | 2 | - | - | - | - | z2w,3=1<br>z4=1<br>z2=1 | Gd1Sn2Zn1 |
| 129.2.22 | 41255 | 60 | 0.0062 | -0.0193 | 2,81 | 5 | 2,81 | 2 | - | - | - | - | z4=1<br>z2=1 | Gd1Ru1Si1 |
| 129.2.23 | 80200 | 64 | 0.0101 | -0.0461 | 2,81 | 7 | 2,81 | 2 | - | - | - | - | z2w,3=1<br>z4=3<br>z2=1 | Fe1O1P1Pr1 |
| 129.2.24 | 85854 | 52 | 0.0033 | -0.0401 | 2,81 | 5 | 2,81 | 2 | - | - | - | - | z4=1<br>z2=1 | Fe1Nd1Si1 |
| 129.2.25 | 57235 | 52 | 0.0071 | -0.0404 | 2,81 | 5 | 2,81 | 2 | - | - | - | - | z4=1<br>z2=1 | Fe1Nd1Si1 |
| 129.2.26 | 57234 | 52 | 0.0059 | -0.0434 | 2,81 | 5 | 2,81 | 2 | - | - | - | - | z4=1<br>z2=1 | Fe1Nd1Si1 |
| 129.2.27 | 95081 | 52 | 0.0054 | -0.2851 | 2,81 | 6 | 2,81 | 2 | - | - | - | - | z2w,3=1<br>z4=1<br>z2=1 | Eu1P1Sn1 |
| 129.2.28 | 88118 | 60 | 0.0040 | -0.0299 | 2,81 | 5 | 2,81 | 2 | - | - | - | - | z4=1<br>z2=1 | Er1Si1Ti1 |
| 129.2.29 | 88116 | 56 | 0.0027 | -0.0401 | 2,81 | 5 | 2,81 | 2 | - | - | - | - | z4=1<br>z2=1 | Dy1Si1Ti1 |
| 129.2.30 | 57233 | 64 | 0.0000 | -0.0411 | 2,81 | 7 | 2,81 | 2 | - | - | - | - | z2w,3=1<br>z4=3<br>z2=1 | Dy1Fe1Si1 |
| 129.2.31 | 57218 | 64 | 0.0028 | -0.0395 | 2,81 | 7 | 2,81 | 2 | - | - | - | - | z2w,3=1<br>z4=3<br>z2=1 | Dy1Fe1Si1 |
| 129.2.32 | 428492 | 140 | 0.0115 | -0.9581 | 2,81 | 4 | 2,81 | 2 | - | - | - | - | z4=3<br>z2=1 | Cu4La1P3 |
| 129.2.33 | 153572 | 64 | 0.0473 | -0.8906 | 2,81 | 4 | 2,81 | 2 | - | - | - | - | z4=3<br>z2=1 | Cu1Sb2Y1 |
| 129.2.34 | 79579 | 72 | 0.0047 | -0.1007 | 2,81 | 4 | 2,81 | 2 | - | - | - | - | z4=3<br>z2=1 | Cu1O1P1U1 |
| 129.2.35 | 261277 | 72 | 0.0064 | -0.0902 | 2,81 | 4 | 2,81 | 2 | - | - | - | - | z4=3<br>z2=1 | Cu1O1P1U1 |
| 129.2.36 | 182310 | 72 | 0.0042 | -0.0964 | 2,81 | 4 | 2,81 | 2 | - | - | - | - | z4=3<br>z2=1 | Cu1O1P1U1 |
| 129.2.37 | 182311 | 68 | 0.0112 | -0.2143 | 2,81 | 7 | 2,81 | 2 | - | - | - | - | z2w,3=1<br>z4=3<br>z2=1 | Cu1O1P1Th1 |
| 129.2.38 | 172798 | 68 | 0.0078 | -0.5872 | 2,81 | 7 | 2,81 | 2 | - | - | - | - | z2w,3=1<br>z4=3<br>z2=1 | Cu1F1Se1Sr1 |
| 129.2.39 | 172796 | 68 | 0.0079 | -0.4304 | 2,81 | 7 | 2,81 | 2 | - | - | - | - | z2w,3=1<br>z4=3<br>z2=1 | Cu1F1S1Sr1 |
| 129.2.40 | 67932 | 84 | 0.0000 | -0.1520 | 2,81 | 5 | 2,81 | 2 | - | - | - | - | z4=1<br>z2=1 | Co2P2U1 |
| 129.2.41 | 106477 | 52 | 0.0099 | -0.1454 | 2,81 | 7 | 2,81 | 2 | - | - | - | - | z2w,3=1<br>z4=3<br>z2=1 | Co1Ge1Pr1 |
| 129.2.42 | 106471 | 48 | 0.0040 | -0.0967 | 2,81 | 5 | 2,81 | 2 | - | - | - | - | z4=1<br>z2=1 | Co1Ge1La1 |
| 129.2.43 | 61370 | 80 | 0.0047 | -0.5416 | 2,81 | 6 | 2,81 | 2 | - | - | - | - | z2w,3=1<br>z4=1<br>z2=1 | Ce1Pt2Si2 |

Continued on next page

**TABLE XIV– continued**

| TQC | ICSD | NoE | DG | IndG | SubSG (w/ T) | cl | SubSG (w/o T) | cl | Max. Sub. SG | yes or no | Max. Sub. SG | yes or no | SIs (w/ T) | Stoich. Formula |
|---|---|---|---|---|---|---|---|---|---|---|---|---|---|---|
| 129.2.44 | 40764 | 80 | 0.0066 | -0.1430 | 2,81 | 6 | 2,81 | 2 | - | - | - | - | z2w,3=1<br>z4=1<br>z2=1 | Ce1P2Rh2 |
| 129.2.45 | 419356 | 60 | 0.0024 | -0.0895 | 2,81 | 4 | 2,81 | 2 | - | - | - | - | z4=3<br>z2=1 | Ce1Mn1O1Sb1 |
| 129.2.46 | 106407 | 76 | 0.0061 | -0.7410 | 2,81 | 5 | 2,81 | 2 | - | - | - | - | z4=1<br>z2=1 | Ce1Ir2Sn2 |
| 129.2.47 | 102076 | 32 | 0.0374 | -0.8629 | 2,81 | 7 | 2,81 | 2 | - | - | - | - | z2w,3=1<br>z4=3<br>z2=1 | Cd1Ti1 |
| 129.2.48 | 2027 | 20 | 0.0000 | -3.1594 | 2,81 | 4 | 2,81 | 2 | - | - | - | - | z4=3<br>z2=1 | Ca1Ga1N1 |
| 129.2.49 | 62598 | 40 | 0.0280 | -0.6697 | 2,81 | 7 | 2,81 | 2 | - | - | - | - | z2w,3=1<br>z4=3<br>z2=1 | C2Co1Sc1 |
| 129.2.50 | 85283 | 56 | 0.0232 | -0.6024 | 2,81 | 6 | 2,81 | 2 | - | - | - | - | z2w,3=1<br>z4=1<br>z2=1 | C1B1Ni1Y1 |
| 129.2.51 | 415728 | 44 | 0.0238 | -0.6228 | 2,81 | 4 | 2,81 | 2 | - | - | - | - | z4=3<br>z2=1 | Bi2La1Li1 |
| 129.2.52 | 426715 | 76 | 0.0000 | -0.0303 | 2,81 | 7 | 2,81 | 2 | - | - | - | - | z2w,3=1<br>z4=3<br>z2=1 | Bi1N1Nd2 |
| 129.2.53 | 252466 | 68 | 0.0295 | -0.3696 | 2,81 | 6 | 2,81 | 2 | - | - | - | - | z2w,3=1<br>z4=1<br>z2=1 | Bi1La1O1S2 |
| 129.2.54 | 426713 | 68 | 0.0039 | -0.3759 | 2,81 | 7 | 2,81 | 2 | - | - | - | - | z2w,3=1<br>z4=3<br>z2=1 | Bi1Ce2N1 |
| 129.2.55 | 25337 | 28 | 0.0077 | -1.1849 | 2,81 | 4 | 2,81 | 2 | - | - | - | - | z4=3<br>z2=1 | Be2Ca1Ge2 |
| 129.2.56 | 25339 | 84 | 0.0026 | -0.8538 | 2,81 | 4 | 2,81 | 2 | - | - | - | - | z4=3<br>z2=1 | Ba1Sn2Zn2 |
| 129.2.57 | 42460 | 32 | 0.0108 | -0.4317 | 2,81 | 6 | 2,81 | 2 | - | - | - | - | z2w,3=1<br>z4=1<br>z2=1 | Ba1Ge1Mg1 |
| 129.2.58 | 236308 | 84 | 0.0602 | -1.2378 | 2,81 | 4 | 2,81 | 2 | - | - | - | - | z4=3<br>z2=1 | Ba1Cu2Sb2 |
| 129.2.59 | 428691 | 80 | 0.0324 | -2.3433 | 2,81 | 4 | 2,81 | 2 | - | - | - | - | z4=3<br>z2=1 | Au2Sn2Sr1 |
| 129.2.60 | 427711 | 80 | 0.0280 | -2.3174 | 2,81 | 4 | 2,81 | 2 | - | - | - | - | z4=3<br>z2=1 | Au2Sn2Sr1 |
| 129.2.61 | 107956 | 88 | 0.0000 | -0.0498 | 2,81 | 6 | 2,81 | 2 | - | - | - | - | z2w,3=1<br>z4=1<br>z2=1 | As2Nd1Ni2 |
| 129.2.62 | 412550 | 64 | 0.0537 | -1.0483 | 2,81 | 4 | 2,81 | 2 | - | - | - | - | z4=3<br>z2=1 | As2Cu1Y1 |
| 129.2.63 | 174358 | 92 | 0.0218 | -1.1117 | 2,81 | 4 | 2,81 | 2 | - | - | - | - | z4=3<br>z2=1 | As2Cu1Lu1 |
| 129.2.64 | 94440 | 84 | 0.0000 | -0.0369 | 2,81 | 4 | 2,81 | 2 | - | - | - | - | z4=3<br>z2=1 | As2Cu1Ho1 |
| 129.2.65 | 412178 | 84 | 0.0000 | -0.0376 | 2,81 | 4 | 2,81 | 2 | - | - | - | - | z4=3<br>z2=1 | As2Cu1Ho1 |
| 129.2.66 | 44038 | 52 | 0.0100 | -0.0757 | 2,81 | 4 | 2,81 | 2 | - | - | - | - | z4=3<br>z2=1 | As1Np1Te1 |
| 129.2.67 | 50242 | 64 | 0.0047 | -0.0965 | 2,81 | 4 | 2,81 | 2 | - | - | - | - | z4=3<br>z2=1 | As1Mn1Nd1O1 |

Continued on next page

**TABLE XIV– continued**

| TQC | ICSD | NoE | DG | IndG | SubSG (w/ T) | cl | SubSG (w/o T) | cl | Max. Sub. SG | yes or no | Max. Sub. SG | yes or no | SIs (w/ T) | Stoich. Formula |
|---|---|---|---|---|---|---|---|---|---|---|---|---|---|---|
| 129.2.68 | 262831 | 60 | 0.0500 | -0.6166 | 2,81 | 7 | 2,81 | 2 | - | - | - | - | z2w,3=1<br>z4=3<br>z2=1 | As1La1O1Ru1 |
| 129.2.69 | 90316 | 64 | 0.0075 | -0.0840 | 2,81 | 7 | 2,81 | 2 | - | - | - | - | z2w,3=1<br>z4=3<br>z2=1 | As1Fe1O1Pr1 |
| 129.2.70 | 422000 | 64 | 0.0099 | -0.0796 | 2,81 | 7 | 2,81 | 2 | - | - | - | - | z2w,3=1<br>z4=3<br>z2=1 | As1Fe1O1Pr1 |
| 129.2.71 | 246832 | 64 | 0.0041 | -0.0855 | 2,81 | 7 | 2,81 | 2 | - | - | - | - | z2w,3=1<br>z4=3<br>z2=1 | As1Fe1O1Pr1 |
| 129.2.72 | 162830 | 64 | 0.0066 | -0.0844 | 2,81 | 7 | 2,81 | 2 | - | - | - | - | z2w,3=1<br>z4=3<br>z2=1 | As1Fe1O1Pr1 |
| 129.2.73 | 93238 | 40 | 0.0182 | -0.0465 | 2,81 | 5 | 2,81 | 2 | - | - | - | - | z4=1<br>z2=1 | As1Fe1Mn1 |
| 129.2.74 | 93237 | 40 | 0.0059 | -0.0629 | 2,81 | 5 | 2,81 | 2 | - | - | - | - | z4=1<br>z2=1 | As1Fe1Mn1 |
| 129.2.75 | 180774 | 68 | 0.0092 | -0.0607 | 2,81 | 7 | 2,81 | 2 | - | - | - | - | z2w,3=1<br>z4=3<br>z2=1 | As1Co1Nd1O1 |
| 129.2.76 | 180773 | 68 | 0.0101 | -0.0598 | 2,81 | 7 | 2,81 | 2 | - | - | - | - | z2w,3=1<br>z4=3<br>z2=1 | As1Co1Nd1O1 |
| 129.2.77 | 180772 | 68 | 0.0099 | -0.0591 | 2,81 | 7 | 2,81 | 2 | - | - | - | - | z2w,3=1<br>z4=3<br>z2=1 | As1Co1Nd1O1 |
| 129.2.78 | 169027 | 68 | 0.0092 | -0.0600 | 2,81 | 7 | 2,81 | 2 | - | - | - | - | z2w,3=1<br>z4=3<br>z2=1 | As1Co1Nd1O1 |
| 129.2.79 | 195499 | 60 | 0.0052 | -0.0804 | 2,81 | 4 | 2,81 | 2 | - | - | - | - | z4=3<br>z2=1 | As1Ce1Mn1O1 |
| 129.2.80 | 195498 | 60 | 0.0031 | -0.0733 | 2,81 | 4 | 2,81 | 2 | - | - | - | - | z4=3<br>z2=1 | As1Ce1Mn1O1 |
| 129.2.81 | 194698 | 76 | 0.0099 | -0.7457 | 2,81 | 5 | 2,81 | 2 | - | - | - | - | z4=1<br>z2=1 | Al2Ce1Pd2 |
| 129.2.82 | 191050 | 76 | 0.0090 | -0.6819 | 2,81 | 5 | 2,81 | 2 | - | - | - | - | z4=1<br>z2=1 | Al2Ce1Pd2 |
| 129.2.83 | 190094 | 64 | 0.0670 | -1.0541 | 2,81 | 4 | 2,81 | 2 | - | - | - | - | z4=3<br>z2=1 | Ag1Sb2Y1 |
| 129.2.84 | 251149 | 168 | 0.0124 | -0.4907 | 2,81 | 4 | 2,81 | 2 | - | - | - | - | z4=3<br>z2=1 | Ag1Cu1S1 |
| 129.3.1 | 76035 | 52 | - | - | - | - | - | - | 115 | yes | 115 | yes | - | Te2U1 |
| | | | | | | | | | 113 | yes | 113 | yes | - | |
| | | | | | | | | | 67 | no(2) | 67 | no(2) | z4=3 | |
| | | | | | | | | | 59 | no(2) | 59 | no(2) | z4=3 | |
| 129.3.2 | 43030 | 48 | - | - | - | - | - | - | 115 | no(81) | 115 | no(81) | z2=1 | Sb2U1 |
| | | | | | | | | | 113 | no(81) | 113 | no(81) | z2=1 | |
| | | | | | | | | | 67 | no(2) | 67 | yes | z2w,3=1<br>z4=2 | |
| | | | | | | | | | 59 | no(2) | 59 | yes | z2w,3=1<br>z4=2 | |
| 129.3.3 | 42345 | 48 | - | - | - | - | - | - | 115 | no(81) | 115 | no(81) | z2=1 | Sb2U1 |
| | | | | | | | | | 113 | no(81) | 113 | no(81) | z2=1 | |
| | | | | | | | | | 67 | no(2) | 67 | yes | z2w,3=1<br>z4=2 | |

Continued on next page

**TABLE XIV– continued**

| TQC | ICSD | NoE | DG | IndG | SubSG cl (w/ T) | SubSG cl (w/o T) | Max. Sub. SG | yes or no | Max. Sub. SG | yes or no | SIs (w/ T) | Stoich. Formula |
|---|---|---|---|---|---|---|---|---|---|---|---|---|
| | | | | | | | 59 | no(2) | 59 | yes | z2w,3=1<br>z4=2 | |
| 129.3.4 | 16656 | 44 | - | - | - - | - - | 115 | yes | 115 | yes | - | Sb2Th1 |
| | | | | | | | 113 | yes | 113 | yes | - | |
| | | | | | | | 67 | no(2) | 67 | no(2) | z4=3 | |
| | | | | | | | 59 | no(2) | 59 | no(2) | z4=3 | |
| 129.3.5 | 61145 | 84 | - | - | - - | - - | 115 | no(81) | 115 | no(81) | z2=1 | Pt2Si2U1 |
| | | | | | | | 113 | no(81) | 113 | no(81) | z2=1 | |
| | | | | | | | 67 | no(2) | 67 | yes | z4=2 | |
| | | | | | | | 59 | no(2) | 59 | yes | z4=2 | |
| 129.3.6 | 57472 | 84 | - | - | - - | - - | 115 | yes | 115 | yes | - | Pt2Si2U1 |
| | | | | | | | 113 | yes | 113 | yes | - | |
| | | | | | | | 67 | no(2) | 67 | no(2) | z2w,3=1<br>z4=1 | |
| | | | | | | | 59 | no(2) | 59 | no(2) | z2w,3=1<br>z4=1 | |
| 129.3.7 | 167713 | 84 | - | - | - - | - - | 115 | yes | 115 | yes | - | Pt2Si2U1 |
| | | | | | | | 113 | yes | 113 | yes | - | |
| | | | | | | | 67 | no(2) | 67 | no(2) | z2w,3=1<br>z4=1 | |
| | | | | | | | 59 | no(2) | 59 | no(2) | z2w,3=1<br>z4=1 | |
| 129.3.8 | 167712 | 84 | - | - | - - | - - | 115 | yes | 115 | yes | - | Pt2Si2U1 |
| | | | | | | | 113 | yes | 113 | yes | - | |
| | | | | | | | 67 | no(2) | 67 | no(2) | z2w,3=1<br>z4=1 | |
| | | | | | | | 59 | no(2) | 59 | no(2) | z2w,3=1<br>z4=1 | |
| 129.3.9 | 33587 | 48 | - | - | - - | - - | 115 | no(81) | 115 | no(81) | z2=1 | O1Te1Th1 |
| | | | | | | | 113 | no(81) | 113 | no(81) | z2=1 | |
| | | | | | | | 67 | yes | 67 | yes | - | |
| | | | | | | | 59 | yes | 59 | yes | - | |
| 129.3.10 | 44379 | 60 | - | - | - - | - - | 115 | yes | 115 | yes | - | Np1 |
| | | | | | | | 113 | yes | 113 | yes | - | |
| | | | | | | | 67 | yes | 67 | yes | - | |
| | | | | | | | 59 | yes | 59 | yes | - | |
| 129.3.11 | 20313 | 52 | - | - | - - | - - | 115 | yes | 115 | yes | - | Nd1Te2 |
| | | | | | | | 113 | yes | 113 | yes | - | |
| | | | | | | | 67 | no(2) | 67 | no(2) | z4=3 | |
| | | | | | | | 59 | no(2) | 59 | no(2) | z4=3 | |
| 129.3.12 | 93156 | 68 | - | - | - - | - - | 115 | no(81) | 115 | no(81) | z2=1 | Nd1Pd1Sb2 |
| | | | | | | | 113 | no(81) | 113 | no(81) | z2=1 | |
| | | | | | | | 67 | yes | 67 | yes | - | |
| | | | | | | | 59 | yes | 59 | yes | - | |
| 129.3.13 | 160053 | 84 | - | - | - - | - - | 115 | no(81) | 115 | no(81) | z2=1 | Nd1Ni2Sn2 |
| | | | | | | | 113 | no(81) | 113 | no(81) | z2=1 | |
| | | | | | | | 67 | no(2) | 67 | yes | z4=2 | |
| | | | | | | | 59 | no(2) | 59 | yes | z4=2 | |
| 129.3.14 | 100437 | 32 | - | - | - - | - - | 115 | yes | 115 | yes | - | Mo1N1Ta1 |
| | | | | | | | 113 | yes | 113 | yes | - | |
| | | | | | | | 67 | no(2) | 67 | no(2) | z4=3 | |
| | | | | | | | 59 | no(2) | 59 | no(2) | z4=3 | |
| 129.3.15 | 391427 | 64 | - | - | - - | - - | 115 | no(81) | 115 | no(81) | z2=1 | La1Ni1O1P1 |
| | | | | | | | 113 | no(81) | 113 | no(81) | z2=1 | |
| | | | | | | | 67 | no(2) | 67 | yes | z2w,3=1 | |
| | | | | | | | 59 | no(2) | 59 | yes | z2w,3=1 | |
| 129.3.16 | 249484 | 64 | - | - | - - | - - | 115 | no(81) | 115 | no(81) | z2=1 | La1Ni1O1P1 |
| | | | | | | | 113 | no(81) | 113 | no(81) | z2=1 | |

Continued on next page

**TABLE XIV– continued**

| TQC | ICSD | NoE | DG | IndG | SubSG cl (w/ T) | SubSG cl (w/o T) | Max. Sub. SG | yes or no | Max. Sub. SG | yes or no | SIs (w/ T) | Stoich. Formula |
|---|---|---|---|---|---|---|---|---|---|---|---|---|
| | | | | | | | 67 | no(2) | 67 | yes | z2w,3=1 | |
| | | | | | | | 59 | no(2) | 59 | yes | z2w,3=1 | |
| 129.3.17 | 85849 | 44 | - | - | - - | - - | 115 | yes | 115 | yes | - | La1Mn1Si1 |
| | | | | | | | 113 | yes | 113 | yes | - | |
| | | | | | | | 67 | no(2) | 67 | no(2) | z4=1 | |
| | | | | | | | 59 | no(2) | 59 | no(2) | z4=1 | |
| 129.3.18 | 75047 | 44 | - | - | - - | - - | 115 | yes | 115 | yes | - | La1Mn1Si1 |
| | | | | | | | 113 | yes | 113 | yes | - | |
| | | | | | | | 67 | no(2) | 67 | no(2) | z4=1 | |
| | | | | | | | 59 | no(2) | 59 | no(2) | z4=1 | |
| 129.3.19 | 75046 | 44 | - | - | - - | - - | 115 | yes | 115 | yes | - | La1Mn1Si1 |
| | | | | | | | 113 | yes | 113 | yes | - | |
| | | | | | | | 67 | no(2) | 67 | no(2) | z4=1 | |
| | | | | | | | 59 | no(2) | 59 | no(2) | z4=1 | |
| 129.3.20 | 89097 | 80 | - | - | - - | - - | 115 | yes | 115 | yes | - | Ir2Si2U1 |
| | | | | | | | 113 | yes | 113 | yes | - | |
| | | | | | | | 67 | no(2) | 67 | no(2) | z2w,3=1<br>z4=1 | |
| | | | | | | | 59 | no(2) | 59 | no(2) | z2w,3=1<br>z4=1 | |
| 129.3.21 | 98944 | 52 | - | - | - - | - - | 115 | yes | 115 | yes | - | H1Ce1Co1Ge1 |
| | | | | | | | 113 | yes | 113 | yes | - | |
| | | | | | | | 67 | no(2) | 67 | no(2) | z4=1 | |
| | | | | | | | 59 | no(2) | 59 | no(2) | z4=1 | |
| 129.3.22 | 15575 | 28 | - | - | - - | - - | 115 | no(81) | 115 | no(81) | z2=1 | Ge1Te1Zr1 |
| | | | | | | | 113 | no(81) | 113 | no(81) | z2=1 | |
| | | | | | | | 67 | no(2) | 67 | yes | z2w,3=1<br>z4=2 | |
| | | | | | | | 59 | no(2) | 59 | yes | z2w,3=1<br>z4=2 | |
| 129.3.23 | 85857 | 44 | - | - | - - | - - | 115 | yes | 115 | yes | - | Ge1La1Mn1 |
| | | | | | | | 113 | yes | 113 | yes | - | |
| | | | | | | | 67 | no(2) | 67 | no(2) | z4=1 | |
| | | | | | | | 59 | no(2) | 59 | no(2) | z4=1 | |
| 129.3.24 | 80008 | 44 | - | - | - - | - - | 115 | yes | 115 | yes | - | Ge1La1Mn1 |
| | | | | | | | 113 | yes | 113 | yes | - | |
| | | | | | | | 67 | no(2) | 67 | no(2) | z4=1 | |
| | | | | | | | 59 | no(2) | 59 | no(2) | z4=1 | |
| 129.3.25 | 80007 | 44 | - | - | - - | - - | 115 | yes | 115 | yes | - | Ge1La1Mn1 |
| | | | | | | | 113 | yes | 113 | yes | - | |
| | | | | | | | 67 | no(2) | 67 | no(2) | z4=1 | |
| | | | | | | | 59 | no(2) | 59 | no(2) | z4=1 | |
| 129.3.26 | 290843 | 52 | - | - | - - | - - | 115 | no(81) | 115 | no(81) | z2=1 | Gd1Sb1Sc1 |
| | | | | | | | 113 | no(81) | 113 | no(81) | z2=1 | |
| | | | | | | | 67 | yes | 67 | yes | - | |
| | | | | | | | 59 | yes | 59 | yes | - | |
| 129.3.27 | 6318 | 60 | - | - | - - | - - | 115 | yes | 115 | yes | - | Gd1O2 |
| | | | | | | | 113 | yes | 113 | yes | - | |
| | | | | | | | 67 | no(2) | 67 | no(2) | z4=3 | |
| | | | | | | | 59 | no(2) | 59 | no(2) | z4=3 | |
| 129.3.28 | 44753 | 28 | - | - | - - | - - | 115 | yes | 115 | yes | - | Fe1Te1 |
| | | | | | | | 113 | yes | 113 | yes | - | |
| | | | | | | | 67 | no(2) | 67 | no(2) | z2w,3=1<br>z4=1 | |
| | | | | | | | 59 | no(2) | 59 | no(2) | z2w,3=1<br>z4=1 | |
| 129.3.29 | 977903 | 28 | - | - | - - | - - | 115 | yes | 115 | yes | - | Fe1Se1 |
| | | | | | | | 113 | yes | 113 | yes | - | |

Continued on next page

**TABLE XIV– continued**

| TQC | ICSD | NoE | DG | IndG | SubSG cl (w/ T) | SubSG cl (w/o T) | Max. Sub. SG | yes or no | Max. Sub. SG | yes or no | SIs (w/ T) | Stoich. Formula |
|---|---|---|---|---|---|---|---|---|---|---|---|---|
| | | | | | | | 67 | no(2) | 67 | no(2) | z2w,3=1<br>z4=1 | |
| | | | | | | | 59 | no(2) | 59 | no(2) | z2w,3=1<br>z4=1 | |
| 129.3.30 | 977901 | 28 | - | - | - - | - - | 115 | yes | 115 | yes | - | Fe1Se1 |
| | | | | | | | 113 | yes | 113 | yes | - | |
| | | | | | | | 67 | no(2) | 67 | no(2) | z2w,3=1<br>z4=1 | |
| | | | | | | | 59 | no(2) | 59 | no(2) | z2w,3=1<br>z4=1 | |
| 129.3.31 | 26889 | 28 | - | - | - - | - - | 115 | yes | 115 | yes | - | Fe1Se1 |
| | | | | | | | 113 | yes | 113 | yes | - | |
| | | | | | | | 67 | no(2) | 67 | no(2) | z2w,3=1<br>z4=1 | |
| | | | | | | | 59 | no(2) | 59 | no(2) | z2w,3=1<br>z4=1 | |
| 129.3.32 | 196298 | 28 | - | - | - - | - - | 115 | yes | 115 | yes | - | Fe1Se1 |
| | | | | | | | 113 | yes | 113 | yes | - | |
| | | | | | | | 67 | no(2) | 67 | no(2) | z2w,3=1<br>z4=1 | |
| | | | | | | | 59 | no(2) | 59 | no(2) | z2w,3=1<br>z4=1 | |
| 129.3.33 | 196297 | 28 | - | - | - - | - - | 115 | yes | 115 | yes | - | Fe1Se1 |
| | | | | | | | 113 | yes | 113 | yes | - | |
| | | | | | | | 67 | no(2) | 67 | no(2) | z2w,3=1<br>z4=1 | |
| | | | | | | | 59 | no(2) | 59 | no(2) | z2w,3=1<br>z4=1 | |
| 129.3.34 | 185464 | 28 | - | - | - - | - - | 115 | yes | 115 | yes | - | Fe1Se1 |
| | | | | | | | 113 | yes | 113 | yes | - | |
| | | | | | | | 67 | no(2) | 67 | no(2) | z2w,3=1<br>z4=1 | |
| | | | | | | | 59 | no(2) | 59 | no(2) | z2w,3=1<br>z4=1 | |
| 129.3.35 | 165960 | 28 | - | - | - - | - - | 115 | yes | 115 | yes | - | Fe1Se1 |
| | | | | | | | 113 | yes | 113 | yes | - | |
| | | | | | | | 67 | no(2) | 67 | no(2) | z2w,3=1<br>z4=1 | |
| | | | | | | | 59 | no(2) | 59 | no(2) | z2w,3=1<br>z4=1 | |
| 129.3.36 | 165959 | 28 | - | - | - - | - - | 115 | yes | 115 | yes | - | Fe1Se1 |
| | | | | | | | 113 | yes | 113 | yes | - | |
| | | | | | | | 67 | no(2) | 67 | no(2) | z2w,3=1<br>z4=1 | |
| | | | | | | | 59 | no(2) | 59 | no(2) | z2w,3=1<br>z4=1 | |
| 129.3.37 | 165958 | 28 | - | - | - - | - - | 115 | yes | 115 | yes | - | Fe1Se1 |
| | | | | | | | 113 | yes | 113 | yes | - | |
| | | | | | | | 67 | no(2) | 67 | no(2) | z2w,3=1<br>z4=1 | |
| | | | | | | | 59 | no(2) | 59 | no(2) | z2w,3=1<br>z4=1 | |
| 129.3.38 | 163562 | 28 | - | - | - - | - - | 115 | yes | 115 | yes | - | Fe1Se1 |
| | | | | | | | 113 | yes | 113 | yes | - | |
| | | | | | | | 67 | no(2) | 67 | no(2) | z2w,3=1<br>z4=1 | |
| | | | | | | | 59 | no(2) | 59 | no(2) | z2w,3=1<br>z4=1 | |

Continued on next page

**TABLE XIV– continued**

| TQC | ICSD | NoE | DG | IndG | SubSG cl (w/ T) | SubSG cl (w/o T) | Max. Sub. SG | yes or no | Max. Sub. SG | yes or no | SIs (w/ T) | Stoich. Formula |
|---|---|---|---|---|---|---|---|---|---|---|---|---|
| 129.3.39 | 163561 | 28 | - | - | - - | - - | 115 | yes | 115 | yes | - | Fe1Se1 |
| | | | | | | | 113 | yes | 113 | yes | - | |
| | | | | | | | 67 | no(2) | 67 | no(2) | z2w,3=1 z4=1 | |
| | | | | | | | 59 | no(2) | 59 | no(2) | z2w,3=1 z4=1 | |
| 129.3.40 | 163560 | 28 | - | - | - - | - - | 115 | yes | 115 | yes | - | Fe1Se1 |
| | | | | | | | 113 | yes | 113 | yes | - | |
| | | | | | | | 67 | no(2) | 67 | no(2) | z2w,3=1 z4=1 | |
| | | | | | | | 59 | no(2) | 59 | no(2) | z2w,3=1 z4=1 | |
| 129.3.41 | 163559 | 28 | - | - | - - | - - | 115 | yes | 115 | yes | - | Fe1Se1 |
| | | | | | | | 113 | yes | 113 | yes | - | |
| | | | | | | | 67 | no(2) | 67 | no(2) | z2w,3=1 z4=1 | |
| | | | | | | | 59 | no(2) | 59 | no(2) | z2w,3=1 z4=1 | |
| 129.3.42 | 42327 | 28 | - | - | - - | - - | 115 | yes | 115 | yes | - | Fe1Li1P1 |
| | | | | | | | 113 | yes | 113 | yes | - | |
| | | | | | | | 67 | no(2) | 67 | no(2) | z2w,3=1 z4=1 | |
| | | | | | | | 59 | no(2) | 59 | no(2) | z2w,3=1 z4=1 | |
| 129.3.43 | 247248 | 28 | - | - | - - | - - | 115 | yes | 115 | yes | - | Fe1Li1P1 |
| | | | | | | | 113 | yes | 113 | yes | - | |
| | | | | | | | 67 | no(2) | 67 | no(2) | z2w,3=1 z4=1 | |
| | | | | | | | 59 | no(2) | 59 | no(2) | z2w,3=1 z4=1 | |
| 129.3.44 | 247089 | 28 | - | - | - - | - - | 115 | yes | 115 | yes | - | Fe1Li1P1 |
| | | | | | | | 113 | yes | 113 | yes | - | |
| | | | | | | | 67 | no(2) | 67 | no(2) | z2w,3=1 z4=1 | |
| | | | | | | | 59 | no(2) | 59 | no(2) | z2w,3=1 z4=1 | |
| 129.3.45 | 94474 | 64 | - | - | - - | - - | 115 | yes | 115 | yes | - | F1S1Tb1 |
| | | | | | | | 113 | yes | 113 | yes | - | |
| | | | | | | | 67 | no(2) | 67 | no(2) | z4=3 | |
| | | | | | | | 59 | no(2) | 59 | no(2) | z4=3 | |
| 129.3.46 | 95024 | 60 | - | - | - - | - - | 115 | no(81) | 115 | no(81) | z2=1 | Er1Ge1Ti1 |
| | | | | | | | 113 | no(81) | 113 | no(81) | z2=1 | |
| | | | | | | | 67 | no(2) | 67 | yes | z2w,3=1 z4=2 | |
| | | | | | | | 59 | no(2) | 59 | yes | z2w,3=1 z4=2 | |
| 129.3.47 | 95022 | 60 | - | - | - - | - - | 115 | no(81) | 115 | no(81) | z2=1 | Er1Ge1Ti1 |
| | | | | | | | 113 | no(81) | 113 | no(81) | z2=1 | |
| | | | | | | | 67 | no(2) | 67 | yes | z2w,3=1 z4=2 | |
| | | | | | | | 59 | no(2) | 59 | yes | z2w,3=1 z4=2 | |
| 129.3.48 | 85907 | 56 | - | - | - - | - - | 115 | yes | 115 | yes | - | Dy1Ge1Ti1 |
| | | | | | | | 113 | yes | 113 | yes | - | |
| | | | | | | | 67 | no(2) | 67 | no(2) | z4=1 | |
| | | | | | | | 59 | no(2) | 59 | no(2) | z4=1 | |
| 129.3.49 | 41923 | 84 | - | - | - - | - - | 115 | yes | 115 | yes | - | Cu2Sb2Sr1 |
| | | | | | | | 113 | yes | 113 | yes | - | |

Continued on next page

**TABLE XIV– continued**

| TQC | ICSD | NoE | DG | IndG | SubSG cl (w/ T) | SubSG cl (w/o T) | Max. Sub. SG | yes or no | Max. Sub. SG | yes or no | SIs (w/ T) | Stoich. Formula |
|---|---|---|---|---|---|---|---|---|---|---|---|---|
| | | | | | | | 67 | no(2) | 67 | no(2) | z4=3 | |
| | | | | | | | 59 | no(2) | 59 | no(2) | z4=3 | |
| 129.3.50 | 93339 | 80 | - | - | - - | - - | 115 | yes | 115 | yes | - | Cu1Sb2Tb1 |
| | | | | | | | 113 | yes | 113 | yes | - | |
| | | | | | | | 67 | no(2) | 67 | no(2) | z4=3 | |
| | | | | | | | 59 | no(2) | 59 | no(2) | z4=3 | |
| 129.3.51 | 153569 | 80 | - | - | - - | - - | 115 | yes | 115 | yes | - | Cu1Sb2Tb1 |
| | | | | | | | 113 | yes | 113 | yes | - | |
| | | | | | | | 67 | no(2) | 67 | no(2) | z4=3 | |
| | | | | | | | 59 | no(2) | 59 | no(2) | z4=3 | |
| 129.3.52 | 12155 | 36 | - | - | - - | - - | 115 | yes | 115 | yes | - | Cu1Na1Se1 |
| | | | | | | | 113 | yes | 113 | yes | - | |
| | | | | | | | 67 | no(2) | 67 | no(2) | z4=3 | |
| | | | | | | | 59 | no(2) | 59 | no(2) | z4=3 | |
| 129.3.53 | 78576 | 84 | - | - | - - | - - | 115 | no(81) | 115 | no(81) | z2=1 | Co2P2U1 |
| | | | | | | | 113 | no(81) | 113 | no(81) | z2=1 | |
| | | | | | | | 67 | yes | 67 | yes | - | |
| | | | | | | | 59 | yes | 59 | yes | - | |
| 129.3.54 | 80516 | 64 | - | - | - - | - - | 115 | no(81) | 115 | no(81) | z2=1 | Co1Si1Tb1 |
| | | | | | | | 113 | no(81) | 113 | no(81) | z2=1 | |
| | | | | | | | 67 | no(2) | 67 | yes | z2w,3=1 z4=2 | |
| | | | | | | | 59 | no(2) | 59 | yes | z2w,3=1 z4=2 | |
| 129.3.55 | 80515 | 64 | - | - | - - | - - | 115 | no(81) | 115 | no(81) | z2=1 | Co1Si1Tb1 |
| | | | | | | | 113 | no(81) | 113 | no(81) | z2=1 | |
| | | | | | | | 67 | no(2) | 67 | yes | z2w,3=1 z4=2 | |
| | | | | | | | 59 | no(2) | 59 | yes | z2w,3=1 z4=2 | |
| 129.3.56 | 418 - | 160 | - | - | - - | - - | 115 | yes | 115 | yes | - | Ce2Ru1Zn4 |
| | | | | | | | 113 | yes | 113 | yes | - | |
| | | | | | | | 67 | no(2) | 67 | no(2) | z4=1 | |
| | | | | | | | 59 | no(2) | 59 | no(2) | z4=1 | |
| 129.3.57 | 43675 | 48 | - | - | - - | - - | 115 | no(81) | 115 | no(81) | z2=1 | Ce1Te2 |
| | | | | | | | 113 | no(81) | 113 | no(81) | z2=1 | |
| | | | | | | | 67 | no(2) | 67 | yes | z2w,3=1 z4=2 | |
| | | | | | | | 59 | no(2) | 59 | yes | z2w,3=1 z4=2 | |
| 129.3.58 | 196996 | 48 | - | - | - - | - - | 115 | no(81) | 115 | no(81) | z2=1 | Ce1Te2 |
| | | | | | | | 113 | no(81) | 113 | no(81) | z2=1 | |
| | | | | | | | 67 | no(2) | 67 | yes | z2w,3=1 z4=2 | |
| | | | | | | | 59 | no(2) | 59 | yes | z2w,3=1 z4=2 | |
| 129.3.59 | 41252 | 48 | - | - | - - | - - | 115 | yes | 115 | yes | - | Ce1Ru1Si1 |
| | | | | | | | 113 | yes | 113 | yes | - | |
| | | | | | | | 67 | no(2) | 67 | no(2) | z4=1 | |
| | | | | | | | 59 | no(2) | 59 | no(2) | z4=1 | |
| 129.3.60 | 41256 | 48 | - | - | - - | - - | 115 | yes | 115 | yes | - | Ce1Os1Si1 |
| | | | | | | | 113 | yes | 113 | yes | - | |
| | | | | | | | 67 | no(2) | 67 | no(2) | z4=1 | |
| | | | | | | | 59 | no(2) | 59 | no(2) | z4=1 | |
| 129.3.61 | 107618 | 40 | - | - | - - | - - | 115 | yes | 115 | yes | - | Ce1Ge1Ti1 |
| | | | | | | | 113 | yes | 113 | yes | - | |
| | | | | | | | 67 | no(2) | 67 | no(2) | z2w,3=1 z4=3 | |

Continued on next page

**TABLE XIV– continued**

| TQC | ICSD | NoE | DG | IndG | SubSG cl (w/ T) | SubSG cl (w/o T) | Max. Sub. SG | yes or no | Max. Sub. SG | yes or no | SIs (w/ T) | Stoich. Formula |
|---|---|---|---|---|---|---|---|---|---|---|---|---|
| | | | | | | | 59 | no(2) | 59 | no(2) | z2w,3=1<br>z4=3 | |
| 129.3.62 | 42683 | 48 | - | - | - - | - - | 115 | yes | 115 | yes | - | Ce1Fe1Si1 |
| | | | | | | | 113 | yes | 113 | yes | - | |
| | | | | | | | 67 | no(2) | 67 | no(2) | z4=1 | |
| | | | | | | | 59 | no(2) | 59 | no(2) | z4=1 | |
| 129.3.63 | 418852 | 32 | - | - | - - | - - | 115 | no(81) | 115 | no(81) | z2=1 | Ca1Ge1Ni1 |
| | | | | | | | 113 | no(81) | 113 | no(81) | z2=1 | |
| | | | | | | | 67 | no(2) | 67 | yes | z2w,3=1<br>z4=2 | |
| | | | | | | | 59 | no(2) | 59 | yes | z2w,3=1<br>z4=2 | |
| 129.3.64 | 248425 | 72 | - | - | - - | - - | 115 | yes | 115 | yes | - | C1B1Ni1Tb1 |
| | | | | | | | 113 | yes | 113 | yes | - | |
| | | | | | | | 67 | no(2) | 67 | no(2) | z4=3 | |
| | | | | | | | 59 | no(2) | 59 | no(2) | z4=3 | |
| 129.3.65 | 42339 | 48 | - | - | - - | - - | 115 | no(81) | 115 | no(81) | z2=1 | Bi2U1 |
| | | | | | | | 113 | no(81) | 113 | no(81) | z2=1 | |
| | | | | | | | 67 | no(2) | 67 | yes | z2w,3=1<br>z4=2 | |
| | | | | | | | 59 | no(2) | 59 | yes | z2w,3=1<br>z4=2 | |
| 129.3.66 | 58861 | 44 | - | - | - - | - - | 115 | yes | 115 | yes | - | Bi2Th1 |
| | | | | | | | 113 | yes | 113 | yes | - | |
| | | | | | | | 67 | no(2) | 67 | no(2) | z4=3 | |
| | | | | | | | 59 | no(2) | 59 | no(2) | z4=3 | |
| 129.3.67 | 425950 | 64 | - | - | - - | - - | 115 | no(81) | 115 | no(81) | z2=1 | Bi1La2N1 |
| | | | | | | | 113 | no(81) | 113 | no(81) | z2=1 | |
| | | | | | | | 67 | no(2) | 67 | yes | z2w,3=1<br>z4=2 | |
| | | | | | | | 59 | no(2) | 59 | yes | z2w,3=1<br>z4=2 | |
| 129.3.68 | 58790 | 16 | - | - | - - | - - | 115 | yes | 115 | yes | - | Bi1In1 |
| | | | | | | | 113 | yes | 113 | yes | - | |
| | | | | | | | 67 | no(2) | 67 | no(2) | z2w,3=1<br>z4=3 | |
| | | | | | | | 59 | no(2) | 59 | no(2) | z2w,3=1<br>z4=3 | |
| 129.3.69 | 150595 | 16 | - | - | - - | - - | 115 | no(81) | 115 | no(81) | z2=1 | Bi1In1 |
| | | | | | | | 113 | no(81) | 113 | no(81) | z2=1 | |
| | | | | | | | 67 | yes | 67 | yes | - | |
| | | | | | | | 59 | yes | 59 | yes | - | |
| 129.3.70 | 1255 | 16 | - | - | - - | - - | 115 | no(81) | 115 | no(81) | z2=1 | Bi1In1 |
| | | | | | | | 113 | no(81) | 113 | no(81) | z2=1 | |
| | | | | | | | 67 | yes | 67 | yes | - | |
| | | | | | | | 59 | yes | 59 | yes | - | |
| 129.3.71 | 1254 | 16 | - | - | - - | - - | 115 | no(81) | 115 | no(81) | z2=1 | Bi1In1 |
| | | | | | | | 113 | no(81) | 113 | no(81) | z2=1 | |
| | | | | | | | 67 | yes | 67 | yes | - | |
| | | | | | | | 59 | yes | 59 | yes | - | |
| 129.3.72 | 1253 | 16 | - | - | - - | - - | 115 | no(81) | 115 | no(81) | z2=1 | Bi1In1 |
| | | | | | | | 113 | no(81) | 113 | no(81) | z2=1 | |
| | | | | | | | 67 | yes | 67 | yes | - | |
| | | | | | | | 59 | yes | 59 | yes | - | |
| 129.3.73 | 1252 | 16 | - | - | - - | - - | 115 | no(81) | 115 | no(81) | z2=1 | Bi1In1 |
| | | | | | | | 113 | no(81) | 113 | no(81) | z2=1 | |
| | | | | | | | 67 | yes | 67 | yes | - | |
| | | | | | | | 59 | yes | 59 | yes | - | |

Continued on next page

**TABLE XIV– continued**

| TQC | ICSD | NoE | DG | IndG | SubSG cl (w/ T) | SubSG cl (w/o T) | Max. Sub. SG | yes or no | Max. Sub. SG | yes or no | SIs (w/ T) | Stoich. Formula |
|---|---|---|---|---|---|---|---|---|---|---|---|---|
| 129.3.74 | 1251 | 16 | - | - | - - | - - | 115 | no(81) | 115 | no(81) | z2=1 | Bi1In1 |
| | | | | | | | 113 | no(81) | 113 | no(81) | z2=1 | |
| | | | | | | | 67 | yes | 67 | yes | - | |
| | | | | | | | 59 | yes | 59 | yes | - | |
| 129.3.75 | 107642 | 16 | - | - | - - | - - | 115 | no(81) | 115 | no(81) | z2=1 | Bi1In1 |
| | | | | | | | 113 | no(81) | 113 | no(81) | z2=1 | |
| | | | | | | | 67 | yes | 67 | yes | - | |
| | | | | | | | 59 | yes | 59 | yes | - | |
| 129.3.76 | 107641 | 16 | - | - | - - | - - | 115 | no(81) | 115 | no(81) | z2=1 | Bi1In1 |
| | | | | | | | 113 | no(81) | 113 | no(81) | z2=1 | |
| | | | | | | | 67 | yes | 67 | yes | - | |
| | | | | | | | 59 | yes | 59 | yes | - | |
| 129.3.77 | 107640 | 16 | - | - | - - | - - | 115 | no(81) | 115 | no(81) | z2=1 | Bi1In1 |
| | | | | | | | 113 | no(81) | 113 | no(81) | z2=1 | |
| | | | | | | | 67 | yes | 67 | yes | - | |
| | | | | | | | 59 | yes | 59 | yes | - | |
| 129.3.78 | 106325 | 16 | - | - | - - | - - | 115 | no(81) | 115 | no(81) | z2=1 | Bi1In1 |
| | | | | | | | 113 | no(81) | 113 | no(81) | z2=1 | |
| | | | | | | | 67 | yes | 67 | yes | - | |
| | | | | | | | 59 | yes | 59 | yes | - | |
| 129.3.79 | 187824 | 56 | - | - | - - | - - | 115 | yes | 115 | yes | - | Bi1Cu1O1Te1 |
| | | | | | | | 113 | yes | 113 | yes | - | |
| | | | | | | | 67 | no(2) | 67 | no(2) | z2w,3=1 z4=1 | |
| | | | | | | | 59 | no(2) | 59 | no(2) | z2w,3=1 z4=1 | |
| 129.3.80 | 159475 | 56 | - | - | - - | - - | 115 | yes | 115 | yes | - | Bi1Cu1O1Te1 |
| | | | | | | | 113 | yes | 113 | yes | - | |
| | | | | | | | 67 | no(2) | 67 | no(2) | z2w,3=1 z4=1 | |
| | | | | | | | 59 | no(2) | 59 | no(2) | z2w,3=1 z4=1 | |
| 129.3.81 | 42459 | 32 | - | - | - - | - - | 115 | yes | 115 | yes | - | Ba1Mg1Si1 |
| | | | | | | | 113 | yes | 113 | yes | - | |
| | | | | | | | 67 | no(2) | 67 | no(2) | z2w,3=1 z4=1 | |
| | | | | | | | 59 | no(2) | 59 | no(2) | z2w,3=1 z4=1 | |
| 129.3.82 | 238025 | 80 | - | - | - - | - - | 115 | no(81) | 115 | no(81) | z2=1 | Au2Ba1Sn2 |
| | | | | | | | 113 | no(81) | 113 | no(81) | z2=1 | |
| | | | | | | | 67 | no(2) | 67 | yes | z4=2 | |
| | | | | | | | 59 | no(2) | 59 | yes | z4=2 | |
| 129.3.83 | 238704 | 64 | - | - | - - | - - | 115 | yes | 115 | yes | - | Au1Bi2La1 |
| | | | | | | | 113 | yes | 113 | yes | - | |
| | | | | | | | 67 | no(2) | 67 | no(2) | z4=3 | |
| | | | | | | | 59 | no(2) | 59 | no(2) | z4=3 | |
| 129.3.84 | 42351 | 44 | - | - | - - | - - | 115 | no(81) | 115 | no(81) | z2=1 | As2Th1 |
| | | | | | | | 113 | no(81) | 113 | no(81) | z2=1 | |
| | | | | | | | 67 | no(2) | 67 | yes | z2w,3=1 z4=2 | |
| | | | | | | | 59 | no(2) | 59 | yes | z2w,3=1 z4=2 | |
| 129.3.85 | 196977 | 44 | - | - | - - | - - | 115 | no(81) | 115 | no(81) | z2=1 | As2Th1 |
| | | | | | | | 113 | no(81) | 113 | no(81) | z2=1 | |
| | | | | | | | 67 | no(2) | 67 | yes | z2w,3=1 z4=2 | |
| | | | | | | | 59 | no(2) | 59 | yes | z2w,3=1 z4=2 | |

Continued on next page

**TABLE XIV– continued**

| TQC | ICSD | NoE | DG | IndG | SubSG cl (w/ T) | | SubSG cl (w/o T) | | Max. Sub. SG | yes or no | Max. Sub. SG | yes or no | SIs (w/ T) | Stoich. Formula |
|---|---|---|---|---|---|---|---|---|---|---|---|---|---|---|
| 129.3.86 | 44032 | 68 | - | - | - | - | - | - | 115 | no(81) | 115 | no(81) | z2=1 | As2Ni1U1 |
| | | | | | | | | | 113 | no(81) | 113 | no(81) | z2=1 | |
| | | | | | | | | | 67 | no(2) | 67 | yes | z4=2 | |
| | | | | | | | | | 59 | no(2) | 59 | yes | z4=2 | |
| 129.3.87 | 44031 | 68 | - | - | - | - | - | - | 115 | no(81) | 115 | no(81) | z2=1 | As2Ni1U1 |
| | | | | | | | | | 113 | no(81) | 113 | no(81) | z2=1 | |
| | | | | | | | | | 67 | no(2) | 67 | yes | z4=2 | |
| | | | | | | | | | 59 | no(2) | 59 | yes | z4=2 | |
| 129.3.88 | 66117 | 88 | - | - | - | - | - | - | 115 | yes | 115 | yes | - | As2Nd1Ni2 |
| | | | | | | | | | 113 | yes | 113 | yes | - | |
| | | | | | | | | | 67 | no(2) | 67 | no(2) | z2w,3=1<br>z4=1 | |
| | | | | | | | | | 59 | no(2) | 59 | no(2) | z2w,3=1<br>z4=1 | |
| 129.3.89 | 186993 | 80 | - | - | - | - | - | - | 115 | yes | 115 | yes | - | As2Ce1Ir2 |
| | | | | | | | | | 113 | yes | 113 | yes | - | |
| | | | | | | | | | 67 | no(2) | 67 | no(2) | z2w,3=1<br>z4=3 | |
| | | | | | | | | | 59 | no(2) | 59 | no(2) | z2w,3=1<br>z4=3 | |
| 129.3.90 | 42355 | 28 | - | - | - | - | - | - | 115 | no(81) | 115 | no(81) | z2=1 | As1Si1Ta1 |
| | | | | | | | | | 113 | no(81) | 113 | no(81) | z2=1 | |
| | | | | | | | | | 67 | no(2) | 67 | yes | z2w,3=1<br>z4=2 | |
| | | | | | | | | | 59 | no(2) | 59 | yes | z2w,3=1<br>z4=2 | |
| 129.3.91 | 44037 | 52 | - | - | - | - | - | - | 115 | yes | 115 | yes | - | As1Np1Se1 |
| | | | | | | | | | 113 | yes | 113 | yes | - | |
| | | | | | | | | | 67 | no(2) | 67 | no(2) | z4=3 | |
| | | | | | | | | | 59 | no(2) | 59 | no(2) | z4=3 | |
| 129.3.92 | 44030 | 48 | - | - | - | - | - | - | 115 | yes | 115 | yes | - | As1Ni1Rh1 |
| | | | | | | | | | 113 | yes | 113 | yes | - | |
| | | | | | | | | | 67 | no(2) | 67 | no(2) | z2w,3=1<br>z4=1 | |
| | | | | | | | | | 59 | no(2) | 59 | no(2) | z2w,3=1<br>z4=1 | |
| 129.3.93 | 187596 | 48 | - | - | - | - | - | - | 115 | yes | 115 | yes | - | As1Ni1Rh1 |
| | | | | | | | | | 113 | yes | 113 | yes | - | |
| | | | | | | | | | 67 | no(2) | 67 | no(2) | z2w,3=1<br>z4=3 | |
| | | | | | | | | | 59 | no(2) | 59 | no(2) | z2w,3=1<br>z4=3 | |
| 129.3.94 | 187693 | 44 | - | - | - | - | - | - | 115 | no(81) | 115 | no(81) | z2=1 | As1Nb1Si1 |
| | | | | | | | | | 113 | no(81) | 113 | no(81) | z2=1 | |
| | | | | | | | | | 67 | no(2) | 67 | yes | z2w,3=1<br>z4=2 | |
| | | | | | | | | | 59 | no(2) | 59 | yes | z2w,3=1<br>z4=2 | |
| 129.3.95 | 16320 | 44 | - | - | - | - | - | - | 115 | no(81) | 115 | no(81) | z2=1 | As1Nb1Si1 |
| | | | | | | | | | 113 | no(81) | 113 | no(81) | z2=1 | |
| | | | | | | | | | 67 | no(2) | 67 | yes | z2w,3=1<br>z4=2 | |
| | | | | | | | | | 59 | no(2) | 59 | yes | z2w,3=1<br>z4=2 | |
| 129.3.96 | 163144 | 28 | - | - | - | - | - | - | 115 | yes | 115 | yes | - | As1Fe1Na1 |
| | | | | | | | | | 113 | yes | 113 | yes | - | |
| | | | | | | | | | 67 | no(2) | 67 | no(2) | z2w,3=1<br>z4=1 | |

Continued on next page

**TABLE XIV– continued**

| TQC | ICSD | NoE | DG | IndG | SubSG cl (w/ T) | SubSG cl (w/o T) | Max. Sub. SG | yes or no | Max. Sub. SG | yes or no | SIs (w/ T) | Stoich. Formula |
|---|---|---|---|---|---|---|---|---|---|---|---|---|
| | | | | | | | 59 | no(2) | 59 | no(2) | z2w,3=1<br>z4=1 | |
| 129.3.97 | 163143 | 28 | - | - | - - | - - | 115 | yes | 115 | yes | - | As1Fe1Na1 |
| | | | | | | | 113 | yes | 113 | yes | - | |
| | | | | | | | 67 | no(2) | 67 | no(2) | z2w,3=1<br>z4=1 | |
| | | | | | | | 59 | no(2) | 59 | no(2) | z2w,3=1<br>z4=1 | |
| 129.3.98 | 169175 | 28 | - | - | - - | - - | 115 | yes | 115 | yes | - | As1Fe1Li1 |
| | | | | | | | 113 | yes | 113 | yes | - | |
| | | | | | | | 67 | no(2) | 67 | no(2) | z2w,3=1<br>z4=1 | |
| | | | | | | | 59 | no(2) | 59 | no(2) | z2w,3=1<br>z4=1 | |
| 129.3.99 | 163870 | 28 | - | - | - - | - - | 115 | yes | 115 | yes | - | As1Fe1Li1 |
| | | | | | | | 113 | yes | 113 | yes | - | |
| | | | | | | | 67 | no(2) | 67 | no(2) | z2w,3=1<br>z4=3 | |
| | | | | | | | 59 | no(2) | 59 | no(2) | z2w,3=1<br>z4=3 | |
| 129.3.100 | 162250 | 28 | - | - | - - | - - | 115 | yes | 115 | yes | - | As1Fe1Li1 |
| | | | | | | | 113 | yes | 113 | yes | - | |
| | | | | | | | 67 | no(2) | 67 | no(2) | z2w,3=1<br>z4=3 | |
| | | | | | | | 59 | no(2) | 59 | no(2) | z2w,3=1<br>z4=3 | |
| 129.3.101 | 162056 | 28 | - | - | - - | - - | 115 | yes | 115 | yes | - | As1Fe1Li1 |
| | | | | | | | 113 | yes | 113 | yes | - | |
| | | | | | | | 67 | no(2) | 67 | no(2) | z2w,3=1<br>z4=1 | |
| | | | | | | | 59 | no(2) | 59 | no(2) | z2w,3=1<br>z4=1 | |
| 129.3.102 | 42324 | 36 | - | - | - - | - - | 115 | yes | 115 | yes | - | As1Cu1Mg1 |
| | | | | | | | 113 | yes | 113 | yes | - | |
| | | | | | | | 67 | no(2) | 67 | no(2) | z4=3 | |
| | | | | | | | 59 | no(2) | 59 | no(2) | z4=3 | |
| 129.3.103 | 187150 | 36 | - | - | - - | - - | 115 | yes | 115 | yes | - | Ag1Mg1Sb1 |
| | | | | | | | 113 | yes | 113 | yes | - | |
| | | | | | | | 67 | no(2) | 67 | no(2) | z4=3 | |
| | | | | | | | 59 | no(2) | 59 | no(2) | z4=3 | |
| 129.4.1 | 183309 | 62 | - | - | - - | - - | See details in Table XIX | | | | | Sn2Y1Zn1 |
| 129.4.2 | 163425 | 78 | - | - | - - | - - | See details in Table XIX | | | | | Sn2Tb1Zn1 |
| 129.4.3 | 42337 | 26 | - | - | - - | - - | See details in Table XIX | | | | | Si1V1Zr1 |
| 129.4.4 | 88113 | 38 | - | - | - - | - - | See details in Table XIX | | | | | Si1Ti1Y1 |
| 129.4.5 | 88115 | 54 | - | - | - - | - - | See details in Table XIX | | | | | Si1Tb1Ti1 |
| 129.4.6 | 26274 | 50 | - | - | - - | - - | See details in Table XIX | | | | | Sb1Te1U1 |
| 129.4.7 | 42486 | 50 | - | - | - - | - - | See details in Table XIX | | | | | Sb1Se1U1 |
| 129.4.8 | 409586 | 22 | - | - | - - | - - | See details in Table XIX | | | | | Sb1Sc2 |
| 129.4.9 | 193328 | 54 | - | - | - - | - - | See details in Table XIX | | | | | Sb1Sc1Tb1 |
| 129.4.10 | 41253 | 50 | - | - | - - | - - | See details in Table XIX | | | | | Pr1Ru1Si1 |
| 129.4.11 | 93154 | 66 | - | - | - - | - - | See details in Table XIX | | | | | Pd1Pr1Sb2 |
| 129.4.12 | 63594 | 38 | - | - | - - | - - | See details in Table XIX | | | | | P1Sn1Sr1 |
| 129.4.13 | 42169 | 50 | - | - | - - | - - | See details in Table XIX | | | | | P1Se1U1 |
| 129.4.14 | 77793 | 50 | - | - | - - | - - | See details in Table XIX | | | | | P1S1U1 |
| 129.4.15 | 57162 | 50 | - | - | - - | - - | See details in Table XIX | | | | | P1S1U1 |
| 129.4.16 | 154031 | 46 | - | - | - - | - - | See details in Table XIX | | | | | P1S1Th1 |
| 129.4.17 | 73036 | 86 | - | - | - - | - - | See details in Table XIX | | | | | Np1Pt2Si2 |
| 129.4.18 | 421034 | 54 | - | - | - - | - - | See details in Table XIX | | | | | Np1O1Se1 |

Continued on next page

**TABLE XIV– continued**

| TQC | ICSD | NoE | DG | IndG | SubSG cl (w/ T) | SubSG cl (w/o T) | Max. Sub. SG yes or no | Max. Sub. SG yes or no | SIs (w/ T) | Stoich. Formula |
|---|---|---|---|---|---|---|---|---|---|---|
| 129.4.19 | 421032 | 54 | - | - | - - | - - | See details in Table XIX | | | Np1O1S1 |
| 129.4.20 | 31656 | 54 | - | - | - - | - - | See details in Table XIX | | | Np1O1S1 |
| 129.4.21 | 419354 | 74 | - | - | - - | - - | See details in Table XIX | | | Nd1O1Sb1Zn1 |
| 129.4.22 | 82004 | 86 | - | - | - - | - - | See details in Table XIX | | | Na1Nd1O4Ti1 |
| 129.4.23 | 422049 | 86 | - | - | - - | - - | See details in Table XIX | | | Na1Nd1O4Ti1 |
| 129.4.24 | 87145 | 50 | - | - | - - | - - | See details in Table XIX | | | N1Te1U1 |
| 129.4.25 | 105125 | 50 | - | - | - - | - - | See details in Table XIX | | | N1Te1U1 |
| 129.4.26 | 105124 | 50 | - | - | - - | - - | See details in Table XIX | | | N1Te1U1 |
| 129.4.27 | 76214 | 38 | - | - | - - | - - | See details in Table XIX | | | Mn2Sb1 |
| 129.4.28 | 45688 | 38 | - | - | - - | - - | See details in Table XIX | | | Mn2Sb1 |
| 129.4.29 | 42325 | 38 | - | - | - - | - - | See details in Table XIX | | | Mn2Sb1 |
| 129.4.30 | 415627 | 38 | - | - | - - | - - | See details in Table XIX | | | Mn2Sb1 |
| 129.4.31 | 190141 | 38 | - | - | - - | - - | See details in Table XIX | | | Mn2Sb1 |
| 129.4.32 | 174032 | 38 | - | - | - - | - - | See details in Table XIX | | | Mn2Sb1 |
| 129.4.33 | 66963 | 42 | - | - | - - | - - | See details in Table XIX | | | Mn1Sn1Sr1 |
| 129.4.34 | 66962 | 42 | - | - | - - | - - | See details in Table XIX | | | Mn1Sn1Sr1 |
| 129.4.35 | 66951 | 42 | - | - | - - | - - | See details in Table XIX | | | Mn1Sn1Sr1 |
| 129.4.36 | 41868 | 42 | - | - | - - | - - | See details in Table XIX | | | Mn1Rb1Sb1 |
| 129.4.37 | 89595 | 42 | - | - | - - | - - | See details in Table XIX | | | Mn1P1Rb1 |
| 129.4.38 | 89594 | 42 | - | - | - - | - - | See details in Table XIX | | | Mn1P1Rb1 |
| 129.4.39 | 85851 | 50 | - | - | - - | - - | See details in Table XIX | | | Mn1Nd1Si1 |
| 129.4.40 | 75057 | 50 | - | - | - - | - - | See details in Table XIX | | | Mn1Nd1Si1 |
| 129.4.41 | 75056 | 50 | - | - | - - | - - | See details in Table XIX | | | Mn1Nd1Si1 |
| 129.4.42 | 60735 | 26 | - | - | - - | - - | See details in Table XIX | | | Mn1Na1Sb1 |
| 129.4.43 | 60734 | 26 | - | - | - - | - - | See details in Table XIX | | | Mn1Na1Sb1 |
| 129.4.44 | 60733 | 26 | - | - | - - | - - | See details in Table XIX | | | Mn1Na1Sb1 |
| 129.4.45 | 26462 | 26 | - | - | - - | - - | See details in Table XIX | | | Mn1Na1Sb1 |
| 129.4.46 | 60729 | 26 | - | - | - - | - - | See details in Table XIX | | | Mn1Na1P1 |
| 129.4.47 | 60728 | 26 | - | - | - - | - - | See details in Table XIX | | | Mn1Na1P1 |
| 129.4.48 | 60727 | 26 | - | - | - - | - - | See details in Table XIX | | | Mn1Na1P1 |
| 129.4.49 | 60726 | 26 | - | - | - - | - - | See details in Table XIX | | | Mn1Na1P1 |
| 129.4.50 | 26460 | 26 | - | - | - - | - - | See details in Table XIX | | | Mn1Na1P1 |
| 129.4.51 | 183314 | 90 | - | - | - - | - - | See details in Table XIX | | | Lu1Sn2Zn1 |
| 129.4.52 | 88208 | 66 | - | - | - - | - - | See details in Table XIX | | | Lu1Si1Ti1 |
| 129.4.53 | 36019 | 50 | - | - | - - | - - | See details in Table XIX | | | Li2Pr1Sb2 |
| 129.4.54 | 86775 | 82 | - | - | - - | - - | See details in Table XIX | | | Li2O5Si1V1 |
| 129.4.55 | 59359 | 82 | - | - | - - | - - | See details in Table XIX | | | Li2O5Si1V1 |
| 129.4.56 | 417286 | 82 | - | - | - - | - - | See details in Table XIX | | | Li2O5Si1V1 |
| 129.4.57 | 417285 | 82 | - | - | - - | - - | See details in Table XIX | | | Li2O5Si1V1 |
| 129.4.58 | 417284 | 82 | - | - | - - | - - | See details in Table XIX | | | Li2O5Si1V1 |
| 129.4.59 | 417283 | 82 | - | - | - - | - - | See details in Table XIX | | | Li2O5Si1V1 |
| 129.4.60 | 417282 | 82 | - | - | - - | - - | See details in Table XIX | | | Li2O5Si1V1 |
| 129.4.61 | 417281 | 82 | - | - | - - | - - | See details in Table XIX | | | Li2O5Si1V1 |
| 129.4.62 | 417280 | 82 | - | - | - - | - - | See details in Table XIX | | | Li2O5Si1V1 |
| 129.4.63 | 417279 | 82 | - | - | - - | - - | See details in Table XIX | | | Li2O5Si1V1 |
| 129.4.64 | 417278 | 82 | - | - | - - | - - | See details in Table XIX | | | Li2O5Si1V1 |
| 129.4.65 | 417277 | 82 | - | - | - - | - - | See details in Table XIX | | | Li2O5Si1V1 |
| 129.4.66 | 417276 | 82 | - | - | - - | - - | See details in Table XIX | | | Li2O5Si1V1 |
| 129.4.67 | 26458 | 26 | - | - | - - | - - | See details in Table XIX | | | Li1Mn1P1 |
| 129.4.68 | 43279 | 46 | - | - | - - | - - | See details in Table XIX | | | La1Te2 |
| 129.4.69 | 42342 | 46 | - | - | - - | - - | See details in Table XIX | | | La1Te2 |
| 129.4.70 | 23957 | 46 | - | - | - - | - - | See details in Table XIX | | | La1Te2 |
| 129.4.71 | 85855 | 46 | - | - | - - | - - | See details in Table XIX | | | La1Ru1Si1 |
| 129.4.72 | 41249 | 46 | - | - | - - | - - | See details in Table XIX | | | La1Ru1Si1 |
| 129.4.73 | 189724 | 82 | - | - | - - | - - | See details in Table XIX | | | La1Pd2Sb2 |
| 129.4.74 | 419355 | 58 | - | - | - - | - - | See details in Table XIX | | | La1Mn1O1Sb1 |
| 129.4.75 | 419856 | 118 | - | - | - - | - - | See details in Table XIX | | | K1Na1Nd1O5Ta1 |

Continued on next page

**TABLE XIV– continued**

| TQC | ICSD | NoE | DG | IndG | SubSG (w/ T) | cl | SubSG (w/o T) | cl | Max. Sub. SG | yes or no | Max. Sub. SG | yes or no | SIs (w/ T) | Stoich. Formula |
|---|---|---|---|---|---|---|---|---|---|---|---|---|---|---|
| 129.4.76 | 420038 | 134 | - | - | - | - | - | - | See details in Table XIX | | | | | K1Na1Nb1Nd1O5 |
| 129.4.77 | 89591 | 42 | - | - | - | - | - | - | See details in Table XIX | | | | | K1Mn1Sb1 |
| 129.4.78 | 89590 | 42 | - | - | - | - | - | - | See details in Table XIX | | | | | K1Mn1Sb1 |
| 129.4.79 | 89593 | 42 | - | - | - | - | - | - | See details in Table XIX | | | | | K1Mn1P1 |
| 129.4.80 | 89592 | 42 | - | - | - | - | - | - | See details in Table XIX | | | | | K1Mn1P1 |
| 129.4.81 | 41882 | 42 | - | - | - | - | - | - | See details in Table XIX | | | | | K1Mn1P1 |
| 129.4.82 | 44668 | 74 | - | - | - | - | - | - | See details in Table XIX | | | | | Ir2Si2Y1 |
| 129.4.83 | 188325 | 74 | - | - | - | - | - | - | See details in Table XIX | | | | | Ir2Si2Y1 |
| 129.4.84 | 73035 | 82 | - | - | - | - | - | - | See details in Table XIX | | | | | Ir2Np1Si2 |
| 129.4.85 | 44658 | 74 | - | - | - | - | - | - | See details in Table XIX | | | | | Ir2La1Si2 |
| 129.4.86 | 188327 | 74 | - | - | - | - | - | - | See details in Table XIX | | | | | Ir2La1Si2 |
| 129.4.87 | 38204 | 58 | - | - | - | - | - | - | See details in Table XIX | | | | | I1O1Pu1 |
| 129.4.88 | 31663 | 58 | - | - | - | - | - | - | See details in Table XIX | | | | | I1O1Pu1 |
| 129.4.89 | 183311 | 82 | - | - | - | - | - | - | See details in Table XIX | | | | | Ho1Sn2Zn1 |
| 129.4.90 | 88117 | 58 | - | - | - | - | - | - | See details in Table XIX | | | | | Ho1Si1Ti1 |
| 129.4.91 | 97037 | 258 | - | - | - | - | - | - | See details in Table XIX | | | | | Ho1Ni10Si2 |
| 129.4.92 | 263028 | 34 | - | - | - | - | - | - | See details in Table XIX | | | | | H2Ca2Ge2Ni2 |
| 129.4.93 | 428162 | 50 | - | - | - | - | - | - | See details in Table XIX | | | | | H1Cl1Eu1 |
| 129.4.94 | 161825 | 50 | - | - | - | - | - | - | See details in Table XIX | | | | | H1Ce1Ru1Si1 |
| 129.4.95 | 416423 | 78 | - | - | - | - | - | - | See details in Table XIX | | | | | Ge2La1Pt2 |
| 129.4.96 | 180145 | 78 | - | - | - | - | - | - | See details in Table XIX | | | | | Ge2Ir2Pr1 |
| 129.4.97 | 90233 | 38 | - | - | - | - | - | - | See details in Table XIX | | | | | Ge1Ti1Y1 |
| 129.4.98 | 85904 | 38 | - | - | - | - | - | - | See details in Table XIX | | | | | Ge1Ti1Y1 |
| 129.4.99 | 90393 | 54 | - | - | - | - | - | - | See details in Table XIX | | | | | Ge1Tb1Ti1 |
| 129.4.100 | 85906 | 54 | - | - | - | - | - | - | See details in Table XIX | | | | | Ge1Tb1Ti1 |
| 129.4.101 | 85000 | 26 | - | - | - | - | - | - | See details in Table XIX | | | | | Ge1Sb1Zr1 |
| 129.4.102 | 260557 | 26 | - | - | - | - | - | - | See details in Table XIX | | | | | Ge1Sb1Ti1 |
| 129.4.103 | 66961 | 42 | - | - | - | - | - | - | See details in Table XIX | | | | | Ge1Mn1Sr1 |
| 129.4.104 | 66960 | 42 | - | - | - | - | - | - | See details in Table XIX | | | | | Ge1Mn1Sr1 |
| 129.4.105 | 53707 | 42 | - | - | - | - | - | - | See details in Table XIX | | | | | Ge1Mn1Sr1 |
| 129.4.106 | 80016 | 50 | - | - | - | - | - | - | See details in Table XIX | | | | | Ge1Mn1Nd1 |
| 129.4.107 | 80014 | 50 | - | - | - | - | - | - | See details in Table XIX | | | | | Ge1Mn1Nd1 |
| 129.4.108 | 87117 | 26 | - | - | - | - | - | - | See details in Table XIX | | | | | Ge1Mg1Mn1 |
| 129.4.109 | 66953 | 26 | - | - | - | - | - | - | See details in Table XIX | | | | | Ge1Mg1Mn1 |
| 129.4.110 | 66952 | 26 | - | - | - | - | - | - | See details in Table XIX | | | | | Ge1Mg1Mn1 |
| 129.4.111 | 66948 | 26 | - | - | - | - | - | - | See details in Table XIX | | | | | Ge1Mg1Mn1 |
| 129.4.112 | 88209 | 66 | - | - | - | - | - | - | See details in Table XIX | | | | | Ge1Lu1Ti1 |
| 129.4.113 | 86777 | 82 | - | - | - | - | - | - | See details in Table XIX | | | | | Ge1Li2O5V1 |
| 129.4.114 | 107617 | 38 | - | - | - | - | - | - | See details in Table XIX | | | | | Ge1La1Ti1 |
| 129.4.115 | 85862 | 46 | - | - | - | - | - | - | See details in Table XIX | | | | | Ge1La1Ru1 |
| 129.4.116 | 95023 | 58 | - | - | - | - | - | - | See details in Table XIX | | | | | Ge1Ho1Ti1 |
| 129.4.117 | 95021 | 58 | - | - | - | - | - | - | See details in Table XIX | | | | | Ge1Ho1Ti1 |
| 129.4.118 | 82006 | 94 | - | - | - | - | - | - | See details in Table XIX | | | | | Gd1Na1O4Ti1 |
| 129.4.119 | 88277 | 58 | - | - | - | - | - | - | See details in Table XIX | | | | | Gd1Mn1Si1 |
| 129.4.120 | 85852 | 58 | - | - | - | - | - | - | See details in Table XIX | | | | | Gd1Mn1Si1 |
| 129.4.121 | 419854 | 126 | - | - | - | - | - | - | See details in Table XIX | | | | | Gd1K1Na1O5Ta1 |
| 129.4.122 | 420037 | 142 | - | - | - | - | - | - | See details in Table XIX | | | | | Gd1K1Na1Nb1O5 |
| 129.4.123 | 191057 | 74 | - | - | - | - | - | - | See details in Table XIX | | | | | Ga2La1Pd2 |
| 129.4.124 | 57297 | 62 | - | - | - | - | - | - | See details in Table XIX | | | | | Fe1Si1Tb1 |
| 129.4.125 | 57237 | 62 | - | - | - | - | - | - | See details in Table XIX | | | | | Fe1Si1Tb1 |
| 129.4.126 | 57236 | 62 | - | - | - | - | - | - | See details in Table XIX | | | | | Fe1Si1Tb1 |
| 129.4.127 | 194319 | 38 | - | - | - | - | - | - | See details in Table XIX | | | | | Fe1N1Se1 |
| 129.4.128 | 85853 | 46 | - | - | - | - | - | - | See details in Table XIX | | | | | Fe1La1Si1 |
| 129.4.129 | 76428 | 58 | - | - | - | - | - | - | See details in Table XIX | | | | | F1O1Pu1 |
| 129.4.130 | 108969 | 54 | - | - | - | - | - | - | See details in Table XIX | | | | | F1Nd1Te1 |
| 129.4.131 | 94471 | 54 | - | - | - | - | - | - | See details in Table XIX | | | | | F1Nd1S1 |
| 129.4.132 | 189941 | 58 | - | - | - | - | - | - | See details in Table XIX | | | | | F1Mn1Sb1Sr1 |

Continued on next page

**TABLE XIV– continued**

| TQC | ICSD | NoE | DG | IndG | SubSG cl (w/ T) | SubSG cl (w/o T) | Max. Sub. SG yes or no | Max. Sub. SG yes or no | SIs (w/ T) | Stoich. Formula |
|---|---|---|---|---|---|---|---|---|---|---|
| 129.4.133 | 189939 | 58 | - | - | - - | - - | See details in Table XIX | | | F1Mn1P1Sr1 |
| 129.4.134 | 94473 | 62 | - | - | - - | - - | See details in Table XIX | | | F1Gd1S1 |
| 129.4.135 | 93355 | 62 | - | - | - - | - - | See details in Table XIX | | | F1Gd1S1 |
| 129.4.136 | 88460 | 106 | - | - | - - | - - | See details in Table XIX | | | F1Fe1O3Sr2 |
| 129.4.137 | 162267 | 98 | - | - | - - | - - | See details in Table XIX | | | Eu1Sn2Zn2 |
| 129.4.138 | 427712 | 90 | - | - | - - | - - | See details in Table XIX | | | Eu1Pd2Sn2 |
| 129.4.139 | 51757 | 98 | - | - | - - | - - | See details in Table XIX | | | Eu1Ge2Zn2 |
| 129.4.140 | 94476 | 70 | - | - | - - | - - | See details in Table XIX | | | Er1F1S1 |
| 129.4.141 | 103133 | 82 | - | - | - - | - - | See details in Table XIX | | | Cu3Ti2 |
| 129.4.142 | 42323 | 54 | - | - | - - | - - | See details in Table XIX | | | Cu2Sb1 |
| 129.4.143 | 412295 | 54 | - | - | - - | - - | See details in Table XIX | | | Cu2Sb1 |
| 129.4.144 | 16843 | 54 | - | - | - - | - - | See details in Table XIX | | | Cu2Sb1 |
| 129.4.145 | 103041 | 82 | - | - | - - | - - | See details in Table XIX | | | Cu2La1Sn2 |
| 129.4.146 | 77197 | 98 | - | - | - - | - - | See details in Table XIX | | | Cu2Eu1Sb2 |
| 129.4.147 | 103128 | 30 | - | - | - - | - - | See details in Table XIX | | | Cu1Ti1 |
| 129.4.148 | 53352 | 46 | - | - | - - | - - | See details in Table XIX | | | Cu1Si2Zr1 |
| 129.4.149 | 261278 | 74 | - | - | - - | - - | See details in Table XIX | | | Cu1Np1O1P1 |
| 129.4.150 | 416523 | 74 | - | - | - - | - - | See details in Table XIX | | | Cu1Nd1O1Te1 |
| 129.4.151 | 96346 | 74 | - | - | - - | - - | See details in Table XIX | | | Cu1Nd1O1S1 |
| 129.4.152 | 87174 | 46 | - | - | - - | - - | See details in Table XIX | | | Cu1Hf1Si2 |
| 129.4.153 | 153568 | 78 | - | - | - - | - - | See details in Table XIX | | | Cu1Gd1Sb2 |
| 129.4.154 | 80358 | 82 | - | - | - - | - - | See details in Table XIX | | | Cu1Gd1O1Se1 |
| 129.4.155 | 84963 | 126 | - | - | - - | - - | See details in Table XIX | | | Cu1Fe1O3S1Sr2 |
| 129.4.156 | 195375 | 126 | - | - | - - | - - | See details in Table XIX | | | Cu1Fe1O3S1Sr2 |
| 129.4.157 | 419381 | 82 | - | - | - - | - - | See details in Table XIX | | | Cu1Eu1F1Te1 |
| 129.4.158 | 172799 | 82 | - | - | - - | - - | See details in Table XIX | | | Cu1Eu1F1Se1 |
| 129.4.159 | 172797 | 82 | - | - | - - | - - | See details in Table XIX | | | Cu1Eu1F1S1 |
| 129.4.160 | - 27 | 82 | - | - | - - | - - | See details in Table XIX | | | Cu1Dy1Sb2 |
| 129.4.161 | 80359 | 86 | - | - | - - | - - | See details in Table XIX | | | Cu1Dy1O1Se1 |
| 129.4.162 | 41869 | 42 | - | - | - - | - - | See details in Table XIX | | | Cs1Mn1Sb1 |
| 129.4.163 | 89599 | 42 | - | - | - - | - - | See details in Table XIX | | | Cs1Mn1P1 |
| 129.4.164 | 89598 | 42 | - | - | - - | - - | See details in Table XIX | | | Cs1Mn1P1 |
| 129.4.165 | 84962 | 122 | - | - | - - | - - | See details in Table XIX | | | Cr1Cu1O3S1Sr2 |
| 129.4.166 | 80514 | 54 | - | - | - - | - - | See details in Table XIX | | | Co1Nd1Si1 |
| 129.4.167 | 80203 | 62 | - | - | - - | - - | See details in Table XIX | | | Co1La1O1P1 |
| 129.4.168 | 161911 | 62 | - | - | - - | - - | See details in Table XIX | | | Co1La1O1P1 |
| 129.4.169 | 85861 | 54 | - | - | - - | - - | See details in Table XIX | | | Co1Ge1Nd1 |
| 129.4.170 | 106473 | 54 | - | - | - - | - - | See details in Table XIX | | | Co1Ge1Nd1 |
| 129.4.171 | 418853 | 30 | - | - | - - | - - | See details in Table XIX | | | Co1Ge1Mg1 |
| 129.4.172 | 31661 | 58 | - | - | - - | - - | See details in Table XIX | | | Cl1O1Pu1 |
| 129.4.173 | 237581 | 110 | - | - | - - | - - | See details in Table XIX | | | Cl1Ni1O3Sr2 |
| 129.4.174 | 426777 | 54 | - | - | - - | - - | See details in Table XIX | | | Cl1Nd1Te1 |
| 129.4.175 | 59231 | 54 | - | - | - - | - - | See details in Table XIX | | | Cl1Nd1O1 |
| 129.4.176 | 31665 | 54 | - | - | - - | - - | See details in Table XIX | | | Cl1Nd1O1 |
| 129.4.177 | 50989 | 114 | - | - | - - | - - | See details in Table XIX | | | Cl1Mo1O5P1Rb1 |
| 129.4.178 | 50988 | 114 | - | - | - - | - - | See details in Table XIX | | | Cl1K1Mo1O5P1 |
| 129.4.179 | 77820 | 62 | - | - | - - | - - | See details in Table XIX | | | Cl1Gd1O1 |
| 129.4.180 | 59232 | 62 | - | - | - - | - - | See details in Table XIX | | | Cl1Gd1O1 |
| 129.4.181 | 80798 | 90 | - | - | - - | - - | See details in Table XIX | | | Cl1Fe1O4W1 |
| 129.4.182 | 93508 | 106 | - | - | - - | - - | See details in Table XIX | | | Cl1Fe1O3Sr2 |
| 129.4.183 | 49516 | 90 | - | - | - - | - - | See details in Table XIX | | | Cl1Fe1Mo1O4 |
| 129.4.184 | 202188 | 90 | - | - | - - | - - | See details in Table XIX | | | Cl1Fe1Mo1O4 |
| 129.4.185 | 202187 | 90 | - | - | - - | - - | See details in Table XIX | | | Cl1Fe1Mo1O4 |
| 129.4.186 | 202186 | 90 | - | - | - - | - - | See details in Table XIX | | | Cl1Fe1Mo1O4 |
| 129.4.187 | 108938 | 62 | - | - | - - | - - | See details in Table XIX | | | Cl1Eu1F1 |
| 129.4.188 | 85449 | 70 | - | - | - - | - - | See details in Table XIX | | | Ce1O1Sb1Zn1 |

Continued on next page

**TABLE XIV– continued**

| TQC | ICSD | NoE | DG | IndG | SubSG cl (w/ T) | SubSG cl (w/o T) | Max. Sub. SG yes or no | Max. Sub. SG yes or no | SIs (w/ T) | Stoich. Formula |
|---|---|---|---|---|---|---|---|---|---|---|
| 129.4.189 | 80201 | 62 | - | - | - - | - - | See details in Table XIX | | | Ce1O1P1Ru1 |
| 129.4.190 | 85850 | 46 | - | - | - - | - - | See details in Table XIX | | | Ce1Mn1Si1 |
| 129.4.191 | 75051 | 46 | - | - | - - | - - | See details in Table XIX | | | Ce1Mn1Si1 |
| 129.4.192 | 75050 | 46 | - | - | - - | - - | See details in Table XIX | | | Ce1Mn1Si1 |
| 129.4.193 | 75049 | 46 | - | - | - - | - - | See details in Table XIX | | | Ce1Mn1Si1 |
| 129.4.194 | 75048 | 46 | - | - | - - | - - | See details in Table XIX | | | Ce1Mn1Si1 |
| 129.4.195 | 85858 | 46 | - | - | - - | - - | See details in Table XIX | | | Ce1Ge1Mn1 |
| 129.4.196 | 80010 | 46 | - | - | - - | - - | See details in Table XIX | | | Ce1Ge1Mn1 |
| 129.4.197 | 80009 | 46 | - | - | - - | - - | See details in Table XIX | | | Ce1Ge1Mn1 |
| 129.4.198 | 94469 | 50 | - | - | - - | - - | See details in Table XIX | | | Ce1F1S1 |
| 129.4.199 | 93477 | 50 | - | - | - - | - - | See details in Table XIX | | | Ce1F1S1 |
| 129.4.200 | 31939 | 50 | - | - | - - | - - | See details in Table XIX | | | Ce1F1S1 |
| 129.4.201 | 106383 | 66 | - | - | - - | - - | See details in Table XIX | | | Ce1Cu1Sb2 |
| 129.4.202 | 416521 | 70 | - | - | - - | - - | See details in Table XIX | | | Ce1Cu1O1Te1 |
| 129.4.203 | 96344 | 70 | - | - | - - | - - | See details in Table XIX | | | Ce1Cu1O1S1 |
| 129.4.204 | 106373 | 50 | - | - | - - | - - | See details in Table XIX | | | Ce1Co1Ge1 |
| 129.4.205 | 426282 | 50 | - | - | - - | - - | See details in Table XIX | | | Ce1Cl1Te1 |
| 129.4.206 | 72154 | 50 | - | - | - - | - - | See details in Table XIX | | | Ce1Cl1O1 |
| 129.4.207 | 412069 | 50 | - | - | - - | - - | See details in Table XIX | | | Ce1Cl1O1 |
| 129.4.208 | 169993 | 94 | - | - | - - | - - | See details in Table XIX | | | Ca2Cu1Fe1O3Se1 |
| 129.4.209 | 195376 | 94 | - | - | - - | - - | See details in Table XIX | | | Ca2Cu1Fe1O3S1 |
| 129.4.210 | 169992 | 94 | - | - | - - | - - | See details in Table XIX | | | Ca2Cu1Fe1O3S1 |
| 129.4.211 | 66959 | 26 | - | - | - - | - - | See details in Table XIX | | | Ca1Mn1Sn1 |
| 129.4.212 | 66958 | 26 | - | - | - - | - - | See details in Table XIX | | | Ca1Mn1Sn1 |
| 129.4.213 | 66950 | 26 | - | - | - - | - - | See details in Table XIX | | | Ca1Mn1Sn1 |
| 129.4.214 | 66955 | 26 | - | - | - - | - - | See details in Table XIX | | | Ca1Mn1Si1 |
| 129.4.215 | 66954 | 26 | - | - | - - | - - | See details in Table XIX | | | Ca1Mn1Si1 |
| 129.4.216 | 66949 | 26 | - | - | - - | - - | See details in Table XIX | | | Ca1Mn1Si1 |
| 129.4.217 | 66957 | 26 | - | - | - - | - - | See details in Table XIX | | | Ca1Ge1Mn1 |
| 129.4.218 | 66956 | 26 | - | - | - - | - - | See details in Table XIX | | | Ca1Ge1Mn1 |
| 129.4.219 | 52758 | 26 | - | - | - - | - - | See details in Table XIX | | | Ca1Ge1Mn1 |
| 129.4.220 | 428148 | 30 | - | - | - - | - - | See details in Table XIX | | | Ca1Co1Si1 |
| 129.4.221 | 248426 | 78 | - | - | - - | - - | See details in Table XIX | | | C1B1Er1Ni1 |
| 129.4.222 | 28532 | 74 | - | - | - - | - - | See details in Table XIX | | | Br1O1Yb1 |
| 129.4.223 | 31662 | 58 | - | - | - - | - - | See details in Table XIX | | | Br1O1Pu1 |
| 129.4.224 | 31666 | 54 | - | - | - - | - - | See details in Table XIX | | | Br1Nd1O1 |
| 129.4.225 | 23955 | 54 | - | - | - - | - - | See details in Table XIX | | | Br1Nd1O1 |
| 129.4.226 | 41071 | 62 | - | - | - - | - - | See details in Table XIX | | | Br1Gd1O1 |
| 129.4.227 | 93509 | 106 | - | - | - - | - - | See details in Table XIX | | | Br1Fe1O3Sr2 |
| 129.4.228 | 10454 | 38 | - | - | - - | - - | See details in Table XIX | | | Bi2Ca1Mn1 |
| 129.4.229 | 60738 | 26 | - | - | - - | - - | See details in Table XIX | | | Bi1Mn1Na1 |
| 129.4.230 | 60737 | 26 | - | - | - - | - - | See details in Table XIX | | | Bi1Mn1Na1 |
| 129.4.231 | 60736 | 26 | - | - | - - | - - | See details in Table XIX | | | Bi1Mn1Na1 |
| 129.4.232 | 80 | 70 | - | - | - - | - - | See details in Table XIX | | | Bi1Ce1O1S2 |
| 129.4.233 | 79261 | 94 | - | - | - - | - - | See details in Table XIX | | | Ba2F1In1O3 |
| 129.4.234 | 88950 | 130 | - | - | - - | - - | See details in Table XIX | | | Ba1Mn2O5Y1 |
| 129.4.235 | 88949 | 130 | - | - | - - | - - | See details in Table XIX | | | Ba1Mn2O5Y1 |
| 129.4.236 | 66965 | 42 | - | - | - - | - - | See details in Table XIX | | | Ba1Ge1Mn1 |
| 129.4.237 | 66964 | 42 | - | - | - - | - - | See details in Table XIX | | | Ba1Ge1Mn1 |
| 129.4.238 | 42370 | 42 | - | - | - - | - - | See details in Table XIX | | | Ba1Ge1Mn1 |
| 129.4.239 | 248366 | 58 | - | - | - - | - - | See details in Table XIX | | | Ba1F1Mn1Sb1 |
| 129.4.240 | 248364 | 58 | - | - | - - | - - | See details in Table XIX | | | Ba1F1Mn1Sb1 |
| 129.4.241 | 82639 | 62 | - | - | - - | - - | See details in Table XIX | | | Ba1Co1S2 |
| 129.4.242 | 426033 | 58 | - | - | - - | - - | See details in Table XIX | | | Ba1Bi1F1Mn1 |
| 129.4.243 | 77939 | 58 | - | - | - - | - - | See details in Table XIX | | | B1La1N1Ni1 |
| 129.4.244 | 25535 | 22 | - | - | - - | - - | See details in Table XIX | | | B1Be4 |
| 129.4.245 | 262387 | 94 | - | - | - - | - - | See details in Table XIX | | | Au3Ge1Sr1 |

Continued on next page

**TABLE XIV– continued**

| TQC | ICSD | NoE | DG | IndG | SubSG cl (w/ T) | SubSG cl (w/o T) | Max. Sub. SG yes or no | Max. Sub. SG yes or no | SIs (w/ T) | Stoich. Formula |
|---|---|---|---|---|---|---|---|---|---|---|
| 129.4.246 | 262386 | 94 | - | - | - - | - - | See details in Table XIX | | | Au3Ge1Sr1 |
| 129.4.247 | 262385 | 94 | - | - | - - | - - | See details in Table XIX | | | Au3Ge1Sr1 |
| 129.4.248 | 157378 | 70 | - | - | - - | - - | See details in Table XIX | | | Au1Sb2U1 |
| 129.4.249 | 238705 | 66 | - | - | - - | - - | See details in Table XIX | | | Au1Bi2Ce1 |
| 129.4.250 | 75132 | 142 | - | - | - - | - - | See details in Table XIX | | | As3Cu2O1U2 |
| 129.4.251 | 44035 | 50 | - | - | - - | - - | See details in Table XIX | | | As2Np1 |
| 129.4.252 | 68147 | 82 | - | - | - - | - - | See details in Table XIX | | | As2La1Ni2 |
| 129.4.253 | 186992 | 78 | - | - | - - | - - | See details in Table XIX | | | As2Ir2La1 |
| 129.4.254 | 412193 | 90 | - | - | - - | - - | See details in Table XIX | | | As2Cu1Yb1 |
| 129.4.255 | 40782 | 70 | - | - | - - | - - | See details in Table XIX | | | As2Cu1U1 |
| 129.4.256 | 91305 | 78 | - | - | - - | - - | See details in Table XIX | | | As2Cu1Gd1 |
| 129.4.257 | 412553 | 86 | - | - | - - | - - | See details in Table XIX | | | As2Cu1Er1 |
| 129.4.258 | 412552 | 82 | - | - | - - | - - | See details in Table XIX | | | As2Cu1Dy1 |
| 129.4.259 | 90313 | 66 | - | - | - - | - - | See details in Table XIX | | | As2Co1U1 |
| 129.4.260 | 42353 | 50 | - | - | - - | - - | See details in Table XIX | | | As1Te1U1 |
| 129.4.261 | 87141 | 50 | - | - | - - | - - | See details in Table XIX | | | As1Se1U1 |
| 129.4.262 | 66946 | 50 | - | - | - - | - - | See details in Table XIX | | | As1Se1U1 |
| 129.4.263 | 42485 | 50 | - | - | - - | - - | See details in Table XIX | | | As1Se1U1 |
| 129.4.264 | 87140 | 50 | - | - | - - | - - | See details in Table XIX | | | As1S1U1 |
| 129.4.265 | 66945 | 50 | - | - | - - | - - | See details in Table XIX | | | As1S1U1 |
| 129.4.266 | 42170 | 50 | - | - | - - | - - | See details in Table XIX | | | As1S1U1 |
| 129.4.267 | 85778 | 74 | - | - | - - | - - | See details in Table XIX | | | As1Nd1O1Zn1 |
| 129.4.268 | 262832 | 66 | - | - | - - | - - | See details in Table XIX | | | As1Nd1O1Ru1 |
| 129.4.269 | 42329 | 38 | - | - | - - | - - | See details in Table XIX | | | As1Mn2 |
| 129.4.270 | 415629 | 38 | - | - | - - | - - | See details in Table XIX | | | As1Mn2 |
| 129.4.271 | 44009 | 34 | - | - | - - | - - | See details in Table XIX | | | As1Mn1V1 |
| 129.4.272 | 89597 | 42 | - | - | - - | - - | See details in Table XIX | | | As1Mn1Rb1 |
| 129.4.273 | 89596 | 42 | - | - | - - | - - | See details in Table XIX | | | As1Mn1Rb1 |
| 129.4.274 | 60732 | 26 | - | - | - - | - - | See details in Table XIX | | | As1Mn1Na1 |
| 129.4.275 | 60731 | 26 | - | - | - - | - - | See details in Table XIX | | | As1Mn1Na1 |
| 129.4.276 | 60730 | 26 | - | - | - - | - - | See details in Table XIX | | | As1Mn1Na1 |
| 129.4.277 | 26461 | 26 | - | - | - - | - - | See details in Table XIX | | | As1Mn1Na1 |
| 129.4.278 | 60742 | 26 | - | - | - - | - - | See details in Table XIX | | | As1Li1Mn1 |
| 129.4.279 | 60741 | 26 | - | - | - - | - - | See details in Table XIX | | | As1Li1Mn1 |
| 129.4.280 | 60740 | 26 | - | - | - - | - - | See details in Table XIX | | | As1Li1Mn1 |
| 129.4.281 | 60739 | 26 | - | - | - - | - - | See details in Table XIX | | | As1Li1Mn1 |
| 129.4.282 | 26459 | 26 | - | - | - - | - - | See details in Table XIX | | | As1Li1Mn1 |
| 129.4.283 | 192823 | 26 | - | - | - - | - - | See details in Table XIX | | | As1Li1Mn1 |
| 129.4.284 | 192822 | 26 | - | - | - - | - - | See details in Table XIX | | | As1Li1Mn1 |
| 129.4.285 | 192821 | 26 | - | - | - - | - - | See details in Table XIX | | | As1Li1Mn1 |
| 129.4.286 | 192820 | 26 | - | - | - - | - - | See details in Table XIX | | | As1Li1Mn1 |
| 129.4.287 | 192819 | 26 | - | - | - - | - - | See details in Table XIX | | | As1Li1Mn1 |
| 129.4.288 | 192818 | 26 | - | - | - - | - - | See details in Table XIX | | | As1Li1Mn1 |
| 129.4.289 | 192817 | 26 | - | - | - - | - - | See details in Table XIX | | | As1Li1Mn1 |
| 129.4.290 | 60745 | 42 | - | - | - - | - - | See details in Table XIX | | | As1K1Mn1 |
| 129.4.291 | 60744 | 42 | - | - | - - | - - | See details in Table XIX | | | As1K1Mn1 |
| 129.4.292 | 60743 | 42 | - | - | - - | - - | See details in Table XIX | | | As1K1Mn1 |
| 129.4.293 | 41883 | 42 | - | - | - - | - - | See details in Table XIX | | | As1K1Mn1 |
| 129.4.294 | 420206 | 82 | - | - | - - | - - | See details in Table XIX | | | As1Gd1O1Zn1 |
| 129.4.295 | 262834 | 74 | - | - | - - | - - | See details in Table XIX | | | As1Gd1O1Ru1 |
| 129.4.296 | 415628 | 42 | - | - | - - | - - | See details in Table XIX | | | As1Fe2 |
| 129.4.297 | 16844 | 42 | - | - | - - | - - | See details in Table XIX | | | As1Fe2 |
| 129.4.298 | 422001 | 66 | - | - | - - | - - | See details in Table XIX | | | As1Fe1Nd1O1 |
| 129.4.299 | 163 - | 66 | - | - | - - | - - | See details in Table XIX | | | As1Fe1Nd1O1 |
| 129.4.300 | 425010 | 74 | - | - | - - | - - | See details in Table XIX | | | As1Fe1Gd1O1 |
| 129.4.301 | 425006 | 74 | - | - | - - | - - | See details in Table XIX | | | As1Fe1Gd1O1 |
| 129.4.302 | 422003 | 74 | - | - | - - | - - | See details in Table XIX | | | As1Fe1Gd1O1 |

Continued on next page

**TABLE XIV– continued**

| TQC | ICSD | NoE | DG | IndG | SubSG (w/ T) | cl | SubSG (w/o T) | cl | Max. Sub. SG | yes or no | Max. Sub. SG | yes or no | SIs (w/ T) | Stoich. Formula |
|---|---|---|---|---|---|---|---|---|---|---|---|---|---|---|
| 129.4.303 | 189940 | 58 | - | - | - | - | - | - | See details in Table XIX | | | | | As1F1Mn1Sr1 |
| 129.4.304 | 420203 | 90 | - | - | - | - | - | - | See details in Table XIX | | | | | As1Er1O1Zn1 |
| 129.4.305 | 15142 | 54 | - | - | - | - | - | - | See details in Table XIX | | | | | As1Cu2 |
| 129.4.306 | 423230 | 46 | - | - | - | - | - | - | See details in Table XIX | | | | | As1Cu1Mn1 |
| 129.4.307 | 40573 | 42 | - | - | - | - | - | - | See details in Table XIX | | | | | As1Cs1Mn1 |
| 129.4.308 | 40572 | 42 | - | - | - | - | - | - | See details in Table XIX | | | | | As1Cs1Mn1 |
| 129.4.309 | 43223 | 34 | - | - | - | - | - | - | See details in Table XIX | | | | | As1Cr2 |
| 129.4.310 | 42335 | 34 | - | - | - | - | - | - | See details in Table XIX | | | | | As1Cr2 |
| 129.4.311 | 196921 | 34 | - | - | - | - | - | - | See details in Table XIX | | | | | As1Cr2 |
| 129.4.312 | 420653 | 114 | - | - | - | - | - | - | See details in Table XIX | | | | | As1Cr1Fe1O3Sr2 |
| 129.4.313 | 107932 | 30 | - | - | - | - | - | - | See details in Table XIX | | | | | As1Co1Li1 |
| 129.4.314 | 188947 | 62 | - | - | - | - | - | - | See details in Table XIX | | | | | As1Ce1O1Ru1 |
| 129.4.315 | 182625 | 66 | - | - | - | - | - | - | See details in Table XIX | | | | | As1Ce1Ni1O1 |
| 129.4.316 | 174440 | 66 | - | - | - | - | - | - | See details in Table XIX | | | | | As1Ce1Ni1O1 |
| 129.4.317 | 421999 | 62 | - | - | - | - | - | - | See details in Table XIX | | | | | As1Ce1Fe1O1 |
| 129.4.318 | 188948 | 62 | - | - | - | - | - | - | See details in Table XIX | | | | | As1Ce1Fe1O1 |
| 129.4.319 | 162817 | 62 | - | - | - | - | - | - | See details in Table XIX | | | | | As1Ce1Fe1O1 |
| 129.4.320 | 88268 | 70 | - | - | - | - | - | - | See details in Table XIX | | | | | As1Cd1Ce1O1 |
| 129.4.321 | 248365 | 58 | - | - | - | - | - | - | See details in Table XIX | | | | | As1Ba1F1Mn1 |
| 129.4.322 | 191054 | 74 | - | - | - | - | - | - | See details in Table XIX | | | | | Al2La1Pd2 |
| 129.4.323 | 55507 | 78 | - | - | - | - | - | - | See details in Table XIX | | | | | Al2Au2Eu1 |
| 129.4.324 | 420343 | 190 | - | - | - | - | - | - | See details in Table XIX | | | | | Ag5O4P1Te1 |
| 129.4.325 | 420342 | 190 | - | - | - | - | - | - | See details in Table XIX | | | | | Ag5O4P1Se1 |
| 129.4.326 | 58390 | 30 | - | - | - | - | - | - | See details in Table XIX | | | | | Ag1Zr1 |
| 129.4.327 | 95133 | 82 | - | - | - | - | - | - | See details in Table XIX | | | | | Ag1Dy1Sb2 |
| 129.4.328 | 79450 | 66 | - | - | - | - | - | - | See details in Table XIX | | | | | Ag1Ce1Sb2 |
| 129.4.329 | 79425 | 66 | - | - | - | - | - | - | See details in Table XIX | | | | | Ag1Ce1Sb2 |
| 129.4.330 | 93150 | 66 | - | - | - | - | - | - | See details in Table XIX | | | | | Ag1As2Ce1 |
| 130.4.1 | 71313 | 180 | - | - | - | - | - | - | See details in Table XIX | | | | | Bi2Cu1O4 |
| 130.4.2 | 69308 | 180 | - | - | - | - | - | - | See details in Table XIX | | | | | Bi2Cu1O4 |
| 130.4.3 | 69307 | 180 | - | - | - | - | - | - | See details in Table XIX | | | | | Bi2Cu1O4 |
| 130.4.4 | 68815 | 180 | - | - | - | - | - | - | See details in Table XIX | | | | | Bi2Cu1O4 |
| 130.4.5 | 68814 | 180 | - | - | - | - | - | - | See details in Table XIX | | | | | Bi2Cu1O4 |
| 130.4.6 | 68813 | 180 | - | - | - | - | - | - | See details in Table XIX | | | | | Bi2Cu1O4 |
| 130.4.7 | 68812 | 180 | - | - | - | - | - | - | See details in Table XIX | | | | | Bi2Cu1O4 |
| 130.4.8 | 56390 | 180 | - | - | - | - | - | - | See details in Table XIX | | | | | Bi2Cu1O4 |
| 130.4.9 | 202994 | 180 | - | - | - | - | - | - | See details in Table XIX | | | | | Bi2Cu1O4 |
| 130.4.10 | 15865 | 180 | - | - | - | - | - | - | See details in Table XIX | | | | | Bi2Cu1O4 |
| 131.2.1 | 50987 | 92 | 0.0000 | -1.9521 | 2,81 | 2 | 2,81 | 2 | - | - | - | - | z2w,1=1<br>z2w,2=1<br>z4=3<br>z2=1 | I1Pd1Te1 |
| 131.2.2 | 6159 | 92 | 0.0067 | -0.2727 | 2,81 | 1 | 2,81 | 1 | - | - | - | - | z4=3<br>z2=1 | Ca1O4Pt2 |
| 131.4.1 | 23300 | 50 | - | - | - | - | - | - | See details in Table XIX | | | | | C2B2La1 |
| 135.3.1 | 86492 | 176 | - | - | - | - | - | - | 117 | yes | 117 | yes | - | Ni1O4Sb2 |
| | | | | | | | | | 114 | yes | 114 | yes | - | |
| | | | | | | | | | 66 | no(2) | 66 | no(2) | z4=1 | |
| | | | | | | | | | 55 | no(2) | 55 | no(2) | z4=1 | |
| 135.3.2 | 86491 | 176 | - | - | - | - | - | - | 117 | yes | 117 | yes | - | Ni1O4Sb2 |
| | | | | | | | | | 114 | yes | 114 | yes | - | |
| | | | | | | | | | 66 | no(2) | 66 | no(2) | z4=1 | |
| | | | | | | | | | 55 | no(2) | 55 | no(2) | z4=1 | |
| 135.3.3 | 86490 | 176 | - | - | - | - | - | - | 117 | yes | 117 | yes | - | Ni1O4Sb2 |
| | | | | | | | | | 114 | yes | 114 | yes | - | |
| | | | | | | | | | 66 | no(2) | 66 | no(2) | z4=1 | |
| | | | | | | | | | 55 | no(2) | 55 | no(2) | z4=1 | |

Continued on next page

**TABLE XIV– continued**

| TQC | ICSD | NoE | DG | IndG | SubSG (w/ T) | cl | SubSG (w/o T) | cl | Max. Sub. SG | yes or no | Max. Sub. SG | yes or no | SIs (w/ T) | Stoich. Formula |
|---|---|---|---|---|---|---|---|---|---|---|---|---|---|---|
| 135.3.4 | 86489 | 176 | - | - | - | - | - | - | 117 | yes | 117 | yes | - | Ni1O4Sb2 |
| | | | | | | | | | 114 | yes | 114 | yes | - | |
| | | | | | | | | | 66 | no(2) | 66 | no(2) | z4=1 | |
| | | | | | | | | | 55 | no(2) | 55 | no(2) | z4=1 | |
| 135.3.5 | 41228 | 176 | - | - | - | - | - | - | 117 | yes | 117 | yes | - | Ni1O4Sb2 |
| | | | | | | | | | 114 | yes | 114 | yes | - | |
| | | | | | | | | | 66 | no(2) | 66 | no(2) | z4=1 | |
| | | | | | | | | | 55 | no(2) | 55 | no(2) | z4=1 | |
| 135.3.6 | 41227 | 176 | - | - | - | - | - | - | 117 | yes | 117 | yes | - | Ni1O4Sb2 |
| | | | | | | | | | 114 | yes | 114 | yes | - | |
| | | | | | | | | | 66 | no(2) | 66 | no(2) | z4=1 | |
| | | | | | | | | | 55 | no(2) | 55 | no(2) | z4=1 | |
| 135.3.7 | 41226 | 176 | - | - | - | - | - | - | 117 | yes | 117 | yes | - | Ni1O4Sb2 |
| | | | | | | | | | 114 | yes | 114 | yes | - | |
| | | | | | | | | | 66 | no(2) | 66 | no(2) | z4=1 | |
| | | | | | | | | | 55 | no(2) | 55 | no(2) | z4=1 | |
| 135.3.8 | 41225 | 176 | - | - | - | - | - | - | 117 | yes | 117 | yes | - | Ni1O4Sb2 |
| | | | | | | | | | 114 | yes | 114 | yes | - | |
| | | | | | | | | | 66 | no(2) | 66 | no(2) | z4=1 | |
| | | | | | | | | | 55 | no(2) | 55 | no(2) | z4=1 | |
| 135.4.1 | 262627 | 172 | - | - | - | - | - | - | See details in Table XIX | | | | | Co1O4Sb2 |
| 135.4.2 | 262626 | 172 | - | - | - | - | - | - | See details in Table XIX | | | | | Co1O4Sb2 |
| 135.4.3 | 4287 | 180 | - | - | - | - | - | - | See details in Table XIX | | | | | As2Cu1O4 |
| 136.1.1 | 76166 | 420 | 0.0237 | -0.0727 | 2 | 3 | 65,84 | 3 | - | - | - | - | z4=2 | U1 |
| 136.1.2 | 90270 | 148 | 0.0085 | -0.5731 | 2,81 | 2 | 2,81 | 1 | - | - | - | - | z4=3<br>z2=1 | Ni4P2Yb1 |
| 136.1.3 | 421229 | 128 | 0.0056 | -0.3416 | 2,81 | 2 | 2,81 | 1 | - | - | - | - | z4=3<br>z2=1 | Li2Sr3 |
| 136.1.4 | 93213 | 116 | 0.0120 | -0.0935 | 2,81 | 2 | 2,81 | 1 | - | - | - | - | z4=3<br>z2=1 | Fe2O6Te1 |
| 136.1.5 | 33797 | 204 | 0.0107 | -0.0095 | 2 | 3 | 65,84 | 3 | - | - | - | - | z4=2 | Co2K4O5Rb2 |
| 136.1.6 | 237545 | 156 | 0.0234 | -1.5977 | 2,81 | 1 | 2,81 | 2 | - | - | - | - | z4=1<br>z2=1 | Cd4Pt2Sr1 |
| 136.1.7 | 62553 | 112 | 0.0312 | -0.8195 | 2,81 | 2 | 2,81 | 1 | - | - | - | - | z4=3<br>z2=1 | Ca1Cu4P2 |
| 136.1.8 | 417667 | 168 | 0.0000 | -0.0179 | 2 | 3 | 65,84 | 3 | - | - | - | - | z4=2 | C2Nd2W1 |
| 136.1.9 | 413345 | 160 | 0.0020 | -0.0463 | 2 | 3 | 65,84 | 3 | - | - | - | - | z4=2 | C2Mo1Pr2 |
| 136.1.10 | 260392 | 160 | 0.0257 | -1.0388 | 2,81 | 2 | 2,81 | 1 | - | - | - | - | z4=3<br>z2=1 | Ba1Hg2Tl2 |
| 136.1.11 | 260391 | 160 | 0.0205 | -1.0463 | 2,81 | 2 | 2,81 | 1 | - | - | - | - | z4=3<br>z2=1 | Ba1Hg2Tl2 |
| 136.1.12 | 237544 | 156 | 0.0085 | -1.5967 | 2,81 | 1 | 2,81 | 2 | - | - | - | - | z4=1<br>z2=1 | Ba1Cd4Pt2 |
| 136.1.13 | 173224 | 144 | 0.0000 | -0.0827 | 2,81 | 2 | 2,81 | 1 | - | - | - | - | z4=3<br>z2=1 | As2Er1Ni4 |
| 136.1.14 | 68526 | 140 | 0.0048 | -0.0314 | 2,81 | 2 | 2,81 | 1 | - | - | - | - | z4=3<br>z2=1 | As2Dy1Ni4 |
| 136.3.1 | 76490 | 420 | - | - | - | - | - | - | 118 | no(81) | 118 | no(81) | z2=1 | U1 |
| | | | | | | | | | 113 | no(81) | 113 | no(81) | z2=1 | |
| | | | | | | | | | 65 | no(2) | 65 | no(65) | z4=2 | |
| | | | | | | | | | 58 | no(2) | 58 | yes | z4=2 | |
| 136.3.2 | 43218 | 420 | - | - | - | - | - | - | 118 | no(81) | 118 | no(81) | z2=1 | U1 |
| | | | | | | | | | 113 | no(81) | 113 | no(81) | z2=1 | |
| | | | | | | | | | 65 | yes | 65 | yes | - | |
| | | | | | | | | | 58 | yes | 58 | yes | - | |
| 136.3.3 | 410745 | 288 | - | - | - | - | - | - | 118 | yes | 118 | yes | - | Pb1Pt2Yb2 |
| | | | | | | | | | 113 | yes | 113 | yes | - | |
| | | | | | | | | | 65 | no(2) | 65 | no(2) | z4=3 | |

Continued on next page

**TABLE XIV– continued**

| TQC | ICSD | NoE | DG | IndG | SubSG cl (w/ T) | SubSG cl (w/o T) | Max. Sub. SG | yes or no | Max. Sub. SG | yes or no | SIs (w/ T) | Stoich. Formula |
|---|---|---|---|---|---|---|---|---|---|---|---|---|
| | | | | | | | 58 | no(2) | 58 | no(2) | z4=3 | |
| 136.3.4 | 2576 | 104 | - | - | - - | - - | 113 | yes | 113 | yes | - | O6V2W1 |
| | | | | | | | 94 | yes | 94 | yes | - | |
| | | | | | | | 84 | no(2) | 84 | yes | z2w,1=1 z2w,2=1 | |
| | | | | | | | 65 | no(2) | 65 | no(65) | z2w,1=1 z2w,2=1 z4=2 | |
| | | | | | | | - | - | 7 | no(7) | - | |
| 136.3.5 | 2575 | 104 | - | - | - - | - - | 113 | yes | 113 | yes | - | O6V2W1 |
| | | | | | | | 94 | yes | 94 | yes | - | |
| | | | | | | | 84 | no(2) | 84 | yes | z2w,1=1 z2w,2=1 | |
| | | | | | | | 65 | no(2) | 65 | no(65) | z2w,1=1 z2w,2=1 z4=2 | |
| | | | | | | | - | - | 7 | no(7) | - | |
| 136.3.6 | 34234 | 32 | - | - | - - | - - | 118 | no(81) | 118 | no(81) | z2=1 | O2Pb1 |
| | | | | | | | 113 | no(81) | 113 | no(81) | z2=1 | |
| | | | | | | | 65 | yes | 65 | yes | - | |
| | | | | | | | 58 | yes | 58 | yes | - | |
| 136.3.7 | 23292 | 32 | - | - | - - | - - | 118 | no(81) | 118 | no(81) | z2=1 | O2Pb1 |
| | | | | | | | 113 | no(81) | 113 | no(81) | z2=1 | |
| | | | | | | | 65 | yes | 65 | yes | - | |
| | | | | | | | 58 | yes | 58 | yes | - | |
| 136.3.8 | 40435 | 108 | - | - | - - | - - | 118 | yes | 118 | yes | - | Ni4P2Zr1 |
| | | | | | | | 113 | yes | 113 | yes | - | |
| | | | | | | | 65 | no(2) | 65 | no(2) | z4=3 | |
| | | | | | | | 58 | no(2) | 58 | no(2) | z4=3 | |
| 136.3.9 | 85896 | 128 | - | - | - - | - - | 118 | yes | 118 | yes | - | Ni4P2U1 |
| | | | | | | | 113 | yes | 113 | yes | - | |
| | | | | | | | 65 | no(2) | 65 | no(2) | z4=3 | |
| | | | | | | | 58 | no(2) | 58 | no(2) | z4=3 | |
| 136.3.10 | 405564 | 128 | - | - | - - | - - | 118 | yes | 118 | yes | - | Ni2Sn1Zr2 |
| | | | | | | | 113 | yes | 113 | yes | - | |
| | | | | | | | 65 | yes | 65 | yes | - | |
| | | | | | | | 58 | yes | 58 | yes | - | |
| 136.3.11 | 407549 | 72 | - | - | - - | - - | 118 | no(81) | 118 | no(81) | z2=1 | In3Rh1 |
| | | | | | | | 113 | no(81) | 113 | no(81) | z2=1 | |
| | | | | | | | 65 | no(2) | 65 | no(65) | z4=2 | |
| | | | | | | | 58 | no(2) | 58 | yes | z4=2 | |
| 136.3.12 | 407548 | 72 | - | - | - - | - - | 118 | no(81) | 118 | no(81) | z2=1 | In3Ir1 |
| | | | | | | | 113 | no(81) | 113 | no(81) | z2=1 | |
| | | | | | | | 65 | no(2) | 65 | no(65) | z4=2 | |
| | | | | | | | 58 | no(2) | 58 | yes | z4=2 | |
| 136.3.13 | 404888 | 124 | - | - | - - | - - | 118 | yes | 118 | yes | - | In1Pd2Zr2 |
| | | | | | | | 113 | yes | 113 | yes | - | |
| | | | | | | | 65 | no(2) | 65 | no(65) | z4=2 | |
| | | | | | | | 58 | no(2) | 58 | yes | z4=2 | |
| 136.3.14 | 107332 | 124 | - | - | - - | - - | 118 | yes | 118 | yes | - | In1Pd2Zr2 |
| | | | | | | | 113 | yes | 113 | yes | - | |
| | | | | | | | 65 | no(2) | 65 | no(65) | z4=2 | |
| | | | | | | | 58 | no(2) | 58 | yes | z4=2 | |
| 136.3.15 | 405563 | 124 | - | - | - - | - - | 118 | yes | 118 | yes | - | In1Ni2Zr2 |
| | | | | | | | 113 | yes | 113 | yes | - | |
| | | | | | | | 65 | no(2) | 65 | no(65) | z4=2 | |
| | | | | | | | 58 | no(2) | 58 | yes | z4=2 | |
| 136.3.16 | 404848 | 124 | - | - | - - | - - | 118 | yes | 118 | yes | - | In1Ni2Ti2 |

Continued on next page

**TABLE XIV– continued**

| TQC | ICSD | NoE | DG | IndG | SubSG cl (w/ T) | SubSG cl (w/o T) | Max. Sub. SG | yes or no | Max. Sub. SG | yes or no | SIs (w/ T) | Stoich. Formula |
|---|---|---|---|---|---|---|---|---|---|---|---|---|
| | | | | | | | 113 | yes | 113 | yes | - | |
| | | | | | | | 65 | no(2) | 65 | no(65) | z4=2 | |
| | | | | | | | 58 | no(2) | 58 | yes | z4=2 | |
| 136.3.17 | 107330 | 124 | - | - | - - | - - | 118 | yes | 118 | yes | - | In1Ni2Ti2 |
| | | | | | | | 113 | yes | 113 | yes | - | |
| | | | | | | | 65 | no(2) | 65 | no(65) | z4=2 | |
| | | | | | | | 58 | no(2) | 58 | yes | z4=2 | |
| 136.3.18 | 407546 | 128 | - | - | - - | - - | 118 | yes | 118 | yes | - | Hf2Ni2Sn1 |
| | | | | | | | 113 | yes | 113 | yes | - | |
| | | | | | | | 65 | yes | 65 | yes | - | |
| | | | | | | | 58 | yes | 58 | yes | - | |
| 136.3.19 | 194982 | 108 | - | - | - - | - - | 118 | yes | 118 | yes | - | Fe4Ge2U1 |
| | | | | | | | 113 | yes | 113 | yes | - | |
| | | | | | | | 65 | no(2) | 65 | no(2) | z4=3 | |
| | | | | | | | 58 | no(2) | 58 | no(2) | z4=3 | |
| 136.3.20 | 9166 | 44 | - | - | - - | - - | 118 | no(81) | 118 | no(81) | z2=1 | F2Fe1 |
| | | | | | | | 113 | no(81) | 113 | no(81) | z2=1 | |
| | | | | | | | 65 | yes | 65 | yes | - | |
| | | | | | | | 58 | yes | 58 | yes | - | |
| 136.3.21 | 65702 | 44 | - | - | - - | - - | 118 | no(81) | 118 | no(81) | z2=1 | F2Fe1 |
| | | | | | | | 113 | no(81) | 113 | no(81) | z2=1 | |
| | | | | | | | 65 | yes | 65 | yes | - | |
| | | | | | | | 58 | yes | 58 | yes | - | |
| 136.3.22 | 14143 | 44 | - | - | - - | - - | 118 | no(81) | 118 | no(81) | z2=1 | F2Fe1 |
| | | | | | | | 113 | no(81) | 113 | no(81) | z2=1 | |
| | | | | | | | 65 | yes | 65 | yes | - | |
| | | | | | | | 58 | yes | 58 | yes | - | |
| 136.3.23 | 412062 | 72 | - | - | - - | - - | 118 | yes | 118 | yes | - | Co4In12 |
| | | | | | | | 113 | yes | 113 | yes | - | |
| | | | | | | | 65 | no(2) | 65 | no(2) | z4=1 | |
| | | | | | | | 58 | no(2) | 58 | no(2) | z4=1 | |
| 136.3.24 | 412061 | 72 | - | - | - - | - - | 118 | yes | 118 | yes | - | Co4Ga12 |
| | | | | | | | 113 | yes | 113 | yes | - | |
| | | | | | | | 65 | no(2) | 65 | no(2) | z4=1 | |
| | | | | | | | 58 | no(2) | 58 | no(2) | z4=1 | |
| 136.3.25 | 64680 | 140 | - | - | - - | - - | 118 | yes | 118 | yes | - | Co2Na4O5Rb2 |
| | | | | | | | 113 | yes | 113 | yes | - | |
| | | | | | | | 65 | no(2) | 65 | no(2) | z4=1 | |
| | | | | | | | 58 | no(2) | 58 | no(2) | z4=1 | |
| 136.3.26 | 64678 | 140 | - | - | - - | - - | 118 | yes | 118 | yes | - | Co2K2Na4O5 |
| | | | | | | | 113 | yes | 113 | yes | - | |
| | | | | | | | 65 | no(2) | 65 | no(2) | z4=3 | |
| | | | | | | | 58 | no(2) | 58 | no(2) | z4=3 | |
| 136.3.27 | 72387 | 80 | - | - | - - | - - | 118 | yes | 118 | yes | - | Co1Li1N2Sr2 |
| | | | | | | | 113 | yes | 113 | yes | - | |
| | | | | | | | 65 | no(2) | 65 | no(2) | z4=1 | |
| | | | | | | | 58 | no(2) | 58 | no(2) | z4=1 | |
| 136.3.28 | 407550 | 72 | - | - | - - | - - | 118 | yes | 118 | yes | - | Co1In3 |
| | | | | | | | 113 | yes | 113 | yes | - | |
| | | | | | | | 65 | no(2) | 65 | no(2) | z4=1 | |
| | | | | | | | 58 | no(2) | 58 | no(2) | z4=1 | |
| 136.3.29 | 417668 | 160 | - | - | - - | - - | 118 | no(81) | 118 | no(81) | z2=1 | C2Pr2W1 |
| | | | | | | | 113 | no(81) | 113 | no(81) | z2=1 | |
| | | | | | | | 65 | no(2) | 65 | no(65) | z4=2 | |
| | | | | | | | 58 | no(2) | 58 | yes | z4=2 | |
| 136.3.30 | 417669 | 152 | - | - | - - | - - | 118 | no(81) | 118 | no(81) | z2=1 | C2Ce2W1 |
| | | | | | | | 113 | no(81) | 113 | no(81) | z2=1 | |
| | | | | | | | 65 | no(2) | 65 | no(65) | z4=2 | |

Continued on next page

**TABLE XIV– continued**

| TQC | ICSD | NoE | DG | IndG | SubSG (w/ T) | cl | SubSG (w/o T) | cl | Max. Sub. SG | yes or no | Max. Sub. SG | yes or no | SIs (w/ T) | Stoich. Formula |
|---|---|---|---|---|---|---|---|---|---|---|---|---|---|---|
| | | | | | | | | | 58 | no(2) | 58 | yes | z4=2 | |
| 136.3.31 | 50005 | 96 | - | - | - | - | - | - | 118 | no(81) | 118 | no(81) | z2=1 | Bi2Mg1O6 |
| | | | | | | | | | 113 | no(81) | 113 | no(81) | z2=1 | |
| | | | | | | | | | 65 | yes | 65 | yes | - | |
| | | | | | | | | | 58 | yes | 58 | yes | - | |
| 136.3.32 | 409532 | 296 | - | - | - | - | - | - | 118 | yes | 118 | yes | - | Au2Pb1Yb2 |
| | | | | | | | | | 113 | yes | 113 | yes | - | |
| | | | | | | | | | 65 | no(2) | 65 | no(2) | z4=3 | |
| | | | | | | | | | 58 | no(2) | 58 | no(2) | z4=3 | |
| 136.4.1 | 66665 | 34 | - | - | - | - | - | - | See details in Table XIX | | | | | O2V1 |
| 136.4.2 | 4110 | 34 | - | - | - | - | - | - | See details in Table XIX | | | | | O2V1 |
| 136.4.3 | 34421 | 34 | - | - | - | - | - | - | See details in Table XIX | | | | | O2V1 |
| 136.4.4 | 34420 | 34 | - | - | - | - | - | - | See details in Table XIX | | | | | O2V1 |
| 136.4.5 | 34418 | 34 | - | - | - | - | - | - | See details in Table XIX | | | | | O2V1 |
| 136.4.6 | 237337 | 34 | - | - | - | - | - | - | See details in Table XIX | | | | | O2V1 |
| 136.4.7 | 1504 | 34 | - | - | - | - | - | - | See details in Table XIX | | | | | O2V1 |
| 136.4.8 | 10141 | 34 | - | - | - | - | - | - | See details in Table XIX | | | | | O2V1 |
| 136.4.9 | 154021 | 38 | - | - | - | - | - | - | See details in Table XIX | | | | | O2Re1 |
| 136.4.10 | 189495 | 122 | - | - | - | - | - | - | See details in Table XIX | | | | | Ni4P2Y1 |
| 136.4.11 | 424579 | 98 | - | - | - | - | - | - | See details in Table XIX | | | | | Ni1O2Rb3 |
| 136.4.12 | 75198 | 50 | - | - | - | - | - | - | See details in Table XIX | | | | | Nb1O2 |
| 136.4.13 | 73716 | 38 | - | - | - | - | - | - | See details in Table XIX | | | | | Mn1O2 |
| 136.4.14 | 393 | 38 | - | - | - | - | - | - | See details in Table XIX | | | | | Mn1O2 |
| 136.4.15 | 262579 | 98 | - | - | - | - | - | - | See details in Table XIX | | | | | K3Ni1O2 |
| 136.4.16 | 84577 | 42 | - | - | - | - | - | - | See details in Table XIX | | | | | Ir1O2 |
| 136.4.17 | 261030 | 126 | - | - | - | - | - | - | See details in Table XIX | | | | | Fe4Si2Tm1 |
| 136.4.18 | 247481 | 126 | - | - | - | - | - | - | See details in Table XIX | | | | | Fe4Si2Tm1 |
| 136.4.19 | 68525 | 90 | - | - | - | - | - | - | See details in Table XIX | | | | | Fe4P2Sc1 |
| 136.4.20 | 186045 | 130 | - | - | - | - | - | - | See details in Table XIX | | | | | Fe4Ge2Lu1 |
| 136.4.21 | 290252 | 98 | - | - | - | - | - | - | See details in Table XIX | | | | | F6Li2V1 |
| 136.4.22 | 62769 | 38 | - | - | - | - | - | - | See details in Table XIX | | | | | F2V1 |
| 136.4.23 | 62768 | 38 | - | - | - | - | - | - | See details in Table XIX | | | | | F2V1 |
| 136.4.24 | 38069 | 38 | - | - | - | - | - | - | See details in Table XIX | | | | | F2V1 |
| 136.4.25 | 32552 | 38 | - | - | - | - | - | - | See details in Table XIX | | | | | F2V1 |
| 136.4.26 | 201245 | 38 | - | - | - | - | - | - | See details in Table XIX | | | | | F2V1 |
| 136.4.27 | 9165 | 42 | - | - | - | - | - | - | See details in Table XIX | | | | | F2Mn1 |
| 136.4.28 | 68740 | 42 | - | - | - | - | - | - | See details in Table XIX | | | | | F2Mn1 |
| 136.4.29 | 68737 | 42 | - | - | - | - | - | - | See details in Table XIX | | | | | F2Mn1 |
| 136.4.30 | 68736 | 42 | - | - | - | - | - | - | See details in Table XIX | | | | | F2Mn1 |
| 136.4.31 | 63654 | 42 | - | - | - | - | - | - | See details in Table XIX | | | | | F2Mn1 |
| 136.4.32 | 200642 | 42 | - | - | - | - | - | - | See details in Table XIX | | | | | F2Mn1 |
| 136.4.33 | 200641 | 42 | - | - | - | - | - | - | See details in Table XIX | | | | | F2Mn1 |
| 136.4.34 | 200640 | 42 | - | - | - | - | - | - | See details in Table XIX | | | | | F2Mn1 |
| 136.4.35 | 84789 | 114 | - | - | - | - | - | - | See details in Table XIX | | | | | Cu1O6Sb2 |
| 136.4.36 | 424578 | 98 | - | - | - | - | - | - | See details in Table XIX | | | | | Cs3Ni1O2 |
| 136.4.37 | 203094 | 110 | - | - | - | - | - | - | See details in Table XIX | | | | | Co1O6Sb2 |
| 136.4.38 | 9167 | 46 | - | - | - | - | - | - | See details in Table XIX | | | | | Co1F2 |
| 136.4.39 | 73460 | 46 | - | - | - | - | - | - | See details in Table XIX | | | | | Co1F2 |
| 136.4.40 | 280604 | 46 | - | - | - | - | - | - | See details in Table XIX | | | | | Co1F2 |
| 136.4.41 | 14144 | 46 | - | - | - | - | - | - | See details in Table XIX | | | | | Co1F2 |
| 137.1.1 | 250760 | 120 | 0.0113 | -0.0119 | 2,81 | 1 | 2,81 | 2 | - | - | - | - | z4=3<br>z2=1 | Np2Se5 |
| 137.1.2 | 248874 | 140 | 0.0139 | -0.1043 | 2,81 | 1 | 2,81 | 2 | - | - | - | - | z4=3<br>z2=1 | Ba2S6U1 |
| 137.1.3 | 49631 | 140 | 0.0000 | -0.0384 | 2 | 3 | 0 | 1 | - | - | - | - | z4=2 | B4Er1Rh4 |
| 137.3.1 | 39595 | 164 | - | - | - | - | - | - | 115 | no(81) | 115 | no(81) | z2=1 | Nb4O5 |
| | | | | | | | | | 114 | no(81) | 114 | no(81) | z2=1 | |

Continued on next page

**TABLE XIV– continued**

| TQC | ICSD | NoE | DG | IndG | SubSG (w/ T) | cl | SubSG (w/o T) | cl | Max. Sub. SG | yes or no | Max. Sub. SG | yes or no | SIs (w/ T) | Stoich. Formula |
|---|---|---|---|---|---|---|---|---|---|---|---|---|---|---|
| | | | | | | | | | 68 | yes | 68 | yes | - | |
| | | | | | | | | | 59 | yes | 59 | yes | - | |
| 137.4.1 | 62688 | 78 | - | - | - | - | - | - | See details in Table XIX | | | | | Co1Li6O4 |
| 139.1.1 | 263005 | 92 | 0.0000 | -0.9007 | 87 | 15 | 139 | 6 | - | - | - | - | z8=4 | Si3Y1 |
| 139.1.2 | 42119 | 50 | 0.0250 | -0.1074 | 2 | 12 | 71,69,87 | 5 | - | - | - | - | z2w,1=1<br>z2w,2=1<br>z2w,3=1<br>z4=2<br>z8=2 | Sb1Sr2 |
| 139.1.3 | 416747 | 88 | 0.0049 | -1.4625 | 2 | 3 | 71,69,87 | 2 | - | - | - | - | z2w,1=1<br>z2w,2=1<br>z2w,3=1<br>z4=2<br>z8=6 | Ru1Zn3 |
| 139.1.4 | 249448 | 88 | 0.0049 | -1.4625 | 2 | 3 | 71,69,87 | 2 | - | - | - | - | z2w,1=1<br>z2w,2=1<br>z2w,3=1<br>z4=2<br>z8=6 | Ru1Zn3 |
| 139.1.5 | 155990 | 88 | 0.0047 | -1.4573 | 2 | 3 | 71,69,87 | 2 | - | - | - | - | z2w,1=1<br>z2w,2=1<br>z2w,3=1<br>z4=2<br>z8=6 | Ru1Zn3 |
| 139.1.6 | 251818 | 64 | 0.0318 | -1.9165 | 2 | 12 | 71,69,87 | 5 | - | - | - | - | z2w,1=1<br>z2w,2=1<br>z2w,3=1<br>z4=2<br>z8=2 | Pt2Th1 |
| 139.1.7 | 423291 | 62 | 0.0032 | -0.0628 | 87 | 15 | 139 | 6 | - | - | - | - | z8=4 | Pr2Sb1 |
| 139.1.8 | 247272 | 66 | 0.0139 | -0.4823 | 2,82 | 13 | 2,82 | 8 | - | - | - | - | z4=3<br>z2=1<br>z8=7 | Pd3Tl1 |
| 139.1.9 | 61195 | 40 | 0.0411 | -2.4999 | 2 | 11 | 71,69,87 | 5 | - | - | - | - | z4=2<br>z8=2 | Pd2Sb2Sr1 |
| 139.1.10 | 105758 | 18 | 0.0335 | -1.5430 | 2,82 | 8 | 2,82 | 4 | - | - | - | - | z4=1<br>z2=1<br>z8=5 | Pd1Zr2 |
| 139.1.11 | 42966 | 36 | 0.0068 | -1.8052 | 2,82 | 5 | 2,82 | 1 | - | - | - | - | z4=1<br>z2=1<br>z8=1 | P2Ru2Sr1 |
| 139.1.12 | 50186 | 38 | 0.0279 | -2.0856 | 2,82 | 8 | 2,82 | 4 | - | - | - | - | z4=1<br>z2=1<br>z8=5 | P2Rh2Sr1 |
| 139.1.13 | 30849 | 44 | 0.0030 | -0.7429 | 2,82 | 7 | 2,82 | 4 | - | - | - | - | z2w,1=1<br>z2w,2=1<br>z2w,3=1<br>z4=1<br>z2=1<br>z8=5 | P2Pd2Yb1 |
| 139.1.14 | 36373 | 40 | 0.0555 | -2.1863 | 2,82 | 13 | 2,82 | 8 | - | - | - | - | z4=3<br>z2=1<br>z8=7 | P2Pd2Sr1 |
| 139.1.15 | 71451 | 82 | 0.0106 | -1.9534 | 87 | 15 | 139 | 6 | - | - | - | - | z8=4 | O7Sr3V2 |
| 139.1.16 | 71320 | 82 | 0.0122 | -1.9829 | 87 | 15 | 139 | 6 | - | - | - | - | z8=4 | O7Sr3V2 |

Continued on next page

**TABLE XIV– continued**

| TQC | ICSD | NoE | DG | IndG | SubSG (w/ T) | cl | SubSG (w/o T) | cl | Max. Sub. SG | yes or no | Max. Sub. SG | yes or no | SIs (w/ T) | Stoich. Formula |
|---|---|---|---|---|---|---|---|---|---|---|---|---|---|---|
| 139.1.17 | 56577 | 88 | 0.0049 | -1.0689 | 2 | 12 | 71,69,87 | 5 | - | - | - | - | z2w,1=1<br>z2w,2=1<br>z2w,3=1<br>z4=2<br>z8=2 | O7Ru2Sr3 |
| 139.1.18 | 79214 | 98 | 0.0155 | -0.1871 | 87 | 15 | 139 | 6 | - | - | - | - | z8=4 | O4Te3U4 |
| 139.1.19 | 68211 | 40 | 0.0258 | -1.1771 | 2 | 6 | 0 | 7 | - | - | - | - | z2w,1=1<br>z2w,2=1<br>z2w,3=1 | Ni2Si2Th1 |
| 139.1.20 | 41771 | 70 | 0.0108 | -1.8567 | 2,82 | 8 | 2,82 | 4 | - | - | - | - | z4=1<br>z2=1<br>z8=5 | Ni2Sc1Si3 |
| 139.1.21 | 89560 | 46 | 0.0287 | -0.0479 | 2,82 | 1 | 2,82 | 3 | - | - | - | - | z4=3<br>z2=1<br>z8=3 | Nd2O2Te1 |
| 139.1.22 | 93412 | 42 | 0.0215 | -0.0988 | 2 | 3 | 71,69,87 | 2 | - | - | - | - | z2w,1=1<br>z2w,2=1<br>z2w,3=1<br>z4=2<br>z8=6 | Nd1Ni2Si2 |
| 139.1.23 | 91220 | 26 | 0.0190 | -0.3763 | 87 | 15 | 139 | 6 | - | - | - | - | z8=4 | Na2O1Sb2Ti2 |
| 139.1.24 | 91219 | 26 | 0.0189 | -0.3766 | 87 | 15 | 139 | 6 | - | - | - | - | z8=4 | Na2O1Sb2Ti2 |
| 139.1.25 | 91218 | 26 | 0.0190 | -0.3759 | 87 | 15 | 139 | 6 | - | - | - | - | z8=4 | Na2O1Sb2Ti2 |
| 139.1.26 | 91217 | 26 | 0.0190 | -0.3781 | 87 | 15 | 139 | 6 | - | - | - | - | z8=4 | Na2O1Sb2Ti2 |
| 139.1.27 | 91216 | 26 | 0.0190 | -0.3774 | 87 | 15 | 139 | 6 | - | - | - | - | z8=4 | Na2O1Sb2Ti2 |
| 139.1.28 | 91215 | 26 | 0.0190 | -0.3788 | 87 | 15 | 139 | 6 | - | - | - | - | z8=4 | Na2O1Sb2Ti2 |
| 139.1.29 | 91214 | 26 | 0.0190 | -0.3787 | 87 | 15 | 139 | 6 | - | - | - | - | z8=4 | Na2O1Sb2Ti2 |
| 139.1.30 | 91213 | 26 | 0.0190 | -0.3777 | 87 | 15 | 139 | 6 | - | - | - | - | z8=4 | Na2O1Sb2Ti2 |
| 139.1.31 | 91212 | 26 | 0.0190 | -0.3783 | 87 | 15 | 139 | 6 | - | - | - | - | z8=4 | Na2O1Sb2Ti2 |
| 139.1.32 | 91211 | 26 | 0.0190 | -0.3795 | 87 | 15 | 139 | 6 | - | - | - | - | z8=4 | Na2O1Sb2Ti2 |
| 139.1.33 | 91210 | 26 | 0.0189 | -0.3825 | 87 | 15 | 139 | 6 | - | - | - | - | z8=4 | Na2O1Sb2Ti2 |
| 139.1.34 | 91209 | 26 | 0.0189 | -0.3792 | 87 | 15 | 139 | 6 | - | - | - | - | z8=4 | Na2O1Sb2Ti2 |
| 139.1.35 | 91208 | 26 | 0.0189 | -0.3802 | 87 | 15 | 139 | 6 | - | - | - | - | z8=4 | Na2O1Sb2Ti2 |
| 139.1.36 | 91207 | 26 | 0.0190 | -0.3795 | 87 | 15 | 139 | 6 | - | - | - | - | z8=4 | Na2O1Sb2Ti2 |
| 139.1.37 | 91206 | 26 | 0.0190 | -0.3809 | 87 | 15 | 139 | 6 | - | - | - | - | z8=4 | Na2O1Sb2Ti2 |
| 139.1.38 | 91205 | 26 | 0.0190 | -0.3815 | 87 | 15 | 139 | 6 | - | - | - | - | z8=4 | Na2O1Sb2Ti2 |
| 139.1.39 | 91204 | 26 | 0.0190 | -0.3821 | 87 | 15 | 139 | 6 | - | - | - | - | z8=4 | Na2O1Sb2Ti2 |
| 139.1.40 | 91203 | 26 | 0.0189 | -0.3852 | 87 | 15 | 139 | 6 | - | - | - | - | z8=4 | Na2O1Sb2Ti2 |
| 139.1.41 | 91202 | 26 | 0.0190 | -0.3848 | 87 | 15 | 139 | 6 | - | - | - | - | z8=4 | Na2O1Sb2Ti2 |
| 139.1.42 | 69648 | 26 | 0.0189 | -0.3911 | 87 | 15 | 139 | 6 | - | - | - | - | z8=4 | Na2O1Sb2Ti2 |
| 139.1.43 | 82464 | 102 | 0.0087 | -0.0344 | 2 | 3 | 71,69,87 | 2 | - | - | - | - | z2w,1=1<br>z2w,2=1<br>z2w,3=1<br>z4=2<br>z8=6 | Na2Nd2O10Ti3 |
| 139.1.44 | 75607 | 102 | 0.0065 | -0.0340 | 2 | 3 | 71,69,87 | 2 | - | - | - | - | z2w,1=1<br>z2w,2=1<br>z2w,3=1<br>z4=2<br>z8=6 | Na2Nd2O10Ti3 |
| 139.1.45 | 75149 | 102 | 0.0065 | -0.0340 | 2 | 3 | 71,69,87 | 2 | - | - | - | - | z2w,1=1<br>z2w,2=1<br>z2w,3=1<br>z4=2<br>z8=6 | Na2Nd2O10Ti3 |
| 139.1.46 | 152123 | 50 | 0.0439 | -2.1985 | 87 | 15 | 139 | 6 | - | - | - | - | z8=4 | Mo1O4Sr2 |

Continued on next page

**TABLE XIV– continued**

| TQC | ICSD | NoE | DG | IndG | SubSG (w/ T) | cl | SubSG (w/o T) | cl | Max. Sub. SG | yes or no | Max. Sub. SG | yes or no | SIs (w/ T) | Stoich. Formula |
|---|---|---|---|---|---|---|---|---|---|---|---|---|---|---|
| 139.1.47 | 290825 | 46 | 0.0000 | -0.8023 | 2,82 | 2 | 2,82 | 1 | - | - | - | - | z2w,1=1<br>z2w,2=1<br>z2w,3=1<br>z4=1<br>z2=1<br>z8=1 | Mn2Si2Yb1 |
| 139.1.48 | 290824 | 46 | 0.0000 | -0.8188 | 2,82 | 2 | 2,82 | 1 | - | - | - | - | z2w,1=1<br>z2w,2=1<br>z2w,3=1<br>z4=1<br>z2=1<br>z8=1 | Mn2Si2Yb1 |
| 139.1.49 | 290822 | 46 | 0.0027 | -0.8244 | 2,82 | 2 | 2,82 | 1 | - | - | - | - | z2w,1=1<br>z2w,2=1<br>z2w,3=1<br>z4=1<br>z2=1<br>z8=1 | Mn2Si2Yb1 |
| 139.1.50 | 104941 | 74 | 0.0000 | -0.5638 | 2 | 3 | 71,69,87 | 2 | - | - | - | - | z2w,1=1<br>z2w,2=1<br>z2w,3=1<br>z4=2<br>z8=6 | Mn1Pd3 |
| 139.1.51 | 150228 | 64 | 0.0026 | -0.6166 | 2,82 | 1 | 2,82 | 3 | - | - | - | - | z4=3<br>z2=1<br>z8=3 | Mg1Pd3 |
| 139.1.52 | 42733 | 54 | 0.0200 | -0.5934 | 2,82 | 5 | 2,82 | 1 | - | - | - | - | z4=1<br>z2=1<br>z8=1 | La2Sb1 |
| 139.1.53 | 423290 | 54 | 0.0197 | -0.5888 | 2,82 | 5 | 2,82 | 1 | - | - | - | - | z4=1<br>z2=1<br>z8=1 | La2Sb1 |
| 139.1.54 | 10442 | 54 | 0.0195 | -0.5947 | 2,82 | 5 | 2,82 | 1 | - | - | - | - | z4=1<br>z2=1<br>z8=1 | La2Sb1 |
| 139.1.55 | 40262 | 56 | 0.0440 | -0.9614 | 2,82 | 1 | 2,82 | 3 | - | - | - | - | z4=3<br>z2=1<br>z8=3 | La2O4Pd1 |
| 139.1.56 | 69172 | 56 | 0.0392 | -0.7822 | 87 | 15 | 139 | 6 | - | - | - | - | z8=4 | La2Ni1O4 |
| 139.1.57 | 33536 | 56 | 0.0177 | -1.3018 | 87 | 15 | 139 | 6 | - | - | - | - | z8=4 | La2Ni1O4 |
| 139.1.58 | 181385 | 66 | 0.0184 | -0.3830 | 2 | 12 | 71,69,87 | 5 | - | - | - | - | z2w,1=1<br>z2w,2=1<br>z2w,3=1<br>z4=2<br>z8=2 | La2Mn2O3Se2 |
| 139.1.59 | 427126 | 38 | 0.0102 | -0.5398 | 2,82 | 5 | 2,82 | 1 | - | - | - | - | z4=1<br>z2=1<br>z8=1 | La1Sb1Sc1 |
| 139.1.60 | 421342 | 40 | 0.0506 | -0.1975 | 87 | 15 | 139 | 6 | - | - | - | - | z8=4 | K2Mg5Sn3 |
| 139.1.61 | 420251 | 66 | 0.0281 | -0.7361 | 2,82 | 13 | 2,82 | 8 | - | - | - | - | z4=3<br>z2=1<br>z8=7 | In1Pd3 |
| 139.1.62 | 72190 | 28 | 0.0021 | -0.0167 | 2 | 3 | 71,69,87 | 2 | - | - | - | - | z2w,1=1<br>z2w,2=1<br>z2w,3=1<br>z4=2<br>z8=6 | I2Nd1 |

Continued on next page

**TABLE XIV– continued**

| TQC | ICSD | NoE | DG | IndG | SubSG (w/ T) | cl | SubSG (w/o T) | cl | Max. Sub. SG | yes or no | Max. Sub. SG | yes or no | SIs (w/ T) | Stoich. Formula |
|---|---|---|---|---|---|---|---|---|---|---|---|---|---|---|
| 139.1.63 | 710068 | 132 | 0.0000 | -0.0062 | 2,82 | 14 | 2,82 | 3 | - | - | - | - | z2w,1=1 z2w,2=1 z2w,3=1 z4=3 z2=1 z8=3 | Ho1Si3 |
| 139.1.64 | 427211 | 40 | 0.0289 | -2.1441 | 2,82 | 13 | 2,82 | 8 | - | - | - | - | z4=3 z2=1 z8=7 | Hg2In2Sr1 |
| 139.1.65 | 75936 | 24 | 0.0116 | -1.6543 | 87 | 15 | 139 | 6 | - | - | - | - | z8=4 | Hf3Te2 |
| 139.1.66 | 415033 | 108 | 0.0655 | 0.0655 | 2,82 | 14 | 2,82 | 3 | - | - | - | - | z2w,1=1 z2w,2=1 z2w,3=1 z4=3 z2=1 z8=3 | H2Ba8Bi4O1 |
| 139.1.67 | 411 | 42 | 0.0180 | -0.8078 | 2,82 | 2 | 2,82 | 1 | - | - | - | - | z2w,1=1 z2w,2=1 z2w,3=1 z4=1 z2=1 z8=1 | Ge2Sr1Zn2 |
| 139.1.68 | 53731 | 42 | 0.0345 | -0.1389 | 2 | 3 | 71,69,87 | 2 | - | - | - | - | z2w,1=1 z2w,2=1 z2w,3=1 z4=2 z8=6 | Ge2Nd1Pd2 |
| 139.1.69 | 81753 | 42 | 0.0134 | -0.1102 | 2 | 3 | 71,69,87 | 2 | - | - | - | - | z2w,1=1 z2w,2=1 z2w,3=1 z4=2 z8=6 | Ge2Nd1Ni2 |
| 139.1.70 | 81744 | 36 | 0.0180 | -0.4425 | 2 | 4 | 139 | 6 | - | - | - | - | z2w,1=1 z2w,2=1 z2w,3=1 z8=4 | Ge2Mn2U1 |
| 139.1.71 | 41976 | 34 | 0.0080 | -0.7931 | 2 | 4 | 139 | 6 | - | - | - | - | z2w,1=1 z2w,2=1 z2w,3=1 z8=4 | Ge2Mn2Th1 |
| 139.1.72 | 403 | 32 | 0.0040 | -0.7019 | 2 | 4 | 139 | 6 | - | - | - | - | z2w,1=1 z2w,2=1 z2w,3=1 z8=4 | Ge2Mn2Sr1 |
| 139.1.73 | 165737 | 36 | 0.0418 | -2.0337 | 2,82 | 14 | 2,82 | 3 | - | - | - | - | z2w,1=1 z2w,2=1 z2w,3=1 z4=3 z2=1 z8=3 | Ge2Ir2Sr1 |
| 139.1.74 | 26277 | 48 | 0.0051 | -0.0501 | 2,82 | 13 | 2,82 | 8 | - | - | - | - | z4=3 z2=1 z8=7 | Ge1Te1U1 |
| 139.1.75 | 87316 | 48 | 0.0097 | -0.0613 | 2,82 | 13 | 2,82 | 8 | - | - | - | - | z4=3 z2=1 z8=7 | Ge1Se1U1 |

Continued on next page

**TABLE XIV– continued**

| TQC | ICSD | NoE | DG | IndG | SubSG (w/ T) | cl | SubSG (w/o T) | cl | Max. Sub. SG | yes or no | Max. Sub. SG | yes or no | SIs (w/ T) | Stoich. Formula |
|---|---|---|---|---|---|---|---|---|---|---|---|---|---|---|
| 139.1.76 | 26276 | 48 | 0.0089 | -0.0542 | 2,82 | 13 | 2,82 | 8 | - | - | - | - | z4=3<br>z2=1<br>z8=7 | Ge1Se1U1 |
| 139.1.77 | 89563 | 54 | 0.0000 | -0.0418 | 2,82 | 7 | 2,82 | 4 | - | - | - | - | z2w,1=1<br>z2w,2=1<br>z2w,3=1<br>z4=1<br>z2=1<br>z8=5 | Gd2O2Te1 |
| 139.1.78 | 82466 | 110 | 0.0023 | -0.0347 | 87 | 15 | 139 | 6 | - | - | - | - | z8=4 | Gd2Na2O10Ti3 |
| 139.1.79 | 30844 | 38 | 0.0037 | -0.0654 | 2,82 | 7 | 2,82 | 4 | - | - | - | - | z2w,1=1<br>z2w,2=1<br>z2w,3=1<br>z4=1<br>z2=1<br>z8=5 | Gd1P2Pd2 |
| 139.1.80 | 26823 | 102 | 0.0063 | -0.1468 | 2 | 6 | 0 | 7 | - | - | - | - | z2w,1=1<br>z2w,2=1<br>z2w,3=1 | Gd1Mn12 |
| 139.1.81 | 90392 | 52 | 0.0259 | -0.0184 | 2 | 12 | 71,69,87 | 5 | - | - | - | - | z2w,1=1<br>z2w,2=1<br>z2w,3=1<br>z4=2<br>z8=2 | Gd1Ge1Ti1 |
| 139.1.82 | 68209 | 36 | 0.0541 | -1.0353 | 2,82 | 8 | 2,82 | 4 | - | - | - | - | z4=1<br>z2=1<br>z8=5 | Fe2Si2Th1 |
| 139.1.83 | 10467 | 36 | 0.0108 | -0.8591 | 2,82 | 9 | 2,82 | 8 | - | - | - | - | z2w,1=1<br>z2w,2=1<br>z2w,3=1<br>z4=3<br>z2=1<br>z8=7 | Fe2P2Sr1 |
| 139.1.84 | 74434 | 82 | 0.0091 | -0.4279 | 2,82 | 7 | 2,82 | 4 | - | - | - | - | z2w,1=1<br>z2w,2=1<br>z2w,3=1<br>z4=1<br>z2=1<br>z8=5 | Fe2O6Sr3 |
| 139.1.85 | 249015 | 82 | 0.0066 | -0.4096 | 2,82 | 7 | 2,82 | 4 | - | - | - | - | z2w,1=1<br>z2w,2=1<br>z2w,3=1<br>z4=1<br>z2=1<br>z8=5 | Fe2O6Sr3 |
| 139.1.86 | 249014 | 82 | 0.0073 | -0.4772 | 2,82 | 7 | 2,82 | 4 | - | - | - | - | z2w,1=1<br>z2w,2=1<br>z2w,3=1<br>z4=1<br>z2=1<br>z8=5 | Fe2O6Sr3 |
| 139.1.87 | 81746 | 38 | 0.0102 | -0.2069 | 2,82 | 8 | 2,82 | 4 | - | - | - | - | z4=1<br>z2=1<br>z8=5 | Fe2Ge2Nd1 |
| 139.1.88 | 74421 | 52 | 0.0393 | -0.5229 | 87 | 15 | 139 | 6 | - | - | - | - | z8=4 | Fe1O4Sr2 |
| 139.1.89 | 74420 | 52 | 0.0376 | -0.5176 | 87 | 15 | 139 | 6 | - | - | - | - | z8=4 | Fe1O4Sr2 |
| 139.1.90 | 74419 | 52 | 0.0393 | -0.5231 | 87 | 15 | 139 | 6 | - | - | - | - | z8=4 | Fe1O4Sr2 |
| 139.1.91 | 69849 | 52 | 0.0393 | -0.5231 | 87 | 15 | 139 | 6 | - | - | - | - | z8=4 | Fe1O4Sr2 |

Continued on next page

**TABLE XIV– continued**

| TQC | ICSD | NoE | DG | IndG | SubSG (w/ T) | cl | SubSG (w/o T) | cl | Max. Sub. SG | yes or no | Max. Sub. SG | yes or no | SIs (w/ T) | Stoich. Formula |
|---|---|---|---|---|---|---|---|---|---|---|---|---|---|---|
| 139.1.92 | 69682 | 56 | 0.0151 | -0.7461 | 87 | 15 | 139 | 6 | - | - | - | - | z8=4 | F4Ni1Rb2 |
| 139.1.93 | 73450 | 56 | 0.0055 | -0.7537 | 87 | 15 | 139 | 6 | - | - | - | - | z8=4 | F4K2Ni1 |
| 139.1.94 | 33520 | 56 | 0.0106 | -0.7618 | 87 | 15 | 139 | 6 | - | - | - | - | z8=4 | F4K2Ni1 |
| 139.1.95 | 167014 | 58 | 0.0211 | -0.3917 | 87 | 15 | 139 | 6 | - | - | - | - | z8=4 | F2O1Sb2Sr2Ti2 |
| 139.1.96 | 426633 | 56 | 0.0030 | -0.0347 | 87 | 15 | 139 | 6 | - | - | - | - | z8=4 | Er2Mg2Ru1 |
| 139.1.97 | 60667 | 80 | 0.0031 | -0.1086 | 2,82 | 13 | 2,82 | 8 | - | - | - | - | z4=3<br>z2=1<br>z8=7 | Er1O2S3U2 |
| 139.1.98 | 245051 | 50 | 0.0043 | -0.1821 | 2,82 | 9 | 2,82 | 8 | - | - | - | - | z2w,1=1<br>z2w,2=1<br>z2w,3=1<br>z4=3<br>z2=1<br>z8=7 | Er1Ni2Si2 |
| 139.1.99 | 109428 | 50 | 0.0035 | -0.1989 | 2,82 | 14 | 2,82 | 3 | - | - | - | - | z2w,1=1<br>z2w,2=1<br>z2w,3=1<br>z4=3<br>z2=1<br>z8=3 | Er1Ge2Ni2 |
| 139.1.100 | 30846 | 40 | 0.0030 | -0.0295 | 2,82 | 7 | 2,82 | 4 | - | - | - | - | z2w,1=1<br>z2w,2=1<br>z2w,3=1<br>z4=1<br>z2=1<br>z8=5 | Dy1P2Pd2 |
| 139.1.101 | 88184 | 50 | 0.0157 | -0.0730 | 2,82 | 5 | 2,82 | 1 | - | - | - | - | z4=1<br>z2=1<br>z8=1 | Dy1Ni2P2 |
| 139.1.102 | 88183 | 50 | 0.0152 | -0.0722 | 2,82 | 5 | 2,82 | 1 | - | - | - | - | z4=1<br>z2=1<br>z8=1 | Dy1Ni2P2 |
| 139.1.103 | 246933 | 44 | 0.0438 | -0.2599 | 2 | 6 | 0 | 7 | - | - | - | - | z2w,1=1<br>z2w,2=1<br>z2w,3=1 | Cu2Si2U1 |
| 139.1.104 | 68212 | 42 | 0.0381 | -1.1634 | 2 | 6 | 0 | 7 | - | - | - | - | z2w,1=1<br>z2w,2=1<br>z2w,3=1 | Cu2Si2Th1 |
| 139.1.105 | 88424 | 76 | 0.0181 | -0.2485 | 2,82 | 1 | 2,82 | 3 | - | - | - | - | z4=3<br>z2=1<br>z8=3 | Cu2Ni1O2S2Sr2 |
| 139.1.106 | 86002 | 44 | 0.0830 | -0.0793 | 2 | 6 | 0 | 7 | - | - | - | - | z2w,1=1<br>z2w,2=1<br>z2w,3=1 | Cu2Ge2U1 |
| 139.1.107 | 86001 | 44 | 0.0825 | -0.0796 | 2 | 6 | 0 | 7 | - | - | - | - | z2w,1=1<br>z2w,2=1<br>z2w,3=1 | Cu2Ge2U1 |
| 139.1.108 | 86000 | 44 | 0.0237 | -0.2006 | 2 | 6 | 0 | 7 | - | - | - | - | z2w,1=1<br>z2w,2=1<br>z2w,3=1 | Cu2Ge2U1 |
| 139.1.109 | 85999 | 44 | 0.0238 | -0.1999 | 2 | 6 | 0 | 7 | - | - | - | - | z2w,1=1<br>z2w,2=1<br>z2w,3=1 | Cu2Ge2U1 |
| 139.1.110 | 85998 | 44 | 0.0237 | -0.1994 | 2 | 6 | 0 | 7 | - | - | - | - | z2w,1=1<br>z2w,2=1<br>z2w,3=1 | Cu2Ge2U1 |
| 139.1.111 | 424224 | 40 | 0.0097 | -2.6608 | 2 | 10 | 71,69,87 | 2 | - | - | - | - | z4=2<br>z8=6 | Cu2Ge2Sr1 |

Continued on next page

**TABLE XIV– continued**

| TQC | ICSD | NoE | DG | IndG | SubSG (w/ T) | cl | SubSG (w/o T) | cl | Max. Sub. SG | yes or no | Max. Sub. SG | yes or no | SIs (w/ T) | Stoich. Formula |
|---|---|---|---|---|---|---|---|---|---|---|---|---|---|---|
| 139.1.112 | 410 | 40 | 0.0086 | -2.5074 | 2 | 10 | 71,69,87 | 2 | - | - | - | - | z4=2<br>z8=6 | Cu2Ge2Sr1 |
| 139.1.113 | 154203 | 110 | 0.0058 | -0.4158 | 2 | 4 | 139 | 6 | - | - | - | - | z2w,1=1<br>z2w,2=1<br>z2w,3=1<br>z8=4 | Cu2Fe2O5Se2Sr3 |
| 139.1.114 | 40763 | 70 | 0.0055 | -0.1392 | 2 | 3 | 71,69,87 | 2 | - | - | - | - | z2w,1=1<br>z2w,2=1<br>z2w,3=1<br>z4=2<br>z8=6 | Cu1P2U1 |
| 139.1.115 | 290194 | 38 | 0.0025 | -2.0766 | 2,82 | 13 | 2,82 | 8 | - | - | - | - | z4=3<br>z2=1<br>z8=7 | Cu1Ge1Hf1 |
| 139.1.116 | 67472 | 38 | 0.0103 | -0.1884 | 2,82 | 7 | 2,82 | 4 | - | - | - | - | z2w,1=1<br>z2w,2=1<br>z2w,3=1<br>z4=1<br>z2=1<br>z8=5 | Cr2Gd1Si2 |
| 139.1.117 | 245595 | 50 | 0.0210 | -1.0360 | 87 | 15 | 139 | 6 | - | - | - | - | z8=4 | Cr1O4Sr2 |
| 139.1.118 | 43091 | 106 | 0.0000 | -0.1296 | 2 | 4 | 139 | 6 | - | - | - | - | z2w,1=1<br>z2w,2=1<br>z2w,3=1<br>z8=4 | Co4Ge7U3 |
| 139.1.119 | 290506 | 106 | 0.0093 | -0.0929 | 2 | 4 | 139 | 6 | - | - | - | - | z2w,1=1<br>z2w,2=1<br>z2w,3=1<br>z8=4 | Co4Ge7U3 |
| 139.1.120 | 290505 | 106 | 0.0092 | -0.0930 | 2 | 4 | 139 | 6 | - | - | - | - | z2w,1=1<br>z2w,2=1<br>z2w,3=1<br>z8=4 | Co4Ge7U3 |
| 139.1.121 | 290504 | 106 | 0.0079 | -0.1102 | 2 | 4 | 139 | 6 | - | - | - | - | z2w,1=1<br>z2w,2=1<br>z2w,3=1<br>z8=4 | Co4Ge7U3 |
| 139.1.122 | 42 - 0 | 36 | 0.0314 | -1.4725 | 2,82 | 8 | 2,82 | 4 | - | - | - | - | z4=1<br>z2=1<br>z8=5 | Co2Si2Sr1 |
| 139.1.123 | 73653 | 42 | 0.0067 | -0.1983 | 2,82 | 14 | 2,82 | 3 | - | - | - | - | z2w,1=1<br>z2w,2=1<br>z2w,3=1<br>z4=3<br>z2=1<br>z8=3 | Co2Nd1P2 |
| 139.1.124 | 73652 | 42 | 0.0058 | -0.1933 | 2,82 | 14 | 2,82 | 3 | - | - | - | - | z2w,1=1<br>z2w,2=1<br>z2w,3=1<br>z4=3<br>z2=1<br>z8=3 | Co2Nd1P2 |
| 139.1.125 | 73651 | 42 | 0.0071 | -0.1931 | 2,82 | 14 | 2,82 | 3 | - | - | - | - | z2w,1=1<br>z2w,2=1<br>z2w,3=1<br>z4=3<br>z2=1<br>z8=3 | Co2Nd1P2 |

Continued on next page

**TABLE XIV– continued**

| TQC | ICSD | NoE | DG | IndG | SubSG (w/ T) | cl | SubSG (w/o T) | cl | Max. Sub. SG | yes or no | Max. Sub. SG | yes or no | SIs (w/ T) | Stoich. Formula |
|---|---|---|---|---|---|---|---|---|---|---|---|---|---|---|
| 139.1.126 | 261192 | 70 | 0.0349 | -0.3842 | 2 | 3 | 71,69,87 | 2 | - | - | - | - | z2w,1=1 z2w,2=1 z2w,3=1 z4=2 z8=6 | Co2La2O3Se2 |
| 139.1.127 | 261191 | 70 | 0.0360 | -0.3853 | 2 | 3 | 71,69,87 | 2 | - | - | - | - | z2w,1=1 z2w,2=1 z2w,3=1 z4=2 z8=6 | Co2La2O3Se2 |
| 139.1.128 | 248239 | 70 | 0.0341 | -0.3832 | 2 | 3 | 71,69,87 | 2 | - | - | - | - | z2w,1=1 z2w,2=1 z2w,3=1 z4=2 z8=6 | Co2La2O3Se2 |
| 139.1.129 | 168330 | 70 | 0.0346 | -0.3799 | 2 | 3 | 71,69,87 | 2 | - | - | - | - | z2w,1=1 z2w,2=1 z2w,3=1 z4=2 z8=6 | Co2La2O3Se2 |
| 139.1.130 | 407 | 36 | 0.0201 | -1.3314 | 2,82 | 8 | 2,82 | 4 | - | - | - | - | z4=1 z2=1 z8=5 | Co2Ge2Sr1 |
| 139.1.131 | 33524 | 94 | 0.0115 | -0.6818 | 87 | 15 | 139 | 6 | - | - | - | - | z8=4 | Co2F7K3 |
| 139.1.132 | 30081 | 90 | 0.0053 | -0.3753 | 87 | 15 | 139 | 6 | - | - | - | - | z8=4 | Cl7Mn2Rb3 |
| 139.1.133 | 2241 | 90 | 0.0048 | -0.3580 | 87 | 15 | 139 | 6 | - | - | - | - | z8=4 | Cl7Mn2Rb3 |
| 139.1.134 | 36127 | 90 | 0.0074 | -0.3664 | 87 | 15 | 139 | 6 | - | - | - | - | z8=4 | Cl7K3Mn2 |
| 139.1.135 | 99942 | 90 | 0.0074 | -0.4406 | 87 | 15 | 139 | 6 | - | - | - | - | z8=4 | Cl2Fe2O5Sr3 |
| 139.1.136 | 69871 | 90 | 0.0049 | -0.4273 | 87 | 15 | 139 | 6 | - | - | - | - | z8=4 | Cl2Fe2O5Sr3 |
| 139.1.137 | 33258 | 90 | 0.0075 | -0.4382 | 87 | 15 | 139 | 6 | - | - | - | - | z8=4 | Cl2Fe2O5Sr3 |
| 139.1.138 | 174532 | 84 | 0.0042 | -0.9615 | 87 | 15 | 139 | 6 | - | - | - | - | z8=4 | Cl2Fe2O4Sr3 |
| 139.1.139 | 91749 | 92 | 0.0243 | -0.4429 | 2 | 12 | 71,69,87 | 5 | - | - | - | - | z2w,1=1 z2w,2=1 z2w,3=1 z4=2 z8=2 | Cl2Co2O5Sr3 |
| 139.1.140 | 90123 | 92 | 0.0262 | -0.3341 | 2 | 12 | 71,69,87 | 5 | - | - | - | - | z2w,1=1 z2w,2=1 z2w,3=1 z4=2 z8=2 | Cl2Co2O5Sr3 |
| 139.1.141 | 251170 | 86 | 0.0073 | -0.6762 | 2 | 12 | 71,69,87 | 5 | - | - | - | - | z2w,1=1 z2w,2=1 z2w,3=1 z4=2 z8=2 | Cl2Co2O4Sr3 |
| 139.1.142 | 409587 | 58 | 0.0141 | -0.1148 | 87 | 15 | 139 | 6 | - | - | - | - | z8=4 | Ce2Sb1 |
| 139.1.143 | 150461 | 64 | 0.0109 | -0.2555 | 2,82 | 13 | 2,82 | 8 | - | - | - | - | z4=3 z2=1 z8=7 | Ce2Ga10Pd1 |
| 139.1.144 | 20525 | 64 | 0.0075 | -0.2497 | 2,82 | 13 | 2,82 | 8 | - | - | - | - | z4=3 z2=1 z8=7 | Ce2Ga10Ni1 |
| 139.1.145 | 87159 | 38 | 0.0042 | -0.3345 | 2,82 | 8 | 2,82 | 4 | - | - | - | - | z4=1 z2=1 z8=5 | Ce1Sc1Si1 |

Continued on next page

**TABLE XIV– continued**

| TQC | ICSD | NoE | DG | IndG | SubSG (w/ T) | cl | SubSG (w/o T) | cl | Max. Sub. SG | yes or no | Max. Sub. SG | yes or no | SIs (w/ T) | Stoich. Formula |
|---|---|---|---|---|---|---|---|---|---|---|---|---|---|---|
| 139.1.146 | 290821 | 38 | 0.0154 | -0.2965 | 2,82 | 8 | 2,82 | 4 | - | - | - | - | z4=1<br>z2=1<br>z8=5 | Ce1Sc1Si1 |
| 139.1.147 | 411677 | 38 | 0.0575 | -0.2675 | 87 | 15 | 139 | 6 | - | - | - | - | z8=4 | Ce1Ir2Si2 |
| 139.1.148 | 85893 | 40 | 0.0208 | -0.3028 | 2,82 | 9 | 2,82 | 8 | - | - | - | - | z2w,1=1<br>z2w,2=1<br>z2w,3=1<br>z4=3<br>z2=1<br>z8=7 | Ce1Co2P2 |
| 139.1.149 | 370021 | 40 | 0.0209 | -2.4329 | 2 | 10 | 71,69,87 | 2 | - | - | - | - | z4=2<br>z8=6 | Cd2Ga2Sr1 |
| 139.1.150 | 416139 | 140 | 0.0048 | -2.1345 | 2 | 4 | 139 | 6 | - | - | - | - | z2w,1=1<br>z2w,2=1<br>z2w,3=1<br>z8=4 | Ca2Pd3Sb4 |
| 139.1.151 | 59649 | 34 | 0.0335 | -0.8643 | 87 | 15 | 139 | 6 | - | - | - | - | z8=4 | Ca1Si2Zn2 |
| 139.1.152 | 425469 | 28 | 0.0361 | -1.8630 | 2,82 | 14 | 2,82 | 3 | - | - | - | - | z2w,1=1<br>z2w,2=1<br>z2w,3=1<br>z4=3<br>z2=1<br>z8=3 | Ca1Rh2Si2 |
| 139.1.153 | 50185 | 30 | 0.0647 | -0.4707 | 2 | 4 | 139 | 6 | - | - | - | - | z2w,1=1<br>z2w,2=1<br>z2w,3=1<br>z8=4 | Ca1P2Rh2 |
| 139.1.154 | 36371 | 32 | 0.0599 | -2.2036 | 2,82 | 13 | 2,82 | 8 | - | - | - | - | z4=3<br>z2=1<br>z8=7 | Ca1P2Pd2 |
| 139.1.155 | 30835 | 22 | 0.0211 | -1.4462 | 2,82 | 7 | 2,82 | 4 | - | - | - | - | z2w,1=1<br>z2w,2=1<br>z2w,3=1<br>z4=1<br>z2=1<br>z8=5 | Ca1P2Pd2 |
| 139.1.156 | 10461 | 32 | 0.0269 | -1.9179 | 2,82 | 13 | 2,82 | 8 | - | - | - | - | z4=3<br>z2=1<br>z8=7 | Ca1Ni2P2 |
| 139.1.157 | 414231 | 20 | 0.0235 | -0.9396 | 2 | 4 | 139 | 6 | - | - | - | - | z2w,1=1<br>z2w,2=1<br>z2w,3=1<br>z8=4 | Ca1In1 |
| 139.1.158 | 52761 | 26 | 0.0076 | -0.7757 | 2,82 | 5 | 2,82 | 1 | - | - | - | - | z4=1<br>z2=1<br>z8=1 | Ca1Ge2Ru2 |
| 139.1.159 | 52759 | 30 | 0.0332 | -1.9520 | 2 | 11 | 71,69,87 | 5 | - | - | - | - | z4=2<br>z8=2 | Ca1Ge2Pd2 |
| 139.1.160 | 81755 | 30 | 0.0335 | -1.8938 | 2 | 4 | 139 | 6 | - | - | - | - | z2w,1=1<br>z2w,2=1<br>z2w,3=1<br>z8=4 | Ca1Ge2Ni2 |
| 139.1.161 | 408 | 30 | 0.0348 | -1.9443 | 2 | 4 | 139 | 6 | - | - | - | - | z2w,1=1<br>z2w,2=1<br>z2w,3=1<br>z8=4 | Ca1Ge2Ni2 |

Continued on next page

**TABLE XIV– continued**

| TQC | ICSD | NoE | DG | IndG | SubSG (w/ T) | cl | SubSG (w/o T) | cl | Max. Sub. SG | yes or no | Max. Sub. SG | yes or no | SIs (w/ T) | Stoich. Formula |
|---|---|---|---|---|---|---|---|---|---|---|---|---|---|---|
| 139.1.162 | 81743 | 24 | 0.0033 | -0.6979 | 2 | 6 | 0 | 7 | - | - | - | - | z2w,1=1 z2w,2=1 z2w,3=1 | Ca1Ge2Mn2 |
| 139.1.163 | 402 | 24 | 0.0035 | -0.7130 | 2 | 6 | 0 | 7 | - | - | - | - | z2w,1=1 z2w,2=1 z2w,3=1 | Ca1Ge2Mn2 |
| 139.1.164 | 52755 | 28 | 0.1292 | -2.0104 | 2,82 | 14 | 2,82 | 3 | - | - | - | - | z2w,1=1 z2w,2=1 z2w,3=1 z4=3 z2=1 z8=3 | Ca1Ge2Ir2 |
| 139.1.165 | 10463 | 28 | 0.0030 | -1.2494 | 2,82 | 8 | 2,82 | 4 | - | - | - | - | z4=1 z2=1 z8=5 | Ca1Fe2P2 |
| 139.1.166 | 41756 | 34 | 0.0608 | -0.8306 | 2,82 | 2 | 2,82 | 1 | - | - | - | - | z2w,1=1 z2w,2=1 z2w,3=1 z4=1 z2=1 z8=1 | Ca1Cu2P2 |
| 139.1.167 | 420974 | 28 | 0.0209 | -1.5873 | 2,82 | 14 | 2,82 | 3 | - | - | - | - | z2w,1=1 z2w,2=1 z2w,3=1 z4=3 z2=1 z8=3 | Ca1Co2Si2 |
| 139.1.168 | 85892 | 30 | 0.0344 | -0.2455 | 2 | 4 | 139 | 6 | - | - | - | - | z2w,1=1 z2w,2=1 z2w,3=1 z8=4 | Ca1Co2P2 |
| 139.1.169 | 85891 | 30 | 0.0621 | -0.2920 | 2 | 4 | 139 | 6 | - | - | - | - | z2w,1=1 z2w,2=1 z2w,3=1 z8=4 | Ca1Co2P2 |
| 139.1.170 | 85890 | 30 | 0.0621 | -0.2920 | 2 | 4 | 139 | 6 | - | - | - | - | z2w,1=1 z2w,2=1 z2w,3=1 z8=4 | Ca1Co2P2 |
| 139.1.171 | 81751 | 28 | 0.0176 | -1.4296 | 2,82 | 14 | 2,82 | 3 | - | - | - | - | z2w,1=1 z2w,2=1 z2w,3=1 z4=3 z2=1 z8=3 | Ca1Co2Ge2 |
| 139.1.172 | 406 | 28 | 0.0173 | -1.4035 | 2,82 | 14 | 2,82 | 3 | - | - | - | - | z2w,1=1 z2w,2=1 z2w,3=1 z4=3 z2=1 z8=3 | Ca1Co2Ge2 |
| 139.1.173 | 74670 | 32 | 0.0121 | -0.1637 | 2 | 12 | 71,69,87 | 5 | - | - | - | - | z2w,1=1 z2w,2=1 z2w,3=1 z4=2 z8=2 | C2Yb1 |

Continued on next page

**TABLE XIV– continued**

| TQC | ICSD | NoE | DG | IndG | SubSG (w/ T) | cl | SubSG (w/o T) | cl | Max. Sub. SG | yes or no | Max. Sub. SG | yes or no | SIs (w/ T) | Stoich. Formula |
|---|---|---|---|---|---|---|---|---|---|---|---|---|---|---|
| 139.1.174 | 23686 | 32 | 0.0160 | -0.1728 | 2 | 12 | 71,69,87 | 5 | - | - | - | - | z2w,1=1<br>z2w,2=1<br>z2w,3=1<br>z4=2<br>z8=2 | C2Yb1 |
| 139.1.175 | 26478 | 22 | 0.0163 | -0.3999 | 2,82 | 13 | 2,82 | 8 | - | - | - | - | z4=3<br>z2=1<br>z8=7 | C2U1 |
| 139.1.176 | 202305 | 22 | 0.0217 | -0.3939 | 2,82 | 13 | 2,82 | 8 | - | - | - | - | z4=3<br>z2=1<br>z8=7 | C2U1 |
| 139.1.177 | 63560 | 22 | 0.0072 | -0.1648 | 87 | 15 | 139 | 6 | - | - | - | - | z8=4 | C2Nd1 |
| 139.1.178 | 69888 | 30 | 0.0360 | -0.1621 | 2 | 3 | 71,69,87 | 2 | - | - | - | - | z2w,1=1<br>z2w,2=1<br>z2w,3=1<br>z4=2<br>z8=6 | C2Er1 |
| 139.1.179 | 23690 | 30 | 0.0312 | -0.1671 | 2 | 3 | 71,69,87 | 2 | - | - | - | - | z2w,1=1<br>z2w,2=1<br>z2w,3=1<br>z4=2<br>z8=6 | C2Er1 |
| 139.1.180 | 23689 | 30 | 0.0317 | -0.1638 | 2 | 3 | 71,69,87 | 2 | - | - | - | - | z2w,1=1<br>z2w,2=1<br>z2w,3=1<br>z4=2<br>z8=6 | C2Er1 |
| 139.1.181 | 380471 | 88 | 0.0246 | -0.0811 | 87 | 15 | 139 | 6 | - | - | - | - | z8=4 | C2Cl1N2Ni1Rb1 |
| 139.1.182 | 380469 | 88 | 0.0174 | -0.0678 | 87 | 15 | 139 | 6 | - | - | - | - | z8=4 | C2Cl1Cs1N2Ni1 |
| 139.1.183 | 380470 | 88 | 0.0098 | -0.0813 | 87 | 15 | 139 | 6 | - | - | - | - | z8=4 | C2Br1N2Ni1Rb1 |
| 139.1.184 | 380468 | 88 | 0.0175 | -0.0849 | 87 | 15 | 139 | 6 | - | - | - | - | z8=4 | C2Br1Cs1N2Ni1 |
| 139.1.185 | 79563 | 48 | 0.0216 | -0.2373 | 2 | 4 | 139 | 6 | - | - | - | - | z2w,1=1<br>z2w,2=1<br>z2w,3=1<br>z8=4 | C1B2Gd1Ni2 |
| 139.1.186 | 56651 | 52 | 0.0174 | -0.0597 | 2 | 6 | 0 | 7 | - | - | - | - | z2w,1=1<br>z2w,2=1<br>z2w,3=1 | C1B2Er1Ni2 |
| 139.1.187 | 56596 | 50 | 0.0034 | -0.0755 | 2 | 4 | 139 | 6 | - | - | - | - | z2w,1=1<br>z2w,2=1<br>z2w,3=1<br>z8=4 | C1B2Dy1Ni2 |
| 139.1.188 | 56595 | 50 | 0.0031 | -0.0755 | 2 | 4 | 139 | 6 | - | - | - | - | z2w,1=1<br>z2w,2=1<br>z2w,3=1<br>z8=4 | C1B2Dy1Ni2 |
| 139.1.189 | 56594 | 50 | 0.0028 | -0.0753 | 2 | 4 | 139 | 6 | - | - | - | - | z2w,1=1<br>z2w,2=1<br>z2w,3=1<br>z8=4 | C1B2Dy1Ni2 |
| 139.1.190 | 56593 | 50 | 0.0028 | -0.0750 | 2 | 4 | 139 | 6 | - | - | - | - | z2w,1=1<br>z2w,2=1<br>z2w,3=1<br>z8=4 | C1B2Dy1Ni2 |
| 139.1.191 | 95097 | 40 | 0.0140 | -0.6541 | 2,82 | 5 | 2,82 | 1 | - | - | - | - | z4=1<br>z2=1<br>z8=1 | C1B2Ce1Rh2 |

Continued on next page

**TABLE XIV– continued**

| TQC | ICSD | NoE | DG | IndG | SubSG (w/ T) | cl | SubSG (w/o T) | cl | Max. Sub. SG | yes or no | Max. Sub. SG | yes or no | SIs (w/ T) | Stoich. Formula |
|---|---|---|---|---|---|---|---|---|---|---|---|---|---|---|
| 139.1.192 | 95098 | 40 | 0.0090 | -0.6549 | 2,82 | 5 | 2,82 | 1 | - | - | - | - | z4=1<br>z2=1<br>z8=1 | C1B2Ce1Ir2 |
| 139.1.193 | 32544 | 90 | 0.0000 | -0.3463 | 87 | 15 | 139 | 6 | - | - | - | - | z8=4 | Br7Mn2Rb3 |
| 139.1.194 | 99944 | 90 | 0.0070 | -0.4365 | 87 | 15 | 139 | 6 | - | - | - | - | z8=4 | Br2Fe2O5Sr3 |
| 139.1.195 | 99943 | 90 | 0.0062 | -0.4281 | 87 | 15 | 139 | 6 | - | - | - | - | z8=4 | Br2Fe2O5Sr3 |
| 139.1.196 | 69872 | 90 | 0.0039 | -0.4211 | 87 | 15 | 139 | 6 | - | - | - | - | z8=4 | Br2Fe2O5Sr3 |
| 139.1.197 | 430065 | 58 | 0.0139 | -0.3524 | 87 | 15 | 139 | 6 | - | - | - | - | z8=4 | Bi2F2O1Sr2Ti2 |
| 139.1.198 | 430064 | 58 | 0.0139 | -0.3518 | 87 | 15 | 139 | 6 | - | - | - | - | z8=4 | Bi2F2O1Sr2Ti2 |
| 139.1.199 | 402954 | 84 | 0.0040 | -0.1693 | 2,82 | 8 | 2,82 | 4 | - | - | - | - | z4=1<br>z2=1<br>z8=5 | Bi2Eu4O1 |
| 139.1.200 | 41836 | 50 | 0.0358 | -0.1011 | 2 | 12 | 71,69,87 | 5 | - | - | - | - | z2w,1=1<br>z2w,2=1<br>z2w,3=1<br>z4=2<br>z8=2 | Bi1Sr2 |
| 139.1.201 | 107609 | 22 | 0.0447 | -2.8384 | 2 | 11 | 71,69,87 | 5 | - | - | - | - | z4=2<br>z8=2 | Be1Pd2 |
| 139.1.202 | 58717 | 30 | 0.0190 | -1.4727 | 2,82 | 2 | 2,82 | 1 | - | - | - | - | z2w,1=1<br>z2w,2=1<br>z2w,3=1<br>z4=1<br>z2=1<br>z8=1 | Be12Mo1 |
| 139.1.203 | 237677 | 112 | 0.0000 | -0.5129 | 2,82 | 9 | 2,82 | 8 | - | - | - | - | z2w,1=1<br>z2w,2=1<br>z2w,3=1<br>z4=3<br>z2=1<br>z8=7 | Ba6Ge32Rh8 |
| 139.1.204 | 427108 | 130 | 0.0266 | -0.5958 | 2 | 10 | 71,69,87 | 2 | - | - | - | - | z4=2<br>z8=6 | Ba3Ge16Ir4 |
| 139.1.205 | 194235 | 130 | 0.0135 | -0.5814 | 2 | 10 | 71,69,87 | 2 | - | - | - | - | z4=2<br>z8=6 | Ba3Ge16Ir4 |
| 139.1.206 | 189363 | 130 | 0.0094 | -0.5768 | 2 | 10 | 71,69,87 | 2 | - | - | - | - | z4=2<br>z8=6 | Ba3Ge16Ir4 |
| 139.1.207 | 41837 | 50 | 0.0096 | -0.3249 | 2 | 12 | 71,69,87 | 5 | - | - | - | - | z2w,1=1<br>z2w,2=1<br>z2w,3=1<br>z4=2<br>z8=2 | Ba2Sb1 |
| 139.1.208 | 30084 | 32 | 0.0413 | -1.1890 | 2 | 12 | 71,69,87 | 5 | - | - | - | - | z2w,1=1<br>z2w,2=1<br>z2w,3=1<br>z4=2<br>z8=2 | Ba2Hg1 |
| 139.1.209 | 183149 | 66 | 0.0053 | -0.4001 | 2 | 12 | 71,69,87 | 5 | - | - | - | - | z2w,1=1<br>z2w,2=1<br>z2w,3=1<br>z4=2<br>z8=2 | Ba2F2Mn2O1Se2 |
| 139.1.210 | 71535 | 130 | 0.0086 | -0.0091 | 2 | 12 | 71,69,87 | 5 | - | - | - | - | z2w,1=1<br>z2w,2=1<br>z2w,3=1<br>z4=2<br>z8=2 | Ba2Cu2O12Si4 |
| 139.1.211 | 78593 | 98 | 0.0080 | -2.3086 | 87 | 15 | 139 | 6 | - | - | - | - | z8=4 | Ba2Ca1Cu2O8Tl2 |

Continued on next page

**TABLE XIV– continued**

| TQC | ICSD | NoE | DG | IndG | SubSG (w/ T) | cl | SubSG (w/o T) | cl | Max. Sub. SG | yes or no | Max. Sub. SG | yes or no | SIs (w/ T) | Stoich. Formula |
|---|---|---|---|---|---|---|---|---|---|---|---|---|---|---|
| 139.1.212 | 78592 | 98 | 0.0073 | -2.3425 | 87 | 15 | 139 | 6 | - | - | - | - | z8=4 | Ba2Ca1Cu2O8Tl2 |
| 139.1.213 | 2141 | 50 | 0.0240 | -0.2559 | 2,82 | 8 | 2,82 | 4 | - | - | - | - | z4=1<br>z2=1<br>z8=5 | Ba2Bi1 |
| 139.1.214 | 61197 | 40 | 0.0542 | -2.2285 | 2 | 11 | 71,69,87 | 5 | - | - | - | - | z4=2<br>z8=2 | Ba1Pd2Sb2 |
| 139.1.215 | 50188 | 38 | 0.0231 | -2.2503 | 2,82 | 8 | 2,82 | 4 | - | - | - | - | z4=1<br>z2=1<br>z8=5 | Ba1P2Rh2 |
| 139.1.216 | 419531 | 34 | 0.0031 | -0.5233 | 2 | 4 | 139 | 6 | - | - | - | - | z2w,1=1<br>z2w,2=1<br>z2w,3=1<br>z8=4 | Ba1Mn2Sb2 |
| 139.1.217 | 32019 | 34 | 0.0033 | -0.5206 | 2 | 4 | 139 | 6 | - | - | - | - | z2w,1=1<br>z2w,2=1<br>z2w,3=1<br>z8=4 | Ba1Mn2Sb2 |
| 139.1.218 | 77668 | 34 | 0.0092 | -0.7492 | 2 | 4 | 139 | 6 | - | - | - | - | z2w,1=1<br>z2w,2=1<br>z2w,3=1<br>z8=4 | Ba1Mn2P2 |
| 139.1.219 | 10469 | 34 | 0.0113 | -0.7605 | 2 | 4 | 139 | 6 | - | - | - | - | z2w,1=1<br>z2w,2=1<br>z2w,3=1<br>z8=4 | Ba1Mn2P2 |
| 139.1.220 | 100024 | 54 | 0.0000 | -0.4307 | 2 | 10 | 71,69,87 | 2 | - | - | - | - | z4=2<br>z8=6 | Ba1Mn1Sb2 |
| 139.1.221 | 10468 | 36 | 0.0055 | -0.6767 | 2,82 | 9 | 2,82 | 8 | - | - | - | - | z2w,1=1<br>z2w,2=1<br>z2w,3=1<br>z4=3<br>z2=1<br>z8=7 | Ba1Fe2P2 |
| 139.1.222 | 420975 | 36 | 0.0240 | -1.1085 | 2,82 | 8 | 2,82 | 4 | - | - | - | - | z4=1<br>z2=1<br>z8=5 | Ba1Co2Ge2 |
| 139.1.223 | 174263 | 42 | 0.1354 | -1.0154 | 2,82 | 7 | 2,82 | 4 | - | - | - | - | z2w,1=1<br>z2w,2=1<br>z2w,3=1<br>z4=1<br>z2=1<br>z8=5 | Ba1Cd2Ge2 |
| 139.1.224 | 71642 | 36 | 0.0502 | -0.2482 | 2 | 3 | 71,69,87 | 2 | - | - | - | - | z2w,1=1<br>z2w,2=1<br>z2w,3=1<br>z4=2<br>z8=6 | B2Ir2Zn1 |
| 139.1.225 | 426150 | 80 | 0.0190 | -0.6189 | 2 | 6 | 0 | 7 | - | - | - | - | z2w,1=1<br>z2w,2=1<br>z2w,3=1 | Au4Ce1Zn2 |
| 139.1.226 | 58388 | 46 | 0.0136 | -0.7134 | 2,82 | 2 | 2,82 | 1 | - | - | - | - | z2w,1=1<br>z2w,2=1<br>z2w,3=1<br>z4=1<br>z2=1<br>z8=1 | Au2Yb1 |

Continued on next page

**TABLE XIV– continued**

| TQC | ICSD | NoE | DG | IndG | SubSG (w/ T) | cl | SubSG (w/o T) | cl | Max. Sub. SG | yes or no | Max. Sub. SG | yes or no | SIs (w/ T) | Stoich. Formula |
|---|---|---|---|---|---|---|---|---|---|---|---|---|---|---|
| 139.1.227 | 413 | 40 | 0.0260 | -2.5891 | 2,82 | 13 | 2,82 | 8 | - | - | - | - | z4=3<br>z2=1<br>z8=7 | Au2Si2Sr1 |
| 139.1.228 | 37453 | 82 | 0.0334 | -0.2167 | 87 | 15 | 139 | 6 | - | - | - | - | z8=4 | Au2Cl6Cs2 |
| 139.1.229 | 418530 | 42 | 0.0153 | -0.4704 | 2 | 6 | 0 | 7 | - | - | - | - | z2w,1=1<br>z2w,2=1<br>z2w,3=1 | Au2Ce1Si2 |
| 139.1.230 | 412 | 32 | 0.0559 | -2.8236 | 2,82 | 13 | 2,82 | 8 | - | - | - | - | z4=3<br>z2=1<br>z8=7 | Au2Ca1Si2 |
| 139.1.231 | 263018 | 70 | 0.0041 | -0.3721 | 2 | 6 | 0 | 7 | - | - | - | - | z2w,1=1<br>z2w,2=1<br>z2w,3=1 | As4Ba2Fe2O1Ti2 |
| 139.1.232 | 417002 | 38 | 0.0114 | -1.3270 | 2 | 4 | 139 | 6 | - | - | - | - | z2w,1=1<br>z2w,2=1<br>z2w,3=1<br>z8=4 | As2Rh2Sr1 |
| 139.1.233 | 47170 | 40 | 0.0512 | -1.8283 | 2 | 6 | 0 | 7 | - | - | - | - | z2w,1=1<br>z2w,2=1<br>z2w,3=1 | As2Pd2Sr1 |
| 139.1.234 | 36374 | 40 | 0.0514 | -1.8219 | 2 | 6 | 0 | 7 | - | - | - | - | z2w,1=1<br>z2w,2=1<br>z2w,3=1 | As2Pd2Sr1 |
| 139.1.235 | 402951 | 112 | 0.0112 | -0.1021 | 2,82 | 5 | 2,82 | 1 | - | - | - | - | z4=1<br>z2=1<br>z8=1 | As2O1Yb4 |
| 139.1.236 | 167013 | 58 | 0.0131 | -0.6075 | 2 | 3 | 71,69,87 | 2 | - | - | - | - | z2w,1=1<br>z2w,2=1<br>z2w,3=1<br>z4=2<br>z8=6 | As2F2O1Sr2Ti2 |
| 139.1.237 | 78756 | 42 | 0.0462 | -0.9580 | 2,82 | 2 | 2,82 | 1 | - | - | - | - | z2w,1=1<br>z2w,2=1<br>z2w,3=1<br>z4=1<br>z2=1<br>z8=1 | As2Cu2Sr1 |
| 139.1.238 | 36372 | 32 | 0.0351 | -1.7520 | 2 | 6 | 0 | 7 | - | - | - | - | z2w,1=1<br>z2w,2=1<br>z2w,3=1 | As2Ca1Pd2 |
| 139.1.239 | 416983 | 38 | 0.0057 | -2.0189 | 2,82 | 5 | 2,82 | 1 | - | - | - | - | z4=1<br>z2=1<br>z8=1 | As2Ba1Rh2 |
| 139.1.240 | 41794 | 34 | 0.0100 | -0.6752 | 2 | 4 | 139 | 6 | - | - | - | - | z2w,1=1<br>z2w,2=1<br>z2w,3=1<br>z8=4 | As2Ba1Mn2 |
| 139.1.241 | 236307 | 42 | 0.0446 | -0.7293 | 2,82 | 2 | 2,82 | 1 | - | - | - | - | z2w,1=1<br>z2w,2=1<br>z2w,3=1<br>z4=1<br>z2=1<br>z8=1 | As2Ba1Cu2 |
| 139.1.242 | 92989 | 26 | 0.0342 | -0.6791 | 2 | 3 | 71,69,87 | 2 | - | - | - | - | z2w,1=1<br>z2w,2=1<br>z2w,3=1<br>z4=2<br>z8=6 | As1Ti1Zr1 |

Continued on next page

**TABLE XIV– continued**

| TQC | ICSD | NoE | DG | IndG | SubSG (w/ T) | cl | SubSG (w/o T) | cl | Max. Sub. SG | yes or no | Max. Sub. SG | yes or no | SIs (w/ T) | Stoich. Formula |
|---|---|---|---|---|---|---|---|---|---|---|---|---|---|---|
| 139.1.243 | 42358 | 50 | 0.0290 | -0.4550 | 2 | 12 | 71,69,87 | 5 | - | - | - | - | z2w,1=1<br>z2w,2=1<br>z2w,3=1<br>z4=2<br>z8=2 | As1Ba2 |
| 139.1.244 | 156967 | 66 | 0.0144 | -0.1595 | 2,82 | 2 | 2,82 | 1 | - | - | - | - | z2w,1=1<br>z2w,2=1<br>z2w,3=1<br>z4=1<br>z2=1<br>z8=1 | Al8Cr4Gd1 |
| 139.1.245 | 57545 | 54 | 0.0026 | -0.7073 | 2 | 10 | 71,69,87 | 2 | - | - | - | - | z4=2<br>z8=6 | Al8Ca1Mn4 |
| 139.1.246 | 57539 | 70 | 0.0327 | -2.1874 | 2,82 | 2 | 2,82 | 1 | - | - | - | - | z2w,1=1<br>z2w,2=1<br>z2w,3=1<br>z4=1<br>z2=1<br>z8=1 | Al8Ca1Cu4 |
| 139.1.247 | 95799 | 68 | 0.0000 | -0.0407 | 2,82 | 8 | 2,82 | 4 | - | - | - | - | z4=1<br>z2=1<br>z8=5 | Al4Ge2Ni1Tb2 |
| 139.1.248 | 168816 | 70 | 0.0101 | -0.1899 | 2,82 | 1 | 2,82 | 3 | - | - | - | - | z4=3<br>z2=1<br>z8=3 | Al2Pd5U1 |
| 139.1.249 | 161313 | 70 | 0.0101 | -0.1899 | 2,82 | 1 | 2,82 | 3 | - | - | - | - | z4=3<br>z2=1<br>z8=3 | Al2Pd5U1 |
| 139.1.250 | 166270 | 72 | 0.0368 | -0.6603 | 2 | 10 | 71,69,87 | 2 | - | - | - | - | z4=2<br>z8=6 | Al2Pd5Pu1 |
| 139.1.251 | 58252 | 44 | 0.0034 | -0.0134 | 2,82 | 2 | 2,82 | 1 | - | - | - | - | z2w,1=1<br>z2w,2=1<br>z2w,3=1<br>z4=1<br>z2=1<br>z8=1 | Ag2Er1 |
| 139.3.1 | 710066 | 144 | - | - | - | - | - | - | 121 | yes | 121 | yes | - | Si3Yb1 |
| | | | | | | | | | 119 | yes | 119 | yes | - | |
| | | | | | | | | | 71 | no(2) | 71 | no(2) | z2w,1=1<br>z2w,2=1<br>z2w,3=1<br>z4=1 | |
| | | | | | | | | | 69 | no(2) | 69 | no(2) | z2w,1=1<br>z2w,3=1<br>z4=1 | |
| 139.3.2 | 73599 | 14 | - | - | - | - | - | - | 121 | no(82) | 121 | no(82) | z2=1 | Si2W1 |
| | | | | | | | | | 119 | no(82) | 119 | no(82) | z2=1 | |
| | | | | | | | | | 71 | no(2) | 71 | no(71) | z4=2 | |
| | | | | | | | | | 69 | no(2) | 69 | no(69) | z4=2 | |
| 139.3.3 | 26869 | 14 | - | - | - | - | - | - | 121 | no(82) | 121 | no(82) | z2=1 | Si2W1 |
| | | | | | | | | | 119 | no(82) | 119 | no(82) | z2=1 | |
| | | | | | | | | | 71 | no(2) | 71 | no(71) | z4=2 | |
| | | | | | | | | | 69 | no(2) | 69 | no(69) | z4=2 | |
| 139.3.4 | 40549 | 38 | - | - | - | - | - | - | 121 | no(82) | 121 | no(82) | z2=1 | Ru2Si2U1 |
| | | | | | | | | | 119 | no(82) | 119 | no(82) | z2=1 | |

Continued on next page

**TABLE XIV– continued**

| TQC | ICSD | NoE | DG | IndG | SubSG cl (w/ T) | SubSG cl (w/o T) | Max. Sub. SG | yes or no | Max. Sub. SG | yes or no | SIs (w/ T) | Stoich. Formula |
|---|---|---|---|---|---|---|---|---|---|---|---|---|
| | | | | | | | 71 | no(2) | 71 | no(71) | z2w,1=1<br>z2w,2=1<br>z2w,3=1<br>z4=2 | |
| | | | | | | | 69 | no(2) | 69 | no(69) | z2w,1=1<br>z2w,3=1<br>z4=2 | |
| 139.3.5 | 89564 | 56 | - | - | - - | - - | 121 | yes | 121 | yes | - | O2Tb2Te1 |
| | | | | | | | 119 | yes | 119 | yes | - | |
| | | | | | | | 71 | no(2) | 71 | no(2) | z4=3 | |
| | | | | | | | 69 | no(2) | 69 | no(2) | z4=3 | |
| 139.3.6 | 89559 | 44 | - | - | - - | - - | 121 | yes | 121 | yes | - | O2Pr2Te1 |
| | | | | | | | 119 | yes | 119 | yes | - | |
| | | | | | | | 71 | no(2) | 71 | no(2) | z4=3 | |
| | | | | | | | 69 | no(2) | 69 | no(2) | z4=3 | |
| 139.3.7 | 41617 | 16 | - | - | - - | - - | 121 | no(82) | 121 | no(82) | z2=1 | O1Pd1 |
| | | | | | | | 119 | no(82) | 119 | no(82) | z2=1 | |
| | | | | | | | 71 | yes | 71 | yes | - | |
| | | | | | | | 69 | yes | 69 | yes | - | |
| 139.3.8 | 419760 | 168 | - | - | - - | - - | 121 | no(82) | 121 | no(82) | z2=1 | Ni2Pd1Se2 |
| | | | | | | | 119 | no(82) | 119 | no(82) | z2=1 | |
| | | | | | | | 71 | no(2) | 71 | yes | z2w,1=1<br>z2w,2=1<br>z2w,3=1 | |
| | | | | | | | 69 | no(2) | 69 | yes | z2w,1=1<br>z2w,3=1 | |
| 139.3.9 | 90298 | 38 | - | - | - - | - - | 121 | no(82) | 121 | no(82) | z2=1 | Nd1Ru2Si2 |
| | | | | | | | 119 | no(82) | 119 | no(82) | z2=1 | |
| | | | | | | | 71 | no(2) | 71 | no(71) | z2w,1=1<br>z2w,2=1<br>z2w,3=1<br>z4=2 | |
| | | | | | | | 69 | no(2) | 69 | no(69) | z2w,1=1<br>z2w,3=1<br>z4=2 | |
| 139.3.10 | 30841 | 34 | - | - | - - | - - | 121 | yes | 121 | yes | - | Nd1P2Pd2 |
| | | | | | | | 119 | yes | 119 | yes | - | |
| | | | | | | | 71 | no(2) | 71 | no(2) | z2w,1=1<br>z2w,2=1<br>z2w,3=1<br>z4=1 | |
| | | | | | | | 69 | no(2) | 69 | no(2) | z2w,1=1<br>z2w,3=1<br>z4=1 | |
| 139.3.11 | 423722 | 20 | - | - | - - | - - | 121 | no(82) | 121 | no(82) | z2=1 | N2Sr1 |
| | | | | | | | 119 | no(82) | 119 | no(82) | z2=1 | |
| | | | | | | | 71 | yes | 71 | yes | - | |
| | | | | | | | 69 | yes | 69 | yes | - | |
| 139.3.12 | 411556 | 20 | - | - | - - | - - | 121 | no(82) | 121 | no(82) | z2=1 | N2Sr1 |
| | | | | | | | 119 | no(82) | 119 | no(82) | z2=1 | |
| | | | | | | | 71 | yes | 71 | yes | - | |
| | | | | | | | 69 | yes | 69 | yes | - | |
| 139.3.13 | 85756 | 14 | - | - | - - | - - | 121 | yes | 121 | yes | - | Mo1Si2 |
| | | | | | | | 119 | yes | 119 | yes | - | |
| | | | | | | | 71 | no(2) | 71 | no(2) | z2w,1=1<br>z2w,2=1<br>z2w,3=1<br>z4=1 | |

Continued on next page

**TABLE XIV– continued**

| TQC | ICSD | NoE | DG | IndG | SubSG cl (w/ T) | SubSG cl (w/o T) | Max. Sub. SG | yes or no | Max. Sub. SG | yes or no | SIs (w/ T) | Stoich. Formula |
|---|---|---|---|---|---|---|---|---|---|---|---|---|
| | | | | | | | 69 | no(2) | 69 | no(2) | z2w,1=1<br>z2w,3=1<br>z4=1 | |
| 139.3.14 | 73598 | 14 | - | - | - - | - - | 121 | yes | 121 | yes | - | Mo1Si2 |
| | | | | | | | 119 | yes | 119 | yes | - | |
| | | | | | | | 71 | no(2) | 71 | no(2) | z2w,1=1<br>z2w,2=1<br>z2w,3=1<br>z4=1 | |
| | | | | | | | 69 | no(2) | 69 | no(2) | z2w,1=1<br>z2w,3=1<br>z4=1 | |
| 139.3.15 | 26868 | 14 | - | - | - - | - - | 121 | yes | 121 | yes | - | Mo1Si2 |
| | | | | | | | 119 | yes | 119 | yes | - | |
| | | | | | | | 71 | no(2) | 71 | no(2) | z2w,1=1<br>z2w,2=1<br>z2w,3=1<br>z4=1 | |
| | | | | | | | 69 | no(2) | 69 | no(2) | z2w,1=1<br>z2w,3=1<br>z4=1 | |
| 139.3.16 | 182115 | 14 | - | - | - - | - - | 121 | yes | 121 | yes | - | Mo1Si2 |
| | | | | | | | 119 | yes | 119 | yes | - | |
| | | | | | | | 71 | no(2) | 71 | no(2) | z2w,1=1<br>z2w,2=1<br>z2w,3=1<br>z4=1 | |
| | | | | | | | 69 | no(2) | 69 | no(2) | z2w,1=1<br>z2w,3=1<br>z4=1 | |
| 139.3.17 | 290823 | 46 | - | - | - - | - - | 121 | yes | 121 | yes | - | Mn2Si2Yb1 |
| | | | | | | | 119 | yes | 119 | yes | - | |
| | | | | | | | 71 | no(2) | 71 | no(2) | z4=3 | |
| | | | | | | | 69 | no(2) | 69 | no(2) | z4=3 | |
| 139.3.18 | 104986 | 96 | - | - | - - | - - | 121 | no(82) | 121 | no(82) | z2=1 | Mn12Th1 |
| | | | | | | | 119 | no(82) | 119 | no(82) | z2=1 | |
| | | | | | | | 71 | no(2) | 71 | no(71) | z4=2 | |
| | | | | | | | 69 | no(2) | 69 | no(69) | z4=2 | |
| 139.3.19 | 55549 | 40 | - | - | - - | - - | 121 | yes | 121 | yes | - | Mg1Tb2 |
| | | | | | | | 119 | yes | 119 | yes | - | |
| | | | | | | | 71 | no(2) | 71 | no(2) | z4=1 | |
| | | | | | | | 69 | no(2) | 69 | no(2) | z4=1 | |
| 139.3.20 | 183307 | 146 | - | - | - - | - - | 121 | no(82) | 121 | no(82) | z2=1 | In5La9Ru4 |
| | | | | | | | 119 | no(82) | 119 | no(82) | z2=1 | |
| | | | | | | | 71 | no(2) | 71 | no(71) | z4=2 | |
| | | | | | | | 69 | no(2) | 69 | no(69) | z4=2 | |
| 139.3.21 | 165985 | 38 | - | - | - - | - - | 121 | no(82) | 121 | no(82) | z2=1 | Ge2Pd2Sr1 |
| | | | | | | | 119 | no(82) | 119 | no(82) | z2=1 | |
| | | | | | | | 71 | no(2) | 71 | no(71) | z4=2 | |
| | | | | | | | 69 | no(2) | 69 | no(69) | z4=2 | |
| 139.3.22 | 409 | 38 | - | - | - - | - - | 121 | no(82) | 121 | no(82) | z2=1 | Ge2Ni2Sr1 |
| | | | | | | | 119 | no(82) | 119 | no(82) | z2=1 | |
| | | | | | | | 71 | no(2) | 71 | yes | z2w,1=1<br>z2w,2=1<br>z2w,3=1 | |
| | | | | | | | 69 | no(2) | 69 | yes | z2w,1=1<br>z2w,3=1 | |
| 139.3.23 | 165736 | 38 | - | - | - - | - - | 121 | no(82) | 121 | no(82) | z2=1 | Ge2Ni2Sr1 |

Continued on next page

**TABLE XIV– continued**

| TQC | ICSD | NoE | DG | IndG | SubSG cl (w/ T) | SubSG cl (w/o T) | Max. Sub. SG | yes or no | Max. Sub. SG | yes or no | SIs (w/ T) | Stoich. Formula |
|---|---|---|---|---|---|---|---|---|---|---|---|---|
| | | | | | | | 119 | no(82) | 119 | no(82) | z2=1 | |
| | | | | | | | 71 | no(2) | 71 | yes | z2w,1=1<br>z2w,2=1<br>z2w,3=1 | |
| | | | | | | | 69 | no(2) | 69 | yes | z2w,1=1<br>z2w,3=1 | |
| 139.3.24 | 53737 | 38 | - | - | - - | - - | 121 | no(82) | 121 | no(82) | z2=1 | Ge2Nd1Ru2 |
| | | | | | | | 119 | no(82) | 119 | no(82) | z2=1 | |
| | | | | | | | 71 | no(2) | 71 | no(71) | z2w,1=1<br>z2w,2=1<br>z2w,3=1<br>z4=2 | |
| | | | | | | | 69 | no(2) | 69 | no(69) | z2w,1=1<br>z2w,3=1<br>z4=2 | |
| 139.3.25 | 53735 | 40 | - | - | - - | - - | 121 | no(82) | 121 | no(82) | z2=1 | Ge2Nd1Rh2 |
| | | | | | | | 119 | no(82) | 119 | no(82) | z2=1 | |
| | | | | | | | 71 | yes | 71 | yes | - | |
| | | | | | | | 69 | yes | 69 | yes | - | |
| 139.3.26 | 182419 | 50 | - | - | - - | - - | 121 | yes | 121 | yes | - | Gd1Sc1Si1 |
| | | | | | | | 119 | yes | 119 | yes | - | |
| | | | | | | | 71 | no(2) | 71 | no(2) | z2w,1=1<br>z2w,2=1<br>z2w,3=1<br>z4=3 | |
| | | | | | | | 69 | no(2) | 69 | no(2) | z2w,1=1<br>z2w,3=1<br>z4=3 | |
| 139.3.27 | 89556 | 40 | - | - | - - | - - | 121 | yes | 121 | yes | - | Gd1Mn2Si2 |
| | | | | | | | 119 | yes | 119 | yes | - | |
| | | | | | | | 71 | no(2) | 71 | no(2) | z2w,1=1<br>z2w,2=1<br>z2w,3=1<br>z4=1 | |
| | | | | | | | 69 | no(2) | 69 | no(2) | z2w,1=1<br>z2w,3=1<br>z4=1 | |
| 139.3.28 | 52771 | 46 | - | - | - - | - - | 121 | yes | 121 | yes | - | Gd1Ge2Pd2 |
| | | | | | | | 119 | yes | 119 | yes | - | |
| | | | | | | | 71 | no(2) | 71 | no(2) | z4=1 | |
| | | | | | | | 69 | no(2) | 69 | no(2) | z4=1 | |
| 139.3.29 | 107607 | 78 | - | - | - - | - - | 121 | yes | 121 | yes | - | Ga5Yb1 |
| | | | | | | | 119 | yes | 119 | yes | - | |
| | | | | | | | 71 | no(2) | 71 | no(2) | z2w,1=1<br>z2w,2=1<br>z2w,3=1<br>z4=1 | |
| | | | | | | | 69 | no(2) | 69 | no(2) | z2w,1=1<br>z2w,3=1<br>z4=1 | |
| 139.3.30 | 23227 | 38 | - | - | - - | - - | 121 | yes | 121 | yes | - | Ga5Mg2 |
| | | | | | | | 119 | yes | 119 | yes | - | |
| | | | | | | | 71 | no(2) | 71 | no(2) | z2w,1=1<br>z2w,2=1<br>z2w,3=1<br>z4=1 | |
| | | | | | | | 69 | no(2) | 69 | no(2) | z2w,1=1<br>z2w,3=1<br>z4=1 | |

Continued on next page

**TABLE XIV– continued**

| TQC | ICSD | NoE | DG | IndG | SubSG cl (w/ T) | SubSG cl (w/o T) | Max. Sub. SG | yes or no | Max. Sub. SG | yes or no | SIs (w/ T) | Stoich. Formula |
|---|---|---|---|---|---|---|---|---|---|---|---|---|
| 139.3.31 | 170782 | 22 | - | - | - - | - - | 121 | yes | 121 | yes | - | Ga4Sr1 |
| | | | | | | | 119 | yes | 119 | yes | - | |
| | | | | | | | 71 | no(2) | 71 | no(2) | z4=3 | |
| | | | | | | | 69 | no(2) | 69 | no(2) | z4=3 | |
| 139.3.32 | 402238 | 216 | - | - | - - | - - | 121 | no(82) | 121 | no(82) | z2=1 | Fe4Si7Zr4 |
| | | | | | | | 119 | no(82) | 119 | no(82) | z2=1 | |
| | | | | | | | 71 | no(2) | 71 | yes | z2w,1=1<br>z2w,2=1<br>z2w,3=1 | |
| | | | | | | | 69 | no(2) | 69 | yes | z2w,1=1<br>z2w,3=1 | |
| 139.3.33 | 55548 | 46 | - | - | - - | - - | 121 | yes | 121 | yes | - | Er2Mg1 |
| | | | | | | | 119 | yes | 119 | yes | - | |
| | | | | | | | 71 | no(2) | 71 | no(2) | z4=3 | |
| | | | | | | | 69 | no(2) | 69 | no(2) | z4=3 | |
| 139.3.34 | 30848 | 42 | - | - | - - | - - | 121 | yes | 121 | yes | - | Er1P2Pd2 |
| | | | | | | | 119 | yes | 119 | yes | - | |
| | | | | | | | 71 | no(2) | 71 | no(2) | z2w,1=1<br>z2w,2=1<br>z2w,3=1<br>z4=1 | |
| | | | | | | | 69 | no(2) | 69 | no(2) | z2w,1=1<br>z2w,3=1<br>z4=1 | |
| 139.3.35 | 88188 | 52 | - | - | - - | - - | 121 | yes | 121 | yes | - | Er1Ni2P2 |
| | | | | | | | 119 | yes | 119 | yes | - | |
| | | | | | | | 71 | no(2) | 71 | no(2) | z4=1 | |
| | | | | | | | 69 | no(2) | 69 | no(2) | z4=1 | |
| 139.3.36 | 88187 | 52 | - | - | - - | - - | 121 | yes | 121 | yes | - | Er1Ni2P2 |
| | | | | | | | 119 | yes | 119 | yes | - | |
| | | | | | | | 71 | no(2) | 71 | no(2) | z4=1 | |
| | | | | | | | 69 | no(2) | 69 | no(2) | z4=1 | |
| 139.3.37 | 26826 | 106 | - | - | - - | - - | 121 | yes | 121 | yes | - | Er1Mn12 |
| | | | | | | | 119 | yes | 119 | yes | - | |
| | | | | | | | 71 | no(2) | 71 | no(2) | z4=1 | |
| | | | | | | | 69 | no(2) | 69 | no(2) | z4=1 | |
| 139.3.38 | 103237 | 42 | - | - | - - | - - | 121 | yes | 121 | yes | - | Er1Ga4Ti2 |
| | | | | | | | 119 | yes | 119 | yes | - | |
| | | | | | | | 71 | no(2) | 71 | no(2) | z4=3 | |
| | | | | | | | 69 | no(2) | 69 | no(2) | z4=3 | |
| 139.3.39 | 89565 | 58 | - | - | - - | - - | 121 | yes | 121 | yes | - | Dy2O2Te1 |
| | | | | | | | 119 | yes | 119 | yes | - | |
| | | | | | | | 71 | no(2) | 71 | no(2) | z4=3 | |
| | | | | | | | 69 | no(2) | 69 | no(2) | z4=3 | |
| 139.3.40 | 41864 | 148 | - | - | - - | - - | 121 | yes | 121 | yes | - | D3Pd1Rb3 |
| | | | | | | | 119 | yes | 119 | yes | - | |
| | | | | | | | 71 | no(2) | 71 | no(2) | z2w,1=1<br>z2w,2=1<br>z2w,3=1<br>z4=1 | |
| | | | | | | | 69 | no(2) | 69 | no(2) | z2w,1=1<br>z2w,3=1<br>z4=1 | |
| 139.3.41 | 98243 | 34 | - | - | - - | - - | 121 | yes | 121 | yes | - | Cr2Si2U1 |
| | | | | | | | 119 | yes | 119 | yes | - | |

Continued on next page

**TABLE XIV– continued**

| TQC | ICSD | NoE | DG | IndG | SubSG cl (w/ T) | SubSG cl (w/o T) | Max. Sub. SG | yes or no | Max. Sub. SG | yes or no | SIs (w/ T) | Stoich. Formula |
|---|---|---|---|---|---|---|---|---|---|---|---|---|
| | | | | | | | 71 | no(2) | 71 | no(2) | z2w,1=1<br>z2w,2=1<br>z2w,3=1<br>z4=1 | |
| | | | | | | | 69 | no(2) | 69 | no(2) | z2w,1=1<br>z2w,3=1<br>z4=1 | |
| 139.3.42 | 68207 | 32 | - | - | - - | - - | 121 | yes | 121 | yes | - | Cr2Si2Th1 |
| | | | | | | | 119 | yes | 119 | yes | - | |
| | | | | | | | 71 | no(2) | 71 | no(2) | z2w,1=1<br>z2w,2=1<br>z2w,3=1<br>z4=1 | |
| | | | | | | | 69 | no(2) | 69 | no(2) | z2w,1=1<br>z2w,3=1<br>z4=1 | |
| 139.3.43 | 68210 | 38 | - | - | - - | - - | 121 | yes | 121 | yes | - | Co2Si2Th1 |
| | | | | | | | 119 | yes | 119 | yes | - | |
| | | | | | | | 71 | no(2) | 71 | no(2) | z2w,1=1<br>z2w,2=1<br>z2w,3=1<br>z4=3 | |
| | | | | | | | 69 | no(2) | 69 | no(2) | z2w,1=1<br>z2w,3=1<br>z4=3 | |
| 139.3.44 | 31421 | 40 | - | - | - - | - - | 121 | yes | 121 | yes | - | Co2Nd1Si2 |
| | | | | | | | 119 | yes | 119 | yes | - | |
| | | | | | | | 71 | no(2) | 71 | no(2) | z2w,1=1<br>z2w,2=1<br>z2w,3=1<br>z4=3 | |
| | | | | | | | 69 | no(2) | 69 | no(2) | z2w,1=1<br>z2w,3=1<br>z4=3 | |
| 139.3.45 | 55758 | 48 | - | - | - - | - - | 121 | yes | 121 | yes | - | Co2Er1Si2 |
| | | | | | | | 119 | yes | 119 | yes | - | |
| | | | | | | | 71 | no(2) | 71 | no(2) | z4=1 | |
| | | | | | | | 69 | no(2) | 69 | no(2) | z4=1 | |
| 139.3.46 | 31427 | 48 | - | - | - - | - - | 121 | yes | 121 | yes | - | Co2Er1Si2 |
| | | | | | | | 119 | yes | 119 | yes | - | |
| | | | | | | | 71 | no(2) | 71 | no(2) | z2w,1=1<br>z2w,2=1<br>z2w,3=1<br>z4=3 | |
| | | | | | | | 69 | no(2) | 69 | no(2) | z2w,1=1<br>z2w,3=1<br>z4=3 | |
| 139.3.47 | 31426 | 48 | - | - | - - | - - | 121 | yes | 121 | yes | - | Co2Er1Si2 |
| | | | | | | | 119 | yes | 119 | yes | - | |
| | | | | | | | 71 | no(2) | 71 | no(2) | z4=1 | |
| | | | | | | | 69 | no(2) | 69 | no(2) | z4=1 | |
| 139.3.48 | 170797 | 48 | - | - | - - | - - | 121 | yes | 121 | yes | - | Co2Er1Si2 |
| | | | | | | | 119 | yes | 119 | yes | - | |
| | | | | | | | 71 | no(2) | 71 | no(2) | z4=1 | |
| | | | | | | | 69 | no(2) | 69 | no(2) | z4=1 | |
| 139.3.49 | 171240 | 204 | - | - | - - | - - | 121 | yes | 121 | yes | - | Cl4Cu4S4W1 |
| | | | | | | | 119 | yes | 119 | yes | - | |

Continued on next page

**TABLE XIV– continued**

| TQC | ICSD | NoE | DG | IndG | SubSG cl (w/ T) | SubSG cl (w/o T) | Max. Sub. SG | yes or no | Max. Sub. SG | yes or no | SIs (w/ T) | Stoich. Formula |
|---|---|---|---|---|---|---|---|---|---|---|---|---|
| | | | | | | | 71 | no(2) | 71 | no(71) | z2w,1=1 z2w,2=1 z2w,3=1 z4=2 | |
| | | | | | | | 69 | no(2) | 69 | no(69) | z2w,1=1 z2w,3=1 z4=2 | |
| 139.3.50 | 240484 | 70 | - | - | - - | - - | 121 | yes | 121 | yes | - | Cl4Cs2I2Pd1 |
| | | | | | | | 119 | yes | 119 | yes | - | |
| | | | | | | | 71 | no(2) | 71 | no(2) | z2w,1=1 z2w,2=1 z2w,3=1 z4=1 | |
| | | | | | | | 69 | no(2) | 69 | no(2) | z2w,1=1 z2w,3=1 z4=1 | |
| 139.3.51 | 9858 | 52 | - | - | - - | - - | 121 | no(82) | 121 | no(82) | z2=1 | Cl4Cr1Rb2 |
| | | | | | | | 119 | no(82) | 119 | no(82) | z2=1 | |
| | | | | | | | 71 | no(2) | 71 | no(71) | z2w,1=1 z2w,2=1 z2w,3=1 z4=2 | |
| | | | | | | | 69 | no(2) | 69 | no(69) | z2w,1=1 z2w,3=1 z4=2 | |
| 139.3.52 | 1923 | 124 | - | - | - - | - - | 121 | yes | 121 | yes | - | Cl10K4O1W2 |
| | | | | | | | 119 | yes | 119 | yes | - | |
| | | | | | | | 71 | no(2) | 71 | no(2) | z2w,1=1 z2w,2=1 z2w,3=1 z4=1 | |
| | | | | | | | 69 | no(2) | 69 | no(2) | z2w,1=1 z2w,3=1 z4=1 | |
| 139.3.53 | 89558 | 42 | - | - | - - | - - | 121 | no(82) | 121 | no(82) | z2=1 | Ce2O2Te1 |
| | | | | | | | 119 | no(82) | 119 | no(82) | z2=1 | |
| | | | | | | | 71 | yes | 71 | yes | - | |
| | | | | | | | 69 | yes | 69 | yes | - | |
| 139.3.54 | 40913 | 40 | - | - | - - | - - | 121 | no(82) | 121 | no(82) | z2=1 | Ce1Pd2Si2 |
| | | | | | | | 119 | no(82) | 119 | no(82) | z2=1 | |
| | | | | | | | 71 | no(2) | 71 | no(71) | z2w,1=1 z2w,2=1 z2w,3=1 z4=2 | |
| | | | | | | | 69 | no(2) | 69 | no(69) | z2w,1=1 z2w,3=1 z4=2 | |
| 139.3.55 | 30839 | 32 | - | - | - - | - - | 121 | yes | 121 | yes | - | Ce1P2Pd2 |
| | | | | | | | 119 | yes | 119 | yes | - | |
| | | | | | | | 71 | no(2) | 71 | no(2) | z4=3 | |
| | | | | | | | 69 | no(2) | 69 | no(2) | z4=3 | |
| 139.3.56 | 87320 | 34 | - | - | - - | - - | 121 | yes | 121 | yes | - | Ce1Mn2Si2 |
| | | | | | | | 119 | yes | 119 | yes | - | |
| | | | | | | | 71 | no(2) | 71 | no(2) | z2w,1=1 z2w,2=1 z2w,3=1 z4=1 | |

Continued on next page

**TABLE XIV– continued**

| TQC | ICSD | NoE | DG | IndG | SubSG cl (w/ T) | SubSG cl (w/o T) | Max. Sub. SG | yes or no | Max. Sub. SG | yes or no | SIs (w/ T) | Stoich. Formula |
|---|---|---|---|---|---|---|---|---|---|---|---|---|
| | | | | | | | 69 | no(2) | 69 | no(2) | z2w,1=1<br>z2w,3=1<br>z4=1 | |
| 139.3.57 | 87319 | 34 | - | - | - - | - - | 121 | yes | 121 | yes | - | Ce1Mn2Si2 |
| | | | | | | | 119 | yes | 119 | yes | - | |
| | | | | | | | 71 | no(2) | 71 | no(2) | z2w,1=1<br>z2w,2=1<br>z2w,3=1<br>z4=1 | |
| | | | | | | | 69 | no(2) | 69 | no(2) | z2w,1=1<br>z2w,3=1<br>z4=1 | |
| 139.3.58 | 87318 | 34 | - | - | - - | - - | 121 | yes | 121 | yes | - | Ce1Mn2Si2 |
| | | | | | | | 119 | yes | 119 | yes | - | |
| | | | | | | | 71 | no(2) | 71 | no(2) | z2w,1=1<br>z2w,2=1<br>z2w,3=1<br>z4=1 | |
| | | | | | | | 69 | no(2) | 69 | no(2) | z2w,1=1<br>z2w,3=1<br>z4=1 | |
| 139.3.59 | 425230 | 38 | - | - | - - | - - | 121 | yes | 121 | yes | - | Ce1Ge2Rh2 |
| | | | | | | | 119 | yes | 119 | yes | - | |
| | | | | | | | 71 | no(2) | 71 | no(2) | z2w,1=1<br>z2w,2=1<br>z2w,3=1<br>z4=3 | |
| | | | | | | | 69 | no(2) | 69 | no(2) | z2w,1=1<br>z2w,3=1<br>z4=3 | |
| 139.3.60 | 423721 | 12 | - | - | - - | - - | 121 | no(82) | 121 | no(82) | z2=1 | Ca1N2 |
| | | | | | | | 119 | no(82) | 119 | no(82) | z2=1 | |
| | | | | | | | 71 | yes | 71 | yes | - | |
| | | | | | | | 69 | yes | 69 | yes | - | |
| 139.3.61 | 262736 | 56 | - | - | - - | - - | 121 | no(82) | 121 | no(82) | z2=1 | Ca1Ge3 |
| | | | | | | | 119 | no(82) | 119 | no(82) | z2=1 | |
| | | | | | | | 71 | no(2) | 71 | no(71) | z2w,1=1<br>z2w,2=1<br>z2w,3=1<br>z4=2 | |
| | | | | | | | 69 | no(2) | 69 | no(69) | z2w,1=1<br>z2w,3=1<br>z4=2 | |
| 139.3.62 | 425467 | 26 | - | - | - - | - - | 121 | yes | 121 | yes | - | Ca1Fe2Si2 |
| | | | | | | | 119 | yes | 119 | yes | - | |
| | | | | | | | 71 | no(2) | 71 | no(2) | z4=1 | |
| | | | | | | | 69 | no(2) | 69 | no(2) | z4=1 | |
| 139.3.63 | 81757 | 32 | - | - | - - | - - | 121 | yes | 121 | yes | - | Ca1Cu2Ge2 |
| | | | | | | | 119 | yes | 119 | yes | - | |
| | | | | | | | 71 | no(2) | 71 | no(2) | z4=3 | |
| | | | | | | | 69 | no(2) | 69 | no(2) | z4=3 | |
| 139.3.64 | 74668 | 20 | - | - | - - | - - | 121 | no(82) | 121 | no(82) | z2=1 | C2Ce1 |
| | | | | | | | 119 | no(82) | 119 | no(82) | z2=1 | |
| | | | | | | | 71 | yes | 71 | yes | - | |
| | | | | | | | 69 | yes | 69 | yes | - | |
| 139.3.65 | 63556 | 20 | - | - | - - | - - | 121 | no(82) | 121 | no(82) | z2=1 | C2Ce1 |
| | | | | | | | 119 | no(82) | 119 | no(82) | z2=1 | |
| | | | | | | | 71 | yes | 71 | yes | - | |

Continued on next page

**TABLE XIV– continued**

| TQC | ICSD | NoE | DG | IndG | SubSG cl (w/ T) | SubSG cl (w/o T) | Max. Sub. SG | yes or no | Max. Sub. SG | yes or no | SIs (w/ T) | Stoich. Formula |
|---|---|---|---|---|---|---|---|---|---|---|---|---|
| | | | | | | | 69 | yes | 69 | yes | - | |
| 139.3.66 | 17022 | 20 | - | - | - - | - - | 121 | no(82) | 121 | no(82) | z2=1 | C2Ce1 |
| | | | | | | | 119 | no(82) | 119 | no(82) | z2=1 | |
| | | | | | | | 71 | yes | 71 | yes | - | |
| | | | | | | | 69 | yes | 69 | yes | - | |
| 139.3.67 | 36246 | 32 | - | - | - - | - - | 121 | yes | 121 | yes | - | C2Bi2Ca1O8 |
| | | | | | | | 119 | yes | 119 | yes | - | |
| | | | | | | | 71 | no(2) | 71 | no(2) | z2w,1=1<br>z2w,2=1<br>z2w,3=1<br>z4=3 | |
| | | | | | | | 69 | no(2) | 69 | no(2) | z2w,1=1<br>z2w,3=1<br>z4=3 | |
| 139.3.68 | 56645 | 42 | - | - | - - | - - | 121 | yes | 121 | yes | - | C1B2Ce1Ni2 |
| | | | | | | | 119 | yes | 119 | yes | - | |
| | | | | | | | 71 | no(2) | 71 | no(2) | z2w,1=1<br>z2w,2=1<br>z2w,3=1<br>z4=3 | |
| | | | | | | | 69 | no(2) | 69 | no(2) | z2w,1=1<br>z2w,3=1<br>z4=3 | |
| 139.3.69 | 174269 | 42 | - | - | - - | - - | 121 | yes | 121 | yes | - | Ba1Si2Zn2 |
| | | | | | | | 119 | yes | 119 | yes | - | |
| | | | | | | | 71 | no(2) | 71 | no(2) | z4=3 | |
| | | | | | | | 69 | no(2) | 69 | no(2) | z4=3 | |
| 139.3.70 | 85408 | 40 | - | - | - - | - - | 121 | yes | 121 | yes | - | Ba1Ni2P2 |
| | | | | | | | 119 | yes | 119 | yes | - | |
| | | | | | | | 71 | no(2) | 71 | no(2) | z2w,1=1<br>z2w,2=1<br>z2w,3=1<br>z4=3 | |
| | | | | | | | 69 | no(2) | 69 | no(2) | z2w,1=1<br>z2w,3=1<br>z4=3 | |
| 139.3.71 | 405 | 32 | - | - | - - | - - | 121 | no(82) | 121 | no(82) | z2=1 | Ba1Mn2Sn2 |
| | | | | | | | 119 | no(82) | 119 | no(82) | z2=1 | |
| | | | | | | | 71 | no(2) | 71 | no(71) | z4=2 | |
| | | | | | | | 69 | no(2) | 69 | no(69) | z4=2 | |
| 139.3.72 | 154540 | 94 | - | - | - - | - - | 121 | no(82) | 121 | no(82) | z2=1 | Ba1Mn2Nd2O7 |
| | | | | | | | 119 | no(82) | 119 | no(82) | z2=1 | |
| | | | | | | | 71 | yes | 71 | yes | - | |
| | | | | | | | 69 | yes | 69 | yes | - | |
| 139.3.73 | 154539 | 94 | - | - | - - | - - | 121 | no(82) | 121 | no(82) | z2=1 | Ba1Mn2Nd2O7 |
| | | | | | | | 119 | no(82) | 119 | no(82) | z2=1 | |
| | | | | | | | 71 | yes | 71 | yes | - | |
| | | | | | | | 69 | yes | 69 | yes | - | |
| 139.3.74 | 95757 | 38 | - | - | - - | - - | 121 | no(82) | 121 | no(82) | z2=1 | Ba1Ir2P2 |
| | | | | | | | 119 | no(82) | 119 | no(82) | z2=1 | |
| | | | | | | | 71 | yes | 71 | yes | - | |
| | | | | | | | 69 | yes | 69 | yes | - | |
| 139.3.75 | 414230 | 22 | - | - | - - | - - | 121 | yes | 121 | yes | - | Ba1In4 |
| | | | | | | | 119 | yes | 119 | yes | - | |
| | | | | | | | 71 | no(2) | 71 | no(2) | z4=3 | |
| | | | | | | | 69 | no(2) | 69 | no(2) | z4=3 | |
| 139.3.76 | 404 | 32 | - | - | - - | - - | 121 | no(82) | 121 | no(82) | z2=1 | Ba1Ge2Mn2 |
| | | | | | | | 119 | no(82) | 119 | no(82) | z2=1 | |

Continued on next page

**TABLE XIV– continued**

| TQC | ICSD | NoE | DG | IndG | SubSG cl (w/ T) | SubSG cl (w/o T) | Max. Sub. SG | yes or no | Max. Sub. SG | yes or no | SIs (w/ T) | Stoich. Formula |
|---|---|---|---|---|---|---|---|---|---|---|---|---|
| | | | | | | | 71 | no(2) | 71 | no(71) | z4=2 | |
| | | | | | | | 69 | no(2) | 69 | no(69) | z4=2 | |
| 139.3.77 | 55449 | 22 | - | - | - - | - - | 121 | yes | 121 | yes | - | Ba1Ga4 |
| | | | | | | | 119 | yes | 119 | yes | - | |
| | | | | | | | 71 | no(2) | 71 | no(2) | z4=3 | |
| | | | | | | | 69 | no(2) | 69 | no(2) | z4=3 | |
| 139.3.78 | 20894 | 176 | - | - | - - | - - | 121 | yes | 121 | yes | - | B4Er1Ni1 |
| | | | | | | | 119 | yes | 119 | yes | - | |
| | | | | | | | 71 | no(2) | 71 | no(2) | z4=3 | |
| | | | | | | | 69 | no(2) | 69 | no(2) | z4=3 | |
| 139.3.79 | 391192 | 74 | - | - | - - | - - | 121 | no(82) | 121 | no(82) | z2=1 | B2La3N3Ni2 |
| | | | | | | | 119 | no(82) | 119 | no(82) | z2=1 | |
| | | | | | | | 71 | no(2) | 71 | yes | z2w,1=1<br>z2w,2=1<br>z2w,3=1 | |
| | | | | | | | 69 | no(2) | 69 | yes | z2w,1=1<br>z2w,3=1 | |
| 139.3.80 | 424033 | 70 | - | - | - - | - - | 121 | yes | 121 | yes | - | Au4Ca1Cd2 |
| | | | | | | | 119 | yes | 119 | yes | - | |
| | | | | | | | 71 | no(2) | 71 | no(2) | z2w,1=1<br>z2w,2=1<br>z2w,3=1<br>z4=1 | |
| | | | | | | | 69 | no(2) | 69 | no(2) | z2w,1=1<br>z2w,3=1<br>z4=1 | |
| 139.3.81 | 409552 | 104 | - | - | - - | - - | 121 | yes | 121 | yes | - | Au1O1Rb1 |
| | | | | | | | 119 | yes | 119 | yes | - | |
| | | | | | | | 71 | no(2) | 71 | no(2) | z4=3 | |
| | | | | | | | 69 | no(2) | 69 | no(2) | z4=3 | |
| 139.3.82 | 23005 | 40 | - | - | - - | - - | 121 | yes | 121 | yes | - | As2Ni2Sr1 |
| | | | | | | | 119 | yes | 119 | yes | - | |
| | | | | | | | 71 | no(2) | 71 | no(2) | z2w,1=1<br>z2w,2=1<br>z2w,3=1<br>z4=3 | |
| | | | | | | | 69 | no(2) | 69 | no(2) | z2w,1=1<br>z2w,3=1<br>z4=3 | |
| 139.3.83 | 182330 | 36 | - | - | - - | - - | 121 | yes | 121 | yes | - | As2Fe2Sr1 |
| | | | | | | | 119 | yes | 119 | yes | - | |
| | | | | | | | 71 | no(2) | 71 | no(2) | z2w,1=1<br>z2w,2=1<br>z2w,3=1<br>z4=3 | |
| | | | | | | | 69 | no(2) | 69 | no(2) | z2w,1=1<br>z2w,3=1<br>z4=3 | |
| 139.3.84 | 163209 | 36 | - | - | - - | - - | 121 | yes | 121 | yes | - | As2Fe2Sr1 |
| | | | | | | | 119 | yes | 119 | yes | - | |
| | | | | | | | 71 | no(2) | 71 | no(2) | z2w,1=1<br>z2w,2=1<br>z2w,3=1<br>z4=3 | |
| | | | | | | | 69 | no(2) | 69 | no(2) | z2w,1=1<br>z2w,3=1<br>z4=3 | |
| 139.3.85 | 1222 | 84 | - | - | - - | - - | 121 | no(82) | 121 | no(82) | z2=1 | As2Eu4O1 |

Continued on next page

**TABLE XIV– continued**

| TQC | ICSD | NoE | DG | IndG | SubSG cl (w/ T) | SubSG cl (w/o T) | Max. Sub. SG | yes or no | Max. Sub. SG | yes or no | SIs (w/ T) | Stoich. Formula |
|---|---|---|---|---|---|---|---|---|---|---|---|---|
| | | | | | | | 119 | no(82) | 119 | no(82) | z2=1 | |
| | | | | | | | 71 | no(2) | 71 | no(71) | z4=2 | |
| | | | | | | | 69 | no(2) | 69 | no(69) | z4=2 | |
| 139.3.86 | 68146 | 42 | - | - | - - | - - | 121 | no(82) | 121 | no(82) | z2=1 | As2Ce1Ni2 |
| | | | | | | | 119 | no(82) | 119 | no(82) | z2=1 | |
| | | | | | | | 71 | no(2) | 71 | yes | z2w,1=1 z2w,2=1 z2w,3=1 | |
| | | | | | | | 69 | no(2) | 69 | yes | z2w,1=1 z2w,3=1 | |
| 139.3.87 | 23004 | 32 | - | - | - - | - - | 121 | yes | 121 | yes | - | As2Ca1Ni2 |
| | | | | | | | 119 | yes | 119 | yes | - | |
| | | | | | | | 71 | no(2) | 71 | no(2) | z2w,1=1 z2w,2=1 z2w,3=1 z4=3 | |
| | | | | | | | 69 | no(2) | 69 | no(2) | z2w,1=1 z2w,3=1 z4=3 | |
| 139.3.88 | 61196 | 40 | - | - | - - | - - | 121 | yes | 121 | yes | - | As2Ba1Pd2 |
| | | | | | | | 119 | yes | 119 | yes | - | |
| | | | | | | | 71 | no(2) | 71 | no(2) | z2w,1=1 z2w,2=1 z2w,3=1 z4=3 | |
| | | | | | | | 69 | no(2) | 69 | no(2) | z2w,1=1 z2w,3=1 z4=3 | |
| 139.3.89 | 36377 | 80 | - | - | - - | - - | 121 | no(82) | 121 | no(82) | z2=1 | As2Ba1Pd2 |
| | | | | | | | 119 | no(82) | 119 | no(82) | z2=1 | |
| | | | | | | | 71 | no(2) | 71 | yes | z2w,1=1 z2w,2=1 z2w,3=1 | |
| | | | | | | | 69 | no(2) | 69 | yes | z2w,1=1 z2w,3=1 | |
| 139.3.90 | 164197 | 40 | - | - | - - | - - | 121 | yes | 121 | yes | - | As2Ba1Ni2 |
| | | | | | | | 119 | yes | 119 | yes | - | |
| | | | | | | | 71 | no(2) | 71 | no(2) | z2w,1=1 z2w,2=1 z2w,3=1 z4=3 | |
| | | | | | | | 69 | no(2) | 69 | no(2) | z2w,1=1 z2w,3=1 z4=3 | |
| 139.3.91 | 87139 | 50 | - | - | - - | - - | 121 | yes | 121 | yes | - | As1Te1U1 |
| | | | | | | | 119 | yes | 119 | yes | - | |
| | | | | | | | 71 | no(2) | 71 | no(2) | z2w,1=1 z2w,2=1 z2w,3=1 z4=1 | |
| | | | | | | | 69 | no(2) | 69 | no(2) | z2w,1=1 z2w,3=1 z4=1 | |
| 139.3.92 | 57749 | 76 | - | - | - - | - - | 121 | yes | 121 | yes | - | Al8Dy1Fe4 |
| | | | | | | | 119 | yes | 119 | yes | - | |
| | | | | | | | 71 | no(2) | 71 | no(2) | z4=1 | |
| | | | | | | | 69 | no(2) | 69 | no(2) | z4=1 | |
| 139.3.93 | 161659 | 76 | - | - | - - | - - | 121 | yes | 121 | yes | - | Al8Dy1Fe4 |

Continued on next page

**TABLE XIV– continued**

| TQC | ICSD | NoE | DG | IndG | SubSG (w/ T) | cl | SubSG (w/o T) | cl | Max. Sub. SG | yes or no | Max. Sub. SG | yes or no | SIs (w/ T) | Stoich. Formula |
|---|---|---|---|---|---|---|---|---|---|---|---|---|---|---|
| | | | | | | | | | 119 | yes | 119 | yes | - | |
| | | | | | | | | | 71 | no(2) | 71 | no(2) | z4=1 | |
| | | | | | | | | | 69 | no(2) | 69 | no(2) | z4=1 | |
| 139.3.94 | 57724 | 82 | - | - | - | - | - | - | 121 | no(82) | 121 | no(82) | z2=1 | Al8Cu4U1 |
| | | | | | | | | | 119 | no(82) | 119 | no(82) | z2=1 | |
| | | | | | | | | | 71 | no(2) | 71 | no(71) | z4=2 | |
| | | | | | | | | | 69 | no(2) | 69 | no(69) | z4=2 | |
| 139.3.95 | 456 | 48 | - | - | - | - | - | - | 121 | no(82) | 121 | no(82) | z2=1 | Al4Mo2Yb1 |
| | | | | | | | | | 119 | no(82) | 119 | no(82) | z2=1 | |
| | | | | | | | | | 71 | no(2) | 71 | no(2) | z4=3 | |
| | | | | | | | | | 69 | no(2) | 69 | no(2) | z4=3 | |
| 139.3.96 | 107130 | 26 | - | - | - | - | - | - | 121 | no(82) | 121 | no(82) | z2=1 | Al3Zr1 |
| | | | | | | | | | 119 | no(82) | 119 | no(82) | z2=1 | |
| | | | | | | | | | 71 | no(2) | 71 | no(71) | z4=2 | |
| | | | | | | | | | 69 | no(2) | 69 | no(69) | z4=2 | |
| 139.3.97 | 281665 | 14 | - | - | - | - | - | - | 121 | no(82) | 121 | no(82) | z2=1 | Al3Ta1 |
| | | | | | | | | | 119 | no(82) | 119 | no(82) | z2=1 | |
| | | | | | | | | | 71 | no(2) | 71 | no(71) | z4=2 | |
| | | | | | | | | | 69 | no(2) | 69 | no(69) | z4=2 | |
| 139.3.98 | 153385 | 24 | - | - | - | - | - | - | 121 | yes | 121 | yes | - | Al2Ba1Si2 |
| | | | | | | | | | 119 | yes | 119 | yes | - | |
| | | | | | | | | | 71 | no(2) | 71 | no(2) | z4=3 | |
| | | | | | | | | | 69 | no(2) | 69 | no(2) | z4=3 | |
| 139.3.99 | 414 | 40 | - | - | - | - | - | - | 121 | yes | 121 | yes | - | Ag2Sn2Sr1 |
| | | | | | | | | | 119 | yes | 119 | yes | - | |
| | | | | | | | | | 71 | no(2) | 71 | no(2) | z4=3 | |
| | | | | | | | | | 69 | no(2) | 69 | no(2) | z4=3 | |
| 140.1.1 | 86940 | 72 | 0.0554 | -0.3464 | 87 | 5 | 0 | 2 | - | - | - | - | z8=4 | O3Ru1Sr1 |
| 140.1.2 | 174206 | 68 | 0.0087 | -2.4774 | 87 | 5 | 0 | 2 | - | - | - | - | z8=4 | Mo1O3Sr1 |
| 140.1.3 | 245120 | 28 | 0.0872 | -0.4835 | 2 | 2 | 69,87 | 1 | - | - | - | - | z2w,1=1<br>z2w,2=1<br>z2w,3=1<br>z4=2<br>z8=6 | Mg2Pt1 |
| 140.1.4 | 88093 | 32 | 0.0042 | -0.2836 | 87 | 5 | 0 | 2 | - | - | - | - | z8=4 | Cr1Sb2 |
| 140.1.5 | 86811 | 32 | 0.0252 | 0.0252 | 87 | 5 | 0 | 2 | - | - | - | - | z8=4 | Cr1Sb2 |
| 140.1.6 | 281417 | 104 | 0.0173 | -0.6461 | 2 | 4 | 0 | 2 | - | - | - | - | z2w,1=1<br>z2w,2=1<br>z2w,3=1<br>z8=4 | Cl1Ga4La3 |
| 140.1.7 | 98408 | 236 | 0.0354 | -0.1201 | 2 | 3 | 0 | 2 | - | - | - | - | z2w,1=1<br>z2w,2=1<br>z2w,3=1 | Ce1Ni9Si4 |
| 140.1.8 | 106369 | 172 | 0.0103 | -0.3008 | 2 | 2 | 69,87 | 1 | - | - | - | - | z2w,1=1<br>z2w,2=1<br>z2w,3=1<br>z4=2<br>z8=6 | Cd3Sr5 |
| 140.1.9 | 106344 | 92 | 0.0115 | -0.3831 | 2 | 2 | 69,87 | 1 | - | - | - | - | z2w,1=1<br>z2w,2=1<br>z2w,3=1<br>z4=2<br>z8=6 | Ca5Hg3 |
| 140.1.10 | 39599 | 168 | 0.0047 | -0.2332 | 87 | 5 | 0 | 2 | - | - | - | - | z8=4 | Br5Cd1Cs3 |
| 140.1.11 | 190146 | 260 | 0.0000 | -0.0028 | 2 | 12 | 69,87 | 1 | - | - | - | - | z4=2<br>z8=2 | Bi1Ho5Ni2 |

Continued on next page

**TABLE XIV– continued**

| TQC | ICSD | NoE | DG | IndG | SubSG (w/ T) | cl | SubSG (w/o T) | cl | Max. Sub. SG | yes or no | Max. Sub. SG | yes or no | SIs (w/ T) | Stoich. Formula |
|---|---|---|---|---|---|---|---|---|---|---|---|---|---|---|
| 140.1.12 | 100084 | 124 | 0.0166 | -0.4943 | 2 | 1 | 69,87 | 1 | - | - | - | - | z2w,1=1<br>z2w,2=1<br>z2w,3=1<br>z4=2<br>z8=2 | Ba5Sn3 |
| 140.1.13 | 247208 | 148 | 0.0034 | -0.1300 | 2 | 1 | 69,87 | 1 | - | - | - | - | z2w,1=1<br>z2w,2=1<br>z2w,3=1<br>z4=2<br>z8=2 | B2Cu2O6Sr1 |
| 140.1.14 | 247207 | 148 | 0.0027 | -0.1235 | 2 | 1 | 69,87 | 1 | - | - | - | - | z2w,1=1<br>z2w,2=1<br>z2w,3=1<br>z4=2<br>z8=2 | B2Cu2O6Sr1 |
| 140.2.1 | 1203 | 124 | 0.0067 | -0.2623 | 2,82 | 9 | 2,82 | 4 | - | - | - | - | z4=1<br>z2=1<br>z8=5 | Sn3Sr5 |
| 140.2.2 | 409797 | 28 | 0.0068 | -0.7197 | 2,82 | 6 | 2,82 | 3 | - | - | - | - | z4=3<br>z2=1<br>z8=7 | Sb2Ti1 |
| 140.2.3 | 249576 | 28 | 0.0066 | -0.7137 | 2,82 | 6 | 2,82 | 3 | - | - | - | - | z4=3<br>z2=1<br>z8=7 | Sb2Ti1 |
| 140.2.4 | 16582 | 184 | 0.0028 | -0.0580 | 2,82 | 6 | 2,82 | 3 | - | - | - | - | z4=3<br>z2=1<br>z8=7 | Pu5Si3 |
| 140.2.5 | 10 - 9 | 36 | 0.0080 | -0.2015 | 2,82 | 10 | 2,82 | 4 | - | - | - | - | z2w,1=1<br>z2w,2=1<br>z2w,3=1<br>z4=1<br>z2=1<br>z8=1 | Ni1Zr2 |
| 140.2.6 | 260983 | 188 | 0.0077 | -0.1153 | 2,82 | 7 | 2,82 | 3 | - | - | - | - | z2w,1=1<br>z2w,2=1<br>z2w,3=1<br>z4=3<br>z2=1<br>z8=3 | Ni1U6 |
| 140.2.7 | 409374 | 124 | 0.0130 | -0.3931 | 2,82 | 9 | 2,82 | 4 | - | - | - | - | z4=1<br>z2=1<br>z8=5 | Ge3Sr5 |
| 140.2.8 | 420322 | 216 | 0.0313 | -0.1453 | 2,82 | 6 | 2,82 | 3 | - | - | - | - | z4=3<br>z2=1<br>z8=7 | Cs9O4Sc1 |
| 140.2.9 | 419823 | 216 | 0.0304 | -0.1653 | 2,82 | 6 | 2,82 | 3 | - | - | - | - | z4=3<br>z2=1<br>z8=7 | Cs9In1O4 |
| 140.2.10 | 420320 | 216 | 0.0343 | -0.1555 | 2,82 | 6 | 2,82 | 3 | - | - | - | - | z4=3<br>z2=1<br>z8=7 | Cs9Ga1O4 |
| 140.2.11 | 418133 | 100 | 0.0038 | -0.8804 | 2,82 | 8 | 2,82 | 4 | - | - | - | - | z2w,1=1<br>z2w,2=1<br>z2w,3=1<br>z4=1<br>z2=1<br>z8=5 | Ca1Ni4Sn2 |

Continued on next page

**TABLE XIV– continued**

| TQC | ICSD | NoE | DG | IndG | SubSG (w/ T) | cl | SubSG (w/o T) | cl | Max. Sub. SG | yes or no | Max. Sub. SG | yes or no | SIs (w/ T) | Stoich. Formula |
|---|---|---|---|---|---|---|---|---|---|---|---|---|---|---|
| 140.2.12 | 150146 | 68 | 0.0038 | -0.1834 | 2,82 | 7 | 2,82 | 3 | - | - | - | - | z2w,1=1<br>z2w,2=1<br>z2w,3=1<br>z4=3<br>z2=1<br>z8=3 | Bi2Fe1Ti4 |
| 140.2.13 | 428736 | 168 | 0.0071 | -0.0079 | 2,82 | 7 | 2,82 | 3 | - | - | - | - | z2w,1=1<br>z2w,2=1<br>z2w,3=1<br>z4=3<br>z2=1<br>z8=3 | Al1Eu3O5 |
| 140.2.14 | 420321 | 216 | 0.0347 | -0.2434 | 2,82 | 6 | 2,82 | 3 | - | - | - | - | z4=3<br>z2=1<br>z8=7 | Al1Cs9O4 |
| 140.3.1 | 90899 | 72 | - | - | - | - | - | - | 121 | no(82) | 121 | no(82) | z2=1 | Te1Tl1 |
| | | | | | | | | | 120 | no(82) | 120 | no(82) | z2=1 | |
| | | | | | | | | | 72 | no(2) | 72 | yes | z2w,1=1<br>z2w,2=1<br>z2w,3=1<br>z4=2 | |
| | | | | | | | | | 69 | no(2) | 69 | no(69) | z2w,1=1<br>z2w,3=1<br>z4=2 | |
| 140.3.2 | 90898 | 72 | - | - | - | - | - | - | 121 | no(82) | 121 | no(82) | z2=1 | Te1Tl1 |
| | | | | | | | | | 120 | no(82) | 120 | no(82) | z2=1 | |
| | | | | | | | | | 72 | no(2) | 72 | yes | z2w,1=1<br>z2w,2=1<br>z2w,3=1<br>z4=2 | |
| | | | | | | | | | 69 | no(2) | 69 | no(69) | z2w,1=1<br>z2w,3=1<br>z4=2 | |
| 140.3.3 | 90897 | 72 | - | - | - | - | - | - | 121 | no(82) | 121 | no(82) | z2=1 | Te1Tl1 |
| | | | | | | | | | 120 | no(82) | 120 | no(82) | z2=1 | |
| | | | | | | | | | 72 | no(2) | 72 | yes | z2w,1=1<br>z2w,2=1<br>z2w,3=1<br>z4=2 | |
| | | | | | | | | | 69 | no(2) | 69 | no(69) | z2w,1=1<br>z2w,3=1<br>z4=2 | |
| 140.3.4 | 90896 | 72 | - | - | - | - | - | - | 121 | no(82) | 121 | no(82) | z2=1 | Te1Tl1 |
| | | | | | | | | | 120 | no(82) | 120 | no(82) | z2=1 | |
| | | | | | | | | | 72 | no(2) | 72 | yes | z2w,1=1<br>z2w,2=1<br>z2w,3=1<br>z4=2 | |
| | | | | | | | | | 69 | no(2) | 69 | no(69) | z2w,1=1<br>z2w,3=1<br>z4=2 | |
| 140.3.5 | 90895 | 72 | - | - | - | - | - | - | 121 | no(82) | 121 | no(82) | z2=1 | Te1Tl1 |
| | | | | | | | | | 120 | no(82) | 120 | no(82) | z2=1 | |
| | | | | | | | | | 72 | no(2) | 72 | yes | z2w,1=1<br>z2w,2=1<br>z2w,3=1<br>z4=2 | |

Continued on next page

**TABLE XIV– continued**

| TQC | ICSD | NoE | DG | IndG | SubSG cl (w/ T) | SubSG cl (w/o T) | Max. Sub. SG | yes or no | Max. Sub. SG | yes or no | SIs (w/ T) | Stoich. Formula |
|---|---|---|---|---|---|---|---|---|---|---|---|---|
| | | | | | | | 69 | no(2) | 69 | no(69) | z2w,1=1 z2w,3=1 z4=2 | |
| 140.3.6 | 69027 | 72 | - | - | - - | - - | 121 | no(82) | 121 | no(82) | z2=1 | Te1Tl1 |
| | | | | | | | 120 | no(82) | 120 | no(82) | z2=1 | |
| | | | | | | | 72 | no(2) | 72 | yes | z2w,1=1 z2w,2=1 z2w,3=1 z4=2 | |
| | | | | | | | 69 | no(2) | 69 | no(69) | z2w,1=1 z2w,3=1 z4=2 | |
| 140.3.7 | 52050 | 72 | - | - | - - | - - | 121 | no(82) | 121 | no(82) | z2=1 | Te1Tl1 |
| | | | | | | | 120 | no(82) | 120 | no(82) | z2=1 | |
| | | | | | | | 72 | no(2) | 72 | yes | z2w,1=1 z2w,2=1 z2w,3=1 z4=2 | |
| | | | | | | | 69 | no(2) | 69 | no(69) | z2w,1=1 z2w,3=1 z4=2 | |
| 140.3.8 | 82564 | 68 | - | - | - - | - - | 121 | yes | 121 | yes | - | Sb2Si1V4 |
| | | | | | | | 120 | yes | 120 | yes | - | |
| | | | | | | | 72 | no(2) | 72 | no(2) | z4=3 | |
| | | | | | | | 69 | no(2) | 69 | no(2) | z4=3 | |
| 140.3.9 | 1916 | 208 | - | - | - - | - - | 121 | yes | 121 | yes | - | Pu5Ru3 |
| | | | | | | | 120 | yes | 120 | yes | - | |
| | | | | | | | 72 | no(2) | 72 | no(2) | z4=1 | |
| | | | | | | | 69 | no(2) | 69 | no(2) | z4=1 | |
| 140.3.10 | 418131 | 116 | - | - | - - | - - | 121 | yes | 121 | yes | - | Ni4Sn2Sr1 |
| | | | | | | | 120 | yes | 120 | yes | - | |
| | | | | | | | 72 | no(2) | 72 | no(2) | z2w,1=1 z2w,2=1 z2w,3=1 z4=1 | |
| | | | | | | | 69 | no(2) | 69 | no(2) | z2w,1=1 z2w,3=1 z4=1 | |
| 140.3.11 | 76377 | 84 | - | - | - - | - - | 121 | no(82) | 121 | no(82) | z2=1 | Mo5Si3 |
| | | | | | | | 120 | no(82) | 120 | no(82) | z2=1 | |
| | | | | | | | 72 | yes | 72 | yes | - | |
| | | | | | | | 69 | yes | 69 | yes | - | |
| 140.3.12 | 35756 | 84 | - | - | - - | - - | 121 | no(82) | 121 | no(82) | z2=1 | Mo5Si3 |
| | | | | | | | 120 | no(82) | 120 | no(82) | z2=1 | |
| | | | | | | | 72 | yes | 72 | yes | - | |
| | | | | | | | 69 | yes | 69 | yes | - | |
| 140.3.13 | 412252 | 52 | - | - | - - | - - | 121 | no(82) | 121 | no(82) | z2=1 | Ir1Li1Sn4 |
| | | | | | | | 120 | no(82) | 120 | no(82) | z2=1 | |
| | | | | | | | 72 | yes | 72 | yes | - | |
| | | | | | | | 69 | yes | 69 | yes | - | |
| 140.3.14 | 172149 | 52 | - | - | - - | - - | 121 | no(82) | 121 | no(82) | z2=1 | Ir1Li1Sn4 |
| | | | | | | | 120 | no(82) | 120 | no(82) | z2=1 | |
| | | | | | | | 72 | yes | 72 | yes | - | |
| | | | | | | | 69 | yes | 69 | yes | - | |
| 140.3.15 | 77281 | 84 | - | - | - - | - - | 121 | no(82) | 121 | no(82) | z2=1 | Ge3Mo5 |
| | | | | | | | 120 | no(82) | 120 | no(82) | z2=1 | |
| | | | | | | | 72 | yes | 72 | yes | - | |
| | | | | | | | 69 | yes | 69 | yes | - | |

Continued on next page

**TABLE XIV– continued**

| TQC | ICSD | NoE | DG | IndG | SubSG cl (w/ T) | SubSG cl (w/o T) | Max. Sub. SG | yes or no | Max. Sub. SG | yes or no | SIs (w/ T) | Stoich. Formula |
|---|---|---|---|---|---|---|---|---|---|---|---|---|
| 140.3.16 | 103835 | 148 | - | - | - - | - - | 121 | yes | 121 | yes | - | Ga3Nb5 |
| | | | | | | | 120 | yes | 120 | yes | - | |
| | | | | | | | 72 | no(2) | 72 | no(2) | z2w,1=1<br>z2w,2=1<br>z2w,3=1<br>z4=3 | |
| | | | | | | | 69 | no(2) | 69 | no(2) | z2w,1=1<br>z2w,3=1<br>z4=3 | |
| 140.3.17 | 103678 | 184 | - | - | - - | - - | 121 | yes | 121 | yes | - | Fe1U6 |
| | | | | | | | 120 | yes | 120 | yes | - | |
| | | | | | | | 72 | no(2) | 72 | no(2) | z2w,1=1<br>z2w,2=1<br>z2w,3=1<br>z4=3 | |
| | | | | | | | 69 | no(2) | 69 | no(2) | z2w,1=1<br>z2w,3=1<br>z4=3 | |
| 140.3.18 | 103677 | 184 | - | - | - - | - - | 121 | yes | 121 | yes | - | Fe1U6 |
| | | | | | | | 120 | yes | 120 | yes | - | |
| | | | | | | | 72 | no(2) | 72 | no(2) | z2w,1=1<br>z2w,2=1<br>z2w,3=1<br>z4=3 | |
| | | | | | | | 69 | no(2) | 69 | no(2) | z2w,1=1<br>z2w,3=1<br>z4=3 | |
| 140.3.19 | 103676 | 184 | - | - | - - | - - | 121 | yes | 121 | yes | - | Fe1U6 |
| | | | | | | | 120 | yes | 120 | yes | - | |
| | | | | | | | 72 | no(2) | 72 | no(2) | z2w,1=1<br>z2w,2=1<br>z2w,3=1<br>z4=3 | |
| | | | | | | | 69 | no(2) | 69 | no(2) | z2w,1=1<br>z2w,3=1<br>z4=3 | |
| 140.3.20 | 103675 | 184 | - | - | - - | - - | 121 | yes | 121 | yes | - | Fe1U6 |
| | | | | | | | 120 | yes | 120 | yes | - | |
| | | | | | | | 72 | no(2) | 72 | no(2) | z2w,1=1<br>z2w,2=1<br>z2w,3=1<br>z4=3 | |
| | | | | | | | 69 | no(2) | 69 | no(2) | z2w,1=1<br>z2w,3=1<br>z4=3 | |
| 140.3.21 | 103674 | 184 | - | - | - - | - - | 121 | yes | 121 | yes | - | Fe1U6 |
| | | | | | | | 120 | yes | 120 | yes | - | |
| | | | | | | | 72 | no(2) | 72 | no(2) | z2w,1=1<br>z2w,2=1<br>z2w,3=1<br>z4=3 | |
| | | | | | | | 69 | no(2) | 69 | no(2) | z2w,1=1<br>z2w,3=1<br>z4=3 | |
| 140.3.22 | 409909 | 32 | - | - | - - | - - | 121 | no(82) | 121 | no(82) | z2=1 | Fe1Sn2 |
| | | | | | | | 120 | no(82) | 120 | no(82) | z2=1 | |
| | | | | | | | 72 | yes | 72 | yes | - | |
| | | | | | | | 69 | yes | 69 | yes | - | |

Continued on next page

**TABLE XIV– continued**

| TQC | ICSD | NoE | DG | IndG | SubSG cl (w/ T) | SubSG cl (w/o T) | Max. Sub. SG | yes or no | Max. Sub. SG | yes or no | SIs (w/ T) | Stoich. Formula |
|---|---|---|---|---|---|---|---|---|---|---|---|---|
| 140.3.23 | 247483 | 32 | - | - | - - | - - | 121 | no(82) | 121 | no(82) | z2=1 | Fe1Sn2 |
| | | | | | | | 120 | no(82) | 120 | no(82) | z2=1 | |
| | | | | | | | 72 | yes | 72 | yes | - | |
| | | | | | | | 69 | yes | 69 | yes | - | |
| 140.3.24 | 96143 | 76 | - | - | - - | - - | 121 | yes | 121 | yes | - | Fe1Sb2Ti5 |
| | | | | | | | 120 | yes | 120 | yes | - | |
| | | | | | | | 72 | no(2) | 72 | no(2) | z4=3 | |
| | | | | | | | 69 | no(2) | 69 | no(2) | z4=3 | |
| 140.3.25 | 422455 | 124 | - | - | - - | - - | 121 | yes | 121 | yes | - | Cu4Sn2Sr1 |
| | | | | | | | 120 | yes | 120 | yes | - | |
| | | | | | | | 72 | no(2) | 72 | no(2) | z2w,1=1<br>z2w,2=1<br>z2w,3=1<br>z4=1 | |
| | | | | | | | 69 | no(2) | 69 | no(2) | z2w,1=1<br>z2w,3=1<br>z4=1 | |
| 140.3.26 | 182104 | 124 | - | - | - - | - - | 121 | yes | 121 | yes | - | Cu4Sn2Sr1 |
| | | | | | | | 120 | yes | 120 | yes | - | |
| | | | | | | | 72 | no(2) | 72 | no(2) | z2w,1=1<br>z2w,2=1<br>z2w,3=1<br>z4=1 | |
| | | | | | | | 69 | no(2) | 69 | no(2) | z2w,1=1<br>z2w,3=1<br>z4=1 | |
| 140.3.27 | 260842 | 216 | - | - | - - | - - | 121 | yes | 121 | yes | - | Cs9In1O4 |
| | | | | | | | 120 | yes | 120 | yes | - | |
| | | | | | | | 72 | no(2) | 72 | no(2) | z4=3 | |
| | | | | | | | 69 | no(2) | 69 | no(2) | z4=3 | |
| 140.3.28 | 169765 | 72 | - | - | - - | - - | 121 | no(82) | 121 | no(82) | z2=1 | Cr1F3K1 |
| | | | | | | | 120 | no(82) | 120 | no(82) | z2=1 | |
| | | | | | | | 72 | yes | 72 | yes | - | |
| | | | | | | | 69 | yes | 69 | yes | - | |
| 140.3.29 | 240270 | 156 | - | - | - - | - - | 121 | yes | 121 | yes | - | Ce5O1Pb3 |
| | | | | | | | 120 | yes | 120 | yes | - | |
| | | | | | | | 72 | no(2) | 72 | no(2) | z4=1 | |
| | | | | | | | 69 | no(2) | 69 | no(2) | z4=1 | |
| 140.3.30 | 55584 | 92 | - | - | - - | - - | 121 | no(82) | 121 | no(82) | z2=1 | Ca5Zn3 |
| | | | | | | | 120 | no(82) | 120 | no(82) | z2=1 | |
| | | | | | | | 72 | no(2) | 72 | yes | z2w,1=1<br>z2w,2=1<br>z2w,3=1<br>z4=2 | |
| | | | | | | | 69 | no(2) | 69 | no(69) | z2w,1=1<br>z2w,3=1<br>z4=2 | |
| 140.3.31 | 93699 | 44 | - | - | - - | - - | 121 | no(82) | 121 | no(82) | z2=1 | Ca5Si3 |
| | | | | | | | 120 | no(82) | 120 | no(82) | z2=1 | |
| | | | | | | | 72 | yes | 72 | yes | - | |
| | | | | | | | 69 | yes | 69 | yes | - | |
| 140.3.32 | 93704 | 44 | - | - | - - | - - | 121 | no(82) | 121 | no(82) | z2=1 | Ca5Ge3 |
| | | | | | | | 120 | no(82) | 120 | no(82) | z2=1 | |
| | | | | | | | 72 | yes | 72 | yes | - | |
| | | | | | | | 69 | yes | 69 | yes | - | |
| 140.3.33 | 58668 | 124 | - | - | - - | - - | 121 | no(82) | 121 | no(82) | z2=1 | Ba5Pb3 |
| | | | | | | | 120 | no(82) | 120 | no(82) | z2=1 | |

Continued on next page

**TABLE XIV– continued**

| TQC | ICSD | NoE | DG | IndG | SubSG (w/ T) | cl | SubSG (w/o T) | cl | Max. Sub. SG | yes or no | Max. Sub. SG | yes or no | SIs (w/ T) | Stoich. Formula |
|---|---|---|---|---|---|---|---|---|---|---|---|---|---|---|
| | | | | | | | | | 72 | no(2) | 72 | yes | z2w,1=1 z2w,2=1 z2w,3=1 z4=2 | |
| | | | | | | | | | 69 | no(2) | 69 | no(69) | z2w,1=1 z2w,3=1 z4=2 | |
| 140.3.34 | 73893 | 132 | - | - | - | - | - | - | 121 | yes | 121 | yes | - | Ba3Cr1O5 |
| | | | | | | | | | 120 | yes | 120 | yes | - | |
| | | | | | | | | | 72 | no(2) | 72 | no(2) | z4=3 | |
| | | | | | | | | | 69 | no(2) | 69 | no(2) | z4=3 | |
| 140.4.1 | 40260 | 66 | - | - | - | - | - | - | See details in Table XIX | | | | | Te3Tl5 |
| 140.4.2 | 195541 | 74 | - | - | - | - | - | - | See details in Table XIX | | | | | Si3Ta5 |
| 140.4.3 | 189722 | 74 | - | - | - | - | - | - | See details in Table XIX | | | | | Si3Ta5 |
| 140.4.4 | 95842 | 78 | - | - | - | - | - | - | See details in Table XIX | | | | | Rh1Sb2Ti5 |
| 140.4.5 | 183452 | 70 | - | - | - | - | - | - | See details in Table XIX | | | | | O3Sr1Tc1 |
| 140.4.6 | 189717 | 154 | - | - | - | - | - | - | See details in Table XIX | | | | | Nb5Si3 |
| 140.4.7 | 16499 | 154 | - | - | - | - | - | - | See details in Table XIX | | | | | Nb5Si1Sn2 |
| 140.4.8 | 423763 | 154 | - | - | - | - | - | - | See details in Table XIX | | | | | Nb20Si12 |
| 140.4.9 | 247482 | 30 | - | - | - | - | - | - | See details in Table XIX | | | | | Mn1Sn2 |
| 140.4.10 | 171198 | 30 | - | - | - | - | - | - | See details in Table XIX | | | | | Mn1Sn2 |
| 140.4.11 | 412250 | 50 | - | - | - | - | - | - | See details in Table XIX | | | | | Li1Ru1Sn4 |
| 140.4.12 | 261925 | 166 | - | - | - | - | - | - | See details in Table XIX | | | | | La5Sn1Zn2 |
| 140.4.13 | 95073 | 134 | - | - | - | - | - | - | See details in Table XIX | | | | | La5Si3 |
| 140.4.14 | 98409 | 234 | - | - | - | - | - | - | See details in Table XIX | | | | | La1Ni9Si4 |
| 140.4.15 | 413805 | 118 | - | - | - | - | - | - | See details in Table XIX | | | | | In3Sr5 |
| 140.4.16 | 165190 | 46 | - | - | - | - | - | - | See details in Table XIX | | | | | H1Ca5Sn3 |
| 140.4.17 | 195542 | 74 | - | - | - | - | - | - | See details in Table XIX | | | | | Ge3Ta5 |
| 140.4.18 | 2003 | 154 | - | - | - | - | - | - | See details in Table XIX | | | | | Ge3Nb5 |
| 140.4.19 | 81469 | 98 | - | - | - | - | - | - | See details in Table XIX | | | | | Ge1In4La3 |
| 140.4.20 | 414109 | 50 | - | - | - | - | - | - | See details in Table XIX | | | | | Ga1K1Se4Sn1 |
| 140.4.21 | 79590 | 198 | - | - | - | - | - | - | See details in Table XIX | | | | | Fe9La1Si4 |
| 140.4.22 | 75412 | 74 | - | - | - | - | - | - | See details in Table XIX | | | | | F3K1Mn1 |
| 140.4.23 | 165195 | 138 | - | - | - | - | - | - | See details in Table XIX | | | | | F1Pb3Sr5 |
| 140.4.24 | 408118 | 194 | - | - | - | - | - | - | See details in Table XIX | | | | | Eu5Si3 |
| 140.4.25 | 424907 | 78 | - | - | - | - | - | - | See details in Table XIX | | | | | Eu1O3Ti1 |
| 140.4.26 | 187117 | 78 | - | - | - | - | - | - | See details in Table XIX | | | | | Eu1O3Ti1 |
| 140.4.27 | 187116 | 78 | - | - | - | - | - | - | See details in Table XIX | | | | | Eu1O3Ti1 |
| 140.4.28 | 187113 | 78 | - | - | - | - | - | - | See details in Table XIX | | | | | Eu1O3Ti1 |
| 140.4.29 | 187112 | 78 | - | - | - | - | - | - | See details in Table XIX | | | | | Eu1O3Ti1 |
| 140.4.30 | 187111 | 78 | - | - | - | - | - | - | See details in Table XIX | | | | | Eu1O3Ti1 |
| 140.4.31 | 187110 | 78 | - | - | - | - | - | - | See details in Table XIX | | | | | Eu1O3Ti1 |
| 140.4.32 | 187109 | 78 | - | - | - | - | - | - | See details in Table XIX | | | | | Eu1O3Ti1 |
| 140.4.33 | 187108 | 78 | - | - | - | - | - | - | See details in Table XIX | | | | | Eu1O3Ti1 |
| 140.4.34 | 187107 | 78 | - | - | - | - | - | - | See details in Table XIX | | | | | Eu1O3Ti1 |
| 140.4.35 | 187106 | 78 | - | - | - | - | - | - | See details in Table XIX | | | | | Eu1O3Ti1 |
| 140.4.36 | 81765 | 250 | - | - | - | - | - | - | See details in Table XIX | | | | | Dy5Ni2Sb1 |
| 140.4.37 | 424219 | 250 | - | - | - | - | - | - | See details in Table XIX | | | | | Cu9Sn4Sr1 |
| 140.4.38 | 422456 | 250 | - | - | - | - | - | - | See details in Table XIX | | | | | Cu9Sn4Sr1 |
| 140.4.39 | 182105 | 250 | - | - | - | - | - | - | See details in Table XIX | | | | | Cu9Sn4Sr1 |
| 140.4.40 | 412882 | 250 | - | - | - | - | - | - | See details in Table XIX | | | | | Cu9Si4Sr1 |
| 140.4.41 | 412884 | 250 | - | - | - | - | - | - | See details in Table XIX | | | | | Cu9Ge4Sr1 |
| 140.4.42 | 416796 | 138 | - | - | - | - | - | - | See details in Table XIX | | | | | Cu4Eu1Sn2 |
| 140.4.43 | 159248 | 82 | - | - | - | - | - | - | See details in Table XIX | | | | | Cu1Sb2Ti5 |
| 140.4.44 | 69656 | 82 | - | - | - | - | - | - | See details in Table XIX | | | | | Cu1F3Rb1 |
| 140.4.45 | 9856 | 82 | - | - | - | - | - | - | See details in Table XIX | | | | | Cu1F3K1 |
| 140.4.46 | 8000 | 82 | - | - | - | - | - | - | See details in Table XIX | | | | | Cu1F3K1 |

Continued on next page

**TABLE XIV– continued**

| TQC | ICSD | NoE | DG | IndG | SubSG cl (w/ T) | SubSG cl (w/o T) | Max. Sub. SG yes or no | Max. Sub. SG yes or no | SIs (w/ T) | Stoich. Formula |
|---|---|---|---|---|---|---|---|---|---|---|
| 140.4.47 | 67445 | 82 | - | - | - - | - - | See details in Table XIX | | | Cu1F3K1 |
| 140.4.48 | 67444 | 82 | - | - | - - | - - | See details in Table XIX | | | Cu1F3K1 |
| 140.4.49 | 21109 | 82 | - | - | - - | - - | See details in Table XIX | | | Cu1F3K1 |
| 140.4.50 | 21108 | 82 | - | - | - - | - - | See details in Table XIX | | | Cu1F3K1 |
| 140.4.51 | 182857 | 82 | - | - | - - | - - | See details in Table XIX | | | Cu1F3K1 |
| 140.4.52 | 182856 | 82 | - | - | - - | - - | See details in Table XIX | | | Cu1F3K1 |
| 140.4.53 | 182855 | 82 | - | - | - - | - - | See details in Table XIX | | | Cu1F3K1 |
| 140.4.54 | 182854 | 82 | - | - | - - | - - | See details in Table XIX | | | Cu1F3K1 |
| 140.4.55 | 182853 | 82 | - | - | - - | - - | See details in Table XIX | | | Cu1F3K1 |
| 140.4.56 | 182852 | 82 | - | - | - - | - - | See details in Table XIX | | | Cu1F3K1 |
| 140.4.57 | 182851 | 82 | - | - | - - | - - | See details in Table XIX | | | Cu1F3K1 |
| 140.4.58 | 182850 | 82 | - | - | - - | - - | See details in Table XIX | | | Cu1F3K1 |
| 140.4.59 | 182849 | 82 | - | - | - - | - - | See details in Table XIX | | | Cu1F3K1 |
| 140.4.60 | 158622 | 82 | - | - | - - | - - | See details in Table XIX | | | Cu1F3K1 |
| 140.4.61 | 420323 | 226 | - | - | - - | - - | See details in Table XIX | | | Cs9Fe1O4 |
| 140.4.62 | 403102 | 138 | - | - | - - | - - | See details in Table XIX | | | Cs3I5Mn1 |
| 140.4.63 | 102740 | 34 | - | - | - - | - - | See details in Table XIX | | | Co1Zr2 |
| 140.4.64 | 409910 | 34 | - | - | - - | - - | See details in Table XIX | | | Co1Sn2 |
| 140.4.65 | 247484 | 34 | - | - | - - | - - | See details in Table XIX | | | Co1Sn2 |
| 140.4.66 | 24 | 138 | - | - | - - | - - | See details in Table XIX | | | Cl5Cs3Mn1 |
| 140.4.67 | 16397 | 142 | - | - | - - | - - | See details in Table XIX | | | Cl5Co1Cs3 |
| 140.4.68 | 15245 | 142 | - | - | - - | - - | See details in Table XIX | | | Cl5Co1Cs3 |
| 140.4.69 | 281418 | 110 | - | - | - - | - - | See details in Table XIX | | | Ce3Cl1Ga4 |
| 140.4.70 | 73478 | 102 | - | - | - - | - - | See details in Table XIX | | | Cd1Cu1F4 |
| 140.4.71 | 424221 | 250 | - | - | - - | - - | See details in Table XIX | | | Ca1Cu9Sn4 |
| 140.4.72 | 417718 | 250 | - | - | - - | - - | See details in Table XIX | | | Ca1Cu9Sn4 |
| 140.4.73 | 201135 | 142 | - | - | - - | - - | See details in Table XIX | | | Br5Co1Cs3 |
| 140.4.74 | 150145 | 66 | - | - | - - | - - | See details in Table XIX | | | Bi2Mn1Ti4 |
| 140.4.75 | 150147 | 70 | - | - | - - | - - | See details in Table XIX | | | Bi2Co1Ti4 |
| 140.4.76 | 107217 | 270 | - | - | - - | - - | See details in Table XIX | | | Bi1Er5Pt2 |
| 140.4.77 | 99477 | 94 | - | - | - - | - - | See details in Table XIX | | | Ba1O3Tb1 |
| 140.4.78 | 99476 | 94 | - | - | - - | - - | See details in Table XIX | | | Ba1O3Tb1 |
| 140.4.79 | 99475 | 94 | - | - | - - | - - | See details in Table XIX | | | Ba1O3Tb1 |
| 140.4.80 | 59596 | 94 | - | - | - - | - - | See details in Table XIX | | | Ba1O3Tb1 |
| 140.4.81 | 91229 | 146 | - | - | - - | - - | See details in Table XIX | | | Ba1Mn1Pr2S5 |
| 140.4.82 | 90639 | 138 | - | - | - - | - - | See details in Table XIX | | | Ba1La2Mn1S5 |
| 140.4.83 | 248378 | 74 | - | - | - - | - - | See details in Table XIX | | | Ba1Ir1O3 |
| 140.4.84 | 424220 | 250 | - | - | - - | - - | See details in Table XIX | | | Ba1Cu9Sn4 |
| 140.4.85 | 422457 | 250 | - | - | - - | - - | See details in Table XIX | | | Ba1Cu9Sn4 |
| 140.4.86 | 182106 | 250 | - | - | - - | - - | See details in Table XIX | | | Ba1Cu9Sn4 |
| 140.4.87 | 412883 | 250 | - | - | - - | - - | See details in Table XIX | | | Ba1Cu9Si4 |
| 140.4.88 | 424223 | 250 | - | - | - - | - - | See details in Table XIX | | | Ba1Cu9Ge4 |
| 140.4.89 | 412885 | 250 | - | - | - - | - - | See details in Table XIX | | | Ba1Cu9Ge4 |
| 140.4.90 | 96555 | 154 | - | - | - - | - - | See details in Table XIX | | | Ba1Co1Nd2S5 |
| 140.4.91 | 96554 | 154 | - | - | - - | - - | See details in Table XIX | | | Ba1Co1Nd2S5 |
| 140.4.92 | 95268 | 142 | - | - | - - | - - | See details in Table XIX | | | Ba1Co1La2S5 |
| 140.4.93 | 91228 | 142 | - | - | - - | - - | See details in Table XIX | | | Ba1Ce2Mn1S5 |
| 140.4.94 | 75792 | 46 | - | - | - - | - - | See details in Table XIX | | | B1Ni2 |
| 140.4.95 | 391330 | 38 | - | - | - - | - - | See details in Table XIX | | | B1Fe2 |
| 140.4.96 | 391328 | 38 | - | - | - - | - - | See details in Table XIX | | | B1Fe2 |
| 140.4.97 | 249220 | 118 | - | - | - - | - - | See details in Table XIX | | | Au4In2Rb1 |
| 140.4.98 | 249221 | 118 | - | - | - - | - - | See details in Table XIX | | | Au4In2K1 |
| 140.4.99 | 58416 | 186 | - | - | - - | - - | See details in Table XIX | | | Au3Cd5 |
| 140.4.100 | 58404 | 86 | - | - | - - | - - | See details in Table XIX | | | Au3Ca5 |
| 140.4.101 | 157338 | 234 | - | - | - - | - - | See details in Table XIX | | | Au2Gd5Sb1 |
| 140.4.102 | 413559 | 190 | - | - | - - | - - | See details in Table XIX | | | Al4Br4La5 |
| 140.4.103 | 42517 | 34 | - | - | - - | - - | See details in Table XIX | | | Al2Cu1 |

Continued on next page

**TABLE XIV– continued**

| TQC | ICSD | NoE | DG | IndG | SubSG (w/ T) | cl | SubSG (w/o T) | cl | Max. Sub. SG | yes or no | Max. Sub. SG | yes or no | SIs (w/ T) | Stoich. Formula |
|---|---|---|---|---|---|---|---|---|---|---|---|---|---|---|
| 140.4.104 | 172800 | 34 | - | - | - | - | - | - | See details in Table XIX | | | | | Al2Cu1 |
| 140.4.105 | 245561 | 66 | - | - | - | - | - | - | See details in Table XIX | | | | | Al1Ce1O3 |
| 140.4.106 | 245554 | 66 | - | - | - | - | - | - | See details in Table XIX | | | | | Al1Ce1O3 |
| 140.4.107 | 245265 | 66 | - | - | - | - | - | - | See details in Table XIX | | | | | Al1Ce1O3 |
| 140.4.108 | 245264 | 66 | - | - | - | - | - | - | See details in Table XIX | | | | | Al1Ce1O3 |
| 140.4.109 | 166284 | 66 | - | - | - | - | - | - | See details in Table XIX | | | | | Al1Ce1O3 |
| 140.4.110 | 150277 | 66 | - | - | - | - | - | - | See details in Table XIX | | | | | Al1Ce1O3 |
| 141.1.1 | 423244 | 64 | 0.0310 | -0.3565 | 2,82 | 1 | 2,82 | 1 | - | - | - | - | z4=3<br>z2=1 | Si2Yb1 |
| 141.1.2 | 170062 | 44 | 0.0256 | -1.5584 | 2,82 | 1 | 2,82 | 1 | - | - | - | - | z4=3<br>z2=1 | Sc1Sn2 |
| 141.1.3 | 88368 | 96 | 0.0031 | -0.0215 | 2,82 | 1 | 2,82 | 1 | - | - | - | - | z4=3<br>z2=1 | O4Tb1V1 |
| 141.1.4 | 78078 | 96 | 0.0034 | -0.0213 | 2,82 | 1 | 2,82 | 1 | - | - | - | - | z4=3<br>z2=1 | O4Tb1V1 |
| 141.1.5 | 78076 | 84 | 0.0055 | -0.0200 | 2,82 | 1 | 2,82 | 1 | - | - | - | - | z4=3<br>z2=1 | O4Pr1V1 |
| 141.1.6 | 157327 | 84 | 0.0031 | -0.0380 | 2,82 | 1 | 2,82 | 1 | - | - | - | - | z4=3<br>z2=1 | O4Pr1V1 |
| 141.1.7 | 79759 | 104 | 0.0033 | -0.0120 | 2,82 | 1 | 2,82 | 1 | - | - | - | - | z4=3<br>z2=1 | O4P1Tm1 |
| 141.1.8 | 36053 | 104 | 0.0027 | -0.0161 | 2,82 | 1 | 2,82 | 1 | - | - | - | - | z4=3<br>z2=1 | O4P1Tm1 |
| 141.1.9 | 61683 | 64 | 0.0083 | -0.0094 | 2,82 | 1 | 2,82 | 1 | - | - | - | - | z4=3<br>z2=1 | Na1O2Tb1 |
| 141.1.10 | 61682 | 52 | 0.0160 | -0.0179 | 2,82 | 1 | 2,82 | 1 | - | - | - | - | z4=3<br>z2=1 | Na1O2Pr1 |
| 141.1.11 | 410968 | 40 | 0.0228 | -1.2123 | 2,82 | 1 | 2,82 | 1 | - | - | - | - | z4=3<br>z2=1 | In2Zr1 |
| 141.1.12 | 61422 | 44 | 0.0021 | -1.2675 | 2,82 | 2 | 2,82 | 1 | - | - | - | - | z4=1<br>z2=1 | H1V2 |
| 141.1.13 | 60511 | 44 | 0.0096 | -1.3000 | 2,82 | 2 | 2,82 | 1 | - | - | - | - | z4=1<br>z2=1 | H1V2 |
| 141.1.14 | 103986 | 36 | 0.0861 | -0.1060 | 2,82 | 1 | 2,82 | 1 | - | - | - | - | z4=3<br>z2=1 | Ga2Th1 |
| 141.1.15 | 100566 | 124 | 0.0242 | -0.7641 | 2,82 | 1 | 2,82 | 1 | - | - | - | - | z4=3<br>z2=1 | Cu4O3 |
| 141.1.16 | 35419 | 124 | 0.0032 | -1.4317 | 2,82 | 1 | 2,82 | 1 | - | - | - | - | z4=3<br>z2=1 | Cu2P1Ti1 |
| 141.1.17 | 100796 | 100 | 0.0116 | -0.7288 | 2,82 | 2 | 2,82 | 1 | - | - | - | - | z4=1<br>z2=1 | Cu2Ge1O4 |
| 141.1.18 | 84377 | 92 | 0.0137 | -0.1773 | 2 | 3 | 0 | 2 | - | - | - | - | z4=2 | Cr2Ni1O4 |
| 141.1.19 | 31300 | 92 | 0.0117 | -0.2149 | 2 | 3 | 0 | 2 | - | - | - | - | z4=2 | Cr2Ni1O4 |
| 141.1.20 | 31299 | 92 | 0.0078 | -0.2194 | 2 | 3 | 0 | 2 | - | - | - | - | z4=2 | Cr2Ni1O4 |
| 141.1.21 | 280062 | 92 | 0.0122 | -0.2219 | 2 | 3 | 0 | 2 | - | - | - | - | z4=2 | Cr2Ni1O4 |
| 141.1.22 | 409800 | 288 | 0.0203 | -0.2099 | 2,82 | 1 | 2,82 | 1 | - | - | - | - | z4=3<br>z2=1 | Ce1Zn11 |
| 141.1.23 | 247146 | 24 | 0.0253 | -3.1676 | 2,82 | 1 | 2,82 | 1 | - | - | - | - | z4=3<br>z2=1 | Ag1Li1 |
| 141.1.24 | 247145 | 24 | 0.0253 | -3.1739 | 2,82 | 1 | 2,82 | 1 | - | - | - | - | z4=3<br>z2=1 | Ag1Li1 |
| 141.3.1 | 194521 | 84 | - | - | - | - | - | - | 122 | no(82) | 122 | no(82) | z2=1 | O4Si1U1 |
| | | | | | | | | | 119 | no(82) | 119 | no(82) | z2=1 | |
| | | | | | | | | | 74 | yes | 74 | yes | - | |
| | | | | | | | | | 70 | yes | 70 | yes | - | |
| 141.3.2 | 79755 | 96 | - | - | - | - | - | - | 122 | no(82) | 122 | no(82) | z2=1 | O4P1Tb1 |
| | | | | | | | | | 119 | no(82) | 119 | no(82) | z2=1 | |

Continued on next page

**TABLE XIV– continued**

| TQC | ICSD | NoE | DG | IndG | SubSG cl (w/ T) | SubSG cl (w/o T) | Max. Sub. SG | yes or no | Max. Sub. SG | yes or no | SIs (w/ T) | Stoich. Formula |
|---|---|---|---|---|---|---|---|---|---|---|---|---|
| | | | | | | | 74 | yes | 74 | yes | - | |
| | | | | | | | 70 | yes | 70 | yes | - | |
| 141.3.3 | 35704 | 96 | - | - | - - | - - | 122 | no(82) | 122 | no(82) | z2=1 | O4P1Tb1 |
| | | | | | | | 119 | no(82) | 119 | no(82) | z2=1 | |
| | | | | | | | 74 | yes | 74 | yes | - | |
| | | | | | | | 70 | yes | 70 | yes | - | |
| 141.3.4 | 29316 | 96 | - | - | - - | - - | 122 | no(82) | 122 | no(82) | z2=1 | O4P1Tb1 |
| | | | | | | | 119 | no(82) | 119 | no(82) | z2=1 | |
| | | | | | | | 74 | yes | 74 | yes | - | |
| | | | | | | | 70 | yes | 70 | yes | - | |
| 141.3.5 | 29315 | 96 | - | - | - - | - - | 122 | no(82) | 122 | no(82) | z2=1 | O4P1Tb1 |
| | | | | | | | 119 | no(82) | 119 | no(82) | z2=1 | |
| | | | | | | | 74 | yes | 74 | yes | - | |
| | | | | | | | 70 | yes | 70 | yes | - | |
| 141.3.6 | 246386 | 84 | - | - | - - | - - | 122 | yes | 122 | yes | - | La1Mg1Si2 |
| | | | | | | | 119 | yes | 119 | yes | - | |
| | | | | | | | 74 | no(2) | 74 | no(2) | z4=3 | |
| | | | | | | | 70 | no(2) | 70 | no(2) | z4=3 | |
| 141.3.7 | 152694 | 100 | - | - | - - | - - | 122 | no(82) | 122 | no(82) | z2=1 | Ho1O4V1 |
| | | | | | | | 119 | no(82) | 119 | no(82) | z2=1 | |
| | | | | | | | 74 | yes | 74 | yes | - | |
| | | | | | | | 70 | yes | 70 | yes | - | |
| 141.3.8 | 93788 | 100 | - | - | - - | - - | 122 | no(82) | 122 | no(82) | z2=1 | Cr1Dy1O4 |
| | | | | | | | 119 | no(82) | 119 | no(82) | z2=1 | |
| | | | | | | | 74 | yes | 74 | yes | - | |
| | | | | | | | 70 | yes | 70 | yes | - | |
| 141.3.9 | 9310 | 84 | - | - | - - | - - | 122 | no(82) | 122 | no(82) | z2=1 | Cl4U1 |
| | | | | | | | 119 | no(82) | 119 | no(82) | z2=1 | |
| | | | | | | | 74 | yes | 74 | yes | - | |
| | | | | | | | 70 | yes | 70 | yes | - | |
| 141.3.10 | 62575 | 84 | - | - | - - | - - | 122 | no(82) | 122 | no(82) | z2=1 | Cl4U1 |
| | | | | | | | 119 | no(82) | 119 | no(82) | z2=1 | |
| | | | | | | | 74 | yes | 74 | yes | - | |
| | | | | | | | 70 | yes | 70 | yes | - | |
| 141.3.11 | 62574 | 84 | - | - | - - | - - | 122 | no(82) | 122 | no(82) | z2=1 | Cl4U1 |
| | | | | | | | 119 | no(82) | 119 | no(82) | z2=1 | |
| | | | | | | | 74 | yes | 74 | yes | - | |
| | | | | | | | 70 | yes | 70 | yes | - | |
| 141.3.12 | 2617 | 84 | - | - | - - | - - | 122 | no(82) | 122 | no(82) | z2=1 | Cl4U1 |
| | | | | | | | 119 | no(82) | 119 | no(82) | z2=1 | |
| | | | | | | | 74 | yes | 74 | yes | - | |
| | | | | | | | 70 | yes | 70 | yes | - | |
| 141.3.13 | 202331 | 84 | - | - | - - | - - | 122 | no(82) | 122 | no(82) | z2=1 | Cl4U1 |
| | | | | | | | 119 | no(82) | 119 | no(82) | z2=1 | |
| | | | | | | | 74 | yes | 74 | yes | - | |
| | | | | | | | 70 | yes | 70 | yes | - | |
| 141.3.14 | 74988 | 80 | - | - | - - | - - | 70 | yes | 70 | yes | - | B4Re4 |
| | | | | | | | 46 | yes | 46 | yes | - | |
| | | | | | | | 12 | yes | 12 | yes | - | |
| 141.3.15 | 200230 | 96 | - | - | - - | - - | 122 | no(82) | 122 | no(82) | z2=1 | As1O4Tb1 |
| | | | | | | | 119 | no(82) | 119 | no(82) | z2=1 | |
| | | | | | | | 74 | yes | 74 | yes | - | |
| | | | | | | | 70 | yes | 70 | yes | - | |
| 141.3.16 | 87609 | 124 | - | - | - - | - - | 122 | yes | 122 | yes | - | Ag2Cu2O3 |
| | | | | | | | 119 | yes | 119 | yes | - | |
| | | | | | | | 74 | no(2) | 74 | no(2) | z4=3 | |
| | | | | | | | 70 | no(2) | 70 | no(2) | z4=3 | |
| 141.3.17 | 51672 | 124 | - | - | - - | - - | 122 | yes | 122 | yes | - | Ag2Cu2O3 |

Continued on next page

**TABLE XIV– continued**

| TQC | ICSD | NoE | DG | IndG | SubSG cl (w/ T) | SubSG cl (w/o T) | Max. Sub. SG yes or no | Max. Sub. SG yes or no | SIs (w/ T) | Stoich. Formula |
|---|---|---|---|---|---|---|---|---|---|---|
| | | | | | | | 119 yes | 119 yes | - | |
| | | | | | | | 74 no(2) | 74 no(2) | z4=3 | |
| | | | | | | | 70 no(2) | 70 no(2) | z4=3 | |
| 141.3.18 | 51502 | 124 | - | - | - - | - - | 122 yes | 122 yes | - | Ag2Cu2O3 |
| | | | | | | | 119 yes | 119 yes | - | |
| | | | | | | | 74 no(2) | 74 no(2) | z4=3 | |
| | | | | | | | 70 no(2) | 70 no(2) | z4=3 | |
| 141.3.19 | 165570 | 124 | - | - | - - | - - | 122 yes | 122 yes | - | Ag2Cu2O3 |
| | | | | | | | 119 yes | 119 yes | - | |
| | | | | | | | 74 no(2) | 74 no(2) | z4=3 | |
| | | | | | | | 70 no(2) | 70 no(2) | z4=3 | |
| 141.4.1 | 79760 | 106 | - | - | - - | - - | See details in Table XIX | | | O4P1Yb1 |
| 141.4.2 | 36054 | 106 | - | - | - - | - - | See details in Table XIX | | | O4P1Yb1 |
| 141.4.3 | 78077 | 86 | - | - | - - | - - | See details in Table XIX | | | Nd1O4V1 |
| 141.4.4 | 191450 | 86 | - | - | - - | - - | See details in Table XIX | | | Nd1O4V1 |
| 141.4.5 | 157328 | 86 | - | - | - - | - - | See details in Table XIX | | | Nd1O4V1 |
| 141.4.6 | 68174 | 90 | - | - | - - | - - | See details in Table XIX | | | Mn3O4 |
| 141.4.7 | 107516 | 30 | - | - | - - | - - | See details in Table XIX | | | Li1Sn1 |
| 141.4.8 | 33955 | 74 | - | - | - - | - - | See details in Table XIX | | | Li1O2Yb1 |
| 141.4.9 | 98663 | 50 | - | - | - - | - - | See details in Table XIX | | | Li1N2U1 |
| 141.4.10 | 62536 | 102 | - | - | - - | - - | See details in Table XIX | | | In2N1Sr4 |
| 141.4.11 | 81703 | 94 | - | - | - - | - - | See details in Table XIX | | | Gd1O4V1 |
| 141.4.12 | 238224 | 94 | - | - | - - | - - | See details in Table XIX | | | Gd1O4V1 |
| 141.4.13 | 238223 | 94 | - | - | - - | - - | See details in Table XIX | | | Gd1O4V1 |
| 141.4.14 | 152692 | 94 | - | - | - - | - - | See details in Table XIX | | | Gd1O4V1 |
| 141.4.15 | 97542 | 62 | - | - | - - | - - | See details in Table XIX | | | Gd1Na1O2 |
| 141.4.16 | 22040 | 62 | - | - | - - | - - | See details in Table XIX | | | Gd1Na1O2 |
| 141.4.17 | 80554 | 102 | - | - | - - | - - | See details in Table XIX | | | Fe2O5P1 |
| 141.4.18 | 174085 | 42 | - | - | - - | - - | See details in Table XIX | | | Fe1Li1O2 |
| 141.4.19 | 103402 | 154 | - | - | - - | - - | See details in Table XIX | | | Eu4Ir1 |
| 141.4.20 | 84235 | 102 | - | - | - - | - - | See details in Table XIX | | | Er1O4V1 |
| 141.4.21 | 78080 | 102 | - | - | - - | - - | See details in Table XIX | | | Er1O4V1 |
| 141.4.22 | 79758 | 102 | - | - | - - | - - | See details in Table XIX | | | Er1O4P1 |
| 141.4.23 | 36052 | 102 | - | - | - - | - - | See details in Table XIX | | | Er1O4P1 |
| 141.4.24 | 95381 | 70 | - | - | - - | - - | See details in Table XIX | | | Er1Li1O2 |
| 141.4.25 | 81704 | 98 | - | - | - - | - - | See details in Table XIX | | | Dy1O4V1 |
| 141.4.26 | 26441 | 98 | - | - | - - | - - | See details in Table XIX | | | Dy1O4V1 |
| 141.4.27 | 152693 | 98 | - | - | - - | - - | See details in Table XIX | | | Dy1O4V1 |
| 141.4.28 | 192553 | 98 | - | - | - - | - - | See details in Table XIX | | | Dy1O4P1 |
| 141.4.29 | 88961 | 106 | - | - | - - | - - | See details in Table XIX | | | Cu1O4Rh2 |
| 141.4.30 | 59267 | 106 | - | - | - - | - - | See details in Table XIX | | | Cu1O4Rh2 |
| 141.4.31 | 426084 | 42 | - | - | - - | - - | See details in Table XIX | | | Cu1Li2Sn2 |
| 141.4.32 | 251619 | 42 | - | - | - - | - - | See details in Table XIX | | | Cu1Li2Sn2 |
| 141.4.33 | 75532 | 106 | - | - | - - | - - | See details in Table XIX | | | Cu1Ir2S4 |
| 141.4.34 | 150187 | 18 | - | - | - - | - - | See details in Table XIX | | | Cs1 |
| 141.4.35 | 84378 | 94 | - | - | - - | - - | See details in Table XIX | | | Cr2Cu1O4 |
| 141.4.36 | 80476 | 94 | - | - | - - | - - | See details in Table XIX | | | Cr2Cu1O4 |
| 141.4.37 | 246080 | 94 | - | - | - - | - - | See details in Table XIX | | | Cr2Cu1O4 |
| 141.4.38 | 95971 | 82 | - | - | - - | - - | See details in Table XIX | | | Cr1O4Y1 |
| 141.4.39 | 157933 | 82 | - | - | - - | - - | See details in Table XIX | | | Cr1O4Y1 |
| 141.4.40 | 169874 | 98 | - | - | - - | - - | See details in Table XIX | | | Cr1O4Tb1 |
| 141.4.41 | 166175 | 98 | - | - | - - | - - | See details in Table XIX | | | Cr1O4Tb1 |
| 141.4.42 | 95967 | 110 | - | - | - - | - - | See details in Table XIX | | | Cr1Lu1O4 |
| 141.4.43 | 173263 | 102 | - | - | - - | - - | See details in Table XIX | | | Cr1Ho1O4 |
| 141.4.44 | 169872 | 102 | - | - | - - | - - | See details in Table XIX | | | Cr1Ho1O4 |
| 141.4.45 | 51696 | 94 | - | - | - - | - - | See details in Table XIX | | | Cr1Eu1O4 |
| 141.4.46 | 81931 | 202 | - | - | - - | - - | See details in Table XIX | | | Co9Si2Th1 |

Continued on next page

**TABLE XIV– continued**

| TQC | ICSD | NoE | DG | IndG | SubSG (w/ T) | cl | SubSG (w/o T) | cl | Max. Sub. SG | yes or no | Max. Sub. SG | yes or no | SIs (w/ T) | Stoich. Formula |
|---|---|---|---|---|---|---|---|---|---|---|---|---|---|---|
| 141.4.47 | 9309 | 82 | - | - | - | - | - | - | See details in Table XIX | | | | | Cl4Pa1 |
| 141.4.48 | 9311 | 86 | - | - | - | - | - | - | See details in Table XIX | | | | | Cl4Np1 |
| 141.4.49 | 75482 | 86 | - | - | - | - | - | - | See details in Table XIX | | | | | Cl4Np1 |
| 141.4.50 | 78075 | 82 | - | - | - | - | - | - | See details in Table XIX | | | | | Ce1O4V1 |
| 141.4.51 | 66033 | 82 | - | - | - | - | - | - | See details in Table XIX | | | | | Ce1O4V1 |
| 141.4.52 | 202425 | 82 | - | - | - | - | - | - | See details in Table XIX | | | | | Ce1O4V1 |
| 141.4.53 | 157326 | 82 | - | - | - | - | - | - | See details in Table XIX | | | | | Ce1O4V1 |
| 141.4.54 | 55349 | 42 | - | - | - | - | - | - | See details in Table XIX | | | | | Au1Li2Sn2 |
| 141.4.55 | 171193 | 106 | - | - | - | - | - | - | See details in Table XIX | | | | | As1O4Yb1 |
| 141.4.56 | 200228 | 98 | - | - | - | - | - | - | See details in Table XIX | | | | | As1Dy1O4 |
| 141.4.57 | 16512 | 98 | - | - | - | - | - | - | See details in Table XIX | | | | | As1Dy1O4 |
| 141.4.58 | 426085 | 42 | - | - | - | - | - | - | See details in Table XIX | | | | | Ag1Li2Sn2 |
| 142.3.1 | 174219 | 100 | - | - | - | - | - | - | 70 | yes | 70 | yes | - | Rh1Sn4 |
| | | | | | | | | | 45 | yes | 45 | yes | - | |
| | | | | | | | | | 15 | yes | 15 | yes | - | |
| 142.3.2 | 412573 | 100 | - | - | - | - | - | - | 70 | yes | 70 | yes | - | Ir1Sn4 |
| | | | | | | | | | 45 | yes | 45 | yes | - | |
| | | | | | | | | | 15 | yes | 15 | yes | - | |
| 143.3.1 | 160324 | 126 | - | - | - | - | - | - | 1 | yes | 1 | yes | - | Ni1Ti1 |
| 145.3.1 | 172481 | 144 | - | - | - | - | - | - | 1 | yes | 1 | yes | - | B2Bi1F1O4 |
| 146.3.1 | 237873 | 72 | - | - | - | - | - | - | 1 | yes | 1 | yes | - | Ni3O6Te1 |
| 146.3.2 | 91527 | 88 | - | - | - | - | - | - | 1 | yes | 1 | yes | - | K1Ni1O9P3 |
| 146.3.3 | 26149 | 82 | - | - | - | - | - | - | 1 | yes | 1 | yes | - | H12Ni1O9S1 |
| 146.3.4 | 171013 | 112 | - | - | - | - | - | - | 1 | yes | 1 | yes | - | Cs4Se9Sn3Sr1 |
| 146.3.5 | 33806 | 176 | - | - | - | - | - | - | 1 | yes | 1 | yes | - | Ba3O9Tm4 |
| 147.1.1 | 166370 | 126 | 0.0000 | -0.5340 | 2 | 2 | 2 | 1 | - | - | - | - | z2w,3=1<br>z4=3 | Ni1Ti1 |
| 147.1.2 | 150943 | 126 | 0.0000 | -0.5453 | 2 | 1 | 2 | 1 | - | - | - | - | z4=1 | Ni1Ti1 |
| 147.3.1 | 109513 | 180 | - | - | - | - | - | - | 2 | no(2) | 2 | no(2) | z2w,3=1<br>z4=3 | Cl18Cs2W6 |
| 148.1.1 | 54337 | 104 | 0.0000 | -0.5327 | 2 | 6 | 2 | 1 | - | - | - | - | z4=1 | Pd4Zr3 |
| 148.1.2 | 1096 | 160 | 0.0029 | -0.4285 | 2 | 7 | 0 | 2 | - | - | - | - | z4=2 | P2Pd15 |
| 148.1.3 | 37134 | 96 | 0.0067 | 0.0053 | 2 | 1 | 2 | 1 | - | - | - | - | z4=3 | O12Te1Tl6 |
| 148.1.4 | 166373 | 104 | 0.0289 | -0.3385 | 2 | 4 | 2 | 1 | - | - | - | - | z2w,1=1<br>z2w,2=1<br>z2w,3=1<br>z4=1 | Ni4Ti3 |
| 148.1.5 | 166371 | 104 | 0.0099 | -0.3584 | 2 | 5 | 0 | 2 | - | - | - | - | z2w,1=1<br>z2w,2=1<br>z2w,3=1<br>z4=2 | Ni4Ti3 |
| 148.1.6 | 166369 | 104 | 0.0097 | -0.3235 | 2 | 6 | 2 | 1 | - | - | - | - | z4=1 | Ni4Ti3 |
| 148.1.7 | 171584 | 64 | 0.0719 | -0.0728 | 2 | 5 | 0 | 2 | - | - | - | - | z2w,1=1<br>z2w,2=1<br>z2w,3=1<br>z4=2 | Ni1O3Ti1 |
| 148.1.8 | 59375 | 84 | 0.1701 | 0.0764 | 2 | 2 | 2 | 1 | - | - | - | - | z2w,1=1<br>z2w,2=1<br>z2w,3=1<br>z4=3 | Mo6Te8 |
| 148.1.9 | 46015 | 108 | 0.0407 | -0.4491 | 2 | 1 | 2 | 1 | - | - | - | - | z4=3 | Mo6S8Yb1 |
| 148.1.10 | 46014 | 108 | 0.0340 | -0.4389 | 2 | 1 | 2 | 1 | - | - | - | - | z4=3 | Mo6S8Yb1 |
| 148.1.11 | 46013 | 108 | 0.0480 | -0.2868 | 2 | 1 | 2 | 1 | - | - | - | - | z4=3 | Mo6S8Yb1 |
| 148.1.12 | 201506 | 108 | 0.0497 | -0.3395 | 2 | 1 | 2 | 1 | - | - | - | - | z4=3 | Mo6S8Yb1 |
| 148.1.13 | 62156 | 94 | 0.0443 | -0.2763 | 2 | 1 | 2 | 1 | - | - | - | - | z4=3 | Mo6S8Sr1 |
| 148.1.14 | 68 | 88 | 0.0758 | -0.1623 | 2 | 1 | 2 | 1 | - | - | - | - | z4=3 | Mo3Pb0.5Se4 |

Continued on next page

**TABLE XIV– continued**

| TQC | ICSD | NoE | DG | IndG | SubSG (w/ T) | cl | SubSG (w/o T) | cl | Max. Sub. SG | yes or no | Max. Sub. SG | yes or no | SIs (w/ T) | Stoich. Formula |
|---|---|---|---|---|---|---|---|---|---|---|---|---|---|---|
| 148.1.15 | 89004 | 174 | 0.0771 | -0.0553 | 2 | 4 | 2 | 1 | - | - | - | - | z2w,1=1 z2w,2=1 z2w,3=1 z4=1 | Mo12Rb2Se14 |
| 148.1.16 | 245054 | 126 | 0.0000 | -0.0505 | 2 | 2 | 2 | 1 | - | - | - | - | z2w,1=1 z2w,2=1 z2w,3=1 z4=3 | Mn3O6Te1 |
| 148.1.17 | 183618 | 126 | 0.0000 | -0.0467 | 2 | 2 | 2 | 1 | - | - | - | - | z2w,1=1 z2w,2=1 z2w,3=1 z4=3 | Mn3O6Te1 |
| 148.1.18 | 182436 | 126 | 0.0000 | -0.0486 | 2 | 2 | 2 | 1 | - | - | - | - | z2w,1=1 z2w,2=1 z2w,3=1 z4=3 | Mn3O6Te1 |
| 148.1.19 | 182435 | 126 | 0.0000 | -0.0502 | 2 | 2 | 2 | 1 | - | - | - | - | z2w,1=1 z2w,2=1 z2w,3=1 z4=3 | Mn3O6Te1 |
| 148.1.20 | 60006 | 58 | 0.0097 | -0.1070 | 2 | 5 | 0 | 2 | - | - | - | - | z2w,1=1 z2w,2=1 z2w,3=1 z4=2 | Mn1O3Ti1 |
| 148.1.21 | 247553 | 58 | 0.0000 | -0.0782 | 2 | 2 | 2 | 1 | - | - | - | - | z2w,1=1 z2w,2=1 z2w,3=1 z4=3 | Mn1O3Ti1 |
| 148.1.22 | 171579 | 58 | 0.0095 | -0.1024 | 2 | 5 | 0 | 2 | - | - | - | - | z2w,1=1 z2w,2=1 z2w,3=1 z4=2 | Mn1O3Ti1 |
| 148.1.23 | 73247 | 58 | 0.0214 | -0.0644 | 2 | 2 | 2 | 1 | - | - | - | - | z2w,1=1 z2w,2=1 z2w,3=1 z4=3 | Ge1Mn1O3 |
| 148.1.24 | 73246 | 58 | 0.0228 | -0.0593 | 2 | 2 | 2 | 1 | - | - | - | - | z2w,1=1 z2w,2=1 z2w,3=1 z4=3 | Ge1Mn1O3 |
| 148.1.25 | 73245 | 58 | 0.0209 | -0.0612 | 2 | 2 | 2 | 1 | - | - | - | - | z2w,1=1 z2w,2=1 z2w,3=1 z4=3 | Ge1Mn1O3 |
| 148.1.26 | 73244 | 58 | 0.0222 | -0.0614 | 2 | 2 | 2 | 1 | - | - | - | - | z2w,1=1 z2w,2=1 z2w,3=1 z4=3 | Ge1Mn1O3 |
| 148.1.27 | 73243 | 58 | 0.0219 | -0.0577 | 2 | 2 | 2 | 1 | - | - | - | - | z2w,1=1 z2w,2=1 z2w,3=1 z4=3 | Ge1Mn1O3 |
| 148.1.28 | 69591 | 58 | 0.0224 | -0.0565 | 2 | 2 | 2 | 1 | - | - | - | - | z2w,1=1 z2w,2=1 z2w,3=1 z4=3 | Ge1Mn1O3 |

Continued on next page

**TABLE XIV– continued**

| TQC | ICSD | NoE | DG | IndG | SubSG (w/ T) | cl | SubSG (w/o T) | cl | Max. Sub. SG | yes or no | Max. Sub. SG | yes or no | SIs (w/ T) | Stoich. Formula |
|---|---|---|---|---|---|---|---|---|---|---|---|---|---|---|
| 148.1.29 | 247127 | 198 | 0.0061 | -0.1330 | 2 | 5 | 0 | 2 | - | - | - | - | z2w,1=1<br>z2w,2=1<br>z2w,3=1<br>z4=2 | Cu1O3Si1 |
| 148.1.30 | 84337 | 174 | 0.0190 | -0.0719 | 2 | 2 | 2 | 1 | - | - | - | - | z2w,1=1<br>z2w,2=1<br>z2w,3=1<br>z4=3 | Cs2Mo12Te14 |
| 148.1.31 | 40820 | 174 | 0.1014 | -0.0501 | 2 | 2 | 2 | 1 | - | - | - | - | z2w,1=1<br>z2w,2=1<br>z2w,3=1<br>z4=3 | Cs2Mo12Se14 |
| 148.1.32 | 71020 | 56 | 0.0179 | 0.0007 | 2 | 4 | 2 | 1 | - | - | - | - | z2w,1=1<br>z2w,2=1<br>z2w,3=1<br>z4=1 | Cr2Si2Te6 |
| 148.1.33 | 106520 | 60 | 0.0256 | -0.0978 | 2 | 1 | 2 | 1 | - | - | - | - | z4=3 | Cr2Se3 |
| 148.1.34 | 79270 | 56 | 0.0305 | -0.0080 | 2 | 4 | 2 | 1 | - | - | - | - | z2w,1=1<br>z2w,2=1<br>z2w,3=1<br>z4=1 | Cr2Ge2Te6 |
| 148.1.35 | 79269 | 56 | 0.0000 | -0.0618 | 2 | 4 | 2 | 1 | - | - | - | - | z2w,1=1<br>z2w,2=1<br>z2w,3=1<br>z4=1 | Cr2Ge2Te6 |
| 148.1.36 | 79268 | 56 | 0.0316 | -0.0062 | 2 | 4 | 2 | 1 | - | - | - | - | z2w,1=1<br>z2w,2=1<br>z2w,3=1<br>z4=1 | Cr2Ge2Te6 |
| 148.1.37 | 252042 | 64 | 0.0265 | 0.0077 | 2 | 3 | 0 | 2 | - | - | - | - | z2w,1=1<br>z2w,2=1<br>z2w,3=1 | Cr1Fe1O3 |
| 148.1.38 | 39426 | 50 | 0.0000 | -0.2393 | 2 | 1 | 2 | 1 | - | - | - | - | z4=3 | Cl3Ti1 |
| 148.1.39 | 201976 | 110 | 0.0360 | -0.0346 | 2 | 2 | 2 | 1 | - | - | - | - | z2w,1=1<br>z2w,2=1<br>z2w,3=1<br>z4=3 | Cl12N1Sc7 |
| 148.1.40 | 424476 | 66 | 0.0000 | -0.0350 | 2 | 1 | 2 | 1 | - | - | - | - | z4=3 | Cl12Co1Sc7 |
| 148.1.41 | 183603 | 134 | 0.0943 | -0.2980 | 2 | 1 | 2 | 1 | - | - | - | - | z4=3 | Ce1Pt5Sb1 |
| 148.1.42 | 62388 | 94 | 0.0428 | -0.3946 | 2 | 1 | 2 | 1 | - | - | - | - | z4=3 | Ca1Mo6S8 |
| 148.1.43 | 39242 | 50 | 0.0093 | -0.2084 | 2 | 1 | 2 | 1 | - | - | - | - | z4=3 | Br3Ti1 |
| 148.1.44 | 419636 | 110 | 0.0468 | -0.2875 | 2 | 4 | 2 | 1 | - | - | - | - | z2w,1=1<br>z2w,2=1<br>z2w,3=1<br>z4=1 | Ba6N6O1Re2 |
| 148.1.45 | 419467 | 112 | 0.0031 | -0.1043 | 2 | 1 | 2 | 1 | - | - | - | - | z4=3 | Ba6N6O1Os2 |
| 148.1.46 | 96087 | 88 | 0.1875 | 0.1578 | 2 | 3 | 0 | 2 | - | - | - | - | z2w,1=1<br>z2w,2=1<br>z2w,3=1 | Ba1Ni2O8V2 |
| 148.1.47 | 96086 | 88 | 0.1799 | 0.1546 | 2 | 3 | 0 | 2 | - | - | - | - | z2w,1=1<br>z2w,2=1<br>z2w,3=1 | Ba1Ni2O8V2 |
| 148.1.48 | 85489 | 94 | 0.0455 | -0.2585 | 2 | 1 | 2 | 1 | - | - | - | - | z4=3 | Ba1Mo6S8 |
| 148.1.49 | 65776 | 94 | 0.0474 | -0.1461 | 2 | 1 | 2 | 1 | - | - | - | - | z4=3 | Ba1Mo6S8 |
| 148.1.50 | 62157 | 94 | 0.0469 | -0.1508 | 2 | 1 | 2 | 1 | - | - | - | - | z4=3 | Ba1Mo6S8 |
| 148.1.51 | 59343 | 86 | 0.0111 | -0.1103 | 2 | 3 | 0 | 2 | - | - | - | - | z2w,1=1<br>z2w,2=1<br>z2w,3=1 | Ba1Co2O8P2 |

Continued on next page

**TABLE XIV– continued**

| TQC | ICSD | NoE | DG | IndG | SubSG (w/ T) | cl | SubSG (w/o T) | cl | Max. Sub. SG | yes or no | Max. Sub. SG | yes or no | SIs (w/ T) | Stoich. Formula |
|---|---|---|---|---|---|---|---|---|---|---|---|---|---|---|
| 148.1.52 | 237739 | 190 | 0.0140 | -0.1613 | 2 | 2 | 2 | 1 | - | - | - | - | z2w,1=1<br>z2w,2=1<br>z2w,3=1<br>z4=3 | Au5Mn1Y3 |
| 148.1.53 | 91777 | 68 | 0.0198 | -0.0466 | 2 | 1 | 2 | 1 | - | - | - | - | z4=3 | Ag1Bi1O3 |
| 148.1.54 | 89432 | 68 | 0.0198 | -0.0466 | 2 | 1 | 2 | 1 | - | - | - | - | z4=3 | Ag1Bi1O3 |
| 148.3.1 | 2516 | 176 | - | - | - | - | - | - | 2 | no(2) | 2 | no(2) | z4=3 | Pd4Pu3 |
| 148.3.2 | 200672 | 88 | - | - | - | - | - | - | 2 | no(2) | 2 | no(2) | z4=3 | Mo6S8Sn1 |
| 148.3.3 | 94879 | 168 | - | - | - | - | - | - | 2 | yes | 2 | yes | - | I12La6Os1Sr1 |
| 148.3.4 | 78905 | 62 | - | - | - | - | - | - | 2 | yes | 2 | yes | - | F6Ni1Pt1 |
| 148.3.5 | 201511 | 58 | - | - | - | - | - | - | 2 | yes | 2 | yes | - | Cs1F6Re1 |
| 148.3.6 | 48107 | 62 | - | - | - | - | - | - | 2 | no(2) | 2 | no(2) | z4=3 | Co1O3Ti1 |
| 148.3.7 | 16548 | 62 | - | - | - | - | - | - | 2 | no(2) | 2 | no(2) | z4=3 | Co1O3Ti1 |
| 148.3.8 | 51513 | 162 | - | - | - | - | - | - | 2 | yes | 2 | yes | - | Cl18W6 |
| 148.3.9 | 109512 | 162 | - | - | - | - | - | - | 2 | yes | 2 | yes | - | Cl18W6 |
| 148.3.10 | 100599 | 86 | - | - | - | - | - | - | 2 | no(2) | 2 | yes | z2w,1=1<br>z2w,2=1<br>z2w,3=1<br>z4=2 | Br2Mo6S6 |
| 148.3.11 | 54 - | 100 | - | - | - | - | - | - | 2 | no(2) | 2 | no(2) | z4=1 | Au4Ca3 |
| 150.4.1 | 190109 | 66 | - | - | - | - | - | - | See details in Table XIX | | | | | O8P2Ta1Tl1 |
| 152.4.1 | 65692 | 18 | - | - | - | - | - | - | See details in Table XIX | | | | | Te1 |
| 155.3.1 | 186536 | 76 | - | - | - | - | - | - | 5 | yes | 5 | yes | - | Li4Pt3Si1 |
| 156.3.1 | 110774 | 20 | - | - | - | - | - | - | 8 | yes | 8 | yes | - | Mn1Na1Te2 |
| 156.3.2 | 50818 | 20 | - | - | - | - | - | - | 8 | yes | 8 | yes | - | Mn1Na1Se2 |
| 156.3.3 | 110773 | 20 | - | - | - | - | - | - | 8 | yes | 8 | yes | - | Li1Mn1Te2 |
| 156.3.4 | 50817 | 20 | - | - | - | - | - | - | 8 | yes | 8 | yes | - | Li1Mn1Se2 |
| 156.3.5 | 68645 | 60 | - | - | - | - | - | - | 8 | yes | 8 | yes | - | In2S6Zn3 |
| 159.3.1 | 237948 | 198 | - | - | - | - | - | - | 143 | yes | 143 | yes | - | Ba1Co4O7Y1 |
| 159.3.2 | 172414 | 224 | - | - | - | - | - | - | 9 | yes | 9 | yes | - | Ba1Co4O6.95Yb1 |
| 159.4.1 | 252019 | 46 | - | - | - | - | - | - | See details in Table XIX | | | | | Ir1Li2Si3 |
| 159.4.2 | 237950 | 198 | - | - | - | - | - | - | See details in Table XIX | | | | | Ba1Co4O7Y1 |
| 159.4.3 | 237949 | 198 | - | - | - | - | - | - | See details in Table XIX | | | | | Ba1Co4O7Y1 |
| 160.3.1 | 40054 | 48 | - | - | - | - | - | - | 8 | yes | 8 | yes | - | Ni1S1 |
| 160.3.2 | 40053 | 48 | - | - | - | - | - | - | 8 | yes | 8 | yes | - | Ni1S1 |
| 160.3.3 | 151602 | 48 | - | - | - | - | - | - | 8 | yes | 8 | yes | - | Ni1S1 |
| 160.3.4 | 151601 | 48 | - | - | - | - | - | - | 8 | yes | 8 | yes | - | Ni1S1 |
| 160.3.5 | 151600 | 48 | - | - | - | - | - | - | 8 | yes | 8 | yes | - | Ni1S1 |
| 160.3.6 | 151599 | 48 | - | - | - | - | - | - | 8 | yes | 8 | yes | - | Ni1S1 |
| 160.3.7 | 65725 | 84 | - | - | - | - | - | - | 8 | yes | 8 | yes | - | In2S4Zn1 |
| 160.3.8 | 42667 | 42 | - | - | - | - | - | - | 8 | yes | 8 | yes | - | In2S4Zn1 |
| 160.3.9 | 69571 | 50 | - | - | - | - | - | - | 8 | yes | 8 | yes | - | H6F3O3V1 |
| 160.3.10 | 4084 | 144 | - | - | - | - | - | - | 8 | yes | 8 | yes | - | F4O1U1 |
| 161.3.1 | 200999 | 54 | - | - | - | - | - | - | 146 | yes | 146 | yes | - | Li2O3Re1 |
| 161.3.2 | 35012 | 52 | - | - | - | - | - | - | 9 | yes | 9 | yes | - | Li1O3Re1 |
| 161.3.3 | 171497 | 88 | - | - | - | - | - | - | 9 | yes | 9 | yes | - | Ge3Li8Zn2 |
| 161.3.4 | 153664 | 222 | - | - | - | - | - | - | 1 | yes | 1 | yes | - | As3Yb4 |
| 161.4.1 | 65359 | 58 | - | - | - | - | - | - | See details in Table XIX | | | | | Mn1O3Ti1 |
| 161.4.2 | 190911 | 58 | - | - | - | - | - | - | See details in Table XIX | | | | | Mn1O3Ti1 |
| 161.4.3 | 97591 | 62 | - | - | - | - | - | - | See details in Table XIX | | | | | Bi1Fe1O3 |
| 161.4.4 | 75324 | 62 | - | - | - | - | - | - | See details in Table XIX | | | | | Bi1Fe1O3 |
| 161.4.5 | 51664 | 62 | - | - | - | - | - | - | See details in Table XIX | | | | | Bi1Fe1O3 |
| 161.4.6 | 428380 | 62 | - | - | - | - | - | - | See details in Table XIX | | | | | Bi1Fe1O3 |
| 161.4.7 | 417312 | 62 | - | - | - | - | - | - | See details in Table XIX | | | | | Bi1Fe1O3 |
| 161.4.8 | 417311 | 62 | - | - | - | - | - | - | See details in Table XIX | | | | | Bi1Fe1O3 |
| 161.4.9 | 417310 | 62 | - | - | - | - | - | - | See details in Table XIX | | | | | Bi1Fe1O3 |
| 161.4.10 | 417309 | 62 | - | - | - | - | - | - | See details in Table XIX | | | | | Bi1Fe1O3 |

Continued on next page

**TABLE XIV– continued**

| TQC | ICSD | NoE | DG | IndG | SubSG (w/ T) | cl | SubSG (w/o T) | cl | Max. Sub. SG | yes or no | Max. Sub. SG | yes or no | SIs (w/ T) | Stoich. Formula |
|---|---|---|---|---|---|---|---|---|---|---|---|---|---|---|
| 161.4.11 | 417308 | 62 | - | - | - | - | - | - | See details in Table XIX | | | | | Bi1Fe1O3 |
| 161.4.12 | 417307 | 62 | - | - | - | - | - | - | See details in Table XIX | | | | | Bi1Fe1O3 |
| 161.4.13 | 417306 | 62 | - | - | - | - | - | - | See details in Table XIX | | | | | Bi1Fe1O3 |
| 161.4.14 | 417305 | 62 | - | - | - | - | - | - | See details in Table XIX | | | | | Bi1Fe1O3 |
| 161.4.15 | 417304 | 62 | - | - | - | - | - | - | See details in Table XIX | | | | | Bi1Fe1O3 |
| 161.4.16 | 417303 | 62 | - | - | - | - | - | - | See details in Table XIX | | | | | Bi1Fe1O3 |
| 161.4.17 | 260119 | 62 | - | - | - | - | - | - | See details in Table XIX | | | | | Bi1Fe1O3 |
| 161.4.18 | 260118 | 62 | - | - | - | - | - | - | See details in Table XIX | | | | | Bi1Fe1O3 |
| 161.4.19 | 245827 | 62 | - | - | - | - | - | - | See details in Table XIX | | | | | Bi1Fe1O3 |
| 161.4.20 | 237368 | 62 | - | - | - | - | - | - | See details in Table XIX | | | | | Bi1Fe1O3 |
| 161.4.21 | 237367 | 62 | - | - | - | - | - | - | See details in Table XIX | | | | | Bi1Fe1O3 |
| 161.4.22 | 192511 | 62 | - | - | - | - | - | - | See details in Table XIX | | | | | Bi1Fe1O3 |
| 161.4.23 | 192509 | 62 | - | - | - | - | - | - | See details in Table XIX | | | | | Bi1Fe1O3 |
| 161.4.24 | 191940 | 62 | - | - | - | - | - | - | See details in Table XIX | | | | | Bi1Fe1O3 |
| 161.4.25 | 188396 | 62 | - | - | - | - | - | - | See details in Table XIX | | | | | Bi1Fe1O3 |
| 161.4.26 | 163688 | 62 | - | - | - | - | - | - | See details in Table XIX | | | | | Bi1Fe1O3 |
| 161.4.27 | 163687 | 62 | - | - | - | - | - | - | See details in Table XIX | | | | | Bi1Fe1O3 |
| 161.4.28 | 15299 | 62 | - | - | - | - | - | - | See details in Table XIX | | | | | Bi1Fe1O3 |
| 161.4.29 | 109370 | 62 | - | - | - | - | - | - | See details in Table XIX | | | | | Bi1Fe1O3 |
| 161.4.30 | 153665 | 222 | - | - | - | - | - | - | See details in Table XIX | | | | | As3Yb4 |
| 162.1.1 | 28086 | 60 | 0.0073 | -0.1584 | 2 | 4 | 2 | 1 | - | - | - | - | z2w,3=1 z4=1 | O6U1V2 |
| 162.1.2 | 275 | 62 | 0.0161 | -0.1353 | 2 | 3 | 2 | 1 | - | - | - | - | z4=1 | Cr2O6U1 |
| 162.1.3 | 15133 | 62 | 0.0185 | -0.1280 | 2 | 3 | 2 | 1 | - | - | - | - | z4=1 | Cr2O6U1 |
| 162.1.4 | 37011 | 100 | 0.0079 | -0.3234 | 2 | 2 | 0 | 2 | - | - | - | - | z2w,3=1 z4=2 | Ba1Fe4O8Sr1 |
| 162.1.5 | 26557 | 78 | 0.0095 | -0.4713 | 2 | 1 | 2 | 1 | - | - | - | - | z4=3 | Ag6O2 |
| 162.3.1 | 260305 | 180 | - | - | - | - | - | - | 12 | no(2) | 12 | no(2) | z2w,3=1 z4=3 | Cl18K2W6 |
| 162.3.2 | 245119 | 56 | - | - | - | - | - | - | 12 | yes | 12 | yes | - | As2O6Pd1 |
| 162.3.3 | 195298 | 56 | - | - | - | - | - | - | 12 | yes | 12 | yes | - | As2O6Pd1 |
| 162.3.4 | 187099 | 56 | - | - | - | - | - | - | 12 | yes | 12 | yes | - | As2O6Pd1 |
| 162.3.5 | 187098 | 56 | - | - | - | - | - | - | 12 | yes | 12 | yes | - | As2O6Pd1 |
| 162.3.6 | 80349 | 56 | - | - | - | - | - | - | 12 | yes | 12 | yes | - | As2Ni1O6 |
| 162.3.7 | 30360 | 156 | - | - | - | - | - | - | 12 | yes | 12 | yes | - | Ag1Se2Tl1 |
| 163.1.1 | 71037 | 112 | 0.0052 | -0.1452 | 2 | 2 | 2 | 1 | - | - | - | - | z4=1 | Mo1Ni1P8 |
| 163.1.2 | 411131 | 140 | 0.0134 | -0.0011 | 2 | 1 | 2 | 1 | - | - | - | - | z4=3 | Al1F6Li1Yb1 |
| 163.3.1 | 78748 | 112 | - | - | - | - | - | - | 15 | yes | 15 | yes | - | F6Ga1Li1Pd1 |
| 163.3.2 | 15039 | 120 | - | - | - | - | - | - | 15 | no(2) | 15 | yes | z4=2 | Cr2Te3 |
| 163.3.3 | 73132 | 112 | - | - | - | - | - | - | 15 | yes | 15 | yes | - | Al1F6Li1Pd1 |
| 163.3.4 | 195337 | 160 | - | - | - | - | - | - | 15 | no(2) | 15 | no(2) | z4=1 | Ag1P2Se6Tm1 |
| 163.4.1 | 62035 | 102 | - | - | - | - | - | - | See details in Table XIX | | | | | Cl6Cr1Na3 |
| 163.4.2 | 195335 | 158 | - | - | - | - | - | - | See details in Table XIX | | | | | Ag1Er1P2Se6 |
| 164.1.1 | 657478 | 16 | 0.0000 | -0.7448 | 2 | 7 | 0 | 1 | - | - | - | - | z4=2 | Te2Zr1 |
| 164.1.2 | 653213 | 16 | 0.0000 | -0.7538 | 2 | 7 | 0 | 1 | - | - | - | - | z4=2 | Te2Zr1 |
| 164.1.3 | 80205 | 16 | 0.0961 | -0.5185 | 2 | 5 | 2 | 2 | - | - | - | - | z4=3 | Si1Te2 |
| 164.1.4 | 651764 | 58 | 0.0064 | -0.4025 | 2 | 4 | 0 | 1 | - | - | - | - | z2w,3=1 | Sb2Yb1Zn2 |
| 164.1.5 | 419689 | 58 | 0.0034 | -0.3910 | 2 | 4 | 0 | 1 | - | - | - | - | z2w,3=1 | Sb2Yb1Zn2 |
| 164.1.6 | 69557 | 48 | 0.1460 | -0.1500 | 2 | 5 | 2 | 2 | - | - | - | - | z4=3 | Sb2Te1 |
| 164.1.7 | 41385 | 22 | 0.1115 | -1.2711 | 2 | 5 | 2 | 2 | - | - | - | - | z4=3 | Pt1Te2 |
| 164.1.8 | 41373 | 22 | 0.1008 | -1.2893 | 2 | 5 | 2 | 2 | - | - | - | - | z4=3 | Pt1Te2 |
| 164.1.9 | 83642 | 22 | 0.1760 | -1.3184 | 2 | 5 | 2 | 2 | - | - | - | - | z4=3 | Pd1Te2 |
| 164.1.10 | 42555 | 22 | 0.2185 | -1.2534 | 2 | 5 | 2 | 2 | - | - | - | - | z4=3 | Pd1Te2 |
| 164.1.11 | 42554 | 22 | 0.2108 | -1.2728 | 2 | 5 | 2 | 2 | - | - | - | - | z4=3 | Pd1Te2 |
| 164.1.12 | 41387 | 22 | 0.2106 | -1.2595 | 2 | 5 | 2 | 2 | - | - | - | - | z4=3 | Pd1Te2 |
| 164.1.13 | 100271 | 58 | 0.0150 | -0.1917 | 2 | 2 | 2 | 2 | - | - | - | - | z4=1 | P2Yb1Zn2 |
| 164.1.14 | 43293 | 22 | 0.1344 | -1.4110 | 2 | 5 | 2 | 2 | - | - | - | - | z4=3 | Ni1Te2 |

Continued on next page

**TABLE XIV– continued**

| TQC | ICSD | NoE | DG | IndG | SubSG (w/ T) | cl | SubSG (w/o T) | cl | Max. Sub. SG | yes or no | Max. Sub. SG | yes or no | SIs (w/ T) | Stoich. Formula |
|---|---|---|---|---|---|---|---|---|---|---|---|---|---|---|
| 164.1.15 | 42559 | 22 | 0.1448 | -1.4167 | 2 | 5 | 2 | 2 | - | - | - | - | z4=3 | Ni1Te2 |
| 164.1.16 | 94478 | 46 | 0.0680 | 0.0542 | 2 | 5 | 2 | 2 | - | - | - | - | z4=3 | Nd2S2Te1 |
| 164.1.17 | 32514 | 46 | 0.0048 | 0.0048 | 2 | 5 | 2 | 2 | - | - | - | - | z4=3 | Nd2O3 |
| 164.1.18 | 25806 | 46 | 0.1130 | 0.0946 | 2 | 5 | 2 | 2 | - | - | - | - | z4=3 | Nd2O2Se1 |
| 164.1.19 | 32515 | 46 | 0.1168 | 0.1029 | 2 | 5 | 2 | 2 | - | - | - | - | z4=3 | Nd2O2S1 |
| 164.1.20 | 300245 | 46 | 0.1147 | 0.1025 | 2 | 5 | 2 | 2 | - | - | - | - | z4=3 | Nd2O2S1 |
| 164.1.21 | 380376 | 48 | 0.0000 | -0.1917 | 2 | 3 | 2 | 2 | - | - | - | - | z2w,3=1 z4=1 | Mn2Sb2Yb1 |
| 164.1.22 | 41790 | 34 | 0.0110 | -0.1670 | 2 | 3 | 2 | 2 | - | - | - | - | z2w,3=1 z4=1 | Mn2Sb2Sr1 |
| 164.1.23 | 416094 | 34 | 0.0128 | -0.1636 | 2 | 3 | 2 | 2 | - | - | - | - | z2w,3=1 z4=1 | Mn2Sb2Sr1 |
| 164.1.24 | 77670 | 34 | 0.0156 | -0.2675 | 2 | 3 | 2 | 2 | - | - | - | - | z2w,3=1 z4=1 | Mn2P2Sr1 |
| 164.1.25 | 77669 | 34 | 0.0055 | -0.2161 | 2 | 3 | 2 | 2 | - | - | - | - | z2w,3=1 z4=1 | Mn2P2Sr1 |
| 164.1.26 | 49019 | 34 | 0.0131 | -0.2243 | 2 | 3 | 2 | 2 | - | - | - | - | z2w,3=1 z4=1 | Mn2P2Sr1 |
| 164.1.27 | 412655 | 38 | 0.0145 | -0.0816 | 2 | 2 | 2 | 2 | - | - | - | - | z4=1 | Mg2Sb2Yb1 |
| 164.1.28 | 191298 | 26 | 0.0056 | -0.0182 | 2 | 3 | 2 | 2 | - | - | - | - | z2w,3=1 z4=1 | Li3Pr1Sb2 |
| 164.1.29 | 26018 | 94 | 0.0101 | -0.4888 | 2 | 3 | 2 | 2 | - | - | - | - | z2w,3=1 z4=1 | K3Na1O8S2 |
| 164.1.30 | 26014 | 94 | 0.0124 | -0.4199 | 2 | 3 | 2 | 2 | - | - | - | - | z2w,3=1 z4=1 | K3Na1O8S2 |
| 164.1.31 | 660273 | 84 | 0.0137 | -0.1924 | 2 | 4 | 0 | 1 | - | - | - | - | z2w,3=1 | In2S4Zn1 |
| 164.1.32 | 42668 | 84 | 0.0137 | -0.1924 | 2 | 4 | 0 | 1 | - | - | - | - | z2w,3=1 | In2S4Zn1 |
| 164.1.33 | 25809 | 60 | 0.0266 | 0.0179 | 2 | 7 | 0 | 1 | - | - | - | - | z4=2 | Ho2O2Se1 |
| 164.1.34 | 185808 | 100 | 0.0278 | -0.4511 | 2 | 1 | 0 | 1 | - | - | - | - | z2w,3=1 z4=2 | Hg1Pt2Se3 |
| 164.1.35 | 56153 | 24 | 0.0069 | -0.8854 | 2 | 5 | 2 | 2 | - | - | - | - | z4=3 | H2Cd1O2 |
| 164.1.36 | 41478 | 10 | 0.0900 | -0.8490 | 2 | 3 | 2 | 2 | - | - | - | - | z2w,3=1 z4=1 | H1Si1 |
| 164.1.37 | 42876 | 48 | 0.0625 | -0.1249 | 2 | 5 | 2 | 2 | - | - | - | - | z4=3 | Ge2Sb2Te5 |
| 164.1.38 | 59801 | 44 | 0.0239 | -1.3395 | 2 | 1 | 0 | 1 | - | - | - | - | z2w,3=1 z4=2 | Ge2I2La2 |
| 164.1.39 | 42875 | 66 | 0.1194 | 0.0653 | 2 | 6 | 2 | 2 | - | - | - | - | z2w,3=1 z4=3 | Ge1Sb4Te7 |
| 164.1.40 | 25808 | 54 | 0.0077 | -0.0082 | 2 | 3 | 2 | 2 | - | - | - | - | z2w,3=1 z4=1 | Gd2O2Se1 |
| 164.1.41 | 91111 | 54 | 0.0050 | -0.0006 | 2 | 3 | 2 | 2 | - | - | - | - | z2w,3=1 z4=1 | Gd2O2S1 |
| 164.1.42 | 159909 | 92 | 0.0101 | -0.0111 | 2 | 1 | 0 | 1 | - | - | - | - | z2w,3=1 z4=2 | Fe2K3Na1O8 |
| 164.1.43 | 155025 | 104 | 0.0034 | -0.4471 | 2 | 5 | 2 | 2 | - | - | - | - | z4=3 | Fe2In2Se5 |
| 164.1.44 | 25810 | 62 | 0.0021 | -0.0557 | 2 | 1 | 0 | 1 | - | - | - | - | z2w,3=1 z4=2 | Er2O2Se1 |
| 164.1.45 | 35585 | 36 | 0.0128 | -0.6593 | 2 | 2 | 2 | 2 | - | - | - | - | z4=1 | Cu2P2Zr1 |
| 164.1.46 | 407246 | 46 | 0.0132 | -0.5295 | 2 | 1 | 0 | 1 | - | - | - | - | z2w,3=1 z4=2 | Ce2I2Si2 |
| 164.1.47 | 78989 | 58 | 0.0143 | -0.2671 | 2 | 4 | 0 | 1 | - | - | - | - | z2w,3=1 | Cd2Sb2Yb1 |
| 164.1.48 | 419690 | 58 | 0.0024 | -0.2678 | 2 | 4 | 0 | 1 | - | - | - | - | z2w,3=1 | Cd2Sb2Yb1 |
| 164.1.49 | 248526 | 10 | 0.0140 | -1.2495 | 2 | 4 | 0 | 1 | - | - | - | - | z2w,3=1 | Ca1Si2 |
| 164.1.50 | 248525 | 10 | 0.0140 | -1.2370 | 2 | 4 | 0 | 1 | - | - | - | - | z2w,3=1 | Ca1Si2 |
| 164.1.51 | 248524 | 10 | 0.0142 | -1.2537 | 2 | 4 | 0 | 1 | - | - | - | - | z2w,3=1 | Ca1Si2 |
| 164.1.52 | 248523 | 10 | 0.0142 | -1.2514 | 2 | 4 | 0 | 1 | - | - | - | - | z2w,3=1 | Ca1Si2 |
| 164.1.53 | 248522 | 10 | 0.0142 | -1.3329 | 2 | 4 | 0 | 1 | - | - | - | - | z2w,3=1 | Ca1Si2 |

Continued on next page

**TABLE XIV– continued**

| TQC | ICSD | NoE | DG | IndG | SubSG (w/ T) | cl | SubSG (w/o T) | cl | Max. Sub. SG | yes or no | Max. Sub. SG | yes or no | SIs (w/ T) | Stoich. Formula |
|---|---|---|---|---|---|---|---|---|---|---|---|---|---|---|
| 164.1.54 | 248521 | 10 | 0.0142 | -1.3887 | 2 | 4 | 0 | 1 | - | - | - | - | z2w,3=1 | Ca1Si2 |
| 164.1.55 | 248520 | 10 | 0.0140 | -1.3802 | 2 | 6 | 2 | 2 | - | - | - | - | z2w,3=1<br>z4=3 | Ca1Si2 |
| 164.1.56 | 248519 | 10 | 0.0140 | -1.3731 | 2 | 6 | 2 | 2 | - | - | - | - | z2w,3=1<br>z4=3 | Ca1Si2 |
| 164.1.57 | 41789 | 26 | 0.0056 | -0.1653 | 2 | 3 | 2 | 2 | - | - | - | - | z2w,3=1<br>z4=1 | Ca1Mn2Sb2 |
| 164.1.58 | 416093 | 26 | 0.0073 | -0.1665 | 2 | 3 | 2 | 2 | - | - | - | - | z2w,3=1<br>z4=1 | Ca1Mn2Sb2 |
| 164.1.59 | 163778 | 26 | 0.0076 | -0.1500 | 2 | 3 | 2 | 2 | - | - | - | - | z2w,3=1<br>z4=1 | Ca1Mn2Sb2 |
| 164.1.60 | 49017 | 26 | 0.0122 | -0.2355 | 2 | 3 | 2 | 2 | - | - | - | - | z2w,3=1<br>z4=1 | Ca1Mn2P2 |
| 164.1.61 | 35512 | 64 | 0.0814 | 0.0724 | 2 | 5 | 2 | 2 | - | - | - | - | z4=3 | C6N6Pt1Rb2S6 |
| 164.1.62 | 35511 | 64 | 0.0806 | 0.0588 | 2 | 5 | 2 | 2 | - | - | - | - | z4=3 | C6K2N6Pt1S6 |
| 164.1.63 | 35510 | 56 | 0.0095 | -0.0382 | 2 | 5 | 2 | 2 | - | - | - | - | z4=3 | C6H8N8Pt1S6 |
| 164.1.64 | 5005 | 14 | 0.0352 | -0.3942 | 2 | 5 | 2 | 2 | - | - | - | - | z4=3 | C1V2 |
| 164.1.65 | 409555 | 14 | 0.3520 | -0.3158 | 2 | 5 | 2 | 2 | - | - | - | - | z4=3 | C1Ta2 |
| 164.1.66 | 95113 | 42 | 0.0348 | -1.2693 | 2 | 3 | 2 | 2 | - | - | - | - | z2w,3=1<br>z4=1 | C1Nb2S2 |
| 164.1.67 | 95111 | 42 | 0.0377 | -1.2681 | 2 | 3 | 2 | 2 | - | - | - | - | z2w,3=1<br>z4=1 | C1Nb2S2 |
| 164.1.68 | 95109 | 42 | 0.0346 | -1.2705 | 2 | 3 | 2 | 2 | - | - | - | - | z2w,3=1<br>z4=1 | C1Nb2S2 |
| 164.1.69 | 656705 | 42 | 0.0367 | -1.2662 | 2 | 3 | 2 | 2 | - | - | - | - | z2w,3=1<br>z4=1 | C1Nb2S2 |
| 164.1.70 | 43670 | 30 | 0.0960 | -0.5439 | 2 | 5 | 2 | 2 | - | - | - | - | z4=3 | C1Nb2 |
| 164.1.71 | 82264 | 52 | 0.0130 | -0.0563 | 2 | 6 | 2 | 2 | - | - | - | - | z2w,3=1<br>z4=3 | C1N2O2Pr2 |
| 164.1.72 | 82265 | 54 | 0.0139 | -0.0471 | 2 | 6 | 2 | 2 | - | - | - | - | z2w,3=1<br>z4=3 | C1N2Nd2O2 |
| 164.1.73 | 71625 | 60 | 0.0054 | -0.0335 | 2 | 1 | 0 | 1 | - | - | - | - | z2w,3=1<br>z4=2 | C1F2Ho2 |
| 164.1.74 | 33912 | 54 | 0.0046 | 0.0011 | 2 | 3 | 2 | 2 | - | - | - | - | z2w,3=1<br>z4=1 | C1F2Gd2 |
| 164.1.75 | 42753 | 66 | 0.0174 | -0.2023 | 2 | 3 | 2 | 2 | - | - | - | - | z2w,3=1<br>z4=1 | Bi4Pb1Te7 |
| 164.1.76 | 42707 | 66 | 0.0236 | -0.1775 | 2 | 3 | 2 | 2 | - | - | - | - | z2w,3=1<br>z4=1 | Bi4Pb1Te7 |
| 164.1.77 | 250249 | 66 | 0.1608 | 0.1268 | 2 | 3 | 2 | 2 | - | - | - | - | z2w,3=1<br>z4=1 | Bi4Pb1Te7 |
| 164.1.78 | 42891 | 66 | 0.0400 | -0.1640 | 2 | 3 | 2 | 2 | - | - | - | - | z2w,3=1<br>z4=1 | Bi4Ge1Te7 |
| 164.1.79 | 42708 | 48 | 0.0226 | -0.1596 | 2 | 4 | 0 | 1 | - | - | - | - | z2w,3=1 | Bi2Pb2Te5 |
| 164.1.80 | 30372 | 48 | 0.1138 | 0.1092 | 2 | 5 | 2 | 2 | - | - | - | - | z4=3 | Bi2Pb2Se5 |
| 164.1.81 | 659569 | 16 | 0.1714 | -0.4237 | 2 | 5 | 2 | 2 | - | - | - | - | z4=3 | Bi2Mg3 |
| 164.1.82 | 41791 | 26 | 0.0098 | -0.1563 | 2 | 3 | 2 | 2 | - | - | - | - | z2w,3=1<br>z4=1 | Bi2Ca1Mn2 |
| 164.1.83 | 30525 | 66 | 0.0897 | -0.0434 | 2 | 4 | 0 | 1 | - | - | - | - | z2w,3=1 | Bi1Te1 |
| 164.1.84 | 100654 | 66 | 0.1363 | 0.0343 | 2 | 1 | 0 | 1 | - | - | - | - | z2w,3=1<br>z4=2 | Bi1Te1 |
| 164.1.85 | 99681 | 170 | 0.0107 | -0.0539 | 2 | 1 | 0 | 1 | - | - | - | - | z2w,3=1<br>z4=2 | Ba6Cl2O12Ru3 |
| 164.1.86 | 152482 | 102 | 0.0110 | -0.1739 | 2 | 3 | 2 | 2 | - | - | - | - | z2w,3=1<br>z4=1 | Ba3Ca1O9Ru2 |
| 164.1.87 | 409958 | 18 | 0.1690 | -0.1601 | 2 | 3 | 2 | 2 | - | - | - | - | z2w,3=1<br>z4=1 | Ba1Sn2 |

Continued on next page

**TABLE XIV– continued**

| TQC | ICSD | NoE | DG | IndG | SubSG (w/ T) | cl | SubSG (w/o T) | cl | Max. Sub. SG | yes or no | Max. Sub. SG | yes or no | SIs (w/ T) | Stoich. Formula |
|---|---|---|---|---|---|---|---|---|---|---|---|---|---|---|
| 164.1.88 | 41107 | 68 | 0.0108 | -0.6074 | 2 | 6 | 2 | 2 | - | - | - | - | z2w,3=1 z4=3 | As4Ge1Te7 |
| 164.1.89 | 88231 | 58 | 0.0158 | -0.1833 | 2 | 4 | 0 | 1 | - | - | - | - | z2w,3=1 | As2Yb1Zn2 |
| 164.1.90 | 39512 | 36 | 0.0108 | -0.2481 | 2 | 4 | 0 | 1 | - | - | - | - | z2w,3=1 | As2Mn2Zn1 |
| 164.1.91 | 49020 | 34 | 0.0086 | -0.1993 | 2 | 3 | 2 | 2 | - | - | - | - | z2w,3=1 z4=1 | As2Mn2Sr1 |
| 164.1.92 | 41793 | 34 | 0.0076 | -0.2030 | 2 | 3 | 2 | 2 | - | - | - | - | z2w,3=1 z4=1 | As2Mn2Sr1 |
| 164.1.93 | 63174 | 76 | 0.0144 | -0.7926 | 2 | 5 | 2 | 2 | - | - | - | - | z4=3 | As2Ge5Te8 |
| 164.1.94 | 41792 | 26 | 0.0054 | -0.2105 | 2 | 3 | 2 | 2 | - | - | - | - | z2w,3=1 z4=1 | As2Ca1Mn2 |
| 164.1.95 | 411054 | 38 | 0.0258 | -0.4950 | 2 | 2 | 2 | 2 | - | - | - | - | z4=1 | Al2Si2Yb1 |
| 164.1.96 | 411056 | 38 | 0.0248 | -0.4677 | 2 | 2 | 2 | 2 | - | - | - | - | z4=1 | Al2Ge2Yb1 |
| 164.1.97 | 409458 | 38 | 0.0248 | -0.4699 | 2 | 2 | 2 | 2 | - | - | - | - | z4=1 | Al2Ge2Yb1 |
| 164.1.98 | 412121 | 28 | 0.0048 | -0.0277 | 2 | 2 | 2 | 2 | - | - | - | - | z4=1 | Al2Ge2Nd1 |
| 164.1.99 | 160311 | 36 | 0.0053 | -0.4897 | 2 | 1 | 0 | 1 | - | - | - | - | z2w,3=1 z4=2 | Al2Er1Ge2 |
| 164.1.100 | 50976 | 26 | 0.0227 | -0.3786 | 2 | 2 | 2 | 2 | - | - | - | - | z4=1 | Al2Ce1Ge2 |
| 164.1.101 | 94502 | 44 | 0.0033 | -1.3922 | 2 | 5 | 2 | 2 | - | - | - | - | z4=3 | Al1La1Si2 |
| 164.1.102 | 94501 | 44 | 0.0028 | -1.3847 | 2 | 1 | 0 | 1 | - | - | - | - | z2w,3=1 z4=2 | Al1La1Si2 |
| 164.3.1 | 652385 | 16 | - | - | - | - | - | - | 12 | yes | 12 | yes | - | Si1Te2 |
| 164.3.2 | 12152 | 44 | - | - | - | - | - | - | 12 | yes | 12 | yes | - | Sb2Sr1Zn2 |
| 164.3.3 | 400048 | 64 | - | - | - | - | - | - | 12 | no(2) | 12 | no(2) | z4=3 | O2S1Tm2 |
| 164.3.4 | 24081 | 14 | - | - | - | - | - | - | 12 | yes | 12 | yes | - | O1Ti2 |
| 164.3.5 | 240197 | 52 | - | - | - | - | - | - | 12 | no(2) | 12 | yes | z2w,3=1 z4=2 | Ni3Sb2Zn1 |
| 164.3.6 | 27955 | 64 | - | - | - | - | - | - | 12 | no(2) | 12 | no(2) | z4=3 | Na2O4S1 |
| 164.3.7 | 416038 | 92 | - | - | - | - | - | - | 12 | no(2) | 12 | yes | z2w,3=1 | Na1O8Rb3Ru2 |
| 164.3.8 | 41942 | 44 | - | - | - | - | - | - | 12 | yes | 12 | yes | - | N2S1U2 |
| 164.3.9 | 41941 | 44 | - | - | - | - | - | - | 12 | yes | 12 | yes | - | N2S1U2 |
| 164.3.10 | 416040 | 92 | - | - | - | - | - | - | 12 | no(2) | 12 | yes | z2w,3=1 | K3Na1O8Ru2 |
| 164.3.11 | 107483 | 102 | - | - | - | - | - | - | 12 | yes | 12 | yes | - | Hg7Rb2 |
| 164.3.12 | 107482 | 102 | - | - | - | - | - | - | 12 | no(2) | 12 | no(2) | z4=3 | Hg7K2 |
| 164.3.13 | 429401 | 44 | - | - | - | - | - | - | 12 | yes | 12 | yes | - | H3Eu2I1 |
| 164.3.14 | 89829 | 14 | - | - | - | - | - | - | 12 | no(2) | 12 | no(2) | z4=3 | H2Mg1O2 |
| 164.3.15 | 89828 | 14 | - | - | - | - | - | - | 12 | no(2) | 12 | no(2) | z4=3 | H2Mg1O2 |
| 164.3.16 | 89827 | 14 | - | - | - | - | - | - | 12 | no(2) | 12 | no(2) | z4=3 | H2Mg1O2 |
| 164.3.17 | 89826 | 14 | - | - | - | - | - | - | 12 | no(2) | 12 | no(2) | z4=3 | H2Mg1O2 |
| 164.3.18 | 89825 | 14 | - | - | - | - | - | - | 12 | no(2) | 12 | yes | z2w,3=1 z4=2 | H2Mg1O2 |
| 164.3.19 | 89824 | 14 | - | - | - | - | - | - | 12 | no(2) | 12 | yes | z2w,3=1 z4=2 | H2Mg1O2 |
| 164.3.20 | 89823 | 14 | - | - | - | - | - | - | 12 | no(2) | 12 | yes | z2w,3=1 z4=2 | H2Mg1O2 |
| 164.3.21 | 31053 | 14 | - | - | - | - | - | - | 12 | no(2) | 12 | yes | z2w,3=1 z4=2 | H2Mg1O2 |
| 164.3.22 | 28275 | 14 | - | - | - | - | - | - | 12 | no(2) | 12 | yes | z2w,3=1 z4=2 | H2Mg1O2 |
| 164.3.23 | 91887 | 22 | - | - | - | - | - | - | 12 | yes | 12 | yes | - | H2Ca1O2 |
| 164.3.24 | 91886 | 22 | - | - | - | - | - | - | 12 | yes | 12 | yes | - | H2Ca1O2 |
| 164.3.25 | 91885 | 22 | - | - | - | - | - | - | 12 | yes | 12 | yes | - | H2Ca1O2 |
| 164.3.26 | 91884 | 22 | - | - | - | - | - | - | 12 | yes | 12 | yes | - | H2Ca1O2 |
| 164.3.27 | 91883 | 22 | - | - | - | - | - | - | 12 | yes | 12 | yes | - | H2Ca1O2 |
| 164.3.28 | 91882 | 22 | - | - | - | - | - | - | 12 | yes | 12 | yes | - | H2Ca1O2 |
| 164.3.29 | 56154 | 22 | - | - | - | - | - | - | 12 | yes | 12 | yes | - | H2Ca1O2 |
| 164.3.30 | 53989 | 22 | - | - | - | - | - | - | 12 | yes | 12 | yes | - | H2Ca1O2 |

Continued on next page

**TABLE XIV– continued**

| TQC | ICSD | NoE | DG | IndG | SubSG (w/ T) | cl | SubSG (w/o T) | cl | Max. Sub. SG | yes or no | Max. Sub. SG | yes or no | SIs (w/ T) | Stoich. Formula |
|---|---|---|---|---|---|---|---|---|---|---|---|---|---|---|
| 164.3.31 | 26075 | 60 | - | - | - | - | - | - | 12 | no(2) | 12 | no(2) | z4=1 | H2Al1Br3S1 |
| 164.3.32 | 59230 | 40 | - | - | - | - | - | - | 12 | yes | 12 | yes | - | Ga2Ni1S4 |
| 164.3.33 | 53287 | 36 | - | - | - | - | - | - | 12 | no(2) | 12 | no(2) | z4=3 | Cu2Hf1P2 |
| 164.3.34 | 38219 | 18 | - | - | - | - | - | - | 12 | yes | 12 | yes | - | Cl2Ti1 |
| 164.3.35 | 58976 | 42 | - | - | - | - | - | - | 12 | no(2) | 12 | no(2) | z4=1 | Cd2Gd1 |
| 164.3.36 | 58951 | 36 | - | - | - | - | - | - | 12 | no(2) | 12 | yes | z2w,3=1<br>z4=2 | Cd2Ce1 |
| 164.3.37 | 12150 | 36 | - | - | - | - | - | - | 12 | yes | 12 | yes | - | Ca1Sb2Zn2 |
| 164.3.38 | 38691 | 96 | - | - | - | - | - | - | 12 | no(2) | 12 | yes | z2w,3=1 | C6H6Cu2N8Pt1 |
| 164.3.39 | 280743 | 10 | - | - | - | - | - | - | 12 | no(2) | 12 | no(2) | z2w,3=1<br>z4=3 | C1Sc2 |
| 164.3.40 | 246280 | 104 | - | - | - | - | - | - | 12 | no(2) | 12 | yes | z2w,3=1<br>z4=2 | Ba3Ca1Ir2O9 |
| 164.3.41 | 423734 | 50 | - | - | - | - | - | - | 12 | no(2) | 12 | no(2) | z2w,3=1<br>z4=3 | Ba2Ni3 |
| 164.3.42 | 195973 | 164 | - | - | - | - | - | - | 12 | no(2) | 12 | yes | z4=2 | Ba1Cr1O2.8 |
| 164.3.43 | 100049 | 24 | - | - | - | - | - | - | 12 | no(2) | 12 | no(2) | z4=3 | Ba1Bi2Mg2 |
| 164.3.44 | 58134 | 58 | - | - | - | - | - | - | 12 | no(2) | 12 | yes | z2w,3=1 | Al3Pt2 |
| 164.3.45 | 41180 | 32 | - | - | - | - | - | - | 12 | no(2) | 12 | yes | z2w,3=1<br>z4=2 | Al2Gd1Si2 |
| 164.3.46 | 89553 | 34 | - | - | - | - | - | - | 12 | no(2) | 12 | yes | z2w,3=1<br>z4=2 | Al2Dy1Si2 |
| 164.3.47 | 82674 | 46 | - | - | - | - | - | - | 12 | yes | 12 | yes | - | Al1Ce1Si2 |
| 164.3.48 | 156301 | 46 | - | - | - | - | - | - | 12 | yes | 12 | yes | - | Ag1Te2Tm1 |
| 165.1.1 | 154809 | 84 | 0.0132 | 0.0101 | 2 | 1 | 2 | 1 | - | - | - | - | z4=3 | H3Y1 |
| 165.1.2 | 26776 | 228 | 0.0028 | -0.2735 | 2 | 1 | 2 | 1 | - | - | - | - | z4=3 | As1Cu3 |
| 165.3.1 | 172186 | 232 | - | - | - | - | - | - | 15 | no(2) | 15 | yes | z4=2 | Co4Nb2O9 |
| 165.4.1 | 165378 | 146 | - | - | - | - | - | - | See details in Table XIX | | | | | Cs2Mn1S6Te2 |
| 165.4.2 | 240159 | 66 | - | - | - | - | - | - | See details in Table XIX | | | | | Ca5Cu1Pb3 |
| 166.1.1 | 653496 | 252 | 0.0124 | -0.4052 | 2 | 2 | 2 | 1 | - | - | - | - | z2w,1=1<br>z2w,2=1<br>z2w,3=1<br>z4=1 | Yb2Zn17 |
| 166.1.2 | 653487 | 252 | 0.0126 | -0.4034 | 2 | 2 | 2 | 1 | - | - | - | - | z2w,1=1<br>z2w,2=1<br>z2w,3=1<br>z4=1 | Yb2Zn17 |
| 166.1.3 | 104207 | 252 | 0.0131 | -0.4055 | 2 | 2 | 2 | 1 | - | - | - | - | z2w,1=1<br>z2w,2=1<br>z2w,3=1<br>z4=1 | Yb2Zn17 |
| 166.1.4 | 653467 | 226 | 0.0000 | -0.7387 | 2 | 6 | 2 | 1 | - | - | - | - | z4=3 | Y2Zn17 |
| 166.1.5 | 104239 | 226 | 0.0000 | -0.7378 | 2 | 6 | 2 | 1 | - | - | - | - | z4=3 | Y2Zn17 |
| 166.1.6 | 653389 | 232 | 0.0344 | -0.0885 | 2 | 6 | 2 | 1 | - | - | - | - | z4=3 | U2Zn17 |
| 166.1.7 | 653384 | 232 | 0.0118 | -0.0859 | 2 | 6 | 2 | 1 | - | - | - | - | z4=3 | U2Zn17 |
| 166.1.8 | 106208 | 232 | 0.0055 | -0.0888 | 2 | 6 | 2 | 1 | - | - | - | - | z4=3 | U2Zn17 |
| 166.1.9 | 106207 | 232 | 0.0114 | -0.0780 | 2 | 6 | 2 | 1 | - | - | - | - | z4=3 | U2Zn17 |
| 166.1.10 | 104248 | 232 | 0.0122 | -0.0869 | 2 | 6 | 2 | 1 | - | - | - | - | z4=3 | U2Zn17 |
| 166.1.11 | 653358 | 250 | 0.0086 | -0.1739 | 2 | 7 | 0 | 2 | - | - | - | - | z2w,1=1<br>z2w,2=1<br>z2w,3=1 | Tm2Zn17 |
| 166.1.12 | 104246 | 250 | 0.0085 | -0.1791 | 2 | 7 | 0 | 2 | - | - | - | - | z2w,1=1<br>z2w,2=1<br>z2w,3=1 | Tm2Zn17 |

Continued on next page

**TABLE XIV– continued**

| TQC | ICSD | NoE | DG | IndG | SubSG (w/ T) | cl | SubSG (w/o T) | cl | Max. Sub. SG | yes or no | Max. Sub. SG | yes or no | SIs (w/ T) | Stoich. Formula |
|---|---|---|---|---|---|---|---|---|---|---|---|---|---|---|
| 166.1.13 | 653216 | 54 | 0.0416 | -0.5926 | 2 | 3 | 0 | 2 | - | - | - | - | z2w,1=1<br>z2w,2=1<br>z2w,3=1<br>z4=2 | Te1Zr3 |
| 166.1.14 | 52499 | 12 | 0.2915 | -2.1560 | 2 | 6 | 2 | 1 | - | - | - | - | z4=3 | Te1 |
| 166.1.15 | 410956 | 88 | 0.0302 | -0.9466 | 2 | 1 | 2 | 1 | - | - | - | - | z2w,1=1<br>z2w,2=1<br>z2w,3=1<br>z4=3 | Sn3Sr1 |
| 166.1.16 | 652055 | 40 | 0.0196 | -1.1235 | 2 | 6 | 2 | 1 | - | - | - | - | z4=3 | Se1Ti1 |
| 166.1.17 | 152188 | 58 | 0.0495 | -0.2430 | 2 | 1 | 2 | 1 | - | - | - | - | z2w,1=1<br>z2w,2=1<br>z2w,3=1<br>z4=3 | Sb8Te3 |
| 166.1.18 | 152189 | 98 | 0.0742 | 0.0742 | 2 | 1 | 2 | 1 | - | - | - | - | z2w,1=1<br>z2w,2=1<br>z2w,3=1<br>z4=3 | Sb48Te9 |
| 166.1.19 | 291530 | 28 | 0.2987 | 0.2987 | 2 | 6 | 2 | 1 | - | - | - | - | z4=3 | Sb3Te4.5 |
| 166.1.20 | 651728 | 50 | 0.0126 | -0.0934 | 2 | 5 | 0 | 2 | - | - | - | - | z4=2 | Sb2V3 |
| 166.1.21 | 41814 | 50 | 0.0167 | -0.2798 | 2 | 5 | 0 | 2 | - | - | - | - | z4=2 | Sb2V3 |
| 166.1.22 | 651638 | 28 | 0.2280 | 0.2141 | 2 | 6 | 2 | 1 | - | - | - | - | z4=3 | Sb2Te3 |
| 166.1.23 | 651635 | 28 | 0.2167 | 0.2121 | 2 | 6 | 2 | 1 | - | - | - | - | z4=3 | Sb2Te3 |
| 166.1.24 | 651629 | 28 | 0.2691 | 0.2572 | 2 | 6 | 2 | 1 | - | - | - | - | z4=3 | Sb2Te3 |
| 166.1.25 | 5420 | 28 | 0.2291 | 0.2127 | 2 | 6 | 2 | 1 | - | - | - | - | z4=3 | Sb2Te3 |
| 166.1.26 | 262171 | 28 | 0.2212 | 0.2205 | 2 | 6 | 2 | 1 | - | - | - | - | z4=3 | Sb2Te3 |
| 166.1.27 | 2084 | 28 | 0.2335 | 0.2159 | 2 | 6 | 2 | 1 | - | - | - | - | z4=3 | Sb2Te3 |
| 166.1.28 | 20236 | 28 | 0.2703 | 0.2628 | 2 | 6 | 2 | 1 | - | - | - | - | z4=3 | Sb2Te3 |
| 166.1.29 | 193347 | 28 | 0.2460 | 0.2287 | 2 | 6 | 2 | 1 | - | - | - | - | z4=3 | Sb2Te3 |
| 166.1.30 | 193346 | 28 | 0.2460 | 0.2309 | 2 | 6 | 2 | 1 | - | - | - | - | z4=3 | Sb2Te3 |
| 166.1.31 | 193345 | 28 | 0.2460 | 0.2304 | 2 | 6 | 2 | 1 | - | - | - | - | z4=3 | Sb2Te3 |
| 166.1.32 | 193344 | 28 | 0.2481 | 0.2332 | 2 | 6 | 2 | 1 | - | - | - | - | z4=3 | Sb2Te3 |
| 166.1.33 | 193343 | 28 | 0.2483 | 0.2348 | 2 | 6 | 2 | 1 | - | - | - | - | z4=3 | Sb2Te3 |
| 166.1.34 | 193342 | 28 | 0.2467 | 0.2329 | 2 | 6 | 2 | 1 | - | - | - | - | z4=3 | Sb2Te3 |
| 166.1.35 | 193341 | 28 | 0.2467 | 0.2339 | 2 | 6 | 2 | 1 | - | - | - | - | z4=3 | Sb2Te3 |
| 166.1.36 | 192780 | 28 | 0.2463 | 0.2284 | 2 | 6 | 2 | 1 | - | - | - | - | z4=3 | Sb2Te3 |
| 166.1.37 | 187495 | 28 | 0.0332 | -0.5970 | 2 | 7 | 0 | 2 | - | - | - | - | z2w,1=1<br>z2w,2=1<br>z2w,3=1 | Sb2Te3 |
| 166.1.38 | 185952 | 28 | 0.1807 | 0.1798 | 2 | 6 | 2 | 1 | - | - | - | - | z4=3 | Sb2Te3 |
| 166.1.39 | 185951 | 28 | 0.0299 | 0.0299 | 2 | 6 | 2 | 1 | - | - | - | - | z4=3 | Sb2Te3 |
| 166.1.40 | 185950 | 28 | 0.0795 | 0.0795 | 2 | 6 | 2 | 1 | - | - | - | - | z4=3 | Sb2Te3 |
| 166.1.41 | 185949 | 28 | 0.0958 | 0.0958 | 2 | 6 | 2 | 1 | - | - | - | - | z4=3 | Sb2Te3 |
| 166.1.42 | 185948 | 28 | 0.0695 | 0.0694 | 2 | 6 | 2 | 1 | - | - | - | - | z4=3 | Sb2Te3 |
| 166.1.43 | 185947 | 28 | 0.1574 | 0.1572 | 2 | 6 | 2 | 1 | - | - | - | - | z4=3 | Sb2Te3 |
| 166.1.44 | 185946 | 28 | 0.1859 | 0.1748 | 2 | 6 | 2 | 1 | - | - | - | - | z4=3 | Sb2Te3 |
| 166.1.45 | 185945 | 28 | 0.2361 | 0.2181 | 2 | 6 | 2 | 1 | - | - | - | - | z4=3 | Sb2Te3 |
| 166.1.46 | 657316 | 38 | 0.2263 | 0.0135 | 2 | 2 | 2 | 1 | - | - | - | - | z2w,1=1<br>z2w,2=1<br>z2w,3=1<br>z4=1 | Sb2Sn1Te4 |
| 166.1.47 | 30392 | 38 | 0.1759 | -0.2160 | 2 | 2 | 2 | 1 | - | - | - | - | z2w,1=1<br>z2w,2=1<br>z2w,3=1<br>z4=1 | Sb2Sn1Te4 |
| 166.1.48 | 651529 | 28 | 0.0644 | 0.0644 | 2 | 6 | 2 | 1 | - | - | - | - | z4=3 | Sb2Se2Te1 |
| 166.1.49 | 52295 | 28 | 0.0797 | 0.0797 | 2 | 6 | 2 | 1 | - | - | - | - | z4=3 | Sb2Se2Te1 |

Continued on next page

**TABLE XIV– continued**

| TQC | ICSD | NoE | DG | IndG | SubSG (w/ T) | cl | SubSG (w/o T) | cl | Max. Sub. SG | yes or no | Max. Sub. SG | yes or no | SIs (w/ T) | Stoich. Formula |
|---|---|---|---|---|---|---|---|---|---|---|---|---|---|---|
| 166.1.50 | 87416 | 28 | 0.2458 | -0.1862 | 2 | 6 | 2 | 1 | - | - | - | - | z4=3 | Sb2Se1Te2 |
| 166.1.51 | 2085 | 28 | 0.3352 | 0.3352 | 2 | 6 | 2 | 1 | - | - | - | - | z4=3 | Sb2Se1Te2 |
| 166.1.52 | 651646 | 20 | 0.1588 | 0.0951 | 2 | 4 | 2 | 1 | - | - | - | - | z4=1 | Sb1Te2Tl1 |
| 166.1.53 | 15411 | 20 | 0.1542 | 0.0326 | 2 | 4 | 2 | 1 | - | - | - | - | z4=1 | Sb1Te2Tl1 |
| 166.1.54 | 651506 | 10 | 0.1044 | 0.1044 | 2 | 1 | 2 | 1 | - | - | - | - | z2w,1=1<br>z2w,2=1<br>z2w,3=1<br>z4=3 | Sb1 |
| 166.1.55 | 651498 | 10 | 0.1158 | 0.1158 | 2 | 1 | 2 | 1 | - | - | - | - | z2w,1=1<br>z2w,2=1<br>z2w,3=1<br>z4=3 | Sb1 |
| 166.1.56 | 651497 | 10 | 0.0827 | 0.0827 | 2 | 1 | 2 | 1 | - | - | - | - | z2w,1=1<br>z2w,2=1<br>z2w,3=1<br>z4=3 | Sb1 |
| 166.1.57 | 651491 | 10 | 0.2949 | -0.0021 | 2 | 1 | 2 | 1 | - | - | - | - | z2w,1=1<br>z2w,2=1<br>z2w,3=1<br>z4=3 | Sb1 |
| 166.1.58 | 651490 | 10 | 0.3039 | 0.0592 | 2 | 1 | 2 | 1 | - | - | - | - | z2w,1=1<br>z2w,2=1<br>z2w,3=1<br>z4=3 | Sb1 |
| 166.1.59 | 651489 | 10 | 0.2780 | -0.1268 | 2 | 1 | 2 | 1 | - | - | - | - | z2w,1=1<br>z2w,2=1<br>z2w,3=1<br>z4=3 | Sb1 |
| 166.1.60 | 651488 | 10 | 0.2913 | -0.0816 | 2 | 1 | 2 | 1 | - | - | - | - | z2w,1=1<br>z2w,2=1<br>z2w,3=1<br>z4=3 | Sb1 |
| 166.1.61 | 64697 | 10 | 0.0889 | 0.0889 | 2 | 1 | 2 | 1 | - | - | - | - | z2w,1=1<br>z2w,2=1<br>z2w,3=1<br>z4=3 | Sb1 |
| 166.1.62 | 64696 | 10 | 0.0896 | 0.0896 | 2 | 1 | 2 | 1 | - | - | - | - | z2w,1=1<br>z2w,2=1<br>z2w,3=1<br>z4=3 | Sb1 |
| 166.1.63 | 64695 | 10 | 0.0827 | 0.0827 | 2 | 1 | 2 | 1 | - | - | - | - | z2w,1=1<br>z2w,2=1<br>z2w,3=1<br>z4=3 | Sb1 |
| 166.1.64 | 55402 | 10 | 0.0828 | 0.0828 | 2 | 2 | 2 | 1 | - | - | - | - | z2w,1=1<br>z2w,2=1<br>z2w,3=1<br>z4=1 | Sb1 |
| 166.1.65 | 53795 | 10 | 0.1378 | 0.1378 | 2 | 1 | 2 | 1 | - | - | - | - | z2w,1=1<br>z2w,2=1<br>z2w,3=1<br>z4=3 | Sb1 |
| 166.1.66 | 409754 | 10 | 0.2030 | 0.2030 | 2 | 1 | 2 | 1 | - | - | - | - | z2w,1=1<br>z2w,2=1<br>z2w,3=1<br>z4=3 | Sb1 |

Continued on next page

**TABLE XIV– continued**

| TQC | ICSD | NoE | DG | IndG | SubSG (w/ T) | cl | SubSG (w/o T) | cl | Max. Sub. SG | yes or no | Max. Sub. SG | yes or no | SIs (w/ T) | Stoich. Formula |
|---|---|---|---|---|---|---|---|---|---|---|---|---|---|---|
| 166.1.67 | 161493 | 10 | 0.0867 | 0.0867 | 2 | 2 | 2 | 1 | - | - | - | - | z2w,1=1<br>z2w,2=1<br>z2w,3=1<br>z4=1 | Sb1 |
| 166.1.68 | 24757 | 34 | 0.0080 | -1.7358 | 2 | 6 | 2 | 1 | - | - | - | - | z4=3 | S2Ta1 |
| 166.1.69 | 651179 | 30 | 0.0706 | -1.3124 | 2 | 1 | 2 | 1 | - | - | - | - | z2w,1=1<br>z2w,2=1<br>z2w,3=1<br>z4=3 | S1Ti1 |
| 166.1.70 | 25561 | 30 | 0.0719 | -1.2996 | 2 | 1 | 2 | 1 | - | - | - | - | z2w,1=1<br>z2w,2=1<br>z2w,3=1<br>z4=3 | S1Ti1 |
| 166.1.71 | 184500 | 30 | 0.0698 | -1.2928 | 2 | 1 | 2 | 1 | - | - | - | - | z2w,1=1<br>z2w,2=1<br>z2w,3=1<br>z4=3 | S1Ti1 |
| 166.1.72 | 650000 | 236 | 0.0162 | -0.1716 | 2 | 5 | 0 | 2 | - | - | - | - | z4=2 | Pu2Zn17 |
| 166.1.73 | 41372 | 54 | 0.0966 | -1.9366 | 2 | 7 | 0 | 2 | - | - | - | - | z2w,1=1<br>z2w,2=1<br>z2w,3=1 | Pt3Te4 |
| 166.1.74 | 41371 | 76 | 0.0387 | -1.5958 | 2 | 2 | 2 | 1 | - | - | - | - | z2w,1=1<br>z2w,2=1<br>z2w,3=1<br>z4=1 | Pt2Te3 |
| 166.1.75 | 41370 | 32 | 0.1313 | -2.7130 | 2 | 2 | 2 | 1 | - | - | - | - | z2w,1=1<br>z2w,2=1<br>z2w,3=1<br>z4=1 | Pt1Te1 |
| 166.1.76 | 649479 | 230 | 0.0059 | -0.0963 | 2 | 3 | 0 | 2 | - | - | - | - | z2w,1=1<br>z2w,2=1<br>z2w,3=1<br>z4=2 | Pr2Zn17 |
| 166.1.77 | 649469 | 230 | 0.0058 | -0.0944 | 2 | 3 | 0 | 2 | - | - | - | - | z2w,1=1<br>z2w,2=1<br>z2w,3=1<br>z4=2 | Pr2Zn17 |
| 166.1.78 | 104241 | 230 | 0.0061 | -0.0939 | 2 | 3 | 0 | 2 | - | - | - | - | z2w,1=1<br>z2w,2=1<br>z2w,3=1<br>z4=2 | Pr2Zn17 |
| 166.1.79 | 54338 | 104 | 0.0000 | -0.1121 | 2 | 4 | 2 | 1 | - | - | - | - | z4=1 | Pr1 |
| 166.1.80 | 648763 | 48 | 0.2043 | -0.5233 | 2 | 3 | 0 | 2 | - | - | - | - | z2w,1=1<br>z2w,2=1<br>z2w,3=1<br>z4=2 | Pd3S2Tl2 |
| 166.1.81 | 648762 | 48 | 0.1095 | -0.5146 | 2 | 7 | 0 | 2 | - | - | - | - | z2w,1=1<br>z2w,2=1<br>z2w,3=1 | Pd3S2Tl2 |
| 166.1.82 | 600019 | 10 | 0.0287 | -0.8713 | 2 | 1 | 2 | 1 | - | - | - | - | z2w,1=1<br>z2w,2=1<br>z2w,3=1<br>z4=3 | P1 |
| 166.1.83 | 53301 | 10 | 0.0638 | -1.2231 | 2 | 5 | 148 | 2 | - | - | - | - | z4=2 | P1 |
| 166.1.84 | 416476 | 72 | 0.0024 | -0.0068 | 2 | 6 | 2 | 1 | - | - | - | - | z4=3 | O1P1Pr1Zn1 |

Continued on next page

**TABLE XIV– continued**

| TQC | ICSD | NoE | DG | IndG | SubSG (w/ T) | cl | SubSG (w/o T) | cl | Max. Sub. SG | yes or no | Max. Sub. SG | yes or no | SIs (w/ T) | Stoich. Formula |
|---|---|---|---|---|---|---|---|---|---|---|---|---|---|---|
| 166.1.85 | 647123 | 236 | 0.0116 | -0.1544 | 2 | 3 | 0 | 2 | - | - | - | - | z2w,1=1 z2w,2=1 z2w,3=1 z4=2 | Ni7Yb2 |
| 166.1.86 | 646928 | 188 | 0.0079 | -0.1896 | 2 | 4 | 2 | 1 | - | - | - | - | z4=1 | Ni7Th2 |
| 166.1.87 | 646978 | 110 | 0.0538 | -0.4912 | 2 | 3 | 0 | 2 | - | - | - | - | z2w,1=1 z2w,2=1 z2w,3=1 z4=2 | Ni3Ti0.67Zr0.33 |
| 166.1.88 | 646399 | 48 | 0.2414 | -0.2844 | 2 | 3 | 0 | 2 | - | - | - | - | z2w,1=1 z2w,2=1 z2w,3=1 z4=2 | Ni3S2Tl2 |
| 166.1.89 | 646398 | 48 | 0.0763 | -0.2828 | 2 | 7 | 0 | 2 | - | - | - | - | z2w,1=1 z2w,2=1 z2w,3=1 | Ni3S2Tl2 |
| 166.1.90 | 422328 | 48 | 0.0915 | -0.2706 | 2 | 3 | 0 | 2 | - | - | - | - | z2w,1=1 z2w,2=1 z2w,3=1 z4=2 | Ni3S2Tl2 |
| 166.1.91 | 108893 | 48 | 0.2406 | -0.2853 | 2 | 7 | 0 | 2 | - | - | - | - | z2w,1=1 z2w,2=1 z2w,3=1 | Ni3S2Tl2 |
| 166.1.92 | 654948 | 50 | 0.0163 | -0.4957 | 2 | 3 | 0 | 2 | - | - | - | - | z2w,1=1 z2w,2=1 z2w,3=1 z4=2 | Ni3S2Sn2 |
| 166.1.93 | 646380 | 50 | 0.0596 | -0.5023 | 2 | 7 | 0 | 2 | - | - | - | - | z2w,1=1 z2w,2=1 z2w,3=1 | Ni3S2Sn2 |
| 166.1.94 | 646379 | 50 | 0.0175 | -0.5011 | 2 | 3 | 0 | 2 | - | - | - | - | z2w,1=1 z2w,2=1 z2w,3=1 z4=2 | Ni3S2Sn2 |
| 166.1.95 | 402458 | 50 | 0.0170 | -0.4995 | 2 | 7 | 0 | 2 | - | - | - | - | z2w,1=1 z2w,2=1 z2w,3=1 | Ni3S2Sn2 |
| 166.1.96 | 100217 | 50 | 0.0394 | -0.5294 | 2 | 3 | 0 | 2 | - | - | - | - | z2w,1=1 z2w,2=1 z2w,3=1 z4=2 | Ni3S2Sn2 |
| 166.1.97 | 646211 | 50 | 0.1123 | -0.2424 | 2 | 3 | 0 | 2 | - | - | - | - | z2w,1=1 z2w,2=1 z2w,3=1 z4=2 | Ni3Pb2Se2 |
| 166.1.98 | 646210 | 50 | 0.0457 | -0.2464 | 2 | 3 | 0 | 2 | - | - | - | - | z2w,1=1 z2w,2=1 z2w,3=1 z4=2 | Ni3Pb2Se2 |
| 166.1.99 | 402453 | 50 | 0.0453 | -0.2456 | 2 | 3 | 0 | 2 | - | - | - | - | z2w,1=1 z2w,2=1 z2w,3=1 z4=2 | Ni3Pb2Se2 |
| 166.1.100 | 108892 | 50 | 0.1037 | -0.2712 | 2 | 7 | 0 | 2 | - | - | - | - | z2w,1=1 z2w,2=1 z2w,3=1 | Ni3Pb2Se2 |

Continued on next page

**TABLE XIV– continued**

| TQC | ICSD | NoE | DG | IndG | SubSG (w/ T) | cl | SubSG (w/o T) | cl | Max. Sub. SG | yes or no | Max. Sub. SG | yes or no | SIs (w/ T) | Stoich. Formula |
|---|---|---|---|---|---|---|---|---|---|---|---|---|---|---|
| 166.1.101 | 646207 | 50 | 0.1946 | -0.0939 | 2 | 3 | 0 | 2 | - | - | - | - | z2w,1=1<br>z2w,2=1<br>z2w,3=1<br>z4=2 | Ni3Pb2S2 |
| 166.1.102 | 646206 | 50 | 0.0462 | -0.1397 | 2 | 3 | 0 | 2 | - | - | - | - | z2w,1=1<br>z2w,2=1<br>z2w,3=1<br>z4=2 | Ni3Pb2S2 |
| 166.1.103 | 417632 | 50 | 0.0454 | -0.1414 | 2 | 3 | 0 | 2 | - | - | - | - | z2w,1=1<br>z2w,2=1<br>z2w,3=1<br>z4=2 | Ni3Pb2S2 |
| 166.1.104 | 24176 | 50 | 0.1051 | -0.2099 | 2 | 3 | 0 | 2 | - | - | - | - | z2w,1=1<br>z2w,2=1<br>z2w,3=1<br>z4=2 | Ni3Pb2S2 |
| 166.1.105 | 108895 | 50 | 0.0972 | -0.2004 | 2 | 7 | 0 | 2 | - | - | - | - | z2w,1=1<br>z2w,2=1<br>z2w,3=1 | Ni3Pb2S2 |
| 166.1.106 | 646072 | 232 | 0.0025 | -0.0438 | 2 | 6 | 2 | 1 | - | - | - | - | z4=3 | Nd2Zn17 |
| 166.1.107 | 601895 | 232 | 0.0032 | -0.0453 | 2 | 6 | 2 | 1 | - | - | - | - | z4=3 | Nd2Zn17 |
| 166.1.108 | 104242 | 232 | 0.0043 | -0.0433 | 2 | 6 | 2 | 1 | - | - | - | - | z4=3 | Nd2Zn17 |
| 166.1.109 | 645562 | 162 | 0.0154 | -0.5255 | 2 | 1 | 2 | 1 | - | - | - | - | z2w,1=1<br>z2w,2=1<br>z2w,3=1<br>z4=3 | Nb6Zn7 |
| 166.1.110 | 30364 | 16 | 0.0021 | -0.1102 | 2 | 6 | 2 | 1 | - | - | - | - | z4=3 | N2W1 |
| 166.1.111 | 44926 | 38 | 0.0067 | -0.0910 | 2 | 2 | 2 | 1 | - | - | - | - | z2w,1=1<br>z2w,2=1<br>z2w,3=1<br>z4=1 | Mg2Ni3Si1 |
| 166.1.112 | 55617 | 108 | 0.0000 | -0.2394 | 2 | 2 | 2 | 1 | - | - | - | - | z2w,1=1<br>z2w,2=1<br>z2w,3=1<br>z4=1 | Mg2Nd1Ni9 |
| 166.1.113 | 41369 | 20 | 0.2273 | -1.7604 | 2 | 1 | 2 | 1 | - | - | - | - | z2w,1=1<br>z2w,2=1<br>z2w,3=1<br>z4=3 | Li8Pb3 |
| 166.1.114 | 191137 | 146 | 0.0077 | -0.0139 | 2 | 1 | 2 | 1 | - | - | - | - | z2w,1=1<br>z2w,2=1<br>z2w,3=1<br>z4=3 | La3Mn2O14Sb3 |
| 166.1.115 | 641738 | 114 | 0.0000 | -0.6512 | 2 | 3 | 0 | 2 | - | - | - | - | z2w,1=1<br>z2w,2=1<br>z2w,3=1<br>z4=2 | La1Rh3 |
| 166.1.116 | 641724 | 114 | 0.0026 | -0.6533 | 2 | 3 | 0 | 2 | - | - | - | - | z2w,1=1<br>z2w,2=1<br>z2w,3=1<br>z4=2 | La1Rh3 |
| 166.1.117 | 640862 | 182 | 0.0024 | -0.0996 | 2 | 6 | 2 | 1 | - | - | - | - | z4=3 | Ir7Nd2 |
| 166.1.118 | 657607 | 24 | 0.1739 | -1.0989 | 2 | 2 | 2 | 1 | - | - | - | - | z2w,1=1<br>z2w,2=1<br>z2w,3=1<br>z4=1 | In2Te3 |

Continued on next page

**TABLE XIV– continued**

| TQC | ICSD | NoE | DG | IndG | SubSG (w/ T) | cl | SubSG (w/o T) | cl | Max. Sub. SG | yes or no | Max. Sub. SG | yes or no | SIs (w/ T) | Stoich. Formula |
|---|---|---|---|---|---|---|---|---|---|---|---|---|---|---|
| 166.1.119 | 640609 | 24 | 0.1211 | -0.7799 | 2 | 2 | 2 | 1 | - | - | - | - | z2w,1=1 z2w,2=1 z2w,3=1 z4=1 | In2Te3 |
| 166.1.120 | 640141 | 48 | 0.2552 | -0.3175 | 2 | 3 | 0 | 2 | - | - | - | - | z2w,1=1 z2w,2=1 z2w,3=1 z4=2 | In2Ni3Se2 |
| 166.1.121 | 640140 | 48 | 0.2831 | -0.4190 | 2 | 7 | 0 | 2 | - | - | - | - | z2w,1=1 z2w,2=1 z2w,3=1 | In2Ni3Se2 |
| 166.1.122 | 640135 | 48 | 0.3473 | -0.2473 | 2 | 3 | 0 | 2 | - | - | - | - | z2w,1=1 z2w,2=1 z2w,3=1 z4=2 | In2Ni3S2 |
| 166.1.123 | 640134 | 48 | 0.3458 | -0.3607 | 2 | 7 | 0 | 2 | - | - | - | - | z2w,1=1 z2w,2=1 z2w,3=1 | In2Ni3S2 |
| 166.1.124 | 415258 | 48 | 0.3438 | -0.3612 | 2 | 3 | 0 | 2 | - | - | - | - | z2w,1=1 z2w,2=1 z2w,3=1 z4=2 | In2Ni3S2 |
| 166.1.125 | 108894 | 48 | 0.2098 | -0.3212 | 2 | 7 | 0 | 2 | - | - | - | - | z2w,1=1 z2w,2=1 z2w,3=1 | In2Ni3S2 |
| 166.1.126 | 172602 | 32 | 0.0259 | -0.4097 | 2 | 4 | 2 | 1 | - | - | - | - | z4=1 | I1N1Sr2 |
| 166.1.127 | 654919 | 12 | 0.0252 | -1.7921 | 2 | 6 | 2 | 1 | - | - | - | - | z4=3 | Hg1 |
| 166.1.128 | 53771 | 12 | 0.1928 | -1.7604 | 2 | 7 | 2 | 2 | - | - | - | - | z2w,1=1 z2w,2=1 z2w,3=1 | Hg1 |
| 166.1.129 | 174006 | 12 | 0.0463 | -1.7925 | 2 | 6 | 2 | 1 | - | - | - | - | z4=3 | Hg1 |
| 166.1.130 | 174005 | 12 | 0.0034 | -1.8230 | 2 | 6 | 2 | 1 | - | - | - | - | z4=3 | Hg1 |
| 166.1.131 | 104296 | 12 | 0.0512 | -1.7780 | 2 | 6 | 2 | 1 | - | - | - | - | z4=3 | Hg1 |
| 166.1.132 | 638697 | 102 | 0.1425 | -0.6109 | 2 | 3 | 0 | 2 | - | - | - | - | z2w,1=1 z2w,2=1 z2w,3=1 z4=2 | Hf1Ni3 |
| 166.1.133 | 2414 | 102 | 0.1458 | -0.6061 | 2 | 3 | 0 | 2 | - | - | - | - | z2w,1=1 z2w,2=1 z2w,3=1 z4=2 | Hf1Ni3 |
| 166.1.134 | 171105 | 126 | 0.0216 | -0.1637 | 2 | 1 | 2 | 1 | - | - | - | - | z2w,1=1 z2w,2=1 z2w,3=1 z4=3 | H9Fe3O15S2 |
| 166.1.135 | 15888 | 74 | 0.0017 | -0.0660 | 2 | 5 | 0 | 2 | - | - | - | - | z4=2 | H8N2O8Pb1S2 |
| 166.1.136 | 195640 | 28 | 0.0433 | 0.0407 | 2 | 7 | 0 | 2 | - | - | - | - | z2w,1=1 z2w,2=1 z2w,3=1 | H6Mg1Nb2 |
| 166.1.137 | 67682 | 124 | 0.0702 | -0.1125 | 2 | 1 | 2 | 1 | - | - | - | - | z2w,1=1 z2w,2=1 z2w,3=1 z4=3 | H6Fe3O14Pb1S2 |
| 166.1.138 | 44626 | 106 | 0.0057 | -0.0778 | 2 | 4 | 2 | 1 | - | - | - | - | z4=1 | H6Al3Na1O14S2 |
| 166.1.139 | 24344 | 80 | 0.0058 | -0.1273 | 2 | 4 | 2 | 1 | - | - | - | - | z4=1 | H4Cu1N1O3S1 |
| 166.1.140 | 623755 | 114 | 0.0145 | -0.2591 | 2 | 6 | 2 | 1 | - | - | - | - | z4=3 | H4Co3Y1 |
| 166.1.141 | 623744 | 114 | 0.0146 | -0.2537 | 2 | 6 | 2 | 1 | - | - | - | - | z4=3 | H4Co3Y1 |
| 166.1.142 | 604264 | 114 | 0.0164 | -0.2586 | 2 | 6 | 2 | 1 | - | - | - | - | z4=3 | H4Co3Y1 |

Continued on next page

**TABLE XIV– continued**

| TQC | ICSD | NoE | DG | IndG | SubSG (w/ T) | cl | SubSG (w/o T) | cl | Max. Sub. SG | yes or no | Max. Sub. SG | yes or no | SIs (w/ T) | Stoich. Formula |
|---|---|---|---|---|---|---|---|---|---|---|---|---|---|---|
| 166.1.143 | 602920 | 114 | 0.0164 | -0.2572 | 2 | 6 | 2 | 1 | - | - | - | - | z4=3 | H4Co3Y1 |
| 166.1.144 | 623723 | 120 | 0.0319 | -0.0031 | 2 | 6 | 2 | 1 | - | - | - | - | z4=3 | H4Co3Pr1 |
| 166.1.145 | 623698 | 144 | 0.0031 | -0.0207 | 2 | 4 | 2 | 1 | - | - | - | - | z4=1 | H4Co3Ho1 |
| 166.1.146 | 602925 | 144 | 0.0039 | -0.0203 | 2 | 2 | 2 | 1 | - | - | - | - | z2w,1=1<br>z2w,2=1<br>z2w,3=1<br>z4=1 | H4Co3Ho1 |
| 166.1.147 | 623752 | 114 | 0.0146 | -0.2570 | 2 | 6 | 2 | 1 | - | - | - | - | z4=3 | H3Co3Y1 |
| 166.1.148 | 24685 | 86 | 0.0517 | -0.0922 | 2 | 3 | 0 | 2 | - | - | - | - | z2w,1=1<br>z2w,2=1<br>z2w,3=1<br>z4=2 | H3Cl1Co2O3 |
| 166.1.149 | 155534 | 86 | 0.0465 | -0.0739 | 2 | 3 | 0 | 2 | - | - | - | - | z2w,1=1<br>z2w,2=1<br>z2w,3=1<br>z4=2 | H3Cl1Co2O3 |
| 166.1.150 | 6016 | 44 | 0.0104 | -0.1573 | 2 | 3 | 0 | 2 | - | - | - | - | z2w,1=1<br>z2w,2=1<br>z2w,3=1<br>z4=2 | H2N2Th1 |
| 166.1.151 | 423880 | 108 | 0.0306 | 0.0232 | 2 | 7 | 0 | 2 | - | - | - | - | z2w,1=1<br>z2w,2=1<br>z2w,3=1 | H2Mo1O4Rb2S1 |
| 166.1.152 | 623746 | 114 | 0.0137 | -0.2806 | 2 | 6 | 2 | 1 | - | - | - | - | z4=3 | H1Co3Y1 |
| 166.1.153 | 40981 | 22 | 0.0108 | -0.3801 | 2 | 7 | 0 | 2 | - | - | - | - | z2w,1=1<br>z2w,2=1<br>z2w,3=1 | H1Cl1Sc1 |
| 166.1.154 | 62226 | 66 | 0.1946 | -0.5550 | 2 | 7 | 0 | 2 | - | - | - | - | z2w,1=1<br>z2w,2=1<br>z2w,3=1 | H1Cl1Lu1 |
| 166.1.155 | 911 | 52 | 0.0226 | -0.3726 | 2 | 3 | 0 | 2 | - | - | - | - | z2w,1=1<br>z2w,2=1<br>z2w,3=1<br>z4=2 | H12K2N6Sn1 |
| 166.1.156 | 159891 | 114 | 0.0000 | -0.2001 | 2 | 3 | 0 | 2 | - | - | - | - | z2w,1=1<br>z2w,2=1<br>z2w,3=1<br>z4=2 | H1.24Co3Y1 |
| 166.1.157 | 94220 | 216 | 0.0000 | -0.0779 | 2 | 3 | 0 | 2 | - | - | - | - | z2w,1=1<br>z2w,2=1<br>z2w,3=1<br>z4=2 | Ge2Nd2Zn15 |
| 166.1.158 | 99161 | 16 | 0.0127 | -0.5883 | 2 | 6 | 2 | 1 | - | - | - | - | z4=3 | Ge2Li4Zr1 |
| 166.1.159 | 166193 | 58 | 0.0210 | -0.4065 | 2 | 3 | 0 | 2 | - | - | - | - | z2w,1=1<br>z2w,2=1<br>z2w,3=1<br>z4=2 | Gd1Sn4Ti6 |
| 166.1.160 | 418524 | 82 | 0.0105 | 0.0015 | 2 | 6 | 2 | 1 | - | - | - | - | z4=3 | Gd1O1P1Zn1 |
| 166.1.161 | 55619 | 112 | 0.0111 | -0.1725 | 2 | 2 | 2 | 1 | - | - | - | - | z2w,1=1<br>z2w,2=1<br>z2w,3=1<br>z4=1 | Gd1Mg2Ni9 |
| 166.1.162 | 635712 | 54 | 0.0000 | -0.0213 | 2 | 4 | 2 | 1 | - | - | - | - | z4=1 | Gd1 |
| 166.1.163 | 635711 | 54 | 0.0032 | -0.0283 | 2 | 3 | 0 | 2 | - | - | - | - | z2w,1=1<br>z2w,2=1<br>z2w,3=1<br>z4=2 | Gd1 |
| 166.1.164 | 52517 | 54 | 0.0000 | -0.0213 | 2 | 4 | 2 | 1 | - | - | - | - | z4=1 | Gd1 |

Continued on next page

**TABLE XIV– continued**

| TQC | ICSD | NoE | DG | IndG | SubSG (w/ T) | cl | SubSG (w/o T) | cl | Max. Sub. SG | yes or no | Max. Sub. SG | yes or no | SIs (w/ T) | Stoich. Formula |
|---|---|---|---|---|---|---|---|---|---|---|---|---|---|---|
| 166.1.165 | 38267 | 60 | 0.0030 | -0.6440 | 2 | 1 | 2 | 1 | - | - | - | - | z2w,1=1<br>z2w,2=1<br>z2w,3=1<br>z4=3 | Ga7Rb1 |
| 166.1.166 | 102891 | 60 | 0.0044 | -0.6 - | 2 | 1 | 2 | 1 | - | - | - | - | z2w,1=1<br>z2w,2=1<br>z2w,3=1<br>z4=3 | Ga7Rb1 |
| 166.1.167 | 602235 | 150 | 0.0000 | -0.0048 | 2 | 4 | 2 | 1 | - | - | - | - | z4=1 | Ga3Ho1 |
| 166.1.168 | 657608 | 24 | 0.1844 | -0.7858 | 2 | 2 | 2 | 1 | - | - | - | - | z2w,1=1<br>z2w,2=1<br>z2w,3=1<br>z4=1 | Ga2Te3 |
| 166.1.169 | 635031 | 16 | 0.0571 | -1.4714 | 2 | 6 | 2 | 1 | - | - | - | - | z4=3 | Ga1P1 |
| 166.1.170 | 96491 | 130 | 0.0250 | -0.0510 | 2 | 2 | 2 | 1 | - | - | - | - | z2w,1=1<br>z2w,2=1<br>z2w,3=1<br>z4=1 | Ga1Mn1 |
| 166.1.171 | 35386 | 26 | 0.0578 | 0.0578 | 2 | 2 | 2 | 1 | - | - | - | - | z2w,1=1<br>z2w,2=1<br>z2w,3=1<br>z4=1 | Ga1Ge1Te1 |
| 166.1.172 | 2513 | 66 | 0.0083 | -1.4767 | 2 | 3 | 0 | 2 | - | - | - | - | z2w,1=1<br>z2w,2=1<br>z2w,3=1<br>z4=2 | Ga1 |
| 166.1.173 | 634071 | 92 | 0.0216 | -0.1560 | 2 | 4 | 2 | 1 | - | - | - | - | z4=1 | Fe7W6 |
| 166.1.174 | 634063 | 92 | 0.0216 | -0.1604 | 2 | 4 | 2 | 1 | - | - | - | - | z4=1 | Fe7W6 |
| 166.1.175 | 634060 | 92 | 0.0357 | -0.0795 | 2 | 4 | 2 | 1 | - | - | - | - | z4=1 | Fe7W6 |
| 166.1.176 | 103694 | 92 | 0.0215 | -0.1551 | 2 | 4 | 2 | 1 | - | - | - | - | z4=1 | Fe7W6 |
| 166.1.177 | 632784 | 134 | 0.0134 | -0.0997 | 2 | 3 | 0 | 2 | - | - | - | - | z2w,1=1<br>z2w,2=1<br>z2w,3=1<br>z4=2 | Fe7Nb6 |
| 166.1.178 | 632622 | 92 | 0.0184 | -0.1807 | 2 | 4 | 2 | 1 | - | - | - | - | z4=1 | Fe7Mo6 |
| 166.1.179 | 632620 | 92 | 0.0444 | -0.1768 | 2 | 4 | 2 | 1 | - | - | - | - | z4=1 | Fe7Mo6 |
| 166.1.180 | 41380 | 138 | 0.0023 | -0.0643 | 2 | 3 | 0 | 2 | - | - | - | - | z2w,1=1<br>z2w,2=1<br>z2w,3=1<br>z4=2 | Fe6Ge5Li1 |
| 166.1.181 | 633920 | 108 | 0.0026 | -0.1478 | 2 | 3 | 0 | 2 | - | - | - | - | z2w,1=1<br>z2w,2=1<br>z2w,3=1<br>z4=2 | Fe3Th1 |
| 166.1.182 | 633912 | 108 | 0.0021 | -0.1485 | 2 | 3 | 0 | 2 | - | - | - | - | z2w,1=1<br>z2w,2=1<br>z2w,3=1<br>z4=2 | Fe3Th1 |
| 166.1.183 | 633900 | 108 | 0.0021 | -0.1485 | 2 | 3 | 0 | 2 | - | - | - | - | z2w,1=1<br>z2w,2=1<br>z2w,3=1<br>z4=2 | Fe3Th1 |
| 166.1.184 | 633896 | 108 | 0.0027 | -0.1466 | 2 | 6 | 2 | 1 | - | - | - | - | z4=3 | Fe3Th1 |
| 166.1.185 | 633886 | 108 | 0.0107 | -0.1474 | 2 | 3 | 0 | 2 | - | - | - | - | z2w,1=1<br>z2w,2=1<br>z2w,3=1<br>z4=2 | Fe3Th1 |
| 166.1.186 | 103654 | 108 | 0.0064 | -0.1414 | 2 | 6 | 2 | 1 | - | - | - | - | z4=3 | Fe3Th1 |

Continued on next page

**TABLE XIV– continued**

| TQC | ICSD | NoE | DG | IndG | SubSG (w/ T) | cl | SubSG (w/o T) | cl | Max. Sub. SG | yes or no | Max. Sub. SG | yes or no | SIs (w/ T) | Stoich. Formula |
|---|---|---|---|---|---|---|---|---|---|---|---|---|---|---|
| 166.1.187 | 71 | 64 | 0.0024 | -0.1384 | 2 | 3 | 0 | 2 | - | - | - | - | z2w,1=1<br>z2w,2=1<br>z2w,3=1<br>z4=2 | Fe3Sn2 |
| 166.1.188 | 631893 | 126 | 0.0123 | -0.0865 | 2 | 6 | 2 | 1 | - | - | - | - | z4=3 | Fe3Gd1 |
| 166.1.189 | 631888 | 126 | 0.0123 | -0.0860 | 2 | 6 | 2 | 1 | - | - | - | - | z4=3 | Fe3Gd1 |
| 166.1.190 | 631864 | 126 | 0.0124 | -0.0861 | 2 | 6 | 2 | 1 | - | - | - | - | z4=3 | Fe3Gd1 |
| 166.1.191 | 103479 | 126 | 0.0128 | -0.0862 | 2 | 6 | 2 | 1 | - | - | - | - | z4=3 | Fe3Gd1 |
| 166.1.192 | 4192 | 64 | 0.0219 | -0.4508 | 2 | 4 | 2 | 1 | - | - | - | - | z4=1 | Fe2O4Yb1 |
| 166.1.193 | 195811 | 64 | 0.0085 | -0.4346 | 2 | 4 | 2 | 1 | - | - | - | - | z4=1 | Fe2O4Yb1 |
| 166.1.194 | 195810 | 64 | 0.0347 | -0.4249 | 2 | 4 | 2 | 1 | - | - | - | - | z4=1 | Fe2O4Yb1 |
| 166.1.195 | 195809 | 64 | 0.0081 | -0.3231 | 2 | 3 | 0 | 2 | - | - | - | - | z2w,1=1<br>z2w,2=1<br>z2w,3=1<br>z4=2 | Fe2O4Yb1 |
| 166.1.196 | 29011 | 46 | 0.0090 | -0.1661 | 2 | 6 | 2 | 1 | - | - | - | - | z4=3 | Fe1O2Tl1 |
| 166.1.197 | 631854 | 148 | 0.0000 | -0.0257 | 2 | 6 | 2 | 1 | - | - | - | - | z4=3 | Fe15Ga2Y2 |
| 166.1.198 | 95660 | 54 | 0.0035 | -0.0177 | 2 | 4 | 2 | 1 | - | - | - | - | z4=1 | F1Nd1O1 |
| 166.1.199 | 95659 | 54 | 0.0056 | -0.0149 | 2 | 4 | 2 | 1 | - | - | - | - | z4=1 | F1Nd1O1 |
| 166.1.200 | 95658 | 54 | 0.0045 | -0.0168 | 2 | 4 | 2 | 1 | - | - | - | - | z4=1 | F1Nd1O1 |
| 166.1.201 | 95657 | 54 | 0.0053 | -0.0154 | 2 | 4 | 2 | 1 | - | - | - | - | z4=1 | F1Nd1O1 |
| 166.1.202 | 95656 | 54 | 0.0058 | -0.0151 | 2 | 4 | 2 | 1 | - | - | - | - | z4=1 | F1Nd1O1 |
| 166.1.203 | 95655 | 54 | 0.0000 | -0.0197 | 2 | 4 | 2 | 1 | - | - | - | - | z4=1 | F1Nd1O1 |
| 166.1.204 | 184005 | 54 | 0.0053 | -0.0149 | 2 | 4 | 2 | 1 | - | - | - | - | z4=1 | F1Nd1O1 |
| 166.1.205 | 184011 | 68 | 0.0095 | 0.0006 | 2 | 3 | 0 | 2 | - | - | - | - | z2w,1=1<br>z2w,2=1<br>z2w,3=1<br>z4=2 | F1Ho1O1 |
| 166.1.206 | 202068 | 46 | 0.0109 | 0.0045 | 2 | 1 | 2 | 1 | - | - | - | - | z2w,1=1<br>z2w,2=1<br>z2w,3=1<br>z4=3 | Eu2I1P1 |
| 166.1.207 | 67700 | 62 | 0.0313 | -0.3188 | 2 | 4 | 2 | 1 | - | - | - | - | z4=1 | Er1Fe2O4 |
| 166.1.208 | 184012 | 70 | 0.0216 | -0.0150 | 2 | 3 | 0 | 2 | - | - | - | - | z2w,1=1<br>z2w,2=1<br>z2w,3=1<br>z4=2 | Er1F1O1 |
| 166.1.209 | 630377 | 244 | 0.0015 | -0.0243 | 2 | 3 | 0 | 2 | - | - | - | - | z2w,1=1<br>z2w,2=1<br>z2w,3=1<br>z4=2 | Dy2Zn17 |
| 166.1.210 | 629979 | 220 | 0.0016 | -0.0165 | 2 | 2 | 2 | 1 | - | - | - | - | z2w,1=1<br>z2w,2=1<br>z2w,3=1<br>z4=1 | Dy2Ni7 |
| 166.1.211 | 262377 | 220 | 0.0064 | -0.0179 | 2 | 3 | 0 | 2 | - | - | - | - | z2w,1=1<br>z2w,2=1<br>z2w,3=1<br>z4=2 | Dy2Ni7 |
| 166.1.212 | 630007 | 150 | 0.0037 | -0.0288 | 2 | 2 | 2 | 1 | - | - | - | - | z2w,1=1<br>z2w,2=1<br>z2w,3=1<br>z4=1 | Dy1Ni3 |
| 166.1.213 | 629986 | 150 | 0.0038 | -0.0287 | 2 | 2 | 2 | 1 | - | - | - | - | z2w,1=1<br>z2w,2=1<br>z2w,3=1<br>z4=1 | Dy1Ni3 |

Continued on next page

**TABLE XIV– continued**

| TQC | ICSD | NoE | DG | IndG | SubSG (w/ T) | cl | SubSG (w/o T) | cl | Max. Sub. SG | yes or no | Max. Sub. SG | yes or no | SIs (w/ T) | Stoich. Formula |
|---|---|---|---|---|---|---|---|---|---|---|---|---|---|---|
| 166.1.214 | 629980 | 150 | 0.0026 | -0.0302 | 2 | 2 | 2 | 1 | - | - | - | - | z2w,1=1 z2w,2=1 z2w,3=1 z4=1 | Dy1Ni3 |
| 166.1.215 | 629598 | 132 | 0.0000 | -0.0684 | 2 | 7 | 0 | 2 | - | - | - | - | z2w,1=1 z2w,2=1 z2w,3=1 | Dy1Fe3 |
| 166.1.216 | 184010 | 66 | 0.0045 | -0.0199 | 2 | 3 | 0 | 2 | - | - | - | - | z2w,1=1 z2w,2=1 z2w,3=1 z4=2 | Dy1F1O1 |
| 166.1.217 | 629540 | 60 | 0.0259 | 0.0181 | 2 | 4 | 2 | 1 | - | - | - | - | z4=1 | Dy1 |
| 166.1.218 | 629537 | 60 | 0.0228 | 0.0187 | 2 | 4 | 2 | 1 | - | - | - | - | z4=1 | Dy1 |
| 166.1.219 | 52518 | 60 | 0.0260 | 0.0180 | 2 | 4 | 2 | 1 | - | - | - | - | z4=1 | Dy1 |
| 166.1.220 | 190938 | 26 | 0.1697 | -0.8060 | 2 | 1 | 2 | 1 | - | - | - | - | z2w,1=1 z2w,2=1 z2w,3=1 z4=3 | Cu1N2Ta1 |
| 166.1.221 | 80259 | 36 | 0.1565 | 0.0572 | 2 | 6 | 2 | 1 | - | - | - | - | z4=3 | Cu1I1 |
| 166.1.222 | 80257 | 36 | 0.1300 | 0.0541 | 2 | 6 | 2 | 1 | - | - | - | - | z4=3 | Cu1I1 |
| 166.1.223 | 80253 | 36 | 0.0039 | 0.0039 | 2 | 6 | 2 | 1 | - | - | - | - | z4=3 | Cu1I1 |
| 166.1.224 | 80252 | 36 | 0.0073 | 0.0073 | 2 | 6 | 2 | 1 | - | - | - | - | z4=3 | Cu1I1 |
| 166.1.225 | 410393 | 120 | 0.0141 | -0.0301 | 2 | 3 | 0 | 2 | - | - | - | - | z2w,1=1 z2w,2=1 z2w,3=1 z4=2 | Cs9F27Ni9 |
| 166.1.226 | 78928 | 146 | 0.0064 | 0.0010 | 2 | 3 | 0 | 2 | - | - | - | - | z2w,1=1 z2w,2=1 z2w,3=1 z4=2 | Cs4F12Fe2K1Li1 |
| 166.1.227 | 42152 | 142 | 0.0085 | -0.0399 | 2 | 2 | 2 | 1 | - | - | - | - | z2w,1=1 z2w,2=1 z2w,3=1 z4=1 | Cs2F6Na1Ni1 |
| 166.1.228 | 16255 | 138 | 0.0040 | -0.0030 | 2 | 1 | 2 | 1 | - | - | - | - | z2w,1=1 z2w,2=1 z2w,3=1 z4=3 | Cs2F6Fe1Na1 |
| 166.1.229 | 102864 | 60 | 0.0308 | -0.7931 | 2 | 1 | 2 | 1 | - | - | - | - | z2w,1=1 z2w,2=1 z2w,3=1 z4=3 | Cs1Ga7 |
| 166.1.230 | 625563 | 170 | 0.0027 | -0.4110 | 2 | 4 | 2 | 1 | - | - | - | - | z4=1 | Co7Y2 |
| 166.1.231 | 625438 | 174 | 0.0052 | -0.3486 | 2 | 2 | 2 | 1 | - | - | - | - | z2w,1=1 z2w,2=1 z2w,3=1 z4=1 | Co7Th2 |
| 166.1.232 | 625416 | 174 | 0.0056 | -0.3517 | 2 | 2 | 2 | 1 | - | - | - | - | z2w,1=1 z2w,2=1 z2w,3=1 z4=1 | Co7Th2 |
| 166.1.233 | 184367 | 178 | 0.0294 | -0.0502 | 2 | 3 | 0 | 2 | - | - | - | - | z2w,1=1 z2w,2=1 z2w,3=1 z4=2 | Co7Pr2 |
| 166.1.234 | 623985 | 170 | 0.0054 | -0.2311 | 2 | 4 | 2 | 1 | - | - | - | - | z4=1 | Co7La2 |
| 166.1.235 | 623978 | 170 | 0.0000 | -0.2333 | 2 | 4 | 2 | 1 | - | - | - | - | z4=1 | Co7La2 |

Continued on next page

**TABLE XIV– continued**

| TQC | ICSD | NoE | DG | IndG | SubSG (w/ T) | cl | SubSG (w/o T) | cl | Max. Sub. SG | yes or no | Max. Sub. SG | yes or no | SIs (w/ T) | Stoich. Formula |
|---|---|---|---|---|---|---|---|---|---|---|---|---|---|---|
| 166.1.236 | 623284 | 198 | 0.0079 | -0.0982 | 2 | 2 | 2 | 1 | - | - | - | - | z2w,1=1<br>z2w,2=1<br>z2w,3=1<br>z4=1 | Co7Gd2 |
| 166.1.237 | 658884 | 206 | 0.0038 | -0.0276 | 2 | 6 | 2 | 1 | - | - | - | - | z4=3 | Co7Dy2 |
| 166.1.238 | 656092 | 114 | 0.0000 | -0.3753 | 2 | 6 | 2 | 1 | - | - | - | - | z4=3 | Co3Y1 |
| 166.1.239 | 656079 | 114 | 0.0000 | -0.3420 | 2 | 6 | 2 | 1 | - | - | - | - | z4=3 | Co3Y1 |
| 166.1.240 | 625636 | 114 | 0.0000 | -0.3602 | 2 | 6 | 2 | 1 | - | - | - | - | z4=3 | Co3Y1 |
| 166.1.241 | 625625 | 114 | 0.0000 | -0.3643 | 2 | 6 | 2 | 1 | - | - | - | - | z4=3 | Co3Y1 |
| 166.1.242 | 625605 | 114 | 0.0000 | -0.3607 | 2 | 6 | 2 | 1 | - | - | - | - | z4=3 | Co3Y1 |
| 166.1.243 | 625600 | 114 | 0.0000 | -0.3608 | 2 | 6 | 2 | 1 | - | - | - | - | z4=3 | Co3Y1 |
| 166.1.244 | 625590 | 114 | 0.0000 | -0.3689 | 2 | 6 | 2 | 1 | - | - | - | - | z4=3 | Co3Y1 |
| 166.1.245 | 625580 | 114 | 0.0000 | -0.3612 | 2 | 6 | 2 | 1 | - | - | - | - | z4=3 | Co3Y1 |
| 166.1.246 | 625572 | 114 | 0.0000 | -0.3646 | 2 | 6 | 2 | 1 | - | - | - | - | z4=3 | Co3Y1 |
| 166.1.247 | 625564 | 114 | 0.0000 | -0.3642 | 2 | 6 | 2 | 1 | - | - | - | - | z4=3 | Co3Y1 |
| 166.1.248 | 603098 | 114 | 0.0000 | -0.3451 | 2 | 6 | 2 | 1 | - | - | - | - | z4=3 | Co3Y1 |
| 166.1.249 | 159894 | 114 | 0.0000 | -0.3290 | 2 | 6 | 2 | 1 | - | - | - | - | z4=3 | Co3Y1 |
| 166.1.250 | 159890 | 114 | 0.0000 | -0.3395 | 2 | 6 | 2 | 1 | - | - | - | - | z4=3 | Co3Y1 |
| 166.1.251 | 102729 | 114 | 0.0047 | -0.3989 | 2 | 6 | 2 | 1 | - | - | - | - | z4=3 | Co3Y1 |
| 166.1.252 | 625506 | 150 | 0.0000 | -0.0869 | 2 | 3 | 0 | 2 | - | - | - | - | z2w,1=1<br>z2w,2=1<br>z2w,3=1<br>z4=2 | Co3Tm1 |
| 166.1.253 | 625499 | 150 | 0.0000 | -0.0870 | 2 | 3 | 0 | 2 | - | - | - | - | z2w,1=1<br>z2w,2=1<br>z2w,3=1<br>z4=2 | Co3Tm1 |
| 166.1.254 | 24581 | 96 | 0.0085 | -0.5958 | 2 | 1 | 2 | 1 | - | - | - | - | z2w,1=1<br>z2w,2=1<br>z2w,3=1<br>z4=3 | Co3Ta1 |
| 166.1.255 | 624740 | 120 | 0.0063 | -0.0365 | 2 | 2 | 2 | 1 | - | - | - | - | z2w,1=1<br>z2w,2=1<br>z2w,3=1<br>z4=1 | Co3Pr1 |
| 166.1.256 | 624707 | 120 | 0.0037 | -0.0381 | 2 | 2 | 2 | 1 | - | - | - | - | z2w,1=1<br>z2w,2=1<br>z2w,3=1<br>z4=1 | Co3Pr1 |
| 166.1.257 | 624697 | 120 | 0.0037 | -0.0382 | 2 | 2 | 2 | 1 | - | - | - | - | z2w,1=1<br>z2w,2=1<br>z2w,3=1<br>z4=1 | Co3Pr1 |
| 166.1.258 | 108889 | 120 | 0.0036 | -0.0381 | 2 | 2 | 2 | 1 | - | - | - | - | z2w,1=1<br>z2w,2=1<br>z2w,3=1<br>z4=1 | Co3Pr1 |
| 166.1.259 | 624054 | 156 | 0.0021 | -0.3779 | 2 | 6 | 2 | 1 | - | - | - | - | z4=3 | Co3Lu1 |
| 166.1.260 | 624039 | 156 | 0.0000 | -0.3764 | 2 | 6 | 2 | 1 | - | - | - | - | z4=3 | Co3Lu1 |
| 166.1.261 | 246307 | 14 | 0.0196 | -0.6125 | 2 | 7 | 0 | 2 | - | - | - | - | z2w,1=1<br>z2w,2=1<br>z2w,3=1 | Co1Li1O2 |
| 166.1.262 | 657284 | 204 | 0.0000 | -0.0580 | 2 | 3 | 0 | 2 | - | - | - | - | z2w,1=1<br>z2w,2=1<br>z2w,3=1<br>z4=2 | Co17Gd2N3 |

Continued on next page

**TABLE XIV– continued**

| TQC | ICSD | NoE | DG | IndG | SubSG (w/ T) | cl | SubSG (w/o T) | cl | Max. Sub. SG | yes or no | Max. Sub. SG | yes or no | SIs (w/ T) | Stoich. Formula |
|---|---|---|---|---|---|---|---|---|---|---|---|---|---|---|
| 166.1.263 | 79717 | 198 | 0.0105 | -0.0513 | 2 | 1 | 2 | 1 | - | - | - | - | z2w,1=1 z2w,2=1 z2w,3=1 z4=3 | Co10Ge3O16 |
| 166.1.264 | 401026 | 70 | 0.0104 | 0.0104 | 2 | 1 | 2 | 1 | - | - | - | - | z2w,1=1 z2w,2=1 z2w,3=1 z4=3 | Cl8Na2Ti3 |
| 166.1.265 | 869 | 22 | 0.0642 | -0.2412 | 2 | 7 | 0 | 2 | - | - | - | - | z2w,1=1 z2w,2=1 z2w,3=1 | Cl1Zr1 |
| 166.1.266 | 20145 | 22 | 0.0626 | 0.0379 | 2 | 7 | 0 | 2 | - | - | - | - | z2w,1=1 z2w,2=1 z2w,3=1 | Cl1Zr1 |
| 166.1.267 | 6077 | 74 | 0.0071 | 0.0071 | 2 | 3 | 0 | 2 | - | - | - | - | z2w,1=1 z2w,2=1 z2w,3=1 z4=2 | Cl1O1Yb1 |
| 166.1.268 | 24410 | 36 | 0.0215 | -0.0225 | 2 | 4 | 2 | 1 | - | - | - | - | z4=1 | Cl1La1 |
| 166.1.269 | 154513 | 206 | 0.0222 | -0.5399 | 2 | 1 | 2 | 1 | - | - | - | - | z2w,1=1 z2w,2=1 z2w,3=1 z4=3 | Ce2Cu5Zn2 |
| 166.1.270 | 620668 | 174 | 0.0260 | -0.2752 | 2 | 1 | 2 | 1 | - | - | - | - | z2w,1=1 z2w,2=1 z2w,3=1 z4=3 | Ce2Co7 |
| 166.1.271 | 620636 | 174 | 0.0187 | -0.2736 | 2 | 1 | 2 | 1 | - | - | - | - | z2w,1=1 z2w,2=1 z2w,3=1 z4=3 | Ce2Co7 |
| 166.1.272 | 603443 | 192 | 0.0093 | -0.0365 | 2 | 2 | 2 | 1 | - | - | - | - | z2w,1=1 z2w,2=1 z2w,3=1 z4=1 | Ce2Co5Fe12N3 |
| 166.1.273 | 201355 | 162 | 0.0116 | -0.0177 | 2 | 2 | 2 | 1 | - | - | - | - | z2w,1=1 z2w,2=1 z2w,3=1 z4=1 | Cd1Cs4F12Ni3 |
| 166.1.274 | 166533 | 16 | 0.0057 | -0.5245 | 2 | 4 | 2 | 1 | - | - | - | - | z4=1 | Ca2I1P1 |
| 166.1.275 | 424597 | 38 | 0.0000 | -1.1270 | 2 | 6 | 2 | 1 | - | - | - | - | z4=3 | Ca2Ge1Pd3 |
| 166.1.276 | 52789 | 36 | 0.6137 | -0.7647 | 2 | 4 | 2 | 1 | - | - | - | - | z4=1 | Ca1Si2 |
| 166.1.277 | 409057 | 10 | 0.8336 | -0.5272 | 2 | 1 | 2 | 1 | - | - | - | - | z2w,1=1 z2w,2=1 z2w,3=1 z4=3 | Ca1Si2 |
| 166.1.278 | 32006 | 20 | 0.0527 | -0.9811 | 2 | 4 | 2 | 1 | - | - | - | - | z4=1 | Ca1Si2 |
| 166.1.279 | 248518 | 20 | 0.0113 | -1.1763 | 2 | 4 | 2 | 1 | - | - | - | - | z4=1 | Ca1Si2 |
| 166.1.280 | 248517 | 20 | 0.0153 | -1.6045 | 2 | 4 | 2 | 1 | - | - | - | - | z4=1 | Ca1Si2 |
| 166.1.281 | 248516 | 20 | 0.0338 | -1.5822 | 2 | 4 | 2 | 1 | - | - | - | - | z4=1 | Ca1Si2 |
| 166.1.282 | 248515 | 20 | 0.1331 | -1.3261 | 2 | 4 | 2 | 1 | - | - | - | - | z4=1 | Ca1Si2 |
| 166.1.283 | 193539 | 10 | 0.8486 | -0.5186 | 2 | 1 | 2 | 1 | - | - | - | - | z2w,1=1 z2w,2=1 z2w,3=1 z4=3 | Ca1Si2 |

Continued on next page

**TABLE XIV– continued**

| TQC | ICSD | NoE | DG | IndG | SubSG (w/ T) | cl | SubSG (w/o T) | cl | Max. Sub. SG | yes or no | Max. Sub. SG | yes or no | SIs (w/ T) | Stoich. Formula |
|---|---|---|---|---|---|---|---|---|---|---|---|---|---|---|
| 166.1.284 | 15632 | 10 | 0.8149 | -0.5284 | 2 | 1 | 2 | 1 | - | - | - | - | z2w,1=1 z2w,2=1 z2w,3=1 z4=3 | Ca1Si2 |
| 166.1.285 | 77316 | 36 | 0.2566 | -0.2056 | 2 | 4 | 2 | 1 | - | - | - | - | z4=1 | Ca1Ge2 |
| 166.1.286 | 619304 | 20 | 0.0322 | -0.7062 | 2 | 4 | 2 | 1 | - | - | - | - | z4=1 | Ca1Ge2 |
| 166.1.287 | 245612 | 20 | 0.2534 | -0.4931 | 2 | 4 | 2 | 1 | - | - | - | - | z4=1 | Ca1Ge2 |
| 166.1.288 | 110107 | 20 | 0.0828 | -0.6388 | 2 | 4 | 2 | 1 | - | - | - | - | z4=1 | Ca1Ge2 |
| 166.1.289 | 20822 | 52 | 0.0188 | 0.0188 | 2 | 4 | 2 | 1 | - | - | - | - | z4=1 | C5Ti8 |
| 166.1.290 | 657171 | 186 | 0.0000 | -0.0385 | 2 | 7 | 0 | 2 | - | - | - | - | z2w,1=1 z2w,2=1 z2w,3=1 | C3Fe17Tb2 |
| 166.1.291 | 659087 | 192 | 0.0000 | -0.0434 | 2 | 6 | 2 | 1 | - | - | - | - | z4=3 | C3Er2Fe17 |
| 166.1.292 | 41260 | 40 | 0.0041 | -0.7993 | 2 | 2 | 2 | 1 | - | - | - | - | z2w,1=1 z2w,2=1 z2w,3=1 z4=1 | C3Al6N2 |
| 166.1.293 | 74178 | 146 | 0.0021 | -0.0021 | 2 | 2 | 2 | 1 | - | - | - | - | z2w,1=1 z2w,2=1 z2w,3=1 z4=1 | C2Ba1Ce1F1O6 |
| 166.1.294 | 96188 | 26 | 0.0618 | -0.1194 | 2 | 2 | 2 | 1 | - | - | - | - | z2w,1=1 z2w,2=1 z2w,3=1 z4=1 | C1Y2 |
| 166.1.295 | 22283 | 26 | 0.0525 | -0.1112 | 2 | 2 | 2 | 1 | - | - | - | - | z2w,1=1 z2w,2=1 z2w,3=1 z4=1 | C1Y2 |
| 166.1.296 | 42762 | 42 | 0.0175 | 0.0009 | 2 | 4 | 2 | 1 | - | - | - | - | z4=1 | C1Tb2 |
| 166.1.297 | 42761 | 42 | 0.0165 | 0.0001 | 2 | 4 | 2 | 1 | - | - | - | - | z4=1 | C1Tb2 |
| 166.1.298 | 108187 | 42 | 0.0165 | -0.0002 | 2 | 4 | 2 | 1 | - | - | - | - | z4=1 | C1Tb2 |
| 166.1.299 | 23791 | 26 | 0.0703 | -0.1015 | 2 | 2 | 2 | 1 | - | - | - | - | z2w,1=1 z2w,2=1 z2w,3=1 z4=1 | C1S2Ta2 |
| 166.1.300 | 95112 | 42 | 0.0420 | -0.4349 | 2 | 3 | 0 | 2 | - | - | - | - | z2w,1=1 z2w,2=1 z2w,3=1 z4=2 | C1Nb2S2 |
| 166.1.301 | 95110 | 42 | 0.0858 | 0.0310 | 2 | 3 | 0 | 2 | - | - | - | - | z2w,1=1 z2w,2=1 z2w,3=1 z4=2 | C1Nb2S2 |
| 166.1.302 | 53780 | 8 | 3.8650 | 3.2031 | 2 | 7 | 0 | 2 | - | - | - | - | z2w,1=1 z2w,2=1 z2w,3=1 | C1 |
| 166.1.303 | 31829 | 8 | 3.8349 | 3.3538 | 2 | 7 | 0 | 2 | - | - | - | - | z2w,1=1 z2w,2=1 z2w,3=1 | C1 |
| 166.1.304 | 29123 | 8 | 3.9661 | 3.2600 | 2 | 7 | 0 | 2 | - | - | - | - | z2w,1=1 z2w,2=1 z2w,3=1 | C1 |
| 166.1.305 | 602542 | 72 | 0.0000 | -0.1686 | 2 | 2 | 2 | 1 | - | - | - | - | z2w,1=1 z2w,2=1 z2w,3=1 z4=1 | C1B2U1 |

Continued on next page

**TABLE XIV– continued**

| TQC | ICSD | NoE | DG | IndG | SubSG (w/ T) | cl | SubSG (w/o T) | cl | Max. Sub. SG | yes or no | Max. Sub. SG | yes or no | SIs (w/ T) | Stoich. Formula |
|---|---|---|---|---|---|---|---|---|---|---|---|---|---|---|
| 166.1.306 | 44142 | 72 | 0.0067 | -0.1716 | 2 | 2 | 2 | 1 | - | - | - | - | z2w,1=1 z2w,2=1 z2w,3=1 z4=1 | C1B2U1 |
| 166.1.307 | 238975 | 78 | 0.3343 | 0.2435 | 2 | 6 | 2 | 1 | - | - | - | - | z4=3 | C1B2Pu1 |
| 166.1.308 | 1168 | 22 | 0.0431 | -0.2468 | 2 | 7 | 0 | 2 | - | - | - | - | z2w,1=1 z2w,2=1 z2w,3=1 | Br1Zr1 |
| 166.1.309 | 23354 | 36 | 0.0783 | -0.0873 | 2 | 7 | 0 | 2 | - | - | - | - | z2w,1=1 z2w,2=1 z2w,3=1 | Br1La1 |
| 166.1.310 | 202067 | 46 | 0.0073 | -0.0034 | 2 | 1 | 2 | 1 | - | - | - | - | z2w,1=1 z2w,2=1 z2w,3=1 z4=3 | Br1Eu2P1 |
| 166.1.311 | 95551 | 94 | 0.0930 | 0.0930 | 2 | 2 | 2 | 1 | - | - | - | - | z2w,1=1 z2w,2=1 z2w,3=1 z4=1 | Bi6Pb1Te10 |
| 166.1.312 | 20319 | 38 | 0.0301 | -0.4953 | 2 | 7 | 0 | 2 | - | - | - | - | z2w,1=1 z2w,2=1 z2w,3=1 | Bi4Se3 |
| 166.1.313 | 74348 | 28 | 0.3465 | 0.3191 | 2 | 6 | 2 | 1 | - | - | - | - | z4=3 | Bi2Te3 |
| 166.1.314 | 658764 | 28 | 0.3458 | 0.3051 | 2 | 6 | 2 | 1 | - | - | - | - | z4=3 | Bi2Te3 |
| 166.1.315 | 44983 | 28 | 0.2704 | 0.2443 | 2 | 6 | 2 | 1 | - | - | - | - | z4=3 | Bi2Te3 |
| 166.1.316 | 42546 | 28 | 0.3516 | 0.3308 | 2 | 6 | 2 | 1 | - | - | - | - | z4=3 | Bi2Te3 |
| 166.1.317 | 193340 | 28 | 0.3495 | 0.3254 | 2 | 6 | 2 | 1 | - | - | - | - | z4=3 | Bi2Te3 |
| 166.1.318 | 193339 | 28 | 0.3531 | 0.3297 | 2 | 6 | 2 | 1 | - | - | - | - | z4=3 | Bi2Te3 |
| 166.1.319 | 193338 | 28 | 0.3455 | 0.3203 | 2 | 6 | 2 | 1 | - | - | - | - | z4=3 | Bi2Te3 |
| 166.1.320 | 193337 | 28 | 0.3517 | 0.3270 | 2 | 6 | 2 | 1 | - | - | - | - | z4=3 | Bi2Te3 |
| 166.1.321 | 193336 | 28 | 0.3470 | 0.3210 | 2 | 6 | 2 | 1 | - | - | - | - | z4=3 | Bi2Te3 |
| 166.1.322 | 193335 | 28 | 0.3561 | 0.3323 | 2 | 6 | 2 | 1 | - | - | - | - | z4=3 | Bi2Te3 |
| 166.1.323 | 193334 | 28 | 0.3558 | 0.3331 | 2 | 6 | 2 | 1 | - | - | - | - | z4=3 | Bi2Te3 |
| 166.1.324 | 193333 | 28 | 0.3560 | 0.3323 | 2 | 6 | 2 | 1 | - | - | - | - | z4=3 | Bi2Te3 |
| 166.1.325 | 193332 | 28 | 0.3571 | 0.3347 | 2 | 6 | 2 | 1 | - | - | - | - | z4=3 | Bi2Te3 |
| 166.1.326 | 193331 | 28 | 0.3565 | 0.3327 | 2 | 6 | 2 | 1 | - | - | - | - | z4=3 | Bi2Te3 |
| 166.1.327 | 193330 | 28 | 0.3605 | 0.3368 | 2 | 6 | 2 | 1 | - | - | - | - | z4=3 | Bi2Te3 |
| 166.1.328 | 184631 | 28 | 0.3517 | 0.3283 | 2 | 6 | 2 | 1 | - | - | - | - | z4=3 | Bi2Te3 |
| 166.1.329 | 158366 | 28 | 0.3529 | 0.3314 | 2 | 6 | 2 | 1 | - | - | - | - | z4=3 | Bi2Te3 |
| 166.1.330 | 15753 | 28 | 0.5349 | 0.4625 | 2 | 6 | 2 | 1 | - | - | - | - | z4=3 | Bi2Te3 |
| 166.1.331 | 42545 | 28 | 0.3244 | 0.3067 | 2 | 6 | 2 | 1 | - | - | - | - | z4=3 | Bi2Se3 |
| 166.1.332 | 20385 | 28 | 0.7109 | 0.6852 | 2 | 6 | 2 | 1 | - | - | - | - | z4=3 | Bi2Se3 |
| 166.1.333 | 165226 | 28 | 0.3162 | 0.2974 | 2 | 6 | 2 | 1 | - | - | - | - | z4=3 | Bi2Se3 |
| 166.1.334 | 54838 | 28 | 0.2191 | 0.2191 | 2 | 6 | 2 | 1 | - | - | - | - | z4=3 | Bi2Se2Te1 |
| 166.1.335 | 43512 | 28 | 0.5426 | 0.4645 | 2 | 6 | 2 | 1 | - | - | - | - | z4=3 | Bi2Se1Te2 |
| 166.1.336 | 658633 | 38 | 0.2102 | 0.1268 | 2 | 2 | 2 | 1 | - | - | - | - | z2w,1=1 z2w,2=1 z2w,3=1 z4=1 | Bi2Ge1Te4 |
| 166.1.337 | 30394 | 38 | 0.2122 | 0.1835 | 2 | 2 | 2 | 1 | - | - | - | - | z2w,1=1 z2w,2=1 z2w,3=1 z4=1 | Bi2Ge1Te4 |
| 166.1.338 | 172572 | 20 | 0.0123 | 0.0123 | 2 | 4 | 2 | 1 | - | - | - | - | z4=1 | Bi1S2Tl1 |
| 166.1.339 | 64705 | 10 | 0.1026 | -0.0778 | 2 | 5 | 148 | 2 | - | - | - | - | z4=2 | Bi1 |
| 166.1.340 | 64704 | 10 | 0.1008 | -0.0813 | 2 | 5 | 148 | 2 | - | - | - | - | z4=2 | Bi1 |
| 166.1.341 | 64703 | 10 | 0.0952 | -0.0904 | 2 | 5 | 148 | 2 | - | - | - | - | z4=2 | Bi1 |

Continued on next page

**TABLE XIV– continued**

| TQC | ICSD | NoE | DG | IndG | SubSG (w/ T) | cl | SubSG (w/o T) | cl | Max. Sub. SG | yes or no | Max. Sub. SG | yes or no | SIs (w/ T) | Stoich. Formula |
|---|---|---|---|---|---|---|---|---|---|---|---|---|---|---|
| 166.1.342 | 53797 | 10 | 0.0954 | -0.1945 | 2 | 5 | 148 | 2 | - | - | - | - | z4=2 | Bi1 |
| 166.1.343 | 246663 | 10 | 0.0942 | -0.1228 | 2 | 5 | 148 | 2 | - | - | - | - | z4=2 | Bi1 |
| 166.1.344 | 242004 | 10 | 0.1056 | 0.0498 | 2 | 5 | 148 | 2 | - | - | - | - | z4=2 | Bi1 |
| 166.1.345 | 242003 | 10 | 0.1091 | 0.0580 | 2 | 5 | 148 | 2 | - | - | - | - | z4=2 | Bi1 |
| 166.1.346 | 242002 | 10 | 0.1091 | 0.0531 | 2 | 5 | 148 | 2 | - | - | - | - | z4=2 | Bi1 |
| 166.1.347 | 242001 | 10 | 0.1090 | 0.0561 | 2 | 5 | 148 | 2 | - | - | - | - | z4=2 | Bi1 |
| 166.1.348 | 242000 | 10 | 0.1091 | 0.0549 | 2 | 5 | 148 | 2 | - | - | - | - | z4=2 | Bi1 |
| 166.1.349 | 241999 | 10 | 0.1108 | 0.0569 | 2 | 5 | 148 | 2 | - | - | - | - | z4=2 | Bi1 |
| 166.1.350 | 241984 | 10 | 0.1091 | 0.0580 | 2 | 5 | 148 | 2 | - | - | - | - | z4=2 | Bi1 |
| 166.1.351 | 241983 | 10 | 0.1083 | 0.0034 | 2 | 5 | 148 | 2 | - | - | - | - | z4=2 | Bi1 |
| 166.1.352 | 58724 | 60 | 0.0327 | -1.4726 | 2 | 6 | 2 | 1 | - | - | - | - | z4=3 | Be17Nb2 |
| 166.1.353 | 71755 | 206 | 0.0090 | -0.0660 | 2 | 3 | 0 | 2 | - | - | - | - | z2w,1=1 z2w,2=1 z2w,3=1 z4=2 | Ba7Cl2O15Ru4 |
| 166.1.354 | 73069 | 206 | 0.0157 | -0.0515 | 2 | 3 | 0 | 2 | - | - | - | - | z2w,1=1 z2w,2=1 z2w,3=1 z4=2 | Ba7Br2O15Ru4 |
| 166.1.355 | 99662 | 146 | 0.0085 | -0.1563 | 2 | 1 | 2 | 1 | - | - | - | - | z2w,1=1 z2w,2=1 z2w,3=1 z4=3 | Ba4Mn3O12Pr1 |
| 166.1.356 | 174186 | 158 | 0.0308 | -0.0615 | 2 | 5 | 0 | 2 | - | - | - | - | z4=2 | Ba4Er1O12Ru3 |
| 166.1.357 | 20621 | 42 | 0.2045 | -0.2393 | 2 | 3 | 0 | 2 | - | - | - | - | z2w,1=1 z2w,2=1 z2w,3=1 z4=2 | Ba3O9Sc4 |
| 166.1.358 | 25005 | 144 | 0.0096 | -0.1218 | 2 | 3 | 0 | 2 | - | - | - | - | z2w,1=1 z2w,2=1 z2w,3=1 z4=2 | Ba2Ni1O6Te1 |
| 166.1.359 | 58660 | 54 | 0.0142 | -0.6363 | 2 | 2 | 2 | 1 | - | - | - | - | z2w,1=1 z2w,2=1 z2w,3=1 z4=1 | Ba2Mg17 |
| 166.1.360 | 58665 | 66 | 0.0373 | -0.9065 | 2 | 4 | 2 | 1 | - | - | - | - | z4=1 | Ba1Pb3 |
| 166.1.361 | 419973 | 66 | 0.0000 | -0.9795 | 2 | 4 | 2 | 1 | - | - | - | - | z4=1 | Ba1Pb3 |
| 166.1.362 | 99682 | 108 | 0.1237 | -0.1080 | 2 | 1 | 2 | 1 | - | - | - | - | z2w,1=1 z2w,2=1 z2w,3=1 z4=3 | Ba1O3Ru1 |
| 166.1.363 | 91075 | 108 | 0.1198 | -0.1082 | 2 | 2 | 2 | 1 | - | - | - | - | z2w,1=1 z2w,2=1 z2w,3=1 z4=1 | Ba1O3Ru1 |
| 166.1.364 | 51293 | 108 | 0.1245 | -0.0994 | 2 | 2 | 2 | 1 | - | - | - | - | z2w,1=1 z2w,2=1 z2w,3=1 z4=1 | Ba1O3Ru1 |
| 166.1.365 | 172175 | 108 | 0.0545 | -0.1338 | 2 | 7 | 0 | 2 | - | - | - | - | z2w,1=1 z2w,2=1 z2w,3=1 | Ba1O3Ru1 |
| 166.1.366 | 110770 | 108 | 0.1198 | -0.1082 | 2 | 2 | 2 | 1 | - | - | - | - | z2w,1=1 z2w,2=1 z2w,3=1 z4=1 | Ba1O3Ru1 |

Continued on next page

**TABLE XIV– continued**

| TQC | ICSD | NoE | DG | IndG | SubSG (w/ T) | cl | SubSG (w/o T) | cl | Max. Sub. SG | yes or no | Max. Sub. SG | yes or no | SIs (w/ T) | Stoich. Formula |
|---|---|---|---|---|---|---|---|---|---|---|---|---|---|---|
| 166.1.367 | 10253 | 108 | 0.1359 | -0.1181 | 2 | 2 | 2 | 1 | - | - | - | - | z2w,1=1 z2w,2=1 z2w,3=1 z4=1 | Ba1O3Ru1 |
| 166.1.368 | 20898 | 98 | 0.0000 | -0.5191 | 2 | 1 | 2 | 1 | - | - | - | - | z2w,1=1 z2w,2=1 z2w,3=1 z4=3 | Ba1Ni4O8 |
| 166.1.369 | 657452 | 124 | 0.0040 | -0.0169 | 2 | 5 | 0 | 2 | - | - | - | - | z4=2 | B6Fe2Gd5 |
| 166.1.370 | 657450 | 94 | 0.0040 | -0.0278 | 2 | 2 | 2 | 1 | - | - | - | - | z2w,1=1 z2w,2=1 z2w,3=1 z4=1 | B6Ce5Fe2 |
| 166.1.371 | 39554 | 24 | 0.0430 | -3.0546 | 2 | 2 | 2 | 1 | - | - | - | - | z2w,1=1 z2w,2=1 z2w,3=1 z4=1 | B4Mo2 |
| 166.1.372 | 191908 | 24 | 0.0518 | -3.0483 | 2 | 7 | 0 | 2 | - | - | - | - | z2w,1=1 z2w,2=1 z2w,3=1 | B4Mo2 |
| 166.1.373 | 601082 | 136 | 0.0021 | 0.0003 | 2 | 6 | 2 | 1 | - | - | - | - | z4=3 | B3Co1Dy2.5 |
| 166.1.374 | 44451 | 24 | 0.0422 | -3.0363 | 2 | 2 | 2 | 1 | - | - | - | - | z2w,1=1 z2w,2=1 z2w,3=1 z4=1 | B2Mo1 |
| 166.1.375 | 418398 | 24 | 0.0387 | -3.0572 | 2 | 2 | 2 | 1 | - | - | - | - | z2w,1=1 z2w,2=1 z2w,3=1 z4=1 | B2Mo1 |
| 166.1.376 | 418397 | 24 | 0.0320 | -3.0737 | 2 | 2 | 2 | 1 | - | - | - | - | z2w,1=1 z2w,2=1 z2w,3=1 z4=1 | B2Mo1 |
| 166.1.377 | 40907 | 24 | 0.0449 | -3.0272 | 2 | 2 | 2 | 1 | - | - | - | - | z2w,1=1 z2w,2=1 z2w,3=1 z4=1 | B2Mo1 |
| 166.1.378 | 252125 | 24 | 0.0506 | -2.7496 | 2 | 2 | 2 | 1 | - | - | - | - | z2w,1=1 z2w,2=1 z2w,3=1 z4=1 | B2Mo1 |
| 166.1.379 | 97343 | 150 | 0.0115 | 0.0075 | 2 | 2 | 2 | 1 | - | - | - | - | z2w,1=1 z2w,2=1 z2w,3=1 z4=1 | B2Ce2Ir5 |
| 166.1.380 | 409982 | 54 | 0.0152 | -0.0428 | 2 | 2 | 2 | 1 | - | - | - | - | z2w,1=1 z2w,2=1 z2w,3=1 z4=1 | B1Br1Eu2N2 |
| 166.1.381 | 262037 | 72 | 0.1153 | 0.0021 | 2 | 3 | 0 | 2 | - | - | - | - | z2w,1=1 z2w,2=1 z2w,3=1 z4=2 | As3Cd4Rb1 |
| 166.1.382 | 262035 | 64 | 0.0732 | -0.0687 | 2 | 3 | 0 | 2 | - | - | - | - | z2w,1=1 z2w,2=1 z2w,3=1 z4=2 | As3Cd4Na1 |

Continued on next page

**TABLE XIV– continued**

| TQC | ICSD | NoE | DG | IndG | SubSG (w/ T) | cl | SubSG (w/o T) | cl | Max. Sub. SG | yes or no | Max. Sub. SG | yes or no | SIs (w/ T) | Stoich. Formula |
|---|---|---|---|---|---|---|---|---|---|---|---|---|---|---|
| 166.1.383 | 262032 | 72 | 0.1110 | -0.0201 | 2 | 3 | 0 | 2 | - | - | - | - | z2w,1=1<br>z2w,2=1<br>z2w,3=1<br>z4=2 | As3Cd4K1 |
| 166.1.384 | 68110 | 28 | 0.0933 | 0.0418 | 2 | 6 | 2 | 1 | - | - | - | - | z4=3 | As2Te3 |
| 166.1.385 | 82371 | 28 | 0.0924 | 0.0503 | 2 | 6 | 2 | 1 | - | - | - | - | z4=3 | As2Sn2Sr1 |
| 166.1.386 | 68114 | 68 | 0.0447 | 0.0447 | 2 | 6 | 2 | 1 | - | - | - | - | z4=3 | As2Ge4Te7 |
| 166.1.387 | 68113 | 56 | 0.0288 | -0.5770 | 2 | 6 | 2 | 1 | - | - | - | - | z4=3 | As2Ge3Te6 |
| 166.1.388 | 422526 | 18 | 0.2392 | 0.1058 | 2 | 1 | 2 | 1 | - | - | - | - | z2w,1=1<br>z2w,2=1<br>z2w,3=1<br>z4=3 | As2Ca1Ga2 |
| 166.1.389 | 53794 | 10 | 0.1659 | 0.0236 | 2 | 1 | 2 | 1 | - | - | - | - | z2w,1=1<br>z2w,2=1<br>z2w,3=1<br>z4=3 | As1 |
| 166.1.390 | 16518 | 10 | 0.0607 | 0.0607 | 2 | 1 | 2 | 1 | - | - | - | - | z2w,1=1<br>z2w,2=1<br>z2w,3=1<br>z4=3 | As1 |
| 166.1.391 | 16517 | 10 | 0.0329 | 0.0329 | 2 | 1 | 2 | 1 | - | - | - | - | z2w,1=1<br>z2w,2=1<br>z2w,3=1<br>z4=3 | As1 |
| 166.1.392 | 16516 | 10 | 0.0288 | 0.0288 | 2 | 1 | 2 | 1 | - | - | - | - | z2w,1=1<br>z2w,2=1<br>z2w,3=1<br>z4=3 | As1 |
| 166.1.393 | 281656 | 80 | 0.0299 | -0.2040 | 2 | 6 | 2 | 1 | - | - | - | - | z4=3 | Al6Gd2Pt1Si4 |
| 166.1.394 | 281660 | 84 | 0.0044 | -0.2353 | 2 | 5 | 0 | 2 | - | - | - | - | z4=2 | Al6Dy2Pt1Si4 |
| 166.1.395 | 58175 | 84 | 0.0031 | -0.0282 | 2 | 3 | 0 | 2 | - | - | - | - | z2w,1=1<br>z2w,2=1<br>z2w,3=1<br>z4=2 | Al3Tb1 |
| 166.1.396 | 420092 | 100 | 0.1663 | -0.0321 | 2 | 4 | 2 | 1 | - | - | - | - | z4=1 | Al10Ba7 |
| 166.1.397 | 424311 | 64 | 0.1395 | -0.7315 | 2 | 3 | 0 | 2 | - | - | - | - | z2w,1=1<br>z2w,2=1<br>z2w,3=1<br>z4=2 | Ag4Sb2Sr1 |
| 166.1.398 | 416373 | 44 | 0.1748 | -0.7447 | 2 | 2 | 2 | 1 | - | - | - | - | z2w,1=1<br>z2w,2=1<br>z2w,3=1<br>z4=1 | Ag2Ni1O2 |
| 166.1.399 | 412279 | 44 | 0.1766 | -0.7482 | 2 | 2 | 2 | 1 | - | - | - | - | z2w,1=1<br>z2w,2=1<br>z2w,3=1<br>z4=1 | Ag2Ni1O2 |
| 166.1.400 | 172558 | 44 | 0.1759 | -0.7570 | 2 | 2 | 2 | 1 | - | - | - | - | z2w,1=1<br>z2w,2=1<br>z2w,3=1<br>z4=1 | Ag2Ni1O2 |
| 166.1.401 | 160582 | 44 | 0.1754 | -0.7403 | 2 | 2 | 2 | 1 | - | - | - | - | z2w,1=1<br>z2w,2=1<br>z2w,3=1<br>z4=1 | Ag2Ni1O2 |

Continued on next page

**TABLE XIV– continued**

| TQC | ICSD | NoE | DG | IndG | SubSG (w/ T) | cl | SubSG (w/o T) | cl | Max. Sub. SG | yes or no | Max. Sub. SG | yes or no | SIs (w/ T) | Stoich. Formula |
|---|---|---|---|---|---|---|---|---|---|---|---|---|---|---|
| 166.1.402 | 160581 | 44 | 0.1757 | -0.7406 | 2 | 2 | 2 | 1 | - | - | - | - | z2w,1=1<br>z2w,2=1<br>z2w,3=1<br>z4=1 | Ag2Ni1O2 |
| 166.1.403 | 160580 | 44 | 0.1754 | -0.7419 | 2 | 2 | 2 | 1 | - | - | - | - | z2w,1=1<br>z2w,2=1<br>z2w,3=1<br>z4=1 | Ag2Ni1O2 |
| 166.1.404 | 160579 | 44 | 0.1756 | -0.7432 | 2 | 2 | 2 | 1 | - | - | - | - | z2w,1=1<br>z2w,2=1<br>z2w,3=1<br>z4=1 | Ag2Ni1O2 |
| 166.1.405 | 160578 | 44 | 0.1759 | -0.7453 | 2 | 2 | 2 | 1 | - | - | - | - | z2w,1=1<br>z2w,2=1<br>z2w,3=1<br>z4=1 | Ag2Ni1O2 |
| 166.1.406 | 160577 | 44 | 0.1757 | -0.7469 | 2 | 2 | 2 | 1 | - | - | - | - | z2w,1=1<br>z2w,2=1<br>z2w,3=1<br>z4=1 | Ag2Ni1O2 |
| 166.1.407 | 160576 | 44 | 0.1763 | -0.7474 | 2 | 2 | 2 | 1 | - | - | - | - | z2w,1=1<br>z2w,2=1<br>z2w,3=1<br>z4=1 | Ag2Ni1O2 |
| 166.1.408 | 160575 | 44 | 0.1762 | -0.7478 | 2 | 2 | 2 | 1 | - | - | - | - | z2w,1=1<br>z2w,2=1<br>z2w,3=1<br>z4=1 | Ag2Ni1O2 |
| 166.1.409 | 160574 | 44 | 0.1759 | -0.7495 | 2 | 2 | 2 | 1 | - | - | - | - | z2w,1=1<br>z2w,2=1<br>z2w,3=1<br>z4=1 | Ag2Ni1O2 |
| 166.1.410 | 251620 | 68 | 0.0427 | -2.5145 | 2 | 1 | 2 | 1 | - | - | - | - | z2w,1=1<br>z2w,2=1<br>z2w,3=1<br>z4=3 | Ag1Ge1Li2 |
| 166.3.1 | 653256 | 228 | - | - | - | - | - | - | 12 | no(2) | 12 | no(2) | z4=3 | Th2Zn17 |
| 166.3.2 | 20238 | 228 | - | - | - | - | - | - | 12 | no(2) | 12 | yes | z4=2 | Th2Zn17 |
| 166.3.3 | 653096 | 38 | - | - | - | - | - | - | 12 | yes | 12 | yes | - | Te2Tl1Tm1 |
| 166.3.4 | 52500 | 6 | - | - | - | - | - | - | 12 | yes | 12 | yes | - | Te1 |
| 166.3.5 | 652991 | 242 | - | - | - | - | - | - | 12 | no(2) | 12 | yes | z2w,1=1<br>z2w,2=1<br>z2w,3=1<br>z4=2 | Tb2Zn17 |
| 166.3.6 | 104244 | 242 | - | - | - | - | - | - | 12 | no(2) | 12 | yes | z2w,1=1<br>z2w,2=1<br>z2w,3=1<br>z4=2 | Tb2Zn17 |
| 166.3.7 | 652071 | 38 | - | - | - | - | - | - | 12 | yes | 12 | yes | - | Se2Tl1Tm1 |
| 166.3.8 | 656457 | 6 | - | - | - | - | - | - | 12 | yes | 12 | yes | - | Se1 |
| 166.3.9 | 57181 | 6 | - | - | - | - | - | - | 12 | yes | 12 | yes | - | Se1 |
| 166.3.10 | 649161 | 6 | - | - | - | - | - | - | 12 | yes | 12 | yes | - | Po1 |
| 166.3.11 | 648844 | 48 | - | - | - | - | - | - | 12 | no(2) | 12 | no(2) | z2w,1=1<br>z2w,2=1<br>z2w,3=1<br>z4=1 | Pd3Se2Tl2 |

Continued on next page

**TABLE XIV– continued**

| TQC | ICSD | NoE | DG | IndG | SubSG (w/ T) | cl | SubSG (w/o T) | cl | Max. Sub. SG | yes or no | Max. Sub. SG | yes or no | SIs (w/ T) | Stoich. Formula |
|---|---|---|---|---|---|---|---|---|---|---|---|---|---|---|
| 166.3.12 | 648843 | 48 | - | - | - | - | - | - | 12 | no(2) | 12 | no(2) | z2w,1=1<br>z2w,2=1<br>z2w,3=1<br>z4=3 | Pd3Se2Tl2 |
| 166.3.13 | 648366 | 50 | - | - | - | - | - | - | 12 | no(2) | 12 | yes | z2w,1=1<br>z2w,2=1<br>z2w,3=1<br>z4=2 | Pb2Pd3Se2 |
| 166.3.14 | 648365 | 50 | - | - | - | - | - | - | 12 | no(2) | 12 | yes | z2w,1=1<br>z2w,2=1<br>z2w,3=1 | Pb2Pd3Se2 |
| 166.3.15 | 421874 | 50 | - | - | - | - | - | - | 12 | no(2) | 12 | yes | z2w,1=1<br>z2w,2=1<br>z2w,3=1 | Pb2Pd3Se2 |
| 166.3.16 | 648158 | 38 | - | - | - | - | - | - | 12 | no(2) | 12 | no(2) | z2w,1=1<br>z2w,2=1<br>z2w,3=1<br>z4=3 | P3Sn1 |
| 166.3.17 | 16293 | 38 | - | - | - | - | - | - | 12 | no(2) | 12 | no(2) | z2w,1=1<br>z2w,2=1<br>z2w,3=1<br>z4=3 | P3Sn1 |
| 166.3.18 | 48205 | 162 | - | - | - | - | - | - | 12 | no(2) | 12 | no(2) | z2w,1=1<br>z2w,2=1<br>z2w,3=1<br>z4=3 | O6Pb1Re2 |
| 166.3.19 | 647076 | 184 | - | - | - | - | - | - | 12 | no(2) | 12 | no(2) | z4=1 | Ni7Y2 |
| 166.3.20 | 647060 | 184 | - | - | - | - | - | - | 12 | no(2) | 12 | no(2) | z4=1 | Ni7Y2 |
| 166.3.21 | 647051 | 184 | - | - | - | - | - | - | 12 | no(2) | 12 | no(2) | z4=1 | Ni7Y2 |
| 166.3.22 | 108650 | 184 | - | - | - | - | - | - | 12 | no(2) | 12 | no(2) | z4=1 | Ni7Y2 |
| 166.3.23 | 646229 | 192 | - | - | - | - | - | - | 12 | yes | 12 | yes | - | Ni7Pr2 |
| 166.3.24 | 185343 | 192 | - | - | - | - | - | - | 12 | no(2) | 12 | yes | z4=2 | Ni7Pr2 |
| 166.3.25 | 76730 | 32 | - | - | - | - | - | - | 12 | no(2) | 12 | yes | z4=2 | Ni1Te1 |
| 166.3.26 | 642642 | 254 | - | - | - | - | - | - | 12 | no(2) | 12 | no(2) | z4=3 | Lu2Zn17 |
| 166.3.27 | 104247 | 254 | - | - | - | - | - | - | 12 | no(2) | 12 | no(2) | z4=3 | Lu2Zn17 |
| 166.3.28 | 104760 | 46 | - | - | - | - | - | - | 12 | no(2) | 12 | no(2) | z2w,1=1<br>z2w,2=1<br>z2w,3=1<br>z4=3 | Li5Na2Sn4 |
| 166.3.29 | 642087 | 226 | - | - | - | - | - | - | 12 | no(2) | 12 | no(2) | z4=3 | La2Zn17 |
| 166.3.30 | 104240 | 226 | - | - | - | - | - | - | 12 | no(2) | 12 | no(2) | z4=3 | La2Zn17 |
| 166.3.31 | 641740 | 170 | - | - | - | - | - | - | 12 | no(2) | 12 | yes | z2w,1=1<br>z2w,2=1<br>z2w,3=1<br>z4=2 | La2Rh7 |
| 166.3.32 | 641502 | 184 | - | - | - | - | - | - | 12 | no(2) | 12 | no(2) | z4=1 | La2Ni7 |
| 166.3.33 | 29335 | 82 | - | - | - | - | - | - | 12 | no(2) | 12 | yes | z2w,1=1<br>z2w,2=1<br>z2w,3=1<br>z4=2 | K2O8Pb1S2 |
| 166.3.34 | 640751 | 170 | - | - | - | - | - | - | 12 | yes | 12 | yes | - | Ir7La2 |
| 166.3.35 | 640512 | 24 | - | - | - | - | - | - | 12 | no(2) | 12 | yes | z2w,1=1<br>z2w,2=1<br>z2w,3=1 | In2Se3 |
| 166.3.36 | 640498 | 24 | - | - | - | - | - | - | 12 | no(2) | 12 | yes | z2w,1=1<br>z2w,2=1<br>z2w,3=1 | In2Se3 |

Continued on next page

**TABLE XIV– continued**

| TQC | ICSD | NoE | DG | IndG | SubSG (w/ T) | cl | SubSG (w/o T) | cl | Max. Sub. SG | yes or no | Max. Sub. SG | yes or no | SIs (w/ T) | Stoich. Formula |
|---|---|---|---|---|---|---|---|---|---|---|---|---|---|---|
| 166.3.37 | 640492 | 24 | - | - | - | - | - | - | 12 | no(2) | 12 | yes | z2w,1=1<br>z2w,2=1<br>z2w,3=1 | In2Se3 |
| 166.3.38 | 640489 | 24 | - | - | - | - | - | - | 12 | no(2) | 12 | yes | z2w,1=1<br>z2w,2=1<br>z2w,3=1 | In2Se3 |
| 166.3.39 | 17007 | 24 | - | - | - | - | - | - | 12 | no(2) | 12 | yes | z2w,1=1<br>z2w,2=1<br>z2w,3=1 | In2Se3 |
| 166.3.40 | 640213 | 46 | - | - | - | - | - | - | 12 | no(2) | 12 | no(2) | z2w,1=1<br>z2w,2=1<br>z2w,3=1<br>z4=3 | In1Pb1Rh3S2 |
| 166.3.41 | 5440 | 46 | - | - | - | - | - | - | 12 | no(2) | 12 | no(2) | z2w,1=1<br>z2w,2=1<br>z2w,3=1<br>z4=1 | In1Pb1Rh3S2 |
| 166.3.42 | 639452 | 224 | - | - | - | - | - | - | 12 | no(2) | 12 | yes | z4=2 | Ho2Ni7 |
| 166.3.43 | 639712 | 36 | - | - | - | - | - | - | 12 | yes | 12 | yes | - | Ho1Se2Tl1 |
| 166.3.44 | 106992 | 36 | - | - | - | - | - | - | 12 | yes | 12 | yes | - | Ho1Se2Tl1 |
| 166.3.45 | 639673 | 36 | - | - | - | - | - | - | 12 | yes | 12 | yes | - | Ho1S2Tl1 |
| 166.3.46 | 291908 | 34 | - | - | - | - | - | - | 12 | yes | 12 | yes | - | Ho1Na1O2 |
| 166.3.47 | 639371 | 34 | - | - | - | - | - | - | 12 | yes | 12 | yes | - | Ho1Li1S2 |
| 166.3.48 | 44959 | 34 | - | - | - | - | - | - | 12 | yes | 12 | yes | - | Ho1Li1S2 |
| 166.3.49 | 157720 | 132 | - | - | - | - | - | - | 12 | no(2) | 12 | no(2) | z2w,1=1<br>z2w,2=1<br>z2w,3=1<br>z4=1 | H9Fe3O15S2 |
| 166.3.50 | 40946 | 114 | - | - | - | - | - | - | 12 | no(2) | 12 | yes | z2w,1=1<br>z2w,2=1<br>z2w,3=1 | H6Ba1O14Ta4 |
| 166.3.51 | 27641 | 114 | - | - | - | - | - | - | 12 | no(2) | 12 | no(2) | z2w,1=1<br>z2w,2=1<br>z2w,3=1<br>z4=1 | H6Al3K1O14S2 |
| 166.3.52 | 24157 | 114 | - | - | - | - | - | - | 12 | no(2) | 12 | no(2) | z2w,1=1<br>z2w,2=1<br>z2w,3=1<br>z4=1 | H6Al3K1O14S2 |
| 166.3.53 | 81510 | 68 | - | - | - | - | - | - | 12 | yes | 12 | yes | - | H4.3Dy1Mn2 |
| 166.3.54 | 623738 | 150 | - | - | - | - | - | - | 12 | yes | 12 | yes | - | H3Co3Tm1 |
| 166.3.55 | 623734 | 138 | - | - | - | - | - | - | 12 | no(2) | 12 | no(2) | z4=3 | H3Co3Tb1 |
| 166.3.56 | 656645 | 178 | - | - | - | - | - | - | 12 | no(2) | 12 | no(2) | z4=1 | H3Co17Y2 |
| 166.3.57 | 64956 | 100 | - | - | - | - | - | - | 12 | no(2) | 12 | no(2) | z4=3 | H3Cl1Cu2O3 |
| 166.3.58 | 630464 | 138 | - | - | - | - | - | - | 12 | yes | 12 | yes | - | H3.5Er1Fe3 |
| 166.3.59 | 16294 | 38 | - | - | - | - | - | - | 12 | no(2) | 12 | no(2) | z2w,1=1<br>z2w,2=1<br>z2w,3=1<br>z4=3 | Ge1P3 |
| 166.3.60 | 636517 | 240 | - | - | - | - | - | - | 12 | no(2) | 12 | yes | z4=2 | Gd2Zn17 |
| 166.3.61 | 104152 | 240 | - | - | - | - | - | - | 12 | no(2) | 12 | yes | z4=2 | Gd2Zn17 |
| 166.3.62 | 104151 | 240 | - | - | - | - | - | - | 12 | no(2) | 12 | no(2) | z2w,1=1<br>z2w,2=1<br>z2w,3=1<br>z4=3 | Gd2Zn17 |
| 166.3.63 | 104093 | 212 | - | - | - | - | - | - | 12 | no(2) | 12 | no(2) | z4=3 | Gd2Ni7 |
| 166.3.64 | 636033 | 144 | - | - | - | - | - | - | 12 | yes | 12 | yes | - | Gd1Ni3 |
| 166.3.65 | 636021 | 144 | - | - | - | - | - | - | 12 | yes | 12 | yes | - | Gd1Ni3 |

Continued on next page

**TABLE XIV– continued**

| TQC | ICSD | NoE | DG | IndG | SubSG cl (w/ T) | SubSG cl (w/o T) | Max. Sub. SG | yes or no | Max. Sub. SG | yes or no | SIs (w/ T) | Stoich. Formula |
|---|---|---|---|---|---|---|---|---|---|---|---|---|
| 166.3.66 | 236382 | 144 | - | - | - - | - - | 12 | no(2) | 12 | no(2) | z4=3 | Gd1Ni3 |
| 166.3.67 | 104090 | 144 | - | - | - - | - - | 12 | yes | 12 | yes | - | Gd1Ni3 |
| 166.3.68 | 246480 | 100 | - | - | - - | - - | 12 | no(2) | 12 | no(2) | z2w,1=1 z2w,2=1 z2w,3=1 z4=3 | Ga3Pu1 |
| 166.3.69 | 103940 | 100 | - | - | - - | - - | 12 | no(2) | 12 | no(2) | z2w,1=1 z2w,2=1 z2w,3=1 z4=3 | Ga3Pu1 |
| 166.3.70 | 633791 | 86 | - | - | - - | - - | 12 | no(2) | 12 | no(2) | z4=1 | Fe7Ta6 |
| 166.3.71 | 631751 | 148 | - | - | - - | - - | 12 | no(2) | 12 | yes | z2w,1=1 z2w,2=1 z2w,3=1 | Fe7Ga6 |
| 166.3.72 | 634081 | 158 | - | - | - - | - - | 12 | no(2) | 12 | no(2) | z4=3 | Fe17Y2 |
| 166.3.73 | 291058 | 158 | - | - | - - | - - | 12 | no(2) | 12 | no(2) | z2w,1=1 z2w,2=1 z2w,3=1 z4=3 | Fe17Y2 |
| 166.3.74 | 658466 | 160 | - | - | - - | - - | 12 | no(2) | 12 | no(2) | z4=3 | Fe17Th2 |
| 166.3.75 | 656795 | 160 | - | - | - - | - - | 12 | no(2) | 12 | no(2) | z4=3 | Fe17Th2 |
| 166.3.76 | 633921 | 160 | - | - | - - | - - | 12 | no(2) | 12 | no(2) | z4=3 | Fe17Th2 |
| 166.3.77 | 633916 | 160 | - | - | - - | - - | 12 | no(2) | 12 | no(2) | z4=3 | Fe17Th2 |
| 166.3.78 | 633906 | 160 | - | - | - - | - - | 12 | no(2) | 12 | no(2) | z4=3 | Fe17Th2 |
| 166.3.79 | 633892 | 160 | - | - | - - | - - | 12 | no(2) | 12 | no(2) | z4=3 | Fe17Th2 |
| 166.3.80 | 23224 | 160 | - | - | - - | - - | 12 | no(2) | 12 | no(2) | z4=3 | Fe17Th2 |
| 166.3.81 | 106684 | 160 | - | - | - - | - - | 12 | no(2) | 12 | no(2) | z4=3 | Fe17Th2 |
| 166.3.82 | 633846 | 174 | - | - | - - | - - | 12 | no(2) | 12 | no(2) | z2w,1=1 z2w,2=1 z2w,3=1 z4=1 | Fe17Tb2 |
| 166.3.83 | 633812 | 174 | - | - | - - | - - | 12 | no(2) | 12 | no(2) | z2w,1=1 z2w,2=1 z2w,3=1 z4=1 | Fe17Tb2 |
| 166.3.84 | 657634 | 162 | - | - | - - | - - | 12 | no(2) | 12 | yes | z4=2 | Fe17Pr2 |
| 166.3.85 | 657633 | 162 | - | - | - - | - - | 12 | no(2) | 12 | no(2) | z4=3 | Fe17Pr2 |
| 166.3.86 | 602990 | 162 | - | - | - - | - - | 12 | no(2) | 12 | no(2) | z4=3 | Fe17Pr2 |
| 166.3.87 | 185381 | 162 | - | - | - - | - - | 12 | no(2) | 12 | no(2) | z4=3 | Fe17Pr2 |
| 166.3.88 | 185380 | 162 | - | - | - - | - - | 12 | no(2) | 12 | yes | z4=2 | Fe17Pr2 |
| 166.3.89 | 163873 | 162 | - | - | - - | - - | 12 | no(2) | 12 | no(2) | z4=3 | Fe17Pr2 |
| 166.3.90 | 657210 | 164 | - | - | - - | - - | 12 | no(2) | 12 | no(2) | z4=3 | Fe17Nd2 |
| 166.3.91 | 657082 | 164 | - | - | - - | - - | 12 | no(2) | 12 | no(2) | z4=3 | Fe17Nd2 |
| 166.3.92 | 632854 | 164 | - | - | - - | - - | 12 | yes | 12 | yes | - | Fe17Nd2 |
| 166.3.93 | 632852 | 164 | - | - | - - | - - | 12 | no(2) | 12 | no(2) | z4=3 | Fe17Nd2 |
| 166.3.94 | 632843 | 164 | - | - | - - | - - | 12 | no(2) | 12 | yes | z4=2 | Fe17Nd2 |
| 166.3.95 | 632839 | 164 | - | - | - - | - - | 12 | no(2) | 12 | no(2) | z4=3 | Fe17Nd2 |
| 166.3.96 | 632837 | 164 | - | - | - - | - - | 12 | no(2) | 12 | no(2) | z4=3 | Fe17Nd2 |
| 166.3.97 | 190296 | 164 | - | - | - - | - - | 12 | no(2) | 12 | no(2) | z4=3 | Fe17Nd2 |
| 166.3.98 | 190295 | 164 | - | - | - - | - - | 12 | no(2) | 12 | no(2) | z4=3 | Fe17Nd2 |
| 166.3.99 | 190294 | 164 | - | - | - - | - - | 12 | no(2) | 12 | no(2) | z4=3 | Fe17Nd2 |
| 166.3.100 | 190293 | 164 | - | - | - - | - - | 12 | no(2) | 12 | no(2) | z4=3 | Fe17Nd2 |
| 166.3.101 | 190292 | 164 | - | - | - - | - - | 12 | no(2) | 12 | no(2) | z4=3 | Fe17Nd2 |
| 166.3.102 | 190272 | 164 | - | - | - - | - - | 12 | no(2) | 12 | no(2) | z4=3 | Fe17Nd2 |
| 166.3.103 | 190229 | 164 | - | - | - - | - - | 12 | no(2) | 12 | no(2) | z4=3 | Fe17Nd2 |
| 166.3.104 | 190226 | 164 | - | - | - - | - - | 12 | no(2) | 12 | no(2) | z4=3 | Fe17Nd2 |
| 166.3.105 | 1 - 09 | 164 | - | - | - - | - - | 12 | no(2) | 12 | no(2) | z4=3 | Fe17Nd2 |

Continued on next page

**TABLE XIV– continued**

| TQC | ICSD | NoE | DG | IndG | SubSG (w/ T) | cl | SubSG (w/o T) | cl | Max. Sub. SG | yes or no | Max. Sub. SG | yes or no | SIs (w/ T) | Stoich. Formula |
|---|---|---|---|---|---|---|---|---|---|---|---|---|---|---|
| 166.3.106 | 107530 | 164 | - | - | - | - | - | - | 12 | no(2) | 12 | no(2) | z4=3 | Fe17Nd2 |
| 166.3.107 | 106661 | 164 | - | - | - | - | - | - | 12 | no(2) | 12 | no(2) | z4=3 | Fe17Nd2 |
| 166.3.108 | 603446 | 162 | - | - | - | - | - | - | 12 | no(2) | 12 | no(2) | z4=3 | Fe17N3Pr2 |
| 166.3.109 | 602496 | 172 | - | - | - | - | - | - | 12 | no(2) | 12 | yes | z2w,1=1<br>z2w,2=1<br>z2w,3=1 | Fe17Gd2 |
| 166.3.110 | 184009 | 64 | - | - | - | - | - | - | 12 | no(2) | 12 | no(2) | z2w,1=1<br>z2w,2=1<br>z2w,3=1<br>z4=3 | F1O1Tb1 |
| 166.3.111 | 184008 | 64 | - | - | - | - | - | - | 12 | no(2) | 12 | no(2) | z2w,1=1<br>z2w,2=1<br>z2w,3=1<br>z4=3 | F1O1Tb1 |
| 166.3.112 | 261705 | 140 | - | - | - | - | - | - | 12 | no(2) | 12 | no(2) | z4=1 | Eu1In1Ni5 |
| 166.3.113 | 630816 | 228 | - | - | - | - | - | - | 12 | no(2) | 12 | yes | z2w,1=1<br>z2w,2=1<br>z2w,3=1 | Er2Ni7 |
| 166.3.114 | 630819 | 156 | - | - | - | - | - | - | 12 | no(2) | 12 | yes | z2w,1=1<br>z2w,2=1<br>z2w,3=1 | Er1Ni3 |
| 166.3.115 | 630813 | 156 | - | - | - | - | - | - | 12 | no(2) | 12 | yes | z2w,1=1<br>z2w,2=1<br>z2w,3=1 | Er1Ni3 |
| 166.3.116 | 152898 | 156 | - | - | - | - | - | - | 12 | no(2) | 12 | yes | z2w,1=1<br>z2w,2=1<br>z2w,3=1 | Er1Ni3 |
| 166.3.117 | 152845 | 156 | - | - | - | - | - | - | 12 | no(2) | 12 | yes | z2w,1=1<br>z2w,2=1<br>z2w,3=1 | Er1Ni3 |
| 166.3.118 | 630418 | 138 | - | - | - | - | - | - | 12 | no(2) | 12 | yes | z2w,1=1<br>z2w,2=1<br>z2w,3=1 | Er1Fe3 |
| 166.3.119 | 602052 | 138 | - | - | - | - | - | - | 12 | no(2) | 12 | yes | z2w,1=1<br>z2w,2=1<br>z2w,3=1 | Er1Fe3 |
| 166.3.120 | 103385 | 244 | - | - | - | - | - | - | 12 | no(2) | 12 | no(2) | z4=3 | Dy2Zn17 |
| 166.3.121 | 189470 | 176 | - | - | - | - | - | - | 12 | no(2) | 12 | no(2) | z4=3 | Dy2Fe17 |
| 166.3.122 | 418526 | 86 | - | - | - | - | - | - | 12 | no(2) | 12 | yes | z2w,1=1<br>z2w,2=1<br>z2w,3=1 | Dy1O1P1Zn1 |
| 166.3.123 | 629614 | 132 | - | - | - | - | - | - | 12 | no(2) | 12 | yes | z2w,1=1<br>z2w,2=1<br>z2w,3=1 | Dy1Fe3 |
| 166.3.124 | 629588 | 132 | - | - | - | - | - | - | 12 | no(2) | 12 | yes | z2w,1=1<br>z2w,2=1<br>z2w,3=1 | Dy1Fe3 |
| 166.3.125 | 629573 | 132 | - | - | - | - | - | - | 12 | no(2) | 12 | yes | z2w,1=1<br>z2w,2=1<br>z2w,3=1 | Dy1Fe3 |
| 166.3.126 | 629569 | 132 | - | - | - | - | - | - | 12 | no(2) | 12 | yes | z2w,1=1<br>z2w,2=1<br>z2w,3=1 | Dy1Fe3 |
| 166.3.127 | 629557 | 132 | - | - | - | - | - | - | 12 | no(2) | 12 | yes | z2w,1=1<br>z2w,2=1<br>z2w,3=1 | Dy1Fe3 |

Continued on next page

**TABLE XIV– continued**

| TQC | ICSD | NoE | DG | IndG | SubSG (w/ T) | cl | SubSG (w/o T) | cl | Max. Sub. SG | yes or no | Max. Sub. SG | yes or no | SIs (w/ T) | Stoich. Formula |
|---|---|---|---|---|---|---|---|---|---|---|---|---|---|---|
| 166.3.128 | 180280 | 132 | - | - | - | - | - | - | 12 | no(2) | 12 | yes | z2w,1=1<br>z2w,2=1<br>z2w,3=1 | Dy1Fe3 |
| 166.3.129 | 152847 | 156 | - | - | - | - | - | - | 12 | no(2) | 12 | no(2) | z4=3 | D5Er1Ni3 |
| 166.3.130 | 152846 | 156 | - | - | - | - | - | - | 12 | no(2) | 12 | no(2) | z4=3 | D4Er1Ni3 |
| 166.3.131 | 170228 | 172 | - | - | - | - | - | - | 12 | no(2) | 12 | no(2) | z2w,1=1<br>z2w,2=1<br>z2w,3=1<br>z4=3 | Cu1S4Ti2 |
| 166.3.132 | 80261 | 36 | - | - | - | - | - | - | 12 | no(2) | 12 | no(2) | z4=3 | Cu1I1 |
| 166.3.133 | 80258 | 36 | - | - | - | - | - | - | 12 | no(2) | 12 | no(2) | z4=3 | Cu1I1 |
| 166.3.134 | 80256 | 36 | - | - | - | - | - | - | 12 | no(2) | 12 | no(2) | z4=3 | Cu1I1 |
| 166.3.135 | 80255 | 36 | - | - | - | - | - | - | 12 | no(2) | 12 | no(2) | z4=3 | Cu1I1 |
| 166.3.136 | 80254 | 36 | - | - | - | - | - | - | 12 | no(2) | 12 | no(2) | z4=3 | Cu1I1 |
| 166.3.137 | 78267 | 36 | - | - | - | - | - | - | 12 | no(2) | 12 | no(2) | z4=3 | Cu1I1 |
| 166.3.138 | 78266 | 36 | - | - | - | - | - | - | 12 | no(2) | 12 | no(2) | z4=3 | Cu1I1 |
| 166.3.139 | 163445 | 36 | - | - | - | - | - | - | 12 | no(2) | 12 | no(2) | z4=3 | Cu1I1 |
| 166.3.140 | 163444 | 36 | - | - | - | - | - | - | 12 | no(2) | 12 | no(2) | z4=3 | Cu1I1 |
| 166.3.141 | 10342 | 58 | - | - | - | - | - | - | 12 | yes | 12 | yes | - | Cr1F6Sr1 |
| 166.3.142 | 656093 | 170 | - | - | - | - | - | - | 12 | yes | 12 | yes | - | Co7Y2 |
| 166.3.143 | 625574 | 170 | - | - | - | - | - | - | 12 | yes | 12 | yes | - | Co7Y2 |
| 166.3.144 | 624343 | 182 | - | - | - | - | - | - | 12 | no(2) | 12 | no(2) | z2w,1=1<br>z2w,2=1<br>z2w,3=1<br>z4=1 | Co7Nd2 |
| 166.3.145 | 623266 | 198 | - | - | - | - | - | - | 12 | yes | 12 | yes | - | Co7Gd2 |
| 166.3.146 | 624692 | 120 | - | - | - | - | - | - | 12 | no(2) | 12 | no(2) | z2w,1=1<br>z2w,2=1<br>z2w,3=1<br>z4=1 | Co3Pr1 |
| 166.3.147 | 425137 | 46 | - | - | - | - | - | - | 12 | no(2) | 12 | yes | z4=2 | Co3In1S2Sn1 |
| 166.3.148 | 622335 | 228 | - | - | - | - | - | - | 12 | no(2) | 12 | yes | z4=2 | Ce2Zn17 |
| 166.3.149 | 622324 | 228 | - | - | - | - | - | - | 12 | no(2) | 12 | yes | z4=2 | Ce2Zn17 |
| 166.3.150 | 622318 | 228 | - | - | - | - | - | - | 12 | no(2) | 12 | yes | z4=2 | Ce2Zn17 |
| 166.3.151 | 102308 | 228 | - | - | - | - | - | - | 12 | no(2) | 12 | yes | z4=2 | Ce2Zn17 |
| 166.3.152 | 658442 | 160 | - | - | - | - | - | - | 12 | no(2) | 12 | no(2) | z4=3 | Ce2Fe17 |
| 166.3.153 | 657969 | 160 | - | - | - | - | - | - | 12 | no(2) | 12 | no(2) | z4=3 | Ce2Fe17 |
| 166.3.154 | 657614 | 160 | - | - | - | - | - | - | 12 | no(2) | 12 | no(2) | z4=3 | Ce2Fe17 |
| 166.3.155 | 657080 | 160 | - | - | - | - | - | - | 12 | no(2) | 12 | no(2) | z4=3 | Ce2Fe17 |
| 166.3.156 | 656479 | 160 | - | - | - | - | - | - | 12 | no(2) | 12 | no(2) | z4=3 | Ce2Fe17 |
| 166.3.157 | 656345 | 160 | - | - | - | - | - | - | 12 | no(2) | 12 | no(2) | z4=3 | Ce2Fe17 |
| 166.3.158 | 621011 | 160 | - | - | - | - | - | - | 12 | no(2) | 12 | no(2) | z4=3 | Ce2Fe17 |
| 166.3.159 | 620990 | 160 | - | - | - | - | - | - | 12 | no(2) | 12 | no(2) | z4=3 | Ce2Fe17 |
| 166.3.160 | 603450 | 160 | - | - | - | - | - | - | 12 | no(2) | 12 | no(2) | z4=3 | Ce2Fe17 |
| 166.3.161 | 603331 | 160 | - | - | - | - | - | - | 12 | no(2) | 12 | no(2) | z4=3 | Ce2Fe17 |
| 166.3.162 | 190316 | 160 | - | - | - | - | - | - | 12 | no(2) | 12 | no(2) | z4=3 | Ce2Fe17 |
| 166.3.163 | 190249 | 160 | - | - | - | - | - | - | 12 | no(2) | 12 | no(2) | z4=3 | Ce2Fe17 |
| 166.3.164 | 151866 | 160 | - | - | - | - | - | - | 12 | no(2) | 12 | no(2) | z4=3 | Ce2Fe17 |
| 166.3.165 | 151865 | 160 | - | - | - | - | - | - | 12 | no(2) | 12 | no(2) | z4=3 | Ce2Fe17 |
| 166.3.166 | 151864 | 160 | - | - | - | - | - | - | 12 | no(2) | 12 | no(2) | z4=3 | Ce2Fe17 |
| 166.3.167 | 151863 | 160 | - | - | - | - | - | - | 12 | no(2) | 12 | no(2) | z4=3 | Ce2Fe17 |
| 166.3.168 | 151862 | 160 | - | - | - | - | - | - | 12 | no(2) | 12 | no(2) | z4=3 | Ce2Fe17 |
| 166.3.169 | 109483 | 160 | - | - | - | - | - | - | 12 | no(2) | 12 | no(2) | z4=3 | Ce2Fe17 |
| 166.3.170 | 106387 | 160 | - | - | - | - | - | - | 12 | no(2) | 12 | no(2) | z4=3 | Ce2Fe17 |

Continued on next page

**TABLE XIV– continued**

| TQC | ICSD | NoE | DG | IndG | SubSG cl (w/ T) | SubSG cl (w/o T) | Max. Sub. SG | yes or no | Max. Sub. SG | yes or no | SIs (w/ T) | Stoich. Formula |
|---|---|---|---|---|---|---|---|---|---|---|---|---|
| 166.3.171 | 621019 | 150 | - | - | - - | - - | 12 | no(2) | 12 | no(2) | z2w,1=1<br>z2w,2=1<br>z2w,3=1<br>z4=3 | Ce2Fe15Ga2 |
| 166.3.172 | 21011 | 58 | - | - | - - | - - | 12 | yes | 12 | yes | - | Ce2F4Se1 |
| 166.3.173 | 262033 | 72 | - | - | - - | - - | 12 | no(2) | 12 | no(2) | z4=3 | Cd4K1P3 |
| 166.3.174 | 619443 | 148 | - | - | - - | - - | 12 | no(2) | 12 | no(2) | z2w,1=1<br>z2w,2=1<br>z2w,3=1<br>z4=1 | Ca2Ni7 |
| 166.3.175 | 617904 | 170 | - | - | - - | - - | 12 | yes | 12 | yes | - | C3Fe17Y2 |
| 166.3.176 | 617875 | 194 | - | - | - - | - - | 12 | no(2) | 12 | yes | z4=2 | C3Fe17Tm2 |
| 166.3.177 | 657167 | 174 | - | - | - - | - - | 12 | no(2) | 12 | yes | z2w,1=1<br>z2w,2=1<br>z2w,3=1 | C3Fe17Pr2 |
| 166.3.178 | 657170 | 184 | - | - | - - | - - | 12 | no(2) | 12 | yes | z2w,1=1<br>z2w,2=1<br>z2w,3=1 | C3Fe17Gd2 |
| 166.3.179 | 657166 | 172 | - | - | - - | - - | 12 | no(2) | 12 | yes | z2w,1=1<br>z2w,2=1<br>z2w,3=1 | C3Ce2Fe17 |
| 166.3.180 | 41261 | 56 | - | - | - - | - - | 12 | no(2) | 12 | yes | z2w,1=1<br>z2w,2=1<br>z2w,3=1 | C3Al8N4 |
| 166.3.181 | 68415 | 66 | - | - | - - | - - | 12 | no(2) | 12 | no(2) | z2w,1=1<br>z2w,2=1<br>z2w,3=1<br>z4=3 | C1B2Th1 |
| 166.3.182 | 30526 | 38 | - | - | - - | - - | 12 | no(2) | 12 | yes | z2w,1=1<br>z2w,2=1<br>z2w,3=1 | Bi4Te3 |
| 166.3.183 | 58759 | 42 | - | - | - - | - - | 12 | no(2) | 12 | no(2) | z4=3 | Be17Zr2 |
| 166.3.184 | 109217 | 42 | - | - | - - | - - | 12 | no(2) | 12 | no(2) | z4=3 | Be17Ti2 |
| 166.3.185 | 47133 | 140 | - | - | - - | - - | 12 | no(2) | 12 | no(2) | z2w,1=1<br>z2w,2=1<br>z2w,3=1<br>z4=3 | Ba4O12Ru3Zr1 |
| 166.3.186 | 47132 | 140 | - | - | - - | - - | 12 | no(2) | 12 | no(2) | z2w,1=1<br>z2w,2=1<br>z2w,3=1<br>z4=3 | Ba4O12Ru3Zr1 |
| 166.3.187 | 240501 | 222 | - | - | - - | - - | 12 | no(2) | 12 | no(2) | z4=3 | Ba3Co10O17 |
| 166.3.188 | 150735 | 54 | - | - | - - | - - | 12 | yes | 12 | yes | - | Ba2Mg17 |
| 166.3.189 | 10341 | 58 | - | - | - - | - - | 12 | yes | 12 | yes | - | Ba1Cr1F6 |
| 166.3.190 | 100286 | 148 | - | - | - - | - - | 12 | no(2) | 12 | yes | z4=2 | B6Ni12Sr1 |
| 166.3.191 | 417868 | 144 | - | - | - - | - - | 12 | no(2) | 12 | no(2) | z4=1 | B6Co12Gd1 |
| 166.3.192 | 184644 | 144 | - | - | - - | - - | 12 | no(2) | 12 | yes | z2w,1=1<br>z2w,2=1<br>z2w,3=1<br>z4=2 | B6Co12Gd1 |
| 166.3.193 | 85465 | 140 | - | - | - - | - - | 12 | no(2) | 12 | no(2) | z2w,1=1<br>z2w,2=1<br>z2w,3=1<br>z4=1 | B6Ca1Ni12 |
| 166.3.194 | 36507 | 140 | - | - | - - | - - | 12 | no(2) | 12 | no(2) | z2w,1=1<br>z2w,2=1<br>z2w,3=1<br>z4=1 | B6Ca1Ni12 |

Continued on next page

**TABLE XIV– continued**

| TQC | ICSD | NoE | DG | IndG | SubSG (w/ T) | cl | SubSG (w/o T) | cl | Max. Sub. SG | yes or no | Max. Sub. SG | yes or no | SIs (w/ T) | Stoich. Formula |
|---|---|---|---|---|---|---|---|---|---|---|---|---|---|---|
| 166.3.195 | 100287 | 148 | - | - | - | - | - | - | 12 | no(2) | 12 | no(2) | z2w,1=1 z2w,2=1 z2w,3=1 z4=3 | B6Ba1Ni12 |
| 166.3.196 | 36505 | 82 | - | - | - | - | - | - | 12 | yes | 12 | yes | - | B2Ca3Ni7 |
| 166.3.197 | 710044 | 178 | - | - | - | - | - | - | 12 | no(2) | 12 | yes | z2w,1=1 z2w,2=1 z2w,3=1 z4=2 | Au3Sn2Yb2 |
| 166.3.198 | 262036 | 64 | - | - | - | - | - | - | 12 | no(2) | 12 | no(2) | z4=3 | As3Na1Zn4 |
| 166.3.199 | 89626 | 64 | - | - | - | - | - | - | 12 | no(2) | 12 | no(2) | z4=3 | As2Cu4Sr1 |
| 166.3.200 | 32619 | 56 | - | - | - | - | - | - | 12 | no(2) | 12 | no(2) | z4=3 | As2Ca1Cu4 |
| 166.3.201 | 89628 | 64 | - | - | - | - | - | - | 12 | no(2) | 12 | no(2) | z4=3 | As2Ba1Cu4 |
| 166.3.202 | 166534 | 16 | - | - | - | - | - | - | 12 | no(2) | 12 | no(2) | z4=3 | As1Ca2I1 |
| 166.3.203 | 58219 | 60 | - | - | - | - | - | - | 12 | no(2) | 12 | no(2) | z4=3 | Al3Y1 |
| 166.3.204 | 58218 | 60 | - | - | - | - | - | - | 12 | no(2) | 12 | no(2) | z4=3 | Al3Y1 |
| 166.3.205 | 58217 | 60 | - | - | - | - | - | - | 12 | no(2) | 12 | no(2) | z4=3 | Al3Y1 |
| 166.3.206 | 173990 | 60 | - | - | - | - | - | - | 12 | no(2) | 12 | no(2) | z4=3 | Al3Y1 |
| 166.3.207 | 150556 | 60 | - | - | - | - | - | - | 12 | no(2) | 12 | no(2) | z4=3 | Al3Y1 |
| 166.3.208 | 151457 | 74 | - | - | - | - | - | - | 12 | no(2) | 12 | no(2) | z4=3 | Al1O4V2 |
| 166.3.209 | 604401 | 26 | - | - | - | - | - | - | 12 | yes | 12 | yes | - | Ag1In1Se2 |
| 167.1.1 | 83385 | 132 | 0.0036 | -0.3978 | 2 | 3 | 0 | 2 | - | - | - | - | z4=2 | S3Yb2 |
| 167.1.2 | 75248 | 132 | 0.0225 | -0.3778 | 2 | 3 | 0 | 2 | - | - | - | - | z4=2 | S3Yb2 |
| 167.1.3 | 95761 | 56 | 0.0057 | -0.1918 | 2 | 1 | 2 | 1 | - | - | - | - | z4=1 | O3V2 |
| 167.1.4 | 95760 | 56 | 0.0079 | -0.1455 | 2 | 1 | 2 | 1 | - | - | - | - | z4=1 | O3V2 |
| 167.1.5 | 94768 | 56 | 0.0051 | -0.1214 | 2 | 3 | 0 | 2 | - | - | - | - | z4=2 | O3V2 |
| 167.1.6 | 64796 | 56 | 0.0061 | -0.1196 | 2 | 3 | 0 | 2 | - | - | - | - | z4=2 | O3V2 |
| 167.1.7 | 64786 | 56 | 0.0065 | -0.1207 | 2 | 3 | 0 | 2 | - | - | - | - | z4=2 | O3V2 |
| 167.1.8 | 25780 | 56 | 0.0087 | -0.1461 | 2 | 1 | 2 | 1 | - | - | - | - | z4=1 | O3V2 |
| 167.1.9 | 201109 | 56 | 0.0058 | -0.2096 | 2 | 1 | 2 | 1 | - | - | - | - | z4=1 | O3V2 |
| 167.1.10 | 201108 | 56 | 0.0033 | -0.1336 | 2 | 3 | 0 | 2 | - | - | - | - | z4=2 | O3V2 |
| 167.1.11 | 201107 | 56 | 0.0039 | -0.1227 | 2 | 3 | 0 | 2 | - | - | - | - | z4=2 | O3V2 |
| 167.1.12 | 201106 | 56 | 0.0101 | -0.1289 | 2 | 3 | 0 | 2 | - | - | - | - | z4=2 | O3V2 |
| 167.1.13 | 1874 | 56 | 0.0052 | -0.0996 | 2 | 1 | 2 | 1 | - | - | - | - | z4=1 | O3V2 |
| 167.1.14 | 1873 | 56 | 0.0073 | -0.1046 | 2 | 1 | 2 | 1 | - | - | - | - | z4=1 | O3V2 |
| 167.1.15 | 1872 | 56 | 0.0026 | -0.1048 | 2 | 1 | 2 | 1 | - | - | - | - | z4=1 | O3V2 |
| 167.1.16 | 1870 | 56 | 0.0037 | -0.1176 | 2 | 3 | 0 | 2 | - | - | - | - | z4=2 | O3V2 |
| 167.1.17 | 1869 | 56 | 0.0042 | -0.1237 | 2 | 3 | 0 | 2 | - | - | - | - | z4=2 | O3V2 |
| 167.1.18 | 1473 | 56 | 0.0042 | -0.1237 | 2 | 3 | 0 | 2 | - | - | - | - | z4=2 | O3V2 |
| 167.1.19 | 167061 | 72 | 0.0000 | -0.2454 | 2 | 1 | 2 | 1 | - | - | - | - | z4=1 | La1Mn1O3.31 |
| 167.1.20 | 167062 | 72 | 0.0073 | -0.1362 | 2 | 1 | 2 | 1 | - | - | - | - | z4=1 | La1Mn1O3.29 |
| 167.1.21 | 167065 | 72 | 0.0069 | -0.1030 | 2 | 1 | 2 | 1 | - | - | - | - | z4=1 | La1Mn1O3.26 |
| 167.1.22 | 167064 | 72 | 0.0098 | -0.0679 | 2 | 1 | 2 | 1 | - | - | - | - | z4=1 | La1Mn1O3.26 |
| 167.1.23 | 167066 | 72 | 0.0000 | -0.2171 | 2 | 1 | 2 | 1 | - | - | - | - | z4=1 | La1Mn1O3.22 |
| 167.1.24 | 167068 | 72 | 0.0000 | -0.2903 | 2 | 1 | 2 | 1 | - | - | - | - | z4=1 | La1Mn1O3.15 |
| 167.1.25 | 167069 | 72 | 0.0000 | -0.2020 | 2 | 1 | 2 | 1 | - | - | - | - | z4=1 | La1Mn1O3.13 |
| 167.1.26 | 96070 | 68 | 0.0185 | -0.4017 | 2 | 2 | 2 | 1 | - | - | - | - | z4=3 | Fe2O3 |
| 167.1.27 | 96069 | 68 | 0.0082 | -0.4119 | 2 | 2 | 2 | 1 | - | - | - | - | z4=3 | Fe2O3 |
| 167.1.28 | 88418 | 68 | 0.0145 | -0.3709 | 2 | 2 | 2 | 1 | - | - | - | - | z4=3 | Fe2O3 |
| 167.1.29 | 88417 | 68 | 0.0158 | -0.3604 | 2 | 2 | 2 | 1 | - | - | - | - | z4=3 | Fe2O3 |
| 167.1.30 | 82904 | 68 | 0.0134 | -0.3695 | 2 | 2 | 2 | 1 | - | - | - | - | z4=3 | Fe2O3 |
| 167.1.31 | 82903 | 68 | 0.0137 | -0.3677 | 2 | 2 | 2 | 1 | - | - | - | - | z4=3 | Fe2O3 |
| 167.1.32 | 82902 | 68 | 0.0136 | -0.3678 | 2 | 2 | 2 | 1 | - | - | - | - | z4=3 | Fe2O3 |
| 167.1.33 | 82137 | 68 | 0.0000 | -0.3231 | 2 | 2 | 2 | 1 | - | - | - | - | z4=3 | Fe2O3 |
| 167.1.34 | 82136 | 68 | 0.0000 | -0.4151 | 2 | 2 | 2 | 1 | - | - | - | - | z4=3 | Fe2O3 |
| 167.1.35 | 82135 | 68 | 0.0133 | -0.3233 | 2 | 2 | 2 | 1 | - | - | - | - | z4=3 | Fe2O3 |
| 167.1.36 | 82134 | 68 | 0.0000 | -0.4608 | 2 | 2 | 2 | 1 | - | - | - | - | z4=3 | Fe2O3 |

Continued on next page

**TABLE XIV– continued**

| TQC | ICSD | NoE | DG | IndG | SubSG (w/ T) | cl | SubSG (w/o T) | cl | Max. Sub. SG | yes or no | Max. Sub. SG | yes or no | SIs (w/ T) | Stoich. Formula |
|---|---|---|---|---|---|---|---|---|---|---|---|---|---|---|
| 167.1.37 | 81248 | 68 | 0.0136 | -0.3674 | 2 | 2 | 2 | 1 | - | - | - | - | z4=3 | Fe2O3 |
| 167.1.38 | 71194 | 68 | 0.0137 | -0.3551 | 2 | 2 | 2 | 1 | - | - | - | - | z4=3 | Fe2O3 |
| 167.1.39 | 66756 | 68 | 0.0126 | -0.3779 | 2 | 2 | 2 | 1 | - | - | - | - | z4=3 | Fe2O3 |
| 167.1.40 | 64599 | 68 | 0.0134 | -0.3655 | 2 | 2 | 2 | 1 | - | - | - | - | z4=3 | Fe2O3 |
| 167.1.41 | 56372 | 68 | 0.0118 | -0.3579 | 2 | 2 | 2 | 1 | - | - | - | - | z4=3 | Fe2O3 |
| 167.1.42 | 43465 | 68 | 0.0151 | -0.3193 | 2 | 2 | 2 | 1 | - | - | - | - | z4=3 | Fe2O3 |
| 167.1.43 | 415251 | 68 | 0.0143 | -0.3698 | 2 | 2 | 2 | 1 | - | - | - | - | z4=3 | Fe2O3 |
| 167.1.44 | 40142 | 68 | 0.0136 | -0.3667 | 2 | 2 | 2 | 1 | - | - | - | - | z4=3 | Fe2O3 |
| 167.1.45 | 24791 | 68 | 0.0474 | -0.2940 | 2 | 2 | 2 | 1 | - | - | - | - | z4=3 | Fe2O3 |
| 167.1.46 | 245851 | 68 | 0.0125 | -0.3758 | 2 | 2 | 2 | 1 | - | - | - | - | z4=3 | Fe2O3 |
| 167.1.47 | 24004 | 68 | 0.0617 | -0.1939 | 2 | 2 | 2 | 1 | - | - | - | - | z4=3 | Fe2O3 |
| 167.1.48 | 22505 | 68 | 0.0141 | -0.3659 | 2 | 2 | 2 | 1 | - | - | - | - | z4=3 | Fe2O3 |
| 167.1.49 | 201101 | 68 | 0.0051 | -0.4147 | 2 | 2 | 2 | 1 | - | - | - | - | z4=3 | Fe2O3 |
| 167.1.50 | 201100 | 68 | 0.0190 | -0.3620 | 2 | 2 | 2 | 1 | - | - | - | - | z4=3 | Fe2O3 |
| 167.1.51 | 201099 | 68 | 0.0157 | -0.3509 | 2 | 2 | 2 | 1 | - | - | - | - | z4=3 | Fe2O3 |
| 167.1.52 | 201098 | 68 | 0.0148 | -0.3618 | 2 | 2 | 2 | 1 | - | - | - | - | z4=3 | Fe2O3 |
| 167.1.53 | 201097 | 68 | 0.0150 | -0.3769 | 2 | 2 | 2 | 1 | - | - | - | - | z4=3 | Fe2O3 |
| 167.1.54 | 201096 | 68 | 0.0147 | -0.3647 | 2 | 2 | 2 | 1 | - | - | - | - | z4=3 | Fe2O3 |
| 167.1.55 | 184766 | 68 | 0.0123 | -0.3502 | 2 | 2 | 2 | 1 | - | - | - | - | z4=3 | Fe2O3 |
| 167.1.56 | 182848 | 68 | 0.0143 | -0.4278 | 2 | 2 | 2 | 1 | - | - | - | - | z4=3 | Fe2O3 |
| 167.1.57 | 182847 | 68 | 0.0136 | -0.4305 | 2 | 2 | 2 | 1 | - | - | - | - | z4=3 | Fe2O3 |
| 167.1.58 | 182846 | 68 | 0.0135 | -0.4273 | 2 | 2 | 2 | 1 | - | - | - | - | z4=3 | Fe2O3 |
| 167.1.59 | 182845 | 68 | 0.0166 | -0.4052 | 2 | 2 | 2 | 1 | - | - | - | - | z4=3 | Fe2O3 |
| 167.1.60 | 182844 | 68 | 0.0177 | -0.4062 | 2 | 2 | 2 | 1 | - | - | - | - | z4=3 | Fe2O3 |
| 167.1.61 | 182843 | 68 | 0.0182 | -0.4188 | 2 | 2 | 2 | 1 | - | - | - | - | z4=3 | Fe2O3 |
| 167.1.62 | 182842 | 68 | 0.0194 | -0.3957 | 2 | 2 | 2 | 1 | - | - | - | - | z4=3 | Fe2O3 |
| 167.1.63 | 182841 | 68 | 0.0194 | -0.4024 | 2 | 2 | 2 | 1 | - | - | - | - | z4=3 | Fe2O3 |
| 167.1.64 | 182840 | 68 | 0.0134 | -0.3855 | 2 | 2 | 2 | 1 | - | - | - | - | z4=3 | Fe2O3 |
| 167.1.65 | 182839 | 68 | 0.0130 | -0.3637 | 2 | 2 | 2 | 1 | - | - | - | - | z4=3 | Fe2O3 |
| 167.1.66 | 173655 | 68 | 0.0131 | -0.3798 | 2 | 2 | 2 | 1 | - | - | - | - | z4=3 | Fe2O3 |
| 167.1.67 | 173653 | 68 | 0.0126 | -0.3757 | 2 | 2 | 2 | 1 | - | - | - | - | z4=3 | Fe2O3 |
| 167.1.68 | 161294 | 68 | 0.0030 | -0.3873 | 2 | 2 | 2 | 1 | - | - | - | - | z4=3 | Fe2O3 |
| 167.1.69 | 161293 | 68 | 0.0030 | -0.3874 | 2 | 2 | 2 | 1 | - | - | - | - | z4=3 | Fe2O3 |
| 167.1.70 | 161292 | 68 | 0.0042 | -0.3916 | 2 | 2 | 2 | 1 | - | - | - | - | z4=3 | Fe2O3 |
| 167.1.71 | 161291 | 68 | 0.0062 | -0.3905 | 2 | 2 | 2 | 1 | - | - | - | - | z4=3 | Fe2O3 |
| 167.1.72 | 161290 | 68 | 0.0089 | -0.3882 | 2 | 2 | 2 | 1 | - | - | - | - | z4=3 | Fe2O3 |
| 167.1.73 | 161289 | 68 | 0.0123 | -0.3617 | 2 | 2 | 2 | 1 | - | - | - | - | z4=3 | Fe2O3 |
| 167.1.74 | 161288 | 68 | 0.0125 | -0.3641 | 2 | 2 | 2 | 1 | - | - | - | - | z4=3 | Fe2O3 |
| 167.1.75 | 161287 | 68 | 0.0123 | -0.3620 | 2 | 2 | 2 | 1 | - | - | - | - | z4=3 | Fe2O3 |
| 167.1.76 | 161286 | 68 | 0.0127 | -0.3643 | 2 | 2 | 2 | 1 | - | - | - | - | z4=3 | Fe2O3 |
| 167.1.77 | 161285 | 68 | 0.0126 | -0.3652 | 2 | 2 | 2 | 1 | - | - | - | - | z4=3 | Fe2O3 |
| 167.1.78 | 161284 | 68 | 0.0133 | -0.3680 | 2 | 2 | 2 | 1 | - | - | - | - | z4=3 | Fe2O3 |
| 167.1.79 | 161283 | 68 | 0.0138 | -0.3695 | 2 | 2 | 2 | 1 | - | - | - | - | z4=3 | Fe2O3 |
| 167.1.80 | 15840 | 68 | 0.0137 | -0.3659 | 2 | 2 | 2 | 1 | - | - | - | - | z4=3 | Fe2O3 |
| 167.1.81 | 154191 | 68 | 0.0373 | -0.3600 | 2 | 2 | 2 | 1 | - | - | - | - | z4=3 | Fe2O3 |
| 167.1.82 | 154190 | 68 | 0.0380 | -0.3605 | 2 | 2 | 2 | 1 | - | - | - | - | z4=3 | Fe2O3 |
| 167.1.83 | 173470 | 60 | 0.0000 | -0.3665 | 2 | 1 | 2 | 1 | - | - | - | - | z4=1 | Cr2O3 |
| 167.1.84 | 201966 | 204 | 0.0175 | 0.0037 | 2 | 2 | 2 | 1 | - | - | - | - | z4=3 | Cr2Cs3I9 |
| 167.1.85 | 50809 | 120 | 0.0074 | 0.0060 | 2 | 2 | 2 | 1 | - | - | - | - | z4=3 | Ca3Co1O6Rh1 |
| 167.1.86 | 421391 | 108 | 0.0123 | -0.1053 | 2 | 3 | 0 | 2 | - | - | - | - | z4=2 | C3Cr2N6 |
| 167.1.87 | 60625 | 180 | 0.0039 | -0.1334 | 2 | 2 | 2 | 1 | - | - | - | - | z4=3 | Br6Cd1Rb4 |
| 167.1.88 | 243759 | 200 | 0.0156 | -0.0603 | 2 | 1 | 2 | 1 | - | - | - | - | z4=1 | Ba3O9Os2 |
| 167.1.89 | 97540 | 156 | 0.0063 | -0.0117 | 2 | 2 | 2 | 1 | - | - | - | - | z4=3 | Ba3Cr2S6 |
| 167.1.90 | 402170 | 156 | 0.0563 | -0.7478 | 2 | 1 | 2 | 1 | - | - | - | - | z4=1 | Au3Ga7Yb1 |
| 167.1.91 | 391108 | 148 | 0.0048 | -0.0593 | 2 | 1 | 2 | 1 | - | - | - | - | z4=1 | Al7Au3Dy1 |
| 167.1.92 | 391101 | 132 | 0.0356 | -0.0238 | 2 | 1 | 2 | 1 | - | - | - | - | z4=1 | Al7Au3Ce1 |
| 167.3.1 | 67115 | 128 | - | - | - | - | - | - | 15 | no(2) | 15 | yes | z4=2 | S3Tm2 |

Continued on next page

**TABLE XIV– continued**

| TQC | ICSD | NoE | DG | IndG | SubSG cl (w/ T) | | SubSG cl (w/o T) | | Max. Sub. SG | yes or no | Max. Sub. SG | yes or no | SIs (w/ T) | Stoich. Formula |
|---|---|---|---|---|---|---|---|---|---|---|---|---|---|---|
| 167.3.2 | 15243 | 116 | - | - | - | - | - | - | 15 | no(2) | 15 | no(2) | z4=3 | Rh3Sc1Si7 |
| 167.3.3 | 51054 | 188 | - | - | - | - | - | - | 15 | yes | 15 | yes | - | O6Rh1Sr3Tb1 |
| 167.3.4 | 26490 | 220 | - | - | - | - | - | - | 15 | yes | 15 | yes | - | O6Pt1Sr4 |
| 167.3.5 | 33641 | 56 | - | - | - | - | - | - | 15 | no(2) | 15 | no(2) | z4=3 | O3V2 |
| 167.3.6 | 9658 | 52 | - | - | - | - | - | - | 15 | no(2) | 15 | no(2) | z4=3 | O3Ti2 |
| 167.3.7 | 9656 | 52 | - | - | - | - | - | - | 15 | no(2) | 15 | no(2) | z4=3 | O3Ti2 |
| 167.3.8 | 9655 | 52 | - | - | - | - | - | - | 15 | no(2) | 15 | no(2) | z4=3 | O3Ti2 |
| 167.3.9 | 9654 | 52 | - | - | - | - | - | - | 15 | no(2) | 15 | no(2) | z4=3 | O3Ti2 |
| 167.3.10 | 9648 | 52 | - | - | - | - | - | - | 15 | no(2) | 15 | no(2) | z4=3 | O3Ti2 |
| 167.3.11 | 9647 | 52 | - | - | - | - | - | - | 15 | no(2) | 15 | no(2) | z4=3 | O3Ti2 |
| 167.3.12 | 9646 | 52 | - | - | - | - | - | - | 15 | no(2) | 15 | no(2) | z4=3 | O3Ti2 |
| 167.3.13 | 77696 | 52 | - | - | - | - | - | - | 15 | no(2) | 15 | no(2) | z4=3 | O3Ti2 |
| 167.3.14 | 6095 | 52 | - | - | - | - | - | - | 15 | no(2) | 15 | no(2) | z4=3 | O3Ti2 |
| 167.3.15 | 54175 | 52 | - | - | - | - | - | - | 15 | no(2) | 15 | no(2) | z4=3 | O3Ti2 |
| 167.3.16 | 29286 | 52 | - | - | - | - | - | - | 15 | no(2) | 15 | no(2) | z4=3 | O3Ti2 |
| 167.3.17 | 28354 | 52 | - | - | - | - | - | - | 15 | no(2) | 15 | no(2) | z4=3 | O3Ti2 |
| 167.3.18 | 28353 | 52 | - | - | - | - | - | - | 15 | no(2) | 15 | no(2) | z4=3 | O3Ti2 |
| 167.3.19 | 25779 | 52 | - | - | - | - | - | - | 15 | no(2) | 15 | no(2) | z4=3 | O3Ti2 |
| 167.3.20 | 1462 | 52 | - | - | - | - | - | - | 15 | no(2) | 15 | no(2) | z4=3 | O3Ti2 |
| 167.3.21 | 27023 | 36 | - | - | - | - | - | - | 15 | no(2) | 15 | no(2) | z4=1 | O1Zr3 |
| 167.3.22 | 87773 | 172 | - | - | - | - | - | - | 15 | yes | 15 | yes | - | Ni1O6Pt1Sr3 |
| 167.3.23 | 75335 | 172 | - | - | - | - | - | - | 15 | yes | 15 | yes | - | Ni1O6Pt1Sr3 |
| 167.3.24 | 33661 | 48 | - | - | - | - | - | - | 15 | no(2) | 15 | yes | z4=2 | Li1N1O3 |
| 167.3.25 | 51056 | 192 | - | - | - | - | - | - | 15 | yes | 15 | yes | - | Ho1O6Rh1Sr3 |
| 167.3.26 | 60752 | 148 | - | - | - | - | - | - | 15 | no(2) | 15 | no(2) | z4=1 | H16Cd1Cl6N4 |
| 167.3.27 | 60622 | 148 | - | - | - | - | - | - | 15 | no(2) | 15 | no(2) | z4=1 | H16Cd1Cl6N4 |
| 167.3.28 | 77645 | 160 | - | - | - | - | - | - | 15 | no(2) | 15 | no(2) | z4=3 | Ga5N1Sr6 |
| 167.3.29 | 96074 | 68 | - | - | - | - | - | - | 15 | no(2) | 15 | no(2) | z4=3 | Fe2O3 |
| 167.3.30 | 96073 | 68 | - | - | - | - | - | - | 15 | no(2) | 15 | no(2) | z4=3 | Fe2O3 |
| 167.3.31 | 96072 | 68 | - | - | - | - | - | - | 15 | no(2) | 15 | no(2) | z4=3 | Fe2O3 |
| 167.3.32 | 69167 | 52 | - | - | - | - | - | - | 15 | yes | 15 | yes | - | F3V1 |
| 167.3.33 | 18028 | 48 | - | - | - | - | - | - | 15 | yes | 15 | yes | - | F3In1 |
| 167.3.34 | 9428 | 80 | - | - | - | - | - | - | 15 | yes | 15 | yes | - | Cu1La1O3 |
| 167.3.35 | 90158 | 60 | - | - | - | - | - | - | 15 | no(2) | 15 | yes | z4=2 | Cr2O3 |
| 167.3.36 | 90157 | 60 | - | - | - | - | - | - | 15 | no(2) | 15 | yes | z4=2 | Cr2O3 |
| 167.3.37 | 75577 | 60 | - | - | - | - | - | - | 15 | no(2) | 15 | yes | z4=2 | Cr2O3 |
| 167.3.38 | 33642 | 60 | - | - | - | - | - | - | 15 | no(2) | 15 | no(2) | z4=3 | Cr2O3 |
| 167.3.39 | 29298 | 60 | - | - | - | - | - | - | 15 | no(2) | 15 | yes | z4=2 | Cr2O3 |
| 167.3.40 | 290245 | 60 | - | - | - | - | - | - | 15 | no(2) | 15 | no(2) | z4=1 | Cr2O3 |
| 167.3.41 | 290244 | 60 | - | - | - | - | - | - | 15 | no(2) | 15 | no(2) | z4=1 | Cr2O3 |
| 167.3.42 | 290243 | 60 | - | - | - | - | - | - | 15 | no(2) | 15 | yes | z4=2 | Cr2O3 |
| 167.3.43 | 290242 | 60 | - | - | - | - | - | - | 15 | no(2) | 15 | yes | z4=2 | Cr2O3 |
| 167.3.44 | 290241 | 60 | - | - | - | - | - | - | 15 | no(2) | 15 | yes | z4=2 | Cr2O3 |
| 167.3.45 | 290240 | 60 | - | - | - | - | - | - | 15 | no(2) | 15 | yes | z4=2 | Cr2O3 |
| 167.3.46 | 290239 | 60 | - | - | - | - | - | - | 15 | no(2) | 15 | yes | z4=2 | Cr2O3 |
| 167.3.47 | 290238 | 60 | - | - | - | - | - | - | 15 | no(2) | 15 | yes | z4=2 | Cr2O3 |
| 167.3.48 | 290237 | 60 | - | - | - | - | - | - | 15 | no(2) | 15 | yes | z4=2 | Cr2O3 |
| 167.3.49 | 290236 | 60 | - | - | - | - | - | - | 15 | no(2) | 15 | yes | z4=2 | Cr2O3 |
| 167.3.50 | 290235 | 60 | - | - | - | - | - | - | 15 | no(2) | 15 | yes | z4=2 | Cr2O3 |
| 167.3.51 | 290234 | 60 | - | - | - | - | - | - | 15 | no(2) | 15 | yes | z4=2 | Cr2O3 |
| 167.3.52 | 26791 | 60 | - | - | - | - | - | - | 15 | no(2) | 15 | yes | z4=2 | Cr2O3 |
| 167.3.53 | 261801 | 60 | - | - | - | - | - | - | 15 | no(2) | 15 | yes | z4=2 | Cr2O3 |
| 167.3.54 | 25781 | 60 | - | - | - | - | - | - | 15 | no(2) | 15 | yes | z4=2 | Cr2O3 |
| 167.3.55 | 250078 | 60 | - | - | - | - | - | - | 15 | no(2) | 15 | yes | z4=2 | Cr2O3 |
| 167.3.56 | 202619 | 60 | - | - | - | - | - | - | 15 | no(2) | 15 | yes | z4=2 | Cr2O3 |
| 167.3.57 | 201103 | 60 | - | - | - | - | - | - | 15 | no(2) | 15 | yes | z4=2 | Cr2O3 |
| 167.3.58 | 201102 | 60 | - | - | - | - | - | - | 15 | no(2) | 15 | yes | z4=2 | Cr2O3 |

Continued on next page

**TABLE XIV– continued**

| TQC | ICSD | NoE | DG | IndG | SubSG cl (w/ T) | SubSG cl (w/o T) | Max. Sub. SG | yes or no | Max. Sub. SG | yes or no | SIs (w/ T) | Stoich. Formula |
|---|---|---|---|---|---|---|---|---|---|---|---|---|
| 167.3.59 | 167291 | 60 | - | - | - - | - - | 15 | no(2) | 15 | yes | z4=2 | Cr2O3 |
| 167.3.60 | 167290 | 60 | - | - | - - | - - | 15 | no(2) | 15 | yes | z4=2 | Cr2O3 |
| 167.3.61 | 167289 | 60 | - | - | - - | - - | 15 | no(2) | 15 | yes | z4=2 | Cr2O3 |
| 167.3.62 | 167288 | 60 | - | - | - - | - - | 15 | no(2) | 15 | yes | z4=2 | Cr2O3 |
| 167.3.63 | 167287 | 60 | - | - | - - | - - | 15 | no(2) | 15 | yes | z4=2 | Cr2O3 |
| 167.3.64 | 167286 | 60 | - | - | - - | - - | 15 | no(2) | 15 | yes | z4=2 | Cr2O3 |
| 167.3.65 | 167285 | 60 | - | - | - - | - - | 15 | no(2) | 15 | yes | z4=2 | Cr2O3 |
| 167.3.66 | 167284 | 60 | - | - | - - | - - | 15 | no(2) | 15 | yes | z4=2 | Cr2O3 |
| 167.3.67 | 167283 | 60 | - | - | - - | - - | 15 | no(2) | 15 | yes | z4=2 | Cr2O3 |
| 167.3.68 | 167282 | 60 | - | - | - - | - - | 15 | no(2) | 15 | yes | z4=2 | Cr2O3 |
| 167.3.69 | 167281 | 60 | - | - | - - | - - | 15 | no(2) | 15 | yes | z4=2 | Cr2O3 |
| 167.3.70 | 167280 | 60 | - | - | - - | - - | 15 | no(2) | 15 | yes | z4=2 | Cr2O3 |
| 167.3.71 | 167279 | 60 | - | - | - - | - - | 15 | no(2) | 15 | yes | z4=2 | Cr2O3 |
| 167.3.72 | 167278 | 60 | - | - | - - | - - | 15 | no(2) | 15 | yes | z4=2 | Cr2O3 |
| 167.3.73 | 167277 | 60 | - | - | - - | - - | 15 | no(2) | 15 | yes | z4=2 | Cr2O3 |
| 167.3.74 | 167276 | 60 | - | - | - - | - - | 15 | no(2) | 15 | yes | z4=2 | Cr2O3 |
| 167.3.75 | 167275 | 60 | - | - | - - | - - | 15 | no(2) | 15 | yes | z4=2 | Cr2O3 |
| 167.3.76 | 167274 | 60 | - | - | - - | - - | 15 | no(2) | 15 | yes | z4=2 | Cr2O3 |
| 167.3.77 | 167273 | 60 | - | - | - - | - - | 15 | no(2) | 15 | yes | z4=2 | Cr2O3 |
| 167.3.78 | 167272 | 60 | - | - | - - | - - | 15 | no(2) | 15 | yes | z4=2 | Cr2O3 |
| 167.3.79 | 167271 | 60 | - | - | - - | - - | 15 | no(2) | 15 | yes | z4=2 | Cr2O3 |
| 167.3.80 | 167270 | 60 | - | - | - - | - - | 15 | no(2) | 15 | yes | z4=2 | Cr2O3 |
| 167.3.81 | 167269 | 60 | - | - | - - | - - | 15 | no(2) | 15 | yes | z4=2 | Cr2O3 |
| 167.3.82 | 167268 | 60 | - | - | - - | - - | 15 | no(2) | 15 | yes | z4=2 | Cr2O3 |
| 167.3.83 | 107035 | 60 | - | - | - - | - - | 15 | no(2) | 15 | yes | z4=2 | Cr2O3 |
| 167.3.84 | 416892 | 180 | - | - | - - | - - | 15 | no(2) | 15 | no(2) | z4=3 | C3N6Yb2 |
| 167.3.85 | 192545 | 68 | - | - | - - | - - | 15 | yes | 15 | yes | - | C1Ni3 |
| 167.3.86 | 61067 | 64 | - | - | - - | - - | 15 | yes | 15 | yes | - | C1Ni1O3 |
| 167.3.87 | 173985 | 64 | - | - | - - | - - | 15 | yes | 15 | yes | - | C1Ni1O3 |
| 167.3.88 | 72790 | 264 | - | - | - - | - - | 15 | no(2) | 15 | no(2) | z4=3 | Br9Cs3Ho2 |
| 167.3.89 | 409805 | 160 | - | - | - - | - - | 15 | no(2) | 15 | no(2) | z4=3 | Ba6In5N1 |
| 167.3.90 | 77731 | 160 | - | - | - - | - - | 15 | no(2) | 15 | no(2) | z4=3 | Ba6Ga5N1 |
| 167.3.91 | 30662 | 128 | - | - | - - | - - | 15 | yes | 15 | yes | - | Ba3Ni1O4 |
| 167.3.92 | 45060 | 52 | - | - | - - | - - | 15 | yes | 15 | yes | - | B1O3V1 |
| 167.3.93 | 419132 | 52 | - | - | - - | - - | 15 | yes | 15 | yes | - | B1O3V1 |
| 167.3.94 | 426464 | 244 | - | - | - - | - - | 15 | no(2) | 15 | no(2) | z4=3 | Au6Sr2Zn3 |
| 167.3.95 | 391110 | 156 | - | - | - - | - - | 15 | yes | 15 | yes | - | Al7Au3Yb1 |
| 167.3.96 | 391107 | 136 | - | - | - - | - - | 15 | no(2) | 15 | no(2) | z4=1 | Al7Au3Nd1 |
| 167.3.97 | 391109 | 152 | - | - | - - | - - | 15 | no(2) | 15 | no(2) | z4=1 | Al7Au3Er1 |
| 167.3.98 | 99783 | 48 | - | - | - - | - - | 15 | no(2) | 15 | no(2) | z4=3 | Al2O3 |
| 167.3.99 | 90571 | 68 | - | - | - - | - - | 15 | no(2) | 15 | no(2) | z4=3 | Al1O3Pr1 |
| 167.3.100 | 90570 | 68 | - | - | - - | - - | 15 | no(2) | 15 | no(2) | z4=3 | Al1O3Pr1 |
| 167.3.101 | 90569 | 68 | - | - | - - | - - | 15 | no(2) | 15 | no(2) | z4=3 | Al1O3Pr1 |
| 167.3.102 | 90568 | 68 | - | - | - - | - - | 15 | no(2) | 15 | no(2) | z4=3 | Al1O3Pr1 |
| 167.3.103 | 90567 | 68 | - | - | - - | - - | 15 | no(2) | 15 | no(2) | z4=3 | Al1O3Pr1 |
| 167.3.104 | 90566 | 68 | - | - | - - | - - | 15 | no(2) | 15 | no(2) | z4=3 | Al1O3Pr1 |
| 167.3.105 | 90565 | 68 | - | - | - - | - - | 15 | no(2) | 15 | no(2) | z4=3 | Al1O3Pr1 |
| 167.3.106 | 90560 | 68 | - | - | - - | - - | 15 | no(2) | 15 | no(2) | z4=3 | Al1O3Pr1 |
| 167.3.107 | 90559 | 68 | - | - | - - | - - | 15 | no(2) | 15 | no(2) | z4=3 | Al1O3Pr1 |
| 167.3.108 | 90558 | 68 | - | - | - - | - - | 15 | no(2) | 15 | no(2) | z4=3 | Al1O3Pr1 |
| 167.3.109 | 90557 | 68 | - | - | - - | - - | 15 | no(2) | 15 | no(2) | z4=3 | Al1O3Pr1 |
| 167.3.110 | 165848 | 68 | - | - | - - | - - | 15 | no(2) | 15 | no(2) | z4=3 | Al1O3Pr1 |
| 167.3.111 | 165847 | 68 | - | - | - - | - - | 15 | no(2) | 15 | no(2) | z4=3 | Al1O3Pr1 |
| 167.3.112 | 165846 | 68 | - | - | - - | - - | 15 | no(2) | 15 | no(2) | z4=3 | Al1O3Pr1 |
| 167.3.113 | 165845 | 68 | - | - | - - | - - | 15 | no(2) | 15 | no(2) | z4=3 | Al1O3Pr1 |
| 167.3.114 | 165844 | 68 | - | - | - - | - - | 15 | no(2) | 15 | no(2) | z4=3 | Al1O3Pr1 |
| 167.3.115 | 165843 | 68 | - | - | - - | - - | 15 | no(2) | 15 | no(2) | z4=3 | Al1O3Pr1 |

Continued on next page

**TABLE XIV– continued**

| TQC | ICSD | NoE | DG | IndG | SubSG (w/ T) | cl | SubSG (w/o T) | cl | Max. Sub. SG | yes or no | Max. Sub. SG | yes or no | SIs (w/ T) | Stoich. Formula |
|---|---|---|---|---|---|---|---|---|---|---|---|---|---|---|
| 167.3.116 | 165842 | 68 | - | - | - | - | - | - | 15 | no(2) | 15 | no(2) | z4=3 | Al1O3Pr1 |
| 167.3.117 | 165841 | 68 | - | - | - | - | - | - | 15 | no(2) | 15 | no(2) | z4=3 | Al1O3Pr1 |
| 167.3.118 | 165840 | 68 | - | - | - | - | - | - | 15 | no(2) | 15 | no(2) | z4=3 | Al1O3Pr1 |
| 167.4.1 | 109297 | 218 | - | - | - | - | - | - | See details in Table XIX | | | | | O6Rh1Sr4 |
| 167.4.2 | 90285 | 174 | - | - | - | - | - | - | See details in Table XIX | | | | | O6Rh1Sr3Zn1 |
| 167.4.3 | 51058 | 198 | - | - | - | - | - | - | See details in Table XIX | | | | | O6Rh1Sr3Yb1 |
| 167.4.4 | 81659 | 158 | - | - | - | - | - | - | See details in Table XIX | | | | | Ni1O6Sc1Sr3 |
| 167.4.5 | 51799 | 170 | - | - | - | - | - | - | See details in Table XIX | | | | | Ni1O6Rh1Sr3 |
| 167.4.6 | 69187 | 82 | - | - | - | - | - | - | See details in Table XIX | | | | | Ni1O3Pr1 |
| 167.4.7 | 69186 | 82 | - | - | - | - | - | - | See details in Table XIX | | | | | Ni1O3Pr1 |
| 167.4.8 | 84738 | 154 | - | - | - | - | - | - | See details in Table XIX | | | | | Mg1O6Rh1Sr3 |
| 167.4.9 | 96220 | 150 | - | - | - | - | - | - | See details in Table XIX | | | | | Li1O6Ru1Sr3 |
| 167.4.10 | 50022 | 150 | - | - | - | - | - | - | See details in Table XIX | | | | | Li1O6Ru1Sr3 |
| 167.4.11 | 252432 | 54 | - | - | - | - | - | - | See details in Table XIX | | | | | Li1O3Os1 |
| 167.4.12 | 249150 | 114 | - | - | - | - | - | - | See details in Table XIX | | | | | Ir3Mg1Si7 |
| 167.4.13 | 82660 | 174 | - | - | - | - | - | - | See details in Table XIX | | | | | Ir1O6Sr3Zn1 |
| 167.4.14 | 82659 | 174 | - | - | - | - | - | - | See details in Table XIX | | | | | Ir1O6Sr3Zn1 |
| 167.4.15 | 82658 | 174 | - | - | - | - | - | - | See details in Table XIX | | | | | Ir1O6Sr3Zn1 |
| 167.4.16 | 81903 | 174 | - | - | - | - | - | - | See details in Table XIX | | | | | Ir1O6Sr3Zn1 |
| 167.4.17 | 80287 | 170 | - | - | - | - | - | - | See details in Table XIX | | | | | Ir1Ni1O6Sr3 |
| 167.4.18 | 80286 | 170 | - | - | - | - | - | - | See details in Table XIX | | | | | Ir1Ni1O6Sr3 |
| 167.4.19 | 80285 | 170 | - | - | - | - | - | - | See details in Table XIX | | | | | Ir1Ni1O6Sr3 |
| 167.4.20 | 84737 | 154 | - | - | - | - | - | - | See details in Table XIX | | | | | Ir1Mg1O6Sr3 |
| 167.4.21 | 81660 | 158 | - | - | - | - | - | - | See details in Table XIX | | | | | In1Ni1O6Sr3 |
| 167.4.22 | 50176 | 186 | - | - | - | - | - | - | See details in Table XIX | | | | | Gd1O6Rh1Sr3 |
| 167.4.23 | 236177 | 74 | - | - | - | - | - | - | See details in Table XIX | | | | | Fe1La1O3 |
| 167.4.24 | 236172 | 74 | - | - | - | - | - | - | See details in Table XIX | | | | | Fe1La1O3 |
| 167.4.25 | 236170 | 74 | - | - | - | - | - | - | See details in Table XIX | | | | | Fe1La1O3 |
| 167.4.26 | 236169 | 74 | - | - | - | - | - | - | See details in Table XIX | | | | | Fe1La1O3 |
| 167.4.27 | 67148 | 50 | - | - | - | - | - | - | See details in Table XIX | | | | | F3Ti1 |
| 167.4.28 | 52165 | 50 | - | - | - | - | - | - | See details in Table XIX | | | | | F3Ti1 |
| 167.4.29 | 52162 | 50 | - | - | - | - | - | - | See details in Table XIX | | | | | F3Ti1 |
| 167.4.30 | 52159 | 50 | - | - | - | - | - | - | See details in Table XIX | | | | | F3Ti1 |
| 167.4.31 | 16673 | 58 | - | - | - | - | - | - | See details in Table XIX | | | | | F3Ru1 |
| 167.4.32 | 165400 | 58 | - | - | - | - | - | - | See details in Table XIX | | | | | F3Ru1 |
| 167.4.33 | 16675 | 62 | - | - | - | - | - | - | See details in Table XIX | | | | | F3Pd1 |
| 167.4.34 | 68527 | 54 | - | - | - | - | - | - | See details in Table XIX | | | | | F3Mo1 |
| 167.4.35 | 52170 | 58 | - | - | - | - | - | - | See details in Table XIX | | | | | F3Fe1 |
| 167.4.36 | 52169 | 58 | - | - | - | - | - | - | See details in Table XIX | | | | | F3Fe1 |
| 167.4.37 | 52168 | 58 | - | - | - | - | - | - | See details in Table XIX | | | | | F3Fe1 |
| 167.4.38 | 240402 | 58 | - | - | - | - | - | - | See details in Table XIX | | | | | F3Fe1 |
| 167.4.39 | 240396 | 58 | - | - | - | - | - | - | See details in Table XIX | | | | | F3Fe1 |
| 167.4.40 | 240395 | 58 | - | - | - | - | - | - | See details in Table XIX | | | | | F3Fe1 |
| 167.4.41 | 240394 | 58 | - | - | - | - | - | - | See details in Table XIX | | | | | F3Fe1 |
| 167.4.42 | 51057 | 194 | - | - | - | - | - | - | See details in Table XIX | | | | | Er1O6Rh1Sr3 |
| 167.4.43 | 51055 | 190 | - | - | - | - | - | - | See details in Table XIX | | | | | Dy1O6Rh1Sr3 |
| 167.4.44 | 167590 | 70 | - | - | - | - | - | - | See details in Table XIX | | | | | Cr1La1O3 |
| 167.4.45 | 59973 | 54 | - | - | - | - | - | - | See details in Table XIX | | | | | Cr1F3 |
| 167.4.46 | 59972 | 54 | - | - | - | - | - | - | See details in Table XIX | | | | | Cr1F3 |
| 167.4.47 | 59969 | 54 | - | - | - | - | - | - | See details in Table XIX | | | | | Cr1F3 |
| 167.4.48 | 59968 | 54 | - | - | - | - | - | - | See details in Table XIX | | | | | Cr1F3 |
| 167.4.49 | 189230 | 174 | - | - | - | - | - | - | See details in Table XIX | | | | | Co1O6Sr3Zn1 |
| 167.4.50 | 192479 | 166 | - | - | - | - | - | - | See details in Table XIX | | | | | Co1O6Ru1Sr3 |
| 167.4.51 | 192478 | 166 | - | - | - | - | - | - | See details in Table XIX | | | | | Co1O6Ru1Sr3 |
| 167.4.52 | 290109 | 154 | - | - | - | - | - | - | See details in Table XIX | | | | | Cl6K3Mn1Na1 |
| 167.4.53 | 52832 | 194 | - | - | - | - | - | - | See details in Table XIX | | | | | Cd6Sb5 |
| 167.4.54 | 81904 | 174 | - | - | - | - | - | - | See details in Table XIX | | | | | Cd1Ir1O6Sr3 |

Continued on next page

**TABLE XIV– continued**

| TQC | ICSD | NoE | DG | IndG | SubSG cl (w/ T) | SubSG cl (w/o T) | Max. Sub. SG yes or no | Max. Sub. SG yes or no | SIs (w/ T) | Stoich. Formula |
|---|---|---|---|---|---|---|---|---|---|---|
| 167.4.55 | 81902 | 106 | - | - | - - | - - | See details in Table XIX | | | Ca4Ir1O6 |
| 167.4.56 | 280873 | 106 | - | - | - - | - - | See details in Table XIX | | | Ca4Ir1O6 |
| 167.4.57 | 50021 | 102 | - | - | - - | - - | See details in Table XIX | | | Ca3Na1O6Ru1 |
| 167.4.58 | 50020 | 102 | - | - | - - | - - | See details in Table XIX | | | Ca3Na1O6Ru1 |
| 167.4.59 | 50808 | 122 | - | - | - - | - - | See details in Table XIX | | | Ca3Mn1O6Zn1 |
| 167.4.60 | 96219 | 102 | - | - | - - | - - | See details in Table XIX | | | Ca3Li1O6Ru1 |
| 167.4.61 | 248206 | 102 | - | - | - - | - - | See details in Table XIX | | | Ca3Li1O6Os1 |
| 167.4.62 | 96484 | 118 | - | - | - - | - - | See details in Table XIX | | | Ca3Fe1O6Rh1 |
| 167.4.63 | 50810 | 118 | - | - | - - | - - | See details in Table XIX | | | Ca3Fe1O6Rh1 |
| 167.4.64 | 181449 | 130 | - | - | - - | - - | See details in Table XIX | | | Ca1Ir3Si7 |
| 167.4.65 | 81899 | 154 | - | - | - - | - - | See details in Table XIX | | | Ca1Ir1O6Sr3 |
| 167.4.66 | 80868 | 58 | - | - | - - | - - | See details in Table XIX | | | C1Mn1O3 |
| 167.4.67 | 80867 | 58 | - | - | - - | - - | See details in Table XIX | | | C1Mn1O3 |
| 167.4.68 | 196052 | 58 | - | - | - - | - - | See details in Table XIX | | | C1Mn1O3 |
| 167.4.69 | 196051 | 58 | - | - | - - | - - | See details in Table XIX | | | C1Mn1O3 |
| 167.4.70 | 196049 | 58 | - | - | - - | - - | See details in Table XIX | | | C1Mn1O3 |
| 167.4.71 | 196048 | 58 | - | - | - - | - - | See details in Table XIX | | | C1Mn1O3 |
| 167.4.72 | 196047 | 58 | - | - | - - | - - | See details in Table XIX | | | C1Mn1O3 |
| 167.4.73 | 196046 | 58 | - | - | - - | - - | See details in Table XIX | | | C1Mn1O3 |
| 167.4.74 | 196043 | 58 | - | - | - - | - - | See details in Table XIX | | | C1Mn1O3 |
| 167.4.75 | 196042 | 58 | - | - | - - | - - | See details in Table XIX | | | C1Mn1O3 |
| 167.4.76 | 196041 | 58 | - | - | - - | - - | See details in Table XIX | | | C1Mn1O3 |
| 167.4.77 | 196039 | 58 | - | - | - - | - - | See details in Table XIX | | | C1Mn1O3 |
| 167.4.78 | 196038 | 58 | - | - | - - | - - | See details in Table XIX | | | C1Mn1O3 |
| 167.4.79 | 196037 | 58 | - | - | - - | - - | See details in Table XIX | | | C1Mn1O3 |
| 167.4.80 | 196036 | 58 | - | - | - - | - - | See details in Table XIX | | | C1Mn1O3 |
| 167.4.81 | 196035 | 58 | - | - | - - | - - | See details in Table XIX | | | C1Mn1O3 |
| 167.4.82 | 196034 | 58 | - | - | - - | - - | See details in Table XIX | | | C1Mn1O3 |
| 167.4.83 | 196033 | 58 | - | - | - - | - - | See details in Table XIX | | | C1Mn1O3 |
| 167.4.84 | 196032 | 58 | - | - | - - | - - | See details in Table XIX | | | C1Mn1O3 |
| 167.4.85 | 196031 | 58 | - | - | - - | - - | See details in Table XIX | | | C1Mn1O3 |
| 167.4.86 | 196030 | 58 | - | - | - - | - - | See details in Table XIX | | | C1Mn1O3 |
| 167.4.87 | 196029 | 58 | - | - | - - | - - | See details in Table XIX | | | C1Mn1O3 |
| 167.4.88 | 196022 | 58 | - | - | - - | - - | See details in Table XIX | | | C1Mn1O3 |
| 167.4.89 | 196020 | 58 | - | - | - - | - - | See details in Table XIX | | | C1Mn1O3 |
| 167.4.90 | 196017 | 58 | - | - | - - | - - | See details in Table XIX | | | C1Mn1O3 |
| 167.4.91 | 100677 | 58 | - | - | - - | - - | See details in Table XIX | | | C1Mn1O3 |
| 167.4.92 | 61066 | 62 | - | - | - - | - - | See details in Table XIX | | | C1Co1O3 |
| 167.4.93 | 5396 | 62 | - | - | - - | - - | See details in Table XIX | | | C1Co1O3 |
| 167.4.94 | 5208 | 62 | - | - | - - | - - | See details in Table XIX | | | C1Co1O3 |
| 167.4.95 | 251725 | 174 | - | - | - - | - - | See details in Table XIX | | | Ba3Mn1Se6U1 |
| 167.4.96 | 160141 | 90 | - | - | - - | - - | See details in Table XIX | | | B1O3Yb1 |
| 167.4.97 | 402039 | 50 | - | - | - - | - - | See details in Table XIX | | | B1O3Ti1 |
| 167.4.98 | 34474 | 58 | - | - | - - | - - | See details in Table XIX | | | B1Fe1O3 |
| 167.4.99 | 43311 | 54 | - | - | - - | - - | See details in Table XIX | | | B1Cr1O3 |
| 167.4.100 | 426465 | 190 | - | - | - - | - - | See details in Table XIX | | | Au6Ga3Sr2 |
| 167.4.101 | 391112 | 154 | - | - | - - | - - | See details in Table XIX | | | Al7Au3Tm1 |
| 167.4.102 | 391104 | 134 | - | - | - - | - - | See details in Table XIX | | | Al7Au3Pr1 |
| 167.4.103 | 391105 | 158 | - | - | - - | - - | See details in Table XIX | | | Al7Au3Lu1 |
| 167.4.104 | 391106 | 150 | - | - | - - | - - | See details in Table XIX | | | Al7Au3Ho1 |
| 167.4.105 | 105520 | 42 | - | - | - - | - - | See details in Table XIX | | | Al5Mo1 |
| 167.4.106 | 427021 | 218 | - | - | - - | - - | See details in Table XIX | | | Al3Au6Eu2 |
| 167.4.107 | 90577 | 70 | - | - | - - | - - | See details in Table XIX | | | Al1Nd1O3 |
| 167.4.108 | 90576 | 70 | - | - | - - | - - | See details in Table XIX | | | Al1Nd1O3 |
| 167.4.109 | 90575 | 70 | - | - | - - | - - | See details in Table XIX | | | Al1Nd1O3 |
| 167.4.110 | 90574 | 70 | - | - | - - | - - | See details in Table XIX | | | Al1Nd1O3 |
| 167.4.111 | 90573 | 70 | - | - | - - | - - | See details in Table XIX | | | Al1Nd1O3 |

Continued on next page

**TABLE XIV– continued**

| TQC | ICSD | NoE | DG | IndG | SubSG (w/ T) | cl | SubSG (w/o T) | cl | Max. Sub. SG | yes or no | Max. Sub. SG | yes or no | SIs (w/ T) | Stoich. Formula |
|---|---|---|---|---|---|---|---|---|---|---|---|---|---|---|
| 167.4.112 | 90572 | 70 | - | - | - | - | - | - | See details in Table XIX | | | | | Al1Nd1O3 |
| 167.4.113 | 35444 | 70 | - | - | - | - | - | - | See details in Table XIX | | | | | Al1Nd1O3 |
| 167.4.114 | 249018 | 70 | - | - | - | - | - | - | See details in Table XIX | | | | | Al1Nd1O3 |
| 167.4.115 | 245564 | 66 | - | - | - | - | - | - | See details in Table XIX | | | | | Al1Ce1O3 |
| 167.4.116 | 245563 | 66 | - | - | - | - | - | - | See details in Table XIX | | | | | Al1Ce1O3 |
| 167.4.117 | 245272 | 66 | - | - | - | - | - | - | See details in Table XIX | | | | | Al1Ce1O3 |
| 167.4.118 | 245271 | 66 | - | - | - | - | - | - | See details in Table XIX | | | | | Al1Ce1O3 |
| 167.4.119 | 245269 | 66 | - | - | - | - | - | - | See details in Table XIX | | | | | Al1Ce1O3 |
| 167.4.120 | 245268 | 66 | - | - | - | - | - | - | See details in Table XIX | | | | | Al1Ce1O3 |
| 167.4.121 | 245267 | 66 | - | - | - | - | - | - | See details in Table XIX | | | | | Al1Ce1O3 |
| 167.4.122 | 238056 | 182 | - | - | - | - | - | - | See details in Table XIX | | | | | Ag1Ba3S6U1 |
| 173.3.1 | 96420 | 162 | - | - | - | - | - | - | 1 | yes | 1 | yes | - | O3Rb0.333W1 |
| 173.3.2 | 426496 | 176 | - | - | - | - | - | - | 143 | yes | 143 | yes | - | In1La3Ni1S7 |
| | | | | | | | | | 4 | yes | 4 | yes | - | |
| 173.3.3 | 154261 | 204 | - | - | - | - | - | - | 1 | yes | 1 | yes | - | I3K1Ti1 |
| 173.3.4 | 426885 | 242 | - | - | - | - | - | - | 143 | yes | 143 | yes | - | Er3Ga1Ni1S7 |
| 173.3.5 | 159888 | 240 | - | - | - | - | - | - | 143 | yes | 143 | yes | - | Cu1Ge1Ho3S7 |
| | | | | | | | | | 4 | yes | 4 | yes | - | |
| 173.3.6 | 155941 | 222 | - | - | - | - | - | - | 143 | yes | 143 | yes | - | Cu1Gd3S7Sn1 |
| 173.3.7 | 155080 | 246 | - | - | - | - | - | - | 1 | yes | 1 | yes | - | Cu1Er3Ge1S7 |
| 173.3.8 | 426495 | 174 | - | - | - | - | - | - | 1 | yes | 1 | yes | - | Co1In1La3S7 |
| 173.3.9 | 426677 | 174 | - | - | - | - | - | - | 1 | yes | 1 | yes | - | Co1Ga1La3S7 |
| 173.3.10 | 152546 | 186 | - | - | - | - | - | - | 143 | yes | 143 | yes | - | Ce3Cu1Se7Sn1 |
| 173.4.1 | 426493 | 178 | - | - | - | - | - | - | See details in Table XIX | | | | | Ce3Fe1In1S7 |
| 174.3.1 | 417604 | 152 | - | - | - | - | - | - | 6 | yes | 6 | yes | - | B3Cl1La2 |
| 176.1.1 | 80772 | 104 | 0.0000 | -0.4465 | 174 | 11 | 174 | 2 | - | - | - | - | z6m0=2<br>z12prime=8 | Mn1N3Sr3 |
| 176.1.2 | 251133 | 240 | 0.0784 | -0.0245 | 2,174 | 1 | 2,174 | 1 | - | - | - | - | z4=3<br>z6m0=5<br>z12prime=11 | Ba4Nb2O9 |
| 176.3.1 | 79629 | 72 | - | - | - | - | - | - | 11 | no(2) | 11 | no(2) | z4=3 | Se4Ti3 |
| 176.3.2 | 29336 | 240 | - | - | - | - | - | - | 11 | no(2) | 11 | no(2) | z4=3 | O1Rb6 |
| 176.3.3 | 106615 | 164 | - | - | - | - | - | - | 11 | yes | 11 | yes | - | Er3Ru2 |
| 176.3.4 | 80771 | 104 | - | - | - | - | - | - | 174 | no(174) | 174 | no(174) | z3m0=1 | Ba3Mn1N3 |
| | | | | | | | | | 147 | no(2) | 147 | no(2) | z4=3 | |
| | | | | | | | | | 11 | no(2) | 11 | no(2) | z4=3 | |
| 176.3.5 | 2530 | 168 | - | - | - | - | - | - | 174 | no(174) | 174 | no(174) | z3m0=1 | Ba1Fe4O7 |
| | | | | | | | | | 147 | yes | 147 | yes | - | |
| | | | | | | | | | 11 | yes | 11 | yes | - | |
| 176.3.6 | 2529 | 168 | - | - | - | - | - | - | 174 | no(174) | 174 | no(174) | z3m0=1 | Ba1Fe4O7 |
| | | | | | | | | | 147 | no(2) | 147 | no(2) | z4=1 | |
| | | | | | | | | | 11 | no(2) | 11 | no(2) | z4=1 | |
| 176.3.7 | 44424 | 84 | - | - | - | - | - | - | 11 | no(2) | 11 | no(2) | z4=3 | B4Ir3Sc1 |
| 176.3.8 | 260285 | 44 | - | - | - | - | - | - | 11 | no(2) | 11 | no(2) | z4=3 | B3Pr1 |
| 176.3.9 | 78985 | 176 | - | - | - | - | - | - | 11 | yes | 11 | yes | - | B3Ir7Sn4 |
| 176.3.10 | 79934 | 124 | - | - | - | - | - | - | 11 | no(2) | 11 | yes | z4=2 | As1Nb3Te3 |
| 176.4.1 | 72922 | 78 | - | - | - | - | - | - | See details in Table XIX | | | | | S4V3 |
| 176.4.2 | 72921 | 78 | - | - | - | - | - | - | See details in Table XIX | | | | | S4V3 |
| 176.4.3 | 72920 | 78 | - | - | - | - | - | - | See details in Table XIX | | | | | S4V3 |
| 176.4.4 | 16278 | 126 | - | - | - | - | - | - | See details in Table XIX | | | | | Nb3Se4 |
| 176.4.5 | 53119 | 78 | - | - | - | - | - | - | See details in Table XIX | | | | | Mo6Se6Tl2 |
| 176.4.6 | 23301 | 78 | - | - | - | - | - | - | See details in Table XIX | | | | | Mo6Se6Tl2 |
| 176.4.7 | 90811 | 78 | - | - | - | - | - | - | See details in Table XIX | | | | | Mo3Te3Tl1 |
| 176.4.8 | 30752 | 90 | - | - | - | - | - | - | See details in Table XIX | | | | | K1Mo3S3 |
| 176.4.9 | 53100 | 78 | - | - | - | - | - | - | See details in Table XIX | | | | | In1Mo3Te3 |
| 176.4.10 | 53098 | 78 | - | - | - | - | - | - | See details in Table XIX | | | | | In1Mo3Se3 |
| 176.4.11 | 200095 | 82 | - | - | - | - | - | - | See details in Table XIX | | | | | H3Dy1O3 |

Continued on next page

**TABLE XIV– continued**

| TQC | ICSD | NoE | DG | IndG | SubSG (w/ T) | cl | SubSG (w/o T) | cl | Max. Sub. SG | yes or no | Max. Sub. SG | yes or no | SIs (w/ T) | Stoich. Formula |
|---|---|---|---|---|---|---|---|---|---|---|---|---|---|---|
| 176.4.12 | 185605 | 82 | - | - | - | - | - | - | See details in Table XIX | | | | | H3Dy1O3 |
| 176.4.13 | 100352 | 90 | - | - | - | - | - | - | See details in Table XIX | | | | | Fe3Te3Tl1 |
| 176.4.14 | 154803 | 102 | - | - | - | - | - | - | See details in Table XIX | | | | | Cr1N3Sr3 |
| 176.4.15 | 4060 | 74 | - | - | - | - | - | - | See details in Table XIX | | | | | Cl3Pu1 |
| 176.4.16 | 9536 | 78 | - | - | - | - | - | - | See details in Table XIX | | | | | Cl3Cm1 |
| 176.4.17 | 4070 | 70 | - | - | - | - | - | - | See details in Table XIX | | | | | Br3U1 |
| 176.4.18 | 36502 | 106 | - | - | - | - | - | - | See details in Table XIX | | | | | Ba3Fe1N3 |
| 176.4.19 | 154802 | 102 | - | - | - | - | - | - | See details in Table XIX | | | | | Ba3Cr1N3 |
| 180.3.1 | 156947 | 102 | - | - | - | - | - | - | 21 | yes | 21 | yes | - | B1Pt2Y1 |
| 180.3.2 | 156946 | 102 | - | - | - | - | - | - | 21 | yes | 21 | yes | - | B1Pt2Y1 |
| 180.3.3 | 409979 | 84 | - | - | - | - | - | - | 154 | yes | 154 | yes | - | B1Ir1Mg1 |
| | | | | | | | | | 153 | yes | 153 | yes | - | |
| | | | | | | | | | 21 | yes | 21 | yes | - | |
| 180.4.1 | 77114 | 84 | - | - | - | - | - | - | See details in Table XIX | | | | | Mo1Sn2 |
| 181.3.1 | 173662 | 36 | - | - | - | - | - | - | 21 | yes | 21 | yes | - | Al2W1 |
| 182.3.1 | 88320 | 36 | - | - | - | - | - | - | 20 | yes | 20 | yes | - | O1Zr3 |
| 182.3.2 | 88316 | 36 | - | - | - | - | - | - | 20 | yes | 20 | yes | - | O1Zr3 |
| 182.3.3 | 80930 | 58 | - | - | - | - | - | - | 5 | yes | 5 | yes | - | Fe3N1 |
| | | | | | | | | | 5 | yes | 5 | yes | - | |
| 182.3.4 | 79984 | 58 | - | - | - | - | - | - | 5 | yes | 5 | yes | - | Fe3N1 |
| | | | | | | | | | 5 | yes | 5 | yes | - | |
| 182.3.5 | 79983 | 58 | - | - | - | - | - | - | 5 | yes | 5 | yes | - | Fe3N1 |
| | | | | | | | | | 5 | yes | 5 | yes | - | |
| 182.3.6 | 79982 | 58 | - | - | - | - | - | - | 5 | yes | 5 | yes | - | Fe3N1 |
| | | | | | | | | | 5 | yes | 5 | yes | - | |
| 182.3.7 | 79981 | 58 | - | - | - | - | - | - | 5 | yes | 5 | yes | - | Fe3N1 |
| | | | | | | | | | 5 | yes | 5 | yes | - | |
| 182.3.8 | 61514 | 148 | - | - | - | - | - | - | 150 | yes | 150 | yes | - | Cu5Na1S3 |
| | | | | | | | | | 149 | yes | 149 | yes | - | |
| | | | | | | | | | 20 | yes | 20 | yes | - | |
| 185.3.1 | 79253 | 168 | - | - | - | - | - | - | 36 | yes | 36 | yes | - | Mn1O3Sc1 |
| 185.3.2 | 50694 | 168 | - | - | - | - | - | - | 36 | yes | 36 | yes | - | Mn1O3Sc1 |
| 185.3.3 | 54253 | 96 | - | - | - | - | - | - | 158 | yes | 158 | yes | - | Mg3Pt1 |
| | | | | | | | | | 157 | yes | 157 | yes | - | |
| | | | | | | | | | 36 | yes | 36 | yes | - | |
| 185.3.4 | 409824 | 96 | - | - | - | - | - | - | 158 | yes | 158 | yes | - | Mg3Pd1 |
| | | | | | | | | | 157 | yes | 157 | yes | - | |
| | | | | | | | | | 36 | yes | 36 | yes | - | |
| 185.3.5 | 15056 | 228 | - | - | - | - | - | - | 36 | yes | 36 | yes | - | Cu3P1 |
| 185.3.6 | 167455 | 240 | - | - | - | - | - | - | 36 | yes | 36 | yes | - | Cl3K1Ni1 |
| 185.3.7 | 409223 | 192 | - | - | - | - | - | - | 36 | yes | 36 | yes | - | Bi1K3 |
| 185.3.8 | 97539 | 132 | - | - | - | - | - | - | 36 | yes | 36 | yes | - | Ba3Cr1S5 |
| 185.4.1 | 106871 | 90 | - | - | - | - | - | - | See details in Table XIX | | | | | Ir1Mg3 |
| 185.4.2 | 26455 | 210 | - | - | - | - | - | - | See details in Table XIX | | | | | I3Rb1V1 |
| 185.4.3 | 89996 | 210 | - | - | - | - | - | - | See details in Table XIX | | | | | Ba1Mn1O3 |
| 185.4.4 | 106287 | 102 | - | - | - | - | - | - | See details in Table XIX | | | | | Au1Mg3 |
| 186.3.1 | 106444 | 80 | - | - | - | - | - | - | 36 | yes | 36 | yes | - | Sn1Yb1Zn1 |
| 186.3.2 | 99453 | 88 | - | - | - | - | - | - | 36 | yes | 36 | yes | - | Mo1N1 |
| 186.3.3 | 104802 | 120 | - | - | - | - | - | - | 36 | yes | 36 | yes | - | Lu1Mn5 |
| 186.3.4 | 61069 | 164 | - | - | - | - | - | - | 36 | yes | 36 | yes | - | Fe2Mo3O8 |
| 186.3.5 | 27216 | 164 | - | - | - | - | - | - | 36 | yes | 36 | yes | - | Fe2Mo3O8 |
| 186.3.6 | 87503 | 112 | - | - | - | - | - | - | 36 | yes | 36 | yes | - | Fe1S1 |
| 186.3.7 | 87502 | 112 | - | - | - | - | - | - | 36 | yes | 36 | yes | - | Fe1S1 |
| 186.3.8 | 68848 | 112 | - | - | - | - | - | - | 36 | yes | 36 | yes | - | Fe1S1 |
| 186.3.9 | 68847 | 112 | - | - | - | - | - | - | 36 | yes | 36 | yes | - | Fe1S1 |
| 186.3.10 | 53528 | 28 | - | - | - | - | - | - | 159 | yes | 159 | yes | - | Fe1S1 |
| | | | | | | | | | 156 | yes | 156 | yes | - | |

Continued on next page

**TABLE XIV– continued**

| TQC | ICSD | NoE | DG | IndG | SubSG (w/ T) | cl | SubSG (w/o T) | cl | Max. Sub. SG | yes or no | Max. Sub. SG | yes or no | SIs (w/ T) | Stoich. Formula |
|---|---|---|---|---|---|---|---|---|---|---|---|---|---|---|
| | | | | | | | | | 36 | yes | 36 | yes | - | |
| 186.3.11 | 72303 | 76 | - | - | - | - | - | - | 36 | yes | 36 | yes | - | Fe1Na6S4 |
| 186.3.12 | 424277 | 108 | - | - | - | - | - | - | 36 | yes | 36 | yes | - | Cu4In1Mn1 |
| 186.3.13 | 416544 | 52 | - | - | - | - | - | - | 159 | yes | 159 | yes | - | Cu1Sn1Y1 |
| | | | | | | | | | 156 | yes | 156 | yes | - | |
| | | | | | | | | | 36 | yes | 36 | yes | - | |
| 186.3.14 | 416543 | 36 | - | - | - | - | - | - | 159 | yes | 159 | yes | - | Cu1Sc1Sn1 |
| | | | | | | | | | 156 | yes | 156 | yes | - | |
| | | | | | | | | | 36 | yes | 36 | yes | - | |
| 186.3.15 | 107004 | 52 | - | - | - | - | - | - | 36 | yes | 36 | yes | - | Cu1Pb1Y1 |
| 186.3.16 | 107285 | 76 | - | - | - | - | - | - | 36 | yes | 36 | yes | - | Cu1Pb1Tm1 |
| 186.3.17 | 416546 | 80 | - | - | - | - | - | - | 159 | yes | 159 | yes | - | Cu1Lu1Sn1 |
| | | | | | | | | | 156 | yes | 156 | yes | - | |
| | | | | | | | | | 36 | yes | 36 | yes | - | |
| 186.3.18 | 107286 | 80 | - | - | - | - | - | - | 36 | yes | 36 | yes | - | Cu1Lu1Pb1 |
| 186.3.19 | 107283 | 72 | - | - | - | - | - | - | 36 | yes | 36 | yes | - | Cu1Ho1Pb1 |
| 186.3.20 | 8105 | 72 | - | - | - | - | - | - | 159 | yes | 159 | yes | - | Cr1Cs1I3 |
| | | | | | | | | | 156 | yes | 156 | yes | - | |
| | | | | | | | | | 36 | yes | 36 | yes | - | |
| 186.3.21 | 10206 | 72 | - | - | - | - | - | - | 159 | yes | 159 | yes | - | Cl3Cr1Cs1 |
| | | | | | | | | | 156 | yes | 156 | yes | - | |
| | | | | | | | | | 36 | yes | 36 | yes | - | |
| 186.3.22 | 417716 | 268 | - | - | - | - | - | - | 36 | yes | 36 | yes | - | Ca3Cu8Sn4 |
| 186.3.23 | 106358 | 36 | - | - | - | - | - | - | 36 | yes | 36 | yes | - | Ca1Sn1Zn1 |
| 186.3.24 | 2590 | 72 | - | - | - | - | - | - | 159 | yes | 159 | yes | - | Br3Cr1Cs1 |
| | | | | | | | | | 156 | yes | 156 | yes | - | |
| | | | | | | | | | 36 | yes | 36 | yes | - | |
| 186.3.25 | 416144 | 80 | - | - | - | - | - | - | 36 | yes | 36 | yes | - | Bi1Cu1Yb1 |
| 186.3.26 | 416143 | 80 | - | - | - | - | - | - | 36 | yes | 36 | yes | - | Bi1Cu1Yb1 |
| 186.3.27 | 412207 | 100 | - | - | - | - | - | - | 36 | yes | 36 | yes | - | Au3Li1Sn4 |
| 186.3.28 | 415826 | 52 | - | - | - | - | - | - | 159 | yes | 159 | yes | - | Au1Sn1Y1 |
| | | | | | | | | | 156 | yes | 156 | yes | - | |
| | | | | | | | | | 36 | yes | 36 | yes | - | |
| 186.3.29 | 83984 | 80 | - | - | - | - | - | - | 36 | yes | 36 | yes | - | Au1Sb1Yb1 |
| 186.3.30 | 57240 | 80 | - | - | - | - | - | - | 36 | yes | 36 | yes | - | Au1Sb1Yb1 |
| 186.3.31 | 57022 | 80 | - | - | - | - | - | - | 36 | yes | 36 | yes | - | Au1Sb1Yb1 |
| 186.3.32 | 429889 | 52 | - | - | - | - | - | - | 36 | yes | 36 | yes | - | Au1La1Sn1 |
| 186.3.33 | 405323 | 52 | - | - | - | - | - | - | 36 | yes | 36 | yes | - | Au1Ge1Y1 |
| 186.3.34 | 81733 | 36 | - | - | - | - | - | - | 159 | yes | 159 | yes | - | Au1Ge1Sc1 |
| | | | | | | | | | 156 | yes | 156 | yes | - | |
| | | | | | | | | | 36 | yes | 36 | yes | - | |
| 186.3.35 | 81734 | 80 | - | - | - | - | - | - | 36 | yes | 36 | yes | - | Au1Ge1Lu1 |
| 186.3.36 | 405322 | 52 | - | - | - | - | - | - | 36 | yes | 36 | yes | - | Au1Ge1La1 |
| 186.3.37 | 93436 | 72 | - | - | - | - | - | - | 36 | yes | 36 | yes | - | Au1Ge1Ho1 |
| 186.3.38 | 93435 | 72 | - | - | - | - | - | - | 36 | yes | 36 | yes | - | Au1Ge1Ho1 |
| 186.3.39 | 248685 | 36 | - | - | - | - | - | - | 36 | yes | 36 | yes | - | Au1Bi1Ca1 |
| 186.3.40 | 248683 | 36 | - | - | - | - | - | - | 36 | yes | 36 | yes | - | Au1Bi1Ca1 |
| 186.3.41 | 248681 | 36 | - | - | - | - | - | - | 36 | yes | 36 | yes | - | Au1Bi1Ca1 |
| 186.3.42 | 248679 | 36 | - | - | - | - | - | - | 36 | yes | 36 | yes | - | Au1Bi1Ca1 |
| 186.3.43 | 248677 | 36 | - | - | - | - | - | - | 36 | yes | 36 | yes | - | Au1Bi1Ca1 |
| 186.3.44 | 96763 | 132 | - | - | - | - | - | - | 36 | yes | 36 | yes | - | As2Ni1Tm2 |
| 186.3.45 | 416375 | 52 | - | - | - | - | - | - | 159 | yes | 159 | yes | - | Ag1Sn1Y1 |
| | | | | | | | | | 156 | yes | 156 | yes | - | |
| | | | | | | | | | 36 | yes | 36 | yes | - | |
| 186.3.46 | 245234 | 52 | - | - | - | - | - | - | 159 | yes | 159 | yes | - | Ag1Sn1Y1 |
| | | | | | | | | | 156 | yes | 156 | yes | - | |
| | | | | | | | | | 36 | yes | 36 | yes | - | |

Continued on next page

**TABLE XIV– continued**

| TQC | ICSD | NoE | DG | IndG | SubSG cl (w/ T) | SubSG cl (w/o T) | Max. Sub. SG | yes or no | Max. Sub. SG | yes or no | SIs (w/ T) | Stoich. Formula |
|---|---|---|---|---|---|---|---|---|---|---|---|---|
| 186.3.47 | 417550 | 76 | - | - | - - | - - | 36 | yes | 36 | yes | - | Ag1Sn1Tm1 |
| 186.3.48 | 107105 | 68 | - | - | - - | - - | 36 | yes | 36 | yes | - | Ag1Pb1Tb1 |
| 186.3.49 | 418580 | 80 | - | - | - - | - - | 159 | yes | 159 | yes | - | Ag1Lu1Sn1 |
| | | | | | | | 156 | yes | 156 | yes | - | |
| | | | | | | | 36 | yes | 36 | yes | - | |
| 186.3.50 | 107099 | 52 | - | - | - - | - - | 36 | yes | 36 | yes | - | Ag1La1Pb1 |
| 186.3.51 | 418581 | 72 | - | - | - - | - - | 36 | yes | 36 | yes | - | Ag1Ho1Sn1 |
| 186.3.52 | 107107 | 72 | - | - | - - | - - | 36 | yes | 36 | yes | - | Ag1Ho1Pb1 |
| 186.3.53 | 416283 | 36 | - | - | - - | - - | 36 | yes | 36 | yes | - | Ag1Bi1Ca1 |
| 186.4.1 | 76600 | 58 | - | - | - - | - - | See details in Table XIX | | | | | Nd1Pt1Sb1 |
| 186.4.2 | 405625 | 170 | - | - | - - | - - | See details in Table XIX | | | | | N7Si4Sr1Yb1 |
| 186.4.3 | 40534 | 162 | - | - | - - | - - | See details in Table XIX | | | | | Mo3O8Sc1Zn1 |
| 186.4.4 | 99452 | 88 | - | - | - - | - - | See details in Table XIX | | | | | Mo1N1 |
| 186.4.5 | 262928 | 26 | - | - | - - | - - | See details in Table XIX | | | | | Mn1O1 |
| 186.4.6 | 65451 | 74 | - | - | - - | - - | See details in Table XIX | | | | | Mn1Na6Te4 |
| 186.4.7 | 95067 | 74 | - | - | - - | - - | See details in Table XIX | | | | | Mn1Na6S4 |
| 186.4.8 | 95065 | 74 | - | - | - - | - - | See details in Table XIX | | | | | Mn1Na6S4 |
| 186.4.9 | 65447 | 74 | - | - | - - | - - | See details in Table XIX | | | | | Mn1Na6S4 |
| 186.4.10 | 104681 | 214 | - | - | - - | - - | See details in Table XIX | | | | | La7Ni3 |
| 186.4.11 | 104680 | 214 | - | - | - - | - - | See details in Table XIX | | | | | La7Ni3 |
| 186.4.12 | 391303 | 186 | - | - | - - | - - | See details in Table XIX | | | | | La4Mn1O1Se6 |
| 186.4.13 | 391305 | 186 | - | - | - - | - - | See details in Table XIX | | | | | La4Mn1O1S6 |
| 186.4.14 | 26218 | 42 | - | - | - - | - - | See details in Table XIX | | | | | Ir1Si3 |
| 186.4.15 | 426618 | 142 | - | - | - - | - - | See details in Table XIX | | | | | In1Li2Mo3O8 |
| 186.4.16 | 423281 | 130 | - | - | - - | - - | See details in Table XIX | | | | | H2Fe1K1O6P2 |
| 186.4.17 | 415946 | 66 | - | - | - - | - - | See details in Table XIX | | | | | Gd1Pd1Sb1 |
| 186.4.18 | 261028 | 74 | - | - | - - | - - | See details in Table XIX | | | | | Er1Ga1Zn1 |
| 186.4.19 | 415945 | 70 | - | - | - - | - - | See details in Table XIX | | | | | Dy1Pd1Sb1 |
| 186.4.20 | 422454 | 268 | - | - | - - | - - | See details in Table XIX | | | | | Cu8Sn4Sr3 |
| 186.4.21 | 182103 | 268 | - | - | - - | - - | See details in Table XIX | | | | | Cu8Sn4Sr3 |
| 186.4.22 | 190121 | 38 | - | - | - - | - - | See details in Table XIX | | | | | Cu1Hf1Sn1 |
| 186.4.23 | 414432 | 78 | - | - | - - | - - | See details in Table XIX | | | | | Cu1Ge1Yb1 |
| 186.4.24 | 151171 | 66 | - | - | - - | - - | See details in Table XIX | | | | | Cu1Gd1Sn1 |
| 186.4.25 | 107281 | 66 | - | - | - - | - - | See details in Table XIX | | | | | Cu1Gd1Pb1 |
| 186.4.26 | 107284 | 74 | - | - | - - | - - | See details in Table XIX | | | | | Cu1Er1Pb1 |
| 186.4.27 | 55581 | 70 | - | - | - - | - - | See details in Table XIX | | | | | Cu1Dy1Sn1 |
| 186.4.28 | 405201 | 70 | - | - | - - | - - | See details in Table XIX | | | | | Cu1Dy1Sn1 |
| 186.4.29 | 107274 | 70 | - | - | - - | - - | See details in Table XIX | | | | | Cu1Dy1Pb1 |
| 186.4.30 | 54572 | 242 | - | - | - - | - - | See details in Table XIX | | | | | Co8Sn4Y3 |
| 186.4.31 | 152945 | 242 | - | - | - - | - - | See details in Table XIX | | | | | Co8Sn4Y3 |
| 186.4.32 | 68604 | 78 | - | - | - - | - - | See details in Table XIX | | | | | Co1Na6Se4 |
| 186.4.33 | 37012 | 78 | - | - | - - | - - | See details in Table XIX | | | | | Co1Na6S4 |
| 186.4.34 | 2064 | 82 | - | - | - - | - - | See details in Table XIX | | | | | Cl3Cs1Cu1 |
| 186.4.35 | 102234 | 228 | - | - | - - | - - | See details in Table XIX | | | | | Ce7Ni3 |
| 186.4.36 | 416788 | 54 | - | - | - - | - - | See details in Table XIX | | | | | Ce1Cu1Sn1 |
| 186.4.37 | 156394 | 54 | - | - | - - | - - | See details in Table XIX | | | | | Ce1Cu1Sn1 |
| 186.4.38 | 43476 | 130 | - | - | - - | - - | See details in Table XIX | | | | | C5Al3Zr3 |
| 186.4.39 | 153859 | 66 | - | - | - - | - - | See details in Table XIX | | | | | C4Al3Zr2 |
| 186.4.40 | 66750 | 106 | - | - | - - | - - | See details in Table XIX | | | | | C4Al3U2 |
| 186.4.41 | 43478 | 70 | - | - | - - | - - | See details in Table XIX | | | | | C3Al3U1 |
| 186.4.42 | 405194 | 170 | - | - | - - | - - | See details in Table XIX | | | | | Ba1N7Si4Yb1 |
| 186.4.43 | 95747 | 198 | - | - | - - | - - | See details in Table XIX | | | | | Ba1Co4O7Y1 |
| 186.4.44 | 95746 | 198 | - | - | - - | - - | See details in Table XIX | | | | | Ba1Co4O7Y1 |
| 186.4.45 | 188853 | 198 | - | - | - - | - - | See details in Table XIX | | | | | Ba1Co4O7Y1 |
| 186.4.46 | 168709 | 198 | - | - | - - | - - | See details in Table XIX | | | | | Ba1Co4O7Y1 |
| 186.4.47 | 59794 | 182 | - | - | - - | - - | See details in Table XIX | | | | | Ba1Co4In1O7 |
| 186.4.48 | 413464 | 182 | - | - | - - | - - | See details in Table XIX | | | | | Ba1Co4In1O7 |

Continued on next page

**TABLE XIV– continued**

| TQC | ICSD | NoE | DG | IndG | SubSG (w/ T) | cl | SubSG (w/o T) | cl | Max. Sub. SG | yes or no | Max. Sub. SG | yes or no | SIs (w/ T) | Stoich. Formula |
|---|---|---|---|---|---|---|---|---|---|---|---|---|---|---|
| 186.4.49 | 59781 | 218 | - | - | - | - | - | - | See details in Table XIX | | | | | Ba1Co4Ho1O7 |
| 186.4.50 | 420427 | 218 | - | - | - | - | - | - | See details in Table XIX | | | | | Ba1Co4Ho1O7 |
| 186.4.51 | 420426 | 218 | - | - | - | - | - | - | See details in Table XIX | | | | | Ba1Co4Ho1O7 |
| 186.4.52 | 420425 | 218 | - | - | - | - | - | - | See details in Table XIX | | | | | Ba1Co4Ho1O7 |
| 186.4.53 | 420424 | 218 | - | - | - | - | - | - | See details in Table XIX | | | | | Ba1Co4Ho1O7 |
| 186.4.54 | 420423 | 218 | - | - | - | - | - | - | See details in Table XIX | | | | | Ba1Co4Ho1O7 |
| 186.4.55 | 58597 | 206 | - | - | - | - | - | - | See details in Table XIX | | | | | Au3Sr7 |
| 186.4.56 | 416419 | 66 | - | - | - | - | - | - | See details in Table XIX | | | | | Au1Gd1Sn1 |
| 186.4.57 | 411539 | 54 | - | - | - | - | - | - | See details in Table XIX | | | | | Au1Ce1Sn1 |
| 186.4.58 | 57176 | 54 | - | - | - | - | - | - | See details in Table XIX | | | | | Au1Ce1Ge1 |
| 186.4.59 | 71615 | 58 | - | - | - | - | - | - | See details in Table XIX | | | | | As1Nd1Pd1 |
| 186.4.60 | 107104 | 66 | - | - | - | - | - | - | See details in Table XIX | | | | | Ag1Gd1Pb1 |
| 186.4.61 | 418582 | 70 | - | - | - | - | - | - | See details in Table XIX | | | | | Ag1Dy1Sn1 |
| 186.4.62 | 107100 | 54 | - | - | - | - | - | - | See details in Table XIX | | | | | Ag1Ce1Pb1 |
| 187.3.1 | 410867 | 10 | - | - | - | - | - | - | 38 | yes | 38 | yes | - | Te1Zr1 |
| 187.3.2 | 89610 | 76 | - | - | - | - | - | - | 38 | yes | 38 | yes | - | Mg1P4Rh6 |
| 187.3.3 | 411151 | 16 | - | - | - | - | - | - | 38 | yes | 38 | yes | - | Li1N1Ni1 |
| 187.3.4 | 72375 | 70 | - | - | - | - | - | - | 38 | yes | 38 | yes | - | Cr6P4U1 |
| 187.3.5 | 94412 | 76 | - | - | - | - | - | - | 38 | yes | 38 | yes | - | Co6Mg1P4 |
| 187.3.6 | 260171 | 10 | - | - | - | - | - | - | 38 | yes | 38 | yes | - | C1W1 |
| 187.3.7 | 260168 | 10 | - | - | - | - | - | - | 38 | yes | 38 | yes | - | C1W1 |
| 187.3.8 | 246151 | 10 | - | - | - | - | - | - | 38 | yes | 38 | yes | - | C1W1 |
| 187.3.9 | 246150 | 10 | - | - | - | - | - | - | 38 | yes | 38 | yes | - | C1W1 |
| 187.3.10 | 246149 | 10 | - | - | - | - | - | - | 38 | yes | 38 | yes | - | C1W1 |
| 187.3.11 | 249661 | 178 | - | - | - | - | - | - | 38 | yes | 38 | yes | - | Au9In13Sr4 |
| 187.3.12 | 89613 | 84 | - | - | - | - | - | - | 38 | yes | 38 | yes | - | As4Rh6Sr1 |
| 187.3.13 | 89611 | 76 | - | - | - | - | - | - | 38 | yes | 38 | yes | - | As4Mg1Rh6 |
| 187.3.14 | 89612 | 76 | - | - | - | - | - | - | 38 | yes | 38 | yes | - | As4Ca1Rh6 |
| 188.3.1 | 193216 | 174 | - | - | - | - | - | - | 6 | yes | 6 | yes | - | Co1K1O9P3 |
| | | | | | | | | | 5 | yes | 5 | yes | - | |
| 188.4.1 | 73339 | 50 | - | - | - | - | - | - | See details in Table XIX | | | | | I3Li1Sc1 |
| 189.1.1 | 51718 | 92 | 0.0000 | -1.2322 | 174 | 1 | 174 | 1 | - | - | - | - | z3m0=1 | Si5Yb3 |
| 189.1.2 | 415599 | 92 | 0.0000 | -1.2333 | 174 | 1 | 174 | 1 | - | - | - | - | z3m0=1 | Si5Yb3 |
| 189.1.3 | 408117 | 92 | 0.0000 | -1.2387 | 174 | 1 | 174 | 1 | - | - | - | - | z3m0=1 | Si5Yb3 |
| 189.1.4 | 410418 | 114 | 0.0000 | -1.2043 | 174 | 5 | 174 | 6 | - | - | - | - | z3m0=1<br>z3mpi=2 | Pt1Sn1Yb1 |
| 189.1.5 | 415370 | 78 | 0.0000 | -0.0121 | 174 | 2 | 174 | 2 | - | - | - | - | z3m0=2 | Pr1Rh1Sn1 |
| 189.1.6 | 196273 | 30 | 0.0037 | -1.3373 | 174 | 1 | 174 | 1 | - | - | - | - | z3m0=1 | O1Ti1 |
| 189.1.7 | 183337 | 108 | 0.0000 | -0.0546 | 174 | 2 | 174 | 2 | - | - | - | - | z3m0=2 | Nd1Pd1Zn1 |
| 189.1.8 | 419191 | 84 | 0.0000 | -0.0568 | 174 | 3 | 174 | 4 | - | - | - | - | z3m0=2<br>z3mpi=1 | Nd1Pd1Sn1 |
| 189.1.9 | 414438 | 48 | 0.0223 | -0.4859 | 174 | 1 | 174 | 1 | - | - | - | - | z3m0=1 | Mg1Tl1Y1 |
| 189.1.10 | 54344 | 90 | 0.0046 | -0.2681 | 174 | 1 | 174 | 1 | - | - | - | - | z3m0=1 | Mg1Sn1Yb1 |
| 189.1.11 | 411152 | 50 | 0.0000 | -1.7068 | 174 | 2 | 174 | 2 | - | - | - | - | z3m0=2 | Li5N3Ni3 |
| 189.1.12 | 415594 | 72 | 0.0000 | -1.2755 | 174 | 3 | 174 | 4 | - | - | - | - | z3m0=2<br>z3mpi=1 | La1Rh1Sn1 |
| 189.1.13 | 414439 | 48 | 0.0237 | -0.2227 | 174 | 1 | 174 | 1 | - | - | - | - | z3m0=1 | La1Mg1Tl1 |
| 189.1.14 | 160910 | 48 | 0.0194 | -0.5872 | 174 | 1 | 174 | 1 | - | - | - | - | z3m0=1 | In1Mg1Y1 |
| 189.1.15 | 85966 | 44 | 0.0000 | -1.4246 | 174 | 2 | 174 | 2 | - | - | - | - | z3m0=2 | Hf6Ni1Sb2 |
| 189.1.16 | 409925 | 54 | 0.0000 | -0.2974 | 174 | 1 | 174 | 1 | - | - | - | - | z3m0=1 | Ge1Pd1Ti1 |
| 189.1.17 | 59177 | 54 | 0.0000 | -0.0259 | 174 | 8 | 174 | 5 | - | - | - | - | z3mpi=1 | Ge1Li1Pr1 |
| 189.1.18 | 240950 | 48 | 0.0246 | -0.5016 | 174 | 1 | 174 | 1 | - | - | - | - | z3m0=1 | Ga1Mg1Y1 |
| 189.1.19 | 160909 | 48 | 0.0350 | -0.5083 | 174 | 1 | 174 | 1 | - | - | - | - | z3m0=1 | Ga1Mg1Y1 |
| 189.1.20 | 413276 | 48 | 0.0040 | -0.4296 | 174 | 1 | 174 | 1 | - | - | - | - | z3m0=1 | Ga1La1Mg1 |
| 189.1.21 | 152689 | 144 | 0.0000 | -0.0127 | 174 | 2 | 174 | 2 | - | - | - | - | z3m0=2 | Fe1Ho6Sb2 |
| 189.1.22 | 418214 | 90 | 0.0000 | -0.8628 | 174 | 4 | 174 | 3 | - | - | - | - | z3m0=1<br>z3mpi=1 | Cu1La1Mg4 |

Continued on next page

**TABLE XIV– continued**

| TQC | ICSD | NoE | DG | IndG | SubSG (w/ T) | cl | SubSG (w/o T) | cl | Max. Sub. SG | yes or no | Max. Sub. SG | yes or no | SIs (w/ T) | Stoich. Formula |
|---|---|---|---|---|---|---|---|---|---|---|---|---|---|---|
| 189.1.23 | 428473 | 54 | 0.0114 | -0.6175 | 174 | 1 | 174 | 1 | - | - | - | - | z3m0=1 | Cu1Ge1Sc1 |
| 189.1.24 | 98233 | 108 | 0.0000 | -0.1634 | 174 | 4 | 174 | 3 | - | - | - | - | z3m0=1<br>z3mpi=1 | Ce2Ge3Zn6 |
| 189.1.25 | 415492 | 78 | 0.0033 | -0.1695 | 174 | 2 | 174 | 2 | - | - | - | - | z3m0=2 | Ce1Pt1Sn1 |
| 189.1.26 | 416969 | 78 | 0.0000 | -0.1961 | 174 | 2 | 174 | 2 | - | - | - | - | z3m0=2 | Ce1Pd1Sn1 |
| 189.1.27 | 100002 | 72 | 0.0050 | -0.3740 | 174 | 3 | 174 | 4 | - | - | - | - | z3m0=2<br>z3mpi=1 | Ce1In1Ir1 |
| 189.1.28 | 428851 | 102 | 0.0000 | -0.4226 | 174 | 3 | 174 | 4 | - | - | - | - | z3m0=2<br>z3mpi=1 | Cd1Ce1Pt1 |
| 189.1.29 | 51528 | 74 | 0.0000 | -0.7393 | 174 | 8 | 174 | 5 | - | - | - | - | z3mpi=1 | C2Rb2 |
| 189.1.30 | 413810 | 48 | 0.0103 | 0.0103 | 174 | 1 | 174 | 1 | - | - | - | - | z3m0=1 | Ba1Bi1Na1 |
| 189.1.31 | 261317 | 100 | 0.0364 | -0.5145 | 174 | 1 | 174 | 1 | - | - | - | - | z3m0=1 | Au1Ge2In3Yb3 |
| 189.1.32 | 104011 | 96 | 0.0035 | -0.0437 | 174 | 2 | 174 | 2 | - | - | - | - | z3m0=2 | Au1Gd1In1 |
| 189.1.33 | 192086 | 48 | 0.0482 | -2.7011 | 174 | 3 | 174 | 4 | - | - | - | - | z3m0=2<br>z3mpi=1 | Au1Ga1Mg1 |
| 189.1.34 | 411543 | 102 | 0.0000 | -1.2661 | 174 | 4 | 174 | 3 | - | - | - | - | z3m0=1<br>z3mpi=1 | Au1Cd1La1 |
| 189.1.35 | 810 | 132 | 0.0036 | -0.0313 | 174 | 1 | 174 | 1 | - | - | - | - | z3m0=1 | As2Eu2 |
| 189.1.36 | 193686 | 80 | 0.0000 | -0.0907 | 174 | 1 | 174 | 1 | - | - | - | - | z3m0=1 | Al3Co1Gd3Ge2 |
| 189.1.37 | 54936 | 96 | 0.0000 | -0.0176 | 174 | 1 | 174 | 1 | - | - | - | - | z3m0=1 | Al1Pd1Tb1 |
| 189.1.38 | 160908 | 48 | 0.0000 | -0.6314 | 174 | 2 | 174 | 2 | - | - | - | - | z3m0=2 | Al1Mg1Y1 |
| 189.1.39 | 107275 | 78 | 0.0259 | -0.5029 | 174 | 1 | 174 | 1 | - | - | - | - | z3m0=1 | Ag1Pb1Y1 |
| 189.1.40 | 416377 | 120 | 0.0176 | -0.8320 | 174 | 1 | 174 | 1 | - | - | - | - | z3m0=1 | Ag1Lu1Sn1 |
| 189.1.41 | 245236 | 120 | 0.0173 | -0.8327 | 174 | 1 | 174 | 1 | - | - | - | - | z3m0=1 | Ag1Lu1Sn1 |
| 189.1.42 | 107278 | 120 | 0.0000 | -0.6900 | 174 | 1 | 174 | 1 | - | - | - | - | z3m0=1 | Ag1Lu1Pb1 |
| 189.1.43 | 10016 | 54 | 0.0264 | 0.0264 | 174 | 1 | 174 | 1 | - | - | - | - | z3m0=1 | Ag1Ca1P1 |
| 189.1.44 | 10017 | 54 | 0.0848 | 0.0781 | 174 | 1 | 174 | 1 | - | - | - | - | z3m0=1 | Ag1As1Ca1 |
| 189.3.1 | 50239 | 78 | - | - | - | - | - | - | 38 | yes | 38 | yes | - | Se2U1 |
| 189.3.2 | 82890 | 78 | - | - | - | - | - | - | 38 | yes | 38 | yes | - | S2U1 |
| 189.3.3 | 411824 | 108 | - | - | - | - | - | - | 38 | yes | 38 | yes | - | Rh1Sn1Tm1 |
| 189.3.4 | 157366 | 44 | - | - | - | - | - | - | 38 | yes | 38 | yes | - | Pt1Sb2Zr6 |
| 189.3.5 | 43209 | 72 | - | - | - | - | - | - | 38 | yes | 38 | yes | - | Pd2Si1 |
| 189.3.6 | 408409 | 114 | - | - | - | - | - | - | 38 | yes | 38 | yes | - | Pd1Sn1Yb1 |
| 189.3.7 | 151166 | 114 | - | - | - | - | - | - | 38 | yes | 38 | yes | - | Pd1Sn1Yb1 |
| 189.3.8 | 98972 | 38 | - | - | - | - | - | - | 38 | yes | 38 | yes | - | Os1Sc6Te2 |
| 189.3.9 | 1396 | 30 | - | - | - | - | - | - | 38 | yes | 38 | yes | - | N1Ta1 |
| 189.3.10 | 415509 | 72 | - | - | - | - | - | - | 38 | yes | 38 | yes | - | In1Pt1Y1 |
| 189.3.11 | 411458 | 96 | - | - | - | - | - | - | 38 | yes | 38 | yes | - | In1Pt1Tb1 |
| 189.3.12 | 411459 | 48 | - | - | - | - | - | - | 38 | yes | 38 | yes | - | In1Pt1Sc1 |
| 189.3.13 | 428555 | 164 | - | - | - | - | - | - | 38 | yes | 38 | yes | - | Ho3Rh9Si2Sn3 |
| 189.3.14 | 97355 | 92 | - | - | - | - | - | - | 38 | yes | 38 | yes | - | Ge5Yb3 |
| 189.3.15 | 59586 | 92 | - | - | - | - | - | - | 38 | yes | 38 | yes | - | Ge5Yb3 |
| 189.3.16 | 43236 | 92 | - | - | - | - | - | - | 38 | yes | 38 | yes | - | Ge5Yb3 |
| 189.3.17 | 98234 | 110 | - | - | - | - | - | - | 38 | yes | 38 | yes | - | Ge3Pr2Zn6 |
| 189.3.18 | 98235 | 112 | - | - | - | - | - | - | 38 | yes | 38 | yes | - | Ge3Nd2Zn6 |
| 189.3.19 | 98232 | 106 | - | - | - | - | - | - | 38 | yes | 38 | yes | - | Ge3La2Zn6 |
| 189.3.20 | 53879 | 72 | - | - | - | - | - | - | 38 | yes | 38 | yes | - | Ge1Pd2 |
| 189.3.21 | 52052 | 72 | - | - | - | - | - | - | 38 | yes | 38 | yes | - | Ge1Pd2 |
| 189.3.22 | 183339 | 120 | - | - | - | - | - | - | 38 | yes | 38 | yes | - | Gd1Pd1Zn1 |
| 189.3.23 | 412443 | 90 | - | - | - | - | - | - | 38 | yes | 38 | yes | - | Gd1Mg1Pt1 |
| 189.3.24 | 412441 | 90 | - | - | - | - | - | - | 38 | yes | 38 | yes | - | Gd1Mg1Pd1 |
| 189.3.25 | 414478 | 90 | - | - | - | - | - | - | 38 | yes | 38 | yes | - | Gd1In1Ir1 |
| 189.3.26 | 413278 | 84 | - | - | - | - | - | - | 38 | yes | 38 | yes | - | Ga1Mg1Tm1 |
| 189.3.27 | 413285 | 54 | - | - | - | - | - | - | 38 | yes | 38 | yes | - | Ga1Mg1Pr1 |
| 189.3.28 | 82530 | 44 | - | - | - | - | - | - | 38 | yes | 38 | yes | - | Fe1Te2Zr6 |
| 189.3.29 | 90081 | 38 | - | - | - | - | - | - | 38 | yes | 38 | yes | - | Fe1Sc6Te2 |
| 189.3.30 | 183341 | 126 | - | - | - | - | - | - | 38 | yes | 38 | yes | - | Dy1Pd1Zn1 |

Continued on next page

**TABLE XIV– continued**

| TQC | ICSD | NoE | DG | IndG | SubSG (w/ T) | cl | SubSG (w/o T) | cl | Max. Sub. SG | yes or no | Max. Sub. SG | yes or no | SIs (w/ T) | Stoich. Formula |
|---|---|---|---|---|---|---|---|---|---|---|---|---|---|---|
| 189.3.31 | 414472 | 96 | - | - | - | - | - | - | 38 | yes | 38 | yes | - | Dy1In1Ir1 |
| 189.3.32 | 420208 | 102 | - | - | - | - | - | - | 38 | yes | 38 | yes | - | Ce1Pd1Zn1 |
| 189.3.33 | 412400 | 72 | - | - | - | - | - | - | 38 | yes | 38 | yes | - | Ce1Mg1Pt1 |
| 189.3.34 | 412401 | 72 | - | - | - | - | - | - | 38 | yes | 38 | yes | - | Ce1Mg1Pd1 |
| 189.3.35 | 280338 | 126 | - | - | - | - | - | - | 38 | yes | 38 | yes | - | Ba2Ni1Si3 |
| 189.3.36 | 429502 | 138 | - | - | - | - | - | - | 38 | yes | 38 | yes | - | B5Cu3Pt9 |
| 189.3.37 | 249113 | 108 | - | - | - | - | - | - | 38 | yes | 38 | yes | - | B4In5Ir9 |
| 189.3.38 | 77352 | 80 | - | - | - | - | - | - | 38 | yes | 38 | yes | - | B2Rh6Sn5 |
| 189.3.39 | 78984 | 80 | - | - | - | - | - | - | 38 | yes | 38 | yes | - | B2Ir6Sn5 |
| 189.3.40 | 68091 | 102 | - | - | - | - | - | - | 38 | yes | 38 | yes | - | B1Li1Pt3 |
| 189.3.41 | 76297 | 60 | - | - | - | - | - | - | 38 | yes | 38 | yes | - | Al3Ge2Ni1Y3 |
| 189.3.42 | 261007 | 40 | - | - | - | - | - | - | 38 | yes | 38 | yes | - | Al2Hf6Pt1 |
| 189.3.43 | 402701 | 38 | - | - | - | - | - | - | 38 | yes | 38 | yes | - | Al2Fe1Zr6 |
| 189.3.44 | 107244 | 102 | - | - | - | - | - | - | 38 | yes | 38 | yes | - | Al1Ho1Ni1 |
| 189.3.45 | 107223 | 102 | - | - | - | - | - | - | 38 | yes | 38 | yes | - | Al1Ho1Ni1 |
| 189.3.46 | 416376 | 114 | - | - | - | - | - | - | 38 | yes | 38 | yes | - | Ag1Sn1Tm1 |
| 189.3.47 | 245235 | 114 | - | - | - | - | - | - | 38 | yes | 38 | yes | - | Ag1Sn1Tm1 |
| 190.2.1 | 81072 | 204 | 0.0235 | -0.0935 | 174 | 1 | 174 | 1 | - | - | - | - | z3m0=2 | Ba3Cr2O9W1 |
| 190.2.2 | 73183 | 220 | 0.0284 | -0.1339 | 174 | 1 | 174 | 1 | - | - | - | - | z3m0=2 | Ba3Ca1O9Ru2 |
| 190.3.1 | 68845 | 168 | - | - | - | - | - | - | 40 | yes | 40 | yes | - | Fe1S1 |
| 190.3.2 | 35007 | 168 | - | - | - | - | - | - | 40 | yes | 40 | yes | - | Fe1S1 |
| 190.3.3 | 35005 | 168 | - | - | - | - | - | - | 40 | yes | 40 | yes | - | Fe1S1 |
| 190.3.4 | 31963 | 168 | - | - | - | - | - | - | 40 | yes | 40 | yes | - | Fe1S1 |
| 190.3.5 | 156620 | 168 | - | - | - | - | - | - | 40 | yes | 40 | yes | - | Fe1S1 |
| 190.3.6 | 156619 | 168 | - | - | - | - | - | - | 40 | yes | 40 | yes | - | Fe1S1 |
| 190.3.7 | 156618 | 168 | - | - | - | - | - | - | 40 | yes | 40 | yes | - | Fe1S1 |
| 190.3.8 | 191973 | 184 | - | - | - | - | - | - | 40 | yes | 40 | yes | - | Ce2Ge3Ni1 |
| 190.4.1 | 152133 | 102 | - | - | - | - | - | - | See details in Table XIX | | | | | Rh1Sn1Zr1 |
| 190.4.2 | 411822 | 102 | - | - | - | - | - | - | See details in Table XIX | | | | | Pt1Sc1Sn1 |
| 190.4.3 | 418976 | 102 | - | - | - | - | - | - | See details in Table XIX | | | | | Pd1Sc1Sn1 |
| 190.4.4 | 408901 | 102 | - | - | - | - | - | - | See details in Table XIX | | | | | Ir1Sn1Zr1 |
| 190.4.5 | 107470 | 102 | - | - | - | - | - | - | See details in Table XIX | | | | | Ir1Sn1Zr1 |
| 190.4.6 | 408900 | 102 | - | - | - | - | - | - | See details in Table XIX | | | | | Hf1Rh1Sn1 |
| 190.4.7 | 107474 | 102 | - | - | - | - | - | - | See details in Table XIX | | | | | Hf1Rh1Sn1 |
| 190.4.8 | 410398 | 102 | - | - | - | - | - | - | See details in Table XIX | | | | | Ga1Pt1Zr1 |
| 190.4.9 | 152132 | 102 | - | - | - | - | - | - | See details in Table XIX | | | | | Ga1Pt1Zr1 |
| 190.4.10 | 156268 | 102 | - | - | - | - | - | - | See details in Table XIX | | | | | Ga1Hf1Pt1 |
| 190.4.11 | 408899 | 102 | - | - | - | - | - | - | See details in Table XIX | | | | | Co1Hf1Sn1 |
| 190.4.12 | 107472 | 102 | - | - | - | - | - | - | See details in Table XIX | | | | | Co1Hf1Sn1 |
| 190.4.13 | 156266 | 102 | - | - | - | - | - | - | See details in Table XIX | | | | | Al1Pt1Zr1 |
| 190.4.14 | 261008 | 102 | - | - | - | - | - | - | See details in Table XIX | | | | | Al1Hf1Pt1 |
| 190.4.15 | 156267 | 102 | - | - | - | - | - | - | See details in Table XIX | | | | | Al1Hf1Pt1 |
| 191.1.1 | 173575 | 160 | 0.0051 | -0.8293 | 2,174 | 3 | 2,174 | 10 | - | - | - | - | z4=1<br>z6m0=1<br>z6mpi=2<br>z12=9 | La10O3Si8 |
| 191.1.2 | 417224 | 16 | 0.0037 | -1.7473 | 2,174 | 8 | 10,174 | 14 | - | - | - | - | z2w,3=1<br>z6m0=1<br>z6mpi=5 | Ga2Sr1 |
| 191.1.3 | 260654 | 16 | 0.0038 | -1.7478 | 2,174 | 8 | 10,174 | 14 | - | - | - | - | z2w,3=1<br>z6m0=1<br>z6mpi=5 | Ga2Sr1 |
| 191.1.4 | 246236 | 16 | 0.0037 | -1.7509 | 2,174 | 8 | 10,174 | 14 | - | - | - | - | z2w,3=1<br>z6m0=1<br>z6mpi=5 | Ga2Sr1 |

Continued on next page

**TABLE XIV– continued**

| TQC | ICSD | NoE | DG | IndG | SubSG (w/ T) | cl | SubSG (w/o T) | cl | Max. Sub. SG | yes or no | Max. Sub. SG | yes or no | SIs (w/ T) | Stoich. Formula |
|---|---|---|---|---|---|---|---|---|---|---|---|---|---|---|
| 191.1.5 | 107425 | 24 | 0.0025 | -0.0418 | 2,174 | 65 | 174,65 | 4 | - | - | - | - | z4=2<br>z6m0=2<br>z12=2 | Ga2Gd1 |
| 191.1.6 | 419163 | 68 | 0.0055 | -0.3102 | 2,174 | 6 | 2,174 | 13 | - | - | - | - | z4=1<br>z6m0=5<br>z12=5 | Ce1Pd2Zn3 |
| 191.1.7 | 418614 | 62 | 0.0036 | -2.1470 | 2,174 | 5 | 2,174 | 12 | - | - | - | - | z4=3<br>z6m0=3<br>z6mpi=4<br>z12=7 | Ca1Zn5 |
| 191.1.8 | 107305 | 30 | 0.0351 | -2.5764 | 2,174 | 4 | 2,174 | 11 | - | - | - | - | z2w,3=1<br>z4=1<br>z6m0=2<br>z6mpi=3<br>z12=5 | Ba1Sn5 |
| 191.1.9 | 246238 | 16 | 0.0091 | -1.1254 | 2,174 | 3 | 2,174 | 10 | - | - | - | - | z4=1<br>z6m0=1<br>z6mpi=2<br>z12=9 | Ba1Ga2 |
| 191.1.10 | 97723 | 150 | 0.0000 | -0.0449 | 2,174 | 11 | 10,174 | 9 | - | - | - | - | z2w,3=1<br>z4=2<br>z6m0=1<br>z6mpi=5<br>z12=6 | B4Co11Pr3 |
| 191.1.11 | 97731 | 224 | 0.0000 | -0.0257 | 2,174 | 16 | 2,174 | 8 | - | - | - | - | z2w,3=1<br>z4=3<br>z6m0=2<br>z6mpi=5<br>z12=7 | B3Co7Dy2 |
| 191.1.12 | 78848 | 10 | 0.0115 | -0.7706 | 2,174 | 2 | 10,174 | 7 | - | - | - | - | z2w,3=1<br>z6m0=5<br>z6mpi=3<br>z12=8 | B2Ti1 |
| 191.1.13 | 78847 | 10 | 0.0115 | -0.7850 | 2,174 | 2 | 10,174 | 7 | - | - | - | - | z2w,3=1<br>z6m0=5<br>z6mpi=3<br>z12=8 | B2Ti1 |
| 191.1.14 | 56723 | 10 | 0.0116 | -0.7819 | 2,174 | 2 | 10,174 | 7 | - | - | - | - | z2w,3=1<br>z6m0=5<br>z6mpi=3<br>z12=8 | B2Ti1 |
| 191.1.15 | 184968 | 44 | 0.0059 | -0.6199 | 174 | 37 | 174 | 6 | - | - | - | - | z6mpi=4<br>z12=4 | B2Fe3U1 |
| 191.1.16 | 237660 | 134 | 0.0000 | -0.3834 | 2,174 | 30 | 10,174 | 5 | - | - | - | - | z2w,3=1<br>z4=2<br>z6m0=1<br>z6mpi=3<br>z12=10 | B1Eu1Pt4 |
| 191.1.17 | 97722 | 104 | 0.0117 | -0.1440 | 2,174 | 65 | 174,65 | 4 | - | - | - | - | z4=2<br>z6m0=2<br>z12=2 | B1Co4Pr1 |
| 191.1.18 | 417869 | 114 | 0.0000 | -0.1690 | 2,174 | 60 | 2,174 | 3 | - | - | - | - | z4=3<br>z6m0=1<br>z6mpi=4<br>z12=11 | B1Co4Gd1 |
| 191.1.19 | 36504 | 90 | 0.0031 | -0.8506 | 2 | 1 | 2 | 2 | - | - | - | - | z4=3<br>z6m0=3<br>z12=3 | B1Ca1Ni4 |

Continued on next page

**TABLE XIV– continued**

| TQC | ICSD | NoE | DG | IndG | SubSG (w/ T) | cl | SubSG (w/o T) | cl | Max. Sub. SG | yes or no | Max. Sub. SG | yes or no | SIs (w/ T) | Stoich. Formula |
|---|---|---|---|---|---|---|---|---|---|---|---|---|---|---|
| 191.1.20 | 409394 | 62 | 0.0025 | -0.2354 | 2,174 | 61 | 2,174 | 1 | - | - | - | - | z4=3<br>z6m0=5<br>z6mpi=2<br>z12=7 | Al9Co2Eu1 |
| 191.3.1 | 418615 | 70 | - | - | - | - | - | - | 65 | no(2) | 65 | no(2) | z4=3 | Sr1Zn5 |
| 191.3.2 | 96116 | 156 | - | - | - | - | - | - | 65 | yes | 65 | yes | - | La1Mn3Ni2 |
| 191.3.3 | 260252 | 56 | - | - | - | - | - | - | 65 | no(2) | 65 | no(2) | z2w,3=1<br>z4=1 | In9K1Ni2 |
| 191.3.4 | 424816 | 46 | - | - | - | - | - | - | 189 | no(174) | 189 | no(174) | z3m0=1<br>z3mpi=1 | Ge2Ir3La1 |
| | | | | | | | | | 187 | no(174) | 187 | no(174) | z3m0=1<br>z3mpi=1 | |
| | | | | | | | | | 164 | no(2) | 164 | yes | z2w,3=1 | |
| | | | | | | | | | 162 | no(2) | 162 | yes | z2w,3=1 | |
| | | | | | | | | | 65 | no(2) | 65 | no(10) | z2w,3=1 | |
| 191.3.5 | 107582 | 30 | - | - | - | - | - | - | 189 | no(174) | 189 | no(174) | z3mpi=1 | Ga2Yb1 |
| | | | | | | | | | 187 | no(174) | 187 | no(174) | z3mpi=1 | |
| | | | | | | | | | 164 | no(2) | 164 | yes | z4=2 | |
| | | | | | | | | | 162 | no(2) | 162 | yes | z4=2 | |
| | | | | | | | | | 65 | no(2) | 65 | no(65) | z4=2 | |
| 191.3.6 | 188716 | 74 | - | - | - | - | - | - | 65 | no(2) | 65 | no(2) | z2w,3=1<br>z4=3 | Fe6Ge6Mg1 |
| 191.3.7 | 391246 | 112 | - | - | - | - | - | - | 65 | no(2) | 65 | no(2) | z2w,3=1<br>z4=1 | Eu2Pd1Si3 |
| 191.3.8 | 107032 | 88 | - | - | - | - | - | - | 65 | no(2) | 65 | no(2) | z4=3 | Er1Ge6Mn6 |
| 191.3.9 | 107031 | 88 | - | - | - | - | - | - | 65 | no(2) | 65 | no(2) | z4=3 | Er1Ge6Mn6 |
| 191.3.10 | 107030 | 88 | - | - | - | - | - | - | 65 | no(2) | 65 | no(2) | z4=3 | Er1Ge6Mn6 |
| 191.3.11 | 85876 | 86 | - | - | - | - | - | - | 65 | no(2) | 65 | no(2) | z2w,3=1<br>z4=1 | Dy1Ge6Mn6 |
| 191.3.12 | 85875 | 86 | - | - | - | - | - | - | 65 | no(2) | 65 | no(2) | z2w,3=1<br>z4=1 | Dy1Ge6Mn6 |
| 191.3.13 | 57229 | 86 | - | - | - | - | - | - | 65 | no(2) | 65 | no(2) | z2w,3=1<br>z4=1 | Dy1Ge6Mn6 |
| 191.3.14 | 57228 | 86 | - | - | - | - | - | - | 65 | no(2) | 65 | no(2) | z2w,3=1<br>z4=1 | Dy1Ge6Mn6 |
| 191.3.15 | 158265 | 66 | - | - | - | - | - | - | 65 | no(2) | 65 | no(10) | z2w,3=1<br>z4=2 | Cu5La1 |
| 191.3.16 | 415256 | 82 | - | - | - | - | - | - | 65 | no(2) | 65 | no(10) | z2w,3=1<br>z4=2 | Co6Ge6Mg1 |
| 191.3.17 | 83895 | 90 | - | - | - | - | - | - | 65 | no(2) | 65 | no(2) | z2w,3=1<br>z4=3 | Ce2Co1Si3 |
| 191.3.18 | 189763 | 18 | - | - | - | - | - | - | 189 | no(174) | 189 | no(174) | z3m0=1 | Ce1Ga2 |
| | | | | | | | | | 187 | no(174) | 187 | no(174) | z3m0=1 | |
| | | | | | | | | | 164 | no(2) | 164 | no(2) | z4=1 | |
| | | | | | | | | | 162 | no(2) | 162 | no(2) | z4=1 | |
| | | | | | | | | | 65 | no(2) | 65 | no(2) | z4=1 | |
| 191.3.19 | 173576 | 170 | - | - | - | - | - | - | 65 | no(2) | 65 | no(10) | z2w,3=1<br>z4=2 | Ce10O3Si8 |
| 191.3.20 | 54474 | 52 | - | - | - | - | - | - | 65 | no(2) | 65 | no(2) | z2w,3=1<br>z4=3 | Ca1Ni5 |
| 191.3.21 | 245512 | 174 | - | - | - | - | - | - | 65 | no(2) | 65 | no(2) | z2w,3=1<br>z4=1 | Ca1Ni2Zn3 |
| 191.3.22 | 260652 | 26 | - | - | - | - | - | - | 189 | yes | 189 | yes | - | Ca1Hg2 |
| | | | | | | | | | 187 | yes | 187 | yes | - | |
| | | | | | | | | | 164 | no(2) | 164 | no(2) | z2w,3=1<br>z4=1 | |
| | | | | | | | | | 162 | no(2) | 162 | no(2) | z2w,3=1<br>z4=1 | |

Continued on next page

**TABLE XIV– continued**

| TQC | ICSD | NoE | DG | IndG | SubSG cl (w/ T) | SubSG cl (w/o T) | Max. Sub. SG | yes or no | Max. Sub. SG | yes or no | SIs (w/ T) | Stoich. Formula |
|---|---|---|---|---|---|---|---|---|---|---|---|---|
| | | | | | | | 65 | no(2) | 65 | no(2) | z2w,3=1<br>z4=1 | |
| 191.3.23 | 165559 | 144 | - | - | - - | - - | 65 | no(2) | 65 | no(2) | z2w,3=1<br>z4=1 | Bi2Ga1La1 |
| 191.3.24 | 97726 | 196 | - | - | - - | - - | 189 | no(174) | 189 | no(174) | z3m0=2 | B3Co7Pr2 |
| | | | | | | | 187 | no(174) | 187 | no(174) | z3m0=2 | |
| | | | | | | | 164 | no(2) | 164 | yes | z4=2 | |
| | | | | | | | 162 | no(2) | 162 | yes | z4=2 | |
| | | | | | | | 65 | no(2) | 65 | no(65) | z4=2 | |
| 191.3.25 | 97725 | 196 | - | - | - - | - - | 189 | no(174) | 189 | no(174) | z3m0=2 | B3Co7Pr2 |
| | | | | | | | 187 | no(174) | 187 | no(174) | z3m0=2 | |
| | | | | | | | 164 | no(2) | 164 | no(2) | z2w,3=1<br>z4=3 | |
| | | | | | | | 162 | no(2) | 162 | no(2) | z2w,3=1<br>z4=3 | |
| | | | | | | | 65 | no(2) | 65 | no(2) | z2w,3=1<br>z4=3 | |
| 191.3.26 | 81809 | 12 | - | - | - - | - - | 65 | no(2) | 65 | no(10) | z2w,3=1<br>z4=2 | B2Mo1 |
| 191.3.27 | 81806 | 12 | - | - | - - | - - | 65 | no(2) | 65 | no(10) | z2w,3=1<br>z4=2 | B2Mo1 |
| 191.3.28 | 97342 | 46 | - | - | - - | - - | 65 | no(2) | 65 | no(10) | z2w,3=1<br>z4=2 | B2Ir3Pr1 |
| 191.3.29 | 23655 | 44 | - | - | - - | - - | 65 | no(2) | 65 | no(2) | z2w,3=1<br>z4=1 | B2Co3Y1 |
| 191.3.30 | 246509 | 52 | - | - | - - | - - | 65 | no(2) | 65 | no(2) | z2w,3=1<br>z4=1 | B2Co3Tb1 |
| 191.3.31 | 246508 | 52 | - | - | - - | - - | 65 | no(2) | 65 | no(2) | z2w,3=1<br>z4=1 | B2Co3Tb1 |
| 191.3.32 | 44170 | 54 | - | - | - - | - - | 65 | no(2) | 65 | no(65) | z4=2 | B2Co3Ho1 |
| 191.3.33 | 55454 | 112 | - | - | - - | - - | 189 | no(174) | 189 | no(174) | z3m0=2 | B1Pr1Pt4 |
| | | | | | | | 187 | no(174) | 187 | no(174) | z3m0=2 | |
| | | | | | | | 164 | no(2) | 164 | yes | z2w,3=1 | |
| | | | | | | | 162 | no(2) | 162 | yes | z2w,3=1 | |
| | | | | | | | 65 | no(2) | 65 | no(10) | z2w,3=1 | |
| 191.3.34 | 156975 | 124 | - | - | - - | - - | 65 | no(2) | 65 | no(10) | z2w,3=1 | B1Ni4Tb1 |
| 191.3.35 | 68092 | 68 | - | - | - - | - - | 189 | yes | 189 | yes | - | B1Na1Pt3 |
| | | | | | | | 187 | yes | 187 | yes | - | |
| | | | | | | | 164 | no(2) | 164 | no(2) | z4=3 | |
| | | | | | | | 162 | no(2) | 162 | no(2) | z4=3 | |
| | | | | | | | 65 | no(2) | 65 | no(2) | z4=3 | |
| 191.3.36 | 170617 | 108 | - | - | - - | - - | 65 | no(2) | 65 | no(65) | z4=2 | B1La1Ni4 |
| 191.3.37 | 170609 | 108 | - | - | - - | - - | 65 | no(2) | 65 | no(65) | z4=2 | B1La1Ni4 |
| 191.3.38 | 156976 | 130 | - | - | - - | - - | 65 | no(2) | 65 | no(10) | z2w,3=1 | B1Er1Ni4 |
| 191.3.39 | 97727 | 106 | - | - | - - | - - | 65 | no(2) | 65 | no(65) | z4=2 | B1Co4Nd1 |
| 191.3.40 | 55455 | 110 | - | - | - - | - - | 189 | no(174) | 189 | no(174) | z3mpi=2 | B1Ce1Pt4 |
| | | | | | | | 187 | no(174) | 187 | no(174) | z3mpi=2 | |
| | | | | | | | 164 | no(2) | 164 | yes | z2w,3=1 | |
| | | | | | | | 162 | no(2) | 162 | yes | z2w,3=1 | |
| | | | | | | | 65 | no(2) | 65 | no(10) | z2w,3=1 | |
| 191.3.41 | 55453 | 110 | - | - | - - | - - | 189 | no(174) | 189 | no(174) | z3mpi=2 | B1Ce1Pt4 |
| | | | | | | | 187 | no(174) | 187 | no(174) | z3mpi=2 | |
| | | | | | | | 164 | no(2) | 164 | yes | z2w,3=1 | |
| | | | | | | | 162 | no(2) | 162 | yes | z2w,3=1 | |
| | | | | | | | 65 | no(2) | 65 | no(10) | z2w,3=1 | |
| 191.3.42 | 106289 | 64 | - | - | - - | - - | 65 | no(2) | 65 | no(2) | z2w,3=1<br>z4=3 | Au5Rb1 |
| 191.3.43 | 106286 | 64 | - | - | - - | - - | 65 | no(2) | 65 | no(10) | z2w,3=1 | Au5K1 |

Continued on next page

**TABLE XIV– continued**

| TQC | ICSD | NoE | DG | IndG | SubSG (w/ T) | cl | SubSG (w/o T) | cl | Max. Sub. SG | yes or no | Max. Sub. SG | yes or no | SIs (w/ T) | Stoich. Formula |
|---|---|---|---|---|---|---|---|---|---|---|---|---|---|---|
| 191.3.44 | 260669 | 32 | - | - | - | - | - | - | 189 | no(174) | 189 | no(174) | z3m0=2 z3mpi=1 | Au2Ba1 |
| | | | | | | | | | 187 | no(174) | 187 | no(174) | z3m0=2 z3mpi=1 | |
| | | | | | | | | | 164 | no(2) | 164 | no(2) | z4=3 | |
| | | | | | | | | | 162 | no(2) | 162 | no(2) | z4=3 | |
| | | | | | | | | | 65 | no(2) | 65 | no(2) | z4=3 | |
| 193.1.1 | 150562 | 64 | 0.0101 | -0.6205 | 2,174 | 1 | 2,174 | 2 | - | - | - | - | z4=1 z6m0=1 z12prime=1 | Sn3Zr5 |
| 193.1.2 | 109292 | 64 | 0.0021 | -0.3650 | 2,174 | 1 | 2,174 | 2 | - | - | - | - | z4=1 z6m0=1 z12prime=1 | Sn3Ti5 |
| 193.1.3 | 62591 | 64 | 0.0021 | -0.5893 | 2,174 | 1 | 2,174 | 2 | - | - | - | - | z4=1 z6m0=1 z12prime=1 | Si3Ti5 |
| 193.1.4 | 105649 | 64 | 0.0080 | -0.4920 | 2,174 | 1 | 2,174 | 2 | - | - | - | - | z4=1 z6m0=1 z12prime=1 | Pb3Zr5 |
| 193.1.5 | 154390 | 160 | 0.0000 | -0.0362 | 2,174 | 3 | 2,174 | 1 | - | - | - | - | z4=3 z6m0=5 z12prime=11 | Nb1Sb5U3 |
| 193.1.6 | 99999 | 64 | 0.0093 | -0.0504 | 174 | 2 | 174 | 3 | - | - | - | - | z6m0=2 z12prime=8 | K3N1 |
| 193.1.7 | 44361 | 64 | 0.0059 | -0.7794 | 2,174 | 1 | 2,174 | 2 | - | - | - | - | z4=1 z6m0=1 z12prime=1 | Ge3Hf5 |
| 193.1.8 | 173358 | 136 | 0.0000 | -0.2324 | 2,174 | 3 | 2,174 | 1 | - | - | - | - | z4=3 z6m0=5 z12prime=11 | Bi5Ce3Mn1 |
| 193.3.1 | 106133 | 208 | - | - | - | - | - | - | 63 | yes | 63 | yes | - | Tb5Tl3 |
| 193.3.2 | 107311 | 72 | - | - | - | - | - | - | 63 | no(2) | 63 | yes | z4=2 | Sn4Zr5 |
| 193.3.3 | 106085 | 152 | - | - | - | - | - | - | 63 | no(2) | 63 | yes | z4=2 | Sn4Th5 |
| 193.3.4 | 106084 | 144 | - | - | - | - | - | - | 63 | no(2) | 63 | no(2) | z4=1 | Sn3Th5 |
| 193.3.5 | 189947 | 80 | - | - | - | - | - | - | 63 | no(2) | 63 | no(2) | z4=1 | Sb4Zr5 |
| 193.3.6 | 153609 | 80 | - | - | - | - | - | - | 63 | yes | 63 | yes | - | Sb4Zr5 |
| 193.3.7 | 52314 | 220 | - | - | - | - | - | - | 63 | no(2) | 63 | yes | z4=2 | Sb3Tb5 |
| 193.3.8 | 182209 | 220 | - | - | - | - | - | - | 63 | no(2) | 63 | yes | z4=2 | Sb3Tb5 |
| 193.3.9 | 182208 | 220 | - | - | - | - | - | - | 63 | no(2) | 63 | yes | z4=2 | Sb3Tb5 |
| 193.3.10 | 182207 | 220 | - | - | - | - | - | - | 63 | no(2) | 63 | yes | z4=2 | Sb3Tb5 |
| 193.3.11 | 182205 | 220 | - | - | - | - | - | - | 63 | no(2) | 63 | yes | z4=2 | Sb3Tb5 |
| 193.3.12 | 391476 | 100 | - | - | - | - | - | - | 63 | no(2) | 63 | no(2) | z4=1 | Pt3Zr5 |
| 193.3.13 | 1917 | 220 | - | - | - | - | - | - | 63 | no(2) | 63 | no(2) | z4=3 | Pt3Pu5 |
| 193.3.14 | 105776 | 220 | - | - | - | - | - | - | 63 | no(2) | 63 | no(2) | z4=3 | Pt3Pu5 |
| 193.3.15 | 54293 | 164 | - | - | - | - | - | - | 63 | no(2) | 63 | no(2) | z4=3 | Nd5Sn3 |
| 193.3.16 | 165098 | 164 | - | - | - | - | - | - | 63 | yes | 63 | yes | - | Nb5P1Si3 |
| 193.3.17 | 43020 | 148 | - | - | - | - | - | - | 63 | yes | 63 | yes | - | Mn1Sb5U3 |
| 193.3.18 | 168592 | 140 | - | - | - | - | - | - | 189 | no(174) | 189 | no(174) | z3m0=1 | La6Mn2S10 |
| | | | | | | | | | 188 | no(174) | 188 | no(174) | z3m0=1 | |
| | | | | | | | | | 165 | yes | 165 | yes | - | |
| | | | | | | | | | 162 | yes | 162 | yes | - | |
| | | | | | | | | | 63 | yes | 63 | yes | - | |
| 193.3.19 | 83905 | 124 | - | - | - | - | - | - | 63 | yes | 63 | yes | - | La3Sb5Zr1 |
| 193.3.20 | 80907 | 124 | - | - | - | - | - | - | 63 | no(2) | 63 | no(2) | z4=3 | La3Sb5Ti1 |
| 193.3.21 | 109145 | 68 | - | - | - | - | - | - | 63 | yes | 63 | yes | - | I3Nb1 |
| 193.3.22 | 409778 | 148 | - | - | - | - | - | - | 63 | no(2) | 63 | yes | z4=2 | I1La5Pb3 |
| 193.3.23 | 104320 | 92 | - | - | - | - | - | - | 63 | no(2) | 63 | no(2) | z4=1 | Hg3Mg5 |

Continued on next page

**TABLE XIV– continued**

| TQC | ICSD | NoE | DG | IndG | SubSG cl (w/ T) | | SubSG cl (w/o T) | | Max. Sub. SG | yes or no | Max. Sub. SG | yes or no | SIs (w/ T) | Stoich. Formula |
|---|---|---|---|---|---|---|---|---|---|---|---|---|---|---|
| 193.3.24 | 107313 | 72 | - | - | - | - | - | - | 63 | no(2) | 63 | yes | z4=2 | Hf5Sn4 |
| 193.3.25 | 183487 | 64 | - | - | - | - | - | - | 63 | no(2) | 63 | no(2) | z4=1 | Hf5Sn3 |
| 193.3.26 | 104273 | 64 | - | - | - | - | - | - | 63 | no(2) | 63 | no(2) | z4=1 | Hf5Sn3 |
| 193.3.27 | 53042 | 64 | - | - | - | - | - | - | 63 | no(2) | 63 | no(2) | z4=1 | Hf5Si3 |
| 193.3.28 | 83906 | 124 | - | - | - | - | - | - | 63 | yes | 63 | yes | - | Hf1La3Sb5 |
| 193.3.29 | 83510 | 100 | - | - | - | - | - | - | 63 | no(2) | 63 | no(2) | z4=1 | H6Al2Br1Li1O6 |
| 193.3.30 | 50740 | 132 | - | - | - | - | - | - | 63 | yes | 63 | yes | - | Ge5Ti1U3 |
| 193.3.31 | 56051 | 164 | - | - | - | - | - | - | 63 | no(2) | 63 | no(2) | z4=1 | Ge3U5 |
| 193.3.32 | 91481 | 164 | - | - | - | - | - | - | 63 | no(2) | 63 | yes | z4=2 | Ge3Nd5 |
| 193.3.33 | 43262 | 204 | - | - | - | - | - | - | 63 | no(2) | 63 | no(2) | z4=3 | Gd5Ge3 |
| 193.3.34 | 103997 | 64 | - | - | - | - | - | - | 63 | yes | 63 | yes | - | Ga4Ti5 |
| 193.3.35 | 150544 | 128 | - | - | - | - | - | - | 63 | no(2) | 63 | no(2) | z4=3 | Ga3Y5 |
| 193.3.36 | 103978 | 68 | - | - | - | - | - | - | 63 | no(2) | 63 | yes | z4=2 | Ga3Ta5 |
| 193.3.37 | 150543 | 48 | - | - | - | - | - | - | 63 | no(2) | 63 | no(2) | z4=3 | Ga3Sc5 |
| 193.3.38 | 99977 | 104 | - | - | - | - | - | - | 63 | no(2) | 63 | no(2) | z4=1 | Fe5Si3 |
| 193.3.39 | 99976 | 104 | - | - | - | - | - | - | 63 | no(2) | 63 | no(2) | z4=1 | Fe5Si3 |
| 193.3.40 | 99975 | 104 | - | - | - | - | - | - | 63 | no(2) | 63 | no(2) | z4=1 | Fe5Si3 |
| 193.3.41 | 99974 | 104 | - | - | - | - | - | - | 63 | no(2) | 63 | no(2) | z4=1 | Fe5Si3 |
| 193.3.42 | 99973 | 104 | - | - | - | - | - | - | 63 | no(2) | 63 | no(2) | z4=1 | Fe5Si3 |
| 193.3.43 | 99972 | 104 | - | - | - | - | - | - | 63 | no(2) | 63 | no(2) | z4=1 | Fe5Si3 |
| 193.3.44 | 99971 | 104 | - | - | - | - | - | - | 63 | no(2) | 63 | no(2) | z4=1 | Fe5Si3 |
| 193.3.45 | 99970 | 104 | - | - | - | - | - | - | 63 | no(2) | 63 | no(2) | z4=1 | Fe5Si3 |
| 193.3.46 | 99969 | 104 | - | - | - | - | - | - | 63 | no(2) | 63 | no(2) | z4=1 | Fe5Si3 |
| 193.3.47 | 99968 | 104 | - | - | - | - | - | - | 63 | no(2) | 63 | no(2) | z4=1 | Fe5Si3 |
| 193.3.48 | 99966 | 104 | - | - | - | - | - | - | 63 | yes | 63 | yes | - | Fe5Si3 |
| 193.3.49 | 42585 | 104 | - | - | - | - | - | - | 63 | no(2) | 63 | no(2) | z4=1 | Fe5Si3 |
| 193.3.50 | 166484 | 104 | - | - | - | - | - | - | 63 | no(2) | 63 | no(2) | z4=1 | Fe5Si3 |
| 193.3.51 | 161129 | 104 | - | - | - | - | - | - | 63 | no(2) | 63 | no(2) | z4=1 | Fe5Si3 |
| 193.3.52 | 161128 | 104 | - | - | - | - | - | - | 63 | no(2) | 63 | no(2) | z4=1 | Fe5Si3 |
| 193.3.53 | 173030 | 200 | - | - | - | - | - | - | 63 | no(2) | 63 | no(2) | z4=3 | Eu5Sb3 |
| 193.3.54 | 103371 | 224 | - | - | - | - | - | - | 63 | no(2) | 63 | no(2) | z4=3 | Dy5Sn3 |
| 193.3.55 | 152604 | 244 | - | - | - | - | - | - | 63 | yes | 63 | yes | - | Dy5Ni1Pb3 |
| 193.3.56 | 107005 | 156 | - | - | - | - | - | - | 63 | no(2) | 63 | no(2) | z4=3 | Cu1Pb3Y5 |
| 193.3.57 | 190115 | 80 | - | - | - | - | - | - | 189 | no(174) | 189 | no(174) | z3m0=1 | Cu1In3Zr5 |
| | | | | | | | | | 188 | no(174) | 188 | no(174) | z3m0=1 | |
| | | | | | | | | | 165 | no(2) | 165 | no(2) | z4=1 | |
| | | | | | | | | | 162 | no(2) | 162 | no(2) | z4=1 | |
| | | | | | | | | | 63 | no(2) | 63 | no(2) | z4=1 | |
| 193.3.58 | 53012 | 76 | - | - | - | - | - | - | 63 | yes | 63 | yes | - | Co2N5Ta4 |
| 193.3.59 | 95238 | 148 | - | - | - | - | - | - | 63 | no(2) | 63 | no(2) | z4=3 | Cl1La5Sn3 |
| 193.3.60 | 108250 | 144 | - | - | - | - | - | - | 189 | no(174) | 189 | no(174) | z3m0=1 | Ce5Ge3 |
| | | | | | | | | | 188 | no(174) | 188 | no(174) | z3m0=1 | |
| | | | | | | | | | 165 | no(2) | 165 | yes | z4=2 | |
| | | | | | | | | | 162 | no(2) | 162 | yes | z4=2 | |
| | | | | | | | | | 63 | no(2) | 63 | yes | z4=2 | |
| 193.3.61 | 80184 | 88 | - | - | - | - | - | - | 63 | no(2) | 63 | no(2) | z4=1 | Ca6Mn1N5 |
| 193.3.62 | 42922 | 92 | - | - | - | - | - | - | 63 | yes | 63 | yes | - | C1Ge3Mo5 |
| 193.3.63 | 95239 | 148 | - | - | - | - | - | - | 63 | no(2) | 63 | no(2) | z4=3 | Br1La5Sn3 |
| 193.3.64 | 415727 | 120 | - | - | - | - | - | - | 63 | no(2) | 63 | no(2) | z4=3 | Bi5La3Mg1 |
| 193.3.65 | 173031 | 200 | - | - | - | - | - | - | 63 | no(2) | 63 | no(2) | z4=3 | Bi3Eu5 |
| 193.3.66 | 188688 | 92 | - | - | - | - | - | - | 63 | no(2) | 63 | yes | z4=2 | Bi3Cu1Zr5 |
| 193.3.67 | 415737 | 280 | - | - | - | - | - | - | 63 | no(2) | 63 | yes | z4=2 | B1Lu5Si3 |
| 193.3.68 | 58475 | 252 | - | - | - | - | - | - | 63 | no(2) | 63 | no(2) | z4=3 | Au6Hg5 |
| 193.3.69 | 150475 | 252 | - | - | - | - | - | - | 63 | no(2) | 63 | no(2) | z4=3 | Au6Hg5 |
| 193.3.70 | 2250 | 200 | - | - | - | - | - | - | 63 | no(2) | 63 | no(2) | z4=3 | As3Eu5 |
| 193.3.71 | 167476 | 64 | - | - | - | - | - | - | 63 | no(2) | 63 | no(2) | z4=1 | Al4Zr5 |
| 193.3.72 | 190116 | 80 | - | - | - | - | - | - | 189 | no(174) | 189 | no(174) | z3m0=1 | Ag1In3Zr5 |

Continued on next page

**TABLE XIV– continued**

| TQC | ICSD | NoE | DG | IndG | SubSG (w/ T) | cl | SubSG (w/o T) | cl | Max. Sub. SG | yes or no | Max. Sub. SG | yes or no | SIs (w/ T) | Stoich. Formula |
|---|---|---|---|---|---|---|---|---|---|---|---|---|---|---|
| | | | | | | | | | 188 | no(174) | 188 | no(174) | z3m0=1 | |
| | | | | | | | | | 165 | yes | 165 | yes | - | |
| | | | | | | | | | 162 | yes | 162 | yes | - | |
| | | | | | | | | | 63 | yes | 63 | yes | - | |
| 193.4.1 | 43692 | 134 | - | - | - | - | - | - | See details in Table XIX | | | | | Si3Y5 |
| 193.4.2 | 415741 | 134 | - | - | - | - | - | - | See details in Table XIX | | | | | Si3Y5 |
| 193.4.3 | 104190 | 214 | - | - | - | - | - | - | See details in Table XIX | | | | | Si3Tb5 |
| 193.4.4 | 76159 | 74 | - | - | - | - | - | - | See details in Table XIX | | | | | Si3Ta5 |
| 193.4.5 | 415673 | 142 | - | - | - | - | - | - | See details in Table XIX | | | | | Sb5U3Zr1 |
| 193.4.6 | 43019 | 142 | - | - | - | - | - | - | See details in Table XIX | | | | | Sb5Ti1U3 |
| 193.4.7 | 52407 | 70 | - | - | - | - | - | - | See details in Table XIX | | | | | Sb3Zr5 |
| 193.4.8 | 189949 | 70 | - | - | - | - | - | - | See details in Table XIX | | | | | Sb3Zr5 |
| 193.4.9 | 153607 | 70 | - | - | - | - | - | - | See details in Table XIX | | | | | Sb3Zr5 |
| 193.4.10 | 57167 | 94 | - | - | - | - | - | - | See details in Table XIX | | | | | Sb3Zn1Zr5 |
| 193.4.11 | 57517 | 130 | - | - | - | - | - | - | See details in Table XIX | | | | | Sb3Sr5 |
| 193.4.12 | 42120 | 130 | - | - | - | - | - | - | See details in Table XIX | | | | | Sb3Sr5 |
| 193.4.13 | 57144 | 78 | - | - | - | - | - | - | See details in Table XIX | | | | | Sb3Si1Zr5 |
| 193.4.14 | 153608 | 78 | - | - | - | - | - | - | See details in Table XIX | | | | | Sb3Si1Zr5 |
| 193.4.15 | 168436 | 254 | - | - | - | - | - | - | See details in Table XIX | | | | | Pb3Tm5 |
| 193.4.16 | 105614 | 54 | - | - | - | - | - | - | See details in Table XIX | | | | | Pb3Sc5 |
| 193.4.17 | 24148 | 54 | - | - | - | - | - | - | See details in Table XIX | | | | | P3Ti5 |
| 193.4.18 | 96419 | 162 | - | - | - | - | - | - | See details in Table XIX | | | | | O3Rb0.333W1 |
| 193.4.19 | 152607 | 274 | - | - | - | - | - | - | See details in Table XIX | | | | | Ni1Pb3Tm5 |
| 193.4.20 | 152602 | 174 | - | - | - | - | - | - | See details in Table XIX | | | | | Ni1Pb3Pr5 |
| 193.4.21 | 36277 | 202 | - | - | - | - | - | - | See details in Table XIX | | | | | Nb5O1Pt3 |
| 193.4.22 | 76461 | 110 | - | - | - | - | - | - | See details in Table XIX | | | | | N6Ta5 |
| 193.4.23 | 29329 | 110 | - | - | - | - | - | - | See details in Table XIX | | | | | N6Ta5 |
| 193.4.24 | 166512 | 110 | - | - | - | - | - | - | See details in Table XIX | | | | | N6Ta5 |
| 193.4.25 | 26252 | 190 | - | - | - | - | - | - | See details in Table XIX | | | | | N6Nb5 |
| 193.4.26 | 29285 | 90 | - | - | - | - | - | - | See details in Table XIX | | | | | N1P3V5 |
| 193.4.27 | 42435 | 94 | - | - | - | - | - | - | See details in Table XIX | | | | | Mn5Si3 |
| 193.4.28 | 152600 | 154 | - | - | - | - | - | - | See details in Table XIX | | | | | La5Ni1Pb3 |
| 193.4.29 | 35317 | 50 | - | - | - | - | - | - | See details in Table XIX | | | | | I3Zr1 |
| 193.4.30 | 32704 | 50 | - | - | - | - | - | - | See details in Table XIX | | | | | I3Zr1 |
| 193.4.31 | 23946 | 50 | - | - | - | - | - | - | See details in Table XIX | | | | | I3Zr1 |
| 193.4.32 | 422871 | 50 | - | - | - | - | - | - | See details in Table XIX | | | | | I3Ti1 |
| 193.4.33 | 23172 | 50 | - | - | - | - | - | - | See details in Table XIX | | | | | I3Ti1 |
| 193.4.34 | 415743 | 234 | - | - | - | - | - | - | See details in Table XIX | | | | | Ho5Si3 |
| 193.4.35 | 42913 | 94 | - | - | - | - | - | - | See details in Table XIX | | | | | Hf5Sb3Zn1 |
| 193.4.36 | 23947 | 50 | - | - | - | - | - | - | See details in Table XIX | | | | | Hf1I3 |
| 193.4.37 | 51371 | 182 | - | - | - | - | - | - | See details in Table XIX | | | | | Ge4Np5 |
| 193.4.38 | 57122 | 134 | - | - | - | - | - | - | See details in Table XIX | | | | | Ge3Y5 |
| 193.4.39 | 44504 | 74 | - | - | - | - | - | - | See details in Table XIX | | | | | Ge3V5 |
| 193.4.40 | 185874 | 214 | - | - | - | - | - | - | See details in Table XIX | | | | | Ge3Tb5 |
| 193.4.41 | 56028 | 74 | - | - | - | - | - | - | See details in Table XIX | | | | | Ge3Ta5 |
| 193.4.42 | 76138 | 94 | - | - | - | - | - | - | See details in Table XIX | | | | | Ge3Mn5 |
| 193.4.43 | 108517 | 94 | - | - | - | - | - | - | See details in Table XIX | | | | | Ge3Mn5 |
| 193.4.44 | 43402 | 134 | - | - | - | - | - | - | See details in Table XIX | | | | | Ge3La5 |
| 193.4.45 | 103837 | 154 | - | - | - | - | - | - | See details in Table XIX | | | | | Ga4Nb5 |
| 193.4.46 | 104003 | 58 | - | - | - | - | - | - | See details in Table XIX | | | | | Ga3Ti2Zr3 |
| 193.4.47 | 103730 | 58 | - | - | - | - | - | - | See details in Table XIX | | | | | Ga3Hf5 |
| 193.4.48 | 103733 | 94 | - | - | - | - | - | - | See details in Table XIX | | | | | Ga3Hf3Nb2 |
| 193.4.49 | 106753 | 70 | - | - | - | - | - | - | See details in Table XIX | | | | | Ga1Sn3Zr5 |
| 193.4.50 | 183490 | 70 | - | - | - | - | - | - | See details in Table XIX | | | | | Ga1Hf5Sn3 |
| 193.4.51 | 107310 | 86 | - | - | - | - | - | - | See details in Table XIX | | | | | Cu1Hf5Sn3 |
| 193.4.52 | 107314 | 86 | - | - | - | - | - | - | See details in Table XIX | | | | | Cu1Hf5Pb3 |
| 193.4.53 | 15695 | 66 | - | - | - | - | - | - | See details in Table XIX | | | | | Cs3O1 |

Continued on next page

**TABLE XIV– continued**

| TQC | ICSD | NoE | DG | IndG | SubSG (w/ T) | cl | SubSG (w/o T) | cl | Max. Sub. SG | yes or no | Max. Sub. SG | yes or no | SIs (w/ T) | Stoich. Formula |
|---|---|---|---|---|---|---|---|---|---|---|---|---|---|---|
| 193.4.54 | 35315 | 50 | - | - | - | - | - | - | See details in Table XIX | | | | | Cl3Zr1 |
| 193.4.55 | 32702 | 50 | - | - | - | - | - | - | See details in Table XIX | | | | | Cl3Zr1 |
| 193.4.56 | 23944 | 50 | - | - | - | - | - | - | See details in Table XIX | | | | | Cl3Zr1 |
| 193.4.57 | 23163 | 50 | - | - | - | - | - | - | See details in Table XIX | | | | | Cl3Zr1 |
| 193.4.58 | 26069 | 50 | - | - | - | - | - | - | See details in Table XIX | | | | | Cl3Ti1 |
| 193.4.59 | 414040 | 58 | - | - | - | - | - | - | See details in Table XIX | | | | | Cl3Ru1 |
| 193.4.60 | 25771 | 58 | - | - | - | - | - | - | See details in Table XIX | | | | | Cl3Ru1 |
| 193.4.61 | 22090 | 58 | - | - | - | - | - | - | See details in Table XIX | | | | | Cl3Ru1 |
| 193.4.62 | 42470 | 198 | - | - | - | - | - | - | See details in Table XIX | | | | | Ce1F3 |
| 193.4.63 | 33796 | 90 | - | - | - | - | - | - | See details in Table XIX | | | | | Ca6Fe1N5 |
| 193.4.64 | 173026 | 50 | - | - | - | - | - | - | See details in Table XIX | | | | | Ca5Sb3 |
| 193.4.65 | 4068 | 50 | - | - | - | - | - | - | See details in Table XIX | | | | | Br3Zr1 |
| 193.4.66 | 35316 | 50 | - | - | - | - | - | - | See details in Table XIX | | | | | Br3Zr1 |
| 193.4.67 | 32703 | 50 | - | - | - | - | - | - | See details in Table XIX | | | | | Br3Zr1 |
| 193.4.68 | 23945 | 50 | - | - | - | - | - | - | See details in Table XIX | | | | | Br3Zr1 |
| 193.4.69 | 165302 | 50 | - | - | - | - | - | - | See details in Table XIX | | | | | Br3Zr1 |
| 193.4.70 | 414043 | 58 | - | - | - | - | - | - | See details in Table XIX | | | | | Br3Ru1 |
| 193.4.71 | 413692 | 58 | - | - | - | - | - | - | See details in Table XIX | | | | | Br3Ru1 |
| 193.4.72 | 413691 | 58 | - | - | - | - | - | - | See details in Table XIX | | | | | Br3Ru1 |
| 193.4.73 | 50725 | 154 | - | - | - | - | - | - | See details in Table XIX | | | | | Br1La5Sb3 |
| 193.4.74 | 409631 | 142 | - | - | - | - | - | - | See details in Table XIX | | | | | Br1In3La5 |
| 193.4.75 | 416609 | 122 | - | - | - | - | - | - | See details in Table XIX | | | | | Bi5La3Sc1 |
| 193.4.76 | 173362 | 130 | - | - | - | - | - | - | See details in Table XIX | | | | | Bi5La3Mn1 |
| 193.4.77 | 41838 | 130 | - | - | - | - | - | - | See details in Table XIX | | | | | Bi3Sr5 |
| 193.4.78 | 95241 | 154 | - | - | - | - | - | - | See details in Table XIX | | | | | Bi3Br1La5 |
| 193.4.79 | 74697 | 122 | - | - | - | - | - | - | See details in Table XIX | | | | | Bi0.67S2Ta1 |
| 193.4.80 | 41840 | 130 | - | - | - | - | - | - | See details in Table XIX | | | | | Ba5Sb3 |
| 193.4.81 | 41839 | 130 | - | - | - | - | - | - | See details in Table XIX | | | | | Ba5Bi3 |
| 193.4.82 | 77730 | 70 | - | - | - | - | - | - | See details in Table XIX | | | | | Ba3N1 |
| 193.4.83 | 44067 | 130 | - | - | - | - | - | - | See details in Table XIX | | | | | As3Sr5 |
| 193.4.84 | 43877 | 50 | - | - | - | - | - | - | See details in Table XIX | | | | | As3Ca5 |
| 193.4.85 | 43872 | 130 | - | - | - | - | - | - | See details in Table XIX | | | | | As3Ba5 |
| 193.4.86 | 58235 | 58 | - | - | - | - | - | - | See details in Table XIX | | | | | Al3Zr5 |
| 193.4.87 | 60633 | 58 | - | - | - | - | - | - | See details in Table XIX | | | | | Al3Hf5 |
| 193.4.88 | 57898 | 58 | - | - | - | - | - | - | See details in Table XIX | | | | | Al3Hf5 |
| 194.1.1 | 421015 | 24 | 0.0122 | -4.7636 | 2,174 | 6 | 2,174 | 1 | - | - | - | - | z4=3<br>z6m0=1<br>z12prime=7 | Zn1 |
| 194.1.2 | 421014 | 24 | 0.0060 | -4.7009 | 2,174 | 6 | 2,174 | 1 | - | - | - | - | z4=3<br>z6m0=1<br>z12prime=7 | Zn1 |
| 194.1.3 | 421013 | 24 | 0.0026 | -4.5632 | 2,174 | 6 | 2,174 | 1 | - | - | - | - | z4=3<br>z6m0=1<br>z12prime=7 | Zn1 |
| 194.1.4 | 247162 | 24 | 0.0241 | -4.7510 | 2,174 | 6 | 2,174 | 1 | - | - | - | - | z4=3<br>z6m0=1<br>z12prime=7 | Zn1 |
| 194.1.5 | 247161 | 24 | 0.0514 | -4.7417 | 2,174 | 6 | 2,174 | 1 | - | - | - | - | z4=3<br>z6m0=1<br>z12prime=7 | Zn1 |
| 194.1.6 | 247160 | 24 | 0.0392 | -4.7219 | 2,174 | 6 | 2,174 | 1 | - | - | - | - | z4=3<br>z6m0=1<br>z12prime=7 | Zn1 |
| 194.1.7 | 247159 | 24 | 0.0189 | -4.6926 | 2,174 | 6 | 2,174 | 1 | - | - | - | - | z4=3<br>z6m0=1<br>z12prime=7 | Zn1 |

Continued on next page

**TABLE XIV– continued**

| TQC | ICSD | NoE | DG | IndG | SubSG (w/ T) | cl | SubSG (w/o T) | cl | Max. Sub. SG | yes or no | Max. Sub. SG | yes or no | SIs (w/ T) | Stoich. Formula |
|---|---|---|---|---|---|---|---|---|---|---|---|---|---|---|
| 194.1.8 | 247158 | 24 | 0.0082 | -4.6835 | 2,174 | 6 | 2,174 | 1 | - | - | - | - | z4=3 z6m0=1 z12prime=7 | Zn1 |
| 194.1.9 | 247157 | 24 | 0.0109 | -4.6744 | 2,174 | 6 | 2,174 | 1 | - | - | - | - | z4=3 z6m0=1 z12prime=7 | Zn1 |
| 194.1.10 | 247156 | 24 | 0.0298 | -4.6627 | 2,174 | 6 | 2,174 | 1 | - | - | - | - | z4=3 z6m0=1 z12prime=7 | Zn1 |
| 194.1.11 | 247155 | 24 | 0.0401 | -4.6571 | 2,174 | 6 | 2,174 | 1 | - | - | - | - | z4=3 z6m0=1 z12prime=7 | Zn1 |
| 194.1.12 | 247154 | 24 | 0.0580 | -4.6458 | 2,174 | 6 | 2,174 | 1 | - | - | - | - | z4=3 z6m0=1 z12prime=7 | Zn1 |
| 194.1.13 | 247153 | 24 | 0.0657 | -4.6435 | 2,174 | 6 | 2,174 | 1 | - | - | - | - | z4=3 z6m0=1 z12prime=7 | Zn1 |
| 194.1.14 | 247152 | 24 | 0.0778 | -4.6242 | 2,174 | 6 | 2,174 | 1 | - | - | - | - | z4=3 z6m0=1 z12prime=7 | Zn1 |
| 194.1.15 | 247151 | 24 | 0.0605 | -4.6109 | 2,174 | 6 | 2,174 | 1 | - | - | - | - | z4=3 z6m0=1 z12prime=7 | Zn1 |
| 194.1.16 | 247150 | 24 | 0.0468 | -4.6009 | 2,174 | 6 | 2,174 | 1 | - | - | - | - | z4=3 z6m0=1 z12prime=7 | Zn1 |
| 194.1.17 | 247149 | 24 | 0.0307 | -4.5876 | 2,174 | 6 | 2,174 | 1 | - | - | - | - | z4=3 z6m0=1 z12prime=7 | Zn1 |
| 194.1.18 | 247148 | 24 | 0.0172 | -4.5769 | 2,174 | 6 | 2,174 | 1 | - | - | - | - | z4=3 z6m0=1 z12prime=7 | Zn1 |
| 194.1.19 | 247147 | 24 | 0.0110 | -4.5714 | 2,174 | 6 | 2,174 | 1 | - | - | - | - | z4=3 z6m0=1 z12prime=7 | Zn1 |
| 194.1.20 | 422990 | 72 | 0.0312 | -0.3492 | 174 | 7 | 174 | 2 | - | - | - | - | z6m0=2 z12prime=8 | Mn1O3Y1 |
| 194.1.21 | 422989 | 72 | 0.0319 | -0.3502 | 174 | 7 | 174 | 2 | - | - | - | - | z6m0=2 z12prime=8 | Mn1O3Y1 |
| 194.1.22 | 422982 | 72 | 0.0315 | -0.3479 | 174 | 7 | 174 | 2 | - | - | - | - | z6m0=2 z12prime=8 | Mn1O3Y1 |
| 194.1.23 | 182284 | 72 | 0.0266 | -0.3418 | 174 | 7 | 174 | 2 | - | - | - | - | z6m0=2 z12prime=8 | Mn1O3Y1 |
| 194.1.24 | 181182 | 72 | 0.0680 | -0.3201 | 174 | 7 | 174 | 2 | - | - | - | - | z6m0=2 z12prime=8 | Mn1O3Y1 |
| 194.1.25 | 193568 | 176 | 0.0065 | -0.9872 | 174 | 7 | 174 | 2 | - | - | - | - | z6m0=2 z12prime=8 | Mg1Ni2 |
| 194.1.26 | 171451 | 36 | 0.0182 | -0.3041 | 174 | 7 | 174 | 2 | - | - | - | - | z6m0=2 z12prime=8 | I1La1 |
| 194.1.27 | 163344 | 136 | 0.0000 | -0.0139 | 2,174 | 2 | 174 | 2 | - | - | - | - | z4=2 z6m0=2 z12prime=2 | Gd1Ge1Zn1 |
| 194.1.28 | 417150 | 152 | 0.0028 | -0.2272 | 2,174 | 6 | 2,174 | 1 | - | - | - | - | z4=3 z6m0=1 z12prime=7 | Fe0.25S2Ta1 |

Continued on next page

**TABLE XIV– continued**

| TQC | ICSD | NoE | DG | IndG | SubSG (w/ T) | cl | SubSG (w/o T) | cl | Max. Sub. SG | yes or no | Max. Sub. SG | yes or no | SIs (w/ T) | Stoich. Formula |
|---|---|---|---|---|---|---|---|---|---|---|---|---|---|---|
| 194.1.29 | 249600 | 96 | 0.0062 | -0.0933 | 2,174 | 1 | 2,174 | 1 | - | - | - | - | z4=1 z6m0=1 z12prime=1 | Eu1Ga1Sn1 |
| 194.1.30 | 249599 | 96 | 0.0048 | -0.1035 | 174 | 7 | 174 | 2 | - | - | - | - | z6m0=2 z12prime=8 | Eu1Ga1Ge1 |
| 194.1.31 | 158979 | 64 | 0.0249 | -0.0369 | 2,174 | 5 | 2,174 | 5 | - | - | - | - | z4=3 z6m0=5 z12prime=11 | Cr1Ge3Nd1 |
| 194.1.32 | 420050 | 112 | 0.0101 | -0.0825 | 2 | 3 | 2 | 3 | - | - | - | - | z4=1 z6m0=3 z12prime=9 | Ce1Sn1Zn1 |
| 194.1.33 | 152624 | 112 | 0.0099 | -0.0848 | 2 | 3 | 2 | 3 | - | - | - | - | z4=1 z6m0=3 z12prime=9 | Ce1Sn1Zn1 |
| 194.1.34 | 420186 | 112 | 0.0072 | -0.0954 | 2,174 | 1 | 2,174 | 1 | - | - | - | - | z4=1 z6m0=1 z12prime=1 | Ce1Ge1Zn1 |
| 194.1.35 | 163343 | 112 | 0.0095 | -0.0937 | 2,174 | 1 | 2,174 | 1 | - | - | - | - | z4=1 z6m0=1 z12prime=1 | Ce1Ge1Zn1 |
| 194.1.36 | 425673 | 152 | 0.0023 | -0.0028 | 2,174 | 1 | 2,174 | 1 | - | - | - | - | z4=1 z6m0=1 z12prime=1 | C6Co1N6Pr1 |
| 194.1.37 | 180423 | 48 | 0.0040 | -1.2156 | 2,174 | 5 | 2,174 | 5 | - | - | - | - | z4=3 z6m0=5 z12prime=11 | C2Si1Ti3 |
| 194.1.38 | 180422 | 48 | 0.0042 | -1.2268 | 2,174 | 5 | 2,174 | 5 | - | - | - | - | z4=3 z6m0=5 z12prime=11 | C2Si1Ti3 |
| 194.1.39 | 180421 | 48 | 0.0023 | -1.2373 | 2,174 | 5 | 2,174 | 5 | - | - | - | - | z4=3 z6m0=5 z12prime=11 | C2Si1Ti3 |
| 194.1.40 | 180420 | 48 | 0.0000 | -1.2466 | 2,174 | 5 | 2,174 | 5 | - | - | - | - | z4=3 z6m0=5 z12prime=11 | C2Si1Ti3 |
| 194.1.41 | 180419 | 48 | 0.0000 | -1.2551 | 2,174 | 5 | 2,174 | 5 | - | - | - | - | z4=3 z6m0=5 z12prime=11 | C2Si1Ti3 |
| 194.1.42 | 153264 | 48 | 0.0000 | -1.2692 | 2,174 | 5 | 2,174 | 5 | - | - | - | - | z4=3 z6m0=5 z12prime=11 | C2Si1Ti3 |
| 194.1.43 | 180431 | 48 | 0.0117 | -1.1436 | 2,174 | 5 | 2,174 | 5 | - | - | - | - | z4=3 z6m0=5 z12prime=11 | C2Ge1Ti3 |
| 194.1.44 | 180430 | 48 | 0.0143 | -1.1477 | 2,174 | 5 | 2,174 | 5 | - | - | - | - | z4=3 z6m0=5 z12prime=11 | C2Ge1Ti3 |
| 194.1.45 | 180429 | 48 | 0.0138 | -1.1519 | 2,174 | 5 | 2,174 | 5 | - | - | - | - | z4=3 z6m0=5 z12prime=11 | C2Ge1Ti3 |
| 194.1.46 | 180428 | 48 | 0.0135 | -1.1553 | 2,174 | 5 | 2,174 | 5 | - | - | - | - | z4=3 z6m0=5 z12prime=11 | C2Ge1Ti3 |
| 194.1.47 | 180427 | 48 | 0.0134 | -1.1582 | 2,174 | 5 | 2,174 | 5 | - | - | - | - | z4=3 z6m0=5 z12prime=11 | C2Ge1Ti3 |

Continued on next page

**TABLE XIV– continued**

| TQC | ICSD | NoE | DG | IndG | SubSG (w/ T) | cl | SubSG (w/o T) | cl | Max. Sub. SG | yes or no | Max. Sub. SG | yes or no | SIs (w/ T) | Stoich. Formula |
|---|---|---|---|---|---|---|---|---|---|---|---|---|---|---|
| 194.1.48 | 180426 | 48 | 0.0134 | -1.1596 | 2,174 | 5 | 2,174 | 5 | - | - | - | - | z4=3<br>z6m0=5<br>z12prime=11 | C2Ge1Ti3 |
| 194.1.49 | 180425 | 48 | 0.0137 | -1.1606 | 2,174 | 5 | 2,174 | 5 | - | - | - | - | z4=3<br>z6m0=5<br>z12prime=11 | C2Ge1Ti3 |
| 194.1.50 | 180424 | 48 | 0.0139 | -1.1555 | 2,174 | 5 | 2,174 | 5 | - | - | - | - | z4=3<br>z6m0=5<br>z12prime=11 | C2Ge1Ti3 |
| 194.1.51 | 249399 | 52 | 0.0632 | -0.2488 | 2,174 | 1 | 2,174 | 1 | - | - | - | - | z4=1<br>z6m0=1<br>z12prime=1 | C2Al1Ta3 |
| 194.1.52 | 246019 | 216 | 0.0157 | -0.1708 | 2 | 4 | 0 | 4 | - | - | - | - | z4=2<br>z12prime=6 | Ba1O3Ru1 |
| 194.1.53 | 173866 | 216 | 0.0157 | -0.1708 | 2 | 4 | 0 | 4 | - | - | - | - | z4=2<br>z12prime=6 | Ba1O3Ru1 |
| 194.1.54 | 160287 | 216 | 0.0184 | -0.1720 | 2 | 4 | 0 | 4 | - | - | - | - | z4=2<br>z12prime=6 | Ba1O3Ru1 |
| 194.1.55 | 192940 | 204 | 0.0060 | -0.6106 | 2 | 3 | 2 | 3 | - | - | - | - | z4=1<br>z6m0=3<br>z12prime=9 | Ba1Cr1O3 |
| 194.1.56 | 192939 | 204 | 0.0130 | -0.5264 | 2,174 | 2 | 174 | 2 | - | - | - | - | z4=2<br>z6m0=2<br>z12prime=2 | Ba1Cr1O3 |
| 194.1.57 | 251687 | 112 | 0.0000 | -0.0785 | 2,174 | 1 | 2,174 | 1 | - | - | - | - | z4=1<br>z6m0=1<br>z12prime=1 | Au1Ce1Sb1 |
| 194.3.1 | 152622 | 108 | - | - | - | - | - | - | 190 | yes | 190 | yes | - | Sn1Y1Zn1 |
| | | | | | | | | | 187 | yes | 187 | yes | - | |
| | | | | | | | | | 164 | no(2) | 164 | yes | z4=2 | |
| | | | | | | | | | 163 | no(2) | 163 | yes | z4=2 | |
| | | | | | | | | | 63 | no(2) | 63 | yes | z4=2 | |
| 194.3.2 | 262042 | 36 | - | - | - | - | - | - | 63 | yes | 63 | yes | - | Sc1Te1 |
| 194.3.3 | 427250 | 172 | - | - | - | - | - | - | 63 | no(2) | 63 | no(2) | z4=1 | Rh1Si3Y2 |
| 194.3.4 | 189925 | 176 | - | - | - | - | - | - | 63 | no(2) | 63 | no(2) | z4=3 | Pd3U1 |
| 194.3.5 | 164724 | 24 | - | - | - | - | - | - | 63 | no(2) | 63 | no(2) | z4=3 | O2 |
| 194.3.6 | 246978 | 120 | - | - | - | - | - | - | 63 | no(2) | 63 | no(2) | z4=3 | Mg5Pd2 |
| 194.3.7 | 163193 | 120 | - | - | - | - | - | - | 164 | no(2) | 164 | no(2) | z4=3 | Mg5Pd2 |
| | | | | | | | | | 163 | no(2) | 163 | no(2) | z4=3 | |
| | | | | | | | | | 63 | no(2) | 63 | no(2) | z4=3 | |
| 194.3.8 | 162567 | 112 | - | - | - | - | - | - | 63 | no(2) | 63 | yes | z4=2 | Mg2Yb1 |
| 194.3.9 | 380503 | 108 | - | - | - | - | - | - | 63 | no(2) | 63 | no(2) | z4=3 | Mg2Tm1 |
| 194.3.10 | 262676 | 16 | - | - | - | - | - | - | 63 | no(2) | 63 | no(2) | z4=3 | Mg1Na2Sn1 |
| 194.3.11 | 238202 | 16 | - | - | - | - | - | - | 63 | no(2) | 63 | no(2) | z4=3 | Mg1Na2Pb1 |
| 194.3.12 | 423654 | 196 | - | - | - | - | - | - | 63 | no(2) | 63 | no(2) | z4=3 | K3Na1O9Os2 |
| 194.3.13 | 414235 | 32 | - | - | - | - | - | - | 63 | no(2) | 63 | yes | z4=2 | In2Sr1 |
| 194.3.14 | 260996 | 128 | - | - | - | - | - | - | 63 | no(2) | 63 | no(2) | z4=3 | Hf9Mo4Si1 |
| 194.3.15 | 164891 | 56 | - | - | - | - | - | - | 63 | yes | 63 | yes | - | Ge3La1V1 |
| 194.3.16 | 419660 | 96 | - | - | - | - | - | - | 63 | no(2) | 63 | no(2) | z4=1 | Ge2Ni3Sr1 |
| 194.3.17 | 163345 | 108 | - | - | - | - | - | - | 190 | yes | 190 | yes | - | Ge1Y1Zn1 |
| | | | | | | | | | 187 | yes | 187 | yes | - | |
| | | | | | | | | | 164 | no(2) | 164 | no(2) | z4=3 | |
| | | | | | | | | | 163 | no(2) | 163 | no(2) | z4=3 | |
| | | | | | | | | | 63 | no(2) | 63 | no(2) | z4=3 | |
| 194.3.18 | 419659 | 68 | - | - | - | - | - | - | 190 | no(174) | 190 | no(174) | z3m0=1 | Ge1Ni2Sr1 |
| | | | | | | | | | 187 | no(174) | 187 | no(174) | z3m0=1 | |
| | | | | | | | | | 164 | no(2) | 164 | yes | z4=2 | |
| | | | | | | | | | 163 | no(2) | 163 | yes | z4=2 | |

Continued on next page

**TABLE XIV– continued**

| TQC | ICSD | NoE | DG | IndG | SubSG cl (w/ T) | SubSG cl (w/o T) | Max. Sub. SG | yes or no | Max. Sub. SG | yes or no | SIs (w/ T) | Stoich. Formula |
|---|---|---|---|---|---|---|---|---|---|---|---|---|
| | | | | | | | 63 | no(2) | 63 | yes | z4=2 | |
| 194.3.19 | 420008 | 84 | - | - | - - | - - | 63 | no(2) | 63 | yes | z4=2 | Gd1Pt2Sn1 |
| 194.3.20 | 260580 | 60 | - | - | - - | - - | 63 | no(2) | 63 | no(2) | z4=3 | Ga2Yb1 |
| 194.3.21 | 161592 | 112 | - | - | - - | - - | 63 | no(2) | 63 | no(2) | z4=1 | Fe3Ge1U2 |
| 194.3.22 | 155674 | 80 | - | - | - - | - - | 63 | no(2) | 63 | no(2) | z4=1 | Fe2Ti1 |
| 194.3.23 | 169307 | 28 | - | - | - - | - - | 190 | no(174) | 190 | no(174) | z3m0=1 | Fe1Se1 |
| | | | | | | | 187 | no(174) | 187 | no(174) | z3m0=1 | |
| | | | | | | | 164 | no(2) | 164 | no(2) | z4=3 | |
| | | | | | | | 163 | no(2) | 163 | no(2) | z4=3 | |
| | | | | | | | 63 | no(2) | 63 | no(2) | z4=3 | |
| 194.3.24 | 169302 | 28 | - | - | - - | - - | 190 | no(174) | 190 | no(174) | z3m0=1 | Fe1Se1 |
| | | | | | | | 187 | no(174) | 187 | no(174) | z3m0=1 | |
| | | | | | | | 164 | yes | 164 | yes | - | |
| | | | | | | | 163 | yes | 163 | yes | - | |
| | | | | | | | 63 | yes | 63 | yes | - | |
| 194.3.25 | 169300 | 28 | - | - | - - | - - | 190 | no(174) | 190 | no(174) | z3m0=1 | Fe1Se1 |
| | | | | | | | 187 | no(174) | 187 | no(174) | z3m0=1 | |
| | | | | | | | 164 | yes | 164 | yes | - | |
| | | | | | | | 163 | yes | 163 | yes | - | |
| | | | | | | | 63 | yes | 63 | yes | - | |
| 194.3.26 | 169298 | 28 | - | - | - - | - - | 190 | no(174) | 190 | no(174) | z3m0=1 | Fe1Se1 |
| | | | | | | | 187 | no(174) | 187 | no(174) | z3m0=1 | |
| | | | | | | | 164 | yes | 164 | yes | - | |
| | | | | | | | 163 | yes | 163 | yes | - | |
| | | | | | | | 63 | yes | 63 | yes | - | |
| 194.3.27 | 169296 | 28 | - | - | - - | - - | 190 | no(174) | 190 | no(174) | z3m0=1 | Fe1Se1 |
| | | | | | | | 187 | no(174) | 187 | no(174) | z3m0=1 | |
| | | | | | | | 164 | yes | 164 | yes | - | |
| | | | | | | | 163 | yes | 163 | yes | - | |
| | | | | | | | 63 | yes | 63 | yes | - | |
| 194.3.28 | 169294 | 28 | - | - | - - | - - | 190 | no(174) | 190 | no(174) | z3m0=1 | Fe1Se1 |
| | | | | | | | 187 | no(174) | 187 | no(174) | z3m0=1 | |
| | | | | | | | 164 | yes | 164 | yes | - | |
| | | | | | | | 163 | yes | 163 | yes | - | |
| | | | | | | | 63 | yes | 63 | yes | - | |
| 194.3.29 | 169292 | 28 | - | - | - - | - - | 190 | no(174) | 190 | no(174) | z3m0=1 | Fe1Se1 |
| | | | | | | | 187 | no(174) | 187 | no(174) | z3m0=1 | |
| | | | | | | | 164 | yes | 164 | yes | - | |
| | | | | | | | 163 | yes | 163 | yes | - | |
| | | | | | | | 63 | yes | 63 | yes | - | |
| 194.3.30 | 169290 | 28 | - | - | - - | - - | 190 | no(174) | 190 | no(174) | z3m0=1 | Fe1Se1 |
| | | | | | | | 187 | no(174) | 187 | no(174) | z3m0=1 | |
| | | | | | | | 164 | yes | 164 | yes | - | |
| | | | | | | | 163 | yes | 163 | yes | - | |
| | | | | | | | 63 | yes | 63 | yes | - | |
| 194.3.31 | 169288 | 28 | - | - | - - | - - | 190 | no(174) | 190 | no(174) | z3m0=1 | Fe1Se1 |
| | | | | | | | 187 | no(174) | 187 | no(174) | z3m0=1 | |
| | | | | | | | 164 | yes | 164 | yes | - | |
| | | | | | | | 163 | yes | 163 | yes | - | |
| | | | | | | | 63 | yes | 63 | yes | - | |
| 194.3.32 | 169286 | 28 | - | - | - - | - - | 190 | no(174) | 190 | no(174) | z3m0=1 | Fe1Se1 |
| | | | | | | | 187 | no(174) | 187 | no(174) | z3m0=1 | |
| | | | | | | | 164 | yes | 164 | yes | - | |
| | | | | | | | 163 | yes | 163 | yes | - | |
| | | | | | | | 63 | yes | 63 | yes | - | |
| 194.3.33 | 169284 | 28 | - | - | - - | - - | 190 | no(174) | 190 | no(174) | z3m0=1 | Fe1Se1 |
| | | | | | | | 187 | no(174) | 187 | no(174) | z3m0=1 | |
| | | | | | | | 164 | yes | 164 | yes | - | |

Continued on next page

**TABLE XIV– continued**

| TQC | ICSD | NoE | DG | IndG | SubSG cl (w/ T) | SubSG cl (w/o T) | Max. Sub. SG | yes or no | Max. Sub. SG | yes or no | SIs (w/ T) | Stoich. Formula |
|---|---|---|---|---|---|---|---|---|---|---|---|---|
| | | | | | | | 163 | yes | 163 | yes | - | |
| | | | | | | | 63 | yes | 63 | yes | - | |
| 194.3.34 | 169272 | 28 | - | - | - - | - - | 190 | no(174) | 190 | no(174) | z3m0=1 | Fe1Se1 |
| | | | | | | | 187 | no(174) | 187 | no(174) | z3m0=1 | |
| | | | | | | | 164 | yes | 164 | yes | - | |
| | | | | | | | 163 | yes | 163 | yes | - | |
| | | | | | | | 63 | yes | 63 | yes | - | |
| 194.3.35 | 169270 | 28 | - | - | - - | - - | 190 | no(174) | 190 | no(174) | z3m0=1 | Fe1Se1 |
| | | | | | | | 187 | no(174) | 187 | no(174) | z3m0=1 | |
| | | | | | | | 164 | yes | 164 | yes | - | |
| | | | | | | | 163 | yes | 163 | yes | - | |
| | | | | | | | 63 | yes | 63 | yes | - | |
| 194.3.36 | 169268 | 28 | - | - | - - | - - | 190 | no(174) | 190 | no(174) | z3m0=1 | Fe1Se1 |
| | | | | | | | 187 | no(174) | 187 | no(174) | z3m0=1 | |
| | | | | | | | 164 | yes | 164 | yes | - | |
| | | | | | | | 163 | yes | 163 | yes | - | |
| | | | | | | | 63 | yes | 63 | yes | - | |
| 194.3.37 | 169266 | 28 | - | - | - - | - - | 190 | no(174) | 190 | no(174) | z3m0=1 | Fe1Se1 |
| | | | | | | | 187 | no(174) | 187 | no(174) | z3m0=1 | |
| | | | | | | | 164 | yes | 164 | yes | - | |
| | | | | | | | 163 | yes | 163 | yes | - | |
| | | | | | | | 63 | yes | 63 | yes | - | |
| 194.3.38 | 169264 | 28 | - | - | - - | - - | 190 | no(174) | 190 | no(174) | z3m0=1 | Fe1Se1 |
| | | | | | | | 187 | no(174) | 187 | no(174) | z3m0=1 | |
| | | | | | | | 164 | yes | 164 | yes | - | |
| | | | | | | | 163 | yes | 163 | yes | - | |
| | | | | | | | 63 | yes | 63 | yes | - | |
| 194.3.39 | 169262 | 28 | - | - | - - | - - | 190 | no(174) | 190 | no(174) | z3m0=1 | Fe1Se1 |
| | | | | | | | 187 | no(174) | 187 | no(174) | z3m0=1 | |
| | | | | | | | 164 | yes | 164 | yes | - | |
| | | | | | | | 163 | yes | 163 | yes | - | |
| | | | | | | | 63 | yes | 63 | yes | - | |
| 194.3.40 | 169260 | 28 | - | - | - - | - - | 190 | no(174) | 190 | no(174) | z3m0=1 | Fe1Se1 |
| | | | | | | | 187 | no(174) | 187 | no(174) | z3m0=1 | |
| | | | | | | | 164 | yes | 164 | yes | - | |
| | | | | | | | 163 | yes | 163 | yes | - | |
| | | | | | | | 63 | yes | 63 | yes | - | |
| 194.3.41 | 169258 | 28 | - | - | - - | - - | 190 | no(174) | 190 | no(174) | z3m0=1 | Fe1Se1 |
| | | | | | | | 187 | no(174) | 187 | no(174) | z3m0=1 | |
| | | | | | | | 164 | yes | 164 | yes | - | |
| | | | | | | | 163 | yes | 163 | yes | - | |
| | | | | | | | 63 | yes | 63 | yes | - | |
| 194.3.42 | 169256 | 28 | - | - | - - | - - | 190 | no(174) | 190 | no(174) | z3m0=1 | Fe1Se1 |
| | | | | | | | 187 | no(174) | 187 | no(174) | z3m0=1 | |
| | | | | | | | 164 | yes | 164 | yes | - | |
| | | | | | | | 163 | yes | 163 | yes | - | |
| | | | | | | | 63 | yes | 63 | yes | - | |
| 194.3.43 | 169254 | 28 | - | - | - - | - - | 190 | no(174) | 190 | no(174) | z3m0=1 | Fe1Se1 |
| | | | | | | | 187 | no(174) | 187 | no(174) | z3m0=1 | |
| | | | | | | | 164 | yes | 164 | yes | - | |
| | | | | | | | 163 | yes | 163 | yes | - | |
| | | | | | | | 63 | yes | 63 | yes | - | |
| 194.3.44 | 169252 | 28 | - | - | - - | - - | 190 | no(174) | 190 | no(174) | z3m0=1 | Fe1Se1 |
| | | | | | | | 187 | no(174) | 187 | no(174) | z3m0=1 | |
| | | | | | | | 164 | yes | 164 | yes | - | |
| | | | | | | | 163 | yes | 163 | yes | - | |
| | | | | | | | 63 | yes | 63 | yes | - | |
| 194.3.45 | 409890 | 104 | - | - | - - | - - | 63 | yes | 63 | yes | - | Eu2Ga3Ir1 |

Continued on next page

**TABLE XIV– continued**

| TQC | ICSD | NoE | DG | IndG | SubSG (w/ T) | cl | SubSG (w/o T) | cl | Max. Sub. SG | yes or no | Max. Sub. SG | yes or no | SIs (w/ T) | Stoich. Formula |
|---|---|---|---|---|---|---|---|---|---|---|---|---|---|---|
| 194.3.46 | 187957 | 144 | - | - | - | - | - | - | 63 | no(2) | 63 | no(2) | z4=3 | Dy1In4Pt4 |
| 194.3.47 | 416327 | 244 | - | - | - | - | - | - | 63 | no(2) | 63 | no(2) | z4=3 | Cu9Mg2Tb1 |
| 194.3.48 | 416545 | 52 | - | - | - | - | - | - | 63 | no(2) | 63 | no(2) | z4=3 | Cu1La1Sn1 |
| 194.3.49 | 154258 | 68 | - | - | - | - | - | - | 63 | yes | 63 | yes | - | Cs1I3Ti1 |
| 194.3.50 | 423830 | 80 | - | - | - | - | - | - | 63 | yes | 63 | yes | - | Cs1I3Ni1 |
| 194.3.51 | 423828 | 80 | - | - | - | - | - | - | 63 | yes | 63 | yes | - | Cl3Cs1Ni1 |
| 194.3.52 | 158977 | 60 | - | - | - | - | - | - | 63 | no(2) | 63 | no(2) | z4=1 | Ce1Cr1Ge3 |
| 194.3.53 | 428620 | 308 | - | - | - | - | - | - | 63 | no(2) | 63 | no(2) | z4=3 | Cd3Ru1Y10 |
| 194.3.54 | 418878 | 104 | - | - | - | - | - | - | 63 | yes | 63 | yes | - | Ca1Zn2 |
| 194.3.55 | 172208 | 60 | - | - | - | - | - | - | 190 | no(174) | 190 | no(174) | z3m0=2 | Ca1Na4Sn6 |
| | | | | | | | | | 187 | no(174) | 187 | no(174) | z3m0=2 | |
| | | | | | | | | | 164 | yes | 164 | yes | - | |
| | | | | | | | | | 163 | yes | 163 | yes | - | |
| | | | | | | | | | 63 | yes | 63 | yes | - | |
| 194.3.56 | 414232 | 16 | - | - | - | - | - | - | 63 | no(2) | 63 | no(2) | z4=1 | Ca1In2 |
| 194.3.57 | 417217 | 16 | - | - | - | - | - | - | 63 | no(2) | 63 | no(2) | z4=1 | Ca1Ga2 |
| 194.3.58 | 260645 | 16 | - | - | - | - | - | - | 63 | no(2) | 63 | no(2) | z4=1 | Ca1Ga2 |
| 194.3.59 | 420577 | 104 | - | - | - | - | - | - | 63 | yes | 63 | yes | - | Ca1Cd2 |
| 194.3.60 | 249388 | 48 | - | - | - | - | - | - | 63 | yes | 63 | yes | - | C1N2Ni1 |
| 194.3.61 | 154257 | 68 | - | - | - | - | - | - | 63 | yes | 63 | yes | - | Br3Rb1Ti1 |
| 194.3.62 | 423829 | 80 | - | - | - | - | - | - | 63 | yes | 63 | yes | - | Br3Cs1Ni1 |
| 194.3.63 | 416142 | 80 | - | - | - | - | - | - | 63 | yes | 63 | yes | - | Bi1Cu1Yb1 |
| 194.3.64 | 416141 | 80 | - | - | - | - | - | - | 63 | yes | 63 | yes | - | Bi1Cu1Yb1 |
| 194.3.65 | 172754 | 224 | - | - | - | - | - | - | 63 | no(2) | 63 | no(2) | z4=3 | Ba3O9Ru2Zr1 |
| 194.3.66 | 245252 | 208 | - | - | - | - | - | - | 63 | no(2) | 63 | yes | z4=2 | Ba3Ir2Mg1O9 |
| 194.3.67 | 245251 | 208 | - | - | - | - | - | - | 63 | no(2) | 63 | yes | z4=2 | Ba3Ir2Mg1O9 |
| 194.3.68 | 419206 | 44 | - | - | - | - | - | - | 63 | yes | 63 | yes | - | Ba1Sn3 |
| 194.3.69 | 409902 | 84 | - | - | - | - | - | - | 63 | no(2) | 63 | no(2) | z4=1 | Ba1Li4 |
| 194.3.70 | 261651 | 44 | - | - | - | - | - | - | 63 | yes | 63 | yes | - | Ba1Ge3 |
| 194.3.71 | 154600 | 296 | - | - | - | - | - | - | 63 | no(2) | 63 | no(2) | z4=1 | B2Ni7Yb3 |
| 194.3.72 | 251686 | 108 | - | - | - | - | - | - | 63 | no(2) | 63 | no(2) | z4=3 | Au1La1Sb1 |
| 194.3.73 | 420340 | 52 | - | - | - | - | - | - | 63 | no(2) | 63 | no(2) | z4=3 | Au1Ba1P1 |
| 194.3.74 | 420341 | 52 | - | - | - | - | - | - | 63 | no(2) | 63 | no(2) | z4=3 | As1Au1Ba1 |
| 194.4.1 | 164099 | 6 | - | - | - | - | - | - | See details in Table XIX | | | | | Sc1 |
| 194.4.2 | 164097 | 6 | - | - | - | - | - | - | See details in Table XIX | | | | | Sc1 |
| 194.4.3 | 164095 | 6 | - | - | - | - | - | - | See details in Table XIX | | | | | Sc1 |
| 194.4.4 | 164094 | 6 | - | - | - | - | - | - | See details in Table XIX | | | | | Sc1 |
| 194.4.5 | 164093 | 6 | - | - | - | - | - | - | See details in Table XIX | | | | | Sc1 |
| 194.4.6 | 164092 | 6 | - | - | - | - | - | - | See details in Table XIX | | | | | Sc1 |
| 194.4.7 | 164091 | 6 | - | - | - | - | - | - | See details in Table XIX | | | | | Sc1 |
| 194.4.8 | 164090 | 6 | - | - | - | - | - | - | See details in Table XIX | | | | | Sc1 |
| 194.4.9 | 164089 | 6 | - | - | - | - | - | - | See details in Table XIX | | | | | Sc1 |
| 194.4.10 | 164088 | 6 | - | - | - | - | - | - | See details in Table XIX | | | | | Sc1 |
| 194.4.11 | 162586 | 58 | - | - | - | - | - | - | See details in Table XIX | | | | | Ni2Sb1Zr1 |
| 194.4.12 | 162585 | 58 | - | - | - | - | - | - | See details in Table XIX | | | | | Ni2Sb1Zr1 |
| 194.4.13 | 158485 | 74 | - | - | - | - | - | - | See details in Table XIX | | | | | Ni2Sb1Te2 |
| 194.4.14 | 190179 | 30 | - | - | - | - | - | - | See details in Table XIX | | | | | Ni1Sb1 |
| 194.4.15 | 710050 | 58 | - | - | - | - | - | - | See details in Table XIX | | | | | Nd1P1Pd1 |
| 194.4.16 | 185417 | 210 | - | - | - | - | - | - | See details in Table XIX | | | | | Mn1O3Sr1 |
| 194.4.17 | 237636 | 86 | - | - | - | - | - | - | See details in Table XIX | | | | | Li3Sn4Zn2 |
| 194.4.18 | 246806 | 102 | - | - | - | - | - | - | See details in Table XIX | | | | | La1Si10 |
| 194.4.19 | 154047 | 54 | - | - | - | - | - | - | See details in Table XIX | | | | | Ge3La1Ti1 |
| 194.4.20 | 425591 | 106 | - | - | - | - | - | - | See details in Table XIX | | | | | Ge1Mn3U2 |
| 194.4.21 | 420009 | 82 | - | - | - | - | - | - | See details in Table XIX | | | | | Gd1In1Pt2 |
| 194.4.22 | 156263 | 34 | - | - | - | - | - | - | See details in Table XIX | | | | | Ga1Pt1Ti1 |
| 194.4.23 | 236572 | 126 | - | - | - | - | - | - | See details in Table XIX | | | | | Cu3Sb4V2 |
| 194.4.24 | 422620 | 66 | - | - | - | - | - | - | See details in Table XIX | | | | | Cu1Eu1Sb1 |

Continued on next page

**TABLE XIV– continued**

| TQC | ICSD | NoE | DG | IndG | SubSG cl (w/ T) | SubSG cl (w/o T) | Max. Sub. SG yes or no | Max. Sub. SG yes or no | SIs (w/ T) | Stoich. Formula |
|---|---|---|---|---|---|---|---|---|---|---|
| 194.4.25 | 158978 | 62 | - | - | - - | - - | See details in Table XIX | | | Cr1Ge3Pr1 |
| 194.4.26 | 158976 | 58 | - | - | - - | - - | See details in Table XIX | | | Cr1Ge3La1 |
| 194.4.27 | 418080 | 126 | - | - | - - | - - | See details in Table XIX | | | Co2I1Tb2 |
| 194.4.28 | 416273 | 102 | - | - | - - | - - | See details in Table XIX | | | Co2I1Pr2 |
| 194.4.29 | 416787 | 54 | - | - | - - | - - | See details in Table XIX | | | Ce1Cu1Sn1 |
| 194.4.30 | 416786 | 54 | - | - | - - | - - | See details in Table XIX | | | Ce1Cu1Sn1 |
| 194.4.31 | 173221 | 54 | - | - | - - | - - | See details in Table XIX | | | Ce1Cu1Si1 |
| 194.4.32 | 173219 | 54 | - | - | - - | - - | See details in Table XIX | | | Ce1Cu1Ge1 |
| 194.4.33 | 153107 | 50 | - | - | - - | - - | See details in Table XIX | | | Ca2I1N1 |
| 194.4.34 | 419117 | 30 | - | - | - - | - - | See details in Table XIX | | | C1Ga1Ti2 |
| 194.4.35 | 419116 | 38 | - | - | - - | - - | See details in Table XIX | | | C1Cr2Ga1 |
| 194.4.36 | 249387 | 46 | - | - | - - | - - | See details in Table XIX | | | C1Co1N2 |
| 194.4.37 | 420348 | 74 | - | - | - - | - - | See details in Table XIX | | | C1Br1Pr2 |
| 194.4.38 | 246360 | 206 | - | - | - - | - - | See details in Table XIX | | | Ba3Ir2Na1O9 |
| 194.4.39 | 246359 | 206 | - | - | - - | - - | See details in Table XIX | | | Ba3Ir2Na1O9 |
| 194.4.40 | 174293 | 226 | - | - | - - | - - | See details in Table XIX | | | Ba3Ir2La1O9 |
| 194.4.41 | 157841 | 206 | - | - | - - | - - | See details in Table XIX | | | Ba3In1O9Ru2 |
| 194.4.42 | 154181 | 66 | - | - | - - | - - | See details in Table XIX | | | Ba1S3V1 |
| 194.4.43 | 424878 | 50 | - | - | - - | - - | See details in Table XIX | | | B3Ru2 |
| 194.4.44 | 424879 | 50 | - | - | - - | - - | See details in Table XIX | | | B3Os2 |
| 194.4.45 | 421522 | 26 | - | - | - - | - - | See details in Table XIX | | | B2Re1 |
| 194.4.46 | 165834 | 26 | - | - | - - | - - | See details in Table XIX | | | B2Re1 |
| 194.4.47 | 427220 | 182 | - | - | - - | - - | See details in Table XIX | | | Au5Cs1Ga9 |
| 194.4.48 | 190987 | 62 | - | - | - - | - - | See details in Table XIX | | | Au2Ga2In1 |
| 194.4.49 | 192087 | 54 | - | - | - - | - - | See details in Table XIX | | | Au2Ga1Mg1 |
| 194.4.50 | 163719 | 146 | - | - | - - | - - | See details in Table XIX | | | Au2Ce3Sb3 |
| 194.4.51 | 260109 | 66 | - | - | - - | - - | See details in Table XIX | | | As2Eu1In2 |
| 194.4.52 | 420091 | 110 | - | - | - - | - - | See details in Table XIX | | | Al5Ba4 |
| 194.4.53 | 420090 | 90 | - | - | - - | - - | See details in Table XIX | | | Al5Ba3 |
| 194.4.54 | 415931 | 90 | - | - | - - | - - | See details in Table XIX | | | Al2Ba2Ga1.33 |
| 194.4.55 | 163350 | 34 | - | - | - - | - - | See details in Table XIX | | | Al1Pt1Ti1 |
| 194.4.56 | 172878 | 154 | - | - | - - | - - | See details in Table XIX | | | Ag3Ni2O4 |
| 194.4.57 | 415451 | 66 | - | - | - - | - - | See details in Table XIX | | | Ag1Ni1O2 |
| 194.4.58 | 422621 | 66 | - | - | - - | - - | See details in Table XIX | | | Ag1Eu1Sb1 |
| 197.4.1 | 100098 | 206 | - | - | - cl | - cl | See details in Table XIX | | | La4O19Ru6 |
| 197.4.2 | 100099 | 206 | - | - | - cl | - cl | See details in Table XIX | | | La4O19Os6 |
| 198.4.1 | 79233 | 52 | - | - | - - | - - | See details in Table XIX | | | Rh1Si1 |
| 198.4.2 | 43240 | 52 | - | - | - - | - - | See details in Table XIX | | | Rh1Si1 |
| 198.4.3 | 24719 | 52 | - | - | - - | - - | See details in Table XIX | | | Rh1Si1 |
| 198.4.4 | 69791 | 96 | - | - | - - | - - | See details in Table XIX | | | Pd1Si1Sr1 |
| 198.4.5 | 93906 | 84 | - | - | - - | - - | See details in Table XIX | | | Pd1Sb1Te1 |
| 198.4.6 | 93905 | 84 | - | - | - - | - - | See details in Table XIX | | | Pd1Sb1Se1 |
| 198.4.7 | 187621 | 56 | - | - | - - | - - | See details in Table XIX | | | Ni1Si1 |
| 198.4.8 | 93902 | 84 | - | - | - - | - - | See details in Table XIX | | | Ni1Sb1Se1 |
| 198.4.9 | 93900 | 84 | - | - | - - | - - | See details in Table XIX | | | Ni1S1Sb1 |
| 198.4.10 | 93898 | 84 | - | - | - - | - - | See details in Table XIX | | | Ni1P1S1 |
| 198.4.11 | 71832 | 44 | - | - | - - | - - | See details in Table XIX | | | Mn1Si1 |
| 198.4.12 | 71831 | 44 | - | - | - - | - - | See details in Table XIX | | | Mn1Si1 |
| 198.4.13 | 71830 | 44 | - | - | - - | - - | See details in Table XIX | | | Mn1Si1 |
| 198.4.14 | 155712 | 96 | - | - | - - | - - | See details in Table XIX | | | Ir1P1Sr1 |
| 198.4.15 | 27187 | 96 | - | - | - - | - - | See details in Table XIX | | | Ir1La1Si1 |
| 198.4.16 | 252044 | 44 | - | - | - - | - - | See details in Table XIX | | | Ge1Mn1 |
| 198.4.17 | 252043 | 44 | - | - | - - | - - | See details in Table XIX | | | Ge1Mn1 |
| 198.4.18 | 187195 | 44 | - | - | - - | - - | See details in Table XIX | | | Ge1Mn1 |
| 198.4.19 | 150395 | 136 | - | - | - - | - - | See details in Table XIX | | | Ge1Ir1Mn3 |
| 198.4.20 | 261111 | 52 | - | - | - - | - - | See details in Table XIX | | | Ga1Pd1 |
| 198.4.21 | 401561 | 124 | - | - | - - | - - | See details in Table XIX | | | Eu1Ge1Pt1 |

Continued on next page

**TABLE XIV– continued**

| TQC | ICSD | NoE | DG | IndG | SubSG (w/ T) | cl | SubSG (w/o T) | cl | Max. Sub. SG | yes or no | Max. Sub. SG | yes or no | SIs (w/ T) | Stoich. Formula |
|---|---|---|---|---|---|---|---|---|---|---|---|---|---|---|
| 198.4.22 | 260982 | 52 | - | - | - | - | - | - | See details in Table XIX | | | | | Co1Si1 |
| 198.4.23 | 413854 | 100 | - | - | - | - | - | - | See details in Table XIX | | | | | Ce1Ir1Si1 |
| 198.4.24 | 72640 | 64 | - | - | - | - | - | - | See details in Table XIX | | | | | Ca1Pt1Si1 |
| 198.4.25 | 63555 | 64 | - | - | - | - | - | - | See details in Table XIX | | | | | Ca1Pt1Si1 |
| 198.4.26 | 70060 | 84 | - | - | - | - | - | - | See details in Table XIX | | | | | Bi1Pd1Te1 |
| 198.4.27 | 59191 | 100 | - | - | - | - | - | - | See details in Table XIX | | | | | Ba1P1Pt1 |
| 198.4.28 | 73529 | 96 | - | - | - | - | - | - | See details in Table XIX | | | | | Ba1Ir1P1 |
| 198.4.29 | 93904 | 84 | - | - | - | - | - | - | See details in Table XIX | | | | | As1Pd1Se1 |
| 198.4.30 | 93903 | 84 | - | - | - | - | - | - | See details in Table XIX | | | | | As1Pd1S1 |
| 198.4.31 | 93901 | 84 | - | - | - | - | - | - | See details in Table XIX | | | | | As1Ni1Se1 |
| 198.4.32 | 93899 | 84 | - | - | - | - | - | - | See details in Table XIX | | | | | As1Ni1S1 |
| 198.4.33 | 59192 | 100 | - | - | - | - | - | - | See details in Table XIX | | | | | As1Ba1Pt1 |
| 198.4.34 | 104616 | 188 | - | - | - | - | - | - | See details in Table XIX | | | | | Al1Au4 |
| 199.3.1 | 195668 | 104 | - | - | - | - | - | - | 24 | yes | 24 | yes | - | Bi2Pt3Se2 |
| 199.4.1 | 417634 | 104 | - | - | - | - | - | - | See details in Table XIX | | | | | Bi2Pd3S2 |
| 199.4.2 | 48178 | 180 | - | - | - | - | - | - | See details in Table XIX | | | | | Ba3Cl2Fe2O5 |
| 199.4.3 | 48179 | 180 | - | - | - | - | - | - | See details in Table XIX | | | | | Ba3Br2Fe2O5 |
| 202.4.1 | 179 | 140 | - | - | - | - | - | - | See details in Table XIX | | | | | K2N6Ni1O12Sr1 |
| 202.4.2 | 1937 | 134 | - | - | - | - | - | - | See details in Table XIX | | | | | K2N6Ni1O12Pb1 |
| 204.1.1 | 237751 | 114 | 0.0042 | -0.0406 | 2 | 4 | 2 | 4 | - | - | - | - | z2w,1=1<br>z2w,2=1<br>z2w,3=1<br>z4=3 | Mn3Na1O12V4 |
| 204.1.2 | 195439 | 136 | 0.0153 | -0.2976 | 2 | 4 | 2 | 4 | - | - | - | - | z2w,1=1<br>z2w,2=1<br>z2w,3=1<br>z4=3 | Cu4O12V4 |
| 204.1.3 | 195438 | 136 | 0.0143 | -0.2943 | 2 | 4 | 2 | 4 | - | - | - | - | z2w,1=1<br>z2w,2=1<br>z2w,3=1<br>z4=3 | Cu4O12V4 |
| 204.1.4 | 51893 | 138 | 0.0155 | -0.4631 | 2 | 3 | 0 | 3 | - | - | - | - | z2w,1=1<br>z2w,2=1<br>z2w,3=1 | Cu3Na1O12Ru4 |
| 204.1.5 | 237422 | 160 | 0.0181 | -0.1382 | 2 | 5 | 2 | 1 | - | - | - | - | z2w,1=1<br>z2w,2=1<br>z2w,3=1<br>z4=1 | Cu3Fe4O12Tm1 |
| 204.1.6 | 248615 | 140 | 0.0159 | -0.1626 | 2 | 3 | 0 | 3 | - | - | - | - | z2w,1=1<br>z2w,2=1<br>z2w,3=1 | Cr4Cu3O12Y1 |
| 204.1.7 | 248614 | 140 | 0.0170 | -0.1626 | 2 | 3 | 0 | 3 | - | - | - | - | z2w,1=1<br>z2w,2=1<br>z2w,3=1 | Cr4Cu3O12Y1 |
| 204.1.8 | 248613 | 140 | 0.0164 | -0.1632 | 2 | 3 | 0 | 3 | - | - | - | - | z2w,1=1<br>z2w,2=1<br>z2w,3=1 | Cr4Cu3O12Y1 |
| 204.1.9 | 248612 | 140 | 0.0163 | -0.1635 | 2 | 3 | 0 | 3 | - | - | - | - | z2w,1=1<br>z2w,2=1<br>z2w,3=1 | Cr4Cu3O12Y1 |
| 204.1.10 | 248611 | 140 | 0.0157 | -0.1639 | 2 | 3 | 0 | 3 | - | - | - | - | z2w,1=1<br>z2w,2=1<br>z2w,3=1 | Cr4Cu3O12Y1 |
| 204.1.11 | 248610 | 140 | 0.0122 | -0.1673 | 2 | 3 | 0 | 3 | - | - | - | - | z2w,1=1<br>z2w,2=1<br>z2w,3=1 | Cr4Cu3O12Y1 |

Continued on next page

**TABLE XIV– continued**

| TQC | ICSD | NoE | DG | IndG | SubSG (w/ T) | cl | SubSG (w/o T) | cl | Max. Sub. SG | yes or no | Max. Sub. SG | yes or no | SIs (w/ T) | Stoich. Formula |
|---|---|---|---|---|---|---|---|---|---|---|---|---|---|---|
| 204.1.12 | 248609 | 140 | 0.0104 | -0.1685 | 2 | 3 | 0 | 3 | - | - | - | - | z2w,1=1 z2w,2=1 z2w,3=1 | Cr4Cu3O12Y1 |
| 204.1.13 | 248608 | 140 | 0.0109 | -0.1683 | 2 | 3 | 0 | 3 | - | - | - | - | z2w,1=1 z2w,2=1 z2w,3=1 | Cr4Cu3O12Y1 |
| 204.1.14 | 248607 | 140 | 0.0050 | -0.1572 | 2 | 2 | 71 | 2 | - | - | - | - | z4=2 | Cr4Cu3La1O12 |
| 204.1.15 | 248606 | 140 | 0.0053 | -0.1577 | 2 | 2 | 71 | 2 | - | - | - | - | z4=2 | Cr4Cu3La1O12 |
| 204.1.16 | 248605 | 140 | 0.0055 | -0.1570 | 2 | 2 | 71 | 2 | - | - | - | - | z4=2 | Cr4Cu3La1O12 |
| 204.1.17 | 248604 | 140 | 0.0056 | -0.1578 | 2 | 2 | 71 | 2 | - | - | - | - | z4=2 | Cr4Cu3La1O12 |
| 204.1.18 | 248603 | 140 | 0.0061 | -0.1582 | 2 | 2 | 71 | 2 | - | - | - | - | z4=2 | Cr4Cu3La1O12 |
| 204.1.19 | 248601 | 140 | 0.0069 | -0.1580 | 2 | 2 | 71 | 2 | - | - | - | - | z4=2 | Cr4Cu3La1O12 |
| 204.1.20 | 248600 | 140 | 0.0070 | -0.1575 | 2 | 2 | 71 | 2 | - | - | - | - | z4=2 | Cr4Cu3La1O12 |
| 204.1.21 | 248599 | 140 | 0.0069 | -0.1562 | 2 | 2 | 71 | 2 | - | - | - | - | z4=2 | Cr4Cu3La1O12 |
| 204.1.22 | 248598 | 140 | 0.0071 | -0.1575 | 2 | 2 | 71 | 2 | - | - | - | - | z4=2 | Cr4Cu3La1O12 |
| 204.1.23 | 174551 | 98 | 0.0393 | -0.6960 | 2 | 1 | 2 | 1 | - | - | - | - | z4=1 | Ba1Ge12Pt4 |
| 204.1.24 | 160797 | 98 | 0.0368 | -0.6863 | 2 | 1 | 2 | 1 | - | - | - | - | z4=1 | Ba1Ge12Pt4 |
| 204.4.1 | 77721 | 100 | - | - | - | - | - | - | See details in Table XIX | | | | | O3Re1 |
| 204.4.2 | 77678 | 100 | - | - | - | - | - | - | See details in Table XIX | | | | | O3Re1 |
| 204.4.3 | 5 - 3 | 100 | - | - | - | - | - | - | See details in Table XIX | | | | | O3Re1 |
| 204.4.4 | 5 - 2 | 100 | - | - | - | - | - | - | See details in Table XIX | | | | | O3Re1 |
| 204.4.5 | 5 - 1 | 100 | - | - | - | - | - | - | See details in Table XIX | | | | | O3Re1 |
| 204.4.6 | 55469 | 100 | - | - | - | - | - | - | See details in Table XIX | | | | | O3Re1 |
| 204.4.7 | 55468 | 100 | - | - | - | - | - | - | See details in Table XIX | | | | | O3Re1 |
| 204.4.8 | 55467 | 100 | - | - | - | - | - | - | See details in Table XIX | | | | | O3Re1 |
| 204.4.9 | 55466 | 100 | - | - | - | - | - | - | See details in Table XIX | | | | | O3Re1 |
| 204.4.10 | 55465 | 100 | - | - | - | - | - | - | See details in Table XIX | | | | | O3Re1 |
| 204.4.11 | 55464 | 100 | - | - | - | - | - | - | See details in Table XIX | | | | | O3Re1 |
| 204.4.12 | 55463 | 100 | - | - | - | - | - | - | See details in Table XIX | | | | | O3Re1 |
| 204.4.13 | 55462 | 100 | - | - | - | - | - | - | See details in Table XIX | | | | | O3Re1 |
| 204.4.14 | 54614 | 196 | - | - | - | - | - | - | See details in Table XIX | | | | | Ni6Pb1Y12 |
| 204.4.15 | 79929 | 106 | - | - | - | - | - | - | See details in Table XIX | | | | | Nd1Os4Sb12 |
| 204.4.16 | 183088 | 106 | - | - | - | - | - | - | See details in Table XIX | | | | | Nd1Os4Sb12 |
| 204.4.17 | 19022 | 122 | - | - | - | - | - | - | See details in Table XIX | | | | | Mn7Na1O12 |
| 204.4.18 | 189912 | 102 | - | - | - | - | - | - | See details in Table XIX | | | | | Ge12Pt4U1 |
| 204.4.19 | 189911 | 100 | - | - | - | - | - | - | See details in Table XIX | | | | | Ge12Pt4Th1 |
| 204.4.20 | 174555 | 102 | - | - | - | - | - | - | See details in Table XIX | | | | | Ge12Nd1Pt4 |
| 204.4.21 | 79926 | 106 | - | - | - | - | - | - | See details in Table XIX | | | | | Fe4P12U1 |
| 204.4.22 | 79927 | 106 | - | - | - | - | - | - | See details in Table XIX | | | | | Fe4Nd1Sb12 |
| 204.4.23 | 93364 | 106 | - | - | - | - | - | - | See details in Table XIX | | | | | Fe4Nd1P12 |
| 204.4.24 | 95716 | 138 | - | - | - | - | - | - | See details in Table XIX | | | | | Cu3Na1O12Ru4 |
| 204.4.25 | 38419 | 144 | - | - | - | - | - | - | See details in Table XIX | | | | | Cu3Mn4O12Y1 |
| 204.4.26 | 38418 | 144 | - | - | - | - | - | - | See details in Table XIX | | | | | Cu3Mn4O12Y1 |
| 204.4.27 | 250793 | 132 | - | - | - | - | - | - | See details in Table XIX | | | | | Cu3Mn1O12V4 |
| 204.4.28 | 250792 | 132 | - | - | - | - | - | - | See details in Table XIX | | | | | Cu3Mn1O12V4 |
| 204.4.29 | 51897 | 148 | - | - | - | - | - | - | See details in Table XIX | | | | | Cu3La1O12Ru4 |
| 204.4.30 | 251657 | 152 | - | - | - | - | - | - | See details in Table XIX | | | | | Cu3Ir4La1O12 |
| 204.4.31 | 237418 | 156 | - | - | - | - | - | - | See details in Table XIX | | | | | Cu3Fe4O12Tb1 |
| 204.4.32 | 237424 | 162 | - | - | - | - | - | - | See details in Table XIX | | | | | Cu3Fe4Lu1O12 |
| 204.4.33 | 290766 | 148 | - | - | - | - | - | - | See details in Table XIX | | | | | Cu3Fe4La1O12 |
| 204.4.34 | 290765 | 148 | - | - | - | - | - | - | See details in Table XIX | | | | | Cu3Fe4La1O12 |
| 204.4.35 | 290764 | 148 | - | - | - | - | - | - | See details in Table XIX | | | | | Cu3Fe4La1O12 |
| 204.4.36 | 290763 | 148 | - | - | - | - | - | - | See details in Table XIX | | | | | Cu3Fe4La1O12 |
| 204.4.37 | 290762 | 148 | - | - | - | - | - | - | See details in Table XIX | | | | | Cu3Fe4La1O12 |
| 204.4.38 | 290761 | 148 | - | - | - | - | - | - | See details in Table XIX | | | | | Cu3Fe4La1O12 |
| 204.4.39 | 290760 | 148 | - | - | - | - | - | - | See details in Table XIX | | | | | Cu3Fe4La1O12 |
| 204.4.40 | 290759 | 148 | - | - | - | - | - | - | See details in Table XIX | | | | | Cu3Fe4La1O12 |

Continued on next page

**TABLE XIV– continued**

| TQC | ICSD | NoE | DG | IndG | SubSG (w/ T) | cl | SubSG (w/o T) | cl | Max. Sub. SG | yes or no | Max. Sub. SG | yes or no | SIs (w/ T) | Stoich. Formula |
|---|---|---|---|---|---|---|---|---|---|---|---|---|---|---|
| 204.4.41 | 290758 | 148 | - | - | - | - | - | - | See details in Table XIX | | | | | Cu3Fe4La1O12 |
| 204.4.42 | 290757 | 148 | - | - | - | - | - | - | See details in Table XIX | | | | | Cu3Fe4La1O12 |
| 204.4.43 | 290756 | 148 | - | - | - | - | - | - | See details in Table XIX | | | | | Cu3Fe4La1O12 |
| 204.4.44 | 237412 | 148 | - | - | - | - | - | - | See details in Table XIX | | | | | Cu3Fe4La1O12 |
| 204.4.45 | 237399 | 148 | - | - | - | - | - | - | See details in Table XIX | | | | | Cu3Fe4La1O12 |
| 204.4.46 | 185584 | 148 | - | - | - | - | - | - | See details in Table XIX | | | | | Cu3Fe4La1O12 |
| 204.4.47 | 185583 | 148 | - | - | - | - | - | - | See details in Table XIX | | | | | Cu3Fe4La1O12 |
| 204.4.48 | 237420 | 158 | - | - | - | - | - | - | See details in Table XIX | | | | | Cu3Fe4Ho1O12 |
| 204.4.49 | 237416 | 154 | - | - | - | - | - | - | See details in Table XIX | | | | | Cu3Eu1Fe4O12 |
| 204.4.50 | 237403 | 154 | - | - | - | - | - | - | See details in Table XIX | | | | | Cu3Eu1Fe4O12 |
| 204.4.51 | 87886 | 100 | - | - | - | - | - | - | See details in Table XIX | | | | | Co4Sb12Sn1 |
| 204.4.52 | 92393 | 96 | - | - | - | - | - | - | See details in Table XIX | | | | | Co1P3 |
| 204.4.53 | 174553 | 100 | - | - | - | - | - | - | See details in Table XIX | | | | | Ce1Ge12Pt4 |
| 204.4.54 | 237495 | 124 | - | - | - | - | - | - | See details in Table XIX | | | | | Ca1O12Pd3V4 |
| 204.4.55 | 9188 | 96 | - | - | - | - | - | - | See details in Table XIX | | | | | As3Co1 |
| 205.2.1 | 77109 | 80 | 0.0037 | -0.6199 | 2 | 1 | 2 | 1 | - | - | - | - | z4=1 | Pd1Sb2 |
| 205.2.2 | 300225 | 80 | 0.0231 | -0.1106 | 2 | 1 | 2 | 1 | - | - | - | - | z4=1 | Os1Te2 |
| 205.2.3 | 86354 | 88 | 0.0132 | -1.1174 | 2 | 1 | 2 | 1 | - | - | - | - | z4=1 | Ni1S2 |
| 205.2.4 | 68169 | 88 | 0.0029 | -1.0577 | 2 | 1 | 2 | 1 | - | - | - | - | z4=1 | Ni1S2 |
| 205.2.5 | 68168 | 88 | 0.0022 | -1.0067 | 2 | 1 | 2 | 1 | - | - | - | - | z4=1 | Ni1S2 |
| 205.2.6 | 68167 | 88 | 0.0043 | -1.0673 | 2 | 1 | 2 | 1 | - | - | - | - | z4=1 | Ni1S2 |
| 205.3.1 | 77111 | 80 | - | - | - | - | - | - | 61 | no(2) | 61 | yes | z4=2 | Bi2Pt1 |
| 205.4.1 | 251565 | 84 | - | - | - | - | - | - | See details in Table XIX | | | | | O2Rh1 |
| 205.4.2 | 36550 | 76 | - | - | - | - | - | - | See details in Table XIX | | | | | Mn1S2 |
| 205.4.3 | 36549 | 76 | - | - | - | - | - | - | See details in Table XIX | | | | | Mn1S2 |
| 205.4.4 | 36548 | 76 | - | - | - | - | - | - | See details in Table XIX | | | | | Mn1S2 |
| 205.4.5 | 36 - | 76 | - | - | - | - | - | - | See details in Table XIX | | | | | Mn1S2 |
| 205.4.6 | 36546 | 76 | - | - | - | - | - | - | See details in Table XIX | | | | | Mn1S2 |
| 205.4.7 | 251567 | 84 | - | - | - | - | - | - | See details in Table XIX | | | | | Ir1O2 |
| 205.4.8 | 100567 | 96 | - | - | - | - | - | - | See details in Table XIX | | | | | F2Pd1 |
| 205.4.9 | 100510 | 92 | - | - | - | - | - | - | See details in Table XIX | | | | | Cu1S2 |
| 205.4.10 | 86351 | 84 | - | - | - | - | - | - | See details in Table XIX | | | | | Co1S2 |
| 212.4.1 | 84932 | 140 | - | - | - | - | - | - | See details in Table XIX | | | | | B1Li2Pt3 |
| 212.4.2 | 156466 | 140 | - | - | - | - | - | - | See details in Table XIX | | | | | B1Li2Pt3 |
| 212.4.3 | 84931 | 140 | - | - | - | - | - | - | See details in Table XIX | | | | | B1Li2Pd3 |
| 212.4.4 | 156465 | 140 | - | - | - | - | - | - | See details in Table XIX | | | | | B1Li2Pd3 |
| 213.4.1 | 86170 | 172 | - | - | - | - | - | - | See details in Table XIX | | | | | N1Ni2W3 |
| 213.4.2 | 150314 | 172 | - | - | - | - | - | - | See details in Table XIX | | | | | Mo3N1Pt2 |
| 213.4.3 | 93502 | 172 | - | - | - | - | - | - | See details in Table XIX | | | | | Mo3N1Pd2 |
| 213.4.4 | 96416 | 172 | - | - | - | - | - | - | See details in Table XIX | | | | | Mo3N1Ni2 |
| 213.4.5 | 50815 | 172 | - | - | - | - | - | - | See details in Table XIX | | | | | Mo3N1Ni2 |
| 213.4.6 | 165261 | 172 | - | - | - | - | - | - | See details in Table XIX | | | | | Mo3N1Ni2 |
| 213.4.7 | 260022 | 88 | - | - | - | - | - | - | See details in Table XIX | | | | | Mg3Ru2 |
| 216.2.1 | 260984 | 64 | 0.0255 | -0.0261 | 82 | 1 | 82 | 1 | - | - | - | - | z2=1 | Ni5U1 |
| 216.2.2 | 54390 | 58 | 0.0548 | -0.1291 | 82 | 1 | 82 | 1 | - | - | - | - | z2=1 | Ni4Sn1U1 |
| 216.2.3 | 54387 | 58 | 0.0108 | -0.1633 | 82 | 1 | 82 | 1 | - | - | - | - | z2=1 | Ni4Sn1U1 |
| 216.2.4 | 418013 | 216 | 0.0000 | -0.1017 | 82 | 1 | 82 | 1 | - | - | - | - | z2=1 | La4Mg1Ru1 |
| 216.2.5 | 417519 | 416 | 0.0000 | -0.0067 | 82 | 1 | 82 | 1 | - | - | - | - | z2=1 | In1Rh1Tm4 |
| 216.2.6 | 81917 | 18 | 0.1118 | 0.0460 | 82 | 1 | 82 | 1 | - | - | - | - | z2=1 | Hg1S1 |
| 216.2.7 | 419068 | 368 | 0.0000 | -0.0049 | 82 | 1 | 82 | 1 | - | - | - | - | z2=1 | Cd1Gd4Ru1 |
| 216.2.8 | 415194 | 64 | 0.0220 | -0.0685 | 82 | 1 | 82 | 1 | - | - | - | - | z2=1 | Cd1Ce1Ni4 |
| 216.3.1 | 189201 | 54 | - | - | - | - | - | - | 160 | yes | 160 | yes | - | Ni5Zr1 |
| | | | | | | | | | 22 | yes | 22 | yes | - | |
| 216.3.2 | 250246 | 86 | - | - | - | - | - | - | 160 | yes | 160 | yes | - | Cr4Cu1In1Se8 |
| | | | | | | | | | 22 | yes | 22 | yes | - | |
| 216.3.3 | 182766 | 232 | - | - | - | - | - | - | 119 | yes | 119 | yes | - | Ce4Mg1Ru1 |
| 216.3.4 | 415196 | 78 | - | - | - | - | - | - | 119 | no(82) | 119 | no(82) | z2=1 | Cd1Cu4Er1 |

Continued on next page

**TABLE XIV– continued**

| TQC | ICSD | NoE | DG | IndG | SubSG (w/ T) | cl | SubSG (w/o T) | cl | Max. Sub. SG | yes or no | Max. Sub. SG | yes or no | SIs (w/ T) | Stoich. Formula |
|---|---|---|---|---|---|---|---|---|---|---|---|---|---|---|
| 216.3.5 | 182767 | 272 | - | - | - | - | - | - | 119 | yes | 119 | yes | - | Cd1Ce4Ru1 |
| 216.3.6 | 54465 | 22 | - | - | - | - | - | - | 160 | yes | 160 | yes | - | Au1Mn1Sn1 |
| | | | | | | | | | 22 | yes | 22 | yes | - | |
| 216.4.1 | 194973 | 56 | - | - | - | - | - | - | See details in Table XIX | | | | | Mg1Nd1Ni4 |
| 216.4.2 | 416868 | 130 | - | - | - | - | - | - | See details in Table XIX | | | | | Ir3Mg3Si8 |
| 216.4.3 | 418746 | 220 | - | - | - | - | - | - | See details in Table XIX | | | | | Ir1La4Mg1 |
| 216.4.4 | 162198 | 8 | - | - | - | - | - | - | See details in Table XIX | | | | | In1Sb1 |
| 216.4.5 | 162197 | 8 | - | - | - | - | - | - | See details in Table XIX | | | | | In1Sb1 |
| 216.4.6 | 162196 | 8 | - | - | - | - | - | - | See details in Table XIX | | | | | In1Sb1 |
| 216.4.7 | 201283 | 104 | - | - | - | - | - | - | See details in Table XIX | | | | | I1Nb1Se1 |
| 216.4.8 | 81213 | 36 | - | - | - | - | - | - | See details in Table XIX | | | | | Ho1Ni1Sb1 |
| 216.4.9 | 290038 | 18 | - | - | - | - | - | - | See details in Table XIX | | | | | Hg1Se1 |
| 216.4.10 | 50502 | 72 | - | - | - | - | - | - | See details in Table XIX | | | | | Ge1S8V4 |
| 216.4.11 | 194977 | 60 | - | - | - | - | - | - | See details in Table XIX | | | | | Cu4Mg1Nd1 |
| 216.4.12 | 412294 | 18 | - | - | - | - | - | - | See details in Table XIX | | | | | Cu1Mg1Sb1 |
| 216.4.13 | 159198 | 54 | - | - | - | - | - | - | See details in Table XIX | | | | | Co4Sn1U1 |
| 216.4.14 | 416990 | 220 | - | - | - | - | - | - | See details in Table XIX | | | | | Co1La4Mg1 |
| 216.4.15 | 419070 | 256 | - | - | - | - | - | - | See details in Table XIX | | | | | Cd1Ru1Y4 |
| 216.4.16 | 419069 | 256 | - | - | - | - | - | - | See details in Table XIX | | | | | Cd1La4Ru1 |
| 216.4.17 | 419067 | 260 | - | - | - | - | - | - | See details in Table XIX | | | | | Cd1Co1La4 |
| 216.4.18 | 51139 | 168 | - | - | - | - | - | - | See details in Table XIX | | | | | Au1S1Ta5 |
| 216.4.19 | 248684 | 18 | - | - | - | - | - | - | See details in Table XIX | | | | | Au1Bi1Ca1 |
| 216.4.20 | 248682 | 18 | - | - | - | - | - | - | See details in Table XIX | | | | | Au1Bi1Ca1 |
| 216.4.21 | 248680 | 18 | - | - | - | - | - | - | See details in Table XIX | | | | | Au1Bi1Ca1 |
| 216.4.22 | 248678 | 18 | - | - | - | - | - | - | See details in Table XIX | | | | | Au1Bi1Ca1 |
| 216.4.23 | 248676 | 18 | - | - | - | - | - | - | See details in Table XIX | | | | | Au1Bi1Ca1 |
| 217.1.1 | 427048 | 128 | 0.0107 | -0.2406 | 82 | 1 | 82 | 1 | - | - | - | - | z2=1 | F6Ni3O6Sb4 |
| 217.1.2 | 427047 | 128 | 0.0103 | -0.2387 | 82 | 1 | 82 | 1 | - | - | - | - | z2=1 | F6Ni3O6Sb4 |
| 217.1.3 | 84572 | 230 | 0.0074 | -0.2368 | 82 | 1 | 82 | 1 | - | - | - | - | z2=1 | Cu12S13Sb4 |
| 217.1.4 | 84571 | 230 | 0.0061 | -0.2569 | 82 | 1 | 82 | 1 | - | - | - | - | z2=1 | Cu12S13Sb4 |
| 217.1.5 | 62116 | 230 | 0.0066 | -0.2440 | 82 | 1 | 82 | 1 | - | - | - | - | z2=1 | Cu12S13Sb4 |
| 217.1.6 | 25707 | 230 | 0.0081 | -0.2247 | 82 | 1 | 82 | 1 | - | - | - | - | z2=1 | Cu12S13Sb4 |
| 217.1.7 | 33588 | 224 | 0.0229 | -0.0397 | 82 | 1 | 82 | 1 | - | - | - | - | z2=1 | As4Cu12S12 |
| 217.4.1 | 77887 | 202 | - | - | - | - | - | - | See details in Table XIX | | | | | Pd16S7 |
| 217.4.2 | 410992 | 82 | - | - | - | - | - | - | See details in Table XIX | | | | | Ir4Sn4Sr3 |
| 217.4.3 | 104605 | 300 | - | - | - | - | - | - | See details in Table XIX | | | | | Ir2Zn11 |
| 217.4.4 | 429725 | 82 | - | - | - | - | - | - | See details in Table XIX | | | | | Ge4Ir4Sr3 |
| 217.4.5 | 429512 | 78 | - | - | - | - | - | - | See details in Table XIX | | | | | Ga4Ir4Sr3 |
| 217.4.6 | 200460 | 182 | - | - | - | - | - | - | See details in Table XIX | | | | | F9U2 |
| 217.4.7 | 240667 | 302 | - | - | - | - | - | - | See details in Table XIX | | | | | Cu5Zn8 |
| 217.4.8 | 2092 | 302 | - | - | - | - | - | - | See details in Table XIX | | | | | Cu5Zn8 |
| 217.4.9 | 194374 | 300 | - | - | - | - | - | - | See details in Table XIX | | | | | Co2Zn11 |
| 217.4.10 | 95788 | 58 | - | - | - | - | - | - | See details in Table XIX | | | | | Ca3Ir4Si4 |
| 217.4.11 | 429727 | 58 | - | - | - | - | - | - | See details in Table XIX | | | | | Ca3Ge4Ir4 |
| 217.4.12 | 195500 | 58 | - | - | - | - | - | - | See details in Table XIX | | | | | Ca3Ge4Ir4 |
| 220.2.1 | 155383 | 136 | 0.0000 | -0.1818 | 82 | 1 | 82 | 1 | - | - | - | - | z2=1 | C3Y2 |
| 220.2.2 | 155382 | 136 | 0.0045 | -0.1921 | 82 | 1 | 82 | 1 | - | - | - | - | z2=1 | C3Y2 |
| 220.2.3 | 26477 | 160 | 0.0022 | -0.0294 | 82 | 1 | 82 | 1 | - | - | - | - | z2=1 | C3U2 |
| 220.2.4 | 2447 | 160 | 0.0102 | -0.0010 | 82 | 1 | 82 | 1 | - | - | - | - | z2=1 | C3Nd2 |
| 220.4.1 | 60537 | 132 | - | - | - | - | - | - | See details in Table XIX | | | | | Se4U3 |
| 220.4.2 | 427458 | 144 | - | - | - | - | - | - | See details in Table XIX | | | | | O6Rb4 |
| 220.4.3 | 33961 | 144 | - | - | - | - | - | - | See details in Table XIX | | | | | O6Rb4 |
| 220.4.4 | 33960 | 144 | - | - | - | - | - | - | See details in Table XIX | | | | | O6Rb4 |
| 220.4.5 | 25718 | 144 | - | - | - | - | - | - | See details in Table XIX | | | | | O6Rb4 |
| 220.4.6 | 56782 | 114 | - | - | - | - | - | - | See details in Table XIX | | | | | La3S4 |
| 220.4.7 | 151632 | 150 | - | - | - | - | - | - | See details in Table XIX | | | | | Eu3S4 |
| 220.4.8 | 151631 | 150 | - | - | - | - | - | - | See details in Table XIX | | | | | Eu3S4 |

Continued on next page

**TABLE XIV– continued**

| TQC | ICSD | NoE | DG | IndG | SubSG (w/ T) | cl | SubSG (w/o T) | cl | Max. Sub. SG | yes or no | Max. Sub. SG | yes or no | SIs (w/ T) | Stoich. Formula |
|---|---|---|---|---|---|---|---|---|---|---|---|---|---|---|
| 220.4.9 | 100522 | 150 | - | - | - | - | - | - | See details in Table XIX | | | | | Eu3S4 |
| 220.4.10 | 93872 | 166 | - | - | - | - | - | - | See details in Table XIX | | | | | Co3Sb4Th3 |
| 220.4.11 | 16511 | 176 | - | - | - | - | - | - | See details in Table XIX | | | | | C3Pu2 |
| 220.4.12 | 2445 | 152 | - | - | - | - | - | - | See details in Table XIX | | | | | C3Pr2 |
| 220.4.13 | 96944 | 110 | - | - | - | - | - | - | See details in Table XIX | | | | | Bi3Sr4 |
| 220.4.14 | 107394 | 166 | - | - | - | - | - | - | See details in Table XIX | | | | | Bi3Eu4 |
| 220.4.15 | 96943 | 110 | - | - | - | - | - | - | See details in Table XIX | | | | | Ba4Bi3 |
| 220.4.16 | 10440 | 166 | - | - | - | - | - | - | See details in Table XIX | | | | | As3Eu4 |
| 221.1.1 | 201036 | 40 | 0.0522 | 0.0356 | 2 | 3 | 47,65,83 | 3 | - | - | - | - | z4=2<br>z8=2 | O1Sn1Sr3 |
| 221.1.2 | 181127 | 34 | 0.0697 | -0.1724 | 2 | 7 | 10 | 7 | - | - | - | - | z2w,1=1<br>z2w,2=1<br>z2w,3=1<br>z4=2<br>z4mpi=3<br>z8=6 | Ni3Sn1 |
| 221.1.3 | 163195 | 12 | 0.0625 | -3.4123 | 2,81 | 6 | 2,81 | 6 | - | - | - | - | z4=3<br>z2=1<br>z8=7 | Mg1Pd1 |
| 221.1.4 | 247065 | 38 | 0.0707 | -0.4238 | 2 | 5 | 10 | 5 | - | - | - | - | z2w,1=1<br>z2w,2=1<br>z2w,3=1<br>z4mpi=3 | In1N1Ni3 |
| 221.1.5 | 247273 | 34 | 0.1487 | -0.9244 | 2 | 5 | 10 | 5 | - | - | - | - | z2w,1=1<br>z2w,2=1<br>z2w,3=1<br>z4mpi=3 | H1Pd3Tl1 |
| 221.1.6 | 100533 | 12 | 0.0088 | -0.9107 | 2,81 | 4 | 2,81 | 4 | - | - | - | - | z2w,1=1<br>z2w,2=1<br>z2w,3=1<br>z4=1<br>z4mpi=1<br>z2=1<br>z8=1 | Fe1Ti1 |
| 221.1.7 | 262924 | 34 | 0.0621 | 0.0621 | 2,81 | 2 | 2,81 | 2 | - | - | - | - | z2w,1=1<br>z2w,2=1<br>z2w,3=1<br>z4=3<br>z4mpi=3<br>z2=1<br>z8=3 | Cs1I3Sn1 |
| 221.1.8 | 100787 | 16 | 0.0181 | 0.0181 | 2 | 3 | 47,65,83 | 3 | - | - | - | - | z4=2<br>z8=2 | Ca3O1Sn1 |
| 221.1.9 | 100789 | 16 | 0.0445 | 0.0445 | 2 | 3 | 47,65,83 | 3 | - | - | - | - | z4=2<br>z8=2 | Ca3O1Pb1 |
| 221.1.10 | 402087 | 16 | 0.0031 | 0.0031 | 2 | 3 | 47,65,83 | 3 | - | - | - | - | z4=2<br>z8=2 | Ca3Ge1O1 |
| 221.1.11 | 80320 | 34 | 0.0816 | 0.0782 | 2,81 | 2 | 2,81 | 2 | - | - | - | - | z2w,1=1<br>z2w,2=1<br>z2w,3=1<br>z4=3<br>z4mpi=3<br>z2=1<br>z8=3 | Br3Cs1Ge1 |

Continued on next page

**TABLE XIV– continued**

| TQC | ICSD | NoE | DG | IndG | SubSG (w/ T) | cl | SubSG (w/o T) | cl | Max. Sub. SG | yes or no | Max. Sub. SG | yes or no | SIs (w/ T) | Stoich. Formula |
|---|---|---|---|---|---|---|---|---|---|---|---|---|---|---|
| 221.1.12 | 409846 | 60 | 0.0000 | -0.1051 | 2,81 | 2 | 2,81 | 2 | - | - | - | - | z2w,1=1<br>z2w,2=1<br>z2w,3=1<br>z4=3<br>z4mpi=3<br>z2=1<br>z8=3 | Br3Ce3Ga1 |
| 221.1.13 | 50313 | 28 | 0.0050 | -0.0275 | 2,81 | 1 | 2,81 | 1 | - | - | - | - | z2w,1=1<br>z2w,2=1<br>z2w,3=1<br>z4=1<br>z4mpi=3<br>z2=1<br>z8=5 | B6Sr1 |
| 221.1.14 | 51495 | 28 | 0.0299 | -0.0343 | 2,81 | 1 | 2,81 | 1 | - | - | - | - | z2w,1=1<br>z2w,2=1<br>z2w,3=1<br>z4=1<br>z4mpi=3<br>z2=1<br>z8=5 | B6Ba1 |
| 221.3.1 | 85207 | 12 | - | - | - | - | - | - | 215 | no(81) | 215 | no(81) | z2=1 | Ru1Si1 |
| | | | | | | | | | 200 | no(2) | 200 | no(2) | z4=2 | |
| | | | | | | | | | 166 | no(2) | 166 | no(2) | z4=2 | |
| | | | | | | | | | 115 | no(81) | 115 | no(81) | z2=1 | |
| | | | | | | | | | 65 | no(2) | 65 | no(2) | z4=2 | |
| 221.3.2 | 79230 | 12 | - | - | - | - | - | - | 215 | no(81) | 215 | no(81) | z2=1 | Ru1Si1 |
| | | | | | | | | | 200 | no(2) | 200 | no(2) | z4=2 | |
| | | | | | | | | | 166 | no(2) | 166 | no(2) | z4=2 | |
| | | | | | | | | | 115 | no(81) | 115 | no(81) | z2=1 | |
| | | | | | | | | | 65 | no(2) | 65 | no(2) | z4=2 | |
| 221.3.3 | 413388 | 82 | - | - | - | - | - | - | 215 | yes | 215 | yes | - | O1Sn1Yb3 |
| | | | | | | | | | 200 | no(2) | 200 | no(2) | z4=1 | |
| | | | | | | | | | 166 | no(2) | 166 | no(2) | z4=1 | |
| | | | | | | | | | 115 | yes | 115 | yes | - | |
| | | | | | | | | | 65 | no(2) | 65 | no(2) | z4=1 | |
| 221.3.4 | 96119 | 26 | - | - | - | - | - | - | 215 | no(81) | 215 | no(81) | z2=1 | Mn3Sb1 |
| | | | | | | | | | 200 | yes | 200 | yes | - | |
| | | | | | | | | | 166 | yes | 166 | yes | - | |
| | | | | | | | | | 115 | no(81) | 115 | no(81) | z2=1 | |
| | | | | | | | | | 65 | yes | 65 | yes | - | |
| 221.3.5 | 194422 | 32 | - | - | - | - | - | - | 215 | yes | 215 | yes | - | Ga3Tm1 |
| | | | | | | | | | 200 | no(2) | 200 | no(2) | z2w,1=1<br>z2w,2=1<br>z2w,3=1<br>z4=1 | |
| | | | | | | | | | 166 | no(2) | 166 | no(2) | z2w,1=1<br>z2w,2=1<br>z2w,3=1<br>z4=1 | |
| | | | | | | | | | 115 | yes | 115 | yes | - | |
| | | | | | | | | | 65 | no(2) | 65 | no(2) | z2w,1=1<br>z2w,2=1<br>z2w,3=1<br>z4=1 | |
| 221.3.6 | 79241 | 40 | - | - | - | - | - | - | 215 | yes | 215 | yes | - | C1Ru3Th1 |

Continued on next page

**TABLE XIV– continued**

| TQC | ICSD | NoE | DG | IndG | SubSG cl (w/ T) | SubSG cl (w/o T) | Max. Sub. SG | yes or no | Max. Sub. SG | yes or no | SIs (w/ T) | Stoich. Formula |
|---|---|---|---|---|---|---|---|---|---|---|---|---|
| | | | | | | | 200 | no(2) | 200 | no(2) | z2w,1=1<br>z2w,2=1<br>z2w,3=1<br>z4=1 | |
| | | | | | | | 166 | no(2) | 166 | no(2) | z2w,1=1<br>z2w,2=1<br>z2w,3=1<br>z4=1 | |
| | | | | | | | 115 | yes | 115 | yes | - | |
| | | | | | | | 65 | no(2) | 65 | no(2) | z2w,1=1<br>z2w,2=1<br>z2w,3=1<br>z4=1 | |
| 221.3.7 | 81546 | 30 | - | - | - - | - - | 215 | yes | 215 | yes | - | B6Th1 |
| | | | | | | | 200 | no(2) | 200 | no(2) | z2w,1=1<br>z2w,2=1<br>z2w,3=1<br>z4=1 | |
| | | | | | | | 166 | no(2) | 166 | no(2) | z2w,1=1<br>z2w,2=1<br>z2w,3=1<br>z4=1 | |
| | | | | | | | 115 | yes | 115 | yes | - | |
| | | | | | | | 65 | no(2) | 65 | no(2) | z2w,1=1<br>z2w,2=1<br>z2w,3=1<br>z4=1 | |
| 221.3.8 | 95053 | 30 | - | - | - - | - - | 215 | yes | 215 | yes | - | B6Ce1 |
| | | | | | | | 200 | no(2) | 200 | no(2) | z2w,1=1<br>z2w,2=1<br>z2w,3=1<br>z4=1 | |
| | | | | | | | 166 | no(2) | 166 | no(2) | z2w,1=1<br>z2w,2=1<br>z2w,3=1<br>z4=1 | |
| | | | | | | | 115 | yes | 115 | yes | - | |
| | | | | | | | 65 | no(2) | 65 | no(2) | z2w,1=1<br>z2w,2=1<br>z2w,3=1<br>z4=1 | |
| 221.3.9 | 84561 | 30 | - | - | - - | - - | 215 | yes | 215 | yes | - | B6Ce1 |
| | | | | | | | 200 | no(2) | 200 | no(2) | z2w,1=1<br>z2w,2=1<br>z2w,3=1<br>z4=3 | |
| | | | | | | | 166 | no(2) | 166 | no(2) | z2w,1=1<br>z2w,2=1<br>z2w,3=1<br>z4=3 | |
| | | | | | | | 115 | yes | 115 | yes | - | |
| | | | | | | | 65 | no(2) | 65 | no(2) | z2w,1=1<br>z2w,2=1<br>z2w,3=1<br>z4=3 | |
| 221.3.10 | 67404 | 30 | - | - | - - | - - | 215 | yes | 215 | yes | - | B6Ce1 |

Continued on next page

**TABLE XIV– continued**

| TQC | ICSD | NoE | DG | IndG | SubSG cl (w/ T) | SubSG cl (w/o T) | Max. Sub. SG | yes or no | Max. Sub. SG | yes or no | SIs (w/ T) | Stoich. Formula |
|---|---|---|---|---|---|---|---|---|---|---|---|---|
| | | | | | | | 200 | no(2) | 200 | no(2) | z2w,1=1 z2w,2=1 z2w,3=1 z4=1 | |
| | | | | | | | 166 | no(2) | 166 | no(2) | z2w,1=1 z2w,2=1 z2w,3=1 z4=1 | |
| | | | | | | | 115 | yes | 115 | yes | - | |
| | | | | | | | 65 | no(2) | 65 | no(2) | z2w,1=1 z2w,2=1 z2w,3=1 z4=1 | |
| 221.3.11 | 241004 | 30 | - | - | - - | - - | 215 | yes | 215 | yes | - | B6Ce1 |
| | | | | | | | 200 | no(2) | 200 | no(2) | z2w,1=1 z2w,2=1 z2w,3=1 z4=3 | |
| | | | | | | | 166 | no(2) | 166 | no(2) | z2w,1=1 z2w,2=1 z2w,3=1 z4=3 | |
| | | | | | | | 115 | yes | 115 | yes | - | |
| | | | | | | | 65 | no(2) | 65 | no(2) | z2w,1=1 z2w,2=1 z2w,3=1 z4=3 | |
| 221.3.12 | 194318 | 30 | - | - | - - | - - | 215 | yes | 215 | yes | - | B6Ce1 |
| | | | | | | | 200 | no(2) | 200 | no(2) | z2w,1=1 z2w,2=1 z2w,3=1 z4=1 | |
| | | | | | | | 166 | no(2) | 166 | no(2) | z2w,1=1 z2w,2=1 z2w,3=1 z4=1 | |
| | | | | | | | 115 | yes | 115 | yes | - | |
| | | | | | | | 65 | no(2) | 65 | no(2) | z2w,1=1 z2w,2=1 z2w,3=1 z4=1 | |
| 221.3.13 | 194317 | 30 | - | - | - - | - - | 215 | yes | 215 | yes | - | B6Ce1 |
| | | | | | | | 200 | no(2) | 200 | no(2) | z2w,1=1 z2w,2=1 z2w,3=1 z4=3 | |
| | | | | | | | 166 | no(2) | 166 | no(2) | z2w,1=1 z2w,2=1 z2w,3=1 z4=3 | |
| | | | | | | | 115 | yes | 115 | yes | - | |
| | | | | | | | 65 | no(2) | 65 | no(2) | z2w,1=1 z2w,2=1 z2w,3=1 z4=3 | |
| 221.3.14 | 194316 | 30 | - | - | - - | - - | 215 | yes | 215 | yes | - | B6Ce1 |

Continued on next page

**TABLE XIV– continued**

| TQC | ICSD | NoE | DG | IndG | SubSG cl (w/ T) | SubSG cl (w/o T) | Max. Sub. SG | yes or no | Max. Sub. SG | yes or no | SIs (w/ T) | Stoich. Formula |
|---|---|---|---|---|---|---|---|---|---|---|---|---|
| | | | | | | | 200 | no(2) | 200 | no(2) | z2w,1=1<br>z2w,2=1<br>z2w,3=1<br>z4=3 | |
| | | | | | | | 166 | no(2) | 166 | no(2) | z2w,1=1<br>z2w,2=1<br>z2w,3=1<br>z4=3 | |
| | | | | | | | 115 | yes | 115 | yes | - | |
| | | | | | | | 65 | no(2) | 65 | no(2) | z2w,1=1<br>z2w,2=1<br>z2w,3=1<br>z4=3 | |
| 221.3.15 | 194315 | 30 | - | - | - - | - - | 215 | yes | 215 | yes | - | B6Ce1 |
| | | | | | | | 200 | no(2) | 200 | no(2) | z2w,1=1<br>z2w,2=1<br>z2w,3=1<br>z4=3 | |
| | | | | | | | 166 | no(2) | 166 | no(2) | z2w,1=1<br>z2w,2=1<br>z2w,3=1<br>z4=3 | |
| | | | | | | | 115 | yes | 115 | yes | - | |
| | | | | | | | 65 | no(2) | 65 | no(2) | z2w,1=1<br>z2w,2=1<br>z2w,3=1<br>z4=3 | |
| 221.3.16 | 194314 | 30 | - | - | - - | - - | 215 | yes | 215 | yes | - | B6Ce1 |
| | | | | | | | 200 | no(2) | 200 | no(2) | z2w,1=1<br>z2w,2=1<br>z2w,3=1<br>z4=1 | |
| | | | | | | | 166 | no(2) | 166 | no(2) | z2w,1=1<br>z2w,2=1<br>z2w,3=1<br>z4=1 | |
| | | | | | | | 115 | yes | 115 | yes | - | |
| | | | | | | | 65 | no(2) | 65 | no(2) | z2w,1=1<br>z2w,2=1<br>z2w,3=1<br>z4=1 | |
| 221.3.17 | 194313 | 30 | - | - | - - | - - | 215 | yes | 215 | yes | - | B6Ce1 |
| | | | | | | | 200 | no(2) | 200 | no(2) | z2w,1=1<br>z2w,2=1<br>z2w,3=1<br>z4=3 | |
| | | | | | | | 166 | no(2) | 166 | no(2) | z2w,1=1<br>z2w,2=1<br>z2w,3=1<br>z4=3 | |
| | | | | | | | 115 | yes | 115 | yes | - | |
| | | | | | | | 65 | no(2) | 65 | no(2) | z2w,1=1<br>z2w,2=1<br>z2w,3=1<br>z4=3 | |
| 221.3.18 | 194312 | 30 | - | - | - - | - - | 215 | yes | 215 | yes | - | B6Ce1 |

Continued on next page

**TABLE XIV– continued**

| TQC | ICSD | NoE | DG | IndG | SubSG cl (w/ T) | SubSG cl (w/o T) | Max. Sub. SG | yes or no | Max. Sub. SG | yes or no | SIs (w/ T) | Stoich. Formula |
|---|---|---|---|---|---|---|---|---|---|---|---|---|
| | | | | | | | 200 | no(2) | 200 | no(2) | z2w,1=1 z2w,2=1 z2w,3=1 z4=3 | |
| | | | | | | | 166 | no(2) | 166 | no(2) | z2w,1=1 z2w,2=1 z2w,3=1 z4=3 | |
| | | | | | | | 115 | yes | 115 | yes | - | |
| | | | | | | | 65 | no(2) | 65 | no(2) | z2w,1=1 z2w,2=1 z2w,3=1 z4=3 | |
| 221.3.19 | 194311 | 30 | - | - | - - | - - | 215 | yes | 215 | yes | - | B6Ce1 |
| | | | | | | | 200 | no(2) | 200 | no(2) | z2w,1=1 z2w,2=1 z2w,3=1 z4=3 | |
| | | | | | | | 166 | no(2) | 166 | no(2) | z2w,1=1 z2w,2=1 z2w,3=1 z4=3 | |
| | | | | | | | 115 | yes | 115 | yes | - | |
| | | | | | | | 65 | no(2) | 65 | no(2) | z2w,1=1 z2w,2=1 z2w,3=1 z4=3 | |
| 221.3.20 | 173598 | 30 | - | - | - - | - - | 215 | yes | 215 | yes | - | B6Ce1 |
| | | | | | | | 200 | no(2) | 200 | no(2) | z2w,1=1 z2w,2=1 z2w,3=1 z4=3 | |
| | | | | | | | 166 | no(2) | 166 | no(2) | z2w,1=1 z2w,2=1 z2w,3=1 z4=3 | |
| | | | | | | | 115 | yes | 115 | yes | - | |
| | | | | | | | 65 | no(2) | 65 | no(2) | z2w,1=1 z2w,2=1 z2w,3=1 z4=3 | |
| 221.3.21 | 173597 | 30 | - | - | - - | - - | 215 | yes | 215 | yes | - | B6Ce1 |
| | | | | | | | 200 | no(2) | 200 | no(2) | z2w,1=1 z2w,2=1 z2w,3=1 z4=3 | |
| | | | | | | | 166 | no(2) | 166 | no(2) | z2w,1=1 z2w,2=1 z2w,3=1 z4=3 | |
| | | | | | | | 115 | yes | 115 | yes | - | |
| | | | | | | | 65 | no(2) | 65 | no(2) | z2w,1=1 z2w,2=1 z2w,3=1 z4=3 | |
| 221.4.1 | 86941 | 36 | - | - | - - | - - | See details in Table XIX | | | | | O3Ru1Sr1 |
| 221.4.2 | 82984 | 36 | - | - | - - | - - | See details in Table XIX | | | | | O3Ru1Sr1 |
| 221.4.3 | 82983 | 36 | - | - | - - | - - | See details in Table XIX | | | | | O3Ru1Sr1 |

Continued on next page

**TABLE XIV– continued**

| TQC | ICSD | NoE | DG | IndG | SubSG (w/ T) | cl | SubSG (w/o T) | cl | Max. Sub. SG | yes or no | Max. Sub. SG | yes or no | SIs (w/ T) | Stoich. Formula |
|---|---|---|---|---|---|---|---|---|---|---|---|---|---|---|
| 221.4.4 | 82982 | 36 | - | - | - | - | - | - | See details in Table XIX | | | | | O3Ru1Sr1 |
| 221.4.5 | 82981 | 36 | - | - | - | - | - | - | See details in Table XIX | | | | | O3Ru1Sr1 |
| 221.4.6 | 69360 | 36 | - | - | - | - | - | - | See details in Table XIX | | | | | O3Ru1Sr1 |
| 221.4.7 | 162777 | 36 | - | - | - | - | - | - | See details in Table XIX | | | | | O3Ru1Sr1 |
| 221.4.8 | 413389 | 82 | - | - | - | - | - | - | See details in Table XIX | | | | | O1Pb1Yb3 |
| 221.4.9 | 421118 | 8 | - | - | - | - | - | - | See details in Table XIX | | | | | N1Na3 |
| 221.4.10 | 421117 | 8 | - | - | - | - | - | - | See details in Table XIX | | | | | N1Na3 |
| 221.4.11 | 421116 | 8 | - | - | - | - | - | - | See details in Table XIX | | | | | N1Na3 |
| 221.4.12 | 421115 | 8 | - | - | - | - | - | - | See details in Table XIX | | | | | N1Na3 |
| 221.4.13 | 421113 | 8 | - | - | - | - | - | - | See details in Table XIX | | | | | N1Na3 |
| 221.4.14 | 421112 | 8 | - | - | - | - | - | - | See details in Table XIX | | | | | N1Na3 |
| 221.4.15 | 421111 | 8 | - | - | - | - | - | - | See details in Table XIX | | | | | N1Na3 |
| 221.4.16 | 421110 | 8 | - | - | - | - | - | - | See details in Table XIX | | | | | N1Na3 |
| 221.4.17 | 421109 | 8 | - | - | - | - | - | - | See details in Table XIX | | | | | N1Na3 |
| 221.4.18 | 412313 | 8 | - | - | - | - | - | - | See details in Table XIX | | | | | N1Na3 |
| 221.4.19 | 249566 | 8 | - | - | - | - | - | - | See details in Table XIX | | | | | N1Na3 |
| 221.4.20 | 71994 | 34 | - | - | - | - | - | - | See details in Table XIX | | | | | Mo1O3Sr1 |
| 221.4.21 | 174208 | 34 | - | - | - | - | - | - | See details in Table XIX | | | | | Mo1O3Sr1 |
| 221.4.22 | 174207 | 34 | - | - | - | - | - | - | See details in Table XIX | | | | | Mo1O3Sr1 |
| 221.4.23 | 237710 | 38 | - | - | - | - | - | - | See details in Table XIX | | | | | Mn3N1Zn1 |
| 221.4.24 | 240907 | 48 | - | - | - | - | - | - | See details in Table XIX | | | | | In1O1Pr3 |
| 221.4.25 | 106748 | 120 | - | - | - | - | - | - | See details in Table XIX | | | | | Ga3Ru1Tm1 |
| 221.4.26 | 92335 | 36 | - | - | - | - | - | - | See details in Table XIX | | | | | Fe1O3Sr1 |
| 221.4.27 | 91062 | 36 | - | - | - | - | - | - | See details in Table XIX | | | | | Fe1O3Sr1 |
| 221.4.28 | 73166 | 40 | - | - | - | - | - | - | See details in Table XIX | | | | | F3Pd1Rb1 |
| 221.4.29 | 73167 | 40 | - | - | - | - | - | - | See details in Table XIX | | | | | F3K1Pd1 |
| 221.4.30 | 37255 | 40 | - | - | - | - | - | - | See details in Table XIX | | | | | F3K1Ni1 |
| 221.4.31 | 189369 | 40 | - | - | - | - | - | - | See details in Table XIX | | | | | F3K1Ni1 |
| 221.4.32 | 415804 | 60 | - | - | - | - | - | - | See details in Table XIX | | | | | Eu3In1O1 |
| 221.4.33 | 187860 | 22 | - | - | - | - | - | - | See details in Table XIX | | | | | Cu1Y1 |
| 221.4.34 | 245834 | 34 | - | - | - | - | - | - | See details in Table XIX | | | | | Cr1O3Sr1 |
| 221.4.35 | 194155 | 28 | - | - | - | - | - | - | See details in Table XIX | | | | | Cr1O3Pb1 |
| 221.4.36 | 160196 | 28 | - | - | - | - | - | - | See details in Table XIX | | | | | Cr1O3Pb1 |
| 221.4.37 | 422858 | 44 | - | - | - | - | - | - | See details in Table XIX | | | | | Cd1Co3N1 |
| 221.4.38 | 411448 | 14 | - | - | - | - | - | - | See details in Table XIX | | | | | Ca3N1Tl1 |
| 221.4.39 | 50164 | 16 | - | - | - | - | - | - | See details in Table XIX | | | | | C1Sc3Tl1 |
| 221.4.40 | 50163 | 16 | - | - | - | - | - | - | See details in Table XIX | | | | | C1In1Sc3 |
| 221.4.41 | 80956 | 40 | - | - | - | - | - | - | See details in Table XIX | | | | | C1In1La3 |
| 221.4.42 | 72269 | 32 | - | - | - | - | - | - | See details in Table XIX | | | | | Ba1O3Pb1 |
| 221.4.43 | 51659 | 32 | - | - | - | - | - | - | See details in Table XIX | | | | | Ba1O3Pb1 |
| 221.4.44 | 245607 | 32 | - | - | - | - | - | - | See details in Table XIX | | | | | Ba1O3Pb1 |
| 221.4.45 | 245606 | 32 | - | - | - | - | - | - | See details in Table XIX | | | | | Ba1O3Pb1 |
| 221.4.46 | 43799 | 34 | - | - | - | - | - | - | See details in Table XIX | | | | | Ba1Mo1O3 |
| 221.4.47 | 262132 | 36 | - | - | - | - | - | - | See details in Table XIX | | | | | Ba1Fe1O3 |
| 221.4.48 | 262131 | 36 | - | - | - | - | - | - | See details in Table XIX | | | | | Ba1Fe1O3 |
| 221.4.49 | 72942 | 22 | - | - | - | - | - | - | See details in Table XIX | | | | | Au1Ca3N1 |
| 223.3.1 | 84666 | 136 | - | - | - | - | - | - | 115 | no(81) | 115 | no(81) | z2=1 | Nd1Pd3S4 |
| | | | | | | | | | 112 | no(81) | 112 | no(81) | z2=1 | |
| | | | | | | | | | 66 | yes | 66 | yes | - | |
| | | | | | | | | | 47 | yes | 47 | yes | - | |
| 223.3.2 | 186819 | 112 | - | - | - | - | - | - | 218 | no(81) | 218 | no(81) | z2=1 | Ca1O4Pd3 |
| | | | | | | | | | 200 | yes | 200 | yes | - | |
| | | | | | | | | | 167 | yes | 167 | yes | - | |
| | | | | | | | | | 115 | no(81) | 115 | no(81) | z2=1 | |
| | | | | | | | | | 66 | yes | 66 | yes | - | |
| 223.4.1 | 87329 | 38 | - | - | - | - | - | - | See details in Table XIX | | | | | Si1V3 |
| 223.4.2 | 87328 | 38 | - | - | - | - | - | - | See details in Table XIX | | | | | Si1V3 |

Continued on next page

**TABLE XIV– continued**

| TQC | ICSD | NoE | DG | IndG | SubSG (w/ T) | cl | SubSG (w/o T) | cl | Max. Sub. SG | yes or no | Max. Sub. SG | yes or no | SIs (w/ T) | Stoich. Formula |
|---|---|---|---|---|---|---|---|---|---|---|---|---|---|---|
| 223.4.3 | 32546 | 110 | - | - | - | - | - | - | See details in Table XIX | | | | | Na1O4Pt3 |
| 223.4.4 | 245608 | 110 | - | - | - | - | - | - | See details in Table XIX | | | | | Na1O4Pd3 |
| 223.4.5 | 35755 | 44 | - | - | - | - | - | - | See details in Table XIX | | | | | Mo3Si1 |
| 223.4.6 | 84665 | 130 | - | - | - | - | - | - | See details in Table XIX | | | | | La1Pd3S4 |
| 223.4.7 | 61031 | 130 | - | - | - | - | - | - | See details in Table XIX | | | | | La1Pd3S4 |
| 223.4.8 | 50296 | 42 | - | - | - | - | - | - | See details in Table XIX | | | | | Ir1Ti3 |
| 223.4.9 | 84667 | 142 | - | - | - | - | - | - | See details in Table XIX | | | | | Eu1Pd3S4 |
| 223.4.10 | 53218 | 44 | - | - | - | - | - | - | See details in Table XIX | | | | | Cr3Si1 |
| 223.4.11 | 32509 | 44 | - | - | - | - | - | - | See details in Table XIX | | | | | Cr3Si1 |
| 224.3.1 | 281041 | 56 | - | - | - | - | - | - | 215 | no(81) | 215 | no(81) | z2=1 | Ag2O1 |
| | | | | | | | | | 201 | yes | 201 | yes | - | |
| | | | | | | | | | 166 | yes | 166 | yes | - | |
| | | | | | | | | | 118 | no(81) | 118 | no(81) | z2=1 | |
| | | | | | | | | | 67 | yes | 67 | yes | - | |
| 224.3.2 | 247821 | 56 | - | - | - | - | - | - | 215 | no(81) | 215 | no(81) | z2=1 | Ag2O1 |
| | | | | | | | | | 201 | yes | 201 | yes | - | |
| | | | | | | | | | 166 | yes | 166 | yes | - | |
| | | | | | | | | | 118 | no(81) | 118 | no(81) | z2=1 | |
| | | | | | | | | | 67 | yes | 67 | yes | - | |
| 225.1.1 | 63098 | 10 | 0.0045 | 0.0045 | 87 | 1 | 139 | 1 | - | - | - | - | z8=4 | Pb1Te1 |
| 225.1.2 | 153711 | 10 | 0.0199 | 0.0199 | 87 | 1 | 139 | 1 | - | - | - | - | z8=4 | Pb1Te1 |
| 225.1.3 | 63097 | 10 | 0.1043 | 0.1043 | 87 | 1 | 139 | 1 | - | - | - | - | z8=4 | Pb1Se1 |
| 225.1.4 | 63096 | 10 | 0.1305 | 0.1305 | 87 | 1 | 139 | 1 | - | - | - | - | z8=4 | Pb1Se1 |
| 225.1.5 | 62196 | 10 | 0.1043 | 0.1043 | 87 | 1 | 139 | 1 | - | - | - | - | z8=4 | Pb1Se1 |
| 225.1.6 | 62195 | 10 | 0.1305 | 0.1305 | 87 | 1 | 139 | 1 | - | - | - | - | z8=4 | Pb1Se1 |
| 225.1.7 | 38294 | 10 | 0.0934 | 0.0934 | 87 | 1 | 139 | 1 | - | - | - | - | z8=4 | Pb1Se1 |
| 225.1.8 | 62192 | 10 | 0.0099 | 0.0099 | 87 | 1 | 139 | 1 | - | - | - | - | z8=4 | Pb1S1 |
| 225.1.9 | 62191 | 10 | 0.0204 | 0.0204 | 87 | 1 | 139 | 1 | - | - | - | - | z8=4 | Pb1S1 |
| 225.1.10 | 62190 | 10 | 0.0277 | 0.0277 | 87 | 1 | 139 | 1 | - | - | - | - | z8=4 | Pb1S1 |
| 225.1.11 | 35204 | 26 | 0.1272 | -0.1787 | 87 | 1 | 139 | 1 | - | - | - | - | z8=4 | O2U1 |
| 225.1.12 | 246851 | 26 | 0.1273 | -0.1789 | 87 | 1 | 139 | 1 | - | - | - | - | z8=4 | O2U1 |
| 225.1.13 | 419774 | 248 | 0.0172 | -0.1731 | 2,82 | 4 | 2,82 | 7 | - | - | - | - | z4=1<br>z2=1<br>z8=1 | Ni21P6Sn2 |
| 225.1.14 | 159261 | 200 | 0.0057 | -0.3337 | 2,82 | 2 | 2,82 | 5 | - | - | - | - | z4=1<br>z2=1<br>z8=5 | Mg6Ni16Si7 |
| 225.1.15 | 106715 | 16 | 0.0467 | -4.8500 | 2,82 | 7 | 2,82 | 6 | - | - | - | - | z4=3<br>z2=1<br>z8=7 | Ga2Li1Rh1 |
| 225.1.16 | 160154 | 240 | 0.0263 | -0.0135 | 87 | 1 | 139 | 1 | - | - | - | - | z8=4 | Fe16Si7U6 |
| 225.1.17 | 188351 | 132 | 0.0000 | -0.4299 | 2 | 6 | 69,71,87 | 2 | - | - | - | - | z4=2<br>z8=6 | Cl1Cu6O8Y1 |
| 225.1.18 | 69612 | 124 | 0.0021 | -0.1308 | 2 | 6 | 69,71,87 | 2 | - | - | - | - | z4=2<br>z8=6 | Cl1Cu6In1O8 |
| 225.1.19 | 75013 | 260 | 0.0143 | -0.1316 | 2,82 | 4 | 2,82 | 7 | - | - | - | - | z4=1<br>z2=1<br>z8=1 | Ce3Pd20Si6 |
| 225.1.20 | 237543 | 332 | 0.0253 | -0.6103 | 2,82 | 5 | 2,82 | 3 | - | - | - | - | z4=3<br>z2=1<br>z8=3 | Ca3Cd8Pt4 |
| 225.1.21 | 62667 | 162 | 0.0076 | -0.0983 | 2,82 | 4 | 2,82 | 7 | - | - | - | - | z4=1<br>z2=1<br>z8=1 | C6Cr23 |
| 225.1.22 | 159874 | 8 | 0.0234 | -1.3404 | 2,82 | 7 | 2,82 | 6 | - | - | - | - | z4=3<br>z2=1<br>z8=7 | C1Zr1 |

Continued on next page

**TABLE XIV– continued**

| TQC | ICSD | NoE | DG | IndG | SubSG (w/ T) | cl | SubSG (w/o T) | cl | Max. Sub. SG | yes or no | Max. Sub. SG | yes or no | SIs (w/ T) | Stoich. Formula |
|---|---|---|---|---|---|---|---|---|---|---|---|---|---|---|
| 225.1.23 | 424342 | 10 | 0.0261 | -3.4985 | 2 | 3 | 2 | 4 | - | - | - | - | z4=2<br>z8=2 | C1W1 |
| 225.1.24 | 159873 | 8 | 0.1567 | -1.3558 | 2,82 | 2 | 2,82 | 5 | - | - | - | - | z4=1<br>z2=1<br>z8=5 | C1Hf1 |
| 225.1.25 | 155551 | 78 | 0.0756 | -0.1212 | 2 | 3 | 69,71,87 | 4 | - | - | - | - | z4=2<br>z8=2 | Ba2Nd1O6Ru1 |
| 225.1.26 | 171491 | 84 | 0.0285 | 0.0096 | 2,82 | 5 | 2,82 | 3 | - | - | - | - | z4=3<br>z2=1<br>z8=3 | Ba2Ir1O6Tb1 |
| 225.1.27 | 153057 | 78 | 0.0341 | 0.0120 | 2,82 | 5 | 2,82 | 3 | - | - | - | - | z4=3<br>z2=1<br>z8=3 | Ba2Ir1O6Pr1 |
| 225.1.28 | 59744 | 86 | 0.0472 | 0.0179 | 2 | 6 | 69,71,87 | 2 | - | - | - | - | z4=2<br>z8=6 | Ba2Er1O6Ru1 |
| 225.1.29 | 59743 | 86 | 0.0465 | 0.0179 | 2 | 6 | 69,71,87 | 2 | - | - | - | - | z4=2<br>z8=6 | Ba2Er1O6Ru1 |
| 225.1.30 | 409959 | 234 | 0.0117 | -0.4146 | 87 | 1 | 139 | 1 | - | - | - | - | z8=4 | B6Ni21Sc2 |
| 225.3.1 | 66720 | 30 | - | - | - | - | - | - | 216 | yes | 216 | yes | - | Se1Yb1 |
| | | | | | | | | | 202 | no(2) | 202 | no(2) | z4=1 | |
| | | | | | | | | | 166 | no(2) | 166 | no(2) | z4=1 | |
| | | | | | | | | | 121 | yes | 121 | yes | - | |
| | | | | | | | | | 71 | no(2) | 71 | no(2) | z4=1 | |
| 225.3.2 | 95561 | 266 | - | - | - | - | - | - | 216 | yes | 216 | yes | - | Pd20Si6U3 |
| | | | | | | | | | 202 | no(2) | 202 | no(2) | z4=1 | |
| | | | | | | | | | 166 | no(2) | 166 | no(2) | z4=1 | |
| | | | | | | | | | 121 | yes | 121 | yes | - | |
| | | | | | | | | | 71 | no(2) | 71 | no(2) | z4=1 | |
| 225.3.3 | 93930 | 266 | - | - | - | - | - | - | 216 | yes | 216 | yes | - | Pd20Si6U3 |
| | | | | | | | | | 202 | no(2) | 202 | no(2) | z4=1 | |
| | | | | | | | | | 166 | no(2) | 166 | no(2) | z4=1 | |
| | | | | | | | | | 121 | yes | 121 | yes | - | |
| | | | | | | | | | 71 | no(2) | 71 | no(2) | z4=1 | |
| 225.3.4 | 159262 | 206 | - | - | - | - | - | - | 121 | no(82) | 121 | no(82) | z2=1 | Ni16Sc6Si7 |
| | | | | | | | | | 119 | no(82) | 119 | no(82) | z2=1 | |
| | | | | | | | | | 71 | no(2) | 71 | no(71) | z4=2 | |
| | | | | | | | | | 69 | no(2) | 69 | no(69) | z4=2 | |
| 225.3.5 | 162574 | 230 | - | - | - | - | - | - | 139 | no(2) | 0 | - | z4=2<br>z8=2 | Mn6Ni16Si7 |
| 225.3.6 | 162573 | 230 | - | - | - | - | - | - | 139 | no(2) | 0 | - | z4=2<br>z8=2 | Mn6Ni16Si7 |
| 225.3.7 | 72350 | 236 | - | - | - | - | - | - | 216 | no(82) | 0 | - | z2=1 | Mg3Ni20P6 |
| | | | | | | | | | 202 | yes | 0 | - | - | |
| | | | | | | | | | 166 | yes | 0 | - | - | |
| | | | | | | | | | 121 | no(82) | 0 | - | z2=1 | |
| | | | | | | | | | 71 | yes | 0 | - | - | |
| 225.3.8 | 10438 | 138 | - | - | - | - | - | - | 216 | no(82) | 0 | - | z2=1 | Ir4Sc11 |
| | | | | | | | | | 202 | yes | 0 | - | - | |
| | | | | | | | | | 166 | yes | 0 | - | - | |
| | | | | | | | | | 121 | no(82) | 0 | - | z2=1 | |
| | | | | | | | | | 71 | yes | 0 | - | - | |
| 225.3.9 | 68998 | 246 | - | - | - | - | - | - | 216 | no(82) | 0 | - | z2=1 | In2Ni21P6 |
| | | | | | | | | | 202 | yes | 0 | - | - | |
| | | | | | | | | | 166 | yes | 0 | - | - | |
| | | | | | | | | | 121 | no(82) | 0 | - | z2=1 | |
| | | | | | | | | | 71 | yes | 0 | - | - | |
| 225.3.10 | 412838 | 28 | - | - | - | - | - | - | 216 | no(82) | 0 | - | z2=1 | Fe3Si1 |
| | | | | | | | | | 202 | yes | 0 | - | - | |

Continued on next page

**TABLE XIV– continued**

| TQC | ICSD | NoE | DG | IndG | SubSG cl (w/ T) | SubSG cl (w/o T) | Max. Sub. SG | yes or no | Max. Sub. SG | yes or no | SIs (w/ T) | Stoich. Formula |
|---|---|---|---|---|---|---|---|---|---|---|---|---|
| | | | | | | | 166 | yes | 0 | - | - | |
| | | | | | | | 121 | no(82) | 0 | - | z2=1 | |
| | | | | | | | 71 | yes | 0 | - | - | |
| 225.3.11 | 102989 | 36 | - | - | - - | - - | 216 | no(82) | 0 | - | z2=1 | Cu2In1La1 |
| | | | | | | | 202 | no(2) | 0 | - | z4=2 | |
| | | | | | | | 166 | no(2) | 0 | - | z4=2 | |
| | | | | | | | 121 | no(82) | 0 | - | z2=1 | |
| | | | | | | | 71 | no(2) | 0 | - | z4=2 | |
| 225.3.12 | 174079 | 258 | - | - | - - | - - | 121 | no(82) | 0 | - | z2=1 | Ce2Pd21Si6 |
| | | | | | | | 119 | no(82) | 0 | - | z2=1 | |
| | | | | | | | 71 | yes | 0 | - | - | |
| | | | | | | | 69 | yes | 0 | - | - | |
| 225.3.13 | 249342 | 38 | - | - | - - | - - | 216 | no(82) | 0 | - | z2=1 | Ce1Mg1Zn2 |
| | | | | | | | 202 | no(2) | 0 | - | z4=2 | |
| | | | | | | | 166 | no(2) | 0 | - | z4=2 | |
| | | | | | | | 121 | no(82) | 0 | - | z2=1 | |
| | | | | | | | 71 | no(2) | 0 | - | z4=2 | |
| 225.3.14 | 26476 | 18 | - | - | - - | - - | 121 | yes | 0 | - | - | C1U1 |
| | | | | | | | 119 | yes | 0 | - | - | |
| | | | | | | | 71 | no(2) | 0 | - | z4=1 | |
| | | | | | | | 69 | no(2) | 0 | - | z4=1 | |
| 225.3.15 | 44494 | 8 | - | - | - - | - - | 216 | no(82) | 0 | - | z2=1 | C1Ti1 |
| | | | | | | | 202 | no(2) | 0 | - | z4=2 | |
| | | | | | | | 166 | no(2) | 0 | - | z4=2 | |
| | | | | | | | 121 | no(82) | 0 | - | z2=1 | |
| | | | | | | | 71 | no(2) | 0 | - | z4=2 | |
| 225.3.16 | 159871 | 8 | - | - | - - | - - | 216 | no(82) | 0 | - | z2=1 | C1Ti1 |
| | | | | | | | 202 | no(2) | 0 | - | z4=2 | |
| | | | | | | | 166 | no(2) | 0 | - | z4=2 | |
| | | | | | | | 121 | no(82) | 0 | - | z2=1 | |
| | | | | | | | 71 | no(2) | 0 | - | z4=2 | |
| 225.3.17 | 1546 | 8 | - | - | - - | - - | 216 | no(82) | 0 | - | z2=1 | C1Ti1 |
| | | | | | | | 202 | no(2) | 0 | - | z4=2 | |
| | | | | | | | 166 | no(2) | 0 | - | z4=2 | |
| | | | | | | | 121 | no(82) | 0 | - | z2=1 | |
| | | | | | | | 71 | no(2) | 0 | - | z4=2 | |
| 225.3.18 | 160155 | 244 | - | - | - - | - - | 139 | no(2) | 0 | - | z4=2<br>z8=2 | C1Fe16Si7U6 |
| 225.3.19 | 409960 | 252 | - | - | - - | - - | 139 | no(87) | 0 | - | z8=4 | B6Ni21Zr2 |
| 225.3.20 | 380216 | 234 | - | - | - - | - - | 216 | yes | 0 | - | - | B6In2Ni21 |
| | | | | | | | 202 | no(2) | 0 | - | z4=1 | |
| | | | | | | | 166 | no(2) | 0 | - | z4=1 | |
| | | | | | | | 121 | yes | 0 | - | - | |
| | | | | | | | 71 | no(2) | 0 | - | z4=1 | |
| 225.3.21 | 105524 | 32 | - | - | - - | - - | 216 | yes | 0 | - | - | Al1Nb1Ru2 |
| | | | | | | | 202 | no(2) | 0 | - | z4=1 | |
| | | | | | | | 166 | no(2) | 0 | - | z4=1 | |
| | | | | | | | 121 | yes | 0 | - | - | |
| | | | | | | | 71 | no(2) | 0 | - | z4=1 | |
| 225.4.1 | 261057 | 196 | - | - | - - | - - | See details in Table XIX | | | | | Ru9Sb8Zn7 |
| 225.4.2 | 10439 | 130 | - | - | - - | - - | See details in Table XIX | | | | | Os4Sc11 |
| 225.4.3 | 55456 | 28 | - | - | - - | - - | See details in Table XIX | | | | | O2Pu1 |
| 225.4.4 | 9866 | 16 | - | - | - - | - - | See details in Table XIX | | | | | Ni1O1 |
| 225.4.5 | 260172 | 10 | - | - | - - | - - | See details in Table XIX | | | | | Ni1 |
| 225.4.6 | 260169 | 10 | - | - | - - | - - | See details in Table XIX | | | | | Ni1 |
| 225.4.7 | 159263 | 212 | - | - | - - | - - | See details in Table XIX | | | | | Ni16Si7Ti6 |
| 225.4.8 | 159264 | 266 | - | - | - - | - - | See details in Table XIX | | | | | Nb6Ni16Si7 |
| 225.4.9 | 262367 | 20 | - | - | - - | - - | See details in Table XIX | | | | | N1Np1 |

Continued on next page

**TABLE XIV– continued**

| TQC | ICSD | NoE | DG | IndG | SubSG (w/ T) | cl | SubSG (w/o T) | cl | Max. Sub. SG | yes or no | Max. Sub. SG | yes or no | SIs (w/ T) | Stoich. Formula |
|---|---|---|---|---|---|---|---|---|---|---|---|---|---|---|
| 225.4.10 | 982 | 18 | - | - | - | - | - | - | See details in Table XIX | | | | | N1Nb1 |
| 225.4.11 | 157027 | 72 | - | - | - | - | - | - | See details in Table XIX | | | | | Mo1Ni1O6Sr2 |
| 225.4.12 | 155733 | 72 | - | - | - | - | - | - | See details in Table XIX | | | | | Mo1Ni1O6Sr2 |
| 225.4.13 | 409428 | 8 | - | - | - | - | - | - | See details in Table XIX | | | | | Mg2Pb1 |
| 225.4.14 | 10052 | 6 | - | - | - | - | - | - | See details in Table XIX | | | | | Li3Tl1 |
| 225.4.15 | 427115 | 52 | - | - | - | - | - | - | See details in Table XIX | | | | | I1Na1O6Pb2 |
| 225.4.16 | 63670 | 8 | - | - | - | - | - | - | See details in Table XIX | | | | | H1Br1 |
| 225.4.17 | 94456 | 50 | - | - | - | - | - | - | See details in Table XIX | | | | | F6Ti1Zr1 |
| 225.4.18 | 59102 | 80 | - | - | - | - | - | - | See details in Table XIX | | | | | Cs2Cu1F6K1 |
| 225.4.19 | 409229 | 54 | - | - | - | - | - | - | See details in Table XIX | | | | | Cl6Tl2W1 |
| 225.4.20 | 409632 | 66 | - | - | - | - | - | - | See details in Table XIX | | | | | Cl6Rb2W1 |
| 225.4.21 | 408777 | 54 | - | - | - | - | - | - | See details in Table XIX | | | | | Cl6Mo1Tl2 |
| 225.4.22 | 409840 | 66 | - | - | - | - | - | - | See details in Table XIX | | | | | Cl6K2W1 |
| 225.4.23 | 409674 | 66 | - | - | - | - | - | - | See details in Table XIX | | | | | Cl6Cs2W1 |
| 225.4.24 | 409808 | 66 | - | - | - | - | - | - | See details in Table XIX | | | | | Cl6Cs2Mo1 |
| 225.4.25 | 83258 | 56 | - | - | - | - | - | - | See details in Table XIX | | | | | Ca2O6Pd1W1 |
| 225.4.26 | 151706 | 90 | - | - | - | - | - | - | See details in Table XIX | | | | | C6Cs2Fe1N6Ni1 |
| 225.4.27 | 89531 | 6 | - | - | - | - | - | - | See details in Table XIX | | | | | C2Na2 |
| 225.4.28 | 89536 | 6 | - | - | - | - | - | - | See details in Table XIX | | | | | C2Li2 |
| 225.4.29 | 89532 | 22 | - | - | - | - | - | - | See details in Table XIX | | | | | C2K2 |
| 225.4.30 | 402439 | 66 | - | - | - | - | - | - | See details in Table XIX | | | | | Br6Rb2W1 |
| 225.4.31 | 402441 | 66 | - | - | - | - | - | - | See details in Table XIX | | | | | Br6Cs2W1 |
| 225.4.32 | 94216 | 74 | - | - | - | - | - | - | See details in Table XIX | | | | | Ba2O6Re1Y1 |
| 225.4.33 | 94215 | 74 | - | - | - | - | - | - | See details in Table XIX | | | | | Ba2O6Re1Y1 |
| 225.4.34 | 168216 | 74 | - | - | - | - | - | - | See details in Table XIX | | | | | Ba2O6Re1Y1 |
| 225.4.35 | 168215 | 74 | - | - | - | - | - | - | See details in Table XIX | | | | | Ba2O6Re1Y1 |
| 225.4.36 | 245142 | 80 | - | - | - | - | - | - | See details in Table XIX | | | | | Ba2Ni1O6U1 |
| 225.4.37 | 169535 | 80 | - | - | - | - | - | - | See details in Table XIX | | | | | Ba2Ni1O6U1 |
| 225.4.38 | 98192 | 72 | - | - | - | - | - | - | See details in Table XIX | | | | | Ba2Mo1Ni1O6 |
| 225.4.39 | 26229 | 76 | - | - | - | - | - | - | See details in Table XIX | | | | | Ba2Mo1Nd1O6 |
| 225.4.40 | 158359 | 82 | - | - | - | - | - | - | See details in Table XIX | | | | | Ba2Ho1O6Ta1 |
| 225.4.41 | 84651 | 82 | - | - | - | - | - | - | See details in Table XIX | | | | | Ba2Ho1O6Sb1 |
| 225.4.42 | 150862 | 82 | - | - | - | - | - | - | See details in Table XIX | | | | | Ba2Ho1O6Sb1 |
| 225.4.43 | 236354 | 80 | - | - | - | - | - | - | See details in Table XIX | | | | | Ba2Gd1Mo1O6 |
| 225.4.44 | 236353 | 80 | - | - | - | - | - | - | See details in Table XIX | | | | | Ba2Gd1Mo1O6 |
| 225.4.45 | 192845 | 66 | - | - | - | - | - | - | See details in Table XIX | | | | | Ba2Ca1O6Os1 |
| 225.4.46 | 192844 | 66 | - | - | - | - | - | - | See details in Table XIX | | | | | Ba2Ca1O6Os1 |
| 225.4.47 | 171989 | 66 | - | - | - | - | - | - | See details in Table XIX | | | | | Ba2Ca1O6Os1 |
| 225.4.48 | 171988 | 66 | - | - | - | - | - | - | See details in Table XIX | | | | | Ba2Ca1O6Os1 |
| 225.4.49 | 189924 | 50 | - | - | - | - | - | - | See details in Table XIX | | | | | B12U1 |
| 225.4.50 | 171472 | 26 | - | - | - | - | - | - | See details in Table XIX | | | | | Ag2In1Li1 |
| 226.3.1 | 58751 | 80 | - | - | - | - | - | - | 219 | yes | 0 | - | - | Be13U1 |
| | | | | | | | | | 202 | no(2) | 0 | - | z4=3 | |
| | | | | | | | | | 167 | no(2) | 0 | - | z4=3 | |
| | | | | | | | | | 121 | yes | 0 | - | - | |
| | | | | | | | | | 72 | no(2) | 0 | - | z4=3 | |
| 226.3.2 | 58750 | 80 | - | - | - | - | - | - | 219 | yes | 0 | - | - | Be13U1 |
| | | | | | | | | | 202 | no(2) | 0 | - | z4=3 | |
| | | | | | | | | | 167 | no(2) | 0 | - | z4=3 | |
| | | | | | | | | | 121 | yes | 0 | - | - | |
| | | | | | | | | | 72 | no(2) | 0 | - | z4=3 | |
| 226.3.3 | 58749 | 80 | - | - | - | - | - | - | 219 | yes | 0 | - | - | Be13U1 |
| | | | | | | | | | 202 | no(2) | 0 | - | z4=3 | |
| | | | | | | | | | 167 | no(2) | 0 | - | z4=3 | |
| | | | | | | | | | 121 | yes | 0 | - | - | |
| | | | | | | | | | 72 | no(2) | 0 | - | z4=3 | |
| 226.3.4 | 58748 | 80 | - | - | - | - | - | - | 219 | yes | 0 | - | - | Be13U1 |

Continued on next page

**TABLE XIV– continued**

| TQC | ICSD | NoE | DG | IndG | SubSG (w/ T) | cl | SubSG (w/o T) | cl | Max. Sub. SG | yes or no | Max. Sub. SG | yes or no | SIs (w/ T) | Stoich. Formula |
|---|---|---|---|---|---|---|---|---|---|---|---|---|---|---|
| | | | | | | | | | 202 | no(2) | 0 | - | z4=3 | |
| | | | | | | | | | 167 | no(2) | 0 | - | z4=3 | |
| | | | | | | | | | 121 | yes | 0 | - | - | |
| | | | | | | | | | 72 | no(2) | 0 | - | z4=3 | |
| 226.3.5 | 58747 | 80 | - | - | - | - | - | - | 219 | yes | 0 | - | - | Be13U1 |
| | | | | | | | | | 202 | no(2) | 0 | - | z4=3 | |
| | | | | | | | | | 167 | no(2) | 0 | - | z4=3 | |
| | | | | | | | | | 121 | yes | 0 | - | - | |
| | | | | | | | | | 72 | no(2) | 0 | - | z4=3 | |
| 226.3.6 | 107528 | 80 | - | - | - | - | - | - | 219 | yes | 0 | - | - | Be13U1 |
| | | | | | | | | | 202 | no(2) | 0 | - | z4=3 | |
| | | | | | | | | | 167 | no(2) | 0 | - | z4=3 | |
| | | | | | | | | | 121 | yes | 0 | - | - | |
| | | | | | | | | | 72 | no(2) | 0 | - | z4=3 | |
| 226.4.1 | 415889 | 330 | - | - | - | - | - | - | See details in Table XIX | | | | | Rb1Zn13 |
| 226.4.2 | 415887 | 314 | - | - | - | - | - | - | See details in Table XIX | | | | | Na1Zn13 |
| 226.4.3 | 421842 | 314 | - | - | - | - | - | - | See details in Table XIX | | | | | Li1Zn13 |
| 226.4.4 | 415886 | 330 | - | - | - | - | - | - | See details in Table XIX | | | | | K1Zn13 |
| 226.4.5 | 55509 | 304 | - | - | - | - | - | - | See details in Table XIX | | | | | Co1Cu12La1 |
| 226.4.6 | 415888 | 330 | - | - | - | - | - | - | See details in Table XIX | | | | | Cd13Rb1 |
| 226.4.7 | 415885 | 330 | - | - | - | - | - | - | See details in Table XIX | | | | | Cd13K1 |
| 226.4.8 | 415884 | 330 | - | - | - | - | - | - | See details in Table XIX | | | | | Cd13Cs1 |
| 226.4.9 | 58742 | 76 | - | - | - | - | - | - | See details in Table XIX | | | | | Be13Th1 |
| 226.4.10 | 58740 | 76 | - | - | - | - | - | - | See details in Table XIX | | | | | Be13Th1 |
| 226.4.11 | 834 | 62 | - | - | - | - | - | - | See details in Table XIX | | | | | Be13Sb1 |
| 226.4.12 | 170628 | 76 | - | - | - | - | - | - | See details in Table XIX | | | | | Be13Ce1 |
| 226.4.13 | 154391 | 306 | - | - | - | - | - | - | See details in Table XIX | | | | | Ba1Cu13 |
| 227.1.1 | 72042 | 64 | 0.0180 | -0.1132 | 2,82 | 2 | 2,82 | 2 | - | - | - | - | z4=1<br>z2=1 | S2Ti1 |
| 227.1.2 | 160865 | 148 | 0.0028 | -0.3432 | 2,82 | 2 | 2,82 | 2 | - | - | - | - | z4=1<br>z2=1 | O7V2Y2 |
| 227.1.3 | 151747 | 176 | 0.0028 | -0.0015 | 2 | 3 | 0 | 1 | - | - | - | - | z4=2 | O7Tb2Ti2 |
| 227.1.4 | 429192 | 152 | 0.0000 | -0.0146 | 2 | 3 | 0 | 1 | - | - | - | - | z4=2 | O7Pr2Zr2 |
| 227.1.5 | 261854 | 156 | 0.0149 | 0.0121 | 2,82 | 2 | 2,82 | 2 | - | - | - | - | z4=1<br>z2=1 | Nd2O7Zr2 |
| 227.1.6 | 249974 | 156 | 0.0069 | -0.0295 | 2,82 | 1 | 2,82 | 2 | - | - | - | - | z4=3<br>z2=1 | Nd2O7Zr2 |
| 227.1.7 | 174187 | 156 | 0.0144 | 0.0132 | 2,82 | 2 | 2,82 | 2 | - | - | - | - | z4=1<br>z2=1 | Nd2O7Zr2 |
| 227.1.8 | 153227 | 156 | 0.0143 | 0.0122 | 2,82 | 2 | 2,82 | 2 | - | - | - | - | z4=1<br>z2=1 | Nd2O7Zr2 |
| 227.1.9 | 82958 | 156 | 0.0021 | 0.0021 | 2,82 | 2 | 2,82 | 2 | - | - | - | - | z4=1<br>z2=1 | Nd2O7Sn2 |
| 227.1.10 | 202522 | 152 | 0.0033 | -0.1901 | 2 | 3 | 0 | 1 | - | - | - | - | z4=2 | Mo2O7Y2 |
| 227.1.11 | 83277 | 124 | 0.0062 | -0.1926 | 2,82 | 1 | 2,82 | 2 | - | - | - | - | z4=3<br>z2=1 | Mn2O7Sc2 |
| 227.1.12 | 187996 | 120 | 0.0000 | -0.6027 | 2,82 | 2 | 2,82 | 2 | - | - | - | - | z4=1<br>z2=1 | Mg3Ni2Ti1 |
| 227.1.13 | 184696 | 68 | 0.0122 | -0.5590 | 2,82 | 2 | 2,82 | 2 | - | - | - | - | z4=1<br>z2=1 | Mg1O4Ti2 |
| 227.1.14 | 32688 | 204 | 0.0000 | -0.2974 | 2,82 | 2 | 2,82 | 2 | - | - | - | - | z4=1<br>z2=1 | Lu2O7V2 |
| 227.1.15 | 159290 | 204 | 0.0000 | -0.3781 | 2,82 | 2 | 2,82 | 2 | - | - | - | - | z4=1<br>z2=1 | Lu2O7V2 |
| 227.1.16 | 159286 | 204 | 0.0029 | -0.3794 | 2,82 | 2 | 2,82 | 2 | - | - | - | - | z4=1<br>z2=1 | Lu2O7V2 |
| 227.1.17 | 48128 | 68 | 0.0057 | -0.6646 | 2,82 | 2 | 2,82 | 2 | - | - | - | - | z4=1<br>z2=1 | Li1O2Ti1 |

Continued on next page

**TABLE XIV– continued**

| TQC | ICSD | NoE | DG | IndG | SubSG (w/ T) | cl | SubSG (w/o T) | cl | Max. Sub. SG | yes or no | Max. Sub. SG | yes or no | SIs (w/ T) | Stoich. Formula |
|---|---|---|---|---|---|---|---|---|---|---|---|---|---|---|
| 227.1.18 | 423608 | 16 | 0.0000 | -2.3367 | 2,82 | 1 | 2,82 | 2 | - | - | - | - | z4=3 z2=1 | In2Mg1 |
| 227.1.19 | 420818 | 164 | 0.0466 | -0.2097 | 2,82 | 2 | 2,82 | 2 | - | - | - | - | z4=1 z2=1 | Hg2O7Ru2 |
| 227.1.20 | 159809 | 164 | 0.0446 | -0.1904 | 2,82 | 2 | 2,82 | 2 | - | - | - | - | z4=1 z2=1 | Hg2O7Os2 |
| 227.1.21 | 159808 | 164 | 0.0450 | -0.1906 | 2,82 | 2 | 2,82 | 2 | - | - | - | - | z4=1 z2=1 | Hg2O7Os2 |
| 227.1.22 | 159807 | 164 | 0.0453 | -0.1909 | 2,82 | 2 | 2,82 | 2 | - | - | - | - | z4=1 z2=1 | Hg2O7Os2 |
| 227.1.23 | 159806 | 164 | 0.0455 | -0.1912 | 2,82 | 2 | 2,82 | 2 | - | - | - | - | z4=1 z2=1 | Hg2O7Os2 |
| 227.1.24 | 159805 | 164 | 0.0460 | -0.1904 | 2,82 | 2 | 2,82 | 2 | - | - | - | - | z4=1 z2=1 | Hg2O7Os2 |
| 227.1.25 | 159804 | 164 | 0.0459 | -0.1907 | 2,82 | 2 | 2,82 | 2 | - | - | - | - | z4=1 z2=1 | Hg2O7Os2 |
| 227.1.26 | 159803 | 164 | 0.0459 | -0.1908 | 2,82 | 2 | 2,82 | 2 | - | - | - | - | z4=1 z2=1 | Hg2O7Os2 |
| 227.1.27 | 159802 | 164 | 0.0459 | -0.1909 | 2,82 | 2 | 2,82 | 2 | - | - | - | - | z4=1 z2=1 | Hg2O7Os2 |
| 227.1.28 | 159801 | 164 | 0.0462 | -0.1904 | 2,82 | 2 | 2,82 | 2 | - | - | - | - | z4=1 z2=1 | Hg2O7Os2 |
| 227.1.29 | 159800 | 164 | 0.0462 | -0.1905 | 2,82 | 2 | 2,82 | 2 | - | - | - | - | z4=1 z2=1 | Hg2O7Os2 |
| 227.1.30 | 159799 | 164 | 0.0460 | -0.1909 | 2,82 | 2 | 2,82 | 2 | - | - | - | - | z4=1 z2=1 | Hg2O7Os2 |
| 227.1.31 | 159798 | 164 | 0.0461 | -0.1907 | 2,82 | 2 | 2,82 | 2 | - | - | - | - | z4=1 z2=1 | Hg2O7Os2 |
| 227.1.32 | 159797 | 164 | 0.0462 | -0.1905 | 2,82 | 2 | 2,82 | 2 | - | - | - | - | z4=1 z2=1 | Hg2O7Os2 |
| 227.1.33 | 159796 | 164 | 0.0461 | -0.1909 | 2,82 | 2 | 2,82 | 2 | - | - | - | - | z4=1 z2=1 | Hg2O7Os2 |
| 227.1.34 | 159795 | 164 | 0.0460 | -0.1911 | 2,82 | 2 | 2,82 | 2 | - | - | - | - | z4=1 z2=1 | Hg2O7Os2 |
| 227.1.35 | 159794 | 164 | 0.0463 | -0.1908 | 2,82 | 2 | 2,82 | 2 | - | - | - | - | z4=1 z2=1 | Hg2O7Os2 |
| 227.1.36 | 84753 | 172 | 0.0159 | 0.0159 | 2,82 | 1 | 2,82 | 2 | - | - | - | - | z4=3 z2=1 | Gd2O7Sn2 |
| 227.1.37 | 160866 | 192 | 0.0000 | -0.0671 | 2 | 3 | 0 | 1 | - | - | - | - | z4=2 | Er2O7V2 |
| 227.1.38 | 50047 | 96 | 0.0000 | -0.0886 | 2,82 | 2 | 2,82 | 2 | - | - | - | - | z4=1 z2=1 | Cr2O4Zn1 |
| 227.1.39 | 196119 | 96 | 0.0000 | -0.0904 | 2,82 | 2 | 2,82 | 2 | - | - | - | - | z4=1 z2=1 | Cr2O4Zn1 |
| 227.1.40 | 171899 | 96 | 0.0000 | -0.0630 | 2,82 | 2 | 2,82 | 2 | - | - | - | - | z4=1 z2=1 | Cr2O4Zn1 |
| 227.1.41 | 171896 | 96 | 0.0000 | -0.0657 | 2,82 | 2 | 2,82 | 2 | - | - | - | - | z4=1 z2=1 | Cr2O4Zn1 |
| 227.1.42 | 171894 | 96 | 0.0000 | -0.0713 | 2,82 | 2 | 2,82 | 2 | - | - | - | - | z4=1 z2=1 | Cr2O4Zn1 |
| 227.1.43 | 171893 | 96 | 0.0000 | -0.0670 | 2,82 | 2 | 2,82 | 2 | - | - | - | - | z4=1 z2=1 | Cr2O4Zn1 |
| 227.1.44 | 97201 | 76 | 0.0027 | -0.0849 | 2,82 | 2 | 2,82 | 2 | - | - | - | - | z4=1 z2=1 | Cr2Mg1O4 |
| 227.1.45 | 75623 | 76 | 0.0094 | -0.0680 | 2,82 | 2 | 2,82 | 2 | - | - | - | - | z4=1 z2=1 | Cr2Mg1O4 |
| 227.1.46 | 290599 | 76 | 0.0095 | -0.0685 | 2,82 | 2 | 2,82 | 2 | - | - | - | - | z4=1 z2=1 | Cr2Mg1O4 |

Continued on next page

**TABLE XIV– continued**

| TQC | ICSD | NoE | DG | IndG | SubSG (w/ T) | cl | SubSG (w/o T) | cl | Max. Sub. SG | yes or no | Max. Sub. SG | yes or no | SIs (w/ T) | Stoich. Formula |
|---|---|---|---|---|---|---|---|---|---|---|---|---|---|---|
| 227.1.47 | 171107 | 76 | 0.0085 | -0.0706 | 2,82 | 2 | 2,82 | 2 | - | - | - | - | z4=1<br>z2=1 | Cr2Mg1O4 |
| 227.1.48 | 171106 | 76 | 0.0084 | -0.0711 | 2,82 | 2 | 2,82 | 2 | - | - | - | - | z4=1<br>z2=1 | Cr2Mg1O4 |
| 227.1.49 | 160953 | 76 | 0.0085 | -0.0707 | 2,82 | 2 | 2,82 | 2 | - | - | - | - | z4=1<br>z2=1 | Cr2Mg1O4 |
| 227.1.50 | 402408 | 96 | 0.0054 | -0.0221 | 2,82 | 2 | 2,82 | 2 | - | - | - | - | z4=1<br>z2=1 | Cr2Hg1Se4 |
| 227.1.51 | 183965 | 88 | 0.0075 | -0.1562 | 2 | 3 | 0 | 1 | - | - | - | - | z4=2 | Cr2Fe1O4 |
| 227.1.52 | 183964 | 88 | 0.0091 | -0.1501 | 2 | 3 | 0 | 1 | - | - | - | - | z4=2 | Cr2Fe1O4 |
| 227.1.53 | 183963 | 88 | 0.0085 | -0.1507 | 2 | 3 | 0 | 1 | - | - | - | - | z4=2 | Cr2Fe1O4 |
| 227.1.54 | 183962 | 88 | 0.0067 | -0.1537 | 2 | 3 | 0 | 1 | - | - | - | - | z4=2 | Cr2Fe1O4 |
| 227.1.55 | 183961 | 88 | 0.0073 | -0.1458 | 2 | 3 | 0 | 1 | - | - | - | - | z4=2 | Cr2Fe1O4 |
| 227.1.56 | 183960 | 88 | 0.0045 | -0.1453 | 2 | 3 | 0 | 1 | - | - | - | - | z4=2 | Cr2Fe1O4 |
| 227.1.57 | 171121 | 88 | 0.0067 | -0.1399 | 2 | 3 | 0 | 1 | - | - | - | - | z4=2 | Cr2Fe1O4 |
| 227.1.58 | 859 | 92 | 0.0066 | -0.2996 | 2,82 | 2 | 2,82 | 2 | - | - | - | - | z4=1<br>z2=1 | Co2O4Si1 |
| 227.1.59 | 1 - 30 | 64 | 0.0137 | -0.0139 | 2,82 | 1 | 2,82 | 2 | - | - | - | - | z4=3<br>z2=1 | Co2Nd1 |
| 227.1.60 | 1 - 29 | 64 | 0.0134 | -0.0139 | 2,82 | 1 | 2,82 | 2 | - | - | - | - | z4=3<br>z2=1 | Co2Nd1 |
| 227.1.61 | 1 - 28 | 64 | 0.0133 | -0.0140 | 2,82 | 1 | 2,82 | 2 | - | - | - | - | z4=3<br>z2=1 | Co2Nd1 |
| 227.1.62 | 1 - 27 | 64 | 0.0132 | -0.0141 | 2,82 | 1 | 2,82 | 2 | - | - | - | - | z4=3<br>z2=1 | Co2Nd1 |
| 227.1.63 | 1 - 26 | 64 | 0.0132 | -0.0140 | 2,82 | 1 | 2,82 | 2 | - | - | - | - | z4=3<br>z2=1 | Co2Nd1 |
| 227.1.64 | 1 - 25 | 64 | 0.0132 | -0.0140 | 2,82 | 1 | 2,82 | 2 | - | - | - | - | z4=3<br>z2=1 | Co2Nd1 |
| 227.1.65 | 1 - 24 | 64 | 0.0132 | -0.0141 | 2,82 | 1 | 2,82 | 2 | - | - | - | - | z4=3<br>z2=1 | Co2Nd1 |
| 227.1.66 | 194667 | 80 | 0.0060 | -0.1279 | 2,82 | 1 | 2,82 | 2 | - | - | - | - | z4=3<br>z2=1 | Co2Er1 |
| 227.1.67 | 168600 | 148 | 0.0079 | 0.0057 | 2,82 | 1 | 2,82 | 2 | - | - | - | - | z4=3<br>z2=1 | Ce2O7Zr2 |
| 227.1.68 | 168595 | 148 | 0.0108 | 0.0062 | 2,82 | 1 | 2,82 | 2 | - | - | - | - | z4=3<br>z2=1 | Ce2O7Zr2 |
| 227.1.69 | 155773 | 164 | 0.0230 | -0.1358 | 2,82 | 2 | 2,82 | 2 | - | - | - | - | z4=1<br>z2=1 | Cd2O7Os2 |
| 227.1.70 | 155772 | 164 | 0.0220 | -0.1365 | 2,82 | 2 | 2,82 | 2 | - | - | - | - | z4=1<br>z2=1 | Cd2O7Os2 |
| 227.1.71 | 155771 | 164 | 0.0227 | -0.1366 | 2,82 | 2 | 2,82 | 2 | - | - | - | - | z4=1<br>z2=1 | Cd2O7Os2 |
| 227.1.72 | 155770 | 164 | 0.0220 | -0.1368 | 2,82 | 2 | 2,82 | 2 | - | - | - | - | z4=1<br>z2=1 | Cd2O7Os2 |
| 227.1.73 | 155769 | 164 | 0.0212 | -0.1369 | 2,82 | 2 | 2,82 | 2 | - | - | - | - | z4=1<br>z2=1 | Cd2O7Os2 |
| 227.1.74 | 155768 | 164 | 0.0225 | -0.1368 | 2,82 | 2 | 2,82 | 2 | - | - | - | - | z4=1<br>z2=1 | Cd2O7Os2 |
| 227.1.75 | 155767 | 164 | 0.0214 | -0.1373 | 2,82 | 2 | 2,82 | 2 | - | - | - | - | z4=1<br>z2=1 | Cd2O7Os2 |
| 227.1.76 | 155766 | 164 | 0.0192 | -0.1378 | 2,82 | 2 | 2,82 | 2 | - | - | - | - | z4=1<br>z2=1 | Cd2O7Os2 |
| 227.1.77 | 155765 | 164 | 0.0191 | -0.1380 | 2,82 | 2 | 2,82 | 2 | - | - | - | - | z4=1<br>z2=1 | Cd2O7Os2 |
| 227.1.78 | 155764 | 164 | 0.0186 | -0.1384 | 2,82 | 2 | 2,82 | 2 | - | - | - | - | z4=1<br>z2=1 | Cd2O7Os2 |

Continued on next page

**TABLE XIV– continued**

| TQC | ICSD | NoE | DG | IndG | SubSG (w/ T) | cl | SubSG (w/o T) | cl | Max. Sub. SG | yes or no | Max. Sub. SG | yes or no | SIs (w/ T) | Stoich. Formula |
|---|---|---|---|---|---|---|---|---|---|---|---|---|---|---|
| 227.1.79 | 155763 | 164 | 0.0191 | -0.1384 | 2,82 | 2 | 2,82 | 2 | - | - | - | - | z4=1<br>z2=1 | Cd2O7Os2 |
| 227.1.80 | 155762 | 164 | 0.0188 | -0.1387 | 2,82 | 2 | 2,82 | 2 | - | - | - | - | z4=1<br>z2=1 | Cd2O7Os2 |
| 227.1.81 | 155761 | 164 | 0.0186 | -0.1389 | 2,82 | 2 | 2,82 | 2 | - | - | - | - | z4=1<br>z2=1 | Cd2O7Os2 |
| 227.1.82 | 37408 | 168 | 0.0052 | -0.4626 | 2,82 | 2 | 2,82 | 2 | - | - | - | - | z4=1<br>z2=1 | Cd1Se4Yb2 |
| 227.1.83 | 188292 | 264 | 0.0109 | -0.2229 | 2,82 | 1 | 2,82 | 2 | - | - | - | - | z4=3<br>z2=1 | Cd1Ni1 |
| 227.1.84 | 246498 | 160 | 0.0000 | -0.0224 | 2,82 | 2 | 2,82 | 2 | - | - | - | - | z4=1<br>z2=1 | Cd1Er2S4 |
| 227.1.85 | 78554 | 96 | 0.0065 | -0.0090 | 2,82 | 2 | 2,82 | 2 | - | - | - | - | z4=1<br>z2=1 | Cd1Cr2Se4 |
| 227.1.86 | 39416 | 96 | 0.0065 | -0.0091 | 2,82 | 2 | 2,82 | 2 | - | - | - | - | z4=1<br>z2=1 | Cd1Cr2Se4 |
| 227.1.87 | 39415 | 96 | 0.0142 | 0.0017 | 2,82 | 2 | 2,82 | 2 | - | - | - | - | z4=1<br>z2=1 | Cd1Cr2S4 |
| 227.1.88 | 57512 | 32 | 0.0119 | -2.0510 | 2 | 3 | 0 | 1 | - | - | - | - | z4=2 | Al2Ba1 |
| 227.3.1 | 32689 | 200 | - | - | - | - | - | - | 216 | yes | 0 | - | - | O7V2Yb2 |
| | | | | | | | | | 203 | no(2) | 0 | - | z4=3 | |
| | | | | | | | | | 166 | no(2) | 0 | - | z4=3 | |
| | | | | | | | | | 122 | yes | 0 | - | - | |
| | | | | | | | | | 74 | no(2) | 0 | - | z4=3 | |
| 227.3.2 | 32690 | 196 | - | - | - | - | - | - | 122 | no(82) | 0 | - | z2=1 | O7Tm2V2 |
| | | | | | | | | | 119 | no(82) | 0 | - | z2=1 | |
| | | | | | | | | | 74 | yes | 0 | - | - | |
| | | | | | | | | | 70 | yes | 0 | - | - | |
| 227.3.3 | 82303 | 168 | - | - | - | - | - | - | 216 | yes | 0 | - | - | O7Pr2Ru2 |
| | | | | | | | | | 203 | no(2) | 0 | - | z4=3 | |
| | | | | | | | | | 166 | no(2) | 0 | - | z4=3 | |
| | | | | | | | | | 122 | yes | 0 | - | - | |
| | | | | | | | | | 74 | no(2) | 0 | - | z4=3 | |
| 227.3.4 | 78123 | 168 | - | - | - | - | - | - | 216 | yes | 0 | - | - | O7Pr2Ru2 |
| | | | | | | | | | 203 | no(2) | 0 | - | z4=3 | |
| | | | | | | | | | 166 | no(2) | 0 | - | z4=3 | |
| | | | | | | | | | 122 | yes | 0 | - | - | |
| | | | | | | | | | 74 | no(2) | 0 | - | z4=3 | |
| 227.3.5 | 163397 | 168 | - | - | - | - | - | - | 216 | yes | 0 | - | - | O7Pr2Ru2 |
| | | | | | | | | | 203 | no(2) | 0 | - | z4=3 | |
| | | | | | | | | | 166 | no(2) | 0 | - | z4=3 | |
| | | | | | | | | | 122 | yes | 0 | - | - | |
| | | | | | | | | | 74 | no(2) | 0 | - | z4=3 | |
| 227.3.6 | 15308 | 104 | - | - | - | - | - | - | 216 | yes | 0 | - | - | O6Pb2Sn2 |
| | | | | | | | | | 203 | no(2) | 0 | - | z4=1 | |
| | | | | | | | | | 166 | no(2) | 0 | - | z4=1 | |
| | | | | | | | | | 122 | yes | 0 | - | - | |
| | | | | | | | | | 74 | no(2) | 0 | - | z4=1 | |
| 227.3.7 | 236758 | 168 | - | - | - | - | - | - | 216 | yes | 0 | - | - | Nd2O7Tc2 |
| | | | | | | | | | 203 | no(2) | 0 | - | z4=3 | |
| | | | | | | | | | 166 | no(2) | 0 | - | z4=3 | |
| | | | | | | | | | 122 | yes | 0 | - | - | |
| | | | | | | | | | 74 | no(2) | 0 | - | z4=3 | |
| 227.3.8 | 79327 | 172 | - | - | - | - | - | - | 216 | no(82) | 0 | - | z2=1 | Nd2O7Ru2 |
| | | | | | | | | | 203 | yes | 0 | - | - | |
| | | | | | | | | | 166 | yes | 0 | - | - | |
| | | | | | | | | | 122 | no(82) | 0 | - | z2=1 | |
| | | | | | | | | | 74 | yes | 0 | - | - | |

Continued on next page

**TABLE XIV– continued**

| TQC | ICSD | NoE | DG | IndG | SubSG cl (w/ T) | | SubSG cl (w/o T) | | Max. Sub. SG | yes or no | Max. Sub. SG | yes or no | SIs (w/ T) | Stoich. Formula |
|---|---|---|---|---|---|---|---|---|---|---|---|---|---|---|
| 227.3.9 | 78133 | 172 | - | - | - | - | - | - | 216 | no(82) | 0 | - | z2=1 | Nd2O7Ru2 |
| | | | | | | | | | 203 | yes | 0 | - | - | |
| | | | | | | | | | 166 | yes | 0 | - | - | |
| | | | | | | | | | 122 | no(82) | 0 | - | z2=1 | |
| | | | | | | | | | 74 | yes | 0 | - | - | |
| 227.3.10 | 83269 | 124 | - | - | - | - | - | - | 141 | no(2,82) | 0 | - | z4=1<br>z2=1 | Mn2O7Tl2 |
| 227.3.11 | 83268 | 124 | - | - | - | - | - | - | 141 | no(2,82) | 0 | - | z4=1<br>z2=1 | Mn2O7Tl2 |
| 227.3.12 | 409837 | 112 | - | - | - | - | - | - | 216 | no(82) | 0 | - | z2=1 | Mg2Pd1 |
| | | | | | | | | | 203 | yes | 0 | - | - | |
| | | | | | | | | | 166 | yes | 0 | - | - | |
| | | | | | | | | | 122 | no(82) | 0 | - | z2=1 | |
| | | | | | | | | | 74 | yes | 0 | - | - | |
| 227.3.13 | 160599 | 204 | - | - | - | - | - | - | 141 | no(2) | 0 | - | z4=2 | Lu2O7V2 |
| 227.3.14 | 156436 | 172 | - | - | - | - | - | - | 216 | no(82) | 0 | - | z2=1 | Ir2O7Pr2 |
| | | | | | | | | | 203 | yes | 0 | - | - | |
| | | | | | | | | | 166 | yes | 0 | - | - | |
| | | | | | | | | | 122 | no(82) | 0 | - | z2=1 | |
| | | | | | | | | | 74 | yes | 0 | - | - | |
| 227.3.15 | 55590 | 40 | - | - | - | - | - | - | 141 | no(2,82) | 0 | - | z4=3<br>z2=1 | Fe2Zr1 |
| 227.3.16 | 106701 | 40 | - | - | - | - | - | - | 141 | no(2,82) | 0 | - | z4=3<br>z2=1 | Fe2Zr1 |
| 227.3.17 | 91827 | 104 | - | - | - | - | - | - | 166 | no(2) | 0 | - | z4=3 | Fe2O4Zn1 |
| | | | | | | | | | 70 | no(2) | 0 | - | z4=3 | |
| | | | | | | | | | 46 | yes | 0 | - | - | |
| 227.3.18 | 94995 | 76 | - | - | - | - | - | - | 141 | no(2,82) | 0 | - | z4=3<br>z2=1 | Fe1S4Sc2 |
| 227.3.19 | 100527 | 76 | - | - | - | - | - | - | 141 | no(2,82) | 0 | - | z4=3<br>z2=1 | Fe1S4Sc2 |
| 227.3.20 | 27328 | 172 | - | - | - | - | - | - | 122 | yes | 0 | - | - | F6Hg2Mn2S1 |
| | | | | | | | | | 119 | yes | 0 | - | - | |
| | | | | | | | | | 74 | no(2) | 0 | - | z4=3 | |
| | | | | | | | | | 70 | no(2) | 0 | - | z4=3 | |
| 227.3.21 | 24209 | 188 | - | - | - | - | - | - | 216 | yes | 0 | - | - | Er2O7Ti2 |
| | | | | | | | | | 203 | no(2) | 0 | - | z4=1 | |
| | | | | | | | | | 166 | no(2) | 0 | - | z4=1 | |
| | | | | | | | | | 122 | yes | 0 | - | - | |
| | | | | | | | | | 74 | no(2) | 0 | - | z4=1 | |
| 227.3.22 | 84749 | 188 | - | - | - | - | - | - | 216 | yes | 0 | - | - | Er2O7Sn2 |
| | | | | | | | | | 203 | no(2) | 0 | - | z4=3 | |
| | | | | | | | | | 166 | no(2) | 0 | - | z4=3 | |
| | | | | | | | | | 122 | yes | 0 | - | - | |
| | | | | | | | | | 74 | no(2) | 0 | - | z4=3 | |
| 227.3.23 | 82962 | 188 | - | - | - | - | - | - | 216 | yes | 0 | - | - | Er2O7Sn2 |
| | | | | | | | | | 203 | no(2) | 0 | - | z4=3 | |
| | | | | | | | | | 166 | no(2) | 0 | - | z4=3 | |
| | | | | | | | | | 122 | yes | 0 | - | - | |
| | | | | | | | | | 74 | no(2) | 0 | - | z4=3 | |
| 227.3.24 | 27327 | 188 | - | - | - | - | - | - | 141 | no(2,82) | 0 | - | z4=3<br>z2=1 | Cu2F6Hg2S1 |
| 227.3.25 | 171898 | 96 | - | - | - | - | - | - | 141 | no(2,82) | 0 | - | z4=3<br>z2=1 | Cr2O4Zn1 |
| 227.3.26 | 171897 | 96 | - | - | - | - | - | - | 141 | no(2,82) | 0 | - | z4=3<br>z2=1 | Cr2O4Zn1 |
| 227.3.27 | 171895 | 96 | - | - | - | - | - | - | 141 | no(2,82) | 0 | - | z4=3<br>z2=1 | Cr2O4Zn1 |

Continued on next page

**TABLE XIV– continued**

| TQC | ICSD | NoE | DG | IndG | SubSG (w/ T) | cl | SubSG (w/o T) | cl | Max. Sub. SG | yes or no | Max. Sub. SG | yes or no | SIs (w/ T) | Stoich. Formula |
|---|---|---|---|---|---|---|---|---|---|---|---|---|---|---|
| 227.3.28 | 171892 | 96 | - | - | - | - | - | - | 141 | no(2,82) | 0 | - | z4=3<br>z2=1 | Cr2O4Zn1 |
| 227.3.29 | 171891 | 96 | - | - | - | - | - | - | 141 | no(2,82) | 0 | - | z4=3<br>z2=1 | Cr2O4Zn1 |
| 227.3.30 | 171890 | 96 | - | - | - | - | - | - | 141 | no(2,82) | 0 | - | z4=3<br>z2=1 | Cr2O4Zn1 |
| 227.3.31 | 171889 | 96 | - | - | - | - | - | - | 141 | no(2,82) | 0 | - | z4=3<br>z2=1 | Cr2O4Zn1 |
| 227.3.32 | 43451 | 96 | - | - | - | - | - | - | 141 | no(2,82) | 0 | - | z4=3<br>z2=1 | Cr2Hg1O4 |
| 227.3.33 | 245275 | 96 | - | - | - | - | - | - | 141 | no(2,82) | 0 | - | z4=3<br>z2=1 | Cr2Hg1O4 |
| 227.3.34 | 247302 | 140 | - | - | - | - | - | - | 141 | no(2,82) | 0 | - | z4=1<br>z2=1 | Co2O7Sb2 |
| 227.3.35 | 106482 | 132 | - | - | - | - | - | - | 216 | no(82) | 0 | - | z2=1 | Co1Mg1 |
| | | | | | | | | | 203 | no(2) | 0 | - | z4=2 | |
| | | | | | | | | | 166 | no(2) | 0 | - | z4=2 | |
| | | | | | | | | | 122 | no(82) | 0 | - | z2=1 | |
| | | | | | | | | | 74 | no(2) | 0 | - | z4=2 | |
| 227.3.36 | 102038 | 264 | - | - | - | - | - | - | 216 | yes | 0 | - | - | Cd1Ni1 |
| | | | | | | | | | 203 | no(2) | 0 | - | z4=3 | |
| | | | | | | | | | 166 | no(2) | 0 | - | z4=3 | |
| | | | | | | | | | 122 | yes | 0 | - | - | |
| | | | | | | | | | 74 | no(2) | 0 | - | z4=3 | |
| 227.3.37 | 413207 | 8 | - | - | - | - | - | - | 216 | yes | 0 | - | - | Ca1Li2 |
| | | | | | | | | | 203 | no(2) | 0 | - | z4=3 | |
| | | | | | | | | | 166 | no(2) | 0 | - | z4=3 | |
| | | | | | | | | | 122 | yes | 0 | - | - | |
| | | | | | | | | | 74 | no(2) | 0 | - | z4=3 | |
| 227.3.38 | 67517 | 84 | - | - | - | - | - | - | 166 | no(2) | 0 | - | z4=3 | Br2Mn1 |
| | | | | | | | | | 70 | no(2) | 0 | - | z4=3 | |
| | | | | | | | | | 46 | yes | 0 | - | - | |
| 227.3.39 | 170867 | 80 | - | - | - | - | - | - | 216 | yes | 0 | - | - | Au1In2Na3 |
| | | | | | | | | | 203 | no(2) | 0 | - | z4=3 | |
| | | | | | | | | | 166 | no(2) | 0 | - | z4=3 | |
| | | | | | | | | | 122 | yes | 0 | - | - | |
| | | | | | | | | | 74 | no(2) | 0 | - | z4=3 | |
| 227.3.40 | 187920 | 76 | - | - | - | - | - | - | 141 | no(2,82) | 0 | - | z4=3<br>z2=1 | Al2Fe1O4 |
| 227.3.41 | 105510 | 116 | - | - | - | - | - | - | 166 | no(2) | 0 | - | z4=3 | Al1Mg3Pt2 |
| | | | | | | | | | 70 | no(2) | 0 | - | z4=3 | |
| | | | | | | | | | 46 | yes | 0 | - | - | |
| 227.4.1 | 160600 | 148 | - | - | - | - | - | - | See details in Table XIX | | | | | O7V2Y2 |
| 227.4.2 | 190698 | 196 | - | - | - | - | - | - | See details in Table XIX | | | | | O7Ti2Yb2 |
| 227.4.3 | 190697 | 196 | - | - | - | - | - | - | See details in Table XIX | | | | | O7Ti2Yb2 |
| 227.4.4 | 190696 | 196 | - | - | - | - | - | - | See details in Table XIX | | | | | O7Ti2Yb2 |
| 227.4.5 | 190695 | 196 | - | - | - | - | - | - | See details in Table XIX | | | | | O7Ti2Yb2 |
| 227.4.6 | 190694 | 196 | - | - | - | - | - | - | See details in Table XIX | | | | | O7Ti2Yb2 |
| 227.4.7 | 190688 | 196 | - | - | - | - | - | - | See details in Table XIX | | | | | O7Ti2Yb2 |
| 227.4.8 | 190687 | 196 | - | - | - | - | - | - | See details in Table XIX | | | | | O7Ti2Yb2 |
| 227.4.9 | 190686 | 196 | - | - | - | - | - | - | See details in Table XIX | | | | | O7Ti2Yb2 |
| 227.4.10 | 190685 | 196 | - | - | - | - | - | - | See details in Table XIX | | | | | O7Ti2Yb2 |
| 227.4.11 | 190684 | 196 | - | - | - | - | - | - | See details in Table XIX | | | | | O7Ti2Yb2 |
| 227.4.12 | 186196 | 196 | - | - | - | - | - | - | See details in Table XIX | | | | | O7Ti2Yb2 |
| 227.4.13 | 186195 | 196 | - | - | - | - | - | - | See details in Table XIX | | | | | O7Ti2Yb2 |
| 227.4.14 | 173750 | 196 | - | - | - | - | - | - | See details in Table XIX | | | | | O7Ti2Yb2 |
| 227.4.15 | 161405 | 196 | - | - | - | - | - | - | See details in Table XIX | | | | | O7Ti2Yb2 |
| 227.4.16 | 82964 | 196 | - | - | - | - | - | - | See details in Table XIX | | | | | O7Sn2Yb2 |

Continued on next page

**TABLE XIV– continued**

| TQC | ICSD | NoE | DG | IndG | SubSG cl (w/ T) | SubSG cl (w/o T) | Max. Sub. SG yes or no | Max. Sub. SG yes or no | SIs (w/ T) | Stoich. Formula |
|---|---|---|---|---|---|---|---|---|---|---|
| 227.4.17 | 84750 | 192 | - | - | - - | - - | See details in Table XIX | | | O7Sn2Tm2 |
| 227.4.18 | 82963 | 192 | - | - | - - | - - | See details in Table XIX | | | O7Sn2Tm2 |
| 227.4.19 | 84744 | 152 | - | - | - - | - - | See details in Table XIX | | | O7Pr2Sn2 |
| 227.4.20 | 82957 | 152 | - | - | - - | - - | See details in Table XIX | | | O7Pr2Sn2 |
| 227.4.21 | 173073 | 122 | - | - | - - | - - | See details in Table XIX | | | O6Os2Rb1 |
| 227.4.22 | 173072 | 122 | - | - | - - | - - | See details in Table XIX | | | O6Os2Rb1 |
| 227.4.23 | 157442 | 122 | - | - | - - | - - | See details in Table XIX | | | O6Os2Rb1 |
| 227.4.24 | 157441 | 122 | - | - | - - | - - | See details in Table XIX | | | O6Os2Rb1 |
| 227.4.25 | 157440 | 122 | - | - | - - | - - | See details in Table XIX | | | O6Os2Rb1 |
| 227.4.26 | 157439 | 122 | - | - | - - | - - | See details in Table XIX | | | O6Os2Rb1 |
| 227.4.27 | 157438 | 122 | - | - | - - | - - | See details in Table XIX | | | O6Os2Rb1 |
| 227.4.28 | 157437 | 122 | - | - | - - | - - | See details in Table XIX | | | O6Os2Rb1 |
| 227.4.29 | 157436 | 122 | - | - | - - | - - | See details in Table XIX | | | O6Os2Rb1 |
| 227.4.30 | 157435 | 122 | - | - | - - | - - | See details in Table XIX | | | O6Os2Rb1 |
| 227.4.31 | 99181 | 92 | - | - | - - | - - | See details in Table XIX | | | O4V2Zn1 |
| 227.4.32 | 55442 | 92 | - | - | - - | - - | See details in Table XIX | | | O4V2Zn1 |
| 227.4.33 | 160164 | 156 | - | - | - - | - - | See details in Table XIX | | | Nd2O7Zr2 |
| 227.4.34 | 105162 | 42 | - | - | - - | - - | See details in Table XIX | | | Na1Pt2 |
| 227.4.35 | 54375 | 50 | - | - | - - | - - | See details in Table XIX | | | Mn2Y1 |
| 227.4.36 | 202516 | 156 | - | - | - - | - - | See details in Table XIX | | | Mn2O7Y2 |
| 227.4.37 | 195759 | 204 | - | - | - - | - - | See details in Table XIX | | | Mn2O7Tm2 |
| 227.4.38 | 104935 | 58 | - | - | - - | - - | See details in Table XIX | | | Mn2Np1 |
| 227.4.39 | 94992 | 74 | - | - | - - | - - | See details in Table XIX | | | Mn1S4Sc2 |
| 227.4.40 | 248565 | 72 | - | - | - - | - - | See details in Table XIX | | | Mg1O4V2 |
| 227.4.41 | 89352 | 70 | - | - | - - | - - | See details in Table XIX | | | Li1O4V2 |
| 227.4.42 | 89351 | 70 | - | - | - - | - - | See details in Table XIX | | | Li1O4V2 |
| 227.4.43 | 89350 | 70 | - | - | - - | - - | See details in Table XIX | | | Li1O4V2 |
| 227.4.44 | 89349 | 70 | - | - | - - | - - | See details in Table XIX | | | Li1O4V2 |
| 227.4.45 | 89348 | 70 | - | - | - - | - - | See details in Table XIX | | | Li1O4V2 |
| 227.4.46 | 89347 | 70 | - | - | - - | - - | See details in Table XIX | | | Li1O4V2 |
| 227.4.47 | 89346 | 70 | - | - | - - | - - | See details in Table XIX | | | Li1O4V2 |
| 227.4.48 | 89345 | 70 | - | - | - - | - - | See details in Table XIX | | | Li1O4V2 |
| 227.4.49 | 89344 | 70 | - | - | - - | - - | See details in Table XIX | | | Li1O4V2 |
| 227.4.50 | 418091 | 70 | - | - | - - | - - | See details in Table XIX | | | Li1O4V2 |
| 227.4.51 | 418090 | 70 | - | - | - - | - - | See details in Table XIX | | | Li1O4V2 |
| 227.4.52 | 201818 | 70 | - | - | - - | - - | See details in Table XIX | | | Li1O4V2 |
| 227.4.53 | 170768 | 70 | - | - | - - | - - | See details in Table XIX | | | Li1O4V2 |
| 227.4.54 | 78482 | 66 | - | - | - - | - - | See details in Table XIX | | | Li1O4Ti2 |
| 227.4.55 | 72859 | 66 | - | - | - - | - - | See details in Table XIX | | | Li1O4Ti2 |
| 227.4.56 | 48127 | 66 | - | - | - - | - - | See details in Table XIX | | | Li1O4Ti2 |
| 227.4.57 | 154982 | 66 | - | - | - - | - - | See details in Table XIX | | | Li1O4Ti2 |
| 227.4.58 | 94340 | 78 | - | - | - - | - - | See details in Table XIX | | | Li1Mn2O4 |
| 227.4.59 | 94339 | 78 | - | - | - - | - - | See details in Table XIX | | | Li1Mn2O4 |
| 227.4.60 | 89985 | 78 | - | - | - - | - - | See details in Table XIX | | | Li1Mn2O4 |
| 227.4.61 | 89459 | 78 | - | - | - - | - - | See details in Table XIX | | | Li1Mn2O4 |
| 227.4.62 | 88644 | 78 | - | - | - - | - - | See details in Table XIX | | | Li1Mn2O4 |
| 227.4.63 | 50415 | 78 | - | - | - - | - - | See details in Table XIX | | | Li1Mn2O4 |
| 227.4.64 | 261557 | 78 | - | - | - - | - - | See details in Table XIX | | | Li1Mn2O4 |
| 227.4.65 | 192369 | 78 | - | - | - - | - - | See details in Table XIX | | | Li1Mn2O4 |
| 227.4.66 | 155665 | 78 | - | - | - - | - - | See details in Table XIX | | | Li1Mn2O4 |
| 227.4.67 | 155282 | 78 | - | - | - - | - - | See details in Table XIX | | | Li1Mn2O4 |
| 227.4.68 | 170775 | 30 | - | - | - - | - - | See details in Table XIX | | | La1Mg2 |
| 227.4.69 | 172620 | 122 | - | - | - - | - - | See details in Table XIX | | | K1O6Os2 |
| 227.4.70 | 172619 | 122 | - | - | - - | - - | See details in Table XIX | | | K1O6Os2 |
| 227.4.71 | 172618 | 122 | - | - | - - | - - | See details in Table XIX | | | K1O6Os2 |
| 227.4.72 | 162266 | 122 | - | - | - - | - - | See details in Table XIX | | | K1O6Os2 |
| 227.4.73 | 160698 | 122 | - | - | - - | - - | See details in Table XIX | | | K1O6Os2 |

Continued on next page

**TABLE XIV– continued**

| TQC | ICSD | NoE | DG | IndG | SubSG cl (w/ T) | SubSG cl (w/o T) | Max. Sub. SG yes or no | Max. Sub. SG yes or no | SIs (w/ T) | Stoich. Formula |
|---|---|---|---|---|---|---|---|---|---|---|
| 227.4.74 | 187534 | 164 | - | - | - - | - - | See details in Table XIX | | | Ir2O7Y2 |
| 227.4.75 | 280390 | 28 | - | - | - - | - - | See details in Table XIX | | | Hf1V2 |
| 227.4.76 | 65618 | 196 | - | - | - - | - - | See details in Table XIX | | | Ge2O7Yb2 |
| 227.4.77 | 52179 | 240 | - | - | - - | - - | See details in Table XIX | | | Ge1Ni2Zn3 |
| 227.4.78 | 261759 | 172 | - | - | - - | - - | See details in Table XIX | | | Gd2O7Zr2 |
| 227.4.79 | 261758 | 172 | - | - | - - | - - | See details in Table XIX | | | Gd2O7Zr2 |
| 227.4.80 | 160158 | 172 | - | - | - - | - - | See details in Table XIX | | | Gd2O7Zr2 |
| 227.4.81 | 79332 | 188 | - | - | - - | - - | See details in Table XIX | | | Gd2O7Ru2 |
| 227.4.82 | 170910 | 94 | - | - | - - | - - | See details in Table XIX | | | Fe2Mn1O4 |
| 227.4.83 | 106646 | 128 | - | - | - - | - - | See details in Table XIX | | | Fe1Hf2 |
| 227.4.84 | 4083 | 184 | - | - | - - | - - | See details in Table XIX | | | F6Hg2Ni2S1 |
| 227.4.85 | 166545 | 184 | - | - | - - | - - | See details in Table XIX | | | F6Hg2Ni2S1 |
| 227.4.86 | 27329 | 184 | - | - | - - | - - | See details in Table XIX | | | F6Hg2Ni2O1 |
| 227.4.87 | 202047 | 116 | - | - | - - | - - | See details in Table XIX | | | F3Fe1 |
| 227.4.88 | 156437 | 188 | - | - | - - | - - | See details in Table XIX | | | Eu2Ir2O7 |
| 227.4.89 | 410701 | 70 | - | - | - - | - - | See details in Table XIX | | | Eu1Ir2 |
| 227.4.90 | 202517 | 200 | - | - | - - | - - | See details in Table XIX | | | Er2Mn2O7 |
| 227.4.91 | 78151 | 196 | - | - | - - | - - | See details in Table XIX | | | Dy2O7Ru2 |
| 227.4.92 | 156368 | 188 | - | - | - - | - - | See details in Table XIX | | | Cu2F6Hg2S1 |
| 227.4.93 | 156367 | 188 | - | - | - - | - - | See details in Table XIX | | | Cu2F6Hg2S1 |
| 227.4.94 | 156366 | 188 | - | - | - - | - - | See details in Table XIX | | | Cu2F6Hg2S1 |
| 227.4.95 | 190441 | 90 | - | - | - - | - - | See details in Table XIX | | | Cu1S4V2 |
| 227.4.96 | 10035 | 90 | - | - | - - | - - | See details in Table XIX | | | Cu1S4V2 |
| 227.4.97 | 75531 | 106 | - | - | - - | - - | See details in Table XIX | | | Cu1Ir2S4 |
| 227.4.98 | 4206 | 146 | - | - | - - | - - | See details in Table XIX | | | Cs2O12U4 |
| 227.4.99 | 421173 | 58 | - | - | - - | - - | See details in Table XIX | | | Cs1Fe1O2 |
| 227.4.100 | 174318 | 58 | - | - | - - | - - | See details in Table XIX | | | Cs1Fe1O2 |
| 227.4.101 | 425260 | 134 | - | - | - - | - - | See details in Table XIX | | | Cs1F6Fe2 |
| 227.4.102 | 31298 | 92 | - | - | - - | - - | See details in Table XIX | | | Cr2Ni1O4 |
| 227.4.103 | 280061 | 92 | - | - | - - | - - | See details in Table XIX | | | Cr2Ni1O4 |
| 227.4.104 | 97620 | 94 | - | - | - - | - - | See details in Table XIX | | | Cr2Cu1Se4 |
| 227.4.105 | 87477 | 94 | - | - | - - | - - | See details in Table XIX | | | Cr2Cu1Se4 |
| 227.4.106 | 87476 | 94 | - | - | - - | - - | See details in Table XIX | | | Cr2Cu1Se4 |
| 227.4.107 | 73403 | 94 | - | - | - - | - - | See details in Table XIX | | | Cr2Cu1Se4 |
| 227.4.108 | 291003 | 94 | - | - | - - | - - | See details in Table XIX | | | Cr2Cu1S4 |
| 227.4.109 | 291002 | 94 | - | - | - - | - - | See details in Table XIX | | | Cr2Cu1S4 |
| 227.4.110 | 291001 | 94 | - | - | - - | - - | See details in Table XIX | | | Cr2Cu1S4 |
| 227.4.111 | 291000 | 94 | - | - | - - | - - | See details in Table XIX | | | Cr2Cu1S4 |
| 227.4.112 | 290999 | 94 | - | - | - - | - - | See details in Table XIX | | | Cr2Cu1S4 |
| 227.4.113 | 246081 | 94 | - | - | - - | - - | See details in Table XIX | | | Cr2Cu1O4 |
| 227.4.114 | 63165 | 102 | - | - | - - | - - | See details in Table XIX | | | Co3O4 |
| 227.4.115 | 63164 | 102 | - | - | - - | - - | See details in Table XIX | | | Co3O4 |
| 227.4.116 | 36256 | 102 | - | - | - - | - - | See details in Table XIX | | | Co3O4 |
| 227.4.117 | 290808 | 90 | - | - | - - | - - | See details in Table XIX | | | Co1Cr2S4 |
| 227.4.118 | 180008 | 160 | - | - | - - | - - | See details in Table XIX | | | Cd2O7Tc2 |
| 227.4.119 | 41112 | 168 | - | - | - - | - - | See details in Table XIX | | | Cd1S4Yb2 |
| 227.4.120 | 37409 | 168 | - | - | - - | - - | See details in Table XIX | | | Cd1S4Yb2 |
| 227.4.121 | 156409 | 124 | - | - | - - | - - | See details in Table XIX | | | Ca2O7Ru2 |
| 227.4.122 | 50048 | 114 | - | - | - - | - - | See details in Table XIX | | | Ca2F1O6Ta2 |
| 227.4.123 | 55069 | 38 | - | - | - - | - - | See details in Table XIX | | | Bi2Rb1 |
| 227.4.124 | 55068 | 38 | - | - | - - | - - | See details in Table XIX | | | Bi2K1 |
| 227.4.125 | 237820 | 140 | - | - | - - | - - | See details in Table XIX | | | Bi2Ir2O7 |
| 227.4.126 | 55070 | 38 | - | - | - - | - - | See details in Table XIX | | | Bi2Cs1 |
| 227.4.127 | 108001 | 46 | - | - | - - | - - | See details in Table XIX | | | Au2Na1 |
| 227.4.128 | 57911 | 54 | - | - | - - | - - | See details in Table XIX | | | Al2Ho1 |
| 227.4.129 | 290133 | 78 | - | - | - - | - - | See details in Table XIX | | | Al2Co1O4 |
| 227.4.130 | 170868 | 80 | - | - | - - | - - | See details in Table XIX | | | Ag1In2Na3 |

Continued on next page

**TABLE XIV– continued**

| TQC | ICSD | NoE | DG | IndG | SubSG (w/ T) | cl | SubSG (w/o T) | cl | Max. Sub. SG | yes or no | Max. Sub. SG | yes or no | SIs (w/ T) | Stoich. Formula |
|---|---|---|---|---|---|---|---|---|---|---|---|---|---|---|
| 227.4.131 | 71695 | 94 | - | - | - | - | - | - | See details in Table XIX | | | | | Ag1Cr2Te4 |
| 229.1.1 | 26270 | 112 | 0.0280 | -0.0127 | 2,82 | 1 | 2,82 | 1 | - | - | - | - | z4=1<br>z2=1<br>z8=1 | As7Re3 |
| 229.3.1 | 249996 | 88 | - | - | - | - | - | - | 121 | no(82) | 0 | - | z2=1 | Fe1Ga12Y4 |
| | | | | | | | | | 119 | no(82) | 0 | - | z2=1 | |
| | | | | | | | | | 71 | no(2) | 0 | - | z4=2 | |
| | | | | | | | | | 69 | no(2) | 0 | - | z4=2 | |
| 229.3.2 | 247295 | 160 | - | - | - | - | - | - | 217 | yes | 0 | - | - | Ce4Rh12Si1 |
| | | | | | | | | | 204 | no(2) | 0 | - | z4=1 | |
| | | | | | | | | | 166 | no(2) | 0 | - | z4=1 | |
| | | | | | | | | | 119 | yes | 0 | - | - | |
| | | | | | | | | | 69 | no(2) | 0 | - | z4=1 | |
| 229.3.3 | 417049 | 256 | - | - | - | - | - | - | 217 | no(82) | 0 | - | z2=1 | Cd4Pd13Pr6 |
| | | | | | | | | | 204 | yes | 0 | - | - | |
| | | | | | | | | | 166 | yes | 0 | - | - | |
| | | | | | | | | | 119 | no(82) | 0 | - | z2=1 | |
| | | | | | | | | | 69 | yes | 0 | - | - | |
| 229.4.1 | 426405 | 182 | - | - | - | - | - | - | See details in Table XIX | | | | | Ru7Sb6Yb4 |
| 229.4.2 | 54511 | 104 | - | - | - | - | - | - | See details in Table XIX | | | | | Ru3Sn7 |
| 229.4.3 | 54510 | 104 | - | - | - | - | - | - | See details in Table XIX | | | | | Ru3Sn7 |
| 229.4.4 | 24303 | 106 | - | - | - | - | - | - | See details in Table XIX | | | | | Mo3Sb7 |
| 229.4.5 | 164762 | 106 | - | - | - | - | - | - | See details in Table XIX | | | | | Mo3Sb7 |
| 229.4.6 | 173131 | 6 | - | - | - | - | - | - | See details in Table XIX | | | | | Mo1 |
| 229.4.7 | 173127 | 6 | - | - | - | - | - | - | See details in Table XIX | | | | | Mo1 |
| 229.4.8 | 428097 | 244 | - | - | - | - | - | - | See details in Table XIX | | | | | La6Pd13Zn4 |
| 229.4.9 | 412776 | 110 | - | - | - | - | - | - | See details in Table XIX | | | | | Ir3Sn7 |
| 229.4.10 | 409596 | 102 | - | - | - | - | - | - | See details in Table XIX | | | | | In7Pt3 |
| 229.4.11 | 408314 | 102 | - | - | - | - | - | - | See details in Table XIX | | | | | In7Pd3 |
| 229.4.12 | 408313 | 102 | - | - | - | - | - | - | See details in Table XIX | | | | | Ga7Ni3 |
| 229.4.13 | 153909 | 90 | - | - | - | - | - | - | See details in Table XIX | | | | | Ga12Pt1Y4 |
| 229.4.14 | 249997 | 120 | - | - | - | - | - | - | See details in Table XIX | | | | | Fe1Ga12Tb4 |
| 229.4.15 | 426404 | 174 | - | - | - | - | - | - | See details in Table XIX | | | | | Er4Ru7Sb6 |
| 229.4.16 | 248000 | 132 | - | - | - | - | - | - | See details in Table XIX | | | | | Er4Fe1Ga12 |
| 229.4.17 | 421178 | 244 | - | - | - | - | - | - | See details in Table XIX | | | | | Cd4La6Pd13 |
| 229.4.18 | 195760 | 94 | - | - | - | - | - | - | See details in Table XIX | | | | | Ca5Ge6Pd6 |
| 229.4.19 | 400465 | 108 | - | - | - | - | - | - | See details in Table XIX | | | | | B3Eu4Li1N6 |
| 229.4.20 | 90326 | 142 | - | - | - | - | - | - | See details in Table XIX | | | | | As6Ru7U4 |

The End

TABLE XV: Topological class of materials, per space group (see Appendix[H] for details). For all the topological gapped materials (SEBR and NLC) in each space group in Table[XIV], we group together all the materials whose representations at high symmetry points differ from each other by adding or subtracting double-group EBR (atomic limit representations) ***with*** time-reversal. If compounds only differed by EBRs, they belong to the same strong-topological class ("cl"), ***with*** time-reversal. We list the symmetry group, topological class, TQC number, ICSD number, Stoichiometric Formula and minimal topological symmetry subgroup (this having been obtained in column 6 of Table[XIV]). All the compounds in the same topological class have the same symmetry indicators (SIs). Compounds in different classes will have at least one different value of the SIs. See the list of tables.

| SG | cl | TQC | ICSD | Stoich. Formula | SubSG |
|---|---|---|---|---|---|
| 2 | 1 | 2.2.36 | 79719 | Fe2O14P4Sr1 | 2 |
| | | 2.2.64 | 92165 | Cl2Cu3O6Se2 | 2 |
| | | 2.2.67 | 410649 | Cl18Mo2P2 | 2 |
| | | 2.2.91 | 86400 | As3Ca1 | 2 |
| | | 2.2.92 | 193 | As3Ca1 | 2 |
| | | 2.2.100 | 51507 | Ag1F6Pd1 | 2 |
| 2 | 2 | 2.2.3 | 66584 | O9V5 | 2 |
| | | 2.2.38 | 171431 | Cu4Mo6Se8 | 2 |
| | | 2.2.65 | 427665 | Cl1O4Tm1W1 | 2 |
| | | 2.2.96 | 428633 | As1Bi1Co1O5 | 2 |
| | | 2.2.99 | 421341 | Ag1O4S1 | 2 |
| 2 | 3 | 2.2.15 | 65941 | Mo6S8Sr1 | 2 |
| | | 2.2.17 | 61050 | Mn3Na2O7 | 2 |
| | | 2.2.18 | 59328 | Mn1O7Te1V2 | 2 |
| | | 2.2.34 | 194210 | H12Cl4N4Pt2 | 2 |
| | | 2.2.57 | 73191 | Co2O4Rb5 | 2 |
| | | 2.2.58 | 73190 | Co2O4Rb5 | 2 |
| | | 2.2.63 | 431558 | Cl3Cu3O3Te1 | 2 |
| | | 2.2.84 | 85490 | Ba1Mo6S8 | 2 |
| | | 2.2.85 | 65775 | Ba1Mo6S8 | 2 |
| | | 2.2.98 | 95832 | Ag2F5 | 2 |
| 2 | 4 | 2.2.35 | 240179 | Ga5Rh3 | 2 |
| | | 2.2.94 | 31787 | As1F6N2S3 | 2 |
| 2 | 5 | 2.2.37 | 72390 | Fe1N2Sr2 | 2 |
| | | 2.2.93 | 404850 | As2Cu4O9 | 2 |
| 2 | 6 | 2.2.2 | 41017 | P3Tc2 | 2 |
| | | 2.2.50 | 75006 | Cu1Gd1O8W2 | 2 |
| | | 2.2.90 | 66662 | As3Tc2 | 2 |
| 2 | 7 | 2.2.27 | 8063 | H1As1O5Zn2 | 2 |
| | | 2.2.28 | 37436 | H1As1O5Zn2 | 2 |
| | | 2.2.29 | 34341 | H1As1O5Zn2 | 2 |
| | | 2.2.79 | 196024 | C1Mn1O3 | 2 |
| | | 2.2.89 | 24284 | B2Co2O5 | 2 |
| 2 | 8 | 2.2.49 | 62367 | Cu1I3Na1O9 | 2 |
| | | 2.2.88 | 96562 | B2Co2O5 | 2 |
| 2 | 9 | 2.2.19 | 431529 | Lu1Na1P2S6 | 2 |
| | | 2.2.23 | 167608 | H1Fe1Li1O5P1 | 2 |
| | | 2.2.39 | 190444 | Cu3Mo2O9 | 2 |
| | | 2.2.59 | 167202 | Co1F1Li1O4S1 | 2 |
| | | 2.2.87 | 192060 | Ba1Cu1O7P2 | 2 |
| 2 | 10 | 2.2.8 | 82204 | O4Pb1Pt2 | 2 |
| | | 2.2.9 | 59657 | O4Pb1Pt2 | 2 |
| | | 2.2.21 | 39887 | H2Cu1O5Se1 | 2 |
| | | 2.2.60 | 410945 | Cl9Te1W1 | 2 |
| | | 2.2.81 | 39784 | Br3Ti1 | 2 |
| | | 2.2.95 | 92916 | As1Bi1Ni1O5 | 2 |
| 2 | 11 | 2.2.10 | 281617 | Na1O6Si2Ti1 | 2 |
| | | 2.2.11 | 281616 | Na1O6Si2Ti1 | 2 |

Continued on next column

**TABLE XV– continued**

| SG | cl | TQC | ICSD | Stoich. Formula | SubSG |
|---|---|---|---|---|---|
| | | 2.2.12 | 281615 | Na1O6Si2Ti1 | 2 |
| | | 2.2.13 | 281614 | Na1O6Si2Ti1 | 2 |
| | | 2.2.14 | 281613 | Na1O6Si2Ti1 | 2 |
| | | 2.2.41 | 69759 | Cu1O4W1 | 2 |
| | | 2.2.43 | 24339 | Cu1O4W1 | 2 |
| | | 2.2.44 | 182753 | Cu1O4W1 | 2 |
| | | 2.2.45 | 182752 | Cu1O4W1 | 2 |
| | | 2.2.46 | 182750 | Cu1O4W1 | 2 |
| | | 2.2.47 | 16009 | Cu1O4W1 | 2 |
| | | 2.2.80 | 423462 | Br5W1 | 2 |
| | | 2.2.101 | 174407 | Ag1Cu1F3 | 2 |
| 2 | 12 | 2.2.4 | 80363 | O9P3Ru1 | 2 |
| | | 2.2.5 | 2212 | O7V4 | 2 |
| | | 2.2.22 | 166101 | H2Cu1O5S1 | 2 |
| | | 2.2.73 | 62436 | C2H8Br2N4Se2Te1 | 2 |
| 2 | 13 | 2.2.72 | 62435 | C2H8Cl2N4Se2Te1 | 2 |
| 2 | 14 | 2.2.26 | 60937 | H1As4Cu3Na5O16 | 2 |
| | | 2.2.30 | 8065 | H1As1Ba1Cr2O10 | 2 |
| | | 2.2.33 | 35417 | H12Cl8Hg3Mg1O6 | 2 |
| | | 2.2.40 | 84558 | Cu1O4W1 | 2 |
| | | 2.2.53 | 200533 | Cu1F1Nb1O3 | 2 |
| | | 2.2.70 | 42698 | C4H16Cl2N8Se4Te1 | 2 |
| 2 | 15 | 2.2.56 | 62676 | Cr1Na4O4 | 2 |
| | | 2.2.69 | 74479 | Ca1P3 | 2 |
| 2 | 16 | 2.2.16 | 237587 | Mo3S4 | 2 |
| | | 2.2.68 | 408774 | Cl14Mo4O4Tl2 | 2 |
| 2 | 17 | 2.2.6 | 19012 | O7V4 | 2 |
| | | 2.2.55 | 150974 | Cs1Hg1 | 2 |
| | | 2.2.62 | 84620 | Cl5Mo1 | 2 |
| | | 2.2.76 | 196027 | C1Mn1O3 | 2 |
| | | 2.2.82 | 51779 | Bi3O7Sb1 | 2 |
| | | 2.2.86 | 237763 | Ba1Fe2O8P2 | 2 |
| 2 | 18 | 2.2.61 | 410101 | Cl7Mo1O1Se1 | 2 |
| | | 2.2.66 | 50441 | Cl18Mo2P2 | 2 |
| 2 | 19 | 2.2.1 | 202564 | S1Ta6 | 2 |
| | | 2.2.7 | 61430 | O6Pb3S1 | 2 |
| | | 2.2.25 | 416823 | H1Co1O5P1Zn1 | 2 |
| | | 2.2.32 | 417328 | H12Cs4O6P2S6 | 2 |
| | | 2.2.54 | 62000 | Cs1Hg1 | 2 |
| 2 | 20 | 2.2.52 | 69655 | Cu1F3Na1 | 2 |
| 2 | 21 | 2.2.42 | 4189 | Cu1O4W1 | 2 |
| | | 2.2.48 | 39439 | Cu1Mo1O4 | 2 |
| | | 2.2.71 | 170721 | C3H6As1N1 | 2 |
| 2 | 22 | 2.2.31 | 73197 | H18Al2Cl6N6 | 2 |
| | | 2.2.75 | 196028 | C1Mn1O3 | 2 |
| | | 2.2.77 | 196026 | C1Mn1O3 | 2 |
| | | 2.2.78 | 196025 | C1Mn1O3 | 2 |
| 2 | 24 | 2.2.51 | 64660 | Cu1F6Pt1 | 2 |
| | | 2.2.97 | 422342 | Al2Fe3Si3 | 2 |
| 2 | 25 | 2.2.20 | 419880 | K1O6Os2 | 2 |
| | | 2.2.24 | 62742 | H1Cu2K3O14P4 | 2 |
| 2 | 26 | 2.2.83 | 97768 | Bi1Ni1O3 | 2 |
| 2 | 27 | 2.2.74 | 186554 | C2H3Na1O5 | 2 |
| 10 | 30 | 10.2.1 | 417035 | Co2Gd4Mg3 | 2 |
| 11 | 1 | 11.2.2 | 424108 | Sn2Sr1Zn2 | 2 |
| | | 11.2.14 | 99271 | Ni3Te2 | 2 |
| | | 11.2.48 | 166218 | H2O4P1Rb1 | 2 |
| | | 11.2.55 | 260832 | H2Cl1O2Tb1 | 2 |
| | | 11.2.57 | 6164 | H1O2Tb1 | 2 |
| | | 11.2.58 | 162390 | H1Nd1O2 | 2 |

Continued on next column

**TABLE XV– continued**

| SG | cl | TQC | ICSD | Stoich. Formula | SubSG |
|---|---|---|---|---|---|
| | | 11.2.62 | 156004 | H12Cl8Hg3N2O2 | 2 |
| | | 11.2.70 | 414338 | Cu1Nb1Te2 | 2 |
| | | 11.2.75 | 65346 | Cl1Fe1Li1Mo1O4 | 2 |
| | | 11.2.76 | 54553 | Ce1In2Pt2 | 2 |
| | | 11.2.92 | 410884 | Au2Eu2Sn5 | 2 |
| | | 11.2.93 | 416298 | Au2Ba1Sb2 | 2 |
| | | 11.2.96 | 104626 | Al1Ce3 | 2 |
| | | 11.2.97 | 57494 | Al1Au1 | 2 |
| | | 11.2.98 | 104649 | Al1Au1 | 2 |
| 11 | 2 | 11.2.9 | 52066 | Rh4Si5 | 2 |
| | | 11.2.39 | 437 | H6Br3N1O1Sn1 | 2 |
| | | 11.2.41 | 15199 | H5Fe1O7S1 | 2 |
| | | 11.2.67 | 94536 | Eu1K3O8P2 | 2 |
| | | 11.2.74 | 415798 | Cl2Co2O3Te1 | 2 |
| | | 11.2.79 | 110059 | C4H12Cd1Cl3N1 | 2 |
| | | 11.2.83 | 42679 | Bi1 | 2 |
| | | 11.2.94 | 902 | As1Cl1Co2O4 | 2 |
| 11 | 3 | 11.2.8 | 260980 | S3Tm2 | 2 |
| | | 11.2.12 | 43283 | Pt6Si5 | 2 |
| | | 11.2.35 | 14349 | Mo1Te2 | 2 |
| | | 11.2.49 | 27406 | H2N1Rb1 | 2 |
| | | 11.2.61 | 54147 | H14Cd1Cu4O18S2 | 2 |
| | | 11.2.82 | 416300 | Bi2Pd2Sr1 | 2 |
| | | 11.2.89 | 416299 | Ba1Bi2Pd2 | 2 |
| | | 11.2.91 | 261042 | Au2In4Yb1 | 2 |
| 11 | 4 | 11.2.5 | 52307 | Sb2Sr1 | 2 |
| | | 11.2.13 | 105804 | Pt3Ta1 | 2 |
| | | 11.2.28 | 66316 | Nd1Ru1Si2 | 2 |
| | | 11.2.30 | 54283 | Nb1Pt3 | 2 |
| | | 11.2.36 | 59708 | Li4S4Si1 | 2 |
| | | 11.2.60 | 201682 | H14Cl2N5O2Tc1 | 2 |
| | | 11.2.63 | 63219 | H10Mo3N2O11 | 2 |
| | | 11.2.71 | 69990 | Cu1La2O6Sn1 | 2 |
| | | 11.2.90 | 413519 | Ba1.5Sb3 | 2 |
| 11 | 5 | 11.2.27 | 105414 | Ni1Ti1 | 2 |
| | | 11.2.33 | 62486 | Mo2S3 | 2 |
| | | 11.2.72 | 108896 | Co1O3Sr1 | 2 |
| | | 11.2.88 | 167615 | Ba1Er2O10Si3 | 2 |
| 11 | 6 | 11.2.3 | 42982 | Se3Ta2 | 2 |
| | | 11.2.65 | 180364 | Ga5Pt1 | 2 |
| | | 11.2.87 | 429621 | Ba1Ho4O17Si5 | 2 |
| 11 | 7 | 11.2.6 | 171417 | S3U1 | 2 |
| | | 11.2.15 | 54304 | Ni3Ta1 | 2 |
| | | 11.2.31 | 93368 | Na1O8P1W2 | 2 |
| | | 11.2.47 | 153707 | H3Cl1Cu2O3 | 2 |
| | | 11.2.50 | 169093 | H2K1O4P1 | 2 |
| | | 11.2.52 | 36300 | H2Cs1O4P1 | 2 |
| | | 11.2.53 | 20449 | H2Cl1O2Y1 | 2 |
| | | 11.2.68 | 190487 | D2O4P1Rb1 | 2 |
| | | 11.2.80 | 110260 | C3H10Cl1N1 | 2 |
| | | 11.2.86 | 194985 | Ba1Nd2O10Ti3 | 2 |
| 11 | 8 | 11.2.37 | 68366 | In2S7Sn3 | 2 |
| | | 11.2.40 | 24075 | H5In1O7S1 | 2 |
| | | 11.2.42 | 262728 | H4Co2O10P2Pb1 | 2 |
| | | 11.2.51 | 6163 | H2Ga11Na1O18 | 2 |
| | | 11.2.84 | 81819 | Ba3Ge2N2 | 2 |
| | | 11.2.85 | 409517 | Ba1Sb2 | 2 |
| 11 | 9 | 11.2.16 | 245081 | Ni1Ti1 | 2 |
| | | 11.2.17 | 240195 | Ni1Ti1 | 2 |
| | | 11.2.18 | 166372 | Ni1Ti1 | 2 |

Continued on next column

**TABLE XV– continued**

| SG | cl | TQC | ICSD | Stoich. Formula | SubSG |
|---|---|---|---|---|---|
| | | 11.2.19 | 166367 | Ni1Ti1 | 2 |
| | | 11.2.20 | 164155 | Ni1Ti1 | 2 |
| | | 11.2.21 | 164154 | Ni1Ti1 | 2 |
| | | 11.2.22 | 164153 | Ni1Ti1 | 2 |
| | | 11.2.23 | 164152 | Ni1Ti1 | 2 |
| | | 11.2.24 | 164151 | Ni1Ti1 | 2 |
| | | 11.2.25 | 157983 | Ni1Ti1 | 2 |
| | | 11.2.26 | 105415 | Ni1Ti1 | 2 |
| | | 11.2.64 | 418027 | Ge1Hf1Pd1 | 2 |
| | | 11.2.73 | 411140 | Cl7Cs1Ti2 | 2 |
| | | 11.2.81 | 20741 | C2H8N4O7Se1U1 | 2 |
| 11 | 10 | 11.2.7 | 430 | S3Tm2 | 2 |
| | | 11.2.11 | 31692 | Pu1 | 2 |
| | | 11.2.46 | 70101 | H3Cl1Cu2O3 | 2 |
| | | 11.2.54 | 260836 | H2Cl1O2Tm1 | 2 |
| | | 11.2.56 | 260834 | H2Cl1Ho1O2 | 2 |
| | | 11.2.59 | 424942 | H1F4Mn1Na2O3P1 | 2 |
| | | 11.2.69 | 67816 | Cu4K1Mg1O13V3 | 2 |
| | | 11.2.78 | 62428 | C4H2As1F17N2 | 2 |
| 11 | 11 | 11.2.10 | 24355 | Rh4Si5 | 2 |
| | | 11.2.77 | 862 | Ca1Sb2 | 2 |
| 11 | 12 | 11.2.4 | 43213 | Se3Ta1 | 2 |
| | | 11.2.45 | 62901 | H3Dy3Ge1O8 | 2 |
| | | 11.2.95 | 43456 | Am1 | 2 |
| 11 | 13 | 11.2.32 | 73453 | Mo2S3 | 2 |
| | | 11.2.34 | 16261 | Mo2.06S3 | 2 |
| | | 11.2.44 | 151476 | H4Ag2K4O2S9Sn3 | 2 |
| 11 | 14 | 11.2.1 | 42076 | Te3Zr1 | 2 |
| | | 11.2.29 | 42981 | Nb2Se3 | 2 |
| | | 11.2.38 | 26525 | H8Ba1O7Pt1 | 2 |
| | | 11.2.43 | 60659 | H4Ba1O3 | 2 |
| 11 | 15 | 11.2.66 | 89343 | F3K1Mn1 | 2 |
| 12 | 1 | 12.1.7 | 42077 | P2V1 | 2 |
| | | 12.1.13 | 646547 | Ni1Se4Ti2 | 2 |
| | | 12.1.14 | 646544 | Ni1Se4Ti2 | 2 |
| | | 12.1.18 | 646388 | Ni1S4Ti2 | 2 |
| | | 12.1.23 | 26183 | Nb2Se1 | 2 |
| | | 12.1.27 | 190180 | Nb1Sb2 | 2 |
| | | 12.1.28 | 18144 | Nb1Sb2 | 2 |
| | | 12.1.47 | 165326 | Li1Mo1O2 | 2 |
| | | 12.1.51 | 407248 | I3La5Si5 | 2 |
| | | 12.1.62 | 637671 | Ge2Re1 | 2 |
| | | 12.1.64 | 430563 | Fe5O7 | 2 |
| | | 12.1.68 | 240050 | Fe1Zn13 | 2 |
| | | 12.1.69 | 163222 | Fe1Zn13 | 2 |
| | | 12.1.74 | 633332 | Fe0.5S2Ti1 | 2 |
| | | 12.1.77 | 151123 | Er2O7Si2 | 2 |
| | | 12.1.86 | 26977 | Cr7Se8 | 2 |
| | | 12.1.97 | 74296 | Cr2S8Tl1V3 | 2 |
| | | 12.1.123 | 73096 | Cl1Dy1O3Ti1 | 2 |
| | | 12.1.124 | 424719 | Ce4Ge6Mg7 | 2 |
| | | 12.1.146 | 79394 | C1Fe2Si2Tm2 | 2 |
| | | 12.1.147 | 195010 | C1Fe2Gd2Si2 | 2 |
| | | 12.1.151 | 416907 | C10Cl4La8 | 2 |
| | | 12.1.159 | 65614 | Ba3Cu2O10Pt1Y2 | 2 |
| | | 12.1.163 | 238228 | Ba1Te4U1 | 2 |
| | | 12.1.170 | 170618 | B4La2Ni5 | 2 |
| | | 12.1.171 | 405291 | B4Br5La4 | 2 |
| | | 12.1.173 | 261932 | B2Ni3Zn1 | 2 |
| | | 12.1.174 | 107555 | Au8Ga1K4 | 2 |

Continued on next column

**TABLE XV– continued**

| SG | cl | TQC | ICSD | Stoich. Formula | SubSG |
|---|---|---|---|---|---|
| | | 12.1.191 | 75454 | As2Ba2Mn2O1 | 2 |
| | | 12.1.195 | 27569 | Al4Cl51Zr12 | 2 |
| | | 12.1.197 | 165239 | Al14Ca13 | 2 |
| 12 | 2 | 12.1.8 | 56694 | O5Ti5 | 2 |
| | | 12.1.11 | 50409 | O13V6 | 2 |
| | | 12.1.12 | 15028 | O13V6 | 2 |
| | | 12.1.22 | 41238 | Nb7P4 | 2 |
| | | 12.1.30 | 29383 | N3Na1 | 2 |
| | | 12.1.31 | 29382 | N3Na1 | 2 |
| | | 12.1.32 | 29381 | N3Na1 | 2 |
| | | 12.1.33 | 29380 | N3Na1 | 2 |
| | | 12.1.34 | 29379 | N3Na1 | 2 |
| | | 12.1.35 | 29378 | N3Na1 | 2 |
| | | 12.1.36 | 29377 | N3Na1 | 2 |
| | | 12.1.37 | 29376 | N3Na1 | 2 |
| | | 12.1.38 | 29375 | N3Na1 | 2 |
| | | 12.1.39 | 29374 | N3Na1 | 2 |
| | | 12.1.40 | 29373 | N3Na1 | 2 |
| | | 12.1.41 | 29372 | N3Na1 | 2 |
| | | 12.1.42 | 29371 | N3Na1 | 2 |
| | | 12.1.43 | 29370 | N3Na1 | 2 |
| | | 12.1.45 | 78328 | Mn1Mo1O4 | 2 |
| | | 12.1.48 | 49021 | K1Li1Mn1O2 | 2 |
| | | 12.1.52 | 407247 | I3La4Si4 | 2 |
| | | 12.1.85 | 71624 | Cs2Ir1Na2O4 | 2 |
| | | 12.1.126 | 20762 | C6H5Cl2Tl1 | 2 |
| | | 12.1.144 | 618692 | C3.5Pu2W2 | 2 |
| | | 12.1.149 | 391256 | C1Cl2Eu2N2 | 2 |
| | | 12.1.154 | 261457 | Bi2Pd3Se2 | 2 |
| | | 12.1.161 | 173685 | Ba3Cd2Sb4 | 2 |
| | | 12.1.166 | 173848 | Ba1Ir1O3 | 2 |
| | | 12.1.168 | 61499 | Ba1Bi1O3 | 2 |
| | | 12.1.175 | 425401 | Au1Ce2O1P2 | 2 |
| | | 12.1.176 | 424760 | As4Ba3Zn2 | 2 |
| | | 12.1.177 | 424761 | As4Ba3Cd2 | 2 |
| | | 12.1.187 | 69003 | As2Mn2O7 | 2 |
| | | 12.1.188 | 261143 | As2Mn2O7 | 2 |
| | | 12.1.196 | 40332 | Al1Cu1 | 2 |
| 12 | 3 | 12.1.1 | 187638 | Y1 | 2 |
| | | 12.1.5 | 601347 | Se4Ti3 | 2 |
| | | 12.1.46 | 239971 | Li2Mn1O3 | 2 |
| | | 12.1.55 | 23361 | H3N1Na2S3 | 2 |
| | | 12.1.57 | 422475 | H3Co2Mo2Na1O10 | 2 |
| | | 12.1.76 | 106627 | Eu1In4 | 2 |
| | | 12.1.78 | 419435 | Er2O3 | 2 |
| | | 12.1.83 | 418227 | Dy1P2Pt8 | 2 |
| | | 12.1.92 | 26976 | Cr3Se4 | 2 |
| | | 12.1.102 | 251718 | Cr1Se2 | 2 |
| | | 12.1.103 | 75420 | Cr1S2 | 2 |
| | | 12.1.107 | 821 | Cr1P1S4 | 2 |
| | | 12.1.108 | 56835 | Cr1P1S4 | 2 |
| | | 12.1.111 | 66006 | Co2Na7O6Rb1 | 2 |
| | | 12.1.120 | 427661 | Cl1Gd1O4W1 | 2 |
| | | 12.1.121 | 35292 | Cl1Gd1O4W1 | 2 |
| | | 12.1.122 | 420055 | Cl1Gd1Mo1O4 | 2 |
| | | 12.1.125 | 100827 | Ca2Fe7O11 | 2 |
| | | 12.1.145 | 262897 | C1H3La1O6P2 | 2 |
| | | 12.1.157 | 1559 | Bi1I1 | 2 |
| | | 12.1.169 | 63510 | B4Nd2Ni5 | 2 |
| | | 12.1.182 | 610428 | As3Eu1 | 2 |
| | | 12.1.198 | 429000 | Ag1Ba1S3Tb1 | 2 |

Continued on next column

**TABLE XV– continued**

| SG | cl | TQC | ICSD | Stoich. Formula | SubSG |
|---|---|---|---|---|---|
| 12 | 4 | 12.1.4 | 37122 | Se8Tl1V5 | 2 |
| | | 12.1.15 | 646533 | Ni1Se4Si2 | 2 |
| | | 12.1.17 | 646389 | Ni1S8Ti4 | 2 |
| | | 12.1.56 | 62646 | H3Cu2F2K1O8P2 | 2 |
| | | 12.1.58 | 24759 | H2I2O12Sb8 | 2 |
| | | 12.1.59 | 24760 | H2Br2O12Sb8 | 2 |
| | | 12.1.60 | 16023 | H12N2O16Zn5 | 2 |
| | | 12.1.67 | 150568 | Fe3Se4 | 2 |
| | | 12.1.70 | 633326 | Fe1S8Ti4 | 2 |
| | | 12.1.71 | 53534 | Fe1S8Ti4 | 2 |
| | | 12.1.72 | 42563 | Fe1S4Ti2 | 2 |
| | | 12.1.73 | 103658 | Fe17Th2 | 2 |
| | | 12.1.90 | 626716 | Cr3Se4 | 2 |
| | | 12.1.91 | 626712 | Cr3Se4 | 2 |
| | | 12.1.93 | 81886 | Cr3S4 | 2 |
| | | 12.1.95 | 192944 | Cr3S4 | 2 |
| | | 12.1.99 | 625968 | Cr2Fe1Se4 | 2 |
| | | 12.1.100 | 155121 | Cr2Fe1Se4 | 2 |
| | | 12.1.117 | 183325 | Co1Gd2Si2 | 2 |
| | | 12.1.118 | 184277 | Co1Gd2Ge2 | 2 |
| | | 12.1.119 | 183326 | Co1Gd2Ge2 | 2 |
| | | 12.1.128 | 39297 | C5Nb6 | 2 |
| | | 12.1.129 | 20695 | C5Nb6 | 2 |
| | | 12.1.135 | 88513 | C3Ho2Mo2 | 2 |
| | | 12.1.136 | 88512 | C3Ho2Mo2 | 2 |
| | | 12.1.137 | 88511 | C3Ho2Mo2 | 2 |
| | | 12.1.143 | 417827 | C3Ce2Mo2 | 2 |
| | | 12.1.152 | 617015 | Bi2Rh3Si2 | 2 |
| | | 12.1.153 | 420725 | Bi2Rh3S2 | 2 |
| | | 12.1.155 | 251720 | Bi1Se1 | 2 |
| | | 12.1.156 | 251717 | Bi1Se1 | 2 |
| | | 12.1.165 | 154011 | Ba1Mn2O6Y1 | 2 |
| | | 12.1.167 | 151699 | Ba1Er2F8 | 2 |
| | | 12.1.185 | 69002 | As2Ni2O7 | 2 |
| | | 12.1.192 | 68906 | As1F6N1O2 | 2 |
| | | 12.1.193 | 418388 | Al6O1Si8Sr13 | 2 |
| | | 12.1.194 | 409319 | Al4Na4P12Sr8 | 2 |
| 12 | 5 | 12.1.3 | 41047 | Ta2Te3 | 2 |
| | | 12.1.9 | 34417 | O2V1 | 2 |
| | | 12.1.10 | 34416 | O2V1 | 2 |
| | | 12.1.19 | 646321 | Ni1Rh2Se4 | 2 |
| | | 12.1.20 | 646318 | Ni1Rh2Se4 | 2 |
| | | 12.1.21 | 15214 | Ni1Rh2Se4 | 2 |
| | | 12.1.24 | 77752 | Nb1Sb2 | 2 |
| | | 12.1.25 | 645350 | Nb1Sb2 | 2 |
| | | 12.1.26 | 57430 | Nb1Sb2 | 2 |
| | | 12.1.29 | 645169 | Nb1P2 | 2 |
| | | 12.1.54 | 20755 | H4Cl2Ni1O2 | 2 |
| | | 12.1.61 | 262220 | Ge6La4Mg7 | 2 |
| | | 12.1.65 | 633477 | Fe3Se4 | 2 |
| | | 12.1.66 | 24560 | Fe3Se4 | 2 |
| | | 12.1.75 | 192506 | F5Fe1K2 | 2 |
| | | 12.1.94 | 626598 | Cr3S4 | 2 |
| | | 12.1.96 | 16722 | Cr3S4 | 2 |
| | | 12.1.101 | 625931 | Cr2Fe1S4 | 2 |
| | | 12.1.104 | 626579 | Cr1Rh2Te4 | 2 |
| | | 12.1.105 | 626577 | Cr1Rh2Se4 | 2 |
| | | 12.1.106 | 626541 | Cr1Pr2Se4 | 2 |
| | | 12.1.109 | 173875 | Cr1O4Re1 | 2 |
| | | 12.1.110 | 624485 | Co2Ni1Se4 | 2 |
| | | 12.1.112 | 2371 | Co1Si2Zr2 | 2 |

Continued on next column

**TABLE XV– continued**

| SG | cl | TQC | ICSD | Stoich. Formula | SubSG |
|---|---|---|---|---|---|
| | | 12.1.131 | 619030 | C3U2W2 | 2 |
| | | 12.1.132 | 618707 | C3Re2U2 | 2 |
| | | 12.1.133 | 618357 | C3Mo2U2 | 2 |
| | | 12.1.134 | 618223 | C3Lu2Mo2 | 2 |
| | | 12.1.138 | 602482 | C3Cr2Yb2 | 2 |
| | | 12.1.139 | 617537 | C3Cr2Tb2 | 2 |
| | | 12.1.140 | 617514 | C3Cr2Lu2 | 2 |
| | | 12.1.141 | 617492 | C3Cr2Dy2 | 2 |
| | | 12.1.142 | 43018 | C3Cr2Dy2 | 2 |
| | | 12.1.148 | 67467 | C1Dy2Fe2Si2 | 2 |
| | | 12.1.150 | 603998 | C1.66Mo1U1 | 2 |
| | | 12.1.160 | 69569 | Ba3Cu2Er2O10Pt1 | 2 |
| | | 12.1.162 | 423268 | Ba2Ge4Ni5 | 2 |
| | | 12.1.178 | 43185 | As3W2 | 2 |
| | | 12.1.180 | 43184 | As3Mo2 | 2 |
| | | 12.1.181 | 41015 | As3Mo2 | 2 |
| | | 12.1.183 | 611574 | As2V1 | 2 |
| | | 12.1.184 | 44069 | As2Ta1 | 2 |
| | | 12.1.186 | 81218 | As2Nb1 | 2 |
| | | 12.1.189 | 79441 | As2Gd1Pd3 | 2 |
| 12 | 6 | 12.1.6 | 48176 | Rh1Si2Y2 | 2 |
| | | 12.1.16 | 48004 | Ni1Sc3Si3 | 2 |
| | | 12.1.44 | 151588 | Mn7Na1O12 | 2 |
| | | 12.1.49 | 240135 | In4Sr1 | 2 |
| | | 12.1.50 | 240133 | In4Sr1 | 2 |
| | | 12.1.53 | 83965 | Hf2Te1 | 2 |
| | | 12.1.63 | 183378 | Gd1O6Ru1Sr2 | 2 |
| | | 12.1.79 | 180224 | Dy1Si1 | 2 |
| | | 12.1.80 | 180223 | Dy1Si1 | 2 |
| | | 12.1.81 | 180222 | Dy1Si1 | 2 |
| | | 12.1.82 | 180221 | Dy1Si1 | 2 |
| | | 12.1.87 | 626888 | Cr3Te4 | 2 |
| | | 12.1.88 | 626884 | Cr3Te4 | 2 |
| | | 12.1.89 | 626874 | Cr3Te4 | 2 |
| | | 12.1.114 | 623423 | Co1Ge1 | 2 |
| | | 12.1.115 | 43677 | Co1Ge1 | 2 |
| | | 12.1.116 | 184254 | Co1Ge1 | 2 |
| | | 12.1.158 | 1560 | Bi1Br1 | 2 |
| | | 12.1.164 | 195017 | Ba1Pt2Sb2 | 2 |
| | | 12.1.172 | 615340 | B2Rh3Y1 | 2 |
| | | 12.1.179 | 15049 | As3V4 | 2 |
| 12 | 7 | 12.1.2 | 43412 | Te4Ti3 | 2 |
| | | 12.1.84 | 421466 | Cu4Na1O12Te2Tl3 | 2 |
| | | 12.1.98 | 626640 | Cr2S4Ti1 | 2 |
| | | 12.1.113 | 624952 | Co1Sc2Si2 | 2 |
| | | 12.1.127 | 63503 | C5Nb6 | 2 |
| | | 12.1.130 | 167353 | C5Nb6 | 2 |
| | | 12.1.190 | 81775 | As2Ce1Pd3 | 2 |
| 12 | 8 | 12.2.2 | 420133 | Ru1Sc2Si2 | 2 |
| | | 12.2.20 | 30433 | Ni2O7P2 | 2 |
| | | 12.2.51 | 24430 | H3Al2O7P1 | 2 |
| | | 12.2.60 | 15044 | Fe3Se4 | 2 |
| | | 12.2.61 | 15043 | Fe3Se4 | 2 |
| | | 12.2.63 | 84205 | Fe1Sc2Si2 | 2 |
| | | 12.2.64 | 76348 | Fe1Sc2Si2 | 2 |
| | | 12.2.65 | 633455 | Fe1Sc2Si2 | 2 |
| | | 12.2.66 | 20661 | Fe1Sc2Si2 | 2 |
| | | 12.2.82 | 426133 | Cu1Ge6Yb2 | 2 |
| | | 12.2.85 | 601348 | Cr3Se4 | 2 |
| | | 12.2.97 | 151873 | Cr1Fe2Se4 | 2 |
| | | 12.2.98 | 74542 | Co2O7P2 | 2 |

Continued on next column

**TABLE XV– continued**

| SG | cl | TQC | ICSD | Stoich. Formula | SubSG |
|---|---|---|---|---|---|
| | | 12.2.108 | 417665 | Ce1Ru1Sn1 | 2 |
| | | 12.2.109 | 192243 | Ce1Ru1Sn1 | 2 |
| | | 12.2.110 | 192242 | Ce1Ru1Sn1 | 2 |
| | | 12.2.111 | 189952 | Ce1Ru1Sn1 | 2 |
| | | 12.2.112 | 189412 | Ce1Ru1Sn1 | 2 |
| | | 12.2.113 | 189411 | Ce1Ru1Sn1 | 2 |
| | | 12.2.114 | 2284 | Ce1 | 2 |
| | | 12.2.115 | 620729 | Ce1Co1Ga1 | 2 |
| | | 12.2.116 | 620726 | Ce1Co1Ga1 | 2 |
| | | 12.2.117 | 429884 | Ce1Co1Ga1 | 2 |
| | | 12.2.118 | 187960 | Ce1Co1Ga1 | 2 |
| | | 12.2.134 | 617825 | C1Fe2Si2Y2 | 2 |
| | | 12.2.186 | 69001 | As2Co2O7 | 2 |
| | | 12.2.190 | 20632 | Al1Ce1Co1 | 2 |
| | | 12.2.192 | 27743 | Ag1Er1O8W2 | 2 |
| 12 | 9 | 12.2.3 | 61353 | Pt1Se7Ta2 | 2 |
| | | 12.2.11 | 248719 | O7Si2Tm2 | 2 |
| | | 12.2.14 | 38223 | O3Rb2Te1 | 2 |
| | | 12.2.30 | 280016 | Nb2Te3 | 2 |
| | | 12.2.38 | 428765 | Hg2Sn2Sr1 | 2 |
| | | 12.2.39 | 157861 | Hg1 | 2 |
| | | 12.2.40 | 405208 | Hf7P4 | 2 |
| | | 12.2.41 | 162649 | H8Cu2Mg1O12V2 | 2 |
| | | 12.2.45 | 16130 | H4Cl8K3N1O2Ru2 | 2 |
| | | 12.2.54 | 49911 | H12Cl2N2O6Pt1 | 2 |
| | | 12.2.57 | 183379 | Gd1O6Ru1Sr2 | 2 |
| | | 12.2.58 | 635673 | Ga4Yb1 | 2 |
| | | 12.2.77 | 416629 | Eu2Li1Si3 | 2 |
| | | 12.2.78 | 425141 | Dy3O3Sb1 | 2 |
| | | 12.2.79 | 422453 | Cu2Sn2Sr1 | 2 |
| | | 12.2.80 | 182102 | Cu2Sn2Sr1 | 2 |
| | | 12.2.81 | 79536 | Cu2Na2S4Zr1 | 2 |
| | | 12.2.84 | 626873 | Cr3Te4 | 2 |
| | | 12.2.86 | 73143 | Cr3Rb1S5 | 2 |
| | | 12.2.88 | 626896 | Cr2Te4Ti1 | 2 |
| | | 12.2.103 | 420059 | Cl1Mo1O4Tb1 | 2 |
| | | 12.2.121 | 619274 | Ca1Ga4 | 2 |
| | | 12.2.122 | 58893 | Ca1Ga4 | 2 |
| | | 12.2.123 | 54994 | Ca1Ga4 | 2 |
| | | 12.2.124 | 417717 | Ca1Cu2Sn2 | 2 |
| | | 12.2.126 | 618350 | C3Mo2Tm2 | 2 |
| | | 12.2.128 | 617669 | C3Er2Mo2 | 2 |
| | | 12.2.129 | 617541 | C3Cr2Tm2 | 2 |
| | | 12.2.130 | 617494 | C3Cr2Er2 | 2 |
| | | 12.2.141 | 58840 | Bi2Pd5 | 2 |
| | | 12.2.148 | 616946 | Bi2Pd1 | 2 |
| | | 12.2.149 | 42565 | Bi2Pd1 | 2 |
| | | 12.2.153 | 429459 | Bi1Gd3O3 | 2 |
| | | 12.2.158 | 240016 | Ba3Li4Sn8 | 2 |
| | | 12.2.159 | 62390 | Ba3Cu2O10Pt1Y2 | 2 |
| | | 12.2.164 | 261206 | Ba1Tl4 | 2 |
| | | 12.2.165 | 238227 | Ba1Te4Th1 | 2 |
| | | 12.2.167 | 16634 | Ba1O5Ti2 | 2 |
| | | 12.2.168 | 83618 | Ba1Cu10P4 | 2 |
| | | 12.2.169 | 418886 | Ba11Cd6Sb12 | 2 |
| | | 12.2.181 | 429212 | Au2Ni4Si7 | 2 |
| | | 12.2.187 | 239291 | As1Pd5 | 2 |
| | | 12.2.191 | 416360 | Ag5Cl5Hg4P8 | 2 |
| 12 | 10 | 12.2.16 | 10455 | O13Sr2Ti6 | 2 |
| | | 12.2.21 | 646556 | Ni1Se4V2 | 2 |
| | | 12.2.22 | 646555 | Ni1Se4V2 | 2 |

Continued on next column

**TABLE XV– continued**

| SG | cl | TQC | ICSD | Stoich. Formula | SubSG |
|---|---|---|---|---|---|
| | | 12.2.23 | 23970 | Ni1Se4V2 | 2 |
| | | 12.2.25 | 646403 | Ni1S4V2 | 2 |
| | | 12.2.26 | 35140 | Ni1S4V2 | 2 |
| | | 12.2.44 | 201563 | H4Cl8N1O2Rb3Ru2 | 2 |
| | | 12.2.47 | 35156 | H3Li1O2 | 2 |
| | | 12.2.48 | 30223 | H3Li1O2 | 2 |
| | | 12.2.50 | 159510 | H3Li1O2 | 2 |
| | | 12.2.59 | 165189 | Ga3Sc5 | 2 |
| | | 12.2.67 | 633361 | Fe1S4V2 | 2 |
| | | 12.2.70 | 90460 | Fe1Ge1Yb1 | 2 |
| | | 12.2.71 | 631983 | Fe1Ge1 | 2 |
| | | 12.2.72 | 53458 | Fe1Ge1 | 2 |
| | | 12.2.73 | 53457 | Fe1Ge1 | 2 |
| | | 12.2.74 | 53456 | Fe1Ge1 | 2 |
| | | 12.2.75 | 53455 | Fe1Ge1 | 2 |
| | | 12.2.76 | 43055 | Fe1Ge1 | 2 |
| | | 12.2.87 | 1824 | Cr3Ga4 | 2 |
| | | 12.2.90 | 625971 | Cr2Fe1Se4 | 2 |
| | | 12.2.91 | 626655 | Cr1S4V2 | 2 |
| | | 12.2.94 | 16087 | Cr1O5Pb2 | 2 |
| | | 12.2.119 | 171243 | Ca6Cu2Sn7 | 2 |
| | | 12.2.125 | 236391 | C4Co1Sc3 | 2 |
| | | 12.2.131 | 252737 | C2Ca1 | 2 |
| | | 12.2.136 | 61141 | C10Fe22Mo12 | 2 |
| | | 12.2.137 | 65114 | C10Au2Cl8S8 | 2 |
| | | 12.2.138 | 240479 | Br4Cs2I2Pd1 | 2 |
| | | 12.2.151 | 195678 | Bi2Hg1S4 | 2 |
| | | 12.2.162 | 420834 | Ba2Cd2Sb3 | 2 |
| | | 12.2.189 | 417659 | Al2Ba5Ge7 | 2 |
| 12 | 11 | 12.2.5 | 61005 | Pd1Se6Ta2 | 2 |
| | | 12.2.7 | 41016 | P3Tc2 | 2 |
| | | 12.2.15 | 16779 | O26V12 | 2 |
| | | 12.2.34 | 15615 | Mn1Mo1O4 | 2 |
| | | 12.2.53 | 26919 | H12Cl2Ni1O6 | 2 |
| | | 12.2.83 | 380416 | Cs4Ge2Se6 | 2 |
| | | 12.2.92 | 626543 | Cr1Pr2Se4 | 2 |
| | | 12.2.95 | 79865 | Cr1O4V1 | 2 |
| | | 12.2.107 | 424718 | Ce5Ge8Mg8 | 2 |
| | | 12.2.154 | 164224 | Bi1Fe1O3 | 2 |
| | | 12.2.155 | 164223 | Bi1Fe1O3 | 2 |
| | | 12.2.163 | 80901 | Ba2Bi1O6Tb1 | 2 |
| | | 12.2.170 | 63501 | B4La2Ni5 | 2 |
| | | 12.2.171 | 170608 | B4La2Ni5 | 2 |
| | | 12.2.188 | 417968 | Al2Ge4Sr3 | 2 |
| 12 | 12 | 12.2.4 | 248500 | Pr1 | 2 |
| | | 12.2.6 | 40281 | P4Zr7 | 2 |
| | | 12.2.28 | 193869 | Ni0.5Se2Ti1 | 2 |
| | | 12.2.31 | 192215 | Nb2Pd1S5 | 2 |
| | | 12.2.36 | 180011 | Li0.5O2Ti1 | 2 |
| | | 12.2.46 | 200012 | H4Al1Cr2K1O10 | 2 |
| | | 12.2.62 | 152533 | Fe1Si3Tm3 | 2 |
| | | 12.2.99 | 41026 | Co2Nb4Pd1Se12 | 2 |
| | | 12.2.120 | 85505 | Ca1K2N4O8 | 2 |
| | | 12.2.135 | 617663 | C1Er2Fe2Si2 | 2 |
| | | 12.2.152 | 415138 | Bi1I1Mn1Se2 | 2 |
| | | 12.2.160 | 62391 | Ba3Cu2Ho2O10Pt1 | 2 |
| | | 12.2.173 | 614701 | B2Lu1Rh3 | 2 |
| | | 12.2.177 | 614561 | B2Ir3Sc1 | 2 |
| | | 12.2.185 | 86402 | As3Ba1 | 2 |
| 12 | 13 | 12.2.1 | 187496 | Sb2Te3 | 2 |
| | | 12.2.13 | 75193 | O5Ti3 | 2 |

Continued on next column

**TABLE XV– continued**

| SG | cl | TQC | ICSD | Stoich. Formula | SubSG |
|---|---|---|---|---|---|
| | | 12.2.33 | 20296 | Mn2O7P2 | 2 |
| | | 12.2.56 | 76350 | Ge2Ho2Os1 | 2 |
| | | 12.2.68 | 603354 | Fe1Rh2Se4 | 2 |
| | | 12.2.69 | 632972 | Fe1Ni2Se4 | 2 |
| | | 12.2.127 | 88516 | C3Er2Mo2 | 2 |
| | | 12.2.139 | 172782 | Br1Mn1S2Sb1 | 2 |
| | | 12.2.142 | 616957 | Bi2Pd3Se2 | 2 |
| | | 12.2.144 | 261459 | Bi2Pd3Se2 | 2 |
| | | 12.2.145 | 261458 | Bi2Pd3Se2 | 2 |
| | | 12.2.150 | 1024 | Bi2Ni3Se2 | 2 |
| | | 12.2.157 | 423458 | Ba5Cd2O2Sb4 | 2 |
| | | 12.2.161 | 408938 | Ba2Mg3Si4 | 2 |
| | | 12.2.182 | 425402 | Au1O1P2Pr2 | 2 |
| | | 12.2.184 | 80375 | As3Mn4 | 2 |
| 12 | 14 | 12.2.8 | 74780 | O7Si2Yb2 | 2 |
| | | 12.2.9 | 65363 | O7Si2Yb2 | 2 |
| | | 12.2.10 | 62356 | O7Si2Yb2 | 2 |
| | | 12.2.24 | 35141 | Ni1S8V4 | 2 |
| | | 12.2.27 | 646394 | Ni1S4Ti2 | 2 |
| | | 12.2.35 | 50001 | Mg2Si10Sr11 | 2 |
| | | 12.2.43 | 2559 | H4F6Li2O2Sn1 | 2 |
| | | 12.2.49 | 27761 | H3Li1O2 | 2 |
| | | 12.2.52 | 421319 | H2Bi1Cl3O1 | 2 |
| | | 12.2.89 | 16884 | Cr2Ni1S4 | 2 |
| | | 12.2.93 | 626536 | Cr1Pr2S4 | 2 |
| | | 12.2.101 | 420061 | Cl1Mo1O4Yb1 | 2 |
| | | 12.2.102 | 420060 | Cl1Mo1O4Tm1 | 2 |
| | | 12.2.140 | 427411 | Bi4Pb7Se13 | 2 |
| | | 12.2.143 | 421873 | Bi2Pd3Se2 | 2 |
| | | 12.2.146 | 261456 | Bi2Pd3Se2 | 2 |
| | | 12.2.183 | 41831 | As3Sr1 | 2 |
| 12 | 15 | 12.2.12 | 75194 | O5Ti3 | 2 |
| | | 12.2.17 | 646757 | Ni3Sn4 | 2 |
| | | 12.2.18 | 105364 | Ni3Sn4 | 2 |
| | | 12.2.19 | 105363 | Ni3Sn4 | 2 |
| | | 12.2.29 | 73315 | Nb8Pt1Se20 | 2 |
| | | 12.2.32 | 81059 | Mo1Ni1O4 | 2 |
| | | 12.2.37 | 246807 | La1Si5 | 2 |
| | | 12.2.42 | 33293 | H6Au1Br3N2 | 2 |
| | | 12.2.55 | 200906 | H10F6Mn1O5Zr1 | 2 |
| | | 12.2.96 | 202377 | Cr1Nd1O1S2 | 2 |
| | | 12.2.100 | 408031 | Cl2La3Si3 | 2 |
| | | 12.2.104 | 427663 | Cl1Ho1O4W1 | 2 |
| | | 12.2.105 | 427664 | Cl1Er1O4W1 | 2 |
| | | 12.2.106 | 420053 | Cl1Er1Mo1O4 | 2 |
| | | 12.2.132 | 252734 | C2Ca1 | 2 |
| | | 12.2.133 | 252716 | C2Ca1 | 2 |
| | | 12.2.147 | 616954 | Bi2Pd3S2 | 2 |
| | | 12.2.156 | 246145 | Bi16I4 | 2 |
| | | 12.2.166 | 49000 | Ba1Sb3 | 2 |
| | | 12.2.172 | 615319 | B2Rh3Tb1 | 2 |
| | | 12.2.174 | 614578 | B2Ir3Y1 | 2 |
| | | 12.2.175 | 614571 | B2Ir3Tm1 | 2 |
| | | 12.2.176 | 614566 | B2Ir3Tb1 | 2 |
| | | 12.2.178 | 614531 | B2Ir3Lu1 | 2 |
| | | 12.2.179 | 614472 | B2Ho1Ir3 | 2 |
| | | 12.2.180 | 8058 | Au2P3 | 2 |
| 13 | 1 | 13.2.2 | 415434 | O4Ta1Tb1 | 2 |
| | | 13.2.9 | 415436 | Ho1O4Ta1 | 2 |
| | | 13.2.13 | 67168 | Cl7Cs1Ti2 | 2 |
| 13 | 2 | 13.2.10 | 155660 | Ho1O4Ta1 | 2 |

Continued on next column

**TABLE XV– continued**

| SG | cl | TQC | ICSD | Stoich. Formula | SubSG |
|---|---|---|---|---|---|
| 13 | 3 | 13.2.8 | 170685 | Mn1O4Re1 | 2 |
| 13 | 4 | 13.2.5 | 15852 | Ni1O4W1 | 2 |
| 13 | 5 | 13.2.1 | 415459 | O4Ta1Tm1 | 2 |
| | | 13.2.4 | 16685 | Ni1O4W1 | 2 |
| | | 13.2.11 | 261836 | Ho1Li1O8W2 | 2 |
| | | 13.2.12 | 36623 | Cu1Ho1O8W2 | 2 |
| 13 | 6 | 13.2.7 | 81060 | Mo1Ni1O4 | 2 |
| 13 | 7 | 13.2.3 | 174299 | O4Sn3 | 2 |
| 13 | 15 | 13.2.6 | 27701 | Mo8O23 | 2 |
| 14 | 1 | 14.2.10 | 80000 | Mo3Si4U2 | 2 |
| | | 14.2.15 | 171490 | Ir1O6Sr2Tb1 | 2 |
| | | 14.2.16 | 153133 | Ir1O6Sr2Tb1 | 2 |
| | | 14.2.21 | 38384 | H1Cl1Cu1O1 | 2 |
| | | 14.2.22 | 80002 | Ge4Mo3U2 | 2 |
| | | 14.2.24 | 421335 | Gd1S2 | 2 |
| | | 14.2.28 | 419955 | F1O3Se1Tm1 | 2 |
| | | 14.2.33 | 67379 | Er2O1S2 | 2 |
| | | 14.2.36 | 162113 | Dy1Ni1O3 | 2 |
| | | 14.2.37 | 60653 | Cu2O4Se1 | 2 |
| | | 14.2.40 | 202451 | Cu1O3Te1 | 2 |
| | | 14.2.41 | 260194 | Cu1Np1Se2 | 2 |
| | | 14.2.58 | 54136 | Br3Cr1In1 | 2 |
| | | 14.2.60 | 75019 | Ba2P3Pd1 | 2 |
| | | 14.2.67 | 74948 | As2Ba1Cu2O8 | 2 |
| | | 14.2.69 | 73402 | Ag2Te1 | 2 |
| | | 14.2.70 | 73230 | Ag2Te1 | 2 |
| | | 14.2.71 | 165661 | Ag1Fe2S3 | 2 |
| 14 | 2 | 14.2.3 | 9402 | O4Te1V1 | 2 |
| | | 14.2.6 | 67381 | O1S2Yb2 | 2 |
| | | 14.2.20 | 91088 | H1Cl1Cu1O1 | 2 |
| | | 14.2.27 | 154372 | Fe2K4O5 | 2 |
| | | 14.2.48 | 250364 | Cl4V1 | 2 |
| | | 14.2.50 | 4411 | Cl3Mo1O1 | 2 |
| | | 14.2.51 | 16150 | Cl3Mo1O1 | 2 |
| | | 14.2.65 | 237469 | Au4Ga2Na1 | 2 |
| | | 14.2.68 | 73978 | As1Fe1O4 | 2 |
| 14 | 3 | 14.2.1 | 96154 | Si4Ti3U2 | 2 |
| | | 14.2.14 | 174231 | Ir1P2 | 2 |
| | | 14.2.29 | 412767 | Eu2P2Se6 | 2 |
| | | 14.2.30 | 412763 | Eu2P2S6 | 2 |
| | | 14.2.32 | 401556 | Eu1Ge1Pd1 | 2 |
| | | 14.2.42 | 73398 | Cu1Na1O4P1 | 2 |
| | | 14.2.52 | 27020 | Ce1Se2 | 2 |
| | | 14.2.54 | 238150 | Ca2Fe1O6Os1 | 2 |
| | | 14.2.55 | 238148 | Ca2Fe1O6Os1 | 2 |
| | | 14.2.57 | 86049 | C7Y4 | 2 |
| | | 14.2.61 | 50158 | Ba1Rh2Si2 | 2 |
| | | 14.2.62 | 50160 | Ba1Ga2Pt2 | 2 |
| | | 14.2.66 | 58406 | Au4Ca5 | 2 |
| 14 | 4 | 14.2.4 | 8217 | O2W1 | 2 |
| | | 14.2.5 | 80829 | O2W1 | 2 |
| | | 14.2.8 | 100194 | Ni2O7P2 | 2 |
| | | 14.2.9 | 246365 | Ni1O10P2V2 | 2 |
| | | 14.2.12 | 80830 | Mo1O2 | 2 |
| | | 14.2.13 | 23722 | Mo1O2 | 2 |
| | | 14.2.18 | 84949 | H1Mn1O2 | 2 |
| | | 14.2.26 | 189438 | Fe2O3 | 2 |
| | | 14.2.31 | 416422 | Eu1Ge2Pt2 | 2 |
| | | 14.2.45 | 73738 | Co1Ta1Te2 | 2 |
| | | 14.2.46 | 63553 | Co1Sb2 | 2 |
| | | 14.2.47 | 73739 | Co1Nb1Te2 | 2 |
| | | 14.2.53 | 419346 | Ce1S2 | 2 |
| | | 14.2.63 | 88450 | B1Fe2O4 | 2 |
| | | 14.2.64 | 86862 | B1Fe2O4 | 2 |
| 14 | 5 | 14.2.19 | 98698 | H1Cu1F1O1 | 2 |
| | | 14.2.23 | 416421 | Ge2La1Pt2 | 2 |
| | | 14.2.43 | 411020 | Cu1Ge1O3 | 2 |
| | | 14.2.49 | 422546 | Cl4Hg4I2S1 | 2 |
| | | 14.2.59 | 172071 | Br2Cu2O6Sr2Te1 | 2 |
| 14 | 6 | 14.2.7 | 281705 | Ni3O8P2 | 2 |
| | | 14.2.11 | 291320 | Mo1O6Sr2Yb1 | 2 |
| | | 14.2.17 | 47158 | H2Li1N1O3 | 2 |
| | | 14.2.25 | 62281 | Ga1La1Ni1 | 2 |
| | | 14.2.44 | 59291 | Co2O7P2 | 2 |
| | | 14.2.56 | 69790 | Ca1Pd1Si1 | 2 |
| 14 | 7 | 14.2.2 | 57480 | Re2Si1 | 2 |
| | | 14.2.34 | 69772 | Er1O6Ru1Sr2 | 2 |
| | | 14.2.35 | 69771 | Er1O6Ru1Sr2 | 2 |
| | | 14.2.38 | 1672 | Cu1O4Te1 | 2 |
| | | 14.2.39 | 1671 | Cu1O4Te1 | 2 |
| 15 | 1 | 15.2.34 | 192833 | Ho1O4V1 | 2 |
| | | 15.2.37 | 15271 | H8O6Pt1 | 2 |
| | | 15.2.39 | 20138 | H4Al2Ca1F4O4 | 2 |
| | | 15.2.70 | 56838 | Cl3Cu1Rb1 | 2 |
| | | 15.2.72 | 247292 | Ce4Rh4Si3 | 2 |
| | | 15.2.91 | 86064 | Ba1Cu1N1 | 2 |
| | | 15.2.95 | 251835 | Al1B2La2Os2 | 2 |
| 15 | 2 | 15.2.6 | 35148 | O5Ti3 | 2 |
| | | 15.2.7 | 194464 | O5Ti3 | 2 |
| | | 15.2.9 | 97442 | O3V2 | 2 |
| | | 15.2.10 | 95763 | O3V2 | 2 |
| | | 15.2.11 | 95762 | O3V2 | 2 |
| | | 15.2.27 | 105057 | Mo1P4 | 2 |
| | | 15.2.28 | 49802 | Mn2P12Ti1 | 2 |
| | | 15.2.45 | 430120 | Gd1Ni8Si3 | 2 |
| | | 15.2.49 | 71059 | Fe2Li1N3Sr2 | 2 |
| | | 15.2.53 | 94884 | D7Al1Sr2 | 2 |
| | | 15.2.67 | 2790 | Cr1P4 | 2 |
| | | 15.2.68 | 26814 | Co1Ge1O3 | 2 |
| | | 15.2.69 | 417870 | Cl8Ga2Ni1 | 2 |
| | | 15.2.74 | 82216 | Ca1Ni1O6Si2 | 2 |
| | | 15.2.75 | 69381 | Ca1Ni1O6Si2 | 2 |
| | | 15.2.76 | 202244 | Ca1Ni1O6Si2 | 2 |
| | | 15.2.78 | 423887 | C2Fe5 | 2 |
| | | 15.2.79 | 423886 | C2Fe5 | 2 |
| | | 15.2.80 | 423885 | C2Fe5 | 2 |
| | | 15.2.81 | 183902 | Bi1Mn1O3 | 2 |
| | | 15.2.90 | 71060 | Ba2Fe2Li1N3 | 2 |
| | | 15.2.94 | 417872 | Al2Cl8Ni1 | 2 |
| 15 | 3 | 15.2.4 | 24327 | P2Pt5 | 2 |
| | | 15.2.5 | 261031 | O6Pb2Sb1Tm1 | 2 |
| | | 15.2.13 | 50703 | Ni2O8Te3 | 2 |
| | | 15.2.40 | 14176 | H1Cr1Hg1O4.5 | 2 |
| | | 15.2.58 | 23479 | Cu2O7V2 | 2 |
| | | 15.2.59 | 158375 | Cu2O7V2 | 2 |
| | | 15.2.83 | 174404 | Bi1Cr1O3 | 2 |
| | | 15.2.84 | 160458 | Bi1Cr1O3 | 2 |
| | | 15.2.85 | 160457 | Bi1Cr1O3 | 2 |
| | | 15.2.86 | 160456 | Bi1Cr1O3 | 2 |
| | | 15.2.87 | 160455 | Bi1Cr1O3 | 2 |

Continued on next column

**TABLE XV– continued**

| SG | cl | TQC | ICSD | Stoich. Formula | SubSG |
|---|---|---|---|---|---|
| | | 15.2.89 | 245253 | Ba3Ca1Ir2O9 | 2 |
| | | 15.2.93 | 58403 | Au2Ca5 | 2 |
| 15 | 4 | 15.2.3 | 16041 | S1Ta6 | 2 |
| | | 15.2.8 | 109187 | O4Ta1Tb1 | 2 |
| | | 15.2.14 | 61148 | Ni1Se5Ta2 | 2 |
| | | 15.2.15 | 109181 | Nb1O4Tm1 | 2 |
| | | 15.2.16 | 109178 | Nb1O4Tb1 | 2 |
| | | 15.2.17 | 78342 | Na1O6Si2V1 | 2 |
| | | 15.2.18 | 163133 | Na1O6Si2V1 | 2 |
| | | 15.2.19 | 163132 | Na1O6Si2V1 | 2 |
| | | 15.2.20 | 163131 | Na1O6Si2V1 | 2 |
| | | 15.2.21 | 163130 | Na1O6Si2V1 | 2 |
| | | 15.2.22 | 163129 | Na1O6Si2V1 | 2 |
| | | 15.2.23 | 163128 | Na1O6Si2V1 | 2 |
| | | 15.2.24 | 163127 | Na1O6Si2V1 | 2 |
| | | 15.2.25 | 163126 | Na1O6Si2V1 | 2 |
| | | 15.2.26 | 163125 | Na1O6Si2V1 | 2 |
| | | 15.2.29 | 78721 | Li2O3Ru1 | 2 |
| | | 15.2.31 | 55164 | Li1O6Si2V1 | 2 |
| | | 15.2.32 | 152074 | Li1O6Si2V1 | 2 |
| | | 15.2.35 | 79499 | Ho1O4Ta1 | 2 |
| | | 15.2.36 | 153363 | Ho1O4Ta1 | 2 |
| | | 15.2.42 | 192134 | H12Br3Mg1O6Rb1 | 2 |
| | | 15.2.48 | 154291 | Fe2O5Rb4 | 2 |
| | | 15.2.50 | 41899 | Fe1S5U2 | 2 |
| | | 15.2.65 | 83494 | Cu2Er1O4 | 2 |
| | | 15.2.71 | 189154 | Cl2Cu2O6Pb1Se2 | 2 |
| | | 15.2.82 | 174405 | Bi1Cr1O3 | 2 |
| | | 15.2.88 | 245254 | Ba3Ir2O9Sr1 | 2 |
| | | 15.2.92 | 418385 | B2Ni1O4 | 2 |
| 15 | 5 | 15.2.2 | 185954 | Sb2Te3 | 2 |
| | | 15.2.47 | 24416 | Fe2O5Ti1 | 2 |
| | | 15.2.51 | 15909 | F2Sn1 | 2 |
| | | 15.2.52 | 2739 | Er1Na1O2 | 2 |
| | | 15.2.64 | 83490 | Cu2Gd1O4 | 2 |
| | | 15.2.77 | 158299 | C2H8Cl3Cu1N1 | 2 |
| 15 | 6 | 15.2.33 | 416039 | K3Na1O8Ru2 | 2 |
| | | 15.2.38 | 14329 | H4Ba1O8S2 | 2 |
| | | 15.2.41 | 183508 | H12Cl3Mg1O6Rb1 | 2 |
| | | 15.2.43 | 237747 | Ge2Mn1Na1O6 | 2 |
| | | 15.2.44 | 237746 | Ge2Mn1Na1O6 | 2 |
| | | 15.2.46 | 151766 | Fe7S8 | 2 |
| | | 15.2.60 | 67316 | Cu2O7P2 | 2 |
| | | 15.2.61 | 28279 | Cu2O7P2 | 2 |
| | | 15.2.62 | 157107 | Cu2O7P2 | 2 |
| | | 15.2.63 | 14369 | Cu2O7P2 | 2 |
| | | 15.2.73 | 95789 | Ca2Ir2Si1 | 2 |
| 15 | 7 | 15.2.1 | 187497 | Sb2Te3 | 2 |
| | | 15.2.12 | 6286 | O3V2 | 2 |
| | | 15.2.30 | 180367 | Li2O3Ru1 | 2 |
| | | 15.2.54 | 185090 | Cu6Sn5 | 2 |
| | | 15.2.55 | 185089 | Cu6Sn5 | 2 |
| | | 15.2.56 | 150124 | Cu6Sn5 | 2 |
| | | 15.2.57 | 106530 | Cu6Sn5 | 2 |
| | | 15.2.66 | 83491 | Cu2Dy1O4 | 2 |
| 47 | 1 | 47.2.4 | 63335 | Ba2Cu3O7Y1 | 2 |
| | | 47.2.13 | 71640 | B3Ir4Zn1 | 2 |
| 47 | 2 | 47.2.12 | 187120 | Ba1Co2Nd1O5.5 | 2 |
| 47 | 3 | 47.2.11 | 96377 | Ba1Co2Nd1O5.5 | 2 |
| 47 | 4 | 47.2.10 | 96380 | Ba1Co2Nd1O5.5 | 2 |

Continued on next column

**TABLE XV– continued**

| SG | cl | TQC | ICSD | Stoich. Formula | SubSG |
|---|---|---|---|---|---|
| 47 | 5 | 47.2.7 | 68044 | Ba2Cu3Ho1O7 | 2 |
| 47 | 6 | 47.2.1 | 86427 | Cu1O5Sr2Tl1 | 2 |
| | | 47.2.2 | 390001 | Ba4Cu6O13Y2 | 2 |
| | | 47.2.6 | 73896 | Ba2Cu3O6.5Y1 | 2 |
| 47 | 7 | 47.2.5 | 262809 | Ba2Cu3O6.93Y1 | 2 |
| 47 | 9 | 47.2.8 | 94046 | Ba2Co4O11Tb2 | 2 |
| | | 47.2.9 | 94045 | Ba2Co4O11Tb2 | 2 |
| 47 | 11 | 47.2.3 | 77745 | Ba2Cu3O7Y1 | 2 |
| 51 | 1 | 51.2.4 | 421074 | Ca2Ge2Ni3 | 2 |
| | | 51.2.17 | 9986 | Al4Co1La1 | 2 |
| 51 | 2 | 51.2.12 | 429116 | B1Ru1Ta1 | 2 |
| | | 51.2.13 | 426468 | B1Nb1Ru1 | 2 |
| | | 51.2.16 | 16325 | Au1I1Te2 | 2 |
| 51 | 3 | 51.2.1 | 248052 | La2Se9U2 | 2 |
| | | 51.2.7 | 247569 | Ba1Fe2Gd1O5 | 2 |
| | | 51.2.15 | 249924 | Au3Rb2Tl1 | 2 |
| 51 | 4 | 51.2.5 | 194877 | C3Al1Y3 | 2 |
| | | 51.2.14 | 194435 | B1Nb1Ru1 | 2 |
| 51 | 5 | 51.2.2 | 106800 | In1La1Ni2 | 2 |
| 51 | 6 | 51.2.8 | 96391 | Ba1Fe2Gd1O4.999 | 2 |
| | | 51.2.11 | 107494 | B3Ir7Zn5 | 2 |
| 51 | 7 | 51.2.10 | 107495 | B3Rh7Zn5 | 2 |
| 51 | 8 | 51.2.9 | 99611 | Ba1Dy1Fe2O4.996 | 2 |
| 51 | 9 | 51.2.6 | 69767 | C1B2U1 | 2 |
| 51 | 13 | 51.2.3 | 370040 | Co2In1Tb1 | 2 |
| 53 | 1 | 53.1.2 | 71520 | Fe1Nb1Te2 | 2 |
| 53 | 2 | 53.1.1 | 71063 | Ni2Ta2Te4 | 2 |
| 55 | 1 | 55.2.47 | 419360 | B4Rh1Tm1 | 2 |
| | | 55.2.51 | 190184 | B4Fe1Yb1 | 2 |
| | | 55.2.55 | 417873 | B4Co1Gd1 | 2 |
| | | 55.2.62 | 167534 | Al1B4Lu1 | 2 |
| | | 55.2.63 | 260799 | Al1.33B5.33Lu1.33 | 2 |
| 55 | 2 | 55.2.2 | 249665 | Pr2S5Sn1 | 2 |
| | | 55.2.20 | 193223 | In1La2Si2 | 2 |
| | | 55.2.26 | 82389 | Cu2La2O5 | 2 |
| | | 55.2.27 | 82388 | Cu2La2O5 | 2 |
| | | 55.2.28 | 73399 | Cu2La2O5 | 2 |
| | | 55.2.31 | 2578 | Ce4O4S3 | 2 |
| | | 55.2.35 | 35219 | Ca1Mn1O2.5 | 2 |
| | | 55.2.37 | 23834 | C2B2Sc1 | 2 |
| | | 55.2.38 | 423924 | C1B2Lu1 | 2 |
| | | 55.2.40 | 415576 | Bi6In2Sr5 | 2 |
| | | 55.2.42 | 186442 | Bi2Fe4O9 | 2 |
| | | 55.2.43 | 186441 | Bi2Fe4O9 | 2 |
| | | 55.2.46 | 419358 | B4Rh1Y1 | 2 |
| | | 55.2.59 | 36364 | Al2Bi6Ca5 | 2 |
| 55 | 3 | 55.2.4 | 249666 | Nd2S5Sn1 | 2 |
| | | 55.2.19 | 391127 | In4Rh2Zr5 | 2 |
| | | 55.2.29 | 82019 | Cu1O2.5Sr1 | 2 |
| | | 55.2.48 | 409820 | B4Ni1Yb1 | 2 |
| | | 55.2.58 | 56973 | As1Ga2Rh5 | 2 |
| 55 | 4 | 55.2.5 | 90183 | Mn2O5Sr2 | 2 |
| | | 55.2.21 | 412706 | Hf5In4Rh2 | 2 |
| | | 55.2.34 | 35130 | Ca2Mn2O5 | 2 |
| | | 55.2.39 | 412563 | Bi8Ca3Pd4 | 2 |
| | | 55.2.41 | 262861 | Bi2Fe4O9 | 2 |
| | | 55.2.44 | 186440 | Bi2Fe4O9 | 2 |
| | | 55.2.53 | 81542 | B4Cr1Th1 | 2 |
| | | 55.2.57 | 78821 | B10Fe1Th2 | 2 |
| | | 55.2.60 | 260796 | Al1B4Yb1 | 2 |

Continued on next column

**TABLE XV– continued**

| SG | cl | TQC | ICSD | Stoich. Formula | SubSG |
|---|---|---|---|---|---|
| | | 55.2.61 | 167532 | Al1B4Yb1 | 2 |
| 55 | 5 | 55.2.1 | 404193 | Se2Ti9 | 2 |
| | | 55.2.3 | 97282 | O4Pb3 | 2 |
| | | 55.2.24 | 426620 | F3O1Rb1V1 | 2 |
| | | 55.2.25 | 249403 | F3O1Rb1V1 | 2 |
| | | 55.2.50 | 61039 | B4Mg2Ru5 | 2 |
| | | 55.2.56 | 78823 | B10Ni1Th2 | 2 |
| 55 | 6 | 55.2.32 | 102075 | Cd7Th6 | 2 |
| | | 55.2.33 | 36466 | Ca5Ga2Sb6 | 2 |
| | | 55.2.45 | 419361 | B4Rh1Yb1 | 2 |
| | | 55.2.49 | 425457 | B4Ni1Sc1 | 2 |
| | | 55.2.52 | 409626 | B4Er1V1 | 2 |
| | | 55.2.54 | 250251 | B4Co1Yb1 | 2 |
| 55 | 7 | 55.2.6 | 417818 | Mn2O5Sr2 | 2 |
| | | 55.2.7 | 417817 | Mn2O5Sr2 | 2 |
| | | 55.2.8 | 417816 | Mn2O5Sr2 | 2 |
| | | 55.2.9 | 417815 | Mn2O5Sr2 | 2 |
| | | 55.2.10 | 417814 | Mn2O5Sr2 | 2 |
| | | 55.2.11 | 417813 | Mn2O5Sr2 | 2 |
| | | 55.2.12 | 417812 | Mn2O5Sr2 | 2 |
| | | 55.2.13 | 417811 | Mn2O5Sr2 | 2 |
| | | 55.2.14 | 417810 | Mn2O5Sr2 | 2 |
| | | 55.2.15 | 417809 | Mn2O5Sr2 | 2 |
| | | 55.2.16 | 417808 | Mn2O5Sr2 | 2 |
| | | 55.2.17 | 417807 | Mn2O5Sr2 | 2 |
| | | 55.2.18 | 417806 | Mn2O5Sr2 | 2 |
| | | 55.2.22 | 22046 | Ge1Mn2O4 | 2 |
| | | 55.2.23 | 23226 | Ga2Mg1 | 2 |
| | | 55.2.30 | 171414 | Ce5In4Pt2 | 2 |
| | | 55.2.36 | 84869 | C5Y4 | 2 |
| 57 | 1 | 57.2.7 | 190838 | Ce1Pd2Sn3 | 2 |
| | | 57.2.8 | 93749 | Ca1Mn2O4 | 2 |
| 57 | 2 | 57.2.6 | 76352 | Ce3Ge2Rh2 | 2 |
| 57 | 3 | 57.2.4 | 83907 | Cr1La1Sb3 | 2 |
| | | 57.2.5 | 247291 | Ce3Rh2Si2 | 2 |
| 57 | 4 | 57.2.2 | 85640 | Cr1La1Sb3 | 2 |
| | | 57.2.3 | 85639 | Cr1La1Sb3 | 2 |
| 57 | 5 | 57.2.1 | 20933 | Fe2Hf1Si2 | 2 |
| | | 57.2.9 | 150592 | Al1Dy1 | 2 |
| 58 | 1 | 58.2.1 | 42724 | Ni1Sb2 | 2 |
| | | 58.2.6 | 81572 | C3Al2Th2 | 2 |
| | | 58.2.7 | 425106 | B4Cr1 | 2 |
| | | 58.2.8 | 186851 | B4Cr1 | 2 |
| | | 58.2.9 | 31387 | B2Ni3O6 | 2 |
| | | 58.2.10 | 2016 | B2Ni3O6 | 2 |
| | | 58.2.12 | 42605 | As2Ni1 | 2 |
| 58 | 2 | 58.2.2 | 414575 | Ge6Li2Sr4 | 2 |
| | | 58.2.3 | 409767 | Eu2Ge3Li1 | 2 |
| | | 58.2.11 | 251066 | Au8Sn3Sr3 | 2 |
| 58 | 3 | 58.2.4 | 42720 | Cr1Sb2 | 2 |
| | | 58.2.5 | 42601 | Cr1Sb2 | 2 |
| 59 | 1 | 59.1.3 | 425285 | Ho1I1S1 | 2 |
| | | 59.1.4 | 154397 | Ge5N2Sr6 | 2 |
| | | 59.1.5 | 410357 | Ge1Pr1Pt1 | 2 |
| | | 59.1.10 | 160543 | Dy2Sn5 | 2 |
| | | 59.1.11 | 72376 | Cr6P4U1 | 2 |
| | | 59.1.14 | 154398 | Ba6Ge5N2 | 2 |
| | | 59.1.15 | 421431 | Ba1Ge1Ni2 | 2 |
| | | 59.1.17 | 421524 | B2Ru1 | 2 |
| | | 59.1.18 | 31871 | B2Ru1 | 2 |

Continued on next column

**TABLE XV– continued**

| SG | cl | TQC | ICSD | Stoich. Formula | SubSG |
|---|---|---|---|---|---|
| | | 59.1.19 | 421523 | B2Os1 | 2 |
| 59 | 2 | 59.1.1 | 408080 | Pd1Si1Y1 | 2 |
| | | 59.1.2 | 154823 | Li1Mn1O2 | 2 |
| | | 59.1.6 | 195027 | Fe2O1Se2Sr1 | 2 |
| | | 59.1.7 | 195026 | Fe2O1S2Sr1 | 2 |
| | | 59.1.8 | 195025 | Fe2O1S2Sr1 | 2 |
| | | 59.1.9 | 88203 | Er2Ge5 | 2 |
| | | 59.1.16 | 428170 | Ba1Fe2O1S2 | 2 |
| 59 | 3 | 59.1.12 | 30701 | Co8La1P5 | 2 |
| | | 59.1.13 | 410358 | Ce1Ge1Pt1 | 2 |
| 59 | 4 | 59.2.1 | 187622 | Ni1Si1 | 2 |
| | | 59.2.2 | 95640 | Li1Mn1O2 | 2 |
| | | 59.2.6 | 99042 | Cu2O3Sr1 | 2 |
| | | 59.2.9 | 249645 | Au3K1Sn3 | 2 |
| | | 59.2.10 | 245679 | Au3In3Sr1 | 2 |
| | | 59.2.11 | 107445 | Au1Rb3Sn4 | 2 |
| | | 59.2.12 | 107447 | Au1Pb4Rb3 | 2 |
| | | 59.2.13 | 107448 | Au1Cs3Pb4 | 2 |
| 59 | 5 | 59.2.7 | 155492 | Co1Na0.5O2 | 2 |
| | | 59.2.8 | 413690 | Br3Mo1 | 2 |
| 59 | 6 | 59.2.4 | 81303 | Fe1Hf8Te6 | 2 |
| | | 59.2.5 | 408081 | Er1Pd1Si1 | 2 |
| | | 59.2.14 | 107525 | Al1Eu1 | 2 |
| 59 | 7 | 59.2.3 | 408082 | Gd1Pd1Si1 | 2 |
| 60 | 1 | 60.2.1 | 415270 | O2Pb1 | 2 |
| | | 60.2.2 | 415269 | O2Pb1 | 2 |
| | | 60.2.3 | 415268 | O2Pb1 | 2 |
| | | 60.2.4 | 246148 | C1Mo2 | 2 |
| | | 60.2.5 | 246147 | C1Mo2 | 2 |
| | | 60.2.6 | 246146 | C1Mo2 | 2 |
| | | 60.2.7 | 430298 | Au2Pb1 | 2 |
| | | 61.2.2 | 415968 | Au1Sn2 | 2 |
| 61 | 2 | 61.2.1 | 162404 | Pd1Sn1Te1 | 2 |
| 62 | 1 | 62.2.2 | 54980 | Sr1Zn1 | 2 |
| | | 62.2.3 | 82295 | Se2U1 | 2 |
| | | 62.2.4 | 10001 | Sb2Sr1Zn1 | 2 |
| | | 62.2.6 | 42430 | S1Te1U1 | 2 |
| | | 62.2.10 | 15248 | Rh1Sc1Si2 | 2 |
| | | 62.2.11 | 159305 | Pt1Yb1Zn1 | 2 |
| | | 62.2.13 | 260792 | Pt1Si1Zr1 | 2 |
| | | 62.2.14 | 260794 | Pt1Si1Ti1 | 2 |
| | | 62.2.19 | 159304 | Pd1Yb1Zn1 | 2 |
| | | 62.2.25 | 1918 | Pd1Pu1 | 2 |
| | | 62.2.26 | 400334 | P2Zr1 | 2 |
| | | 62.2.27 | 417911 | P1Pt1Sc1 | 2 |
| | | 62.2.28 | 100624 | O4S1Sr1 | 2 |
| | | 62.2.29 | 415748 | O4Rb2Ru1 | 2 |
| | | 62.2.31 | 185833 | O3Pr1V1 | 2 |
| | | 62.2.32 | 172354 | O3Pr1Rh1 | 2 |
| | | 62.2.33 | 172348 | O3Pr1Rh1 | 2 |
| | | 62.2.49 | 248408 | Ni1S3U1 | 2 |
| | | 62.2.50 | 419759 | Ni1Pd1Te1 | 2 |
| | | 62.2.51 | 50990 | Ni1P1Sc1 | 2 |
| | | 62.2.52 | 189299 | Ni1O3Yb1 | 2 |
| | | 62.2.53 | 72290 | Nd2S3 | 2 |
| | | 62.2.54 | 415953 | Nd1Pt1Sn1 | 2 |
| | | 62.2.55 | 82008 | Nd1O3Ti1 | 2 |
| | | 62.2.56 | 67723 | Nd1Ni1O3 | 2 |
| | | 62.2.57 | 67722 | Nd1Ni1O3 | 2 |
| | | 62.2.58 | 75008 | Nb1P1Zr1 | 2 |

Continued on next column

**TABLE XV– continued**

| SG | cl | TQC | ICSD | Stoich. Formula | SubSG |
|---|---|---|---|---|---|
| | | 62.2.59 | 49728 | Nb1Ni1P1 | 2 |
| | | 62.2.61 | 43399 | Mn1P1 | 2 |
| | | 62.2.62 | 43246 | Mn1P1 | 2 |
| | | 62.2.64 | 97084 | Mn1O3Pr1 | 2 |
| | | 62.2.65 | 97083 | Mn1O3Pr1 | 2 |
| | | 62.2.66 | 97082 | Mn1O3Pr1 | 2 |
| | | 62.2.67 | 84922 | Mn1O3Pr1 | 2 |
| | | 62.2.68 | 426614 | Mg1Pt1Sr1 | 2 |
| | | 62.2.72 | 417912 | La1P1Pt1 | 2 |
| | | 62.2.84 | 240201 | In3Ir1 | 2 |
| | | 62.2.85 | 412913 | In1Rh1Yb1 | 2 |
| | | 62.2.87 | 99546 | Ho1O3Sc1 | 2 |
| | | 62.2.91 | 42912 | Ge3Pt2 | 2 |
| | | 62.2.92 | 16822 | Ge2Mo1 | 2 |
| | | 62.2.93 | 429881 | Ge2Li3Na1 | 2 |
| | | 62.2.94 | 280691 | Ge2In1Li1Sr2 | 2 |
| | | 62.2.96 | 421092 | Ge1Rh1Ta1 | 2 |
| | | 62.2.97 | 260793 | Ge1Pt1Zr1 | 2 |
| | | 62.2.100 | 409926 | Ge1Pd1Ti1 | 2 |
| | | 62.2.101 | 413387 | Ge1O1Yb3 | 2 |
| | | 62.2.106 | 280201 | Gd2O1Se2 | 2 |
| | | 62.2.109 | 103904 | Ga2Pd5 | 2 |
| | | 62.2.110 | 420711 | Ga1Ho1O3 | 2 |
| | | 62.2.112 | 169317 | Fe1Se1 | 2 |
| | | 62.2.114 | 169314 | Fe1Se1 | 2 |
| | | 62.2.118 | 169305 | Fe1Se1 | 2 |
| | | 62.2.119 | 169282 | Fe1Se1 | 2 |
| | | 62.2.120 | 169281 | Fe1Se1 | 2 |
| | | 62.2.121 | 169280 | Fe1Se1 | 2 |
| | | 62.2.122 | 169279 | Fe1Se1 | 2 |
| | | 62.2.123 | 169278 | Fe1Se1 | 2 |
| | | 62.2.124 | 169277 | Fe1Se1 | 2 |
| | | 62.2.127 | 169274 | Fe1Se1 | 2 |
| | | 62.2.128 | 169273 | Fe1Se1 | 2 |
| | | 62.2.131 | 35008 | Fe1S1 | 2 |
| | | 62.2.133 | 153236 | Fe1Nd1O3 | 2 |
| | | 62.2.134 | 32756 | Fe1K2O4 | 2 |
| | | 62.2.135 | 2876 | Fe1K2O4 | 2 |
| | | 62.2.136 | 240891 | F2K1N1O1 | 2 |
| | | 62.2.137 | 2635 | F1Se1Y1 | 2 |
| | | 62.2.141 | 423115 | Er1Pt1Zn1 | 2 |
| | | 62.2.142 | 27282 | Er1Fe1O3 | 2 |
| | | 62.2.143 | 419 | Dy2Se3 | 2 |
| | | 62.2.144 | 80768 | Dy2S3 | 2 |
| | | 62.2.145 | 66532 | Dy2S3 | 2 |
| | | 62.2.146 | 66531 | Dy2S3 | 2 |
| | | 62.2.147 | 36603 | Dy2O1S2 | 2 |
| | | 62.2.151 | 86391 | Cu1Sc1Si1 | 2 |
| | | 62.2.152 | 172850 | Cu1S2Tm1 | 2 |
| | | 62.2.154 | 33799 | Cs2O4Ru1 | 2 |
| | | 62.2.155 | 33861 | Cs2Fe1O4 | 2 |
| | | 62.2.156 | 192388 | Cr2Fe1O4 | 2 |
| | | 62.2.162 | 251083 | Cr1Er1O3 | 2 |
| | | 62.2.163 | 183759 | Cr1Er1O3 | 2 |
| | | 62.2.174 | 49726 | Co1P1Zr1 | 2 |
| | | 62.2.177 | 172046 | Co1O3Tb1 | 2 |
| | | 62.2.178 | 88738 | Co1O3Pr1 | 2 |
| | | 62.2.179 | 94437 | Co1O2Rb3 | 2 |
| | | 62.2.183 | 172048 | Co1Ho1O3 | 2 |
| | | 62.2.184 | 102428 | Co1Ga1Gd1 | 2 |
| | | 62.2.185 | 48206 | Cl2Nd1 | 2 |

Continued on next column

**TABLE XV– continued**

| SG | cl | TQC | ICSD | Stoich. Formula | SubSG |
|---|---|---|---|---|---|
| | | 62.2.191 | 409972 | Ce1P2Pt4 | 2 |
| | | 62.2.192 | 54733 | Ce1In4Pd2 | 2 |
| | | 62.2.195 | 237728 | Cd1Pt1Sr1 | 2 |
| | | 62.2.196 | 12159 | Cd1Na1Sb1 | 2 |
| | | 62.2.198 | 426616 | Ca1Mg1Pd1 | 2 |
| | | 62.2.199 | 410841 | Ca1In1Rh1 | 2 |
| | | 62.2.202 | 426610 | Ca1Cd1Pt1 | 2 |
| | | 62.2.203 | 426611 | Ca1Cd1Pd1 | 2 |
| | | 62.2.204 | 68116 | C2U1W1 | 2 |
| | | 62.2.205 | 16549 | C2Mo1U1 | 2 |
| | | 62.2.208 | 247128 | C1H2Cl1Cu1N2 | 2 |
| | | 62.2.209 | 236383 | Bi3Co1 | 2 |
| | | 62.2.210 | 193272 | Bi3Co1 | 2 |
| | | 62.2.211 | 428110 | Bi1Li1Yb1 | 2 |
| | | 62.2.215 | 424857 | Ba1Hg6 | 2 |
| | | 62.2.217 | 424315 | Ba1Fe2Se3 | 2 |
| | | 62.2.218 | 424314 | Ba1Fe2Se3 | 2 |
| | | 62.2.219 | 290595 | Ba1Fe2Se3 | 2 |
| | | 62.2.220 | 185512 | Ba1Fe2Se3 | 2 |
| | | 62.2.222 | 421745 | B1O3Pr1 | 2 |
| | | 62.2.223 | 42894 | B1Co1Mo1 | 2 |
| | | 62.2.224 | 50842 | B1Ca2F1N2 | 2 |
| | | 62.2.225 | 263057 | Au3Ga2Rb1 | 2 |
| | | 62.2.226 | 263058 | Au3Cs1Ga2 | 2 |
| | | 62.2.227 | 418239 | Au3Ca3In1 | 2 |
| | | 62.2.232 | 411544 | Au1Cd1Eu1 | 2 |
| | | 62.2.235 | 168665 | As2Zr1 | 2 |
| | | 62.2.239 | 412296 | As1Cu1Mg1 | 2 |
| | | 62.2.240 | 406953 | As1Co1Hf1 | 2 |
| | | 62.2.245 | 9571 | As1Cd1Na1 | 2 |
| | | 62.2.246 | 249802 | As1.01Ce1Se0.99 | 2 |
| | | 62.2.253 | 83983 | Ag1Sb1Yb1 | 2 |
| | | 62.2.255 | 98736 | Ag1As2Pr1 | 2 |
| | | 62.2.256 | 420611 | Ag1As2Pr1 | 2 |
| | | 62.2.257 | 412816 | Ag1As2Pr1 | 2 |
| 62 | 2 | 62.2.1 | 418619 | Sr1Zn5 | 2 |
| | | 62.2.7 | 30398 | Ru4Si3 | 2 |
| | | 62.2.9 | 421255 | Rh1Sc3Si3 | 2 |
| | | 62.2.12 | 412012 | Pt1Sn1Sr1 | 2 |
| | | 62.2.20 | 408408 | Pd1Sn1Yb1 | 2 |
| | | 62.2.21 | 151167 | Pd1Sn1Yb1 | 2 |
| | | 62.2.36 | 64952 | Ni2O4Si1 | 2 |
| | | 62.2.37 | 35675 | Ni2O4Si1 | 2 |
| | | 62.2.38 | 202376 | Ni2O4Si1 | 2 |
| | | 62.2.39 | 100645 | Ni2O4Si1 | 2 |
| | | 62.2.40 | 100644 | Ni2O4Si1 | 2 |
| | | 62.2.41 | 100643 | Ni2O4Si1 | 2 |
| | | 62.2.42 | 100642 | Ni2O4Si1 | 2 |
| | | 62.2.43 | 105403 | Ni1Th1 | 2 |
| | | 62.2.44 | 54360 | Ni1Sn1Yb1 | 2 |
| | | 62.2.45 | 18188 | Ni1Si1Ti1 | 2 |
| | | 62.2.63 | 185418 | Mn1O3V1 | 2 |
| | | 62.2.69 | 97765 | Li1Ni1O4P1 | 2 |
| | | 62.2.70 | 72929 | Li1Ni1O4P1 | 2 |
| | | 62.2.71 | 402760 | Li1Ni1O4P1 | 2 |
| | | 62.2.73 | 415749 | K2O4Ru1 | 2 |
| | | 62.2.74 | 427252 | Ir1Sn1Y1 | 2 |
| | | 62.2.75 | 427253 | Ir1Si1Y1 | 2 |
| | | 62.2.77 | 421256 | Ir1Sc3Si3 | 2 |
| | | 62.2.78 | 420416 | Ir1Sc1Si1 | 2 |
| | | 62.2.79 | 422623 | Ir1Sb1Yb1 | 2 |

Continued on next column

**TABLE XV– continued**

| SG | cl | TQC | ICSD | Stoich. Formula | SubSG |
|---|---|---|---|---|---|
| | | 62.2.83 | 413857 | Ir1Lu1Si1 | 2 |
| | | 62.2.88 | 165196 | Ho1Mn1O3 | 2 |
| | | 62.2.89 | 82727 | Hf3P2 | 2 |
| | | 62.2.98 | 419135 | Ge1Pt1Yb1 | 2 |
| | | 62.2.102 | 237302 | Ge1Mg1Ni1 | 2 |
| | | 62.2.103 | 428468 | Ge1Ir1Sc1 | 2 |
| | | 62.2.105 | 15534 | Gd3Ni1Si2 | 2 |
| | | 62.2.108 | 419793 | Gd1Ge1Ru1 | 2 |
| | | 62.2.111 | 55537 | Fe1Si1U1 | 2 |
| | | 62.2.113 | 169316 | Fe1Se1 | 2 |
| | | 62.2.115 | 169312 | Fe1Se1 | 2 |
| | | 62.2.116 | 169310 | Fe1Se1 | 2 |
| | | 62.2.117 | 169308 | Fe1Se1 | 2 |
| | | 62.2.125 | 169276 | Fe1Se1 | 2 |
| | | 62.2.126 | 169275 | Fe1Se1 | 2 |
| | | 62.2.129 | 35010 | Fe1S1 | 2 |
| | | 62.2.130 | 35009 | Fe1S1 | 2 |
| | | 62.2.139 | 405049 | Eu1In1Pt1 | 2 |
| | | 62.2.140 | 404044 | Eu1In1Pd1 | 2 |
| | | 62.2.164 | 65751 | Co2O4Si1 | 2 |
| | | 62.2.165 | 6248 | Co2O4Si1 | 2 |
| | | 62.2.166 | 50072 | Co2O4Si1 | 2 |
| | | 62.2.167 | 260092 | Co2O4Si1 | 2 |
| | | 62.2.168 | 260091 | Co2O4Si1 | 2 |
| | | 62.2.169 | 202255 | Co2O4Si1 | 2 |
| | | 62.2.170 | 200705 | Co2O4Si1 | 2 |
| | | 62.2.171 | 192245 | Co1Y3 | 2 |
| | | 62.2.172 | 192244 | Co1Y3 | 2 |
| | | 62.2.187 | 428424 | Ce2Te3 | 2 |
| | | 62.2.188 | 106433 | Ce1Si1 | 2 |
| | | 62.2.189 | 418546 | Ce1Pt1Zn1 | 2 |
| | | 62.2.190 | 420207 | Ce1Pd1Zn1 | 2 |
| | | 62.2.197 | 72642 | Ca1Pt1Si1 | 2 |
| | | 62.2.200 | 106343 | Ca1Ge1Pt1 | 2 |
| | | 62.2.201 | 291004 | Ca1Fe2O4 | 2 |
| | | 62.2.212 | 52698 | Ba2Si1 | 2 |
| | | 62.2.213 | 416561 | Ba2Pb1 | 2 |
| | | 62.2.216 | 237304 | Ba1Ge1Ni1 | 2 |
| | | 62.2.229 | 249562 | Au2Ba1In2 | 2 |
| | | 62.2.230 | 391422 | Au1In1Sr1 | 2 |
| | | 62.2.231 | 420674 | Au1Eu1Zn1 | 2 |
| | | 62.2.233 | 408579 | Au1Ca1In1 | 2 |
| | | 62.2.234 | 106273 | Au1Ca1Ga1 | 2 |
| | | 62.2.237 | 51201 | As1Li1Ni1O4 | 2 |
| | | 62.2.238 | 245184 | As1Li1Ni1O4 | 2 |
| | | 62.2.248 | 160051 | Al1Ce1Ru1 | 2 |
| | | 62.2.251 | 370027 | Al1Au1Yb1 | 2 |
| | | 62.2.252 | 370015 | Al1Au1Ca1 | 2 |
| 62 | 3 | 62.2.5 | 402457 | S3Tm2 | 2 |
| | | 62.2.8 | 421663 | Ru2Sn2Zn3 | 2 |
| | | 62.2.15 | 79238 | Pt1Si1 | 2 |
| | | 62.2.16 | 2623 | Pt1Si1 | 2 |
| | | 62.2.17 | 105785 | Pt1Sc2 | 2 |
| | | 62.2.18 | 158364 | Pd2Sn1 | 2 |
| | | 62.2.22 | 109415 | Pd1Sn1 | 2 |
| | | 62.2.23 | 79237 | Pd1Si1 | 2 |
| | | 62.2.24 | 15016 | Pd1Si1 | 2 |
| | | 62.2.30 | 100625 | O4Pb1S1 | 2 |
| | | 62.2.34 | 413386 | O1Si1Yb3 | 2 |
| | | 62.2.35 | 74785 | Ni2P2Th1 | 2 |
| | | 62.2.46 | 95449 | Ni1Si1 | 2 |

Continued on next column

**TABLE XV– continued**

| SG | cl | TQC | ICSD | Stoich. Formula | SubSG |
|---|---|---|---|---|---|
| | | 62.2.47 | 95448 | Ni1Si1 | 2 |
| | | 62.2.48 | 190105 | Ni1Si1 | 2 |
| | | 62.2.60 | 41149 | Mn1Rh1Si1 | 2 |
| | | 62.2.76 | 411884 | Ir1Si1Ta1 | 2 |
| | | 62.2.80 | 411882 | Ir1Nb1Si1 | 2 |
| | | 62.2.81 | 246548 | Ir1Mn1Si1 | 2 |
| | | 62.2.82 | 246547 | Ir1Mn1Si1 | 2 |
| | | 62.2.86 | 72291 | Ho2S3 | 2 |
| | | 62.2.90 | 260839 | H2Cl1O2Tm1 | 2 |
| | | 62.2.95 | 44860 | Ge1Rh2 | 2 |
| | | 62.2.99 | 2624 | Ge1Pt1 | 2 |
| | | 62.2.104 | 411883 | Ge1Ir1Nb1 | 2 |
| | | 62.2.107 | 1375 | Gd1P1S1 | 2 |
| | | 62.2.132 | 49729 | Fe1P1Ta1 | 2 |
| | | 62.2.138 | 195382 | F1Ho1Se1 | 2 |
| | | 62.2.148 | 180220 | Dy1Si1 | 2 |
| | | 62.2.149 | 280091 | Dy1Fe1O3 | 2 |
| | | 62.2.150 | 27280 | Dy1Fe1O3 | 2 |
| | | 62.2.153 | 172848 | Cu1Ho1S2 | 2 |
| | | 62.2.157 | 80332 | Cr1Nd1O3 | 2 |
| | | 62.2.158 | 79349 | Cr1Nd1O3 | 2 |
| | | 62.2.159 | 156326 | Cr1Nd1O3 | 2 |
| | | 62.2.160 | 156322 | Cr1Nd1O3 | 2 |
| | | 62.2.161 | 156320 | Cr1Nd1O3 | 2 |
| | | 62.2.173 | 409847 | Co1Si1V1 | 2 |
| | | 62.2.175 | 43401 | Co1P1 | 2 |
| | | 62.2.176 | 43249 | Co1P1 | 2 |
| | | 62.2.180 | 2421 | Co1Mo1P1 | 2 |
| | | 62.2.181 | 150531 | Co1La3 | 2 |
| | | 62.2.182 | 73212 | Co1K3O2 | 2 |
| | | 62.2.186 | 157332 | Ce3In2Ru2 | 2 |
| | | 62.2.193 | 99240 | Ce1In2Pd3 | 2 |
| | | 62.2.194 | 150105 | Ce1Cu6 | 2 |
| | | 62.2.206 | 151477 | C2Cr3 | 2 |
| | | 62.2.207 | 15086 | C2Cr3 | 2 |
| | | 62.2.214 | 90336 | Ba1Ho2Te4 | 2 |
| | | 62.2.221 | 41893 | B2Co1Nb1 | 2 |
| | | 62.2.228 | 249563 | Au2In2Sr1 | 2 |
| | | 62.2.236 | 42446 | As1V1 | 2 |
| | | 62.2.241 | 48030 | As1Co1 | 2 |
| | | 62.2.242 | 48029 | As1Co1 | 2 |
| | | 62.2.243 | 48028 | As1Co1 | 2 |
| | | 62.2.244 | 48026 | As1Co1 | 2 |
| | | 62.2.247 | 150173 | Al1Pr1Pt1 | 2 |
| | | 62.2.249 | 55623 | Al1Ce1Rh1 | 2 |
| | | 62.2.250 | 160052 | Al1Ce1Rh1 | 2 |
| | | 62.2.254 | 420615 | Ag1As2Tb1 | 2 |
| 63 | 1 | 63.1.2 | 106074 | Sn1Sr1 | 2 |
| | | 63.1.5 | 78997 | Si1Sr1 | 2 |
| | | 63.1.9 | 99253 | O7Re1Tb3 | 2 |
| | | 63.1.10 | 96410 | O7Pr3Re1 | 2 |
| | | 63.1.20 | 165363 | In5Y3 | 2 |
| | | 63.1.29 | 26225 | H5Li4Rh1 | 2 |
| | | 63.1.33 | 170875 | Gd3O7Os1 | 2 |
| | | 63.1.34 | 423067 | Ga4Gd1Ni1 | 2 |
| | | 63.1.36 | 261037 | Fe1Se3U1 | 2 |
| | | 63.1.37 | 245105 | Eu3O7Re1 | 2 |
| | | 63.1.38 | 400536 | Eu2O7Ta2 | 2 |
| | | 63.1.41 | 80501 | Er3Ge4 | 2 |
| | | 63.1.47 | 57225 | Dy1Ge3 | 2 |
| | | 63.1.55 | 429089 | Cs1Sc1Se5U1 | 2 |

Continued on next column

**TABLE XV– continued**

| SG | cl | TQC | ICSD | Stoich. Formula | SubSG |
|---|---|---|---|---|---|
| | | 63.1.60 | 236871 | Cr2Mg1O4 | 2 |
| | | 63.1.65 | 63036 | Ce2Ge3Li2 | 2 |
| | | 63.1.66 | 262430 | Ce2Ge3Li2 | 2 |
| | | 63.1.68 | 421810 | Cd2F1Sb5Sr5 | 2 |
| | | 63.1.70 | 96228 | Ca2In1N1 | 2 |
| | | 63.1.71 | 165193 | Ca1Sn1 | 2 |
| | | 63.1.75 | 16356 | Ca1Fe5O7 | 2 |
| | | 63.1.77 | 421809 | Ba5Cd2F1Sb5 | 2 |
| | | 63.1.78 | 411979 | Ba3Pb5 | 2 |
| | | 63.1.79 | 165184 | Ba3Pb5 | 2 |
| | | 63.1.82 | 78998 | Ba1Si1 | 2 |
| | | 63.1.84 | 58664 | Ba1Pb1 | 2 |
| | | 63.1.86 | 290596 | Ba1Fe2Se3 | 2 |
| | | 63.1.87 | 290594 | Ba1Fe2Se3 | 2 |
| | | 63.1.88 | 16307 | Ba1Fe2S3 | 2 |
| | | 63.1.91 | 420401 | B4Fe1O7 | 2 |
| | | 63.1.93 | 237221 | Au1Eu1In4 | 2 |
| | | 63.1.98 | 423614 | Al2Ge1La2 | 2 |
| | | 63.1.99 | 99138 | Al1Si1Y1 | 2 |
| | | 63.1.100 | 415373 | Al1Lu1Si1 | 2 |
| | | 63.1.101 | 76295 | Al1Ge1Y1 | 2 |
| | | 63.1.102 | 426322 | Al1Fe1.5Si2 | 2 |
| 63 | 2 | 63.1.1 | 68416 | Te3U1 | 2 |
| | | 63.1.3 | 416971 | Si6Sr1 | 2 |
| | | 63.1.4 | 240779 | Si6Sr1 | 2 |
| | | 63.1.12 | 86102 | Mg1O5V2 | 2 |
| | | 63.1.13 | 50981 | Mg1O5V2 | 2 |
| | | 63.1.14 | 50979 | Mg1O5V2 | 2 |
| | | 63.1.15 | 250752 | La2Sn1Zn5 | 2 |
| | | 63.1.16 | 250756 | La2Pb1Zn5 | 2 |
| | | 63.1.17 | 423611 | La2Li2Si3 | 2 |
| | | 63.1.18 | 410732 | La1Rh1Sn2 | 2 |
| | | 63.1.19 | 38311 | La1Ni5P3 | 2 |
| | | 63.1.30 | 262431 | Ge3La2Li2 | 2 |
| | | 63.1.35 | 189439 | Fe2O3 | 2 |
| | | 63.1.42 | 171141 | Er1Ni1Si2 | 2 |
| | | 63.1.48 | 59858 | Cu3Ho2Rb1S5 | 2 |
| | | 63.1.49 | 174371 | Cu3Ho2K1Se5 | 2 |
| | | 63.1.52 | 95028 | Cu1Ho2K1Se4 | 2 |
| | | 63.1.53 | 97561 | Cu1Ho2K1S4 | 2 |
| | | 63.1.57 | 413677 | Cs1Cu3Ho2Se5 | 2 |
| | | 63.1.59 | 413674 | Cs1Cu3Dy2Se5 | 2 |
| | | 63.1.61 | 20952 | Co3Ga9Ho2 | 2 |
| | | 63.1.62 | 427218 | Ce3Ge3Mn2 | 2 |
| | | 63.1.67 | 99133 | Ce1Rh2Si1 | 2 |
| | | 63.1.72 | 410774 | Ca1Pd1Sn2 | 2 |
| | | 63.1.74 | 408881 | Ca1In2Pd1 | 2 |
| | | 63.1.76 | 75133 | C1Fe2Si1Th1 | 2 |
| | | 63.1.80 | 710014 | Ba1Si6 | 2 |
| | | 63.1.81 | 245295 | Ba1Si6 | 2 |
| | | 63.1.85 | 251583 | Ba1Ge6 | 2 |
| | | 63.1.89 | 64596 | B7Fe1Y3 | 2 |
| | | 63.1.90 | 91243 | B4Mg1Os3 | 2 |
| | | 63.1.92 | 418162 | B2Cr2Ir1 | 2 |
| | | 63.1.96 | 58157 | Al6Ru1 | 2 |
| | | 63.1.103 | 93684 | Ag3Cs1Se5Tb2 | 2 |
| 63 | 3 | 63.1.6 | 87915 | Pd3Te2 | 2 |
| | | 63.1.7 | 77902 | Pd3Te2 | 2 |
| | | 63.1.8 | 411798 | Pd1Sn2Yb1 | 2 |
| | | 63.1.11 | 96149 | Np1Si1 | 2 |
| | | 63.1.21 | 165362 | In5La3 | 2 |

Continued on next column

**TABLE XV– continued**

| SG | cl | TQC | ICSD | Stoich. Formula | SubSG |
|---|---|---|---|---|---|
| | | 63.1.22 | 418178 | In4Ni1Sr1 | 2 |
| | | 63.1.23 | 424100 | In2Pt1Yb1 | 2 |
| | | 63.1.24 | 410437 | In2Pd1Yb1 | 2 |
| | | 63.1.25 | 414077 | In2Ni1Yb1 | 2 |
| | | 63.1.26 | 412431 | In2La1Rh1 | 2 |
| | | 63.1.27 | 414479 | In2Ir1La1 | 2 |
| | | 63.1.28 | 200067 | Hf3Ni2Si3 | 2 |
| | | 63.1.31 | 42375 | Ge1Nd1 | 2 |
| | | 63.1.32 | 42374 | Ge1Nd1 | 2 |
| | | 63.1.39 | 40812 | Eu1Ni5P3 | 2 |
| | | 63.1.40 | 237971 | Eu1In4Ir1 | 2 |
| | | 63.1.43 | 423066 | Er1Ga4Ni1 | 2 |
| | | 63.1.44 | 152349 | Dy1Sn2 | 2 |
| | | 63.1.45 | 180214 | Dy1Si1 | 2 |
| | | 63.1.46 | 154699 | Dy1Si1 | 2 |
| | | 63.1.50 | 423716 | Cu3Er2S5Tl1 | 2 |
| | | 63.1.51 | 174372 | Cu3Er2K1Se5 | 2 |
| | | 63.1.54 | 106440 | Cu1Eu1In4 | 2 |
| | | 63.1.56 | 413682 | Cs1Cu3Se5Yb2 | 2 |
| | | 63.1.58 | 412074 | Cs1Cu3Er2S5 | 2 |
| | | 63.1.63 | 250753 | Ce2Sn1Zn5 | 2 |
| | | 63.1.64 | 250757 | Ce2Pb1Zn5 | 2 |
| | | 63.1.69 | 424107 | Ca2Sn6Zn3 | 2 |
| | | 63.1.73 | 408880 | Ca1In2Pt1 | 2 |
| | | 63.1.83 | 427286 | Ba1Ru3Sn6 | 2 |
| | | 63.1.94 | 33916 | As3Eu1Ni5 | 2 |
| | | 63.1.95 | 104653 | Al9Co3Y2 | 2 |
| | | 63.1.97 | 370014 | Al2Ni1Yb1 | 2 |
| 63 | 4 | 63.2.32 | 15051 | Hf1Te5 | 2 |
| | | 63.2.33 | 171235 | Ge6Pr2Zn3 | 2 |
| | | 63.2.34 | 171234 | Ge6Nd2Zn3 | 2 |
| | | 63.2.35 | 171232 | Ge6La2Zn3 | 2 |
| | | 63.2.40 | 104087 | Gd1Ni1 | 2 |
| | | 63.2.43 | 262740 | Ga2Ni1Yb1 | 2 |
| | | 63.2.44 | 262739 | Ga2Ni1Yb1 | 2 |
| | | 63.2.45 | 262738 | Ga2Ni1Yb1 | 2 |
| | | 63.2.46 | 262737 | Ga2Ni1Yb1 | 2 |
| | | 63.2.78 | 58948 | Ca3Zn1 | 2 |
| | | 63.2.95 | 429723 | Ba2Ge4Ir4Mg1 | 2 |
| | | 63.2.109 | 15032 | As3Nb4 | 2 |
| | | 63.2.112 | 415372 | Al1Si1Tm1 | 2 |
| | | 63.2.113 | 54470 | Al1La1 | 2 |
| | | 63.2.120 | 238167 | Ag1.2Cs0.4Er0.8Te2 | 2 |
| 63 | 5 | 63.2.3 | 411981 | Sn5Sr3 | 2 |
| | | 63.2.5 | 54756 | Sn4Sr1 | 2 |
| | | 63.2.6 | 281669 | Sn4Sr1 | 2 |
| | | 63.2.8 | 95179 | Si2Zr1 | 2 |
| | | 63.2.10 | 22235 | Sb2Yb1 | 2 |
| | | 63.2.20 | 71692 | Nd1Ni1Si2 | 2 |
| | | 63.2.21 | 71691 | Nd1Ni1Si2 | 2 |
| | | 63.2.25 | 418180 | In4Pt1Sr1 | 2 |
| | | 63.2.27 | 418179 | In4Pd1Sr1 | 2 |
| | | 63.2.28 | 415937 | In4Pb1Sr3 | 2 |
| | | 63.2.38 | 240728 | Ge1Na2Zn1 | 2 |
| | | 63.2.54 | 410700 | Eu1Ir1Sn2 | 2 |
| | | 63.2.56 | 404599 | Eu1Ge2Ir1 | 2 |
| | | 63.2.60 | 59896 | Cu1Gd2Rb1S4 | 2 |
| | | 63.2.61 | 416795 | Cu1Eu1Sn2 | 2 |
| | | 63.2.65 | 193548 | Cr2Mg1O4 | 2 |
| | | 63.2.66 | 192389 | Cr2Fe1O4 | 2 |
| | | 63.2.68 | 102378 | Co2Eu1Ga3 | 2 |

Continued on next column

**TABLE XV– continued**

| SG | cl | TQC | ICSD | Stoich. Formula | SubSG |
|---|---|---|---|---|---|
| | | 63.2.70 | 42711 | Ce1Ni1Si2 | 2 |
| | | 63.2.73 | 150132 | Ce1In2Pd1 | 2 |
| | | 63.2.74 | 106394 | Ce1Ge2Ni1 | 2 |
| | | 63.2.76 | 425064 | Cd2Nd1Pd1 | 2 |
| | | 63.2.81 | 410890 | Ca1In4Pd1 | 2 |
| | | 63.2.82 | 106347 | Ca1In4Ni1 | 2 |
| | | 63.2.87 | 159578 | C1Gd1Ru2Si1 | 2 |
| | | 63.2.90 | 411622 | C1B3Nb3 | 2 |
| | | 63.2.92 | 411980 | Ba3Sn5 | 2 |
| | | 63.2.93 | 167166 | Ba3Sn5 | 2 |
| | | 63.2.94 | 167163 | Ba3Ge5 | 2 |
| | | 63.2.96 | 418610 | Ba1Zn5 | 2 |
| | | 63.2.106 | 411390 | Au1Eu1In2 | 2 |
| | | 63.2.107 | 107515 | Au1Co1Na4O5 | 2 |
| | | 63.2.111 | 153384 | Al2Ba1Si2 | 2 |
| | | 63.2.114 | 251809 | Al1B1W1 | 2 |
| | | 63.2.115 | 251808 | Al1B1Mo1 | 2 |
| | | 63.2.118 | 152354 | Al10Fe2Th1 | 2 |
| | | 63.2.119 | 59912 | Al10Ce1Ru2 | 2 |
| 63 | 6 | 63.2.4 | 165183 | Sn5Sr3 | 2 |
| | | 63.2.9 | 20284 | Si1Zr1 | 2 |
| | | 63.2.13 | 165577 | Pd1Sr1Tl2 | 2 |
| | | 63.2.14 | 78010 | O7Pr3Sb1 | 2 |
| | | 63.2.18 | 63084 | Nd3O7Ru1 | 2 |
| | | 63.2.19 | 190762 | Nd3O7Ru1 | 2 |
| | | 63.2.22 | 425493 | Mg2Pd1Sr1 | 2 |
| | | 63.2.24 | 412908 | In4Pt1Yb1 | 2 |
| | | 63.2.26 | 411398 | In4Pd1Yb1 | 2 |
| | | 63.2.29 | 410400 | In2Pt1Sr1 | 2 |
| | | 63.2.30 | 410402 | In2Pd1Sr1 | 2 |
| | | 63.2.31 | 190810 | Hf2Pt3 | 2 |
| | | 63.2.36 | 56047 | Ge2Th1 | 2 |
| | | 63.2.37 | 240343 | Ge2Ni1Sr1 | 2 |
| | | 63.2.41 | 423070 | Ga4Ni1Yb1 | 2 |
| | | 63.2.42 | 154676 | Ga4Ni1Yb1 | 2 |
| | | 63.2.48 | 2795 | Ga1 | 2 |
| | | 63.2.50 | 98 | Fe1S3U1 | 2 |
| | | 63.2.51 | 261040 | Fe1S3U1 | 2 |
| | | 63.2.52 | 261039 | Fe1S3U1 | 2 |
| | | 63.2.53 | 261038 | Fe1S3U1 | 2 |
| | | 63.2.55 | 411268 | Eu1In2Rh1 | 2 |
| | | 63.2.57 | 425774 | Er3Ge3Ru2 | 2 |
| | | 63.2.58 | 55524 | Er1Ge3 | 2 |
| | | 63.2.59 | 55523 | Er1Ge3 | 2 |
| | | 63.2.63 | 96226 | Cs1Cu3Se5Tb2 | 2 |
| | | 63.2.64 | 413680 | Cs1Cu3Se5Tb2 | 2 |
| | | 63.2.67 | 20951 | Co3Ga9Y2 | 2 |
| | | 63.2.69 | 246169 | Ce1Pt1Si2 | 2 |
| | | 63.2.72 | 157363 | Ce1In4Pd1 | 2 |
| | | 63.2.75 | 425494 | Cd2Pd1Sr1 | 2 |
| | | 63.2.77 | 421432 | Ca4Ge3Ni4 | 2 |
| | | 63.2.83 | 240342 | Ca1Ge2Ni1 | 2 |
| | | 63.2.84 | 710075 | Ca1Ga2Ni1 | 2 |
| | | 63.2.85 | 16355 | Ca1Fe4O6 | 2 |
| | | 63.2.88 | 50633 | C1Dy1Ru2Si1 | 2 |
| | | 63.2.89 | 40773 | C1Dy1Fe2Si1 | 2 |
| | | 63.2.91 | 107610 | Be1Pd3 | 2 |
| | | 63.2.97 | 405111 | Ba1Pd1Sb2 | 2 |
| | | 63.2.99 | 411228 | Ba1In2Pt1 | 2 |
| | | 63.2.100 | 414228 | Ba1In1 | 2 |
| | | 63.2.101 | 429115 | Ba1Cd2Pt1 | 2 |

Continued on next column

**TABLE XV– continued**

| SG | cl | TQC | ICSD | Stoich. Formula | SubSG |
|---|---|---|---|---|---|
| | | 63.2.104 | 43662 | B1Re3 | 2 |
| | | 63.2.105 | 107505 | Au1In2Na1 | 2 |
| | | 63.2.108 | 290262 | As4Cd5Rb2 | 2 |
| | | 63.2.110 | 405110 | As2Ba1Pd1 | 2 |
| | | 63.2.116 | 182114 | Al10Ru2U1 | 2 |
| | | 63.2.117 | 99179 | Al10Fe2U1 | 2 |
| 63 | 7 | 63.2.1 | 43419 | U1 | 2 |
| | | 63.2.2 | 43339 | U1 | 2 |
| | | 63.2.7 | 73619 | Si3Tb2 | 2 |
| | | 63.2.11 | 163454 | Rh1Zr1 | 2 |
| | | 63.2.12 | 55546 | Pd1Zr1 | 2 |
| | | 63.2.15 | 96322 | Ni1Zr1 | 2 |
| | | 63.2.16 | 105478 | Ni1Zr1 | 2 |
| | | 63.2.17 | 71964 | Ni1O1Zr3 | 2 |
| | | 63.2.23 | 163453 | Ir1Zr1 | 2 |
| | | 63.2.39 | 163917 | Gd3Ge4 | 2 |
| | | 63.2.47 | 423623 | Ga1La1 | 2 |
| | | 63.2.49 | 290249 | Fe4O5 | 2 |
| | | 63.2.62 | 413681 | Cs1Cu3Se5Tm2 | 2 |
| | | 63.2.71 | 154440 | Ce1Ni1 | 2 |
| | | 63.2.79 | 78996 | Ca1Si1 | 2 |
| | | 63.2.80 | 190659 | Ca1Si1 | 2 |
| | | 63.2.86 | 260482 | C1P1Ru2Tb1 | 2 |
| | | 63.2.98 | 14110 | Ba1Ni2Si2 | 2 |
| | | 63.2.102 | 39663 | B7W1Y3 | 2 |
| | | 63.2.103 | 62282 | B7Co4Nb3 | 2 |
| 64 | 1 | 64.1.1 | 411192 | Sn3Ti2 | 2 |
| | | 64.1.2 | 410902 | Sn3Ti2 | 2 |
| | | 64.1.26 | 68347 | Ag1Fe2S3 | 2 |
| 64 | 2 | 64.1.4 | 417180 | P1 | 2 |
| | | 64.1.5 | 23836 | P1 | 2 |
| | | 64.1.6 | 150873 | P1 | 2 |
| | | 64.1.8 | 79878 | K1Na2Ni1O2 | 2 |
| | | 64.1.11 | 150569 | Ga1Sn2V2 | 2 |
| | | 64.1.12 | 106752 | Ga1Sn2V2 | 2 |
| | | 64.1.19 | 58636 | Ba2Bi4Cd3 | 2 |
| | | 64.1.20 | 416447 | Ba2Bi4Cd3 | 2 |
| | | 64.1.21 | 416446 | Ba2Bi4Cd3 | 2 |
| | | 64.1.22 | 416445 | Ba2Bi4Cd3 | 2 |
| | | 64.1.23 | 416444 | Ba2Bi4Cd3 | 2 |
| | | 64.1.25 | 63708 | As2Ni3O8 | 2 |
| 64 | 3 | 64.1.18 | 186936 | Ba2Ge4Pd5 | 2 |
| 64 | 4 | 64.1.7 | 77831 | La1S1 | 2 |
| | | 64.1.10 | 90341 | Ge2Ni1 | 2 |
| | | 64.1.15 | 424937 | Co1Cs1O2 | 2 |
| | | 64.1.16 | 40372 | Ca5Ga2N4 | 2 |
| | | 64.1.17 | 103352 | Ca2Mn1O3.5 | 2 |
| 64 | 5 | 64.1.13 | 71196 | Co2Nb2Te4 | 2 |
| | | 64.1.14 | 71519 | Co1Nb1Te2 | 2 |
| | | 64.1.24 | 412877 | Ba1Fe1K1O3 | 2 |
| 64 | 6 | 64.1.9 | 40442 | Hf2Ni3Si4 | 2 |
| 64 | 7 | 64.1.3 | 413279 | Pd8Sn24 | 2 |
| 65 | 1 | 65.1.9 | 60646 | Ge3Ni1Y1 | 2 |
| | | 65.1.12 | 261512 | Fe2O5Sr3 | 2 |
| | | 65.1.16 | 418215 | Cu1Mg4Tb1 | 2 |
| | | 65.1.22 | 78621 | Ba2Cu4O8Yb1 | 2 |
| | | 65.1.23 | 154010 | Ba1Mn2O6Tb1 | 2 |
| | | 65.1.24 | 106291 | Au7Rb3 | 2 |
| | | 65.1.25 | 106290 | Au7Rb3 | 2 |
| | | 65.1.26 | 418529 | Au4Ce1Si2 | 2 |

Continued on next column

**TABLE XV– continued**

| SG | cl | TQC | ICSD | Stoich. Formula | SubSG |
|---|---|---|---|---|---|
| 65 | 2 | 65.1.1 | 87827 | Ni1Si3Y1 | 2 |
| | | 65.1.7 | 415513 | In16Pt7Tb2 | 2 |
| | | 65.1.13 | 238922 | Fe2Ge7U3 | 2 |
| | | 65.1.21 | 413212 | Br2Ce4Ga5 | 2 |
| 65 | 3 | 65.1.2 | 78760 | Mn2S4Sn1 | 2 |
| | | 65.1.5 | 167637 | In9Ni4Y11 | 2 |
| | | 65.1.11 | 20337 | Fe2Si7U3 | 2 |
| | | 65.1.18 | 63667 | Ce3Sn7 | 2 |
| | | 65.1.19 | 261436 | Cd1Cu2Nd2 | 2 |
| 65 | 4 | 65.1.3 | 240761 | Mg3Ni2Tb2 | 2 |
| | | 65.1.6 | 167624 | In9La11Ni4 | 2 |
| | | 65.1.14 | 174439 | Er3Sn7 | 2 |
| | | 65.1.15 | 165783 | Dy2Ni1Sn6 | 2 |
| | | 65.1.17 | 424317 | Ce7Ge10Li8 | 2 |
| 65 | 5 | 65.1.4 | 261871 | La2Sn5 | 2 |
| | | 65.1.10 | 421933 | Gd2Mg3Ni2 | 2 |
| | | 65.1.20 | 419476 | C3B2Tb2 | 2 |
| 65 | 6 | 65.1.8 | 54612 | Ho2Ni2Pb1 | 2 |
| 65 | 7 | 65.2.1 | 15927 | Mg1O3V1 | 2 |
| | | 65.2.3 | 106923 | Lu1Ni1Sn4 | 2 |
| | | 65.2.9 | 40487 | Ge2Lu1Mn1 | 2 |
| | | 65.2.12 | 171189 | Er1Ni1Si3 | 2 |
| | | 65.2.13 | 154835 | Dy1Fe4Ge2 | 2 |
| | | 65.2.18 | 174528 | Ce2Si7 | 2 |
| | | 65.2.26 | 68002 | B6Lu2Ni3 | 2 |
| | | 65.2.29 | 260798 | Al1B4Yb1 | 2 |
| | | 65.2.30 | 167533 | Al1B4Yb1 | 2 |
| 65 | 8 | 65.2.28 | 413191 | Al5Br2La4 | 2 |
| 65 | 9 | 65.2.4 | 83885 | Ge2U1 | 2 |
| | | 65.2.5 | 82702 | Ge2U1 | 2 |
| | | 65.2.6 | 236995 | Ge2U1 | 2 |
| | | 65.2.14 | 419475 | Cu1Mg4Y1 | 2 |
| | | 65.2.27 | 417990 | B4Rh8Zn5 | 2 |
| 65 | 10 | 65.2.2 | 104809 | Lu2Ni1Sn6 | 2 |
| | | 65.2.7 | 56046 | Ge2Th1 | 2 |
| | | 65.2.20 | 185392 | Br2Cd2Hg2S2 | 2 |
| | | 65.2.25 | 39394 | B6Ni3Yb2 | 2 |
| 65 | 11 | 65.2.10 | 104134 | Gd3Sn7 | 2 |
| | | 65.2.21 | 391060 | Ba3Ge4 | 2 |
| | | 65.2.22 | 75687 | Ba2Cu4Er1O8 | 2 |
| | | 65.2.23 | 74127 | Ba2Cu4Er1O8 | 2 |
| | | 65.2.24 | 67642 | Ba2Cu4Er1O8 | 2 |
| 65 | 12 | 65.2.8 | 193219 | Ge2Pb1Sr2 | 2 |
| | | 65.2.17 | 16497 | Ce3Ni2Si8 | 2 |
| 65 | 13 | 65.2.11 | 91064 | Fe4O11Sr4 | 2 |
| | | 65.2.16 | 72872 | Co1O4Re1 | 2 |
| | | 65.2.31 | 167535 | Al1B4Lu1 | 2 |
| 65 | 15 | 65.2.15 | 418093 | Co2Ge7U3 | 2 |
| | | 65.2.19 | 39221 | C4Er4Ni13 | 2 |
| 67 | 1 | 67.2.1 | 167666 | Al2Cu1Ir1 | 2 |
| | | 69.1.5 | 414574 | Ge6Li2Sr4 | 2 |
| | | 69.1.18 | 410522 | Ag2Ca4Si6 | 2 |
| 69 | 2 | 69.1.14 | 72328 | Ba2Cu3Na1O6 | 2 |
| | | 69.1.15 | 59589 | Ba2Cu3Na1O6 | 2 |
| | | 69.1.16 | 401239 | Ba2Cu3Li1O6 | 2 |
| 69 | 3 | 69.1.6 | 68676 | Cu2O3Sr2 | 2 |
| | | 69.1.12 | 73160 | Ba2Na1Ni3O6 | 2 |
| | | 69.1.13 | 59588 | Ba2Na1Ni3O6 | 2 |
| 69 | 4 | 69.1.1 | 262559 | Pt3Rb2Se6U1 | 2 |
| | | 69.1.2 | 262561 | Pd3Rb2Se6U1 | 2 |

Continued on next column

**TABLE XV– continued**

| SG | cl | TQC | ICSD | Stoich. Formula | SubSG |
|---|---|---|---|---|---|
| | | 69.1.3 | 262562 | Pd3Rb2S6U1 | 2 |
| | | 69.1.4 | 262563 | K2Pd3S6U1 | 2 |
| | | 69.1.7 | 262555 | Cs2Pt3Se6U1 | 2 |
| | | 69.1.9 | 262557 | Cs2Pd3Se6U1 | 2 |
| | | 69.1.10 | 261242 | Cs2Pd3Se6U1 | 2 |
| | | 69.1.11 | 262558 | Cs2Pd3S6U1 | 2 |
| 69 | 11 | 69.1.8 | 262556 | Cs2Pt3S6U1 | 2 |
| | | 69.1.17 | 174521 | Au1Eu2Ge3 | 2 |
| 70 | 1 | 70.1.4 | 106552 | Nb1Sn2 | 2 |
| | | 70.1.5 | 429511 | Ga2Rh2Sr1 | 2 |
| | | 70.1.14 | 77143 | Ba1Ge2Ru2 | 2 |
| | | 70.1.15 | 40620 | B2Gd1Ir2 | 2 |
| | | 70.1.16 | 40799 | B2Ce1Ru2 | 2 |
| | | 70.1.17 | 40618 | B2Ce1Ir2 | 2 |
| | | 70.1.18 | 49737 | B2Ca1Rh2 | 2 |
| | | 70.1.19 | 49738 | B2Ca1Ir2 | 2 |
| | | 70.1.20 | 75028 | B1Ir1Li1 | 2 |
| | | 70.1.21 | 75027 | B1Ir1Li1 | 2 |
| | | 70.1.23 | 410521 | Ag2Eu4Si6 | 2 |
| 70 | 2 | 70.1.1 | 1089 | Si2Ti1 | 2 |
| | | 70.1.2 | 97545 | Sb1Sn1Ti1 | 2 |
| | | 70.1.3 | 88369 | O4Tb1V1 | 2 |
| | | 70.1.6 | 89074 | Cr2O4Si1 | 2 |
| | | 70.1.7 | 89073 | Cr2O4Si1 | 2 |
| | | 70.1.8 | 89072 | Cr2O4Si1 | 2 |
| | | 70.1.9 | 89071 | Cr2O4Si1 | 2 |
| | | 70.1.10 | 89070 | Cr2O4Si1 | 2 |
| | | 70.1.11 | 89069 | Cr2O4Si1 | 2 |
| | | 70.1.12 | 75639 | Cr2O4Si1 | 2 |
| | | 70.1.13 | 96610 | Cl2Cu2O1 | 2 |
| | | 70.1.22 | 418163 | B1Cr2 | 2 |
| 71 | 1 | 71.1.21 | 56747 | H1Li1Nd2O3 | 2 |
| | | 71.1.22 | 56746 | H1Li1Nd2O3 | 2 |
| | | 71.1.27 | 262435 | Ge4Li4Nd3 | 2 |
| | | 71.1.32 | 300260 | Ge2Ho1Pt1 | 2 |
| | | 71.1.49 | 151160 | Ce3Cu4Sn4 | 2 |
| | | 71.1.65 | 72631 | Ba1Ni1O5Tm2 | 2 |
| | | 71.1.66 | 71415 | Ba1Ni1O5Tm2 | 2 |
| | | 71.1.67 | 67831 | Ba1Ni1O5Tm2 | 2 |
| | | 71.1.68 | 62761 | Ba1Ni1O5Tm2 | 2 |
| | | 71.1.69 | 62426 | Ba1Ni1O5Tm2 | 2 |
| | | 71.1.72 | 72626 | Ba1Nd2Ni1O5 | 2 |
| | | 71.1.73 | 62627 | Ba1Nd2Ni1O5 | 2 |
| | | 71.1.86 | 81543 | B6Cr2Th1 | 2 |
| | | 71.1.87 | 81544 | B6Ce1Cr2 | 2 |
| | | 71.1.92 | 281420 | As1Te1Ti1 | 2 |
| | | 71.1.105 | 417491 | Ag4Ce3Ge4 | 2 |
| 71 | 2 | 71.1.28 | 170342 | Ge2Ni1Yb1 | 2 |
| | | 71.1.29 | 170343 | Ge2Lu1Ni1 | 2 |
| | | 71.1.31 | 88227 | Ge2Ho1Pt1 | 2 |
| | | 71.1.37 | 83190 | Cu8Ge8Nd6 | 2 |
| | | 71.1.38 | 102956 | Cu8Gd6Sn8 | 2 |
| | | 71.1.43 | 42717 | Cu4Gd3Ge4 | 2 |
| | | 71.1.46 | 247288 | Ce3Rh4Si4 | 2 |
| | | 71.1.56 | 40950 | C4Co1Sc3 | 2 |
| | | 71.1.63 | 290306 | Ba3Cd1Hg10 | 2 |
| | | 71.1.77 | 72630 | Ba1Er2Ni1O5 | 2 |
| | | 71.1.78 | 69324 | Ba1Er2Ni1O5 | 2 |
| | | 71.1.79 | 69323 | Ba1Er2Ni1O5 | 2 |
| | | 71.1.80 | 69315 | Ba1Er2Ni1O5 | 2 |

Continued on next column

**TABLE XV– continued**

| SG | cl | TQC | ICSD | Stoich. Formula | SubSG |
|---|---|---|---|---|---|
| | | 71.1.81 | 62760 | Ba1Er2Ni1O5 | 2 |
| | | 71.1.82 | 62425 | Ba1Er2Ni1O5 | 2 |
| | | 71.1.83 | 85046 | Ba1Dy2Ni1O5 | 2 |
| | | 71.1.84 | 72627 | Ba1Dy2Ni1O5 | 2 |
| | | 71.1.103 | 163921 | Al16Ce2Pt9 | 2 |
| | | 71.1.104 | 157372 | Ag4Gd3Sn4 | 2 |
| 71 | 3 | 71.1.20 | 416817 | La2Ni2Zn1 | 2 |
| | | 71.1.23 | 56745 | H1Ce2Li1O3 | 2 |
| | | 71.1.24 | 710017 | Ge5Nd1 | 2 |
| | | 71.1.25 | 246814 | Ge5Nd1 | 2 |
| | | 71.1.33 | 174530 | Gd1Si5 | 2 |
| | | 71.1.34 | 710016 | Gd1Ge5 | 2 |
| | | 71.1.36 | 187334 | Fe4Ge4U3 | 2 |
| | | 71.1.51 | 174529 | Ce1Si5 | 2 |
| | | 71.1.52 | 246812 | Ce1Ge5 | 2 |
| | | 71.1.101 | 100128 | Al2Ba3Si2 | 2 |
| | | 71.1.102 | 52612 | Al2Ba3Ge2 | 2 |
| 71 | 4 | 71.1.18 | 85406 | Ni2P2Sr1 | 2 |
| | | 71.1.35 | 154471 | Fe6Ga6U1 | 2 |
| | | 71.1.39 | 413485 | Cu6Sn5Yb3 | 2 |
| | | 71.1.40 | 426132 | Cu4Ge4Yb3 | 2 |
| | | 71.1.47 | 174076 | Ce3Pd4Si4 | 2 |
| | | 71.1.60 | 238028 | C1Co2Pr2Si1 | 2 |
| | | 71.1.61 | 418403 | C1B4Dy2 | 2 |
| | | 71.1.90 | 407576 | B2N4Nd3 | 2 |
| | | 71.1.91 | 68610 | B2Ce3N4 | 2 |
| | | 71.1.93 | 58046 | Al5Ni2Pr1 | 2 |
| | | 71.1.95 | 9564 | Al2Sn2Sr3 | 2 |
| | | 71.1.96 | 100127 | Al2Si2Sr3 | 2 |
| | | 71.1.97 | 52617 | Al2Ge2Sr3 | 2 |
| | | 71.1.98 | 100126 | Al2Ca3Si2 | 2 |
| | | 71.1.100 | 9565 | Al2Ba3Sn2 | 2 |
| 71 | 5 | 71.1.17 | 416820 | Ni2Tb2Zn1 | 2 |
| | | 71.1.19 | 25307 | Li4Sb4Sr3 | 2 |
| | | 71.1.45 | 201112 | Cs1Fe1S2 | 2 |
| | | 71.1.48 | 262434 | Ce3Ge4Li4 | 2 |
| | | 71.1.50 | 416816 | Ce2Ni2Zn1 | 2 |
| | | 71.1.53 | 429175 | Ca2Mg1Pt2 | 2 |
| | | 71.1.54 | 429174 | Ca2Mg1Pd2 | 2 |
| | | 71.1.55 | 425471 | Ca2Cd1Pd2 | 2 |
| | | 71.1.59 | 94741 | C2Bi2Ca1O8 | 2 |
| | | 71.1.62 | 428114 | Ba3Li4Sb4 | 2 |
| | | 71.1.64 | 428115 | Ba3Bi4Li4 | 2 |
| | | 71.1.71 | 75035 | Ba1Ni1O5Pr2 | 2 |
| | | 71.1.74 | 68425 | Ba1Gd2Ni1O5 | 2 |
| | | 71.1.75 | 62758 | Ba1Gd2Ni1O5 | 2 |
| | | 71.1.76 | 62423 | Ba1Gd2Ni1O5 | 2 |
| | | 71.1.85 | 167153 | B6Fe2U1 | 2 |
| | | 71.1.94 | 418387 | Al4O1Si6Sr10 | 2 |
| | | 71.1.99 | 52616 | Al2Ca3Ge2 | 2 |
| 71 | 6 | 71.1.26 | 411182 | Ge4Pd4Yb3 | 2 |
| | | 71.1.30 | 76509 | Ge2Ir1Y1 | 2 |
| | | 71.1.44 | 55650 | Cu4Er3Si4 | 2 |
| | | 71.1.88 | 40791 | B4Cr3 | 2 |
| | | 71.1.89 | 423380 | B4Cr1 | 2 |
| 71 | 7 | 71.1.1 | 82655 | Te2U1 | 2 |
| | | 71.1.2 | 82654 | Te2U1 | 2 |
| | | 71.1.3 | 82653 | Te2U1 | 2 |
| | | 71.1.4 | 82652 | Te2U1 | 2 |
| | | 71.1.5 | 82651 | Te2U1 | 2 |

Continued on next column

**TABLE XV– continued**

| SG | cl | TQC | ICSD | Stoich. Formula | SubSG |
|---|---|---|---|---|---|
| | | 71.1.6 | 82650 | Te2U1 | 2 |
| | | 71.1.7 | 82649 | Te2U1 | 2 |
| | | 71.1.8 | 82648 | Te2U1 | 2 |
| | | 71.1.9 | 82647 | Te2U1 | 2 |
| | | 71.1.10 | 82646 | Te2U1 | 2 |
| | | 71.1.11 | 82645 | Te2U1 | 2 |
| | | 71.1.12 | 82644 | Te2U1 | 2 |
| | | 71.1.13 | 82643 | Te2U1 | 2 |
| | | 71.1.14 | 82642 | Te2U1 | 2 |
| | | 71.1.15 | 63170 | Te2U1 | 2 |
| | | 71.1.16 | 40694 | Te2U1 | 2 |
| | | 71.1.41 | 82565 | Cu4Ge4U3 | 2 |
| | | 71.1.42 | 81142 | Cu4Ge4U3 | 2 |
| | | 71.1.57 | 236393 | C4Co1Sc3 | 2 |
| | | 71.1.58 | 173167 | C4Co1Sc3 | 2 |
| | | 71.1.70 | 66078 | Ba1Ni1O5Tb2 | 2 |
| 72 | 1 | 72.1.1 | 84206 | Ni3Si5Y2 | 2 |
| | | 72.1.2 | 428478 | Eu2Pt3Si5 | 2 |
| 72 | 2 | 72.2.5 | 54107 | Cr4Sc2Si5 | 2 |
| | | 72.2.7 | 28232 | Br2Hg5O4 | 2 |
| 72 | 3 | 72.2.2 | 416512 | Ni1Pd2Te2 | 2 |
| | | 72.2.3 | 416515 | Fe1Pd2Se2 | 2 |
| 72 | 4 | 72.2.1 | 170063 | Pb5Sc6 | 2 |
| 72 | 5 | 72.2.4 | 68792 | Er2Ru3Si5 | 2 |
| | | 72.2.6 | 40758 | C4Fe1Tm2 | 2 |
| 74 | 1 | 74.2.6 | 93229 | Fe2Mo4O7 | 2 |
| | | 74.2.12 | 97094 | Ca2O7Os2 | 2 |
| | | 74.2.13 | 97093 | Ca2O7Os2 | 2 |
| | | 74.2.14 | 97092 | Ca2O7Os2 | 2 |
| | | 74.2.15 | 97091 | Ca2O7Os2 | 2 |
| | | 74.2.18 | 414229 | Ba1In2 | 2 |
| | | 74.2.25 | 417227 | Ag1Ca1Zn1 | 2 |
| 74 | 2 | 74.2.2 | 429139 | Ga2La1Rh3 | 2 |
| | | 74.2.22 | 55542 | Au2Ca1 | 2 |
| | | 74.2.24 | 417225 | Ag2Ca1 | 2 |
| 74 | 3 | 74.2.7 | 74288 | Cs1F5Ni1Pd1 | 2 |
| | | 74.2.11 | 260381 | Cd2Sr1 | 2 |
| | | 74.2.16 | 418971 | Ca1Zn2 | 2 |
| | | 74.2.17 | 420576 | Ca1Cd2 | 2 |
| | | 74.2.21 | 260668 | Ba1Cd2 | 2 |
| 74 | 4 | 74.2.1 | 106234 | Yb1Zn2 | 2 |
| | | 74.2.19 | 251586 | Ba1Ge5 | 2 |
| | | 74.2.20 | 194880 | Ba1Ge5 | 2 |
| | | 74.2.23 | 72289 | Al1F7Na2Ni1 | 2 |
| 74 | 5 | 74.2.3 | 238760 | Fe2O4Si1 | 2 |
| | | 74.2.4 | 238759 | Fe2O4Si1 | 2 |
| | | 74.2.5 | 238758 | Fe2O4Si1 | 2 |
| | | 74.2.9 | 246171 | Ce3Pt5Si1 | 2 |
| | | 74.2.10 | 93244 | Ce3Pd5Si1 | 2 |
| 74 | 6 | 74.2.8 | 10156 | Co2O4Si1 | 2 |
| 82 | 1 | 82.2.1 | 81539 | Te1Zr3 | 82 |
| | | 82.2.2 | 51048 | Te1Zr3 | 82 |
| | | 82.2.3 | 195057 | Sb1Zr3 | 82 |
| | | 82.2.4 | 98373 | Ni3P1 | 82 |
| | | 82.2.5 | 23612 | In1P1S4 | 82 |
| | | 82.2.6 | 56027 | Ge1Ta3 | 82 |
| | | 82.2.7 | 43365 | Fe3P1 | 82 |
| | | 82.2.8 | 23560 | Cr3P1 | 82 |
| | | 83.2.1 | 106354 | Ca7Ni4Sn13 | 2,81 |
| | | 84.2.1 | 74390 | P2S6U1 | 2 |

Continued on next column

**TABLE XV– continued**

| SG | cl | TQC | ICSD | Stoich. Formula | SubSG |
|---|---|---|---|---|---|
| 85 | 6 | 85.2.1 | 33884 | As2Cl4F6 | 2,81 |
| 87 | 1 | 87.1.3 | 55075 | Se4Ti5 | 2,82 |
| | | 87.1.11 | 66017 | As4Ba1Cu8 | 2,82 |
| 87 | 2 | 87.1.10 | 43186 | As4Mo5 | 2,82 |
| 87 | 3 | 87.1.8 | 60822 | C4Cr4U1 | 2,82 |
| | | 87.1.9 | 40319 | C4Cr4U1 | 2,82 |
| 87 | 4 | 87.1.1 | 51047 | Te4Zr5 | 2,82 |
| | | 87.1.2 | 409337 | Te4Zr5 | 2,82 |
| | | 87.1.6 | 57076 | Ge4Li4Yb5 | 2,82 |
| 87 | 5 | 87.1.4 | 157026 | Mo1Ni1O6Sr2 | 87 |
| | | 87.1.5 | 157024 | Mo1Ni1O6Sr2 | 87 |
| | | 87.1.7 | 173488 | Co1O6Re1Sr2 | 87 |
| 113 | 1 | 113.2.1 | 246245 | Ca1Fe1O2 | 81 |
| | | 113.2.2 | 39120 | Be2Gd2Ge1O7 | 81 |
| | | 113.2.3 | 160121 | Ag1Tb1Te2 | 81 |
| | | 113.2.4 | 154793 | Ag1Tb1Te2 | 81 |
| | | 115.1.1 | 93242 | Cu1Sb3Zr2 | 81 |
| | | 115.1.2 | 426313 | Cu1Sb3Zr2 | 81 |
| | | 115.1.3 | 195058 | Cu1Sb3Zr2 | 81 |
| | | 115.1.4 | 93241 | Cu1Sb3Ti2 | 81 |
| | | 115.1.5 | 93243 | Cu1Hf2Sb3 | 81 |
| | | 121.1.1 | 237706 | Mo3P1 | 82 |
| | | 121.1.2 | 104607 | Ir1Zr3 | 82 |
| | | 121.1.3 | 400652 | Cu3Sb1Se4 | 82 |
| | | 121.1.4 | 412239 | Cu3S4Sb1 | 82 |
| | | 121.1.5 | 2857 | Cu3S4Sb1 | 82 |
| | | 121.1.6 | 95119 | Cu2Hg1Se4Sn1 | 82 |
| | | 121.1.7 | 152761 | Cu2Ge1Hg1Se4 | 82 |
| | | 121.1.8 | 85126 | Cu2Fe1Se4Sn1 | 82 |
| | | 121.1.9 | 165094 | Cd1Cu2Ge1Te4 | 82 |
| | | 122.1.1 | 248539 | O6Rb4 | 82 |
| | | 122.1.2 | 659 | F5U1 | 82 |
| | | 122.1.3 | 16427 | F5U1 | 82 |
| | | 122.1.4 | 16426 | F5U1 | 82 |
| | | 122.1.5 | 16425 | F5U1 | 82 |
| | | 122.1.6 | 16424 | F5U1 | 82 |
| | | 122.1.7 | 84027 | Cu1In1Se2 | 82 |
| | | 122.1.8 | 73351 | Cu1In1Se2 | 82 |
| | | 122.1.9 | 154409 | Cu1In1Se2 | 82 |
| | | 122.1.10 | 154408 | Cu1In1Se2 | 82 |
| | | 122.1.11 | 154407 | Cu1In1Se2 | 82 |
| | | 122.1.12 | 154406 | Cu1In1Se2 | 82 |
| | | 122.1.13 | 154405 | Cu1In1Se2 | 82 |
| | | 122.1.14 | 68325 | As2Sn1Zn1 | 82 |
| | | 122.1.15 | 42098 | As2Cd1Ge1 | 82 |
| | | 123.1.17 | 74163 | Ba2Cu2O7Tl1Y1 | 2 |
| | | 123.1.24 | 170662 | Ag1Sb1Te2 | 2 |
| 123 | 2 | 123.1.1 | 193701 | In5Rh1U1 | 2 |
| | | 123.1.2 | 103639 | Ga5Ir1U1 | 2 |
| | | 123.1.5 | 102455 | Co1Ga5U1 | 2 |
| | | 123.1.22 | 262038 | As3Rb1Zn4 | 2 |
| | | 123.1.23 | 262030 | As3Cs1Zn4 | 2 |
| 123 | 3 | 123.1.21 | 165631 | Ba1Ge3Mg4 | 2 |
| 123 | 4 | 123.1.20 | 237789 | Ba1O1Sb2Ti2 | 2 |
| 123 | 5 | 123.1.4 | 106640 | Fe1Ga8Tm2 | 2 |
| | | 123.1.18 | 430061 | Ba1O1Sb2Ti2 | 2 |
| | | 123.1.19 | 430060 | Ba1O1Sb2Ti2 | 2 |
| 123 | 6 | 123.1.16 | 68047 | Ba2Cu3Ho1O6 | 2 |
| 123 | 7 | 123.1.7 | 150226 | Ce1In5Rh1 | 83 |
| | | 123.1.8 | 110780 | Ce1In5Rh1 | 83 |

Continued on next column

**TABLE XV– continued**

| SG | cl | TQC | ICSD | Stoich. Formula | SubSG |
|---|---|---|---|---|---|
| | | 123.1.9 | 110779 | Ce1In5Rh1 | 83 |
| | | 123.1.10 | 110778 | Ce1In5Rh1 | 83 |
| | | 123.1.11 | 110777 | Ce1In5Rh1 | 83 |
| | | 123.1.12 | 150225 | Ce1In5Ir1 | 83 |
| | | 123.1.13 | 92435 | Bi2I1O4Tm1 | 83 |
| | | 123.1.14 | 92408 | Bi2Cl1O4Tm1 | 83 |
| | | 123.1.15 | 92421 | Bi2Br1O4Tm1 | 83 |
| 123 | 8 | 123.1.3 | 5257 | Fe1Si2 | 2 |
| 123 | 9 | 123.2.31 | 251834 | Al2B1La1Os2 | 2,81 |
| 123 | 10 | 123.2.30 | 174342 | Al2B1La1Ru2 | 2,81 |
| 123 | 11 | 123.2.2 | 428960 | Mg1Ru1Sc2 | 2,81 |
| | | 123.2.29 | 169074 | As2Ba1O1Ti2 | 2,81 |
| 123 | 12 | 123.2.28 | 165630 | Ba1Mg4Si3 | 2,81 |
| 123 | 13 | 123.2.27 | 64639 | Ba2Cu2La1O8Ta1 | 2,81 |
| 123 | 14 | 123.2.16 | 84739 | Ba2Cu3O6Y1 | 2,81 |
| | | 123.2.17 | 72254 | Ba2Cu3O6Y1 | 2,81 |
| | | 123.2.18 | 68290 | Ba2Cu3O6Y1 | 2,81 |
| | | 123.2.19 | 68046 | Ba2Cu3O6Y1 | 2,81 |
| | | 123.2.20 | 68008 | Ba2Cu3O6Y1 | 2,81 |
| | | 123.2.21 | 67016 | Ba2Cu3O6Y1 | 2,81 |
| | | 123.2.22 | 66903 | Ba2Cu3O6Y1 | 2,81 |
| | | 123.2.23 | 63269 | Ba2Cu3O6Y1 | 2,81 |
| | | 123.2.24 | 203116 | Ba2Cu3O6Y1 | 2,81 |
| | | 123.2.25 | 98114 | Ba2Cu3Lu1O6 | 2,81 |
| | | 123.2.26 | 98113 | Ba2Cu3Lu1O6 | 2,81 |
| 123 | 15 | 123.2.7 | 240161 | Ce1Ga6Pd1 | 2,81 |
| 123 | 16 | 123.2.5 | 74164 | Cu2O7Sr2Tl1Y1 | 2,81 |
| | | 123.2.8 | 102108 | Ce1Co1In5 | 2,81 |
| 123 | 17 | 123.2.3 | 157999 | Fe5Si3U1 | 2,81 |
| 123 | 24 | 123.2.9 | 92425 | Bi2I1O4Pr1 | 2,81 |
| | | 123.2.10 | 92403 | Bi2Cl1O4Tb1 | 2,81 |
| | | 123.2.11 | 92398 | Bi2Cl1O4Pr1 | 2,81 |
| | | 123.2.12 | 92401 | Bi2Cl1Eu1O4 | 2,81 |
| | | 123.2.13 | 92416 | Bi2Br1O4Tb1 | 2,81 |
| | | 123.2.14 | 92411 | Bi2Br1O4Pr1 | 2,81 |
| 123 | 31 | 123.2.1 | 98585 | Nd1Ni1O2 | 2,81 |
| 123 | 38 | 123.2.6 | 85244 | Cu2Gd1O8Ru1Sr2 | 2,81 |
| 123 | 43 | 123.2.4 | 427123 | Er2Mg3Ru1 | 2,81 |
| 123 | 47 | 123.2.15 | 424316 | Ba3O5Te1Yb2 | 2,81 |
| 123 | 56 | 123.1.6 | 192621 | Ce2In8Pt1 | 2 |
| 125 | 1 | 125.1.1 | 412794 | Pr1Rb1Te4 | 2 |
| | | 125.1.2 | 391204 | K1Pr1Te4 | 2 |
| | | 125.1.3 | 152562 | Ce1Ga6 | 2 |
| | | 125.1.4 | 102167 | Ce1Ga6 | 2 |
| 125 | 2 | 125.2.1 | 400461 | H8I1N4P1 | 2,81 |
| 127 | 1 | 127.2.37 | 107439 | Al1Pt3 | 2,81 |
| 127 | 2 | 127.2.1 | 105627 | Pb3Sr2 | 2 |
| 127 | 3 | 127.2.35 | 182659 | B4Ho1 | 83 |
| 127 | 4 | 127.2.3 | 391441 | Mg1Si2Yb2 | 2,81 |
| | | 127.2.5 | 415117 | Dy2Mg1Si2 | 2,81 |
| | | 127.2.6 | 423452 | Dy2Ge2Mg1 | 2,81 |
| | | 127.2.7 | 262925 | Cs1I3Sn1 | 2,81 |
| | | 127.2.13 | 107389 | Cd1La2Rh2 | 2,81 |
| | | 127.2.14 | 424179 | Cd1Ge2Yb2 | 2,81 |
| | | 127.2.17 | 424177 | Cd1Er2Ge2 | 2,81 |
| | | 127.2.18 | 424175 | Cd1Dy2Ge2 | 2,81 |
| | | 127.2.33 | 26777 | B4Y1 | 2,81 |
| | | 127.2.34 | 2360 | B4La1 | 2,81 |
| 127 | 5 | 127.2.32 | 67222 | B4Y1 | 2,81 |
| 127 | 6 | 127.2.30 | 404497 | Ba1Ho2O5Pd1 | 2,81 |

Continued on next column

**TABLE XV– continued**

| SG | cl | TQC | ICSD | Stoich. Formula | SubSG |
|---|---|---|---|---|---|
| 127 | 7 | 127.2.25 | 94036 | C2B2Ce1 | 2 |
| | | 127.2.26 | 88561 | C2B2Ce1 | 2 |
| | | 127.2.27 | 88560 | C2B2Ce1 | 2 |
| | | 127.2.28 | 280180 | C2B2Ce1 | 2 |
| | | 127.2.31 | 404496 | Ba1Dy2O5Pd1 | 2 |
| 127 | 8 | 127.2.21 | 39204 | C5Ru6Th2 | 2,81 |
| 127 | 9 | 127.2.20 | 79239 | C5Ru6Th2 | 2,81 |
| 127 | 10 | 127.2.15 | 424169 | Cd1Ge2Y2 | 2,81 |
| 127 | 11 | 127.2.8 | 413849 | Ce2Ge2Mg1 | 2,81 |
| 127 | 12 | 127.2.4 | 26200 | Li2Sn5 | 2 |
| 127 | 13 | 127.2.24 | 94040 | C2B2Dy1 | 2 |
| 127 | 17 | 127.2.22 | 427156 | C2B2Gd1 | 2 |
| | | 127.2.23 | 94042 | C2B2Er1 | 2 |
| 127 | 19 | 127.2.36 | 51434 | B2Rh5Sc3 | 2,81 |
| 127 | 20 | 127.2.29 | 202173 | Ba1O5Pd1Tb2 | 2 |
| 127 | 21 | 127.2.19 | 193226 | Ca2Ge2Sn1 | 2,81 |
| 127 | 23 | 127.2.11 | 157385 | Cd1Pd2Pr2 | 2,81 |
| | | 127.2.12 | 414596 | Cd1Ni2Pr2 | 2,81 |
| 127 | 26 | 127.2.16 | 424178 | Cd1Ge2Tm2 | 2,81 |
| 127 | 28 | 127.2.2 | 427282 | Ni2Sn1Tb2 | 2,81 |
| | | 127.2.9 | 157389 | Cd1Pd2Yb2 | 2,81 |
| | | 127.2.10 | 156063 | Cd1Pd2Yb2 | 2,81 |
| 128 | 1 | 128.1.1 | 425655 | B2Nb2Os1 | 2 |
| | | 129.1.1 | 80190 | Sn1Te1Zr1 | 2 |
| | | 129.1.2 | 26273 | Sn1Te1U1 | 2 |
| | | 129.1.4 | 25734 | Si1Te1Zr1 | 2 |
| | | 129.1.5 | 15572 | Si1Te1Zr1 | 2 |
| | | 129.1.6 | 25733 | Se1Si1Zr1 | 2 |
| | | 129.1.7 | 15571 | Se1Si1Zr1 | 2 |
| | | 129.1.8 | 25732 | S1Si1Zr1 | 2 |
| | | 129.1.9 | 15570 | S1Si1Zr1 | 2 |
| | | 129.1.10 | 15569 | S1Si1Zr1 | 2 |
| | | 129.1.11 | 26271 | S1Si1U1 | 2 |
| | | 129.1.17 | 77853 | P2U1 | 2 |
| | | 129.1.18 | 25731 | O1Si1Zr1 | 2 |
| | | 129.1.19 | 85856 | Nd1Ru1Si1 | 2 |
| | | 129.1.20 | 41250 | Nd1Ru1Si1 | 2 |
| | | 129.1.21 | 76235 | Mn1Si1Y1 | 2 |
| | | 129.1.23 | 15420 | I1O1Tm1 | 2 |
| | | 129.1.24 | 25737 | Hf1Si1Te1 | 2 |
| | | 129.1.25 | 25736 | Hf1Se1Si1 | 2 |
| | | 129.1.26 | 25735 | Hf1S1Si1 | 2 |
| | | 129.1.27 | 25740 | Ge1Te1Zr1 | 2 |
| | | 129.1.28 | 25739 | Ge1Se1Zr1 | 2 |
| | | 129.1.29 | 15574 | Ge1Se1Zr1 | 2 |
| | | 129.1.30 | 25738 | Ge1S1Zr1 | 2 |
| | | 129.1.31 | 15573 | Ge1S1Zr1 | 2 |
| | | 129.1.32 | 87315 | Ge1S1U1 | 2 |
| | | 129.1.33 | 26272 | Ge1S1U1 | 2 |
| | | 129.1.34 | 85863 | Ge1Nd1Ru1 | 2 |
| | | 129.1.35 | 25743 | Ge1Hf1Te1 | 2 |
| | | 129.1.36 | 25741 | Ge1Hf1Se1 | 2 |
| | | 129.1.37 | 25742 | Ge1Hf1S1 | 2 |
| | | 129.1.38 | 106758 | Gd1Pt2Si2 | 2 |
| | | 129.1.39 | 180602 | Fe1Te1 | 2 |
| | | 129.1.40 | 169271 | Fe1Se1 | 2 |
| | | 129.1.41 | 169269 | Fe1Se1 | 2 |
| | | 129.1.42 | 169267 | Fe1Se1 | 2 |
| | | 129.1.43 | 169265 | Fe1Se1 | 2 |
| | | 129.1.44 | 169263 | Fe1Se1 | 2 |

Continued on next column

**TABLE XV– continued**

| SG | cl | TQC | ICSD | Stoich. Formula | SubSG |
|---|---|---|---|---|---|
| | | 129.1.45 | 169261 | Fe1Se1 | 2 |
| | | 129.1.46 | 169259 | Fe1Se1 | 2 |
| | | 129.1.47 | 169257 | Fe1Se1 | 2 |
| | | 129.1.48 | 169255 | Fe1Se1 | 2 |
| | | 129.1.49 | 169253 | Fe1Se1 | 2 |
| | | 129.1.52 | 59594 | Cu1P1Si1Zr1 | 2 |
| | | 129.1.54 | 27230 | Cs1F4Ti1 | 2 |
| | | 129.1.55 | 62696 | Cs1F4Fe1 | 2 |
| | | 129.1.56 | 6043 | Cs1F4Fe1 | 2 |
| | | 129.1.58 | 78578 | Co2P2U1 | 2 |
| | | 129.1.59 | 15770 | Co1K1O2 | 2 |
| | | 129.1.60 | 97019 | Cl1Mn1O3Sr2 | 2 |
| | | 129.1.61 | 97018 | Cl1Mn1O3Sr2 | 2 |
| | | 129.1.62 | 96044 | Cl1Mn1O3Sr2 | 2 |
| | | 129.1.63 | 94745 | Cl1Mn1O3Sr2 | 2 |
| | | 129.1.64 | 76171 | Cl1Ho1O1 | 2 |
| | | 129.1.65 | 91750 | Cl1Co1O3Sr2 | 2 |
| | | 129.1.66 | 90122 | Cl1Co1O3Sr2 | 2 |
| | | 129.1.68 | 411678 | Ce1Ir2Si2 | 2 |
| | | 129.1.69 | 61371 | Ce1Ge2Pt2 | 2 |
| | | 129.1.72 | 92319 | Br1Nd1O3Te1 | 2 |
| | | 129.1.81 | 66765 | As2U1 | 2 |
| | | 129.1.82 | 44081 | As2U1 | 2 |
| | | 129.1.83 | 107523 | As2U1 | 2 |
| | | 129.1.89 | 422004 | As1Fe1O1Tb1 | 2 |
| | | 129.1.90 | 42165 | As1Cu1Si1Zr1 | 2 |
| | | 129.1.91 | 423103 | As1Cu1Ge1Zr1 | 2 |
| | | 129.1.92 | 42750 | Al1Ge1Mn1 | 2 |
| | | 129.1.93 | 190195 | Ag1Pr1Sb2 | 2 |
| 129 | 2 | 129.1.3 | 74522 | Si1Te1Zr1 | 2 |
| | | 129.1.22 | 73039 | Ir2Pu1Si2 | 2 |
| | | 129.1.50 | 94475 | F1Ho1S1 | 2 |
| | | 129.1.53 | 91199 | Cu1Mn1O3S1Sr2 | 2 |
| | | 129.1.57 | 16496 | Cr1N1Nb1 | 2 |
| | | 129.1.70 | 106393 | Ce1Ge2Ir2 | 2 |
| | | 129.1.73 | 92318 | Bi1I1O3Te1 | 2 |
| | | 129.1.74 | 246945 | Ba1Ni1S2 | 2 |
| | | 129.1.75 | 15289 | Ba1Ni1S2 | 2 |
| | | 129.1.76 | 41915 | Ba1Mg2Sn2 | 2 |
| | | 129.1.77 | 25338 | Ba1Mg2Pb2 | 2 |
| | | 129.1.79 | 238706 | Au1Bi2Pr1 | 2 |
| | | 129.1.80 | 238707 | Au1Bi2Nd1 | 2 |
| | | 129.1.87 | 246061 | As1La1Ni1O1 | 2 |
| | | 129.1.88 | 248695 | As1Fe1O3Sr2V1 | 2 |
| 129 | 3 | 129.1.12 | 50351 | Ru1Sb2U1 | 2 |
| | | 129.1.13 | 50350 | Ru1Sb2U1 | 2 |
| | | 129.1.14 | 61194 | Pd2Sb2Sr1 | 2 |
| | | 129.1.15 | 50349 | Pd1Sb2U1 | 2 |
| | | 129.1.16 | 50348 | Pd1Sb2U1 | 2 |
| | | 129.1.51 | 42969 | Er1Se2 | 2 |
| | | 129.1.67 | 84529 | Ce1Ni1Sb2 | 2 |
| | | 129.1.71 | 191053 | Ce1Ga2Pd2 | 2 |
| | | 129.1.78 | 370004 | Au2Ga2Sr1 | 2 |
| | | 129.1.84 | 56976 | As2Pd1U1 | 2 |
| | | 129.1.85 | 412554 | As2Cu1Tm1 | 2 |
| | | 129.1.86 | 412551 | As2Cu1Tb1 | 2 |
| 129 | 4 | 129.2.1 | 52412 | Sc1 | 2,81 |
| | | 129.2.2 | 54346 | Pt2Si2U1 | 2,81 |
| | | 129.2.3 | 38176 | O1Te1U1 | 2,81 |
| | | 129.2.4 | 22204 | O1Te1U1 | 2,81 |
| | | 129.2.6 | 74786 | Ni2P2Th1 | 2,81 |

Continued on next column

**TABLE XV– continued**

| SG | cl | TQC | ICSD | Stoich. Formula | SubSG |
|---|---|---|---|---|---|
| | | 129.2.11 | 50243 | Mn1Nd1O1Sb1 | 2,81 |
| | | 129.2.12 | 50241 | Mn1Nd1O1P1 | 2,81 |
| | | 129.2.13 | 423892 | La1N1 | 2,81 |
| | | 129.2.15 | 200099 | I1Np1O1 | 2,81 |
| | | 129.2.16 | 87913 | Ge1Ni1Pd1 | 2,81 |
| | | 129.2.17 | 42369 | Ge1Ni1Pd1 | 2,81 |
| | | 129.2.32 | 428492 | Cu4La1P3 | 2,81 |
| | | 129.2.33 | 153572 | Cu1Sb2Y1 | 2,81 |
| | | 129.2.34 | 79579 | Cu1O1P1U1 | 2,81 |
| | | 129.2.35 | 261277 | Cu1O1P1U1 | 2,81 |
| | | 129.2.36 | 182310 | Cu1O1P1U1 | 2,81 |
| | | 129.2.45 | 419356 | Ce1Mn1O1Sb1 | 2,81 |
| | | 129.2.48 | 2027 | Ca1Ga1N1 | 2,81 |
| | | 129.2.51 | 415728 | Bi2La1Li1 | 2,81 |
| | | 129.2.55 | 25337 | Be2Ca1Ge2 | 2,81 |
| | | 129.2.56 | 25339 | Ba1Sn2Zn2 | 2,81 |
| | | 129.2.58 | 236308 | Ba1Cu2Sb2 | 2,81 |
| | | 129.2.59 | 428691 | Au2Sn2Sr1 | 2,81 |
| | | 129.2.60 | 427711 | Au2Sn2Sr1 | 2,81 |
| | | 129.2.62 | 412550 | As2Cu1Y1 | 2,81 |
| | | 129.2.63 | 174358 | As2Cu1Lu1 | 2,81 |
| | | 129.2.64 | 94440 | As2Cu1Ho1 | 2,81 |
| | | 129.2.65 | 412178 | As2Cu1Ho1 | 2,81 |
| | | 129.2.66 | 44038 | As1Np1Te1 | 2,81 |
| | | 129.2.67 | 50242 | As1Mn1Nd1O1 | 2,81 |
| | | 129.2.79 | 195499 | As1Ce1Mn1O1 | 2,81 |
| | | 129.2.80 | 195498 | As1Ce1Mn1O1 | 2,81 |
| | | 129.2.83 | 190094 | Ag1Sb2Y1 | 2,81 |
| | | 129.2.84 | 251149 | Ag1Cu1S1 | 2,81 |
| 129 | 5 | 129.2.7 | 54276 | Mn1Si1Tb1 | 2,81 |
| | | 129.2.22 | 41255 | Gd1Ru1Si1 | 2,81 |
| | | 129.2.24 | 85854 | Fe1Nd1Si1 | 2,81 |
| | | 129.2.25 | 57235 | Fe1Nd1Si1 | 2,81 |
| | | 129.2.26 | 57234 | Fe1Nd1Si1 | 2,81 |
| | | 129.2.28 | 88118 | Er1Si1Ti1 | 2,81 |
| | | 129.2.29 | 88116 | Dy1Si1Ti1 | 2,81 |
| | | 129.2.40 | 67932 | Co2P2U1 | 2,81 |
| | | 129.2.42 | 106471 | Co1Ge1La1 | 2,81 |
| | | 129.2.46 | 106407 | Ce1Ir2Sn2 | 2,81 |
| | | 129.2.73 | 93238 | As1Fe1Mn1 | 2,81 |
| | | 129.2.74 | 93237 | As1Fe1Mn1 | 2,81 |
| | | 129.2.81 | 194698 | Al2Ce1Pd2 | 2,81 |
| | | 129.2.82 | 191050 | Al2Ce1Pd2 | 2,81 |
| 129 | 6 | 129.2.14 | 62716 | Ir2Si2U1 | 2,81 |
| | | 129.2.21 | 183310 | Gd1Sn2Zn1 | 2,81 |
| | | 129.2.27 | 95081 | Eu1P1Sn1 | 2,81 |
| | | 129.2.43 | 61370 | Ce1Pt2Si2 | 2,81 |
| | | 129.2.44 | 40764 | Ce1P2Rh2 | 2,81 |
| | | 129.2.50 | 85283 | C1B1Ni1Y1 | 2,81 |
| | | 129.2.53 | 252466 | Bi1La1O1S2 | 2,81 |
| | | 129.2.57 | 42460 | Ba1Ge1Mg1 | 2,81 |
| | | 129.2.61 | 107956 | As2Nd1Ni2 | 2,81 |
| 129 | 7 | 129.2.5 | 419464 | O1Os1P1Pr1 | 2,81 |
| | | 129.2.8 | 75054 | Mn1Pr1Si1 | 2,81 |
| | | 129.2.9 | 75053 | Mn1Pr1Si1 | 2,81 |
| | | 129.2.10 | 75052 | Mn1Pr1Si1 | 2,81 |
| | | 129.2.18 | 85859 | Ge1Mn1Pr1 | 2,81 |
| | | 129.2.19 | 80012 | Ge1Mn1Pr1 | 2,81 |
| | | 129.2.20 | 80011 | Ge1Mn1Pr1 | 2,81 |
| | | 129.2.23 | 80200 | Fe1O1P1Pr1 | 2,81 |
| | | 129.2.30 | 57233 | Dy1Fe1Si1 | 2,81 |

Continued on next column

**TABLE XV– continued**

| SG | cl | TQC | ICSD | Stoich. Formula | SubSG |
|---|---|---|---|---|---|
| | | 129.2.31 | 57218 | Dy1Fe1Si1 | 2,81 |
| | | 129.2.37 | 182311 | Cu1O1P1Th1 | 2,81 |
| | | 129.2.38 | 172798 | Cu1F1Se1Sr1 | 2,81 |
| | | 129.2.39 | 172796 | Cu1F1S1Sr1 | 2,81 |
| | | 129.2.41 | 106477 | Co1Ge1Pr1 | 2,81 |
| | | 129.2.47 | 102076 | Cd1Ti1 | 2,81 |
| | | 129.2.49 | 62598 | C2Co1Sc1 | 2,81 |
| | | 129.2.52 | 426715 | Bi1N1Nd2 | 2,81 |
| | | 129.2.54 | 426713 | Bi1Ce2N1 | 2,81 |
| | | 129.2.68 | 262831 | As1La1O1Ru1 | 2,81 |
| | | 129.2.69 | 90316 | As1Fe1O1Pr1 | 2,81 |
| | | 129.2.70 | 422000 | As1Fe1O1Pr1 | 2,81 |
| | | 129.2.71 | 246832 | As1Fe1O1Pr1 | 2,81 |
| | | 129.2.72 | 162830 | As1Fe1O1Pr1 | 2,81 |
| | | 129.2.75 | 180774 | As1Co1Nd1O1 | 2,81 |
| | | 129.2.76 | 180773 | As1Co1Nd1O1 | 2,81 |
| | | 129.2.77 | 180772 | As1Co1Nd1O1 | 2,81 |
| | | 129.2.78 | 169027 | As1Co1Nd1O1 | 2,81 |
| 131 | 1 | 131.2.2 | 6159 | Ca1O4Pt2 | 2,81 |
| 131 | 2 | 131.2.1 | 50987 | I1Pd1Te1 | 2,81 |
| 136 | 1 | 136.1.6 | 237545 | Cd4Pt2Sr1 | 2,81 |
| | | 136.1.12 | 237544 | Ba1Cd4Pt2 | 2,81 |
| 136 | 2 | 136.1.2 | 90270 | Ni4P2Yb1 | 2,81 |
| | | 136.1.3 | 421229 | Li2Sr3 | 2,81 |
| | | 136.1.4 | 93213 | Fe2O6Te1 | 2,81 |
| | | 136.1.7 | 62553 | Ca1Cu4P2 | 2,81 |
| | | 136.1.10 | 260392 | Ba1Hg2Tl2 | 2,81 |
| | | 136.1.11 | 260391 | Ba1Hg2Tl2 | 2,81 |
| | | 136.1.13 | 173224 | As2Er1Ni4 | 2,81 |
| | | 136.1.14 | 68526 | As2Dy1Ni4 | 2,81 |
| 136 | 3 | 136.1.1 | 76166 | U1 | 2 |
| | | 136.1.5 | 33797 | Co2K4O5Rb2 | 2 |
| | | 136.1.8 | 417667 | C2Nd2W1 | 2 |
| | | 136.1.9 | 413345 | C2Mo1Pr2 | 2 |
| 137 | 1 | 137.1.1 | 250760 | Np2Se5 | 2,81 |
| | | 137.1.2 | 248874 | Ba2S6U1 | 2,81 |
| 137 | 3 | 137.1.3 | 49631 | B4Er1Rh4 | 2 |
| 139 | 1 | 139.1.21 | 89560 | Nd2O2Te1 | 2,82 |
| | | 139.1.51 | 150228 | Mg1Pd3 | 2,82 |
| | | 139.1.55 | 40262 | La2O4Pd1 | 2,82 |
| | | 139.1.105 | 88424 | Cu2Ni1O2S2Sr2 | 2,82 |
| | | 139.1.248 | 168816 | Al2Pd5U1 | 2,82 |
| | | 139.1.249 | 161313 | Al2Pd5U1 | 2,82 |
| 139 | 2 | 139.1.47 | 290825 | Mn2Si2Yb1 | 2,82 |
| | | 139.1.48 | 290824 | Mn2Si2Yb1 | 2,82 |
| | | 139.1.49 | 290822 | Mn2Si2Yb1 | 2,82 |
| | | 139.1.67 | 411 | Ge2Sr1Zn2 | 2,82 |
| | | 139.1.166 | 41756 | Ca1Cu2P2 | 2,82 |
| | | 139.1.202 | 58717 | Be12Mo1 | 2,82 |
| | | 139.1.226 | 58388 | Au2Yb1 | 2,82 |
| | | 139.1.237 | 78756 | As2Cu2Sr1 | 2,82 |
| | | 139.1.241 | 236307 | As2Ba1Cu2 | 2,82 |
| | | 139.1.244 | 156967 | Al8Cr4Gd1 | 2,82 |
| | | 139.1.246 | 57539 | Al8Ca1Cu4 | 2,82 |
| | | 139.1.251 | 58252 | Ag2Er1 | 2,82 |
| 139 | 3 | 139.1.3 | 416747 | Ru1Zn3 | 2 |
| | | 139.1.4 | 249448 | Ru1Zn3 | 2 |
| | | 139.1.5 | 155990 | Ru1Zn3 | 2 |
| | | 139.1.22 | 93412 | Nd1Ni2Si2 | 2 |
| | | 139.1.43 | 82464 | Na2Nd2O10Ti3 | 2 |

Continued on next column

**TABLE XV– continued**

| SG | cl | TQC | ICSD | Stoich. Formula | SubSG |
|---|---|---|---|---|---|
| | | 139.1.44 | 75607 | Na2Nd2O10Ti3 | 2 |
| | | 139.1.45 | 75149 | Na2Nd2O10Ti3 | 2 |
| | | 139.1.50 | 104941 | Mn1Pd3 | 2 |
| | | 139.1.62 | 72190 | I2Nd1 | 2 |
| | | 139.1.68 | 53731 | Ge2Nd1Pd2 | 2 |
| | | 139.1.69 | 81753 | Ge2Nd1Ni2 | 2 |
| | | 139.1.114 | 40763 | Cu1P2U1 | 2 |
| | | 139.1.126 | 261192 | Co2La2O3Se2 | 2 |
| | | 139.1.127 | 261191 | Co2La2O3Se2 | 2 |
| | | 139.1.128 | 248239 | Co2La2O3Se2 | 2 |
| | | 139.1.129 | 168330 | Co2La2O3Se2 | 2 |
| | | 139.1.178 | 69888 | C2Er1 | 2 |
| | | 139.1.179 | 23690 | C2Er1 | 2 |
| | | 139.1.180 | 23689 | C2Er1 | 2 |
| | | 139.1.224 | 71642 | B2Ir2Zn1 | 2 |
| | | 139.1.236 | 167013 | As2F2O1Sr2Ti2 | 2 |
| | | 139.1.242 | 92989 | As1Ti1Zr1 | 2 |
| 139 | 4 | 139.1.70 | 81744 | Ge2Mn2U1 | 2 |
| | | 139.1.71 | 41976 | Ge2Mn2Th1 | 2 |
| | | 139.1.72 | 403 | Ge2Mn2Sr1 | 2 |
| | | 139.1.113 | 154203 | Cu2Fe2O5Se2Sr3 | 2 |
| | | 139.1.118 | 43091 | Co4Ge7U3 | 2 |
| | | 139.1.119 | 290506 | Co4Ge7U3 | 2 |
| | | 139.1.120 | 290505 | Co4Ge7U3 | 2 |
| | | 139.1.121 | 290504 | Co4Ge7U3 | 2 |
| | | 139.1.150 | 416139 | Ca2Pd3Sb4 | 2 |
| | | 139.1.153 | 50185 | Ca1P2Rh2 | 2 |
| | | 139.1.157 | 414231 | Ca1In1 | 2 |
| | | 139.1.160 | 81755 | Ca1Ge2Ni2 | 2 |
| | | 139.1.161 | 408 | Ca1Ge2Ni2 | 2 |
| | | 139.1.168 | 85892 | Ca1Co2P2 | 2 |
| | | 139.1.169 | 85891 | Ca1Co2P2 | 2 |
| | | 139.1.170 | 85890 | Ca1Co2P2 | 2 |
| | | 139.1.185 | 79563 | C1B2Gd1Ni2 | 2 |
| | | 139.1.187 | 56596 | C1B2Dy1Ni2 | 2 |
| | | 139.1.188 | 56595 | C1B2Dy1Ni2 | 2 |
| | | 139.1.189 | 56594 | C1B2Dy1Ni2 | 2 |
| | | 139.1.190 | 56593 | C1B2Dy1Ni2 | 2 |
| | | 139.1.216 | 419531 | Ba1Mn2Sb2 | 2 |
| | | 139.1.217 | 32019 | Ba1Mn2Sb2 | 2 |
| | | 139.1.218 | 77668 | Ba1Mn2P2 | 2 |
| | | 139.1.219 | 10469 | Ba1Mn2P2 | 2 |
| | | 139.1.232 | 417002 | As2Rh2Sr1 | 2 |
| | | 139.1.240 | 41794 | As2Ba1Mn2 | 2 |
| 139 | 5 | 139.1.11 | 42966 | P2Ru2Sr1 | 2,82 |
| | | 139.1.52 | 42733 | La2Sb1 | 2,82 |
| | | 139.1.53 | 423290 | La2Sb1 | 2,82 |
| | | 139.1.54 | 10442 | La2Sb1 | 2,82 |
| | | 139.1.59 | 427126 | La1Sb1Sc1 | 2,82 |
| | | 139.1.101 | 88184 | Dy1Ni2P2 | 2,82 |
| | | 139.1.102 | 88183 | Dy1Ni2P2 | 2,82 |
| | | 139.1.158 | 52761 | Ca1Ge2Ru2 | 2,82 |
| | | 139.1.191 | 95097 | C1B2Ce1Rh2 | 2,82 |
| | | 139.1.192 | 95098 | C1B2Ce1Ir2 | 2,82 |
| | | 139.1.235 | 402951 | As2O1Yb4 | 2,82 |
| | | 139.1.239 | 416983 | As2Ba1Rh2 | 2,82 |
| 139 | 6 | 139.1.19 | 68211 | Ni2Si2Th1 | 2 |
| | | 139.1.80 | 26823 | Gd1Mn12 | 2 |
| | | 139.1.103 | 246933 | Cu2Si2U1 | 2 |
| | | 139.1.104 | 68212 | Cu2Si2Th1 | 2 |
| | | 139.1.106 | 86002 | Cu2Ge2U1 | 2 |

Continued on next column

**TABLE XV– continued**

| SG | cl | TQC | ICSD | Stoich. Formula | SubSG |
|---|---|---|---|---|---|
| | | 139.1.107 | 86001 | Cu2Ge2U1 | 2 |
| | | 139.1.108 | 86000 | Cu2Ge2U1 | 2 |
| | | 139.1.109 | 85999 | Cu2Ge2U1 | 2 |
| | | 139.1.110 | 85998 | Cu2Ge2U1 | 2 |
| | | 139.1.162 | 81743 | Ca1Ge2Mn2 | 2 |
| | | 139.1.163 | 402 | Ca1Ge2Mn2 | 2 |
| | | 139.1.186 | 56651 | C1B2Er1Ni2 | 2 |
| | | 139.1.225 | 426150 | Au4Ce1Zn2 | 2 |
| | | 139.1.229 | 418530 | Au2Ce1Si2 | 2 |
| | | 139.1.231 | 263018 | As4Ba2Fe2O1Ti2 | 2 |
| | | 139.1.233 | 47170 | As2Pd2Sr1 | 2 |
| | | 139.1.234 | 36374 | As2Pd2Sr1 | 2 |
| | | 139.1.238 | 36372 | As2Ca1Pd2 | 2 |
| 139 | 7 | 139.1.13 | 30849 | P2Pd2Yb1 | 2,82 |
| | | 139.1.77 | 89563 | Gd2O2Te1 | 2,82 |
| | | 139.1.79 | 30844 | Gd1P2Pd2 | 2,82 |
| | | 139.1.84 | 74434 | Fe2O6Sr3 | 2,82 |
| | | 139.1.85 | 249015 | Fe2O6Sr3 | 2,82 |
| | | 139.1.86 | 249014 | Fe2O6Sr3 | 2,82 |
| | | 139.1.100 | 30846 | Dy1P2Pd2 | 2,82 |
| | | 139.1.116 | 67472 | Cr2Gd1Si2 | 2,82 |
| | | 139.1.155 | 30835 | Ca1P2Pd2 | 2,82 |
| | | 139.1.223 | 174263 | Ba1Cd2Ge2 | 2,82 |
| 139 | 8 | 139.1.10 | 105758 | Pd1Zr2 | 2,82 |
| | | 139.1.12 | 50186 | P2Rh2Sr1 | 2,82 |
| | | 139.1.20 | 41771 | Ni2Sc1Si3 | 2,82 |
| | | 139.1.82 | 68209 | Fe2Si2Th1 | 2,82 |
| | | 139.1.87 | 81746 | Fe2Ge2Nd1 | 2,82 |
| | | 139.1.122 | 425470 | Co2Si2Sr1 | 2,82 |
| | | 139.1.130 | 407 | Co2Ge2Sr1 | 2,82 |
| | | 139.1.145 | 87159 | Ce1Sc1Si1 | 2,82 |
| | | 139.1.146 | 290821 | Ce1Sc1Si1 | 2,82 |
| | | 139.1.165 | 10463 | Ca1Fe2P2 | 2,82 |
| | | 139.1.199 | 402954 | Bi2Eu4O1 | 2,82 |
| | | 139.1.213 | 2141 | Ba2Bi1 | 2,82 |
| | | 139.1.215 | 50188 | Ba1P2Rh2 | 2,82 |
| | | 139.1.222 | 420975 | Ba1Co2Ge2 | 2,82 |
| | | 139.1.247 | 95799 | Al4Ge2Ni1Tb2 | 2,82 |
| 139 | 9 | 139.1.83 | 10467 | Fe2P2Sr1 | 2,82 |
| | | 139.1.98 | 245051 | Er1Ni2Si2 | 2,82 |
| | | 139.1.148 | 85893 | Ce1Co2P2 | 2,82 |
| | | 139.1.203 | 237677 | Ba6Ge32Rh8 | 2,82 |
| | | 139.1.221 | 10468 | Ba1Fe2P2 | 2,82 |
| 139 | 10 | 139.1.111 | 424224 | Cu2Ge2Sr1 | 2 |
| | | 139.1.112 | 410 | Cu2Ge2Sr1 | 2 |
| | | 139.1.149 | 370021 | Cd2Ga2Sr1 | 2 |
| | | 139.1.204 | 427108 | Ba3Ge16Ir4 | 2 |
| | | 139.1.205 | 194235 | Ba3Ge16Ir4 | 2 |
| | | 139.1.206 | 189363 | Ba3Ge16Ir4 | 2 |
| | | 139.1.220 | 100024 | Ba1Mn1Sb2 | 2 |
| | | 139.1.245 | 57545 | Al8Ca1Mn4 | 2 |
| | | 139.1.250 | 166270 | Al2Pd5Pu1 | 2 |
| 139 | 11 | 139.1.9 | 61195 | Pd2Sb2Sr1 | 2 |
| | | 139.1.159 | 52759 | Ca1Ge2Pd2 | 2 |
| | | 139.1.201 | 107609 | Be1Pd2 | 2 |
| | | 139.1.214 | 61197 | Ba1Pd2Sb2 | 2 |
| 139 | 12 | 139.1.2 | 42119 | Sb1Sr2 | 2 |
| | | 139.1.6 | 251818 | Pt2Th1 | 2 |
| | | 139.1.17 | 56577 | O7Ru2Sr3 | 2 |
| | | 139.1.58 | 181385 | La2Mn2O3Se2 | 2 |

Continued on next column

**TABLE XV– continued**

| SG | cl | TQC | ICSD | Stoich. Formula | SubSG |
|---|---|---|---|---|---|
| | | 139.1.81 | 90392 | Gd1Ge1Ti1 | 2 |
| | | 139.1.139 | 91749 | Cl2Co2O5Sr3 | 2 |
| | | 139.1.140 | 90123 | Cl2Co2O5Sr3 | 2 |
| | | 139.1.141 | 251170 | Cl2Co2O4Sr3 | 2 |
| | | 139.1.173 | 74670 | C2Yb1 | 2 |
| | | 139.1.174 | 23686 | C2Yb1 | 2 |
| | | 139.1.200 | 41836 | Bi1Sr2 | 2 |
| | | 139.1.207 | 41837 | Ba2Sb1 | 2 |
| | | 139.1.208 | 30084 | Ba2Hg1 | 2 |
| | | 139.1.209 | 183149 | Ba2F2Mn2O1Se2 | 2 |
| | | 139.1.210 | 71535 | Ba2Cu2O12Si4 | 2 |
| | | 139.1.243 | 42358 | As1Ba2 | 2 |
| 139 | 13 | 139.1.8 | 247272 | Pd3Tl1 | 2,82 |
| | | 139.1.14 | 36373 | P2Pd2Sr1 | 2,82 |
| | | 139.1.61 | 420251 | In1Pd3 | 2,82 |
| | | 139.1.64 | 427211 | Hg2In2Sr1 | 2,82 |
| | | 139.1.74 | 26277 | Ge1Te1U1 | 2,82 |
| | | 139.1.75 | 87316 | Ge1Se1U1 | 2,82 |
| | | 139.1.76 | 26276 | Ge1Se1U1 | 2,82 |
| | | 139.1.97 | 60667 | Er1O2S3U2 | 2,82 |
| | | 139.1.115 | 290194 | Cu1Ge1Hf1 | 2,82 |
| | | 139.1.143 | 150461 | Ce2Ga10Pd1 | 2,82 |
| | | 139.1.144 | 20525 | Ce2Ga10Ni1 | 2,82 |
| | | 139.1.154 | 36371 | Ca1P2Pd2 | 2,82 |
| | | 139.1.156 | 10461 | Ca1Ni2P2 | 2,82 |
| | | 139.1.175 | 26478 | C2U1 | 2,82 |
| | | 139.1.176 | 202305 | C2U1 | 2,82 |
| | | 139.1.227 | 413 | Au2Si2Sr1 | 2,82 |
| | | 139.1.230 | 412 | Au2Ca1Si2 | 2,82 |
| 139 | 14 | 139.1.63 | 710068 | Ho1Si3 | 2,82 |
| | | 139.1.66 | 415033 | H2Ba8Bi4O1 | 2,82 |
| | | 139.1.73 | 165737 | Ge2Ir2Sr1 | 2,82 |
| | | 139.1.99 | 109428 | Er1Ge2Ni2 | 2,82 |
| | | 139.1.123 | 73653 | Co2Nd1P2 | 2,82 |
| | | 139.1.124 | 73652 | Co2Nd1P2 | 2,82 |
| | | 139.1.125 | 73651 | Co2Nd1P2 | 2,82 |
| | | 139.1.152 | 425469 | Ca1Rh2Si2 | 2,82 |
| | | 139.1.164 | 52755 | Ca1Ge2Ir2 | 2,82 |
| | | 139.1.167 | 420974 | Ca1Co2Si2 | 2,82 |
| | | 139.1.171 | 81751 | Ca1Co2Ge2 | 2,82 |
| | | 139.1.172 | 406 | Ca1Co2Ge2 | 2,82 |
| 139 | 15 | 139.1.1 | 263005 | Si3Y1 | 87 |
| | | 139.1.7 | 423291 | Pr2Sb1 | 87 |
| | | 139.1.15 | 71451 | O7Sr3V2 | 87 |
| | | 139.1.16 | 71320 | O7Sr3V2 | 87 |
| | | 139.1.18 | 79214 | O4Te3U4 | 87 |
| | | 139.1.23 | 91220 | Na2O1Sb2Ti2 | 87 |
| | | 139.1.24 | 91219 | Na2O1Sb2Ti2 | 87 |
| | | 139.1.25 | 91218 | Na2O1Sb2Ti2 | 87 |
| | | 139.1.26 | 91217 | Na2O1Sb2Ti2 | 87 |
| | | 139.1.27 | 91216 | Na2O1Sb2Ti2 | 87 |
| | | 139.1.28 | 91215 | Na2O1Sb2Ti2 | 87 |
| | | 139.1.29 | 91214 | Na2O1Sb2Ti2 | 87 |
| | | 139.1.30 | 91213 | Na2O1Sb2Ti2 | 87 |
| | | 139.1.31 | 91212 | Na2O1Sb2Ti2 | 87 |
| | | 139.1.32 | 91211 | Na2O1Sb2Ti2 | 87 |
| | | 139.1.33 | 91210 | Na2O1Sb2Ti2 | 87 |
| | | 139.1.34 | 91209 | Na2O1Sb2Ti2 | 87 |
| | | 139.1.35 | 91208 | Na2O1Sb2Ti2 | 87 |
| | | 139.1.36 | 91207 | Na2O1Sb2Ti2 | 87 |
| | | 139.1.37 | 91206 | Na2O1Sb2Ti2 | 87 |
| | | 139.1.38 | 91205 | Na2O1Sb2Ti2 | 87 |
| | | 139.1.39 | 91204 | Na2O1Sb2Ti2 | 87 |
| | | 139.1.40 | 91203 | Na2O1Sb2Ti2 | 87 |
| | | 139.1.41 | 91202 | Na2O1Sb2Ti2 | 87 |
| | | 139.1.42 | 69648 | Na2O1Sb2Ti2 | 87 |
| | | 139.1.46 | 152123 | Mo1O4Sr2 | 87 |
| | | 139.1.56 | 69172 | La2Ni1O4 | 87 |
| | | 139.1.57 | 33536 | La2Ni1O4 | 87 |
| | | 139.1.60 | 421342 | K2Mg5Sn3 | 87 |
| | | 139.1.65 | 75936 | Hf3Te2 | 87 |
| | | 139.1.78 | 82466 | Gd2Na2O10Ti3 | 87 |
| | | 139.1.88 | 74421 | Fe1O4Sr2 | 87 |
| | | 139.1.89 | 74420 | Fe1O4Sr2 | 87 |
| | | 139.1.90 | 74419 | Fe1O4Sr2 | 87 |
| | | 139.1.91 | 69849 | Fe1O4Sr2 | 87 |
| | | 139.1.92 | 69682 | F4Ni1Rb2 | 87 |
| | | 139.1.93 | 73450 | F4K2Ni1 | 87 |
| | | 139.1.94 | 33520 | F4K2Ni1 | 87 |
| | | 139.1.95 | 167014 | F2O1Sb2Sr2Ti2 | 87 |
| | | 139.1.96 | 426633 | Er2Mg2Ru1 | 87 |
| | | 139.1.117 | 245595 | Cr1O4Sr2 | 87 |
| | | 139.1.131 | 33524 | Co2F7K3 | 87 |
| | | 139.1.132 | 30081 | Cl7Mn2Rb3 | 87 |
| | | 139.1.133 | 2241 | Cl7Mn2Rb3 | 87 |
| | | 139.1.134 | 36127 | Cl7K3Mn2 | 87 |
| | | 139.1.135 | 99942 | Cl2Fe2O5Sr3 | 87 |
| | | 139.1.136 | 69871 | Cl2Fe2O5Sr3 | 87 |
| | | 139.1.137 | 33258 | Cl2Fe2O5Sr3 | 87 |
| | | 139.1.138 | 174532 | Cl2Fe2O4Sr3 | 87 |
| | | 139.1.142 | 409587 | Ce2Sb1 | 87 |
| | | 139.1.147 | 411677 | Ce1Ir2Si2 | 87 |
| | | 139.1.151 | 59649 | Ca1Si2Zn2 | 87 |
| | | 139.1.177 | 63560 | C2Nd1 | 87 |
| | | 139.1.181 | 380471 | C2Cl1N2Ni1Rb1 | 87 |
| | | 139.1.182 | 380469 | C2Cl1Cs1N2Ni1 | 87 |
| | | 139.1.183 | 380470 | C2Br1N2Ni1Rb1 | 87 |
| | | 139.1.184 | 380468 | C2Br1Cs1N2Ni1 | 87 |
| | | 139.1.193 | 32544 | Br7Mn2Rb3 | 87 |
| | | 139.1.194 | 99944 | Br2Fe2O5Sr3 | 87 |
| | | 139.1.195 | 99943 | Br2Fe2O5Sr3 | 87 |
| | | 139.1.196 | 69872 | Br2Fe2O5Sr3 | 87 |
| | | 139.1.197 | 430065 | Bi2F2O1Sr2Ti2 | 87 |
| | | 139.1.198 | 430064 | Bi2F2O1Sr2Ti2 | 87 |
| | | 139.1.211 | 78593 | Ba2Ca1Cu2O8Tl2 | 87 |
| | | 139.1.212 | 78592 | Ba2Ca1Cu2O8Tl2 | 87 |
| | | 139.1.228 | 37453 | Au2Cl6Cs2 | 87 |
| 140 | 1 | 140.1.12 | 100084 | Ba5Sn3 | 2 |
| | | 140.1.13 | 247208 | B2Cu2O6Sr1 | 2 |
| | | 140.1.14 | 247207 | B2Cu2O6Sr1 | 2 |
| 140 | 2 | 140.1.3 | 245120 | Mg2Pt1 | 2 |
| | | 140.1.8 | 106369 | Cd3Sr5 | 2 |
| | | 140.1.9 | 106344 | Ca5Hg3 | 2 |
| 140 | 3 | 140.1.7 | 98408 | Ce1Ni9Si4 | 2 |
| 140 | 4 | 140.1.6 | 281417 | Cl1Ga4La3 | 2 |
| 140 | 5 | 140.1.1 | 86940 | O3Ru1Sr1 | 87 |
| | | 140.1.2 | 174206 | Mo1O3Sr1 | 87 |
| | | 140.1.4 | 88093 | Cr1Sb2 | 87 |
| | | 140.1.5 | 86811 | Cr1Sb2 | 87 |
| | | 140.1.10 | 39599 | Br5Cd1Cs3 | 87 |
| 140 | 6 | 140.2.2 | 409797 | Sb2Ti1 | 2,82 |

Continued on next column

**TABLE XV– continued**

| SG | cl | TQC | ICSD | Stoich. Formula | SubSG |
|---|---|---|---|---|---|
| | | 140.2.3 | 249576 | Sb2Ti1 | 2,82 |
| | | 140.2.4 | 16582 | Pu5Si3 | 2,82 |
| | | 140.2.8 | 420322 | Cs9O4Sc1 | 2,82 |
| | | 140.2.9 | 419823 | Cs9In1O4 | 2,82 |
| | | 140.2.10 | 420320 | Cs9Ga1O4 | 2,82 |
| | | 140.2.14 | 420321 | Al1Cs9O4 | 2,82 |
| 140 | 7 | 140.2.6 | 260983 | Ni1U6 | 2,82 |
| | | 140.2.12 | 150146 | Bi2Fe1Ti4 | 2,82 |
| | | 140.2.13 | 428736 | Al1Eu3O5 | 2,82 |
| 140 | 8 | 140.2.11 | 418133 | Ca1Ni4Sn2 | 2,82 |
| 140 | 9 | 140.2.1 | 1203 | Sn3Sr5 | 2,82 |
| | | 140.2.7 | 409374 | Ge3Sr5 | 2,82 |
| 140 | 10 | 140.2.5 | 105479 | Ni1Zr2 | 2,82 |
| 140 | 12 | 140.1.11 | 190146 | Bi1Ho5Ni2 | 2 |
| 141 | 1 | 141.1.1 | 423244 | Si2Yb1 | 2,82 |
| | | 141.1.2 | 170062 | Sc1Sn2 | 2,82 |
| | | 141.1.3 | 88368 | O4Tb1V1 | 2,82 |
| | | 141.1.4 | 78078 | O4Tb1V1 | 2,82 |
| | | 141.1.5 | 78076 | O4Pr1V1 | 2,82 |
| | | 141.1.6 | 157327 | O4Pr1V1 | 2,82 |
| | | 141.1.7 | 79759 | O4P1Tm1 | 2,82 |
| | | 141.1.8 | 36053 | O4P1Tm1 | 2,82 |
| | | 141.1.9 | 61683 | Na1O2Tb1 | 2,82 |
| | | 141.1.10 | 61682 | Na1O2Pr1 | 2,82 |
| | | 141.1.11 | 410968 | In2Zr1 | 2,82 |
| | | 141.1.14 | 103986 | Ga2Th1 | 2,82 |
| | | 141.1.15 | 100566 | Cu4O3 | 2,82 |
| | | 141.1.16 | 35419 | Cu2P1Ti1 | 2,82 |
| | | 141.1.22 | 409800 | Ce1Zn11 | 2,82 |
| | | 141.1.23 | 247146 | Ag1Li1 | 2,82 |
| | | 141.1.24 | 247145 | Ag1Li1 | 2,82 |
| 141 | 2 | 141.1.12 | 61422 | H1V2 | 2,82 |
| | | 141.1.13 | 60511 | H1V2 | 2,82 |
| | | 141.1.17 | 100796 | Cu2Ge1O4 | 2,82 |
| 141 | 3 | 141.1.18 | 84377 | Cr2Ni1O4 | 2 |
| | | 141.1.19 | 31300 | Cr2Ni1O4 | 2 |
| | | 141.1.20 | 31299 | Cr2Ni1O4 | 2 |
| | | 141.1.21 | 280062 | Cr2Ni1O4 | 2 |
| 147 | 1 | 147.1.2 | 150943 | Ni1Ti1 | 2 |
| 147 | 2 | 147.1.1 | 166370 | Ni1Ti1 | 2 |
| 148 | 1 | 148.1.3 | 37134 | O12Te1Tl6 | 2 |
| | | 148.1.9 | 46015 | Mo6S8Yb1 | 2 |
| | | 148.1.10 | 46014 | Mo6S8Yb1 | 2 |
| | | 148.1.11 | 46013 | Mo6S8Yb1 | 2 |
| | | 148.1.12 | 201506 | Mo6S8Yb1 | 2 |
| | | 148.1.13 | 62156 | Mo6S8Sr1 | 2 |
| | | 148.1.14 | 68 | Mo3Pb0.5Se4 | 2 |
| | | 148.1.33 | 106520 | Cr2Se3 | 2 |
| | | 148.1.38 | 39426 | Cl3Ti1 | 2 |
| | | 148.1.40 | 424476 | Cl12Co1Sc7 | 2 |
| | | 148.1.41 | 183603 | Ce1Pt5Sb1 | 2 |
| | | 148.1.42 | 62388 | Ca1Mo6S8 | 2 |
| | | 148.1.43 | 39242 | Br3Ti1 | 2 |
| | | 148.1.45 | 419467 | Ba6N6O1Os2 | 2 |
| | | 148.1.48 | 85489 | Ba1Mo6S8 | 2 |
| | | 148.1.49 | 65776 | Ba1Mo6S8 | 2 |
| | | 148.1.50 | 62157 | Ba1Mo6S8 | 2 |
| | | 148.1.53 | 91777 | Ag1Bi1O3 | 2 |
| | | 148.1.54 | 89432 | Ag1Bi1O3 | 2 |
| 148 | 2 | 148.1.8 | 59375 | Mo6Te8 | 2 |

Continued on next column

**TABLE XV– continued**

| SG | cl | TQC | ICSD | Stoich. Formula | SubSG |
|---|---|---|---|---|---|
| | | 148.1.16 | 245054 | Mn3O6Te1 | 2 |
| | | 148.1.17 | 183618 | Mn3O6Te1 | 2 |
| | | 148.1.18 | 182436 | Mn3O6Te1 | 2 |
| | | 148.1.19 | 182435 | Mn3O6Te1 | 2 |
| | | 148.1.21 | 247553 | Mn1O3Ti1 | 2 |
| | | 148.1.23 | 73247 | Ge1Mn1O3 | 2 |
| | | 148.1.24 | 73246 | Ge1Mn1O3 | 2 |
| | | 148.1.25 | 73245 | Ge1Mn1O3 | 2 |
| | | 148.1.26 | 73244 | Ge1Mn1O3 | 2 |
| | | 148.1.27 | 73243 | Ge1Mn1O3 | 2 |
| | | 148.1.28 | 69591 | Ge1Mn1O3 | 2 |
| | | 148.1.30 | 84337 | Cs2Mo12Te14 | 2 |
| | | 148.1.31 | 40820 | Cs2Mo12Se14 | 2 |
| | | 148.1.39 | 201976 | Cl12N1Sc7 | 2 |
| | | 148.1.52 | 237739 | Au5Mn1Y3 | 2 |
| 148 | 3 | 148.1.37 | 252042 | Cr1Fe1O3 | 2 |
| | | 148.1.46 | 96087 | Ba1Ni2O8V2 | 2 |
| | | 148.1.47 | 96086 | Ba1Ni2O8V2 | 2 |
| | | 148.1.51 | 59343 | Ba1Co2O8P2 | 2 |
| 148 | 4 | 148.1.4 | 166373 | Ni4Ti3 | 2 |
| | | 148.1.15 | 89004 | Mo12Rb2Se14 | 2 |
| | | 148.1.32 | 71020 | Cr2Si2Te6 | 2 |
| | | 148.1.34 | 79270 | Cr2Ge2Te6 | 2 |
| | | 148.1.35 | 79269 | Cr2Ge2Te6 | 2 |
| | | 148.1.36 | 79268 | Cr2Ge2Te6 | 2 |
| | | 148.1.44 | 419636 | Ba6N6O1Re2 | 2 |
| 148 | 5 | 148.1.5 | 166371 | Ni4Ti3 | 2 |
| | | 148.1.7 | 171584 | Ni1O3Ti1 | 2 |
| | | 148.1.20 | 60006 | Mn1O3Ti1 | 2 |
| | | 148.1.22 | 171579 | Mn1O3Ti1 | 2 |
| | | 148.1.29 | 247127 | Cu1O3Si1 | 2 |
| 148 | 6 | 148.1.1 | 54337 | Pd4Zr3 | 2 |
| | | 148.1.6 | 166369 | Ni4Ti3 | 2 |
| 148 | 7 | 148.1.2 | 1096 | P2Pd15 | 2 |
| 162 | 1 | 162.1.5 | 26557 | Ag6O2 | 2 |
| 162 | 2 | 162.1.4 | 37011 | Ba1Fe4O8Sr1 | 2 |
| 162 | 3 | 162.1.2 | 275 | Cr2O6U1 | 2 |
| | | 162.1.3 | 15133 | Cr2O6U1 | 2 |
| 162 | 4 | 162.1.1 | 28086 | O6U1V2 | 2 |
| 163 | 1 | 163.1.2 | 411131 | Al1F6Li1Yb1 | 2 |
| 163 | 2 | 163.1.1 | 71037 | Mo1Ni1P8 | 2 |
| 164 | 1 | 164.1.34 | 185808 | Hg1Pt2Se3 | 2 |
| | | 164.1.38 | 59801 | Ge2I2La2 | 2 |
| | | 164.1.42 | 159909 | Fe2K3Na1O8 | 2 |
| | | 164.1.44 | 25810 | Er2O2Se1 | 2 |
| | | 164.1.46 | 407246 | Ce2I2Si2 | 2 |
| | | 164.1.73 | 71625 | C1F2Ho2 | 2 |
| | | 164.1.84 | 100654 | Bi1Te1 | 2 |
| | | 164.1.85 | 99681 | Ba6Cl2O12Ru3 | 2 |
| | | 164.1.99 | 160311 | Al2Er1Ge2 | 2 |
| | | 164.1.102 | 94501 | Al1La1Si2 | 2 |
| 164 | 2 | 164.1.13 | 100271 | P2Yb1Zn2 | 2 |
| | | 164.1.27 | 412655 | Mg2Sb2Yb1 | 2 |
| | | 164.1.45 | 35585 | Cu2P2Zr1 | 2 |
| | | 164.1.95 | 411054 | Al2Si2Yb1 | 2 |
| | | 164.1.96 | 411056 | Al2Ge2Yb1 | 2 |
| | | 164.1.97 | 409458 | Al2Ge2Yb1 | 2 |
| | | 164.1.98 | 412121 | Al2Ge2Nd1 | 2 |
| | | 164.1.100 | 50976 | Al2Ce1Ge2 | 2 |
| 164 | 3 | 164.1.21 | 380376 | Mn2Sb2Yb1 | 2 |

Continued on next column

**TABLE XV– continued**

| SG | cl | TQC | ICSD | Stoich. Formula | SubSG |
|---|---|---|---|---|---|
| | | 164.1.22 | 41790 | Mn2Sb2Sr1 | 2 |
| | | 164.1.23 | 416094 | Mn2Sb2Sr1 | 2 |
| | | 164.1.24 | 77670 | Mn2P2Sr1 | 2 |
| | | 164.1.25 | 77669 | Mn2P2Sr1 | 2 |
| | | 164.1.26 | 49019 | Mn2P2Sr1 | 2 |
| | | 164.1.28 | 191298 | Li3Pr1Sb2 | 2 |
| | | 164.1.29 | 26018 | K3Na1O8S2 | 2 |
| | | 164.1.30 | 26014 | K3Na1O8S2 | 2 |
| | | 164.1.36 | 41478 | H1Si1 | 2 |
| | | 164.1.40 | 25808 | Gd2O2Se1 | 2 |
| | | 164.1.41 | 91111 | Gd2O2S1 | 2 |
| | | 164.1.57 | 41789 | Ca1Mn2Sb2 | 2 |
| | | 164.1.58 | 416093 | Ca1Mn2Sb2 | 2 |
| | | 164.1.59 | 163778 | Ca1Mn2Sb2 | 2 |
| | | 164.1.60 | 49017 | Ca1Mn2P2 | 2 |
| | | 164.1.66 | 95113 | C1Nb2S2 | 2 |
| | | 164.1.67 | 95111 | C1Nb2S2 | 2 |
| | | 164.1.68 | 95109 | C1Nb2S2 | 2 |
| | | 164.1.69 | 656705 | C1Nb2S2 | 2 |
| | | 164.1.74 | 33912 | C1F2Gd2 | 2 |
| | | 164.1.75 | 42753 | Bi4Pb1Te7 | 2 |
| | | 164.1.76 | 42707 | Bi4Pb1Te7 | 2 |
| | | 164.1.77 | 250249 | Bi4Pb1Te7 | 2 |
| | | 164.1.78 | 42891 | Bi4Ge1Te7 | 2 |
| | | 164.1.82 | 41791 | Bi2Ca1Mn2 | 2 |
| | | 164.1.86 | 152482 | Ba3Ca1O9Ru2 | 2 |
| | | 164.1.87 | 409958 | Ba1Sn2 | 2 |
| | | 164.1.91 | 49020 | As2Mn2Sr1 | 2 |
| | | 164.1.92 | 41793 | As2Mn2Sr1 | 2 |
| | | 164.1.94 | 41792 | As2Ca1Mn2 | 2 |
| 164 | 4 | 164.1.4 | 651764 | Sb2Yb1Zn2 | 2 |
| | | 164.1.5 | 419689 | Sb2Yb1Zn2 | 2 |
| | | 164.1.31 | 660273 | In2S4Zn1 | 2 |
| | | 164.1.32 | 42668 | In2S4Zn1 | 2 |
| | | 164.1.47 | 78989 | Cd2Sb2Yb1 | 2 |
| | | 164.1.48 | 419690 | Cd2Sb2Yb1 | 2 |
| | | 164.1.49 | 248526 | Ca1Si2 | 2 |
| | | 164.1.50 | 248525 | Ca1Si2 | 2 |
| | | 164.1.51 | 248524 | Ca1Si2 | 2 |
| | | 164.1.52 | 248523 | Ca1Si2 | 2 |
| | | 164.1.53 | 248522 | Ca1Si2 | 2 |
| | | 164.1.54 | 248521 | Ca1Si2 | 2 |
| | | 164.1.79 | 42708 | Bi2Pb2Te5 | 2 |
| | | 164.1.83 | 30525 | Bi1Te1 | 2 |
| | | 164.1.89 | 88231 | As2Yb1Zn2 | 2 |
| | | 164.1.90 | 39512 | As2Mn2Zn1 | 2 |
| 164 | 5 | 164.1.3 | 80205 | Si1Te2 | 2 |
| | | 164.1.6 | 69557 | Sb2Te1 | 2 |
| | | 164.1.7 | 41385 | Pt1Te2 | 2 |
| | | 164.1.8 | 41373 | Pt1Te2 | 2 |
| | | 164.1.9 | 83642 | Pd1Te2 | 2 |
| | | 164.1.10 | 42555 | Pd1Te2 | 2 |
| | | 164.1.11 | 42554 | Pd1Te2 | 2 |
| | | 164.1.12 | 41387 | Pd1Te2 | 2 |
| | | 164.1.14 | 43293 | Ni1Te2 | 2 |
| | | 164.1.15 | 42559 | Ni1Te2 | 2 |
| | | 164.1.16 | 94478 | Nd2S2Te1 | 2 |
| | | 164.1.17 | 32514 | Nd2O3 | 2 |
| | | 164.1.18 | 25806 | Nd2O2Se1 | 2 |
| | | 164.1.19 | 32515 | Nd2O2S1 | 2 |
| | | 164.1.20 | 300245 | Nd2O2S1 | 2 |

Continued on next column

**TABLE XV– continued**

| SG | cl | TQC | ICSD | Stoich. Formula | SubSG |
|---|---|---|---|---|---|
| | | 164.1.35 | 56153 | H2Cd1O2 | 2 |
| | | 164.1.37 | 42876 | Ge2Sb2Te5 | 2 |
| | | 164.1.43 | 155025 | Fe2In2Se5 | 2 |
| | | 164.1.61 | 35512 | C6N6Pt1Rb2S6 | 2 |
| | | 164.1.62 | 35511 | C6K2N6Pt1S6 | 2 |
| | | 164.1.63 | 35510 | C6H8N8Pt1S6 | 2 |
| | | 164.1.64 | 5005 | C1V2 | 2 |
| | | 164.1.65 | 409555 | C1Ta2 | 2 |
| | | 164.1.70 | 43670 | C1Nb2 | 2 |
| | | 164.1.80 | 30372 | Bi2Pb2Se5 | 2 |
| | | 164.1.81 | 659569 | Bi2Mg3 | 2 |
| | | 164.1.93 | 63174 | As2Ge5Te8 | 2 |
| | | 164.1.101 | 94502 | Al1La1Si2 | 2 |
| 164 | 6 | 164.1.39 | 42875 | Ge1Sb4Te7 | 2 |
| | | 164.1.55 | 248520 | Ca1Si2 | 2 |
| | | 164.1.56 | 248519 | Ca1Si2 | 2 |
| | | 164.1.71 | 82264 | C1N2O2Pr2 | 2 |
| | | 164.1.72 | 82265 | C1N2Nd2O2 | 2 |
| | | 164.1.88 | 41107 | As4Ge1Te7 | 2 |
| 164 | 7 | 164.1.1 | 657478 | Te2Zr1 | 2 |
| | | 164.1.2 | 653213 | Te2Zr1 | 2 |
| | | 164.1.33 | 25809 | Ho2O2Se1 | 2 |
| 165 | 1 | 165.1.1 | 154809 | H3Y1 | 2 |
| | | 165.1.2 | 26776 | As1Cu3 | 2 |
| | | 166.1.15 | 410956 | Sn3Sr1 | 2 |
| | | 166.1.17 | 152188 | Sb8Te3 | 2 |
| | | 166.1.18 | 152189 | Sb48Te9 | 2 |
| | | 166.1.54 | 651506 | Sb1 | 2 |
| | | 166.1.55 | 651498 | Sb1 | 2 |
| | | 166.1.56 | 651497 | Sb1 | 2 |
| | | 166.1.57 | 651491 | Sb1 | 2 |
| | | 166.1.58 | 651490 | Sb1 | 2 |
| | | 166.1.59 | 651489 | Sb1 | 2 |
| | | 166.1.60 | 651488 | Sb1 | 2 |
| | | 166.1.61 | 64697 | Sb1 | 2 |
| | | 166.1.62 | 64696 | Sb1 | 2 |
| | | 166.1.63 | 64695 | Sb1 | 2 |
| | | 166.1.65 | 53795 | Sb1 | 2 |
| | | 166.1.66 | 409754 | Sb1 | 2 |
| | | 166.1.69 | 651179 | S1Ti1 | 2 |
| | | 166.1.70 | 25561 | S1Ti1 | 2 |
| | | 166.1.71 | 184500 | S1Ti1 | 2 |
| | | 166.1.82 | 600019 | P1 | 2 |
| | | 166.1.109 | 645562 | Nb6Zn7 | 2 |
| | | 166.1.113 | 41369 | Li8Pb3 | 2 |
| | | 166.1.114 | 191137 | La3Mn2O14Sb3 | 2 |
| | | 166.1.134 | 171105 | H9Fe3O15S2 | 2 |
| | | 166.1.137 | 67682 | H6Fe3O14Pb1S2 | 2 |
| | | 166.1.165 | 38267 | Ga7Rb1 | 2 |
| | | 166.1.166 | 102891 | Ga7Rb1 | 2 |
| | | 166.1.206 | 202068 | Eu2I1P1 | 2 |
| | | 166.1.220 | 190938 | Cu1N2Ta1 | 2 |
| | | 166.1.228 | 16255 | Cs2F6Fe1Na1 | 2 |
| | | 166.1.229 | 102864 | Cs1Ga7 | 2 |
| | | 166.1.254 | 24581 | Co3Ta1 | 2 |
| | | 166.1.263 | 79717 | Co10Ge3O16 | 2 |
| | | 166.1.264 | 401026 | Cl8Na2Ti3 | 2 |
| | | 166.1.269 | 154513 | Ce2Cu5Zn2 | 2 |
| | | 166.1.270 | 620668 | Ce2Co7 | 2 |
| | | 166.1.271 | 620636 | Ce2Co7 | 2 |
| | | 166.1.277 | 409057 | Ca1Si2 | 2 |

Continued on next column

**TABLE XV– continued**

| SG | cl | TQC | ICSD | Stoich. Formula | SubSG |
|---|---|---|---|---|---|
| | | 166.1.283 | 193539 | Ca1Si2 | 2 |
| | | 166.1.284 | 15632 | Ca1Si2 | 2 |
| | | 166.1.310 | 202067 | Br1Eu2P1 | 2 |
| | | 166.1.355 | 99662 | Ba4Mn3O12Pr1 | 2 |
| | | 166.1.362 | 99682 | Ba1O3Ru1 | 2 |
| | | 166.1.368 | 20898 | Ba1Ni4O8 | 2 |
| | | 166.1.388 | 422526 | As2Ca1Ga2 | 2 |
| | | 166.1.389 | 53794 | As1 | 2 |
| | | 166.1.390 | 16518 | As1 | 2 |
| | | 166.1.391 | 16517 | As1 | 2 |
| | | 166.1.392 | 16516 | As1 | 2 |
| | | 166.1.410 | 251620 | Ag1Ge1Li2 | 2 |
| 166 | 2 | 166.1.1 | 653496 | Yb2Zn17 | 2 |
| | | 166.1.2 | 653487 | Yb2Zn17 | 2 |
| | | 166.1.3 | 104207 | Yb2Zn17 | 2 |
| | | 166.1.46 | 657316 | Sb2Sn1Te4 | 2 |
| | | 166.1.47 | 30392 | Sb2Sn1Te4 | 2 |
| | | 166.1.64 | 55402 | Sb1 | 2 |
| | | 166.1.67 | 161493 | Sb1 | 2 |
| | | 166.1.74 | 41371 | Pt2Te3 | 2 |
| | | 166.1.75 | 41370 | Pt1Te1 | 2 |
| | | 166.1.111 | 44926 | Mg2Ni3Si1 | 2 |
| | | 166.1.112 | 55617 | Mg2Nd1Ni9 | 2 |
| | | 166.1.118 | 657607 | In2Te3 | 2 |
| | | 166.1.119 | 640609 | In2Te3 | 2 |
| | | 166.1.146 | 602925 | H4Co3Ho1 | 2 |
| | | 166.1.161 | 55619 | Gd1Mg2Ni9 | 2 |
| | | 166.1.168 | 657608 | Ga2Te3 | 2 |
| | | 166.1.170 | 96491 | Ga1Mn1 | 2 |
| | | 166.1.171 | 35386 | Ga1Ge1Te1 | 2 |
| | | 166.1.210 | 629979 | Dy2Ni7 | 2 |
| | | 166.1.212 | 630007 | Dy1Ni3 | 2 |
| | | 166.1.213 | 629986 | Dy1Ni3 | 2 |
| | | 166.1.214 | 629980 | Dy1Ni3 | 2 |
| | | 166.1.227 | 42152 | Cs2F6Na1Ni1 | 2 |
| | | 166.1.231 | 625438 | Co7Th2 | 2 |
| | | 166.1.232 | 625416 | Co7Th2 | 2 |
| | | 166.1.236 | 623284 | Co7Gd2 | 2 |
| | | 166.1.255 | 624740 | Co3Pr1 | 2 |
| | | 166.1.256 | 624707 | Co3Pr1 | 2 |
| | | 166.1.257 | 624697 | Co3Pr1 | 2 |
| | | 166.1.258 | 108889 | Co3Pr1 | 2 |
| | | 166.1.272 | 603443 | Ce2Co5Fe12N3 | 2 |
| | | 166.1.273 | 201355 | Cd1Cs4F12Ni3 | 2 |
| | | 166.1.292 | 41260 | C3Al6N2 | 2 |
| | | 166.1.293 | 74178 | C2Ba1Ce1F1O6 | 2 |
| | | 166.1.294 | 96188 | C1Y2 | 2 |
| | | 166.1.295 | 22283 | C1Y2 | 2 |
| | | 166.1.299 | 23791 | C1S2Ta2 | 2 |
| | | 166.1.305 | 602542 | C1B2U1 | 2 |
| | | 166.1.306 | 44142 | C1B2U1 | 2 |
| | | 166.1.311 | 95551 | Bi6Pb1Te10 | 2 |
| | | 166.1.336 | 658633 | Bi2Ge1Te4 | 2 |
| | | 166.1.337 | 30394 | Bi2Ge1Te4 | 2 |
| | | 166.1.359 | 58660 | Ba2Mg17 | 2 |
| | | 166.1.363 | 91075 | Ba1O3Ru1 | 2 |
| | | 166.1.364 | 51293 | Ba1O3Ru1 | 2 |
| | | 166.1.366 | 110770 | Ba1O3Ru1 | 2 |
| | | 166.1.367 | 10253 | Ba1O3Ru1 | 2 |
| | | 166.1.370 | 657450 | B6Ce5Fe2 | 2 |
| | | 166.1.371 | 39554 | B4Mo2 | 2 |

Continued on next column

**TABLE XV– continued**

| SG | cl | TQC | ICSD | Stoich. Formula | SubSG |
|---|---|---|---|---|---|
| | | 166.1.374 | 44451 | B2Mo1 | 2 |
| | | 166.1.375 | 418398 | B2Mo1 | 2 |
| | | 166.1.376 | 418397 | B2Mo1 | 2 |
| | | 166.1.377 | 40907 | B2Mo1 | 2 |
| | | 166.1.378 | 252125 | B2Mo1 | 2 |
| | | 166.1.379 | 97343 | B2Ce2Ir5 | 2 |
| | | 166.1.380 | 409982 | B1Br1Eu2N2 | 2 |
| | | 166.1.398 | 416373 | Ag2Ni1O2 | 2 |
| | | 166.1.399 | 412279 | Ag2Ni1O2 | 2 |
| | | 166.1.400 | 172558 | Ag2Ni1O2 | 2 |
| | | 166.1.401 | 160582 | Ag2Ni1O2 | 2 |
| | | 166.1.402 | 160581 | Ag2Ni1O2 | 2 |
| | | 166.1.403 | 160580 | Ag2Ni1O2 | 2 |
| | | 166.1.404 | 160579 | Ag2Ni1O2 | 2 |
| | | 166.1.405 | 160578 | Ag2Ni1O2 | 2 |
| | | 166.1.406 | 160577 | Ag2Ni1O2 | 2 |
| | | 166.1.407 | 160576 | Ag2Ni1O2 | 2 |
| | | 166.1.408 | 160575 | Ag2Ni1O2 | 2 |
| | | 166.1.409 | 160574 | Ag2Ni1O2 | 2 |
| 166 | 3 | 166.1.13 | 653216 | Te1Zr3 | 2 |
| | | 166.1.76 | 649479 | Pr2Zn17 | 2 |
| | | 166.1.77 | 649469 | Pr2Zn17 | 2 |
| | | 166.1.78 | 104241 | Pr2Zn17 | 2 |
| | | 166.1.80 | 648763 | Pd3S2Tl2 | 2 |
| | | 166.1.85 | 647123 | Ni7Yb2 | 2 |
| | | 166.1.87 | 646978 | Ni3Ti0.67Zr0.33 | 2 |
| | | 166.1.88 | 646399 | Ni3S2Tl2 | 2 |
| | | 166.1.90 | 422328 | Ni3S2Tl2 | 2 |
| | | 166.1.92 | 654948 | Ni3S2Sn2 | 2 |
| | | 166.1.94 | 646379 | Ni3S2Sn2 | 2 |
| | | 166.1.96 | 100217 | Ni3S2Sn2 | 2 |
| | | 166.1.97 | 646211 | Ni3Pb2Se2 | 2 |
| | | 166.1.98 | 646210 | Ni3Pb2Se2 | 2 |
| | | 166.1.99 | 402453 | Ni3Pb2Se2 | 2 |
| | | 166.1.101 | 646207 | Ni3Pb2S2 | 2 |
| | | 166.1.102 | 646206 | Ni3Pb2S2 | 2 |
| | | 166.1.103 | 417632 | Ni3Pb2S2 | 2 |
| | | 166.1.104 | 24176 | Ni3Pb2S2 | 2 |
| | | 166.1.115 | 641738 | La1Rh3 | 2 |
| | | 166.1.116 | 641724 | La1Rh3 | 2 |
| | | 166.1.120 | 640141 | In2Ni3Se2 | 2 |
| | | 166.1.122 | 640135 | In2Ni3S2 | 2 |
| | | 166.1.124 | 415258 | In2Ni3S2 | 2 |
| | | 166.1.132 | 638697 | Hf1Ni3 | 2 |
| | | 166.1.133 | 2414 | Hf1Ni3 | 2 |
| | | 166.1.148 | 24685 | H3Cl1Co2O3 | 2 |
| | | 166.1.149 | 155534 | H3Cl1Co2O3 | 2 |
| | | 166.1.150 | 6016 | H2N2Th1 | 2 |
| | | 166.1.155 | 911 | H12K2N6Sn1 | 2 |
| | | 166.1.156 | 159891 | H1.24Co3Y1 | 2 |
| | | 166.1.157 | 94220 | Ge2Nd2Zn15 | 2 |
| | | 166.1.159 | 166193 | Gd1Sn4Ti6 | 2 |
| | | 166.1.163 | 635711 | Gd1 | 2 |
| | | 166.1.172 | 2513 | Ga1 | 2 |
| | | 166.1.177 | 632784 | Fe7Nb6 | 2 |
| | | 166.1.180 | 41380 | Fe6Ge5Li1 | 2 |
| | | 166.1.181 | 633920 | Fe3Th1 | 2 |
| | | 166.1.182 | 633912 | Fe3Th1 | 2 |
| | | 166.1.183 | 633900 | Fe3Th1 | 2 |
| | | 166.1.185 | 633886 | Fe3Th1 | 2 |
| | | 166.1.187 | 71 | Fe3Sn2 | 2 |

Continued on next column

**TABLE XV– continued**

| SG | cl | TQC | ICSD | Stoich. Formula | SubSG |
|---|---|---|---|---|---|
| | | 166.1.195 | 195809 | Fe2O4Yb1 | 2 |
| | | 166.1.205 | 184011 | F1Ho1O1 | 2 |
| | | 166.1.208 | 184012 | Er1F1O1 | 2 |
| | | 166.1.209 | 630377 | Dy2Zn17 | 2 |
| | | 166.1.211 | 262377 | Dy2Ni7 | 2 |
| | | 166.1.216 | 184010 | Dy1F1O1 | 2 |
| | | 166.1.225 | 410393 | Cs9F27Ni9 | 2 |
| | | 166.1.226 | 78928 | Cs4F12Fe2K1Li1 | 2 |
| | | 166.1.233 | 184367 | Co7Pr2 | 2 |
| | | 166.1.252 | 625506 | Co3Tm1 | 2 |
| | | 166.1.253 | 625499 | Co3Tm1 | 2 |
| | | 166.1.262 | 657284 | Co17Gd2N3 | 2 |
| | | 166.1.267 | 6077 | Cl1O1Yb1 | 2 |
| | | 166.1.300 | 95112 | C1Nb2S2 | 2 |
| | | 166.1.301 | 95110 | C1Nb2S2 | 2 |
| | | 166.1.353 | 71755 | Ba7Cl2O15Ru4 | 2 |
| | | 166.1.354 | 73069 | Ba7Br2O15Ru4 | 2 |
| | | 166.1.357 | 20621 | Ba3O9Sc4 | 2 |
| | | 166.1.358 | 25005 | Ba2Ni1O6Te1 | 2 |
| | | 166.1.381 | 262037 | As3Cd4Rb1 | 2 |
| | | 166.1.382 | 262035 | As3Cd4Na1 | 2 |
| | | 166.1.383 | 262032 | As3Cd4K1 | 2 |
| | | 166.1.395 | 58175 | Al3Tb1 | 2 |
| | | 166.1.397 | 424311 | Ag4Sb2Sr1 | 2 |
| 166 | 4 | 166.1.52 | 651646 | Sb1Te2Tl1 | 2 |
| | | 166.1.53 | 15411 | Sb1Te2Tl1 | 2 |
| | | 166.1.79 | 54338 | Pr1 | 2 |
| | | 166.1.86 | 646928 | Ni7Th2 | 2 |
| | | 166.1.126 | 172602 | I1N1Sr2 | 2 |
| | | 166.1.138 | 44626 | H6Al3Na1O14S2 | 2 |
| | | 166.1.139 | 24344 | H4Cu1N1O3S1 | 2 |
| | | 166.1.145 | 623698 | H4Co3Ho1 | 2 |
| | | 166.1.162 | 635712 | Gd1 | 2 |
| | | 166.1.164 | 52517 | Gd1 | 2 |
| | | 166.1.167 | 602235 | Ga3Ho1 | 2 |
| | | 166.1.173 | 634071 | Fe7W6 | 2 |
| | | 166.1.174 | 634063 | Fe7W6 | 2 |
| | | 166.1.175 | 634060 | Fe7W6 | 2 |
| | | 166.1.176 | 103694 | Fe7W6 | 2 |
| | | 166.1.178 | 632622 | Fe7Mo6 | 2 |
| | | 166.1.179 | 632620 | Fe7Mo6 | 2 |
| | | 166.1.192 | 4192 | Fe2O4Yb1 | 2 |
| | | 166.1.193 | 195811 | Fe2O4Yb1 | 2 |
| | | 166.1.194 | 195810 | Fe2O4Yb1 | 2 |
| | | 166.1.198 | 95660 | F1Nd1O1 | 2 |
| | | 166.1.199 | 95659 | F1Nd1O1 | 2 |
| | | 166.1.200 | 95658 | F1Nd1O1 | 2 |
| | | 166.1.201 | 95657 | F1Nd1O1 | 2 |
| | | 166.1.202 | 95656 | F1Nd1O1 | 2 |
| | | 166.1.203 | 95655 | F1Nd1O1 | 2 |
| | | 166.1.204 | 184005 | F1Nd1O1 | 2 |
| | | 166.1.207 | 67700 | Er1Fe2O4 | 2 |
| | | 166.1.217 | 629540 | Dy1 | 2 |
| | | 166.1.218 | 629537 | Dy1 | 2 |
| | | 166.1.219 | 52518 | Dy1 | 2 |
| | | 166.1.230 | 625563 | Co7Y2 | 2 |
| | | 166.1.234 | 623985 | Co7La2 | 2 |
| | | 166.1.235 | 623978 | Co7La2 | 2 |
| | | 166.1.268 | 24410 | Cl1La1 | 2 |
| | | 166.1.274 | 166533 | Ca2I1P1 | 2 |
| | | 166.1.276 | 52789 | Ca1Si2 | 2 |

Continued on next column

**TABLE XV– continued**

| SG | cl | TQC | ICSD | Stoich. Formula | SubSG |
|---|---|---|---|---|---|
| | | 166.1.278 | 32006 | Ca1Si2 | 2 |
| | | 166.1.279 | 248518 | Ca1Si2 | 2 |
| | | 166.1.280 | 248517 | Ca1Si2 | 2 |
| | | 166.1.281 | 248516 | Ca1Si2 | 2 |
| | | 166.1.282 | 248515 | Ca1Si2 | 2 |
| | | 166.1.285 | 77316 | Ca1Ge2 | 2 |
| | | 166.1.286 | 619304 | Ca1Ge2 | 2 |
| | | 166.1.287 | 245612 | Ca1Ge2 | 2 |
| | | 166.1.288 | 110107 | Ca1Ge2 | 2 |
| | | 166.1.289 | 20822 | C5Ti8 | 2 |
| | | 166.1.296 | 42762 | C1Tb2 | 2 |
| | | 166.1.297 | 42761 | C1Tb2 | 2 |
| | | 166.1.298 | 108187 | C1Tb2 | 2 |
| | | 166.1.338 | 172572 | Bi1S2Tl1 | 2 |
| | | 166.1.360 | 58665 | Ba1Pb3 | 2 |
| | | 166.1.361 | 419973 | Ba1Pb3 | 2 |
| | | 166.1.396 | 420092 | Al10Ba7 | 2 |
| 166 | 5 | 166.1.20 | 651728 | Sb2V3 | 2 |
| | | 166.1.21 | 41814 | Sb2V3 | 2 |
| | | 166.1.72 | 650000 | Pu2Zn17 | 2 |
| | | 166.1.83 | 53301 | P1 | 2 |
| | | 166.1.135 | 15888 | H8N2O8Pb1S2 | 2 |
| | | 166.1.339 | 64705 | Bi1 | 2 |
| | | 166.1.340 | 64704 | Bi1 | 2 |
| | | 166.1.341 | 64703 | Bi1 | 2 |
| | | 166.1.342 | 53797 | Bi1 | 2 |
| | | 166.1.343 | 246663 | Bi1 | 2 |
| | | 166.1.344 | 242004 | Bi1 | 2 |
| | | 166.1.345 | 242003 | Bi1 | 2 |
| | | 166.1.346 | 242002 | Bi1 | 2 |
| | | 166.1.347 | 242001 | Bi1 | 2 |
| | | 166.1.348 | 242000 | Bi1 | 2 |
| | | 166.1.349 | 241999 | Bi1 | 2 |
| | | 166.1.350 | 241984 | Bi1 | 2 |
| | | 166.1.351 | 241983 | Bi1 | 2 |
| | | 166.1.356 | 174186 | Ba4Er1O12Ru3 | 2 |
| | | 166.1.369 | 657452 | B6Fe2Gd5 | 2 |
| | | 166.1.394 | 281660 | Al6Dy2Pt1Si4 | 2 |
| 166 | 6 | 166.1.4 | 653467 | Y2Zn17 | 2 |
| | | 166.1.5 | 104239 | Y2Zn17 | 2 |
| | | 166.1.6 | 653389 | U2Zn17 | 2 |
| | | 166.1.7 | 653384 | U2Zn17 | 2 |
| | | 166.1.8 | 106208 | U2Zn17 | 2 |
| | | 166.1.9 | 106207 | U2Zn17 | 2 |
| | | 166.1.10 | 104248 | U2Zn17 | 2 |
| | | 166.1.14 | 52499 | Te1 | 2 |
| | | 166.1.16 | 652055 | Se1Ti1 | 2 |
| | | 166.1.19 | 291530 | Sb3Te4.5 | 2 |
| | | 166.1.22 | 651638 | Sb2Te3 | 2 |
| | | 166.1.23 | 651635 | Sb2Te3 | 2 |
| | | 166.1.24 | 651629 | Sb2Te3 | 2 |
| | | 166.1.25 | 5420 | Sb2Te3 | 2 |
| | | 166.1.26 | 262171 | Sb2Te3 | 2 |
| | | 166.1.27 | 2084 | Sb2Te3 | 2 |
| | | 166.1.28 | 20236 | Sb2Te3 | 2 |
| | | 166.1.29 | 193347 | Sb2Te3 | 2 |
| | | 166.1.30 | 193346 | Sb2Te3 | 2 |
| | | 166.1.31 | 193345 | Sb2Te3 | 2 |
| | | 166.1.32 | 193344 | Sb2Te3 | 2 |
| | | 166.1.33 | 193343 | Sb2Te3 | 2 |
| | | 166.1.34 | 193342 | Sb2Te3 | 2 |

Continued on next column

**TABLE XV– continued**

| SG | cl | TQC | ICSD | Stoich. Formula | SubSG |
|---|---|---|---|---|---|
| | | 166.1.35 | 193341 | Sb2Te3 | 2 |
| | | 166.1.36 | 192780 | Sb2Te3 | 2 |
| | | 166.1.38 | 185952 | Sb2Te3 | 2 |
| | | 166.1.39 | 185951 | Sb2Te3 | 2 |
| | | 166.1.40 | 185950 | Sb2Te3 | 2 |
| | | 166.1.41 | 185949 | Sb2Te3 | 2 |
| | | 166.1.42 | 185948 | Sb2Te3 | 2 |
| | | 166.1.43 | 185947 | Sb2Te3 | 2 |
| | | 166.1.44 | 185946 | Sb2Te3 | 2 |
| | | 166.1.45 | 185945 | Sb2Te3 | 2 |
| | | 166.1.48 | 651529 | Sb2Se2Te1 | 2 |
| | | 166.1.49 | 52295 | Sb2Se2Te1 | 2 |
| | | 166.1.50 | 87416 | Sb2Se1Te2 | 2 |
| | | 166.1.51 | 2085 | Sb2Se1Te2 | 2 |
| | | 166.1.68 | 24757 | S2Ta1 | 2 |
| | | 166.1.84 | 416476 | O1P1Pr1Zn1 | 2 |
| | | 166.1.106 | 646072 | Nd2Zn17 | 2 |
| | | 166.1.107 | 601895 | Nd2Zn17 | 2 |
| | | 166.1.108 | 104242 | Nd2Zn17 | 2 |
| | | 166.1.110 | 30364 | N2W1 | 2 |
| | | 166.1.117 | 640862 | Ir7Nd2 | 2 |
| | | 166.1.127 | 654919 | Hg1 | 2 |
| | | 166.1.129 | 174006 | Hg1 | 2 |
| | | 166.1.130 | 174005 | Hg1 | 2 |
| | | 166.1.131 | 104296 | Hg1 | 2 |
| | | 166.1.140 | 623755 | H4Co3Y1 | 2 |
| | | 166.1.141 | 623744 | H4Co3Y1 | 2 |
| | | 166.1.142 | 604264 | H4Co3Y1 | 2 |
| | | 166.1.143 | 602920 | H4Co3Y1 | 2 |
| | | 166.1.144 | 623723 | H4Co3Pr1 | 2 |
| | | 166.1.147 | 623752 | H3Co3Y1 | 2 |
| | | 166.1.152 | 623746 | H1Co3Y1 | 2 |
| | | 166.1.158 | 99161 | Ge2Li4Zr1 | 2 |
| | | 166.1.160 | 418524 | Gd1O1P1Zn1 | 2 |
| | | 166.1.169 | 635031 | Ga1P1 | 2 |
| | | 166.1.184 | 633896 | Fe3Th1 | 2 |
| | | 166.1.186 | 103654 | Fe3Th1 | 2 |
| | | 166.1.188 | 631893 | Fe3Gd1 | 2 |
| | | 166.1.189 | 631888 | Fe3Gd1 | 2 |
| | | 166.1.190 | 631864 | Fe3Gd1 | 2 |
| | | 166.1.191 | 103479 | Fe3Gd1 | 2 |
| | | 166.1.196 | 29011 | Fe1O2Tl1 | 2 |
| | | 166.1.197 | 631854 | Fe15Ga2Y2 | 2 |
| | | 166.1.221 | 80259 | Cu1I1 | 2 |
| | | 166.1.222 | 80257 | Cu1I1 | 2 |
| | | 166.1.223 | 80253 | Cu1I1 | 2 |
| | | 166.1.224 | 80252 | Cu1I1 | 2 |
| | | 166.1.237 | 658884 | Co7Dy2 | 2 |
| | | 166.1.238 | 656092 | Co3Y1 | 2 |
| | | 166.1.239 | 656079 | Co3Y1 | 2 |
| | | 166.1.240 | 625636 | Co3Y1 | 2 |
| | | 166.1.241 | 625625 | Co3Y1 | 2 |
| | | 166.1.242 | 625605 | Co3Y1 | 2 |
| | | 166.1.243 | 625600 | Co3Y1 | 2 |
| | | 166.1.244 | 625590 | Co3Y1 | 2 |
| | | 166.1.245 | 625580 | Co3Y1 | 2 |
| | | 166.1.246 | 625572 | Co3Y1 | 2 |
| | | 166.1.247 | 625564 | Co3Y1 | 2 |
| | | 166.1.248 | 603098 | Co3Y1 | 2 |
| | | 166.1.249 | 159894 | Co3Y1 | 2 |
| | | 166.1.250 | 159890 | Co3Y1 | 2 |

Continued on next column

**TABLE XV– continued**

| SG | cl | TQC | ICSD | Stoich. Formula | SubSG |
|---|---|---|---|---|---|
| | | 166.1.251 | 102729 | Co3Y1 | 2 |
| | | 166.1.259 | 624054 | Co3Lu1 | 2 |
| | | 166.1.260 | 624039 | Co3Lu1 | 2 |
| | | 166.1.275 | 424597 | Ca2Ge1Pd3 | 2 |
| | | 166.1.291 | 659087 | C3Er2Fe17 | 2 |
| | | 166.1.307 | 238975 | C1B2Pu1 | 2 |
| | | 166.1.313 | 74348 | Bi2Te3 | 2 |
| | | 166.1.314 | 658764 | Bi2Te3 | 2 |
| | | 166.1.315 | 44983 | Bi2Te3 | 2 |
| | | 166.1.316 | 42546 | Bi2Te3 | 2 |
| | | 166.1.317 | 193340 | Bi2Te3 | 2 |
| | | 166.1.318 | 193339 | Bi2Te3 | 2 |
| | | 166.1.319 | 193338 | Bi2Te3 | 2 |
| | | 166.1.320 | 193337 | Bi2Te3 | 2 |
| | | 166.1.321 | 193336 | Bi2Te3 | 2 |
| | | 166.1.322 | 193335 | Bi2Te3 | 2 |
| | | 166.1.323 | 193334 | Bi2Te3 | 2 |
| | | 166.1.324 | 193333 | Bi2Te3 | 2 |
| | | 166.1.325 | 193332 | Bi2Te3 | 2 |
| | | 166.1.326 | 193331 | Bi2Te3 | 2 |
| | | 166.1.327 | 193330 | Bi2Te3 | 2 |
| | | 166.1.328 | 184631 | Bi2Te3 | 2 |
| | | 166.1.329 | 158366 | Bi2Te3 | 2 |
| | | 166.1.330 | 15753 | Bi2Te3 | 2 |
| | | 166.1.331 | 42545 | Bi2Se3 | 2 |
| | | 166.1.332 | 20385 | Bi2Se3 | 2 |
| | | 166.1.333 | 165226 | Bi2Se3 | 2 |
| | | 166.1.334 | 54838 | Bi2Se2Te1 | 2 |
| | | 166.1.335 | 43512 | Bi2Se1Te2 | 2 |
| | | 166.1.352 | 58724 | Be17Nb2 | 2 |
| | | 166.1.373 | 601082 | B3Co1Dy2.5 | 2 |
| | | 166.1.384 | 68110 | As2Te3 | 2 |
| | | 166.1.385 | 82371 | As2Sn2Sr1 | 2 |
| | | 166.1.386 | 68114 | As2Ge4Te7 | 2 |
| | | 166.1.387 | 68113 | As2Ge3Te6 | 2 |
| | | 166.1.393 | 281656 | Al6Gd2Pt1Si4 | 2 |
| 166 | 7 | 166.1.11 | 653358 | Tm2Zn17 | 2 |
| | | 166.1.12 | 104246 | Tm2Zn17 | 2 |
| | | 166.1.37 | 187495 | Sb2Te3 | 2 |
| | | 166.1.73 | 41372 | Pt3Te4 | 2 |
| | | 166.1.81 | 648762 | Pd3S2Tl2 | 2 |
| | | 166.1.89 | 646398 | Ni3S2Tl2 | 2 |
| | | 166.1.91 | 108893 | Ni3S2Tl2 | 2 |
| | | 166.1.93 | 646380 | Ni3S2Sn2 | 2 |
| | | 166.1.95 | 402458 | Ni3S2Sn2 | 2 |
| | | 166.1.100 | 108892 | Ni3Pb2Se2 | 2 |
| | | 166.1.105 | 108895 | Ni3Pb2S2 | 2 |
| | | 166.1.121 | 640140 | In2Ni3Se2 | 2 |
| | | 166.1.123 | 640134 | In2Ni3S2 | 2 |
| | | 166.1.125 | 108894 | In2Ni3S2 | 2 |
| | | 166.1.128 | 53771 | Hg1 | 2 |
| | | 166.1.136 | 195640 | H6Mg1Nb2 | 2 |
| | | 166.1.151 | 423880 | H2Mo1O4Rb2S1 | 2 |
| | | 166.1.153 | 40981 | H1Cl1Sc1 | 2 |
| | | 166.1.154 | 62226 | H1Cl1Lu1 | 2 |
| | | 166.1.215 | 629598 | Dy1Fe3 | 2 |
| | | 166.1.261 | 246307 | Co1Li1O2 | 2 |
| | | 166.1.265 | 869 | Cl1Zr1 | 2 |
| | | 166.1.266 | 20145 | Cl1Zr1 | 2 |
| | | 166.1.290 | 657171 | C3Fe17Tb2 | 2 |
| | | 166.1.302 | 53780 | C1 | 2 |

Continued on next column

**TABLE XV– continued**

| SG | cl | TQC | ICSD | Stoich. Formula | SubSG |
|---|---|---|---|---|---|
| | | 166.1.303 | 31829 | C1 | 2 |
| | | 166.1.304 | 29123 | C1 | 2 |
| | | 166.1.308 | 1168 | Br1Zr1 | 2 |
| | | 166.1.309 | 23354 | Br1La1 | 2 |
| | | 166.1.312 | 20319 | Bi4Se3 | 2 |
| | | 166.1.365 | 172175 | Ba1O3Ru1 | 2 |
| | | 166.1.372 | 191908 | B4Mo2 | 2 |
| 167 | 1 | 167.1.3 | 95761 | O3V2 | 2 |
| | | 167.1.4 | 95760 | O3V2 | 2 |
| | | 167.1.8 | 25780 | O3V2 | 2 |
| | | 167.1.9 | 201109 | O3V2 | 2 |
| | | 167.1.13 | 1874 | O3V2 | 2 |
| | | 167.1.14 | 1873 | O3V2 | 2 |
| | | 167.1.15 | 1872 | O3V2 | 2 |
| | | 167.1.19 | 167061 | La1Mn1O3.31 | 2 |
| | | 167.1.20 | 167062 | La1Mn1O3.29 | 2 |
| | | 167.1.21 | 167065 | La1Mn1O3.26 | 2 |
| | | 167.1.22 | 167064 | La1Mn1O3.26 | 2 |
| | | 167.1.23 | 167066 | La1Mn1O3.22 | 2 |
| | | 167.1.24 | 167068 | La1Mn1O3.15 | 2 |
| | | 167.1.25 | 167069 | La1Mn1O3.13 | 2 |
| | | 167.1.83 | 173470 | Cr2O3 | 2 |
| | | 167.1.88 | 243759 | Ba3O9Os2 | 2 |
| | | 167.1.90 | 402170 | Au3Ga7Yb1 | 2 |
| | | 167.1.91 | 391108 | Al7Au3Dy1 | 2 |
| | | 167.1.92 | 391101 | Al7Au3Ce1 | 2 |
| 167 | 2 | 167.1.26 | 96070 | Fe2O3 | 2 |
| | | 167.1.27 | 96069 | Fe2O3 | 2 |
| | | 167.1.28 | 88418 | Fe2O3 | 2 |
| | | 167.1.29 | 88417 | Fe2O3 | 2 |
| | | 167.1.30 | 82904 | Fe2O3 | 2 |
| | | 167.1.31 | 82903 | Fe2O3 | 2 |
| | | 167.1.32 | 82902 | Fe2O3 | 2 |
| | | 167.1.33 | 82137 | Fe2O3 | 2 |
| | | 167.1.34 | 82136 | Fe2O3 | 2 |
| | | 167.1.35 | 82135 | Fe2O3 | 2 |
| | | 167.1.36 | 82134 | Fe2O3 | 2 |
| | | 167.1.37 | 81248 | Fe2O3 | 2 |
| | | 167.1.38 | 71194 | Fe2O3 | 2 |
| | | 167.1.39 | 66756 | Fe2O3 | 2 |
| | | 167.1.40 | 64599 | Fe2O3 | 2 |
| | | 167.1.41 | 56372 | Fe2O3 | 2 |
| | | 167.1.42 | 43465 | Fe2O3 | 2 |
| | | 167.1.43 | 415251 | Fe2O3 | 2 |
| | | 167.1.44 | 40142 | Fe2O3 | 2 |
| | | 167.1.45 | 24791 | Fe2O3 | 2 |
| | | 167.1.46 | 245851 | Fe2O3 | 2 |
| | | 167.1.47 | 24004 | Fe2O3 | 2 |
| | | 167.1.48 | 22505 | Fe2O3 | 2 |
| | | 167.1.49 | 201101 | Fe2O3 | 2 |
| | | 167.1.50 | 201100 | Fe2O3 | 2 |
| | | 167.1.51 | 201099 | Fe2O3 | 2 |
| | | 167.1.52 | 201098 | Fe2O3 | 2 |
| | | 167.1.53 | 201097 | Fe2O3 | 2 |
| | | 167.1.54 | 201096 | Fe2O3 | 2 |
| | | 167.1.55 | 184766 | Fe2O3 | 2 |
| | | 167.1.56 | 182848 | Fe2O3 | 2 |
| | | 167.1.57 | 182847 | Fe2O3 | 2 |
| | | 167.1.58 | 182846 | Fe2O3 | 2 |
| | | 167.1.59 | 182845 | Fe2O3 | 2 |
| | | 167.1.60 | 182844 | Fe2O3 | 2 |

Continued on next column

**TABLE XV– continued**

| SG | cl | TQC | ICSD | Stoich. Formula | SubSG |
|---|---|---|---|---|---|
| | | 167.1.61 | 182843 | Fe2O3 | 2 |
| | | 167.1.62 | 182842 | Fe2O3 | 2 |
| | | 167.1.63 | 182841 | Fe2O3 | 2 |
| | | 167.1.64 | 182840 | Fe2O3 | 2 |
| | | 167.1.65 | 182839 | Fe2O3 | 2 |
| | | 167.1.66 | 173655 | Fe2O3 | 2 |
| | | 167.1.67 | 173653 | Fe2O3 | 2 |
| | | 167.1.68 | 161294 | Fe2O3 | 2 |
| | | 167.1.69 | 161293 | Fe2O3 | 2 |
| | | 167.1.70 | 161292 | Fe2O3 | 2 |
| | | 167.1.71 | 161291 | Fe2O3 | 2 |
| | | 167.1.72 | 161290 | Fe2O3 | 2 |
| | | 167.1.73 | 161289 | Fe2O3 | 2 |
| | | 167.1.74 | 161288 | Fe2O3 | 2 |
| | | 167.1.75 | 161287 | Fe2O3 | 2 |
| | | 167.1.76 | 161286 | Fe2O3 | 2 |
| | | 167.1.77 | 161285 | Fe2O3 | 2 |
| | | 167.1.78 | 161284 | Fe2O3 | 2 |
| | | 167.1.79 | 161283 | Fe2O3 | 2 |
| | | 167.1.80 | 15840 | Fe2O3 | 2 |
| | | 167.1.81 | 154191 | Fe2O3 | 2 |
| | | 167.1.82 | 154190 | Fe2O3 | 2 |
| | | 167.1.84 | 201966 | Cr2Cs3I9 | 2 |
| | | 167.1.85 | 50809 | Ca3Co1O6Rh1 | 2 |
| | | 167.1.87 | 60625 | Br6Cd1Rb4 | 2 |
| | | 167.1.89 | 97540 | Ba3Cr2S6 | 2 |
| 167 | 3 | 167.1.1 | 83385 | S3Yb2 | 2 |
| | | 167.1.2 | 75248 | S3Yb2 | 2 |
| | | 167.1.5 | 94768 | O3V2 | 2 |
| | | 167.1.6 | 64796 | O3V2 | 2 |
| | | 167.1.7 | 64786 | O3V2 | 2 |
| | | 167.1.10 | 201108 | O3V2 | 2 |
| | | 167.1.11 | 201107 | O3V2 | 2 |
| | | 167.1.12 | 201106 | O3V2 | 2 |
| | | 167.1.16 | 1870 | O3V2 | 2 |
| | | 167.1.17 | 1869 | O3V2 | 2 |
| | | 167.1.18 | 1473 | O3V2 | 2 |
| | | 167.1.86 | 421391 | C3Cr2N6 | 2 |
| 176 | 1 | 176.1.2 | 251133 | Ba4Nb2O9 | 2,174 |
| 176 | 11 | 176.1.1 | 80772 | Mn1N3Sr3 | 174 |
| 189 | 1 | 189.1.1 | 51718 | Si5Yb3 | 174 |
| | | 189.1.2 | 415599 | Si5Yb3 | 174 |
| | | 189.1.3 | 408117 | Si5Yb3 | 174 |
| | | 189.1.6 | 196273 | O1Ti1 | 174 |
| | | 189.1.9 | 414438 | Mg1Tl1Y1 | 174 |
| | | 189.1.10 | 54344 | Mg1Sn1Yb1 | 174 |
| | | 189.1.13 | 414439 | La1Mg1Tl1 | 174 |
| | | 189.1.14 | 160910 | In1Mg1Y1 | 174 |
| | | 189.1.16 | 409925 | Ge1Pd1Ti1 | 174 |
| | | 189.1.18 | 240950 | Ga1Mg1Y1 | 174 |
| | | 189.1.19 | 160909 | Ga1Mg1Y1 | 174 |
| | | 189.1.20 | 413276 | Ga1La1Mg1 | 174 |
| | | 189.1.23 | 428473 | Cu1Ge1Sc1 | 174 |
| | | 189.1.30 | 413810 | Ba1Bi1Na1 | 174 |
| | | 189.1.31 | 261317 | Au1Ge2In3Yb3 | 174 |
| | | 189.1.35 | 810 | As2Eu2 | 174 |
| | | 189.1.36 | 193686 | Al3Co1Gd3Ge2 | 174 |
| | | 189.1.37 | 54936 | Al1Pd1Tb1 | 174 |
| | | 189.1.39 | 107275 | Ag1Pb1Y1 | 174 |
| | | 189.1.40 | 416377 | Ag1Lu1Sn1 | 174 |
| | | 189.1.41 | 245236 | Ag1Lu1Sn1 | 174 |

Continued on next column

**TABLE XV– continued**

| SG | cl | TQC | ICSD | Stoich. Formula | SubSG |
|---|---|---|---|---|---|
| | | 189.1.42 | 107278 | Ag1Lu1Pb1 | 174 |
| | | 189.1.43 | 10016 | Ag1Ca1P1 | 174 |
| | | 189.1.44 | 10017 | Ag1As1Ca1 | 174 |
| 189 | 2 | 189.1.5 | 415370 | Pr1Rh1Sn1 | 174 |
| | | 189.1.7 | 183337 | Nd1Pd1Zn1 | 174 |
| | | 189.1.11 | 411152 | Li5N3Ni3 | 174 |
| | | 189.1.15 | 85966 | Hf6Ni1Sb2 | 174 |
| | | 189.1.21 | 152689 | Fe1Ho6Sb2 | 174 |
| | | 189.1.25 | 415492 | Ce1Pt1Sn1 | 174 |
| | | 189.1.26 | 416969 | Ce1Pd1Sn1 | 174 |
| | | 189.1.32 | 104011 | Au1Gd1In1 | 174 |
| | | 189.1.38 | 160908 | Al1Mg1Y1 | 174 |
| 189 | 3 | 189.1.8 | 419191 | Nd1Pd1Sn1 | 174 |
| | | 189.1.12 | 415594 | La1Rh1Sn1 | 174 |
| | | 189.1.27 | 100002 | Ce1In1Ir1 | 174 |
| | | 189.1.28 | 428851 | Cd1Ce1Pt1 | 174 |
| | | 189.1.33 | 192086 | Au1Ga1Mg1 | 174 |
| 189 | 4 | 189.1.22 | 418214 | Cu1La1Mg4 | 174 |
| | | 189.1.24 | 98233 | Ce2Ge3Zn6 | 174 |
| | | 189.1.34 | 411543 | Au1Cd1La1 | 174 |
| 189 | 5 | 189.1.4 | 410418 | Pt1Sn1Yb1 | 174 |
| 189 | 8 | 189.1.17 | 59177 | Ge1Li1Pr1 | 174 |
| | | 189.1.29 | 51528 | C2Rb2 | 174 |
| 190 | 1 | 190.2.1 | 81072 | Ba3Cr2O9W1 | 174 |
| | | 190.2.2 | 73183 | Ba3Ca1O9Ru2 | 174 |
| | | 191.1.19 | 36504 | B1Ca1Ni4 | 2 |
| 191 | 2 | 191.1.12 | 78848 | B2Ti1 | 2,174 |
| | | 191.1.13 | 78847 | B2Ti1 | 2,174 |
| | | 191.1.14 | 56723 | B2Ti1 | 2,174 |
| 191 | 3 | 191.1.1 | 173575 | La10O3Si8 | 2,174 |
| | | 191.1.9 | 246238 | Ba1Ga2 | 2,174 |
| 191 | 4 | 191.1.8 | 107305 | Ba1Sn5 | 2,174 |
| 191 | 5 | 191.1.7 | 418614 | Ca1Zn5 | 2,174 |
| 191 | 6 | 191.1.6 | 419163 | Ce1Pd2Zn3 | 2,174 |
| 191 | 8 | 191.1.2 | 417224 | Ga2Sr1 | 2,174 |
| | | 191.1.3 | 260654 | Ga2Sr1 | 2,174 |
| | | 191.1.4 | 246236 | Ga2Sr1 | 2,174 |
| 191 | 11 | 191.1.10 | 97723 | B4Co11Pr3 | 2,174 |
| 191 | 16 | 191.1.11 | 97731 | B3Co7Dy2 | 2,174 |
| 191 | 30 | 191.1.16 | 237660 | B1Eu1Pt4 | 2,174 |
| 191 | 37 | 191.1.15 | 184968 | B2Fe3U1 | 174 |
| 191 | 60 | 191.1.18 | 417869 | B1Co4Gd1 | 2,174 |
| 191 | 61 | 191.1.20 | 409394 | Al9Co2Eu1 | 2,174 |
| 191 | 65 | 191.1.5 | 107425 | Ga2Gd1 | 2,174 |
| | | 191.1.17 | 97722 | B1Co4Pr1 | 2,174 |
| 193 | 1 | 193.1.1 | 150562 | Sn3Zr5 | 2,174 |
| | | 193.1.2 | 109292 | Sn3Ti5 | 2,174 |
| | | 193.1.3 | 62591 | Si3Ti5 | 2,174 |
| | | 193.1.4 | 105649 | Pb3Zr5 | 2,174 |
| | | 193.1.7 | 44361 | Ge3Hf5 | 2,174 |
| 193 | 2 | 193.1.6 | 99999 | K3N1 | 174 |
| 193 | 3 | 193.1.5 | 154390 | Nb1Sb5U3 | 2,174 |
| | | 193.1.8 | 173358 | Bi5Ce3Mn1 | 2,174 |
| 194 | 1 | 194.1.29 | 249600 | Eu1Ga1Sn1 | 2,174 |
| | | 194.1.34 | 420186 | Ce1Ge1Zn1 | 2,174 |
| | | 194.1.35 | 163343 | Ce1Ge1Zn1 | 2,174 |
| | | 194.1.36 | 425673 | C6Co1N6Pr1 | 2,174 |
| | | 194.1.51 | 249399 | C2Al1Ta3 | 2,174 |
| | | 194.1.57 | 251687 | Au1Ce1Sb1 | 2,174 |
| 194 | 2 | 194.1.27 | 163344 | Gd1Ge1Zn1 | 2,174 |

Continued on next column

**TABLE XV– continued**

| SG | cl | TQC | ICSD | Stoich. Formula | SubSG |
|---|---|---|---|---|---|
| | | 194.1.56 | 192939 | Ba1Cr1O3 | 2,174 |
| 194 | 3 | 194.1.32 | 420050 | Ce1Sn1Zn1 | 2 |
| | | 194.1.33 | 152624 | Ce1Sn1Zn1 | 2 |
| | | 194.1.55 | 192940 | Ba1Cr1O3 | 2 |
| 194 | 4 | 194.1.52 | 246019 | Ba1O3Ru1 | 2 |
| | | 194.1.53 | 173866 | Ba1O3Ru1 | 2 |
| | | 194.1.54 | 160287 | Ba1O3Ru1 | 2 |
| 194 | 5 | 194.1.31 | 158979 | Cr1Ge3Nd1 | 2,174 |
| | | 194.1.37 | 180423 | C2Si1Ti3 | 2,174 |
| | | 194.1.38 | 180422 | C2Si1Ti3 | 2,174 |
| | | 194.1.39 | 180421 | C2Si1Ti3 | 2,174 |
| | | 194.1.40 | 180420 | C2Si1Ti3 | 2,174 |
| | | 194.1.41 | 180419 | C2Si1Ti3 | 2,174 |
| | | 194.1.42 | 153264 | C2Si1Ti3 | 2,174 |
| | | 194.1.43 | 180431 | C2Ge1Ti3 | 2,174 |
| | | 194.1.44 | 180430 | C2Ge1Ti3 | 2,174 |
| | | 194.1.45 | 180429 | C2Ge1Ti3 | 2,174 |
| | | 194.1.46 | 180428 | C2Ge1Ti3 | 2,174 |
| | | 194.1.47 | 180427 | C2Ge1Ti3 | 2,174 |
| | | 194.1.48 | 180426 | C2Ge1Ti3 | 2,174 |
| | | 194.1.49 | 180425 | C2Ge1Ti3 | 2,174 |
| | | 194.1.50 | 180424 | C2Ge1Ti3 | 2,174 |
| 194 | 6 | 194.1.1 | 421015 | Zn1 | 2,174 |
| | | 194.1.2 | 421014 | Zn1 | 2,174 |
| | | 194.1.3 | 421013 | Zn1 | 2,174 |
| | | 194.1.4 | 247162 | Zn1 | 2,174 |
| | | 194.1.5 | 247161 | Zn1 | 2,174 |
| | | 194.1.6 | 247160 | Zn1 | 2,174 |
| | | 194.1.7 | 247159 | Zn1 | 2,174 |
| | | 194.1.8 | 247158 | Zn1 | 2,174 |
| | | 194.1.9 | 247157 | Zn1 | 2,174 |
| | | 194.1.10 | 247156 | Zn1 | 2,174 |
| | | 194.1.11 | 247155 | Zn1 | 2,174 |
| | | 194.1.12 | 247154 | Zn1 | 2,174 |
| | | 194.1.13 | 247153 | Zn1 | 2,174 |
| | | 194.1.14 | 247152 | Zn1 | 2,174 |
| | | 194.1.15 | 247151 | Zn1 | 2,174 |
| | | 194.1.16 | 247150 | Zn1 | 2,174 |
| | | 194.1.17 | 247149 | Zn1 | 2,174 |
| | | 194.1.18 | 247148 | Zn1 | 2,174 |
| | | 194.1.19 | 247147 | Zn1 | 2,174 |
| | | 194.1.28 | 417150 | Fe0.25S2Ta1 | 2,174 |
| 194 | 7 | 194.1.20 | 422990 | Mn1O3Y1 | 174 |
| | | 194.1.21 | 422989 | Mn1O3Y1 | 174 |
| | | 194.1.22 | 422982 | Mn1O3Y1 | 174 |
| | | 194.1.23 | 182284 | Mn1O3Y1 | 174 |
| | | 194.1.24 | 181182 | Mn1O3Y1 | 174 |
| | | 194.1.25 | 193568 | Mg1Ni2 | 174 |
| | | 194.1.26 | 171451 | I1La1 | 174 |
| | | 194.1.30 | 249599 | Eu1Ga1Ge1 | 174 |
| 204 | 1 | 204.1.23 | 174551 | Ba1Ge12Pt4 | 2 |
| | | 204.1.24 | 160797 | Ba1Ge12Pt4 | 2 |
| 204 | 2 | 204.1.14 | 248607 | Cr4Cu3La1O12 | 2 |
| | | 204.1.15 | 248606 | Cr4Cu3La1O12 | 2 |
| | | 204.1.16 | 248605 | Cr4Cu3La1O12 | 2 |
| | | 204.1.17 | 248604 | Cr4Cu3La1O12 | 2 |
| | | 204.1.18 | 248603 | Cr4Cu3La1O12 | 2 |
| | | 204.1.19 | 248601 | Cr4Cu3La1O12 | 2 |
| | | 204.1.20 | 248600 | Cr4Cu3La1O12 | 2 |
| | | 204.1.21 | 248599 | Cr4Cu3La1O12 | 2 |

Continued on next column

**TABLE XV– continued**

| SG | cl | TQC | ICSD | Stoich. Formula | SubSG |
|---|---|---|---|---|---|
| | | 204.1.22 | 248598 | Cr4Cu3La1O12 | 2 |
| 204 | 3 | 204.1.4 | 51893 | Cu3Na1O12Ru4 | 2 |
| | | 204.1.6 | 248615 | Cr4Cu3O12Y1 | 2 |
| | | 204.1.7 | 248614 | Cr4Cu3O12Y1 | 2 |
| | | 204.1.8 | 248613 | Cr4Cu3O12Y1 | 2 |
| | | 204.1.9 | 248612 | Cr4Cu3O12Y1 | 2 |
| | | 204.1.10 | 248611 | Cr4Cu3O12Y1 | 2 |
| | | 204.1.11 | 248610 | Cr4Cu3O12Y1 | 2 |
| | | 204.1.12 | 248609 | Cr4Cu3O12Y1 | 2 |
| | | 204.1.13 | 248608 | Cr4Cu3O12Y1 | 2 |
| 204 | 4 | 204.1.1 | 237751 | Mn3Na1O12V4 | 2 |
| | | 204.1.2 | 195439 | Cu4O12V4 | 2 |
| | | 204.1.3 | 195438 | Cu4O12V4 | 2 |
| 204 | 5 | 204.1.5 | 237422 | Cu3Fe4O12Tm1 | 2 |
| 205 | 1 | 205.2.1 | 77109 | Pd1Sb2 | 2 |
| | | 205.2.2 | 300225 | Os1Te2 | 2 |
| | | 205.2.3 | 86354 | Ni1S2 | 2 |
| | | 205.2.4 | 68169 | Ni1S2 | 2 |
| | | 205.2.5 | 68168 | Ni1S2 | 2 |
| | | 205.2.6 | 68167 | Ni1S2 | 2 |
| | | 216.2.1 | 260984 | Ni5U1 | 82 |
| | | 216.2.2 | 54390 | Ni4Sn1U1 | 82 |
| | | 216.2.3 | 54387 | Ni4Sn1U1 | 82 |
| | | 216.2.4 | 418013 | La4Mg1Ru1 | 82 |
| | | 216.2.5 | 417519 | In1Rh1Tm4 | 82 |
| | | 216.2.6 | 81917 | Hg1S1 | 82 |
| | | 216.2.7 | 419068 | Cd1Gd4Ru1 | 82 |
| | | 216.2.8 | 415194 | Cd1Ce1Ni4 | 82 |
| | | 217.1.1 | 427048 | F6Ni3O6Sb4 | 82 |
| | | 217.1.2 | 427047 | F6Ni3O6Sb4 | 82 |
| | | 217.1.3 | 84572 | Cu12S13Sb4 | 82 |
| | | 217.1.4 | 84571 | Cu12S13Sb4 | 82 |
| | | 217.1.5 | 62116 | Cu12S13Sb4 | 82 |
| | | 217.1.6 | 25707 | Cu12S13Sb4 | 82 |
| | | 217.1.7 | 33588 | As4Cu12S12 | 82 |
| | | 220.2.1 | 155383 | C3Y2 | 82 |
| | | 220.2.2 | 155382 | C3Y2 | 82 |
| | | 220.2.3 | 26477 | C3U2 | 82 |
| | | 220.2.4 | 2447 | C3Nd2 | 82 |
| | | 221.1.13 | 50313 | B6Sr1 | 2,81 |
| | | 221.1.14 | 51495 | B6Ba1 | 2,81 |
| 221 | 2 | 221.1.7 | 262924 | Cs1I3Sn1 | 2,81 |
| | | 221.1.11 | 80320 | Br3Cs1Ge1 | 2,81 |
| | | 221.1.12 | 409846 | Br3Ce3Ga1 | 2,81 |
| 221 | 3 | 221.1.1 | 201036 | O1Sn1Sr3 | 2 |
| | | 221.1.8 | 100787 | Ca3O1Sn1 | 2 |
| | | 221.1.9 | 100789 | Ca3O1Pb1 | 2 |
| | | 221.1.10 | 402087 | Ca3Ge1O1 | 2 |
| 221 | 4 | 221.1.6 | 100533 | Fe1Ti1 | 2,81 |
| 221 | 5 | 221.1.4 | 247065 | In1N1Ni3 | 2 |
| | | 221.1.5 | 247273 | H1Pd3Tl1 | 2 |
| 221 | 6 | 221.1.3 | 163195 | Mg1Pd1 | 2,81 |
| 221 | 7 | 221.1.2 | 181127 | Ni3Sn1 | 2 |
| 225 | 1 | 225.1.1 | 63098 | Pb1Te1 | 87 |
| | | 225.1.2 | 153711 | Pb1Te1 | 87 |
| | | 225.1.3 | 63097 | Pb1Se1 | 87 |
| | | 225.1.4 | 63096 | Pb1Se1 | 87 |
| | | 225.1.5 | 62196 | Pb1Se1 | 87 |
| | | 225.1.6 | 62195 | Pb1Se1 | 87 |
| | | 225.1.7 | 38294 | Pb1Se1 | 87 |

Continued on next column

**TABLE XV– continued**

| SG | cl | TQC | ICSD | Stoich. Formula | SubSG |
|---|---|---|---|---|---|
| | | 225.1.8 | 62192 | Pb1S1 | 87 |
| | | 225.1.9 | 62191 | Pb1S1 | 87 |
| | | 225.1.10 | 62190 | Pb1S1 | 87 |
| | | 225.1.11 | 35204 | O2U1 | 87 |
| | | 225.1.12 | 246851 | O2U1 | 87 |
| | | 225.1.16 | 160154 | Fe16Si7U6 | 87 |
| | | 225.1.30 | 409959 | B6Ni21Sc2 | 87 |
| 225 | 2 | 225.1.14 | 159261 | Mg6Ni16Si7 | 2,82 |
| | | 225.1.24 | 159873 | C1Hf1 | 2,82 |
| 225 | 3 | 225.1.23 | 424342 | C1W1 | 2 |
| | | 225.1.25 | 155551 | Ba2Nd1O6Ru1 | 2 |
| 225 | 4 | 225.1.13 | 419774 | Ni21P6Sn2 | 2,82 |
| | | 225.1.19 | 75013 | Ce3Pd20Si6 | 2,82 |
| | | 225.1.21 | 62667 | C6Cr23 | 2,82 |
| 225 | 5 | 225.1.20 | 237543 | Ca3Cd8Pt4 | 2,82 |
| | | 225.1.26 | 171491 | Ba2Ir1O6Tb1 | 2,82 |
| | | 225.1.27 | 153057 | Ba2Ir1O6Pr1 | 2,82 |
| 225 | 6 | 225.1.17 | 188351 | Cl1Cu6O8Y1 | 2 |
| | | 225.1.18 | 69612 | Cl1Cu6In1O8 | 2 |
| | | 225.1.28 | 59744 | Ba2Er1O6Ru1 | 2 |
| | | 225.1.29 | 59743 | Ba2Er1O6Ru1 | 2 |
| 225 | 7 | 225.1.15 | 106715 | Ga2Li1Rh1 | 2,82 |
| | | 225.1.22 | 159874 | C1Zr1 | 2,82 |
| 227 | 1 | 227.1.6 | 249974 | Nd2O7Zr2 | 2,82 |
| | | 227.1.11 | 83277 | Mn2O7Sc2 | 2,82 |
| | | 227.1.18 | 423608 | In2Mg1 | 2,82 |
| | | 227.1.36 | 84753 | Gd2O7Sn2 | 2,82 |
| | | 227.1.59 | 154730 | Co2Nd1 | 2,82 |
| | | 227.1.60 | 154729 | Co2Nd1 | 2,82 |
| | | 227.1.61 | 154728 | Co2Nd1 | 2,82 |
| | | 227.1.62 | 154727 | Co2Nd1 | 2,82 |
| | | 227.1.63 | 154726 | Co2Nd1 | 2,82 |
| | | 227.1.64 | 154725 | Co2Nd1 | 2,82 |
| | | 227.1.65 | 154724 | Co2Nd1 | 2,82 |
| | | 227.1.66 | 194667 | Co2Er1 | 2,82 |
| | | 227.1.67 | 168600 | Ce2O7Zr2 | 2,82 |
| | | 227.1.68 | 168595 | Ce2O7Zr2 | 2,82 |
| | | 227.1.83 | 188292 | Cd1Ni1 | 2,82 |
| 227 | 2 | 227.1.1 | 72042 | S2Ti1 | 2,82 |
| | | 227.1.2 | 160865 | O7V2Y2 | 2,82 |
| | | 227.1.5 | 261854 | Nd2O7Zr2 | 2,82 |
| | | 227.1.7 | 174187 | Nd2O7Zr2 | 2,82 |
| | | 227.1.8 | 153227 | Nd2O7Zr2 | 2,82 |
| | | 227.1.9 | 82958 | Nd2O7Sn2 | 2,82 |
| | | 227.1.12 | 187996 | Mg3Ni2Ti1 | 2,82 |
| | | 227.1.13 | 184696 | Mg1O4Ti2 | 2,82 |
| | | 227.1.14 | 32688 | Lu2O7V2 | 2,82 |
| | | 227.1.15 | 159290 | Lu2O7V2 | 2,82 |
| | | 227.1.16 | 159286 | Lu2O7V2 | 2,82 |
| | | 227.1.17 | 48128 | Li1O2Ti1 | 2,82 |
| | | 227.1.19 | 420818 | Hg2O7Ru2 | 2,82 |
| | | 227.1.20 | 159809 | Hg2O7Os2 | 2,82 |
| | | 227.1.21 | 159808 | Hg2O7Os2 | 2,82 |
| | | 227.1.22 | 159807 | Hg2O7Os2 | 2,82 |
| | | 227.1.23 | 159806 | Hg2O7Os2 | 2,82 |
| | | 227.1.24 | 159805 | Hg2O7Os2 | 2,82 |
| | | 227.1.25 | 159804 | Hg2O7Os2 | 2,82 |
| | | 227.1.26 | 159803 | Hg2O7Os2 | 2,82 |
| | | 227.1.27 | 159802 | Hg2O7Os2 | 2,82 |
| | | 227.1.28 | 159801 | Hg2O7Os2 | 2,82 |

Continued on next column

**TABLE XV– continued**

| SG | cl | TQC | ICSD | Stoich. Formula | SubSG |
|---|---|---|---|---|---|
| | | 227.1.29 | 159800 | Hg2O7Os2 | 2,82 |
| | | 227.1.30 | 159799 | Hg2O7Os2 | 2,82 |
| | | 227.1.31 | 159798 | Hg2O7Os2 | 2,82 |
| | | 227.1.32 | 159797 | Hg2O7Os2 | 2,82 |
| | | 227.1.33 | 159796 | Hg2O7Os2 | 2,82 |
| | | 227.1.34 | 159795 | Hg2O7Os2 | 2,82 |
| | | 227.1.35 | 159794 | Hg2O7Os2 | 2,82 |
| | | 227.1.38 | 50047 | Cr2O4Zn1 | 2,82 |
| | | 227.1.39 | 196119 | Cr2O4Zn1 | 2,82 |
| | | 227.1.40 | 171899 | Cr2O4Zn1 | 2,82 |
| | | 227.1.41 | 171896 | Cr2O4Zn1 | 2,82 |
| | | 227.1.42 | 171894 | Cr2O4Zn1 | 2,82 |
| | | 227.1.43 | 171893 | Cr2O4Zn1 | 2,82 |
| | | 227.1.44 | 97201 | Cr2Mg1O4 | 2,82 |
| | | 227.1.45 | 75623 | Cr2Mg1O4 | 2,82 |
| | | 227.1.46 | 290599 | Cr2Mg1O4 | 2,82 |
| | | 227.1.47 | 171107 | Cr2Mg1O4 | 2,82 |
| | | 227.1.48 | 171106 | Cr2Mg1O4 | 2,82 |
| | | 227.1.49 | 160953 | Cr2Mg1O4 | 2,82 |
| | | 227.1.50 | 402408 | Cr2Hg1Se4 | 2,82 |
| | | 227.1.58 | 859 | Co2O4Si1 | 2,82 |
| | | 227.1.69 | 155773 | Cd2O7Os2 | 2,82 |
| | | 227.1.70 | 155772 | Cd2O7Os2 | 2,82 |
| | | 227.1.71 | 155771 | Cd2O7Os2 | 2,82 |
| | | 227.1.72 | 155770 | Cd2O7Os2 | 2,82 |
| | | 227.1.73 | 155769 | Cd2O7Os2 | 2,82 |
| | | 227.1.74 | 155768 | Cd2O7Os2 | 2,82 |
| | | 227.1.75 | 155767 | Cd2O7Os2 | 2,82 |
| | | 227.1.76 | 155766 | Cd2O7Os2 | 2,82 |
| | | 227.1.77 | 155765 | Cd2O7Os2 | 2,82 |
| | | 227.1.78 | 155764 | Cd2O7Os2 | 2,82 |
| | | 227.1.79 | 155763 | Cd2O7Os2 | 2,82 |
| | | 227.1.80 | 155762 | Cd2O7Os2 | 2,82 |
| | | 227.1.81 | 155761 | Cd2O7Os2 | 2,82 |
| | | 227.1.82 | 37408 | Cd1Se4Yb2 | 2,82 |
| | | 227.1.84 | 246498 | Cd1Er2S4 | 2,82 |
| | | 227.1.85 | 78554 | Cd1Cr2Se4 | 2,82 |
| | | 227.1.86 | 39416 | Cd1Cr2Se4 | 2,82 |
| | | 227.1.87 | 39415 | Cd1Cr2S4 | 2,82 |
| 227 | 3 | 227.1.3 | 151747 | O7Tb2Ti2 | 2 |
| | | 227.1.4 | 429192 | O7Pr2Zr2 | 2 |
| | | 227.1.10 | 202522 | Mo2O7Y2 | 2 |
| | | 227.1.37 | 160866 | Er2O7V2 | 2 |
| | | 227.1.51 | 183965 | Cr2Fe1O4 | 2 |
| | | 227.1.52 | 183964 | Cr2Fe1O4 | 2 |
| | | 227.1.53 | 183963 | Cr2Fe1O4 | 2 |
| | | 227.1.54 | 183962 | Cr2Fe1O4 | 2 |
| | | 227.1.55 | 183961 | Cr2Fe1O4 | 2 |
| | | 227.1.56 | 183960 | Cr2Fe1O4 | 2 |
| | | 227.1.57 | 171121 | Cr2Fe1O4 | 2 |
| | | 227.1.88 | 57512 | Al2Ba1 | 2 |
| 229 | 1 | 229.1.1 | 26270 | As7Re3 | 2,82 |

The End

TABLE XVI: Topological class of materials, per space group (see Appendix[H] for details). For the topological gapped materials (SEBR and NLC) in each space group in Table[XIV], we group together all the materials whose representations at high symmetry points differ from each other by adding or subtracting double-group EBR (atomic limit representations) ***without*** time-reversal. If compounds only differed by EBRs, they belong to the same strong-topological class ("cl"), ***without*** time-reversal. We list the symmetry group, topological class, TQC number, ICSD number, Stoichiometric Formula and minimal topological symmetry subgroup (this having been obtained in column 7 of Table[XIV]). A "0" in the last column means that none of its subgroups is topological. See the list of tables.

| SG | cl | TQC | ICSD | Stoich. Formula | SubSG |
|---|---|---|---|---|---|
| 2 | 1 | 2.2.2 | 41017 | P3Tc2 | 2 |
| | | 2.2.3 | 66584 | O9V5 | 2 |
| | | 2.2.10 | 281617 | Na1O6Si2Ti1 | 2 |
| | | 2.2.11 | 281616 | Na1O6Si2Ti1 | 2 |
| | | 2.2.12 | 281615 | Na1O6Si2Ti1 | 2 |
| | | 2.2.13 | 281614 | Na1O6Si2Ti1 | 2 |
| | | 2.2.14 | 281613 | Na1O6Si2Ti1 | 2 |
| | | 2.2.15 | 65941 | Mo6S8Sr1 | 2 |
| | | 2.2.16 | 237587 | Mo3S4 | 2 |
| | | 2.2.17 | 61050 | Mn3Na2O7 | 2 |
| | | 2.2.18 | 59328 | Mn1O7Te1V2 | 2 |
| | | 2.2.20 | 419880 | K1O6Os2 | 2 |
| | | 2.2.24 | 62742 | H1Cu2K3O14P4 | 2 |
| | | 2.2.26 | 60937 | H1As4Cu3Na5O16 | 2 |
| | | 2.2.27 | 8063 | H1As1O5Zn2 | 2 |
| | | 2.2.28 | 37436 | H1As1O5Zn2 | 2 |
| | | 2.2.29 | 34341 | H1As1O5Zn2 | 2 |
| | | 2.2.30 | 8065 | H1As1Ba1Cr2O10 | 2 |
| | | 2.2.33 | 35417 | H12Cl8Hg3Mg1O6 | 2 |
| | | 2.2.34 | 194210 | H12Cl4N4Pt2 | 2 |
| | | 2.2.36 | 79719 | Fe2O14P4Sr1 | 2 |
| | | 2.2.38 | 171431 | Cu4Mo6Se8 | 2 |
| | | 2.2.40 | 84558 | Cu1O4W1 | 2 |
| | | 2.2.41 | 69759 | Cu1O4W1 | 2 |
| | | 2.2.43 | 24339 | Cu1O4W1 | 2 |
| | | 2.2.44 | 182753 | Cu1O4W1 | 2 |
| | | 2.2.45 | 182752 | Cu1O4W1 | 2 |
| | | 2.2.46 | 182750 | Cu1O4W1 | 2 |
| | | 2.2.47 | 16009 | Cu1O4W1 | 2 |
| | | 2.2.49 | 62367 | Cu1I3Na1O9 | 2 |
| | | 2.2.50 | 75006 | Cu1Gd1O8W2 | 2 |
| | | 2.2.51 | 64660 | Cu1F6Pt1 | 2 |
| | | 2.2.53 | 200533 | Cu1F1Nb1O3 | 2 |
| | | 2.2.56 | 62676 | Cr1Na4O4 | 2 |
| | | 2.2.57 | 73191 | Co2O4Rb5 | 2 |
| | | 2.2.58 | 73190 | Co2O4Rb5 | 2 |
| | | 2.2.61 | 410101 | Cl7Mo1O1Se1 | 2 |
| | | 2.2.63 | 431558 | Cl3Cu3O3Te1 | 2 |
| | | 2.2.64 | 92165 | Cl2Cu3O6Se2 | 2 |
| | | 2.2.65 | 427665 | Cl1O4Tm1W1 | 2 |
| | | 2.2.66 | 50441 | Cl18Mo2P2 | 2 |
| | | 2.2.67 | 410649 | Cl18Mo2P2 | 2 |
| | | 2.2.68 | 408774 | Cl14Mo4O4Tl2 | 2 |
| | | 2.2.69 | 74479 | Ca1P3 | 2 |
| | | 2.2.70 | 42698 | C4H16Cl2N8Se4Te1 | 2 |
| | | 2.2.79 | 196024 | C1Mn1O3 | 2 |
| | | 2.2.80 | 423462 | Br5W1 | 2 |
| | | 2.2.83 | 97768 | Bi1Ni1O3 | 2 |
| | | 2.2.84 | 85490 | Ba1Mo6S8 | 2 |

Continued on next column

**TABLE XVI– continued**

| SG | cl | TQC | ICSD | Stoich. Formula | SubSG |
|---|---|---|---|---|---|
| | | 2.2.85 | 65775 | Ba1Mo6S8 | 2 |
| | | 2.2.88 | 96562 | B2Co2O5 | 2 |
| | | 2.2.89 | 24284 | B2Co2O5 | 2 |
| | | 2.2.90 | 66662 | As3Tc2 | 2 |
| | | 2.2.91 | 86400 | As3Ca1 | 2 |
| | | 2.2.92 | 193 | As3Ca1 | 2 |
| | | 2.2.96 | 428633 | As1Bi1Co1O5 | 2 |
| | | 2.2.97 | 422342 | Al2Fe3Si3 | 2 |
| | | 2.2.98 | 95832 | Ag2F5 | 2 |
| | | 2.2.99 | 421341 | Ag1O4S1 | 2 |
| | | 2.2.100 | 51507 | Ag1F6Pd1 | 2 |
| | | 2.2.101 | 174407 | Ag1Cu1F3 | 2 |
| 2 | 2 | 2.2.1 | 202564 | S1Ta6 | 0 |
| | | 2.2.4 | 80363 | O9P3Ru1 | 0 |
| | | 2.2.5 | 2212 | O7V4 | 0 |
| | | 2.2.6 | 19012 | O7V4 | 0 |
| | | 2.2.7 | 61430 | O6Pb3S1 | 0 |
| | | 2.2.8 | 82204 | O4Pb1Pt2 | 0 |
| | | 2.2.9 | 59657 | O4Pb1Pt2 | 0 |
| | | 2.2.19 | 431529 | Lu1Na1P2S6 | 0 |
| | | 2.2.21 | 39887 | H2Cu1O5Se1 | 0 |
| | | 2.2.22 | 166101 | H2Cu1O5S1 | 0 |
| | | 2.2.23 | 167608 | H1Fe1Li1O5P1 | 0 |
| | | 2.2.25 | 416823 | H1Co1O5P1Zn1 | 0 |
| | | 2.2.31 | 73197 | H18Al2Cl6N6 | 0 |
| | | 2.2.32 | 417328 | H12Cs4O6P2S6 | 0 |
| | | 2.2.35 | 240179 | Ga5Rh3 | 0 |
| | | 2.2.37 | 72390 | Fe1N2Sr2 | 0 |
| | | 2.2.39 | 190444 | Cu3Mo2O9 | 0 |
| | | 2.2.42 | 4189 | Cu1O4W1 | 0 |
| | | 2.2.48 | 39439 | Cu1Mo1O4 | 0 |
| | | 2.2.52 | 69655 | Cu1F3Na1 | 0 |
| | | 2.2.54 | 62000 | Cs1Hg1 | 0 |
| | | 2.2.55 | 150974 | Cs1Hg1 | 0 |
| | | 2.2.59 | 167202 | Co1F1Li1O4S1 | 0 |
| | | 2.2.60 | 410945 | Cl9Te1W1 | 0 |
| | | 2.2.62 | 84620 | Cl5Mo1 | 0 |
| | | 2.2.71 | 170721 | C3H6As1N1 | 0 |
| | | 2.2.72 | 62435 | C2H8Cl2N4Se2Te1 | 0 |
| | | 2.2.73 | 62436 | C2H8Br2N4Se2Te1 | 0 |
| | | 2.2.74 | 186554 | C2H3Na1O5 | 0 |
| | | 2.2.75 | 196028 | C1Mn1O3 | 0 |
| | | 2.2.76 | 196027 | C1Mn1O3 | 0 |
| | | 2.2.77 | 196026 | C1Mn1O3 | 0 |
| | | 2.2.78 | 196025 | C1Mn1O3 | 0 |
| | | 2.2.81 | 39784 | Br3Ti1 | 0 |
| | | 2.2.82 | 51779 | Bi3O7Sb1 | 0 |
| | | 2.2.86 | 237763 | Ba1Fe2O8P2 | 0 |
| | | 2.2.87 | 192060 | Ba1Cu1O7P2 | 0 |
| | | 2.2.93 | 404850 | As2Cu4O9 | 0 |
| | | 2.2.94 | 31787 | As1F6N2S3 | 0 |
| | | 2.2.95 | 92916 | As1Bi1Ni1O5 | 0 |
| 10 | 1 | 10.2.1 | 417035 | Co2Gd4Mg3 | 2 |
| | | 11.2.1 | 42076 | Te3Zr1 | 2 |
| | | 11.2.2 | 424108 | Sn2Sr1Zn2 | 2 |
| | | 11.2.3 | 42982 | Se3Ta2 | 2 |
| | | 11.2.9 | 52066 | Rh4Si5 | 2 |
| | | 11.2.10 | 24355 | Rh4Si5 | 2 |
| | | 11.2.14 | 99271 | Ni3Te2 | 2 |
| | | 11.2.16 | 245081 | Ni1Ti1 | 2 |
| | | 11.2.17 | 240195 | Ni1Ti1 | 2 |

Continued on next column

**TABLE XVI– continued**

| SG | cl | TQC | ICSD | Stoich. Formula | SubSG |
|---|---|---|---|---|---|
| | | 11.2.18 | 166372 | Ni1Ti1 | 2 |
| | | 11.2.19 | 166367 | Ni1Ti1 | 2 |
| | | 11.2.20 | 164155 | Ni1Ti1 | 2 |
| | | 11.2.21 | 164154 | Ni1Ti1 | 2 |
| | | 11.2.22 | 164153 | Ni1Ti1 | 2 |
| | | 11.2.23 | 164152 | Ni1Ti1 | 2 |
| | | 11.2.24 | 164151 | Ni1Ti1 | 2 |
| | | 11.2.25 | 157983 | Ni1Ti1 | 2 |
| | | 11.2.26 | 105415 | Ni1Ti1 | 2 |
| | | 11.2.27 | 105414 | Ni1Ti1 | 2 |
| | | 11.2.29 | 42981 | Nb2Se3 | 2 |
| | | 11.2.32 | 73453 | Mo2S3 | 2 |
| | | 11.2.33 | 62486 | Mo2S3 | 2 |
| | | 11.2.34 | 16261 | Mo2.06S3 | 2 |
| | | 11.2.38 | 26525 | H8Ba1O7Pt1 | 2 |
| | | 11.2.39 | 437 | H6Br3N1O1Sn1 | 2 |
| | | 11.2.41 | 15199 | H5Fe1O7S1 | 2 |
| | | 11.2.43 | 60659 | H4Ba1O3 | 2 |
| | | 11.2.44 | 151476 | H4Ag2K4O2S9Sn3 | 2 |
| | | 11.2.48 | 166218 | H2O4P1Rb1 | 2 |
| | | 11.2.55 | 260832 | H2Cl1O2Tb1 | 2 |
| | | 11.2.57 | 6164 | H1O2Tb1 | 2 |
| | | 11.2.58 | 162390 | H1Nd1O2 | 2 |
| | | 11.2.62 | 156004 | H12Cl8Hg3N2O2 | 2 |
| | | 11.2.64 | 418027 | Ge1Hf1Pd1 | 2 |
| | | 11.2.65 | 180364 | Ga5Pt1 | 2 |
| | | 11.2.67 | 94536 | Eu1K3O8P2 | 2 |
| | | 11.2.70 | 414338 | Cu1Nb1Te2 | 2 |
| | | 11.2.72 | 108896 | Co1O3Sr1 | 2 |
| | | 11.2.73 | 411140 | Cl7Cs1Ti2 | 2 |
| | | 11.2.74 | 415798 | Cl2Co2O3Te1 | 2 |
| | | 11.2.75 | 65346 | Cl1Fe1Li1Mo1O4 | 2 |
| | | 11.2.76 | 54553 | Ce1In2Pt2 | 2 |
| | | 11.2.77 | 862 | Ca1Sb2 | 2 |
| | | 11.2.79 | 110059 | C4H12Cd1Cl3N1 | 2 |
| | | 11.2.81 | 20741 | C2H8N4O7Se1U1 | 2 |
| | | 11.2.83 | 42679 | Bi1 | 2 |
| | | 11.2.87 | 429621 | Ba1Ho4O17Si5 | 2 |
| | | 11.2.88 | 167615 | Ba1Er2O10Si3 | 2 |
| | | 11.2.92 | 410884 | Au2Eu2Sn5 | 2 |
| | | 11.2.93 | 416298 | Au2Ba1Sb2 | 2 |
| | | 11.2.94 | 902 | As1Cl1Co2O4 | 2 |
| | | 11.2.96 | 104626 | Al1Ce3 | 2 |
| | | 11.2.97 | 57494 | Al1Au1 | 2 |
| | | 11.2.98 | 104649 | Al1Au1 | 2 |
| 11 | 2 | 11.2.4 | 43213 | Se3Ta1 | 0 |
| | | 11.2.5 | 52307 | Sb2Sr1 | 0 |
| | | 11.2.6 | 171417 | S3U1 | 0 |
| | | 11.2.7 | 430 | S3Tm2 | 0 |
| | | 11.2.8 | 260980 | S3Tm2 | 0 |
| | | 11.2.11 | 31692 | Pu1 | 0 |
| | | 11.2.12 | 43283 | Pt6Si5 | 0 |
| | | 11.2.13 | 105804 | Pt3Ta1 | 0 |
| | | 11.2.15 | 54304 | Ni3Ta1 | 0 |
| | | 11.2.28 | 66316 | Nd1Ru1Si2 | 0 |
| | | 11.2.30 | 54283 | Nb1Pt3 | 0 |
| | | 11.2.31 | 93368 | Na1O8P1W2 | 0 |
| | | 11.2.35 | 14349 | Mo1Te2 | 0 |
| | | 11.2.36 | 59708 | Li4S4Si1 | 0 |
| | | 11.2.37 | 68366 | In2S7Sn3 | 0 |
| | | 11.2.40 | 24075 | H5In1O7S1 | 0 |

Continued on next column

**TABLE XVI– continued**

| SG | cl | TQC | ICSD | Stoich. Formula | SubSG |
|---|---|---|---|---|---|
| | | 11.2.42 | 262728 | H4Co2O10P2Pb1 | 0 |
| | | 11.2.45 | 62901 | H3Dy3Ge1O8 | 0 |
| | | 11.2.46 | 70101 | H3Cl1Cu2O3 | 0 |
| | | 11.2.47 | 153707 | H3Cl1Cu2O3 | 0 |
| | | 11.2.49 | 27406 | H2N1Rb1 | 0 |
| | | 11.2.50 | 169093 | H2K1O4P1 | 0 |
| | | 11.2.51 | 6163 | H2Ga11Na1O18 | 0 |
| | | 11.2.52 | 36300 | H2Cs1O4P1 | 0 |
| | | 11.2.53 | 20449 | H2Cl1O2Y1 | 0 |
| | | 11.2.54 | 260836 | H2Cl1O2Tm1 | 0 |
| | | 11.2.56 | 260834 | H2Cl1Ho1O2 | 0 |
| | | 11.2.59 | 424942 | H1F4Mn1Na2O3P1 | 0 |
| | | 11.2.60 | 201682 | H14Cl2N5O2Tc1 | 0 |
| | | 11.2.61 | 54147 | H14Cd1Cu4O18S2 | 0 |
| | | 11.2.63 | 63219 | H10Mo3N2O11 | 0 |
| | | 11.2.66 | 89343 | F3K1Mn1 | 0 |
| | | 11.2.68 | 190487 | D2O4P1Rb1 | 0 |
| | | 11.2.69 | 67816 | Cu4K1Mg1O13V3 | 0 |
| | | 11.2.71 | 69990 | Cu1La2O6Sn1 | 0 |
| | | 11.2.78 | 62428 | C4H2As1F17N2 | 0 |
| | | 11.2.80 | 110260 | C3H10Cl1N1 | 0 |
| | | 11.2.82 | 416300 | Bi2Pd2Sr1 | 0 |
| | | 11.2.84 | 81819 | Ba3Ge2N2 | 0 |
| | | 11.2.85 | 409517 | Ba1Sb2 | 0 |
| | | 11.2.86 | 194985 | Ba1Nd2O10Ti3 | 0 |
| | | 11.2.89 | 416299 | Ba1Bi2Pd2 | 0 |
| | | 11.2.90 | 413519 | Ba1.5Sb3 | 0 |
| | | 11.2.91 | 261042 | Au2In4Yb1 | 0 |
| | | 11.2.95 | 43456 | Am1 | 0 |
| 12 | 1 | 12.1.1 | 187638 | Y1 | 0 |
| | | 12.1.2 | 43412 | Te4Ti3 | 0 |
| | | 12.1.3 | 41047 | Ta2Te3 | 0 |
| | | 12.1.4 | 37122 | Se8Tl1V5 | 0 |
| | | 12.1.5 | 601347 | Se4Ti3 | 0 |
| | | 12.1.6 | 48176 | Rh1Si2Y2 | 0 |
| | | 12.1.7 | 42077 | P2V1 | 0 |
| | | 12.1.8 | 56694 | O5Ti5 | 0 |
| | | 12.1.9 | 34417 | O2V1 | 0 |
| | | 12.1.10 | 34416 | O2V1 | 0 |
| | | 12.1.11 | 50409 | O13V6 | 0 |
| | | 12.1.12 | 15028 | O13V6 | 0 |
| | | 12.1.13 | 646547 | Ni1Se4Ti2 | 0 |
| | | 12.1.14 | 646544 | Ni1Se4Ti2 | 0 |
| | | 12.1.15 | 646533 | Ni1Se4Si2 | 0 |
| | | 12.1.16 | 48004 | Ni1Sc3Si3 | 0 |
| | | 12.1.17 | 646389 | Ni1S8Ti4 | 0 |
| | | 12.1.18 | 646388 | Ni1S4Ti2 | 0 |
| | | 12.1.19 | 646321 | Ni1Rh2Se4 | 0 |
| | | 12.1.20 | 646318 | Ni1Rh2Se4 | 0 |
| | | 12.1.21 | 15214 | Ni1Rh2Se4 | 0 |
| | | 12.1.22 | 41238 | Nb7P4 | 0 |
| | | 12.1.23 | 26183 | Nb2Se1 | 0 |
| | | 12.1.24 | 77752 | Nb1Sb2 | 0 |
| | | 12.1.25 | 645350 | Nb1Sb2 | 0 |
| | | 12.1.26 | 57430 | Nb1Sb2 | 0 |
| | | 12.1.27 | 190180 | Nb1Sb2 | 0 |
| | | 12.1.28 | 18144 | Nb1Sb2 | 0 |
| | | 12.1.29 | 645169 | Nb1P2 | 0 |
| | | 12.1.30 | 29383 | N3Na1 | 0 |
| | | 12.1.31 | 29382 | N3Na1 | 0 |
| | | 12.1.32 | 29381 | N3Na1 | 0 |

Continued on next column

TABLE XVI– continued

| SG | cl | TQC | ICSD | Stoich. Formula | SubSG |
|---|---|---|---|---|---|
| | | 12.1.33 | 29380 | N3Na1 | 0 |
| | | 12.1.34 | 29379 | N3Na1 | 0 |
| | | 12.1.35 | 29378 | N3Na1 | 0 |
| | | 12.1.36 | 29377 | N3Na1 | 0 |
| | | 12.1.37 | 29376 | N3Na1 | 0 |
| | | 12.1.38 | 29375 | N3Na1 | 0 |
| | | 12.1.39 | 29374 | N3Na1 | 0 |
| | | 12.1.40 | 29373 | N3Na1 | 0 |
| | | 12.1.41 | 29372 | N3Na1 | 0 |
| | | 12.1.42 | 29371 | N3Na1 | 0 |
| | | 12.1.43 | 29370 | N3Na1 | 0 |
| | | 12.1.44 | 151588 | Mn7Na1O12 | 0 |
| | | 12.1.45 | 78328 | Mn1Mo1O4 | 0 |
| | | 12.1.46 | 239971 | Li2Mn1O3 | 0 |
| | | 12.1.47 | 165326 | Li1Mo1O2 | 0 |
| | | 12.1.48 | 49021 | K1Li1Mn1O2 | 0 |
| | | 12.1.49 | 240135 | In4Sr1 | 0 |
| | | 12.1.50 | 240133 | In4Sr1 | 0 |
| | | 12.1.51 | 407248 | I3La5Si5 | 0 |
| | | 12.1.52 | 407247 | I3La4Si4 | 0 |
| | | 12.1.53 | 83965 | Hf2Te1 | 0 |
| | | 12.1.54 | 20755 | H4Cl2Ni1O2 | 0 |
| | | 12.1.55 | 23361 | H3N1Na2S3 | 0 |
| | | 12.1.56 | 62646 | H3Cu2F2K1O8P2 | 0 |
| | | 12.1.57 | 422475 | H3Co2Mo2Na1O10 | 0 |
| | | 12.1.58 | 24759 | H2I2O12Sb8 | 0 |
| | | 12.1.59 | 24760 | H2Br2O12Sb8 | 0 |
| | | 12.1.60 | 16023 | H12N2O16Zn5 | 0 |
| | | 12.1.61 | 262220 | Ge6La4Mg7 | 0 |
| | | 12.1.62 | 637671 | Ge2Re1 | 0 |
| | | 12.1.63 | 183378 | Gd1O6Ru1Sr2 | 0 |
| | | 12.1.64 | 430563 | Fe5O7 | 0 |
| | | 12.1.65 | 633477 | Fe3Se4 | 0 |
| | | 12.1.66 | 24560 | Fe3Se4 | 0 |
| | | 12.1.67 | 150568 | Fe3Se4 | 0 |
| | | 12.1.68 | 240050 | Fe1Zn13 | 0 |
| | | 12.1.69 | 163222 | Fe1Zn13 | 0 |
| | | 12.1.70 | 633326 | Fe1S8Ti4 | 0 |
| | | 12.1.71 | 53534 | Fe1S8Ti4 | 0 |
| | | 12.1.72 | 42563 | Fe1S4Ti2 | 0 |
| | | 12.1.73 | 103658 | Fe17Th2 | 0 |
| | | 12.1.74 | 633332 | Fe0.5S2Ti1 | 0 |
| | | 12.1.75 | 192506 | F5Fe1K2 | 0 |
| | | 12.1.76 | 106627 | Eu1In4 | 0 |
| | | 12.1.77 | 151123 | Er2O7Si2 | 0 |
| | | 12.1.78 | 419435 | Er2O3 | 0 |
| | | 12.1.79 | 180224 | Dy1Si1 | 0 |
| | | 12.1.80 | 180223 | Dy1Si1 | 0 |
| | | 12.1.81 | 180222 | Dy1Si1 | 0 |
| | | 12.1.82 | 180221 | Dy1Si1 | 0 |
| | | 12.1.83 | 418227 | Dy1P2Pt8 | 0 |
| | | 12.1.84 | 421466 | Cu4Na1O12Te2Tl3 | 0 |
| | | 12.1.85 | 71624 | Cs2Ir1Na2O4 | 0 |
| | | 12.1.86 | 26977 | Cr7Se8 | 0 |
| | | 12.1.87 | 626888 | Cr3Te4 | 0 |
| | | 12.1.88 | 626884 | Cr3Te4 | 0 |
| | | 12.1.89 | 626874 | Cr3Te4 | 0 |
| | | 12.1.90 | 626716 | Cr3Se4 | 0 |
| | | 12.1.91 | 626712 | Cr3Se4 | 0 |
| | | 12.1.92 | 26976 | Cr3Se4 | 0 |
| | | 12.1.93 | 81886 | Cr3S4 | 0 |

Continued on next column

TABLE XVI– continued

| SG | cl | TQC | ICSD | Stoich. Formula | SubSG |
|---|---|---|---|---|---|
| | | 12.1.94 | 626598 | Cr3S4 | 0 |
| | | 12.1.95 | 192944 | Cr3S4 | 0 |
| | | 12.1.96 | 16722 | Cr3S4 | 0 |
| | | 12.1.97 | 74296 | Cr2S8Tl1V3 | 0 |
| | | 12.1.98 | 626640 | Cr2S4Ti1 | 0 |
| | | 12.1.99 | 625968 | Cr2Fe1Se4 | 0 |
| | | 12.1.100 | 155121 | Cr2Fe1Se4 | 0 |
| | | 12.1.101 | 625931 | Cr2Fe1S4 | 0 |
| | | 12.1.102 | 251718 | Cr1Se2 | 0 |
| | | 12.1.103 | 75420 | Cr1S2 | 0 |
| | | 12.1.104 | 626579 | Cr1Rh2Te4 | 0 |
| | | 12.1.105 | 626577 | Cr1Rh2Se4 | 0 |
| | | 12.1.106 | 626541 | Cr1Pr2Se4 | 0 |
| | | 12.1.107 | 821 | Cr1P1S4 | 0 |
| | | 12.1.108 | 56835 | Cr1P1S4 | 0 |
| | | 12.1.109 | 173875 | Cr1O4Re1 | 0 |
| | | 12.1.110 | 624485 | Co2Ni1Se4 | 0 |
| | | 12.1.111 | 66006 | Co2Na7O6Rb1 | 0 |
| | | 12.1.112 | 2371 | Co1Si2Zr2 | 0 |
| | | 12.1.113 | 624952 | Co1Sc2Si2 | 0 |
| | | 12.1.114 | 623423 | Co1Ge1 | 0 |
| | | 12.1.115 | 43677 | Co1Ge1 | 0 |
| | | 12.1.116 | 184254 | Co1Ge1 | 0 |
| | | 12.1.117 | 183325 | Co1Gd2Si2 | 0 |
| | | 12.1.118 | 184277 | Co1Gd2Ge2 | 0 |
| | | 12.1.119 | 183326 | Co1Gd2Ge2 | 0 |
| | | 12.1.120 | 427661 | Cl1Gd1O4W1 | 0 |
| | | 12.1.121 | 35292 | Cl1Gd1O4W1 | 0 |
| | | 12.1.122 | 420055 | Cl1Gd1Mo1O4 | 0 |
| | | 12.1.123 | 73096 | Cl1Dy1O3Ti1 | 0 |
| | | 12.1.124 | 424719 | Ce4Ge6Mg7 | 0 |
| | | 12.1.125 | 100827 | Ca2Fe7O11 | 0 |
| | | 12.1.126 | 20762 | C6H5Cl2Tl1 | 0 |
| | | 12.1.127 | 63503 | C5Nb6 | 0 |
| | | 12.1.128 | 39297 | C5Nb6 | 0 |
| | | 12.1.129 | 20695 | C5Nb6 | 0 |
| | | 12.1.130 | 167353 | C5Nb6 | 0 |
| | | 12.1.131 | 619030 | C3U2W2 | 0 |
| | | 12.1.132 | 618707 | C3Re2U2 | 0 |
| | | 12.1.133 | 618357 | C3Mo2U2 | 0 |
| | | 12.1.134 | 618223 | C3Lu2Mo2 | 0 |
| | | 12.1.135 | 88513 | C3Ho2Mo2 | 0 |
| | | 12.1.136 | 88512 | C3Ho2Mo2 | 0 |
| | | 12.1.137 | 88511 | C3Ho2Mo2 | 0 |
| | | 12.1.138 | 602482 | C3Cr2Yb2 | 0 |
| | | 12.1.139 | 617537 | C3Cr2Tb2 | 0 |
| | | 12.1.140 | 617514 | C3Cr2Lu2 | 0 |
| | | 12.1.141 | 617492 | C3Cr2Dy2 | 0 |
| | | 12.1.142 | 43018 | C3Cr2Dy2 | 0 |
| | | 12.1.143 | 417827 | C3Ce2Mo2 | 0 |
| | | 12.1.144 | 618692 | C3.5Pu2W2 | 0 |
| | | 12.1.145 | 262897 | C1H3La1O6P2 | 0 |
| | | 12.1.146 | 79394 | C1Fe2Si2Tm2 | 0 |
| | | 12.1.147 | 195010 | C1Fe2Gd2Si2 | 0 |
| | | 12.1.148 | 67467 | C1Dy2Fe2Si2 | 0 |
| | | 12.1.149 | 391256 | C1Cl2Eu2N2 | 0 |
| | | 12.1.150 | 603998 | C1.66Mo1U1 | 0 |
| | | 12.1.151 | 416907 | C10Cl4La8 | 0 |
| | | 12.1.152 | 617015 | Bi2Rh3Si2 | 0 |
| | | 12.1.153 | 420725 | Bi2Rh3S2 | 0 |
| | | 12.1.154 | 261457 | Bi2Pd3Se2 | 0 |

Continued on next column

**TABLE XVI– continued**

| SG | cl | TQC | ICSD | Stoich. Formula | SubSG |
|---|---|---|---|---|---|
| | | 12.1.155 | 251720 | Bi1Se1 | 0 |
| | | 12.1.156 | 251717 | Bi1Se1 | 0 |
| | | 12.1.157 | 1559 | Bi1I1 | 0 |
| | | 12.1.158 | 1560 | Bi1Br1 | 0 |
| | | 12.1.159 | 65614 | Ba3Cu2O10Pt1Y2 | 0 |
| | | 12.1.160 | 69569 | Ba3Cu2Er2O10Pt1 | 0 |
| | | 12.1.161 | 173685 | Ba3Cd2Sb4 | 0 |
| | | 12.1.162 | 423268 | Ba2Ge4Ni5 | 0 |
| | | 12.1.163 | 238228 | Ba1Te4U1 | 0 |
| | | 12.1.164 | 195017 | Ba1Pt2Sb2 | 0 |
| | | 12.1.165 | 154011 | Ba1Mn2O6Y1 | 0 |
| | | 12.1.166 | 173848 | Ba1Ir1O3 | 0 |
| | | 12.1.167 | 151699 | Ba1Er2F8 | 0 |
| | | 12.1.168 | 61499 | Ba1Bi1O3 | 0 |
| | | 12.1.169 | 63510 | B4Nd2Ni5 | 0 |
| | | 12.1.170 | 170618 | B4La2Ni5 | 0 |
| | | 12.1.171 | 405291 | B4Br5La4 | 0 |
| | | 12.1.172 | 615340 | B2Rh3Y1 | 0 |
| | | 12.1.173 | 261932 | B2Ni3Zn1 | 0 |
| | | 12.1.174 | 107555 | Au8Ga1K4 | 0 |
| | | 12.1.175 | 425401 | Au1Ce2O1P2 | 0 |
| | | 12.1.176 | 424760 | As4Ba3Zn2 | 0 |
| | | 12.1.177 | 424761 | As4Ba3Cd2 | 0 |
| | | 12.1.178 | 43185 | As3W2 | 0 |
| | | 12.1.179 | 15049 | As3V4 | 0 |
| | | 12.1.180 | 43184 | As3Mo2 | 0 |
| | | 12.1.181 | 41015 | As3Mo2 | 0 |
| | | 12.1.182 | 610428 | As3Eu1 | 0 |
| | | 12.1.183 | 611574 | As2V1 | 0 |
| | | 12.1.184 | 44069 | As2Ta1 | 0 |
| | | 12.1.185 | 69002 | As2Ni2O7 | 0 |
| | | 12.1.186 | 81218 | As2Nb1 | 0 |
| | | 12.1.187 | 69003 | As2Mn2O7 | 0 |
| | | 12.1.188 | 261143 | As2Mn2O7 | 0 |
| | | 12.1.189 | 79441 | As2Gd1Pd3 | 0 |
| | | 12.1.190 | 81775 | As2Ce1Pd3 | 0 |
| | | 12.1.191 | 75454 | As2Ba2Mn2O1 | 0 |
| | | 12.1.192 | 68906 | As1F6N1O2 | 0 |
| | | 12.1.193 | 418388 | Al6O1Si8Sr13 | 0 |
| | | 12.1.194 | 409319 | Al4Na4P12Sr8 | 0 |
| | | 12.1.195 | 27569 | Al4Cl51Zr12 | 0 |
| | | 12.1.196 | 40332 | Al1Cu1 | 0 |
| | | 12.1.197 | 165239 | Al14Ca13 | 0 |
| | | 12.1.198 | 429000 | Ag1Ba1S3Tb1 | 0 |
| 12 | 2 | 12.2.1 | 187496 | Sb2Te3 | 2 |
| | | 12.2.2 | 420133 | Ru1Sc2Si2 | 2 |
| | | 12.2.3 | 61353 | Pt1Se7Ta2 | 2 |
| | | 12.2.4 | 248500 | Pr1 | 2 |
| | | 12.2.5 | 61005 | Pd1Se6Ta2 | 2 |
| | | 12.2.6 | 40281 | P4Zr7 | 2 |
| | | 12.2.7 | 41016 | P3Tc2 | 2 |
| | | 12.2.8 | 74780 | O7Si2Yb2 | 2 |
| | | 12.2.9 | 65363 | O7Si2Yb2 | 2 |
| | | 12.2.10 | 62356 | O7Si2Yb2 | 2 |
| | | 12.2.11 | 248719 | O7Si2Tm2 | 2 |
| | | 12.2.12 | 75194 | O5Ti3 | 2 |
| | | 12.2.13 | 75193 | O5Ti3 | 2 |
| | | 12.2.14 | 38223 | O3Rb2Te1 | 2 |
| | | 12.2.15 | 16779 | O26V12 | 2 |
| | | 12.2.16 | 10455 | O13Sr2Ti6 | 2 |
| | | 12.2.17 | 646757 | Ni3Sn4 | 2 |

Continued on next column

**TABLE XVI– continued**

| SG | cl | TQC | ICSD | Stoich. Formula | SubSG |
|---|---|---|---|---|---|
| | | 12.2.18 | 105364 | Ni3Sn4 | 2 |
| | | 12.2.19 | 105363 | Ni3Sn4 | 2 |
| | | 12.2.20 | 30433 | Ni2O7P2 | 2 |
| | | 12.2.21 | 646556 | Ni1Se4V2 | 2 |
| | | 12.2.22 | 646555 | Ni1Se4V2 | 2 |
| | | 12.2.23 | 23970 | Ni1Se4V2 | 2 |
| | | 12.2.24 | 35141 | Ni1S8V4 | 2 |
| | | 12.2.25 | 646403 | Ni1S4V2 | 2 |
| | | 12.2.26 | 35140 | Ni1S4V2 | 2 |
| | | 12.2.27 | 646394 | Ni1S4Ti2 | 2 |
| | | 12.2.28 | 193869 | Ni0.5Se2Ti1 | 2 |
| | | 12.2.29 | 73315 | Nb8Pt1Se20 | 2 |
| | | 12.2.30 | 280016 | Nb2Te3 | 2 |
| | | 12.2.31 | 192215 | Nb2Pd1S5 | 2 |
| | | 12.2.32 | 81059 | Mo1Ni1O4 | 2 |
| | | 12.2.33 | 20296 | Mn2O7P2 | 2 |
| | | 12.2.34 | 15615 | Mn1Mo1O4 | 2 |
| | | 12.2.35 | 50001 | Mg2Si10Sr11 | 2 |
| | | 12.2.36 | 180011 | Li0.5O2Ti1 | 2 |
| | | 12.2.37 | 246807 | La1Si5 | 2 |
| | | 12.2.38 | 428765 | Hg2Sn2Sr1 | 2 |
| | | 12.2.39 | 157861 | Hg1 | 2 |
| | | 12.2.40 | 405208 | Hf7P4 | 2 |
| | | 12.2.41 | 162649 | H8Cu2Mg1O12V2 | 2 |
| | | 12.2.42 | 33293 | H6Au1Br3N2 | 2 |
| | | 12.2.43 | 2559 | H4F6Li2O2Sn1 | 2 |
| | | 12.2.44 | 201563 | H4Cl8N1O2Rb3Ru2 | 2 |
| | | 12.2.45 | 16130 | H4Cl8K3N1O2Ru2 | 2 |
| | | 12.2.46 | 200012 | H4Al1Cr2K1O10 | 2 |
| | | 12.2.47 | 35156 | H3Li1O2 | 2 |
| | | 12.2.48 | 30223 | H3Li1O2 | 2 |
| | | 12.2.49 | 27761 | H3Li1O2 | 2 |
| | | 12.2.50 | 159510 | H3Li1O2 | 2 |
| | | 12.2.51 | 24430 | H3Al2O7P1 | 2 |
| | | 12.2.52 | 421319 | H2Bi1Cl3O1 | 2 |
| | | 12.2.53 | 26919 | H12Cl2Ni1O6 | 2 |
| | | 12.2.54 | 49911 | H12Cl2N2O6Pt1 | 2 |
| | | 12.2.55 | 200906 | H10F6Mn1O5Zr1 | 2 |
| | | 12.2.56 | 76350 | Ge2Ho2Os1 | 2 |
| | | 12.2.57 | 183379 | Gd1O6Ru1Sr2 | 2 |
| | | 12.2.58 | 635673 | Ga4Yb1 | 2 |
| | | 12.2.59 | 165189 | Ga3Sc5 | 2 |
| | | 12.2.60 | 15044 | Fe3Se4 | 2 |
| | | 12.2.61 | 15043 | Fe3Se4 | 2 |
| | | 12.2.62 | 152533 | Fe1Si3Tm3 | 2 |
| | | 12.2.63 | 84205 | Fe1Sc2Si2 | 2 |
| | | 12.2.64 | 76348 | Fe1Sc2Si2 | 2 |
| | | 12.2.65 | 633455 | Fe1Sc2Si2 | 2 |
| | | 12.2.66 | 20661 | Fe1Sc2Si2 | 2 |
| | | 12.2.67 | 633361 | Fe1S4V2 | 2 |
| | | 12.2.68 | 603354 | Fe1Rh2Se4 | 2 |
| | | 12.2.69 | 632972 | Fe1Ni2Se4 | 2 |
| | | 12.2.70 | 90460 | Fe1Ge1Yb1 | 2 |
| | | 12.2.71 | 631983 | Fe1Ge1 | 2 |
| | | 12.2.72 | 53458 | Fe1Ge1 | 2 |
| | | 12.2.73 | 53457 | Fe1Ge1 | 2 |
| | | 12.2.74 | 53456 | Fe1Ge1 | 2 |
| | | 12.2.75 | 53455 | Fe1Ge1 | 2 |
| | | 12.2.76 | 43055 | Fe1Ge1 | 2 |
| | | 12.2.77 | 416629 | Eu2Li1Si3 | 2 |
| | | 12.2.78 | 425141 | Dy3O3Sb1 | 2 |

Continued on next column

**TABLE XVI– continued**

| SG | cl | TQC | ICSD | Stoich. Formula | SubSG |
|---|---|---|---|---|---|
| | | 12.2.79 | 422453 | Cu2Sn2Sr1 | 2 |
| | | 12.2.80 | 182102 | Cu2Sn2Sr1 | 2 |
| | | 12.2.81 | 79536 | Cu2Na2S4Zr1 | 2 |
| | | 12.2.82 | 426133 | Cu1Ge6Yb2 | 2 |
| | | 12.2.83 | 380416 | Cs4Ge2Se6 | 2 |
| | | 12.2.84 | 626873 | Cr3Te4 | 2 |
| | | 12.2.85 | 601348 | Cr3Se4 | 2 |
| | | 12.2.86 | 73143 | Cr3Rb1S5 | 2 |
| | | 12.2.87 | 1824 | Cr3Ga4 | 2 |
| | | 12.2.88 | 626896 | Cr2Te4Ti1 | 2 |
| | | 12.2.89 | 16884 | Cr2Ni1S4 | 2 |
| | | 12.2.90 | 625971 | Cr2Fe1Se4 | 2 |
| | | 12.2.91 | 626655 | Cr1S4V2 | 2 |
| | | 12.2.92 | 626543 | Cr1Pr2Se4 | 2 |
| | | 12.2.93 | 626536 | Cr1Pr2S4 | 2 |
| | | 12.2.94 | 16087 | Cr1O5Pb2 | 2 |
| | | 12.2.95 | 79865 | Cr1O4V1 | 2 |
| | | 12.2.96 | 202377 | Cr1Nd1O1S2 | 2 |
| | | 12.2.97 | 151873 | Cr1Fe2Se4 | 2 |
| | | 12.2.98 | 74542 | Co2O7P2 | 2 |
| | | 12.2.99 | 41026 | Co2Nb4Pd1Se12 | 2 |
| | | 12.2.100 | 408031 | Cl2La3Si3 | 2 |
| | | 12.2.101 | 420061 | Cl1Mo1O4Yb1 | 2 |
| | | 12.2.102 | 420060 | Cl1Mo1O4Tm1 | 2 |
| | | 12.2.103 | 420059 | Cl1Mo1O4Tb1 | 2 |
| | | 12.2.104 | 427663 | Cl1Ho1O4W1 | 2 |
| | | 12.2.105 | 427664 | Cl1Er1O4W1 | 2 |
| | | 12.2.106 | 420053 | Cl1Er1Mo1O4 | 2 |
| | | 12.2.107 | 424718 | Ce5Ge8Mg8 | 2 |
| | | 12.2.108 | 417665 | Ce1Ru1Sn1 | 2 |
| | | 12.2.109 | 192243 | Ce1Ru1Sn1 | 2 |
| | | 12.2.110 | 192242 | Ce1Ru1Sn1 | 2 |
| | | 12.2.111 | 189952 | Ce1Ru1Sn1 | 2 |
| | | 12.2.112 | 189412 | Ce1Ru1Sn1 | 2 |
| | | 12.2.113 | 189411 | Ce1Ru1Sn1 | 2 |
| | | 12.2.114 | 2284 | Ce1 | 2 |
| | | 12.2.115 | 620729 | Ce1Co1Ga1 | 2 |
| | | 12.2.116 | 620726 | Ce1Co1Ga1 | 2 |
| | | 12.2.117 | 429884 | Ce1Co1Ga1 | 2 |
| | | 12.2.118 | 187960 | Ce1Co1Ga1 | 2 |
| | | 12.2.119 | 171243 | Ca6Cu2Sn7 | 2 |
| | | 12.2.120 | 85505 | Ca1K2N4O8 | 2 |
| | | 12.2.121 | 619274 | Ca1Ga4 | 2 |
| | | 12.2.122 | 58893 | Ca1Ga4 | 2 |
| | | 12.2.123 | 54994 | Ca1Ga4 | 2 |
| | | 12.2.124 | 417717 | Ca1Cu2Sn2 | 2 |
| | | 12.2.125 | 236391 | C4Co1Sc3 | 2 |
| | | 12.2.126 | 618350 | C3Mo2Tm2 | 2 |
| | | 12.2.127 | 88516 | C3Er2Mo2 | 2 |
| | | 12.2.128 | 617669 | C3Er2Mo2 | 2 |
| | | 12.2.129 | 617541 | C3Cr2Tm2 | 2 |
| | | 12.2.130 | 617494 | C3Cr2Er2 | 2 |
| | | 12.2.131 | 252737 | C2Ca1 | 2 |
| | | 12.2.132 | 252734 | C2Ca1 | 2 |
| | | 12.2.133 | 252716 | C2Ca1 | 2 |
| | | 12.2.134 | 617825 | C1Fe2Si2Y2 | 2 |
| | | 12.2.135 | 617663 | C1Er2Fe2Si2 | 2 |
| | | 12.2.136 | 61141 | C10Fe22Mo12 | 2 |
| | | 12.2.137 | 65114 | C10Au2Cl8S8 | 2 |
| | | 12.2.138 | 240479 | Br4Cs2I2Pd1 | 2 |
| | | 12.2.139 | 172782 | Br1Mn1S2Sb1 | 2 |

Continued on next column

**TABLE XVI– continued**

| SG | cl | TQC | ICSD | Stoich. Formula | SubSG |
|---|---|---|---|---|---|
| | | 12.2.140 | 427411 | Bi4Pb7Se13 | 2 |
| | | 12.2.141 | 58840 | Bi2Pd5 | 2 |
| | | 12.2.142 | 616957 | Bi2Pd3Se2 | 2 |
| | | 12.2.143 | 421873 | Bi2Pd3Se2 | 2 |
| | | 12.2.144 | 261459 | Bi2Pd3Se2 | 2 |
| | | 12.2.145 | 261458 | Bi2Pd3Se2 | 2 |
| | | 12.2.146 | 261456 | Bi2Pd3Se2 | 2 |
| | | 12.2.147 | 616954 | Bi2Pd3S2 | 2 |
| | | 12.2.148 | 616946 | Bi2Pd1 | 2 |
| | | 12.2.149 | 42565 | Bi2Pd1 | 2 |
| | | 12.2.150 | 1024 | Bi2Ni3Se2 | 2 |
| | | 12.2.151 | 195678 | Bi2Hg1S4 | 2 |
| | | 12.2.152 | 415138 | Bi1I1Mn1Se2 | 2 |
| | | 12.2.153 | 429459 | Bi1Gd3O3 | 2 |
| | | 12.2.154 | 164224 | Bi1Fe1O3 | 2 |
| | | 12.2.155 | 164223 | Bi1Fe1O3 | 2 |
| | | 12.2.156 | 246145 | Bi16I4 | 2 |
| | | 12.2.157 | 423458 | Ba5Cd2O2Sb4 | 2 |
| | | 12.2.158 | 240016 | Ba3Li4Sn8 | 2 |
| | | 12.2.159 | 62390 | Ba3Cu2O10Pt1Y2 | 2 |
| | | 12.2.160 | 62391 | Ba3Cu2Ho2O10Pt1 | 2 |
| | | 12.2.161 | 408938 | Ba2Mg3Si4 | 2 |
| | | 12.2.162 | 420834 | Ba2Cd2Sb3 | 2 |
| | | 12.2.163 | 80901 | Ba2Bi1O6Tb1 | 2 |
| | | 12.2.164 | 261206 | Ba1Tl4 | 2 |
| | | 12.2.165 | 238227 | Ba1Te4Th1 | 2 |
| | | 12.2.166 | 49000 | Ba1Sb3 | 2 |
| | | 12.2.167 | 16634 | Ba1O5Ti2 | 2 |
| | | 12.2.168 | 83618 | Ba1Cu10P4 | 2 |
| | | 12.2.169 | 418886 | Ba11Cd6Sb12 | 2 |
| | | 12.2.170 | 63501 | B4La2Ni5 | 2 |
| | | 12.2.171 | 170608 | B4La2Ni5 | 2 |
| | | 12.2.172 | 615319 | B2Rh3Tb1 | 2 |
| | | 12.2.173 | 614701 | B2Lu1Rh3 | 2 |
| | | 12.2.174 | 614578 | B2Ir3Y1 | 2 |
| | | 12.2.175 | 614571 | B2Ir3Tm1 | 2 |
| | | 12.2.176 | 614566 | B2Ir3Tb1 | 2 |
| | | 12.2.177 | 614561 | B2Ir3Sc1 | 2 |
| | | 12.2.178 | 614531 | B2Ir3Lu1 | 2 |
| | | 12.2.179 | 614472 | B2Ho1Ir3 | 2 |
| | | 12.2.180 | 8058 | Au2P3 | 2 |
| | | 12.2.181 | 429212 | Au2Ni4Si7 | 2 |
| | | 12.2.182 | 425402 | Au1O1P2Pr2 | 2 |
| | | 12.2.183 | 41831 | As3Sr1 | 2 |
| | | 12.2.184 | 80375 | As3Mn4 | 2 |
| | | 12.2.185 | 86402 | As3Ba1 | 2 |
| | | 12.2.186 | 69001 | As2Co2O7 | 2 |
| | | 12.2.187 | 239291 | As1Pd5 | 2 |
| | | 12.2.188 | 417968 | Al2Ge4Sr3 | 2 |
| | | 12.2.189 | 417659 | Al2Ba5Ge7 | 2 |
| | | 12.2.190 | 20632 | Al1Ce1Co1 | 2 |
| | | 12.2.191 | 416360 | Ag5Cl5Hg4P8 | 2 |
| | | 12.2.192 | 27743 | Ag1Er1O8W2 | 2 |
| 13 | 1 | 13.2.2 | 415434 | O4Ta1Tb1 | 0 |
| | | 13.2.3 | 174299 | O4Sn3 | 0 |
| | | 13.2.7 | 81060 | Mo1Ni1O4 | 0 |
| | | 13.2.9 | 415436 | Ho1O4Ta1 | 0 |
| | | 13.2.10 | 155660 | Ho1O4Ta1 | 0 |
| | | 13.2.13 | 67168 | Cl7Cs1Ti2 | 0 |
| 13 | 2 | 13.2.1 | 415459 | O4Ta1Tm1 | 2 |
| | | 13.2.4 | 16685 | Ni1O4W1 | 2 |

Continued on next column

**TABLE XVI– continued**

| SG | cl | TQC | ICSD | Stoich. Formula | SubSG |
|---|---|---|---|---|---|
| | | 13.2.5 | 15852 | Ni1O4W1 | 2 |
| | | 13.2.6 | 27701 | Mo8O23 | 2 |
| | | 13.2.8 | 170685 | Mn1O4Re1 | 2 |
| | | 13.2.11 | 261836 | Ho1Li1O8W2 | 2 |
| | | 13.2.12 | 36623 | Cu1Ho1O8W2 | 2 |
| 14 | 1 | 14.2.1 | 96154 | Si4Ti3U2 | 2 |
| | | 14.2.4 | 8217 | O2W1 | 2 |
| | | 14.2.5 | 80829 | O2W1 | 2 |
| | | 14.2.7 | 281705 | Ni3O8P2 | 2 |
| | | 14.2.8 | 100194 | Ni2O7P2 | 2 |
| | | 14.2.9 | 246365 | Ni1O10P2V2 | 2 |
| | | 14.2.10 | 80000 | Mo3Si4U2 | 2 |
| | | 14.2.11 | 291320 | Mo1O6Sr2Yb1 | 2 |
| | | 14.2.12 | 80830 | Mo1O2 | 2 |
| | | 14.2.13 | 23722 | Mo1O2 | 2 |
| | | 14.2.14 | 174231 | Ir1P2 | 2 |
| | | 14.2.15 | 171490 | Ir1O6Sr2Tb1 | 2 |
| | | 14.2.16 | 153133 | Ir1O6Sr2Tb1 | 2 |
| | | 14.2.17 | 47158 | H2Li1N1O3 | 2 |
| | | 14.2.18 | 84949 | H1Mn1O2 | 2 |
| | | 14.2.21 | 38384 | H1Cl1Cu1O1 | 2 |
| | | 14.2.22 | 80002 | Ge4Mo3U2 | 2 |
| | | 14.2.24 | 421335 | Gd1S2 | 2 |
| | | 14.2.25 | 62281 | Ga1La1Ni1 | 2 |
| | | 14.2.26 | 189438 | Fe2O3 | 2 |
| | | 14.2.28 | 419955 | F1O3Se1Tm1 | 2 |
| | | 14.2.29 | 412767 | Eu2P2Se6 | 2 |
| | | 14.2.30 | 412763 | Eu2P2S6 | 2 |
| | | 14.2.31 | 416422 | Eu1Ge2Pt2 | 2 |
| | | 14.2.32 | 401556 | Eu1Ge1Pd1 | 2 |
| | | 14.2.33 | 67379 | Er2O1S2 | 2 |
| | | 14.2.36 | 162113 | Dy1Ni1O3 | 2 |
| | | 14.2.37 | 60653 | Cu2O4Se1 | 2 |
| | | 14.2.40 | 202451 | Cu1O3Te1 | 2 |
| | | 14.2.41 | 260194 | Cu1Np1Se2 | 2 |
| | | 14.2.42 | 73398 | Cu1Na1O4P1 | 2 |
| | | 14.2.44 | 59291 | Co2O7P2 | 2 |
| | | 14.2.45 | 73738 | Co1Ta1Te2 | 2 |
| | | 14.2.46 | 63553 | Co1Sb2 | 2 |
| | | 14.2.47 | 73739 | Co1Nb1Te2 | 2 |
| | | 14.2.52 | 27020 | Ce1Se2 | 2 |
| | | 14.2.53 | 419346 | Ce1S2 | 2 |
| | | 14.2.54 | 238150 | Ca2Fe1O6Os1 | 2 |
| | | 14.2.55 | 238148 | Ca2Fe1O6Os1 | 2 |
| | | 14.2.56 | 69790 | Ca1Pd1Si1 | 2 |
| | | 14.2.57 | 86049 | C7Y4 | 2 |
| | | 14.2.58 | 54136 | Br3Cr1In1 | 2 |
| | | 14.2.60 | 75019 | Ba2P3Pd1 | 2 |
| | | 14.2.61 | 50158 | Ba1Rh2Si2 | 2 |
| | | 14.2.62 | 50160 | Ba1Ga2Pt2 | 2 |
| | | 14.2.63 | 88450 | B1Fe2O4 | 2 |
| | | 14.2.64 | 86862 | B1Fe2O4 | 2 |
| | | 14.2.66 | 58406 | Au4Ca5 | 2 |
| | | 14.2.67 | 74948 | As2Ba1Cu2O8 | 2 |
| | | 14.2.69 | 73402 | Ag2Te1 | 2 |
| | | 14.2.70 | 73230 | Ag2Te1 | 2 |
| | | 14.2.71 | 165661 | Ag1Fe2S3 | 2 |
| 14 | 2 | 14.2.2 | 57480 | Re2Si1 | 0 |
| | | 14.2.3 | 9402 | O4Te1V1 | 0 |
| | | 14.2.6 | 67381 | O1S2Yb2 | 0 |
| | | 14.2.19 | 98698 | H1Cu1F1O1 | 0 |

Continued on next column

**TABLE XVI– continued**

| SG | cl | TQC | ICSD | Stoich. Formula | SubSG |
|---|---|---|---|---|---|
| | | 14.2.20 | 91088 | H1Cl1Cu1O1 | 0 |
| | | 14.2.23 | 416421 | Ge2La1Pt2 | 0 |
| | | 14.2.27 | 154372 | Fe2K4O5 | 0 |
| | | 14.2.34 | 69772 | Er1O6Ru1Sr2 | 0 |
| | | 14.2.35 | 69771 | Er1O6Ru1Sr2 | 0 |
| | | 14.2.38 | 1672 | Cu1O4Te1 | 0 |
| | | 14.2.39 | 1671 | Cu1O4Te1 | 0 |
| | | 14.2.43 | 411020 | Cu1Ge1O3 | 0 |
| | | 14.2.48 | 250364 | Cl4V1 | 0 |
| | | 14.2.49 | 422546 | Cl4Hg4I2S1 | 0 |
| | | 14.2.50 | 4411 | Cl3Mo1O1 | 0 |
| | | 14.2.51 | 16150 | Cl3Mo1O1 | 0 |
| | | 14.2.59 | 172071 | Br2Cu2O6Sr2Te1 | 0 |
| | | 14.2.65 | 237469 | Au4Ga2Na1 | 0 |
| | | 14.2.68 | 73978 | As1Fe1O4 | 0 |
| 15 | 1 | 15.2.2 | 185954 | Sb2Te3 | 2 |
| | | 15.2.3 | 16041 | S1Ta6 | 2 |
| | | 15.2.6 | 35148 | O5Ti3 | 2 |
| | | 15.2.7 | 194464 | O5Ti3 | 2 |
| | | 15.2.8 | 109187 | O4Ta1Tb1 | 2 |
| | | 15.2.9 | 97442 | O3V2 | 2 |
| | | 15.2.10 | 95763 | O3V2 | 2 |
| | | 15.2.11 | 95762 | O3V2 | 2 |
| | | 15.2.14 | 61148 | Ni1Se5Ta2 | 2 |
| | | 15.2.15 | 109181 | Nb1O4Tm1 | 2 |
| | | 15.2.16 | 109178 | Nb1O4Tb1 | 2 |
| | | 15.2.17 | 78342 | Na1O6Si2V1 | 2 |
| | | 15.2.18 | 163133 | Na1O6Si2V1 | 2 |
| | | 15.2.19 | 163132 | Na1O6Si2V1 | 2 |
| | | 15.2.20 | 163131 | Na1O6Si2V1 | 2 |
| | | 15.2.21 | 163130 | Na1O6Si2V1 | 2 |
| | | 15.2.22 | 163129 | Na1O6Si2V1 | 2 |
| | | 15.2.23 | 163128 | Na1O6Si2V1 | 2 |
| | | 15.2.24 | 163127 | Na1O6Si2V1 | 2 |
| | | 15.2.25 | 163126 | Na1O6Si2V1 | 2 |
| | | 15.2.26 | 163125 | Na1O6Si2V1 | 2 |
| | | 15.2.27 | 105057 | Mo1P4 | 2 |
| | | 15.2.28 | 49802 | Mn2P12Ti1 | 2 |
| | | 15.2.29 | 78721 | Li2O3Ru1 | 2 |
| | | 15.2.31 | 55164 | Li1O6Si2V1 | 2 |
| | | 15.2.32 | 152074 | Li1O6Si2V1 | 2 |
| | | 15.2.34 | 192833 | Ho1O4V1 | 2 |
| | | 15.2.35 | 79499 | Ho1O4Ta1 | 2 |
| | | 15.2.36 | 153363 | Ho1O4Ta1 | 2 |
| | | 15.2.37 | 15271 | H8O6Pt1 | 2 |
| | | 15.2.39 | 20138 | H4Al2Ca1F4O4 | 2 |
| | | 15.2.42 | 192134 | H12Br3Mg1O6Rb1 | 2 |
| | | 15.2.45 | 430120 | Gd1Ni8Si3 | 2 |
| | | 15.2.47 | 24416 | Fe2O5Ti1 | 2 |
| | | 15.2.48 | 154291 | Fe2O5Rb4 | 2 |
| | | 15.2.49 | 71059 | Fe2Li1N3Sr2 | 2 |
| | | 15.2.50 | 41899 | Fe1S5U2 | 2 |
| | | 15.2.51 | 15909 | F2Sn1 | 2 |
| | | 15.2.52 | 2739 | Er1Na1O2 | 2 |
| | | 15.2.53 | 94884 | D7Al1Sr2 | 2 |
| | | 15.2.64 | 83490 | Cu2Gd1O4 | 2 |
| | | 15.2.65 | 83494 | Cu2Er1O4 | 2 |
| | | 15.2.67 | 2790 | Cr1P4 | 2 |
| | | 15.2.68 | 26814 | Co1Ge1O3 | 2 |
| | | 15.2.69 | 417870 | Cl8Ga2Ni1 | 2 |
| | | 15.2.70 | 56838 | Cl3Cu1Rb1 | 2 |

Continued on next column

**TABLE XVI– continued**

| SG | cl | TQC | ICSD | Stoich. Formula | SubSG |
|---|---|---|---|---|---|
| | | 15.2.71 | 189154 | Cl2Cu2O6Pb1Se2 | 2 |
| | | 15.2.72 | 247292 | Ce4Rh4Si3 | 2 |
| | | 15.2.74 | 82216 | Ca1Ni1O6Si2 | 2 |
| | | 15.2.75 | 69381 | Ca1Ni1O6Si2 | 2 |
| | | 15.2.76 | 202244 | Ca1Ni1O6Si2 | 2 |
| | | 15.2.77 | 158299 | C2H8Cl3Cu1N1 | 2 |
| | | 15.2.78 | 423887 | C2Fe5 | 2 |
| | | 15.2.79 | 423886 | C2Fe5 | 2 |
| | | 15.2.80 | 423885 | C2Fe5 | 2 |
| | | 15.2.81 | 183902 | Bi1Mn1O3 | 2 |
| | | 15.2.82 | 174405 | Bi1Cr1O3 | 2 |
| | | 15.2.88 | 245254 | Ba3Ir2O9Sr1 | 2 |
| | | 15.2.90 | 71060 | Ba2Fe2Li1N3 | 2 |
| | | 15.2.91 | 86064 | Ba1Cu1N1 | 2 |
| | | 15.2.92 | 418385 | B2Ni1O4 | 2 |
| | | 15.2.94 | 417872 | Al2Cl8Ni1 | 2 |
| | | 15.2.95 | 251835 | Al1B2La2Os2 | 2 |
| 15 | 2 | 15.2.1 | 187497 | Sb2Te3 | 0 |
| | | 15.2.4 | 24327 | P2Pt5 | 0 |
| | | 15.2.5 | 261031 | O6Pb2Sb1Tm1 | 0 |
| | | 15.2.12 | 6286 | O3V2 | 0 |
| | | 15.2.13 | 50703 | Ni2O8Te3 | 0 |
| | | 15.2.30 | 180367 | Li2O3Ru1 | 0 |
| | | 15.2.33 | 416039 | K3Na1O8Ru2 | 0 |
| | | 15.2.38 | 14329 | H4Ba1O8S2 | 0 |
| | | 15.2.40 | 14176 | H1Cr1Hg1O4.5 | 0 |
| | | 15.2.41 | 183508 | H12Cl3Mg1O6Rb1 | 0 |
| | | 15.2.43 | 237747 | Ge2Mn1Na1O6 | 0 |
| | | 15.2.44 | 237746 | Ge2Mn1Na1O6 | 0 |
| | | 15.2.46 | 151766 | Fe7S8 | 0 |
| | | 15.2.54 | 185090 | Cu6Sn5 | 0 |
| | | 15.2.55 | 185089 | Cu6Sn5 | 0 |
| | | 15.2.56 | 150124 | Cu6Sn5 | 0 |
| | | 15.2.57 | 106530 | Cu6Sn5 | 0 |
| | | 15.2.58 | 23479 | Cu2O7V2 | 0 |
| | | 15.2.59 | 158375 | Cu2O7V2 | 0 |
| | | 15.2.60 | 67316 | Cu2O7P2 | 0 |
| | | 15.2.61 | 28279 | Cu2O7P2 | 0 |
| | | 15.2.62 | 157107 | Cu2O7P2 | 0 |
| | | 15.2.63 | 14369 | Cu2O7P2 | 0 |
| | | 15.2.66 | 83491 | Cu2Dy1O4 | 0 |
| | | 15.2.73 | 95789 | Ca2Ir2Si1 | 0 |
| | | 15.2.83 | 174404 | Bi1Cr1O3 | 0 |
| | | 15.2.84 | 160458 | Bi1Cr1O3 | 0 |
| | | 15.2.85 | 160457 | Bi1Cr1O3 | 0 |
| | | 15.2.86 | 160456 | Bi1Cr1O3 | 0 |
| | | 15.2.87 | 160455 | Bi1Cr1O3 | 0 |
| | | 15.2.89 | 245253 | Ba3Ca1Ir2O9 | 0 |
| | | 15.2.93 | 58403 | Au2Ca5 | 0 |
| 47 | 1 | 47.2.4 | 63335 | Ba2Cu3O7Y1 | 10 |
| | | 47.2.13 | 71640 | B3Ir4Zn1 | 10 |
| 47 | 2 | 47.2.12 | 187120 | Ba1Co2Nd1O5.5 | 10,10 |
| 47 | 3 | 47.2.11 | 96377 | Ba1Co2Nd1O5.5 | 10,10 |
| 47 | 4 | 47.2.10 | 96380 | Ba1Co2Nd1O5.5 | 10,10 |
| 47 | 5 | 47.2.8 | 94046 | Ba2Co4O11Tb2 | 2 |
| | | 47.2.9 | 94045 | Ba2Co4O11Tb2 | 2 |
| 47 | 6 | 47.2.7 | 68044 | Ba2Cu3Ho1O7 | 10,10 |
| 47 | 7 | 47.2.1 | 86427 | Cu1O5Sr2Tl1 | 2 |
| | | 47.2.2 | 390001 | Ba4Cu6O13Y2 | 2 |
| | | 47.2.6 | 73896 | Ba2Cu3O6.5Y1 | 2 |

Continued on next column

**TABLE XVI– continued**

| SG | cl | TQC | ICSD | Stoich. Formula | SubSG |
|---|---|---|---|---|---|
| 47 | 8 | 47.2.5 | 262809 | Ba2Cu3O6.93Y1 | 10,10,10 |
| 47 | 9 | 47.2.3 | 77745 | Ba2Cu3O7Y1 | 2 |
| 51 | 1 | 51.2.4 | 421074 | Ca2Ge2Ni3 | 2 |
| | | 51.2.6 | 69767 | C1B2U1 | 2 |
| | | 51.2.17 | 9986 | Al4Co1La1 | 2 |
| 51 | 2 | 51.2.1 | 248052 | La2Se9U2 | 2 |
| | | 51.2.7 | 247569 | Ba1Fe2Gd1O5 | 2 |
| | | 51.2.10 | 107495 | B3Rh7Zn5 | 2 |
| | | 51.2.12 | 429116 | B1Ru1Ta1 | 2 |
| | | 51.2.13 | 426468 | B1Nb1Ru1 | 2 |
| | | 51.2.15 | 249924 | Au3Rb2Tl1 | 2 |
| | | 51.2.16 | 16325 | Au1I1Te2 | 2 |
| 51 | 3 | 51.2.3 | 370040 | Co2In1Tb1 | 10 |
| | | 51.2.5 | 194877 | C3Al1Y3 | 10 |
| | | 51.2.8 | 96391 | Ba1Fe2Gd1O4.999 | 10 |
| | | 51.2.11 | 107494 | B3Ir7Zn5 | 10 |
| | | 51.2.14 | 194435 | B1Nb1Ru1 | 10 |
| 51 | 4 | 51.2.2 | 106800 | In1La1Ni2 | 0 |
| | | 51.2.9 | 99611 | Ba1Dy1Fe2O4.996 | 0 |
| 53 | 1 | 53.1.1 | 71063 | Ni2Ta2Te4 | 2 |
| | | 53.1.2 | 71520 | Fe1Nb1Te2 | 2 |
| | | 55.2.1 | 404193 | Se2Ti9 | 2 |
| | | 55.2.3 | 97282 | O4Pb3 | 2 |
| | | 55.2.24 | 426620 | F3O1Rb1V1 | 2 |
| | | 55.2.25 | 249403 | F3O1Rb1V1 | 2 |
| | | 55.2.47 | 419360 | B4Rh1Tm1 | 2 |
| | | 55.2.50 | 61039 | B4Mg2Ru5 | 2 |
| | | 55.2.51 | 190184 | B4Fe1Yb1 | 2 |
| | | 55.2.55 | 417873 | B4Co1Gd1 | 2 |
| | | 55.2.56 | 78823 | B10Ni1Th2 | 2 |
| | | 55.2.62 | 167534 | Al1B4Lu1 | 2 |
| | | 55.2.63 | 260799 | Al1.33B5.33Lu1.33 | 2 |
| 55 | 2 | 55.2.5 | 90183 | Mn2O5Sr2 | 10 |
| | | 55.2.6 | 417818 | Mn2O5Sr2 | 10 |
| | | 55.2.7 | 417817 | Mn2O5Sr2 | 10 |
| | | 55.2.8 | 417816 | Mn2O5Sr2 | 10 |
| | | 55.2.9 | 417815 | Mn2O5Sr2 | 10 |
| | | 55.2.10 | 417814 | Mn2O5Sr2 | 10 |
| | | 55.2.11 | 417813 | Mn2O5Sr2 | 10 |
| | | 55.2.12 | 417812 | Mn2O5Sr2 | 10 |
| | | 55.2.13 | 417811 | Mn2O5Sr2 | 10 |
| | | 55.2.14 | 417810 | Mn2O5Sr2 | 10 |
| | | 55.2.15 | 417809 | Mn2O5Sr2 | 10 |
| | | 55.2.16 | 417808 | Mn2O5Sr2 | 10 |
| | | 55.2.17 | 417807 | Mn2O5Sr2 | 10 |
| | | 55.2.18 | 417806 | Mn2O5Sr2 | 10 |
| | | 55.2.21 | 412706 | Hf5In4Rh2 | 10 |
| | | 55.2.22 | 22046 | Ge1Mn2O4 | 10 |
| | | 55.2.23 | 23226 | Ga2Mg1 | 10 |
| | | 55.2.30 | 171414 | Ce5In4Pt2 | 10 |
| | | 55.2.34 | 35130 | Ca2Mn2O5 | 10 |
| | | 55.2.36 | 84869 | C5Y4 | 10 |
| | | 55.2.39 | 412563 | Bi8Ca3Pd4 | 10 |
| | | 55.2.41 | 262861 | Bi2Fe4O9 | 10 |
| | | 55.2.44 | 186440 | Bi2Fe4O9 | 10 |
| | | 55.2.53 | 81542 | B4Cr1Th1 | 10 |
| | | 55.2.57 | 78821 | B10Fe1Th2 | 10 |
| | | 55.2.60 | 260796 | Al1B4Yb1 | 10 |
| | | 55.2.61 | 167532 | Al1B4Yb1 | 10 |
| 55 | 3 | 55.2.2 | 249665 | Pr2S5Sn1 | 2 |

Continued on next column

**TABLE XVI– continued**

| SG | cl | TQC | ICSD | Stoich. Formula | SubSG |
|---|---|---|---|---|---|
| | | 55.2.4 | 249666 | Nd2S5Sn1 | 2 |
| | | 55.2.19 | 391127 | In4Rh2Zr5 | 2 |
| | | 55.2.20 | 193223 | In1La2Si2 | 2 |
| | | 55.2.26 | 82389 | Cu2La2O5 | 2 |
| | | 55.2.27 | 82388 | Cu2La2O5 | 2 |
| | | 55.2.28 | 73399 | Cu2La2O5 | 2 |
| | | 55.2.29 | 82019 | Cu1O2.5Sr1 | 2 |
| | | 55.2.31 | 2578 | Ce4O4S3 | 2 |
| | | 55.2.35 | 35219 | Ca1Mn1O2.5 | 2 |
| | | 55.2.37 | 23834 | C2B2Sc1 | 2 |
| | | 55.2.38 | 423924 | C1B2Lu1 | 2 |
| | | 55.2.40 | 415576 | Bi6In2Sr5 | 2 |
| | | 55.2.42 | 186442 | Bi2Fe4O9 | 2 |
| | | 55.2.43 | 186441 | Bi2Fe4O9 | 2 |
| | | 55.2.46 | 419358 | B4Rh1Y1 | 2 |
| | | 55.2.48 | 409820 | B4Ni1Yb1 | 2 |
| | | 55.2.58 | 56973 | As1Ga2Rh5 | 2 |
| | | 55.2.59 | 36364 | Al2Bi6Ca5 | 2 |
| 55 | 4 | 55.2.32 | 102075 | Cd7Th6 | 0 |
| | | 55.2.33 | 36466 | Ca5Ga2Sb6 | 0 |
| | | 55.2.45 | 419361 | B4Rh1Yb1 | 0 |
| | | 55.2.49 | 425457 | B4Ni1Sc1 | 0 |
| | | 55.2.52 | 409626 | B4Er1V1 | 0 |
| | | 55.2.54 | 250251 | B4Co1Yb1 | 0 |
| 57 | 1 | 57.2.1 | 20933 | Fe2Hf1Si2 | 2 |
| | | 57.2.7 | 190838 | Ce1Pd2Sn3 | 2 |
| | | 57.2.8 | 93749 | Ca1Mn2O4 | 2 |
| | | 57.2.9 | 150592 | Al1Dy1 | 2 |
| 57 | 2 | 57.2.2 | 85640 | Cr1La1Sb3 | 0 |
| | | 57.2.3 | 85639 | Cr1La1Sb3 | 0 |
| | | 57.2.4 | 83907 | Cr1La1Sb3 | 0 |
| | | 57.2.5 | 247291 | Ce3Rh2Si2 | 0 |
| | | 57.2.6 | 76352 | Ce3Ge2Rh2 | 0 |
| 58 | 1 | 58.2.1 | 42724 | Ni1Sb2 | 2 |
| | | 58.2.2 | 414575 | Ge6Li2Sr4 | 2 |
| | | 58.2.3 | 409767 | Eu2Ge3Li1 | 2 |
| | | 58.2.6 | 81572 | C3Al2Th2 | 2 |
| | | 58.2.7 | 425106 | B4Cr1 | 2 |
| | | 58.2.8 | 186851 | B4Cr1 | 2 |
| | | 58.2.9 | 31387 | B2Ni3O6 | 2 |
| | | 58.2.10 | 2016 | B2Ni3O6 | 2 |
| | | 58.2.11 | 251066 | Au8Sn3Sr3 | 2 |
| | | 58.2.12 | 42605 | As2Ni1 | 2 |
| 58 | 2 | 58.2.4 | 42720 | Cr1Sb2 | 0 |
| | | 58.2.5 | 42601 | Cr1Sb2 | 0 |
| 59 | 1 | 59.1.1 | 408080 | Pd1Si1Y1 | 0 |
| | | 59.1.2 | 154823 | Li1Mn1O2 | 0 |
| | | 59.1.3 | 425285 | Ho1I1S1 | 0 |
| | | 59.1.4 | 154397 | Ge5N2Sr6 | 0 |
| | | 59.1.5 | 410357 | Ge1Pr1Pt1 | 0 |
| | | 59.1.6 | 195027 | Fe2O1Se2Sr1 | 0 |
| | | 59.1.7 | 195026 | Fe2O1S2Sr1 | 0 |
| | | 59.1.8 | 195025 | Fe2O1S2Sr1 | 0 |
| | | 59.1.9 | 88203 | Er2Ge5 | 0 |
| | | 59.1.10 | 160543 | Dy2Sn5 | 0 |
| | | 59.1.11 | 72376 | Cr6P4U1 | 0 |
| | | 59.1.12 | 30701 | Co8La1P5 | 0 |
| | | 59.1.13 | 410358 | Ce1Ge1Pt1 | 0 |
| | | 59.1.14 | 154398 | Ba6Ge5N2 | 0 |
| | | 59.1.15 | 421431 | Ba1Ge1Ni2 | 0 |

Continued on next column

**TABLE XVI– continued**

| SG | cl | TQC | ICSD | Stoich. Formula | SubSG |
|---|---|---|---|---|---|
| | | 59.1.16 | 428170 | Ba1Fe2O1S2 | 0 |
| | | 59.1.17 | 421524 | B2Ru1 | 0 |
| | | 59.1.18 | 31871 | B2Ru1 | 0 |
| | | 59.1.19 | 421523 | B2Os1 | 0 |
| 59 | 2 | 59.2.1 | 187622 | Ni1Si1 | 2 |
| | | 59.2.2 | 95640 | Li1Mn1O2 | 2 |
| | | 59.2.3 | 408082 | Gd1Pd1Si1 | 2 |
| | | 59.2.4 | 81303 | Fe1Hf8Te6 | 2 |
| | | 59.2.5 | 408081 | Er1Pd1Si1 | 2 |
| | | 59.2.6 | 99042 | Cu2O3Sr1 | 2 |
| | | 59.2.7 | 155492 | Co1Na0.5O2 | 2 |
| | | 59.2.8 | 413690 | Br3Mo1 | 2 |
| | | 59.2.9 | 249645 | Au3K1Sn3 | 2 |
| | | 59.2.10 | 245679 | Au3In3Sr1 | 2 |
| | | 59.2.11 | 107445 | Au1Rb3Sn4 | 2 |
| | | 59.2.12 | 107447 | Au1Pb4Rb3 | 2 |
| | | 59.2.13 | 107448 | Au1Cs3Pb4 | 2 |
| | | 59.2.14 | 107525 | Al1Eu1 | 2 |
| 60 | 1 | 60.2.1 | 415270 | O2Pb1 | 2 |
| | | 60.2.2 | 415269 | O2Pb1 | 2 |
| | | 60.2.3 | 415268 | O2Pb1 | 2 |
| | | 60.2.4 | 246148 | C1Mo2 | 2 |
| | | 60.2.5 | 246147 | C1Mo2 | 2 |
| | | 60.2.6 | 246146 | C1Mo2 | 2 |
| | | 60.2.7 | 430298 | Au2Pb1 | 2 |
| | | 61.2.1 | 162404 | Pd1Sn1Te1 | 2 |
| | | 61.2.2 | 415968 | Au1Sn2 | 2 |
| | | 62.2.1 | 418619 | Sr1Zn5 | 2 |
| | | 62.2.2 | 54980 | Sr1Zn1 | 2 |
| | | 62.2.3 | 82295 | Se2U1 | 2 |
| | | 62.2.4 | 10001 | Sb2Sr1Zn1 | 2 |
| | | 62.2.6 | 42430 | S1Te1U1 | 2 |
| | | 62.2.7 | 30398 | Ru4Si3 | 2 |
| | | 62.2.9 | 421255 | Rh1Sc3Si3 | 2 |
| | | 62.2.10 | 15248 | Rh1Sc1Si2 | 2 |
| | | 62.2.11 | 159305 | Pt1Yb1Zn1 | 2 |
| | | 62.2.12 | 412012 | Pt1Sn1Sr1 | 2 |
| | | 62.2.13 | 260792 | Pt1Si1Zr1 | 2 |
| | | 62.2.14 | 260794 | Pt1Si1Ti1 | 2 |
| | | 62.2.19 | 159304 | Pd1Yb1Zn1 | 2 |
| | | 62.2.20 | 408408 | Pd1Sn1Yb1 | 2 |
| | | 62.2.21 | 151167 | Pd1Sn1Yb1 | 2 |
| | | 62.2.25 | 1918 | Pd1Pu1 | 2 |
| | | 62.2.26 | 400334 | P2Zr1 | 2 |
| | | 62.2.27 | 417911 | P1Pt1Sc1 | 2 |
| | | 62.2.28 | 100624 | O4S1Sr1 | 2 |
| | | 62.2.29 | 415748 | O4Rb2Ru1 | 2 |
| | | 62.2.31 | 185833 | O3Pr1V1 | 2 |
| | | 62.2.32 | 172354 | O3Pr1Rh1 | 2 |
| | | 62.2.33 | 172348 | O3Pr1Rh1 | 2 |
| | | 62.2.36 | 64952 | Ni2O4Si1 | 2 |
| | | 62.2.37 | 35675 | Ni2O4Si1 | 2 |
| | | 62.2.38 | 202376 | Ni2O4Si1 | 2 |
| | | 62.2.39 | 100645 | Ni2O4Si1 | 2 |
| | | 62.2.40 | 100644 | Ni2O4Si1 | 2 |
| | | 62.2.41 | 100643 | Ni2O4Si1 | 2 |
| | | 62.2.42 | 100642 | Ni2O4Si1 | 2 |
| | | 62.2.43 | 105403 | Ni1Th1 | 2 |
| | | 62.2.44 | 54360 | Ni1Sn1Yb1 | 2 |
| | | 62.2.45 | 18188 | Ni1Si1Ti1 | 2 |
| | | 62.2.49 | 248408 | Ni1S3U1 | 2 |

Continued on next column

**TABLE XVI– continued**

| SG | cl | TQC | ICSD | Stoich. Formula | SubSG |
|---|---|---|---|---|---|
| | | 62.2.50 | 419759 | Ni1Pd1Te1 | 2 |
| | | 62.2.51 | 50990 | Ni1P1Sc1 | 2 |
| | | 62.2.52 | 189299 | Ni1O3Yb1 | 2 |
| | | 62.2.53 | 72290 | Nd2S3 | 2 |
| | | 62.2.54 | 415953 | Nd1Pt1Sn1 | 2 |
| | | 62.2.55 | 82008 | Nd1O3Ti1 | 2 |
| | | 62.2.56 | 67723 | Nd1Ni1O3 | 2 |
| | | 62.2.57 | 67722 | Nd1Ni1O3 | 2 |
| | | 62.2.58 | 75008 | Nb1P1Zr1 | 2 |
| | | 62.2.59 | 49728 | Nb1Ni1P1 | 2 |
| | | 62.2.61 | 43399 | Mn1P1 | 2 |
| | | 62.2.62 | 43246 | Mn1P1 | 2 |
| | | 62.2.63 | 185418 | Mn1O3V1 | 2 |
| | | 62.2.64 | 97084 | Mn1O3Pr1 | 2 |
| | | 62.2.65 | 97083 | Mn1O3Pr1 | 2 |
| | | 62.2.66 | 97082 | Mn1O3Pr1 | 2 |
| | | 62.2.67 | 84922 | Mn1O3Pr1 | 2 |
| | | 62.2.68 | 426614 | Mg1Pt1Sr1 | 2 |
| | | 62.2.69 | 97765 | Li1Ni1O4P1 | 2 |
| | | 62.2.70 | 72929 | Li1Ni1O4P1 | 2 |
| | | 62.2.71 | 402760 | Li1Ni1O4P1 | 2 |
| | | 62.2.72 | 417912 | La1P1Pt1 | 2 |
| | | 62.2.73 | 415749 | K2O4Ru1 | 2 |
| | | 62.2.74 | 427252 | Ir1Sn1Y1 | 2 |
| | | 62.2.75 | 427253 | Ir1Si1Y1 | 2 |
| | | 62.2.77 | 421256 | Ir1Sc3Si3 | 2 |
| | | 62.2.78 | 420416 | Ir1Sc1Si1 | 2 |
| | | 62.2.79 | 422623 | Ir1Sb1Yb1 | 2 |
| | | 62.2.83 | 413857 | Ir1Lu1Si1 | 2 |
| | | 62.2.84 | 240201 | In3Ir1 | 2 |
| | | 62.2.85 | 412913 | In1Rh1Yb1 | 2 |
| | | 62.2.87 | 99546 | Ho1O3Sc1 | 2 |
| | | 62.2.88 | 165196 | Ho1Mn1O3 | 2 |
| | | 62.2.89 | 82727 | Hf3P2 | 2 |
| | | 62.2.91 | 42912 | Ge3Pt2 | 2 |
| | | 62.2.92 | 16822 | Ge2Mo1 | 2 |
| | | 62.2.93 | 429881 | Ge2Li3Na1 | 2 |
| | | 62.2.94 | 280691 | Ge2In1Li1Sr2 | 2 |
| | | 62.2.96 | 421092 | Ge1Rh1Ta1 | 2 |
| | | 62.2.97 | 260793 | Ge1Pt1Zr1 | 2 |
| | | 62.2.98 | 419135 | Ge1Pt1Yb1 | 2 |
| | | 62.2.100 | 409926 | Ge1Pd1Ti1 | 2 |
| | | 62.2.101 | 413387 | Ge1O1Yb3 | 2 |
| | | 62.2.102 | 237302 | Ge1Mg1Ni1 | 2 |
| | | 62.2.103 | 428468 | Ge1Ir1Sc1 | 2 |
| | | 62.2.105 | 15534 | Gd3Ni1Si2 | 2 |
| | | 62.2.106 | 280201 | Gd2O1Se2 | 2 |
| | | 62.2.108 | 419793 | Gd1Ge1Ru1 | 2 |
| | | 62.2.109 | 103904 | Ga2Pd5 | 2 |
| | | 62.2.110 | 420711 | Ga1Ho1O3 | 2 |
| | | 62.2.111 | 55537 | Fe1Si1U1 | 2 |
| | | 62.2.112 | 169317 | Fe1Se1 | 2 |
| | | 62.2.113 | 169316 | Fe1Se1 | 2 |
| | | 62.2.114 | 169314 | Fe1Se1 | 2 |
| | | 62.2.115 | 169312 | Fe1Se1 | 2 |
| | | 62.2.116 | 169310 | Fe1Se1 | 2 |
| | | 62.2.117 | 169308 | Fe1Se1 | 2 |
| | | 62.2.118 | 169305 | Fe1Se1 | 2 |
| | | 62.2.119 | 169282 | Fe1Se1 | 2 |
| | | 62.2.120 | 169281 | Fe1Se1 | 2 |
| | | 62.2.121 | 169280 | Fe1Se1 | 2 |

Continued on next column

**TABLE XVI– continued**

| SG | cl | TQC | ICSD | Stoich. Formula | SubSG |
|---|---|---|---|---|---|
| | | 62.2.122 | 169279 | Fe1Se1 | 2 |
| | | 62.2.123 | 169278 | Fe1Se1 | 2 |
| | | 62.2.124 | 169277 | Fe1Se1 | 2 |
| | | 62.2.125 | 169276 | Fe1Se1 | 2 |
| | | 62.2.126 | 169275 | Fe1Se1 | 2 |
| | | 62.2.127 | 169274 | Fe1Se1 | 2 |
| | | 62.2.128 | 169273 | Fe1Se1 | 2 |
| | | 62.2.129 | 35010 | Fe1S1 | 2 |
| | | 62.2.130 | 35009 | Fe1S1 | 2 |
| | | 62.2.131 | 35008 | Fe1S1 | 2 |
| | | 62.2.133 | 153236 | Fe1Nd1O3 | 2 |
| | | 62.2.134 | 32756 | Fe1K2O4 | 2 |
| | | 62.2.135 | 2876 | Fe1K2O4 | 2 |
| | | 62.2.136 | 240891 | F2K1N1O1 | 2 |
| | | 62.2.137 | 2635 | F1Se1Y1 | 2 |
| | | 62.2.139 | 405049 | Eu1In1Pt1 | 2 |
| | | 62.2.140 | 404044 | Eu1In1Pd1 | 2 |
| | | 62.2.141 | 423115 | Er1Pt1Zn1 | 2 |
| | | 62.2.142 | 27282 | Er1Fe1O3 | 2 |
| | | 62.2.143 | 419 | Dy2Se3 | 2 |
| | | 62.2.144 | 80768 | Dy2S3 | 2 |
| | | 62.2.145 | 66532 | Dy2S3 | 2 |
| | | 62.2.146 | 66531 | Dy2S3 | 2 |
| | | 62.2.147 | 36603 | Dy2O1S2 | 2 |
| | | 62.2.151 | 86391 | Cu1Sc1Si1 | 2 |
| | | 62.2.152 | 172850 | Cu1S2Tm1 | 2 |
| | | 62.2.154 | 33799 | Cs2O4Ru1 | 2 |
| | | 62.2.155 | 33861 | Cs2Fe1O4 | 2 |
| | | 62.2.156 | 192388 | Cr2Fe1O4 | 2 |
| | | 62.2.162 | 251083 | Cr1Er1O3 | 2 |
| | | 62.2.163 | 183759 | Cr1Er1O3 | 2 |
| | | 62.2.164 | 65751 | Co2O4Si1 | 2 |
| | | 62.2.165 | 6248 | Co2O4Si1 | 2 |
| | | 62.2.166 | 50072 | Co2O4Si1 | 2 |
| | | 62.2.167 | 260092 | Co2O4Si1 | 2 |
| | | 62.2.168 | 260091 | Co2O4Si1 | 2 |
| | | 62.2.169 | 202255 | Co2O4Si1 | 2 |
| | | 62.2.170 | 200705 | Co2O4Si1 | 2 |
| | | 62.2.171 | 192245 | Co1Y3 | 2 |
| | | 62.2.172 | 192244 | Co1Y3 | 2 |
| | | 62.2.174 | 49726 | Co1P1Zr1 | 2 |
| | | 62.2.177 | 172046 | Co1O3Tb1 | 2 |
| | | 62.2.178 | 88738 | Co1O3Pr1 | 2 |
| | | 62.2.179 | 94437 | Co1O2Rb3 | 2 |
| | | 62.2.183 | 172048 | Co1Ho1O3 | 2 |
| | | 62.2.184 | 102428 | Co1Ga1Gd1 | 2 |
| | | 62.2.185 | 48206 | Cl2Nd1 | 2 |
| | | 62.2.187 | 428424 | Ce2Te3 | 2 |
| | | 62.2.188 | 106433 | Ce1Si1 | 2 |
| | | 62.2.189 | 418546 | Ce1Pt1Zn1 | 2 |
| | | 62.2.190 | 420207 | Ce1Pd1Zn1 | 2 |
| | | 62.2.191 | 409972 | Ce1P2Pt4 | 2 |
| | | 62.2.192 | 54733 | Ce1In4Pd2 | 2 |
| | | 62.2.195 | 237728 | Cd1Pt1Sr1 | 2 |
| | | 62.2.196 | 12159 | Cd1Na1Sb1 | 2 |
| | | 62.2.197 | 72642 | Ca1Pt1Si1 | 2 |
| | | 62.2.198 | 426616 | Ca1Mg1Pd1 | 2 |
| | | 62.2.199 | 410841 | Ca1In1Rh1 | 2 |
| | | 62.2.200 | 106343 | Ca1Ge1Pt1 | 2 |
| | | 62.2.201 | 291004 | Ca1Fe2O4 | 2 |
| | | 62.2.202 | 426610 | Ca1Cd1Pt1 | 2 |

Continued on next column

**TABLE XVI– continued**

| SG | cl | TQC | ICSD | Stoich. Formula | SubSG |
|---|---|---|---|---|---|
| | | 62.2.203 | 426611 | Ca1Cd1Pd1 | 2 |
| | | 62.2.204 | 68116 | C2U1W1 | 2 |
| | | 62.2.205 | 16549 | C2Mo1U1 | 2 |
| | | 62.2.208 | 247128 | C1H2Cl1Cu1N2 | 2 |
| | | 62.2.209 | 236383 | Bi3Co1 | 2 |
| | | 62.2.210 | 193272 | Bi3Co1 | 2 |
| | | 62.2.211 | 428110 | Bi1Li1Yb1 | 2 |
| | | 62.2.212 | 52698 | Ba2Si1 | 2 |
| | | 62.2.213 | 416561 | Ba2Pb1 | 2 |
| | | 62.2.215 | 424857 | Ba1Hg6 | 2 |
| | | 62.2.216 | 237304 | Ba1Ge1Ni1 | 2 |
| | | 62.2.217 | 424315 | Ba1Fe2Se3 | 2 |
| | | 62.2.218 | 424314 | Ba1Fe2Se3 | 2 |
| | | 62.2.219 | 290595 | Ba1Fe2Se3 | 2 |
| | | 62.2.220 | 185512 | Ba1Fe2Se3 | 2 |
| | | 62.2.222 | 421745 | B1O3Pr1 | 2 |
| | | 62.2.223 | 42894 | B1Co1Mo1 | 2 |
| | | 62.2.224 | 50842 | B1Ca2F1N2 | 2 |
| | | 62.2.225 | 263057 | Au3Ga2Rb1 | 2 |
| | | 62.2.226 | 263058 | Au3Cs1Ga2 | 2 |
| | | 62.2.227 | 418239 | Au3Ca3In1 | 2 |
| | | 62.2.229 | 249562 | Au2Ba1In2 | 2 |
| | | 62.2.230 | 391422 | Au1In1Sr1 | 2 |
| | | 62.2.231 | 420674 | Au1Eu1Zn1 | 2 |
| | | 62.2.232 | 411544 | Au1Cd1Eu1 | 2 |
| | | 62.2.233 | 408579 | Au1Ca1In1 | 2 |
| | | 62.2.234 | 106273 | Au1Ca1Ga1 | 2 |
| | | 62.2.235 | 168665 | As2Zr1 | 2 |
| | | 62.2.237 | 51201 | As1Li1Ni1O4 | 2 |
| | | 62.2.238 | 245184 | As1Li1Ni1O4 | 2 |
| | | 62.2.239 | 412296 | As1Cu1Mg1 | 2 |
| | | 62.2.240 | 406953 | As1Co1Hf1 | 2 |
| | | 62.2.245 | 9571 | As1Cd1Na1 | 2 |
| | | 62.2.246 | 249802 | As1.01Ce1Se0.99 | 2 |
| | | 62.2.248 | 160051 | Al1Ce1Ru1 | 2 |
| | | 62.2.251 | 370027 | Al1Au1Yb1 | 2 |
| | | 62.2.252 | 370015 | Al1Au1Ca1 | 2 |
| | | 62.2.253 | 83983 | Ag1Sb1Yb1 | 2 |
| | | 62.2.255 | 98736 | Ag1As2Pr1 | 2 |
| | | 62.2.256 | 420611 | Ag1As2Pr1 | 2 |
| | | 62.2.257 | 412816 | Ag1As2Pr1 | 2 |
| 62 | 2 | 62.2.5 | 402457 | S3Tm2 | 0 |
| | | 62.2.8 | 421663 | Ru2Sn2Zn3 | 0 |
| | | 62.2.15 | 79238 | Pt1Si1 | 0 |
| | | 62.2.16 | 2623 | Pt1Si1 | 0 |
| | | 62.2.17 | 105785 | Pt1Sc2 | 0 |
| | | 62.2.18 | 158364 | Pd2Sn1 | 0 |
| | | 62.2.22 | 109415 | Pd1Sn1 | 0 |
| | | 62.2.23 | 79237 | Pd1Si1 | 0 |
| | | 62.2.24 | 15016 | Pd1Si1 | 0 |
| | | 62.2.30 | 100625 | O4Pb1S1 | 0 |
| | | 62.2.34 | 413386 | O1Si1Yb3 | 0 |
| | | 62.2.35 | 74785 | Ni2P2Th1 | 0 |
| | | 62.2.46 | 95449 | Ni1Si1 | 0 |
| | | 62.2.47 | 95448 | Ni1Si1 | 0 |
| | | 62.2.48 | 190105 | Ni1Si1 | 0 |
| | | 62.2.60 | 41149 | Mn1Rh1Si1 | 0 |
| | | 62.2.76 | 411884 | Ir1Si1Ta1 | 0 |
| | | 62.2.80 | 411882 | Ir1Nb1Si1 | 0 |
| | | 62.2.81 | 246548 | Ir1Mn1Si1 | 0 |
| | | 62.2.82 | 246547 | Ir1Mn1Si1 | 0 |

Continued on next column

**TABLE XVI– continued**

| SG | cl | TQC | ICSD | Stoich. Formula | SubSG |
|---|---|---|---|---|---|
| | | 62.2.86 | 72291 | Ho2S3 | 0 |
| | | 62.2.90 | 260839 | H2Cl1O2Tm1 | 0 |
| | | 62.2.95 | 44860 | Ge1Rh2 | 0 |
| | | 62.2.99 | 2624 | Ge1Pt1 | 0 |
| | | 62.2.104 | 411883 | Ge1Ir1Nb1 | 0 |
| | | 62.2.107 | 1375 | Gd1P1S1 | 0 |
| | | 62.2.132 | 49729 | Fe1P1Ta1 | 0 |
| | | 62.2.138 | 195382 | F1Ho1Se1 | 0 |
| | | 62.2.148 | 180220 | Dy1Si1 | 0 |
| | | 62.2.149 | 280091 | Dy1Fe1O3 | 0 |
| | | 62.2.150 | 27280 | Dy1Fe1O3 | 0 |
| | | 62.2.153 | 172848 | Cu1Ho1S2 | 0 |
| | | 62.2.157 | 80332 | Cr1Nd1O3 | 0 |
| | | 62.2.158 | 79349 | Cr1Nd1O3 | 0 |
| | | 62.2.159 | 156326 | Cr1Nd1O3 | 0 |
| | | 62.2.160 | 156322 | Cr1Nd1O3 | 0 |
| | | 62.2.161 | 156320 | Cr1Nd1O3 | 0 |
| | | 62.2.173 | 409847 | Co1Si1V1 | 0 |
| | | 62.2.175 | 43401 | Co1P1 | 0 |
| | | 62.2.176 | 43249 | Co1P1 | 0 |
| | | 62.2.180 | 2421 | Co1Mo1P1 | 0 |
| | | 62.2.181 | 150531 | Co1La3 | 0 |
| | | 62.2.182 | 73212 | Co1K3O2 | 0 |
| | | 62.2.186 | 157332 | Ce3In2Ru2 | 0 |
| | | 62.2.193 | 99240 | Ce1In2Pd3 | 0 |
| | | 62.2.194 | 150105 | Ce1Cu6 | 0 |
| | | 62.2.206 | 151477 | C2Cr3 | 0 |
| | | 62.2.207 | 15086 | C2Cr3 | 0 |
| | | 62.2.214 | 90336 | Ba1Ho2Te4 | 0 |
| | | 62.2.221 | 41893 | B2Co1Nb1 | 0 |
| | | 62.2.228 | 249563 | Au2In2Sr1 | 0 |
| | | 62.2.236 | 42446 | As1V1 | 0 |
| | | 62.2.241 | 48030 | As1Co1 | 0 |
| | | 62.2.242 | 48029 | As1Co1 | 0 |
| | | 62.2.243 | 48028 | As1Co1 | 0 |
| | | 62.2.244 | 48026 | As1Co1 | 0 |
| | | 62.2.247 | 150173 | Al1Pr1Pt1 | 0 |
| | | 62.2.249 | 55623 | Al1Ce1Rh1 | 0 |
| | | 62.2.250 | 160052 | Al1Ce1Rh1 | 0 |
| | | 62.2.254 | 420615 | Ag1As2Tb1 | 0 |
| 63 | 1 | 63.1.1 | 68416 | Te3U1 | 0 |
| | | 63.1.2 | 106074 | Sn1Sr1 | 0 |
| | | 63.1.3 | 416971 | Si6Sr1 | 0 |
| | | 63.1.4 | 240779 | Si6Sr1 | 0 |
| | | 63.1.5 | 78997 | Si1Sr1 | 0 |
| | | 63.1.6 | 87915 | Pd3Te2 | 0 |
| | | 63.1.7 | 77902 | Pd3Te2 | 0 |
| | | 63.1.8 | 411798 | Pd1Sn2Yb1 | 0 |
| | | 63.1.9 | 99253 | O7Re1Tb3 | 0 |
| | | 63.1.10 | 96410 | O7Pr3Re1 | 0 |
| | | 63.1.11 | 96149 | Np1Si1 | 0 |
| | | 63.1.12 | 86102 | Mg1O5V2 | 0 |
| | | 63.1.13 | 50981 | Mg1O5V2 | 0 |
| | | 63.1.14 | 50979 | Mg1O5V2 | 0 |
| | | 63.1.15 | 250752 | La2Sn1Zn5 | 0 |
| | | 63.1.16 | 250756 | La2Pb1Zn5 | 0 |
| | | 63.1.17 | 423611 | La2Li2Si3 | 0 |
| | | 63.1.18 | 410732 | La1Rh1Sn2 | 0 |
| | | 63.1.19 | 38311 | La1Ni5P3 | 0 |
| | | 63.1.20 | 165363 | In5Y3 | 0 |
| | | 63.1.21 | 165362 | In5La3 | 0 |

Continued on next column

**TABLE XVI– continued**

| SG | cl | TQC | ICSD | Stoich. Formula | SubSG |
|---|---|---|---|---|---|
| | | 63.1.22 | 418178 | In4Ni1Sr1 | 0 |
| | | 63.1.23 | 424100 | In2Pt1Yb1 | 0 |
| | | 63.1.24 | 410437 | In2Pd1Yb1 | 0 |
| | | 63.1.25 | 414077 | In2Ni1Yb1 | 0 |
| | | 63.1.26 | 412431 | In2La1Rh1 | 0 |
| | | 63.1.27 | 414479 | In2Ir1La1 | 0 |
| | | 63.1.28 | 200067 | Hf3Ni2Si3 | 0 |
| | | 63.1.29 | 26225 | H5Li4Rh1 | 0 |
| | | 63.1.30 | 262431 | Ge3La2Li2 | 0 |
| | | 63.1.31 | 42375 | Ge1Nd1 | 0 |
| | | 63.1.32 | 42374 | Ge1Nd1 | 0 |
| | | 63.1.33 | 170875 | Gd3O7Os1 | 0 |
| | | 63.1.34 | 423067 | Ga4Gd1Ni1 | 0 |
| | | 63.1.35 | 189439 | Fe2O3 | 0 |
| | | 63.1.36 | 261037 | Fe1Se3U1 | 0 |
| | | 63.1.37 | 245105 | Eu3O7Re1 | 0 |
| | | 63.1.38 | 400536 | Eu2O7Ta2 | 0 |
| | | 63.1.39 | 40812 | Eu1Ni5P3 | 0 |
| | | 63.1.40 | 237971 | Eu1In4Ir1 | 0 |
| | | 63.1.41 | 80501 | Er3Ge4 | 0 |
| | | 63.1.42 | 171141 | Er1Ni1Si2 | 0 |
| | | 63.1.43 | 423066 | Er1Ga4Ni1 | 0 |
| | | 63.1.44 | 152349 | Dy1Sn2 | 0 |
| | | 63.1.45 | 180214 | Dy1Si1 | 0 |
| | | 63.1.46 | 154699 | Dy1Si1 | 0 |
| | | 63.1.47 | 57225 | Dy1Ge3 | 0 |
| | | 63.1.48 | 59858 | Cu3Ho2Rb1S5 | 0 |
| | | 63.1.49 | 174371 | Cu3Ho2K1Se5 | 0 |
| | | 63.1.50 | 423716 | Cu3Er2S5Tl1 | 0 |
| | | 63.1.51 | 174372 | Cu3Er2K1Se5 | 0 |
| | | 63.1.52 | 95028 | Cu1Ho2K1Se4 | 0 |
| | | 63.1.53 | 97561 | Cu1Ho2K1S4 | 0 |
| | | 63.1.54 | 106440 | Cu1Eu1In4 | 0 |
| | | 63.1.55 | 429089 | Cs1Sc1Se5U1 | 0 |
| | | 63.1.56 | 413682 | Cs1Cu3Se5Yb2 | 0 |
| | | 63.1.57 | 413677 | Cs1Cu3Ho2Se5 | 0 |
| | | 63.1.58 | 412074 | Cs1Cu3Er2S5 | 0 |
| | | 63.1.59 | 413674 | Cs1Cu3Dy2Se5 | 0 |
| | | 63.1.60 | 236871 | Cr2Mg1O4 | 0 |
| | | 63.1.61 | 20952 | Co3Ga9Ho2 | 0 |
| | | 63.1.62 | 427218 | Ce3Ge3Mn2 | 0 |
| | | 63.1.63 | 250753 | Ce2Sn1Zn5 | 0 |
| | | 63.1.64 | 250757 | Ce2Pb1Zn5 | 0 |
| | | 63.1.65 | 63036 | Ce2Ge3Li2 | 0 |
| | | 63.1.66 | 262430 | Ce2Ge3Li2 | 0 |
| | | 63.1.67 | 99133 | Ce1Rh2Si1 | 0 |
| | | 63.1.68 | 421810 | Cd2F1Sb5Sr5 | 0 |
| | | 63.1.69 | 424107 | Ca2Sn6Zn3 | 0 |
| | | 63.1.70 | 96228 | Ca2In1N1 | 0 |
| | | 63.1.71 | 165193 | Ca1Sn1 | 0 |
| | | 63.1.72 | 410774 | Ca1Pd1Sn2 | 0 |
| | | 63.1.73 | 408880 | Ca1In2Pt1 | 0 |
| | | 63.1.74 | 408881 | Ca1In2Pd1 | 0 |
| | | 63.1.75 | 16356 | Ca1Fe5O7 | 0 |
| | | 63.1.76 | 75133 | C1Fe2Si1Th1 | 0 |
| | | 63.1.77 | 421809 | Ba5Cd2F1Sb5 | 0 |
| | | 63.1.78 | 411979 | Ba3Pb5 | 0 |
| | | 63.1.79 | 165184 | Ba3Pb5 | 0 |
| | | 63.1.80 | 710014 | Ba1Si6 | 0 |
| | | 63.1.81 | 245295 | Ba1Si6 | 0 |
| | | 63.1.82 | 78998 | Ba1Si1 | 0 |

Continued on next column

| SG | cl | TQC | ICSD | Stoich. Formula | SubSG |
|---|---|---|---|---|---|
| | | 63.1.83 | 427286 | Ba1Ru3Sn6 | 0 |
| | | 63.1.84 | 58664 | Ba1Pb1 | 0 |
| | | 63.1.85 | 251583 | Ba1Ge6 | 0 |
| | | 63.1.86 | 290596 | Ba1Fe2Se3 | 0 |
| | | 63.1.87 | 290594 | Ba1Fe2Se3 | 0 |
| | | 63.1.88 | 16307 | Ba1Fe2S3 | 0 |
| | | 63.1.89 | 64596 | B7Fe1Y3 | 0 |
| | | 63.1.90 | 91243 | B4Mg1Os3 | 0 |
| | | 63.1.91 | 420401 | B4Fe1O7 | 0 |
| | | 63.1.92 | 418162 | B2Cr2Ir1 | 0 |
| | | 63.1.93 | 237221 | Au1Eu1In4 | 0 |
| | | 63.1.94 | 33916 | As3Eu1Ni5 | 0 |
| | | 63.1.95 | 104653 | Al9Co3Y2 | 0 |
| | | 63.1.96 | 58157 | Al6Ru1 | 0 |
| | | 63.1.97 | 370014 | Al2Ni1Yb1 | 0 |
| | | 63.1.98 | 423614 | Al2Ge1La2 | 0 |
| | | 63.1.99 | 99138 | Al1Si1Y1 | 0 |
| | | 63.1.100 | 415373 | Al1Lu1Si1 | 0 |
| | | 63.1.101 | 76295 | Al1Ge1Y1 | 0 |
| | | 63.1.102 | 426322 | Al1Fe1.5Si2 | 0 |
| | | 63.1.103 | 93684 | Ag3Cs1Se5Tb2 | 0 |
| 63 | 2 | 63.2.1 | 43419 | U1 | 2 |
| | | 63.2.2 | 43339 | U1 | 2 |
| | | 63.2.3 | 411981 | Sn5Sr3 | 2 |
| | | 63.2.4 | 165183 | Sn5Sr3 | 2 |
| | | 63.2.5 | 54756 | Sn4Sr1 | 2 |
| | | 63.2.6 | 281669 | Sn4Sr1 | 2 |
| | | 63.2.7 | 73619 | Si3Tb2 | 2 |
| | | 63.2.8 | 95179 | Si2Zr1 | 2 |
| | | 63.2.9 | 20284 | Si1Zr1 | 2 |
| | | 63.2.10 | 22235 | Sb2Yb1 | 2 |
| | | 63.2.11 | 163454 | Rh1Zr1 | 2 |
| | | 63.2.12 | 55546 | Pd1Zr1 | 2 |
| | | 63.2.13 | 165577 | Pd1Sr1Tl2 | 2 |
| | | 63.2.14 | 78010 | O7Pr3Sb1 | 2 |
| | | 63.2.15 | 96322 | Ni1Zr1 | 2 |
| | | 63.2.16 | 105478 | Ni1Zr1 | 2 |
| | | 63.2.17 | 71964 | Ni1O1Zr3 | 2 |
| | | 63.2.18 | 63084 | Nd3O7Ru1 | 2 |
| | | 63.2.19 | 190762 | Nd3O7Ru1 | 2 |
| | | 63.2.20 | 71692 | Nd1Ni1Si2 | 2 |
| | | 63.2.21 | 71691 | Nd1Ni1Si2 | 2 |
| | | 63.2.22 | 425493 | Mg2Pd1Sr1 | 2 |
| | | 63.2.23 | 163453 | Ir1Zr1 | 2 |
| | | 63.2.24 | 412908 | In4Pt1Yb1 | 2 |
| | | 63.2.25 | 418180 | In4Pt1Sr1 | 2 |
| | | 63.2.26 | 411398 | In4Pd1Yb1 | 2 |
| | | 63.2.27 | 418179 | In4Pd1Sr1 | 2 |
| | | 63.2.28 | 415937 | In4Pb1Sr3 | 2 |
| | | 63.2.29 | 410400 | In2Pt1Sr1 | 2 |
| | | 63.2.30 | 410402 | In2Pd1Sr1 | 2 |
| | | 63.2.31 | 190810 | Hf2Pt3 | 2 |
| | | 63.2.32 | 15051 | Hf1Te5 | 2 |
| | | 63.2.33 | 171235 | Ge6Pr2Zn3 | 2 |
| | | 63.2.34 | 171234 | Ge6Nd2Zn3 | 2 |
| | | 63.2.35 | 171232 | Ge6La2Zn3 | 2 |
| | | 63.2.36 | 56047 | Ge2Th1 | 2 |
| | | 63.2.37 | 240343 | Ge2Ni1Sr1 | 2 |
| | | 63.2.38 | 240728 | Ge1Na2Zn1 | 2 |
| | | 63.2.39 | 163917 | Gd3Ge4 | 2 |
| | | 63.2.40 | 104087 | Gd1Ni1 | 2 |

Continued on next column

**TABLE XVI– continued**

| SG | cl | TQC | ICSD | Stoich. Formula | SubSG |
|---|---|---|---|---|---|
| | | 63.2.41 | 423070 | Ga4Ni1Yb1 | 2 |
| | | 63.2.42 | 154676 | Ga4Ni1Yb1 | 2 |
| | | 63.2.43 | 262740 | Ga2Ni1Yb1 | 2 |
| | | 63.2.44 | 262739 | Ga2Ni1Yb1 | 2 |
| | | 63.2.45 | 262738 | Ga2Ni1Yb1 | 2 |
| | | 63.2.46 | 262737 | Ga2Ni1Yb1 | 2 |
| | | 63.2.47 | 423623 | Ga1La1 | 2 |
| | | 63.2.48 | 2795 | Ga1 | 2 |
| | | 63.2.49 | 290249 | Fe4O5 | 2 |
| | | 63.2.50 | 98 | Fe1S3U1 | 2 |
| | | 63.2.51 | 261040 | Fe1S3U1 | 2 |
| | | 63.2.52 | 261039 | Fe1S3U1 | 2 |
| | | 63.2.53 | 261038 | Fe1S3U1 | 2 |
| | | 63.2.54 | 410700 | Eu1Ir1Sn2 | 2 |
| | | 63.2.55 | 411268 | Eu1In2Rh1 | 2 |
| | | 63.2.56 | 404599 | Eu1Ge2Ir1 | 2 |
| | | 63.2.57 | 425774 | Er3Ge3Ru2 | 2 |
| | | 63.2.58 | 55524 | Er1Ge3 | 2 |
| | | 63.2.59 | 55523 | Er1Ge3 | 2 |
| | | 63.2.60 | 59896 | Cu1Gd2Rb1S4 | 2 |
| | | 63.2.61 | 416795 | Cu1Eu1Sn2 | 2 |
| | | 63.2.62 | 413681 | Cs1Cu3Se5Tm2 | 2 |
| | | 63.2.63 | 96226 | Cs1Cu3Se5Tb2 | 2 |
| | | 63.2.64 | 413680 | Cs1Cu3Se5Tb2 | 2 |
| | | 63.2.65 | 193548 | Cr2Mg1O4 | 2 |
| | | 63.2.66 | 192389 | Cr2Fe1O4 | 2 |
| | | 63.2.67 | 20951 | Co3Ga9Y2 | 2 |
| | | 63.2.68 | 102378 | Co2Eu1Ga3 | 2 |
| | | 63.2.69 | 246169 | Ce1Pt1Si2 | 2 |
| | | 63.2.70 | 42711 | Ce1Ni1Si2 | 2 |
| | | 63.2.71 | 154440 | Ce1Ni1 | 2 |
| | | 63.2.72 | 157363 | Ce1In4Pd1 | 2 |
| | | 63.2.73 | 150132 | Ce1In2Pd1 | 2 |
| | | 63.2.74 | 106394 | Ce1Ge2Ni1 | 2 |
| | | 63.2.75 | 425494 | Cd2Pd1Sr1 | 2 |
| | | 63.2.76 | 425064 | Cd2Nd1Pd1 | 2 |
| | | 63.2.77 | 421432 | Ca4Ge3Ni4 | 2 |
| | | 63.2.78 | 58948 | Ca3Zn1 | 2 |
| | | 63.2.79 | 78996 | Ca1Si1 | 2 |
| | | 63.2.80 | 190659 | Ca1Si1 | 2 |
| | | 63.2.81 | 410890 | Ca1In4Pd1 | 2 |
| | | 63.2.82 | 106347 | Ca1In4Ni1 | 2 |
| | | 63.2.83 | 240342 | Ca1Ge2Ni1 | 2 |
| | | 63.2.84 | 710075 | Ca1Ga2Ni1 | 2 |
| | | 63.2.85 | 16355 | Ca1Fe4O6 | 2 |
| | | 63.2.86 | 260482 | C1P1Ru2Tb1 | 2 |
| | | 63.2.87 | 159578 | C1Gd1Ru2Si1 | 2 |
| | | 63.2.88 | 50633 | C1Dy1Ru2Si1 | 2 |
| | | 63.2.89 | 40773 | C1Dy1Fe2Si1 | 2 |
| | | 63.2.90 | 411622 | C1B3Nb3 | 2 |
| | | 63.2.91 | 107610 | Be1Pd3 | 2 |
| | | 63.2.92 | 411980 | Ba3Sn5 | 2 |
| | | 63.2.93 | 167166 | Ba3Sn5 | 2 |
| | | 63.2.94 | 167163 | Ba3Ge5 | 2 |
| | | 63.2.95 | 429723 | Ba2Ge4Ir4Mg1 | 2 |
| | | 63.2.96 | 418610 | Ba1Zn5 | 2 |
| | | 63.2.97 | 405111 | Ba1Pd1Sb2 | 2 |
| | | 63.2.98 | 14110 | Ba1Ni2Si2 | 2 |
| | | 63.2.99 | 411228 | Ba1In2Pt1 | 2 |
| | | 63.2.100 | 414228 | Ba1In1 | 2 |
| | | 63.2.101 | 429115 | Ba1Cd2Pt1 | 2 |

Continued on next column

**TABLE XVI– continued**

| SG | cl | TQC | ICSD | Stoich. Formula | SubSG |
|---|---|---|---|---|---|
| | | 63.2.102 | 39663 | B7W1Y3 | 2 |
| | | 63.2.103 | 62282 | B7Co4Nb3 | 2 |
| | | 63.2.104 | 43662 | B1Re3 | 2 |
| | | 63.2.105 | 107505 | Au1In2Na1 | 2 |
| | | 63.2.106 | 411390 | Au1Eu1In2 | 2 |
| | | 63.2.107 | 107515 | Au1Co1Na4O5 | 2 |
| | | 63.2.108 | 290262 | As4Cd5Rb2 | 2 |
| | | 63.2.109 | 15032 | As3Nb4 | 2 |
| | | 63.2.110 | 405110 | As2Ba1Pd1 | 2 |
| | | 63.2.111 | 153384 | Al2Ba1Si2 | 2 |
| | | 63.2.112 | 415372 | Al1Si1Tm1 | 2 |
| | | 63.2.113 | 54470 | Al1La1 | 2 |
| | | 63.2.114 | 251809 | Al1B1W1 | 2 |
| | | 63.2.115 | 251808 | Al1B1Mo1 | 2 |
| | | 63.2.116 | 182114 | Al10Ru2U1 | 2 |
| | | 63.2.117 | 99179 | Al10Fe2U1 | 2 |
| | | 63.2.118 | 152354 | Al10Fe2Th1 | 2 |
| | | 63.2.119 | 59912 | Al10Ce1Ru2 | 2 |
| | | 63.2.120 | 238167 | Ag1.2Cs0.4Er0.8Te2 | 2 |
| 64 | 1 | 64.1.1 | 411192 | Sn3Ti2 | 2 |
| | | 64.1.2 | 410902 | Sn3Ti2 | 2 |
| | | 64.1.3 | 413279 | Pd8Sn24 | 2 |
| | | 64.1.4 | 417180 | P1 | 2 |
| | | 64.1.5 | 23836 | P1 | 2 |
| | | 64.1.6 | 150873 | P1 | 2 |
| | | 64.1.8 | 79878 | K1Na2Ni1O2 | 2 |
| | | 64.1.9 | 40442 | Hf2Ni3Si4 | 2 |
| | | 64.1.11 | 150569 | Ga1Sn2V2 | 2 |
| | | 64.1.12 | 106752 | Ga1Sn2V2 | 2 |
| | | 64.1.19 | 58636 | Ba2Bi4Cd3 | 2 |
| | | 64.1.20 | 416447 | Ba2Bi4Cd3 | 2 |
| | | 64.1.21 | 416446 | Ba2Bi4Cd3 | 2 |
| | | 64.1.22 | 416445 | Ba2Bi4Cd3 | 2 |
| | | 64.1.23 | 416444 | Ba2Bi4Cd3 | 2 |
| | | 64.1.25 | 63708 | As2Ni3O8 | 2 |
| | | 64.1.26 | 68347 | Ag1Fe2S3 | 2 |
| 64 | 2 | 64.1.7 | 77831 | La1S1 | 0 |
| | | 64.1.10 | 90341 | Ge2Ni1 | 0 |
| | | 64.1.13 | 71196 | Co2Nb2Te4 | 0 |
| | | 64.1.14 | 71519 | Co1Nb1Te2 | 0 |
| | | 64.1.15 | 424937 | Co1Cs1O2 | 0 |
| | | 64.1.16 | 40372 | Ca5Ga2N4 | 0 |
| | | 64.1.17 | 103352 | Ca2Mn1O3.5 | 0 |
| | | 64.1.18 | 186936 | Ba2Ge4Pd5 | 0 |
| | | 64.1.24 | 412877 | Ba1Fe1K1O3 | 0 |
| 65 | 1 | 65.1.8 | 54612 | Ho2Ni2Pb1 | 65 |
| | | 65.1.9 | 60646 | Ge3Ni1Y1 | 65 |
| | | 65.1.12 | 261512 | Fe2O5Sr3 | 65 |
| | | 65.1.16 | 418215 | Cu1Mg4Tb1 | 65 |
| | | 65.1.22 | 78621 | Ba2Cu4O8Yb1 | 65 |
| | | 65.1.23 | 154010 | Ba1Mn2O6Tb1 | 65 |
| | | 65.1.24 | 106291 | Au7Rb3 | 65 |
| | | 65.1.25 | 106290 | Au7Rb3 | 65 |
| | | 65.1.26 | 418529 | Au4Ce1Si2 | 65 |
| 65 | 2 | 65.1.1 | 87827 | Ni1Si3Y1 | 10 |
| | | 65.1.4 | 261871 | La2Sn5 | 10 |
| | | 65.1.7 | 415513 | In16Pt7Tb2 | 10 |
| | | 65.1.10 | 421933 | Gd2Mg3Ni2 | 10 |
| | | 65.1.13 | 238922 | Fe2Ge7U3 | 10 |
| | | 65.1.20 | 419476 | C3B2Tb2 | 10 |
| | | 65.1.21 | 413212 | Br2Ce4Ga5 | 10 |

Continued on next column

**TABLE XVI– continued**

| SG | cl | TQC | ICSD | Stoich. Formula | SubSG |
|---|---|---|---|---|---|
| 65 | 3 | 65.1.2 | 78760 | Mn2S4Sn1 | 10 |
| | | 65.1.5 | 167637 | In9Ni4Y11 | 10 |
| | | 65.1.11 | 20337 | Fe2Si7U3 | 10 |
| | | 65.1.18 | 63667 | Ce3Sn7 | 10 |
| | | 65.1.19 | 261436 | Cd1Cu2Nd2 | 10 |
| 65 | 4 | 65.1.3 | 240761 | Mg3Ni2Tb2 | 0 |
| | | 65.1.6 | 167624 | In9La11Ni4 | 0 |
| | | 65.1.14 | 174439 | Er3Sn7 | 0 |
| | | 65.1.15 | 165783 | Dy2Ni1Sn6 | 0 |
| | | 65.1.17 | 424317 | Ce7Ge10Li8 | 0 |
| 65 | 5 | 65.2.11 | 91064 | Fe4O11Sr4 | 2 |
| | | 65.2.15 | 418093 | Co2Ge7U3 | 2 |
| | | 65.2.16 | 72872 | Co1O4Re1 | 2 |
| | | 65.2.19 | 39221 | C4Er4Ni13 | 2 |
| | | 65.2.31 | 167535 | Al1B4Lu1 | 2 |
| 65 | 6 | 65.2.1 | 15927 | Mg1O3V1 | 2 |
| | | 65.2.3 | 106923 | Lu1Ni1Sn4 | 2 |
| | | 65.2.4 | 83885 | Ge2U1 | 2 |
| | | 65.2.5 | 82702 | Ge2U1 | 2 |
| | | 65.2.6 | 236995 | Ge2U1 | 2 |
| | | 65.2.9 | 40487 | Ge2Lu1Mn1 | 2 |
| | | 65.2.12 | 171189 | Er1Ni1Si3 | 2 |
| | | 65.2.13 | 154835 | Dy1Fe4Ge2 | 2 |
| | | 65.2.14 | 419475 | Cu1Mg4Y1 | 2 |
| | | 65.2.18 | 174528 | Ce2Si7 | 2 |
| | | 65.2.26 | 68002 | B6Lu2Ni3 | 2 |
| | | 65.2.27 | 417990 | B4Rh8Zn5 | 2 |
| | | 65.2.29 | 260798 | Al1B4Yb1 | 2 |
| | | 65.2.30 | 167533 | Al1B4Yb1 | 2 |
| 65 | 7 | 65.2.2 | 104809 | Lu2Ni1Sn6 | 2 |
| | | 65.2.7 | 56046 | Ge2Th1 | 2 |
| | | 65.2.20 | 185392 | Br2Cd2Hg2S2 | 2 |
| | | 65.2.25 | 39394 | B6Ni3Yb2 | 2 |
| | | 65.2.28 | 413191 | Al5Br2La4 | 2 |
| 65 | 8 | 65.2.8 | 193219 | Ge2Pb1Sr2 | 2 |
| | | 65.2.10 | 104134 | Gd3Sn7 | 2 |
| | | 65.2.17 | 16497 | Ce3Ni2Si8 | 2 |
| | | 65.2.21 | 391060 | Ba3Ge4 | 2 |
| | | 65.2.22 | 75687 | Ba2Cu4Er1O8 | 2 |
| | | 65.2.23 | 74127 | Ba2Cu4Er1O8 | 2 |
| | | 65.2.24 | 67642 | Ba2Cu4Er1O8 | 2 |
| 67 | 1 | 67.2.1 | 167666 | Al2Cu1Ir1 | 2 |
| | | 69.1.1 | 262559 | Pt3Rb2Se6U1 | 69 |
| | | 69.1.2 | 262561 | Pd3Rb2Se6U1 | 69 |
| | | 69.1.3 | 262562 | Pd3Rb2S6U1 | 69 |
| | | 69.1.4 | 262563 | K2Pd3S6U1 | 69 |
| | | 69.1.5 | 414574 | Ge6Li2Sr4 | 69 |
| | | 69.1.7 | 262555 | Cs2Pt3Se6U1 | 69 |
| | | 69.1.9 | 262557 | Cs2Pd3Se6U1 | 69 |
| | | 69.1.10 | 261242 | Cs2Pd3Se6U1 | 69 |
| | | 69.1.11 | 262558 | Cs2Pd3S6U1 | 69 |
| | | 69.1.18 | 410522 | Ag2Ca4Si6 | 69 |
| 69 | 2 | 69.1.6 | 68676 | Cu2O3Sr2 | 2 |
| | | 69.1.8 | 262556 | Cs2Pt3S6U1 | 2 |
| | | 69.1.12 | 73160 | Ba2Na1Ni3O6 | 2 |
| | | 69.1.13 | 59588 | Ba2Na1Ni3O6 | 2 |
| | | 69.1.14 | 72328 | Ba2Cu3Na1O6 | 2 |
| | | 69.1.15 | 59589 | Ba2Cu3Na1O6 | 2 |
| | | 69.1.16 | 401239 | Ba2Cu3Li1O6 | 2 |
| | | 69.1.17 | 174521 | Au1Eu2Ge3 | 2 |

Continued on next column

**TABLE XVI– continued**

| SG | cl | TQC | ICSD | Stoich. Formula | SubSG |
|---|---|---|---|---|---|
| 70 | 1 | 70.1.1 | 1089 | Si2Ti1 | 2 |
| | | 70.1.2 | 97545 | Sb1Sn1Ti1 | 2 |
| | | 70.1.3 | 88369 | O4Tb1V1 | 2 |
| | | 70.1.4 | 106552 | Nb1Sn2 | 2 |
| | | 70.1.5 | 429511 | Ga2Rh2Sr1 | 2 |
| | | 70.1.6 | 89074 | Cr2O4Si1 | 2 |
| | | 70.1.7 | 89073 | Cr2O4Si1 | 2 |
| | | 70.1.8 | 89072 | Cr2O4Si1 | 2 |
| | | 70.1.9 | 89071 | Cr2O4Si1 | 2 |
| | | 70.1.10 | 89070 | Cr2O4Si1 | 2 |
| | | 70.1.11 | 89069 | Cr2O4Si1 | 2 |
| | | 70.1.12 | 75639 | Cr2O4Si1 | 2 |
| | | 70.1.13 | 96610 | Cl2Cu2O1 | 2 |
| | | 70.1.14 | 77143 | Ba1Ge2Ru2 | 2 |
| | | 70.1.15 | 40620 | B2Gd1Ir2 | 2 |
| | | 70.1.16 | 40799 | B2Ce1Ru2 | 2 |
| | | 70.1.17 | 40618 | B2Ce1Ir2 | 2 |
| | | 70.1.18 | 49737 | B2Ca1Rh2 | 2 |
| | | 70.1.19 | 49738 | B2Ca1Ir2 | 2 |
| | | 70.1.20 | 75028 | B1Ir1Li1 | 2 |
| | | 70.1.21 | 75027 | B1Ir1Li1 | 2 |
| | | 70.1.22 | 418163 | B1Cr2 | 2 |
| | | 70.1.23 | 410521 | Ag2Eu4Si6 | 2 |
| | | 71.1.18 | 85406 | Ni2P2Sr1 | 2 |
| | | 71.1.21 | 56747 | H1Li1Nd2O3 | 2 |
| | | 71.1.22 | 56746 | H1Li1Nd2O3 | 2 |
| | | 71.1.27 | 262435 | Ge4Li4Nd3 | 2 |
| | | 71.1.32 | 300260 | Ge2Ho1Pt1 | 2 |
| | | 71.1.35 | 154471 | Fe6Ga6U1 | 2 |
| | | 71.1.39 | 413485 | Cu6Sn5Yb3 | 2 |
| | | 71.1.40 | 426132 | Cu4Ge4Yb3 | 2 |
| | | 71.1.47 | 174076 | Ce3Pd4Si4 | 2 |
| | | 71.1.49 | 151160 | Ce3Cu4Sn4 | 2 |
| | | 71.1.60 | 238028 | C1Co2Pr2Si1 | 2 |
| | | 71.1.61 | 418403 | C1B4Dy2 | 2 |
| | | 71.1.65 | 72631 | Ba1Ni1O5Tm2 | 2 |
| | | 71.1.66 | 71415 | Ba1Ni1O5Tm2 | 2 |
| | | 71.1.67 | 67831 | Ba1Ni1O5Tm2 | 2 |
| | | 71.1.68 | 62761 | Ba1Ni1O5Tm2 | 2 |
| | | 71.1.69 | 62426 | Ba1Ni1O5Tm2 | 2 |
| | | 71.1.72 | 72626 | Ba1Nd2Ni1O5 | 2 |
| | | 71.1.73 | 62627 | Ba1Nd2Ni1O5 | 2 |
| | | 71.1.86 | 81543 | B6Cr2Th1 | 2 |
| | | 71.1.87 | 81544 | B6Ce1Cr2 | 2 |
| | | 71.1.90 | 407576 | B2N4Nd3 | 2 |
| | | 71.1.91 | 68610 | B2Ce3N4 | 2 |
| | | 71.1.92 | 281420 | As1Te1Ti1 | 2 |
| | | 71.1.93 | 58046 | Al5Ni2Pr1 | 2 |
| | | 71.1.95 | 9564 | Al2Sn2Sr3 | 2 |
| | | 71.1.96 | 100127 | Al2Si2Sr3 | 2 |
| | | 71.1.97 | 52617 | Al2Ge2Sr3 | 2 |
| | | 71.1.98 | 100126 | Al2Ca3Si2 | 2 |
| | | 71.1.100 | 9565 | Al2Ba3Sn2 | 2 |
| | | 71.1.105 | 417491 | Ag4Ce3Ge4 | 2 |
| 71 | 2 | 71.1.28 | 170342 | Ge2Ni1Yb1 | 0 |
| | | 71.1.29 | 170343 | Ge2Lu1Ni1 | 0 |
| | | 71.1.31 | 88227 | Ge2Ho1Pt1 | 0 |
| | | 71.1.37 | 83190 | Cu8Ge8Nd6 | 0 |
| | | 71.1.38 | 102956 | Cu8Gd6Sn8 | 0 |
| | | 71.1.43 | 42717 | Cu4Gd3Ge4 | 0 |
| | | 71.1.46 | 247288 | Ce3Rh4Si4 | 0 |

Continued on next column

**TABLE XVI– continued**

| SG | cl | TQC | ICSD | Stoich. Formula | SubSG |
|---|---|---|---|---|---|
| | | 71.1.56 | 40950 | C4Co1Sc3 | 0 |
| | | 71.1.63 | 290306 | Ba3Cd1Hg10 | 0 |
| | | 71.1.77 | 72630 | Ba1Er2Ni1O5 | 0 |
| | | 71.1.78 | 69324 | Ba1Er2Ni1O5 | 0 |
| | | 71.1.79 | 69323 | Ba1Er2Ni1O5 | 0 |
| | | 71.1.80 | 69315 | Ba1Er2Ni1O5 | 0 |
| | | 71.1.81 | 62760 | Ba1Er2Ni1O5 | 0 |
| | | 71.1.82 | 62425 | Ba1Er2Ni1O5 | 0 |
| | | 71.1.83 | 85046 | Ba1Dy2Ni1O5 | 0 |
| | | 71.1.84 | 72627 | Ba1Dy2Ni1O5 | 0 |
| | | 71.1.103 | 163921 | Al16Ce2Pt9 | 0 |
| | | 71.1.104 | 157372 | Ag4Gd3Sn4 | 0 |
| 71 | 3 | 71.1.1 | 82655 | Te2U1 | 2 |
| | | 71.1.2 | 82654 | Te2U1 | 2 |
| | | 71.1.3 | 82653 | Te2U1 | 2 |
| | | 71.1.4 | 82652 | Te2U1 | 2 |
| | | 71.1.5 | 82651 | Te2U1 | 2 |
| | | 71.1.6 | 82650 | Te2U1 | 2 |
| | | 71.1.7 | 82649 | Te2U1 | 2 |
| | | 71.1.8 | 82648 | Te2U1 | 2 |
| | | 71.1.9 | 82647 | Te2U1 | 2 |
| | | 71.1.10 | 82646 | Te2U1 | 2 |
| | | 71.1.11 | 82645 | Te2U1 | 2 |
| | | 71.1.12 | 82644 | Te2U1 | 2 |
| | | 71.1.13 | 82643 | Te2U1 | 2 |
| | | 71.1.14 | 82642 | Te2U1 | 2 |
| | | 71.1.15 | 63170 | Te2U1 | 2 |
| | | 71.1.16 | 40694 | Te2U1 | 2 |
| | | 71.1.20 | 416817 | La2Ni2Zn1 | 2 |
| | | 71.1.23 | 56745 | H1Ce2Li1O3 | 2 |
| | | 71.1.24 | 710017 | Ge5Nd1 | 2 |
| | | 71.1.25 | 246814 | Ge5Nd1 | 2 |
| | | 71.1.33 | 174530 | Gd1Si5 | 2 |
| | | 71.1.34 | 710016 | Gd1Ge5 | 2 |
| | | 71.1.36 | 187334 | Fe4Ge4U3 | 2 |
| | | 71.1.41 | 82565 | Cu4Ge4U3 | 2 |
| | | 71.1.42 | 81142 | Cu4Ge4U3 | 2 |
| | | 71.1.51 | 174529 | Ce1Si5 | 2 |
| | | 71.1.52 | 246812 | Ce1Ge5 | 2 |
| | | 71.1.57 | 236393 | C4Co1Sc3 | 2 |
| | | 71.1.58 | 173167 | C4Co1Sc3 | 2 |
| | | 71.1.70 | 66078 | Ba1Ni1O5Tb2 | 2 |
| | | 71.1.101 | 100128 | Al2Ba3Si2 | 2 |
| | | 71.1.102 | 52612 | Al2Ba3Ge2 | 2 |
| 71 | 4 | 71.1.17 | 416820 | Ni2Tb2Zn1 | 71 |
| | | 71.1.19 | 25307 | Li4Sb4Sr3 | 71 |
| | | 71.1.26 | 411182 | Ge4Pd4Yb3 | 71 |
| | | 71.1.30 | 76509 | Ge2Ir1Y1 | 71 |
| | | 71.1.44 | 55650 | Cu4Er3Si4 | 71 |
| | | 71.1.45 | 201112 | Cs1Fe1S2 | 71 |
| | | 71.1.48 | 262434 | Ce3Ge4Li4 | 71 |
| | | 71.1.50 | 416816 | Ce2Ni2Zn1 | 71 |
| | | 71.1.53 | 429175 | Ca2Mg1Pt2 | 71 |
| | | 71.1.54 | 429174 | Ca2Mg1Pd2 | 71 |
| | | 71.1.55 | 425471 | Ca2Cd1Pd2 | 71 |
| | | 71.1.59 | 94741 | C2Bi2Ca1O8 | 71 |
| | | 71.1.62 | 428114 | Ba3Li4Sb4 | 71 |
| | | 71.1.64 | 428115 | Ba3Bi4Li4 | 71 |
| | | 71.1.71 | 75035 | Ba1Ni1O5Pr2 | 71 |
| | | 71.1.74 | 68425 | Ba1Gd2Ni1O5 | 71 |
| | | 71.1.75 | 62758 | Ba1Gd2Ni1O5 | 71 |

Continued on next column

**TABLE XVI– continued**

| SG | cl | TQC | ICSD | Stoich. Formula | SubSG |
|---|---|---|---|---|---|
| | | 71.1.76 | 62423 | Ba1Gd2Ni1O5 | 71 |
| | | 71.1.85 | 167153 | B6Fe2U1 | 71 |
| | | 71.1.88 | 40791 | B4Cr3 | 71 |
| | | 71.1.89 | 423380 | B4Cr1 | 71 |
| | | 71.1.94 | 418387 | Al4O1Si6Sr10 | 71 |
| | | 71.1.99 | 52616 | Al2Ca3Ge2 | 71 |
| 72 | 1 | 72.1.1 | 84206 | Ni3Si5Y2 | 0 |
| | | 72.1.2 | 428478 | Eu2Pt3Si5 | 0 |
| 72 | 2 | 72.2.1 | 170063 | Pb5Sc6 | 2 |
| | | 72.2.2 | 416512 | Ni1Pd2Te2 | 2 |
| | | 72.2.3 | 416515 | Fe1Pd2Se2 | 2 |
| | | 72.2.4 | 68792 | Er2Ru3Si5 | 2 |
| | | 72.2.5 | 54107 | Cr4Sc2Si5 | 2 |
| | | 72.2.6 | 40758 | C4Fe1Tm2 | 2 |
| | | 72.2.7 | 28232 | Br2Hg5O4 | 2 |
| 74 | 1 | 74.2.2 | 429139 | Ga2La1Rh3 | 2 |
| | | 74.2.3 | 238760 | Fe2O4Si1 | 2 |
| | | 74.2.4 | 238759 | Fe2O4Si1 | 2 |
| | | 74.2.5 | 238758 | Fe2O4Si1 | 2 |
| | | 74.2.6 | 93229 | Fe2Mo4O7 | 2 |
| | | 74.2.9 | 246171 | Ce3Pt5Si1 | 2 |
| | | 74.2.10 | 93244 | Ce3Pd5Si1 | 2 |
| | | 74.2.12 | 97094 | Ca2O7Os2 | 2 |
| | | 74.2.13 | 97093 | Ca2O7Os2 | 2 |
| | | 74.2.14 | 97092 | Ca2O7Os2 | 2 |
| | | 74.2.15 | 97091 | Ca2O7Os2 | 2 |
| | | 74.2.18 | 414229 | Ba1In2 | 2 |
| | | 74.2.22 | 55542 | Au2Ca1 | 2 |
| | | 74.2.24 | 417225 | Ag2Ca1 | 2 |
| | | 74.2.25 | 417227 | Ag1Ca1Zn1 | 2 |
| 74 | 2 | 74.2.1 | 106234 | Yb1Zn2 | 0 |
| | | 74.2.7 | 74288 | Cs1F5Ni1Pd1 | 0 |
| | | 74.2.8 | 10156 | Co2O4Si1 | 0 |
| | | 74.2.11 | 260381 | Cd2Sr1 | 0 |
| | | 74.2.16 | 418971 | Ca1Zn2 | 0 |
| | | 74.2.17 | 420576 | Ca1Cd2 | 0 |
| | | 74.2.19 | 251586 | Ba1Ge5 | 0 |
| | | 74.2.20 | 194880 | Ba1Ge5 | 0 |
| | | 74.2.21 | 260668 | Ba1Cd2 | 0 |
| | | 74.2.23 | 72289 | Al1F7Na2Ni1 | 0 |
| 82 | 1 | 82.2.1 | 81539 | Te1Zr3 | 82 |
| | | 82.2.2 | 51048 | Te1Zr3 | 82 |
| | | 82.2.3 | 195057 | Sb1Zr3 | 82 |
| | | 82.2.4 | 98373 | Ni3P1 | 82 |
| | | 82.2.5 | 23612 | In1P1S4 | 82 |
| | | 82.2.6 | 56027 | Ge1Ta3 | 82 |
| | | 82.2.7 | 43365 | Fe3P1 | 82 |
| | | 82.2.8 | 23560 | Cr3P1 | 82 |
| | | 83.2.1 | 106354 | Ca7Ni4Sn13 | 2,81 |
| | | 84.2.1 | 74390 | P2S6U1 | 0 |
| | | 85.2.1 | 33884 | As2Cl4F6 | 2,81 |
| | | 87.1.1 | 51047 | Te4Zr5 | 2,82 |
| | | 87.1.2 | 409337 | Te4Zr5 | 2,82 |
| | | 87.1.3 | 55075 | Se4Ti5 | 2,82 |
| | | 87.1.6 | 57076 | Ge4Li4Yb5 | 2,82 |
| | | 87.1.11 | 66017 | As4Ba1Cu8 | 2,82 |
| 87 | 2 | 87.1.8 | 60822 | C4Cr4U1 | 2,82 |
| | | 87.1.9 | 40319 | C4Cr4U1 | 2,82 |
| | | 87.1.10 | 43186 | As4Mo5 | 2,82 |
| 87 | 3 | 87.1.4 | 157026 | Mo1Ni1O6Sr2 | 0 |

Continued on next column

**TABLE XVI– continued**

| SG | cl | TQC | ICSD | Stoich. Formula | SubSG |
|---|---|---|---|---|---|
| | | 87.1.5 | 157024 | Mo1Ni1O6Sr2 | 0 |
| | | 87.1.7 | 173488 | Co1O6Re1Sr2 | 0 |
| 113 | 1 | 113.2.1 | 246245 | Ca1Fe1O2 | 81 |
| | | 113.2.2 | 39120 | Be2Gd2Ge1O7 | 81 |
| | | 113.2.3 | 160121 | Ag1Tb1Te2 | 81 |
| | | 113.2.4 | 154793 | Ag1Tb1Te2 | 81 |
| | | 115.1.1 | 93242 | Cu1Sb3Zr2 | 81 |
| | | 115.1.2 | 426313 | Cu1Sb3Zr2 | 81 |
| | | 115.1.3 | 195058 | Cu1Sb3Zr2 | 81 |
| | | 115.1.4 | 93241 | Cu1Sb3Ti2 | 81 |
| | | 115.1.5 | 93243 | Cu1Hf2Sb3 | 81 |
| | | 121.1.1 | 237706 | Mo3P1 | 82 |
| | | 121.1.2 | 104607 | Ir1Zr3 | 82 |
| | | 121.1.3 | 400652 | Cu3Sb1Se4 | 82 |
| | | 121.1.4 | 412239 | Cu3S4Sb1 | 82 |
| | | 121.1.5 | 2857 | Cu3S4Sb1 | 82 |
| | | 121.1.6 | 95119 | Cu2Hg1Se4Sn1 | 82 |
| | | 121.1.7 | 152761 | Cu2Ge1Hg1Se4 | 82 |
| | | 121.1.8 | 85126 | Cu2Fe1Se4Sn1 | 82 |
| | | 121.1.9 | 165094 | Cd1Cu2Ge1Te4 | 82 |
| | | 122.1.1 | 248539 | O6Rb4 | 82 |
| | | 122.1.2 | 659 | F5U1 | 82 |
| | | 122.1.3 | 16427 | F5U1 | 82 |
| | | 122.1.4 | 16426 | F5U1 | 82 |
| | | 122.1.5 | 16425 | F5U1 | 82 |
| | | 122.1.6 | 16424 | F5U1 | 82 |
| | | 122.1.7 | 84027 | Cu1In1Se2 | 82 |
| | | 122.1.8 | 73351 | Cu1In1Se2 | 82 |
| | | 122.1.9 | 154409 | Cu1In1Se2 | 82 |
| | | 122.1.10 | 154408 | Cu1In1Se2 | 82 |
| | | 122.1.11 | 154407 | Cu1In1Se2 | 82 |
| | | 122.1.12 | 154406 | Cu1In1Se2 | 82 |
| | | 122.1.13 | 154405 | Cu1In1Se2 | 82 |
| | | 122.1.14 | 68325 | As2Sn1Zn1 | 82 |
| | | 122.1.15 | 42098 | As2Cd1Ge1 | 82 |
| | | 123.1.17 | 74163 | Ba2Cu2O7Tl1Y1 | 10,65,83 |
| | | 123.1.24 | 170662 | Ag1Sb1Te2 | 10,65,83 |
| 123 | 2 | 123.1.1 | 193701 | In5Rh1U1 | 10 |
| | | 123.1.2 | 103639 | Ga5Ir1U1 | 10 |
| | | 123.1.5 | 102455 | Co1Ga5U1 | 10 |
| | | 123.1.22 | 262038 | As3Rb1Zn4 | 10 |
| | | 123.1.23 | 262030 | As3Cs1Zn4 | 10 |
| 123 | 3 | 123.1.21 | 165631 | Ba1Ge3Mg4 | 10 |
| 123 | 4 | 123.1.20 | 237789 | Ba1O1Sb2Ti2 | 10,10 |
| 123 | 5 | 123.1.4 | 106640 | Fe1Ga8Tm2 | 10,10 |
| | | 123.1.18 | 430061 | Ba1O1Sb2Ti2 | 10,10 |
| | | 123.1.19 | 430060 | Ba1O1Sb2Ti2 | 10,10 |
| 123 | 6 | 123.1.16 | 68047 | Ba2Cu3Ho1O6 | 10 |
| 123 | 7 | 123.1.7 | 150226 | Ce1In5Rh1 | 83 |
| | | 123.1.8 | 110780 | Ce1In5Rh1 | 83 |
| | | 123.1.9 | 110779 | Ce1In5Rh1 | 83 |
| | | 123.1.10 | 110778 | Ce1In5Rh1 | 83 |
| | | 123.1.11 | 110777 | Ce1In5Rh1 | 83 |
| | | 123.1.12 | 150225 | Ce1In5Ir1 | 83 |
| | | 123.1.13 | 92435 | Bi2I1O4Tm1 | 83 |
| | | 123.1.14 | 92408 | Bi2Cl1O4Tm1 | 83 |
| | | 123.1.15 | 92421 | Bi2Br1O4Tm1 | 83 |
| 123 | 8 | 123.1.6 | 192621 | Ce2In8Pt1 | 10,83 |
| 123 | 9 | 123.1.3 | 5257 | Fe1Si2 | 10 |
| 123 | 10 | 123.2.31 | 251834 | Al2B1La1Os2 | 2,81 |

Continued on next column

**TABLE XVI– continued**

| SG | cl | TQC | ICSD | Stoich. Formula | SubSG |
|---|---|---|---|---|---|
| 123 | 11 | 123.2.30 | 174342 | Al2B1La1Ru2 | 2,81 |
| 123 | 12 | 123.2.2 | 428960 | Mg1Ru1Sc2 | 2,81 |
| | | 123.2.29 | 169074 | As2Ba1O1Ti2 | 2,81 |
| 123 | 13 | 123.2.28 | 165630 | Ba1Mg4Si3 | 2,81 |
| 123 | 14 | 123.2.27 | 64639 | Ba2Cu2La1O8Ta1 | 2,81 |
| 123 | 15 | 123.2.16 | 84739 | Ba2Cu3O6Y1 | 2,81 |
| | | 123.2.17 | 72254 | Ba2Cu3O6Y1 | 2,81 |
| | | 123.2.18 | 68290 | Ba2Cu3O6Y1 | 2,81 |
| | | 123.2.19 | 68046 | Ba2Cu3O6Y1 | 2,81 |
| | | 123.2.20 | 68008 | Ba2Cu3O6Y1 | 2,81 |
| | | 123.2.21 | 67016 | Ba2Cu3O6Y1 | 2,81 |
| | | 123.2.22 | 66903 | Ba2Cu3O6Y1 | 2,81 |
| | | 123.2.23 | 63269 | Ba2Cu3O6Y1 | 2,81 |
| | | 123.2.24 | 203116 | Ba2Cu3O6Y1 | 2,81 |
| | | 123.2.25 | 98114 | Ba2Cu3Lu1O6 | 2,81 |
| | | 123.2.26 | 98113 | Ba2Cu3Lu1O6 | 2,81 |
| 123 | 16 | 123.2.15 | 424316 | Ba3O5Te1Yb2 | 2,81 |
| 123 | 17 | 123.2.9 | 92425 | Bi2I1O4Pr1 | 2,81 |
| | | 123.2.10 | 92403 | Bi2Cl1O4Tb1 | 2,81 |
| | | 123.2.11 | 92398 | Bi2Cl1O4Pr1 | 2,81 |
| | | 123.2.12 | 92401 | Bi2Cl1Eu1O4 | 2,81 |
| | | 123.2.13 | 92416 | Bi2Br1O4Tb1 | 2,81 |
| | | 123.2.14 | 92411 | Bi2Br1O4Pr1 | 2,81 |
| 123 | 18 | 123.2.5 | 74164 | Cu2O7Sr2Tl1Y1 | 2,81 |
| | | 123.2.8 | 102108 | Ce1Co1In5 | 2,81 |
| 123 | 19 | 123.2.7 | 240161 | Ce1Ga6Pd1 | 2,81 |
| 123 | 20 | 123.2.6 | 85244 | Cu2Gd1O8Ru1Sr2 | 2,81 |
| 123 | 21 | 123.2.4 | 427123 | Er2Mg3Ru1 | 2,81 |
| 123 | 22 | 123.2.3 | 157999 | Fe5Si3U1 | 2,81 |
| 123 | 23 | 123.2.1 | 98585 | Nd1Ni1O2 | 2,81 |
| 125 | 1 | 125.1.1 | 412794 | Pr1Rb1Te4 | 0 |
| | | 125.1.2 | 391204 | K1Pr1Te4 | 0 |
| | | 125.1.3 | 152562 | Ce1Ga6 | 0 |
| | | 125.1.4 | 102167 | Ce1Ga6 | 0 |
| 125 | 2 | 125.2.1 | 400461 | H8I1N4P1 | 2,81 |
| 127 | 1 | 127.2.3 | 391441 | Mg1Si2Yb2 | 2,81 |
| | | 127.2.5 | 415117 | Dy2Mg1Si2 | 2,81 |
| | | 127.2.6 | 423452 | Dy2Ge2Mg1 | 2,81 |
| | | 127.2.7 | 262925 | Cs1I3Sn1 | 2,81 |
| | | 127.2.13 | 107389 | Cd1La2Rh2 | 2,81 |
| | | 127.2.14 | 424179 | Cd1Ge2Yb2 | 2,81 |
| | | 127.2.17 | 424177 | Cd1Er2Ge2 | 2,81 |
| | | 127.2.18 | 424175 | Cd1Dy2Ge2 | 2,81 |
| | | 127.2.33 | 26777 | B4Y1 | 2,81 |
| | | 127.2.34 | 2360 | B4La1 | 2,81 |
| | | 127.2.37 | 107439 | Al1Pt3 | 2,81 |
| 127 | 2 | 127.2.19 | 193226 | Ca2Ge2Sn1 | 2,81 |
| | | 127.2.36 | 51434 | B2Rh5Sc3 | 2,81 |
| 127 | 3 | 127.2.35 | 182659 | B4Ho1 | 0 |
| 127 | 4 | 127.2.8 | 413849 | Ce2Ge2Mg1 | 2,81 |
| | | 127.2.32 | 67222 | B4Y1 | 2,81 |
| 127 | 5 | 127.2.25 | 94036 | C2B2Ce1 | 10 |
| | | 127.2.26 | 88561 | C2B2Ce1 | 10 |
| | | 127.2.27 | 88560 | C2B2Ce1 | 10 |
| | | 127.2.28 | 280180 | C2B2Ce1 | 10 |
| | | 127.2.31 | 404496 | Ba1Dy2O5Pd1 | 10 |
| 127 | 6 | 127.2.11 | 157385 | Cd1Pd2Pr2 | 2,81 |
| | | 127.2.12 | 414596 | Cd1Ni2Pr2 | 2,81 |
| | | 127.2.30 | 404497 | Ba1Ho2O5Pd1 | 2,81 |
| 127 | 7 | 127.2.29 | 202173 | Ba1O5Pd1Tb2 | 10 |

Continued on next column

**TABLE XVI– continued**

| SG | cl | TQC | ICSD | Stoich. Formula | SubSG |
|---|---|---|---|---|---|
| 127 | 8 | 127.2.24 | 94040 | C2B2Dy1 | 10 |
| 127 | 9 | 127.2.22 | 427156 | C2B2Gd1 | 10 |
| | | 127.2.23 | 94042 | C2B2Er1 | 10 |
| 127 | 10 | 127.2.20 | 79239 | C5Ru6Th2 | 2,81 |
| | | 127.2.21 | 39204 | C5Ru6Th2 | 2,81 |
| 127 | 11 | 127.2.16 | 424178 | Cd1Ge2Tm2 | 2,81 |
| 127 | 12 | 127.2.15 | 424169 | Cd1Ge2Y2 | 2,81 |
| 127 | 13 | 127.2.2 | 427282 | Ni2Sn1Tb2 | 2,81 |
| | | 127.2.9 | 157389 | Cd1Pd2Yb2 | 2,81 |
| | | 127.2.10 | 156063 | Cd1Pd2Yb2 | 2,81 |
| 127 | 14 | 127.2.4 | 26200 | Li2Sn5 | 65,83 |
| 127 | 15 | 127.2.1 | 105627 | Pb3Sr2 | 65,83 |
| 128 | 1 | 128.1.1 | 425655 | B2Nb2Os1 | 83 |
| | | 129.1.1 | 80190 | Sn1Te1Zr1 | 0 |
| | | 129.1.2 | 26273 | Sn1Te1U1 | 0 |
| | | 129.1.3 | 74522 | Si1Te1Zr1 | 0 |
| | | 129.1.4 | 25734 | Si1Te1Zr1 | 0 |
| | | 129.1.5 | 15572 | Si1Te1Zr1 | 0 |
| | | 129.1.6 | 25733 | Se1Si1Zr1 | 0 |
| | | 129.1.7 | 15571 | Se1Si1Zr1 | 0 |
| | | 129.1.8 | 25732 | S1Si1Zr1 | 0 |
| | | 129.1.9 | 15570 | S1Si1Zr1 | 0 |
| | | 129.1.10 | 15569 | S1Si1Zr1 | 0 |
| | | 129.1.11 | 26271 | S1Si1U1 | 0 |
| | | 129.1.12 | 50351 | Ru1Sb2U1 | 0 |
| | | 129.1.13 | 50350 | Ru1Sb2U1 | 0 |
| | | 129.1.14 | 61194 | Pd2Sb2Sr1 | 0 |
| | | 129.1.15 | 50349 | Pd1Sb2U1 | 0 |
| | | 129.1.16 | 50348 | Pd1Sb2U1 | 0 |
| | | 129.1.17 | 77853 | P2U1 | 0 |
| | | 129.1.18 | 25731 | O1Si1Zr1 | 0 |
| | | 129.1.19 | 85856 | Nd1Ru1Si1 | 0 |
| | | 129.1.20 | 41250 | Nd1Ru1Si1 | 0 |
| | | 129.1.21 | 76235 | Mn1Si1Y1 | 0 |
| | | 129.1.22 | 73039 | Ir2Pu1Si2 | 0 |
| | | 129.1.23 | 15420 | I1O1Tm1 | 0 |
| | | 129.1.24 | 25737 | Hf1Si1Te1 | 0 |
| | | 129.1.25 | 25736 | Hf1Se1Si1 | 0 |
| | | 129.1.26 | 25735 | Hf1S1Si1 | 0 |
| | | 129.1.27 | 25740 | Ge1Te1Zr1 | 0 |
| | | 129.1.28 | 25739 | Ge1Se1Zr1 | 0 |
| | | 129.1.29 | 15574 | Ge1Se1Zr1 | 0 |
| | | 129.1.30 | 25738 | Ge1S1Zr1 | 0 |
| | | 129.1.31 | 15573 | Ge1S1Zr1 | 0 |
| | | 129.1.32 | 87315 | Ge1S1U1 | 0 |
| | | 129.1.33 | 26272 | Ge1S1U1 | 0 |
| | | 129.1.34 | 85863 | Ge1Nd1Ru1 | 0 |
| | | 129.1.35 | 25743 | Ge1Hf1Te1 | 0 |
| | | 129.1.36 | 25741 | Ge1Hf1Se1 | 0 |
| | | 129.1.37 | 25742 | Ge1Hf1S1 | 0 |
| | | 129.1.38 | 106758 | Gd1Pt2Si2 | 0 |
| | | 129.1.39 | 180602 | Fe1Te1 | 0 |
| | | 129.1.40 | 169271 | Fe1Se1 | 0 |
| | | 129.1.41 | 169269 | Fe1Se1 | 0 |
| | | 129.1.42 | 169267 | Fe1Se1 | 0 |
| | | 129.1.43 | 169265 | Fe1Se1 | 0 |
| | | 129.1.44 | 169263 | Fe1Se1 | 0 |
| | | 129.1.45 | 169261 | Fe1Se1 | 0 |
| | | 129.1.46 | 169259 | Fe1Se1 | 0 |
| | | 129.1.47 | 169257 | Fe1Se1 | 0 |

Continued on next column

**TABLE XVI– continued**

| SG | cl | TQC | ICSD | Stoich. Formula | SubSG |
|---|---|---|---|---|---|
| | | 129.1.48 | 169255 | Fe1Se1 | 0 |
| | | 129.1.49 | 169253 | Fe1Se1 | 0 |
| | | 129.1.50 | 94475 | F1Ho1S1 | 0 |
| | | 129.1.51 | 42969 | Er1Se2 | 0 |
| | | 129.1.52 | 59594 | Cu1P1Si1Zr1 | 0 |
| | | 129.1.53 | 91199 | Cu1Mn1O3S1Sr2 | 0 |
| | | 129.1.54 | 27230 | Cs1F4Ti1 | 0 |
| | | 129.1.55 | 62696 | Cs1F4Fe1 | 0 |
| | | 129.1.56 | 6043 | Cs1F4Fe1 | 0 |
| | | 129.1.57 | 16496 | Cr1N1Nb1 | 0 |
| | | 129.1.58 | 78578 | Co2P2U1 | 0 |
| | | 129.1.59 | 15770 | Co1K1O2 | 0 |
| | | 129.1.60 | 97019 | Cl1Mn1O3Sr2 | 0 |
| | | 129.1.61 | 97018 | Cl1Mn1O3Sr2 | 0 |
| | | 129.1.62 | 96044 | Cl1Mn1O3Sr2 | 0 |
| | | 129.1.63 | 94745 | Cl1Mn1O3Sr2 | 0 |
| | | 129.1.64 | 76171 | Cl1Ho1O1 | 0 |
| | | 129.1.65 | 91750 | Cl1Co1O3Sr2 | 0 |
| | | 129.1.66 | 90122 | Cl1Co1O3Sr2 | 0 |
| | | 129.1.67 | 84529 | Ce1Ni1Sb2 | 0 |
| | | 129.1.68 | 411678 | Ce1Ir2Si2 | 0 |
| | | 129.1.69 | 61371 | Ce1Ge2Pt2 | 0 |
| | | 129.1.70 | 106393 | Ce1Ge2Ir2 | 0 |
| | | 129.1.71 | 191053 | Ce1Ga2Pd2 | 0 |
| | | 129.1.72 | 92319 | Br1Nd1O3Te1 | 0 |
| | | 129.1.73 | 92318 | Bi1I1O3Te1 | 0 |
| | | 129.1.74 | 246945 | Ba1Ni1S2 | 0 |
| | | 129.1.75 | 15289 | Ba1Ni1S2 | 0 |
| | | 129.1.76 | 41915 | Ba1Mg2Sn2 | 0 |
| | | 129.1.77 | 25338 | Ba1Mg2Pb2 | 0 |
| | | 129.1.78 | 370004 | Au2Ga2Sr1 | 0 |
| | | 129.1.79 | 238706 | Au1Bi2Pr1 | 0 |
| | | 129.1.80 | 238707 | Au1Bi2Nd1 | 0 |
| | | 129.1.81 | 66765 | As2U1 | 0 |
| | | 129.1.82 | 44081 | As2U1 | 0 |
| | | 129.1.83 | 107523 | As2U1 | 0 |
| | | 129.1.84 | 56976 | As2Pd1U1 | 0 |
| | | 129.1.85 | 412554 | As2Cu1Tm1 | 0 |
| | | 129.1.86 | 412551 | As2Cu1Tb1 | 0 |
| | | 129.1.87 | 246061 | As1La1Ni1O1 | 0 |
| | | 129.1.88 | 248695 | As1Fe1O3Sr2V1 | 0 |
| | | 129.1.89 | 422004 | As1Fe1O1Tb1 | 0 |
| | | 129.1.90 | 42165 | As1Cu1Si1Zr1 | 0 |
| | | 129.1.91 | 423103 | As1Cu1Ge1Zr1 | 0 |
| | | 129.1.92 | 42750 | Al1Ge1Mn1 | 0 |
| | | 129.1.93 | 190195 | Ag1Pr1Sb2 | 0 |
| 129 | 2 | 129.2.1 | 52412 | Sc1 | 2,81 |
| | | 129.2.2 | 54346 | Pt2Si2U1 | 2,81 |
| | | 129.2.3 | 38176 | O1Te1U1 | 2,81 |
| | | 129.2.4 | 22204 | O1Te1U1 | 2,81 |
| | | 129.2.5 | 419464 | O1Os1P1Pr1 | 2,81 |
| | | 129.2.6 | 74786 | Ni2P2Th1 | 2,81 |
| | | 129.2.7 | 54276 | Mn1Si1Tb1 | 2,81 |
| | | 129.2.8 | 75054 | Mn1Pr1Si1 | 2,81 |
| | | 129.2.9 | 75053 | Mn1Pr1Si1 | 2,81 |
| | | 129.2.10 | 75052 | Mn1Pr1Si1 | 2,81 |
| | | 129.2.11 | 50243 | Mn1Nd1O1Sb1 | 2,81 |
| | | 129.2.12 | 50241 | Mn1Nd1O1P1 | 2,81 |
| | | 129.2.13 | 423892 | La1N1 | 2,81 |
| | | 129.2.14 | 62716 | Ir2Si2U1 | 2,81 |
| | | 129.2.15 | 200099 | I1Np1O1 | 2,81 |

Continued on next column

**TABLE XVI– continued**

| SG | cl | TQC | ICSD | Stoich. Formula | SubSG |
|---|---|---|---|---|---|
| | | 129.2.16 | 87913 | Ge1Ni1Pd1 | 2,81 |
| | | 129.2.17 | 42369 | Ge1Ni1Pd1 | 2,81 |
| | | 129.2.18 | 85859 | Ge1Mn1Pr1 | 2,81 |
| | | 129.2.19 | 80012 | Ge1Mn1Pr1 | 2,81 |
| | | 129.2.20 | 80011 | Ge1Mn1Pr1 | 2,81 |
| | | 129.2.21 | 183310 | Gd1Sn2Zn1 | 2,81 |
| | | 129.2.22 | 41255 | Gd1Ru1Si1 | 2,81 |
| | | 129.2.23 | 80200 | Fe1O1P1Pr1 | 2,81 |
| | | 129.2.24 | 85854 | Fe1Nd1Si1 | 2,81 |
| | | 129.2.25 | 57235 | Fe1Nd1Si1 | 2,81 |
| | | 129.2.26 | 57234 | Fe1Nd1Si1 | 2,81 |
| | | 129.2.27 | 95081 | Eu1P1Sn1 | 2,81 |
| | | 129.2.28 | 88118 | Er1Si1Ti1 | 2,81 |
| | | 129.2.29 | 88116 | Dy1Si1Ti1 | 2,81 |
| | | 129.2.30 | 57233 | Dy1Fe1Si1 | 2,81 |
| | | 129.2.31 | 57218 | Dy1Fe1Si1 | 2,81 |
| | | 129.2.32 | 428492 | Cu4La1P3 | 2,81 |
| | | 129.2.33 | 153572 | Cu1Sb2Y1 | 2,81 |
| | | 129.2.34 | 79579 | Cu1O1P1U1 | 2,81 |
| | | 129.2.35 | 261277 | Cu1O1P1U1 | 2,81 |
| | | 129.2.36 | 182310 | Cu1O1P1U1 | 2,81 |
| | | 129.2.37 | 182311 | Cu1O1P1Th1 | 2,81 |
| | | 129.2.38 | 172798 | Cu1F1Se1Sr1 | 2,81 |
| | | 129.2.39 | 172796 | Cu1F1S1Sr1 | 2,81 |
| | | 129.2.40 | 67932 | Co2P2U1 | 2,81 |
| | | 129.2.41 | 106477 | Co1Ge1Pr1 | 2,81 |
| | | 129.2.42 | 106471 | Co1Ge1La1 | 2,81 |
| | | 129.2.43 | 61370 | Ce1Pt2Si2 | 2,81 |
| | | 129.2.44 | 40764 | Ce1P2Rh2 | 2,81 |
| | | 129.2.45 | 419356 | Ce1Mn1O1Sb1 | 2,81 |
| | | 129.2.46 | 106407 | Ce1Ir2Sn2 | 2,81 |
| | | 129.2.47 | 102076 | Cd1Ti1 | 2,81 |
| | | 129.2.48 | 2027 | Ca1Ga1N1 | 2,81 |
| | | 129.2.49 | 62598 | C2Co1Sc1 | 2,81 |
| | | 129.2.50 | 85283 | C1B1Ni1Y1 | 2,81 |
| | | 129.2.51 | 415728 | Bi2La1Li1 | 2,81 |
| | | 129.2.52 | 426715 | Bi1N1Nd2 | 2,81 |
| | | 129.2.53 | 252466 | Bi1La1O1S2 | 2,81 |
| | | 129.2.54 | 426713 | Bi1Ce2N1 | 2,81 |
| | | 129.2.55 | 25337 | Be2Ca1Ge2 | 2,81 |
| | | 129.2.56 | 25339 | Ba1Sn2Zn2 | 2,81 |
| | | 129.2.57 | 42460 | Ba1Ge1Mg1 | 2,81 |
| | | 129.2.58 | 236308 | Ba1Cu2Sb2 | 2,81 |
| | | 129.2.59 | 428691 | Au2Sn2Sr1 | 2,81 |
| | | 129.2.60 | 427711 | Au2Sn2Sr1 | 2,81 |
| | | 129.2.61 | 107956 | As2Nd1Ni2 | 2,81 |
| | | 129.2.62 | 412550 | As2Cu1Y1 | 2,81 |
| | | 129.2.63 | 174358 | As2Cu1Lu1 | 2,81 |
| | | 129.2.64 | 94440 | As2Cu1Ho1 | 2,81 |
| | | 129.2.65 | 412178 | As2Cu1Ho1 | 2,81 |
| | | 129.2.66 | 44038 | As1Np1Te1 | 2,81 |
| | | 129.2.67 | 50242 | As1Mn1Nd1O1 | 2,81 |
| | | 129.2.68 | 262831 | As1La1O1Ru1 | 2,81 |
| | | 129.2.69 | 90316 | As1Fe1O1Pr1 | 2,81 |
| | | 129.2.70 | 422000 | As1Fe1O1Pr1 | 2,81 |
| | | 129.2.71 | 246832 | As1Fe1O1Pr1 | 2,81 |
| | | 129.2.72 | 162830 | As1Fe1O1Pr1 | 2,81 |
| | | 129.2.73 | 93238 | As1Fe1Mn1 | 2,81 |
| | | 129.2.74 | 93237 | As1Fe1Mn1 | 2,81 |
| | | 129.2.75 | 180774 | As1Co1Nd1O1 | 2,81 |
| | | 129.2.76 | 180773 | As1Co1Nd1O1 | 2,81 |

Continued on next column

**TABLE XVI– continued**

| SG | cl | TQC | ICSD | Stoich. Formula | SubSG |
|---|---|---|---|---|---|
| | | 129.2.77 | 180772 | As1Co1Nd1O1 | 2,81 |
| | | 129.2.78 | 169027 | As1Co1Nd1O1 | 2,81 |
| | | 129.2.79 | 195499 | As1Ce1Mn1O1 | 2,81 |
| | | 129.2.80 | 195498 | As1Ce1Mn1O1 | 2,81 |
| | | 129.2.81 | 194698 | Al2Ce1Pd2 | 2,81 |
| | | 129.2.82 | 191050 | Al2Ce1Pd2 | 2,81 |
| | | 129.2.83 | 190094 | Ag1Sb2Y1 | 2,81 |
| | | 129.2.84 | 251149 | Ag1Cu1S1 | 2,81 |
| 131 | 1 | 131.2.2 | 6159 | Ca1O4Pt2 | 2,81 |
| 131 | 2 | 131.2.1 | 50987 | I1Pd1Te1 | 2,81 |
| 136 | 1 | 136.1.2 | 90270 | Ni4P2Yb1 | 2,81 |
| | | 136.1.3 | 421229 | Li2Sr3 | 2,81 |
| | | 136.1.4 | 93213 | Fe2O6Te1 | 2,81 |
| | | 136.1.7 | 62553 | Ca1Cu4P2 | 2,81 |
| | | 136.1.10 | 260392 | Ba1Hg2Tl2 | 2,81 |
| | | 136.1.11 | 260391 | Ba1Hg2Tl2 | 2,81 |
| | | 136.1.13 | 173224 | As2Er1Ni4 | 2,81 |
| | | 136.1.14 | 68526 | As2Dy1Ni4 | 2,81 |
| 136 | 2 | 136.1.6 | 237545 | Cd4Pt2Sr1 | 2,81 |
| | | 136.1.12 | 237544 | Ba1Cd4Pt2 | 2,81 |
| 136 | 3 | 136.1.1 | 76166 | U1 | 65,84 |
| | | 136.1.5 | 33797 | Co2K4O5Rb2 | 65,84 |
| | | 136.1.8 | 417667 | C2Nd2W1 | 65,84 |
| | | 136.1.9 | 413345 | C2Mo1Pr2 | 65,84 |
| 137 | 1 | 137.1.3 | 49631 | B4Er1Rh4 | 0 |
| 137 | 2 | 137.1.1 | 250760 | Np2Se5 | 2,81 |
| | | 137.1.2 | 248874 | Ba2S6U1 | 2,81 |
| 139 | 1 | 139.1.11 | 42966 | P2Ru2Sr1 | 2,82 |
| | | 139.1.47 | 290825 | Mn2Si2Yb1 | 2,82 |
| | | 139.1.48 | 290824 | Mn2Si2Yb1 | 2,82 |
| | | 139.1.49 | 290822 | Mn2Si2Yb1 | 2,82 |
| | | 139.1.52 | 42733 | La2Sb1 | 2,82 |
| | | 139.1.53 | 423290 | La2Sb1 | 2,82 |
| | | 139.1.54 | 10442 | La2Sb1 | 2,82 |
| | | 139.1.59 | 427126 | La1Sb1Sc1 | 2,82 |
| | | 139.1.67 | 411 | Ge2Sr1Zn2 | 2,82 |
| | | 139.1.101 | 88184 | Dy1Ni2P2 | 2,82 |
| | | 139.1.102 | 88183 | Dy1Ni2P2 | 2,82 |
| | | 139.1.158 | 52761 | Ca1Ge2Ru2 | 2,82 |
| | | 139.1.166 | 41756 | Ca1Cu2P2 | 2,82 |
| | | 139.1.191 | 95097 | C1B2Ce1Rh2 | 2,82 |
| | | 139.1.192 | 95098 | C1B2Ce1Ir2 | 2,82 |
| | | 139.1.202 | 58717 | Be12Mo1 | 2,82 |
| | | 139.1.226 | 58388 | Au2Yb1 | 2,82 |
| | | 139.1.235 | 402951 | As2O1Yb4 | 2,82 |
| | | 139.1.237 | 78756 | As2Cu2Sr1 | 2,82 |
| | | 139.1.239 | 416983 | As2Ba1Rh2 | 2,82 |
| | | 139.1.241 | 236307 | As2Ba1Cu2 | 2,82 |
| | | 139.1.244 | 156967 | Al8Cr4Gd1 | 2,82 |
| | | 139.1.246 | 57539 | Al8Ca1Cu4 | 2,82 |
| | | 139.1.251 | 58252 | Ag2Er1 | 2,82 |
| 139 | 2 | 139.1.3 | 416747 | Ru1Zn3 | 71,69,87 |
| | | 139.1.4 | 249448 | Ru1Zn3 | 71,69,87 |
| | | 139.1.5 | 155990 | Ru1Zn3 | 71,69,87 |
| | | 139.1.22 | 93412 | Nd1Ni2Si2 | 71,69,87 |
| | | 139.1.43 | 82464 | Na2Nd2O10Ti3 | 71,69,87 |
| | | 139.1.44 | 75607 | Na2Nd2O10Ti3 | 71,69,87 |
| | | 139.1.45 | 75149 | Na2Nd2O10Ti3 | 71,69,87 |
| | | 139.1.50 | 104941 | Mn1Pd3 | 71,69,87 |
| | | 139.1.62 | 72190 | I2Nd1 | 71,69,87 |

Continued on next column

**TABLE XVI– continued**

| SG | cl | TQC | ICSD | Stoich. Formula | SubSG |
|---|---|---|---|---|---|
| | | 139.1.68 | 53731 | Ge2Nd1Pd2 | 71,69,87 |
| | | 139.1.69 | 81753 | Ge2Nd1Ni2 | 71,69,87 |
| | | 139.1.111 | 424224 | Cu2Ge2Sr1 | 71,69,87 |
| | | 139.1.112 | 410 | Cu2Ge2Sr1 | 71,69,87 |
| | | 139.1.114 | 40763 | Cu1P2U1 | 71,69,87 |
| | | 139.1.126 | 261192 | Co2La2O3Se2 | 71,69,87 |
| | | 139.1.127 | 261191 | Co2La2O3Se2 | 71,69,87 |
| | | 139.1.128 | 248239 | Co2La2O3Se2 | 71,69,87 |
| | | 139.1.129 | 168330 | Co2La2O3Se2 | 71,69,87 |
| | | 139.1.149 | 370021 | Cd2Ga2Sr1 | 71,69,87 |
| | | 139.1.178 | 69888 | C2Er1 | 71,69,87 |
| | | 139.1.179 | 23690 | C2Er1 | 71,69,87 |
| | | 139.1.180 | 23689 | C2Er1 | 71,69,87 |
| | | 139.1.204 | 427108 | Ba3Ge16Ir4 | 71,69,87 |
| | | 139.1.205 | 194235 | Ba3Ge16Ir4 | 71,69,87 |
| | | 139.1.206 | 189363 | Ba3Ge16Ir4 | 71,69,87 |
| | | 139.1.220 | 100024 | Ba1Mn1Sb2 | 71,69,87 |
| | | 139.1.224 | 71642 | B2Ir2Zn1 | 71,69,87 |
| | | 139.1.236 | 167013 | As2F2O1Sr2Ti2 | 71,69,87 |
| | | 139.1.242 | 92989 | As1Ti1Zr1 | 71,69,87 |
| | | 139.1.245 | 57545 | Al8Ca1Mn4 | 71,69,87 |
| | | 139.1.250 | 166270 | Al2Pd5Pu1 | 71,69,87 |
| 139 | 3 | 139.1.21 | 89560 | Nd2O2Te1 | 2,82 |
| | | 139.1.51 | 150228 | Mg1Pd3 | 2,82 |
| | | 139.1.55 | 40262 | La2O4Pd1 | 2,82 |
| | | 139.1.63 | 710068 | Ho1Si3 | 2,82 |
| | | 139.1.66 | 415033 | H2Ba8Bi4O1 | 2,82 |
| | | 139.1.73 | 165737 | Ge2Ir2Sr1 | 2,82 |
| | | 139.1.99 | 109428 | Er1Ge2Ni2 | 2,82 |
| | | 139.1.105 | 88424 | Cu2Ni1O2S2Sr2 | 2,82 |
| | | 139.1.123 | 73653 | Co2Nd1P2 | 2,82 |
| | | 139.1.124 | 73652 | Co2Nd1P2 | 2,82 |
| | | 139.1.125 | 73651 | Co2Nd1P2 | 2,82 |
| | | 139.1.152 | 425469 | Ca1Rh2Si2 | 2,82 |
| | | 139.1.164 | 52755 | Ca1Ge2Ir2 | 2,82 |
| | | 139.1.167 | 420974 | Ca1Co2Si2 | 2,82 |
| | | 139.1.171 | 81751 | Ca1Co2Ge2 | 2,82 |
| | | 139.1.172 | 406 | Ca1Co2Ge2 | 2,82 |
| | | 139.1.248 | 168816 | Al2Pd5U1 | 2,82 |
| | | 139.1.249 | 161313 | Al2Pd5U1 | 2,82 |
| 139 | 4 | 139.1.10 | 105758 | Pd1Zr2 | 2,82 |
| | | 139.1.12 | 50186 | P2Rh2Sr1 | 2,82 |
| | | 139.1.13 | 30849 | P2Pd2Yb1 | 2,82 |
| | | 139.1.20 | 41771 | Ni2Sc1Si3 | 2,82 |
| | | 139.1.77 | 89563 | Gd2O2Te1 | 2,82 |
| | | 139.1.79 | 30844 | Gd1P2Pd2 | 2,82 |
| | | 139.1.82 | 68209 | Fe2Si2Th1 | 2,82 |
| | | 139.1.84 | 74434 | Fe2O6Sr3 | 2,82 |
| | | 139.1.85 | 249015 | Fe2O6Sr3 | 2,82 |
| | | 139.1.86 | 249014 | Fe2O6Sr3 | 2,82 |
| | | 139.1.87 | 81746 | Fe2Ge2Nd1 | 2,82 |
| | | 139.1.100 | 30846 | Dy1P2Pd2 | 2,82 |
| | | 139.1.116 | 67472 | Cr2Gd1Si2 | 2,82 |
| | | 139.1.122 | 425470 | Co2Si2Sr1 | 2,82 |
| | | 139.1.130 | 407 | Co2Ge2Sr1 | 2,82 |
| | | 139.1.145 | 87159 | Ce1Sc1Si1 | 2,82 |
| | | 139.1.146 | 290821 | Ce1Sc1Si1 | 2,82 |
| | | 139.1.155 | 30835 | Ca1P2Pd2 | 2,82 |
| | | 139.1.165 | 10463 | Ca1Fe2P2 | 2,82 |
| | | 139.1.199 | 402954 | Bi2Eu4O1 | 2,82 |
| | | 139.1.213 | 2141 | Ba2Bi1 | 2,82 |

Continued on next column

**TABLE XVI– continued**

| SG | cl | TQC | ICSD | Stoich. Formula | SubSG |
|---|---|---|---|---|---|
| | | 139.1.215 | 50188 | Ba1P2Rh2 | 2,82 |
| | | 139.1.222 | 420975 | Ba1Co2Ge2 | 2,82 |
| | | 139.1.223 | 174263 | Ba1Cd2Ge2 | 2,82 |
| | | 139.1.247 | 95799 | Al4Ge2Ni1Tb2 | 2,82 |
| 139 | 5 | 139.1.2 | 42119 | Sb1Sr2 | 71,69,87 |
| | | 139.1.6 | 251818 | Pt2Th1 | 71,69,87 |
| | | 139.1.9 | 61195 | Pd2Sb2Sr1 | 71,69,87 |
| | | 139.1.17 | 56577 | O7Ru2Sr3 | 71,69,87 |
| | | 139.1.58 | 181385 | La2Mn2O3Se2 | 71,69,87 |
| | | 139.1.81 | 90392 | Gd1Ge1Ti1 | 71,69,87 |
| | | 139.1.139 | 91749 | Cl2Co2O5Sr3 | 71,69,87 |
| | | 139.1.140 | 90123 | Cl2Co2O5Sr3 | 71,69,87 |
| | | 139.1.141 | 251170 | Cl2Co2O4Sr3 | 71,69,87 |
| | | 139.1.159 | 52759 | Ca1Ge2Pd2 | 71,69,87 |
| | | 139.1.173 | 74670 | C2Yb1 | 71,69,87 |
| | | 139.1.174 | 23686 | C2Yb1 | 71,69,87 |
| | | 139.1.200 | 41836 | Bi1Sr2 | 71,69,87 |
| | | 139.1.201 | 107609 | Be1Pd2 | 71,69,87 |
| | | 139.1.207 | 41837 | Ba2Sb1 | 71,69,87 |
| | | 139.1.208 | 30084 | Ba2Hg1 | 71,69,87 |
| | | 139.1.209 | 183149 | Ba2F2Mn2O1Se2 | 71,69,87 |
| | | 139.1.210 | 71535 | Ba2Cu2O12Si4 | 71,69,87 |
| | | 139.1.214 | 61197 | Ba1Pd2Sb2 | 71,69,87 |
| | | 139.1.243 | 42358 | As1Ba2 | 71,69,87 |
| 139 | 6 | 139.1.1 | 263005 | Si3Y1 | 139 |
| | | 139.1.7 | 423291 | Pr2Sb1 | 139 |
| | | 139.1.15 | 71451 | O7Sr3V2 | 139 |
| | | 139.1.16 | 71320 | O7Sr3V2 | 139 |
| | | 139.1.18 | 79214 | O4Te3U4 | 139 |
| | | 139.1.23 | 91220 | Na2O1Sb2Ti2 | 139 |
| | | 139.1.24 | 91219 | Na2O1Sb2Ti2 | 139 |
| | | 139.1.25 | 91218 | Na2O1Sb2Ti2 | 139 |
| | | 139.1.26 | 91217 | Na2O1Sb2Ti2 | 139 |
| | | 139.1.27 | 91216 | Na2O1Sb2Ti2 | 139 |
| | | 139.1.28 | 91215 | Na2O1Sb2Ti2 | 139 |
| | | 139.1.29 | 91214 | Na2O1Sb2Ti2 | 139 |
| | | 139.1.30 | 91213 | Na2O1Sb2Ti2 | 139 |
| | | 139.1.31 | 91212 | Na2O1Sb2Ti2 | 139 |
| | | 139.1.32 | 91211 | Na2O1Sb2Ti2 | 139 |
| | | 139.1.33 | 91210 | Na2O1Sb2Ti2 | 139 |
| | | 139.1.34 | 91209 | Na2O1Sb2Ti2 | 139 |
| | | 139.1.35 | 91208 | Na2O1Sb2Ti2 | 139 |
| | | 139.1.36 | 91207 | Na2O1Sb2Ti2 | 139 |
| | | 139.1.37 | 91206 | Na2O1Sb2Ti2 | 139 |
| | | 139.1.38 | 91205 | Na2O1Sb2Ti2 | 139 |
| | | 139.1.39 | 91204 | Na2O1Sb2Ti2 | 139 |
| | | 139.1.40 | 91203 | Na2O1Sb2Ti2 | 139 |
| | | 139.1.41 | 91202 | Na2O1Sb2Ti2 | 139 |
| | | 139.1.42 | 69648 | Na2O1Sb2Ti2 | 139 |
| | | 139.1.46 | 152123 | Mo1O4Sr2 | 139 |
| | | 139.1.56 | 69172 | La2Ni1O4 | 139 |
| | | 139.1.57 | 33536 | La2Ni1O4 | 139 |
| | | 139.1.60 | 421342 | K2Mg5Sn3 | 139 |
| | | 139.1.65 | 75936 | Hf3Te2 | 139 |
| | | 139.1.70 | 81744 | Ge2Mn2U1 | 139 |
| | | 139.1.71 | 41976 | Ge2Mn2Th1 | 139 |
| | | 139.1.72 | 403 | Ge2Mn2Sr1 | 139 |
| | | 139.1.78 | 82466 | Gd2Na2O10Ti3 | 139 |
| | | 139.1.88 | 74421 | Fe1O4Sr2 | 139 |
| | | 139.1.89 | 74420 | Fe1O4Sr2 | 139 |
| | | 139.1.90 | 74419 | Fe1O4Sr2 | 139 |

Continued on next column

**TABLE XVI– continued**

| SG | cl | TQC | ICSD | Stoich. Formula | SubSG |
|---|---|---|---|---|---|
| | | 139.1.91 | 69849 | Fe1O4Sr2 | 139 |
| | | 139.1.92 | 69682 | F4Ni1Rb2 | 139 |
| | | 139.1.93 | 73450 | F4K2Ni1 | 139 |
| | | 139.1.94 | 33520 | F4K2Ni1 | 139 |
| | | 139.1.95 | 167014 | F2O1Sb2Sr2Ti2 | 139 |
| | | 139.1.96 | 426633 | Er2Mg2Ru1 | 139 |
| | | 139.1.113 | 154203 | Cu2Fe2O5Se2Sr3 | 139 |
| | | 139.1.117 | 245595 | Cr1O4Sr2 | 139 |
| | | 139.1.118 | 43091 | Co4Ge7U3 | 139 |
| | | 139.1.119 | 290506 | Co4Ge7U3 | 139 |
| | | 139.1.120 | 290505 | Co4Ge7U3 | 139 |
| | | 139.1.121 | 290504 | Co4Ge7U3 | 139 |
| | | 139.1.131 | 33524 | Co2F7K3 | 139 |
| | | 139.1.132 | 30081 | Cl7Mn2Rb3 | 139 |
| | | 139.1.133 | 2241 | Cl7Mn2Rb3 | 139 |
| | | 139.1.134 | 36127 | Cl7K3Mn2 | 139 |
| | | 139.1.135 | 99942 | Cl2Fe2O5Sr3 | 139 |
| | | 139.1.136 | 69871 | Cl2Fe2O5Sr3 | 139 |
| | | 139.1.137 | 33258 | Cl2Fe2O5Sr3 | 139 |
| | | 139.1.138 | 174532 | Cl2Fe2O4Sr3 | 139 |
| | | 139.1.142 | 409587 | Ce2Sb1 | 139 |
| | | 139.1.147 | 411677 | Ce1Ir2Si2 | 139 |
| | | 139.1.150 | 416139 | Ca2Pd3Sb4 | 139 |
| | | 139.1.151 | 59649 | Ca1Si2Zn2 | 139 |
| | | 139.1.153 | 50185 | Ca1P2Rh2 | 139 |
| | | 139.1.157 | 414231 | Ca1In1 | 139 |
| | | 139.1.160 | 81755 | Ca1Ge2Ni2 | 139 |
| | | 139.1.161 | 408 | Ca1Ge2Ni2 | 139 |
| | | 139.1.168 | 85892 | Ca1Co2P2 | 139 |
| | | 139.1.169 | 85891 | Ca1Co2P2 | 139 |
| | | 139.1.170 | 85890 | Ca1Co2P2 | 139 |
| | | 139.1.177 | 63560 | C2Nd1 | 139 |
| | | 139.1.181 | 380471 | C2Cl1N2Ni1Rb1 | 139 |
| | | 139.1.182 | 380469 | C2Cl1Cs1N2Ni1 | 139 |
| | | 139.1.183 | 380470 | C2Br1N2Ni1Rb1 | 139 |
| | | 139.1.184 | 380468 | C2Br1Cs1N2Ni1 | 139 |
| | | 139.1.185 | 79563 | C1B2Gd1Ni2 | 139 |
| | | 139.1.187 | 56596 | C1B2Dy1Ni2 | 139 |
| | | 139.1.188 | 56595 | C1B2Dy1Ni2 | 139 |
| | | 139.1.189 | 56594 | C1B2Dy1Ni2 | 139 |
| | | 139.1.190 | 56593 | C1B2Dy1Ni2 | 139 |
| | | 139.1.193 | 32544 | Br7Mn2Rb3 | 139 |
| | | 139.1.194 | 99944 | Br2Fe2O5Sr3 | 139 |
| | | 139.1.195 | 99943 | Br2Fe2O5Sr3 | 139 |
| | | 139.1.196 | 69872 | Br2Fe2O5Sr3 | 139 |
| | | 139.1.197 | 430065 | Bi2F2O1Sr2Ti2 | 139 |
| | | 139.1.198 | 430064 | Bi2F2O1Sr2Ti2 | 139 |
| | | 139.1.211 | 78593 | Ba2Ca1Cu2O8Tl2 | 139 |
| | | 139.1.212 | 78592 | Ba2Ca1Cu2O8Tl2 | 139 |
| | | 139.1.216 | 419531 | Ba1Mn2Sb2 | 139 |
| | | 139.1.217 | 32019 | Ba1Mn2Sb2 | 139 |
| | | 139.1.218 | 77668 | Ba1Mn2P2 | 139 |
| | | 139.1.219 | 10469 | Ba1Mn2P2 | 139 |
| | | 139.1.228 | 37453 | Au2Cl6Cs2 | 139 |
| | | 139.1.232 | 417002 | As2Rh2Sr1 | 139 |
| | | 139.1.240 | 41794 | As2Ba1Mn2 | 139 |
| 139 | 7 | 139.1.19 | 68211 | Ni2Si2Th1 | 0 |
| | | 139.1.80 | 26823 | Gd1Mn12 | 0 |
| | | 139.1.103 | 246933 | Cu2Si2U1 | 0 |
| | | 139.1.104 | 68212 | Cu2Si2Th1 | 0 |
| | | 139.1.106 | 86002 | Cu2Ge2U1 | 0 |

Continued on next column

**TABLE XVI– continued**

| SG | cl | TQC | ICSD | Stoich. Formula | SubSG |
|---|---|---|---|---|---|
| | | 139.1.107 | 86001 | Cu2Ge2U1 | 0 |
| | | 139.1.108 | 86000 | Cu2Ge2U1 | 0 |
| | | 139.1.109 | 85999 | Cu2Ge2U1 | 0 |
| | | 139.1.110 | 85998 | Cu2Ge2U1 | 0 |
| | | 139.1.162 | 81743 | Ca1Ge2Mn2 | 0 |
| | | 139.1.163 | 402 | Ca1Ge2Mn2 | 0 |
| | | 139.1.186 | 56651 | C1B2Er1Ni2 | 0 |
| | | 139.1.225 | 426150 | Au4Ce1Zn2 | 0 |
| | | 139.1.229 | 418530 | Au2Ce1Si2 | 0 |
| | | 139.1.231 | 263018 | As4Ba2Fe2O1Ti2 | 0 |
| | | 139.1.233 | 47170 | As2Pd2Sr1 | 0 |
| | | 139.1.234 | 36374 | As2Pd2Sr1 | 0 |
| | | 139.1.238 | 36372 | As2Ca1Pd2 | 0 |
| 139 | 8 | 139.1.8 | 247272 | Pd3Tl1 | 2,82 |
| | | 139.1.14 | 36373 | P2Pd2Sr1 | 2,82 |
| | | 139.1.61 | 420251 | In1Pd3 | 2,82 |
| | | 139.1.64 | 427211 | Hg2In2Sr1 | 2,82 |
| | | 139.1.74 | 26277 | Ge1Te1U1 | 2,82 |
| | | 139.1.75 | 87316 | Ge1Se1U1 | 2,82 |
| | | 139.1.76 | 26276 | Ge1Se1U1 | 2,82 |
| | | 139.1.83 | 10467 | Fe2P2Sr1 | 2,82 |
| | | 139.1.97 | 60667 | Er1O2S3U2 | 2,82 |
| | | 139.1.98 | 245051 | Er1Ni2Si2 | 2,82 |
| | | 139.1.115 | 290194 | Cu1Ge1Hf1 | 2,82 |
| | | 139.1.143 | 150461 | Ce2Ga10Pd1 | 2,82 |
| | | 139.1.144 | 20525 | Ce2Ga10Ni1 | 2,82 |
| | | 139.1.148 | 85893 | Ce1Co2P2 | 2,82 |
| | | 139.1.154 | 36371 | Ca1P2Pd2 | 2,82 |
| | | 139.1.156 | 10461 | Ca1Ni2P2 | 2,82 |
| | | 139.1.175 | 26478 | C2U1 | 2,82 |
| | | 139.1.176 | 202305 | C2U1 | 2,82 |
| | | 139.1.203 | 237677 | Ba6Ge32Rh8 | 2,82 |
| | | 139.1.221 | 10468 | Ba1Fe2P2 | 2,82 |
| | | 139.1.227 | 413 | Au2Si2Sr1 | 2,82 |
| | | 139.1.230 | 412 | Au2Ca1Si2 | 2,82 |
| 140 | 1 | 140.1.3 | 245120 | Mg2Pt1 | 69,87 |
| | | 140.1.8 | 106369 | Cd3Sr5 | 69,87 |
| | | 140.1.9 | 106344 | Ca5Hg3 | 69,87 |
| | | 140.1.11 | 190146 | Bi1Ho5Ni2 | 69,87 |
| | | 140.1.12 | 100084 | Ba5Sn3 | 69,87 |
| | | 140.1.13 | 247208 | B2Cu2O6Sr1 | 69,87 |
| | | 140.1.14 | 247207 | B2Cu2O6Sr1 | 69,87 |
| 140 | 2 | 140.1.1 | 86940 | O3Ru1Sr1 | 0 |
| | | 140.1.2 | 174206 | Mo1O3Sr1 | 0 |
| | | 140.1.4 | 88093 | Cr1Sb2 | 0 |
| | | 140.1.5 | 86811 | Cr1Sb2 | 0 |
| | | 140.1.6 | 281417 | Cl1Ga4La3 | 0 |
| | | 140.1.7 | 98408 | Ce1Ni9Si4 | 0 |
| | | 140.1.10 | 39599 | Br5Cd1Cs3 | 0 |
| 140 | 3 | 140.2.2 | 409797 | Sb2Ti1 | 2,82 |
| | | 140.2.3 | 249576 | Sb2Ti1 | 2,82 |
| | | 140.2.4 | 16582 | Pu5Si3 | 2,82 |
| | | 140.2.6 | 260983 | Ni1U6 | 2,82 |
| | | 140.2.8 | 420322 | Cs9O4Sc1 | 2,82 |
| | | 140.2.9 | 419823 | Cs9In1O4 | 2,82 |
| | | 140.2.10 | 420320 | Cs9Ga1O4 | 2,82 |
| | | 140.2.12 | 150146 | Bi2Fe1Ti4 | 2,82 |
| | | 140.2.13 | 428736 | Al1Eu3O5 | 2,82 |
| | | 140.2.14 | 420321 | Al1Cs9O4 | 2,82 |
| 140 | 4 | 140.2.1 | 1203 | Sn3Sr5 | 2,82 |

Continued on next column

**TABLE XVI– continued**

| SG | cl | TQC | ICSD | Stoich. Formula | SubSG |
|---|---|---|---|---|---|
| | | 140.2.5 | 105479 | Ni1Zr2 | 2,82 |
| | | 140.2.7 | 409374 | Ge3Sr5 | 2,82 |
| | | 140.2.11 | 418133 | Ca1Ni4Sn2 | 2,82 |
| 141 | 1 | 141.1.1 | 423244 | Si2Yb1 | 2,82 |
| | | 141.1.2 | 170062 | Sc1Sn2 | 2,82 |
| | | 141.1.3 | 88368 | O4Tb1V1 | 2,82 |
| | | 141.1.4 | 78078 | O4Tb1V1 | 2,82 |
| | | 141.1.5 | 78076 | O4Pr1V1 | 2,82 |
| | | 141.1.6 | 157327 | O4Pr1V1 | 2,82 |
| | | 141.1.7 | 79759 | O4P1Tm1 | 2,82 |
| | | 141.1.8 | 36053 | O4P1Tm1 | 2,82 |
| | | 141.1.9 | 61683 | Na1O2Tb1 | 2,82 |
| | | 141.1.10 | 61682 | Na1O2Pr1 | 2,82 |
| | | 141.1.11 | 410968 | In2Zr1 | 2,82 |
| | | 141.1.12 | 61422 | H1V2 | 2,82 |
| | | 141.1.13 | 60511 | H1V2 | 2,82 |
| | | 141.1.14 | 103986 | Ga2Th1 | 2,82 |
| | | 141.1.15 | 100566 | Cu4O3 | 2,82 |
| | | 141.1.16 | 35419 | Cu2P1Ti1 | 2,82 |
| | | 141.1.17 | 100796 | Cu2Ge1O4 | 2,82 |
| | | 141.1.22 | 409800 | Ce1Zn11 | 2,82 |
| | | 141.1.23 | 247146 | Ag1Li1 | 2,82 |
| | | 141.1.24 | 247145 | Ag1Li1 | 2,82 |
| 141 | 2 | 141.1.18 | 84377 | Cr2Ni1O4 | 0 |
| | | 141.1.19 | 31300 | Cr2Ni1O4 | 0 |
| | | 141.1.20 | 31299 | Cr2Ni1O4 | 0 |
| | | 141.1.21 | 280062 | Cr2Ni1O4 | 0 |
| 147 | 1 | 147.1.1 | 166370 | Ni1Ti1 | 2 |
| | | 147.1.2 | 150943 | Ni1Ti1 | 2 |
| | | 148.1.1 | 54337 | Pd4Zr3 | 2 |
| | | 148.1.3 | 37134 | O12Te1Tl6 | 2 |
| | | 148.1.4 | 166373 | Ni4Ti3 | 2 |
| | | 148.1.6 | 166369 | Ni4Ti3 | 2 |
| | | 148.1.8 | 59375 | Mo6Te8 | 2 |
| | | 148.1.9 | 46015 | Mo6S8Yb1 | 2 |
| | | 148.1.10 | 46014 | Mo6S8Yb1 | 2 |
| | | 148.1.11 | 46013 | Mo6S8Yb1 | 2 |
| | | 148.1.12 | 201506 | Mo6S8Yb1 | 2 |
| | | 148.1.13 | 62156 | Mo6S8Sr1 | 2 |
| | | 148.1.14 | 68 | Mo3Pb0.5Se4 | 2 |
| | | 148.1.15 | 89004 | Mo12Rb2Se14 | 2 |
| | | 148.1.16 | 245054 | Mn3O6Te1 | 2 |
| | | 148.1.17 | 183618 | Mn3O6Te1 | 2 |
| | | 148.1.18 | 182436 | Mn3O6Te1 | 2 |
| | | 148.1.19 | 182435 | Mn3O6Te1 | 2 |
| | | 148.1.21 | 247553 | Mn1O3Ti1 | 2 |
| | | 148.1.23 | 73247 | Ge1Mn1O3 | 2 |
| | | 148.1.24 | 73246 | Ge1Mn1O3 | 2 |
| | | 148.1.25 | 73245 | Ge1Mn1O3 | 2 |
| | | 148.1.26 | 73244 | Ge1Mn1O3 | 2 |
| | | 148.1.27 | 73243 | Ge1Mn1O3 | 2 |
| | | 148.1.28 | 69591 | Ge1Mn1O3 | 2 |
| | | 148.1.30 | 84337 | Cs2Mo12Te14 | 2 |
| | | 148.1.31 | 40820 | Cs2Mo12Se14 | 2 |
| | | 148.1.32 | 71020 | Cr2Si2Te6 | 2 |
| | | 148.1.33 | 106520 | Cr2Se3 | 2 |
| | | 148.1.34 | 79270 | Cr2Ge2Te6 | 2 |
| | | 148.1.35 | 79269 | Cr2Ge2Te6 | 2 |
| | | 148.1.36 | 79268 | Cr2Ge2Te6 | 2 |
| | | 148.1.38 | 39426 | Cl3Ti1 | 2 |
| | | 148.1.39 | 201976 | Cl12N1Sc7 | 2 |

Continued on next column

**TABLE XVI– continued**

| SG | cl | TQC | ICSD | Stoich. Formula | SubSG |
|---|---|---|---|---|---|
| | | 148.1.40 | 424476 | Cl12Co1Sc7 | 2 |
| | | 148.1.41 | 183603 | Ce1Pt5Sb1 | 2 |
| | | 148.1.42 | 62388 | Ca1Mo6S8 | 2 |
| | | 148.1.43 | 39242 | Br3Ti1 | 2 |
| | | 148.1.44 | 419636 | Ba6N6O1Re2 | 2 |
| | | 148.1.45 | 419467 | Ba6N6O1Os2 | 2 |
| | | 148.1.48 | 85489 | Ba1Mo6S8 | 2 |
| | | 148.1.49 | 65776 | Ba1Mo6S8 | 2 |
| | | 148.1.50 | 62157 | Ba1Mo6S8 | 2 |
| | | 148.1.52 | 237739 | Au5Mn1Y3 | 2 |
| | | 148.1.53 | 91777 | Ag1Bi1O3 | 2 |
| | | 148.1.54 | 89432 | Ag1Bi1O3 | 2 |
| 148 | 2 | 148.1.2 | 1096 | P2Pd15 | 0 |
| | | 148.1.5 | 166371 | Ni4Ti3 | 0 |
| | | 148.1.7 | 171584 | Ni1O3Ti1 | 0 |
| | | 148.1.20 | 60006 | Mn1O3Ti1 | 0 |
| | | 148.1.22 | 171579 | Mn1O3Ti1 | 0 |
| | | 148.1.29 | 247127 | Cu1O3Si1 | 0 |
| | | 148.1.37 | 252042 | Cr1Fe1O3 | 0 |
| | | 148.1.46 | 96087 | Ba1Ni2O8V2 | 0 |
| | | 148.1.47 | 96086 | Ba1Ni2O8V2 | 0 |
| | | 148.1.51 | 59343 | Ba1Co2O8P2 | 0 |
| 162 | 1 | 162.1.1 | 28086 | O6U1V2 | 2 |
| | | 162.1.2 | 275 | Cr2O6U1 | 2 |
| | | 162.1.3 | 15133 | Cr2O6U1 | 2 |
| | | 162.1.5 | 26557 | Ag6O2 | 2 |
| 162 | 2 | 162.1.4 | 37011 | Ba1Fe4O8Sr1 | 0 |
| 163 | 1 | 163.1.1 | 71037 | Mo1Ni1P8 | 2 |
| | | 163.1.2 | 411131 | Al1F6Li1Yb1 | 2 |
| | | 164.1.1 | 657478 | Te2Zr1 | 0 |
| | | 164.1.2 | 653213 | Te2Zr1 | 0 |
| | | 164.1.4 | 651764 | Sb2Yb1Zn2 | 0 |
| | | 164.1.5 | 419689 | Sb2Yb1Zn2 | 0 |
| | | 164.1.31 | 660273 | In2S4Zn1 | 0 |
| | | 164.1.32 | 42668 | In2S4Zn1 | 0 |
| | | 164.1.33 | 25809 | Ho2O2Se1 | 0 |
| | | 164.1.34 | 185808 | Hg1Pt2Se3 | 0 |
| | | 164.1.38 | 59801 | Ge2I2La2 | 0 |
| | | 164.1.42 | 159909 | Fe2K3Na1O8 | 0 |
| | | 164.1.44 | 25810 | Er2O2Se1 | 0 |
| | | 164.1.46 | 407246 | Ce2I2Si2 | 0 |
| | | 164.1.47 | 78989 | Cd2Sb2Yb1 | 0 |
| | | 164.1.48 | 419690 | Cd2Sb2Yb1 | 0 |
| | | 164.1.49 | 248526 | Ca1Si2 | 0 |
| | | 164.1.50 | 248525 | Ca1Si2 | 0 |
| | | 164.1.51 | 248524 | Ca1Si2 | 0 |
| | | 164.1.52 | 248523 | Ca1Si2 | 0 |
| | | 164.1.53 | 248522 | Ca1Si2 | 0 |
| | | 164.1.54 | 248521 | Ca1Si2 | 0 |
| | | 164.1.73 | 71625 | C1F2Ho2 | 0 |
| | | 164.1.79 | 42708 | Bi2Pb2Te5 | 0 |
| | | 164.1.83 | 30525 | Bi1Te1 | 0 |
| | | 164.1.84 | 100654 | Bi1Te1 | 0 |
| | | 164.1.85 | 99681 | Ba6Cl2O12Ru3 | 0 |
| | | 164.1.89 | 88231 | As2Yb1Zn2 | 0 |
| | | 164.1.90 | 39512 | As2Mn2Zn1 | 0 |
| | | 164.1.99 | 160311 | Al2Er1Ge2 | 0 |
| | | 164.1.102 | 94501 | Al1La1Si2 | 0 |
| 164 | 2 | 164.1.3 | 80205 | Si1Te2 | 2 |
| | | 164.1.6 | 69557 | Sb2Te1 | 2 |

Continued on next column

**TABLE XVI– continued**

| SG | cl | TQC | ICSD | Stoich. Formula | SubSG |
|---|---|---|---|---|---|
| | | 164.1.7 | 41385 | Pt1Te2 | 2 |
| | | 164.1.8 | 41373 | Pt1Te2 | 2 |
| | | 164.1.9 | 83642 | Pd1Te2 | 2 |
| | | 164.1.10 | 42555 | Pd1Te2 | 2 |
| | | 164.1.11 | 42554 | Pd1Te2 | 2 |
| | | 164.1.12 | 41387 | Pd1Te2 | 2 |
| | | 164.1.13 | 100271 | P2Yb1Zn2 | 2 |
| | | 164.1.14 | 43293 | Ni1Te2 | 2 |
| | | 164.1.15 | 42559 | Ni1Te2 | 2 |
| | | 164.1.16 | 94478 | Nd2S2Te1 | 2 |
| | | 164.1.17 | 32514 | Nd2O3 | 2 |
| | | 164.1.18 | 25806 | Nd2O2Se1 | 2 |
| | | 164.1.19 | 32515 | Nd2O2S1 | 2 |
| | | 164.1.20 | 300245 | Nd2O2S1 | 2 |
| | | 164.1.21 | 380376 | Mn2Sb2Yb1 | 2 |
| | | 164.1.22 | 41790 | Mn2Sb2Sr1 | 2 |
| | | 164.1.23 | 416094 | Mn2Sb2Sr1 | 2 |
| | | 164.1.24 | 77670 | Mn2P2Sr1 | 2 |
| | | 164.1.25 | 77669 | Mn2P2Sr1 | 2 |
| | | 164.1.26 | 49019 | Mn2P2Sr1 | 2 |
| | | 164.1.27 | 412655 | Mg2Sb2Yb1 | 2 |
| | | 164.1.28 | 191298 | Li3Pr1Sb2 | 2 |
| | | 164.1.29 | 26018 | K3Na1O8S2 | 2 |
| | | 164.1.30 | 26014 | K3Na1O8S2 | 2 |
| | | 164.1.35 | 56153 | H2Cd1O2 | 2 |
| | | 164.1.36 | 41478 | H1Si1 | 2 |
| | | 164.1.37 | 42876 | Ge2Sb2Te5 | 2 |
| | | 164.1.39 | 42875 | Ge1Sb4Te7 | 2 |
| | | 164.1.40 | 25808 | Gd2O2Se1 | 2 |
| | | 164.1.41 | 91111 | Gd2O2S1 | 2 |
| | | 164.1.43 | 155025 | Fe2In2Se5 | 2 |
| | | 164.1.45 | 35585 | Cu2P2Zr1 | 2 |
| | | 164.1.55 | 248520 | Ca1Si2 | 2 |
| | | 164.1.56 | 248519 | Ca1Si2 | 2 |
| | | 164.1.57 | 41789 | Ca1Mn2Sb2 | 2 |
| | | 164.1.58 | 416093 | Ca1Mn2Sb2 | 2 |
| | | 164.1.59 | 163778 | Ca1Mn2Sb2 | 2 |
| | | 164.1.60 | 49017 | Ca1Mn2P2 | 2 |
| | | 164.1.61 | 35512 | C6N6Pt1Rb2S6 | 2 |
| | | 164.1.62 | 35511 | C6K2N6Pt1S6 | 2 |
| | | 164.1.63 | 35510 | C6H8N8Pt1S6 | 2 |
| | | 164.1.64 | 5005 | C1V2 | 2 |
| | | 164.1.65 | 409555 | C1Ta2 | 2 |
| | | 164.1.66 | 95113 | C1Nb2S2 | 2 |
| | | 164.1.67 | 95111 | C1Nb2S2 | 2 |
| | | 164.1.68 | 95109 | C1Nb2S2 | 2 |
| | | 164.1.69 | 656705 | C1Nb2S2 | 2 |
| | | 164.1.70 | 43670 | C1Nb2 | 2 |
| | | 164.1.71 | 82264 | C1N2O2Pr2 | 2 |
| | | 164.1.72 | 82265 | C1N2Nd2O2 | 2 |
| | | 164.1.74 | 33912 | C1F2Gd2 | 2 |
| | | 164.1.75 | 42753 | Bi4Pb1Te7 | 2 |
| | | 164.1.76 | 42707 | Bi4Pb1Te7 | 2 |
| | | 164.1.77 | 250249 | Bi4Pb1Te7 | 2 |
| | | 164.1.78 | 42891 | Bi4Ge1Te7 | 2 |
| | | 164.1.80 | 30372 | Bi2Pb2Se5 | 2 |
| | | 164.1.81 | 659569 | Bi2Mg3 | 2 |
| | | 164.1.82 | 41791 | Bi2Ca1Mn2 | 2 |
| | | 164.1.86 | 152482 | Ba3Ca1O9Ru2 | 2 |
| | | 164.1.87 | 409958 | Ba1Sn2 | 2 |
| | | 164.1.88 | 41107 | As4Ge1Te7 | 2 |

Continued on next column

**TABLE XVI– continued**

| SG | cl | TQC | ICSD | Stoich. Formula | SubSG |
|---|---|---|---|---|---|
| | | 164.1.91 | 49020 | As2Mn2Sr1 | 2 |
| | | 164.1.92 | 41793 | As2Mn2Sr1 | 2 |
| | | 164.1.93 | 63174 | As2Ge5Te8 | 2 |
| | | 164.1.94 | 41792 | As2Ca1Mn2 | 2 |
| | | 164.1.95 | 411054 | Al2Si2Yb1 | 2 |
| | | 164.1.96 | 411056 | Al2Ge2Yb1 | 2 |
| | | 164.1.97 | 409458 | Al2Ge2Yb1 | 2 |
| | | 164.1.98 | 412121 | Al2Ge2Nd1 | 2 |
| | | 164.1.100 | 50976 | Al2Ce1Ge2 | 2 |
| | | 164.1.101 | 94502 | Al1La1Si2 | 2 |
| 165 | 1 | 165.1.1 | 154809 | H3Y1 | 2 |
| | | 165.1.2 | 26776 | As1Cu3 | 2 |
| | | 166.1.1 | 653496 | Yb2Zn17 | 2 |
| | | 166.1.2 | 653487 | Yb2Zn17 | 2 |
| | | 166.1.3 | 104207 | Yb2Zn17 | 2 |
| | | 166.1.4 | 653467 | Y2Zn17 | 2 |
| | | 166.1.5 | 104239 | Y2Zn17 | 2 |
| | | 166.1.6 | 653389 | U2Zn17 | 2 |
| | | 166.1.7 | 653384 | U2Zn17 | 2 |
| | | 166.1.8 | 106208 | U2Zn17 | 2 |
| | | 166.1.9 | 106207 | U2Zn17 | 2 |
| | | 166.1.10 | 104248 | U2Zn17 | 2 |
| | | 166.1.14 | 52499 | Te1 | 2 |
| | | 166.1.15 | 410956 | Sn3Sr1 | 2 |
| | | 166.1.16 | 652055 | Se1Ti1 | 2 |
| | | 166.1.17 | 152188 | Sb8Te3 | 2 |
| | | 166.1.18 | 152189 | Sb48Te9 | 2 |
| | | 166.1.19 | 291530 | Sb3Te4.5 | 2 |
| | | 166.1.22 | 651638 | Sb2Te3 | 2 |
| | | 166.1.23 | 651635 | Sb2Te3 | 2 |
| | | 166.1.24 | 651629 | Sb2Te3 | 2 |
| | | 166.1.25 | 5420 | Sb2Te3 | 2 |
| | | 166.1.26 | 262171 | Sb2Te3 | 2 |
| | | 166.1.27 | 2084 | Sb2Te3 | 2 |
| | | 166.1.28 | 20236 | Sb2Te3 | 2 |
| | | 166.1.29 | 193347 | Sb2Te3 | 2 |
| | | 166.1.30 | 193346 | Sb2Te3 | 2 |
| | | 166.1.31 | 193345 | Sb2Te3 | 2 |
| | | 166.1.32 | 193344 | Sb2Te3 | 2 |
| | | 166.1.33 | 193343 | Sb2Te3 | 2 |
| | | 166.1.34 | 193342 | Sb2Te3 | 2 |
| | | 166.1.35 | 193341 | Sb2Te3 | 2 |
| | | 166.1.36 | 192780 | Sb2Te3 | 2 |
| | | 166.1.38 | 185952 | Sb2Te3 | 2 |
| | | 166.1.39 | 185951 | Sb2Te3 | 2 |
| | | 166.1.40 | 185950 | Sb2Te3 | 2 |
| | | 166.1.41 | 185949 | Sb2Te3 | 2 |
| | | 166.1.42 | 185948 | Sb2Te3 | 2 |
| | | 166.1.43 | 185947 | Sb2Te3 | 2 |
| | | 166.1.44 | 185946 | Sb2Te3 | 2 |
| | | 166.1.45 | 185945 | Sb2Te3 | 2 |
| | | 166.1.46 | 657316 | Sb2Sn1Te4 | 2 |
| | | 166.1.47 | 30392 | Sb2Sn1Te4 | 2 |
| | | 166.1.48 | 651529 | Sb2Se2Te1 | 2 |
| | | 166.1.49 | 52295 | Sb2Se2Te1 | 2 |
| | | 166.1.50 | 87416 | Sb2Se1Te2 | 2 |
| | | 166.1.51 | 2085 | Sb2Se1Te2 | 2 |
| | | 166.1.52 | 651646 | Sb1Te2Tl1 | 2 |
| | | 166.1.53 | 15411 | Sb1Te2Tl1 | 2 |
| | | 166.1.54 | 651506 | Sb1 | 2 |
| | | 166.1.55 | 651498 | Sb1 | 2 |

Continued on next column

**TABLE XVI– continued**

| SG | cl | TQC | ICSD | Stoich. Formula | SubSG |
|---|---|---|---|---|---|
| | | 166.1.56 | 651497 | Sb1 | 2 |
| | | 166.1.57 | 651491 | Sb1 | 2 |
| | | 166.1.58 | 651490 | Sb1 | 2 |
| | | 166.1.59 | 651489 | Sb1 | 2 |
| | | 166.1.60 | 651488 | Sb1 | 2 |
| | | 166.1.61 | 64697 | Sb1 | 2 |
| | | 166.1.62 | 64696 | Sb1 | 2 |
| | | 166.1.63 | 64695 | Sb1 | 2 |
| | | 166.1.64 | 55402 | Sb1 | 2 |
| | | 166.1.65 | 53795 | Sb1 | 2 |
| | | 166.1.66 | 409754 | Sb1 | 2 |
| | | 166.1.67 | 161493 | Sb1 | 2 |
| | | 166.1.68 | 24757 | S2Ta1 | 2 |
| | | 166.1.69 | 651179 | S1Ti1 | 2 |
| | | 166.1.70 | 25561 | S1Ti1 | 2 |
| | | 166.1.71 | 184500 | S1Ti1 | 2 |
| | | 166.1.74 | 41371 | Pt2Te3 | 2 |
| | | 166.1.75 | 41370 | Pt1Te1 | 2 |
| | | 166.1.79 | 54338 | Pr1 | 2 |
| | | 166.1.82 | 600019 | P1 | 2 |
| | | 166.1.84 | 416476 | O1P1Pr1Zn1 | 2 |
| | | 166.1.86 | 646928 | Ni7Th2 | 2 |
| | | 166.1.106 | 646072 | Nd2Zn17 | 2 |
| | | 166.1.107 | 601895 | Nd2Zn17 | 2 |
| | | 166.1.108 | 104242 | Nd2Zn17 | 2 |
| | | 166.1.109 | 645562 | Nb6Zn7 | 2 |
| | | 166.1.110 | 30364 | N2W1 | 2 |
| | | 166.1.111 | 44926 | Mg2Ni3Si1 | 2 |
| | | 166.1.112 | 55617 | Mg2Nd1Ni9 | 2 |
| | | 166.1.113 | 41369 | Li8Pb3 | 2 |
| | | 166.1.114 | 191137 | La3Mn2O14Sb3 | 2 |
| | | 166.1.117 | 640862 | Ir7Nd2 | 2 |
| | | 166.1.118 | 657607 | In2Te3 | 2 |
| | | 166.1.119 | 640609 | In2Te3 | 2 |
| | | 166.1.126 | 172602 | I1N1Sr2 | 2 |
| | | 166.1.127 | 654919 | Hg1 | 2 |
| | | 166.1.129 | 174006 | Hg1 | 2 |
| | | 166.1.130 | 174005 | Hg1 | 2 |
| | | 166.1.131 | 104296 | Hg1 | 2 |
| | | 166.1.134 | 171105 | H9Fe3O15S2 | 2 |
| | | 166.1.137 | 67682 | H6Fe3O14Pb1S2 | 2 |
| | | 166.1.138 | 44626 | H6Al3Na1O14S2 | 2 |
| | | 166.1.139 | 24344 | H4Cu1N1O3S1 | 2 |
| | | 166.1.140 | 623755 | H4Co3Y1 | 2 |
| | | 166.1.141 | 623744 | H4Co3Y1 | 2 |
| | | 166.1.142 | 604264 | H4Co3Y1 | 2 |
| | | 166.1.143 | 602920 | H4Co3Y1 | 2 |
| | | 166.1.144 | 623723 | H4Co3Pr1 | 2 |
| | | 166.1.145 | 623698 | H4Co3Ho1 | 2 |
| | | 166.1.146 | 602925 | H4Co3Ho1 | 2 |
| | | 166.1.147 | 623752 | H3Co3Y1 | 2 |
| | | 166.1.152 | 623746 | H1Co3Y1 | 2 |
| | | 166.1.158 | 99161 | Ge2Li4Zr1 | 2 |
| | | 166.1.160 | 418524 | Gd1O1P1Zn1 | 2 |
| | | 166.1.161 | 55619 | Gd1Mg2Ni9 | 2 |
| | | 166.1.162 | 635712 | Gd1 | 2 |
| | | 166.1.164 | 52517 | Gd1 | 2 |
| | | 166.1.165 | 38267 | Ga7Rb1 | 2 |
| | | 166.1.166 | 102891 | Ga7Rb1 | 2 |
| | | 166.1.167 | 602235 | Ga3Ho1 | 2 |
| | | 166.1.168 | 657608 | Ga2Te3 | 2 |
| | | 166.1.169 | 635031 | Ga1P1 | 2 |
| | | 166.1.170 | 96491 | Ga1Mn1 | 2 |
| | | 166.1.171 | 35386 | Ga1Ge1Te1 | 2 |
| | | 166.1.173 | 634071 | Fe7W6 | 2 |
| | | 166.1.174 | 634063 | Fe7W6 | 2 |
| | | 166.1.175 | 634060 | Fe7W6 | 2 |
| | | 166.1.176 | 103694 | Fe7W6 | 2 |
| | | 166.1.178 | 632622 | Fe7Mo6 | 2 |
| | | 166.1.179 | 632620 | Fe7Mo6 | 2 |
| | | 166.1.184 | 633896 | Fe3Th1 | 2 |
| | | 166.1.186 | 103654 | Fe3Th1 | 2 |
| | | 166.1.188 | 631893 | Fe3Gd1 | 2 |
| | | 166.1.189 | 631888 | Fe3Gd1 | 2 |
| | | 166.1.190 | 631864 | Fe3Gd1 | 2 |
| | | 166.1.191 | 103479 | Fe3Gd1 | 2 |
| | | 166.1.192 | 4192 | Fe2O4Yb1 | 2 |
| | | 166.1.193 | 195811 | Fe2O4Yb1 | 2 |
| | | 166.1.194 | 195810 | Fe2O4Yb1 | 2 |
| | | 166.1.196 | 29011 | Fe1O2Tl1 | 2 |
| | | 166.1.197 | 631854 | Fe15Ga2Y2 | 2 |
| | | 166.1.198 | 95660 | F1Nd1O1 | 2 |
| | | 166.1.199 | 95659 | F1Nd1O1 | 2 |
| | | 166.1.200 | 95658 | F1Nd1O1 | 2 |
| | | 166.1.201 | 95657 | F1Nd1O1 | 2 |
| | | 166.1.202 | 95656 | F1Nd1O1 | 2 |
| | | 166.1.203 | 95655 | F1Nd1O1 | 2 |
| | | 166.1.204 | 184005 | F1Nd1O1 | 2 |
| | | 166.1.206 | 202068 | Eu2I1P1 | 2 |
| | | 166.1.207 | 67700 | Er1Fe2O4 | 2 |
| | | 166.1.210 | 629979 | Dy2Ni7 | 2 |
| | | 166.1.212 | 630007 | Dy1Ni3 | 2 |
| | | 166.1.213 | 629986 | Dy1Ni3 | 2 |
| | | 166.1.214 | 629980 | Dy1Ni3 | 2 |
| | | 166.1.217 | 629540 | Dy1 | 2 |
| | | 166.1.218 | 629537 | Dy1 | 2 |
| | | 166.1.219 | 52518 | Dy1 | 2 |
| | | 166.1.220 | 190938 | Cu1N2Ta1 | 2 |
| | | 166.1.221 | 80259 | Cu1I1 | 2 |
| | | 166.1.222 | 80257 | Cu1I1 | 2 |
| | | 166.1.223 | 80253 | Cu1I1 | 2 |
| | | 166.1.224 | 80252 | Cu1I1 | 2 |
| | | 166.1.227 | 42152 | Cs2F6Na1Ni1 | 2 |
| | | 166.1.228 | 16255 | Cs2F6Fe1Na1 | 2 |
| | | 166.1.229 | 102864 | Cs1Ga7 | 2 |
| | | 166.1.230 | 625563 | Co7Y2 | 2 |
| | | 166.1.231 | 625438 | Co7Th2 | 2 |
| | | 166.1.232 | 625416 | Co7Th2 | 2 |
| | | 166.1.234 | 623985 | Co7La2 | 2 |
| | | 166.1.235 | 623978 | Co7La2 | 2 |
| | | 166.1.236 | 623284 | Co7Gd2 | 2 |
| | | 166.1.237 | 658884 | Co7Dy2 | 2 |
| | | 166.1.238 | 656092 | Co3Y1 | 2 |
| | | 166.1.239 | 656079 | Co3Y1 | 2 |
| | | 166.1.240 | 625636 | Co3Y1 | 2 |
| | | 166.1.241 | 625625 | Co3Y1 | 2 |
| | | 166.1.242 | 625605 | Co3Y1 | 2 |
| | | 166.1.243 | 625600 | Co3Y1 | 2 |
| | | 166.1.244 | 625590 | Co3Y1 | 2 |
| | | 166.1.245 | 625580 | Co3Y1 | 2 |
| | | 166.1.246 | 625572 | Co3Y1 | 2 |
| | | 166.1.247 | 625564 | Co3Y1 | 2 |

Continued on next column

**TABLE XVI– continued**

| SG | cl | TQC | ICSD | Stoich. Formula | SubSG |
|---|---|---|---|---|---|
| | | 166.1.248 | 603098 | Co3Y1 | 2 |
| | | 166.1.249 | 159894 | Co3Y1 | 2 |
| | | 166.1.250 | 159890 | Co3Y1 | 2 |
| | | 166.1.251 | 102729 | Co3Y1 | 2 |
| | | 166.1.254 | 24581 | Co3Ta1 | 2 |
| | | 166.1.255 | 624740 | Co3Pr1 | 2 |
| | | 166.1.256 | 624707 | Co3Pr1 | 2 |
| | | 166.1.257 | 624697 | Co3Pr1 | 2 |
| | | 166.1.258 | 108889 | Co3Pr1 | 2 |
| | | 166.1.259 | 624054 | Co3Lu1 | 2 |
| | | 166.1.260 | 624039 | Co3Lu1 | 2 |
| | | 166.1.263 | 79717 | Co10Ge3O16 | 2 |
| | | 166.1.264 | 401026 | Cl8Na2Ti3 | 2 |
| | | 166.1.268 | 24410 | Cl1La1 | 2 |
| | | 166.1.269 | 154513 | Ce2Cu5Zn2 | 2 |
| | | 166.1.270 | 620668 | Ce2Co7 | 2 |
| | | 166.1.271 | 620636 | Ce2Co7 | 2 |
| | | 166.1.272 | 603443 | Ce2Co5Fe12N3 | 2 |
| | | 166.1.273 | 201355 | Cd1Cs4F12Ni3 | 2 |
| | | 166.1.274 | 166533 | Ca2I1P1 | 2 |
| | | 166.1.275 | 424597 | Ca2Ge1Pd3 | 2 |
| | | 166.1.276 | 52789 | Ca1Si2 | 2 |
| | | 166.1.277 | 409057 | Ca1Si2 | 2 |
| | | 166.1.278 | 32006 | Ca1Si2 | 2 |
| | | 166.1.279 | 248518 | Ca1Si2 | 2 |
| | | 166.1.280 | 248517 | Ca1Si2 | 2 |
| | | 166.1.281 | 248516 | Ca1Si2 | 2 |
| | | 166.1.282 | 248515 | Ca1Si2 | 2 |
| | | 166.1.283 | 193539 | Ca1Si2 | 2 |
| | | 166.1.284 | 15632 | Ca1Si2 | 2 |
| | | 166.1.285 | 77316 | Ca1Ge2 | 2 |
| | | 166.1.286 | 619304 | Ca1Ge2 | 2 |
| | | 166.1.287 | 245612 | Ca1Ge2 | 2 |
| | | 166.1.288 | 110107 | Ca1Ge2 | 2 |
| | | 166.1.289 | 20822 | C5Ti8 | 2 |
| | | 166.1.291 | 659087 | C3Er2Fe17 | 2 |
| | | 166.1.292 | 41260 | C3Al6N2 | 2 |
| | | 166.1.293 | 74178 | C2Ba1Ce1F1O6 | 2 |
| | | 166.1.294 | 96188 | C1Y2 | 2 |
| | | 166.1.295 | 22283 | C1Y2 | 2 |
| | | 166.1.296 | 42762 | C1Tb2 | 2 |
| | | 166.1.297 | 42761 | C1Tb2 | 2 |
| | | 166.1.298 | 108187 | C1Tb2 | 2 |
| | | 166.1.299 | 23791 | C1S2Ta2 | 2 |
| | | 166.1.305 | 602542 | C1B2U1 | 2 |
| | | 166.1.306 | 44142 | C1B2U1 | 2 |
| | | 166.1.307 | 238975 | C1B2Pu1 | 2 |
| | | 166.1.310 | 202067 | Br1Eu2P1 | 2 |
| | | 166.1.311 | 95551 | Bi6Pb1Te10 | 2 |
| | | 166.1.313 | 74348 | Bi2Te3 | 2 |
| | | 166.1.314 | 658764 | Bi2Te3 | 2 |
| | | 166.1.315 | 44983 | Bi2Te3 | 2 |
| | | 166.1.316 | 42546 | Bi2Te3 | 2 |
| | | 166.1.317 | 193340 | Bi2Te3 | 2 |
| | | 166.1.318 | 193339 | Bi2Te3 | 2 |
| | | 166.1.319 | 193338 | Bi2Te3 | 2 |
| | | 166.1.320 | 193337 | Bi2Te3 | 2 |
| | | 166.1.321 | 193336 | Bi2Te3 | 2 |
| | | 166.1.322 | 193335 | Bi2Te3 | 2 |
| | | 166.1.323 | 193334 | Bi2Te3 | 2 |
| | | 166.1.324 | 193333 | Bi2Te3 | 2 |
| | | 166.1.325 | 193332 | Bi2Te3 | 2 |
| | | 166.1.326 | 193331 | Bi2Te3 | 2 |
| | | 166.1.327 | 193330 | Bi2Te3 | 2 |
| | | 166.1.328 | 184631 | Bi2Te3 | 2 |
| | | 166.1.329 | 158366 | Bi2Te3 | 2 |
| | | 166.1.330 | 15753 | Bi2Te3 | 2 |
| | | 166.1.331 | 42545 | Bi2Se3 | 2 |
| | | 166.1.332 | 20385 | Bi2Se3 | 2 |
| | | 166.1.333 | 165226 | Bi2Se3 | 2 |
| | | 166.1.334 | 54838 | Bi2Se2Te1 | 2 |
| | | 166.1.335 | 43512 | Bi2Se1Te2 | 2 |
| | | 166.1.336 | 658633 | Bi2Ge1Te4 | 2 |
| | | 166.1.337 | 30394 | Bi2Ge1Te4 | 2 |
| | | 166.1.338 | 172572 | Bi1S2Tl1 | 2 |
| | | 166.1.352 | 58724 | Be17Nb2 | 2 |
| | | 166.1.355 | 99662 | Ba4Mn3O12Pr1 | 2 |
| | | 166.1.359 | 58660 | Ba2Mg17 | 2 |
| | | 166.1.360 | 58665 | Ba1Pb3 | 2 |
| | | 166.1.361 | 419973 | Ba1Pb3 | 2 |
| | | 166.1.362 | 99682 | Ba1O3Ru1 | 2 |
| | | 166.1.363 | 91075 | Ba1O3Ru1 | 2 |
| | | 166.1.364 | 51293 | Ba1O3Ru1 | 2 |
| | | 166.1.366 | 110770 | Ba1O3Ru1 | 2 |
| | | 166.1.367 | 10253 | Ba1O3Ru1 | 2 |
| | | 166.1.368 | 20898 | Ba1Ni4O8 | 2 |
| | | 166.1.370 | 657450 | B6Ce5Fe2 | 2 |
| | | 166.1.371 | 39554 | B4Mo2 | 2 |
| | | 166.1.373 | 601082 | B3Co1Dy2.5 | 2 |
| | | 166.1.374 | 44451 | B2Mo1 | 2 |
| | | 166.1.375 | 418398 | B2Mo1 | 2 |
| | | 166.1.376 | 418397 | B2Mo1 | 2 |
| | | 166.1.377 | 40907 | B2Mo1 | 2 |
| | | 166.1.378 | 252125 | B2Mo1 | 2 |
| | | 166.1.379 | 97343 | B2Ce2Ir5 | 2 |
| | | 166.1.380 | 409982 | B1Br1Eu2N2 | 2 |
| | | 166.1.384 | 68110 | As2Te3 | 2 |
| | | 166.1.385 | 82371 | As2Sn2Sr1 | 2 |
| | | 166.1.386 | 68114 | As2Ge4Te7 | 2 |
| | | 166.1.387 | 68113 | As2Ge3Te6 | 2 |
| | | 166.1.388 | 422526 | As2Ca1Ga2 | 2 |
| | | 166.1.389 | 53794 | As1 | 2 |
| | | 166.1.390 | 16518 | As1 | 2 |
| | | 166.1.391 | 16517 | As1 | 2 |
| | | 166.1.392 | 16516 | As1 | 2 |
| | | 166.1.393 | 281656 | Al6Gd2Pt1Si4 | 2 |
| | | 166.1.396 | 420092 | Al10Ba7 | 2 |
| | | 166.1.398 | 416373 | Ag2Ni1O2 | 2 |
| | | 166.1.399 | 412279 | Ag2Ni1O2 | 2 |
| | | 166.1.400 | 172558 | Ag2Ni1O2 | 2 |
| | | 166.1.401 | 160582 | Ag2Ni1O2 | 2 |
| | | 166.1.402 | 160581 | Ag2Ni1O2 | 2 |
| | | 166.1.403 | 160580 | Ag2Ni1O2 | 2 |
| | | 166.1.404 | 160579 | Ag2Ni1O2 | 2 |
| | | 166.1.405 | 160578 | Ag2Ni1O2 | 2 |
| | | 166.1.406 | 160577 | Ag2Ni1O2 | 2 |
| | | 166.1.407 | 160576 | Ag2Ni1O2 | 2 |
| | | 166.1.408 | 160575 | Ag2Ni1O2 | 2 |
| | | 166.1.409 | 160574 | Ag2Ni1O2 | 2 |
| | | 166.1.410 | 251620 | Ag1Ge1Li2 | 2 |
| 166 | 2 | 166.1.11 | 653358 | Tm2Zn17 | 0 |
| | | 166.1.12 | 104246 | Tm2Zn17 | 0 |

Continued on next column

**TABLE XVI– continued**

| SG | cl | TQC | ICSD | Stoich. Formula | SubSG |
|---|---|---|---|---|---|
| | | 166.1.13 | 653216 | Te1Zr3 | 0 |
| | | 166.1.20 | 651728 | Sb2V3 | 0 |
| | | 166.1.21 | 41814 | Sb2V3 | 0 |
| | | 166.1.37 | 187495 | Sb2Te3 | 0 |
| | | 166.1.72 | 650000 | Pu2Zn17 | 0 |
| | | 166.1.73 | 41372 | Pt3Te4 | 0 |
| | | 166.1.76 | 649479 | Pr2Zn17 | 0 |
| | | 166.1.77 | 649469 | Pr2Zn17 | 0 |
| | | 166.1.78 | 104241 | Pr2Zn17 | 0 |
| | | 166.1.80 | 648763 | Pd3S2Tl2 | 0 |
| | | 166.1.81 | 648762 | Pd3S2Tl2 | 0 |
| | | 166.1.83 | 53301 | P1 | 148 |
| | | 166.1.85 | 647123 | Ni7Yb2 | 0 |
| | | 166.1.87 | 646978 | Ni3Ti0.67Zr0.33 | 0 |
| | | 166.1.88 | 646399 | Ni3S2Tl2 | 0 |
| | | 166.1.89 | 646398 | Ni3S2Tl2 | 0 |
| | | 166.1.90 | 422328 | Ni3S2Tl2 | 0 |
| | | 166.1.91 | 108893 | Ni3S2Tl2 | 0 |
| | | 166.1.92 | 654948 | Ni3S2Sn2 | 0 |
| | | 166.1.93 | 646380 | Ni3S2Sn2 | 0 |
| | | 166.1.94 | 646379 | Ni3S2Sn2 | 0 |
| | | 166.1.95 | 402458 | Ni3S2Sn2 | 0 |
| | | 166.1.96 | 100217 | Ni3S2Sn2 | 0 |
| | | 166.1.97 | 646211 | Ni3Pb2Se2 | 0 |
| | | 166.1.98 | 646210 | Ni3Pb2Se2 | 0 |
| | | 166.1.99 | 402453 | Ni3Pb2Se2 | 0 |
| | | 166.1.100 | 108892 | Ni3Pb2Se2 | 0 |
| | | 166.1.101 | 646207 | Ni3Pb2S2 | 0 |
| | | 166.1.102 | 646206 | Ni3Pb2S2 | 0 |
| | | 166.1.103 | 417632 | Ni3Pb2S2 | 0 |
| | | 166.1.104 | 24176 | Ni3Pb2S2 | 0 |
| | | 166.1.105 | 108895 | Ni3Pb2S2 | 0 |
| | | 166.1.115 | 641738 | La1Rh3 | 0 |
| | | 166.1.116 | 641724 | La1Rh3 | 0 |
| | | 166.1.120 | 640141 | In2Ni3Se2 | 0 |
| | | 166.1.121 | 640140 | In2Ni3Se2 | 0 |
| | | 166.1.122 | 640135 | In2Ni3S2 | 0 |
| | | 166.1.123 | 640134 | In2Ni3S2 | 0 |
| | | 166.1.124 | 415258 | In2Ni3S2 | 0 |
| | | 166.1.125 | 108894 | In2Ni3S2 | 0 |
| | | 166.1.128 | 53771 | Hg1 | 2 |
| | | 166.1.132 | 638697 | Hf1Ni3 | 0 |
| | | 166.1.133 | 2414 | Hf1Ni3 | 0 |
| | | 166.1.135 | 15888 | H8N2O8Pb1S2 | 0 |
| | | 166.1.136 | 195640 | H6Mg1Nb2 | 0 |
| | | 166.1.148 | 24685 | H3Cl1Co2O3 | 0 |
| | | 166.1.149 | 155534 | H3Cl1Co2O3 | 0 |
| | | 166.1.150 | 6016 | H2N2Th1 | 0 |
| | | 166.1.151 | 423880 | H2Mo1O4Rb2S1 | 0 |
| | | 166.1.153 | 40981 | H1Cl1Sc1 | 0 |
| | | 166.1.154 | 62226 | H1Cl1Lu1 | 0 |
| | | 166.1.155 | 911 | H12K2N6Sn1 | 0 |
| | | 166.1.156 | 159891 | H1.24Co3Y1 | 0 |
| | | 166.1.157 | 94220 | Ge2Nd2Zn15 | 0 |
| | | 166.1.159 | 166193 | Gd1Sn4Ti6 | 0 |
| | | 166.1.163 | 635711 | Gd1 | 0 |
| | | 166.1.172 | 2513 | Ga1 | 0 |
| | | 166.1.177 | 632784 | Fe7Nb6 | 0 |
| | | 166.1.180 | 41380 | Fe6Ge5Li1 | 0 |
| | | 166.1.181 | 633920 | Fe3Th1 | 0 |
| | | 166.1.182 | 633912 | Fe3Th1 | 0 |

Continued on next column

**TABLE XVI– continued**

| SG | cl | TQC | ICSD | Stoich. Formula | SubSG |
|---|---|---|---|---|---|
| | | 166.1.183 | 633900 | Fe3Th1 | 0 |
| | | 166.1.185 | 633886 | Fe3Th1 | 0 |
| | | 166.1.187 | 71 | Fe3Sn2 | 0 |
| | | 166.1.195 | 195809 | Fe2O4Yb1 | 0 |
| | | 166.1.205 | 184011 | F1Ho1O1 | 0 |
| | | 166.1.208 | 184012 | Er1F1O1 | 0 |
| | | 166.1.209 | 630377 | Dy2Zn17 | 0 |
| | | 166.1.211 | 262377 | Dy2Ni7 | 0 |
| | | 166.1.215 | 629598 | Dy1Fe3 | 0 |
| | | 166.1.216 | 184010 | Dy1F1O1 | 0 |
| | | 166.1.225 | 410393 | Cs9F27Ni9 | 0 |
| | | 166.1.226 | 78928 | Cs4F12Fe2K1Li1 | 0 |
| | | 166.1.233 | 184367 | Co7Pr2 | 0 |
| | | 166.1.252 | 625506 | Co3Tm1 | 0 |
| | | 166.1.253 | 625499 | Co3Tm1 | 0 |
| | | 166.1.261 | 246307 | Co1Li1O2 | 0 |
| | | 166.1.262 | 657284 | Co17Gd2N3 | 0 |
| | | 166.1.265 | 869 | Cl1Zr1 | 0 |
| | | 166.1.266 | 20145 | Cl1Zr1 | 0 |
| | | 166.1.267 | 6077 | Cl1O1Yb1 | 0 |
| | | 166.1.290 | 657171 | C3Fe17Tb2 | 0 |
| | | 166.1.300 | 95112 | C1Nb2S2 | 0 |
| | | 166.1.301 | 95110 | C1Nb2S2 | 0 |
| | | 166.1.302 | 53780 | C1 | 0 |
| | | 166.1.303 | 31829 | C1 | 0 |
| | | 166.1.304 | 29123 | C1 | 0 |
| | | 166.1.308 | 1168 | Br1Zr1 | 0 |
| | | 166.1.309 | 23354 | Br1La1 | 0 |
| | | 166.1.312 | 20319 | Bi4Se3 | 0 |
| | | 166.1.339 | 64705 | Bi1 | 148 |
| | | 166.1.340 | 64704 | Bi1 | 148 |
| | | 166.1.341 | 64703 | Bi1 | 148 |
| | | 166.1.342 | 53797 | Bi1 | 148 |
| | | 166.1.343 | 246663 | Bi1 | 148 |
| | | 166.1.344 | 242004 | Bi1 | 148 |
| | | 166.1.345 | 242003 | Bi1 | 148 |
| | | 166.1.346 | 242002 | Bi1 | 148 |
| | | 166.1.347 | 242001 | Bi1 | 148 |
| | | 166.1.348 | 242000 | Bi1 | 148 |
| | | 166.1.349 | 241999 | Bi1 | 148 |
| | | 166.1.350 | 241984 | Bi1 | 148 |
| | | 166.1.351 | 241983 | Bi1 | 148 |
| | | 166.1.353 | 71755 | Ba7Cl2O15Ru4 | 0 |
| | | 166.1.354 | 73069 | Ba7Br2O15Ru4 | 0 |
| | | 166.1.356 | 174186 | Ba4Er1O12Ru3 | 0 |
| | | 166.1.357 | 20621 | Ba3O9Sc4 | 0 |
| | | 166.1.358 | 25005 | Ba2Ni1O6Te1 | 0 |
| | | 166.1.365 | 172175 | Ba1O3Ru1 | 0 |
| | | 166.1.369 | 657452 | B6Fe2Gd5 | 0 |
| | | 166.1.372 | 191908 | B4Mo2 | 0 |
| | | 166.1.381 | 262037 | As3Cd4Rb1 | 0 |
| | | 166.1.382 | 262035 | As3Cd4Na1 | 0 |
| | | 166.1.383 | 262032 | As3Cd4K1 | 0 |
| | | 166.1.394 | 281660 | Al6Dy2Pt1Si4 | 0 |
| | | 166.1.395 | 58175 | Al3Tb1 | 0 |
| | | 166.1.397 | 424311 | Ag4Sb2Sr1 | 0 |
| 167 | 1 | 167.1.3 | 95761 | O3V2 | 2 |
| | | 167.1.4 | 95760 | O3V2 | 2 |
| | | 167.1.8 | 25780 | O3V2 | 2 |
| | | 167.1.9 | 201109 | O3V2 | 2 |
| | | 167.1.13 | 1874 | O3V2 | 2 |

Continued on next column

**TABLE XVI– continued**

| SG | cl | TQC | ICSD | Stoich. Formula | SubSG |
|---|---|---|---|---|---|
| | | 167.1.14 | 1873 | O3V2 | 2 |
| | | 167.1.15 | 1872 | O3V2 | 2 |
| | | 167.1.19 | 167061 | La1Mn1O3.31 | 2 |
| | | 167.1.20 | 167062 | La1Mn1O3.29 | 2 |
| | | 167.1.21 | 167065 | La1Mn1O3.26 | 2 |
| | | 167.1.22 | 167064 | La1Mn1O3.26 | 2 |
| | | 167.1.23 | 167066 | La1Mn1O3.22 | 2 |
| | | 167.1.24 | 167068 | La1Mn1O3.15 | 2 |
| | | 167.1.25 | 167069 | La1Mn1O3.13 | 2 |
| | | 167.1.26 | 96070 | Fe2O3 | 2 |
| | | 167.1.27 | 96069 | Fe2O3 | 2 |
| | | 167.1.28 | 88418 | Fe2O3 | 2 |
| | | 167.1.29 | 88417 | Fe2O3 | 2 |
| | | 167.1.30 | 82904 | Fe2O3 | 2 |
| | | 167.1.31 | 82903 | Fe2O3 | 2 |
| | | 167.1.32 | 82902 | Fe2O3 | 2 |
| | | 167.1.33 | 82137 | Fe2O3 | 2 |
| | | 167.1.34 | 82136 | Fe2O3 | 2 |
| | | 167.1.35 | 82135 | Fe2O3 | 2 |
| | | 167.1.36 | 82134 | Fe2O3 | 2 |
| | | 167.1.37 | 81248 | Fe2O3 | 2 |
| | | 167.1.38 | 71194 | Fe2O3 | 2 |
| | | 167.1.39 | 66756 | Fe2O3 | 2 |
| | | 167.1.40 | 64599 | Fe2O3 | 2 |
| | | 167.1.41 | 56372 | Fe2O3 | 2 |
| | | 167.1.42 | 43465 | Fe2O3 | 2 |
| | | 167.1.43 | 415251 | Fe2O3 | 2 |
| | | 167.1.44 | 40142 | Fe2O3 | 2 |
| | | 167.1.45 | 24791 | Fe2O3 | 2 |
| | | 167.1.46 | 245851 | Fe2O3 | 2 |
| | | 167.1.47 | 24004 | Fe2O3 | 2 |
| | | 167.1.48 | 22505 | Fe2O3 | 2 |
| | | 167.1.49 | 201101 | Fe2O3 | 2 |
| | | 167.1.50 | 201100 | Fe2O3 | 2 |
| | | 167.1.51 | 201099 | Fe2O3 | 2 |
| | | 167.1.52 | 201098 | Fe2O3 | 2 |
| | | 167.1.53 | 201097 | Fe2O3 | 2 |
| | | 167.1.54 | 201096 | Fe2O3 | 2 |
| | | 167.1.55 | 184766 | Fe2O3 | 2 |
| | | 167.1.56 | 182848 | Fe2O3 | 2 |
| | | 167.1.57 | 182847 | Fe2O3 | 2 |
| | | 167.1.58 | 182846 | Fe2O3 | 2 |
| | | 167.1.59 | 182845 | Fe2O3 | 2 |
| | | 167.1.60 | 182844 | Fe2O3 | 2 |
| | | 167.1.61 | 182843 | Fe2O3 | 2 |
| | | 167.1.62 | 182842 | Fe2O3 | 2 |
| | | 167.1.63 | 182841 | Fe2O3 | 2 |
| | | 167.1.64 | 182840 | Fe2O3 | 2 |
| | | 167.1.65 | 182839 | Fe2O3 | 2 |
| | | 167.1.66 | 173655 | Fe2O3 | 2 |
| | | 167.1.67 | 173653 | Fe2O3 | 2 |
| | | 167.1.68 | 161294 | Fe2O3 | 2 |
| | | 167.1.69 | 161293 | Fe2O3 | 2 |
| | | 167.1.70 | 161292 | Fe2O3 | 2 |
| | | 167.1.71 | 161291 | Fe2O3 | 2 |
| | | 167.1.72 | 161290 | Fe2O3 | 2 |
| | | 167.1.73 | 161289 | Fe2O3 | 2 |
| | | 167.1.74 | 161288 | Fe2O3 | 2 |
| | | 167.1.75 | 161287 | Fe2O3 | 2 |
| | | 167.1.76 | 161286 | Fe2O3 | 2 |
| | | 167.1.77 | 161285 | Fe2O3 | 2 |

Continued on next column

**TABLE XVI– continued**

| SG | cl | TQC | ICSD | Stoich. Formula | SubSG |
|---|---|---|---|---|---|
| | | 167.1.78 | 161284 | Fe2O3 | 2 |
| | | 167.1.79 | 161283 | Fe2O3 | 2 |
| | | 167.1.80 | 15840 | Fe2O3 | 2 |
| | | 167.1.81 | 154191 | Fe2O3 | 2 |
| | | 167.1.82 | 154190 | Fe2O3 | 2 |
| | | 167.1.83 | 173470 | Cr2O3 | 2 |
| | | 167.1.84 | 201966 | Cr2Cs3I9 | 2 |
| | | 167.1.85 | 50809 | Ca3Co1O6Rh1 | 2 |
| | | 167.1.87 | 60625 | Br6Cd1Rb4 | 2 |
| | | 167.1.88 | 243759 | Ba3O9Os2 | 2 |
| | | 167.1.89 | 97540 | Ba3Cr2S6 | 2 |
| | | 167.1.90 | 402170 | Au3Ga7Yb1 | 2 |
| | | 167.1.91 | 391108 | Al7Au3Dy1 | 2 |
| | | 167.1.92 | 391101 | Al7Au3Ce1 | 2 |
| 167 | 2 | 167.1.1 | 83385 | S3Yb2 | 0 |
| | | 167.1.2 | 75248 | S3Yb2 | 0 |
| | | 167.1.5 | 94768 | O3V2 | 0 |
| | | 167.1.6 | 64796 | O3V2 | 0 |
| | | 167.1.7 | 64786 | O3V2 | 0 |
| | | 167.1.10 | 201108 | O3V2 | 0 |
| | | 167.1.11 | 201107 | O3V2 | 0 |
| | | 167.1.12 | 201106 | O3V2 | 0 |
| | | 167.1.16 | 1870 | O3V2 | 0 |
| | | 167.1.17 | 1869 | O3V2 | 0 |
| | | 167.1.18 | 1473 | O3V2 | 0 |
| | | 167.1.86 | 421391 | C3Cr2N6 | 0 |
| 176 | 1 | 176.1.2 | 251133 | Ba4Nb2O9 | 2,174 |
| 176 | 2 | 176.1.1 | 80772 | Mn1N3Sr3 | 174 |
| 189 | 1 | 189.1.1 | 51718 | Si5Yb3 | 174 |
| | | 189.1.2 | 415599 | Si5Yb3 | 174 |
| | | 189.1.3 | 408117 | Si5Yb3 | 174 |
| | | 189.1.6 | 196273 | O1Ti1 | 174 |
| | | 189.1.9 | 414438 | Mg1Tl1Y1 | 174 |
| | | 189.1.10 | 54344 | Mg1Sn1Yb1 | 174 |
| | | 189.1.13 | 414439 | La1Mg1Tl1 | 174 |
| | | 189.1.14 | 160910 | In1Mg1Y1 | 174 |
| | | 189.1.16 | 409925 | Ge1Pd1Ti1 | 174 |
| | | 189.1.18 | 240950 | Ga1Mg1Y1 | 174 |
| | | 189.1.19 | 160909 | Ga1Mg1Y1 | 174 |
| | | 189.1.20 | 413276 | Ga1La1Mg1 | 174 |
| | | 189.1.23 | 428473 | Cu1Ge1Sc1 | 174 |
| | | 189.1.30 | 413810 | Ba1Bi1Na1 | 174 |
| | | 189.1.31 | 261317 | Au1Ge2In3Yb3 | 174 |
| | | 189.1.35 | 810 | As2Eu2 | 174 |
| | | 189.1.36 | 193686 | Al3Co1Gd3Ge2 | 174 |
| | | 189.1.37 | 54936 | Al1Pd1Tb1 | 174 |
| | | 189.1.39 | 107275 | Ag1Pb1Y1 | 174 |
| | | 189.1.40 | 416377 | Ag1Lu1Sn1 | 174 |
| | | 189.1.41 | 245236 | Ag1Lu1Sn1 | 174 |
| | | 189.1.42 | 107278 | Ag1Lu1Pb1 | 174 |
| | | 189.1.43 | 10016 | Ag1Ca1P1 | 174 |
| | | 189.1.44 | 10017 | Ag1As1Ca1 | 174 |
| 189 | 2 | 189.1.5 | 415370 | Pr1Rh1Sn1 | 174 |
| | | 189.1.7 | 183337 | Nd1Pd1Zn1 | 174 |
| | | 189.1.11 | 411152 | Li5N3Ni3 | 174 |
| | | 189.1.15 | 85966 | Hf6Ni1Sb2 | 174 |
| | | 189.1.21 | 152689 | Fe1Ho6Sb2 | 174 |
| | | 189.1.25 | 415492 | Ce1Pt1Sn1 | 174 |
| | | 189.1.26 | 416969 | Ce1Pd1Sn1 | 174 |
| | | 189.1.32 | 104011 | Au1Gd1In1 | 174 |

Continued on next column

**TABLE XVI– continued**

| SG | cl | TQC | ICSD | Stoich. Formula | SubSG |
|---|---|---|---|---|---|
| | | 189.1.38 | 160908 | Al1Mg1Y1 | 174 |
| 189 | 3 | 189.1.22 | 418214 | Cu1La1Mg4 | 174 |
| | | 189.1.24 | 98233 | Ce2Ge3Zn6 | 174 |
| | | 189.1.34 | 411543 | Au1Cd1La1 | 174 |
| 189 | 4 | 189.1.8 | 419191 | Nd1Pd1Sn1 | 174 |
| | | 189.1.12 | 415594 | La1Rh1Sn1 | 174 |
| | | 189.1.27 | 100002 | Ce1In1Ir1 | 174 |
| | | 189.1.28 | 428851 | Cd1Ce1Pt1 | 174 |
| | | 189.1.33 | 192086 | Au1Ga1Mg1 | 174 |
| 189 | 5 | 189.1.17 | 59177 | Ge1Li1Pr1 | 174 |
| | | 189.1.29 | 51528 | C2Rb2 | 174 |
| 189 | 6 | 189.1.4 | 410418 | Pt1Sn1Yb1 | 174 |
| 190 | 1 | 190.2.1 | 81072 | Ba3Cr2O9W1 | 174 |
| | | 190.2.2 | 73183 | Ba3Ca1O9Ru2 | 174 |
| 191 | 1 | 191.1.20 | 409394 | Al9Co2Eu1 | 2,174 |
| 191 | 2 | 191.1.19 | 36504 | B1Ca1Ni4 | 2 |
| 191 | 3 | 191.1.18 | 417869 | B1Co4Gd1 | 2,174 |
| 191 | 4 | 191.1.5 | 107425 | Ga2Gd1 | 174,65 |
| | | 191.1.17 | 97722 | B1Co4Pr1 | 174,65 |
| 191 | 5 | 191.1.16 | 237660 | B1Eu1Pt4 | 10,174 |
| 191 | 6 | 191.1.15 | 184968 | B2Fe3U1 | 174 |
| 191 | 7 | 191.1.12 | 78848 | B2Ti1 | 10,174 |
| | | 191.1.13 | 78847 | B2Ti1 | 10,174 |
| | | 191.1.14 | 56723 | B2Ti1 | 10,174 |
| 191 | 8 | 191.1.11 | 97731 | B3Co7Dy2 | 2,174 |
| 191 | 9 | 191.1.10 | 97723 | B4Co11Pr3 | 10,174 |
| 191 | 10 | 191.1.1 | 173575 | La10O3Si8 | 2,174 |
| | | 191.1.9 | 246238 | Ba1Ga2 | 2,174 |
| 191 | 11 | 191.1.8 | 107305 | Ba1Sn5 | 2,174 |
| 191 | 12 | 191.1.7 | 418614 | Ca1Zn5 | 2,174 |
| 191 | 13 | 191.1.6 | 419163 | Ce1Pd2Zn3 | 2,174 |
| 191 | 14 | 191.1.2 | 417224 | Ga2Sr1 | 10,174 |
| | | 191.1.3 | 260654 | Ga2Sr1 | 10,174 |
| | | 191.1.4 | 246236 | Ga2Sr1 | 10,174 |
| 193 | 1 | 193.1.5 | 154390 | Nb1Sb5U3 | 2,174 |
| | | 193.1.8 | 173358 | Bi5Ce3Mn1 | 2,174 |
| 193 | 2 | 193.1.1 | 150562 | Sn3Zr5 | 2,174 |
| | | 193.1.2 | 109292 | Sn3Ti5 | 2,174 |
| | | 193.1.3 | 62591 | Si3Ti5 | 2,174 |
| | | 193.1.4 | 105649 | Pb3Zr5 | 2,174 |
| | | 193.1.7 | 44361 | Ge3Hf5 | 2,174 |
| 193 | 3 | 193.1.6 | 99999 | K3N1 | 174 |
| 194 | 1 | 194.1.1 | 421015 | Zn1 | 2,174 |
| | | 194.1.2 | 421014 | Zn1 | 2,174 |
| | | 194.1.3 | 421013 | Zn1 | 2,174 |
| | | 194.1.4 | 247162 | Zn1 | 2,174 |
| | | 194.1.5 | 247161 | Zn1 | 2,174 |
| | | 194.1.6 | 247160 | Zn1 | 2,174 |
| | | 194.1.7 | 247159 | Zn1 | 2,174 |
| | | 194.1.8 | 247158 | Zn1 | 2,174 |
| | | 194.1.9 | 247157 | Zn1 | 2,174 |
| | | 194.1.10 | 247156 | Zn1 | 2,174 |
| | | 194.1.11 | 247155 | Zn1 | 2,174 |
| | | 194.1.12 | 247154 | Zn1 | 2,174 |
| | | 194.1.13 | 247153 | Zn1 | 2,174 |
| | | 194.1.14 | 247152 | Zn1 | 2,174 |
| | | 194.1.15 | 247151 | Zn1 | 2,174 |
| | | 194.1.16 | 247150 | Zn1 | 2,174 |
| | | 194.1.17 | 247149 | Zn1 | 2,174 |
| | | 194.1.18 | 247148 | Zn1 | 2,174 |

Continued on next column

**TABLE XVI– continued**

| SG | cl | TQC | ICSD | Stoich. Formula | SubSG |
|---|---|---|---|---|---|
| | | 194.1.19 | 247147 | Zn1 | 2,174 |
| | | 194.1.28 | 417150 | Fe0.25S2Ta1 | 2,174 |
| | | 194.1.29 | 249600 | Eu1Ga1Sn1 | 2,174 |
| | | 194.1.34 | 420186 | Ce1Ge1Zn1 | 2,174 |
| | | 194.1.35 | 163343 | Ce1Ge1Zn1 | 2,174 |
| | | 194.1.36 | 425673 | C6Co1N6Pr1 | 2,174 |
| | | 194.1.51 | 249399 | C2Al1Ta3 | 2,174 |
| | | 194.1.57 | 251687 | Au1Ce1Sb1 | 2,174 |
| 194 | 2 | 194.1.20 | 422990 | Mn1O3Y1 | 174 |
| | | 194.1.21 | 422989 | Mn1O3Y1 | 174 |
| | | 194.1.22 | 422982 | Mn1O3Y1 | 174 |
| | | 194.1.23 | 182284 | Mn1O3Y1 | 174 |
| | | 194.1.24 | 181182 | Mn1O3Y1 | 174 |
| | | 194.1.25 | 193568 | Mg1Ni2 | 174 |
| | | 194.1.26 | 171451 | I1La1 | 174 |
| | | 194.1.27 | 163344 | Gd1Ge1Zn1 | 174 |
| | | 194.1.30 | 249599 | Eu1Ga1Ge1 | 174 |
| | | 194.1.56 | 192939 | Ba1Cr1O3 | 174 |
| 194 | 3 | 194.1.32 | 420050 | Ce1Sn1Zn1 | 2 |
| | | 194.1.33 | 152624 | Ce1Sn1Zn1 | 2 |
| | | 194.1.55 | 192940 | Ba1Cr1O3 | 2 |
| 194 | 4 | 194.1.52 | 246019 | Ba1O3Ru1 | 0 |
| | | 194.1.53 | 173866 | Ba1O3Ru1 | 0 |
| | | 194.1.54 | 160287 | Ba1O3Ru1 | 0 |
| 194 | 5 | 194.1.31 | 158979 | Cr1Ge3Nd1 | 2,174 |
| | | 194.1.37 | 180423 | C2Si1Ti3 | 2,174 |
| | | 194.1.38 | 180422 | C2Si1Ti3 | 2,174 |
| | | 194.1.39 | 180421 | C2Si1Ti3 | 2,174 |
| | | 194.1.40 | 180420 | C2Si1Ti3 | 2,174 |
| | | 194.1.41 | 180419 | C2Si1Ti3 | 2,174 |
| | | 194.1.42 | 153264 | C2Si1Ti3 | 2,174 |
| | | 194.1.43 | 180431 | C2Ge1Ti3 | 2,174 |
| | | 194.1.44 | 180430 | C2Ge1Ti3 | 2,174 |
| | | 194.1.45 | 180429 | C2Ge1Ti3 | 2,174 |
| | | 194.1.46 | 180428 | C2Ge1Ti3 | 2,174 |
| | | 194.1.47 | 180427 | C2Ge1Ti3 | 2,174 |
| | | 194.1.48 | 180426 | C2Ge1Ti3 | 2,174 |
| | | 194.1.49 | 180425 | C2Ge1Ti3 | 2,174 |
| | | 194.1.50 | 180424 | C2Ge1Ti3 | 2,174 |
| 204 | 1 | 204.1.5 | 237422 | Cu3Fe4O12Tm1 | 2 |
| | | 204.1.23 | 174551 | Ba1Ge12Pt4 | 2 |
| | | 204.1.24 | 160797 | Ba1Ge12Pt4 | 2 |
| 204 | 2 | 204.1.14 | 248607 | Cr4Cu3La1O12 | 71 |
| | | 204.1.15 | 248606 | Cr4Cu3La1O12 | 71 |
| | | 204.1.16 | 248605 | Cr4Cu3La1O12 | 71 |
| | | 204.1.17 | 248604 | Cr4Cu3La1O12 | 71 |
| | | 204.1.18 | 248603 | Cr4Cu3La1O12 | 71 |
| | | 204.1.19 | 248601 | Cr4Cu3La1O12 | 71 |
| | | 204.1.20 | 248600 | Cr4Cu3La1O12 | 71 |
| | | 204.1.21 | 248599 | Cr4Cu3La1O12 | 71 |
| | | 204.1.22 | 248598 | Cr4Cu3La1O12 | 71 |
| 204 | 3 | 204.1.4 | 51893 | Cu3Na1O12Ru4 | 0 |
| | | 204.1.6 | 248615 | Cr4Cu3O12Y1 | 0 |
| | | 204.1.7 | 248614 | Cr4Cu3O12Y1 | 0 |
| | | 204.1.8 | 248613 | Cr4Cu3O12Y1 | 0 |
| | | 204.1.9 | 248612 | Cr4Cu3O12Y1 | 0 |
| | | 204.1.10 | 248611 | Cr4Cu3O12Y1 | 0 |
| | | 204.1.11 | 248610 | Cr4Cu3O12Y1 | 0 |
| | | 204.1.12 | 248609 | Cr4Cu3O12Y1 | 0 |
| | | 204.1.13 | 248608 | Cr4Cu3O12Y1 | 0 |

Continued on next column

**TABLE XVI– continued**

| SG | cl | TQC | ICSD | Stoich. Formula | SubSG |
|---|---|---|---|---|---|
| 204 | 4 | 204.1.1 | 237751 | Mn3Na1O12V4 | 2 |
| | | 204.1.2 | 195439 | Cu4O12V4 | 2 |
| | | 204.1.3 | 195438 | Cu4O12V4 | 2 |
| 205 | 1 | 205.2.1 | 77109 | Pd1Sb2 | 2 |
| | | 205.2.2 | 300225 | Os1Te2 | 2 |
| | | 205.2.3 | 86354 | Ni1S2 | 2 |
| | | 205.2.4 | 68169 | Ni1S2 | 2 |
| | | 205.2.5 | 68168 | Ni1S2 | 2 |
| | | 205.2.6 | 68167 | Ni1S2 | 2 |
| | | 216.2.1 | 260984 | Ni5U1 | 82 |
| | | 216.2.2 | 54390 | Ni4Sn1U1 | 82 |
| | | 216.2.3 | 54387 | Ni4Sn1U1 | 82 |
| | | 216.2.4 | 418013 | La4Mg1Ru1 | 82 |
| | | 216.2.5 | 417519 | In1Rh1Tm4 | 82 |
| | | 216.2.6 | 81917 | Hg1S1 | 82 |
| | | 216.2.7 | 419068 | Cd1Gd4Ru1 | 82 |
| | | 216.2.8 | 415194 | Cd1Ce1Ni4 | 82 |
| | | 217.1.1 | 427048 | F6Ni3O6Sb4 | 82 |
| | | 217.1.2 | 427047 | F6Ni3O6Sb4 | 82 |
| | | 217.1.3 | 84572 | Cu12S13Sb4 | 82 |
| | | 217.1.4 | 84571 | Cu12S13Sb4 | 82 |
| | | 217.1.5 | 62116 | Cu12S13Sb4 | 82 |
| | | 217.1.6 | 25707 | Cu12S13Sb4 | 82 |
| | | 217.1.7 | 33588 | As4Cu12S12 | 82 |
| | | 220.2.1 | 155383 | C3Y2 | 82 |
| | | 220.2.2 | 155382 | C3Y2 | 82 |
| | | 220.2.3 | 26477 | C3U2 | 82 |
| | | 220.2.4 | 2447 | C3Nd2 | 82 |
| | | 221.1.13 | 50313 | B6Sr1 | 2,81 |
| | | 221.1.14 | 51495 | B6Ba1 | 2,81 |
| 221 | 2 | 221.1.7 | 262924 | Cs1I3Sn1 | 2,81 |
| | | 221.1.11 | 80320 | Br3Cs1Ge1 | 2,81 |
| | | 221.1.12 | 409846 | Br3Ce3Ga1 | 2,81 |
| 221 | 3 | 221.1.1 | 201036 | O1Sn1Sr3 | 47,65,83 |
| | | 221.1.8 | 100787 | Ca3O1Sn1 | 47,65,83 |
| | | 221.1.9 | 100789 | Ca3O1Pb1 | 47,65,83 |
| | | 221.1.10 | 402087 | Ca3Ge1O1 | 47,65,83 |
| 221 | 4 | 221.1.6 | 100533 | Fe1Ti1 | 2,81 |
| 221 | 5 | 221.1.4 | 247065 | In1N1Ni3 | 10 |
| | | 221.1.5 | 247273 | H1Pd3Tl1 | 10 |
| 221 | 6 | 221.1.3 | 163195 | Mg1Pd1 | 2,81 |
| 221 | 7 | 221.1.2 | 181127 | Ni3Sn1 | 10 |
| 225 | 1 | 225.1.1 | 63098 | Pb1Te1 | 139 |
| | | 225.1.2 | 153711 | Pb1Te1 | 139 |
| | | 225.1.3 | 63097 | Pb1Se1 | 139 |
| | | 225.1.4 | 63096 | Pb1Se1 | 139 |
| | | 225.1.5 | 62196 | Pb1Se1 | 139 |
| | | 225.1.6 | 62195 | Pb1Se1 | 139 |
| | | 225.1.7 | 38294 | Pb1Se1 | 139 |
| | | 225.1.8 | 62192 | Pb1S1 | 139 |
| | | 225.1.9 | 62191 | Pb1S1 | 139 |
| | | 225.1.10 | 62190 | Pb1S1 | 139 |
| | | 225.1.11 | 35204 | O2U1 | 139 |
| | | 225.1.12 | 246851 | O2U1 | 139 |
| | | 225.1.16 | 160154 | Fe16Si7U6 | 139 |
| | | 225.1.30 | 409959 | B6Ni21Sc2 | 139 |
| 225 | 2 | 225.1.17 | 188351 | Cl1Cu6O8Y1 | 69,71,87 |
| | | 225.1.18 | 69612 | Cl1Cu6In1O8 | 69,71,87 |
| | | 225.1.28 | 59744 | Ba2Er1O6Ru1 | 69,71,87 |
| | | 225.1.29 | 59743 | Ba2Er1O6Ru1 | 69,71,87 |

Continued on next column

**TABLE XVI– continued**

| SG | cl | TQC | ICSD | Stoich. Formula | SubSG |
|---|---|---|---|---|---|
| 225 | 3 | 225.1.20 | 237543 | Ca3Cd8Pt4 | 2,82 |
| | | 225.1.26 | 171491 | Ba2Ir1O6Tb1 | 2,82 |
| | | 225.1.27 | 153057 | Ba2Ir1O6Pr1 | 2,82 |
| 225 | 4 | 225.1.23 | 424342 | C1W1 | 2 |
| | | 225.1.25 | 155551 | Ba2Nd1O6Ru1 | 69,71,87 |
| 225 | 5 | 225.1.14 | 159261 | Mg6Ni16Si7 | 2,82 |
| | | 225.1.24 | 159873 | C1Hf1 | 2,82 |
| 225 | 6 | 225.1.15 | 106715 | Ga2Li1Rh1 | 2,82 |
| | | 225.1.22 | 159874 | C1Zr1 | 2,82 |
| 225 | 7 | 225.1.13 | 419774 | Ni21P6Sn2 | 2,82 |
| | | 225.1.19 | 75013 | Ce3Pd20Si6 | 2,82 |
| | | 225.1.21 | 62667 | C6Cr23 | 2,82 |
| 227 | 1 | 227.1.3 | 151747 | O7Tb2Ti2 | 0 |
| | | 227.1.4 | 429192 | O7Pr2Zr2 | 0 |
| | | 227.1.10 | 202522 | Mo2O7Y2 | 0 |
| | | 227.1.37 | 160866 | Er2O7V2 | 0 |
| | | 227.1.51 | 183965 | Cr2Fe1O4 | 0 |
| | | 227.1.52 | 183964 | Cr2Fe1O4 | 0 |
| | | 227.1.53 | 183963 | Cr2Fe1O4 | 0 |
| | | 227.1.54 | 183962 | Cr2Fe1O4 | 0 |
| | | 227.1.55 | 183961 | Cr2Fe1O4 | 0 |
| | | 227.1.56 | 183960 | Cr2Fe1O4 | 0 |
| | | 227.1.57 | 171121 | Cr2Fe1O4 | 0 |
| | | 227.1.88 | 57512 | Al2Ba1 | 0 |
| 227 | 2 | 227.1.1 | 72042 | S2Ti1 | 2,82 |
| | | 227.1.2 | 160865 | O7V2Y2 | 2,82 |
| | | 227.1.5 | 261854 | Nd2O7Zr2 | 2,82 |
| | | 227.1.6 | 249974 | Nd2O7Zr2 | 2,82 |
| | | 227.1.7 | 174187 | Nd2O7Zr2 | 2,82 |
| | | 227.1.8 | 153227 | Nd2O7Zr2 | 2,82 |
| | | 227.1.9 | 82958 | Nd2O7Sn2 | 2,82 |
| | | 227.1.11 | 83277 | Mn2O7Sc2 | 2,82 |
| | | 227.1.12 | 187996 | Mg3Ni2Ti1 | 2,82 |
| | | 227.1.13 | 184696 | Mg1O4Ti2 | 2,82 |
| | | 227.1.14 | 32688 | Lu2O7V2 | 2,82 |
| | | 227.1.15 | 159290 | Lu2O7V2 | 2,82 |
| | | 227.1.16 | 159286 | Lu2O7V2 | 2,82 |
| | | 227.1.17 | 48128 | Li1O2Ti1 | 2,82 |
| | | 227.1.18 | 423608 | In2Mg1 | 2,82 |
| | | 227.1.19 | 420818 | Hg2O7Ru2 | 2,82 |
| | | 227.1.20 | 159809 | Hg2O7Os2 | 2,82 |
| | | 227.1.21 | 159808 | Hg2O7Os2 | 2,82 |
| | | 227.1.22 | 159807 | Hg2O7Os2 | 2,82 |
| | | 227.1.23 | 159806 | Hg2O7Os2 | 2,82 |
| | | 227.1.24 | 159805 | Hg2O7Os2 | 2,82 |
| | | 227.1.25 | 159804 | Hg2O7Os2 | 2,82 |
| | | 227.1.26 | 159803 | Hg2O7Os2 | 2,82 |
| | | 227.1.27 | 159802 | Hg2O7Os2 | 2,82 |
| | | 227.1.28 | 159801 | Hg2O7Os2 | 2,82 |
| | | 227.1.29 | 159800 | Hg2O7Os2 | 2,82 |
| | | 227.1.30 | 159799 | Hg2O7Os2 | 2,82 |
| | | 227.1.31 | 159798 | Hg2O7Os2 | 2,82 |
| | | 227.1.32 | 159797 | Hg2O7Os2 | 2,82 |
| | | 227.1.33 | 159796 | Hg2O7Os2 | 2,82 |
| | | 227.1.34 | 159795 | Hg2O7Os2 | 2,82 |
| | | 227.1.35 | 159794 | Hg2O7Os2 | 2,82 |
| | | 227.1.36 | 84753 | Gd2O7Sn2 | 2,82 |
| | | 227.1.38 | 50047 | Cr2O4Zn1 | 2,82 |
| | | 227.1.39 | 196119 | Cr2O4Zn1 | 2,82 |
| | | 227.1.40 | 171899 | Cr2O4Zn1 | 2,82 |

Continued on next column

**TABLE XVI– continued**

| SG | cl | TQC | ICSD | Stoich. Formula | SubSG |
|---|---|---|---|---|---|
| | | 227.1.41 | 171896 | Cr2O4Zn1 | 2,82 |
| | | 227.1.42 | 171894 | Cr2O4Zn1 | 2,82 |
| | | 227.1.43 | 171893 | Cr2O4Zn1 | 2,82 |
| | | 227.1.44 | 97201 | Cr2Mg1O4 | 2,82 |
| | | 227.1.45 | 75623 | Cr2Mg1O4 | 2,82 |
| | | 227.1.46 | 290599 | Cr2Mg1O4 | 2,82 |
| | | 227.1.47 | 171107 | Cr2Mg1O4 | 2,82 |
| | | 227.1.48 | 171106 | Cr2Mg1O4 | 2,82 |
| | | 227.1.49 | 160953 | Cr2Mg1O4 | 2,82 |
| | | 227.1.50 | 402408 | Cr2Hg1Se4 | 2,82 |
| | | 227.1.58 | 859 | Co2O4Si1 | 2,82 |
| | | 227.1.59 | 154730 | Co2Nd1 | 2,82 |
| | | 227.1.60 | 154729 | Co2Nd1 | 2,82 |
| | | 227.1.61 | 154728 | Co2Nd1 | 2,82 |
| | | 227.1.62 | 154727 | Co2Nd1 | 2,82 |
| | | 227.1.63 | 154726 | Co2Nd1 | 2,82 |
| | | 227.1.64 | 154725 | Co2Nd1 | 2,82 |
| | | 227.1.65 | 154724 | Co2Nd1 | 2,82 |
| | | 227.1.66 | 194667 | Co2Er1 | 2,82 |
| | | 227.1.67 | 168600 | Ce2O7Zr2 | 2,82 |
| | | 227.1.68 | 168595 | Ce2O7Zr2 | 2,82 |
| | | 227.1.69 | 155773 | Cd2O7Os2 | 2,82 |
| | | 227.1.70 | 155772 | Cd2O7Os2 | 2,82 |
| | | 227.1.71 | 155771 | Cd2O7Os2 | 2,82 |
| | | 227.1.72 | 155770 | Cd2O7Os2 | 2,82 |
| | | 227.1.73 | 155769 | Cd2O7Os2 | 2,82 |
| | | 227.1.74 | 155768 | Cd2O7Os2 | 2,82 |
| | | 227.1.75 | 155767 | Cd2O7Os2 | 2,82 |
| | | 227.1.76 | 155766 | Cd2O7Os2 | 2,82 |
| | | 227.1.77 | 155765 | Cd2O7Os2 | 2,82 |
| | | 227.1.78 | 155764 | Cd2O7Os2 | 2,82 |
| | | 227.1.79 | 155763 | Cd2O7Os2 | 2,82 |
| | | 227.1.80 | 155762 | Cd2O7Os2 | 2,82 |
| | | 227.1.81 | 155761 | Cd2O7Os2 | 2,82 |
| | | 227.1.82 | 37408 | Cd1Se4Yb2 | 2,82 |
| | | 227.1.83 | 188292 | Cd1Ni1 | 2,82 |
| | | 227.1.84 | 246498 | Cd1Er2S4 | 2,82 |
| | | 227.1.85 | 78554 | Cd1Cr2Se4 | 2,82 |
| | | 227.1.86 | 39416 | Cd1Cr2Se4 | 2,82 |
| | | 227.1.87 | 39415 | Cd1Cr2S4 | 2,82 |
| 229 | 1 | 229.1.1 | 26270 | As7Re3 | 2,82 |

The End

TABLE XVII: A full list of all our $f$-electron topological compounds. We give, in this order: the TQC number used for fast referencing with other tables, the ICSD database number, the Number of Electrons (filling), the smallest Direct Gap calculated by our *ab-initio* methods, and the stoichiometric formula. See the list of tables.

| TQC | ICSD | NoE | DG | Stoich. Formula |
|---|---|---|---|---|
| 11.2.f1 | 93367 | 372 | 0.0036 | Er2S3 |
| 11.2.f2 | 73558 | 248 | 0.0056 | Er2S3 |
| 12.1.f1 | 418227 | 110 | 0.0062 | Dy1P2Pt8 |
| 12.1.f2 | 420055 | 110 | 0.0098 | Cl1Gd1Mo1O4 |
| 12.2.f1 | 43336 | 272 | 0.0011 | Pu1 |
| 12.2.f2 | 62228 | 52 | 0.0368 | Nd2O4 |
| 12.2.f3 | 88516 | 68 | 0.0259 | C3Er2Mo2 |
| 12.2.f4 | 67467 | 68 | 0.0020 | C1Dy2Fe2Si2 |
| 12.2.f5 | 75074 | 122 | 0.0220 | Ag1Ba1Er1S3 |
| 13.4.f1 | 415433 | 94 | ∅ | Gd1O4Ta1 |
| 13.4.f2 | 155659 | 94 | ∅ | Gd1O4Ta1 |
| 13.4.f3 | 56734 | 158 | ∅ | Al2Cl8Eu1 |
| 14.2.f1 | 164826 | 168 | 0.0150 | Nd1Ni1O3 |
| 14.2.f2 | 421335 | 120 | 0.0068 | Gd1S2 |
| 14.2.f3 | 416422 | 180 | 0.0120 | Eu1Ge2Pt2 |
| 14.2.f4 | 36604 | 232 | 0.0248 | Dy2O1S2 |
| 14.2.f5 | 291317 | 164 | 0.0068 | Dy1Mo1O6Sr2 |
| 14.2.f6 | 150110 | 320 | 0.0097 | Cu6Nd1 |
| 14.2.f7 | 99678 | 148 | 0.0323 | Cu1Nd1Se2 |
| 14.2.f8 | 90643 | 156 | 0.0520 | Ba2Nd1O6Ru1 |
| 14.2.f9 | 90641 | 156 | 0.0064 | Ba2Nd1O6Ru1 |
| 14.3.f1 | 238697 | 180 | ∅ | O4P1Pu1 |
| 14.3.f2 | 95380 | 140 | ∅ | Er1Li1O2 |
| 14.3.f3 | 249707 | 196 | ∅ | As1Eu1Rb1S3 |
| 15.2.f1 | 430120 | 220 | 0.0138 | Gd1Ni8Si3 |
| 15.2.f2 | 2739 | 140 | 0.0038 | Er1Na1O2 |
| 15.2.f3 | 415206 | 284 | 0.0384 | Dy2Ge3Pt9 |
| 15.2.f4 | 83494 | 136 | 0.0227 | Cu2Er1O4 |
| 15.2.f5 | 83491 | 132 | 0.0043 | Cu2Dy1O4 |
| 15.2.f6 | 402192 | 152 | 0.0094 | Br5Dy2Li1 |
| 15.2.f7 | 93811 | 228 | 0.0127 | Ba3Nd1O9Ru2 |
| 15.2.f8 | 93810 | 228 | 0.0304 | Ba3Nd1O9Ru2 |
| 15.4.f1 | 87720 | 174 | ∅ | Gd1K1O8W2 |
| 15.4.f2 | 76349 | 194 | ∅ | Co3Lu2Si5 |
| 15.4.f3 | 407226 | 142 | ∅ | C2Gd1K1O6 |
| 15.4.f4 | 407832 | 126 | ∅ | Ba1Cl5Gd1 |
| 19.3.f1 | 412745 | 236 | ∅ | Cs1Er1S4Si1 |
| 36.4.f1 | 262892 | 146 | ∅ | Ba2Dy1In1Se5 |
| 40.3.f1 | 280199 | 94 | ∅ | Eu1Li1P1Se4 |
| 47.2.f1 | 89221 | 108 | 0.0134 | Ba2Cu3O7Pr1 |
| 51.2.f1 | 80274 | 116 | 0.0888 | Nd1Pd1Sb1 |
| 51.2.f2 | 370040 | 80 | 0.0100 | Co2In1Tb1 |
| 51.2.f3 | 247569 | 148 | 0.0107 | Ba1Fe2Gd1O5 |
| 55.2.f1 | 423924 | 140 | 0.0859 | C1B2Lu1 |
| 55.2.f2 | 409626 | 156 | 0.0127 | B4Er1V1 |
| 55.2.f3 | 419359 | 172 | 0.0044 | B4Er1Rh1 |
| 55.2.f4 | 417873 | 156 | 0.0142 | B4Co1Gd1 |
| 55.2.f5 | 167534 | 160 | 0.2306 | Al1B4Lu1 |
| 55.4.f1 | 190739 | 294 | ∅ | In4Ni2Tm5 |
| 55.4.f2 | 165139 | 314 | ∅ | In4Lu5Pd2 |
| 55.4.f3 | 106807 | 314 | ∅ | In4Lu5Ni2 |
| 57.2.f1 | 57098 | 264 | 0.0112 | Ge2Nd3Ru2 |
| 57.3.f1 | 81537 | 188 | ∅ | Gd1Na1O4Ti1 |
| 58.2.f1 | 41773 | 200 | 0.0739 | Nd1Re4Si2 |
| 58.2.f2 | 156069 | 124 | 0.0112 | Er1Fe4Ge2 |
| 58.4.f1 | 186046 | 130 | ∅ | Fe4Ge2Lu1 |

Continued on next column

**TABLE XVII– continued**

| TQC | ICSD | NoE | DG | Stoich. Formula |
|---|---|---|---|---|
| 59.2.f1 | 408082 | 256 | 0.0071 | Gd1Pd1Si1 |
| 59.2.f2 | 408081 | 288 | 0.0069 | Er1Pd1Si1 |
| 59.2.f3 | 68074 | 228 | 0.0338 | Co8Eu1P5 |
| 59.4.f1 | 415947 | 278 | ∅ | Gd3Pd6Sb5 |
| 59.4.f2 | 416462 | 62 | ∅ | Gd1I1S1 |
| 59.4.f3 | 50194 | 70 | ∅ | Er1I1Se1 |
| 59.4.f4 | 79107 | 66 | ∅ | Dy1I1S1 |
| 59.4.f5 | 21009 | 70 | ∅ | Cl1Er1S1 |
| 59.4.f6 | 79106 | 66 | ∅ | Br1Dy1S1 |
| 59.4.f7 | 245680 | 118 | ∅ | Au3Eu1In3 |
| 62.2.f1 | 67111 | 328 | 0.0026 | S4Tm2Zn1 |
| 62.2.f2 | 402457 | 256 | 0.0048 | S3Tm2 |
| 62.2.f3 | 1918 | 104 | 0.1222 | Pd1Pu1 |
| 62.2.f4 | 246552 | 320 | 0.0105 | O4Sr1Tm2 |
| 62.2.f5 | 155481 | 168 | 0.0126 | O3Tb1V1 |
| 62.2.f6 | 172358 | 184 | 0.0204 | O3Rh1Tb1 |
| 62.2.f7 | 195600 | 136 | 0.0021 | O3Pr1Sc1 |
| 62.2.f8 | 172354 | 160 | 0.0249 | O3Pr1Rh1 |
| 62.2.f9 | 172348 | 160 | 0.0249 | O3Pr1Rh1 |
| 62.2.f10 | 415953 | 112 | 0.0245 | Nd1Pt1Sn1 |
| 62.2.f11 | 155746 | 168 | 0.0254 | Nd1Ni1O3 |
| 62.2.f12 | 413857 | 152 | 0.1504 | Ir1Lu1Si1 |
| 62.2.f13 | 260839 | 176 | 0.0161 | H2Cl1O2Tm1 |
| 62.2.f14 | 84094 | 120 | 0.0143 | Ge1Mn1Tb1 |
| 62.2.f15 | 15534 | 288 | 0.0224 | Gd3Ni1Si2 |
| 62.2.f16 | 22223 | 216 | 0.0054 | Gd2S3 |
| 62.2.f17 | 280201 | 216 | 0.0066 | Gd2O1Se2 |
| 62.2.f18 | 27275 | 160 | 0.0696 | Fe1Nd1O3 |
| 62.2.f19 | 156444 | 160 | 0.0134 | Fe1Nd1O3 |
| 62.2.f20 | 153442 | 160 | 0.0414 | Fe1Nd1O3 |
| 62.2.f21 | 153441 | 160 | 0.1150 | Fe1Nd1O3 |
| 62.2.f22 | 150360 | 176 | 0.0193 | Fe1Gd1O3 |
| 62.2.f23 | 95492 | 168 | 0.0021 | Eu1Mn1O3 |
| 62.2.f24 | 262205 | 104 | 0.0024 | Eu1Ge2Li1 |
| 62.2.f25 | 191795 | 200 | 0.0223 | Er1Ni1O3 |
| 62.2.f26 | 27282 | 192 | 0.0055 | Er1Fe1O3 |
| 62.2.f27 | 419 | 232 | 0.0023 | Dy2Se3 |
| 62.2.f28 | 80768 | 232 | 0.0037 | Dy2S3 |
| 62.2.f29 | 66532 | 232 | 0.0056 | Dy2S3 |
| 62.2.f30 | 183814 | 240 | 0.0234 | Cu1Eu1Nd1S3 |
| 62.2.f31 | 156329 | 152 | 0.0114 | Cr1Nd1O3 |
| 62.2.f32 | 156324 | 152 | 0.0079 | Cr1Nd1O3 |
| 62.2.f33 | 156319 | 152 | 0.0337 | Cr1Nd1O3 |
| 62.2.f34 | 9985 | 168 | 0.0152 | Cr1Gd1O3 |
| 62.2.f35 | 183759 | 184 | 0.0141 | Cr1Er1O3 |
| 62.2.f36 | 172050 | 200 | 0.0246 | Co1O3Tm1 |
| 62.2.f37 | 172045 | 160 | 0.0265 | Co1O3Pr1 |
| 62.2.f38 | 164817 | 160 | 0.0860 | Co1O3Pr1 |
| 62.2.f39 | 173604 | 272 | 0.0050 | Cl3Pr3Ru1 |
| 62.2.f40 | 90338 | 320 | 0.0027 | Ba1Te4Tm2 |
| 62.2.f41 | 260611 | 320 | 0.0202 | Ba1Se4Tm2 |
| 62.2.f42 | 427492 | 320 | 0.0115 | Ba1O4Tm2 |
| 62.2.f43 | 96546 | 248 | 0.0091 | Ba1Nd2O4 |
| 62.2.f44 | 78486 | 248 | 0.0052 | Ba1Nd2O4 |
| 62.2.f45 | 90337 | 312 | 0.0063 | Ba1Er2Te4 |
| 62.2.f46 | 260610 | 312 | 0.0058 | Ba1Er2Se4 |
| 62.2.f47 | 188656 | 312 | 0.0136 | Ba1Er2S4 |
| 62.2.f48 | 249532 | 192 | 0.0029 | Au2In4Nd1 |
| 62.2.f49 | 411544 | 160 | 0.0048 | Au1Cd1Eu1 |
| 62.2.f50 | 420615 | 160 | 0.0036 | Ag1As2Tb1 |
| 62.2.f51 | 98736 | 136 | 0.0228 | Ag1As2Pr1 |

Continued on next column

**TABLE XVII– continued**

| TQC | ICSD | NoE | DG | Stoich. Formula |
|---|---|---|---|---|
| 62.2.f52 | 420611 | 136 | 0.0102 | Ag1As2Pr1 |
| 62.2.f53 | 412816 | 136 | 0.0275 | Ag1As2Pr1 |
| 62.3.f1 | 51842 | 132 | ∅ | Rh1Sb1Tb1 |
| 62.3.f2 | 417533 | 132 | ∅ | Pt1Sn1Tb1 |
| 62.3.f3 | 183345 | 180 | ∅ | Pd1Tm1Zn1 |
| 62.3.f4 | 185834 | 148 | ∅ | Nd1O3V1 |
| 62.3.f5 | 167138 | 316 | ∅ | Lu1Ni5Sn1 |
| 62.3.f6 | 412912 | 148 | ∅ | In1Lu1Rh1 |
| 62.3.f7 | 420232 | 156 | ∅ | Ge1Lu1Pt1 |
| 62.3.f8 | 88247 | 156 | ∅ | Ge1Lu1Ni1 |
| 62.3.f9 | 419895 | 156 | ∅ | Gd1Rh1Zn1 |
| 62.3.f10 | 85496 | 188 | ∅ | Gd1Ge1Na1O4 |
| 62.3.f11 | 66694 | 188 | ∅ | Eu1Se3Zr1 |
| 62.3.f12 | 192313 | 188 | ∅ | Eu1O3Zr1 |
| 62.3.f13 | 40393 | 180 | ∅ | Er1O3V1 |
| 62.3.f14 | 185838 | 180 | ∅ | Er1O3V1 |
| 62.3.f15 | 54199 | 204 | ∅ | Er1Ge1Na1O4 |
| 62.3.f16 | 416219 | 332 | ∅ | Dy3N1S3 |
| 62.3.f17 | 422562 | 132 | ∅ | Dy1Li1O2 |
| 62.3.f18 | 429061 | 164 | ∅ | Dy1Ga1O3 |
| 62.3.f19 | 103089 | 316 | ∅ | Cu6Pr1 |
| 62.3.f20 | 150365 | 252 | ∅ | Cu1Eu2S3 |
| 62.3.f21 | 189144 | 236 | ∅ | Cu1Eu1Pr1S3 |
| 62.3.f22 | 412861 | 212 | ∅ | Cl1Er1O3Se1 |
| 62.3.f23 | 428850 | 156 | ∅ | Cd1Eu1Pt1 |
| 62.3.f24 | 100774 | 156 | ∅ | B2Lu1Ru1 |
| 62.3.f25 | 409908 | 148 | ∅ | As1Er1Pd1 |
| 62.3.f26 | 150354 | 156 | ∅ | Al1Gd1O3 |
| 62.3.f27 | 150353 | 156 | ∅ | Al1Gd1O3 |
| 62.3.f28 | 150352 | 156 | ∅ | Al1Gd1O3 |
| 62.3.f29 | 150349 | 156 | ∅ | Al1Gd1O3 |
| 62.3.f30 | 150348 | 156 | ∅ | Al1Gd1O3 |
| 63.1.f1 | 350 | 196 | 0.0107 | Pd5Pu3 |
| 63.1.f2 | 96410 | 176 | 0.0788 | O7Pr3Re1 |
| 63.1.f3 | 150528 | 52 | 0.0669 | Ni1Pu1 |
| 63.1.f4 | 250755 | 184 | 0.0065 | Nd2Sn1Zn5 |
| 63.1.f5 | 78007 | 188 | 0.0023 | Nb1O7Pr3 |
| 63.1.f6 | 42375 | 36 | 0.0263 | Ge1Nd1 |
| 63.1.f7 | 42374 | 36 | 0.0244 | Ge1Nd1 |
| 63.1.f8 | 88248 | 312 | 0.0082 | Ge1Lu1Ni1 |
| 63.1.f9 | 400536 | 172 | 0.0036 | Eu2O7Ta2 |
| 63.1.f10 | 425774 | 188 | 0.0024 | Er3Ge3Ru2 |
| 63.1.f11 | 415373 | 64 | 0.4603 | Al1Lu1Si1 |
| 63.2.f1 | 73619 | 100 | 0.0073 | Si3Tb2 |
| 63.2.f2 | 250754 | 180 | 0.0028 | Pr2Sn1Zn5 |
| 63.2.f3 | 250758 | 180 | 0.0362 | Pb1Pr2Zn5 |
| 63.2.f4 | 78009 | 172 | 0.0030 | O7Pr3Sb1 |
| 63.2.f5 | 191961 | 172 | 0.0132 | O7Pr3Sb1 |
| 63.2.f6 | 71691 | 64 | 0.0699 | Nd1Ni1Si2 |
| 63.2.f7 | 237376 | 68 | 0.0294 | Mg5Nd1Ni1 |
| 63.2.f8 | 171235 | 172 | 0.0154 | Ge6Pr2Zn3 |
| 63.2.f9 | 171234 | 176 | 0.0245 | Ge6Nd2Zn3 |
| 63.2.f10 | 170875 | 208 | 0.0095 | Gd3O7Os1 |
| 63.2.f11 | 416379 | 176 | 0.0136 | Eu2Ga9Rh3 |
| 63.2.f12 | 416378 | 176 | 0.0382 | Eu2Ga9Ir3 |
| 63.2.f13 | 410700 | 68 | 0.0368 | Eu1Ir1Sn2 |
| 63.2.f14 | 411268 | 64 | 0.0198 | Eu1In2Rh1 |
| 63.2.f15 | 150369 | 284 | 0.0052 | Er5Ni2Te2 |
| 63.2.f16 | 55524 | 68 | 0.0198 | Er1Ge3 |
| 63.2.f17 | 55523 | 68 | 0.0207 | Er1Ge3 |
| 63.2.f18 | 42380 | 52 | 0.0061 | Er1Ge1 |

Continued on next column

**TABLE XVII– continued**

| TQC | ICSD | NoE | DG | Stoich. Formula |
|---|---|---|---|---|
| 63.2.f19 | 425772 | 176 | 0.0135 | Dy3Ge3Ru2 |
| 63.2.f20 | 154699 | 48 | 0.0123 | Dy1Si1 |
| 63.2.f21 | 85841 | 76 | 0.0088 | Dy1Ni1Si2 |
| 63.2.f22 | 85840 | 76 | 0.0059 | Dy1Ni1Si2 |
| 63.2.f23 | 423716 | 220 | 0.0128 | Cu3Er2S5Tl1 |
| 63.2.f24 | 97560 | 164 | 0.0097 | Cu1K1S4Tb2 |
| 63.2.f25 | 59896 | 160 | 0.0055 | Cu1Gd2Rb1S4 |
| 63.2.f26 | 95029 | 176 | 0.0157 | Cu1Er2K1Se4 |
| 63.2.f27 | 413680 | 220 | 0.0055 | Cs1Cu3Se5Tb2 |
| 63.2.f28 | 413676 | 216 | 0.0147 | Cs1Cu3Gd2Se5 |
| 63.2.f29 | 421087 | 144 | 0.0122 | Cs1Cu1Nd2Se4 |
| 63.2.f30 | 40773 | 88 | 0.0048 | C1Dy1Fe2Si1 |
| 63.2.f31 | 411390 | 68 | 0.0144 | Au1Eu1In2 |
| 63.2.f32 | 415372 | 60 | 0.0061 | Al1Si1Tm1 |
| 63.2.f33 | 186914 | 128 | 0.0259 | Al10Gd1Ru2 |
| 63.4.f1 | 412465 | 62 | ∅ | Pr1Se1Te2 |
| 63.4.f2 | 166874 | 62 | ∅ | Mg1Ni1Tb1 |
| 63.4.f3 | 97005 | 362 | ∅ | Lu7Te1 |
| 63.4.f4 | 426769 | 66 | ∅ | Lu1Sn2 |
| 63.4.f5 | 169059 | 70 | ∅ | Ge3Tm1 |
| 63.4.f6 | 425776 | 206 | ∅ | Ge3Lu3Ru2 |
| 63.4.f7 | 88246 | 214 | ∅ | Ge3Lu3Ni2 |
| 63.4.f8 | 412709 | 210 | ∅ | Ge3Ir2Lu3 |
| 63.4.f9 | 159585 | 62 | ∅ | Ge2Tm1 |
| 63.4.f10 | 42377 | 34 | ∅ | Ge1Pr1 |
| 63.4.f11 | 188506 | 202 | ∅ | Gd3O7Sb1 |
| 63.4.f12 | 423069 | 90 | ∅ | Ga4Ni1Tm1 |
| 63.4.f13 | 423068 | 82 | ∅ | Ga4Ni1Tb1 |
| 63.4.f14 | 245230 | 142 | ∅ | Eu3Ge5 |
| 63.4.f15 | 416579 | 82 | ∅ | Eu1Si6 |
| 63.4.f16 | 416577 | 82 | ∅ | Eu1Si6 |
| 63.4.f17 | 391366 | 66 | ∅ | Eu1Pd1Tl2 |
| 63.4.f18 | 193724 | 78 | ∅ | Eu1In4Pt1 |
| 63.4.f19 | 411391 | 78 | ∅ | Eu1In4Pd1 |
| 63.4.f20 | 404251 | 78 | ∅ | Eu1In4Ni1 |
| 63.4.f21 | 402571 | 66 | ∅ | Eu1In2Ni1 |
| 63.4.f22 | 55672 | 138 | ∅ | Cu1Eu1S3Tm1 |
| 63.4.f23 | 40064 | 82 | ∅ | Cl3Dy1 |
| 63.4.f24 | 260483 | 86 | ∅ | C1Gd1Os2P1 |
| 63.4.f25 | 78176 | 122 | ∅ | Ba1Cu1Er1S3 |
| 63.4.f26 | 62630 | 186 | ∅ | B7Cr1Er3 |
| 63.4.f27 | 154855 | 82 | ∅ | Al4Ni1Tb1 |
| 63.4.f28 | 186913 | 118 | ∅ | Al10Pr1Ru2 |
| 63.4.f29 | 238041 | 142 | ∅ | Al10Lu1Os2 |
| 64.4.f1 | 261995 | 318 | ∅ | Au2Lu5Te2 |
| 65.1.f1 | 415513 | 156 | 0.0790 | In16Pt7Tb2 |
| 65.1.f2 | 174439 | 94 | 0.0041 | Er3Sn7 |
| 65.1.f3 | 418215 | 152 | 0.0063 | Cu1Mg4Tb1 |
| 65.1.f4 | 261436 | 62 | 0.0231 | Cd1Cu2Nd2 |
| 65.1.f5 | 419476 | 56 | 0.0225 | C3B2Tb2 |
| 65.1.f6 | 78623 | 134 | 0.0482 | Ba2Cu4Er1O8 |
| 65.1.f7 | 67641 | 134 | 0.0546 | Ba2Cu4Er1O8 |
| 65.1.f8 | 78625 | 132 | 0.0141 | Ba2Cu4Dy1O8 |
| 65.2.f1 | 104809 | 84 | 0.0550 | Lu2Ni1Sn6 |
| 65.2.f2 | 106923 | 102 | 0.0889 | Lu1Ni1Sn4 |
| 65.2.f3 | 40487 | 240 | 0.0215 | Ge2Lu1Mn1 |
| 65.2.f4 | 104134 | 82 | 0.0715 | Gd3Sn7 |
| 65.2.f5 | 171189 | 88 | 0.0073 | Er1Ni1Si3 |
| 65.2.f6 | 165783 | 74 | 0.0058 | Dy2Ni1Sn6 |
| 65.2.f7 | 412576 | 158 | 0.0079 | Dy2In16Pt7 |
| 65.2.f8 | 154010 | 158 | 0.0042 | Ba1Mn2O6Tb1 |

Continued on next column

**TABLE XVII– continued**

| TQC | ICSD | NoE | DG | Stoich. Formula |
|---|---|---|---|---|
| 65.2.f9 | 68002 | 98 | 0.0200 | B6Lu2Ni3 |
| 67.4.f1 | 59306 | 142 | ∅ | Ba1F6Tb1 |
| 70.4.f1 | 40619 | 82 | ∅ | B2Eu1Ir2 |
| 71.1.f1 | 246814 | 34 | 0.0409 | Ge5Nd1 |
| 71.1.f2 | 95144 | 148 | 0.0056 | Ge2Pt1Tb1 |
| 71.1.f3 | 83190 | 102 | 0.0169 | Cu8Ge8Nd6 |
| 71.1.f4 | 238028 | 52 | 0.2836 | C1Co2Pr2Si1 |
| 71.1.f5 | 238027 | 54 | 0.1246 | C1Co2Nd2Si1 |
| 71.1.f6 | 157373 | 120 | 0.0045 | Ag4Dy3Sn4 |
| 72.2.f1 | 40758 | 140 | 0.0060 | C4Fe1Tm2 |
| 72.4.f1 | 165350 | 222 | ∅ | Eu1K4P2Se8 |
| 72.4.f2 | 279621 | 222 | ∅ | Eu1K4P2S8 |
| 72.4.f3 | 73169 | 134 | ∅ | C4Er2Mn1 |
| 74.2.f1 | 411845 | 64 | 0.0329 | Eu1Ga1Zn1 |
| 74.4.f1 | 79599 | 70 | ∅ | Ge1Pr1Pt1 |
| 74.4.f2 | 50592 | 66 | ∅ | Gd1Pd1Si1 |
| 74.4.f3 | 104006 | 58 | ∅ | Ga2Tm1 |
| 74.4.f4 | 62924 | 114 | ∅ | Er1Rh3Si2 |
| 74.4.f5 | 77728 | 14 | ∅ | Er1Pd1Si1 |
| 74.4.f6 | 20660 | 266 | ∅ | Co3Ga9Nd1 |
| 74.4.f7 | 78818 | 58 | ∅ | Ag1Eu1Ge1 |
| 84.4.f1 | 59987 | 146 | ∅ | Cd1F6Tb1 |
| 113.2.f1 | 262242 | 172 | 0.0056 | Eu2Ga2Ge1S7 |
| 113.2.f2 | 160121 | 84 | 0.0121 | Ag1Tb1Te2 |
| 113.2.f3 | 154793 | 84 | 0.0112 | Ag1Tb1Te2 |
| 113.4.f1 | 160122 | 86 | ∅ | Ag1Dy1Te2 |
| 113.4.f2 | 154794 | 86 | ∅ | Ag1Dy1Te2 |
| 123.1.f1 | 106640 | 78 | 0.0021 | Fe1Ga8Tm2 |
| 123.2.f1 | 98585 | 36 | 0.5439 | Nd1Ni1O2 |
| 123.2.f2 | 427123 | 58 | 0.0261 | Er2Mg3Ru1 |
| 123.2.f3 | 92430 | 60 | 0.0059 | Bi2I1O4Tb1 |
| 123.2.f4 | 92403 | 60 | 0.0057 | Bi2Cl1O4Tb1 |
| 123.2.f5 | 92401 | 58 | 0.0659 | Bi2Cl1Eu1O4 |
| 123.2.f6 | 92416 | 60 | 0.0583 | Bi2Br1O4Tb1 |
| 123.2.f7 | 92411 | 54 | 0.0093 | Bi2Br1O4Pr1 |
| 123.2.f8 | 98114 | 114 | 0.0730 | Ba2Cu3Lu1O6 |
| 123.2.f9 | 98113 | 114 | 0.0724 | Ba2Cu3Lu1O6 |
| 123.2.f10 | 79534 | 154 | 0.0430 | Ba2Cu2O11Tb2Ti2 |
| 123.2.f11 | 66812 | 150 | 0.0131 | Ba2Cu2Eu2O11Ti2 |
| 123.2.f12 | 154107 | 78 | 0.0664 | Ba1Co2Nd1O6 |
| 123.2.f13 | 422759 | 76 | 0.0825 | Ba1Co2Gd1O5 |
| 123.3.f1 | 92435 | 64 | ∅ | Bi2I1O4Tm1 |
| 123.3.f2 | 92408 | 64 | ∅ | Bi2Cl1O4Tm1 |
| 123.3.f3 | 92421 | 64 | ∅ | Bi2Br1O4Tm1 |
| 125.3.f1 | 183718 | 144 | ∅ | Ga12Pd1Pr2 |
| 125.3.f2 | 167197 | 144 | ∅ | Ga12Ni1Pr2 |
| 125.3.f3 | 183719 | 148 | ∅ | Ga12Nd2Pd1 |
| 125.3.f4 | 167198 | 148 | ∅ | Ga12Nd2Ni1 |
| 125.4.f1 | 412792 | 94 | ∅ | K1Nd1Te4 |
| 125.4.f2 | 391206 | 102 | ∅ | Gd1Rb1Te4 |
| 125.4.f3 | 59298 | 102 | ∅ | Cs1Gd1Te4 |
| 126.3.f1 | 421051 | 152 | ∅ | Mo2O8Pr1Tl1 |
| 126.3.f2 | 152176 | 152 | ∅ | Mo2O8Pr1Tl1 |
| 127.2.f1 | 99199 | 140 | 0.0034 | Pb1Pd2Tm2 |
| 127.2.f2 | 425432 | 140 | 0.0032 | Ni2Sn1Tm2 |
| 127.2.f3 | 427282 | 124 | 0.0203 | Ni2Sn1Tb2 |
| 127.2.f4 | 192644 | 224 | 0.0175 | In2Ni9Sm1 |
| 127.2.f5 | 413850 | 72 | 0.0023 | Ge2Mg1Pr2 |
| 127.2.f6 | 260384 | 200 | 0.0238 | Ga10Os3Tm2 |
| 127.2.f7 | 107023 | 200 | 0.0227 | Ga10Os3Tm2 |
| 127.2.f8 | 425431 | 136 | 0.0066 | Er2Ni2Sn1 |
| 127.2.f9 | 260383 | 196 | 0.0125 | Er2Ga10Os3 |
| 127.2.f10 | 192648 | 232 | 0.0099 | Dy1In2Ni9 |
| 127.2.f11 | 157385 | 116 | 0.0242 | Cd1Pd2Pr2 |
| 127.2.f12 | 424178 | 132 | 0.0052 | Cd1Ge2Tm2 |
| 127.2.f13 | 424172 | 104 | 0.0088 | Cd1Ge2Sm2 |
| 127.2.f14 | 424177 | 128 | 0.0097 | Cd1Er2Ge2 |
| 127.2.f15 | 424175 | 120 | 0.0036 | Cd1Dy2Ge2 |
| 127.2.f16 | 427156 | 64 | 0.0190 | C2B2Gd1 |
| 127.2.f17 | 94042 | 72 | 0.0259 | C2B2Er1 |
| 127.2.f18 | 94040 | 68 | 0.0032 | C2B2Dy1 |
| 127.2.f19 | 404496 | 180 | 0.0041 | Ba1Dy2O5Pd1 |
| 127.3.f1 | 99192 | 100 | ∅ | Pb1Pd2Pr2 |
| 127.3.f2 | 412728 | 120 | ∅ | Mg1Pd2Tb2 |
| 127.3.f3 | 411080 | 96 | ∅ | Mg1Ni2Pr2 |
| 127.3.f4 | 107328 | 96 | ∅ | Mg1Ni2Pr2 |
| 127.3.f5 | 411079 | 100 | ∅ | Mg1Nd2Ni2 |
| 127.3.f6 | 99200 | 148 | ∅ | Lu2Pb1Pd2 |
| 127.3.f7 | 192643 | 220 | ∅ | In2Nd1Ni9 |
| 127.3.f8 | 189071 | 220 | ∅ | In2Nd1Ni9 |
| 127.3.f9 | 423454 | 112 | ∅ | Ge2Mg1Tm2 |
| 127.3.f10 | 423455 | 120 | ∅ | Ge2Lu2Mg1 |
| 127.3.f11 | 99194 | 120 | ∅ | Gd2Pb1Pd2 |
| 127.3.f12 | 99198 | 136 | ∅ | Er2Pb1Pd2 |
| 127.3.f13 | 423456 | 108 | ∅ | Er2Ge2Mg1 |
| 127.3.f14 | 427283 | 128 | ∅ | Dy2Ni2Sn1 |
| 127.3.f15 | 425067 | 272 | ∅ | Cd2Cu9Pr1 |
| 127.3.f16 | 414596 | 116 | ∅ | Cd1Ni2Pr2 |
| 127.3.f17 | 424174 | 116 | ∅ | Cd1Ge2Tb2 |
| 127.3.f18 | 424171 | 96 | ∅ | Cd1Ge2Nd2 |
| 127.3.f19 | 157387 | 136 | ∅ | Cd1Gd2Pd2 |
| 127.3.f20 | 157388 | 144 | ∅ | Cd1Dy2Pd2 |
| 127.3.f21 | 419851 | 140 | ∅ | Cd1Cu2Gd2 |
| 127.3.f22 | 202172 | 152 | ∅ | Ba1O5Pd1Pr2 |
| 127.3.f23 | 62609 | 156 | ∅ | Ba1Nd2O5Pd1 |
| 127.3.f24 | 404495 | 172 | ∅ | Ba1Gd2O5Pd1 |
| 127.3.f25 | 93225 | 104 | ∅ | B4Nd1 |
| 127.3.f26 | 154656 | 104 | ∅ | B4Nd1 |
| 127.3.f27 | 68650 | 136 | ∅ | B4Er1 |
| 127.3.f28 | 44214 | 136 | ∅ | B4Er1 |
| 127.4.f1 | 192642 | 218 | ∅ | In2Ni9Pr1 |
| 127.4.f2 | 54488 | 138 | ∅ | In1Pd2Tm2 |
| 127.4.f3 | 54480 | 98 | ∅ | In1Pd2Pr2 |
| 127.4.f4 | 54490 | 146 | ∅ | In1Lu2Pd2 |
| 127.4.f5 | 172489 | 86 | ∅ | Ge2In1Sm2 |
| 127.4.f6 | 414072 | 74 | ∅ | Ge2In1Pr2 |
| 127.4.f7 | 414071 | 78 | ∅ | Ge2In1Nd2 |
| 127.4.f8 | 107343 | 126 | ∅ | Dy2In1Pd2 |
| 127.4.f9 | 172492 | 102 | ∅ | Dy2Ge2In1 |
| 127.4.f10 | 94043 | 74 | ∅ | C2B2Tm1 |
| 127.4.f11 | 94044 | 78 | ∅ | C2B2Lu1 |
| 127.4.f12 | 172061 | 102 | ∅ | Al1Dy2Ge2 |
| 129.1.f1 | 50243 | 64 | 0.1368 | Mn1Nd1O1Sb1 |
| 129.1.f2 | 50241 | 64 | 0.1103 | Mn1Nd1O1P1 |
| 129.1.f3 | 186991 | 88 | 0.0210 | Ir2P2Sm1 |
| 129.1.f4 | 85863 | 52 | 0.0378 | Ge1Nd1Ru1 |
| 129.1.f5 | 85859 | 48 | 0.5305 | Ge1Mn1Pr1 |
| 129.1.f6 | 183312 | 84 | 0.0024 | Er1Sn2Zn1 |
| 129.1.f7 | 96345 | 72 | 0.1479 | Cu1O1Pr1S1 |
| 129.1.f8 | 412554 | 88 | 0.0215 | As2Cu1Tm1 |
| 129.1.f9 | 412551 | 80 | 0.0125 | As2Cu1Tb1 |

Continued on next column

Continued on next column

**TABLE XVII– continued**

| TQC | ICSD | NoE | DG | Stoich. Formula |
|---|---|---|---|---|
| 129.1.f10 | 90316 | 64 | 0.0703 | As1Fe1O1Pr1 |
| 129.1.f11 | 180774 | 68 | 1.0339 | As1Co1Nd1O1 |
| 129.2.f1 | 183310 | 76 | 0.0140 | Gd1Sn2Zn1 |
| 129.2.f2 | 95081 | 52 | 0.0033 | Eu1P1Sn1 |
| 129.2.f3 | 421440 | 56 | 0.0150 | Dy1Sb1Sc1 |
| 129.2.f4 | 421439 | 56 | 0.0140 | Dy1Sb1Sc1 |
| 129.2.f5 | 174358 | 92 | 0.0742 | As2Cu1Lu1 |
| 129.2.f6 | 422000 | 64 | 0.0939 | As1Fe1O1Pr1 |
| 129.3.f1 | 261768 | 168 | ∅ | H1Nd1O1 |
| 129.3.f2 | 95024 | 60 | ∅ | Er1Ge1Ti1 |
| 129.3.f3 | 95022 | 60 | ∅ | Er1Ge1Ti1 |
| 129.3.f4 | 238706 | 68 | ∅ | Au1Bi2Pr1 |
| 129.3.f5 | 238707 | 68 | ∅ | Au1Bi2Nd1 |
| 129.3.f6 | 107956 | 88 | ∅ | As2Nd1Ni2 |
| 129.3.f7 | 162830 | 64 | ∅ | As1Fe1O1Pr1 |
| 129.3.f8 | 180772 | 68 | ∅ | As1Co1Nd1O1 |
| 129.4.f1 | 163425 | 78 | ∅ | Sn2Tb1Zn1 |
| 129.4.f2 | 82004 | 86 | ∅ | Na1Nd1O4Ti1 |
| 129.4.f3 | 422049 | 86 | ∅ | Na1Nd1O4Ti1 |
| 129.4.f4 | 183314 | 90 | ∅ | Lu1Sn2Zn1 |
| 129.4.f5 | 419856 | 118 | ∅ | K1Na1Nd1O5Ta1 |
| 129.4.f6 | 82006 | 94 | ∅ | Gd1Na1O4Ti1 |
| 129.4.f7 | 419854 | 126 | ∅ | Gd1K1Na1O5Ta1 |
| 129.4.f8 | 420037 | 142 | ∅ | Gd1K1Na1Nb1O5 |
| 129.4.f9 | 94473 | 62 | ∅ | F1Gd1S1 |
| 129.4.f10 | 93355 | 62 | ∅ | F1Gd1S1 |
| 129.4.f11 | 162267 | 98 | ∅ | Eu1Sn2Zn2 |
| 129.4.f12 | 427712 | 90 | ∅ | Eu1Pd2Sn2 |
| 129.4.f13 | 51757 | 98 | ∅ | Eu1Ge2Zn2 |
| 129.4.f14 | 94476 | 70 | ∅ | Er1F1S1 |
| 129.4.f15 | 80358 | 82 | ∅ | Cu1Gd1O1Se1 |
| 129.4.f16 | 80359 | 86 | ∅ | Cu1Dy1O1Se1 |
| 129.4.f17 | 59232 | 62 | ∅ | Cl1Gd1O1 |
| 129.4.f18 | 248426 | 78 | ∅ | C1B1Er1Ni1 |
| 129.4.f19 | 412553 | 86 | ∅ | As2Cu1Er1 |
| 129.4.f20 | 412552 | 82 | ∅ | As2Cu1Dy1 |
| 129.4.f21 | 420206 | 82 | ∅ | As1Gd1O1Zn1 |
| 129.4.f22 | 425010 | 74 | ∅ | As1Fe1Gd1O1 |
| 129.4.f23 | 425006 | 74 | ∅ | As1Fe1Gd1O1 |
| 129.4.f24 | 422003 | 74 | ∅ | As1Fe1Gd1O1 |
| 129.4.f25 | 420203 | 90 | ∅ | As1Er1O1Zn1 |
| 129.4.f26 | 55507 | 90 | ∅ | Al2Au2Eu1 |
| 136.1.f1 | 154834 | 120 | 0.0104 | Dy1Fe4Ge2 |
| 136.3.f1 | 63543 | 160 | ∅ | Cl3Tb1 |
| 136.3.f2 | 106282 | 280 | ∅ | Au2Er2Sn1 |
| 139.1.f1 | 411149 | 76 | 0.0027 | O5Pr2S2Ti2 |
| 139.1.f2 | 409859 | 66 | 0.1589 | Nd2Sb1 |
| 139.1.f3 | 181937 | 70 | 0.0030 | Mn2O3Pr2Se2 |
| 139.1.f4 | 93205 | 40 | 0.0274 | Ge2Nd1Rh2 |
| 139.1.f5 | 182419 | 50 | 0.0178 | Gd1Sc1Si1 |
| 139.1.f6 | 90392 | 52 | 0.0625 | Gd1Ge1Ti1 |
| 139.1.f7 | 240091 | 42 | 0.0117 | Ga2Pd2Sm1 |
| 139.1.f8 | 181168 | 72 | 0.2020 | Fe2O3Pr2S2 |
| 139.1.f9 | 73653 | 42 | 0.0495 | Co2Nd1P2 |
| 139.1.f10 | 73651 | 42 | 0.0274 | Co2Nd1P2 |
| 139.3.f1 | 89564 | 56 | ∅ | O2Tb2Te1 |
| 139.3.f2 | 89559 | 44 | ∅ | O2Pr2Te1 |
| 139.3.f3 | 90298 | 38 | ∅ | Nd1Ru2Si2 |
| 139.3.f4 | 93412 | 42 | ∅ | Nd1Ni2Si2 |
| 139.3.f5 | 263006 | 92 | ∅ | Lu1Si3 |
| 139.3.f6 | 53735 | 40 | ∅ | Ge2Nd1Rh2 |

Continued on next column

**TABLE XVII– continued**

| TQC | ICSD | NoE | DG | Stoich. Formula |
|---|---|---|---|---|
| 139.3.f7 | 89556 | 40 | ∅ | Gd1Mn2Si2 |
| 139.3.f8 | 81746 | 38 | ∅ | Fe2Ge2Nd1 |
| 139.3.f9 | 60667 | 80 | ∅ | Er1O2S3U2 |
| 139.3.f10 | 88188 | 52 | ∅ | Er1Ni2P2 |
| 139.3.f11 | 88187 | 52 | ∅ | Er1Ni2P2 |
| 139.3.f12 | 103237 | 42 | ∅ | Er1Ga4Ti2 |
| 139.3.f13 | 89565 | 58 | ∅ | Dy2O2Te1 |
| 139.3.f14 | 55759 | 48 | ∅ | Co2Er1Si2 |
| 139.3.f15 | 58252 | 44 | ∅ | Ag2Er1 |
| 140.1.f1 | 424222 | 264 | 0.0218 | Cu9Eu1Sn4 |
| 140.1.f2 | 416794 | 264 | 0.0150 | Cu9Eu1Sn4 |
| 140.1.f3 | 41629 | 240 | 0.0036 | Cu2Eu1Nb1O8Sr2 |
| 140.1.f4 | 90161 | 232 | 0.0365 | Co9Si4Tb1 |
| 140.2.f1 | 156956 | 244 | 0.0144 | Au2Bi1Tb5 |
| 140.2.f2 | 428736 | 168 | 0.0126 | Al1Eu3O5 |
| 140.3.f1 | 1916 | 208 | ∅ | Pu5Ru3 |
| 140.3.f2 | 87452 | 300 | ∅ | Lu5Pd2Sb1 |
| 140.3.f3 | 72092 | 232 | ∅ | Co9Si4Tb1 |
| 140.4.f1 | 424907 | 78 | ∅ | Eu1O3Ti1 |
| 140.4.f2 | 290422 | 78 | ∅ | Eu1O3Ti1 |
| 140.4.f3 | 187116 | 78 | ∅ | Eu1O3Ti1 |
| 140.4.f4 | 187113 | 78 | ∅ | Eu1O3Ti1 |
| 140.4.f5 | 187112 | 78 | ∅ | Eu1O3Ti1 |
| 140.4.f6 | 187111 | 78 | ∅ | Eu1O3Ti1 |
| 140.4.f7 | 187110 | 78 | ∅ | Eu1O3Ti1 |
| 140.4.f8 | 187109 | 78 | ∅ | Eu1O3Ti1 |
| 140.4.f9 | 187108 | 78 | ∅ | Eu1O3Ti1 |
| 140.4.f10 | 187107 | 78 | ∅ | Eu1O3Ti1 |
| 140.4.f11 | 187106 | 78 | ∅ | Eu1O3Ti1 |
| 140.4.f12 | 416796 | 138 | ∅ | Cu4Eu1Sn2 |
| 140.4.f13 | 99476 | 94 | ∅ | Ba1O3Tb1 |
| 140.4.f14 | 99475 | 94 | ∅ | Ba1O3Tb1 |
| 140.4.f15 | 59596 | 94 | ∅ | Ba1O3Tb1 |
| 163.3.f1 | 420304 | 160 | ∅ | Ag1P2Se6Tm1 |
| 163.3.f2 | 195337 | 160 | ∅ | Ag1P2Se6Tm1 |
| 164.3.f1 | 400048 | 64 | ∅ | O2S1Tm2 |
| 164.3.f2 | 41180 | 32 | ∅ | Al2Gd1Si2 |
| 164.3.f3 | 156301 | 46 | ∅ | Ag1Te2Tm1 |
| 166.3.f1 | 417868 | 144 | ∅ | B6Co12Gd1 |
| 167.3.f1 | 67115 | 128 | ∅ | S3Tm2 |
| 167.3.f2 | 391109 | 152 | ∅ | Al7Au3Er1 |
| 176.3.f1 | 34735 | 80 | ∅ | H3O3Tb1 |
| 186.3.f1 | 104802 | 120 | ∅ | Lu1Mn5 |
| 186.3.f2 | 107285 | 76 | ∅ | Cu1Pb1Tm1 |
| 186.3.f3 | 416546 | 80 | ∅ | Cu1Lu1Sn1 |
| 186.3.f4 | 107286 | 80 | ∅ | Cu1Lu1Pb1 |
| 186.3.f5 | 81734 | 80 | ∅ | Au1Ge1Lu1 |
| 186.3.f6 | 96763 | 132 | ∅ | As2Ni1Tm2 |
| 186.3.f7 | 417550 | 76 | ∅ | Ag1Sn1Tm1 |
| 186.3.f8 | 418580 | 80 | ∅ | Ag1Lu1Sn1 |
| 189.1.f1 | 415370 | 78 | 0.0052 | Pr1Rh1Sn1 |
| 189.1.f2 | 412356 | 78 | 0.0167 | Mg1Nd1Pt1 |
| 189.1.f3 | 59177 | 54 | 0.0052 | Ge1Li1Pr1 |
| 189.1.f4 | 810 | 132 | 0.0301 | As2Eu2 |
| 189.3.f1 | 415952 | 84 | ∅ | Nd1Pt1Sn1 |
| 189.3.f2 | 419191 | 84 | ∅ | Nd1Pd1Sn1 |
| 189.3.f3 | 51845 | 102 | ∅ | Bi1Dy1Rh1 |
| 189.3.f4 | 245235 | 114 | ∅ | Ag1Sn1Tm1 |
| 191.3.f1 | 107031 | 88 | ∅ | Er1Ge6Mn6 |
| 191.3.f2 | 85875 | 86 | ∅ | Dy1Ge6Mn6 |

Continued on next column

**TABLE XVII– continued**

| TQC | ICSD | NoE | DG | Stoich. Formula |
|---|---|---|---|---|
| 191.3.f3 | 57229 | 86 | ∅ | Dy1Ge6Mn6 |
| 191.3.f4 | 57228 | 86 | ∅ | Dy1Ge6Mn6 |
| 191.3.f5 | 92053 | 254 | ∅ | B6Co19Pr5 |
| 191.3.f6 | 97725 | 196 | ∅ | B3Co7Pr2 |
| 191.3.f7 | 97342 | 46 | ∅ | B2Ir3Pr1 |
| 191.3.f8 | 246509 | 52 | ∅ | B2Co3Tb1 |
| 191.3.f9 | 246508 | 52 | ∅ | B2Co3Tb1 |
| 191.3.f10 | 156975 | 124 | ∅ | B1Ni4Tb1 |
| 191.3.f11 | 156974 | 122 | ∅ | B1Gd1Ni4 |
| 191.3.f12 | 237660 | 120 | ∅ | B1Eu1Pt4 |
| 191.3.f13 | 156976 | 130 | ∅ | B1Er1Ni4 |
| 191.3.f14 | 97722 | 104 | ∅ | B1Co4Pr1 |
| 191.3.f15 | 417869 | 114 | ∅ | B1Co4Gd1 |
| 191.3.f16 | 409394 | 62 | ∅ | Al9Co2Eu1 |
| 193.3.f1 | 415737 | 280 | ∅ | B1Lu5Si3 |
| 194.3.f1 | 164893 | 60 | ∅ | Ge3Pr1V1 |
| 194.3.f2 | 416327 | 244 | ∅ | Cu9Mg2Tb1 |
| 204.4.f1 | 183089 | 108 | ∅ | Os4Sb12Sm1 |
| 204.4.f2 | 156468 | 108 | ∅ | Os4Sb12Sm1 |
| 204.4.f3 | 156470 | 106 | ∅ | Nd1Os4Sb12 |
| 204.4.f4 | 174555 | 102 | ∅ | Ge12Nd1Pt4 |
| 204.4.f5 | 237413 | 150 | ∅ | Cu3Fe4O12Pr1 |
| 204.4.f6 | 237424 | 162 | ∅ | Cu3Fe4Lu1O12 |
| 204.4.f7 | 237416 | 154 | ∅ | Cu3Eu1Fe4O12 |
| 204.4.f8 | 237403 | 154 | ∅ | Cu3Eu1Fe4O12 |
| 216.4.f1 | 194973 | 56 | ∅ | Mg1Nd1Ni4 |
| 216.4.f2 | 194977 | 60 | ∅ | Cu4Mg1Nd1 |
| 216.4.f3 | 416420 | 38 | ∅ | Au1Sn1Tm1 |
| 221.3.f1 | 194422 | 32 | ∅ | Ga3Tm1 |
| 221.4.f1 | 106748 | 120 | ∅ | Ga3Ru1Tm1 |
| 223.3.f1 | 84666 | 136 | ∅ | Nd1Pd3S4 |
| 223.4.f1 | 84667 | 142 | ∅ | Eu1Pd3S4 |
| 225.4.f1 | 55456 | 28 | ∅ | O2Pu1 |
| 225.4.f2 | 26229 | 76 | ∅ | Ba2Mo1Nd1O6 |
| 225.4.f3 | 236354 | 80 | ∅ | Ba2Gd1Mo1O6 |
| 225.4.f4 | 236353 | 80 | ∅ | Ba2Gd1Mo1O6 |
| 226.4.f1 | 108888 | 90 | ∅ | Be13Tb1 |
| 226.4.f2 | 58707 | 92 | ∅ | Be13Dy1 |
| 229.3.f1 | 417049 | 256 | ∅ | Cd4Pd13Pr6 |
| 229.4.f1 | 248000 | 132 | ∅ | Er4Fe1Ga12 |
| 229.4.f2 | 424496 | 122 | ∅ | Cr1Dy4Ga12 |

The End

TABLE XVIII: The full list of inversion-based higher order topological insulators (*i.e.* z4=2) from Table[XIV], and their symmetry indicators (SIs). The indicators that are not written out are *zero*. See the list of tables.

| TQC | ICSD | Stoich. Formula | SIs |
|---|---|---|---|
| 2.2.1 | 202564 | S1Ta6 | z2w,3=1<br>z4=2 |
| 2.2.4 | 80363 | O9P3Ru1 | z2w,2=1<br>z2w,3=1<br>z4=2 |
| 2.2.5 | 2212 | O7V4 | z2w,2=1<br>z2w,3=1<br>z4=2 |
| 2.2.7 | 61430 | O6Pb3S1 | z2w,3=1<br>z4=2 |
| 2.2.19 | 431529 | Lu1Na1P2S6 | z2w,1=1<br>z4=2 |
| 2.2.22 | 166101 | H2Cu1O5S1 | z2w,2=1<br>z2w,3=1<br>z4=2 |
| 2.2.23 | 167608 | H1Fe1Li1O5P1 | z2w,1=1<br>z4=2 |
| 2.2.25 | 416823 | H1Co1O5P1Zn1 | z2w,3=1<br>z4=2 |
| 2.2.32 | 417328 | H12Cs4O6P2S6 | z2w,3=1<br>z4=2 |
| 2.2.37 | 72390 | Fe1N2Sr2 | z2w,1=1<br>z2w,2=1<br>z2w,3=1<br>z4=2 |
| 2.2.39 | 190444 | Cu3Mo2O9 | z2w,1=1<br>z4=2 |
| 2.2.52 | 69655 | Cu1F3Na1 | z4=2 |
| 2.2.54 | 62000 | Cs1Hg1 | z2w,3=1<br>z4=2 |
| 2.2.59 | 167202 | Co1F1Li1O4S1 | z2w,1=1<br>z4=2 |
| 2.2.73 | 62436 | C2H8Br2N4Se2Te1 | z2w,2=1<br>z2w,3=1<br>z4=2 |
| 2.2.74 | 186554 | C2H3Na1O5 | z2w,2=1<br>z4=2 |
| 2.2.87 | 192060 | Ba1Cu1O7P2 | z2w,1=1<br>z4=2 |
| 2.2.93 | 404850 | As2Cu4O9 | z2w,1=1<br>z2w,2=1<br>z2w,3=1<br>z4=2 |
| 11.2.4 | 43213 | Se3Ta1 | z2w,1=1<br>z2w,3=1<br>z4=2 |
| 11.2.5 | 52307 | Sb2Sr1 | z2w,1=1<br>z4=2 |
| 11.2.6 | 171417 | S3U1 | z2w,3=1<br>z4=2 |
| 11.2.8 | 260980 | S3Tm2 | z4=2 |
| 11.2.12 | 43283 | Pt6Si5 | z4=2 |
| 11.2.13 | 105804 | Pt3Ta1 | z2w,1=1<br>z4=2 |
| 11.2.15 | 54304 | Ni3Ta1 | z2w,3=1<br>z4=2 |

Continued on next column

**TABLE XVIII– continued**

| TQC | ICSD | Stoich. Formula | SIs |
|---|---|---|---|
| 11.2.28 | 66316 | Nd1Ru1Si2 | z2w,1=1<br>z4=2 |
| 11.2.30 | 54283 | Nb1Pt3 | z2w,1=1<br>z4=2 |
| 11.2.31 | 93368 | Na1O8P1W2 | z2w,3=1<br>z4=2 |
| 11.2.35 | 14349 | Mo1Te2 | z4=2 |
| 11.2.36 | 59708 | Li4S4Si1 | z2w,1=1<br>z4=2 |
| 11.2.45 | 62901 | H3Dy3Ge1O8 | z2w,1=1<br>z2w,3=1<br>z4=2 |
| 11.2.47 | 153707 | H3Cl1Cu2O3 | z2w,3=1<br>z4=2 |
| 11.2.49 | 27406 | H2N1Rb1 | z4=2 |
| 11.2.50 | 169093 | H2K1O4P1 | z2w,3=1<br>z4=2 |
| 11.2.52 | 36300 | H2Cs1O4P1 | z2w,3=1<br>z4=2 |
| 11.2.53 | 20449 | H2Cl1O2Y1 | z2w,3=1<br>z4=2 |
| 11.2.60 | 201682 | H14Cl2N5O2Tc1 | z2w,1=1<br>z4=2 |
| 11.2.61 | 54147 | H14Cd1Cu4O18S2 | z4=2 |
| 11.2.63 | 63219 | H10Mo3N2O11 | z2w,1=1<br>z4=2 |
| 11.2.68 | 190487 | D2O4P1Rb1 | z2w,3=1<br>z4=2 |
| 11.2.71 | 69990 | Cu1La2O6Sn1 | z2w,1=1<br>z4=2 |
| 11.2.80 | 110260 | C3H10Cl1N1 | z2w,3=1<br>z4=2 |
| 11.2.82 | 416300 | Bi2Pd2Sr1 | z4=2 |
| 11.2.86 | 194985 | Ba1Nd2O10Ti3 | z2w,3=1<br>z4=2 |
| 11.2.89 | 416299 | Ba1Bi2Pd2 | z4=2 |
| 11.2.90 | 413519 | Ba1.5Sb3 | z2w,1=1<br>z4=2 |
| 11.2.91 | 261042 | Au2In4Yb1 | z4=2 |
| 11.2.95 | 43456 | Am1 | z2w,1=1<br>z2w,3=1<br>z4=2 |
| 12.1.1 | 187638 | Y1 | z2w,3=1<br>z4=2 |
| 12.1.3 | 41047 | Ta2Te3 | z2w,1=1<br>z2w,2=1<br>z2w,3=1<br>z4=2 |
| 12.1.5 | 601347 | Se4Ti3 | z2w,3=1<br>z4=2 |
| 12.1.6 | 48176 | Rh1Si2Y2 | z4=2 |
| 12.1.8 | 56694 | O5Ti5 | z2w,1=1<br>z2w,2=1<br>z4=2 |
| 12.1.9 | 34417 | O2V1 | z2w,1=1<br>z2w,2=1<br>z2w,3=1<br>z4=2 |
| 12.1.10 | 34416 | O2V1 | z2w,1=1<br>z2w,2=1 |

Continued on next column

**TABLE XVIII– continued**

| TQC | ICSD | Stoich. Formula | SIs |
|---|---|---|---|
| | | | z2w,3=1<br>z4=2 |
| 12.1.11 | 50409 | O13V6 | z2w,1=1<br>z2w,2=1<br>z4=2 |
| 12.1.12 | 15028 | O13V6 | z2w,1=1<br>z2w,2=1<br>z4=2 |
| 12.1.16 | 48004 | Ni1Sc3Si3 | z4=2 |
| 12.1.19 | 646321 | Ni1Rh2Se4 | z2w,1=1<br>z2w,2=1<br>z2w,3=1<br>z4=2 |
| 12.1.20 | 646318 | Ni1Rh2Se4 | z2w,1=1<br>z2w,2=1<br>z2w,3=1<br>z4=2 |
| 12.1.21 | 15214 | Ni1Rh2Se4 | z2w,1=1<br>z2w,2=1<br>z2w,3=1<br>z4=2 |
| 12.1.22 | 41238 | Nb7P4 | z2w,1=1<br>z2w,2=1<br>z4=2 |
| 12.1.24 | 77752 | Nb1Sb2 | z2w,1=1<br>z2w,2=1<br>z2w,3=1<br>z4=2 |
| 12.1.25 | 645350 | Nb1Sb2 | z2w,1=1<br>z2w,2=1<br>z2w,3=1<br>z4=2 |
| 12.1.26 | 57430 | Nb1Sb2 | z2w,1=1<br>z2w,2=1<br>z2w,3=1<br>z4=2 |
| 12.1.29 | 645169 | Nb1P2 | z2w,1=1<br>z2w,2=1<br>z2w,3=1<br>z4=2 |
| 12.1.30 | 29383 | N3Na1 | z2w,1=1<br>z2w,2=1<br>z4=2 |
| 12.1.31 | 29382 | N3Na1 | z2w,1=1<br>z2w,2=1<br>z4=2 |
| 12.1.32 | 29381 | N3Na1 | z2w,1=1<br>z2w,2=1<br>z4=2 |
| 12.1.33 | 29380 | N3Na1 | z2w,1=1<br>z2w,2=1<br>z4=2 |
| 12.1.34 | 29379 | N3Na1 | z2w,1=1<br>z2w,2=1<br>z4=2 |
| 12.1.35 | 29378 | N3Na1 | z2w,1=1<br>z2w,2=1<br>z4=2 |
| 12.1.36 | 29377 | N3Na1 | z2w,1=1<br>z2w,2=1 |

Continued on next column

**TABLE XVIII– continued**

| TQC | ICSD | Stoich. Formula | SIs |
|---|---|---|---|
| | | | z4=2 |
| 12.1.37 | 29376 | N3Na1 | z2w,1=1<br>z2w,2=1<br>z4=2 |
| 12.1.38 | 29375 | N3Na1 | z2w,1=1<br>z2w,2=1<br>z4=2 |
| 12.1.39 | 29374 | N3Na1 | z2w,1=1<br>z2w,2=1<br>z4=2 |
| 12.1.40 | 29373 | N3Na1 | z2w,1=1<br>z2w,2=1<br>z4=2 |
| 12.1.41 | 29372 | N3Na1 | z2w,1=1<br>z2w,2=1<br>z4=2 |
| 12.1.42 | 29371 | N3Na1 | z2w,1=1<br>z2w,2=1<br>z4=2 |
| 12.1.43 | 29370 | N3Na1 | z2w,1=1<br>z2w,2=1<br>z4=2 |
| 12.1.44 | 151588 | Mn7Na1O12 | z4=2 |
| 12.1.45 | 78328 | Mn1Mo1O4 | z2w,1=1<br>z2w,2=1<br>z4=2 |
| 12.1.46 | 239971 | Li2Mn1O3 | z2w,3=1<br>z4=2 |
| 12.1.48 | 49021 | K1Li1Mn1O2 | z2w,1=1<br>z2w,2=1<br>z4=2 |
| 12.1.49 | 240135 | In4Sr1 | z4=2 |
| 12.1.50 | 240133 | In4Sr1 | z4=2 |
| 12.1.52 | 407247 | I3La4Si4 | z2w,1=1<br>z2w,2=1<br>z4=2 |
| 12.1.53 | 83965 | Hf2Te1 | z4=2 |
| 12.1.54 | 20755 | H4Cl2Ni1O2 | z2w,1=1<br>z2w,2=1<br>z2w,3=1<br>z4=2 |
| 12.1.55 | 23361 | H3N1Na2S3 | z2w,3=1<br>z4=2 |
| 12.1.57 | 422475 | H3Co2Mo2Na1O10 | z2w,3=1<br>z4=2 |
| 12.1.61 | 262220 | Ge6La4Mg7 | z2w,1=1<br>z2w,2=1<br>z2w,3=1<br>z4=2 |
| 12.1.63 | 183378 | Gd1O6Ru1Sr2 | z4=2 |
| 12.1.65 | 633477 | Fe3Se4 | z2w,1=1<br>z2w,2=1<br>z2w,3=1<br>z4=2 |
| 12.1.66 | 24560 | Fe3Se4 | z2w,1=1<br>z2w,2=1<br>z2w,3=1<br>z4=2 |
| 12.1.75 | 192506 | F5Fe1K2 | z2w,1=1<br>z2w,2=1 |

Continued on next column

**TABLE XVIII– continued**

| TQC | ICSD | Stoich. Formula | SIs |
|---|---|---|---|
| | | | z2w,3=1<br>z4=2 |
| 12.1.76 | 106627 | Eu1In4 | z2w,3=1<br>z4=2 |
| 12.1.78 | 419435 | Er2O3 | z2w,3=1<br>z4=2 |
| 12.1.79 | 180224 | Dy1Si1 | z4=2 |
| 12.1.80 | 180223 | Dy1Si1 | z4=2 |
| 12.1.81 | 180222 | Dy1Si1 | z4=2 |
| 12.1.82 | 180221 | Dy1Si1 | z4=2 |
| 12.1.83 | 418227 | Dy1P2Pt8 | z2w,3=1<br>z4=2 |
| 12.1.85 | 71624 | Cs2Ir1Na2O4 | z2w,1=1<br>z2w,2=1<br>z4=2 |
| 12.1.87 | 626888 | Cr3Te4 | z4=2 |
| 12.1.88 | 626884 | Cr3Te4 | z4=2 |
| 12.1.89 | 626874 | Cr3Te4 | z4=2 |
| 12.1.92 | 26976 | Cr3Se4 | z2w,3=1<br>z4=2 |
| 12.1.94 | 626598 | Cr3S4 | z2w,1=1<br>z2w,2=1<br>z2w,3=1<br>z4=2 |
| 12.1.96 | 16722 | Cr3S4 | z2w,1=1<br>z2w,2=1<br>z2w,3=1<br>z4=2 |
| 12.1.101 | 625931 | Cr2Fe1S4 | z2w,1=1<br>z2w,2=1<br>z2w,3=1<br>z4=2 |
| 12.1.102 | 251718 | Cr1Se2 | z2w,3=1<br>z4=2 |
| 12.1.103 | 75420 | Cr1S2 | z2w,3=1<br>z4=2 |
| 12.1.104 | 626579 | Cr1Rh2Te4 | z2w,1=1<br>z2w,2=1<br>z2w,3=1<br>z4=2 |
| 12.1.105 | 626577 | Cr1Rh2Se4 | z2w,1=1<br>z2w,2=1<br>z2w,3=1<br>z4=2 |
| 12.1.106 | 626541 | Cr1Pr2Se4 | z2w,1=1<br>z2w,2=1<br>z2w,3=1<br>z4=2 |
| 12.1.107 | 821 | Cr1P1S4 | z2w,3=1<br>z4=2 |
| 12.1.108 | 56835 | Cr1P1S4 | z2w,3=1<br>z4=2 |
| 12.1.109 | 173875 | Cr1O4Re1 | z2w,1=1<br>z2w,2=1<br>z2w,3=1<br>z4=2 |
| 12.1.110 | 624485 | Co2Ni1Se4 | z2w,1=1<br>z2w,2=1<br>z2w,3=1<br>z4=2 |

Continued on next column

**TABLE XVIII– continued**

| TQC | ICSD | Stoich. Formula | SIs |
|---|---|---|---|
| 12.1.111 | 66006 | Co2Na7O6Rb1 | z2w,3=1<br>z4=2 |
| 12.1.112 | 2371 | Co1Si2Zr2 | z2w,1=1<br>z2w,2=1<br>z2w,3=1<br>z4=2 |
| 12.1.114 | 623423 | Co1Ge1 | z4=2 |
| 12.1.115 | 43677 | Co1Ge1 | z4=2 |
| 12.1.116 | 184254 | Co1Ge1 | z4=2 |
| 12.1.120 | 427661 | Cl1Gd1O4W1 | z2w,3=1<br>z4=2 |
| 12.1.121 | 35292 | Cl1Gd1O4W1 | z2w,3=1<br>z4=2 |
| 12.1.122 | 420055 | Cl1Gd1Mo1O4 | z2w,3=1<br>z4=2 |
| 12.1.125 | 100827 | Ca2Fe7O11 | z2w,3=1<br>z4=2 |
| 12.1.126 | 20762 | C6H5Cl2Tl1 | z2w,1=1<br>z2w,2=1<br>z4=2 |
| 12.1.131 | 619030 | C3U2W2 | z2w,1=1<br>z2w,2=1<br>z2w,3=1<br>z4=2 |
| 12.1.132 | 618707 | C3Re2U2 | z2w,1=1<br>z2w,2=1<br>z2w,3=1<br>z4=2 |
| 12.1.133 | 618357 | C3Mo2U2 | z2w,1=1<br>z2w,2=1<br>z2w,3=1<br>z4=2 |
| 12.1.134 | 618223 | C3Lu2Mo2 | z2w,1=1<br>z2w,2=1<br>z2w,3=1<br>z4=2 |
| 12.1.138 | 602482 | C3Cr2Yb2 | z2w,1=1<br>z2w,2=1<br>z2w,3=1<br>z4=2 |
| 12.1.139 | 617537 | C3Cr2Tb2 | z2w,1=1<br>z2w,2=1<br>z2w,3=1<br>z4=2 |
| 12.1.140 | 617514 | C3Cr2Lu2 | z2w,1=1<br>z2w,2=1<br>z2w,3=1<br>z4=2 |
| 12.1.141 | 617492 | C3Cr2Dy2 | z2w,1=1<br>z2w,2=1<br>z2w,3=1<br>z4=2 |
| 12.1.142 | 43018 | C3Cr2Dy2 | z2w,1=1<br>z2w,2=1<br>z2w,3=1<br>z4=2 |
| 12.1.144 | 618692 | C3.5Pu2W2 | z2w,1=1<br>z2w,2=1<br>z4=2 |
| 12.1.145 | 262897 | C1H3La1O6P2 | z2w,3=1 |

Continued on next column

**TABLE XVIII– continued**

| TQC | ICSD | Stoich. Formula | SIs |
|---|---|---|---|
| | | | z4=2 |
| 12.1.148 | 67467 | C1Dy2Fe2Si2 | z2w,1=1<br>z2w,2=1<br>z2w,3=1<br>z4=2 |
| 12.1.149 | 391256 | C1Cl2Eu2N2 | z2w,1=1<br>z2w,2=1<br>z4=2 |
| 12.1.150 | 603998 | C1.66Mo1U1 | z2w,1=1<br>z2w,2=1<br>z2w,3=1<br>z4=2 |
| 12.1.154 | 261457 | Bi2Pd3Se2 | z2w,1=1<br>z2w,2=1<br>z4=2 |
| 12.1.157 | 1559 | Bi1I1 | z2w,3=1<br>z4=2 |
| 12.1.158 | 1560 | Bi1Br1 | z4=2 |
| 12.1.160 | 69569 | Ba3Cu2Er2O10Pt1 | z2w,1=1<br>z2w,2=1<br>z2w,3=1<br>z4=2 |
| 12.1.161 | 173685 | Ba3Cd2Sb4 | z2w,1=1<br>z2w,2=1<br>z4=2 |
| 12.1.162 | 423268 | Ba2Ge4Ni5 | z2w,1=1<br>z2w,2=1<br>z2w,3=1<br>z4=2 |
| 12.1.164 | 195017 | Ba1Pt2Sb2 | z4=2 |
| 12.1.166 | 173848 | Ba1Ir1O3 | z2w,1=1<br>z2w,2=1<br>z4=2 |
| 12.1.168 | 61499 | Ba1Bi1O3 | z2w,1=1<br>z2w,2=1<br>z4=2 |
| 12.1.169 | 63510 | B4Nd2Ni5 | z2w,3=1<br>z4=2 |
| 12.1.172 | 615340 | B2Rh3Y1 | z4=2 |
| 12.1.175 | 425401 | Au1Ce2O1P2 | z2w,1=1<br>z2w,2=1<br>z4=2 |
| 12.1.176 | 424760 | As4Ba3Zn2 | z2w,1=1<br>z2w,2=1<br>z4=2 |
| 12.1.177 | 424761 | As4Ba3Cd2 | z2w,1=1<br>z2w,2=1<br>z4=2 |
| 12.1.178 | 43185 | As3W2 | z2w,1=1<br>z2w,2=1<br>z2w,3=1<br>z4=2 |
| 12.1.179 | 15049 | As3V4 | z4=2 |
| 12.1.180 | 43184 | As3Mo2 | z2w,1=1<br>z2w,2=1<br>z2w,3=1<br>z4=2 |
| 12.1.181 | 41015 | As3Mo2 | z2w,1=1<br>z2w,2=1<br>z2w,3=1 |

Continued on next column

**TABLE XVIII– continued**

| TQC | ICSD | Stoich. Formula | SIs |
|---|---|---|---|
| | | | z4=2 |
| 12.1.182 | 610428 | As3Eu1 | z2w,3=1<br>z4=2 |
| 12.1.183 | 611574 | As2V1 | z2w,1=1<br>z2w,2=1<br>z2w,3=1<br>z4=2 |
| 12.1.184 | 44069 | As2Ta1 | z2w,1=1<br>z2w,2=1<br>z2w,3=1<br>z4=2 |
| 12.1.186 | 81218 | As2Nb1 | z2w,1=1<br>z2w,2=1<br>z2w,3=1<br>z4=2 |
| 12.1.187 | 69003 | As2Mn2O7 | z2w,1=1<br>z2w,2=1<br>z4=2 |
| 12.1.188 | 261143 | As2Mn2O7 | z2w,1=1<br>z2w,2=1<br>z4=2 |
| 12.1.189 | 79441 | As2Gd1Pd3 | z2w,1=1<br>z2w,2=1<br>z2w,3=1<br>z4=2 |
| 12.1.196 | 40332 | Al1Cu1 | z2w,1=1<br>z2w,2=1<br>z4=2 |
| 12.1.198 | 429000 | Ag1Ba1S3Tb1 | z2w,3=1<br>z4=2 |
| 13.2.2 | 415434 | O4Ta1Tb1 | z2w,1=1<br>z4=2 |
| 13.2.9 | 415436 | Ho1O4Ta1 | z2w,1=1<br>z4=2 |
| 13.2.13 | 67168 | Cl7Cs1Ti2 | z2w,1=1<br>z4=2 |
| 14.2.2 | 57480 | Re2Si1 | z4=2 |
| 14.2.19 | 98698 | H1Cu1F1O1 | z2w,1=1<br>z4=2 |
| 14.2.23 | 416421 | Ge2La1Pt2 | z2w,1=1<br>z4=2 |
| 14.2.34 | 69772 | Er1O6Ru1Sr2 | z4=2 |
| 14.2.35 | 69771 | Er1O6Ru1Sr2 | z4=2 |
| 14.2.38 | 1672 | Cu1O4Te1 | z4=2 |
| 14.2.39 | 1671 | Cu1O4Te1 | z4=2 |
| 14.2.43 | 411020 | Cu1Ge1O3 | z2w,1=1<br>z4=2 |
| 14.2.49 | 422546 | Cl4Hg4I2S1 | z2w,1=1<br>z4=2 |
| 14.2.59 | 172071 | Br2Cu2O6Sr2Te1 | z2w,1=1<br>z4=2 |
| 15.2.1 | 187497 | Sb2Te3 | z4=2 |
| 15.2.12 | 6286 | O3V2 | z4=2 |
| 15.2.30 | 180367 | Li2O3Ru1 | z4=2 |
| 15.2.33 | 416039 | K3Na1O8Ru2 | z2w,1=1<br>z2w,2=1<br>z4=2 |
| 15.2.38 | 14329 | H4Ba1O8S2 | z2w,1=1<br>z2w,2=1<br>z4=2 |

Continued on next column

**TABLE XVIII– continued**

| TQC | ICSD | Stoich. Formula | SIs |
|---|---|---|---|
| 15.2.41 | 183508 | H12Cl3Mg1O6Rb1 | z2w,1=1<br>z2w,2=1<br>z4=2 |
| 15.2.43 | 237747 | Ge2Mn1Na1O6 | z2w,1=1<br>z2w,2=1<br>z4=2 |
| 15.2.44 | 237746 | Ge2Mn1Na1O6 | z2w,1=1<br>z2w,2=1<br>z4=2 |
| 15.2.46 | 151766 | Fe7S8 | z2w,1=1<br>z2w,2=1<br>z4=2 |
| 15.2.54 | 185090 | Cu6Sn5 | z4=2 |
| 15.2.55 | 185089 | Cu6Sn5 | z4=2 |
| 15.2.56 | 150124 | Cu6Sn5 | z4=2 |
| 15.2.57 | 106530 | Cu6Sn5 | z4=2 |
| 15.2.60 | 67316 | Cu2O7P2 | z2w,1=1<br>z2w,2=1<br>z4=2 |
| 15.2.61 | 28279 | Cu2O7P2 | z2w,1=1<br>z2w,2=1<br>z4=2 |
| 15.2.62 | 157107 | Cu2O7P2 | z2w,1=1<br>z2w,2=1<br>z4=2 |
| 15.2.63 | 14369 | Cu2O7P2 | z2w,1=1<br>z2w,2=1<br>z4=2 |
| 15.2.66 | 83491 | Cu2Dy1O4 | z4=2 |
| 15.2.73 | 95789 | Ca2Ir2Si1 | z2w,1=1<br>z2w,2=1<br>z4=2 |
| 47.2.4 | 63335 | Ba2Cu3O7Y1 | z2w,2=1<br>z4=2 |
| 47.2.10 | 96380 | Ba1Co2Nd1O5.5 | z2w,1=1<br>z2w,2=1<br>z4=2 |
| 47.2.12 | 187120 | Ba1Co2Nd1O5.5 | z2w,2=1<br>z2w,3=1<br>z4=2 |
| 47.2.13 | 71640 | B3Ir4Zn1 | z2w,2=1<br>z4=2 |
| 51.2.2 | 106800 | In1La1Ni2 | z2w,3=1<br>z4=2 |
| 51.2.3 | 370040 | Co2In1Tb1 | z2w,2=1<br>z4=2 |
| 51.2.5 | 194877 | C3Al1Y3 | z2w,2=1<br>z2w,3=1<br>z4=2 |
| 51.2.9 | 99611 | Ba1Dy1Fe2O4.996 | z4=2 |
| 51.2.14 | 194435 | B1Nb1Ru1 | z2w,2=1<br>z2w,3=1<br>z4=2 |
| 55.2.6 | 417818 | Mn2O5Sr2 | z2w,3=1<br>z4=2 |
| 55.2.7 | 417817 | Mn2O5Sr2 | z2w,3=1<br>z4=2 |
| 55.2.8 | 417816 | Mn2O5Sr2 | z2w,3=1<br>z4=2 |
| 55.2.9 | 417815 | Mn2O5Sr2 | z2w,3=1 |

Continued on next column

**TABLE XVIII– continued**

| TQC | ICSD | Stoich. Formula | SIs |
|---|---|---|---|
| | | | z4=2 |
| 55.2.10 | 417814 | Mn2O5Sr2 | z2w,3=1<br>z4=2 |
| 55.2.11 | 417813 | Mn2O5Sr2 | z2w,3=1<br>z4=2 |
| 55.2.12 | 417812 | Mn2O5Sr2 | z2w,3=1<br>z4=2 |
| 55.2.13 | 417811 | Mn2O5Sr2 | z2w,3=1<br>z4=2 |
| 55.2.14 | 417810 | Mn2O5Sr2 | z2w,3=1<br>z4=2 |
| 55.2.15 | 417809 | Mn2O5Sr2 | z2w,3=1<br>z4=2 |
| 55.2.16 | 417808 | Mn2O5Sr2 | z2w,3=1<br>z4=2 |
| 55.2.17 | 417807 | Mn2O5Sr2 | z2w,3=1<br>z4=2 |
| 55.2.18 | 417806 | Mn2O5Sr2 | z2w,3=1<br>z4=2 |
| 55.2.22 | 22046 | Ge1Mn2O4 | z2w,3=1<br>z4=2 |
| 55.2.23 | 23226 | Ga2Mg1 | z2w,3=1<br>z4=2 |
| 55.2.30 | 171414 | Ce5In4Pt2 | z2w,3=1<br>z4=2 |
| 55.2.32 | 102075 | Cd7Th6 | z4=2 |
| 55.2.33 | 36466 | Ca5Ga2Sb6 | z4=2 |
| 55.2.36 | 84869 | C5Y4 | z2w,3=1<br>z4=2 |
| 55.2.45 | 419361 | B4Rh1Yb1 | z4=2 |
| 55.2.49 | 425457 | B4Ni1Sc1 | z4=2 |
| 55.2.52 | 409626 | B4Er1V1 | z4=2 |
| 55.2.54 | 250251 | B4Co1Yb1 | z4=2 |
| 57.2.2 | 85640 | Cr1La1Sb3 | z4=2 |
| 57.2.3 | 85639 | Cr1La1Sb3 | z4=2 |
| 57.2.6 | 76352 | Ce3Ge2Rh2 | z2w,1=1<br>z4=2 |
| 58.2.4 | 42720 | Cr1Sb2 | z4=2 |
| 58.2.5 | 42601 | Cr1Sb2 | z4=2 |
| 59.1.3 | 425285 | Ho1I1S1 | z2w,3=1<br>z4=2 |
| 59.1.4 | 154397 | Ge5N2Sr6 | z2w,3=1<br>z4=2 |
| 59.1.5 | 410357 | Ge1Pr1Pt1 | z2w,3=1<br>z4=2 |
| 59.1.10 | 160543 | Dy2Sn5 | z2w,3=1<br>z4=2 |
| 59.1.11 | 72376 | Cr6P4U1 | z2w,3=1<br>z4=2 |
| 59.1.12 | 30701 | Co8La1P5 | z4=2 |
| 59.1.13 | 410358 | Ce1Ge1Pt1 | z4=2 |
| 59.1.14 | 154398 | Ba6Ge5N2 | z2w,3=1<br>z4=2 |
| 59.1.15 | 421431 | Ba1Ge1Ni2 | z2w,3=1<br>z4=2 |
| 59.1.17 | 421524 | B2Ru1 | z2w,3=1<br>z4=2 |
| 59.1.18 | 31871 | B2Ru1 | z2w,3=1<br>z4=2 |

Continued on next column

**TABLE XVIII– continued**

| TQC | ICSD | Stoich. Formula | SIs |
|---|---|---|---|
| 59.1.19 | 421523 | B2Os1 | z2w,3=1<br>z4=2 |
| 62.2.5 | 402457 | S3Tm2 | z4=2 |
| 62.2.8 | 421663 | Ru2Sn2Zn3 | z4=2 |
| 62.2.15 | 79238 | Pt1Si1 | z4=2 |
| 62.2.16 | 2623 | Pt1Si1 | z4=2 |
| 62.2.17 | 105785 | Pt1Sc2 | z4=2 |
| 62.2.18 | 158364 | Pd2Sn1 | z4=2 |
| 62.2.22 | 109415 | Pd1Sn1 | z4=2 |
| 62.2.23 | 79237 | Pd1Si1 | z4=2 |
| 62.2.24 | 15016 | Pd1Si1 | z4=2 |
| 62.2.30 | 100625 | O4Pb1S1 | z4=2 |
| 62.2.34 | 413386 | O1Si1Yb3 | z4=2 |
| 62.2.35 | 74785 | Ni2P2Th1 | z4=2 |
| 62.2.46 | 95449 | Ni1Si1 | z4=2 |
| 62.2.47 | 95448 | Ni1Si1 | z4=2 |
| 62.2.48 | 190105 | Ni1Si1 | z4=2 |
| 62.2.60 | 41149 | Mn1Rh1Si1 | z4=2 |
| 62.2.76 | 411884 | Ir1Si1Ta1 | z4=2 |
| 62.2.80 | 411882 | Ir1Nb1Si1 | z4=2 |
| 62.2.81 | 246548 | Ir1Mn1Si1 | z4=2 |
| 62.2.82 | 246547 | Ir1Mn1Si1 | z4=2 |
| 62.2.86 | 72291 | Ho2S3 | z4=2 |
| 62.2.90 | 260839 | H2Cl1O2Tm1 | z4=2 |
| 62.2.95 | 44860 | Ge1Rh2 | z4=2 |
| 62.2.99 | 2624 | Ge1Pt1 | z4=2 |
| 62.2.104 | 411883 | Ge1Ir1Nb1 | z4=2 |
| 62.2.107 | 1375 | Gd1P1S1 | z4=2 |
| 62.2.132 | 49729 | Fe1P1Ta1 | z4=2 |
| 62.2.138 | 195382 | F1Ho1Se1 | z4=2 |
| 62.2.148 | 180220 | Dy1Si1 | z4=2 |
| 62.2.149 | 280091 | Dy1Fe1O3 | z4=2 |
| 62.2.150 | 27280 | Dy1Fe1O3 | z4=2 |
| 62.2.153 | 172848 | Cu1Ho1S2 | z4=2 |
| 62.2.157 | 80332 | Cr1Nd1O3 | z4=2 |
| 62.2.158 | 79349 | Cr1Nd1O3 | z4=2 |
| 62.2.159 | 156326 | Cr1Nd1O3 | z4=2 |
| 62.2.160 | 156322 | Cr1Nd1O3 | z4=2 |
| 62.2.161 | 156320 | Cr1Nd1O3 | z4=2 |
| 62.2.173 | 409847 | Co1Si1V1 | z4=2 |
| 62.2.175 | 43401 | Co1P1 | z4=2 |
| 62.2.176 | 43249 | Co1P1 | z4=2 |
| 62.2.180 | 2421 | Co1Mo1P1 | z4=2 |
| 62.2.181 | 150531 | Co1La3 | z4=2 |
| 62.2.182 | 73212 | Co1K3O2 | z4=2 |
| 62.2.186 | 157332 | Ce3In2Ru2 | z4=2 |
| 62.2.193 | 99240 | Ce1In2Pd3 | z4=2 |
| 62.2.194 | 150105 | Ce1Cu6 | z4=2 |
| 62.2.206 | 151477 | C2Cr3 | z4=2 |
| 62.2.207 | 15086 | C2Cr3 | z4=2 |
| 62.2.214 | 90336 | Ba1Ho2Te4 | z4=2 |
| 62.2.221 | 41893 | B2Co1Nb1 | z4=2 |
| 62.2.228 | 249563 | Au2In2Sr1 | z4=2 |
| 62.2.236 | 42446 | As1V1 | z4=2 |
| 62.2.241 | 48030 | As1Co1 | z4=2 |
| 62.2.242 | 48029 | As1Co1 | z4=2 |
| 62.2.243 | 48028 | As1Co1 | z4=2 |
| 62.2.244 | 48026 | As1Co1 | z4=2 |
| 62.2.247 | 150173 | Al1Pr1Pt1 | z4=2 |

Continued on next column

**TABLE XVIII– continued**

| TQC | ICSD | Stoich. Formula | SIs |
|---|---|---|---|
| 62.2.249 | 55623 | Al1Ce1Rh1 | z4=2 |
| 62.2.250 | 160052 | Al1Ce1Rh1 | z4=2 |
| 62.2.254 | 420615 | Ag1As2Tb1 | z4=2 |
| 63.1.2 | 106074 | Sn1Sr1 | z2w,1=1<br>z2w,2=1<br>z4=2 |
| 63.1.5 | 78997 | Si1Sr1 | z2w,1=1<br>z2w,2=1<br>z4=2 |
| 63.1.6 | 87915 | Pd3Te2 | z4=2 |
| 63.1.7 | 77902 | Pd3Te2 | z4=2 |
| 63.1.8 | 411798 | Pd1Sn2Yb1 | z4=2 |
| 63.1.9 | 99253 | O7Re1Tb3 | z2w,1=1<br>z2w,2=1<br>z4=2 |
| 63.1.10 | 96410 | O7Pr3Re1 | z2w,1=1<br>z2w,2=1<br>z4=2 |
| 63.1.11 | 96149 | Np1Si1 | z4=2 |
| 63.1.20 | 165363 | In5Y3 | z2w,1=1<br>z2w,2=1<br>z4=2 |
| 63.1.21 | 165362 | In5La3 | z4=2 |
| 63.1.22 | 418178 | In4Ni1Sr1 | z4=2 |
| 63.1.23 | 424100 | In2Pt1Yb1 | z4=2 |
| 63.1.24 | 410437 | In2Pd1Yb1 | z4=2 |
| 63.1.25 | 414077 | In2Ni1Yb1 | z4=2 |
| 63.1.26 | 412431 | In2La1Rh1 | z4=2 |
| 63.1.27 | 414479 | In2Ir1La1 | z4=2 |
| 63.1.28 | 200067 | Hf3Ni2Si3 | z4=2 |
| 63.1.29 | 26225 | H5Li4Rh1 | z2w,1=1<br>z2w,2=1<br>z4=2 |
| 63.1.31 | 42375 | Ge1Nd1 | z4=2 |
| 63.1.32 | 42374 | Ge1Nd1 | z4=2 |
| 63.1.33 | 170875 | Gd3O7Os1 | z2w,1=1<br>z2w,2=1<br>z4=2 |
| 63.1.34 | 423067 | Ga4Gd1Ni1 | z2w,1=1<br>z2w,2=1<br>z4=2 |
| 63.1.36 | 261037 | Fe1Se3U1 | z2w,1=1<br>z2w,2=1<br>z4=2 |
| 63.1.37 | 245105 | Eu3O7Re1 | z2w,1=1<br>z2w,2=1<br>z4=2 |
| 63.1.38 | 400536 | Eu2O7Ta2 | z2w,1=1<br>z2w,2=1<br>z4=2 |
| 63.1.39 | 40812 | Eu1Ni5P3 | z4=2 |
| 63.1.40 | 237971 | Eu1In4Ir1 | z4=2 |
| 63.1.41 | 80501 | Er3Ge4 | z2w,1=1<br>z2w,2=1<br>z4=2 |
| 63.1.43 | 423066 | Er1Ga4Ni1 | z4=2 |
| 63.1.44 | 152349 | Dy1Sn2 | z4=2 |
| 63.1.45 | 180214 | Dy1Si1 | z4=2 |
| 63.1.46 | 154699 | Dy1Si1 | z4=2 |

Continued on next column

**TABLE XVIII– continued**

| TQC | ICSD | Stoich. Formula | SIs |
|---|---|---|---|
| 63.1.47 | 57225 | Dy1Ge3 | z2w,1=1<br>z2w,2=1<br>z4=2 |
| 63.1.50 | 423716 | Cu3Er2S5Tl1 | z4=2 |
| 63.1.51 | 174372 | Cu3Er2K1Se5 | z4=2 |
| 63.1.54 | 106440 | Cu1Eu1In4 | z4=2 |
| 63.1.55 | 429089 | Cs1Sc1Se5U1 | z2w,1=1<br>z2w,2=1<br>z4=2 |
| 63.1.56 | 413682 | Cs1Cu3Se5Yb2 | z4=2 |
| 63.1.58 | 412074 | Cs1Cu3Er2S5 | z4=2 |
| 63.1.60 | 236871 | Cr2Mg1O4 | z2w,1=1<br>z2w,2=1<br>z4=2 |
| 63.1.63 | 250753 | Ce2Sn1Zn5 | z4=2 |
| 63.1.64 | 250757 | Ce2Pb1Zn5 | z4=2 |
| 63.1.65 | 63036 | Ce2Ge3Li2 | z2w,1=1<br>z2w,2=1<br>z4=2 |
| 63.1.66 | 262430 | Ce2Ge3Li2 | z2w,1=1<br>z2w,2=1<br>z4=2 |
| 63.1.68 | 421810 | Cd2F1Sb5Sr5 | z2w,1=1<br>z2w,2=1<br>z4=2 |
| 63.1.69 | 424107 | Ca2Sn6Zn3 | z4=2 |
| 63.1.70 | 96228 | Ca2In1N1 | z2w,1=1<br>z2w,2=1<br>z4=2 |
| 63.1.71 | 165193 | Ca1Sn1 | z2w,1=1<br>z2w,2=1<br>z4=2 |
| 63.1.73 | 408880 | Ca1In2Pt1 | z4=2 |
| 63.1.75 | 16356 | Ca1Fe5O7 | z2w,1=1<br>z2w,2=1<br>z4=2 |
| 63.1.77 | 421809 | Ba5Cd2F1Sb5 | z2w,1=1<br>z2w,2=1<br>z4=2 |
| 63.1.78 | 411979 | Ba3Pb5 | z2w,1=1<br>z2w,2=1<br>z4=2 |
| 63.1.79 | 165184 | Ba3Pb5 | z2w,1=1<br>z2w,2=1<br>z4=2 |
| 63.1.82 | 78998 | Ba1Si1 | z2w,1=1<br>z2w,2=1<br>z4=2 |
| 63.1.83 | 427286 | Ba1Ru3Sn6 | z4=2 |
| 63.1.84 | 58664 | Ba1Pb1 | z2w,1=1<br>z2w,2=1<br>z4=2 |
| 63.1.86 | 290596 | Ba1Fe2Se3 | z2w,1=1<br>z2w,2=1<br>z4=2 |
| 63.1.87 | 290594 | Ba1Fe2Se3 | z2w,1=1<br>z2w,2=1<br>z4=2 |
| 63.1.88 | 16307 | Ba1Fe2S3 | z2w,1=1<br>z2w,2=1 |

Continued on next column

**TABLE XVIII– continued**

| TQC | ICSD | Stoich. Formula | SIs |
|---|---|---|---|
| | | | z4=2 |
| 63.1.91 | 420401 | B4Fe1O7 | z2w,1=1<br>z2w,2=1<br>z4=2 |
| 63.1.93 | 237221 | Au1Eu1In4 | z2w,1=1<br>z2w,2=1<br>z4=2 |
| 63.1.94 | 33916 | As3Eu1Ni5 | z4=2 |
| 63.1.95 | 104653 | Al9Co3Y2 | z4=2 |
| 63.1.97 | 370014 | Al2Ni1Yb1 | z4=2 |
| 63.1.98 | 423614 | Al2Ge1La2 | z2w,1=1<br>z2w,2=1<br>z4=2 |
| 63.1.99 | 99138 | Al1Si1Y1 | z2w,1=1<br>z2w,2=1<br>z4=2 |
| 63.1.100 | 415373 | Al1Lu1Si1 | z2w,1=1<br>z2w,2=1<br>z4=2 |
| 63.1.101 | 76295 | Al1Ge1Y1 | z2w,1=1<br>z2w,2=1<br>z4=2 |
| 63.1.102 | 426322 | Al1Fe1.5Si2 | z2w,1=1<br>z2w,2=1<br>z4=2 |
| 64.1.7 | 77831 | La1S1 | z2w,1=1<br>z2w,2=1<br>z4=2 |
| 64.1.10 | 90341 | Ge2Ni1 | z2w,1=1<br>z2w,2=1<br>z4=2 |
| 64.1.13 | 71196 | Co2Nb2Te4 | z4=2 |
| 64.1.14 | 71519 | Co1Nb1Te2 | z4=2 |
| 64.1.15 | 424937 | Co1Cs1O2 | z2w,1=1<br>z2w,2=1<br>z4=2 |
| 64.1.16 | 40372 | Ca5Ga2N4 | z2w,1=1<br>z2w,2=1<br>z4=2 |
| 64.1.17 | 103352 | Ca2Mn1O3.5 | z2w,1=1<br>z2w,2=1<br>z4=2 |
| 64.1.24 | 412877 | Ba1Fe1K1O3 | z4=2 |
| 65.1.1 | 87827 | Ni1Si3Y1 | z2w,3=1<br>z4=2 |
| 65.1.4 | 261871 | La2Sn5 | z2w,1=1<br>z2w,2=1<br>z2w,3=1<br>z4=2 |
| 65.1.7 | 415513 | In16Pt7Tb2 | z2w,3=1<br>z4=2 |
| 65.1.8 | 54612 | Ho2Ni2Pb1 | z4=2 |
| 65.1.9 | 60646 | Ge3Ni1Y1 | z2w,1=1<br>z2w,2=1<br>z4=2 |
| 65.1.10 | 421933 | Gd2Mg3Ni2 | z2w,1=1<br>z2w,2=1<br>z2w,3=1<br>z4=2 |
| 65.1.12 | 261512 | Fe2O5Sr3 | z2w,1=1 |

Continued on next column

**TABLE XVIII– continued**

| TQC | ICSD | Stoich. Formula | SIs |
|---|---|---|---|
| | | | z2w,2=1<br>z4=2 |
| 65.1.13 | 238922 | Fe2Ge7U3 | z2w,3=1<br>z4=2 |
| 65.1.16 | 418215 | Cu1Mg4Tb1 | z2w,1=1<br>z2w,2=1<br>z4=2 |
| 65.1.20 | 419476 | C3B2Tb2 | z2w,1=1<br>z2w,2=1<br>z2w,3=1<br>z4=2 |
| 65.1.21 | 413212 | Br2Ce4Ga5 | z2w,3=1<br>z4=2 |
| 65.1.22 | 78621 | Ba2Cu4O8Yb1 | z2w,1=1<br>z2w,2=1<br>z4=2 |
| 65.1.23 | 154010 | Ba1Mn2O6Tb1 | z2w,1=1<br>z2w,2=1<br>z4=2 |
| 65.1.24 | 106291 | Au7Rb3 | z2w,1=1<br>z2w,2=1<br>z4=2 |
| 65.1.25 | 106290 | Au7Rb3 | z2w,1=1<br>z2w,2=1<br>z4=2 |
| 65.1.26 | 418529 | Au4Ce1Si2 | z2w,1=1<br>z2w,2=1<br>z4=2 |
| 69.1.1 | 262559 | Pt3Rb2Se6U1 | z2w,1=1<br>z2w,3=1<br>z4=2 |
| 69.1.2 | 262561 | Pd3Rb2Se6U1 | z2w,1=1<br>z2w,3=1<br>z4=2 |
| 69.1.3 | 262562 | Pd3Rb2S6U1 | z2w,1=1<br>z2w,3=1<br>z4=2 |
| 69.1.4 | 262563 | K2Pd3S6U1 | z2w,1=1<br>z2w,3=1<br>z4=2 |
| 69.1.5 | 414574 | Ge6Li2Sr4 | z2w,1=1<br>z2w,2=1<br>z4=2 |
| 69.1.7 | 262555 | Cs2Pt3Se6U1 | z2w,1=1<br>z2w,3=1<br>z4=2 |
| 69.1.9 | 262557 | Cs2Pd3Se6U1 | z2w,1=1<br>z2w,3=1<br>z4=2 |
| 69.1.10 | 261242 | Cs2Pd3Se6U1 | z2w,1=1<br>z2w,3=1<br>z4=2 |
| 69.1.11 | 262558 | Cs2Pd3S6U1 | z2w,1=1<br>z2w,3=1<br>z4=2 |
| 69.1.18 | 410522 | Ag2Ca4Si6 | z2w,1=1<br>z2w,2=1<br>z4=2 |
| 71.1.17 | 416820 | Ni2Tb2Zn1 | z2w,1=1<br>z2w,2=1 |

Continued on next column

**TABLE XVIII– continued**

| TQC | ICSD | Stoich. Formula | SIs |
|---|---|---|---|
| | | | z2w,3=1<br>z4=2 |
| 71.1.19 | 25307 | Li4Sb4Sr3 | z2w,1=1<br>z2w,2=1<br>z2w,3=1<br>z4=2 |
| 71.1.26 | 411182 | Ge4Pd4Yb3 | z4=2 |
| 71.1.30 | 76509 | Ge2Ir1Y1 | z4=2 |
| 71.1.44 | 55650 | Cu4Er3Si4 | z4=2 |
| 71.1.45 | 201112 | Cs1Fe1S2 | z2w,1=1<br>z2w,2=1<br>z2w,3=1<br>z4=2 |
| 71.1.48 | 262434 | Ce3Ge4Li4 | z2w,1=1<br>z2w,2=1<br>z2w,3=1<br>z4=2 |
| 71.1.50 | 416816 | Ce2Ni2Zn1 | z2w,1=1<br>z2w,2=1<br>z2w,3=1<br>z4=2 |
| 71.1.53 | 429175 | Ca2Mg1Pt2 | z2w,1=1<br>z2w,2=1<br>z2w,3=1<br>z4=2 |
| 71.1.54 | 429174 | Ca2Mg1Pd2 | z2w,1=1<br>z2w,2=1<br>z2w,3=1<br>z4=2 |
| 71.1.55 | 425471 | Ca2Cd1Pd2 | z2w,1=1<br>z2w,2=1<br>z2w,3=1<br>z4=2 |
| 71.1.59 | 94741 | C2Bi2Ca1O8 | z2w,1=1<br>z2w,2=1<br>z2w,3=1<br>z4=2 |
| 71.1.62 | 428114 | Ba3Li4Sb4 | z2w,1=1<br>z2w,2=1<br>z2w,3=1<br>z4=2 |
| 71.1.64 | 428115 | Ba3Bi4Li4 | z2w,1=1<br>z2w,2=1<br>z2w,3=1<br>z4=2 |
| 71.1.71 | 75035 | Ba1Ni1O5Pr2 | z2w,1=1<br>z2w,2=1<br>z2w,3=1<br>z4=2 |
| 71.1.74 | 68425 | Ba1Gd2Ni1O5 | z2w,1=1<br>z2w,2=1<br>z2w,3=1<br>z4=2 |
| 71.1.75 | 62758 | Ba1Gd2Ni1O5 | z2w,1=1<br>z2w,2=1<br>z2w,3=1<br>z4=2 |
| 71.1.76 | 62423 | Ba1Gd2Ni1O5 | z2w,1=1<br>z2w,2=1<br>z2w,3=1 |

Continued on next column

**TABLE XVIII– continued**

| TQC | ICSD | Stoich. Formula | SIs |
|---|---|---|---|
| | | | z4=2 |
| 71.1.85 | 167153 | B6Fe2U1 | z2w,1=1 z2w,2=1 z2w,3=1 z4=2 |
| 71.1.88 | 40791 | B4Cr3 | z4=2 |
| 71.1.89 | 423380 | B4Cr1 | z4=2 |
| 71.1.94 | 418387 | Al4O1Si6Sr10 | z2w,1=1 z2w,2=1 z2w,3=1 z4=2 |
| 71.1.99 | 52616 | Al2Ca3Ge2 | z2w,1=1 z2w,2=1 z2w,3=1 z4=2 |
| 72.1.1 | 84206 | Ni3Si5Y2 | z2w,1=1 z2w,2=1 z2w,3=1 z4=2 |
| 72.1.2 | 428478 | Eu2Pt3Si5 | z2w,1=1 z2w,2=1 z2w,3=1 z4=2 |
| 74.2.1 | 106234 | Yb1Zn2 | z2w,1=1 z2w,2=1 z2w,3=1 z4=2 |
| 74.2.8 | 10156 | Co2O4Si1 | z4=2 |
| 74.2.19 | 251586 | Ba1Ge5 | z2w,1=1 z2w,2=1 z2w,3=1 z4=2 |
| 74.2.20 | 194880 | Ba1Ge5 | z2w,1=1 z2w,2=1 z2w,3=1 z4=2 |
| 74.2.23 | 72289 | Al1F7Na2Ni1 | z2w,1=1 z2w,2=1 z2w,3=1 z4=2 |
| 123.1.1 | 193701 | In5Rh1U1 | z2w,3=1 z4=2 z4mpi=3 z8=6 |
| 123.1.2 | 103639 | Ga5Ir1U1 | z2w,3=1 z4=2 z4mpi=3 z8=6 |
| 123.1.3 | 5257 | Fe1Si2 | z2w,3=1 z4=2 z4mpi=1 z8=2 |
| 123.1.4 | 106640 | Fe1Ga8Tm2 | z2w,1=1 z2w,2=1 z2w,3=1 z4=2 z4mpi=3 z8=2 |
| 123.1.5 | 102455 | Co1Ga5U1 | z2w,3=1 z4=2 |

Continued on next column

**TABLE XVIII– continued**

| TQC | ICSD | Stoich. Formula | SIs |
|---|---|---|---|
| | | | z4mpi=3 z8=6 |
| 123.1.17 | 74163 | Ba2Cu2O7Tl1Y1 | z2w,1=1 z2w,2=1 z4=2 z4mpi=2 z8=6 |
| 123.1.18 | 430061 | Ba1O1Sb2Ti2 | z2w,1=1 z2w,2=1 z2w,3=1 z4=2 z4mpi=3 z8=2 |
| 123.1.19 | 430060 | Ba1O1Sb2Ti2 | z2w,1=1 z2w,2=1 z2w,3=1 z4=2 z4mpi=3 z8=2 |
| 123.1.22 | 262038 | As3Rb1Zn4 | z2w,3=1 z4=2 z4mpi=3 z8=6 |
| 123.1.23 | 262030 | As3Cs1Zn4 | z2w,3=1 z4=2 z4mpi=3 z8=6 |
| 123.1.24 | 170662 | Ag1Sb1Te2 | z2w,1=1 z2w,2=1 z4=2 z4mpi=2 z8=6 |
| 127.2.1 | 105627 | Pb3Sr2 | z4=2 z4mpi=2 z8=2 |
| 127.2.4 | 26200 | Li2Sn5 | z4=2 z8=6 |
| 127.2.22 | 427156 | C2B2Gd1 | z2w,3=1 z4=2 z4mpi=1 z8=6 |
| 127.2.23 | 94042 | C2B2Er1 | z2w,3=1 z4=2 z4mpi=1 z8=6 |
| 127.2.24 | 94040 | C2B2Dy1 | z2w,3=1 z4=2 z4mpi=3 z8=2 |
| 128.1.1 | 425655 | B2Nb2Os1 | z4=2 z8=2 |
| 129.1.1 | 80190 | Sn1Te1Zr1 | z2w,3=1 z4=2 |
| 129.1.2 | 26273 | Sn1Te1U1 | z2w,3=1 z4=2 |
| 129.1.4 | 25734 | Si1Te1Zr1 | z2w,3=1 z4=2 |
| 129.1.5 | 15572 | Si1Te1Zr1 | z2w,3=1 z4=2 |
| 129.1.6 | 25733 | Se1Si1Zr1 | z2w,3=1 |

Continued on next column

**TABLE XVIII– continued**

| TQC | ICSD | Stoich. Formula | SIs |
|---|---|---|---|
| | | | z4=2 |
| 129.1.7 | 15571 | Se1Si1Zr1 | z2w,3=1<br>z4=2 |
| 129.1.8 | 25732 | S1Si1Zr1 | z2w,3=1<br>z4=2 |
| 129.1.9 | 15570 | S1Si1Zr1 | z2w,3=1<br>z4=2 |
| 129.1.10 | 15569 | S1Si1Zr1 | z2w,3=1<br>z4=2 |
| 129.1.11 | 26271 | S1Si1U1 | z2w,3=1<br>z4=2 |
| 129.1.12 | 50351 | Ru1Sb2U1 | z4=2 |
| 129.1.13 | 50350 | Ru1Sb2U1 | z4=2 |
| 129.1.14 | 61194 | Pd2Sb2Sr1 | z4=2 |
| 129.1.15 | 50349 | Pd1Sb2U1 | z4=2 |
| 129.1.16 | 50348 | Pd1Sb2U1 | z4=2 |
| 129.1.17 | 77853 | P2U1 | z2w,3=1<br>z4=2 |
| 129.1.18 | 25731 | O1Si1Zr1 | z2w,3=1<br>z4=2 |
| 129.1.19 | 85856 | Nd1Ru1Si1 | z2w,3=1<br>z4=2 |
| 129.1.20 | 41250 | Nd1Ru1Si1 | z2w,3=1<br>z4=2 |
| 129.1.21 | 76235 | Mn1Si1Y1 | z2w,3=1<br>z4=2 |
| 129.1.23 | 15420 | I1O1Tm1 | z2w,3=1<br>z4=2 |
| 129.1.24 | 25737 | Hf1Si1Te1 | z2w,3=1<br>z4=2 |
| 129.1.25 | 25736 | Hf1Se1Si1 | z2w,3=1<br>z4=2 |
| 129.1.26 | 25735 | Hf1S1Si1 | z2w,3=1<br>z4=2 |
| 129.1.27 | 25740 | Ge1Te1Zr1 | z2w,3=1<br>z4=2 |
| 129.1.28 | 25739 | Ge1Se1Zr1 | z2w,3=1<br>z4=2 |
| 129.1.29 | 15574 | Ge1Se1Zr1 | z2w,3=1<br>z4=2 |
| 129.1.30 | 25738 | Ge1S1Zr1 | z2w,3=1<br>z4=2 |
| 129.1.31 | 15573 | Ge1S1Zr1 | z2w,3=1<br>z4=2 |
| 129.1.32 | 87315 | Ge1S1U1 | z2w,3=1<br>z4=2 |
| 129.1.33 | 26272 | Ge1S1U1 | z2w,3=1<br>z4=2 |
| 129.1.34 | 85863 | Ge1Nd1Ru1 | z2w,3=1<br>z4=2 |
| 129.1.35 | 25743 | Ge1Hf1Te1 | z2w,3=1<br>z4=2 |
| 129.1.36 | 25741 | Ge1Hf1Se1 | z2w,3=1<br>z4=2 |
| 129.1.37 | 25742 | Ge1Hf1S1 | z2w,3=1<br>z4=2 |
| 129.1.38 | 106758 | Gd1Pt2Si2 | z2w,3=1<br>z4=2 |
| 129.1.39 | 180602 | Fe1Te1 | z2w,3=1 |

Continued on next column

**TABLE XVIII– continued**

| TQC | ICSD | Stoich. Formula | SIs |
|---|---|---|---|
| | | | z4=2 |
| 129.1.40 | 169271 | Fe1Se1 | z2w,3=1<br>z4=2 |
| 129.1.41 | 169269 | Fe1Se1 | z2w,3=1<br>z4=2 |
| 129.1.42 | 169267 | Fe1Se1 | z2w,3=1<br>z4=2 |
| 129.1.43 | 169265 | Fe1Se1 | z2w,3=1<br>z4=2 |
| 129.1.44 | 169263 | Fe1Se1 | z2w,3=1<br>z4=2 |
| 129.1.45 | 169261 | Fe1Se1 | z2w,3=1<br>z4=2 |
| 129.1.46 | 169259 | Fe1Se1 | z2w,3=1<br>z4=2 |
| 129.1.47 | 169257 | Fe1Se1 | z2w,3=1<br>z4=2 |
| 129.1.48 | 169255 | Fe1Se1 | z2w,3=1<br>z4=2 |
| 129.1.49 | 169253 | Fe1Se1 | z2w,3=1<br>z4=2 |
| 129.1.51 | 42969 | Er1Se2 | z4=2 |
| 129.1.52 | 59594 | Cu1P1Si1Zr1 | z2w,3=1<br>z4=2 |
| 129.1.54 | 27230 | Cs1F4Ti1 | z2w,3=1<br>z4=2 |
| 129.1.55 | 62696 | Cs1F4Fe1 | z2w,3=1<br>z4=2 |
| 129.1.56 | 6043 | Cs1F4Fe1 | z2w,3=1<br>z4=2 |
| 129.1.58 | 78578 | Co2P2U1 | z2w,3=1<br>z4=2 |
| 129.1.59 | 15770 | Co1K1O2 | z2w,3=1<br>z4=2 |
| 129.1.60 | 97019 | Cl1Mn1O3Sr2 | z2w,3=1<br>z4=2 |
| 129.1.61 | 97018 | Cl1Mn1O3Sr2 | z2w,3=1<br>z4=2 |
| 129.1.62 | 96044 | Cl1Mn1O3Sr2 | z2w,3=1<br>z4=2 |
| 129.1.63 | 94745 | Cl1Mn1O3Sr2 | z2w,3=1<br>z4=2 |
| 129.1.64 | 76171 | Cl1Ho1O1 | z2w,3=1<br>z4=2 |
| 129.1.65 | 91750 | Cl1Co1O3Sr2 | z2w,3=1<br>z4=2 |
| 129.1.66 | 90122 | Cl1Co1O3Sr2 | z2w,3=1<br>z4=2 |
| 129.1.67 | 84529 | Ce1Ni1Sb2 | z4=2 |
| 129.1.68 | 411678 | Ce1Ir2Si2 | z2w,3=1<br>z4=2 |
| 129.1.69 | 61371 | Ce1Ge2Pt2 | z2w,3=1<br>z4=2 |
| 129.1.71 | 191053 | Ce1Ga2Pd2 | z4=2 |
| 129.1.72 | 92319 | Br1Nd1O3Te1 | z2w,3=1<br>z4=2 |
| 129.1.78 | 370004 | Au2Ga2Sr1 | z4=2 |
| 129.1.81 | 66765 | As2U1 | z2w,3=1<br>z4=2 |

Continued on next column

**TABLE XVIII– continued**

| TQC | ICSD | Stoich. Formula | SIs |
|---|---|---|---|
| 129.1.82 | 44081 | As2U1 | z2w,3=1<br>z4=2 |
| 129.1.83 | 107523 | As2U1 | z2w,3=1<br>z4=2 |
| 129.1.84 | 56976 | As2Pd1U1 | z4=2 |
| 129.1.85 | 412554 | As2Cu1Tm1 | z4=2 |
| 129.1.86 | 412551 | As2Cu1Tb1 | z4=2 |
| 129.1.89 | 422004 | As1Fe1O1Tb1 | z2w,3=1<br>z4=2 |
| 129.1.90 | 42165 | As1Cu1Si1Zr1 | z2w,3=1<br>z4=2 |
| 129.1.91 | 423103 | As1Cu1Ge1Zr1 | z2w,3=1<br>z4=2 |
| 129.1.92 | 42750 | Al1Ge1Mn1 | z2w,3=1<br>z4=2 |
| 129.1.93 | 190195 | Ag1Pr1Sb2 | z2w,3=1<br>z4=2 |
| 136.1.1 | 76166 | U1 | z4=2 |
| 136.1.5 | 33797 | Co2K4O5Rb2 | z4=2 |
| 136.1.8 | 417667 | C2Nd2W1 | z4=2 |
| 136.1.9 | 413345 | C2Mo1Pr2 | z4=2 |
| 137.1.3 | 49631 | B4Er1Rh4 | z4=2 |
| 139.1.2 | 42119 | Sb1Sr2 | z2w,1=1<br>z2w,2=1<br>z2w,3=1<br>z4=2<br>z8=2 |
| 139.1.3 | 416747 | Ru1Zn3 | z2w,1=1<br>z2w,2=1<br>z2w,3=1<br>z4=2<br>z8=6 |
| 139.1.4 | 249448 | Ru1Zn3 | z2w,1=1<br>z2w,2=1<br>z2w,3=1<br>z4=2<br>z8=6 |
| 139.1.5 | 155990 | Ru1Zn3 | z2w,1=1<br>z2w,2=1<br>z2w,3=1<br>z4=2<br>z8=6 |
| 139.1.6 | 251818 | Pt2Th1 | z2w,1=1<br>z2w,2=1<br>z2w,3=1<br>z4=2<br>z8=2 |
| 139.1.9 | 61195 | Pd2Sb2Sr1 | z4=2<br>z8=2 |
| 139.1.17 | 56577 | O7Ru2Sr3 | z2w,1=1<br>z2w,2=1<br>z2w,3=1<br>z4=2<br>z8=2 |
| 139.1.22 | 93412 | Nd1Ni2Si2 | z2w,1=1<br>z2w,2=1<br>z2w,3=1<br>z4=2<br>z8=6 |
| 139.1.43 | 82464 | Na2Nd2O10Ti3 | z2w,1=1 |

Continued on next column

**TABLE XVIII– continued**

| TQC | ICSD | Stoich. Formula | SIs |
|---|---|---|---|
| | | | z2w,2=1<br>z2w,3=1<br>z4=2<br>z8=6 |
| 139.1.44 | 75607 | Na2Nd2O10Ti3 | z2w,1=1<br>z2w,2=1<br>z2w,3=1<br>z4=2<br>z8=6 |
| 139.1.45 | 75149 | Na2Nd2O10Ti3 | z2w,1=1<br>z2w,2=1<br>z2w,3=1<br>z4=2<br>z8=6 |
| 139.1.50 | 104941 | Mn1Pd3 | z2w,1=1<br>z2w,2=1<br>z2w,3=1<br>z4=2<br>z8=6 |
| 139.1.58 | 181385 | La2Mn2O3Se2 | z2w,1=1<br>z2w,2=1<br>z2w,3=1<br>z4=2<br>z8=2 |
| 139.1.62 | 72190 | I2Nd1 | z2w,1=1<br>z2w,2=1<br>z2w,3=1<br>z4=2<br>z8=6 |
| 139.1.68 | 53731 | Ge2Nd1Pd2 | z2w,1=1<br>z2w,2=1<br>z2w,3=1<br>z4=2<br>z8=6 |
| 139.1.69 | 81753 | Ge2Nd1Ni2 | z2w,1=1<br>z2w,2=1<br>z2w,3=1<br>z4=2<br>z8=6 |
| 139.1.81 | 90392 | Gd1Ge1Ti1 | z2w,1=1<br>z2w,2=1<br>z2w,3=1<br>z4=2<br>z8=2 |
| 139.1.111 | 424224 | Cu2Ge2Sr1 | z4=2<br>z8=6 |
| 139.1.112 | 410 | Cu2Ge2Sr1 | z4=2<br>z8=6 |
| 139.1.114 | 40763 | Cu1P2U1 | z2w,1=1<br>z2w,2=1<br>z2w,3=1<br>z4=2<br>z8=6 |
| 139.1.126 | 261192 | Co2La2O3Se2 | z2w,1=1<br>z2w,2=1<br>z2w,3=1<br>z4=2<br>z8=6 |
| 139.1.127 | 261191 | Co2La2O3Se2 | z2w,1=1<br>z2w,2=1 |

Continued on next column

**TABLE XVIII– continued**

| TQC | ICSD | Stoich. Formula | SIs |
|---|---|---|---|
| | | | z2w,3=1<br>z4=2<br>z8=6 |
| 139.1.128 | 248239 | Co2La2O3Se2 | z2w,1=1<br>z2w,2=1<br>z2w,3=1<br>z4=2<br>z8=6 |
| 139.1.129 | 168330 | Co2La2O3Se2 | z2w,1=1<br>z2w,2=1<br>z2w,3=1<br>z4=2<br>z8=6 |
| 139.1.139 | 91749 | Cl2Co2O5Sr3 | z2w,1=1<br>z2w,2=1<br>z2w,3=1<br>z4=2<br>z8=2 |
| 139.1.140 | 90123 | Cl2Co2O5Sr3 | z2w,1=1<br>z2w,2=1<br>z2w,3=1<br>z4=2<br>z8=2 |
| 139.1.141 | 251170 | Cl2Co2O4Sr3 | z2w,1=1<br>z2w,2=1<br>z2w,3=1<br>z4=2<br>z8=2 |
| 139.1.149 | 370021 | Cd2Ga2Sr1 | z4=2<br>z8=6 |
| 139.1.159 | 52759 | Ca1Ge2Pd2 | z4=2<br>z8=2 |
| 139.1.173 | 74670 | C2Yb1 | z2w,1=1<br>z2w,2=1<br>z2w,3=1<br>z4=2<br>z8=2 |
| 139.1.174 | 23686 | C2Yb1 | z2w,1=1<br>z2w,2=1<br>z2w,3=1<br>z4=2<br>z8=2 |
| 139.1.178 | 69888 | C2Er1 | z2w,1=1<br>z2w,2=1<br>z2w,3=1<br>z4=2<br>z8=6 |
| 139.1.179 | 23690 | C2Er1 | z2w,1=1<br>z2w,2=1<br>z2w,3=1<br>z4=2<br>z8=6 |
| 139.1.180 | 23689 | C2Er1 | z2w,1=1<br>z2w,2=1<br>z2w,3=1<br>z4=2<br>z8=6 |
| 139.1.200 | 41836 | Bi1Sr2 | z2w,1=1<br>z2w,2=1<br>z2w,3=1 |

Continued on next column

**TABLE XVIII– continued**

| TQC | ICSD | Stoich. Formula | SIs |
|---|---|---|---|
| | | | z4=2<br>z8=2 |
| 139.1.201 | 107609 | Be1Pd2 | z4=2<br>z8=2 |
| 139.1.204 | 427108 | Ba3Ge16Ir4 | z4=2<br>z8=6 |
| 139.1.205 | 194235 | Ba3Ge16Ir4 | z4=2<br>z8=6 |
| 139.1.206 | 189363 | Ba3Ge16Ir4 | z4=2<br>z8=6 |
| 139.1.207 | 41837 | Ba2Sb1 | z2w,1=1<br>z2w,2=1<br>z2w,3=1<br>z4=2<br>z8=2 |
| 139.1.208 | 30084 | Ba2Hg1 | z2w,1=1<br>z2w,2=1<br>z2w,3=1<br>z4=2<br>z8=2 |
| 139.1.209 | 183149 | Ba2F2Mn2O1Se2 | z2w,1=1<br>z2w,2=1<br>z2w,3=1<br>z4=2<br>z8=2 |
| 139.1.210 | 71535 | Ba2Cu2O12Si4 | z2w,1=1<br>z2w,2=1<br>z2w,3=1<br>z4=2<br>z8=2 |
| 139.1.214 | 61197 | Ba1Pd2Sb2 | z4=2<br>z8=2 |
| 139.1.220 | 100024 | Ba1Mn1Sb2 | z4=2<br>z8=6 |
| 139.1.224 | 71642 | B2Ir2Zn1 | z2w,1=1<br>z2w,2=1<br>z2w,3=1<br>z4=2<br>z8=6 |
| 139.1.236 | 167013 | As2F2O1Sr2Ti2 | z2w,1=1<br>z2w,2=1<br>z2w,3=1<br>z4=2<br>z8=6 |
| 139.1.242 | 92989 | As1Ti1Zr1 | z2w,1=1<br>z2w,2=1<br>z2w,3=1<br>z4=2<br>z8=6 |
| 139.1.243 | 42358 | As1Ba2 | z2w,1=1<br>z2w,2=1<br>z2w,3=1<br>z4=2<br>z8=2 |
| 139.1.245 | 57545 | Al8Ca1Mn4 | z4=2<br>z8=6 |
| 139.1.250 | 166270 | Al2Pd5Pu1 | z4=2<br>z8=6 |
| 140.1.3 | 245120 | Mg2Pt1 | z2w,1=1<br>z2w,2=1 |

Continued on next column

**TABLE XVIII– continued**

| TQC | ICSD | Stoich. Formula | SIs |
|---|---|---|---|
| | | | z2w,3=1<br>z4=2<br>z8=6 |
| 140.1.8 | 106369 | Cd3Sr5 | z2w,1=1<br>z2w,2=1<br>z2w,3=1<br>z4=2<br>z8=6 |
| 140.1.9 | 106344 | Ca5Hg3 | z2w,1=1<br>z2w,2=1<br>z2w,3=1<br>z4=2<br>z8=6 |
| 140.1.11 | 190146 | Bi1Ho5Ni2 | z4=2<br>z8=2 |
| 140.1.12 | 100084 | Ba5Sn3 | z2w,1=1<br>z2w,2=1<br>z2w,3=1<br>z4=2<br>z8=2 |
| 140.1.13 | 247208 | B2Cu2O6Sr1 | z2w,1=1<br>z2w,2=1<br>z2w,3=1<br>z4=2<br>z8=2 |
| 140.1.14 | 247207 | B2Cu2O6Sr1 | z2w,1=1<br>z2w,2=1<br>z2w,3=1<br>z4=2<br>z8=2 |
| 141.1.18 | 84377 | Cr2Ni1O4 | z4=2 |
| 141.1.19 | 31300 | Cr2Ni1O4 | z4=2 |
| 141.1.20 | 31299 | Cr2Ni1O4 | z4=2 |
| 141.1.21 | 280062 | Cr2Ni1O4 | z4=2 |
| 148.1.2 | 1096 | P2Pd15 | z4=2 |
| 148.1.5 | 166371 | Ni4Ti3 | z2w,1=1<br>z2w,2=1<br>z2w,3=1<br>z4=2 |
| 148.1.7 | 171584 | Ni1O3Ti1 | z2w,1=1<br>z2w,2=1<br>z2w,3=1<br>z4=2 |
| 148.1.20 | 60006 | Mn1O3Ti1 | z2w,1=1<br>z2w,2=1<br>z2w,3=1<br>z4=2 |
| 148.1.22 | 171579 | Mn1O3Ti1 | z2w,1=1<br>z2w,2=1<br>z2w,3=1<br>z4=2 |
| 148.1.29 | 247127 | Cu1O3Si1 | z2w,1=1<br>z2w,2=1<br>z2w,3=1<br>z4=2 |
| 162.1.4 | 37011 | Ba1Fe4O8Sr1 | z2w,3=1<br>z4=2 |
| 164.1.1 | 657478 | Te2Zr1 | z4=2 |
| 164.1.2 | 653213 | Te2Zr1 | z4=2 |
| 164.1.33 | 25809 | Ho2O2Se1 | z4=2 |

Continued on next column

**TABLE XVIII– continued**

| TQC | ICSD | Stoich. Formula | SIs |
|---|---|---|---|
| 164.1.34 | 185808 | Hg1Pt2Se3 | z2w,3=1<br>z4=2 |
| 164.1.38 | 59801 | Ge2I2La2 | z2w,3=1<br>z4=2 |
| 164.1.42 | 159909 | Fe2K3Na1O8 | z2w,3=1<br>z4=2 |
| 164.1.44 | 25810 | Er2O2Se1 | z2w,3=1<br>z4=2 |
| 164.1.46 | 407246 | Ce2I2Si2 | z2w,3=1<br>z4=2 |
| 164.1.73 | 71625 | C1F2Ho2 | z2w,3=1<br>z4=2 |
| 164.1.84 | 100654 | Bi1Te1 | z2w,3=1<br>z4=2 |
| 164.1.85 | 99681 | Ba6Cl2O12Ru3 | z2w,3=1<br>z4=2 |
| 164.1.99 | 160311 | Al2Er1Ge2 | z2w,3=1<br>z4=2 |
| 164.1.102 | 94501 | Al1La1Si2 | z2w,3=1<br>z4=2 |
| 166.1.13 | 653216 | Te1Zr3 | z2w,1=1<br>z2w,2=1<br>z2w,3=1<br>z4=2 |
| 166.1.20 | 651728 | Sb2V3 | z4=2 |
| 166.1.21 | 41814 | Sb2V3 | z4=2 |
| 166.1.72 | 650000 | Pu2Zn17 | z4=2 |
| 166.1.76 | 649479 | Pr2Zn17 | z2w,1=1<br>z2w,2=1<br>z2w,3=1<br>z4=2 |
| 166.1.77 | 649469 | Pr2Zn17 | z2w,1=1<br>z2w,2=1<br>z2w,3=1<br>z4=2 |
| 166.1.78 | 104241 | Pr2Zn17 | z2w,1=1<br>z2w,2=1<br>z2w,3=1<br>z4=2 |
| 166.1.80 | 648763 | Pd3S2Tl2 | z2w,1=1<br>z2w,2=1<br>z2w,3=1<br>z4=2 |
| 166.1.83 | 53301 | P1 | z4=2 |
| 166.1.85 | 647123 | Ni7Yb2 | z2w,1=1<br>z2w,2=1<br>z2w,3=1<br>z4=2 |
| 166.1.87 | 646978 | Ni3Ti0.67Zr0.33 | z2w,1=1<br>z2w,2=1<br>z2w,3=1<br>z4=2 |
| 166.1.88 | 646399 | Ni3S2Tl2 | z2w,1=1<br>z2w,2=1<br>z2w,3=1<br>z4=2 |
| 166.1.90 | 422328 | Ni3S2Tl2 | z2w,1=1<br>z2w,2=1<br>z2w,3=1<br>z4=2 |

Continued on next column

**TABLE XVIII– continued**

| TQC | ICSD | Stoich. Formula | SIs |
|---|---|---|---|
| 166.1.92 | 654948 | Ni3S2Sn2 | z2w,1=1<br>z2w,2=1<br>z2w,3=1<br>z4=2 |
| 166.1.94 | 646379 | Ni3S2Sn2 | z2w,1=1<br>z2w,2=1<br>z2w,3=1<br>z4=2 |
| 166.1.96 | 100217 | Ni3S2Sn2 | z2w,1=1<br>z2w,2=1<br>z2w,3=1<br>z4=2 |
| 166.1.97 | 646211 | Ni3Pb2Se2 | z2w,1=1<br>z2w,2=1<br>z2w,3=1<br>z4=2 |
| 166.1.98 | 646210 | Ni3Pb2Se2 | z2w,1=1<br>z2w,2=1<br>z2w,3=1<br>z4=2 |
| 166.1.99 | 402453 | Ni3Pb2Se2 | z2w,1=1<br>z2w,2=1<br>z2w,3=1<br>z4=2 |
| 166.1.101 | 646207 | Ni3Pb2S2 | z2w,1=1<br>z2w,2=1<br>z2w,3=1<br>z4=2 |
| 166.1.102 | 646206 | Ni3Pb2S2 | z2w,1=1<br>z2w,2=1<br>z2w,3=1<br>z4=2 |
| 166.1.103 | 417632 | Ni3Pb2S2 | z2w,1=1<br>z2w,2=1<br>z2w,3=1<br>z4=2 |
| 166.1.104 | 24176 | Ni3Pb2S2 | z2w,1=1<br>z2w,2=1<br>z2w,3=1<br>z4=2 |
| 166.1.115 | 641738 | La1Rh3 | z2w,1=1<br>z2w,2=1<br>z2w,3=1<br>z4=2 |
| 166.1.116 | 641724 | La1Rh3 | z2w,1=1<br>z2w,2=1<br>z2w,3=1<br>z4=2 |
| 166.1.120 | 640141 | In2Ni3Se2 | z2w,1=1<br>z2w,2=1<br>z2w,3=1<br>z4=2 |
| 166.1.122 | 640135 | In2Ni3S2 | z2w,1=1<br>z2w,2=1<br>z2w,3=1<br>z4=2 |
| 166.1.124 | 415258 | In2Ni3S2 | z2w,1=1<br>z2w,2=1<br>z2w,3=1<br>z4=2 |

Continued on next column

**TABLE XVIII– continued**

| TQC | ICSD | Stoich. Formula | SIs |
|---|---|---|---|
| 166.1.132 | 638697 | Hf1Ni3 | z2w,1=1<br>z2w,2=1<br>z2w,3=1<br>z4=2 |
| 166.1.133 | 2414 | Hf1Ni3 | z2w,1=1<br>z2w,2=1<br>z2w,3=1<br>z4=2 |
| 166.1.135 | 15888 | H8N2O8Pb1S2 | z4=2 |
| 166.1.148 | 24685 | H3Cl1Co2O3 | z2w,1=1<br>z2w,2=1<br>z2w,3=1<br>z4=2 |
| 166.1.149 | 155534 | H3Cl1Co2O3 | z2w,1=1<br>z2w,2=1<br>z2w,3=1<br>z4=2 |
| 166.1.150 | 6016 | H2N2Th1 | z2w,1=1<br>z2w,2=1<br>z2w,3=1<br>z4=2 |
| 166.1.155 | 911 | H12K2N6Sn1 | z2w,1=1<br>z2w,2=1<br>z2w,3=1<br>z4=2 |
| 166.1.156 | 159891 | H1.24Co3Y1 | z2w,1=1<br>z2w,2=1<br>z2w,3=1<br>z4=2 |
| 166.1.157 | 94220 | Ge2Nd2Zn15 | z2w,1=1<br>z2w,2=1<br>z2w,3=1<br>z4=2 |
| 166.1.159 | 166193 | Gd1Sn4Ti6 | z2w,1=1<br>z2w,2=1<br>z2w,3=1<br>z4=2 |
| 166.1.163 | 635711 | Gd1 | z2w,1=1<br>z2w,2=1<br>z2w,3=1<br>z4=2 |
| 166.1.172 | 2513 | Ga1 | z2w,1=1<br>z2w,2=1<br>z2w,3=1<br>z4=2 |
| 166.1.177 | 632784 | Fe7Nb6 | z2w,1=1<br>z2w,2=1<br>z2w,3=1<br>z4=2 |
| 166.1.180 | 41380 | Fe6Ge5Li1 | z2w,1=1<br>z2w,2=1<br>z2w,3=1<br>z4=2 |
| 166.1.181 | 633920 | Fe3Th1 | z2w,1=1<br>z2w,2=1<br>z2w,3=1<br>z4=2 |
| 166.1.182 | 633912 | Fe3Th1 | z2w,1=1<br>z2w,2=1<br>z2w,3=1 |

Continued on next column

**TABLE XVIII– continued**

| TQC | ICSD | Stoich. Formula | SIs |
|---|---|---|---|
| | | | z4=2 |
| 166.1.183 | 633900 | Fe3Th1 | z2w,1=1<br>z2w,2=1<br>z2w,3=1<br>z4=2 |
| 166.1.185 | 633886 | Fe3Th1 | z2w,1=1<br>z2w,2=1<br>z2w,3=1<br>z4=2 |
| 166.1.187 | 71 | Fe3Sn2 | z2w,1=1<br>z2w,2=1<br>z2w,3=1<br>z4=2 |
| 166.1.195 | 195809 | Fe2O4Yb1 | z2w,1=1<br>z2w,2=1<br>z2w,3=1<br>z4=2 |
| 166.1.205 | 184011 | F1Ho1O1 | z2w,1=1<br>z2w,2=1<br>z2w,3=1<br>z4=2 |
| 166.1.208 | 184012 | Er1F1O1 | z2w,1=1<br>z2w,2=1<br>z2w,3=1<br>z4=2 |
| 166.1.209 | 630377 | Dy2Zn17 | z2w,1=1<br>z2w,2=1<br>z2w,3=1<br>z4=2 |
| 166.1.211 | 262377 | Dy2Ni7 | z2w,1=1<br>z2w,2=1<br>z2w,3=1<br>z4=2 |
| 166.1.216 | 184010 | Dy1F1O1 | z2w,1=1<br>z2w,2=1<br>z2w,3=1<br>z4=2 |
| 166.1.225 | 410393 | Cs9F27Ni9 | z2w,1=1<br>z2w,2=1<br>z2w,3=1<br>z4=2 |
| 166.1.226 | 78928 | Cs4F12Fe2K1Li1 | z2w,1=1<br>z2w,2=1<br>z2w,3=1<br>z4=2 |
| 166.1.233 | 184367 | Co7Pr2 | z2w,1=1<br>z2w,2=1<br>z2w,3=1<br>z4=2 |
| 166.1.252 | 625506 | Co3Tm1 | z2w,1=1<br>z2w,2=1<br>z2w,3=1<br>z4=2 |
| 166.1.253 | 625499 | Co3Tm1 | z2w,1=1<br>z2w,2=1<br>z2w,3=1<br>z4=2 |
| 166.1.262 | 657284 | Co17Gd2N3 | z2w,1=1<br>z2w,2=1<br>z2w,3=1 |

Continued on next column

**TABLE XVIII– continued**

| TQC | ICSD | Stoich. Formula | SIs |
|---|---|---|---|
| | | | z4=2 |
| 166.1.267 | 6077 | Cl1O1Yb1 | z2w,1=1<br>z2w,2=1<br>z2w,3=1<br>z4=2 |
| 166.1.300 | 95112 | C1Nb2S2 | z2w,1=1<br>z2w,2=1<br>z2w,3=1<br>z4=2 |
| 166.1.301 | 95110 | C1Nb2S2 | z2w,1=1<br>z2w,2=1<br>z2w,3=1<br>z4=2 |
| 166.1.339 | 64705 | Bi1 | z4=2 |
| 166.1.340 | 64704 | Bi1 | z4=2 |
| 166.1.341 | 64703 | Bi1 | z4=2 |
| 166.1.342 | 53797 | Bi1 | z4=2 |
| 166.1.343 | 246663 | Bi1 | z4=2 |
| 166.1.344 | 242004 | Bi1 | z4=2 |
| 166.1.345 | 242003 | Bi1 | z4=2 |
| 166.1.346 | 242002 | Bi1 | z4=2 |
| 166.1.347 | 242001 | Bi1 | z4=2 |
| 166.1.348 | 242000 | Bi1 | z4=2 |
| 166.1.349 | 241999 | Bi1 | z4=2 |
| 166.1.350 | 241984 | Bi1 | z4=2 |
| 166.1.351 | 241983 | Bi1 | z4=2 |
| 166.1.353 | 71755 | Ba7Cl2O15Ru4 | z2w,1=1<br>z2w,2=1<br>z2w,3=1<br>z4=2 |
| 166.1.354 | 73069 | Ba7Br2O15Ru4 | z2w,1=1<br>z2w,2=1<br>z2w,3=1<br>z4=2 |
| 166.1.356 | 174186 | Ba4Er1O12Ru3 | z4=2 |
| 166.1.357 | 20621 | Ba3O9Sc4 | z2w,1=1<br>z2w,2=1<br>z2w,3=1<br>z4=2 |
| 166.1.358 | 25005 | Ba2Ni1O6Te1 | z2w,1=1<br>z2w,2=1<br>z2w,3=1<br>z4=2 |
| 166.1.369 | 657452 | B6Fe2Gd5 | z4=2 |
| 166.1.381 | 262037 | As3Cd4Rb1 | z2w,1=1<br>z2w,2=1<br>z2w,3=1<br>z4=2 |
| 166.1.382 | 262035 | As3Cd4Na1 | z2w,1=1<br>z2w,2=1<br>z2w,3=1<br>z4=2 |
| 166.1.383 | 262032 | As3Cd4K1 | z2w,1=1<br>z2w,2=1<br>z2w,3=1<br>z4=2 |
| 166.1.394 | 281660 | Al6Dy2Pt1Si4 | z4=2 |
| 166.1.395 | 58175 | Al3Tb1 | z2w,1=1<br>z2w,2=1<br>z2w,3=1 |

Continued on next column

**TABLE XVIII– continued**

| TQC | ICSD | Stoich. Formula | SIs |
|---|---|---|---|
| | | | z4=2 |
| 166.1.397 | 424311 | Ag4Sb2Sr1 | z2w,1=1<br>z2w,2=1<br>z2w,3=1<br>z4=2 |
| 167.1.1 | 83385 | S3Yb2 | z4=2 |
| 167.1.2 | 75248 | S3Yb2 | z4=2 |
| 167.1.5 | 94768 | O3V2 | z4=2 |
| 167.1.6 | 64796 | O3V2 | z4=2 |
| 167.1.7 | 64786 | O3V2 | z4=2 |
| 167.1.10 | 201108 | O3V2 | z4=2 |
| 167.1.11 | 201107 | O3V2 | z4=2 |
| 167.1.12 | 201106 | O3V2 | z4=2 |
| 167.1.16 | 1870 | O3V2 | z4=2 |
| 167.1.17 | 1869 | O3V2 | z4=2 |
| 167.1.18 | 1473 | O3V2 | z4=2 |
| 167.1.86 | 421391 | C3Cr2N6 | z4=2 |
| 191.1.5 | 107425 | Ga2Gd1 | z4=2<br>z6m0=2<br>z12=2 |
| 191.1.10 | 97723 | B4Co11Pr3 | z2w,3=1<br>z4=2<br>z6m0=1<br>z6mpi=5<br>z12=6 |
| 191.1.16 | 237660 | B1Eu1Pt4 | z2w,3=1<br>z4=2<br>z6m0=1<br>z6mpi=3<br>z12=10 |
| 191.1.17 | 97722 | B1Co4Pr1 | z4=2<br>z6m0=2<br>z12=2 |
| 194.1.27 | 163344 | Gd1Ge1Zn1 | z4=2<br>z6m0=2<br>z12prime=2 |
| 194.1.52 | 246019 | Ba1O3Ru1 | z4=2<br>z12prime=6 |
| 194.1.53 | 173866 | Ba1O3Ru1 | z4=2<br>z12prime=6 |
| 194.1.54 | 160287 | Ba1O3Ru1 | z4=2<br>z12prime=6 |
| 194.1.56 | 192939 | Ba1Cr1O3 | z4=2<br>z6m0=2<br>z12prime=2 |
| 204.1.14 | 248607 | Cr4Cu3La1O12 | z4=2 |
| 204.1.15 | 248606 | Cr4Cu3La1O12 | z4=2 |
| 204.1.16 | 248605 | Cr4Cu3La1O12 | z4=2 |
| 204.1.17 | 248604 | Cr4Cu3La1O12 | z4=2 |
| 204.1.18 | 248603 | Cr4Cu3La1O12 | z4=2 |
| 204.1.19 | 248601 | Cr4Cu3La1O12 | z4=2 |
| 204.1.20 | 248600 | Cr4Cu3La1O12 | z4=2 |
| 204.1.21 | 248599 | Cr4Cu3La1O12 | z4=2 |
| 204.1.22 | 248598 | Cr4Cu3La1O12 | z4=2 |
| 221.1.1 | 201036 | O1Sn1Sr3 | z4=2<br>z8=2 |
| 221.1.2 | 181127 | Ni3Sn1 | z2w,1=1<br>z2w,2=1<br>z2w,3=1 |

Continued on next column

**TABLE XVIII– continued**

| TQC | ICSD | Stoich. Formula | SIs |
|---|---|---|---|
| | | | z4=2<br>z4mpi=3<br>z8=6 |
| 221.1.8 | 100787 | Ca3O1Sn1 | z4=2<br>z8=2 |
| 221.1.9 | 100789 | Ca3O1Pb1 | z4=2<br>z8=2 |
| 221.1.10 | 402087 | Ca3Ge1O1 | z4=2<br>z8=2 |
| 225.1.17 | 188351 | Cl1Cu6O8Y1 | z4=2<br>z8=6 |
| 225.1.18 | 69612 | Cl1Cu6In1O8 | z4=2<br>z8=6 |
| 225.1.23 | 424342 | C1W1 | z4=2<br>z8=2 |
| 225.1.25 | 155551 | Ba2Nd1O6Ru1 | z4=2<br>z8=2 |
| 225.1.28 | 59744 | Ba2Er1O6Ru1 | z4=2<br>z8=6 |
| 225.1.29 | 59743 | Ba2Er1O6Ru1 | z4=2<br>z8=6 |
| 227.1.3 | 151747 | O7Tb2Ti2 | z4=2 |
| 227.1.4 | 429192 | O7Pr2Zr2 | z4=2 |
| 227.1.10 | 202522 | Mo2O7Y2 | z4=2 |
| 227.1.37 | 160866 | Er2O7V2 | z4=2 |
| 227.1.51 | 183965 | Cr2Fe1O4 | z4=2 |
| 227.1.52 | 183964 | Cr2Fe1O4 | z4=2 |
| 227.1.53 | 183963 | Cr2Fe1O4 | z4=2 |
| 227.1.54 | 183962 | Cr2Fe1O4 | z4=2 |
| 227.1.55 | 183961 | Cr2Fe1O4 | z4=2 |
| 227.1.56 | 183960 | Cr2Fe1O4 | z4=2 |
| 227.1.57 | 171121 | Cr2Fe1O4 | z4=2 |
| 227.1.88 | 57512 | Al2Ba1 | z4=2 |

The End

TABLE XIX: The computed symmetry indicators (SIs) for the the insulators obtained by gapping enforced semimetals ESFDs in Table [XIV]. The indicators that are not written out are *zero*. See the list of tables.

| TQC | ICSD SSG | Stoich. Formula SIs |
|---|---|---|
| 11.4.1 | 166509 | S3Ta1 |
| | 2 | z2w,2=1 z4=3 |
| | 2 | z2w,1=1 z2w,2=1 z2w,3=1 z4=3 |
| | 2 | z2w,1=1 z2w,3=1 z4=2 |
| | 2 | z2w,2=1 z2w,3=1 z4=3 |
| | 2 | z2w,3=1 z4=2 |
| | 2 | z2w,1=1 z4=2 |
| | 2 | z2w,1=1 z2w,2=1 z4=1 |
| | 2 | z2w,1=1 z2w,2=1 z4=3 |
| | 2 | z2w,1=1 z4=2 |
| | 2 | z2w,3=1 z4=2 |
| | 2 | z2w,2=1 z2w,3=1 z4=1 |
| | 2 | z2w,1=1 z2w,3=1 z4=2 |
| | 2 | z2w,1=1 z2w,2=1 z2w,3=1 z4=1 |
| | 2 | z2w,2=1 z4=1 |
| 11.4.2 | 15251 | S3Ta1 |
| | 2 | z2w,2=1 z4=3 |
| | 2 | z2w,1=1 z2w,2=1 z2w,3=1 z4=3 |
| | 2 | z2w,1=1 z2w,3=1 z4=2 |
| | 2 | z2w,2=1 z2w,3=1 z4=3 |
| | 2 | z2w,3=1 z4=2 |
| | 2 | z2w,1=1 z4=2 |
| | 2 | z2w,1=1 z2w,2=1 z4=1 |
| | 2 | z2w,1=1 z2w,2=1 z4=3 |
| | 2 | z2w,1=1 z4=2 |
| | 2 | z2w,3=1 z4=2 |
| | 2 | z2w,2=1 z2w,3=1 z4=1 |
| | 2 | z2w,1=1 z2w,3=1 z4=2 |
| | 2 | z2w,1=1 z2w,2=1 z2w,3=1 z4=1 |
| | 2 | z2w,2=1 z4=1 |
| 11.4.3 | 99170 | Pr1Ru1Si2 |
| | 2 | z2w,1=1 z4=3 |
| | 2 | z2w,1=1 z2w,2=1 z4=2 |
| | 2 | z2w,2=1 z2w,3=1 z4=2 |
| | 2 | z2w,3=1 z4=1 |
| | 2 | z2w,1=1 z2w,2=1 z2w,3=1 z4=2 |
| | 2 | z2w,1=1 z2w,3=1 z4=1 |
| | 2 | z4=1 |
| | 2 | z2w,2=1 |
| | 2 | z2w,2=1 z4=2 |
| | 2 | z4=1 |
| | 2 | z2w,1=1 z2w,3=1 z4=1 |
| | 2 | z2w,1=1 z2w,2=1 z2w,3=1 |
| | 2 | z2w,3=1 z4=1 |
| | 2 | z2w,2=1 z2w,3=1 |
| | 2 | z2w,1=1 z2w,2=1 |
| | 2 | z2w,1=1 z4=3 |
| 11.4.4 | 409179 | P5Yb1 |
| | 2 | z2w,3=1 z4=3 |
| | 2 | z2w,2=1 z2w,3=1 z4=2 |
| | 2 | z2w,1=1 z2w,2=1 z4=2 |
| | 2 | z2w,1=1 z4=1 |
| | 2 | z2w,2=1 z4=2 |
| | 2 | z4=1 |
| | 2 | z2w,1=1 z2w,3=1 z4=1 |
| | 2 | z2w,1=1 z2w,2=1 z2w,3=1 |
| | 2 | z2w,1=1 z2w,2=1 z2w,3=1 z4=2 |

Continued on next column

**TABLE XIX– continued**

| TQC | ICSD SSG | Stoich. Formula SIs |
|---|---|---|
| | 2 | z2w,1=1 z2w,3=1 z4=1 |
| | 2 | z4=1 |
| | 2 | z2w,2=1 |
| | 2 | z2w,1=1 z4=1 |
| | 2 | z2w,1=1 z2w,2=1 |
| | 2 | z2w,2=1 z2w,3=1 |
| | 2 | z2w,3=1 z4=3 |
| 11.4.5 | 1239 | O2Rb9 |
| | 2 | z4=3 |
| | 2 | z2w,2=1 z4=2 |
| | 2 | z2w,1=1 z2w,2=1 z2w,3=1 z4=2 |
| | 2 | z2w,1=1 z2w,3=1 z4=1 |
| | 2 | z2w,2=1 z2w,3=1 z4=2 |
| | 2 | z2w,3=1 z4=1 |
| | 2 | z2w,1=1 z4=1 |
| | 2 | z2w,1=1 z2w,2=1 |
| | 2 | z2w,1=1 z2w,2=1 z4=2 |
| | 2 | z2w,1=1 z4=1 |
| | 2 | z2w,3=1 z4=1 |
| | 2 | z2w,2=1 z2w,3=1 |
| | 2 | z2w,1=1 z2w,3=1 z4=1 |
| | 2 | z2w,1=1 z2w,2=1 z2w,3=1 |
| | 2 | z2w,2=1 |
| | 2 | z4=3 |
| 11.4.6 | 424526 | Np1Se3 |
| | 2 | z2w,1=1 z2w,3=1 z4=1 |
| | 2 | z2w,1=1 z2w,2=1 z2w,3=1 |
| | 2 | z2w,2=1 |
| | 2 | z4=3 |
| | 2 | z2w,1=1 z2w,2=1 |
| | 2 | z2w,1=1 z4=3 |
| | 2 | z2w,3=1 z4=3 |
| | 2 | z2w,2=1 z2w,3=1 z4=2 |
| | 2 | z2w,2=1 z2w,3=1 |
| | 2 | z2w,3=1 z4=3 |
| | 2 | z2w,1=1 z4=3 |
| | 2 | z2w,1=1 z2w,2=1 z4=2 |
| | 2 | z4=3 |
| | 2 | z2w,2=1 z4=2 |
| | 2 | z2w,1=1 z2w,2=1 z2w,3=1 z4=2 |
| | 2 | z2w,1=1 z2w,3=1 z4=1 |
| 11.4.7 | 82605 | Ni2Ta1Te3 |
| | 2 | z2w,1=1 z4=3 |
| | 2 | z2w,1=1 z2w,2=1 z4=2 |
| | 2 | z2w,2=1 z2w,3=1 z4=2 |
| | 2 | z2w,3=1 z4=1 |
| | 2 | z2w,1=1 z2w,2=1 z2w,3=1 z4=2 |
| | 2 | z2w,1=1 z2w,3=1 z4=1 |
| | 2 | z4=1 |
| | 2 | z2w,2=1 |
| | 2 | z2w,2=1 z4=2 |
| | 2 | z4=1 |
| | 2 | z2w,1=1 z2w,3=1 z4=1 |
| | 2 | z2w,1=1 z2w,2=1 z2w,3=1 |
| | 2 | z2w,3=1 z4=1 |
| | 2 | z2w,2=1 z2w,3=1 |
| | 2 | z2w,1=1 z2w,2=1 |
| | 2 | z2w,1=1 z4=3 |
| 11.4.8 | 84826 | Ni1P1Zr2 |
| | 2 | z2w,1=1 z4=2 |

Continued on next column

**TABLE XIX– continued**

| TQC | ICSD SSG | Stoich. Formula SIs |
|---|---|---|
| | 2 | z2w,1=1 z2w,2=1 z4=1 |
| | 2 | z2w,2=1 z2w,3=1 z4=1 |
| | 2 | z2w,3=1 |
| | 2 | z2w,1=1 z2w,2=1 z2w,3=1 z4=1 |
| | 2 | z2w,1=1 z2w,3=1 |
| | 2 | z2w,2=1 z4=3 |
| | 2 | z2w,2=1 z4=1 |
| | 2 | z2w,1=1 z2w,3=1 |
| | 2 | z2w,1=1 z2w,2=1 z2w,3=1 z4=3 |
| | 2 | z2w,3=1 |
| | 2 | z2w,2=1 z2w,3=1 z4=3 |
| | 2 | z2w,1=1 z2w,2=1 z4=3 |
| | 2 | z2w,1=1 z4=2 |
| 11.4.9 | 71498 | Nb1Se3 |
| | 2 | z2w,3=1 z4=2 |
| | 2 | z2w,2=1 z2w,3=1 z4=1 |
| | 2 | z2w,1=1 z2w,2=1 z4=1 |
| | 2 | z2w,1=1 |
| | 2 | z2w,2=1 z4=1 |
| | 2 | z2w,1=1 z2w,3=1 |
| | 2 | z2w,1=1 z2w,2=1 z2w,3=1 z4=3 |
| | 2 | z2w,1=1 z2w,2=1 z2w,3=1 z4=1 |
| | 2 | z2w,1=1 z2w,3=1 |
| | 2 | z2w,2=1 z4=3 |
| | 2 | z2w,1=1 |
| | 2 | z2w,1=1 z2w,2=1 z4=3 |
| | 2 | z2w,2=1 z2w,3=1 z4=3 |
| | 2 | z2w,3=1 z4=2 |
| 11.4.10 | 30002 | Nb1Se3 |
| | 2 | z2w,1=1 z2w,3=1 z4=2 |
| | 2 | z2w,1=1 z2w,2=1 z2w,3=1 z4=1 |
| | 2 | z2w,2=1 z4=1 |
| | 2 | z2w,1=1 z2w,2=1 z4=1 |
| | 2 | z2w,1=1 |
| | 2 | z2w,3=1 |
| | 2 | z2w,2=1 z2w,3=1 z4=3 |
| | 2 | z2w,2=1 z2w,3=1 z4=1 |
| | 2 | z2w,3=1 |
| | 2 | z2w,1=1 |
| | 2 | z2w,1=1 z2w,2=1 z4=3 |
| | 2 | z2w,2=1 z4=3 |
| | 2 | z2w,1=1 z2w,2=1 z2w,3=1 z4=3 |
| | 2 | z2w,1=1 z2w,3=1 z4=2 |
| 11.4.11 | 97682 | Mo2S2Sb1 |
| | 2 | z2w,3=1 z4=3 |
| | 2 | z2w,2=1 z2w,3=1 z4=2 |
| | 2 | z2w,1=1 z2w,2=1 z4=2 |
| | 2 | z2w,1=1 z4=1 |
| | 2 | z2w,2=1 z4=2 |
| | 2 | z4=1 |
| | 2 | z2w,1=1 z2w,3=1 z4=1 |
| | 2 | z2w,1=1 z2w,2=1 z2w,3=1 |
| | 2 | z2w,1=1 z2w,2=1 z2w,3=1 z4=2 |
| | 2 | z2w,1=1 z2w,3=1 z4=1 |
| | 2 | z4=1 |
| | 2 | z2w,2=1 |
| | 2 | z2w,1=1 z4=1 |
| | 2 | z2w,1=1 z2w,2=1 |
| | 2 | z2w,2=1 z2w,3=1 |
| | 2 | z2w,3=1 z4=3 |

Continued on next column

**TABLE XIX– continued**

| TQC | ICSD SSG | Stoich. Formula SIs |
|---|---|---|
| 11.4.12 | 104785 | Li7Sn3 |
| | 2 | z2w,2=1 z4=3 |
| | 2 | z2w,1=1 z2w,2=1 z2w,3=1 z4=3 |
| | 2 | z2w,1=1 z2w,3=1 z4=2 |
| | 2 | z2w,2=1 z2w,3=1 z4=3 |
| | 2 | z2w,3=1 z4=2 |
| | 2 | z2w,1=1 z4=2 |
| | 2 | z2w,1=1 z2w,2=1 z4=1 |
| | 2 | z2w,1=1 z2w,2=1 z4=3 |
| | 2 | z2w,1=1 z4=2 |
| | 2 | z2w,3=1 z4=2 |
| | 2 | z2w,2=1 z2w,3=1 z4=1 |
| | 2 | z2w,1=1 z2w,3=1 z4=2 |
| | 2 | z2w,1=1 z2w,2=1 z2w,3=1 z4=1 |
| | 2 | z2w,2=1 z4=1 |
| 11.4.13 | 67845 | Li4O8V3 |
| | 2 | z2w,1=1 z2w,3=1 z4=2 |
| | 2 | z2w,1=1 z2w,2=1 z2w,3=1 z4=1 |
| | 2 | z2w,2=1 z4=1 |
| | 2 | z2w,1=1 z2w,2=1 z4=1 |
| | 2 | z2w,1=1 |
| | 2 | z2w,3=1 |
| | 2 | z2w,2=1 z2w,3=1 z4=3 |
| | 2 | z2w,2=1 z2w,3=1 z4=1 |
| | 2 | z2w,3=1 |
| | 2 | z2w,1=1 |
| | 2 | z2w,1=1 z2w,2=1 z4=3 |
| | 2 | z2w,2=1 z4=3 |
| | 2 | z2w,1=1 z2w,2=1 z2w,3=1 z4=3 |
| | 2 | z2w,1=1 z2w,3=1 z4=2 |
| 11.4.14 | 71960 | La1Ru1Si2 |
| | 2 | z4=2 |
| | 2 | z2w,2=1 z4=1 |
| | 2 | z2w,1=1 z2w,2=1 z2w,3=1 z4=1 |
| | 2 | z2w,1=1 z2w,3=1 |
| | 2 | z2w,2=1 z2w,3=1 z4=1 |
| | 2 | z2w,3=1 |
| | 2 | z2w,1=1 |
| | 2 | z2w,1=1 z2w,2=1 z4=3 |
| | 2 | z2w,1=1 z2w,2=1 z4=1 |
| | 2 | z2w,1=1 |
| | 2 | z2w,3=1 |
| | 2 | z2w,2=1 z2w,3=1 z4=3 |
| | 2 | z2w,1=1 z2w,3=1 |
| | 2 | z2w,1=1 z2w,2=1 z2w,3=1 z4=3 |
| | 2 | z2w,2=1 z4=3 |
| | 2 | z4=2 |
| 11.4.15 | 402125 | In5S7Tl1 |
| | 2 | z2w,1=1 z2w,3=1 z4=1 |
| | 2 | z2w,1=1 z2w,2=1 z2w,3=1 |
| | 2 | z2w,2=1 |
| | 2 | z4=3 |
| | 2 | z2w,1=1 z2w,2=1 |
| | 2 | z2w,1=1 z4=3 |
| | 2 | z2w,3=1 z4=3 |
| | 2 | z2w,2=1 z2w,3=1 z4=2 |
| | 2 | z2w,2=1 z2w,3=1 |
| | 2 | z2w,3=1 z4=3 |
| | 2 | z2w,1=1 z4=3 |
| | 2 | z2w,1=1 z2w,2=1 z4=2 |

Continued on next column

**TABLE XIX– continued**

| TQC | ICSD SSG | Stoich. Formula SIs |
|---|---|---|
| | 2 | z4=3 |
| | 2 | z2w,2=1 z4=2 |
| | 2 | z2w,1=1 z2w,2=1 z2w,3=1 z4=2 |
| | 2 | z2w,1=1 z2w,3=1 z4=1 |
| 11.4.16 | 71524 | I3Ir1Y3 |
| | 2 | z4=1 |
| | 2 | z2w,2=1 |
| | 2 | z2w,1=1 z2w,2=1 z2w,3=1 |
| | 2 | z2w,1=1 z2w,3=1 z4=3 |
| | 2 | z2w,2=1 z2w,3=1 |
| | 2 | z2w,3=1 z4=3 |
| | 2 | z2w,1=1 z4=3 |
| | 2 | z2w,1=1 z2w,2=1 z4=2 |
| | 2 | z2w,1=1 z2w,2=1 |
| | 2 | z2w,1=1 z4=3 |
| | 2 | z2w,3=1 z4=3 |
| | 2 | z2w,2=1 z2w,3=1 z4=2 |
| | 2 | z2w,1=1 z2w,3=1 z4=3 |
| | 2 | z2w,1=1 z2w,2=1 z2w,3=1 z4=2 |
| | 2 | z2w,2=1 z4=2 |
| | 2 | z4=1 |
| 11.4.17 | 84828 | Hf2Ni1P1 |
| | 2 | z2w,1=1 z4=2 |
| | 2 | z2w,1=1 z2w,2=1 z4=1 |
| | 2 | z2w,2=1 z2w,3=1 z4=1 |
| | 2 | z2w,3=1 |
| | 2 | z2w,1=1 z2w,2=1 z2w,3=1 z4=1 |
| | 2 | z2w,1=1 z2w,3=1 |
| | 2 | z2w,2=1 z4=3 |
| | 2 | z2w,2=1 z4=1 |
| | 2 | z2w,1=1 z2w,3=1 |
| | 2 | z2w,1=1 z2w,2=1 z2w,3=1 z4=3 |
| | 2 | z2w,3=1 |
| | 2 | z2w,2=1 z2w,3=1 z4=3 |
| | 2 | z2w,1=1 z2w,2=1 z4=3 |
| | 2 | z2w,1=1 z4=2 |
| 11.4.18 | 405616 | Hf2Ni1P1 |
| | 2 | z2w,1=1 z4=2 |
| | 2 | z2w,1=1 z2w,2=1 z4=1 |
| | 2 | z2w,2=1 z2w,3=1 z4=1 |
| | 2 | z2w,3=1 |
| | 2 | z2w,1=1 z2w,2=1 z2w,3=1 z4=1 |
| | 2 | z2w,1=1 z2w,3=1 |
| | 2 | z2w,2=1 z4=3 |
| | 2 | z2w,2=1 z4=1 |
| | 2 | z2w,1=1 z2w,3=1 |
| | 2 | z2w,1=1 z2w,2=1 z2w,3=1 z4=3 |
| | 2 | z2w,3=1 |
| | 2 | z2w,2=1 z2w,3=1 z4=3 |
| | 2 | z2w,1=1 z2w,2=1 z4=3 |
| | 2 | z2w,1=1 z4=2 |
| 11.4.19 | 248158 | H6Cu1N1Na3O13P2 |
| | 2 | z2w,3=1 z4=2 |
| | 2 | z2w,2=1 z2w,3=1 z4=1 |
| | 2 | z2w,1=1 z2w,2=1 z4=1 |
| | 2 | z2w,1=1 |
| | 2 | z2w,2=1 z4=1 |
| | 2 | z2w,1=1 z2w,3=1 |
| | 2 | z2w,1=1 z2w,2=1 z2w,3=1 z4=3 |
| | 2 | z2w,1=1 z2w,2=1 z2w,3=1 z4=1 |

Continued on next column

**TABLE XIX– continued**

| TQC | ICSD SSG | Stoich. Formula SIs |
|---|---|---|
| | 2 | z2w,1=1 z2w,3=1 |
| | 2 | z2w,2=1 z4=3 |
| | 2 | z2w,1=1 |
| | 2 | z2w,1=1 z2w,2=1 z4=3 |
| | 2 | z2w,2=1 z2w,3=1 z4=3 |
| | 2 | z2w,3=1 z4=2 |
| 11.4.20 | 48018 | H6Cl2Cu4O6 |
| | 2 | z2w,3=1 z4=2 |
| | 2 | z2w,2=1 z2w,3=1 z4=1 |
| | 2 | z2w,1=1 z2w,2=1 z4=1 |
| | 2 | z2w,1=1 |
| | 2 | z2w,2=1 z4=1 |
| | 2 | z2w,1=1 z2w,3=1 |
| | 2 | z2w,1=1 z2w,2=1 z2w,3=1 z4=3 |
| | 2 | z2w,1=1 z2w,2=1 z2w,3=1 z4=1 |
| | 2 | z2w,1=1 z2w,3=1 |
| | 2 | z2w,2=1 z4=3 |
| | 2 | z2w,1=1 |
| | 2 | z2w,1=1 z2w,2=1 z4=3 |
| | 2 | z2w,2=1 z2w,3=1 z4=3 |
| | 2 | z2w,3=1 z4=2 |
| 11.4.21 | 16780 | H6Br2Cu4O6 |
| | 2 | z2w,2=1 z4=3 |
| | 2 | z2w,1=1 z2w,2=1 z2w,3=1 z4=3 |
| | 2 | z2w,1=1 z2w,3=1 z4=2 |
| | 2 | z2w,2=1 z2w,3=1 z4=3 |
| | 2 | z2w,3=1 z4=2 |
| | 2 | z2w,1=1 z4=2 |
| | 2 | z2w,1=1 z2w,2=1 z4=1 |
| | 2 | z2w,1=1 z2w,2=1 z4=3 |
| | 2 | z2w,1=1 z4=2 |
| | 2 | z2w,3=1 z4=2 |
| | 2 | z2w,2=1 z2w,3=1 z4=1 |
| | 2 | z2w,1=1 z2w,3=1 z4=2 |
| | 2 | z2w,1=1 z2w,2=1 z2w,3=1 z4=1 |
| | 2 | z2w,2=1 z4=1 |
| 11.4.22 | 250377 | H5Ga1N1O5P1 |
| | 2 | z2w,2=1 z4=3 |
| | 2 | z2w,1=1 z2w,2=1 z2w,3=1 z4=3 |
| | 2 | z2w,1=1 z2w,3=1 z4=2 |
| | 2 | z2w,2=1 z2w,3=1 z4=3 |
| | 2 | z2w,3=1 z4=2 |
| | 2 | z2w,1=1 z4=2 |
| | 2 | z2w,1=1 z2w,2=1 z4=1 |
| | 2 | z2w,1=1 z2w,2=1 z4=3 |
| | 2 | z2w,1=1 z4=2 |
| | 2 | z2w,3=1 z4=2 |
| | 2 | z2w,2=1 z2w,3=1 z4=1 |
| | 2 | z2w,1=1 z2w,3=1 z4=2 |
| | 2 | z2w,1=1 z2w,2=1 z2w,3=1 z4=1 |
| | 2 | z2w,2=1 z4=1 |
| 11.4.25 | 239167 | H4Cu5K2O19S3Te1 |
| | 2 | z2w,3=1 z4=2 |
| | 2 | z2w,2=1 z2w,3=1 z4=1 |
| | 2 | z2w,1=1 z2w,2=1 z4=1 |
| | 2 | z2w,1=1 |
| | 2 | z2w,2=1 z4=1 |
| | 2 | z2w,1=1 z2w,3=1 |
| | 2 | z2w,1=1 z2w,2=1 z2w,3=1 z4=3 |
| | 2 | z2w,1=1 z2w,2=1 z2w,3=1 z4=1 |

Continued on next column

**TABLE XIX– continued**

| TQC | ICSD<br>SSG | Stoich. Formula<br>SIs |
|---|---|---|
| | 2 | z2w,1=1 z2w,3=1 |
| | 2 | z2w,2=1 z4=3 |
| | 2 | z2w,1=1 |
| | 2 | z2w,1=1 z2w,2=1 z4=3 |
| | 2 | z2w,2=1 z2w,3=1 z4=3 |
| | 2 | z2w,3=1 z4=2 |
| 11.4.26 | 183584 | H4Cu3Mo1O8 |
| | 2 | z2w,1=1 z2w,3=1 |
| | 2 | z2w,1=1 z2w,2=1 z2w,3=1 z4=3 |
| | 2 | z2w,2=1 z4=3 |
| | 2 | z4=2 |
| | 2 | z2w,1=1 z2w,2=1 z4=3 |
| | 2 | z2w,1=1 z4=2 |
| | 2 | z2w,3=1 z4=2 |
| | 2 | z2w,2=1 z2w,3=1 z4=1 |
| | 2 | z2w,2=1 z2w,3=1 z4=3 |
| | 2 | z2w,3=1 z4=2 |
| | 2 | z2w,1=1 z4=2 |
| | 2 | z2w,1=1 z2w,2=1 z4=1 |
| | 2 | z4=2 |
| | 2 | z2w,2=1 z4=1 |
| | 2 | z2w,1=1 z2w,2=1 z2w,3=1 z4=1 |
| | 2 | z2w,1=1 z2w,3=1 |
| 11.4.27 | 262400 | H3O8V1Y3 |
| | 2 | z2w,1=1 z2w,3=1 z4=2 |
| | 2 | z2w,1=1 z2w,2=1 z2w,3=1 z4=1 |
| | 2 | z2w,2=1 z4=1 |
| | 2 | z2w,1=1 z2w,2=1 z4=1 |
| | 2 | z2w,1=1 |
| | 2 | z2w,3=1 |
| | 2 | z2w,2=1 z2w,3=1 z4=3 |
| | 2 | z2w,2=1 z2w,3=1 z4=1 |
| | 2 | z2w,3=1 |
| | 2 | z2w,1=1 |
| | 2 | z2w,1=1 z2w,2=1 z4=3 |
| | 2 | z2w,2=1 z4=3 |
| | 2 | z2w,1=1 z2w,2=1 z2w,3=1 z4=3 |
| | 2 | z2w,1=1 z2w,3=1 z4=2 |
| 11.4.28 | 36417 | H3O14Sb5 |
| | 2 | z2w,2=1 z4=3 |
| | 2 | z2w,1=1 z2w,2=1 z2w,3=1 z4=3 |
| | 2 | z2w,1=1 z2w,3=1 z4=2 |
| | 2 | z2w,2=1 z2w,3=1 z4=3 |
| | 2 | z2w,3=1 z4=2 |
| | 2 | z2w,1=1 z4=2 |
| | 2 | z2w,1=1 z2w,2=1 z4=1 |
| | 2 | z2w,1=1 z2w,2=1 z4=3 |
| | 2 | z2w,1=1 z4=2 |
| | 2 | z2w,3=1 z4=2 |
| | 2 | z2w,2=1 z2w,3=1 z4=1 |
| | 2 | z2w,1=1 z2w,3=1 z4=2 |
| | 2 | z2w,1=1 z2w,2=1 z2w,3=1 z4=1 |
| | 2 | z2w,2=1 z4=1 |
| 11.4.29 | 38152 | H3Cu2N1O6 |
| | 2 | z2w,3=1 z4=2 |
| | 2 | z2w,2=1 z2w,3=1 z4=1 |
| | 2 | z2w,1=1 z2w,2=1 z4=1 |
| | 2 | z2w,1=1 |
| | 2 | z2w,2=1 z4=1 |
| | 2 | z2w,1=1 z2w,3=1 |

Continued on next column

**TABLE XIX– continued**

| TQC | ICSD<br>SSG | Stoich. Formula<br>SIs |
|---|---|---|
| | 2 | z2w,1=1 z2w,2=1 z2w,3=1 z4=3 |
| | 2 | z2w,1=1 z2w,2=1 z2w,3=1 z4=1 |
| | 2 | z2w,1=1 z2w,3=1 |
| | 2 | z2w,2=1 z4=3 |
| | 2 | z2w,1=1 |
| | 2 | z2w,1=1 z2w,2=1 z4=3 |
| | 2 | z2w,2=1 z2w,3=1 z4=3 |
| | 2 | z2w,3=1 z4=2 |
| 11.4.30 | 27668 | H3Cu2N1O6 |
| | 2 | z2w,2=1 z4=3 |
| | 2 | z2w,1=1 z2w,2=1 z2w,3=1 z4=3 |
| | 2 | z2w,1=1 z2w,3=1 z4=2 |
| | 2 | z2w,2=1 z2w,3=1 z4=3 |
| | 2 | z2w,3=1 z4=2 |
| | 2 | z2w,1=1 z4=2 |
| | 2 | z2w,1=1 z2w,2=1 z4=1 |
| | 2 | z2w,1=1 z2w,2=1 z4=3 |
| | 2 | z2w,1=1 z4=2 |
| | 2 | z2w,3=1 z4=2 |
| | 2 | z2w,2=1 z2w,3=1 z4=1 |
| | 2 | z2w,1=1 z2w,3=1 z4=2 |
| | 2 | z2w,1=1 z2w,2=1 z2w,3=1 z4=1 |
| | 2 | z2w,2=1 z4=1 |
| 11.4.31 | 252429 | H3Cl1Cu1O3Zn1 |
| | 2 | z2w,1=1 z4=2 |
| | 2 | z2w,1=1 z2w,2=1 z4=1 |
| | 2 | z2w,2=1 z2w,3=1 z4=1 |
| | 2 | z2w,3=1 |
| | 2 | z2w,1=1 z2w,2=1 z2w,3=1 z4=1 |
| | 2 | z2w,1=1 z2w,3=1 |
| | 2 | z2w,2=1 z4=3 |
| | 2 | z2w,2=1 z4=1 |
| | 2 | z2w,1=1 z2w,3=1 |
| | 2 | z2w,1=1 z2w,2=1 z2w,3=1 z4=3 |
| | 2 | z2w,3=1 |
| | 2 | z2w,2=1 z2w,3=1 z4=3 |
| | 2 | z2w,1=1 z2w,2=1 z4=3 |
| | 2 | z2w,1=1 z4=2 |
| 11.4.32 | 413481 | H3Ca1Cs1O12Se3 |
| | 2 | z2w,1=1 z2w,3=1 z4=2 |
| | 2 | z2w,1=1 z2w,2=1 z2w,3=1 z4=1 |
| | 2 | z2w,2=1 z4=1 |
| | 2 | z2w,1=1 z2w,2=1 z4=1 |
| | 2 | z2w,1=1 |
| | 2 | z2w,3=1 |
| | 2 | z2w,2=1 z2w,3=1 z4=3 |
| | 2 | z2w,2=1 z2w,3=1 z4=1 |
| | 2 | z2w,3=1 |
| | 2 | z2w,1=1 |
| | 2 | z2w,1=1 z2w,2=1 z4=3 |
| | 2 | z2w,2=1 z4=3 |
| | 2 | z2w,1=1 z2w,2=1 z2w,3=1 z4=3 |
| | 2 | z2w,1=1 z2w,3=1 z4=2 |
| 11.4.33 | 27493 | H3Br1Cu2O3 |
| | 2 | z2w,2=1 z4=3 |
| | 2 | z2w,1=1 z2w,2=1 z2w,3=1 z4=3 |
| | 2 | z2w,1=1 z2w,3=1 z4=2 |
| | 2 | z2w,2=1 z2w,3=1 z4=3 |
| | 2 | z2w,3=1 z4=2 |
| | 2 | z2w,1=1 z4=2 |

Continued on next column

**TABLE XIX– continued**

| TQC | ICSD SSG | Stoich. Formula SIs |
|---|---|---|
| | 2 | z2w,1=1 z2w,2=1 z4=1 |
| | 2 | z2w,1=1 z2w,2=1 z4=3 |
| | 2 | z2w,1=1 z4=2 |
| | 2 | z2w,3=1 z4=2 |
| | 2 | z2w,2=1 z2w,3=1 z4=1 |
| | 2 | z2w,1=1 z2w,3=1 z4=2 |
| | 2 | z2w,1=1 z2w,2=1 z2w,3=1 z4=1 |
| | 2 | z2w,2=1 z4=1 |
| 11.4.34 | 172060 | H2Na3O6Os1 |
| | 2 | z2w,1=1 z4=3 |
| | 2 | z2w,1=1 z2w,2=1 z4=2 |
| | 2 | z2w,2=1 z2w,3=1 z4=2 |
| | 2 | z2w,3=1 z4=1 |
| | 2 | z2w,1=1 z2w,2=1 z2w,3=1 z4=2 |
| | 2 | z2w,1=1 z2w,3=1 z4=1 |
| | 2 | z4=1 |
| | 2 | z2w,2=1 |
| | 2 | z2w,2=1 z4=2 |
| | 2 | z4=1 |
| | 2 | z2w,1=1 z2w,3=1 z4=1 |
| | 2 | z2w,1=1 z2w,2=1 z2w,3=1 |
| | 2 | z2w,3=1 z4=1 |
| | 2 | z2w,2=1 z2w,3=1 |
| | 2 | z2w,1=1 z2w,2=1 |
| | 2 | z2w,1=1 z4=3 |
| 11.4.35 | 159479 | H2K1O9S2Ti1 |
| | 2 | z2w,2=1 z4=3 |
| | 2 | z2w,1=1 z2w,2=1 z2w,3=1 z4=3 |
| | 2 | z2w,1=1 z2w,3=1 z4=2 |
| | 2 | z2w,2=1 z2w,3=1 z4=3 |
| | 2 | z2w,3=1 z4=2 |
| | 2 | z2w,1=1 z4=2 |
| | 2 | z2w,1=1 z2w,2=1 z4=1 |
| | 2 | z2w,1=1 z2w,2=1 z4=3 |
| | 2 | z2w,1=1 z4=2 |
| | 2 | z2w,3=1 z4=2 |
| | 2 | z2w,2=1 z2w,3=1 z4=1 |
| | 2 | z2w,1=1 z2w,3=1 z4=2 |
| | 2 | z2w,1=1 z2w,2=1 z2w,3=1 z4=1 |
| | 2 | z2w,2=1 z4=1 |
| 11.4.36 | 68173 | H2Cu1O6Pb1S1 |
| | 2 | z2w,2=1 z4=3 |
| | 2 | z2w,1=1 z2w,2=1 z2w,3=1 z4=3 |
| | 2 | z2w,1=1 z2w,3=1 z4=2 |
| | 2 | z2w,2=1 z2w,3=1 z4=3 |
| | 2 | z2w,3=1 z4=2 |
| | 2 | z2w,1=1 z4=2 |
| | 2 | z2w,1=1 z2w,2=1 z4=1 |
| | 2 | z2w,1=1 z2w,2=1 z4=3 |
| | 2 | z2w,1=1 z4=2 |
| | 2 | z2w,3=1 z4=2 |
| | 2 | z2w,2=1 z2w,3=1 z4=1 |
| | 2 | z2w,1=1 z2w,3=1 z4=2 |
| | 2 | z2w,1=1 z2w,2=1 z2w,3=1 z4=1 |
| | 2 | z2w,2=1 z4=1 |
| 11.4.37 | 25714 | H2Cu1O6Pb1S1 |
| | 2 | z2w,2=1 z4=3 |
| | 2 | z2w,1=1 z2w,2=1 z2w,3=1 z4=3 |
| | 2 | z2w,1=1 z2w,3=1 z4=2 |
| | 2 | z2w,2=1 z2w,3=1 z4=3 |

Continued on next column

**TABLE XIX– continued**

| TQC | ICSD SSG | Stoich. Formula SIs |
|---|---|---|
| | 2 | z2w,3=1 z4=2 |
| | 2 | z2w,1=1 z4=2 |
| | 2 | z2w,1=1 z2w,2=1 z4=1 |
| | 2 | z2w,1=1 z2w,2=1 z4=3 |
| | 2 | z2w,1=1 z4=2 |
| | 2 | z2w,3=1 z4=2 |
| | 2 | z2w,2=1 z2w,3=1 z4=1 |
| | 2 | z2w,1=1 z2w,3=1 z4=2 |
| | 2 | z2w,1=1 z2w,2=1 z2w,3=1 z4=1 |
| | 2 | z2w,2=1 z4=1 |
| 11.4.38 | 15915 | H2Cu1O6Pb1S1 |
| | 2 | z2w,2=1 z4=3 |
| | 2 | z2w,1=1 z2w,2=1 z2w,3=1 z4=3 |
| | 2 | z2w,1=1 z2w,3=1 z4=2 |
| | 2 | z2w,2=1 z2w,3=1 z4=3 |
| | 2 | z2w,3=1 z4=2 |
| | 2 | z2w,1=1 z4=2 |
| | 2 | z2w,1=1 z2w,2=1 z4=1 |
| | 2 | z2w,1=1 z2w,2=1 z4=3 |
| | 2 | z2w,1=1 z4=2 |
| | 2 | z2w,3=1 z4=2 |
| | 2 | z2w,2=1 z2w,3=1 z4=1 |
| | 2 | z2w,1=1 z2w,3=1 z4=2 |
| | 2 | z2w,1=1 z2w,2=1 z2w,3=1 z4=1 |
| | 2 | z2w,2=1 z4=1 |
| 11.4.39 | 15435 | H2Cu1O6Pb1S1 |
| | 2 | z2w,2=1 z4=3 |
| | 2 | z2w,1=1 z2w,2=1 z2w,3=1 z4=3 |
| | 2 | z2w,1=1 z2w,3=1 z4=2 |
| | 2 | z2w,2=1 z2w,3=1 z4=3 |
| | 2 | z2w,3=1 z4=2 |
| | 2 | z2w,1=1 z4=2 |
| | 2 | z2w,1=1 z2w,2=1 z4=1 |
| | 2 | z2w,1=1 z2w,2=1 z4=3 |
| | 2 | z2w,1=1 z4=2 |
| | 2 | z2w,3=1 z4=2 |
| | 2 | z2w,2=1 z2w,3=1 z4=1 |
| | 2 | z2w,1=1 z2w,3=1 z4=2 |
| | 2 | z2w,1=1 z2w,2=1 z2w,3=1 z4=1 |
| | 2 | z2w,2=1 z4=1 |
| 11.4.40 | 200123 | H2Cl2Cu1O4Pb3 |
| | 2 | z2w,2=1 z4=3 |
| | 2 | z2w,1=1 z2w,2=1 z2w,3=1 z4=3 |
| | 2 | z2w,1=1 z2w,3=1 z4=2 |
| | 2 | z2w,2=1 z2w,3=1 z4=3 |
| | 2 | z2w,3=1 z4=2 |
| | 2 | z2w,1=1 z4=2 |
| | 2 | z2w,1=1 z2w,2=1 z4=1 |
| | 2 | z2w,1=1 z2w,2=1 z4=3 |
| | 2 | z2w,1=1 z4=2 |
| | 2 | z2w,3=1 z4=2 |
| | 2 | z2w,2=1 z2w,3=1 z4=1 |
| | 2 | z2w,1=1 z2w,3=1 z4=2 |
| | 2 | z2w,1=1 z2w,2=1 z2w,3=1 z4=1 |
| | 2 | z2w,2=1 z4=1 |
| 11.4.41 | 260837 | H2Cl1O2Yb1 |
| | 2 | z2w,1=1 z4=2 |
| | 2 | z2w,1=1 z2w,2=1 z4=1 |
| | 2 | z2w,2=1 z2w,3=1 z4=1 |
| | 2 | z2w,3=1 |

Continued on next column

**TABLE XIX– continued**

| TQC | ICSD SSG | Stoich. Formula SIs |
|---|---|---|
| | 2 | z2w,1=1 z2w,2=1 z2w,3=1 z4=1 |
| | 2 | z2w,1=1 z2w,3=1 |
| | 2 | z2w,2=1 z4=3 |
| | 2 | z2w,2=1 z4=1 |
| | 2 | z2w,1=1 z2w,3=1 |
| | 2 | z2w,1=1 z2w,2=1 z2w,3=1 z4=3 |
| | 2 | z2w,3=1 |
| | 2 | z2w,2=1 z2w,3=1 z4=3 |
| | 2 | z2w,1=1 z2w,2=1 z4=3 |
| | 2 | z2w,1=1 z4=2 |
| 11.4.42 | 260831 | H2Cl1Gd1O2 |
| | 2 | z2w,3=1 z4=2 |
| | 2 | z2w,2=1 z2w,3=1 z4=1 |
| | 2 | z2w,1=1 z2w,2=1 z4=1 |
| | 2 | z2w,1=1 |
| | 2 | z2w,2=1 z4=1 |
| | 2 | z2w,1=1 z2w,3=1 |
| | 2 | z2w,1=1 z2w,2=1 z2w,3=1 z4=3 |
| | 2 | z2w,1=1 z2w,2=1 z2w,3=1 z4=1 |
| | 2 | z2w,1=1 z2w,3=1 |
| | 2 | z2w,2=1 z4=3 |
| | 2 | z2w,1=1 |
| | 2 | z2w,1=1 z2w,2=1 z4=3 |
| | 2 | z2w,2=1 z2w,3=1 z4=3 |
| | 2 | z2w,3=1 z4=2 |
| 11.4.43 | 260833 | H2Cl1Dy1O2 |
| | 2 | z2w,2=1 z4=3 |
| | 2 | z2w,1=1 z2w,2=1 z2w,3=1 z4=3 |
| | 2 | z2w,1=1 z2w,3=1 z4=2 |
| | 2 | z2w,2=1 z2w,3=1 z4=3 |
| | 2 | z2w,3=1 z4=2 |
| | 2 | z2w,1=1 z4=2 |
| | 2 | z2w,1=1 z2w,2=1 z4=1 |
| | 2 | z2w,1=1 z2w,2=1 z4=3 |
| | 2 | z2w,1=1 z4=2 |
| | 2 | z2w,3=1 z4=2 |
| | 2 | z2w,2=1 z2w,3=1 z4=1 |
| | 2 | z2w,1=1 z2w,3=1 z4=2 |
| | 2 | z2w,1=1 z2w,2=1 z2w,3=1 z4=1 |
| | 2 | z2w,2=1 z4=1 |
| 11.4.44 | 26341 | H1O2Yb1 |
| | 2 | z2w,1=1 z4=2 |
| | 2 | z2w,1=1 z2w,2=1 z4=1 |
| | 2 | z2w,2=1 z2w,3=1 z4=1 |
| | 2 | z2w,3=1 |
| | 2 | z2w,1=1 z2w,2=1 z2w,3=1 z4=1 |
| | 2 | z2w,1=1 z2w,3=1 |
| | 2 | z2w,2=1 z4=3 |
| | 2 | z2w,2=1 z4=1 |
| | 2 | z2w,1=1 z2w,3=1 |
| | 2 | z2w,1=1 z2w,2=1 z2w,3=1 z4=3 |
| | 2 | z2w,3=1 |
| | 2 | z2w,2=1 z2w,3=1 z4=3 |
| | 2 | z2w,1=1 z2w,2=1 z4=3 |
| | 2 | z2w,1=1 z4=2 |
| 11.4.45 | 60675 | H1La1O2 |
| | 2 | z2w,3=1 z4=2 |
| | 2 | z2w,2=1 z2w,3=1 z4=1 |
| | 2 | z2w,1=1 z2w,2=1 z4=1 |
| | 2 | z2w,1=1 |

Continued on next column

**TABLE XIX– continued**

| TQC | ICSD SSG | Stoich. Formula SIs |
|---|---|---|
| | 2 | z2w,2=1 z4=1 |
| | 2 | z2w,1=1 z2w,3=1 |
| | 2 | z2w,1=1 z2w,2=1 z2w,3=1 z4=3 |
| | 2 | z2w,1=1 z2w,2=1 z2w,3=1 z4=1 |
| | 2 | z2w,1=1 z2w,3=1 |
| | 2 | z2w,2=1 z4=3 |
| | 2 | z2w,1=1 |
| | 2 | z2w,1=1 z2w,2=1 z4=3 |
| | 2 | z2w,2=1 z2w,3=1 z4=3 |
| | 2 | z2w,3=1 z4=2 |
| 11.4.46 | 82535 | H1K1O7Se2V1 |
| | 2 | z2w,2=1 z4=3 |
| | 2 | z2w,1=1 z2w,2=1 z2w,3=1 z4=3 |
| | 2 | z2w,1=1 z2w,3=1 z4=2 |
| | 2 | z2w,2=1 z2w,3=1 z4=3 |
| | 2 | z2w,3=1 z4=2 |
| | 2 | z2w,1=1 z4=2 |
| | 2 | z2w,1=1 z2w,2=1 z4=1 |
| | 2 | z2w,1=1 z2w,2=1 z4=3 |
| | 2 | z2w,1=1 z4=2 |
| | 2 | z2w,3=1 z4=2 |
| | 2 | z2w,2=1 z2w,3=1 z4=1 |
| | 2 | z2w,1=1 z2w,3=1 z4=2 |
| | 2 | z2w,1=1 z2w,2=1 z2w,3=1 z4=1 |
| | 2 | z2w,2=1 z4=1 |
| 11.4.47 | 35408 | H1F3Sr1 |
| | 2 | z2w,1=1 z4=1 |
| | 2 | z2w,1=1 z2w,2=1 |
| | 2 | z2w,2=1 z2w,3=1 |
| | 2 | z2w,3=1 z4=3 |
| | 2 | z2w,1=1 z2w,2=1 z2w,3=1 |
| | 2 | z2w,1=1 z2w,3=1 z4=3 |
| | 2 | z4=3 |
| | 2 | z2w,2=1 z4=2 |
| | 2 | z2w,2=1 |
| | 2 | z4=3 |
| | 2 | z2w,1=1 z2w,3=1 z4=3 |
| | 2 | z2w,1=1 z2w,2=1 z2w,3=1 z4=2 |
| | 2 | z2w,3=1 z4=3 |
| | 2 | z2w,2=1 z2w,3=1 z4=2 |
| | 2 | z2w,1=1 z2w,2=1 z4=2 |
| | 2 | z2w,1=1 z4=1 |
| 11.4.48 | 36582 | H1Cu1O5Se1Tl1 |
| | 2 | z2w,3=1 z4=2 |
| | 2 | z2w,2=1 z2w,3=1 z4=1 |
| | 2 | z2w,1=1 z2w,2=1 z4=1 |
| | 2 | z2w,1=1 |
| | 2 | z2w,2=1 z4=1 |
| | 2 | z2w,1=1 z2w,3=1 |
| | 2 | z2w,1=1 z2w,2=1 z2w,3=1 z4=3 |
| | 2 | z2w,1=1 z2w,2=1 z2w,3=1 z4=1 |
| | 2 | z2w,1=1 z2w,3=1 |
| | 2 | z2w,2=1 z4=3 |
| | 2 | z2w,1=1 |
| | 2 | z2w,1=1 z2w,2=1 z4=3 |
| | 2 | z2w,2=1 z2w,3=1 z4=3 |
| | 2 | z2w,3=1 z4=2 |
| 11.4.49 | 36581 | H1Cu1O5S1Tl1 |
| | 2 | z2w,2=1 z4=3 |
| | 2 | z2w,1=1 z2w,2=1 z2w,3=1 z4=3 |

Continued on next column

TABLE XIX– continued

| TQC | ICSD SSG | Stoich. Formula SIs |
|---|---|---|
| | 2 | z2w,1=1 z2w,3=1 z4=2 |
| | 2 | z2w,2=1 z2w,3=1 z4=3 |
| | 2 | z2w,3=1 z4=2 |
| | 2 | z2w,1=1 z4=2 |
| | 2 | z2w,1=1 z2w,2=1 z4=1 |
| | 2 | z2w,1=1 z2w,2=1 z4=3 |
| | 2 | z2w,1=1 z4=2 |
| | 2 | z2w,3=1 z4=2 |
| | 2 | z2w,2=1 z2w,3=1 z4=1 |
| | 2 | z2w,1=1 z2w,3=1 z4=2 |
| | 2 | z2w,1=1 z2w,2=1 z2w,3=1 z4=1 |
| | 2 | z2w,2=1 z4=1 |
| 11.4.50 | 4333 | H1Ca1Na1O4Si1 |
| | 2 | z2w,1=1 z4=2 |
| | 2 | z2w,1=1 z2w,2=1 z4=1 |
| | 2 | z2w,2=1 z2w,3=1 z4=1 |
| | 2 | z2w,3=1 |
| | 2 | z2w,1=1 z2w,2=1 z2w,3=1 z4=1 |
| | 2 | z2w,1=1 z2w,3=1 |
| | 2 | z2w,2=1 z4=3 |
| | 2 | z2w,2=1 z4=1 |
| | 2 | z2w,1=1 z2w,3=1 |
| | 2 | z2w,1=1 z2w,2=1 z2w,3=1 z4=3 |
| | 2 | z2w,3=1 |
| | 2 | z2w,2=1 z2w,3=1 z4=3 |
| | 2 | z2w,1=1 z2w,2=1 z4=3 |
| | 2 | z2w,1=1 z4=2 |
| 11.4.51 | 35409 | H1Ba1F3 |
| | 2 | z2w,1=1 z4=1 |
| | 2 | z2w,1=1 z2w,2=1 |
| | 2 | z2w,2=1 z2w,3=1 |
| | 2 | z2w,3=1 z4=3 |
| | 2 | z2w,1=1 z2w,2=1 z2w,3=1 |
| | 2 | z2w,1=1 z2w,3=1 z4=3 |
| | 2 | z4=3 |
| | 2 | z2w,2=1 z4=2 |
| | 2 | z2w,2=1 |
| | 2 | z4=3 |
| | 2 | z2w,1=1 z2w,3=1 z4=3 |
| | 2 | z2w,1=1 z2w,2=1 z2w,3=1 z4=2 |
| | 2 | z2w,3=1 z4=3 |
| | 2 | z2w,2=1 z2w,3=1 z4=2 |
| | 2 | z2w,1=1 z2w,2=1 z4=2 |
| | 2 | z2w,1=1 z4=1 |
| 11.4.52 | 420467 | H1B2F10Nd1 |
| | 2 | z2w,1=1 z4=2 |
| | 2 | z2w,1=1 z2w,2=1 z4=1 |
| | 2 | z2w,2=1 z2w,3=1 z4=1 |
| | 2 | z2w,3=1 |
| | 2 | z2w,1=1 z2w,2=1 z2w,3=1 z4=1 |
| | 2 | z2w,1=1 z2w,3=1 |
| | 2 | z2w,2=1 z4=3 |
| | 2 | z2w,2=1 z4=1 |
| | 2 | z2w,1=1 z2w,3=1 |
| | 2 | z2w,1=1 z2w,2=1 z2w,3=1 z4=3 |
| | 2 | z2w,3=1 |
| | 2 | z2w,2=1 z2w,3=1 z4=3 |
| | 2 | z2w,1=1 z2w,2=1 z4=3 |
| | 2 | z2w,1=1 z4=2 |
| 11.4.53 | 180447 | H1B1Ca1O3 |

Continued on next column

TABLE XIX– continued

| TQC | ICSD SSG | Stoich. Formula SIs |
|---|---|---|
| | 2 | z2w,1=1 z4=1 |
| | 2 | z2w,1=1 z2w,2=1 |
| | 2 | z2w,2=1 z2w,3=1 |
| | 2 | z2w,3=1 z4=3 |
| | 2 | z2w,1=1 z2w,2=1 z2w,3=1 |
| | 2 | z2w,1=1 z2w,3=1 z4=3 |
| | 2 | z4=3 |
| | 2 | z2w,2=1 z4=2 |
| | 2 | z2w,2=1 |
| | 2 | z4=3 |
| | 2 | z2w,1=1 z2w,3=1 z4=3 |
| | 2 | z2w,1=1 z2w,2=1 z2w,3=1 z4=2 |
| | 2 | z2w,3=1 z4=3 |
| | 2 | z2w,2=1 z2w,3=1 z4=2 |
| | 2 | z2w,1=1 z2w,2=1 z4=2 |
| | 2 | z2w,1=1 z4=1 |
| 11.4.54 | 67314 | H1As1Cl2O3Pb2 |
| | 2 | z2w,3=1 z4=2 |
| | 2 | z2w,2=1 z2w,3=1 z4=1 |
| | 2 | z2w,1=1 z2w,2=1 z4=1 |
| | 2 | z2w,1=1 |
| | 2 | z2w,2=1 z4=1 |
| | 2 | z2w,1=1 z2w,3=1 |
| | 2 | z2w,1=1 z2w,2=1 z2w,3=1 z4=3 |
| | 2 | z2w,1=1 z2w,2=1 z2w,3=1 z4=1 |
| | 2 | z2w,1=1 z2w,3=1 |
| | 2 | z2w,2=1 z4=3 |
| | 2 | z2w,1=1 |
| | 2 | z2w,1=1 z2w,2=1 z4=3 |
| | 2 | z2w,2=1 z2w,3=1 z4=3 |
| | 2 | z2w,3=1 z4=2 |
| 11.4.55 | 188917 | H1As1Cl2O3Pb2 |
| | 2 | z2w,3=1 z4=2 |
| | 2 | z2w,2=1 z2w,3=1 z4=1 |
| | 2 | z2w,1=1 z2w,2=1 z4=1 |
| | 2 | z2w,1=1 |
| | 2 | z2w,2=1 z4=1 |
| | 2 | z2w,1=1 z2w,3=1 |
| | 2 | z2w,1=1 z2w,2=1 z2w,3=1 z4=3 |
| | 2 | z2w,1=1 z2w,2=1 z2w,3=1 z4=1 |
| | 2 | z2w,1=1 z2w,3=1 |
| | 2 | z2w,2=1 z4=3 |
| | 2 | z2w,1=1 |
| | 2 | z2w,1=1 z2w,2=1 z4=3 |
| | 2 | z2w,2=1 z2w,3=1 z4=3 |
| | 2 | z2w,3=1 z4=2 |
| 11.4.56 | 423907 | H1Al4B3O14U1 |
| | 2 | z2w,1=1 z2w,3=1 z4=3 |
| | 2 | z2w,1=1 z2w,2=1 z2w,3=1 z4=2 |
| | 2 | z2w,2=1 z4=2 |
| | 2 | z4=1 |
| | 2 | z2w,1=1 z2w,2=1 z4=2 |
| | 2 | z2w,1=1 z4=1 |
| | 2 | z2w,3=1 z4=1 |
| | 2 | z2w,2=1 z2w,3=1 |
| | 2 | z2w,2=1 z2w,3=1 z4=2 |
| | 2 | z2w,3=1 z4=1 |
| | 2 | z2w,1=1 z4=1 |
| | 2 | z2w,1=1 z2w,2=1 |
| | 2 | z4=1 |

Continued on next column

**TABLE XIX– continued**

| TQC | ICSD SSG | Stoich. Formula SIs |
|---|---|---|
| | 2 | z2w,2=1 |
| | 2 | z2w,1=1 z2w,2=1 z2w,3=1 |
| | 2 | z2w,1=1 z2w,3=1 z4=3 |
| 11.4.57 | 68016 | H18Br7N6Ru2Zn1 |
| | 2 | z2w,2=1 z4=3 |
| | 2 | z2w,1=1 z2w,2=1 z2w,3=1 z4=3 |
| | 2 | z2w,1=1 z2w,3=1 z4=2 |
| | 2 | z2w,2=1 z2w,3=1 z4=3 |
| | 2 | z2w,3=1 z4=2 |
| | 2 | z2w,1=1 z4=2 |
| | 2 | z2w,1=1 z2w,2=1 z4=1 |
| | 2 | z2w,1=1 z2w,2=1 z4=3 |
| | 2 | z2w,1=1 z4=2 |
| | 2 | z2w,3=1 z4=2 |
| | 2 | z2w,2=1 z2w,3=1 z4=1 |
| | 2 | z2w,1=1 z2w,3=1 z4=2 |
| | 2 | z2w,1=1 z2w,2=1 z2w,3=1 z4=1 |
| | 2 | z2w,2=1 z4=1 |
| 11.4.58 | 63297 | H18Br7N6Ru2Zn1 |
| | 2 | z2w,2=1 z4=3 |
| | 2 | z2w,1=1 z2w,2=1 z2w,3=1 z4=3 |
| | 2 | z2w,1=1 z2w,3=1 z4=2 |
| | 2 | z2w,2=1 z2w,3=1 z4=3 |
| | 2 | z2w,3=1 z4=2 |
| | 2 | z2w,1=1 z4=2 |
| | 2 | z2w,1=1 z2w,2=1 z4=1 |
| | 2 | z2w,1=1 z2w,2=1 z4=3 |
| | 2 | z2w,1=1 z4=2 |
| | 2 | z2w,3=1 z4=2 |
| | 2 | z2w,2=1 z2w,3=1 z4=1 |
| | 2 | z2w,1=1 z2w,3=1 z4=2 |
| | 2 | z2w,1=1 z2w,2=1 z2w,3=1 z4=1 |
| | 2 | z2w,2=1 z4=1 |
| 11.4.59 | 100691 | H11Cs4O18P3Te1 |
| | 2 | z2w,3=1 z4=1 |
| | 2 | z2w,2=1 z2w,3=1 |
| | 2 | z2w,1=1 z2w,2=1 |
| | 2 | z2w,1=1 z4=3 |
| | 2 | z2w,2=1 |
| | 2 | z4=3 |
| | 2 | z2w,1=1 z2w,3=1 z4=3 |
| | 2 | z2w,1=1 z2w,2=1 z2w,3=1 z4=2 |
| | 2 | z2w,1=1 z2w,2=1 z2w,3=1 |
| | 2 | z2w,1=1 z2w,3=1 z4=3 |
| | 2 | z4=3 |
| | 2 | z2w,2=1 z4=2 |
| | 2 | z2w,1=1 z4=3 |
| | 2 | z2w,1=1 z2w,2=1 z4=2 |
| | 2 | z2w,2=1 z2w,3=1 z4=2 |
| | 2 | z2w,3=1 z4=1 |
| 11.4.60 | 76293 | Ge6Pt4Y3 |
| | 2 | z2w,3=1 z4=1 |
| | 2 | z2w,2=1 z2w,3=1 |
| | 2 | z2w,1=1 z2w,2=1 |
| | 2 | z2w,1=1 z4=3 |
| | 2 | z2w,2=1 |
| | 2 | z4=3 |
| | 2 | z2w,1=1 z2w,3=1 z4=3 |
| | 2 | z2w,1=1 z2w,2=1 z2w,3=1 z4=2 |
| | 2 | z2w,1=1 z2w,2=1 z2w,3=1 |

Continued on next column

**TABLE XIX– continued**

| TQC | ICSD SSG | Stoich. Formula SIs |
|---|---|---|
| | 2 | z2w,1=1 z2w,3=1 z4=3 |
| | 2 | z4=3 |
| | 2 | z2w,2=1 z4=2 |
| | 2 | z2w,1=1 z4=3 |
| | 2 | z2w,1=1 z2w,2=1 z4=2 |
| | 2 | z2w,2=1 z2w,3=1 z4=2 |
| | 2 | z2w,3=1 z4=1 |
| 11.4.61 | 114 | Gd1P5 |
| | 2 | z2w,1=1 z2w,3=1 z4=2 |
| | 2 | z2w,1=1 z2w,2=1 z2w,3=1 z4=1 |
| | 2 | z2w,2=1 z4=1 |
| | 2 | z2w,1=1 z2w,2=1 z4=1 |
| | 2 | z2w,1=1 |
| | 2 | z2w,3=1 |
| | 2 | z2w,2=1 z2w,3=1 z4=3 |
| | 2 | z2w,2=1 z2w,3=1 z4=1 |
| | 2 | z2w,3=1 |
| | 2 | z2w,1=1 |
| | 2 | z2w,1=1 z2w,2=1 z4=3 |
| | 2 | z2w,2=1 z4=3 |
| | 2 | z2w,1=1 z2w,2=1 z2w,3=1 z4=3 |
| | 2 | z2w,1=1 z2w,3=1 z4=2 |
| 11.4.62 | 260894 | Gd1K3O8P2 |
| | 2 | z2w,1=1 z4=2 |
| | 2 | z2w,1=1 z2w,2=1 z4=1 |
| | 2 | z2w,2=1 z2w,3=1 z4=1 |
| | 2 | z2w,3=1 |
| | 2 | z2w,1=1 z2w,2=1 z2w,3=1 z4=1 |
| | 2 | z2w,1=1 z2w,3=1 |
| | 2 | z2w,2=1 z4=3 |
| | 2 | z2w,2=1 z4=1 |
| | 2 | z2w,1=1 z2w,3=1 |
| | 2 | z2w,1=1 z2w,2=1 z2w,3=1 z4=3 |
| | 2 | z2w,3=1 |
| | 2 | z2w,2=1 z2w,3=1 z4=3 |
| | 2 | z2w,1=1 z2w,2=1 z4=3 |
| | 2 | z2w,1=1 z4=2 |
| 11.4.63 | 66513 | Fe1Ta1Te3 |
| | 2 | z2w,3=1 z4=1 |
| | 2 | z2w,2=1 z2w,3=1 |
| | 2 | z2w,1=1 z2w,2=1 |
| | 2 | z2w,1=1 z4=3 |
| | 2 | z2w,2=1 |
| | 2 | z4=3 |
| | 2 | z2w,1=1 z2w,3=1 z4=3 |
| | 2 | z2w,1=1 z2w,2=1 z2w,3=1 z4=2 |
| | 2 | z2w,1=1 z2w,2=1 z2w,3=1 |
| | 2 | z2w,1=1 z2w,3=1 z4=3 |
| | 2 | z4=3 |
| | 2 | z2w,2=1 z4=2 |
| | 2 | z2w,1=1 z4=3 |
| | 2 | z2w,1=1 z2w,2=1 z4=2 |
| | 2 | z2w,2=1 z2w,3=1 z4=2 |
| | 2 | z2w,3=1 z4=1 |
| 11.4.64 | 166586 | F5K1N1O1Os1 |
| | 2 | z2w,2=1 z4=3 |
| | 2 | z2w,1=1 z2w,2=1 z2w,3=1 z4=3 |
| | 2 | z2w,1=1 z2w,3=1 z4=2 |
| | 2 | z2w,2=1 z2w,3=1 z4=3 |
| | 2 | z2w,3=1 z4=2 |

Continued on next column

**TABLE XIX– continued**

| TQC | ICSD SSG | Stoich. Formula SIs |
|---|---|---|
| | 2 | z2w,1=1 z4=2 |
| | 2 | z2w,1=1 z2w,2=1 z4=1 |
| | 2 | z2w,1=1 z2w,2=1 z4=3 |
| | 2 | z2w,1=1 z4=2 |
| | 2 | z2w,3=1 z4=2 |
| | 2 | z2w,2=1 z2w,3=1 z4=1 |
| | 2 | z2w,1=1 z2w,3=1 z4=2 |
| | 2 | z2w,1=1 z2w,2=1 z2w,3=1 z4=1 |
| | 2 | z2w,2=1 z4=1 |
| 11.4.65 | 424958 | F3Fe1Na2O4S1 |
| | 2 | z2w,3=1 z4=2 |
| | 2 | z2w,2=1 z2w,3=1 z4=1 |
| | 2 | z2w,1=1 z2w,2=1 z4=1 |
| | 2 | z2w,1=1 |
| | 2 | z2w,2=1 z4=1 |
| | 2 | z2w,1=1 z2w,3=1 |
| | 2 | z2w,1=1 z2w,2=1 z2w,3=1 z4=3 |
| | 2 | z2w,1=1 z2w,2=1 z2w,3=1 z4=1 |
| | 2 | z2w,1=1 z2w,3=1 |
| | 2 | z2w,2=1 z4=3 |
| | 2 | z2w,1=1 |
| | 2 | z2w,1=1 z2w,2=1 z4=3 |
| | 2 | z2w,2=1 z2w,3=1 z4=3 |
| | 2 | z2w,3=1 z4=2 |
| 11.4.66 | 10081 | Eu1Sb2 |
| | 2 | z2w,1=1 z2w,3=1 |
| | 2 | z2w,1=1 z2w,2=1 z2w,3=1 z4=3 |
| | 2 | z2w,2=1 z4=3 |
| | 2 | z4=2 |
| | 2 | z2w,1=1 z2w,2=1 z4=3 |
| | 2 | z2w,1=1 z4=2 |
| | 2 | z2w,3=1 z4=2 |
| | 2 | z2w,2=1 z2w,3=1 z4=1 |
| | 2 | z2w,2=1 z2w,3=1 z4=3 |
| | 2 | z2w,3=1 z4=2 |
| | 2 | z2w,1=1 z4=2 |
| | 2 | z2w,1=1 z2w,2=1 z4=1 |
| | 2 | z4=2 |
| | 2 | z2w,2=1 z4=1 |
| | 2 | z2w,1=1 z2w,2=1 z2w,3=1 z4=1 |
| | 2 | z2w,1=1 z2w,3=1 |
| 11.4.67 | 280200 | Eu1K1P1Se4 |
| | 2 | z2w,1=1 z4=1 |
| | 2 | z2w,1=1 z2w,2=1 |
| | 2 | z2w,2=1 z2w,3=1 |
| | 2 | z2w,3=1 z4=3 |
| | 2 | z2w,1=1 z2w,2=1 z2w,3=1 |
| | 2 | z2w,1=1 z2w,3=1 z4=3 |
| | 2 | z4=3 |
| | 2 | z2w,2=1 z4=2 |
| | 2 | z2w,2=1 |
| | 2 | z4=3 |
| | 2 | z2w,1=1 z2w,3=1 z4=3 |
| | 2 | z2w,1=1 z2w,2=1 z2w,3=1 z4=2 |
| | 2 | z2w,3=1 z4=3 |
| | 2 | z2w,2=1 z2w,3=1 z4=2 |
| | 2 | z2w,1=1 z2w,2=1 z4=2 |
| | 2 | z2w,1=1 z4=1 |
| 11.4.68 | 90775 | Er1Na1O8S2 |
| | 2 | z2w,1=1 z2w,3=1 z4=2 |

Continued on next column

**TABLE XIX– continued**

| TQC | ICSD SSG | Stoich. Formula SIs |
|---|---|---|
| | 2 | z2w,1=1 z2w,2=1 z2w,3=1 z4=1 |
| | 2 | z2w,2=1 z4=1 |
| | 2 | z2w,1=1 z2w,2=1 z4=1 |
| | 2 | z2w,1=1 |
| | 2 | z2w,3=1 |
| | 2 | z2w,2=1 z2w,3=1 z4=3 |
| | 2 | z2w,2=1 z2w,3=1 z4=1 |
| | 2 | z2w,3=1 |
| | 2 | z2w,1=1 |
| | 2 | z2w,1=1 z2w,2=1 z4=3 |
| | 2 | z2w,2=1 z4=3 |
| | 2 | z2w,1=1 z2w,2=1 z2w,3=1 z4=3 |
| | 2 | z2w,1=1 z2w,3=1 z4=2 |
| 11.4.69 | 33963 | D1Rb1S1 |
| | 2 | z2w,3=1 z4=2 |
| | 2 | z2w,2=1 z2w,3=1 z4=1 |
| | 2 | z2w,1=1 z2w,2=1 z4=1 |
| | 2 | z2w,1=1 |
| | 2 | z2w,2=1 z4=1 |
| | 2 | z2w,1=1 z2w,3=1 |
| | 2 | z2w,1=1 z2w,2=1 z2w,3=1 z4=3 |
| | 2 | z2w,1=1 z2w,2=1 z2w,3=1 z4=1 |
| | 2 | z2w,1=1 z2w,3=1 |
| | 2 | z2w,2=1 z4=3 |
| | 2 | z2w,1=1 |
| | 2 | z2w,1=1 z2w,2=1 z4=3 |
| | 2 | z2w,2=1 z2w,3=1 z4=3 |
| | 2 | z2w,3=1 z4=2 |
| 11.4.70 | 426627 | Cu3Ge1O10P2 |
| | 2 | z2w,1=1 z2w,3=1 z4=3 |
| | 2 | z2w,1=1 z2w,2=1 z2w,3=1 z4=2 |
| | 2 | z2w,2=1 z4=2 |
| | 2 | z4=1 |
| | 2 | z2w,1=1 z2w,2=1 z4=2 |
| | 2 | z2w,1=1 z4=1 |
| | 2 | z2w,3=1 z4=1 |
| | 2 | z2w,2=1 z2w,3=1 |
| | 2 | z2w,2=1 z2w,3=1 z4=2 |
| | 2 | z2w,3=1 z4=1 |
| | 2 | z2w,1=1 z4=1 |
| | 2 | z2w,1=1 z2w,2=1 |
| | 2 | z4=1 |
| | 2 | z2w,2=1 |
| | 2 | z2w,1=1 z2w,2=1 z2w,3=1 |
| | 2 | z2w,1=1 z2w,3=1 z4=3 |
| 11.4.71 | 167596 | Cu1Zr1 |
| | 2 | z2w,3=1 z4=3 |
| | 2 | z2w,2=1 z2w,3=1 z4=2 |
| | 2 | z2w,1=1 z2w,2=1 z4=2 |
| | 2 | z2w,1=1 z4=1 |
| | 2 | z2w,2=1 z4=2 |
| | 2 | z4=1 |
| | 2 | z2w,1=1 z2w,3=1 z4=1 |
| | 2 | z2w,1=1 z2w,2=1 z2w,3=1 |
| | 2 | z2w,1=1 z2w,2=1 z2w,3=1 z4=2 |
| | 2 | z2w,1=1 z2w,3=1 z4=1 |
| | 2 | z4=1 |
| | 2 | z2w,2=1 |
| | 2 | z2w,1=1 z4=1 |
| | 2 | z2w,1=1 z2w,2=1 |

Continued on next column

TABLE XIX– continued

| TQC | ICSD SSG | Stoich. Formula SIs |
|---|---|---|
| | 2 | z2w,2=1 z2w,3=1 |
| | 2 | z2w,3=1 z4=3 |
| 11.4.72 | 422033 | Cu1N4O10 |
| | 2 | z2w,1=1 z4=2 |
| | 2 | z2w,1=1 z2w,2=1 z4=1 |
| | 2 | z2w,2=1 z2w,3=1 z4=1 |
| | 2 | z2w,3=1 |
| | 2 | z2w,1=1 z2w,2=1 z2w,3=1 z4=1 |
| | 2 | z2w,1=1 z2w,3=1 |
| | 2 | z2w,2=1 z4=3 |
| | 2 | z2w,2=1 z4=1 |
| | 2 | z2w,1=1 z2w,3=1 |
| | 2 | z2w,1=1 z2w,2=1 z2w,3=1 z4=3 |
| | 2 | z2w,3=1 |
| | 2 | z2w,2=1 z2w,3=1 z4=3 |
| | 2 | z2w,1=1 z2w,2=1 z4=3 |
| | 2 | z2w,1=1 z4=2 |
| 11.4.73 | 2232 | Cu1I2O6 |
| | 2 | z2w,1=1 z2w,3=1 z4=2 |
| | 2 | z2w,1=1 z2w,2=1 z2w,3=1 z4=1 |
| | 2 | z2w,2=1 z4=1 |
| | 2 | z2w,1=1 z2w,2=1 z4=1 |
| | 2 | z2w,1=1 |
| | 2 | z2w,3=1 |
| | 2 | z2w,2=1 z2w,3=1 z4=3 |
| | 2 | z2w,2=1 z2w,3=1 z4=1 |
| | 2 | z2w,3=1 |
| | 2 | z2w,1=1 |
| | 2 | z2w,1=1 z2w,2=1 z4=3 |
| | 2 | z2w,2=1 z4=3 |
| | 2 | z2w,1=1 z2w,2=1 z2w,3=1 z4=3 |
| | 2 | z2w,1=1 z2w,3=1 z4=2 |
| 11.4.74 | 60759 | Cu1F2O3W1 |
| | 2 | z2w,1=1 z4=2 |
| | 2 | z2w,1=1 z2w,2=1 z4=1 |
| | 2 | z2w,2=1 z2w,3=1 z4=1 |
| | 2 | z2w,3=1 |
| | 2 | z2w,1=1 z2w,2=1 z2w,3=1 z4=1 |
| | 2 | z2w,1=1 z2w,3=1 |
| | 2 | z2w,2=1 z4=3 |
| | 2 | z2w,2=1 z4=1 |
| | 2 | z2w,1=1 z2w,3=1 |
| | 2 | z2w,1=1 z2w,2=1 z2w,3=1 z4=3 |
| | 2 | z2w,3=1 |
| | 2 | z2w,2=1 z2w,3=1 z4=3 |
| | 2 | z2w,1=1 z2w,2=1 z4=3 |
| | 2 | z2w,1=1 z4=2 |
| 11.4.75 | 182754 | Cs2Cu3F12Zr1 |
| | 2 | z2w,2=1 z4=3 |
| | 2 | z2w,1=1 z2w,2=1 z2w,3=1 z4=3 |
| | 2 | z2w,1=1 z2w,3=1 z4=2 |
| | 2 | z2w,2=1 z2w,3=1 z4=3 |
| | 2 | z2w,3=1 z4=2 |
| | 2 | z2w,1=1 z4=2 |
| | 2 | z2w,1=1 z2w,2=1 z4=1 |
| | 2 | z2w,1=1 z2w,2=1 z4=3 |
| | 2 | z2w,1=1 z4=2 |
| | 2 | z2w,3=1 z4=2 |
| | 2 | z2w,2=1 z2w,3=1 z4=1 |
| | 2 | z2w,1=1 z2w,3=1 z4=2 |

Continued on next column

TABLE XIX– continued

| TQC | ICSD SSG | Stoich. Formula SIs |
|---|---|---|
| | 2 | z2w,1=1 z2w,2=1 z2w,3=1 z4=1 |
| | 2 | z2w,2=1 z4=1 |
| 11.4.76 | 166587 | Cs1F5N1O1Os1 |
| | 2 | z2w,2=1 z4=3 |
| | 2 | z2w,1=1 z2w,2=1 z2w,3=1 z4=3 |
| | 2 | z2w,1=1 z2w,3=1 z4=2 |
| | 2 | z2w,2=1 z2w,3=1 z4=3 |
| | 2 | z2w,3=1 z4=2 |
| | 2 | z2w,1=1 z4=2 |
| | 2 | z2w,1=1 z2w,2=1 z4=1 |
| | 2 | z2w,1=1 z2w,2=1 z4=3 |
| | 2 | z2w,1=1 z4=2 |
| | 2 | z2w,3=1 z4=2 |
| | 2 | z2w,2=1 z2w,3=1 z4=1 |
| | 2 | z2w,1=1 z2w,3=1 z4=2 |
| | 2 | z2w,1=1 z2w,2=1 z2w,3=1 z4=1 |
| | 2 | z2w,2=1 z4=1 |
| 11.4.77 | 409488 | Cr5P3Zr1 |
| | 2 | z2w,1=1 z2w,3=1 z4=2 |
| | 2 | z2w,1=1 z2w,2=1 z2w,3=1 z4=1 |
| | 2 | z2w,2=1 z4=1 |
| | 2 | z2w,1=1 z2w,2=1 z4=1 |
| | 2 | z2w,1=1 |
| | 2 | z2w,3=1 |
| | 2 | z2w,2=1 z2w,3=1 z4=3 |
| | 2 | z2w,2=1 z2w,3=1 z4=1 |
| | 2 | z2w,3=1 |
| | 2 | z2w,1=1 |
| | 2 | z2w,1=1 z2w,2=1 z4=3 |
| | 2 | z2w,2=1 z4=3 |
| | 2 | z2w,1=1 z2w,2=1 z2w,3=1 z4=3 |
| | 2 | z2w,1=1 z2w,3=1 z4=2 |
| 11.4.78 | 35044 | Cr2Nb2Se10 |
| | 2 | z2w,1=1 z4=2 |
| | 2 | z2w,1=1 z2w,2=1 z4=1 |
| | 2 | z2w,2=1 z2w,3=1 z4=1 |
| | 2 | z2w,3=1 |
| | 2 | z2w,1=1 z2w,2=1 z2w,3=1 z4=1 |
| | 2 | z2w,1=1 z2w,3=1 |
| | 2 | z2w,2=1 z4=3 |
| | 2 | z2w,2=1 z4=1 |
| | 2 | z2w,1=1 z2w,3=1 |
| | 2 | z2w,1=1 z2w,2=1 z2w,3=1 z4=3 |
| | 2 | z2w,3=1 |
| | 2 | z2w,2=1 z2w,3=1 z4=3 |
| | 2 | z2w,1=1 z2w,2=1 z4=3 |
| | 2 | z2w,1=1 z4=2 |
| 11.4.79 | 1029 | Co1I4Rb2 |
| | 2 | z2w,1=1 z2w,3=1 z4=2 |
| | 2 | z2w,1=1 z2w,2=1 z2w,3=1 z4=1 |
| | 2 | z2w,2=1 z4=1 |
| | 2 | z2w,1=1 z2w,2=1 z4=1 |
| | 2 | z2w,1=1 |
| | 2 | z2w,3=1 |
| | 2 | z2w,2=1 z2w,3=1 z4=3 |
| | 2 | z2w,2=1 z2w,3=1 z4=1 |
| | 2 | z2w,3=1 |
| | 2 | z2w,1=1 |
| | 2 | z2w,1=1 z2w,2=1 z4=3 |
| | 2 | z2w,2=1 z4=3 |

Continued on next column

**TABLE XIX– continued**

| TQC | ICSD<br>SSG | Stoich. Formula<br>SIs |
|---|---|---|
| | 2 | z2w,1=1 z2w,2=1 z2w,3=1 z4=3 |
| | 2 | z2w,1=1 z2w,3=1 z4=2 |
| 11.4.80 | 99178 | Ce1P5 |
| | 2 | z2w,2=1 z4=3 |
| | 2 | z2w,1=1 z2w,2=1 z2w,3=1 z4=3 |
| | 2 | z2w,1=1 z2w,3=1 z4=2 |
| | 2 | z2w,2=1 z2w,3=1 z4=3 |
| | 2 | z2w,3=1 z4=2 |
| | 2 | z2w,1=1 z4=2 |
| | 2 | z2w,1=1 z2w,2=1 z4=1 |
| | 2 | z2w,1=1 z2w,2=1 z4=3 |
| | 2 | z2w,1=1 z4=2 |
| | 2 | z2w,3=1 z4=2 |
| | 2 | z2w,2=1 z2w,3=1 z4=1 |
| | 2 | z2w,1=1 z2w,3=1 z4=2 |
| | 2 | z2w,1=1 z2w,2=1 z2w,3=1 z4=1 |
| | 2 | z2w,2=1 z4=1 |
| 11.4.81 | 409181 | Ce1P5 |
| | 2 | z2w,3=1 z4=2 |
| | 2 | z2w,2=1 z2w,3=1 z4=1 |
| | 2 | z2w,1=1 z2w,2=1 z4=1 |
| | 2 | z2w,1=1 |
| | 2 | z2w,2=1 z4=1 |
| | 2 | z2w,1=1 z2w,3=1 |
| | 2 | z2w,1=1 z2w,2=1 z2w,3=1 z4=3 |
| | 2 | z2w,1=1 z2w,2=1 z2w,3=1 z4=1 |
| | 2 | z2w,1=1 z2w,3=1 |
| | 2 | z2w,2=1 z4=3 |
| | 2 | z2w,1=1 |
| | 2 | z2w,1=1 z2w,2=1 z4=3 |
| | 2 | z2w,2=1 z2w,3=1 z4=3 |
| | 2 | z2w,3=1 z4=2 |
| 11.4.82 | 413847 | Ca4Ir8P7 |
| | 2 | z2w,3=1 z4=3 |
| | 2 | z2w,2=1 z2w,3=1 z4=2 |
| | 2 | z2w,1=1 z2w,2=1 z4=2 |
| | 2 | z2w,1=1 z4=1 |
| | 2 | z2w,2=1 z4=2 |
| | 2 | z4=1 |
| | 2 | z2w,1=1 z2w,3=1 z4=1 |
| | 2 | z2w,1=1 z2w,2=1 z2w,3=1 |
| | 2 | z2w,1=1 z2w,2=1 z2w,3=1 z4=2 |
| | 2 | z2w,1=1 z2w,3=1 z4=1 |
| | 2 | z4=1 |
| | 2 | z2w,2=1 |
| | 2 | z2w,1=1 z4=1 |
| | 2 | z2w,1=1 z2w,2=1 |
| | 2 | z2w,2=1 z2w,3=1 |
| | 2 | z2w,3=1 z4=3 |
| 11.4.83 | 183583 | Ca1O3Rh1 |
| | 2 | z2w,1=1 z4=1 |
| | 2 | z2w,1=1 z2w,2=1 |
| | 2 | z2w,2=1 z2w,3=1 |
| | 2 | z2w,3=1 z4=3 |
| | 2 | z2w,1=1 z2w,2=1 z2w,3=1 |
| | 2 | z2w,1=1 z2w,3=1 z4=3 |
| | 2 | z4=3 |
| | 2 | z2w,2=1 z4=2 |
| | 2 | z2w,2=1 |
| | 2 | z4=3 |

Continued on next column

**TABLE XIX– continued**

| TQC | ICSD<br>SSG | Stoich. Formula<br>SIs |
|---|---|---|
| | 2 | z2w,1=1 z2w,3=1 z4=3 |
| | 2 | z2w,1=1 z2w,2=1 z2w,3=1 z4=2 |
| | 2 | z2w,3=1 z4=3 |
| | 2 | z2w,2=1 z2w,3=1 z4=2 |
| | 2 | z2w,1=1 z2w,2=1 z4=2 |
| | 2 | z2w,1=1 z4=1 |
| 11.4.84 | 58880 | Ca1Cu1 |
| | 2 | z2w,3=1 z4=3 |
| | 2 | z2w,2=1 z2w,3=1 z4=2 |
| | 2 | z2w,1=1 z2w,2=1 z4=2 |
| | 2 | z2w,1=1 z4=1 |
| | 2 | z2w,2=1 z4=2 |
| | 2 | z4=1 |
| | 2 | z2w,1=1 z2w,3=1 z4=1 |
| | 2 | z2w,1=1 z2w,2=1 z2w,3=1 |
| | 2 | z2w,1=1 z2w,2=1 z2w,3=1 z4=2 |
| | 2 | z2w,1=1 z2w,3=1 z4=1 |
| | 2 | z4=1 |
| | 2 | z2w,2=1 |
| | 2 | z2w,1=1 z4=1 |
| | 2 | z2w,1=1 z2w,2=1 |
| | 2 | z2w,2=1 z2w,3=1 |
| | 2 | z2w,3=1 z4=3 |
| 11.4.85 | 248211 | C6H7N1O14U2 |
| | 2 | z2w,3=1 z4=2 |
| | 2 | z2w,2=1 z2w,3=1 z4=1 |
| | 2 | z2w,1=1 z2w,2=1 z4=1 |
| | 2 | z2w,1=1 |
| | 2 | z2w,2=1 z4=1 |
| | 2 | z2w,1=1 z2w,3=1 |
| | 2 | z2w,1=1 z2w,2=1 z2w,3=1 z4=3 |
| | 2 | z2w,1=1 z2w,2=1 z2w,3=1 z4=1 |
| | 2 | z2w,1=1 z2w,3=1 |
| | 2 | z2w,2=1 z4=3 |
| | 2 | z2w,1=1 |
| | 2 | z2w,1=1 z2w,2=1 z4=3 |
| | 2 | z2w,2=1 z2w,3=1 z4=3 |
| | 2 | z2w,3=1 z4=2 |
| 11.4.86 | 1138 | C4H10Cs1O13Ti1 |
| | 2 | z2w,1=1 z2w,3=1 z4=2 |
| | 2 | z2w,1=1 z2w,2=1 z2w,3=1 z4=1 |
| | 2 | z2w,2=1 z4=1 |
| | 2 | z2w,1=1 z2w,2=1 z4=1 |
| | 2 | z2w,1=1 |
| | 2 | z2w,3=1 |
| | 2 | z2w,2=1 z2w,3=1 z4=3 |
| | 2 | z2w,2=1 z2w,3=1 z4=1 |
| | 2 | z2w,3=1 |
| | 2 | z2w,1=1 |
| | 2 | z2w,1=1 z2w,2=1 z4=3 |
| | 2 | z2w,2=1 z4=3 |
| | 2 | z2w,1=1 z2w,2=1 z2w,3=1 z4=3 |
| | 2 | z2w,1=1 z2w,3=1 z4=2 |
| 11.4.87 | 237796 | C1Cu1Na3O7P1 |
| | 2 | z2w,1=1 z2w,3=1 z4=3 |
| | 2 | z2w,1=1 z2w,2=1 z2w,3=1 z4=2 |
| | 2 | z2w,2=1 z4=2 |
| | 2 | z4=1 |
| | 2 | z2w,1=1 z2w,2=1 z4=2 |
| | 2 | z2w,1=1 z4=1 |

Continued on next column

TABLE XIX– continued

| TQC | ICSD SSG | Stoich. Formula SIs |
|---|---|---|
| | 2 | z2w,3=1 z4=1 |
| | 2 | z2w,2=1 z2w,3=1 |
| | 2 | z2w,2=1 z2w,3=1 z4=2 |
| | 2 | z2w,3=1 z4=1 |
| | 2 | z2w,1=1 z4=1 |
| | 2 | z2w,1=1 z2w,2=1 |
| | 2 | z4=1 |
| | 2 | z2w,2=1 |
| | 2 | z2w,1=1 z2w,2=1 z2w,3=1 |
| | 2 | z2w,1=1 z2w,3=1 z4=3 |
| 11.4.88 | 237797 | C1Co1Na3O7P1 |
| | 2 | z2w,1=1 z4=3 |
| | 2 | z2w,1=1 z2w,2=1 z4=2 |
| | 2 | z2w,2=1 z2w,3=1 z4=2 |
| | 2 | z2w,3=1 z4=1 |
| | 2 | z2w,1=1 z2w,2=1 z2w,3=1 z4=2 |
| | 2 | z2w,1=1 z2w,3=1 z4=1 |
| | 2 | z4=1 |
| | 2 | z2w,2=1 |
| | 2 | z2w,2=1 z4=2 |
| | 2 | z4=1 |
| | 2 | z2w,1=1 z2w,3=1 z4=1 |
| | 2 | z2w,1=1 z2w,2=1 z2w,3=1 |
| | 2 | z2w,3=1 z4=1 |
| | 2 | z2w,2=1 z2w,3=1 |
| | 2 | z2w,1=1 z2w,2=1 |
| | 2 | z2w,1=1 z4=3 |
| 11.4.89 | 20270 | C1Br1Hg1N1S1 |
| | 2 | z2w,1=1 z4=2 |
| | 2 | z2w,1=1 z2w,2=1 z4=1 |
| | 2 | z2w,2=1 z2w,3=1 z4=1 |
| | 2 | z2w,3=1 |
| | 2 | z2w,1=1 z2w,2=1 z2w,3=1 z4=1 |
| | 2 | z2w,1=1 z2w,3=1 |
| | 2 | z2w,2=1 z4=3 |
| | 2 | z2w,2=1 z4=1 |
| | 2 | z2w,1=1 z2w,3=1 |
| | 2 | z2w,1=1 z2w,2=1 z2w,3=1 z4=3 |
| | 2 | z2w,3=1 |
| | 2 | z2w,2=1 z2w,3=1 z4=3 |
| | 2 | z2w,1=1 z2w,2=1 z4=3 |
| | 2 | z2w,1=1 z4=2 |
| 11.4.90 | 151019 | C10H21I3N4 |
| | 2 | z2w,2=1 z4=3 |
| | 2 | z2w,1=1 z2w,2=1 z2w,3=1 z4=3 |
| | 2 | z2w,1=1 z2w,3=1 z4=2 |
| | 2 | z2w,2=1 z2w,3=1 z4=3 |
| | 2 | z2w,3=1 z4=2 |
| | 2 | z2w,1=1 z4=2 |
| | 2 | z2w,1=1 z2w,2=1 z4=1 |
| | 2 | z2w,1=1 z2w,2=1 z4=3 |
| | 2 | z2w,1=1 z4=2 |
| | 2 | z2w,3=1 z4=2 |
| | 2 | z2w,2=1 z2w,3=1 z4=1 |
| | 2 | z2w,1=1 z2w,3=1 z4=2 |
| | 2 | z2w,1=1 z2w,2=1 z2w,3=1 z4=1 |
| | 2 | z2w,2=1 z4=1 |
| 11.4.91 | 410948 | Br9Te1W1 |
| | 2 | z2w,1=1 z4=1 |
| | 2 | z2w,1=1 z2w,2=1 |

Continued on next column

TABLE XIX– continued

| TQC | ICSD SSG | Stoich. Formula SIs |
|---|---|---|
| | 2 | z2w,2=1 z2w,3=1 |
| | 2 | z2w,3=1 z4=3 |
| | 2 | z2w,1=1 z2w,2=1 z2w,3=1 |
| | 2 | z2w,1=1 z2w,3=1 z4=3 |
| | 2 | z4=3 |
| | 2 | z2w,2=1 z4=2 |
| | 2 | z2w,2=1 |
| | 2 | z4=3 |
| | 2 | z2w,1=1 z2w,3=1 z4=3 |
| | 2 | z2w,1=1 z2w,2=1 z2w,3=1 z4=2 |
| | 2 | z2w,3=1 z4=3 |
| | 2 | z2w,2=1 z2w,3=1 z4=2 |
| | 2 | z2w,1=1 z2w,2=1 z4=2 |
| | 2 | z2w,1=1 z4=1 |
| 11.4.92 | 39257 | Bi9I2 |
| | 2 | z2w,3=1 z4=3 |
| | 2 | z2w,2=1 z2w,3=1 z4=2 |
| | 2 | z2w,1=1 z2w,2=1 z4=2 |
| | 2 | z2w,1=1 z4=1 |
| | 2 | z2w,2=1 z4=2 |
| | 2 | z4=1 |
| | 2 | z2w,1=1 z2w,3=1 z4=1 |
| | 2 | z2w,1=1 z2w,2=1 z2w,3=1 |
| | 2 | z2w,1=1 z2w,2=1 z2w,3=1 z4=2 |
| | 2 | z2w,1=1 z2w,3=1 z4=1 |
| | 2 | z4=1 |
| | 2 | z2w,2=1 |
| | 2 | z2w,1=1 z4=1 |
| | 2 | z2w,1=1 z2w,2=1 |
| | 2 | z2w,2=1 z2w,3=1 |
| | 2 | z2w,3=1 z4=3 |
| 11.4.93 | 162834 | Bi1Fe1O3 |
| | 2 | z2w,1=1 z2w,3=1 z4=1 |
| | 2 | z2w,1=1 z2w,2=1 z2w,3=1 |
| | 2 | z2w,2=1 |
| | 2 | z4=3 |
| | 2 | z2w,1=1 z2w,2=1 |
| | 2 | z2w,1=1 z4=3 |
| | 2 | z2w,3=1 z4=3 |
| | 2 | z2w,2=1 z2w,3=1 z4=2 |
| | 2 | z2w,2=1 z2w,3=1 |
| | 2 | z2w,3=1 z4=3 |
| | 2 | z2w,1=1 z4=3 |
| | 2 | z2w,1=1 z2w,2=1 z4=2 |
| | 2 | z4=3 |
| | 2 | z2w,2=1 z4=2 |
| | 2 | z2w,1=1 z2w,2=1 z2w,3=1 z4=2 |
| | 2 | z2w,1=1 z2w,3=1 z4=1 |
| 11.4.94 | 79480 | Ba2Cl1F7Mn1Ni1 |
| | 2 | z2w,3=1 z4=2 |
| | 2 | z2w,2=1 z2w,3=1 z4=1 |
| | 2 | z2w,1=1 z2w,2=1 z4=1 |
| | 2 | z2w,1=1 |
| | 2 | z2w,2=1 z4=1 |
| | 2 | z2w,1=1 z2w,3=1 |
| | 2 | z2w,1=1 z2w,2=1 z2w,3=1 z4=3 |
| | 2 | z2w,1=1 z2w,2=1 z2w,3=1 z4=1 |
| | 2 | z2w,1=1 z2w,3=1 |
| | 2 | z2w,2=1 z4=3 |
| | 2 | z2w,1=1 |

Continued on next column

**TABLE XIX– continued**

| TQC | ICSD SSG | Stoich. Formula SIs |
|---|---|---|
| | 2 | z2w,1=1 z2w,2=1 z4=3 |
| | 2 | z2w,2=1 z2w,3=1 z4=3 |
| | 2 | z2w,3=1 z4=2 |
| 11.4.95 | 82063 | Ba1O8V3 |
| | 2 | z2w,1=1 z2w,3=1 z4=1 |
| | 2 | z2w,1=1 z2w,2=1 z2w,3=1 |
| | 2 | z2w,2=1 |
| | 2 | z4=3 |
| | 2 | z2w,1=1 z2w,2=1 |
| | 2 | z2w,1=1 z4=3 |
| | 2 | z2w,3=1 z4=3 |
| | 2 | z2w,2=1 z2w,3=1 z4=2 |
| | 2 | z2w,2=1 z2w,3=1 |
| | 2 | z2w,3=1 z4=3 |
| | 2 | z2w,1=1 z4=3 |
| | 2 | z2w,1=1 z2w,2=1 z4=2 |
| | 2 | z4=3 |
| | 2 | z2w,2=1 z4=2 |
| | 2 | z2w,1=1 z2w,2=1 z2w,3=1 z4=2 |
| | 2 | z2w,1=1 z2w,3=1 z4=1 |
| 11.4.96 | 65932 | B3Er1Mo1 |
| | 2 | z2w,1=1 z2w,3=1 z4=1 |
| | 2 | z2w,1=1 z2w,2=1 z2w,3=1 |
| | 2 | z2w,2=1 |
| | 2 | z4=3 |
| | 2 | z2w,1=1 z2w,2=1 |
| | 2 | z2w,1=1 z4=3 |
| | 2 | z2w,3=1 z4=3 |
| | 2 | z2w,2=1 z2w,3=1 z4=2 |
| | 2 | z2w,2=1 z2w,3=1 |
| | 2 | z2w,3=1 z4=3 |
| | 2 | z2w,1=1 z4=3 |
| | 2 | z2w,1=1 z2w,2=1 z4=2 |
| | 2 | z4=3 |
| | 2 | z2w,2=1 z4=2 |
| | 2 | z2w,1=1 z2w,2=1 z2w,3=1 z4=2 |
| | 2 | z2w,1=1 z2w,3=1 z4=1 |
| 11.4.97 | 65934 | B2Ba2Cl1O6Yb1 |
| | 2 | z2w,1=1 z2w,3=1 z4=1 |
| | 2 | z2w,1=1 z2w,2=1 z2w,3=1 |
| | 2 | z2w,2=1 |
| | 2 | z4=3 |
| | 2 | z2w,1=1 z2w,2=1 |
| | 2 | z2w,1=1 z4=3 |
| | 2 | z2w,3=1 z4=3 |
| | 2 | z2w,2=1 z2w,3=1 z4=2 |
| | 2 | z2w,2=1 z2w,3=1 |
| | 2 | z2w,3=1 z4=3 |
| | 2 | z2w,1=1 z4=3 |
| | 2 | z2w,1=1 z2w,2=1 z4=2 |
| | 2 | z4=3 |
| | 2 | z2w,2=1 z4=2 |
| | 2 | z2w,1=1 z2w,2=1 z2w,3=1 z4=2 |
| | 2 | z2w,1=1 z2w,3=1 z4=1 |
| 11.4.98 | 262073 | B2Ba2Cl1Gd1O6 |
| | 2 | z2w,1=1 z2w,3=1 z4=2 |
| | 2 | z2w,1=1 z2w,2=1 z2w,3=1 z4=1 |
| | 2 | z2w,2=1 z4=1 |
| | 2 | z2w,1=1 z2w,2=1 z4=1 |
| | 2 | z2w,1=1 |

Continued on next column

**TABLE XIX– continued**

| TQC | ICSD SSG | Stoich. Formula SIs |
|---|---|---|
| | 2 | z2w,3=1 |
| | 2 | z2w,2=1 z2w,3=1 z4=3 |
| | 2 | z2w,2=1 z2w,3=1 z4=1 |
| | 2 | z2w,3=1 |
| | 2 | z2w,1=1 |
| | 2 | z2w,1=1 z2w,2=1 z4=3 |
| | 2 | z2w,2=1 z4=3 |
| | 2 | z2w,1=1 z2w,2=1 z2w,3=1 z4=3 |
| | 2 | z2w,1=1 z2w,3=1 z4=2 |
| 11.4.99 | 413846 | As7Ca4Ir8 |
| | 2 | z4=1 |
| | 2 | z2w,2=1 |
| | 2 | z2w,1=1 z2w,2=1 z2w,3=1 |
| | 2 | z2w,1=1 z2w,3=1 z4=3 |
| | 2 | z2w,2=1 z2w,3=1 |
| | 2 | z2w,3=1 z4=3 |
| | 2 | z2w,1=1 z4=3 |
| | 2 | z2w,1=1 z2w,2=1 z4=2 |
| | 2 | z2w,1=1 z2w,2=1 |
| | 2 | z2w,1=1 z4=3 |
| | 2 | z2w,3=1 z4=3 |
| | 2 | z2w,2=1 z2w,3=1 z4=2 |
| | 2 | z2w,1=1 z2w,3=1 z4=3 |
| | 2 | z2w,1=1 z2w,2=1 z2w,3=1 z4=2 |
| | 2 | z2w,2=1 z4=2 |
| | 2 | z4=1 |
| 11.4.100 | 427443 | As3Ca1Fe5 |
| | 2 | z2w,3=1 z4=3 |
| | 2 | z2w,2=1 z2w,3=1 z4=2 |
| | 2 | z2w,1=1 z2w,2=1 z4=2 |
| | 2 | z2w,1=1 z4=1 |
| | 2 | z2w,2=1 z4=2 |
| | 2 | z4=1 |
| | 2 | z2w,1=1 z2w,3=1 z4=1 |
| | 2 | z2w,1=1 z2w,2=1 z2w,3=1 |
| | 2 | z2w,1=1 z2w,2=1 z2w,3=1 z4=2 |
| | 2 | z2w,1=1 z2w,3=1 z4=1 |
| | 2 | z4=1 |
| | 2 | z2w,2=1 |
| | 2 | z2w,1=1 z4=1 |
| | 2 | z2w,1=1 z2w,2=1 |
| | 2 | z2w,2=1 z2w,3=1 |
| | 2 | z2w,3=1 z4=3 |
| 11.4.101 | 420736 | As1Eu1K1S4 |
| | 2 | z4=3 |
| | 2 | z2w,2=1 z4=2 |
| | 2 | z2w,1=1 z2w,2=1 z2w,3=1 z4=2 |
| | 2 | z2w,1=1 z2w,3=1 z4=1 |
| | 2 | z2w,2=1 z2w,3=1 z4=2 |
| | 2 | z2w,3=1 z4=1 |
| | 2 | z2w,1=1 z4=1 |
| | 2 | z2w,1=1 z2w,2=1 |
| | 2 | z2w,1=1 z2w,2=1 z4=2 |
| | 2 | z2w,1=1 z4=1 |
| | 2 | z2w,3=1 z4=1 |
| | 2 | z2w,2=1 z2w,3=1 |
| | 2 | z2w,1=1 z2w,3=1 z4=1 |
| | 2 | z2w,1=1 z2w,2=1 z2w,3=1 |
| | 2 | z2w,2=1 |
| | 2 | z4=3 |

Continued on next column

**TABLE XIX– continued**

| TQC | ICSD<br>SSG | Stoich. Formula<br>SIs |
|---|---|---|
| 13.4.1 | 415460 | O4Ta1Yb1 |
| | 2 | z2w,1=1 z4=1 |
| | 2 | z2w,2=1 z2w,3=1 |
| | 2 | z2w,1=1 z2w,2=1 z2w,3=1 |
| | 2 | z4=3 |
| | 2 | z2w,1=1 z2w,3=1 |
| | 2 | z2w,2=1 z4=3 |
| | 2 | z2w,1=1 z2w,2=1 z4=3 |
| | 2 | z2w,3=1 z4=2 |
| | 2 | z2w,3=1 |
| | 2 | z2w,1=1 z2w,2=1 z4=3 |
| | 2 | z2w,2=1 z4=3 |
| | 2 | z2w,1=1 z2w,3=1 z4=2 |
| | 2 | z4=3 |
| | 2 | z2w,1=1 z2w,2=1 z2w,3=1 z4=2 |
| | 2 | z2w,2=1 z2w,3=1 z4=2 |
| | 2 | z2w,1=1 z4=1 |
| 13.4.2 | 261838 | Li1O8W2Yb1 |
| | 2 | z2w,1=1 z2w,2=1 z2w,3=1 z4=3 |
| | 2 | z2w,2=1 z2w,3=1 z4=3 |
| | 2 | z2w,1=1 z4=2 |
| | 2 | z2w,3=1 z4=3 |
| | 2 | z2w,1=1 z2w,2=1 z4=2 |
| | 2 | z2w,2=1 z4=2 |
| | 2 | z2w,1=1 z2w,3=1 z4=1 |
| | 2 | z2w,1=1 z2w,3=1 z4=3 |
| | 2 | z2w,2=1 z4=2 |
| | 2 | z2w,1=1 z2w,2=1 z4=2 |
| | 2 | z2w,3=1 z4=1 |
| | 2 | z2w,1=1 z4=2 |
| | 2 | z2w,2=1 z2w,3=1 z4=1 |
| | 2 | z2w,1=1 z2w,2=1 z2w,3=1 z4=1 |
| 13.4.3 | 16163 | Li1O8W2Yb1 |
| | 2 | z2w,2=1 z4=1 |
| | 2 | z2w,1=1 z2w,3=1 |
| | 2 | z2w,3=1 |
| | 2 | z2w,1=1 z2w,2=1 z4=3 |
| | 2 | z2w,2=1 z2w,3=1 |
| | 2 | z2w,1=1 z4=3 |
| | 2 | z4=3 |
| | 2 | z2w,1=1 z2w,2=1 z2w,3=1 z4=2 |
| | 2 | z2w,1=1 z2w,2=1 z2w,3=1 |
| | 2 | z4=3 |
| | 2 | z2w,1=1 z4=3 |
| | 2 | z2w,2=1 z2w,3=1 z4=2 |
| | 2 | z2w,1=1 z2w,2=1 z4=3 |
| | 2 | z2w,3=1 z4=2 |
| | 2 | z2w,1=1 z2w,3=1 z4=2 |
| | 2 | z2w,2=1 z4=1 |
| 13.4.4 | 154808 | H6Cu1O5Sr1 |
| | 2 | z2w,1=1 z2w,2=1 z2w,3=1 z4=3 |
| | 2 | z2w,2=1 z2w,3=1 z4=3 |
| | 2 | z2w,1=1 z4=2 |
| | 2 | z2w,3=1 z4=3 |
| | 2 | z2w,1=1 z2w,2=1 z4=2 |
| | 2 | z2w,2=1 z4=2 |
| | 2 | z2w,1=1 z2w,3=1 z4=1 |
| | 2 | z2w,1=1 z2w,3=1 z4=3 |
| | 2 | z2w,2=1 z4=2 |
| | 2 | z2w,1=1 z2w,2=1 z4=2 |

Continued on next column

**TABLE XIX– continued**

| TQC | ICSD<br>SSG | Stoich. Formula<br>SIs |
|---|---|---|
| | 2 | z2w,3=1 z4=1 |
| | 2 | z2w,1=1 z4=2 |
| | 2 | z2w,2=1 z2w,3=1 z4=1 |
| | 2 | z2w,1=1 z2w,2=1 z2w,3=1 z4=1 |
| 13.4.5 | 20264 | Gd1O4Ta1 |
| | 2 | z2w,1=1 z2w,2=1 z4=2 |
| | 2 | z2w,3=1 z4=1 |
| | 2 | z2w,1=1 z2w,3=1 z4=1 |
| | 2 | z2w,2=1 |
| | 2 | z2w,1=1 z2w,2=1 z2w,3=1 z4=1 |
| | 2 | z2w,1=1 |
| | 2 | z2w,2=1 z2w,3=1 z4=3 |
| | 2 | z2w,2=1 z2w,3=1 z4=1 |
| | 2 | z2w,1=1 |
| | 2 | z2w,1=1 z2w,2=1 z2w,3=1 z4=3 |
| | 2 | z2w,2=1 |
| | 2 | z2w,1=1 z2w,3=1 z4=3 |
| | 2 | z2w,3=1 z4=3 |
| | 2 | z2w,1=1 z2w,2=1 z4=2 |
| 13.4.6 | 155659 | Gd1O4Ta1 |
| | 2 | z4=3 |
| | 2 | z2w,1=1 z2w,2=1 z2w,3=1 z4=2 |
| | 2 | z2w,2=1 z2w,3=1 z4=2 |
| | 2 | z2w,1=1 z4=1 |
| | 2 | z2w,3=1 z4=2 |
| | 2 | z2w,1=1 z2w,2=1 z4=1 |
| | 2 | z2w,2=1 z4=1 |
| | 2 | z2w,1=1 z2w,3=1 |
| | 2 | z2w,1=1 z2w,3=1 z4=2 |
| | 2 | z2w,2=1 z4=1 |
| | 2 | z2w,1=1 z2w,2=1 z4=1 |
| | 2 | z2w,3=1 |
| | 2 | z2w,1=1 z4=1 |
| | 2 | z2w,2=1 z2w,3=1 |
| | 2 | z2w,1=1 z2w,2=1 z2w,3=1 |
| | 2 | z4=3 |
| 13.4.7 | 161847 | Fe1Na1O8W2 |
| | 2 | z2w,1=1 z2w,2=1 z4=2 |
| | 2 | z2w,3=1 z4=1 |
| | 2 | z2w,1=1 z2w,3=1 z4=1 |
| | 2 | z2w,2=1 |
| | 2 | z2w,1=1 z2w,2=1 z2w,3=1 z4=1 |
| | 2 | z2w,1=1 |
| | 2 | z2w,2=1 z2w,3=1 z4=3 |
| | 2 | z2w,2=1 z2w,3=1 z4=1 |
| | 2 | z2w,1=1 |
| | 2 | z2w,1=1 z2w,2=1 z2w,3=1 z4=3 |
| | 2 | z2w,2=1 |
| | 2 | z2w,1=1 z2w,3=1 z4=3 |
| | 2 | z2w,3=1 z4=3 |
| | 2 | z2w,1=1 z2w,2=1 z4=2 |
| 13.4.8 | 415437 | Er1O4Ta1 |
| | 2 | z4=3 |
| | 2 | z2w,1=1 z2w,2=1 z2w,3=1 z4=2 |
| | 2 | z2w,2=1 z2w,3=1 z4=2 |
| | 2 | z2w,1=1 z4=1 |
| | 2 | z2w,3=1 z4=2 |
| | 2 | z2w,1=1 z2w,2=1 z4=1 |
| | 2 | z2w,2=1 z4=1 |
| | 2 | z2w,1=1 z2w,3=1 |

Continued on next column

**TABLE XIX– continued**

| TQC | ICSD<br>SSG | Stoich. Formula<br>SIs |
|---|---|---|
| | 2 | z2w,1=1 z2w,3=1 z4=2 |
| | 2 | z2w,2=1 z4=1 |
| | 2 | z2w,1=1 z2w,2=1 z4=1 |
| | 2 | z2w,3=1 |
| | 2 | z2w,1=1 z4=1 |
| | 2 | z2w,2=1 z2w,3=1 |
| | 2 | z2w,1=1 z2w,2=1 z2w,3=1 |
| | 2 | z4=3 |
| 13.4.9 | 20262 | Er1O4Ta1 |
| | 2 | z2w,2=1 z4=2 |
| | 2 | z2w,1=1 z2w,3=1 z4=1 |
| | 2 | z2w,3=1 z4=1 |
| | 2 | z2w,1=1 z2w,2=1 |
| | 2 | z2w,2=1 z2w,3=1 z4=1 |
| | 2 | z2w,1=1 |
| | 2 | z2w,1=1 z2w,2=1 z2w,3=1 z4=3 |
| | 2 | z2w,1=1 z2w,2=1 z2w,3=1 z4=1 |
| | 2 | z2w,1=1 |
| | 2 | z2w,2=1 z2w,3=1 z4=3 |
| | 2 | z2w,1=1 z2w,2=1 |
| | 2 | z2w,3=1 z4=3 |
| | 2 | z2w,1=1 z2w,3=1 z4=3 |
| | 2 | z2w,2=1 z4=2 |
| 13.4.10 | 155661 | Er1O4Ta1 |
| | 2 | z4=2 |
| | 2 | z2w,1=1 z2w,2=1 z2w,3=1 z4=1 |
| | 2 | z2w,2=1 z2w,3=1 z4=1 |
| | 2 | z2w,1=1 |
| | 2 | z2w,3=1 z4=1 |
| | 2 | z2w,1=1 z2w,2=1 |
| | 2 | z2w,2=1 |
| | 2 | z2w,1=1 z2w,3=1 z4=3 |
| | 2 | z2w,1=1 z2w,3=1 z4=1 |
| | 2 | z2w,2=1 |
| | 2 | z2w,1=1 z2w,2=1 |
| | 2 | z2w,3=1 z4=3 |
| | 2 | z2w,1=1 |
| | 2 | z2w,2=1 z2w,3=1 z4=3 |
| | 2 | z2w,1=1 z2w,2=1 z2w,3=1 z4=3 |
| | 2 | z4=2 |
| 13.4.11 | 261837 | Er1Li1O8W2 |
| | 2 | z2w,2=1 z4=2 |
| | 2 | z2w,1=1 z2w,3=1 z4=1 |
| | 2 | z2w,3=1 z4=1 |
| | 2 | z2w,1=1 z2w,2=1 |
| | 2 | z2w,2=1 z2w,3=1 z4=1 |
| | 2 | z2w,1=1 |
| | 2 | z2w,1=1 z2w,2=1 z2w,3=1 z4=3 |
| | 2 | z2w,1=1 z2w,2=1 z2w,3=1 z4=1 |
| | 2 | z2w,1=1 |
| | 2 | z2w,2=1 z2w,3=1 z4=3 |
| | 2 | z2w,1=1 z2w,2=1 |
| | 2 | z2w,3=1 z4=3 |
| | 2 | z2w,1=1 z2w,3=1 z4=3 |
| | 2 | z2w,2=1 z4=2 |
| 13.4.12 | 415435 | Dy1O4Ta1 |
| | 2 | z2w,1=1 z2w,2=1 z2w,3=1 z4=3 |
| | 2 | z2w,2=1 z2w,3=1 z4=3 |
| | 2 | z2w,1=1 z4=2 |
| | 2 | z2w,3=1 z4=3 |

Continued on next column

**TABLE XIX– continued**

| TQC | ICSD<br>SSG | Stoich. Formula<br>SIs |
|---|---|---|
| | 2 | z2w,1=1 z2w,2=1 z4=2 |
| | 2 | z2w,2=1 z4=2 |
| | 2 | z2w,1=1 z2w,3=1 z4=1 |
| | 2 | z2w,1=1 z2w,3=1 z4=3 |
| | 2 | z2w,2=1 z4=2 |
| | 2 | z2w,1=1 z2w,2=1 z4=2 |
| | 2 | z2w,3=1 z4=1 |
| | 2 | z2w,1=1 z4=2 |
| | 2 | z2w,2=1 z2w,3=1 z4=1 |
| | 2 | z2w,1=1 z2w,2=1 z2w,3=1 z4=1 |
| 13.4.13 | 20263 | Dy1O4Ta1 |
| | 2 | z2w,2=1 z4=1 |
| | 2 | z2w,1=1 z2w,3=1 |
| | 2 | z2w,3=1 |
| | 2 | z2w,1=1 z2w,2=1 z4=3 |
| | 2 | z2w,2=1 z2w,3=1 |
| | 2 | z2w,1=1 z4=3 |
| | 2 | z4=3 |
| | 2 | z2w,1=1 z2w,2=1 z2w,3=1 z4=2 |
| | 2 | z2w,1=1 z2w,2=1 z2w,3=1 |
| | 2 | z4=3 |
| | 2 | z2w,1=1 z4=3 |
| | 2 | z2w,2=1 z2w,3=1 z4=2 |
| | 2 | z2w,1=1 z2w,2=1 z4=3 |
| | 2 | z2w,3=1 z4=2 |
| | 2 | z2w,1=1 z2w,3=1 z4=2 |
| | 2 | z2w,2=1 z4=1 |
| 13.4.14 | 182751 | Cu1O4W1 |
| | 2 | z2w,1=1 z2w,2=1 z2w,3=1 z4=3 |
| | 2 | z2w,2=1 z2w,3=1 z4=3 |
| | 2 | z2w,1=1 z4=2 |
| | 2 | z2w,3=1 z4=3 |
| | 2 | z2w,1=1 z2w,2=1 z4=2 |
| | 2 | z2w,2=1 z4=2 |
| | 2 | z2w,1=1 z2w,3=1 z4=1 |
| | 2 | z2w,1=1 z2w,3=1 z4=3 |
| | 2 | z2w,2=1 z4=2 |
| | 2 | z2w,1=1 z2w,2=1 z4=2 |
| | 2 | z2w,3=1 z4=1 |
| | 2 | z2w,1=1 z4=2 |
| | 2 | z2w,2=1 z2w,3=1 z4=1 |
| | 2 | z2w,1=1 z2w,2=1 z2w,3=1 z4=1 |
| 13.4.15 | 73748 | Cu1Er1O8W2 |
| | 2 | z2w,1=1 z4=1 |
| | 2 | z2w,2=1 z2w,3=1 |
| | 2 | z2w,1=1 z2w,2=1 z2w,3=1 |
| | 2 | z4=3 |
| | 2 | z2w,1=1 z2w,3=1 |
| | 2 | z2w,2=1 z4=3 |
| | 2 | z2w,1=1 z2w,2=1 z4=3 |
| | 2 | z2w,3=1 z4=2 |
| | 2 | z2w,3=1 |
| | 2 | z2w,1=1 z2w,2=1 z4=3 |
| | 2 | z2w,2=1 z4=3 |
| | 2 | z2w,1=1 z2w,3=1 z4=2 |
| | 2 | z4=3 |
| | 2 | z2w,1=1 z2w,2=1 z2w,3=1 z4=2 |
| | 2 | z2w,2=1 z2w,3=1 z4=2 |
| | 2 | z2w,1=1 z4=1 |
| 13.4.16 | 192430 | Cs1Gd1Mo2O8 |

Continued on next column

**TABLE XIX– continued**

| TQC | ICSD SSG | Stoich. Formula SIs |
|---|---|---|
| | 2 | z2w,1=1 z2w,2=1 z2w,3=1 z4=3 |
| | 2 | z2w,2=1 z2w,3=1 z4=3 |
| | 2 | z2w,1=1 z4=2 |
| | 2 | z2w,3=1 z4=3 |
| | 2 | z2w,1=1 z2w,2=1 z4=2 |
| | 2 | z2w,2=1 z4=2 |
| | 2 | z2w,1=1 z2w,3=1 z4=1 |
| | 2 | z2w,1=1 z2w,3=1 z4=3 |
| | 2 | z2w,2=1 z4=2 |
| | 2 | z2w,1=1 z2w,2=1 z4=2 |
| | 2 | z2w,3=1 z4=1 |
| | 2 | z2w,1=1 z4=2 |
| | 2 | z2w,2=1 z2w,3=1 z4=1 |
| | 2 | z2w,1=1 z2w,2=1 z2w,3=1 z4=1 |
| 13.4.17 | 15851 | Co1O4W1 |
| | 2 | z2w,2=1 z4=1 |
| | 2 | z2w,1=1 z2w,3=1 |
| | 2 | z2w,3=1 |
| | 2 | z2w,1=1 z2w,2=1 z4=3 |
| | 2 | z2w,2=1 z2w,3=1 |
| | 2 | z2w,1=1 z4=3 |
| | 2 | z4=3 |
| | 2 | z2w,1=1 z2w,2=1 z2w,3=1 z4=2 |
| | 2 | z2w,1=1 z2w,2=1 z2w,3=1 |
| | 2 | z4=3 |
| | 2 | z2w,1=1 z4=3 |
| | 2 | z2w,2=1 z2w,3=1 z4=2 |
| | 2 | z2w,1=1 z2w,2=1 z4=3 |
| | 2 | z2w,3=1 z4=2 |
| | 2 | z2w,1=1 z2w,3=1 z4=2 |
| | 2 | z2w,2=1 z4=1 |
| 13.4.18 | 253697 | Co1Na4O6Te1 |
| | 2 | z2w,1=1 z2w,2=1 z4=2 |
| | 2 | z2w,3=1 z4=1 |
| | 2 | z2w,1=1 z2w,3=1 z4=1 |
| | 2 | z2w,2=1 |
| | 2 | z2w,1=1 z2w,2=1 z2w,3=1 z4=1 |
| | 2 | z2w,1=1 |
| | 2 | z2w,2=1 z2w,3=1 z4=3 |
| | 2 | z2w,2=1 z2w,3=1 z4=1 |
| | 2 | z2w,1=1 |
| | 2 | z2w,1=1 z2w,2=1 z2w,3=1 z4=3 |
| | 2 | z2w,2=1 |
| | 2 | z2w,1=1 z2w,3=1 z4=3 |
| | 2 | z2w,3=1 z4=3 |
| | 2 | z2w,1=1 z2w,2=1 z4=2 |
| 13.4.19 | 81061 | Co1Mo1O4 |
| | 2 | z4=3 |
| | 2 | z2w,1=1 z2w,2=1 z2w,3=1 z4=2 |
| | 2 | z2w,2=1 z2w,3=1 z4=2 |
| | 2 | z2w,1=1 z4=1 |
| | 2 | z2w,3=1 z4=2 |
| | 2 | z2w,1=1 z2w,2=1 z4=1 |
| | 2 | z2w,2=1 z4=1 |
| | 2 | z2w,1=1 z2w,3=1 |
| | 2 | z2w,1=1 z2w,3=1 z4=2 |
| | 2 | z2w,2=1 z4=1 |
| | 2 | z2w,1=1 z2w,2=1 z4=1 |
| | 2 | z2w,3=1 |
| | 2 | z2w,1=1 z4=1 |

Continued on next column

**TABLE XIX– continued**

| TQC | ICSD SSG | Stoich. Formula SIs |
|---|---|---|
| | 2 | z2w,2=1 z2w,3=1 |
| | 2 | z2w,1=1 z2w,2=1 z2w,3=1 |
| | 2 | z4=3 |
| 13.4.20 | 281235 | Co1Mo1O4 |
| | 2 | z2w,2=1 z4=1 |
| | 2 | z2w,1=1 z2w,3=1 |
| | 2 | z2w,3=1 |
| | 2 | z2w,1=1 z2w,2=1 z4=3 |
| | 2 | z2w,2=1 z2w,3=1 |
| | 2 | z2w,1=1 z4=3 |
| | 2 | z4=3 |
| | 2 | z2w,1=1 z2w,2=1 z2w,3=1 z4=2 |
| | 2 | z2w,1=1 z2w,2=1 z2w,3=1 |
| | 2 | z4=3 |
| | 2 | z2w,1=1 z4=3 |
| | 2 | z2w,2=1 z2w,3=1 z4=2 |
| | 2 | z2w,1=1 z2w,2=1 z4=3 |
| | 2 | z2w,3=1 z4=2 |
| | 2 | z2w,1=1 z2w,3=1 z4=2 |
| | 2 | z2w,2=1 z4=1 |
| 13.4.21 | 82364 | Cl4Er1Na1 |
| | 2 | z4=3 |
| | 2 | z2w,1=1 z2w,2=1 z2w,3=1 z4=2 |
| | 2 | z2w,2=1 z2w,3=1 z4=2 |
| | 2 | z2w,1=1 z4=1 |
| | 2 | z2w,3=1 z4=2 |
| | 2 | z2w,1=1 z2w,2=1 z4=1 |
| | 2 | z2w,2=1 z4=1 |
| | 2 | z2w,1=1 z2w,3=1 |
| | 2 | z2w,1=1 z2w,3=1 z4=2 |
| | 2 | z2w,2=1 z4=1 |
| | 2 | z2w,1=1 z2w,2=1 z4=1 |
| | 2 | z2w,3=1 |
| | 2 | z2w,1=1 z4=1 |
| | 2 | z2w,2=1 z2w,3=1 |
| | 2 | z2w,1=1 z2w,2=1 z2w,3=1 |
| | 2 | z4=3 |
| 13.4.22 | 251426 | C6Li1N4 |
| | 2 | z2w,1=1 z2w,2=1 z2w,3=1 z4=3 |
| | 2 | z2w,2=1 z2w,3=1 z4=3 |
| | 2 | z2w,1=1 z4=2 |
| | 2 | z2w,3=1 z4=3 |
| | 2 | z2w,1=1 z2w,2=1 z4=2 |
| | 2 | z2w,2=1 z4=2 |
| | 2 | z2w,1=1 z2w,3=1 z4=1 |
| | 2 | z2w,1=1 z2w,3=1 z4=3 |
| | 2 | z2w,2=1 z4=2 |
| | 2 | z2w,1=1 z2w,2=1 z4=2 |
| | 2 | z2w,3=1 z4=1 |
| | 2 | z2w,1=1 z4=2 |
| | 2 | z2w,2=1 z2w,3=1 z4=1 |
| | 2 | z2w,1=1 z2w,2=1 z2w,3=1 z4=1 |
| 13.4.23 | 251425 | C6Li1N4 |
| | 2 | z2w,1=1 z2w,2=1 z2w,3=1 z4=3 |
| | 2 | z2w,2=1 z2w,3=1 z4=3 |
| | 2 | z2w,1=1 z4=2 |
| | 2 | z2w,3=1 z4=3 |
| | 2 | z2w,1=1 z2w,2=1 z4=2 |
| | 2 | z2w,2=1 z4=2 |
| | 2 | z2w,1=1 z2w,3=1 z4=1 |

Continued on next column

**TABLE XIX– continued**

| TQC | ICSD SSG | Stoich. Formula SIs |
|---|---|---|
| | 2 | z2w,1=1 z2w,3=1 z4=3 |
| | 2 | z2w,2=1 z4=2 |
| | 2 | z2w,1=1 z2w,2=1 z4=2 |
| | 2 | z2w,3=1 z4=1 |
| | 2 | z2w,1=1 z4=2 |
| | 2 | z2w,2=1 z2w,3=1 z4=1 |
| | 2 | z2w,1=1 z2w,2=1 z2w,3=1 z4=1 |
| 13.4.24 | 251423 | C6Li1N4 |
| | 2 | z2w,1=1 z2w,2=1 z2w,3=1 z4=3 |
| | 2 | z2w,2=1 z2w,3=1 z4=3 |
| | 2 | z2w,1=1 z4=2 |
| | 2 | z2w,3=1 z4=3 |
| | 2 | z2w,1=1 z2w,2=1 z4=2 |
| | 2 | z2w,2=1 z4=2 |
| | 2 | z2w,1=1 z2w,3=1 z4=1 |
| | 2 | z2w,1=1 z2w,3=1 z4=3 |
| | 2 | z2w,2=1 z4=2 |
| | 2 | z2w,1=1 z2w,2=1 z4=2 |
| | 2 | z2w,3=1 z4=1 |
| | 2 | z2w,1=1 z4=2 |
| | 2 | z2w,2=1 z2w,3=1 z4=1 |
| | 2 | z2w,1=1 z2w,2=1 z2w,3=1 z4=1 |
| 13.4.25 | 251421 | C6Li1N4 |
| | 2 | z2w,2=1 z4=2 |
| | 2 | z2w,1=1 z2w,3=1 z4=1 |
| | 2 | z2w,3=1 z4=1 |
| | 2 | z2w,1=1 z2w,2=1 |
| | 2 | z2w,2=1 z2w,3=1 z4=1 |
| | 2 | z2w,1=1 |
| | 2 | z2w,1=1 z2w,2=1 z2w,3=1 z4=3 |
| | 2 | z2w,1=1 z2w,2=1 z2w,3=1 z4=1 |
| | 2 | z2w,1=1 |
| | 2 | z2w,2=1 z2w,3=1 z4=3 |
| | 2 | z2w,1=1 z2w,2=1 |
| | 2 | z2w,3=1 z4=3 |
| | 2 | z2w,1=1 z2w,3=1 z4=3 |
| | 2 | z2w,2=1 z4=2 |
| 13.4.26 | 251420 | C6Li1N4 |
| | 2 | z2w,2=1 z4=2 |
| | 2 | z2w,1=1 z2w,3=1 z4=1 |
| | 2 | z2w,3=1 z4=1 |
| | 2 | z2w,1=1 z2w,2=1 |
| | 2 | z2w,2=1 z2w,3=1 z4=1 |
| | 2 | z2w,1=1 |
| | 2 | z2w,1=1 z2w,2=1 z2w,3=1 z4=3 |
| | 2 | z2w,1=1 z2w,2=1 z2w,3=1 z4=1 |
| | 2 | z2w,1=1 |
| | 2 | z2w,2=1 z2w,3=1 z4=3 |
| | 2 | z2w,1=1 z2w,2=1 |
| | 2 | z2w,3=1 z4=3 |
| | 2 | z2w,1=1 z2w,3=1 z4=3 |
| | 2 | z2w,2=1 z4=2 |
| 13.4.27 | 251419 | C6Li1N4 |
| | 2 | z2w,1=1 z2w,2=1 z2w,3=1 z4=3 |
| | 2 | z2w,2=1 z2w,3=1 z4=3 |
| | 2 | z2w,1=1 z4=2 |
| | 2 | z2w,3=1 z4=3 |
| | 2 | z2w,1=1 z2w,2=1 z4=2 |
| | 2 | z2w,2=1 z4=2 |
| | 2 | z2w,1=1 z2w,3=1 z4=1 |

Continued on next column

**TABLE XIX– continued**

| TQC | ICSD SSG | Stoich. Formula SIs |
|---|---|---|
| | 2 | z2w,1=1 z2w,3=1 z4=3 |
| | 2 | z2w,2=1 z4=2 |
| | 2 | z2w,1=1 z2w,2=1 z4=2 |
| | 2 | z2w,3=1 z4=1 |
| | 2 | z2w,1=1 z4=2 |
| | 2 | z2w,2=1 z2w,3=1 z4=1 |
| | 2 | z2w,1=1 z2w,2=1 z2w,3=1 z4=1 |
| 13.4.28 | 251418 | C6Li1N4 |
| | 2 | z2w,1=1 z2w,2=1 z2w,3=1 z4=3 |
| | 2 | z2w,2=1 z2w,3=1 z4=3 |
| | 2 | z2w,1=1 z4=2 |
| | 2 | z2w,3=1 z4=3 |
| | 2 | z2w,1=1 z2w,2=1 z4=2 |
| | 2 | z2w,2=1 z4=2 |
| | 2 | z2w,1=1 z2w,3=1 z4=1 |
| | 2 | z2w,1=1 z2w,3=1 z4=3 |
| | 2 | z2w,2=1 z4=2 |
| | 2 | z2w,1=1 z2w,2=1 z4=2 |
| | 2 | z2w,3=1 z4=1 |
| | 2 | z2w,1=1 z4=2 |
| | 2 | z2w,2=1 z2w,3=1 z4=1 |
| | 2 | z2w,1=1 z2w,2=1 z2w,3=1 z4=1 |
| 13.4.29 | 56734 | Al2Cl8Eu1 |
| | 2 | z2w,2=1 z4=1 |
| | 2 | z2w,1=1 z2w,3=1 |
| | 2 | z2w,3=1 |
| | 2 | z2w,1=1 z2w,2=1 z4=3 |
| | 2 | z2w,2=1 z2w,3=1 |
| | 2 | z2w,1=1 z4=3 |
| | 2 | z4=3 |
| | 2 | z2w,1=1 z2w,2=1 z2w,3=1 z4=2 |
| | 2 | z2w,1=1 z2w,2=1 z2w,3=1 |
| | 2 | z4=3 |
| | 2 | z2w,1=1 z4=3 |
| | 2 | z2w,2=1 z2w,3=1 z4=2 |
| | 2 | z2w,1=1 z2w,2=1 z4=3 |
| | 2 | z2w,3=1 z4=2 |
| | 2 | z2w,1=1 z2w,3=1 z4=2 |
| | 2 | z2w,2=1 z4=1 |
| 14.4.1 | 193236 | O6Sr2Ta1Yb1 |
| | 2 | z2w,2=1 z2w,3=1 z4=2 |
| | 2 | z2w,3=1 z4=1 |
| | 2 | z2w,1=1 z2w,3=1 z4=1 |
| | 2 | z2w,1=1 z2w,2=1 z2w,3=1 |
| | 2 | z2w,2=1 z4=1 |
| | 2 | z2w,1=1 |
| | 2 | z2w,1=1 z2w,2=1 z4=3 |
| | 2 | z2w,1=1 z2w,2=1 z4=1 |
| | 2 | z2w,1=1 |
| | 2 | z2w,2=1 z4=3 |
| | 2 | z2w,1=1 z2w,2=1 z2w,3=1 |
| | 2 | z2w,1=1 z2w,3=1 z4=3 |
| | 2 | z2w,3=1 z4=3 |
| | 2 | z2w,2=1 z2w,3=1 z4=2 |
| 14.4.2 | 49500 | O6Ru1Sr2Y1 |
| | 2 | z2w,2=1 z2w,3=1 z4=2 |
| | 2 | z2w,3=1 z4=1 |
| | 2 | z2w,1=1 z2w,3=1 z4=1 |
| | 2 | z2w,1=1 z2w,2=1 z2w,3=1 |
| | 2 | z2w,2=1 z4=1 |

Continued on next column

**TABLE XIX– continued**

| TQC | ICSD SSG | Stoich. Formula SIs |
|---|---|---|
| | 2 | z2w,1=1 |
| | 2 | z2w,1=1 z2w,2=1 z4=3 |
| | 2 | z2w,1=1 z2w,2=1 z4=1 |
| | 2 | z2w,1=1 |
| | 2 | z2w,2=1 z4=3 |
| | 2 | z2w,1=1 z2w,2=1 z2w,3=1 |
| | 2 | z2w,1=1 z2w,3=1 z4=3 |
| | 2 | z2w,3=1 z4=3 |
| | 2 | z2w,2=1 z2w,3=1 z4=2 |
| 14.4.3 | 238814 | O6Ru1Sr2Y1 |
| | 2 | z2w,1=1 |
| | 2 | z2w,1=1 z2w,2=1 z4=3 |
| | 2 | z2w,2=1 z4=3 |
| | 2 | z4=2 |
| | 2 | z2w,1=1 z2w,3=1 z4=3 |
| | 2 | z2w,1=1 z2w,2=1 z2w,3=1 z4=2 |
| | 2 | z2w,2=1 z2w,3=1 z4=2 |
| | 2 | z2w,3=1 z4=1 |
| | 2 | z2w,3=1 z4=3 |
| | 2 | z2w,2=1 z2w,3=1 z4=2 |
| | 2 | z2w,1=1 z2w,2=1 z2w,3=1 z4=2 |
| | 2 | z2w,1=1 z2w,3=1 z4=1 |
| | 2 | z4=2 |
| | 2 | z2w,2=1 z4=1 |
| | 2 | z2w,1=1 z2w,2=1 z4=1 |
| | 2 | z2w,1=1 |
| 14.4.4 | 192766 | O6Ru1Sr2Y1 |
| | 2 | z2w,1=1 z4=2 |
| | 2 | z2w,1=1 z2w,2=1 z4=1 |
| | 2 | z2w,2=1 z4=1 |
| | 2 | z2w,1=1 z2w,3=1 z4=1 |
| | 2 | z2w,1=1 z2w,2=1 z2w,3=1 |
| | 2 | z2w,2=1 z2w,3=1 |
| | 2 | z2w,3=1 z4=3 |
| | 2 | z2w,3=1 z4=1 |
| | 2 | z2w,2=1 z2w,3=1 |
| | 2 | z2w,1=1 z2w,2=1 z2w,3=1 |
| | 2 | z2w,1=1 z2w,3=1 z4=3 |
| | 2 | z2w,2=1 z4=3 |
| | 2 | z2w,1=1 z2w,2=1 z4=3 |
| | 2 | z2w,1=1 z4=2 |
| 14.4.5 | 93361 | O6Ru1Sr2Tb1 |
| | 2 | z2w,2=1 z2w,3=1 |
| | 2 | z2w,3=1 z4=3 |
| | 2 | z2w,1=1 z2w,3=1 z4=3 |
| | 2 | z2w,1=1 z2w,2=1 z2w,3=1 z4=2 |
| | 2 | z2w,2=1 z4=3 |
| | 2 | z4=2 |
| | 2 | z2w,1=1 z4=2 |
| | 2 | z2w,1=1 z2w,2=1 z4=1 |
| | 2 | z2w,1=1 z2w,2=1 z4=3 |
| | 2 | z2w,1=1 z4=2 |
| | 2 | z4=2 |
| | 2 | z2w,2=1 z4=1 |
| | 2 | z2w,1=1 z2w,2=1 z2w,3=1 z4=2 |
| | 2 | z2w,1=1 z2w,3=1 z4=1 |
| | 2 | z2w,3=1 z4=1 |
| | 2 | z2w,2=1 z2w,3=1 |
| 14.4.6 | 290794 | O6Ru1Sr2Tb1 |
| | 2 | z2w,2=1 z2w,3=1 |

Continued on next column

**TABLE XIX– continued**

| TQC | ICSD SSG | Stoich. Formula SIs |
|---|---|---|
| | 2 | z2w,3=1 z4=3 |
| | 2 | z2w,1=1 z2w,3=1 z4=3 |
| | 2 | z2w,1=1 z2w,2=1 z2w,3=1 z4=2 |
| | 2 | z2w,2=1 z4=3 |
| | 2 | z4=2 |
| | 2 | z2w,1=1 z4=2 |
| | 2 | z2w,1=1 z2w,2=1 z4=1 |
| | 2 | z2w,1=1 z2w,2=1 z4=3 |
| | 2 | z2w,1=1 z4=2 |
| | 2 | z4=2 |
| | 2 | z2w,2=1 z4=1 |
| | 2 | z2w,1=1 z2w,2=1 z2w,3=1 z4=2 |
| | 2 | z2w,1=1 z2w,3=1 z4=1 |
| | 2 | z2w,3=1 z4=1 |
| | 2 | z2w,2=1 z2w,3=1 |
| 14.4.7 | 153132 | O6Ru1Sr2Tb1 |
| | 2 | z2w,1=1 z2w,2=1 z2w,3=1 |
| | 2 | z2w,1=1 z2w,3=1 z4=3 |
| | 2 | z2w,3=1 z4=3 |
| | 2 | z2w,2=1 z2w,3=1 z4=2 |
| | 2 | z2w,1=1 z2w,2=1 z4=3 |
| | 2 | z2w,1=1 z4=2 |
| | 2 | z4=2 |
| | 2 | z2w,2=1 z4=1 |
| | 2 | z2w,2=1 z4=3 |
| | 2 | z4=2 |
| | 2 | z2w,1=1 z4=2 |
| | 2 | z2w,1=1 z2w,2=1 z4=1 |
| | 2 | z2w,2=1 z2w,3=1 z4=2 |
| | 2 | z2w,3=1 z4=1 |
| | 2 | z2w,1=1 z2w,3=1 z4=1 |
| | 2 | z2w,1=1 z2w,2=1 z2w,3=1 |
| 14.4.8 | 245796 | O6Re2Sc1 |
| | 2 | z2w,2=1 z4=3 |
| | 2 | z2w,1=1 z2w,2=1 z4=3 |
| | 2 | z2w,1=1 z4=2 |
| | 2 | z2w,3=1 z4=3 |
| | 2 | z2w,2=1 z2w,3=1 z4=2 |
| | 2 | z2w,1=1 z2w,2=1 z2w,3=1 z4=2 |
| | 2 | z2w,1=1 z2w,3=1 z4=1 |
| | 2 | z2w,1=1 z2w,3=1 z4=3 |
| | 2 | z2w,1=1 z2w,2=1 z2w,3=1 z4=2 |
| | 2 | z2w,2=1 z2w,3=1 z4=2 |
| | 2 | z2w,3=1 z4=1 |
| | 2 | z2w,1=1 z4=2 |
| | 2 | z2w,1=1 z2w,2=1 z4=1 |
| | 2 | z2w,2=1 z4=1 |
| 14.4.9 | 238315 | O6Os1Sr2Y1 |
| | 2 | z2w,1=1 z2w,2=1 z2w,3=1 z4=2 |
| | 2 | z2w,1=1 z2w,3=1 z4=1 |
| | 2 | z2w,3=1 z4=1 |
| | 2 | z2w,2=1 z2w,3=1 |
| | 2 | z2w,1=1 z2w,2=1 z4=1 |
| | 2 | z2w,1=1 |
| | 2 | z2w,2=1 z4=3 |
| | 2 | z2w,2=1 z4=1 |
| | 2 | z2w,1=1 |
| | 2 | z2w,1=1 z2w,2=1 z4=3 |
| | 2 | z2w,2=1 z2w,3=1 |
| | 2 | z2w,3=1 z4=3 |

Continued on next column

**TABLE XIX– continued**

| TQC | ICSD<br>SSG | Stoich. Formula<br>SIs |
|---|---|---|
| | 2 | z2w,1=1 z2w,3=1 z4=3 |
| | 2 | z2w,1=1 z2w,2=1 z2w,3=1 z4=2 |
| 14.4.10 | 238314 | O6Os1Sr2Y1 |
| | 2 | z2w,2=1 z4=3 |
| | 2 | z2w,1=1 z2w,2=1 z4=3 |
| | 2 | z2w,1=1 z4=2 |
| | 2 | z2w,3=1 z4=3 |
| | 2 | z2w,2=1 z2w,3=1 z4=2 |
| | 2 | z2w,1=1 z2w,2=1 z2w,3=1 z4=2 |
| | 2 | z2w,1=1 z2w,3=1 z4=1 |
| | 2 | z2w,1=1 z2w,3=1 z4=3 |
| | 2 | z2w,1=1 z2w,2=1 z2w,3=1 z4=2 |
| | 2 | z2w,2=1 z2w,3=1 z4=2 |
| | 2 | z2w,3=1 z4=1 |
| | 2 | z2w,1=1 z4=2 |
| | 2 | z2w,1=1 z2w,2=1 z4=1 |
| | 2 | z2w,2=1 z4=1 |
| 14.4.11 | 238313 | O6Os1Sr2Y1 |
| | 2 | z2w,1=1 z2w,2=1 z2w,3=1 z4=2 |
| | 2 | z2w,1=1 z2w,3=1 z4=1 |
| | 2 | z2w,3=1 z4=1 |
| | 2 | z2w,2=1 z2w,3=1 |
| | 2 | z2w,1=1 z2w,2=1 z4=1 |
| | 2 | z2w,1=1 |
| | 2 | z2w,2=1 z4=3 |
| | 2 | z2w,2=1 z4=1 |
| | 2 | z2w,1=1 |
| | 2 | z2w,1=1 z2w,2=1 z4=3 |
| | 2 | z2w,2=1 z2w,3=1 |
| | 2 | z2w,3=1 z4=3 |
| | 2 | z2w,1=1 z2w,3=1 z4=3 |
| | 2 | z2w,1=1 z2w,2=1 z2w,3=1 z4=2 |
| 14.4.12 | 238312 | O6Os1Sr2Y1 |
| | 2 | z2w,1=1 z2w,2=1 z2w,3=1 z4=2 |
| | 2 | z2w,1=1 z2w,3=1 z4=1 |
| | 2 | z2w,3=1 z4=1 |
| | 2 | z2w,2=1 z2w,3=1 |
| | 2 | z2w,1=1 z2w,2=1 z4=1 |
| | 2 | z2w,1=1 |
| | 2 | z2w,2=1 z4=3 |
| | 2 | z2w,2=1 z4=1 |
| | 2 | z2w,1=1 |
| | 2 | z2w,1=1 z2w,2=1 z4=3 |
| | 2 | z2w,2=1 z2w,3=1 |
| | 2 | z2w,3=1 z4=3 |
| | 2 | z2w,1=1 z2w,3=1 z4=3 |
| | 2 | z2w,1=1 z2w,2=1 z2w,3=1 z4=2 |
| 14.4.13 | 238311 | O6Os1Sr2Y1 |
| | 2 | z2w,1=1 z2w,2=1 z2w,3=1 z4=2 |
| | 2 | z2w,1=1 z2w,3=1 z4=1 |
| | 2 | z2w,3=1 z4=1 |
| | 2 | z2w,2=1 z2w,3=1 |
| | 2 | z2w,1=1 z2w,2=1 z4=1 |
| | 2 | z2w,1=1 |
| | 2 | z2w,2=1 z4=3 |
| | 2 | z2w,2=1 z4=1 |
| | 2 | z2w,1=1 |
| | 2 | z2w,1=1 z2w,2=1 z4=3 |
| | 2 | z2w,2=1 z2w,3=1 |
| | 2 | z2w,3=1 z4=3 |

Continued on next column

**TABLE XIX– continued**

| TQC | ICSD<br>SSG | Stoich. Formula<br>SIs |
|---|---|---|
| | 2 | z2w,1=1 z2w,3=1 z4=3 |
| | 2 | z2w,1=1 z2w,2=1 z2w,3=1 z4=2 |
| 14.4.14 | 193235 | Nb1O6Sr2Yb1 |
| | 2 | z2w,2=1 z2w,3=1 z4=2 |
| | 2 | z2w,3=1 z4=1 |
| | 2 | z2w,1=1 z2w,3=1 z4=1 |
| | 2 | z2w,1=1 z2w,2=1 z2w,3=1 |
| | 2 | z2w,2=1 z4=1 |
| | 2 | z2w,1=1 |
| | 2 | z2w,1=1 z2w,2=1 z4=3 |
| | 2 | z2w,1=1 z2w,2=1 z4=1 |
| | 2 | z2w,1=1 |
| | 2 | z2w,2=1 z4=3 |
| | 2 | z2w,1=1 z2w,2=1 z2w,3=1 |
| | 2 | z2w,1=1 z2w,3=1 z4=3 |
| | 2 | z2w,3=1 z4=3 |
| | 2 | z2w,2=1 z2w,3=1 z4=2 |
| 14.4.15 | 280907 | Mn1P2Rb2S6 |
| | 2 | z2w,1=1 z2w,2=1 z2w,3=1 z4=2 |
| | 2 | z2w,1=1 z2w,3=1 z4=1 |
| | 2 | z2w,3=1 z4=1 |
| | 2 | z2w,2=1 z2w,3=1 |
| | 2 | z2w,1=1 z2w,2=1 z4=1 |
| | 2 | z2w,1=1 |
| | 2 | z2w,2=1 z4=3 |
| | 2 | z2w,2=1 z4=1 |
| | 2 | z2w,1=1 |
| | 2 | z2w,1=1 z2w,2=1 z4=3 |
| | 2 | z2w,2=1 z2w,3=1 |
| | 2 | z2w,3=1 z4=3 |
| | 2 | z2w,1=1 z2w,3=1 z4=3 |
| | 2 | z2w,1=1 z2w,2=1 z2w,3=1 z4=2 |
| 14.4.16 | 69801 | Mn1O6Se2 |
| | 2 | z2w,2=1 z4=3 |
| | 2 | z2w,1=1 z2w,2=1 z4=3 |
| | 2 | z2w,1=1 z4=2 |
| | 2 | z2w,3=1 z4=3 |
| | 2 | z2w,2=1 z2w,3=1 z4=2 |
| | 2 | z2w,1=1 z2w,2=1 z2w,3=1 z4=2 |
| | 2 | z2w,1=1 z2w,3=1 z4=1 |
| | 2 | z2w,1=1 z2w,3=1 z4=3 |
| | 2 | z2w,1=1 z2w,2=1 z2w,3=1 z4=2 |
| | 2 | z2w,2=1 z2w,3=1 z4=2 |
| | 2 | z2w,3=1 z4=1 |
| | 2 | z2w,1=1 z4=2 |
| | 2 | z2w,1=1 z2w,2=1 z4=1 |
| | 2 | z2w,2=1 z4=1 |
| 14.4.17 | 202867 | Lu1O6Ru1Sr2 |
| | 2 | z2w,1=1 z4=2 |
| | 2 | z2w,1=1 z2w,2=1 z4=1 |
| | 2 | z2w,2=1 z4=1 |
| | 2 | z2w,1=1 z2w,3=1 z4=1 |
| | 2 | z2w,1=1 z2w,2=1 z2w,3=1 |
| | 2 | z2w,2=1 z2w,3=1 |
| | 2 | z2w,3=1 z4=3 |
| | 2 | z2w,3=1 z4=1 |
| | 2 | z2w,2=1 z2w,3=1 |
| | 2 | z2w,1=1 z2w,2=1 z2w,3=1 |
| | 2 | z2w,1=1 z2w,3=1 z4=3 |
| | 2 | z2w,2=1 z4=3 |

Continued on next column

**TABLE XIX– continued**

| TQC | ICSD SSG | Stoich. Formula SIs |
|---|---|---|
| | 2 | z2w,1=1 z2w,2=1 z4=3 |
| | 2 | z2w,1=1 z4=2 |
| 14.4.18 | 202866 | Lu1O6Ru1Sr2 |
| | 2 | z2w,1=1 z4=2 |
| | 2 | z2w,1=1 z2w,2=1 z4=1 |
| | 2 | z2w,2=1 z4=1 |
| | 2 | z2w,1=1 z2w,3=1 z4=1 |
| | 2 | z2w,1=1 z2w,2=1 z2w,3=1 |
| | 2 | z2w,2=1 z2w,3=1 |
| | 2 | z2w,3=1 z4=3 |
| | 2 | z2w,3=1 z4=1 |
| | 2 | z2w,2=1 z2w,3=1 |
| | 2 | z2w,1=1 z2w,2=1 z2w,3=1 |
| | 2 | z2w,1=1 z2w,3=1 z4=3 |
| | 2 | z2w,2=1 z4=3 |
| | 2 | z2w,1=1 z2w,2=1 z4=3 |
| | 2 | z2w,1=1 z4=2 |
| 14.4.19 | 88523 | La2O6Rh1Zn1 |
| | 2 | z2w,1=1 z4=2 |
| | 2 | z2w,1=1 z2w,2=1 z4=1 |
| | 2 | z2w,2=1 z4=1 |
| | 2 | z2w,1=1 z2w,3=1 z4=1 |
| | 2 | z2w,1=1 z2w,2=1 z2w,3=1 |
| | 2 | z2w,2=1 z2w,3=1 |
| | 2 | z2w,3=1 z4=3 |
| | 2 | z2w,3=1 z4=1 |
| | 2 | z2w,2=1 z2w,3=1 |
| | 2 | z2w,1=1 z2w,2=1 z2w,3=1 |
| | 2 | z2w,1=1 z2w,3=1 z4=3 |
| | 2 | z2w,2=1 z4=3 |
| | 2 | z2w,1=1 z2w,2=1 z4=3 |
| | 2 | z2w,1=1 z4=2 |
| 14.4.20 | 88522 | La2Mg1O6Rh1 |
| | 2 | z2w,1=1 z4=2 |
| | 2 | z2w,1=1 z2w,2=1 z4=1 |
| | 2 | z2w,2=1 z4=1 |
| | 2 | z2w,1=1 z2w,3=1 z4=1 |
| | 2 | z2w,1=1 z2w,2=1 z2w,3=1 |
| | 2 | z2w,2=1 z2w,3=1 |
| | 2 | z2w,3=1 z4=3 |
| | 2 | z2w,3=1 z4=1 |
| | 2 | z2w,2=1 z2w,3=1 |
| | 2 | z2w,1=1 z2w,2=1 z2w,3=1 |
| | 2 | z2w,1=1 z2w,3=1 z4=3 |
| | 2 | z2w,2=1 z4=3 |
| | 2 | z2w,1=1 z2w,2=1 z4=3 |
| | 2 | z2w,1=1 z4=2 |
| 14.4.21 | 97487 | La2Li1O6Ru1 |
| | 2 | z2w,1=1 z4=2 |
| | 2 | z2w,1=1 z2w,2=1 z4=1 |
| | 2 | z2w,2=1 z4=1 |
| | 2 | z2w,1=1 z2w,3=1 z4=1 |
| | 2 | z2w,1=1 z2w,2=1 z2w,3=1 |
| | 2 | z2w,2=1 z2w,3=1 |
| | 2 | z2w,3=1 z4=3 |
| | 2 | z2w,3=1 z4=1 |
| | 2 | z2w,2=1 z2w,3=1 |
| | 2 | z2w,1=1 z2w,2=1 z2w,3=1 |
| | 2 | z2w,1=1 z2w,3=1 z4=3 |
| | 2 | z2w,2=1 z4=3 |

Continued on next column

**TABLE XIX– continued**

| TQC | ICSD SSG | Stoich. Formula SIs |
|---|---|---|
| | 2 | z2w,1=1 z2w,2=1 z4=3 |
| | 2 | z2w,1=1 z4=2 |
| 14.4.22 | 81882 | Ir1La2O6Zn1 |
| | 2 | z2w,1=1 z4=2 |
| | 2 | z2w,1=1 z2w,2=1 z4=1 |
| | 2 | z2w,2=1 z4=1 |
| | 2 | z2w,1=1 z2w,3=1 z4=1 |
| | 2 | z2w,1=1 z2w,2=1 z2w,3=1 |
| | 2 | z2w,2=1 z2w,3=1 |
| | 2 | z2w,3=1 z4=3 |
| | 2 | z2w,3=1 z4=1 |
| | 2 | z2w,2=1 z2w,3=1 |
| | 2 | z2w,1=1 z2w,2=1 z2w,3=1 |
| | 2 | z2w,1=1 z2w,3=1 z4=3 |
| | 2 | z2w,2=1 z4=3 |
| | 2 | z2w,1=1 z2w,2=1 z4=3 |
| | 2 | z2w,1=1 z4=2 |
| 14.4.23 | 79492 | Ir1La2Mg1O6 |
| | 2 | z2w,2=1 z2w,3=1 z4=2 |
| | 2 | z2w,3=1 z4=1 |
| | 2 | z2w,1=1 z2w,3=1 z4=1 |
| | 2 | z2w,1=1 z2w,2=1 z2w,3=1 |
| | 2 | z2w,2=1 z4=1 |
| | 2 | z2w,1=1 |
| | 2 | z2w,1=1 z2w,2=1 z4=3 |
| | 2 | z2w,1=1 z2w,2=1 z4=1 |
| | 2 | z2w,1=1 |
| | 2 | z2w,2=1 z4=3 |
| | 2 | z2w,1=1 z2w,2=1 z2w,3=1 |
| | 2 | z2w,1=1 z2w,3=1 z4=3 |
| | 2 | z2w,3=1 z4=3 |
| | 2 | z2w,2=1 z2w,3=1 z4=2 |
| 14.4.24 | 238318 | In1O6Os1Sr2 |
| | 2 | z2w,1=1 z2w,2=1 z2w,3=1 z4=2 |
| | 2 | z2w,1=1 z2w,3=1 z4=1 |
| | 2 | z2w,3=1 z4=1 |
| | 2 | z2w,2=1 z2w,3=1 |
| | 2 | z2w,1=1 z2w,2=1 z4=1 |
| | 2 | z2w,1=1 |
| | 2 | z2w,2=1 z4=3 |
| | 2 | z2w,2=1 z4=1 |
| | 2 | z2w,1=1 |
| | 2 | z2w,1=1 z2w,2=1 z4=3 |
| | 2 | z2w,2=1 z2w,3=1 |
| | 2 | z2w,3=1 z4=3 |
| | 2 | z2w,1=1 z2w,3=1 z4=3 |
| | 2 | z2w,1=1 z2w,2=1 z2w,3=1 z4=2 |
| 14.4.25 | 238317 | In1O6Os1Sr2 |
| | 2 | z2w,1=1 z2w,2=1 z2w,3=1 z4=2 |
| | 2 | z2w,1=1 z2w,3=1 z4=1 |
| | 2 | z2w,3=1 z4=1 |
| | 2 | z2w,2=1 z2w,3=1 |
| | 2 | z2w,1=1 z2w,2=1 z4=1 |
| | 2 | z2w,1=1 |
| | 2 | z2w,2=1 z4=3 |
| | 2 | z2w,2=1 z4=1 |
| | 2 | z2w,1=1 |
| | 2 | z2w,1=1 z2w,2=1 z4=3 |
| | 2 | z2w,2=1 z2w,3=1 |
| | 2 | z2w,3=1 z4=3 |

Continued on next column

**TABLE XIX– continued**

| TQC | ICSD SSG | Stoich. Formula SIs |
|---|---|---|
| | 2 | z2w,1=1 z2w,3=1 z4=3 |
| | 2 | z2w,1=1 z2w,2=1 z2w,3=1 z4=2 |
| 14.4.26 | 65510 | I6K2Tc1 |
| | 2 | z2w,2=1 z4=3 |
| | 2 | z2w,1=1 z2w,2=1 z4=3 |
| | 2 | z2w,1=1 z4=2 |
| | 2 | z2w,3=1 z4=3 |
| | 2 | z2w,2=1 z2w,3=1 z4=2 |
| | 2 | z2w,1=1 z2w,2=1 z2w,3=1 z4=2 |
| | 2 | z2w,1=1 z2w,3=1 z4=1 |
| | 2 | z2w,1=1 z2w,3=1 z4=3 |
| | 2 | z2w,1=1 z2w,2=1 z2w,3=1 z4=2 |
| | 2 | z2w,2=1 z2w,3=1 z4=2 |
| | 2 | z2w,3=1 z4=1 |
| | 2 | z2w,1=1 z4=2 |
| | 2 | z2w,1=1 z2w,2=1 z4=1 |
| | 2 | z2w,2=1 z4=1 |
| 14.4.27 | 67555 | H8Br2Cu1O4 |
| | 2 | z2w,2=1 z4=3 |
| | 2 | z2w,1=1 z2w,2=1 z4=3 |
| | 2 | z2w,1=1 z4=2 |
| | 2 | z2w,3=1 z4=3 |
| | 2 | z2w,2=1 z2w,3=1 z4=2 |
| | 2 | z2w,1=1 z2w,2=1 z2w,3=1 z4=2 |
| | 2 | z2w,1=1 z2w,3=1 z4=1 |
| | 2 | z2w,1=1 z2w,3=1 z4=3 |
| | 2 | z2w,1=1 z2w,2=1 z2w,3=1 z4=2 |
| | 2 | z2w,2=1 z2w,3=1 z4=2 |
| | 2 | z2w,3=1 z4=1 |
| | 2 | z2w,1=1 z4=2 |
| | 2 | z2w,1=1 z2w,2=1 z4=1 |
| | 2 | z2w,2=1 z4=1 |
| 14.4.28 | 280918 | H4Cu1O4P2 |
| | 2 | z2w,2=1 z4=3 |
| | 2 | z2w,1=1 z2w,2=1 z4=3 |
| | 2 | z2w,1=1 z4=2 |
| | 2 | z2w,3=1 z4=3 |
| | 2 | z2w,2=1 z2w,3=1 z4=2 |
| | 2 | z2w,1=1 z2w,2=1 z2w,3=1 z4=2 |
| | 2 | z2w,1=1 z2w,3=1 z4=1 |
| | 2 | z2w,1=1 z2w,3=1 z4=3 |
| | 2 | z2w,1=1 z2w,2=1 z2w,3=1 z4=2 |
| | 2 | z2w,2=1 z2w,3=1 z4=2 |
| | 2 | z2w,3=1 z4=1 |
| | 2 | z2w,1=1 z4=2 |
| | 2 | z2w,1=1 z2w,2=1 z4=1 |
| | 2 | z2w,2=1 z4=1 |
| 14.4.29 | 280917 | H4Cu1O4P2 |
| | 2 | z2w,2=1 z4=3 |
| | 2 | z2w,1=1 z2w,2=1 z4=3 |
| | 2 | z2w,1=1 z4=2 |
| | 2 | z2w,3=1 z4=3 |
| | 2 | z2w,2=1 z2w,3=1 z4=2 |
| | 2 | z2w,1=1 z2w,2=1 z2w,3=1 z4=2 |
| | 2 | z2w,1=1 z2w,3=1 z4=1 |
| | 2 | z2w,1=1 z2w,3=1 z4=3 |
| | 2 | z2w,1=1 z2w,2=1 z2w,3=1 z4=2 |
| | 2 | z2w,2=1 z2w,3=1 z4=2 |
| | 2 | z2w,3=1 z4=1 |
| | 2 | z2w,1=1 z4=2 |

Continued on next column

**TABLE XIX– continued**

| TQC | ICSD SSG | Stoich. Formula SIs |
|---|---|---|
| | 2 | z2w,1=1 z2w,2=1 z4=1 |
| | 2 | z2w,2=1 z4=1 |
| 14.4.30 | 98699 | H2Cu3F4O2 |
| | 2 | z2w,2=1 z2w,3=1 z4=2 |
| | 2 | z2w,3=1 z4=1 |
| | 2 | z2w,1=1 z2w,3=1 z4=1 |
| | 2 | z2w,1=1 z2w,2=1 z2w,3=1 |
| | 2 | z2w,2=1 z4=1 |
| | 2 | z2w,1=1 |
| | 2 | z2w,1=1 z2w,2=1 z4=3 |
| | 2 | z2w,1=1 z2w,2=1 z4=1 |
| | 2 | z2w,1=1 |
| | 2 | z2w,2=1 z4=3 |
| | 2 | z2w,1=1 z2w,2=1 z2w,3=1 |
| | 2 | z2w,1=1 z2w,3=1 z4=3 |
| | 2 | z2w,3=1 z4=3 |
| | 2 | z2w,2=1 z2w,3=1 z4=2 |
| 14.4.31 | 290251 | F6Li2V1 |
| | 2 | z2w,2=1 z4=3 |
| | 2 | z2w,1=1 z2w,2=1 z4=3 |
| | 2 | z2w,1=1 z4=2 |
| | 2 | z2w,3=1 z4=3 |
| | 2 | z2w,2=1 z2w,3=1 z4=2 |
| | 2 | z2w,1=1 z2w,2=1 z2w,3=1 z4=2 |
| | 2 | z2w,1=1 z2w,3=1 z4=1 |
| | 2 | z2w,1=1 z2w,3=1 z4=3 |
| | 2 | z2w,1=1 z2w,2=1 z2w,3=1 z4=2 |
| | 2 | z2w,2=1 z2w,3=1 z4=2 |
| | 2 | z2w,3=1 z4=1 |
| | 2 | z2w,1=1 z4=2 |
| | 2 | z2w,1=1 z2w,2=1 z4=1 |
| | 2 | z2w,2=1 z4=1 |
| 14.4.33 | 420874 | Eu2Ir1Mg1O6 |
| | 2 | z2w,1=1 z2w,2=1 z2w,3=1 z4=2 |
| | 2 | z2w,1=1 z2w,3=1 z4=1 |
| | 2 | z2w,3=1 z4=1 |
| | 2 | z2w,2=1 z2w,3=1 |
| | 2 | z2w,1=1 z2w,2=1 z4=1 |
| | 2 | z2w,1=1 |
| | 2 | z2w,2=1 z4=3 |
| | 2 | z2w,2=1 z4=1 |
| | 2 | z2w,1=1 |
| | 2 | z2w,1=1 z2w,2=1 z4=3 |
| | 2 | z2w,2=1 z2w,3=1 |
| | 2 | z2w,3=1 z4=3 |
| | 2 | z2w,1=1 z2w,3=1 z4=3 |
| | 2 | z2w,1=1 z2w,2=1 z2w,3=1 z4=2 |
| 14.4.34 | 80576 | Cu1O6Sb2 |
| | 2 | z2w,1=1 z2w,2=1 z2w,3=1 z4=2 |
| | 2 | z2w,1=1 z2w,3=1 z4=1 |
| | 2 | z2w,3=1 z4=1 |
| | 2 | z2w,2=1 z2w,3=1 |
| | 2 | z2w,1=1 z2w,2=1 z4=1 |
| | 2 | z2w,1=1 |
| | 2 | z2w,2=1 z4=3 |
| | 2 | z2w,2=1 z4=1 |
| | 2 | z2w,1=1 |
| | 2 | z2w,1=1 z2w,2=1 z4=3 |
| | 2 | z2w,2=1 z2w,3=1 |
| | 2 | z2w,3=1 z4=3 |

Continued on next column

**TABLE XIX– continued**

| TQC | ICSD SSG | Stoich. Formula SIs |
|---|---|---|
| | 2 | z2w,1=1 z2w,3=1 z4=3 |
| | 2 | z2w,1=1 z2w,2=1 z2w,3=1 z4=2 |
| 14.4.35 | 36071 | Cu1O4U1 |
| | 2 | z2w,1=1 z2w,2=1 z2w,3=1 z4=2 |
| | 2 | z2w,1=1 z2w,3=1 z4=1 |
| | 2 | z2w,3=1 z4=1 |
| | 2 | z2w,2=1 z2w,3=1 |
| | 2 | z2w,1=1 z2w,2=1 z4=1 |
| | 2 | z2w,1=1 |
| | 2 | z2w,2=1 z4=3 |
| | 2 | z2w,2=1 z4=1 |
| | 2 | z2w,1=1 |
| | 2 | z2w,1=1 z2w,2=1 z4=3 |
| | 2 | z2w,2=1 z2w,3=1 |
| | 2 | z2w,3=1 z4=3 |
| | 2 | z2w,1=1 z2w,3=1 z4=3 |
| | 2 | z2w,1=1 z2w,2=1 z2w,3=1 z4=2 |
| 14.4.36 | 9790 | Cu1F2 |
| | 2 | z2w,2=1 z4=3 |
| | 2 | z2w,1=1 z2w,2=1 z4=3 |
| | 2 | z2w,1=1 z4=2 |
| | 2 | z2w,3=1 z4=3 |
| | 2 | z2w,2=1 z2w,3=1 z4=2 |
| | 2 | z2w,1=1 z2w,2=1 z2w,3=1 z4=2 |
| | 2 | z2w,1=1 z2w,3=1 z4=1 |
| | 2 | z2w,1=1 z2w,3=1 z4=3 |
| | 2 | z2w,1=1 z2w,2=1 z2w,3=1 z4=2 |
| | 2 | z2w,2=1 z2w,3=1 z4=2 |
| | 2 | z2w,3=1 z4=1 |
| | 2 | z2w,1=1 z4=2 |
| | 2 | z2w,1=1 z2w,2=1 z4=1 |
| | 2 | z2w,2=1 z4=1 |
| 14.4.37 | 9789 | Cu1F2 |
| | 2 | z2w,2=1 z4=3 |
| | 2 | z2w,1=1 z2w,2=1 z4=3 |
| | 2 | z2w,1=1 z4=2 |
| | 2 | z2w,3=1 z4=3 |
| | 2 | z2w,2=1 z2w,3=1 z4=2 |
| | 2 | z2w,1=1 z2w,2=1 z2w,3=1 z4=2 |
| | 2 | z2w,1=1 z2w,3=1 z4=1 |
| | 2 | z2w,1=1 z2w,3=1 z4=3 |
| | 2 | z2w,1=1 z2w,2=1 z2w,3=1 z4=2 |
| | 2 | z2w,2=1 z2w,3=1 z4=2 |
| | 2 | z2w,3=1 z4=1 |
| | 2 | z2w,1=1 z4=2 |
| | 2 | z2w,1=1 z2w,2=1 z4=1 |
| | 2 | z2w,2=1 z4=1 |
| 14.4.38 | 9788 | Cu1F2 |
| | 2 | z2w,2=1 z4=3 |
| | 2 | z2w,1=1 z2w,2=1 z4=3 |
| | 2 | z2w,1=1 z4=2 |
| | 2 | z2w,3=1 z4=3 |
| | 2 | z2w,2=1 z2w,3=1 z4=2 |
| | 2 | z2w,1=1 z2w,2=1 z2w,3=1 z4=2 |
| | 2 | z2w,1=1 z2w,3=1 z4=1 |
| | 2 | z2w,1=1 z2w,3=1 z4=3 |
| | 2 | z2w,1=1 z2w,2=1 z2w,3=1 z4=2 |
| | 2 | z2w,2=1 z2w,3=1 z4=2 |
| | 2 | z2w,3=1 z4=1 |
| | 2 | z2w,1=1 z4=2 |

Continued on next column

**TABLE XIX– continued**

| TQC | ICSD SSG | Stoich. Formula SIs |
|---|---|---|
| | 2 | z2w,1=1 z2w,2=1 z4=1 |
| | 2 | z2w,2=1 z4=1 |
| 14.4.39 | 71833 | Cu1F2 |
| | 2 | z2w,2=1 z4=3 |
| | 2 | z2w,1=1 z2w,2=1 z4=3 |
| | 2 | z2w,1=1 z4=2 |
| | 2 | z2w,3=1 z4=3 |
| | 2 | z2w,2=1 z2w,3=1 z4=2 |
| | 2 | z2w,1=1 z2w,2=1 z2w,3=1 z4=2 |
| | 2 | z2w,1=1 z2w,3=1 z4=1 |
| | 2 | z2w,1=1 z2w,3=1 z4=3 |
| | 2 | z2w,1=1 z2w,2=1 z2w,3=1 z4=2 |
| | 2 | z2w,2=1 z2w,3=1 z4=2 |
| | 2 | z2w,3=1 z4=1 |
| | 2 | z2w,1=1 z4=2 |
| | 2 | z2w,1=1 z2w,2=1 z4=1 |
| | 2 | z2w,2=1 z4=1 |
| 14.4.40 | 6024 | Cu1F2 |
| | 2 | z2w,2=1 z4=3 |
| | 2 | z2w,1=1 z2w,2=1 z4=3 |
| | 2 | z2w,1=1 z4=2 |
| | 2 | z2w,3=1 z4=3 |
| | 2 | z2w,2=1 z2w,3=1 z4=2 |
| | 2 | z2w,1=1 z2w,2=1 z2w,3=1 z4=2 |
| | 2 | z2w,1=1 z2w,3=1 z4=1 |
| | 2 | z2w,1=1 z2w,3=1 z4=3 |
| | 2 | z2w,1=1 z2w,2=1 z2w,3=1 z4=2 |
| | 2 | z2w,2=1 z2w,3=1 z4=2 |
| | 2 | z2w,3=1 z4=1 |
| | 2 | z2w,1=1 z4=2 |
| | 2 | z2w,1=1 z2w,2=1 z4=1 |
| | 2 | z2w,2=1 z4=1 |
| 14.4.41 | 173093 | Cu0.5O5P1Ti1 |
| | 2 | z2w,2=1 z4=3 |
| | 2 | z2w,1=1 z2w,2=1 z4=3 |
| | 2 | z2w,1=1 z4=2 |
| | 2 | z2w,3=1 z4=3 |
| | 2 | z2w,2=1 z2w,3=1 z4=2 |
| | 2 | z2w,1=1 z2w,2=1 z2w,3=1 z4=2 |
| | 2 | z2w,1=1 z2w,3=1 z4=1 |
| | 2 | z2w,1=1 z2w,3=1 z4=3 |
| | 2 | z2w,1=1 z2w,2=1 z2w,3=1 z4=2 |
| | 2 | z2w,2=1 z2w,3=1 z4=2 |
| | 2 | z2w,3=1 z4=1 |
| | 2 | z2w,1=1 z4=2 |
| | 2 | z2w,1=1 z2w,2=1 z4=1 |
| | 2 | z2w,2=1 z4=1 |
| 14.4.42 | 195658 | Co1O6Sr2Te1 |
| | 2 | z2w,1=1 z2w,2=1 z2w,3=1 z4=2 |
| | 2 | z2w,1=1 z2w,3=1 z4=1 |
| | 2 | z2w,3=1 z4=1 |
| | 2 | z2w,2=1 z2w,3=1 |
| | 2 | z2w,1=1 z2w,2=1 z4=1 |
| | 2 | z2w,1=1 |
| | 2 | z2w,2=1 z4=3 |
| | 2 | z2w,2=1 z4=1 |
| | 2 | z2w,1=1 |
| | 2 | z2w,1=1 z2w,2=1 z4=3 |
| | 2 | z2w,2=1 z2w,3=1 |
| | 2 | z2w,3=1 z4=3 |

Continued on next column

**TABLE XIX– continued**

| TQC | ICSD SSG | Stoich. Formula SIs |
|---|---|---|
| | 2 | z2w,1=1 z2w,3=1 z4=3 |
| | 2 | z2w,1=1 z2w,2=1 z2w,3=1 z4=2 |
| 14.4.43 | 195651 | Co1O6Sr2Te1 |
| | 2 | z2w,2=1 z4=3 |
| | 2 | z2w,1=1 z2w,2=1 z4=3 |
| | 2 | z2w,1=1 z4=2 |
| | 2 | z2w,3=1 z4=3 |
| | 2 | z2w,2=1 z2w,3=1 z4=2 |
| | 2 | z2w,1=1 z2w,2=1 z2w,3=1 z4=2 |
| | 2 | z2w,1=1 z2w,3=1 z4=1 |
| | 2 | z2w,1=1 z2w,3=1 z4=3 |
| | 2 | z2w,1=1 z2w,2=1 z2w,3=1 z4=2 |
| | 2 | z2w,2=1 z2w,3=1 z4=2 |
| | 2 | z2w,3=1 z4=1 |
| | 2 | z2w,1=1 z4=2 |
| | 2 | z2w,1=1 z2w,2=1 z4=1 |
| | 2 | z2w,2=1 z4=1 |
| 14.4.44 | 195650 | Co1O6Sr2Te1 |
| | 2 | z2w,1=1 z2w,2=1 z2w,3=1 z4=2 |
| | 2 | z2w,1=1 z2w,3=1 z4=1 |
| | 2 | z2w,3=1 z4=1 |
| | 2 | z2w,2=1 z2w,3=1 |
| | 2 | z2w,1=1 z2w,2=1 z4=1 |
| | 2 | z2w,1=1 |
| | 2 | z2w,2=1 z4=3 |
| | 2 | z2w,2=1 z4=1 |
| | 2 | z2w,1=1 |
| | 2 | z2w,1=1 z2w,2=1 z4=3 |
| | 2 | z2w,2=1 z2w,3=1 |
| | 2 | z2w,3=1 z4=3 |
| | 2 | z2w,1=1 z2w,3=1 z4=3 |
| | 2 | z2w,1=1 z2w,2=1 z2w,3=1 z4=2 |
| 14.4.45 | 153002 | Co1O6Sr2Te1 |
| | 2 | z2w,2=1 z2w,3=1 z4=2 |
| | 2 | z2w,3=1 z4=1 |
| | 2 | z2w,1=1 z2w,3=1 z4=1 |
| | 2 | z2w,1=1 z2w,2=1 z2w,3=1 |
| | 2 | z2w,2=1 z4=1 |
| | 2 | z2w,1=1 |
| | 2 | z2w,1=1 z2w,2=1 z4=3 |
| | 2 | z2w,1=1 z2w,2=1 z4=1 |
| | 2 | z2w,1=1 |
| | 2 | z2w,2=1 z4=3 |
| | 2 | z2w,1=1 z2w,2=1 z2w,3=1 |
| | 2 | z2w,1=1 z2w,3=1 z4=3 |
| | 2 | z2w,3=1 z4=3 |
| | 2 | z2w,2=1 z2w,3=1 z4=2 |
| 14.4.46 | 169193 | Co1O6Pb2Te1 |
| | 2 | z2w,2=1 z4=3 |
| | 2 | z2w,1=1 z2w,2=1 z4=3 |
| | 2 | z2w,1=1 z4=2 |
| | 2 | z2w,3=1 z4=3 |
| | 2 | z2w,2=1 z2w,3=1 z4=2 |
| | 2 | z2w,1=1 z2w,2=1 z2w,3=1 z4=2 |
| | 2 | z2w,1=1 z2w,3=1 z4=1 |
| | 2 | z2w,1=1 z2w,3=1 z4=3 |
| | 2 | z2w,1=1 z2w,2=1 z2w,3=1 z4=2 |
| | 2 | z2w,2=1 z2w,3=1 z4=2 |
| | 2 | z2w,3=1 z4=1 |
| | 2 | z2w,1=1 z4=2 |

Continued on next column

| TQC | ICSD SSG | Stoich. Formula SIs |
|---|---|---|
| | 2 | z2w,1=1 z2w,2=1 z4=1 |
| | 2 | z2w,2=1 z4=1 |
| 14.4.47 | 411351 | Cl4F6Ir1 |
| | 2 | z2w,2=1 z4=3 |
| | 2 | z2w,1=1 z2w,2=1 z4=3 |
| | 2 | z2w,1=1 z4=2 |
| | 2 | z2w,3=1 z4=3 |
| | 2 | z2w,2=1 z2w,3=1 z4=2 |
| | 2 | z2w,1=1 z2w,2=1 z2w,3=1 z4=2 |
| | 2 | z2w,1=1 z2w,3=1 z4=1 |
| | 2 | z2w,1=1 z2w,3=1 z4=3 |
| | 2 | z2w,1=1 z2w,2=1 z2w,3=1 z4=2 |
| | 2 | z2w,2=1 z2w,3=1 z4=2 |
| | 2 | z2w,3=1 z4=1 |
| | 2 | z2w,1=1 z4=2 |
| | 2 | z2w,1=1 z2w,2=1 z4=1 |
| | 2 | z2w,2=1 z4=1 |
| 14.4.48 | 150697 | Ca2Fe1O6Re1 |
| | 2 | z2w,2=1 z2w,3=1 z4=2 |
| | 2 | z2w,3=1 z4=1 |
| | 2 | z2w,1=1 z2w,3=1 z4=1 |
| | 2 | z2w,1=1 z2w,2=1 z2w,3=1 |
| | 2 | z2w,2=1 z4=1 |
| | 2 | z2w,1=1 |
| | 2 | z2w,1=1 z2w,2=1 z4=3 |
| | 2 | z2w,1=1 z2w,2=1 z4=1 |
| | 2 | z2w,1=1 |
| | 2 | z2w,2=1 z4=3 |
| | 2 | z2w,1=1 z2w,2=1 z2w,3=1 |
| | 2 | z2w,1=1 z2w,3=1 z4=3 |
| | 2 | z2w,3=1 z4=3 |
| | 2 | z2w,2=1 z2w,3=1 z4=2 |
| 14.4.49 | 98732 | Ca2Co1O6W1 |
| | 2 | z2w,2=1 z2w,3=1 z4=2 |
| | 2 | z2w,3=1 z4=1 |
| | 2 | z2w,1=1 z2w,3=1 z4=1 |
| | 2 | z2w,1=1 z2w,2=1 z2w,3=1 |
| | 2 | z2w,2=1 z4=1 |
| | 2 | z2w,1=1 |
| | 2 | z2w,1=1 z2w,2=1 z4=3 |
| | 2 | z2w,1=1 z2w,2=1 z4=1 |
| | 2 | z2w,1=1 |
| | 2 | z2w,2=1 z4=3 |
| | 2 | z2w,1=1 z2w,2=1 z2w,3=1 |
| | 2 | z2w,1=1 z2w,3=1 z4=3 |
| | 2 | z2w,3=1 z4=3 |
| | 2 | z2w,2=1 z2w,3=1 z4=2 |
| 14.4.50 | 417591 | C2H2Cu3O8 |
| | 2 | z2w,1=1 z2w,2=1 z2w,3=1 z4=1 |
| | 2 | z2w,1=1 z2w,3=1 |
| | 2 | z2w,3=1 |
| | 2 | z2w,2=1 z2w,3=1 z4=3 |
| | 2 | z2w,1=1 z2w,2=1 |
| | 2 | z2w,1=1 z4=3 |
| | 2 | z4=3 |
| | 2 | z2w,2=1 z4=2 |
| | 2 | z2w,2=1 |
| | 2 | z4=3 |
| | 2 | z2w,1=1 z4=3 |
| | 2 | z2w,1=1 z2w,2=1 z4=2 |

Continued on next column

**TABLE XIX– continued**

| TQC | ICSD SSG | Stoich. Formula SIs |
|---|---|---|
| | 2 | z2w,2=1 z2w,3=1 z4=3 |
| | 2 | z2w,3=1 z4=2 |
| | 2 | z2w,1=1 z2w,3=1 z4=2 |
| | 2 | z2w,1=1 z2w,2=1 z2w,3=1 z4=1 |
| 14.4.51 | 2934 | C2H2Cu3O8 |
| | 2 | z2w,1=1 z4=1 |
| | 2 | z2w,1=1 z2w,2=1 |
| | 2 | z2w,2=1 |
| | 2 | z4=3 |
| | 2 | z2w,1=1 z2w,3=1 |
| | 2 | z2w,1=1 z2w,2=1 z2w,3=1 z4=3 |
| | 2 | z2w,2=1 z2w,3=1 z4=3 |
| | 2 | z2w,3=1 z4=2 |
| | 2 | z2w,3=1 |
| | 2 | z2w,2=1 z2w,3=1 z4=3 |
| | 2 | z2w,1=1 z2w,2=1 z2w,3=1 z4=3 |
| | 2 | z2w,1=1 z2w,3=1 z4=2 |
| | 2 | z4=3 |
| | 2 | z2w,2=1 z4=2 |
| | 2 | z2w,1=1 z2w,2=1 z4=2 |
| | 2 | z2w,1=1 z4=1 |
| 14.4.52 | 182324 | C2H2Cu3O8 |
| | 2 | z2w,1=1 z4=2 |
| | 2 | z2w,1=1 z2w,2=1 z4=1 |
| | 2 | z2w,2=1 z4=1 |
| | 2 | z2w,1=1 z2w,3=1 z4=1 |
| | 2 | z2w,1=1 z2w,2=1 z2w,3=1 |
| | 2 | z2w,2=1 z2w,3=1 |
| | 2 | z2w,3=1 z4=3 |
| | 2 | z2w,3=1 z4=1 |
| | 2 | z2w,2=1 z2w,3=1 |
| | 2 | z2w,1=1 z2w,2=1 z2w,3=1 |
| | 2 | z2w,1=1 z2w,3=1 z4=3 |
| | 2 | z2w,2=1 z4=3 |
| | 2 | z2w,1=1 z2w,2=1 z4=3 |
| | 2 | z2w,1=1 z4=2 |
| 14.4.53 | 158577 | C2H2Cu3O8 |
| | 2 | z2w,1=1 z4=1 |
| | 2 | z2w,1=1 z2w,2=1 |
| | 2 | z2w,2=1 |
| | 2 | z4=3 |
| | 2 | z2w,1=1 z2w,3=1 |
| | 2 | z2w,1=1 z2w,2=1 z2w,3=1 z4=3 |
| | 2 | z2w,2=1 z2w,3=1 z4=3 |
| | 2 | z2w,3=1 z4=2 |
| | 2 | z2w,3=1 |
| | 2 | z2w,2=1 z2w,3=1 z4=3 |
| | 2 | z2w,1=1 z2w,2=1 z2w,3=1 z4=3 |
| | 2 | z2w,1=1 z2w,3=1 z4=2 |
| | 2 | z4=3 |
| | 2 | z2w,2=1 z4=2 |
| | 2 | z2w,1=1 z2w,2=1 z4=2 |
| | 2 | z2w,1=1 z4=1 |
| 14.4.54 | 60858 | C2Cu1Na2O6 |
| | 2 | z2w,2=1 z4=3 |
| | 2 | z2w,1=1 z2w,2=1 z4=3 |
| | 2 | z2w,1=1 z4=2 |
| | 2 | z2w,3=1 z4=3 |
| | 2 | z2w,2=1 z2w,3=1 z4=2 |
| | 2 | z2w,1=1 z2w,2=1 z2w,3=1 z4=2 |

Continued on next column

**TABLE XIX– continued**

| TQC | ICSD SSG | Stoich. Formula SIs |
|---|---|---|
| | 2 | z2w,1=1 z2w,3=1 z4=1 |
| | 2 | z2w,1=1 z2w,3=1 z4=3 |
| | 2 | z2w,1=1 z2w,2=1 z2w,3=1 z4=2 |
| | 2 | z2w,2=1 z2w,3=1 z4=2 |
| | 2 | z2w,3=1 z4=1 |
| | 2 | z2w,1=1 z4=2 |
| | 2 | z2w,1=1 z2w,2=1 z4=1 |
| | 2 | z2w,2=1 z4=1 |
| 14.4.55 | 426579 | C2Cu1Na2O6 |
| | 2 | z2w,2=1 z4=3 |
| | 2 | z2w,1=1 z2w,2=1 z4=3 |
| | 2 | z2w,1=1 z4=2 |
| | 2 | z2w,3=1 z4=3 |
| | 2 | z2w,2=1 z2w,3=1 z4=2 |
| | 2 | z2w,1=1 z2w,2=1 z2w,3=1 z4=2 |
| | 2 | z2w,1=1 z2w,3=1 z4=1 |
| | 2 | z2w,1=1 z2w,3=1 z4=3 |
| | 2 | z2w,1=1 z2w,2=1 z2w,3=1 z4=2 |
| | 2 | z2w,2=1 z2w,3=1 z4=2 |
| | 2 | z2w,3=1 z4=1 |
| | 2 | z2w,1=1 z4=2 |
| | 2 | z2w,1=1 z2w,2=1 z4=1 |
| | 2 | z2w,2=1 z4=1 |
| 14.4.56 | 10355 | C2Cu1Na2O6 |
| | 2 | z2w,2=1 z4=3 |
| | 2 | z2w,1=1 z2w,2=1 z4=3 |
| | 2 | z2w,1=1 z4=2 |
| | 2 | z2w,3=1 z4=3 |
| | 2 | z2w,2=1 z2w,3=1 z4=2 |
| | 2 | z2w,1=1 z2w,2=1 z2w,3=1 z4=2 |
| | 2 | z2w,1=1 z2w,3=1 z4=1 |
| | 2 | z2w,1=1 z2w,3=1 z4=3 |
| | 2 | z2w,1=1 z2w,2=1 z2w,3=1 z4=2 |
| | 2 | z2w,2=1 z2w,3=1 z4=2 |
| | 2 | z2w,3=1 z4=1 |
| | 2 | z2w,1=1 z4=2 |
| | 2 | z2w,1=1 z2w,2=1 z4=1 |
| | 2 | z2w,2=1 z4=1 |
| 14.4.57 | 155317 | B2Cu1O6Pb2 |
| | 2 | z2w,1=1 z2w,2=1 z2w,3=1 z4=2 |
| | 2 | z2w,1=1 z2w,3=1 z4=1 |
| | 2 | z2w,3=1 z4=1 |
| | 2 | z2w,2=1 z2w,3=1 |
| | 2 | z2w,1=1 z2w,2=1 z4=1 |
| | 2 | z2w,1=1 |
| | 2 | z2w,2=1 z4=3 |
| | 2 | z2w,2=1 z4=1 |
| | 2 | z2w,1=1 |
| | 2 | z2w,1=1 z2w,2=1 z4=3 |
| | 2 | z2w,2=1 z2w,3=1 |
| | 2 | z2w,3=1 z4=3 |
| | 2 | z2w,1=1 z2w,3=1 z4=3 |
| | 2 | z2w,1=1 z2w,2=1 z2w,3=1 z4=2 |
| 14.4.58 | 89620 | Au3F8 |
| | 2 | z2w,2=1 z2w,3=1 z4=2 |
| | 2 | z2w,3=1 z4=1 |
| | 2 | z2w,1=1 z2w,3=1 z4=1 |
| | 2 | z2w,1=1 z2w,2=1 z2w,3=1 |
| | 2 | z2w,2=1 z4=1 |
| | 2 | z2w,1=1 |

Continued on next column

**TABLE XIX– continued**

| TQC | ICSD SSG | Stoich. Formula SIs |
|---|---|---|
| | 2 | z2w,1=1 z2w,2=1 z4=3 |
| | 2 | z2w,1=1 z2w,2=1 z4=1 |
| | 2 | z2w,1=1 |
| | 2 | z2w,2=1 z4=3 |
| | 2 | z2w,1=1 z2w,2=1 z2w,3=1 |
| | 2 | z2w,1=1 z2w,3=1 z4=3 |
| | 2 | z2w,3=1 z4=3 |
| | 2 | z2w,2=1 z2w,3=1 z4=2 |
| 14.4.59 | 58154 | Al9Rh2 |
| | 2 | z4=3 |
| | 2 | z2w,2=1 z4=2 |
| | 2 | z2w,1=1 z2w,2=1 z4=2 |
| | 2 | z2w,1=1 z4=1 |
| | 2 | z2w,3=1 z4=2 |
| | 2 | z2w,2=1 z2w,3=1 z4=1 |
| | 2 | z2w,1=1 z2w,2=1 z2w,3=1 z4=1 |
| | 2 | z2w,1=1 z2w,3=1 |
| | 2 | z2w,1=1 z2w,3=1 z4=2 |
| | 2 | z2w,1=1 z2w,2=1 z2w,3=1 z4=1 |
| | 2 | z2w,2=1 z2w,3=1 z4=1 |
| | 2 | z2w,3=1 |
| | 2 | z2w,1=1 z4=1 |
| | 2 | z2w,1=1 z2w,2=1 |
| | 2 | z2w,2=1 |
| | 2 | z4=3 |
| 14.4.60 | 414304 | Al9Rh2 |
| | 2 | z2w,1=1 z4=2 |
| | 2 | z2w,1=1 z2w,2=1 z4=1 |
| | 2 | z2w,2=1 z4=1 |
| | 2 | z2w,1=1 z2w,3=1 z4=1 |
| | 2 | z2w,1=1 z2w,2=1 z2w,3=1 |
| | 2 | z2w,2=1 z2w,3=1 |
| | 2 | z2w,3=1 z4=3 |
| | 2 | z2w,3=1 z4=1 |
| | 2 | z2w,2=1 z2w,3=1 |
| | 2 | z2w,1=1 z2w,2=1 z2w,3=1 |
| | 2 | z2w,1=1 z2w,3=1 z4=3 |
| | 2 | z2w,2=1 z4=3 |
| | 2 | z2w,1=1 z2w,2=1 z4=3 |
| | 2 | z2w,1=1 z4=2 |
| 14.4.61 | 414302 | Al9Ir2 |
| | 2 | z2w,1=1 z4=2 |
| | 2 | z2w,1=1 z2w,2=1 z4=1 |
| | 2 | z2w,2=1 z4=1 |
| | 2 | z2w,1=1 z2w,3=1 z4=1 |
| | 2 | z2w,1=1 z2w,2=1 z2w,3=1 |
| | 2 | z2w,2=1 z2w,3=1 |
| | 2 | z2w,3=1 z4=3 |
| | 2 | z2w,3=1 z4=1 |
| | 2 | z2w,2=1 z2w,3=1 |
| | 2 | z2w,1=1 z2w,2=1 z2w,3=1 |
| | 2 | z2w,1=1 z2w,3=1 z4=3 |
| | 2 | z2w,2=1 z4=3 |
| | 2 | z2w,1=1 z2w,2=1 z4=3 |
| | 2 | z2w,1=1 z4=2 |
| 14.4.62 | 59225 | Ag3O4 |
| | 2 | z2w,2=1 z2w,3=1 z4=1 |
| | 2 | z2w,3=1 |
| | 2 | z2w,1=1 z2w,3=1 |
| | 2 | z2w,1=1 z2w,2=1 z2w,3=1 z4=3 |

Continued on next column

**TABLE XIX– continued**

| TQC | ICSD SSG | Stoich. Formula SIs |
|---|---|---|
| | 2 | z2w,2=1 |
| | 2 | z4=3 |
| | 2 | z2w,1=1 z4=3 |
| | 2 | z2w,1=1 z2w,2=1 z4=2 |
| | 2 | z2w,1=1 z2w,2=1 |
| | 2 | z2w,1=1 z4=3 |
| | 2 | z4=3 |
| | 2 | z2w,2=1 z4=2 |
| | 2 | z2w,1=1 z2w,2=1 z2w,3=1 z4=3 |
| | 2 | z2w,1=1 z2w,3=1 z4=2 |
| | 2 | z2w,3=1 z4=2 |
| | 2 | z2w,2=1 z2w,3=1 z4=1 |
| 14.4.63 | 202218 | Ag3O4 |
| | 2 | z2w,2=1 z2w,3=1 z4=1 |
| | 2 | z2w,3=1 |
| | 2 | z2w,1=1 z2w,3=1 |
| | 2 | z2w,1=1 z2w,2=1 z2w,3=1 z4=3 |
| | 2 | z2w,2=1 |
| | 2 | z4=3 |
| | 2 | z2w,1=1 z4=3 |
| | 2 | z2w,1=1 z2w,2=1 z4=2 |
| | 2 | z2w,1=1 z2w,2=1 |
| | 2 | z2w,1=1 z4=3 |
| | 2 | z4=3 |
| | 2 | z2w,2=1 z4=2 |
| | 2 | z2w,1=1 z2w,2=1 z2w,3=1 z4=3 |
| | 2 | z2w,1=1 z2w,3=1 z4=2 |
| | 2 | z2w,3=1 z4=2 |
| | 2 | z2w,2=1 z2w,3=1 z4=1 |
| 14.4.64 | 421461 | Ag1F4K2 |
| | 2 | z2w,2=1 z4=3 |
| | 2 | z2w,1=1 z2w,2=1 z4=3 |
| | 2 | z2w,1=1 z4=2 |
| | 2 | z2w,3=1 z4=3 |
| | 2 | z2w,2=1 z2w,3=1 z4=2 |
| | 2 | z2w,1=1 z2w,2=1 z2w,3=1 z4=2 |
| | 2 | z2w,1=1 z2w,3=1 z4=1 |
| | 2 | z2w,1=1 z2w,3=1 z4=3 |
| | 2 | z2w,1=1 z2w,2=1 z2w,3=1 z4=2 |
| | 2 | z2w,2=1 z2w,3=1 z4=2 |
| | 2 | z2w,3=1 z4=1 |
| | 2 | z2w,1=1 z4=2 |
| | 2 | z2w,1=1 z2w,2=1 z4=1 |
| | 2 | z2w,2=1 z4=1 |
| 15.4.1 | 40760 | Rh3Si5Y2 |
| | 2 | z2w,1=1 z2w,3=1 z4=2 |
| | 2 | z2w,2=1 z4=1 |
| | 2 | z2w,1=1 z4=1 |
| | 2 | z2w,2=1 z2w,3=1 |
| 15.4.2 | 74837 | Rb1Te6 |
| | 2 | z2w,1=1 z2w,3=1 z4=3 |
| | 2 | z2w,2=1 z4=2 |
| | 2 | z2w,1=1 z4=2 |
| | 2 | z2w,2=1 z2w,3=1 z4=1 |
| 15.4.3 | 39762 | O7P2V1 |
| | 2 | z2w,2=1 z2w,3=1 z4=1 |
| | 2 | z2w,1=1 |
| | 2 | z2w,2=1 |
| | 2 | z2w,1=1 z2w,3=1 z4=3 |

Continued on next column

**TABLE XIX– continued**

| TQC | ICSD<br>SSG | Stoich. Formula<br>SIs |
|---|---|---|
| 15.4.4 | 160944 | O7P2V1 |
| | 2 | z2w,2=1 z2w,3=1 z4=1 |
| | 2 | z2w,1=1 |
| | 2 | z2w,2=1 |
| | 2 | z2w,1=1 z2w,3=1 z4=3 |
| 15.4.5 | 160943 | O7P2V1 |
| | 2 | z2w,1=1 z2w,3=1 z4=1 |
| | 2 | z2w,2=1 |
| | 2 | z2w,1=1 |
| | 2 | z2w,2=1 z2w,3=1 z4=3 |
| 15.4.6 | 160942 | O7P2V1 |
| | 2 | z2w,1=1 z2w,3=1 z4=1 |
| | 2 | z2w,2=1 |
| | 2 | z2w,1=1 |
| | 2 | z2w,2=1 z2w,3=1 z4=3 |
| 15.4.7 | 51023 | O6Re2Sb1 |
| | 2 | z2w,1=1 z2w,3=1 |
| | 2 | z2w,2=1 z4=3 |
| | 2 | z2w,1=1 z4=3 |
| | 2 | z2w,2=1 z2w,3=1 z4=2 |
| 15.4.8 | 32587 | O5V3 |
| | 2 | z2w,2=1 z2w,3=1 z4=1 |
| | 2 | z2w,1=1 |
| | 2 | z2w,2=1 |
| | 2 | z2w,1=1 z2w,3=1 z4=3 |
| 15.4.9 | 27800 | O5Sb2V1 |
| | 2 | z2w,2=1 z2w,3=1 z4=1 |
| | 2 | z2w,1=1 |
| | 2 | z2w,2=1 |
| | 2 | z2w,1=1 z2w,3=1 z4=3 |
| 15.4.10 | 2292 | O5Sb2V1 |
| | 2 | z2w,2=1 z2w,3=1 z4=1 |
| | 2 | z2w,1=1 |
| | 2 | z2w,2=1 |
| | 2 | z2w,1=1 z2w,3=1 z4=3 |
| 15.4.11 | 95210 | O4Ta1Yb1 |
| | 2 | z2w,1=1 z2w,3=1 z4=2 |
| | 2 | z2w,2=1 z4=1 |
| | 2 | z2w,1=1 z4=1 |
| | 2 | z2w,2=1 z2w,3=1 |
| 15.4.12 | 430121 | Ni8Si3Tb1 |
| | 2 | z2w,2=1 z2w,3=1 z4=1 |
| | 2 | z2w,1=1 |
| | 2 | z2w,2=1 |
| | 2 | z2w,1=1 z2w,3=1 z4=3 |
| 15.4.13 | 281609 | Na1O6Si2Ti1 |
| | 2 | z2w,1=1 z2w,3=1 |
| | 2 | z2w,2=1 z4=3 |
| | 2 | z2w,1=1 z4=3 |
| | 2 | z2w,2=1 z2w,3=1 z4=2 |
| 15.4.14 | 281607 | Na1O6Si2Ti1 |
| | 2 | z2w,1=1 z2w,3=1 |
| | 2 | z2w,2=1 z4=3 |
| | 2 | z2w,1=1 z4=3 |
| | 2 | z2w,2=1 z2w,3=1 z4=2 |
| 15.4.15 | 281606 | Na1O6Si2Ti1 |
| | 2 | z2w,1=1 z2w,3=1 |
| | 2 | z2w,2=1 z4=3 |
| | 2 | z2w,1=1 z4=3 |

Continued on next column

**TABLE XIX– continued**

| TQC | ICSD<br>SSG | Stoich. Formula<br>SIs |
|---|---|---|
| | 2 | z2w,2=1 z2w,3=1 z4=2 |
| 15.4.16 | 281604 | Na1O6Si2Ti1 |
| | 2 | z2w,1=1 z2w,3=1 |
| | 2 | z2w,2=1 z4=3 |
| | 2 | z2w,1=1 z4=3 |
| | 2 | z2w,2=1 z2w,3=1 z4=2 |
| 15.4.17 | 248826 | Na0.5O2V1 |
| | 2 | z2w,1=1 z2w,3=1 z4=2 |
| | 2 | z2w,2=1 z4=1 |
| | 2 | z2w,1=1 z4=1 |
| | 2 | z2w,2=1 z2w,3=1 |
| 15.4.18 | 425216 | Mn3O4Rb2 |
| | 2 | z2w,2=1 z2w,3=1 z4=2 |
| | 2 | z2w,1=1 z4=1 |
| | 2 | z2w,2=1 z4=1 |
| | 2 | z2w,1=1 z2w,3=1 |
| 15.4.19 | 280877 | K1O8W2Yb1 |
| | 2 | z2w,1=1 z2w,3=1 z4=2 |
| | 2 | z2w,2=1 z4=1 |
| | 2 | z2w,1=1 z4=1 |
| | 2 | z2w,2=1 z2w,3=1 |
| 15.4.20 | 87184 | K1O2 |
| | 2 | z2w,1=1 z2w,3=1 z4=2 |
| | 2 | z2w,2=1 z4=1 |
| | 2 | z2w,1=1 z4=1 |
| | 2 | z2w,2=1 z2w,3=1 |
| 15.4.21 | 104493 | Ir2Lu5 |
| | 2 | z2w,1=1 z2w,3=1 z4=2 |
| | 2 | z2w,2=1 z4=1 |
| | 2 | z2w,1=1 z4=1 |
| | 2 | z2w,2=1 z2w,3=1 |
| 15.4.22 | 187813 | Ir1O3Sr1 |
| | 2 | z2w,1=1 z2w,3=1 z4=3 |
| | 2 | z2w,2=1 z4=2 |
| | 2 | z2w,1=1 z4=2 |
| | 2 | z2w,2=1 z2w,3=1 z4=1 |
| 15.4.23 | 391433 | In1Pt2Sr2 |
| | 2 | z2w,2=1 z2w,3=1 z4=2 |
| | 2 | z2w,1=1 z4=1 |
| | 2 | z2w,2=1 z4=1 |
| | 2 | z2w,1=1 z2w,3=1 |
| 15.4.24 | 391432 | In1Pd2Sr2 |
| | 2 | z2w,1=1 z2w,3=1 z4=2 |
| | 2 | z2w,2=1 z4=1 |
| | 2 | z2w,1=1 z4=1 |
| | 2 | z2w,2=1 z2w,3=1 |
| 15.4.25 | 412275 | I4La1Sr1 |
| | 2 | z2w,2=1 z2w,3=1 z4=3 |
| | 2 | z2w,1=1 z4=2 |
| | 2 | z2w,2=1 z4=2 |
| | 2 | z2w,1=1 z2w,3=1 z4=1 |
| 15.4.26 | 16738 | H5Ca1O6P1 |
| | 2 | z2w,2=1 z2w,3=1 z4=2 |
| | 2 | z2w,1=1 z4=1 |
| | 2 | z2w,2=1 z4=1 |
| | 2 | z2w,1=1 z2w,3=1 |
| 15.4.27 | 16727 | H4O6V2 |
| | 2 | z2w,1=1 z2w,3=1 z4=1 |
| | 2 | z2w,2=1 |

Continued on next column

**TABLE XIX– continued**

| TQC | ICSD SSG | Stoich. Formula SIs |
|---|---|---|
| | 2 | z2w,1=1 |
| | 2 | z2w,2=1 z2w,3=1 z4=3 |
| 15.4.28 | 412876 | H4Cl6N1W1 |
| | 2 | z2w,1=1 z2w,3=1 z4=2 |
| | 2 | z2w,2=1 z4=1 |
| | 2 | z2w,1=1 z4=1 |
| | 2 | z2w,2=1 z2w,3=1 |
| 15.4.29 | 195740 | H3K2N1Se2 |
| | 2 | z2w,2=1 z2w,3=1 z4=2 |
| | 2 | z2w,1=1 z4=1 |
| | 2 | z2w,2=1 z4=1 |
| | 2 | z2w,1=1 z2w,3=1 |
| 15.4.30 | 16628 | H1K2N1O6S2 |
| | 2 | z2w,1=1 z2w,3=1 z4=1 |
| | 2 | z2w,2=1 |
| | 2 | z2w,1=1 |
| | 2 | z2w,2=1 z2w,3=1 z4=3 |
| 15.4.31 | 15499 | H1Cs1F2O6S2 |
| | 2 | z2w,1=1 z2w,3=1 |
| | 2 | z2w,2=1 z4=3 |
| | 2 | z2w,1=1 z4=3 |
| | 2 | z2w,2=1 z2w,3=1 z4=2 |
| 15.4.32 | 109186 | Gd1O4Ta1 |
| | 2 | z2w,2=1 z2w,3=1 z4=3 |
| | 2 | z2w,1=1 z4=2 |
| | 2 | z2w,2=1 z4=2 |
| | 2 | z2w,1=1 z2w,3=1 z4=1 |
| 15.4.33 | 40781 | Fe1Rb1Se2 |
| | 2 | z2w,1=1 z2w,3=1 z4=3 |
| | 2 | z2w,2=1 z4=2 |
| | 2 | z2w,1=1 z4=2 |
| | 2 | z2w,2=1 z2w,3=1 z4=1 |
| 15.4.34 | 202385 | Fe1Rb1S2 |
| | 2 | z2w,1=1 z2w,3=1 z4=3 |
| | 2 | z2w,2=1 z4=2 |
| | 2 | z2w,1=1 z4=2 |
| | 2 | z2w,2=1 z2w,3=1 z4=1 |
| 15.4.35 | 202384 | Fe1Rb1S2 |
| | 2 | z2w,1=1 z2w,3=1 z4=3 |
| | 2 | z2w,2=1 z4=2 |
| | 2 | z2w,1=1 z4=2 |
| | 2 | z2w,2=1 z2w,3=1 z4=1 |
| 15.4.36 | 202381 | Fe1Rb1S2 |
| | 2 | z2w,2=1 z2w,3=1 z4=3 |
| | 2 | z2w,1=1 z4=2 |
| | 2 | z2w,2=1 z4=2 |
| | 2 | z2w,1=1 z2w,3=1 z4=1 |
| 15.4.37 | 9671 | Fe1Na1O6Si2 |
| | 2 | z2w,1=1 z2w,3=1 z4=2 |
| | 2 | z2w,2=1 z4=1 |
| | 2 | z2w,1=1 z4=1 |
| | 2 | z2w,2=1 z2w,3=1 |
| 15.4.38 | 180313 | Fe1Na1O6Si2 |
| | 2 | z2w,2=1 z2w,3=1 z4=2 |
| | 2 | z2w,1=1 z4=1 |
| | 2 | z2w,2=1 z4=1 |
| | 2 | z2w,1=1 z2w,3=1 |
| 15.4.39 | 180312 | Fe1Na1O6Si2 |
| | 2 | z2w,2=1 z2w,3=1 z4=2 |

Continued on next column

**TABLE XIX– continued**

| TQC | ICSD SSG | Stoich. Formula SIs |
|---|---|---|
| | 2 | z2w,1=1 z4=1 |
| | 2 | z2w,2=1 z4=1 |
| | 2 | z2w,1=1 z2w,3=1 |
| 15.4.40 | 180311 | Fe1Na1O6Si2 |
| | 2 | z2w,2=1 z2w,3=1 z4=2 |
| | 2 | z2w,1=1 z4=1 |
| | 2 | z2w,2=1 z4=1 |
| | 2 | z2w,1=1 z2w,3=1 |
| 15.4.41 | 161982 | Fe1Na1O6Si2 |
| | 2 | z2w,1=1 z2w,3=1 z4=3 |
| | 2 | z2w,2=1 z4=2 |
| | 2 | z2w,1=1 z4=2 |
| | 2 | z2w,2=1 z2w,3=1 z4=1 |
| 15.4.42 | 161980 | Fe1Na1O6Si2 |
| | 2 | z2w,2=1 z2w,3=1 z4=2 |
| | 2 | z2w,1=1 z4=1 |
| | 2 | z2w,2=1 z4=1 |
| | 2 | z2w,1=1 z2w,3=1 |
| 15.4.43 | 161979 | Fe1Na1O6Si2 |
| | 2 | z2w,2=1 z2w,3=1 z4=2 |
| | 2 | z2w,1=1 z4=1 |
| | 2 | z2w,2=1 z4=1 |
| | 2 | z2w,1=1 z2w,3=1 |
| 15.4.44 | 161978 | Fe1Na1O6Si2 |
| | 2 | z2w,2=1 z2w,3=1 z4=2 |
| | 2 | z2w,1=1 z4=1 |
| | 2 | z2w,2=1 z4=1 |
| | 2 | z2w,1=1 z2w,3=1 |
| 15.4.45 | 161977 | Fe1Na1O6Si2 |
| | 2 | z2w,2=1 z2w,3=1 z4=2 |
| | 2 | z2w,1=1 z4=1 |
| | 2 | z2w,2=1 z4=1 |
| | 2 | z2w,1=1 z2w,3=1 |
| 15.4.46 | 161976 | Fe1Na1O6Si2 |
| | 2 | z2w,1=1 z2w,3=1 z4=2 |
| | 2 | z2w,2=1 z4=1 |
| | 2 | z2w,1=1 z4=1 |
| | 2 | z2w,2=1 z2w,3=1 |
| 15.4.47 | 161975 | Fe1Na1O6Si2 |
| | 2 | z2w,2=1 z2w,3=1 z4=2 |
| | 2 | z2w,1=1 z4=1 |
| | 2 | z2w,2=1 z4=1 |
| | 2 | z2w,1=1 z2w,3=1 |
| 15.4.48 | 158082 | Fe1Na1O6Si2 |
| | 2 | z2w,1=1 z2w,3=1 z4=2 |
| | 2 | z2w,2=1 z4=1 |
| | 2 | z2w,1=1 z4=1 |
| | 2 | z2w,2=1 z2w,3=1 |
| 15.4.49 | 157733 | Fe1Na1O6Si2 |
| | 2 | z2w,2=1 z2w,3=1 z4=2 |
| | 2 | z2w,1=1 z4=1 |
| | 2 | z2w,2=1 z4=1 |
| | 2 | z2w,1=1 z2w,3=1 |
| 15.4.50 | 156561 | Fe1Na1O6Si2 |
| | 2 | z2w,1=1 z2w,3=1 z4=2 |
| | 2 | z2w,2=1 z4=1 |
| | 2 | z2w,1=1 z4=1 |
| | 2 | z2w,2=1 z2w,3=1 |
| 15.4.51 | 156560 | Fe1Na1O6Si2 |

Continued on next column

**TABLE XIX– continued**

| TQC | ICSD<br>SSG | Stoich. Formula<br>SIs |
|---|---|---|
| | 2 | z2w,1=1 z2w,3=1 z4=2 |
| | 2 | z2w,2=1 z4=1 |
| | 2 | z2w,1=1 z4=1 |
| | 2 | z2w,2=1 z2w,3=1 |
| 15.4.52 | 10219 | Fe1Na1O6Si2 |
| | 2 | z2w,1=1 z2w,3=1 z4=2 |
| | 2 | z2w,2=1 z4=1 |
| | 2 | z2w,1=1 z4=1 |
| | 2 | z2w,2=1 z2w,3=1 |
| 15.4.53 | 262314 | Fe1Li1O8W2 |
| | 2 | z2w,1=1 z2w,3=1 z4=2 |
| | 2 | z2w,2=1 z4=1 |
| | 2 | z2w,1=1 z4=1 |
| | 2 | z2w,2=1 z2w,3=1 |
| 15.4.54 | 9669 | Fe1Li1O6Si2 |
| | 2 | z2w,2=1 z2w,3=1 z4=2 |
| | 2 | z2w,1=1 z4=1 |
| | 2 | z2w,2=1 z4=1 |
| | 2 | z2w,1=1 z2w,3=1 |
| 15.4.55 | 158582 | Fe1Li1O6Si2 |
| | 2 | z2w,2=1 z2w,3=1 z4=2 |
| | 2 | z2w,1=1 z4=1 |
| | 2 | z2w,2=1 z4=1 |
| | 2 | z2w,1=1 z2w,3=1 |
| 15.4.56 | 158581 | Fe1Li1O6Si2 |
| | 2 | z2w,1=1 z2w,3=1 z4=2 |
| | 2 | z2w,2=1 z4=1 |
| | 2 | z2w,1=1 z4=1 |
| | 2 | z2w,2=1 z2w,3=1 |
| 15.4.57 | 40780 | Fe1K1Se2 |
| | 2 | z2w,2=1 z2w,3=1 z4=3 |
| | 2 | z2w,1=1 z4=2 |
| | 2 | z2w,2=1 z4=2 |
| | 2 | z2w,1=1 z2w,3=1 z4=1 |
| 15.4.58 | 68383 | Fe1K1S2 |
| | 2 | z2w,2=1 z2w,3=1 z4=3 |
| | 2 | z2w,1=1 z4=2 |
| | 2 | z2w,2=1 z4=2 |
| | 2 | z2w,1=1 z2w,3=1 z4=1 |
| 15.4.59 | 202383 | Fe1K1S2 |
| | 2 | z2w,2=1 z2w,3=1 z4=3 |
| | 2 | z2w,1=1 z4=2 |
| | 2 | z2w,2=1 z4=2 |
| | 2 | z2w,1=1 z2w,3=1 z4=1 |
| 15.4.60 | 202382 | Fe1K1S2 |
| | 2 | z2w,1=1 z2w,3=1 z4=3 |
| | 2 | z2w,2=1 z4=2 |
| | 2 | z2w,1=1 z4=2 |
| | 2 | z2w,2=1 z2w,3=1 z4=1 |
| 15.4.61 | 202380 | Fe1K1S2 |
| | 2 | z2w,2=1 z2w,3=1 z4=3 |
| | 2 | z2w,1=1 z4=2 |
| | 2 | z2w,2=1 z4=2 |
| | 2 | z2w,1=1 z2w,3=1 z4=1 |
| 15.4.62 | 167454 | Fe1K1S2 |
| | 2 | z2w,1=1 z2w,3=1 z4=3 |
| | 2 | z2w,2=1 z4=2 |
| | 2 | z2w,1=1 z4=2 |
| | 2 | z2w,2=1 z2w,3=1 z4=1 |

Continued on next column

**TABLE XIX– continued**

| TQC | ICSD<br>SSG | Stoich. Formula<br>SIs |
|---|---|---|
| 15.4.63 | 167453 | Fe1K1S2 |
| | 2 | z2w,2=1 z2w,3=1 z4=3 |
| | 2 | z2w,1=1 z4=2 |
| | 2 | z2w,2=1 z4=2 |
| | 2 | z2w,1=1 z2w,3=1 z4=1 |
| 15.4.64 | 15557 | Fe1K1S2 |
| | 2 | z2w,2=1 z2w,3=1 z4=3 |
| | 2 | z2w,1=1 z4=2 |
| | 2 | z2w,2=1 z4=2 |
| | 2 | z2w,1=1 z2w,3=1 z4=1 |
| 15.4.65 | 180310 | Fe1Ge2Na1O6 |
| | 2 | z2w,2=1 z2w,3=1 z4=3 |
| | 2 | z2w,1=1 z4=2 |
| | 2 | z2w,2=1 z4=2 |
| | 2 | z2w,1=1 z2w,3=1 z4=1 |
| 15.4.66 | 180308 | Fe1Ge2Na1O6 |
| | 2 | z2w,2=1 z2w,3=1 z4=3 |
| | 2 | z2w,1=1 z4=2 |
| | 2 | z2w,2=1 z4=2 |
| | 2 | z2w,1=1 z2w,3=1 z4=1 |
| 15.4.67 | 180307 | Fe1Ge2Na1O6 |
| | 2 | z2w,2=1 z2w,3=1 z4=3 |
| | 2 | z2w,1=1 z4=2 |
| | 2 | z2w,2=1 z4=2 |
| | 2 | z2w,1=1 z2w,3=1 z4=1 |
| 15.4.68 | 180306 | Fe1Ge2Na1O6 |
| | 2 | z2w,2=1 z2w,3=1 z4=3 |
| | 2 | z2w,1=1 z4=2 |
| | 2 | z2w,2=1 z4=2 |
| | 2 | z2w,1=1 z2w,3=1 z4=1 |
| 15.4.69 | 169480 | Fe1Ge2Li1O6 |
| | 2 | z2w,2=1 z2w,3=1 z4=2 |
| | 2 | z2w,1=1 z4=1 |
| | 2 | z2w,2=1 z4=1 |
| | 2 | z2w,1=1 z2w,3=1 |
| 15.4.70 | 169479 | Fe1Ge2Li1O6 |
| | 2 | z2w,2=1 z2w,3=1 z4=2 |
| | 2 | z2w,1=1 z4=1 |
| | 2 | z2w,2=1 z4=1 |
| | 2 | z2w,1=1 z2w,3=1 |
| 15.4.71 | 162951 | Fe1Ge2Li1O6 |
| | 2 | z2w,1=1 z2w,3=1 z4=3 |
| | 2 | z2w,2=1 z4=2 |
| | 2 | z2w,1=1 z4=2 |
| | 2 | z2w,2=1 z2w,3=1 z4=1 |
| 15.4.72 | 162950 | Fe1Ge2Li1O6 |
| | 2 | z2w,1=1 z2w,3=1 z4=3 |
| | 2 | z2w,2=1 z4=2 |
| | 2 | z2w,1=1 z4=2 |
| | 2 | z2w,2=1 z2w,3=1 z4=1 |
| 15.4.73 | 162949 | Fe1Ge2Li1O6 |
| | 2 | z2w,1=1 z2w,3=1 z4=3 |
| | 2 | z2w,2=1 z4=2 |
| | 2 | z2w,1=1 z4=2 |
| | 2 | z2w,2=1 z2w,3=1 z4=1 |
| 15.4.74 | 162948 | Fe1Ge2Li1O6 |
| | 2 | z2w,1=1 z2w,3=1 z4=3 |
| | 2 | z2w,2=1 z4=2 |
| | 2 | z2w,1=1 z4=2 |

Continued on next column

**TABLE XIX– continued**

| TQC | ICSD SSG | Stoich. Formula SIs |
|---|---|---|
| | 2 | z2w,2=1 z2w,3=1 z4=1 |
| 15.4.75 | 251694 | F3Nb2O2 |
| | 2 | z2w,2=1 z2w,3=1 |
| | 2 | z2w,1=1 z4=3 |
| | 2 | z2w,2=1 z4=3 |
| | 2 | z2w,1=1 z2w,3=1 z4=2 |
| 15.4.76 | 262447 | Eu3Ga2P4 |
| | 2 | z2w,1=1 z2w,3=1 z4=2 |
| | 2 | z2w,2=1 z4=1 |
| | 2 | z2w,1=1 z4=1 |
| | 2 | z2w,2=1 z2w,3=1 |
| 15.4.77 | 106624 | Eu3Ga2 |
| | 2 | z2w,2=1 z2w,3=1 |
| | 2 | z2w,1=1 z4=3 |
| | 2 | z2w,2=1 z4=3 |
| | 2 | z2w,1=1 z2w,3=1 z4=2 |
| 15.4.78 | 423578 | Eu2In1Pt2 |
| | 2 | z2w,2=1 z2w,3=1 z4=2 |
| | 2 | z2w,1=1 z4=1 |
| | 2 | z2w,2=1 z4=1 |
| | 2 | z2w,1=1 z2w,3=1 |
| 15.4.79 | 423577 | Eu2In1Pd2 |
| | 2 | z2w,2=1 z2w,3=1 |
| | 2 | z2w,1=1 z4=3 |
| | 2 | z2w,2=1 z4=3 |
| | 2 | z2w,1=1 z2w,3=1 z4=2 |
| 15.4.80 | 153364 | Er1O4Ta1 |
| | 2 | z2w,2=1 z2w,3=1 z4=1 |
| | 2 | z2w,1=1 |
| | 2 | z2w,2=1 |
| | 2 | z2w,1=1 z2w,3=1 z4=3 |
| 15.4.81 | 109189 | Er1O4Ta1 |
| | 2 | z2w,1=1 z2w,3=1 |
| | 2 | z2w,2=1 z4=3 |
| | 2 | z2w,1=1 z4=3 |
| | 2 | z2w,2=1 z2w,3=1 z4=2 |
| 15.4.82 | 109304 | Er1Nb1O4 |
| | 2 | z2w,2=1 z2w,3=1 |
| | 2 | z2w,1=1 z4=3 |
| | 2 | z2w,2=1 z4=3 |
| | 2 | z2w,1=1 z2w,3=1 z4=2 |
| 15.4.83 | 109180 | Er1Nb1O4 |
| | 2 | z2w,1=1 z2w,3=1 |
| | 2 | z2w,2=1 z4=3 |
| | 2 | z2w,1=1 z4=3 |
| | 2 | z2w,2=1 z2w,3=1 z4=2 |
| 15.4.84 | 157832 | Er1K1O8W2 |
| | 2 | z2w,2=1 z2w,3=1 z4=2 |
| | 2 | z2w,1=1 z4=1 |
| | 2 | z2w,2=1 z4=1 |
| | 2 | z2w,1=1 z2w,3=1 |
| 15.4.85 | 109188 | Dy1O4Ta1 |
| | 2 | z2w,2=1 z2w,3=1 z4=3 |
| | 2 | z2w,1=1 z4=2 |
| | 2 | z2w,2=1 z4=2 |
| | 2 | z2w,1=1 z2w,3=1 z4=1 |
| 15.4.86 | 109305 | Dy1Nb1O4 |
| | 2 | z2w,2=1 z2w,3=1 z4=1 |
| | 2 | z2w,1=1 |

Continued on next column

**TABLE XIX– continued**

| TQC | ICSD SSG | Stoich. Formula SIs |
|---|---|---|
| | 2 | z2w,2=1 |
| | 2 | z2w,1=1 z2w,3=1 z4=3 |
| 15.4.87 | 109179 | Dy1Nb1O4 |
| | 2 | z2w,2=1 z2w,3=1 z4=3 |
| | 2 | z2w,1=1 z4=2 |
| | 2 | z2w,2=1 z4=2 |
| | 2 | z2w,1=1 z2w,3=1 z4=1 |
| 15.4.88 | 202587 | Cu3F7Na1 |
| | 2 | z2w,2=1 z2w,3=1 z4=3 |
| | 2 | z2w,1=1 z4=2 |
| | 2 | z2w,2=1 z4=2 |
| | 2 | z2w,1=1 z2w,3=1 z4=1 |
| 15.4.89 | 83493 | Cu2O4Y1 |
| | 2 | z2w,1=1 z2w,3=1 z4=1 |
| | 2 | z2w,2=1 |
| | 2 | z2w,1=1 |
| | 2 | z2w,2=1 z2w,3=1 z4=3 |
| 15.4.90 | 83486 | Cu2La1O4 |
| | 2 | z2w,2=1 z2w,3=1 z4=1 |
| | 2 | z2w,1=1 |
| | 2 | z2w,2=1 |
| | 2 | z2w,1=1 z2w,3=1 z4=3 |
| 15.4.91 | 83489 | Cu2Eu1O4 |
| | 2 | z2w,1=1 z2w,3=1 z4=1 |
| | 2 | z2w,2=1 |
| | 2 | z2w,1=1 |
| | 2 | z2w,2=1 z2w,3=1 z4=3 |
| 15.4.92 | 72721 | Cu1O6Pt1Sr3 |
| | 2 | z2w,2=1 z2w,3=1 z4=2 |
| | 2 | z2w,1=1 z4=1 |
| | 2 | z2w,2=1 z4=1 |
| | 2 | z2w,1=1 z2w,3=1 |
| 15.4.93 | 66347 | Cu1O6Pt1Sr3 |
| | 2 | z2w,2=1 z2w,3=1 z4=2 |
| | 2 | z2w,1=1 z4=1 |
| | 2 | z2w,2=1 z4=1 |
| | 2 | z2w,1=1 z2w,3=1 |
| 15.4.94 | 603 | Cu1O5Se2 |
| | 2 | z2w,1=1 z2w,3=1 z4=3 |
| | 2 | z2w,2=1 z4=2 |
| | 2 | z2w,1=1 z4=2 |
| | 2 | z2w,2=1 z2w,3=1 z4=1 |
| 15.4.95 | 245059 | Cu1O5Se2 |
| | 2 | z2w,1=1 z2w,3=1 z4=1 |
| | 2 | z2w,2=1 |
| | 2 | z2w,1=1 |
| | 2 | z2w,2=1 z2w,3=1 z4=3 |
| 15.4.96 | 69094 | Cu1O1 |
| | 2 | z2w,1=1 z2w,3=1 |
| | 2 | z2w,2=1 z4=3 |
| | 2 | z2w,1=1 z4=3 |
| | 2 | z2w,2=1 z2w,3=1 z4=2 |
| 15.4.97 | 67850 | Cu1O1 |
| | 2 | z2w,2=1 z2w,3=1 |
| | 2 | z2w,1=1 z4=3 |
| | 2 | z2w,2=1 z4=3 |
| | 2 | z2w,1=1 z2w,3=1 z4=2 |
| 15.4.98 | 291393 | Cu1O1 |
| | 2 | z2w,1=1 z2w,3=1 |

Continued on next column

**TABLE XIX– continued**

| TQC | ICSD SSG | Stoich. Formula SIs |
|---|---|---|
| | 2 | z2w,2=1 z4=3 |
| | 2 | z2w,1=1 z4=3 |
| | 2 | z2w,2=1 z2w,3=1 z4=2 |
| 15.4.99 | 291390 | Cu1O1 |
| | 2 | z2w,1=1 z2w,3=1 |
| | 2 | z2w,2=1 z4=3 |
| | 2 | z2w,1=1 z4=3 |
| | 2 | z2w,2=1 z2w,3=1 z4=2 |
| 15.4.100 | 291387 | Cu1O1 |
| | 2 | z2w,1=1 z2w,3=1 |
| | 2 | z2w,2=1 z4=3 |
| | 2 | z2w,1=1 z4=3 |
| | 2 | z2w,2=1 z2w,3=1 z4=2 |
| 15.4.101 | 160630 | Cu1O1 |
| | 2 | z2w,1=1 z2w,3=1 |
| | 2 | z2w,2=1 z4=3 |
| | 2 | z2w,1=1 z4=3 |
| | 2 | z2w,2=1 z2w,3=1 z4=2 |
| 15.4.102 | 16025 | Cu1O1 |
| | 2 | z2w,1=1 z2w,3=1 |
| | 2 | z2w,2=1 z4=3 |
| | 2 | z2w,1=1 z4=3 |
| | 2 | z2w,2=1 z2w,3=1 z4=2 |
| 15.4.103 | 80418 | Cu1Na2O7P2 |
| | 2 | z2w,2=1 z2w,3=1 z4=3 |
| | 2 | z2w,1=1 z4=2 |
| | 2 | z2w,2=1 z4=2 |
| | 2 | z2w,1=1 z2w,3=1 z4=1 |
| 15.4.104 | 79868 | Cu1Na2O7P2 |
| | 2 | z2w,2=1 z2w,3=1 z4=3 |
| | 2 | z2w,1=1 z4=2 |
| | 2 | z2w,2=1 z4=2 |
| | 2 | z2w,1=1 z2w,3=1 z4=1 |
| 15.4.105 | 72485 | Cu1Li2O7P2 |
| | 2 | z2w,1=1 z2w,3=1 z4=1 |
| | 2 | z2w,2=1 |
| | 2 | z2w,1=1 |
| | 2 | z2w,2=1 z2w,3=1 z4=3 |
| 15.4.106 | 160321 | Cu1Li2O7P2 |
| | 2 | z2w,2=1 z2w,3=1 z4=3 |
| | 2 | z2w,1=1 z4=2 |
| | 2 | z2w,2=1 z4=2 |
| | 2 | z2w,1=1 z2w,3=1 z4=1 |
| 15.4.107 | 262273 | Cu1F1Na1O4S1 |
| | 2 | z2w,2=1 z2w,3=1 z4=3 |
| | 2 | z2w,1=1 z4=2 |
| | 2 | z2w,2=1 z4=2 |
| | 2 | z2w,1=1 z2w,3=1 z4=1 |
| 15.4.108 | 97262 | Cr1Na1O6Si2 |
| | 2 | z2w,1=1 z2w,3=1 z4=3 |
| | 2 | z2w,2=1 z4=2 |
| | 2 | z2w,1=1 z4=2 |
| | 2 | z2w,2=1 z2w,3=1 z4=1 |
| 15.4.109 | 97258 | Cr1Na1O6Si2 |
| | 2 | z2w,1=1 z2w,3=1 z4=3 |
| | 2 | z2w,2=1 z4=2 |
| | 2 | z2w,1=1 z4=2 |
| | 2 | z2w,2=1 z2w,3=1 z4=1 |
| 15.4.110 | 97257 | Cr1Na1O6Si2 |

Continued on next column

**TABLE XIX– continued**

| TQC | ICSD SSG | Stoich. Formula SIs |
|---|---|---|
| | 2 | z2w,1=1 z2w,3=1 z4=3 |
| | 2 | z2w,2=1 z4=2 |
| | 2 | z2w,1=1 z4=2 |
| | 2 | z2w,2=1 z2w,3=1 z4=1 |
| 15.4.111 | 9670 | Cr1Na1O6Si2 |
| | 2 | z2w,2=1 z2w,3=1 z4=2 |
| | 2 | z2w,1=1 z4=1 |
| | 2 | z2w,2=1 z4=1 |
| | 2 | z2w,1=1 z2w,3=1 |
| 15.4.112 | 192422 | Cr1Na1O6Si2 |
| | 2 | z2w,1=1 z2w,3=1 z4=3 |
| | 2 | z2w,2=1 z4=2 |
| | 2 | z2w,1=1 z4=2 |
| | 2 | z2w,2=1 z2w,3=1 z4=1 |
| 15.4.113 | 192420 | Cr1Na1O6Si2 |
| | 2 | z2w,2=1 z2w,3=1 z4=3 |
| | 2 | z2w,1=1 z4=2 |
| | 2 | z2w,2=1 z4=2 |
| | 2 | z2w,1=1 z2w,3=1 z4=1 |
| 15.4.114 | 192419 | Cr1Na1O6Si2 |
| | 2 | z2w,1=1 z2w,3=1 z4=3 |
| | 2 | z2w,2=1 z4=2 |
| | 2 | z2w,1=1 z4=2 |
| | 2 | z2w,2=1 z2w,3=1 z4=1 |
| 15.4.115 | 192418 | Cr1Na1O6Si2 |
| | 2 | z2w,2=1 z2w,3=1 z4=2 |
| | 2 | z2w,1=1 z4=1 |
| | 2 | z2w,2=1 z4=1 |
| | 2 | z2w,1=1 z2w,3=1 |
| 15.4.116 | 10240 | Cr1Na1O6Si2 |
| | 2 | z2w,2=1 z2w,3=1 z4=2 |
| | 2 | z2w,1=1 z4=1 |
| | 2 | z2w,2=1 z4=1 |
| | 2 | z2w,1=1 z2w,3=1 |
| 15.4.117 | 10239 | Cr1Na1O6Si2 |
| | 2 | z2w,1=1 z2w,3=1 z4=2 |
| | 2 | z2w,2=1 z4=1 |
| | 2 | z2w,1=1 z4=1 |
| | 2 | z2w,2=1 z2w,3=1 |
| 15.4.118 | 152072 | Cr1Li1O6Si2 |
| | 2 | z2w,2=1 z2w,3=1 z4=3 |
| | 2 | z2w,1=1 z4=2 |
| | 2 | z2w,2=1 z4=2 |
| | 2 | z2w,1=1 z2w,3=1 z4=1 |
| 15.4.119 | 152071 | Cr1Li1O6Si2 |
| | 2 | z2w,1=1 z2w,3=1 z4=3 |
| | 2 | z2w,2=1 z4=2 |
| | 2 | z2w,1=1 z4=2 |
| | 2 | z2w,2=1 z2w,3=1 z4=1 |
| 15.4.120 | 260079 | Cr1Ge2Na1O6 |
| | 2 | z2w,1=1 z2w,3=1 z4=1 |
| | 2 | z2w,2=1 |
| | 2 | z2w,1=1 |
| | 2 | z2w,2=1 z2w,3=1 z4=3 |
| 15.4.121 | 194813 | Cr1Ge2Li1O6 |
| | 2 | z2w,1=1 z2w,3=1 z4=1 |
| | 2 | z2w,2=1 |
| | 2 | z2w,1=1 |
| | 2 | z2w,2=1 z2w,3=1 z4=3 |

Continued on next column

**TABLE XIX– continued**

| TQC | ICSD<br>SSG | Stoich. Formula<br>SIs |
|---|---|---|
| 15.4.122 | 194812 | Cr1Ge2Li1O6 |
| | 2 | z2w,2=1 z2w,3=1 z4=1 |
| | 2 | z2w,1=1 |
| | 2 | z2w,2=1 |
| | 2 | z2w,1=1 z2w,3=1 z4=3 |
| 15.4.123 | 174432 | Cr1Ge2Li1O6 |
| | 2 | z2w,1=1 z2w,3=1 z4=3 |
| | 2 | z2w,2=1 z4=2 |
| | 2 | z2w,1=1 z4=2 |
| | 2 | z2w,2=1 z2w,3=1 z4=1 |
| 15.4.124 | 414127 | Co2Na7O6 |
| | 2 | z2w,1=1 z2w,3=1 z4=3 |
| | 2 | z2w,2=1 z4=2 |
| | 2 | z2w,1=1 z4=2 |
| | 2 | z2w,2=1 z2w,3=1 z4=1 |
| 15.4.125 | 157364 | Co2Ga7Mg3 |
| | 2 | z2w,2=1 z2w,3=1 |
| | 2 | z2w,1=1 z4=3 |
| | 2 | z2w,2=1 z4=3 |
| | 2 | z2w,1=1 z2w,3=1 z4=2 |
| 15.4.126 | 290052 | Co1F1Na1O4S1 |
| | 2 | z2w,1=1 z2w,3=1 z4=3 |
| | 2 | z2w,2=1 z4=2 |
| | 2 | z2w,1=1 z4=2 |
| | 2 | z2w,2=1 z2w,3=1 z4=1 |
| 15.4.127 | 262275 | Co1F1Na1O4S1 |
| | 2 | z2w,1=1 z2w,3=1 z4=3 |
| | 2 | z2w,2=1 z4=2 |
| | 2 | z2w,1=1 z4=2 |
| | 2 | z2w,2=1 z2w,3=1 z4=1 |
| 15.4.128 | 1242 | Cl6Cs1W1 |
| | 2 | z2w,1=1 z2w,3=1 z4=2 |
| | 2 | z2w,2=1 z4=1 |
| | 2 | z2w,1=1 z4=1 |
| | 2 | z2w,2=1 z2w,3=1 |
| 15.4.129 | 418203 | Ca2In1Pt2 |
| | 2 | z2w,2=1 z2w,3=1 z4=2 |
| | 2 | z2w,1=1 z4=1 |
| | 2 | z2w,2=1 z4=1 |
| | 2 | z2w,1=1 z2w,3=1 |
| 15.4.130 | 418202 | Ca2In1Pd2 |
| | 2 | z2w,1=1 z2w,3=1 z4=2 |
| | 2 | z2w,2=1 z4=1 |
| | 2 | z2w,1=1 z4=1 |
| | 2 | z2w,2=1 z2w,3=1 |
| 15.4.131 | 429177 | Ca2Ga1Pt2 |
| | 2 | z2w,2=1 z2w,3=1 z4=2 |
| | 2 | z2w,1=1 z4=1 |
| | 2 | z2w,2=1 z4=1 |
| | 2 | z2w,1=1 z2w,3=1 |
| 15.4.132 | 429176 | Ca2Ga1Pd2 |
| | 2 | z2w,1=1 z2w,3=1 z4=2 |
| | 2 | z2w,2=1 z4=1 |
| | 2 | z2w,1=1 z4=1 |
| | 2 | z2w,2=1 z2w,3=1 |
| 15.4.133 | 170824 | Ca1Cu1Ge2O6 |
| | 2 | z2w,1=1 z2w,3=1 z4=3 |
| | 2 | z2w,2=1 z4=2 |
| | 2 | z2w,1=1 z4=2 |

Continued on next column

**TABLE XIX– continued**

| TQC | ICSD<br>SSG | Stoich. Formula<br>SIs |
|---|---|---|
| | 2 | z2w,2=1 z2w,3=1 z4=1 |
| 15.4.134 | 170823 | Ca1Cu1Ge2O6 |
| | 2 | z2w,1=1 z2w,3=1 z4=3 |
| | 2 | z2w,2=1 z4=2 |
| | 2 | z2w,1=1 z4=2 |
| | 2 | z2w,2=1 z2w,3=1 z4=1 |
| 15.4.135 | 170822 | Ca1Cu1Ge2O6 |
| | 2 | z2w,1=1 z2w,3=1 z4=3 |
| | 2 | z2w,2=1 z4=2 |
| | 2 | z2w,1=1 z4=2 |
| | 2 | z2w,2=1 z2w,3=1 z4=1 |
| 15.4.136 | 170821 | Ca1Cu1Ge2O6 |
| | 2 | z2w,1=1 z2w,3=1 z4=3 |
| | 2 | z2w,2=1 z4=2 |
| | 2 | z2w,1=1 z4=2 |
| | 2 | z2w,2=1 z2w,3=1 z4=1 |
| 15.4.137 | 82212 | Ca1Co1O6Si2 |
| | 2 | z2w,2=1 z2w,3=1 z4=2 |
| | 2 | z2w,1=1 z4=1 |
| | 2 | z2w,2=1 z4=1 |
| | 2 | z2w,1=1 z2w,3=1 |
| 15.4.138 | 246209 | Ca1Co1O6Si2 |
| | 2 | z2w,2=1 z2w,3=1 z4=2 |
| | 2 | z2w,1=1 z4=1 |
| | 2 | z2w,2=1 z4=1 |
| | 2 | z2w,1=1 z2w,3=1 |
| 15.4.139 | 415406 | C4Cu1O4 |
| | 2 | z2w,1=1 z2w,3=1 z4=3 |
| | 2 | z2w,2=1 z4=2 |
| | 2 | z2w,1=1 z4=2 |
| | 2 | z2w,2=1 z2w,3=1 z4=1 |
| 15.4.140 | 174608 | C4Cu1Hg1N4S4 |
| | 2 | z2w,1=1 z2w,3=1 z4=1 |
| | 2 | z2w,2=1 |
| | 2 | z2w,1=1 |
| | 2 | z2w,2=1 z2w,3=1 z4=3 |
| 15.4.141 | 240902 | C2H4Mn1O6 |
| | 2 | z2w,1=1 z2w,3=1 |
| | 2 | z2w,2=1 z4=3 |
| | 2 | z2w,1=1 z4=3 |
| | 2 | z2w,2=1 z2w,3=1 z4=2 |
| 15.4.142 | 407226 | C2Gd1K1O6 |
| | 2 | z2w,2=1 z2w,3=1 z4=1 |
| | 2 | z2w,1=1 |
| | 2 | z2w,2=1 |
| | 2 | z2w,1=1 z2w,3=1 z4=3 |
| 15.4.143 | 380495 | C2Eu1 |
| | 2 | z2w,1=1 z2w,3=1 z4=2 |
| | 2 | z2w,2=1 z4=1 |
| | 2 | z2w,1=1 z4=1 |
| | 2 | z2w,2=1 z2w,3=1 |
| 15.4.144 | 419171 | C2Cl1La2 |
| | 2 | z2w,1=1 z2w,3=1 z4=2 |
| | 2 | z2w,2=1 z4=1 |
| | 2 | z2w,1=1 z4=1 |
| | 2 | z2w,2=1 z2w,3=1 |
| 15.4.145 | 109643 | C1H1Na1O2 |
| | 2 | z2w,1=1 z2w,3=1 |
| | 2 | z2w,2=1 z4=3 |

Continued on next column

**TABLE XIX– continued**

| TQC | ICSD SSG | Stoich. Formula SIs |
|---|---|---|
| | 2 | z2w,1=1 z4=3 |
| | 2 | z2w,2=1 z2w,3=1 z4=2 |
| 15.4.146 | 240017 | Bi2Cs3 |
| | 2 | z2w,1=1 z2w,3=1 |
| | 2 | z2w,2=1 z4=3 |
| | 2 | z2w,1=1 z4=3 |
| | 2 | z2w,2=1 z2w,3=1 z4=2 |
| 15.4.147 | 181425 | Bi1O6Re2 |
| | 2 | z2w,2=1 z2w,3=1 |
| | 2 | z2w,1=1 z4=3 |
| | 2 | z2w,2=1 z4=3 |
| | 2 | z2w,1=1 z2w,3=1 z4=2 |
| 15.4.148 | 246363 | Ba3Ir2Na1O9 |
| | 2 | z2w,2=1 z2w,3=1 z4=2 |
| | 2 | z2w,1=1 z4=1 |
| | 2 | z2w,2=1 z4=1 |
| | 2 | z2w,1=1 z2w,3=1 |
| 15.4.149 | 246362 | Ba3Ir2Na1O9 |
| | 2 | z2w,2=1 z2w,3=1 |
| | 2 | z2w,1=1 z4=3 |
| | 2 | z2w,2=1 z4=3 |
| | 2 | z2w,1=1 z2w,3=1 z4=2 |
| 15.4.150 | 246361 | Ba3Ir2Na1O9 |
| | 2 | z2w,1=1 z2w,3=1 z4=2 |
| | 2 | z2w,2=1 z4=1 |
| | 2 | z2w,1=1 z4=1 |
| | 2 | z2w,2=1 z2w,3=1 |
| 15.4.151 | 72448 | Ba3Bi1O9Ru2 |
| | 2 | z2w,1=1 z2w,3=1 z4=2 |
| | 2 | z2w,2=1 z4=1 |
| | 2 | z2w,1=1 z4=1 |
| | 2 | z2w,2=1 z2w,3=1 |
| 15.4.152 | 174292 | Ba3Bi1Ir2O9 |
| | 2 | z2w,1=1 z2w,3=1 z4=2 |
| | 2 | z2w,2=1 z4=1 |
| | 2 | z2w,1=1 z4=1 |
| | 2 | z2w,2=1 z2w,3=1 |
| 15.4.153 | 20905 | Ba2Cu1O7Si2 |
| | 2 | z2w,1=1 z2w,3=1 z4=1 |
| | 2 | z2w,2=1 |
| | 2 | z2w,1=1 |
| | 2 | z2w,2=1 z2w,3=1 z4=3 |
| 15.4.154 | 173850 | Ba1Ir1O3 |
| | 2 | z2w,2=1 z2w,3=1 |
| | 2 | z2w,1=1 z4=3 |
| | 2 | z2w,2=1 z4=3 |
| | 2 | z2w,1=1 z2w,3=1 z4=2 |
| 15.4.155 | 202388 | Ba1Cu1O6Se2 |
| | 2 | z2w,1=1 z2w,3=1 z4=3 |
| | 2 | z2w,2=1 z4=2 |
| | 2 | z2w,1=1 z4=2 |
| | 2 | z2w,2=1 z2w,3=1 z4=1 |
| 15.4.156 | 407832 | Ba1Cl5Gd1 |
| | 2 | z2w,2=1 z2w,3=1 z4=3 |
| | 2 | z2w,1=1 z4=2 |
| | 2 | z2w,2=1 z4=2 |
| | 2 | z2w,1=1 z2w,3=1 z4=1 |
| 15.4.157 | 24308 | B3Ni4 |
| | 2 | z2w,1=1 z2w,3=1 z4=1 |

Continued on next column

**TABLE XIX– continued**

| TQC | ICSD SSG | Stoich. Formula SIs |
|---|---|---|
| | 2 | z2w,2=1 |
| | 2 | z2w,1=1 |
| | 2 | z2w,2=1 z2w,3=1 z4=3 |
| 15.4.158 | 187403 | B3Ni4 |
| | 2 | z2w,1=1 z2w,3=1 z4=1 |
| | 2 | z2w,2=1 |
| | 2 | z2w,1=1 |
| | 2 | z2w,2=1 z2w,3=1 z4=3 |
| 15.4.159 | 150561 | B3Ni4 |
| | 2 | z2w,1=1 z2w,3=1 z4=3 |
| | 2 | z2w,2=1 z4=2 |
| | 2 | z2w,1=1 z4=2 |
| | 2 | z2w,2=1 z2w,3=1 z4=1 |
| 15.4.160 | 39487 | B2Ho1Ni2 |
| | 2 | z2w,2=1 z2w,3=1 z4=1 |
| | 2 | z2w,1=1 |
| | 2 | z2w,2=1 |
| | 2 | z2w,1=1 z2w,3=1 z4=3 |
| 15.4.161 | 2125 | Au6Pr1 |
| | 2 | z2w,1=1 z2w,3=1 z4=3 |
| | 2 | z2w,2=1 z4=2 |
| | 2 | z2w,1=1 z4=2 |
| | 2 | z2w,2=1 z2w,3=1 z4=1 |
| 15.4.162 | 423784 | As4Eu3Ga2 |
| | 2 | z2w,2=1 z2w,3=1 |
| | 2 | z2w,1=1 z4=3 |
| | 2 | z2w,2=1 z4=3 |
| | 2 | z2w,1=1 z2w,3=1 z4=2 |
| 15.4.163 | 415951 | Al2Cl8V1 |
| | 2 | z2w,1=1 z2w,3=1 z4=3 |
| | 2 | z2w,2=1 z4=2 |
| | 2 | z2w,1=1 z4=2 |
| | 2 | z2w,2=1 z2w,3=1 z4=1 |
| 15.4.164 | 95801 | Al1Gd2Ge2 |
| | 2 | z2w,2=1 z2w,3=1 z4=3 |
| | 2 | z2w,1=1 z4=2 |
| | 2 | z2w,2=1 z4=2 |
| | 2 | z2w,1=1 z2w,3=1 z4=1 |
| 15.4.165 | 412545 | Al1Co2Pr2 |
| | 2 | z2w,1=1 z2w,3=1 |
| | 2 | z2w,2=1 z4=3 |
| | 2 | z2w,1=1 z4=3 |
| | 2 | z2w,2=1 z2w,3=1 z4=2 |
| 15.4.166 | 428674 | Al1Ca2Pt2 |
| | 2 | z2w,2=1 z2w,3=1 z4=2 |
| | 2 | z2w,1=1 z4=1 |
| | 2 | z2w,2=1 z4=1 |
| | 2 | z2w,1=1 z2w,3=1 |
| 15.4.167 | 428673 | Al1Ca2Pd2 |
| | 2 | z2w,1=1 z2w,3=1 z4=2 |
| | 2 | z2w,2=1 z4=1 |
| | 2 | z2w,1=1 z4=1 |
| | 2 | z2w,2=1 z2w,3=1 |
| 15.4.168 | 416931 | Ag4Cu1O6Te1 |
| | 2 | z2w,2=1 z2w,3=1 z4=3 |
| | 2 | z2w,1=1 z4=2 |
| | 2 | z2w,2=1 z4=2 |
| | 2 | z2w,1=1 z2w,3=1 z4=1 |
| 15.4.169 | 158336 | Ag2Cu2O3 |

Continued on next column

**TABLE XIX– continued**

| TQC | ICSD SSG | Stoich. Formula SIs |
|---|---|---|
| | 2 | z2w,1=1 z2w,3=1 z4=2 |
| | 2 | z2w,2=1 z4=1 |
| | 2 | z2w,1=1 z4=1 |
| | 2 | z2w,2=1 z2w,3=1 |
| 51.4.1 | 417836 | In4Ir1Sr1 |
| | 10 | z2w,2=1 z4=3 |
| | 10 | z2w,1=1 z2w,2=1 |
| | 10 | z2w,1=1 z2w,2=1 z2w,3=1 |
| | 10 | z2w,2=1 z2w,3=1 z4=1 |
| | 10 | z2w,1=1 |
| | 10 | z4=1 |
| | 10 | z2w,3=1 z4=1 |
| | 10 | z2w,1=1 z2w,3=1 z4=2 |
| | 10 | z2w,1=1 z2w,3=1 |
| | 10 | z2w,3=1 z4=1 |
| | 10 | z4=1 |
| | 10 | z2w,1=1 z4=2 |
| | 10 | z2w,2=1 z2w,3=1 z4=1 |
| | 10 | z2w,1=1 z2w,2=1 z2w,3=1 z4=2 |
| | 10 | z2w,1=1 z2w,2=1 z4=2 |
| | 10 | z2w,2=1 z4=3 |
| 51.4.2 | 280921 | H4Cu1O4P2 |
| | 10 | z2w,1=1 z4=1 |
| | 10 | z2w,1=1 z2w,3=1 z4=1 |
| | 10 | z2w,3=1 z4=2 |
| | 10 | z2w,1=1 z2w,2=1 z4=1 |
| | 10 | z2w,2=1 z4=2 |
| | 10 | z2w,2=1 z2w,3=1 z4=2 |
| | 10 | z2w,1=1 z2w,2=1 z2w,3=1 z4=3 |
| | 10 | z2w,1=1 z2w,2=1 z2w,3=1 z4=1 |
| | 10 | z2w,2=1 z2w,3=1 z4=2 |
| | 10 | z2w,2=1 z4=2 |
| | 10 | z2w,1=1 z2w,2=1 z4=3 |
| | 10 | z2w,3=1 z4=2 |
| | 10 | z2w,1=1 z2w,3=1 z4=3 |
| | 10 | z2w,1=1 z4=3 |
| 51.4.3 | 89669 | Cu1O3Si1 |
| | 10 | z2w,2=1 z4=2 |
| | 10 | z2w,1=1 z2w,2=1 z4=3 |
| | 10 | z2w,1=1 z2w,2=1 z2w,3=1 z4=3 |
| | 10 | z2w,2=1 z2w,3=1 |
| | 10 | z2w,1=1 z4=3 |
| | 10 | z2w,3=1 |
| | 10 | z2w,1=1 z2w,3=1 z4=1 |
| | 10 | z2w,1=1 z2w,3=1 z4=3 |
| | 10 | z2w,3=1 |
| | 10 | z2w,1=1 z4=1 |
| | 10 | z2w,2=1 z2w,3=1 |
| | 10 | z2w,1=1 z2w,2=1 z2w,3=1 z4=1 |
| | 10 | z2w,1=1 z2w,2=1 z4=1 |
| | 10 | z2w,2=1 z4=2 |
| 51.4.4 | 89407 | Cu1Ge1O3 |
| | 10 | z2w,2=1 z4=2 |
| | 10 | z2w,1=1 z2w,2=1 z4=3 |
| | 10 | z2w,1=1 z2w,2=1 z2w,3=1 z4=3 |
| | 10 | z2w,2=1 z2w,3=1 |
| | 10 | z2w,1=1 z4=3 |
| | 10 | z2w,3=1 |
| | 10 | z2w,1=1 z2w,3=1 z4=1 |
| | 10 | z2w,1=1 z2w,3=1 z4=3 |

Continued on next column

**TABLE XIX– continued**

| TQC | ICSD SSG | Stoich. Formula SIs |
|---|---|---|
| | 10 | z2w,3=1 |
| | 10 | z2w,1=1 z4=1 |
| | 10 | z2w,2=1 z2w,3=1 |
| | 10 | z2w,1=1 z2w,2=1 z2w,3=1 z4=1 |
| | 10 | z2w,1=1 z2w,2=1 z4=1 |
| | 10 | z2w,2=1 z4=2 |
| 51.4.5 | 89406 | Cu1Ge1O3 |
| | 10 | z2w,2=1 z4=2 |
| | 10 | z2w,1=1 z2w,2=1 z4=3 |
| | 10 | z2w,1=1 z2w,2=1 z2w,3=1 z4=3 |
| | 10 | z2w,2=1 z2w,3=1 |
| | 10 | z2w,1=1 z4=3 |
| | 10 | z2w,3=1 |
| | 10 | z2w,1=1 z2w,3=1 z4=1 |
| | 10 | z2w,1=1 z2w,3=1 z4=3 |
| | 10 | z2w,3=1 |
| | 10 | z2w,1=1 z4=1 |
| | 10 | z2w,2=1 z2w,3=1 |
| | 10 | z2w,1=1 z2w,2=1 z2w,3=1 z4=1 |
| | 10 | z2w,1=1 z2w,2=1 z4=1 |
| | 10 | z2w,2=1 z4=2 |
| 51.4.6 | 89405 | Cu1Ge1O3 |
| | 10 | z2w,2=1 z4=2 |
| | 10 | z2w,1=1 z2w,2=1 z4=3 |
| | 10 | z2w,1=1 z2w,2=1 z2w,3=1 z4=3 |
| | 10 | z2w,2=1 z2w,3=1 |
| | 10 | z2w,1=1 z4=3 |
| | 10 | z2w,3=1 |
| | 10 | z2w,1=1 z2w,3=1 z4=1 |
| | 10 | z2w,1=1 z2w,3=1 z4=3 |
| | 10 | z2w,3=1 |
| | 10 | z2w,1=1 z4=1 |
| | 10 | z2w,2=1 z2w,3=1 |
| | 10 | z2w,1=1 z2w,2=1 z2w,3=1 z4=1 |
| | 10 | z2w,1=1 z2w,2=1 z4=1 |
| | 10 | z2w,2=1 z4=2 |
| 51.4.7 | 89404 | Cu1Ge1O3 |
| | 10 | z2w,2=1 z4=2 |
| | 10 | z2w,1=1 z2w,2=1 z4=3 |
| | 10 | z2w,1=1 z2w,2=1 z2w,3=1 z4=3 |
| | 10 | z2w,2=1 z2w,3=1 |
| | 10 | z2w,1=1 z4=3 |
| | 10 | z2w,3=1 |
| | 10 | z2w,1=1 z2w,3=1 z4=1 |
| | 10 | z2w,1=1 z2w,3=1 z4=3 |
| | 10 | z2w,3=1 |
| | 10 | z2w,1=1 z4=1 |
| | 10 | z2w,2=1 z2w,3=1 |
| | 10 | z2w,1=1 z2w,2=1 z2w,3=1 z4=1 |
| | 10 | z2w,1=1 z2w,2=1 z4=1 |
| | 10 | z2w,2=1 z4=2 |
| 51.4.8 | 89403 | Cu1Ge1O3 |
| | 10 | z2w,2=1 z4=2 |
| | 10 | z2w,1=1 z2w,2=1 z4=3 |
| | 10 | z2w,1=1 z2w,2=1 z2w,3=1 z4=3 |
| | 10 | z2w,2=1 z2w,3=1 |
| | 10 | z2w,1=1 z4=3 |
| | 10 | z2w,3=1 |
| | 10 | z2w,1=1 z2w,3=1 z4=1 |
| | 10 | z2w,1=1 z2w,3=1 z4=3 |

Continued on next column

**TABLE XIX– continued**

| TQC | ICSD SSG | Stoich. Formula SIs |
|---|---|---|
| | 10 | z2w,3=1 |
| | 10 | z2w,1=1 z4=1 |
| | 10 | z2w,2=1 z2w,3=1 |
| | 10 | z2w,1=1 z2w,2=1 z2w,3=1 z4=1 |
| | 10 | z2w,1=1 z2w,2=1 z4=1 |
| | 10 | z2w,2=1 z4=2 |
| 51.4.9 | 89402 | Cu1Ge1O3 |
| | 10 | z2w,2=1 z4=2 |
| | 10 | z2w,1=1 z2w,2=1 z4=3 |
| | 10 | z2w,1=1 z2w,2=1 z2w,3=1 z4=3 |
| | 10 | z2w,2=1 z2w,3=1 |
| | 10 | z2w,1=1 z4=3 |
| | 10 | z2w,3=1 |
| | 10 | z2w,1=1 z2w,3=1 z4=1 |
| | 10 | z2w,1=1 z2w,3=1 z4=3 |
| | 10 | z2w,3=1 |
| | 10 | z2w,1=1 z4=1 |
| | 10 | z2w,2=1 z2w,3=1 |
| | 10 | z2w,1=1 z2w,2=1 z2w,3=1 z4=1 |
| | 10 | z2w,1=1 z2w,2=1 z4=1 |
| | 10 | z2w,2=1 z4=2 |
| 51.4.10 | 89401 | Cu1Ge1O3 |
| | 10 | z2w,2=1 z4=2 |
| | 10 | z2w,1=1 z2w,2=1 z4=3 |
| | 10 | z2w,1=1 z2w,2=1 z2w,3=1 z4=3 |
| | 10 | z2w,2=1 z2w,3=1 |
| | 10 | z2w,1=1 z4=3 |
| | 10 | z2w,3=1 |
| | 10 | z2w,1=1 z2w,3=1 z4=1 |
| | 10 | z2w,1=1 z2w,3=1 z4=3 |
| | 10 | z2w,3=1 |
| | 10 | z2w,1=1 z4=1 |
| | 10 | z2w,2=1 z2w,3=1 |
| | 10 | z2w,1=1 z2w,2=1 z2w,3=1 z4=1 |
| | 10 | z2w,1=1 z2w,2=1 z4=1 |
| | 10 | z2w,2=1 z4=2 |
| 51.4.11 | 89400 | Cu1Ge1O3 |
| | 10 | z2w,2=1 z4=2 |
| | 10 | z2w,1=1 z2w,2=1 z4=3 |
| | 10 | z2w,1=1 z2w,2=1 z2w,3=1 z4=3 |
| | 10 | z2w,2=1 z2w,3=1 |
| | 10 | z2w,1=1 z4=3 |
| | 10 | z2w,3=1 |
| | 10 | z2w,1=1 z2w,3=1 z4=1 |
| | 10 | z2w,1=1 z2w,3=1 z4=3 |
| | 10 | z2w,3=1 |
| | 10 | z2w,1=1 z4=1 |
| | 10 | z2w,2=1 z2w,3=1 |
| | 10 | z2w,1=1 z2w,2=1 z2w,3=1 z4=1 |
| | 10 | z2w,1=1 z2w,2=1 z4=1 |
| | 10 | z2w,2=1 z4=2 |
| 51.4.12 | 56700 | Cu1Ge1O3 |
| | 10 | z2w,2=1 z4=2 |
| | 10 | z2w,1=1 z2w,2=1 z4=3 |
| | 10 | z2w,1=1 z2w,2=1 z2w,3=1 z4=3 |
| | 10 | z2w,2=1 z2w,3=1 |
| | 10 | z2w,1=1 z4=3 |
| | 10 | z2w,3=1 |
| | 10 | z2w,1=1 z2w,3=1 z4=1 |
| | 10 | z2w,1=1 z2w,3=1 z4=3 |

Continued on next column

**TABLE XIX– continued**

| TQC | ICSD SSG | Stoich. Formula SIs |
|---|---|---|
| | 10 | z2w,3=1 |
| | 10 | z2w,1=1 z4=1 |
| | 10 | z2w,2=1 z2w,3=1 |
| | 10 | z2w,1=1 z2w,2=1 z2w,3=1 z4=1 |
| | 10 | z2w,1=1 z2w,2=1 z4=1 |
| | 10 | z2w,2=1 z4=2 |
| 51.4.13 | 56699 | Cu1Ge1O3 |
| | 10 | z2w,2=1 z4=2 |
| | 10 | z2w,1=1 z2w,2=1 z4=3 |
| | 10 | z2w,1=1 z2w,2=1 z2w,3=1 z4=3 |
| | 10 | z2w,2=1 z2w,3=1 |
| | 10 | z2w,1=1 z4=3 |
| | 10 | z2w,3=1 |
| | 10 | z2w,1=1 z2w,3=1 z4=1 |
| | 10 | z2w,1=1 z2w,3=1 z4=3 |
| | 10 | z2w,3=1 |
| | 10 | z2w,1=1 z4=1 |
| | 10 | z2w,2=1 z2w,3=1 |
| | 10 | z2w,1=1 z2w,2=1 z2w,3=1 z4=1 |
| | 10 | z2w,1=1 z2w,2=1 z4=1 |
| | 10 | z2w,2=1 z4=2 |
| 51.4.14 | 50331 | Cu1Ge1O3 |
| | 10 | z2w,2=1 z4=2 |
| | 10 | z2w,1=1 z2w,2=1 z4=3 |
| | 10 | z2w,1=1 z2w,2=1 z2w,3=1 z4=3 |
| | 10 | z2w,2=1 z2w,3=1 |
| | 10 | z2w,1=1 z4=3 |
| | 10 | z2w,3=1 |
| | 10 | z2w,1=1 z2w,3=1 z4=1 |
| | 10 | z2w,1=1 z2w,3=1 z4=3 |
| | 10 | z2w,3=1 |
| | 10 | z2w,1=1 z4=1 |
| | 10 | z2w,2=1 z2w,3=1 |
| | 10 | z2w,1=1 z2w,2=1 z2w,3=1 z4=1 |
| | 10 | z2w,1=1 z2w,2=1 z4=1 |
| | 10 | z2w,2=1 z4=2 |
| 51.4.15 | 50330 | Cu1Ge1O3 |
| | 10 | z2w,2=1 z4=2 |
| | 10 | z2w,1=1 z2w,2=1 z4=3 |
| | 10 | z2w,1=1 z2w,2=1 z2w,3=1 z4=3 |
| | 10 | z2w,2=1 z2w,3=1 |
| | 10 | z2w,1=1 z4=3 |
| | 10 | z2w,3=1 |
| | 10 | z2w,1=1 z2w,3=1 z4=1 |
| | 10 | z2w,1=1 z2w,3=1 z4=3 |
| | 10 | z2w,3=1 |
| | 10 | z2w,1=1 z4=1 |
| | 10 | z2w,2=1 z2w,3=1 |
| | 10 | z2w,1=1 z2w,2=1 z2w,3=1 z4=1 |
| | 10 | z2w,1=1 z2w,2=1 z4=1 |
| | 10 | z2w,2=1 z4=2 |
| 51.4.16 | 50329 | Cu1Ge1O3 |
| | 10 | z2w,2=1 z4=2 |
| | 10 | z2w,1=1 z2w,2=1 z4=3 |
| | 10 | z2w,1=1 z2w,2=1 z2w,3=1 z4=3 |
| | 10 | z2w,2=1 z2w,3=1 |
| | 10 | z2w,1=1 z4=3 |
| | 10 | z2w,3=1 |
| | 10 | z2w,1=1 z2w,3=1 z4=1 |
| | 10 | z2w,1=1 z2w,3=1 z4=3 |

Continued on next column

**TABLE XIX– continued**

| TQC | ICSD SSG | Stoich. Formula SIs |
|---|---|---|
| | 10 | z2w,3=1 |
| | 10 | z2w,1=1 z4=1 |
| | 10 | z2w,2=1 z2w,3=1 |
| | 10 | z2w,1=1 z2w,2=1 z2w,3=1 z4=1 |
| | 10 | z2w,1=1 z2w,2=1 z4=1 |
| | 10 | z2w,2=1 z4=2 |
| 51.4.17 | 411019 | Cu1Ge1O3 |
| | 10 | z2w,2=1 z2w,3=1 z4=2 |
| | 10 | z2w,1=1 z2w,2=1 z2w,3=1 z4=3 |
| | 10 | z2w,1=1 z2w,2=1 z4=3 |
| | 10 | z2w,2=1 |
| | 10 | z2w,1=1 z2w,3=1 z4=3 |
| | 10 | z2w,3=1 |
| | 10 | z2w,1=1 z4=1 |
| | 10 | z2w,1=1 z4=3 |
| | 10 | z2w,3=1 |
| | 10 | z2w,1=1 z2w,3=1 z4=1 |
| | 10 | z2w,2=1 |
| | 10 | z2w,1=1 z2w,2=1 z4=1 |
| | 10 | z2w,1=1 z2w,2=1 z2w,3=1 z4=1 |
| | 10 | z2w,2=1 z2w,3=1 z4=2 |
| 51.4.18 | 411018 | Cu1Ge1O3 |
| | 10 | z2w,2=1 z2w,3=1 z4=2 |
| | 10 | z2w,1=1 z2w,2=1 z2w,3=1 z4=3 |
| | 10 | z2w,1=1 z2w,2=1 z4=3 |
| | 10 | z2w,2=1 |
| | 10 | z2w,1=1 z2w,3=1 z4=3 |
| | 10 | z2w,3=1 |
| | 10 | z2w,1=1 z4=1 |
| | 10 | z2w,1=1 z4=3 |
| | 10 | z2w,3=1 |
| | 10 | z2w,1=1 z2w,3=1 z4=1 |
| | 10 | z2w,2=1 |
| | 10 | z2w,1=1 z2w,2=1 z4=1 |
| | 10 | z2w,1=1 z2w,2=1 z2w,3=1 z4=1 |
| | 10 | z2w,2=1 z2w,3=1 z4=2 |
| 51.4.19 | 411017 | Cu1Ge1O3 |
| | 10 | z2w,2=1 z2w,3=1 z4=2 |
| | 10 | z2w,1=1 z2w,2=1 z2w,3=1 z4=3 |
| | 10 | z2w,1=1 z2w,2=1 z4=3 |
| | 10 | z2w,2=1 |
| | 10 | z2w,1=1 z2w,3=1 z4=3 |
| | 10 | z2w,3=1 |
| | 10 | z2w,1=1 z4=1 |
| | 10 | z2w,1=1 z4=3 |
| | 10 | z2w,3=1 |
| | 10 | z2w,1=1 z2w,3=1 z4=1 |
| | 10 | z2w,2=1 |
| | 10 | z2w,1=1 z2w,2=1 z4=1 |
| | 10 | z2w,1=1 z2w,2=1 z2w,3=1 z4=1 |
| | 10 | z2w,2=1 z2w,3=1 z4=2 |
| 51.4.20 | 25754 | Cu1Ge1O3 |
| | 10 | z2w,2=1 z4=2 |
| | 10 | z2w,1=1 z2w,2=1 z4=3 |
| | 10 | z2w,1=1 z2w,2=1 z2w,3=1 z4=3 |
| | 10 | z2w,2=1 z2w,3=1 |
| | 10 | z2w,1=1 z4=3 |
| | 10 | z2w,3=1 |
| | 10 | z2w,1=1 z2w,3=1 z4=1 |
| | 10 | z2w,1=1 z2w,3=1 z4=3 |

Continued on next column

**TABLE XIX– continued**

| TQC | ICSD SSG | Stoich. Formula SIs |
|---|---|---|
| | 10 | z2w,3=1 |
| | 10 | z2w,1=1 z4=1 |
| | 10 | z2w,2=1 z2w,3=1 |
| | 10 | z2w,1=1 z2w,2=1 z2w,3=1 z4=1 |
| | 10 | z2w,1=1 z2w,2=1 z4=1 |
| | 10 | z2w,2=1 z4=2 |
| 51.4.21 | 62330 | Cu1Fe2Ge2 |
| | 10 | z2w,2=1 z2w,3=1 z4=1 |
| | 10 | z2w,1=1 z2w,2=1 z2w,3=1 z4=2 |
| | 10 | z2w,1=1 z2w,2=1 z4=2 |
| | 10 | z2w,2=1 z4=3 |
| | 10 | z2w,1=1 z2w,3=1 z4=2 |
| | 10 | z2w,3=1 z4=3 |
| | 10 | z4=3 |
| | 10 | z2w,1=1 |
| | 10 | z2w,1=1 z4=2 |
| | 10 | z4=3 |
| | 10 | z2w,3=1 z4=3 |
| | 10 | z2w,1=1 z2w,3=1 |
| | 10 | z2w,2=1 z4=3 |
| | 10 | z2w,1=1 z2w,2=1 |
| | 10 | z2w,1=1 z2w,2=1 z2w,3=1 |
| | 10 | z2w,2=1 z2w,3=1 z4=1 |
| 51.4.22 | 172189 | Cs1Te5U1Zr1 |
| | 10 | z2w,2=1 z2w,3=1 z4=1 |
| | 10 | z2w,1=1 z2w,2=1 z2w,3=1 z4=2 |
| | 10 | z2w,1=1 z2w,2=1 z4=2 |
| | 10 | z2w,2=1 z4=3 |
| | 10 | z2w,1=1 z2w,3=1 z4=2 |
| | 10 | z2w,3=1 z4=3 |
| | 10 | z4=3 |
| | 10 | z2w,1=1 |
| | 10 | z2w,1=1 z4=2 |
| | 10 | z4=3 |
| | 10 | z2w,3=1 z4=3 |
| | 10 | z2w,1=1 z2w,3=1 |
| | 10 | z2w,2=1 z4=3 |
| | 10 | z2w,1=1 z2w,2=1 |
| | 10 | z2w,1=1 z2w,2=1 z2w,3=1 |
| | 10 | z2w,2=1 z2w,3=1 z4=1 |
| 51.4.23 | 79885 | Cs1Te5Ti1U1 |
| | 10 | z2w,2=1 z2w,3=1 z4=1 |
| | 10 | z2w,1=1 z2w,2=1 z2w,3=1 z4=2 |
| | 10 | z2w,1=1 z2w,2=1 z4=2 |
| | 10 | z2w,2=1 z4=3 |
| | 10 | z2w,1=1 z2w,3=1 z4=2 |
| | 10 | z2w,3=1 z4=3 |
| | 10 | z4=3 |
| | 10 | z2w,1=1 |
| | 10 | z2w,1=1 z4=2 |
| | 10 | z4=3 |
| | 10 | z2w,3=1 z4=3 |
| | 10 | z2w,1=1 z2w,3=1 |
| | 10 | z2w,2=1 z4=3 |
| | 10 | z2w,1=1 z2w,2=1 |
| | 10 | z2w,1=1 z2w,2=1 z2w,3=1 |
| | 10 | z2w,2=1 z2w,3=1 z4=1 |
| 51.4.24 | 62856 | Co2Cu1Ge2 |
| | 10 | z4=1 |
| | 10 | z2w,1=1 z4=2 |

Continued on next column

**TABLE XIX– continued**

| TQC | ICSD SSG | Stoich. Formula SIs |
|---|---|---|
| | 10 | z2w,1=1 z2w,3=1 z4=2 |
| | 10 | z2w,3=1 z4=3 |
| | 10 | z2w,1=1 z2w,2=1 z4=2 |
| | 10 | z2w,2=1 z4=3 |
| | 10 | z2w,2=1 z2w,3=1 z4=3 |
| | 10 | z2w,1=1 z2w,2=1 z2w,3=1 |
| | 10 | z2w,1=1 z2w,2=1 z2w,3=1 z4=2 |
| | 10 | z2w,2=1 z2w,3=1 z4=3 |
| | 10 | z2w,2=1 z4=3 |
| | 10 | z2w,1=1 z2w,2=1 |
| | 10 | z2w,3=1 z4=3 |
| | 10 | z2w,1=1 z2w,3=1 |
| | 10 | z2w,1=1 |
| | 10 | z4=1 |
| 51.4.25 | 410891 | Ca1In4Rh1 |
| | 10 | z2w,2=1 |
| | 10 | z2w,1=1 z2w,2=1 z4=1 |
| | 10 | z2w,1=1 z2w,2=1 z2w,3=1 z4=1 |
| | 10 | z2w,2=1 z2w,3=1 z4=2 |
| | 10 | z2w,1=1 z4=1 |
| | 10 | z4=2 |
| | 10 | z2w,3=1 z4=2 |
| | 10 | z2w,1=1 z2w,3=1 z4=3 |
| | 10 | z2w,1=1 z2w,3=1 z4=1 |
| | 10 | z2w,3=1 z4=2 |
| | 10 | z4=2 |
| | 10 | z2w,1=1 z4=3 |
| | 10 | z2w,2=1 z2w,3=1 z4=2 |
| | 10 | z2w,1=1 z2w,2=1 z2w,3=1 z4=3 |
| | 10 | z2w,1=1 z2w,2=1 z4=3 |
| | 10 | z2w,2=1 |
| 51.4.26 | 410892 | Ca1In4Ir1 |
| | 10 | z2w,2=1 z4=3 |
| | 10 | z2w,1=1 z2w,2=1 |
| | 10 | z2w,1=1 z2w,2=1 z2w,3=1 |
| | 10 | z2w,2=1 z2w,3=1 z4=1 |
| | 10 | z2w,1=1 |
| | 10 | z4=1 |
| | 10 | z2w,3=1 z4=1 |
| | 10 | z2w,1=1 z2w,3=1 z4=2 |
| | 10 | z2w,1=1 z2w,3=1 |
| | 10 | z2w,3=1 z4=1 |
| | 10 | z4=1 |
| | 10 | z2w,1=1 z4=2 |
| | 10 | z2w,2=1 z2w,3=1 z4=1 |
| | 10 | z2w,1=1 z2w,2=1 z2w,3=1 z4=2 |
| | 10 | z2w,1=1 z2w,2=1 z4=2 |
| | 10 | z2w,2=1 z4=3 |
| 51.4.27 | 173616 | Ba1In4Ir1 |
| | 10 | z4=2 |
| | 10 | z2w,1=1 z4=3 |
| | 10 | z2w,1=1 z2w,3=1 z4=3 |
| | 10 | z2w,3=1 |
| | 10 | z2w,1=1 z2w,2=1 z4=3 |
| | 10 | z2w,2=1 |
| | 10 | z2w,2=1 z2w,3=1 |
| | 10 | z2w,1=1 z2w,2=1 z2w,3=1 z4=1 |
| | 10 | z2w,1=1 z2w,2=1 z2w,3=1 z4=3 |
| | 10 | z2w,2=1 z2w,3=1 |
| | 10 | z2w,2=1 |

Continued on next column

**TABLE XIX– continued**

| TQC | ICSD SSG | Stoich. Formula SIs |
|---|---|---|
| | 10 | z2w,1=1 z2w,2=1 z4=1 |
| | 10 | z2w,3=1 |
| | 10 | z2w,1=1 z2w,3=1 z4=1 |
| | 10 | z2w,1=1 z4=1 |
| | 10 | z4=2 |
| 51.4.28 | 281204 | Ba1Fe2O5Y1 |
| | 10 | z2w,2=1 z2w,3=1 z4=1 |
| | 10 | z2w,1=1 z2w,2=1 z2w,3=1 z4=2 |
| | 10 | z2w,1=1 z2w,2=1 z4=2 |
| | 10 | z2w,2=1 z4=3 |
| | 10 | z2w,1=1 z2w,3=1 z4=2 |
| | 10 | z2w,3=1 z4=3 |
| | 10 | z4=3 |
| | 10 | z2w,1=1 |
| | 10 | z2w,1=1 z4=2 |
| | 10 | z4=3 |
| | 10 | z2w,3=1 z4=3 |
| | 10 | z2w,1=1 z2w,3=1 |
| | 10 | z2w,2=1 z4=3 |
| | 10 | z2w,1=1 z2w,2=1 |
| | 10 | z2w,1=1 z2w,2=1 z2w,3=1 |
| | 10 | z2w,2=1 z2w,3=1 z4=1 |
| 51.4.29 | 281203 | Ba1Fe2O5Y1 |
| | 10 | z4=2 |
| | 10 | z2w,1=1 z4=3 |
| | 10 | z2w,1=1 z2w,3=1 z4=3 |
| | 10 | z2w,3=1 |
| | 10 | z2w,1=1 z2w,2=1 z4=3 |
| | 10 | z2w,2=1 |
| | 10 | z2w,2=1 z2w,3=1 |
| | 10 | z2w,1=1 z2w,2=1 z2w,3=1 z4=1 |
| | 10 | z2w,1=1 z2w,2=1 z2w,3=1 z4=3 |
| | 10 | z2w,2=1 z2w,3=1 |
| | 10 | z2w,2=1 |
| | 10 | z2w,1=1 z2w,2=1 z4=1 |
| | 10 | z2w,3=1 |
| | 10 | z2w,1=1 z2w,3=1 z4=1 |
| | 10 | z2w,1=1 z4=1 |
| | 10 | z4=2 |
| 51.4.30 | 154678 | Al4Co1Nd1 |
| | 10 | z2w,3=1 z4=3 |
| | 10 | z2w,1=1 z2w,3=1 |
| | 10 | z2w,1=1 |
| | 10 | z4=1 |
| | 10 | z2w,1=1 z2w,2=1 z2w,3=1 |
| | 10 | z2w,2=1 z2w,3=1 z4=1 |
| | 10 | z2w,2=1 z4=1 |
| | 10 | z2w,1=1 z2w,2=1 z4=2 |
| | 10 | z2w,1=1 z2w,2=1 |
| | 10 | z2w,2=1 z4=1 |
| | 10 | z2w,2=1 z2w,3=1 z4=1 |
| | 10 | z2w,1=1 z2w,2=1 z2w,3=1 z4=2 |
| | 10 | z4=1 |
| | 10 | z2w,1=1 z4=2 |
| | 10 | z2w,1=1 z2w,3=1 z4=2 |
| | 10 | z2w,3=1 z4=3 |
| 55.4.1 | 90366 | O4Pr1Sr2 |
| | 10 | z2w,3=1 z4=1 |
| | 10 | z2w,1=1 z4=1 |
| | 10 | z2w,1=1 z2w,3=1 z4=2 |

Continued on next column

**TABLE XIX– continued**

| TQC | ICSD<br>SSG | Stoich. Formula<br>SIs |
|---|---|---|
| | 10 | z2w,1=1 z2w,2=1 z4=1 |
| | 10 | z2w,1=1 z2w,2=1 z2w,3=1 z4=2 |
| | 10 | z2w,2=1 z4=2 |
| | 10 | z2w,2=1 z2w,3=1 z4=3 |
| | 10 | z2w,2=1 z2w,3=1 z4=1 |
| | 10 | z2w,2=1 z4=2 |
| | 10 | z2w,1=1 z2w,2=1 z2w,3=1 z4=2 |
| | 10 | z2w,1=1 z2w,2=1 z4=3 |
| | 10 | z2w,1=1 z2w,3=1 z4=2 |
| | 10 | z2w,1=1 z4=3 |
| | 10 | z2w,3=1 z4=3 |
| 55.4.2 | 237190 | O4Pr1Sr2 |
| | 10 | z2w,3=1 z4=1 |
| | 10 | z2w,1=1 z4=1 |
| | 10 | z2w,1=1 z2w,3=1 z4=2 |
| | 10 | z2w,1=1 z2w,2=1 z4=1 |
| | 10 | z2w,1=1 z2w,2=1 z2w,3=1 z4=2 |
| | 10 | z2w,2=1 z4=2 |
| | 10 | z2w,2=1 z2w,3=1 z4=3 |
| | 10 | z2w,2=1 z2w,3=1 z4=1 |
| | 10 | z2w,2=1 z4=2 |
| | 10 | z2w,1=1 z2w,2=1 z2w,3=1 z4=2 |
| | 10 | z2w,1=1 z2w,2=1 z4=3 |
| | 10 | z2w,1=1 z2w,3=1 z4=2 |
| | 10 | z2w,1=1 z4=3 |
| | 10 | z2w,3=1 z4=3 |
| 55.4.3 | 414506 | In4Rh2Sc5 |
| | 10 | z2w,1=1 z2w,3=1 z4=2 |
| | 10 | z2w,1=1 z4=3 |
| | 10 | z2w,3=1 z4=3 |
| | 10 | z2w,2=1 z2w,3=1 z4=3 |
| | 10 | z2w,2=1 |
| | 10 | z2w,1=1 z2w,2=1 z2w,3=1 |
| | 10 | z2w,1=1 z2w,2=1 z4=1 |
| | 10 | z2w,1=1 z2w,2=1 z4=3 |
| | 10 | z2w,1=1 z2w,2=1 z2w,3=1 |
| | 10 | z2w,2=1 |
| | 10 | z2w,2=1 z2w,3=1 z4=1 |
| | 10 | z2w,3=1 z4=1 |
| | 10 | z2w,1=1 z4=1 |
| | 10 | z2w,1=1 z2w,3=1 z4=2 |
| 55.4.4 | 165133 | In4Pd2Y5 |
| | 10 | z4=1 |
| | 10 | z2w,3=1 z4=2 |
| | 10 | z2w,1=1 z4=2 |
| | 10 | z2w,1=1 z2w,3=1 z4=3 |
| | 10 | z2w,1=1 z2w,2=1 z4=2 |
| | 10 | z2w,1=1 z2w,2=1 z2w,3=1 z4=3 |
| | 10 | z2w,2=1 z4=3 |
| | 10 | z2w,2=1 z2w,3=1 |
| | 10 | z2w,2=1 z2w,3=1 z4=2 |
| | 10 | z2w,2=1 z4=3 |
| | 10 | z2w,1=1 z2w,2=1 z2w,3=1 z4=3 |
| | 10 | z2w,1=1 z2w,2=1 |
| | 10 | z2w,1=1 z2w,3=1 z4=3 |
| | 10 | z2w,1=1 |
| | 10 | z2w,3=1 |
| | 10 | z4=1 |
| 55.4.5 | 414505 | In4Ni2Sc5 |
| | 10 | z2w,3=1 z4=1 |

Continued on next column

**TABLE XIX– continued**

| TQC | ICSD<br>SSG | Stoich. Formula<br>SIs |
|---|---|---|
| | 10 | z2w,1=1 z4=1 |
| | 10 | z2w,1=1 z2w,3=1 z4=2 |
| | 10 | z2w,1=1 z2w,2=1 z4=1 |
| | 10 | z2w,1=1 z2w,2=1 z2w,3=1 z4=2 |
| | 10 | z2w,2=1 z4=2 |
| | 10 | z2w,2=1 z2w,3=1 z4=3 |
| | 10 | z2w,2=1 z2w,3=1 z4=1 |
| | 10 | z2w,2=1 z4=2 |
| | 10 | z2w,1=1 z2w,2=1 z2w,3=1 z4=2 |
| | 10 | z2w,1=1 z2w,2=1 z4=3 |
| | 10 | z2w,1=1 z2w,3=1 z4=2 |
| | 10 | z2w,1=1 z4=3 |
| | 10 | z2w,3=1 z4=3 |
| 55.4.6 | 165139 | In4Lu5Pd2 |
| | 10 | z2w,3=1 z4=1 |
| | 10 | z2w,1=1 z4=1 |
| | 10 | z2w,1=1 z2w,3=1 z4=2 |
| | 10 | z2w,1=1 z2w,2=1 z4=1 |
| | 10 | z2w,1=1 z2w,2=1 z2w,3=1 z4=2 |
| | 10 | z2w,2=1 z4=2 |
| | 10 | z2w,2=1 z2w,3=1 z4=3 |
| | 10 | z2w,2=1 z2w,3=1 z4=1 |
| | 10 | z2w,2=1 z4=2 |
| | 10 | z2w,1=1 z2w,2=1 z2w,3=1 z4=2 |
| | 10 | z2w,1=1 z2w,2=1 z4=3 |
| | 10 | z2w,1=1 z2w,3=1 z4=2 |
| | 10 | z2w,1=1 z4=3 |
| | 10 | z2w,3=1 z4=3 |
| 55.4.7 | 106807 | In4Lu5Ni2 |
| | 10 | z4=3 |
| | 10 | z2w,3=1 |
| | 10 | z2w,1=1 |
| | 10 | z2w,1=1 z2w,3=1 z4=1 |
| | 10 | z2w,1=1 z2w,2=1 |
| | 10 | z2w,1=1 z2w,2=1 z2w,3=1 z4=1 |
| | 10 | z2w,2=1 z4=1 |
| | 10 | z2w,2=1 z2w,3=1 z4=2 |
| | 10 | z2w,2=1 z2w,3=1 |
| | 10 | z2w,2=1 z4=1 |
| | 10 | z2w,1=1 z2w,2=1 z2w,3=1 z4=1 |
| | 10 | z2w,1=1 z2w,2=1 z4=2 |
| | 10 | z2w,1=1 z2w,3=1 z4=1 |
| | 10 | z2w,1=1 z4=2 |
| | 10 | z2w,3=1 z4=2 |
| | 10 | z4=3 |
| 55.4.8 | 417906 | In3Pd5 |
| | 10 | z2w,2=1 |
| | 10 | z2w,2=1 z2w,3=1 z4=1 |
| | 10 | z2w,1=1 z2w,2=1 z4=1 |
| | 10 | z2w,1=1 z2w,2=1 z2w,3=1 z4=2 |
| | 10 | z2w,1=1 z4=1 |
| | 10 | z2w,1=1 z2w,3=1 z4=2 |
| | 10 | z4=2 |
| | 10 | z2w,3=1 z4=3 |
| | 10 | z2w,3=1 z4=1 |
| | 10 | z4=2 |
| | 10 | z2w,1=1 z2w,3=1 z4=2 |
| | 10 | z2w,1=1 z4=3 |
| | 10 | z2w,1=1 z2w,2=1 z2w,3=1 z4=2 |
| | 10 | z2w,1=1 z2w,2=1 z4=3 |

Continued on next column

**TABLE XIX– continued**

| TQC | ICSD SSG | Stoich. Formula SIs |
|---|---|---|
| | 10 | z2w,2=1 z2w,3=1 z4=3 |
| | 10 | z2w,2=1 |
| 55.4.9 | 10004 | Cu4Na3S4 |
| | 10 | z2w,2=1 z4=2 |
| | 10 | z2w,2=1 z2w,3=1 z4=3 |
| | 10 | z2w,1=1 z2w,2=1 z4=3 |
| | 10 | z2w,1=1 z2w,2=1 z2w,3=1 |
| | 10 | z2w,1=1 z4=3 |
| | 10 | z2w,1=1 z2w,3=1 |
| | 10 | z2w,3=1 z4=1 |
| | 10 | z2w,3=1 z4=3 |
| | 10 | z2w,1=1 z2w,3=1 |
| | 10 | z2w,1=1 z4=1 |
| | 10 | z2w,1=1 z2w,2=1 z2w,3=1 |
| | 10 | z2w,1=1 z2w,2=1 z4=1 |
| | 10 | z2w,2=1 z2w,3=1 z4=1 |
| | 10 | z2w,2=1 z4=2 |
| 55.4.10 | 201979 | Cl6N1Sc4 |
| | 10 | z2w,3=1 z4=1 |
| | 10 | z2w,1=1 z4=1 |
| | 10 | z2w,1=1 z2w,3=1 z4=2 |
| | 10 | z2w,1=1 z2w,2=1 z4=1 |
| | 10 | z2w,1=1 z2w,2=1 z2w,3=1 z4=2 |
| | 10 | z2w,2=1 z4=2 |
| | 10 | z2w,2=1 z2w,3=1 z4=3 |
| | 10 | z2w,2=1 z2w,3=1 z4=1 |
| | 10 | z2w,2=1 z4=2 |
| | 10 | z2w,1=1 z2w,2=1 z2w,3=1 z4=2 |
| | 10 | z2w,1=1 z2w,2=1 z4=3 |
| | 10 | z2w,1=1 z2w,3=1 z4=2 |
| | 10 | z2w,1=1 z4=3 |
| | 10 | z2w,3=1 z4=3 |
| 55.4.11 | 201978 | Cl6N1Sc4 |
| | 10 | z2w,3=1 z4=1 |
| | 10 | z2w,1=1 z4=1 |
| | 10 | z2w,1=1 z2w,3=1 z4=2 |
| | 10 | z2w,1=1 z2w,2=1 z4=1 |
| | 10 | z2w,1=1 z2w,2=1 z2w,3=1 z4=2 |
| | 10 | z2w,2=1 z4=2 |
| | 10 | z2w,2=1 z2w,3=1 z4=3 |
| | 10 | z2w,2=1 z2w,3=1 z4=1 |
| | 10 | z2w,2=1 z4=2 |
| | 10 | z2w,1=1 z2w,2=1 z2w,3=1 z4=2 |
| | 10 | z2w,1=1 z2w,2=1 z4=3 |
| | 10 | z2w,1=1 z2w,3=1 z4=2 |
| | 10 | z2w,1=1 z4=3 |
| | 10 | z2w,3=1 z4=3 |
| 55.4.12 | 9136 | Cl4Mn1Na2 |
| | 10 | z2w,3=1 z4=1 |
| | 10 | z2w,1=1 z4=1 |
| | 10 | z2w,1=1 z2w,3=1 z4=2 |
| | 10 | z2w,1=1 z2w,2=1 z4=1 |
| | 10 | z2w,1=1 z2w,2=1 z2w,3=1 z4=2 |
| | 10 | z2w,2=1 z4=2 |
| | 10 | z2w,2=1 z2w,3=1 z4=3 |
| | 10 | z2w,2=1 z2w,3=1 z4=1 |
| | 10 | z2w,2=1 z4=2 |
| | 10 | z2w,1=1 z2w,2=1 z2w,3=1 z4=2 |
| | 10 | z2w,1=1 z2w,2=1 z4=3 |
| | 10 | z2w,1=1 z2w,3=1 z4=2 |

Continued on next column

**TABLE XIX– continued**

| TQC | ICSD SSG | Stoich. Formula SIs |
|---|---|---|
| | 10 | z2w,1=1 z4=3 |
| | 10 | z2w,3=1 z4=3 |
| 55.4.13 | 414344 | C2Ge2La5 |
| | 10 | z2w,1=1 z2w,3=1 z4=2 |
| | 10 | z2w,1=1 z4=3 |
| | 10 | z2w,3=1 z4=3 |
| | 10 | z2w,2=1 z2w,3=1 z4=3 |
| | 10 | z2w,2=1 |
| | 10 | z2w,1=1 z2w,2=1 z2w,3=1 |
| | 10 | z2w,1=1 z2w,2=1 z4=1 |
| | 10 | z2w,1=1 z2w,2=1 z4=3 |
| | 10 | z2w,1=1 z2w,2=1 z2w,3=1 |
| | 10 | z2w,2=1 |
| | 10 | z2w,2=1 z2w,3=1 z4=1 |
| | 10 | z2w,3=1 z4=1 |
| | 10 | z2w,1=1 z4=1 |
| | 10 | z2w,1=1 z2w,3=1 z4=2 |
| 55.4.14 | 417442 | B2Li2Rh3 |
| | 10 | z2w,2=1 z4=1 |
| | 10 | z2w,2=1 z2w,3=1 z4=2 |
| | 10 | z2w,1=1 z2w,2=1 z4=2 |
| | 10 | z2w,1=1 z2w,2=1 z2w,3=1 z4=3 |
| | 10 | z2w,1=1 z4=2 |
| | 10 | z2w,1=1 z2w,3=1 z4=3 |
| | 10 | z4=3 |
| | 10 | z2w,3=1 |
| | 10 | z2w,3=1 z4=2 |
| | 10 | z4=3 |
| | 10 | z2w,1=1 z2w,3=1 z4=3 |
| | 10 | z2w,1=1 |
| | 10 | z2w,1=1 z2w,2=1 z2w,3=1 z4=3 |
| | 10 | z2w,1=1 z2w,2=1 |
| | 10 | z2w,2=1 z2w,3=1 |
| | 10 | z2w,2=1 z4=1 |
| 55.4.15 | 201977 | B1Cl6Sc4 |
| | 10 | z2w,1=1 z2w,3=1 z4=2 |
| | 10 | z2w,1=1 z4=3 |
| | 10 | z2w,3=1 z4=3 |
| | 10 | z2w,2=1 z2w,3=1 z4=3 |
| | 10 | z2w,2=1 |
| | 10 | z2w,1=1 z2w,2=1 z2w,3=1 |
| | 10 | z2w,1=1 z2w,2=1 z4=1 |
| | 10 | z2w,1=1 z2w,2=1 z4=3 |
| | 10 | z2w,1=1 z2w,2=1 z2w,3=1 |
| | 10 | z2w,2=1 |
| | 10 | z2w,2=1 z2w,3=1 z4=1 |
| | 10 | z2w,3=1 z4=1 |
| | 10 | z2w,1=1 z4=1 |
| | 10 | z2w,1=1 z2w,3=1 z4=2 |
| 55.4.16 | 78822 | B10Co1Th2 |
| | 10 | z2w,1=1 z2w,3=1 |
| | 10 | z2w,1=1 z4=1 |
| | 10 | z2w,3=1 z4=1 |
| | 10 | z4=2 |
| | 10 | z2w,2=1 z2w,3=1 z4=1 |
| | 10 | z2w,2=1 z4=2 |
| | 10 | z2w,1=1 z2w,2=1 z2w,3=1 z4=2 |
| | 10 | z2w,1=1 z2w,2=1 z4=3 |
| | 10 | z2w,1=1 z2w,2=1 z4=1 |
| | 10 | z2w,1=1 z2w,2=1 z2w,3=1 z4=2 |

Continued on next column

**TABLE XIX– continued**

| TQC | ICSD SSG | Stoich. Formula SIs |
|---|---|---|
| | 10 | z2w,2=1 z4=2 |
| | 10 | z2w,2=1 z2w,3=1 z4=3 |
| | 10 | z4=2 |
| | 10 | z2w,3=1 z4=3 |
| | 10 | z2w,1=1 z4=3 |
| | 10 | z2w,1=1 z2w,3=1 |
| 55.4.17 | 58186 | Al7Th2 |
| | 10 | z2w,3=1 z4=1 |
| | 10 | z2w,1=1 z4=1 |
| | 10 | z2w,1=1 z2w,3=1 z4=2 |
| | 10 | z2w,1=1 z2w,2=1 z4=1 |
| | 10 | z2w,1=1 z2w,2=1 z2w,3=1 z4=2 |
| | 10 | z2w,2=1 z4=2 |
| | 10 | z2w,2=1 z2w,3=1 z4=3 |
| | 10 | z2w,2=1 z2w,3=1 z4=1 |
| | 10 | z2w,2=1 z4=2 |
| | 10 | z2w,1=1 z2w,2=1 z2w,3=1 z4=2 |
| | 10 | z2w,1=1 z2w,2=1 z4=3 |
| | 10 | z2w,1=1 z2w,3=1 z4=2 |
| | 10 | z2w,1=1 z4=3 |
| | 10 | z2w,3=1 z4=3 |
| 55.4.18 | 55579 | Al3Pt5 |
| | 10 | z2w,2=1 |
| | 10 | z2w,2=1 z2w,3=1 z4=1 |
| | 10 | z2w,1=1 z2w,2=1 z4=1 |
| | 10 | z2w,1=1 z2w,2=1 z2w,3=1 z4=2 |
| | 10 | z2w,1=1 z4=1 |
| | 10 | z2w,1=1 z2w,3=1 z4=2 |
| | 10 | z4=2 |
| | 10 | z2w,3=1 z4=3 |
| | 10 | z2w,3=1 z4=1 |
| | 10 | z4=2 |
| | 10 | z2w,1=1 z2w,3=1 z4=2 |
| | 10 | z2w,1=1 z4=3 |
| | 10 | z2w,1=1 z2w,2=1 z2w,3=1 z4=2 |
| | 10 | z2w,1=1 z2w,2=1 z4=3 |
| | 10 | z2w,2=1 z2w,3=1 z4=3 |
| | 10 | z2w,2=1 |
| 58.4.1 | 48007 | Li1O2Ru1 |
| | 10 | z2w,2=1 z4=1 |
| | 10 | z2w,3=1 z4=1 |
| | 10 | z2w,2=1 z2w,3=1 z4=2 |
| | 10 | z2w,1=1 z4=1 |
| | 10 | z2w,1=1 z2w,2=1 z4=2 |
| | 10 | z2w,1=1 z2w,3=1 z4=2 |
| | 10 | z2w,1=1 z2w,2=1 z2w,3=1 z4=3 |
| | 10 | z2w,1=1 z2w,2=1 z2w,3=1 z4=1 |
| | 10 | z2w,1=1 z2w,3=1 z4=2 |
| | 10 | z2w,1=1 z2w,2=1 z4=2 |
| | 10 | z2w,1=1 z4=3 |
| | 10 | z2w,2=1 z2w,3=1 z4=2 |
| | 10 | z2w,3=1 z4=3 |
| | 10 | z2w,2=1 z4=3 |
| 58.4.2 | 186047 | Fe4Ge2Lu1 |
| | 10 | z2w,2=1 z2w,3=1 |
| | 10 | z2w,3=1 z4=1 |
| | 10 | z2w,2=1 z4=1 |
| | 10 | z4=2 |
| | 10 | z2w,1=1 z2w,2=1 z2w,3=1 z4=1 |
| | 10 | z2w,1=1 z2w,3=1 z4=2 |
| | 10 | z2w,1=1 z2w,2=1 z4=2 |
| | 10 | z2w,1=1 z4=3 |
| | 10 | z2w,1=1 z4=1 |
| | 10 | z2w,1=1 z2w,2=1 z4=2 |
| | 10 | z2w,1=1 z2w,3=1 z4=2 |
| | 10 | z2w,1=1 z2w,2=1 z2w,3=1 z4=3 |
| | 10 | z4=2 |
| | 10 | z2w,2=1 z4=3 |
| | 10 | z2w,3=1 z4=3 |
| | 10 | z2w,2=1 z2w,3=1 |
| 58.4.3 | 186046 | Fe4Ge2Lu1 |
| | 10 | z2w,2=1 z2w,3=1 |
| | 10 | z2w,3=1 z4=1 |
| | 10 | z2w,2=1 z4=1 |
| | 10 | z4=2 |
| | 10 | z2w,1=1 z2w,2=1 z2w,3=1 z4=1 |
| | 10 | z2w,1=1 z2w,3=1 z4=2 |
| | 10 | z2w,1=1 z2w,2=1 z4=2 |
| | 10 | z2w,1=1 z4=3 |
| | 10 | z2w,1=1 z4=1 |
| | 10 | z2w,1=1 z2w,2=1 z4=2 |
| | 10 | z2w,1=1 z2w,3=1 z4=2 |
| | 10 | z2w,1=1 z2w,2=1 z2w,3=1 z4=3 |
| | 10 | z4=2 |
| | 10 | z2w,2=1 z4=3 |
| | 10 | z2w,3=1 z4=3 |
| | 10 | z2w,2=1 z2w,3=1 |
| 58.4.4 | 170634 | Eu3In2P4 |
| | 10 | z2w,1=1 z2w,2=1 z4=2 |
| | 10 | z2w,1=1 z4=3 |
| | 10 | z2w,1=1 z2w,2=1 z2w,3=1 z4=3 |
| | 10 | z2w,1=1 z2w,3=1 |
| | 10 | z2w,2=1 z4=3 |
| | 10 | z2w,2=1 z2w,3=1 |
| | 10 | z2w,3=1 z4=1 |
| | 10 | z2w,3=1 z4=3 |
| | 10 | z2w,2=1 z2w,3=1 |
| | 10 | z2w,2=1 z4=1 |
| | 10 | z2w,1=1 z2w,3=1 |
| | 10 | z2w,1=1 z2w,2=1 z2w,3=1 z4=1 |
| | 10 | z2w,1=1 z4=1 |
| | 10 | z2w,1=1 z2w,2=1 z4=2 |
| 58.4.5 | 2598 | Co3O8U2 |
| | 10 | z2w,2=1 z2w,3=1 z4=1 |
| | 10 | z2w,3=1 z4=2 |
| | 10 | z2w,2=1 z4=2 |
| | 10 | z4=3 |
| | 10 | z2w,1=1 z2w,2=1 z2w,3=1 z4=2 |
| | 10 | z2w,1=1 z2w,3=1 z4=3 |
| | 10 | z2w,1=1 z2w,2=1 z4=3 |
| | 10 | z2w,1=1 |
| | 10 | z2w,1=1 z4=2 |
| | 10 | z2w,1=1 z2w,2=1 z4=3 |
| | 10 | z2w,1=1 z2w,3=1 z4=3 |
| | 10 | z2w,1=1 z2w,2=1 z2w,3=1 |
| | 10 | z4=3 |
| | 10 | z2w,2=1 |
| | 10 | z2w,3=1 |
| | 10 | z2w,2=1 z2w,3=1 z4=1 |
| 58.4.6 | 86115 | Co1Te2 |

Continued on next column

**TABLE XIX– continued**

| TQC | ICSD SSG | Stoich. Formula SIs |
|---|---|---|
| | 10 | z2w,2=1 z4=1 |
| | 10 | z2w,3=1 z4=1 |
| | 10 | z2w,2=1 z2w,3=1 z4=2 |
| | 10 | z2w,1=1 z4=1 |
| | 10 | z2w,1=1 z2w,2=1 z4=2 |
| | 10 | z2w,1=1 z2w,3=1 z4=2 |
| | 10 | z2w,1=1 z2w,2=1 z2w,3=1 z4=3 |
| | 10 | z2w,1=1 z2w,2=1 z2w,3=1 z4=1 |
| | 10 | z2w,1=1 z2w,3=1 z4=2 |
| | 10 | z2w,1=1 z2w,2=1 z4=2 |
| | 10 | z2w,1=1 z4=3 |
| | 10 | z2w,2=1 z2w,3=1 z4=2 |
| | 10 | z2w,3=1 z4=3 |
| | 10 | z2w,2=1 z4=3 |
| 58.4.7 | 92076 | C4Mn1N6 |
| | 10 | z2w,2=1 z2w,3=1 z4=2 |
| | 10 | z2w,3=1 z4=3 |
| | 10 | z2w,2=1 z4=3 |
| | 10 | z2w,1=1 z2w,2=1 z2w,3=1 z4=3 |
| | 10 | z2w,1=1 z2w,3=1 |
| | 10 | z2w,1=1 z2w,2=1 |
| | 10 | z2w,1=1 z4=1 |
| | 10 | z2w,1=1 z4=3 |
| | 10 | z2w,1=1 z2w,2=1 |
| | 10 | z2w,1=1 z2w,3=1 |
| | 10 | z2w,1=1 z2w,2=1 z2w,3=1 z4=1 |
| | 10 | z2w,2=1 z4=1 |
| | 10 | z2w,3=1 z4=1 |
| | 10 | z2w,2=1 z2w,3=1 z4=2 |
| 58.4.8 | 85618 | C4Co1N6 |
| | 10 | z2w,2=1 z4=1 |
| | 10 | z2w,3=1 z4=1 |
| | 10 | z2w,2=1 z2w,3=1 z4=2 |
| | 10 | z2w,1=1 z4=1 |
| | 10 | z2w,1=1 z2w,2=1 z4=2 |
| | 10 | z2w,1=1 z2w,3=1 z4=2 |
| | 10 | z2w,1=1 z2w,2=1 z2w,3=1 z4=3 |
| | 10 | z2w,1=1 z2w,2=1 z2w,3=1 z4=1 |
| | 10 | z2w,1=1 z2w,3=1 z4=2 |
| | 10 | z2w,1=1 z2w,2=1 z4=2 |
| | 10 | z2w,1=1 z4=3 |
| | 10 | z2w,2=1 z2w,3=1 z4=2 |
| | 10 | z2w,3=1 z4=3 |
| | 10 | z2w,2=1 z4=3 |
| 58.4.9 | 422726 | C2H2Co1N4 |
| | 10 | z2w,2=1 z4=1 |
| | 10 | z2w,3=1 z4=1 |
| | 10 | z2w,2=1 z2w,3=1 z4=2 |
| | 10 | z2w,1=1 z4=1 |
| | 10 | z2w,1=1 z2w,2=1 z4=2 |
| | 10 | z2w,1=1 z2w,3=1 z4=2 |
| | 10 | z2w,1=1 z2w,2=1 z2w,3=1 z4=3 |
| | 10 | z2w,1=1 z2w,2=1 z2w,3=1 z4=1 |
| | 10 | z2w,1=1 z2w,3=1 z4=2 |
| | 10 | z2w,1=1 z2w,2=1 z4=2 |
| | 10 | z2w,1=1 z4=3 |
| | 10 | z2w,2=1 z2w,3=1 z4=2 |
| | 10 | z2w,3=1 z4=3 |
| | 10 | z2w,2=1 z4=3 |
| 58.4.10 | 236414 | C2H2Co1N4 |

Continued on next column

**TABLE XIX– continued**

| TQC | ICSD SSG | Stoich. Formula SIs |
|---|---|---|
| | 10 | z2w,2=1 z4=1 |
| | 10 | z2w,3=1 z4=1 |
| | 10 | z2w,2=1 z2w,3=1 z4=2 |
| | 10 | z2w,1=1 z4=1 |
| | 10 | z2w,1=1 z2w,2=1 z4=2 |
| | 10 | z2w,1=1 z2w,3=1 z4=2 |
| | 10 | z2w,1=1 z2w,2=1 z2w,3=1 z4=3 |
| | 10 | z2w,1=1 z2w,2=1 z2w,3=1 z4=1 |
| | 10 | z2w,1=1 z2w,3=1 z4=2 |
| | 10 | z2w,1=1 z2w,2=1 z4=2 |
| | 10 | z2w,1=1 z4=3 |
| | 10 | z2w,2=1 z2w,3=1 z4=2 |
| | 10 | z2w,3=1 z4=3 |
| | 10 | z2w,2=1 z4=3 |
| 58.4.11 | 236413 | C2H2Co1N4 |
| | 10 | z2w,1=1 z2w,2=1 z4=2 |
| | 10 | z2w,1=1 z4=3 |
| | 10 | z2w,1=1 z2w,2=1 z2w,3=1 z4=3 |
| | 10 | z2w,1=1 z2w,3=1 |
| | 10 | z2w,2=1 z4=3 |
| | 10 | z2w,2=1 z2w,3=1 |
| | 10 | z2w,3=1 z4=1 |
| | 10 | z2w,3=1 z4=3 |
| | 10 | z2w,2=1 z2w,3=1 |
| | 10 | z2w,2=1 z4=1 |
| | 10 | z2w,1=1 z2w,3=1 |
| | 10 | z2w,1=1 z2w,2=1 z2w,3=1 z4=1 |
| | 10 | z2w,1=1 z4=1 |
| | 10 | z2w,1=1 z2w,2=1 z4=2 |
| 58.4.12 | 194807 | B4Mn1 |
| | 10 | z2w,1=1 z2w,3=1 z4=2 |
| | 10 | z2w,1=1 z2w,2=1 z2w,3=1 z4=3 |
| | 10 | z2w,1=1 z4=3 |
| | 10 | z2w,1=1 z2w,2=1 |
| | 10 | z2w,3=1 z4=3 |
| | 10 | z2w,2=1 z2w,3=1 |
| | 10 | z2w,2=1 z4=1 |
| | 10 | z2w,2=1 z4=3 |
| | 10 | z2w,2=1 z2w,3=1 |
| | 10 | z2w,3=1 z4=1 |
| | 10 | z2w,1=1 z2w,2=1 |
| | 10 | z2w,1=1 z4=1 |
| | 10 | z2w,1=1 z2w,2=1 z2w,3=1 z4=1 |
| | 10 | z2w,1=1 z2w,3=1 z4=2 |
| 58.4.13 | 236773 | B2Co3O6 |
| | 10 | z4=1 |
| | 10 | z2w,2=1 z4=2 |
| | 10 | z2w,3=1 z4=2 |
| | 10 | z2w,2=1 z2w,3=1 z4=3 |
| | 10 | z2w,1=1 z4=2 |
| | 10 | z2w,1=1 z2w,2=1 z4=3 |
| | 10 | z2w,1=1 z2w,3=1 z4=3 |
| | 10 | z2w,1=1 z2w,2=1 z2w,3=1 |
| | 10 | z2w,1=1 z2w,2=1 z2w,3=1 z4=2 |
| | 10 | z2w,1=1 z2w,3=1 z4=3 |
| | 10 | z2w,1=1 z2w,2=1 z4=3 |
| | 10 | z2w,1=1 |
| | 10 | z2w,2=1 z2w,3=1 z4=3 |
| | 10 | z2w,3=1 |
| | 10 | z2w,2=1 |

Continued on next column

**TABLE XIX– continued**

| TQC | ICSD SSG | Stoich. Formula SIs |
|---|---|---|
| | 10 | z4=1 |
| 58.4.14 | 10487 | B1Pd2 |
| | 10 | z2w,1=1 z2w,2=1 z4=2 |
| | 10 | z2w,1=1 z4=3 |
| | 10 | z2w,1=1 z2w,2=1 z2w,3=1 z4=3 |
| | 10 | z2w,1=1 z2w,3=1 |
| | 10 | z2w,2=1 z4=3 |
| | 10 | z2w,2=1 z2w,3=1 |
| | 10 | z2w,3=1 z4=1 |
| | 10 | z2w,3=1 z4=3 |
| | 10 | z2w,2=1 z2w,3=1 |
| | 10 | z2w,2=1 z4=1 |
| | 10 | z2w,1=1 z2w,3=1 |
| | 10 | z2w,1=1 z2w,2=1 z2w,3=1 z4=1 |
| | 10 | z2w,1=1 z4=1 |
| | 10 | z2w,1=1 z2w,2=1 z4=2 |
| 58.4.15 | 57497 | Al1Au2 |
| | 10 | z2w,2=1 z2w,3=1 z4=2 |
| | 10 | z2w,3=1 z4=3 |
| | 10 | z2w,2=1 z4=3 |
| | 10 | z2w,1=1 z2w,2=1 z2w,3=1 z4=3 |
| | 10 | z2w,1=1 z2w,3=1 |
| | 10 | z2w,1=1 z2w,2=1 |
| | 10 | z2w,1=1 z4=1 |
| | 10 | z2w,1=1 z4=3 |
| | 10 | z2w,1=1 z2w,2=1 |
| | 10 | z2w,1=1 z2w,3=1 |
| | 10 | z2w,1=1 z2w,2=1 z2w,3=1 z4=1 |
| | 10 | z2w,2=1 z4=1 |
| | 10 | z2w,3=1 z4=1 |
| | 10 | z2w,2=1 z2w,3=1 z4=2 |
| 59.4.1 | 241187 | O7Sr1V3 |
| | 2 | z4=3 |
| | 2 | z2w,2=1 z4=2 |
| | 2 | z2w,1=1 z4=2 |
| | 2 | z2w,1=1 z2w,2=1 z4=1 |
| | 2 | z2w,1=1 z2w,3=1 z4=2 |
| | 2 | z2w,1=1 z2w,2=1 z2w,3=1 z4=1 |
| | 2 | z2w,3=1 z4=1 |
| | 2 | z2w,2=1 z2w,3=1 |
| | 2 | z2w,2=1 z2w,3=1 z4=2 |
| | 2 | z2w,3=1 z4=1 |
| | 2 | z2w,1=1 z2w,2=1 z2w,3=1 z4=1 |
| | 2 | z2w,1=1 z2w,3=1 |
| | 2 | z2w,1=1 z2w,2=1 z4=1 |
| | 2 | z2w,1=1 |
| | 2 | z2w,2=1 |
| | 2 | z4=3 |
| | 2 | z2w,1=1 z2w,2=1 z4=2 |
| | 2 | z2w,1=1 z4=1 |
| | 2 | z2w,2=1 z4=1 |
| | 2 | z2w,2=1 z2w,3=1 z4=1 |
| | 2 | z2w,3=1 |
| | 2 | z2w,1=1 z2w,2=1 z2w,3=1 |
| | 2 | z2w,1=1 z2w,3=1 z4=3 |
| | 2 | z2w,1=1 z2w,3=1 z4=1 |
| | 2 | z2w,1=1 z2w,2=1 z2w,3=1 |
| | 2 | z2w,3=1 |
| | 2 | z2w,2=1 z2w,3=1 z4=3 |
| | 2 | z2w,2=1 z4=3 |

Continued on next column

**TABLE XIX– continued**

| TQC | ICSD SSG | Stoich. Formula SIs |
|---|---|---|
| | 2 | z2w,1=1 z4=3 |
| | 2 | z2w,1=1 z2w,2=1 z4=2 |
| | 2 | z2w,1=1 z2w,2=1 z2w,3=1 z4=2 |
| | 2 | z2w,1=1 z2w,3=1 z4=1 |
| | 2 | z2w,2=1 z2w,3=1 z4=1 |
| | 2 | z2w,3=1 |
| | 2 | z2w,2=1 z4=1 |
| | 2 | z2w,1=1 z2w,2=1 |
| | 2 | z2w,1=1 z4=3 |
| | 2 | z2w,1=1 z4=1 |
| | 2 | z2w,1=1 z2w,2=1 |
| | 2 | z2w,2=1 z4=3 |
| | 2 | z2w,3=1 |
| | 2 | z2w,2=1 z2w,3=1 z4=3 |
| | 2 | z2w,1=1 z2w,3=1 z4=3 |
| | 2 | z2w,1=1 z2w,2=1 z2w,3=1 z4=2 |
| | 2 | z2w,3=1 z4=1 |
| | 2 | z2w,2=1 z2w,3=1 |
| | 2 | z2w,1=1 z2w,3=1 |
| | 2 | z2w,1=1 z2w,2=1 z2w,3=1 z4=3 |
| | 2 | z2w,1=1 |
| | 2 | z2w,1=1 z2w,2=1 z4=3 |
| | 2 | z4=3 |
| | 2 | z2w,2=1 z4=2 |
| | 2 | z2w,2=1 |
| | 2 | z4=3 |
| | 2 | z2w,1=1 z2w,2=1 z4=3 |
| | 2 | z2w,1=1 z4=2 |
| | 2 | z2w,1=1 z2w,2=1 z2w,3=1 z4=3 |
| | 2 | z2w,1=1 z2w,3=1 z4=2 |
| | 2 | z2w,2=1 z2w,3=1 z4=2 |
| | 2 | z2w,3=1 z4=1 |
| 59.4.2 | 241186 | O7Sr1V3 |
| | 2 | z4=3 |
| | 2 | z2w,2=1 z4=2 |
| | 2 | z2w,1=1 z4=2 |
| | 2 | z2w,1=1 z2w,2=1 z4=1 |
| | 2 | z2w,1=1 z2w,3=1 z4=2 |
| | 2 | z2w,1=1 z2w,2=1 z2w,3=1 z4=1 |
| | 2 | z2w,3=1 z4=1 |
| | 2 | z2w,2=1 z2w,3=1 |
| | 2 | z2w,2=1 z2w,3=1 z4=2 |
| | 2 | z2w,3=1 z4=1 |
| | 2 | z2w,1=1 z2w,2=1 z2w,3=1 z4=1 |
| | 2 | z2w,1=1 z2w,3=1 |
| | 2 | z2w,1=1 z2w,2=1 z4=1 |
| | 2 | z2w,1=1 |
| | 2 | z2w,2=1 |
| | 2 | z4=3 |
| | 2 | z2w,1=1 z2w,2=1 z4=2 |
| | 2 | z2w,1=1 z4=1 |
| | 2 | z2w,2=1 z4=1 |
| | 2 | z2w,2=1 z2w,3=1 z4=1 |
| | 2 | z2w,3=1 |
| | 2 | z2w,1=1 z2w,2=1 z2w,3=1 |
| | 2 | z2w,1=1 z2w,3=1 z4=3 |
| | 2 | z2w,1=1 z2w,3=1 z4=1 |
| | 2 | z2w,1=1 z2w,2=1 z2w,3=1 |
| | 2 | z2w,3=1 |
| | 2 | z2w,2=1 z2w,3=1 z4=3 |

Continued on next column

**TABLE XIX– continued**

| TQC | ICSD SSG | Stoich. Formula SIs |
|---|---|---|
| | 2 | z2w,2=1 z4=3 |
| | 2 | z2w,1=1 z4=3 |
| | 2 | z2w,1=1 z2w,2=1 z4=2 |
| | 2 | z2w,1=1 z2w,2=1 z2w,3=1 z4=2 |
| | 2 | z2w,1=1 z2w,3=1 z4=1 |
| | 2 | z2w,2=1 z2w,3=1 z4=1 |
| | 2 | z2w,3=1 |
| | 2 | z2w,2=1 z4=1 |
| | 2 | z2w,1=1 z2w,2=1 |
| | 2 | z2w,1=1 z4=3 |
| | 2 | z2w,1=1 z4=1 |
| | 2 | z2w,1=1 z2w,2=1 |
| | 2 | z2w,2=1 z4=3 |
| | 2 | z2w,3=1 |
| | 2 | z2w,2=1 z2w,3=1 z4=3 |
| | 2 | z2w,1=1 z2w,3=1 z4=3 |
| | 2 | z2w,1=1 z2w,2=1 z2w,3=1 z4=2 |
| | 2 | z2w,3=1 z4=1 |
| | 2 | z2w,2=1 z2w,3=1 |
| | 2 | z2w,1=1 z2w,3=1 |
| | 2 | z2w,1=1 z2w,2=1 z2w,3=1 z4=3 |
| | 2 | z2w,1=1 |
| | 2 | z2w,1=1 z2w,2=1 z4=3 |
| | 2 | z4=3 |
| | 2 | z2w,2=1 z4=2 |
| | 2 | z2w,2=1 |
| | 2 | z4=3 |
| | 2 | z2w,1=1 z2w,2=1 z4=3 |
| | 2 | z2w,1=1 z4=2 |
| | 2 | z2w,1=1 z2w,2=1 z2w,3=1 z4=3 |
| | 2 | z2w,1=1 z2w,3=1 z4=2 |
| | 2 | z2w,2=1 z2w,3=1 z4=2 |
| | 2 | z2w,3=1 z4=1 |
| 59.4.3 | 241185 | O7Sr1V3 |
| | 2 | z4=3 |
| | 2 | z2w,2=1 z4=2 |
| | 2 | z2w,1=1 z4=2 |
| | 2 | z2w,1=1 z2w,2=1 z4=1 |
| | 2 | z2w,1=1 z2w,3=1 z4=2 |
| | 2 | z2w,1=1 z2w,2=1 z2w,3=1 z4=1 |
| | 2 | z2w,3=1 z4=1 |
| | 2 | z2w,2=1 z2w,3=1 |
| | 2 | z2w,2=1 z2w,3=1 z4=2 |
| | 2 | z2w,3=1 z4=1 |
| | 2 | z2w,1=1 z2w,2=1 z2w,3=1 z4=1 |
| | 2 | z2w,1=1 z2w,3=1 |
| | 2 | z2w,1=1 z2w,2=1 z4=1 |
| | 2 | z2w,1=1 |
| | 2 | z2w,2=1 |
| | 2 | z4=3 |
| | 2 | z2w,1=1 z2w,2=1 z4=2 |
| | 2 | z2w,1=1 z4=1 |
| | 2 | z2w,2=1 z4=1 |
| | 2 | z2w,2=1 z2w,3=1 z4=1 |
| | 2 | z2w,3=1 |
| | 2 | z2w,1=1 z2w,2=1 z2w,3=1 |
| | 2 | z2w,1=1 z2w,3=1 z4=3 |
| | 2 | z2w,1=1 z2w,3=1 z4=1 |
| | 2 | z2w,1=1 z2w,2=1 z2w,3=1 |
| | 2 | z2w,3=1 |

Continued on next column

**TABLE XIX– continued**

| TQC | ICSD SSG | Stoich. Formula SIs |
|---|---|---|
| | 2 | z2w,2=1 z2w,3=1 z4=3 |
| | 2 | z2w,2=1 z4=3 |
| | 2 | z2w,1=1 z4=3 |
| | 2 | z2w,1=1 z2w,2=1 z4=2 |
| | 2 | z2w,1=1 z2w,2=1 z2w,3=1 z4=2 |
| | 2 | z2w,1=1 z2w,3=1 z4=1 |
| | 2 | z2w,2=1 z2w,3=1 z4=1 |
| | 2 | z2w,3=1 |
| | 2 | z2w,2=1 z4=1 |
| | 2 | z2w,1=1 z2w,2=1 |
| | 2 | z2w,1=1 z4=3 |
| | 2 | z2w,1=1 z4=1 |
| | 2 | z2w,1=1 z2w,2=1 |
| | 2 | z2w,2=1 z4=3 |
| | 2 | z2w,3=1 |
| | 2 | z2w,2=1 z2w,3=1 z4=3 |
| | 2 | z2w,1=1 z2w,3=1 z4=3 |
| | 2 | z2w,1=1 z2w,2=1 z2w,3=1 z4=2 |
| | 2 | z2w,3=1 z4=1 |
| | 2 | z2w,2=1 z2w,3=1 |
| | 2 | z2w,1=1 z2w,3=1 |
| | 2 | z2w,1=1 z2w,2=1 z2w,3=1 z4=3 |
| | 2 | z2w,1=1 |
| | 2 | z2w,1=1 z2w,2=1 z4=3 |
| | 2 | z4=3 |
| | 2 | z2w,2=1 z4=2 |
| | 2 | z2w,2=1 |
| | 2 | z4=3 |
| | 2 | z2w,1=1 z2w,2=1 z4=3 |
| | 2 | z2w,1=1 z4=2 |
| | 2 | z2w,1=1 z2w,2=1 z2w,3=1 z4=3 |
| | 2 | z2w,1=1 z2w,3=1 z4=2 |
| | 2 | z2w,2=1 z2w,3=1 z4=2 |
| | 2 | z2w,3=1 z4=1 |
| 59.4.4 | 241184 | O7Sr1V3 |
| | 2 | z4=3 |
| | 2 | z2w,2=1 z4=2 |
| | 2 | z2w,1=1 z4=2 |
| | 2 | z2w,1=1 z2w,2=1 z4=1 |
| | 2 | z2w,1=1 z2w,3=1 z4=2 |
| | 2 | z2w,1=1 z2w,2=1 z2w,3=1 z4=1 |
| | 2 | z2w,3=1 z4=1 |
| | 2 | z2w,2=1 z2w,3=1 |
| | 2 | z2w,2=1 z2w,3=1 z4=2 |
| | 2 | z2w,3=1 z4=1 |
| | 2 | z2w,1=1 z2w,2=1 z2w,3=1 z4=1 |
| | 2 | z2w,1=1 z2w,3=1 |
| | 2 | z2w,1=1 z2w,2=1 z4=1 |
| | 2 | z2w,1=1 |
| | 2 | z2w,2=1 |
| | 2 | z4=3 |
| | 2 | z2w,1=1 z2w,2=1 z4=2 |
| | 2 | z2w,1=1 z4=1 |
| | 2 | z2w,2=1 z4=1 |
| | 2 | z2w,2=1 z2w,3=1 z4=1 |
| | 2 | z2w,3=1 |
| | 2 | z2w,1=1 z2w,2=1 z2w,3=1 |
| | 2 | z2w,1=1 z2w,3=1 z4=3 |
| | 2 | z2w,1=1 z2w,3=1 z4=1 |
| | 2 | z2w,1=1 z2w,2=1 z2w,3=1 |

Continued on next column

**TABLE XIX– continued**

| TQC | ICSD SSG | Stoich. Formula SIs |
|---|---|---|
| | 2 | z2w,3=1 |
| | 2 | z2w,2=1 z2w,3=1 z4=3 |
| | 2 | z2w,2=1 z4=3 |
| | 2 | z2w,1=1 z4=3 |
| | 2 | z2w,1=1 z2w,2=1 z4=2 |
| | 2 | z2w,1=1 z2w,2=1 z2w,3=1 z4=2 |
| | 2 | z2w,1=1 z2w,3=1 z4=1 |
| | 2 | z2w,2=1 z2w,3=1 z4=1 |
| | 2 | z2w,3=1 |
| | 2 | z2w,2=1 z4=1 |
| | 2 | z2w,1=1 z2w,2=1 |
| | 2 | z2w,1=1 z4=3 |
| | 2 | z2w,1=1 z4=1 |
| | 2 | z2w,1=1 z2w,2=1 |
| | 2 | z2w,2=1 z4=3 |
| | 2 | z2w,3=1 |
| | 2 | z2w,2=1 z2w,3=1 z4=3 |
| | 2 | z2w,1=1 z2w,3=1 z4=3 |
| | 2 | z2w,1=1 z2w,2=1 z2w,3=1 z4=2 |
| | 2 | z2w,3=1 z4=1 |
| | 2 | z2w,2=1 z2w,3=1 |
| | 2 | z2w,1=1 z2w,3=1 |
| | 2 | z2w,1=1 z2w,2=1 z2w,3=1 z4=3 |
| | 2 | z2w,1=1 |
| | 2 | z2w,1=1 z2w,2=1 z4=3 |
| | 2 | z4=3 |
| | 2 | z2w,2=1 z4=2 |
| | 2 | z2w,2=1 |
| | 2 | z4=3 |
| | 2 | z2w,1=1 z2w,2=1 z4=3 |
| | 2 | z2w,1=1 z4=2 |
| | 2 | z2w,1=1 z2w,2=1 z2w,3=1 z4=3 |
| | 2 | z2w,1=1 z2w,3=1 z4=2 |
| | 2 | z2w,2=1 z2w,3=1 z4=2 |
| | 2 | z2w,3=1 z4=1 |
| 59.4.5 | 241183 | O7Sr1V3 |
| | 2 | z4=3 |
| | 2 | z2w,2=1 z4=2 |
| | 2 | z2w,1=1 z4=2 |
| | 2 | z2w,1=1 z2w,2=1 z4=1 |
| | 2 | z2w,1=1 z2w,3=1 z4=2 |
| | 2 | z2w,1=1 z2w,2=1 z2w,3=1 z4=1 |
| | 2 | z2w,3=1 z4=1 |
| | 2 | z2w,2=1 z2w,3=1 |
| | 2 | z2w,2=1 z2w,3=1 z4=2 |
| | 2 | z2w,3=1 z4=1 |
| | 2 | z2w,1=1 z2w,2=1 z2w,3=1 z4=1 |
| | 2 | z2w,1=1 z2w,3=1 |
| | 2 | z2w,1=1 z2w,2=1 z4=1 |
| | 2 | z2w,1=1 |
| | 2 | z2w,2=1 |
| | 2 | z4=3 |
| | 2 | z2w,1=1 z2w,2=1 z4=2 |
| | 2 | z2w,1=1 z4=1 |
| | 2 | z2w,2=1 z4=1 |
| | 2 | z2w,2=1 z2w,3=1 z4=1 |
| | 2 | z2w,3=1 |
| | 2 | z2w,1=1 z2w,2=1 z2w,3=1 |
| | 2 | z2w,1=1 z2w,3=1 z4=3 |
| | 2 | z2w,1=1 z2w,3=1 z4=1 |

Continued on next column

**TABLE XIX– continued**

| TQC | ICSD SSG | Stoich. Formula SIs |
|---|---|---|
| | 2 | z2w,1=1 z2w,2=1 z2w,3=1 |
| | 2 | z2w,3=1 |
| | 2 | z2w,2=1 z2w,3=1 z4=3 |
| | 2 | z2w,2=1 z4=3 |
| | 2 | z2w,1=1 z4=3 |
| | 2 | z2w,1=1 z2w,2=1 z4=2 |
| | 2 | z2w,1=1 z2w,2=1 z2w,3=1 z4=2 |
| | 2 | z2w,1=1 z2w,3=1 z4=1 |
| | 2 | z2w,2=1 z2w,3=1 z4=1 |
| | 2 | z2w,3=1 |
| | 2 | z2w,2=1 z4=1 |
| | 2 | z2w,1=1 z2w,2=1 |
| | 2 | z2w,1=1 z4=3 |
| | 2 | z2w,1=1 z4=1 |
| | 2 | z2w,1=1 z2w,2=1 |
| | 2 | z2w,2=1 z4=3 |
| | 2 | z2w,3=1 |
| | 2 | z2w,2=1 z2w,3=1 z4=3 |
| | 2 | z2w,1=1 z2w,3=1 z4=3 |
| | 2 | z2w,1=1 z2w,2=1 z2w,3=1 z4=2 |
| | 2 | z2w,3=1 z4=1 |
| | 2 | z2w,2=1 z2w,3=1 |
| | 2 | z2w,1=1 z2w,3=1 |
| | 2 | z2w,1=1 z2w,2=1 z2w,3=1 z4=3 |
| | 2 | z2w,1=1 |
| | 2 | z2w,1=1 z2w,2=1 z4=3 |
| | 2 | z4=3 |
| | 2 | z2w,2=1 z4=2 |
| | 2 | z2w,2=1 |
| | 2 | z4=3 |
| | 2 | z2w,1=1 z2w,2=1 z4=3 |
| | 2 | z2w,1=1 z4=2 |
| | 2 | z2w,1=1 z2w,2=1 z2w,3=1 z4=3 |
| | 2 | z2w,1=1 z2w,3=1 z4=2 |
| | 2 | z2w,2=1 z2w,3=1 z4=2 |
| | 2 | z2w,3=1 z4=1 |
| 59.4.6 | 241182 | O7Sr1V3 |
| | 2 | z4=3 |
| | 2 | z2w,2=1 z4=2 |
| | 2 | z2w,1=1 z4=2 |
| | 2 | z2w,1=1 z2w,2=1 z4=1 |
| | 2 | z2w,1=1 z2w,3=1 z4=2 |
| | 2 | z2w,1=1 z2w,2=1 z2w,3=1 z4=1 |
| | 2 | z2w,3=1 z4=1 |
| | 2 | z2w,2=1 z2w,3=1 |
| | 2 | z2w,2=1 z2w,3=1 z4=2 |
| | 2 | z2w,3=1 z4=1 |
| | 2 | z2w,1=1 z2w,2=1 z2w,3=1 z4=1 |
| | 2 | z2w,1=1 z2w,3=1 |
| | 2 | z2w,1=1 z2w,2=1 z4=1 |
| | 2 | z2w,1=1 |
| | 2 | z2w,2=1 |
| | 2 | z4=3 |
| | 2 | z2w,1=1 z2w,2=1 z4=2 |
| | 2 | z2w,1=1 z4=1 |
| | 2 | z2w,2=1 z4=1 |
| | 2 | z2w,2=1 z2w,3=1 z4=1 |
| | 2 | z2w,3=1 |
| | 2 | z2w,1=1 z2w,2=1 z2w,3=1 |
| | 2 | z2w,1=1 z2w,3=1 z4=3 |

Continued on next column

**TABLE XIX– continued**

| TQC | ICSD SSG | Stoich. Formula SIs |
|---|---|---|
| | 2 | z2w,1=1 z2w,3=1 z4=1 |
| | 2 | z2w,1=1 z2w,2=1 z2w,3=1 |
| | 2 | z2w,3=1 |
| | 2 | z2w,2=1 z2w,3=1 z4=3 |
| | 2 | z2w,2=1 z4=3 |
| | 2 | z2w,1=1 z4=3 |
| | 2 | z2w,1=1 z2w,2=1 z4=2 |
| | 2 | z2w,1=1 z2w,2=1 z2w,3=1 z4=2 |
| | 2 | z2w,1=1 z2w,3=1 z4=1 |
| | 2 | z2w,2=1 z2w,3=1 z4=1 |
| | 2 | z2w,3=1 |
| | 2 | z2w,2=1 z4=1 |
| | 2 | z2w,1=1 z2w,2=1 |
| | 2 | z2w,1=1 z4=3 |
| | 2 | z2w,1=1 z4=1 |
| | 2 | z2w,1=1 z2w,2=1 |
| | 2 | z2w,2=1 z4=3 |
| | 2 | z2w,3=1 |
| | 2 | z2w,2=1 z2w,3=1 z4=3 |
| | 2 | z2w,1=1 z2w,3=1 z4=3 |
| | 2 | z2w,1=1 z2w,2=1 z2w,3=1 z4=2 |
| | 2 | z2w,3=1 z4=1 |
| | 2 | z2w,2=1 z2w,3=1 |
| | 2 | z2w,1=1 z2w,3=1 |
| | 2 | z2w,1=1 z2w,2=1 z2w,3=1 z4=3 |
| | 2 | z2w,1=1 |
| | 2 | z2w,1=1 z2w,2=1 z4=3 |
| | 2 | z4=3 |
| | 2 | z2w,2=1 z4=2 |
| | 2 | z2w,2=1 |
| | 2 | z4=3 |
| | 2 | z2w,1=1 z2w,2=1 z4=3 |
| | 2 | z2w,1=1 z4=2 |
| | 2 | z2w,1=1 z2w,2=1 z2w,3=1 z4=3 |
| | 2 | z2w,1=1 z2w,3=1 z4=2 |
| | 2 | z2w,2=1 z2w,3=1 z4=2 |
| | 2 | z2w,3=1 z4=1 |
| 59.4.7 | 241181 | O7Sr1V3 |
| | 2 | z4=3 |
| | 2 | z2w,2=1 z4=2 |
| | 2 | z2w,1=1 z4=2 |
| | 2 | z2w,1=1 z2w,2=1 z4=1 |
| | 2 | z2w,1=1 z2w,3=1 z4=2 |
| | 2 | z2w,1=1 z2w,2=1 z2w,3=1 z4=1 |
| | 2 | z2w,3=1 z4=1 |
| | 2 | z2w,2=1 z2w,3=1 |
| | 2 | z2w,2=1 z2w,3=1 z4=2 |
| | 2 | z2w,3=1 z4=1 |
| | 2 | z2w,1=1 z2w,2=1 z2w,3=1 z4=1 |
| | 2 | z2w,1=1 z2w,3=1 |
| | 2 | z2w,1=1 z2w,2=1 z4=1 |
| | 2 | z2w,1=1 |
| | 2 | z2w,2=1 |
| | 2 | z4=3 |
| | 2 | z2w,1=1 z2w,2=1 z4=2 |
| | 2 | z2w,1=1 z4=1 |
| | 2 | z2w,2=1 z4=1 |
| | 2 | z2w,2=1 z2w,3=1 z4=1 |
| | 2 | z2w,3=1 |
| | 2 | z2w,1=1 z2w,2=1 z2w,3=1 |

Continued on next column

**TABLE XIX– continued**

| TQC | ICSD SSG | Stoich. Formula SIs |
|---|---|---|
| | 2 | z2w,1=1 z2w,3=1 z4=3 |
| | 2 | z2w,1=1 z2w,3=1 z4=1 |
| | 2 | z2w,1=1 z2w,2=1 z2w,3=1 |
| | 2 | z2w,3=1 |
| | 2 | z2w,2=1 z2w,3=1 z4=3 |
| | 2 | z2w,2=1 z4=3 |
| | 2 | z2w,1=1 z4=3 |
| | 2 | z2w,1=1 z2w,2=1 z4=2 |
| | 2 | z2w,1=1 z2w,2=1 z2w,3=1 z4=2 |
| | 2 | z2w,1=1 z2w,3=1 z4=1 |
| | 2 | z2w,2=1 z2w,3=1 z4=1 |
| | 2 | z2w,3=1 |
| | 2 | z2w,2=1 z4=1 |
| | 2 | z2w,1=1 z2w,2=1 |
| | 2 | z2w,1=1 z4=3 |
| | 2 | z2w,1=1 z4=1 |
| | 2 | z2w,1=1 z2w,2=1 |
| | 2 | z2w,2=1 z4=3 |
| | 2 | z2w,3=1 |
| | 2 | z2w,2=1 z2w,3=1 z4=3 |
| | 2 | z2w,1=1 z2w,3=1 z4=3 |
| | 2 | z2w,1=1 z2w,2=1 z2w,3=1 z4=2 |
| | 2 | z2w,3=1 z4=1 |
| | 2 | z2w,2=1 z2w,3=1 |
| | 2 | z2w,1=1 z2w,3=1 |
| | 2 | z2w,1=1 z2w,2=1 z2w,3=1 z4=3 |
| | 2 | z2w,1=1 |
| | 2 | z2w,1=1 z2w,2=1 z4=3 |
| | 2 | z4=3 |
| | 2 | z2w,2=1 z4=2 |
| | 2 | z2w,2=1 |
| | 2 | z4=3 |
| | 2 | z2w,1=1 z2w,2=1 z4=3 |
| | 2 | z2w,1=1 z4=2 |
| | 2 | z2w,1=1 z2w,2=1 z2w,3=1 z4=3 |
| | 2 | z2w,1=1 z2w,3=1 z4=2 |
| | 2 | z2w,2=1 z2w,3=1 z4=2 |
| | 2 | z2w,3=1 z4=1 |
| 59.4.8 | 241180 | O7Sr1V3 |
| | 2 | z4=3 |
| | 2 | z2w,2=1 z4=2 |
| | 2 | z2w,1=1 z4=2 |
| | 2 | z2w,1=1 z2w,2=1 z4=1 |
| | 2 | z2w,1=1 z2w,3=1 z4=2 |
| | 2 | z2w,1=1 z2w,2=1 z2w,3=1 z4=1 |
| | 2 | z2w,3=1 z4=1 |
| | 2 | z2w,2=1 z2w,3=1 |
| | 2 | z2w,2=1 z2w,3=1 z4=2 |
| | 2 | z2w,3=1 z4=1 |
| | 2 | z2w,1=1 z2w,2=1 z2w,3=1 z4=1 |
| | 2 | z2w,1=1 z2w,3=1 |
| | 2 | z2w,1=1 z2w,2=1 z4=1 |
| | 2 | z2w,1=1 |
| | 2 | z2w,2=1 |
| | 2 | z4=3 |
| | 2 | z2w,1=1 z2w,2=1 z4=2 |
| | 2 | z2w,1=1 z4=1 |
| | 2 | z2w,2=1 z4=1 |
| | 2 | z2w,2=1 z2w,3=1 z4=1 |
| | 2 | z2w,3=1 |

Continued on next column

**TABLE XIX– continued**

| TQC | ICSD SSG | Stoich. Formula SIs |
|---|---|---|
| | 2 | z2w,1=1 z2w,2=1 z2w,3=1 |
| | 2 | z2w,1=1 z2w,3=1 z4=3 |
| | 2 | z2w,1=1 z2w,3=1 z4=1 |
| | 2 | z2w,1=1 z2w,2=1 z2w,3=1 |
| | 2 | z2w,3=1 |
| | 2 | z2w,2=1 z2w,3=1 z4=3 |
| | 2 | z2w,2=1 z4=3 |
| | 2 | z2w,1=1 z4=3 |
| | 2 | z2w,1=1 z2w,2=1 z4=2 |
| | 2 | z2w,1=1 z2w,2=1 z2w,3=1 z4=2 |
| | 2 | z2w,1=1 z2w,3=1 z4=1 |
| | 2 | z2w,2=1 z2w,3=1 z4=1 |
| | 2 | z2w,3=1 |
| | 2 | z2w,2=1 z4=1 |
| | 2 | z2w,1=1 z2w,2=1 |
| | 2 | z2w,1=1 z4=3 |
| | 2 | z2w,1=1 z4=1 |
| | 2 | z2w,1=1 z2w,2=1 |
| | 2 | z2w,2=1 z4=3 |
| | 2 | z2w,3=1 |
| | 2 | z2w,2=1 z2w,3=1 z4=3 |
| | 2 | z2w,1=1 z2w,3=1 z4=3 |
| | 2 | z2w,1=1 z2w,2=1 z2w,3=1 z4=2 |
| | 2 | z2w,3=1 z4=1 |
| | 2 | z2w,2=1 z2w,3=1 |
| | 2 | z2w,1=1 z2w,3=1 |
| | 2 | z2w,1=1 z2w,2=1 z2w,3=1 z4=3 |
| | 2 | z2w,1=1 |
| | 2 | z2w,1=1 z2w,2=1 z4=3 |
| | 2 | z4=3 |
| | 2 | z2w,2=1 z4=2 |
| | 2 | z2w,2=1 |
| | 2 | z4=3 |
| | 2 | z2w,1=1 z2w,2=1 z4=3 |
| | 2 | z2w,1=1 z4=2 |
| | 2 | z2w,1=1 z2w,2=1 z2w,3=1 z4=3 |
| | 2 | z2w,1=1 z2w,3=1 z4=2 |
| | 2 | z2w,2=1 z2w,3=1 z4=2 |
| | 2 | z2w,3=1 z4=1 |
| 59.4.9 | 241179 | O7Sr1V3 |
| | 2 | z4=3 |
| | 2 | z2w,2=1 z4=2 |
| | 2 | z2w,1=1 z4=2 |
| | 2 | z2w,1=1 z2w,2=1 z4=1 |
| | 2 | z2w,1=1 z2w,3=1 z4=2 |
| | 2 | z2w,1=1 z2w,2=1 z2w,3=1 z4=1 |
| | 2 | z2w,3=1 z4=1 |
| | 2 | z2w,2=1 z2w,3=1 |
| | 2 | z2w,2=1 z2w,3=1 z4=2 |
| | 2 | z2w,3=1 z4=1 |
| | 2 | z2w,1=1 z2w,2=1 z2w,3=1 z4=1 |
| | 2 | z2w,1=1 z2w,3=1 |
| | 2 | z2w,1=1 z2w,2=1 z4=1 |
| | 2 | z2w,1=1 |
| | 2 | z2w,2=1 |
| | 2 | z4=3 |
| | 2 | z2w,1=1 z2w,2=1 z4=2 |
| | 2 | z2w,1=1 z4=1 |
| | 2 | z2w,2=1 z4=1 |
| | 2 | z2w,2=1 z2w,3=1 z4=1 |

Continued on next column

**TABLE XIX– continued**

| TQC | ICSD SSG | Stoich. Formula SIs |
|---|---|---|
| | 2 | z2w,3=1 |
| | 2 | z2w,1=1 z2w,2=1 z2w,3=1 |
| | 2 | z2w,1=1 z2w,3=1 z4=3 |
| | 2 | z2w,1=1 z2w,3=1 z4=1 |
| | 2 | z2w,1=1 z2w,2=1 z2w,3=1 |
| | 2 | z2w,3=1 |
| | 2 | z2w,2=1 z2w,3=1 z4=3 |
| | 2 | z2w,2=1 z4=3 |
| | 2 | z2w,1=1 z4=3 |
| | 2 | z2w,1=1 z2w,2=1 z4=2 |
| | 2 | z2w,1=1 z2w,2=1 z2w,3=1 z4=2 |
| | 2 | z2w,1=1 z2w,3=1 z4=1 |
| | 2 | z2w,2=1 z2w,3=1 z4=1 |
| | 2 | z2w,3=1 |
| | 2 | z2w,2=1 z4=1 |
| | 2 | z2w,1=1 z2w,2=1 |
| | 2 | z2w,1=1 z4=3 |
| | 2 | z2w,1=1 z4=1 |
| | 2 | z2w,1=1 z2w,2=1 |
| | 2 | z2w,2=1 z4=3 |
| | 2 | z2w,3=1 |
| | 2 | z2w,2=1 z2w,3=1 z4=3 |
| | 2 | z2w,1=1 z2w,3=1 z4=3 |
| | 2 | z2w,1=1 z2w,2=1 z2w,3=1 z4=2 |
| | 2 | z2w,3=1 z4=1 |
| | 2 | z2w,2=1 z2w,3=1 |
| | 2 | z2w,1=1 z2w,3=1 |
| | 2 | z2w,1=1 z2w,2=1 z2w,3=1 z4=3 |
| | 2 | z2w,1=1 |
| | 2 | z2w,1=1 z2w,2=1 z4=3 |
| | 2 | z4=3 |
| | 2 | z2w,2=1 z4=2 |
| | 2 | z2w,2=1 |
| | 2 | z4=3 |
| | 2 | z2w,1=1 z2w,2=1 z4=3 |
| | 2 | z2w,1=1 z4=2 |
| | 2 | z2w,1=1 z2w,2=1 z2w,3=1 z4=3 |
| | 2 | z2w,1=1 z2w,3=1 z4=2 |
| | 2 | z2w,2=1 z2w,3=1 z4=2 |
| | 2 | z2w,3=1 z4=1 |
| 59.4.10 | 241178 | O7Sr1V3 |
| | 2 | z4=3 |
| | 2 | z2w,2=1 z4=2 |
| | 2 | z2w,1=1 z4=2 |
| | 2 | z2w,1=1 z2w,2=1 z4=1 |
| | 2 | z2w,1=1 z2w,3=1 z4=2 |
| | 2 | z2w,1=1 z2w,2=1 z2w,3=1 z4=1 |
| | 2 | z2w,3=1 z4=1 |
| | 2 | z2w,2=1 z2w,3=1 |
| | 2 | z2w,2=1 z2w,3=1 z4=2 |
| | 2 | z2w,3=1 z4=1 |
| | 2 | z2w,1=1 z2w,2=1 z2w,3=1 z4=1 |
| | 2 | z2w,1=1 z2w,3=1 |
| | 2 | z2w,1=1 z2w,2=1 z4=1 |
| | 2 | z2w,1=1 |
| | 2 | z2w,2=1 |
| | 2 | z4=3 |
| | 2 | z2w,1=1 z2w,2=1 z4=2 |
| | 2 | z2w,1=1 z4=1 |
| | 2 | z2w,2=1 z4=1 |

Continued on next column

**TABLE XIX– continued**

| TQC | ICSD<br>SSG | Stoich. Formula<br>SIs |
|---|---|---|
| | 2 | z2w,2=1 z2w,3=1 z4=1 |
| | 2 | z2w,3=1 |
| | 2 | z2w,1=1 z2w,2=1 z2w,3=1 |
| | 2 | z2w,1=1 z2w,3=1 z4=3 |
| | 2 | z2w,1=1 z2w,3=1 z4=1 |
| | 2 | z2w,1=1 z2w,2=1 z2w,3=1 |
| | 2 | z2w,3=1 |
| | 2 | z2w,2=1 z2w,3=1 z4=3 |
| | 2 | z2w,2=1 z4=3 |
| | 2 | z2w,1=1 z4=3 |
| | 2 | z2w,1=1 z2w,2=1 z4=2 |
| | 2 | z2w,1=1 z2w,2=1 z2w,3=1 z4=2 |
| | 2 | z2w,1=1 z2w,3=1 z4=1 |
| | 2 | z2w,2=1 z2w,3=1 z4=1 |
| | 2 | z2w,3=1 |
| | 2 | z2w,2=1 z4=1 |
| | 2 | z2w,1=1 z2w,2=1 |
| | 2 | z2w,1=1 z4=3 |
| | 2 | z2w,1=1 z4=1 |
| | 2 | z2w,1=1 z2w,2=1 |
| | 2 | z2w,2=1 z4=3 |
| | 2 | z2w,3=1 |
| | 2 | z2w,2=1 z2w,3=1 z4=3 |
| | 2 | z2w,1=1 z2w,3=1 z4=3 |
| | 2 | z2w,1=1 z2w,2=1 z2w,3=1 z4=2 |
| | 2 | z2w,3=1 z4=1 |
| | 2 | z2w,2=1 z2w,3=1 |
| | 2 | z2w,1=1 z2w,3=1 |
| | 2 | z2w,1=1 z2w,2=1 z2w,3=1 z4=3 |
| | 2 | z2w,1=1 |
| | 2 | z2w,1=1 z2w,2=1 z4=3 |
| | 2 | z4=3 |
| | 2 | z2w,2=1 z4=2 |
| | 2 | z2w,2=1 |
| | 2 | z4=3 |
| | 2 | z2w,1=1 z2w,2=1 z4=3 |
| | 2 | z2w,1=1 z4=2 |
| | 2 | z2w,1=1 z2w,2=1 z2w,3=1 z4=3 |
| | 2 | z2w,1=1 z2w,3=1 z4=2 |
| | 2 | z2w,2=1 z2w,3=1 z4=2 |
| | 2 | z2w,3=1 z4=1 |
| 59.4.11 | 241177 | O7Sr1V3 |
| | 2 | z4=3 |
| | 2 | z2w,2=1 z4=2 |
| | 2 | z2w,1=1 z4=2 |
| | 2 | z2w,1=1 z2w,2=1 z4=1 |
| | 2 | z2w,1=1 z2w,3=1 z4=2 |
| | 2 | z2w,1=1 z2w,2=1 z2w,3=1 z4=1 |
| | 2 | z2w,3=1 z4=1 |
| | 2 | z2w,2=1 z2w,3=1 |
| | 2 | z2w,2=1 z2w,3=1 z4=2 |
| | 2 | z2w,3=1 z4=1 |
| | 2 | z2w,1=1 z2w,2=1 z2w,3=1 z4=1 |
| | 2 | z2w,1=1 z2w,3=1 |
| | 2 | z2w,1=1 z2w,2=1 z4=1 |
| | 2 | z2w,1=1 |
| | 2 | z2w,2=1 |
| | 2 | z4=3 |
| | 2 | z2w,1=1 z2w,2=1 z4=2 |
| | 2 | z2w,1=1 z4=1 |

Continued on next column

**TABLE XIX– continued**

| TQC | ICSD<br>SSG | Stoich. Formula<br>SIs |
|---|---|---|
| | 2 | z2w,2=1 z4=1 |
| | 2 | z2w,2=1 z2w,3=1 z4=1 |
| | 2 | z2w,3=1 |
| | 2 | z2w,1=1 z2w,2=1 z2w,3=1 |
| | 2 | z2w,1=1 z2w,3=1 z4=3 |
| | 2 | z2w,1=1 z2w,3=1 z4=1 |
| | 2 | z2w,1=1 z2w,2=1 z2w,3=1 |
| | 2 | z2w,3=1 |
| | 2 | z2w,2=1 z2w,3=1 z4=3 |
| | 2 | z2w,2=1 z4=3 |
| | 2 | z2w,1=1 z4=3 |
| | 2 | z2w,1=1 z2w,2=1 z4=2 |
| | 2 | z2w,1=1 z2w,2=1 z2w,3=1 z4=2 |
| | 2 | z2w,1=1 z2w,3=1 z4=1 |
| | 2 | z2w,2=1 z2w,3=1 z4=1 |
| | 2 | z2w,3=1 |
| | 2 | z2w,2=1 z4=1 |
| | 2 | z2w,1=1 z2w,2=1 |
| | 2 | z2w,1=1 z4=3 |
| | 2 | z2w,1=1 z4=1 |
| | 2 | z2w,1=1 z2w,2=1 |
| | 2 | z2w,2=1 z4=3 |
| | 2 | z2w,3=1 |
| | 2 | z2w,2=1 z2w,3=1 z4=3 |
| | 2 | z2w,1=1 z2w,3=1 z4=3 |
| | 2 | z2w,1=1 z2w,2=1 z2w,3=1 z4=2 |
| | 2 | z2w,3=1 z4=1 |
| | 2 | z2w,2=1 z2w,3=1 |
| | 2 | z2w,1=1 z2w,3=1 |
| | 2 | z2w,1=1 z2w,2=1 z2w,3=1 z4=3 |
| | 2 | z2w,1=1 |
| | 2 | z2w,1=1 z2w,2=1 z4=3 |
| | 2 | z4=3 |
| | 2 | z2w,2=1 z4=2 |
| | 2 | z2w,2=1 |
| | 2 | z4=3 |
| | 2 | z2w,1=1 z2w,2=1 z4=3 |
| | 2 | z2w,1=1 z4=2 |
| | 2 | z2w,1=1 z2w,2=1 z2w,3=1 z4=3 |
| | 2 | z2w,1=1 z2w,3=1 z4=2 |
| | 2 | z2w,2=1 z2w,3=1 z4=2 |
| | 2 | z2w,3=1 z4=1 |
| 59.4.12 | 241176 | O7Sr1V3 |
| | 2 | z4=3 |
| | 2 | z2w,2=1 z4=2 |
| | 2 | z2w,1=1 z4=2 |
| | 2 | z2w,1=1 z2w,2=1 z4=1 |
| | 2 | z2w,1=1 z2w,3=1 z4=2 |
| | 2 | z2w,1=1 z2w,2=1 z2w,3=1 z4=1 |
| | 2 | z2w,3=1 z4=1 |
| | 2 | z2w,2=1 z2w,3=1 |
| | 2 | z2w,2=1 z2w,3=1 z4=2 |
| | 2 | z2w,3=1 z4=1 |
| | 2 | z2w,1=1 z2w,2=1 z2w,3=1 z4=1 |
| | 2 | z2w,1=1 z2w,3=1 |
| | 2 | z2w,1=1 z2w,2=1 z4=1 |
| | 2 | z2w,1=1 |
| | 2 | z2w,2=1 |
| | 2 | z4=3 |
| | 2 | z2w,1=1 z2w,2=1 z4=2 |

Continued on next column

**TABLE XIX– continued**

| TQC | ICSD<br>SSG | Stoich. Formula<br>SIs |
|---|---|---|
| | 2 | z2w,1=1 z4=1 |
| | 2 | z2w,2=1 z4=1 |
| | 2 | z2w,2=1 z2w,3=1 z4=1 |
| | 2 | z2w,3=1 |
| | 2 | z2w,1=1 z2w,2=1 z2w,3=1 |
| | 2 | z2w,1=1 z2w,3=1 z4=3 |
| | 2 | z2w,1=1 z2w,3=1 z4=1 |
| | 2 | z2w,1=1 z2w,2=1 z2w,3=1 |
| | 2 | z2w,3=1 |
| | 2 | z2w,2=1 z2w,3=1 z4=3 |
| | 2 | z2w,2=1 z4=3 |
| | 2 | z2w,1=1 z4=3 |
| | 2 | z2w,1=1 z2w,2=1 z4=2 |
| | 2 | z2w,1=1 z2w,2=1 z2w,3=1 z4=2 |
| | 2 | z2w,1=1 z2w,3=1 z4=1 |
| | 2 | z2w,2=1 z2w,3=1 z4=1 |
| | 2 | z2w,3=1 |
| | 2 | z2w,2=1 z4=1 |
| | 2 | z2w,1=1 z2w,2=1 |
| | 2 | z2w,1=1 z4=3 |
| | 2 | z2w,1=1 z4=1 |
| | 2 | z2w,1=1 z2w,2=1 |
| | 2 | z2w,2=1 z4=3 |
| | 2 | z2w,3=1 |
| | 2 | z2w,2=1 z2w,3=1 z4=3 |
| | 2 | z2w,1=1 z2w,3=1 z4=3 |
| | 2 | z2w,1=1 z2w,2=1 z2w,3=1 z4=2 |
| | 2 | z2w,3=1 z4=1 |
| | 2 | z2w,2=1 z2w,3=1 |
| | 2 | z2w,1=1 z2w,3=1 |
| | 2 | z2w,1=1 z2w,2=1 z2w,3=1 z4=3 |
| | 2 | z2w,1=1 |
| | 2 | z2w,1=1 z2w,2=1 z4=3 |
| | 2 | z4=3 |
| | 2 | z2w,2=1 z4=2 |
| | 2 | z2w,2=1 |
| | 2 | z4=3 |
| | 2 | z2w,1=1 z2w,2=1 z4=3 |
| | 2 | z2w,1=1 z4=2 |
| | 2 | z2w,1=1 z2w,2=1 z2w,3=1 z4=3 |
| | 2 | z2w,1=1 z2w,3=1 z4=2 |
| | 2 | z2w,2=1 z2w,3=1 z4=2 |
| | 2 | z2w,3=1 z4=1 |
| 59.4.13 | 88354 | Na1O5V2 |
| | 2 | z4=3 |
| | 2 | z2w,2=1 z4=2 |
| | 2 | z2w,1=1 z4=2 |
| | 2 | z2w,1=1 z2w,2=1 z4=1 |
| | 2 | z2w,1=1 z2w,3=1 z4=2 |
| | 2 | z2w,1=1 z2w,2=1 z2w,3=1 z4=1 |
| | 2 | z2w,3=1 z4=1 |
| | 2 | z2w,2=1 z2w,3=1 |
| | 2 | z2w,2=1 z2w,3=1 z4=2 |
| | 2 | z2w,3=1 z4=1 |
| | 2 | z2w,1=1 z2w,2=1 z2w,3=1 z4=1 |
| | 2 | z2w,1=1 z2w,3=1 |
| | 2 | z2w,1=1 z2w,2=1 z4=1 |
| | 2 | z2w,1=1 |
| | 2 | z2w,2=1 |
| | 2 | z4=3 |

Continued on next column

**TABLE XIX– continued**

| TQC | ICSD<br>SSG | Stoich. Formula<br>SIs |
|---|---|---|
| | 2 | z2w,1=1 z2w,2=1 z4=2 |
| | 2 | z2w,1=1 z4=1 |
| | 2 | z2w,2=1 z4=1 |
| | 2 | z2w,2=1 z2w,3=1 z4=1 |
| | 2 | z2w,3=1 |
| | 2 | z2w,1=1 z2w,2=1 z2w,3=1 |
| | 2 | z2w,1=1 z2w,3=1 z4=3 |
| | 2 | z2w,1=1 z2w,3=1 z4=1 |
| | 2 | z2w,1=1 z2w,2=1 z2w,3=1 |
| | 2 | z2w,3=1 |
| | 2 | z2w,2=1 z2w,3=1 z4=3 |
| | 2 | z2w,2=1 z4=3 |
| | 2 | z2w,1=1 z4=3 |
| | 2 | z2w,1=1 z2w,2=1 z4=2 |
| | 2 | z2w,1=1 z2w,2=1 z2w,3=1 z4=2 |
| | 2 | z2w,1=1 z2w,3=1 z4=1 |
| | 2 | z2w,2=1 z2w,3=1 z4=1 |
| | 2 | z2w,3=1 |
| | 2 | z2w,2=1 z4=1 |
| | 2 | z2w,1=1 z2w,2=1 |
| | 2 | z2w,1=1 z4=3 |
| | 2 | z2w,1=1 z4=1 |
| | 2 | z2w,1=1 z2w,2=1 |
| | 2 | z2w,2=1 z4=3 |
| | 2 | z2w,3=1 |
| | 2 | z2w,2=1 z2w,3=1 z4=3 |
| | 2 | z2w,1=1 z2w,3=1 z4=3 |
| | 2 | z2w,1=1 z2w,2=1 z2w,3=1 z4=2 |
| | 2 | z2w,3=1 z4=1 |
| | 2 | z2w,2=1 z2w,3=1 |
| | 2 | z2w,1=1 z2w,3=1 |
| | 2 | z2w,1=1 z2w,2=1 z2w,3=1 z4=3 |
| | 2 | z2w,1=1 |
| | 2 | z2w,1=1 z2w,2=1 z4=3 |
| | 2 | z4=3 |
| | 2 | z2w,2=1 z4=2 |
| | 2 | z2w,2=1 |
| | 2 | z4=3 |
| | 2 | z2w,1=1 z2w,2=1 z4=3 |
| | 2 | z2w,1=1 z4=2 |
| | 2 | z2w,1=1 z2w,2=1 z2w,3=1 z4=3 |
| | 2 | z2w,1=1 z2w,3=1 z4=2 |
| | 2 | z2w,2=1 z2w,3=1 z4=2 |
| | 2 | z2w,3=1 z4=1 |
| 59.4.14 | 59345 | Na1O5V2 |
| | 2 | z4=3 |
| | 2 | z2w,2=1 z4=2 |
| | 2 | z2w,1=1 z4=2 |
| | 2 | z2w,1=1 z2w,2=1 z4=1 |
| | 2 | z2w,1=1 z2w,3=1 z4=2 |
| | 2 | z2w,1=1 z2w,2=1 z2w,3=1 z4=1 |
| | 2 | z2w,3=1 z4=1 |
| | 2 | z2w,2=1 z2w,3=1 |
| | 2 | z2w,2=1 z2w,3=1 z4=2 |
| | 2 | z2w,3=1 z4=1 |
| | 2 | z2w,1=1 z2w,2=1 z2w,3=1 z4=1 |
| | 2 | z2w,1=1 z2w,3=1 |
| | 2 | z2w,1=1 z2w,2=1 z4=1 |
| | 2 | z2w,1=1 |
| | 2 | z2w,2=1 |

Continued on next column

**TABLE XIX– continued**

| TQC | ICSD SSG | Stoich. Formula SIs |
|---|---|---|
| | 2 | z4=3 |
| | 2 | z2w,1=1 z2w,2=1 z4=2 |
| | 2 | z2w,1=1 z4=1 |
| | 2 | z2w,2=1 z4=1 |
| | 2 | z2w,2=1 z2w,3=1 z4=1 |
| | 2 | z2w,3=1 |
| | 2 | z2w,1=1 z2w,2=1 z2w,3=1 |
| | 2 | z2w,1=1 z2w,3=1 z4=3 |
| | 2 | z2w,1=1 z2w,3=1 z4=1 |
| | 2 | z2w,1=1 z2w,2=1 z2w,3=1 |
| | 2 | z2w,3=1 |
| | 2 | z2w,2=1 z2w,3=1 z4=3 |
| | 2 | z2w,2=1 z4=3 |
| | 2 | z2w,1=1 z4=3 |
| | 2 | z2w,1=1 z2w,2=1 z4=2 |
| | 2 | z2w,1=1 z2w,2=1 z2w,3=1 z4=2 |
| | 2 | z2w,1=1 z2w,3=1 z4=1 |
| | 2 | z2w,2=1 z2w,3=1 z4=1 |
| | 2 | z2w,3=1 |
| | 2 | z2w,2=1 z4=1 |
| | 2 | z2w,1=1 z2w,2=1 |
| | 2 | z2w,1=1 z4=3 |
| | 2 | z2w,1=1 z4=1 |
| | 2 | z2w,1=1 z2w,2=1 |
| | 2 | z2w,2=1 z4=3 |
| | 2 | z2w,3=1 |
| | 2 | z2w,2=1 z2w,3=1 z4=3 |
| | 2 | z2w,1=1 z2w,3=1 z4=3 |
| | 2 | z2w,1=1 z2w,2=1 z2w,3=1 z4=2 |
| | 2 | z2w,3=1 z4=1 |
| | 2 | z2w,2=1 z2w,3=1 |
| | 2 | z2w,1=1 z2w,3=1 |
| | 2 | z2w,1=1 z2w,2=1 z2w,3=1 z4=3 |
| | 2 | z2w,1=1 |
| | 2 | z2w,1=1 z2w,2=1 z4=3 |
| | 2 | z4=3 |
| | 2 | z2w,2=1 z4=2 |
| | 2 | z2w,2=1 |
| | 2 | z4=3 |
| | 2 | z2w,1=1 z2w,2=1 z4=3 |
| | 2 | z2w,1=1 z4=2 |
| | 2 | z2w,1=1 z2w,2=1 z2w,3=1 z4=3 |
| | 2 | z2w,1=1 z2w,3=1 z4=2 |
| | 2 | z2w,2=1 z2w,3=1 z4=2 |
| | 2 | z2w,3=1 z4=1 |
| 59.4.15 | 409210 | Na1O5V2 |
| | 2 | z2w,3=1 z4=2 |
| | 2 | z2w,2=1 z2w,3=1 z4=1 |
| | 2 | z2w,1=1 z2w,3=1 z4=1 |
| | 2 | z2w,1=1 z2w,2=1 z2w,3=1 |
| | 2 | z2w,1=1 z4=1 |
| | 2 | z2w,1=1 z2w,2=1 |
| | 2 | z2w,2=1 z4=3 |
| | 2 | z2w,2=1 z4=1 |
| | 2 | z2w,1=1 z2w,2=1 |
| | 2 | z2w,1=1 z4=3 |
| | 2 | z2w,1=1 z2w,2=1 z2w,3=1 |
| | 2 | z2w,1=1 z2w,3=1 z4=3 |
| | 2 | z2w,2=1 z2w,3=1 z4=3 |
| | 2 | z2w,3=1 z4=2 |

Continued on next column

**TABLE XIX– continued**

| TQC | ICSD SSG | Stoich. Formula SIs |
|---|---|---|
| | 2 | z2w,1=1 z2w,2=1 z2w,3=1 z4=1 |
| | 2 | z2w,1=1 z2w,3=1 |
| | 2 | z2w,2=1 z2w,3=1 |
| | 2 | z2w,3=1 z4=3 |
| | 2 | z2w,2=1 |
| | 2 | z4=3 |
| | 2 | z2w,1=1 z2w,2=1 z4=3 |
| | 2 | z2w,1=1 z4=2 |
| | 2 | z2w,1=1 |
| | 2 | z2w,1=1 z2w,2=1 z4=3 |
| | 2 | z4=3 |
| | 2 | z2w,2=1 z4=2 |
| | 2 | z2w,3=1 z4=3 |
| | 2 | z2w,2=1 z2w,3=1 z4=2 |
| | 2 | z2w,1=1 z2w,3=1 z4=2 |
| | 2 | z2w,1=1 z2w,2=1 z2w,3=1 z4=1 |
| | 2 | z2w,1=1 z2w,2=1 z4=1 |
| | 2 | z2w,1=1 |
| | 2 | z2w,2=1 |
| | 2 | z4=3 |
| | 2 | z2w,2=1 z2w,3=1 |
| | 2 | z2w,3=1 z4=3 |
| | 2 | z2w,1=1 z2w,2=1 z2w,3=1 z4=3 |
| | 2 | z2w,1=1 z2w,3=1 z4=2 |
| | 2 | z2w,1=1 z2w,3=1 |
| | 2 | z2w,1=1 z2w,2=1 z2w,3=1 z4=3 |
| | 2 | z2w,3=1 z4=3 |
| | 2 | z2w,2=1 z2w,3=1 z4=2 |
| | 2 | z4=3 |
| | 2 | z2w,2=1 z4=2 |
| | 2 | z2w,1=1 z4=2 |
| | 2 | z2w,1=1 z2w,2=1 z4=1 |
| | 2 | z2w,2=1 z4=3 |
| | 2 | z2w,1=1 z4=3 |
| | 2 | z2w,1=1 z2w,2=1 z4=2 |
| | 2 | z2w,1=1 z2w,3=1 z4=3 |
| | 2 | z2w,1=1 z2w,2=1 z2w,3=1 z4=2 |
| | 2 | z2w,3=1 z4=2 |
| | 2 | z2w,2=1 z2w,3=1 z4=1 |
| | 2 | z2w,2=1 z2w,3=1 z4=3 |
| | 2 | z2w,3=1 z4=2 |
| | 2 | z2w,1=1 z2w,2=1 z2w,3=1 z4=2 |
| | 2 | z2w,1=1 z2w,3=1 z4=1 |
| | 2 | z2w,1=1 z2w,2=1 z4=2 |
| | 2 | z2w,1=1 z4=1 |
| | 2 | z2w,2=1 z4=1 |
| 59.4.16 | 251182 | Na1O5V2 |
| | 2 | z4=3 |
| | 2 | z2w,2=1 z4=2 |
| | 2 | z2w,1=1 z4=2 |
| | 2 | z2w,1=1 z2w,2=1 z4=1 |
| | 2 | z2w,1=1 z2w,3=1 z4=2 |
| | 2 | z2w,1=1 z2w,2=1 z2w,3=1 z4=1 |
| | 2 | z2w,3=1 z4=1 |
| | 2 | z2w,2=1 z2w,3=1 |
| | 2 | z2w,2=1 z2w,3=1 z4=2 |
| | 2 | z2w,3=1 z4=1 |
| | 2 | z2w,1=1 z2w,2=1 z2w,3=1 z4=1 |
| | 2 | z2w,1=1 z2w,3=1 |
| | 2 | z2w,1=1 z2w,2=1 z4=1 |

Continued on next column

**TABLE XIX– continued**

| TQC | ICSD SSG | Stoich. Formula SIs |
|---|---|---|
| | 2 | z2w,1=1 |
| | 2 | z2w,2=1 |
| | 2 | z4=3 |
| | 2 | z2w,1=1 z2w,2=1 z4=2 |
| | 2 | z2w,1=1 z4=1 |
| | 2 | z2w,2=1 z4=1 |
| | 2 | z2w,2=1 z2w,3=1 z4=1 |
| | 2 | z2w,3=1 |
| | 2 | z2w,1=1 z2w,2=1 z2w,3=1 |
| | 2 | z2w,1=1 z2w,3=1 z4=3 |
| | 2 | z2w,1=1 z2w,3=1 z4=1 |
| | 2 | z2w,1=1 z2w,2=1 z2w,3=1 |
| | 2 | z2w,3=1 |
| | 2 | z2w,2=1 z2w,3=1 z4=3 |
| | 2 | z2w,2=1 z4=3 |
| | 2 | z2w,1=1 z4=3 |
| | 2 | z2w,1=1 z2w,2=1 z4=2 |
| | 2 | z2w,1=1 z2w,2=1 z2w,3=1 z4=2 |
| | 2 | z2w,1=1 z2w,3=1 z4=1 |
| | 2 | z2w,2=1 z2w,3=1 z4=1 |
| | 2 | z2w,3=1 |
| | 2 | z2w,2=1 z4=1 |
| | 2 | z2w,1=1 z2w,2=1 |
| | 2 | z2w,1=1 z4=3 |
| | 2 | z2w,1=1 z4=1 |
| | 2 | z2w,1=1 z2w,2=1 |
| | 2 | z2w,2=1 z4=3 |
| | 2 | z2w,3=1 |
| | 2 | z2w,2=1 z2w,3=1 z4=3 |
| | 2 | z2w,1=1 z2w,3=1 z4=3 |
| | 2 | z2w,1=1 z2w,2=1 z2w,3=1 z4=2 |
| | 2 | z2w,3=1 z4=1 |
| | 2 | z2w,2=1 z2w,3=1 |
| | 2 | z2w,1=1 z2w,3=1 |
| | 2 | z2w,1=1 z2w,2=1 z2w,3=1 z4=3 |
| | 2 | z2w,1=1 |
| | 2 | z2w,1=1 z2w,2=1 z4=3 |
| | 2 | z4=3 |
| | 2 | z2w,2=1 z4=2 |
| | 2 | z2w,2=1 |
| | 2 | z4=3 |
| | 2 | z2w,1=1 z2w,2=1 z4=3 |
| | 2 | z2w,1=1 z4=2 |
| | 2 | z2w,1=1 z2w,2=1 z2w,3=1 z4=3 |
| | 2 | z2w,1=1 z2w,3=1 z4=2 |
| | 2 | z2w,2=1 z2w,3=1 z4=2 |
| | 2 | z2w,3=1 z4=1 |
| 59.4.17 | 20825 | Mo2Ni6P3 |
| | 2 | z4=2 |
| | 2 | z2w,2=1 z4=1 |
| | 2 | z2w,1=1 z4=1 |
| | 2 | z2w,1=1 z2w,2=1 |
| | 2 | z2w,1=1 z2w,3=1 z4=1 |
| | 2 | z2w,1=1 z2w,2=1 z2w,3=1 |
| | 2 | z2w,3=1 |
| | 2 | z2w,2=1 z2w,3=1 z4=3 |
| | 2 | z2w,2=1 z2w,3=1 z4=1 |
| | 2 | z2w,3=1 |
| | 2 | z2w,1=1 z2w,2=1 z2w,3=1 |
| | 2 | z2w,1=1 z2w,3=1 z4=3 |

Continued on next column

**TABLE XIX– continued**

| TQC | ICSD SSG | Stoich. Formula SIs |
|---|---|---|
| | 2 | z2w,1=1 z2w,2=1 |
| | 2 | z2w,1=1 z4=3 |
| | 2 | z2w,2=1 z4=3 |
| | 2 | z4=2 |
| | 2 | z2w,1=1 z2w,2=1 z4=1 |
| | 2 | z2w,1=1 |
| | 2 | z2w,2=1 |
| | 2 | z4=3 |
| | 2 | z2w,2=1 z2w,3=1 |
| | 2 | z2w,3=1 z4=3 |
| | 2 | z2w,1=1 z2w,2=1 z2w,3=1 z4=3 |
| | 2 | z2w,1=1 z2w,3=1 z4=2 |
| | 2 | z2w,1=1 z2w,3=1 |
| | 2 | z2w,1=1 z2w,2=1 z2w,3=1 z4=3 |
| | 2 | z2w,3=1 z4=3 |
| | 2 | z2w,2=1 z2w,3=1 z4=2 |
| | 2 | z4=3 |
| | 2 | z2w,2=1 z4=2 |
| | 2 | z2w,1=1 z4=2 |
| | 2 | z2w,1=1 z2w,2=1 z4=1 |
| | 2 | z2w,1=1 z2w,2=1 z2w,3=1 z4=1 |
| | 2 | z2w,1=1 z2w,3=1 |
| | 2 | z2w,2=1 z2w,3=1 |
| | 2 | z2w,3=1 z4=3 |
| | 2 | z2w,2=1 |
| | 2 | z4=3 |
| | 2 | z2w,1=1 z2w,2=1 z4=3 |
| | 2 | z2w,1=1 z4=2 |
| | 2 | z2w,1=1 |
| | 2 | z2w,1=1 z2w,2=1 z4=3 |
| | 2 | z4=3 |
| | 2 | z2w,2=1 z4=2 |
| | 2 | z2w,3=1 z4=3 |
| | 2 | z2w,2=1 z2w,3=1 z4=2 |
| | 2 | z2w,1=1 z2w,3=1 z4=2 |
| | 2 | z2w,1=1 z2w,2=1 z2w,3=1 z4=1 |
| | 2 | z2w,3=1 |
| | 2 | z2w,2=1 z2w,3=1 z4=3 |
| | 2 | z2w,1=1 z2w,3=1 z4=3 |
| | 2 | z2w,1=1 z2w,2=1 z2w,3=1 z4=2 |
| | 2 | z2w,1=1 z4=3 |
| | 2 | z2w,1=1 z2w,2=1 z4=2 |
| | 2 | z4=2 |
| | 2 | z2w,2=1 z4=1 |
| | 2 | z2w,2=1 z4=3 |
| | 2 | z4=2 |
| | 2 | z2w,1=1 z2w,2=1 z4=2 |
| | 2 | z2w,1=1 z4=1 |
| | 2 | z2w,1=1 z2w,2=1 z2w,3=1 z4=2 |
| | 2 | z2w,1=1 z2w,3=1 z4=1 |
| | 2 | z2w,2=1 z2w,3=1 z4=1 |
| | 2 | z2w,3=1 |
| 59.4.18 | 88639 | Li1O5V2 |
| | 2 | z2w,3=1 z4=2 |
| | 2 | z2w,2=1 z2w,3=1 z4=1 |
| | 2 | z2w,1=1 z2w,3=1 z4=1 |
| | 2 | z2w,1=1 z2w,2=1 z2w,3=1 |
| | 2 | z2w,1=1 z4=1 |
| | 2 | z2w,1=1 z2w,2=1 |
| | 2 | z2w,2=1 z4=3 |

Continued on next column

**TABLE XIX– continued**

| TQC | ICSD SSG | Stoich. Formula SIs |
|---|---|---|
| | 2 | z2w,2=1 z4=1 |
| | 2 | z2w,1=1 z2w,2=1 |
| | 2 | z2w,1=1 z4=3 |
| | 2 | z2w,1=1 z2w,2=1 z2w,3=1 |
| | 2 | z2w,1=1 z2w,3=1 z4=3 |
| | 2 | z2w,2=1 z2w,3=1 z4=3 |
| | 2 | z2w,3=1 z4=2 |
| | 2 | z2w,1=1 z2w,2=1 z2w,3=1 z4=1 |
| | 2 | z2w,1=1 z2w,3=1 |
| | 2 | z2w,2=1 z2w,3=1 |
| | 2 | z2w,3=1 z4=3 |
| | 2 | z2w,2=1 |
| | 2 | z4=3 |
| | 2 | z2w,1=1 z2w,2=1 z4=3 |
| | 2 | z2w,1=1 z4=2 |
| | 2 | z2w,1=1 |
| | 2 | z2w,1=1 z2w,2=1 z4=3 |
| | 2 | z4=3 |
| | 2 | z2w,2=1 z4=2 |
| | 2 | z2w,3=1 z4=3 |
| | 2 | z2w,2=1 z2w,3=1 z4=2 |
| | 2 | z2w,1=1 z2w,3=1 z4=2 |
| | 2 | z2w,1=1 z2w,2=1 z2w,3=1 z4=1 |
| | 2 | z2w,1=1 z2w,2=1 z4=1 |
| | 2 | z2w,1=1 |
| | 2 | z2w,2=1 |
| | 2 | z4=3 |
| | 2 | z2w,2=1 z2w,3=1 |
| | 2 | z2w,3=1 z4=3 |
| | 2 | z2w,1=1 z2w,2=1 z2w,3=1 z4=3 |
| | 2 | z2w,1=1 z2w,3=1 z4=2 |
| | 2 | z2w,1=1 z2w,3=1 |
| | 2 | z2w,1=1 z2w,2=1 z2w,3=1 z4=3 |
| | 2 | z2w,3=1 z4=3 |
| | 2 | z2w,2=1 z2w,3=1 z4=2 |
| | 2 | z4=3 |
| | 2 | z2w,2=1 z4=2 |
| | 2 | z2w,1=1 z4=2 |
| | 2 | z2w,1=1 z2w,2=1 z4=1 |
| | 2 | z2w,2=1 z4=3 |
| | 2 | z2w,1=1 z4=3 |
| | 2 | z2w,1=1 z2w,2=1 z4=2 |
| | 2 | z2w,1=1 z2w,3=1 z4=3 |
| | 2 | z2w,1=1 z2w,2=1 z2w,3=1 z4=2 |
| | 2 | z2w,3=1 z4=2 |
| | 2 | z2w,2=1 z2w,3=1 z4=1 |
| | 2 | z2w,2=1 z2w,3=1 z4=3 |
| | 2 | z2w,3=1 z4=2 |
| | 2 | z2w,1=1 z2w,2=1 z2w,3=1 z4=2 |
| | 2 | z2w,1=1 z2w,3=1 z4=1 |
| | 2 | z2w,1=1 z2w,2=1 z4=2 |
| | 2 | z2w,1=1 z4=1 |
| | 2 | z2w,2=1 z4=1 |
| 59.4.19 | 107524 | In6La1Ni3 |
| | 2 | z2w,3=1 z4=1 |
| | 2 | z2w,2=1 z2w,3=1 |
| | 2 | z2w,1=1 z2w,3=1 |
| | 2 | z2w,1=1 z2w,2=1 z2w,3=1 z4=3 |
| | 2 | z2w,1=1 |
| | 2 | z2w,1=1 z2w,2=1 z4=3 |

Continued on next column

**TABLE XIX– continued**

| TQC | ICSD SSG | Stoich. Formula SIs |
|---|---|---|
| | 2 | z4=3 |
| | 2 | z2w,2=1 z4=2 |
| | 2 | z2w,2=1 |
| | 2 | z4=3 |
| | 2 | z2w,1=1 z2w,2=1 z4=3 |
| | 2 | z2w,1=1 z4=2 |
| | 2 | z2w,1=1 z2w,2=1 z2w,3=1 z4=3 |
| | 2 | z2w,1=1 z2w,3=1 z4=2 |
| | 2 | z2w,2=1 z2w,3=1 z4=2 |
| | 2 | z2w,3=1 z4=1 |
| | 2 | z2w,1=1 z2w,2=1 z2w,3=1 |
| | 2 | z2w,1=1 z2w,3=1 z4=3 |
| | 2 | z2w,2=1 z2w,3=1 z4=3 |
| | 2 | z2w,3=1 z4=2 |
| | 2 | z2w,2=1 z4=3 |
| | 2 | z4=2 |
| | 2 | z2w,1=1 z2w,2=1 z4=2 |
| | 2 | z2w,1=1 z4=1 |
| | 2 | z2w,1=1 z4=3 |
| | 2 | z2w,1=1 z2w,2=1 z4=2 |
| | 2 | z4=2 |
| | 2 | z2w,2=1 z4=1 |
| | 2 | z2w,3=1 z4=2 |
| | 2 | z2w,2=1 z2w,3=1 z4=1 |
| | 2 | z2w,1=1 z2w,3=1 z4=1 |
| | 2 | z2w,1=1 z2w,2=1 z2w,3=1 |
| | 2 | z2w,1=1 z2w,2=1 |
| | 2 | z2w,1=1 z4=3 |
| | 2 | z2w,2=1 z4=3 |
| | 2 | z4=2 |
| | 2 | z2w,2=1 z2w,3=1 z4=3 |
| | 2 | z2w,3=1 z4=2 |
| | 2 | z2w,1=1 z2w,2=1 z2w,3=1 z4=2 |
| | 2 | z2w,1=1 z2w,3=1 z4=1 |
| | 2 | z2w,1=1 z2w,3=1 z4=3 |
| | 2 | z2w,1=1 z2w,2=1 z2w,3=1 z4=2 |
| | 2 | z2w,3=1 z4=2 |
| | 2 | z2w,2=1 z2w,3=1 z4=1 |
| | 2 | z4=2 |
| | 2 | z2w,2=1 z4=1 |
| | 2 | z2w,1=1 z4=1 |
| | 2 | z2w,1=1 z2w,2=1 |
| | 2 | z4=3 |
| | 2 | z2w,2=1 z4=2 |
| | 2 | z2w,1=1 z4=2 |
| | 2 | z2w,1=1 z2w,2=1 z4=1 |
| | 2 | z2w,1=1 z2w,3=1 z4=2 |
| | 2 | z2w,1=1 z2w,2=1 z2w,3=1 z4=1 |
| | 2 | z2w,3=1 z4=1 |
| | 2 | z2w,2=1 z2w,3=1 |
| | 2 | z2w,2=1 z2w,3=1 z4=2 |
| | 2 | z2w,3=1 z4=1 |
| | 2 | z2w,1=1 z2w,2=1 z2w,3=1 z4=1 |
| | 2 | z2w,1=1 z2w,3=1 |
| | 2 | z2w,1=1 z2w,2=1 z4=1 |
| | 2 | z2w,1=1 |
| | 2 | z2w,2=1 |
| | 2 | z4=3 |
| 59.4.20 | 416462 | Gd1I1S1 |
| | 2 | z2w,3=1 z4=2 |

Continued on next column

**TABLE XIX– continued**

| TQC | ICSD SSG | Stoich. Formula SIs |
|---|---|---|
| | 2 | z2w,2=1 z2w,3=1 z4=1 |
| | 2 | z2w,1=1 z2w,3=1 z4=1 |
| | 2 | z2w,1=1 z2w,2=1 z2w,3=1 |
| | 2 | z2w,1=1 z4=1 |
| | 2 | z2w,1=1 z2w,2=1 |
| | 2 | z2w,2=1 z4=3 |
| | 2 | z2w,2=1 z4=1 |
| | 2 | z2w,1=1 z2w,2=1 |
| | 2 | z2w,1=1 z4=3 |
| | 2 | z2w,1=1 z2w,2=1 z2w,3=1 |
| | 2 | z2w,1=1 z2w,3=1 z4=3 |
| | 2 | z2w,2=1 z2w,3=1 z4=3 |
| | 2 | z2w,3=1 z4=2 |
| | 2 | z2w,1=1 z2w,2=1 z2w,3=1 z4=1 |
| | 2 | z2w,1=1 z2w,3=1 |
| | 2 | z2w,2=1 z2w,3=1 |
| | 2 | z2w,3=1 z4=3 |
| | 2 | z2w,2=1 |
| | 2 | z4=3 |
| | 2 | z2w,1=1 z2w,2=1 z4=3 |
| | 2 | z2w,1=1 z4=2 |
| | 2 | z2w,1=1 |
| | 2 | z2w,1=1 z2w,2=1 z4=3 |
| | 2 | z4=3 |
| | 2 | z2w,2=1 z4=2 |
| | 2 | z2w,3=1 z4=3 |
| | 2 | z2w,2=1 z2w,3=1 z4=2 |
| | 2 | z2w,1=1 z2w,3=1 z4=2 |
| | 2 | z2w,1=1 z2w,2=1 z2w,3=1 z4=1 |
| | 2 | z2w,1=1 z2w,2=1 z4=1 |
| | 2 | z2w,1=1 |
| | 2 | z2w,2=1 |
| | 2 | z4=3 |
| | 2 | z2w,2=1 z2w,3=1 |
| | 2 | z2w,3=1 z4=3 |
| | 2 | z2w,1=1 z2w,2=1 z2w,3=1 z4=3 |
| | 2 | z2w,1=1 z2w,3=1 z4=2 |
| | 2 | z2w,1=1 z2w,3=1 |
| | 2 | z2w,1=1 z2w,2=1 z2w,3=1 z4=3 |
| | 2 | z2w,3=1 z4=3 |
| | 2 | z2w,2=1 z2w,3=1 z4=2 |
| | 2 | z4=3 |
| | 2 | z2w,2=1 z4=2 |
| | 2 | z2w,1=1 z4=2 |
| | 2 | z2w,1=1 z2w,2=1 z4=1 |
| | 2 | z2w,2=1 z4=3 |
| | 2 | z2w,1=1 z4=3 |
| | 2 | z2w,1=1 z2w,2=1 z4=2 |
| | 2 | z2w,1=1 z2w,3=1 z4=3 |
| | 2 | z2w,1=1 z2w,2=1 z2w,3=1 z4=2 |
| | 2 | z2w,3=1 z4=2 |
| | 2 | z2w,2=1 z2w,3=1 z4=1 |
| | 2 | z2w,2=1 z2w,3=1 z4=3 |
| | 2 | z2w,3=1 z4=2 |
| | 2 | z2w,1=1 z2w,2=1 z2w,3=1 z4=2 |
| | 2 | z2w,1=1 z2w,3=1 z4=1 |
| | 2 | z2w,1=1 z2w,2=1 z4=2 |
| | 2 | z2w,1=1 z4=1 |
| | 2 | z2w,2=1 z4=1 |
| 59.4.21 | 50194 | Er1I1Se1 |

Continued on next column

**TABLE XIX– continued**

| TQC | ICSD SSG | Stoich. Formula SIs |
|---|---|---|
| | 2 | z2w,2=1 z4=3 |
| | 2 | z2w,1=1 z4=3 |
| | 2 | z2w,1=1 z2w,2=1 z4=2 |
| | 2 | z2w,1=1 z2w,3=1 z4=3 |
| | 2 | z2w,1=1 z2w,2=1 z2w,3=1 z4=2 |
| | 2 | z2w,3=1 z4=2 |
| | 2 | z2w,2=1 z2w,3=1 z4=1 |
| | 2 | z2w,2=1 z2w,3=1 z4=3 |
| | 2 | z2w,3=1 z4=2 |
| | 2 | z2w,1=1 z2w,2=1 z2w,3=1 z4=2 |
| | 2 | z2w,1=1 z2w,3=1 z4=1 |
| | 2 | z2w,1=1 z2w,2=1 z4=2 |
| | 2 | z2w,1=1 z4=1 |
| | 2 | z2w,2=1 z4=1 |
| | 2 | z2w,1=1 z2w,2=1 z4=3 |
| | 2 | z2w,1=1 z4=2 |
| | 2 | z2w,2=1 z4=2 |
| | 2 | z4=1 |
| | 2 | z2w,2=1 z2w,3=1 z4=2 |
| | 2 | z2w,3=1 z4=1 |
| | 2 | z2w,1=1 z2w,2=1 z2w,3=1 z4=1 |
| | 2 | z2w,1=1 z2w,3=1 |
| | 2 | z2w,1=1 z2w,3=1 z4=2 |
| | 2 | z2w,1=1 z2w,2=1 z2w,3=1 z4=1 |
| | 2 | z2w,3=1 z4=1 |
| | 2 | z2w,2=1 z2w,3=1 |
| | 2 | z4=1 |
| | 2 | z2w,2=1 |
| | 2 | z2w,1=1 |
| | 2 | z2w,1=1 z2w,2=1 z4=3 |
| | 2 | z2w,1=1 z2w,2=1 z2w,3=1 z4=3 |
| | 2 | z2w,1=1 z2w,3=1 z4=2 |
| | 2 | z2w,2=1 z2w,3=1 z4=2 |
| | 2 | z2w,3=1 z4=1 |
| | 2 | z2w,2=1 z4=2 |
| | 2 | z4=1 |
| | 2 | z2w,1=1 z2w,2=1 z4=1 |
| | 2 | z2w,1=1 |
| | 2 | z2w,1=1 z4=2 |
| | 2 | z2w,1=1 z2w,2=1 z4=1 |
| | 2 | z4=1 |
| | 2 | z2w,2=1 |
| | 2 | z2w,3=1 z4=1 |
| | 2 | z2w,2=1 z2w,3=1 |
| | 2 | z2w,1=1 z2w,3=1 |
| | 2 | z2w,1=1 z2w,2=1 z2w,3=1 z4=3 |
| | 2 | z2w,3=1 z4=2 |
| | 2 | z2w,2=1 z2w,3=1 z4=1 |
| | 2 | z2w,1=1 z2w,3=1 z4=1 |
| | 2 | z2w,1=1 z2w,2=1 z2w,3=1 |
| | 2 | z2w,1=1 z4=1 |
| | 2 | z2w,1=1 z2w,2=1 |
| | 2 | z2w,2=1 z4=3 |
| | 2 | z2w,2=1 z4=1 |
| | 2 | z2w,1=1 z2w,2=1 |
| | 2 | z2w,1=1 z4=3 |
| | 2 | z2w,1=1 z2w,2=1 z2w,3=1 |
| | 2 | z2w,1=1 z2w,3=1 z4=3 |
| | 2 | z2w,2=1 z2w,3=1 z4=3 |
| | 2 | z2w,3=1 z4=2 |

Continued on next column

**TABLE XIX– continued**

| TQC | ICSD SSG | Stoich. Formula SIs |
|---|---|---|
| 59.4.22 | 79107 | Dy1I1S1 |
| | 2 | z2w,3=1 z4=2 |
| | 2 | z2w,2=1 z2w,3=1 z4=1 |
| | 2 | z2w,1=1 z2w,3=1 z4=1 |
| | 2 | z2w,1=1 z2w,2=1 z2w,3=1 |
| | 2 | z2w,1=1 z4=1 |
| | 2 | z2w,1=1 z2w,2=1 |
| | 2 | z2w,2=1 z4=3 |
| | 2 | z2w,2=1 z4=1 |
| | 2 | z2w,1=1 z2w,2=1 |
| | 2 | z2w,1=1 z4=3 |
| | 2 | z2w,1=1 z2w,2=1 z2w,3=1 |
| | 2 | z2w,1=1 z2w,3=1 z4=3 |
| | 2 | z2w,2=1 z2w,3=1 z4=3 |
| | 2 | z2w,3=1 z4=2 |
| | 2 | z2w,1=1 z2w,2=1 z2w,3=1 z4=1 |
| | 2 | z2w,1=1 z2w,3=1 |
| | 2 | z2w,2=1 z2w,3=1 |
| | 2 | z2w,3=1 z4=3 |
| | 2 | z2w,2=1 |
| | 2 | z4=3 |
| | 2 | z2w,1=1 z2w,2=1 z4=3 |
| | 2 | z2w,1=1 z4=2 |
| | 2 | z2w,1=1 |
| | 2 | z2w,1=1 z2w,2=1 z4=3 |
| | 2 | z4=3 |
| | 2 | z2w,2=1 z4=2 |
| | 2 | z2w,3=1 z4=3 |
| | 2 | z2w,2=1 z2w,3=1 z4=2 |
| | 2 | z2w,1=1 z2w,3=1 z4=2 |
| | 2 | z2w,1=1 z2w,2=1 z2w,3=1 z4=1 |
| | 2 | z2w,1=1 z2w,2=1 z4=1 |
| | 2 | z2w,1=1 |
| | 2 | z2w,2=1 |
| | 2 | z4=3 |
| | 2 | z2w,2=1 z2w,3=1 |
| | 2 | z2w,3=1 z4=3 |
| | 2 | z2w,1=1 z2w,2=1 z2w,3=1 z4=3 |
| | 2 | z2w,1=1 z2w,3=1 z4=2 |
| | 2 | z2w,1=1 z2w,3=1 |
| | 2 | z2w,1=1 z2w,2=1 z2w,3=1 z4=3 |
| | 2 | z2w,3=1 z4=3 |
| | 2 | z2w,2=1 z2w,3=1 z4=2 |
| | 2 | z4=3 |
| | 2 | z2w,2=1 z4=2 |
| | 2 | z2w,1=1 z4=2 |
| | 2 | z2w,1=1 z2w,2=1 z4=1 |
| | 2 | z2w,2=1 z4=3 |
| | 2 | z2w,1=1 z4=3 |
| | 2 | z2w,1=1 z2w,2=1 z4=2 |
| | 2 | z2w,1=1 z2w,3=1 z4=3 |
| | 2 | z2w,1=1 z2w,2=1 z2w,3=1 z4=2 |
| | 2 | z2w,3=1 z4=2 |
| | 2 | z2w,2=1 z2w,3=1 z4=1 |
| | 2 | z2w,2=1 z2w,3=1 z4=3 |
| | 2 | z2w,3=1 z4=2 |
| | 2 | z2w,1=1 z2w,2=1 z2w,3=1 z4=2 |
| | 2 | z2w,1=1 z2w,3=1 z4=1 |
| | 2 | z2w,1=1 z2w,2=1 z4=2 |
| | 2 | z2w,1=1 z4=1 |

Continued on next column

**TABLE XIX– continued**

| TQC | ICSD SSG | Stoich. Formula SIs |
|---|---|---|
| | 2 | z2w,2=1 z4=1 |
| 59.4.23 | 93966 | Cu1Te1 |
| | 2 | z4=3 |
| | 2 | z2w,2=1 z4=2 |
| | 2 | z2w,1=1 z4=2 |
| | 2 | z2w,1=1 z2w,2=1 z4=1 |
| | 2 | z2w,1=1 z2w,3=1 z4=2 |
| | 2 | z2w,1=1 z2w,2=1 z2w,3=1 z4=1 |
| | 2 | z2w,3=1 z4=1 |
| | 2 | z2w,2=1 z2w,3=1 |
| | 2 | z2w,2=1 z2w,3=1 z4=2 |
| | 2 | z2w,3=1 z4=1 |
| | 2 | z2w,1=1 z2w,2=1 z2w,3=1 z4=1 |
| | 2 | z2w,1=1 z2w,3=1 |
| | 2 | z2w,1=1 z2w,2=1 z4=1 |
| | 2 | z2w,1=1 |
| | 2 | z2w,2=1 |
| | 2 | z4=3 |
| | 2 | z2w,1=1 z2w,2=1 z4=2 |
| | 2 | z2w,1=1 z4=1 |
| | 2 | z2w,2=1 z4=1 |
| | 2 | z2w,2=1 z2w,3=1 z4=1 |
| | 2 | z2w,3=1 |
| | 2 | z2w,1=1 z2w,2=1 z2w,3=1 |
| | 2 | z2w,1=1 z2w,3=1 z4=3 |
| | 2 | z2w,1=1 z2w,3=1 z4=1 |
| | 2 | z2w,1=1 z2w,2=1 z2w,3=1 |
| | 2 | z2w,3=1 |
| | 2 | z2w,2=1 z2w,3=1 z4=3 |
| | 2 | z2w,2=1 z4=3 |
| | 2 | z2w,1=1 z4=3 |
| | 2 | z2w,1=1 z2w,2=1 z4=2 |
| | 2 | z2w,1=1 z2w,2=1 z2w,3=1 z4=2 |
| | 2 | z2w,1=1 z2w,3=1 z4=1 |
| | 2 | z2w,2=1 z2w,3=1 z4=1 |
| | 2 | z2w,3=1 |
| | 2 | z2w,2=1 z4=1 |
| | 2 | z2w,1=1 z2w,2=1 |
| | 2 | z2w,1=1 z4=3 |
| | 2 | z2w,1=1 z4=1 |
| | 2 | z2w,1=1 z2w,2=1 |
| | 2 | z2w,2=1 z4=3 |
| | 2 | z2w,3=1 |
| | 2 | z2w,2=1 z2w,3=1 z4=3 |
| | 2 | z2w,1=1 z2w,3=1 z4=3 |
| | 2 | z2w,1=1 z2w,2=1 z2w,3=1 z4=2 |
| | 2 | z2w,3=1 z4=1 |
| | 2 | z2w,2=1 z2w,3=1 |
| | 2 | z2w,1=1 z2w,3=1 |
| | 2 | z2w,1=1 z2w,2=1 z2w,3=1 z4=3 |
| | 2 | z2w,1=1 |
| | 2 | z2w,1=1 z2w,2=1 z4=3 |
| | 2 | z4=3 |
| | 2 | z2w,2=1 z4=2 |
| | 2 | z2w,2=1 |
| | 2 | z4=3 |
| | 2 | z2w,1=1 z2w,2=1 z4=3 |
| | 2 | z2w,1=1 z4=2 |
| | 2 | z2w,1=1 z2w,2=1 z2w,3=1 z4=3 |
| | 2 | z2w,1=1 z2w,3=1 z4=2 |

Continued on next column

**TABLE XIX– continued**

| TQC | ICSD SSG | Stoich. Formula SIs |
|---|---|---|
| | 2 | z2w,2=1 z2w,3=1 z4=2 |
| | 2 | z2w,3=1 z4=1 |
| 59.4.24 | 33996 | Co1Cu2O3 |
| | 2 | z2w,3=1 z4=3 |
| | 2 | z2w,2=1 z2w,3=1 z4=2 |
| | 2 | z2w,1=1 z2w,3=1 z4=2 |
| | 2 | z2w,1=1 z2w,2=1 z2w,3=1 z4=1 |
| | 2 | z2w,1=1 z4=2 |
| | 2 | z2w,1=1 z2w,2=1 z4=1 |
| | 2 | z4=1 |
| | 2 | z2w,2=1 |
| | 2 | z2w,2=1 z4=2 |
| | 2 | z4=1 |
| | 2 | z2w,1=1 z2w,2=1 z4=1 |
| | 2 | z2w,1=1 |
| | 2 | z2w,1=1 z2w,2=1 z2w,3=1 z4=1 |
| | 2 | z2w,1=1 z2w,3=1 |
| | 2 | z2w,2=1 z2w,3=1 |
| | 2 | z2w,3=1 z4=3 |
| | 2 | z2w,1=1 z2w,2=1 z2w,3=1 z4=2 |
| | 2 | z2w,1=1 z2w,3=1 z4=1 |
| | 2 | z2w,2=1 z2w,3=1 z4=1 |
| | 2 | z2w,3=1 |
| | 2 | z2w,2=1 z4=1 |
| | 2 | z2w,1=1 z2w,2=1 |
| | 2 | z2w,1=1 z4=3 |
| | 2 | z2w,1=1 z4=1 |
| | 2 | z2w,1=1 z2w,2=1 |
| | 2 | z2w,2=1 z4=3 |
| | 2 | z2w,3=1 |
| | 2 | z2w,2=1 z2w,3=1 z4=3 |
| | 2 | z2w,1=1 z2w,3=1 z4=3 |
| | 2 | z2w,1=1 z2w,2=1 z2w,3=1 z4=2 |
| | 2 | z2w,1=1 z2w,2=1 z4=2 |
| | 2 | z2w,1=1 z4=1 |
| | 2 | z2w,2=1 z4=1 |
| | 2 | z2w,2=1 z2w,3=1 z4=1 |
| | 2 | z2w,3=1 |
| | 2 | z2w,1=1 z2w,2=1 z2w,3=1 |
| | 2 | z2w,1=1 z2w,3=1 z4=3 |
| | 2 | z2w,1=1 z2w,3=1 z4=1 |
| | 2 | z2w,1=1 z2w,2=1 z2w,3=1 |
| | 2 | z2w,3=1 |
| | 2 | z2w,2=1 z2w,3=1 z4=3 |
| | 2 | z2w,2=1 z4=3 |
| | 2 | z2w,1=1 z4=3 |
| | 2 | z2w,1=1 z2w,2=1 z4=2 |
| | 2 | z4=1 |
| | 2 | z2w,2=1 |
| | 2 | z2w,1=1 |
| | 2 | z2w,1=1 z2w,2=1 z4=3 |
| | 2 | z2w,1=1 z2w,3=1 |
| | 2 | z2w,1=1 z2w,2=1 z2w,3=1 z4=3 |
| | 2 | z2w,3=1 z4=3 |
| | 2 | z2w,2=1 z2w,3=1 z4=2 |
| | 2 | z2w,2=1 z2w,3=1 |
| | 2 | z2w,3=1 z4=3 |
| | 2 | z2w,1=1 z2w,2=1 z2w,3=1 z4=3 |
| | 2 | z2w,1=1 z2w,3=1 z4=2 |
| | 2 | z2w,1=1 z2w,2=1 z4=3 |

Continued on next column

**TABLE XIX– continued**

| TQC | ICSD SSG | Stoich. Formula SIs |
|---|---|---|
| | 2 | z2w,1=1 z4=2 |
| | 2 | z2w,2=1 z4=2 |
| | 2 | z4=1 |
| 59.4.25 | 39314 | Cl1O1Ti1 |
| | 2 | z2w,3=1 z4=2 |
| | 2 | z2w,2=1 z2w,3=1 z4=1 |
| | 2 | z2w,1=1 z2w,3=1 z4=1 |
| | 2 | z2w,1=1 z2w,2=1 z2w,3=1 |
| | 2 | z2w,1=1 z4=1 |
| | 2 | z2w,1=1 z2w,2=1 |
| | 2 | z2w,2=1 z4=3 |
| | 2 | z2w,2=1 z4=1 |
| | 2 | z2w,1=1 z2w,2=1 |
| | 2 | z2w,1=1 z4=3 |
| | 2 | z2w,1=1 z2w,2=1 z2w,3=1 |
| | 2 | z2w,1=1 z2w,3=1 z4=3 |
| | 2 | z2w,2=1 z2w,3=1 z4=3 |
| | 2 | z2w,3=1 z4=2 |
| | 2 | z2w,1=1 z2w,2=1 z2w,3=1 z4=1 |
| | 2 | z2w,1=1 z2w,3=1 |
| | 2 | z2w,2=1 z2w,3=1 |
| | 2 | z2w,3=1 z4=3 |
| | 2 | z2w,2=1 |
| | 2 | z4=3 |
| | 2 | z2w,1=1 z2w,2=1 z4=3 |
| | 2 | z2w,1=1 z4=2 |
| | 2 | z2w,1=1 |
| | 2 | z2w,1=1 z2w,2=1 z4=3 |
| | 2 | z4=3 |
| | 2 | z2w,2=1 z4=2 |
| | 2 | z2w,3=1 z4=3 |
| | 2 | z2w,2=1 z2w,3=1 z4=2 |
| | 2 | z2w,1=1 z2w,3=1 z4=2 |
| | 2 | z2w,1=1 z2w,2=1 z2w,3=1 z4=1 |
| | 2 | z2w,1=1 z2w,2=1 z4=1 |
| | 2 | z2w,1=1 |
| | 2 | z2w,2=1 |
| | 2 | z4=3 |
| | 2 | z2w,2=1 z2w,3=1 |
| | 2 | z2w,3=1 z4=3 |
| | 2 | z2w,1=1 z2w,2=1 z2w,3=1 z4=3 |
| | 2 | z2w,1=1 z2w,3=1 z4=2 |
| | 2 | z2w,1=1 z2w,3=1 |
| | 2 | z2w,1=1 z2w,2=1 z2w,3=1 z4=3 |
| | 2 | z2w,3=1 z4=3 |
| | 2 | z2w,2=1 z2w,3=1 z4=2 |
| | 2 | z4=3 |
| | 2 | z2w,2=1 z4=2 |
| | 2 | z2w,1=1 z4=2 |
| | 2 | z2w,1=1 z2w,2=1 z4=1 |
| | 2 | z2w,2=1 z4=3 |
| | 2 | z2w,1=1 z4=3 |
| | 2 | z2w,1=1 z2w,2=1 z4=2 |
| | 2 | z2w,1=1 z2w,3=1 z4=3 |
| | 2 | z2w,1=1 z2w,2=1 z2w,3=1 z4=2 |
| | 2 | z2w,3=1 z4=2 |
| | 2 | z2w,2=1 z2w,3=1 z4=1 |
| | 2 | z2w,2=1 z2w,3=1 z4=3 |
| | 2 | z2w,3=1 z4=2 |
| | 2 | z2w,1=1 z2w,2=1 z2w,3=1 z4=2 |

Continued on next column

**TABLE XIX– continued**

| TQC | ICSD SSG | Stoich. Formula SIs |
|---|---|---|
| | 2 | z2w,1=1 z2w,3=1 z4=1 |
| | 2 | z2w,1=1 z2w,2=1 z4=2 |
| | 2 | z2w,1=1 z4=1 |
| | 2 | z2w,2=1 z4=1 |
| 59.4.26 | 155833 | Cl1O1Ti1 |
| | 2 | z2w,3=1 z4=3 |
| | 2 | z2w,2=1 z2w,3=1 z4=2 |
| | 2 | z2w,1=1 z2w,3=1 z4=2 |
| | 2 | z2w,1=1 z2w,2=1 z2w,3=1 z4=1 |
| | 2 | z2w,1=1 z4=2 |
| | 2 | z2w,1=1 z2w,2=1 z4=1 |
| | 2 | z4=1 |
| | 2 | z2w,2=1 |
| | 2 | z2w,2=1 z4=2 |
| | 2 | z4=1 |
| | 2 | z2w,1=1 z2w,2=1 z4=1 |
| | 2 | z2w,1=1 |
| | 2 | z2w,1=1 z2w,2=1 z2w,3=1 z4=1 |
| | 2 | z2w,1=1 z2w,3=1 |
| | 2 | z2w,2=1 z2w,3=1 |
| | 2 | z2w,3=1 z4=3 |
| | 2 | z2w,1=1 z2w,2=1 z2w,3=1 z4=2 |
| | 2 | z2w,1=1 z2w,3=1 z4=1 |
| | 2 | z2w,2=1 z2w,3=1 z4=1 |
| | 2 | z2w,3=1 |
| | 2 | z2w,2=1 z4=1 |
| | 2 | z2w,1=1 z2w,2=1 |
| | 2 | z2w,1=1 z4=3 |
| | 2 | z2w,1=1 z4=1 |
| | 2 | z2w,1=1 z2w,2=1 |
| | 2 | z2w,2=1 z4=3 |
| | 2 | z2w,3=1 |
| | 2 | z2w,2=1 z2w,3=1 z4=3 |
| | 2 | z2w,1=1 z2w,3=1 z4=3 |
| | 2 | z2w,1=1 z2w,2=1 z2w,3=1 z4=2 |
| | 2 | z2w,1=1 z2w,2=1 z4=2 |
| | 2 | z2w,1=1 z4=1 |
| | 2 | z2w,2=1 z4=1 |
| | 2 | z2w,2=1 z2w,3=1 z4=1 |
| | 2 | z2w,3=1 |
| | 2 | z2w,1=1 z2w,2=1 z2w,3=1 |
| | 2 | z2w,1=1 z2w,3=1 z4=3 |
| | 2 | z2w,1=1 z2w,3=1 z4=1 |
| | 2 | z2w,1=1 z2w,2=1 z2w,3=1 |
| | 2 | z2w,3=1 |
| | 2 | z2w,2=1 z2w,3=1 z4=3 |
| | 2 | z2w,2=1 z4=3 |
| | 2 | z2w,1=1 z4=3 |
| | 2 | z2w,1=1 z2w,2=1 z4=2 |
| | 2 | z4=1 |
| | 2 | z2w,2=1 |
| | 2 | z2w,1=1 |
| | 2 | z2w,1=1 z2w,2=1 z4=3 |
| | 2 | z2w,1=1 z2w,3=1 |
| | 2 | z2w,1=1 z2w,2=1 z2w,3=1 z4=3 |
| | 2 | z2w,3=1 z4=3 |
| | 2 | z2w,2=1 z2w,3=1 z4=2 |
| | 2 | z2w,2=1 z2w,3=1 |
| | 2 | z2w,3=1 z4=3 |
| | 2 | z2w,1=1 z2w,2=1 z2w,3=1 z4=3 |

Continued on next column

**TABLE XIX– continued**

| TQC | ICSD SSG | Stoich. Formula SIs |
|---|---|---|
| | 2 | z2w,1=1 z2w,3=1 z4=2 |
| | 2 | z2w,1=1 z2w,2=1 z4=3 |
| | 2 | z2w,1=1 z4=2 |
| | 2 | z2w,2=1 z4=2 |
| | 2 | z4=1 |
| 59.4.27 | 261526 | Cl1Li1N1Ti1 |
| | 2 | z2w,3=1 z4=2 |
| | 2 | z2w,2=1 z2w,3=1 z4=1 |
| | 2 | z2w,1=1 z2w,3=1 z4=1 |
| | 2 | z2w,1=1 z2w,2=1 z2w,3=1 |
| | 2 | z2w,1=1 z4=1 |
| | 2 | z2w,1=1 z2w,2=1 |
| | 2 | z2w,2=1 z4=3 |
| | 2 | z2w,2=1 z4=1 |
| | 2 | z2w,1=1 z2w,2=1 |
| | 2 | z2w,1=1 z4=3 |
| | 2 | z2w,1=1 z2w,2=1 z2w,3=1 |
| | 2 | z2w,1=1 z2w,3=1 z4=3 |
| | 2 | z2w,2=1 z2w,3=1 z4=3 |
| | 2 | z2w,3=1 z4=2 |
| | 2 | z2w,1=1 z2w,2=1 z2w,3=1 z4=1 |
| | 2 | z2w,1=1 z2w,3=1 |
| | 2 | z2w,2=1 z2w,3=1 |
| | 2 | z2w,3=1 z4=3 |
| | 2 | z2w,2=1 |
| | 2 | z4=3 |
| | 2 | z2w,1=1 z2w,2=1 z4=3 |
| | 2 | z2w,1=1 z4=2 |
| | 2 | z2w,1=1 |
| | 2 | z2w,1=1 z2w,2=1 z4=3 |
| | 2 | z4=3 |
| | 2 | z2w,2=1 z4=2 |
| | 2 | z2w,3=1 z4=3 |
| | 2 | z2w,2=1 z2w,3=1 z4=2 |
| | 2 | z2w,1=1 z2w,3=1 z4=2 |
| | 2 | z2w,1=1 z2w,2=1 z2w,3=1 z4=1 |
| | 2 | z2w,1=1 z2w,2=1 z4=1 |
| | 2 | z2w,1=1 |
| | 2 | z2w,2=1 |
| | 2 | z4=3 |
| | 2 | z2w,2=1 z2w,3=1 |
| | 2 | z2w,3=1 z4=3 |
| | 2 | z2w,1=1 z2w,2=1 z2w,3=1 z4=3 |
| | 2 | z2w,1=1 z2w,3=1 z4=2 |
| | 2 | z2w,1=1 z2w,3=1 |
| | 2 | z2w,1=1 z2w,2=1 z2w,3=1 z4=3 |
| | 2 | z2w,3=1 z4=3 |
| | 2 | z2w,2=1 z2w,3=1 z4=2 |
| | 2 | z4=3 |
| | 2 | z2w,2=1 z4=2 |
| | 2 | z2w,1=1 z4=2 |
| | 2 | z2w,1=1 z2w,2=1 z4=1 |
| | 2 | z2w,2=1 z4=3 |
| | 2 | z2w,1=1 z4=3 |
| | 2 | z2w,1=1 z2w,2=1 z4=2 |
| | 2 | z2w,1=1 z2w,3=1 z4=3 |
| | 2 | z2w,1=1 z2w,2=1 z2w,3=1 z4=2 |
| | 2 | z2w,3=1 z4=2 |
| | 2 | z2w,2=1 z2w,3=1 z4=1 |
| | 2 | z2w,2=1 z2w,3=1 z4=3 |

Continued on next column

**TABLE XIX– continued**

| TQC | ICSD SSG | Stoich. Formula SIs |
|---|---|---|
| | 2 | z2w,3=1 z4=2 |
| | 2 | z2w,1=1 z2w,2=1 z2w,3=1 z4=2 |
| | 2 | z2w,1=1 z2w,3=1 z4=1 |
| | 2 | z2w,1=1 z2w,2=1 z4=2 |
| | 2 | z2w,1=1 z4=1 |
| | 2 | z2w,2=1 z4=1 |
| 59.4.28 | 40963 | Cl1Fe1O1 |
| | 2 | z2w,2=1 z4=3 |
| | 2 | z2w,1=1 z4=3 |
| | 2 | z2w,1=1 z2w,2=1 z4=2 |
| | 2 | z2w,1=1 z2w,3=1 z4=3 |
| | 2 | z2w,1=1 z2w,2=1 z2w,3=1 z4=2 |
| | 2 | z2w,3=1 z4=2 |
| | 2 | z2w,2=1 z2w,3=1 z4=1 |
| | 2 | z2w,2=1 z2w,3=1 z4=3 |
| | 2 | z2w,3=1 z4=2 |
| | 2 | z2w,1=1 z2w,2=1 z2w,3=1 z4=2 |
| | 2 | z2w,1=1 z2w,3=1 z4=1 |
| | 2 | z2w,1=1 z2w,2=1 z4=2 |
| | 2 | z2w,1=1 z4=1 |
| | 2 | z2w,2=1 z4=1 |
| | 2 | z2w,1=1 z2w,2=1 z4=3 |
| | 2 | z2w,1=1 z4=2 |
| | 2 | z2w,2=1 z4=2 |
| | 2 | z4=1 |
| | 2 | z2w,2=1 z2w,3=1 z4=2 |
| | 2 | z2w,3=1 z4=1 |
| | 2 | z2w,1=1 z2w,2=1 z2w,3=1 z4=1 |
| | 2 | z2w,1=1 z2w,3=1 |
| | 2 | z2w,1=1 z2w,3=1 z4=2 |
| | 2 | z2w,1=1 z2w,2=1 z2w,3=1 z4=1 |
| | 2 | z2w,3=1 z4=1 |
| | 2 | z2w,2=1 z2w,3=1 |
| | 2 | z4=1 |
| | 2 | z2w,2=1 |
| | 2 | z2w,1=1 |
| | 2 | z2w,1=1 z2w,2=1 z4=3 |
| | 2 | z2w,1=1 z2w,2=1 z2w,3=1 z4=3 |
| | 2 | z2w,1=1 z2w,3=1 z4=2 |
| | 2 | z2w,2=1 z2w,3=1 z4=2 |
| | 2 | z2w,3=1 z4=1 |
| | 2 | z2w,2=1 z4=2 |
| | 2 | z4=1 |
| | 2 | z2w,1=1 z2w,2=1 z4=1 |
| | 2 | z2w,1=1 |
| | 2 | z2w,1=1 z4=2 |
| | 2 | z2w,1=1 z2w,2=1 z4=1 |
| | 2 | z4=1 |
| | 2 | z2w,2=1 |
| | 2 | z2w,3=1 z4=1 |
| | 2 | z2w,2=1 z2w,3=1 |
| | 2 | z2w,1=1 z2w,3=1 |
| | 2 | z2w,1=1 z2w,2=1 z2w,3=1 z4=3 |
| | 2 | z2w,3=1 z4=2 |
| | 2 | z2w,2=1 z2w,3=1 z4=1 |
| | 2 | z2w,1=1 z2w,3=1 z4=1 |
| | 2 | z2w,1=1 z2w,2=1 z2w,3=1 |
| | 2 | z2w,1=1 z4=1 |
| | 2 | z2w,1=1 z2w,2=1 |
| | 2 | z2w,2=1 z4=3 |

Continued on next column

**TABLE XIX– continued**

| TQC | ICSD SSG | Stoich. Formula SIs |
|---|---|---|
| | 2 | z2w,2=1 z4=1 |
| | 2 | z2w,1=1 z2w,2=1 |
| | 2 | z2w,1=1 z4=3 |
| | 2 | z2w,1=1 z2w,2=1 z2w,3=1 |
| | 2 | z2w,1=1 z2w,3=1 z4=3 |
| | 2 | z2w,2=1 z2w,3=1 z4=3 |
| | 2 | z2w,3=1 z4=2 |
| 59.4.29 | 16013 | Cl1Fe1O1 |
| | 2 | z2w,2=1 z4=3 |
| | 2 | z2w,1=1 z4=3 |
| | 2 | z2w,1=1 z2w,2=1 z4=2 |
| | 2 | z2w,1=1 z2w,3=1 z4=3 |
| | 2 | z2w,1=1 z2w,2=1 z2w,3=1 z4=2 |
| | 2 | z2w,3=1 z4=2 |
| | 2 | z2w,2=1 z2w,3=1 z4=1 |
| | 2 | z2w,2=1 z2w,3=1 z4=3 |
| | 2 | z2w,3=1 z4=2 |
| | 2 | z2w,1=1 z2w,2=1 z2w,3=1 z4=2 |
| | 2 | z2w,1=1 z2w,3=1 z4=1 |
| | 2 | z2w,1=1 z2w,2=1 z4=2 |
| | 2 | z2w,1=1 z4=1 |
| | 2 | z2w,2=1 z4=1 |
| | 2 | z2w,1=1 z2w,2=1 z4=3 |
| | 2 | z2w,1=1 z4=2 |
| | 2 | z2w,2=1 z4=2 |
| | 2 | z4=1 |
| | 2 | z2w,2=1 z2w,3=1 z4=2 |
| | 2 | z2w,3=1 z4=1 |
| | 2 | z2w,1=1 z2w,2=1 z2w,3=1 z4=1 |
| | 2 | z2w,1=1 z2w,3=1 |
| | 2 | z2w,1=1 z2w,3=1 z4=2 |
| | 2 | z2w,1=1 z2w,2=1 z2w,3=1 z4=1 |
| | 2 | z2w,3=1 z4=1 |
| | 2 | z2w,2=1 z2w,3=1 |
| | 2 | z4=1 |
| | 2 | z2w,2=1 |
| | 2 | z2w,1=1 |
| | 2 | z2w,1=1 z2w,2=1 z4=3 |
| | 2 | z2w,1=1 z2w,2=1 z2w,3=1 z4=3 |
| | 2 | z2w,1=1 z2w,3=1 z4=2 |
| | 2 | z2w,2=1 z2w,3=1 z4=2 |
| | 2 | z2w,3=1 z4=1 |
| | 2 | z2w,2=1 z4=2 |
| | 2 | z4=1 |
| | 2 | z2w,1=1 z2w,2=1 z4=1 |
| | 2 | z2w,1=1 |
| | 2 | z2w,1=1 z4=2 |
| | 2 | z2w,1=1 z2w,2=1 z4=1 |
| | 2 | z4=1 |
| | 2 | z2w,2=1 |
| | 2 | z2w,3=1 z4=1 |
| | 2 | z2w,2=1 z2w,3=1 |
| | 2 | z2w,1=1 z2w,3=1 |
| | 2 | z2w,1=1 z2w,2=1 z2w,3=1 z4=3 |
| | 2 | z2w,3=1 z4=2 |
| | 2 | z2w,2=1 z2w,3=1 z4=1 |
| | 2 | z2w,1=1 z2w,3=1 z4=1 |
| | 2 | z2w,1=1 z2w,2=1 z2w,3=1 |
| | 2 | z2w,1=1 z4=1 |
| | 2 | z2w,1=1 z2w,2=1 |

Continued on next column

**TABLE XIX– continued**

| TQC | ICSD SSG | Stoich. Formula SIs |
|---|---|---|
| | 2 | z2w,2=1 z4=3 |
| | 2 | z2w,2=1 z4=1 |
| | 2 | z2w,1=1 z2w,2=1 |
| | 2 | z2w,1=1 z4=3 |
| | 2 | z2w,1=1 z2w,2=1 z2w,3=1 |
| | 2 | z2w,1=1 z2w,3=1 z4=3 |
| | 2 | z2w,2=1 z2w,3=1 z4=3 |
| | 2 | z2w,3=1 z4=2 |
| 59.4.30 | 21009 | Cl1Er1S1 |
| | 2 | z2w,3=1 z4=3 |
| | 2 | z2w,2=1 z2w,3=1 z4=2 |
| | 2 | z2w,1=1 z2w,3=1 z4=2 |
| | 2 | z2w,1=1 z2w,2=1 z2w,3=1 z4=1 |
| | 2 | z2w,1=1 z4=2 |
| | 2 | z2w,1=1 z2w,2=1 z4=1 |
| | 2 | z4=1 |
| | 2 | z2w,2=1 |
| | 2 | z2w,2=1 z4=2 |
| | 2 | z4=1 |
| | 2 | z2w,1=1 z2w,2=1 z4=1 |
| | 2 | z2w,1=1 |
| | 2 | z2w,1=1 z2w,2=1 z2w,3=1 z4=1 |
| | 2 | z2w,1=1 z2w,3=1 |
| | 2 | z2w,2=1 z2w,3=1 |
| | 2 | z2w,3=1 z4=3 |
| | 2 | z2w,1=1 z2w,2=1 z2w,3=1 z4=2 |
| | 2 | z2w,1=1 z2w,3=1 z4=1 |
| | 2 | z2w,2=1 z2w,3=1 z4=1 |
| | 2 | z2w,3=1 |
| | 2 | z2w,2=1 z4=1 |
| | 2 | z2w,1=1 z2w,2=1 |
| | 2 | z2w,1=1 z4=3 |
| | 2 | z2w,1=1 z4=1 |
| | 2 | z2w,1=1 z2w,2=1 |
| | 2 | z2w,2=1 z4=3 |
| | 2 | z2w,3=1 |
| | 2 | z2w,2=1 z2w,3=1 z4=3 |
| | 2 | z2w,1=1 z2w,3=1 z4=3 |
| | 2 | z2w,1=1 z2w,2=1 z2w,3=1 z4=2 |
| | 2 | z2w,1=1 z2w,2=1 z4=2 |
| | 2 | z2w,1=1 z4=1 |
| | 2 | z2w,2=1 z4=1 |
| | 2 | z2w,2=1 z2w,3=1 z4=1 |
| | 2 | z2w,3=1 |
| | 2 | z2w,1=1 z2w,2=1 z2w,3=1 |
| | 2 | z2w,1=1 z2w,3=1 z4=3 |
| | 2 | z2w,1=1 z2w,3=1 z4=1 |
| | 2 | z2w,1=1 z2w,2=1 z2w,3=1 |
| | 2 | z2w,3=1 |
| | 2 | z2w,2=1 z2w,3=1 z4=3 |
| | 2 | z2w,2=1 z4=3 |
| | 2 | z2w,1=1 z4=3 |
| | 2 | z2w,1=1 z2w,2=1 z4=2 |
| | 2 | z4=1 |
| | 2 | z2w,2=1 |
| | 2 | z2w,1=1 |
| | 2 | z2w,1=1 z2w,2=1 z4=3 |
| | 2 | z2w,1=1 z2w,3=1 |
| | 2 | z2w,1=1 z2w,2=1 z2w,3=1 z4=3 |
| | 2 | z2w,3=1 z4=3 |

Continued on next column

**TABLE XIX– continued**

| TQC | ICSD SSG | Stoich. Formula SIs |
|---|---|---|
| | 2 | z2w,2=1 z2w,3=1 z4=2 |
| | 2 | z2w,2=1 z2w,3=1 |
| | 2 | z2w,3=1 z4=3 |
| | 2 | z2w,1=1 z2w,2=1 z2w,3=1 z4=3 |
| | 2 | z2w,1=1 z2w,3=1 z4=2 |
| | 2 | z2w,1=1 z2w,2=1 z4=3 |
| | 2 | z2w,1=1 z4=2 |
| | 2 | z2w,2=1 z4=2 |
| | 2 | z4=1 |
| 59.4.31 | 161723 | Cl1Cu3O8Se2Y1 |
| | 2 | z4=1 |
| | 2 | z2w,2=1 |
| | 2 | z2w,1=1 |
| | 2 | z2w,1=1 z2w,2=1 z4=3 |
| | 2 | z2w,1=1 z2w,3=1 |
| | 2 | z2w,1=1 z2w,2=1 z2w,3=1 z4=3 |
| | 2 | z2w,3=1 z4=3 |
| | 2 | z2w,2=1 z2w,3=1 z4=2 |
| | 2 | z2w,2=1 z2w,3=1 |
| | 2 | z2w,3=1 z4=3 |
| | 2 | z2w,1=1 z2w,2=1 z2w,3=1 z4=3 |
| | 2 | z2w,1=1 z2w,3=1 z4=2 |
| | 2 | z2w,1=1 z2w,2=1 z4=3 |
| | 2 | z2w,1=1 z4=2 |
| | 2 | z2w,2=1 z4=2 |
| | 2 | z4=1 |
| | 2 | z2w,1=1 z2w,2=1 |
| | 2 | z2w,1=1 z4=3 |
| | 2 | z2w,2=1 z4=3 |
| | 2 | z4=2 |
| | 2 | z2w,2=1 z2w,3=1 z4=3 |
| | 2 | z2w,3=1 z4=2 |
| | 2 | z2w,1=1 z2w,2=1 z2w,3=1 z4=2 |
| | 2 | z2w,1=1 z2w,3=1 z4=1 |
| | 2 | z2w,1=1 z2w,3=1 z4=3 |
| | 2 | z2w,1=1 z2w,2=1 z2w,3=1 z4=2 |
| | 2 | z2w,3=1 z4=2 |
| | 2 | z2w,2=1 z2w,3=1 z4=1 |
| | 2 | z4=2 |
| | 2 | z2w,2=1 z4=1 |
| | 2 | z2w,1=1 z4=1 |
| | 2 | z2w,1=1 z2w,2=1 |
| | 2 | z2w,1=1 z2w,2=1 z2w,3=1 |
| | 2 | z2w,1=1 z2w,3=1 z4=3 |
| | 2 | z2w,2=1 z2w,3=1 z4=3 |
| | 2 | z2w,3=1 z4=2 |
| | 2 | z2w,2=1 z4=3 |
| | 2 | z4=2 |
| | 2 | z2w,1=1 z2w,2=1 z4=2 |
| | 2 | z2w,1=1 z4=1 |
| | 2 | z2w,1=1 z4=3 |
| | 2 | z2w,1=1 z2w,2=1 z4=2 |
| | 2 | z4=2 |
| | 2 | z2w,2=1 z4=1 |
| | 2 | z2w,3=1 z4=2 |
| | 2 | z2w,2=1 z2w,3=1 z4=1 |
| | 2 | z2w,1=1 z2w,3=1 z4=1 |
| | 2 | z2w,1=1 z2w,2=1 z2w,3=1 |
| | 2 | z2w,3=1 z4=3 |
| | 2 | z2w,2=1 z2w,3=1 z4=2 |

Continued on next column

**TABLE XIX– continued**

| TQC | ICSD<br>SSG | Stoich. Formula<br>SIs |
|---|---|---|
| | 2 | z2w,1=1 z2w,3=1 z4=2 |
| | 2 | z2w,1=1 z2w,2=1 z2w,3=1 z4=1 |
| | 2 | z2w,1=1 z4=2 |
| | 2 | z2w,1=1 z2w,2=1 z4=1 |
| | 2 | z4=1 |
| | 2 | z2w,2=1 |
| | 2 | z2w,2=1 z4=2 |
| | 2 | z4=1 |
| | 2 | z2w,1=1 z2w,2=1 z4=1 |
| | 2 | z2w,1=1 |
| | 2 | z2w,1=1 z2w,2=1 z2w,3=1 z4=1 |
| | 2 | z2w,1=1 z2w,3=1 |
| | 2 | z2w,2=1 z2w,3=1 |
| | 2 | z2w,3=1 z4=3 |
| 59.4.32 | 4086 | Cl1Cr1O1 |
| | 2 | z4=3 |
| | 2 | z2w,2=1 z4=2 |
| | 2 | z2w,1=1 z4=2 |
| | 2 | z2w,1=1 z2w,2=1 z4=1 |
| | 2 | z2w,1=1 z2w,3=1 z4=2 |
| | 2 | z2w,1=1 z2w,2=1 z2w,3=1 z4=1 |
| | 2 | z2w,3=1 z4=1 |
| | 2 | z2w,2=1 z2w,3=1 |
| | 2 | z2w,2=1 z2w,3=1 z4=2 |
| | 2 | z2w,3=1 z4=1 |
| | 2 | z2w,1=1 z2w,2=1 z2w,3=1 z4=1 |
| | 2 | z2w,1=1 z2w,3=1 |
| | 2 | z2w,1=1 z2w,2=1 z4=1 |
| | 2 | z2w,1=1 |
| | 2 | z2w,2=1 |
| | 2 | z4=3 |
| | 2 | z2w,1=1 z2w,2=1 z4=2 |
| | 2 | z2w,1=1 z4=1 |
| | 2 | z2w,2=1 z4=1 |
| | 2 | z2w,2=1 z2w,3=1 z4=1 |
| | 2 | z2w,3=1 |
| | 2 | z2w,1=1 z2w,2=1 z2w,3=1 |
| | 2 | z2w,1=1 z2w,3=1 z4=3 |
| | 2 | z2w,1=1 z2w,3=1 z4=1 |
| | 2 | z2w,1=1 z2w,2=1 z2w,3=1 |
| | 2 | z2w,3=1 |
| | 2 | z2w,2=1 z2w,3=1 z4=3 |
| | 2 | z2w,2=1 z4=3 |
| | 2 | z2w,1=1 z4=3 |
| | 2 | z2w,1=1 z2w,2=1 z4=2 |
| | 2 | z2w,1=1 z2w,2=1 z2w,3=1 z4=2 |
| | 2 | z2w,1=1 z2w,3=1 z4=1 |
| | 2 | z2w,2=1 z2w,3=1 z4=1 |
| | 2 | z2w,3=1 |
| | 2 | z2w,2=1 z4=1 |
| | 2 | z2w,1=1 z2w,2=1 |
| | 2 | z2w,1=1 z4=3 |
| | 2 | z2w,1=1 z4=1 |
| | 2 | z2w,1=1 z2w,2=1 |
| | 2 | z2w,2=1 z4=3 |
| | 2 | z2w,3=1 |
| | 2 | z2w,2=1 z2w,3=1 z4=3 |
| | 2 | z2w,1=1 z2w,3=1 z4=3 |
| | 2 | z2w,1=1 z2w,2=1 z2w,3=1 z4=2 |
| | 2 | z2w,3=1 z4=1 |

Continued on next column

**TABLE XIX– continued**

| TQC | ICSD<br>SSG | Stoich. Formula<br>SIs |
|---|---|---|
| | 2 | z2w,2=1 z2w,3=1 |
| | 2 | z2w,1=1 z2w,3=1 |
| | 2 | z2w,1=1 z2w,2=1 z2w,3=1 z4=3 |
| | 2 | z2w,1=1 |
| | 2 | z2w,1=1 z2w,2=1 z4=3 |
| | 2 | z4=3 |
| | 2 | z2w,2=1 z4=2 |
| | 2 | z2w,2=1 |
| | 2 | z4=3 |
| | 2 | z2w,1=1 z2w,2=1 z4=3 |
| | 2 | z2w,1=1 z4=2 |
| | 2 | z2w,1=1 z2w,2=1 z2w,3=1 z4=3 |
| | 2 | z2w,1=1 z2w,3=1 z4=2 |
| | 2 | z2w,2=1 z2w,3=1 z4=2 |
| | 2 | z2w,3=1 z4=1 |
| 59.4.33 | 80005 | Ce3Pd6Sb5 |
| | 2 | z2w,3=1 z4=1 |
| | 2 | z2w,2=1 z2w,3=1 |
| | 2 | z2w,1=1 z2w,3=1 |
| | 2 | z2w,1=1 z2w,2=1 z2w,3=1 z4=3 |
| | 2 | z2w,1=1 |
| | 2 | z2w,1=1 z2w,2=1 z4=3 |
| | 2 | z4=3 |
| | 2 | z2w,2=1 z4=2 |
| | 2 | z2w,2=1 |
| | 2 | z4=3 |
| | 2 | z2w,1=1 z2w,2=1 z4=3 |
| | 2 | z2w,1=1 z4=2 |
| | 2 | z2w,1=1 z2w,2=1 z2w,3=1 z4=3 |
| | 2 | z2w,1=1 z2w,3=1 z4=2 |
| | 2 | z2w,2=1 z2w,3=1 z4=2 |
| | 2 | z2w,3=1 z4=1 |
| | 2 | z2w,1=1 z2w,2=1 z2w,3=1 |
| | 2 | z2w,1=1 z2w,3=1 z4=3 |
| | 2 | z2w,2=1 z2w,3=1 z4=3 |
| | 2 | z2w,3=1 z4=2 |
| | 2 | z2w,2=1 z4=3 |
| | 2 | z4=2 |
| | 2 | z2w,1=1 z2w,2=1 z4=2 |
| | 2 | z2w,1=1 z4=1 |
| | 2 | z2w,1=1 z4=3 |
| | 2 | z2w,1=1 z2w,2=1 z4=2 |
| | 2 | z4=2 |
| | 2 | z2w,2=1 z4=1 |
| | 2 | z2w,3=1 z4=2 |
| | 2 | z2w,2=1 z2w,3=1 z4=1 |
| | 2 | z2w,1=1 z2w,3=1 z4=1 |
| | 2 | z2w,1=1 z2w,2=1 z2w,3=1 |
| | 2 | z2w,1=1 z2w,2=1 |
| | 2 | z2w,1=1 z4=3 |
| | 2 | z2w,2=1 z4=3 |
| | 2 | z4=2 |
| | 2 | z2w,2=1 z2w,3=1 z4=3 |
| | 2 | z2w,3=1 z4=2 |
| | 2 | z2w,1=1 z2w,2=1 z2w,3=1 z4=2 |
| | 2 | z2w,1=1 z2w,3=1 z4=1 |
| | 2 | z2w,1=1 z2w,3=1 z4=3 |
| | 2 | z2w,1=1 z2w,2=1 z2w,3=1 z4=2 |
| | 2 | z2w,3=1 z4=2 |
| | 2 | z2w,2=1 z2w,3=1 z4=1 |

Continued on next column

**TABLE XIX– continued**

| TQC | ICSD SSG | Stoich. Formula SIs |
|---|---|---|
| | 2 | z4=2 |
| | 2 | z2w,2=1 z4=1 |
| | 2 | z2w,1=1 z4=1 |
| | 2 | z2w,1=1 z2w,2=1 |
| | 2 | z4=3 |
| | 2 | z2w,2=1 z4=2 |
| | 2 | z2w,1=1 z4=2 |
| | 2 | z2w,1=1 z2w,2=1 z4=1 |
| | 2 | z2w,1=1 z2w,3=1 z4=2 |
| | 2 | z2w,1=1 z2w,2=1 z2w,3=1 z4=1 |
| | 2 | z2w,3=1 z4=1 |
| | 2 | z2w,2=1 z2w,3=1 |
| | 2 | z2w,2=1 z2w,3=1 z4=2 |
| | 2 | z2w,3=1 z4=1 |
| | 2 | z2w,1=1 z2w,2=1 z2w,3=1 z4=1 |
| | 2 | z2w,1=1 z2w,3=1 |
| | 2 | z2w,1=1 z2w,2=1 z4=1 |
| | 2 | z2w,1=1 |
| | 2 | z2w,2=1 |
| | 2 | z4=3 |
| 59.4.34 | 79106 | Br1Dy1S1 |
| | 2 | z2w,3=1 z4=2 |
| | 2 | z2w,2=1 z2w,3=1 z4=1 |
| | 2 | z2w,1=1 z2w,3=1 z4=1 |
| | 2 | z2w,1=1 z2w,2=1 z2w,3=1 |
| | 2 | z2w,1=1 z4=1 |
| | 2 | z2w,1=1 z2w,2=1 |
| | 2 | z2w,2=1 z4=3 |
| | 2 | z2w,2=1 z4=1 |
| | 2 | z2w,1=1 z2w,2=1 |
| | 2 | z2w,1=1 z4=3 |
| | 2 | z2w,1=1 z2w,2=1 z2w,3=1 |
| | 2 | z2w,1=1 z2w,3=1 z4=3 |
| | 2 | z2w,2=1 z2w,3=1 z4=3 |
| | 2 | z2w,3=1 z4=2 |
| | 2 | z2w,1=1 z2w,2=1 z2w,3=1 z4=1 |
| | 2 | z2w,1=1 z2w,3=1 |
| | 2 | z2w,2=1 z2w,3=1 |
| | 2 | z2w,3=1 z4=3 |
| | 2 | z2w,2=1 |
| | 2 | z4=3 |
| | 2 | z2w,1=1 z2w,2=1 z4=3 |
| | 2 | z2w,1=1 z4=2 |
| | 2 | z2w,1=1 |
| | 2 | z2w,1=1 z2w,2=1 z4=3 |
| | 2 | z4=3 |
| | 2 | z2w,2=1 z4=2 |
| | 2 | z2w,3=1 z4=3 |
| | 2 | z2w,2=1 z2w,3=1 z4=2 |
| | 2 | z2w,1=1 z2w,3=1 z4=2 |
| | 2 | z2w,1=1 z2w,2=1 z2w,3=1 z4=1 |
| | 2 | z2w,1=1 z2w,2=1 z4=1 |
| | 2 | z2w,1=1 |
| | 2 | z2w,2=1 |
| | 2 | z4=3 |
| | 2 | z2w,2=1 z2w,3=1 |
| | 2 | z2w,3=1 z4=3 |
| | 2 | z2w,1=1 z2w,2=1 z2w,3=1 z4=3 |
| | 2 | z2w,1=1 z2w,3=1 z4=2 |
| | 2 | z2w,1=1 z2w,3=1 |

Continued on next column

**TABLE XIX– continued**

| TQC | ICSD SSG | Stoich. Formula SIs |
|---|---|---|
| | 2 | z2w,1=1 z2w,2=1 z2w,3=1 z4=3 |
| | 2 | z2w,3=1 z4=3 |
| | 2 | z2w,2=1 z2w,3=1 z4=2 |
| | 2 | z4=3 |
| | 2 | z2w,2=1 z4=2 |
| | 2 | z2w,1=1 z4=2 |
| | 2 | z2w,1=1 z2w,2=1 z4=1 |
| | 2 | z2w,2=1 z4=3 |
| | 2 | z2w,1=1 z4=3 |
| | 2 | z2w,1=1 z2w,2=1 z4=2 |
| | 2 | z2w,1=1 z2w,3=1 z4=3 |
| | 2 | z2w,1=1 z2w,2=1 z2w,3=1 z4=2 |
| | 2 | z2w,3=1 z4=2 |
| | 2 | z2w,2=1 z2w,3=1 z4=1 |
| | 2 | z2w,2=1 z2w,3=1 z4=3 |
| | 2 | z2w,3=1 z4=2 |
| | 2 | z2w,1=1 z2w,2=1 z2w,3=1 z4=2 |
| | 2 | z2w,1=1 z2w,3=1 z4=1 |
| | 2 | z2w,1=1 z2w,2=1 z4=2 |
| | 2 | z2w,1=1 z4=1 |
| | 2 | z2w,2=1 z4=1 |
| 59.4.35 | 69661 | Br1Cr1S1 |
| | 2 | z4=3 |
| | 2 | z2w,2=1 z4=2 |
| | 2 | z2w,1=1 z4=2 |
| | 2 | z2w,1=1 z2w,2=1 z4=1 |
| | 2 | z2w,1=1 z2w,3=1 z4=2 |
| | 2 | z2w,1=1 z2w,2=1 z2w,3=1 z4=1 |
| | 2 | z2w,3=1 z4=1 |
| | 2 | z2w,2=1 z2w,3=1 |
| | 2 | z2w,2=1 z2w,3=1 z4=2 |
| | 2 | z2w,3=1 z4=1 |
| | 2 | z2w,1=1 z2w,2=1 z2w,3=1 z4=1 |
| | 2 | z2w,1=1 z2w,3=1 |
| | 2 | z2w,1=1 z2w,2=1 z4=1 |
| | 2 | z2w,1=1 |
| | 2 | z2w,2=1 |
| | 2 | z4=3 |
| | 2 | z2w,1=1 z2w,2=1 z4=2 |
| | 2 | z2w,1=1 z4=1 |
| | 2 | z2w,2=1 z4=1 |
| | 2 | z2w,2=1 z2w,3=1 z4=1 |
| | 2 | z2w,3=1 |
| | 2 | z2w,1=1 z2w,2=1 z2w,3=1 |
| | 2 | z2w,1=1 z2w,3=1 z4=3 |
| | 2 | z2w,1=1 z2w,3=1 z4=1 |
| | 2 | z2w,1=1 z2w,2=1 z2w,3=1 |
| | 2 | z2w,3=1 |
| | 2 | z2w,2=1 z2w,3=1 z4=3 |
| | 2 | z2w,2=1 z4=3 |
| | 2 | z2w,1=1 z4=3 |
| | 2 | z2w,1=1 z2w,2=1 z4=2 |
| | 2 | z2w,1=1 z2w,2=1 z2w,3=1 z4=2 |
| | 2 | z2w,1=1 z2w,3=1 z4=1 |
| | 2 | z2w,2=1 z2w,3=1 z4=1 |
| | 2 | z2w,3=1 |
| | 2 | z2w,2=1 z4=1 |
| | 2 | z2w,1=1 z2w,2=1 |
| | 2 | z2w,1=1 z4=3 |
| | 2 | z2w,1=1 z4=1 |

Continued on next column

**TABLE XIX– continued**

| TQC | ICSD SSG | Stoich. Formula SIs |
|---|---|---|
| | 2 | z2w,1=1 z2w,2=1 |
| | 2 | z2w,2=1 z4=3 |
| | 2 | z2w,3=1 |
| | 2 | z2w,2=1 z2w,3=1 z4=3 |
| | 2 | z2w,1=1 z2w,3=1 z4=3 |
| | 2 | z2w,1=1 z2w,2=1 z2w,3=1 z4=2 |
| | 2 | z2w,3=1 z4=1 |
| | 2 | z2w,2=1 z2w,3=1 |
| | 2 | z2w,1=1 z2w,3=1 |
| | 2 | z2w,1=1 z2w,2=1 z2w,3=1 z4=3 |
| | 2 | z2w,1=1 |
| | 2 | z2w,1=1 z2w,2=1 z4=3 |
| | 2 | z4=3 |
| | 2 | z2w,2=1 z4=2 |
| | 2 | z2w,2=1 |
| | 2 | z4=3 |
| | 2 | z2w,1=1 z2w,2=1 z4=3 |
| | 2 | z2w,1=1 z4=2 |
| | 2 | z2w,1=1 z2w,2=1 z2w,3=1 z4=3 |
| | 2 | z2w,1=1 z2w,3=1 z4=2 |
| | 2 | z2w,2=1 z2w,3=1 z4=2 |
| | 2 | z2w,3=1 z4=1 |
| 59.4.36 | 69660 | Br1Cr1S1 |
| | 2 | z4=3 |
| | 2 | z2w,2=1 z4=2 |
| | 2 | z2w,1=1 z4=2 |
| | 2 | z2w,1=1 z2w,2=1 z4=1 |
| | 2 | z2w,1=1 z2w,3=1 z4=2 |
| | 2 | z2w,1=1 z2w,2=1 z2w,3=1 z4=1 |
| | 2 | z2w,3=1 z4=1 |
| | 2 | z2w,2=1 z2w,3=1 |
| | 2 | z2w,2=1 z2w,3=1 z4=2 |
| | 2 | z2w,3=1 z4=1 |
| | 2 | z2w,1=1 z2w,2=1 z2w,3=1 z4=1 |
| | 2 | z2w,1=1 z2w,3=1 |
| | 2 | z2w,1=1 z2w,2=1 z4=1 |
| | 2 | z2w,1=1 |
| | 2 | z2w,2=1 |
| | 2 | z4=3 |
| | 2 | z2w,1=1 z2w,2=1 z4=2 |
| | 2 | z2w,1=1 z4=1 |
| | 2 | z2w,2=1 z4=1 |
| | 2 | z2w,2=1 z2w,3=1 z4=1 |
| | 2 | z2w,3=1 |
| | 2 | z2w,1=1 z2w,2=1 z2w,3=1 |
| | 2 | z2w,1=1 z2w,3=1 z4=3 |
| | 2 | z2w,1=1 z2w,3=1 z4=1 |
| | 2 | z2w,1=1 z2w,2=1 z2w,3=1 |
| | 2 | z2w,3=1 |
| | 2 | z2w,2=1 z2w,3=1 z4=3 |
| | 2 | z2w,2=1 z4=3 |
| | 2 | z2w,1=1 z4=3 |
| | 2 | z2w,1=1 z2w,2=1 z4=2 |
| | 2 | z2w,1=1 z2w,2=1 z2w,3=1 z4=2 |
| | 2 | z2w,1=1 z2w,3=1 z4=1 |
| | 2 | z2w,2=1 z2w,3=1 z4=1 |
| | 2 | z2w,3=1 |
| | 2 | z2w,2=1 z4=1 |
| | 2 | z2w,1=1 z2w,2=1 |
| | 2 | z2w,1=1 z4=3 |

Continued on next column

**TABLE XIX– continued**

| TQC | ICSD SSG | Stoich. Formula SIs |
|---|---|---|
| | 2 | z2w,1=1 z4=1 |
| | 2 | z2w,1=1 z2w,2=1 |
| | 2 | z2w,2=1 z4=3 |
| | 2 | z2w,3=1 |
| | 2 | z2w,2=1 z2w,3=1 z4=3 |
| | 2 | z2w,1=1 z2w,3=1 z4=3 |
| | 2 | z2w,1=1 z2w,2=1 z2w,3=1 z4=2 |
| | 2 | z2w,3=1 z4=1 |
| | 2 | z2w,2=1 z2w,3=1 |
| | 2 | z2w,1=1 z2w,3=1 |
| | 2 | z2w,1=1 z2w,2=1 z2w,3=1 z4=3 |
| | 2 | z2w,1=1 |
| | 2 | z2w,1=1 z2w,2=1 z4=3 |
| | 2 | z4=3 |
| | 2 | z2w,2=1 z4=2 |
| | 2 | z2w,2=1 |
| | 2 | z4=3 |
| | 2 | z2w,1=1 z2w,2=1 z4=3 |
| | 2 | z2w,1=1 z4=2 |
| | 2 | z2w,1=1 z2w,2=1 z2w,3=1 z4=3 |
| | 2 | z2w,1=1 z2w,3=1 z4=2 |
| | 2 | z2w,2=1 z2w,3=1 z4=2 |
| | 2 | z2w,3=1 z4=1 |
| 59.4.37 | 69659 | Br1Cr1S1 |
| | 2 | z2w,3=1 z4=2 |
| | 2 | z2w,2=1 z2w,3=1 z4=1 |
| | 2 | z2w,1=1 z2w,3=1 z4=1 |
| | 2 | z2w,1=1 z2w,2=1 z2w,3=1 |
| | 2 | z2w,1=1 z4=1 |
| | 2 | z2w,1=1 z2w,2=1 |
| | 2 | z2w,2=1 z4=3 |
| | 2 | z2w,2=1 z4=1 |
| | 2 | z2w,1=1 z2w,2=1 |
| | 2 | z2w,1=1 z4=3 |
| | 2 | z2w,1=1 z2w,2=1 z2w,3=1 |
| | 2 | z2w,1=1 z2w,3=1 z4=3 |
| | 2 | z2w,2=1 z2w,3=1 z4=3 |
| | 2 | z2w,3=1 z4=2 |
| | 2 | z2w,1=1 z2w,2=1 z2w,3=1 z4=1 |
| | 2 | z2w,1=1 z2w,3=1 |
| | 2 | z2w,2=1 z2w,3=1 |
| | 2 | z2w,3=1 z4=3 |
| | 2 | z2w,2=1 |
| | 2 | z4=3 |
| | 2 | z2w,1=1 z2w,2=1 z4=3 |
| | 2 | z2w,1=1 z4=2 |
| | 2 | z2w,1=1 |
| | 2 | z2w,1=1 z2w,2=1 z4=3 |
| | 2 | z4=3 |
| | 2 | z2w,2=1 z4=2 |
| | 2 | z2w,3=1 z4=3 |
| | 2 | z2w,2=1 z2w,3=1 z4=2 |
| | 2 | z2w,1=1 z2w,3=1 z4=2 |
| | 2 | z2w,1=1 z2w,2=1 z2w,3=1 z4=1 |
| | 2 | z2w,1=1 z2w,2=1 z4=1 |
| | 2 | z2w,1=1 |
| | 2 | z2w,2=1 |
| | 2 | z4=3 |
| | 2 | z2w,2=1 z2w,3=1 |
| | 2 | z2w,3=1 z4=3 |

Continued on next column

**TABLE XIX– continued**

| TQC | ICSD SSG | Stoich. Formula SIs |
|---|---|---|
| | 2 | z2w,1=1 z2w,2=1 z2w,3=1 z4=3 |
| | 2 | z2w,1=1 z2w,3=1 z4=2 |
| | 2 | z2w,1=1 z2w,3=1 |
| | 2 | z2w,1=1 z2w,2=1 z2w,3=1 z4=3 |
| | 2 | z2w,3=1 z4=3 |
| | 2 | z2w,2=1 z2w,3=1 z4=2 |
| | 2 | z4=3 |
| | 2 | z2w,2=1 z4=2 |
| | 2 | z2w,1=1 z4=2 |
| | 2 | z2w,1=1 z2w,2=1 z4=1 |
| | 2 | z2w,2=1 z4=3 |
| | 2 | z2w,1=1 z4=3 |
| | 2 | z2w,1=1 z2w,2=1 z4=2 |
| | 2 | z2w,1=1 z2w,3=1 z4=3 |
| | 2 | z2w,1=1 z2w,2=1 z2w,3=1 z4=2 |
| | 2 | z2w,3=1 z4=2 |
| | 2 | z2w,2=1 z2w,3=1 z4=1 |
| | 2 | z2w,2=1 z2w,3=1 z4=3 |
| | 2 | z2w,3=1 z4=2 |
| | 2 | z2w,1=1 z2w,2=1 z2w,3=1 z4=2 |
| | 2 | z2w,1=1 z2w,3=1 z4=1 |
| | 2 | z2w,1=1 z2w,2=1 z4=2 |
| | 2 | z2w,1=1 z4=1 |
| | 2 | z2w,2=1 z4=1 |
| 59.4.38 | 238106 | Bi5K6Zn1 |
| | 2 | z2w,3=1 z4=2 |
| | 2 | z2w,2=1 z2w,3=1 z4=1 |
| | 2 | z2w,1=1 z2w,3=1 z4=1 |
| | 2 | z2w,1=1 z2w,2=1 z2w,3=1 |
| | 2 | z2w,1=1 z4=1 |
| | 2 | z2w,1=1 z2w,2=1 |
| | 2 | z2w,2=1 z4=3 |
| | 2 | z2w,2=1 z4=1 |
| | 2 | z2w,1=1 z2w,2=1 |
| | 2 | z2w,1=1 z4=3 |
| | 2 | z2w,1=1 z2w,2=1 z2w,3=1 |
| | 2 | z2w,1=1 z2w,3=1 z4=3 |
| | 2 | z2w,2=1 z2w,3=1 z4=3 |
| | 2 | z2w,3=1 z4=2 |
| | 2 | z2w,1=1 z2w,2=1 z2w,3=1 z4=1 |
| | 2 | z2w,1=1 z2w,3=1 |
| | 2 | z2w,2=1 z2w,3=1 |
| | 2 | z2w,3=1 z4=3 |
| | 2 | z2w,2=1 |
| | 2 | z4=3 |
| | 2 | z2w,1=1 z2w,2=1 z4=3 |
| | 2 | z2w,1=1 z4=2 |
| | 2 | z2w,1=1 |
| | 2 | z2w,1=1 z2w,2=1 z4=3 |
| | 2 | z4=3 |
| | 2 | z2w,2=1 z4=2 |
| | 2 | z2w,3=1 z4=3 |
| | 2 | z2w,2=1 z2w,3=1 z4=2 |
| | 2 | z2w,1=1 z2w,3=1 z4=2 |
| | 2 | z2w,1=1 z2w,2=1 z2w,3=1 z4=1 |
| | 2 | z2w,1=1 z2w,2=1 z4=1 |
| | 2 | z2w,1=1 |
| | 2 | z2w,2=1 |
| | 2 | z4=3 |
| | 2 | z2w,2=1 z2w,3=1 |

Continued on next column

**TABLE XIX– continued**

| TQC | ICSD SSG | Stoich. Formula SIs |
|---|---|---|
| | 2 | z2w,3=1 z4=3 |
| | 2 | z2w,1=1 z2w,2=1 z2w,3=1 z4=3 |
| | 2 | z2w,1=1 z2w,3=1 z4=2 |
| | 2 | z2w,1=1 z2w,3=1 |
| | 2 | z2w,1=1 z2w,2=1 z2w,3=1 z4=3 |
| | 2 | z2w,3=1 z4=3 |
| | 2 | z2w,2=1 z2w,3=1 z4=2 |
| | 2 | z4=3 |
| | 2 | z2w,2=1 z4=2 |
| | 2 | z2w,1=1 z4=2 |
| | 2 | z2w,1=1 z2w,2=1 z4=1 |
| | 2 | z2w,2=1 z4=3 |
| | 2 | z2w,1=1 z4=3 |
| | 2 | z2w,1=1 z2w,2=1 z4=2 |
| | 2 | z2w,1=1 z2w,3=1 z4=3 |
| | 2 | z2w,1=1 z2w,2=1 z2w,3=1 z4=2 |
| | 2 | z2w,3=1 z4=2 |
| | 2 | z2w,2=1 z2w,3=1 z4=1 |
| | 2 | z2w,2=1 z2w,3=1 z4=3 |
| | 2 | z2w,3=1 z4=2 |
| | 2 | z2w,1=1 z2w,2=1 z2w,3=1 z4=2 |
| | 2 | z2w,1=1 z2w,3=1 z4=1 |
| | 2 | z2w,1=1 z2w,2=1 z4=2 |
| | 2 | z2w,1=1 z4=1 |
| | 2 | z2w,2=1 z4=1 |
| 59.4.39 | 69385 | Bi1Cl1Cu3O8Se2 |
| | 2 | z2w,2=1 z4=3 |
| | 2 | z2w,1=1 z4=3 |
| | 2 | z2w,1=1 z2w,2=1 z4=2 |
| | 2 | z2w,1=1 z2w,3=1 z4=3 |
| | 2 | z2w,1=1 z2w,2=1 z2w,3=1 z4=2 |
| | 2 | z2w,3=1 z4=2 |
| | 2 | z2w,2=1 z2w,3=1 z4=1 |
| | 2 | z2w,2=1 z2w,3=1 z4=3 |
| | 2 | z2w,3=1 z4=2 |
| | 2 | z2w,1=1 z2w,2=1 z2w,3=1 z4=2 |
| | 2 | z2w,1=1 z2w,3=1 z4=1 |
| | 2 | z2w,1=1 z2w,2=1 z4=2 |
| | 2 | z2w,1=1 z4=1 |
| | 2 | z2w,2=1 z4=1 |
| | 2 | z2w,1=1 z2w,2=1 z4=3 |
| | 2 | z2w,1=1 z4=2 |
| | 2 | z2w,2=1 z4=2 |
| | 2 | z4=1 |
| | 2 | z2w,2=1 z2w,3=1 z4=2 |
| | 2 | z2w,3=1 z4=1 |
| | 2 | z2w,1=1 z2w,2=1 z2w,3=1 z4=1 |
| | 2 | z2w,1=1 z2w,3=1 |
| | 2 | z2w,1=1 z2w,3=1 z4=2 |
| | 2 | z2w,1=1 z2w,2=1 z2w,3=1 z4=1 |
| | 2 | z2w,3=1 z4=1 |
| | 2 | z2w,2=1 z2w,3=1 |
| | 2 | z4=1 |
| | 2 | z2w,2=1 |
| | 2 | z2w,1=1 |
| | 2 | z2w,1=1 z2w,2=1 z4=3 |
| | 2 | z2w,1=1 z2w,2=1 z2w,3=1 z4=3 |
| | 2 | z2w,1=1 z2w,3=1 z4=2 |
| | 2 | z2w,2=1 z2w,3=1 z4=2 |
| | 2 | z2w,3=1 z4=1 |

Continued on next column

**TABLE XIX– continued**

| TQC | ICSD SSG | Stoich. Formula SIs |
|---|---|---|
| | 2 | z2w,2=1 z4=2 |
| | 2 | z4=1 |
| | 2 | z2w,1=1 z2w,2=1 z4=1 |
| | 2 | z2w,1=1 |
| | 2 | z2w,1=1 z4=2 |
| | 2 | z2w,1=1 z2w,2=1 z4=1 |
| | 2 | z4=1 |
| | 2 | z2w,2=1 |
| | 2 | z2w,3=1 z4=1 |
| | 2 | z2w,2=1 z2w,3=1 |
| | 2 | z2w,1=1 z2w,3=1 |
| | 2 | z2w,1=1 z2w,2=1 z2w,3=1 z4=3 |
| | 2 | z2w,3=1 z4=2 |
| | 2 | z2w,2=1 z2w,3=1 z4=1 |
| | 2 | z2w,1=1 z2w,3=1 z4=1 |
| | 2 | z2w,1=1 z2w,2=1 z2w,3=1 |
| | 2 | z2w,1=1 z4=1 |
| | 2 | z2w,1=1 z2w,2=1 |
| | 2 | z2w,2=1 z4=3 |
| | 2 | z2w,2=1 z4=1 |
| | 2 | z2w,1=1 z2w,2=1 |
| | 2 | z2w,1=1 z4=3 |
| | 2 | z2w,1=1 z2w,2=1 z2w,3=1 |
| | 2 | z2w,1=1 z2w,3=1 z4=3 |
| | 2 | z2w,2=1 z2w,3=1 z4=3 |
| | 2 | z2w,3=1 z4=2 |
| 59.4.40 | 54190 | Bi1Cl1Cu3O8Se2 |
| | 2 | z2w,2=1 z4=3 |
| | 2 | z2w,1=1 z4=3 |
| | 2 | z2w,1=1 z2w,2=1 z4=2 |
| | 2 | z2w,1=1 z2w,3=1 z4=3 |
| | 2 | z2w,1=1 z2w,2=1 z2w,3=1 z4=2 |
| | 2 | z2w,3=1 z4=2 |
| | 2 | z2w,2=1 z2w,3=1 z4=1 |
| | 2 | z2w,2=1 z2w,3=1 z4=3 |
| | 2 | z2w,3=1 z4=2 |
| | 2 | z2w,1=1 z2w,2=1 z2w,3=1 z4=2 |
| | 2 | z2w,1=1 z2w,3=1 z4=1 |
| | 2 | z2w,1=1 z2w,2=1 z4=2 |
| | 2 | z2w,1=1 z4=1 |
| | 2 | z2w,2=1 z4=1 |
| | 2 | z2w,1=1 z2w,2=1 z4=3 |
| | 2 | z2w,1=1 z4=2 |
| | 2 | z2w,2=1 z4=2 |
| | 2 | z4=1 |
| | 2 | z2w,2=1 z2w,3=1 z4=2 |
| | 2 | z2w,3=1 z4=1 |
| | 2 | z2w,1=1 z2w,2=1 z2w,3=1 z4=1 |
| | 2 | z2w,1=1 z2w,3=1 |
| | 2 | z2w,1=1 z2w,3=1 z4=2 |
| | 2 | z2w,1=1 z2w,2=1 z2w,3=1 z4=1 |
| | 2 | z2w,3=1 z4=1 |
| | 2 | z2w,2=1 z2w,3=1 |
| | 2 | z4=1 |
| | 2 | z2w,2=1 |
| | 2 | z2w,1=1 |
| | 2 | z2w,1=1 z2w,2=1 z4=3 |
| | 2 | z2w,1=1 z2w,2=1 z2w,3=1 z4=3 |
| | 2 | z2w,1=1 z2w,3=1 z4=2 |
| | 2 | z2w,2=1 z2w,3=1 z4=2 |

Continued on next column

| TQC | ICSD SSG | Stoich. Formula SIs |
|---|---|---|
| | 2 | z2w,3=1 z4=1 |
| | 2 | z2w,2=1 z4=2 |
| | 2 | z4=1 |
| | 2 | z2w,1=1 z2w,2=1 z4=1 |
| | 2 | z2w,1=1 |
| | 2 | z2w,1=1 z4=2 |
| | 2 | z2w,1=1 z2w,2=1 z4=1 |
| | 2 | z4=1 |
| | 2 | z2w,2=1 |
| | 2 | z2w,3=1 z4=1 |
| | 2 | z2w,2=1 z2w,3=1 |
| | 2 | z2w,1=1 z2w,3=1 |
| | 2 | z2w,1=1 z2w,2=1 z2w,3=1 z4=3 |
| | 2 | z2w,3=1 z4=2 |
| | 2 | z2w,2=1 z2w,3=1 z4=1 |
| | 2 | z2w,1=1 z2w,3=1 z4=1 |
| | 2 | z2w,1=1 z2w,2=1 z2w,3=1 |
| | 2 | z2w,1=1 z4=1 |
| | 2 | z2w,1=1 z2w,2=1 |
| | 2 | z2w,2=1 z4=3 |
| | 2 | z2w,2=1 z4=1 |
| | 2 | z2w,1=1 z2w,2=1 |
| | 2 | z2w,1=1 z4=3 |
| | 2 | z2w,1=1 z2w,2=1 z2w,3=1 |
| | 2 | z2w,1=1 z2w,3=1 z4=3 |
| | 2 | z2w,2=1 z2w,3=1 z4=3 |
| | 2 | z2w,3=1 z4=2 |
| 59.4.41 | 280759 | Bi1Br1Cu3O8Se2 |
| | 2 | z2w,2=1 z4=3 |
| | 2 | z2w,1=1 z4=3 |
| | 2 | z2w,1=1 z2w,2=1 z4=2 |
| | 2 | z2w,1=1 z2w,3=1 z4=3 |
| | 2 | z2w,1=1 z2w,2=1 z2w,3=1 z4=2 |
| | 2 | z2w,3=1 z4=2 |
| | 2 | z2w,2=1 z2w,3=1 z4=1 |
| | 2 | z2w,2=1 z2w,3=1 z4=3 |
| | 2 | z2w,3=1 z4=2 |
| | 2 | z2w,1=1 z2w,2=1 z2w,3=1 z4=2 |
| | 2 | z2w,1=1 z2w,3=1 z4=1 |
| | 2 | z2w,1=1 z2w,2=1 z4=2 |
| | 2 | z2w,1=1 z4=1 |
| | 2 | z2w,2=1 z4=1 |
| | 2 | z2w,1=1 z2w,2=1 z4=3 |
| | 2 | z2w,1=1 z4=2 |
| | 2 | z2w,2=1 z4=2 |
| | 2 | z4=1 |
| | 2 | z2w,2=1 z2w,3=1 z4=2 |
| | 2 | z2w,3=1 z4=1 |
| | 2 | z2w,1=1 z2w,2=1 z2w,3=1 z4=1 |
| | 2 | z2w,1=1 z2w,3=1 |
| | 2 | z2w,1=1 z2w,3=1 z4=2 |
| | 2 | z2w,1=1 z2w,2=1 z2w,3=1 z4=1 |
| | 2 | z2w,3=1 z4=1 |
| | 2 | z2w,2=1 z2w,3=1 |
| | 2 | z4=1 |
| | 2 | z2w,2=1 |
| | 2 | z2w,1=1 |
| | 2 | z2w,1=1 z2w,2=1 z4=3 |
| | 2 | z2w,1=1 z2w,2=1 z2w,3=1 z4=3 |
| | 2 | z2w,1=1 z2w,3=1 z4=2 |

Continued on next column

**TABLE XIX– continued**

| TQC | ICSD SSG | Stoich. Formula SIs |
|---|---|---|
| | 2 | z2w,2=1 z2w,3=1 z4=2 |
| | 2 | z2w,3=1 z4=1 |
| | 2 | z2w,2=1 z4=2 |
| | 2 | z4=1 |
| | 2 | z2w,1=1 z2w,2=1 z4=1 |
| | 2 | z2w,1=1 |
| | 2 | z2w,1=1 z4=2 |
| | 2 | z2w,1=1 z2w,2=1 z4=1 |
| | 2 | z4=1 |
| | 2 | z2w,2=1 |
| | 2 | z2w,3=1 z4=1 |
| | 2 | z2w,2=1 z2w,3=1 |
| | 2 | z2w,1=1 z2w,3=1 |
| | 2 | z2w,1=1 z2w,2=1 z2w,3=1 z4=3 |
| | 2 | z2w,3=1 z4=2 |
| | 2 | z2w,2=1 z2w,3=1 z4=1 |
| | 2 | z2w,1=1 z2w,3=1 z4=1 |
| | 2 | z2w,1=1 z2w,2=1 z2w,3=1 |
| | 2 | z2w,1=1 z4=1 |
| | 2 | z2w,1=1 z2w,2=1 |
| | 2 | z2w,2=1 z4=3 |
| | 2 | z2w,2=1 z4=1 |
| | 2 | z2w,1=1 z2w,2=1 |
| | 2 | z2w,1=1 z4=3 |
| | 2 | z2w,1=1 z2w,2=1 z2w,3=1 |
| | 2 | z2w,1=1 z2w,3=1 z4=3 |
| | 2 | z2w,2=1 z2w,3=1 z4=3 |
| | 2 | z2w,3=1 z4=2 |
| 59.4.42 | 54011 | B3Ge2Ni8 |
| | 2 | z2w,3=1 z4=2 |
| | 2 | z2w,2=1 z2w,3=1 z4=1 |
| | 2 | z2w,1=1 z2w,3=1 z4=1 |
| | 2 | z2w,1=1 z2w,2=1 z2w,3=1 |
| | 2 | z2w,1=1 z4=1 |
| | 2 | z2w,1=1 z2w,2=1 |
| | 2 | z2w,2=1 z4=3 |
| | 2 | z2w,2=1 z4=1 |
| | 2 | z2w,1=1 z2w,2=1 |
| | 2 | z2w,1=1 z4=3 |
| | 2 | z2w,1=1 z2w,2=1 z2w,3=1 |
| | 2 | z2w,1=1 z2w,3=1 z4=3 |
| | 2 | z2w,2=1 z2w,3=1 z4=3 |
| | 2 | z2w,3=1 z4=2 |
| | 2 | z2w,1=1 z2w,2=1 z2w,3=1 z4=1 |
| | 2 | z2w,1=1 z2w,3=1 |
| | 2 | z2w,2=1 z2w,3=1 |
| | 2 | z2w,3=1 z4=3 |
| | 2 | z2w,2=1 |
| | 2 | z4=3 |
| | 2 | z2w,1=1 z2w,2=1 z4=3 |
| | 2 | z2w,1=1 z4=2 |
| | 2 | z2w,1=1 |
| | 2 | z2w,1=1 z2w,2=1 z4=3 |
| | 2 | z4=3 |
| | 2 | z2w,2=1 z4=2 |
| | 2 | z2w,3=1 z4=3 |
| | 2 | z2w,2=1 z2w,3=1 z4=2 |
| | 2 | z2w,1=1 z2w,3=1 z4=2 |
| | 2 | z2w,1=1 z2w,2=1 z2w,3=1 z4=1 |
| | 2 | z2w,1=1 z2w,2=1 z4=1 |

Continued on next column

**TABLE XIX– continued**

| TQC | ICSD SSG | Stoich. Formula SIs |
|---|---|---|
| | 2 | z2w,1=1 |
| | 2 | z2w,2=1 |
| | 2 | z4=3 |
| | 2 | z2w,2=1 z2w,3=1 |
| | 2 | z2w,3=1 z4=3 |
| | 2 | z2w,1=1 z2w,2=1 z2w,3=1 z4=3 |
| | 2 | z2w,1=1 z2w,3=1 z4=2 |
| | 2 | z2w,1=1 z2w,3=1 |
| | 2 | z2w,1=1 z2w,2=1 z2w,3=1 z4=3 |
| | 2 | z2w,3=1 z4=3 |
| | 2 | z2w,2=1 z2w,3=1 z4=2 |
| | 2 | z4=3 |
| | 2 | z2w,2=1 z4=2 |
| | 2 | z2w,1=1 z4=2 |
| | 2 | z2w,1=1 z2w,2=1 z4=1 |
| | 2 | z2w,2=1 z4=3 |
| | 2 | z2w,1=1 z4=3 |
| | 2 | z2w,1=1 z2w,2=1 z4=2 |
| | 2 | z2w,1=1 z2w,3=1 z4=3 |
| | 2 | z2w,1=1 z2w,2=1 z2w,3=1 z4=2 |
| | 2 | z2w,3=1 z4=2 |
| | 2 | z2w,2=1 z2w,3=1 z4=1 |
| | 2 | z2w,2=1 z2w,3=1 z4=3 |
| | 2 | z2w,3=1 z4=2 |
| | 2 | z2w,1=1 z2w,2=1 z2w,3=1 z4=2 |
| | 2 | z2w,1=1 z2w,3=1 z4=1 |
| | 2 | z2w,1=1 z2w,2=1 z4=2 |
| | 2 | z2w,1=1 z4=1 |
| | 2 | z2w,2=1 z4=1 |
| 59.4.43 | 245680 | Au3Eu1In3 |
| | 2 | z2w,3=1 z4=1 |
| | 2 | z2w,2=1 z2w,3=1 |
| | 2 | z2w,1=1 z2w,3=1 |
| | 2 | z2w,1=1 z2w,2=1 z2w,3=1 z4=3 |
| | 2 | z2w,1=1 |
| | 2 | z2w,1=1 z2w,2=1 z4=3 |
| | 2 | z4=3 |
| | 2 | z2w,2=1 z4=2 |
| | 2 | z2w,2=1 |
| | 2 | z4=3 |
| | 2 | z2w,1=1 z2w,2=1 z4=3 |
| | 2 | z2w,1=1 z4=2 |
| | 2 | z2w,1=1 z2w,2=1 z2w,3=1 z4=3 |
| | 2 | z2w,1=1 z2w,3=1 z4=2 |
| | 2 | z2w,2=1 z2w,3=1 z4=2 |
| | 2 | z2w,3=1 z4=1 |
| | 2 | z2w,1=1 z2w,2=1 z2w,3=1 |
| | 2 | z2w,1=1 z2w,3=1 z4=3 |
| | 2 | z2w,2=1 z2w,3=1 z4=3 |
| | 2 | z2w,3=1 z4=2 |
| | 2 | z2w,2=1 z4=3 |
| | 2 | z4=2 |
| | 2 | z2w,1=1 z2w,2=1 z4=2 |
| | 2 | z2w,1=1 z4=1 |
| | 2 | z2w,1=1 z4=3 |
| | 2 | z2w,1=1 z2w,2=1 z4=2 |
| | 2 | z4=2 |
| | 2 | z2w,2=1 z4=1 |
| | 2 | z2w,3=1 z4=2 |
| | 2 | z2w,2=1 z2w,3=1 z4=1 |

Continued on next column

**TABLE XIX– continued**

| TQC | ICSD SSG | Stoich. Formula SIs |
|---|---|---|
| | 2 | z2w,1=1 z2w,3=1 z4=1 |
| | 2 | z2w,1=1 z2w,2=1 z2w,3=1 |
| | 2 | z2w,1=1 z2w,2=1 |
| | 2 | z2w,1=1 z4=3 |
| | 2 | z2w,2=1 z4=3 |
| | 2 | z4=2 |
| | 2 | z2w,2=1 z2w,3=1 z4=3 |
| | 2 | z2w,3=1 z4=2 |
| | 2 | z2w,1=1 z2w,2=1 z2w,3=1 z4=2 |
| | 2 | z2w,1=1 z2w,3=1 z4=1 |
| | 2 | z2w,1=1 z2w,3=1 z4=3 |
| | 2 | z2w,1=1 z2w,2=1 z2w,3=1 z4=2 |
| | 2 | z2w,3=1 z4=2 |
| | 2 | z2w,2=1 z2w,3=1 z4=1 |
| | 2 | z4=2 |
| | 2 | z2w,2=1 z4=1 |
| | 2 | z2w,1=1 z4=1 |
| | 2 | z2w,1=1 z2w,2=1 |
| | 2 | z4=3 |
| | 2 | z2w,2=1 z4=2 |
| | 2 | z2w,1=1 z4=2 |
| | 2 | z2w,1=1 z2w,2=1 z4=1 |
| | 2 | z2w,1=1 z2w,3=1 z4=2 |
| | 2 | z2w,1=1 z2w,2=1 z2w,3=1 z4=1 |
| | 2 | z2w,3=1 z4=1 |
| | 2 | z2w,2=1 z2w,3=1 |
| | 2 | z2w,2=1 z2w,3=1 z4=2 |
| | 2 | z2w,3=1 z4=1 |
| | 2 | z2w,1=1 z2w,2=1 z2w,3=1 z4=1 |
| | 2 | z2w,1=1 z2w,3=1 |
| | 2 | z2w,1=1 z2w,2=1 z4=1 |
| | 2 | z2w,1=1 |
| | 2 | z2w,2=1 |
| | 2 | z4=3 |
| 59.4.44 | 92440 | As1Pd1Zr1 |
| | 2 | z2w,3=1 z4=3 |
| | 2 | z2w,2=1 z2w,3=1 z4=2 |
| | 2 | z2w,1=1 z2w,3=1 z4=2 |
| | 2 | z2w,1=1 z2w,2=1 z2w,3=1 z4=1 |
| | 2 | z2w,1=1 z4=2 |
| | 2 | z2w,1=1 z2w,2=1 z4=1 |
| | 2 | z4=1 |
| | 2 | z2w,2=1 |
| | 2 | z2w,2=1 z4=2 |
| | 2 | z4=1 |
| | 2 | z2w,1=1 z2w,2=1 z4=1 |
| | 2 | z2w,1=1 |
| | 2 | z2w,1=1 z2w,2=1 z2w,3=1 z4=1 |
| | 2 | z2w,1=1 z2w,3=1 |
| | 2 | z2w,2=1 z2w,3=1 |
| | 2 | z2w,3=1 z4=3 |
| | 2 | z2w,1=1 z2w,2=1 z2w,3=1 z4=2 |
| | 2 | z2w,1=1 z2w,3=1 z4=1 |
| | 2 | z2w,2=1 z2w,3=1 z4=1 |
| | 2 | z2w,3=1 |
| | 2 | z2w,2=1 z4=1 |
| | 2 | z2w,1=1 z2w,2=1 |
| | 2 | z2w,1=1 z4=3 |
| | 2 | z2w,1=1 z4=1 |
| | 2 | z2w,1=1 z2w,2=1 |

Continued on next column

**TABLE XIX– continued**

| TQC | ICSD SSG | Stoich. Formula SIs |
|---|---|---|
| | 2 | z2w,2=1 z4=3 |
| | 2 | z2w,3=1 |
| | 2 | z2w,2=1 z2w,3=1 z4=3 |
| | 2 | z2w,1=1 z2w,3=1 z4=3 |
| | 2 | z2w,1=1 z2w,2=1 z2w,3=1 z4=2 |
| | 2 | z2w,1=1 z2w,2=1 z4=2 |
| | 2 | z2w,1=1 z4=1 |
| | 2 | z2w,2=1 z4=1 |
| | 2 | z2w,2=1 z2w,3=1 z4=1 |
| | 2 | z2w,3=1 |
| | 2 | z2w,1=1 z2w,2=1 z2w,3=1 |
| | 2 | z2w,1=1 z2w,3=1 z4=3 |
| | 2 | z2w,1=1 z2w,3=1 z4=1 |
| | 2 | z2w,1=1 z2w,2=1 z2w,3=1 |
| | 2 | z2w,3=1 |
| | 2 | z2w,2=1 z2w,3=1 z4=3 |
| | 2 | z2w,2=1 z4=3 |
| | 2 | z2w,1=1 z4=3 |
| | 2 | z2w,1=1 z2w,2=1 z4=2 |
| | 2 | z4=1 |
| | 2 | z2w,2=1 |
| | 2 | z2w,1=1 |
| | 2 | z2w,1=1 z2w,2=1 z4=3 |
| | 2 | z2w,1=1 z2w,3=1 |
| | 2 | z2w,1=1 z2w,2=1 z2w,3=1 z4=3 |
| | 2 | z2w,3=1 z4=3 |
| | 2 | z2w,2=1 z2w,3=1 z4=2 |
| | 2 | z2w,2=1 z2w,3=1 |
| | 2 | z2w,3=1 z4=3 |
| | 2 | z2w,1=1 z2w,2=1 z2w,3=1 z4=3 |
| | 2 | z2w,1=1 z2w,3=1 z4=2 |
| | 2 | z2w,1=1 z2w,2=1 z4=3 |
| | 2 | z2w,1=1 z4=2 |
| | 2 | z2w,2=1 z4=2 |
| | 2 | z4=1 |
| 59.4.45 | 2721 | Ag3Sn1 |
| | 2 | z2w,3=1 z4=2 |
| | 2 | z2w,2=1 z2w,3=1 z4=1 |
| | 2 | z2w,1=1 z2w,3=1 z4=1 |
| | 2 | z2w,1=1 z2w,2=1 z2w,3=1 |
| | 2 | z2w,1=1 z4=1 |
| | 2 | z2w,1=1 z2w,2=1 |
| | 2 | z2w,2=1 z4=3 |
| | 2 | z2w,2=1 z4=1 |
| | 2 | z2w,1=1 z2w,2=1 |
| | 2 | z2w,1=1 z4=3 |
| | 2 | z2w,1=1 z2w,2=1 z2w,3=1 |
| | 2 | z2w,1=1 z2w,3=1 z4=3 |
| | 2 | z2w,2=1 z2w,3=1 z4=3 |
| | 2 | z2w,3=1 z4=2 |
| | 2 | z2w,1=1 z2w,2=1 z2w,3=1 z4=1 |
| | 2 | z2w,1=1 z2w,3=1 |
| | 2 | z2w,2=1 z2w,3=1 |
| | 2 | z2w,3=1 z4=3 |
| | 2 | z2w,2=1 |
| | 2 | z4=3 |
| | 2 | z2w,1=1 z2w,2=1 z4=3 |
| | 2 | z2w,1=1 z4=2 |
| | 2 | z2w,1=1 |
| | 2 | z2w,1=1 z2w,2=1 z4=3 |

Continued on next column

**TABLE XIX– continued**

| TQC | ICSD SSG | Stoich. Formula SIs |
|---|---|---|
| | 2 | z4=3 |
| | 2 | z2w,2=1 z4=2 |
| | 2 | z2w,3=1 z4=3 |
| | 2 | z2w,2=1 z2w,3=1 z4=2 |
| | 2 | z2w,1=1 z2w,3=1 z4=2 |
| | 2 | z2w,1=1 z2w,2=1 z2w,3=1 z4=1 |
| | 2 | z2w,1=1 z2w,2=1 z4=1 |
| | 2 | z2w,1=1 |
| | 2 | z2w,2=1 |
| | 2 | z4=3 |
| | 2 | z2w,2=1 z2w,3=1 |
| | 2 | z2w,3=1 z4=3 |
| | 2 | z2w,1=1 z2w,2=1 z2w,3=1 z4=3 |
| | 2 | z2w,1=1 z2w,3=1 z4=2 |
| | 2 | z2w,1=1 z2w,3=1 |
| | 2 | z2w,1=1 z2w,2=1 z2w,3=1 z4=3 |
| | 2 | z2w,3=1 z4=3 |
| | 2 | z2w,2=1 z2w,3=1 z4=2 |
| | 2 | z4=3 |
| | 2 | z2w,2=1 z4=2 |
| | 2 | z2w,1=1 z4=2 |
| | 2 | z2w,1=1 z2w,2=1 z4=1 |
| | 2 | z2w,2=1 z4=3 |
| | 2 | z2w,1=1 z4=3 |
| | 2 | z2w,1=1 z2w,2=1 z4=2 |
| | 2 | z2w,1=1 z2w,3=1 z4=3 |
| | 2 | z2w,1=1 z2w,2=1 z2w,3=1 z4=2 |
| | 2 | z2w,3=1 z4=2 |
| | 2 | z2w,2=1 z2w,3=1 z4=1 |
| | 2 | z2w,2=1 z2w,3=1 z4=3 |
| | 2 | z2w,3=1 z4=2 |
| | 2 | z2w,1=1 z2w,2=1 z2w,3=1 z4=2 |
| | 2 | z2w,1=1 z2w,3=1 z4=1 |
| | 2 | z2w,1=1 z2w,2=1 z4=2 |
| | 2 | z2w,1=1 z4=1 |
| | 2 | z2w,2=1 z4=1 |
| 59.4.46 | 154084 | Ag3Sn1 |
| | 2 | z4=3 |
| | 2 | z2w,2=1 z4=2 |
| | 2 | z2w,1=1 z4=2 |
| | 2 | z2w,1=1 z2w,2=1 z4=1 |
| | 2 | z2w,1=1 z2w,3=1 z4=2 |
| | 2 | z2w,1=1 z2w,2=1 z2w,3=1 z4=1 |
| | 2 | z2w,3=1 z4=1 |
| | 2 | z2w,2=1 z2w,3=1 |
| | 2 | z2w,2=1 z2w,3=1 z4=2 |
| | 2 | z2w,3=1 z4=1 |
| | 2 | z2w,1=1 z2w,2=1 z2w,3=1 z4=1 |
| | 2 | z2w,1=1 z2w,3=1 |
| | 2 | z2w,1=1 z2w,2=1 z4=1 |
| | 2 | z2w,1=1 |
| | 2 | z2w,2=1 |
| | 2 | z4=3 |
| | 2 | z2w,1=1 z2w,2=1 z4=2 |
| | 2 | z2w,1=1 z4=1 |
| | 2 | z2w,2=1 z4=1 |
| | 2 | z2w,2=1 z2w,3=1 z4=1 |
| | 2 | z2w,3=1 |
| | 2 | z2w,1=1 z2w,2=1 z2w,3=1 |
| | 2 | z2w,1=1 z2w,3=1 z4=3 |

Continued on next column

**TABLE XIX– continued**

| TQC | ICSD SSG | Stoich. Formula SIs |
|---|---|---|
| | 2 | z2w,1=1 z2w,3=1 z4=1 |
| | 2 | z2w,1=1 z2w,2=1 z2w,3=1 |
| | 2 | z2w,3=1 |
| | 2 | z2w,2=1 z2w,3=1 z4=3 |
| | 2 | z2w,2=1 z4=3 |
| | 2 | z2w,1=1 z4=3 |
| | 2 | z2w,1=1 z2w,2=1 z4=2 |
| | 2 | z2w,1=1 z2w,2=1 z2w,3=1 z4=2 |
| | 2 | z2w,1=1 z2w,3=1 z4=1 |
| | 2 | z2w,2=1 z2w,3=1 z4=1 |
| | 2 | z2w,3=1 |
| | 2 | z2w,2=1 z4=1 |
| | 2 | z2w,1=1 z2w,2=1 |
| | 2 | z2w,1=1 z4=3 |
| | 2 | z2w,1=1 z4=1 |
| | 2 | z2w,1=1 z2w,2=1 |
| | 2 | z2w,2=1 z4=3 |
| | 2 | z2w,3=1 |
| | 2 | z2w,2=1 z2w,3=1 z4=3 |
| | 2 | z2w,1=1 z2w,3=1 z4=3 |
| | 2 | z2w,1=1 z2w,2=1 z2w,3=1 z4=2 |
| | 2 | z2w,3=1 z4=1 |
| | 2 | z2w,2=1 z2w,3=1 |
| | 2 | z2w,1=1 z2w,3=1 |
| | 2 | z2w,1=1 z2w,2=1 z2w,3=1 z4=3 |
| | 2 | z2w,1=1 |
| | 2 | z2w,1=1 z2w,2=1 z4=3 |
| | 2 | z4=3 |
| | 2 | z2w,2=1 z4=2 |
| | 2 | z2w,2=1 |
| | 2 | z4=3 |
| | 2 | z2w,1=1 z2w,2=1 z4=3 |
| | 2 | z2w,1=1 z4=2 |
| | 2 | z2w,1=1 z2w,2=1 z2w,3=1 z4=3 |
| | 2 | z2w,1=1 z2w,3=1 z4=2 |
| | 2 | z2w,2=1 z2w,3=1 z4=2 |
| | 2 | z2w,3=1 z4=1 |
| 63.4.1 | 422877 | Sr3Tl5 |
| | 12 | z2w,1=1 z2w,2=1 |
| | 12 | z4=2 |
| | 12 | z2w,1=1 z2w,2=1 z2w,3=1 z4=1 |
| | 12 | z2w,3=1 z4=3 |
| | 12 | z2w,3=1 z4=1 |
| | 12 | z2w,1=1 z2w,2=1 z2w,3=1 z4=3 |
| | 12 | z4=2 |
| | 12 | z2w,1=1 z2w,2=1 |
| 63.4.2 | 380325 | Sr3Tl5 |
| | 12 | z2w,1=1 z2w,2=1 |
| | 12 | z4=2 |
| | 12 | z2w,1=1 z2w,2=1 z2w,3=1 z4=1 |
| | 12 | z2w,3=1 z4=3 |
| | 12 | z2w,3=1 z4=1 |
| | 12 | z2w,1=1 z2w,2=1 z2w,3=1 z4=3 |
| | 12 | z4=2 |
| | 12 | z2w,1=1 z2w,2=1 |
| 63.4.3 | 416771 | Sn2Y1 |
| | 12 | z2w,1=1 z2w,2=1 |
| | 12 | z4=2 |
| | 12 | z2w,1=1 z2w,2=1 z2w,3=1 z4=1 |
| | 12 | z2w,3=1 z4=3 |

Continued on next column

**TABLE XIX– continued**

| TQC | ICSD SSG | Stoich. Formula SIs |
|---|---|---|
| | 12 | z2w,3=1 z4=1 |
| | 12 | z2w,1=1 z2w,2=1 z2w,3=1 z4=3 |
| | 12 | z4=2 |
| | 12 | z2w,1=1 z2w,2=1 |
| 63.4.4 | 2239 | S3Sc1U1 |
| | 12 | z4=3 |
| | 12 | z2w,1=1 z2w,2=1 z4=1 |
| | 12 | z2w,3=1 |
| | 12 | z2w,1=1 z2w,2=1 z2w,3=1 z4=2 |
| | 12 | z2w,1=1 z2w,2=1 z2w,3=1 |
| | 12 | z2w,3=1 z4=2 |
| | 12 | z2w,1=1 z2w,2=1 z4=1 |
| | 12 | z4=3 |
| 63.4.5 | 63105 | Rh5Zr3 |
| | 12 | z2w,1=1 z2w,2=1 z4=2 |
| | 12 | z2w,3=1 z4=1 |
| | 12 | z2w,1=1 z2w,2=1 z2w,3=1 z4=3 |
| | 12 | z2w,1=1 z2w,2=1 z2w,3=1 z4=1 |
| | 12 | z2w,3=1 z4=3 |
| | 12 | z2w,1=1 z2w,2=1 z4=2 |
| 63.4.6 | 62932 | Rh5Zr3 |
| | 12 | z2w,1=1 z2w,2=1 z4=2 |
| | 12 | z2w,3=1 z4=1 |
| | 12 | z2w,1=1 z2w,2=1 z2w,3=1 z4=3 |
| | 12 | z2w,1=1 z2w,2=1 z2w,3=1 z4=1 |
| | 12 | z2w,3=1 z4=3 |
| | 12 | z2w,1=1 z2w,2=1 z4=2 |
| 63.4.7 | 14382 | Rh3Te2 |
| | 12 | z4=2 |
| | 12 | z2w,1=1 z2w,2=1 |
| | 12 | z2w,3=1 z4=3 |
| | 12 | z2w,1=1 z2w,2=1 z2w,3=1 z4=1 |
| | 12 | z2w,1=1 z2w,2=1 z2w,3=1 z4=3 |
| | 12 | z2w,3=1 z4=1 |
| | 12 | z2w,1=1 z2w,2=1 |
| | 12 | z4=2 |
| 63.4.8 | 420355 | Pt8Sc4Si4 |
| | 12 | z2w,1=1 z2w,2=1 |
| | 12 | z4=2 |
| | 12 | z2w,1=1 z2w,2=1 z2w,3=1 z4=1 |
| | 12 | z2w,3=1 z4=3 |
| | 12 | z2w,3=1 z4=1 |
| | 12 | z2w,1=1 z2w,2=1 z2w,3=1 z4=3 |
| | 12 | z4=2 |
| | 12 | z2w,1=1 z2w,2=1 |
| 63.4.9 | 71292 | Pt1Ta1Te5 |
| | 12 | z2w,1=1 z2w,2=1 z4=2 |
| | 12 | z2w,3=1 z4=1 |
| | 12 | z2w,1=1 z2w,2=1 z2w,3=1 z4=3 |
| | 12 | z2w,1=1 z2w,2=1 z2w,3=1 z4=1 |
| | 12 | z2w,3=1 z4=3 |
| | 12 | z2w,1=1 z2w,2=1 z4=2 |
| 63.4.10 | 424600 | Pt1Sc1Zn1 |
| | 12 | z2w,1=1 z2w,2=1 z4=2 |
| | 12 | z2w,3=1 z4=1 |
| | 12 | z2w,1=1 z2w,2=1 z2w,3=1 z4=3 |
| | 12 | z2w,1=1 z2w,2=1 z2w,3=1 z4=1 |
| | 12 | z2w,3=1 z4=3 |
| | 12 | z2w,1=1 z2w,2=1 z4=2 |

Continued on next column

**TABLE XIX– continued**

| TQC | ICSD SSG | Stoich. Formula SIs |
|---|---|---|
| 63.4.11 | 412465 | Pr1Se1Te2 |
| | 12 | z4=3 |
| | 12 | z2w,1=1 z2w,2=1 z4=1 |
| | 12 | z2w,3=1 |
| | 12 | z2w,1=1 z2w,2=1 z2w,3=1 z4=2 |
| | 12 | z2w,1=1 z2w,2=1 z2w,3=1 |
| | 12 | z2w,3=1 z4=2 |
| | 12 | z2w,1=1 z2w,2=1 z4=1 |
| | 12 | z4=3 |
| 63.4.12 | 35086 | Pd2Si3Y3 |
| | 12 | z4=2 |
| | 12 | z2w,1=1 z2w,2=1 |
| | 12 | z2w,3=1 z4=3 |
| | 12 | z2w,1=1 z2w,2=1 z2w,3=1 z4=1 |
| | 12 | z2w,1=1 z2w,2=1 z2w,3=1 z4=3 |
| | 12 | z2w,3=1 z4=1 |
| | 12 | z2w,1=1 z2w,2=1 |
| | 12 | z4=2 |
| 63.4.13 | 424601 | Pd1Sc1Zn1 |
| | 12 | z2w,1=1 z2w,2=1 z4=2 |
| | 12 | z2w,3=1 z4=1 |
| | 12 | z2w,1=1 z2w,2=1 z2w,3=1 z4=3 |
| | 12 | z2w,1=1 z2w,2=1 z2w,3=1 z4=1 |
| | 12 | z2w,3=1 z4=3 |
| | 12 | z2w,1=1 z2w,2=1 z4=2 |
| 63.4.14 | 36630 | P5Pd5Sr4 |
| | 12 | z2w,1=1 z2w,2=1 z4=2 |
| | 12 | z2w,3=1 z4=1 |
| | 12 | z2w,1=1 z2w,2=1 z2w,3=1 z4=3 |
| | 12 | z2w,1=1 z2w,2=1 z2w,3=1 z4=1 |
| | 12 | z2w,3=1 z4=3 |
| | 12 | z2w,1=1 z2w,2=1 z4=2 |
| 63.4.15 | 192737 | O7Ru1Tb3 |
| | 12 | z2w,1=1 z2w,2=1 z4=1 |
| | 12 | z4=3 |
| | 12 | z2w,1=1 z2w,2=1 z2w,3=1 z4=2 |
| | 12 | z2w,3=1 |
| | 12 | z2w,3=1 z4=2 |
| | 12 | z2w,1=1 z2w,2=1 z2w,3=1 |
| | 12 | z4=3 |
| | 12 | z2w,1=1 z2w,2=1 z4=1 |
| 63.4.16 | 187410 | O7Os1Tb3 |
| | 12 | z4=1 |
| | 12 | z2w,1=1 z2w,2=1 z4=3 |
| | 12 | z2w,3=1 z4=2 |
| | 12 | z2w,1=1 z2w,2=1 z2w,3=1 |
| | 12 | z2w,1=1 z2w,2=1 z2w,3=1 z4=2 |
| | 12 | z2w,3=1 |
| | 12 | z2w,1=1 z2w,2=1 z4=3 |
| | 12 | z4=1 |
| 63.4.17 | 36520 | O4P1Ti1 |
| | 12 | z2w,1=1 z2w,2=1 z4=2 |
| | 12 | z2w,1=1 z2w,2=1 z2w,3=1 z4=3 |
| | 12 | z2w,3=1 z4=1 |
| | 12 | z2w,3=1 z4=3 |
| | 12 | z2w,1=1 z2w,2=1 z2w,3=1 z4=1 |
| | 12 | z2w,1=1 z2w,2=1 z4=2 |
| 63.4.18 | 83361 | O14Si4Sr2V2 |
| | 12 | z2w,1=1 z2w,2=1 z4=2 |

Continued on next column

TABLE XIX– continued

| TQC | ICSD<br>SSG | Stoich. Formula<br>SIs |
|---|---|---|
| | 12 | z2w,1=1 z2w,2=1 z2w,3=1 z4=3 |
| | 12 | z2w,3=1 z4=1 |
| | 12 | z2w,3=1 z4=3 |
| | 12 | z2w,1=1 z2w,2=1 z2w,3=1 z4=1 |
| | 12 | z2w,1=1 z2w,2=1 z4=2 |
| 63.4.19 | 33914 | Ni5P3Sr1 |
| | 12 | z4=2 |
| | 12 | z2w,1=1 z2w,2=1 |
| | 12 | z2w,3=1 z4=3 |
| | 12 | z2w,1=1 z2w,2=1 z2w,3=1 z4=1 |
| | 12 | z2w,1=1 z2w,2=1 z2w,3=1 z4=3 |
| | 12 | z2w,3=1 z4=1 |
| | 12 | z2w,1=1 z2w,2=1 |
| | 12 | z4=2 |
| 63.4.20 | 65760 | Ni2Sc3Si3 |
| | 12 | z2w,1=1 z2w,2=1 |
| | 12 | z4=2 |
| | 12 | z2w,1=1 z2w,2=1 z2w,3=1 z4=1 |
| | 12 | z2w,3=1 z4=3 |
| | 12 | z2w,3=1 z4=1 |
| | 12 | z2w,1=1 z2w,2=1 z2w,3=1 z4=3 |
| | 12 | z4=2 |
| | 12 | z2w,1=1 z2w,2=1 |
| 63.4.21 | 73314 | Ni1Ta1Te5 |
| | 12 | z2w,1=1 z2w,2=1 z4=2 |
| | 12 | z2w,3=1 z4=1 |
| | 12 | z2w,1=1 z2w,2=1 z2w,3=1 z4=3 |
| | 12 | z2w,1=1 z2w,2=1 z2w,3=1 z4=1 |
| | 12 | z2w,3=1 z4=3 |
| | 12 | z2w,1=1 z2w,2=1 z4=2 |
| 63.4.22 | 71694 | Ni1Pr1Si2 |
| | 12 | z2w,1=1 z2w,2=1 z4=2 |
| | 12 | z2w,3=1 z4=1 |
| | 12 | z2w,1=1 z2w,2=1 z2w,3=1 z4=3 |
| | 12 | z2w,1=1 z2w,2=1 z2w,3=1 z4=1 |
| | 12 | z2w,3=1 z4=3 |
| | 12 | z2w,1=1 z2w,2=1 z4=2 |
| 63.4.23 | 99823 | Ni1Pr1 |
| | 12 | z4=1 |
| | 12 | z2w,1=1 z2w,2=1 z4=3 |
| | 12 | z2w,3=1 z4=2 |
| | 12 | z2w,1=1 z2w,2=1 z2w,3=1 |
| | 12 | z2w,1=1 z2w,2=1 z2w,3=1 z4=2 |
| | 12 | z2w,3=1 |
| | 12 | z2w,1=1 z2w,2=1 z4=3 |
| | 12 | z4=1 |
| 63.4.24 | 73316 | Nb1Ni1Te5 |
| | 12 | z2w,1=1 z2w,2=1 z4=2 |
| | 12 | z2w,3=1 z4=1 |
| | 12 | z2w,1=1 z2w,2=1 z2w,3=1 z4=3 |
| | 12 | z2w,1=1 z2w,2=1 z2w,3=1 z4=1 |
| | 12 | z2w,3=1 z4=3 |
| | 12 | z2w,1=1 z2w,2=1 z4=2 |
| 63.4.25 | 105166 | Na9Sn4 |
| | 12 | z2w,1=1 z2w,2=1 z4=2 |
| | 12 | z2w,3=1 z4=1 |
| | 12 | z2w,1=1 z2w,2=1 z2w,3=1 z4=3 |
| | 12 | z2w,1=1 z2w,2=1 z2w,3=1 z4=1 |
| | 12 | z2w,3=1 z4=3 |

Continued on next column

| TQC | ICSD<br>SSG | Stoich. Formula<br>SIs |
|---|---|---|
| | 12 | z2w,1=1 z2w,2=1 z4=2 |
| 63.4.26 | 237248 | Na1Si6 |
| | 12 | z4=1 |
| | 12 | z2w,1=1 z2w,2=1 z4=3 |
| | 12 | z2w,3=1 z4=2 |
| | 12 | z2w,1=1 z2w,2=1 z2w,3=1 |
| | 12 | z2w,1=1 z2w,2=1 z2w,3=1 z4=2 |
| | 12 | z2w,3=1 |
| | 12 | z2w,1=1 z2w,2=1 z4=3 |
| | 12 | z4=1 |
| 63.4.27 | 84911 | N5Na1Ta3 |
| | 12 | z4=3 |
| | 12 | z2w,1=1 z2w,2=1 z4=1 |
| | 12 | z2w,3=1 |
| | 12 | z2w,1=1 z2w,2=1 z2w,3=1 z4=2 |
| | 12 | z2w,1=1 z2w,2=1 z2w,3=1 |
| | 12 | z2w,3=1 z4=2 |
| | 12 | z2w,1=1 z2w,2=1 z4=1 |
| | 12 | z4=3 |
| 63.4.28 | 99582 | Mn1Se3Th1 |
| | 12 | z2w,1=1 z2w,2=1 z4=2 |
| | 12 | z2w,1=1 z2w,2=1 z2w,3=1 z4=3 |
| | 12 | z2w,3=1 z4=1 |
| | 12 | z2w,3=1 z4=3 |
| | 12 | z2w,1=1 z2w,2=1 z2w,3=1 z4=1 |
| | 12 | z2w,1=1 z2w,2=1 z4=2 |
| 63.4.29 | 166874 | Mg1Ni1Tb1 |
| | 12 | z2w,1=1 z2w,2=1 z4=1 |
| | 12 | z4=3 |
| | 12 | z2w,1=1 z2w,2=1 z2w,3=1 z4=2 |
| | 12 | z2w,3=1 |
| | 12 | z2w,3=1 z4=2 |
| | 12 | z2w,1=1 z2w,2=1 z2w,3=1 |
| | 12 | z4=3 |
| | 12 | z2w,1=1 z2w,2=1 z4=1 |
| 63.4.30 | 426769 | Lu1Sn2 |
| | 12 | z4=2 |
| | 12 | z2w,1=1 z2w,2=1 |
| | 12 | z2w,3=1 z4=3 |
| | 12 | z2w,1=1 z2w,2=1 z2w,3=1 z4=1 |
| | 12 | z2w,1=1 z2w,2=1 z2w,3=1 z4=3 |
| | 12 | z2w,3=1 z4=1 |
| | 12 | z2w,1=1 z2w,2=1 |
| | 12 | z4=2 |
| 63.4.31 | 249621 | Lu1Nd1Se3 |
| | 12 | z4=3 |
| | 12 | z2w,1=1 z2w,2=1 z4=1 |
| | 12 | z2w,3=1 |
| | 12 | z2w,1=1 z2w,2=1 z2w,3=1 z4=2 |
| | 12 | z2w,1=1 z2w,2=1 z2w,3=1 |
| | 12 | z2w,3=1 z4=2 |
| | 12 | z2w,1=1 z2w,2=1 z4=1 |
| | 12 | z4=3 |
| 63.4.32 | 50982 | Li1O5V2 |
| | 12 | z2w,1=1 z2w,2=1 z4=1 |
| | 12 | z4=3 |
| | 12 | z2w,1=1 z2w,2=1 z2w,3=1 z4=2 |
| | 12 | z2w,3=1 |
| | 12 | z2w,3=1 z4=2 |

Continued on next column

**TABLE XIX– continued**

| TQC | ICSD SSG | Stoich. Formula SIs |
|---|---|---|
| | 12 | z2w,1=1 z2w,2=1 z2w,3=1 |
| | 12 | z4=3 |
| | 12 | z2w,1=1 z2w,2=1 z4=1 |
| 63.4.33 | 50980 | Li1O5V2 |
| | 12 | z2w,1=1 z2w,2=1 z4=1 |
| | 12 | z4=3 |
| | 12 | z2w,1=1 z2w,2=1 z2w,3=1 z4=2 |
| | 12 | z2w,3=1 |
| | 12 | z2w,3=1 z4=2 |
| | 12 | z2w,1=1 z2w,2=1 z2w,3=1 |
| | 12 | z4=3 |
| | 12 | z2w,1=1 z2w,2=1 z4=1 |
| 63.4.34 | 201974 | Li1O5V2 |
| | 12 | z4=3 |
| | 12 | z2w,1=1 z2w,2=1 z4=1 |
| | 12 | z2w,3=1 |
| | 12 | z2w,1=1 z2w,2=1 z2w,3=1 z4=2 |
| | 12 | z2w,1=1 z2w,2=1 z2w,3=1 |
| | 12 | z2w,3=1 z4=2 |
| | 12 | z2w,1=1 z2w,2=1 z4=1 |
| | 12 | z4=3 |
| 63.4.35 | 281581 | Li1Mn1O4V1 |
| | 12 | z2w,1=1 z2w,2=1 z4=2 |
| | 12 | z2w,1=1 z2w,2=1 z2w,3=1 z4=3 |
| | 12 | z2w,3=1 z4=1 |
| | 12 | z2w,3=1 z4=3 |
| | 12 | z2w,1=1 z2w,2=1 z2w,3=1 z4=1 |
| | 12 | z2w,1=1 z2w,2=1 z4=2 |
| 63.4.36 | 165185 | La3Sn5 |
| | 12 | z2w,1=1 z2w,2=1 z4=1 |
| | 12 | z4=3 |
| | 12 | z2w,1=1 z2w,2=1 z2w,3=1 z4=2 |
| | 12 | z2w,3=1 |
| | 12 | z2w,3=1 z4=2 |
| | 12 | z2w,1=1 z2w,2=1 z2w,3=1 |
| | 12 | z4=3 |
| | 12 | z2w,1=1 z2w,2=1 z4=1 |
| 63.4.37 | 423418 | La3Sn4 |
| | 12 | z2w,1=1 z2w,2=1 z4=2 |
| | 12 | z2w,3=1 z4=1 |
| | 12 | z2w,1=1 z2w,2=1 z2w,3=1 z4=3 |
| | 12 | z2w,1=1 z2w,2=1 z2w,3=1 z4=1 |
| | 12 | z2w,3=1 z4=3 |
| | 12 | z2w,1=1 z2w,2=1 z4=2 |
| 63.4.38 | 416718 | La3O7Ru1 |
| | 12 | z2w,1=1 z2w,2=1 z4=2 |
| | 12 | z2w,3=1 z4=1 |
| | 12 | z2w,1=1 z2w,2=1 z2w,3=1 z4=3 |
| | 12 | z2w,1=1 z2w,2=1 z2w,3=1 z4=1 |
| | 12 | z2w,3=1 z4=3 |
| | 12 | z2w,1=1 z2w,2=1 z4=2 |
| 63.4.39 | 416818 | La3Ni3Zn1 |
| | 12 | z2w,1=1 z2w,2=1 z4=3 |
| | 12 | z4=1 |
| | 12 | z2w,1=1 z2w,2=1 z2w,3=1 |
| | 12 | z2w,3=1 z4=2 |
| | 12 | z2w,3=1 |
| | 12 | z2w,1=1 z2w,2=1 z2w,3=1 z4=2 |
| | 12 | z4=1 |

Continued on next column

**TABLE XIX– continued**

| TQC | ICSD SSG | Stoich. Formula SIs |
|---|---|---|
| | 12 | z2w,1=1 z2w,2=1 z4=3 |
| 63.4.40 | 91142 | La3Ni2O7 |
| | 12 | z2w,1=1 z2w,2=1 z4=2 |
| | 12 | z2w,1=1 z2w,2=1 z2w,3=1 z4=3 |
| | 12 | z2w,3=1 z4=1 |
| | 12 | z2w,3=1 z4=3 |
| | 12 | z2w,1=1 z2w,2=1 z2w,3=1 z4=1 |
| | 12 | z2w,1=1 z2w,2=1 z4=2 |
| 63.4.41 | 91141 | La3Ni2O7 |
| | 12 | z2w,1=1 z2w,2=1 z4=2 |
| | 12 | z2w,3=1 z4=1 |
| | 12 | z2w,1=1 z2w,2=1 z2w,3=1 z4=3 |
| | 12 | z2w,1=1 z2w,2=1 z2w,3=1 z4=1 |
| | 12 | z2w,3=1 z4=3 |
| | 12 | z2w,1=1 z2w,2=1 z4=2 |
| 63.4.42 | 262792 | La1Zn4 |
| | 12 | z4=1 |
| | 12 | z2w,1=1 z2w,2=1 z4=3 |
| | 12 | z2w,3=1 z4=2 |
| | 12 | z2w,1=1 z2w,2=1 z2w,3=1 |
| | 12 | z2w,1=1 z2w,2=1 z2w,3=1 z4=2 |
| | 12 | z2w,3=1 |
| | 12 | z2w,1=1 z2w,2=1 z4=3 |
| | 12 | z4=1 |
| 63.4.43 | 423421 | La1Sn1 |
| | 12 | z2w,1=1 z2w,2=1 z4=2 |
| | 12 | z2w,1=1 z2w,2=1 z2w,3=1 z4=3 |
| | 12 | z2w,3=1 z4=1 |
| | 12 | z2w,3=1 z4=3 |
| | 12 | z2w,1=1 z2w,2=1 z2w,3=1 z4=1 |
| | 12 | z2w,1=1 z2w,2=1 z4=2 |
| 63.4.44 | 413173 | La1Se1Te2 |
| | 12 | z2w,1=1 z2w,2=1 z4=3 |
| | 12 | z4=1 |
| | 12 | z2w,1=1 z2w,2=1 z2w,3=1 |
| | 12 | z2w,3=1 z4=2 |
| | 12 | z2w,3=1 |
| | 12 | z2w,1=1 z2w,2=1 z2w,3=1 z4=2 |
| | 12 | z4=1 |
| | 12 | z2w,1=1 z2w,2=1 z4=3 |
| 63.4.45 | 413171 | La1Se1Te2 |
| | 12 | z2w,1=1 z2w,2=1 z4=3 |
| | 12 | z4=1 |
| | 12 | z2w,1=1 z2w,2=1 z2w,3=1 |
| | 12 | z2w,3=1 z4=2 |
| | 12 | z2w,3=1 |
| | 12 | z2w,1=1 z2w,2=1 z2w,3=1 z4=2 |
| | 12 | z4=1 |
| | 12 | z2w,1=1 z2w,2=1 z4=3 |
| 63.4.46 | 158253 | La1Mg2Pd1 |
| | 12 | z4=3 |
| | 12 | z2w,1=1 z2w,2=1 z4=1 |
| | 12 | z2w,3=1 |
| | 12 | z2w,1=1 z2w,2=1 z2w,3=1 z4=2 |
| | 12 | z2w,1=1 z2w,2=1 z2w,3=1 |
| | 12 | z2w,3=1 z4=2 |
| | 12 | z2w,1=1 z2w,2=1 z4=1 |
| | 12 | z4=3 |
| 63.4.47 | 96152 | La1Mg2Ni1 |

Continued on next column

**TABLE XIX– continued**

| TQC | ICSD SSG | Stoich. Formula SIs |
|---|---|---|
| | 12 | z4=3 |
| | 12 | z2w,1=1 z2w,2=1 z4=1 |
| | 12 | z2w,3=1 |
| | 12 | z2w,1=1 z2w,2=1 z2w,3=1 z4=2 |
| | 12 | z2w,1=1 z2w,2=1 z2w,3=1 |
| | 12 | z2w,3=1 z4=2 |
| | 12 | z2w,1=1 z2w,2=1 z4=1 |
| | 12 | z4=3 |
| 63.4.48 | 64625 | K4P3 |
| | 12 | z2w,1=1 z2w,2=1 z4=1 |
| | 12 | z4=3 |
| | 12 | z2w,1=1 z2w,2=1 z2w,3=1 z4=2 |
| | 12 | z2w,3=1 |
| | 12 | z2w,3=1 z4=2 |
| | 12 | z2w,1=1 z2w,2=1 z2w,3=1 |
| | 12 | z4=3 |
| | 12 | z2w,1=1 z2w,2=1 z4=1 |
| 63.4.49 | 245610 | K3O4Pd2 |
| | 12 | z2w,1=1 z2w,2=1 z4=2 |
| | 12 | z2w,1=1 z2w,2=1 z2w,3=1 z4=3 |
| | 12 | z2w,3=1 z4=1 |
| | 12 | z2w,3=1 z4=3 |
| | 12 | z2w,1=1 z2w,2=1 z2w,3=1 z4=1 |
| | 12 | z2w,1=1 z2w,2=1 z4=2 |
| 63.4.50 | 14157 | K3Ni2O4 |
| | 12 | z2w,1=1 z2w,2=1 z4=2 |
| | 12 | z2w,1=1 z2w,2=1 z2w,3=1 z4=3 |
| | 12 | z2w,3=1 z4=1 |
| | 12 | z2w,3=1 z4=3 |
| | 12 | z2w,1=1 z2w,2=1 z2w,3=1 z4=1 |
| | 12 | z2w,1=1 z2w,2=1 z4=2 |
| 63.4.51 | 165188 | In7K2Sr1 |
| | 12 | z2w,1=1 z2w,2=1 z4=1 |
| | 12 | z4=3 |
| | 12 | z2w,1=1 z2w,2=1 z2w,3=1 z4=2 |
| | 12 | z2w,3=1 |
| | 12 | z2w,3=1 z4=2 |
| | 12 | z2w,1=1 z2w,2=1 z2w,3=1 |
| | 12 | z4=3 |
| | 12 | z2w,1=1 z2w,2=1 z4=1 |
| 63.4.52 | 411746 | In4La1Ni1 |
| | 12 | z2w,1=1 z2w,2=1 z4=1 |
| | 12 | z4=3 |
| | 12 | z2w,1=1 z2w,2=1 z2w,3=1 z4=2 |
| | 12 | z2w,3=1 |
| | 12 | z2w,3=1 z4=2 |
| | 12 | z2w,1=1 z2w,2=1 z2w,3=1 |
| | 12 | z4=3 |
| | 12 | z2w,1=1 z2w,2=1 z4=1 |
| 63.4.53 | 152048 | In2Rh1Yb1 |
| | 12 | z4=2 |
| | 12 | z2w,1=1 z2w,2=1 |
| | 12 | z2w,3=1 z4=3 |
| | 12 | z2w,1=1 z2w,2=1 z2w,3=1 z4=1 |
| | 12 | z2w,1=1 z2w,2=1 z2w,3=1 z4=3 |
| | 12 | z2w,3=1 z4=1 |
| | 12 | z2w,1=1 z2w,2=1 |
| | 12 | z4=2 |
| 63.4.54 | 413459 | In2Ni1Y1 |

Continued on next column

**TABLE XIX– continued**

| TQC | ICSD SSG | Stoich. Formula SIs |
|---|---|---|
| | 12 | z2w,1=1 z2w,2=1 |
| | 12 | z4=2 |
| | 12 | z2w,1=1 z2w,2=1 z2w,3=1 z4=1 |
| | 12 | z2w,3=1 z4=3 |
| | 12 | z2w,3=1 z4=1 |
| | 12 | z2w,1=1 z2w,2=1 z2w,3=1 z4=3 |
| | 12 | z4=2 |
| | 12 | z2w,1=1 z2w,2=1 |
| 63.4.55 | 413448 | In2La1Pd1 |
| | 12 | z2w,1=1 z2w,2=1 |
| | 12 | z4=2 |
| | 12 | z2w,1=1 z2w,2=1 z2w,3=1 z4=1 |
| | 12 | z2w,3=1 z4=3 |
| | 12 | z2w,3=1 z4=1 |
| | 12 | z2w,1=1 z2w,2=1 z2w,3=1 z4=3 |
| | 12 | z4=2 |
| | 12 | z2w,1=1 z2w,2=1 |
| 63.4.56 | 410401 | In2Ir1Sr1 |
| | 12 | z4=2 |
| | 12 | z2w,1=1 z2w,2=1 |
| | 12 | z2w,3=1 z4=3 |
| | 12 | z2w,1=1 z2w,2=1 z2w,3=1 z4=1 |
| | 12 | z2w,1=1 z2w,2=1 z2w,3=1 z4=3 |
| | 12 | z2w,3=1 z4=1 |
| | 12 | z2w,1=1 z2w,2=1 |
| | 12 | z4=2 |
| 63.4.57 | 201170 | I3U1 |
| | 12 | z2w,1=1 z2w,2=1 z4=2 |
| | 12 | z2w,1=1 z2w,2=1 z2w,3=1 z4=3 |
| | 12 | z2w,3=1 z4=1 |
| | 12 | z2w,3=1 z4=3 |
| | 12 | z2w,1=1 z2w,2=1 z2w,3=1 z4=1 |
| | 12 | z2w,1=1 z2w,2=1 z4=2 |
| 63.4.58 | 72343 | I3K1Tm1 |
| | 12 | z2w,1=1 z2w,2=1 z4=2 |
| | 12 | z2w,3=1 z4=1 |
| | 12 | z2w,1=1 z2w,2=1 z2w,3=1 z4=3 |
| | 12 | z2w,1=1 z2w,2=1 z2w,3=1 z4=1 |
| | 12 | z2w,3=1 z4=3 |
| | 12 | z2w,1=1 z2w,2=1 z4=2 |
| 63.4.59 | 154700 | Ho1Si1 |
| | 12 | z4=1 |
| | 12 | z2w,1=1 z2w,2=1 z4=3 |
| | 12 | z2w,3=1 z4=2 |
| | 12 | z2w,1=1 z2w,2=1 z2w,3=1 |
| | 12 | z2w,1=1 z2w,2=1 z2w,3=1 z4=2 |
| | 12 | z2w,3=1 |
| | 12 | z2w,1=1 z2w,2=1 z4=3 |
| | 12 | z4=1 |
| 63.4.60 | 83917 | Ho1Ni1Si2 |
| | 12 | z2w,1=1 z2w,2=1 |
| | 12 | z4=2 |
| | 12 | z2w,1=1 z2w,2=1 z2w,3=1 z4=1 |
| | 12 | z2w,3=1 z4=3 |
| | 12 | z2w,3=1 z4=1 |
| | 12 | z2w,1=1 z2w,2=1 z2w,3=1 z4=3 |
| | 12 | z4=2 |
| | 12 | z2w,1=1 z2w,2=1 |
| 63.4.61 | 83916 | Ho1Ni1Si2 |

Continued on next column

TABLE XIX– continued

| TQC | ICSD SSG | Stoich. Formula SIs |
|---|---|---|
| | 12 | z2w,1=1 z2w,2=1 |
| | 12 | z4=2 |
| | 12 | z2w,1=1 z2w,2=1 z2w,3=1 z4=1 |
| | 12 | z2w,3=1 z4=3 |
| | 12 | z2w,3=1 z4=1 |
| | 12 | z2w,1=1 z2w,2=1 z2w,3=1 z4=3 |
| | 12 | z4=2 |
| | 12 | z2w,1=1 z2w,2=1 |
| 63.4.62 | 107591 | Hf1Sb1 |
| | 12 | z4=3 |
| | 12 | z2w,1=1 z2w,2=1 z4=1 |
| | 12 | z2w,3=1 |
| | 12 | z2w,1=1 z2w,2=1 z2w,3=1 z4=2 |
| | 12 | z2w,1=1 z2w,2=1 z2w,3=1 |
| | 12 | z2w,3=1 z4=2 |
| | 12 | z2w,1=1 z2w,2=1 z4=1 |
| | 12 | z4=3 |
| 63.4.63 | 180189 | H6Cl2Cu1N2 |
| | 12 | z4=1 |
| | 12 | z2w,1=1 z2w,2=1 z4=3 |
| | 12 | z2w,3=1 z4=2 |
| | 12 | z2w,1=1 z2w,2=1 z2w,3=1 |
| | 12 | z2w,1=1 z2w,2=1 z2w,3=1 z4=2 |
| | 12 | z2w,3=1 |
| | 12 | z2w,1=1 z2w,2=1 z4=3 |
| | 12 | z4=1 |
| 63.4.64 | 23318 | H2Cl5Cs2Fe1O1 |
| | 12 | z4=3 |
| | 12 | z2w,1=1 z2w,2=1 z4=1 |
| | 12 | z2w,3=1 |
| | 12 | z2w,1=1 z2w,2=1 z2w,3=1 z4=2 |
| | 12 | z2w,1=1 z2w,2=1 z2w,3=1 |
| | 12 | z2w,3=1 z4=2 |
| | 12 | z2w,1=1 z2w,2=1 z4=1 |
| | 12 | z4=3 |
| 63.4.65 | 409766 | H2Cl5Cr1Cs2O1 |
| | 12 | z4=3 |
| | 12 | z2w,1=1 z2w,2=1 z4=1 |
| | 12 | z2w,3=1 |
| | 12 | z2w,1=1 z2w,2=1 z2w,3=1 z4=2 |
| | 12 | z2w,1=1 z2w,2=1 z2w,3=1 |
| | 12 | z2w,3=1 z4=2 |
| | 12 | z2w,1=1 z2w,2=1 z4=1 |
| | 12 | z4=3 |
| 63.4.66 | 88105 | Ge4Y3 |
| | 12 | z2w,1=1 z2w,2=1 z4=2 |
| | 12 | z2w,3=1 z4=1 |
| | 12 | z2w,1=1 z2w,2=1 z2w,3=1 z4=3 |
| | 12 | z2w,1=1 z2w,2=1 z2w,3=1 z4=1 |
| | 12 | z2w,3=1 z4=3 |
| | 12 | z2w,1=1 z2w,2=1 z4=2 |
| 63.4.67 | 88075 | Ge3Y1 |
| | 12 | z2w,1=1 z2w,2=1 z4=2 |
| | 12 | z2w,1=1 z2w,2=1 z2w,3=1 z4=3 |
| | 12 | z2w,3=1 z4=1 |
| | 12 | z2w,3=1 z4=3 |
| | 12 | z2w,1=1 z2w,2=1 z2w,3=1 z4=1 |
| | 12 | z2w,1=1 z2w,2=1 z4=2 |
| 63.4.68 | 169059 | Ge3Tm1 |

Continued on next column

TABLE XIX– continued

| TQC | ICSD SSG | Stoich. Formula SIs |
|---|---|---|
| | 12 | z2w,1=1 z2w,2=1 z4=2 |
| | 12 | z2w,1=1 z2w,2=1 z2w,3=1 z4=3 |
| | 12 | z2w,3=1 z4=1 |
| | 12 | z2w,3=1 z4=3 |
| | 12 | z2w,1=1 z2w,2=1 z2w,3=1 z4=1 |
| | 12 | z2w,1=1 z2w,2=1 z4=2 |
| 63.4.69 | 425769 | Ge3Ru2Y3 |
| | 12 | z4=1 |
| | 12 | z2w,1=1 z2w,2=1 z4=3 |
| | 12 | z2w,3=1 z4=2 |
| | 12 | z2w,1=1 z2w,2=1 z2w,3=1 |
| | 12 | z2w,1=1 z2w,2=1 z2w,3=1 z4=2 |
| | 12 | z2w,3=1 |
| | 12 | z2w,1=1 z2w,2=1 z4=3 |
| | 12 | z4=1 |
| 63.4.70 | 425775 | Ge3Ru2Tm3 |
| | 12 | z2w,1=1 z2w,2=1 z4=1 |
| | 12 | z4=3 |
| | 12 | z2w,1=1 z2w,2=1 z2w,3=1 z4=2 |
| | 12 | z2w,3=1 |
| | 12 | z2w,3=1 z4=2 |
| | 12 | z2w,1=1 z2w,2=1 z2w,3=1 |
| | 12 | z4=3 |
| | 12 | z2w,1=1 z2w,2=1 z4=1 |
| 63.4.71 | 88246 | Ge3Lu3Ni2 |
| | 12 | z4=2 |
| | 12 | z2w,1=1 z2w,2=1 |
| | 12 | z2w,3=1 z4=3 |
| | 12 | z2w,1=1 z2w,2=1 z2w,3=1 z4=1 |
| | 12 | z2w,1=1 z2w,2=1 z2w,3=1 z4=3 |
| | 12 | z2w,3=1 z4=1 |
| | 12 | z2w,1=1 z2w,2=1 |
| | 12 | z4=2 |
| 63.4.72 | 412709 | Ge3Ir2Lu3 |
| | 12 | z2w,1=1 z2w,2=1 z4=3 |
| | 12 | z4=1 |
| | 12 | z2w,1=1 z2w,2=1 z2w,3=1 |
| | 12 | z2w,3=1 z4=2 |
| | 12 | z2w,3=1 |
| | 12 | z2w,1=1 z2w,2=1 z2w,3=1 z4=2 |
| | 12 | z4=1 |
| | 12 | z2w,1=1 z2w,2=1 z4=3 |
| 63.4.73 | 57071 | Ge3Ho3Ru2 |
| | 12 | z4=3 |
| | 12 | z2w,1=1 z2w,2=1 z4=1 |
| | 12 | z2w,3=1 |
| | 12 | z2w,1=1 z2w,2=1 z2w,3=1 z4=2 |
| | 12 | z2w,1=1 z2w,2=1 z2w,3=1 |
| | 12 | z2w,3=1 z4=2 |
| | 12 | z2w,1=1 z2w,2=1 z4=1 |
| | 12 | z4=3 |
| 63.4.74 | 425773 | Ge3Ho3Ru2 |
| | 12 | z2w,1=1 z2w,2=1 z4=1 |
| | 12 | z4=3 |
| | 12 | z2w,1=1 z2w,2=1 z2w,3=1 z4=2 |
| | 12 | z2w,3=1 |
| | 12 | z2w,3=1 z4=2 |
| | 12 | z2w,1=1 z2w,2=1 z2w,3=1 |
| | 12 | z4=3 |

Continued on next column

**TABLE XIX– continued**

| TQC | ICSD<br>SSG | Stoich. Formula<br>SIs |
|---|---|---|
| | 12 | z2w,1=1 z2w,2=1 z4=1 |
| 63.4.75 | 82875 | Ge3Ho3Ir2 |
| | 12 | z2w,1=1 z2w,2=1 z4=2 |
| | 12 | z2w,1=1 z2w,2=1 z2w,3=1 z4=3 |
| | 12 | z2w,3=1 z4=1 |
| | 12 | z2w,3=1 z4=3 |
| | 12 | z2w,1=1 z2w,2=1 z2w,3=1 z4=1 |
| | 12 | z2w,1=1 z2w,2=1 z4=2 |
| 63.4.76 | 159585 | Ge2Tm1 |
| | 12 | z2w,1=1 z2w,2=1 |
| | 12 | z4=2 |
| | 12 | z2w,1=1 z2w,2=1 z2w,3=1 z4=1 |
| | 12 | z2w,3=1 z4=3 |
| | 12 | z2w,3=1 z4=1 |
| | 12 | z2w,1=1 z2w,2=1 z2w,3=1 z4=3 |
| | 12 | z4=2 |
| | 12 | z2w,1=1 z2w,2=1 |
| 63.4.77 | 189136 | Ge2Ni1Pr1 |
| | 12 | z2w,1=1 z2w,2=1 z4=2 |
| | 12 | z2w,3=1 z4=1 |
| | 12 | z2w,1=1 z2w,2=1 z2w,3=1 z4=3 |
| | 12 | z2w,1=1 z2w,2=1 z2w,3=1 z4=1 |
| | 12 | z2w,3=1 z4=3 |
| | 12 | z2w,1=1 z2w,2=1 z4=2 |
| 63.4.78 | 85600 | Ge2La1Ni1 |
| | 12 | z2w,1=1 z2w,2=1 z4=1 |
| | 12 | z4=3 |
| | 12 | z2w,1=1 z2w,2=1 z2w,3=1 z4=2 |
| | 12 | z2w,3=1 |
| | 12 | z2w,3=1 z4=2 |
| | 12 | z2w,1=1 z2w,2=1 z2w,3=1 |
| | 12 | z4=3 |
| | 12 | z2w,1=1 z2w,2=1 z4=1 |
| 63.4.79 | 42377 | Ge1Pr1 |
| | 12 | z4=1 |
| | 12 | z2w,1=1 z2w,2=1 z4=3 |
| | 12 | z2w,3=1 z4=2 |
| | 12 | z2w,1=1 z2w,2=1 z2w,3=1 |
| | 12 | z2w,1=1 z2w,2=1 z2w,3=1 z4=2 |
| | 12 | z2w,3=1 |
| | 12 | z2w,1=1 z2w,2=1 z4=3 |
| | 12 | z4=1 |
| 63.4.80 | 42376 | Ge1Pr1 |
| | 12 | z4=1 |
| | 12 | z2w,1=1 z2w,2=1 z4=3 |
| | 12 | z2w,3=1 z4=2 |
| | 12 | z2w,1=1 z2w,2=1 z2w,3=1 |
| | 12 | z2w,1=1 z2w,2=1 z2w,3=1 z4=2 |
| | 12 | z2w,3=1 |
| | 12 | z2w,1=1 z2w,2=1 z4=3 |
| | 12 | z4=1 |
| 63.4.81 | 53647 | Ge1Ho1 |
| | 12 | z4=1 |
| | 12 | z2w,1=1 z2w,2=1 z4=3 |
| | 12 | z2w,3=1 z4=2 |
| | 12 | z2w,1=1 z2w,2=1 z2w,3=1 |
| | 12 | z2w,1=1 z2w,2=1 z2w,3=1 z4=2 |
| | 12 | z2w,3=1 |
| | 12 | z2w,1=1 z2w,2=1 z4=3 |

Continued on next column

**TABLE XIX– continued**

| TQC | ICSD<br>SSG | Stoich. Formula<br>SIs |
|---|---|---|
| | 12 | z4=1 |
| 63.4.82 | 188506 | Gd3O7Sb1 |
| | 12 | z2w,1=1 z2w,2=1 z4=1 |
| | 12 | z4=3 |
| | 12 | z2w,1=1 z2w,2=1 z2w,3=1 z4=2 |
| | 12 | z2w,3=1 |
| | 12 | z2w,3=1 z4=2 |
| | 12 | z2w,1=1 z2w,2=1 z2w,3=1 |
| | 12 | z4=3 |
| | 12 | z2w,1=1 z2w,2=1 z4=1 |
| 63.4.83 | 190729 | Gd3Nb1O7 |
| | 12 | z2w,1=1 z2w,2=1 z4=1 |
| | 12 | z4=3 |
| | 12 | z2w,1=1 z2w,2=1 z2w,3=1 z4=2 |
| | 12 | z2w,3=1 |
| | 12 | z2w,3=1 z4=2 |
| | 12 | z2w,1=1 z2w,2=1 z2w,3=1 |
| | 12 | z4=3 |
| | 12 | z2w,1=1 z2w,2=1 z4=1 |
| 63.4.84 | 411392 | Gd1In2Rh1 |
| | 12 | z2w,1=1 z2w,2=1 z4=1 |
| | 12 | z4=3 |
| | 12 | z2w,1=1 z2w,2=1 z2w,3=1 z4=2 |
| | 12 | z2w,3=1 |
| | 12 | z2w,3=1 z4=2 |
| | 12 | z2w,1=1 z2w,2=1 z2w,3=1 |
| | 12 | z4=3 |
| | 12 | z2w,1=1 z2w,2=1 z4=1 |
| 63.4.85 | 423074 | Ga4Ni1Y1 |
| | 12 | z2w,1=1 z2w,2=1 z4=1 |
| | 12 | z4=3 |
| | 12 | z2w,1=1 z2w,2=1 z2w,3=1 z4=2 |
| | 12 | z2w,3=1 |
| | 12 | z2w,3=1 z4=2 |
| | 12 | z2w,1=1 z2w,2=1 z2w,3=1 |
| | 12 | z4=3 |
| | 12 | z2w,1=1 z2w,2=1 z4=1 |
| 63.4.86 | 423069 | Ga4Ni1Tm1 |
| | 12 | z4=2 |
| | 12 | z2w,1=1 z2w,2=1 |
| | 12 | z2w,3=1 z4=3 |
| | 12 | z2w,1=1 z2w,2=1 z2w,3=1 z4=1 |
| | 12 | z2w,1=1 z2w,2=1 z2w,3=1 z4=3 |
| | 12 | z2w,3=1 z4=1 |
| | 12 | z2w,1=1 z2w,2=1 |
| | 12 | z4=2 |
| 63.4.87 | 423068 | Ga4Ni1Tb1 |
| | 12 | z2w,1=1 z2w,2=1 z4=1 |
| | 12 | z4=3 |
| | 12 | z2w,1=1 z2w,2=1 z2w,3=1 z4=2 |
| | 12 | z2w,3=1 |
| | 12 | z2w,3=1 z4=2 |
| | 12 | z2w,1=1 z2w,2=1 z2w,3=1 |
| | 12 | z4=3 |
| | 12 | z2w,1=1 z2w,2=1 z4=1 |
| 63.4.88 | 423079 | Ga4Ho1Ni1 |
| | 12 | z4=1 |
| | 12 | z2w,1=1 z2w,2=1 z4=3 |
| | 12 | z2w,3=1 z4=2 |

Continued on next column

**TABLE XIX– continued**

| TQC | ICSD SSG | Stoich. Formula SIs |
|---|---|---|
| | 12 | z2w,1=1 z2w,2=1 z2w,3=1 |
| | 12 | z2w,1=1 z2w,2=1 z2w,3=1 z4=2 |
| | 12 | z2w,3=1 |
| | 12 | z2w,1=1 z2w,2=1 z4=3 |
| | 12 | z4=1 |
| 63.4.89 | 192026 | Ga3Pd1Sr1 |
| | 12 | z4=1 |
| | 12 | z2w,1=1 z2w,2=1 z4=3 |
| | 12 | z2w,3=1 z4=2 |
| | 12 | z2w,1=1 z2w,2=1 z2w,3=1 |
| | 12 | z2w,1=1 z2w,2=1 z2w,3=1 z4=2 |
| | 12 | z2w,3=1 |
| | 12 | z2w,1=1 z2w,2=1 z4=3 |
| | 12 | z4=1 |
| 63.4.90 | 103895 | Ga2Ni1Y1 |
| | 12 | z2w,1=1 z2w,2=1 |
| | 12 | z4=2 |
| | 12 | z2w,1=1 z2w,2=1 z2w,3=1 z4=1 |
| | 12 | z2w,3=1 z4=3 |
| | 12 | z2w,3=1 z4=1 |
| | 12 | z2w,1=1 z2w,2=1 z2w,3=1 z4=3 |
| | 12 | z4=2 |
| | 12 | z2w,1=1 z2w,2=1 |
| 63.4.91 | 413557 | Ga2Ir1 |
| | 12 | z4=3 |
| | 12 | z2w,1=1 z2w,2=1 z4=1 |
| | 12 | z2w,3=1 |
| | 12 | z2w,1=1 z2w,2=1 z2w,3=1 z4=2 |
| | 12 | z2w,1=1 z2w,2=1 z2w,3=1 |
| | 12 | z2w,3=1 z4=2 |
| | 12 | z2w,1=1 z2w,2=1 z4=1 |
| | 12 | z4=3 |
| 63.4.92 | 100051 | Ga1Li2 |
| | 12 | z4=3 |
| | 12 | z2w,1=1 z2w,2=1 z4=1 |
| | 12 | z2w,3=1 |
| | 12 | z2w,1=1 z2w,2=1 z2w,3=1 z4=2 |
| | 12 | z2w,1=1 z2w,2=1 z2w,3=1 |
| | 12 | z2w,3=1 z4=2 |
| | 12 | z2w,1=1 z2w,2=1 z4=1 |
| | 12 | z4=3 |
| 63.4.93 | 98966 | Fe2Te2Y5 |
| | 12 | z2w,1=1 z2w,2=1 z4=2 |
| | 12 | z2w,1=1 z2w,2=1 z2w,3=1 z4=3 |
| | 12 | z2w,3=1 z4=1 |
| | 12 | z2w,3=1 z4=3 |
| | 12 | z2w,1=1 z2w,2=1 z2w,3=1 z4=1 |
| | 12 | z2w,1=1 z2w,2=1 z4=2 |
| 63.4.94 | 166049 | Fe2S3Tl1 |
| | 12 | z2w,1=1 z2w,2=1 z4=2 |
| | 12 | z2w,1=1 z2w,2=1 z2w,3=1 z4=3 |
| | 12 | z2w,3=1 z4=1 |
| | 12 | z2w,3=1 z4=3 |
| | 12 | z2w,1=1 z2w,2=1 z2w,3=1 z4=1 |
| | 12 | z2w,1=1 z2w,2=1 z4=2 |
| 63.4.95 | 81548 | Fe2Rb1Te3 |
| | 12 | z2w,1=1 z2w,2=1 z4=2 |
| | 12 | z2w,1=1 z2w,2=1 z2w,3=1 z4=3 |
| | 12 | z2w,3=1 z4=1 |

Continued on next column

**TABLE XIX– continued**

| TQC | ICSD SSG | Stoich. Formula SIs |
|---|---|---|
| | 12 | z2w,3=1 z4=3 |
| | 12 | z2w,1=1 z2w,2=1 z2w,3=1 z4=1 |
| | 12 | z2w,1=1 z2w,2=1 z4=2 |
| 63.4.96 | 81547 | Fe2Rb1Se3 |
| | 12 | z2w,1=1 z2w,2=1 z4=2 |
| | 12 | z2w,1=1 z2w,2=1 z2w,3=1 z4=3 |
| | 12 | z2w,3=1 z4=1 |
| | 12 | z2w,3=1 z4=3 |
| | 12 | z2w,1=1 z2w,2=1 z2w,3=1 z4=1 |
| | 12 | z2w,1=1 z2w,2=1 z4=2 |
| 63.4.97 | 99505 | Fe2Rb1S3 |
| | 12 | z2w,1=1 z2w,2=1 z4=2 |
| | 12 | z2w,1=1 z2w,2=1 z2w,3=1 z4=3 |
| | 12 | z2w,3=1 z4=1 |
| | 12 | z2w,3=1 z4=3 |
| | 12 | z2w,1=1 z2w,2=1 z2w,3=1 z4=1 |
| | 12 | z2w,1=1 z2w,2=1 z4=2 |
| 63.4.98 | 424672 | Fe2K1Se3 |
| | 12 | z2w,1=1 z2w,2=1 z4=2 |
| | 12 | z2w,3=1 z4=1 |
| | 12 | z2w,1=1 z2w,2=1 z2w,3=1 z4=3 |
| | 12 | z2w,1=1 z2w,2=1 z2w,3=1 z4=1 |
| | 12 | z2w,3=1 z4=3 |
| | 12 | z2w,1=1 z2w,2=1 z4=2 |
| 63.4.99 | 424671 | Fe2K1Se3 |
| | 12 | z2w,1=1 z2w,2=1 z4=2 |
| | 12 | z2w,1=1 z2w,2=1 z2w,3=1 z4=3 |
| | 12 | z2w,3=1 z4=1 |
| | 12 | z2w,3=1 z4=3 |
| | 12 | z2w,1=1 z2w,2=1 z2w,3=1 z4=1 |
| | 12 | z2w,1=1 z2w,2=1 z4=2 |
| 63.4.100 | 99507 | Fe2K1S3 |
| | 12 | z2w,1=1 z2w,2=1 z4=2 |
| | 12 | z2w,1=1 z2w,2=1 z2w,3=1 z4=3 |
| | 12 | z2w,3=1 z4=1 |
| | 12 | z2w,3=1 z4=3 |
| | 12 | z2w,1=1 z2w,2=1 z2w,3=1 z4=1 |
| | 12 | z2w,1=1 z2w,2=1 z4=2 |
| 63.4.101 | 100180 | Fe2K1S3 |
| | 12 | z2w,1=1 z2w,2=1 z4=2 |
| | 12 | z2w,1=1 z2w,2=1 z2w,3=1 z4=3 |
| | 12 | z2w,3=1 z4=1 |
| | 12 | z2w,3=1 z4=3 |
| | 12 | z2w,1=1 z2w,2=1 z2w,3=1 z4=1 |
| | 12 | z2w,1=1 z2w,2=1 z4=2 |
| 63.4.102 | 155065 | Fe1O4P1 |
| | 12 | z2w,1=1 z2w,2=1 z4=2 |
| | 12 | z2w,1=1 z2w,2=1 z2w,3=1 z4=3 |
| | 12 | z2w,3=1 z4=1 |
| | 12 | z2w,3=1 z4=3 |
| | 12 | z2w,1=1 z2w,2=1 z2w,3=1 z4=1 |
| | 12 | z2w,1=1 z2w,2=1 z4=2 |
| 63.4.103 | 60615 | F4Fe1K1 |
| | 12 | z2w,1=1 z2w,2=1 z4=2 |
| | 12 | z2w,1=1 z2w,2=1 z2w,3=1 z4=3 |
| | 12 | z2w,3=1 z4=1 |
| | 12 | z2w,3=1 z4=3 |
| | 12 | z2w,1=1 z2w,2=1 z2w,3=1 z4=1 |
| | 12 | z2w,1=1 z2w,2=1 z4=2 |

Continued on next column

TABLE XIX– continued

| TQC | ICSD SSG | Stoich. Formula SIs |
|---|---|---|
| 63.4.104 | 93668 | Eu3O7Ru1 |
| | 12 | z2w,1=1 z2w,2=1 z4=2 |
| | 12 | z2w,3=1 z4=1 |
| | 12 | z2w,1=1 z2w,2=1 z2w,3=1 z4=3 |
| | 12 | z2w,1=1 z2w,2=1 z2w,3=1 z4=1 |
| | 12 | z2w,3=1 z4=3 |
| | 12 | z2w,1=1 z2w,2=1 z4=2 |
| 63.4.105 | 416436 | Eu3Ge5 |
| | 12 | z2w,1=1 z2w,2=1 z4=2 |
| | 12 | z2w,3=1 z4=1 |
| | 12 | z2w,1=1 z2w,2=1 z2w,3=1 z4=3 |
| | 12 | z2w,1=1 z2w,2=1 z2w,3=1 z4=1 |
| | 12 | z2w,3=1 z4=3 |
| | 12 | z2w,1=1 z2w,2=1 z4=2 |
| 63.4.106 | 245230 | Eu3Ge5 |
| | 12 | z2w,1=1 z2w,2=1 z4=2 |
| | 12 | z2w,3=1 z4=1 |
| | 12 | z2w,1=1 z2w,2=1 z2w,3=1 z4=3 |
| | 12 | z2w,1=1 z2w,2=1 z2w,3=1 z4=1 |
| | 12 | z2w,3=1 z4=3 |
| | 12 | z2w,1=1 z2w,2=1 z4=2 |
| 63.4.107 | 245229 | Eu3Ge5 |
| | 12 | z2w,1=1 z2w,2=1 z4=2 |
| | 12 | z2w,3=1 z4=1 |
| | 12 | z2w,1=1 z2w,2=1 z2w,3=1 z4=3 |
| | 12 | z2w,1=1 z2w,2=1 z2w,3=1 z4=1 |
| | 12 | z2w,3=1 z4=3 |
| | 12 | z2w,1=1 z2w,2=1 z4=2 |
| 63.4.108 | 416579 | Eu1Si6 |
| | 12 | z2w,1=1 z2w,2=1 z4=1 |
| | 12 | z4=3 |
| | 12 | z2w,1=1 z2w,2=1 z2w,3=1 z4=2 |
| | 12 | z2w,3=1 |
| | 12 | z2w,3=1 z4=2 |
| | 12 | z2w,1=1 z2w,2=1 z2w,3=1 |
| | 12 | z4=3 |
| | 12 | z2w,1=1 z2w,2=1 z4=1 |
| 63.4.109 | 416578 | Eu1Si6 |
| | 12 | z2w,1=1 z2w,2=1 z4=1 |
| | 12 | z4=3 |
| | 12 | z2w,1=1 z2w,2=1 z2w,3=1 z4=2 |
| | 12 | z2w,3=1 |
| | 12 | z2w,3=1 z4=2 |
| | 12 | z2w,1=1 z2w,2=1 z2w,3=1 |
| | 12 | z4=3 |
| | 12 | z2w,1=1 z2w,2=1 z4=1 |
| 63.4.110 | 416577 | Eu1Si6 |
| | 12 | z2w,1=1 z2w,2=1 z4=1 |
| | 12 | z4=3 |
| | 12 | z2w,1=1 z2w,2=1 z2w,3=1 z4=2 |
| | 12 | z2w,3=1 |
| | 12 | z2w,3=1 z4=2 |
| | 12 | z2w,1=1 z2w,2=1 z2w,3=1 |
| | 12 | z4=3 |
| | 12 | z2w,1=1 z2w,2=1 z4=1 |
| 63.4.111 | 416576 | Eu1Si6 |
| | 12 | z2w,1=1 z2w,2=1 z4=1 |
| | 12 | z4=3 |
| | 12 | z2w,1=1 z2w,2=1 z2w,3=1 z4=2 |
| | 12 | z2w,3=1 |
| | 12 | z2w,3=1 z4=2 |
| | 12 | z2w,1=1 z2w,2=1 z2w,3=1 |
| | 12 | z4=3 |
| | 12 | z2w,1=1 z2w,2=1 z4=1 |
| 63.4.112 | 193724 | Eu1In4Pt1 |
| | 12 | z2w,1=1 z2w,2=1 z4=1 |
| | 12 | z4=3 |
| | 12 | z2w,1=1 z2w,2=1 z2w,3=1 z4=2 |
| | 12 | z2w,3=1 |
| | 12 | z2w,3=1 z4=2 |
| | 12 | z2w,1=1 z2w,2=1 z2w,3=1 |
| | 12 | z4=3 |
| | 12 | z2w,1=1 z2w,2=1 z4=1 |
| 63.4.113 | 411391 | Eu1In4Pd1 |
| | 12 | z4=3 |
| | 12 | z2w,1=1 z2w,2=1 z4=1 |
| | 12 | z2w,3=1 |
| | 12 | z2w,1=1 z2w,2=1 z2w,3=1 z4=2 |
| | 12 | z2w,1=1 z2w,2=1 z2w,3=1 |
| | 12 | z2w,3=1 z4=2 |
| | 12 | z2w,1=1 z2w,2=1 z4=1 |
| | 12 | z4=3 |
| 63.4.114 | 404251 | Eu1In4Ni1 |
| | 12 | z2w,1=1 z2w,2=1 z4=1 |
| | 12 | z4=3 |
| | 12 | z2w,1=1 z2w,2=1 z2w,3=1 z4=2 |
| | 12 | z2w,3=1 |
| | 12 | z2w,3=1 z4=2 |
| | 12 | z2w,1=1 z2w,2=1 z2w,3=1 |
| | 12 | z4=3 |
| | 12 | z2w,1=1 z2w,2=1 z4=1 |
| 63.4.115 | 410438 | Eu1In2Pd1 |
| | 12 | z2w,1=1 z2w,2=1 |
| | 12 | z4=2 |
| | 12 | z2w,1=1 z2w,2=1 z2w,3=1 z4=1 |
| | 12 | z2w,3=1 z4=3 |
| | 12 | z2w,3=1 z4=1 |
| | 12 | z2w,1=1 z2w,2=1 z2w,3=1 z4=3 |
| | 12 | z4=2 |
| | 12 | z2w,1=1 z2w,2=1 |
| 63.4.116 | 402571 | Eu1In2Ni1 |
| | 12 | z4=2 |
| | 12 | z2w,1=1 z2w,2=1 |
| | 12 | z2w,3=1 z4=3 |
| | 12 | z2w,1=1 z2w,2=1 z2w,3=1 z4=1 |
| | 12 | z2w,1=1 z2w,2=1 z2w,3=1 z4=3 |
| | 12 | z2w,3=1 z4=1 |
| | 12 | z2w,1=1 z2w,2=1 |
| | 12 | z4=2 |
| 63.4.117 | 42420 | Eu1Ge3Mg3 |
| | 12 | z2w,1=1 z2w,2=1 z4=2 |
| | 12 | z2w,1=1 z2w,2=1 z2w,3=1 z4=3 |
| | 12 | z2w,3=1 z4=1 |
| | 12 | z2w,3=1 z4=3 |
| | 12 | z2w,1=1 z2w,2=1 z2w,3=1 z4=1 |
| | 12 | z2w,1=1 z2w,2=1 z4=2 |
| 63.4.118 | 71204 | Cu5K2Te5 |
| | 12 | z2w,1=1 z2w,2=1 z4=1 |

Continued on next column

**TABLE XIX– continued**

| TQC | ICSD<br>SSG | Stoich. Formula<br>SIs |
|---|---|---|
| | 12 | z4=3 |
| | 12 | z2w,1=1 z2w,2=1 z2w,3=1 z4=2 |
| | 12 | z2w,3=1 |
| | 12 | z2w,3=1 z4=2 |
| | 12 | z2w,1=1 z2w,2=1 z2w,3=1 |
| | 12 | z4=3 |
| | 12 | z2w,1=1 z2w,2=1 z4=1 |
| 63.4.119 | 54129 | Cu4O8 |
| | 12 | z2w,1=1 z2w,2=1 z4=2 |
| | 12 | z2w,3=1 z4=1 |
| | 12 | z2w,1=1 z2w,2=1 z2w,3=1 z4=3 |
| | 12 | z2w,1=1 z2w,2=1 z2w,3=1 z4=1 |
| | 12 | z2w,3=1 z4=3 |
| | 12 | z2w,1=1 z2w,2=1 z4=2 |
| 63.4.120 | 249533 | Cu3K2S5U1 |
| | 12 | z2w,1=1 z2w,2=1 z4=2 |
| | 12 | z2w,3=1 z4=1 |
| | 12 | z2w,1=1 z2w,2=1 z2w,3=1 z4=3 |
| | 12 | z2w,1=1 z2w,2=1 z2w,3=1 z4=1 |
| | 12 | z2w,3=1 z4=3 |
| | 12 | z2w,1=1 z2w,2=1 z4=2 |
| 63.4.121 | 380487 | Cu2K3O4 |
| | 12 | z2w,1=1 z2w,2=1 z4=2 |
| | 12 | z2w,1=1 z2w,2=1 z2w,3=1 z4=3 |
| | 12 | z2w,3=1 z4=1 |
| | 12 | z2w,3=1 z4=3 |
| | 12 | z2w,1=1 z2w,2=1 z2w,3=1 z4=1 |
| | 12 | z2w,1=1 z2w,2=1 z4=2 |
| 63.4.122 | 82330 | Cu1Se1 |
| | 12 | z2w,1=1 z2w,2=1 z4=3 |
| | 12 | z4=1 |
| | 12 | z2w,1=1 z2w,2=1 z2w,3=1 |
| | 12 | z2w,3=1 z4=2 |
| | 12 | z2w,3=1 |
| | 12 | z2w,1=1 z2w,2=1 z2w,3=1 z4=2 |
| | 12 | z4=1 |
| | 12 | z2w,1=1 z2w,2=1 z4=3 |
| 63.4.123 | 83051 | Cu1O2Sr1 |
| | 12 | z2w,1=1 z2w,2=1 z4=2 |
| | 12 | z2w,3=1 z4=1 |
| | 12 | z2w,1=1 z2w,2=1 z2w,3=1 z4=3 |
| | 12 | z2w,1=1 z2w,2=1 z2w,3=1 z4=1 |
| | 12 | z2w,3=1 z4=3 |
| | 12 | z2w,1=1 z2w,2=1 z4=2 |
| 63.4.124 | 77293 | Cu1O2Sr1 |
| | 12 | z2w,1=1 z2w,2=1 z4=2 |
| | 12 | z2w,1=1 z2w,2=1 z2w,3=1 z4=3 |
| | 12 | z2w,3=1 z4=1 |
| | 12 | z2w,3=1 z4=3 |
| | 12 | z2w,1=1 z2w,2=1 z2w,3=1 z4=1 |
| | 12 | z2w,1=1 z2w,2=1 z4=2 |
| 63.4.125 | 77291 | Cu1O2Sr1 |
| | 12 | z2w,1=1 z2w,2=1 z4=2 |
| | 12 | z2w,3=1 z4=1 |
| | 12 | z2w,1=1 z2w,2=1 z2w,3=1 z4=3 |
| | 12 | z2w,1=1 z2w,2=1 z2w,3=1 z4=1 |
| | 12 | z2w,3=1 z4=3 |
| | 12 | z2w,1=1 z2w,2=1 z4=2 |
| 63.4.126 | 202992 | Cu1O2Sr1 |

Continued on next column

**TABLE XIX– continued**

| TQC | ICSD<br>SSG | Stoich. Formula<br>SIs |
|---|---|---|
| | 12 | z2w,1=1 z2w,2=1 z4=2 |
| | 12 | z2w,3=1 z4=1 |
| | 12 | z2w,1=1 z2w,2=1 z2w,3=1 z4=3 |
| | 12 | z2w,1=1 z2w,2=1 z2w,3=1 z4=1 |
| | 12 | z2w,3=1 z4=3 |
| | 12 | z2w,1=1 z2w,2=1 z4=2 |
| 63.4.127 | 150913 | Cu1O2 |
| | 12 | z2w,1=1 z2w,2=1 z4=2 |
| | 12 | z2w,3=1 z4=1 |
| | 12 | z2w,1=1 z2w,2=1 z2w,3=1 z4=3 |
| | 12 | z2w,1=1 z2w,2=1 z2w,3=1 z4=1 |
| | 12 | z2w,3=1 z4=3 |
| | 12 | z2w,1=1 z2w,2=1 z4=2 |
| 63.4.128 | 249967 | Cu1Np1Rb1S3 |
| | 12 | z2w,1=1 z2w,2=1 z4=2 |
| | 12 | z2w,1=1 z2w,2=1 z2w,3=1 z4=3 |
| | 12 | z2w,3=1 z4=1 |
| | 12 | z2w,3=1 z4=3 |
| | 12 | z2w,1=1 z2w,2=1 z2w,3=1 z4=1 |
| | 12 | z2w,1=1 z2w,2=1 z4=2 |
| 63.4.129 | 249966 | Cu1K1Np1S3 |
| | 12 | z2w,1=1 z2w,2=1 z4=2 |
| | 12 | z2w,1=1 z2w,2=1 z2w,3=1 z4=3 |
| | 12 | z2w,3=1 z4=1 |
| | 12 | z2w,3=1 z4=3 |
| | 12 | z2w,1=1 z2w,2=1 z2w,3=1 z4=1 |
| | 12 | z2w,1=1 z2w,2=1 z4=2 |
| 63.4.130 | 84200 | Cu1Ge4Zr2 |
| | 12 | z4=2 |
| | 12 | z2w,1=1 z2w,2=1 |
| | 12 | z2w,3=1 z4=3 |
| | 12 | z2w,1=1 z2w,2=1 z2w,3=1 z4=1 |
| | 12 | z2w,1=1 z2w,2=1 z2w,3=1 z4=3 |
| | 12 | z2w,3=1 z4=1 |
| | 12 | z2w,1=1 z2w,2=1 |
| | 12 | z4=2 |
| 63.4.131 | 54101 | Cs2Cu5Se4 |
| | 12 | z4=3 |
| | 12 | z2w,1=1 z2w,2=1 z4=1 |
| | 12 | z2w,3=1 |
| | 12 | z2w,1=1 z2w,2=1 z2w,3=1 z4=2 |
| | 12 | z2w,1=1 z2w,2=1 z2w,3=1 |
| | 12 | z2w,3=1 z4=2 |
| | 12 | z2w,1=1 z2w,2=1 z4=1 |
| | 12 | z4=3 |
| 63.4.132 | 281446 | Cs1Gd1Hg1Se3 |
| | 12 | z2w,1=1 z2w,2=1 z4=2 |
| | 12 | z2w,1=1 z2w,2=1 z2w,3=1 z4=3 |
| | 12 | z2w,3=1 z4=1 |
| | 12 | z2w,3=1 z4=3 |
| | 12 | z2w,1=1 z2w,2=1 z2w,3=1 z4=1 |
| | 12 | z2w,1=1 z2w,2=1 z4=2 |
| 63.4.133 | 81549 | Cs1Fe2Se3 |
| | 12 | z2w,1=1 z2w,2=1 z4=2 |
| | 12 | z2w,1=1 z2w,2=1 z2w,3=1 z4=3 |
| | 12 | z2w,3=1 z4=1 |
| | 12 | z2w,3=1 z4=3 |
| | 12 | z2w,1=1 z2w,2=1 z2w,3=1 z4=1 |
| | 12 | z2w,1=1 z2w,2=1 z4=2 |

Continued on next column

**TABLE XIX– continued**

| TQC | ICSD<br>SSG | Stoich. Formula<br>SIs |
|---|---|---|
| 63.4.134 | 187535 | Cs1Fe2Se3 |
| | 12 | z2w,1=1 z2w,2=1 z4=2 |
| | 12 | z2w,1=1 z2w,2=1 z2w,3=1 z4=3 |
| | 12 | z2w,3=1 z4=1 |
| | 12 | z2w,3=1 z4=3 |
| | 12 | z2w,1=1 z2w,2=1 z2w,3=1 z4=1 |
| | 12 | z2w,1=1 z2w,2=1 z4=2 |
| 63.4.135 | 99504 | Cs1Fe2S3 |
| | 12 | z2w,1=1 z2w,2=1 z4=2 |
| | 12 | z2w,1=1 z2w,2=1 z2w,3=1 z4=3 |
| | 12 | z2w,3=1 z4=1 |
| | 12 | z2w,3=1 z4=3 |
| | 12 | z2w,1=1 z2w,2=1 z2w,3=1 z4=1 |
| | 12 | z2w,1=1 z2w,2=1 z4=2 |
| 63.4.136 | 280844 | Cs1Er1Se3Zn1 |
| | 12 | z2w,1=1 z2w,2=1 z4=2 |
| | 12 | z2w,1=1 z2w,2=1 z2w,3=1 z4=3 |
| | 12 | z2w,3=1 z4=1 |
| | 12 | z2w,3=1 z4=3 |
| | 12 | z2w,1=1 z2w,2=1 z2w,3=1 z4=1 |
| | 12 | z2w,1=1 z2w,2=1 z4=2 |
| 63.4.137 | 280842 | Cs1Dy1Se3Zn1 |
| | 12 | z2w,1=1 z2w,2=1 z4=2 |
| | 12 | z2w,1=1 z2w,2=1 z2w,3=1 z4=3 |
| | 12 | z2w,3=1 z4=1 |
| | 12 | z2w,3=1 z4=3 |
| | 12 | z2w,1=1 z2w,2=1 z2w,3=1 z4=1 |
| | 12 | z2w,1=1 z2w,2=1 z4=2 |
| 63.4.138 | 249968 | Cs1Cu1Np1S3 |
| | 12 | z2w,1=1 z2w,2=1 z4=2 |
| | 12 | z2w,1=1 z2w,2=1 z2w,3=1 z4=3 |
| | 12 | z2w,3=1 z4=1 |
| | 12 | z2w,3=1 z4=3 |
| | 12 | z2w,1=1 z2w,2=1 z2w,3=1 z4=1 |
| | 12 | z2w,1=1 z2w,2=1 z4=2 |
| 63.4.139 | 62159 | Cr1O4P1 |
| | 12 | z2w,1=1 z2w,2=1 z4=2 |
| | 12 | z2w,1=1 z2w,2=1 z2w,3=1 z4=3 |
| | 12 | z2w,3=1 z4=1 |
| | 12 | z2w,3=1 z4=3 |
| | 12 | z2w,1=1 z2w,2=1 z2w,3=1 z4=1 |
| | 12 | z2w,1=1 z2w,2=1 z4=2 |
| 63.4.140 | 60825 | Cr1Cu1O4 |
| | 12 | z2w,1=1 z2w,2=1 z4=2 |
| | 12 | z2w,1=1 z2w,2=1 z2w,3=1 z4=3 |
| | 12 | z2w,3=1 z4=1 |
| | 12 | z2w,3=1 z4=3 |
| | 12 | z2w,1=1 z2w,2=1 z2w,3=1 z4=1 |
| | 12 | z2w,1=1 z2w,2=1 z4=2 |
| 63.4.141 | 183216 | Cr1Cu1O4 |
| | 12 | z2w,1=1 z2w,2=1 z4=2 |
| | 12 | z2w,1=1 z2w,2=1 z2w,3=1 z4=3 |
| | 12 | z2w,3=1 z4=1 |
| | 12 | z2w,3=1 z4=3 |
| | 12 | z2w,1=1 z2w,2=1 z2w,3=1 z4=1 |
| | 12 | z2w,1=1 z2w,2=1 z4=2 |
| 63.4.142 | 102490 | Co5Ga1Y5 |
| | 12 | z2w,1=1 z2w,2=1 z4=2 |
| | 12 | z2w,3=1 z4=1 |

Continued on next column

**TABLE XIX– continued**

| TQC | ICSD<br>SSG | Stoich. Formula<br>SIs |
|---|---|---|
| | 12 | z2w,1=1 z2w,2=1 z2w,3=1 z4=3 |
| | 12 | z2w,1=1 z2w,2=1 z2w,3=1 z4=1 |
| | 12 | z2w,3=1 z4=3 |
| | 12 | z2w,1=1 z2w,2=1 z4=2 |
| 63.4.143 | 10044 | Co3Ga1Y3 |
| | 12 | z2w,1=1 z2w,2=1 z4=2 |
| | 12 | z2w,1=1 z2w,2=1 z2w,3=1 z4=3 |
| | 12 | z2w,3=1 z4=1 |
| | 12 | z2w,3=1 z4=3 |
| | 12 | z2w,1=1 z2w,2=1 z2w,3=1 z4=1 |
| | 12 | z2w,1=1 z2w,2=1 z4=2 |
| 63.4.144 | 41745 | Co2Sc3Si3 |
| | 12 | z2w,1=1 z2w,2=1 z4=3 |
| | 12 | z4=1 |
| | 12 | z2w,1=1 z2w,2=1 z2w,3=1 |
| | 12 | z2w,3=1 z4=2 |
| | 12 | z2w,3=1 |
| | 12 | z2w,1=1 z2w,2=1 z2w,3=1 z4=2 |
| | 12 | z4=1 |
| | 12 | z2w,1=1 z2w,2=1 z4=3 |
| 63.4.145 | 106479 | Co2Ge3Sc3 |
| | 12 | z2w,1=1 z2w,2=1 z4=3 |
| | 12 | z4=1 |
| | 12 | z2w,1=1 z2w,2=1 z2w,3=1 |
| | 12 | z2w,3=1 z4=2 |
| | 12 | z2w,3=1 |
| | 12 | z2w,1=1 z2w,2=1 z2w,3=1 z4=2 |
| | 12 | z4=1 |
| | 12 | z2w,1=1 z2w,2=1 z4=3 |
| 63.4.146 | 33736 | Co1O4S1 |
| | 12 | z2w,1=1 z2w,2=1 z4=2 |
| | 12 | z2w,1=1 z2w,2=1 z2w,3=1 z4=3 |
| | 12 | z2w,3=1 z4=1 |
| | 12 | z2w,3=1 z4=3 |
| | 12 | z2w,1=1 z2w,2=1 z2w,3=1 z4=1 |
| | 12 | z2w,1=1 z2w,2=1 z4=2 |
| 63.4.147 | 55670 | Co1Ga4Th1 |
| | 12 | z4=3 |
| | 12 | z2w,1=1 z2w,2=1 z4=1 |
| | 12 | z2w,3=1 |
| | 12 | z2w,1=1 z2w,2=1 z2w,3=1 z4=2 |
| | 12 | z2w,1=1 z2w,2=1 z2w,3=1 |
| | 12 | z2w,3=1 z4=2 |
| | 12 | z2w,1=1 z2w,2=1 z4=1 |
| | 12 | z4=3 |
| 63.4.148 | 417507 | Co1Cs1Ho1Se3 |
| | 12 | z2w,1=1 z2w,2=1 z4=2 |
| | 12 | z2w,1=1 z2w,2=1 z2w,3=1 z4=3 |
| | 12 | z2w,3=1 z4=1 |
| | 12 | z2w,3=1 z4=3 |
| | 12 | z2w,1=1 z2w,2=1 z2w,3=1 z4=1 |
| | 12 | z2w,1=1 z2w,2=1 z4=2 |
| 63.4.149 | 169508 | Ce1Rh1Si2 |
| | 12 | z2w,1=1 z2w,2=1 |
| | 12 | z4=2 |
| | 12 | z2w,1=1 z2w,2=1 z2w,3=1 z4=1 |
| | 12 | z2w,3=1 z4=3 |
| | 12 | z2w,3=1 z4=1 |
| | 12 | z2w,1=1 z2w,2=1 z2w,3=1 z4=3 |

Continued on next column

**TABLE XIX– continued**

| TQC | ICSD<br>SSG | Stoich. Formula<br>SIs |
|---|---|---|
| | 12 | z4=2 |
| | 12 | z2w,1=1 z2w,2=1 |
| 63.4.150 | 106411 | Ce1Mn1Si2 |
| | 12 | z2w,1=1 z2w,2=1 z4=2 |
| | 12 | z2w,3=1 z4=1 |
| | 12 | z2w,1=1 z2w,2=1 z2w,3=1 z4=3 |
| | 12 | z2w,1=1 z2w,2=1 z2w,3=1 z4=1 |
| | 12 | z2w,3=1 z4=3 |
| | 12 | z2w,1=1 z2w,2=1 z4=2 |
| 63.4.151 | 106410 | Ce1Mn1Si2 |
| | 12 | z2w,1=1 z2w,2=1 z4=2 |
| | 12 | z2w,3=1 z4=1 |
| | 12 | z2w,1=1 z2w,2=1 z2w,3=1 z4=3 |
| | 12 | z2w,1=1 z2w,2=1 z2w,3=1 z4=1 |
| | 12 | z2w,3=1 z4=3 |
| | 12 | z2w,1=1 z2w,2=1 z4=2 |
| 63.4.152 | 102212 | Ce1Li1Sn2 |
| | 12 | z2w,1=1 z2w,2=1 |
| | 12 | z4=2 |
| | 12 | z2w,1=1 z2w,2=1 z2w,3=1 z4=1 |
| | 12 | z2w,3=1 z4=3 |
| | 12 | z2w,3=1 z4=1 |
| | 12 | z2w,1=1 z2w,2=1 z2w,3=1 z4=3 |
| | 12 | z4=2 |
| | 12 | z2w,1=1 z2w,2=1 |
| 63.4.153 | 414474 | Ce1In2Ir1 |
| | 12 | z2w,1=1 z2w,2=1 z4=2 |
| | 12 | z2w,3=1 z4=1 |
| | 12 | z2w,1=1 z2w,2=1 z2w,3=1 z4=3 |
| | 12 | z2w,1=1 z2w,2=1 z2w,3=1 z4=1 |
| | 12 | z2w,3=1 z4=3 |
| | 12 | z2w,1=1 z2w,2=1 z4=2 |
| 63.4.154 | 427867 | Ce1Ga3Pd1 |
| | 12 | z2w,1=1 z2w,2=1 z4=2 |
| | 12 | z2w,3=1 z4=1 |
| | 12 | z2w,1=1 z2w,2=1 z2w,3=1 z4=3 |
| | 12 | z2w,1=1 z2w,2=1 z2w,3=1 z4=1 |
| | 12 | z2w,3=1 z4=3 |
| | 12 | z2w,1=1 z2w,2=1 z4=2 |
| 63.4.155 | 16501 | Ce1Co1Si2 |
| | 12 | z2w,1=1 z2w,2=1 |
| | 12 | z4=2 |
| | 12 | z2w,1=1 z2w,2=1 z2w,3=1 z4=1 |
| | 12 | z2w,3=1 z4=3 |
| | 12 | z2w,3=1 z4=1 |
| | 12 | z2w,1=1 z2w,2=1 z2w,3=1 z4=3 |
| | 12 | z4=2 |
| | 12 | z2w,1=1 z2w,2=1 |
| 63.4.156 | 417029 | Cd2La1Pd1 |
| | 12 | z4=3 |
| | 12 | z2w,1=1 z2w,2=1 z4=1 |
| | 12 | z2w,3=1 |
| | 12 | z2w,1=1 z2w,2=1 z2w,3=1 z4=2 |
| | 12 | z2w,1=1 z2w,2=1 z2w,3=1 |
| | 12 | z2w,3=1 z4=2 |
| | 12 | z2w,1=1 z2w,2=1 z4=1 |
| | 12 | z4=3 |
| 63.4.157 | 99139 | Cd2Cu1Er1 |
| | 12 | z2w,1=1 z2w,2=1 z4=1 |

Continued on next column

**TABLE XIX– continued**

| TQC | ICSD<br>SSG | Stoich. Formula<br>SIs |
|---|---|---|
| | 12 | z4=3 |
| | 12 | z2w,1=1 z2w,2=1 z2w,3=1 z4=2 |
| | 12 | z2w,3=1 |
| | 12 | z2w,3=1 z4=2 |
| | 12 | z2w,1=1 z2w,2=1 z2w,3=1 |
| | 12 | z4=3 |
| | 12 | z2w,1=1 z2w,2=1 z4=1 |
| 63.4.158 | 281437 | Cd1Cs1Gd1Se3 |
| | 12 | z2w,1=1 z2w,2=1 z4=2 |
| | 12 | z2w,1=1 z2w,2=1 z2w,3=1 z4=3 |
| | 12 | z2w,3=1 z4=1 |
| | 12 | z2w,3=1 z4=3 |
| | 12 | z2w,1=1 z2w,2=1 z2w,3=1 z4=1 |
| | 12 | z2w,1=1 z2w,2=1 z4=2 |
| 63.4.159 | 173313 | Cd1Cs1Er1Te3 |
| | 12 | z2w,1=1 z2w,2=1 z4=2 |
| | 12 | z2w,1=1 z2w,2=1 z2w,3=1 z4=3 |
| | 12 | z2w,3=1 z4=1 |
| | 12 | z2w,3=1 z4=3 |
| | 12 | z2w,1=1 z2w,2=1 z2w,3=1 z4=1 |
| | 12 | z2w,1=1 z2w,2=1 z4=2 |
| 63.4.160 | 281434 | Cd1Ce1Cs1Se3 |
| | 12 | z2w,1=1 z2w,2=1 z4=2 |
| | 12 | z2w,1=1 z2w,2=1 z2w,3=1 z4=3 |
| | 12 | z2w,3=1 z4=1 |
| | 12 | z2w,3=1 z4=3 |
| | 12 | z2w,1=1 z2w,2=1 z2w,3=1 z4=1 |
| | 12 | z2w,1=1 z2w,2=1 z4=2 |
| 63.4.161 | 79095 | Ca4P5Pd5 |
| | 12 | z2w,1=1 z2w,2=1 z4=2 |
| | 12 | z2w,1=1 z2w,2=1 z2w,3=1 z4=3 |
| | 12 | z2w,3=1 z4=1 |
| | 12 | z2w,3=1 z4=3 |
| | 12 | z2w,1=1 z2w,2=1 z2w,3=1 z4=1 |
| | 12 | z2w,1=1 z2w,2=1 z4=2 |
| 63.4.162 | 380324 | Ca3Tl5 |
| | 12 | z4=1 |
| | 12 | z2w,1=1 z2w,2=1 z4=3 |
| | 12 | z2w,3=1 z4=2 |
| | 12 | z2w,1=1 z2w,2=1 z2w,3=1 |
| | 12 | z2w,1=1 z2w,2=1 z2w,3=1 z4=2 |
| | 12 | z2w,3=1 |
| | 12 | z2w,1=1 z2w,2=1 z4=3 |
| | 12 | z4=1 |
| 63.4.163 | 58894 | Ca3Ga5 |
| | 12 | z2w,1=1 z2w,2=1 z4=2 |
| | 12 | z2w,3=1 z4=1 |
| | 12 | z2w,1=1 z2w,2=1 z2w,3=1 z4=3 |
| | 12 | z2w,1=1 z2w,2=1 z2w,3=1 z4=1 |
| | 12 | z2w,3=1 z4=3 |
| | 12 | z2w,1=1 z2w,2=1 z4=2 |
| 63.4.164 | 410775 | Ca1Rh1Sn2 |
| | 12 | z2w,1=1 z2w,2=1 |
| | 12 | z4=2 |
| | 12 | z2w,1=1 z2w,2=1 z2w,3=1 z4=1 |
| | 12 | z2w,3=1 z4=3 |
| | 12 | z2w,3=1 z4=1 |
| | 12 | z2w,1=1 z2w,2=1 z2w,3=1 z4=3 |
| | 12 | z4=2 |

Continued on next column

**TABLE XIX– continued**

| TQC | ICSD SSG | Stoich. Formula SIs |
|---|---|---|
| | 12 | z2w,1=1 z2w,2=1 |
| 63.4.165 | 410773 | Ca1Ir1Sn2 |
| | 12 | z2w,1=1 z2w,2=1 |
| | 12 | z4=2 |
| | 12 | z2w,1=1 z2w,2=1 z2w,3=1 z4=1 |
| | 12 | z2w,3=1 z4=3 |
| | 12 | z2w,3=1 z4=1 |
| | 12 | z2w,1=1 z2w,2=1 z2w,3=1 z4=3 |
| | 12 | z4=2 |
| | 12 | z2w,1=1 z2w,2=1 |
| 63.4.166 | 420479 | Ca1Ir1O3 |
| | 12 | z2w,1=1 z2w,2=1 z4=2 |
| | 12 | z2w,1=1 z2w,2=1 z2w,3=1 z4=3 |
| | 12 | z2w,3=1 z4=1 |
| | 12 | z2w,3=1 z4=3 |
| | 12 | z2w,1=1 z2w,2=1 z2w,3=1 z4=1 |
| | 12 | z2w,1=1 z2w,2=1 z4=2 |
| 63.4.167 | 420478 | Ca1Ir1O3 |
| | 12 | z2w,1=1 z2w,2=1 z4=2 |
| | 12 | z2w,1=1 z2w,2=1 z2w,3=1 z4=3 |
| | 12 | z2w,3=1 z4=1 |
| | 12 | z2w,3=1 z4=3 |
| | 12 | z2w,1=1 z2w,2=1 z2w,3=1 z4=1 |
| | 12 | z2w,1=1 z2w,2=1 z4=2 |
| 63.4.168 | 182246 | Ca1Ir1O3 |
| | 12 | z2w,1=1 z2w,2=1 z4=2 |
| | 12 | z2w,1=1 z2w,2=1 z2w,3=1 z4=3 |
| | 12 | z2w,3=1 z4=1 |
| | 12 | z2w,3=1 z4=3 |
| | 12 | z2w,1=1 z2w,2=1 z2w,3=1 z4=1 |
| | 12 | z2w,1=1 z2w,2=1 z4=2 |
| 63.4.169 | 180015 | Ca1Ir1O3 |
| | 12 | z2w,1=1 z2w,2=1 z4=2 |
| | 12 | z2w,1=1 z2w,2=1 z2w,3=1 z4=3 |
| | 12 | z2w,3=1 z4=1 |
| | 12 | z2w,3=1 z4=3 |
| | 12 | z2w,1=1 z2w,2=1 z2w,3=1 z4=1 |
| | 12 | z2w,1=1 z2w,2=1 z4=2 |
| 63.4.170 | 160816 | Ca1Ir1O3 |
| | 12 | z2w,1=1 z2w,2=1 z4=2 |
| | 12 | z2w,1=1 z2w,2=1 z2w,3=1 z4=3 |
| | 12 | z2w,3=1 z4=1 |
| | 12 | z2w,3=1 z4=3 |
| | 12 | z2w,1=1 z2w,2=1 z2w,3=1 z4=1 |
| | 12 | z2w,1=1 z2w,2=1 z4=2 |
| 63.4.171 | 159038 | Ca1Ir1O3 |
| | 12 | z2w,1=1 z2w,2=1 z4=2 |
| | 12 | z2w,1=1 z2w,2=1 z2w,3=1 z4=3 |
| | 12 | z2w,3=1 z4=1 |
| | 12 | z2w,3=1 z4=3 |
| | 12 | z2w,1=1 z2w,2=1 z2w,3=1 z4=1 |
| | 12 | z2w,1=1 z2w,2=1 z4=2 |
| 63.4.172 | 159037 | Ca1Ir1O3 |
| | 12 | z2w,1=1 z2w,2=1 z4=2 |
| | 12 | z2w,1=1 z2w,2=1 z2w,3=1 z4=3 |
| | 12 | z2w,3=1 z4=1 |
| | 12 | z2w,3=1 z4=3 |
| | 12 | z2w,1=1 z2w,2=1 z2w,3=1 z4=1 |
| | 12 | z2w,1=1 z2w,2=1 z4=2 |

Continued on next column

| TQC | ICSD SSG | Stoich. Formula SIs |
|---|---|---|
| 63.4.173 | 159036 | Ca1Ir1O3 |
| | 12 | z2w,1=1 z2w,2=1 z4=2 |
| | 12 | z2w,3=1 z4=1 |
| | 12 | z2w,1=1 z2w,2=1 z2w,3=1 z4=3 |
| | 12 | z2w,1=1 z2w,2=1 z2w,3=1 z4=1 |
| | 12 | z2w,3=1 z4=3 |
| | 12 | z2w,1=1 z2w,2=1 z4=2 |
| 63.4.174 | 159035 | Ca1Ir1O3 |
| | 12 | z2w,1=1 z2w,2=1 z4=2 |
| | 12 | z2w,1=1 z2w,2=1 z2w,3=1 z4=3 |
| | 12 | z2w,3=1 z4=1 |
| | 12 | z2w,3=1 z4=3 |
| | 12 | z2w,1=1 z2w,2=1 z2w,3=1 z4=1 |
| | 12 | z2w,1=1 z2w,2=1 z4=2 |
| 63.4.175 | 159034 | Ca1Ir1O3 |
| | 12 | z2w,1=1 z2w,2=1 z4=2 |
| | 12 | z2w,1=1 z2w,2=1 z2w,3=1 z4=3 |
| | 12 | z2w,3=1 z4=1 |
| | 12 | z2w,3=1 z4=3 |
| | 12 | z2w,1=1 z2w,2=1 z2w,3=1 z4=1 |
| | 12 | z2w,1=1 z2w,2=1 z4=2 |
| 63.4.176 | 159033 | Ca1Ir1O3 |
| | 12 | z2w,1=1 z2w,2=1 z4=2 |
| | 12 | z2w,3=1 z4=1 |
| | 12 | z2w,1=1 z2w,2=1 z2w,3=1 z4=3 |
| | 12 | z2w,1=1 z2w,2=1 z2w,3=1 z4=1 |
| | 12 | z2w,3=1 z4=3 |
| | 12 | z2w,1=1 z2w,2=1 z4=2 |
| 63.4.177 | 159029 | Ca1Ir1O3 |
| | 12 | z2w,1=1 z2w,2=1 z4=2 |
| | 12 | z2w,3=1 z4=1 |
| | 12 | z2w,1=1 z2w,2=1 z2w,3=1 z4=3 |
| | 12 | z2w,1=1 z2w,2=1 z2w,3=1 z4=1 |
| | 12 | z2w,3=1 z4=3 |
| | 12 | z2w,1=1 z2w,2=1 z4=2 |
| 63.4.178 | 159028 | Ca1Ir1O3 |
| | 12 | z2w,1=1 z2w,2=1 z4=2 |
| | 12 | z2w,3=1 z4=1 |
| | 12 | z2w,1=1 z2w,2=1 z2w,3=1 z4=3 |
| | 12 | z2w,1=1 z2w,2=1 z2w,3=1 z4=1 |
| | 12 | z2w,3=1 z4=3 |
| | 12 | z2w,1=1 z2w,2=1 z4=2 |
| 63.4.179 | 159027 | Ca1Ir1O3 |
| | 12 | z2w,1=1 z2w,2=1 z4=2 |
| | 12 | z2w,3=1 z4=1 |
| | 12 | z2w,1=1 z2w,2=1 z2w,3=1 z4=3 |
| | 12 | z2w,1=1 z2w,2=1 z2w,3=1 z4=1 |
| | 12 | z2w,3=1 z4=3 |
| | 12 | z2w,1=1 z2w,2=1 z4=2 |
| 63.4.180 | 58898 | Ca1Ga3Ni2 |
| | 12 | z4=2 |
| | 12 | z2w,1=1 z2w,2=1 |
| | 12 | z2w,3=1 z4=3 |
| | 12 | z2w,1=1 z2w,2=1 z2w,3=1 z4=1 |
| | 12 | z2w,1=1 z2w,2=1 z2w,3=1 z4=3 |
| | 12 | z2w,3=1 z4=1 |
| | 12 | z2w,1=1 z2w,2=1 |
| | 12 | z4=2 |
| 63.4.181 | 419100 | Ca1Ga1 |

Continued on next column

**TABLE XIX– continued**

| TQC | ICSD SSG | Stoich. Formula SIs |
|---|---|---|
| | 12 | z2w,1=1 z2w,2=1 z4=2 |
| | 12 | z2w,1=1 z2w,2=1 z2w,3=1 z4=3 |
| | 12 | z2w,3=1 z4=1 |
| | 12 | z2w,3=1 z4=3 |
| | 12 | z2w,1=1 z2w,2=1 z2w,3=1 z4=1 |
| | 12 | z2w,1=1 z2w,2=1 z4=2 |
| 63.4.182 | 260641 | Ca1Ga1 |
| | 12 | z2w,1=1 z2w,2=1 z4=2 |
| | 12 | z2w,3=1 z4=1 |
| | 12 | z2w,1=1 z2w,2=1 z2w,3=1 z4=3 |
| | 12 | z2w,1=1 z2w,2=1 z2w,3=1 z4=1 |
| | 12 | z2w,3=1 z4=3 |
| | 12 | z2w,1=1 z2w,2=1 z4=2 |
| 63.4.183 | 106332 | Ca1Cu1In2 |
| | 12 | z2w,1=1 z2w,2=1 |
| | 12 | z4=2 |
| | 12 | z2w,1=1 z2w,2=1 z2w,3=1 z4=1 |
| | 12 | z2w,3=1 z4=3 |
| | 12 | z2w,3=1 z4=1 |
| | 12 | z2w,1=1 z2w,2=1 z2w,3=1 z4=3 |
| | 12 | z4=2 |
| | 12 | z2w,1=1 z2w,2=1 |
| 63.4.184 | 411623 | C2B3Nb4 |
| | 12 | z2w,1=1 z2w,2=1 z4=2 |
| | 12 | z2w,3=1 z4=1 |
| | 12 | z2w,1=1 z2w,2=1 z2w,3=1 z4=3 |
| | 12 | z2w,1=1 z2w,2=1 z2w,3=1 z4=1 |
| | 12 | z2w,3=1 z4=3 |
| | 12 | z2w,1=1 z2w,2=1 z4=2 |
| 63.4.185 | 50632 | C1Re2Si1Tb1 |
| | 12 | z2w,1=1 z2w,2=1 z4=1 |
| | 12 | z4=3 |
| | 12 | z2w,1=1 z2w,2=1 z2w,3=1 z4=2 |
| | 12 | z2w,3=1 |
| | 12 | z2w,3=1 z4=2 |
| | 12 | z2w,1=1 z2w,2=1 z2w,3=1 |
| | 12 | z4=3 |
| | 12 | z2w,1=1 z2w,2=1 z4=1 |
| 63.4.186 | 50634 | C1Ho1Ru2Si1 |
| | 12 | z4=2 |
| | 12 | z2w,1=1 z2w,2=1 |
| | 12 | z2w,3=1 z4=3 |
| | 12 | z2w,1=1 z2w,2=1 z2w,3=1 z4=1 |
| | 12 | z2w,1=1 z2w,2=1 z2w,3=1 z4=3 |
| | 12 | z2w,3=1 z4=1 |
| | 12 | z2w,1=1 z2w,2=1 |
| | 12 | z4=2 |
| 63.4.187 | 260483 | C1Gd1Os2P1 |
| | 12 | z2w,1=1 z2w,2=1 |
| | 12 | z4=2 |
| | 12 | z2w,1=1 z2w,2=1 z2w,3=1 z4=1 |
| | 12 | z2w,3=1 z4=3 |
| | 12 | z2w,3=1 z4=1 |
| | 12 | z2w,1=1 z2w,2=1 z2w,3=1 z4=3 |
| | 12 | z4=2 |
| | 12 | z2w,1=1 z2w,2=1 |
| 63.4.188 | 415220 | C1Cu1N2 |
| | 12 | z2w,1=1 z2w,2=1 z4=2 |
| | 12 | z2w,1=1 z2w,2=1 z2w,3=1 z4=3 |

Continued on next column

**TABLE XIX– continued**

| TQC | ICSD SSG | Stoich. Formula SIs |
|---|---|---|
| | 12 | z2w,3=1 z4=1 |
| | 12 | z2w,3=1 z4=3 |
| | 12 | z2w,1=1 z2w,2=1 z2w,3=1 z4=1 |
| | 12 | z2w,1=1 z2w,2=1 z4=2 |
| 63.4.189 | 161461 | C1Cu1N2 |
| | 12 | z2w,1=1 z2w,2=1 z4=2 |
| | 12 | z2w,1=1 z2w,2=1 z2w,3=1 z4=3 |
| | 12 | z2w,3=1 z4=1 |
| | 12 | z2w,3=1 z4=3 |
| | 12 | z2w,1=1 z2w,2=1 z2w,3=1 z4=1 |
| | 12 | z2w,1=1 z2w,2=1 z4=2 |
| 63.4.190 | 161460 | C1Cu1N2 |
| | 12 | z2w,1=1 z2w,2=1 z4=2 |
| | 12 | z2w,1=1 z2w,2=1 z2w,3=1 z4=3 |
| | 12 | z2w,3=1 z4=1 |
| | 12 | z2w,3=1 z4=3 |
| | 12 | z2w,1=1 z2w,2=1 z2w,3=1 z4=1 |
| | 12 | z2w,1=1 z2w,2=1 z4=2 |
| 63.4.191 | 260480 | C1Ce1P1Ru2 |
| | 12 | z4=2 |
| | 12 | z2w,1=1 z2w,2=1 |
| | 12 | z2w,3=1 z4=3 |
| | 12 | z2w,1=1 z2w,2=1 z2w,3=1 z4=1 |
| | 12 | z2w,1=1 z2w,2=1 z2w,3=1 z4=3 |
| | 12 | z2w,3=1 z4=1 |
| | 12 | z2w,1=1 z2w,2=1 |
| | 12 | z4=2 |
| 63.4.192 | 73828 | C1B1U1 |
| | 12 | z2w,1=1 z2w,2=1 z4=2 |
| | 12 | z2w,3=1 z4=1 |
| | 12 | z2w,1=1 z2w,2=1 z2w,3=1 z4=3 |
| | 12 | z2w,1=1 z2w,2=1 z2w,3=1 z4=1 |
| | 12 | z2w,3=1 z4=3 |
| | 12 | z2w,1=1 z2w,2=1 z4=2 |
| 63.4.193 | 108984 | Br3Cm1 |
| | 12 | z2w,1=1 z2w,2=1 z4=1 |
| | 12 | z4=3 |
| | 12 | z2w,1=1 z2w,2=1 z2w,3=1 z4=2 |
| | 12 | z2w,3=1 |
| | 12 | z2w,3=1 z4=2 |
| | 12 | z2w,1=1 z2w,2=1 z2w,3=1 |
| | 12 | z4=3 |
| | 12 | z2w,1=1 z2w,2=1 z4=1 |
| 63.4.194 | 1107 | Br1O2U1 |
| | 12 | z2w,1=1 z2w,2=1 z4=2 |
| | 12 | z2w,1=1 z2w,2=1 z2w,3=1 z4=3 |
| | 12 | z2w,3=1 z4=1 |
| | 12 | z2w,3=1 z4=3 |
| | 12 | z2w,1=1 z2w,2=1 z2w,3=1 z4=1 |
| | 12 | z2w,1=1 z2w,2=1 z4=2 |
| 63.4.195 | 409756 | Bi1Zr1 |
| | 12 | z4=3 |
| | 12 | z2w,1=1 z2w,2=1 z4=1 |
| | 12 | z2w,3=1 |
| | 12 | z2w,1=1 z2w,2=1 z2w,3=1 z4=2 |
| | 12 | z2w,1=1 z2w,2=1 z2w,3=1 |
| | 12 | z2w,3=1 z4=2 |
| | 12 | z2w,1=1 z2w,2=1 z4=1 |
| | 12 | z4=3 |

Continued on next column

**TABLE XIX– continued**

| TQC | ICSD SSG | Stoich. Formula SIs |
|---|---|---|
| 63.4.196 | 50826 | Ba3Cu1O9Ru2 |
| | 12 | z2w,1=1 z2w,2=1 z4=2 |
| | 12 | z2w,1=1 z2w,2=1 z2w,3=1 z4=3 |
| | 12 | z2w,3=1 z4=1 |
| | 12 | z2w,3=1 z4=3 |
| | 12 | z2w,1=1 z2w,2=1 z2w,3=1 z4=1 |
| | 12 | z2w,1=1 z2w,2=1 z4=2 |
| 63.4.197 | 50825 | Ba3Cu1O9Ru2 |
| | 12 | z2w,1=1 z2w,2=1 z4=2 |
| | 12 | z2w,1=1 z2w,2=1 z2w,3=1 z4=3 |
| | 12 | z2w,3=1 z4=1 |
| | 12 | z2w,3=1 z4=3 |
| | 12 | z2w,1=1 z2w,2=1 z2w,3=1 z4=1 |
| | 12 | z2w,1=1 z2w,2=1 z4=2 |
| 63.4.198 | 63228 | Ba1S3V1 |
| | 12 | z2w,1=1 z2w,2=1 z4=2 |
| | 12 | z2w,1=1 z2w,2=1 z2w,3=1 z4=3 |
| | 12 | z2w,3=1 z4=1 |
| | 12 | z2w,3=1 z4=3 |
| | 12 | z2w,1=1 z2w,2=1 z2w,3=1 z4=1 |
| | 12 | z2w,1=1 z2w,2=1 z4=2 |
| 63.4.199 | 58669 | Ba1Pb6Rh3 |
| | 12 | z4=2 |
| | 12 | z2w,1=1 z2w,2=1 |
| | 12 | z2w,3=1 z4=3 |
| | 12 | z2w,1=1 z2w,2=1 z2w,3=1 z4=1 |
| | 12 | z2w,1=1 z2w,2=1 z2w,3=1 z4=3 |
| | 12 | z2w,3=1 z4=1 |
| | 12 | z2w,1=1 z2w,2=1 |
| | 12 | z4=2 |
| 63.4.200 | 195765 | Ba1La1Si2 |
| | 12 | z2w,1=1 z2w,2=1 z4=2 |
| | 12 | z2w,1=1 z2w,2=1 z2w,3=1 z4=3 |
| | 12 | z2w,3=1 z4=1 |
| | 12 | z2w,3=1 z4=3 |
| | 12 | z2w,1=1 z2w,2=1 z2w,3=1 z4=1 |
| | 12 | z2w,1=1 z2w,2=1 z4=2 |
| 63.4.201 | 411232 | Ba1In2Rh1 |
| | 12 | z2w,1=1 z2w,2=1 |
| | 12 | z4=2 |
| | 12 | z2w,1=1 z2w,2=1 z2w,3=1 z4=1 |
| | 12 | z2w,3=1 z4=3 |
| | 12 | z2w,3=1 z4=1 |
| | 12 | z2w,1=1 z2w,2=1 z2w,3=1 z4=3 |
| | 12 | z4=2 |
| | 12 | z2w,1=1 z2w,2=1 |
| 63.4.202 | 414434 | Ba1In2Ir1 |
| | 12 | z4=2 |
| | 12 | z2w,1=1 z2w,2=1 |
| | 12 | z2w,3=1 z4=3 |
| | 12 | z2w,1=1 z2w,2=1 z2w,3=1 z4=1 |
| | 12 | z2w,1=1 z2w,2=1 z2w,3=1 z4=3 |
| | 12 | z2w,3=1 z4=1 |
| | 12 | z2w,1=1 z2w,2=1 |
| | 12 | z4=2 |
| 63.4.203 | 58648 | Ba1Cu1Sn2 |
| | 12 | z4=3 |
| | 12 | z2w,1=1 z2w,2=1 z4=1 |
| | 12 | z2w,3=1 |
| | 12 | z2w,1=1 z2w,2=1 z2w,3=1 z4=2 |
| | 12 | z2w,1=1 z2w,2=1 z2w,3=1 |
| | 12 | z2w,3=1 z4=2 |
| | 12 | z2w,1=1 z2w,2=1 z4=1 |
| | 12 | z4=3 |
| 63.4.204 | 429001 | Ba1Cu1Gd1Te3 |
| | 12 | z2w,1=1 z2w,2=1 z4=2 |
| | 12 | z2w,1=1 z2w,2=1 z2w,3=1 z4=3 |
| | 12 | z2w,3=1 z4=1 |
| | 12 | z2w,3=1 z4=3 |
| | 12 | z2w,1=1 z2w,2=1 z2w,3=1 z4=1 |
| | 12 | z2w,1=1 z2w,2=1 z4=2 |
| 63.4.205 | 280698 | Ba1Cu1Gd1Se3 |
| | 12 | z2w,1=1 z2w,2=1 z4=2 |
| | 12 | z2w,3=1 z4=1 |
| | 12 | z2w,1=1 z2w,2=1 z2w,3=1 z4=3 |
| | 12 | z2w,1=1 z2w,2=1 z2w,3=1 z4=1 |
| | 12 | z2w,3=1 z4=3 |
| | 12 | z2w,1=1 z2w,2=1 z4=2 |
| 63.4.206 | 78176 | Ba1Cu1Er1S3 |
| | 12 | z2w,1=1 z2w,2=1 z4=2 |
| | 12 | z2w,1=1 z2w,2=1 z2w,3=1 z4=3 |
| | 12 | z2w,3=1 z4=1 |
| | 12 | z2w,3=1 z4=3 |
| | 12 | z2w,1=1 z2w,2=1 z2w,3=1 z4=1 |
| | 12 | z2w,1=1 z2w,2=1 z4=2 |
| 63.4.207 | 428663 | Ba1Co1O1S1 |
| | 12 | z2w,1=1 z2w,2=1 z4=2 |
| | 12 | z2w,1=1 z2w,2=1 z2w,3=1 z4=3 |
| | 12 | z2w,3=1 z4=1 |
| | 12 | z2w,3=1 z4=3 |
| | 12 | z2w,1=1 z2w,2=1 z2w,3=1 z4=1 |
| | 12 | z2w,1=1 z2w,2=1 z4=2 |
| 63.4.208 | 428516 | Ba1Co1O1S1 |
| | 12 | z2w,1=1 z2w,2=1 z4=2 |
| | 12 | z2w,3=1 z4=1 |
| | 12 | z2w,1=1 z2w,2=1 z2w,3=1 z4=3 |
| | 12 | z2w,1=1 z2w,2=1 z2w,3=1 z4=1 |
| | 12 | z2w,3=1 z4=3 |
| | 12 | z2w,1=1 z2w,2=1 z4=2 |
| 63.4.209 | 64595 | B7Re1Y3 |
| | 12 | z2w,1=1 z2w,2=1 |
| | 12 | z4=2 |
| | 12 | z2w,1=1 z2w,2=1 z2w,3=1 z4=1 |
| | 12 | z2w,3=1 z4=3 |
| | 12 | z2w,3=1 z4=1 |
| | 12 | z2w,1=1 z2w,2=1 z2w,3=1 z4=3 |
| | 12 | z4=2 |
| | 12 | z2w,1=1 z2w,2=1 |
| 63.4.210 | 62630 | B7Cr1Er3 |
| | 12 | z4=2 |
| | 12 | z2w,1=1 z2w,2=1 |
| | 12 | z2w,3=1 z4=3 |
| | 12 | z2w,1=1 z2w,2=1 z2w,3=1 z4=1 |
| | 12 | z2w,1=1 z2w,2=1 z2w,3=1 z4=3 |
| | 12 | z2w,3=1 z4=1 |
| | 12 | z2w,1=1 z2w,2=1 |
| | 12 | z4=2 |
| 63.4.211 | 91244 | B4Os3Sc1 |

Continued on next column

**TABLE XIX– continued**

| TQC | ICSD SSG | Stoich. Formula SIs |
|---|---|---|
| | 12 | z4=1 |
| | 12 | z2w,1=1 z2w,2=1 z4=3 |
| | 12 | z2w,3=1 z4=2 |
| | 12 | z2w,1=1 z2w,2=1 z2w,3=1 |
| | 12 | z2w,1=1 z2w,2=1 z2w,3=1 z4=2 |
| | 12 | z2w,3=1 |
| | 12 | z2w,1=1 z2w,2=1 z4=3 |
| | 12 | z4=1 |
| 63.4.212 | 391409 | B4Mn1O7 |
| | 12 | z2w,1=1 z2w,2=1 z4=2 |
| | 12 | z2w,1=1 z2w,2=1 z2w,3=1 z4=3 |
| | 12 | z2w,3=1 z4=1 |
| | 12 | z2w,3=1 z4=3 |
| | 12 | z2w,1=1 z2w,2=1 z2w,3=1 z4=1 |
| | 12 | z2w,1=1 z2w,2=1 z4=2 |
| 63.4.213 | 391408 | B4Cu1O7 |
| | 12 | z2w,1=1 z2w,2=1 z4=2 |
| | 12 | z2w,3=1 z4=1 |
| | 12 | z2w,1=1 z2w,2=1 z2w,3=1 z4=3 |
| | 12 | z2w,1=1 z2w,2=1 z2w,3=1 z4=1 |
| | 12 | z2w,3=1 z4=3 |
| | 12 | z2w,1=1 z2w,2=1 z4=2 |
| 63.4.214 | 420402 | B4Co1O7 |
| | 12 | z2w,1=1 z2w,2=1 z4=2 |
| | 12 | z2w,1=1 z2w,2=1 z2w,3=1 z4=3 |
| | 12 | z2w,3=1 z4=1 |
| | 12 | z2w,3=1 z4=3 |
| | 12 | z2w,1=1 z2w,2=1 z2w,3=1 z4=1 |
| | 12 | z2w,1=1 z2w,2=1 z4=2 |
| 63.4.215 | 79258 | B3V2 |
| | 12 | z2w,1=1 z2w,2=1 z4=2 |
| | 12 | z2w,3=1 z4=1 |
| | 12 | z2w,1=1 z2w,2=1 z2w,3=1 z4=3 |
| | 12 | z2w,1=1 z2w,2=1 z2w,3=1 z4=1 |
| | 12 | z2w,3=1 z4=3 |
| | 12 | z2w,1=1 z2w,2=1 z4=2 |
| 63.4.216 | 71643 | B1Ir2Zn2 |
| | 12 | z2w,1=1 z2w,2=1 z4=2 |
| | 12 | z2w,3=1 z4=1 |
| | 12 | z2w,1=1 z2w,2=1 z2w,3=1 z4=3 |
| | 12 | z2w,1=1 z2w,2=1 z2w,3=1 z4=1 |
| | 12 | z2w,3=1 z4=3 |
| | 12 | z2w,1=1 z2w,2=1 z4=2 |
| 63.4.217 | 44249 | B1Cr1 |
| | 12 | z2w,1=1 z2w,2=1 |
| | 12 | z4=2 |
| | 12 | z2w,1=1 z2w,2=1 z2w,3=1 z4=1 |
| | 12 | z2w,3=1 z4=3 |
| | 12 | z2w,3=1 z4=1 |
| | 12 | z2w,1=1 z2w,2=1 z2w,3=1 z4=3 |
| | 12 | z4=2 |
| | 12 | z2w,1=1 z2w,2=1 |
| 63.4.218 | 410436 | Au1In2Yb1 |
| | 12 | z2w,1=1 z2w,2=1 |
| | 12 | z4=2 |
| | 12 | z2w,1=1 z2w,2=1 z2w,3=1 z4=1 |
| | 12 | z2w,3=1 z4=3 |
| | 12 | z2w,3=1 z4=1 |
| | 12 | z2w,1=1 z2w,2=1 z2w,3=1 z4=3 |

Continued on next column

**TABLE XIX– continued**

| TQC | ICSD SSG | Stoich. Formula SIs |
|---|---|---|
| | 12 | z4=2 |
| | 12 | z2w,1=1 z2w,2=1 |
| 63.4.219 | 408882 | Au1Ca1In2 |
| | 12 | z2w,1=1 z2w,2=1 |
| | 12 | z4=2 |
| | 12 | z2w,1=1 z2w,2=1 z2w,3=1 z4=1 |
| | 12 | z2w,3=1 z4=3 |
| | 12 | z2w,3=1 z4=1 |
| | 12 | z2w,1=1 z2w,2=1 z2w,3=1 z4=3 |
| | 12 | z4=2 |
| | 12 | z2w,1=1 z2w,2=1 |
| 63.4.220 | 88718 | Au1Ba1Gd1Se3 |
| | 12 | z2w,1=1 z2w,2=1 z4=2 |
| | 12 | z2w,1=1 z2w,2=1 z2w,3=1 z4=3 |
| | 12 | z2w,3=1 z4=1 |
| | 12 | z2w,3=1 z4=3 |
| | 12 | z2w,1=1 z2w,2=1 z2w,3=1 z4=1 |
| | 12 | z2w,1=1 z2w,2=1 z4=2 |
| 63.4.221 | 33915 | As3Ni5Sr1 |
| | 12 | z4=2 |
| | 12 | z2w,1=1 z2w,2=1 |
| | 12 | z2w,3=1 z4=3 |
| | 12 | z2w,1=1 z2w,2=1 z2w,3=1 z4=1 |
| | 12 | z2w,1=1 z2w,2=1 z2w,3=1 z4=3 |
| | 12 | z2w,3=1 z4=1 |
| | 12 | z2w,1=1 z2w,2=1 |
| | 12 | z4=2 |
| 63.4.222 | 57973 | Al6Mn1 |
| | 12 | z2w,1=1 z2w,2=1 z4=1 |
| | 12 | z4=3 |
| | 12 | z2w,1=1 z2w,2=1 z2w,3=1 z4=2 |
| | 12 | z2w,3=1 |
| | 12 | z2w,3=1 z4=2 |
| | 12 | z2w,1=1 z2w,2=1 z2w,3=1 |
| | 12 | z4=3 |
| | 12 | z2w,1=1 z2w,2=1 z4=1 |
| 63.4.223 | 154855 | Al4Ni1Tb1 |
| | 12 | z4=3 |
| | 12 | z2w,1=1 z2w,2=1 z4=1 |
| | 12 | z2w,3=1 |
| | 12 | z2w,1=1 z2w,2=1 z2w,3=1 z4=2 |
| | 12 | z2w,1=1 z2w,2=1 z2w,3=1 |
| | 12 | z2w,3=1 z4=2 |
| | 12 | z2w,1=1 z2w,2=1 z4=1 |
| | 12 | z4=3 |
| 63.4.224 | 370012 | Al2Ni1Sc1 |
| | 12 | z4=2 |
| | 12 | z2w,1=1 z2w,2=1 |
| | 12 | z2w,3=1 z4=3 |
| | 12 | z2w,1=1 z2w,2=1 z2w,3=1 z4=1 |
| | 12 | z2w,1=1 z2w,2=1 z2w,3=1 z4=3 |
| | 12 | z2w,3=1 z4=1 |
| | 12 | z2w,1=1 z2w,2=1 |
| | 12 | z4=2 |
| 63.4.225 | 415062 | Al2Cu1Mg1 |
| | 12 | z4=3 |
| | 12 | z2w,1=1 z2w,2=1 z4=1 |
| | 12 | z2w,3=1 |
| | 12 | z2w,1=1 z2w,2=1 z2w,3=1 z4=2 |

Continued on next column

**TABLE XIX– continued**

| TQC | ICSD<br>SSG | Stoich. Formula<br>SIs |
|---|---|---|
| | 12 | z2w,1=1 z2w,2=1 z2w,3=1 |
| | 12 | z2w,3=1 z4=2 |
| | 12 | z2w,1=1 z2w,2=1 z4=1 |
| | 12 | z4=3 |
| 63.4.226 | 370003 | Ag7Yb2 |
| | 12 | z2w,1=1 z2w,2=1 |
| | 12 | z4=2 |
| | 12 | z2w,1=1 z2w,2=1 z2w,3=1 z4=1 |
| | 12 | z2w,3=1 z4=3 |
| | 12 | z2w,3=1 z4=1 |
| | 12 | z2w,1=1 z2w,2=1 z2w,3=1 z4=3 |
| | 12 | z4=2 |
| | 12 | z2w,1=1 z2w,2=1 |
| 63.4.227 | 55510 | Ag7Ca2 |
| | 12 | z4=3 |
| | 12 | z2w,1=1 z2w,2=1 z4=1 |
| | 12 | z2w,3=1 |
| | 12 | z2w,1=1 z2w,2=1 z2w,3=1 z4=2 |
| | 12 | z2w,1=1 z2w,2=1 z2w,3=1 |
| | 12 | z2w,3=1 z4=2 |
| | 12 | z2w,1=1 z2w,2=1 z4=1 |
| | 12 | z4=3 |
| 63.4.228 | 247421 | Ag1Ce1Si2 |
| | 12 | z4=3 |
| | 12 | z2w,1=1 z2w,2=1 z4=1 |
| | 12 | z2w,3=1 |
| | 12 | z2w,1=1 z2w,2=1 z2w,3=1 z4=2 |
| | 12 | z2w,1=1 z2w,2=1 z2w,3=1 |
| | 12 | z2w,3=1 z4=2 |
| | 12 | z2w,1=1 z2w,2=1 z4=1 |
| | 12 | z4=3 |
| 63.4.229 | 57353 | Ag1Ca1 |
| | 12 | z2w,1=1 z2w,2=1 z4=2 |
| | 12 | z2w,3=1 z4=1 |
| | 12 | z2w,1=1 z2w,2=1 z2w,3=1 z4=3 |
| | 12 | z2w,1=1 z2w,2=1 z2w,3=1 z4=1 |
| | 12 | z2w,3=1 z4=3 |
| | 12 | z2w,1=1 z2w,2=1 z4=2 |
| 63.4.230 | 54979 | Ag1Ca1 |
| | 12 | z2w,1=1 z2w,2=1 z4=2 |
| | 12 | z2w,3=1 z4=1 |
| | 12 | z2w,1=1 z2w,2=1 z2w,3=1 z4=3 |
| | 12 | z2w,1=1 z2w,2=1 z2w,3=1 z4=1 |
| | 12 | z2w,3=1 z4=3 |
| | 12 | z2w,1=1 z2w,2=1 z4=2 |
| 64.4.1 | 78300 | Cu1La2O4 |
| | 12 | z2w,1=1 z2w,2=1 z4=2 |
| | 12 | z2w,1=1 z2w,2=1 z2w,3=1 z4=3 |
| | 12 | z2w,3=1 z4=1 |
| | 12 | z2w,3=1 z4=3 |
| | 12 | z2w,1=1 z2w,2=1 z2w,3=1 z4=1 |
| | 12 | z2w,1=1 z2w,2=1 z4=2 |
| 64.4.2 | 78299 | Cu1La2O4 |
| | 12 | z2w,1=1 z2w,2=1 z4=2 |
| | 12 | z2w,3=1 z4=1 |
| | 12 | z2w,1=1 z2w,2=1 z2w,3=1 z4=3 |
| | 12 | z2w,1=1 z2w,2=1 z2w,3=1 z4=1 |
| | 12 | z2w,3=1 z4=3 |
| | 12 | z2w,1=1 z2w,2=1 z4=2 |

Continued on next column

**TABLE XIX– continued**

| TQC | ICSD<br>SSG | Stoich. Formula<br>SIs |
|---|---|---|
| 64.4.3 | 78298 | Cu1La2O4 |
| | 12 | z2w,1=1 z2w,2=1 z4=2 |
| | 12 | z2w,3=1 z4=1 |
| | 12 | z2w,1=1 z2w,2=1 z2w,3=1 z4=3 |
| | 12 | z2w,1=1 z2w,2=1 z2w,3=1 z4=1 |
| | 12 | z2w,3=1 z4=3 |
| | 12 | z2w,1=1 z2w,2=1 z4=2 |
| 64.4.4 | 78297 | Cu1La2O4 |
| | 12 | z2w,1=1 z2w,2=1 z4=2 |
| | 12 | z2w,1=1 z2w,2=1 z2w,3=1 z4=3 |
| | 12 | z2w,3=1 z4=1 |
| | 12 | z2w,3=1 z4=3 |
| | 12 | z2w,1=1 z2w,2=1 z2w,3=1 z4=1 |
| | 12 | z2w,1=1 z2w,2=1 z4=2 |
| 64.4.5 | 78296 | Cu1La2O4 |
| | 12 | z2w,1=1 z2w,2=1 z4=2 |
| | 12 | z2w,3=1 z4=1 |
| | 12 | z2w,1=1 z2w,2=1 z2w,3=1 z4=3 |
| | 12 | z2w,1=1 z2w,2=1 z2w,3=1 z4=1 |
| | 12 | z2w,3=1 z4=3 |
| | 12 | z2w,1=1 z2w,2=1 z4=2 |
| 64.4.6 | 78295 | Cu1La2O4 |
| | 12 | z2w,1=1 z2w,2=1 z4=2 |
| | 12 | z2w,3=1 z4=1 |
| | 12 | z2w,1=1 z2w,2=1 z2w,3=1 z4=3 |
| | 12 | z2w,1=1 z2w,2=1 z2w,3=1 z4=1 |
| | 12 | z2w,3=1 z4=3 |
| | 12 | z2w,1=1 z2w,2=1 z4=2 |
| 64.4.7 | 78294 | Cu1La2O4 |
| | 12 | z2w,1=1 z2w,2=1 z4=2 |
| | 12 | z2w,3=1 z4=1 |
| | 12 | z2w,1=1 z2w,2=1 z2w,3=1 z4=3 |
| | 12 | z2w,1=1 z2w,2=1 z2w,3=1 z4=1 |
| | 12 | z2w,3=1 z4=3 |
| | 12 | z2w,1=1 z2w,2=1 z4=2 |
| 64.4.8 | 78293 | Cu1La2O4 |
| | 12 | z2w,1=1 z2w,2=1 z4=2 |
| | 12 | z2w,1=1 z2w,2=1 z2w,3=1 z4=3 |
| | 12 | z2w,3=1 z4=1 |
| | 12 | z2w,3=1 z4=3 |
| | 12 | z2w,1=1 z2w,2=1 z2w,3=1 z4=1 |
| | 12 | z2w,1=1 z2w,2=1 z4=2 |
| 64.4.9 | 78292 | Cu1La2O4 |
| | 12 | z2w,1=1 z2w,2=1 z4=2 |
| | 12 | z2w,3=1 z4=1 |
| | 12 | z2w,1=1 z2w,2=1 z2w,3=1 z4=3 |
| | 12 | z2w,1=1 z2w,2=1 z2w,3=1 z4=1 |
| | 12 | z2w,3=1 z4=3 |
| | 12 | z2w,1=1 z2w,2=1 z4=2 |
| 64.4.10 | 78291 | Cu1La2O4 |
| | 12 | z2w,1=1 z2w,2=1 z4=2 |
| | 12 | z2w,1=1 z2w,2=1 z2w,3=1 z4=3 |
| | 12 | z2w,3=1 z4=1 |
| | 12 | z2w,3=1 z4=3 |
| | 12 | z2w,1=1 z2w,2=1 z2w,3=1 z4=1 |
| | 12 | z2w,1=1 z2w,2=1 z4=2 |
| 64.4.11 | 78290 | Cu1La2O4 |
| | 12 | z2w,1=1 z2w,2=1 z4=2 |
| | 12 | z2w,1=1 z2w,2=1 z2w,3=1 z4=3 |

Continued on next column

**TABLE XIX– continued**

| TQC | ICSD<br>SSG | Stoich. Formula<br>SIs |
|---|---|---|
| | 12 | z2w,3=1 z4=1 |
| | 12 | z2w,3=1 z4=3 |
| | 12 | z2w,1=1 z2w,2=1 z2w,3=1 z4=1 |
| | 12 | z2w,1=1 z2w,2=1 z4=2 |
| 64.4.12 | 78289 | Cu1La2O4 |
| | 12 | z2w,1=1 z2w,2=1 z4=2 |
| | 12 | z2w,3=1 z4=1 |
| | 12 | z2w,1=1 z2w,2=1 z2w,3=1 z4=3 |
| | 12 | z2w,1=1 z2w,2=1 z2w,3=1 z4=1 |
| | 12 | z2w,3=1 z4=3 |
| | 12 | z2w,1=1 z2w,2=1 z4=2 |
| 64.4.13 | 78288 | Cu1La2O4 |
| | 12 | z2w,1=1 z2w,2=1 z4=2 |
| | 12 | z2w,3=1 z4=1 |
| | 12 | z2w,1=1 z2w,2=1 z2w,3=1 z4=3 |
| | 12 | z2w,1=1 z2w,2=1 z2w,3=1 z4=1 |
| | 12 | z2w,3=1 z4=3 |
| | 12 | z2w,1=1 z2w,2=1 z4=2 |
| 64.4.14 | 78287 | Cu1La2O4 |
| | 12 | z2w,1=1 z2w,2=1 z4=2 |
| | 12 | z2w,1=1 z2w,2=1 z2w,3=1 z4=3 |
| | 12 | z2w,3=1 z4=1 |
| | 12 | z2w,3=1 z4=3 |
| | 12 | z2w,1=1 z2w,2=1 z2w,3=1 z4=1 |
| | 12 | z2w,1=1 z2w,2=1 z4=2 |
| 64.4.15 | 78237 | Cu1La2O4 |
| | 12 | z2w,1=1 z2w,2=1 z4=2 |
| | 12 | z2w,1=1 z2w,2=1 z2w,3=1 z4=3 |
| | 12 | z2w,3=1 z4=1 |
| | 12 | z2w,3=1 z4=3 |
| | 12 | z2w,1=1 z2w,2=1 z2w,3=1 z4=1 |
| | 12 | z2w,1=1 z2w,2=1 z4=2 |
| 64.4.16 | 78194 | Cu1La2O4 |
| | 12 | z2w,1=1 z2w,2=1 z4=2 |
| | 12 | z2w,1=1 z2w,2=1 z2w,3=1 z4=3 |
| | 12 | z2w,3=1 z4=1 |
| | 12 | z2w,3=1 z4=3 |
| | 12 | z2w,1=1 z2w,2=1 z2w,3=1 z4=1 |
| | 12 | z2w,1=1 z2w,2=1 z4=2 |
| 64.4.17 | 75891 | Cu1La2O4 |
| | 12 | z2w,1=1 z2w,2=1 z4=2 |
| | 12 | z2w,3=1 z4=1 |
| | 12 | z2w,1=1 z2w,2=1 z2w,3=1 z4=3 |
| | 12 | z2w,1=1 z2w,2=1 z2w,3=1 z4=1 |
| | 12 | z2w,3=1 z4=3 |
| | 12 | z2w,1=1 z2w,2=1 z4=2 |
| 64.4.18 | 74150 | Cu1La2O4 |
| | 12 | z2w,1=1 z2w,2=1 z4=2 |
| | 12 | z2w,3=1 z4=1 |
| | 12 | z2w,1=1 z2w,2=1 z2w,3=1 z4=3 |
| | 12 | z2w,1=1 z2w,2=1 z2w,3=1 z4=1 |
| | 12 | z2w,3=1 z4=3 |
| | 12 | z2w,1=1 z2w,2=1 z4=2 |
| 64.4.19 | 74145 | Cu1La2O4 |
| | 12 | z2w,1=1 z2w,2=1 z4=2 |
| | 12 | z2w,1=1 z2w,2=1 z2w,3=1 z4=3 |
| | 12 | z2w,3=1 z4=1 |
| | 12 | z2w,3=1 z4=3 |
| | 12 | z2w,1=1 z2w,2=1 z2w,3=1 z4=1 |

Continued on next column

**TABLE XIX– continued**

| TQC | ICSD<br>SSG | Stoich. Formula<br>SIs |
|---|---|---|
| | 12 | z2w,1=1 z2w,2=1 z4=2 |
| 64.4.20 | 69312 | Cu1La2O4 |
| | 12 | z2w,1=1 z2w,2=1 z4=2 |
| | 12 | z2w,1=1 z2w,2=1 z2w,3=1 z4=3 |
| | 12 | z2w,3=1 z4=1 |
| | 12 | z2w,3=1 z4=3 |
| | 12 | z2w,1=1 z2w,2=1 z2w,3=1 z4=1 |
| | 12 | z2w,1=1 z2w,2=1 z4=2 |
| 64.4.21 | 69311 | Cu1La2O4 |
| | 12 | z2w,1=1 z2w,2=1 z4=2 |
| | 12 | z2w,3=1 z4=1 |
| | 12 | z2w,1=1 z2w,2=1 z2w,3=1 z4=3 |
| | 12 | z2w,1=1 z2w,2=1 z2w,3=1 z4=1 |
| | 12 | z2w,3=1 z4=3 |
| | 12 | z2w,1=1 z2w,2=1 z4=2 |
| 64.4.22 | 50265 | Cu1La2O4 |
| | 12 | z2w,1=1 z2w,2=1 z4=2 |
| | 12 | z2w,3=1 z4=1 |
| | 12 | z2w,1=1 z2w,2=1 z2w,3=1 z4=3 |
| | 12 | z2w,1=1 z2w,2=1 z2w,3=1 z4=1 |
| | 12 | z2w,3=1 z4=3 |
| | 12 | z2w,1=1 z2w,2=1 z4=2 |
| 64.4.23 | 261661 | Cu1La2O4 |
| | 12 | z2w,1=1 z2w,2=1 z4=2 |
| | 12 | z2w,1=1 z2w,2=1 z2w,3=1 z4=3 |
| | 12 | z2w,3=1 z4=1 |
| | 12 | z2w,3=1 z4=3 |
| | 12 | z2w,1=1 z2w,2=1 z2w,3=1 z4=1 |
| | 12 | z2w,1=1 z2w,2=1 z4=2 |
| 64.4.24 | 173895 | Cu1La2O4 |
| | 12 | z2w,1=1 z2w,2=1 z4=2 |
| | 12 | z2w,1=1 z2w,2=1 z2w,3=1 z4=3 |
| | 12 | z2w,3=1 z4=1 |
| | 12 | z2w,3=1 z4=3 |
| | 12 | z2w,1=1 z2w,2=1 z2w,3=1 z4=1 |
| | 12 | z2w,1=1 z2w,2=1 z4=2 |
| 64.4.25 | 173893 | Cu1La2O4 |
| | 12 | z2w,1=1 z2w,2=1 z4=2 |
| | 12 | z2w,1=1 z2w,2=1 z2w,3=1 z4=3 |
| | 12 | z2w,3=1 z4=1 |
| | 12 | z2w,3=1 z4=3 |
| | 12 | z2w,1=1 z2w,2=1 z2w,3=1 z4=1 |
| | 12 | z2w,1=1 z2w,2=1 z4=2 |
| 64.4.26 | 173891 | Cu1La2O4 |
| | 12 | z2w,1=1 z2w,2=1 z4=2 |
| | 12 | z2w,1=1 z2w,2=1 z2w,3=1 z4=3 |
| | 12 | z2w,3=1 z4=1 |
| | 12 | z2w,3=1 z4=3 |
| | 12 | z2w,1=1 z2w,2=1 z2w,3=1 z4=1 |
| | 12 | z2w,1=1 z2w,2=1 z4=2 |
| 64.4.27 | 202724 | Ba2Cu1O6Tl2 |
| | 12 | z2w,1=1 z2w,2=1 z4=2 |
| | 12 | z2w,3=1 z4=1 |
| | 12 | z2w,1=1 z2w,2=1 z2w,3=1 z4=3 |
| | 12 | z2w,1=1 z2w,2=1 z2w,3=1 z4=1 |
| | 12 | z2w,3=1 z4=3 |
| | 12 | z2w,1=1 z2w,2=1 z4=2 |
| 64.4.28 | 202722 | Ba2Cu1O6Tl2 |
| | 12 | z2w,1=1 z2w,2=1 z4=2 |

Continued on next column

**TABLE XIX– continued**

| TQC | ICSD<br>SSG | Stoich. Formula<br>SIs |
|---|---|---|
| | 12 | z2w,1=1 z2w,2=1 z2w,3=1 z4=3 |
| | 12 | z2w,3=1 z4=1 |
| | 12 | z2w,3=1 z4=3 |
| | 12 | z2w,1=1 z2w,2=1 z2w,3=1 z4=1 |
| | 12 | z2w,1=1 z2w,2=1 z4=2 |
| 64.4.29 | 76359 | Al3Ge4La2 |
| | 12 | z2w,1=1 z2w,2=1 z4=2 |
| | 12 | z2w,1=1 z2w,2=1 z2w,3=1 z4=3 |
| | 12 | z2w,3=1 z4=1 |
| | 12 | z2w,3=1 z4=3 |
| | 12 | z2w,1=1 z2w,2=1 z2w,3=1 z4=1 |
| | 12 | z2w,1=1 z2w,2=1 z4=2 |
| 64.4.30 | 50977 | Al3Ce2Ge4 |
| | 12 | z4=3 |
| | 12 | z2w,1=1 z2w,2=1 z4=1 |
| | 12 | z2w,3=1 |
| | 12 | z2w,1=1 z2w,2=1 z2w,3=1 z4=2 |
| | 12 | z2w,1=1 z2w,2=1 z2w,3=1 |
| | 12 | z2w,3=1 z4=2 |
| | 12 | z2w,1=1 z2w,2=1 z4=1 |
| | 12 | z4=3 |
| 67.4.1 | 411905 | Ba2Na1O1 |
| | 12 | z4=3 |
| | 12 | z2w,1=1 z2w,2=1 z4=1 |
| | 12 | z2w,1=1 z2w,2=1 z4=1 |
| | 12 | z4=3 |
| | 12 | z2w,3=1 z4=1 |
| | 12 | z2w,1=1 z2w,2=1 z2w,3=1 z4=3 |
| | 12 | z2w,1=1 z2w,2=1 z2w,3=1 z4=3 |
| | 12 | z2w,3=1 z4=1 |
| 67.4.2 | 59306 | Ba1F6Tb1 |
| | 12 | z2w,3=1 z4=2 |
| | 12 | z2w,1=1 z2w,2=1 z2w,3=1 |
| | 12 | z2w,1=1 z2w,2=1 z2w,3=1 |
| | 12 | z2w,3=1 z4=2 |
| | 12 | z2w,1=1 z2w,2=1 z4=2 |
| | 12 | z2w,1=1 z2w,2=1 z4=2 |
| 70.4.1 | 73610 | O10U1V3 |
| | 2 | z2w,1=1 z4=2 |
| | 2 | z2w,2=1 z4=1 |
| | 2 | z2w,3=1 z4=1 |
| | 2 | z2w,1=1 z2w,2=1 z2w,3=1 |
| | 2 | z2w,1=1 z2w,2=1 z2w,3=1 z4=1 |
| | 2 | z2w,3=1 |
| | 2 | z2w,2=1 |
| | 2 | z2w,1=1 z4=3 |
| 70.4.2 | 84406 | Nb3O10U1 |
| | 2 | z2w,1=1 |
| | 2 | z2w,2=1 z4=3 |
| | 2 | z2w,3=1 z4=3 |
| | 2 | z2w,1=1 z2w,2=1 z2w,3=1 z4=2 |
| | 2 | z2w,1=1 z2w,2=1 z2w,3=1 z4=3 |
| | 2 | z2w,3=1 z4=2 |
| | 2 | z2w,2=1 z4=2 |
| | 2 | z2w,1=1 z4=1 |
| 70.4.3 | 94466 | Fe2K3O4 |
| | 2 | z2w,1=1 |
| | 2 | z2w,2=1 z4=3 |
| | 2 | z2w,3=1 z4=3 |

Continued on next column

**TABLE XIX– continued**

| TQC | ICSD<br>SSG | Stoich. Formula<br>SIs |
|---|---|---|
| | 2 | z2w,1=1 z2w,2=1 z2w,3=1 z4=2 |
| | 2 | z2w,1=1 z2w,2=1 z2w,3=1 z4=3 |
| | 2 | z2w,3=1 z4=2 |
| | 2 | z2w,2=1 z4=2 |
| | 2 | z2w,1=1 z4=1 |
| 70.4.4 | 40621 | B2Ir2Y1 |
| | 2 | z2w,1=1 |
| | 2 | z2w,2=1 z4=3 |
| | 2 | z2w,3=1 z4=3 |
| | 2 | z2w,1=1 z2w,2=1 z2w,3=1 z4=2 |
| | 2 | z2w,1=1 z2w,2=1 z2w,3=1 z4=3 |
| | 2 | z2w,3=1 z4=2 |
| | 2 | z2w,2=1 z4=2 |
| | 2 | z2w,1=1 z4=1 |
| 70.4.5 | 40619 | B2Eu1Ir2 |
| | 2 | z2w,1=1 |
| | 2 | z2w,2=1 z4=3 |
| | 2 | z2w,3=1 z4=3 |
| | 2 | z2w,1=1 z2w,2=1 z2w,3=1 z4=2 |
| | 2 | z2w,1=1 z2w,2=1 z2w,3=1 z4=3 |
| | 2 | z2w,3=1 z4=2 |
| | 2 | z2w,2=1 z4=2 |
| | 2 | z2w,1=1 z4=1 |
| 72.4.1 | 65460 | Mn1Rb2Te2 |
| | 12 | z2w,1=1 z2w,2=1 z4=2 |
| | 12 | z2w,1=1 z2w,2=1 z4=2 |
| 72.4.2 | 65457 | Mn1Rb2Se2 |
| | 12 | z2w,1=1 z2w,2=1 z4=2 |
| | 12 | z2w,1=1 z2w,2=1 z4=2 |
| 72.4.3 | 65454 | Mn1Rb2S2 |
| | 12 | z2w,1=1 z2w,2=1 z4=2 |
| | 12 | z2w,1=1 z2w,2=1 z4=2 |
| 72.4.4 | 40427 | La2Rh3Si5 |
| | 12 | z4=1 |
| | 12 | z2w,1=1 z2w,2=1 z4=3 |
| | 12 | z2w,1=1 z2w,2=1 z4=3 |
| | 12 | z4=1 |
| 72.4.5 | 65459 | K2Mn1Te2 |
| | 12 | z2w,1=1 z2w,2=1 z4=2 |
| | 12 | z2w,1=1 z2w,2=1 z4=2 |
| 72.4.6 | 65456 | K2Mn1Se2 |
| | 12 | z2w,1=1 z2w,2=1 z4=2 |
| | 12 | z2w,1=1 z2w,2=1 z4=2 |
| 72.4.7 | 65453 | K2Mn1S2 |
| | 12 | z2w,1=1 z2w,2=1 z4=2 |
| | 12 | z2w,1=1 z2w,2=1 z4=2 |
| 72.4.8 | 86502 | Ge5La2Rh3 |
| | 12 | z4=1 |
| | 12 | z2w,1=1 z2w,2=1 z4=3 |
| | 12 | z2w,1=1 z2w,2=1 z4=3 |
| | 12 | z4=1 |
| 72.4.9 | 86503 | Ge5Ir3Y2 |
| | 12 | z2w,1=1 z2w,2=1 z4=2 |
| | 12 | z2w,1=1 z2w,2=1 z4=2 |
| 72.4.10 | 86499 | Ge5Ir3La2 |
| | 12 | z4=1 |
| | 12 | z2w,1=1 z2w,2=1 z4=3 |
| | 12 | z2w,1=1 z2w,2=1 z4=3 |
| | 12 | z4=1 |

Continued on next column

**TABLE XIX– continued**

| TQC | ICSD SSG | Stoich. Formula SIs |
|---|---|---|
| 72.4.11 | 40462 | Ge2Mn5 |
| | 12 | z2w,1=1 z2w,2=1 z4=2 |
| | 12 | z2w,1=1 z2w,2=1 z4=2 |
| 72.4.12 | 68523 | Fe1Li3N2 |
| | 12 | z2w,1=1 z2w,2=1 z4=2 |
| | 12 | z2w,1=1 z2w,2=1 z4=2 |
| 72.4.13 | 78933 | Cs2Mn3Te4 |
| | 12 | z4=2 |
| | 12 | z2w,1=1 z2w,2=1 |
| | 12 | z2w,1=1 z2w,2=1 |
| | 12 | z4=2 |
| 72.4.14 | 59276 | Cs2Mn3Te4 |
| | 12 | z4=2 |
| | 12 | z2w,1=1 z2w,2=1 |
| | 12 | z2w,1=1 z2w,2=1 |
| | 12 | z4=2 |
| 72.4.15 | 78932 | Cs2Mn3Se4 |
| | 12 | z4=2 |
| | 12 | z2w,1=1 z2w,2=1 |
| | 12 | z2w,1=1 z2w,2=1 |
| | 12 | z4=2 |
| 72.4.16 | 65461 | Cs2Mn1Te2 |
| | 12 | z2w,1=1 z2w,2=1 z4=2 |
| | 12 | z2w,1=1 z2w,2=1 z4=2 |
| 72.4.17 | 65458 | Cs2Mn1Se2 |
| | 12 | z2w,1=1 z2w,2=1 z4=2 |
| | 12 | z2w,1=1 z2w,2=1 z4=2 |
| 72.4.18 | 65455 | Cs2Mn1S2 |
| | 12 | z2w,1=1 z2w,2=1 z4=2 |
| | 12 | z2w,1=1 z2w,2=1 z4=2 |
| 72.4.19 | 65254 | Co3Cs2S4 |
| | 12 | z4=1 |
| | 12 | z2w,1=1 z2w,2=1 z4=3 |
| | 12 | z2w,1=1 z2w,2=1 z4=3 |
| | 12 | z4=1 |
| 72.4.20 | 67391 | Co1Rb2Se2 |
| | 12 | z2w,1=1 z2w,2=1 z4=2 |
| | 12 | z2w,1=1 z2w,2=1 z4=2 |
| 72.4.21 | 67388 | Co1Rb2S2 |
| | 12 | z2w,1=1 z2w,2=1 z4=2 |
| | 12 | z2w,1=1 z2w,2=1 z4=2 |
| 72.4.22 | 416514 | Co1Pd2Te2 |
| | 12 | z2w,1=1 z2w,2=1 z4=2 |
| | 12 | z2w,1=1 z2w,2=1 z4=2 |
| 72.4.23 | 416513 | Co1Pd2Se2 |
| | 12 | z2w,1=1 z2w,2=1 z4=3 |
| | 12 | z4=1 |
| | 12 | z4=1 |
| | 12 | z2w,1=1 z2w,2=1 z4=3 |
| 72.4.24 | 67386 | Co1Na2S2 |
| | 12 | z2w,1=1 z2w,2=1 z4=2 |
| | 12 | z2w,1=1 z2w,2=1 z4=2 |
| 72.4.25 | 67390 | Co1K2Se2 |
| | 12 | z2w,1=1 z2w,2=1 z4=2 |
| | 12 | z2w,1=1 z2w,2=1 z4=2 |
| 72.4.26 | 67387 | Co1K2S2 |
| | 12 | z2w,1=1 z2w,2=1 z4=2 |
| | 12 | z2w,1=1 z2w,2=1 z4=2 |
| 72.4.27 | 67392 | Co1Cs2Se2 |

Continued on next column

**TABLE XIX– continued**

| TQC | ICSD SSG | Stoich. Formula SIs |
|---|---|---|
| | 12 | z2w,1=1 z2w,2=1 z4=2 |
| | 12 | z2w,1=1 z2w,2=1 z4=2 |
| 72.4.28 | 67389 | Co1Cs2S2 |
| | 12 | z2w,1=1 z2w,2=1 z4=2 |
| | 12 | z2w,1=1 z2w,2=1 z4=2 |
| 72.4.29 | 73169 | C4Er2Mn1 |
| | 12 | z2w,1=1 z2w,2=1 z4=2 |
| | 12 | z2w,1=1 z2w,2=1 z4=2 |
| 72.4.30 | 69494 | Ba2Cu3P4 |
| | 12 | z2w,1=1 z2w,2=1 z4=3 |
| | 12 | z4=1 |
| | 12 | z4=1 |
| | 12 | z2w,1=1 z2w,2=1 z4=3 |
| 72.4.31 | 26253 | As2Cu5 |
| | 12 | z4=2 |
| | 12 | z2w,1=1 z2w,2=1 |
| | 12 | z2w,1=1 z2w,2=1 |
| | 12 | z4=2 |
| 74.4.1 | 236759 | O3Sr1Tc1 |
| | 12 | z2w,1=1 z2w,2=1 z2w,3=1 z4=3 |
| | 12 | z2w,3=1 z4=1 |
| | 12 | z2w,1=1 z2w,2=1 z2w,3=1 z4=2 |
| | 12 | z2w,3=1 |
| 74.4.2 | 183451 | O3Sr1Tc1 |
| | 12 | z2w,1=1 z2w,2=1 z2w,3=1 z4=3 |
| | 12 | z2w,3=1 z4=1 |
| | 12 | z2w,1=1 z2w,2=1 z2w,3=1 z4=2 |
| | 12 | z2w,3=1 |
| 74.4.3 | 260033 | La3Pd5Si1 |
| | 12 | z2w,3=1 |
| | 12 | z2w,1=1 z2w,2=1 z2w,3=1 z4=2 |
| | 12 | z2w,3=1 z4=1 |
| | 12 | z2w,1=1 z2w,2=1 z2w,3=1 z4=3 |
| 74.4.4 | 200066 | La1Re2Si2 |
| | 12 | z2w,3=1 z4=1 |
| | 12 | z2w,1=1 z2w,2=1 z2w,3=1 z4=3 |
| | 12 | z2w,3=1 z4=2 |
| | 12 | z2w,1=1 z2w,2=1 z2w,3=1 |
| 74.4.5 | 408698 | Hg2Rb1 |
| | 12 | z2w,3=1 z4=3 |
| | 12 | z2w,1=1 z2w,2=1 z2w,3=1 z4=1 |
| | 12 | z2w,3=1 |
| | 12 | z2w,1=1 z2w,2=1 z2w,3=1 z4=2 |
| 74.4.6 | 195614 | Ge2Na1Pd3 |
| | 12 | z2w,1=1 z2w,2=1 z2w,3=1 |
| | 12 | z2w,3=1 z4=2 |
| | 12 | z2w,1=1 z2w,2=1 z2w,3=1 z4=3 |
| | 12 | z2w,3=1 z4=1 |
| 74.4.7 | 104006 | Ga2Tm1 |
| | 12 | z2w,1=1 z2w,2=1 z2w,3=1 z4=2 |
| | 12 | z2w,3=1 |
| | 12 | z2w,1=1 z2w,2=1 z2w,3=1 z4=1 |
| | 12 | z2w,3=1 z4=3 |
| 74.4.8 | 104005 | Ga2Tm1 |
| | 12 | z2w,1=1 z2w,2=1 z2w,3=1 |
| | 12 | z2w,3=1 z4=2 |
| | 12 | z2w,1=1 z2w,2=1 z2w,3=1 z4=3 |
| | 12 | z2w,3=1 z4=1 |
| 74.4.9 | 202872 | F7Fe1Na2Ni1 |

Continued on next column

**TABLE XIX– continued**

| TQC | ICSD SSG | Stoich. Formula SIs |
|---|---|---|
| | 12 | z2w,3=1 z4=2 |
| | 12 | z2w,1=1 z2w,2=1 z2w,3=1 |
| | 12 | z2w,3=1 z4=3 |
| | 12 | z2w,1=1 z2w,2=1 z2w,3=1 z4=1 |
| 74.4.10 | 202871 | F7Fe1Na2Ni1 |
| | 12 | z2w,3=1 z4=2 |
| | 12 | z2w,1=1 z2w,2=1 z2w,3=1 |
| | 12 | z2w,3=1 z4=3 |
| | 12 | z2w,1=1 z2w,2=1 z2w,3=1 z4=1 |
| 74.4.11 | 409889 | Eu1Ga2 |
| | 12 | z2w,3=1 z4=1 |
| | 12 | z2w,1=1 z2w,2=1 z2w,3=1 z4=3 |
| | 12 | z2w,3=1 z4=2 |
| | 12 | z2w,1=1 z2w,2=1 z2w,3=1 |
| 74.4.12 | 36535 | Cu2O2Tl1 |
| | 12 | z2w,1=1 z2w,2=1 z2w,3=1 z4=2 |
| | 12 | z2w,3=1 |
| | 12 | z2w,1=1 z2w,2=1 z2w,3=1 z4=1 |
| | 12 | z2w,3=1 z4=3 |
| 74.4.13 | 408695 | Cs1Hg2 |
| | 12 | z2w,1=1 z2w,2=1 z2w,3=1 z4=3 |
| | 12 | z2w,3=1 z4=1 |
| | 12 | z2w,1=1 z2w,2=1 z2w,3=1 z4=2 |
| | 12 | z2w,3=1 |
| 74.4.14 | 79221 | Ce1Ir3Si2 |
| | 12 | z2w,3=1 z4=3 |
| | 12 | z2w,1=1 z2w,2=1 z2w,3=1 z4=1 |
| | 12 | z2w,3=1 |
| | 12 | z2w,1=1 z2w,2=1 z2w,3=1 z4=2 |
| 74.4.15 | 40664 | Br4Li2Mn1 |
| | 12 | z2w,1=1 z2w,2=1 z2w,3=1 z4=1 |
| | 12 | z2w,3=1 z4=3 |
| | 12 | z2w,1=1 z2w,2=1 z2w,3=1 |
| | 12 | z2w,3=1 z4=2 |
| 74.4.16 | 249923 | Au5K3Tl1 |
| | 12 | z2w,3=1 |
| | 12 | z2w,1=1 z2w,2=1 z2w,3=1 z4=2 |
| | 12 | z2w,3=1 z4=1 |
| | 12 | z2w,1=1 z2w,2=1 z2w,3=1 z4=3 |
| 74.4.17 | 249922 | Au5In1K3 |
| | 12 | z2w,3=1 z4=1 |
| | 12 | z2w,1=1 z2w,2=1 z2w,3=1 z4=3 |
| | 12 | z2w,3=1 z4=2 |
| | 12 | z2w,1=1 z2w,2=1 z2w,3=1 |
| 74.4.18 | 413972 | As1Cu1F7 |
| | 12 | z2w,3=1 z4=2 |
| | 12 | z2w,1=1 z2w,2=1 z2w,3=1 |
| | 12 | z2w,3=1 z4=3 |
| | 12 | z2w,1=1 z2w,2=1 z2w,3=1 z4=1 |
| 74.4.19 | 245266 | Al1Ce1O3 |
| | 12 | z2w,3=1 z4=3 |
| | 12 | z2w,1=1 z2w,2=1 z2w,3=1 z4=1 |
| | 12 | z2w,3=1 |
| | 12 | z2w,1=1 z2w,2=1 z2w,3=1 z4=2 |
| 84.4.1 | 59987 | Cd1F6Tb1 |
| | 81 | z2=1 |
| | 81 | z2=1 |
| | 10 | z2w,1=1 z2w,2=1 z4=3 |
| | 10 | z2w,1=1 |

Continued on next column

**TABLE XIX– continued**

| TQC | ICSD SSG | Stoich. Formula SIs |
|---|---|---|
| | 10 | z2w,3=1 |
| | 10 | z2w,2=1 z2w,3=1 z4=1 |
| 85.4.1 | 195053 | Ni1Sb9Zr5 |
| | 81 | z2=1 |
| | 81 | z2=1 |
| | 2 | z4=1 |
| | 2 | z2w,2=1 |
| | 2 | z2w,1=1 |
| | 2 | z2w,1=1 z2w,2=1 z4=3 |
| | 2 | z2w,1=1 z2w,3=1 |
| | 2 | z2w,1=1 z2w,2=1 z2w,3=1 z4=3 |
| | 2 | z2w,3=1 z4=3 |
| | 2 | z2w,2=1 z2w,3=1 z4=2 |
| | 2 | z2w,2=1 z2w,3=1 |
| | 2 | z2w,3=1 z4=3 |
| | 2 | z2w,1=1 z2w,2=1 z2w,3=1 z4=3 |
| | 2 | z2w,1=1 z2w,3=1 z4=2 |
| | 2 | z2w,1=1 z2w,2=1 z4=3 |
| | 2 | z2w,1=1 z4=2 |
| | 2 | z2w,2=1 z4=2 |
| | 2 | z4=1 |
| | 2 | z2w,1=1 z2w,2=1 |
| | 2 | z2w,1=1 z4=3 |
| | 2 | z2w,2=1 z4=3 |
| | 2 | z4=2 |
| | 2 | z2w,2=1 z2w,3=1 z4=3 |
| | 2 | z2w,3=1 z4=2 |
| | 2 | z2w,1=1 z2w,2=1 z2w,3=1 z4=2 |
| | 2 | z2w,1=1 z2w,3=1 z4=1 |
| | 2 | z2w,1=1 z2w,3=1 z4=3 |
| | 2 | z2w,1=1 z2w,2=1 z2w,3=1 z4=2 |
| | 2 | z2w,3=1 z4=2 |
| | 2 | z2w,2=1 z2w,3=1 z4=1 |
| | 2 | z4=2 |
| | 2 | z2w,2=1 z4=1 |
| | 2 | z2w,1=1 z4=1 |
| | 2 | z2w,1=1 z2w,2=1 |
| | 2 | z2w,1=1 z2w,2=1 z2w,3=1 |
| | 2 | z2w,1=1 z2w,3=1 z4=3 |
| | 2 | z2w,2=1 z2w,3=1 z4=3 |
| | 2 | z2w,3=1 z4=2 |
| | 2 | z2w,2=1 z4=3 |
| | 2 | z4=2 |
| | 2 | z2w,1=1 z2w,2=1 z4=2 |
| | 2 | z2w,1=1 z4=1 |
| | 2 | z2w,1=1 z4=3 |
| | 2 | z2w,1=1 z2w,2=1 z4=2 |
| | 2 | z4=2 |
| | 2 | z2w,2=1 z4=1 |
| | 2 | z2w,3=1 z4=2 |
| | 2 | z2w,2=1 z2w,3=1 z4=1 |
| | 2 | z2w,1=1 z2w,3=1 z4=1 |
| | 2 | z2w,1=1 z2w,2=1 z2w,3=1 |
| | 2 | z2w,3=1 z4=3 |
| | 2 | z2w,2=1 z2w,3=1 z4=2 |
| | 2 | z2w,1=1 z2w,3=1 z4=2 |
| | 2 | z2w,1=1 z2w,2=1 z2w,3=1 z4=1 |
| | 2 | z2w,1=1 z4=2 |
| | 2 | z2w,1=1 z2w,2=1 z4=1 |
| | 2 | z4=1 |

Continued on next column

**TABLE XIX– continued**

| TQC | ICSD SSG | Stoich. Formula SIs |
|---|---|---|
| | 2 | z2w,2=1 |
| | 2 | z2w,2=1 z4=2 |
| | 2 | z4=1 |
| | 2 | z2w,1=1 z2w,2=1 z4=1 |
| | 2 | z2w,1=1 |
| | 2 | z2w,1=1 z2w,2=1 z2w,3=1 z4=1 |
| | 2 | z2w,1=1 z2w,3=1 |
| | 2 | z2w,2=1 z2w,3=1 |
| | 2 | z2w,3=1 z4=3 |
| 85.4.2 | 27315 | Mo1O5V1 |
| | 81 | z2=1 |
| | 81 | z2=1 |
| | 2 | z4=3 |
| | 2 | z2w,2=1 z4=2 |
| | 2 | z2w,1=1 z4=2 |
| | 2 | z2w,1=1 z2w,2=1 z4=1 |
| | 2 | z2w,1=1 z2w,3=1 z4=2 |
| | 2 | z2w,1=1 z2w,2=1 z2w,3=1 z4=1 |
| | 2 | z2w,3=1 z4=1 |
| | 2 | z2w,2=1 z2w,3=1 |
| | 2 | z2w,2=1 z2w,3=1 z4=2 |
| | 2 | z2w,3=1 z4=1 |
| | 2 | z2w,1=1 z2w,2=1 z2w,3=1 z4=1 |
| | 2 | z2w,1=1 z2w,3=1 |
| | 2 | z2w,1=1 z2w,2=1 z4=1 |
| | 2 | z2w,1=1 |
| | 2 | z2w,2=1 |
| | 2 | z4=3 |
| | 2 | z2w,1=1 z2w,2=1 z4=2 |
| | 2 | z2w,1=1 z4=1 |
| | 2 | z2w,2=1 z4=1 |
| | 2 | z2w,2=1 z2w,3=1 z4=1 |
| | 2 | z2w,3=1 |
| | 2 | z2w,1=1 z2w,2=1 z2w,3=1 |
| | 2 | z2w,1=1 z2w,3=1 z4=3 |
| | 2 | z2w,1=1 z2w,3=1 z4=1 |
| | 2 | z2w,1=1 z2w,2=1 z2w,3=1 |
| | 2 | z2w,3=1 |
| | 2 | z2w,2=1 z2w,3=1 z4=3 |
| | 2 | z2w,2=1 z4=3 |
| | 2 | z2w,1=1 z4=3 |
| | 2 | z2w,1=1 z2w,2=1 z4=2 |
| | 2 | z2w,1=1 z2w,2=1 z2w,3=1 z4=2 |
| | 2 | z2w,1=1 z2w,3=1 z4=1 |
| | 2 | z2w,2=1 z2w,3=1 z4=1 |
| | 2 | z2w,3=1 |
| | 2 | z2w,2=1 z4=1 |
| | 2 | z2w,1=1 z2w,2=1 |
| | 2 | z2w,1=1 z4=3 |
| | 2 | z2w,1=1 z4=1 |
| | 2 | z2w,1=1 z2w,2=1 |
| | 2 | z2w,2=1 z4=3 |
| | 2 | z2w,3=1 |
| | 2 | z2w,2=1 z2w,3=1 z4=3 |
| | 2 | z2w,1=1 z2w,3=1 z4=3 |
| | 2 | z2w,1=1 z2w,2=1 z2w,3=1 z4=2 |
| | 2 | z2w,3=1 z4=1 |
| | 2 | z2w,2=1 z2w,3=1 |
| | 2 | z2w,1=1 z2w,3=1 |
| | 2 | z2w,1=1 z2w,2=1 z2w,3=1 z4=3 |

Continued on next column

**TABLE XIX– continued**

| TQC | ICSD SSG | Stoich. Formula SIs |
|---|---|---|
| | 2 | z2w,1=1 |
| | 2 | z2w,1=1 z2w,2=1 z4=3 |
| | 2 | z4=3 |
| | 2 | z2w,2=1 z4=2 |
| | 2 | z2w,2=1 |
| | 2 | z4=3 |
| | 2 | z2w,1=1 z2w,2=1 z4=3 |
| | 2 | z2w,1=1 z4=2 |
| | 2 | z2w,1=1 z2w,2=1 z2w,3=1 z4=3 |
| | 2 | z2w,1=1 z2w,3=1 z4=2 |
| | 2 | z2w,2=1 z2w,3=1 z4=2 |
| | 2 | z2w,3=1 z4=1 |
| 85.4.3 | 36095 | Mo1O5P1 |
| | 81 | z2=1 |
| | 81 | z2=1 |
| | 2 | z4=3 |
| | 2 | z2w,2=1 z4=2 |
| | 2 | z2w,1=1 z4=2 |
| | 2 | z2w,1=1 z2w,2=1 z4=1 |
| | 2 | z2w,1=1 z2w,3=1 z4=2 |
| | 2 | z2w,1=1 z2w,2=1 z2w,3=1 z4=1 |
| | 2 | z2w,3=1 z4=1 |
| | 2 | z2w,2=1 z2w,3=1 |
| | 2 | z2w,2=1 z2w,3=1 z4=2 |
| | 2 | z2w,3=1 z4=1 |
| | 2 | z2w,1=1 z2w,2=1 z2w,3=1 z4=1 |
| | 2 | z2w,1=1 z2w,3=1 |
| | 2 | z2w,1=1 z2w,2=1 z4=1 |
| | 2 | z2w,1=1 |
| | 2 | z2w,2=1 |
| | 2 | z4=3 |
| | 2 | z2w,1=1 z2w,2=1 z4=2 |
| | 2 | z2w,1=1 z4=1 |
| | 2 | z2w,2=1 z4=1 |
| | 2 | z2w,2=1 z2w,3=1 z4=1 |
| | 2 | z2w,3=1 |
| | 2 | z2w,1=1 z2w,2=1 z2w,3=1 |
| | 2 | z2w,1=1 z2w,3=1 z4=3 |
| | 2 | z2w,1=1 z2w,3=1 z4=1 |
| | 2 | z2w,1=1 z2w,2=1 z2w,3=1 |
| | 2 | z2w,3=1 |
| | 2 | z2w,2=1 z2w,3=1 z4=3 |
| | 2 | z2w,2=1 z4=3 |
| | 2 | z2w,1=1 z4=3 |
| | 2 | z2w,1=1 z2w,2=1 z4=2 |
| | 2 | z2w,1=1 z2w,2=1 z2w,3=1 z4=2 |
| | 2 | z2w,1=1 z2w,3=1 z4=1 |
| | 2 | z2w,2=1 z2w,3=1 z4=1 |
| | 2 | z2w,3=1 |
| | 2 | z2w,2=1 z4=1 |
| | 2 | z2w,1=1 z2w,2=1 |
| | 2 | z2w,1=1 z4=3 |
| | 2 | z2w,1=1 z4=1 |
| | 2 | z2w,1=1 z2w,2=1 |
| | 2 | z2w,2=1 z4=3 |
| | 2 | z2w,3=1 |
| | 2 | z2w,2=1 z2w,3=1 z4=3 |
| | 2 | z2w,1=1 z2w,3=1 z4=3 |
| | 2 | z2w,1=1 z2w,2=1 z2w,3=1 z4=2 |
| | 2 | z2w,3=1 z4=1 |

Continued on next column

TABLE XIX– continued

| TQC | ICSD SSG | Stoich. Formula SIs |
|---|---|---|
| | 2 | z2w,2=1 z2w,3=1 |
| | 2 | z2w,1=1 z2w,3=1 |
| | 2 | z2w,1=1 z2w,2=1 z2w,3=1 z4=3 |
| | 2 | z2w,1=1 |
| | 2 | z2w,1=1 z2w,2=1 z4=3 |
| | 2 | z4=3 |
| | 2 | z2w,2=1 z4=2 |
| | 2 | z2w,2=1 |
| | 2 | z4=3 |
| | 2 | z2w,1=1 z2w,2=1 z4=3 |
| | 2 | z2w,1=1 z4=2 |
| | 2 | z2w,1=1 z2w,2=1 z2w,3=1 z4=3 |
| | 2 | z2w,1=1 z2w,3=1 z4=2 |
| | 2 | z2w,2=1 z2w,3=1 z4=2 |
| | 2 | z2w,3=1 z4=1 |
| 85.4.4 | 24894 | Mo1O5P1 |
| | 81 | z2=1 |
| | 81 | z2=1 |
| | 2 | z4=3 |
| | 2 | z2w,2=1 z4=2 |
| | 2 | z2w,1=1 z4=2 |
| | 2 | z2w,1=1 z2w,2=1 z4=1 |
| | 2 | z2w,1=1 z2w,3=1 z4=2 |
| | 2 | z2w,1=1 z2w,2=1 z2w,3=1 z4=1 |
| | 2 | z2w,3=1 z4=1 |
| | 2 | z2w,2=1 z2w,3=1 |
| | 2 | z2w,2=1 z2w,3=1 z4=2 |
| | 2 | z2w,3=1 z4=1 |
| | 2 | z2w,1=1 z2w,2=1 z2w,3=1 z4=1 |
| | 2 | z2w,1=1 z2w,3=1 |
| | 2 | z2w,1=1 z2w,2=1 z4=1 |
| | 2 | z2w,1=1 |
| | 2 | z2w,2=1 |
| | 2 | z4=3 |
| | 2 | z2w,1=1 z2w,2=1 z4=2 |
| | 2 | z2w,1=1 z4=1 |
| | 2 | z2w,2=1 z4=1 |
| | 2 | z2w,2=1 z2w,3=1 z4=1 |
| | 2 | z2w,3=1 |
| | 2 | z2w,1=1 z2w,2=1 z2w,3=1 |
| | 2 | z2w,1=1 z2w,3=1 z4=3 |
| | 2 | z2w,1=1 z2w,3=1 z4=1 |
| | 2 | z2w,1=1 z2w,2=1 z2w,3=1 |
| | 2 | z2w,3=1 |
| | 2 | z2w,2=1 z2w,3=1 z4=3 |
| | 2 | z2w,2=1 z4=3 |
| | 2 | z2w,1=1 z4=3 |
| | 2 | z2w,1=1 z2w,2=1 z4=2 |
| | 2 | z2w,1=1 z2w,2=1 z2w,3=1 z4=2 |
| | 2 | z2w,1=1 z2w,3=1 z4=1 |
| | 2 | z2w,2=1 z2w,3=1 z4=1 |
| | 2 | z2w,3=1 |
| | 2 | z2w,2=1 z4=1 |
| | 2 | z2w,1=1 z2w,2=1 |
| | 2 | z2w,1=1 z4=3 |
| | 2 | z2w,1=1 z4=1 |
| | 2 | z2w,1=1 z2w,2=1 |
| | 2 | z2w,2=1 z4=3 |
| | 2 | z2w,3=1 |
| | 2 | z2w,2=1 z2w,3=1 z4=3 |

Continued on next column

TABLE XIX– continued

| TQC | ICSD SSG | Stoich. Formula SIs |
|---|---|---|
| | 2 | z2w,1=1 z2w,3=1 z4=3 |
| | 2 | z2w,1=1 z2w,2=1 z2w,3=1 z4=2 |
| | 2 | z2w,3=1 z4=1 |
| | 2 | z2w,2=1 z2w,3=1 |
| | 2 | z2w,1=1 z2w,3=1 |
| | 2 | z2w,1=1 z2w,2=1 z2w,3=1 z4=3 |
| | 2 | z2w,1=1 |
| | 2 | z2w,1=1 z2w,2=1 z4=3 |
| | 2 | z4=3 |
| | 2 | z2w,2=1 z4=2 |
| | 2 | z2w,2=1 |
| | 2 | z4=3 |
| | 2 | z2w,1=1 z2w,2=1 z4=3 |
| | 2 | z2w,1=1 z4=2 |
| | 2 | z2w,1=1 z2w,2=1 z2w,3=1 z4=3 |
| | 2 | z2w,1=1 z2w,3=1 z4=2 |
| | 2 | z2w,2=1 z2w,3=1 z4=2 |
| | 2 | z2w,3=1 z4=1 |
| 85.4.5 | 413979 | Hf5Sb9 |
| | 81 | z2=1 |
| | 81 | z2=1 |
| | 2 | z4=1 |
| | 2 | z2w,2=1 |
| | 2 | z2w,1=1 |
| | 2 | z2w,1=1 z2w,2=1 z4=3 |
| | 2 | z2w,1=1 z2w,3=1 |
| | 2 | z2w,1=1 z2w,2=1 z2w,3=1 z4=3 |
| | 2 | z2w,3=1 z4=3 |
| | 2 | z2w,2=1 z2w,3=1 z4=2 |
| | 2 | z2w,2=1 z2w,3=1 |
| | 2 | z2w,3=1 z4=3 |
| | 2 | z2w,1=1 z2w,2=1 z2w,3=1 z4=3 |
| | 2 | z2w,1=1 z2w,3=1 z4=2 |
| | 2 | z2w,1=1 z2w,2=1 z4=3 |
| | 2 | z2w,1=1 z4=2 |
| | 2 | z2w,2=1 z4=2 |
| | 2 | z4=1 |
| | 2 | z2w,1=1 z2w,2=1 |
| | 2 | z2w,1=1 z4=3 |
| | 2 | z2w,2=1 z4=3 |
| | 2 | z4=2 |
| | 2 | z2w,2=1 z2w,3=1 z4=3 |
| | 2 | z2w,3=1 z4=2 |
| | 2 | z2w,1=1 z2w,2=1 z2w,3=1 z4=2 |
| | 2 | z2w,1=1 z2w,3=1 z4=1 |
| | 2 | z2w,1=1 z2w,3=1 z4=3 |
| | 2 | z2w,1=1 z2w,2=1 z2w,3=1 z4=2 |
| | 2 | z2w,3=1 z4=2 |
| | 2 | z2w,2=1 z2w,3=1 z4=1 |
| | 2 | z4=2 |
| | 2 | z2w,2=1 z4=1 |
| | 2 | z2w,1=1 z4=1 |
| | 2 | z2w,1=1 z2w,2=1 |
| | 2 | z2w,1=1 z2w,2=1 z2w,3=1 |
| | 2 | z2w,1=1 z2w,3=1 z4=3 |
| | 2 | z2w,2=1 z2w,3=1 z4=3 |
| | 2 | z2w,3=1 z4=2 |
| | 2 | z2w,2=1 z4=3 |
| | 2 | z4=2 |
| | 2 | z2w,1=1 z2w,2=1 z4=2 |

Continued on next column

**TABLE XIX– continued**

| TQC | ICSD SSG | Stoich. Formula SIs |
|---|---|---|
| | 2 | z2w,1=1 z4=1 |
| | 2 | z2w,1=1 z4=3 |
| | 2 | z2w,1=1 z2w,2=1 z4=2 |
| | 2 | z4=2 |
| | 2 | z2w,2=1 z4=1 |
| | 2 | z2w,3=1 z4=2 |
| | 2 | z2w,2=1 z2w,3=1 z4=1 |
| | 2 | z2w,1=1 z2w,3=1 z4=1 |
| | 2 | z2w,1=1 z2w,2=1 z2w,3=1 |
| | 2 | z2w,3=1 z4=3 |
| | 2 | z2w,2=1 z2w,3=1 z4=2 |
| | 2 | z2w,1=1 z2w,3=1 z4=2 |
| | 2 | z2w,1=1 z2w,2=1 z2w,3=1 z4=1 |
| | 2 | z2w,1=1 z4=2 |
| | 2 | z2w,1=1 z2w,2=1 z4=1 |
| | 2 | z4=1 |
| | 2 | z2w,2=1 |
| | 2 | z2w,2=1 z4=2 |
| | 2 | z4=1 |
| | 2 | z2w,1=1 z2w,2=1 z4=1 |
| | 2 | z2w,1=1 |
| | 2 | z2w,1=1 z2w,2=1 z2w,3=1 z4=1 |
| | 2 | z2w,1=1 z2w,3=1 |
| | 2 | z2w,2=1 z2w,3=1 |
| | 2 | z2w,3=1 z4=3 |
| 85.4.6 | 56963 | H4B1Cl1Cu1O4 |
| | 81 | z2=1 |
| | 81 | z2=1 |
| | 2 | z4=3 |
| | 2 | z2w,2=1 z4=2 |
| | 2 | z2w,1=1 z4=2 |
| | 2 | z2w,1=1 z2w,2=1 z4=1 |
| | 2 | z2w,1=1 z2w,3=1 z4=2 |
| | 2 | z2w,1=1 z2w,2=1 z2w,3=1 z4=1 |
| | 2 | z2w,3=1 z4=1 |
| | 2 | z2w,2=1 z2w,3=1 |
| | 2 | z2w,2=1 z2w,3=1 z4=2 |
| | 2 | z2w,3=1 z4=1 |
| | 2 | z2w,1=1 z2w,2=1 z2w,3=1 z4=1 |
| | 2 | z2w,1=1 z2w,3=1 |
| | 2 | z2w,1=1 z2w,2=1 z4=1 |
| | 2 | z2w,1=1 |
| | 2 | z2w,2=1 |
| | 2 | z4=3 |
| | 2 | z2w,1=1 z2w,2=1 z4=2 |
| | 2 | z2w,1=1 z4=1 |
| | 2 | z2w,2=1 z4=1 |
| | 2 | z2w,2=1 z2w,3=1 z4=1 |
| | 2 | z2w,3=1 |
| | 2 | z2w,1=1 z2w,2=1 z2w,3=1 |
| | 2 | z2w,1=1 z2w,3=1 z4=3 |
| | 2 | z2w,1=1 z2w,3=1 z4=1 |
| | 2 | z2w,1=1 z2w,2=1 z2w,3=1 |
| | 2 | z2w,3=1 |
| | 2 | z2w,2=1 z2w,3=1 z4=3 |
| | 2 | z2w,2=1 z4=3 |
| | 2 | z2w,1=1 z4=3 |
| | 2 | z2w,1=1 z2w,2=1 z4=2 |
| | 2 | z2w,1=1 z2w,2=1 z2w,3=1 z4=2 |
| | 2 | z2w,1=1 z2w,3=1 z4=1 |

Continued on next column

**TABLE XIX– continued**

| TQC | ICSD SSG | Stoich. Formula SIs |
|---|---|---|
| | 2 | z2w,2=1 z2w,3=1 z4=1 |
| | 2 | z2w,3=1 |
| | 2 | z2w,2=1 z4=1 |
| | 2 | z2w,1=1 z2w,2=1 |
| | 2 | z2w,1=1 z4=3 |
| | 2 | z2w,1=1 z4=1 |
| | 2 | z2w,1=1 z2w,2=1 |
| | 2 | z2w,2=1 z4=3 |
| | 2 | z2w,3=1 |
| | 2 | z2w,2=1 z2w,3=1 z4=3 |
| | 2 | z2w,1=1 z2w,3=1 z4=3 |
| | 2 | z2w,1=1 z2w,2=1 z2w,3=1 z4=2 |
| | 2 | z2w,3=1 z4=1 |
| | 2 | z2w,2=1 z2w,3=1 |
| | 2 | z2w,1=1 z2w,3=1 |
| | 2 | z2w,1=1 z2w,2=1 z2w,3=1 z4=3 |
| | 2 | z2w,1=1 |
| | 2 | z2w,1=1 z2w,2=1 z4=3 |
| | 2 | z4=3 |
| | 2 | z2w,2=1 z4=2 |
| | 2 | z2w,2=1 |
| | 2 | z4=3 |
| | 2 | z2w,1=1 z2w,2=1 z4=3 |
| | 2 | z2w,1=1 z4=2 |
| | 2 | z2w,1=1 z2w,2=1 z2w,3=1 z4=3 |
| | 2 | z2w,1=1 z2w,3=1 z4=2 |
| | 2 | z2w,2=1 z2w,3=1 z4=2 |
| | 2 | z2w,3=1 z4=1 |
| 85.4.7 | 80646 | Ag1B1F5 |
| | 81 | z2=1 |
| | 81 | z2=1 |
| | 2 | z4=3 |
| | 2 | z2w,2=1 z4=2 |
| | 2 | z2w,1=1 z4=2 |
| | 2 | z2w,1=1 z2w,2=1 z4=1 |
| | 2 | z2w,1=1 z2w,3=1 z4=2 |
| | 2 | z2w,1=1 z2w,2=1 z2w,3=1 z4=1 |
| | 2 | z2w,3=1 z4=1 |
| | 2 | z2w,2=1 z2w,3=1 |
| | 2 | z2w,2=1 z2w,3=1 z4=2 |
| | 2 | z2w,3=1 z4=1 |
| | 2 | z2w,1=1 z2w,2=1 z2w,3=1 z4=1 |
| | 2 | z2w,1=1 z2w,3=1 |
| | 2 | z2w,1=1 z2w,2=1 z4=1 |
| | 2 | z2w,1=1 |
| | 2 | z2w,2=1 |
| | 2 | z4=3 |
| | 2 | z2w,1=1 z2w,2=1 z4=2 |
| | 2 | z2w,1=1 z4=1 |
| | 2 | z2w,2=1 z4=1 |
| | 2 | z2w,2=1 z2w,3=1 z4=1 |
| | 2 | z2w,3=1 |
| | 2 | z2w,1=1 z2w,2=1 z2w,3=1 |
| | 2 | z2w,1=1 z2w,3=1 z4=3 |
| | 2 | z2w,1=1 z2w,3=1 z4=1 |
| | 2 | z2w,1=1 z2w,2=1 z2w,3=1 |
| | 2 | z2w,3=1 |
| | 2 | z2w,2=1 z2w,3=1 z4=3 |
| | 2 | z2w,2=1 z4=3 |
| | 2 | z2w,1=1 z4=3 |

Continued on next column

**TABLE XIX– continued**

| TQC | ICSD SSG | Stoich. Formula SIs |
|---|---|---|
| | 2 | z2w,1=1 z2w,2=1 z4=2 |
| | 2 | z2w,1=1 z2w,2=1 z2w,3=1 z4=2 |
| | 2 | z2w,1=1 z2w,3=1 z4=1 |
| | 2 | z2w,2=1 z2w,3=1 z4=1 |
| | 2 | z2w,3=1 |
| | 2 | z2w,2=1 z4=1 |
| | 2 | z2w,1=1 z2w,2=1 |
| | 2 | z2w,1=1 z4=3 |
| | 2 | z2w,1=1 z4=1 |
| | 2 | z2w,1=1 z2w,2=1 |
| | 2 | z2w,2=1 z4=3 |
| | 2 | z2w,3=1 |
| | 2 | z2w,2=1 z2w,3=1 z4=3 |
| | 2 | z2w,1=1 z2w,3=1 z4=3 |
| | 2 | z2w,1=1 z2w,2=1 z2w,3=1 z4=2 |
| | 2 | z2w,3=1 z4=1 |
| | 2 | z2w,2=1 z2w,3=1 |
| | 2 | z2w,1=1 z2w,3=1 |
| | 2 | z2w,1=1 z2w,2=1 z2w,3=1 z4=3 |
| | 2 | z2w,1=1 |
| | 2 | z2w,1=1 z2w,2=1 z4=3 |
| | 2 | z4=3 |
| | 2 | z2w,2=1 z4=2 |
| | 2 | z2w,2=1 |
| | 2 | z4=3 |
| | 2 | z2w,1=1 z2w,2=1 z4=3 |
| | 2 | z2w,1=1 z4=2 |
| | 2 | z2w,1=1 z2w,2=1 z2w,3=1 z4=3 |
| | 2 | z2w,1=1 z2w,3=1 z4=2 |
| | 2 | z2w,2=1 z2w,3=1 z4=2 |
| | 2 | z2w,3=1 z4=1 |
| 85.4.8 | 80645 | Ag1B1F5 |
| | 81 | z2=1 |
| | 81 | z2=1 |
| | 2 | z4=3 |
| | 2 | z2w,2=1 z4=2 |
| | 2 | z2w,1=1 z4=2 |
| | 2 | z2w,1=1 z2w,2=1 z4=1 |
| | 2 | z2w,1=1 z2w,3=1 z4=2 |
| | 2 | z2w,1=1 z2w,2=1 z2w,3=1 z4=1 |
| | 2 | z2w,3=1 z4=1 |
| | 2 | z2w,2=1 z2w,3=1 |
| | 2 | z2w,2=1 z2w,3=1 z4=2 |
| | 2 | z2w,3=1 z4=1 |
| | 2 | z2w,1=1 z2w,2=1 z2w,3=1 z4=1 |
| | 2 | z2w,1=1 z2w,3=1 |
| | 2 | z2w,1=1 z2w,2=1 z4=1 |
| | 2 | z2w,1=1 |
| | 2 | z2w,2=1 |
| | 2 | z4=3 |
| | 2 | z2w,1=1 z2w,2=1 z4=2 |
| | 2 | z2w,1=1 z4=1 |
| | 2 | z2w,2=1 z4=1 |
| | 2 | z2w,2=1 z2w,3=1 z4=1 |
| | 2 | z2w,3=1 |
| | 2 | z2w,1=1 z2w,2=1 z2w,3=1 |
| | 2 | z2w,1=1 z2w,3=1 z4=3 |
| | 2 | z2w,1=1 z2w,3=1 z4=1 |
| | 2 | z2w,1=1 z2w,2=1 z2w,3=1 |
| | 2 | z2w,3=1 |

Continued on next column

**TABLE XIX– continued**

| TQC | ICSD SSG | Stoich. Formula SIs |
|---|---|---|
| | 2 | z2w,2=1 z2w,3=1 z4=3 |
| | 2 | z2w,2=1 z4=3 |
| | 2 | z2w,1=1 z4=3 |
| | 2 | z2w,1=1 z2w,2=1 z4=2 |
| | 2 | z2w,1=1 z2w,2=1 z2w,3=1 z4=2 |
| | 2 | z2w,1=1 z2w,3=1 z4=1 |
| | 2 | z2w,2=1 z2w,3=1 z4=1 |
| | 2 | z2w,3=1 |
| | 2 | z2w,2=1 z4=1 |
| | 2 | z2w,1=1 z2w,2=1 |
| | 2 | z2w,1=1 z4=3 |
| | 2 | z2w,1=1 z4=1 |
| | 2 | z2w,1=1 z2w,2=1 |
| | 2 | z2w,2=1 z4=3 |
| | 2 | z2w,3=1 |
| | 2 | z2w,2=1 z2w,3=1 z4=3 |
| | 2 | z2w,1=1 z2w,3=1 z4=3 |
| | 2 | z2w,1=1 z2w,2=1 z2w,3=1 z4=2 |
| | 2 | z2w,3=1 z4=1 |
| | 2 | z2w,2=1 z2w,3=1 |
| | 2 | z2w,1=1 z2w,3=1 |
| | 2 | z2w,1=1 z2w,2=1 z2w,3=1 z4=3 |
| | 2 | z2w,1=1 |
| | 2 | z2w,1=1 z2w,2=1 z4=3 |
| | 2 | z4=3 |
| | 2 | z2w,2=1 z4=2 |
| | 2 | z2w,2=1 |
| | 2 | z4=3 |
| | 2 | z2w,1=1 z2w,2=1 z4=3 |
| | 2 | z2w,1=1 z4=2 |
| | 2 | z2w,1=1 z2w,2=1 z2w,3=1 z4=3 |
| | 2 | z2w,1=1 z2w,3=1 z4=2 |
| | 2 | z2w,2=1 z2w,3=1 z4=2 |
| | 2 | z2w,3=1 z4=1 |
| 86.4.1 | 8078 | B4La1Ru4 |
| | 81 | z2=1 |
| | 81 | z2=1 |
| | 2 | z2w,2=1 z4=3 |
| | 2 | z2w,1=1 z4=3 |
| | 2 | z2w,1=1 z2w,2=1 z4=2 |
| | 2 | z2w,1=1 z2w,3=1 z4=3 |
| | 2 | z2w,1=1 z2w,2=1 z2w,3=1 z4=2 |
| | 2 | z2w,3=1 z4=2 |
| | 2 | z2w,2=1 z2w,3=1 z4=1 |
| | 2 | z2w,2=1 z2w,3=1 z4=3 |
| | 2 | z2w,3=1 z4=2 |
| | 2 | z2w,1=1 z2w,2=1 z2w,3=1 z4=2 |
| | 2 | z2w,1=1 z2w,3=1 z4=1 |
| | 2 | z2w,1=1 z2w,2=1 z4=2 |
| | 2 | z2w,1=1 z4=1 |
| | 2 | z2w,2=1 z4=1 |
| | 2 | z2w,3=1 z4=3 |
| | 2 | z2w,2=1 z2w,3=1 z4=2 |
| | 2 | z2w,1=1 z2w,3=1 z4=2 |
| | 2 | z2w,1=1 z2w,2=1 z2w,3=1 z4=1 |
| | 2 | z2w,1=1 z4=2 |
| | 2 | z2w,1=1 z2w,2=1 z4=1 |
| | 2 | z4=1 |
| | 2 | z2w,2=1 |
| | 2 | z2w,2=1 z4=2 |

Continued on next column

**TABLE XIX– continued**

| TQC | ICSD<br>SSG | Stoich. Formula<br>SIs |
|---|---|---|
| | 2 | z4=1 |
| | 2 | z2w,1=1 z2w,2=1 z4=1 |
| | 2 | z2w,1=1 |
| | 2 | z2w,1=1 z2w,2=1 z2w,3=1 z4=1 |
| | 2 | z2w,1=1 z2w,3=1 |
| | 2 | z2w,2=1 z2w,3=1 |
| | 2 | z2w,3=1 z4=3 |
| | 2 | z2w,1=1 z2w,2=1 z4=3 |
| | 2 | z2w,1=1 z4=2 |
| | 2 | z2w,2=1 z4=2 |
| | 2 | z4=1 |
| | 2 | z2w,2=1 z2w,3=1 z4=2 |
| | 2 | z2w,3=1 z4=1 |
| | 2 | z2w,1=1 z2w,2=1 z2w,3=1 z4=1 |
| | 2 | z2w,1=1 z2w,3=1 |
| | 2 | z2w,1=1 z2w,3=1 z4=2 |
| | 2 | z2w,1=1 z2w,2=1 z2w,3=1 z4=1 |
| | 2 | z2w,3=1 z4=1 |
| | 2 | z2w,2=1 z2w,3=1 |
| | 2 | z4=1 |
| | 2 | z2w,2=1 |
| | 2 | z2w,1=1 |
| | 2 | z2w,1=1 z2w,2=1 z4=3 |
| | 2 | z2w,1=1 z2w,2=1 z2w,3=1 z4=2 |
| | 2 | z2w,1=1 z2w,3=1 z4=1 |
| | 2 | z2w,2=1 z2w,3=1 z4=1 |
| | 2 | z2w,3=1 |
| | 2 | z2w,2=1 z4=1 |
| | 2 | z2w,1=1 z2w,2=1 |
| | 2 | z2w,1=1 z4=3 |
| | 2 | z2w,1=1 z4=1 |
| | 2 | z2w,1=1 z2w,2=1 |
| | 2 | z2w,2=1 z4=3 |
| | 2 | z2w,3=1 |
| | 2 | z2w,2=1 z2w,3=1 z4=3 |
| | 2 | z2w,1=1 z2w,3=1 z4=3 |
| | 2 | z2w,1=1 z2w,2=1 z2w,3=1 z4=2 |
| 86.4.2 | 100717 | B4Ir4La1 |
| | 81 | z2=1 |
| | 81 | z2=1 |
| | 2 | z2w,2=1 z4=3 |
| | 2 | z2w,1=1 z4=3 |
| | 2 | z2w,1=1 z2w,2=1 z4=2 |
| | 2 | z2w,1=1 z2w,3=1 z4=3 |
| | 2 | z2w,1=1 z2w,2=1 z2w,3=1 z4=2 |
| | 2 | z2w,3=1 z4=2 |
| | 2 | z2w,2=1 z2w,3=1 z4=1 |
| | 2 | z2w,2=1 z2w,3=1 z4=3 |
| | 2 | z2w,3=1 z4=2 |
| | 2 | z2w,1=1 z2w,2=1 z2w,3=1 z4=2 |
| | 2 | z2w,1=1 z2w,3=1 z4=1 |
| | 2 | z2w,1=1 z2w,2=1 z4=2 |
| | 2 | z2w,1=1 z4=1 |
| | 2 | z2w,2=1 z4=1 |
| | 2 | z2w,3=1 z4=3 |
| | 2 | z2w,2=1 z2w,3=1 z4=2 |
| | 2 | z2w,1=1 z2w,3=1 z4=2 |
| | 2 | z2w,1=1 z2w,2=1 z2w,3=1 z4=1 |
| | 2 | z2w,1=1 z4=2 |
| | 2 | z2w,1=1 z2w,2=1 z4=1 |

Continued on next column

**TABLE XIX– continued**

| TQC | ICSD<br>SSG | Stoich. Formula<br>SIs |
|---|---|---|
| | 2 | z4=1 |
| | 2 | z2w,2=1 |
| | 2 | z2w,2=1 z4=2 |
| | 2 | z4=1 |
| | 2 | z2w,1=1 z2w,2=1 z4=1 |
| | 2 | z2w,1=1 |
| | 2 | z2w,1=1 z2w,2=1 z2w,3=1 z4=1 |
| | 2 | z2w,1=1 z2w,3=1 |
| | 2 | z2w,2=1 z2w,3=1 |
| | 2 | z2w,3=1 z4=3 |
| | 2 | z2w,1=1 z2w,2=1 z4=3 |
| | 2 | z2w,1=1 z4=2 |
| | 2 | z2w,2=1 z4=2 |
| | 2 | z4=1 |
| | 2 | z2w,2=1 z2w,3=1 z4=2 |
| | 2 | z2w,3=1 z4=1 |
| | 2 | z2w,1=1 z2w,2=1 z2w,3=1 z4=1 |
| | 2 | z2w,1=1 z2w,3=1 |
| | 2 | z2w,1=1 z2w,3=1 z4=2 |
| | 2 | z2w,1=1 z2w,2=1 z2w,3=1 z4=1 |
| | 2 | z2w,3=1 z4=1 |
| | 2 | z2w,2=1 z2w,3=1 |
| | 2 | z4=1 |
| | 2 | z2w,2=1 |
| | 2 | z2w,1=1 |
| | 2 | z2w,1=1 z2w,2=1 z4=3 |
| | 2 | z2w,1=1 z2w,2=1 z2w,3=1 z4=2 |
| | 2 | z2w,1=1 z2w,3=1 z4=1 |
| | 2 | z2w,2=1 z2w,3=1 z4=1 |
| | 2 | z2w,3=1 |
| | 2 | z2w,2=1 z4=1 |
| | 2 | z2w,1=1 z2w,2=1 |
| | 2 | z2w,1=1 z4=3 |
| | 2 | z2w,1=1 z4=1 |
| | 2 | z2w,1=1 z2w,2=1 |
| | 2 | z2w,2=1 z4=3 |
| | 2 | z2w,3=1 |
| | 2 | z2w,2=1 z2w,3=1 z4=3 |
| | 2 | z2w,1=1 z2w,3=1 z4=3 |
| | 2 | z2w,1=1 z2w,2=1 z2w,3=1 z4=2 |
| 88.4.1 | 78846 | O4Pd2Y1 |
| | 2 | z2w,1=1 z4=1 |
| | 2 | z2w,1=1 z2w,2=1 |
| | 2 | z2w,3=1 |
| | 2 | z2w,2=1 z2w,3=1 z4=3 |
| | 2 | z2w,2=1 z2w,3=1 |
| | 2 | z2w,3=1 z4=3 |
| | 2 | z2w,1=1 z2w,2=1 z4=3 |
| | 2 | z2w,1=1 z4=2 |
| 88.4.2 | 79340 | La1O4Pd2 |
| | 2 | z2w,1=1 z4=1 |
| | 2 | z2w,1=1 z2w,2=1 |
| | 2 | z2w,3=1 |
| | 2 | z2w,2=1 z2w,3=1 z4=3 |
| | 2 | z2w,2=1 z2w,3=1 |
| | 2 | z2w,3=1 z4=3 |
| | 2 | z2w,1=1 z2w,2=1 z4=3 |
| | 2 | z2w,1=1 z4=2 |
| 88.4.3 | 9914 | F4Li1Yb1 |
| | 2 | z2w,1=1 |

Continued on next column

**TABLE XIX– continued**

| TQC | ICSD SSG | Stoich. Formula SIs |
|---|---|---|
| | 2 | z2w,1=1 z2w,2=1 z4=3 |
| | 2 | z2w,3=1 z4=3 |
| | 2 | z2w,2=1 z2w,3=1 z4=2 |
| | 2 | z2w,2=1 z2w,3=1 z4=3 |
| | 2 | z2w,3=1 z4=2 |
| | 2 | z2w,1=1 z2w,2=1 z4=2 |
| | 2 | z2w,1=1 z4=1 |
| 88.4.4 | 429261 | Eu1O4W1 |
| | 2 | z2w,1=1 |
| | 2 | z2w,1=1 z2w,2=1 z4=3 |
| | 2 | z2w,3=1 z4=3 |
| | 2 | z2w,2=1 z2w,3=1 z4=2 |
| | 2 | z2w,2=1 z2w,3=1 z4=3 |
| | 2 | z2w,3=1 z4=2 |
| | 2 | z2w,1=1 z2w,2=1 z4=2 |
| | 2 | z2w,1=1 z4=1 |
| 88.4.5 | 173264 | Cr1Ho1O4 |
| | 2 | z2w,1=1 |
| | 2 | z2w,1=1 z2w,2=1 z4=3 |
| | 2 | z2w,3=1 z4=3 |
| | 2 | z2w,2=1 z2w,3=1 z4=2 |
| | 2 | z2w,2=1 z2w,3=1 z4=3 |
| | 2 | z2w,3=1 z4=2 |
| | 2 | z2w,1=1 z2w,2=1 z4=2 |
| | 2 | z2w,1=1 z4=1 |
| 88.4.6 | 169873 | Cr1Ho1O4 |
| | 2 | z2w,1=1 |
| | 2 | z2w,1=1 z2w,2=1 z4=3 |
| | 2 | z2w,3=1 z4=3 |
| | 2 | z2w,2=1 z2w,3=1 z4=2 |
| | 2 | z2w,2=1 z2w,3=1 z4=3 |
| | 2 | z2w,3=1 z4=2 |
| | 2 | z2w,1=1 z2w,2=1 z4=2 |
| | 2 | z2w,1=1 z4=1 |
| 88.4.7 | 38326 | Cl4Gd1Li1 |
| | 2 | z2w,1=1 |
| | 2 | z2w,1=1 z2w,2=1 z4=3 |
| | 2 | z2w,3=1 z4=3 |
| | 2 | z2w,2=1 z2w,3=1 z4=2 |
| | 2 | z2w,2=1 z2w,3=1 z4=3 |
| | 2 | z2w,3=1 z4=2 |
| | 2 | z2w,1=1 z2w,2=1 z4=2 |
| | 2 | z2w,1=1 z4=1 |
| 113.4.1 | 54208 | Ta30 |
| | 81 | z2=1 |
| | 81 | z2=1 |
| 113.4.2 | 280872 | Ta30 |
| | 81 | z2=1 |
| | 81 | z2=1 |
| 113.4.3 | 250214 | Ta30 |
| | 81 | z2=1 |
| | 81 | z2=1 |
| 113.4.4 | 54206 | Ta22 |
| | 81 | z2=1 |
| | 81 | z2=1 |
| 113.4.5 | 96594 | Ta1 |
| | 81 | z2=1 |
| | 81 | z2=1 |
| 113.4.6 | 89533 | P6Rh7Sr2 |

Continued on next column

**TABLE XIX– continued**

| TQC | ICSD SSG | Stoich. Formula SIs |
|---|---|---|
| | 81 | z2=1 |
| | 81 | z2=1 |
| 113.4.7 | 408045 | Na1Sn5 |
| | 81 | z2=1 |
| | 81 | z2=1 |
| 113.4.8 | 291413 | Mn1Mo1O6Te1 |
| | 81 | z2=1 |
| | 81 | z2=1 |
| 113.4.9 | 239 | Cu3Se2 |
| | 81 | z2=1 |
| | 81 | z2=1 |
| 113.4.10 | 191282 | Cu1K2O7P2 |
| | 81 | z2=1 |
| | 81 | z2=1 |
| 113.4.11 | 262399 | Co1Ge2O7Sr2 |
| | 81 | z2=1 |
| | 81 | z2=1 |
| 113.4.12 | 262398 | Co1Ge2O7Sr2 |
| | 81 | z2=1 |
| | 81 | z2=1 |
| 113.4.13 | 94139 | Ca2Co1O7Si2 |
| | 81 | z2=1 |
| | 81 | z2=1 |
| 113.4.14 | 94137 | Ca2Co1O7Si2 |
| | 81 | z2=1 |
| | 81 | z2=1 |
| 113.4.15 | 94136 | Ca2Co1O7Si2 |
| | 81 | z2=1 |
| | 81 | z2=1 |
| 113.4.16 | 94132 | Ca2Co1O7Si2 |
| | 81 | z2=1 |
| | 81 | z2=1 |
| 113.4.17 | 72543 | Ca2Co1O7Si2 |
| | 81 | z2=1 |
| | 81 | z2=1 |
| 113.4.18 | 186949 | Ca2Co1O7Si2 |
| | 81 | z2=1 |
| | 81 | z2=1 |
| 113.4.19 | 97762 | Ba2Cu1O7Si2 |
| | 81 | z2=1 |
| | 81 | z2=1 |
| 113.4.20 | 281293 | Ba2Co1O7Si2 |
| | 81 | z2=1 |
| | 81 | z2=1 |
| 113.4.21 | 423351 | Ba2Co1Ge2O7 |
| | 81 | z2=1 |
| | 81 | z2=1 |
| 113.4.22 | 423350 | Ba2Co1Ge2O7 |
| | 81 | z2=1 |
| | 81 | z2=1 |
| 113.4.23 | 290484 | Ba2Co1Ge2O7 |
| | 81 | z2=1 |
| | 81 | z2=1 |
| 113.4.24 | 290483 | Ba2Co1Ge2O7 |
| | 81 | z2=1 |
| | 81 | z2=1 |
| 113.4.25 | 154791 | Ag1Er1Te2 |
| | 81 | z2=1 |
| | 81 | z2=1 |

Continued on next column

**TABLE XIX– continued**

| TQC | ICSD SSG | Stoich. Formula SIs |
|---|---|---|
| 113.4.26 | 160122 | Ag1Dy1Te2 |
| | 81 | z2=1 |
| | 81 | z2=1 |
| 113.4.27 | 154794 | Ag1Dy1Te2 |
| | 81 | z2=1 |
| | 81 | z2=1 |
| 124.4.1 | 62049 | Ta1Te4 |
| | 83 | z2w,1=1 z2w,2=1 z4=3 z4mpi=2 z2=1 z8=3 |
| | 83 | z4=1 z2=1 z8=1 |
| | 83 | z2w,1=1 z2w,2=1 z2w,3=1 z4mpi=3 z8=4 |
| | 83 | z2w,3=1 z4=2 z4mpi=1 z8=2 |
| | 83 | z2w,3=1 z4mpi=3 |
| | 83 | z2w,1=1 z2w,2=1 z2w,3=1 z4=2 z4mpi=1 z8=6 |
| | 83 | z4=1 z2=1 z8=1 |
| | 83 | z2w,1=1 z2w,2=1 z4=3 z4mpi=2 z2=1 z8=7 |
| 124.4.2 | 75321 | Nb1Te4 |
| | 83 | z2w,1=1 z2w,2=1 z4=3 z4mpi=2 z2=1 z8=3 |
| | 83 | z4=1 z2=1 z8=1 |
| | 83 | z2w,1=1 z2w,2=1 z2w,3=1 z4mpi=3 z8=4 |
| | 83 | z2w,3=1 z4=2 z4mpi=1 z8=2 |
| | 83 | z2w,3=1 z4mpi=3 |
| | 83 | z2w,1=1 z2w,2=1 z2w,3=1 z4=2 z4mpi=1 z8=6 |
| | 83 | z4=1 z2=1 z8=1 |
| | 83 | z2w,1=1 z2w,2=1 z4=3 z4mpi=2 z2=1 z8=7 |
| 124.4.3 | 75319 | Nb1Te4 |
| | 83 | z2w,1=1 z2w,2=1 z4=3 z4mpi=2 z2=1 z8=3 |
| | 83 | z4=1 z2=1 z8=1 |
| | 83 | z2w,1=1 z2w,2=1 z2w,3=1 z4mpi=3 z8=4 |
| | 83 | z2w,3=1 z4=2 z4mpi=1 z8=2 |
| | 83 | z2w,3=1 z4mpi=3 |
| | 83 | z2w,1=1 z2w,2=1 z2w,3=1 z4=2 z4mpi=1 z8=6 |
| | 83 | z4=1 z2=1 z8=1 |
| | 83 | z2w,1=1 z2w,2=1 z4=3 z4mpi=2 z2=1 z8=7 |
| 124.4.4 | 65130 | Nb1Te4 |
| | 83 | z2w,1=1 z2w,2=1 z4=3 z4mpi=2 z2=1 z8=3 |
| | 83 | z4=1 z2=1 z8=1 |
| | 83 | z2w,1=1 z2w,2=1 z2w,3=1 z4mpi=3 z8=4 |
| | 83 | z2w,3=1 z4=2 z4mpi=1 z8=2 |
| | 83 | z2w,3=1 z4mpi=3 |
| | 83 | z2w,1=1 z2w,2=1 z2w,3=1 z4=2 z4mpi=1 z8=6 |
| | 83 | z4=1 z2=1 z8=1 |
| | 83 | z2w,1=1 z2w,2=1 z4=3 z4mpi=2 z2=1 z8=7 |
| 124.4.5 | 61113 | Nb1Te4 |

Continued on next column

**TABLE XIX– continued**

| TQC | ICSD SSG | Stoich. Formula SIs |
|---|---|---|
| | 81 | z2=1 |
| | 81 | z2=1 |
| | 10 | z2w,1=1 z2w,3=1 z4=1 |
| | 10 | z2w,1=1 z2w,2=1 z4=2 |
| | 10 | z2w,2=1 z2w,3=1 z4=2 |
| | 10 | z4=3 |
| | 10 | z2w,1=1 z2w,2=1 z2w,3=1 z4=2 |
| | 10 | z2w,1=1 z4=3 |
| | 10 | z2w,3=1 z4=3 |
| | 10 | z2w,2=1 |
| | 10 | z2w,2=1 z4=2 |
| | 10 | z2w,3=1 z4=3 |
| | 10 | z2w,1=1 z4=3 |
| | 10 | z2w,1=1 z2w,2=1 z2w,3=1 |
| | 10 | z4=3 |
| | 10 | z2w,2=1 z2w,3=1 |
| | 10 | z2w,1=1 z2w,2=1 |
| | 10 | z2w,1=1 z2w,3=1 z4=1 |
| 124.4.6 | 60599 | Nb1Te4 |
| | 83 | z2w,1=1 z2w,2=1 z4=3 z4mpi=2 z2=1 z8=3 |
| | 83 | z4=1 z2=1 z8=1 |
| | 83 | z2w,1=1 z2w,2=1 z2w,3=1 z4mpi=3 z8=4 |
| | 83 | z2w,3=1 z4=2 z4mpi=1 z8=2 |
| | 83 | z2w,3=1 z4mpi=3 |
| | 83 | z2w,1=1 z2w,2=1 z2w,3=1 z4=2 z4mpi=1 z8=6 |
| | 83 | z4=1 z2=1 z8=1 |
| | 83 | z2w,1=1 z2w,2=1 z4=3 z4mpi=2 z2=1 z8=7 |
| 124.4.7 | 60598 | Nb1Te4 |
| | 83 | z2w,1=1 z2w,2=1 z4=3 z4mpi=2 z2=1 z8=3 |
| | 83 | z4=1 z2=1 z8=1 |
| | 83 | z2w,1=1 z2w,2=1 z2w,3=1 z4mpi=3 z8=4 |
| | 83 | z2w,3=1 z4=2 z4mpi=1 z8=2 |
| | 83 | z2w,3=1 z4mpi=3 |
| | 83 | z2w,1=1 z2w,2=1 z2w,3=1 z4=2 z4mpi=1 z8=6 |
| | 83 | z4=1 z2=1 z8=1 |
| | 83 | z2w,1=1 z2w,2=1 z4=3 z4mpi=2 z2=1 z8=7 |
| 125.4.1 | 412793 | Nd1Rb1Te4 |
| | 111 | z2=1 |
| | 111 | z2=1 |
| | 12 | z2w,3=1 z4=1 |
| | 12 | z2w,1=1 z2w,2=1 z2w,3=1 z4=3 |
| | 12 | z2w,1=1 z2w,2=1 z2w,3=1 z4=3 |
| | 12 | z2w,3=1 z4=1 |
| | 12 | z2w,1=1 z2w,2=1 z2w,3=1 z4=2 |
| | 12 | z2w,3=1 |
| | 12 | z2w,3=1 |
| | 12 | z2w,1=1 z2w,2=1 z2w,3=1 z4=2 |
| | 12 | z2w,1=1 z2w,2=1 z4=2 |
| | 12 | z2w,1=1 z2w,2=1 z4=2 |
| | 12 | z4=3 |
| | 12 | z2w,1=1 z2w,2=1 z4=1 |
| | 12 | z2w,1=1 z2w,2=1 z4=1 |

Continued on next column

**TABLE XIX– continued**

| TQC | ICSD SSG | Stoich. Formula SIs |
|---|---|---|
| | 12 | z4=3 |
| 125.4.2 | 412792 | K1Nd1Te4 |
| | 111 | z2=1 |
| | 111 | z2=1 |
| | 12 | z2w,3=1 z4=1 |
| | 12 | z2w,1=1 z2w,2=1 z2w,3=1 z4=3 |
| | 12 | z2w,1=1 z2w,2=1 z2w,3=1 z4=3 |
| | 12 | z2w,3=1 z4=1 |
| | 12 | z2w,1=1 z2w,2=1 z2w,3=1 z4=2 |
| | 12 | z2w,3=1 |
| | 12 | z2w,3=1 |
| | 12 | z2w,1=1 z2w,2=1 z2w,3=1 z4=2 |
| | 12 | z2w,1=1 z2w,2=1 z4=2 |
| | 12 | z2w,1=1 z2w,2=1 z4=2 |
| | 12 | z4=3 |
| | 12 | z2w,1=1 z2w,2=1 z4=1 |
| | 12 | z2w,1=1 z2w,2=1 z4=1 |
| | 12 | z4=3 |
| 125.4.3 | 391206 | Gd1Rb1Te4 |
| | 111 | z2=1 |
| | 111 | z2=1 |
| | 12 | z2w,3=1 z4=1 |
| | 12 | z2w,1=1 z2w,2=1 z2w,3=1 z4=3 |
| | 12 | z2w,1=1 z2w,2=1 z2w,3=1 z4=3 |
| | 12 | z2w,3=1 z4=1 |
| | 12 | z2w,1=1 z2w,2=1 z2w,3=1 z4=2 |
| | 12 | z2w,3=1 |
| | 12 | z2w,3=1 |
| | 12 | z2w,1=1 z2w,2=1 z2w,3=1 z4=2 |
| | 12 | z2w,1=1 z2w,2=1 z4=2 |
| | 12 | z2w,1=1 z2w,2=1 z4=2 |
| | 12 | z4=3 |
| | 12 | z2w,1=1 z2w,2=1 z4=1 |
| | 12 | z2w,1=1 z2w,2=1 z4=1 |
| | 12 | z4=3 |
| 125.4.4 | 391205 | Gd1K1Te4 |
| | 111 | z2=1 |
| | 111 | z2=1 |
| | 12 | z2w,3=1 z4=1 |
| | 12 | z2w,1=1 z2w,2=1 z2w,3=1 z4=3 |
| | 12 | z2w,1=1 z2w,2=1 z2w,3=1 z4=3 |
| | 12 | z2w,3=1 z4=1 |
| | 12 | z2w,1=1 z2w,2=1 z2w,3=1 z4=2 |
| | 12 | z2w,3=1 |
| | 12 | z2w,3=1 |
| | 12 | z2w,1=1 z2w,2=1 z2w,3=1 z4=2 |
| | 12 | z2w,1=1 z2w,2=1 z4=2 |
| | 12 | z2w,1=1 z2w,2=1 z4=2 |
| | 12 | z4=3 |
| | 12 | z2w,1=1 z2w,2=1 z4=1 |
| | 12 | z2w,1=1 z2w,2=1 z4=1 |
| | 12 | z4=3 |
| 125.4.5 | 59298 | Cs1Gd1Te4 |
| | 111 | z2=1 |
| | 111 | z2=1 |
| | 12 | z2w,3=1 z4=1 |
| | 12 | z2w,1=1 z2w,2=1 z2w,3=1 z4=3 |
| | 12 | z2w,1=1 z2w,2=1 z2w,3=1 z4=3 |
| | 12 | z2w,3=1 z4=1 |
| | 12 | z2w,1=1 z2w,2=1 z2w,3=1 z4=2 |

Continued on next column

**TABLE XIX– continued**

| TQC | ICSD SSG | Stoich. Formula SIs |
|---|---|---|
| | 12 | z2w,3=1 |
| | 12 | z2w,3=1 |
| | 12 | z2w,1=1 z2w,2=1 z2w,3=1 z4=2 |
| | 12 | z2w,1=1 z2w,2=1 z4=2 |
| | 12 | z2w,1=1 z2w,2=1 z4=2 |
| | 12 | z4=3 |
| | 12 | z2w,1=1 z2w,2=1 z4=1 |
| | 12 | z2w,1=1 z2w,2=1 z4=1 |
| | 12 | z4=3 |
| 125.4.6 | 161768 | Ce2Cu1Ga12 |
| | 111 | z2=1 |
| | 111 | z2=1 |
| | 12 | z2w,3=1 z4=3 |
| | 12 | z2w,1=1 z2w,2=1 z2w,3=1 z4=1 |
| | 12 | z2w,1=1 z2w,2=1 z2w,3=1 z4=1 |
| | 12 | z2w,3=1 z4=3 |
| | 12 | z2w,1=1 z2w,2=1 z2w,3=1 |
| | 12 | z2w,3=1 z4=2 |
| | 12 | z2w,3=1 z4=2 |
| | 12 | z2w,1=1 z2w,2=1 z2w,3=1 |
| | 12 | z2w,1=1 z2w,2=1 |
| | 12 | z4=2 |
| | 12 | z4=2 |
| | 12 | z2w,1=1 z2w,2=1 |
| | 12 | z4=1 |
| | 12 | z2w,1=1 z2w,2=1 z4=3 |
| | 12 | z2w,1=1 z2w,2=1 z4=3 |
| | 12 | z4=1 |
| 125.4.7 | 161767 | Ce2Cu1Ga12 |
| | 111 | z2=1 |
| | 111 | z2=1 |
| | 12 | z2w,3=1 z4=2 |
| | 12 | z2w,1=1 z2w,2=1 z2w,3=1 |
| | 12 | z2w,1=1 z2w,2=1 z2w,3=1 |
| | 12 | z2w,3=1 z4=2 |
| | 12 | z2w,1=1 z2w,2=1 z2w,3=1 z4=3 |
| | 12 | z2w,3=1 z4=1 |
| | 12 | z2w,3=1 z4=1 |
| | 12 | z2w,1=1 z2w,2=1 z2w,3=1 z4=3 |
| | 12 | z2w,1=1 z2w,2=1 z4=3 |
| | 12 | z4=1 |
| | 12 | z4=1 |
| | 12 | z2w,1=1 z2w,2=1 z4=3 |
| | 12 | z2w,1=1 z2w,2=1 z4=2 |
| | 12 | z2w,1=1 z2w,2=1 z4=2 |
| 125.4.8 | 67656 | Ce1K1Se4 |
| | 111 | z2=1 |
| | 111 | z2=1 |
| | 12 | z2w,1=1 z2w,2=1 z4=2 |
| | 12 | z2w,1=1 z2w,2=1 z4=2 |
| | 12 | z2w,1=1 z2w,2=1 z4=1 |
| | 12 | z4=3 |
| | 12 | z4=3 |
| | 12 | z2w,1=1 z2w,2=1 z4=1 |
| | 12 | z2w,1=1 z2w,2=1 z2w,3=1 z4=1 |
| | 12 | z2w,3=1 z4=3 |
| | 12 | z2w,3=1 z4=3 |
| | 12 | z2w,1=1 z2w,2=1 z2w,3=1 z4=1 |
| | 12 | z2w,3=1 z4=2 |
| | 12 | z2w,1=1 z2w,2=1 z2w,3=1 |

Continued on next column

**TABLE XIX– continued**

| TQC | ICSD<br>SSG | Stoich. Formula<br>SIs |
|---|---|---|
| | 12 | z2w,1=1 z2w,2=1 z2w,3=1 |
| | 12 | z2w,3=1 z4=2 |
| 126.4.1 | 161960 | Mo2Nd1O8Tl1 |
| | 81 | z2=1 |
| | 81 | z2=1 |
| | 81 | z2=1 |
| | 81 | z2=1 |
| | 2 | z2w,2=1 z4=3 |
| | 2 | z2w,1=1 z4=3 |
| | 2 | z2w,1=1 z2w,2=1 z4=2 |
| | 2 | z2w,1=1 z2w,3=1 z4=3 |
| | 2 | z2w,1=1 z2w,2=1 z2w,3=1 z4=2 |
| | 2 | z2w,3=1 z4=2 |
| | 2 | z2w,2=1 z2w,3=1 z4=1 |
| | 2 | z2w,2=1 z2w,3=1 z4=3 |
| | 2 | z2w,3=1 z4=2 |
| | 2 | z2w,1=1 z2w,2=1 z2w,3=1 z4=2 |
| | 2 | z2w,1=1 z2w,3=1 z4=1 |
| | 2 | z2w,1=1 z2w,2=1 z4=2 |
| | 2 | z2w,1=1 z4=1 |
| | 2 | z2w,2=1 z4=1 |
| | 2 | z2w,3=1 z4=3 |
| | 2 | z2w,2=1 z2w,3=1 z4=2 |
| | 2 | z2w,1=1 z2w,3=1 z4=2 |
| | 2 | z2w,1=1 z2w,2=1 z2w,3=1 z4=1 |
| | 2 | z2w,1=1 z4=2 |
| | 2 | z2w,1=1 z2w,2=1 z4=1 |
| | 2 | z4=1 |
| | 2 | z2w,2=1 |
| | 2 | z2w,2=1 z4=2 |
| | 2 | z4=1 |
| | 2 | z2w,1=1 z2w,2=1 z4=1 |
| | 2 | z2w,1=1 |
| | 2 | z2w,1=1 z2w,2=1 z2w,3=1 z4=1 |
| | 2 | z2w,1=1 z2w,3=1 |
| | 2 | z2w,2=1 z2w,3=1 |
| | 2 | z2w,3=1 z4=3 |
| | 2 | z2w,1=1 z2w,2=1 z4=3 |
| | 2 | z2w,1=1 z4=2 |
| | 2 | z2w,2=1 z4=2 |
| | 2 | z4=1 |
| | 2 | z2w,2=1 z2w,3=1 z4=2 |
| | 2 | z2w,3=1 z4=1 |
| | 2 | z2w,1=1 z2w,2=1 z2w,3=1 z4=1 |
| | 2 | z2w,1=1 z2w,3=1 |
| | 2 | z2w,1=1 z2w,3=1 z4=2 |
| | 2 | z2w,1=1 z2w,2=1 z2w,3=1 z4=1 |
| | 2 | z2w,3=1 z4=1 |
| | 2 | z2w,2=1 z2w,3=1 |
| | 2 | z4=1 |
| | 2 | z2w,2=1 |
| | 2 | z2w,1=1 |
| | 2 | z2w,1=1 z2w,2=1 z4=3 |
| | 2 | z2w,1=1 z2w,2=1 z2w,3=1 z4=2 |
| | 2 | z2w,1=1 z2w,3=1 z4=1 |
| | 2 | z2w,2=1 z2w,3=1 z4=1 |
| | 2 | z2w,3=1 |
| | 2 | z2w,2=1 z4=1 |
| | 2 | z2w,1=1 z2w,2=1 |
| | 2 | z2w,1=1 z4=3 |

Continued on next column

**TABLE XIX– continued**

| TQC | ICSD<br>SSG | Stoich. Formula<br>SIs |
|---|---|---|
| | 2 | z2w,1=1 z4=1 |
| | 2 | z2w,1=1 z2w,2=1 |
| | 2 | z2w,2=1 z4=3 |
| | 2 | z2w,3=1 |
| | 2 | z2w,2=1 z2w,3=1 z4=3 |
| | 2 | z2w,1=1 z2w,3=1 z4=3 |
| | 2 | z2w,1=1 z2w,2=1 z2w,3=1 z4=2 |
| | 2 | z2w,1=1 z2w,2=1 z2w,3=1 z4=3 |
| | 2 | z2w,1=1 z2w,3=1 z4=2 |
| | 2 | z2w,2=1 z2w,3=1 z4=2 |
| | 2 | z2w,3=1 z4=1 |
| | 2 | z2w,2=1 z4=2 |
| | 2 | z4=1 |
| | 2 | z2w,1=1 z2w,2=1 z4=1 |
| | 2 | z2w,1=1 |
| | 2 | z2w,1=1 z4=2 |
| | 2 | z2w,1=1 z2w,2=1 z4=1 |
| | 2 | z4=1 |
| | 2 | z2w,2=1 |
| | 2 | z2w,3=1 z4=1 |
| | 2 | z2w,2=1 z2w,3=1 |
| | 2 | z2w,1=1 z2w,3=1 |
| | 2 | z2w,1=1 z2w,2=1 z2w,3=1 z4=3 |
| | 2 | z2w,1=1 z2w,2=1 z4=2 |
| | 2 | z2w,1=1 z4=1 |
| | 2 | z2w,2=1 z4=1 |
| | 2 | z2w,2=1 z2w,3=1 z4=1 |
| | 2 | z2w,3=1 |
| | 2 | z2w,1=1 z2w,2=1 z2w,3=1 |
| | 2 | z2w,1=1 z2w,3=1 z4=3 |
| | 2 | z2w,1=1 z2w,3=1 z4=1 |
| | 2 | z2w,1=1 z2w,2=1 z2w,3=1 |
| | 2 | z2w,3=1 |
| | 2 | z2w,2=1 z2w,3=1 z4=3 |
| | 2 | z2w,2=1 z4=3 |
| | 2 | z2w,1=1 z4=3 |
| | 2 | z2w,1=1 z2w,2=1 z4=2 |
| | 2 | z2w,3=1 z4=2 |
| | 2 | z2w,2=1 z2w,3=1 z4=1 |
| | 2 | z2w,1=1 z2w,3=1 z4=1 |
| | 2 | z2w,1=1 z2w,2=1 z2w,3=1 |
| | 2 | z2w,1=1 z4=1 |
| | 2 | z2w,1=1 z2w,2=1 |
| | 2 | z2w,2=1 z4=3 |
| | 2 | z2w,2=1 z4=1 |
| | 2 | z2w,1=1 z2w,2=1 |
| | 2 | z2w,1=1 z4=3 |
| | 2 | z2w,1=1 z2w,2=1 z2w,3=1 |
| | 2 | z2w,1=1 z2w,3=1 z4=3 |
| | 2 | z2w,2=1 z2w,3=1 z4=3 |
| | 2 | z2w,3=1 z4=2 |
| | 2 | z4=1 |
| | 2 | z2w,2=1 |
| | 2 | z2w,1=1 |
| | 2 | z2w,1=1 z2w,2=1 z4=3 |
| | 2 | z2w,1=1 z2w,3=1 |
| | 2 | z2w,1=1 z2w,2=1 z2w,3=1 z4=3 |
| | 2 | z2w,3=1 z4=3 |
| | 2 | z2w,2=1 z2w,3=1 z4=2 |
| | 2 | z2w,2=1 z2w,3=1 |

Continued on next column

**TABLE XIX– continued**

| TQC | ICSD SSG | Stoich. Formula SIs |
|---|---|---|
| | 2 | z2w,3=1 z4=3 |
| | 2 | z2w,1=1 z2w,2=1 z2w,3=1 z4=3 |
| | 2 | z2w,1=1 z2w,3=1 z4=2 |
| | 2 | z2w,1=1 z2w,2=1 z4=3 |
| | 2 | z2w,1=1 z4=2 |
| | 2 | z2w,2=1 z4=2 |
| | 2 | z4=1 |
| 127.4.1 | 40962 | Mo4Na1O6 |
| | 81 | z2=1 |
| | 81 | z2=1 |
| | 65 | z2w,1=1 z2w,2=1 z4=3 |
| | 65 | z4=1 |
| | 65 | z4=1 |
| | 65 | z2w,1=1 z2w,2=1 z4=3 |
| 127.4.2 | 420085 | Li1Rh2Si2 |
| | 81 | z2=1 |
| | 81 | z2=1 |
| | 65 | z4=1 |
| | 65 | z2w,1=1 z2w,2=1 z4=3 |
| | 65 | z2w,1=1 z2w,2=1 z4=3 |
| | 65 | z4=1 |
| 127.4.3 | 159116 | La7Ni2Zn1 |
| | 83 | z2w,3=1 z4=3 z4mpi=3 z2=1 z8=3 |
| | 83 | z2w,1=1 z2w,2=1 z2w,3=1 z4=1 z4mpi=3 z2=1 z8=1 |
| | 83 | z2w,1=1 z2w,2=1 z2w,3=1 z4=1 z4mpi=1 z2=1 z8=1 |
| | 83 | z2w,3=1 z4=3 z4mpi=1 z2=1 z8=7 |
| | 83 | z2w,3=1 z4=3 z4mpi=1 z2=1 z8=7 |
| | 83 | z2w,1=1 z2w,2=1 z2w,3=1 z4=1 z4mpi=1 z2=1 z8=5 |
| | 83 | z2w,1=1 z2w,2=1 z2w,3=1 z4=1 z4mpi=3 z2=1 z8=5 |
| | 83 | z2w,3=1 z4=3 z4mpi=3 z2=1 z8=3 |
| | 65 | z4=1 |
| | 65 | z2w,1=1 z2w,2=1 z4=3 |
| | 65 | z2w,1=1 z2w,2=1 z4=3 |
| | 65 | z4=1 |
| 127.4.4 | 68533 | K1Mo4O6 |
| | 81 | z2=1 |
| | 81 | z2=1 |
| | 65 | z2w,1=1 z2w,2=1 z4=2 |
| | 65 | z2w,1=1 z2w,2=1 z4=2 |
| 127.4.5 | 401730 | In5Ti2 |
| | 83 | z4=1 z4mpi=2 z2=1 z8=5 |
| | 83 | z2w,1=1 z2w,2=1 z4=3 z4mpi=2 z2=1 z8=3 |
| | 83 | z2w,1=1 z2w,2=1 z4=3 z2=1 z8=3 |
| | 83 | z4=1 z2=1 z8=1 |
| | 83 | z4=1 z2=1 z8=1 |
| | 83 | z2w,1=1 z2w,2=1 z4=3 z2=1 z8=7 |
| | 83 | z2w,1=1 z2w,2=1 z4=3 z4mpi=2 z2=1 z8=7 |
| | 83 | z4=1 z4mpi=2 z2=1 z8=5 |
| | 65 | z2w,3=1 z4=3 |
| | 65 | z2w,1=1 z2w,2=1 z2w,3=1 z4=1 |
| | 65 | z2w,1=1 z2w,2=1 z2w,3=1 z4=1 |
| | 65 | z2w,3=1 z4=3 |
| 127.4.6 | 192651 | In2Ni9Y1 |

Continued on next column

**TABLE XIX– continued**

| TQC | ICSD SSG | Stoich. Formula SIs |
|---|---|---|
| | 83 | z4mpi=2 z8=4 |
| | 83 | z2w,1=1 z2w,2=1 z4=2 z4mpi=2 z8=2 |
| | 83 | z2w,1=1 z2w,2=1 z4=2 z8=2 |
| | 83 | z2w,1=1 z2w,2=1 z4=2 z8=6 |
| | 83 | z2w,1=1 z2w,2=1 z4=2 z4mpi=2 z8=6 |
| | 83 | z4mpi=2 z8=4 |
| | 65 | z2w,3=1 z4=2 |
| | 65 | z2w,1=1 z2w,2=1 z2w,3=1 |
| | 65 | z2w,1=1 z2w,2=1 z2w,3=1 |
| | 65 | z2w,3=1 z4=2 |
| 127.4.7 | 189069 | In2Ni9Pr1 |
| | 81 | z2=1 |
| | 81 | z2=1 |
| | 65 | z2w,1=1 z2w,2=1 z4=1 |
| | 65 | z4=3 |
| | 65 | z4=3 |
| | 65 | z2w,1=1 z2w,2=1 z4=1 |
| 127.4.8 | 192640 | In2La1Ni9 |
| | 83 | z2w,1=1 z2w,2=1 z4=3 z4mpi=2 z2=1 z8=3 |
| | 83 | z4=1 z4mpi=2 z2=1 z8=1 |
| | 83 | z4=1 z2=1 z8=1 |
| | 83 | z2w,1=1 z2w,2=1 z4=3 z2=1 z8=7 |
| | 83 | z2w,1=1 z2w,2=1 z4=3 z2=1 z8=3 |
| | 83 | z4=1 z2=1 z8=1 |
| | 83 | z4=1 z4mpi=2 z2=1 z8=1 |
| | 83 | z2w,1=1 z2w,2=1 z4=3 z4mpi=2 z2=1 z8=7 |
| | 65 | z2w,1=1 z2w,2=1 z2w,3=1 z4=3 |
| | 65 | z2w,3=1 z4=1 |
| | 65 | z2w,3=1 z4=1 |
| | 65 | z2w,1=1 z2w,2=1 z2w,3=1 z4=3 |
| 127.4.9 | 54484 | In1Pd2Tb2 |
| | 81 | z2=1 |
| | 81 | z2=1 |
| | 65 | z2w,3=1 z4=1 |
| | 65 | z2w,1=1 z2w,2=1 z2w,3=1 z4=3 |
| | 65 | z2w,1=1 z2w,2=1 z2w,3=1 z4=3 |
| | 65 | z2w,3=1 z4=1 |
| 127.4.10 | 54480 | In1Pd2Pr2 |
| | 81 | z2=1 |
| | 81 | z2=1 |
| | 65 | z2w,3=1 |
| | 65 | z2w,1=1 z2w,2=1 z2w,3=1 z4=2 |
| | 65 | z2w,1=1 z2w,2=1 z2w,3=1 z4=2 |
| | 65 | z2w,3=1 |
| 127.4.11 | 107372 | In1Ni2Zr2 |
| | 81 | z2=1 |
| | 81 | z2=1 |
| | 65 | z2w,1=1 z2w,2=1 z2w,3=1 z4=3 |
| | 65 | z2w,3=1 z4=1 |
| | 65 | z2w,3=1 z4=1 |
| | 65 | z2w,1=1 z2w,2=1 z2w,3=1 z4=3 |
| 127.4.12 | 107333 | In1Ni2Sc2 |
| | 81 | z2=1 |
| | 81 | z2=1 |
| | 65 | z2w,3=1 z4=1 |

Continued on next column

**TABLE XIX– continued**

| TQC | ICSD SSG | Stoich. Formula SIs |
|---|---|---|
| | 65 | z2w,1=1 z2w,2=1 z2w,3=1 z4=3 |
| | 65 | z2w,1=1 z2w,2=1 z2w,3=1 z4=3 |
| | 65 | z2w,3=1 z4=1 |
| 127.4.13 | 54490 | In1Lu2Pd2 |
| | 81 | z2=1 |
| | 81 | z2=1 |
| | 65 | z2w,3=1 z4=1 |
| | 65 | z2w,1=1 z2w,2=1 z2w,3=1 z4=3 |
| | 65 | z2w,1=1 z2w,2=1 z2w,3=1 z4=3 |
| | 65 | z2w,3=1 z4=1 |
| 127.4.14 | 404531 | Hf2In5 |
| | 81 | z2=1 |
| | 81 | z2=1 |
| | 65 | z2w,3=1 z4=3 |
| | 65 | z2w,1=1 z2w,2=1 z2w,3=1 z4=1 |
| | 65 | z2w,1=1 z2w,2=1 z2w,3=1 z4=1 |
| | 65 | z2w,3=1 z4=3 |
| 127.4.15 | 402570 | Ge2In1Yb2 |
| | 83 | z2w,1=1 z2w,2=1 z2w,3=1 z4=3 z4mpi=3 z2=1 z8=3 |
| | 83 | z2w,3=1 z4=1 z4mpi=3 z2=1 z8=1 |
| | 83 | z2w,3=1 z4=1 z4mpi=1 z2=1 z8=1 |
| | 83 | z2w,1=1 z2w,2=1 z2w,3=1 z4=3 z4mpi=1 z2=1 z8=7 |
| | 83 | z2w,1=1 z2w,2=1 z2w,3=1 z4=3 z4mpi=1 z2=1 z8=3 |
| | 83 | z2w,3=1 z4=1 z4mpi=1 z2=1 z8=1 |
| | 83 | z2w,3=1 z4=1 z4mpi=3 z2=1 z8=1 |
| | 83 | z2w,1=1 z2w,2=1 z2w,3=1 z4=3 z4mpi=3 z2=1 z8=7 |
| | 65 | z2w,1=1 z2w,2=1 z4=3 |
| | 65 | z4=1 |
| | 65 | z4=1 |
| | 65 | z2w,1=1 z2w,2=1 z4=3 |
| 127.4.16 | 172494 | Ge2In1Yb2 |
| | 83 | z2w,3=1 z4=3 z4mpi=1 z2=1 z8=3 |
| | 83 | z2w,1=1 z2w,2=1 z2w,3=1 z4=1 z4mpi=1 z2=1 z8=1 |
| | 83 | z2w,1=1 z2w,2=1 z2w,3=1 z4=1 z4mpi=3 z2=1 z8=1 |
| | 83 | z2w,3=1 z4=3 z4mpi=3 z2=1 z8=7 |
| | 83 | z2w,3=1 z4=3 z4mpi=3 z2=1 z8=7 |
| | 83 | z2w,1=1 z2w,2=1 z2w,3=1 z4=1 z4mpi=3 z2=1 z8=5 |
| | 83 | z2w,1=1 z2w,2=1 z2w,3=1 z4=1 z4mpi=1 z2=1 z8=5 |
| | 83 | z2w,3=1 z4=3 z4mpi=1 z2=1 z8=3 |
| | 65 | z4=1 |
| | 65 | z2w,1=1 z2w,2=1 z4=3 |
| | 65 | z2w,1=1 z2w,2=1 z4=3 |
| | 65 | z4=1 |
| 127.4.17 | 172491 | Ge2In1Tb2 |
| | 81 | z2=1 |
| | 81 | z2=1 |
| | 65 | z2w,1=1 z2w,2=1 |
| | 65 | z4=2 |
| | 65 | z4=2 |
| | 65 | z2w,1=1 z2w,2=1 |
| 127.4.18 | 414072 | Ge2In1Pr2 |

Continued on next column

**TABLE XIX– continued**

| TQC | ICSD SSG | Stoich. Formula SIs |
|---|---|---|
| | 81 | z2=1 |
| | 81 | z2=1 |
| | 65 | z2w,1=1 z2w,2=1 |
| | 65 | z4=2 |
| | 65 | z4=2 |
| | 65 | z2w,1=1 z2w,2=1 |
| 127.4.19 | 414071 | Ge2In1Nd2 |
| | 83 | z2w,1=1 z2w,2=1 z4=3 z2=1 z8=7 |
| | 83 | z4=1 z2=1 z8=5 |
| | 83 | z4=1 z4mpi=2 z2=1 z8=5 |
| | 83 | z2w,1=1 z2w,2=1 z4=3 z4mpi=2 z2=1 z8=3 |
| | 83 | z2w,1=1 z2w,2=1 z4=3 z4mpi=2 z2=1 z8=7 |
| | 83 | z4=1 z4mpi=2 z2=1 z8=5 |
| | 83 | z4=1 z2=1 z8=5 |
| | 83 | z2w,1=1 z2w,2=1 z4=3 z2=1 z8=3 |
| | 65 | z2w,1=1 z2w,2=1 z2w,3=1 z4=3 |
| | 65 | z2w,3=1 z4=1 |
| | 65 | z2w,3=1 z4=1 |
| | 65 | z2w,1=1 z2w,2=1 z2w,3=1 z4=3 |
| 127.4.20 | 87511 | Ge2In1La2 |
| | 83 | z2w,1=1 z2w,2=1 z4=2 z4mpi=2 z8=2 |
| | 83 | z4mpi=2 |
| | 83 | z2w,1=1 z2w,2=1 z4=2 z8=6 |
| | 83 | z2w,1=1 z2w,2=1 z4=2 z8=2 |
| | 83 | z4mpi=2 |
| | 83 | z2w,1=1 z2w,2=1 z4=2 z4mpi=2 z8=6 |
| | 65 | z2w,1=1 z2w,2=1 z2w,3=1 z4=2 |
| | 65 | z2w,3=1 |
| | 65 | z2w,3=1 |
| | 65 | z2w,1=1 z2w,2=1 z2w,3=1 z4=2 |
| 127.4.21 | 172493 | Ge2Ho2In1 |
| | 81 | z2=1 |
| | 81 | z2=1 |
| | 65 | z2w,1=1 z2w,2=1 z4=1 |
| | 65 | z4=3 |
| | 65 | z4=3 |
| | 65 | z2w,1=1 z2w,2=1 z4=1 |
| 127.4.22 | 249632 | Ga5Mn2 |
| | 83 | z2w,1=1 z2w,2=1 z4=2 z8=6 |
| | 83 | z8=4 |
| | 83 | z4mpi=2 z8=4 |
| | 83 | z2w,1=1 z2w,2=1 z4=2 z4mpi=2 z8=2 |
| | 83 | z2w,1=1 z2w,2=1 z4=2 z4mpi=2 z8=6 |
| | 83 | z4mpi=2 z8=4 |
| | 83 | z8=4 |
| | 83 | z2w,1=1 z2w,2=1 z4=2 z8=2 |
| | 65 | z2w,1=1 z2w,2=1 z2w,3=1 z4=2 |
| | 65 | z2w,3=1 |
| | 65 | z2w,3=1 |
| | 65 | z2w,1=1 z2w,2=1 z2w,3=1 z4=2 |
| 127.4.23 | 85499 | F4Fe1Rb1 |
| | 83 | z2w,1=1 z2w,2=1 z4=2 z8=6 |
| | 83 | z2w,1=1 z2w,2=1 z4=2 z4mpi=2 z8=6 |

Continued on next column

**TABLE XIX– continued**

| TQC | ICSD SSG | Stoich. Formula SIs |
|---|---|---|
| | 83 | z4mpi=2 z8=4 |
| | 83 | z4mpi=2 z8=4 |
| | 83 | z2w,1=1 z2w,2=1 z4=2 z4mpi=2 z8=2 |
| | 83 | z2w,1=1 z2w,2=1 z4=2 z8=2 |
| | 65 | z2w,3=1 z4=2 |
| | 65 | z2w,1=1 z2w,2=1 z2w,3=1 |
| | 65 | z2w,1=1 z2w,2=1 z2w,3=1 |
| | 65 | z2w,3=1 z4=2 |
| 127.4.24 | 54487 | Er2In1Pd2 |
| | 81 | z2=1 |
| | 81 | z2=1 |
| | 65 | z2w,1=1 z2w,2=1 z4=3 |
| | 65 | z4=1 |
| | 65 | z4=1 |
| | 65 | z2w,1=1 z2w,2=1 z4=3 |
| 127.4.25 | 107343 | Dy2In1Pd2 |
| | 83 | z4=3 z4mpi=2 z2=1 z8=3 |
| | 83 | z2w,1=1 z2w,2=1 z4=1 z4mpi=2 z2=1 z8=1 |
| | 83 | z2w,1=1 z2w,2=1 z4=1 z2=1 z8=1 |
| | 83 | z4=3 z2=1 z8=7 |
| | 83 | z4=3 z2=1 z8=7 |
| | 83 | z2w,1=1 z2w,2=1 z4=1 z2=1 z8=5 |
| | 83 | z2w,1=1 z2w,2=1 z4=1 z4mpi=2 z2=1 z8=5 |
| | 83 | z4=3 z4mpi=2 z2=1 z8=3 |
| | 65 | z2w,3=1 z4=1 |
| | 65 | z2w,1=1 z2w,2=1 z2w,3=1 z4=3 |
| | 65 | z2w,1=1 z2w,2=1 z2w,3=1 z4=3 |
| | 65 | z2w,3=1 z4=1 |
| 127.4.26 | 411708 | Cu2In1La2 |
| | 81 | z2=1 |
| | 81 | z2=1 |
| | 65 | z2w,3=1 |
| | 65 | z2w,1=1 z2w,2=1 z2w,3=1 z4=2 |
| | 65 | z2w,1=1 z2w,2=1 z2w,3=1 z4=2 |
| | 65 | z2w,3=1 |
| 127.4.27 | 21111 | Cu1F3K1 |
| | 81 | z2=1 |
| | 81 | z2=1 |
| | 65 | z2w,1=1 z2w,2=1 z4=2 |
| | 65 | z2w,1=1 z2w,2=1 z4=2 |
| 127.4.28 | 21110 | Cu1F3K1 |
| | 81 | z2=1 |
| | 81 | z2=1 |
| | 65 | z2w,1=1 z2w,2=1 z4=2 |
| | 65 | z2w,1=1 z2w,2=1 z4=2 |
| 127.4.29 | 404847 | Co2In1Zr2 |
| | 81 | z2=1 |
| | 81 | z2=1 |
| | 65 | z2w,3=1 z4=1 |
| | 65 | z2w,1=1 z2w,2=1 z2w,3=1 z4=3 |
| | 65 | z2w,1=1 z2w,2=1 z2w,3=1 z4=3 |
| | 65 | z2w,3=1 z4=1 |
| 127.4.30 | 107331 | Co2In1Zr2 |
| | 81 | z2=1 |
| | 81 | z2=1 |
| | 65 | z2w,3=1 z4=1 |

Continued on next column

**TABLE XIX– continued**

| TQC | ICSD SSG | Stoich. Formula SIs |
|---|---|---|
| | 65 | z2w,1=1 z2w,2=1 z2w,3=1 z4=3 |
| | 65 | z2w,1=1 z2w,2=1 z2w,3=1 z4=3 |
| | 65 | z2w,3=1 z4=1 |
| 127.4.31 | 410293 | Ce2In1Pt2 |
| | 81 | z2=1 |
| | 81 | z2=1 |
| | 65 | z2w,3=1 |
| | 65 | z2w,1=1 z2w,2=1 z2w,3=1 z4=2 |
| | 65 | z2w,1=1 z2w,2=1 z2w,3=1 z4=2 |
| | 65 | z2w,3=1 |
| 127.4.32 | 150615 | Ce2In1Ni2 |
| | 81 | z2=1 |
| | 81 | z2=1 |
| | 65 | z2w,3=1 |
| | 65 | z2w,1=1 z2w,2=1 z2w,3=1 z4=2 |
| | 65 | z2w,1=1 z2w,2=1 z2w,3=1 z4=2 |
| | 65 | z2w,3=1 |
| 127.4.33 | 415108 | Ce2Ge2In1 |
| | 83 | z2w,1=1 z2w,2=1 z4=3 z2=1 z8=7 |
| | 83 | z4=1 z2=1 z8=5 |
| | 83 | z4=1 z4mpi=2 z2=1 z8=5 |
| | 83 | z2w,1=1 z2w,2=1 z4=3 z4mpi=2 z2=1 z8=3 |
| | 83 | z2w,1=1 z2w,2=1 z4=3 z4mpi=2 z2=1 z8=7 |
| | 83 | z4=1 z4mpi=2 z2=1 z8=5 |
| | 83 | z4=1 z2=1 z8=5 |
| | 83 | z2w,1=1 z2w,2=1 z4=3 z2=1 z8=3 |
| | 65 | z2w,1=1 z2w,2=1 z2w,3=1 z4=3 |
| | 65 | z2w,3=1 z4=1 |
| | 65 | z2w,3=1 z4=1 |
| | 65 | z2w,1=1 z2w,2=1 z2w,3=1 z4=3 |
| 127.4.34 | 424134 | Ca1Cd2Cu9 |
| | 83 | z2w,1=1 z2w,2=1 z2w,3=1 z4mpi=1 |
| | 83 | z2w,3=1 z4=2 z4mpi=1 z8=6 |
| | 83 | z2w,3=1 z4=2 z4mpi=3 z8=6 |
| | 83 | z2w,1=1 z2w,2=1 z2w,3=1 z4mpi=3 z8=4 |
| | 83 | z2w,1=1 z2w,2=1 z2w,3=1 z4mpi=3 |
| | 83 | z2w,3=1 z4=2 z4mpi=3 z8=6 |
| | 83 | z2w,3=1 z4=2 z4mpi=1 z8=6 |
| | 83 | z2w,1=1 z2w,2=1 z2w,3=1 z4mpi=1 z8=4 |
| | 65 | z2w,1=1 z2w,2=1 |
| | 65 | z4=2 |
| | 65 | z4=2 |
| | 65 | z2w,1=1 z2w,2=1 |
| 127.4.35 | 427155 | C2B2Y1 |
| | 81 | z2=1 |
| | 81 | z2=1 |
| | 65 | z2w,3=1 z4=2 |
| | 65 | z2w,1=1 z2w,2=1 z2w,3=1 |
| | 65 | z2w,1=1 z2w,2=1 z2w,3=1 |
| | 65 | z2w,3=1 z4=2 |
| 127.4.36 | 94043 | C2B2Tm1 |
| | 83 | z2w,3=1 z4=2 z4mpi=3 z8=2 |
| | 83 | z2w,1=1 z2w,2=1 z2w,3=1 z4mpi=3 |
| | 83 | z2w,1=1 z2w,2=1 z2w,3=1 z4mpi=1 |
| | 83 | z2w,3=1 z4=2 z4mpi=1 z8=6 |

Continued on next column

**TABLE XIX– continued**

| TQC | ICSD SSG | Stoich. Formula SIs |
|---|---|---|
| | 83 | z2w,3=1 z4=2 z4mpi=1 z8=6 |
| | 83 | z2w,1=1 z2w,2=1 z2w,3=1 z4mpi=1 z8=4 |
| | 83 | z2w,1=1 z2w,2=1 z2w,3=1 z4mpi=3 z8=4 |
| | 83 | z2w,3=1 z4=2 z4mpi=3 z8=2 |
| | 65 | z2w,1=1 z2w,2=1 z4=2 |
| | 65 | z2w,1=1 z2w,2=1 z4=2 |
| 127.4.37 | 94039 | C2B2Tb1 |
| | 83 | z2w,3=1 z4=2 z4mpi=1 z8=6 |
| | 83 | z2w,1=1 z2w,2=1 z2w,3=1 z4mpi=1 z8=4 |
| | 83 | z2w,1=1 z2w,2=1 z2w,3=1 z4mpi=3 z8=4 |
| | 83 | z2w,3=1 z4=2 z4mpi=3 z8=2 |
| | 83 | z2w,3=1 z4=2 z4mpi=3 z8=2 |
| | 83 | z2w,1=1 z2w,2=1 z2w,3=1 z4mpi=3 |
| | 83 | z2w,1=1 z2w,2=1 z2w,3=1 z4mpi=1 |
| | 83 | z2w,3=1 z4=2 z4mpi=1 z8=6 |
| | 65 | z2w,1=1 z2w,2=1 z4=2 |
| | 65 | z2w,1=1 z2w,2=1 z4=2 |
| 127.4.38 | 94037 | C2B2Pr1 |
| | 83 | z2w,3=1 z4=2 z4mpi=1 z8=6 |
| | 83 | z2w,1=1 z2w,2=1 z2w,3=1 z4mpi=1 z8=4 |
| | 83 | z2w,1=1 z2w,2=1 z2w,3=1 z4mpi=3 z8=4 |
| | 83 | z2w,3=1 z4=2 z4mpi=3 z8=2 |
| | 83 | z2w,3=1 z4=2 z4mpi=3 z8=2 |
| | 83 | z2w,1=1 z2w,2=1 z2w,3=1 z4mpi=3 |
| | 83 | z2w,1=1 z2w,2=1 z2w,3=1 z4mpi=1 |
| | 83 | z2w,3=1 z4=2 z4mpi=1 z8=6 |
| | 65 | z2w,1=1 z2w,2=1 z4=2 |
| | 65 | z2w,1=1 z2w,2=1 z4=2 |
| 127.4.39 | 94044 | C2B2Lu1 |
| | 81 | z2=1 |
| | 81 | z2=1 |
| | 65 | z2w,3=1 z4=2 |
| | 65 | z2w,1=1 z2w,2=1 z2w,3=1 |
| | 65 | z2w,1=1 z2w,2=1 z2w,3=1 |
| | 65 | z2w,3=1 z4=2 |
| 127.4.40 | 94035 | C2B2La1 |
| | 81 | z2=1 |
| | 81 | z2=1 |
| | 65 | z2w,3=1 z4=2 |
| | 65 | z2w,1=1 z2w,2=1 z2w,3=1 |
| | 65 | z2w,1=1 z2w,2=1 z2w,3=1 |
| | 65 | z2w,3=1 z4=2 |
| 127.4.41 | 63497 | Ba1Cu1Nd2O5 |
| | 83 | z2w,3=1 z4mpi=3 |
| | 83 | z2w,1=1 z2w,2=1 z2w,3=1 z4=2 z4mpi=3 z8=6 |
| | 83 | z2w,1=1 z2w,2=1 z2w,3=1 z4=2 z4mpi=1 z8=6 |
| | 83 | z2w,3=1 z4mpi=1 z8=4 |
| | 83 | z2w,3=1 z4mpi=1 z8=4 |
| | 83 | z2w,1=1 z2w,2=1 z2w,3=1 z4=2 z4mpi=1 z8=2 |
| | 83 | z2w,1=1 z2w,2=1 z2w,3=1 z4=2 z4mpi=3 z8=2 |

Continued on next column

**TABLE XIX– continued**

| TQC | ICSD SSG | Stoich. Formula SIs |
|---|---|---|
| | 83 | z2w,3=1 z4mpi=3 |
| | 65 | z4=2 |
| | 65 | z2w,1=1 z2w,2=1 |
| | 65 | z2w,1=1 z2w,2=1 |
| | 65 | z4=2 |
| 127.4.42 | 88317 | B2V3 |
| | 81 | z2=1 |
| | 81 | z2=1 |
| | 65 | z2w,3=1 z4=2 |
| | 65 | z2w,1=1 z2w,2=1 z2w,3=1 |
| | 65 | z2w,1=1 z2w,2=1 z2w,3=1 |
| | 65 | z2w,3=1 z4=2 |
| 127.4.43 | 427424 | B2Ru5Ta3 |
| | 83 | z2w,1=1 z2w,2=1 z2w,3=1 z4mpi=1 z8=4 |
| | 83 | z2w,3=1 z4=2 z4mpi=1 z8=2 |
| | 83 | z2w,3=1 z4=2 z4mpi=3 z8=2 |
| | 83 | z2w,1=1 z2w,2=1 z2w,3=1 z4mpi=3 |
| | 83 | z2w,1=1 z2w,2=1 z2w,3=1 z4mpi=3 z8=4 |
| | 83 | z2w,3=1 z4=2 z4mpi=3 z8=2 |
| | 83 | z2w,3=1 z4=2 z4mpi=1 z8=2 |
| | 83 | z2w,1=1 z2w,2=1 z2w,3=1 z4mpi=1 |
| | 65 | z2w,1=1 z2w,2=1 |
| | 65 | z4=2 |
| | 65 | z4=2 |
| | 65 | z2w,1=1 z2w,2=1 |
| 127.4.44 | 423469 | B2Nb3Ru5 |
| | 83 | z2w,1=1 z2w,2=1 z2w,3=1 z4mpi=1 z8=4 |
| | 83 | z2w,3=1 z4=2 z4mpi=1 z8=2 |
| | 83 | z2w,3=1 z4=2 z4mpi=3 z8=2 |
| | 83 | z2w,1=1 z2w,2=1 z2w,3=1 z4mpi=3 |
| | 83 | z2w,1=1 z2w,2=1 z2w,3=1 z4mpi=3 z8=4 |
| | 83 | z2w,3=1 z4=2 z4mpi=3 z8=2 |
| | 83 | z2w,3=1 z4=2 z4mpi=1 z8=2 |
| | 83 | z2w,1=1 z2w,2=1 z2w,3=1 z4mpi=1 |
| | 65 | z2w,1=1 z2w,2=1 |
| | 65 | z4=2 |
| | 65 | z4=2 |
| | 65 | z2w,1=1 z2w,2=1 |
| 127.4.45 | 69487 | B2Ir5Mg2Si1 |
| | 81 | z2=1 |
| | 81 | z2=1 |
| | 65 | z2w,1=1 z2w,2=1 |
| | 65 | z4=2 |
| | 65 | z4=2 |
| | 65 | z2w,1=1 z2w,2=1 |
| 127.4.46 | 44316 | B2Hf3Ir5 |
| | 81 | z2=1 |
| | 81 | z2=1 |
| | 65 | z4=1 |
| | 65 | z2w,1=1 z2w,2=1 z4=3 |
| | 65 | z2w,1=1 z2w,2=1 z4=3 |
| | 65 | z4=1 |
| 127.4.47 | 51432 | B2Ge1Mg2Rh5 |
| | 81 | z2=1 |
| | 81 | z2=1 |

Continued on next column

**TABLE XIX– continued**

| TQC | ICSD SSG | Stoich. Formula SIs |
|---|---|---|
| | 65 | z2w,3=1 z4=2 |
| | 65 | z2w,1=1 z2w,2=1 z2w,3=1 |
| | 65 | z2w,1=1 z2w,2=1 z2w,3=1 |
| | 65 | z2w,3=1 z4=2 |
| 127.4.48 | 51437 | B2Fe1Rh5Sc2 |
| | 81 | z2=1 |
| | 81 | z2=1 |
| | 65 | z2w,1=1 z2w,2=1 z2w,3=1 z4=2 |
| | 65 | z2w,3=1 |
| | 65 | z2w,3=1 |
| | 65 | z2w,1=1 z2w,2=1 z2w,3=1 z4=2 |
| 127.4.49 | 51435 | B2Cr1Rh5Sc2 |
| | 81 | z2=1 |
| | 81 | z2=1 |
| | 65 | z2w,1=1 z2w,2=1 z2w,3=1 z4=2 |
| | 65 | z2w,3=1 |
| | 65 | z2w,3=1 |
| | 65 | z2w,1=1 z2w,2=1 z2w,3=1 z4=2 |
| 127.4.50 | 262043 | Au2Y3 |
| | 83 | z2w,3=1 z4=2 z4mpi=3 z8=6 |
| | 83 | z2w,1=1 z2w,2=1 z2w,3=1 z4mpi=3 z8=4 |
| | 83 | z2w,1=1 z2w,2=1 z2w,3=1 z4mpi=1 z8=4 |
| | 83 | z2w,3=1 z4=2 z4mpi=1 z8=2 |
| | 83 | z2w,3=1 z4=2 z4mpi=1 z8=2 |
| | 83 | z2w,1=1 z2w,2=1 z2w,3=1 z4mpi=1 |
| | 83 | z2w,1=1 z2w,2=1 z2w,3=1 z4mpi=3 |
| | 83 | z2w,3=1 z4=2 z4mpi=3 z8=6 |
| | 65 | z2w,1=1 z2w,2=1 z4=2 |
| | 65 | z2w,1=1 z2w,2=1 z4=2 |
| 127.4.51 | 92455 | Al1Si2Yb2 |
| | 83 | z2w,3=1 z4=3 z4mpi=1 z2=1 z8=3 |
| | 83 | z2w,1=1 z2w,2=1 z2w,3=1 z4=1 z4mpi=1 z2=1 z8=1 |
| | 83 | z2w,1=1 z2w,2=1 z2w,3=1 z4=1 z4mpi=3 z2=1 z8=1 |
| | 83 | z2w,3=1 z4=3 z4mpi=3 z2=1 z8=7 |
| | 83 | z2w,3=1 z4=3 z4mpi=3 z2=1 z8=7 |
| | 83 | z2w,1=1 z2w,2=1 z2w,3=1 z4=1 z4mpi=3 z2=1 z8=5 |
| | 83 | z2w,1=1 z2w,2=1 z2w,3=1 z4=1 z4mpi=1 z2=1 z8=5 |
| | 83 | z2w,3=1 z4=3 z4mpi=1 z2=1 z8=3 |
| | 65 | z4=1 |
| | 65 | z2w,1=1 z2w,2=1 z4=3 |
| | 65 | z2w,1=1 z2w,2=1 z4=3 |
| | 65 | z4=1 |
| 128.4.1 | 245749 | Cl6K2Nb1 |
| | 81 | z2=1 |
| | 81 | z2=1 |
| | 10 | z2w,1=1 z4=1 |
| | 10 | z2w,3=1 z4=1 |
| | 10 | z2w,1=1 z2w,3=1 z4=2 |
| | 10 | z2w,2=1 z4=1 |
| | 10 | z2w,1=1 z2w,2=1 z4=2 |
| | 10 | z2w,2=1 z2w,3=1 z4=2 |
| | 10 | z2w,1=1 z2w,2=1 z2w,3=1 z4=3 |
| | 10 | z2w,1=1 z2w,2=1 z2w,3=1 z4=1 |

Continued on next column

**TABLE XIX– continued**

| TQC | ICSD SSG | Stoich. Formula SIs |
|---|---|---|
| | 10 | z2w,2=1 z2w,3=1 z4=2 |
| | 10 | z2w,1=1 z2w,2=1 z4=2 |
| | 10 | z2w,2=1 z4=3 |
| | 10 | z2w,1=1 z2w,3=1 z4=2 |
| | 10 | z2w,3=1 z4=3 |
| | 10 | z2w,1=1 z4=3 |
| 129.4.1 | 183309 | Sn2Y1Zn1 |
| | 115 | z2=1 |
| | 115 | z2=1 |
| | 12 | z2w,3=1 |
| | 12 | z2w,1=1 z2w,2=1 z2w,3=1 z4=2 |
| | 12 | z2w,1=1 z2w,2=1 z2w,3=1 z4=2 |
| | 12 | z2w,3=1 |
| | 12 | z2w,1=1 z2w,2=1 z2w,3=1 z4=1 |
| | 12 | z2w,3=1 z4=3 |
| | 12 | z2w,3=1 z4=3 |
| | 12 | z2w,1=1 z2w,2=1 z2w,3=1 z4=1 |
| | 12 | z2w,1=1 z2w,2=1 z4=1 |
| | 12 | z4=3 |
| | 12 | z4=3 |
| | 12 | z2w,1=1 z2w,2=1 z4=1 |
| | 12 | z4=2 |
| | 12 | z2w,1=1 z2w,2=1 |
| | 12 | z2w,1=1 z2w,2=1 |
| | 12 | z4=2 |
| 129.4.2 | 163425 | Sn2Tb1Zn1 |
| | 115 | z2=1 |
| | 115 | z2=1 |
| | 12 | z2w,3=1 z4=3 |
| | 12 | z2w,1=1 z2w,2=1 z2w,3=1 z4=1 |
| | 12 | z2w,1=1 z2w,2=1 z2w,3=1 z4=1 |
| | 12 | z2w,3=1 z4=3 |
| | 12 | z2w,1=1 z2w,2=1 z2w,3=1 |
| | 12 | z2w,3=1 z4=2 |
| | 12 | z2w,3=1 z4=2 |
| | 12 | z2w,1=1 z2w,2=1 z2w,3=1 |
| | 12 | z2w,1=1 z2w,2=1 |
| | 12 | z4=2 |
| | 12 | z4=2 |
| | 12 | z2w,1=1 z2w,2=1 |
| | 12 | z4=1 |
| | 12 | z2w,1=1 z2w,2=1 z4=3 |
| | 12 | z2w,1=1 z2w,2=1 z4=3 |
| | 12 | z4=1 |
| 129.4.3 | 42337 | Si1V1Zr1 |
| | 115 | z2=1 |
| | 115 | z2=1 |
| | 12 | z2w,1=1 z2w,2=1 z4=2 |
| | 12 | z2w,1=1 z2w,2=1 z4=2 |
| | 12 | z2w,1=1 z2w,2=1 z4=1 |
| | 12 | z4=3 |
| | 12 | z4=3 |
| | 12 | z2w,1=1 z2w,2=1 z4=1 |
| | 12 | z2w,1=1 z2w,2=1 z2w,3=1 z4=1 |
| | 12 | z2w,3=1 z4=3 |
| | 12 | z2w,3=1 z4=3 |
| | 12 | z2w,1=1 z2w,2=1 z2w,3=1 z4=1 |
| | 12 | z2w,3=1 z4=2 |
| | 12 | z2w,1=1 z2w,2=1 z2w,3=1 |
| | 12 | z2w,1=1 z2w,2=1 z2w,3=1 |

Continued on next column

**TABLE XIX– continued**

| TQC | ICSD SSG | Stoich. Formula SIs |
|---|---|---|
| | 12 | z2w,3=1 z4=2 |
| 129.4.4 | 88113 | Si1Ti1Y1 |
| | 115 | z2=1 |
| | 115 | z2=1 |
| | 12 | z2w,3=1 z4=2 |
| | 12 | z2w,1=1 z2w,2=1 z2w,3=1 |
| | 12 | z2w,1=1 z2w,2=1 z2w,3=1 |
| | 12 | z2w,3=1 z4=2 |
| | 12 | z2w,1=1 z2w,2=1 z2w,3=1 z4=3 |
| | 12 | z2w,3=1 z4=1 |
| | 12 | z2w,3=1 z4=1 |
| | 12 | z2w,1=1 z2w,2=1 z2w,3=1 z4=3 |
| | 12 | z2w,1=1 z2w,2=1 z4=3 |
| | 12 | z4=1 |
| | 12 | z4=1 |
| | 12 | z2w,1=1 z2w,2=1 z4=3 |
| | 12 | z2w,1=1 z2w,2=1 z4=2 |
| | 12 | z2w,1=1 z2w,2=1 z4=2 |
| 129.4.5 | 88115 | Si1Tb1Ti1 |
| | 115 | z2=1 |
| | 115 | z2=1 |
| | 12 | z4=1 |
| | 12 | z2w,1=1 z2w,2=1 z4=3 |
| | 12 | z2w,1=1 z2w,2=1 z4=3 |
| | 12 | z4=1 |
| | 12 | z2w,1=1 z2w,2=1 z4=2 |
| | 12 | z2w,1=1 z2w,2=1 z4=2 |
| | 12 | z2w,1=1 z2w,2=1 z2w,3=1 z4=2 |
| | 12 | z2w,3=1 |
| | 12 | z2w,3=1 |
| | 12 | z2w,1=1 z2w,2=1 z2w,3=1 z4=2 |
| | 12 | z2w,3=1 z4=3 |
| | 12 | z2w,1=1 z2w,2=1 z2w,3=1 z4=1 |
| | 12 | z2w,1=1 z2w,2=1 z2w,3=1 z4=1 |
| | 12 | z2w,3=1 z4=3 |
| 129.4.6 | 26274 | Sb1Te1U1 |
| | 115 | z2=1 |
| | 115 | z2=1 |
| | 12 | z4=3 |
| | 12 | z2w,1=1 z2w,2=1 z4=1 |
| | 12 | z2w,1=1 z2w,2=1 z4=1 |
| | 12 | z4=3 |
| | 12 | z2w,1=1 z2w,2=1 |
| | 12 | z4=2 |
| | 12 | z4=2 |
| | 12 | z2w,1=1 z2w,2=1 |
| | 12 | z2w,1=1 z2w,2=1 z2w,3=1 |
| | 12 | z2w,3=1 z4=2 |
| | 12 | z2w,3=1 z4=2 |
| | 12 | z2w,1=1 z2w,2=1 z2w,3=1 |
| | 12 | z2w,3=1 z4=1 |
| | 12 | z2w,1=1 z2w,2=1 z2w,3=1 z4=3 |
| | 12 | z2w,1=1 z2w,2=1 z2w,3=1 z4=3 |
| | 12 | z2w,3=1 z4=1 |
| 129.4.7 | 42486 | Sb1Se1U1 |
| | 115 | z2=1 |
| | 115 | z2=1 |
| | 12 | z4=3 |
| | 12 | z2w,1=1 z2w,2=1 z4=1 |
| | 12 | z2w,1=1 z2w,2=1 z4=1 |

Continued on next column

**TABLE XIX– continued**

| TQC | ICSD SSG | Stoich. Formula SIs |
|---|---|---|
| | 12 | z4=3 |
| | 12 | z2w,1=1 z2w,2=1 |
| | 12 | z4=2 |
| | 12 | z4=2 |
| | 12 | z2w,1=1 z2w,2=1 |
| | 12 | z2w,1=1 z2w,2=1 z2w,3=1 |
| | 12 | z2w,3=1 z4=2 |
| | 12 | z2w,3=1 z4=2 |
| | 12 | z2w,1=1 z2w,2=1 z2w,3=1 |
| | 12 | z2w,3=1 z4=1 |
| | 12 | z2w,1=1 z2w,2=1 z2w,3=1 z4=3 |
| | 12 | z2w,1=1 z2w,2=1 z2w,3=1 z4=3 |
| | 12 | z2w,3=1 z4=1 |
| 129.4.8 | 409586 | Sb1Sc2 |
| | 115 | z2=1 |
| | 115 | z2=1 |
| | 12 | z4=1 |
| | 12 | z2w,1=1 z2w,2=1 z4=3 |
| | 12 | z2w,1=1 z2w,2=1 z4=3 |
| | 12 | z4=1 |
| | 12 | z2w,1=1 z2w,2=1 z4=2 |
| | 12 | z2w,1=1 z2w,2=1 z4=2 |
| | 12 | z2w,1=1 z2w,2=1 z2w,3=1 z4=2 |
| | 12 | z2w,3=1 |
| | 12 | z2w,3=1 |
| | 12 | z2w,1=1 z2w,2=1 z2w,3=1 z4=2 |
| | 12 | z2w,3=1 z4=3 |
| | 12 | z2w,1=1 z2w,2=1 z2w,3=1 z4=1 |
| | 12 | z2w,1=1 z2w,2=1 z2w,3=1 z4=1 |
| | 12 | z2w,3=1 z4=3 |
| 129.4.9 | 193328 | Sb1Sc1Tb1 |
| | 115 | z2=1 |
| | 115 | z2=1 |
| | 12 | z4=1 |
| | 12 | z2w,1=1 z2w,2=1 z4=3 |
| | 12 | z2w,1=1 z2w,2=1 z4=3 |
| | 12 | z4=1 |
| | 12 | z2w,1=1 z2w,2=1 z4=2 |
| | 12 | z2w,1=1 z2w,2=1 z4=2 |
| | 12 | z2w,1=1 z2w,2=1 z2w,3=1 z4=2 |
| | 12 | z2w,3=1 |
| | 12 | z2w,3=1 |
| | 12 | z2w,1=1 z2w,2=1 z2w,3=1 z4=2 |
| | 12 | z2w,3=1 z4=3 |
| | 12 | z2w,1=1 z2w,2=1 z2w,3=1 z4=1 |
| | 12 | z2w,1=1 z2w,2=1 z2w,3=1 z4=1 |
| | 12 | z2w,3=1 z4=3 |
| 129.4.10 | 41253 | Pr1Ru1Si1 |
| | 115 | z2=1 |
| | 115 | z2=1 |
| | 12 | z2w,3=1 z4=3 |
| | 12 | z2w,1=1 z2w,2=1 z2w,3=1 z4=1 |
| | 12 | z2w,1=1 z2w,2=1 z2w,3=1 z4=1 |
| | 12 | z2w,3=1 z4=3 |
| | 12 | z2w,1=1 z2w,2=1 z2w,3=1 |
| | 12 | z2w,3=1 z4=2 |
| | 12 | z2w,3=1 z4=2 |
| | 12 | z2w,1=1 z2w,2=1 z2w,3=1 |
| | 12 | z2w,1=1 z2w,2=1 |
| | 12 | z4=2 |

Continued on next column

**TABLE XIX– continued**

| TQC | ICSD SSG | Stoich. Formula SIs |
|---|---|---|
| | 12 | z4=2 |
| | 12 | z2w,1=1 z2w,2=1 |
| | 12 | z4=1 |
| | 12 | z2w,1=1 z2w,2=1 z4=3 |
| | 12 | z2w,1=1 z2w,2=1 z4=3 |
| | 12 | z4=1 |
| 129.4.11 | 93154 | Pd1Pr1Sb2 |
| | 115 | z2=1 |
| | 115 | z2=1 |
| | 12 | z2w,3=1 z4=1 |
| | 12 | z2w,1=1 z2w,2=1 z2w,3=1 z4=3 |
| | 12 | z2w,1=1 z2w,2=1 z2w,3=1 z4=3 |
| | 12 | z2w,3=1 z4=1 |
| | 12 | z2w,1=1 z2w,2=1 z2w,3=1 z4=2 |
| | 12 | z2w,3=1 |
| | 12 | z2w,3=1 |
| | 12 | z2w,1=1 z2w,2=1 z2w,3=1 z4=2 |
| | 12 | z2w,1=1 z2w,2=1 z4=2 |
| | 12 | z2w,1=1 z2w,2=1 z4=2 |
| | 12 | z4=3 |
| | 12 | z2w,1=1 z2w,2=1 z4=1 |
| | 12 | z2w,1=1 z2w,2=1 z4=1 |
| | 12 | z4=3 |
| 129.4.12 | 63594 | P1Sn1Sr1 |
| | 115 | z2=1 |
| | 115 | z2=1 |
| | 12 | z2w,1=1 z2w,2=1 z4=2 |
| | 12 | z2w,1=1 z2w,2=1 z4=2 |
| | 12 | z2w,1=1 z2w,2=1 z4=1 |
| | 12 | z4=3 |
| | 12 | z4=3 |
| | 12 | z2w,1=1 z2w,2=1 z4=1 |
| | 12 | z2w,1=1 z2w,2=1 z2w,3=1 z4=1 |
| | 12 | z2w,3=1 z4=3 |
| | 12 | z2w,3=1 z4=3 |
| | 12 | z2w,1=1 z2w,2=1 z2w,3=1 z4=1 |
| | 12 | z2w,3=1 z4=2 |
| | 12 | z2w,1=1 z2w,2=1 z2w,3=1 |
| | 12 | z2w,1=1 z2w,2=1 z2w,3=1 |
| | 12 | z2w,3=1 z4=2 |
| 129.4.13 | 42169 | P1Se1U1 |
| | 115 | z2=1 |
| | 115 | z2=1 |
| | 12 | z4=3 |
| | 12 | z2w,1=1 z2w,2=1 z4=1 |
| | 12 | z2w,1=1 z2w,2=1 z4=1 |
| | 12 | z4=3 |
| | 12 | z2w,1=1 z2w,2=1 |
| | 12 | z4=2 |
| | 12 | z4=2 |
| | 12 | z2w,1=1 z2w,2=1 |
| | 12 | z2w,1=1 z2w,2=1 z2w,3=1 |
| | 12 | z2w,3=1 z4=2 |
| | 12 | z2w,3=1 z4=2 |
| | 12 | z2w,1=1 z2w,2=1 z2w,3=1 |
| | 12 | z2w,3=1 z4=1 |
| | 12 | z2w,1=1 z2w,2=1 z2w,3=1 z4=3 |
| | 12 | z2w,1=1 z2w,2=1 z2w,3=1 z4=3 |
| | 12 | z2w,3=1 z4=1 |
| 129.4.14 | 77793 | P1S1U1 |

Continued on next column

**TABLE XIX– continued**

| TQC | ICSD SSG | Stoich. Formula SIs |
|---|---|---|
| | 115 | z2=1 |
| | 115 | z2=1 |
| | 12 | z4=3 |
| | 12 | z2w,1=1 z2w,2=1 z4=1 |
| | 12 | z2w,1=1 z2w,2=1 z4=1 |
| | 12 | z4=3 |
| | 12 | z2w,1=1 z2w,2=1 |
| | 12 | z4=2 |
| | 12 | z4=2 |
| | 12 | z2w,1=1 z2w,2=1 |
| | 12 | z2w,1=1 z2w,2=1 z2w,3=1 |
| | 12 | z2w,3=1 z4=2 |
| | 12 | z2w,3=1 z4=2 |
| | 12 | z2w,1=1 z2w,2=1 z2w,3=1 |
| | 12 | z2w,3=1 z4=1 |
| | 12 | z2w,1=1 z2w,2=1 z2w,3=1 z4=3 |
| | 12 | z2w,1=1 z2w,2=1 z2w,3=1 z4=3 |
| | 12 | z2w,3=1 z4=1 |
| 129.4.15 | 57162 | P1S1U1 |
| | 115 | z2=1 |
| | 115 | z2=1 |
| | 12 | z4=3 |
| | 12 | z2w,1=1 z2w,2=1 z4=1 |
| | 12 | z2w,1=1 z2w,2=1 z4=1 |
| | 12 | z4=3 |
| | 12 | z2w,1=1 z2w,2=1 |
| | 12 | z4=2 |
| | 12 | z4=2 |
| | 12 | z2w,1=1 z2w,2=1 |
| | 12 | z2w,1=1 z2w,2=1 z2w,3=1 |
| | 12 | z2w,3=1 z4=2 |
| | 12 | z2w,3=1 z4=2 |
| | 12 | z2w,1=1 z2w,2=1 z2w,3=1 |
| | 12 | z2w,3=1 z4=1 |
| | 12 | z2w,1=1 z2w,2=1 z2w,3=1 z4=3 |
| | 12 | z2w,1=1 z2w,2=1 z2w,3=1 z4=3 |
| | 12 | z2w,3=1 z4=1 |
| 129.4.16 | 154031 | P1S1Th1 |
| | 115 | z2=1 |
| | 115 | z2=1 |
| | 12 | z2w,1=1 z2w,2=1 z4=2 |
| | 12 | z2w,1=1 z2w,2=1 z4=2 |
| | 12 | z2w,1=1 z2w,2=1 z4=1 |
| | 12 | z4=3 |
| | 12 | z4=3 |
| | 12 | z2w,1=1 z2w,2=1 z4=1 |
| | 12 | z2w,1=1 z2w,2=1 z2w,3=1 z4=1 |
| | 12 | z2w,3=1 z4=3 |
| | 12 | z2w,3=1 z4=3 |
| | 12 | z2w,1=1 z2w,2=1 z2w,3=1 z4=1 |
| | 12 | z2w,3=1 z4=2 |
| | 12 | z2w,1=1 z2w,2=1 z2w,3=1 |
| | 12 | z2w,1=1 z2w,2=1 z2w,3=1 |
| | 12 | z2w,3=1 z4=2 |
| 129.4.17 | 73036 | Np1Pt2Si2 |
| | 115 | z2=1 |
| | 115 | z2=1 |
| | 12 | z4=2 |
| | 12 | z2w,1=1 z2w,2=1 |
| | 12 | z2w,1=1 z2w,2=1 |

Continued on next column

**TABLE XIX– continued**

| TQC | ICSD SSG | Stoich. Formula SIs |
|---|---|---|
| | 12 | z4=2 |
| | 12 | z2w,1=1 z2w,2=1 z4=3 |
| | 12 | z4=1 |
| | 12 | z4=1 |
| | 12 | z2w,1=1 z2w,2=1 z4=3 |
| | 12 | z2w,1=1 z2w,2=1 z2w,3=1 z4=3 |
| | 12 | z2w,3=1 z4=1 |
| | 12 | z2w,3=1 z4=1 |
| | 12 | z2w,1=1 z2w,2=1 z2w,3=1 z4=3 |
| | 12 | z2w,3=1 |
| | 12 | z2w,1=1 z2w,2=1 z2w,3=1 z4=2 |
| | 12 | z2w,1=1 z2w,2=1 z2w,3=1 z4=2 |
| | 12 | z2w,3=1 |
| 129.4.18 | 421034 | Np1O1Se1 |
| | 115 | z2=1 |
| | 115 | z2=1 |
| | 12 | z2w,1=1 z2w,2=1 z4=2 |
| | 12 | z2w,1=1 z2w,2=1 z4=2 |
| | 12 | z2w,1=1 z2w,2=1 z4=1 |
| | 12 | z4=3 |
| | 12 | z4=3 |
| | 12 | z2w,1=1 z2w,2=1 z4=1 |
| | 12 | z2w,1=1 z2w,2=1 z2w,3=1 z4=1 |
| | 12 | z2w,3=1 z4=3 |
| | 12 | z2w,3=1 z4=3 |
| | 12 | z2w,1=1 z2w,2=1 z2w,3=1 z4=1 |
| | 12 | z2w,3=1 z4=2 |
| | 12 | z2w,1=1 z2w,2=1 z2w,3=1 |
| | 12 | z2w,1=1 z2w,2=1 z2w,3=1 |
| | 12 | z2w,3=1 z4=2 |
| 129.4.19 | 421032 | Np1O1S1 |
| | 115 | z2=1 |
| | 115 | z2=1 |
| | 12 | z2w,1=1 z2w,2=1 z4=2 |
| | 12 | z2w,1=1 z2w,2=1 z4=2 |
| | 12 | z2w,1=1 z2w,2=1 z4=1 |
| | 12 | z4=3 |
| | 12 | z4=3 |
| | 12 | z2w,1=1 z2w,2=1 z4=1 |
| | 12 | z2w,1=1 z2w,2=1 z2w,3=1 z4=1 |
| | 12 | z2w,3=1 z4=3 |
| | 12 | z2w,3=1 z4=3 |
| | 12 | z2w,1=1 z2w,2=1 z2w,3=1 z4=1 |
| | 12 | z2w,3=1 z4=2 |
| | 12 | z2w,1=1 z2w,2=1 z2w,3=1 |
| | 12 | z2w,1=1 z2w,2=1 z2w,3=1 |
| | 12 | z2w,3=1 z4=2 |
| 129.4.20 | 31656 | Np1O1S1 |
| | 115 | z2=1 |
| | 115 | z2=1 |
| | 12 | z2w,1=1 z2w,2=1 z4=2 |
| | 12 | z2w,1=1 z2w,2=1 z4=2 |
| | 12 | z2w,1=1 z2w,2=1 z4=1 |
| | 12 | z4=3 |
| | 12 | z4=3 |
| | 12 | z2w,1=1 z2w,2=1 z4=1 |
| | 12 | z2w,1=1 z2w,2=1 z2w,3=1 z4=1 |
| | 12 | z2w,3=1 z4=3 |
| | 12 | z2w,3=1 z4=3 |
| | 12 | z2w,1=1 z2w,2=1 z2w,3=1 z4=1 |

Continued on next column

**TABLE XIX– continued**

| TQC | ICSD SSG | Stoich. Formula SIs |
|---|---|---|
| | 12 | z2w,3=1 z4=2 |
| | 12 | z2w,1=1 z2w,2=1 z2w,3=1 |
| | 12 | z2w,1=1 z2w,2=1 z2w,3=1 |
| | 12 | z2w,3=1 z4=2 |
| 129.4.21 | 419354 | Nd1O1Sb1Zn1 |
| | 115 | z2=1 |
| | 115 | z2=1 |
| | 12 | z2w,1=1 z2w,2=1 z4=2 |
| | 12 | z2w,1=1 z2w,2=1 z4=2 |
| | 12 | z2w,1=1 z2w,2=1 z4=1 |
| | 12 | z4=3 |
| | 12 | z4=3 |
| | 12 | z2w,1=1 z2w,2=1 z4=1 |
| | 12 | z2w,1=1 z2w,2=1 z2w,3=1 z4=1 |
| | 12 | z2w,3=1 z4=3 |
| | 12 | z2w,3=1 z4=3 |
| | 12 | z2w,1=1 z2w,2=1 z2w,3=1 z4=1 |
| | 12 | z2w,3=1 z4=2 |
| | 12 | z2w,1=1 z2w,2=1 z2w,3=1 |
| | 12 | z2w,1=1 z2w,2=1 z2w,3=1 |
| | 12 | z2w,3=1 z4=2 |
| 129.4.22 | 82004 | Na1Nd1O4Ti1 |
| | 115 | z2=1 |
| | 115 | z2=1 |
| | 12 | z2w,1=1 z2w,2=1 z4=2 |
| | 12 | z2w,1=1 z2w,2=1 z4=2 |
| | 12 | z2w,1=1 z2w,2=1 z4=1 |
| | 12 | z4=3 |
| | 12 | z4=3 |
| | 12 | z2w,1=1 z2w,2=1 z4=1 |
| | 12 | z2w,1=1 z2w,2=1 z2w,3=1 z4=1 |
| | 12 | z2w,3=1 z4=3 |
| | 12 | z2w,3=1 z4=3 |
| | 12 | z2w,1=1 z2w,2=1 z2w,3=1 z4=1 |
| | 12 | z2w,3=1 z4=2 |
| | 12 | z2w,1=1 z2w,2=1 z2w,3=1 |
| | 12 | z2w,1=1 z2w,2=1 z2w,3=1 |
| | 12 | z2w,3=1 z4=2 |
| 129.4.23 | 422049 | Na1Nd1O4Ti1 |
| | 115 | z2=1 |
| | 115 | z2=1 |
| | 12 | z2w,1=1 z2w,2=1 z4=2 |
| | 12 | z2w,1=1 z2w,2=1 z4=2 |
| | 12 | z2w,1=1 z2w,2=1 z4=1 |
| | 12 | z4=3 |
| | 12 | z4=3 |
| | 12 | z2w,1=1 z2w,2=1 z4=1 |
| | 12 | z2w,1=1 z2w,2=1 z2w,3=1 z4=1 |
| | 12 | z2w,3=1 z4=3 |
| | 12 | z2w,3=1 z4=3 |
| | 12 | z2w,1=1 z2w,2=1 z2w,3=1 z4=1 |
| | 12 | z2w,3=1 z4=2 |
| | 12 | z2w,1=1 z2w,2=1 z2w,3=1 |
| | 12 | z2w,1=1 z2w,2=1 z2w,3=1 |
| | 12 | z2w,3=1 z4=2 |
| 129.4.24 | 87145 | N1Te1U1 |
| | 115 | z2=1 |
| | 115 | z2=1 |
| | 12 | z2w,1=1 z2w,2=1 z4=2 |
| | 12 | z2w,1=1 z2w,2=1 z4=2 |

Continued on next column

**TABLE XIX– continued**

| TQC | ICSD<br>SSG | Stoich. Formula<br>SIs |
|---|---|---|
| | 12 | z2w,1=1 z2w,2=1 z4=1 |
| | 12 | z4=3 |
| | 12 | z4=3 |
| | 12 | z2w,1=1 z2w,2=1 z4=1 |
| | 12 | z2w,1=1 z2w,2=1 z2w,3=1 z4=1 |
| | 12 | z2w,3=1 z4=3 |
| | 12 | z2w,3=1 z4=3 |
| | 12 | z2w,1=1 z2w,2=1 z2w,3=1 z4=1 |
| | 12 | z2w,3=1 z4=2 |
| | 12 | z2w,1=1 z2w,2=1 z2w,3=1 |
| | 12 | z2w,1=1 z2w,2=1 z2w,3=1 |
| | 12 | z2w,3=1 z4=2 |
| 129.4.25 | 105125 | N1Te1U1 |
| | 115 | z2=1 |
| | 115 | z2=1 |
| | 12 | z2w,1=1 z2w,2=1 z4=2 |
| | 12 | z2w,1=1 z2w,2=1 z4=2 |
| | 12 | z2w,1=1 z2w,2=1 z4=1 |
| | 12 | z4=3 |
| | 12 | z4=3 |
| | 12 | z2w,1=1 z2w,2=1 z4=1 |
| | 12 | z2w,1=1 z2w,2=1 z2w,3=1 z4=1 |
| | 12 | z2w,3=1 z4=3 |
| | 12 | z2w,3=1 z4=3 |
| | 12 | z2w,1=1 z2w,2=1 z2w,3=1 z4=1 |
| | 12 | z2w,3=1 z4=2 |
| | 12 | z2w,1=1 z2w,2=1 z2w,3=1 |
| | 12 | z2w,1=1 z2w,2=1 z2w,3=1 |
| | 12 | z2w,3=1 z4=2 |
| 129.4.26 | 105124 | N1Te1U1 |
| | 115 | z2=1 |
| | 115 | z2=1 |
| | 12 | z2w,1=1 z2w,2=1 z4=2 |
| | 12 | z2w,1=1 z2w,2=1 z4=2 |
| | 12 | z2w,1=1 z2w,2=1 z4=1 |
| | 12 | z4=3 |
| | 12 | z4=3 |
| | 12 | z2w,1=1 z2w,2=1 z4=1 |
| | 12 | z2w,1=1 z2w,2=1 z2w,3=1 z4=1 |
| | 12 | z2w,3=1 z4=3 |
| | 12 | z2w,3=1 z4=3 |
| | 12 | z2w,1=1 z2w,2=1 z2w,3=1 z4=1 |
| | 12 | z2w,3=1 z4=2 |
| | 12 | z2w,1=1 z2w,2=1 z2w,3=1 |
| | 12 | z2w,1=1 z2w,2=1 z2w,3=1 |
| | 12 | z2w,3=1 z4=2 |
| 129.4.27 | 76214 | Mn2Sb1 |
| | 115 | z2=1 |
| | 115 | z2=1 |
| | 12 | z4=3 |
| | 12 | z2w,1=1 z2w,2=1 z4=1 |
| | 12 | z2w,1=1 z2w,2=1 z4=1 |
| | 12 | z4=3 |
| | 12 | z2w,1=1 z2w,2=1 |
| | 12 | z4=2 |
| | 12 | z4=2 |
| | 12 | z2w,1=1 z2w,2=1 |
| | 12 | z2w,1=1 z2w,2=1 z2w,3=1 |
| | 12 | z2w,3=1 z4=2 |
| | 12 | z2w,3=1 z4=2 |

Continued on next column

**TABLE XIX– continued**

| TQC | ICSD<br>SSG | Stoich. Formula<br>SIs |
|---|---|---|
| | 12 | z2w,1=1 z2w,2=1 z2w,3=1 |
| | 12 | z2w,3=1 z4=1 |
| | 12 | z2w,1=1 z2w,2=1 z2w,3=1 z4=3 |
| | 12 | z2w,1=1 z2w,2=1 z2w,3=1 z4=3 |
| | 12 | z2w,3=1 z4=1 |
| 129.4.28 | 45688 | Mn2Sb1 |
| | 115 | z2=1 |
| | 115 | z2=1 |
| | 12 | z4=3 |
| | 12 | z2w,1=1 z2w,2=1 z4=1 |
| | 12 | z2w,1=1 z2w,2=1 z4=1 |
| | 12 | z4=3 |
| | 12 | z2w,1=1 z2w,2=1 |
| | 12 | z4=2 |
| | 12 | z4=2 |
| | 12 | z2w,1=1 z2w,2=1 |
| | 12 | z2w,1=1 z2w,2=1 z2w,3=1 |
| | 12 | z2w,3=1 z4=2 |
| | 12 | z2w,3=1 z4=2 |
| | 12 | z2w,1=1 z2w,2=1 z2w,3=1 |
| | 12 | z2w,3=1 z4=1 |
| | 12 | z2w,1=1 z2w,2=1 z2w,3=1 z4=3 |
| | 12 | z2w,1=1 z2w,2=1 z2w,3=1 z4=3 |
| | 12 | z2w,3=1 z4=1 |
| 129.4.29 | 42325 | Mn2Sb1 |
| | 115 | z2=1 |
| | 115 | z2=1 |
| | 12 | z4=3 |
| | 12 | z2w,1=1 z2w,2=1 z4=1 |
| | 12 | z2w,1=1 z2w,2=1 z4=1 |
| | 12 | z4=3 |
| | 12 | z2w,1=1 z2w,2=1 |
| | 12 | z4=2 |
| | 12 | z4=2 |
| | 12 | z2w,1=1 z2w,2=1 |
| | 12 | z2w,1=1 z2w,2=1 z2w,3=1 |
| | 12 | z2w,3=1 z4=2 |
| | 12 | z2w,3=1 z4=2 |
| | 12 | z2w,1=1 z2w,2=1 z2w,3=1 |
| | 12 | z2w,3=1 z4=1 |
| | 12 | z2w,1=1 z2w,2=1 z2w,3=1 z4=3 |
| | 12 | z2w,1=1 z2w,2=1 z2w,3=1 z4=3 |
| | 12 | z2w,3=1 z4=1 |
| 129.4.30 | 415627 | Mn2Sb1 |
| | 115 | z2=1 |
| | 115 | z2=1 |
| | 12 | z4=3 |
| | 12 | z2w,1=1 z2w,2=1 z4=1 |
| | 12 | z2w,1=1 z2w,2=1 z4=1 |
| | 12 | z4=3 |
| | 12 | z2w,1=1 z2w,2=1 |
| | 12 | z4=2 |
| | 12 | z4=2 |
| | 12 | z2w,1=1 z2w,2=1 |
| | 12 | z2w,1=1 z2w,2=1 z2w,3=1 |
| | 12 | z2w,3=1 z4=2 |
| | 12 | z2w,3=1 z4=2 |
| | 12 | z2w,1=1 z2w,2=1 z2w,3=1 |
| | 12 | z2w,3=1 z4=1 |
| | 12 | z2w,1=1 z2w,2=1 z2w,3=1 z4=3 |

Continued on next column

**TABLE XIX– continued**

| TQC | ICSD SSG | Stoich. Formula SIs |
|---|---|---|
| | 12 | z2w,1=1 z2w,2=1 z2w,3=1 z4=3 |
| | 12 | z2w,3=1 z4=1 |
| 129.4.31 | 190141 | Mn2Sb1 |
| | 115 | z2=1 |
| | 115 | z2=1 |
| | 12 | z4=3 |
| | 12 | z2w,1=1 z2w,2=1 z4=1 |
| | 12 | z2w,1=1 z2w,2=1 z4=1 |
| | 12 | z4=3 |
| | 12 | z2w,1=1 z2w,2=1 |
| | 12 | z4=2 |
| | 12 | z4=2 |
| | 12 | z2w,1=1 z2w,2=1 |
| | 12 | z2w,1=1 z2w,2=1 z2w,3=1 |
| | 12 | z2w,3=1 z4=2 |
| | 12 | z2w,3=1 z4=2 |
| | 12 | z2w,1=1 z2w,2=1 z2w,3=1 |
| | 12 | z2w,3=1 z4=1 |
| | 12 | z2w,1=1 z2w,2=1 z2w,3=1 z4=3 |
| | 12 | z2w,1=1 z2w,2=1 z2w,3=1 z4=3 |
| | 12 | z2w,3=1 z4=1 |
| 129.4.32 | 174032 | Mn2Sb1 |
| | 115 | z2=1 |
| | 115 | z2=1 |
| | 12 | z4=3 |
| | 12 | z2w,1=1 z2w,2=1 z4=1 |
| | 12 | z2w,1=1 z2w,2=1 z4=1 |
| | 12 | z4=3 |
| | 12 | z2w,1=1 z2w,2=1 |
| | 12 | z4=2 |
| | 12 | z4=2 |
| | 12 | z2w,1=1 z2w,2=1 |
| | 12 | z2w,1=1 z2w,2=1 z2w,3=1 |
| | 12 | z2w,3=1 z4=2 |
| | 12 | z2w,3=1 z4=2 |
| | 12 | z2w,1=1 z2w,2=1 z2w,3=1 |
| | 12 | z2w,3=1 z4=1 |
| | 12 | z2w,1=1 z2w,2=1 z2w,3=1 z4=3 |
| | 12 | z2w,1=1 z2w,2=1 z2w,3=1 z4=3 |
| | 12 | z2w,3=1 z4=1 |
| 129.4.33 | 66963 | Mn1Sn1Sr1 |
| | 115 | z2=1 |
| | 115 | z2=1 |
| | 12 | z2w,1=1 z2w,2=1 z4=2 |
| | 12 | z2w,1=1 z2w,2=1 z4=2 |
| | 12 | z2w,1=1 z2w,2=1 z4=1 |
| | 12 | z4=3 |
| | 12 | z4=3 |
| | 12 | z2w,1=1 z2w,2=1 z4=1 |
| | 12 | z2w,1=1 z2w,2=1 z2w,3=1 z4=1 |
| | 12 | z2w,3=1 z4=3 |
| | 12 | z2w,3=1 z4=3 |
| | 12 | z2w,1=1 z2w,2=1 z2w,3=1 z4=1 |
| | 12 | z2w,3=1 z4=2 |
| | 12 | z2w,1=1 z2w,2=1 z2w,3=1 |
| | 12 | z2w,1=1 z2w,2=1 z2w,3=1 |
| | 12 | z2w,3=1 z4=2 |
| 129.4.34 | 66962 | Mn1Sn1Sr1 |
| | 115 | z2=1 |
| | 115 | z2=1 |

Continued on next column

**TABLE XIX– continued**

| TQC | ICSD SSG | Stoich. Formula SIs |
|---|---|---|
| | 12 | z2w,1=1 z2w,2=1 z4=2 |
| | 12 | z2w,1=1 z2w,2=1 z4=2 |
| | 12 | z2w,1=1 z2w,2=1 z4=1 |
| | 12 | z4=3 |
| | 12 | z4=3 |
| | 12 | z2w,1=1 z2w,2=1 z4=1 |
| | 12 | z2w,1=1 z2w,2=1 z2w,3=1 z4=1 |
| | 12 | z2w,3=1 z4=3 |
| | 12 | z2w,3=1 z4=3 |
| | 12 | z2w,1=1 z2w,2=1 z2w,3=1 z4=1 |
| | 12 | z2w,3=1 z4=2 |
| | 12 | z2w,1=1 z2w,2=1 z2w,3=1 |
| | 12 | z2w,1=1 z2w,2=1 z2w,3=1 |
| | 12 | z2w,3=1 z4=2 |
| 129.4.35 | 66951 | Mn1Sn1Sr1 |
| | 115 | z2=1 |
| | 115 | z2=1 |
| | 12 | z4=1 |
| | 12 | z2w,1=1 z2w,2=1 z4=3 |
| | 12 | z2w,1=1 z2w,2=1 z4=3 |
| | 12 | z4=1 |
| | 12 | z2w,1=1 z2w,2=1 z4=2 |
| | 12 | z2w,1=1 z2w,2=1 z4=2 |
| | 12 | z2w,1=1 z2w,2=1 z2w,3=1 z4=2 |
| | 12 | z2w,3=1 |
| | 12 | z2w,3=1 |
| | 12 | z2w,1=1 z2w,2=1 z2w,3=1 z4=2 |
| | 12 | z2w,3=1 z4=3 |
| | 12 | z2w,1=1 z2w,2=1 z2w,3=1 z4=1 |
| | 12 | z2w,1=1 z2w,2=1 z2w,3=1 z4=1 |
| | 12 | z2w,3=1 z4=3 |
| 129.4.36 | 41868 | Mn1Rb1Sb1 |
| | 115 | z2=1 |
| | 115 | z2=1 |
| | 12 | z2w,3=1 z4=2 |
| | 12 | z2w,1=1 z2w,2=1 z2w,3=1 |
| | 12 | z2w,1=1 z2w,2=1 z2w,3=1 |
| | 12 | z2w,3=1 z4=2 |
| | 12 | z2w,1=1 z2w,2=1 z2w,3=1 z4=3 |
| | 12 | z2w,3=1 z4=1 |
| | 12 | z2w,3=1 z4=1 |
| | 12 | z2w,1=1 z2w,2=1 z2w,3=1 z4=3 |
| | 12 | z2w,1=1 z2w,2=1 z4=3 |
| | 12 | z4=1 |
| | 12 | z4=1 |
| | 12 | z2w,1=1 z2w,2=1 z4=3 |
| | 12 | z2w,1=1 z2w,2=1 z4=2 |
| | 12 | z2w,1=1 z2w,2=1 z4=2 |
| 129.4.37 | 89595 | Mn1P1Rb1 |
| | 115 | z2=1 |
| | 115 | z2=1 |
| | 12 | z2w,1=1 z2w,2=1 z4=2 |
| | 12 | z2w,1=1 z2w,2=1 z4=2 |
| | 12 | z2w,1=1 z2w,2=1 z4=1 |
| | 12 | z4=3 |
| | 12 | z4=3 |
| | 12 | z2w,1=1 z2w,2=1 z4=1 |
| | 12 | z2w,1=1 z2w,2=1 z2w,3=1 z4=1 |
| | 12 | z2w,3=1 z4=3 |
| | 12 | z2w,3=1 z4=3 |

Continued on next column

**TABLE XIX– continued**

| TQC | ICSD SSG | Stoich. Formula SIs |
|---|---|---|
| | 12 | z2w,1=1 z2w,2=1 z2w,3=1 z4=1 |
| | 12 | z2w,3=1 z4=2 |
| | 12 | z2w,1=1 z2w,2=1 z2w,3=1 |
| | 12 | z2w,1=1 z2w,2=1 z2w,3=1 |
| | 12 | z2w,3=1 z4=2 |
| 129.4.38 | 89594 | Mn1P1Rb1 |
| | 115 | z2=1 |
| | 115 | z2=1 |
| | 12 | z2w,3=1 z4=3 |
| | 12 | z2w,1=1 z2w,2=1 z2w,3=1 z4=1 |
| | 12 | z2w,1=1 z2w,2=1 z2w,3=1 z4=1 |
| | 12 | z2w,3=1 z4=3 |
| | 12 | z2w,1=1 z2w,2=1 z2w,3=1 |
| | 12 | z2w,3=1 z4=2 |
| | 12 | z2w,3=1 z4=2 |
| | 12 | z2w,1=1 z2w,2=1 z2w,3=1 |
| | 12 | z2w,1=1 z2w,2=1 |
| | 12 | z4=2 |
| | 12 | z4=2 |
| | 12 | z2w,1=1 z2w,2=1 |
| | 12 | z4=1 |
| | 12 | z2w,1=1 z2w,2=1 z4=3 |
| | 12 | z2w,1=1 z2w,2=1 z4=3 |
| | 12 | z4=1 |
| 129.4.39 | 85851 | Mn1Nd1Si1 |
| | 115 | z2=1 |
| | 115 | z2=1 |
| | 12 | z4=1 |
| | 12 | z2w,1=1 z2w,2=1 z4=3 |
| | 12 | z2w,1=1 z2w,2=1 z4=3 |
| | 12 | z4=1 |
| | 12 | z2w,1=1 z2w,2=1 z4=2 |
| | 12 | z2w,1=1 z2w,2=1 z4=2 |
| | 12 | z2w,1=1 z2w,2=1 z2w,3=1 z4=2 |
| | 12 | z2w,3=1 |
| | 12 | z2w,3=1 |
| | 12 | z2w,1=1 z2w,2=1 z2w,3=1 z4=2 |
| | 12 | z2w,3=1 z4=3 |
| | 12 | z2w,1=1 z2w,2=1 z2w,3=1 z4=1 |
| | 12 | z2w,1=1 z2w,2=1 z2w,3=1 z4=1 |
| | 12 | z2w,3=1 z4=3 |
| 129.4.40 | 75057 | Mn1Nd1Si1 |
| | 115 | z2=1 |
| | 115 | z2=1 |
| | 12 | z4=1 |
| | 12 | z2w,1=1 z2w,2=1 z4=3 |
| | 12 | z2w,1=1 z2w,2=1 z4=3 |
| | 12 | z4=1 |
| | 12 | z2w,1=1 z2w,2=1 z4=2 |
| | 12 | z2w,1=1 z2w,2=1 z4=2 |
| | 12 | z2w,1=1 z2w,2=1 z2w,3=1 z4=2 |
| | 12 | z2w,3=1 |
| | 12 | z2w,3=1 |
| | 12 | z2w,1=1 z2w,2=1 z2w,3=1 z4=2 |
| | 12 | z2w,3=1 z4=3 |
| | 12 | z2w,1=1 z2w,2=1 z2w,3=1 z4=1 |
| | 12 | z2w,1=1 z2w,2=1 z2w,3=1 z4=1 |
| | 12 | z2w,3=1 z4=3 |
| 129.4.41 | 75056 | Mn1Nd1Si1 |
| | 115 | z2=1 |

Continued on next column

**TABLE XIX– continued**

| TQC | ICSD SSG | Stoich. Formula SIs |
|---|---|---|
| | 115 | z2=1 |
| | 12 | z4=1 |
| | 12 | z2w,1=1 z2w,2=1 z4=3 |
| | 12 | z2w,1=1 z2w,2=1 z4=3 |
| | 12 | z4=1 |
| | 12 | z2w,1=1 z2w,2=1 z4=2 |
| | 12 | z2w,1=1 z2w,2=1 z4=2 |
| | 12 | z2w,1=1 z2w,2=1 z2w,3=1 z4=2 |
| | 12 | z2w,3=1 |
| | 12 | z2w,3=1 |
| | 12 | z2w,1=1 z2w,2=1 z2w,3=1 z4=2 |
| | 12 | z2w,3=1 z4=3 |
| | 12 | z2w,1=1 z2w,2=1 z2w,3=1 z4=1 |
| | 12 | z2w,1=1 z2w,2=1 z2w,3=1 z4=1 |
| | 12 | z2w,3=1 z4=3 |
| 129.4.42 | 60735 | Mn1Na1Sb1 |
| | 115 | z2=1 |
| | 115 | z2=1 |
| | 12 | z4=3 |
| | 12 | z2w,1=1 z2w,2=1 z4=1 |
| | 12 | z2w,1=1 z2w,2=1 z4=1 |
| | 12 | z4=3 |
| | 12 | z2w,1=1 z2w,2=1 |
| | 12 | z4=2 |
| | 12 | z4=2 |
| | 12 | z2w,1=1 z2w,2=1 |
| | 12 | z2w,1=1 z2w,2=1 z2w,3=1 |
| | 12 | z2w,3=1 z4=2 |
| | 12 | z2w,3=1 z4=2 |
| | 12 | z2w,1=1 z2w,2=1 z2w,3=1 |
| | 12 | z2w,3=1 z4=1 |
| | 12 | z2w,1=1 z2w,2=1 z2w,3=1 z4=3 |
| | 12 | z2w,1=1 z2w,2=1 z2w,3=1 z4=3 |
| | 12 | z2w,3=1 z4=1 |
| 129.4.43 | 60734 | Mn1Na1Sb1 |
| | 115 | z2=1 |
| | 115 | z2=1 |
| | 12 | z4=3 |
| | 12 | z2w,1=1 z2w,2=1 z4=1 |
| | 12 | z2w,1=1 z2w,2=1 z4=1 |
| | 12 | z4=3 |
| | 12 | z2w,1=1 z2w,2=1 |
| | 12 | z4=2 |
| | 12 | z4=2 |
| | 12 | z2w,1=1 z2w,2=1 |
| | 12 | z2w,1=1 z2w,2=1 z2w,3=1 |
| | 12 | z2w,3=1 z4=2 |
| | 12 | z2w,3=1 z4=2 |
| | 12 | z2w,1=1 z2w,2=1 z2w,3=1 |
| | 12 | z2w,3=1 z4=1 |
| | 12 | z2w,1=1 z2w,2=1 z2w,3=1 z4=3 |
| | 12 | z2w,1=1 z2w,2=1 z2w,3=1 z4=3 |
| | 12 | z2w,3=1 z4=1 |
| 129.4.44 | 60733 | Mn1Na1Sb1 |
| | 115 | z2=1 |
| | 115 | z2=1 |
| | 12 | z4=3 |
| | 12 | z2w,1=1 z2w,2=1 z4=1 |
| | 12 | z2w,1=1 z2w,2=1 z4=1 |
| | 12 | z4=3 |

Continued on next column

**TABLE XIX– continued**

| TQC | ICSD SSG | Stoich. Formula SIs |
|---|---|---|
| | 12 | z2w,1=1 z2w,2=1 |
| | 12 | z4=2 |
| | 12 | z4=2 |
| | 12 | z2w,1=1 z2w,2=1 |
| | 12 | z2w,1=1 z2w,2=1 z2w,3=1 |
| | 12 | z2w,3=1 z4=2 |
| | 12 | z2w,3=1 z4=2 |
| | 12 | z2w,1=1 z2w,2=1 z2w,3=1 |
| | 12 | z2w,3=1 z4=1 |
| | 12 | z2w,1=1 z2w,2=1 z2w,3=1 z4=3 |
| | 12 | z2w,1=1 z2w,2=1 z2w,3=1 z4=3 |
| | 12 | z2w,3=1 z4=1 |
| 129.4.45 | 26462 | Mn1Na1Sb1 |
| | 115 | z2=1 |
| | 115 | z2=1 |
| | 12 | z4=3 |
| | 12 | z2w,1=1 z2w,2=1 z4=1 |
| | 12 | z2w,1=1 z2w,2=1 z4=1 |
| | 12 | z4=3 |
| | 12 | z2w,1=1 z2w,2=1 |
| | 12 | z4=2 |
| | 12 | z4=2 |
| | 12 | z2w,1=1 z2w,2=1 |
| | 12 | z2w,1=1 z2w,2=1 z2w,3=1 |
| | 12 | z2w,3=1 z4=2 |
| | 12 | z2w,3=1 z4=2 |
| | 12 | z2w,1=1 z2w,2=1 z2w,3=1 |
| | 12 | z2w,3=1 z4=1 |
| | 12 | z2w,1=1 z2w,2=1 z2w,3=1 z4=3 |
| | 12 | z2w,1=1 z2w,2=1 z2w,3=1 z4=3 |
| | 12 | z2w,3=1 z4=1 |
| 129.4.46 | 60729 | Mn1Na1P1 |
| | 115 | z2=1 |
| | 115 | z2=1 |
| | 12 | z4=3 |
| | 12 | z2w,1=1 z2w,2=1 z4=1 |
| | 12 | z2w,1=1 z2w,2=1 z4=1 |
| | 12 | z4=3 |
| | 12 | z2w,1=1 z2w,2=1 |
| | 12 | z4=2 |
| | 12 | z4=2 |
| | 12 | z2w,1=1 z2w,2=1 |
| | 12 | z2w,1=1 z2w,2=1 z2w,3=1 |
| | 12 | z2w,3=1 z4=2 |
| | 12 | z2w,3=1 z4=2 |
| | 12 | z2w,1=1 z2w,2=1 z2w,3=1 |
| | 12 | z2w,3=1 z4=1 |
| | 12 | z2w,1=1 z2w,2=1 z2w,3=1 z4=3 |
| | 12 | z2w,1=1 z2w,2=1 z2w,3=1 z4=3 |
| | 12 | z2w,3=1 z4=1 |
| 129.4.47 | 60728 | Mn1Na1P1 |
| | 115 | z2=1 |
| | 115 | z2=1 |
| | 12 | z4=3 |
| | 12 | z2w,1=1 z2w,2=1 z4=1 |
| | 12 | z2w,1=1 z2w,2=1 z4=1 |
| | 12 | z4=3 |
| | 12 | z2w,1=1 z2w,2=1 |
| | 12 | z4=2 |
| | 12 | z4=2 |

Continued on next column

**TABLE XIX– continued**

| TQC | ICSD SSG | Stoich. Formula SIs |
|---|---|---|
| | 12 | z2w,1=1 z2w,2=1 |
| | 12 | z2w,1=1 z2w,2=1 z2w,3=1 |
| | 12 | z2w,3=1 z4=2 |
| | 12 | z2w,3=1 z4=2 |
| | 12 | z2w,1=1 z2w,2=1 z2w,3=1 |
| | 12 | z2w,3=1 z4=1 |
| | 12 | z2w,1=1 z2w,2=1 z2w,3=1 z4=3 |
| | 12 | z2w,1=1 z2w,2=1 z2w,3=1 z4=3 |
| | 12 | z2w,3=1 z4=1 |
| 129.4.48 | 60727 | Mn1Na1P1 |
| | 115 | z2=1 |
| | 115 | z2=1 |
| | 12 | z4=3 |
| | 12 | z2w,1=1 z2w,2=1 z4=1 |
| | 12 | z2w,1=1 z2w,2=1 z4=1 |
| | 12 | z4=3 |
| | 12 | z2w,1=1 z2w,2=1 |
| | 12 | z4=2 |
| | 12 | z4=2 |
| | 12 | z2w,1=1 z2w,2=1 |
| | 12 | z2w,1=1 z2w,2=1 z2w,3=1 |
| | 12 | z2w,3=1 z4=2 |
| | 12 | z2w,3=1 z4=2 |
| | 12 | z2w,1=1 z2w,2=1 z2w,3=1 |
| | 12 | z2w,3=1 z4=1 |
| | 12 | z2w,1=1 z2w,2=1 z2w,3=1 z4=3 |
| | 12 | z2w,1=1 z2w,2=1 z2w,3=1 z4=3 |
| | 12 | z2w,3=1 z4=1 |
| 129.4.49 | 60726 | Mn1Na1P1 |
| | 115 | z2=1 |
| | 115 | z2=1 |
| | 12 | z4=3 |
| | 12 | z2w,1=1 z2w,2=1 z4=1 |
| | 12 | z2w,1=1 z2w,2=1 z4=1 |
| | 12 | z4=3 |
| | 12 | z2w,1=1 z2w,2=1 |
| | 12 | z4=2 |
| | 12 | z4=2 |
| | 12 | z2w,1=1 z2w,2=1 |
| | 12 | z2w,1=1 z2w,2=1 z2w,3=1 |
| | 12 | z2w,3=1 z4=2 |
| | 12 | z2w,3=1 z4=2 |
| | 12 | z2w,1=1 z2w,2=1 z2w,3=1 |
| | 12 | z2w,3=1 z4=1 |
| | 12 | z2w,1=1 z2w,2=1 z2w,3=1 z4=3 |
| | 12 | z2w,1=1 z2w,2=1 z2w,3=1 z4=3 |
| | 12 | z2w,3=1 z4=1 |
| 129.4.50 | 26460 | Mn1Na1P1 |
| | 115 | z2=1 |
| | 115 | z2=1 |
| | 12 | z4=3 |
| | 12 | z2w,1=1 z2w,2=1 z4=1 |
| | 12 | z2w,1=1 z2w,2=1 z4=1 |
| | 12 | z4=3 |
| | 12 | z2w,1=1 z2w,2=1 |
| | 12 | z4=2 |
| | 12 | z4=2 |
| | 12 | z2w,1=1 z2w,2=1 |
| | 12 | z2w,1=1 z2w,2=1 z2w,3=1 |
| | 12 | z2w,3=1 z4=2 |

Continued on next column

**TABLE XIX– continued**

| TQC | ICSD<br>SSG | Stoich. Formula<br>SIs |
|---|---|---|
| | 12 | z2w,3=1 z4=2 |
| | 12 | z2w,1=1 z2w,2=1 z2w,3=1 |
| | 12 | z2w,3=1 z4=1 |
| | 12 | z2w,1=1 z2w,2=1 z2w,3=1 z4=3 |
| | 12 | z2w,1=1 z2w,2=1 z2w,3=1 z4=3 |
| | 12 | z2w,3=1 z4=1 |
| 129.4.51 | 183314 | Lu1Sn2Zn1 |
| | 115 | z2=1 |
| | 115 | z2=1 |
| | 12 | z4=3 |
| | 12 | z2w,1=1 z2w,2=1 z4=1 |
| | 12 | z2w,1=1 z2w,2=1 z4=1 |
| | 12 | z4=3 |
| | 12 | z2w,1=1 z2w,2=1 |
| | 12 | z4=2 |
| | 12 | z4=2 |
| | 12 | z2w,1=1 z2w,2=1 |
| | 12 | z2w,1=1 z2w,2=1 z2w,3=1 |
| | 12 | z2w,3=1 z4=2 |
| | 12 | z2w,3=1 z4=2 |
| | 12 | z2w,1=1 z2w,2=1 z2w,3=1 |
| | 12 | z2w,3=1 z4=1 |
| | 12 | z2w,1=1 z2w,2=1 z2w,3=1 z4=3 |
| | 12 | z2w,1=1 z2w,2=1 z2w,3=1 z4=3 |
| | 12 | z2w,3=1 z4=1 |
| 129.4.52 | 88208 | Lu1Si1Ti1 |
| | 115 | z2=1 |
| | 115 | z2=1 |
| | 12 | z2w,3=1 z4=2 |
| | 12 | z2w,1=1 z2w,2=1 z2w,3=1 |
| | 12 | z2w,1=1 z2w,2=1 z2w,3=1 |
| | 12 | z2w,3=1 z4=2 |
| | 12 | z2w,1=1 z2w,2=1 z2w,3=1 z4=3 |
| | 12 | z2w,3=1 z4=1 |
| | 12 | z2w,3=1 z4=1 |
| | 12 | z2w,1=1 z2w,2=1 z2w,3=1 z4=3 |
| | 12 | z2w,1=1 z2w,2=1 z4=3 |
| | 12 | z4=1 |
| | 12 | z4=1 |
| | 12 | z2w,1=1 z2w,2=1 z4=3 |
| | 12 | z2w,1=1 z2w,2=1 z4=2 |
| | 12 | z2w,1=1 z2w,2=1 z4=2 |
| 129.4.53 | 36019 | Li2Pr1Sb2 |
| | 115 | z2=1 |
| | 115 | z2=1 |
| | 12 | z2w,3=1 |
| | 12 | z2w,1=1 z2w,2=1 z2w,3=1 z4=2 |
| | 12 | z2w,1=1 z2w,2=1 z2w,3=1 z4=2 |
| | 12 | z2w,3=1 |
| | 12 | z2w,1=1 z2w,2=1 z2w,3=1 z4=1 |
| | 12 | z2w,3=1 z4=3 |
| | 12 | z2w,3=1 z4=3 |
| | 12 | z2w,1=1 z2w,2=1 z2w,3=1 z4=1 |
| | 12 | z2w,1=1 z2w,2=1 z4=1 |
| | 12 | z4=3 |
| | 12 | z4=3 |
| | 12 | z2w,1=1 z2w,2=1 z4=1 |
| | 12 | z4=2 |
| | 12 | z2w,1=1 z2w,2=1 |
| | 12 | z2w,1=1 z2w,2=1 |

Continued on next column

**TABLE XIX– continued**

| TQC | ICSD<br>SSG | Stoich. Formula<br>SIs |
|---|---|---|
| | 12 | z4=2 |
| 129.4.54 | 86775 | Li2O5Si1V1 |
| | 115 | z2=1 |
| | 115 | z2=1 |
| | 12 | z4=1 |
| | 12 | z2w,1=1 z2w,2=1 z4=3 |
| | 12 | z2w,1=1 z2w,2=1 z4=3 |
| | 12 | z4=1 |
| | 12 | z2w,1=1 z2w,2=1 z4=2 |
| | 12 | z2w,1=1 z2w,2=1 z4=2 |
| | 12 | z2w,1=1 z2w,2=1 z2w,3=1 z4=2 |
| | 12 | z2w,3=1 |
| | 12 | z2w,3=1 |
| | 12 | z2w,1=1 z2w,2=1 z2w,3=1 z4=2 |
| | 12 | z2w,3=1 z4=3 |
| | 12 | z2w,1=1 z2w,2=1 z2w,3=1 z4=1 |
| | 12 | z2w,1=1 z2w,2=1 z2w,3=1 z4=1 |
| | 12 | z2w,3=1 z4=3 |
| 129.4.55 | 59359 | Li2O5Si1V1 |
| | 115 | z2=1 |
| | 115 | z2=1 |
| | 12 | z2w,3=1 z4=2 |
| | 12 | z2w,1=1 z2w,2=1 z2w,3=1 |
| | 12 | z2w,1=1 z2w,2=1 z2w,3=1 |
| | 12 | z2w,3=1 z4=2 |
| | 12 | z2w,1=1 z2w,2=1 z2w,3=1 z4=3 |
| | 12 | z2w,3=1 z4=1 |
| | 12 | z2w,3=1 z4=1 |
| | 12 | z2w,1=1 z2w,2=1 z2w,3=1 z4=3 |
| | 12 | z2w,1=1 z2w,2=1 z4=3 |
| | 12 | z4=1 |
| | 12 | z4=1 |
| | 12 | z2w,1=1 z2w,2=1 z4=3 |
| | 12 | z2w,1=1 z2w,2=1 z4=2 |
| | 12 | z2w,1=1 z2w,2=1 z4=2 |
| 129.4.56 | 417286 | Li2O5Si1V1 |
| | 115 | z2=1 |
| | 115 | z2=1 |
| | 12 | z4=1 |
| | 12 | z2w,1=1 z2w,2=1 z4=3 |
| | 12 | z2w,1=1 z2w,2=1 z4=3 |
| | 12 | z4=1 |
| | 12 | z2w,1=1 z2w,2=1 z4=2 |
| | 12 | z2w,1=1 z2w,2=1 z4=2 |
| | 12 | z2w,1=1 z2w,2=1 z2w,3=1 z4=2 |
| | 12 | z2w,3=1 |
| | 12 | z2w,3=1 |
| | 12 | z2w,1=1 z2w,2=1 z2w,3=1 z4=2 |
| | 12 | z2w,3=1 z4=3 |
| | 12 | z2w,1=1 z2w,2=1 z2w,3=1 z4=1 |
| | 12 | z2w,1=1 z2w,2=1 z2w,3=1 z4=1 |
| | 12 | z2w,3=1 z4=3 |
| 129.4.57 | 417285 | Li2O5Si1V1 |
| | 115 | z2=1 |
| | 115 | z2=1 |
| | 12 | z4=1 |
| | 12 | z2w,1=1 z2w,2=1 z4=3 |
| | 12 | z2w,1=1 z2w,2=1 z4=3 |
| | 12 | z4=1 |
| | 12 | z2w,1=1 z2w,2=1 z4=2 |

Continued on next column

**TABLE XIX– continued**

| TQC | ICSD SSG | Stoich. Formula SIs |
|---|---|---|
| | 12 | z2w,1=1 z2w,2=1 z4=2 |
| | 12 | z2w,1=1 z2w,2=1 z2w,3=1 z4=2 |
| | 12 | z2w,3=1 |
| | 12 | z2w,3=1 |
| | 12 | z2w,1=1 z2w,2=1 z2w,3=1 z4=2 |
| | 12 | z2w,3=1 z4=3 |
| | 12 | z2w,1=1 z2w,2=1 z2w,3=1 z4=1 |
| | 12 | z2w,1=1 z2w,2=1 z2w,3=1 z4=1 |
| | 12 | z2w,3=1 z4=3 |
| 129.4.58 | 417284 | Li2O5Si1V1 |
| | 115 | z2=1 |
| | 115 | z2=1 |
| | 12 | z4=1 |
| | 12 | z2w,1=1 z2w,2=1 z4=3 |
| | 12 | z2w,1=1 z2w,2=1 z4=3 |
| | 12 | z4=1 |
| | 12 | z2w,1=1 z2w,2=1 z4=2 |
| | 12 | z2w,1=1 z2w,2=1 z4=2 |
| | 12 | z2w,1=1 z2w,2=1 z2w,3=1 z4=2 |
| | 12 | z2w,3=1 |
| | 12 | z2w,3=1 |
| | 12 | z2w,1=1 z2w,2=1 z2w,3=1 z4=2 |
| | 12 | z2w,3=1 z4=3 |
| | 12 | z2w,1=1 z2w,2=1 z2w,3=1 z4=1 |
| | 12 | z2w,1=1 z2w,2=1 z2w,3=1 z4=1 |
| | 12 | z2w,3=1 z4=3 |
| 129.4.59 | 417283 | Li2O5Si1V1 |
| | 115 | z2=1 |
| | 115 | z2=1 |
| | 12 | z4=1 |
| | 12 | z2w,1=1 z2w,2=1 z4=3 |
| | 12 | z2w,1=1 z2w,2=1 z4=3 |
| | 12 | z4=1 |
| | 12 | z2w,1=1 z2w,2=1 z4=2 |
| | 12 | z2w,1=1 z2w,2=1 z4=2 |
| | 12 | z2w,1=1 z2w,2=1 z2w,3=1 z4=2 |
| | 12 | z2w,3=1 |
| | 12 | z2w,3=1 |
| | 12 | z2w,1=1 z2w,2=1 z2w,3=1 z4=2 |
| | 12 | z2w,3=1 z4=3 |
| | 12 | z2w,1=1 z2w,2=1 z2w,3=1 z4=1 |
| | 12 | z2w,1=1 z2w,2=1 z2w,3=1 z4=1 |
| | 12 | z2w,3=1 z4=3 |
| 129.4.60 | 417282 | Li2O5Si1V1 |
| | 115 | z2=1 |
| | 115 | z2=1 |
| | 12 | z4=1 |
| | 12 | z2w,1=1 z2w,2=1 z4=3 |
| | 12 | z2w,1=1 z2w,2=1 z4=3 |
| | 12 | z4=1 |
| | 12 | z2w,1=1 z2w,2=1 z4=2 |
| | 12 | z2w,1=1 z2w,2=1 z4=2 |
| | 12 | z2w,1=1 z2w,2=1 z2w,3=1 z4=2 |
| | 12 | z2w,3=1 |
| | 12 | z2w,3=1 |
| | 12 | z2w,1=1 z2w,2=1 z2w,3=1 z4=2 |
| | 12 | z2w,3=1 z4=3 |
| | 12 | z2w,1=1 z2w,2=1 z2w,3=1 z4=1 |
| | 12 | z2w,1=1 z2w,2=1 z2w,3=1 z4=1 |
| | 12 | z2w,3=1 z4=3 |

Continued on next column

**TABLE XIX– continued**

| TQC | ICSD SSG | Stoich. Formula SIs |
|---|---|---|
| 129.4.61 | 417281 | Li2O5Si1V1 |
| | 115 | z2=1 |
| | 115 | z2=1 |
| | 12 | z4=1 |
| | 12 | z2w,1=1 z2w,2=1 z4=3 |
| | 12 | z2w,1=1 z2w,2=1 z4=3 |
| | 12 | z4=1 |
| | 12 | z2w,1=1 z2w,2=1 z4=2 |
| | 12 | z2w,1=1 z2w,2=1 z4=2 |
| | 12 | z2w,1=1 z2w,2=1 z2w,3=1 z4=2 |
| | 12 | z2w,3=1 |
| | 12 | z2w,3=1 |
| | 12 | z2w,1=1 z2w,2=1 z2w,3=1 z4=2 |
| | 12 | z2w,3=1 z4=3 |
| | 12 | z2w,1=1 z2w,2=1 z2w,3=1 z4=1 |
| | 12 | z2w,1=1 z2w,2=1 z2w,3=1 z4=1 |
| | 12 | z2w,3=1 z4=3 |
| 129.4.62 | 417280 | Li2O5Si1V1 |
| | 115 | z2=1 |
| | 115 | z2=1 |
| | 12 | z4=1 |
| | 12 | z2w,1=1 z2w,2=1 z4=3 |
| | 12 | z2w,1=1 z2w,2=1 z4=3 |
| | 12 | z4=1 |
| | 12 | z2w,1=1 z2w,2=1 z4=2 |
| | 12 | z2w,1=1 z2w,2=1 z4=2 |
| | 12 | z2w,1=1 z2w,2=1 z2w,3=1 z4=2 |
| | 12 | z2w,3=1 |
| | 12 | z2w,3=1 |
| | 12 | z2w,1=1 z2w,2=1 z2w,3=1 z4=2 |
| | 12 | z2w,3=1 z4=3 |
| | 12 | z2w,1=1 z2w,2=1 z2w,3=1 z4=1 |
| | 12 | z2w,1=1 z2w,2=1 z2w,3=1 z4=1 |
| | 12 | z2w,3=1 z4=3 |
| 129.4.63 | 417279 | Li2O5Si1V1 |
| | 115 | z2=1 |
| | 115 | z2=1 |
| | 12 | z4=1 |
| | 12 | z2w,1=1 z2w,2=1 z4=3 |
| | 12 | z2w,1=1 z2w,2=1 z4=3 |
| | 12 | z4=1 |
| | 12 | z2w,1=1 z2w,2=1 z4=2 |
| | 12 | z2w,1=1 z2w,2=1 z4=2 |
| | 12 | z2w,1=1 z2w,2=1 z2w,3=1 z4=2 |
| | 12 | z2w,3=1 |
| | 12 | z2w,3=1 |
| | 12 | z2w,1=1 z2w,2=1 z2w,3=1 z4=2 |
| | 12 | z2w,3=1 z4=3 |
| | 12 | z2w,1=1 z2w,2=1 z2w,3=1 z4=1 |
| | 12 | z2w,1=1 z2w,2=1 z2w,3=1 z4=1 |
| | 12 | z2w,3=1 z4=3 |
| 129.4.64 | 417278 | Li2O5Si1V1 |
| | 115 | z2=1 |
| | 115 | z2=1 |
| | 12 | z4=1 |
| | 12 | z2w,1=1 z2w,2=1 z4=3 |
| | 12 | z2w,1=1 z2w,2=1 z4=3 |
| | 12 | z4=1 |
| | 12 | z2w,1=1 z2w,2=1 z4=2 |
| | 12 | z2w,1=1 z2w,2=1 z4=2 |

Continued on next column

**TABLE XIX– continued**

| TQC | ICSD<br>SSG | Stoich. Formula<br>SIs |
|---|---|---|
| | 12 | z2w,1=1 z2w,2=1 z2w,3=1 z4=2 |
| | 12 | z2w,3=1 |
| | 12 | z2w,3=1 |
| | 12 | z2w,1=1 z2w,2=1 z2w,3=1 z4=2 |
| | 12 | z2w,3=1 z4=3 |
| | 12 | z2w,1=1 z2w,2=1 z2w,3=1 z4=1 |
| | 12 | z2w,1=1 z2w,2=1 z2w,3=1 z4=1 |
| | 12 | z2w,3=1 z4=3 |
| 129.4.65 | 417277 | Li2O5Si1V1 |
| | 115 | z2=1 |
| | 115 | z2=1 |
| | 12 | z2w,3=1 z4=2 |
| | 12 | z2w,1=1 z2w,2=1 z2w,3=1 |
| | 12 | z2w,1=1 z2w,2=1 z2w,3=1 |
| | 12 | z2w,3=1 z4=2 |
| | 12 | z2w,1=1 z2w,2=1 z2w,3=1 z4=3 |
| | 12 | z2w,3=1 z4=1 |
| | 12 | z2w,3=1 z4=1 |
| | 12 | z2w,1=1 z2w,2=1 z2w,3=1 z4=3 |
| | 12 | z2w,1=1 z2w,2=1 z4=3 |
| | 12 | z4=1 |
| | 12 | z4=1 |
| | 12 | z2w,1=1 z2w,2=1 z4=3 |
| | 12 | z2w,1=1 z2w,2=1 z4=2 |
| | 12 | z2w,1=1 z2w,2=1 z4=2 |
| 129.4.66 | 417276 | Li2O5Si1V1 |
| | 115 | z2=1 |
| | 115 | z2=1 |
| | 12 | z4=1 |
| | 12 | z2w,1=1 z2w,2=1 z4=3 |
| | 12 | z2w,1=1 z2w,2=1 z4=3 |
| | 12 | z4=1 |
| | 12 | z2w,1=1 z2w,2=1 z4=2 |
| | 12 | z2w,1=1 z2w,2=1 z4=2 |
| | 12 | z2w,1=1 z2w,2=1 z2w,3=1 z4=2 |
| | 12 | z2w,3=1 |
| | 12 | z2w,3=1 |
| | 12 | z2w,1=1 z2w,2=1 z2w,3=1 z4=2 |
| | 12 | z2w,3=1 z4=3 |
| | 12 | z2w,1=1 z2w,2=1 z2w,3=1 z4=1 |
| | 12 | z2w,1=1 z2w,2=1 z2w,3=1 z4=1 |
| | 12 | z2w,3=1 z4=3 |
| 129.4.67 | 26458 | Li1Mn1P1 |
| | 115 | z2=1 |
| | 115 | z2=1 |
| | 12 | z2w,3=1 z4=2 |
| | 12 | z2w,1=1 z2w,2=1 z2w,3=1 |
| | 12 | z2w,1=1 z2w,2=1 z2w,3=1 |
| | 12 | z2w,3=1 z4=2 |
| | 12 | z2w,1=1 z2w,2=1 z2w,3=1 z4=3 |
| | 12 | z2w,3=1 z4=1 |
| | 12 | z2w,3=1 z4=1 |
| | 12 | z2w,1=1 z2w,2=1 z2w,3=1 z4=3 |
| | 12 | z2w,1=1 z2w,2=1 z4=3 |
| | 12 | z4=1 |
| | 12 | z4=1 |
| | 12 | z2w,1=1 z2w,2=1 z4=3 |
| | 12 | z2w,1=1 z2w,2=1 z4=2 |
| | 12 | z2w,1=1 z2w,2=1 z4=2 |
| 129.4.68 | 43279 | La1Te2 |

Continued on next column

**TABLE XIX– continued**

| TQC | ICSD<br>SSG | Stoich. Formula<br>SIs |
|---|---|---|
| | 115 | z2=1 |
| | 115 | z2=1 |
| | 12 | z2w,1=1 z2w,2=1 z4=2 |
| | 12 | z2w,1=1 z2w,2=1 z4=2 |
| | 12 | z2w,1=1 z2w,2=1 z4=1 |
| | 12 | z4=3 |
| | 12 | z4=3 |
| | 12 | z2w,1=1 z2w,2=1 z4=1 |
| | 12 | z2w,1=1 z2w,2=1 z2w,3=1 z4=1 |
| | 12 | z2w,3=1 z4=3 |
| | 12 | z2w,3=1 z4=3 |
| | 12 | z2w,1=1 z2w,2=1 z2w,3=1 z4=1 |
| | 12 | z2w,3=1 z4=2 |
| | 12 | z2w,1=1 z2w,2=1 z2w,3=1 |
| | 12 | z2w,1=1 z2w,2=1 z2w,3=1 |
| | 12 | z2w,3=1 z4=2 |
| 129.4.69 | 42342 | La1Te2 |
| | 115 | z2=1 |
| | 115 | z2=1 |
| | 12 | z2w,1=1 z2w,2=1 z4=2 |
| | 12 | z2w,1=1 z2w,2=1 z4=2 |
| | 12 | z2w,1=1 z2w,2=1 z4=1 |
| | 12 | z4=3 |
| | 12 | z4=3 |
| | 12 | z2w,1=1 z2w,2=1 z4=1 |
| | 12 | z2w,1=1 z2w,2=1 z2w,3=1 z4=1 |
| | 12 | z2w,3=1 z4=3 |
| | 12 | z2w,3=1 z4=3 |
| | 12 | z2w,1=1 z2w,2=1 z2w,3=1 z4=1 |
| | 12 | z2w,3=1 z4=2 |
| | 12 | z2w,1=1 z2w,2=1 z2w,3=1 |
| | 12 | z2w,1=1 z2w,2=1 z2w,3=1 |
| | 12 | z2w,3=1 z4=2 |
| 129.4.70 | 23957 | La1Te2 |
| | 115 | z2=1 |
| | 115 | z2=1 |
| | 12 | z2w,1=1 z2w,2=1 z4=2 |
| | 12 | z2w,1=1 z2w,2=1 z4=2 |
| | 12 | z2w,1=1 z2w,2=1 z4=1 |
| | 12 | z4=3 |
| | 12 | z4=3 |
| | 12 | z2w,1=1 z2w,2=1 z4=1 |
| | 12 | z2w,1=1 z2w,2=1 z2w,3=1 z4=1 |
| | 12 | z2w,3=1 z4=3 |
| | 12 | z2w,3=1 z4=3 |
| | 12 | z2w,1=1 z2w,2=1 z2w,3=1 z4=1 |
| | 12 | z2w,3=1 z4=2 |
| | 12 | z2w,1=1 z2w,2=1 z2w,3=1 |
| | 12 | z2w,1=1 z2w,2=1 z2w,3=1 |
| | 12 | z2w,3=1 z4=2 |
| 129.4.71 | 85855 | La1Ru1Si1 |
| | 115 | z2=1 |
| | 115 | z2=1 |
| | 12 | z2w,1=1 z2w,2=1 z4=2 |
| | 12 | z2w,1=1 z2w,2=1 z4=2 |
| | 12 | z2w,1=1 z2w,2=1 z4=1 |
| | 12 | z4=3 |
| | 12 | z4=3 |
| | 12 | z2w,1=1 z2w,2=1 z4=1 |
| | 12 | z2w,1=1 z2w,2=1 z2w,3=1 z4=1 |

Continued on next column

TABLE XIX– continued

| TQC | ICSD SSG | Stoich. Formula SIs |
|---|---|---|
| | 12 | z2w,3=1 z4=3 |
| | 12 | z2w,3=1 z4=3 |
| | 12 | z2w,1=1 z2w,2=1 z2w,3=1 z4=1 |
| | 12 | z2w,3=1 z4=2 |
| | 12 | z2w,1=1 z2w,2=1 z2w,3=1 |
| | 12 | z2w,1=1 z2w,2=1 z2w,3=1 |
| | 12 | z2w,3=1 z4=2 |
| 129.4.72 | 41249 | La1Ru1Si1 |
| | 115 | z2=1 |
| | 115 | z2=1 |
| | 12 | z2w,1=1 z2w,2=1 z4=2 |
| | 12 | z2w,1=1 z2w,2=1 z4=2 |
| | 12 | z2w,1=1 z2w,2=1 z4=1 |
| | 12 | z4=3 |
| | 12 | z4=3 |
| | 12 | z2w,1=1 z2w,2=1 z4=1 |
| | 12 | z2w,1=1 z2w,2=1 z2w,3=1 z4=1 |
| | 12 | z2w,3=1 z4=3 |
| | 12 | z2w,3=1 z4=3 |
| | 12 | z2w,1=1 z2w,2=1 z2w,3=1 z4=1 |
| | 12 | z2w,3=1 z4=2 |
| | 12 | z2w,1=1 z2w,2=1 z2w,3=1 |
| | 12 | z2w,1=1 z2w,2=1 z2w,3=1 |
| | 12 | z2w,3=1 z4=2 |
| 129.4.73 | 189724 | La1Pd2Sb2 |
| | 115 | z2=1 |
| | 115 | z2=1 |
| | 12 | z4=3 |
| | 12 | z2w,1=1 z2w,2=1 z4=1 |
| | 12 | z2w,1=1 z2w,2=1 z4=1 |
| | 12 | z4=3 |
| | 12 | z2w,1=1 z2w,2=1 |
| | 12 | z4=2 |
| | 12 | z4=2 |
| | 12 | z2w,1=1 z2w,2=1 |
| | 12 | z2w,1=1 z2w,2=1 z2w,3=1 |
| | 12 | z2w,3=1 z4=2 |
| | 12 | z2w,3=1 z4=2 |
| | 12 | z2w,1=1 z2w,2=1 z2w,3=1 |
| | 12 | z2w,3=1 z4=1 |
| | 12 | z2w,1=1 z2w,2=1 z2w,3=1 z4=3 |
| | 12 | z2w,1=1 z2w,2=1 z2w,3=1 z4=3 |
| | 12 | z2w,3=1 z4=1 |
| 129.4.74 | 419355 | La1Mn1O1Sb1 |
| | 115 | z2=1 |
| | 115 | z2=1 |
| | 12 | z2w,3=1 z4=1 |
| | 12 | z2w,1=1 z2w,2=1 z2w,3=1 z4=3 |
| | 12 | z2w,1=1 z2w,2=1 z2w,3=1 z4=3 |
| | 12 | z2w,3=1 z4=1 |
| | 12 | z2w,1=1 z2w,2=1 z2w,3=1 z4=2 |
| | 12 | z2w,3=1 |
| | 12 | z2w,3=1 |
| | 12 | z2w,1=1 z2w,2=1 z2w,3=1 z4=2 |
| | 12 | z2w,1=1 z2w,2=1 z4=2 |
| | 12 | z2w,1=1 z2w,2=1 z4=2 |
| | 12 | z4=3 |
| | 12 | z2w,1=1 z2w,2=1 z4=1 |
| | 12 | z2w,1=1 z2w,2=1 z4=1 |
| | 12 | z4=3 |

Continued on next column

TABLE XIX– continued

| TQC | ICSD SSG | Stoich. Formula SIs |
|---|---|---|
| 129.4.75 | 419856 | K1Na1Nd1O5Ta1 |
| | 115 | z2=1 |
| | 115 | z2=1 |
| | 12 | z2w,3=1 z4=1 |
| | 12 | z2w,1=1 z2w,2=1 z2w,3=1 z4=3 |
| | 12 | z2w,1=1 z2w,2=1 z2w,3=1 z4=3 |
| | 12 | z2w,3=1 z4=1 |
| | 12 | z2w,1=1 z2w,2=1 z2w,3=1 z4=2 |
| | 12 | z2w,3=1 |
| | 12 | z2w,3=1 |
| | 12 | z2w,1=1 z2w,2=1 z2w,3=1 z4=2 |
| | 12 | z2w,1=1 z2w,2=1 z4=2 |
| | 12 | z2w,1=1 z2w,2=1 z4=2 |
| | 12 | z4=3 |
| | 12 | z2w,1=1 z2w,2=1 z4=1 |
| | 12 | z2w,1=1 z2w,2=1 z4=1 |
| | 12 | z4=3 |
| 129.4.76 | 420038 | K1Na1Nb1Nd1O5 |
| | 115 | z2=1 |
| | 115 | z2=1 |
| | 12 | z2w,3=1 z4=1 |
| | 12 | z2w,1=1 z2w,2=1 z2w,3=1 z4=3 |
| | 12 | z2w,1=1 z2w,2=1 z2w,3=1 z4=3 |
| | 12 | z2w,3=1 z4=1 |
| | 12 | z2w,1=1 z2w,2=1 z2w,3=1 z4=2 |
| | 12 | z2w,3=1 |
| | 12 | z2w,3=1 |
| | 12 | z2w,1=1 z2w,2=1 z2w,3=1 z4=2 |
| | 12 | z2w,1=1 z2w,2=1 z4=2 |
| | 12 | z2w,1=1 z2w,2=1 z4=2 |
| | 12 | z4=3 |
| | 12 | z2w,1=1 z2w,2=1 z4=1 |
| | 12 | z2w,1=1 z2w,2=1 z4=1 |
| | 12 | z4=3 |
| 129.4.77 | 89591 | K1Mn1Sb1 |
| | 115 | z2=1 |
| | 115 | z2=1 |
| | 12 | z4=3 |
| | 12 | z2w,1=1 z2w,2=1 z4=1 |
| | 12 | z2w,1=1 z2w,2=1 z4=1 |
| | 12 | z4=3 |
| | 12 | z2w,1=1 z2w,2=1 |
| | 12 | z4=2 |
| | 12 | z4=2 |
| | 12 | z2w,1=1 z2w,2=1 |
| | 12 | z2w,1=1 z2w,2=1 z2w,3=1 |
| | 12 | z2w,3=1 z4=2 |
| | 12 | z2w,3=1 z4=2 |
| | 12 | z2w,1=1 z2w,2=1 z2w,3=1 |
| | 12 | z2w,3=1 z4=1 |
| | 12 | z2w,1=1 z2w,2=1 z2w,3=1 z4=3 |
| | 12 | z2w,1=1 z2w,2=1 z2w,3=1 z4=3 |
| | 12 | z2w,3=1 z4=1 |
| 129.4.78 | 89590 | K1Mn1Sb1 |
| | 115 | z2=1 |
| | 115 | z2=1 |
| | 12 | z2w,3=1 z4=2 |
| | 12 | z2w,1=1 z2w,2=1 z2w,3=1 |
| | 12 | z2w,1=1 z2w,2=1 z2w,3=1 |
| | 12 | z2w,3=1 z4=2 |

Continued on next column

**TABLE XIX– continued**

| TQC | ICSD SSG | Stoich. Formula SIs |
|---|---|---|
| | 12 | z2w,1=1 z2w,2=1 z2w,3=1 z4=3 |
| | 12 | z2w,3=1 z4=1 |
| | 12 | z2w,3=1 z4=1 |
| | 12 | z2w,1=1 z2w,2=1 z2w,3=1 z4=3 |
| | 12 | z2w,1=1 z2w,2=1 z4=3 |
| | 12 | z4=1 |
| | 12 | z4=1 |
| | 12 | z2w,1=1 z2w,2=1 z4=3 |
| | 12 | z2w,1=1 z2w,2=1 z4=2 |
| | 12 | z2w,1=1 z2w,2=1 z4=2 |
| 129.4.79 | 89593 | K1Mn1P1 |
| | 115 | z2=1 |
| | 115 | z2=1 |
| | 12 | z4=3 |
| | 12 | z2w,1=1 z2w,2=1 z4=1 |
| | 12 | z2w,1=1 z2w,2=1 z4=1 |
| | 12 | z4=3 |
| | 12 | z2w,1=1 z2w,2=1 |
| | 12 | z4=2 |
| | 12 | z4=2 |
| | 12 | z2w,1=1 z2w,2=1 |
| | 12 | z2w,1=1 z2w,2=1 z2w,3=1 |
| | 12 | z2w,3=1 z4=2 |
| | 12 | z2w,3=1 z4=2 |
| | 12 | z2w,1=1 z2w,2=1 z2w,3=1 |
| | 12 | z2w,3=1 z4=1 |
| | 12 | z2w,1=1 z2w,2=1 z2w,3=1 z4=3 |
| | 12 | z2w,1=1 z2w,2=1 z2w,3=1 z4=3 |
| | 12 | z2w,3=1 z4=1 |
| 129.4.80 | 89592 | K1Mn1P1 |
| | 115 | z2=1 |
| | 115 | z2=1 |
| | 12 | z4=3 |
| | 12 | z2w,1=1 z2w,2=1 z4=1 |
| | 12 | z2w,1=1 z2w,2=1 z4=1 |
| | 12 | z4=3 |
| | 12 | z2w,1=1 z2w,2=1 |
| | 12 | z4=2 |
| | 12 | z4=2 |
| | 12 | z2w,1=1 z2w,2=1 |
| | 12 | z2w,1=1 z2w,2=1 z2w,3=1 |
| | 12 | z2w,3=1 z4=2 |
| | 12 | z2w,3=1 z4=2 |
| | 12 | z2w,1=1 z2w,2=1 z2w,3=1 |
| | 12 | z2w,3=1 z4=1 |
| | 12 | z2w,1=1 z2w,2=1 z2w,3=1 z4=3 |
| | 12 | z2w,1=1 z2w,2=1 z2w,3=1 z4=3 |
| | 12 | z2w,3=1 z4=1 |
| 129.4.81 | 41882 | K1Mn1P1 |
| | 115 | z2=1 |
| | 115 | z2=1 |
| | 12 | z4=3 |
| | 12 | z2w,1=1 z2w,2=1 z4=1 |
| | 12 | z2w,1=1 z2w,2=1 z4=1 |
| | 12 | z4=3 |
| | 12 | z2w,1=1 z2w,2=1 |
| | 12 | z4=2 |
| | 12 | z4=2 |
| | 12 | z2w,1=1 z2w,2=1 |
| | 12 | z2w,1=1 z2w,2=1 z2w,3=1 |

Continued on next column

**TABLE XIX– continued**

| TQC | ICSD SSG | Stoich. Formula SIs |
|---|---|---|
| | 12 | z2w,3=1 z4=2 |
| | 12 | z2w,3=1 z4=2 |
| | 12 | z2w,1=1 z2w,2=1 z2w,3=1 |
| | 12 | z2w,3=1 z4=1 |
| | 12 | z2w,1=1 z2w,2=1 z2w,3=1 z4=3 |
| | 12 | z2w,1=1 z2w,2=1 z2w,3=1 z4=3 |
| | 12 | z2w,3=1 z4=1 |
| 129.4.82 | 44668 | Ir2Si2Y1 |
| | 115 | z2=1 |
| | 115 | z2=1 |
| | 12 | z4=2 |
| | 12 | z2w,1=1 z2w,2=1 |
| | 12 | z2w,1=1 z2w,2=1 |
| | 12 | z4=2 |
| | 12 | z2w,1=1 z2w,2=1 z4=3 |
| | 12 | z4=1 |
| | 12 | z4=1 |
| | 12 | z2w,1=1 z2w,2=1 z4=3 |
| | 12 | z2w,1=1 z2w,2=1 z2w,3=1 z4=3 |
| | 12 | z2w,3=1 z4=1 |
| | 12 | z2w,3=1 z4=1 |
| | 12 | z2w,1=1 z2w,2=1 z2w,3=1 z4=3 |
| | 12 | z2w,3=1 |
| | 12 | z2w,1=1 z2w,2=1 z2w,3=1 z4=2 |
| | 12 | z2w,1=1 z2w,2=1 z2w,3=1 z4=2 |
| | 12 | z2w,3=1 |
| 129.4.83 | 188325 | Ir2Si2Y1 |
| | 115 | z2=1 |
| | 115 | z2=1 |
| | 12 | z4=2 |
| | 12 | z2w,1=1 z2w,2=1 |
| | 12 | z2w,1=1 z2w,2=1 |
| | 12 | z4=2 |
| | 12 | z2w,1=1 z2w,2=1 z4=3 |
| | 12 | z4=1 |
| | 12 | z4=1 |
| | 12 | z2w,1=1 z2w,2=1 z4=3 |
| | 12 | z2w,1=1 z2w,2=1 z2w,3=1 z4=3 |
| | 12 | z2w,3=1 z4=1 |
| | 12 | z2w,3=1 z4=1 |
| | 12 | z2w,1=1 z2w,2=1 z2w,3=1 z4=3 |
| | 12 | z2w,3=1 |
| | 12 | z2w,1=1 z2w,2=1 z2w,3=1 z4=2 |
| | 12 | z2w,1=1 z2w,2=1 z2w,3=1 z4=2 |
| | 12 | z2w,3=1 |
| 129.4.84 | 73035 | Ir2Np1Si2 |
| | 115 | z2=1 |
| | 115 | z2=1 |
| | 12 | z4=2 |
| | 12 | z2w,1=1 z2w,2=1 |
| | 12 | z2w,1=1 z2w,2=1 |
| | 12 | z4=2 |
| | 12 | z2w,1=1 z2w,2=1 z4=3 |
| | 12 | z4=1 |
| | 12 | z4=1 |
| | 12 | z2w,1=1 z2w,2=1 z4=3 |
| | 12 | z2w,1=1 z2w,2=1 z2w,3=1 z4=3 |
| | 12 | z2w,3=1 z4=1 |
| | 12 | z2w,3=1 z4=1 |
| | 12 | z2w,1=1 z2w,2=1 z2w,3=1 z4=3 |

Continued on next column

**TABLE XIX– continued**

| TQC | ICSD SSG | Stoich. Formula SIs |
|---|---|---|
| | 12 | z2w,3=1 |
| | 12 | z2w,1=1 z2w,2=1 z2w,3=1 z4=2 |
| | 12 | z2w,1=1 z2w,2=1 z2w,3=1 z4=2 |
| | 12 | z2w,3=1 |
| 129.4.85 | 44658 | Ir2La1Si2 |
| | 115 | z2=1 |
| | 115 | z2=1 |
| | 12 | z2w,3=1 z4=3 |
| | 12 | z2w,1=1 z2w,2=1 z2w,3=1 z4=1 |
| | 12 | z2w,1=1 z2w,2=1 z2w,3=1 z4=1 |
| | 12 | z2w,3=1 z4=3 |
| | 12 | z2w,1=1 z2w,2=1 z2w,3=1 |
| | 12 | z2w,3=1 z4=2 |
| | 12 | z2w,3=1 z4=2 |
| | 12 | z2w,1=1 z2w,2=1 z2w,3=1 |
| | 12 | z2w,1=1 z2w,2=1 |
| | 12 | z4=2 |
| | 12 | z4=2 |
| | 12 | z2w,1=1 z2w,2=1 |
| | 12 | z4=1 |
| | 12 | z2w,1=1 z2w,2=1 z4=3 |
| | 12 | z2w,1=1 z2w,2=1 z4=3 |
| | 12 | z4=1 |
| 129.4.86 | 188327 | Ir2La1Si2 |
| | 115 | z2=1 |
| | 115 | z2=1 |
| | 12 | z4=2 |
| | 12 | z2w,1=1 z2w,2=1 |
| | 12 | z2w,1=1 z2w,2=1 |
| | 12 | z4=2 |
| | 12 | z2w,1=1 z2w,2=1 z4=3 |
| | 12 | z4=1 |
| | 12 | z4=1 |
| | 12 | z2w,1=1 z2w,2=1 z4=3 |
| | 12 | z2w,1=1 z2w,2=1 z2w,3=1 z4=3 |
| | 12 | z2w,3=1 z4=1 |
| | 12 | z2w,3=1 z4=1 |
| | 12 | z2w,1=1 z2w,2=1 z2w,3=1 z4=3 |
| | 12 | z2w,3=1 |
| | 12 | z2w,1=1 z2w,2=1 z2w,3=1 z4=2 |
| | 12 | z2w,1=1 z2w,2=1 z2w,3=1 z4=2 |
| | 12 | z2w,3=1 |
| 129.4.87 | 38204 | I1O1Pu1 |
| | 115 | z2=1 |
| | 115 | z2=1 |
| | 12 | z2w,1=1 z2w,2=1 z4=2 |
| | 12 | z2w,1=1 z2w,2=1 z4=2 |
| | 12 | z2w,1=1 z2w,2=1 z4=1 |
| | 12 | z4=3 |
| | 12 | z4=3 |
| | 12 | z2w,1=1 z2w,2=1 z4=1 |
| | 12 | z2w,1=1 z2w,2=1 z2w,3=1 z4=1 |
| | 12 | z2w,3=1 z4=3 |
| | 12 | z2w,3=1 z4=3 |
| | 12 | z2w,1=1 z2w,2=1 z2w,3=1 z4=1 |
| | 12 | z2w,3=1 z4=2 |
| | 12 | z2w,1=1 z2w,2=1 z2w,3=1 |
| | 12 | z2w,1=1 z2w,2=1 z2w,3=1 |
| | 12 | z2w,3=1 z4=2 |
| 129.4.88 | 31663 | I1O1Pu1 |

Continued on next column

**TABLE XIX– continued**

| TQC | ICSD SSG | Stoich. Formula SIs |
|---|---|---|
| | 115 | z2=1 |
| | 115 | z2=1 |
| | 12 | z2w,1=1 z2w,2=1 z4=2 |
| | 12 | z2w,1=1 z2w,2=1 z4=2 |
| | 12 | z2w,1=1 z2w,2=1 z4=1 |
| | 12 | z4=3 |
| | 12 | z4=3 |
| | 12 | z2w,1=1 z2w,2=1 z4=1 |
| | 12 | z2w,1=1 z2w,2=1 z2w,3=1 z4=1 |
| | 12 | z2w,3=1 z4=3 |
| | 12 | z2w,3=1 z4=3 |
| | 12 | z2w,1=1 z2w,2=1 z2w,3=1 z4=1 |
| | 12 | z2w,3=1 z4=2 |
| | 12 | z2w,1=1 z2w,2=1 z2w,3=1 |
| | 12 | z2w,1=1 z2w,2=1 z2w,3=1 |
| | 12 | z2w,3=1 z4=2 |
| 129.4.89 | 183311 | Ho1Sn2Zn1 |
| | 115 | z2=1 |
| | 115 | z2=1 |
| | 12 | z2w,3=1 |
| | 12 | z2w,1=1 z2w,2=1 z2w,3=1 z4=2 |
| | 12 | z2w,1=1 z2w,2=1 z2w,3=1 z4=2 |
| | 12 | z2w,3=1 |
| | 12 | z2w,1=1 z2w,2=1 z2w,3=1 z4=1 |
| | 12 | z2w,3=1 z4=3 |
| | 12 | z2w,3=1 z4=3 |
| | 12 | z2w,1=1 z2w,2=1 z2w,3=1 z4=1 |
| | 12 | z2w,1=1 z2w,2=1 z4=1 |
| | 12 | z4=3 |
| | 12 | z4=3 |
| | 12 | z2w,1=1 z2w,2=1 z4=1 |
| | 12 | z4=2 |
| | 12 | z2w,1=1 z2w,2=1 |
| | 12 | z2w,1=1 z2w,2=1 |
| | 12 | z4=2 |
| 129.4.90 | 88117 | Ho1Si1Ti1 |
| | 115 | z2=1 |
| | 115 | z2=1 |
| | 12 | z4=2 |
| | 12 | z2w,1=1 z2w,2=1 |
| | 12 | z2w,1=1 z2w,2=1 |
| | 12 | z4=2 |
| | 12 | z2w,1=1 z2w,2=1 z4=3 |
| | 12 | z4=1 |
| | 12 | z4=1 |
| | 12 | z2w,1=1 z2w,2=1 z4=3 |
| | 12 | z2w,1=1 z2w,2=1 z2w,3=1 z4=3 |
| | 12 | z2w,3=1 z4=1 |
| | 12 | z2w,3=1 z4=1 |
| | 12 | z2w,1=1 z2w,2=1 z2w,3=1 z4=3 |
| | 12 | z2w,3=1 |
| | 12 | z2w,1=1 z2w,2=1 z2w,3=1 z4=2 |
| | 12 | z2w,1=1 z2w,2=1 z2w,3=1 z4=2 |
| | 12 | z2w,3=1 |
| 129.4.91 | 97037 | Ho1Ni10Si2 |
| | 115 | z2=1 |
| | 115 | z2=1 |
| | 12 | z2w,3=1 z4=2 |
| | 12 | z2w,1=1 z2w,2=1 z2w,3=1 |
| | 12 | z2w,1=1 z2w,2=1 z2w,3=1 |

Continued on next column

TABLE XIX– continued

| TQC | ICSD SSG | Stoich. Formula SIs |
|---|---|---|
| | 12 | z2w,3=1 z4=2 |
| | 12 | z2w,1=1 z2w,2=1 z2w,3=1 z4=3 |
| | 12 | z2w,3=1 z4=1 |
| | 12 | z2w,3=1 z4=1 |
| | 12 | z2w,1=1 z2w,2=1 z2w,3=1 z4=3 |
| | 12 | z2w,1=1 z2w,2=1 z4=3 |
| | 12 | z4=1 |
| | 12 | z4=1 |
| | 12 | z2w,1=1 z2w,2=1 z4=3 |
| | 12 | z2w,1=1 z2w,2=1 z4=2 |
| | 12 | z2w,1=1 z2w,2=1 z4=2 |
| 129.4.92 | 263028 | H2Ca2Ge2Ni2 |
| | 115 | z2=1 |
| | 115 | z2=1 |
| | 12 | z4=1 |
| | 12 | z2w,1=1 z2w,2=1 z4=3 |
| | 12 | z2w,1=1 z2w,2=1 z4=3 |
| | 12 | z4=1 |
| | 12 | z2w,1=1 z2w,2=1 z4=2 |
| | 12 | z2w,1=1 z2w,2=1 z4=2 |
| | 12 | z2w,1=1 z2w,2=1 z2w,3=1 z4=2 |
| | 12 | z2w,3=1 |
| | 12 | z2w,3=1 |
| | 12 | z2w,1=1 z2w,2=1 z2w,3=1 z4=2 |
| | 12 | z2w,3=1 z4=3 |
| | 12 | z2w,1=1 z2w,2=1 z2w,3=1 z4=1 |
| | 12 | z2w,1=1 z2w,2=1 z2w,3=1 z4=1 |
| | 12 | z2w,3=1 z4=3 |
| 129.4.93 | 428162 | H1Cl1Eu1 |
| | 115 | z2=1 |
| | 115 | z2=1 |
| | 12 | z2w,1=1 z2w,2=1 z4=2 |
| | 12 | z2w,1=1 z2w,2=1 z4=2 |
| | 12 | z2w,1=1 z2w,2=1 z4=1 |
| | 12 | z4=3 |
| | 12 | z4=3 |
| | 12 | z2w,1=1 z2w,2=1 z4=1 |
| | 12 | z2w,1=1 z2w,2=1 z2w,3=1 z4=1 |
| | 12 | z2w,3=1 z4=3 |
| | 12 | z2w,3=1 z4=3 |
| | 12 | z2w,1=1 z2w,2=1 z2w,3=1 z4=1 |
| | 12 | z2w,3=1 z4=2 |
| | 12 | z2w,1=1 z2w,2=1 z2w,3=1 |
| | 12 | z2w,1=1 z2w,2=1 z2w,3=1 |
| | 12 | z2w,3=1 z4=2 |
| 129.4.94 | 161825 | H1Ce1Ru1Si1 |
| | 115 | z2=1 |
| | 115 | z2=1 |
| | 12 | z2w,3=1 z4=2 |
| | 12 | z2w,1=1 z2w,2=1 z2w,3=1 |
| | 12 | z2w,1=1 z2w,2=1 z2w,3=1 |
| | 12 | z2w,3=1 z4=2 |
| | 12 | z2w,1=1 z2w,2=1 z2w,3=1 z4=3 |
| | 12 | z2w,3=1 z4=1 |
| | 12 | z2w,3=1 z4=1 |
| | 12 | z2w,1=1 z2w,2=1 z2w,3=1 z4=3 |
| | 12 | z2w,1=1 z2w,2=1 z4=3 |
| | 12 | z4=1 |
| | 12 | z4=1 |
| | 12 | z2w,1=1 z2w,2=1 z4=3 |

Continued on next column

TABLE XIX– continued

| TQC | ICSD SSG | Stoich. Formula SIs |
|---|---|---|
| | 12 | z2w,1=1 z2w,2=1 z4=2 |
| | 12 | z2w,1=1 z2w,2=1 z4=2 |
| 129.4.95 | 416423 | Ge2La1Pt2 |
| | 115 | z2=1 |
| | 115 | z2=1 |
| | 12 | z4=2 |
| | 12 | z2w,1=1 z2w,2=1 |
| | 12 | z2w,1=1 z2w,2=1 |
| | 12 | z4=2 |
| | 12 | z2w,1=1 z2w,2=1 z4=3 |
| | 12 | z4=1 |
| | 12 | z4=1 |
| | 12 | z2w,1=1 z2w,2=1 z4=3 |
| | 12 | z2w,1=1 z2w,2=1 z2w,3=1 z4=3 |
| | 12 | z2w,3=1 z4=1 |
| | 12 | z2w,3=1 z4=1 |
| | 12 | z2w,1=1 z2w,2=1 z2w,3=1 z4=3 |
| | 12 | z2w,3=1 |
| | 12 | z2w,1=1 z2w,2=1 z2w,3=1 z4=2 |
| | 12 | z2w,1=1 z2w,2=1 z2w,3=1 z4=2 |
| | 12 | z2w,3=1 |
| 129.4.96 | 180145 | Ge2Ir2Pr1 |
| | 115 | z2=1 |
| | 115 | z2=1 |
| | 12 | z2w,3=1 z4=1 |
| | 12 | z2w,1=1 z2w,2=1 z2w,3=1 z4=3 |
| | 12 | z2w,1=1 z2w,2=1 z2w,3=1 z4=3 |
| | 12 | z2w,3=1 z4=1 |
| | 12 | z2w,1=1 z2w,2=1 z2w,3=1 z4=2 |
| | 12 | z2w,3=1 |
| | 12 | z2w,3=1 |
| | 12 | z2w,1=1 z2w,2=1 z2w,3=1 z4=2 |
| | 12 | z2w,1=1 z2w,2=1 z4=2 |
| | 12 | z2w,1=1 z2w,2=1 z4=2 |
| | 12 | z4=3 |
| | 12 | z2w,1=1 z2w,2=1 z4=1 |
| | 12 | z2w,1=1 z2w,2=1 z4=1 |
| | 12 | z4=3 |
| 129.4.97 | 90233 | Ge1Ti1Y1 |
| | 115 | z2=1 |
| | 115 | z2=1 |
| | 12 | z2w,3=1 z4=2 |
| | 12 | z2w,1=1 z2w,2=1 z2w,3=1 |
| | 12 | z2w,1=1 z2w,2=1 z2w,3=1 |
| | 12 | z2w,3=1 z4=2 |
| | 12 | z2w,1=1 z2w,2=1 z2w,3=1 z4=3 |
| | 12 | z2w,3=1 z4=1 |
| | 12 | z2w,3=1 z4=1 |
| | 12 | z2w,1=1 z2w,2=1 z2w,3=1 z4=3 |
| | 12 | z2w,1=1 z2w,2=1 z4=3 |
| | 12 | z4=1 |
| | 12 | z4=1 |
| | 12 | z2w,1=1 z2w,2=1 z4=3 |
| | 12 | z2w,1=1 z2w,2=1 z4=2 |
| | 12 | z2w,1=1 z2w,2=1 z4=2 |
| 129.4.98 | 85904 | Ge1Ti1Y1 |
| | 115 | z2=1 |
| | 115 | z2=1 |
| | 12 | z2w,3=1 z4=2 |
| | 12 | z2w,1=1 z2w,2=1 z2w,3=1 |

Continued on next column

**TABLE XIX– continued**

| TQC | ICSD<br>SSG | Stoich. Formula<br>SIs |
|---|---|---|
| | 12 | z2w,1=1 z2w,2=1 z2w,3=1 |
| | 12 | z2w,3=1 z4=2 |
| | 12 | z2w,1=1 z2w,2=1 z2w,3=1 z4=3 |
| | 12 | z2w,3=1 z4=1 |
| | 12 | z2w,3=1 z4=1 |
| | 12 | z2w,1=1 z2w,2=1 z2w,3=1 z4=3 |
| | 12 | z2w,1=1 z2w,2=1 z4=3 |
| | 12 | z4=1 |
| | 12 | z4=1 |
| | 12 | z2w,1=1 z2w,2=1 z4=3 |
| | 12 | z2w,1=1 z2w,2=1 z4=2 |
| | 12 | z2w,1=1 z2w,2=1 z4=2 |
| 129.4.99 | 90393 | Ge1Tb1Ti1 |
| | 115 | z2=1 |
| | 115 | z2=1 |
| | 12 | z4=1 |
| | 12 | z2w,1=1 z2w,2=1 z4=3 |
| | 12 | z2w,1=1 z2w,2=1 z4=3 |
| | 12 | z4=1 |
| | 12 | z2w,1=1 z2w,2=1 z4=2 |
| | 12 | z2w,1=1 z2w,2=1 z4=2 |
| | 12 | z2w,1=1 z2w,2=1 z2w,3=1 z4=2 |
| | 12 | z2w,3=1 |
| | 12 | z2w,3=1 |
| | 12 | z2w,1=1 z2w,2=1 z2w,3=1 z4=2 |
| | 12 | z2w,3=1 z4=3 |
| | 12 | z2w,1=1 z2w,2=1 z2w,3=1 z4=1 |
| | 12 | z2w,1=1 z2w,2=1 z2w,3=1 z4=1 |
| | 12 | z2w,3=1 z4=3 |
| 129.4.100 | 85906 | Ge1Tb1Ti1 |
| | 115 | z2=1 |
| | 115 | z2=1 |
| | 12 | z2w,3=1 z4=2 |
| | 12 | z2w,1=1 z2w,2=1 z2w,3=1 |
| | 12 | z2w,1=1 z2w,2=1 z2w,3=1 |
| | 12 | z2w,3=1 z4=2 |
| | 12 | z2w,1=1 z2w,2=1 z2w,3=1 z4=3 |
| | 12 | z2w,3=1 z4=1 |
| | 12 | z2w,3=1 z4=1 |
| | 12 | z2w,1=1 z2w,2=1 z2w,3=1 z4=3 |
| | 12 | z2w,1=1 z2w,2=1 z4=3 |
| | 12 | z4=1 |
| | 12 | z4=1 |
| | 12 | z2w,1=1 z2w,2=1 z4=3 |
| | 12 | z2w,1=1 z2w,2=1 z4=2 |
| | 12 | z2w,1=1 z2w,2=1 z4=2 |
| 129.4.101 | 85000 | Ge1Sb1Zr1 |
| | 115 | z2=1 |
| | 115 | z2=1 |
| | 12 | z2w,3=1 |
| | 12 | z2w,1=1 z2w,2=1 z2w,3=1 z4=2 |
| | 12 | z2w,1=1 z2w,2=1 z2w,3=1 z4=2 |
| | 12 | z2w,3=1 |
| | 12 | z2w,1=1 z2w,2=1 z2w,3=1 z4=1 |
| | 12 | z2w,3=1 z4=3 |
| | 12 | z2w,3=1 z4=3 |
| | 12 | z2w,1=1 z2w,2=1 z2w,3=1 z4=1 |
| | 12 | z2w,1=1 z2w,2=1 z4=1 |
| | 12 | z4=3 |
| | 12 | z4=3 |

Continued on next column

**TABLE XIX– continued**

| TQC | ICSD<br>SSG | Stoich. Formula<br>SIs |
|---|---|---|
| | 12 | z2w,1=1 z2w,2=1 z4=1 |
| | 12 | z4=2 |
| | 12 | z2w,1=1 z2w,2=1 |
| | 12 | z2w,1=1 z2w,2=1 |
| | 12 | z4=2 |
| 129.4.102 | 260557 | Ge1Sb1Ti1 |
| | 115 | z2=1 |
| | 115 | z2=1 |
| | 12 | z4=3 |
| | 12 | z2w,1=1 z2w,2=1 z4=1 |
| | 12 | z2w,1=1 z2w,2=1 z4=1 |
| | 12 | z4=3 |
| | 12 | z2w,1=1 z2w,2=1 |
| | 12 | z4=2 |
| | 12 | z4=2 |
| | 12 | z2w,1=1 z2w,2=1 |
| | 12 | z2w,1=1 z2w,2=1 z2w,3=1 |
| | 12 | z2w,3=1 z4=2 |
| | 12 | z2w,3=1 z4=2 |
| | 12 | z2w,1=1 z2w,2=1 z2w,3=1 |
| | 12 | z2w,3=1 z4=1 |
| | 12 | z2w,1=1 z2w,2=1 z2w,3=1 z4=3 |
| | 12 | z2w,1=1 z2w,2=1 z2w,3=1 z4=3 |
| | 12 | z2w,3=1 z4=1 |
| 129.4.103 | 66961 | Ge1Mn1Sr1 |
| | 115 | z2=1 |
| | 115 | z2=1 |
| | 12 | z2w,1=1 z2w,2=1 z4=2 |
| | 12 | z2w,1=1 z2w,2=1 z4=2 |
| | 12 | z2w,1=1 z2w,2=1 z4=1 |
| | 12 | z4=3 |
| | 12 | z4=3 |
| | 12 | z2w,1=1 z2w,2=1 z4=1 |
| | 12 | z2w,1=1 z2w,2=1 z2w,3=1 z4=1 |
| | 12 | z2w,3=1 z4=3 |
| | 12 | z2w,3=1 z4=3 |
| | 12 | z2w,1=1 z2w,2=1 z2w,3=1 z4=1 |
| | 12 | z2w,3=1 z4=2 |
| | 12 | z2w,1=1 z2w,2=1 z2w,3=1 |
| | 12 | z2w,1=1 z2w,2=1 z2w,3=1 |
| | 12 | z2w,3=1 z4=2 |
| 129.4.104 | 66960 | Ge1Mn1Sr1 |
| | 115 | z2=1 |
| | 115 | z2=1 |
| | 12 | z2w,1=1 z2w,2=1 z4=2 |
| | 12 | z2w,1=1 z2w,2=1 z4=2 |
| | 12 | z2w,1=1 z2w,2=1 z4=1 |
| | 12 | z4=3 |
| | 12 | z4=3 |
| | 12 | z2w,1=1 z2w,2=1 z4=1 |
| | 12 | z2w,1=1 z2w,2=1 z2w,3=1 z4=1 |
| | 12 | z2w,3=1 z4=3 |
| | 12 | z2w,3=1 z4=3 |
| | 12 | z2w,1=1 z2w,2=1 z2w,3=1 z4=1 |
| | 12 | z2w,3=1 z4=2 |
| | 12 | z2w,1=1 z2w,2=1 z2w,3=1 |
| | 12 | z2w,1=1 z2w,2=1 z2w,3=1 |
| | 12 | z2w,3=1 z4=2 |
| 129.4.105 | 53707 | Ge1Mn1Sr1 |
| | 115 | z2=1 |

Continued on next column

**TABLE XIX– continued**

| TQC | ICSD SSG | Stoich. Formula SIs |
|---|---|---|
| | 115 | z2=1 |
| | 12 | z2w,1=1 z2w,2=1 z4=2 |
| | 12 | z2w,1=1 z2w,2=1 z4=2 |
| | 12 | z2w,1=1 z2w,2=1 z4=1 |
| | 12 | z4=3 |
| | 12 | z4=3 |
| | 12 | z2w,1=1 z2w,2=1 z4=1 |
| | 12 | z2w,1=1 z2w,2=1 z2w,3=1 z4=1 |
| | 12 | z2w,3=1 z4=3 |
| | 12 | z2w,3=1 z4=3 |
| | 12 | z2w,1=1 z2w,2=1 z2w,3=1 z4=1 |
| | 12 | z2w,3=1 z4=2 |
| | 12 | z2w,1=1 z2w,2=1 z2w,3=1 |
| | 12 | z2w,1=1 z2w,2=1 z2w,3=1 |
| | 12 | z2w,3=1 z4=2 |
| 129.4.106 | 80016 | Ge1Mn1Nd1 |
| | 115 | z2=1 |
| | 115 | z2=1 |
| | 12 | z4=1 |
| | 12 | z2w,1=1 z2w,2=1 z4=3 |
| | 12 | z2w,1=1 z2w,2=1 z4=3 |
| | 12 | z4=1 |
| | 12 | z2w,1=1 z2w,2=1 z4=2 |
| | 12 | z2w,1=1 z2w,2=1 z4=2 |
| | 12 | z2w,1=1 z2w,2=1 z2w,3=1 z4=2 |
| | 12 | z2w,3=1 |
| | 12 | z2w,3=1 |
| | 12 | z2w,1=1 z2w,2=1 z2w,3=1 z4=2 |
| | 12 | z2w,3=1 z4=3 |
| | 12 | z2w,1=1 z2w,2=1 z2w,3=1 z4=1 |
| | 12 | z2w,1=1 z2w,2=1 z2w,3=1 z4=1 |
| | 12 | z2w,3=1 z4=3 |
| 129.4.107 | 80014 | Ge1Mn1Nd1 |
| | 115 | z2=1 |
| | 115 | z2=1 |
| | 12 | z4=1 |
| | 12 | z2w,1=1 z2w,2=1 z4=3 |
| | 12 | z2w,1=1 z2w,2=1 z4=3 |
| | 12 | z4=1 |
| | 12 | z2w,1=1 z2w,2=1 z4=2 |
| | 12 | z2w,1=1 z2w,2=1 z4=2 |
| | 12 | z2w,1=1 z2w,2=1 z2w,3=1 z4=2 |
| | 12 | z2w,3=1 |
| | 12 | z2w,3=1 |
| | 12 | z2w,1=1 z2w,2=1 z2w,3=1 z4=2 |
| | 12 | z2w,3=1 z4=3 |
| | 12 | z2w,1=1 z2w,2=1 z2w,3=1 z4=1 |
| | 12 | z2w,1=1 z2w,2=1 z2w,3=1 z4=1 |
| | 12 | z2w,3=1 z4=3 |
| 129.4.108 | 87117 | Ge1Mg1Mn1 |
| | 115 | z2=1 |
| | 115 | z2=1 |
| | 12 | z2w,3=1 |
| | 12 | z2w,1=1 z2w,2=1 z2w,3=1 z4=2 |
| | 12 | z2w,1=1 z2w,2=1 z2w,3=1 z4=2 |
| | 12 | z2w,3=1 |
| | 12 | z2w,1=1 z2w,2=1 z2w,3=1 z4=1 |
| | 12 | z2w,3=1 z4=3 |
| | 12 | z2w,3=1 z4=3 |
| | 12 | z2w,1=1 z2w,2=1 z2w,3=1 z4=1 |

Continued on next column

**TABLE XIX– continued**

| TQC | ICSD SSG | Stoich. Formula SIs |
|---|---|---|
| | 12 | z2w,1=1 z2w,2=1 z4=1 |
| | 12 | z4=3 |
| | 12 | z4=3 |
| | 12 | z2w,1=1 z2w,2=1 z4=1 |
| | 12 | z4=2 |
| | 12 | z2w,1=1 z2w,2=1 |
| | 12 | z2w,1=1 z2w,2=1 |
| | 12 | z4=2 |
| 129.4.109 | 66953 | Ge1Mg1Mn1 |
| | 115 | z2=1 |
| | 115 | z2=1 |
| | 12 | z4=3 |
| | 12 | z2w,1=1 z2w,2=1 z4=1 |
| | 12 | z2w,1=1 z2w,2=1 z4=1 |
| | 12 | z4=3 |
| | 12 | z2w,1=1 z2w,2=1 |
| | 12 | z4=2 |
| | 12 | z4=2 |
| | 12 | z2w,1=1 z2w,2=1 |
| | 12 | z2w,1=1 z2w,2=1 z2w,3=1 |
| | 12 | z2w,3=1 z4=2 |
| | 12 | z2w,3=1 z4=2 |
| | 12 | z2w,1=1 z2w,2=1 z2w,3=1 |
| | 12 | z2w,3=1 z4=1 |
| | 12 | z2w,1=1 z2w,2=1 z2w,3=1 z4=3 |
| | 12 | z2w,1=1 z2w,2=1 z2w,3=1 z4=3 |
| | 12 | z2w,3=1 z4=1 |
| 129.4.110 | 66952 | Ge1Mg1Mn1 |
| | 115 | z2=1 |
| | 115 | z2=1 |
| | 12 | z4=3 |
| | 12 | z2w,1=1 z2w,2=1 z4=1 |
| | 12 | z2w,1=1 z2w,2=1 z4=1 |
| | 12 | z4=3 |
| | 12 | z2w,1=1 z2w,2=1 |
| | 12 | z4=2 |
| | 12 | z4=2 |
| | 12 | z2w,1=1 z2w,2=1 |
| | 12 | z2w,1=1 z2w,2=1 z2w,3=1 |
| | 12 | z2w,3=1 z4=2 |
| | 12 | z2w,3=1 z4=2 |
| | 12 | z2w,1=1 z2w,2=1 z2w,3=1 |
| | 12 | z2w,3=1 z4=1 |
| | 12 | z2w,1=1 z2w,2=1 z2w,3=1 z4=3 |
| | 12 | z2w,1=1 z2w,2=1 z2w,3=1 z4=3 |
| | 12 | z2w,3=1 z4=1 |
| 129.4.111 | 66948 | Ge1Mg1Mn1 |
| | 115 | z2=1 |
| | 115 | z2=1 |
| | 12 | z2w,3=1 |
| | 12 | z2w,1=1 z2w,2=1 z2w,3=1 z4=2 |
| | 12 | z2w,1=1 z2w,2=1 z2w,3=1 z4=2 |
| | 12 | z2w,3=1 |
| | 12 | z2w,1=1 z2w,2=1 z2w,3=1 z4=1 |
| | 12 | z2w,3=1 z4=3 |
| | 12 | z2w,3=1 z4=3 |
| | 12 | z2w,1=1 z2w,2=1 z2w,3=1 z4=1 |
| | 12 | z2w,1=1 z2w,2=1 z4=1 |
| | 12 | z4=3 |
| | 12 | z4=3 |

Continued on next column

TABLE XIX– continued

| TQC | ICSD SSG | Stoich. Formula SIs |
|---|---|---|
| | 12 | z2w,1=1 z2w,2=1 z4=1 |
| | 12 | z4=2 |
| | 12 | z2w,1=1 z2w,2=1 |
| | 12 | z2w,1=1 z2w,2=1 |
| | 12 | z4=2 |
| 129.4.112 | 88209 | Ge1Lu1Ti1 |
| | 115 | z2=1 |
| | 115 | z2=1 |
| | 12 | z2w,3=1 z4=2 |
| | 12 | z2w,1=1 z2w,2=1 z2w,3=1 |
| | 12 | z2w,1=1 z2w,2=1 z2w,3=1 |
| | 12 | z2w,3=1 z4=2 |
| | 12 | z2w,1=1 z2w,2=1 z2w,3=1 z4=3 |
| | 12 | z2w,3=1 z4=1 |
| | 12 | z2w,3=1 z4=1 |
| | 12 | z2w,1=1 z2w,2=1 z2w,3=1 z4=3 |
| | 12 | z2w,1=1 z2w,2=1 z4=3 |
| | 12 | z4=1 |
| | 12 | z4=1 |
| | 12 | z2w,1=1 z2w,2=1 z4=3 |
| | 12 | z2w,1=1 z2w,2=1 z4=2 |
| | 12 | z2w,1=1 z2w,2=1 z4=2 |
| 129.4.113 | 86777 | Ge1Li2O5V1 |
| | 115 | z2=1 |
| | 115 | z2=1 |
| | 12 | z4=1 |
| | 12 | z2w,1=1 z2w,2=1 z4=3 |
| | 12 | z2w,1=1 z2w,2=1 z4=3 |
| | 12 | z4=1 |
| | 12 | z2w,1=1 z2w,2=1 z4=2 |
| | 12 | z2w,1=1 z2w,2=1 z4=2 |
| | 12 | z2w,1=1 z2w,2=1 z2w,3=1 z4=2 |
| | 12 | z2w,3=1 |
| | 12 | z2w,3=1 |
| | 12 | z2w,1=1 z2w,2=1 z2w,3=1 z4=2 |
| | 12 | z2w,3=1 z4=3 |
| | 12 | z2w,1=1 z2w,2=1 z2w,3=1 z4=1 |
| | 12 | z2w,1=1 z2w,2=1 z2w,3=1 z4=1 |
| | 12 | z2w,3=1 z4=3 |
| 129.4.114 | 107617 | Ge1La1Ti1 |
| | 115 | z2=1 |
| | 115 | z2=1 |
| | 12 | z4=1 |
| | 12 | z2w,1=1 z2w,2=1 z4=3 |
| | 12 | z2w,1=1 z2w,2=1 z4=3 |
| | 12 | z4=1 |
| | 12 | z2w,1=1 z2w,2=1 z4=2 |
| | 12 | z2w,1=1 z2w,2=1 z4=2 |
| | 12 | z2w,1=1 z2w,2=1 z2w,3=1 z4=2 |
| | 12 | z2w,3=1 |
| | 12 | z2w,3=1 |
| | 12 | z2w,1=1 z2w,2=1 z2w,3=1 z4=2 |
| | 12 | z2w,3=1 z4=3 |
| | 12 | z2w,1=1 z2w,2=1 z2w,3=1 z4=1 |
| | 12 | z2w,1=1 z2w,2=1 z2w,3=1 z4=1 |
| | 12 | z2w,3=1 z4=3 |
| 129.4.115 | 85862 | Ge1La1Ru1 |
| | 115 | z2=1 |
| | 115 | z2=1 |
| | 12 | z2w,1=1 z2w,2=1 z4=2 |

Continued on next column

| TQC | ICSD SSG | Stoich. Formula SIs |
|---|---|---|
| | 12 | z2w,1=1 z2w,2=1 z4=2 |
| | 12 | z2w,1=1 z2w,2=1 z4=1 |
| | 12 | z4=3 |
| | 12 | z4=3 |
| | 12 | z2w,1=1 z2w,2=1 z4=1 |
| | 12 | z2w,1=1 z2w,2=1 z2w,3=1 z4=1 |
| | 12 | z2w,3=1 z4=3 |
| | 12 | z2w,3=1 z4=3 |
| | 12 | z2w,1=1 z2w,2=1 z2w,3=1 z4=1 |
| | 12 | z2w,3=1 z4=2 |
| | 12 | z2w,1=1 z2w,2=1 z2w,3=1 |
| | 12 | z2w,1=1 z2w,2=1 z2w,3=1 |
| | 12 | z2w,3=1 z4=2 |
| 129.4.116 | 95023 | Ge1Ho1Ti1 |
| | 115 | z2=1 |
| | 115 | z2=1 |
| | 12 | z2w,3=1 z4=3 |
| | 12 | z2w,1=1 z2w,2=1 z2w,3=1 z4=1 |
| | 12 | z2w,1=1 z2w,2=1 z2w,3=1 z4=1 |
| | 12 | z2w,3=1 z4=3 |
| | 12 | z2w,1=1 z2w,2=1 z2w,3=1 |
| | 12 | z2w,3=1 z4=2 |
| | 12 | z2w,3=1 z4=2 |
| | 12 | z2w,1=1 z2w,2=1 z2w,3=1 |
| | 12 | z2w,1=1 z2w,2=1 |
| | 12 | z4=2 |
| | 12 | z4=2 |
| | 12 | z2w,1=1 z2w,2=1 |
| | 12 | z4=1 |
| | 12 | z2w,1=1 z2w,2=1 z4=3 |
| | 12 | z2w,1=1 z2w,2=1 z4=3 |
| | 12 | z4=1 |
| 129.4.117 | 95021 | Ge1Ho1Ti1 |
| | 115 | z2=1 |
| | 115 | z2=1 |
| | 12 | z2w,3=1 z4=3 |
| | 12 | z2w,1=1 z2w,2=1 z2w,3=1 z4=1 |
| | 12 | z2w,1=1 z2w,2=1 z2w,3=1 z4=1 |
| | 12 | z2w,3=1 z4=3 |
| | 12 | z2w,1=1 z2w,2=1 z2w,3=1 |
| | 12 | z2w,3=1 z4=2 |
| | 12 | z2w,3=1 z4=2 |
| | 12 | z2w,1=1 z2w,2=1 z2w,3=1 |
| | 12 | z2w,1=1 z2w,2=1 |
| | 12 | z4=2 |
| | 12 | z4=2 |
| | 12 | z2w,1=1 z2w,2=1 |
| | 12 | z4=1 |
| | 12 | z2w,1=1 z2w,2=1 z4=3 |
| | 12 | z2w,1=1 z2w,2=1 z4=3 |
| | 12 | z4=1 |
| 129.4.118 | 82006 | Gd1Na1O4Ti1 |
| | 115 | z2=1 |
| | 115 | z2=1 |
| | 12 | z2w,3=1 z4=2 |
| | 12 | z2w,1=1 z2w,2=1 z2w,3=1 |
| | 12 | z2w,1=1 z2w,2=1 z2w,3=1 |
| | 12 | z2w,3=1 z4=2 |
| | 12 | z2w,1=1 z2w,2=1 z2w,3=1 z4=3 |
| | 12 | z2w,3=1 z4=1 |

Continued on next column

**TABLE XIX– continued**

| TQC | ICSD<br>SSG | Stoich. Formula<br>SIs |
|---|---|---|
| | 12 | z2w,3=1 z4=1 |
| | 12 | z2w,1=1 z2w,2=1 z2w,3=1 z4=3 |
| | 12 | z2w,1=1 z2w,2=1 z4=3 |
| | 12 | z4=1 |
| | 12 | z4=1 |
| | 12 | z2w,1=1 z2w,2=1 z4=3 |
| | 12 | z2w,1=1 z2w,2=1 z4=2 |
| | 12 | z2w,1=1 z2w,2=1 z4=2 |
| 129.4.119 | 88277 | Gd1Mn1Si1 |
| | 115 | z2=1 |
| | 115 | z2=1 |
| | 12 | z4=1 |
| | 12 | z2w,1=1 z2w,2=1 z4=3 |
| | 12 | z2w,1=1 z2w,2=1 z4=3 |
| | 12 | z4=1 |
| | 12 | z2w,1=1 z2w,2=1 z4=2 |
| | 12 | z2w,1=1 z2w,2=1 z4=2 |
| | 12 | z2w,1=1 z2w,2=1 z2w,3=1 z4=2 |
| | 12 | z2w,3=1 |
| | 12 | z2w,3=1 |
| | 12 | z2w,1=1 z2w,2=1 z2w,3=1 z4=2 |
| | 12 | z2w,3=1 z4=3 |
| | 12 | z2w,1=1 z2w,2=1 z2w,3=1 z4=1 |
| | 12 | z2w,1=1 z2w,2=1 z2w,3=1 z4=1 |
| | 12 | z2w,3=1 z4=3 |
| 129.4.120 | 85852 | Gd1Mn1Si1 |
| | 115 | z2=1 |
| | 115 | z2=1 |
| | 12 | z4=1 |
| | 12 | z2w,1=1 z2w,2=1 z4=3 |
| | 12 | z2w,1=1 z2w,2=1 z4=3 |
| | 12 | z4=1 |
| | 12 | z2w,1=1 z2w,2=1 z4=2 |
| | 12 | z2w,1=1 z2w,2=1 z4=2 |
| | 12 | z2w,1=1 z2w,2=1 z2w,3=1 z4=2 |
| | 12 | z2w,3=1 |
| | 12 | z2w,3=1 |
| | 12 | z2w,1=1 z2w,2=1 z2w,3=1 z4=2 |
| | 12 | z2w,3=1 z4=3 |
| | 12 | z2w,1=1 z2w,2=1 z2w,3=1 z4=1 |
| | 12 | z2w,1=1 z2w,2=1 z2w,3=1 z4=1 |
| | 12 | z2w,3=1 z4=3 |
| 129.4.121 | 419854 | Gd1K1Na1O5Ta1 |
| | 115 | z2=1 |
| | 115 | z2=1 |
| | 12 | z2w,3=1 z4=1 |
| | 12 | z2w,1=1 z2w,2=1 z2w,3=1 z4=3 |
| | 12 | z2w,1=1 z2w,2=1 z2w,3=1 z4=3 |
| | 12 | z2w,3=1 z4=1 |
| | 12 | z2w,1=1 z2w,2=1 z2w,3=1 z4=2 |
| | 12 | z2w,3=1 |
| | 12 | z2w,3=1 |
| | 12 | z2w,1=1 z2w,2=1 z2w,3=1 z4=2 |
| | 12 | z2w,1=1 z2w,2=1 z4=2 |
| | 12 | z2w,1=1 z2w,2=1 z4=2 |
| | 12 | z4=3 |
| | 12 | z2w,1=1 z2w,2=1 z4=1 |
| | 12 | z2w,1=1 z2w,2=1 z4=1 |
| | 12 | z4=3 |
| 129.4.122 | 420037 | Gd1K1Na1Nb1O5 |

Continued on next column

**TABLE XIX– continued**

| TQC | ICSD<br>SSG | Stoich. Formula<br>SIs |
|---|---|---|
| | 115 | z2=1 |
| | 115 | z2=1 |
| | 12 | z2w,3=1 z4=1 |
| | 12 | z2w,1=1 z2w,2=1 z2w,3=1 z4=3 |
| | 12 | z2w,1=1 z2w,2=1 z2w,3=1 z4=3 |
| | 12 | z2w,3=1 z4=1 |
| | 12 | z2w,1=1 z2w,2=1 z2w,3=1 z4=2 |
| | 12 | z2w,3=1 |
| | 12 | z2w,3=1 |
| | 12 | z2w,1=1 z2w,2=1 z2w,3=1 z4=2 |
| | 12 | z2w,1=1 z2w,2=1 z4=2 |
| | 12 | z2w,1=1 z2w,2=1 z4=2 |
| | 12 | z4=3 |
| | 12 | z2w,1=1 z2w,2=1 z4=1 |
| | 12 | z2w,1=1 z2w,2=1 z4=1 |
| | 12 | z4=3 |
| 129.4.123 | 191057 | Ga2La1Pd2 |
| | 115 | z2=1 |
| | 115 | z2=1 |
| | 12 | z4=2 |
| | 12 | z2w,1=1 z2w,2=1 |
| | 12 | z2w,1=1 z2w,2=1 |
| | 12 | z4=2 |
| | 12 | z2w,1=1 z2w,2=1 z4=3 |
| | 12 | z4=1 |
| | 12 | z4=1 |
| | 12 | z2w,1=1 z2w,2=1 z4=3 |
| | 12 | z2w,1=1 z2w,2=1 z2w,3=1 z4=3 |
| | 12 | z2w,3=1 z4=1 |
| | 12 | z2w,3=1 z4=1 |
| | 12 | z2w,1=1 z2w,2=1 z2w,3=1 z4=3 |
| | 12 | z2w,3=1 |
| | 12 | z2w,1=1 z2w,2=1 z2w,3=1 z4=2 |
| | 12 | z2w,1=1 z2w,2=1 z2w,3=1 z4=2 |
| | 12 | z2w,3=1 |
| 129.4.124 | 57297 | Fe1Si1Tb1 |
| | 115 | z2=1 |
| | 115 | z2=1 |
| | 12 | z2w,1=1 z2w,2=1 z4=2 |
| | 12 | z2w,1=1 z2w,2=1 z4=2 |
| | 12 | z2w,1=1 z2w,2=1 z4=1 |
| | 12 | z4=3 |
| | 12 | z4=3 |
| | 12 | z2w,1=1 z2w,2=1 z4=1 |
| | 12 | z2w,1=1 z2w,2=1 z2w,3=1 z4=1 |
| | 12 | z2w,3=1 z4=3 |
| | 12 | z2w,3=1 z4=3 |
| | 12 | z2w,1=1 z2w,2=1 z2w,3=1 z4=1 |
| | 12 | z2w,3=1 z4=2 |
| | 12 | z2w,1=1 z2w,2=1 z2w,3=1 |
| | 12 | z2w,1=1 z2w,2=1 z2w,3=1 |
| | 12 | z2w,3=1 z4=2 |
| 129.4.125 | 57237 | Fe1Si1Tb1 |
| | 115 | z2=1 |
| | 115 | z2=1 |
| | 12 | z2w,1=1 z2w,2=1 z4=2 |
| | 12 | z2w,1=1 z2w,2=1 z4=2 |
| | 12 | z2w,1=1 z2w,2=1 z4=1 |
| | 12 | z4=3 |
| | 12 | z4=3 |

Continued on next column

**TABLE XIX– continued**

| TQC | ICSD SSG | Stoich. Formula SIs |
|---|---|---|
| | 12 | z2w,1=1 z2w,2=1 z4=1 |
| | 12 | z2w,1=1 z2w,2=1 z2w,3=1 z4=1 |
| | 12 | z2w,3=1 z4=3 |
| | 12 | z2w,3=1 z4=3 |
| | 12 | z2w,1=1 z2w,2=1 z2w,3=1 z4=1 |
| | 12 | z2w,3=1 z4=2 |
| | 12 | z2w,1=1 z2w,2=1 z2w,3=1 |
| | 12 | z2w,1=1 z2w,2=1 z2w,3=1 |
| | 12 | z2w,3=1 z4=2 |
| 129.4.126 | 57236 | Fe1Si1Tb1 |
| | 115 | z2=1 |
| | 115 | z2=1 |
| | 12 | z2w,1=1 z2w,2=1 z4=2 |
| | 12 | z2w,1=1 z2w,2=1 z4=2 |
| | 12 | z2w,1=1 z2w,2=1 z4=1 |
| | 12 | z4=3 |
| | 12 | z4=3 |
| | 12 | z2w,1=1 z2w,2=1 z4=1 |
| | 12 | z2w,1=1 z2w,2=1 z2w,3=1 z4=1 |
| | 12 | z2w,3=1 z4=3 |
| | 12 | z2w,3=1 z4=3 |
| | 12 | z2w,1=1 z2w,2=1 z2w,3=1 z4=1 |
| | 12 | z2w,3=1 z4=2 |
| | 12 | z2w,1=1 z2w,2=1 z2w,3=1 |
| | 12 | z2w,1=1 z2w,2=1 z2w,3=1 |
| | 12 | z2w,3=1 z4=2 |
| 129.4.127 | 194319 | Fe1N1Se1 |
| | 115 | z2=1 |
| | 115 | z2=1 |
| | 12 | z4=3 |
| | 12 | z2w,1=1 z2w,2=1 z4=1 |
| | 12 | z2w,1=1 z2w,2=1 z4=1 |
| | 12 | z4=3 |
| | 12 | z2w,1=1 z2w,2=1 |
| | 12 | z4=2 |
| | 12 | z4=2 |
| | 12 | z2w,1=1 z2w,2=1 |
| | 12 | z2w,1=1 z2w,2=1 z2w,3=1 |
| | 12 | z2w,3=1 z4=2 |
| | 12 | z2w,3=1 z4=2 |
| | 12 | z2w,1=1 z2w,2=1 z2w,3=1 |
| | 12 | z2w,3=1 z4=1 |
| | 12 | z2w,1=1 z2w,2=1 z2w,3=1 z4=3 |
| | 12 | z2w,1=1 z2w,2=1 z2w,3=1 z4=3 |
| | 12 | z2w,3=1 z4=1 |
| 129.4.128 | 85853 | Fe1La1Si1 |
| | 115 | z2=1 |
| | 115 | z2=1 |
| | 12 | z2w,1=1 z2w,2=1 z4=2 |
| | 12 | z2w,1=1 z2w,2=1 z4=2 |
| | 12 | z2w,1=1 z2w,2=1 z4=1 |
| | 12 | z4=3 |
| | 12 | z4=3 |
| | 12 | z2w,1=1 z2w,2=1 z4=1 |
| | 12 | z2w,1=1 z2w,2=1 z2w,3=1 z4=1 |
| | 12 | z2w,3=1 z4=3 |
| | 12 | z2w,3=1 z4=3 |
| | 12 | z2w,1=1 z2w,2=1 z2w,3=1 z4=1 |
| | 12 | z2w,3=1 z4=2 |
| | 12 | z2w,1=1 z2w,2=1 z2w,3=1 |

Continued on next column

**TABLE XIX– continued**

| TQC | ICSD SSG | Stoich. Formula SIs |
|---|---|---|
| | 12 | z2w,1=1 z2w,2=1 z2w,3=1 |
| | 12 | z2w,3=1 z4=2 |
| 129.4.129 | 76428 | F1O1Pu1 |
| | 115 | z2=1 |
| | 115 | z2=1 |
| | 12 | z2w,3=1 z4=1 |
| | 12 | z2w,1=1 z2w,2=1 z2w,3=1 z4=3 |
| | 12 | z2w,1=1 z2w,2=1 z2w,3=1 z4=3 |
| | 12 | z2w,3=1 z4=1 |
| | 12 | z2w,1=1 z2w,2=1 z2w,3=1 z4=2 |
| | 12 | z2w,3=1 |
| | 12 | z2w,3=1 |
| | 12 | z2w,1=1 z2w,2=1 z2w,3=1 z4=2 |
| | 12 | z2w,1=1 z2w,2=1 z4=2 |
| | 12 | z2w,1=1 z2w,2=1 z4=2 |
| | 12 | z4=3 |
| | 12 | z2w,1=1 z2w,2=1 z4=1 |
| | 12 | z2w,1=1 z2w,2=1 z4=1 |
| | 12 | z4=3 |
| 129.4.130 | 108969 | F1Nd1Te1 |
| | 115 | z2=1 |
| | 115 | z2=1 |
| | 12 | z2w,3=1 z4=1 |
| | 12 | z2w,1=1 z2w,2=1 z2w,3=1 z4=3 |
| | 12 | z2w,1=1 z2w,2=1 z2w,3=1 z4=3 |
| | 12 | z2w,3=1 z4=1 |
| | 12 | z2w,1=1 z2w,2=1 z2w,3=1 z4=2 |
| | 12 | z2w,3=1 |
| | 12 | z2w,3=1 |
| | 12 | z2w,1=1 z2w,2=1 z2w,3=1 z4=2 |
| | 12 | z2w,1=1 z2w,2=1 z4=2 |
| | 12 | z2w,1=1 z2w,2=1 z4=2 |
| | 12 | z4=3 |
| | 12 | z2w,1=1 z2w,2=1 z4=1 |
| | 12 | z2w,1=1 z2w,2=1 z4=1 |
| | 12 | z4=3 |
| 129.4.131 | 94471 | F1Nd1S1 |
| | 115 | z2=1 |
| | 115 | z2=1 |
| | 12 | z2w,3=1 z4=1 |
| | 12 | z2w,1=1 z2w,2=1 z2w,3=1 z4=3 |
| | 12 | z2w,1=1 z2w,2=1 z2w,3=1 z4=3 |
| | 12 | z2w,3=1 z4=1 |
| | 12 | z2w,1=1 z2w,2=1 z2w,3=1 z4=2 |
| | 12 | z2w,3=1 |
| | 12 | z2w,3=1 |
| | 12 | z2w,1=1 z2w,2=1 z2w,3=1 z4=2 |
| | 12 | z2w,1=1 z2w,2=1 z4=2 |
| | 12 | z2w,1=1 z2w,2=1 z4=2 |
| | 12 | z4=3 |
| | 12 | z2w,1=1 z2w,2=1 z4=1 |
| | 12 | z2w,1=1 z2w,2=1 z4=1 |
| | 12 | z4=3 |
| 129.4.132 | 189941 | F1Mn1Sb1Sr1 |
| | 115 | z2=1 |
| | 115 | z2=1 |
| | 12 | z2w,3=1 z4=1 |
| | 12 | z2w,1=1 z2w,2=1 z2w,3=1 z4=3 |
| | 12 | z2w,1=1 z2w,2=1 z2w,3=1 z4=3 |
| | 12 | z2w,3=1 z4=1 |

Continued on next column

**TABLE XIX– continued**

| TQC | ICSD SSG | Stoich. Formula SIs |
|---|---|---|
| | 12 | z2w,1=1 z2w,2=1 z2w,3=1 z4=2 |
| | 12 | z2w,3=1 |
| | 12 | z2w,3=1 |
| | 12 | z2w,1=1 z2w,2=1 z2w,3=1 z4=2 |
| | 12 | z2w,1=1 z2w,2=1 z4=2 |
| | 12 | z2w,1=1 z2w,2=1 z4=2 |
| | 12 | z4=3 |
| | 12 | z2w,1=1 z2w,2=1 z4=1 |
| | 12 | z2w,1=1 z2w,2=1 z4=1 |
| | 12 | z4=3 |
| 129.4.133 | 189939 | F1Mn1P1Sr1 |
| | 115 | z2=1 |
| | 115 | z2=1 |
| | 12 | z2w,3=1 |
| | 12 | z2w,1=1 z2w,2=1 z2w,3=1 z4=2 |
| | 12 | z2w,1=1 z2w,2=1 z2w,3=1 z4=2 |
| | 12 | z2w,3=1 |
| | 12 | z2w,1=1 z2w,2=1 z2w,3=1 z4=1 |
| | 12 | z2w,3=1 z4=3 |
| | 12 | z2w,3=1 z4=3 |
| | 12 | z2w,1=1 z2w,2=1 z2w,3=1 z4=1 |
| | 12 | z2w,1=1 z2w,2=1 z4=1 |
| | 12 | z4=3 |
| | 12 | z4=3 |
| | 12 | z2w,1=1 z2w,2=1 z4=1 |
| | 12 | z4=2 |
| | 12 | z2w,1=1 z2w,2=1 |
| | 12 | z2w,1=1 z2w,2=1 |
| | 12 | z4=2 |
| 129.4.134 | 94473 | F1Gd1S1 |
| | 115 | z2=1 |
| | 115 | z2=1 |
| | 12 | z2w,3=1 z4=1 |
| | 12 | z2w,1=1 z2w,2=1 z2w,3=1 z4=3 |
| | 12 | z2w,1=1 z2w,2=1 z2w,3=1 z4=3 |
| | 12 | z2w,3=1 z4=1 |
| | 12 | z2w,1=1 z2w,2=1 z2w,3=1 z4=2 |
| | 12 | z2w,3=1 |
| | 12 | z2w,3=1 |
| | 12 | z2w,1=1 z2w,2=1 z2w,3=1 z4=2 |
| | 12 | z2w,1=1 z2w,2=1 z4=2 |
| | 12 | z2w,1=1 z2w,2=1 z4=2 |
| | 12 | z4=3 |
| | 12 | z2w,1=1 z2w,2=1 z4=1 |
| | 12 | z2w,1=1 z2w,2=1 z4=1 |
| | 12 | z4=3 |
| 129.4.135 | 93355 | F1Gd1S1 |
| | 115 | z2=1 |
| | 115 | z2=1 |
| | 12 | z2w,3=1 z4=1 |
| | 12 | z2w,1=1 z2w,2=1 z2w,3=1 z4=3 |
| | 12 | z2w,1=1 z2w,2=1 z2w,3=1 z4=3 |
| | 12 | z2w,3=1 z4=1 |
| | 12 | z2w,1=1 z2w,2=1 z2w,3=1 z4=2 |
| | 12 | z2w,3=1 |
| | 12 | z2w,3=1 |
| | 12 | z2w,1=1 z2w,2=1 z2w,3=1 z4=2 |
| | 12 | z2w,1=1 z2w,2=1 z4=2 |
| | 12 | z2w,1=1 z2w,2=1 z4=2 |
| | 12 | z4=3 |
| | 12 | z2w,1=1 z2w,2=1 z4=1 |
| | 12 | z2w,1=1 z2w,2=1 z4=1 |
| | 12 | z4=3 |
| 129.4.136 | 88460 | F1Fe1O3Sr2 |
| | 115 | z2=1 |
| | 115 | z2=1 |
| | 12 | z2w,1=1 z2w,2=1 z4=2 |
| | 12 | z2w,1=1 z2w,2=1 z4=2 |
| | 12 | z2w,1=1 z2w,2=1 z4=1 |
| | 12 | z4=3 |
| | 12 | z4=3 |
| | 12 | z2w,1=1 z2w,2=1 z4=1 |
| | 12 | z2w,1=1 z2w,2=1 z2w,3=1 z4=1 |
| | 12 | z2w,3=1 z4=3 |
| | 12 | z2w,3=1 z4=3 |
| | 12 | z2w,1=1 z2w,2=1 z2w,3=1 z4=1 |
| | 12 | z2w,3=1 z4=2 |
| | 12 | z2w,1=1 z2w,2=1 z2w,3=1 |
| | 12 | z2w,1=1 z2w,2=1 z2w,3=1 |
| | 12 | z2w,3=1 z4=2 |
| 129.4.137 | 162267 | Eu1Sn2Zn2 |
| | 115 | z2=1 |
| | 115 | z2=1 |
| | 12 | z2w,3=1 z4=1 |
| | 12 | z2w,1=1 z2w,2=1 z2w,3=1 z4=3 |
| | 12 | z2w,1=1 z2w,2=1 z2w,3=1 z4=3 |
| | 12 | z2w,3=1 z4=1 |
| | 12 | z2w,1=1 z2w,2=1 z2w,3=1 z4=2 |
| | 12 | z2w,3=1 |
| | 12 | z2w,3=1 |
| | 12 | z2w,1=1 z2w,2=1 z2w,3=1 z4=2 |
| | 12 | z2w,1=1 z2w,2=1 z4=2 |
| | 12 | z2w,1=1 z2w,2=1 z4=2 |
| | 12 | z4=3 |
| | 12 | z2w,1=1 z2w,2=1 z4=1 |
| | 12 | z2w,1=1 z2w,2=1 z4=1 |
| | 12 | z4=3 |
| 129.4.138 | 427712 | Eu1Pd2Sn2 |
| | 115 | z2=1 |
| | 115 | z2=1 |
| | 12 | z4=3 |
| | 12 | z2w,1=1 z2w,2=1 z4=1 |
| | 12 | z2w,1=1 z2w,2=1 z4=1 |
| | 12 | z4=3 |
| | 12 | z2w,1=1 z2w,2=1 |
| | 12 | z4=2 |
| | 12 | z4=2 |
| | 12 | z2w,1=1 z2w,2=1 |
| | 12 | z2w,1=1 z2w,2=1 z2w,3=1 |
| | 12 | z2w,3=1 z4=2 |
| | 12 | z2w,3=1 z4=2 |
| | 12 | z2w,1=1 z2w,2=1 z2w,3=1 |
| | 12 | z2w,3=1 z4=1 |
| | 12 | z2w,1=1 z2w,2=1 z2w,3=1 z4=3 |
| | 12 | z2w,1=1 z2w,2=1 z2w,3=1 z4=3 |
| | 12 | z2w,3=1 z4=1 |
| 129.4.139 | 51757 | Eu1Ge2Zn2 |
| | 115 | z2=1 |
| | 115 | z2=1 |
| | 12 | z4=3 |

Continued on next column

Continued on next column

**TABLE XIX– continued**

| TQC | ICSD SSG | Stoich. Formula SIs |
|---|---|---|
| | 12 | z2w,1=1 z2w,2=1 z4=1 |
| | 12 | z2w,1=1 z2w,2=1 z4=1 |
| | 12 | z4=3 |
| | 12 | z2w,1=1 z2w,2=1 |
| | 12 | z4=2 |
| | 12 | z4=2 |
| | 12 | z2w,1=1 z2w,2=1 |
| | 12 | z2w,1=1 z2w,2=1 z2w,3=1 |
| | 12 | z2w,3=1 z4=2 |
| | 12 | z2w,3=1 z4=2 |
| | 12 | z2w,1=1 z2w,2=1 z2w,3=1 |
| | 12 | z2w,3=1 z4=1 |
| | 12 | z2w,1=1 z2w,2=1 z2w,3=1 z4=3 |
| | 12 | z2w,1=1 z2w,2=1 z2w,3=1 z4=3 |
| | 12 | z2w,3=1 z4=1 |
| 129.4.140 | 94476 | Er1F1S1 |
| | 115 | z2=1 |
| | 115 | z2=1 |
| | 12 | z2w,3=1 z4=1 |
| | 12 | z2w,1=1 z2w,2=1 z2w,3=1 z4=3 |
| | 12 | z2w,1=1 z2w,2=1 z2w,3=1 z4=3 |
| | 12 | z2w,3=1 z4=1 |
| | 12 | z2w,1=1 z2w,2=1 z2w,3=1 z4=2 |
| | 12 | z2w,3=1 |
| | 12 | z2w,3=1 |
| | 12 | z2w,1=1 z2w,2=1 z2w,3=1 z4=2 |
| | 12 | z2w,1=1 z2w,2=1 z4=2 |
| | 12 | z2w,1=1 z2w,2=1 z4=2 |
| | 12 | z4=3 |
| | 12 | z2w,1=1 z2w,2=1 z4=1 |
| | 12 | z2w,1=1 z2w,2=1 z4=1 |
| | 12 | z4=3 |
| 129.4.141 | 103133 | Cu3Ti2 |
| | 115 | z2=1 |
| | 115 | z2=1 |
| | 12 | z2w,1=1 z2w,2=1 z4=2 |
| | 12 | z2w,1=1 z2w,2=1 z4=2 |
| | 12 | z2w,1=1 z2w,2=1 z4=1 |
| | 12 | z4=3 |
| | 12 | z4=3 |
| | 12 | z2w,1=1 z2w,2=1 z4=1 |
| | 12 | z2w,1=1 z2w,2=1 z2w,3=1 z4=1 |
| | 12 | z2w,3=1 z4=3 |
| | 12 | z2w,3=1 z4=3 |
| | 12 | z2w,1=1 z2w,2=1 z2w,3=1 z4=1 |
| | 12 | z2w,3=1 z4=2 |
| | 12 | z2w,1=1 z2w,2=1 z2w,3=1 |
| | 12 | z2w,1=1 z2w,2=1 z2w,3=1 |
| | 12 | z2w,3=1 z4=2 |
| 129.4.142 | 42323 | Cu2Sb1 |
| | 115 | z2=1 |
| | 115 | z2=1 |
| | 12 | z2w,1=1 z2w,2=1 z4=2 |
| | 12 | z2w,1=1 z2w,2=1 z4=2 |
| | 12 | z2w,1=1 z2w,2=1 z4=1 |
| | 12 | z4=3 |
| | 12 | z4=3 |
| | 12 | z2w,1=1 z2w,2=1 z4=1 |
| | 12 | z2w,1=1 z2w,2=1 z2w,3=1 z4=1 |
| | 12 | z2w,3=1 z4=3 |

Continued on next column

**TABLE XIX– continued**

| TQC | ICSD SSG | Stoich. Formula SIs |
|---|---|---|
| | 12 | z2w,3=1 z4=3 |
| | 12 | z2w,1=1 z2w,2=1 z2w,3=1 z4=1 |
| | 12 | z2w,3=1 z4=2 |
| | 12 | z2w,1=1 z2w,2=1 z2w,3=1 |
| | 12 | z2w,1=1 z2w,2=1 z2w,3=1 |
| | 12 | z2w,3=1 z4=2 |
| 129.4.143 | 412295 | Cu2Sb1 |
| | 115 | z2=1 |
| | 115 | z2=1 |
| | 12 | z2w,1=1 z2w,2=1 z4=2 |
| | 12 | z2w,1=1 z2w,2=1 z4=2 |
| | 12 | z2w,1=1 z2w,2=1 z4=1 |
| | 12 | z4=3 |
| | 12 | z4=3 |
| | 12 | z2w,1=1 z2w,2=1 z4=1 |
| | 12 | z2w,1=1 z2w,2=1 z2w,3=1 z4=1 |
| | 12 | z2w,3=1 z4=3 |
| | 12 | z2w,3=1 z4=3 |
| | 12 | z2w,1=1 z2w,2=1 z2w,3=1 z4=1 |
| | 12 | z2w,3=1 z4=2 |
| | 12 | z2w,1=1 z2w,2=1 z2w,3=1 |
| | 12 | z2w,1=1 z2w,2=1 z2w,3=1 |
| | 12 | z2w,3=1 z4=2 |
| 129.4.144 | 16843 | Cu2Sb1 |
| | 115 | z2=1 |
| | 115 | z2=1 |
| | 12 | z2w,1=1 z2w,2=1 z4=2 |
| | 12 | z2w,1=1 z2w,2=1 z4=2 |
| | 12 | z2w,1=1 z2w,2=1 z4=1 |
| | 12 | z4=3 |
| | 12 | z4=3 |
| | 12 | z2w,1=1 z2w,2=1 z4=1 |
| | 12 | z2w,1=1 z2w,2=1 z2w,3=1 z4=1 |
| | 12 | z2w,3=1 z4=3 |
| | 12 | z2w,3=1 z4=3 |
| | 12 | z2w,1=1 z2w,2=1 z2w,3=1 z4=1 |
| | 12 | z2w,3=1 z4=2 |
| | 12 | z2w,1=1 z2w,2=1 z2w,3=1 |
| | 12 | z2w,1=1 z2w,2=1 z2w,3=1 |
| | 12 | z2w,3=1 z4=2 |
| 129.4.145 | 103041 | Cu2La1Sn2 |
| | 115 | z2=1 |
| | 115 | z2=1 |
| | 12 | z4=3 |
| | 12 | z2w,1=1 z2w,2=1 z4=1 |
| | 12 | z2w,1=1 z2w,2=1 z4=1 |
| | 12 | z4=3 |
| | 12 | z2w,1=1 z2w,2=1 |
| | 12 | z4=2 |
| | 12 | z4=2 |
| | 12 | z2w,1=1 z2w,2=1 |
| | 12 | z2w,1=1 z2w,2=1 z2w,3=1 |
| | 12 | z2w,3=1 z4=2 |
| | 12 | z2w,3=1 z4=2 |
| | 12 | z2w,1=1 z2w,2=1 z2w,3=1 |
| | 12 | z2w,3=1 z4=1 |
| | 12 | z2w,1=1 z2w,2=1 z2w,3=1 z4=3 |
| | 12 | z2w,1=1 z2w,2=1 z2w,3=1 z4=3 |
| | 12 | z2w,3=1 z4=1 |
| 129.4.146 | 77197 | Cu2Eu1Sb2 |

Continued on next column

**TABLE XIX– continued**

| TQC | ICSD SSG | Stoich. Formula SIs |
|---|---|---|
| | 115 | z2=1 |
| | 115 | z2=1 |
| | 12 | z2w,3=1 z4=1 |
| | 12 | z2w,1=1 z2w,2=1 z2w,3=1 z4=3 |
| | 12 | z2w,1=1 z2w,2=1 z2w,3=1 z4=3 |
| | 12 | z2w,3=1 z4=1 |
| | 12 | z2w,1=1 z2w,2=1 z2w,3=1 z4=2 |
| | 12 | z2w,3=1 |
| | 12 | z2w,3=1 |
| | 12 | z2w,1=1 z2w,2=1 z2w,3=1 z4=2 |
| | 12 | z2w,1=1 z2w,2=1 z4=2 |
| | 12 | z2w,1=1 z2w,2=1 z4=2 |
| | 12 | z4=3 |
| | 12 | z2w,1=1 z2w,2=1 z4=1 |
| | 12 | z2w,1=1 z2w,2=1 z4=1 |
| | 12 | z4=3 |
| 129.4.147 | 103128 | Cu1Ti1 |
| | 115 | z2=1 |
| | 115 | z2=1 |
| | 12 | z2w,1=1 z2w,2=1 z4=2 |
| | 12 | z2w,1=1 z2w,2=1 z4=2 |
| | 12 | z2w,1=1 z2w,2=1 z4=1 |
| | 12 | z4=3 |
| | 12 | z4=3 |
| | 12 | z2w,1=1 z2w,2=1 z4=1 |
| | 12 | z2w,1=1 z2w,2=1 z2w,3=1 z4=1 |
| | 12 | z2w,3=1 z4=3 |
| | 12 | z2w,3=1 z4=3 |
| | 12 | z2w,1=1 z2w,2=1 z2w,3=1 z4=1 |
| | 12 | z2w,3=1 z4=2 |
| | 12 | z2w,1=1 z2w,2=1 z2w,3=1 |
| | 12 | z2w,1=1 z2w,2=1 z2w,3=1 |
| | 12 | z2w,3=1 z4=2 |
| 129.4.148 | 53352 | Cu1Si2Zr1 |
| | 115 | z2=1 |
| | 115 | z2=1 |
| | 12 | z4=3 |
| | 12 | z2w,1=1 z2w,2=1 z4=1 |
| | 12 | z2w,1=1 z2w,2=1 z4=1 |
| | 12 | z4=3 |
| | 12 | z2w,1=1 z2w,2=1 |
| | 12 | z4=2 |
| | 12 | z4=2 |
| | 12 | z2w,1=1 z2w,2=1 |
| | 12 | z2w,1=1 z2w,2=1 z2w,3=1 |
| | 12 | z2w,3=1 z4=2 |
| | 12 | z2w,3=1 z4=2 |
| | 12 | z2w,1=1 z2w,2=1 z2w,3=1 |
| | 12 | z2w,3=1 z4=1 |
| | 12 | z2w,1=1 z2w,2=1 z2w,3=1 z4=3 |
| | 12 | z2w,1=1 z2w,2=1 z2w,3=1 z4=3 |
| | 12 | z2w,3=1 z4=1 |
| 129.4.149 | 261278 | Cu1Np1O1P1 |
| | 115 | z2=1 |
| | 115 | z2=1 |
| | 12 | z2w,1=1 z2w,2=1 z4=2 |
| | 12 | z2w,1=1 z2w,2=1 z4=2 |
| | 12 | z2w,1=1 z2w,2=1 z4=1 |
| | 12 | z4=3 |
| | 12 | z4=3 |

Continued on next column

**TABLE XIX– continued**

| TQC | ICSD SSG | Stoich. Formula SIs |
|---|---|---|
| | 12 | z2w,1=1 z2w,2=1 z4=1 |
| | 12 | z2w,1=1 z2w,2=1 z2w,3=1 z4=1 |
| | 12 | z2w,3=1 z4=3 |
| | 12 | z2w,3=1 z4=3 |
| | 12 | z2w,1=1 z2w,2=1 z2w,3=1 z4=1 |
| | 12 | z2w,3=1 z4=2 |
| | 12 | z2w,1=1 z2w,2=1 z2w,3=1 |
| | 12 | z2w,1=1 z2w,2=1 z2w,3=1 |
| | 12 | z2w,3=1 z4=2 |
| 129.4.150 | 416523 | Cu1Nd1O1Te1 |
| | 115 | z2=1 |
| | 115 | z2=1 |
| | 12 | z2w,1=1 z2w,2=1 z4=2 |
| | 12 | z2w,1=1 z2w,2=1 z4=2 |
| | 12 | z2w,1=1 z2w,2=1 z4=1 |
| | 12 | z4=3 |
| | 12 | z4=3 |
| | 12 | z2w,1=1 z2w,2=1 z4=1 |
| | 12 | z2w,1=1 z2w,2=1 z2w,3=1 z4=1 |
| | 12 | z2w,3=1 z4=3 |
| | 12 | z2w,3=1 z4=3 |
| | 12 | z2w,1=1 z2w,2=1 z2w,3=1 z4=1 |
| | 12 | z2w,3=1 z4=2 |
| | 12 | z2w,1=1 z2w,2=1 z2w,3=1 |
| | 12 | z2w,1=1 z2w,2=1 z2w,3=1 |
| | 12 | z2w,3=1 z4=2 |
| 129.4.151 | 96346 | Cu1Nd1O1S1 |
| | 115 | z2=1 |
| | 115 | z2=1 |
| | 12 | z2w,1=1 z2w,2=1 z4=2 |
| | 12 | z2w,1=1 z2w,2=1 z4=2 |
| | 12 | z2w,1=1 z2w,2=1 z4=1 |
| | 12 | z4=3 |
| | 12 | z4=3 |
| | 12 | z2w,1=1 z2w,2=1 z4=1 |
| | 12 | z2w,1=1 z2w,2=1 z2w,3=1 z4=1 |
| | 12 | z2w,3=1 z4=3 |
| | 12 | z2w,3=1 z4=3 |
| | 12 | z2w,1=1 z2w,2=1 z2w,3=1 z4=1 |
| | 12 | z2w,3=1 z4=2 |
| | 12 | z2w,1=1 z2w,2=1 z2w,3=1 |
| | 12 | z2w,1=1 z2w,2=1 z2w,3=1 |
| | 12 | z2w,3=1 z4=2 |
| 129.4.152 | 87174 | Cu1Hf1Si2 |
| | 115 | z2=1 |
| | 115 | z2=1 |
| | 12 | z4=3 |
| | 12 | z2w,1=1 z2w,2=1 z4=1 |
| | 12 | z2w,1=1 z2w,2=1 z4=1 |
| | 12 | z4=3 |
| | 12 | z2w,1=1 z2w,2=1 |
| | 12 | z4=2 |
| | 12 | z4=2 |
| | 12 | z2w,1=1 z2w,2=1 |
| | 12 | z2w,1=1 z2w,2=1 z2w,3=1 |
| | 12 | z2w,3=1 z4=2 |
| | 12 | z2w,3=1 z4=2 |
| | 12 | z2w,1=1 z2w,2=1 z2w,3=1 |
| | 12 | z2w,3=1 z4=1 |
| | 12 | z2w,1=1 z2w,2=1 z2w,3=1 z4=3 |

Continued on next column

**TABLE XIX– continued**

| TQC | ICSD SSG | Stoich. Formula SIs |
|---|---|---|
| | 12 | z2w,1=1 z2w,2=1 z2w,3=1 z4=3 |
| | 12 | z2w,3=1 z4=1 |
| 129.4.153 | 153568 | Cu1Gd1Sb2 |
| | 115 | z2=1 |
| | 115 | z2=1 |
| | 12 | z2w,1=1 z2w,2=1 z4=2 |
| | 12 | z2w,1=1 z2w,2=1 z4=2 |
| | 12 | z2w,1=1 z2w,2=1 z4=1 |
| | 12 | z4=3 |
| | 12 | z4=3 |
| | 12 | z2w,1=1 z2w,2=1 z4=1 |
| | 12 | z2w,1=1 z2w,2=1 z2w,3=1 z4=1 |
| | 12 | z2w,3=1 z4=3 |
| | 12 | z2w,3=1 z4=3 |
| | 12 | z2w,1=1 z2w,2=1 z2w,3=1 z4=1 |
| | 12 | z2w,3=1 z4=2 |
| | 12 | z2w,1=1 z2w,2=1 z2w,3=1 |
| | 12 | z2w,1=1 z2w,2=1 z2w,3=1 |
| | 12 | z2w,3=1 z4=2 |
| 129.4.154 | 80358 | Cu1Gd1O1Se1 |
| | 115 | z2=1 |
| | 115 | z2=1 |
| | 12 | z2w,1=1 z2w,2=1 z4=2 |
| | 12 | z2w,1=1 z2w,2=1 z4=2 |
| | 12 | z2w,1=1 z2w,2=1 z4=1 |
| | 12 | z4=3 |
| | 12 | z4=3 |
| | 12 | z2w,1=1 z2w,2=1 z4=1 |
| | 12 | z2w,1=1 z2w,2=1 z2w,3=1 z4=1 |
| | 12 | z2w,3=1 z4=3 |
| | 12 | z2w,3=1 z4=3 |
| | 12 | z2w,1=1 z2w,2=1 z2w,3=1 z4=1 |
| | 12 | z2w,3=1 z4=2 |
| | 12 | z2w,1=1 z2w,2=1 z2w,3=1 |
| | 12 | z2w,1=1 z2w,2=1 z2w,3=1 |
| | 12 | z2w,3=1 z4=2 |
| 129.4.155 | 84963 | Cu1Fe1O3S1Sr2 |
| | 115 | z2=1 |
| | 115 | z2=1 |
| | 12 | z2w,3=1 z4=1 |
| | 12 | z2w,1=1 z2w,2=1 z2w,3=1 z4=3 |
| | 12 | z2w,1=1 z2w,2=1 z2w,3=1 z4=3 |
| | 12 | z2w,3=1 z4=1 |
| | 12 | z2w,1=1 z2w,2=1 z2w,3=1 z4=2 |
| | 12 | z2w,3=1 |
| | 12 | z2w,3=1 |
| | 12 | z2w,1=1 z2w,2=1 z2w,3=1 z4=2 |
| | 12 | z2w,1=1 z2w,2=1 z4=2 |
| | 12 | z2w,1=1 z2w,2=1 z4=2 |
| | 12 | z4=3 |
| | 12 | z2w,1=1 z2w,2=1 z4=1 |
| | 12 | z2w,1=1 z2w,2=1 z4=1 |
| | 12 | z4=3 |
| 129.4.156 | 195375 | Cu1Fe1O3S1Sr2 |
| | 115 | z2=1 |
| | 115 | z2=1 |
| | 12 | z2w,3=1 z4=1 |
| | 12 | z2w,1=1 z2w,2=1 z2w,3=1 z4=3 |
| | 12 | z2w,1=1 z2w,2=1 z2w,3=1 z4=3 |
| | 12 | z2w,3=1 z4=1 |
| | 12 | z2w,1=1 z2w,2=1 z2w,3=1 z4=2 |
| | 12 | z2w,3=1 |
| | 12 | z2w,3=1 |
| | 12 | z2w,1=1 z2w,2=1 z2w,3=1 z4=2 |
| | 12 | z2w,1=1 z2w,2=1 z4=2 |
| | 12 | z2w,1=1 z2w,2=1 z4=2 |
| | 12 | z4=3 |
| | 12 | z2w,1=1 z2w,2=1 z4=1 |
| | 12 | z2w,1=1 z2w,2=1 z4=1 |
| | 12 | z4=3 |
| 129.4.157 | 419381 | Cu1Eu1F1Te1 |
| | 115 | z2=1 |
| | 115 | z2=1 |
| | 12 | z2w,3=1 z4=2 |
| | 12 | z2w,1=1 z2w,2=1 z2w,3=1 |
| | 12 | z2w,1=1 z2w,2=1 z2w,3=1 |
| | 12 | z2w,3=1 z4=2 |
| | 12 | z2w,1=1 z2w,2=1 z2w,3=1 z4=3 |
| | 12 | z2w,3=1 z4=1 |
| | 12 | z2w,3=1 z4=1 |
| | 12 | z2w,1=1 z2w,2=1 z2w,3=1 z4=3 |
| | 12 | z2w,1=1 z2w,2=1 z4=3 |
| | 12 | z4=1 |
| | 12 | z4=1 |
| | 12 | z2w,1=1 z2w,2=1 z4=3 |
| | 12 | z2w,1=1 z2w,2=1 z4=2 |
| | 12 | z2w,1=1 z2w,2=1 z4=2 |
| 129.4.158 | 172799 | Cu1Eu1F1Se1 |
| | 115 | z2=1 |
| | 115 | z2=1 |
| | 12 | z4=1 |
| | 12 | z2w,1=1 z2w,2=1 z4=3 |
| | 12 | z2w,1=1 z2w,2=1 z4=3 |
| | 12 | z4=1 |
| | 12 | z2w,1=1 z2w,2=1 z4=2 |
| | 12 | z2w,1=1 z2w,2=1 z4=2 |
| | 12 | z2w,1=1 z2w,2=1 z2w,3=1 z4=2 |
| | 12 | z2w,3=1 |
| | 12 | z2w,3=1 |
| | 12 | z2w,1=1 z2w,2=1 z2w,3=1 z4=2 |
| | 12 | z2w,3=1 z4=3 |
| | 12 | z2w,1=1 z2w,2=1 z2w,3=1 z4=1 |
| | 12 | z2w,1=1 z2w,2=1 z2w,3=1 z4=1 |
| | 12 | z2w,3=1 z4=3 |
| 129.4.159 | 172797 | Cu1Eu1F1S1 |
| | 115 | z2=1 |
| | 115 | z2=1 |
| | 12 | z4=1 |
| | 12 | z2w,1=1 z2w,2=1 z4=3 |
| | 12 | z2w,1=1 z2w,2=1 z4=3 |
| | 12 | z4=1 |
| | 12 | z2w,1=1 z2w,2=1 z4=2 |
| | 12 | z2w,1=1 z2w,2=1 z4=2 |
| | 12 | z2w,1=1 z2w,2=1 z2w,3=1 z4=2 |
| | 12 | z2w,3=1 |
| | 12 | z2w,3=1 |
| | 12 | z2w,1=1 z2w,2=1 z2w,3=1 z4=2 |
| | 12 | z2w,3=1 z4=3 |
| | 12 | z2w,1=1 z2w,2=1 z2w,3=1 z4=1 |
| | 12 | z2w,1=1 z2w,2=1 z2w,3=1 z4=1 |

Continued on next column

**TABLE XIX– continued**

| TQC | ICSD SSG | Stoich. Formula SIs |
|---|---|---|
| | 12 | z2w,3=1 z4=3 |
| 129.4.160 | 54727 | Cu1Dy1Sb2 |
| | 115 | z2=1 |
| | 115 | z2=1 |
| | 12 | z2w,1=1 z2w,2=1 z4=2 |
| | 12 | z2w,1=1 z2w,2=1 z4=2 |
| | 12 | z2w,1=1 z2w,2=1 z4=1 |
| | 12 | z4=3 |
| | 12 | z4=3 |
| | 12 | z2w,1=1 z2w,2=1 z4=1 |
| | 12 | z2w,1=1 z2w,2=1 z2w,3=1 z4=1 |
| | 12 | z2w,3=1 z4=3 |
| | 12 | z2w,3=1 z4=3 |
| | 12 | z2w,1=1 z2w,2=1 z2w,3=1 z4=1 |
| | 12 | z2w,3=1 z4=2 |
| | 12 | z2w,1=1 z2w,2=1 z2w,3=1 |
| | 12 | z2w,1=1 z2w,2=1 z2w,3=1 |
| | 12 | z2w,3=1 z4=2 |
| 129.4.161 | 80359 | Cu1Dy1O1Se1 |
| | 115 | z2=1 |
| | 115 | z2=1 |
| | 12 | z2w,3=1 |
| | 12 | z2w,1=1 z2w,2=1 z2w,3=1 z4=2 |
| | 12 | z2w,1=1 z2w,2=1 z2w,3=1 z4=2 |
| | 12 | z2w,3=1 |
| | 12 | z2w,1=1 z2w,2=1 z2w,3=1 z4=1 |
| | 12 | z2w,3=1 z4=3 |
| | 12 | z2w,3=1 z4=3 |
| | 12 | z2w,1=1 z2w,2=1 z2w,3=1 z4=1 |
| | 12 | z2w,1=1 z2w,2=1 z4=1 |
| | 12 | z4=3 |
| | 12 | z4=3 |
| | 12 | z2w,1=1 z2w,2=1 z4=1 |
| | 12 | z4=2 |
| | 12 | z2w,1=1 z2w,2=1 |
| | 12 | z2w,1=1 z2w,2=1 |
| | 12 | z4=2 |
| 129.4.162 | 41869 | Cs1Mn1Sb1 |
| | 115 | z2=1 |
| | 115 | z2=1 |
| | 12 | z2w,3=1 z4=2 |
| | 12 | z2w,1=1 z2w,2=1 z2w,3=1 |
| | 12 | z2w,1=1 z2w,2=1 z2w,3=1 |
| | 12 | z2w,3=1 z4=2 |
| | 12 | z2w,1=1 z2w,2=1 z2w,3=1 z4=3 |
| | 12 | z2w,3=1 z4=1 |
| | 12 | z2w,3=1 z4=1 |
| | 12 | z2w,1=1 z2w,2=1 z2w,3=1 z4=3 |
| | 12 | z2w,1=1 z2w,2=1 z4=3 |
| | 12 | z4=1 |
| | 12 | z4=1 |
| | 12 | z2w,1=1 z2w,2=1 z4=3 |
| | 12 | z2w,1=1 z2w,2=1 z4=2 |
| | 12 | z2w,1=1 z2w,2=1 z4=2 |
| 129.4.163 | 89599 | Cs1Mn1P1 |
| | 115 | z2=1 |
| | 115 | z2=1 |
| | 12 | z2w,3=1 z4=3 |
| | 12 | z2w,1=1 z2w,2=1 z2w,3=1 z4=1 |
| | 12 | z2w,1=1 z2w,2=1 z2w,3=1 z4=1 |

Continued on next column

**TABLE XIX– continued**

| TQC | ICSD SSG | Stoich. Formula SIs |
|---|---|---|
| | 12 | z2w,3=1 z4=3 |
| | 12 | z2w,1=1 z2w,2=1 z2w,3=1 |
| | 12 | z2w,3=1 z4=2 |
| | 12 | z2w,3=1 z4=2 |
| | 12 | z2w,1=1 z2w,2=1 z2w,3=1 |
| | 12 | z2w,1=1 z2w,2=1 |
| | 12 | z4=2 |
| | 12 | z4=2 |
| | 12 | z2w,1=1 z2w,2=1 |
| | 12 | z4=1 |
| | 12 | z2w,1=1 z2w,2=1 z4=3 |
| | 12 | z2w,1=1 z2w,2=1 z4=3 |
| | 12 | z4=1 |
| 129.4.164 | 89598 | Cs1Mn1P1 |
| | 115 | z2=1 |
| | 115 | z2=1 |
| | 12 | z2w,3=1 z4=3 |
| | 12 | z2w,1=1 z2w,2=1 z2w,3=1 z4=1 |
| | 12 | z2w,1=1 z2w,2=1 z2w,3=1 z4=1 |
| | 12 | z2w,3=1 z4=3 |
| | 12 | z2w,1=1 z2w,2=1 z2w,3=1 |
| | 12 | z2w,3=1 z4=2 |
| | 12 | z2w,3=1 z4=2 |
| | 12 | z2w,1=1 z2w,2=1 z2w,3=1 |
| | 12 | z2w,1=1 z2w,2=1 |
| | 12 | z4=2 |
| | 12 | z4=2 |
| | 12 | z2w,1=1 z2w,2=1 |
| | 12 | z4=1 |
| | 12 | z2w,1=1 z2w,2=1 z4=3 |
| | 12 | z2w,1=1 z2w,2=1 z4=3 |
| | 12 | z4=1 |
| 129.4.165 | 84962 | Cr1Cu1O3S1Sr2 |
| | 115 | z2=1 |
| | 115 | z2=1 |
| | 12 | z2w,3=1 z4=1 |
| | 12 | z2w,1=1 z2w,2=1 z2w,3=1 z4=3 |
| | 12 | z2w,1=1 z2w,2=1 z2w,3=1 z4=3 |
| | 12 | z2w,3=1 z4=1 |
| | 12 | z2w,1=1 z2w,2=1 z2w,3=1 z4=2 |
| | 12 | z2w,3=1 |
| | 12 | z2w,3=1 |
| | 12 | z2w,1=1 z2w,2=1 z2w,3=1 z4=2 |
| | 12 | z2w,1=1 z2w,2=1 z4=2 |
| | 12 | z2w,1=1 z2w,2=1 z4=2 |
| | 12 | z4=3 |
| | 12 | z2w,1=1 z2w,2=1 z4=1 |
| | 12 | z2w,1=1 z2w,2=1 z4=1 |
| | 12 | z4=3 |
| 129.4.166 | 80514 | Co1Nd1Si1 |
| | 115 | z2=1 |
| | 115 | z2=1 |
| | 12 | z2w,3=1 z4=3 |
| | 12 | z2w,1=1 z2w,2=1 z2w,3=1 z4=1 |
| | 12 | z2w,1=1 z2w,2=1 z2w,3=1 z4=1 |
| | 12 | z2w,3=1 z4=3 |
| | 12 | z2w,1=1 z2w,2=1 z2w,3=1 |
| | 12 | z2w,3=1 z4=2 |
| | 12 | z2w,3=1 z4=2 |
| | 12 | z2w,1=1 z2w,2=1 z2w,3=1 |

Continued on next column

TABLE XIX– continued

| TQC | ICSD SSG | Stoich. Formula SIs |
|---|---|---|
| | 12 | z2w,1=1 z2w,2=1 |
| | 12 | z4=2 |
| | 12 | z4=2 |
| | 12 | z2w,1=1 z2w,2=1 |
| | 12 | z4=1 |
| | 12 | z2w,1=1 z2w,2=1 z4=3 |
| | 12 | z2w,1=1 z2w,2=1 z4=3 |
| | 12 | z4=1 |
| 129.4.167 | 80203 | Co1La1O1P1 |
| | 115 | z2=1 |
| | 115 | z2=1 |
| | 12 | z4=1 |
| | 12 | z2w,1=1 z2w,2=1 z4=3 |
| | 12 | z2w,1=1 z2w,2=1 z4=3 |
| | 12 | z4=1 |
| | 12 | z2w,1=1 z2w,2=1 z4=2 |
| | 12 | z2w,1=1 z2w,2=1 z4=2 |
| | 12 | z2w,1=1 z2w,2=1 z2w,3=1 z4=2 |
| | 12 | z2w,3=1 |
| | 12 | z2w,3=1 |
| | 12 | z2w,1=1 z2w,2=1 z2w,3=1 z4=2 |
| | 12 | z2w,3=1 z4=3 |
| | 12 | z2w,1=1 z2w,2=1 z2w,3=1 z4=1 |
| | 12 | z2w,1=1 z2w,2=1 z2w,3=1 z4=1 |
| | 12 | z2w,3=1 z4=3 |
| 129.4.168 | 161911 | Co1La1O1P1 |
| | 115 | z2=1 |
| | 115 | z2=1 |
| | 12 | z4=1 |
| | 12 | z2w,1=1 z2w,2=1 z4=3 |
| | 12 | z2w,1=1 z2w,2=1 z4=3 |
| | 12 | z4=1 |
| | 12 | z2w,1=1 z2w,2=1 z4=2 |
| | 12 | z2w,1=1 z2w,2=1 z4=2 |
| | 12 | z2w,1=1 z2w,2=1 z2w,3=1 z4=2 |
| | 12 | z2w,3=1 |
| | 12 | z2w,3=1 |
| | 12 | z2w,1=1 z2w,2=1 z2w,3=1 z4=2 |
| | 12 | z2w,3=1 z4=3 |
| | 12 | z2w,1=1 z2w,2=1 z2w,3=1 z4=1 |
| | 12 | z2w,1=1 z2w,2=1 z2w,3=1 z4=1 |
| | 12 | z2w,3=1 z4=3 |
| 129.4.169 | 85861 | Co1Ge1Nd1 |
| | 115 | z2=1 |
| | 115 | z2=1 |
| | 12 | z2w,3=1 z4=3 |
| | 12 | z2w,1=1 z2w,2=1 z2w,3=1 z4=1 |
| | 12 | z2w,1=1 z2w,2=1 z2w,3=1 z4=1 |
| | 12 | z2w,3=1 z4=3 |
| | 12 | z2w,1=1 z2w,2=1 z2w,3=1 |
| | 12 | z2w,3=1 z4=2 |
| | 12 | z2w,3=1 z4=2 |
| | 12 | z2w,1=1 z2w,2=1 z2w,3=1 |
| | 12 | z2w,1=1 z2w,2=1 |
| | 12 | z4=2 |
| | 12 | z4=2 |
| | 12 | z2w,1=1 z2w,2=1 |
| | 12 | z4=1 |
| | 12 | z2w,1=1 z2w,2=1 z4=3 |
| | 12 | z2w,1=1 z2w,2=1 z4=3 |

Continued on next column

TABLE XIX– continued

| TQC | ICSD SSG | Stoich. Formula SIs |
|---|---|---|
| | 12 | z4=1 |
| 129.4.170 | 106473 | Co1Ge1Nd1 |
| | 115 | z2=1 |
| | 115 | z2=1 |
| | 12 | z2w,3=1 z4=3 |
| | 12 | z2w,1=1 z2w,2=1 z2w,3=1 z4=1 |
| | 12 | z2w,1=1 z2w,2=1 z2w,3=1 z4=1 |
| | 12 | z2w,3=1 z4=3 |
| | 12 | z2w,1=1 z2w,2=1 z2w,3=1 |
| | 12 | z2w,3=1 z4=2 |
| | 12 | z2w,3=1 z4=2 |
| | 12 | z2w,1=1 z2w,2=1 z2w,3=1 |
| | 12 | z2w,1=1 z2w,2=1 |
| | 12 | z4=2 |
| | 12 | z4=2 |
| | 12 | z2w,1=1 z2w,2=1 |
| | 12 | z4=1 |
| | 12 | z2w,1=1 z2w,2=1 z4=3 |
| | 12 | z2w,1=1 z2w,2=1 z4=3 |
| | 12 | z4=1 |
| 129.4.171 | 418853 | Co1Ge1Mg1 |
| | 115 | z2=1 |
| | 115 | z2=1 |
| | 12 | z2w,3=1 z4=2 |
| | 12 | z2w,1=1 z2w,2=1 z2w,3=1 |
| | 12 | z2w,1=1 z2w,2=1 z2w,3=1 |
| | 12 | z2w,3=1 z4=2 |
| | 12 | z2w,1=1 z2w,2=1 z2w,3=1 z4=3 |
| | 12 | z2w,3=1 z4=1 |
| | 12 | z2w,3=1 z4=1 |
| | 12 | z2w,1=1 z2w,2=1 z2w,3=1 z4=3 |
| | 12 | z2w,1=1 z2w,2=1 z4=3 |
| | 12 | z4=1 |
| | 12 | z4=1 |
| | 12 | z2w,1=1 z2w,2=1 z4=3 |
| | 12 | z2w,1=1 z2w,2=1 z4=2 |
| | 12 | z2w,1=1 z2w,2=1 z4=2 |
| 129.4.172 | 31661 | Cl1O1Pu1 |
| | 115 | z2=1 |
| | 115 | z2=1 |
| | 12 | z2w,1=1 z2w,2=1 z4=2 |
| | 12 | z2w,1=1 z2w,2=1 z4=2 |
| | 12 | z2w,1=1 z2w,2=1 z4=1 |
| | 12 | z4=3 |
| | 12 | z4=3 |
| | 12 | z2w,1=1 z2w,2=1 z4=1 |
| | 12 | z2w,1=1 z2w,2=1 z2w,3=1 z4=1 |
| | 12 | z2w,3=1 z4=3 |
| | 12 | z2w,3=1 z4=3 |
| | 12 | z2w,1=1 z2w,2=1 z2w,3=1 z4=1 |
| | 12 | z2w,3=1 z4=2 |
| | 12 | z2w,1=1 z2w,2=1 z2w,3=1 |
| | 12 | z2w,1=1 z2w,2=1 z2w,3=1 |
| | 12 | z2w,3=1 z4=2 |
| 129.4.173 | 237581 | Cl1Ni1O3Sr2 |
| | 115 | z2=1 |
| | 115 | z2=1 |
| | 12 | z2w,1=1 z2w,2=1 z4=2 |
| | 12 | z2w,1=1 z2w,2=1 z4=2 |
| | 12 | z2w,1=1 z2w,2=1 z4=1 |

Continued on next column

TABLE XIX– continued

| TQC | ICSD SSG | Stoich. Formula SIs |
|---|---|---|
| | 12 | z4=3 |
| | 12 | z4=3 |
| | 12 | z2w,1=1 z2w,2=1 z4=1 |
| | 12 | z2w,1=1 z2w,2=1 z2w,3=1 z4=1 |
| | 12 | z2w,3=1 z4=3 |
| | 12 | z2w,3=1 z4=3 |
| | 12 | z2w,1=1 z2w,2=1 z2w,3=1 z4=1 |
| | 12 | z2w,3=1 z4=2 |
| | 12 | z2w,1=1 z2w,2=1 z2w,3=1 |
| | 12 | z2w,1=1 z2w,2=1 z2w,3=1 |
| | 12 | z2w,3=1 z4=2 |
| 129.4.174 | 426777 | Cl1Nd1Te1 |
| | 115 | z2=1 |
| | 115 | z2=1 |
| | 12 | z2w,3=1 z4=1 |
| | 12 | z2w,1=1 z2w,2=1 z2w,3=1 z4=3 |
| | 12 | z2w,1=1 z2w,2=1 z2w,3=1 z4=3 |
| | 12 | z2w,3=1 z4=1 |
| | 12 | z2w,1=1 z2w,2=1 z2w,3=1 z4=2 |
| | 12 | z2w,3=1 |
| | 12 | z2w,3=1 |
| | 12 | z2w,1=1 z2w,2=1 z2w,3=1 z4=2 |
| | 12 | z2w,1=1 z2w,2=1 z4=2 |
| | 12 | z2w,1=1 z2w,2=1 z4=2 |
| | 12 | z4=3 |
| | 12 | z2w,1=1 z2w,2=1 z4=1 |
| | 12 | z2w,1=1 z2w,2=1 z4=1 |
| | 12 | z4=3 |
| 129.4.175 | 59231 | Cl1Nd1O1 |
| | 115 | z2=1 |
| | 115 | z2=1 |
| | 12 | z2w,1=1 z2w,2=1 z4=2 |
| | 12 | z2w,1=1 z2w,2=1 z4=2 |
| | 12 | z2w,1=1 z2w,2=1 z4=1 |
| | 12 | z4=3 |
| | 12 | z4=3 |
| | 12 | z2w,1=1 z2w,2=1 z4=1 |
| | 12 | z2w,1=1 z2w,2=1 z2w,3=1 z4=1 |
| | 12 | z2w,3=1 z4=3 |
| | 12 | z2w,3=1 z4=3 |
| | 12 | z2w,1=1 z2w,2=1 z2w,3=1 z4=1 |
| | 12 | z2w,3=1 z4=2 |
| | 12 | z2w,1=1 z2w,2=1 z2w,3=1 |
| | 12 | z2w,1=1 z2w,2=1 z2w,3=1 |
| | 12 | z2w,3=1 z4=2 |
| 129.4.176 | 31665 | Cl1Nd1O1 |
| | 115 | z2=1 |
| | 115 | z2=1 |
| | 12 | z2w,1=1 z2w,2=1 z4=2 |
| | 12 | z2w,1=1 z2w,2=1 z4=2 |
| | 12 | z2w,1=1 z2w,2=1 z4=1 |
| | 12 | z4=3 |
| | 12 | z4=3 |
| | 12 | z2w,1=1 z2w,2=1 z4=1 |
| | 12 | z2w,1=1 z2w,2=1 z2w,3=1 z4=1 |
| | 12 | z2w,3=1 z4=3 |
| | 12 | z2w,3=1 z4=3 |
| | 12 | z2w,1=1 z2w,2=1 z2w,3=1 z4=1 |
| | 12 | z2w,3=1 z4=2 |
| | 12 | z2w,1=1 z2w,2=1 z2w,3=1 |

Continued on next column

TABLE XIX– continued

| TQC | ICSD SSG | Stoich. Formula SIs |
|---|---|---|
| | 12 | z2w,1=1 z2w,2=1 z2w,3=1 |
| | 12 | z2w,3=1 z4=2 |
| 129.4.177 | 50989 | Cl1Mo1O5P1Rb1 |
| | 115 | z2=1 |
| | 115 | z2=1 |
| | 12 | z2w,1=1 z2w,2=1 z4=2 |
| | 12 | z2w,1=1 z2w,2=1 z4=2 |
| | 12 | z2w,1=1 z2w,2=1 z4=1 |
| | 12 | z4=3 |
| | 12 | z4=3 |
| | 12 | z2w,1=1 z2w,2=1 z4=1 |
| | 12 | z2w,1=1 z2w,2=1 z2w,3=1 z4=1 |
| | 12 | z2w,3=1 z4=3 |
| | 12 | z2w,3=1 z4=3 |
| | 12 | z2w,1=1 z2w,2=1 z2w,3=1 z4=1 |
| | 12 | z2w,3=1 z4=2 |
| | 12 | z2w,1=1 z2w,2=1 z2w,3=1 |
| | 12 | z2w,1=1 z2w,2=1 z2w,3=1 |
| | 12 | z2w,3=1 z4=2 |
| 129.4.178 | 50988 | Cl1K1Mo1O5P1 |
| | 115 | z2=1 |
| | 115 | z2=1 |
| | 12 | z2w,1=1 z2w,2=1 z4=2 |
| | 12 | z2w,1=1 z2w,2=1 z4=2 |
| | 12 | z2w,1=1 z2w,2=1 z4=1 |
| | 12 | z4=3 |
| | 12 | z4=3 |
| | 12 | z2w,1=1 z2w,2=1 z4=1 |
| | 12 | z2w,1=1 z2w,2=1 z2w,3=1 z4=1 |
| | 12 | z2w,3=1 z4=3 |
| | 12 | z2w,3=1 z4=3 |
| | 12 | z2w,1=1 z2w,2=1 z2w,3=1 z4=1 |
| | 12 | z2w,3=1 z4=2 |
| | 12 | z2w,1=1 z2w,2=1 z2w,3=1 |
| | 12 | z2w,1=1 z2w,2=1 z2w,3=1 |
| | 12 | z2w,3=1 z4=2 |
| 129.4.179 | 77820 | Cl1Gd1O1 |
| | 115 | z2=1 |
| | 115 | z2=1 |
| | 12 | z2w,1=1 z2w,2=1 z4=2 |
| | 12 | z2w,1=1 z2w,2=1 z4=2 |
| | 12 | z2w,1=1 z2w,2=1 z4=1 |
| | 12 | z4=3 |
| | 12 | z4=3 |
| | 12 | z2w,1=1 z2w,2=1 z4=1 |
| | 12 | z2w,1=1 z2w,2=1 z2w,3=1 z4=1 |
| | 12 | z2w,3=1 z4=3 |
| | 12 | z2w,3=1 z4=3 |
| | 12 | z2w,1=1 z2w,2=1 z2w,3=1 z4=1 |
| | 12 | z2w,3=1 z4=2 |
| | 12 | z2w,1=1 z2w,2=1 z2w,3=1 |
| | 12 | z2w,1=1 z2w,2=1 z2w,3=1 |
| | 12 | z2w,3=1 z4=2 |
| 129.4.180 | 59232 | Cl1Gd1O1 |
| | 115 | z2=1 |
| | 115 | z2=1 |
| | 12 | z2w,1=1 z2w,2=1 z4=2 |
| | 12 | z2w,1=1 z2w,2=1 z4=2 |
| | 12 | z2w,1=1 z2w,2=1 z4=1 |
| | 12 | z4=3 |

Continued on next column

**TABLE XIX– continued**

| TQC | ICSD SSG | Stoich. Formula SIs |
|---|---|---|
| | 12 | z4=3 |
| | 12 | z2w,1=1 z2w,2=1 z4=1 |
| | 12 | z2w,1=1 z2w,2=1 z2w,3=1 z4=1 |
| | 12 | z2w,3=1 z4=3 |
| | 12 | z2w,3=1 z4=3 |
| | 12 | z2w,1=1 z2w,2=1 z2w,3=1 z4=1 |
| | 12 | z2w,3=1 z4=2 |
| | 12 | z2w,1=1 z2w,2=1 z2w,3=1 |
| | 12 | z2w,1=1 z2w,2=1 z2w,3=1 |
| | 12 | z2w,3=1 z4=2 |
| 129.4.181 | 80798 | Cl1Fe1O4W1 |
| | 115 | z2=1 |
| | 115 | z2=1 |
| | 12 | z2w,1=1 z2w,2=1 z4=2 |
| | 12 | z2w,1=1 z2w,2=1 z4=2 |
| | 12 | z2w,1=1 z2w,2=1 z4=1 |
| | 12 | z4=3 |
| | 12 | z4=3 |
| | 12 | z2w,1=1 z2w,2=1 z4=1 |
| | 12 | z2w,1=1 z2w,2=1 z2w,3=1 z4=1 |
| | 12 | z2w,3=1 z4=3 |
| | 12 | z2w,3=1 z4=3 |
| | 12 | z2w,1=1 z2w,2=1 z2w,3=1 z4=1 |
| | 12 | z2w,3=1 z4=2 |
| | 12 | z2w,1=1 z2w,2=1 z2w,3=1 |
| | 12 | z2w,1=1 z2w,2=1 z2w,3=1 |
| | 12 | z2w,3=1 z4=2 |
| 129.4.182 | 93508 | Cl1Fe1O3Sr2 |
| | 115 | z2=1 |
| | 115 | z2=1 |
| | 12 | z2w,1=1 z2w,2=1 z4=2 |
| | 12 | z2w,1=1 z2w,2=1 z4=2 |
| | 12 | z2w,1=1 z2w,2=1 z4=1 |
| | 12 | z4=3 |
| | 12 | z4=3 |
| | 12 | z2w,1=1 z2w,2=1 z4=1 |
| | 12 | z2w,1=1 z2w,2=1 z2w,3=1 z4=1 |
| | 12 | z2w,3=1 z4=3 |
| | 12 | z2w,3=1 z4=3 |
| | 12 | z2w,1=1 z2w,2=1 z2w,3=1 z4=1 |
| | 12 | z2w,3=1 z4=2 |
| | 12 | z2w,1=1 z2w,2=1 z2w,3=1 |
| | 12 | z2w,1=1 z2w,2=1 z2w,3=1 |
| | 12 | z2w,3=1 z4=2 |
| 129.4.183 | 49516 | Cl1Fe1Mo1O4 |
| | 115 | z2=1 |
| | 115 | z2=1 |
| | 12 | z2w,1=1 z2w,2=1 z4=2 |
| | 12 | z2w,1=1 z2w,2=1 z4=2 |
| | 12 | z2w,1=1 z2w,2=1 z4=1 |
| | 12 | z4=3 |
| | 12 | z4=3 |
| | 12 | z2w,1=1 z2w,2=1 z4=1 |
| | 12 | z2w,1=1 z2w,2=1 z2w,3=1 z4=1 |
| | 12 | z2w,3=1 z4=3 |
| | 12 | z2w,3=1 z4=3 |
| | 12 | z2w,1=1 z2w,2=1 z2w,3=1 z4=1 |
| | 12 | z2w,3=1 z4=2 |
| | 12 | z2w,1=1 z2w,2=1 z2w,3=1 |
| | 12 | z2w,1=1 z2w,2=1 z2w,3=1 |

Continued on next column

**TABLE XIX– continued**

| TQC | ICSD SSG | Stoich. Formula SIs |
|---|---|---|
| | 12 | z2w,3=1 z4=2 |
| 129.4.184 | 202188 | Cl1Fe1Mo1O4 |
| | 115 | z2=1 |
| | 115 | z2=1 |
| | 12 | z2w,1=1 z2w,2=1 z4=2 |
| | 12 | z2w,1=1 z2w,2=1 z4=2 |
| | 12 | z2w,1=1 z2w,2=1 z4=1 |
| | 12 | z4=3 |
| | 12 | z4=3 |
| | 12 | z2w,1=1 z2w,2=1 z4=1 |
| | 12 | z2w,1=1 z2w,2=1 z2w,3=1 z4=1 |
| | 12 | z2w,3=1 z4=3 |
| | 12 | z2w,3=1 z4=3 |
| | 12 | z2w,1=1 z2w,2=1 z2w,3=1 z4=1 |
| | 12 | z2w,3=1 z4=2 |
| | 12 | z2w,1=1 z2w,2=1 z2w,3=1 |
| | 12 | z2w,1=1 z2w,2=1 z2w,3=1 |
| | 12 | z2w,3=1 z4=2 |
| 129.4.185 | 202187 | Cl1Fe1Mo1O4 |
| | 115 | z2=1 |
| | 115 | z2=1 |
| | 12 | z2w,1=1 z2w,2=1 z4=2 |
| | 12 | z2w,1=1 z2w,2=1 z4=2 |
| | 12 | z2w,1=1 z2w,2=1 z4=1 |
| | 12 | z4=3 |
| | 12 | z4=3 |
| | 12 | z2w,1=1 z2w,2=1 z4=1 |
| | 12 | z2w,1=1 z2w,2=1 z2w,3=1 z4=1 |
| | 12 | z2w,3=1 z4=3 |
| | 12 | z2w,3=1 z4=3 |
| | 12 | z2w,1=1 z2w,2=1 z2w,3=1 z4=1 |
| | 12 | z2w,3=1 z4=2 |
| | 12 | z2w,1=1 z2w,2=1 z2w,3=1 |
| | 12 | z2w,1=1 z2w,2=1 z2w,3=1 |
| | 12 | z2w,3=1 z4=2 |
| 129.4.186 | 202186 | Cl1Fe1Mo1O4 |
| | 115 | z2=1 |
| | 115 | z2=1 |
| | 12 | z2w,1=1 z2w,2=1 z4=2 |
| | 12 | z2w,1=1 z2w,2=1 z4=2 |
| | 12 | z2w,1=1 z2w,2=1 z4=1 |
| | 12 | z4=3 |
| | 12 | z4=3 |
| | 12 | z2w,1=1 z2w,2=1 z4=1 |
| | 12 | z2w,1=1 z2w,2=1 z2w,3=1 z4=1 |
| | 12 | z2w,3=1 z4=3 |
| | 12 | z2w,3=1 z4=3 |
| | 12 | z2w,1=1 z2w,2=1 z2w,3=1 z4=1 |
| | 12 | z2w,3=1 z4=2 |
| | 12 | z2w,1=1 z2w,2=1 z2w,3=1 |
| | 12 | z2w,1=1 z2w,2=1 z2w,3=1 |
| | 12 | z2w,3=1 z4=2 |
| 129.4.187 | 108938 | Cl1Eu1F1 |
| | 115 | z2=1 |
| | 115 | z2=1 |
| | 12 | z4=1 |
| | 12 | z2w,1=1 z2w,2=1 z4=3 |
| | 12 | z2w,1=1 z2w,2=1 z4=3 |
| | 12 | z4=1 |
| | 12 | z2w,1=1 z2w,2=1 z4=2 |

Continued on next column

**TABLE XIX– continued**

| TQC | ICSD SSG | Stoich. Formula SIs |
|---|---|---|
| | 12 | z2w,1=1 z2w,2=1 z4=2 |
| | 12 | z2w,1=1 z2w,2=1 z2w,3=1 z4=2 |
| | 12 | z2w,3=1 |
| | 12 | z2w,3=1 |
| | 12 | z2w,1=1 z2w,2=1 z2w,3=1 z4=2 |
| | 12 | z2w,3=1 z4=3 |
| | 12 | z2w,1=1 z2w,2=1 z2w,3=1 z4=1 |
| | 12 | z2w,1=1 z2w,2=1 z2w,3=1 z4=1 |
| | 12 | z2w,3=1 z4=3 |
| 129.4.188 | 85449 | Ce1O1Sb1Zn1 |
| | 115 | z2=1 |
| | 115 | z2=1 |
| | 12 | z2w,3=1 z4=2 |
| | 12 | z2w,1=1 z2w,2=1 z2w,3=1 |
| | 12 | z2w,1=1 z2w,2=1 z2w,3=1 |
| | 12 | z2w,3=1 z4=2 |
| | 12 | z2w,1=1 z2w,2=1 z2w,3=1 z4=3 |
| | 12 | z2w,3=1 z4=1 |
| | 12 | z2w,3=1 z4=1 |
| | 12 | z2w,1=1 z2w,2=1 z2w,3=1 z4=3 |
| | 12 | z2w,1=1 z2w,2=1 z4=3 |
| | 12 | z4=1 |
| | 12 | z4=1 |
| | 12 | z2w,1=1 z2w,2=1 z4=3 |
| | 12 | z2w,1=1 z2w,2=1 z4=2 |
| | 12 | z2w,1=1 z2w,2=1 z4=2 |
| 129.4.189 | 80201 | Ce1O1P1Ru1 |
| | 115 | z2=1 |
| | 115 | z2=1 |
| | 12 | z2w,1=1 z2w,2=1 z4=2 |
| | 12 | z2w,1=1 z2w,2=1 z4=2 |
| | 12 | z2w,1=1 z2w,2=1 z4=1 |
| | 12 | z4=3 |
| | 12 | z4=3 |
| | 12 | z2w,1=1 z2w,2=1 z4=1 |
| | 12 | z2w,1=1 z2w,2=1 z2w,3=1 z4=1 |
| | 12 | z2w,3=1 z4=3 |
| | 12 | z2w,3=1 z4=3 |
| | 12 | z2w,1=1 z2w,2=1 z2w,3=1 z4=1 |
| | 12 | z2w,3=1 z4=2 |
| | 12 | z2w,1=1 z2w,2=1 z2w,3=1 |
| | 12 | z2w,1=1 z2w,2=1 z2w,3=1 |
| | 12 | z2w,3=1 z4=2 |
| 129.4.190 | 85850 | Ce1Mn1Si1 |
| | 115 | z2=1 |
| | 115 | z2=1 |
| | 12 | z2w,3=1 |
| | 12 | z2w,1=1 z2w,2=1 z2w,3=1 z4=2 |
| | 12 | z2w,1=1 z2w,2=1 z2w,3=1 z4=2 |
| | 12 | z2w,3=1 |
| | 12 | z2w,1=1 z2w,2=1 z2w,3=1 z4=1 |
| | 12 | z2w,3=1 z4=3 |
| | 12 | z2w,3=1 z4=3 |
| | 12 | z2w,1=1 z2w,2=1 z2w,3=1 z4=1 |
| | 12 | z2w,1=1 z2w,2=1 z4=1 |
| | 12 | z4=3 |
| | 12 | z4=3 |
| | 12 | z2w,1=1 z2w,2=1 z4=1 |
| | 12 | z4=2 |
| | 12 | z2w,1=1 z2w,2=1 |

Continued on next column

**TABLE XIX– continued**

| TQC | ICSD SSG | Stoich. Formula SIs |
|---|---|---|
| | 12 | z2w,1=1 z2w,2=1 |
| | 12 | z4=2 |
| 129.4.191 | 75051 | Ce1Mn1Si1 |
| | 115 | z2=1 |
| | 115 | z2=1 |
| | 12 | z2w,3=1 z4=3 |
| | 12 | z2w,1=1 z2w,2=1 z2w,3=1 z4=1 |
| | 12 | z2w,1=1 z2w,2=1 z2w,3=1 z4=1 |
| | 12 | z2w,3=1 z4=3 |
| | 12 | z2w,1=1 z2w,2=1 z2w,3=1 |
| | 12 | z2w,3=1 z4=2 |
| | 12 | z2w,3=1 z4=2 |
| | 12 | z2w,1=1 z2w,2=1 z2w,3=1 |
| | 12 | z2w,1=1 z2w,2=1 |
| | 12 | z4=2 |
| | 12 | z4=2 |
| | 12 | z2w,1=1 z2w,2=1 |
| | 12 | z4=1 |
| | 12 | z2w,1=1 z2w,2=1 z4=3 |
| | 12 | z2w,1=1 z2w,2=1 z4=3 |
| | 12 | z4=1 |
| 129.4.192 | 75050 | Ce1Mn1Si1 |
| | 115 | z2=1 |
| | 115 | z2=1 |
| | 12 | z2w,3=1 z4=3 |
| | 12 | z2w,1=1 z2w,2=1 z2w,3=1 z4=1 |
| | 12 | z2w,1=1 z2w,2=1 z2w,3=1 z4=1 |
| | 12 | z2w,3=1 z4=3 |
| | 12 | z2w,1=1 z2w,2=1 z2w,3=1 |
| | 12 | z2w,3=1 z4=2 |
| | 12 | z2w,3=1 z4=2 |
| | 12 | z2w,1=1 z2w,2=1 z2w,3=1 |
| | 12 | z2w,1=1 z2w,2=1 |
| | 12 | z4=2 |
| | 12 | z4=2 |
| | 12 | z2w,1=1 z2w,2=1 |
| | 12 | z4=1 |
| | 12 | z2w,1=1 z2w,2=1 z4=3 |
| | 12 | z2w,1=1 z2w,2=1 z4=3 |
| | 12 | z4=1 |
| 129.4.193 | 75049 | Ce1Mn1Si1 |
| | 115 | z2=1 |
| | 115 | z2=1 |
| | 12 | z2w,3=1 z4=3 |
| | 12 | z2w,1=1 z2w,2=1 z2w,3=1 z4=1 |
| | 12 | z2w,1=1 z2w,2=1 z2w,3=1 z4=1 |
| | 12 | z2w,3=1 z4=3 |
| | 12 | z2w,1=1 z2w,2=1 z2w,3=1 |
| | 12 | z2w,3=1 z4=2 |
| | 12 | z2w,3=1 z4=2 |
| | 12 | z2w,1=1 z2w,2=1 z2w,3=1 |
| | 12 | z2w,1=1 z2w,2=1 |
| | 12 | z4=2 |
| | 12 | z4=2 |
| | 12 | z2w,1=1 z2w,2=1 |
| | 12 | z4=1 |
| | 12 | z2w,1=1 z2w,2=1 z4=3 |
| | 12 | z2w,1=1 z2w,2=1 z4=3 |
| | 12 | z4=1 |
| 129.4.194 | 75048 | Ce1Mn1Si1 |

Continued on next column

**TABLE XIX– continued**

| TQC | ICSD SSG | Stoich. Formula SIs |
|---|---|---|
| | 115 | z2=1 |
| | 115 | z2=1 |
| | 12 | z2w,3=1 z4=3 |
| | 12 | z2w,1=1 z2w,2=1 z2w,3=1 z4=1 |
| | 12 | z2w,1=1 z2w,2=1 z2w,3=1 z4=1 |
| | 12 | z2w,3=1 z4=3 |
| | 12 | z2w,1=1 z2w,2=1 z2w,3=1 |
| | 12 | z2w,3=1 z4=2 |
| | 12 | z2w,3=1 z4=2 |
| | 12 | z2w,1=1 z2w,2=1 z2w,3=1 |
| | 12 | z2w,1=1 z2w,2=1 |
| | 12 | z4=2 |
| | 12 | z4=2 |
| | 12 | z2w,1=1 z2w,2=1 |
| | 12 | z4=1 |
| | 12 | z2w,1=1 z2w,2=1 z4=3 |
| | 12 | z2w,1=1 z2w,2=1 z4=3 |
| | 12 | z4=1 |
| 129.4.195 | 85858 | Ce1Ge1Mn1 |
| | 115 | z2=1 |
| | 115 | z2=1 |
| | 12 | z2w,3=1 |
| | 12 | z2w,1=1 z2w,2=1 z2w,3=1 z4=2 |
| | 12 | z2w,1=1 z2w,2=1 z2w,3=1 z4=2 |
| | 12 | z2w,3=1 |
| | 12 | z2w,1=1 z2w,2=1 z2w,3=1 z4=1 |
| | 12 | z2w,3=1 z4=3 |
| | 12 | z2w,3=1 z4=3 |
| | 12 | z2w,1=1 z2w,2=1 z2w,3=1 z4=1 |
| | 12 | z2w,1=1 z2w,2=1 z4=1 |
| | 12 | z4=3 |
| | 12 | z4=3 |
| | 12 | z2w,1=1 z2w,2=1 z4=1 |
| | 12 | z4=2 |
| | 12 | z2w,1=1 z2w,2=1 |
| | 12 | z2w,1=1 z2w,2=1 |
| | 12 | z4=2 |
| 129.4.196 | 80010 | Ce1Ge1Mn1 |
| | 115 | z2=1 |
| | 115 | z2=1 |
| | 12 | z2w,3=1 |
| | 12 | z2w,1=1 z2w,2=1 z2w,3=1 z4=2 |
| | 12 | z2w,1=1 z2w,2=1 z2w,3=1 z4=2 |
| | 12 | z2w,3=1 |
| | 12 | z2w,1=1 z2w,2=1 z2w,3=1 z4=1 |
| | 12 | z2w,3=1 z4=3 |
| | 12 | z2w,3=1 z4=3 |
| | 12 | z2w,1=1 z2w,2=1 z2w,3=1 z4=1 |
| | 12 | z2w,1=1 z2w,2=1 z4=1 |
| | 12 | z4=3 |
| | 12 | z4=3 |
| | 12 | z2w,1=1 z2w,2=1 z4=1 |
| | 12 | z4=2 |
| | 12 | z2w,1=1 z2w,2=1 |
| | 12 | z2w,1=1 z2w,2=1 |
| | 12 | z4=2 |
| 129.4.197 | 80009 | Ce1Ge1Mn1 |
| | 115 | z2=1 |
| | 115 | z2=1 |
| | 12 | z2w,3=1 |

Continued on next column

**TABLE XIX– continued**

| TQC | ICSD SSG | Stoich. Formula SIs |
|---|---|---|
| | 12 | z2w,1=1 z2w,2=1 z2w,3=1 z4=2 |
| | 12 | z2w,1=1 z2w,2=1 z2w,3=1 z4=2 |
| | 12 | z2w,3=1 |
| | 12 | z2w,1=1 z2w,2=1 z2w,3=1 z4=1 |
| | 12 | z2w,3=1 z4=3 |
| | 12 | z2w,3=1 z4=3 |
| | 12 | z2w,1=1 z2w,2=1 z2w,3=1 z4=1 |
| | 12 | z2w,1=1 z2w,2=1 z4=1 |
| | 12 | z4=3 |
| | 12 | z4=3 |
| | 12 | z2w,1=1 z2w,2=1 z4=1 |
| | 12 | z4=2 |
| | 12 | z2w,1=1 z2w,2=1 |
| | 12 | z2w,1=1 z2w,2=1 |
| | 12 | z4=2 |
| 129.4.198 | 94469 | Ce1F1S1 |
| | 115 | z2=1 |
| | 115 | z2=1 |
| | 12 | z4=1 |
| | 12 | z2w,1=1 z2w,2=1 z4=3 |
| | 12 | z2w,1=1 z2w,2=1 z4=3 |
| | 12 | z4=1 |
| | 12 | z2w,1=1 z2w,2=1 z4=2 |
| | 12 | z2w,1=1 z2w,2=1 z4=2 |
| | 12 | z2w,1=1 z2w,2=1 z2w,3=1 z4=2 |
| | 12 | z2w,3=1 |
| | 12 | z2w,3=1 |
| | 12 | z2w,1=1 z2w,2=1 z2w,3=1 z4=2 |
| | 12 | z2w,3=1 z4=3 |
| | 12 | z2w,1=1 z2w,2=1 z2w,3=1 z4=1 |
| | 12 | z2w,1=1 z2w,2=1 z2w,3=1 z4=1 |
| | 12 | z2w,3=1 z4=3 |
| 129.4.199 | 93477 | Ce1F1S1 |
| | 115 | z2=1 |
| | 115 | z2=1 |
| | 12 | z4=1 |
| | 12 | z2w,1=1 z2w,2=1 z4=3 |
| | 12 | z2w,1=1 z2w,2=1 z4=3 |
| | 12 | z4=1 |
| | 12 | z2w,1=1 z2w,2=1 z4=2 |
| | 12 | z2w,1=1 z2w,2=1 z4=2 |
| | 12 | z2w,1=1 z2w,2=1 z2w,3=1 z4=2 |
| | 12 | z2w,3=1 |
| | 12 | z2w,3=1 |
| | 12 | z2w,1=1 z2w,2=1 z2w,3=1 z4=2 |
| | 12 | z2w,3=1 z4=3 |
| | 12 | z2w,1=1 z2w,2=1 z2w,3=1 z4=1 |
| | 12 | z2w,1=1 z2w,2=1 z2w,3=1 z4=1 |
| | 12 | z2w,3=1 z4=3 |
| 129.4.200 | 31939 | Ce1F1S1 |
| | 115 | z2=1 |
| | 115 | z2=1 |
| | 12 | z2w,1=1 z2w,2=1 z4=2 |
| | 12 | z2w,1=1 z2w,2=1 z4=2 |
| | 12 | z2w,1=1 z2w,2=1 z4=1 |
| | 12 | z4=3 |
| | 12 | z4=3 |
| | 12 | z2w,1=1 z2w,2=1 z4=1 |
| | 12 | z2w,1=1 z2w,2=1 z2w,3=1 z4=1 |
| | 12 | z2w,3=1 z4=3 |

Continued on next column

**TABLE XIX– continued**

| TQC | ICSD<br>SSG | Stoich. Formula<br>SIs |
|---|---|---|
| | 12 | z2w,3=1 z4=3 |
| | 12 | z2w,1=1 z2w,2=1 z2w,3=1 z4=1 |
| | 12 | z2w,3=1 z4=2 |
| | 12 | z2w,1=1 z2w,2=1 z2w,3=1 |
| | 12 | z2w,1=1 z2w,2=1 z2w,3=1 |
| | 12 | z2w,3=1 z4=2 |
| 129.4.201 | 106383 | Ce1Cu1Sb2 |
| | 115 | z2=1 |
| | 115 | z2=1 |
| | 12 | z2w,3=1 z4=1 |
| | 12 | z2w,1=1 z2w,2=1 z2w,3=1 z4=3 |
| | 12 | z2w,1=1 z2w,2=1 z2w,3=1 z4=3 |
| | 12 | z2w,3=1 z4=1 |
| | 12 | z2w,1=1 z2w,2=1 z2w,3=1 z4=2 |
| | 12 | z2w,3=1 |
| | 12 | z2w,3=1 |
| | 12 | z2w,1=1 z2w,2=1 z2w,3=1 z4=2 |
| | 12 | z2w,1=1 z2w,2=1 z4=2 |
| | 12 | z2w,1=1 z2w,2=1 z4=2 |
| | 12 | z4=3 |
| | 12 | z2w,1=1 z2w,2=1 z4=1 |
| | 12 | z2w,1=1 z2w,2=1 z4=1 |
| | 12 | z4=3 |
| 129.4.202 | 416521 | Ce1Cu1O1Te1 |
| | 115 | z2=1 |
| | 115 | z2=1 |
| | 12 | z4=1 |
| | 12 | z2w,1=1 z2w,2=1 z4=3 |
| | 12 | z2w,1=1 z2w,2=1 z4=3 |
| | 12 | z4=1 |
| | 12 | z2w,1=1 z2w,2=1 z4=2 |
| | 12 | z2w,1=1 z2w,2=1 z4=2 |
| | 12 | z2w,1=1 z2w,2=1 z2w,3=1 z4=2 |
| | 12 | z2w,3=1 |
| | 12 | z2w,3=1 |
| | 12 | z2w,1=1 z2w,2=1 z2w,3=1 z4=2 |
| | 12 | z2w,3=1 z4=3 |
| | 12 | z2w,1=1 z2w,2=1 z2w,3=1 z4=1 |
| | 12 | z2w,1=1 z2w,2=1 z2w,3=1 z4=1 |
| | 12 | z2w,3=1 z4=3 |
| 129.4.203 | 96344 | Ce1Cu1O1S1 |
| | 115 | z2=1 |
| | 115 | z2=1 |
| | 12 | z2w,3=1 z4=2 |
| | 12 | z2w,1=1 z2w,2=1 z2w,3=1 |
| | 12 | z2w,1=1 z2w,2=1 z2w,3=1 |
| | 12 | z2w,3=1 z4=2 |
| | 12 | z2w,1=1 z2w,2=1 z2w,3=1 z4=3 |
| | 12 | z2w,3=1 z4=1 |
| | 12 | z2w,3=1 z4=1 |
| | 12 | z2w,1=1 z2w,2=1 z2w,3=1 z4=3 |
| | 12 | z2w,1=1 z2w,2=1 z4=3 |
| | 12 | z4=1 |
| | 12 | z4=1 |
| | 12 | z2w,1=1 z2w,2=1 z4=3 |
| | 12 | z2w,1=1 z2w,2=1 z4=2 |
| | 12 | z2w,1=1 z2w,2=1 z4=2 |
| 129.4.204 | 106373 | Ce1Co1Ge1 |
| | 115 | z2=1 |
| | 115 | z2=1 |

Continued on next column

**TABLE XIX– continued**

| TQC | ICSD<br>SSG | Stoich. Formula<br>SIs |
|---|---|---|
| | 12 | z2w,3=1 z4=3 |
| | 12 | z2w,1=1 z2w,2=1 z2w,3=1 z4=1 |
| | 12 | z2w,1=1 z2w,2=1 z2w,3=1 z4=1 |
| | 12 | z2w,3=1 z4=3 |
| | 12 | z2w,1=1 z2w,2=1 z2w,3=1 |
| | 12 | z2w,3=1 z4=2 |
| | 12 | z2w,3=1 z4=2 |
| | 12 | z2w,1=1 z2w,2=1 z2w,3=1 |
| | 12 | z2w,1=1 z2w,2=1 |
| | 12 | z4=2 |
| | 12 | z4=2 |
| | 12 | z2w,1=1 z2w,2=1 |
| | 12 | z4=1 |
| | 12 | z2w,1=1 z2w,2=1 z4=3 |
| | 12 | z2w,1=1 z2w,2=1 z4=3 |
| | 12 | z4=1 |
| 129.4.205 | 426282 | Ce1Cl1Te1 |
| | 115 | z2=1 |
| | 115 | z2=1 |
| | 12 | z4=1 |
| | 12 | z2w,1=1 z2w,2=1 z4=3 |
| | 12 | z2w,1=1 z2w,2=1 z4=3 |
| | 12 | z4=1 |
| | 12 | z2w,1=1 z2w,2=1 z4=2 |
| | 12 | z2w,1=1 z2w,2=1 z4=2 |
| | 12 | z2w,1=1 z2w,2=1 z2w,3=1 z4=2 |
| | 12 | z2w,3=1 |
| | 12 | z2w,3=1 |
| | 12 | z2w,1=1 z2w,2=1 z2w,3=1 z4=2 |
| | 12 | z2w,3=1 z4=3 |
| | 12 | z2w,1=1 z2w,2=1 z2w,3=1 z4=1 |
| | 12 | z2w,1=1 z2w,2=1 z2w,3=1 z4=1 |
| | 12 | z2w,3=1 z4=3 |
| 129.4.206 | 72154 | Ce1Cl1O1 |
| | 115 | z2=1 |
| | 115 | z2=1 |
| | 12 | z4=1 |
| | 12 | z2w,1=1 z2w,2=1 z4=3 |
| | 12 | z2w,1=1 z2w,2=1 z4=3 |
| | 12 | z4=1 |
| | 12 | z2w,1=1 z2w,2=1 z4=2 |
| | 12 | z2w,1=1 z2w,2=1 z4=2 |
| | 12 | z2w,1=1 z2w,2=1 z2w,3=1 z4=2 |
| | 12 | z2w,3=1 |
| | 12 | z2w,3=1 |
| | 12 | z2w,1=1 z2w,2=1 z2w,3=1 z4=2 |
| | 12 | z2w,3=1 z4=3 |
| | 12 | z2w,1=1 z2w,2=1 z2w,3=1 z4=1 |
| | 12 | z2w,1=1 z2w,2=1 z2w,3=1 z4=1 |
| | 12 | z2w,3=1 z4=3 |
| 129.4.207 | 412069 | Ce1Cl1O1 |
| | 115 | z2=1 |
| | 115 | z2=1 |
| | 12 | z4=1 |
| | 12 | z2w,1=1 z2w,2=1 z4=3 |
| | 12 | z2w,1=1 z2w,2=1 z4=3 |
| | 12 | z4=1 |
| | 12 | z2w,1=1 z2w,2=1 z4=2 |
| | 12 | z2w,1=1 z2w,2=1 z4=2 |
| | 12 | z2w,1=1 z2w,2=1 z2w,3=1 z4=2 |

Continued on next column

**TABLE XIX– continued**

| TQC | ICSD SSG | Stoich. Formula SIs |
|---|---|---|
| | 12 | z2w,3=1 |
| | 12 | z2w,3=1 |
| | 12 | z2w,1=1 z2w,2=1 z2w,3=1 z4=2 |
| | 12 | z2w,3=1 z4=3 |
| | 12 | z2w,1=1 z2w,2=1 z2w,3=1 z4=1 |
| | 12 | z2w,1=1 z2w,2=1 z2w,3=1 z4=1 |
| | 12 | z2w,3=1 z4=3 |
| 129.4.208 | 169993 | Ca2Cu1Fe1O3Se1 |
| | 115 | z2=1 |
| | 115 | z2=1 |
| | 12 | z2w,3=1 z4=1 |
| | 12 | z2w,1=1 z2w,2=1 z2w,3=1 z4=3 |
| | 12 | z2w,1=1 z2w,2=1 z2w,3=1 z4=3 |
| | 12 | z2w,3=1 z4=1 |
| | 12 | z2w,1=1 z2w,2=1 z2w,3=1 z4=2 |
| | 12 | z2w,3=1 |
| | 12 | z2w,3=1 |
| | 12 | z2w,1=1 z2w,2=1 z2w,3=1 z4=2 |
| | 12 | z2w,1=1 z2w,2=1 z4=2 |
| | 12 | z2w,1=1 z2w,2=1 z4=2 |
| | 12 | z4=3 |
| | 12 | z2w,1=1 z2w,2=1 z4=1 |
| | 12 | z2w,1=1 z2w,2=1 z4=1 |
| | 12 | z4=3 |
| 129.4.209 | 195376 | Ca2Cu1Fe1O3S1 |
| | 115 | z2=1 |
| | 115 | z2=1 |
| | 12 | z2w,3=1 z4=1 |
| | 12 | z2w,1=1 z2w,2=1 z2w,3=1 z4=3 |
| | 12 | z2w,1=1 z2w,2=1 z2w,3=1 z4=3 |
| | 12 | z2w,3=1 z4=1 |
| | 12 | z2w,1=1 z2w,2=1 z2w,3=1 z4=2 |
| | 12 | z2w,3=1 |
| | 12 | z2w,3=1 |
| | 12 | z2w,1=1 z2w,2=1 z2w,3=1 z4=2 |
| | 12 | z2w,1=1 z2w,2=1 z4=2 |
| | 12 | z2w,1=1 z2w,2=1 z4=2 |
| | 12 | z4=3 |
| | 12 | z2w,1=1 z2w,2=1 z4=1 |
| | 12 | z2w,1=1 z2w,2=1 z4=1 |
| | 12 | z4=3 |
| 129.4.210 | 169992 | Ca2Cu1Fe1O3S1 |
| | 115 | z2=1 |
| | 115 | z2=1 |
| | 12 | z2w,3=1 z4=1 |
| | 12 | z2w,1=1 z2w,2=1 z2w,3=1 z4=3 |
| | 12 | z2w,1=1 z2w,2=1 z2w,3=1 z4=3 |
| | 12 | z2w,3=1 z4=1 |
| | 12 | z2w,1=1 z2w,2=1 z2w,3=1 z4=2 |
| | 12 | z2w,3=1 |
| | 12 | z2w,3=1 |
| | 12 | z2w,1=1 z2w,2=1 z2w,3=1 z4=2 |
| | 12 | z2w,1=1 z2w,2=1 z4=2 |
| | 12 | z2w,1=1 z2w,2=1 z4=2 |
| | 12 | z4=3 |
| | 12 | z2w,1=1 z2w,2=1 z4=1 |
| | 12 | z2w,1=1 z2w,2=1 z4=1 |
| | 12 | z4=3 |
| 129.4.211 | 66959 | Ca1Mn1Sn1 |
| | 115 | z2=1 |

Continued on next column

**TABLE XIX– continued**

| TQC | ICSD SSG | Stoich. Formula SIs |
|---|---|---|
| | 115 | z2=1 |
| | 12 | z4=3 |
| | 12 | z2w,1=1 z2w,2=1 z4=1 |
| | 12 | z2w,1=1 z2w,2=1 z4=1 |
| | 12 | z4=3 |
| | 12 | z2w,1=1 z2w,2=1 |
| | 12 | z4=2 |
| | 12 | z4=2 |
| | 12 | z2w,1=1 z2w,2=1 |
| | 12 | z2w,1=1 z2w,2=1 z2w,3=1 |
| | 12 | z2w,3=1 z4=2 |
| | 12 | z2w,3=1 z4=2 |
| | 12 | z2w,1=1 z2w,2=1 z2w,3=1 |
| | 12 | z2w,3=1 z4=1 |
| | 12 | z2w,1=1 z2w,2=1 z2w,3=1 z4=3 |
| | 12 | z2w,1=1 z2w,2=1 z2w,3=1 z4=3 |
| | 12 | z2w,3=1 z4=1 |
| 129.4.212 | 66958 | Ca1Mn1Sn1 |
| | 115 | z2=1 |
| | 115 | z2=1 |
| | 12 | z4=3 |
| | 12 | z2w,1=1 z2w,2=1 z4=1 |
| | 12 | z2w,1=1 z2w,2=1 z4=1 |
| | 12 | z4=3 |
| | 12 | z2w,1=1 z2w,2=1 |
| | 12 | z4=2 |
| | 12 | z4=2 |
| | 12 | z2w,1=1 z2w,2=1 |
| | 12 | z2w,1=1 z2w,2=1 z2w,3=1 |
| | 12 | z2w,3=1 z4=2 |
| | 12 | z2w,3=1 z4=2 |
| | 12 | z2w,1=1 z2w,2=1 z2w,3=1 |
| | 12 | z2w,3=1 z4=1 |
| | 12 | z2w,1=1 z2w,2=1 z2w,3=1 z4=3 |
| | 12 | z2w,1=1 z2w,2=1 z2w,3=1 z4=3 |
| | 12 | z2w,3=1 z4=1 |
| 129.4.213 | 66950 | Ca1Mn1Sn1 |
| | 115 | z2=1 |
| | 115 | z2=1 |
| | 12 | z2w,3=1 |
| | 12 | z2w,1=1 z2w,2=1 z2w,3=1 z4=2 |
| | 12 | z2w,1=1 z2w,2=1 z2w,3=1 z4=2 |
| | 12 | z2w,3=1 |
| | 12 | z2w,1=1 z2w,2=1 z2w,3=1 z4=1 |
| | 12 | z2w,3=1 z4=3 |
| | 12 | z2w,3=1 z4=3 |
| | 12 | z2w,1=1 z2w,2=1 z2w,3=1 z4=1 |
| | 12 | z2w,1=1 z2w,2=1 z4=1 |
| | 12 | z4=3 |
| | 12 | z4=3 |
| | 12 | z2w,1=1 z2w,2=1 z4=1 |
| | 12 | z4=2 |
| | 12 | z2w,1=1 z2w,2=1 |
| | 12 | z2w,1=1 z2w,2=1 |
| | 12 | z4=2 |
| 129.4.214 | 66955 | Ca1Mn1Si1 |
| | 115 | z2=1 |
| | 115 | z2=1 |
| | 12 | z4=3 |
| | 12 | z2w,1=1 z2w,2=1 z4=1 |

Continued on next column

**TABLE XIX– continued**

| TQC | ICSD<br>SSG | Stoich. Formula<br>SIs |
|---|---|---|
| | 12 | z2w,1=1 z2w,2=1 z4=1 |
| | 12 | z4=3 |
| | 12 | z2w,1=1 z2w,2=1 |
| | 12 | z4=2 |
| | 12 | z4=2 |
| | 12 | z2w,1=1 z2w,2=1 |
| | 12 | z2w,1=1 z2w,2=1 z2w,3=1 |
| | 12 | z2w,3=1 z4=2 |
| | 12 | z2w,3=1 z4=2 |
| | 12 | z2w,1=1 z2w,2=1 z2w,3=1 |
| | 12 | z2w,3=1 z4=1 |
| | 12 | z2w,1=1 z2w,2=1 z2w,3=1 z4=3 |
| | 12 | z2w,1=1 z2w,2=1 z2w,3=1 z4=3 |
| | 12 | z2w,3=1 z4=1 |
| 129.4.215 | 66954 | Ca1Mn1Si1 |
| | 115 | z2=1 |
| | 115 | z2=1 |
| | 12 | z4=3 |
| | 12 | z2w,1=1 z2w,2=1 z4=1 |
| | 12 | z2w,1=1 z2w,2=1 z4=1 |
| | 12 | z4=3 |
| | 12 | z2w,1=1 z2w,2=1 |
| | 12 | z4=2 |
| | 12 | z4=2 |
| | 12 | z2w,1=1 z2w,2=1 |
| | 12 | z2w,1=1 z2w,2=1 z2w,3=1 |
| | 12 | z2w,3=1 z4=2 |
| | 12 | z2w,3=1 z4=2 |
| | 12 | z2w,1=1 z2w,2=1 z2w,3=1 |
| | 12 | z2w,3=1 z4=1 |
| | 12 | z2w,1=1 z2w,2=1 z2w,3=1 z4=3 |
| | 12 | z2w,1=1 z2w,2=1 z2w,3=1 z4=3 |
| | 12 | z2w,3=1 z4=1 |
| 129.4.216 | 66949 | Ca1Mn1Si1 |
| | 115 | z2=1 |
| | 115 | z2=1 |
| | 12 | z2w,3=1 |
| | 12 | z2w,1=1 z2w,2=1 z2w,3=1 z4=2 |
| | 12 | z2w,1=1 z2w,2=1 z2w,3=1 z4=2 |
| | 12 | z2w,3=1 |
| | 12 | z2w,1=1 z2w,2=1 z2w,3=1 z4=1 |
| | 12 | z2w,3=1 z4=3 |
| | 12 | z2w,3=1 z4=3 |
| | 12 | z2w,1=1 z2w,2=1 z2w,3=1 z4=1 |
| | 12 | z2w,1=1 z2w,2=1 z4=1 |
| | 12 | z4=3 |
| | 12 | z4=3 |
| | 12 | z2w,1=1 z2w,2=1 z4=1 |
| | 12 | z4=2 |
| | 12 | z2w,1=1 z2w,2=1 |
| | 12 | z2w,1=1 z2w,2=1 |
| | 12 | z4=2 |
| 129.4.217 | 66957 | Ca1Ge1Mn1 |
| | 115 | z2=1 |
| | 115 | z2=1 |
| | 12 | z4=3 |
| | 12 | z2w,1=1 z2w,2=1 z4=1 |
| | 12 | z2w,1=1 z2w,2=1 z4=1 |
| | 12 | z4=3 |
| | 12 | z2w,1=1 z2w,2=1 |

Continued on next column

**TABLE XIX– continued**

| TQC | ICSD<br>SSG | Stoich. Formula<br>SIs |
|---|---|---|
| | 12 | z4=2 |
| | 12 | z4=2 |
| | 12 | z2w,1=1 z2w,2=1 |
| | 12 | z2w,1=1 z2w,2=1 z2w,3=1 |
| | 12 | z2w,3=1 z4=2 |
| | 12 | z2w,3=1 z4=2 |
| | 12 | z2w,1=1 z2w,2=1 z2w,3=1 |
| | 12 | z2w,3=1 z4=1 |
| | 12 | z2w,1=1 z2w,2=1 z2w,3=1 z4=3 |
| | 12 | z2w,1=1 z2w,2=1 z2w,3=1 z4=3 |
| | 12 | z2w,3=1 z4=1 |
| 129.4.218 | 66956 | Ca1Ge1Mn1 |
| | 115 | z2=1 |
| | 115 | z2=1 |
| | 12 | z4=3 |
| | 12 | z2w,1=1 z2w,2=1 z4=1 |
| | 12 | z2w,1=1 z2w,2=1 z4=1 |
| | 12 | z4=3 |
| | 12 | z2w,1=1 z2w,2=1 |
| | 12 | z4=2 |
| | 12 | z4=2 |
| | 12 | z2w,1=1 z2w,2=1 |
| | 12 | z2w,1=1 z2w,2=1 z2w,3=1 |
| | 12 | z2w,3=1 z4=2 |
| | 12 | z2w,3=1 z4=2 |
| | 12 | z2w,1=1 z2w,2=1 z2w,3=1 |
| | 12 | z2w,3=1 z4=1 |
| | 12 | z2w,1=1 z2w,2=1 z2w,3=1 z4=3 |
| | 12 | z2w,1=1 z2w,2=1 z2w,3=1 z4=3 |
| | 12 | z2w,3=1 z4=1 |
| 129.4.219 | 52758 | Ca1Ge1Mn1 |
| | 115 | z2=1 |
| | 115 | z2=1 |
| | 12 | z4=3 |
| | 12 | z2w,1=1 z2w,2=1 z4=1 |
| | 12 | z2w,1=1 z2w,2=1 z4=1 |
| | 12 | z4=3 |
| | 12 | z2w,1=1 z2w,2=1 |
| | 12 | z4=2 |
| | 12 | z4=2 |
| | 12 | z2w,1=1 z2w,2=1 |
| | 12 | z2w,1=1 z2w,2=1 z2w,3=1 |
| | 12 | z2w,3=1 z4=2 |
| | 12 | z2w,3=1 z4=2 |
| | 12 | z2w,1=1 z2w,2=1 z2w,3=1 |
| | 12 | z2w,3=1 z4=1 |
| | 12 | z2w,1=1 z2w,2=1 z2w,3=1 z4=3 |
| | 12 | z2w,1=1 z2w,2=1 z2w,3=1 z4=3 |
| | 12 | z2w,3=1 z4=1 |
| 129.4.220 | 428148 | Ca1Co1Si1 |
| | 115 | z2=1 |
| | 115 | z2=1 |
| | 12 | z2w,3=1 z4=2 |
| | 12 | z2w,1=1 z2w,2=1 z2w,3=1 |
| | 12 | z2w,1=1 z2w,2=1 z2w,3=1 |
| | 12 | z2w,3=1 z4=2 |
| | 12 | z2w,1=1 z2w,2=1 z2w,3=1 z4=3 |
| | 12 | z2w,3=1 z4=1 |
| | 12 | z2w,3=1 z4=1 |
| | 12 | z2w,1=1 z2w,2=1 z2w,3=1 z4=3 |

Continued on next column

**TABLE XIX– continued**

| TQC | ICSD SSG | Stoich. Formula SIs |
|---|---|---|
| | 12 | z2w,1=1 z2w,2=1 z4=3 |
| | 12 | z4=1 |
| | 12 | z4=1 |
| | 12 | z2w,1=1 z2w,2=1 z4=3 |
| | 12 | z2w,1=1 z2w,2=1 z4=2 |
| | 12 | z2w,1=1 z2w,2=1 z4=2 |
| 129.4.221 | 248426 | C1B1Er1Ni1 |
| | 115 | z2=1 |
| | 115 | z2=1 |
| | 12 | z2w,3=1 |
| | 12 | z2w,1=1 z2w,2=1 z2w,3=1 z4=2 |
| | 12 | z2w,1=1 z2w,2=1 z2w,3=1 z4=2 |
| | 12 | z2w,3=1 |
| | 12 | z2w,1=1 z2w,2=1 z2w,3=1 z4=1 |
| | 12 | z2w,3=1 z4=3 |
| | 12 | z2w,3=1 z4=3 |
| | 12 | z2w,1=1 z2w,2=1 z2w,3=1 z4=1 |
| | 12 | z2w,1=1 z2w,2=1 z4=1 |
| | 12 | z4=3 |
| | 12 | z4=3 |
| | 12 | z2w,1=1 z2w,2=1 z4=1 |
| | 12 | z4=2 |
| | 12 | z2w,1=1 z2w,2=1 |
| | 12 | z2w,1=1 z2w,2=1 |
| | 12 | z4=2 |
| 129.4.222 | 28532 | Br1O1Yb1 |
| | 115 | z2=1 |
| | 115 | z2=1 |
| | 12 | z2w,1=1 z2w,2=1 z4=2 |
| | 12 | z2w,1=1 z2w,2=1 z4=2 |
| | 12 | z2w,1=1 z2w,2=1 z4=1 |
| | 12 | z4=3 |
| | 12 | z4=3 |
| | 12 | z2w,1=1 z2w,2=1 z4=1 |
| | 12 | z2w,1=1 z2w,2=1 z2w,3=1 z4=1 |
| | 12 | z2w,3=1 z4=3 |
| | 12 | z2w,3=1 z4=3 |
| | 12 | z2w,1=1 z2w,2=1 z2w,3=1 z4=1 |
| | 12 | z2w,3=1 z4=2 |
| | 12 | z2w,1=1 z2w,2=1 z2w,3=1 |
| | 12 | z2w,1=1 z2w,2=1 z2w,3=1 |
| | 12 | z2w,3=1 z4=2 |
| 129.4.223 | 31662 | Br1O1Pu1 |
| | 115 | z2=1 |
| | 115 | z2=1 |
| | 12 | z2w,1=1 z2w,2=1 z4=2 |
| | 12 | z2w,1=1 z2w,2=1 z4=2 |
| | 12 | z2w,1=1 z2w,2=1 z4=1 |
| | 12 | z4=3 |
| | 12 | z4=3 |
| | 12 | z2w,1=1 z2w,2=1 z4=1 |
| | 12 | z2w,1=1 z2w,2=1 z2w,3=1 z4=1 |
| | 12 | z2w,3=1 z4=3 |
| | 12 | z2w,3=1 z4=3 |
| | 12 | z2w,1=1 z2w,2=1 z2w,3=1 z4=1 |
| | 12 | z2w,3=1 z4=2 |
| | 12 | z2w,1=1 z2w,2=1 z2w,3=1 |
| | 12 | z2w,1=1 z2w,2=1 z2w,3=1 |
| | 12 | z2w,3=1 z4=2 |
| 129.4.224 | 31666 | Br1Nd1O1 |

Continued on next column

**TABLE XIX– continued**

| TQC | ICSD SSG | Stoich. Formula SIs |
|---|---|---|
| | 115 | z2=1 |
| | 115 | z2=1 |
| | 12 | z2w,1=1 z2w,2=1 z4=2 |
| | 12 | z2w,1=1 z2w,2=1 z4=2 |
| | 12 | z2w,1=1 z2w,2=1 z4=1 |
| | 12 | z4=3 |
| | 12 | z4=3 |
| | 12 | z2w,1=1 z2w,2=1 z4=1 |
| | 12 | z2w,1=1 z2w,2=1 z2w,3=1 z4=1 |
| | 12 | z2w,3=1 z4=3 |
| | 12 | z2w,3=1 z4=3 |
| | 12 | z2w,1=1 z2w,2=1 z2w,3=1 z4=1 |
| | 12 | z2w,3=1 z4=2 |
| | 12 | z2w,1=1 z2w,2=1 z2w,3=1 |
| | 12 | z2w,1=1 z2w,2=1 z2w,3=1 |
| | 12 | z2w,3=1 z4=2 |
| 129.4.225 | 23955 | Br1Nd1O1 |
| | 115 | z2=1 |
| | 115 | z2=1 |
| | 12 | z2w,1=1 z2w,2=1 z4=2 |
| | 12 | z2w,1=1 z2w,2=1 z4=2 |
| | 12 | z2w,1=1 z2w,2=1 z4=1 |
| | 12 | z4=3 |
| | 12 | z4=3 |
| | 12 | z2w,1=1 z2w,2=1 z4=1 |
| | 12 | z2w,1=1 z2w,2=1 z2w,3=1 z4=1 |
| | 12 | z2w,3=1 z4=3 |
| | 12 | z2w,3=1 z4=3 |
| | 12 | z2w,1=1 z2w,2=1 z2w,3=1 z4=1 |
| | 12 | z2w,3=1 z4=2 |
| | 12 | z2w,1=1 z2w,2=1 z2w,3=1 |
| | 12 | z2w,1=1 z2w,2=1 z2w,3=1 |
| | 12 | z2w,3=1 z4=2 |
| 129.4.226 | 41071 | Br1Gd1O1 |
| | 115 | z2=1 |
| | 115 | z2=1 |
| | 12 | z2w,1=1 z2w,2=1 z4=2 |
| | 12 | z2w,1=1 z2w,2=1 z4=2 |
| | 12 | z2w,1=1 z2w,2=1 z4=1 |
| | 12 | z4=3 |
| | 12 | z4=3 |
| | 12 | z2w,1=1 z2w,2=1 z4=1 |
| | 12 | z2w,1=1 z2w,2=1 z2w,3=1 z4=1 |
| | 12 | z2w,3=1 z4=3 |
| | 12 | z2w,3=1 z4=3 |
| | 12 | z2w,1=1 z2w,2=1 z2w,3=1 z4=1 |
| | 12 | z2w,3=1 z4=2 |
| | 12 | z2w,1=1 z2w,2=1 z2w,3=1 |
| | 12 | z2w,1=1 z2w,2=1 z2w,3=1 |
| | 12 | z2w,3=1 z4=2 |
| 129.4.227 | 93509 | Br1Fe1O3Sr2 |
| | 115 | z2=1 |
| | 115 | z2=1 |
| | 12 | z2w,1=1 z2w,2=1 z4=2 |
| | 12 | z2w,1=1 z2w,2=1 z4=2 |
| | 12 | z2w,1=1 z2w,2=1 z4=1 |
| | 12 | z4=3 |
| | 12 | z4=3 |
| | 12 | z2w,1=1 z2w,2=1 z4=1 |
| | 12 | z2w,1=1 z2w,2=1 z2w,3=1 z4=1 |

Continued on next column

**TABLE XIX– continued**

| TQC | ICSD SSG | Stoich. Formula SIs |
|---|---|---|
| | 12 | z2w,3=1 z4=3 |
| | 12 | z2w,3=1 z4=3 |
| | 12 | z2w,1=1 z2w,2=1 z2w,3=1 z4=1 |
| | 12 | z2w,3=1 z4=2 |
| | 12 | z2w,1=1 z2w,2=1 z2w,3=1 |
| | 12 | z2w,1=1 z2w,2=1 z2w,3=1 |
| | 12 | z2w,3=1 z4=2 |
| 129.4.228 | 10454 | Bi2Ca1Mn1 |
| | 115 | z2=1 |
| | 115 | z2=1 |
| | 12 | z4=2 |
| | 12 | z2w,1=1 z2w,2=1 |
| | 12 | z2w,1=1 z2w,2=1 |
| | 12 | z4=2 |
| | 12 | z2w,1=1 z2w,2=1 z4=3 |
| | 12 | z4=1 |
| | 12 | z4=1 |
| | 12 | z2w,1=1 z2w,2=1 z4=3 |
| | 12 | z2w,1=1 z2w,2=1 z2w,3=1 z4=3 |
| | 12 | z2w,3=1 z4=1 |
| | 12 | z2w,3=1 z4=1 |
| | 12 | z2w,1=1 z2w,2=1 z2w,3=1 z4=3 |
| | 12 | z2w,3=1 |
| | 12 | z2w,1=1 z2w,2=1 z2w,3=1 z4=2 |
| | 12 | z2w,1=1 z2w,2=1 z2w,3=1 z4=2 |
| | 12 | z2w,3=1 |
| 129.4.229 | 60738 | Bi1Mn1Na1 |
| | 115 | z2=1 |
| | 115 | z2=1 |
| | 12 | z4=3 |
| | 12 | z2w,1=1 z2w,2=1 z4=1 |
| | 12 | z2w,1=1 z2w,2=1 z4=1 |
| | 12 | z4=3 |
| | 12 | z2w,1=1 z2w,2=1 |
| | 12 | z4=2 |
| | 12 | z4=2 |
| | 12 | z2w,1=1 z2w,2=1 |
| | 12 | z2w,1=1 z2w,2=1 z2w,3=1 |
| | 12 | z2w,3=1 z4=2 |
| | 12 | z2w,3=1 z4=2 |
| | 12 | z2w,1=1 z2w,2=1 z2w,3=1 |
| | 12 | z2w,3=1 z4=1 |
| | 12 | z2w,1=1 z2w,2=1 z2w,3=1 z4=3 |
| | 12 | z2w,1=1 z2w,2=1 z2w,3=1 z4=3 |
| | 12 | z2w,3=1 z4=1 |
| 129.4.230 | 60737 | Bi1Mn1Na1 |
| | 115 | z2=1 |
| | 115 | z2=1 |
| | 12 | z4=3 |
| | 12 | z2w,1=1 z2w,2=1 z4=1 |
| | 12 | z2w,1=1 z2w,2=1 z4=1 |
| | 12 | z4=3 |
| | 12 | z2w,1=1 z2w,2=1 |
| | 12 | z4=2 |
| | 12 | z4=2 |
| | 12 | z2w,1=1 z2w,2=1 |
| | 12 | z2w,1=1 z2w,2=1 z2w,3=1 |
| | 12 | z2w,3=1 z4=2 |
| | 12 | z2w,3=1 z4=2 |
| | 12 | z2w,1=1 z2w,2=1 z2w,3=1 |

Continued on next column

**TABLE XIX– continued**

| TQC | ICSD SSG | Stoich. Formula SIs |
|---|---|---|
| | 12 | z2w,3=1 z4=1 |
| | 12 | z2w,1=1 z2w,2=1 z2w,3=1 z4=3 |
| | 12 | z2w,1=1 z2w,2=1 z2w,3=1 z4=3 |
| | 12 | z2w,3=1 z4=1 |
| 129.4.231 | 60736 | Bi1Mn1Na1 |
| | 115 | z2=1 |
| | 115 | z2=1 |
| | 12 | z4=3 |
| | 12 | z2w,1=1 z2w,2=1 z4=1 |
| | 12 | z2w,1=1 z2w,2=1 z4=1 |
| | 12 | z4=3 |
| | 12 | z2w,1=1 z2w,2=1 |
| | 12 | z4=2 |
| | 12 | z4=2 |
| | 12 | z2w,1=1 z2w,2=1 |
| | 12 | z2w,1=1 z2w,2=1 z2w,3=1 |
| | 12 | z2w,3=1 z4=2 |
| | 12 | z2w,3=1 z4=2 |
| | 12 | z2w,1=1 z2w,2=1 z2w,3=1 |
| | 12 | z2w,3=1 z4=1 |
| | 12 | z2w,1=1 z2w,2=1 z2w,3=1 z4=3 |
| | 12 | z2w,1=1 z2w,2=1 z2w,3=1 z4=3 |
| | 12 | z2w,3=1 z4=1 |
| 129.4.232 | 80 | Bi1Ce1O1S2 |
| | 115 | z2=1 |
| | 115 | z2=1 |
| | 12 | z2w,3=1 z4=1 |
| | 12 | z2w,1=1 z2w,2=1 z2w,3=1 z4=3 |
| | 12 | z2w,1=1 z2w,2=1 z2w,3=1 z4=3 |
| | 12 | z2w,3=1 z4=1 |
| | 12 | z2w,1=1 z2w,2=1 z2w,3=1 z4=2 |
| | 12 | z2w,3=1 |
| | 12 | z2w,3=1 |
| | 12 | z2w,1=1 z2w,2=1 z2w,3=1 z4=2 |
| | 12 | z2w,1=1 z2w,2=1 z4=2 |
| | 12 | z2w,1=1 z2w,2=1 z4=2 |
| | 12 | z4=3 |
| | 12 | z2w,1=1 z2w,2=1 z4=1 |
| | 12 | z2w,1=1 z2w,2=1 z4=1 |
| | 12 | z4=3 |
| 129.4.233 | 79261 | Ba2F1In1O3 |
| | 115 | z2=1 |
| | 115 | z2=1 |
| | 12 | z2w,3=1 z4=2 |
| | 12 | z2w,1=1 z2w,2=1 z2w,3=1 |
| | 12 | z2w,1=1 z2w,2=1 z2w,3=1 |
| | 12 | z2w,3=1 z4=2 |
| | 12 | z2w,1=1 z2w,2=1 z2w,3=1 z4=3 |
| | 12 | z2w,3=1 z4=1 |
| | 12 | z2w,3=1 z4=1 |
| | 12 | z2w,1=1 z2w,2=1 z2w,3=1 z4=3 |
| | 12 | z2w,1=1 z2w,2=1 z4=3 |
| | 12 | z4=1 |
| | 12 | z4=1 |
| | 12 | z2w,1=1 z2w,2=1 z4=3 |
| | 12 | z2w,1=1 z2w,2=1 z4=2 |
| | 12 | z2w,1=1 z2w,2=1 z4=2 |
| 129.4.234 | 88950 | Ba1Mn2O5Y1 |
| | 115 | z2=1 |
| | 115 | z2=1 |

Continued on next column

**TABLE XIX– continued**

| TQC | ICSD SSG | Stoich. Formula SIs |
|---|---|---|
| | 12 | z2w,3=1 z4=2 |
| | 12 | z2w,1=1 z2w,2=1 z2w,3=1 |
| | 12 | z2w,1=1 z2w,2=1 z2w,3=1 |
| | 12 | z2w,3=1 z4=2 |
| | 12 | z2w,1=1 z2w,2=1 z2w,3=1 z4=3 |
| | 12 | z2w,3=1 z4=1 |
| | 12 | z2w,3=1 z4=1 |
| | 12 | z2w,1=1 z2w,2=1 z2w,3=1 z4=3 |
| | 12 | z2w,1=1 z2w,2=1 z4=3 |
| | 12 | z4=1 |
| | 12 | z4=1 |
| | 12 | z2w,1=1 z2w,2=1 z4=3 |
| | 12 | z2w,1=1 z2w,2=1 z4=2 |
| | 12 | z2w,1=1 z2w,2=1 z4=2 |
| 129.4.235 | 88949 | Ba1Mn2O5Y1 |
| | 115 | z2=1 |
| | 115 | z2=1 |
| | 12 | z2w,3=1 z4=1 |
| | 12 | z2w,1=1 z2w,2=1 z2w,3=1 z4=3 |
| | 12 | z2w,1=1 z2w,2=1 z2w,3=1 z4=3 |
| | 12 | z2w,3=1 z4=1 |
| | 12 | z2w,1=1 z2w,2=1 z2w,3=1 z4=2 |
| | 12 | z2w,3=1 |
| | 12 | z2w,3=1 |
| | 12 | z2w,1=1 z2w,2=1 z2w,3=1 z4=2 |
| | 12 | z2w,1=1 z2w,2=1 z4=2 |
| | 12 | z2w,1=1 z2w,2=1 z4=2 |
| | 12 | z4=3 |
| | 12 | z2w,1=1 z2w,2=1 z4=1 |
| | 12 | z2w,1=1 z2w,2=1 z4=1 |
| | 12 | z4=3 |
| 129.4.236 | 66965 | Ba1Ge1Mn1 |
| | 115 | z2=1 |
| | 115 | z2=1 |
| | 12 | z2w,1=1 z2w,2=1 z4=2 |
| | 12 | z2w,1=1 z2w,2=1 z4=2 |
| | 12 | z2w,1=1 z2w,2=1 z4=1 |
| | 12 | z4=3 |
| | 12 | z4=3 |
| | 12 | z2w,1=1 z2w,2=1 z4=1 |
| | 12 | z2w,1=1 z2w,2=1 z2w,3=1 z4=1 |
| | 12 | z2w,3=1 z4=3 |
| | 12 | z2w,3=1 z4=3 |
| | 12 | z2w,1=1 z2w,2=1 z2w,3=1 z4=1 |
| | 12 | z2w,3=1 z4=2 |
| | 12 | z2w,1=1 z2w,2=1 z2w,3=1 |
| | 12 | z2w,1=1 z2w,2=1 z2w,3=1 |
| | 12 | z2w,3=1 z4=2 |
| 129.4.237 | 66964 | Ba1Ge1Mn1 |
| | 115 | z2=1 |
| | 115 | z2=1 |
| | 12 | z2w,1=1 z2w,2=1 z4=2 |
| | 12 | z2w,1=1 z2w,2=1 z4=2 |
| | 12 | z2w,1=1 z2w,2=1 z4=1 |
| | 12 | z4=3 |
| | 12 | z4=3 |
| | 12 | z2w,1=1 z2w,2=1 z4=1 |
| | 12 | z2w,1=1 z2w,2=1 z2w,3=1 z4=1 |
| | 12 | z2w,3=1 z4=3 |
| | 12 | z2w,3=1 z4=3 |

Continued on next column

**TABLE XIX– continued**

| TQC | ICSD SSG | Stoich. Formula SIs |
|---|---|---|
| | 12 | z2w,1=1 z2w,2=1 z2w,3=1 z4=1 |
| | 12 | z2w,3=1 z4=2 |
| | 12 | z2w,1=1 z2w,2=1 z2w,3=1 |
| | 12 | z2w,1=1 z2w,2=1 z2w,3=1 |
| | 12 | z2w,3=1 z4=2 |
| 129.4.238 | 42370 | Ba1Ge1Mn1 |
| | 115 | z2=1 |
| | 115 | z2=1 |
| | 12 | z2w,3=1 z4=2 |
| | 12 | z2w,1=1 z2w,2=1 z2w,3=1 |
| | 12 | z2w,1=1 z2w,2=1 z2w,3=1 |
| | 12 | z2w,3=1 z4=2 |
| | 12 | z2w,1=1 z2w,2=1 z2w,3=1 z4=3 |
| | 12 | z2w,3=1 z4=1 |
| | 12 | z2w,3=1 z4=1 |
| | 12 | z2w,1=1 z2w,2=1 z2w,3=1 z4=3 |
| | 12 | z2w,1=1 z2w,2=1 z4=3 |
| | 12 | z4=1 |
| | 12 | z4=1 |
| | 12 | z2w,1=1 z2w,2=1 z4=3 |
| | 12 | z2w,1=1 z2w,2=1 z4=2 |
| | 12 | z2w,1=1 z2w,2=1 z4=2 |
| 129.4.239 | 248366 | Ba1F1Mn1Sb1 |
| | 115 | z2=1 |
| | 115 | z2=1 |
| | 12 | z2w,1=1 z2w,2=1 z4=2 |
| | 12 | z2w,1=1 z2w,2=1 z4=2 |
| | 12 | z2w,1=1 z2w,2=1 z4=1 |
| | 12 | z4=3 |
| | 12 | z4=3 |
| | 12 | z2w,1=1 z2w,2=1 z4=1 |
| | 12 | z2w,1=1 z2w,2=1 z2w,3=1 z4=1 |
| | 12 | z2w,3=1 z4=3 |
| | 12 | z2w,3=1 z4=3 |
| | 12 | z2w,1=1 z2w,2=1 z2w,3=1 z4=1 |
| | 12 | z2w,3=1 z4=2 |
| | 12 | z2w,1=1 z2w,2=1 z2w,3=1 |
| | 12 | z2w,1=1 z2w,2=1 z2w,3=1 |
| | 12 | z2w,3=1 z4=2 |
| 129.4.240 | 248364 | Ba1F1Mn1Sb1 |
| | 115 | z2=1 |
| | 115 | z2=1 |
| | 12 | z2w,1=1 z2w,2=1 z4=2 |
| | 12 | z2w,1=1 z2w,2=1 z4=2 |
| | 12 | z2w,1=1 z2w,2=1 z4=1 |
| | 12 | z4=3 |
| | 12 | z4=3 |
| | 12 | z2w,1=1 z2w,2=1 z4=1 |
| | 12 | z2w,1=1 z2w,2=1 z2w,3=1 z4=1 |
| | 12 | z2w,3=1 z4=3 |
| | 12 | z2w,3=1 z4=3 |
| | 12 | z2w,1=1 z2w,2=1 z2w,3=1 z4=1 |
| | 12 | z2w,3=1 z4=2 |
| | 12 | z2w,1=1 z2w,2=1 z2w,3=1 |
| | 12 | z2w,1=1 z2w,2=1 z2w,3=1 |
| | 12 | z2w,3=1 z4=2 |
| 129.4.241 | 82639 | Ba1Co1S2 |
| | 115 | z2=1 |
| | 115 | z2=1 |
| | 12 | z4=3 |

Continued on next column

**TABLE XIX– continued**

| TQC | ICSD SSG | Stoich. Formula SIs |
|---|---|---|
| | 12 | z2w,1=1 z2w,2=1 z4=1 |
| | 12 | z2w,1=1 z2w,2=1 z4=1 |
| | 12 | z4=3 |
| | 12 | z2w,1=1 z2w,2=1 |
| | 12 | z4=2 |
| | 12 | z4=2 |
| | 12 | z2w,1=1 z2w,2=1 |
| | 12 | z2w,1=1 z2w,2=1 z2w,3=1 |
| | 12 | z2w,3=1 z4=2 |
| | 12 | z2w,3=1 z4=2 |
| | 12 | z2w,1=1 z2w,2=1 z2w,3=1 |
| | 12 | z2w,3=1 z4=1 |
| | 12 | z2w,1=1 z2w,2=1 z2w,3=1 z4=3 |
| | 12 | z2w,1=1 z2w,2=1 z2w,3=1 z4=3 |
| | 12 | z2w,3=1 z4=1 |
| 129.4.242 | 426033 | Ba1Bi1F1Mn1 |
| | 115 | z2=1 |
| | 115 | z2=1 |
| | 12 | z2w,1=1 z2w,2=1 z4=2 |
| | 12 | z2w,1=1 z2w,2=1 z4=2 |
| | 12 | z2w,1=1 z2w,2=1 z4=1 |
| | 12 | z4=3 |
| | 12 | z4=3 |
| | 12 | z2w,1=1 z2w,2=1 z4=1 |
| | 12 | z2w,1=1 z2w,2=1 z2w,3=1 z4=1 |
| | 12 | z2w,3=1 z4=3 |
| | 12 | z2w,3=1 z4=3 |
| | 12 | z2w,1=1 z2w,2=1 z2w,3=1 z4=1 |
| | 12 | z2w,3=1 z4=2 |
| | 12 | z2w,1=1 z2w,2=1 z2w,3=1 |
| | 12 | z2w,1=1 z2w,2=1 z2w,3=1 |
| | 12 | z2w,3=1 z4=2 |
| 129.4.243 | 77939 | B1La1N1Ni1 |
| | 115 | z2=1 |
| | 115 | z2=1 |
| | 12 | z4=1 |
| | 12 | z2w,1=1 z2w,2=1 z4=3 |
| | 12 | z2w,1=1 z2w,2=1 z4=3 |
| | 12 | z4=1 |
| | 12 | z2w,1=1 z2w,2=1 z4=2 |
| | 12 | z2w,1=1 z2w,2=1 z4=2 |
| | 12 | z2w,1=1 z2w,2=1 z2w,3=1 z4=2 |
| | 12 | z2w,3=1 |
| | 12 | z2w,3=1 |
| | 12 | z2w,1=1 z2w,2=1 z2w,3=1 z4=2 |
| | 12 | z2w,3=1 z4=3 |
| | 12 | z2w,1=1 z2w,2=1 z2w,3=1 z4=1 |
| | 12 | z2w,1=1 z2w,2=1 z2w,3=1 z4=1 |
| | 12 | z2w,3=1 z4=3 |
| 129.4.244 | 25535 | B1Be4 |
| | 115 | z2=1 |
| | 115 | z2=1 |
| | 12 | z2w,3=1 |
| | 12 | z2w,1=1 z2w,2=1 z2w,3=1 z4=2 |
| | 12 | z2w,1=1 z2w,2=1 z2w,3=1 z4=2 |
| | 12 | z2w,3=1 |
| | 12 | z2w,1=1 z2w,2=1 z2w,3=1 z4=1 |
| | 12 | z2w,3=1 z4=3 |
| | 12 | z2w,3=1 z4=3 |
| | 12 | z2w,1=1 z2w,2=1 z2w,3=1 z4=1 |

Continued on next column

**TABLE XIX– continued**

| TQC | ICSD SSG | Stoich. Formula SIs |
|---|---|---|
| | 12 | z2w,1=1 z2w,2=1 z4=1 |
| | 12 | z4=3 |
| | 12 | z4=3 |
| | 12 | z2w,1=1 z2w,2=1 z4=1 |
| | 12 | z4=2 |
| | 12 | z2w,1=1 z2w,2=1 |
| | 12 | z2w,1=1 z2w,2=1 |
| | 12 | z4=2 |
| 129.4.245 | 262387 | Au3Ge1Sr1 |
| | 115 | z2=1 |
| | 115 | z2=1 |
| | 12 | z4=1 |
| | 12 | z2w,1=1 z2w,2=1 z4=3 |
| | 12 | z2w,1=1 z2w,2=1 z4=3 |
| | 12 | z4=1 |
| | 12 | z2w,1=1 z2w,2=1 z4=2 |
| | 12 | z2w,1=1 z2w,2=1 z4=2 |
| | 12 | z2w,1=1 z2w,2=1 z2w,3=1 z4=2 |
| | 12 | z2w,3=1 |
| | 12 | z2w,3=1 |
| | 12 | z2w,1=1 z2w,2=1 z2w,3=1 z4=2 |
| | 12 | z2w,3=1 z4=3 |
| | 12 | z2w,1=1 z2w,2=1 z2w,3=1 z4=1 |
| | 12 | z2w,1=1 z2w,2=1 z2w,3=1 z4=1 |
| | 12 | z2w,3=1 z4=3 |
| 129.4.246 | 262386 | Au3Ge1Sr1 |
| | 115 | z2=1 |
| | 115 | z2=1 |
| | 12 | z4=1 |
| | 12 | z2w,1=1 z2w,2=1 z4=3 |
| | 12 | z2w,1=1 z2w,2=1 z4=3 |
| | 12 | z4=1 |
| | 12 | z2w,1=1 z2w,2=1 z4=2 |
| | 12 | z2w,1=1 z2w,2=1 z4=2 |
| | 12 | z2w,1=1 z2w,2=1 z2w,3=1 z4=2 |
| | 12 | z2w,3=1 |
| | 12 | z2w,3=1 |
| | 12 | z2w,1=1 z2w,2=1 z2w,3=1 z4=2 |
| | 12 | z2w,3=1 z4=3 |
| | 12 | z2w,1=1 z2w,2=1 z2w,3=1 z4=1 |
| | 12 | z2w,1=1 z2w,2=1 z2w,3=1 z4=1 |
| | 12 | z2w,3=1 z4=3 |
| 129.4.247 | 262385 | Au3Ge1Sr1 |
| | 115 | z2=1 |
| | 115 | z2=1 |
| | 12 | z4=1 |
| | 12 | z2w,1=1 z2w,2=1 z4=3 |
| | 12 | z2w,1=1 z2w,2=1 z4=3 |
| | 12 | z4=1 |
| | 12 | z2w,1=1 z2w,2=1 z4=2 |
| | 12 | z2w,1=1 z2w,2=1 z4=2 |
| | 12 | z2w,1=1 z2w,2=1 z2w,3=1 z4=2 |
| | 12 | z2w,3=1 |
| | 12 | z2w,3=1 |
| | 12 | z2w,1=1 z2w,2=1 z2w,3=1 z4=2 |
| | 12 | z2w,3=1 z4=3 |
| | 12 | z2w,1=1 z2w,2=1 z2w,3=1 z4=1 |
| | 12 | z2w,1=1 z2w,2=1 z2w,3=1 z4=1 |
| | 12 | z2w,3=1 z4=3 |
| 129.4.248 | 157378 | Au1Sb2U1 |

Continued on next column

**TABLE XIX– continued**

| TQC | ICSD<br>SSG | Stoich. Formula<br>SIs |
|---|---|---|
| | 115 | z2=1 |
| | 115 | z2=1 |
| | 12 | z4=3 |
| | 12 | z2w,1=1 z2w,2=1 z4=1 |
| | 12 | z2w,1=1 z2w,2=1 z4=1 |
| | 12 | z4=3 |
| | 12 | z2w,1=1 z2w,2=1 |
| | 12 | z4=2 |
| | 12 | z4=2 |
| | 12 | z2w,1=1 z2w,2=1 |
| | 12 | z2w,1=1 z2w,2=1 z2w,3=1 |
| | 12 | z2w,3=1 z4=2 |
| | 12 | z2w,3=1 z4=2 |
| | 12 | z2w,1=1 z2w,2=1 z2w,3=1 |
| | 12 | z2w,3=1 z4=1 |
| | 12 | z2w,1=1 z2w,2=1 z2w,3=1 z4=3 |
| | 12 | z2w,1=1 z2w,2=1 z2w,3=1 z4=3 |
| | 12 | z2w,3=1 z4=1 |
| 129.4.249 | 238705 | Au1Bi2Ce1 |
| | 115 | z2=1 |
| | 115 | z2=1 |
| | 12 | z2w,3=1 z4=1 |
| | 12 | z2w,1=1 z2w,2=1 z2w,3=1 z4=3 |
| | 12 | z2w,1=1 z2w,2=1 z2w,3=1 z4=3 |
| | 12 | z2w,3=1 z4=1 |
| | 12 | z2w,1=1 z2w,2=1 z2w,3=1 z4=2 |
| | 12 | z2w,3=1 |
| | 12 | z2w,3=1 |
| | 12 | z2w,1=1 z2w,2=1 z2w,3=1 z4=2 |
| | 12 | z2w,1=1 z2w,2=1 z4=2 |
| | 12 | z2w,1=1 z2w,2=1 z4=2 |
| | 12 | z4=3 |
| | 12 | z2w,1=1 z2w,2=1 z4=1 |
| | 12 | z2w,1=1 z2w,2=1 z4=1 |
| | 12 | z4=3 |
| 129.4.250 | 75132 | As3Cu2O1U2 |
| | 115 | z2=1 |
| | 115 | z2=1 |
| | 12 | z2w,3=1 |
| | 12 | z2w,1=1 z2w,2=1 z2w,3=1 z4=2 |
| | 12 | z2w,1=1 z2w,2=1 z2w,3=1 z4=2 |
| | 12 | z2w,3=1 |
| | 12 | z2w,1=1 z2w,2=1 z2w,3=1 z4=1 |
| | 12 | z2w,3=1 z4=3 |
| | 12 | z2w,3=1 z4=3 |
| | 12 | z2w,1=1 z2w,2=1 z2w,3=1 z4=1 |
| | 12 | z2w,1=1 z2w,2=1 z4=1 |
| | 12 | z4=3 |
| | 12 | z4=3 |
| | 12 | z2w,1=1 z2w,2=1 z4=1 |
| | 12 | z4=2 |
| | 12 | z2w,1=1 z2w,2=1 |
| | 12 | z2w,1=1 z2w,2=1 |
| | 12 | z4=2 |
| 129.4.251 | 44035 | As2Np1 |
| | 115 | z2=1 |
| | 115 | z2=1 |
| | 12 | z4=3 |
| | 12 | z2w,1=1 z2w,2=1 z4=1 |
| | 12 | z2w,1=1 z2w,2=1 z4=1 |

Continued on next column

**TABLE XIX– continued**

| TQC | ICSD<br>SSG | Stoich. Formula<br>SIs |
|---|---|---|
| | 12 | z4=3 |
| | 12 | z2w,1=1 z2w,2=1 |
| | 12 | z4=2 |
| | 12 | z4=2 |
| | 12 | z2w,1=1 z2w,2=1 |
| | 12 | z2w,1=1 z2w,2=1 z2w,3=1 |
| | 12 | z2w,3=1 z4=2 |
| | 12 | z2w,3=1 z4=2 |
| | 12 | z2w,1=1 z2w,2=1 z2w,3=1 |
| | 12 | z2w,3=1 z4=1 |
| | 12 | z2w,1=1 z2w,2=1 z2w,3=1 z4=3 |
| | 12 | z2w,1=1 z2w,2=1 z2w,3=1 z4=3 |
| | 12 | z2w,3=1 z4=1 |
| 129.4.252 | 68147 | As2La1Ni2 |
| | 115 | z2=1 |
| | 115 | z2=1 |
| | 12 | z2w,3=1 |
| | 12 | z2w,1=1 z2w,2=1 z2w,3=1 z4=2 |
| | 12 | z2w,1=1 z2w,2=1 z2w,3=1 z4=2 |
| | 12 | z2w,3=1 |
| | 12 | z2w,1=1 z2w,2=1 z2w,3=1 z4=1 |
| | 12 | z2w,3=1 z4=3 |
| | 12 | z2w,3=1 z4=3 |
| | 12 | z2w,1=1 z2w,2=1 z2w,3=1 z4=1 |
| | 12 | z2w,1=1 z2w,2=1 z4=1 |
| | 12 | z4=3 |
| | 12 | z4=3 |
| | 12 | z2w,1=1 z2w,2=1 z4=1 |
| | 12 | z4=2 |
| | 12 | z2w,1=1 z2w,2=1 |
| | 12 | z2w,1=1 z2w,2=1 |
| | 12 | z4=2 |
| 129.4.253 | 186992 | As2Ir2La1 |
| | 115 | z2=1 |
| | 115 | z2=1 |
| | 12 | z2w,3=1 z4=3 |
| | 12 | z2w,1=1 z2w,2=1 z2w,3=1 z4=1 |
| | 12 | z2w,1=1 z2w,2=1 z2w,3=1 z4=1 |
| | 12 | z2w,3=1 z4=3 |
| | 12 | z2w,1=1 z2w,2=1 z2w,3=1 |
| | 12 | z2w,3=1 z4=2 |
| | 12 | z2w,3=1 z4=2 |
| | 12 | z2w,1=1 z2w,2=1 z2w,3=1 |
| | 12 | z2w,1=1 z2w,2=1 |
| | 12 | z4=2 |
| | 12 | z4=2 |
| | 12 | z2w,1=1 z2w,2=1 |
| | 12 | z4=1 |
| | 12 | z2w,1=1 z2w,2=1 z4=3 |
| | 12 | z2w,1=1 z2w,2=1 z4=3 |
| | 12 | z4=1 |
| 129.4.254 | 412193 | As2Cu1Yb1 |
| | 115 | z2=1 |
| | 115 | z2=1 |
| | 12 | z4=3 |
| | 12 | z2w,1=1 z2w,2=1 z4=1 |
| | 12 | z2w,1=1 z2w,2=1 z4=1 |
| | 12 | z4=3 |
| | 12 | z2w,1=1 z2w,2=1 |
| | 12 | z4=2 |

Continued on next column

**TABLE XIX– continued**

| TQC | ICSD SSG | Stoich. Formula SIs |
|---|---|---|
| | 12 | z4=2 |
| | 12 | z2w,1=1 z2w,2=1 |
| | 12 | z2w,1=1 z2w,2=1 z2w,3=1 |
| | 12 | z2w,3=1 z4=2 |
| | 12 | z2w,3=1 z4=2 |
| | 12 | z2w,1=1 z2w,2=1 z2w,3=1 |
| | 12 | z2w,3=1 z4=1 |
| | 12 | z2w,1=1 z2w,2=1 z2w,3=1 z4=3 |
| | 12 | z2w,1=1 z2w,2=1 z2w,3=1 z4=3 |
| | 12 | z2w,3=1 z4=1 |
| 129.4.255 | 40782 | As2Cu1U1 |
| | 115 | z2=1 |
| | 115 | z2=1 |
| | 12 | z4=3 |
| | 12 | z2w,1=1 z2w,2=1 z4=1 |
| | 12 | z2w,1=1 z2w,2=1 z4=1 |
| | 12 | z4=3 |
| | 12 | z2w,1=1 z2w,2=1 |
| | 12 | z4=2 |
| | 12 | z4=2 |
| | 12 | z2w,1=1 z2w,2=1 |
| | 12 | z2w,1=1 z2w,2=1 z2w,3=1 |
| | 12 | z2w,3=1 z4=2 |
| | 12 | z2w,3=1 z4=2 |
| | 12 | z2w,1=1 z2w,2=1 z2w,3=1 |
| | 12 | z2w,3=1 z4=1 |
| | 12 | z2w,1=1 z2w,2=1 z2w,3=1 z4=3 |
| | 12 | z2w,1=1 z2w,2=1 z2w,3=1 z4=3 |
| | 12 | z2w,3=1 z4=1 |
| 129.4.256 | 91305 | As2Cu1Gd1 |
| | 115 | z2=1 |
| | 115 | z2=1 |
| | 12 | z4=3 |
| | 12 | z2w,1=1 z2w,2=1 z4=1 |
| | 12 | z2w,1=1 z2w,2=1 z4=1 |
| | 12 | z4=3 |
| | 12 | z2w,1=1 z2w,2=1 |
| | 12 | z4=2 |
| | 12 | z4=2 |
| | 12 | z2w,1=1 z2w,2=1 |
| | 12 | z2w,1=1 z2w,2=1 z2w,3=1 |
| | 12 | z2w,3=1 z4=2 |
| | 12 | z2w,3=1 z4=2 |
| | 12 | z2w,1=1 z2w,2=1 z2w,3=1 |
| | 12 | z2w,3=1 z4=1 |
| | 12 | z2w,1=1 z2w,2=1 z2w,3=1 z4=3 |
| | 12 | z2w,1=1 z2w,2=1 z2w,3=1 z4=3 |
| | 12 | z2w,3=1 z4=1 |
| 129.4.257 | 412553 | As2Cu1Er1 |
| | 115 | z2=1 |
| | 115 | z2=1 |
| | 12 | z4=3 |
| | 12 | z2w,1=1 z2w,2=1 z4=1 |
| | 12 | z2w,1=1 z2w,2=1 z4=1 |
| | 12 | z4=3 |
| | 12 | z2w,1=1 z2w,2=1 |
| | 12 | z4=2 |
| | 12 | z4=2 |
| | 12 | z2w,1=1 z2w,2=1 |
| | 12 | z2w,1=1 z2w,2=1 z2w,3=1 |

Continued on next column

**TABLE XIX– continued**

| TQC | ICSD SSG | Stoich. Formula SIs |
|---|---|---|
| | 12 | z2w,3=1 z4=2 |
| | 12 | z2w,3=1 z4=2 |
| | 12 | z2w,1=1 z2w,2=1 z2w,3=1 |
| | 12 | z2w,3=1 z4=1 |
| | 12 | z2w,1=1 z2w,2=1 z2w,3=1 z4=3 |
| | 12 | z2w,1=1 z2w,2=1 z2w,3=1 z4=3 |
| | 12 | z2w,3=1 z4=1 |
| 129.4.258 | 412552 | As2Cu1Dy1 |
| | 115 | z2=1 |
| | 115 | z2=1 |
| | 12 | z2w,3=1 |
| | 12 | z2w,1=1 z2w,2=1 z2w,3=1 z4=2 |
| | 12 | z2w,1=1 z2w,2=1 z2w,3=1 z4=2 |
| | 12 | z2w,3=1 |
| | 12 | z2w,1=1 z2w,2=1 z2w,3=1 z4=1 |
| | 12 | z2w,3=1 z4=3 |
| | 12 | z2w,3=1 z4=3 |
| | 12 | z2w,1=1 z2w,2=1 z2w,3=1 z4=1 |
| | 12 | z2w,1=1 z2w,2=1 z4=1 |
| | 12 | z4=3 |
| | 12 | z4=3 |
| | 12 | z2w,1=1 z2w,2=1 z4=1 |
| | 12 | z4=2 |
| | 12 | z2w,1=1 z2w,2=1 |
| | 12 | z2w,1=1 z2w,2=1 |
| | 12 | z4=2 |
| 129.4.259 | 90313 | As2Co1U1 |
| | 115 | z2=1 |
| | 115 | z2=1 |
| | 12 | z4=3 |
| | 12 | z2w,1=1 z2w,2=1 z4=1 |
| | 12 | z2w,1=1 z2w,2=1 z4=1 |
| | 12 | z4=3 |
| | 12 | z2w,1=1 z2w,2=1 |
| | 12 | z4=2 |
| | 12 | z4=2 |
| | 12 | z2w,1=1 z2w,2=1 |
| | 12 | z2w,1=1 z2w,2=1 z2w,3=1 |
| | 12 | z2w,3=1 z4=2 |
| | 12 | z2w,3=1 z4=2 |
| | 12 | z2w,1=1 z2w,2=1 z2w,3=1 |
| | 12 | z2w,3=1 z4=1 |
| | 12 | z2w,1=1 z2w,2=1 z2w,3=1 z4=3 |
| | 12 | z2w,1=1 z2w,2=1 z2w,3=1 z4=3 |
| | 12 | z2w,3=1 z4=1 |
| 129.4.260 | 42353 | As1Te1U1 |
| | 115 | z2=1 |
| | 115 | z2=1 |
| | 12 | z4=3 |
| | 12 | z2w,1=1 z2w,2=1 z4=1 |
| | 12 | z2w,1=1 z2w,2=1 z4=1 |
| | 12 | z4=3 |
| | 12 | z2w,1=1 z2w,2=1 |
| | 12 | z4=2 |
| | 12 | z4=2 |
| | 12 | z2w,1=1 z2w,2=1 |
| | 12 | z2w,1=1 z2w,2=1 z2w,3=1 |
| | 12 | z2w,3=1 z4=2 |
| | 12 | z2w,3=1 z4=2 |
| | 12 | z2w,1=1 z2w,2=1 z2w,3=1 |

Continued on next column

**TABLE XIX– continued**

| TQC | ICSD<br>SSG | Stoich. Formula<br>SIs |
|---|---|---|
| | 12 | z2w,3=1 z4=1 |
| | 12 | z2w,1=1 z2w,2=1 z2w,3=1 z4=3 |
| | 12 | z2w,1=1 z2w,2=1 z2w,3=1 z4=3 |
| | 12 | z2w,3=1 z4=1 |
| 129.4.261 | 87141 | As1Se1U1 |
| | 115 | z2=1 |
| | 115 | z2=1 |
| | 12 | z4=3 |
| | 12 | z2w,1=1 z2w,2=1 z4=1 |
| | 12 | z2w,1=1 z2w,2=1 z4=1 |
| | 12 | z4=3 |
| | 12 | z2w,1=1 z2w,2=1 |
| | 12 | z4=2 |
| | 12 | z4=2 |
| | 12 | z2w,1=1 z2w,2=1 |
| | 12 | z2w,1=1 z2w,2=1 z2w,3=1 |
| | 12 | z2w,3=1 z4=2 |
| | 12 | z2w,3=1 z4=2 |
| | 12 | z2w,1=1 z2w,2=1 z2w,3=1 |
| | 12 | z2w,3=1 z4=1 |
| | 12 | z2w,1=1 z2w,2=1 z2w,3=1 z4=3 |
| | 12 | z2w,1=1 z2w,2=1 z2w,3=1 z4=3 |
| | 12 | z2w,3=1 z4=1 |
| 129.4.262 | 66946 | As1Se1U1 |
| | 115 | z2=1 |
| | 115 | z2=1 |
| | 12 | z2w,3=1 |
| | 12 | z2w,1=1 z2w,2=1 z2w,3=1 z4=2 |
| | 12 | z2w,1=1 z2w,2=1 z2w,3=1 z4=2 |
| | 12 | z2w,3=1 |
| | 12 | z2w,1=1 z2w,2=1 z2w,3=1 z4=1 |
| | 12 | z2w,3=1 z4=3 |
| | 12 | z2w,3=1 z4=3 |
| | 12 | z2w,1=1 z2w,2=1 z2w,3=1 z4=1 |
| | 12 | z2w,1=1 z2w,2=1 z4=1 |
| | 12 | z4=3 |
| | 12 | z4=3 |
| | 12 | z2w,1=1 z2w,2=1 z4=1 |
| | 12 | z4=2 |
| | 12 | z2w,1=1 z2w,2=1 |
| | 12 | z2w,1=1 z2w,2=1 |
| | 12 | z4=2 |
| 129.4.263 | 42485 | As1Se1U1 |
| | 115 | z2=1 |
| | 115 | z2=1 |
| | 12 | z4=3 |
| | 12 | z2w,1=1 z2w,2=1 z4=1 |
| | 12 | z2w,1=1 z2w,2=1 z4=1 |
| | 12 | z4=3 |
| | 12 | z2w,1=1 z2w,2=1 |
| | 12 | z4=2 |
| | 12 | z4=2 |
| | 12 | z2w,1=1 z2w,2=1 |
| | 12 | z2w,1=1 z2w,2=1 z2w,3=1 |
| | 12 | z2w,3=1 z4=2 |
| | 12 | z2w,3=1 z4=2 |
| | 12 | z2w,1=1 z2w,2=1 z2w,3=1 |
| | 12 | z2w,3=1 z4=1 |
| | 12 | z2w,1=1 z2w,2=1 z2w,3=1 z4=3 |
| | 12 | z2w,1=1 z2w,2=1 z2w,3=1 z4=3 |
| | 12 | z2w,3=1 z4=1 |
| 129.4.264 | 87140 | As1S1U1 |
| | 115 | z2=1 |
| | 115 | z2=1 |
| | 12 | z4=3 |
| | 12 | z2w,1=1 z2w,2=1 z4=1 |
| | 12 | z2w,1=1 z2w,2=1 z4=1 |
| | 12 | z4=3 |
| | 12 | z2w,1=1 z2w,2=1 |
| | 12 | z4=2 |
| | 12 | z4=2 |
| | 12 | z2w,1=1 z2w,2=1 |
| | 12 | z2w,1=1 z2w,2=1 z2w,3=1 |
| | 12 | z2w,3=1 z4=2 |
| | 12 | z2w,3=1 z4=2 |
| | 12 | z2w,1=1 z2w,2=1 z2w,3=1 |
| | 12 | z2w,3=1 z4=1 |
| | 12 | z2w,1=1 z2w,2=1 z2w,3=1 z4=3 |
| | 12 | z2w,1=1 z2w,2=1 z2w,3=1 z4=3 |
| | 12 | z2w,3=1 z4=1 |
| 129.4.265 | 66945 | As1S1U1 |
| | 115 | z2=1 |
| | 115 | z2=1 |
| | 12 | z2w,3=1 |
| | 12 | z2w,1=1 z2w,2=1 z2w,3=1 z4=2 |
| | 12 | z2w,1=1 z2w,2=1 z2w,3=1 z4=2 |
| | 12 | z2w,3=1 |
| | 12 | z2w,1=1 z2w,2=1 z2w,3=1 z4=1 |
| | 12 | z2w,3=1 z4=3 |
| | 12 | z2w,3=1 z4=3 |
| | 12 | z2w,1=1 z2w,2=1 z2w,3=1 z4=1 |
| | 12 | z2w,1=1 z2w,2=1 z4=1 |
| | 12 | z4=3 |
| | 12 | z4=3 |
| | 12 | z2w,1=1 z2w,2=1 z4=1 |
| | 12 | z4=2 |
| | 12 | z2w,1=1 z2w,2=1 |
| | 12 | z2w,1=1 z2w,2=1 |
| | 12 | z4=2 |
| 129.4.266 | 42170 | As1S1U1 |
| | 115 | z2=1 |
| | 115 | z2=1 |
| | 12 | z4=3 |
| | 12 | z2w,1=1 z2w,2=1 z4=1 |
| | 12 | z2w,1=1 z2w,2=1 z4=1 |
| | 12 | z4=3 |
| | 12 | z2w,1=1 z2w,2=1 |
| | 12 | z4=2 |
| | 12 | z4=2 |
| | 12 | z2w,1=1 z2w,2=1 |
| | 12 | z2w,1=1 z2w,2=1 z2w,3=1 |
| | 12 | z2w,3=1 z4=2 |
| | 12 | z2w,3=1 z4=2 |
| | 12 | z2w,1=1 z2w,2=1 z2w,3=1 |
| | 12 | z2w,3=1 z4=1 |
| | 12 | z2w,1=1 z2w,2=1 z2w,3=1 z4=3 |
| | 12 | z2w,1=1 z2w,2=1 z2w,3=1 z4=3 |
| | 12 | z2w,3=1 z4=1 |
| 129.4.267 | 85778 | As1Nd1O1Zn1 |
| | 115 | z2=1 |

Continued on next column

**TABLE XIX– continued**

| TQC | ICSD SSG | Stoich. Formula SIs |
|---|---|---|
| | 115 | z2=1 |
| | 12 | z2w,1=1 z2w,2=1 z4=2 |
| | 12 | z2w,1=1 z2w,2=1 z4=2 |
| | 12 | z2w,1=1 z2w,2=1 z4=1 |
| | 12 | z4=3 |
| | 12 | z4=3 |
| | 12 | z2w,1=1 z2w,2=1 z4=1 |
| | 12 | z2w,1=1 z2w,2=1 z2w,3=1 z4=1 |
| | 12 | z2w,3=1 z4=3 |
| | 12 | z2w,3=1 z4=3 |
| | 12 | z2w,1=1 z2w,2=1 z2w,3=1 z4=1 |
| | 12 | z2w,3=1 z4=2 |
| | 12 | z2w,1=1 z2w,2=1 z2w,3=1 |
| | 12 | z2w,1=1 z2w,2=1 z2w,3=1 |
| | 12 | z2w,3=1 z4=2 |
| 129.4.268 | 262832 | As1Nd1O1Ru1 |
| | 115 | z2=1 |
| | 115 | z2=1 |
| | 12 | z2w,3=1 z4=2 |
| | 12 | z2w,1=1 z2w,2=1 z2w,3=1 |
| | 12 | z2w,1=1 z2w,2=1 z2w,3=1 |
| | 12 | z2w,3=1 z4=2 |
| | 12 | z2w,1=1 z2w,2=1 z2w,3=1 z4=3 |
| | 12 | z2w,3=1 z4=1 |
| | 12 | z2w,3=1 z4=1 |
| | 12 | z2w,1=1 z2w,2=1 z2w,3=1 z4=3 |
| | 12 | z2w,1=1 z2w,2=1 z4=3 |
| | 12 | z4=1 |
| | 12 | z4=1 |
| | 12 | z2w,1=1 z2w,2=1 z4=3 |
| | 12 | z2w,1=1 z2w,2=1 z4=2 |
| | 12 | z2w,1=1 z2w,2=1 z4=2 |
| 129.4.269 | 42329 | As1Mn2 |
| | 115 | z2=1 |
| | 115 | z2=1 |
| | 12 | z2w,3=1 z4=3 |
| | 12 | z2w,1=1 z2w,2=1 z2w,3=1 z4=1 |
| | 12 | z2w,1=1 z2w,2=1 z2w,3=1 z4=1 |
| | 12 | z2w,3=1 z4=3 |
| | 12 | z2w,1=1 z2w,2=1 z2w,3=1 |
| | 12 | z2w,3=1 z4=2 |
| | 12 | z2w,3=1 z4=2 |
| | 12 | z2w,1=1 z2w,2=1 z2w,3=1 |
| | 12 | z2w,1=1 z2w,2=1 |
| | 12 | z4=2 |
| | 12 | z4=2 |
| | 12 | z2w,1=1 z2w,2=1 |
| | 12 | z4=1 |
| | 12 | z2w,1=1 z2w,2=1 z4=3 |
| | 12 | z2w,1=1 z2w,2=1 z4=3 |
| | 12 | z4=1 |
| 129.4.270 | 415629 | As1Mn2 |
| | 115 | z2=1 |
| | 115 | z2=1 |
| | 12 | z2w,3=1 z4=3 |
| | 12 | z2w,1=1 z2w,2=1 z2w,3=1 z4=1 |
| | 12 | z2w,1=1 z2w,2=1 z2w,3=1 z4=1 |
| | 12 | z2w,3=1 z4=3 |
| | 12 | z2w,1=1 z2w,2=1 z2w,3=1 |
| | 12 | z2w,3=1 z4=2 |

Continued on next column

**TABLE XIX– continued**

| TQC | ICSD SSG | Stoich. Formula SIs |
|---|---|---|
| | 12 | z2w,3=1 z4=2 |
| | 12 | z2w,1=1 z2w,2=1 z2w,3=1 |
| | 12 | z2w,1=1 z2w,2=1 |
| | 12 | z4=2 |
| | 12 | z4=2 |
| | 12 | z2w,1=1 z2w,2=1 |
| | 12 | z4=1 |
| | 12 | z2w,1=1 z2w,2=1 z4=3 |
| | 12 | z2w,1=1 z2w,2=1 z4=3 |
| | 12 | z4=1 |
| 129.4.271 | 44009 | As1Mn1V1 |
| | 115 | z2=1 |
| | 115 | z2=1 |
| | 12 | z4=1 |
| | 12 | z2w,1=1 z2w,2=1 z4=3 |
| | 12 | z2w,1=1 z2w,2=1 z4=3 |
| | 12 | z4=1 |
| | 12 | z2w,1=1 z2w,2=1 z4=2 |
| | 12 | z2w,1=1 z2w,2=1 z4=2 |
| | 12 | z2w,1=1 z2w,2=1 z2w,3=1 z4=2 |
| | 12 | z2w,3=1 |
| | 12 | z2w,3=1 |
| | 12 | z2w,1=1 z2w,2=1 z2w,3=1 z4=2 |
| | 12 | z2w,3=1 z4=3 |
| | 12 | z2w,1=1 z2w,2=1 z2w,3=1 z4=1 |
| | 12 | z2w,1=1 z2w,2=1 z2w,3=1 z4=1 |
| | 12 | z2w,3=1 z4=3 |
| 129.4.272 | 89597 | As1Mn1Rb1 |
| | 115 | z2=1 |
| | 115 | z2=1 |
| | 12 | z2w,1=1 z2w,2=1 z4=2 |
| | 12 | z2w,1=1 z2w,2=1 z4=2 |
| | 12 | z2w,1=1 z2w,2=1 z4=1 |
| | 12 | z4=3 |
| | 12 | z4=3 |
| | 12 | z2w,1=1 z2w,2=1 z4=1 |
| | 12 | z2w,1=1 z2w,2=1 z2w,3=1 z4=1 |
| | 12 | z2w,3=1 z4=3 |
| | 12 | z2w,3=1 z4=3 |
| | 12 | z2w,1=1 z2w,2=1 z2w,3=1 z4=1 |
| | 12 | z2w,3=1 z4=2 |
| | 12 | z2w,1=1 z2w,2=1 z2w,3=1 |
| | 12 | z2w,1=1 z2w,2=1 z2w,3=1 |
| | 12 | z2w,3=1 z4=2 |
| 129.4.273 | 89596 | As1Mn1Rb1 |
| | 115 | z2=1 |
| | 115 | z2=1 |
| | 12 | z2w,1=1 z2w,2=1 z4=2 |
| | 12 | z2w,1=1 z2w,2=1 z4=2 |
| | 12 | z2w,1=1 z2w,2=1 z4=1 |
| | 12 | z4=3 |
| | 12 | z4=3 |
| | 12 | z2w,1=1 z2w,2=1 z4=1 |
| | 12 | z2w,1=1 z2w,2=1 z2w,3=1 z4=1 |
| | 12 | z2w,3=1 z4=3 |
| | 12 | z2w,3=1 z4=3 |
| | 12 | z2w,1=1 z2w,2=1 z2w,3=1 z4=1 |
| | 12 | z2w,3=1 z4=2 |
| | 12 | z2w,1=1 z2w,2=1 z2w,3=1 |
| | 12 | z2w,1=1 z2w,2=1 z2w,3=1 |

Continued on next column

**TABLE XIX– continued**

| TQC | ICSD SSG | Stoich. Formula SIs |
|---|---|---|
| | 12 | z2w,3=1 z4=2 |
| 129.4.274 | 60732 | As1Mn1Na1 |
| | 115 | z2=1 |
| | 115 | z2=1 |
| | 12 | z4=3 |
| | 12 | z2w,1=1 z2w,2=1 z4=1 |
| | 12 | z2w,1=1 z2w,2=1 z4=1 |
| | 12 | z4=3 |
| | 12 | z2w,1=1 z2w,2=1 |
| | 12 | z4=2 |
| | 12 | z4=2 |
| | 12 | z2w,1=1 z2w,2=1 |
| | 12 | z2w,1=1 z2w,2=1 z2w,3=1 |
| | 12 | z2w,3=1 z4=2 |
| | 12 | z2w,3=1 z4=2 |
| | 12 | z2w,1=1 z2w,2=1 z2w,3=1 |
| | 12 | z2w,3=1 z4=1 |
| | 12 | z2w,1=1 z2w,2=1 z2w,3=1 z4=3 |
| | 12 | z2w,1=1 z2w,2=1 z2w,3=1 z4=3 |
| | 12 | z2w,3=1 z4=1 |
| 129.4.275 | 60731 | As1Mn1Na1 |
| | 115 | z2=1 |
| | 115 | z2=1 |
| | 12 | z4=3 |
| | 12 | z2w,1=1 z2w,2=1 z4=1 |
| | 12 | z2w,1=1 z2w,2=1 z4=1 |
| | 12 | z4=3 |
| | 12 | z2w,1=1 z2w,2=1 |
| | 12 | z4=2 |
| | 12 | z4=2 |
| | 12 | z2w,1=1 z2w,2=1 |
| | 12 | z2w,1=1 z2w,2=1 z2w,3=1 |
| | 12 | z2w,3=1 z4=2 |
| | 12 | z2w,3=1 z4=2 |
| | 12 | z2w,1=1 z2w,2=1 z2w,3=1 |
| | 12 | z2w,3=1 z4=1 |
| | 12 | z2w,1=1 z2w,2=1 z2w,3=1 z4=3 |
| | 12 | z2w,1=1 z2w,2=1 z2w,3=1 z4=3 |
| | 12 | z2w,3=1 z4=1 |
| 129.4.276 | 60730 | As1Mn1Na1 |
| | 115 | z2=1 |
| | 115 | z2=1 |
| | 12 | z4=3 |
| | 12 | z2w,1=1 z2w,2=1 z4=1 |
| | 12 | z2w,1=1 z2w,2=1 z4=1 |
| | 12 | z4=3 |
| | 12 | z2w,1=1 z2w,2=1 |
| | 12 | z4=2 |
| | 12 | z4=2 |
| | 12 | z2w,1=1 z2w,2=1 |
| | 12 | z2w,1=1 z2w,2=1 z2w,3=1 |
| | 12 | z2w,3=1 z4=2 |
| | 12 | z2w,3=1 z4=2 |
| | 12 | z2w,1=1 z2w,2=1 z2w,3=1 |
| | 12 | z2w,3=1 z4=1 |
| | 12 | z2w,1=1 z2w,2=1 z2w,3=1 z4=3 |
| | 12 | z2w,1=1 z2w,2=1 z2w,3=1 z4=3 |
| | 12 | z2w,3=1 z4=1 |
| 129.4.277 | 26461 | As1Mn1Na1 |
| | 115 | z2=1 |
| | 115 | z2=1 |
| | 12 | z4=3 |
| | 12 | z2w,1=1 z2w,2=1 z4=1 |
| | 12 | z2w,1=1 z2w,2=1 z4=1 |
| | 12 | z4=3 |
| | 12 | z2w,1=1 z2w,2=1 |
| | 12 | z4=2 |
| | 12 | z4=2 |
| | 12 | z2w,1=1 z2w,2=1 |
| | 12 | z2w,1=1 z2w,2=1 z2w,3=1 |
| | 12 | z2w,3=1 z4=2 |
| | 12 | z2w,3=1 z4=2 |
| | 12 | z2w,1=1 z2w,2=1 z2w,3=1 |
| | 12 | z2w,3=1 z4=1 |
| | 12 | z2w,1=1 z2w,2=1 z2w,3=1 z4=3 |
| | 12 | z2w,1=1 z2w,2=1 z2w,3=1 z4=3 |
| | 12 | z2w,3=1 z4=1 |
| 129.4.278 | 60742 | As1Li1Mn1 |
| | 115 | z2=1 |
| | 115 | z2=1 |
| | 12 | z2w,3=1 z4=2 |
| | 12 | z2w,1=1 z2w,2=1 z2w,3=1 |
| | 12 | z2w,1=1 z2w,2=1 z2w,3=1 |
| | 12 | z2w,3=1 z4=2 |
| | 12 | z2w,1=1 z2w,2=1 z2w,3=1 z4=3 |
| | 12 | z2w,3=1 z4=1 |
| | 12 | z2w,3=1 z4=1 |
| | 12 | z2w,1=1 z2w,2=1 z2w,3=1 z4=3 |
| | 12 | z2w,1=1 z2w,2=1 z4=3 |
| | 12 | z4=1 |
| | 12 | z4=1 |
| | 12 | z2w,1=1 z2w,2=1 z4=3 |
| | 12 | z2w,1=1 z2w,2=1 z4=2 |
| | 12 | z2w,1=1 z2w,2=1 z4=2 |
| 129.4.279 | 60741 | As1Li1Mn1 |
| | 115 | z2=1 |
| | 115 | z2=1 |
| | 12 | z2w,3=1 z4=2 |
| | 12 | z2w,1=1 z2w,2=1 z2w,3=1 |
| | 12 | z2w,1=1 z2w,2=1 z2w,3=1 |
| | 12 | z2w,3=1 z4=2 |
| | 12 | z2w,1=1 z2w,2=1 z2w,3=1 z4=3 |
| | 12 | z2w,3=1 z4=1 |
| | 12 | z2w,3=1 z4=1 |
| | 12 | z2w,1=1 z2w,2=1 z2w,3=1 z4=3 |
| | 12 | z2w,1=1 z2w,2=1 z4=3 |
| | 12 | z4=1 |
| | 12 | z4=1 |
| | 12 | z2w,1=1 z2w,2=1 z4=3 |
| | 12 | z2w,1=1 z2w,2=1 z4=2 |
| | 12 | z2w,1=1 z2w,2=1 z4=2 |
| 129.4.280 | 60740 | As1Li1Mn1 |
| | 115 | z2=1 |
| | 115 | z2=1 |
| | 12 | z2w,3=1 z4=2 |
| | 12 | z2w,1=1 z2w,2=1 z2w,3=1 |
| | 12 | z2w,1=1 z2w,2=1 z2w,3=1 |
| | 12 | z2w,3=1 z4=2 |
| | 12 | z2w,1=1 z2w,2=1 z2w,3=1 z4=3 |
| | 12 | z2w,3=1 z4=1 |

Continued on next column

**TABLE XIX– continued**

| TQC | ICSD SSG | Stoich. Formula SIs |
|---|---|---|
| | 12 | z2w,3=1 z4=1 |
| | 12 | z2w,1=1 z2w,2=1 z2w,3=1 z4=3 |
| | 12 | z2w,1=1 z2w,2=1 z4=3 |
| | 12 | z4=1 |
| | 12 | z4=1 |
| | 12 | z2w,1=1 z2w,2=1 z4=3 |
| | 12 | z2w,1=1 z2w,2=1 z4=2 |
| | 12 | z2w,1=1 z2w,2=1 z4=2 |
| 129.4.281 | 60739 | As1Li1Mn1 |
| | 115 | z2=1 |
| | 115 | z2=1 |
| | 12 | z2w,3=1 z4=2 |
| | 12 | z2w,1=1 z2w,2=1 z2w,3=1 |
| | 12 | z2w,1=1 z2w,2=1 z2w,3=1 |
| | 12 | z2w,3=1 z4=2 |
| | 12 | z2w,1=1 z2w,2=1 z2w,3=1 z4=3 |
| | 12 | z2w,3=1 z4=1 |
| | 12 | z2w,3=1 z4=1 |
| | 12 | z2w,1=1 z2w,2=1 z2w,3=1 z4=3 |
| | 12 | z2w,1=1 z2w,2=1 z4=3 |
| | 12 | z4=1 |
| | 12 | z4=1 |
| | 12 | z2w,1=1 z2w,2=1 z4=3 |
| | 12 | z2w,1=1 z2w,2=1 z4=2 |
| | 12 | z2w,1=1 z2w,2=1 z4=2 |
| 129.4.282 | 26459 | As1Li1Mn1 |
| | 115 | z2=1 |
| | 115 | z2=1 |
| | 12 | z2w,3=1 z4=2 |
| | 12 | z2w,1=1 z2w,2=1 z2w,3=1 |
| | 12 | z2w,1=1 z2w,2=1 z2w,3=1 |
| | 12 | z2w,3=1 z4=2 |
| | 12 | z2w,1=1 z2w,2=1 z2w,3=1 z4=3 |
| | 12 | z2w,3=1 z4=1 |
| | 12 | z2w,3=1 z4=1 |
| | 12 | z2w,1=1 z2w,2=1 z2w,3=1 z4=3 |
| | 12 | z2w,1=1 z2w,2=1 z4=3 |
| | 12 | z4=1 |
| | 12 | z4=1 |
| | 12 | z2w,1=1 z2w,2=1 z4=3 |
| | 12 | z2w,1=1 z2w,2=1 z4=2 |
| | 12 | z2w,1=1 z2w,2=1 z4=2 |
| 129.4.283 | 192823 | As1Li1Mn1 |
| | 115 | z2=1 |
| | 115 | z2=1 |
| | 12 | z2w,3=1 z4=2 |
| | 12 | z2w,1=1 z2w,2=1 z2w,3=1 |
| | 12 | z2w,1=1 z2w,2=1 z2w,3=1 |
| | 12 | z2w,3=1 z4=2 |
| | 12 | z2w,1=1 z2w,2=1 z2w,3=1 z4=3 |
| | 12 | z2w,3=1 z4=1 |
| | 12 | z2w,3=1 z4=1 |
| | 12 | z2w,1=1 z2w,2=1 z2w,3=1 z4=3 |
| | 12 | z2w,1=1 z2w,2=1 z4=3 |
| | 12 | z4=1 |
| | 12 | z4=1 |
| | 12 | z2w,1=1 z2w,2=1 z4=3 |
| | 12 | z2w,1=1 z2w,2=1 z4=2 |
| | 12 | z2w,1=1 z2w,2=1 z4=2 |
| 129.4.284 | 192822 | As1Li1Mn1 |

Continued on next column

**TABLE XIX– continued**

| TQC | ICSD SSG | Stoich. Formula SIs |
|---|---|---|
| | 115 | z2=1 |
| | 115 | z2=1 |
| | 12 | z2w,3=1 z4=2 |
| | 12 | z2w,1=1 z2w,2=1 z2w,3=1 |
| | 12 | z2w,1=1 z2w,2=1 z2w,3=1 |
| | 12 | z2w,3=1 z4=2 |
| | 12 | z2w,1=1 z2w,2=1 z2w,3=1 z4=3 |
| | 12 | z2w,3=1 z4=1 |
| | 12 | z2w,3=1 z4=1 |
| | 12 | z2w,1=1 z2w,2=1 z2w,3=1 z4=3 |
| | 12 | z2w,1=1 z2w,2=1 z4=3 |
| | 12 | z4=1 |
| | 12 | z4=1 |
| | 12 | z2w,1=1 z2w,2=1 z4=3 |
| | 12 | z2w,1=1 z2w,2=1 z4=2 |
| | 12 | z2w,1=1 z2w,2=1 z4=2 |
| 129.4.285 | 192821 | As1Li1Mn1 |
| | 115 | z2=1 |
| | 115 | z2=1 |
| | 12 | z2w,3=1 z4=2 |
| | 12 | z2w,1=1 z2w,2=1 z2w,3=1 |
| | 12 | z2w,1=1 z2w,2=1 z2w,3=1 |
| | 12 | z2w,3=1 z4=2 |
| | 12 | z2w,1=1 z2w,2=1 z2w,3=1 z4=3 |
| | 12 | z2w,3=1 z4=1 |
| | 12 | z2w,3=1 z4=1 |
| | 12 | z2w,1=1 z2w,2=1 z2w,3=1 z4=3 |
| | 12 | z2w,1=1 z2w,2=1 z4=3 |
| | 12 | z4=1 |
| | 12 | z4=1 |
| | 12 | z2w,1=1 z2w,2=1 z4=3 |
| | 12 | z2w,1=1 z2w,2=1 z4=2 |
| | 12 | z2w,1=1 z2w,2=1 z4=2 |
| 129.4.286 | 192820 | As1Li1Mn1 |
| | 115 | z2=1 |
| | 115 | z2=1 |
| | 12 | z2w,3=1 z4=2 |
| | 12 | z2w,1=1 z2w,2=1 z2w,3=1 |
| | 12 | z2w,1=1 z2w,2=1 z2w,3=1 |
| | 12 | z2w,3=1 z4=2 |
| | 12 | z2w,1=1 z2w,2=1 z2w,3=1 z4=3 |
| | 12 | z2w,3=1 z4=1 |
| | 12 | z2w,3=1 z4=1 |
| | 12 | z2w,1=1 z2w,2=1 z2w,3=1 z4=3 |
| | 12 | z2w,1=1 z2w,2=1 z4=3 |
| | 12 | z4=1 |
| | 12 | z4=1 |
| | 12 | z2w,1=1 z2w,2=1 z4=3 |
| | 12 | z2w,1=1 z2w,2=1 z4=2 |
| | 12 | z2w,1=1 z2w,2=1 z4=2 |
| 129.4.287 | 192819 | As1Li1Mn1 |
| | 115 | z2=1 |
| | 115 | z2=1 |
| | 12 | z2w,3=1 z4=2 |
| | 12 | z2w,1=1 z2w,2=1 z2w,3=1 |
| | 12 | z2w,1=1 z2w,2=1 z2w,3=1 |
| | 12 | z2w,3=1 z4=2 |
| | 12 | z2w,1=1 z2w,2=1 z2w,3=1 z4=3 |
| | 12 | z2w,3=1 z4=1 |
| | 12 | z2w,3=1 z4=1 |

Continued on next column

**TABLE XIX– continued**

| TQC | ICSD SSG | Stoich. Formula SIs |
|---|---|---|
| | 12 | z2w,1=1 z2w,2=1 z2w,3=1 z4=3 |
| | 12 | z2w,1=1 z2w,2=1 z4=3 |
| | 12 | z4=1 |
| | 12 | z4=1 |
| | 12 | z2w,1=1 z2w,2=1 z4=3 |
| | 12 | z2w,1=1 z2w,2=1 z4=2 |
| | 12 | z2w,1=1 z2w,2=1 z4=2 |
| 129.4.288 | 192818 | As1Li1Mn1 |
| | 115 | z2=1 |
| | 115 | z2=1 |
| | 12 | z2w,3=1 z4=2 |
| | 12 | z2w,1=1 z2w,2=1 z2w,3=1 |
| | 12 | z2w,1=1 z2w,2=1 z2w,3=1 |
| | 12 | z2w,3=1 z4=2 |
| | 12 | z2w,1=1 z2w,2=1 z2w,3=1 z4=3 |
| | 12 | z2w,3=1 z4=1 |
| | 12 | z2w,3=1 z4=1 |
| | 12 | z2w,1=1 z2w,2=1 z2w,3=1 z4=3 |
| | 12 | z2w,1=1 z2w,2=1 z4=3 |
| | 12 | z4=1 |
| | 12 | z4=1 |
| | 12 | z2w,1=1 z2w,2=1 z4=3 |
| | 12 | z2w,1=1 z2w,2=1 z4=2 |
| | 12 | z2w,1=1 z2w,2=1 z4=2 |
| 129.4.289 | 192817 | As1Li1Mn1 |
| | 115 | z2=1 |
| | 115 | z2=1 |
| | 12 | z2w,3=1 z4=2 |
| | 12 | z2w,1=1 z2w,2=1 z2w,3=1 |
| | 12 | z2w,1=1 z2w,2=1 z2w,3=1 |
| | 12 | z2w,3=1 z4=2 |
| | 12 | z2w,1=1 z2w,2=1 z2w,3=1 z4=3 |
| | 12 | z2w,3=1 z4=1 |
| | 12 | z2w,3=1 z4=1 |
| | 12 | z2w,1=1 z2w,2=1 z2w,3=1 z4=3 |
| | 12 | z2w,1=1 z2w,2=1 z4=3 |
| | 12 | z4=1 |
| | 12 | z4=1 |
| | 12 | z2w,1=1 z2w,2=1 z4=3 |
| | 12 | z2w,1=1 z2w,2=1 z4=2 |
| | 12 | z2w,1=1 z2w,2=1 z4=2 |
| 129.4.290 | 60745 | As1K1Mn1 |
| | 115 | z2=1 |
| | 115 | z2=1 |
| | 12 | z4=3 |
| | 12 | z2w,1=1 z2w,2=1 z4=1 |
| | 12 | z2w,1=1 z2w,2=1 z4=1 |
| | 12 | z4=3 |
| | 12 | z2w,1=1 z2w,2=1 |
| | 12 | z4=2 |
| | 12 | z4=2 |
| | 12 | z2w,1=1 z2w,2=1 |
| | 12 | z2w,1=1 z2w,2=1 z2w,3=1 |
| | 12 | z2w,3=1 z4=2 |
| | 12 | z2w,3=1 z4=2 |
| | 12 | z2w,1=1 z2w,2=1 z2w,3=1 |
| | 12 | z2w,3=1 z4=1 |
| | 12 | z2w,1=1 z2w,2=1 z2w,3=1 z4=3 |
| | 12 | z2w,1=1 z2w,2=1 z2w,3=1 z4=3 |
| | 12 | z2w,3=1 z4=1 |

Continued on next column

**TABLE XIX– continued**

| TQC | ICSD SSG | Stoich. Formula SIs |
|---|---|---|
| 129.4.291 | 60744 | As1K1Mn1 |
| | 115 | z2=1 |
| | 115 | z2=1 |
| | 12 | z4=3 |
| | 12 | z2w,1=1 z2w,2=1 z4=1 |
| | 12 | z2w,1=1 z2w,2=1 z4=1 |
| | 12 | z4=3 |
| | 12 | z2w,1=1 z2w,2=1 |
| | 12 | z4=2 |
| | 12 | z4=2 |
| | 12 | z2w,1=1 z2w,2=1 |
| | 12 | z2w,1=1 z2w,2=1 z2w,3=1 |
| | 12 | z2w,3=1 z4=2 |
| | 12 | z2w,3=1 z4=2 |
| | 12 | z2w,1=1 z2w,2=1 z2w,3=1 |
| | 12 | z2w,3=1 z4=1 |
| | 12 | z2w,1=1 z2w,2=1 z2w,3=1 z4=3 |
| | 12 | z2w,1=1 z2w,2=1 z2w,3=1 z4=3 |
| | 12 | z2w,3=1 z4=1 |
| 129.4.292 | 60743 | As1K1Mn1 |
| | 115 | z2=1 |
| | 115 | z2=1 |
| | 12 | z4=3 |
| | 12 | z2w,1=1 z2w,2=1 z4=1 |
| | 12 | z2w,1=1 z2w,2=1 z4=1 |
| | 12 | z4=3 |
| | 12 | z2w,1=1 z2w,2=1 |
| | 12 | z4=2 |
| | 12 | z4=2 |
| | 12 | z2w,1=1 z2w,2=1 |
| | 12 | z2w,1=1 z2w,2=1 z2w,3=1 |
| | 12 | z2w,3=1 z4=2 |
| | 12 | z2w,3=1 z4=2 |
| | 12 | z2w,1=1 z2w,2=1 z2w,3=1 |
| | 12 | z2w,3=1 z4=1 |
| | 12 | z2w,1=1 z2w,2=1 z2w,3=1 z4=3 |
| | 12 | z2w,1=1 z2w,2=1 z2w,3=1 z4=3 |
| | 12 | z2w,3=1 z4=1 |
| 129.4.293 | 41883 | As1K1Mn1 |
| | 115 | z2=1 |
| | 115 | z2=1 |
| | 12 | z4=3 |
| | 12 | z2w,1=1 z2w,2=1 z4=1 |
| | 12 | z2w,1=1 z2w,2=1 z4=1 |
| | 12 | z4=3 |
| | 12 | z2w,1=1 z2w,2=1 |
| | 12 | z4=2 |
| | 12 | z4=2 |
| | 12 | z2w,1=1 z2w,2=1 |
| | 12 | z2w,1=1 z2w,2=1 z2w,3=1 |
| | 12 | z2w,3=1 z4=2 |
| | 12 | z2w,3=1 z4=2 |
| | 12 | z2w,1=1 z2w,2=1 z2w,3=1 |
| | 12 | z2w,3=1 z4=1 |
| | 12 | z2w,1=1 z2w,2=1 z2w,3=1 z4=3 |
| | 12 | z2w,1=1 z2w,2=1 z2w,3=1 z4=3 |
| | 12 | z2w,3=1 z4=1 |
| 129.4.294 | 420206 | As1Gd1O1Zn1 |
| | 115 | z2=1 |
| | 115 | z2=1 |

Continued on next column

**TABLE XIX– continued**

| TQC | ICSD SSG | Stoich. Formula SIs |
|---|---|---|
| | 12 | z4=3 |
| | 12 | z2w,1=1 z2w,2=1 z4=1 |
| | 12 | z2w,1=1 z2w,2=1 z4=1 |
| | 12 | z4=3 |
| | 12 | z2w,1=1 z2w,2=1 |
| | 12 | z4=2 |
| | 12 | z4=2 |
| | 12 | z2w,1=1 z2w,2=1 |
| | 12 | z2w,1=1 z2w,2=1 z2w,3=1 |
| | 12 | z2w,3=1 z4=2 |
| | 12 | z2w,3=1 z4=2 |
| | 12 | z2w,1=1 z2w,2=1 z2w,3=1 |
| | 12 | z2w,3=1 z4=1 |
| | 12 | z2w,1=1 z2w,2=1 z2w,3=1 z4=3 |
| | 12 | z2w,1=1 z2w,2=1 z2w,3=1 z4=3 |
| | 12 | z2w,3=1 z4=1 |
| 129.4.295 | 262834 | As1Gd1O1Ru1 |
| | 115 | z2=1 |
| | 115 | z2=1 |
| | 12 | z2w,3=1 z4=2 |
| | 12 | z2w,1=1 z2w,2=1 z2w,3=1 |
| | 12 | z2w,1=1 z2w,2=1 z2w,3=1 |
| | 12 | z2w,3=1 z4=2 |
| | 12 | z2w,1=1 z2w,2=1 z2w,3=1 z4=3 |
| | 12 | z2w,3=1 z4=1 |
| | 12 | z2w,3=1 z4=1 |
| | 12 | z2w,1=1 z2w,2=1 z2w,3=1 z4=3 |
| | 12 | z2w,1=1 z2w,2=1 z4=3 |
| | 12 | z4=1 |
| | 12 | z4=1 |
| | 12 | z2w,1=1 z2w,2=1 z4=3 |
| | 12 | z2w,1=1 z2w,2=1 z4=2 |
| | 12 | z2w,1=1 z2w,2=1 z4=2 |
| 129.4.296 | 415628 | As1Fe2 |
| | 115 | z2=1 |
| | 115 | z2=1 |
| | 12 | z2w,3=1 z4=3 |
| | 12 | z2w,1=1 z2w,2=1 z2w,3=1 z4=1 |
| | 12 | z2w,1=1 z2w,2=1 z2w,3=1 z4=1 |
| | 12 | z2w,3=1 z4=3 |
| | 12 | z2w,1=1 z2w,2=1 z2w,3=1 |
| | 12 | z2w,3=1 z4=2 |
| | 12 | z2w,3=1 z4=2 |
| | 12 | z2w,1=1 z2w,2=1 z2w,3=1 |
| | 12 | z2w,1=1 z2w,2=1 |
| | 12 | z4=2 |
| | 12 | z4=2 |
| | 12 | z2w,1=1 z2w,2=1 |
| | 12 | z4=1 |
| | 12 | z2w,1=1 z2w,2=1 z4=3 |
| | 12 | z2w,1=1 z2w,2=1 z4=3 |
| | 12 | z4=1 |
| 129.4.297 | 16844 | As1Fe2 |
| | 115 | z2=1 |
| | 115 | z2=1 |
| | 12 | z2w,3=1 z4=3 |
| | 12 | z2w,1=1 z2w,2=1 z2w,3=1 z4=1 |
| | 12 | z2w,1=1 z2w,2=1 z2w,3=1 z4=1 |
| | 12 | z2w,3=1 z4=3 |
| | 12 | z2w,1=1 z2w,2=1 z2w,3=1 |

Continued on next column

**TABLE XIX– continued**

| TQC | ICSD SSG | Stoich. Formula SIs |
|---|---|---|
| | 12 | z2w,3=1 z4=2 |
| | 12 | z2w,3=1 z4=2 |
| | 12 | z2w,1=1 z2w,2=1 z2w,3=1 |
| | 12 | z2w,1=1 z2w,2=1 |
| | 12 | z4=2 |
| | 12 | z4=2 |
| | 12 | z2w,1=1 z2w,2=1 |
| | 12 | z4=1 |
| | 12 | z2w,1=1 z2w,2=1 z4=3 |
| | 12 | z2w,1=1 z2w,2=1 z4=3 |
| | 12 | z4=1 |
| 129.4.298 | 422001 | As1Fe1Nd1O1 |
| | 115 | z2=1 |
| | 115 | z2=1 |
| | 12 | z2w,3=1 |
| | 12 | z2w,1=1 z2w,2=1 z2w,3=1 z4=2 |
| | 12 | z2w,1=1 z2w,2=1 z2w,3=1 z4=2 |
| | 12 | z2w,3=1 |
| | 12 | z2w,1=1 z2w,2=1 z2w,3=1 z4=1 |
| | 12 | z2w,3=1 z4=3 |
| | 12 | z2w,3=1 z4=3 |
| | 12 | z2w,1=1 z2w,2=1 z2w,3=1 z4=1 |
| | 12 | z2w,1=1 z2w,2=1 z4=1 |
| | 12 | z4=3 |
| | 12 | z4=3 |
| | 12 | z2w,1=1 z2w,2=1 z4=1 |
| | 12 | z4=2 |
| | 12 | z2w,1=1 z2w,2=1 |
| | 12 | z2w,1=1 z2w,2=1 |
| | 12 | z4=2 |
| 129.4.299 | 163547 | As1Fe1Nd1O1 |
| | 115 | z2=1 |
| | 115 | z2=1 |
| | 12 | z2w,3=1 |
| | 12 | z2w,1=1 z2w,2=1 z2w,3=1 z4=2 |
| | 12 | z2w,1=1 z2w,2=1 z2w,3=1 z4=2 |
| | 12 | z2w,3=1 |
| | 12 | z2w,1=1 z2w,2=1 z2w,3=1 z4=1 |
| | 12 | z2w,3=1 z4=3 |
| | 12 | z2w,3=1 z4=3 |
| | 12 | z2w,1=1 z2w,2=1 z2w,3=1 z4=1 |
| | 12 | z2w,1=1 z2w,2=1 z4=1 |
| | 12 | z4=3 |
| | 12 | z4=3 |
| | 12 | z2w,1=1 z2w,2=1 z4=1 |
| | 12 | z4=2 |
| | 12 | z2w,1=1 z2w,2=1 |
| | 12 | z2w,1=1 z2w,2=1 |
| | 12 | z4=2 |
| 129.4.300 | 425010 | As1Fe1Gd1O1 |
| | 115 | z2=1 |
| | 115 | z2=1 |
| | 12 | z4=3 |
| | 12 | z2w,1=1 z2w,2=1 z4=1 |
| | 12 | z2w,1=1 z2w,2=1 z4=1 |
| | 12 | z4=3 |
| | 12 | z2w,1=1 z2w,2=1 |
| | 12 | z4=2 |
| | 12 | z4=2 |
| | 12 | z2w,1=1 z2w,2=1 |

Continued on next column

**TABLE XIX– continued**

| TQC | ICSD SSG | Stoich. Formula SIs |
|---|---|---|
| | 12 | z2w,1=1 z2w,2=1 z2w,3=1 |
| | 12 | z2w,3=1 z4=2 |
| | 12 | z2w,3=1 z4=2 |
| | 12 | z2w,1=1 z2w,2=1 z2w,3=1 |
| | 12 | z2w,3=1 z4=1 |
| | 12 | z2w,1=1 z2w,2=1 z2w,3=1 z4=3 |
| | 12 | z2w,1=1 z2w,2=1 z2w,3=1 z4=3 |
| | 12 | z2w,3=1 z4=1 |
| 129.4.301 | 425006 | As1Fe1Gd1O1 |
| | 115 | z2=1 |
| | 115 | z2=1 |
| | 12 | z4=3 |
| | 12 | z2w,1=1 z2w,2=1 z4=1 |
| | 12 | z2w,1=1 z2w,2=1 z4=1 |
| | 12 | z4=3 |
| | 12 | z2w,1=1 z2w,2=1 |
| | 12 | z4=2 |
| | 12 | z4=2 |
| | 12 | z2w,1=1 z2w,2=1 |
| | 12 | z2w,1=1 z2w,2=1 z2w,3=1 |
| | 12 | z2w,3=1 z4=2 |
| | 12 | z2w,3=1 z4=2 |
| | 12 | z2w,1=1 z2w,2=1 z2w,3=1 |
| | 12 | z2w,3=1 z4=1 |
| | 12 | z2w,1=1 z2w,2=1 z2w,3=1 z4=3 |
| | 12 | z2w,1=1 z2w,2=1 z2w,3=1 z4=3 |
| | 12 | z2w,3=1 z4=1 |
| 129.4.302 | 422003 | As1Fe1Gd1O1 |
| | 115 | z2=1 |
| | 115 | z2=1 |
| | 12 | z4=3 |
| | 12 | z2w,1=1 z2w,2=1 z4=1 |
| | 12 | z2w,1=1 z2w,2=1 z4=1 |
| | 12 | z4=3 |
| | 12 | z2w,1=1 z2w,2=1 |
| | 12 | z4=2 |
| | 12 | z4=2 |
| | 12 | z2w,1=1 z2w,2=1 |
| | 12 | z2w,1=1 z2w,2=1 z2w,3=1 |
| | 12 | z2w,3=1 z4=2 |
| | 12 | z2w,3=1 z4=2 |
| | 12 | z2w,1=1 z2w,2=1 z2w,3=1 |
| | 12 | z2w,3=1 z4=1 |
| | 12 | z2w,1=1 z2w,2=1 z2w,3=1 z4=3 |
| | 12 | z2w,1=1 z2w,2=1 z2w,3=1 z4=3 |
| | 12 | z2w,3=1 z4=1 |
| 129.4.303 | 189940 | As1F1Mn1Sr1 |
| | 115 | z2=1 |
| | 115 | z2=1 |
| | 12 | z2w,3=1 |
| | 12 | z2w,1=1 z2w,2=1 z2w,3=1 z4=2 |
| | 12 | z2w,1=1 z2w,2=1 z2w,3=1 z4=2 |
| | 12 | z2w,3=1 |
| | 12 | z2w,1=1 z2w,2=1 z2w,3=1 z4=1 |
| | 12 | z2w,3=1 z4=3 |
| | 12 | z2w,3=1 z4=3 |
| | 12 | z2w,1=1 z2w,2=1 z2w,3=1 z4=1 |
| | 12 | z2w,1=1 z2w,2=1 z4=1 |
| | 12 | z4=3 |
| | 12 | z4=3 |

Continued on next column

**TABLE XIX– continued**

| TQC | ICSD SSG | Stoich. Formula SIs |
|---|---|---|
| | 12 | z2w,1=1 z2w,2=1 z4=1 |
| | 12 | z4=2 |
| | 12 | z2w,1=1 z2w,2=1 |
| | 12 | z2w,1=1 z2w,2=1 |
| | 12 | z4=2 |
| 129.4.304 | 420203 | As1Er1O1Zn1 |
| | 115 | z2=1 |
| | 115 | z2=1 |
| | 12 | z2w,3=1 |
| | 12 | z2w,1=1 z2w,2=1 z2w,3=1 z4=2 |
| | 12 | z2w,1=1 z2w,2=1 z2w,3=1 z4=2 |
| | 12 | z2w,3=1 |
| | 12 | z2w,1=1 z2w,2=1 z2w,3=1 z4=1 |
| | 12 | z2w,3=1 z4=3 |
| | 12 | z2w,3=1 z4=3 |
| | 12 | z2w,1=1 z2w,2=1 z2w,3=1 z4=1 |
| | 12 | z2w,1=1 z2w,2=1 z4=1 |
| | 12 | z4=3 |
| | 12 | z4=3 |
| | 12 | z2w,1=1 z2w,2=1 z4=1 |
| | 12 | z4=2 |
| | 12 | z2w,1=1 z2w,2=1 |
| | 12 | z2w,1=1 z2w,2=1 |
| | 12 | z4=2 |
| 129.4.305 | 15142 | As1Cu2 |
| | 115 | z2=1 |
| | 115 | z2=1 |
| | 12 | z4=3 |
| | 12 | z2w,1=1 z2w,2=1 z4=1 |
| | 12 | z2w,1=1 z2w,2=1 z4=1 |
| | 12 | z4=3 |
| | 12 | z2w,1=1 z2w,2=1 |
| | 12 | z4=2 |
| | 12 | z4=2 |
| | 12 | z2w,1=1 z2w,2=1 |
| | 12 | z2w,1=1 z2w,2=1 z2w,3=1 |
| | 12 | z2w,3=1 z4=2 |
| | 12 | z2w,3=1 z4=2 |
| | 12 | z2w,1=1 z2w,2=1 z2w,3=1 |
| | 12 | z2w,3=1 z4=1 |
| | 12 | z2w,1=1 z2w,2=1 z2w,3=1 z4=3 |
| | 12 | z2w,1=1 z2w,2=1 z2w,3=1 z4=3 |
| | 12 | z2w,3=1 z4=1 |
| 129.4.306 | 423230 | As1Cu1Mn1 |
| | 115 | z2=1 |
| | 115 | z2=1 |
| | 12 | z4=2 |
| | 12 | z2w,1=1 z2w,2=1 |
| | 12 | z2w,1=1 z2w,2=1 |
| | 12 | z4=2 |
| | 12 | z2w,1=1 z2w,2=1 z4=3 |
| | 12 | z4=1 |
| | 12 | z4=1 |
| | 12 | z2w,1=1 z2w,2=1 z4=3 |
| | 12 | z2w,1=1 z2w,2=1 z2w,3=1 z4=3 |
| | 12 | z2w,3=1 z4=1 |
| | 12 | z2w,3=1 z4=1 |
| | 12 | z2w,1=1 z2w,2=1 z2w,3=1 z4=3 |
| | 12 | z2w,3=1 |
| | 12 | z2w,1=1 z2w,2=1 z2w,3=1 z4=2 |

Continued on next column

**TABLE XIX– continued**

| TQC | ICSD<br>SSG | Stoich. Formula<br>SIs |
|---|---|---|
| | 12 | z2w,1=1 z2w,2=1 z2w,3=1 z4=2 |
| | 12 | z2w,3=1 |
| 129.4.307 | 40573 | As1Cs1Mn1 |
| | 115 | z2=1 |
| | 115 | z2=1 |
| | 12 | z2w,3=1 z4=3 |
| | 12 | z2w,1=1 z2w,2=1 z2w,3=1 z4=1 |
| | 12 | z2w,1=1 z2w,2=1 z2w,3=1 z4=1 |
| | 12 | z2w,3=1 z4=3 |
| | 12 | z2w,1=1 z2w,2=1 z2w,3=1 |
| | 12 | z2w,3=1 z4=2 |
| | 12 | z2w,3=1 z4=2 |
| | 12 | z2w,1=1 z2w,2=1 z2w,3=1 |
| | 12 | z2w,1=1 z2w,2=1 |
| | 12 | z4=2 |
| | 12 | z4=2 |
| | 12 | z2w,1=1 z2w,2=1 |
| | 12 | z4=1 |
| | 12 | z2w,1=1 z2w,2=1 z4=3 |
| | 12 | z2w,1=1 z2w,2=1 z4=3 |
| | 12 | z4=1 |
| 129.4.308 | 40572 | As1Cs1Mn1 |
| | 115 | z2=1 |
| | 115 | z2=1 |
| | 12 | z2w,3=1 z4=3 |
| | 12 | z2w,1=1 z2w,2=1 z2w,3=1 z4=1 |
| | 12 | z2w,1=1 z2w,2=1 z2w,3=1 z4=1 |
| | 12 | z2w,3=1 z4=3 |
| | 12 | z2w,1=1 z2w,2=1 z2w,3=1 |
| | 12 | z2w,3=1 z4=2 |
| | 12 | z2w,3=1 z4=2 |
| | 12 | z2w,1=1 z2w,2=1 z2w,3=1 |
| | 12 | z2w,1=1 z2w,2=1 |
| | 12 | z4=2 |
| | 12 | z4=2 |
| | 12 | z2w,1=1 z2w,2=1 |
| | 12 | z4=1 |
| | 12 | z2w,1=1 z2w,2=1 z4=3 |
| | 12 | z2w,1=1 z2w,2=1 z4=3 |
| | 12 | z4=1 |
| 129.4.309 | 43223 | As1Cr2 |
| | 115 | z2=1 |
| | 115 | z2=1 |
| | 12 | z4=1 |
| | 12 | z2w,1=1 z2w,2=1 z4=3 |
| | 12 | z2w,1=1 z2w,2=1 z4=3 |
| | 12 | z4=1 |
| | 12 | z2w,1=1 z2w,2=1 z4=2 |
| | 12 | z2w,1=1 z2w,2=1 z4=2 |
| | 12 | z2w,1=1 z2w,2=1 z2w,3=1 z4=2 |
| | 12 | z2w,3=1 |
| | 12 | z2w,3=1 |
| | 12 | z2w,1=1 z2w,2=1 z2w,3=1 z4=2 |
| | 12 | z2w,3=1 z4=3 |
| | 12 | z2w,1=1 z2w,2=1 z2w,3=1 z4=1 |
| | 12 | z2w,1=1 z2w,2=1 z2w,3=1 z4=1 |
| | 12 | z2w,3=1 z4=3 |
| 129.4.310 | 42335 | As1Cr2 |
| | 115 | z2=1 |
| | 115 | z2=1 |

Continued on next column

**TABLE XIX– continued**

| TQC | ICSD<br>SSG | Stoich. Formula<br>SIs |
|---|---|---|
| | 12 | z4=1 |
| | 12 | z2w,1=1 z2w,2=1 z4=3 |
| | 12 | z2w,1=1 z2w,2=1 z4=3 |
| | 12 | z4=1 |
| | 12 | z2w,1=1 z2w,2=1 z4=2 |
| | 12 | z2w,1=1 z2w,2=1 z4=2 |
| | 12 | z2w,1=1 z2w,2=1 z2w,3=1 z4=2 |
| | 12 | z2w,3=1 |
| | 12 | z2w,3=1 |
| | 12 | z2w,1=1 z2w,2=1 z2w,3=1 z4=2 |
| | 12 | z2w,3=1 z4=3 |
| | 12 | z2w,1=1 z2w,2=1 z2w,3=1 z4=1 |
| | 12 | z2w,1=1 z2w,2=1 z2w,3=1 z4=1 |
| | 12 | z2w,3=1 z4=3 |
| 129.4.311 | 196921 | As1Cr2 |
| | 115 | z2=1 |
| | 115 | z2=1 |
| | 12 | z4=1 |
| | 12 | z2w,1=1 z2w,2=1 z4=3 |
| | 12 | z2w,1=1 z2w,2=1 z4=3 |
| | 12 | z4=1 |
| | 12 | z2w,1=1 z2w,2=1 z4=2 |
| | 12 | z2w,1=1 z2w,2=1 z4=2 |
| | 12 | z2w,1=1 z2w,2=1 z2w,3=1 z4=2 |
| | 12 | z2w,3=1 |
| | 12 | z2w,3=1 |
| | 12 | z2w,1=1 z2w,2=1 z2w,3=1 z4=2 |
| | 12 | z2w,3=1 z4=3 |
| | 12 | z2w,1=1 z2w,2=1 z2w,3=1 z4=1 |
| | 12 | z2w,1=1 z2w,2=1 z2w,3=1 z4=1 |
| | 12 | z2w,3=1 z4=3 |
| 129.4.312 | 420653 | As1Cr1Fe1O3Sr2 |
| | 115 | z2=1 |
| | 115 | z2=1 |
| | 12 | z2w,3=1 z4=1 |
| | 12 | z2w,1=1 z2w,2=1 z2w,3=1 z4=3 |
| | 12 | z2w,1=1 z2w,2=1 z2w,3=1 z4=3 |
| | 12 | z2w,3=1 z4=1 |
| | 12 | z2w,1=1 z2w,2=1 z2w,3=1 z4=2 |
| | 12 | z2w,3=1 |
| | 12 | z2w,3=1 |
| | 12 | z2w,1=1 z2w,2=1 z2w,3=1 z4=2 |
| | 12 | z2w,1=1 z2w,2=1 z4=2 |
| | 12 | z2w,1=1 z2w,2=1 z4=2 |
| | 12 | z4=3 |
| | 12 | z2w,1=1 z2w,2=1 z4=1 |
| | 12 | z2w,1=1 z2w,2=1 z4=1 |
| | 12 | z4=3 |
| 129.4.313 | 107932 | As1Co1Li1 |
| | 115 | z2=1 |
| | 115 | z2=1 |
| | 12 | z2w,3=1 z4=2 |
| | 12 | z2w,1=1 z2w,2=1 z2w,3=1 |
| | 12 | z2w,1=1 z2w,2=1 z2w,3=1 |
| | 12 | z2w,3=1 z4=2 |
| | 12 | z2w,1=1 z2w,2=1 z2w,3=1 z4=3 |
| | 12 | z2w,3=1 z4=1 |
| | 12 | z2w,3=1 z4=1 |
| | 12 | z2w,1=1 z2w,2=1 z2w,3=1 z4=3 |
| | 12 | z2w,1=1 z2w,2=1 z4=3 |

Continued on next column

TABLE XIX– continued

| TQC | ICSD SSG | Stoich. Formula SIs |
|---|---|---|
| | 12 | z4=1 |
| | 12 | z4=1 |
| | 12 | z2w,1=1 z2w,2=1 z4=3 |
| | 12 | z2w,1=1 z2w,2=1 z4=2 |
| | 12 | z2w,1=1 z2w,2=1 z4=2 |
| 129.4.314 | 188947 | As1Ce1O1Ru1 |
| | 115 | z2=1 |
| | 115 | z2=1 |
| | 12 | z2w,1=1 z2w,2=1 z4=2 |
| | 12 | z2w,1=1 z2w,2=1 z4=2 |
| | 12 | z2w,1=1 z2w,2=1 z4=1 |
| | 12 | z4=3 |
| | 12 | z4=3 |
| | 12 | z2w,1=1 z2w,2=1 z4=1 |
| | 12 | z2w,1=1 z2w,2=1 z2w,3=1 z4=1 |
| | 12 | z2w,3=1 z4=3 |
| | 12 | z2w,3=1 z4=3 |
| | 12 | z2w,1=1 z2w,2=1 z2w,3=1 z4=1 |
| | 12 | z2w,3=1 z4=2 |
| | 12 | z2w,1=1 z2w,2=1 z2w,3=1 |
| | 12 | z2w,1=1 z2w,2=1 z2w,3=1 |
| | 12 | z2w,3=1 z4=2 |
| 129.4.315 | 182625 | As1Ce1Ni1O1 |
| | 115 | z2=1 |
| | 115 | z2=1 |
| | 12 | z2w,1=1 z2w,2=1 z4=2 |
| | 12 | z2w,1=1 z2w,2=1 z4=2 |
| | 12 | z2w,1=1 z2w,2=1 z4=1 |
| | 12 | z4=3 |
| | 12 | z4=3 |
| | 12 | z2w,1=1 z2w,2=1 z4=1 |
| | 12 | z2w,1=1 z2w,2=1 z2w,3=1 z4=1 |
| | 12 | z2w,3=1 z4=3 |
| | 12 | z2w,3=1 z4=3 |
| | 12 | z2w,1=1 z2w,2=1 z2w,3=1 z4=1 |
| | 12 | z2w,3=1 z4=2 |
| | 12 | z2w,1=1 z2w,2=1 z2w,3=1 |
| | 12 | z2w,1=1 z2w,2=1 z2w,3=1 |
| | 12 | z2w,3=1 z4=2 |
| 129.4.316 | 174440 | As1Ce1Ni1O1 |
| | 115 | z2=1 |
| | 115 | z2=1 |
| | 12 | z2w,1=1 z2w,2=1 z4=2 |
| | 12 | z2w,1=1 z2w,2=1 z4=2 |
| | 12 | z2w,1=1 z2w,2=1 z4=1 |
| | 12 | z4=3 |
| | 12 | z4=3 |
| | 12 | z2w,1=1 z2w,2=1 z4=1 |
| | 12 | z2w,1=1 z2w,2=1 z2w,3=1 z4=1 |
| | 12 | z2w,3=1 z4=3 |
| | 12 | z2w,3=1 z4=3 |
| | 12 | z2w,1=1 z2w,2=1 z2w,3=1 z4=1 |
| | 12 | z2w,3=1 z4=2 |
| | 12 | z2w,1=1 z2w,2=1 z2w,3=1 |
| | 12 | z2w,1=1 z2w,2=1 z2w,3=1 |
| | 12 | z2w,3=1 z4=2 |
| 129.4.317 | 421999 | As1Ce1Fe1O1 |
| | 115 | z2=1 |
| | 115 | z2=1 |
| | 12 | z4=1 |

Continued on next column

TABLE XIX– continued

| TQC | ICSD SSG | Stoich. Formula SIs |
|---|---|---|
| | 12 | z2w,1=1 z2w,2=1 z4=3 |
| | 12 | z2w,1=1 z2w,2=1 z4=3 |
| | 12 | z4=1 |
| | 12 | z2w,1=1 z2w,2=1 z4=2 |
| | 12 | z2w,1=1 z2w,2=1 z4=2 |
| | 12 | z2w,1=1 z2w,2=1 z2w,3=1 z4=2 |
| | 12 | z2w,3=1 |
| | 12 | z2w,3=1 |
| | 12 | z2w,1=1 z2w,2=1 z2w,3=1 z4=2 |
| | 12 | z2w,3=1 z4=3 |
| | 12 | z2w,1=1 z2w,2=1 z2w,3=1 z4=1 |
| | 12 | z2w,1=1 z2w,2=1 z2w,3=1 z4=1 |
| | 12 | z2w,3=1 z4=3 |
| 129.4.318 | 188948 | As1Ce1Fe1O1 |
| | 115 | z2=1 |
| | 115 | z2=1 |
| | 12 | z4=1 |
| | 12 | z2w,1=1 z2w,2=1 z4=3 |
| | 12 | z2w,1=1 z2w,2=1 z4=3 |
| | 12 | z4=1 |
| | 12 | z2w,1=1 z2w,2=1 z4=2 |
| | 12 | z2w,1=1 z2w,2=1 z4=2 |
| | 12 | z2w,1=1 z2w,2=1 z2w,3=1 z4=2 |
| | 12 | z2w,3=1 |
| | 12 | z2w,3=1 |
| | 12 | z2w,1=1 z2w,2=1 z2w,3=1 z4=2 |
| | 12 | z2w,3=1 z4=3 |
| | 12 | z2w,1=1 z2w,2=1 z2w,3=1 z4=1 |
| | 12 | z2w,1=1 z2w,2=1 z2w,3=1 z4=1 |
| | 12 | z2w,3=1 z4=3 |
| 129.4.319 | 162817 | As1Ce1Fe1O1 |
| | 115 | z2=1 |
| | 115 | z2=1 |
| | 12 | z4=1 |
| | 12 | z2w,1=1 z2w,2=1 z4=3 |
| | 12 | z2w,1=1 z2w,2=1 z4=3 |
| | 12 | z4=1 |
| | 12 | z2w,1=1 z2w,2=1 z4=2 |
| | 12 | z2w,1=1 z2w,2=1 z4=2 |
| | 12 | z2w,1=1 z2w,2=1 z2w,3=1 z4=2 |
| | 12 | z2w,3=1 |
| | 12 | z2w,3=1 |
| | 12 | z2w,1=1 z2w,2=1 z2w,3=1 z4=2 |
| | 12 | z2w,3=1 z4=3 |
| | 12 | z2w,1=1 z2w,2=1 z2w,3=1 z4=1 |
| | 12 | z2w,1=1 z2w,2=1 z2w,3=1 z4=1 |
| | 12 | z2w,3=1 z4=3 |
| 129.4.320 | 88268 | As1Cd1Ce1O1 |
| | 115 | z2=1 |
| | 115 | z2=1 |
| | 12 | z2w,3=1 z4=2 |
| | 12 | z2w,1=1 z2w,2=1 z2w,3=1 |
| | 12 | z2w,1=1 z2w,2=1 z2w,3=1 |
| | 12 | z2w,3=1 z4=2 |
| | 12 | z2w,1=1 z2w,2=1 z2w,3=1 z4=3 |
| | 12 | z2w,3=1 z4=1 |
| | 12 | z2w,3=1 z4=1 |
| | 12 | z2w,1=1 z2w,2=1 z2w,3=1 z4=3 |
| | 12 | z2w,1=1 z2w,2=1 z4=3 |
| | 12 | z4=1 |

Continued on next column

**TABLE XIX– continued**

| TQC | ICSD SSG | Stoich. Formula SIs |
|---|---|---|
| | 12 | z4=1 |
| | 12 | z2w,1=1 z2w,2=1 z4=3 |
| | 12 | z2w,1=1 z2w,2=1 z4=2 |
| | 12 | z2w,1=1 z2w,2=1 z4=2 |
| 129.4.321 | 248365 | As1Ba1F1Mn1 |
| | 115 | z2=1 |
| | 115 | z2=1 |
| | 12 | z2w,1=1 z2w,2=1 z4=2 |
| | 12 | z2w,1=1 z2w,2=1 z4=2 |
| | 12 | z2w,1=1 z2w,2=1 z4=1 |
| | 12 | z4=3 |
| | 12 | z4=3 |
| | 12 | z2w,1=1 z2w,2=1 z4=1 |
| | 12 | z2w,1=1 z2w,2=1 z2w,3=1 z4=1 |
| | 12 | z2w,3=1 z4=3 |
| | 12 | z2w,3=1 z4=3 |
| | 12 | z2w,1=1 z2w,2=1 z2w,3=1 z4=1 |
| | 12 | z2w,3=1 z4=2 |
| | 12 | z2w,1=1 z2w,2=1 z2w,3=1 |
| | 12 | z2w,1=1 z2w,2=1 z2w,3=1 |
| | 12 | z2w,3=1 z4=2 |
| 129.4.322 | 191054 | Al2La1Pd2 |
| | 115 | z2=1 |
| | 115 | z2=1 |
| | 12 | z4=2 |
| | 12 | z2w,1=1 z2w,2=1 |
| | 12 | z2w,1=1 z2w,2=1 |
| | 12 | z4=2 |
| | 12 | z2w,1=1 z2w,2=1 z4=3 |
| | 12 | z4=1 |
| | 12 | z4=1 |
| | 12 | z2w,1=1 z2w,2=1 z4=3 |
| | 12 | z2w,1=1 z2w,2=1 z2w,3=1 z4=3 |
| | 12 | z2w,3=1 z4=1 |
| | 12 | z2w,3=1 z4=1 |
| | 12 | z2w,1=1 z2w,2=1 z2w,3=1 z4=3 |
| | 12 | z2w,3=1 |
| | 12 | z2w,1=1 z2w,2=1 z2w,3=1 z4=2 |
| | 12 | z2w,1=1 z2w,2=1 z2w,3=1 z4=2 |
| | 12 | z2w,3=1 |
| 129.4.323 | 55507 | Al2Au2Eu1 |
| | 115 | z2=1 |
| | 115 | z2=1 |
| | 12 | z4=3 |
| | 12 | z2w,1=1 z2w,2=1 z4=1 |
| | 12 | z2w,1=1 z2w,2=1 z4=1 |
| | 12 | z4=3 |
| | 12 | z2w,1=1 z2w,2=1 |
| | 12 | z4=2 |
| | 12 | z4=2 |
| | 12 | z2w,1=1 z2w,2=1 |
| | 12 | z2w,1=1 z2w,2=1 z2w,3=1 |
| | 12 | z2w,3=1 z4=2 |
| | 12 | z2w,3=1 z4=2 |
| | 12 | z2w,1=1 z2w,2=1 z2w,3=1 |
| | 12 | z2w,3=1 z4=1 |
| | 12 | z2w,1=1 z2w,2=1 z2w,3=1 z4=3 |
| | 12 | z2w,1=1 z2w,2=1 z2w,3=1 z4=3 |
| | 12 | z2w,3=1 z4=1 |
| 129.4.324 | 420343 | Ag5O4P1Te1 |
| | 115 | z2=1 |
| | 115 | z2=1 |
| | 12 | z2w,1=1 z2w,2=1 z4=2 |
| | 12 | z2w,1=1 z2w,2=1 z4=2 |
| | 12 | z2w,1=1 z2w,2=1 z4=1 |
| | 12 | z4=3 |
| | 12 | z4=3 |
| | 12 | z2w,1=1 z2w,2=1 z4=1 |
| | 12 | z2w,1=1 z2w,2=1 z2w,3=1 z4=1 |
| | 12 | z2w,3=1 z4=3 |
| | 12 | z2w,3=1 z4=3 |
| | 12 | z2w,1=1 z2w,2=1 z2w,3=1 z4=1 |
| | 12 | z2w,3=1 z4=2 |
| | 12 | z2w,1=1 z2w,2=1 z2w,3=1 |
| | 12 | z2w,1=1 z2w,2=1 z2w,3=1 |
| | 12 | z2w,3=1 z4=2 |
| 129.4.325 | 420342 | Ag5O4P1Se1 |
| | 115 | z2=1 |
| | 115 | z2=1 |
| | 12 | z2w,1=1 z2w,2=1 z4=2 |
| | 12 | z2w,1=1 z2w,2=1 z4=2 |
| | 12 | z2w,1=1 z2w,2=1 z4=1 |
| | 12 | z4=3 |
| | 12 | z4=3 |
| | 12 | z2w,1=1 z2w,2=1 z4=1 |
| | 12 | z2w,1=1 z2w,2=1 z2w,3=1 z4=1 |
| | 12 | z2w,3=1 z4=3 |
| | 12 | z2w,3=1 z4=3 |
| | 12 | z2w,1=1 z2w,2=1 z2w,3=1 z4=1 |
| | 12 | z2w,3=1 z4=2 |
| | 12 | z2w,1=1 z2w,2=1 z2w,3=1 |
| | 12 | z2w,1=1 z2w,2=1 z2w,3=1 |
| | 12 | z2w,3=1 z4=2 |
| 129.4.326 | 58390 | Ag1Zr1 |
| | 115 | z2=1 |
| | 115 | z2=1 |
| | 12 | z2w,1=1 z2w,2=1 z4=2 |
| | 12 | z2w,1=1 z2w,2=1 z4=2 |
| | 12 | z2w,1=1 z2w,2=1 z4=1 |
| | 12 | z4=3 |
| | 12 | z4=3 |
| | 12 | z2w,1=1 z2w,2=1 z4=1 |
| | 12 | z2w,1=1 z2w,2=1 z2w,3=1 z4=1 |
| | 12 | z2w,3=1 z4=3 |
| | 12 | z2w,3=1 z4=3 |
| | 12 | z2w,1=1 z2w,2=1 z2w,3=1 z4=1 |
| | 12 | z2w,3=1 z4=2 |
| | 12 | z2w,1=1 z2w,2=1 z2w,3=1 |
| | 12 | z2w,1=1 z2w,2=1 z2w,3=1 |
| | 12 | z2w,3=1 z4=2 |
| 129.4.327 | 95133 | Ag1Dy1Sb2 |
| | 115 | z2=1 |
| | 115 | z2=1 |
| | 12 | z4=3 |
| | 12 | z2w,1=1 z2w,2=1 z4=1 |
| | 12 | z2w,1=1 z2w,2=1 z4=1 |
| | 12 | z4=3 |
| | 12 | z2w,1=1 z2w,2=1 |
| | 12 | z4=2 |
| | 12 | z4=2 |

Continued on next column

**TABLE XIX– continued**

| TQC | ICSD<br>SSG | Stoich. Formula<br>SIs |
|---|---|---|
| | 12 | z2w,1=1 z2w,2=1 |
| | 12 | z2w,1=1 z2w,2=1 z2w,3=1 |
| | 12 | z2w,3=1 z4=2 |
| | 12 | z2w,3=1 z4=2 |
| | 12 | z2w,1=1 z2w,2=1 z2w,3=1 |
| | 12 | z2w,3=1 z4=1 |
| | 12 | z2w,1=1 z2w,2=1 z2w,3=1 z4=3 |
| | 12 | z2w,1=1 z2w,2=1 z2w,3=1 z4=3 |
| | 12 | z2w,3=1 z4=1 |
| 129.4.328 | 79450 | Ag1Ce1Sb2 |
| | 115 | z2=1 |
| | 115 | z2=1 |
| | 12 | z2w,1=1 z2w,2=1 z4=2 |
| | 12 | z2w,1=1 z2w,2=1 z4=2 |
| | 12 | z2w,1=1 z2w,2=1 z4=1 |
| | 12 | z4=3 |
| | 12 | z4=3 |
| | 12 | z2w,1=1 z2w,2=1 z4=1 |
| | 12 | z2w,1=1 z2w,2=1 z2w,3=1 z4=1 |
| | 12 | z2w,3=1 z4=3 |
| | 12 | z2w,3=1 z4=3 |
| | 12 | z2w,1=1 z2w,2=1 z2w,3=1 z4=1 |
| | 12 | z2w,3=1 z4=2 |
| | 12 | z2w,1=1 z2w,2=1 z2w,3=1 |
| | 12 | z2w,1=1 z2w,2=1 z2w,3=1 |
| | 12 | z2w,3=1 z4=2 |
| 129.4.329 | 79425 | Ag1Ce1Sb2 |
| | 115 | z2=1 |
| | 115 | z2=1 |
| | 12 | z2w,1=1 z2w,2=1 z4=2 |
| | 12 | z2w,1=1 z2w,2=1 z4=2 |
| | 12 | z2w,1=1 z2w,2=1 z4=1 |
| | 12 | z4=3 |
| | 12 | z4=3 |
| | 12 | z2w,1=1 z2w,2=1 z4=1 |
| | 12 | z2w,1=1 z2w,2=1 z2w,3=1 z4=1 |
| | 12 | z2w,3=1 z4=3 |
| | 12 | z2w,3=1 z4=3 |
| | 12 | z2w,1=1 z2w,2=1 z2w,3=1 z4=1 |
| | 12 | z2w,3=1 z4=2 |
| | 12 | z2w,1=1 z2w,2=1 z2w,3=1 |
| | 12 | z2w,1=1 z2w,2=1 z2w,3=1 |
| | 12 | z2w,3=1 z4=2 |
| 129.4.330 | 93150 | Ag1As2Ce1 |
| | 115 | z2=1 |
| | 115 | z2=1 |
| | 12 | z2w,3=1 |
| | 12 | z2w,1=1 z2w,2=1 z2w,3=1 z4=2 |
| | 12 | z2w,1=1 z2w,2=1 z2w,3=1 z4=2 |
| | 12 | z2w,3=1 |
| | 12 | z2w,1=1 z2w,2=1 z2w,3=1 z4=1 |
| | 12 | z2w,3=1 z4=3 |
| | 12 | z2w,3=1 z4=3 |
| | 12 | z2w,1=1 z2w,2=1 z2w,3=1 z4=1 |
| | 12 | z2w,1=1 z2w,2=1 z4=1 |
| | 12 | z4=3 |
| | 12 | z4=3 |
| | 12 | z2w,1=1 z2w,2=1 z4=1 |
| | 12 | z4=2 |
| | 12 | z2w,1=1 z2w,2=1 |

Continued on next column

**TABLE XIX– continued**

| TQC | ICSD<br>SSG | Stoich. Formula<br>SIs |
|---|---|---|
| | 12 | z2w,1=1 z2w,2=1 |
| | 12 | z4=2 |
| 130.4.1 | 71313 | Bi2Cu1O4 |
| | 116 | z2=1 |
| | 85 | z4=3 z2=1 |
| | 68 | z4=3 |
| 130.4.2 | 69308 | Bi2Cu1O4 |
| | 116 | z2=1 |
| | 85 | z4=3 z2=1 |
| | 68 | z4=3 |
| 130.4.3 | 69307 | Bi2Cu1O4 |
| | 116 | z2=1 |
| | 85 | z4=3 z2=1 |
| | 68 | z4=3 |
| 130.4.4 | 68815 | Bi2Cu1O4 |
| | 116 | z2=1 |
| | 85 | z4=3 z2=1 |
| | 68 | z4=3 |
| 130.4.5 | 68814 | Bi2Cu1O4 |
| | 116 | z2=1 |
| | 85 | z4=3 z2=1 |
| | 68 | z4=3 |
| 130.4.6 | 68813 | Bi2Cu1O4 |
| | 116 | z2=1 |
| | 85 | z4=3 z2=1 |
| | 68 | z4=3 |
| 130.4.7 | 68812 | Bi2Cu1O4 |
| | 116 | z2=1 |
| | 85 | z4=3 z2=1 |
| | 68 | z4=3 |
| 130.4.8 | 56390 | Bi2Cu1O4 |
| | 116 | z2=1 |
| | 85 | z4=3 z2=1 |
| | 68 | z4=3 |
| 130.4.9 | 202994 | Bi2Cu1O4 |
| | 116 | z2=1 |
| | 85 | z4=3 z2=1 |
| | 68 | z4=3 |
| 130.4.10 | 15865 | Bi2Cu1O4 |
| | 116 | z2=1 |
| | 85 | z4=3 z2=1 |
| | 68 | z4=3 |
| 131.4.1 | 23300 | C2B2La1 |
| | 115 | z2=1 |
| | 115 | z2=1 |
| | 47 | z2w,1=1 z2w,3=1 z4=1 |
| | 47 | z2w,1=1 z4=2 |
| | 47 | z2w,2=1 z4=2 |
| | 47 | z2w,2=1 z2w,3=1 z4=3 |
| 135.4.1 | 262627 | Co1O4Sb2 |
| | 114 | z2=1 |
| | 114 | z2=1 |
| | 66 | z2w,1=1 z2w,2=1 z4=1 |
| | 55 | z2w,3=1 z4=2 |
| | 55 | z2w,3=1 |
| 135.4.2 | 262626 | Co1O4Sb2 |
| | 114 | z2=1 |
| | 114 | z2=1 |
| | 66 | z2w,1=1 z2w,2=1 z4=1 |

Continued on next column

**TABLE XIX– continued**

| TQC | ICSD<br>SSG | Stoich. Formula<br>SIs |
|---|---|---|
| | 55 | z2w,3=1 z4=2 |
| | 55 | z2w,3=1 |
| 135.4.3 | 4287 | As2Cu1O4 |
| | 114 | z2=1 |
| | 114 | z2=1 |
| | 84 | z2w,1=1 z2w,2=1 |
| | 66 | z2w,1=1 z2w,2=1 z4=2 |
| | 55 | z2w,3=1 z4=3 |
| | 55 | z2w,3=1 z4=1 |
| 136.4.1 | 66665 | O2V1 |
| | 81 | z2=1 |
| | 81 | z2=1 |
| | 81 | z2=1 |
| | 81 | z2=1 |
| | 65 | z2w,1=1 z2w,2=1 z4=2 |
| | 65 | z2w,3=1 z4=1 |
| | 65 | z2w,1=1 z2w,2=1 z2w,3=1 z4=3 |
| | 65 | z2w,1=1 z2w,2=1 z2w,3=1 z4=1 |
| | 65 | z2w,3=1 z4=3 |
| | 65 | z2w,1=1 z2w,2=1 z4=2 |
| 136.4.2 | 4110 | O2V1 |
| | 81 | z2=1 |
| | 81 | z2=1 |
| | 81 | z2=1 |
| | 81 | z2=1 |
| | 65 | z2w,1=1 z2w,2=1 z4=2 |
| | 65 | z2w,3=1 z4=1 |
| | 65 | z2w,1=1 z2w,2=1 z2w,3=1 z4=3 |
| | 65 | z2w,1=1 z2w,2=1 z2w,3=1 z4=1 |
| | 65 | z2w,3=1 z4=3 |
| | 65 | z2w,1=1 z2w,2=1 z4=2 |
| 136.4.3 | 34421 | O2V1 |
| | 81 | z2=1 |
| | 81 | z2=1 |
| | 81 | z2=1 |
| | 81 | z2=1 |
| | 65 | z2w,1=1 z2w,2=1 z4=2 |
| | 65 | z2w,3=1 z4=1 |
| | 65 | z2w,1=1 z2w,2=1 z2w,3=1 z4=3 |
| | 65 | z2w,1=1 z2w,2=1 z2w,3=1 z4=1 |
| | 65 | z2w,3=1 z4=3 |
| | 65 | z2w,1=1 z2w,2=1 z4=2 |
| 136.4.4 | 34420 | O2V1 |
| | 81 | z2=1 |
| | 81 | z2=1 |
| | 81 | z2=1 |
| | 81 | z2=1 |
| | 65 | z2w,1=1 z2w,2=1 z4=2 |
| | 65 | z2w,3=1 z4=1 |
| | 65 | z2w,1=1 z2w,2=1 z2w,3=1 z4=3 |
| | 65 | z2w,1=1 z2w,2=1 z2w,3=1 z4=1 |
| | 65 | z2w,3=1 z4=3 |
| | 65 | z2w,1=1 z2w,2=1 z4=2 |
| 136.4.5 | 34418 | O2V1 |
| | 81 | z2=1 |
| | 81 | z2=1 |
| | 81 | z2=1 |
| | 81 | z2=1 |
| | 65 | z2w,1=1 z2w,2=1 z4=2 |

Continued on next column

**TABLE XIX– continued**

| TQC | ICSD<br>SSG | Stoich. Formula<br>SIs |
|---|---|---|
| | 65 | z2w,3=1 z4=1 |
| | 65 | z2w,1=1 z2w,2=1 z2w,3=1 z4=3 |
| | 65 | z2w,1=1 z2w,2=1 z2w,3=1 z4=1 |
| | 65 | z2w,3=1 z4=3 |
| | 65 | z2w,1=1 z2w,2=1 z4=2 |
| 136.4.6 | 237337 | O2V1 |
| | 81 | z2=1 |
| | 81 | z2=1 |
| | 81 | z2=1 |
| | 81 | z2=1 |
| | 65 | z2w,1=1 z2w,2=1 z4=2 |
| | 65 | z2w,3=1 z4=1 |
| | 65 | z2w,1=1 z2w,2=1 z2w,3=1 z4=3 |
| | 65 | z2w,1=1 z2w,2=1 z2w,3=1 z4=1 |
| | 65 | z2w,3=1 z4=3 |
| | 65 | z2w,1=1 z2w,2=1 z4=2 |
| 136.4.7 | 1504 | O2V1 |
| | 81 | z2=1 |
| | 81 | z2=1 |
| | 81 | z2=1 |
| | 81 | z2=1 |
| | 65 | z2w,1=1 z2w,2=1 z4=2 |
| | 65 | z2w,3=1 z4=1 |
| | 65 | z2w,1=1 z2w,2=1 z2w,3=1 z4=3 |
| | 65 | z2w,1=1 z2w,2=1 z2w,3=1 z4=1 |
| | 65 | z2w,3=1 z4=3 |
| | 65 | z2w,1=1 z2w,2=1 z4=2 |
| 136.4.8 | 10141 | O2V1 |
| | 81 | z2=1 |
| | 81 | z2=1 |
| | 81 | z2=1 |
| | 81 | z2=1 |
| | 65 | z2w,1=1 z2w,2=1 z4=2 |
| | 65 | z2w,3=1 z4=1 |
| | 65 | z2w,1=1 z2w,2=1 z2w,3=1 z4=3 |
| | 65 | z2w,1=1 z2w,2=1 z2w,3=1 z4=1 |
| | 65 | z2w,3=1 z4=3 |
| | 65 | z2w,1=1 z2w,2=1 z4=2 |
| 136.4.9 | 154021 | O2Re1 |
| | 81 | z2=1 |
| | 81 | z2=1 |
| | 81 | z2=1 |
| | 81 | z2=1 |
| | 65 | z2w,1=1 z2w,2=1 z4=2 |
| | 65 | z2w,3=1 z4=1 |
| | 65 | z2w,1=1 z2w,2=1 z2w,3=1 z4=3 |
| | 65 | z2w,1=1 z2w,2=1 z2w,3=1 z4=1 |
| | 65 | z2w,3=1 z4=3 |
| | 65 | z2w,1=1 z2w,2=1 z4=2 |
| 136.4.10 | 189495 | Ni4P2Y1 |
| | 81 | z2=1 |
| | 81 | z2=1 |
| | 81 | z2=1 |
| | 81 | z2=1 |
| | 65 | z4=3 |
| | 65 | z2w,1=1 z2w,2=1 z4=1 |
| | 65 | z2w,3=1 |
| | 65 | z2w,1=1 z2w,2=1 z2w,3=1 z4=2 |
| | 65 | z2w,1=1 z2w,2=1 z2w,3=1 |

Continued on next column

**TABLE XIX– continued**

| TQC | ICSD SSG | Stoich. Formula SIs |
|---|---|---|
| | 65 | z2w,3=1 z4=2 |
| | 65 | z2w,1=1 z2w,2=1 z4=1 |
| | 65 | z4=3 |
| 136.4.11 | 424579 | Ni1O2Rb3 |
| | 81 | z2=1 |
| | 81 | z2=1 |
| | 81 | z2=1 |
| | 81 | z2=1 |
| | 65 | z2w,1=1 z2w,2=1 z4=2 |
| | 65 | z2w,3=1 z4=1 |
| | 65 | z2w,1=1 z2w,2=1 z2w,3=1 z4=3 |
| | 65 | z2w,1=1 z2w,2=1 z2w,3=1 z4=1 |
| | 65 | z2w,3=1 z4=3 |
| | 65 | z2w,1=1 z2w,2=1 z4=2 |
| 136.4.12 | 75198 | Nb1O2 |
| | 81 | z2=1 |
| | 81 | z2=1 |
| | 81 | z2=1 |
| | 81 | z2=1 |
| | 65 | z2w,1=1 z2w,2=1 z4=2 |
| | 65 | z2w,1=1 z2w,2=1 z2w,3=1 z4=3 |
| | 65 | z2w,3=1 z4=1 |
| | 65 | z2w,3=1 z4=3 |
| | 65 | z2w,1=1 z2w,2=1 z2w,3=1 z4=1 |
| | 65 | z2w,1=1 z2w,2=1 z4=2 |
| 136.4.13 | 73716 | Mn1O2 |
| | 81 | z2=1 |
| | 81 | z2=1 |
| | 81 | z2=1 |
| | 81 | z2=1 |
| | 65 | z2w,1=1 z2w,2=1 z4=2 |
| | 65 | z2w,3=1 z4=1 |
| | 65 | z2w,1=1 z2w,2=1 z2w,3=1 z4=3 |
| | 65 | z2w,1=1 z2w,2=1 z2w,3=1 z4=1 |
| | 65 | z2w,3=1 z4=3 |
| | 65 | z2w,1=1 z2w,2=1 z4=2 |
| 136.4.14 | 393 | Mn1O2 |
| | 81 | z2=1 |
| | 81 | z2=1 |
| | 81 | z2=1 |
| | 81 | z2=1 |
| | 65 | z2w,1=1 z2w,2=1 z4=2 |
| | 65 | z2w,3=1 z4=1 |
| | 65 | z2w,1=1 z2w,2=1 z2w,3=1 z4=3 |
| | 65 | z2w,1=1 z2w,2=1 z2w,3=1 z4=1 |
| | 65 | z2w,3=1 z4=3 |
| | 65 | z2w,1=1 z2w,2=1 z4=2 |
| 136.4.15 | 262579 | K3Ni1O2 |
| | 81 | z2=1 |
| | 81 | z2=1 |
| | 81 | z2=1 |
| | 81 | z2=1 |
| | 65 | z2w,1=1 z2w,2=1 z4=2 |
| | 65 | z2w,3=1 z4=1 |
| | 65 | z2w,1=1 z2w,2=1 z2w,3=1 z4=3 |
| | 65 | z2w,1=1 z2w,2=1 z2w,3=1 z4=1 |
| | 65 | z2w,3=1 z4=3 |
| | 65 | z2w,1=1 z2w,2=1 z4=2 |
| 136.4.16 | 84577 | Ir1O2 |

Continued on next column

**TABLE XIX– continued**

| TQC | ICSD SSG | Stoich. Formula SIs |
|---|---|---|
| | 81 | z2=1 |
| | 81 | z2=1 |
| | 81 | z2=1 |
| | 81 | z2=1 |
| | 65 | z2w,1=1 z2w,2=1 z4=2 |
| | 65 | z2w,3=1 z4=1 |
| | 65 | z2w,1=1 z2w,2=1 z2w,3=1 z4=3 |
| | 65 | z2w,1=1 z2w,2=1 z2w,3=1 z4=1 |
| | 65 | z2w,3=1 z4=3 |
| | 65 | z2w,1=1 z2w,2=1 z4=2 |
| 136.4.17 | 261030 | Fe4Si2Tm1 |
| | 81 | z2=1 |
| | 81 | z2=1 |
| | 81 | z2=1 |
| | 81 | z2=1 |
| | 65 | z2w,1=1 z2w,2=1 z4=1 |
| | 65 | z4=3 |
| | 65 | z2w,1=1 z2w,2=1 z2w,3=1 z4=2 |
| | 65 | z2w,3=1 |
| | 65 | z2w,3=1 z4=2 |
| | 65 | z2w,1=1 z2w,2=1 z2w,3=1 |
| | 65 | z4=3 |
| | 65 | z2w,1=1 z2w,2=1 z4=1 |
| 136.4.18 | 247481 | Fe4Si2Tm1 |
| | 81 | z2=1 |
| | 81 | z2=1 |
| | 81 | z2=1 |
| | 81 | z2=1 |
| | 65 | z2w,1=1 z2w,2=1 z4=1 |
| | 65 | z4=3 |
| | 65 | z2w,1=1 z2w,2=1 z2w,3=1 z4=2 |
| | 65 | z2w,3=1 |
| | 65 | z2w,3=1 z4=2 |
| | 65 | z2w,1=1 z2w,2=1 z2w,3=1 |
| | 65 | z4=3 |
| | 65 | z2w,1=1 z2w,2=1 z4=1 |
| 136.4.19 | 68525 | Fe4P2Sc1 |
| | 81 | z2=1 |
| | 81 | z2=1 |
| | 81 | z2=1 |
| | 81 | z2=1 |
| | 65 | z4=3 |
| | 65 | z2w,1=1 z2w,2=1 z4=1 |
| | 65 | z2w,3=1 |
| | 65 | z2w,1=1 z2w,2=1 z2w,3=1 z4=2 |
| | 65 | z2w,1=1 z2w,2=1 z2w,3=1 |
| | 65 | z2w,3=1 z4=2 |
| | 65 | z2w,1=1 z2w,2=1 z4=1 |
| | 65 | z4=3 |
| 136.4.20 | 186045 | Fe4Ge2Lu1 |
| | 81 | z2=1 |
| | 81 | z2=1 |
| | 81 | z2=1 |
| | 81 | z2=1 |
| | 65 | z2w,1=1 z2w,2=1 z4=3 |
| | 65 | z4=1 |
| | 65 | z2w,1=1 z2w,2=1 z2w,3=1 |
| | 65 | z2w,3=1 z4=2 |
| | 65 | z2w,3=1 |
| | 65 | z2w,1=1 z2w,2=1 z2w,3=1 z4=2 |

Continued on next column

TABLE XIX– continued

| TQC | ICSD<br>SSG | Stoich. Formula<br>SIs |
|---|---|---|
| | 65 | z4=1 |
| | 65 | z2w,1=1 z2w,2=1 z4=3 |
| 136.4.21 | 290252 | F6Li2V1 |
| | 81 | z2=1 |
| | 81 | z2=1 |
| | 81 | z2=1 |
| | 81 | z2=1 |
| | 65 | z2w,1=1 z2w,2=1 z4=2 |
| | 65 | z2w,1=1 z2w,2=1 z2w,3=1 z4=3 |
| | 65 | z2w,3=1 z4=1 |
| | 65 | z2w,3=1 z4=3 |
| | 65 | z2w,1=1 z2w,2=1 z2w,3=1 z4=1 |
| | 65 | z2w,1=1 z2w,2=1 z4=2 |
| 136.4.22 | 62769 | F2V1 |
| | 81 | z2=1 |
| | 81 | z2=1 |
| | 81 | z2=1 |
| | 81 | z2=1 |
| | 65 | z2w,1=1 z2w,2=1 z4=2 |
| | 65 | z2w,1=1 z2w,2=1 z2w,3=1 z4=3 |
| | 65 | z2w,3=1 z4=1 |
| | 65 | z2w,3=1 z4=3 |
| | 65 | z2w,1=1 z2w,2=1 z2w,3=1 z4=1 |
| | 65 | z2w,1=1 z2w,2=1 z4=2 |
| 136.4.23 | 62768 | F2V1 |
| | 81 | z2=1 |
| | 81 | z2=1 |
| | 81 | z2=1 |
| | 81 | z2=1 |
| | 65 | z2w,1=1 z2w,2=1 z4=2 |
| | 65 | z2w,1=1 z2w,2=1 z2w,3=1 z4=3 |
| | 65 | z2w,3=1 z4=1 |
| | 65 | z2w,3=1 z4=3 |
| | 65 | z2w,1=1 z2w,2=1 z2w,3=1 z4=1 |
| | 65 | z2w,1=1 z2w,2=1 z4=2 |
| 136.4.24 | 38069 | F2V1 |
| | 81 | z2=1 |
| | 81 | z2=1 |
| | 81 | z2=1 |
| | 81 | z2=1 |
| | 65 | z2w,1=1 z2w,2=1 z4=2 |
| | 65 | z2w,3=1 z4=1 |
| | 65 | z2w,1=1 z2w,2=1 z2w,3=1 z4=3 |
| | 65 | z2w,1=1 z2w,2=1 z2w,3=1 z4=1 |
| | 65 | z2w,3=1 z4=3 |
| | 65 | z2w,1=1 z2w,2=1 z4=2 |
| 136.4.25 | 32552 | F2V1 |
| | 81 | z2=1 |
| | 81 | z2=1 |
| | 81 | z2=1 |
| | 81 | z2=1 |
| | 65 | z2w,1=1 z2w,2=1 z4=2 |
| | 65 | z2w,1=1 z2w,2=1 z2w,3=1 z4=3 |
| | 65 | z2w,3=1 z4=1 |
| | 65 | z2w,3=1 z4=3 |
| | 65 | z2w,1=1 z2w,2=1 z2w,3=1 z4=1 |
| | 65 | z2w,1=1 z2w,2=1 z4=2 |
| 136.4.26 | 201245 | F2V1 |
| | 81 | z2=1 |

Continued on next column

TABLE XIX– continued

| TQC | ICSD<br>SSG | Stoich. Formula<br>SIs |
|---|---|---|
| | 81 | z2=1 |
| | 81 | z2=1 |
| | 81 | z2=1 |
| | 65 | z2w,1=1 z2w,2=1 z4=2 |
| | 65 | z2w,1=1 z2w,2=1 z2w,3=1 z4=3 |
| | 65 | z2w,3=1 z4=1 |
| | 65 | z2w,3=1 z4=3 |
| | 65 | z2w,1=1 z2w,2=1 z2w,3=1 z4=1 |
| | 65 | z2w,1=1 z2w,2=1 z4=2 |
| 136.4.27 | 9165 | F2Mn1 |
| | 81 | z2=1 |
| | 81 | z2=1 |
| | 81 | z2=1 |
| | 81 | z2=1 |
| | 65 | z2w,1=1 z2w,2=1 z4=2 |
| | 65 | z2w,3=1 z4=1 |
| | 65 | z2w,1=1 z2w,2=1 z2w,3=1 z4=3 |
| | 65 | z2w,1=1 z2w,2=1 z2w,3=1 z4=1 |
| | 65 | z2w,3=1 z4=3 |
| | 65 | z2w,1=1 z2w,2=1 z4=2 |
| 136.4.28 | 68740 | F2Mn1 |
| | 81 | z2=1 |
| | 81 | z2=1 |
| | 81 | z2=1 |
| | 81 | z2=1 |
| | 65 | z2w,1=1 z2w,2=1 z4=2 |
| | 65 | z2w,3=1 z4=1 |
| | 65 | z2w,1=1 z2w,2=1 z2w,3=1 z4=3 |
| | 65 | z2w,1=1 z2w,2=1 z2w,3=1 z4=1 |
| | 65 | z2w,3=1 z4=3 |
| | 65 | z2w,1=1 z2w,2=1 z4=2 |
| 136.4.29 | 68737 | F2Mn1 |
| | 81 | z2=1 |
| | 81 | z2=1 |
| | 81 | z2=1 |
| | 81 | z2=1 |
| | 65 | z2w,1=1 z2w,2=1 z4=2 |
| | 65 | z2w,3=1 z4=1 |
| | 65 | z2w,1=1 z2w,2=1 z2w,3=1 z4=3 |
| | 65 | z2w,1=1 z2w,2=1 z2w,3=1 z4=1 |
| | 65 | z2w,3=1 z4=3 |
| | 65 | z2w,1=1 z2w,2=1 z4=2 |
| 136.4.30 | 68736 | F2Mn1 |
| | 81 | z2=1 |
| | 81 | z2=1 |
| | 81 | z2=1 |
| | 81 | z2=1 |
| | 65 | z2w,1=1 z2w,2=1 z4=2 |
| | 65 | z2w,3=1 z4=1 |
| | 65 | z2w,1=1 z2w,2=1 z2w,3=1 z4=3 |
| | 65 | z2w,1=1 z2w,2=1 z2w,3=1 z4=1 |
| | 65 | z2w,3=1 z4=3 |
| | 65 | z2w,1=1 z2w,2=1 z4=2 |
| 136.4.31 | 63654 | F2Mn1 |
| | 81 | z2=1 |
| | 81 | z2=1 |
| | 81 | z2=1 |
| | 81 | z2=1 |
| | 65 | z2w,1=1 z2w,2=1 z4=2 |

Continued on next column

**TABLE XIX– continued**

| TQC | ICSD<br>SSG | Stoich. Formula<br>SIs |
|---|---|---|
| | 65 | z2w,3=1 z4=1 |
| | 65 | z2w,1=1 z2w,2=1 z2w,3=1 z4=3 |
| | 65 | z2w,1=1 z2w,2=1 z2w,3=1 z4=1 |
| | 65 | z2w,3=1 z4=3 |
| | 65 | z2w,1=1 z2w,2=1 z4=2 |
| 136.4.32 | 200642 | F2Mn1 |
| | 81 | z2=1 |
| | 81 | z2=1 |
| | 81 | z2=1 |
| | 81 | z2=1 |
| | 65 | z2w,1=1 z2w,2=1 z4=2 |
| | 65 | z2w,3=1 z4=1 |
| | 65 | z2w,1=1 z2w,2=1 z2w,3=1 z4=3 |
| | 65 | z2w,1=1 z2w,2=1 z2w,3=1 z4=1 |
| | 65 | z2w,3=1 z4=3 |
| | 65 | z2w,1=1 z2w,2=1 z4=2 |
| 136.4.33 | 200641 | F2Mn1 |
| | 81 | z2=1 |
| | 81 | z2=1 |
| | 81 | z2=1 |
| | 81 | z2=1 |
| | 65 | z2w,1=1 z2w,2=1 z4=2 |
| | 65 | z2w,3=1 z4=1 |
| | 65 | z2w,1=1 z2w,2=1 z2w,3=1 z4=3 |
| | 65 | z2w,1=1 z2w,2=1 z2w,3=1 z4=1 |
| | 65 | z2w,3=1 z4=3 |
| | 65 | z2w,1=1 z2w,2=1 z4=2 |
| 136.4.34 | 200640 | F2Mn1 |
| | 81 | z2=1 |
| | 81 | z2=1 |
| | 81 | z2=1 |
| | 81 | z2=1 |
| | 65 | z2w,1=1 z2w,2=1 z4=2 |
| | 65 | z2w,3=1 z4=1 |
| | 65 | z2w,1=1 z2w,2=1 z2w,3=1 z4=3 |
| | 65 | z2w,1=1 z2w,2=1 z2w,3=1 z4=1 |
| | 65 | z2w,3=1 z4=3 |
| | 65 | z2w,1=1 z2w,2=1 z4=2 |
| 136.4.35 | 84789 | Cu1O6Sb2 |
| | 81 | z2=1 |
| | 81 | z2=1 |
| | 81 | z2=1 |
| | 81 | z2=1 |
| | 65 | z2w,1=1 z2w,2=1 z4=2 |
| | 65 | z2w,3=1 z4=1 |
| | 65 | z2w,1=1 z2w,2=1 z2w,3=1 z4=3 |
| | 65 | z2w,1=1 z2w,2=1 z2w,3=1 z4=1 |
| | 65 | z2w,3=1 z4=3 |
| | 65 | z2w,1=1 z2w,2=1 z4=2 |
| 136.4.36 | 424578 | Cs3Ni1O2 |
| | 81 | z2=1 |
| | 81 | z2=1 |
| | 81 | z2=1 |
| | 81 | z2=1 |
| | 65 | z2w,1=1 z2w,2=1 z4=2 |
| | 65 | z2w,3=1 z4=1 |
| | 65 | z2w,1=1 z2w,2=1 z2w,3=1 z4=3 |
| | 65 | z2w,1=1 z2w,2=1 z2w,3=1 z4=1 |
| | 65 | z2w,3=1 z4=3 |

Continued on next column

**TABLE XIX– continued**

| TQC | ICSD<br>SSG | Stoich. Formula<br>SIs |
|---|---|---|
| | 65 | z2w,1=1 z2w,2=1 z4=2 |
| 136.4.37 | 203094 | Co1O6Sb2 |
| | 81 | z2=1 |
| | 81 | z2=1 |
| | 81 | z2=1 |
| | 81 | z2=1 |
| | 65 | z2w,1=1 z2w,2=1 z4=2 |
| | 65 | z2w,3=1 z4=1 |
| | 65 | z2w,1=1 z2w,2=1 z2w,3=1 z4=3 |
| | 65 | z2w,1=1 z2w,2=1 z2w,3=1 z4=1 |
| | 65 | z2w,3=1 z4=3 |
| | 65 | z2w,1=1 z2w,2=1 z4=2 |
| 136.4.38 | 9167 | Co1F2 |
| | 81 | z2=1 |
| | 81 | z2=1 |
| | 81 | z2=1 |
| | 81 | z2=1 |
| | 65 | z2w,1=1 z2w,2=1 z4=2 |
| | 65 | z2w,3=1 z4=1 |
| | 65 | z2w,1=1 z2w,2=1 z2w,3=1 z4=3 |
| | 65 | z2w,1=1 z2w,2=1 z2w,3=1 z4=1 |
| | 65 | z2w,3=1 z4=3 |
| | 65 | z2w,1=1 z2w,2=1 z4=2 |
| 136.4.39 | 73460 | Co1F2 |
| | 81 | z2=1 |
| | 81 | z2=1 |
| | 81 | z2=1 |
| | 81 | z2=1 |
| | 65 | z2w,1=1 z2w,2=1 z4=2 |
| | 65 | z2w,3=1 z4=1 |
| | 65 | z2w,1=1 z2w,2=1 z2w,3=1 z4=3 |
| | 65 | z2w,1=1 z2w,2=1 z2w,3=1 z4=1 |
| | 65 | z2w,3=1 z4=3 |
| | 65 | z2w,1=1 z2w,2=1 z4=2 |
| 136.4.40 | 280604 | Co1F2 |
| | 81 | z2=1 |
| | 81 | z2=1 |
| | 81 | z2=1 |
| | 81 | z2=1 |
| | 65 | z2w,1=1 z2w,2=1 z4=2 |
| | 65 | z2w,3=1 z4=1 |
| | 65 | z2w,1=1 z2w,2=1 z2w,3=1 z4=3 |
| | 65 | z2w,1=1 z2w,2=1 z2w,3=1 z4=1 |
| | 65 | z2w,3=1 z4=3 |
| | 65 | z2w,1=1 z2w,2=1 z4=2 |
| 136.4.41 | 14144 | Co1F2 |
| | 81 | z2=1 |
| | 81 | z2=1 |
| | 81 | z2=1 |
| | 81 | z2=1 |
| | 65 | z2w,1=1 z2w,2=1 z4=2 |
| | 65 | z2w,3=1 z4=1 |
| | 65 | z2w,1=1 z2w,2=1 z2w,3=1 z4=3 |
| | 65 | z2w,1=1 z2w,2=1 z2w,3=1 z4=1 |
| | 65 | z2w,3=1 z4=3 |
| | 65 | z2w,1=1 z2w,2=1 z4=2 |
| 137.4.1 | 62688 | Co1Li6O4 |
| | 115 | z2=1 |
| | 115 | z2=1 |

Continued on next column

**TABLE XIX– continued**

| TQC | ICSD SSG | Stoich. Formula SIs |
|---|---|---|
| | 2 | z2w,2=1 z4=3 |
| | 2 | z2w,1=1 z4=3 |
| | 2 | z2w,1=1 z2w,2=1 z4=2 |
| | 2 | z2w,1=1 z2w,3=1 z4=3 |
| | 2 | z2w,1=1 z2w,2=1 z2w,3=1 z4=2 |
| | 2 | z2w,3=1 z4=2 |
| | 2 | z2w,2=1 z2w,3=1 z4=1 |
| | 2 | z2w,2=1 z2w,3=1 z4=3 |
| | 2 | z2w,3=1 z4=2 |
| | 2 | z2w,1=1 z2w,2=1 z2w,3=1 z4=2 |
| | 2 | z2w,1=1 z2w,3=1 z4=1 |
| | 2 | z2w,1=1 z2w,2=1 z4=2 |
| | 2 | z2w,1=1 z4=1 |
| | 2 | z2w,2=1 z4=1 |
| | 2 | z2w,3=1 z4=3 |
| | 2 | z2w,2=1 z2w,3=1 z4=2 |
| | 2 | z2w,1=1 z2w,3=1 z4=2 |
| | 2 | z2w,1=1 z2w,2=1 z2w,3=1 z4=1 |
| | 2 | z2w,1=1 z4=2 |
| | 2 | z2w,1=1 z2w,2=1 z4=1 |
| | 2 | z4=1 |
| | 2 | z2w,2=1 |
| | 2 | z2w,2=1 z4=2 |
| | 2 | z4=1 |
| | 2 | z2w,1=1 z2w,2=1 z4=1 |
| | 2 | z2w,1=1 |
| | 2 | z2w,1=1 z2w,2=1 z2w,3=1 z4=1 |
| | 2 | z2w,1=1 z2w,3=1 |
| | 2 | z2w,2=1 z2w,3=1 |
| | 2 | z2w,3=1 z4=3 |
| | 2 | z2w,1=1 z2w,2=1 z4=3 |
| | 2 | z2w,1=1 z4=2 |
| | 2 | z2w,2=1 z4=2 |
| | 2 | z4=1 |
| | 2 | z2w,2=1 z2w,3=1 z4=2 |
| | 2 | z2w,3=1 z4=1 |
| | 2 | z2w,1=1 z2w,2=1 z2w,3=1 z4=1 |
| | 2 | z2w,1=1 z2w,3=1 |
| | 2 | z2w,1=1 z2w,3=1 z4=2 |
| | 2 | z2w,1=1 z2w,2=1 z2w,3=1 z4=1 |
| | 2 | z2w,3=1 z4=1 |
| | 2 | z2w,2=1 z2w,3=1 |
| | 2 | z4=1 |
| | 2 | z2w,2=1 |
| | 2 | z2w,1=1 |
| | 2 | z2w,1=1 z2w,2=1 z4=3 |
| | 2 | z2w,1=1 z2w,2=1 z2w,3=1 z4=2 |
| | 2 | z2w,1=1 z2w,3=1 z4=1 |
| | 2 | z2w,2=1 z2w,3=1 z4=1 |
| | 2 | z2w,3=1 |
| | 2 | z2w,2=1 z4=1 |
| | 2 | z2w,1=1 z2w,2=1 |
| | 2 | z2w,1=1 z4=3 |
| | 2 | z2w,1=1 z4=1 |
| | 2 | z2w,1=1 z2w,2=1 |
| | 2 | z2w,2=1 z4=3 |
| | 2 | z2w,3=1 |
| | 2 | z2w,2=1 z2w,3=1 z4=3 |
| | 2 | z2w,1=1 z2w,3=1 z4=3 |
| | 2 | z2w,1=1 z2w,2=1 z2w,3=1 z4=2 |

Continued on next column

| TQC | ICSD SSG | Stoich. Formula SIs |
|---|---|---|
| | 2 | z2w,1=1 z2w,2=1 z2w,3=1 z4=3 |
| | 2 | z2w,1=1 z2w,3=1 z4=2 |
| | 2 | z2w,2=1 z2w,3=1 z4=2 |
| | 2 | z2w,3=1 z4=1 |
| | 2 | z2w,2=1 z4=2 |
| | 2 | z4=1 |
| | 2 | z2w,1=1 z2w,2=1 z4=1 |
| | 2 | z2w,1=1 |
| | 2 | z2w,1=1 z4=2 |
| | 2 | z2w,1=1 z2w,2=1 z4=1 |
| | 2 | z4=1 |
| | 2 | z2w,2=1 |
| | 2 | z2w,3=1 z4=1 |
| | 2 | z2w,2=1 z2w,3=1 |
| | 2 | z2w,1=1 z2w,3=1 |
| | 2 | z2w,1=1 z2w,2=1 z2w,3=1 z4=3 |
| | 2 | z2w,1=1 z2w,2=1 z4=2 |
| | 2 | z2w,1=1 z4=1 |
| | 2 | z2w,2=1 z4=1 |
| | 2 | z2w,2=1 z2w,3=1 z4=1 |
| | 2 | z2w,3=1 |
| | 2 | z2w,1=1 z2w,2=1 z2w,3=1 |
| | 2 | z2w,1=1 z2w,3=1 z4=3 |
| | 2 | z2w,1=1 z2w,3=1 z4=1 |
| | 2 | z2w,1=1 z2w,2=1 z2w,3=1 |
| | 2 | z2w,3=1 |
| | 2 | z2w,2=1 z2w,3=1 z4=3 |
| | 2 | z2w,2=1 z4=3 |
| | 2 | z2w,1=1 z4=3 |
| | 2 | z2w,1=1 z2w,2=1 z4=2 |
| | 2 | z2w,3=1 z4=2 |
| | 2 | z2w,2=1 z2w,3=1 z4=1 |
| | 2 | z2w,1=1 z2w,3=1 z4=1 |
| | 2 | z2w,1=1 z2w,2=1 z2w,3=1 |
| | 2 | z2w,1=1 z4=1 |
| | 2 | z2w,1=1 z2w,2=1 |
| | 2 | z2w,2=1 z4=3 |
| | 2 | z2w,2=1 z4=1 |
| | 2 | z2w,1=1 z2w,2=1 |
| | 2 | z2w,1=1 z4=3 |
| | 2 | z2w,1=1 z2w,2=1 z2w,3=1 |
| | 2 | z2w,1=1 z2w,3=1 z4=3 |
| | 2 | z2w,2=1 z2w,3=1 z4=3 |
| | 2 | z2w,3=1 z4=2 |
| | 2 | z4=1 |
| | 2 | z2w,2=1 |
| | 2 | z2w,1=1 |
| | 2 | z2w,1=1 z2w,2=1 z4=3 |
| | 2 | z2w,1=1 z2w,3=1 |
| | 2 | z2w,1=1 z2w,2=1 z2w,3=1 z4=3 |
| | 2 | z2w,3=1 z4=3 |
| | 2 | z2w,2=1 z2w,3=1 z4=2 |
| | 2 | z2w,2=1 z2w,3=1 |
| | 2 | z2w,3=1 z4=3 |
| | 2 | z2w,1=1 z2w,2=1 z2w,3=1 z4=3 |
| | 2 | z2w,1=1 z2w,3=1 z4=2 |
| | 2 | z2w,1=1 z2w,2=1 z4=3 |
| | 2 | z2w,1=1 z4=2 |
| | 2 | z2w,2=1 z4=2 |
| | 2 | z4=1 |

Continued on next column

**TABLE XIX– continued**

| TQC | ICSD<br>SSG | Stoich. Formula<br>SIs |
|---|---|---|
| 140.4.1 | 40260 | Te3Tl5 |
| | 121 | z2=1 |
| | 121 | z2=1 |
| | 87 | z8=4 |
| | 87 | z8=4 |
| | 69 | z2w,1=1 z2w,3=1 z4=2 |
| | 69 | z2w,1=1 z2w,3=1 z4=2 |
| 140.4.2 | 195541 | Si3Ta5 |
| | 82 | z2=1 |
| | 82 | z2=1 |
| | 69 | z2w,1=1 z2w,3=1 z4=2 |
| | 69 | z2w,1=1 z2w,3=1 z4=2 |
| 140.4.3 | 189722 | Si3Ta5 |
| | 121 | z2=1 |
| | 121 | z2=1 |
| | 87 | z2w,1=1 z2w,2=1 z2w,3=1 z4=2 z8=6 |
| | 87 | z2w,1=1 z2w,2=1 z2w,3=1 z4=2 z8=2 |
| | 87 | z2w,1=1 z2w,2=1 z2w,3=1 z4=2 z8=2 |
| | 87 | z2w,1=1 z2w,2=1 z2w,3=1 z4=2 z8=6 |
| 140.4.4 | 95842 | Rh1Sb2Ti5 |
| | 121 | z2=1 |
| | 121 | z2=1 |
| | 69 | z2w,1=1 z2w,3=1 z4=1 |
| | 69 | z2w,1=1 z2w,3=1 z4=1 |
| 140.4.5 | 183452 | O3Sr1Tc1 |
| | 121 | z2=1 |
| | 121 | z2=1 |
| | 87 | z8=4 |
| | 87 | z8=4 |
| | 69 | z2w,1=1 z2w,3=1 z4=2 |
| | 69 | z2w,1=1 z2w,3=1 z4=2 |
| 140.4.6 | 189717 | Nb5Si3 |
| | 121 | z2=1 |
| | 121 | z2=1 |
| | 87 | z2w,1=1 z2w,2=1 z2w,3=1 z4=2 z8=6 |
| | 87 | z2w,1=1 z2w,2=1 z2w,3=1 z4=2 z8=2 |
| | 87 | z2w,1=1 z2w,2=1 z2w,3=1 z4=2 z8=2 |
| | 87 | z2w,1=1 z2w,2=1 z2w,3=1 z4=2 z8=6 |
| 140.4.7 | 16499 | Nb5Si1Sn2 |
| | 121 | z2=1 |
| | 121 | z2=1 |
| | 69 | z2w,1=1 z2w,3=1 z4=1 |
| | 69 | z2w,1=1 z2w,3=1 z4=1 |
| 140.4.8 | 423763 | Nb20Si12 |
| | 82 | z2=1 |
| | 82 | z2=1 |
| | 69 | z2w,1=1 z2w,3=1 z4=2 |
| | 69 | z2w,1=1 z2w,3=1 z4=2 |
| 140.4.9 | 247482 | Mn1Sn2 |
| | 121 | z2=1 |
| | 121 | z2=1 |
| | 69 | z2w,1=1 z2w,3=1 z4=1 |
| | 69 | z2w,1=1 z2w,3=1 z4=1 |
| 140.4.10 | 171198 | Mn1Sn2 |
| | 121 | z2=1 |
| | 121 | z2=1 |
| | 69 | z2w,1=1 z2w,3=1 z4=1 |
| | 69 | z2w,1=1 z2w,3=1 z4=1 |
| 140.4.11 | 412250 | Li1Ru1Sn4 |

Continued on next column

**TABLE XIX– continued**

| TQC | ICSD<br>SSG | Stoich. Formula<br>SIs |
|---|---|---|
| | 87 | z4=3 z2=1 z8=7 |
| | 87 | z4=3 z2=1 z8=3 |
| | 87 | z4=3 z2=1 z8=3 |
| | 87 | z4=3 z2=1 z8=7 |
| | 69 | z2w,1=1 z2w,3=1 z4=1 |
| | 69 | z2w,1=1 z2w,3=1 z4=1 |
| 140.4.12 | 261925 | La5Sn1Zn2 |
| | 121 | z2=1 |
| | 121 | z2=1 |
| | 87 | z2w,1=1 z2w,2=1 z2w,3=1 z4=2 z8=6 |
| | 87 | z2w,1=1 z2w,2=1 z2w,3=1 z4=2 z8=2 |
| | 87 | z2w,1=1 z2w,2=1 z2w,3=1 z4=2 z8=2 |
| | 87 | z2w,1=1 z2w,2=1 z2w,3=1 z4=2 z8=6 |
| 140.4.13 | 95073 | La5Si3 |
| | 82 | z2=1 |
| | 82 | z2=1 |
| | 69 | z2w,1=1 z2w,3=1 |
| | 69 | z2w,1=1 z2w,3=1 |
| 140.4.14 | 98409 | La1Ni9Si4 |
| | 87 | z4=1 z2=1 z8=1 |
| | 87 | z4=1 z2=1 z8=5 |
| | 87 | z4=1 z2=1 z8=5 |
| | 87 | z4=1 z2=1 z8=1 |
| | 69 | z2w,1=1 z2w,3=1 z4=3 |
| | 69 | z2w,1=1 z2w,3=1 z4=3 |
| 140.4.15 | 413805 | In3Sr5 |
| | 121 | z2=1 |
| | 121 | z2=1 |
| | 87 | z2w,1=1 z2w,2=1 z2w,3=1 |
| | 87 | z2w,1=1 z2w,2=1 z2w,3=1 z8=4 |
| | 87 | z2w,1=1 z2w,2=1 z2w,3=1 z8=4 |
| | 87 | z2w,1=1 z2w,2=1 z2w,3=1 |
| | 69 | z4=2 |
| | 69 | z4=2 |
| 140.4.16 | 165190 | H1Ca5Sn3 |
| | 121 | z2=1 |
| | 121 | z2=1 |
| | 87 | z2w,1=1 z2w,2=1 z2w,3=1 z4=2 z8=6 |
| | 87 | z2w,1=1 z2w,2=1 z2w,3=1 z4=2 z8=2 |
| | 87 | z2w,1=1 z2w,2=1 z2w,3=1 z4=2 z8=2 |
| | 87 | z2w,1=1 z2w,2=1 z2w,3=1 z4=2 z8=6 |
| 140.4.17 | 195542 | Ge3Ta5 |
| | 82 | z2=1 |
| | 82 | z2=1 |
| | 69 | z2w,1=1 z2w,3=1 z4=2 |
| | 69 | z2w,1=1 z2w,3=1 z4=2 |
| 140.4.18 | 2003 | Ge3Nb5 |
| | 121 | z2=1 |
| | 121 | z2=1 |
| | 87 | z2w,1=1 z2w,2=1 z2w,3=1 z4=2 z8=6 |
| | 87 | z2w,1=1 z2w,2=1 z2w,3=1 z4=2 z8=2 |
| | 87 | z2w,1=1 z2w,2=1 z2w,3=1 z4=2 z8=2 |
| | 87 | z2w,1=1 z2w,2=1 z2w,3=1 z4=2 z8=6 |
| 140.4.19 | 81469 | Ge1In4La3 |
| | 121 | z2=1 |
| | 121 | z2=1 |
| | 87 | z2w,1=1 z2w,2=1 z2w,3=1 z8=4 |
| | 87 | z2w,1=1 z2w,2=1 z2w,3=1 |
| | 87 | z2w,1=1 z2w,2=1 z2w,3=1 |

Continued on next column

**TABLE XIX– continued**

| TQC | ICSD SSG | Stoich. Formula SIs |
|---|---|---|
| | 87 | z2w,1=1 z2w,2=1 z2w,3=1 z8=4 |
| | 69 | z4=2 |
| | 69 | z4=2 |
| 140.4.20 | 414109 | Ga1K1Se4Sn1 |
| | 87 | z2w,1=1 z2w,2=1 z2w,3=1 z4=1 z2=1 z8=1 |
| | 87 | z2w,1=1 z2w,2=1 z2w,3=1 z4=1 z2=1 z8=5 |
| | 87 | z2w,1=1 z2w,2=1 z2w,3=1 z4=1 z2=1 z8=5 |
| | 87 | z2w,1=1 z2w,2=1 z2w,3=1 z4=1 z2=1 z8=1 |
| | 69 | z4=3 |
| | 69 | z4=3 |
| 140.4.21 | 79590 | Fe9La1Si4 |
| | 121 | z2=1 |
| | 121 | z2=1 |
| | 69 | z2w,1=1 z2w,3=1 z4=1 |
| | 69 | z2w,1=1 z2w,3=1 z4=1 |
| 140.4.22 | 75412 | F3K1Mn1 |
| | 121 | z2=1 |
| | 121 | z2=1 |
| | 87 | z8=4 |
| | 87 | z8=4 |
| | 69 | z2w,1=1 z2w,3=1 z4=2 |
| | 69 | z2w,1=1 z2w,3=1 z4=2 |
| 140.4.23 | 165195 | F1Pb3Sr5 |
| | 121 | z2=1 |
| | 121 | z2=1 |
| | 87 | z2w,1=1 z2w,2=1 z2w,3=1 z4=2 z8=6 |
| | 87 | z2w,1=1 z2w,2=1 z2w,3=1 z4=2 z8=2 |
| | 87 | z2w,1=1 z2w,2=1 z2w,3=1 z4=2 z8=2 |
| | 87 | z2w,1=1 z2w,2=1 z2w,3=1 z4=2 z8=6 |
| 140.4.24 | 408118 | Eu5Si3 |
| | 121 | z2=1 |
| | 121 | z2=1 |
| | 87 | z4=2 z8=2 |
| | 87 | z4=2 z8=6 |
| | 87 | z4=2 z8=6 |
| | 87 | z4=2 z8=2 |
| | 69 | z2w,1=1 z2w,3=1 |
| | 69 | z2w,1=1 z2w,3=1 |
| 140.4.25 | 424907 | Eu1O3Ti1 |
| | 121 | z2=1 |
| | 121 | z2=1 |
| | 87 | z4=2 z8=6 |
| | 87 | z4=2 z8=2 |
| | 87 | z4=2 z8=2 |
| | 87 | z4=2 z8=6 |
| | 69 | z2w,1=1 z2w,3=1 |
| | 69 | z2w,1=1 z2w,3=1 |
| 140.4.26 | 187117 | Eu1O3Ti1 |
| | 121 | z2=1 |
| | 121 | z2=1 |
| | 87 | z4=2 z8=6 |
| | 87 | z4=2 z8=2 |
| | 87 | z4=2 z8=2 |
| | 87 | z4=2 z8=6 |
| | 69 | z2w,1=1 z2w,3=1 |

Continued on next column

**TABLE XIX– continued**

| TQC | ICSD SSG | Stoich. Formula SIs |
|---|---|---|
| | 69 | z2w,1=1 z2w,3=1 |
| 140.4.27 | 187116 | Eu1O3Ti1 |
| | 121 | z2=1 |
| | 121 | z2=1 |
| | 87 | z8=4 |
| | 87 | z8=4 |
| | 69 | z2w,1=1 z2w,3=1 z4=2 |
| | 69 | z2w,1=1 z2w,3=1 z4=2 |
| 140.4.28 | 187113 | Eu1O3Ti1 |
| | 121 | z2=1 |
| | 121 | z2=1 |
| | 87 | z4=2 z8=6 |
| | 87 | z4=2 z8=2 |
| | 87 | z4=2 z8=2 |
| | 87 | z4=2 z8=6 |
| | 69 | z2w,1=1 z2w,3=1 |
| | 69 | z2w,1=1 z2w,3=1 |
| 140.4.29 | 187112 | Eu1O3Ti1 |
| | 121 | z2=1 |
| | 121 | z2=1 |
| | 87 | z8=4 |
| | 87 | z8=4 |
| | 69 | z2w,1=1 z2w,3=1 z4=2 |
| | 69 | z2w,1=1 z2w,3=1 z4=2 |
| 140.4.30 | 187111 | Eu1O3Ti1 |
| | 121 | z2=1 |
| | 121 | z2=1 |
| | 87 | z4=2 z8=6 |
| | 87 | z4=2 z8=2 |
| | 87 | z4=2 z8=2 |
| | 87 | z4=2 z8=6 |
| | 69 | z2w,1=1 z2w,3=1 |
| | 69 | z2w,1=1 z2w,3=1 |
| 140.4.31 | 187110 | Eu1O3Ti1 |
| | 121 | z2=1 |
| | 121 | z2=1 |
| | 87 | z4=2 z8=6 |
| | 87 | z4=2 z8=2 |
| | 87 | z4=2 z8=2 |
| | 87 | z4=2 z8=6 |
| | 69 | z2w,1=1 z2w,3=1 |
| | 69 | z2w,1=1 z2w,3=1 |
| 140.4.32 | 187109 | Eu1O3Ti1 |
| | 121 | z2=1 |
| | 121 | z2=1 |
| | 87 | z4=2 z8=6 |
| | 87 | z4=2 z8=2 |
| | 87 | z4=2 z8=2 |
| | 87 | z4=2 z8=6 |
| | 69 | z2w,1=1 z2w,3=1 |
| | 69 | z2w,1=1 z2w,3=1 |
| 140.4.33 | 187108 | Eu1O3Ti1 |
| | 121 | z2=1 |
| | 121 | z2=1 |
| | 87 | z4=2 z8=6 |
| | 87 | z4=2 z8=2 |
| | 87 | z4=2 z8=2 |
| | 87 | z4=2 z8=6 |
| | 69 | z2w,1=1 z2w,3=1 |

Continued on next column

**TABLE XIX– continued**

| TQC | ICSD<br>SSG | Stoich. Formula<br>SIs |
|---|---|---|
| | 69 | z2w,1=1 z2w,3=1 |
| 140.4.34 | 187107 | Eu1O3Ti1 |
| | 121 | z2=1 |
| | 121 | z2=1 |
| | 87 | z4=2 z8=6 |
| | 87 | z4=2 z8=2 |
| | 87 | z4=2 z8=2 |
| | 87 | z4=2 z8=6 |
| | 69 | z2w,1=1 z2w,3=1 |
| | 69 | z2w,1=1 z2w,3=1 |
| 140.4.35 | 187106 | Eu1O3Ti1 |
| | 121 | z2=1 |
| | 121 | z2=1 |
| | 87 | z4=2 z8=6 |
| | 87 | z4=2 z8=2 |
| | 87 | z4=2 z8=2 |
| | 87 | z4=2 z8=6 |
| | 69 | z2w,1=1 z2w,3=1 |
| | 69 | z2w,1=1 z2w,3=1 |
| 140.4.36 | 81765 | Dy5Ni2Sb1 |
| | 82 | z2=1 |
| | 82 | z2=1 |
| | 69 | z2w,1=1 z2w,3=1 z4=2 |
| | 69 | z2w,1=1 z2w,3=1 z4=2 |
| 140.4.37 | 424219 | Cu9Sn4Sr1 |
| | 87 | z2w,1=1 z2w,2=1 z2w,3=1 z4=1 z2=1 z8=1 |
| | 87 | z2w,1=1 z2w,2=1 z2w,3=1 z4=1 z2=1 z8=5 |
| | 87 | z2w,1=1 z2w,2=1 z2w,3=1 z4=1 z2=1 z8=5 |
| | 87 | z2w,1=1 z2w,2=1 z2w,3=1 z4=1 z2=1 z8=1 |
| | 69 | z4=3 |
| | 69 | z4=3 |
| 140.4.38 | 422456 | Cu9Sn4Sr1 |
| | 87 | z2w,1=1 z2w,2=1 z2w,3=1 z4=1 z2=1 z8=1 |
| | 87 | z2w,1=1 z2w,2=1 z2w,3=1 z4=1 z2=1 z8=5 |
| | 87 | z2w,1=1 z2w,2=1 z2w,3=1 z4=1 z2=1 z8=5 |
| | 87 | z2w,1=1 z2w,2=1 z2w,3=1 z4=1 z2=1 z8=1 |
| | 69 | z4=3 |
| | 69 | z4=3 |
| 140.4.39 | 182105 | Cu9Sn4Sr1 |
| | 87 | z2w,1=1 z2w,2=1 z2w,3=1 z4=1 z2=1 z8=1 |
| | 87 | z2w,1=1 z2w,2=1 z2w,3=1 z4=1 z2=1 z8=5 |
| | 87 | z2w,1=1 z2w,2=1 z2w,3=1 z4=1 z2=1 z8=5 |
| | 87 | z2w,1=1 z2w,2=1 z2w,3=1 z4=1 z2=1 z8=1 |
| | 69 | z4=3 |
| | 69 | z4=3 |
| 140.4.40 | 412882 | Cu9Si4Sr1 |
| | 87 | z2w,1=1 z2w,2=1 z2w,3=1 z4=1 z2=1 z8=1 |

Continued on next column

**TABLE XIX– continued**

| TQC | ICSD<br>SSG | Stoich. Formula<br>SIs |
|---|---|---|
| | 87 | z2w,1=1 z2w,2=1 z2w,3=1 z4=1 z2=1 z8=5 |
| | 87 | z2w,1=1 z2w,2=1 z2w,3=1 z4=1 z2=1 z8=5 |
| | 87 | z2w,1=1 z2w,2=1 z2w,3=1 z4=1 z2=1 z8=1 |
| | 69 | z4=3 |
| | 69 | z4=3 |
| 140.4.41 | 412884 | Cu9Ge4Sr1 |
| | 87 | z2w,1=1 z2w,2=1 z2w,3=1 z4=1 z2=1 z8=1 |
| | 87 | z2w,1=1 z2w,2=1 z2w,3=1 z4=1 z2=1 z8=5 |
| | 87 | z2w,1=1 z2w,2=1 z2w,3=1 z4=1 z2=1 z8=5 |
| | 87 | z2w,1=1 z2w,2=1 z2w,3=1 z4=1 z2=1 z8=1 |
| | 69 | z4=3 |
| | 69 | z4=3 |
| 140.4.42 | 416796 | Cu4Eu1Sn2 |
| | 87 | z2w,1=1 z2w,2=1 z2w,3=1 z4=3 z2=1 z8=7 |
| | 87 | z2w,1=1 z2w,2=1 z2w,3=1 z4=3 z2=1 z8=3 |
| | 87 | z2w,1=1 z2w,2=1 z2w,3=1 z4=3 z2=1 z8=3 |
| | 87 | z2w,1=1 z2w,2=1 z2w,3=1 z4=3 z2=1 z8=7 |
| | 69 | z4=1 |
| | 69 | z4=1 |
| 140.4.43 | 159248 | Cu1Sb2Ti5 |
| | 121 | z2=1 |
| | 121 | z2=1 |
| | 69 | z2w,1=1 z2w,3=1 z4=3 |
| | 69 | z2w,1=1 z2w,3=1 z4=3 |
| 140.4.44 | 69656 | Cu1F3Rb1 |
| | 82 | z2=1 |
| | 82 | z2=1 |
| | 12 | z2w,1=1 z2w,2=1 z4=2 |
| | 12 | z2w,1=1 z2w,2=1 z4=2 |
| | 12 | z2w,1=1 z2w,2=1 z4=2 |
| | 12 | z2w,1=1 z2w,2=1 z4=2 |
| 140.4.45 | 9856 | Cu1F3K1 |
| | 82 | z2=1 |
| | 82 | z2=1 |
| | 12 | z2w,1=1 z2w,2=1 z4=2 |
| | 12 | z2w,1=1 z2w,2=1 z4=2 |
| | 12 | z2w,1=1 z2w,2=1 z4=2 |
| | 12 | z2w,1=1 z2w,2=1 z4=2 |
| 140.4.46 | 8000 | Cu1F3K1 |
| | 82 | z2=1 |
| | 82 | z2=1 |
| | 12 | z2w,1=1 z2w,2=1 z4=2 |
| | 12 | z2w,1=1 z2w,2=1 z4=2 |
| | 12 | z2w,1=1 z2w,2=1 z4=2 |
| | 12 | z2w,1=1 z2w,2=1 z4=2 |
| 140.4.47 | 67445 | Cu1F3K1 |
| | 82 | z2=1 |
| | 82 | z2=1 |

Continued on next column

**TABLE XIX– continued**

| TQC | ICSD<br>SSG | Stoich. Formula<br>SIs |
|---|---|---|
| | 12 | z2w,1=1 z2w,2=1 z4=2 |
| | 12 | z2w,1=1 z2w,2=1 z4=2 |
| | 12 | z2w,1=1 z2w,2=1 z4=2 |
| | 12 | z2w,1=1 z2w,2=1 z4=2 |
| 140.4.48 | 67444 | Cu1F3K1 |
| | 82 | z2=1 |
| | 82 | z2=1 |
| | 12 | z2w,1=1 z2w,2=1 z4=2 |
| | 12 | z2w,1=1 z2w,2=1 z4=2 |
| | 12 | z2w,1=1 z2w,2=1 z4=2 |
| | 12 | z2w,1=1 z2w,2=1 z4=2 |
| 140.4.49 | 21109 | Cu1F3K1 |
| | 82 | z2=1 |
| | 82 | z2=1 |
| | 12 | z2w,1=1 z2w,2=1 z4=2 |
| | 12 | z2w,1=1 z2w,2=1 z4=2 |
| | 12 | z2w,1=1 z2w,2=1 z4=2 |
| | 12 | z2w,1=1 z2w,2=1 z4=2 |
| 140.4.50 | 21108 | Cu1F3K1 |
| | 82 | z2=1 |
| | 82 | z2=1 |
| | 12 | z2w,1=1 z2w,2=1 z4=2 |
| | 12 | z2w,1=1 z2w,2=1 z4=2 |
| | 12 | z2w,1=1 z2w,2=1 z4=2 |
| | 12 | z2w,1=1 z2w,2=1 z4=2 |
| 140.4.51 | 182857 | Cu1F3K1 |
| | 82 | z2=1 |
| | 82 | z2=1 |
| | 12 | z2w,1=1 z2w,2=1 z4=2 |
| | 12 | z2w,1=1 z2w,2=1 z4=2 |
| | 12 | z2w,1=1 z2w,2=1 z4=2 |
| | 12 | z2w,1=1 z2w,2=1 z4=2 |
| 140.4.52 | 182856 | Cu1F3K1 |
| | 82 | z2=1 |
| | 82 | z2=1 |
| | 12 | z2w,1=1 z2w,2=1 z4=2 |
| | 12 | z2w,1=1 z2w,2=1 z4=2 |
| | 12 | z2w,1=1 z2w,2=1 z4=2 |
| | 12 | z2w,1=1 z2w,2=1 z4=2 |
| 140.4.53 | 182855 | Cu1F3K1 |
| | 82 | z2=1 |
| | 82 | z2=1 |
| | 12 | z2w,1=1 z2w,2=1 z4=2 |
| | 12 | z2w,1=1 z2w,2=1 z4=2 |
| | 12 | z2w,1=1 z2w,2=1 z4=2 |
| | 12 | z2w,1=1 z2w,2=1 z4=2 |
| 140.4.54 | 182854 | Cu1F3K1 |
| | 82 | z2=1 |
| | 82 | z2=1 |
| | 12 | z2w,1=1 z2w,2=1 z4=2 |
| | 12 | z2w,1=1 z2w,2=1 z4=2 |
| | 12 | z2w,1=1 z2w,2=1 z4=2 |
| | 12 | z2w,1=1 z2w,2=1 z4=2 |
| 140.4.55 | 182853 | Cu1F3K1 |
| | 82 | z2=1 |
| | 82 | z2=1 |
| | 12 | z2w,1=1 z2w,2=1 z4=2 |
| | 12 | z2w,1=1 z2w,2=1 z4=2 |
| | 12 | z2w,1=1 z2w,2=1 z4=2 |

Continued on next column

**TABLE XIX– continued**

| TQC | ICSD<br>SSG | Stoich. Formula<br>SIs |
|---|---|---|
| | 12 | z2w,1=1 z2w,2=1 z4=2 |
| 140.4.56 | 182852 | Cu1F3K1 |
| | 82 | z2=1 |
| | 82 | z2=1 |
| | 12 | z2w,1=1 z2w,2=1 z4=2 |
| | 12 | z2w,1=1 z2w,2=1 z4=2 |
| | 12 | z2w,1=1 z2w,2=1 z4=2 |
| | 12 | z2w,1=1 z2w,2=1 z4=2 |
| 140.4.57 | 182851 | Cu1F3K1 |
| | 82 | z2=1 |
| | 82 | z2=1 |
| | 12 | z2w,1=1 z2w,2=1 z4=2 |
| | 12 | z2w,1=1 z2w,2=1 z4=2 |
| | 12 | z2w,1=1 z2w,2=1 z4=2 |
| | 12 | z2w,1=1 z2w,2=1 z4=2 |
| 140.4.58 | 182850 | Cu1F3K1 |
| | 82 | z2=1 |
| | 82 | z2=1 |
| | 12 | z2w,1=1 z2w,2=1 z4=2 |
| | 12 | z2w,1=1 z2w,2=1 z4=2 |
| | 12 | z2w,1=1 z2w,2=1 z4=2 |
| | 12 | z2w,1=1 z2w,2=1 z4=2 |
| 140.4.59 | 182849 | Cu1F3K1 |
| | 82 | z2=1 |
| | 82 | z2=1 |
| | 12 | z2w,1=1 z2w,2=1 z4=2 |
| | 12 | z2w,1=1 z2w,2=1 z4=2 |
| | 12 | z2w,1=1 z2w,2=1 z4=2 |
| | 12 | z2w,1=1 z2w,2=1 z4=2 |
| 140.4.60 | 158622 | Cu1F3K1 |
| | 82 | z2=1 |
| | 82 | z2=1 |
| | 12 | z2w,1=1 z2w,2=1 z4=2 |
| | 12 | z2w,1=1 z2w,2=1 z4=2 |
| | 12 | z2w,1=1 z2w,2=1 z4=2 |
| | 12 | z2w,1=1 z2w,2=1 z4=2 |
| 140.4.61 | 420323 | Cs9Fe1O4 |
| | 121 | z2=1 |
| | 121 | z2=1 |
| | 87 | z8=4 |
| | 87 | z8=4 |
| | 69 | z2w,1=1 z2w,3=1 z4=2 |
| | 69 | z2w,1=1 z2w,3=1 z4=2 |
| 140.4.62 | 403102 | Cs3I5Mn1 |
| | 82 | z2=1 |
| | 82 | z2=1 |
| | 12 | z2w,1=1 z2w,2=1 z4=2 |
| | 12 | z2w,1=1 z2w,2=1 z4=2 |
| | 12 | z2w,1=1 z2w,2=1 z4=2 |
| | 12 | z2w,1=1 z2w,2=1 z4=2 |
| 140.4.63 | 102740 | Co1Zr2 |
| | 121 | z2=1 |
| | 121 | z2=1 |
| | 69 | z2w,1=1 z2w,3=1 z4=1 |
| | 69 | z2w,1=1 z2w,3=1 z4=1 |
| 140.4.64 | 409910 | Co1Sn2 |
| | 121 | z2=1 |
| | 121 | z2=1 |
| | 69 | z4=3 |

Continued on next column

**TABLE XIX– continued**

| TQC | ICSD SSG | Stoich. Formula SIs |
|---|---|---|
| | 69 | z4=3 |
| 140.4.65 | 247484 | Co1Sn2 |
| | 121 | z2=1 |
| | 121 | z2=1 |
| | 69 | z4=3 |
| | 69 | z4=3 |
| 140.4.66 | 24 | Cl5Cs3Mn1 |
| | 82 | z2=1 |
| | 82 | z2=1 |
| | 12 | z2w,1=1 z2w,2=1 z4=2 |
| | 12 | z2w,1=1 z2w,2=1 z4=2 |
| | 12 | z2w,1=1 z2w,2=1 z4=2 |
| | 12 | z2w,1=1 z2w,2=1 z4=2 |
| 140.4.67 | 16397 | Cl5Co1Cs3 |
| | 121 | z2=1 |
| | 121 | z2=1 |
| | 87 | z2w,1=1 z2w,2=1 z2w,3=1 z4=2 z8=6 |
| | 87 | z2w,1=1 z2w,2=1 z2w,3=1 z4=2 z8=2 |
| | 87 | z2w,1=1 z2w,2=1 z2w,3=1 z4=2 z8=2 |
| | 87 | z2w,1=1 z2w,2=1 z2w,3=1 z4=2 z8=6 |
| 140.4.68 | 15245 | Cl5Co1Cs3 |
| | 121 | z2=1 |
| | 121 | z2=1 |
| | 87 | z2w,1=1 z2w,2=1 z2w,3=1 z4=2 z8=6 |
| | 87 | z2w,1=1 z2w,2=1 z2w,3=1 z4=2 z8=2 |
| | 87 | z2w,1=1 z2w,2=1 z2w,3=1 z4=2 z8=2 |
| | 87 | z2w,1=1 z2w,2=1 z2w,3=1 z4=2 z8=6 |
| 140.4.69 | 281418 | Ce3Cl1Ga4 |
| | 121 | z2=1 |
| | 121 | z2=1 |
| | 69 | z4=1 |
| | 69 | z4=1 |
| 140.4.70 | 73478 | Cd1Cu1F4 |
| | 121 | z2=1 |
| | 121 | z2=1 |
| | 87 | z2w,1=1 z2w,2=1 z2w,3=1 z4=2 z8=6 |
| | 87 | z2w,1=1 z2w,2=1 z2w,3=1 z4=2 z8=2 |
| | 87 | z2w,1=1 z2w,2=1 z2w,3=1 z4=2 z8=2 |
| | 87 | z2w,1=1 z2w,2=1 z2w,3=1 z4=2 z8=6 |
| 140.4.71 | 424221 | Ca1Cu9Sn4 |
| | 87 | z2w,1=1 z2w,2=1 z2w,3=1 z4=1 z2=1 z8=1 |
| | 87 | z2w,1=1 z2w,2=1 z2w,3=1 z4=1 z2=1 z8=5 |
| | 87 | z2w,1=1 z2w,2=1 z2w,3=1 z4=1 z2=1 z8=5 |
| | 87 | z2w,1=1 z2w,2=1 z2w,3=1 z4=1 z2=1 z8=1 |
| | 69 | z4=3 |
| | 69 | z4=3 |
| 140.4.72 | 417718 | Ca1Cu9Sn4 |
| | 87 | z2w,1=1 z2w,2=1 z2w,3=1 z4=1 z2=1 z8=1 |
| | 87 | z2w,1=1 z2w,2=1 z2w,3=1 z4=1 z2=1 z8=5 |
| | 87 | z2w,1=1 z2w,2=1 z2w,3=1 z4=1 z2=1 z8=5 |
| | 87 | z2w,1=1 z2w,2=1 z2w,3=1 z4=1 z2=1 z8=1 |

Continued on next column

**TABLE XIX– continued**

| TQC | ICSD SSG | Stoich. Formula SIs |
|---|---|---|
| | 69 | z4=3 |
| | 69 | z4=3 |
| 140.4.73 | 201135 | Br5Co1Cs3 |
| | 82 | z2=1 |
| | 82 | z2=1 |
| | 12 | z2w,1=1 z2w,2=1 z4=2 |
| | 12 | z2w,1=1 z2w,2=1 z4=2 |
| | 12 | z2w,1=1 z2w,2=1 z4=2 |
| | 12 | z2w,1=1 z2w,2=1 z4=2 |
| 140.4.74 | 150145 | Bi2Mn1Ti4 |
| | 87 | z2w,1=1 z2w,2=1 z2w,3=1 z4=3 z2=1 z8=3 |
| | 87 | z2w,1=1 z2w,2=1 z2w,3=1 z4=3 z2=1 z8=7 |
| | 87 | z2w,1=1 z2w,2=1 z2w,3=1 z4=3 z2=1 z8=7 |
| | 87 | z2w,1=1 z2w,2=1 z2w,3=1 z4=3 z2=1 z8=3 |
| | 69 | z4=1 |
| | 69 | z4=1 |
| 140.4.75 | 150147 | Bi2Co1Ti4 |
| | 87 | z2w,1=1 z2w,2=1 z2w,3=1 z4=3 z2=1 z8=7 |
| | 87 | z2w,1=1 z2w,2=1 z2w,3=1 z4=3 z2=1 z8=3 |
| | 87 | z2w,1=1 z2w,2=1 z2w,3=1 z4=3 z2=1 z8=3 |
| | 87 | z2w,1=1 z2w,2=1 z2w,3=1 z4=3 z2=1 z8=7 |
| | 69 | z4=1 |
| | 69 | z4=1 |
| 140.4.76 | 107217 | Bi1Er5Pt2 |
| | 87 | z4=1 z2=1 z8=1 |
| | 87 | z4=1 z2=1 z8=5 |
| | 87 | z4=1 z2=1 z8=5 |
| | 87 | z4=1 z2=1 z8=1 |
| | 69 | z2w,1=1 z2w,3=1 z4=3 |
| | 69 | z2w,1=1 z2w,3=1 z4=3 |
| 140.4.77 | 99477 | Ba1O3Tb1 |
| | 121 | z2=1 |
| | 121 | z2=1 |
| | 87 | z8=4 |
| | 87 | z8=4 |
| | 69 | z2w,1=1 z2w,3=1 z4=2 |
| | 69 | z2w,1=1 z2w,3=1 z4=2 |
| 140.4.78 | 99476 | Ba1O3Tb1 |
| | 121 | z2=1 |
| | 121 | z2=1 |
| | 87 | z8=4 |
| | 87 | z8=4 |
| | 69 | z2w,1=1 z2w,3=1 z4=2 |
| | 69 | z2w,1=1 z2w,3=1 z4=2 |
| 140.4.79 | 99475 | Ba1O3Tb1 |
| | 121 | z2=1 |
| | 121 | z2=1 |
| | 87 | z8=4 |
| | 87 | z8=4 |
| | 69 | z2w,1=1 z2w,3=1 z4=2 |
| | 69 | z2w,1=1 z2w,3=1 z4=2 |

Continued on next column

**TABLE XIX– continued**

| TQC | ICSD SSG | Stoich. Formula SIs |
|---|---|---|
| 140.4.80 | 59596 | Ba1O3Tb1 |
| | 121 | z2=1 |
| | 121 | z2=1 |
| | 87 | z8=4 |
| | 87 | z8=4 |
| | 69 | z2w,1=1 z2w,3=1 z4=2 |
| | 69 | z2w,1=1 z2w,3=1 z4=2 |
| 140.4.81 | 91229 | Ba1Mn1Pr2S5 |
| | 82 | z2=1 |
| | 82 | z2=1 |
| | 12 | z2w,1=1 z2w,2=1 z4=2 |
| | 12 | z2w,1=1 z2w,2=1 z4=2 |
| | 12 | z2w,1=1 z2w,2=1 z4=2 |
| | 12 | z2w,1=1 z2w,2=1 z4=2 |
| 140.4.82 | 90639 | Ba1La2Mn1S5 |
| | 121 | z2=1 |
| | 121 | z2=1 |
| | 69 | z2w,1=1 z2w,3=1 z4=1 |
| | 69 | z2w,1=1 z2w,3=1 z4=1 |
| 140.4.83 | 248378 | Ba1Ir1O3 |
| | 121 | z2=1 |
| | 121 | z2=1 |
| | 87 | z8=4 |
| | 87 | z8=4 |
| | 69 | z2w,1=1 z2w,3=1 z4=2 |
| | 69 | z2w,1=1 z2w,3=1 z4=2 |
| 140.4.84 | 424220 | Ba1Cu9Sn4 |
| | 87 | z2w,1=1 z2w,2=1 z2w,3=1 z4=1 z2=1 z8=1 |
| | 87 | z2w,1=1 z2w,2=1 z2w,3=1 z4=1 z2=1 z8=5 |
| | 87 | z2w,1=1 z2w,2=1 z2w,3=1 z4=1 z2=1 z8=5 |
| | 87 | z2w,1=1 z2w,2=1 z2w,3=1 z4=1 z2=1 z8=1 |
| | 69 | z4=3 |
| | 69 | z4=3 |
| 140.4.85 | 422457 | Ba1Cu9Sn4 |
| | 87 | z2w,1=1 z2w,2=1 z2w,3=1 z4=1 z2=1 z8=1 |
| | 87 | z2w,1=1 z2w,2=1 z2w,3=1 z4=1 z2=1 z8=5 |
| | 87 | z2w,1=1 z2w,2=1 z2w,3=1 z4=1 z2=1 z8=5 |
| | 87 | z2w,1=1 z2w,2=1 z2w,3=1 z4=1 z2=1 z8=1 |
| | 69 | z4=3 |
| | 69 | z4=3 |
| 140.4.86 | 182106 | Ba1Cu9Sn4 |
| | 87 | z2w,1=1 z2w,2=1 z2w,3=1 z4=1 z2=1 z8=1 |
| | 87 | z2w,1=1 z2w,2=1 z2w,3=1 z4=1 z2=1 z8=5 |
| | 87 | z2w,1=1 z2w,2=1 z2w,3=1 z4=1 z2=1 z8=5 |
| | 87 | z2w,1=1 z2w,2=1 z2w,3=1 z4=1 z2=1 z8=1 |
| | 69 | z4=3 |
| | 69 | z4=3 |

Continued on next column

**TABLE XIX– continued**

| TQC | ICSD SSG | Stoich. Formula SIs |
|---|---|---|
| 140.4.87 | 412883 | Ba1Cu9Si4 |
| | 87 | z2w,1=1 z2w,2=1 z2w,3=1 z4=1 z2=1 z8=1 |
| | 87 | z2w,1=1 z2w,2=1 z2w,3=1 z4=1 z2=1 z8=5 |
| | 87 | z2w,1=1 z2w,2=1 z2w,3=1 z4=1 z2=1 z8=5 |
| | 87 | z2w,1=1 z2w,2=1 z2w,3=1 z4=1 z2=1 z8=1 |
| | 69 | z4=3 |
| | 69 | z4=3 |
| 140.4.88 | 424223 | Ba1Cu9Ge4 |
| | 87 | z2w,1=1 z2w,2=1 z2w,3=1 z4=1 z2=1 z8=1 |
| | 87 | z2w,1=1 z2w,2=1 z2w,3=1 z4=1 z2=1 z8=5 |
| | 87 | z2w,1=1 z2w,2=1 z2w,3=1 z4=1 z2=1 z8=5 |
| | 87 | z2w,1=1 z2w,2=1 z2w,3=1 z4=1 z2=1 z8=1 |
| | 69 | z4=3 |
| | 69 | z4=3 |
| 140.4.89 | 412885 | Ba1Cu9Ge4 |
| | 87 | z2w,1=1 z2w,2=1 z2w,3=1 z4=1 z2=1 z8=1 |
| | 87 | z2w,1=1 z2w,2=1 z2w,3=1 z4=1 z2=1 z8=5 |
| | 87 | z2w,1=1 z2w,2=1 z2w,3=1 z4=1 z2=1 z8=5 |
| | 87 | z2w,1=1 z2w,2=1 z2w,3=1 z4=1 z2=1 z8=1 |
| | 69 | z4=3 |
| | 69 | z4=3 |
| 140.4.90 | 96555 | Ba1Co1Nd2S5 |
| | 82 | z2=1 |
| | 82 | z2=1 |
| | 12 | z2w,1=1 z2w,2=1 z4=2 |
| | 12 | z2w,1=1 z2w,2=1 z4=2 |
| | 12 | z2w,1=1 z2w,2=1 z4=2 |
| | 12 | z2w,1=1 z2w,2=1 z4=2 |
| 140.4.91 | 96554 | Ba1Co1Nd2S5 |
| | 82 | z2=1 |
| | 82 | z2=1 |
| | 12 | z2w,1=1 z2w,2=1 z4=2 |
| | 12 | z2w,1=1 z2w,2=1 z4=2 |
| | 12 | z2w,1=1 z2w,2=1 z4=2 |
| | 12 | z2w,1=1 z2w,2=1 z4=2 |
| 140.4.92 | 95268 | Ba1Co1La2S5 |
| | 82 | z2=1 |
| | 82 | z2=1 |
| | 69 | z2w,1=1 z2w,3=1 z4=2 |
| | 69 | z2w,1=1 z2w,3=1 z4=2 |
| 140.4.93 | 91228 | Ba1Ce2Mn1S5 |
| | 121 | z2=1 |
| | 121 | z2=1 |
| | 69 | z2w,1=1 z2w,3=1 z4=1 |
| | 69 | z2w,1=1 z2w,3=1 z4=1 |
| 140.4.94 | 75792 | B1Ni2 |
| | 121 | z2=1 |

Continued on next column

**TABLE XIX– continued**

| TQC | ICSD SSG | Stoich. Formula SIs |
|---|---|---|
| | 121 | z2=1 |
| | 87 | z8=4 |
| | 87 | z8=4 |
| | 69 | z2w,1=1 z2w,3=1 z4=2 |
| | 69 | z2w,1=1 z2w,3=1 z4=2 |
| 140.4.95 | 391330 | B1Fe2 |
| | 87 | z2w,1=1 z2w,2=1 z2w,3=1 z4=3 z2=1 z8=7 |
| | 87 | z2w,1=1 z2w,2=1 z2w,3=1 z4=3 z2=1 z8=3 |
| | 87 | z2w,1=1 z2w,2=1 z2w,3=1 z4=3 z2=1 z8=3 |
| | 87 | z2w,1=1 z2w,2=1 z2w,3=1 z4=3 z2=1 z8=7 |
| | 69 | z4=1 |
| | 69 | z4=1 |
| 140.4.96 | 391328 | B1Fe2 |
| | 87 | z2w,1=1 z2w,2=1 z2w,3=1 z4=3 z2=1 z8=7 |
| | 87 | z2w,1=1 z2w,2=1 z2w,3=1 z4=3 z2=1 z8=3 |
| | 87 | z2w,1=1 z2w,2=1 z2w,3=1 z4=3 z2=1 z8=3 |
| | 87 | z2w,1=1 z2w,2=1 z2w,3=1 z4=3 z2=1 z8=7 |
| | 69 | z4=1 |
| | 69 | z4=1 |
| 140.4.97 | 249220 | Au4In2Rb1 |
| | 121 | z2=1 |
| | 121 | z2=1 |
| | 69 | z2w,1=1 z2w,3=1 z4=1 |
| | 69 | z2w,1=1 z2w,3=1 z4=1 |
| 140.4.98 | 249221 | Au4In2K1 |
| | 87 | z4=3 z2=1 z8=7 |
| | 87 | z4=3 z2=1 z8=3 |
| | 87 | z4=3 z2=1 z8=3 |
| | 87 | z4=3 z2=1 z8=7 |
| | 69 | z2w,1=1 z2w,3=1 z4=1 |
| | 69 | z2w,1=1 z2w,3=1 z4=1 |
| 140.4.99 | 58416 | Au3Cd5 |
| | 121 | z2=1 |
| | 121 | z2=1 |
| | 69 | z2w,1=1 z2w,3=1 z4=3 |
| | 69 | z2w,1=1 z2w,3=1 z4=3 |
| 140.4.100 | 58404 | Au3Ca5 |
| | 82 | z2=1 |
| | 82 | z2=1 |
| | 69 | z2w,1=1 z2w,3=1 z4=2 |
| | 69 | z2w,1=1 z2w,3=1 z4=2 |
| 140.4.101 | 157338 | Au2Gd5Sb1 |
| | 121 | z2=1 |
| | 121 | z2=1 |
| | 69 | z4=1 |
| | 69 | z4=1 |
| 140.4.102 | 413559 | Al4Br4La5 |
| | 82 | z2=1 |
| | 82 | z2=1 |
| | 69 | z2w,1=1 z2w,3=1 |
| | 69 | z2w,1=1 z2w,3=1 |

Continued on next column

**TABLE XIX– continued**

| TQC | ICSD SSG | Stoich. Formula SIs |
|---|---|---|
| 140.4.103 | 42517 | Al2Cu1 |
| | 87 | z2w,1=1 z2w,2=1 z2w,3=1 z4=1 z2=1 z8=5 |
| | 87 | z2w,1=1 z2w,2=1 z2w,3=1 z4=1 z2=1 z8=1 |
| | 87 | z2w,1=1 z2w,2=1 z2w,3=1 z4=1 z2=1 z8=1 |
| | 87 | z2w,1=1 z2w,2=1 z2w,3=1 z4=1 z2=1 z8=5 |
| | 69 | z4=3 |
| | 69 | z4=3 |
| 140.4.104 | 172800 | Al2Cu1 |
| | 87 | z2w,1=1 z2w,2=1 z2w,3=1 z4=1 z2=1 z8=5 |
| | 87 | z2w,1=1 z2w,2=1 z2w,3=1 z4=1 z2=1 z8=1 |
| | 87 | z2w,1=1 z2w,2=1 z2w,3=1 z4=1 z2=1 z8=1 |
| | 87 | z2w,1=1 z2w,2=1 z2w,3=1 z4=1 z2=1 z8=5 |
| | 69 | z4=3 |
| | 69 | z4=3 |
| 140.4.105 | 245561 | Al1Ce1O3 |
| | 87 | z4=3 z2=1 z8=7 |
| | 87 | z4=3 z2=1 z8=3 |
| | 87 | z4=3 z2=1 z8=3 |
| | 87 | z4=3 z2=1 z8=7 |
| | 69 | z2w,1=1 z2w,3=1 z4=1 |
| | 69 | z2w,1=1 z2w,3=1 z4=1 |
| 140.4.106 | 245554 | Al1Ce1O3 |
| | 87 | z4=3 z2=1 z8=7 |
| | 87 | z4=3 z2=1 z8=3 |
| | 87 | z4=3 z2=1 z8=3 |
| | 87 | z4=3 z2=1 z8=7 |
| | 69 | z2w,1=1 z2w,3=1 z4=1 |
| | 69 | z2w,1=1 z2w,3=1 z4=1 |
| 140.4.107 | 245265 | Al1Ce1O3 |
| | 87 | z4=3 z2=1 z8=7 |
| | 87 | z4=3 z2=1 z8=3 |
| | 87 | z4=3 z2=1 z8=3 |
| | 87 | z4=3 z2=1 z8=7 |
| | 69 | z2w,1=1 z2w,3=1 z4=1 |
| | 69 | z2w,1=1 z2w,3=1 z4=1 |
| 140.4.108 | 245264 | Al1Ce1O3 |
| | 87 | z4=3 z2=1 z8=7 |
| | 87 | z4=3 z2=1 z8=3 |
| | 87 | z4=3 z2=1 z8=3 |
| | 87 | z4=3 z2=1 z8=7 |
| | 69 | z2w,1=1 z2w,3=1 z4=1 |
| | 69 | z2w,1=1 z2w,3=1 z4=1 |
| 140.4.109 | 166284 | Al1Ce1O3 |
| | 87 | z4=3 z2=1 z8=7 |
| | 87 | z4=3 z2=1 z8=3 |
| | 87 | z4=3 z2=1 z8=3 |
| | 87 | z4=3 z2=1 z8=7 |
| | 69 | z2w,1=1 z2w,3=1 z4=1 |
| | 69 | z2w,1=1 z2w,3=1 z4=1 |
| 140.4.110 | 150277 | Al1Ce1O3 |
| | 87 | z4=3 z2=1 z8=7 |

Continued on next column

**TABLE XIX– continued**

| TQC | ICSD SSG | Stoich. Formula SIs |
|---|---|---|
| | 87 | z4=3 z2=1 z8=3 |
| | 87 | z4=3 z2=1 z8=3 |
| | 87 | z4=3 z2=1 z8=7 |
| | 69 | z2w,1=1 z2w,3=1 z4=1 |
| | 69 | z2w,1=1 z2w,3=1 z4=1 |
| 141.4.1 | 79760 | O4P1Yb1 |
| | 12 | z2w,3=1 |
| | 12 | z2w,1=1 z2w,2=1 z2w,3=1 z4=2 |
| | 12 | z2w,1=1 z2w,2=1 z2w,3=1 z4=1 |
| | 12 | z2w,3=1 z4=3 |
| 141.4.2 | 36054 | O4P1Yb1 |
| | 12 | z2w,3=1 z4=2 |
| | 12 | z2w,1=1 z2w,2=1 z2w,3=1 |
| | 12 | z2w,1=1 z2w,2=1 z2w,3=1 z4=3 |
| | 12 | z2w,3=1 z4=1 |
| 141.4.3 | 78077 | Nd1O4V1 |
| | 12 | z2w,3=1 z4=1 |
| | 12 | z2w,1=1 z2w,2=1 z2w,3=1 z4=3 |
| | 12 | z2w,1=1 z2w,2=1 z2w,3=1 z4=2 |
| | 12 | z2w,3=1 |
| 141.4.4 | 191450 | Nd1O4V1 |
| | 12 | z2w,3=1 z4=1 |
| | 12 | z2w,1=1 z2w,2=1 z2w,3=1 z4=3 |
| | 12 | z2w,1=1 z2w,2=1 z2w,3=1 z4=2 |
| | 12 | z2w,3=1 |
| 141.4.5 | 157328 | Nd1O4V1 |
| | 12 | z2w,3=1 z4=1 |
| | 12 | z2w,1=1 z2w,2=1 z2w,3=1 z4=3 |
| | 12 | z2w,1=1 z2w,2=1 z2w,3=1 z4=2 |
| | 12 | z2w,3=1 |
| 141.4.6 | 68174 | Mn3O4 |
| | 12 | z2w,3=1 z4=2 |
| | 12 | z2w,1=1 z2w,2=1 z2w,3=1 |
| | 12 | z2w,1=1 z2w,2=1 z2w,3=1 z4=3 |
| | 12 | z2w,3=1 z4=1 |
| 141.4.7 | 107516 | Li1Sn1 |
| | 12 | z2w,3=1 |
| | 12 | z2w,1=1 z2w,2=1 z2w,3=1 z4=2 |
| | 12 | z2w,1=1 z2w,2=1 z2w,3=1 z4=1 |
| | 12 | z2w,3=1 z4=3 |
| 141.4.8 | 33955 | Li1O2Yb1 |
| | 12 | z2w,3=1 z4=1 |
| | 12 | z2w,1=1 z2w,2=1 z2w,3=1 z4=3 |
| | 12 | z2w,1=1 z2w,2=1 z2w,3=1 z4=2 |
| | 12 | z2w,3=1 |
| 141.4.9 | 98663 | Li1N2U1 |
| | 12 | z2w,3=1 z4=3 |
| | 12 | z2w,1=1 z2w,2=1 z2w,3=1 z4=1 |
| | 12 | z2w,1=1 z2w,2=1 z2w,3=1 |
| | 12 | z2w,3=1 z4=2 |
| 141.4.10 | 62536 | In2N1Sr4 |
| | 12 | z2w,3=1 z4=2 |
| | 12 | z2w,1=1 z2w,2=1 z2w,3=1 |
| | 12 | z2w,1=1 z2w,2=1 z2w,3=1 z4=3 |
| | 12 | z2w,3=1 z4=1 |
| 141.4.11 | 81703 | Gd1O4V1 |
| | 12 | z2w,3=1 z4=3 |
| | 12 | z2w,1=1 z2w,2=1 z2w,3=1 z4=1 |
| | 12 | z2w,1=1 z2w,2=1 z2w,3=1 |

Continued on next column

**TABLE XIX– continued**

| TQC | ICSD SSG | Stoich. Formula SIs |
|---|---|---|
| | 12 | z2w,3=1 z4=2 |
| 141.4.12 | 238224 | Gd1O4V1 |
| | 12 | z2w,3=1 z4=2 |
| | 12 | z2w,1=1 z2w,2=1 z2w,3=1 |
| | 12 | z2w,1=1 z2w,2=1 z2w,3=1 z4=3 |
| | 12 | z2w,3=1 z4=1 |
| 141.4.13 | 238223 | Gd1O4V1 |
| | 12 | z2w,3=1 z4=2 |
| | 12 | z2w,1=1 z2w,2=1 z2w,3=1 |
| | 12 | z2w,1=1 z2w,2=1 z2w,3=1 z4=3 |
| | 12 | z2w,3=1 z4=1 |
| 141.4.14 | 152692 | Gd1O4V1 |
| | 12 | z2w,3=1 z4=1 |
| | 12 | z2w,1=1 z2w,2=1 z2w,3=1 z4=3 |
| | 12 | z2w,1=1 z2w,2=1 z2w,3=1 z4=2 |
| | 12 | z2w,3=1 |
| 141.4.15 | 97542 | Gd1Na1O2 |
| | 12 | z2w,3=1 z4=1 |
| | 12 | z2w,1=1 z2w,2=1 z2w,3=1 z4=3 |
| | 12 | z2w,1=1 z2w,2=1 z2w,3=1 z4=2 |
| | 12 | z2w,3=1 |
| 141.4.16 | 22040 | Gd1Na1O2 |
| | 12 | z2w,3=1 z4=1 |
| | 12 | z2w,1=1 z2w,2=1 z2w,3=1 z4=3 |
| | 12 | z2w,1=1 z2w,2=1 z2w,3=1 z4=2 |
| | 12 | z2w,3=1 |
| 141.4.17 | 80554 | Fe2O5P1 |
| | 12 | z2w,3=1 |
| | 12 | z2w,1=1 z2w,2=1 z2w,3=1 z4=2 |
| | 12 | z2w,1=1 z2w,2=1 z2w,3=1 z4=1 |
| | 12 | z2w,3=1 z4=3 |
| 141.4.18 | 174085 | Fe1Li1O2 |
| | 12 | z2w,3=1 z4=2 |
| | 12 | z2w,1=1 z2w,2=1 z2w,3=1 |
| | 12 | z2w,1=1 z2w,2=1 z2w,3=1 z4=3 |
| | 12 | z2w,3=1 z4=1 |
| 141.4.19 | 103402 | Eu4Ir1 |
| | 12 | z2w,3=1 z4=3 |
| | 12 | z2w,1=1 z2w,2=1 z2w,3=1 z4=1 |
| | 12 | z2w,1=1 z2w,2=1 z2w,3=1 |
| | 12 | z2w,3=1 z4=2 |
| 141.4.20 | 84235 | Er1O4V1 |
| | 12 | z2w,3=1 z4=1 |
| | 12 | z2w,1=1 z2w,2=1 z2w,3=1 z4=3 |
| | 12 | z2w,1=1 z2w,2=1 z2w,3=1 z4=2 |
| | 12 | z2w,3=1 |
| 141.4.21 | 78080 | Er1O4V1 |
| | 12 | z2w,3=1 z4=1 |
| | 12 | z2w,1=1 z2w,2=1 z2w,3=1 z4=3 |
| | 12 | z2w,1=1 z2w,2=1 z2w,3=1 z4=2 |
| | 12 | z2w,3=1 |
| 141.4.22 | 79758 | Er1O4P1 |
| | 12 | z2w,3=1 z4=1 |
| | 12 | z2w,1=1 z2w,2=1 z2w,3=1 z4=3 |
| | 12 | z2w,1=1 z2w,2=1 z2w,3=1 z4=2 |
| | 12 | z2w,3=1 |
| 141.4.23 | 36052 | Er1O4P1 |
| | 12 | z2w,3=1 z4=1 |
| | 12 | z2w,1=1 z2w,2=1 z2w,3=1 z4=3 |

Continued on next column

**TABLE XIX– continued**

| TQC | ICSD<br>SSG | Stoich. Formula<br>SIs |
|---|---|---|
| | 12 | z2w,1=1 z2w,2=1 z2w,3=1 z4=2 |
| | 12 | z2w,3=1 |
| 141.4.24 | 95381 | Er1Li1O2 |
| | 12 | z2w,3=1 z4=1 |
| | 12 | z2w,1=1 z2w,2=1 z2w,3=1 z4=3 |
| | 12 | z2w,1=1 z2w,2=1 z2w,3=1 z4=2 |
| | 12 | z2w,3=1 |
| 141.4.25 | 81704 | Dy1O4V1 |
| | 12 | z2w,3=1 |
| | 12 | z2w,1=1 z2w,2=1 z2w,3=1 z4=2 |
| | 12 | z2w,1=1 z2w,2=1 z2w,3=1 z4=1 |
| | 12 | z2w,3=1 z4=3 |
| 141.4.26 | 26441 | Dy1O4V1 |
| | 12 | z2w,3=1 |
| | 12 | z2w,1=1 z2w,2=1 z2w,3=1 z4=2 |
| | 12 | z2w,1=1 z2w,2=1 z2w,3=1 z4=1 |
| | 12 | z2w,3=1 z4=3 |
| 141.4.27 | 152693 | Dy1O4V1 |
| | 12 | z2w,3=1 z4=3 |
| | 12 | z2w,1=1 z2w,2=1 z2w,3=1 z4=1 |
| | 12 | z2w,1=1 z2w,2=1 z2w,3=1 |
| | 12 | z2w,3=1 z4=2 |
| 141.4.28 | 192553 | Dy1O4P1 |
| | 12 | z2w,3=1 z4=3 |
| | 12 | z2w,1=1 z2w,2=1 z2w,3=1 z4=1 |
| | 12 | z2w,1=1 z2w,2=1 z2w,3=1 |
| | 12 | z2w,3=1 z4=2 |
| 141.4.29 | 88961 | Cu1O4Rh2 |
| | 12 | z2w,3=1 z4=2 |
| | 12 | z2w,1=1 z2w,2=1 z2w,3=1 |
| | 12 | z2w,1=1 z2w,2=1 z2w,3=1 z4=3 |
| | 12 | z2w,3=1 z4=1 |
| 141.4.30 | 59267 | Cu1O4Rh2 |
| | 12 | z2w,3=1 |
| | 12 | z2w,1=1 z2w,2=1 z2w,3=1 z4=2 |
| | 12 | z2w,1=1 z2w,2=1 z2w,3=1 z4=1 |
| | 12 | z2w,3=1 z4=3 |
| 141.4.31 | 426084 | Cu1Li2Sn2 |
| | 12 | z2w,3=1 z4=3 |
| | 12 | z2w,1=1 z2w,2=1 z2w,3=1 z4=1 |
| | 12 | z2w,1=1 z2w,2=1 z2w,3=1 |
| | 12 | z2w,3=1 z4=2 |
| 141.4.32 | 251619 | Cu1Li2Sn2 |
| | 12 | z2w,3=1 z4=3 |
| | 12 | z2w,1=1 z2w,2=1 z2w,3=1 z4=1 |
| | 12 | z2w,1=1 z2w,2=1 z2w,3=1 |
| | 12 | z2w,3=1 z4=2 |
| 141.4.33 | 75532 | Cu1Ir2S4 |
| | 12 | z2w,3=1 z4=1 |
| | 12 | z2w,1=1 z2w,2=1 z2w,3=1 z4=3 |
| | 12 | z2w,1=1 z2w,2=1 z2w,3=1 z4=2 |
| | 12 | z2w,3=1 |
| 141.4.34 | 150187 | Cs1 |
| | 12 | z2w,3=1 z4=3 |
| | 12 | z2w,1=1 z2w,2=1 z2w,3=1 z4=1 |
| | 12 | z2w,1=1 z2w,2=1 z2w,3=1 |
| | 12 | z2w,3=1 z4=2 |
| 141.4.35 | 84378 | Cr2Cu1O4 |
| | 12 | z2w,3=1 z4=3 |

Continued on next column

**TABLE XIX– continued**

| TQC | ICSD<br>SSG | Stoich. Formula<br>SIs |
|---|---|---|
| | 12 | z2w,1=1 z2w,2=1 z2w,3=1 z4=1 |
| | 12 | z2w,1=1 z2w,2=1 z2w,3=1 |
| | 12 | z2w,3=1 z4=2 |
| 141.4.36 | 80476 | Cr2Cu1O4 |
| | 12 | z2w,3=1 z4=3 |
| | 12 | z2w,1=1 z2w,2=1 z2w,3=1 z4=1 |
| | 12 | z2w,1=1 z2w,2=1 z2w,3=1 |
| | 12 | z2w,3=1 z4=2 |
| 141.4.37 | 246080 | Cr2Cu1O4 |
| | 12 | z2w,3=1 z4=2 |
| | 12 | z2w,1=1 z2w,2=1 z2w,3=1 |
| | 12 | z2w,1=1 z2w,2=1 z2w,3=1 z4=3 |
| | 12 | z2w,3=1 z4=1 |
| 141.4.38 | 95971 | Cr1O4Y1 |
| | 12 | z2w,3=1 z4=2 |
| | 12 | z2w,1=1 z2w,2=1 z2w,3=1 |
| | 12 | z2w,1=1 z2w,2=1 z2w,3=1 z4=3 |
| | 12 | z2w,3=1 z4=1 |
| 141.4.39 | 157933 | Cr1O4Y1 |
| | 12 | z2w,3=1 z4=2 |
| | 12 | z2w,1=1 z2w,2=1 z2w,3=1 |
| | 12 | z2w,1=1 z2w,2=1 z2w,3=1 z4=3 |
| | 12 | z2w,3=1 z4=1 |
| 141.4.40 | 169874 | Cr1O4Tb1 |
| | 12 | z2w,3=1 z4=1 |
| | 12 | z2w,1=1 z2w,2=1 z2w,3=1 z4=3 |
| | 12 | z2w,1=1 z2w,2=1 z2w,3=1 z4=2 |
| | 12 | z2w,3=1 |
| 141.4.41 | 166175 | Cr1O4Tb1 |
| | 12 | z2w,3=1 z4=1 |
| | 12 | z2w,1=1 z2w,2=1 z2w,3=1 z4=3 |
| | 12 | z2w,1=1 z2w,2=1 z2w,3=1 z4=2 |
| | 12 | z2w,3=1 |
| 141.4.42 | 95967 | Cr1Lu1O4 |
| | 12 | z2w,3=1 z4=2 |
| | 12 | z2w,1=1 z2w,2=1 z2w,3=1 |
| | 12 | z2w,1=1 z2w,2=1 z2w,3=1 z4=3 |
| | 12 | z2w,3=1 z4=1 |
| 141.4.43 | 173263 | Cr1Ho1O4 |
| | 12 | z2w,3=1 z4=1 |
| | 12 | z2w,1=1 z2w,2=1 z2w,3=1 z4=3 |
| | 12 | z2w,1=1 z2w,2=1 z2w,3=1 z4=2 |
| | 12 | z2w,3=1 |
| 141.4.44 | 169872 | Cr1Ho1O4 |
| | 12 | z2w,3=1 z4=1 |
| | 12 | z2w,1=1 z2w,2=1 z2w,3=1 z4=3 |
| | 12 | z2w,1=1 z2w,2=1 z2w,3=1 z4=2 |
| | 12 | z2w,3=1 |
| 141.4.45 | 51696 | Cr1Eu1O4 |
| | 12 | z2w,3=1 |
| | 12 | z2w,1=1 z2w,2=1 z2w,3=1 z4=2 |
| | 12 | z2w,1=1 z2w,2=1 z2w,3=1 z4=1 |
| | 12 | z2w,3=1 z4=3 |
| 141.4.46 | 81931 | Co9Si2Th1 |
| | 12 | z2w,3=1 z4=2 |
| | 12 | z2w,1=1 z2w,2=1 z2w,3=1 |
| | 12 | z2w,1=1 z2w,2=1 z2w,3=1 z4=3 |
| | 12 | z2w,3=1 z4=1 |
| 141.4.47 | 9309 | Cl4Pa1 |

Continued on next column

**TABLE XIX– continued**

| TQC | ICSD<br>SSG | Stoich. Formula<br>SIs |
|---|---|---|
| | 12 | z2w,3=1 z4=2 |
| | 12 | z2w,1=1 z2w,2=1 z2w,3=1 |
| | 12 | z2w,1=1 z2w,2=1 z2w,3=1 z4=3 |
| | 12 | z2w,3=1 z4=1 |
| 141.4.48 | 9311 | Cl4Np1 |
| | 12 | z2w,3=1 z4=1 |
| | 12 | z2w,1=1 z2w,2=1 z2w,3=1 z4=3 |
| | 12 | z2w,1=1 z2w,2=1 z2w,3=1 z4=2 |
| | 12 | z2w,3=1 |
| 141.4.49 | 75482 | Cl4Np1 |
| | 12 | z2w,3=1 z4=1 |
| | 12 | z2w,1=1 z2w,2=1 z2w,3=1 z4=3 |
| | 12 | z2w,1=1 z2w,2=1 z2w,3=1 z4=2 |
| | 12 | z2w,3=1 |
| 141.4.50 | 78075 | Ce1O4V1 |
| | 12 | z2w,3=1 z4=3 |
| | 12 | z2w,1=1 z2w,2=1 z2w,3=1 z4=1 |
| | 12 | z2w,1=1 z2w,2=1 z2w,3=1 |
| | 12 | z2w,3=1 z4=2 |
| 141.4.51 | 66033 | Ce1O4V1 |
| | 12 | z2w,3=1 z4=1 |
| | 12 | z2w,1=1 z2w,2=1 z2w,3=1 z4=3 |
| | 12 | z2w,1=1 z2w,2=1 z2w,3=1 z4=2 |
| | 12 | z2w,3=1 |
| 141.4.52 | 202425 | Ce1O4V1 |
| | 12 | z2w,3=1 z4=1 |
| | 12 | z2w,1=1 z2w,2=1 z2w,3=1 z4=3 |
| | 12 | z2w,1=1 z2w,2=1 z2w,3=1 z4=2 |
| | 12 | z2w,3=1 |
| 141.4.53 | 157326 | Ce1O4V1 |
| | 12 | z2w,3=1 z4=1 |
| | 12 | z2w,1=1 z2w,2=1 z2w,3=1 z4=3 |
| | 12 | z2w,1=1 z2w,2=1 z2w,3=1 z4=2 |
| | 12 | z2w,3=1 |
| 141.4.54 | 55349 | Au1Li2Sn2 |
| | 12 | z2w,3=1 z4=3 |
| | 12 | z2w,1=1 z2w,2=1 z2w,3=1 z4=1 |
| | 12 | z2w,1=1 z2w,2=1 z2w,3=1 |
| | 12 | z2w,3=1 z4=2 |
| 141.4.55 | 171193 | As1O4Yb1 |
| | 12 | z2w,3=1 z4=1 |
| | 12 | z2w,1=1 z2w,2=1 z2w,3=1 z4=3 |
| | 12 | z2w,1=1 z2w,2=1 z2w,3=1 z4=2 |
| | 12 | z2w,3=1 |
| 141.4.56 | 200228 | As1Dy1O4 |
| | 12 | z2w,3=1 z4=1 |
| | 12 | z2w,1=1 z2w,2=1 z2w,3=1 z4=3 |
| | 12 | z2w,1=1 z2w,2=1 z2w,3=1 z4=2 |
| | 12 | z2w,3=1 |
| 141.4.57 | 16512 | As1Dy1O4 |
| | 12 | z2w,3=1 z4=1 |
| | 12 | z2w,1=1 z2w,2=1 z2w,3=1 z4=3 |
| | 12 | z2w,1=1 z2w,2=1 z2w,3=1 z4=2 |
| | 12 | z2w,3=1 |
| 141.4.58 | 426085 | Ag1Li2Sn2 |
| | 12 | z2w,3=1 z4=1 |
| | 12 | z2w,1=1 z2w,2=1 z2w,3=1 z4=3 |
| | 12 | z2w,1=1 z2w,2=1 z2w,3=1 z4=2 |
| | 12 | z2w,3=1 |

Continued on next column

**TABLE XIX– continued**

| TQC | ICSD<br>SSG | Stoich. Formula<br>SIs |
|---|---|---|
| 163.4.1 | 62035 | Cl6Cr1Na3 |
| | 147 | z2w,3=1 z4=2 |
| | 147 | z4=3 |
| | 147 | z4=3 |
| | 147 | z2w,3=1 |
| 163.4.2 | 195335 | Ag1Er1P2Se6 |
| | 147 | z2w,3=1 |
| | 147 | z4=1 |
| | 147 | z4=1 |
| | 147 | z2w,3=1 z4=2 |
| 165.4.1 | 165378 | Cs2Mn1S6Te2 |
| | 147 | z2w,3=1 z4=3 |
| | 147 | z2w,3=1 z4=1 |
| 165.4.2 | 240159 | Ca5Cu1Pb3 |
| | 2 | z4=3 |
| | 2 | z2w,1=1 z2w,3=1 z4=2 |
| | 2 | z2w,2=1 z2w,3=1 z4=2 |
| | 2 | z2w,1=1 z2w,2=1 z4=1 |
| | 2 | z2w,1=1 z2w,2=1 z2w,3=1 z4=2 |
| | 2 | z2w,2=1 z4=1 |
| | 2 | z2w,1=1 z4=1 |
| | 2 | z2w,3=1 |
| | 2 | z2w,3=1 z4=2 |
| | 2 | z2w,1=1 z4=1 |
| | 2 | z2w,2=1 z4=1 |
| | 2 | z2w,1=1 z2w,2=1 z2w,3=1 |
| | 2 | z2w,1=1 z2w,2=1 z4=1 |
| | 2 | z2w,2=1 z2w,3=1 |
| | 2 | z2w,1=1 z2w,3=1 |
| | 2 | z4=3 |
| 167.4.1 | 109297 | O6Rh1Sr4 |
| | 148 | z2w,1=1 z2w,2=1 z2w,3=1 z4=2 |
| | 148 | z4=3 |
| | 148 | z4=3 |
| | 148 | z2w,1=1 z2w,2=1 z2w,3=1 |
| 167.4.2 | 90285 | O6Rh1Sr3Zn1 |
| | 148 | z2w,1=1 z2w,2=1 z2w,3=1 z4=3 |
| | 148 | z2w,1=1 z2w,2=1 z2w,3=1 z4=1 |
| 167.4.3 | 51058 | O6Rh1Sr3Yb1 |
| | 148 | z2w,1=1 z2w,2=1 z2w,3=1 z4=3 |
| | 148 | z2w,1=1 z2w,2=1 z2w,3=1 z4=1 |
| 167.4.4 | 81659 | Ni1O6Sc1Sr3 |
| | 148 | z2w,1=1 z2w,2=1 z2w,3=1 z4=3 |
| | 148 | z2w,1=1 z2w,2=1 z2w,3=1 z4=1 |
| 167.4.5 | 51799 | Ni1O6Rh1Sr3 |
| | 2 | z2w,3=1 z4=3 |
| | 2 | z2w,2=1 z4=3 |
| | 2 | z2w,2=1 z2w,3=1 z4=2 |
| | 2 | z2w,1=1 z4=3 |
| | 2 | z2w,1=1 z2w,3=1 z4=2 |
| | 2 | z2w,1=1 z2w,2=1 z4=2 |
| | 2 | z2w,1=1 z2w,2=1 z2w,3=1 z4=1 |
| | 2 | z2w,1=1 z2w,2=1 z2w,3=1 z4=3 |
| | 2 | z2w,1=1 z2w,2=1 z4=2 |
| | 2 | z2w,1=1 z2w,3=1 z4=2 |
| | 2 | z2w,1=1 z4=1 |
| | 2 | z2w,2=1 z2w,3=1 z4=2 |
| | 2 | z2w,2=1 z4=1 |
| | 2 | z2w,3=1 z4=1 |

Continued on next column

**TABLE XIX– continued**

| TQC | ICSD SSG | Stoich. Formula SIs |
|---|---|---|
| 167.4.6 | 69187 | Ni1O3Pr1 |
| | 2 | z4=3 |
| | 2 | z2w,3=1 z4=2 |
| | 2 | z2w,2=1 z4=2 |
| | 2 | z2w,2=1 z2w,3=1 z4=1 |
| | 2 | z2w,1=1 z4=2 |
| | 2 | z2w,1=1 z2w,3=1 z4=1 |
| | 2 | z2w,1=1 z2w,2=1 z4=1 |
| | 2 | z2w,1=1 z2w,2=1 z2w,3=1 |
| | 2 | z2w,1=1 z2w,2=1 z2w,3=1 z4=2 |
| | 2 | z2w,1=1 z2w,2=1 z4=1 |
| | 2 | z2w,1=1 z2w,3=1 z4=1 |
| | 2 | z2w,1=1 |
| | 2 | z2w,2=1 z2w,3=1 z4=1 |
| | 2 | z2w,2=1 |
| | 2 | z2w,3=1 |
| | 2 | z4=3 |
| 167.4.7 | 69186 | Ni1O3Pr1 |
| | 2 | z4=3 |
| | 2 | z2w,3=1 z4=2 |
| | 2 | z2w,2=1 z4=2 |
| | 2 | z2w,2=1 z2w,3=1 z4=1 |
| | 2 | z2w,1=1 z4=2 |
| | 2 | z2w,1=1 z2w,3=1 z4=1 |
| | 2 | z2w,1=1 z2w,2=1 z4=1 |
| | 2 | z2w,1=1 z2w,2=1 z2w,3=1 |
| | 2 | z2w,1=1 z2w,2=1 z2w,3=1 z4=2 |
| | 2 | z2w,1=1 z2w,2=1 z4=1 |
| | 2 | z2w,1=1 z2w,3=1 z4=1 |
| | 2 | z2w,1=1 |
| | 2 | z2w,2=1 z2w,3=1 z4=1 |
| | 2 | z2w,2=1 |
| | 2 | z2w,3=1 |
| | 2 | z4=3 |
| 167.4.8 | 84738 | Mg1O6Rh1Sr3 |
| | 148 | z2w,1=1 z2w,2=1 z2w,3=1 z4=3 |
| | 148 | z2w,1=1 z2w,2=1 z2w,3=1 z4=1 |
| 167.4.9 | 96220 | Li1O6Ru1Sr3 |
| | 148 | z2w,1=1 z2w,2=1 z2w,3=1 z4=3 |
| | 148 | z2w,1=1 z2w,2=1 z2w,3=1 z4=1 |
| 167.4.10 | 50022 | Li1O6Ru1Sr3 |
| | 148 | z2w,1=1 z2w,2=1 z2w,3=1 z4=3 |
| | 148 | z2w,1=1 z2w,2=1 z2w,3=1 z4=1 |
| 167.4.11 | 252432 | Li1O3Os1 |
| | 148 | z2w,1=1 z2w,2=1 z2w,3=1 z4=3 |
| | 148 | z2w,1=1 z2w,2=1 z2w,3=1 z4=1 |
| 167.4.12 | 249150 | Ir3Mg1Si7 |
| | 148 | z2w,1=1 z2w,2=1 z2w,3=1 z4=3 |
| | 148 | z2w,1=1 z2w,2=1 z2w,3=1 z4=1 |
| 167.4.13 | 82660 | Ir1O6Sr3Zn1 |
| | 148 | z2w,1=1 z2w,2=1 z2w,3=1 z4=3 |
| | 148 | z2w,1=1 z2w,2=1 z2w,3=1 z4=1 |
| 167.4.14 | 82659 | Ir1O6Sr3Zn1 |
| | 148 | z2w,1=1 z2w,2=1 z2w,3=1 z4=3 |
| | 148 | z2w,1=1 z2w,2=1 z2w,3=1 z4=1 |
| 167.4.15 | 82658 | Ir1O6Sr3Zn1 |
| | 148 | z2w,1=1 z2w,2=1 z2w,3=1 z4=3 |
| | 148 | z2w,1=1 z2w,2=1 z2w,3=1 z4=1 |
| 167.4.16 | 81903 | Ir1O6Sr3Zn1 |
| | 148 | z2w,1=1 z2w,2=1 z2w,3=1 z4=3 |
| | 148 | z2w,1=1 z2w,2=1 z2w,3=1 z4=1 |
| 167.4.17 | 80287 | Ir1Ni1O6Sr3 |
| | 148 | z2w,1=1 z2w,2=1 z2w,3=1 |
| | 148 | z4=1 |
| | 148 | z4=1 |
| | 148 | z2w,1=1 z2w,2=1 z2w,3=1 z4=2 |
| 167.4.18 | 80286 | Ir1Ni1O6Sr3 |
| | 148 | z2w,1=1 z2w,2=1 z2w,3=1 |
| | 148 | z4=1 |
| | 148 | z4=1 |
| | 148 | z2w,1=1 z2w,2=1 z2w,3=1 z4=2 |
| 167.4.19 | 80285 | Ir1Ni1O6Sr3 |
| | 148 | z2w,1=1 z2w,2=1 z2w,3=1 |
| | 148 | z4=1 |
| | 148 | z4=1 |
| | 148 | z2w,1=1 z2w,2=1 z2w,3=1 z4=2 |
| 167.4.20 | 84737 | Ir1Mg1O6Sr3 |
| | 148 | z2w,1=1 z2w,2=1 z2w,3=1 z4=3 |
| | 148 | z2w,1=1 z2w,2=1 z2w,3=1 z4=1 |
| 167.4.21 | 81660 | In1Ni1O6Sr3 |
| | 148 | z2w,1=1 z2w,2=1 z2w,3=1 z4=3 |
| | 148 | z2w,1=1 z2w,2=1 z2w,3=1 z4=1 |
| 167.4.22 | 50176 | Gd1O6Rh1Sr3 |
| | 148 | z2w,1=1 z2w,2=1 z2w,3=1 |
| | 148 | z4=1 |
| | 148 | z4=1 |
| | 148 | z2w,1=1 z2w,2=1 z2w,3=1 z4=2 |
| 167.4.23 | 236177 | Fe1La1O3 |
| | 148 | z2w,1=1 z2w,2=1 z2w,3=1 z4=3 |
| | 148 | z2w,1=1 z2w,2=1 z2w,3=1 z4=1 |
| 167.4.24 | 236172 | Fe1La1O3 |
| | 148 | z2w,1=1 z2w,2=1 z2w,3=1 z4=3 |
| | 148 | z2w,1=1 z2w,2=1 z2w,3=1 z4=1 |
| 167.4.25 | 236170 | Fe1La1O3 |
| | 148 | z2w,1=1 z2w,2=1 z2w,3=1 z4=3 |
| | 148 | z2w,1=1 z2w,2=1 z2w,3=1 z4=1 |
| 167.4.26 | 236169 | Fe1La1O3 |
| | 148 | z2w,1=1 z2w,2=1 z2w,3=1 z4=3 |
| | 148 | z2w,1=1 z2w,2=1 z2w,3=1 z4=1 |
| 167.4.27 | 67148 | F3Ti1 |
| | 2 | z2w,3=1 z4=3 |
| | 2 | z2w,2=1 z4=3 |
| | 2 | z2w,2=1 z2w,3=1 z4=2 |
| | 2 | z2w,1=1 z4=3 |
| | 2 | z2w,1=1 z2w,3=1 z4=2 |
| | 2 | z2w,1=1 z2w,2=1 z4=2 |
| | 2 | z2w,1=1 z2w,2=1 z2w,3=1 z4=1 |
| | 2 | z2w,1=1 z2w,2=1 z2w,3=1 z4=3 |
| | 2 | z2w,1=1 z2w,2=1 z4=2 |
| | 2 | z2w,1=1 z2w,3=1 z4=2 |
| | 2 | z2w,1=1 z4=1 |
| | 2 | z2w,2=1 z2w,3=1 z4=2 |
| | 2 | z2w,2=1 z4=1 |
| | 2 | z2w,3=1 z4=1 |
| 167.4.28 | 52165 | F3Ti1 |
| | 148 | z2w,1=1 z2w,2=1 z2w,3=1 z4=3 |
| | 148 | z2w,1=1 z2w,2=1 z2w,3=1 z4=1 |
| 167.4.29 | 52162 | F3Ti1 |

Continued on next column

**TABLE XIX– continued**

| TQC | ICSD<br>SSG | Stoich. Formula<br>SIs |
|---|---|---|
| | 148 | z2w,1=1 z2w,2=1 z2w,3=1 z4=3 |
| | 148 | z2w,1=1 z2w,2=1 z2w,3=1 z4=1 |
| 167.4.30 | 52159 | F3Ti1 |
| | 2 | z2w,3=1 z4=3 |
| | 2 | z2w,2=1 z4=3 |
| | 2 | z2w,2=1 z2w,3=1 z4=2 |
| | 2 | z2w,1=1 z4=3 |
| | 2 | z2w,1=1 z2w,3=1 z4=2 |
| | 2 | z2w,1=1 z2w,2=1 z4=2 |
| | 2 | z2w,1=1 z2w,2=1 z2w,3=1 z4=1 |
| | 2 | z2w,1=1 z2w,2=1 z2w,3=1 z4=3 |
| | 2 | z2w,1=1 z2w,2=1 z4=2 |
| | 2 | z2w,1=1 z2w,3=1 z4=2 |
| | 2 | z2w,1=1 z4=1 |
| | 2 | z2w,2=1 z2w,3=1 z4=2 |
| | 2 | z2w,2=1 z4=1 |
| | 2 | z2w,3=1 z4=1 |
| 167.4.31 | 16673 | F3Ru1 |
| | 148 | z2w,1=1 z2w,2=1 z2w,3=1 z4=3 |
| | 148 | z2w,1=1 z2w,2=1 z2w,3=1 z4=1 |
| 167.4.32 | 165400 | F3Ru1 |
| | 148 | z2w,1=1 z2w,2=1 z2w,3=1 z4=3 |
| | 148 | z2w,1=1 z2w,2=1 z2w,3=1 z4=1 |
| 167.4.33 | 16675 | F3Pd1 |
| | 2 | z2w,3=1 z4=3 |
| | 2 | z2w,2=1 z4=3 |
| | 2 | z2w,2=1 z2w,3=1 z4=2 |
| | 2 | z2w,1=1 z4=3 |
| | 2 | z2w,1=1 z2w,3=1 z4=2 |
| | 2 | z2w,1=1 z2w,2=1 z4=2 |
| | 2 | z2w,1=1 z2w,2=1 z2w,3=1 z4=1 |
| | 2 | z2w,1=1 z2w,2=1 z2w,3=1 z4=3 |
| | 2 | z2w,1=1 z2w,2=1 z4=2 |
| | 2 | z2w,1=1 z2w,3=1 z4=2 |
| | 2 | z2w,1=1 z4=1 |
| | 2 | z2w,2=1 z2w,3=1 z4=2 |
| | 2 | z2w,2=1 z4=1 |
| | 2 | z2w,3=1 z4=1 |
| 167.4.34 | 68527 | F3Mo1 |
| | 148 | z2w,1=1 z2w,2=1 z2w,3=1 z4=3 |
| | 148 | z2w,1=1 z2w,2=1 z2w,3=1 z4=1 |
| 167.4.35 | 52170 | F3Fe1 |
| | 148 | z2w,1=1 z2w,2=1 z2w,3=1 z4=3 |
| | 148 | z2w,1=1 z2w,2=1 z2w,3=1 z4=1 |
| 167.4.36 | 52169 | F3Fe1 |
| | 148 | z2w,1=1 z2w,2=1 z2w,3=1 z4=3 |
| | 148 | z2w,1=1 z2w,2=1 z2w,3=1 z4=1 |
| 167.4.37 | 52168 | F3Fe1 |
| | 148 | z2w,1=1 z2w,2=1 z2w,3=1 z4=3 |
| | 148 | z2w,1=1 z2w,2=1 z2w,3=1 z4=1 |
| 167.4.38 | 240402 | F3Fe1 |
| | 148 | z2w,1=1 z2w,2=1 z2w,3=1 z4=3 |
| | 148 | z2w,1=1 z2w,2=1 z2w,3=1 z4=1 |
| 167.4.39 | 240396 | F3Fe1 |
| | 148 | z2w,1=1 z2w,2=1 z2w,3=1 z4=3 |
| | 148 | z2w,1=1 z2w,2=1 z2w,3=1 z4=1 |
| 167.4.40 | 240395 | F3Fe1 |
| | 148 | z2w,1=1 z2w,2=1 z2w,3=1 z4=3 |
| | 148 | z2w,1=1 z2w,2=1 z2w,3=1 z4=1 |

Continued on next column

**TABLE XIX– continued**

| TQC | ICSD<br>SSG | Stoich. Formula<br>SIs |
|---|---|---|
| 167.4.41 | 240394 | F3Fe1 |
| | 148 | z2w,1=1 z2w,2=1 z2w,3=1 z4=3 |
| | 148 | z2w,1=1 z2w,2=1 z2w,3=1 z4=1 |
| 167.4.42 | 51057 | Er1O6Rh1Sr3 |
| | 148 | z2w,1=1 z2w,2=1 z2w,3=1 z4=3 |
| | 148 | z2w,1=1 z2w,2=1 z2w,3=1 z4=1 |
| 167.4.43 | 51055 | Dy1O6Rh1Sr3 |
| | 148 | z2w,1=1 z2w,2=1 z2w,3=1 z4=3 |
| | 148 | z2w,1=1 z2w,2=1 z2w,3=1 z4=1 |
| 167.4.44 | 167590 | Cr1La1O3 |
| | 148 | z2w,1=1 z2w,2=1 z2w,3=1 z4=3 |
| | 148 | z2w,1=1 z2w,2=1 z2w,3=1 z4=1 |
| 167.4.45 | 59973 | Cr1F3 |
| | 148 | z2w,1=1 z2w,2=1 z2w,3=1 z4=3 |
| | 148 | z2w,1=1 z2w,2=1 z2w,3=1 z4=1 |
| 167.4.46 | 59972 | Cr1F3 |
| | 148 | z2w,1=1 z2w,2=1 z2w,3=1 z4=3 |
| | 148 | z2w,1=1 z2w,2=1 z2w,3=1 z4=1 |
| 167.4.47 | 59969 | Cr1F3 |
| | 148 | z2w,1=1 z2w,2=1 z2w,3=1 z4=3 |
| | 148 | z2w,1=1 z2w,2=1 z2w,3=1 z4=1 |
| 167.4.48 | 59968 | Cr1F3 |
| | 148 | z2w,1=1 z2w,2=1 z2w,3=1 z4=3 |
| | 148 | z2w,1=1 z2w,2=1 z2w,3=1 z4=1 |
| 167.4.49 | 189230 | Co1O6Sr3Zn1 |
| | 148 | z2w,1=1 z2w,2=1 z2w,3=1 z4=3 |
| | 148 | z2w,1=1 z2w,2=1 z2w,3=1 z4=1 |
| 167.4.50 | 192479 | Co1O6Ru1Sr3 |
| | 148 | z2w,1=1 z2w,2=1 z2w,3=1 z4=3 |
| | 148 | z2w,1=1 z2w,2=1 z2w,3=1 z4=1 |
| 167.4.51 | 192478 | Co1O6Ru1Sr3 |
| | 148 | z2w,1=1 z2w,2=1 z2w,3=1 z4=3 |
| | 148 | z2w,1=1 z2w,2=1 z2w,3=1 z4=1 |
| 167.4.52 | 290109 | Cl6K3Mn1Na1 |
| | 148 | z2w,1=1 z2w,2=1 z2w,3=1 z4=3 |
| | 148 | z2w,1=1 z2w,2=1 z2w,3=1 z4=1 |
| 167.4.53 | 52832 | Cd6Sb5 |
| | 148 | z2w,1=1 z2w,2=1 z2w,3=1 z4=1 |
| | 148 | z4=2 |
| | 148 | z4=2 |
| | 148 | z2w,1=1 z2w,2=1 z2w,3=1 z4=3 |
| 167.4.54 | 81904 | Cd1Ir1O6Sr3 |
| | 148 | z2w,1=1 z2w,2=1 z2w,3=1 z4=3 |
| | 148 | z2w,1=1 z2w,2=1 z2w,3=1 z4=1 |
| 167.4.55 | 81902 | Ca4Ir1O6 |
| | 148 | z2w,1=1 z2w,2=1 z2w,3=1 z4=3 |
| | 148 | z2w,1=1 z2w,2=1 z2w,3=1 z4=1 |
| 167.4.56 | 280873 | Ca4Ir1O6 |
| | 148 | z2w,1=1 z2w,2=1 z2w,3=1 z4=3 |
| | 148 | z2w,1=1 z2w,2=1 z2w,3=1 z4=1 |
| 167.4.57 | 50021 | Ca3Na1O6Ru1 |
| | 148 | z2w,1=1 z2w,2=1 z2w,3=1 z4=3 |
| | 148 | z2w,1=1 z2w,2=1 z2w,3=1 z4=1 |
| 167.4.58 | 50020 | Ca3Na1O6Ru1 |
| | 148 | z2w,1=1 z2w,2=1 z2w,3=1 z4=3 |
| | 148 | z2w,1=1 z2w,2=1 z2w,3=1 z4=1 |
| 167.4.59 | 50808 | Ca3Mn1O6Zn1 |
| | 148 | z2w,1=1 z2w,2=1 z2w,3=1 z4=3 |
| | 148 | z2w,1=1 z2w,2=1 z2w,3=1 z4=1 |

Continued on next column

**TABLE XIX– continued**

| TQC | ICSD SSG | Stoich. Formula SIs |
|---|---|---|
| 167.4.60 | 96219 | Ca3Li1O6Ru1 |
| | 148 | z2w,1=1 z2w,2=1 z2w,3=1 z4=3 |
| | 148 | z2w,1=1 z2w,2=1 z2w,3=1 z4=1 |
| 167.4.61 | 248206 | Ca3Li1O6Os1 |
| | 148 | z2w,1=1 z2w,2=1 z2w,3=1 z4=3 |
| | 148 | z2w,1=1 z2w,2=1 z2w,3=1 z4=1 |
| 167.4.62 | 96484 | Ca3Fe1O6Rh1 |
| | 148 | z2w,1=1 z2w,2=1 z2w,3=1 z4=3 |
| | 148 | z2w,1=1 z2w,2=1 z2w,3=1 z4=1 |
| 167.4.63 | 50810 | Ca3Fe1O6Rh1 |
| | 148 | z2w,1=1 z2w,2=1 z2w,3=1 z4=3 |
| | 148 | z2w,1=1 z2w,2=1 z2w,3=1 z4=1 |
| 167.4.64 | 181449 | Ca1Ir3Si7 |
| | 148 | z2w,1=1 z2w,2=1 z2w,3=1 z4=3 |
| | 148 | z2w,1=1 z2w,2=1 z2w,3=1 z4=1 |
| 167.4.65 | 81899 | Ca1Ir1O6Sr3 |
| | 148 | z2w,1=1 z2w,2=1 z2w,3=1 z4=3 |
| | 148 | z2w,1=1 z2w,2=1 z2w,3=1 z4=1 |
| 167.4.66 | 80868 | C1Mn1O3 |
| | 148 | z2w,1=1 z2w,2=1 z2w,3=1 z4=3 |
| | 148 | z2w,1=1 z2w,2=1 z2w,3=1 z4=1 |
| 167.4.67 | 80867 | C1Mn1O3 |
| | 148 | z2w,1=1 z2w,2=1 z2w,3=1 z4=3 |
| | 148 | z2w,1=1 z2w,2=1 z2w,3=1 z4=1 |
| 167.4.68 | 196052 | C1Mn1O3 |
| | 148 | z2w,1=1 z2w,2=1 z2w,3=1 z4=3 |
| | 148 | z2w,1=1 z2w,2=1 z2w,3=1 z4=1 |
| 167.4.69 | 196051 | C1Mn1O3 |
| | 148 | z2w,1=1 z2w,2=1 z2w,3=1 z4=3 |
| | 148 | z2w,1=1 z2w,2=1 z2w,3=1 z4=1 |
| 167.4.70 | 196049 | C1Mn1O3 |
| | 148 | z2w,1=1 z2w,2=1 z2w,3=1 z4=3 |
| | 148 | z2w,1=1 z2w,2=1 z2w,3=1 z4=1 |
| 167.4.71 | 196048 | C1Mn1O3 |
| | 148 | z2w,1=1 z2w,2=1 z2w,3=1 z4=3 |
| | 148 | z2w,1=1 z2w,2=1 z2w,3=1 z4=1 |
| 167.4.72 | 196047 | C1Mn1O3 |
| | 148 | z2w,1=1 z2w,2=1 z2w,3=1 z4=3 |
| | 148 | z2w,1=1 z2w,2=1 z2w,3=1 z4=1 |
| 167.4.73 | 196046 | C1Mn1O3 |
| | 148 | z2w,1=1 z2w,2=1 z2w,3=1 z4=3 |
| | 148 | z2w,1=1 z2w,2=1 z2w,3=1 z4=1 |
| 167.4.74 | 196043 | C1Mn1O3 |
| | 148 | z2w,1=1 z2w,2=1 z2w,3=1 z4=3 |
| | 148 | z2w,1=1 z2w,2=1 z2w,3=1 z4=1 |
| 167.4.75 | 196042 | C1Mn1O3 |
| | 148 | z2w,1=1 z2w,2=1 z2w,3=1 z4=3 |
| | 148 | z2w,1=1 z2w,2=1 z2w,3=1 z4=1 |
| 167.4.76 | 196041 | C1Mn1O3 |
| | 148 | z2w,1=1 z2w,2=1 z2w,3=1 z4=3 |
| | 148 | z2w,1=1 z2w,2=1 z2w,3=1 z4=1 |
| 167.4.77 | 196039 | C1Mn1O3 |
| | 148 | z2w,1=1 z2w,2=1 z2w,3=1 z4=3 |
| | 148 | z2w,1=1 z2w,2=1 z2w,3=1 z4=1 |
| 167.4.78 | 196038 | C1Mn1O3 |
| | 148 | z2w,1=1 z2w,2=1 z2w,3=1 z4=3 |
| | 148 | z2w,1=1 z2w,2=1 z2w,3=1 z4=1 |
| 167.4.79 | 196037 | C1Mn1O3 |
| | 148 | z2w,1=1 z2w,2=1 z2w,3=1 z4=3 |

Continued on next column

**TABLE XIX– continued**

| TQC | ICSD SSG | Stoich. Formula SIs |
|---|---|---|
| | 148 | z2w,1=1 z2w,2=1 z2w,3=1 z4=1 |
| 167.4.80 | 196036 | C1Mn1O3 |
| | 148 | z2w,1=1 z2w,2=1 z2w,3=1 z4=3 |
| | 148 | z2w,1=1 z2w,2=1 z2w,3=1 z4=1 |
| 167.4.81 | 196035 | C1Mn1O3 |
| | 148 | z2w,1=1 z2w,2=1 z2w,3=1 z4=3 |
| | 148 | z2w,1=1 z2w,2=1 z2w,3=1 z4=1 |
| 167.4.82 | 196034 | C1Mn1O3 |
| | 148 | z2w,1=1 z2w,2=1 z2w,3=1 z4=3 |
| | 148 | z2w,1=1 z2w,2=1 z2w,3=1 z4=1 |
| 167.4.83 | 196033 | C1Mn1O3 |
| | 148 | z2w,1=1 z2w,2=1 z2w,3=1 z4=3 |
| | 148 | z2w,1=1 z2w,2=1 z2w,3=1 z4=1 |
| 167.4.84 | 196032 | C1Mn1O3 |
| | 148 | z2w,1=1 z2w,2=1 z2w,3=1 z4=3 |
| | 148 | z2w,1=1 z2w,2=1 z2w,3=1 z4=1 |
| 167.4.85 | 196031 | C1Mn1O3 |
| | 148 | z2w,1=1 z2w,2=1 z2w,3=1 z4=3 |
| | 148 | z2w,1=1 z2w,2=1 z2w,3=1 z4=1 |
| 167.4.86 | 196030 | C1Mn1O3 |
| | 148 | z2w,1=1 z2w,2=1 z2w,3=1 z4=3 |
| | 148 | z2w,1=1 z2w,2=1 z2w,3=1 z4=1 |
| 167.4.87 | 196029 | C1Mn1O3 |
| | 148 | z2w,1=1 z2w,2=1 z2w,3=1 z4=3 |
| | 148 | z2w,1=1 z2w,2=1 z2w,3=1 z4=1 |
| 167.4.88 | 196022 | C1Mn1O3 |
| | 148 | z2w,1=1 z2w,2=1 z2w,3=1 z4=3 |
| | 148 | z2w,1=1 z2w,2=1 z2w,3=1 z4=1 |
| 167.4.89 | 196020 | C1Mn1O3 |
| | 148 | z2w,1=1 z2w,2=1 z2w,3=1 z4=3 |
| | 148 | z2w,1=1 z2w,2=1 z2w,3=1 z4=1 |
| 167.4.90 | 196017 | C1Mn1O3 |
| | 148 | z2w,1=1 z2w,2=1 z2w,3=1 z4=3 |
| | 148 | z2w,1=1 z2w,2=1 z2w,3=1 z4=1 |
| 167.4.91 | 100677 | C1Mn1O3 |
| | 148 | z2w,1=1 z2w,2=1 z2w,3=1 z4=3 |
| | 148 | z2w,1=1 z2w,2=1 z2w,3=1 z4=1 |
| 167.4.92 | 61066 | C1Co1O3 |
| | 2 | z2w,3=1 z4=3 |
| | 2 | z2w,2=1 z4=3 |
| | 2 | z2w,2=1 z2w,3=1 z4=2 |
| | 2 | z2w,1=1 z4=3 |
| | 2 | z2w,1=1 z2w,3=1 z4=2 |
| | 2 | z2w,1=1 z2w,2=1 z4=2 |
| | 2 | z2w,1=1 z2w,2=1 z2w,3=1 z4=1 |
| | 2 | z2w,1=1 z2w,2=1 z2w,3=1 z4=3 |
| | 2 | z2w,1=1 z2w,2=1 z4=2 |
| | 2 | z2w,1=1 z2w,3=1 z4=2 |
| | 2 | z2w,1=1 z4=1 |
| | 2 | z2w,2=1 z2w,3=1 z4=2 |
| | 2 | z2w,2=1 z4=1 |
| | 2 | z2w,3=1 z4=1 |
| 167.4.93 | 5396 | C1Co1O3 |
| | 2 | z2w,3=1 z4=3 |
| | 2 | z2w,2=1 z4=3 |
| | 2 | z2w,2=1 z2w,3=1 z4=2 |
| | 2 | z2w,1=1 z4=3 |
| | 2 | z2w,1=1 z2w,3=1 z4=2 |
| | 2 | z2w,1=1 z2w,2=1 z4=2 |

Continued on next column

**TABLE XIX– continued**

| TQC | ICSD<br>SSG | Stoich. Formula<br>SIs |
|---|---|---|
| | 2 | z2w,1=1 z2w,2=1 z2w,3=1 z4=1 |
| | 2 | z2w,1=1 z2w,2=1 z2w,3=1 z4=3 |
| | 2 | z2w,1=1 z2w,2=1 z4=2 |
| | 2 | z2w,1=1 z2w,3=1 z4=2 |
| | 2 | z2w,1=1 z4=1 |
| | 2 | z2w,2=1 z2w,3=1 z4=2 |
| | 2 | z2w,2=1 z4=1 |
| | 2 | z2w,3=1 z4=1 |
| 167.4.94 | 5208 | C1Co1O3 |
| | 2 | z2w,3=1 z4=3 |
| | 2 | z2w,2=1 z4=3 |
| | 2 | z2w,2=1 z2w,3=1 z4=2 |
| | 2 | z2w,1=1 z4=3 |
| | 2 | z2w,1=1 z2w,3=1 z4=2 |
| | 2 | z2w,1=1 z2w,2=1 z4=2 |
| | 2 | z2w,1=1 z2w,2=1 z2w,3=1 z4=1 |
| | 2 | z2w,1=1 z2w,2=1 z2w,3=1 z4=3 |
| | 2 | z2w,1=1 z2w,2=1 z4=2 |
| | 2 | z2w,1=1 z2w,3=1 z4=2 |
| | 2 | z2w,1=1 z4=1 |
| | 2 | z2w,2=1 z2w,3=1 z4=2 |
| | 2 | z2w,2=1 z4=1 |
| | 2 | z2w,3=1 z4=1 |
| 167.4.95 | 251725 | Ba3Mn1Se6U1 |
| | 148 | z2w,1=1 z2w,2=1 z2w,3=1 z4=3 |
| | 148 | z2w,1=1 z2w,2=1 z2w,3=1 z4=1 |
| 167.4.96 | 160141 | B1O3Yb1 |
| | 148 | z2w,1=1 z2w,2=1 z2w,3=1 z4=3 |
| | 148 | z2w,1=1 z2w,2=1 z2w,3=1 z4=1 |
| 167.4.97 | 402039 | B1O3Ti1 |
| | 148 | z2w,1=1 z2w,2=1 z2w,3=1 z4=3 |
| | 148 | z2w,1=1 z2w,2=1 z2w,3=1 z4=1 |
| 167.4.98 | 34474 | B1Fe1O3 |
| | 148 | z2w,1=1 z2w,2=1 z2w,3=1 z4=3 |
| | 148 | z2w,1=1 z2w,2=1 z2w,3=1 z4=1 |
| 167.4.99 | 43311 | B1Cr1O3 |
| | 148 | z2w,1=1 z2w,2=1 z2w,3=1 z4=3 |
| | 148 | z2w,1=1 z2w,2=1 z2w,3=1 z4=1 |
| 167.4.100 | 426465 | Au6Ga3Sr2 |
| | 148 | z2w,1=1 z2w,2=1 z2w,3=1 z4=2 |
| | 148 | z4=3 |
| | 148 | z4=3 |
| | 148 | z2w,1=1 z2w,2=1 z2w,3=1 |
| 167.4.101 | 391112 | Al7Au3Tm1 |
| | 148 | z2w,1=1 z2w,2=1 z2w,3=1 |
| | 148 | z4=1 |
| | 148 | z4=1 |
| | 148 | z2w,1=1 z2w,2=1 z2w,3=1 z4=2 |
| 167.4.102 | 391104 | Al7Au3Pr1 |
| | 148 | z2w,1=1 z2w,2=1 z2w,3=1 |
| | 148 | z4=1 |
| | 148 | z4=1 |
| | 148 | z2w,1=1 z2w,2=1 z2w,3=1 z4=2 |
| 167.4.103 | 391105 | Al7Au3Lu1 |
| | 148 | z2w,1=1 z2w,2=1 z2w,3=1 z4=3 |
| | 148 | z2w,1=1 z2w,2=1 z2w,3=1 z4=1 |
| 167.4.104 | 391106 | Al7Au3Ho1 |
| | 2 | z4=1 |
| | 2 | z2w,3=1 |

Continued on next column

**TABLE XIX– continued**

| TQC | ICSD<br>SSG | Stoich. Formula<br>SIs |
|---|---|---|
| | 2 | z2w,2=1 |
| | 2 | z2w,2=1 z2w,3=1 z4=3 |
| | 2 | z2w,1=1 |
| | 2 | z2w,1=1 z2w,3=1 z4=3 |
| | 2 | z2w,1=1 z2w,2=1 z4=3 |
| | 2 | z2w,1=1 z2w,2=1 z2w,3=1 z4=2 |
| | 2 | z2w,1=1 z2w,2=1 z2w,3=1 |
| | 2 | z2w,1=1 z2w,2=1 z4=3 |
| | 2 | z2w,1=1 z2w,3=1 z4=3 |
| | 2 | z2w,1=1 z4=2 |
| | 2 | z2w,2=1 z2w,3=1 z4=3 |
| | 2 | z2w,2=1 z4=2 |
| | 2 | z2w,3=1 z4=2 |
| | 2 | z4=1 |
| 167.4.105 | 105520 | Al5Mo1 |
| | 2 | z4=1 |
| | 2 | z2w,3=1 |
| | 2 | z2w,2=1 |
| | 2 | z2w,2=1 z2w,3=1 z4=3 |
| | 2 | z2w,1=1 |
| | 2 | z2w,1=1 z2w,3=1 z4=3 |
| | 2 | z2w,1=1 z2w,2=1 z4=3 |
| | 2 | z2w,1=1 z2w,2=1 z2w,3=1 z4=2 |
| | 2 | z2w,1=1 z2w,2=1 z2w,3=1 |
| | 2 | z2w,1=1 z2w,2=1 z4=3 |
| | 2 | z2w,1=1 z2w,3=1 z4=3 |
| | 2 | z2w,1=1 z4=2 |
| | 2 | z2w,2=1 z2w,3=1 z4=3 |
| | 2 | z2w,2=1 z4=2 |
| | 2 | z2w,3=1 z4=2 |
| | 2 | z4=1 |
| 167.4.106 | 427021 | Al3Au6Eu2 |
| | 148 | z2w,1=1 z2w,2=1 z2w,3=1 z4=2 |
| | 148 | z4=3 |
| | 148 | z4=3 |
| | 148 | z2w,1=1 z2w,2=1 z2w,3=1 |
| 167.4.107 | 90577 | Al1Nd1O3 |
| | 148 | z2w,1=1 z2w,2=1 z2w,3=1 z4=3 |
| | 148 | z2w,1=1 z2w,2=1 z2w,3=1 z4=1 |
| 167.4.108 | 90576 | Al1Nd1O3 |
| | 148 | z2w,1=1 z2w,2=1 z2w,3=1 z4=3 |
| | 148 | z2w,1=1 z2w,2=1 z2w,3=1 z4=1 |
| 167.4.109 | 90575 | Al1Nd1O3 |
| | 148 | z2w,1=1 z2w,2=1 z2w,3=1 z4=3 |
| | 148 | z2w,1=1 z2w,2=1 z2w,3=1 z4=1 |
| 167.4.110 | 90574 | Al1Nd1O3 |
| | 148 | z2w,1=1 z2w,2=1 z2w,3=1 z4=3 |
| | 148 | z2w,1=1 z2w,2=1 z2w,3=1 z4=1 |
| 167.4.111 | 90573 | Al1Nd1O3 |
| | 148 | z2w,1=1 z2w,2=1 z2w,3=1 z4=3 |
| | 148 | z2w,1=1 z2w,2=1 z2w,3=1 z4=1 |
| 167.4.112 | 90572 | Al1Nd1O3 |
| | 148 | z2w,1=1 z2w,2=1 z2w,3=1 z4=3 |
| | 148 | z2w,1=1 z2w,2=1 z2w,3=1 z4=1 |
| 167.4.113 | 35444 | Al1Nd1O3 |
| | 148 | z2w,1=1 z2w,2=1 z2w,3=1 z4=3 |
| | 148 | z2w,1=1 z2w,2=1 z2w,3=1 z4=1 |
| 167.4.114 | 249018 | Al1Nd1O3 |
| | 148 | z2w,1=1 z2w,2=1 z2w,3=1 z4=3 |

Continued on next column

**TABLE XIX– continued**

| TQC | ICSD<br>SSG | Stoich. Formula<br>SIs |
|---|---|---|
| | 148 | z2w,1=1 z2w,2=1 z2w,3=1 z4=1 |
| 167.4.115 | 245564 | Al1Ce1O3 |
| | 148 | z2w,1=1 z2w,2=1 z2w,3=1 z4=2 |
| | 148 | z4=3 |
| | 148 | z4=3 |
| | 148 | z2w,1=1 z2w,2=1 z2w,3=1 |
| 167.4.116 | 245563 | Al1Ce1O3 |
| | 148 | z2w,1=1 z2w,2=1 z2w,3=1 z4=2 |
| | 148 | z4=3 |
| | 148 | z4=3 |
| | 148 | z2w,1=1 z2w,2=1 z2w,3=1 |
| 167.4.117 | 245272 | Al1Ce1O3 |
| | 148 | z2w,1=1 z2w,2=1 z2w,3=1 z4=2 |
| | 148 | z4=3 |
| | 148 | z4=3 |
| | 148 | z2w,1=1 z2w,2=1 z2w,3=1 |
| 167.4.118 | 245271 | Al1Ce1O3 |
| | 148 | z2w,1=1 z2w,2=1 z2w,3=1 z4=2 |
| | 148 | z4=3 |
| | 148 | z4=3 |
| | 148 | z2w,1=1 z2w,2=1 z2w,3=1 |
| 167.4.119 | 245269 | Al1Ce1O3 |
| | 148 | z2w,1=1 z2w,2=1 z2w,3=1 z4=2 |
| | 148 | z4=3 |
| | 148 | z4=3 |
| | 148 | z2w,1=1 z2w,2=1 z2w,3=1 |
| 167.4.120 | 245268 | Al1Ce1O3 |
| | 148 | z2w,1=1 z2w,2=1 z2w,3=1 z4=2 |
| | 148 | z4=3 |
| | 148 | z4=3 |
| | 148 | z2w,1=1 z2w,2=1 z2w,3=1 |
| 167.4.121 | 245267 | Al1Ce1O3 |
| | 148 | z2w,1=1 z2w,2=1 z2w,3=1 z4=2 |
| | 148 | z4=3 |
| | 148 | z4=3 |
| | 148 | z2w,1=1 z2w,2=1 z2w,3=1 |
| 167.4.122 | 238056 | Ag1Ba3S6U1 |
| | 148 | z2w,1=1 z2w,2=1 z2w,3=1 z4=3 |
| | 148 | z2w,1=1 z2w,2=1 z2w,3=1 z4=1 |
| 176.4.1 | 72922 | S4V3 |
| | 2 | z4=3 |
| | 2 | z2w,1=1 z2w,3=1 z4=2 |
| | 2 | z2w,2=1 z2w,3=1 z4=2 |
| | 2 | z2w,1=1 z2w,2=1 z4=1 |
| | 2 | z2w,1=1 z2w,2=1 z2w,3=1 z4=2 |
| | 2 | z2w,2=1 z4=1 |
| | 2 | z2w,1=1 z4=1 |
| | 2 | z2w,3=1 |
| | 2 | z2w,3=1 z4=2 |
| | 2 | z2w,1=1 z4=1 |
| | 2 | z2w,2=1 z4=1 |
| | 2 | z2w,1=1 z2w,2=1 z2w,3=1 |
| | 2 | z2w,1=1 z2w,2=1 z4=1 |
| | 2 | z2w,2=1 z2w,3=1 |
| | 2 | z2w,1=1 z2w,3=1 |
| | 2 | z4=3 |
| 176.4.2 | 72921 | S4V3 |
| | 2 | z4=3 |
| | 2 | z2w,1=1 z2w,3=1 z4=2 |

Continued on next column

**TABLE XIX– continued**

| TQC | ICSD<br>SSG | Stoich. Formula<br>SIs |
|---|---|---|
| | 2 | z2w,2=1 z2w,3=1 z4=2 |
| | 2 | z2w,1=1 z2w,2=1 z4=1 |
| | 2 | z2w,1=1 z2w,2=1 z2w,3=1 z4=2 |
| | 2 | z2w,2=1 z4=1 |
| | 2 | z2w,1=1 z4=1 |
| | 2 | z2w,3=1 |
| | 2 | z2w,3=1 z4=2 |
| | 2 | z2w,1=1 z4=1 |
| | 2 | z2w,2=1 z4=1 |
| | 2 | z2w,1=1 z2w,2=1 z2w,3=1 |
| | 2 | z2w,1=1 z2w,2=1 z4=1 |
| | 2 | z2w,2=1 z2w,3=1 |
| | 2 | z2w,1=1 z2w,3=1 |
| | 2 | z4=3 |
| 176.4.3 | 72920 | S4V3 |
| | 2 | z4=3 |
| | 2 | z2w,1=1 z2w,3=1 z4=2 |
| | 2 | z2w,2=1 z2w,3=1 z4=2 |
| | 2 | z2w,1=1 z2w,2=1 z4=1 |
| | 2 | z2w,1=1 z2w,2=1 z2w,3=1 z4=2 |
| | 2 | z2w,2=1 z4=1 |
| | 2 | z2w,1=1 z4=1 |
| | 2 | z2w,3=1 |
| | 2 | z2w,3=1 z4=2 |
| | 2 | z2w,1=1 z4=1 |
| | 2 | z2w,2=1 z4=1 |
| | 2 | z2w,1=1 z2w,2=1 z2w,3=1 |
| | 2 | z2w,1=1 z2w,2=1 z4=1 |
| | 2 | z2w,2=1 z2w,3=1 |
| | 2 | z2w,1=1 z2w,3=1 |
| | 2 | z4=3 |
| 176.4.4 | 16278 | Nb3Se4 |
| | 2 | z4=3 |
| | 2 | z2w,1=1 z2w,3=1 z4=2 |
| | 2 | z2w,2=1 z2w,3=1 z4=2 |
| | 2 | z2w,1=1 z2w,2=1 z4=1 |
| | 2 | z2w,1=1 z2w,2=1 z2w,3=1 z4=2 |
| | 2 | z2w,2=1 z4=1 |
| | 2 | z2w,1=1 z4=1 |
| | 2 | z2w,3=1 |
| | 2 | z2w,3=1 z4=2 |
| | 2 | z2w,1=1 z4=1 |
| | 2 | z2w,2=1 z4=1 |
| | 2 | z2w,1=1 z2w,2=1 z2w,3=1 |
| | 2 | z2w,1=1 z2w,2=1 z4=1 |
| | 2 | z2w,2=1 z2w,3=1 |
| | 2 | z2w,1=1 z2w,3=1 |
| | 2 | z4=3 |
| 176.4.5 | 53119 | Mo6Se6Tl2 |
| | 147 | z2w,3=1 z4=3 |
| | 147 | z2w,3=1 z4=1 |
| 176.4.6 | 23301 | Mo6Se6Tl2 |
| | 147 | z2w,3=1 z4=3 |
| | 147 | z2w,3=1 z4=1 |
| 176.4.7 | 90811 | Mo3Te3Tl1 |
| | 147 | z2w,3=1 z4=3 |
| | 147 | z2w,3=1 z4=1 |
| 176.4.8 | 30752 | K1Mo3S3 |
| | 147 | z2w,3=1 z4=3 |

Continued on next column

**TABLE XIX– continued**

| TQC | ICSD SSG | Stoich. Formula SIs |
|---|---|---|
| | 147 | z2w,3=1 z4=1 |
| 176.4.9 | 53100 | In1Mo3Te3 |
| | 147 | z2w,3=1 z4=3 |
| | 147 | z2w,3=1 z4=1 |
| 176.4.10 | 53098 | In1Mo3Se3 |
| | 147 | z2w,3=1 z4=3 |
| | 147 | z2w,3=1 z4=1 |
| 176.4.11 | 200095 | H3Dy1O3 |
| | 147 | z2w,3=1 z4=2 |
| | 147 | z4=3 |
| | 147 | z4=3 |
| | 147 | z2w,3=1 |
| 176.4.12 | 185605 | H3Dy1O3 |
| | 147 | z2w,3=1 z4=2 |
| | 147 | z4=3 |
| | 147 | z4=3 |
| | 147 | z2w,3=1 |
| 176.4.13 | 100352 | Fe3Te3Tl1 |
| | 2 | z4=1 |
| | 2 | z2w,1=1 z2w,3=1 |
| | 2 | z2w,2=1 z2w,3=1 |
| | 2 | z2w,1=1 z2w,2=1 z4=3 |
| | 2 | z2w,1=1 z2w,2=1 z2w,3=1 |
| | 2 | z2w,2=1 z4=3 |
| | 2 | z2w,1=1 z4=3 |
| | 2 | z2w,3=1 z4=2 |
| | 2 | z2w,3=1 |
| | 2 | z2w,1=1 z4=3 |
| | 2 | z2w,2=1 z4=3 |
| | 2 | z2w,1=1 z2w,2=1 z2w,3=1 z4=2 |
| | 2 | z2w,1=1 z2w,2=1 z4=3 |
| | 2 | z2w,2=1 z2w,3=1 z4=2 |
| | 2 | z2w,1=1 z2w,3=1 z4=2 |
| | 2 | z4=1 |
| 176.4.14 | 154803 | Cr1N3Sr3 |
| | 147 | z2w,3=1 z4=3 |
| | 147 | z2w,3=1 z4=1 |
| 176.4.15 | 4060 | Cl3Pu1 |
| | 2 | z4=3 |
| | 2 | z2w,1=1 z2w,3=1 z4=2 |
| | 2 | z2w,2=1 z2w,3=1 z4=2 |
| | 2 | z2w,1=1 z2w,2=1 z4=1 |
| | 2 | z2w,1=1 z2w,2=1 z2w,3=1 z4=2 |
| | 2 | z2w,2=1 z4=1 |
| | 2 | z2w,1=1 z4=1 |
| | 2 | z2w,3=1 |
| | 2 | z2w,3=1 z4=2 |
| | 2 | z2w,1=1 z4=1 |
| | 2 | z2w,2=1 z4=1 |
| | 2 | z2w,1=1 z2w,2=1 z2w,3=1 |
| | 2 | z2w,1=1 z2w,2=1 z4=1 |
| | 2 | z2w,2=1 z2w,3=1 |
| | 2 | z2w,1=1 z2w,3=1 |
| | 2 | z4=3 |
| 176.4.16 | 9536 | Cl3Cm1 |
| | 147 | z2w,3=1 z4=2 |
| | 147 | z4=3 |
| | 147 | z4=3 |
| | 147 | z2w,3=1 |

Continued on next column

**TABLE XIX– continued**

| TQC | ICSD SSG | Stoich. Formula SIs |
|---|---|---|
| 176.4.17 | 4070 | Br3U1 |
| | 147 | z2w,3=1 z4=2 |
| | 147 | z4=3 |
| | 147 | z4=3 |
| | 147 | z2w,3=1 |
| 176.4.18 | 36502 | Ba3Fe1N3 |
| | 147 | z2w,3=1 z4=2 |
| | 147 | z4=3 |
| | 147 | z4=3 |
| | 147 | z2w,3=1 |
| 176.4.19 | 154802 | Ba3Cr1N3 |
| | 147 | z2w,3=1 z4=3 |
| | 147 | z2w,3=1 z4=1 |
| 193.4.1 | 43692 | Si3Y5 |
| | 189 | z3m0=2 z3mpi=1 |
| | 189 | z3m0=2 z3mpi=2 |
| | 162 | z4=3 |
| | 162 | z2w,3=1 z4=2 |
| | 162 | z2w,3=1 |
| | 162 | z4=3 |
| 193.4.2 | 415741 | Si3Y5 |
| | 189 | z3m0=2 z3mpi=1 |
| | 189 | z3m0=2 z3mpi=2 |
| | 162 | z4=2 |
| | 162 | z2w,3=1 z4=1 |
| | 162 | z2w,3=1 z4=3 |
| | 162 | z4=2 |
| 193.4.3 | 104190 | Si3Tb5 |
| | 189 | z3m0=1 z3mpi=2 |
| | 189 | z3m0=1 z3mpi=1 |
| | 162 | z4=3 |
| | 162 | z2w,3=1 z4=2 |
| | 162 | z2w,3=1 |
| | 162 | z4=3 |
| 193.4.4 | 76159 | Si3Ta5 |
| | 189 | z3m0=1 |
| | 189 | z3m0=1 z3mpi=2 |
| | 189 | z3m0=1 z3mpi=1 |
| | 189 | z3m0=1 |
| | 162 | z2w,3=1 z4=3 |
| | 162 | z2w,3=1 z4=1 |
| 193.4.5 | 415673 | Sb5U3Zr1 |
| | 12 | z2w,1=1 z2w,2=1 z4=2 |
| | 12 | z2w,3=1 z4=3 |
| | 12 | z2w,1=1 z2w,2=1 z2w,3=1 z4=1 |
| | 12 | z2w,1=1 z2w,2=1 z2w,3=1 z4=3 |
| | 12 | z2w,3=1 z4=1 |
| | 12 | z2w,1=1 z2w,2=1 z4=2 |
| 193.4.6 | 43019 | Sb5Ti1U3 |
| | 12 | z2w,1=1 z2w,2=1 |
| | 12 | z4=2 |
| | 12 | z2w,3=1 z4=1 |
| | 12 | z2w,1=1 z2w,2=1 z2w,3=1 z4=3 |
| | 12 | z2w,1=1 z2w,2=1 z2w,3=1 z4=1 |
| | 12 | z2w,3=1 z4=3 |
| | 12 | z4=2 |
| | 12 | z2w,1=1 z2w,2=1 |
| 193.4.7 | 52407 | Sb3Zr5 |
| | 12 | z2w,1=1 z2w,2=1 z4=1 |

Continued on next column

**TABLE XIX– continued**

| TQC | ICSD SSG | Stoich. Formula SIs |
|---|---|---|
| | 12 | z4=3 |
| | 12 | z2w,3=1 z4=2 |
| | 12 | z2w,1=1 z2w,2=1 z2w,3=1 |
| | 12 | z2w,1=1 z2w,2=1 z2w,3=1 z4=2 |
| | 12 | z2w,3=1 |
| | 12 | z4=3 |
| | 12 | z2w,1=1 z2w,2=1 z4=1 |
| 193.4.8 | 189949 | Sb3Zr5 |
| | 12 | z2w,1=1 z2w,2=1 z4=1 |
| | 12 | z4=3 |
| | 12 | z2w,3=1 z4=2 |
| | 12 | z2w,1=1 z2w,2=1 z2w,3=1 |
| | 12 | z2w,1=1 z2w,2=1 z2w,3=1 z4=2 |
| | 12 | z2w,3=1 |
| | 12 | z4=3 |
| | 12 | z2w,1=1 z2w,2=1 z4=1 |
| 193.4.9 | 153607 | Sb3Zr5 |
| | 12 | z2w,1=1 z2w,2=1 z4=1 |
| | 12 | z4=3 |
| | 12 | z2w,3=1 z4=2 |
| | 12 | z2w,1=1 z2w,2=1 z2w,3=1 |
| | 12 | z2w,1=1 z2w,2=1 z2w,3=1 z4=2 |
| | 12 | z2w,3=1 |
| | 12 | z4=3 |
| | 12 | z2w,1=1 z2w,2=1 z4=1 |
| 193.4.10 | 57167 | Sb3Zn1Zr5 |
| | 189 | z3m0=1 |
| | 189 | z3m0=1 z3mpi=2 |
| | 189 | z3m0=1 z3mpi=1 |
| | 189 | z3m0=1 |
| | 162 | z4=3 |
| | 162 | z2w,3=1 z4=2 |
| | 162 | z2w,3=1 |
| | 162 | z4=3 |
| 193.4.11 | 57517 | Sb3Sr5 |
| | 189 | z3mpi=2 |
| | 189 | z3mpi=1 |
| | 162 | z2w,3=1 z4=3 |
| | 162 | z2w,3=1 z4=1 |
| 193.4.12 | 42120 | Sb3Sr5 |
| | 189 | z3mpi=2 |
| | 189 | z3mpi=1 |
| | 162 | z2w,3=1 z4=3 |
| | 162 | z2w,3=1 z4=1 |
| 193.4.13 | 57144 | Sb3Si1Zr5 |
| | 12 | z2w,1=1 z2w,2=1 z4=2 |
| | 12 | z2w,3=1 z4=3 |
| | 12 | z2w,1=1 z2w,2=1 z2w,3=1 z4=1 |
| | 12 | z2w,1=1 z2w,2=1 z2w,3=1 z4=3 |
| | 12 | z2w,3=1 z4=1 |
| | 12 | z2w,1=1 z2w,2=1 z4=2 |
| 193.4.14 | 153608 | Sb3Si1Zr5 |
| | 12 | z2w,1=1 z2w,2=1 z4=2 |
| | 12 | z2w,3=1 z4=3 |
| | 12 | z2w,1=1 z2w,2=1 z2w,3=1 z4=1 |
| | 12 | z2w,1=1 z2w,2=1 z2w,3=1 z4=3 |
| | 12 | z2w,3=1 z4=1 |
| | 12 | z2w,1=1 z2w,2=1 z4=2 |
| 193.4.15 | 168436 | Pb3Tm5 |

Continued on next column

**TABLE XIX– continued**

| TQC | ICSD SSG | Stoich. Formula SIs |
|---|---|---|
| | 12 | z2w,1=1 z2w,2=1 z4=2 |
| | 12 | z2w,3=1 z4=3 |
| | 12 | z2w,1=1 z2w,2=1 z2w,3=1 z4=1 |
| | 12 | z2w,1=1 z2w,2=1 z2w,3=1 z4=3 |
| | 12 | z2w,3=1 z4=1 |
| | 12 | z2w,1=1 z2w,2=1 z4=2 |
| 193.4.16 | 105614 | Pb3Sc5 |
| | 12 | z2w,1=1 z2w,2=1 |
| | 12 | z4=2 |
| | 12 | z2w,3=1 z4=1 |
| | 12 | z2w,1=1 z2w,2=1 z2w,3=1 z4=3 |
| | 12 | z2w,1=1 z2w,2=1 z2w,3=1 z4=1 |
| | 12 | z2w,3=1 z4=3 |
| | 12 | z4=2 |
| | 12 | z2w,1=1 z2w,2=1 |
| 193.4.17 | 24148 | P3Ti5 |
| | 189 | z3m0=2 z3mpi=1 |
| | 189 | z3m0=2 z3mpi=2 |
| | 162 | z2w,3=1 z4=3 |
| | 162 | z2w,3=1 z4=1 |
| 193.4.18 | 96419 | O3Rb0.333W1 |
| | 12 | z2w,1=1 z2w,2=1 z4=1 |
| | 12 | z4=3 |
| | 12 | z2w,3=1 z4=2 |
| | 12 | z2w,1=1 z2w,2=1 z2w,3=1 |
| | 12 | z2w,1=1 z2w,2=1 z2w,3=1 z4=2 |
| | 12 | z2w,3=1 |
| | 12 | z4=3 |
| | 12 | z2w,1=1 z2w,2=1 z4=1 |
| 193.4.19 | 152607 | Ni1Pb3Tm5 |
| | 12 | z2w,1=1 z2w,2=1 z4=1 |
| | 12 | z4=3 |
| | 12 | z2w,3=1 z4=2 |
| | 12 | z2w,1=1 z2w,2=1 z2w,3=1 |
| | 12 | z2w,1=1 z2w,2=1 z2w,3=1 z4=2 |
| | 12 | z2w,3=1 |
| | 12 | z4=3 |
| | 12 | z2w,1=1 z2w,2=1 z4=1 |
| 193.4.20 | 152602 | Ni1Pb3Pr5 |
| | 12 | z2w,1=1 z2w,2=1 z4=2 |
| | 12 | z2w,3=1 z4=3 |
| | 12 | z2w,1=1 z2w,2=1 z2w,3=1 z4=1 |
| | 12 | z2w,1=1 z2w,2=1 z2w,3=1 z4=3 |
| | 12 | z2w,3=1 z4=1 |
| | 12 | z2w,1=1 z2w,2=1 z4=2 |
| 193.4.21 | 36277 | Nb5O1Pt3 |
| | 12 | z2w,1=1 z2w,2=1 z4=2 |
| | 12 | z2w,3=1 z4=3 |
| | 12 | z2w,1=1 z2w,2=1 z2w,3=1 z4=1 |
| | 12 | z2w,1=1 z2w,2=1 z2w,3=1 z4=3 |
| | 12 | z2w,3=1 z4=1 |
| | 12 | z2w,1=1 z2w,2=1 z4=2 |
| 193.4.22 | 76461 | N6Ta5 |
| | 189 | z3mpi=2 |
| | 189 | z3mpi=1 |
| | 162 | z4=2 |
| | 162 | z2w,3=1 z4=1 |
| | 162 | z2w,3=1 z4=3 |
| | 162 | z4=2 |

Continued on next column

**TABLE XIX– continued**

| TQC | ICSD SSG | Stoich. Formula SIs |
|---|---|---|
| 193.4.23 | 29329 | N6Ta5 |
| | 189 | z3mpi=2 |
| | 189 | z3mpi=1 |
| | 162 | z4=2 |
| | 162 | z2w,3=1 z4=1 |
| | 162 | z2w,3=1 z4=3 |
| | 162 | z4=2 |
| 193.4.24 | 166512 | N6Ta5 |
| | 189 | z3mpi=2 |
| | 189 | z3mpi=1 |
| | 162 | z4=2 |
| | 162 | z2w,3=1 z4=1 |
| | 162 | z2w,3=1 z4=3 |
| | 162 | z4=2 |
| 193.4.25 | 26252 | N6Nb5 |
| | 189 | z3mpi=2 |
| | 189 | z3mpi=1 |
| | 162 | z4=2 |
| | 162 | z2w,3=1 z4=1 |
| | 162 | z2w,3=1 z4=3 |
| | 162 | z4=2 |
| 193.4.26 | 29285 | N1P3V5 |
| | 189 | z3m0=2 z3mpi=2 |
| | 189 | z3m0=2 z3mpi=1 |
| | 162 | z2w,3=1 z4=3 |
| | 162 | z2w,3=1 z4=1 |
| 193.4.27 | 42435 | Mn5Si3 |
| | 189 | z3m0=2 z3mpi=1 |
| | 189 | z3m0=2 z3mpi=2 |
| | 162 | z4=1 |
| | 162 | z2w,3=1 |
| | 162 | z2w,3=1 z4=2 |
| | 162 | z4=1 |
| 193.4.28 | 152600 | La5Ni1Pb3 |
| | 189 | z3mpi=1 |
| | 189 | z3mpi=2 |
| | 162 | z4=3 |
| | 162 | z2w,3=1 z4=2 |
| | 162 | z2w,3=1 |
| | 162 | z4=3 |
| 193.4.29 | 35317 | I3Zr1 |
| | 12 | z2w,1=1 z2w,2=1 z4=2 |
| | 12 | z2w,3=1 z4=3 |
| | 12 | z2w,1=1 z2w,2=1 z2w,3=1 z4=1 |
| | 12 | z2w,1=1 z2w,2=1 z2w,3=1 z4=3 |
| | 12 | z2w,3=1 z4=1 |
| | 12 | z2w,1=1 z2w,2=1 z4=2 |
| 193.4.30 | 32704 | I3Zr1 |
| | 12 | z2w,1=1 z2w,2=1 z4=2 |
| | 12 | z2w,3=1 z4=3 |
| | 12 | z2w,1=1 z2w,2=1 z2w,3=1 z4=1 |
| | 12 | z2w,1=1 z2w,2=1 z2w,3=1 z4=3 |
| | 12 | z2w,3=1 z4=1 |
| | 12 | z2w,1=1 z2w,2=1 z4=2 |
| 193.4.31 | 23946 | I3Zr1 |
| | 12 | z2w,1=1 z2w,2=1 z4=2 |
| | 12 | z2w,3=1 z4=3 |
| | 12 | z2w,1=1 z2w,2=1 z2w,3=1 z4=1 |
| | 12 | z2w,1=1 z2w,2=1 z2w,3=1 z4=3 |

Continued on next column

**TABLE XIX– continued**

| TQC | ICSD SSG | Stoich. Formula SIs |
|---|---|---|
| | 12 | z2w,3=1 z4=1 |
| | 12 | z2w,1=1 z2w,2=1 z4=2 |
| 193.4.32 | 422871 | I3Ti1 |
| | 12 | z2w,1=1 z2w,2=1 z4=2 |
| | 12 | z2w,3=1 z4=3 |
| | 12 | z2w,1=1 z2w,2=1 z2w,3=1 z4=1 |
| | 12 | z2w,1=1 z2w,2=1 z2w,3=1 z4=3 |
| | 12 | z2w,3=1 z4=1 |
| | 12 | z2w,1=1 z2w,2=1 z4=2 |
| 193.4.33 | 23172 | I3Ti1 |
| | 12 | z2w,1=1 z2w,2=1 z4=2 |
| | 12 | z2w,3=1 z4=3 |
| | 12 | z2w,1=1 z2w,2=1 z2w,3=1 z4=1 |
| | 12 | z2w,1=1 z2w,2=1 z2w,3=1 z4=3 |
| | 12 | z2w,3=1 z4=1 |
| | 12 | z2w,1=1 z2w,2=1 z4=2 |
| 193.4.34 | 415743 | Ho5Si3 |
| | 189 | z3mpi=2 |
| | 189 | z3mpi=1 |
| | 162 | z4=3 |
| | 162 | z2w,3=1 z4=2 |
| | 162 | z2w,3=1 |
| | 162 | z4=3 |
| 193.4.35 | 42913 | Hf5Sb3Zn1 |
| | 189 | z3m0=1 |
| | 189 | z3m0=1 z3mpi=2 |
| | 189 | z3m0=1 z3mpi=1 |
| | 189 | z3m0=1 |
| | 162 | z4=3 |
| | 162 | z2w,3=1 z4=2 |
| | 162 | z2w,3=1 |
| | 162 | z4=3 |
| 193.4.36 | 23947 | Hf1I3 |
| | 12 | z2w,1=1 z2w,2=1 z4=2 |
| | 12 | z2w,3=1 z4=3 |
| | 12 | z2w,1=1 z2w,2=1 z2w,3=1 z4=1 |
| | 12 | z2w,1=1 z2w,2=1 z2w,3=1 z4=3 |
| | 12 | z2w,3=1 z4=1 |
| | 12 | z2w,1=1 z2w,2=1 z4=2 |
| 193.4.37 | 51371 | Ge4Np5 |
| | 12 | z2w,1=1 z2w,2=1 z4=2 |
| | 12 | z2w,3=1 z4=3 |
| | 12 | z2w,1=1 z2w,2=1 z2w,3=1 z4=1 |
| | 12 | z2w,1=1 z2w,2=1 z2w,3=1 z4=3 |
| | 12 | z2w,3=1 z4=1 |
| | 12 | z2w,1=1 z2w,2=1 z4=2 |
| 193.4.38 | 57122 | Ge3Y5 |
| | 189 | z3m0=2 z3mpi=1 |
| | 189 | z3m0=2 z3mpi=2 |
| | 162 | z4=3 |
| | 162 | z2w,3=1 z4=2 |
| | 162 | z2w,3=1 |
| | 162 | z4=3 |
| 193.4.39 | 44504 | Ge3V5 |
| | 189 | z3m0=1 |
| | 189 | z3m0=1 z3mpi=2 |
| | 189 | z3m0=1 z3mpi=1 |
| | 189 | z3m0=1 |
| | 162 | z2w,3=1 z4=3 |

Continued on next column

**TABLE XIX– continued**

| TQC | ICSD SSG | Stoich. Formula SIs |
|---|---|---|
| | 162 | z2w,3=1 z4=1 |
| 193.4.40 | 185874 | Ge3Tb5 |
| | 12 | z2w,1=1 z2w,2=1 z4=1 |
| | 12 | z4=3 |
| | 12 | z2w,3=1 z4=2 |
| | 12 | z2w,1=1 z2w,2=1 z2w,3=1 |
| | 12 | z2w,1=1 z2w,2=1 z2w,3=1 z4=2 |
| | 12 | z2w,3=1 |
| | 12 | z4=3 |
| | 12 | z2w,1=1 z2w,2=1 z4=1 |
| 193.4.41 | 56028 | Ge3Ta5 |
| | 189 | z3m0=2 |
| | 189 | z3m0=2 z3mpi=2 |
| | 189 | z3m0=2 z3mpi=1 |
| | 189 | z3m0=2 |
| | 162 | z2w,3=1 z4=3 |
| | 162 | z2w,3=1 z4=1 |
| 193.4.42 | 76138 | Ge3Mn5 |
| | 12 | z2w,1=1 z2w,2=1 z4=1 |
| | 12 | z4=3 |
| | 12 | z2w,3=1 z4=2 |
| | 12 | z2w,1=1 z2w,2=1 z2w,3=1 |
| | 12 | z2w,1=1 z2w,2=1 z2w,3=1 z4=2 |
| | 12 | z2w,3=1 |
| | 12 | z4=3 |
| | 12 | z2w,1=1 z2w,2=1 z4=1 |
| 193.4.43 | 108517 | Ge3Mn5 |
| | 12 | z2w,1=1 z2w,2=1 z4=1 |
| | 12 | z4=3 |
| | 12 | z2w,3=1 z4=2 |
| | 12 | z2w,1=1 z2w,2=1 z2w,3=1 |
| | 12 | z2w,1=1 z2w,2=1 z2w,3=1 z4=2 |
| | 12 | z2w,3=1 |
| | 12 | z4=3 |
| | 12 | z2w,1=1 z2w,2=1 z4=1 |
| 193.4.44 | 43402 | Ge3La5 |
| | 189 | z3m0=2 z3mpi=1 |
| | 189 | z3m0=2 z3mpi=2 |
| | 162 | z4=3 |
| | 162 | z2w,3=1 z4=2 |
| | 162 | z2w,3=1 |
| | 162 | z4=3 |
| 193.4.45 | 103837 | Ga4Nb5 |
| | 189 | z3m0=1 |
| | 189 | z3m0=1 z3mpi=2 |
| | 189 | z3m0=1 z3mpi=1 |
| | 189 | z3m0=1 |
| | 162 | z4=2 |
| | 162 | z2w,3=1 z4=1 |
| | 162 | z2w,3=1 z4=3 |
| | 162 | z4=2 |
| 193.4.46 | 104003 | Ga3Ti2Zr3 |
| | 189 | z3m0=2 z3mpi=2 |
| | 189 | z3m0=2 z3mpi=1 |
| | 162 | z4=3 |
| | 162 | z2w,3=1 z4=2 |
| | 162 | z2w,3=1 |
| | 162 | z4=3 |
| 193.4.47 | 103730 | Ga3Hf5 |

Continued on next column

**TABLE XIX– continued**

| TQC | ICSD SSG | Stoich. Formula SIs |
|---|---|---|
| | 12 | z2w,1=1 z2w,2=1 z4=1 |
| | 12 | z4=3 |
| | 12 | z2w,3=1 z4=2 |
| | 12 | z2w,1=1 z2w,2=1 z2w,3=1 |
| | 12 | z2w,1=1 z2w,2=1 z2w,3=1 z4=2 |
| | 12 | z2w,3=1 |
| | 12 | z4=3 |
| | 12 | z2w,1=1 z2w,2=1 z4=1 |
| 193.4.48 | 103733 | Ga3Hf3Nb2 |
| | 189 | z3m0=2 |
| | 189 | z3m0=2 z3mpi=2 |
| | 189 | z3m0=2 z3mpi=1 |
| | 189 | z3m0=2 |
| | 162 | z4=1 |
| | 162 | z2w,3=1 |
| | 162 | z2w,3=1 z4=2 |
| | 162 | z4=1 |
| 193.4.49 | 106753 | Ga1Sn3Zr5 |
| | 12 | z2w,1=1 z2w,2=1 z4=3 |
| | 12 | z4=1 |
| | 12 | z2w,3=1 |
| | 12 | z2w,1=1 z2w,2=1 z2w,3=1 z4=2 |
| | 12 | z2w,1=1 z2w,2=1 z2w,3=1 |
| | 12 | z2w,3=1 z4=2 |
| | 12 | z4=1 |
| | 12 | z2w,1=1 z2w,2=1 z4=3 |
| 193.4.50 | 183490 | Ga1Hf5Sn3 |
| | 12 | z2w,1=1 z2w,2=1 z4=3 |
| | 12 | z4=1 |
| | 12 | z2w,3=1 |
| | 12 | z2w,1=1 z2w,2=1 z2w,3=1 z4=2 |
| | 12 | z2w,1=1 z2w,2=1 z2w,3=1 |
| | 12 | z2w,3=1 z4=2 |
| | 12 | z4=1 |
| | 12 | z2w,1=1 z2w,2=1 z4=3 |
| 193.4.51 | 107310 | Cu1Hf5Sn3 |
| | 12 | z2w,1=1 z2w,2=1 z4=3 |
| | 12 | z4=1 |
| | 12 | z2w,3=1 |
| | 12 | z2w,1=1 z2w,2=1 z2w,3=1 z4=2 |
| | 12 | z2w,1=1 z2w,2=1 z2w,3=1 |
| | 12 | z2w,3=1 z4=2 |
| | 12 | z4=1 |
| | 12 | z2w,1=1 z2w,2=1 z4=3 |
| 193.4.52 | 107314 | Cu1Hf5Pb3 |
| | 12 | z2w,1=1 z2w,2=1 z4=3 |
| | 12 | z4=1 |
| | 12 | z2w,3=1 |
| | 12 | z2w,1=1 z2w,2=1 z2w,3=1 z4=2 |
| | 12 | z2w,1=1 z2w,2=1 z2w,3=1 |
| | 12 | z2w,3=1 z4=2 |
| | 12 | z4=1 |
| | 12 | z2w,1=1 z2w,2=1 z4=3 |
| 193.4.53 | 15695 | Cs3O1 |
| | 189 | z3m0=1 |
| | 189 | z3m0=1 z3mpi=2 |
| | 189 | z3m0=1 z3mpi=1 |
| | 189 | z3m0=1 |
| | 162 | z2w,3=1 z4=3 |

Continued on next column

**TABLE XIX– continued**

| TQC | ICSD SSG | Stoich. Formula SIs |
|---|---|---|
| | 162 | z2w,3=1 z4=1 |
| 193.4.54 | 35315 | Cl3Zr1 |
| | 189 | z3mpi=2 |
| | 189 | z3mpi=1 |
| | 162 | z2w,3=1 z4=3 |
| | 162 | z2w,3=1 z4=1 |
| 193.4.55 | 32702 | Cl3Zr1 |
| | 189 | z3mpi=2 |
| | 189 | z3mpi=1 |
| | 162 | z2w,3=1 z4=3 |
| | 162 | z2w,3=1 z4=1 |
| 193.4.56 | 23944 | Cl3Zr1 |
| | 189 | z3mpi=2 |
| | 189 | z3mpi=1 |
| | 162 | z2w,3=1 z4=3 |
| | 162 | z2w,3=1 z4=1 |
| 193.4.57 | 23163 | Cl3Zr1 |
| | 189 | z3mpi=2 |
| | 189 | z3mpi=1 |
| | 162 | z2w,3=1 z4=3 |
| | 162 | z2w,3=1 z4=1 |
| 193.4.58 | 26069 | Cl3Ti1 |
| | 12 | z2w,1=1 z2w,2=1 z4=2 |
| | 12 | z2w,3=1 z4=3 |
| | 12 | z2w,1=1 z2w,2=1 z2w,3=1 z4=1 |
| | 12 | z2w,1=1 z2w,2=1 z2w,3=1 z4=3 |
| | 12 | z2w,3=1 z4=1 |
| | 12 | z2w,1=1 z2w,2=1 z4=2 |
| 193.4.59 | 414040 | Cl3Ru1 |
| | 189 | z3mpi=2 |
| | 189 | z3mpi=1 |
| | 162 | z2w,3=1 z4=3 |
| | 162 | z2w,3=1 z4=1 |
| 193.4.60 | 25771 | Cl3Ru1 |
| | 189 | z3mpi=2 |
| | 189 | z3mpi=1 |
| | 162 | z2w,3=1 z4=3 |
| | 162 | z2w,3=1 z4=1 |
| 193.4.61 | 22090 | Cl3Ru1 |
| | 189 | z3mpi=2 |
| | 189 | z3mpi=1 |
| | 162 | z2w,3=1 z4=3 |
| | 162 | z2w,3=1 z4=1 |
| 193.4.62 | 42470 | Ce1F3 |
| | 12 | z2w,1=1 z2w,2=1 z4=1 |
| | 12 | z4=3 |
| | 12 | z2w,3=1 z4=2 |
| | 12 | z2w,1=1 z2w,2=1 z2w,3=1 |
| | 12 | z2w,1=1 z2w,2=1 z2w,3=1 z4=2 |
| | 12 | z2w,3=1 |
| | 12 | z4=3 |
| | 12 | z2w,1=1 z2w,2=1 z4=1 |
| 193.4.63 | 33796 | Ca6Fe1N5 |
| | 162 | z2w,3=1 z4=3 |
| | 162 | z2w,3=1 z4=1 |
| 193.4.64 | 173026 | Ca5Sb3 |
| | 189 | z3mpi=2 |
| | 189 | z3mpi=1 |
| | 162 | z4=3 |

Continued on next column

**TABLE XIX– continued**

| TQC | ICSD SSG | Stoich. Formula SIs |
|---|---|---|
| | 162 | z2w,3=1 z4=2 |
| | 162 | z2w,3=1 |
| | 162 | z4=3 |
| 193.4.65 | 4068 | Br3Zr1 |
| | 12 | z2w,1=1 z2w,2=1 z4=2 |
| | 12 | z2w,3=1 z4=3 |
| | 12 | z2w,1=1 z2w,2=1 z2w,3=1 z4=1 |
| | 12 | z2w,1=1 z2w,2=1 z2w,3=1 z4=3 |
| | 12 | z2w,3=1 z4=1 |
| | 12 | z2w,1=1 z2w,2=1 z4=2 |
| 193.4.66 | 35316 | Br3Zr1 |
| | 12 | z2w,1=1 z2w,2=1 z4=2 |
| | 12 | z2w,3=1 z4=3 |
| | 12 | z2w,1=1 z2w,2=1 z2w,3=1 z4=1 |
| | 12 | z2w,1=1 z2w,2=1 z2w,3=1 z4=3 |
| | 12 | z2w,3=1 z4=1 |
| | 12 | z2w,1=1 z2w,2=1 z4=2 |
| 193.4.67 | 32703 | Br3Zr1 |
| | 12 | z2w,1=1 z2w,2=1 z4=2 |
| | 12 | z2w,3=1 z4=3 |
| | 12 | z2w,1=1 z2w,2=1 z2w,3=1 z4=1 |
| | 12 | z2w,1=1 z2w,2=1 z2w,3=1 z4=3 |
| | 12 | z2w,3=1 z4=1 |
| | 12 | z2w,1=1 z2w,2=1 z4=2 |
| 193.4.68 | 23945 | Br3Zr1 |
| | 12 | z2w,1=1 z2w,2=1 z4=2 |
| | 12 | z2w,3=1 z4=3 |
| | 12 | z2w,1=1 z2w,2=1 z2w,3=1 z4=1 |
| | 12 | z2w,1=1 z2w,2=1 z2w,3=1 z4=3 |
| | 12 | z2w,3=1 z4=1 |
| | 12 | z2w,1=1 z2w,2=1 z4=2 |
| 193.4.69 | 165302 | Br3Zr1 |
| | 12 | z2w,1=1 z2w,2=1 z4=2 |
| | 12 | z2w,3=1 z4=3 |
| | 12 | z2w,1=1 z2w,2=1 z2w,3=1 z4=1 |
| | 12 | z2w,1=1 z2w,2=1 z2w,3=1 z4=3 |
| | 12 | z2w,3=1 z4=1 |
| | 12 | z2w,1=1 z2w,2=1 z4=2 |
| 193.4.70 | 414043 | Br3Ru1 |
| | 189 | z3mpi=2 |
| | 189 | z3mpi=1 |
| | 162 | z2w,3=1 z4=3 |
| | 162 | z2w,3=1 z4=1 |
| 193.4.71 | 413692 | Br3Ru1 |
| | 189 | z3mpi=2 |
| | 189 | z3mpi=1 |
| | 162 | z2w,3=1 z4=3 |
| | 162 | z2w,3=1 z4=1 |
| 193.4.72 | 413691 | Br3Ru1 |
| | 189 | z3mpi=2 |
| | 189 | z3mpi=1 |
| | 162 | z2w,3=1 z4=3 |
| | 162 | z2w,3=1 z4=1 |
| 193.4.73 | 50725 | Br1La5Sb3 |
| | 12 | z2w,1=1 z2w,2=1 |
| | 12 | z4=2 |
| | 12 | z2w,3=1 z4=1 |
| | 12 | z2w,1=1 z2w,2=1 z2w,3=1 z4=3 |
| | 12 | z2w,1=1 z2w,2=1 z2w,3=1 z4=1 |

Continued on next column

**TABLE XIX– continued**

| TQC | ICSD SSG | Stoich. Formula SIs |
|---|---|---|
| | 12 | z2w,3=1 z4=3 |
| | 12 | z4=2 |
| | 12 | z2w,1=1 z2w,2=1 |
| 193.4.74 | 409631 | Br1In3La5 |
| | 12 | z2w,1=1 z2w,2=1 z4=3 |
| | 12 | z4=1 |
| | 12 | z2w,3=1 |
| | 12 | z2w,1=1 z2w,2=1 z2w,3=1 z4=2 |
| | 12 | z2w,1=1 z2w,2=1 z2w,3=1 |
| | 12 | z2w,3=1 z4=2 |
| | 12 | z4=1 |
| | 12 | z2w,1=1 z2w,2=1 z4=3 |
| 193.4.75 | 416609 | Bi5La3Sc1 |
| | 12 | z2w,1=1 z2w,2=1 z4=1 |
| | 12 | z4=3 |
| | 12 | z2w,3=1 z4=2 |
| | 12 | z2w,1=1 z2w,2=1 z2w,3=1 |
| | 12 | z2w,1=1 z2w,2=1 z2w,3=1 z4=2 |
| | 12 | z2w,3=1 |
| | 12 | z4=3 |
| | 12 | z2w,1=1 z2w,2=1 z4=1 |
| 193.4.76 | 173362 | Bi5La3Mn1 |
| | 12 | z2w,1=1 z2w,2=1 z4=1 |
| | 12 | z4=3 |
| | 12 | z2w,3=1 z4=2 |
| | 12 | z2w,1=1 z2w,2=1 z2w,3=1 |
| | 12 | z2w,1=1 z2w,2=1 z2w,3=1 z4=2 |
| | 12 | z2w,3=1 |
| | 12 | z4=3 |
| | 12 | z2w,1=1 z2w,2=1 z4=1 |
| 193.4.77 | 41838 | Bi3Sr5 |
| | 189 | z3mpi=2 |
| | 189 | z3mpi=1 |
| | 162 | z4=3 |
| | 162 | z2w,3=1 z4=2 |
| | 162 | z2w,3=1 |
| | 162 | z4=3 |
| 193.4.78 | 95241 | Bi3Br1La5 |
| | 12 | z2w,1=1 z2w,2=1 z4=3 |
| | 12 | z4=1 |
| | 12 | z2w,3=1 |
| | 12 | z2w,1=1 z2w,2=1 z2w,3=1 z4=2 |
| | 12 | z2w,1=1 z2w,2=1 z2w,3=1 |
| | 12 | z2w,3=1 z4=2 |
| | 12 | z4=1 |
| | 12 | z2w,1=1 z2w,2=1 z4=3 |
| 193.4.79 | 74697 | Bi0.67S2Ta1 |
| | 162 | z4=2 |
| | 162 | z2w,3=1 z4=1 |
| | 162 | z2w,3=1 z4=3 |
| | 162 | z4=2 |
| 193.4.80 | 41840 | Ba5Sb3 |
| | 189 | z3mpi=2 |
| | 189 | z3mpi=1 |
| | 162 | z2w,3=1 z4=3 |
| | 162 | z2w,3=1 z4=1 |
| 193.4.81 | 41839 | Ba5Bi3 |
| | 189 | z3mpi=2 |
| | 189 | z3mpi=1 |

Continued on next column

**TABLE XIX– continued**

| TQC | ICSD SSG | Stoich. Formula SIs |
|---|---|---|
| | 162 | z4=3 |
| | 162 | z2w,3=1 z4=2 |
| | 162 | z2w,3=1 |
| | 162 | z4=3 |
| 193.4.82 | 77730 | Ba3N1 |
| | 189 | z3mpi=1 |
| | 189 | z3mpi=2 |
| | 162 | z2w,3=1 z4=3 |
| | 162 | z2w,3=1 z4=1 |
| 193.4.83 | 44067 | As3Sr5 |
| | 189 | z3mpi=2 |
| | 189 | z3mpi=1 |
| | 162 | z2w,3=1 z4=3 |
| | 162 | z2w,3=1 z4=1 |
| 193.4.84 | 43877 | As3Ca5 |
| | 189 | z3mpi=2 |
| | 189 | z3mpi=1 |
| | 162 | z2w,3=1 z4=3 |
| | 162 | z2w,3=1 z4=1 |
| 193.4.85 | 43872 | As3Ba5 |
| | 189 | z3mpi=2 |
| | 189 | z3mpi=1 |
| | 162 | z2w,3=1 z4=3 |
| | 162 | z2w,3=1 z4=1 |
| 193.4.86 | 58235 | Al3Zr5 |
| | 12 | z2w,1=1 z2w,2=1 z4=1 |
| | 12 | z4=3 |
| | 12 | z2w,3=1 z4=2 |
| | 12 | z2w,1=1 z2w,2=1 z2w,3=1 |
| | 12 | z2w,1=1 z2w,2=1 z2w,3=1 z4=2 |
| | 12 | z2w,3=1 |
| | 12 | z4=3 |
| | 12 | z2w,1=1 z2w,2=1 z4=1 |
| 193.4.87 | 60633 | Al3Hf5 |
| | 12 | z2w,1=1 z2w,2=1 z4=1 |
| | 12 | z4=3 |
| | 12 | z2w,3=1 z4=2 |
| | 12 | z2w,1=1 z2w,2=1 z2w,3=1 |
| | 12 | z2w,1=1 z2w,2=1 z2w,3=1 z4=2 |
| | 12 | z2w,3=1 |
| | 12 | z4=3 |
| | 12 | z2w,1=1 z2w,2=1 z4=1 |
| 193.4.88 | 57898 | Al3Hf5 |
| | 12 | z2w,1=1 z2w,2=1 |
| | 12 | z4=2 |
| | 12 | z2w,3=1 z4=1 |
| | 12 | z2w,1=1 z2w,2=1 z2w,3=1 z4=3 |
| | 12 | z2w,1=1 z2w,2=1 z2w,3=1 z4=1 |
| | 12 | z2w,3=1 z4=3 |
| | 12 | z4=2 |
| | 12 | z2w,1=1 z2w,2=1 |
| 194.4.1 | 164099 | Sc1 |
| | 164 | z2w,3=1 z4=3 |
| | 164 | z2w,3=1 z4=1 |
| 194.4.2 | 164097 | Sc1 |
| | 164 | z2w,3=1 z4=3 |
| | 164 | z2w,3=1 z4=1 |
| 194.4.3 | 164095 | Sc1 |
| | 164 | z2w,3=1 z4=3 |

Continued on next column

**TABLE XIX– continued**

| TQC | ICSD SSG | Stoich. Formula SIs |
|---|---|---|
| | 164 | z2w,3=1 z4=1 |
| 194.4.4 | 164094 | Sc1 |
| | 164 | z2w,3=1 z4=3 |
| | 164 | z2w,3=1 z4=1 |
| 194.4.5 | 164093 | Sc1 |
| | 164 | z2w,3=1 z4=3 |
| | 164 | z2w,3=1 z4=1 |
| 194.4.6 | 164092 | Sc1 |
| | 164 | z2w,3=1 z4=3 |
| | 164 | z2w,3=1 z4=1 |
| 194.4.7 | 164091 | Sc1 |
| | 164 | z2w,3=1 z4=3 |
| | 164 | z2w,3=1 z4=1 |
| 194.4.8 | 164090 | Sc1 |
| | 164 | z2w,3=1 z4=3 |
| | 164 | z2w,3=1 z4=1 |
| 194.4.9 | 164089 | Sc1 |
| | 164 | z2w,3=1 z4=3 |
| | 164 | z2w,3=1 z4=1 |
| 194.4.10 | 164088 | Sc1 |
| | 164 | z2w,3=1 z4=3 |
| | 164 | z2w,3=1 z4=1 |
| 194.4.11 | 162586 | Ni2Sb1Zr1 |
| | 164 | z4=2 |
| | 164 | z2w,3=1 z4=1 |
| | 164 | z2w,3=1 z4=3 |
| | 164 | z4=2 |
| 194.4.12 | 162585 | Ni2Sb1Zr1 |
| | 164 | z4=2 |
| | 164 | z2w,3=1 z4=1 |
| | 164 | z2w,3=1 z4=3 |
| | 164 | z4=2 |
| 194.4.13 | 158485 | Ni2Sb1Te2 |
| | 164 | z4=2 |
| | 164 | z2w,3=1 z4=1 |
| | 164 | z2w,3=1 z4=3 |
| | 164 | z4=2 |
| 194.4.14 | 190179 | Ni1Sb1 |
| | 12 | z2w,1=1 z2w,2=1 z4=1 |
| | 12 | z4=3 |
| | 12 | z2w,3=1 z4=2 |
| | 12 | z2w,1=1 z2w,2=1 z2w,3=1 |
| | 12 | z2w,1=1 z2w,2=1 z2w,3=1 z4=2 |
| | 12 | z2w,3=1 |
| | 12 | z4=3 |
| | 12 | z2w,1=1 z2w,2=1 z4=1 |
| 194.4.15 | 710050 | Nd1P1Pd1 |
| | 164 | z2w,3=1 z4=3 |
| | 164 | z2w,3=1 z4=1 |
| 194.4.16 | 185417 | Mn1O3Sr1 |
| | 164 | z4=3 |
| | 164 | z2w,3=1 z4=2 |
| | 164 | z2w,3=1 |
| | 164 | z4=3 |
| 194.4.17 | 237636 | Li3Sn4Zn2 |
| | 164 | z4=2 |
| | 164 | z2w,3=1 z4=1 |
| | 164 | z2w,3=1 z4=3 |
| | 164 | z4=2 |

Continued on next column

**TABLE XIX– continued**

| TQC | ICSD SSG | Stoich. Formula SIs |
|---|---|---|
| 194.4.18 | 246806 | La1Si10 |
| | 164 | z4=3 |
| | 164 | z2w,3=1 z4=2 |
| | 164 | z2w,3=1 |
| | 164 | z4=3 |
| 194.4.19 | 154047 | Ge3La1Ti1 |
| | 164 | z4=1 |
| | 164 | z2w,3=1 |
| | 164 | z2w,3=1 z4=2 |
| | 164 | z4=1 |
| 194.4.20 | 425591 | Ge1Mn3U2 |
| | 12 | z2w,1=1 z2w,2=1 z4=2 |
| | 12 | z2w,3=1 z4=3 |
| | 12 | z2w,1=1 z2w,2=1 z2w,3=1 z4=1 |
| | 12 | z2w,1=1 z2w,2=1 z2w,3=1 z4=3 |
| | 12 | z2w,3=1 z4=1 |
| | 12 | z2w,1=1 z2w,2=1 z4=2 |
| 194.4.21 | 420009 | Gd1In1Pt2 |
| | 12 | z2w,1=1 z2w,2=1 z4=1 |
| | 12 | z4=3 |
| | 12 | z2w,3=1 z4=2 |
| | 12 | z2w,1=1 z2w,2=1 z2w,3=1 |
| | 12 | z2w,1=1 z2w,2=1 z2w,3=1 z4=2 |
| | 12 | z2w,3=1 |
| | 12 | z4=3 |
| | 12 | z2w,1=1 z2w,2=1 z4=1 |
| 194.4.22 | 156263 | Ga1Pt1Ti1 |
| | 164 | z4=2 |
| | 164 | z2w,3=1 z4=1 |
| | 164 | z2w,3=1 z4=3 |
| | 164 | z4=2 |
| 194.4.23 | 236572 | Cu3Sb4V2 |
| | 12 | z2w,1=1 z2w,2=1 z4=1 |
| | 12 | z4=3 |
| | 12 | z2w,3=1 z4=2 |
| | 12 | z2w,1=1 z2w,2=1 z2w,3=1 |
| | 12 | z2w,1=1 z2w,2=1 z2w,3=1 z4=2 |
| | 12 | z2w,3=1 |
| | 12 | z4=3 |
| | 12 | z2w,1=1 z2w,2=1 z4=1 |
| 194.4.24 | 422620 | Cu1Eu1Sb1 |
| | 164 | z4=1 |
| | 164 | z2w,3=1 |
| | 164 | z2w,3=1 z4=2 |
| | 164 | z4=1 |
| 194.4.25 | 158978 | Cr1Ge3Pr1 |
| | 164 | z2w,3=1 z4=3 |
| | 164 | z2w,3=1 z4=1 |
| 194.4.26 | 158976 | Cr1Ge3La1 |
| | 164 | z2w,3=1 z4=3 |
| | 164 | z2w,3=1 z4=1 |
| 194.4.27 | 418080 | Co2I1Tb2 |
| | 164 | z4=3 |
| | 164 | z2w,3=1 z4=2 |
| | 164 | z2w,3=1 |
| | 164 | z4=3 |
| 194.4.28 | 416273 | Co2I1Pr2 |
| | 164 | z4=2 |
| | 164 | z2w,3=1 z4=1 |

Continued on next column

**TABLE XIX– continued**

| TQC | ICSD<br>SSG | Stoich. Formula<br>SIs |
|---|---|---|
| | 164 | z2w,3=1 z4=3 |
| | 164 | z4=2 |
| 194.4.29 | 416787 | Ce1Cu1Sn1 |
| | 12 | z2w,1=1 z2w,2=1 z4=3 |
| | 12 | z4=1 |
| | 12 | z2w,3=1 |
| | 12 | z2w,1=1 z2w,2=1 z2w,3=1 z4=2 |
| | 12 | z2w,1=1 z2w,2=1 z2w,3=1 |
| | 12 | z2w,3=1 z4=2 |
| | 12 | z4=1 |
| | 12 | z2w,1=1 z2w,2=1 z4=3 |
| 194.4.30 | 416786 | Ce1Cu1Sn1 |
| | 12 | z2w,1=1 z2w,2=1 z4=3 |
| | 12 | z4=1 |
| | 12 | z2w,3=1 |
| | 12 | z2w,1=1 z2w,2=1 z2w,3=1 z4=2 |
| | 12 | z2w,1=1 z2w,2=1 z2w,3=1 |
| | 12 | z2w,3=1 z4=2 |
| | 12 | z4=1 |
| | 12 | z2w,1=1 z2w,2=1 z4=3 |
| 194.4.31 | 173221 | Ce1Cu1Si1 |
| | 12 | z2w,1=1 z2w,2=1 z4=3 |
| | 12 | z4=1 |
| | 12 | z2w,3=1 |
| | 12 | z2w,1=1 z2w,2=1 z2w,3=1 z4=2 |
| | 12 | z2w,1=1 z2w,2=1 z2w,3=1 |
| | 12 | z2w,3=1 z4=2 |
| | 12 | z4=1 |
| | 12 | z2w,1=1 z2w,2=1 z4=3 |
| 194.4.32 | 173219 | Ce1Cu1Ge1 |
| | 12 | z2w,1=1 z2w,2=1 z4=3 |
| | 12 | z4=1 |
| | 12 | z2w,3=1 |
| | 12 | z2w,1=1 z2w,2=1 z2w,3=1 z4=2 |
| | 12 | z2w,1=1 z2w,2=1 z2w,3=1 |
| | 12 | z2w,3=1 z4=2 |
| | 12 | z4=1 |
| | 12 | z2w,1=1 z2w,2=1 z4=3 |
| 194.4.33 | 153107 | Ca2I1N1 |
| | 164 | z2w,3=1 z4=3 |
| | 164 | z2w,3=1 z4=1 |
| 194.4.34 | 419117 | C1Ga1Ti2 |
| | 164 | z4=3 |
| | 164 | z2w,3=1 z4=2 |
| | 164 | z2w,3=1 |
| | 164 | z4=3 |
| 194.4.35 | 419116 | C1Cr2Ga1 |
| | 164 | z4=3 |
| | 164 | z2w,3=1 z4=2 |
| | 164 | z2w,3=1 |
| | 164 | z4=3 |
| 194.4.36 | 249387 | C1Co1N2 |
| | 12 | z2w,1=1 z2w,2=1 z4=2 |
| | 12 | z2w,3=1 z4=3 |
| | 12 | z2w,1=1 z2w,2=1 z2w,3=1 z4=1 |
| | 12 | z2w,1=1 z2w,2=1 z2w,3=1 z4=3 |
| | 12 | z2w,3=1 z4=1 |
| | 12 | z2w,1=1 z2w,2=1 z4=2 |
| 194.4.37 | 420348 | C1Br1Pr2 |

Continued on next column

**TABLE XIX– continued**

| TQC | ICSD<br>SSG | Stoich. Formula<br>SIs |
|---|---|---|
| | 164 | z4=2 |
| | 164 | z2w,3=1 z4=1 |
| | 164 | z2w,3=1 z4=3 |
| | 164 | z4=2 |
| 194.4.38 | 246360 | Ba3Ir2Na1O9 |
| | 12 | z2w,1=1 z2w,2=1 |
| | 12 | z4=2 |
| | 12 | z2w,3=1 z4=1 |
| | 12 | z2w,1=1 z2w,2=1 z2w,3=1 z4=3 |
| | 12 | z2w,1=1 z2w,2=1 z2w,3=1 z4=1 |
| | 12 | z2w,3=1 z4=3 |
| | 12 | z4=2 |
| | 12 | z2w,1=1 z2w,2=1 |
| 194.4.39 | 246359 | Ba3Ir2Na1O9 |
| | 12 | z2w,1=1 z2w,2=1 |
| | 12 | z4=2 |
| | 12 | z2w,3=1 z4=1 |
| | 12 | z2w,1=1 z2w,2=1 z2w,3=1 z4=3 |
| | 12 | z2w,1=1 z2w,2=1 z2w,3=1 z4=1 |
| | 12 | z2w,3=1 z4=3 |
| | 12 | z4=2 |
| | 12 | z2w,1=1 z2w,2=1 |
| 194.4.40 | 174293 | Ba3Ir2La1O9 |
| | 12 | z2w,1=1 z2w,2=1 z4=1 |
| | 12 | z4=3 |
| | 12 | z2w,3=1 z4=2 |
| | 12 | z2w,1=1 z2w,2=1 z2w,3=1 |
| | 12 | z2w,1=1 z2w,2=1 z2w,3=1 z4=2 |
| | 12 | z2w,3=1 |
| | 12 | z4=3 |
| | 12 | z2w,1=1 z2w,2=1 z4=1 |
| 194.4.41 | 157841 | Ba3In1O9Ru2 |
| | 164 | z4=3 |
| | 164 | z2w,3=1 z4=2 |
| | 164 | z2w,3=1 |
| | 164 | z4=3 |
| 194.4.42 | 154181 | Ba1S3V1 |
| | 12 | z2w,1=1 z2w,2=1 z4=2 |
| | 12 | z2w,3=1 z4=3 |
| | 12 | z2w,1=1 z2w,2=1 z2w,3=1 z4=1 |
| | 12 | z2w,1=1 z2w,2=1 z2w,3=1 z4=3 |
| | 12 | z2w,3=1 z4=1 |
| | 12 | z2w,1=1 z2w,2=1 z4=2 |
| 194.4.43 | 424878 | B3Ru2 |
| | 12 | z2w,1=1 z2w,2=1 z4=1 |
| | 12 | z4=3 |
| | 12 | z2w,3=1 z4=2 |
| | 12 | z2w,1=1 z2w,2=1 z2w,3=1 |
| | 12 | z2w,1=1 z2w,2=1 z2w,3=1 z4=2 |
| | 12 | z2w,3=1 |
| | 12 | z4=3 |
| | 12 | z2w,1=1 z2w,2=1 z4=1 |
| 194.4.44 | 424879 | B3Os2 |
| | 164 | z4=3 |
| | 164 | z2w,3=1 z4=2 |
| | 164 | z2w,3=1 |
| | 164 | z4=3 |
| 194.4.45 | 421522 | B2Re1 |
| | 164 | z4=3 |

Continued on next column

**TABLE XIX– continued**

| TQC | ICSD SSG | Stoich. Formula SIs |
|---|---|---|
| | 164 | z2w,3=1 z4=2 |
| | 164 | z2w,3=1 |
| | 164 | z4=3 |
| 194.4.46 | 165834 | B2Re1 |
| | 164 | z4=3 |
| | 164 | z2w,3=1 z4=2 |
| | 164 | z2w,3=1 |
| | 164 | z4=3 |
| 194.4.47 | 427220 | Au5Cs1Ga9 |
| | 12 | z2w,1=1 z2w,2=1 z4=1 |
| | 12 | z4=3 |
| | 12 | z2w,3=1 z4=2 |
| | 12 | z2w,1=1 z2w,2=1 z2w,3=1 |
| | 12 | z2w,1=1 z2w,2=1 z2w,3=1 z4=2 |
| | 12 | z2w,3=1 |
| | 12 | z4=3 |
| | 12 | z2w,1=1 z2w,2=1 z4=1 |
| 194.4.48 | 190987 | Au2Ga2In1 |
| | 164 | z4=3 |
| | 164 | z2w,3=1 z4=2 |
| | 164 | z2w,3=1 |
| | 164 | z4=3 |
| 194.4.49 | 192087 | Au2Ga1Mg1 |
| | 164 | z4=2 |
| | 164 | z2w,3=1 z4=1 |
| | 164 | z2w,3=1 z4=3 |
| | 164 | z4=2 |
| 194.4.50 | 163719 | Au2Ce3Sb3 |
| | 164 | z4=1 |
| | 164 | z2w,3=1 |
| | 164 | z2w,3=1 z4=2 |
| | 164 | z4=1 |
| 194.4.51 | 260109 | As2Eu1In2 |
| | 164 | z2w,3=1 z4=3 |
| | 164 | z2w,3=1 z4=1 |
| 194.4.52 | 420091 | Al5Ba4 |
| | 12 | z2w,1=1 z2w,2=1 z4=2 |
| | 12 | z2w,3=1 z4=3 |
| | 12 | z2w,1=1 z2w,2=1 z2w,3=1 z4=1 |
| | 12 | z2w,1=1 z2w,2=1 z2w,3=1 z4=3 |
| | 12 | z2w,3=1 z4=1 |
| | 12 | z2w,1=1 z2w,2=1 z4=2 |
| 194.4.53 | 420090 | Al5Ba3 |
| | 164 | z2w,3=1 z4=3 |
| | 164 | z2w,3=1 z4=1 |
| 194.4.54 | 415931 | Al2Ba2Ga1.33 |
| | 164 | z2w,3=1 z4=3 |
| | 164 | z2w,3=1 z4=1 |
| 194.4.55 | 163350 | Al1Pt1Ti1 |
| | 164 | z4=2 |
| | 164 | z2w,3=1 z4=1 |
| | 164 | z2w,3=1 z4=3 |
| | 164 | z4=2 |
| 194.4.56 | 172878 | Ag3Ni2O4 |
| | 164 | z4=3 |
| | 164 | z2w,3=1 z4=2 |
| | 164 | z2w,3=1 |
| | 164 | z4=3 |
| 194.4.57 | 415451 | Ag1Ni1O2 |

Continued on next column

**TABLE XIX– continued**

| TQC | ICSD SSG | Stoich. Formula SIs |
|---|---|---|
| | 12 | z2w,1=1 z2w,2=1 z4=2 |
| | 12 | z2w,3=1 z4=3 |
| | 12 | z2w,1=1 z2w,2=1 z2w,3=1 z4=1 |
| | 12 | z2w,1=1 z2w,2=1 z2w,3=1 z4=3 |
| | 12 | z2w,3=1 z4=1 |
| | 12 | z2w,1=1 z2w,2=1 z4=2 |
| 194.4.58 | 422621 | Ag1Eu1Sb1 |
| | 12 | z2w,1=1 z2w,2=1 z4=3 |
| | 12 | z4=1 |
| | 12 | z2w,3=1 |
| | 12 | z2w,1=1 z2w,2=1 z2w,3=1 z4=2 |
| | 12 | z2w,1=1 z2w,2=1 z2w,3=1 |
| | 12 | z2w,3=1 z4=2 |
| | 12 | z4=1 |
| | 12 | z2w,1=1 z2w,2=1 z4=3 |
| 204.4.1 | 77721 | O3Re1 |
| | 148 | z2w,1=1 z2w,2=1 z2w,3=1 |
| | 71 | z2w,1=1 z2w,2=1 z2w,3=1 |
| 204.4.2 | 77678 | O3Re1 |
| | 148 | z2w,1=1 z2w,2=1 z2w,3=1 |
| | 71 | z2w,1=1 z2w,2=1 z2w,3=1 |
| 204.4.3 | 55473 | O3Re1 |
| | 148 | z2w,1=1 z2w,2=1 z2w,3=1 |
| | 71 | z2w,1=1 z2w,2=1 z2w,3=1 |
| 204.4.4 | 55472 | O3Re1 |
| | 148 | z2w,1=1 z2w,2=1 z2w,3=1 |
| | 71 | z2w,1=1 z2w,2=1 z2w,3=1 |
| 204.4.5 | 55471 | O3Re1 |
| | 148 | z2w,1=1 z2w,2=1 z2w,3=1 |
| | 71 | z2w,1=1 z2w,2=1 z2w,3=1 |
| 204.4.6 | 55469 | O3Re1 |
| | 148 | z2w,1=1 z2w,2=1 z2w,3=1 |
| | 71 | z2w,1=1 z2w,2=1 z2w,3=1 |
| 204.4.7 | 55468 | O3Re1 |
| | 148 | z2w,1=1 z2w,2=1 z2w,3=1 |
| | 71 | z2w,1=1 z2w,2=1 z2w,3=1 |
| 204.4.8 | 55467 | O3Re1 |
| | 148 | z2w,1=1 z2w,2=1 z2w,3=1 |
| | 71 | z2w,1=1 z2w,2=1 z2w,3=1 |
| 204.4.9 | 55466 | O3Re1 |
| | 148 | z2w,1=1 z2w,2=1 z2w,3=1 |
| | 71 | z2w,1=1 z2w,2=1 z2w,3=1 |
| 204.4.10 | 55465 | O3Re1 |
| | 148 | z2w,1=1 z2w,2=1 z2w,3=1 |
| | 71 | z2w,1=1 z2w,2=1 z2w,3=1 |
| 204.4.11 | 55464 | O3Re1 |
| | 148 | z2w,1=1 z2w,2=1 z2w,3=1 |
| | 71 | z2w,1=1 z2w,2=1 z2w,3=1 |
| 204.4.12 | 55463 | O3Re1 |
| | 148 | z2w,1=1 z2w,2=1 z2w,3=1 |
| | 71 | z2w,1=1 z2w,2=1 z2w,3=1 |
| 204.4.13 | 55462 | O3Re1 |
| | 148 | z2w,1=1 z2w,2=1 z2w,3=1 |
| | 71 | z2w,1=1 z2w,2=1 z2w,3=1 |
| 204.4.14 | 54614 | Ni6Pb1Y12 |
| | 148 | z4=1 |
| | 71 | z4=1 |
| 204.4.15 | 79929 | Nd1Os4Sb12 |
| | 148 | z4=1 |

Continued on next column

**TABLE XIX– continued**

| TQC | ICSD<br>SSG | Stoich. Formula<br>SIs |
|---|---|---|
| | 71 | z4=1 |
| 204.4.16 | 183088 | Nd1Os4Sb12 |
| | 148 | z4=1 |
| | 71 | z4=1 |
| 204.4.17 | 19022 | Mn7Na1O12 |
| | 148 | z4=2 |
| | 148 | z4=2 |
| | 71 | z4=2 |
| 204.4.18 | 189912 | Ge12Pt4U1 |
| | 148 | z4=3 |
| | 71 | z4=3 |
| 204.4.19 | 189911 | Ge12Pt4Th1 |
| | 148 | z2w,1=1 z2w,2=1 z2w,3=1 |
| | 148 | z2w,1=1 z2w,2=1 z2w,3=1 |
| | 71 | z2w,1=1 z2w,2=1 z2w,3=1 |
| 204.4.20 | 174555 | Ge12Nd1Pt4 |
| | 148 | z2w,1=1 z2w,2=1 z2w,3=1 z4=1 |
| | 71 | z2w,1=1 z2w,2=1 z2w,3=1 z4=1 |
| 204.4.22 | 79927 | Fe4Nd1Sb12 |
| | 148 | z4=1 |
| | 71 | z4=1 |
| 204.4.24 | 95716 | Cu3Na1O12Ru4 |
| | 148 | z2w,1=1 z2w,2=1 z2w,3=1 |
| | 71 | z2w,1=1 z2w,2=1 z2w,3=1 |
| 204.4.25 | 38419 | Cu3Mn4O12Y1 |
| | 148 | z4=2 |
| | 71 | z4=2 |
| 204.4.26 | 38418 | Cu3Mn4O12Y1 |
| | 148 | z4=2 |
| | 71 | z4=2 |
| 204.4.27 | 250793 | Cu3Mn1O12V4 |
| | 71 | z2w,1=1 z2w,2=1 z2w,3=1 z4=3 |
| 204.4.28 | 250792 | Cu3Mn1O12V4 |
| | 71 | z2w,1=1 z2w,2=1 z2w,3=1 z4=3 |
| 204.4.29 | 51897 | Cu3La1O12Ru4 |
| | 148 | z4=1 |
| | 148 | z4=1 |
| | 71 | z4=1 |
| 204.4.30 | 251657 | Cu3Ir4La1O12 |
| | 148 | z2w,1=1 z2w,2=1 z2w,3=1 z4=1 |
| | 71 | z2w,1=1 z2w,2=1 z2w,3=1 z4=1 |
| 204.4.31 | 237418 | Cu3Fe4O12Tb1 |
| | 148 | z2w,1=1 z2w,2=1 z2w,3=1 z4=3 |
| | 71 | z2w,1=1 z2w,2=1 z2w,3=1 z4=3 |
| 204.4.32 | 237424 | Cu3Fe4Lu1O12 |
| | 148 | z2w,1=1 z2w,2=1 z2w,3=1 |
| | 148 | z2w,1=1 z2w,2=1 z2w,3=1 |
| | 71 | z2w,1=1 z2w,2=1 z2w,3=1 |
| 204.4.33 | 290766 | Cu3Fe4La1O12 |
| | 148 | z2w,1=1 z2w,2=1 z2w,3=1 |
| | 148 | z2w,1=1 z2w,2=1 z2w,3=1 |
| | 71 | z2w,1=1 z2w,2=1 z2w,3=1 |
| 204.4.34 | 290765 | Cu3Fe4La1O12 |
| | 148 | z2w,1=1 z2w,2=1 z2w,3=1 |
| | 148 | z2w,1=1 z2w,2=1 z2w,3=1 |
| | 71 | z2w,1=1 z2w,2=1 z2w,3=1 |
| 204.4.35 | 290764 | Cu3Fe4La1O12 |
| | 148 | z2w,1=1 z2w,2=1 z2w,3=1 |
| | 148 | z2w,1=1 z2w,2=1 z2w,3=1 |

Continued on next column

**TABLE XIX– continued**

| TQC | ICSD<br>SSG | Stoich. Formula<br>SIs |
|---|---|---|
| | 71 | z2w,1=1 z2w,2=1 z2w,3=1 |
| 204.4.36 | 290763 | Cu3Fe4La1O12 |
| | 148 | z2w,1=1 z2w,2=1 z2w,3=1 |
| | 148 | z2w,1=1 z2w,2=1 z2w,3=1 |
| | 71 | z2w,1=1 z2w,2=1 z2w,3=1 |
| 204.4.37 | 290762 | Cu3Fe4La1O12 |
| | 148 | z2w,1=1 z2w,2=1 z2w,3=1 |
| | 148 | z2w,1=1 z2w,2=1 z2w,3=1 |
| | 71 | z2w,1=1 z2w,2=1 z2w,3=1 |
| 204.4.38 | 290761 | Cu3Fe4La1O12 |
| | 148 | z2w,1=1 z2w,2=1 z2w,3=1 |
| | 148 | z2w,1=1 z2w,2=1 z2w,3=1 |
| | 71 | z2w,1=1 z2w,2=1 z2w,3=1 |
| 204.4.39 | 290760 | Cu3Fe4La1O12 |
| | 148 | z2w,1=1 z2w,2=1 z2w,3=1 |
| | 148 | z2w,1=1 z2w,2=1 z2w,3=1 |
| | 71 | z2w,1=1 z2w,2=1 z2w,3=1 |
| 204.4.40 | 290759 | Cu3Fe4La1O12 |
| | 148 | z2w,1=1 z2w,2=1 z2w,3=1 |
| | 148 | z2w,1=1 z2w,2=1 z2w,3=1 |
| | 71 | z2w,1=1 z2w,2=1 z2w,3=1 |
| 204.4.41 | 290758 | Cu3Fe4La1O12 |
| | 148 | z2w,1=1 z2w,2=1 z2w,3=1 |
| | 148 | z2w,1=1 z2w,2=1 z2w,3=1 |
| | 71 | z2w,1=1 z2w,2=1 z2w,3=1 |
| 204.4.42 | 290757 | Cu3Fe4La1O12 |
| | 148 | z2w,1=1 z2w,2=1 z2w,3=1 |
| | 148 | z2w,1=1 z2w,2=1 z2w,3=1 |
| | 71 | z2w,1=1 z2w,2=1 z2w,3=1 |
| 204.4.43 | 290756 | Cu3Fe4La1O12 |
| | 148 | z2w,1=1 z2w,2=1 z2w,3=1 |
| | 148 | z2w,1=1 z2w,2=1 z2w,3=1 |
| | 71 | z2w,1=1 z2w,2=1 z2w,3=1 |
| 204.4.44 | 237412 | Cu3Fe4La1O12 |
| | 148 | z2w,1=1 z2w,2=1 z2w,3=1 |
| | 148 | z2w,1=1 z2w,2=1 z2w,3=1 |
| | 71 | z2w,1=1 z2w,2=1 z2w,3=1 |
| 204.4.45 | 237399 | Cu3Fe4La1O12 |
| | 148 | z2w,1=1 z2w,2=1 z2w,3=1 |
| | 148 | z2w,1=1 z2w,2=1 z2w,3=1 |
| | 71 | z2w,1=1 z2w,2=1 z2w,3=1 |
| 204.4.46 | 185584 | Cu3Fe4La1O12 |
| | 148 | z2w,1=1 z2w,2=1 z2w,3=1 |
| | 148 | z2w,1=1 z2w,2=1 z2w,3=1 |
| | 71 | z2w,1=1 z2w,2=1 z2w,3=1 |
| 204.4.47 | 185583 | Cu3Fe4La1O12 |
| | 148 | z2w,1=1 z2w,2=1 z2w,3=1 |
| | 71 | z2w,1=1 z2w,2=1 z2w,3=1 |
| 204.4.48 | 237420 | Cu3Fe4Ho1O12 |
| | 148 | z2w,1=1 z2w,2=1 z2w,3=1 z4=1 |
| | 71 | z2w,1=1 z2w,2=1 z2w,3=1 z4=1 |
| 204.4.49 | 237416 | Cu3Eu1Fe4O12 |
| | 148 | z2w,1=1 z2w,2=1 z2w,3=1 |
| | 71 | z2w,1=1 z2w,2=1 z2w,3=1 |
| 204.4.50 | 237403 | Cu3Eu1Fe4O12 |
| | 148 | z2w,1=1 z2w,2=1 z2w,3=1 |
| | 71 | z2w,1=1 z2w,2=1 z2w,3=1 |
| 204.4.51 | 87886 | Co4Sb12Sn1 |
| | 148 | z4=3 |

Continued on next column

**TABLE XIX– continued**

| TQC | ICSD<br>SSG | Stoich. Formula<br>SIs |
|---|---|---|
| | 71 | z4=3 |
| 204.4.52 | 92393<br>148<br>71 | Co1P3<br>z4=3<br>z4=3 |
| 204.4.53 | 174553<br>148<br>71 | Ce1Ge12Pt4<br>z2w,1=1 z2w,2=1 z2w,3=1<br>z2w,1=1 z2w,2=1 z2w,3=1 |
| 204.4.54 | 237495<br>148<br>71 | Ca1O12Pd3V4<br>z2w,1=1 z2w,2=1 z2w,3=1<br>z2w,1=1 z2w,2=1 z2w,3=1 |
| 204.4.55 | 9188<br>148<br>71 | As3Co1<br>z4=3<br>z4=3 |
| 205.4.1 | 251565<br>148 | O2Rh1<br>z2w,1=1 z2w,2=1 z2w,3=1 z4=2 |
| 205.4.2 | 36550<br>148<br>148 | Mn1S2<br>z2w,1=1 z2w,2=1 z2w,3=1 z4=2<br>z2w,1=1 z2w,2=1 z2w,3=1 z4=2 |
| 205.4.3 | 36549<br>148<br>148 | Mn1S2<br>z2w,1=1 z2w,2=1 z2w,3=1 z4=2<br>z2w,1=1 z2w,2=1 z2w,3=1 z4=2 |
| 205.4.4 | 36548<br>148<br>148 | Mn1S2<br>z2w,1=1 z2w,2=1 z2w,3=1 z4=2<br>z2w,1=1 z2w,2=1 z2w,3=1 z4=2 |
| 205.4.5 | 36547<br>148<br>148 | Mn1S2<br>z2w,1=1 z2w,2=1 z2w,3=1 z4=2<br>z2w,1=1 z2w,2=1 z2w,3=1 z4=2 |
| 205.4.6 | 36546<br>148<br>148 | Mn1S2<br>z2w,1=1 z2w,2=1 z2w,3=1 z4=2<br>z2w,1=1 z2w,2=1 z2w,3=1 z4=2 |
| 205.4.7 | 251567<br>148 | Ir1O2<br>z2w,1=1 z2w,2=1 z2w,3=1 z4=2 |
| 205.4.9 | 100510<br>148 | Cu1S2<br>z2w,1=1 z2w,2=1 z2w,3=1 z4=2 |
| 205.4.10 | 86351<br>148 | Co1S2<br>z2w,1=1 z2w,2=1 z2w,3=1 z4=2 |
| 216.4.1 | 194973<br>119 | Mg1Nd1Ni4<br>z2=1 |
| 216.4.2 | 416868<br>119 | Ir3Mg3Si8<br>z2=1 |
| 216.4.3 | 418746<br>119 | Ir1La4Mg1<br>z2=1 |
| 216.4.4 | 162198<br>119 | In1Sb1<br>z2=1 |
| 216.4.5 | 162197<br>119 | In1Sb1<br>z2=1 |
| 216.4.6 | 162196<br>119 | In1Sb1<br>z2=1 |
| 216.4.7 | 201283<br>119 | I1Nb1Se1<br>z2=1 |
| 216.4.8 | 81213<br>119 | Ho1Ni1Sb1<br>z2=1 |
| 216.4.9 | 290038<br>119 | Hg1Se1<br>z2=1 |
| 216.4.10 | 50502<br>119 | Ge1S8V4<br>z2=1 |
| 216.4.11 | 194977<br>119 | Cu4Mg1Nd1<br>z2=1 |
| 216.4.12 | 412294 | Cu1Mg1Sb1 |

Continued on next column

**TABLE XIX– continued**

| TQC | ICSD<br>SSG | Stoich. Formula<br>SIs |
|---|---|---|
| | 119 | z2=1 |
| 216.4.13 | 159198<br>119 | Co4Sn1U1<br>z2=1 |
| 216.4.14 | 416990<br>119 | Co1La4Mg1<br>z2=1 |
| 216.4.15 | 419070<br>119 | Cd1Ru1Y4<br>z2=1 |
| 216.4.16 | 419069<br>119 | Cd1La4Ru1<br>z2=1 |
| 216.4.17 | 419067<br>119 | Cd1Co1La4<br>z2=1 |
| 216.4.19 | 248684<br>119 | Au1Bi1Ca1<br>z2=1 |
| 216.4.20 | 248682<br>119 | Au1Bi1Ca1<br>z2=1 |
| 216.4.21 | 248680<br>119 | Au1Bi1Ca1<br>z2=1 |
| 216.4.22 | 248678<br>119 | Au1Bi1Ca1<br>z2=1 |
| 216.4.23 | 248676<br>119 | Au1Bi1Ca1<br>z2=1 |
| 217.4.1 | 77887<br>121<br>121 | Pd16S7<br>z2=1<br>z2=1 |
| 217.4.2 | 410992<br>121<br>121<br>121<br>121 | Ir4Sn4Sr3<br>z2=1<br>z2=1<br>z2=1<br>z2=1 |
| 217.4.3 | 104605<br>121 | Ir2Zn11<br>z2=1 |
| 217.4.4 | 429725<br>121<br>121<br>121<br>121 | Ge4Ir4Sr3<br>z2=1<br>z2=1<br>z2=1<br>z2=1 |
| 217.4.5 | 429512<br>121 | Ga4Ir4Sr3<br>z2=1 |
| 217.4.6 | 200460<br>121 | F9U2<br>z2=1 |
| 217.4.7 | 240667<br>121 | Cu5Zn8<br>z2=1 |
| 217.4.8 | 2092<br>121 | Cu5Zn8<br>z2=1 |
| 217.4.9 | 194374<br>121<br>121<br>121<br>121 | Co2Zn11<br>z2=1<br>z2=1<br>z2=1<br>z2=1 |
| 217.4.10 | 95788<br>121<br>121 | Ca3Ir4Si4<br>z2=1<br>z2=1 |
| 217.4.11 | 429727<br>121<br>121<br>121<br>121 | Ca3Ge4Ir4<br>z2=1<br>z2=1<br>z2=1<br>z2=1 |
| 217.4.12 | 195500<br>121 | Ca3Ge4Ir4<br>z2=1 |

Continued on next column

**TABLE XIX– continued**

| TQC | ICSD<br>SSG | Stoich. Formula<br>SIs |
|---|---|---|
| | 121 | z2=1 |
| | 121 | z2=1 |
| | 121 | z2=1 |
| 220.4.1 | 60537 | Se4U3 |
| | 82 | z2=1 |
| | 82 | z2=1 |
| | 82 | z2=1 |
| | 82 | z2=1 |
| 220.4.2 | 427458 | O6Rb4 |
| | 122 | z2=1 |
| 220.4.3 | 33961 | O6Rb4 |
| | 122 | z2=1 |
| 220.4.4 | 33960 | O6Rb4 |
| | 122 | z2=1 |
| 220.4.5 | 25718 | O6Rb4 |
| | 122 | z2=1 |
| 220.4.6 | 56782 | La3S4 |
| | 82 | z2=1 |
| | 82 | z2=1 |
| | 82 | z2=1 |
| | 82 | z2=1 |
| 220.4.7 | 151632 | Eu3S4 |
| | 82 | z2=1 |
| | 82 | z2=1 |
| | 82 | z2=1 |
| | 82 | z2=1 |
| 220.4.8 | 151631 | Eu3S4 |
| | 82 | z2=1 |
| | 82 | z2=1 |
| | 82 | z2=1 |
| | 82 | z2=1 |
| 220.4.9 | 100522 | Eu3S4 |
| | 82 | z2=1 |
| | 82 | z2=1 |
| | 82 | z2=1 |
| | 82 | z2=1 |
| 220.4.11 | 16511 | C3Pu2 |
| | 122 | z2=1 |
| 220.4.13 | 96944 | Bi3Sr4 |
| | 82 | z2=1 |
| | 82 | z2=1 |
| | 82 | z2=1 |
| | 82 | z2=1 |
| | 82 | z2=1 |
| | 82 | z2=1 |
| | 82 | z2=1 |
| | 82 | z2=1 |
| 220.4.14 | 107394 | Bi3Eu4 |
| | 82 | z2=1 |
| | 82 | z2=1 |
| 220.4.15 | 96943 | Ba4Bi3 |
| | 82 | z2=1 |
| | 82 | z2=1 |
| | 82 | z2=1 |
| | 82 | z2=1 |
| | 82 | z2=1 |
| | 82 | z2=1 |
| | 82 | z2=1 |
| | 82 | z2=1 |

Continued on next column

**TABLE XIX– continued**

| TQC | ICSD<br>SSG | Stoich. Formula<br>SIs |
|---|---|---|
| 220.4.16 | 10440 | As3Eu4 |
| | 82 | z2=1 |
| | 82 | z2=1 |
| 221.4.1 | 86941 | O3Ru1Sr1 |
| | 166 | z4=3 |
| | 123 | z4=3 z4mpi=2 z2=1 z8=3 |
| 221.4.2 | 82984 | O3Ru1Sr1 |
| | 166 | z4=3 |
| | 123 | z4=3 z4mpi=2 z2=1 z8=7 |
| 221.4.3 | 82983 | O3Ru1Sr1 |
| | 166 | z4=3 |
| | 123 | z4=3 z4mpi=2 z2=1 z8=7 |
| 221.4.4 | 82982 | O3Ru1Sr1 |
| | 166 | z4=3 |
| | 123 | z4=3 z4mpi=2 z2=1 z8=7 |
| 221.4.5 | 82981 | O3Ru1Sr1 |
| | 166 | z4=3 |
| | 123 | z4=3 z4mpi=2 z2=1 z8=7 |
| 221.4.6 | 69360 | O3Ru1Sr1 |
| | 166 | z4=3 |
| | 123 | z4=3 z4mpi=2 z2=1 z8=7 |
| 221.4.7 | 162777 | O3Ru1Sr1 |
| | 166 | z4=3 |
| | 123 | z4=3 z4mpi=2 z2=1 z8=7 |
| 221.4.8 | 413389 | O1Pb1Yb3 |
| | 166 | z4=1 |
| | 123 | z4=1 z2=1 z8=1 |
| 221.4.9 | 421118 | N1Na3 |
| | 166 | z4=3 |
| | 123 | z4=3 z2=1 z8=7 |
| 221.4.10 | 421117 | N1Na3 |
| | 166 | z4=3 |
| | 123 | z4=3 z2=1 z8=7 |
| 221.4.11 | 421116 | N1Na3 |
| | 166 | z4=3 |
| | 123 | z4=3 z2=1 z8=7 |
| 221.4.12 | 421115 | N1Na3 |
| | 166 | z4=3 |
| | 123 | z4=3 z2=1 z8=7 |
| 221.4.13 | 421113 | N1Na3 |
| | 166 | z4=3 |
| | 123 | z4=3 z2=1 z8=7 |
| 221.4.14 | 421112 | N1Na3 |
| | 166 | z4=3 |
| | 123 | z4=3 z2=1 z8=7 |
| 221.4.15 | 421111 | N1Na3 |
| | 166 | z4=3 |
| | 123 | z4=3 z2=1 z8=7 |
| 221.4.16 | 421110 | N1Na3 |
| | 166 | z4=3 |
| | 123 | z4=3 z2=1 z8=7 |
| 221.4.17 | 421109 | N1Na3 |
| | 166 | z4=3 |
| | 123 | z4=3 z2=1 z8=7 |
| 221.4.18 | 412313 | N1Na3 |
| | 166 | z4=3 |
| | 123 | z4=3 z2=1 z8=7 |
| 221.4.19 | 249566 | N1Na3 |
| | 166 | z4=3 |

Continued on next column

**TABLE XIX– continued**

| TQC | ICSD SSG | Stoich. Formula SIs |
|---|---|---|
| | 123 | z4=3 z2=1 z8=7 |
| 221.4.20 | 71994 | Mo1O3Sr1 |
| | 123 | z4mpi=2 |
| 221.4.21 | 174208 | Mo1O3Sr1 |
| | 123 | z4mpi=2 z8=4 |
| 221.4.22 | 174207 | Mo1O3Sr1 |
| | 123 | z4mpi=2 z8=4 |
| 221.4.23 | 237710 | Mn3N1Zn1 |
| | 166 | z4=3 |
| | 123 | z4=3 z4mpi=2 z2=1 z8=7 |
| 221.4.24 | 240907 | In1O1Pr3 |
| | 115 | z2=1 |
| | 111 | z2=1 |
| 221.4.25 | 106748 | Ga3Ru1Tm1 |
| | 166 | z4=2 |
| | 166 | z4=2 |
| | 115 | z2=1 |
| | 115 | z2=1 |
| | 111 | z2=1 |
| | 111 | z2=1 |
| | 65 | z4=2 |
| | 47 | z4=2 |
| 221.4.26 | 92335 | Fe1O3Sr1 |
| | 166 | z4=3 |
| | 123 | z4=3 z4mpi=2 z2=1 z8=7 |
| 221.4.27 | 91062 | Fe1O3Sr1 |
| | 166 | z4=3 |
| | 123 | z4=3 z4mpi=2 z2=1 z8=3 |
| 221.4.28 | 73166 | F3Pd1Rb1 |
| | 123 | z4mpi=2 z8=4 |
| 221.4.29 | 73167 | F3K1Pd1 |
| | 123 | z4mpi=2 z8=4 |
| 221.4.30 | 37255 | F3K1Ni1 |
| | 123 | z4mpi=2 z8=4 |
| 221.4.31 | 189369 | F3K1Ni1 |
| | 123 | z4mpi=2 |
| 221.4.32 | 415804 | Eu3In1O1 |
| | 166 | z4=2 |
| | 123 | z4=2 z4mpi=2 z8=2 |
| 221.4.33 | 187860 | Cu1Y1 |
| | 166 | z2w,1=1 z2w,2=1 z2w,3=1 z4=2 |
| | 166 | z2w,1=1 z2w,2=1 z2w,3=1 z4=2 |
| | 123 | z2w,1=1 z2w,2=1 z2w,3=1 z4=2 z4mpi=1 z8=2 |
| 221.4.34 | 245834 | Cr1O3Sr1 |
| | 123 | z4mpi=2 z8=4 |
| 221.4.35 | 194155 | Cr1O3Pb1 |
| | 115 | z2=1 |
| | 111 | z2=1 |
| 221.4.36 | 160196 | Cr1O3Pb1 |
| | 115 | z2=1 |
| | 111 | z2=1 |
| 221.4.37 | 422858 | Cd1Co3N1 |
| | 166 | z4=2 |
| | 123 | z4=2 z8=6 |
| 221.4.38 | 411448 | Ca3N1Tl1 |
| | 166 | z4=2 |
| | 115 | z2=1 |
| | 111 | z2=1 |

Continued on next column

**TABLE XIX– continued**

| TQC | ICSD SSG | Stoich. Formula SIs |
|---|---|---|
| | 65 | z4=2 |
| | 47 | z4=2 |
| 221.4.39 | 50164 | C1Sc3Tl1 |
| | 166 | z4=1 |
| | 123 | z4=1 z2=1 z8=1 |
| 221.4.40 | 50163 | C1In1Sc3 |
| | 166 | z4=1 |
| | 123 | z4=1 z2=1 z8=1 |
| 221.4.41 | 80956 | C1In1La3 |
| | 166 | z4=3 |
| | 115 | z2=1 |
| | 111 | z2=1 |
| | 65 | z4=3 |
| | 47 | z4=3 |
| 221.4.42 | 72269 | Ba1O3Pb1 |
| | 166 | z4=3 |
| | 123 | z4=3 z2=1 z8=7 |
| 221.4.43 | 51659 | Ba1O3Pb1 |
| | 166 | z4=2 |
| | 166 | z4=2 |
| | 123 | z4=2 z4mpi=2 z8=6 |
| 221.4.44 | 245607 | Ba1O3Pb1 |
| | 166 | z4=3 |
| | 123 | z4=3 z2=1 z8=7 |
| 221.4.45 | 245606 | Ba1O3Pb1 |
| | 166 | z4=3 |
| | 123 | z4=3 z2=1 z8=7 |
| 221.4.46 | 43799 | Ba1Mo1O3 |
| | 123 | z4mpi=2 z8=4 |
| 221.4.47 | 262132 | Ba1Fe1O3 |
| | 166 | z4=3 |
| | 166 | z4=3 |
| | 123 | z4=3 z4mpi=2 z2=1 z8=7 |
| 221.4.48 | 262131 | Ba1Fe1O3 |
| | 166 | z4=3 |
| | 166 | z4=3 |
| | 123 | z4=3 z4mpi=2 z2=1 z8=7 |
| 221.4.49 | 72942 | Au1Ca3N1 |
| | 166 | z4=3 |
| | 123 | z4=3 z2=1 z8=7 |
| 223.4.1 | 87329 | Si1V3 |
| | 148 | z2w,1=1 z2w,2=1 z2w,3=1 |
| | 148 | z4=1 |
| | 148 | z4=1 |
| | 148 | z2w,1=1 z2w,2=1 z2w,3=1 z4=2 |
| | 115 | z2=1 |
| | 115 | z2=1 |
| | 115 | z2=1 |
| | 115 | z2=1 |
| | 47 | z4=1 |
| | 47 | z2w,3=1 z4=2 |
| | 47 | z2w,2=1 z4=2 |
| | 47 | z2w,2=1 z2w,3=1 z4=3 |
| | 47 | z2w,1=1 z4=2 |
| | 47 | z2w,1=1 z2w,3=1 z4=3 |
| | 47 | z2w,1=1 z2w,2=1 z4=3 |
| | 47 | z2w,1=1 z2w,2=1 z2w,3=1 |
| | 47 | z2w,1=1 z2w,2=1 z2w,3=1 z4=2 |
| | 47 | z2w,1=1 z2w,2=1 z4=3 |

Continued on next column

TABLE XIX– continued

| TQC | ICSD SSG | Stoich. Formula SIs |
|---|---|---|
| | 47 | z2w,1=1 z2w,3=1 z4=3 |
| | 47 | z2w,1=1 |
| | 47 | z2w,2=1 z2w,3=1 z4=3 |
| | 47 | z2w,2=1 |
| | 47 | z2w,3=1 |
| | 47 | z4=1 |
| 223.4.2 | 87328 | Si1V3 |
| | 148 | z2w,1=1 z2w,2=1 z2w,3=1 |
| | 148 | z4=1 |
| | 148 | z4=1 |
| | 148 | z2w,1=1 z2w,2=1 z2w,3=1 z4=2 |
| | 115 | z2=1 |
| | 115 | z2=1 |
| | 115 | z2=1 |
| | 115 | z2=1 |
| | 47 | z4=1 |
| | 47 | z2w,3=1 z4=2 |
| | 47 | z2w,2=1 z4=2 |
| | 47 | z2w,2=1 z2w,3=1 z4=3 |
| | 47 | z2w,1=1 z4=2 |
| | 47 | z2w,1=1 z2w,3=1 z4=3 |
| | 47 | z2w,1=1 z2w,2=1 z4=3 |
| | 47 | z2w,1=1 z2w,2=1 z2w,3=1 |
| | 47 | z2w,1=1 z2w,2=1 z2w,3=1 z4=2 |
| | 47 | z2w,1=1 z2w,2=1 z4=3 |
| | 47 | z2w,1=1 z2w,3=1 z4=3 |
| | 47 | z2w,1=1 |
| | 47 | z2w,2=1 z2w,3=1 z4=3 |
| | 47 | z2w,2=1 |
| | 47 | z2w,3=1 |
| | 47 | z4=1 |
| 223.4.3 | 32546 | Na1O4Pt3 |
| | 148 | z2w,1=1 z2w,2=1 z2w,3=1 z4=2 |
| | 148 | z4=3 |
| | 148 | z4=3 |
| | 148 | z2w,1=1 z2w,2=1 z2w,3=1 |
| | 115 | z2=1 |
| | 115 | z2=1 |
| | 115 | z2=1 |
| | 115 | z2=1 |
| | 47 | z4=3 |
| | 47 | z2w,3=1 |
| | 47 | z2w,2=1 |
| | 47 | z2w,2=1 z2w,3=1 z4=1 |
| | 47 | z2w,1=1 |
| | 47 | z2w,1=1 z2w,3=1 z4=1 |
| | 47 | z2w,1=1 z2w,2=1 z4=1 |
| | 47 | z2w,1=1 z2w,2=1 z2w,3=1 z4=2 |
| | 47 | z2w,1=1 z2w,2=1 z2w,3=1 |
| | 47 | z2w,1=1 z2w,2=1 z4=1 |
| | 47 | z2w,1=1 z2w,3=1 z4=1 |
| | 47 | z2w,1=1 z4=2 |
| | 47 | z2w,2=1 z2w,3=1 z4=1 |
| | 47 | z2w,2=1 z4=2 |
| | 47 | z2w,3=1 z4=2 |
| | 47 | z4=3 |
| 223.4.4 | 245608 | Na1O4Pd3 |
| | 148 | z4=2 |
| | 148 | z2w,1=1 z2w,2=1 z2w,3=1 z4=3 |
| | 148 | z2w,1=1 z2w,2=1 z2w,3=1 z4=3 |

Continued on next column

TABLE XIX– continued

| TQC | ICSD SSG | Stoich. Formula SIs |
|---|---|---|
| | 148 | z2w,1=1 z2w,2=1 z2w,3=1 z4=3 |
| | 148 | z2w,1=1 z2w,2=1 z2w,3=1 z4=1 |
| | 148 | z2w,1=1 z2w,2=1 z2w,3=1 z4=1 |
| | 148 | z2w,1=1 z2w,2=1 z2w,3=1 z4=1 |
| | 148 | z4=2 |
| | 148 | z4=2 |
| | 148 | z2w,1=1 z2w,2=1 z2w,3=1 z4=3 |
| | 148 | z2w,1=1 z2w,2=1 z2w,3=1 z4=3 |
| | 148 | z2w,1=1 z2w,2=1 z2w,3=1 z4=1 |
| | 148 | z2w,1=1 z2w,2=1 z2w,3=1 z4=3 |
| | 148 | z2w,1=1 z2w,2=1 z2w,3=1 z4=1 |
| | 148 | z2w,1=1 z2w,2=1 z2w,3=1 z4=1 |
| | 148 | z4=2 |
| | 115 | z2=1 |
| | 115 | z2=1 |
| | 115 | z2=1 |
| | 115 | z2=1 |
| | 115 | z2=1 |
| | 115 | z2=1 |
| | 115 | z2=1 |
| | 115 | z2=1 |
| | 47 | z2w,1=1 z2w,2=1 z2w,3=1 z4=3 |
| | 47 | z2w,1=1 z2w,2=1 |
| | 47 | z2w,1=1 z2w,3=1 |
| | 47 | z2w,1=1 z4=1 |
| | 47 | z2w,2=1 z2w,3=1 |
| | 47 | z2w,2=1 z4=1 |
| | 47 | z2w,3=1 z4=1 |
| | 47 | z4=2 |
| | 47 | z2w,3=1 z4=1 |
| | 47 | z2w,2=1 z4=1 |
| | 47 | z2w,2=1 z2w,3=1 z4=2 |
| | 47 | z2w,1=1 z4=1 |
| | 47 | z2w,1=1 z2w,3=1 z4=2 |
| | 47 | z2w,1=1 z2w,2=1 z4=2 |
| | 47 | z2w,1=1 z2w,2=1 z2w,3=1 z4=3 |
| | 47 | z2w,1=1 z2w,2=1 z2w,3=1 z4=1 |
| | 47 | z2w,1=1 z2w,2=1 z4=2 |
| | 47 | z2w,1=1 z2w,3=1 z4=2 |
| | 47 | z2w,1=1 z4=3 |
| | 47 | z2w,2=1 z2w,3=1 z4=2 |
| | 47 | z2w,2=1 z4=3 |
| | 47 | z2w,3=1 z4=3 |
| | 47 | z4=2 |
| | 47 | z2w,3=1 z4=3 |
| | 47 | z2w,2=1 z4=3 |
| | 47 | z2w,2=1 z2w,3=1 |
| | 47 | z2w,1=1 z4=3 |
| | 47 | z2w,1=1 z2w,3=1 |
| | 47 | z2w,1=1 z2w,2=1 |
| | 47 | z2w,1=1 z2w,2=1 z2w,3=1 z4=1 |
| 223.4.5 | 35755 | Mo3Si1 |
| | 167 | z4=1 |
| | 131 | z4=1 z2=1 |
| 223.4.6 | 84665 | La1Pd3S4 |
| | 148 | z2w,1=1 z2w,2=1 z2w,3=1 z4=3 |
| | 148 | z2w,1=1 z2w,2=1 z2w,3=1 z4=1 |
| | 115 | z2=1 |
| | 115 | z2=1 |
| | 47 | z2w,3=1 z4=1 |

Continued on next column

**TABLE XIX– continued**

| TQC | ICSD SSG | Stoich. Formula SIs |
|---|---|---|
| | 47 | z2w,2=1 z4=1 |
| | 47 | z2w,2=1 z2w,3=1 z4=2 |
| | 47 | z2w,1=1 z4=1 |
| | 47 | z2w,1=1 z2w,3=1 z4=2 |
| | 47 | z2w,1=1 z2w,2=1 z4=2 |
| | 47 | z2w,1=1 z2w,2=1 z2w,3=1 z4=3 |
| | 47 | z2w,1=1 z2w,2=1 z2w,3=1 z4=1 |
| | 47 | z2w,1=1 z2w,2=1 z4=2 |
| | 47 | z2w,1=1 z2w,3=1 z4=2 |
| | 47 | z2w,1=1 z4=3 |
| | 47 | z2w,2=1 z2w,3=1 z4=2 |
| | 47 | z2w,2=1 z4=3 |
| | 47 | z2w,3=1 z4=3 |
| 223.4.7 | 61031 | La1Pd3S4 |
| | 148 | z2w,1=1 z2w,2=1 z2w,3=1 z4=3 |
| | 148 | z2w,1=1 z2w,2=1 z2w,3=1 z4=1 |
| | 115 | z2=1 |
| | 115 | z2=1 |
| | 47 | z2w,3=1 z4=1 |
| | 47 | z2w,2=1 z4=1 |
| | 47 | z2w,2=1 z2w,3=1 z4=2 |
| | 47 | z2w,1=1 z4=1 |
| | 47 | z2w,1=1 z2w,3=1 z4=2 |
| | 47 | z2w,1=1 z2w,2=1 z4=2 |
| | 47 | z2w,1=1 z2w,2=1 z2w,3=1 z4=3 |
| | 47 | z2w,1=1 z2w,2=1 z2w,3=1 z4=1 |
| | 47 | z2w,1=1 z2w,2=1 z4=2 |
| | 47 | z2w,1=1 z2w,3=1 z4=2 |
| | 47 | z2w,1=1 z4=3 |
| | 47 | z2w,2=1 z2w,3=1 z4=2 |
| | 47 | z2w,2=1 z4=3 |
| | 47 | z2w,3=1 z4=3 |
| 223.4.8 | 50296 | Ir1Ti3 |
| | 148 | z4=2 |
| | 148 | z2w,1=1 z2w,2=1 z2w,3=1 z4=3 |
| | 148 | z2w,1=1 z2w,2=1 z2w,3=1 z4=3 |
| | 148 | z2w,1=1 z2w,2=1 z2w,3=1 z4=1 |
| | 148 | z2w,1=1 z2w,2=1 z2w,3=1 z4=3 |
| | 148 | z2w,1=1 z2w,2=1 z2w,3=1 z4=1 |
| | 148 | z2w,1=1 z2w,2=1 z2w,3=1 z4=1 |
| | 148 | z4=2 |
| | 148 | z4=2 |
| | 148 | z2w,1=1 z2w,2=1 z2w,3=1 z4=3 |
| | 148 | z2w,1=1 z2w,2=1 z2w,3=1 z4=3 |
| | 148 | z2w,1=1 z2w,2=1 z2w,3=1 z4=1 |
| | 148 | z2w,1=1 z2w,2=1 z2w,3=1 z4=1 |
| | 148 | z4=2 |
| | 115 | z2=1 |
| | 115 | z2=1 |
| | 115 | z2=1 |
| | 115 | z2=1 |
| | 115 | z2=1 |
| | 115 | z2=1 |
| | 115 | z2=1 |
| | 115 | z2=1 |
| | 47 | z2w,1=1 z2w,2=1 z2w,3=1 z4=3 |
| | 47 | z2w,1=1 z2w,2=1 |
| | 47 | z2w,1=1 z2w,3=1 |
| | 47 | z2w,1=1 z4=1 |
| | 47 | z2w,2=1 z2w,3=1 |

Continued on next column

**TABLE XIX– continued**

| TQC | ICSD SSG | Stoich. Formula SIs |
|---|---|---|
| | 47 | z2w,2=1 z4=1 |
| | 47 | z2w,3=1 z4=1 |
| | 47 | z4=2 |
| | 47 | z2w,3=1 z4=1 |
| | 47 | z2w,2=1 z4=1 |
| | 47 | z2w,2=1 z2w,3=1 z4=2 |
| | 47 | z2w,1=1 z4=1 |
| | 47 | z2w,1=1 z2w,3=1 z4=2 |
| | 47 | z2w,1=1 z2w,2=1 z4=2 |
| | 47 | z2w,1=1 z2w,2=1 z2w,3=1 z4=3 |
| | 47 | z2w,1=1 z2w,2=1 z2w,3=1 z4=1 |
| | 47 | z2w,1=1 z2w,2=1 z4=2 |
| | 47 | z2w,1=1 z2w,3=1 z4=2 |
| | 47 | z2w,1=1 z4=3 |
| | 47 | z2w,2=1 z2w,3=1 z4=2 |
| | 47 | z2w,2=1 z4=3 |
| | 47 | z2w,3=1 z4=3 |
| | 47 | z4=2 |
| | 47 | z2w,3=1 z4=3 |
| | 47 | z2w,2=1 z4=3 |
| | 47 | z2w,2=1 z2w,3=1 |
| | 47 | z2w,1=1 z4=3 |
| | 47 | z2w,1=1 z2w,3=1 |
| | 47 | z2w,1=1 z2w,2=1 |
| | 47 | z2w,1=1 z2w,2=1 z2w,3=1 z4=1 |
| 223.4.9 | 84667 | Eu1Pd3S4 |
| | 200 | z2w,1=1 z2w,2=1 z2w,3=1 z4=3 |
| | 200 | z2w,1=1 z2w,2=1 z2w,3=1 z4=1 |
| | 115 | z2=1 |
| | 115 | z2=1 |
| 223.4.10 | 53218 | Cr3Si1 |
| | 167 | z4=1 |
| | 131 | z4=1 z2=1 |
| 223.4.11 | 32509 | Cr3Si1 |
| | 167 | z4=1 |
| | 131 | z4=1 z2=1 |
| 225.4.1 | 261057 | Ru9Sb8Zn7 |
| | 166 | z4=1 |
| | 139 | z4=1 z2=1 z8=5 |
| 225.4.2 | 10439 | Os4Sc11 |
| | 139 | z8=4 |
| 225.4.3 | 55456 | O2Pu1 |
| | 139 | z8=4 |
| 225.4.4 | 9866 | Ni1O1 |
| | 121 | z2=1 |
| | 119 | z2=1 |
| 225.4.5 | 260172 | Ni1 |
| | 139 | z8=4 |
| 225.4.6 | 260169 | Ni1 |
| | 139 | z8=4 |
| 225.4.7 | 159263 | Ni16Si7Ti6 |
| | 121 | z2=1 |
| | 119 | z2=1 |
| 225.4.8 | 159264 | Nb6Ni16Si7 |
| | 166 | z4=1 |
| | 121 | z2=1 |
| | 119 | z2=1 |
| | 71 | z4=1 |
| | 69 | z4=1 |

Continued on next column

**TABLE XIX– continued**

| TQC | ICSD<br>SSG | Stoich. Formula<br>SIs |
|---|---|---|
| 225.4.9 | 262367 | N1Np1 |
| | 166 | z4=1 |
| | 139 | z4=1 z2=1 z8=1 |
| 225.4.10 | 982 | N1Nb1 |
| | 139 | z8=4 |
| 225.4.11 | 157027 | Mo1Ni1O6Sr2 |
| | 121 | z2=1 |
| | 119 | z2=1 |
| 225.4.12 | 155733 | Mo1Ni1O6Sr2 |
| | 121 | z2=1 |
| | 119 | z2=1 |
| 225.4.13 | 409428 | Mg2Pb1 |
| | 166 | z4=3 |
| | 139 | z4=3 z2=1 z8=7 |
| 225.4.14 | 10052 | Li3Tl1 |
| | 139 | z8=4 |
| 225.4.15 | 427115 | I1Na1O6Pb2 |
| | 121 | z2=1 |
| | 119 | z2=1 |
| 225.4.16 | 63670 | H1Br1 |
| | 166 | z4=3 |
| | 139 | z4=3 z2=1 z8=7 |
| 225.4.17 | 94456 | F6Ti1Zr1 |
| | 139 | z8=4 |
| 225.4.18 | 59102 | Cs2Cu1F6K1 |
| | 139 | z8=4 |
| 225.4.19 | 409229 | Cl6Tl2W1 |
| | 139 | z8=4 |
| 225.4.20 | 409632 | Cl6Rb2W1 |
| | 139 | z8=4 |
| 225.4.21 | 408777 | Cl6Mo1Tl2 |
| | 139 | z8=4 |
| 225.4.22 | 409840 | Cl6K2W1 |
| | 139 | z8=4 |
| 225.4.23 | 409674 | Cl6Cs2W1 |
| | 139 | z8=4 |
| 225.4.24 | 409808 | Cl6Cs2Mo1 |
| | 139 | z8=4 |
| 225.4.25 | 83258 | Ca2O6Pd1W1 |
| | 121 | z2=1 |
| | 119 | z2=1 |
| 225.4.26 | 151706 | C6Cs2Fe1N6Ni1 |
| | 121 | z2=1 |
| | 119 | z2=1 |
| 225.4.27 | 89531 | C2Na2 |
| | 139 | z8=4 |
| 225.4.28 | 89536 | C2Li2 |
| | 139 | z8=4 |
| 225.4.29 | 89532 | C2K2 |
| | 121 | z2=1 |
| | 119 | z2=1 |
| 225.4.30 | 402439 | Br6Rb2W1 |
| | 139 | z8=4 |
| 225.4.31 | 402441 | Br6Cs2W1 |
| | 139 | z8=4 |
| 225.4.32 | 94216 | Ba2O6Re1Y1 |
| | 139 | z8=4 |
| 225.4.33 | 94215 | Ba2O6Re1Y1 |
| | 139 | z8=4 |

Continued on next column

| TQC | ICSD<br>SSG | Stoich. Formula<br>SIs |
|---|---|---|
| 225.4.34 | 168216 | Ba2O6Re1Y1 |
| | 139 | z8=4 |
| 225.4.35 | 168215 | Ba2O6Re1Y1 |
| | 139 | z8=4 |
| 225.4.36 | 245142 | Ba2Ni1O6U1 |
| | 139 | z8=4 |
| 225.4.37 | 169535 | Ba2Ni1O6U1 |
| | 139 | z8=4 |
| 225.4.38 | 98192 | Ba2Mo1Ni1O6 |
| | 121 | z2=1 |
| | 119 | z2=1 |
| 225.4.39 | 26229 | Ba2Mo1Nd1O6 |
| | 139 | z8=4 |
| 225.4.40 | 158359 | Ba2Ho1O6Ta1 |
| | 121 | z2=1 |
| | 119 | z2=1 |
| 225.4.41 | 84651 | Ba2Ho1O6Sb1 |
| | 121 | z2=1 |
| | 119 | z2=1 |
| 225.4.42 | 150862 | Ba2Ho1O6Sb1 |
| | 121 | z2=1 |
| | 119 | z2=1 |
| 225.4.43 | 236354 | Ba2Gd1Mo1O6 |
| | 121 | z2=1 |
| | 119 | z2=1 |
| 225.4.44 | 236353 | Ba2Gd1Mo1O6 |
| | 121 | z2=1 |
| | 119 | z2=1 |
| 225.4.45 | 192845 | Ba2Ca1O6Os1 |
| | 139 | z8=4 |
| 225.4.46 | 192844 | Ba2Ca1O6Os1 |
| | 139 | z8=4 |
| 225.4.47 | 171989 | Ba2Ca1O6Os1 |
| | 139 | z8=4 |
| 225.4.48 | 171988 | Ba2Ca1O6Os1 |
| | 139 | z8=4 |
| 225.4.49 | 189924 | B12U1 |
| | 166 | z4=2 |
| | 139 | z4=2 z8=2 |
| 225.4.50 | 171472 | Ag2In1Li1 |
| | 166 | z4=2 |
| | 139 | z4=2 z8=2 |
| 226.4.1 | 415889 | Rb1Zn13 |
| | 148 | z2w,1=1 z2w,2=1 z2w,3=1 z4=3 |
| | 148 | z2w,1=1 z2w,2=1 z2w,3=1 z4=1 |
| | 82 | z2=1 |
| | 82 | z2=1 |
| | 12 | z2w,1=1 z2w,2=1 z4=2 |
| | 12 | z2w,1=1 z2w,2=1 z4=2 |
| 226.4.2 | 415887 | Na1Zn13 |
| | 148 | z2w,1=1 z2w,2=1 z2w,3=1 z4=3 |
| | 148 | z2w,1=1 z2w,2=1 z2w,3=1 z4=1 |
| | 82 | z2=1 |
| | 82 | z2=1 |
| | 12 | z2w,1=1 z2w,2=1 z4=2 |
| | 12 | z2w,1=1 z2w,2=1 z4=2 |
| 226.4.3 | 421842 | Li1Zn13 |
| | 148 | z2w,1=1 z2w,2=1 z2w,3=1 z4=3 |
| | 148 | z2w,1=1 z2w,2=1 z2w,3=1 z4=1 |

Continued on next column

**TABLE XIX– continued**

| TQC | ICSD<br>SSG | Stoich. Formula<br>SIs |
|---|---|---|
| | 82 | z2=1 |
| | 82 | z2=1 |
| | 12 | z2w,1=1 z2w,2=1 z4=2 |
| | 12 | z2w,1=1 z2w,2=1 z4=2 |
| 226.4.4 | 415886 | K1Zn13 |
| | 148 | z2w,1=1 z2w,2=1 z2w,3=1 z4=3 |
| | 148 | z2w,1=1 z2w,2=1 z2w,3=1 z4=1 |
| | 82 | z2=1 |
| | 82 | z2=1 |
| | 12 | z2w,1=1 z2w,2=1 z4=2 |
| | 12 | z2w,1=1 z2w,2=1 z4=2 |
| 226.4.5 | 55509 | Co1Cu12La1 |
| | 167 | z4=1 |
| | 140 | z4=1 z2=1 z8=1 |
| 226.4.6 | 415888 | Cd13Rb1 |
| | 148 | z2w,1=1 z2w,2=1 z2w,3=1 z4=3 |
| | 148 | z2w,1=1 z2w,2=1 z2w,3=1 z4=1 |
| | 82 | z2=1 |
| | 82 | z2=1 |
| | 12 | z2w,1=1 z2w,2=1 z4=2 |
| | 12 | z2w,1=1 z2w,2=1 z4=2 |
| 226.4.7 | 415885 | Cd13K1 |
| | 82 | z2=1 |
| | 82 | z2=1 |
| | 12 | z2w,1=1 z2w,2=1 z4=2 |
| | 12 | z2w,1=1 z2w,2=1 z4=2 |
| 226.4.8 | 415884 | Cd13Cs1 |
| | 148 | z2w,1=1 z2w,2=1 z2w,3=1 z4=3 |
| | 148 | z2w,1=1 z2w,2=1 z2w,3=1 z4=1 |
| | 82 | z2=1 |
| | 82 | z2=1 |
| | 12 | z2w,1=1 z2w,2=1 z4=2 |
| | 12 | z2w,1=1 z2w,2=1 z4=2 |
| 226.4.9 | 58742 | Be13Th1 |
| | 140 | z4=3 z2=1 z8=7 |
| 226.4.10 | 58740 | Be13Th1 |
| | 140 | z4=3 z2=1 z8=7 |
| 226.4.11 | 834 | Be13Sb1 |
| | 87 | z2w,1=1 z2w,2=1 z2w,3=1 z4=1 z2=1 z8=1 |
| | 87 | z2w,1=1 z2w,2=1 z2w,3=1 z4=1 z2=1 z8=5 |
| | 87 | z2w,1=1 z2w,2=1 z2w,3=1 z4=1 z2=1 z8=5 |
| | 87 | z2w,1=1 z2w,2=1 z2w,3=1 z4=1 z2=1 z8=1 |
| | 69 | z4=3 |
| | 69 | z4=3 |
| 226.4.12 | 170628 | Be13Ce1 |
| | 167 | z4=2 |
| | 140 | z4=2 z8=2 |
| 226.4.13 | 154391 | Ba1Cu13 |
| | 148 | z2w,1=1 z2w,2=1 z2w,3=1 z4=3 |
| | 148 | z2w,1=1 z2w,2=1 z2w,3=1 z4=1 |
| | 121 | z2=1 |
| | 121 | z2=1 |
| | 12 | z2w,1=1 z2w,2=1 z4=2 |
| | 12 | z2w,1=1 z2w,2=1 z4=2 |
| 227.4.1 | 160600 | O7V2Y2 |

Continued on next column

**TABLE XIX– continued**

| TQC | ICSD<br>SSG | Stoich. Formula<br>SIs |
|---|---|---|
| | 166 | z4=2 |
| | 141 | z4=2 |
| 227.4.2 | 190698 | O7Ti2Yb2 |
| | 166 | z4=1 |
| | 141 | z4=1 z2=1 |
| 227.4.3 | 190697 | O7Ti2Yb2 |
| | 166 | z4=1 |
| | 141 | z4=1 z2=1 |
| 227.4.4 | 190696 | O7Ti2Yb2 |
| | 166 | z4=1 |
| | 141 | z4=1 z2=1 |
| 227.4.5 | 190695 | O7Ti2Yb2 |
| | 166 | z4=1 |
| | 141 | z4=1 z2=1 |
| 227.4.6 | 190694 | O7Ti2Yb2 |
| | 166 | z4=1 |
| | 141 | z4=1 z2=1 |
| 227.4.7 | 190688 | O7Ti2Yb2 |
| | 166 | z4=1 |
| | 141 | z4=1 z2=1 |
| 227.4.8 | 190687 | O7Ti2Yb2 |
| | 166 | z4=1 |
| | 141 | z4=1 z2=1 |
| 227.4.9 | 190686 | O7Ti2Yb2 |
| | 166 | z4=1 |
| | 141 | z4=1 z2=1 |
| 227.4.10 | 190685 | O7Ti2Yb2 |
| | 166 | z4=1 |
| | 141 | z4=1 z2=1 |
| 227.4.11 | 190684 | O7Ti2Yb2 |
| | 166 | z4=1 |
| | 141 | z4=1 z2=1 |
| 227.4.12 | 186196 | O7Ti2Yb2 |
| | 166 | z4=1 |
| | 141 | z4=1 z2=1 |
| 227.4.13 | 186195 | O7Ti2Yb2 |
| | 166 | z4=1 |
| | 141 | z4=1 z2=1 |
| 227.4.14 | 173750 | O7Ti2Yb2 |
| | 166 | z4=1 |
| | 141 | z4=1 z2=1 |
| 227.4.15 | 161405 | O7Ti2Yb2 |
| | 166 | z4=1 |
| | 141 | z4=1 z2=1 |
| 227.4.16 | 82964 | O7Sn2Yb2 |
| | 166 | z4=1 |
| | 141 | z4=1 z2=1 |
| 227.4.17 | 84750 | O7Sn2Tm2 |
| | 166 | z4=2 |
| | 141 | z4=2 |
| 227.4.18 | 82963 | O7Sn2Tm2 |
| | 166 | z4=2 |
| | 141 | z4=2 |
| 227.4.19 | 84744 | O7Pr2Sn2 |
| | 166 | z4=2 |
| | 141 | z4=2 |
| 227.4.20 | 82957 | O7Pr2Sn2 |
| | 166 | z4=2 |
| | 141 | z4=2 |

Continued on next column

TABLE XIX– continued

| TQC | ICSD<br>SSG | Stoich. Formula<br>SIs |
|---|---|---|
| 227.4.21 | 173073 | O6Os2Rb1 |
| | 166 | z2w,1=1 z2w,2=1 z2w,3=1 z4=1 |
| | 166 | z2w,1=1 z2w,2=1 z2w,3=1 |
| 227.4.22 | 173072 | O6Os2Rb1 |
| | 166 | z2w,1=1 z2w,2=1 z2w,3=1 z4=1 |
| | 166 | z2w,1=1 z2w,2=1 z2w,3=1 |
| 227.4.23 | 157442 | O6Os2Rb1 |
| | 166 | z2w,1=1 z2w,2=1 z2w,3=1 z4=1 |
| | 166 | z2w,1=1 z2w,2=1 z2w,3=1 |
| 227.4.24 | 157441 | O6Os2Rb1 |
| | 166 | z2w,1=1 z2w,2=1 z2w,3=1 z4=1 |
| | 166 | z2w,1=1 z2w,2=1 z2w,3=1 |
| 227.4.25 | 157440 | O6Os2Rb1 |
| | 166 | z2w,1=1 z2w,2=1 z2w,3=1 z4=1 |
| | 166 | z2w,1=1 z2w,2=1 z2w,3=1 |
| 227.4.26 | 157439 | O6Os2Rb1 |
| | 166 | z2w,1=1 z2w,2=1 z2w,3=1 z4=1 |
| | 166 | z2w,1=1 z2w,2=1 z2w,3=1 |
| 227.4.27 | 157438 | O6Os2Rb1 |
| | 166 | z2w,1=1 z2w,2=1 z2w,3=1 z4=1 |
| | 166 | z2w,1=1 z2w,2=1 z2w,3=1 |
| 227.4.28 | 157437 | O6Os2Rb1 |
| | 166 | z2w,1=1 z2w,2=1 z2w,3=1 z4=1 |
| | 166 | z2w,1=1 z2w,2=1 z2w,3=1 |
| 227.4.29 | 157436 | O6Os2Rb1 |
| | 166 | z2w,1=1 z2w,2=1 z2w,3=1 z4=1 |
| | 166 | z2w,1=1 z2w,2=1 z2w,3=1 |
| 227.4.30 | 157435 | O6Os2Rb1 |
| | 166 | z2w,1=1 z2w,2=1 z2w,3=1 z4=1 |
| | 166 | z2w,1=1 z2w,2=1 z2w,3=1 |
| 227.4.31 | 99181 | O4V2Zn1 |
| | 122 | z2=1 |
| | 119 | z2=1 |
| 227.4.32 | 55442 | O4V2Zn1 |
| | 122 | z2=1 |
| | 119 | z2=1 |
| 227.4.33 | 160164 | Nd2O7Zr2 |
| | 166 | z4=2 |
| | 141 | z4=2 |
| 227.4.34 | 105162 | Na1Pt2 |
| | 166 | z2w,1=1 z2w,2=1 z2w,3=1 |
| | 166 | z2w,1=1 z2w,2=1 z2w,3=1 z4=3 |
| 227.4.35 | 54375 | Mn2Y1 |
| | 166 | z2w,1=1 z2w,2=1 z2w,3=1 z4=3 |
| | 166 | z2w,1=1 z2w,2=1 z2w,3=1 z4=2 |
| 227.4.36 | 202516 | Mn2O7Y2 |
| | 166 | z4=3 |
| | 141 | z4=3 z2=1 |
| 227.4.37 | 195759 | Mn2O7Tm2 |
| | 166 | z4=2 |
| | 141 | z4=2 |
| 227.4.38 | 104935 | Mn2Np1 |
| | 166 | z2w,1=1 z2w,2=1 z2w,3=1 z4=2 |
| | 166 | z2w,1=1 z2w,2=1 z2w,3=1 z4=1 |
| 227.4.39 | 94992 | Mn1S4Sc2 |
| | 166 | z2w,1=1 z2w,2=1 z2w,3=1 z4=3 |
| | 166 | z2w,1=1 z2w,2=1 z2w,3=1 z4=2 |
| 227.4.40 | 248565 | Mg1O4V2 |
| | 122 | z2=1 |

Continued on next column

TABLE XIX– continued

| TQC | ICSD<br>SSG | Stoich. Formula<br>SIs |
|---|---|---|
| | 119 | z2=1 |
| 227.4.41 | 89352 | Li1O4V2 |
| | 166 | z2w,1=1 z2w,2=1 z2w,3=1 z4=2 |
| | 166 | z2w,1=1 z2w,2=1 z2w,3=1 z4=1 |
| 227.4.42 | 89351 | Li1O4V2 |
| | 166 | z2w,1=1 z2w,2=1 z2w,3=1 z4=2 |
| | 166 | z2w,1=1 z2w,2=1 z2w,3=1 z4=1 |
| 227.4.43 | 89350 | Li1O4V2 |
| | 166 | z2w,1=1 z2w,2=1 z2w,3=1 z4=2 |
| | 166 | z2w,1=1 z2w,2=1 z2w,3=1 z4=1 |
| 227.4.44 | 89349 | Li1O4V2 |
| | 166 | z2w,1=1 z2w,2=1 z2w,3=1 z4=2 |
| | 166 | z2w,1=1 z2w,2=1 z2w,3=1 z4=1 |
| 227.4.45 | 89348 | Li1O4V2 |
| | 166 | z2w,1=1 z2w,2=1 z2w,3=1 z4=2 |
| | 166 | z2w,1=1 z2w,2=1 z2w,3=1 z4=1 |
| 227.4.46 | 89347 | Li1O4V2 |
| | 166 | z2w,1=1 z2w,2=1 z2w,3=1 z4=2 |
| | 166 | z2w,1=1 z2w,2=1 z2w,3=1 z4=1 |
| 227.4.47 | 89346 | Li1O4V2 |
| | 166 | z2w,1=1 z2w,2=1 z2w,3=1 z4=2 |
| | 166 | z2w,1=1 z2w,2=1 z2w,3=1 z4=1 |
| 227.4.48 | 89345 | Li1O4V2 |
| | 166 | z2w,1=1 z2w,2=1 z2w,3=1 z4=2 |
| | 166 | z2w,1=1 z2w,2=1 z2w,3=1 z4=1 |
| 227.4.49 | 89344 | Li1O4V2 |
| | 166 | z2w,1=1 z2w,2=1 z2w,3=1 z4=2 |
| | 166 | z2w,1=1 z2w,2=1 z2w,3=1 z4=1 |
| 227.4.50 | 418091 | Li1O4V2 |
| | 166 | z2w,1=1 z2w,2=1 z2w,3=1 z4=2 |
| | 166 | z2w,1=1 z2w,2=1 z2w,3=1 z4=1 |
| 227.4.51 | 418090 | Li1O4V2 |
| | 166 | z2w,1=1 z2w,2=1 z2w,3=1 z4=2 |
| | 166 | z2w,1=1 z2w,2=1 z2w,3=1 z4=1 |
| 227.4.52 | 201818 | Li1O4V2 |
| | 166 | z2w,1=1 z2w,2=1 z2w,3=1 z4=2 |
| | 166 | z2w,1=1 z2w,2=1 z2w,3=1 z4=1 |
| 227.4.53 | 170768 | Li1O4V2 |
| | 166 | z2w,1=1 z2w,2=1 z2w,3=1 z4=2 |
| | 166 | z2w,1=1 z2w,2=1 z2w,3=1 z4=1 |
| 227.4.54 | 78482 | Li1O4Ti2 |
| | 166 | z2w,1=1 z2w,2=1 z2w,3=1 z4=1 |
| | 166 | z2w,1=1 z2w,2=1 z2w,3=1 |
| 227.4.55 | 72859 | Li1O4Ti2 |
| | 166 | z2w,1=1 z2w,2=1 z2w,3=1 z4=1 |
| | 166 | z2w,1=1 z2w,2=1 z2w,3=1 |
| 227.4.56 | 48127 | Li1O4Ti2 |
| | 166 | z2w,1=1 z2w,2=1 z2w,3=1 z4=1 |
| | 166 | z2w,1=1 z2w,2=1 z2w,3=1 |
| 227.4.57 | 154982 | Li1O4Ti2 |
| | 166 | z2w,1=1 z2w,2=1 z2w,3=1 z4=1 |
| | 166 | z2w,1=1 z2w,2=1 z2w,3=1 |
| 227.4.58 | 94340 | Li1Mn2O4 |
| | 166 | z2w,1=1 z2w,2=1 z2w,3=1 |
| | 166 | z2w,1=1 z2w,2=1 z2w,3=1 z4=3 |
| 227.4.59 | 94339 | Li1Mn2O4 |
| | 166 | z2w,1=1 z2w,2=1 z2w,3=1 z4=2 |
| | 166 | z2w,1=1 z2w,2=1 z2w,3=1 z4=1 |
| 227.4.60 | 89985 | Li1Mn2O4 |

Continued on next column

**TABLE XIX– continued**

| TQC | ICSD<br>SSG | Stoich. Formula<br>SIs |
|---|---|---|
| | 12 | z2w,1=1 z2w,2=1 z2w,3=1 |
| | 12 | z2w,3=1 z4=2 |
| | 12 | z2w,3=1 z4=1 |
| | 12 | z2w,1=1 z2w,2=1 z2w,3=1 z4=3 |
| 227.4.61 | 89459 | Li1Mn2O4 |
| | 12 | z2w,1=1 z2w,2=1 z2w,3=1 |
| | 12 | z2w,3=1 z4=2 |
| | 12 | z2w,3=1 z4=1 |
| | 12 | z2w,1=1 z2w,2=1 z2w,3=1 z4=3 |
| 227.4.62 | 88644 | Li1Mn2O4 |
| | 12 | z2w,1=1 z2w,2=1 z2w,3=1 |
| | 12 | z2w,3=1 z4=2 |
| | 12 | z2w,3=1 z4=1 |
| | 12 | z2w,1=1 z2w,2=1 z2w,3=1 z4=3 |
| 227.4.63 | 50415 | Li1Mn2O4 |
| | 12 | z2w,1=1 z2w,2=1 z2w,3=1 |
| | 12 | z2w,3=1 z4=2 |
| | 12 | z2w,3=1 z4=1 |
| | 12 | z2w,1=1 z2w,2=1 z2w,3=1 z4=3 |
| 227.4.64 | 261557 | Li1Mn2O4 |
| | 12 | z2w,1=1 z2w,2=1 z2w,3=1 |
| | 12 | z2w,3=1 z4=2 |
| | 12 | z2w,3=1 z4=1 |
| | 12 | z2w,1=1 z2w,2=1 z2w,3=1 z4=3 |
| 227.4.65 | 192369 | Li1Mn2O4 |
| | 12 | z2w,1=1 z2w,2=1 z2w,3=1 |
| | 12 | z2w,3=1 z4=2 |
| | 12 | z2w,3=1 z4=1 |
| | 12 | z2w,1=1 z2w,2=1 z2w,3=1 z4=3 |
| 227.4.66 | 155665 | Li1Mn2O4 |
| | 12 | z2w,1=1 z2w,2=1 z2w,3=1 |
| | 12 | z2w,3=1 z4=2 |
| | 12 | z2w,3=1 z4=1 |
| | 12 | z2w,1=1 z2w,2=1 z2w,3=1 z4=3 |
| 227.4.67 | 155282 | Li1Mn2O4 |
| | 166 | z2w,1=1 z2w,2=1 z2w,3=1 z4=2 |
| | 166 | z2w,1=1 z2w,2=1 z2w,3=1 z4=1 |
| 227.4.68 | 170775 | La1Mg2 |
| | 166 | z2w,1=1 z2w,2=1 z2w,3=1 z4=3 |
| | 166 | z2w,1=1 z2w,2=1 z2w,3=1 z4=2 |
| 227.4.69 | 172620 | K1O6Os2 |
| | 166 | z2w,1=1 z2w,2=1 z2w,3=1 z4=1 |
| | 166 | z2w,1=1 z2w,2=1 z2w,3=1 |
| 227.4.70 | 172619 | K1O6Os2 |
| | 166 | z2w,1=1 z2w,2=1 z2w,3=1 z4=1 |
| | 166 | z2w,1=1 z2w,2=1 z2w,3=1 |
| 227.4.71 | 172618 | K1O6Os2 |
| | 166 | z2w,1=1 z2w,2=1 z2w,3=1 z4=1 |
| | 166 | z2w,1=1 z2w,2=1 z2w,3=1 |
| 227.4.72 | 162266 | K1O6Os2 |
| | 166 | z2w,1=1 z2w,2=1 z2w,3=1 z4=1 |
| | 166 | z2w,1=1 z2w,2=1 z2w,3=1 |
| 227.4.73 | 160698 | K1O6Os2 |
| | 166 | z2w,1=1 z2w,2=1 z2w,3=1 z4=1 |
| | 166 | z2w,1=1 z2w,2=1 z2w,3=1 |
| 227.4.74 | 187534 | Ir2O7Y2 |
| | 166 | z4=3 |
| | 141 | z4=3 z2=1 |
| 227.4.75 | 280390 | Hf1V2 |

Continued on next column

**TABLE XIX– continued**

| TQC | ICSD<br>SSG | Stoich. Formula<br>SIs |
|---|---|---|
| | 166 | z4=3 |
| | 141 | z4=3 z2=1 |
| 227.4.76 | 65618 | Ge2O7Yb2 |
| | 166 | z4=1 |
| | 141 | z4=1 z2=1 |
| 227.4.77 | 52179 | Ge1Ni2Zn3 |
| | 166 | z4=2 |
| | 141 | z4=2 |
| 227.4.78 | 261759 | Gd2O7Zr2 |
| | 122 | z2=1 |
| | 119 | z2=1 |
| 227.4.79 | 261758 | Gd2O7Zr2 |
| | 122 | z2=1 |
| | 119 | z2=1 |
| 227.4.80 | 160158 | Gd2O7Zr2 |
| | 122 | z2=1 |
| | 119 | z2=1 |
| 227.4.81 | 79332 | Gd2O7Ru2 |
| | 122 | z2=1 |
| | 119 | z2=1 |
| 227.4.82 | 170910 | Fe2Mn1O4 |
| | 166 | z2w,1=1 z2w,2=1 z2w,3=1 z4=2 |
| | 166 | z2w,1=1 z2w,2=1 z2w,3=1 z4=1 |
| 227.4.83 | 106646 | Fe1Hf2 |
| | 166 | z4=3 |
| | 141 | z4=3 z2=1 |
| 227.4.84 | 4083 | F6Hg2Ni2S1 |
| | 122 | z2=1 |
| | 119 | z2=1 |
| 227.4.85 | 166545 | F6Hg2Ni2S1 |
| | 122 | z2=1 |
| | 119 | z2=1 |
| 227.4.86 | 27329 | F6Hg2Ni2O1 |
| | 122 | z2=1 |
| | 119 | z2=1 |
| 227.4.87 | 202047 | F3Fe1 |
| | 166 | z4=3 |
| | 141 | z4=3 z2=1 |
| 227.4.88 | 156437 | Eu2Ir2O7 |
| | 166 | z4=3 |
| | 141 | z4=3 z2=1 |
| 227.4.89 | 410701 | Eu1Ir2 |
| | 12 | z2w,1=1 z2w,2=1 z2w,3=1 |
| | 12 | z2w,3=1 z4=2 |
| | 12 | z2w,3=1 z4=1 |
| | 12 | z2w,1=1 z2w,2=1 z2w,3=1 z4=3 |
| 227.4.90 | 202517 | Er2Mn2O7 |
| | 122 | z2=1 |
| | 119 | z2=1 |
| 227.4.91 | 78151 | Dy2O7Ru2 |
| | 122 | z2=1 |
| | 119 | z2=1 |
| 227.4.92 | 156368 | Cu2F6Hg2S1 |
| | 166 | z4=3 |
| | 141 | z4=3 z2=1 |
| 227.4.93 | 156367 | Cu2F6Hg2S1 |
| | 166 | z4=3 |
| | 141 | z4=3 z2=1 |
| 227.4.94 | 156366 | Cu2F6Hg2S1 |

Continued on next column

**TABLE XIX– continued**

| TQC | ICSD<br>SSG | Stoich. Formula<br>SIs |
|---|---|---|
| | 166 | z4=3 |
| | 141 | z4=3 z2=1 |
| 227.4.95 | 190441 | Cu1S4V2 |
| | 166 | z2w,1=1 z2w,2=1 z2w,3=1 |
| | 166 | z2w,1=1 z2w,2=1 z2w,3=1 z4=3 |
| 227.4.96 | 10035 | Cu1S4V2 |
| | 166 | z2w,1=1 z2w,2=1 z2w,3=1 |
| | 166 | z2w,1=1 z2w,2=1 z2w,3=1 z4=3 |
| 227.4.97 | 75531 | Cu1Ir2S4 |
| | 166 | z2w,1=1 z2w,2=1 z2w,3=1 z4=3 |
| | 166 | z2w,1=1 z2w,2=1 z2w,3=1 z4=2 |
| 227.4.98 | 4206 | Cs2O12U4 |
| | 166 | z2w,1=1 z2w,2=1 z2w,3=1 z4=2 |
| | 166 | z2w,1=1 z2w,2=1 z2w,3=1 z4=1 |
| 227.4.99 | 421173 | Cs1Fe1O2 |
| | 166 | z2w,1=1 z2w,2=1 z2w,3=1 z4=1 |
| | 166 | z2w,1=1 z2w,2=1 z2w,3=1 |
| 227.4.100 | 174318 | Cs1Fe1O2 |
| | 166 | z2w,1=1 z2w,2=1 z2w,3=1 z4=1 |
| | 166 | z2w,1=1 z2w,2=1 z2w,3=1 |
| 227.4.101 | 425260 | Cs1F6Fe2 |
| | 166 | z2w,1=1 z2w,2=1 z2w,3=1 z4=2 |
| | 166 | z2w,1=1 z2w,2=1 z2w,3=1 z4=1 |
| 227.4.102 | 31298 | Cr2Ni1O4 |
| | 166 | z4=2 |
| | 141 | z4=2 |
| 227.4.103 | 280061 | Cr2Ni1O4 |
| | 166 | z4=2 |
| | 141 | z4=2 |
| 227.4.104 | 97620 | Cr2Cu1Se4 |
| | 166 | z2w,1=1 z2w,2=1 z2w,3=1 z4=3 |
| | 166 | z2w,1=1 z2w,2=1 z2w,3=1 z4=2 |
| 227.4.105 | 87477 | Cr2Cu1Se4 |
| | 166 | z2w,1=1 z2w,2=1 z2w,3=1 z4=3 |
| | 166 | z2w,1=1 z2w,2=1 z2w,3=1 z4=2 |
| 227.4.106 | 87476 | Cr2Cu1Se4 |
| | 166 | z2w,1=1 z2w,2=1 z2w,3=1 z4=3 |
| | 166 | z2w,1=1 z2w,2=1 z2w,3=1 z4=2 |
| 227.4.107 | 73403 | Cr2Cu1Se4 |
| | 166 | z2w,1=1 z2w,2=1 z2w,3=1 z4=3 |
| | 166 | z2w,1=1 z2w,2=1 z2w,3=1 z4=2 |
| 227.4.108 | 291003 | Cr2Cu1S4 |
| | 166 | z2w,1=1 z2w,2=1 z2w,3=1 z4=2 |
| | 166 | z2w,1=1 z2w,2=1 z2w,3=1 z4=1 |
| 227.4.109 | 291002 | Cr2Cu1S4 |
| | 166 | z2w,1=1 z2w,2=1 z2w,3=1 |
| | 166 | z2w,1=1 z2w,2=1 z2w,3=1 z4=3 |
| 227.4.110 | 291001 | Cr2Cu1S4 |
| | 166 | z2w,1=1 z2w,2=1 z2w,3=1 |
| | 166 | z2w,1=1 z2w,2=1 z2w,3=1 z4=3 |
| 227.4.111 | 291000 | Cr2Cu1S4 |
| | 166 | z2w,1=1 z2w,2=1 z2w,3=1 |
| | 166 | z2w,1=1 z2w,2=1 z2w,3=1 z4=3 |
| 227.4.112 | 290999 | Cr2Cu1S4 |
| | 166 | z2w,1=1 z2w,2=1 z2w,3=1 |
| | 166 | z2w,1=1 z2w,2=1 z2w,3=1 z4=3 |
| 227.4.113 | 246081 | Cr2Cu1O4 |
| | 166 | z2w,1=1 z2w,2=1 z2w,3=1 z4=1 |
| | 166 | z2w,1=1 z2w,2=1 z2w,3=1 |

Continued on next column

**TABLE XIX– continued**

| TQC | ICSD<br>SSG | Stoich. Formula<br>SIs |
|---|---|---|
| 227.4.114 | 63165 | Co3O4 |
| | 166 | z2w,1=1 z2w,2=1 z2w,3=1 |
| | 166 | z2w,1=1 z2w,2=1 z2w,3=1 z4=3 |
| 227.4.115 | 63164 | Co3O4 |
| | 166 | z2w,1=1 z2w,2=1 z2w,3=1 |
| | 166 | z2w,1=1 z2w,2=1 z2w,3=1 z4=3 |
| 227.4.116 | 36256 | Co3O4 |
| | 166 | z2w,1=1 z2w,2=1 z2w,3=1 |
| | 166 | z2w,1=1 z2w,2=1 z2w,3=1 z4=3 |
| 227.4.117 | 290808 | Co1Cr2S4 |
| | 166 | z2w,1=1 z2w,2=1 z2w,3=1 z4=1 |
| | 166 | z2w,1=1 z2w,2=1 z2w,3=1 |
| 227.4.118 | 180008 | Cd2O7Tc2 |
| | 166 | z4=2 |
| | 141 | z4=2 |
| 227.4.119 | 41112 | Cd1S4Yb2 |
| | 166 | z4=1 |
| | 141 | z4=1 z2=1 |
| 227.4.120 | 37409 | Cd1S4Yb2 |
| | 166 | z4=1 |
| | 141 | z4=1 z2=1 |
| 227.4.121 | 156409 | Ca2O7Ru2 |
| | 166 | z4=1 |
| | 141 | z4=1 z2=1 |
| 227.4.122 | 50048 | Ca2F1O6Ta2 |
| | 166 | z2w,1=1 z2w,2=1 z2w,3=1 z4=3 |
| | 166 | z2w,1=1 z2w,2=1 z2w,3=1 z4=2 |
| 227.4.123 | 55069 | Bi2Rb1 |
| | 166 | z2w,1=1 z2w,2=1 z2w,3=1 z4=2 |
| | 166 | z2w,1=1 z2w,2=1 z2w,3=1 z4=1 |
| 227.4.124 | 55068 | Bi2K1 |
| | 166 | z2w,1=1 z2w,2=1 z2w,3=1 z4=2 |
| | 166 | z2w,1=1 z2w,2=1 z2w,3=1 z4=1 |
| 227.4.125 | 237820 | Bi2Ir2O7 |
| | 166 | z4=1 |
| | 141 | z4=1 z2=1 |
| 227.4.126 | 55070 | Bi2Cs1 |
| | 166 | z2w,1=1 z2w,2=1 z2w,3=1 z4=2 |
| | 166 | z2w,1=1 z2w,2=1 z2w,3=1 z4=1 |
| 227.4.127 | 108001 | Au2Na1 |
| | 166 | z2w,1=1 z2w,2=1 z2w,3=1 |
| | 166 | z2w,1=1 z2w,2=1 z2w,3=1 z4=3 |
| 227.4.128 | 57911 | Al2Ho1 |
| | 12 | z2w,1=1 z2w,2=1 z2w,3=1 z4=1 |
| | 12 | z2w,3=1 z4=3 |
| | 12 | z2w,3=1 z4=2 |
| | 12 | z2w,1=1 z2w,2=1 z2w,3=1 |
| 227.4.129 | 290133 | Al2Co1O4 |
| | 166 | z2w,1=1 z2w,2=1 z2w,3=1 z4=2 |
| | 166 | z2w,1=1 z2w,2=1 z2w,3=1 z4=1 |
| 227.4.130 | 170868 | Ag1In2Na3 |
| | 122 | z2=1 |
| | 119 | z2=1 |
| 227.4.131 | 71695 | Ag1Cr2Te4 |
| | 166 | z2w,1=1 z2w,2=1 z2w,3=1 z4=3 |
| | 166 | z2w,1=1 z2w,2=1 z2w,3=1 z4=2 |
| 229.4.1 | 426405 | Ru7Sb6Yb4 |
| | 121 | z2=1 |
| | 121 | z2=1 |

Continued on next column

**TABLE XIX– continued**

| TQC | ICSD<br>SSG | Stoich. Formula<br>SIs |
|---|---|---|
| | 119 | z2=1 |
| | 119 | z2=1 |
| | 71 | z2w,1=1 z2w,2=1 z2w,3=1 z4=2 |
| 229.4.2 | 54511 | Ru3Sn7 |
| | 166 | z2w,1=1 z2w,2=1 z2w,3=1 z4=3 |
| | 139 | z2w,1=1 z2w,2=1 z2w,3=1 z4=3 z2=1 z8=7 |
| 229.4.3 | 54510 | Ru3Sn7 |
| | 166 | z2w,1=1 z2w,2=1 z2w,3=1 z4=3 |
| | 139 | z2w,1=1 z2w,2=1 z2w,3=1 z4=3 z2=1 z8=7 |
| 229.4.4 | 24303 | Mo3Sb7 |
| | 166 | z2w,1=1 z2w,2=1 z2w,3=1 |
| | 166 | z2w,1=1 z2w,2=1 z2w,3=1 |
| | 139 | z2w,1=1 z2w,2=1 z2w,3=1 |
| 229.4.5 | 164762 | Mo3Sb7 |
| | 166 | z2w,1=1 z2w,2=1 z2w,3=1 |
| | 166 | z2w,1=1 z2w,2=1 z2w,3=1 |
| | 139 | z2w,1=1 z2w,2=1 z2w,3=1 |
| 229.4.6 | 173131 | Mo1 |
| | 166 | z2w,1=1 z2w,2=1 z2w,3=1 z4=2 |
| | 139 | z2w,1=1 z2w,2=1 z2w,3=1 z4=2 z8=6 |
| 229.4.7 | 173127 | Mo1 |
| | 166 | z2w,1=1 z2w,2=1 z2w,3=1 z4=2 |
| | 139 | z2w,1=1 z2w,2=1 z2w,3=1 z4=2 z8=6 |
| 229.4.8 | 428097 | La6Pd13Zn4 |
| | 166 | z4=3 |
| | 139 | z4=3 z2=1 z8=3 |
| 229.4.9 | 412776 | Ir3Sn7 |
| | 166 | z4=1 |
| | 139 | z4=1 z2=1 z8=5 |
| 229.4.10 | 409596 | In7Pt3 |
| | 166 | z4=1 |
| | 166 | z4=1 |
| | 139 | z4=1 z2=1 z8=1 |
| | 139 | z4=1 z2=1 z8=5 |
| | 139 | z4=1 z2=1 z8=1 |
| | 139 | z4=1 z2=1 z8=5 |
| 229.4.11 | 408314 | In7Pd3 |
| | 166 | z4=1 |
| | 139 | z4=1 z2=1 z8=1 |
| | 139 | z4=1 z2=1 z8=5 |
| 229.4.12 | 408313 | Ga7Ni3 |
| | 166 | z4=1 |
| | 166 | z4=1 |
| | 139 | z4=1 z2=1 z8=1 |
| | 139 | z4=1 z2=1 z8=5 |
| | 139 | z4=1 z2=1 z8=1 |
| | 139 | z4=1 z2=1 z8=5 |
| 229.4.13 | 153909 | Ga12Pt1Y4 |
| | 166 | z4=2 |
| | 139 | z4=2 z8=2 |
| 229.4.14 | 249997 | Fe1Ga12Tb4 |
| | 166 | z2w,1=1 z2w,2=1 z2w,3=1 z4=2 |
| | 139 | z2w,1=1 z2w,2=1 z2w,3=1 z4=2 z8=6 |
| | 139 | z2w,1=1 z2w,2=1 z2w,3=1 z4=2 z8=2 |
| 229.4.15 | 426404 | Er4Ru7Sb6 |
| | 121 | z2=1 |
| | 121 | z2=1 |

Continued on next column

**TABLE XIX– continued**

| TQC | ICSD<br>SSG | Stoich. Formula<br>SIs |
|---|---|---|
| | 119 | z2=1 |
| | 119 | z2=1 |
| | 71 | z4=3 |
| 229.4.16 | 248000 | Er4Fe1Ga12 |
| | 166 | z4=2 |
| | 139 | z4=2 z8=6 |
| | 139 | z4=2 z8=2 |
| 229.4.17 | 421178 | Cd4La6Pd13 |
| | 166 | z4=3 |
| | 139 | z4=3 z2=1 z8=3 |
| 229.4.18 | 195760 | Ca5Ge6Pd6 |
| | 166 | z2w,1=1 z2w,2=1 z2w,3=1 z4=2 |
| | 139 | z2w,1=1 z2w,2=1 z2w,3=1 z4=2 z8=6 |
| 229.4.19 | 400465 | B3Eu4Li1N6 |
| | 166 | z2w,1=1 z2w,2=1 z2w,3=1 z4=2 |
| | 166 | z2w,1=1 z2w,2=1 z2w,3=1 z4=2 |
| | 139 | z2w,1=1 z2w,2=1 z2w,3=1 z4=2 z8=2 |
| 229.4.20 | 90326 | As6Ru7U4 |
| | 166 | z2w,1=1 z2w,2=1 z2w,3=1 z4=3 |
| | 139 | z2w,1=1 z2w,2=1 z2w,3=1 z4=3 z2=1 z8=7 |

The End

TABLE XX: Miscellaneous Chemical Information about the compounds in our "high-quality" list. We classify them as (A): compound is perfect and can be made easily. (B): compound looks good but maybe no stable in air. (C): compound can be AFM Ferro or MOTT. (D): compound has to be check - DFT problem! Non transition metal radical such as N3-, or (O2)2- or (N2O4) or La+1 or La+2. (E): compound is radio active and mostly magnetic - spin polarized calculation necessary. Miscellaneous other comments are added; also see Appendix R. See the list of tables.

| TQC | ICSD | Stoich. Formula | SubSG |
|---|---|---|---|
| 2.2.1 | 202564 | S1Ta6 | A |
| 2.2.2 | 41017 | P3Tc2 | E |
| 2.2.3 | 66584 | O9V5 | A or C |
| 2.2.4 | 80363 | O9P3Ru1 | A or C |
| 2.2.5 | 2212 | O7V4 | C |
| 2.2.6 | 19012 | O7V4 | A or C |
| 2.2.9 | 59657 | O4Pb1Pt2 | C |
| 2.2.16 | 237587 | Mo3S4 | And |
| 2.2.35 | 240179 | Ga5Rh3 | A |
| 2.2.40 | 84558 | Cu1O4W1 | C |
| 2.2.41 | 69759 | Cu1O4W1 | C |
| 2.2.48 | 39439 | Cu1Mo1O4 | C |
| 2.2.55 | 150974 | Cs1Hg1 | A or B since Hg might evaporate |
| 2.2.62 | 84620 | Cl5Mo1 | A |
| 2.2.80 | 423462 | Br5W1 | A or C |
| 2.2.81 | 39784 | Br3Ti1 | A or C |
| 2.2.84 | 85490 | Ba1Mo6S8 | A |
| 2.2.85 | 65775 | Ba1Mo6S8 | A |
| 2.2.88 | 96562 | B2Co2O5 | C |
| 2.2.90 | 66662 | As3Tc2 | E |
| 2.2.91 | 86400 | As3Ca1 | A |
| 2.2.98 | 95832 | Ag2F5 | C |
| 2.2.99 | 421341 | Ag1O4S1 | C |
| 2.2.100 | 51507 | Ag1F6Pd1 | A or C |
| 4.3.2 | 53666 | Ge2La1Pt2 | A |
| 4.3.7 | 4327 | Cu1I2O6 | A or C |
| 7.3.1 | 281242 | I6K2Re1 | A |
| 7.3.2 | 414303 | Ga9Ir2 | A |
| 9.3.1 | 69758 | Cu1O1 | A or C |
| 9.3.2 | 69757 | Cu1O1 | A or C |
| 9.3.7 | 192517 | Bi1Fe1O3 | A or C the multiferroic !!! |
| 9.3.8 | 192515 | Bi1Fe1O3 | A or C |
| 11.2.1 | 42076 | Te3Zr1 | A |
| 11.2.2 | 424108 | Sn2Sr1Zn2 | A |
| 11.2.5 | 52307 | Sb2Sr1 | A |
| 11.2.6 | 171417 | S3U1 | E |
| 11.2.10 | 24355 | Rh4Si5 | A |
| 11.2.14 | 99271 | Ni3Te2 | A or C |
| 11.2.15 | 54304 | Ni3Ta1 | A or C |
| 11.2.16 | 245081 | Ni1Ti1 | A or C |
| 11.2.18 | 166372 | Ni1Ti1 | A or C |
| 11.2.19 | 166367 | Ni1Ti1 | A or C |
| 11.2.20 | 164155 | Ni1Ti1 | A or C |
| 11.2.21 | 164154 | Ni1Ti1 | A or C |
| 11.2.22 | 164153 | Ni1Ti1 | A or C |
| 11.2.23 | 164152 | Ni1Ti1 | A or C |
| 11.2.24 | 164151 | Ni1Ti1 | A or C |
| 11.2.26 | 105415 | Ni1Ti1 | A or C |
| 11.2.31 | 93368 | Na1O8P1W2 | A or C |
| 11.2.32 | 73453 | Mo2S3 | A |
| 11.2.33 | 62486 | Mo2S3 | A |

Continued on next column

**TABLE XX– continued**

| TQC | ICSD | Stoich. Formula | SubSG |
|---|---|---|---|
| 11.2.46 | 70101 | H3Cl1Cu2O3 | A or C |
| 11.2.47 | 153707 | H3Cl1Cu2O3 | A or C |
| 11.2.56 | 260834 | H2Cl1Ho1O2 | A or C |
| 11.2.64 | 418027 | Ge1Hf1Pd1 | A |
| 11.2.66 | 89343 | F3K1Mn1 | A or C |
| 11.2.70 | 414338 | Cu1Nb1Te2 | A or C |
| 11.2.73 | 411140 | Cl7Cs1Ti2 | A or C |
| 11.2.74 | 415798 | Cl2Co2O3Te1 | A or C |
| 11.2.76 | 54553 | Ce1In2Pt2 | A or C |
| 11.2.77 | 862 | Ca1Sb2 | A |
| 11.2.82 | 416300 | Bi2Pd2Sr1 | A |
| 11.2.84 | 81819 | Ba3Ge2N2 | A |
| 11.2.85 | 409517 | Ba1Sb2 | A |
| 11.2.89 | 416299 | Ba1Bi2Pd2 | A |
| 11.2.93 | 416298 | Au2Ba1Sb2 | A |
| 11.2.94 | 902 | As1Cl1Co2O4 | A or C |
| 11.2.97 | 57494 | Al1Au1 | A |
| 11.2.98 | 104649 | Al1Au1 | A |
| 11.4.1 | 166509 | S3Ta1 | A |
| 11.4.2 | 15251 | S3Ta1 | A |
| 11.4.5 | 1239 | O2Rb9 | B |
| 11.4.6 | 424526 | Np1Se3 | A or C |
| 11.4.7 | 82605 | Ni2Ta1Te3 | A evt A or C |
| 11.4.8 | 84826 | Ni1P1Zr2 | A or C |
| 11.4.10 | 30002 | Nb1Se3 | A |
| 11.4.11 | 97682 | Mo2S2Sb1 | A |
| 11.4.14 | 71960 | La1Ru1Si2 | A |
| 11.4.16 | 71524 | I3Ir1Y3 | A |
| 11.4.17 | 84828 | Hf2Ni1P1 | A or C |
| 11.4.18 | 405616 | Hf2Ni1P1 | A or C |
| 11.4.34 | 172060 | H2Na3O6Os1 | ? |
| 11.4.60 | 76293 | Ge6Pt4Y3 | A |
| 11.4.63 | 66513 | Fe1Ta1Te3 | A or C |
| 11.4.72 | 422033 | Cu1N4O10 | A or C |
| 11.4.77 | 409488 | Cr5P3Zr1 | A or C |
| 11.4.78 | 35044 | Cr2Nb2Se10 | A or C |
| 11.4.80 | 99178 | Ce1P5 | A or C ? |
| 11.4.81 | 409181 | Ce1P5 | A or C |
| 11.4.83 | 183583 | Ca1O3Rh1 | A |
| 11.4.84 | 58880 | Ca1Cu1 | A or C |
| 11.4.87 | 237796 | C1Cu1Na3O7P1 | A or C |
| 11.4.88 | 237797 | C1Co1Na3O7P1 | A or C |
| 11.4.91 | 410948 | Br9Te1W1 | A |
| 11.4.92 | 39257 | Bi9I2 | A |
| 11.4.93 | 162834 | Bi1Fe1O3 | A or C |
| 11.4.100 | 427443 | As3Ca1Fe5 | A or C |
| 12.1.3 | 41047 | Ta2Te3 | A |
| 12.1.6 | 48176 | Rh1Si2Y2 | A |
| 12.1.7 | 42077 | P2V1 | A |
| 12.1.9 | 34417 | O2V1 | A or C |
| 12.1.10 | 34416 | O2V1 | A evt A or C |
| 12.1.11 | 50409 | O13V6 | A or C |
| 12.1.12 | 15028 | O13V6 | A or C |
| 12.1.16 | 48004 | Ni1Sc3Si3 | A |
| 12.1.22 | 41238 | Nb7P4 | A or C |
| 12.1.24 | 77752 | Nb1Sb2 | A |
| 12.1.26 | 57430 | Nb1Sb2 | A |
| 12.1.27 | 190180 | Nb1Sb2 | A |
| 12.1.30 | 29383 | N3Na1 | colorless salt and explosive |
| 12.1.31 | 29382 | N3Na1 | D |

Continued on next column

**TABLE XX– continued**

| TQC | ICSD | Stoich. Formula | SubSG |
|---|---|---|---|
| 12.1.32 | 29381 | N3Na1 | explosive and white D because N3- is radical |
| 12.1.33 | 29380 | N3Na1 | explosive and white D because N3- is radical |
| 12.1.34 | 29379 | N3Na1 | colorless salt and explosive |
| 12.1.35 | 29378 | N3Na1 | colorless salt and explosive |
| 12.1.36 | 29377 | N3Na1 | A |
| 12.1.37 | 29376 | N3Na1 | D |
| 12.1.38 | 29375 | N3Na1 | colorless salt and explosive |
| 12.1.39 | 29374 | N3Na1 | colorless salt or explosive |
| 12.1.40 | 29373 | N3Na1 | D |
| 12.1.41 | 29372 | N3Na1 | D |
| 12.1.42 | 29371 | N3Na1 | D |
| 12.1.43 | 29370 | N3Na1 | A |
| 12.1.44 | 151588 | Mn7Na1O12 | A or C |
| 12.1.47 | 165326 | Li1Mo1O2 | A |
| 12.1.48 | 49021 | K1Li1Mn1O2 | A or C |
| 12.1.52 | 407247 | I3La4Si4 | A |
| 12.1.53 | 83965 | Hf2Te1 | A |
| 12.1.84 | 421466 | Cu4Na1O12Te2Tl3 | A or C |
| 12.1.85 | 71624 | Cs2Ir1Na2O4 | A |
| 12.1.93 | 81886 | Cr3S4 | A or C |
| 12.1.95 | 192944 | Cr3S4 | A or C |
| 12.1.97 | 74296 | Cr2S8Tl1V3 | A or C |
| 12.1.103 | 75420 | Cr1S2 | A or C |
| 12.1.107 | 821 | Cr1P1S4 | A or C |
| 12.1.108 | 56835 | Cr1P1S4 | A or C |
| 12.1.111 | 66006 | Co2Na7O6Rb1 | A or C |
| 12.1.124 | 424719 | Ce4Ge6Mg7 | A or C |
| 12.1.127 | 63503 | C5Nb6 | A |
| 12.1.135 | 88513 | C3Ho2Mo2 | A or C |
| 12.1.136 | 88512 | C3Ho2Mo2 | A or C |
| 12.1.137 | 88511 | C3Ho2Mo2 | A or C |
| 12.1.143 | 417827 | C3Ce2Mo2 | A or C |
| 12.1.157 | 1559 | Bi1I1 | A |
| 12.1.158 | 1560 | Bi1Br1 | A |
| 12.1.161 | 173685 | Ba3Cd2Sb4 | A |
| 12.1.165 | 154011 | Ba1Mn2O6Y1 | A or C |
| 12.1.166 | 173848 | Ba1Ir1O3 | A |
| 12.1.168 | 61499 | Ba1Bi1O3 | A |
| 12.1.173 | 261932 | B2Ni3Zn1 | A or C |
| 12.1.175 | 425401 | Au1Ce2O1P2 | A or C |
| 12.1.176 | 424760 | As4Ba3Zn2 | A |
| 12.1.177 | 424761 | As4Ba3Cd2 | A |
| 12.1.178 | 43185 | As3W2 | A |
| 12.1.179 | 15049 | As3V4 | A or C |
| 12.1.180 | 43184 | As3Mo2 | A |
| 12.1.181 | 41015 | As3Mo2 | A |
| 12.1.184 | 44069 | As2Ta1 | A |
| 12.1.185 | 69002 | As2Ni2O7 | A or C |
| 12.1.186 | 81218 | As2Nb1 | A |
| 12.1.188 | 261143 | As2Mn2O7 | A or C |
| 12.1.190 | 81775 | As2Ce1Pd3 | A or C |
| 12.1.191 | 75454 | As2Ba2Mn2O1 | A or C |
| 12.1.192 | 68906 | As1F6N1O2 | A |
| 12.2.2 | 420133 | Ru1Sc2Si2 | A |
| 12.2.3 | 61353 | Pt1Se7Ta2 | A |
| 12.2.5 | 61005 | Pd1Se6Ta2 | A |
| 12.2.6 | 40281 | P4Zr7 | A |
| 12.2.7 | 41016 | P3Tc2 | E |
| 12.2.12 | 75194 | O5Ti3 | A or C |
| 12.2.13 | 75193 | O5Ti3 | A or C |
| 12.2.29 | 73315 | Nb8Pt1Se20 | A |
| 12.2.30 | 280016 | Nb2Te3 | A |
| 12.2.32 | 81059 | Mo1Ni1O4 | A or C |
| 12.2.34 | 15615 | Mn1Mo1O4 | A or C |
| 12.2.35 | 50001 | Mg2Si10Sr11 | A |
| 12.2.37 | 246807 | La1Si5 | A |
| 12.2.38 | 428765 | Hg2Sn2Sr1 | A |
| 12.2.40 | 405208 | Hf7P4 | A |
| 12.2.56 | 76350 | Ge2Ho2Os1 | A or C |
| 12.2.59 | 165189 | Ga3Sc5 | A |
| 12.2.60 | 15044 | Fe3Se4 | A or C |
| 12.2.61 | 15043 | Fe3Se4 | A or C |
| 12.2.63 | 84205 | Fe1Sc2Si2 | A or C |
| 12.2.64 | 76348 | Fe1Sc2Si2 | A or C |
| 12.2.66 | 20661 | Fe1Sc2Si2 | A or C |
| 12.2.76 | 43055 | Fe1Ge1 | A |
| 12.2.79 | 422453 | Cu2Sn2Sr1 | A |
| 12.2.80 | 182102 | Cu2Sn2Sr1 | A |
| 12.2.81 | 79536 | Cu2Na2S4Zr1 | A or C |
| 12.2.86 | 73143 | Cr3Rb1S5 | A or C |
| 12.2.87 | 1824 | Cr3Ga4 | A or C |
| 12.2.97 | 151873 | Cr1Fe2Se4 | C AFM? |
| 12.2.99 | 41026 | Co2Nb4Pd1Se12 | A or C |
| 12.2.100 | 408031 | Cl2La3Si3 | A |
| 12.2.104 | 427663 | Cl1Ho1O4W1 | A or C |
| 12.2.107 | 424718 | Ce5Ge8Mg8 | A or C |
| 12.2.108 | 417665 | Ce1Ru1Sn1 | A or C |
| 12.2.109 | 192243 | Ce1Ru1Sn1 | A or C |
| 12.2.110 | 192242 | Ce1Ru1Sn1 | A or C |
| 12.2.120 | 85505 | Ca1K2N4O8 | check might be also a radical like N3- |
| 12.2.123 | 54994 | Ca1Ga4 | A |
| 12.2.124 | 417717 | Ca1Cu2Sn2 | xv |
| 12.2.125 | 236391 | C4Co1Sc3 | A or C |
| 12.2.136 | 61141 | C10Fe22Mo12 | A or C |
| 12.2.137 | 65114 | C10Au2Cl8S8 | ? |
| 12.2.138 | 240479 | Br4Cs2I2Pd1 | A |
| 12.2.140 | 427411 | Bi4Pb7Se13 | A |
| 12.2.141 | 58840 | Bi2Pd5 | A |
| 12.2.143 | 421873 | Bi2Pd3Se2 | A |
| 12.2.150 | 1024 | Bi2Ni3Se2 | A or C |
| 12.2.151 | 195678 | Bi2Hg1S4 | B Hg evaporates |
| 12.2.152 | 415138 | Bi1I1Mn1Se2 | A or C |
| 12.2.154 | 164224 | Bi1Fe1O3 | A or C |
| 12.2.155 | 164223 | Bi1Fe1O3 | A or C |
| 12.2.156 | 246145 | Bi16I4 | A |
| 12.2.157 | 423458 | Ba5Cd2O2Sb4 | A |
| 12.2.159 | 62390 | Ba3Cu2O10Pt1Y2 | A or C |
| 12.2.160 | 62391 | Ba3Cu2Ho2O10Pt1 | A or C |
| 12.2.161 | 408938 | Ba2Mg3Si4 | A |
| 12.2.162 | 420834 | Ba2Cd2Sb3 | A |
| 12.2.164 | 261206 | Ba1Tl4 | colorless salt and explosive |
| 12.2.165 | 238227 | Ba1Te4Th1 | A or E |
| 12.2.166 | 49000 | Ba1Sb3 | A |
| 12.2.168 | 83618 | Ba1Cu10P4 | A or C |

Continued on next column

Continued on next column

**TABLE XX– continued**

| TQC | ICSD | Stoich. Formula | SubSG |
|---|---|---|---|
| 12.2.169 | 418886 | Ba11Cd6Sb12 | A |
| 12.2.170 | 63501 | B4La2Ni5 | A or C |
| 12.2.180 | 8058 | Au2P3 | A |
| 12.2.181 | 429212 | Au2Ni4Si7 | A |
| 12.2.183 | 41831 | As3Sr1 | A |
| 12.2.184 | 80375 | As3Mn4 | A or C |
| 12.2.185 | 86402 | As3Ba1 | A |
| 12.2.186 | 69001 | As2Co2O7 | A or C |
| 12.2.188 | 417968 | Al2Ge4Sr3 | A |
| 12.2.189 | 417659 | Al2Ba5Ge7 | A |
| 12.2.190 | 20632 | Al1Ce1Co1 | A or C |
| 13.2.8 | 170685 | Mn1O4Re1 | A or C |
| 14.2.2 | 57480 | Re2Si1 | A |
| 14.2.4 | 8217 | O2W1 | A or C |
| 14.2.5 | 80829 | O2W1 | A or C photocatalyst and large gap |
| 14.2.12 | 80830 | Mo1O2 | A or C large gap |
| 14.2.13 | 23722 | Mo1O2 | A or C large gap |
| 14.2.14 | 174231 | Ir1P2 | A |
| 14.2.26 | 189438 | Fe2O3 | A or C |
| 14.2.46 | 63553 | Co1Sb2 | A or C |
| 14.2.57 | 86049 | C7Y4 | A |
| 14.2.66 | 58406 | Au4Ca5 | A |
| 14.2.69 | 73402 | Ag2Te1 | A |
| 14.2.70 | 73230 | Ag2Te1 | A |
| 14.3.2 | 26902 | P1Pd6 | A |
| 14.3.19 | 189806 | Bi1 | A |
| 15.2.3 | 16041 | S1Ta6 | A |
| 15.2.4 | 24327 | P2Pt5 | A |
| 15.2.6 | 35148 | O5Ti3 | A or C |
| 15.2.7 | 194464 | O5Ti3 | A or C |
| 15.2.9 | 97442 | O3V2 | A or C |
| 15.2.10 | 95763 | O3V2 | A or C |
| 15.2.11 | 95762 | O3V2 | A or C |
| 15.2.12 | 6286 | O3V2 | A or C |
| 15.2.13 | 50703 | Ni2O8Te3 | A or C |
| 15.2.14 | 61148 | Ni1Se5Ta2 | A |
| 15.2.17 | 78342 | Na1O6Si2V1 | A or C |
| 15.2.19 | 163132 | Na1O6Si2V1 | A or C |
| 15.2.20 | 163131 | Na1O6Si2V1 | A or C |
| 15.2.21 | 163130 | Na1O6Si2V1 | A or C |
| 15.2.22 | 163129 | Na1O6Si2V1 | A or C |
| 15.2.23 | 163128 | Na1O6Si2V1 | A |
| 15.2.24 | 163127 | Na1O6Si2V1 | A or C |
| 15.2.26 | 163125 | Na1O6Si2V1 | A or C |
| 15.2.28 | 49802 | Mn2P12Ti1 | A or C |
| 15.2.29 | 78721 | Li2O3Ru1 | A |
| 15.2.30 | 180367 | Li2O3Ru1 | A or C MIT |
| 15.2.31 | 55164 | Li1O6Si2V1 | A or C |
| 15.2.33 | 416039 | K3Na1O8Ru2 | A or C MAGNETIC with gap |
| 15.2.35 | 79499 | Ho1O4Ta1 | A or C |
| 15.2.36 | 153363 | Ho1O4Ta1 | A or C |
| 15.2.46 | 151766 | Fe7S8 | A or C |
| 15.2.47 | 24416 | Fe2O5Ti1 | A or C |
| 15.2.48 | 154291 | Fe2O5Rb4 | A or C |
| 15.2.49 | 71059 | Fe2Li1N3Sr2 | A or C |
| 15.2.50 | 41899 | Fe1S5U2 | A or C and E |
| 15.2.53 | 94884 | D7Al1Sr2 | ? |
| 15.2.57 | 106530 | Cu6Sn5 | A or C |
| 15.2.58 | 23479 | Cu2O7V2 | A or C |

Continued on next column

**TABLE XX– continued**

| TQC | ICSD | Stoich. Formula | SubSG |
|---|---|---|---|
| 15.2.59 | 158375 | Cu2O7V2 | A or C |
| 15.2.60 | 67316 | Cu2O7P2 | A or C |
| 15.2.61 | 28279 | Cu2O7P2 | A or C |
| 15.2.62 | 157107 | Cu2O7P2 | Copper diphosphate should be insulating A or C |
| 15.2.63 | 14369 | Cu2O7P2 | A or C |
| 15.2.67 | 2790 | Cr1P4 | A or C |
| 15.2.68 | 26814 | Co1Ge1O3 | A or C |
| 15.2.69 | 417870 | Cl8Ga2Ni1 | A or C |
| 15.2.70 | 56838 | Cl3Cu1Rb1 | A or C AFM at 10K a |
| 15.2.71 | 189154 | Cl2Cu2O6Pb1Se2 | A or C |
| 15.2.73 | 95789 | Ca2Ir2Si1 | A |
| 15.2.74 | 82216 | Ca1Ni1O6Si2 | A or C |
| 15.2.75 | 69381 | Ca1Ni1O6Si2 | A or C |
| 15.2.78 | 423887 | C2Fe5 | A or C |
| 15.2.79 | 423886 | C2Fe5 | A or C |
| 15.2.80 | 423885 | C2Fe5 | A or C |
| 15.2.81 | 183902 | Bi1Mn1O3 | A or C |
| 15.2.84 | 160458 | Bi1Cr1O3 | A or C |
| 15.2.86 | 160456 | Bi1Cr1O3 | A or C |
| 15.2.88 | 245254 | Ba3Ir2O9Sr1 | A or C |
| 15.2.89 | 245253 | Ba3Ca1Ir2O9 | A or C |
| 15.2.90 | 71060 | Ba2Fe2Li1N3 | A or C |
| 15.2.91 | 86064 | Ba1Cu1N1 | A or C |
| 15.2.92 | 418385 | B2Ni1O4 | A or C |
| 15.2.93 | 58403 | Au2Ca5 | A |
| 15.2.94 | 417872 | Al2Cl8Ni1 | A or C |
| 15.2.95 | 251835 | Al1B2La2Os2 | A |
| 15.4.1 | 40760 | Rh3Si5Y2 | A |
| 15.4.2 | 74837 | Rb1Te6 | B |
| 15.4.7 | 51023 | O6Re2Sb1 | A |
| 15.4.13 | 281609 | Na1O6Si2Ti1 | A or C |
| 15.4.14 | 281607 | Na1O6Si2Ti1 | A or C Silicate wide band gap |
| 15.4.15 | 281606 | Na1O6Si2Ti1 | A or C |
| 15.4.16 | 281604 | Na1O6Si2Ti1 | A or C |
| 15.4.18 | 425216 | Mn3O4Rb2 | A or C |
| 15.4.20 | 87184 | K1O2 | (O2)- is a MOTT insulator yellow powder |
| 15.4.23 | 391433 | In1Pt2Sr2 | A |
| 15.4.24 | 391432 | In1Pd2Sr2 | A |
| 15.4.33 | 40781 | Fe1Rb1Se2 | A (or C) superconductor with Rb1-x |
| 15.4.34 | 202385 | Fe1Rb1S2 | A or C |
| 15.4.35 | 202384 | Fe1Rb1S2 | A or C |
| 15.4.36 | 202381 | Fe1Rb1S2 | A or C |
| 15.4.37 | 9671 | Fe1Na1O6Si2 | A or C |
| 15.4.38 | 180313 | Fe1Na1O6Si2 | A or C |
| 15.4.41 | 161982 | Fe1Na1O6Si2 | A or C |
| 15.4.42 | 161980 | Fe1Na1O6Si2 | A or C |
| 15.4.43 | 161979 | Fe1Na1O6Si2 | A or C |
| 15.4.46 | 161976 | Fe1Na1O6Si2 | A or C |
| 15.4.48 | 158082 | Fe1Na1O6Si2 | A or C |
| 15.4.52 | 10219 | Fe1Na1O6Si2 | A or C |
| 15.4.54 | 9669 | Fe1Li1O6Si2 | A or C |
| 15.4.57 | 40780 | Fe1K1Se2 | A (or C) superconductor with K1-x |
| 15.4.58 | 68383 | Fe1K1S2 | A or C |
| 15.4.59 | 202383 | Fe1K1S2 | A or C |
| 15.4.60 | 202382 | Fe1K1S2 | A or C |

Continued on next column

**TABLE XX– continued**

| TQC | ICSD | Stoich. Formula | SubSG |
|---|---|---|---|
| 15.4.61 | 202380 | Fe1K1S2 | A or C |
| 15.4.68 | 180306 | Fe1Ge2Na1O6 | A or C |
| 15.4.88 | 202587 | Cu3F7Na1 | A or C Flouride |
| 15.4.89 | 83493 | Cu2O4Y1 | A or C |
| 15.4.92 | 72721 | Cu1O6Pt1Sr3 | A or C |
| 15.4.94 | 603 | Cu1O5Se2 | A or C |
| 15.4.95 | 245059 | Cu1O5Se2 | A or C |
| 15.4.96 | 69094 | Cu1O1 | A or C |
| 15.4.97 | 67850 | Cu1O1 | Cu2+ black oxide AFM Mott |
| 15.4.98 | 291393 | Cu1O1 | Cu2+ black oxide AFM Mott |
| 15.4.99 | 291390 | Cu1O1 | Cu2+ black oxide AFM Mott |
| 15.4.100 | 291387 | Cu1O1 | A or C |
| 15.4.102 | 16025 | Cu1O1 | A or C |
| 15.4.106 | 160321 | Cu1Li2O7P2 | A or C |
| 15.4.108 | 97262 | Cr1Na1O6Si2 | A or C |
| 15.4.109 | 97258 | Cr1Na1O6Si2 | A or C |
| 15.4.110 | 97257 | Cr1Na1O6Si2 | A or C |
| 15.4.111 | 9670 | Cr1Na1O6Si2 | A or C |
| 15.4.115 | 192418 | Cr1Na1O6Si2 | A or C |
| 15.4.116 | 10240 | Cr1Na1O6Si2 | A or C |
| 15.4.124 | 414127 | Co2Na7O6 | A or C |
| 15.4.125 | 157364 | Co2Ga7Mg3 | A |
| 15.4.128 | 1242 | Cl6Cs1W1 | A or C |
| 15.4.129 | 418203 | Ca2In1Pt2 | A |
| 15.4.130 | 418202 | Ca2In1Pd2 | A |
| 15.4.131 | 429177 | Ca2Ga1Pt2 | A |
| 15.4.132 | 429176 | Ca2Ga1Pd2 | A |
| 15.4.134 | 170823 | Ca1Cu1Ge2O6 | A or C |
| 15.4.135 | 170822 | Ca1Cu1Ge2O6 | A or C |
| 15.4.136 | 170821 | Ca1Cu1Ge2O6 | A or C |
| 15.4.137 | 82212 | Ca1Co1O6Si2 | A or C |
| 15.4.139 | 415406 | C4Cu1O4 | A or C |
| 15.4.141 | 240902 | C2H4Mn1O6 | A or C |
| 15.4.144 | 419171 | C2Cl1La2 | C Mott insulator because La+ instead of La3+ |
| 15.4.146 | 240017 | Bi2Cs3 | A |
| 15.4.147 | 181425 | Bi1O6Re2 | A with a small probability of MOTT |
| 15.4.148 | 246363 | Ba3Ir2Na1O9 | A or C |
| 15.4.149 | 246362 | Ba3Ir2Na1O9 | A or C |
| 15.4.150 | 246361 | Ba3Ir2Na1O9 | A or C |
| 15.4.151 | 72448 | Ba3Bi1O9Ru2 | A or C |
| 15.4.153 | 20905 | Ba2Cu1O7Si2 | A or C Silicate with a wide band gap |
| 15.4.154 | 173850 | Ba1Ir1O3 | A or C |
| 15.4.155 | 202388 | Ba1Cu1O6Se2 | A or C |
| 15.4.157 | 24308 | B3Ni4 | A |
| 15.4.158 | 187403 | B3Ni4 | A |
| 15.4.160 | 39487 | B2Ho1Ni2 | A or C |
| 15.4.166 | 428674 | Al1Ca2Pt2 | A |
| 15.4.167 | 428673 | Al1Ca2Pd2 | A |
| 15.4.168 | 416931 | Ag4Cu1O6Te1 | A or C |
| 15.4.169 | 158336 | Ag2Cu2O3 | A or C probably AFM |
| 19.3.9 | 187197 | Ge1Mn1 | is magnetic and a Skyrmion system FeSi is the non magnetic coun |

Continued on next column

**TABLE XX– continued**

| TQC | ICSD | Stoich. Formula | SubSG |
|---|---|---|---|
| 19.3.10 | 187196 | Ge1Mn1 | is magnetic and a Skyrmion system FeSi is the non magnetic cou |
| 19.3.11 | 35451 | Cu1Na1O4P1 | A or C is a phosphate should be insulating |
| 19.3.13 | 425075 | Ce2Ga3Ru2 | A |
| 20.3.1 | 63230 | Ba1S3V1 | A |
| 26.4.2 | 75401 | C1Cu1F1K1O3 | A or C |
| 26.4.3 | 65082 | Bi1Cu4Fe1Pb1S6 | A or C |
| 33.3.6 | 15057 | Fe1P1 | looks A like non magnetic |
| 33.3.7 | 23822 | Fe1O3Y1 | A or C |
| 33.3.8 | 68984 | Fe1Na1O2 | A or C |
| 33.3.9 | 33772 | Fe1Na1O2 | A or C |
| 33.3.10 | 33771 | Fe1Na1O2 | A or C |
| 33.3.11 | 33770 | Fe1Na1O2 | A or C |
| 33.3.12 | 33767 | Fe1Na1O2 | A or C |
| 33.3.13 | 33766 | Fe1Na1O2 | A or C |
| 33.3.14 | 33765 | Fe1Na1O2 | A or C |
| 33.3.15 | 33764 | Fe1Na1O2 | A or C |
| 33.3.16 | 33763 | Fe1Na1O2 | A or C |
| 33.3.18 | 186309 | Fe1Na1O2 | A or C |
| 33.3.20 | 240332 | F5O1Os1 | Os+7 which is d1 therefore can be MOTT A or C |
| 33.3.21 | 89904 | Cu1O4P1Rb1 | A or C |
| 36.4.3 | 188064 | Ni1P1 | A |
| 36.4.5 | 68459 | H2Cu1O2 | C MOTT |
| 36.4.7 | 159032 | Ca1Ir1O3 | A or C C MOTT and magnetic |
| 36.4.10 | 63229 | Ba1S3V1 | V+4 d1 might be MOTT A or C |
| 36.4.11 | 52692 | Ba1S3V1 | V+4 d1 might be MOTT A or C |
| 36.4.12 | 154182 | Ba1S3V1 | V+4 d1 might be MOTT A or C |
| 36.4.15 | 66120 | As1Ni1 | A |
| 43.4.1 | 20612 | Ge5Y3 | A |
| 43.4.2 | 200779 | C2Cu1K2O6 | A or C |
| 43.4.3 | 200695 | C2Cu1K2O6 | A or C |
| 43.4.4 | 98667 | Ba5P9 | A |
| 47.2.1 | 86427 | Cu1O5Sr2Tl1 | A or C more A when doped metallic high Tc - cool C when und |
| 47.2.2 | 390001 | Ba4Cu6O13Y2 | A or C |
| 47.2.3 | 77745 | Ba2Cu3O7Y1 | more A when doped metallic high Tc - cool C when undoped an |
| 47.2.4 | 63335 | Ba2Cu3O7Y1 | A or C more A when doped metallic high Tc - cool C when und |
| 47.2.7 | 68044 | Ba2Cu3Ho1O7 | A or C |
| 47.2.13 | 71640 | B3Ir4Zn1 | A |
| 51.2.1 | 248052 | La2Se9U2 | E |
| 51.2.2 | 106800 | In1La1Ni2 | A |
| 51.2.4 | 421074 | Ca2Ge2Ni3 | A |
| 51.2.5 | 194877 | C3Al1Y3 | A |
| 51.2.6 | 69767 | C1B2U1 | E and spin polarized |
| 51.2.10 | 107495 | B3Rh7Zn5 | A |
| 51.2.11 | 107494 | B3Ir7Zn5 | A |
| 51.2.12 | 429116 | B1Ru1Ta1 | A |

Continued on next column

**TABLE XX– continued**

| TQC | ICSD | Stoich. Formula | SubSG |
|---|---|---|---|
| 51.2.13 | 426468 | B1Nb1Ru1 | A |
| 51.2.14 | 194435 | B1Nb1Ru1 | A |
| 51.2.15 | 249924 | Au3Rb2Tl1 | B |
| 51.2.16 | 16325 | Au1I1Te2 | A |
| 51.2.17 | 9986 | Al4Co1La1 | A or C |
| 51.4.1 | 417836 | In4Ir1Sr1 | A |
| 51.4.2 | 280921 | H4Cu1O4P2 | A or C insulator |
| 51.4.3 | 89669 | Cu1O3Si1 | A or C spin Peierls ? with AFM |
| 51.4.4 | 89407 | Cu1Ge1O3 | A or C |
| 51.4.5 | 89406 | Cu1Ge1O3 | A or C |
| 51.4.6 | 89405 | Cu1Ge1O3 | A or C |
| 51.4.7 | 89404 | Cu1Ge1O3 | A or C |
| 51.4.8 | 89403 | Cu1Ge1O3 | A or C |
| 51.4.9 | 89402 | Cu1Ge1O3 | A or C |
| 51.4.10 | 89401 | Cu1Ge1O3 | A or C spin Peierls with AFM |
| 51.4.11 | 89400 | Cu1Ge1O3 | A or C |
| 51.4.12 | 56700 | Cu1Ge1O3 | A or C |
| 51.4.13 | 56699 | Cu1Ge1O3 | A or C |
| 51.4.14 | 50331 | Cu1Ge1O3 | A or C |
| 51.4.15 | 50330 | Cu1Ge1O3 | A or C |
| 51.4.16 | 50329 | Cu1Ge1O3 | A or C |
| 51.4.17 | 411019 | Cu1Ge1O3 | A or C |
| 51.4.18 | 411018 | Cu1Ge1O3 | A or C |
| 51.4.19 | 411017 | Cu1Ge1O3 | A or C |
| 51.4.20 | 25754 | Cu1Ge1O3 | A or C |
| 51.4.21 | 62330 | Cu1Fe2Ge2 | AFM up to 175 K A or C |
| 51.4.24 | 62856 | Co2Cu1Ge2 | A or C |
| 51.4.25 | 410891 | Ca1In4Rh1 | A |
| 51.4.26 | 410892 | Ca1In4Ir1 | A |
| 51.4.27 | 173616 | Ba1In4Ir1 | A |
| 51.4.28 | 281204 | Ba1Fe2O5Y1 | A or C |
| 51.4.29 | 281203 | Ba1Fe2O5Y1 | A or C |
| 52.3.1 | 402064 | Br4Fe1Tl1 | A or C |
| 52.3.2 | 164987 | Bi3Sr2 | A |
| 52.3.3 | 106329 | Bi3Sr2 | A |
| 52.3.4 | 410665 | Ag2Bi1O3 | Hourglass nodal arXiv:1805.08489 |
| 53.1.1 | 71063 | Ni2Ta2Te4 | A |
| 55.2.1 | 404193 | Se2Ti9 | A but metallic |
| 55.2.3 | 97282 | O4Pb3 | A or C if charge ordered and gap ¿1eV |
| 55.2.5 | 90183 | Mn2O5Sr2 | A or C |
| 55.2.6 | 417818 | Mn2O5Sr2 | A or C |
| 55.2.7 | 417817 | Mn2O5Sr2 | A or C |
| 55.2.8 | 417816 | Mn2O5Sr2 | A or C |
| 55.2.9 | 417815 | Mn2O5Sr2 | A or C |
| 55.2.10 | 417814 | Mn2O5Sr2 | A or C |
| 55.2.11 | 417813 | Mn2O5Sr2 | A or C insulting manganite AFM |
| 55.2.12 | 417812 | Mn2O5Sr2 | A or C |
| 55.2.13 | 417811 | Mn2O5Sr2 | A or C |
| 55.2.14 | 417810 | Mn2O5Sr2 | A or C |
| 55.2.15 | 417809 | Mn2O5Sr2 | A or C |
| 55.2.16 | 417808 | Mn2O5Sr2 | A or C |
| 55.2.17 | 417807 | Mn2O5Sr2 | A or C |
| 55.2.18 | 417806 | Mn2O5Sr2 | A or C |
| 55.2.19 | 391127 | In4Rh2Zr5 | A |
| 55.2.20 | 193223 | In1La2Si2 | A |

Continued on next column

**TABLE XX– continued**

| TQC | ICSD | Stoich. Formula | SubSG |
|---|---|---|---|
| 55.2.21 | 412706 | Hf5In4Rh2 | A |
| 55.2.23 | 23226 | Ga2Mg1 | A |
| 55.2.24 | 426620 | F3O1Rb1V1 | A or C |
| 55.2.25 | 249403 | F3O1Rb1V1 | A or C |
| 55.2.30 | 171414 | Ce5In4Pt2 | A even Ce should be checked |
| 55.2.31 | 2578 | Ce4O4S3 | A or C |
| 55.2.32 | 102075 | Cd7Th6 | E or A |
| 55.2.33 | 36466 | Ca5Ga2Sb6 | A |
| 55.2.34 | 35130 | Ca2Mn2O5 | A or C |
| 55.2.36 | 84869 | C5Y4 | A |
| 55.2.37 | 23834 | C2B2Sc1 | D |
| 55.2.39 | 412563 | Bi8Ca3Pd4 | A |
| 55.2.40 | 415576 | Bi6In2Sr5 | A |
| 55.2.41 | 262861 | Bi2Fe4O9 | A or C |
| 55.2.42 | 186442 | Bi2Fe4O9 | A or C |
| 55.2.43 | 186441 | Bi2Fe4O9 | A or C |
| 55.2.46 | 419358 | B4Rh1Y1 | A |
| 55.2.49 | 425457 | B4Ni1Sc1 | A |
| 55.2.50 | 61039 | B4Mg2Ru5 | A |
| 55.2.53 | 81542 | B4Cr1Th1 | A or C and E |
| 55.2.56 | 78823 | B10Ni1Th2 | E or A |
| 55.2.57 | 78821 | B10Fe1Th2 | E or A |
| 55.2.58 | 56973 | As1Ga2Rh5 | A |
| 55.2.59 | 36364 | Al2Bi6Ca5 | A |
| 55.4.3 | 414506 | In4Rh2Sc5 | A |
| 55.4.4 | 165133 | In4Pd2Y5 | A |
| 55.4.5 | 414505 | In4Ni2Sc5 | A |
| 55.4.8 | 417906 | In3Pd5 | A |
| 55.4.9 | 10004 | Cu4Na3S4 | A or C |
| 55.4.10 | 201979 | Cl6N1Sc4 | A or C |
| 55.4.11 | 201978 | Cl6N1Sc4 | A or C |
| 55.4.14 | 417442 | B2Li2Rh3 | A |
| 55.4.15 | 201977 | B1Cl6Sc4 | A or C |
| 55.4.16 | 78822 | B10Co1Th2 | E and A or C |
| 55.4.17 | 58186 | Al7Th2 | A and E |
| 55.4.18 | 55579 | Al3Pt5 | A |
| 57.2.1 | 20933 | Fe2Hf1Si2 | A or C |
| 57.2.2 | 85640 | Cr1La1Sb3 | A or C |
| 57.2.3 | 85639 | Cr1La1Sb3 | A or C |
| 57.2.4 | 83907 | Cr1La1Sb3 | A or C |
| 57.2.5 | 247291 | Ce3Rh2Si2 | A or C |
| 57.2.6 | 76352 | Ce3Ge2Rh2 | A or C |
| 57.2.7 | 190838 | Ce1Pd2Sn3 | A or C |
| 57.2.8 | 93749 | Ca1Mn2O4 | A or C |
| 57.3.1 | 30725 | Rh2Si2Y3 | A or C |
| 57.3.2 | 188903 | Mn3O4 | A or C |
| 57.3.3 | 102123 | In1Th1 | E |
| 57.3.4 | 20939 | Ga2La3Ni2 | A |
| 57.3.5 | 402545 | Fe2Sc1Si2 | A or C |
| 58.2.1 | 42724 | Ni1Sb2 | A |
| 58.2.2 | 414575 | Ge6Li2Sr4 | A |
| 58.2.4 | 42720 | Cr1Sb2 | A or C but AFM and superintersting properties such as a super l |
| 58.2.5 | 42601 | Cr1Sb2 | or C but AFM and superintersting properties such as a super la |
| 58.2.6 | 81572 | C3Al2Th2 | E |
| 58.2.7 | 425106 | B4Cr1 | A metallic low spin |
| 58.2.8 | 186851 | B4Cr1 | A or C |

Continued on next column

**TABLE XX– continued**

| TQC | ICSD | Stoich. Formula | SubSG |
|---|---|---|---|
| 58.2.9 | 31387 | B2Ni3O6 | C magnetic with a large gap ¿ 3 eV |
| 58.2.10 | 2016 | B2Ni3O6 | A or C |
| 58.2.11 | 251066 | Au8Sn3Sr3 | A |
| 58.2.12 | 42605 | As2Ni1 | A |
| 58.3.1 | 402015 | Re4Si2Th1 | E A or C |
| 58.3.2 | 42721 | Cr1Sb2 | or C but AFM and superintersting properties such as a super la |
| 58.4.1 | 48007 | Li1O2Ru1 | A |
| 58.4.5 | 2598 | Co3O8U2 | E A or C |
| 58.4.6 | 86115 | Co1Te2 | A |
| 58.4.12 | 194807 | B4Mn1 | A bur spin polarized calculations metallic ground state with |
| 58.4.14 | 10487 | B1Pd2 | A |
| 58.4.15 | 57497 | Al1Au2 | A |
| 59.1.1 | 408080 | Pd1Si1Y1 | A |
| 59.1.3 | 425285 | Ho1I1S1 | A or C Ho! |
| 59.1.4 | 154397 | Ge5N2Sr6 | A |
| 59.1.6 | 195027 | Fe2O1Se2Sr1 | A or C |
| 59.1.7 | 195026 | Fe2O1S2Sr1 | A or C |
| 59.1.8 | 195025 | Fe2O1S2Sr1 | A or C |
| 59.1.12 | 30701 | Co8La1P5 | A Pauli para |
| 59.1.14 | 154398 | Ba6Ge5N2 | A |
| 59.1.15 | 421431 | Ba1Ge1Ni2 | A |
| 59.1.16 | 428170 | Ba1Fe2O1S2 | A or C |
| 59.1.17 | 421524 | B2Ru1 | A |
| 59.1.18 | 31871 | B2Ru1 | A |
| 59.1.19 | 421523 | B2Os1 | A |
| 59.2.1 | 187622 | Ni1Si1 | A nice band structure and non magnetic |
| 59.2.4 | 81303 | Fe1Hf8Te6 | A or C |
| 59.2.6 | 99042 | Cu2O3Sr1 | A or C |
| 59.2.8 | 413690 | Br3Mo1 | A see also Phys. Rev. Lett. 120 |
| 59.2.9 | 249645 | Au3K1Sn3 | A |
| 59.2.10 | 245679 | Au3In3Sr1 | A |
| 59.2.11 | 107445 | Au1Rb3Sn4 | B |
| 59.2.12 | 107447 | Au1Pb4Rb3 | B |
| 59.2.13 | 107448 | Au1Cs3Pb4 | B |
| 59.4.1 | 241187 | O7Sr1V3 | A or C |
| 59.4.2 | 241186 | O7Sr1V3 | A or C |
| 59.4.3 | 241185 | O7Sr1V3 | A or C |
| 59.4.4 | 241184 | O7Sr1V3 | TN=34.3 K A or C gap 2 |
| 59.4.5 | 241183 | O7Sr1V3 | TN=34.3 K A or C gap 2 |
| 59.4.6 | 241182 | O7Sr1V3 | TN=34.3 K A or C |
| 59.4.7 | 241181 | O7Sr1V3 | TN=34.3 K A or C gap 2 |
| 59.4.8 | 241180 | O7Sr1V3 | A or C |
| 59.4.9 | 241179 | O7Sr1V3 | TN=34.3 K A or C gap 2 |
| 59.4.10 | 241178 | O7Sr1V3 | TN=34.3 K A or C |
| 59.4.11 | 241177 | O7Sr1V3 | A or C V+4 d1 |
| 59.4.12 | 241176 | O7Sr1V3 | TN=34.3 K A or C gap 2 |
| 59.4.13 | 88354 | Na1O5V2 | A or C |
| 59.4.14 | 59345 | Na1O5V2 | A or C |
| 59.4.15 | 409210 | Na1O5V2 | A or C |

Continued on next column

**TABLE XX– continued**

| TQC | ICSD | Stoich. Formula | SubSG |
|---|---|---|---|
| 59.4.16 | 251182 | Na1O5V2 | A or C |
| 59.4.17 | 20825 | Mo2Ni6P3 | A |
| 59.4.18 | 88639 | Li1O5V2 | A or C |
| 59.4.19 | 107524 | In6La1Ni3 | A |
| 59.4.23 | 93966 | Cu1Te1 | A superintersting for hydrodynamics |
| 59.4.24 | 33996 | Co1Cu2O3 | A or C |
| 59.4.25 | 39314 | Cl1O1Ti1 | A or C Ti2+ d2 configuration |
| 59.4.26 | 155833 | Cl1O1Ti1 | A or C |
| 59.4.27 | 261526 | Cl1Li1N1Ti1 | A or C |
| 59.4.28 | 40963 | Cl1Fe1O1 | A or C |
| 59.4.29 | 16013 | Cl1Fe1O1 | A or C |
| 59.4.31 | 161723 | Cl1Cu3O8Se2Y1 | A or C |
| 59.4.32 | 4086 | Cl1Cr1O1 | A or C |
| 59.4.33 | 80005 | Ce3Pd6Sb5 | A but Ce |
| 59.4.35 | 69661 | Br1Cr1S1 | A or C |
| 59.4.36 | 69660 | Br1Cr1S1 | A or C |
| 59.4.37 | 69659 | Br1Cr1S1 | A or C |
| 59.4.38 | 238106 | Bi5K6Zn1 | A |
| 59.4.39 | 69385 | Bi1Cl1Cu3O8Se2 | A or C |
| 59.4.40 | 54190 | Bi1Cl1Cu3O8Se2 | A or C |
| 59.4.41 | 280759 | Bi1Br1Cu3O8Se2 | A or C |
| 59.4.42 | 54011 | B3Ge2Ni8 | A |
| 59.4.44 | 92440 | As1Pd1Zr1 | A |
| 59.4.45 | 2721 | Ag3Sn1 | A |
| 59.4.46 | 154084 | Ag3Sn1 | A |
| 60.2.1 | 415270 | O2Pb1 | A |
| 60.2.2 | 415269 | O2Pb1 | with hybrid it is n-type semiconductor with a small indirect b |
| 60.2.3 | 415268 | O2Pb1 | A |
| 60.2.4 | 246148 | C1Mo2 | A |
| 60.2.5 | 246147 | C1Mo2 | A |
| 60.2.6 | 246146 | C1Mo2 | A |
| 60.2.7 | 430298 | Au2Pb1 | A |
| 60.3.1 | 40944 | Cl3Cu1Rb1 | A or C |
| 60.3.2 | 15522 | Cl3Cu1Rb1 | A or C |
| 61.2.1 | 162404 | Pd1Sn1Te1 | A |
| 61.2.2 | 415968 | Au1Sn2 | A |
| 61.3.1 | 9922 | Ag1F2 | A or C |
| 61.3.2 | 66014 | Ag1F2 | A or C |
| 61.3.3 | 6277 | Ag1F2 | A or C |
| 62.2.1 | 418619 | Sr1Zn5 | A |
| 62.2.2 | 54980 | Sr1Zn1 | A |
| 62.2.3 | 82295 | Se2U1 | E |
| 62.2.4 | 10001 | Sb2Sr1Zn1 | A |
| 62.2.6 | 42430 | S1Te1U1 | E |
| 62.2.7 | 30398 | Ru4Si3 | A |
| 62.2.8 | 421663 | Ru2Sn2Zn3 | A |
| 62.2.9 | 421255 | Rh1Sc3Si3 | A |
| 62.2.10 | 15248 | Rh1Sc1Si2 | B |
| 62.2.12 | 412012 | Pt1Sn1Sr1 | A |
| 62.2.13 | 260792 | Pt1Si1Zr1 | A |
| 62.2.14 | 260794 | Pt1Si1Ti1 | A |
| 62.2.15 | 79238 | Pt1Si1 | A |
| 62.2.16 | 2623 | Pt1Si1 | A |
| 62.2.17 | 105785 | Pt1Sc2 | A |
| 62.2.18 | 158364 | Pd2Sn1 | A |
| 62.2.22 | 109415 | Pd1Sn1 | A |
| 62.2.23 | 79237 | Pd1Si1 | A |

Continued on next column

**TABLE XX– continued**

| TQC | ICSD | Stoich. Formula | SubSG |
|---|---|---|---|
| 62.2.24 | 15016 | Pd1Si1 | A |
| 62.2.26 | 400334 | P2Zr1 | A. We just made the crystal - |
| 62.2.27 | 417911 | P1Pt1Sc1 | A |
| 62.2.28 | 100624 | O4S1Sr1 | D ? Seems to have a large band gap |
| 62.2.29 | 415748 | O4Rb2Ru1 | A or C |
| 62.2.30 | 100625 | O4Pb1S1 | A |
| 62.2.35 | 74785 | Ni2P2Th1 | E |
| 62.2.36 | 64952 | Ni2O4Si1 | Silicate wide band A or C |
| 62.2.37 | 35675 | Ni2O4Si1 | A or C |
| 62.2.38 | 202376 | Ni2O4Si1 | Silcate A or C |
| 62.2.39 | 100645 | Ni2O4Si1 | A or C |
| 62.2.40 | 100644 | Ni2O4Si1 | A or C spinel with anti-ferromagnetic ordering transition (TN |
| 62.2.41 | 100643 | Ni2O4Si1 | A |
| 62.2.42 | 100642 | Ni2O4Si1 | A |
| 62.2.43 | 105403 | Ni1Th1 | E |
| 62.2.45 | 18188 | Ni1Si1Ti1 | A |
| 62.2.46 | 95449 | Ni1Si1 | A |
| 62.2.47 | 95448 | Ni1Si1 | A |
| 62.2.48 | 190105 | Ni1Si1 | A |
| 62.2.49 | 248408 | Ni1S3U1 | E |
| 62.2.50 | 419759 | Ni1Pd1Te1 | A |
| 62.2.51 | 50990 | Ni1P1Sc1 | A |
| 62.2.58 | 75008 | Nb1P1Zr1 | A |
| 62.2.59 | 49728 | Nb1Ni1P1 | A |
| 62.2.60 | 41149 | Mn1Rh1Si1 | A or C |
| 62.2.61 | 43399 | Mn1P1 | A or C |
| 62.2.62 | 43246 | Mn1P1 | A or C MnP exhibits complex noncollinear ferromagnetic features |
| 62.2.63 | 185418 | Mn1O3V1 | A or C |
| 62.2.68 | 426614 | Mg1Pt1Sr1 | A |
| 62.2.69 | 97765 | Li1Ni1O4P1 | Phospate A or C |
| 62.2.70 | 72929 | Li1Ni1O4P1 | A |
| 62.2.71 | 402760 | Li1Ni1O4P1 | A or C |
| 62.2.72 | 417912 | La1P1Pt1 | A |
| 62.2.73 | 415749 | K2O4Ru1 | A or C |
| 62.2.74 | 427252 | Ir1Sn1Y1 | A |
| 62.2.75 | 427253 | Ir1Si1Y1 | A |
| 62.2.76 | 411884 | Ir1Si1Ta1 | A |
| 62.2.77 | 421256 | Ir1Sc3Si3 | A |
| 62.2.78 | 420416 | Ir1Sc1Si1 | A |
| 62.2.80 | 411882 | Ir1Nb1Si1 | A |
| 62.2.81 | 246548 | Ir1Mn1Si1 | A |
| 62.2.82 | 246547 | Ir1Mn1Si1 | A |
| 62.2.84 | 240201 | In3Ir1 | A |
| 62.2.86 | 72291 | Ho2S3 | A or C check Y2S3 or La2S3 for all Lathanides |
| 62.2.87 | 99546 | Ho1O3Sc1 | A or C |
| 62.2.88 | 165196 | Ho1Mn1O3 | A or C |
| 62.2.89 | 82727 | Hf3P2 | A |
| 62.2.91 | 42912 | Ge3Pt2 | A |
| 62.2.92 | 16822 | Ge2Mo1 | A |
| 62.2.93 | 429881 | Ge2Li3Na1 | A or C |
| 62.2.94 | 280691 | Ge2In1Li1Sr2 | A |
| 62.2.95 | 44860 | Ge1Rh2 | A |
| 62.2.96 | 421092 | Ge1Rh1Ta1 | A |

Continued on next column

**TABLE XX– continued**

| TQC | ICSD | Stoich. Formula | SubSG |
|---|---|---|---|
| 62.2.97 | 260793 | Ge1Pt1Zr1 | A |
| 62.2.99 | 2624 | Ge1Pt1 | A |
| 62.2.100 | 409926 | Ge1Pd1Ti1 | A |
| 62.2.102 | 237302 | Ge1Mg1Ni1 | A |
| 62.2.103 | 428468 | Ge1Ir1Sc1 | A |
| 62.2.104 | 411883 | Ge1Ir1Nb1 | A |
| 62.2.109 | 103904 | Ga2Pd5 | A |
| 62.2.112 | 169317 | Fe1Se1 | A superconductor |
| 62.2.113 | 169316 | Fe1Se1 | A |
| 62.2.114 | 169314 | Fe1Se1 | A |
| 62.2.115 | 169312 | Fe1Se1 | A |
| 62.2.116 | 169310 | Fe1Se1 | A |
| 62.2.117 | 169308 | Fe1Se1 | A |
| 62.2.118 | 169305 | Fe1Se1 | A |
| 62.2.119 | 169282 | Fe1Se1 | A |
| 62.2.120 | 169281 | Fe1Se1 | A since non magnetic |
| 62.2.121 | 169280 | Fe1Se1 | A |
| 62.2.122 | 169279 | Fe1Se1 | A |
| 62.2.123 | 169278 | Fe1Se1 | A |
| 62.2.124 | 169277 | Fe1Se1 | A |
| 62.2.125 | 169276 | Fe1Se1 | A |
| 62.2.126 | 169275 | Fe1Se1 | A dot |
| 62.2.127 | 169274 | Fe1Se1 | A |
| 62.2.128 | 169273 | Fe1Se1 | A |
| 62.2.129 | 35010 | Fe1S1 | A or C |
| 62.2.130 | 35009 | Fe1S1 | A or C |
| 62.2.131 | 35008 | Fe1S1 | A or C |
| 62.2.132 | 49729 | Fe1P1Ta1 | A |
| 62.2.134 | 32756 | Fe1K2O4 | A or C |
| 62.2.135 | 2876 | Fe1K2O4 | A or C |
| 62.2.136 | 240891 | F2K1N1O1 | A or C should be an isolator |
| 62.2.137 | 2635 | F1Se1Y1 | A |
| 62.2.151 | 86391 | Cu1Sc1Si1 | A |
| 62.2.153 | 172848 | Cu1Ho1S2 | A or C |
| 62.2.154 | 33799 | Cs2O4Ru1 | B |
| 62.2.155 | 33861 | Cs2Fe1O4 | A or C |
| 62.2.156 | 192388 | Cr2Fe1O4 | A or C |
| 62.2.164 | 65751 | Co2O4Si1 | A or C |
| 62.2.165 | 6248 | Co2O4Si1 | other spinel gap more than 2 eV |
| 62.2.166 | 50072 | Co2O4Si1 | A or C gap more than 2 eV |
| 62.2.167 | 260092 | Co2O4Si1 | Another spinel ig gap more than 2 eV |
| 62.2.168 | 260091 | Co2O4Si1 | A or C silicate large band gap |
| 62.2.169 | 202255 | Co2O4Si1 | A or C gap more than 2 eV |
| 62.2.170 | 200705 | Co2O4Si1 | Cobalt spinel silicate gap more than 2 eV |
| 62.2.171 | 192245 | Co1Y3 | A or C |
| 62.2.172 | 192244 | Co1Y3 | A or C |
| 62.2.173 | 409847 | Co1Si1V1 | A or C |
| 62.2.174 | 49726 | Co1P1Zr1 | A |
| 62.2.175 | 43401 | Co1P1 | A |
| 62.2.176 | 43249 | Co1P1 | A |
| 62.2.180 | 2421 | Co1Mo1P1 | A or C |
| 62.2.181 | 150531 | Co1La3 | A |
| 62.2.183 | 172048 | Co1Ho1O3 | A or C because spin polarized Ce |

Continued on next column

**TABLE XX– continued**

| TQC | ICSD | Stoich. Formula | SubSG |
|---|---|---|---|
| 62.2.186 | 157332 | Ce3In2Ru2 | A or C |
| 62.2.187 | 428424 | Ce2Te3 | A or C |
| 62.2.189 | 418546 | Ce1Pt1Zn1 | A |
| 62.2.190 | 420207 | Ce1Pd1Zn1 | A or C |
| 62.2.191 | 409972 | Ce1P2Pt4 | A |
| 62.2.192 | 54733 | Ce1In4Pd2 | A or C because Ce |
| 62.2.193 | 99240 | Ce1In2Pd3 | A |
| 62.2.194 | 150105 | Ce1Cu6 | A or C |
| 62.2.195 | 237728 | Cd1Pt1Sr1 | A |
| 62.2.196 | 12159 | Cd1Na1Sb1 | A |
| 62.2.197 | 72642 | Ca1Pt1Si1 | A |
| 62.2.198 | 426616 | Ca1Mg1Pd1 | A |
| 62.2.199 | 410841 | Ca1In1Rh1 | A |
| 62.2.200 | 106343 | Ca1Ge1Pt1 | A |
| 62.2.201 | 291004 | Ca1Fe2O4 | A or C |
| 62.2.202 | 426610 | Ca1Cd1Pt1 | A |
| 62.2.203 | 426611 | Ca1Cd1Pd1 | A |
| 62.2.204 | 68116 | C2U1W1 | E and C |
| 62.2.205 | 16549 | C2Mo1U1 | E |
| 62.2.206 | 151477 | C2Cr3 | A or C |
| 62.2.207 | 15086 | C2Cr3 | E |
| 62.2.208 | 247128 | C1H2Cl1Cu1N2 | looks like an insulator A or C |
| 62.2.209 | 236383 | Bi3Co1 | A |
| 62.2.210 | 193272 | Bi3Co1 | A |
| 62.2.212 | 52698 | Ba2Si1 | A |
| 62.2.213 | 416561 | Ba2Pb1 | A |
| 62.2.215 | 424857 | Ba1Hg6 | A |
| 62.2.216 | 237304 | Ba1Ge1Ni1 | A |
| 62.2.217 | 424315 | Ba1Fe2Se3 | C AFM |
| 62.2.218 | 424314 | Ba1Fe2Se3 | C BaFe2S4 are linear chain antiferromagnets exhibiting semico |
| 62.2.219 | 290595 | Ba1Fe2Se3 | C BaFe2S4 are linear chain antiferromagnets exhibiting semicon |
| 62.2.220 | 185512 | Ba1Fe2Se3 | ABaFe2S4 are linear chain antiferromagnets exhibiting semicond |
| 62.2.221 | 41893 | B2Co1Nb1 | A |
| 62.2.223 | 42894 | B1Co1Mo1 | A evt or C |
| 62.2.224 | 50842 | B1Ca2F1N2 | A or C insulator |
| 62.2.225 | 263057 | Au3Ga2Rb1 | A |
| 62.2.226 | 263058 | Au3Cs1Ga2 | B |
| 62.2.227 | 418239 | Au3Ca3In1 | A |
| 62.2.228 | 249563 | Au2In2Sr1 | A |
| 62.2.229 | 249562 | Au2Ba1In2 | A |
| 62.2.230 | 391422 | Au1In1Sr1 | A |
| 62.2.233 | 408579 | Au1Ca1In1 | A |
| 62.2.234 | 106273 | Au1Ca1Ga1 | A |
| 62.2.235 | 168665 | As2Zr1 | A |
| 62.2.236 | 42446 | As1V1 | V3+ d2 A or C |
| 62.2.237 | 51201 | As1Li1Ni1O4 | A or C |
| 62.2.238 | 245184 | As1Li1Ni1O4 | A or C |
| 62.2.239 | 412296 | As1Cu1Mg1 | A |
| 62.2.240 | 406953 | As1Co1Hf1 | A |
| 62.2.241 | 48030 | As1Co1 | A |
| 62.2.242 | 48029 | As1Co1 | A |
| 62.2.243 | 48028 | As1Co1 | A or C |
| 62.2.244 | 48026 | As1Co1 | A |
| 62.2.245 | 9571 | As1Cd1Na1 | A |

Continued on next column

| TQC | ICSD | Stoich. Formula | SubSG |
|---|---|---|---|
| 62.2.248 | 160051 | Al1Ce1Ru1 | A or C |
| 62.2.249 | 55623 | Al1Ce1Rh1 | A or C |
| 62.2.250 | 160052 | Al1Ce1Rh1 | A or C Ce |
| 62.2.252 | 370015 | Al1Au1Ca1 | A |
| 62.3.5 | 421254 | Ru1Sc3Si3 | A |
| 62.3.6 | 420417 | Ru1Sc1Si1 | A |
| 62.3.7 | 24354 | Rh3Si4 | A |
| 62.3.8 | 42960 | Rh2Sb1 | A |
| 62.3.10 | 202438 | Rh1S3U1 | E |
| 62.3.11 | 8016 | Pt1Y3 | A |
| 62.3.12 | 417531 | Pt1Sn1Y1 | A |
| 62.3.13 | 421257 | Pt1Sc3Si3 | A |
| 62.3.14 | 79597 | Pt1Sc1Si1 | A |
| 62.3.15 | 35087 | Pd2Si1Y1 | A |
| 62.3.16 | 183334 | Pd1Y1Zn1 | A |
| 62.3.17 | 424012 | Pd1Sn1Y1 | A |
| 62.3.18 | 420418 | Pd1Sc1Si1 | A |
| 62.3.19 | 92441 | Pd1Sb1Zr1 | A |
| 62.3.21 | 42056 | P1W1 | A |
| 62.3.22 | 77856 | P1V2 | A |
| 62.3.23 | 30727 | P1Ru1Zr1 | A |
| 62.3.24 | 48123 | P1Re2 | A |
| 62.3.25 | 186202 | P1Re2 | A |
| 62.3.31 | 183450 | O3Sr1Tc1 | E |
| 62.3.32 | 66346 | Ni2Ta1Te2 | A |
| 62.3.33 | 86174 | Ni2Se1Ta1Te1 | A |
| 62.3.35 | 105379 | Ni1Sn1Y1 | A |
| 62.3.37 | 79598 | Ni1Si1Y1 | A |
| 62.3.38 | 418977 | Ni1Sc1Sn1 | A |
| 62.3.39 | 105338 | Ni1Sc1Sn1 | A |
| 62.3.40 | 41800 | Ni1Sc1Si1 | A |
| 62.3.41 | 408195 | Ni1Sb1Zr1 | A |
| 62.3.42 | 427254 | Ni1Pb1Y1 | A |
| 62.3.44 | 93507 | Ni1O3Tl1 | A or C |
| 62.3.46 | 23281 | Nb2P5 | A |
| 62.3.47 | 68103 | Nb1Pd1Te5 | C MOTT and magnetic |
| 62.3.48 | 416177 | Nb1Pd1Si1 | A |
| 62.3.49 | 88723 | Nb1O3Sr1 | insulator |
| 62.3.50 | 20633 | Nb1Ni1P2 | A |
| 62.3.51 | 100824 | Nb1Ni1P2 | A |
| 62.3.52 | 202906 | Na1O4Ti2 | A or C |
| 62.3.53 | 172608 | Na1O4Ru2 | A |
| 62.3.54 | 170598 | Na1O4Rh2 | A |
| 62.3.58 | 55622 | Mn3Sn2 | A or C |
| 62.3.59 | 150139 | Mn3Sn2 | A or C ferrimagnetic metal spin polarized calculations is neces |
| 62.3.61 | 85947 | Mn1P1Ta1 | A or C |
| 62.3.63 | 78331 | Mn1O2 | A or C |
| 62.3.65 | 68280 | Mn1Nb1P1 | A or C |
| 62.3.66 | 74784 | Mg1Ni2P1 | A |
| 62.3.67 | 189097 | Li1Mn1O4P1 | A or C |
| 62.3.68 | 427930 | La2Ni1Sb1 | A |
| 62.3.69 | 423625 | La1Si1 | A |
| 62.3.71 | 99274 | La1Pt1Sn1 | A |
| 62.3.72 | 416896 | La1Pt1Sn1 | A |
| 62.3.73 | 183335 | La1Pd1Zn1 | A |
| 62.3.74 | 422484 | La1Pd1Sn1 | A |
| 62.3.76 | 98414 | La1O3Ti1 | A or C TI3+ |
| 62.3.78 | 63575 | La1O3Ti1 | A or C |
| 62.3.79 | 63574 | La1O3Ti1 | Ti3+ d1 A or C |

Continued on next column

**TABLE XX– continued**

| TQC | ICSD | Stoich. Formula | SubSG |
|---|---|---|---|
| 62.3.80 | 63573 | La1O3Ti1 | A |
| 62.3.82 | 63571 | La1O3Ti1 | A or C TI3+ D1 |
| 62.3.84 | 42731 | Ir3Si4 | A |
| 62.3.86 | 79236 | Ir1Si1 | A |
| 62.3.87 | 410704 | In1Pt1Sr1 | A |
| 62.3.88 | 417907 | In1Pd2 | A |
| 62.3.92 | 24876 | Ho1Ni1O3 | A or C |
| 62.3.93 | 427001 | Hg6K1 | A |
| 62.3.94 | 68790 | Hf1Rh1Si1 | A |
| 62.3.95 | 237382 | H2Cl2Co1O1 | A or C |
| 62.3.96 | 245057 | H1Fe1O2 | A or C |
| 62.3.97 | 41804 | H1Cu1I1O4 | A or C looks like an insulator |
| 62.3.98 | 429637 | Ge2Rh4Y1 | A |
| 62.3.100 | 162582 | Ge2Li1Sr1 | A |
| 62.3.101 | 20520 | Ge2La3Ni1 | A or C |
| 62.3.104 | 44861 | Ge1Rh1 | A |
| 62.3.105 | 428474 | Ge1Pt1Sc1 | A |
| 62.3.108 | 427255 | Ge1Ni1Y1 | A |
| 62.3.109 | 428471 | Ge1Ni1Sc1 | A |
| 62.3.110 | 186940 | Ge1Nb1Pt1 | A |
| 62.3.111 | 425590 | Ge1Mn1U1 | E |
| 62.3.114 | 189328 | Ge1La1Pd2 | A |
| 62.3.115 | 423624 | Ge1La1 | A |
| 62.3.117 | 57070 | Ge1Ho1Ru1 | A or C |
| 62.3.122 | 409939 | Ga1Pd2 | A |
| 62.3.124 | 84203 | Fe1Sc1Si1 | A |
| 62.3.125 | 76346 | Fe1Sc1Si1 | A or C |
| 62.3.127 | 43400 | Fe1P1 | A or C |
| 62.3.128 | 43248 | Fe1P1 | A or C |
| 62.3.129 | 92199 | Fe1O4P1 | A or C |
| 62.3.132 | 43260 | Fe1O3Y1 | A or C |
| 62.3.134 | 171654 | F6Ir1 | A |
| 62.3.135 | 16872 | F5O1Os1 | Os7+ d1 |
| 62.3.163 | 107269 | Cu5Sn1Th1 | E |
| 62.3.164 | 71683 | Cu2Li1O2 | A or C |
| 62.3.165 | 71221 | Cu2Li1O2 | A or C Cu1 |
| 62.3.167 | 190123 | Cu1Sn1Zr1 | A |
| 62.3.168 | 97747 | Cu1O3Se1 | A or C skyrmions insulator |
| 62.3.170 | 498 | Cu1O3Se1 | A or C |
| 62.3.171 | 67531 | Cu1Fe2S3 | C AFM |
| 62.3.172 | 67530 | Cu1Fe2S3 | C AFM |
| 62.3.173 | 67529 | Cu1Fe2S3 | C AFM |
| 62.3.174 | 42104 | Cu1Fe2S3 | C AFM |
| 62.3.175 | 14255 | Cu1Fe2S3 | C AFM |
| 62.3.179 | 43247 | Cr1P1 | A or C |
| 62.3.182 | 94380 | Co2P1 | A or C ferrimagnetic metal spin polarized calculations is neces |
| 62.3.183 | 94379 | Co2P1 | A or C magnetic spin polarized |
| 62.3.185 | 49727 | Co1Nb1P1 | A or C |
| 62.3.186 | 87422 | Co1Li1O4P1 | A or C |
| 62.3.187 | 400625 | Co1Li1O4P1 | C multiferroic strongly anisotropic AFM LiCoPO4(TN=21.7 K) |
| 62.3.188 | 291401 | Co1Li1O4P1 | AC multiferroic strongly anisotropic AFM LiCoPO4(TN=21.7 K) |

Continued on next column

**TABLE XX– continued**

| TQC | ICSD | Stoich. Formula | SubSG |
|---|---|---|---|
| 62.3.189 | 247498 | Co1Li1O4P1 | C multiferroic strongly anisotropic AFM LiCoPO4(TN=21.7 K) |
| 62.3.190 | 247497 | Co1Li1O4P1 | C multiferroic strongly anisotropic AFM LiCoPO4(TN=21.7 K) Oliv |
| 62.3.191 | 247496 | Co1Li1O4P1 | A or C |
| 62.3.192 | 173773 | Co1Li1O4P1 | C multiferroic strongly anisotropic AFM LiCoPO4(TN=21.7 K) |
| 62.3.195 | 72321 | Co1F3Na1 | A or C magnetic |
| 62.3.196 | 87902 | Co1Cs2I4 | A or C |
| 62.3.197 | 87901 | Co1Cs2I4 | A or C |
| 62.3.198 | 87898 | Co1Cs2I4 | D |
| 62.3.199 | 87897 | Co1Cs2I4 | A or C |
| 62.3.200 | 9468 | Cl4Cs2Cu1 | A or C |
| 62.3.201 | 91188 | Cl4Cs2Cu1 | A or C |
| 62.3.202 | 91187 | Cl4Cs2Cu1 | A or C |
| 62.3.203 | 91186 | Cl4Cs2Cu1 | A or C |
| 62.3.204 | 15699 | Cl4Cs2Cu1 | A or C |
| 62.3.213 | 99834 | Ce1O3V1 | A or C |
| 62.3.214 | 185832 | Ce1O3V1 | A or C |
| 62.3.215 | 165721 | Ce1O3V1 | A or C |
| 62.3.216 | 162748 | Ce1O3V1 | A or C |
| 62.3.218 | 262201 | Ce1Ge2Li1 | A or C because Ce |
| 62.3.219 | 414509 | Ce1Ge1Ir1 | A |
| 62.3.221 | 150269 | Ce1Ga1Ni1 | A or C |
| 62.3.222 | 426281 | Ce1Cl1Te1 | A or C |
| 62.3.223 | 180009 | Cd1O3Tc1 | E |
| 62.3.225 | 156887 | Ca5Ir1 | A |
| 62.3.226 | 413803 | Ca2In1 | A |
| 62.3.227 | 60830 | Ca1Pt1Sb1 | A |
| 62.3.230 | 261342 | Ca1O3Tc1 | E |
| 62.3.231 | 261341 | Ca1O3Tc1 | A or C |
| 62.3.232 | 164775 | Ca1O3Rh1 | A |
| 62.3.233 | 85569 | Ca1Nb1O3 | A or C Nb4+ d1 |
| 62.3.235 | 165699 | Ca1Mn1O3 | A or C |
| 62.3.237 | 161466 | Ca1Mn1O3 | A or C |
| 62.3.238 | 161465 | Ca1Mn1O3 | A or C |
| 62.3.240 | 182245 | Ca1Ir1O3 | C MOTT and magnetic |
| 62.3.241 | 106348 | Ca1In1Pt1 | A |
| 62.3.242 | 106340 | Ca1Ga1Pt1 | A |
| 62.3.243 | 23742 | C2U1V1 | E |
| 62.3.244 | 69098 | C2Re1Y2 | A |
| 62.3.247 | 26024 | Br4Cs2Cu1 | A or C |
| 62.3.248 | 10185 | Br4Cs2Cu1 | B |
| 62.3.250 | 58821 | Bi3Ni1 | A |
| 62.3.251 | 391336 | Bi3Ni1 | A |
| 62.3.255 | 172156 | Bi1Cl1Mn1S2 | A or C |
| 62.3.257 | 261714 | Ba2In1 | A |
| 62.3.258 | 160295 | Ba2Co1S3 | A or C |
| 62.3.259 | 160294 | Ba2Co1S3 | A or C |
| 62.3.260 | 100435 | B2Ni1Ta1 | A |
| 62.3.262 | 247263 | B1Pd4Sr1 | A |
| 62.3.263 | 391331 | B1Fe1 | A or C |
| 62.3.264 | 391329 | B1Fe1 | A magnetic spin polarized |
| 62.3.265 | 247264 | B1Ba1Pd4 | A |
| 62.3.266 | 249529 | Au2In4La1 | A |
| 62.3.267 | 237467 | Au2Ga4Na1 | A |

Continued on next column

**TABLE XX– continued**

| TQC | ICSD | Stoich. Formula | SubSG |
|---|---|---|---|
| 62.3.269 | 262044 | Au1Y2 | A |
| 62.3.274 | 426612 | Au1Cd1Sr1 | A |
| 62.3.275 | 58401 | Au1Ca3 | A |
| 62.3.276 | 420574 | Au1Ca1Cd1 | A |
| 62.3.277 | 419559 | Au1Ba1 | A |
| 62.3.278 | 38412 | As2Nb1Ni1 | A |
| 62.3.279 | 48036 | As1Fe1 | A or C |
| 62.3.280 | 48035 | As1Fe1 | A or C |
| 62.3.281 | 48034 | As1Fe1 | A or C |
| 62.3.282 | 48033 | As1Fe1 | A or C |
| 62.3.283 | 48031 | As1Fe1 | A or C spin polarized band structure |
| 62.3.287 | 195014 | As1Cr1 | Superconductivity in the vicinity of anti-ferromagnetic order in |
| 62.3.288 | 195013 | As1Cr1 | Superconductivity in the vicinity of anti-ferromagnetic order in |
| 62.3.289 | 245183 | As1Co1Li1O4 | C Olivine older calcuation LDA +U |
| 62.3.290 | 155305 | As1Co1Li1O4 | A orC multi-ferroic strongly anisotropic AFM LiCoPO4(TN=21.7 K) |
| 62.3.292 | 245328 | Al1Pd5 | A |
| 62.3.294 | 78969 | Al1Ge3Y2 | A |
| 62.3.303 | 150172 | Al1Ce1Pt1 | A or C Ce |
| 62.3.304 | 104635 | Al1Ce1Pt1 | A or C |
| 62.3.305 | 261896 | Al1Ca1Pt1 | A |
| 62.3.306 | 57496 | Al1Au2 | A |
| 62.3.307 | 57342 | Ag1Ba1 | A |
| 62.3.308 | 260623 | Ag1Ba1 | A |
| 63.1.1 | 68416 | Te3U1 | E spin polarized |
| 63.1.2 | 106074 | Sn1Sr1 | A |
| 63.1.3 | 416971 | Si6Sr1 | A |
| 63.1.4 | 240779 | Si6Sr1 | A |
| 63.1.5 | 78997 | Si1Sr1 | A |
| 63.1.6 | 87915 | Pd3Te2 | A |
| 63.1.7 | 77902 | Pd3Te2 | A |
| 63.1.11 | 96149 | Np1Si1 | E |
| 63.1.12 | 86102 | Mg1O5V2 | A |
| 63.1.13 | 50981 | Mg1O5V2 | E |
| 63.1.14 | 50979 | Mg1O5V2 | E |
| 63.1.15 | 250752 | La2Sn1Zn5 | A |
| 63.1.16 | 250756 | La2Pb1Zn5 | A |
| 63.1.17 | 423611 | La2Li2Si3 | A |
| 63.1.18 | 410732 | La1Rh1Sn2 | A |
| 63.1.19 | 38311 | La1Ni5P3 | A |
| 63.1.20 | 165363 | In5Y3 | A |
| 63.1.21 | 165362 | In5La3 | A |
| 63.1.22 | 418178 | In4Ni1Sr1 | A |
| 63.1.26 | 412431 | In2La1Rh1 | A |
| 63.1.27 | 414479 | In2Ir1La1 | A |
| 63.1.28 | 200067 | Hf3Ni2Si3 | A |
| 63.1.29 | 26225 | H5Li4Rh1 | A |
| 63.1.30 | 262431 | Ge3La2Li2 | A |
| 63.1.35 | 189439 | Fe2O3 | A |
| 63.1.52 | 95028 | Cu1Ho2K1Se4 | C AFM ? |
| 63.1.53 | 97561 | Cu1Ho2K1S4 | E |

Continued on next column

**TABLE XX– continued**

| TQC | ICSD | Stoich. Formula | SubSG |
|---|---|---|---|
| 63.1.55 | 429089 | Cs1Sc1Se5U1 | E spin polarized |
| 63.1.57 | 413677 | Cs1Cu3Ho2Se5 | spin polarized |
| 63.1.60 | 236871 | Cr2Mg1O4 | A |
| 63.1.61 | 20952 | Co3Ga9Ho2 | spin polarized |
| 63.1.63 | 250753 | Ce2Sn1Zn5 | A |
| 63.1.64 | 250757 | Ce2Pb1Zn5 | A |
| 63.1.65 | 63036 | Ce2Ge3Li2 | spin polarized |
| 63.1.66 | 262430 | Ce2Ge3Li2 | spin polarized |
| 63.1.68 | 421810 | Cd2F1Sb5Sr5 | A |
| 63.1.69 | 424107 | Ca2Sn6Zn3 | A |
| 63.1.70 | 96228 | Ca2In1N1 | A |
| 63.1.71 | 165193 | Ca1Sn1 | A |
| 63.1.72 | 410774 | Ca1Pd1Sn2 | A |
| 63.1.73 | 408880 | Ca1In2Pt1 | A |
| 63.1.74 | 408881 | Ca1In2Pd1 | A |
| 63.1.75 | 16356 | Ca1Fe5O7 | spin polarized |
| 63.1.76 | 75133 | C1Fe2Si1Th1 | E |
| 63.1.77 | 421809 | Ba5Cd2F1Sb5 | A |
| 63.1.78 | 411979 | Ba3Pb5 | A |
| 63.1.79 | 165184 | Ba3Pb5 | A |
| 63.1.80 | 710014 | Ba1Si6 | A |
| 63.1.81 | 245295 | Ba1Si6 | A |
| 63.1.82 | 78998 | Ba1Si1 | A |
| 63.1.83 | 427286 | Ba1Ru3Sn6 | A |
| 63.1.84 | 58664 | Ba1Pb1 | A |
| 63.1.85 | 251583 | Ba1Ge6 | A |
| 63.1.86 | 290596 | Ba1Fe2Se3 | C AFM |
| 63.1.87 | 290594 | Ba1Fe2Se3 | C AFM spin polarized |
| 63.1.88 | 16307 | Ba1Fe2S3 | BaFe2S4 are linear chain antiferromagnets exhibiting semiconduc |
| 63.1.89 | 64596 | B7Fe1Y3 | A |
| 63.1.90 | 91243 | B4Mg1Os3 | A |
| 63.1.92 | 418162 | B2Cr2Ir1 | A |
| 63.1.95 | 104653 | Al9Co3Y2 | A |
| 63.1.96 | 58157 | Al6Ru1 | A |
| 63.1.98 | 423614 | Al2Ge1La2 | A |
| 63.1.99 | 99138 | Al1Si1Y1 | A |
| 63.1.101 | 76295 | Al1Ge1Y1 | A |
| 63.2.1 | 43419 | U1 | E |
| 63.2.2 | 43339 | U1 | E Uranium spin polarized |
| 63.2.3 | 411981 | Sn5Sr3 | A |
| 63.2.4 | 165183 | Sn5Sr3 | A |
| 63.2.5 | 54756 | Sn4Sr1 | A |
| 63.2.6 | 281669 | Sn4Sr1 | A |
| 63.2.8 | 95179 | Si2Zr1 | A |
| 63.2.9 | 20284 | Si1Zr1 | A |
| 63.2.11 | 163454 | Rh1Zr1 | A |
| 63.2.12 | 55546 | Pd1Zr1 | A |
| 63.2.13 | 165577 | Pd1Sr1Tl2 | A |
| 63.2.15 | 96322 | Ni1Zr1 | A |
| 63.2.16 | 105478 | Ni1Zr1 | A |
| 63.2.17 | 71964 | Ni1O1Zr3 | E spin polarized |
| 63.2.22 | 425493 | Mg2Pd1Sr1 | A |
| 63.2.23 | 163453 | Ir1Zr1 | A |
| 63.2.25 | 418180 | In4Pt1Sr1 | A |
| 63.2.27 | 418179 | In4Pd1Sr1 | A |
| 63.2.28 | 415937 | In4Pb1Sr3 | A |
| 63.2.29 | 410400 | In2Pt1Sr1 | A |
| 63.2.30 | 410402 | In2Pd1Sr1 | A |

Continued on next column

**TABLE XX– continued**

| TQC | ICSD | Stoich. Formula | SubSG |
|---|---|---|---|
| 63.2.31 | 190810 | Hf2Pt3 | A |
| 63.2.32 | 15051 | Hf1Te5 | A |
| 63.2.35 | 171232 | Ge6La2Zn3 | A |
| 63.2.36 | 56047 | Ge2Th1 | E |
| 63.2.37 | 240343 | Ge2Ni1Sr1 | A |
| 63.2.38 | 240728 | Ge1Na2Zn1 | A |
| 63.2.47 | 423623 | Ga1La1 | A |
| 63.2.48 | 2795 | Ga1 | A |
| 63.2.49 | 290249 | Fe4O5 | C spin polarized |
| 63.2.50 | 98 | Fe1S3U1 | E |
| 63.2.51 | 261040 | Fe1S3U1 | E magnetic |
| 63.2.52 | 261039 | Fe1S3U1 | E spin polarized |
| 63.2.53 | 261038 | Fe1S3U1 | E |
| 63.2.65 | 193548 | Cr2Mg1O4 | C spin polarized |
| 63.2.66 | 192389 | Cr2Fe1O4 | A |
| 63.2.67 | 20951 | Co3Ga9Y2 | C spin polarized |
| 63.2.69 | 246169 | Ce1Pt1Si2 | A |
| 63.2.70 | 42711 | Ce1Ni1Si2 | A |
| 63.2.72 | 157363 | Ce1In4Pd1 | A |
| 63.2.73 | 150132 | Ce1In2Pd1 | A |
| 63.2.74 | 106394 | Ce1Ge2Ni1 | A |
| 63.2.75 | 425494 | Cd2Pd1Sr1 | A |
| 63.2.77 | 421432 | Ca4Ge3Ni4 | A |
| 63.2.78 | 58948 | Ca3Zn1 | A |
| 63.2.79 | 78996 | Ca1Si1 | A |
| 63.2.80 | 190659 | Ca1Si1 | A |
| 63.2.81 | 410890 | Ca1In4Pd1 | A |
| 63.2.82 | 106347 | Ca1In4Ni1 | A |
| 63.2.83 | 240342 | Ca1Ge2Ni1 | A |
| 63.2.84 | 710075 | Ca1Ga2Ni1 | A |
| 63.2.85 | 16355 | Ca1Fe4O6 | spin polarized |
| 63.2.90 | 411622 | C1B3Nb3 | A |
| 63.2.91 | 107610 | Be1Pd3 | A |
| 63.2.92 | 411980 | Ba3Sn5 | A |
| 63.2.93 | 167166 | Ba3Sn5 | A |
| 63.2.94 | 167163 | Ba3Ge5 | A |
| 63.2.95 | 429723 | Ba2Ge4Ir4Mg1 | A |
| 63.2.96 | 418610 | Ba1Zn5 | A |
| 63.2.97 | 405111 | Ba1Pd1Sb2 | A |
| 63.2.98 | 14110 | Ba1Ni2Si2 | A |
| 63.2.99 | 411228 | Ba1In2Pt1 | A |
| 63.2.100 | 414228 | Ba1In1 | A |
| 63.2.101 | 429115 | Ba1Cd2Pt1 | A |
| 63.2.102 | 39663 | B7W1Y3 | A |
| 63.2.103 | 62282 | B7Co4Nb3 | A |
| 63.2.104 | 43662 | B1Re3 | A |
| 63.2.105 | 107505 | Au1In2Na1 | A |
| 63.2.108 | 290262 | As4Cd5Rb2 | A |
| 63.2.109 | 15032 | As3Nb4 | A |
| 63.2.110 | 405110 | As2Ba1Pd1 | A |
| 63.2.111 | 153384 | Al2Ba1Si2 | A |
| 63.2.113 | 54470 | Al1La1 | A |
| 63.2.114 | 251809 | Al1B1W1 | A |
| 63.2.115 | 251808 | Al1B1Mo1 | A |
| 63.2.116 | 182114 | Al10Ru2U1 | A |
| 63.2.117 | 99179 | Al10Fe2U1 | E |
| 63.2.118 | 152354 | Al10Fe2Th1 | E spin polarized |
| 63.2.119 | 59912 | Al10Ce1Ru2 | A |
| 63.4.1 | 422877 | Sr3Tl5 | A |
| 63.4.2 | 380325 | Sr3Tl5 | A |
| 63.4.3 | 416771 | Sn2Y1 | A |

Continued on next column

**TABLE XX– continued**

| TQC | ICSD | Stoich. Formula | SubSG |
|---|---|---|---|
| 63.4.5 | 63105 | Rh5Zr3 | A |
| 63.4.6 | 62932 | Rh5Zr3 | A |
| 63.4.7 | 14382 | Rh3Te2 | A |
| 63.4.8 | 420355 | Pt8Sc4Si4 | A |
| 63.4.9 | 71292 | Pt1Ta1Te5 | A |
| 63.4.10 | 424600 | Pt1Sc1Zn1 | A |
| 63.4.12 | 35086 | Pd2Si3Y3 | A |
| 63.4.13 | 424601 | Pd1Sc1Zn1 | A |
| 63.4.14 | 36630 | P5Pd5Sr4 | A |
| 63.4.17 | 36520 | O4P1Ti1 | C spin polarized |
| 63.4.18 | 83361 | O14Si4Sr2V2 | Silicate wide band gap |
| 63.4.19 | 33914 | Ni5P3Sr1 | A |
| 63.4.20 | 65760 | Ni2Sc3Si3 | A |
| 63.4.21 | 73314 | Ni1Ta1Te5 | A |
| 63.4.24 | 73316 | Nb1Ni1Te5 | A |
| 63.4.25 | 105166 | Na9Sn4 | A |
| 63.4.26 | 237248 | Na1Si6 | A |
| 63.4.27 | 84911 | N5Na1Ta3 | A |
| 63.4.32 | 50982 | Li1O5V2 | C probably MOTT |
| 63.4.33 | 50980 | Li1O5V2 | C spin polarized |
| 63.4.34 | 201974 | Li1O5V2 | C spin polarized |
| 63.4.35 | 281581 | Li1Mn1O4V1 | spin polarized |
| 63.4.36 | 165185 | La3Sn5 | A |
| 63.4.37 | 423418 | La3Sn4 | A |
| 63.4.39 | 416818 | La3Ni3Zn1 | A |
| 63.4.40 | 91142 | La3Ni2O7 | spin polarized |
| 63.4.41 | 91141 | La3Ni2O7 | C spin polarized |
| 63.4.42 | 262792 | La1Zn4 | A |
| 63.4.43 | 423421 | La1Sn1 | A |
| 63.4.44 | 413173 | La1Se1Te2 | A |
| 63.4.45 | 413171 | La1Se1Te2 | A |
| 63.4.46 | 158253 | La1Mg2Pd1 | A |
| 63.4.47 | 96152 | La1Mg2Ni1 | A |
| 63.4.48 | 64625 | K4P3 | A |
| 63.4.50 | 14157 | K3Ni2O4 | C spin polarized |
| 63.4.51 | 165188 | In7K2Sr1 | A |
| 63.4.52 | 411746 | In4La1Ni1 | A |
| 63.4.54 | 413459 | In2Ni1Y1 | A |
| 63.4.55 | 413448 | In2La1Pd1 | A |
| 63.4.56 | 410401 | In2Ir1Sr1 | A |
| 63.4.59 | 154700 | Ho1Si1 | C spin pol |
| 63.4.60 | 83917 | Ho1Ni1Si2 | C magnetic spin polarized |
| 63.4.61 | 83916 | Ho1Ni1Si2 | E |
| 63.4.62 | 107591 | Hf1Sb1 | A |
| 63.4.63 | 180189 | H6Cl2Cu1N2 | A |
| 63.4.64 | 23318 | H2Cl5Cs2Fe1O1 | C magnetic with large band gap |
| 63.4.65 | 409766 | H2Cl5Cr1Cs2O1 | C magnetic with large band gap |
| 63.4.66 | 88105 | Ge4Y3 | A |
| 63.4.67 | 88075 | Ge3Y1 | A |
| 63.4.69 | 425769 | Ge3Ru2Y3 | A |
| 63.4.73 | 57071 | Ge3Ho3Ru2 | C Ho |
| 63.4.75 | 82875 | Ge3Ho3Ir2 | C AFM ? |
| 63.4.78 | 85600 | Ge2La1Ni1 | A |
| 63.4.81 | 53647 | Ge1Ho1 | C spin pol |
| 63.4.85 | 423074 | Ga4Ni1Y1 | A |
| 63.4.88 | 423079 | Ga4Ho1Ni1 | A |
| 63.4.89 | 192026 | Ga3Pd1Sr1 | A |
| 63.4.90 | 103895 | Ga2Ni1Y1 | C MOTT and magnetic |

Continued on next column

**TABLE XX– continued**

| TQC | ICSD | Stoich. Formula | SubSG |
|---|---|---|---|
| 63.4.91 | 413557 | Ga2Ir1 | A |
| 63.4.92 | 100051 | Ga1Li2 | A |
| 63.4.93 | 98966 | Fe2Te2Y5 | C spin polarized |
| 63.4.94 | 166049 | Fe2S3Tl1 | C AFM? |
| 63.4.95 | 81548 | Fe2Rb1Te3 | C magnetic spin polarized |
| 63.4.96 | 81547 | Fe2Rb1Se3 | C magnetic |
| 63.4.98 | 424672 | Fe2K1Se3 | C AFM |
| 63.4.99 | 424671 | Fe2K1Se3 | C stripe antiferromagnetic order in KFe2Se3 |
| 63.4.100 | 99507 | Fe2K1S3 | C AFM? |
| 63.4.101 | 100180 | Fe2K1S3 | C AFM ? |
| 63.4.118 | 71204 | Cu5K2Te5 | C spin polarized |
| 63.4.119 | 54129 | Cu4O8 | A |
| 63.4.121 | 380487 | Cu2K3O4 | C spin polarized |
| 63.4.122 | 82330 | Cu1Se1 | A or C |
| 63.4.123 | 83051 | Cu1O2Sr1 | C spin polarized |
| 63.4.124 | 77293 | Cu1O2Sr1 | C MOTT? Mit gap |
| 63.4.125 | 77291 | Cu1O2Sr1 | C spin polarized |
| 63.4.126 | 202992 | Cu1O2Sr1 | C spin polarized |
| 63.4.127 | 150913 | Cu1O2 | C spin polarized |
| 63.4.130 | 84200 | Cu1Ge4Zr2 | A |
| 63.4.131 | 54101 | Cs2Cu5Se4 | C magnetic with large band gap |
| 63.4.134 | 187535 | Cs1Fe2Se3 | C magnetic spin polarized |
| 63.4.135 | 99504 | Cs1Fe2S3 | C magnetic spin polarized |
| 63.4.141 | 183216 | Cr1Cu1O4 | C magnetic with large band gap |
| 63.4.142 | 102490 | Co5Ga1Y5 | magnetic |
| 63.4.144 | 41745 | Co2Sc3Si3 | A |
| 63.4.145 | 106479 | Co2Ge3Sc3 | C MOTT and magnetic |
| 63.4.147 | 55670 | Co1Ga4Th1 | E |
| 63.4.148 | 417507 | Co1Cs1Ho1Se3 | C magnetic |
| 63.4.149 | 169508 | Ce1Rh1Si2 | C Ce |
| 63.4.150 | 106411 | Ce1Mn1Si2 | spin polarized Ce and Mn |
| 63.4.151 | 106410 | Ce1Mn1Si2 | magnetic C spin pol |
| 63.4.152 | 102212 | Ce1Li1Sn2 | A |
| 63.4.153 | 414474 | Ce1In2Ir1 | A |
| 63.4.154 | 427867 | Ce1Ga3Pd1 | spin polarized |
| 63.4.155 | 16501 | Ce1Co1Si2 | C Ce an dCo |
| 63.4.156 | 417029 | Cd2La1Pd1 | A |
| 63.4.161 | 79095 | Ca4P5Pd5 | A |
| 63.4.162 | 380324 | Ca3Tl5 | A |
| 63.4.163 | 58894 | Ca3Ga5 | A |
| 63.4.164 | 410775 | Ca1Rh1Sn2 | A |
| 63.4.165 | 410773 | Ca1Ir1Sn2 | A |
| 63.4.166 | 420479 | Ca1Ir1O3 | C MOTT and magnetic |
| 63.4.167 | 420478 | Ca1Ir1O3 | C MOTT and magnetic |
| 63.4.168 | 182246 | Ca1Ir1O3 | C MOTT and magnetic |
| 63.4.169 | 180015 | Ca1Ir1O3 | C MOTT and magnetic |
| 63.4.170 | 160816 | Ca1Ir1O3 | C MOTT and magnetic |
| 63.4.171 | 159038 | Ca1Ir1O3 | C MOTT and magnetic |
| 63.4.172 | 159037 | Ca1Ir1O3 | C MOTT and magnetic |
| 63.4.173 | 159036 | Ca1Ir1O3 | C MOTT and magnetic |
| 63.4.174 | 159035 | Ca1Ir1O3 | C MOTT and magnetic |
| 63.4.175 | 159034 | Ca1Ir1O3 | A |
| 63.4.176 | 159033 | Ca1Ir1O3 | C MOTT and magnetic |
| 63.4.177 | 159029 | Ca1Ir1O3 | C MOTT and magnetic |
| 63.4.178 | 159028 | Ca1Ir1O3 | C MOTT and magnetic |
| 63.4.179 | 159027 | Ca1Ir1O3 | C MOTT and magnetic |
| 63.4.180 | 58898 | Ca1Ga3Ni2 | A |
| 63.4.181 | 419100 | Ca1Ga1 | A |
| 63.4.182 | 260641 | Ca1Ga1 | A |
| 63.4.183 | 106332 | Ca1Cu1In2 | A |
| 63.4.184 | 411623 | C2B3Nb4 | A |
| 63.4.186 | 50634 | C1Ho1Ru2Si1 | C spin polarized |
| 63.4.188 | 415220 | C1Cu1N2 | C probably MOTT |
| 63.4.189 | 161461 | C1Cu1N2 | A or C |
| 63.4.190 | 161460 | C1Cu1N2 | C probably MOTT |
| 63.4.191 | 260480 | C1Ce1P1Ru2 | E |
| 63.4.192 | 73828 | C1B1U1 | E |
| 63.4.193 | 108984 | Br3Cm1 | E |
| 63.4.194 | 1107 | Br1O2U1 | E spin polarized |
| 63.4.195 | 409756 | Bi1Zr1 | A |
| 63.4.196 | 50826 | Ba3Cu1O9Ru2 | A |
| 63.4.197 | 50825 | Ba3Cu1O9Ru2 | C spin polarized |
| 63.4.198 | 63228 | Ba1S3V1 | A |
| 63.4.199 | 58669 | Ba1Pb6Rh3 | A |
| 63.4.200 | 195765 | Ba1La1Si2 | A |
| 63.4.201 | 411232 | Ba1In2Rh1 | A |
| 63.4.202 | 414434 | Ba1In2Ir1 | A |
| 63.4.203 | 58648 | Ba1Cu1Sn2 | A |
| 63.4.207 | 428663 | Ba1Co1O1S1 | C spin polarized |
| 63.4.208 | 428516 | Ba1Co1O1S1 | C magnetic with large band gap |
| 63.4.209 | 64595 | B7Re1Y3 | A |
| 63.4.211 | 91244 | B4Os3Sc1 | A |
| 63.4.213 | 391408 | B4Cu1O7 | C spin polarized |
| 63.4.214 | 420402 | B4Co1O7 | C spin polarized large gap |
| 63.4.215 | 79258 | B3V2 | A |
| 63.4.216 | 71643 | B1Ir2Zn2 | A |
| 63.4.217 | 44249 | B1Cr1 | A |
| 63.4.219 | 408882 | Au1Ca1In2 | A |
| 63.4.221 | 33915 | As3Ni5Sr1 | A |
| 63.4.222 | 57973 | Al6Mn1 | A |
| 63.4.224 | 370012 | Al2Ni1Sc1 | A |
| 63.4.225 | 415062 | Al2Cu1Mg1 | A |
| 63.4.227 | 55510 | Ag7Ca2 | A |
| 63.4.228 | 247421 | Ag1Ce1Si2 | C spin pol |
| 63.4.229 | 57353 | Ag1Ca1 | A |
| 63.4.230 | 54979 | Ag1Ca1 | A |
| 64.1.1 | 411192 | Sn3Ti2 | A |
| 64.1.2 | 410902 | Sn3Ti2 | A |
| 64.1.3 | 413279 | Pd8Sn24 | A |
| 64.1.4 | 417180 | P1 | A |
| 64.1.5 | 23836 | P1 | A |
| 64.1.6 | 150873 | P1 | A |
| 64.1.7 | 77831 | La1S1 | A |
| 64.1.8 | 79878 | K1Na2Ni1O2 | C spin polarized |
| 64.1.9 | 40442 | Hf2Ni3Si4 | A |
| 64.1.10 | 90341 | Ge2Ni1 | A |
| 64.1.11 | 150569 | Ga1Sn2V2 | A |
| 64.1.12 | 106752 | Ga1Sn2V2 | A |
| 64.1.13 | 71196 | Co2Nb2Te2 | A |
| 64.1.14 | 71519 | Co1Nb1Te2 | A |
| 64.1.15 | 424937 | Co1Cs1O2 | C spin polarized |
| 64.1.16 | 40372 | Ca5Ga2N4 | A |
| 64.1.18 | 186936 | Ba2Ge4Pd5 | A |

Continued on next column

**TABLE XX– continued**

| TQC | ICSD | Stoich. Formula | SubSG |
|---|---|---|---|
| 64.1.19 | 58636 | Ba2Bi4Cd3 | A |
| 64.1.20 | 416447 | Ba2Bi4Cd3 | A |
| 64.1.21 | 416446 | Ba2Bi4Cd3 | A |
| 64.1.22 | 416445 | Ba2Bi4Cd3 | A |
| 64.1.23 | 416444 | Ba2Bi4Cd3 | A |
| 64.1.25 | 63708 | As2Ni3O8 | C spin polarized |
| 64.4.4 | 78297 | Cu1La2O4 | A high Tc if doped otherwise C MOTT |
| 64.4.5 | 78296 | Cu1La2O4 | A high Tc if doped otherwise C MOTT |
| 64.4.12 | 78289 | Cu1La2O4 | A high Tc if doped otherwise C MOTT |
| 64.4.13 | 78288 | Cu1La2O4 | A high Tc if doped otherwise C MOTT |
| 64.4.15 | 78237 | Cu1La2O4 | A high Tc if doped otherwise C MOTT |
| 64.4.27 | 202724 | Ba2Cu1O6Tl2 | A high Tc if doped otherwise C MOTT |
| 64.4.28 | 202722 | Ba2Cu1O6Tl2 | A high Tc if doped otherwise C MOTT |
| 64.4.29 | 76359 | Al3Ge4La2 | A |
| 64.4.30 | 50977 | Al3Ce2Ge4 | C spin polarized Ce |
| 65.1.1 | 87827 | Ni1Si3Y1 | A |
| 65.1.2 | 78760 | Mn2S4Sn1 | A |
| 65.1.6 | 167624 | In9La11Ni4 | A |
| 65.1.8 | 54612 | Ho2Ni2Pb1 | C spin polarized Ho |
| 65.1.9 | 60646 | Ge3Ni1Y1 | A |
| 65.1.11 | 20337 | Fe2Si7U3 | C spin polarized Fe U |
| 65.1.12 | 261512 | Fe2O5Sr3 | C spin polarized |
| 65.1.17 | 424317 | Ce7Ge10Li8 | C spin polarized Ce |
| 65.1.18 | 63667 | Ce3Sn7 | Ce C needs spin polarisation or U |
| 65.1.24 | 106291 | Au7Rb3 | A |
| 65.1.25 | 106290 | Au7Rb3 | A |
| 65.1.26 | 418529 | Au4Ce1Si2 | C spin polarized Ce |
| 65.2.1 | 15927 | Mg1O3V1 | C spin polarized V4+ d1 |
| 65.2.4 | 83885 | Ge2U1 | E |
| 65.2.5 | 82702 | Ge2U1 | E C spin polarized U |
| 65.2.6 | 236995 | Ge2U1 | C spin polarized U |
| 65.2.7 | 56046 | Ge2Th1 | E magnetic |
| 65.2.8 | 193219 | Ge2Pb1Sr2 | A |
| 65.2.11 | 91064 | Fe4O11Sr4 | C spin polarized |
| 65.2.14 | 419475 | Cu1Mg4Y1 | A |
| 65.2.16 | 72872 | Co1O4Re1 | C spin polarized |
| 65.2.18 | 174528 | Ce2Si7 | C spin polarized Ce |
| 65.2.20 | 185392 | Br2Cd2Hg2S2 | A |
| 65.2.21 | 391060 | Ba3Ge4 | A |
| 65.2.27 | 417990 | B4Rh8Zn5 | A |
| 65.2.28 | 413191 | Al5Br2La4 | A |
| 67.2.1 | 167666 | Al2Cu1Ir1 | A |
| 67.4.1 | 411905 | Ba2Na1O1 | A sub oxide - interesting band structure maybe hybride |
| 69.1.1 | 262559 | Pt3Rb2Se6U1 | E |
| 69.1.2 | 262561 | Pd3Rb2Se6U1 | EC spin polarized U |
| 69.1.3 | 262562 | Pd3Rb2S6U1 | E C spin polarized U |
| 69.1.4 | 262563 | K2Pd3S6U1 | E |
| 69.1.5 | 414574 | Ge6Li2Sr4 | A |
| 69.1.6 | 68676 | Cu2O3Sr2 | C spin polarized |
| 69.1.7 | 262555 | Cs2Pt3Se6U1 | E |
| 69.1.8 | 262556 | Cs2Pt3S6U1 | E |

Continued on next column

**TABLE XX– continued**

| TQC | ICSD | Stoich. Formula | SubSG |
|---|---|---|---|
| 69.1.9 | 262557 | Cs2Pd3Se6U1 | EC spin polarized U |
| 69.1.10 | 261242 | Cs2Pd3Se6U1 | EC spin polarized U |
| 69.1.11 | 262558 | Cs2Pd3S6U1 | C spin polarized U |
| 69.1.12 | 73160 | Ba2Na1Ni3O6 | C spin polarized large band gap |
| 69.1.13 | 59588 | Ba2Na1Ni3O6 | C spin polarized large band gap |
| 69.1.14 | 72328 | Ba2Cu3Na1O6 | C spin polarized |
| 69.1.15 | 59589 | Ba2Cu3Na1O6 | C spin polarized magnetic small band gap |
| 69.1.16 | 401239 | Ba2Cu3Li1O6 | C spin polarized Cu |
| 69.1.18 | 410522 | Ag2Ca4Si6 | A |
| 70.1.1 | 1089 | Si2Ti1 | A |
| 70.1.2 | 97545 | Sb1Sn1Ti1 | A |
| 70.1.4 | 106552 | Nb1Sn2 | A |
| 70.1.5 | 429511 | Ga2Rh2Sr1 | A |
| 70.1.7 | 89073 | Cr2O4Si1 | C magnetic with large band gap |
| 70.1.9 | 89071 | Cr2O4Si1 | C magnetic with large band gap |
| 70.1.13 | 96610 | Cl2Cu2O1 | C spin polarized |
| 70.1.14 | 77143 | Ba1Ge2Ru2 | A |
| 70.1.16 | 40799 | B2Ce1Ru2 | A |
| 70.1.17 | 40618 | B2Ce1Ir2 | C spin polarized Ce |
| 70.1.18 | 49737 | B2Ca1Rh2 | A |
| 70.1.19 | 49738 | B2Ca1Ir2 | A |
| 70.1.20 | 75028 | B1Ir1Li1 | A |
| 70.1.21 | 75027 | B1Ir1Li1 | A |
| 70.1.22 | 418163 | B1Cr2 | A |
| 70.4.1 | 73610 | O10U1V3 | C spin polarized U |
| 70.4.2 | 84406 | Nb3O10U1 | C spin polarized U |
| 70.4.3 | 94466 | Fe2K3O4 | Fe2K1S3 C AFM? |
| 70.4.4 | 40621 | B2Ir2Y1 | A |
| 71.1.1 | 82655 | Te2U1 | C spin polarized U |
| 71.1.2 | 82654 | Te2U1 | C spin polarized U |
| 71.1.3 | 82653 | Te2U1 | C spin polarized U |
| 71.1.4 | 82652 | Te2U1 | C spin polarized U |
| 71.1.5 | 82651 | Te2U1 | C spin polarized U |
| 71.1.6 | 82650 | Te2U1 | C spin polarized U |
| 71.1.7 | 82649 | Te2U1 | C spin polarized U |
| 71.1.8 | 82648 | Te2U1 | C spin polarized U |
| 71.1.9 | 82647 | Te2U1 | C spin polarized U |
| 71.1.10 | 82646 | Te2U1 | C spin polarized U |
| 71.1.11 | 82645 | Te2U1 | C spin polarized U |
| 71.1.12 | 82644 | Te2U1 | C spin polarized U |
| 71.1.13 | 82643 | Te2U1 | C spin polarized U |
| 71.1.14 | 82642 | Te2U1 | C spin polarized U |
| 71.1.15 | 63170 | Te2U1 | C spin polarized U |
| 71.1.16 | 40694 | Te2U1 | C spin polarized U |
| 71.1.20 | 416817 | La2Ni2Zn1 | A |
| 71.1.23 | 56745 | H1Ce2Li1O3 | C spin polarized Ce |
| 71.1.32 | 300260 | Ge2Ho1Pt1 | spin polarized |
| 71.1.50 | 416816 | Ce2Ni2Zn1 | A |
| 71.1.51 | 174529 | Ce1Si5 | C spin polarized Ce |
| 71.1.52 | 246812 | Ce1Ge5 | C spin polarized Ce |
| 71.1.53 | 429175 | Ca2Mg1Pt2 | A |
| 71.1.54 | 429174 | Ca2Mg1Pd2 | A |
| 71.1.55 | 425471 | Ca2Cd1Pd2 | A |
| 71.1.56 | 40950 | C4Co1Sc3 | A CDW and superconductivity |

Continued on next column

**TABLE XX– continued**

| TQC | ICSD | Stoich. Formula | SubSG |
|---|---|---|---|
| 71.1.57 | 236393 | C4Co1Sc3 | A CDW and superconductivity |
| 71.1.58 | 173167 | C4Co1Sc3 | A |
| 71.1.85 | 167153 | B6Fe2U1 | E spin polarized U |
| 71.1.86 | 81543 | B6Cr2Th1 | E spin polarized Th |
| 71.1.87 | 81544 | B6Ce1Cr2 | C spin polarized Ce Cr |
| 71.1.88 | 40791 | B4Cr3 | A |
| 71.1.89 | 423380 | B4Cr1 | A |
| 71.1.92 | 281420 | As1Te1Ti1 | A |
| 71.1.95 | 9564 | Al2Sn2Sr3 | A |
| 71.1.96 | 100127 | Al2Si2Sr3 | A |
| 71.1.97 | 52617 | Al2Ge2Sr3 | A |
| 71.1.98 | 100126 | Al2Ca3Si2 | C spin polarized U |
| 71.1.99 | 52616 | Al2Ca3Ge2 | A |
| 71.1.100 | 9565 | Al2Ba3Sn2 | A |
| 71.1.101 | 100128 | Al2Ba3Si2 | A |
| 71.1.102 | 52612 | Al2Ba3Ge2 | A |
| 72.1.1 | 84206 | Ni3Si5Y2 | A |
| 72.2.1 | 170063 | Pb5Sc6 | A |
| 72.2.2 | 416512 | Ni1Pd2Te2 | A Pauli paramagnetism and metallic |
| 72.2.3 | 416515 | Fe1Pd2Se2 | A or C |
| 72.2.5 | 54107 | Cr4Sc2Si5 | A but messy band structure |
| 72.2.7 | 28232 | Br2Hg5O4 | A |
| 72.4.1 | 65460 | Mn1Rb2Te2 | C magnetic spin polarized |
| 72.4.2 | 65457 | Mn1Rb2Se2 | C magnetic spin polarized |
| 72.4.3 | 65454 | Mn1Rb2S2 | C magnetic spin polarized |
| 72.4.4 | 40427 | La2Rh3Si5 | A nonmagnetic metals |
| 72.4.5 | 65459 | K2Mn1Te2 | C magnetic spin polarized |
| 72.4.6 | 65456 | K2Mn1Se2 | C magnetic spin polarized |
| 72.4.7 | 65453 | K2Mn1S2 | C magnetic spin polarized |
| 72.4.8 | 86502 | Ge5La2Rh3 | A non magnetic metallic |
| 72.4.9 | 86503 | Ge5Ir3Y2 | A |
| 72.4.10 | 86499 | Ge5Ir3La2 | A |
| 72.4.11 | 40462 | Ge2Mn5 | C magnetic spin polarized |
| 72.4.12 | 68523 | Fe1Li3N2 | A or C para band structure has a peak at EF |
| 72.4.13 | 78933 | Cs2Mn3Te4 | C magnetic spin polarized |
| 72.4.14 | 59276 | Cs2Mn3Te4 | C magnetic spin polarized |
| 72.4.15 | 78932 | Cs2Mn3Se4 | C magnetic spin polarized |
| 72.4.16 | 65461 | Cs2Mn1Te2 | C magnetic spin polarized |
| 72.4.17 | 65458 | Cs2Mn1Se2 | C magnetic spin polarized |
| 72.4.18 | 65455 | Cs2Mn1S2 | C magnetic spin polarized |
| 72.4.19 | 65254 | Co3Cs2S4 | C magnetic spin polarized |
| 72.4.20 | 67391 | Co1Rb2Se2 | C magnetic spin polarized |

Continued on next column

**TABLE XX– continued**

| TQC | ICSD | Stoich. Formula | SubSG |
|---|---|---|---|
| 72.4.21 | 67388 | Co1Rb2S2 | C magnetic spin polarized |
| 72.4.22 | 416514 | Co1Pd2Te2 | A metallic |
| 72.4.23 | 416513 | Co1Pd2Se2 | A metallic |
| 72.4.24 | 67386 | Co1Na2S2 | C magnetic spin polarized |
| 72.4.25 | 67390 | Co1K2Se2 | C magnetic spin polarized |
| 72.4.26 | 67387 | Co1K2S2 | C magnetic spin polarized |
| 72.4.27 | 67392 | Co1Cs2Se2 | C ferro spin polarized |
| 72.4.28 | 67389 | Co1Cs2S2 | C magnetic spin polarized |
| 72.4.30 | 69494 | Ba2Cu3P4 | A |
| 72.4.31 | 26253 | As2Cu5 | A |
| 74.2.2 | 429139 | Ga2La1Rh3 | A |
| 74.2.3 | 238760 | Fe2O4Si1 | C magnetic with large band gap |
| 74.2.4 | 238759 | Fe2O4Si1 | C magnetic with large band gap |
| 74.2.6 | 93229 | Fe2Mo4O7 | C spin polarized magnetic |
| 74.2.7 | 74288 | Cs1F5Ni1Pd1 | B |
| 74.2.8 | 10156 | Co2O4Si1 | C spin polarized |
| 74.2.9 | 246171 | Ce3Pt5Si1 | C spin polarized Ce |
| 74.2.10 | 93244 | Ce3Pd5Si1 | A Ce caution |
| 74.2.11 | 260381 | Cd2Sr1 | A |
| 74.2.12 | 97094 | Ca2O7Os2 | C ferrimagnetic insulating state at 327 K |
| 74.2.13 | 97093 | Ca2O7Os2 | C ferrimagnetic insulating state at 327 K |
| 74.2.14 | 97092 | Ca2O7Os2 | C ferrimagnetic insulating state at 327 K |
| 74.2.15 | 97091 | Ca2O7Os2 | C ferrimagnetic insulating state at 327 K |
| 74.2.16 | 418971 | Ca1Zn2 | A |
| 74.2.17 | 420576 | Ca1Cd2 | A |
| 74.2.18 | 414229 | Ba1In2 | A |
| 74.2.19 | 251586 | Ba1Ge5 | A |
| 74.2.20 | 194880 | Ba1Ge5 | A |
| 74.2.21 | 260668 | Ba1Cd2 | A |
| 74.2.22 | 55542 | Au2Ca1 | A |
| 74.2.24 | 417225 | Ag2Ca1 | A |
| 74.2.25 | 417227 | Ag1Ca1Zn1 | A |
| 74.4.1 | 236759 | O3Sr1Tc1 | C spin polarized Tc |
| 74.4.2 | 183451 | O3Sr1Tc1 | C spin polarized |
| 74.4.3 | 260033 | La3Pd5Si1 | A |
| 74.4.4 | 200066 | La1Re2Si2 | A |
| 74.4.5 | 408698 | Hg2Rb1 | B |
| 74.4.6 | 195614 | Ge2Na1Pd3 | A |
| 74.4.12 | 36535 | Cu2O2Tl1 | C spin polarized |
| 74.4.13 | 408695 | Cs1Hg2 | B |
| 74.4.14 | 79221 | Ce1Ir3Si2 | C spin polarized |
| 74.4.15 | 40664 | Br4Li2Mn1 | C spin polarized large band gap |
| 74.4.16 | 249923 | Au5K3Tl1 | A |
| 74.4.17 | 249922 | Au5In1K3 | A |
| 74.4.18 | 413972 | As1Cu1F7 | C spin polarized |
| 82.2.1 | 81539 | Te1Zr3 | A |
| 82.2.2 | 51048 | Te1Zr3 | A |
| 82.2.3 | 195057 | Sb1Zr3 | A |

Continued on next column

**TABLE XX– continued**

| TQC | ICSD | Stoich. Formula | SubSG |
|---|---|---|---|
| 82.2.4 | 98373 | Ni3P1 | A |
| 82.2.5 | 23612 | In1P1S4 | A |
| 82.2.6 | 56027 | Ge1Ta3 | A |
| 82.2.7 | 43365 | Fe3P1 | C spin polarized |
| 82.2.8 | 23560 | Cr3P1 | A |
| 83.2.1 | 106354 | Ca7Ni4Sn13 | A Pauli paramagnetic with simple metallic behavior |
| 83.3.1 | 42959 | C4U1W4 | C spin polarized U |
| 83.3.2 | 39222 | C4U1W4 | C spin polarized U |
| 83.3.3 | 999 | As2V3 | A |
| 84.2.1 | 74390 | P2S6U1 | C spin polarized U |
| 84.3.1 | 155153 | Fe1O4Sb2 | C spin polarized |
| 85.4.1 | 195053 | Ni1Sb9Zr5 | A |
| 85.4.2 | 27315 | Mo1O5V1 | C spin polarized |
| 85.4.3 | 36095 | Mo1O5P1 | C spin polarized |
| 85.4.4 | 24894 | Mo1O5P1 | C spin polarized |
| 85.4.5 | 413979 | Hf5Sb9 | A |
| 85.4.6 | 56963 | H4B1Cl1Cu1O4 | C spin polarized |
| 85.4.7 | 80646 | Ag1B1F5 | C spin polarized MOTT |
| 85.4.8 | 80645 | Ag1B1F5 | C MOTT |
| 86.4.1 | 8078 | B4La1Ru4 | A |
| 86.4.2 | 100717 | B4Ir4La1 | A |
| 87.1.3 | 55075 | Se4Ti5 | A |
| 87.1.8 | 60822 | C4Cr4U1 | C spin polarized U Cr |
| 87.1.9 | 40319 | C4Cr4U1 | C spin polarized U |
| 87.1.10 | 43186 | As4Mo5 | A |
| 87.3.2 | 194696 | Mg1O6Os1Sr2 | C spin polarized |
| 87.3.3 | 190088 | In4Pd17Se4 | A |
| 87.3.4 | 421410 | Ba9In4 | A |
| 87.3.8 | 238713 | Ba1Fe2Se4 | C AFM |
| 87.3.9 | 23081 | Ba1Fe2S4 | C BaFe2S4 are linear chain antiferromagnets exhibiting semicond |
| 87.3.10 | 66016 | Ba1Cu8P4 | A |
| 88.3.1 | 16189 | O4Pb1W1 | A |
| 88.4.1 | 78846 | O4Pd2Y1 | A |
| 88.4.2 | 79340 | La1O4Pd2 | A |
| 88.4.5 | 173264 | Cr1Ho1O4 | C spin polarized Ho |
| 88.4.6 | 169873 | Cr1Ho1O4 | C spin polarized Ho |
| 92.3.1 | 33775 | Fe1Na1O2 | C spin polarized |
| 92.3.2 | 33774 | Fe1Na1O2 | C spin polarized Fe magnetic |
| 92.3.3 | 33773 | Fe1Na1O2 | C spin polarized |
| 92.3.4 | 33769 | Fe1Na1O2 | C spin polarized |
| 92.3.5 | 33768 | Fe1Na1O2 | C spin polarized |
| 92.4.1 | 424580 | Ni1O2Rb3 | C spin polarized |
| 92.4.2 | 262578 | K3Ni1O2 | C spin polarized |
| 92.4.3 | 262577 | K3Ni1O2 | C spin polarized |
| 97.3.1 | 35190 | I1Se8Ta2 | A |
| 99.3.2 | 87466 | Ba1Cu1Fe1O5Y1 | C spin polarized |
| 99.3.3 | 71355 | Ba1Cu1Fe1O5Y1 | C spin polarized |
| 102.4.1 | 202705 | F6Fe2Li1 | C spin polarized FLOURIDES ARE AS MOTT AS OXIDES |
| 103.4.1 | 62050 | Ta1Te4 | A |
| 103.4.2 | 65129 | Nb1Te4 | A |
| 103.4.3 | 65128 | Nb1Te4 | A |
| 106.3.1 | 190731 | Cu1O4Sb2 | C spin polarized |
| 107.3.1 | 161866 | Ce1Ru1Si3 | C spin polarized |
| 107.3.2 | 58662 | Ba1Ni1Sn3 | A |

Continued on next column

**TABLE XX– continued**

| TQC | ICSD | Stoich. Formula | SubSG |
|---|---|---|---|
| 107.3.3 | 174261 | Ba1Ge3Pd1 | A |
| 109.4.2 | 414514 | Ir1La1P1 | A |
| 109.4.3 | 95191 | As1La1Rh1 | A |
| 113.2.1 | 246245 | Ca1Fe1O2 | C spin polarized |
| 113.4.1 | 54208 | Ta30 | A |
| 113.4.2 | 280872 | Ta30 | A |
| 113.4.3 | 250214 | Ta30 | A |
| 113.4.4 | 54206 | Ta22 | A |
| 113.4.5 | 96594 | Ta1 | A |
| 113.4.6 | 89533 | P6Rh7Sr2 | A |
| 113.4.7 | 408045 | Na1Sn5 | A |
| 113.4.9 | 239 | Cu3Se2 | C spin polarized small band gap AFM mixed valent |
| 113.4.10 | 191282 | Cu1K2O7P2 | C spin polarized |
| 113.4.12 | 262398 | Co1Ge2O7Sr2 | C spin polarized |
| 113.4.14 | 94137 | Ca2Co1O7Si2 | C spin polarized |
| 113.4.20 | 281293 | Ba2Co1O7Si2 | A |
| 113.4.21 | 423351 | Ba2Co1Ge2O7 | C spin polarized |
| 113.4.22 | 423350 | Ba2Co1Ge2O7 | C spin polarized |
| 115.1.1 | 93242 | Cu1Sb3Zr2 | A |
| 115.1.2 | 426313 | Cu1Sb3Zr2 | A |
| 115.1.3 | 195058 | Cu1Sb3Zr2 | A |
| 115.1.4 | 93241 | Cu1Sb3Ti2 | A |
| 115.1.5 | 93243 | Cu1Hf2Sb3 | A |
| 120.3.1 | 58522 | Au4K1Sn2 | A |
| 121.1.1 | 237706 | Mo3P1 | A |
| 121.1.3 | 400652 | Cu3Sb1Se4 | C spin polarized |
| 121.1.4 | 412239 | Cu3S4Sb1 | C spin polarized |
| 121.1.5 | 2857 | Cu3S4Sb1 | C spin polarized |
| 121.1.6 | 95119 | Cu2Hg1Se4Sn1 | C spin polarized |
| 121.1.7 | 152761 | Cu2Ge1Hg1Se4 | C spin polarized |
| 121.1.8 | 85126 | Cu2Fe1Se4Sn1 | C spin polarized |
| 121.1.9 | 165094 | Cd1Cu2Ge1Te4 | C spin polarized |
| 122.4.1 | 80095 | Cu1Fe1S2 | C MAGNETIC semi-conductor - TI??? |
| 122.4.2 | 80094 | Cu1Fe1S2 | C MAGNETIC semi-conductor - TI??? |
| 122.4.3 | 261883 | Cu1Fe1S2 | C MAGNETIC semi-conductor - TI??? |
| 122.4.4 | 261882 | Cu1Fe1S2 | C MAGNETIC semi-conductor - TI??? |
| 122.4.5 | 2518 | Cu1Fe1S2 | C MAGNETIC semi-conductor - TI??? |
| 122.4.6 | 156643 | Ag1Fe1S2 | C MAGNETIC semi-conductor - TI??? |
| 123.1.3 | 5257 | Fe1Si2 | A !!! Non magnetic berry phase - sounds interesting |
| 123.1.6 | 192621 | Ce2In8Pt1 | C spin polarized |
| 123.1.17 | 74163 | Ba2Cu2O7Tl1Y1 | A when doped HIGH TC |
| 123.1.21 | 165631 | Ba1Ge3Mg4 | A |
| 123.1.24 | 170662 | Ag1Sb1Te2 | A |
| 123.2.3 | 157999 | Fe5Si3U1 | C spin polarized |
| 123.2.5 | 74164 | Cu2O7Sr2Tl1Y1 | A when doped HIGH TC |
| 123.2.7 | 240161 | Ce1Ga6Pd1 | C because Ce |
| 123.2.16 | 84739 | Ba2Cu3O6Y1 | A when doped HIGH TC |

Continued on next column

**TABLE XX– continued**

| TQC | ICSD | Stoich. Formula | SubSG |
|---|---|---|---|
| 123.2.17 | 72254 | Ba2Cu3O6Y1 | A when doped HIGH TC |
| 123.2.18 | 68290 | Ba2Cu3O6Y1 | Awhen doped HIGH TC |
| 123.2.19 | 68046 | Ba2Cu3O6Y1 | A when doped HIGH TC |
| 123.2.20 | 68008 | Ba2Cu3O6Y1 | A when doped HIGH TC |
| 123.2.21 | 67016 | Ba2Cu3O6Y1 | A when doped HIGH TC |
| 123.2.22 | 66903 | Ba2Cu3O6Y1 | A when doped HIGH TC |
| 123.2.23 | 63269 | Ba2Cu3O6Y1 | A when doped HIGH TC |
| 123.2.24 | 203116 | Ba2Cu3O6Y1 | A when doped HIGH TC |
| 123.2.27 | 64639 | Ba2Cu2La1O8Ta1 | A when doped HIGH TC |
| 123.2.28 | 165630 | Ba1Mg4Si3 | A |
| 123.2.29 | 169074 | As2Ba1O1Ti2 | Ti3+ C spin polarized |
| 123.2.30 | 174342 | Al2B1La1Ru2 | A |
| 123.2.31 | 251834 | Al2B1La1Os2 | A |
| 123.3.1 | 79355 | Nb4O6Sr1 | C spin polarized |
| 123.3.4 | 290932 | In2Pd8Se1 | A |
| 123.3.5 | 236215 | Hg1Pd5Se1 | A |
| 123.3.6 | 187377 | Hg1Pd5Se1 | A |
| 123.3.7 | 418606 | Fe1O2Sr1 | C spin polarized |
| 123.3.8 | 418605 | Fe1O2Sr1 | C spin polarized |
| 123.3.9 | 418603 | Fe1O2Sr1 | C spin polarized |
| 123.3.10 | 173434 | Fe1O2Sr1 | C spin polarized |
| 123.3.11 | 88880 | F1K1Nb4O5 | A or C |
| 123.3.15 | 247864 | Ce2In8Pd1 | C spin polarized |
| 123.3.16 | 173438 | Ca1Fe1O2 | C spin polarized |
| 123.3.21 | 430063 | Ba1Bi2O1Ti2 | C AFM but also superconductor |
| 123.3.22 | 79256 | As2Ba1Cu6 | A or C |
| 123.3.23 | 193115 | As1In1Pt5 | A |
| 123.3.24 | 193117 | As1In1Pd5 | A |
| 127.2.1 | 105627 | Pb3Sr2 | A |
| 127.2.4 | 26200 | Li2Sn5 | A |
| 127.2.7 | 262925 | Cs1I3Sn1 | B |
| 127.2.8 | 413849 | Ce2Ge2Mg1 | C spin polarized |
| 127.2.13 | 107389 | Cd1La2Rh2 | A |
| 127.2.15 | 424169 | Cd1Ge2Y2 | A |
| 127.2.19 | 193226 | Ca2Ge2Sn1 | C spin polarized |
| 127.2.20 | 79239 | C5Ru6Th2 | C spin polarized |
| 127.2.21 | 39204 | C5Ru6Th2 | C spin polarized |
| 127.2.25 | 94036 | C2B2Ce1 | C spin polarized |
| 127.2.26 | 88561 | C2B2Ce1 | C spin polarized |
| 127.2.27 | 88560 | C2B2Ce1 | C spin polarized |
| 127.2.28 | 280180 | C2B2Ce1 | C spin polarized |
| 127.2.30 | 404497 | Ba1Ho2O5Pd1 | C spin polarized |
| 127.2.32 | 67222 | B4Y1 | A |
| 127.2.33 | 26777 | B4Y1 | A |
| 127.2.34 | 2360 | B4La1 | A |
| 127.2.35 | 182659 | B4Ho1 | A |
| 127.2.36 | 51434 | B2Rh5Sc3 | A |
| 127.2.37 | 107439 | Al1Pt3 | A |
| 127.3.1 | 73695 | Si2U3 | E C spin polarized |
| 127.3.2 | 54347 | Rh2Sn1U2 | E C spin polarized |
| 127.3.3 | 99189 | Pb1Pd2Y2 | A |
| 127.3.6 | 54348 | Ni2Sc2Sn1 | A |

Continued on next column

**TABLE XX– continued**

| TQC | ICSD | Stoich. Formula | SubSG |
|---|---|---|---|
| 127.3.7 | 81707 | Ni2P4Zr9 | C spin polarized |
| 127.3.8 | 92839 | Mo4O6Sn1 | A small band semiconductor |
| 127.3.15 | 99190 | La2Pb1Pd2 | A |
| 127.3.16 | 411081 | La2Mg1Ni2 | A |
| 127.3.17 | 107327 | La2Mg1Ni2 | A |
| 127.3.20 | 99197 | Ho2Pb1Pd2 | C spin polarized |
| 127.3.21 | 107371 | Hg2Sr3 | A |
| 127.3.22 | 107096 | Hf2Ni2Sn1 | A |
| 127.3.25 | 423453 | Ge2Ho2Mg1 | C spin polarized |
| 127.3.28 | 103988 | Ga2Th3 | C spin polarized |
| 127.3.29 | 260213 | Ga2Mg1Sc2 | A |
| 127.3.33 | 419472 | Cu2Mg1Y2 | A |
| 127.3.34 | 411711 | Cu2Mg1Y2 | A |
| 127.3.35 | 411709 | Cu2La2Mg1 | A |
| 127.3.36 | 99191 | Ce2Pb1Pd2 | C spin polarized |
| 127.3.37 | 411012 | Ce2Mg1Ni2 | C spin polarized |
| 127.3.38 | 411013 | Ce2Cu2Mg1 | C spin polarized |
| 127.3.39 | 107322 | Ce2Cu2Mg1 | C spin polarized |
| 127.3.46 | 411009 | Cd1Ce2Pt2 | C spin polarized |
| 127.3.47 | 411010 | Cd1Ce2Pd2 | C spin polarized |
| 127.3.48 | 195139 | C4Al3Y5 | A |
| 127.3.49 | 67376 | C3La2Ni5 | A |
| 127.3.50 | 62277 | C3La2Ni5 | C spin polarized |
| 127.3.54 | 81545 | B4Th1 | C spin polarized |
| 127.3.58 | 417745 | B4Ce1 | C spin polarized |
| 127.3.59 | 51438 | B2Ga1Rh5Sc2 | A |
| 127.3.60 | 409531 | Au2Ca2Pb1 | A |
| 127.4.1 | 40962 | Mo4Na1O6 | A |
| 127.4.2 | 420085 | Li1Rh2Si2 | A |
| 127.4.3 | 159116 | La7Ni2Zn1 | A |
| 127.4.4 | 68533 | K1Mo4O6 | A |
| 127.4.5 | 401730 | In5Ti2 | A |
| 127.4.6 | 192651 | In2Ni9Y1 | A |
| 127.4.8 | 192640 | In2La1Ni9 | A |
| 127.4.11 | 107372 | In1Ni2Zr2 | A |
| 127.4.12 | 107333 | In1Ni2Sc2 | A |
| 127.4.14 | 404531 | Hf2In5 | A |
| 127.4.20 | 87511 | Ge2In1La2 | A |
| 127.4.21 | 172493 | Ge2Ho2In1 | C spin polarized |
| 127.4.22 | 249632 | Ga5Mn2 | C spin polarized |
| 127.4.23 | 85499 | F4Fe1Rb1 | C spin polarized |
| 127.4.26 | 411708 | Cu2In1La2 | A |
| 127.4.27 | 21111 | Cu1F3K1 | C spin polarized |
| 127.4.28 | 21110 | Cu1F3K1 | C spin polarized |
| 127.4.29 | 404847 | Co2In1Zr2 | A |
| 127.4.30 | 107331 | Co2In1Zr2 | A |
| 127.4.31 | 410293 | Ce2In1Pt2 | C spin polarized |
| 127.4.32 | 150615 | Ce2In1Ni2 | A |
| 127.4.33 | 415108 | Ce2Ge2In1 | C spin polarized |
| 127.4.34 | 424134 | Ca1Cd2Cu9 | A |
| 127.4.35 | 427155 | C2B2Y1 | A |
| 127.4.40 | 94035 | C2B2La1 | A interesting band structure |
| 127.4.42 | 88317 | B2V3 | A |
| 127.4.43 | 427424 | B2Ru5Ta3 | A |
| 127.4.44 | 423469 | B2Nb3Ru5 | A |
| 127.4.45 | 69487 | B2Ir5Mg2Si1 | A |
| 127.4.46 | 44316 | B2Hf3Ir5 | A |
| 127.4.47 | 51432 | B2Ge1Mg2Rh5 | A |
| 127.4.49 | 51435 | B2Cr1Rh5Sc2 | C spin polarized |

Continued on next column

**TABLE XX– continued**

| TQC | ICSD | Stoich. Formula | SubSG |
|---|---|---|---|
| 127.4.50 | 262043 | Au2Y3 | A |
| 128.1.1 | 425655 | B2Nb2Os1 | A |
| 128.3.1 | 38143 | Cr1I6Tl4 | C spin polarized |
| 128.3.2 | 429117 | B2Os1Ta2 | A |
| 128.4.1 | 245749 | Cl6K2Nb1 | C spin polarized Nb d1 |
| 129.1.1 | 80190 | Sn1Te1Zr1 | A |
| 129.1.3 | 74522 | Si1Te1Zr1 | A |
| 129.1.14 | 61194 | Pd2Sb2Sr1 | A |
| 129.1.32 | 87315 | Ge1S1U1 | C spin polarized U |
| 129.1.40 | 169271 | Fe1Se1 | A Fe high TC |
| 129.1.41 | 169269 | Fe1Se1 | A Fe high TC |
| 129.1.42 | 169267 | Fe1Se1 | A Fe high TC |
| 129.1.43 | 169265 | Fe1Se1 | A Fe high TC |
| 129.1.44 | 169263 | Fe1Se1 | A Fe high TC |
| 129.1.45 | 169261 | Fe1Se1 | A Fe high TC |
| 129.1.46 | 169259 | Fe1Se1 | A Fe high TC |
| 129.1.47 | 169257 | Fe1Se1 | A Fe high TC |
| 129.1.48 | 169255 | Fe1Se1 | A Fe high TC |
| 129.1.49 | 169253 | Fe1Se1 | A Fe high TC |
| 129.1.50 | 94475 | F1Ho1S1 | C spin polarized |
| 129.1.52 | 59594 | Cu1P1Si1Zr1 | A |
| 129.1.54 | 27230 | Cs1F4Ti1 | C spin polarized Ti d1 |
| 129.1.60 | 97019 | Cl1Mn1O3Sr2 | C spin polarized |
| 129.1.61 | 97018 | Cl1Mn1O3Sr2 | C spin polarized |
| 129.1.62 | 96044 | Cl1Mn1O3Sr2 | C spin polarized |
| 129.1.63 | 94745 | Cl1Mn1O3Sr2 | C spin polarized |
| 129.1.68 | 411678 | Ce1Ir2Si2 | A |
| 129.1.69 | 61371 | Ce1Ge2Pt2 | C spin polarized |
| 129.1.73 | 92318 | Bi1I1O3Te1 | C spin polarized |
| 129.1.74 | 246945 | Ba1Ni1S2 | A |
| 129.1.75 | 15289 | Ba1Ni1S2 | A |
| 129.1.78 | 370004 | Au2Ga2Sr1 | A |
| 129.1.84 | 56976 | As2Pd1U1 | C spin polarized U |
| 129.1.87 | 246061 | As1La1Ni1O1 | A |
| 129.1.90 | 42165 | As1Cu1Si1Zr1 | A |
| 129.2.6 | 74786 | Ni2P2Th1 | EC spin polarized |
| 129.2.13 | 423892 | La1N1 | A |
| 129.2.16 | 87913 | Ge1Ni1Pd1 | A |
| 129.2.32 | 428492 | Cu4La1P3 | A |
| 129.2.34 | 79579 | Cu1O1P1U1 | C spin polarized U |
| 129.2.37 | 182311 | Cu1O1P1Th1 | E C spin polarized |
| 129.2.38 | 172798 | Cu1F1Se1Sr1 | check a 2 eV gap semi-conductor TI???? |
| 129.2.39 | 172796 | Cu1F1S1Sr1 | C spin polarized |
| 129.2.43 | 61370 | Ce1Pt2Si2 | C spin polarized |
| 129.2.44 | 40764 | Ce1P2Rh2 | C spin polarized |
| 129.2.48 | 2027 | Ca1Ga1N1 | A |
| 129.2.49 | 62598 | C2Co1Sc1 | A nice bona structure |
| 129.2.50 | 85283 | C1B1Ni1Y1 | A |
| 129.2.51 | 415728 | Bi2La1Li1 | A |
| 129.2.58 | 236308 | Ba1Cu2Sb2 | A |
| 129.2.59 | 428691 | Au2Sn2Sr1 | A |
| 129.2.60 | 427711 | Au2Sn2Sr1 | A |
| 129.2.62 | 412550 | As2Cu1Y1 | A |
| 129.2.64 | 94440 | As2Cu1Ho1 | C spin polarized |
| 129.2.65 | 412178 | As2Cu1Ho1 | C spin polarized |
| 129.2.81 | 194698 | Al2Ce1Pd2 | C spin polarized |
| 129.2.83 | 190094 | Ag1Sb2Y1 | A |
| 129.3.15 | 391427 | La1Ni1O1P1 | A |
| 129.3.16 | 249484 | La1Ni1O1P1 | C spin polarized |
| 129.3.21 | 98944 | H1Ce1Co1Ge1 | C spin polarized |

Continued on next column

**TABLE XX– continued**

| TQC | ICSD | Stoich. Formula | SubSG |
|---|---|---|---|
| 129.3.23 | 85857 | Ge1La1Mn1 | A |
| 129.3.29 | 977903 | Fe1Se1 | A Fe high TC |
| 129.3.30 | 977901 | Fe1Se1 | A Fe high TC |
| 129.3.34 | 185464 | Fe1Se1 | A Fe high TC |
| 129.3.38 | 163562 | Fe1Se1 | A Fe high TC |
| 129.3.39 | 163561 | Fe1Se1 | A Fe high TC |
| 129.3.40 | 163560 | Fe1Se1 | A Fe high TC |
| 129.3.41 | 163559 | Fe1Se1 | A Fe high TC |
| 129.3.49 | 41923 | Cu2Sb2Sr1 | A |
| 129.3.52 | 12155 | Cu1Na1Se1 | A |
| 129.3.56 | 418547 | Ce2Ru1Zn4 | A |
| 129.3.63 | 418852 | Ca1Ge1Ni1 | A |
| 129.3.67 | 425950 | Bi1La2N1 | A |
| 129.3.75 | 107642 | Bi1In1 | A |
| 129.3.76 | 107641 | Bi1In1 | A |
| 129.3.77 | 107640 | Bi1In1 | A |
| 129.3.79 | 187824 | Bi1Cu1O1Te1 | C spin polarized |
| 129.3.80 | 159475 | Bi1Cu1O1Te1 | C spin polarized |
| 129.3.82 | 238025 | Au2Ba1Sn2 | A |
| 129.3.83 | 238704 | Au1Bi2La1 | A |
| 129.3.86 | 44032 | As2Ni1U1 | C spin polarized |
| 129.3.89 | 186993 | As2Ce1Ir2 | C spin polarized |
| 129.3.93 | 187596 | As1Ni1Rh1 | A |
| 129.3.95 | 16320 | As1Nb1Si1 | A |
| 129.3.96 | 163144 | As1Fe1Na1 | A Fe high TC |
| 129.3.97 | 163143 | As1Fe1Na1 | A Fe high TC |
| 129.3.98 | 169175 | As1Fe1Li1 | A Fe high TC |
| 129.3.100 | 162250 | As1Fe1Li1 | A pnictide Superconductor |
| 129.4.1 | 183309 | Sn2Y1Zn1 | A |
| 129.4.8 | 409586 | Sb1Sc2 | A |
| 129.4.12 | 63594 | P1Sn1Sr1 | A |
| 129.4.18 | 421034 | Np1O1Se1 | C spin polarized |
| 129.4.19 | 421032 | Np1O1S1 | A |
| 129.4.30 | 415627 | Mn2Sb1 | C spin polarized |
| 129.4.33 | 66963 | Mn1Sn1Sr1 | C spin polarized |
| 129.4.34 | 66962 | Mn1Sn1Sr1 | C spin polarized |
| 129.4.35 | 66951 | Mn1Sn1Sr1 | A |
| 129.4.36 | 41868 | Mn1Rb1Sb1 | C spin polarized |
| 129.4.37 | 89595 | Mn1P1Rb1 | C spin polarized |
| 129.4.38 | 89594 | Mn1P1Rb1 | C spin polarized |
| 129.4.42 | 60735 | Mn1Na1Sb1 | C spin polarized |
| 129.4.43 | 60734 | Mn1Na1Sb1 | C spin polarized |
| 129.4.44 | 60733 | Mn1Na1Sb1 | C spin polarized |
| 129.4.46 | 60729 | Mn1Na1P1 | C spin polarized |
| 129.4.47 | 60728 | Mn1Na1P1 | C spin polarized |
| 129.4.48 | 60727 | Mn1Na1P1 | C spin polarized |
| 129.4.49 | 60726 | Mn1Na1P1 | C spin polarized |
| 129.4.54 | 86775 | Li2O5Si1V1 | C spin polarized |
| 129.4.55 | 59359 | Li2O5Si1V1 | C spin polarized |
| 129.4.56 | 417286 | Li2O5Si1V1 | C spin polarized |
| 129.4.57 | 417285 | Li2O5Si1V1 | C spin polarized |
| 129.4.58 | 417284 | Li2O5Si1V1 | V4+ d1 C spin polarized |
| 129.4.59 | 417283 | Li2O5Si1V1 | C spin polarized |
| 129.4.60 | 417282 | Li2O5Si1V1 | C spin polarized |
| 129.4.61 | 417281 | Li2O5Si1V1 | C spin polarized |
| 129.4.62 | 417280 | Li2O5Si1V1 | C spin polarized |
| 129.4.63 | 417279 | Li2O5Si1V1 | A |
| 129.4.64 | 417278 | Li2O5Si1V1 | C spin polarized |
| 129.4.65 | 417277 | Li2O5Si1V1 | C spin polarized |
| 129.4.66 | 417276 | Li2O5Si1V1 | C spin polarized |

Continued on next column

**TABLE XX– continued**

| TQC | ICSD | Stoich. Formula | SubSG |
|---|---|---|---|
| 129.4.70 | 23957 | La1Te2 | A |
| 129.4.71 | 85855 | La1Ru1Si1 | A |
| 129.4.73 | 189724 | La1Pd2Sb2 | A |
| 129.4.74 | 419355 | La1Mn1O1Sb1 | A |
| 129.4.77 | 89591 | K1Mn1Sb1 | C spin polarized |
| 129.4.78 | 89590 | K1Mn1Sb1 | C spin polarized |
| 129.4.79 | 89593 | K1Mn1P1 | C spin polarized |
| 129.4.80 | 89592 | K1Mn1P1 | A |
| 129.4.81 | 41882 | K1Mn1P1 | C spin polarized |
| 129.4.82 | 44668 | Ir2Si2Y1 | A |
| 129.4.89 | 183311 | Ho1Sn2Zn1 | A |
| 129.4.91 | 97037 | Ho1Ni10Si2 | C spin polarized |
| 129.4.92 | 263028 | H2Ca2Ge2Ni2 | A |
| 129.4.94 | 161825 | H1Ce1Ru1Si1 | C spin polarized |
| 129.4.95 | 416423 | Ge2La1Pt2 | C spin polarized |
| 129.4.101 | 85000 | Ge1Sb1Zr1 | A |
| 129.4.102 | 260557 | Ge1Sb1Ti1 | A |
| 129.4.103 | 66961 | Ge1Mn1Sr1 | C spin polarized |
| 129.4.104 | 66960 | Ge1Mn1Sr1 | C spin polarized |
| 129.4.105 | 53707 | Ge1Mn1Sr1 | C spin polarized |
| 129.4.109 | 66953 | Ge1Mg1Mn1 | C spin polarized |
| 129.4.110 | 66952 | Ge1Mg1Mn1 | C spin polarized |
| 129.4.111 | 66948 | Ge1Mg1Mn1 | C spin polarized |
| 129.4.113 | 86777 | Ge1Li2O5V1 | C spin polarized |
| 129.4.115 | 85862 | Ge1La1Ru1 | A |
| 129.4.116 | 95023 | Ge1Ho1Ti1 | C spin polarized |
| 129.4.117 | 95021 | Ge1Ho1Ti1 | C spin polarized |
| 129.4.127 | 194319 | Fe1N1Se1 | C spin polarized |
| 129.4.128 | 85853 | Fe1La1Si1 | A |
| 129.4.132 | 189941 | F1Mn1Sb1Sr1 | C spin polarized |
| 129.4.133 | 189939 | F1Mn1P1Sr1 | C spin polarized |
| 129.4.136 | 88460 | F1Fe1O3Sr2 | C spin polarized |
| 129.4.143 | 412295 | Cu2Sb1 | A |
| 129.4.145 | 103041 | Cu2La1Sn2 | C spin polarized |
| 129.4.149 | 261278 | Cu1Np1O1P1 | C spin polarized |
| 129.4.162 | 41869 | Cs1Mn1Sb1 | C spin polarized |
| 129.4.163 | 89599 | Cs1Mn1P1 | A |
| 129.4.164 | 89598 | Cs1Mn1P1 | C spin polarized |
| 129.4.167 | 80203 | Co1La1O1P1 | C spin polarized |
| 129.4.171 | 418853 | Co1Ge1Mg1 | C spin polarized |
| 129.4.173 | 237581 | Cl1Ni1O3Sr2 | C spin polarized |
| 129.4.178 | 50988 | Cl1K1Mo1O5P1 | C spin polarized |
| 129.4.181 | 80798 | Cl1Fe1O4W1 | C spin polarized |
| 129.4.182 | 93508 | Cl1Fe1O3Sr2 | C spin polarized |
| 129.4.183 | 49516 | Cl1Fe1Mo1O4 | C spin polarized |
| 129.4.184 | 202188 | Cl1Fe1Mo1O4 | C spin polarized |
| 129.4.185 | 202187 | Cl1Fe1Mo1O4 | C spin polarized |
| 129.4.186 | 202186 | Cl1Fe1Mo1O4 | C spin polarized |
| 129.4.188 | 85449 | Ce1O1Sb1Zn1 | C spin polarized |
| 129.4.189 | 80201 | Ce1O1P1Ru1 | C spin polarized |
| 129.4.190 | 85850 | Ce1Mn1Si1 | C spin polarized |
| 129.4.195 | 85858 | Ce1Ge1Mn1 | C spin polarized |
| 129.4.198 | 94469 | Ce1F1S1 | Ce3+ d1 C spin polarized |
| 129.4.199 | 93477 | Ce1F1S1 | C spin polarized |
| 129.4.202 | 416521 | Ce1Cu1O1Te1 | A |
| 129.4.203 | 96344 | Ce1Cu1O1S1 | C spin polarized |
| 129.4.205 | 426282 | Ce1Cl1Te1 | C spin polarized |
| 129.4.206 | 72154 | Ce1Cl1O1 | C spin polarized |
| 129.4.207 | 412069 | Ce1Cl1O1 | A |
| 129.4.211 | 66959 | Ca1Mn1Sn1 | A |

Continued on next column

**TABLE XX– continued**

| TQC | ICSD | Stoich. Formula | SubSG |
|---|---|---|---|
| 129.4.212 | 66958 | Ca1Mn1Sn1 | C spin polarized |
| 129.4.213 | 66950 | Ca1Mn1Sn1 | C spin polarized |
| 129.4.214 | 66955 | Ca1Mn1Si1 | A |
| 129.4.215 | 66954 | Ca1Mn1Si1 | C spin polarized |
| 129.4.216 | 66949 | Ca1Mn1Si1 | C spin polarized |
| 129.4.217 | 66957 | Ca1Ge1Mn1 | C spin polarized |
| 129.4.218 | 66956 | Ca1Ge1Mn1 | C spin polarized |
| 129.4.219 | 52758 | Ca1Ge1Mn1 | C spin polarized |
| 129.4.220 | 428148 | Ca1Co1Si1 | C spin polarized |
| 129.4.227 | 93509 | Br1Fe1O3Sr2 | C spin polarized |
| 129.4.228 | 10454 | Bi2Ca1Mn1 | C spin polarized |
| 129.4.229 | 60738 | Bi1Mn1Na1 | C spin polarized |
| 129.4.230 | 60737 | Bi1Mn1Na1 | C spin polarized |
| 129.4.231 | 60736 | Bi1Mn1Na1 | A |
| 129.4.232 | 80 | Bi1Ce1O1S2 | C spin polarized |
| 129.4.234 | 88950 | Ba1Mn2O5Y1 | C spin polarized |
| 129.4.235 | 88949 | Ba1Mn2O5Y1 | A |
| 129.4.236 | 66965 | Ba1Ge1Mn1 | C spin polarized |
| 129.4.237 | 66964 | Ba1Ge1Mn1 | A |
| 129.4.238 | 42370 | Ba1Ge1Mn1 | C spin polarized |
| 129.4.239 | 248366 | Ba1F1Mn1Sb1 | C spin polarized |
| 129.4.240 | 248364 | Ba1F1Mn1Sb1 | C spin polarized |
| 129.4.241 | 82639 | Ba1Co1S2 | C spin polarized ?????????? |
| 129.4.242 | 426033 | Ba1Bi1F1Mn1 | C spin polarized |
| 129.4.245 | 262387 | Au3Ge1Sr1 | C spin polarized |
| 129.4.246 | 262386 | Au3Ge1Sr1 | C spin polarized |
| 129.4.247 | 262385 | Au3Ge1Sr1 | A |
| 129.4.249 | 238705 | Au1Bi2Ce1 | C spin polarized |
| 129.4.250 | 75132 | As3Cu2O1U2 | C spin polarized |
| 129.4.252 | 68147 | As2La1Ni2 | C spin polarized |
| 129.4.253 | 186992 | As2Ir2La1 | A |
| 129.4.255 | 40782 | As2Cu1U1 | A |
| 129.4.259 | 90313 | As2Co1U1 | C spin polarized |
| 129.4.261 | 87141 | As1Se1U1 | C spin polarized |
| 129.4.262 | 66946 | As1Se1U1 | C spin polarized |
| 129.4.264 | 87140 | As1S1U1 | C spin polarized |
| 129.4.265 | 66945 | As1S1U1 | C spin polarized |
| 129.4.270 | 415629 | As1Mn2 | C spin polarized |
| 129.4.272 | 89597 | As1Mn1Rb1 | A |
| 129.4.273 | 89596 | As1Mn1Rb1 | C spin polarized |
| 129.4.274 | 60732 | As1Mn1Na1 | C spin polarized |
| 129.4.275 | 60731 | As1Mn1Na1 | C spin polarized |
| 129.4.276 | 60730 | As1Mn1Na1 | C spin polarized |
| 129.4.278 | 60742 | As1Li1Mn1 | A |
| 129.4.279 | 60741 | As1Li1Mn1 | C spin polarized |
| 129.4.280 | 60740 | As1Li1Mn1 | C spin polarized |
| 129.4.281 | 60739 | As1Li1Mn1 | C spin polarized |
| 129.4.290 | 60745 | As1K1Mn1 | C spin polarized |
| 129.4.291 | 60744 | As1K1Mn1 | C spin polarized |
| 129.4.292 | 60743 | As1K1Mn1 | C spin polarized |
| 129.4.293 | 41883 | As1K1Mn1 | A |
| 129.4.296 | 415628 | As1Fe2 | C spin polarized |
| 129.4.303 | 189940 | As1F1Mn1Sr1 | C spin polarized |
| 129.4.305 | 15142 | As1Cu2 | A |
| 129.4.306 | 423230 | As1Cu1Mn1 | C spin polarized |
| 129.4.307 | 40573 | As1Cs1Mn1 | C spin polarized |
| 129.4.308 | 40572 | As1Cs1Mn1 | C spin polarized |
| 129.4.316 | 174440 | As1Ce1Ni1O1 | A |
| 129.4.317 | 421999 | As1Ce1Fe1O1 | A superconductor |
| 129.4.319 | 162817 | As1Ce1Fe1O1 | C spin polarized |

Continued on next column

**TABLE XX– continued**

| TQC | ICSD | Stoich. Formula | SubSG |
|---|---|---|---|
| 129.4.320 | 88268 | As1Cd1Ce1O1 | A |
| 129.4.321 | 248365 | As1Ba1F1Mn1 | A |
| 129.4.328 | 79450 | Ag1Ce1Sb2 | A |
| 129.4.329 | 79425 | Ag1Ce1Sb2 | C spin polarized |
| 129.4.330 | 93150 | Ag1As2Ce1 | C spin polarized |
| 130.4.1 | 71313 | Bi2Cu1O4 | A with band gap |
| 130.4.2 | 69308 | Bi2Cu1O4 | A with band gap |
| 130.4.3 | 69307 | Bi2Cu1O4 | A with band gap |
| 130.4.4 | 68815 | Bi2Cu1O4 | A with band gap |
| 130.4.5 | 68814 | Bi2Cu1O4 | A with band gap |
| 130.4.6 | 68813 | Bi2Cu1O4 | A with band gap |
| 130.4.7 | 68812 | Bi2Cu1O4 | A with band gap |
| 130.4.8 | 56390 | Bi2Cu1O4 | A with band gap |
| 130.4.9 | 202994 | Bi2Cu1O4 | A with band gap |
| 130.4.10 | 15865 | Bi2Cu1O4 | A with band gap |
| 131.2.1 | 50987 | I1Pd1Te1 | A |
| 131.2.2 | 6159 | Ca1O4Pt2 | A |
| 131.4.1 | 23300 | C2B2La1 | A |
| 135.3.1 | 86492 | Ni1O4Sb2 | C spin polarized |
| 135.3.2 | 86491 | Ni1O4Sb2 | C spin polarized |
| 135.3.3 | 86490 | Ni1O4Sb2 | C spin polarized |
| 135.3.4 | 86489 | Ni1O4Sb2 | C spin polarized |
| 135.3.5 | 41228 | Ni1O4Sb2 | C spin polarized |
| 135.3.6 | 41227 | Ni1O4Sb2 | C spin polarized |
| 135.3.7 | 41226 | Ni1O4Sb2 | C spin polarized |
| 135.3.8 | 41225 | Ni1O4Sb2 | C spin polarized |
| 135.4.1 | 262627 | Co1O4Sb2 | C spin polarized |
| 135.4.2 | 262626 | Co1O4Sb2 | C spin polarized |
| 135.4.3 | 4287 | As2Cu1O4 | A |
| 136.1.3 | 421229 | Li2Sr3 | A |
| 136.1.4 | 93213 | Fe2O6Te1 | C magnetic with large band gap |
| 136.1.5 | 33797 | Co2K4O5Rb2 | C spin polarized |
| 136.1.6 | 237545 | Cd4Pt2Sr1 | A |
| 136.1.7 | 62553 | Ca1Cu4P2 | A |
| 136.1.10 | 260392 | Ba1Hg2Tl2 | A |
| 136.1.11 | 260391 | Ba1Hg2Tl2 | A |
| 136.1.12 | 237544 | Ba1Cd4Pt2 | A |
| 136.3.1 | 76490 | U1 | C spin polarized |
| 136.3.2 | 43218 | U1 | C spin polarized |
| 136.3.4 | 2576 | O6V2W1 | C spin polarized |
| 136.3.5 | 2575 | O6V2W1 | V d2 C spin polarized |
| 136.3.6 | 34234 | O2Pb1 | A |
| 136.3.7 | 23292 | O2Pb1 | A |
| 136.3.8 | 40435 | Ni4P2Zr1 | A |
| 136.3.9 | 85896 | Ni4P2U1 | C spin polarized |
| 136.3.10 | 405564 | Ni2Sn1Zr2 | A |
| 136.3.11 | 407549 | In3Rh1 | A |
| 136.3.12 | 407548 | In3Ir1 | A |
| 136.3.13 | 404888 | In1Pd2Zr2 | A |
| 136.3.14 | 107332 | In1Pd2Zr2 | A |
| 136.3.15 | 405563 | In1Ni2Zr2 | A |
| 136.3.16 | 404848 | In1Ni2Ti2 | A |
| 136.3.17 | 107330 | In1Ni2Ti2 | A |
| 136.3.18 | 407546 | Hf2Ni2Sn1 | C spin polarized |
| 136.3.19 | 194982 | Fe4Ge2U1 | C spin polarized |
| 136.3.23 | 412062 | Co4In12 | A |
| 136.3.24 | 412061 | Co4Ga12 | A |
| 136.3.25 | 64680 | Co2Na4O5Rb2 | C spin polarized |
| 136.3.26 | 64678 | Co2K2Na4O5 | C spin polarized |
| 136.3.27 | 72387 | Co1Li1N2Sr2 | C spin polarized |

Continued on next column

**TABLE XX– continued**

| TQC | ICSD | Stoich. Formula | SubSG |
|---|---|---|---|
| 136.3.28 | 407550 | Co1In3 | A |
| 136.3.30 | 417669 | C2Ce2W1 | C spin polarized |
| 136.3.31 | 50005 | Bi2Mg1O6 | A |
| 136.4.1 | 66665 | O2V1 | C spin polarized |
| 136.4.2 | 4110 | O2V1 | C spin polarized |
| 136.4.3 | 34421 | O2V1 | C spin polarized |
| 136.4.4 | 34420 | O2V1 | C spin polarized |
| 136.4.5 | 34418 | O2V1 | C spin polarized |
| 136.4.6 | 237337 | O2V1 | C spin polarized |
| 136.4.7 | 1504 | O2V1 | C spin polarized |
| 136.4.8 | 10141 | O2V1 | C spin polarized |
| 136.4.9 | 154021 | O2Re1 | A |
| 136.4.10 | 189495 | Ni4P2Y1 | A |
| 136.4.11 | 424579 | Ni1O2Rb3 | C spin polarized |
| 136.4.12 | 75198 | Nb1O2 | A |
| 136.4.13 | 73716 | Mn1O2 | C spin polarized |
| 136.4.14 | 393 | Mn1O2 | C spin polarized |
| 136.4.15 | 262579 | K3Ni1O2 | C spin polarized |
| 136.4.16 | 84577 | Ir1O2 | A |
| 136.4.19 | 68525 | Fe4P2Sc1 | A |
| 136.4.21 | 290252 | F6Li2V1 | C spin polarized |
| 136.4.23 | 62768 | F2V1 | C spin polarized |
| 136.4.24 | 38069 | F2V1 | C spin polarized |
| 136.4.25 | 32552 | F2V1 | C spin polarized |
| 136.4.26 | 201245 | F2V1 | C spin polarized |
| 136.4.27 | 9165 | F2Mn1 | C spin polarized |
| 136.4.28 | 68740 | F2Mn1 | C spin polarized |
| 136.4.29 | 68737 | F2Mn1 | C spin polarized |
| 136.4.30 | 68736 | F2Mn1 | C spin polarized |
| 136.4.31 | 63654 | F2Mn1 | C spin polarized |
| 136.4.32 | 200642 | F2Mn1 | C spin polarized |
| 136.4.33 | 200641 | F2Mn1 | C spin polarized |
| 136.4.34 | 200640 | F2Mn1 | C spin polarized |
| 136.4.35 | 84789 | Cu1O6Sb2 | C spin polarized long range magnetic ordering in CuSb2O6 |
| 136.4.36 | 424578 | Cs3Ni1O2 | C spin polarized |
| 136.4.37 | 203094 | Co1O6Sb2 | C spin polarized |
| 136.4.38 | 9167 | Co1F2 | C spin polarized |
| 136.4.39 | 73460 | Co1F2 | C spin polarized |
| 136.4.40 | 280604 | Co1F2 | C spin polarized |
| 136.4.41 | 14144 | Co1F2 | C spin polarized |
| 137.1.1 | 250760 | Np2Se5 | C spin polarized |
| 137.1.2 | 248874 | Ba2S6U1 | C spin polarized |
| 137.3.1 | 39595 | Nb4O5 | A |
| 137.4.1 | 62688 | Co1Li6O4 | C spin polarized |
| 139.1.2 | 42119 | Sb1Sr2 | A |
| 139.1.6 | 251818 | Pt2Th1 | C spin polarized |
| 139.1.8 | 247272 | Pd3Tl1 | A |
| 139.1.9 | 61195 | Pd2Sb2Sr1 | A |
| 139.1.10 | 105758 | Pd1Zr2 | A |
| 139.1.11 | 42966 | P2Ru2Sr1 | A |
| 139.1.12 | 50186 | P2Rh2Sr1 | A |
| 139.1.19 | 68211 | Ni2Si2Th1 | C spin polarized |
| 139.1.23 | 91220 | Na2O1Sb2Ti2 | A |
| 139.1.24 | 91219 | Na2O1Sb2Ti2 | A |
| 139.1.25 | 91218 | Na2O1Sb2Ti2 | A |
| 139.1.26 | 91217 | Na2O1Sb2Ti2 | A |
| 139.1.27 | 91216 | Na2O1Sb2Ti2 | A |
| 139.1.28 | 91215 | Na2O1Sb2Ti2 | A |
| 139.1.29 | 91214 | Na2O1Sb2Ti2 | A superconductor |

Continued on next column

**TABLE XX– continued**

| TQC | ICSD | Stoich. Formula | SubSG |
|---|---|---|---|
| 139.1.30 | 91213 | Na2O1Sb2Ti2 | A |
| 139.1.31 | 91212 | Na2O1Sb2Ti2 | A |
| 139.1.32 | 91211 | Na2O1Sb2Ti2 | A |
| 139.1.33 | 91210 | Na2O1Sb2Ti2 | A |
| 139.1.34 | 91209 | Na2O1Sb2Ti2 | A |
| 139.1.35 | 91208 | Na2O1Sb2Ti2 | A |
| 139.1.36 | 91207 | Na2O1Sb2Ti2 | A |
| 139.1.37 | 91206 | Na2O1Sb2Ti2 | A |
| 139.1.38 | 91205 | Na2O1Sb2Ti2 | A |
| 139.1.39 | 91204 | Na2O1Sb2Ti2 | A |
| 139.1.40 | 91203 | Na2O1Sb2Ti2 | A |
| 139.1.41 | 91202 | Na2O1Sb2Ti2 | A |
| 139.1.42 | 69648 | Na2O1Sb2Ti2 | A |
| 139.1.46 | 152123 | Mo1O4Sr2 | A |
| 139.1.51 | 150228 | Mg1Pd3 | A |
| 139.1.52 | 42733 | La2Sb1 | A |
| 139.1.53 | 423290 | La2Sb1 | A |
| 139.1.54 | 10442 | La2Sb1 | A |
| 139.1.56 | 69172 | La2Ni1O4 | C spin polarized |
| 139.1.58 | 181385 | La2Mn2O3Se2 | C spin polarized |
| 139.1.59 | 427126 | La1Sb1Sc1 | A |
| 139.1.60 | 421342 | K2Mg5Sn3 | A |
| 139.1.61 | 420251 | In1Pd3 | A |
| 139.1.64 | 427211 | Hg2In2Sr1 | A |
| 139.1.73 | 165737 | Ge2Ir2Sr1 | A |
| 139.1.75 | 87316 | Ge1Se1U1 | C spin polarized |
| 139.1.82 | 68209 | Fe2Si2Th1 | C spin polarized |
| 139.1.83 | 10467 | Fe2P2Sr1 | A |
| 139.1.88 | 74421 | Fe1O4Sr2 | C spin polarized |
| 139.1.89 | 74420 | Fe1O4Sr2 | C spin polarized |
| 139.1.90 | 74419 | Fe1O4Sr2 | C spin polarized |
| 139.1.91 | 69849 | Fe1O4Sr2 | C spin polarized |
| 139.1.92 | 69682 | F4Ni1Rb2 | C spin polarized |
| 139.1.93 | 73450 | F4K2Ni1 | C spin polarized |
| 139.1.106 | 86002 | Cu2Ge2U1 | C spin polarized |
| 139.1.107 | 86001 | Cu2Ge2U1 | C spin polarized |
| 139.1.108 | 86000 | Cu2Ge2U1 | C spin polarized |
| 139.1.109 | 85999 | Cu2Ge2U1 | C spin polarized |
| 139.1.111 | 424224 | Cu2Ge2Sr1 | A |
| 139.1.115 | 290194 | Cu1Ge1Hf1 | A |
| 139.1.117 | 245595 | Cr1O4Sr2 | C spin polarized |
| 139.1.122 | 425470 | Co2Si2Sr1 | A |
| 139.1.135 | 99942 | Cl2Fe2O5Sr3 | C spin polarized |
| 139.1.137 | 33258 | Cl2Fe2O5Sr3 | C magnetic with large band gap |
| 139.1.142 | 409587 | Ce2Sb1 | C spin polarized |
| 139.1.146 | 290821 | Ce1Sc1Si1 | C spin polarized |
| 139.1.147 | 411677 | Ce1Ir2Si2 | A |
| 139.1.148 | 85893 | Ce1Co2P2 | A |
| 139.1.149 | 370021 | Cd2Ga2Sr1 | A |
| 139.1.151 | 59649 | Ca1Si2Zn2 | A |
| 139.1.152 | 425469 | Ca1Rh2Si2 | A |
| 139.1.153 | 50185 | Ca1P2Rh2 | A |
| 139.1.156 | 10461 | Ca1Ni2P2 | A |
| 139.1.157 | 414231 | Ca1In1 | A |
| 139.1.158 | 52761 | Ca1Ge2Ru2 | A |
| 139.1.160 | 81755 | Ca1Ge2Ni2 | A |
| 139.1.162 | 81743 | Ca1Ge2Mn2 | C spin polarized |
| 139.1.164 | 52755 | Ca1Ge2Ir2 | A |
| 139.1.165 | 10463 | Ca1Fe2P2 | A |
| 139.1.167 | 420974 | Ca1Co2Si2 | A |

Continued on next column

**TABLE XX– continued**

| TQC | ICSD | Stoich. Formula | SubSG |
|---|---|---|---|
| 139.1.168 | 85892 | Ca1Co2P2 | A |
| 139.1.169 | 85891 | Ca1Co2P2 | A |
| 139.1.170 | 85890 | Ca1Co2P2 | A |
| 139.1.171 | 81751 | Ca1Co2Ge2 | A |
| 139.1.175 | 26478 | C2U1 | C spin polarized |
| 139.1.176 | 202305 | C2U1 | C spin polarized |
| 139.1.191 | 95097 | C1B2Ce1Rh2 | C spin polarized |
| 139.1.192 | 95098 | C1B2Ce1Ir2 | A |
| 139.1.201 | 107609 | Be1Pd2 | A |
| 139.1.202 | 58717 | Be12Mo1 | B |
| 139.1.213 | 2141 | Ba2Bi1 | A |
| 139.1.214 | 61197 | Ba1Pd2Sb2 | A |
| 139.1.215 | 50188 | Ba1P2Rh2 | A |
| 139.1.216 | 419531 | Ba1Mn2Sb2 | C spin polarized |
| 139.1.219 | 10469 | Ba1Mn2P2 | C spin polarized |
| 139.1.221 | 10468 | Ba1Fe2P2 | A |
| 139.1.222 | 420975 | Ba1Co2Ge2 | A |
| 139.1.223 | 174263 | Ba1Cd2Ge2 | A |
| 139.1.224 | 71642 | B2Ir2Zn1 | A |
| 139.1.229 | 418530 | Au2Ce1Si2 | C spin polarized |
| 139.1.232 | 417002 | As2Rh2Sr1 | A |
| 139.1.233 | 47170 | As2Pd2Sr1 | A |
| 139.1.237 | 78756 | As2Cu2Sr1 | A |
| 139.1.239 | 416983 | As2Ba1Rh2 | A |
| 139.1.240 | 41794 | As2Ba1Mn2 | C spin polarized |
| 139.1.241 | 236307 | As2Ba1Cu2 | A |
| 139.1.242 | 92989 | As1Ti1Zr1 | A |
| 139.3.2 | 73599 | Si2W1 | A |
| 139.3.4 | 40549 | Ru2Si2U1 | C spin polarized |
| 139.3.8 | 419760 | Ni2Pd1Se2 | A |
| 139.3.11 | 423722 | N2Sr1 | A |
| 139.3.12 | 411556 | N2Sr1 | A |
| 139.3.13 | 85756 | Mo1Si2 | A |
| 139.3.14 | 73598 | Mo1Si2 | A |
| 139.3.16 | 182115 | Mo1Si2 | A |
| 139.3.18 | 104986 | Mn12Th1 | A |
| 139.3.20 | 183307 | In5La9Ru4 | A |
| 139.3.21 | 165985 | Ge2Pd2Sr1 | A |
| 139.3.23 | 165736 | Ge2Ni2Sr1 | A |
| 139.3.30 | 23227 | Ga5Mg2 | A |
| 139.3.31 | 170782 | Ga4Sr1 | A |
| 139.3.32 | 402238 | Fe4Si7Zr4 | A |
| 139.3.41 | 98243 | Cr2Si2U1 | C spin polarized |
| 139.3.42 | 68207 | Cr2Si2Th1 | C spin polarized |
| 139.3.43 | 68210 | Co2Si2Th1 | C spin polarized |
| 139.3.49 | 171240 | Cl4Cu4S4W1 | A |
| 139.3.50 | 240484 | Cl4Cs2I2Pd1 | A |
| 139.3.52 | 1923 | Cl10K4O1W2 | C spin polarized |
| 139.3.53 | 89558 | Ce2O2Te1 | C spin polarized |
| 139.3.56 | 87320 | Ce1Mn2Si2 | C spin polarized |
| 139.3.57 | 87319 | Ce1Mn2Si2 | C spin polarized |
| 139.3.58 | 87318 | Ce1Mn2Si2 | C spin polarized |
| 139.3.59 | 425230 | Ce1Ge2Rh2 | C spin polarized |
| 139.3.60 | 423721 | Ca1N2 | A |
| 139.3.61 | 262736 | Ca1Ge3 | A |
| 139.3.62 | 425467 | Ca1Fe2Si2 | A high TC super |
| 139.3.63 | 81757 | Ca1Cu2Ge2 | A |
| 139.3.64 | 74668 | C2Ce1 | C spin polarized |
| 139.3.65 | 63556 | C2Ce1 | C spin polarized |
| 139.3.66 | 17022 | C2Ce1 | C spin polarized |
| 139.3.69 | 174269 | Ba1Si2Zn2 | A |

Continued on next column

**TABLE XX– continued**

| TQC | ICSD | Stoich. Formula | SubSG |
|---|---|---|---|
| 139.3.70 | 85408 | Ba1Ni2P2 | A |
| 139.3.74 | 95757 | Ba1Ir2P2 | A |
| 139.3.75 | 414230 | Ba1In4 | A |
| 139.3.77 | 55449 | Ba1Ga4 | A |
| 139.3.80 | 424033 | Au4Ca1Cd2 | A |
| 139.3.81 | 409552 | Au1O1Rb1 | A |
| 139.3.82 | 23005 | As2Ni2Sr1 | A |
| 139.3.83 | 182330 | As2Fe2Sr1 | A high TC superconductor |
| 139.3.84 | 163209 | As2Fe2Sr1 | A high TC with doping |
| 139.3.86 | 68146 | As2Ce1Ni2 | C spin polarized |
| 139.3.87 | 23004 | As2Ca1Ni2 | A |
| 139.3.88 | 61196 | As2Ba1Pd2 | A |
| 139.3.90 | 164197 | As2Ba1Ni2 | A |
| 139.3.91 | 87139 | As1Te1U1 | C spin polarized |
| 139.3.94 | 57724 | Al8Cu4U1 | C spin polarized |
| 139.3.97 | 281665 | Al3Ta1 | A |
| 139.3.98 | 153385 | Al2Ba1Si2 | A |
| 140.1.1 | 86940 | O3Ru1Sr1 | A |
| 140.1.2 | 174206 | Mo1O3Sr1 | A |
| 140.1.3 | 245120 | Mg2Pt1 | A |
| 140.1.4 | 88093 | Cr1Sb2 | C spin polarized |
| 140.1.5 | 86811 | Cr1Sb2 | C spin polarized |
| 140.1.6 | 281417 | Cl1Ga4La3 | A |
| 140.1.8 | 106369 | Cd3Sr5 | A |
| 140.1.9 | 106344 | Ca5Hg3 | A |
| 140.1.11 | 190146 | Bi1Ho5Ni2 | C spin polarized |
| 140.1.12 | 100084 | Ba5Sn3 | A |
| 140.1.13 | 247208 | B2Cu2O6Sr1 | C spin polarized |
| 140.1.14 | 247207 | B2Cu2O6Sr1 | C spin polarized |
| 140.2.1 | 1203 | Sn3Sr5 | A |
| 140.2.2 | 409797 | Sb2Ti1 | A |
| 140.2.3 | 249576 | Sb2Ti1 | A |
| 140.2.5 | 105479 | Ni1Zr2 | A |
| 140.2.7 | 409374 | Ge3Sr5 | A |
| 140.2.8 | 420322 | Cs9O4Sc1 | A |
| 140.2.9 | 419823 | Cs9In1O4 | A |
| 140.2.10 | 420320 | Cs9Ga1O4 | A |
| 140.2.11 | 418133 | Ca1Ni4Sn2 | A |
| 140.2.12 | 150146 | Bi2Fe1Ti4 | A |
| 140.2.14 | 420321 | Al1Cs9O4 | A |
| 140.3.1 | 90899 | Te1Tl1 | A |
| 140.3.2 | 90898 | Te1Tl1 | A |
| 140.3.3 | 90897 | Te1Tl1 | A |
| 140.3.4 | 90896 | Te1Tl1 | A |
| 140.3.5 | 90895 | Te1Tl1 | A |
| 140.3.6 | 69027 | Te1Tl1 | A |
| 140.3.7 | 52050 | Te1Tl1 | A |
| 140.3.8 | 82564 | Sb2Si1V4 | A |
| 140.3.10 | 418131 | Ni4Sn2Sr1 | A |
| 140.3.11 | 76377 | Mo5Si3 | A |
| 140.3.12 | 35756 | Mo5Si3 | A |
| 140.3.13 | 412252 | Ir1Li1Sn4 | A |
| 140.3.14 | 172149 | Ir1Li1Sn4 | A |
| 140.3.15 | 77281 | Ge3Mo5 | A |
| 140.3.16 | 103835 | Ga3Nb5 | A |
| 140.3.18 | 103677 | Fe1U6 | C spin polarized |
| 140.3.22 | 409909 | Fe1Sn2 | A |
| 140.3.23 | 247483 | Fe1Sn2 | A |
| 140.3.24 | 96143 | Fe1Sb2Ti5 | A |
| 140.3.25 | 422455 | Cu4Sn2Sr1 | A |

Continued on next column

**TABLE XX– continued**

| TQC | ICSD | Stoich. Formula | SubSG |
|---|---|---|---|
| 140.3.26 | 182104 | Cu4Sn2Sr1 | A |
| 140.3.27 | 260842 | Cs9In1O4 | A |
| 140.3.29 | 240270 | Ce5O1Pb3 | C spin polarized |
| 140.3.30 | 55584 | Ca5Zn3 | A |
| 140.3.31 | 93699 | Ca5Si3 | A |
| 140.3.32 | 93704 | Ca5Ge3 | A |
| 140.3.33 | 58668 | Ba5Pb3 | A |
| 140.3.34 | 73893 | Ba3Cr1O5 | C spin polarized |
| 140.4.1 | 40260 | Te3Tl5 | A |
| 140.4.2 | 195541 | Si3Ta5 | A |
| 140.4.3 | 189722 | Si3Ta5 | A |
| 140.4.4 | 95842 | Rh1Sb2Ti5 | A |
| 140.4.5 | 183452 | O3Sr1Tc1 | C spin polarized |
| 140.4.6 | 189717 | Nb5Si3 | A |
| 140.4.7 | 16499 | Nb5Si1Sn2 | A |
| 140.4.8 | 423763 | Nb20Si12 | A |
| 140.4.9 | 247482 | Mn1Sn2 | C spin polarized |
| 140.4.10 | 171198 | Mn1Sn2 | A or C |
| 140.4.11 | 412250 | Li1Ru1Sn4 | A |
| 140.4.12 | 261925 | La5Sn1Zn2 | A |
| 140.4.13 | 95073 | La5Si3 | A |
| 140.4.14 | 98409 | La1Ni9Si4 | A |
| 140.4.15 | 413805 | In3Sr5 | A |
| 140.4.16 | 165190 | H1Ca5Sn3 | A |
| 140.4.17 | 195542 | Ge3Ta5 | A |
| 140.4.18 | 2003 | Ge3Nb5 | A |
| 140.4.19 | 81469 | Ge1In4La3 | A |
| 140.4.20 | 414109 | Ga1K1Se4Sn1 | A |
| 140.4.21 | 79590 | Fe9La1Si4 | C spin polarized ferri |
| 140.4.22 | 75412 | F3K1Mn1 | C spin polarized |
| 140.4.23 | 165195 | F1Pb3Sr5 | A |
| 140.4.37 | 424219 | Cu9Sn4Sr1 | A or C |
| 140.4.38 | 422456 | Cu9Sn4Sr1 | A |
| 140.4.39 | 182105 | Cu9Sn4Sr1 | A or C is the same as C spin polarized |
| 140.4.40 | 412882 | Cu9Si4Sr1 | A |
| 140.4.41 | 412884 | Cu9Ge4Sr1 | A |
| 140.4.43 | 159248 | Cu1Sb2Ti5 | A |
| 140.4.44 | 69656 | Cu1F3Rb1 | C spin polarized |
| 140.4.45 | 9856 | Cu1F3K1 | C spin polarized |
| 140.4.46 | 8000 | Cu1F3K1 | C spin polarized |
| 140.4.47 | 67445 | Cu1F3K1 | C spin polarized |
| 140.4.48 | 67444 | Cu1F3K1 | C spin polarized |
| 140.4.49 | 21109 | Cu1F3K1 | C spin polarized |
| 140.4.50 | 21108 | Cu1F3K1 | C spin polarized |
| 140.4.51 | 182857 | Cu1F3K1 | C spin polarized |
| 140.4.52 | 182856 | Cu1F3K1 | A or C |
| 140.4.53 | 182855 | Cu1F3K1 | C spin polarized |
| 140.4.54 | 182854 | Cu1F3K1 | C spin polarized |
| 140.4.55 | 182853 | Cu1F3K1 | C spin polarized |
| 140.4.56 | 182852 | Cu1F3K1 | C spin polarized |
| 140.4.57 | 182851 | Cu1F3K1 | A or C |
| 140.4.58 | 182850 | Cu1F3K1 | C spin polarized |
| 140.4.59 | 182849 | Cu1F3K1 | C spin polarized |
| 140.4.60 | 158622 | Cu1F3K1 | C spin polarized |
| 140.4.61 | 420323 | Cs9Fe1O4 | A or C |
| 140.4.62 | 403102 | Cs3I5Mn1 | C spin polarized |
| 140.4.63 | 102740 | Co1Zr2 | A |
| 140.4.64 | 409910 | Co1Sn2 | A |
| 140.4.65 | 247484 | Co1Sn2 | A |
| 140.4.67 | 16397 | Cl5Co1Cs3 | C spin polarized |

Continued on next column

**TABLE XX– continued**

| TQC | ICSD | Stoich. Formula | SubSG |
|---|---|---|---|
| 140.4.68 | 15245 | Cl5Co1Cs3 | C spin polarized |
| 140.4.69 | 281418 | Ce3Cl1Ga4 | C spin polarized |
| 140.4.70 | 73478 | Cd1Cu1F4 | C spin polarized |
| 140.4.71 | 424221 | Ca1Cu9Sn4 | A |
| 140.4.72 | 417718 | Ca1Cu9Sn4 | A |
| 140.4.73 | 201135 | Br5Co1Cs3 | A or C |
| 140.4.74 | 150145 | Bi2Mn1Ti4 | A or C |
| 140.4.75 | 150147 | Bi2Co1Ti4 | A |
| 140.4.82 | 90639 | Ba1La2Mn1S5 | C spin polarized |
| 140.4.83 | 248378 | Ba1Ir1O3 | A or C MOTT and magnetic ? |
| 140.4.84 | 424220 | Ba1Cu9Sn4 | A or C |
| 140.4.85 | 422457 | Ba1Cu9Sn4 | A |
| 140.4.86 | 182106 | Ba1Cu9Sn4 | A |
| 140.4.87 | 412883 | Ba1Cu9Si4 | A |
| 140.4.88 | 424223 | Ba1Cu9Ge4 | A |
| 140.4.89 | 412885 | Ba1Cu9Ge4 | A or C |
| 140.4.92 | 95268 | Ba1Co1La2S5 | A or C |
| 140.4.94 | 75792 | B1Ni2 | A or C |
| 140.4.95 | 391330 | B1Fe2 | C spin polarized ferri |
| 140.4.96 | 391328 | B1Fe2 | A or C |
| 140.4.97 | 249220 | Au4In2Rb1 | A |
| 140.4.98 | 249221 | Au4In2K1 | A |
| 140.4.99 | 58416 | Au3Cd5 | A |
| 140.4.100 | 58404 | Au3Ca5 | A |
| 140.4.102 | 413559 | Al4Br4La5 | A |
| 140.4.103 | 42517 | Al2Cu1 | A |
| 140.4.104 | 172800 | Al2Cu1 | A |
| 140.4.105 | 245561 | Al1Ce1O3 | C spin polarized |
| 140.4.106 | 245554 | Al1Ce1O3 | C spin polarized |
| 140.4.107 | 245265 | Al1Ce1O3 | A or C |
| 140.4.108 | 245264 | Al1Ce1O3 | C spin polarized |
| 140.4.109 | 166284 | Al1Ce1O3 | A or C |
| 140.4.110 | 150277 | Al1Ce1O3 | C spin polarized |
| 141.4.6 | 68174 | Mn3O4 | C spin polarized |
| 141.4.7 | 107516 | Li1Sn1 | A |
| 141.4.9 | 98663 | Li1N2U1 | C spin polarized |
| 141.4.10 | 62536 | In2N1Sr4 | A |
| 141.4.17 | 80554 | Fe2O5P1 | C magnetic spin polarized |
| 141.4.18 | 174085 | Fe1Li1O2 | C spin polarized |
| 141.4.29 | 88961 | Cu1O4Rh2 | A |
| 141.4.30 | 59267 | Cu1O4Rh2 | C spin polarized |
| 141.4.31 | 426084 | Cu1Li2Sn2 | A |
| 141.4.32 | 251619 | Cu1Li2Sn2 | A |
| 141.4.33 | 75532 | Cu1Ir2S4 | A |
| 141.4.34 | 150187 | Cs1 | A |
| 141.4.35 | 84378 | Cr2Cu1O4 | C spin polarized |
| 141.4.36 | 80476 | Cr2Cu1O4 | C spin polarized |
| 141.4.37 | 246080 | Cr2Cu1O4 | C spin polarized |
| 141.4.38 | 95971 | Cr1O4Y1 | C spin polarized |
| 141.4.39 | 157933 | Cr1O4Y1 | C spin polarized |
| 141.4.43 | 173263 | Cr1Ho1O4 | C spin polarized |
| 141.4.44 | 169872 | Cr1Ho1O4 | C spin polarized |
| 141.4.46 | 81931 | Co9Si2Th1 | C spin polarized |
| 141.4.47 | 9309 | Cl4Pa1 | C spin polarized |
| 141.4.48 | 9311 | Cl4Np1 | C spin polarized |
| 141.4.49 | 75482 | Cl4Np1 | C spin polarized |
| 141.4.50 | 78075 | Ce1O4V1 | C spin polarized |
| 141.4.51 | 66033 | Ce1O4V1 | C spin polarized |
| 141.4.52 | 202425 | Ce1O4V1 | C spin polarized |

Continued on next column

**TABLE XX– continued**

| TQC | ICSD | Stoich. Formula | SubSG |
|---|---|---|---|
| 141.4.53 | 157326 | Ce1O4V1 | C spin polarized |
| 141.4.54 | 55349 | Au1Li2Sn2 | A |
| 141.4.58 | 426085 | Ag1Li2Sn2 | A |
| 142.3.1 | 174219 | Rh1Sn4 | A |
| 142.3.2 | 412573 | Ir1Sn4 | A |
| 146.3.1 | 237873 | Ni3O6Te1 | A or C MOTT and magnetic long-range antiferromagnetic order has |
| 146.3.3 | 26149 | H12Ni1O9S1 | A or C MOTT and magnetic long-range antiferromagnetic order has |
| 146.3.4 | 171013 | Cs4Se9Sn3Sr1 | A or C MOTT and magnetic long-range antiferromagnetic order has |
| 147.3.1 | 109513 | Cl18Cs2W6 | A or C MOTT and magnetic long-range antiferromagnetic order has |
| 148.1.3 | 37134 | O12Te1Tl6 | A |
| 148.1.7 | 171584 | Ni1O3Ti1 | C magnetic insulator |
| 148.1.8 | 59375 | Mo6Te8 | A |
| 148.1.10 | 46014 | Mo6S8Yb1 | A or CYb3+ magnetic moments perpendicularly to the c axis below |
| 148.1.11 | 46013 | Mo6S8Yb1 | A or CYb3+ magnetic moments perpendicularly to the c axis below |
| 148.1.12 | 201506 | Mo6S8Yb1 | A or CYb3+ magnetic moments perpendicularly to the c axis below |
| 148.1.13 | 62156 | Mo6S8Sr1 | C spin polarized |
| 148.1.18 | 182436 | Mn3O6Te1 | C spin polarized |
| 148.1.21 | 247553 | Mn1O3Ti1 | magnetic insulator |
| 148.1.23 | 73247 | Ge1Mn1O3 | A magnetic isolator |
| 148.1.24 | 73246 | Ge1Mn1O3 | C magnetic insulator |
| 148.1.25 | 73245 | Ge1Mn1O3 | C spin polarized |
| 148.1.26 | 73244 | Ge1Mn1O3 | C magnetic |
| 148.1.27 | 73243 | Ge1Mn1O3 | A |
| 148.1.28 | 69591 | Ge1Mn1O3 | C spin polarized |
| 148.1.33 | 106520 | Cr2Se3 | looks like stoner A or C than spin polarized |
| 148.1.34 | 79270 | Cr2Ge2Te6 | C spin polarized ferri |
| 148.1.36 | 79268 | Cr2Ge2Te6 | A |
| 148.1.37 | 252042 | Cr1Fe1O3 | C spin polarized |
| 148.1.38 | 39426 | Cl3Ti1 | Ti d1 C spin polarized |
| 148.1.42 | 62388 | Ca1Mo6S8 | A |
| 148.1.43 | 39242 | Br3Ti1 | A |
| 148.1.48 | 85489 | Ba1Mo6S8A | or C MOTT Chevrel |
| 148.1.49 | 65776 | Ba1Mo6S8 | A or C MOTT |
| 148.1.50 | 62157 | Ba1Mo6S8 | A or C MOTT |
| 148.1.53 | 91777 | Ag1Bi1O3 | A |
| 148.1.54 | 89432 | Ag1Bi1O3 | A |
| 148.3.2 | 200672 | Mo6S8Sn1 | A |
| 148.3.6 | 48107 | Co1O3Ti1 | C spin polarized |
| 148.3.7 | 16548 | Co1O3Ti1 | C spin polarized |
| 148.3.8 | 51513 | Cl18W6 | A |
| 148.3.9 | 109512 | Cl18W6 | A |
| 148.3.10 | 100599 | Br2Mo6S6 | A superconductor |
| 148.3.11 | 54547 | Au4Ca3 | A |

Continued on next column

**TABLE XX– continued**

| TQC | ICSD | Stoich. Formula | SubSG |
|---|---|---|---|
| 152.4.1 | 65692 | Te1 | A |
| 155.3.1 | 186536 | Li4Pt3Si1 | A |
| 156.3.1 | 110774 | Mn1Na1Te2 | C spin polarized |
| 156.3.2 | 50818 | Mn1Na1Se2 | C spin polarized |
| 156.3.3 | 110773 | Li1Mn1Te2 | C spin polarized |
| 156.3.4 | 50817 | Li1Mn1Se2 | C spin polarized |
| 159.3.1 | 237948 | Ba1Co4O7Y1 | C spin polarized |
| 159.4.1 | 252019 | Ir1Li2Si3 | A |
| 159.4.2 | 237950 | Ba1Co4O7Y1 | C spin polarized |
| 159.4.3 | 237949 | Ba1Co4O7Y1 | C spin polarized |
| 160.3.1 | 40054 | Ni1S1 | C spin polarized |
| 160.3.2 | 40053 | Ni1S1 | C spin polarized |
| 160.3.3 | 151602 | Ni1S1 | C spin polarized |
| 160.3.4 | 151601 | Ni1S1 | C spin polarized |
| 160.3.5 | 151600 | Ni1S1 | C spin polarized |
| 160.3.6 | 151599 | Ni1S1 | C spin polarized |
| 160.3.7 | 65725 | In2S4Zn1 | A |
| 160.3.8 | 42667 | In2S4Zn1 | A |
| 160.3.10 | 4084 | F4O1U1 | C spin polarized |
| 161.3.1 | 200999 | Li2O3Re1 | A |
| 161.3.2 | 35012 | Li1O3Re1 | A |
| 161.3.3 | 171497 | Ge3Li8Zn2 | A |
| 161.4.3 | 97591 | Bi1Fe1O3 | C spin polarized |
| 161.4.4 | 75324 | Bi1Fe1O3 | C spin polarized |
| 161.4.5 | 51664 | Bi1Fe1O3 | C spin polarized |
| 161.4.6 | 428380 | Bi1Fe1O3 | C spin polarized |
| 161.4.7 | 417312 | Bi1Fe1O3 | C spin polarized |
| 161.4.8 | 417311 | Bi1Fe1O3 | C spin polarized |
| 161.4.9 | 417310 | Bi1Fe1O3 | C spin polarized |
| 161.4.11 | 417308 | Bi1Fe1O3 | C spin polarized |
| 161.4.12 | 417307 | Bi1Fe1O3 | C spin polarized |
| 161.4.13 | 417306 | Bi1Fe1O3 | C spin polarized |
| 161.4.14 | 417305 | Bi1Fe1O3 | C spin polarized |
| 161.4.15 | 417304 | Bi1Fe1O3 | C spin polarized |
| 161.4.16 | 417303 | Bi1Fe1O3 | C spin polarized |
| 161.4.17 | 260119 | Bi1Fe1O3 | C spin polarized |
| 161.4.18 | 260118 | Bi1Fe1O3 | C spin polarized |
| 161.4.19 | 245827 | Bi1Fe1O3 | C spin polarized |
| 161.4.20 | 237368 | Bi1Fe1O3 | C spin polarized |
| 161.4.21 | 237367 | Bi1Fe1O3 | C spin polarized |
| 161.4.22 | 192511 | Bi1Fe1O3 | C spin polarized |
| 161.4.23 | 192509 | Bi1Fe1O3 | C spin polarized |
| 161.4.24 | 191940 | Bi1Fe1O3 | C spin polarized |
| 161.4.25 | 188396 | Bi1Fe1O3 | C spin polarized |
| 161.4.26 | 163688 | Bi1Fe1O3 | C spin polarized |
| 161.4.27 | 163687 | Bi1Fe1O3 | C spin polarized |
| 161.4.28 | 15299 | Bi1Fe1O3 | C spin polarized |
| 161.4.29 | 109370 | Bi1Fe1O3 | C spin polarized |
| 162.1.2 | 275 | Cr2O6U1 | C spin polarized |
| 162.3.1 | 260305 | Cl18K2W6 | A |
| 162.3.2 | 245119 | As2O6Pd1 | A |
| 162.3.3 | 195298 | As2O6Pd1 | A |
| 162.3.4 | 187099 | As2O6Pd1 | A |
| 162.3.5 | 187098 | As2O6Pd1 | A |
| 162.3.6 | 80349 | As2Ni1O6 | C spin polarized |
| 162.3.7 | 30360 | Ag1Se2Tl1 | A |
| 163.3.2 | 15039 | Cr2Te3 | C spin polarized |
| 163.4.1 | 62035 | Cl6Cr1Na3 | C spin polarized |
| 164.1.6 | 69557 | Sb2Te1 | A |
| 164.1.9 | 83642 | Pd1Te2 | A |
| 164.1.22 | 41790 | Mn2Sb2Sr1 | C spin polarized |

Continued on next column

**TABLE XX– continued**

| TQC | ICSD | Stoich. Formula | SubSG |
|---|---|---|---|
| 164.1.23 | 416094 | Mn2Sb2Sr1 | C spin polarized |
| 164.1.24 | 77670 | Mn2P2Sr1 | C spin polarized |
| 164.1.25 | 77669 | Mn2P2Sr1 | C spin polarized |
| 164.1.26 | 49019 | Mn2P2Sr1 | C spin polarized |
| 164.1.45 | 35585 | Cu2P2Zr1 | A |
| 164.1.57 | 41789 | Ca1Mn2Sb2 | C spin polarized |
| 164.1.58 | 416093 | Ca1Mn2Sb2 | C spin polarized |
| 164.1.59 | 163778 | Ca1Mn2Sb2 | C spin polarized |
| 164.1.60 | 49017 | Ca1Mn2P2 | C spin polarized |
| 164.1.65 | 409555 | C1Ta2 | A |
| 164.1.66 | 95113 | C1Nb2S2 | A |
| 164.1.67 | 95111 | C1Nb2S2 | A |
| 164.1.68 | 95109 | C1Nb2S2 | A |
| 164.1.73 | 71625 | C1F2Ho2 | C spin polarized |
| 164.1.77 | 250249 | Bi4Pb1Te7 | A |
| 164.1.82 | 41791 | Bi2Ca1Mn2 | C spin polarized |
| 164.1.83 | 30525 | Bi1Te1 | A |
| 164.1.84 | 100654 | Bi1Te1 | A |
| 164.1.87 | 409958 | Ba1Sn2 | A |
| 164.1.91 | 49020 | As2Mn2Sr1 | C spin polarized |
| 164.1.92 | 41793 | As2Mn2Sr1 | C spin polarized |
| 164.1.93 | 63174 | As2Ge5Te8 | A |
| 164.1.94 | 41792 | As2Ca1Mn2 | C spin polarized |
| 164.1.100 | 50976 | Al2Ce1Ge2 | A |
| 164.1.102 | 94501 | Al1La1Si2 | A |
| 164.3.2 | 12152 | Sb2Sr1Zn2 | A |
| 164.3.5 | 240197 | Ni3Sb2Zn1 | A |
| 164.3.7 | 416038 | Na1O8Rb3Ru2 | A |
| 164.3.10 | 416040 | K3Na1O8Ru2 | A |
| 164.3.11 | 107483 | Hg7Rb2 | A |
| 164.3.12 | 107482 | Hg7K2 | A |
| 164.3.32 | 59230 | Ga2Ni1S4 | A |
| 164.3.33 | 53287 | Cu2Hf1P2 | A |
| 164.3.37 | 12150 | Ca1Sb2Zn2 | A |
| 164.3.39 | 280743 | C1Sc2 | A |
| 164.3.40 | 246280 | Ba3Ca1Ir2O9 | C spin polarized |
| 164.3.41 | 423734 | Ba2Ni3 | A |
| 164.3.43 | 100049 | Ba1Bi2Mg2 | A |
| 164.3.47 | 82674 | Al1Ce1Si2 | C spin polarized |
| 165.1.1 | 154809 | H3Y1 | A but with wide gap please check |
| 165.3.1 | 172186 | Co4Nb2O9 | C spin polarized |
| 165.4.1 | 165378 | Cs2Mn1S6Te2 | C spin polarized |
| 165.4.2 | 240159 | Ca5Cu1Pb3 | A |
| 166.1.15 | 410956 | Sn3Sr1 | A |
| 166.1.17 | 152188 | Sb8Te3 | A |
| 166.1.18 | 152189 | Sb48Te9 | A |
| 166.1.27 | 2084 | Sb2Te3 | A |
| 166.1.29 | 193347 | Sb2Te3 | A |
| 166.1.30 | 193346 | Sb2Te3 | A |
| 166.1.31 | 193345 | Sb2Te3 | A |
| 166.1.32 | 193344 | Sb2Te3 | A |
| 166.1.33 | 193343 | Sb2Te3 | A |
| 166.1.34 | 193342 | Sb2Te3 | A |
| 166.1.35 | 193341 | Sb2Te3 | A |
| 166.1.36 | 192780 | Sb2Te3 | A |
| 166.1.49 | 52295 | Sb2Se2Te1 | A |
| 166.1.51 | 2085 | Sb2Se1Te2 | A |
| 166.1.61 | 64697 | Sb1 | A |
| 166.1.62 | 64696 | Sb1 | A |
| 166.1.63 | 64695 | Sb1 | A |

Continued on next column

**TABLE XX– continued**

| TQC | ICSD | Stoich. Formula | SubSG |
|---|---|---|---|
| 166.1.64 | 55402 | Sb1 | A |
| 166.1.66 | 409754 | Sb1 | A |
| 166.1.67 | 161493 | Sb1 | A |
| 166.1.90 | 422328 | Ni3S2Tl2 | A |
| 166.1.95 | 402458 | Ni3S2Sn2 | A shandite - really please check |
| 166.1.99 | 402453 | Ni3Pb2Se2 | A |
| 166.1.103 | 417632 | Ni3Pb2S2 | A |
| 166.1.124 | 415258 | In2Ni3S2 | A |
| 166.1.126 | 172602 | I1N1Sr2 | A |
| 166.1.153 | 40981 | H1Cl1Sc1 | A |
| 166.1.158 | 99161 | Ge2Li4Zr1 | A |
| 166.1.166 | 102891 | Ga7Rb1 | A |
| 166.1.171 | 35386 | Ga1Ge1Te1 | A |
| 166.1.172 | 2513 | Ga1 | A |
| 166.1.187 | 71 | Fe3Sn2 | C spin polarized |
| 166.1.205 | 184011 | F1Ho1O1 | C spin polarized |
| 166.1.221 | 80259 | Cu1I1 | C spin polarized |
| 166.1.229 | 102864 | Cs1Ga7 | B |
| 166.1.249 | 159894 | Co3Y1 | A |
| 166.1.250 | 159890 | Co3Y1 | A |
| 166.1.261 | 246307 | Co1Li1O2 | C spin polarized |
| 166.1.265 | 869 | Cl1Zr1 | D DFT PROBLEM!!! |
| 166.1.268 | 24410 | Cl1La1 | D DFT problem |
| 166.1.277 | 409057 | Ca1Si2 | A |
| 166.1.283 | 193539 | Ca1Si2 | A |
| 166.1.287 | 245612 | Ca1Ge2 | A |
| 166.1.293 | 74178 | C2Ba1Ce1F1O6 | C spin polarized |
| 166.1.294 | 96188 | C1Y2 | A superconductor up to 18K |
| 166.1.295 | 22283 | C1Y2 | A |
| 166.1.300 | 95112 | C1Nb2S2 | A |
| 166.1.301 | 95110 | C1Nb2S2 | A |
| 166.1.308 | 1168 | Br1Zr1 | D DFT PROBLEM!!! |
| 166.1.309 | 23354 | Br1La1 | D DFT PROBLEM!!! |
| 166.1.311 | 95551 | Bi6Pb1Te10 | A |
| 166.1.313 | 74348 | Bi2Te3 | A |
| 166.1.317 | 193340 | Bi2Te3 | A |
| 166.1.318 | 193339 | Bi2Te3 | A |
| 166.1.319 | 193338 | Bi2Te3 | A |
| 166.1.320 | 193337 | Bi2Te3 | A |
| 166.1.321 | 193336 | Bi2Te3 | A |
| 166.1.322 | 193335 | Bi2Te3 | A |
| 166.1.323 | 193334 | Bi2Te3 | A |
| 166.1.324 | 193333 | Bi2Te3 | A |
| 166.1.325 | 193332 | Bi2Te3 | A |
| 166.1.326 | 193331 | Bi2Te3 | A |
| 166.1.327 | 193330 | Bi2Te3 | A |
| 166.1.328 | 184631 | Bi2Te3 | A |
| 166.1.333 | 165226 | Bi2Se3 | A |
| 166.1.334 | 54838 | Bi2Se2Te1 | A |
| 166.1.335 | 43512 | Bi2Se1Te2 | A |
| 166.1.339 | 64705 | Bi1 | A |
| 166.1.340 | 64704 | Bi1 | A |
| 166.1.341 | 64703 | Bi1 | A |
| 166.1.352 | 58724 | Be17Nb2 | A |
| 166.1.359 | 58660 | Ba2Mg17 | A |
| 166.1.361 | 419973 | Ba1Pb3 | A |
| 166.1.371 | 39554 | B4Mo2 | A |
| 166.1.375 | 418398 | B2Mo1 | A |
| 166.1.376 | 418397 | B2Mo1 | A |

Continued on next column

**TABLE XX– continued**

| TQC | ICSD | Stoich. Formula | SubSG |
|---|---|---|---|
| 166.1.377 | 40907 | B2Mo1 | A |
| 166.1.381 | 262037 | As3Cd4Rb1 | A |
| 166.1.382 | 262035 | As3Cd4Na1 | A |
| 166.1.383 | 262032 | As3Cd4K1 | A |
| 166.1.385 | 82371 | As2Sn2Sr1 | A |
| 166.1.388 | 422526 | As2Ca1Ga2 | A |
| 166.1.390 | 16518 | As1 | A |
| 166.1.391 | 16517 | As1 | A |
| 166.1.392 | 16516 | As1 | A |
| 166.1.396 | 420092 | Al10Ba7 | A |
| 166.1.397 | 424311 | Ag4Sb2Sr1 | A |
| 166.1.398 | 416373 | Ag2Ni1O2 | A |
| 166.1.399 | 412279 | Ag2Ni1O2 | A |
| 166.1.400 | 172558 | Ag2Ni1O2 | A |
| 166.1.410 | 251620 | Ag1Ge1Li2 | A |
| 166.3.28 | 104760 | Li5Na2Sn4 | A |
| 166.3.167 | 151863 | Ce2Fe17 | C spin polarized |
| 166.3.168 | 151862 | Ce2Fe17 | C spin polarized |
| 166.3.169 | 109483 | Ce2Fe17 | C spin polarized |
| 166.3.187 | 240501 | Ba3Co10O17 | C spin polarized |
| 167.1.1 | 83385 | S3Yb2 | A |
| 167.1.2 | 75248 | S3Yb2 | A |
| 167.1.3 | 95761 | O3V2 | C spin polarized |
| 167.1.4 | 95760 | O3V2 | C magnetic with large band gap |
| 167.1.5 | 94768 | O3V2 | C magnetic with large band gap |
| 167.1.7 | 64786 | O3V2 | C magnetic with large band gap |
| 167.1.12 | 201106 | O3V2 | C magnetic with large band gap |
| 167.1.13 | 1874 | O3V2 | C magnetic with large band gap |
| 167.1.14 | 1873 | O3V2 | C magnetic with large band gap |
| 167.1.16 | 1870 | O3V2 | C magnetic with large band gap |
| 167.1.17 | 1869 | O3V2 | C magnetic with large band gap |
| 167.1.18 | 1473 | O3V2 | C magnetic with large band gap |
| 167.1.28 | 88418 | Fe2O3 | C magnetic with large band gap |
| 167.1.29 | 88417 | Fe2O3 | C magnetic with large band gap |
| 167.1.30 | 82904 | Fe2O3 | C magnetic with large band gap |
| 167.1.31 | 82903 | Fe2O3 | C magnetic with large band gap |
| 167.1.32 | 82902 | Fe2O3 | C magnetic with large band gap |
| 167.1.33 | 82137 | Fe2O3 | C magnetic with large band gap |
| 167.1.34 | 82136 | Fe2O3 | C magnetic with large band gap |
| 167.1.35 | 82135 | Fe2O3 | C magnetic with large band gap |
| 167.1.36 | 82134 | Fe2O3 | C magnetic with large band gap |
| 167.1.37 | 81248 | Fe2O3 | C magnetic with large band gap |

Continued on next column

**TABLE XX– continued**

| TQC | ICSD | Stoich. Formula | SubSG |
|---|---|---|---|
| 167.1.38 | 71194 | Fe2O3 | C magnetic with large band gap |
| 167.1.40 | 64599 | Fe2O3 | C magnetic with large band gap |
| 167.1.41 | 56372 | Fe2O3 | C magnetic with large band gap |
| 167.1.43 | 415251 | Fe2O3 | C magnetic with large band gap |
| 167.1.44 | 40142 | Fe2O3 | C magnetic with large band gap |
| 167.1.46 | 245851 | Fe2O3 | C magnetic with large band gap |
| 167.1.54 | 201096 | Fe2O3 | C magnetic with large band gap |
| 167.1.66 | 173655 | Fe2O3 | C magnetic with large band gap |
| 167.1.67 | 173653 | Fe2O3 | C magnetic with large band gap |
| 167.1.80 | 15840 | Fe2O3 | C magnetic with large band gap |
| 167.1.81 | 154191 | Fe2O3 | C magnetic with large band gap |
| 167.1.82 | 154190 | Fe2O3 | C magnetic with large band gap |
| 167.1.83 | 173470 | Cr2O3 | C magnetic with large band gap |
| 167.1.84 | 201966 | Cr2Cs3I9 | C spin polarized |
| 167.1.89 | 97540 | Ba3Cr2S6 | C spin polarized |
| 167.1.90 | 402170 | Au3Ga7Yb1 | A |
| 167.1.92 | 391101 | Al7Au3Ce1 | A |
| 167.3.22 | 87773 | Ni1O6Pt1Sr3 | A |
| 167.3.28 | 77645 | Ga5N1Sr6 | A |
| 167.3.34 | 9428 | Cu1La1O3 | C spin polarized |
| 167.3.37 | 75577 | Cr2O3 | C magnetic with large band gap |
| 167.3.40 | 290245 | Cr2O3 | C wide band gap spin polarized gap ¿ 2 eV |
| 167.3.41 | 290244 | Cr2O3 | C magnetic with large band gap |
| 167.3.42 | 290243 | Cr2O3 | C magnetic with large band gap |
| 167.3.43 | 290242 | Cr2O3 | C magnetic with large band gap |
| 167.3.44 | 290241 | Cr2O3 | C magnetic with large band gap |
| 167.3.45 | 290240 | Cr2O3 | C magnetic with large band gap |
| 167.3.46 | 290239 | Cr2O3 | C magnetic with large band gap |
| 167.3.47 | 290238 | Cr2O3 | C magnetic with large band gap |
| 167.3.48 | 290237 | Cr2O3 | C magnetic with large band gap |
| 167.3.49 | 290236 | Cr2O3 | C magnetic with large band gap |
| 167.3.50 | 290235 | Cr2O3 | C magnetic with large band gap |
| 167.3.51 | 290234 | Cr2O3 | C magnetic with large band gap |
| 167.3.53 | 261801 | Cr2O3 | C magnetic with large band gap |

Continued on next column

**TABLE XX– continued**

| TQC | ICSD | Stoich. Formula | SubSG |
|---|---|---|---|
| 167.3.55 | 250078 | Cr2O3 | C magnetic with large band gap |
| 167.3.59 | 167291 | Cr2O3 | C magnetic with large band gap |
| 167.3.60 | 167290 | Cr2O3 | C magnetic with large band gap |
| 167.3.61 | 167289 | Cr2O3 | C magnetic with large band gap |
| 167.3.62 | 167288 | Cr2O3 | C magnetic with large band gap |
| 167.3.63 | 167287 | Cr2O3 | C magnetic with large band gap |
| 167.3.64 | 167286 | Cr2O3 | C magnetic with large band gap |
| 167.3.65 | 167285 | Cr2O3 | C magnetic with large band gap |
| 167.3.66 | 167284 | Cr2O3 | C magnetic with large band gap |
| 167.3.67 | 167283 | Cr2O3 | C magnetic with large band gap |
| 167.3.68 | 167282 | Cr2O3 | C magnetic with large band gap |
| 167.3.69 | 167281 | Cr2O3 | C magnetic with large band gap |
| 167.3.71 | 167279 | Cr2O3 | C magnetic with large band gap |
| 167.3.72 | 167278 | Cr2O3 | C magnetic with large band gap |
| 167.3.73 | 167277 | Cr2O3 | C magnetic with large band gap |
| 167.3.74 | 167276 | Cr2O3 | C magnetic with large band gap |
| 167.3.75 | 167275 | Cr2O3 | C magnetic with large band gap |
| 167.3.76 | 167274 | Cr2O3 | C magnetic with large band gap |
| 167.3.77 | 167273 | Cr2O3 | C magnetic with large band gap |
| 167.3.78 | 167272 | Cr2O3 | C magnetic with large band gap |
| 167.3.79 | 167271 | Cr2O3 | C magnetic with large band gap |
| 167.3.80 | 167270 | Cr2O3 | C magnetic with large band gap |
| 167.3.81 | 167269 | Cr2O3 | C magnetic with large band gap |
| 167.3.82 | 167268 | Cr2O3 | C magnetic with large band gap |
| 167.3.84 | 416892 | C3N6Yb2 | D wide band gap spin polarized carbonate CN2- red orange cryst |
| 167.3.86 | 61067 | C1Ni1O3 | C spin polarized |
| 167.3.88 | 72790 | Br9Cs3Ho2 | C spin polarized |
| 167.3.89 | 409805 | Ba6In5N1 | A |
| 167.3.90 | 77731 | Ba6Ga5N1 | A |
| 167.3.94 | 426464 | Au6Sr2Zn3 | A |
| 167.3.95 | 391110 | Al7Au3Yb1 | A |
| 167.4.2 | 90285 | O6Rh1Sr3Zn1 | A |
| 167.4.3 | 51058 | O6Rh1Sr3Yb1 | hex perovskite eve magnetic |
| 167.4.4 | 81659 | Ni1O6Sc1Sr3 | C spin polarized |

Continued on next column

TABLE XX– continued

| TQC | ICSD | Stoich. Formula | SubSG |
|---|---|---|---|
| 167.4.5 | 51799 | Ni1O6Rh1Sr3 | A |
| 167.4.8 | 84738 | Mg1O6Rh1Sr3 | A |
| 167.4.12 | 249150 | Ir3Mg1Si7 | A |
| 167.4.14 | 82659 | Ir1O6Sr3Zn1 | A |
| 167.4.16 | 81903 | Ir1O6Sr3Zn1 | A |
| 167.4.17 | 80287 | Ir1Ni1O6Sr3 | C spin polarized |
| 167.4.18 | 80286 | Ir1Ni1O6Sr3 | C spin polarized |
| 167.4.19 | 80285 | Ir1Ni1O6Sr3 | A |
| 167.4.27 | 67148 | F3Ti1 | C Ti3+ d1 |
| 167.4.34 | 68527 | F3Mo1 | A |
| 167.4.44 | 167590 | Cr1La1O3 | C spin polarized |
| 167.4.47 | 59969 | Cr1F3 | C wide band gap spin polarized |
| 167.4.49 | 189230 | Co1O6Sr3Zn1 | C spin polarized |
| 167.4.50 | 192479 | Co1O6Ru1Sr3 | C spin polarized |
| 167.4.51 | 192478 | Co1O6Ru1Sr3 | A |
| 167.4.54 | 81904 | Cd1Ir1O6Sr3 | A |
| 167.4.55 | 81902 | Ca4Ir1O6 | A |
| 167.4.56 | 280873 | Ca4Ir1O6 | A |
| 167.4.63 | 50810 | Ca3Fe1O6Rh1 | C spin polarized |
| 167.4.64 | 181449 | Ca1Ir3Si7 | A |
| 167.4.65 | 81899 | Ca1Ir1O6Sr3 | A |
| 167.4.70 | 196049 | C1Mn1O3 | C spin polarized |
| 167.4.88 | 196022 | C1Mn1O3 | C wide band gap spin polarized carbonate CO3 2- |
| 167.4.92 | 61066 | C1Co1O3 | C spin polarized |
| 167.4.96 | 160141 | B1O3Yb1 | semiconducting borate rare earth Yb needs spin polarized calcul |
| 167.4.104 | 391106 | Al7Au3Ho1 | C magnetic with large band gap |
| 167.4.115 | 245564 | Al1Ce1O3 | C spin polarized |
| 167.4.116 | 245563 | Al1Ce1O3 | C spin polarized |
| 167.4.117 | 245272 | Al1Ce1O3 | C spin polarized |
| 167.4.118 | 245271 | Al1Ce1O3 | C spin polarized |
| 167.4.119 | 245269 | Al1Ce1O3 | C spin polarized |
| 167.4.120 | 245268 | Al1Ce1O3 | C spin polarized |
| 167.4.121 | 245267 | Al1Ce1O3 | C spin polarized |
| 167.4.122 | 238056 | Ag1Ba3S6U1 | C spin polarized |
| 173.3.3 | 154261 | I3K1Ti1 | C spin polarized d2 |
| 173.4.1 | 426493 | Ce3Fe1In1S7 | A |
| 174.3.1 | 417604 | B3Cl1La2 | A |
| 176.1.1 | 80772 | Mn1N3Sr3 | C spin polarized |
| 176.3.1 | 79629 | Se4Ti3 | A |
| 176.3.4 | 80771 | Ba3Mn1N3 | C spin polarized |
| 176.3.7 | 44424 | B4Ir3Sc1 | A |
| 176.3.9 | 78985 | B3Ir7Sn4 | A |
| 176.3.10 | 79934 | As1Nb3Te3 | A |
| 176.4.1 | 72922 | S4V3 | A or C |
| 176.4.3 | 72920 | S4V3 | A |
| 176.4.4 | 16278 | Nb3Se4 | A |
| 176.4.5 | 53119 | Mo6Se6Tl2 | A Chevrel |
| 176.4.6 | 23301 | Mo6Se6Tl2 | A |
| 176.4.7 | 90811 | Mo3Te3Tl1 | A |
| 176.4.8 | 30752 | K1Mo3S3 | A |
| 176.4.9 | 53100 | In1Mo3Te3 | A |
| 176.4.10 | 53098 | In1Mo3Se3 | A |
| 176.4.13 | 100352 | Fe3Te3Tl1 | A |
| 176.4.14 | 154803 | Cr1N3Sr3 | A ??? |
| 176.4.16 | 9536 | Cl3Cm1 | C spin polarized |

Continued on next column

TABLE XX– continued

| TQC | ICSD | Stoich. Formula | SubSG |
|---|---|---|---|
| 176.4.18 | 36502 | Ba3Fe1N3 | A |
| 176.4.19 | 154802 | Ba3Cr1N3 | C spin polarized |
| 180.3.1 | 156947 | B1Pt2Y1 | A |
| 180.3.2 | 156946 | B1Pt2Y1 | A |
| 180.3.3 | 409979 | B1Ir1Mg1 | A |
| 180.4.1 | 77114 | Mo1Sn2 | A |
| 181.3.1 | 173662 | Al2W1 | A |
| 182.3.2 | 88316 | O1Zr3 | A |
| 182.3.3 | 80930 | Fe3N1 | C spin polarized |
| 182.3.4 | 79984 | Fe3N1 | C spin polarized |
| 182.3.5 | 79983 | Fe3N1 | C spin polarized |
| 182.3.6 | 79982 | Fe3N1 | C spin polarized |
| 182.3.7 | 79981 | Fe3N1 | C spin polarized |
| 182.3.8 | 61514 | Cu5Na1S3 | A |
| 185.3.1 | 79253 | Mn1O3Sc1 | C spin polarized |
| 185.3.2 | 50694 | Mn1O3Sc1 | C spin polarized |
| 185.3.3 | 54253 | Mg3Pt1 | A |
| 185.3.4 | 409824 | Mg3Pd1 | A |
| 185.3.5 | 15056 | Cu3P1 | A |
| 185.3.6 | 167455 | Cl3K1Ni1 | C spin polarized |
| 185.3.7 | 409223 | Bi1K3 | A |
| 185.3.8 | 97539 | Ba3Cr1S5 | A |
| 185.4.1 | 106871 | Ir1Mg3 | A |
| 185.4.2 | 26455 | I3Rb1V1 | A |
| 185.4.3 | 89996 | Ba1Mn1O3 | C spin polarized |
| 185.4.4 | 106287 | Au1Mg3 | A |
| 186.3.1 | 106444 | Sn1Yb1Zn1 | A |
| 186.3.2 | 99453 | Mo1N1 | A |
| 186.3.4 | 61069 | Fe2Mo3O8 | A |
| 186.3.5 | 27216 | Fe2Mo3O8 | A |
| 186.3.6 | 87503 | Fe1S1 | A |
| 186.3.7 | 87502 | Fe1S1 | A |
| 186.3.8 | 68848 | Fe1S1 | A |
| 186.3.9 | 68847 | Fe1S1 | A |
| 186.3.10 | 53528 | Fe1S1 | A |
| 186.3.11 | 72303 | Fe1Na6S4 | A |
| 186.3.12 | 424277 | Cu4In1Mn1 | C spin polarized |
| 186.3.13 | 416544 | Cu1Sn1Y1 | A |
| 186.3.14 | 416543 | Cu1Sc1Sn1 | A |
| 186.3.15 | 107004 | Cu1Pb1Y1 | A |
| 186.3.19 | 107283 | Cu1Ho1Pb1 | C spin polarized |
| 186.3.20 | 8105 | Cr1Cs1I3 | A |
| 186.3.21 | 10206 | Cl3Cr1Cs1 | A |
| 186.3.23 | 106358 | Ca1Sn1Zn1 | A |
| 186.3.24 | 2590 | Br3Cr1Cs1 | A |
| 186.3.25 | 416144 | Bi1Cu1Yb1 | A |
| 186.3.26 | 416143 | Bi1Cu1Yb1 | A |
| 186.3.27 | 412207 | Au3Li1Sn4 | A |
| 186.3.28 | 415826 | Au1Sn1Y1 | A |
| 186.3.29 | 83984 | Au1Sb1Yb1 | A |
| 186.3.30 | 57240 | Au1Sb1Yb1 | A |
| 186.3.31 | 57022 | Au1Sb1Yb1 | A |
| 186.3.32 | 429889 | Au1La1Sn1 | A |
| 186.3.33 | 405323 | Au1Ge1Y1 | A |
| 186.3.34 | 81733 | Au1Ge1Sc1 | A |
| 186.3.36 | 405322 | Au1Ge1La1 | A |
| 186.3.37 | 93436 | Au1Ge1Ho1 | A |
| 186.3.38 | 93435 | Au1Ge1Ho1 | C spin polarized |
| 186.3.39 | 248685 | Au1Bi1Ca1 | A |
| 186.3.40 | 248683 | Au1Bi1Ca1 | A |
| 186.3.41 | 248681 | Au1Bi1Ca1 | A |

Continued on next column

**TABLE XX– continued**

| TQC | ICSD | Stoich. Formula | SubSG |
|---|---|---|---|
| 186.3.42 | 248679 | Au1Bi1Ca1 | A |
| 186.3.43 | 248677 | Au1Bi1Ca1 | A |
| 186.3.45 | 416375 | Ag1Sn1Y1 | A |
| 186.3.46 | 245234 | Ag1Sn1Y1 | A |
| 186.3.50 | 107099 | Ag1La1Pb1 | A |
| 186.3.51 | 418581 | Ag1Ho1Sn1 | C spin polarized |
| 186.3.53 | 416283 | Ag1Bi1Ca1 | A |
| 186.4.2 | 405625 | N7Si4Sr1Yb1 | A |
| 186.4.3 | 40534 | Mo3O8Sc1Zn1 | A |
| 186.4.4 | 99452 | Mo1N1 | A |
| 186.4.5 | 262928 | Mn1O1 | C spin polarized |
| 186.4.6 | 65451 | Mn1Na6Te4 | C spin polarized |
| 186.4.7 | 95067 | Mn1Na6S4 | C spin polarized |
| 186.4.8 | 95065 | Mn1Na6S4 | C spin polarized |
| 186.4.9 | 65447 | Mn1Na6S4 | C spin polarized |
| 186.4.11 | 104680 | La7Ni3 | A |
| 186.4.13 | 391305 | La4Mn1O1S6 | C spin polarized |
| 186.4.14 | 26218 | Ir1Si3 | A |
| 186.4.15 | 426618 | In1Li2Mo3O8 | A |
| 186.4.16 | 423281 | H2Fe1K1O6P2 | A |
| 186.4.20 | 422454 | Cu8Sn4Sr3 | A |
| 186.4.21 | 182103 | Cu8Sn4Sr3 | A |
| 186.4.22 | 190121 | Cu1Hf1Sn1 | A |
| 186.4.23 | 414432 | Cu1Ge1Yb1 | A |
| 186.4.30 | 54572 | Co8Sn4Y3 | A |
| 186.4.31 | 152945 | Co8Sn4Y3 | A |
| 186.4.32 | 68604 | Co1Na6Se4 | A |
| 186.4.33 | 37012 | Co1Na6S4 | A |
| 186.4.34 | 2064 | Cl3Cs1Cu1 | A |
| 186.4.35 | 102234 | Ce7Ni3 | C spin polarized |
| 186.4.36 | 416788 | Ce1Cu1Sn1 | C spin polarized |
| 186.4.37 | 156394 | Ce1Cu1Sn1 | A |
| 186.4.38 | 43476 | C5Al3Zr3 | A |
| 186.4.39 | 153859 | C4Al3Zr2 | A |
| 186.4.40 | 66750 | C4Al3U2 | C spin polarized |
| 186.4.41 | 43478 | C3Al3U1 | C spin polarized |
| 186.4.42 | 405194 | Ba1N7Si4Yb1 | A |
| 186.4.43 | 95747 | Ba1Co4O7Y1 | C spin polarized |
| 186.4.44 | 95746 | Ba1Co4O7Y1 | C spin polarized |
| 186.4.45 | 188853 | Ba1Co4O7Y1 | C spin polarized |
| 186.4.46 | 168709 | Ba1Co4O7Y1 | A |
| 186.4.47 | 59794 | Ba1Co4In1O7 | C spin polarized |
| 186.4.48 | 413464 | Ba1Co4In1O7 | C spin polarized |
| 186.4.49 | 59781 | Ba1Co4Ho1O7 | C spin polarized |
| 186.4.50 | 420427 | Ba1Co4Ho1O7 | C spin polarized |
| 186.4.51 | 420426 | Ba1Co4Ho1O7 | C spin polarized |
| 186.4.52 | 420425 | Ba1Co4Ho1O7 | C spin polarized |
| 186.4.53 | 420424 | Ba1Co4Ho1O7 | C spin polarized |
| 186.4.54 | 420423 | Ba1Co4Ho1O7 | C spin polarized |
| 186.4.55 | 58597 | Au3Sr7 | A |
| 186.4.57 | 411539 | Au1Ce1Sn1 | C spin polarized |
| 186.4.58 | 57176 | Au1Ce1Ge1 | C spin polarized |
| 186.4.62 | 107100 | Ag1Ce1Pb1 | C spin polarized |
| 187.3.1 | 410867 | Te1Zr1 | D |
| 187.3.2 | 89610 | Mg1P4Rh6 | A |
| 187.3.3 | 411151 | Li1N1Ni1 | A |
| 187.3.4 | 72375 | Cr6P4U1 | C spin polarized |
| 187.3.5 | 94412 | Co6Mg1P4 | A |
| 187.3.6 | 260171 | C1W1 | A |
| 187.3.7 | 260168 | C1W1 | A |
| 187.3.8 | 246151 | C1W1 | A |
| 187.3.9 | 246150 | C1W1 | A |
| 187.3.10 | 246149 | C1W1 | A |
| 187.3.11 | 249661 | Au9In13Sr4 | A |
| 187.3.12 | 89613 | As4Rh6Sr1 | A |
| 187.3.13 | 89611 | As4Mg1Rh6 | A |
| 187.3.14 | 89612 | As4Ca1Rh6 | A |
| 188.3.1 | 193216 | Co1K1O9P3 | A |
| 188.4.1 | 73339 | I3Li1Sc1 | A |
| 189.1.1 | 51718 | Si5Yb3 | A |
| 189.1.2 | 415599 | Si5Yb3 | A |
| 189.1.3 | 408117 | Si5Yb3 | A |
| 189.1.4 | 410418 | Pt1Sn1Yb1 | A |
| 189.1.6 | 196273 | O1Ti1 | A |
| 189.1.9 | 414438 | Mg1Tl1Y1 | A |
| 189.1.10 | 54344 | Mg1Sn1Yb1 | A |
| 189.1.11 | 411152 | Li5N3Ni3 | A |
| 189.1.12 | 415594 | La1Rh1Sn1 | A |
| 189.1.13 | 414439 | La1Mg1Tl1 | A |
| 189.1.14 | 160910 | In1Mg1Y1 | A |
| 189.1.15 | 85966 | Hf6Ni1Sb2 | A |
| 189.1.16 | 409925 | Ge1Pd1Ti1 | A |
| 189.1.18 | 240950 | Ga1Mg1Y1 | A |
| 189.1.19 | 160909 | Ga1Mg1Y1 | A |
| 189.1.20 | 413276 | Ga1La1Mg1 | A |
| 189.1.23 | 428473 | Cu1Ge1Sc1 | A |
| 189.1.24 | 98233 | Ce2Ge3Zn6 | A |
| 189.1.25 | 415492 | Ce1Pt1Sn1 | C spin polarized |
| 189.1.26 | 416969 | Ce1Pd1Sn1 | C spin polarized |
| 189.1.27 | 100002 | Ce1In1Ir1 | A |
| 189.1.30 | 413810 | Ba1Bi1Na1 | A |
| 189.1.31 | 261317 | Au1Ge2In3Yb3 | A |
| 189.1.33 | 192086 | Au1Ga1Mg1 | A |
| 189.1.34 | 411543 | Au1Cd1La1 | A |
| 189.1.38 | 160908 | Al1Mg1Y1 | A |
| 189.1.39 | 107275 | Ag1Pb1Y1 | A |
| 189.1.43 | 10016 | Ag1Ca1P1 | A |
| 189.1.44 | 10017 | Ag1As1Ca1 | A |
| 189.3.1 | 50239 | Se2U1 | C spin polarized |
| 189.3.2 | 82890 | S2U1 | C spin polarized |
| 189.3.4 | 157366 | Pt1Sb2Zr6 | A |
| 189.3.5 | 43209 | Pd2Si1 | A |
| 189.3.6 | 408409 | Pd1Sn1Yb1 | A |
| 189.3.7 | 151166 | Pd1Sn1Yb1 | A |
| 189.3.8 | 98972 | Os1Sc6Te2 | A |
| 189.3.9 | 1396 | N1Ta1 | A |
| 189.3.10 | 415509 | In1Pt1Y1 | A |
| 189.3.12 | 411459 | In1Pt1Sc1 | A |
| 189.3.14 | 97355 | Ge5Yb3 | A |
| 189.3.15 | 59586 | Ge5Yb3 | A |
| 189.3.16 | 43236 | Ge5Yb3 | A |
| 189.3.19 | 98232 | Ge3La2Zn6 | A |
| 189.3.20 | 53879 | Ge1Pd2 | A |
| 189.3.21 | 52052 | Ge1Pd2 | A |
| 189.3.28 | 82530 | Fe1Te2Zr6 | A |
| 189.3.29 | 90081 | Fe1Sc6Te2 | A |
| 189.3.32 | 420208 | Ce1Pd1Zn1 | C spin polarized |
| 189.3.33 | 412400 | Ce1Mg1Pt1 | C spin polarized |
| 189.3.34 | 412401 | Ce1Mg1Pd1 | C spin polarized |
| 189.3.35 | 280338 | Ba2Ni1Si3 | A |
| 189.3.36 | 429502 | B5Cu3Pt9 | A |
| 189.3.37 | 249113 | B4In5Ir9 | A |

Continued on next column

**TABLE XX– continued**

| TQC | ICSD | Stoich. Formula | SubSG |
|---|---|---|---|
| 189.3.38 | 77352 | B2Rh6Sn5 | A |
| 189.3.39 | 78984 | B2Ir6Sn5 | A |
| 189.3.40 | 68091 | B1Li1Pt3 | A |
| 189.3.41 | 76297 | Al3Ge2Ni1Y3 | A |
| 189.3.42 | 261007 | Al2Hf6Pt1 | A |
| 189.3.43 | 402701 | Al2Fe1Zr6 | A |
| 189.3.44 | 107244 | Al1Ho1Ni1 | C spin polarized |
| 189.3.45 | 107223 | Al1Ho1Ni1 | C spin polarized |
| 190.2.1 | 81072 | Ba3Cr2O9W1 | C spin polarized |
| 190.2.2 | 73183 | Ba3Ca1O9Ru2 | A |
| 190.3.1 | 68845 | Fe1S1 | A |
| 190.3.2 | 35007 | Fe1S1 | A |
| 190.3.3 | 35005 | Fe1S1 | A |
| 190.3.4 | 31963 | Fe1S1 | A |
| 190.3.5 | 156620 | Fe1S1 | A |
| 190.3.6 | 156619 | Fe1S1 | A |
| 190.3.7 | 156618 | Fe1S1 | A |
| 190.3.8 | 191973 | Ce2Ge3Ni1 | C spin polarized |
| 190.4.1 | 152133 | Rh1Sn1Zr1 | A |
| 190.4.3 | 418976 | Pd1Sc1Sn1 | A |
| 190.4.4 | 408901 | Ir1Sn1Zr1 | A |
| 190.4.5 | 107470 | Ir1Sn1Zr1 | A |
| 190.4.6 | 408900 | Hf1Rh1Sn1 | A |
| 190.4.7 | 107474 | Hf1Rh1Sn1 | A |
| 190.4.8 | 410398 | Ga1Pt1Zr1 | A |
| 190.4.9 | 152132 | Ga1Pt1Zr1 | A |
| 190.4.10 | 156268 | Ga1Hf1Pt1 | A |
| 190.4.11 | 408899 | Co1Hf1Sn1 | A |
| 190.4.12 | 107472 | Co1Hf1Sn1 | A |
| 190.4.13 | 156266 | Al1Pt1Zr1 | A |
| 190.4.14 | 261008 | Al1Hf1Pt1 | A |
| 190.4.15 | 156267 | Al1Hf1Pt1 | A |
| 191.1.2 | 417224 | Ga2Sr1 | A |
| 191.1.3 | 260654 | Ga2Sr1 | A |
| 191.1.4 | 246236 | Ga2Sr1 | A |
| 191.1.6 | 419163 | Ce1Pd2Zn3 | C spin polarized |
| 191.1.7 | 418614 | Ca1Zn5 | A |
| 191.1.8 | 107305 | Ba1Sn5 | A |
| 191.1.9 | 246238 | Ba1Ga2 | A |
| 191.1.12 | 78848 | B2Ti1 | A |
| 191.1.13 | 78847 | B2Ti1 | A |
| 191.1.14 | 56723 | B2Ti1 | A |
| 191.1.15 | 184968 | B2Fe3U1 | C spin polarized |
| 191.3.1 | 418615 | Sr1Zn5 | A |
| 191.3.2 | 96116 | La1Mn3Ni2 | C spin polarized |
| 191.3.3 | 260252 | In9K1Ni2 | A |
| 191.3.4 | 424816 | Ge2Ir3La1 | A |
| 191.3.5 | 107582 | Ga2Yb1 | A |
| 191.3.6 | 188716 | Fe6Ge6Mg1 | A |
| 191.3.15 | 158265 | Cu5La1 | A |
| 191.3.16 | 415256 | Co6Ge6Mg1 | A |
| 191.3.17 | 83895 | Ce2Co1Si3 | C spin polarized |
| 191.3.18 | 189763 | Ce1Ga2 | C spin polarized |
| 191.3.19 | 173576 | Ce10O3Si8 | C spin polarized |
| 191.3.20 | 54474 | Ca1Ni5 | A |
| 191.3.23 | 165559 | Bi2Ga1La1 | A |
| 191.3.26 | 81809 | B2Mo1 | A |
| 191.3.27 | 81806 | B2Mo1 | A |
| 191.3.29 | 23655 | B2Co3Y1 | A |
| 191.3.32 | 44170 | B2Co3Ho1 | C spin polarized |
| 191.3.35 | 68092 | B1Na1Pt3 | A |

Continued on next column

**TABLE XX– continued**

| TQC | ICSD | Stoich. Formula | SubSG |
|---|---|---|---|
| 191.3.36 | 170617 | B1La1Ni4 | A |
| 191.3.37 | 170609 | B1La1Ni4 | A |
| 191.3.41 | 55453 | B1Ce1Pt4 | C spin polarized |
| 191.3.42 | 106289 | Au5Rb1 | A |
| 191.3.43 | 106286 | Au5K1 | A |
| 191.3.44 | 260669 | Au2Ba1 | A |
| 193.3.6 | 153609 | Sb4Zr5 | A |
| 193.3.12 | 391476 | Pt3Zr5 | A |
| 193.3.16 | 165098 | Nb5P1Si3 | A |
| 193.3.19 | 83905 | La3Sb5Zr1 | A |
| 193.3.20 | 80907 | La3Sb5Ti1 | A |
| 193.3.22 | 409778 | I1La5Pb3 | A |
| 193.3.25 | 183487 | Hf5Sn3 | A |
| 193.3.28 | 83906 | Hf1La3Sb5 | A |
| 193.3.30 | 50740 | Ge5Ti1U3 | C spin polarized |
| 193.3.56 | 107005 | Cu1Pb3Y5 | A |
| 193.3.59 | 95238 | Cl1La5Sn3 | A |
| 193.3.61 | 80184 | Ca6Mn1N5 | C spin polarized |
| 193.3.63 | 95239 | Br1La5Sn3 | A |
| 193.3.64 | 415727 | Bi5La3Mg1 | A |
| 193.3.68 | 58475 | Au6Hg5 | A |
| 193.3.69 | 150475 | Au6Hg5 | A |
| 193.4.2 | 415741 | Si3Y5 | A |
| 193.4.5 | 415673 | Sb5U3Zr1 | C spin polarized |
| 193.4.6 | 43019 | Sb5Ti1U3 | C spin polarized |
| 193.4.9 | 153607 | Sb3Zr5 | A |
| 193.4.10 | 57167 | Sb3Zn1Zr5 | A |
| 193.4.11 | 57517 | Sb3Sr5 | A |
| 193.4.12 | 42120 | Sb3Sr5 | A |
| 193.4.13 | 57144 | Sb3Si1Zr5 | A |
| 193.4.14 | 153608 | Sb3Si1Zr5 | A |
| 193.4.28 | 152600 | La5Ni1Pb3 | A |
| 193.4.32 | 422871 | I3Ti1 | A |
| 193.4.34 | 415743 | Ho5Si3 | C spin polarized |
| 193.4.37 | 51371 | Ge4Np5 | C spin polarized |
| 193.4.50 | 183490 | Ga1Hf5Sn3 | A |
| 193.4.59 | 414040 | Cl3Ru1 | A |
| 193.4.63 | 33796 | Ca6Fe1N5 | A |
| 193.4.64 | 173026 | Ca5Sb3 | A |
| 193.4.65 | 4068 | Br3Zr1 | A |
| 193.4.69 | 165302 | Br3Zr1 | A |
| 193.4.70 | 414043 | Br3Ru1 | A |
| 193.4.71 | 413692 | Br3Ru1 | A |
| 193.4.72 | 413691 | Br3Ru1 | A |
| 193.4.73 | 50725 | Br1La5Sb3 | A |
| 193.4.74 | 409631 | Br1In3La5 | A |
| 193.4.75 | 416609 | Bi5La3Sc1 | A |
| 193.4.76 | 173362 | Bi5La3Mn1 | C spin polarized |
| 193.4.78 | 95241 | Bi3Br1La5 | A |
| 194.1.1 | 421015 | Zn1 | A |
| 194.1.2 | 421014 | Zn1 | A |
| 194.1.3 | 421013 | Zn1 | A |
| 194.1.4 | 247162 | Zn1 | A |
| 194.1.5 | 247161 | Zn1 | A |
| 194.1.6 | 247160 | Zn1 | A |
| 194.1.7 | 247159 | Zn1 | A |
| 194.1.8 | 247158 | Zn1 | A |
| 194.1.9 | 247157 | Zn1 | A |
| 194.1.10 | 247156 | Zn1 | A |
| 194.1.11 | 247155 | Zn1 | A |
| 194.1.12 | 247154 | Zn1 | A |

Continued on next column

**TABLE XX– continued**

| TQC | ICSD | Stoich. Formula | SubSG |
|---|---|---|---|
| 194.1.13 | 247153 | Zn1 | A |
| 194.1.14 | 247152 | Zn1 | A |
| 194.1.15 | 247151 | Zn1 | A |
| 194.1.16 | 247150 | Zn1 | A |
| 194.1.17 | 247149 | Zn1 | A |
| 194.1.18 | 247148 | Zn1 | A |
| 194.1.19 | 247147 | Zn1 | A |
| 194.1.20 | 422990 | Mn1O3Y1 | C spin polarized |
| 194.1.21 | 422989 | Mn1O3Y1 | C spin polarized |
| 194.1.22 | 422982 | Mn1O3Y1 | C spin polarized |
| 194.1.23 | 182284 | Mn1O3Y1 | C spin polarized |
| 194.1.24 | 181182 | Mn1O3Y1 | C spin polarized |
| 194.1.26 | 171451 | I1La1 | A |
| 194.1.27 | 163344 | Gd1Ge1Zn1 | C spin polarized |
| 194.1.36 | 425673 | C6Co1N6Pr1 | A or C |
| 194.1.37 | 180423 | C2Si1Ti3 | A |
| 194.1.38 | 180422 | C2Si1Ti3 | A |
| 194.1.39 | 180421 | C2Si1Ti3 | A |
| 194.1.40 | 180420 | C2Si1Ti3 | A |
| 194.1.41 | 180419 | C2Si1Ti3 | A |
| 194.1.42 | 153264 | C2Si1Ti3 | A |
| 194.1.43 | 180431 | C2Ge1Ti3 | A |
| 194.1.44 | 180430 | C2Ge1Ti3 | A |
| 194.1.45 | 180429 | C2Ge1Ti3 | A |
| 194.1.46 | 180428 | C2Ge1Ti3 | A |
| 194.1.47 | 180427 | C2Ge1Ti3 | A |
| 194.1.48 | 180426 | C2Ge1Ti3 | A |
| 194.1.49 | 180425 | C2Ge1Ti3 | A |
| 194.1.50 | 180424 | C2Ge1Ti3 | A |
| 194.1.51 | 249399 | C2Al1Ta3 | A |
| 194.3.1 | 152622 | Sn1Y1Zn1 | A |
| 194.3.2 | 262042 | Sc1Te1 | A |
| 194.3.3 | 427250 | Rh1Si3Y2 | A |
| 194.3.4 | 189925 | Pd3U1 | C spin polarized |
| 194.3.5 | 164724 | O2 | Aha solid superconducting O2 or AfM |
| 194.3.6 | 246978 | Mg5Pd2 | A |
| 194.3.7 | 163193 | Mg5Pd2 | A |
| 194.3.8 | 162567 | Mg2Yb1 | A |
| 194.3.10 | 262676 | Mg1Na2Sn1 | A |
| 194.3.11 | 238202 | Mg1Na2Pb1 | A |
| 194.3.12 | 423654 | K3Na1O9Os2 | A |
| 194.3.13 | 414235 | In2Sr1 | A |
| 194.3.14 | 260996 | Hf9Mo4Si1 | A |
| 194.3.15 | 164891 | Ge3La1V1 | A |
| 194.3.16 | 419660 | Ge2Ni3Sr1 | A |
| 194.3.17 | 163345 | Ge1Y1Zn1 | A |
| 194.3.18 | 419659 | Ge1Ni2Sr1 | A |
| 194.3.19 | 420008 | Gd1Pt2Sn1 | C spin polarized |
| 194.3.20 | 260580 | Ga2Yb1 | A |
| 194.3.22 | 155674 | Fe2Ti1 | A |
| 194.3.23 | 169307 | Fe1Se1 | A |
| 194.3.24 | 169302 | Fe1Se1 | A |
| 194.3.25 | 169300 | Fe1Se1 | A |
| 194.3.26 | 169298 | Fe1Se1 | A |
| 194.3.27 | 169296 | Fe1Se1 | A |
| 194.3.28 | 169294 | Fe1Se1 | A |
| 194.3.29 | 169292 | Fe1Se1 | A |
| 194.3.30 | 169290 | Fe1Se1 | A |
| 194.3.31 | 169288 | Fe1Se1 | A |
| 194.3.32 | 169286 | Fe1Se1 | A |

Continued on next column

**TABLE XX– continued**

| TQC | ICSD | Stoich. Formula | SubSG |
|---|---|---|---|
| 194.3.33 | 169284 | Fe1Se1 | A |
| 194.3.34 | 169272 | Fe1Se1 | A |
| 194.3.35 | 169270 | Fe1Se1 | A |
| 194.3.36 | 169268 | Fe1Se1 | A |
| 194.3.37 | 169266 | Fe1Se1 | A |
| 194.3.38 | 169264 | Fe1Se1 | A |
| 194.3.39 | 169262 | Fe1Se1 | A |
| 194.3.40 | 169260 | Fe1Se1 | A |
| 194.3.41 | 169258 | Fe1Se1 | A |
| 194.3.42 | 169256 | Fe1Se1 | A |
| 194.3.43 | 169254 | Fe1Se1 | A |
| 194.3.44 | 169252 | Fe1Se1 | A |
| 194.3.45 | 409890 | Eu2Ga3Ir1 | A |
| 194.3.47 | 416327 | Cu9Mg2Tb1 | A |
| 194.3.48 | 416545 | Cu1La1Sn1 | A |
| 194.3.49 | 154258 | Cs1I3Ti1 | A |
| 194.3.50 | 423830 | Cs1I3Ni1 | A |
| 194.3.51 | 423828 | Cl3Cs1Ni1 | C spin polarized |
| 194.3.53 | 428620 | Cd3Ru1Y10 | A |
| 194.3.54 | 418878 | Ca1Zn2 | A |
| 194.3.55 | 172208 | Ca1Na4Sn6 | A |
| 194.3.56 | 414232 | Ca1In2 | A |
| 194.3.57 | 417217 | Ca1Ga2 | A |
| 194.3.58 | 260645 | Ca1Ga2 | A |
| 194.3.59 | 420577 | Ca1Cd2 | A |
| 194.3.60 | 249388 | C1N2Ni1 | A |
| 194.3.61 | 154257 | Br3Rb1Ti1 | A |
| 194.3.62 | 423829 | Br3Cs1Ni1 | A |
| 194.3.63 | 416142 | Bi1Cu1Yb1 | A |
| 194.3.64 | 416141 | Bi1Cu1Yb1 | A |
| 194.3.65 | 172754 | Ba3O9Ru2Zr1 | A |
| 194.3.66 | 245252 | Ba3Ir2Mg1O9 | A |
| 194.3.67 | 245251 | Ba3Ir2Mg1O9 | A |
| 194.3.68 | 419206 | Ba1Sn3 | A |
| 194.3.69 | 409902 | Ba1Li4 | A |
| 194.3.70 | 261651 | Ba1Ge3 | A |
| 194.3.71 | 154600 | B2Ni7Yb3 | A |
| 194.3.72 | 251686 | Au1La1Sb1 | A |
| 194.3.73 | 420340 | Au1Ba1P1 | A |
| 194.3.74 | 420341 | As1Au1Ba1 | A |
| 194.4.1 | 164099 | Sc1 | A |
| 194.4.2 | 164097 | Sc1 | A |
| 194.4.3 | 164095 | Sc1 | A |
| 194.4.4 | 164094 | Sc1 | A |
| 194.4.5 | 164093 | Sc1 | A |
| 194.4.6 | 164092 | Sc1 | A |
| 194.4.7 | 164091 | Sc1 | A |
| 194.4.8 | 164090 | Sc1 | A |
| 194.4.9 | 164089 | Sc1 | A |
| 194.4.10 | 164088 | Sc1 | A |
| 194.4.11 | 162586 | Ni2Sb1Zr1 | A |
| 194.4.12 | 162585 | Ni2Sb1Zr1 | A |
| 194.4.13 | 158485 | Ni2Sb1Te2 | A |
| 194.4.14 | 190179 | Ni1Sb1 | A |
| 194.4.15 | 710050 | Nd1P1Pd1 | A |
| 194.4.16 | 185417 | Mn1O3Sr1 | C spin polarized |
| 194.4.17 | 237636 | Li3Sn4Zn2 | A |
| 194.4.18 | 246806 | La1Si10 | A |
| 194.4.19 | 154047 | Ge3La1Ti1 | A |
| 194.4.20 | 425591 | Ge1Mn3U2 | C spin polarized |
| 194.4.21 | 420009 | Gd1In1Pt2 | A |

Continued on next column

**TABLE XX– continued**

| TQC | ICSD | Stoich. Formula | SubSG |
|---|---|---|---|
| 194.4.22 | 156263 | Ga1Pt1Ti1 | A |
| 194.4.23 | 236572 | Cu3Sb4V2 | A |
| 194.4.24 | 422620 | Cu1Eu1Sb1 | C spin polarized |
| 194.4.25 | 158978 | Cr1Ge3Pr1 | C spin polarized |
| 194.4.26 | 158976 | Cr1Ge3La1 | A |
| 194.4.27 | 418080 | Co2I1Tb2 | C spin polarized |
| 194.4.28 | 416273 | Co2I1Pr2 | C spin polarized |
| 194.4.29 | 416787 | Ce1Cu1Sn1 | A |
| 194.4.30 | 416786 | Ce1Cu1Sn1 | C spin polarized |
| 194.4.31 | 173221 | Ce1Cu1Si1 | C spin polarized |
| 194.4.32 | 173219 | Ce1Cu1Ge1 | A |
| 194.4.33 | 153107 | Ca2I1N1 | A |
| 194.4.34 | 419117 | C1Ga1Ti2 | A |
| 194.4.35 | 419116 | C1Cr2Ga1 | A |
| 194.4.36 | 249387 | C1Co1N2 | C spin polarized |
| 194.4.37 | 420348 | C1Br1Pr2 | A |
| 194.4.38 | 246360 | Ba3Ir2Na1O9 | A |
| 194.4.39 | 246359 | Ba3Ir2Na1O9 | C spin polarized |
| 194.4.40 | 174293 | Ba3Ir2La1O9 | A or C |
| 194.4.41 | 157841 | Ba3In1O9Ru2 | A |
| 194.4.42 | 154181 | Ba1S3V1 | A |
| 194.4.43 | 424878 | B3Ru2 | A |
| 194.4.44 | 424879 | B3Os2 | A |
| 194.4.45 | 421522 | B2Re1 | A |
| 194.4.46 | 165834 | B2Re1 | A |
| 194.4.47 | 427220 | Au5Cs1Ga9 | A |
| 194.4.48 | 190987 | Au2Ga2In1 | A |
| 194.4.49 | 192087 | Au2Ga1Mg1 | A |
| 194.4.50 | 163719 | Au2Ce3Sb3 | C spin polarized |
| 194.4.51 | 260109 | As2Eu1In2 | C spin polarized |
| 194.4.52 | 420091 | Al5Ba4 | A |
| 194.4.53 | 420090 | Al5Ba3 | A |
| 194.4.55 | 163350 | Al1Pt1Ti1 | A |
| 194.4.56 | 172878 | Ag3Ni2O4 | A |
| 194.4.57 | 415451 | Ag1Ni1O2 | A |
| 194.4.58 | 422621 | Ag1Eu1Sb1 | C spin polarized |
| 198.4.1 | 79233 | Rh1Si1 | chiral |
| 198.4.2 | 43240 | Rh1Si1 | A |
| 198.4.3 | 24719 | Rh1Si1 | A chiral |
| 198.4.4 | 69791 | Pd1Si1Sr1 | A |
| 198.4.5 | 93906 | Pd1Sb1Te1 | A |
| 198.4.6 | 93905 | Pd1Sb1Se1 | A |
| 198.4.7 | 187621 | Ni1Si1 | A |
| 198.4.8 | 93902 | Ni1Sb1Se1 | A |
| 198.4.9 | 93900 | Ni1S1Sb1 | A |
| 198.4.10 | 93898 | Ni1P1S1 | A |
| 198.4.11 | 71832 | Mn1Si1 | C magnetic spin polarized |
| 198.4.12 | 71831 | Mn1Si1 | magnetic skyrmion spin polarized |
| 198.4.13 | 71830 | Mn1Si1 | magnetic skyrmion |
| 198.4.14 | 155712 | Ir1P1Sr1 | A |
| 198.4.15 | 27187 | Ir1La1Si1 | A |
| 198.4.16 | 252044 | Ge1Mn1 | C magnetic spin polarized |
| 198.4.17 | 252043 | Ge1Mn1 | magnetic skyrmion |
| 198.4.18 | 187195 | Ge1Mn1 | magnetic skyrmion |
| 198.4.19 | 150395 | Ge1Ir1Mn3 | C magnetic spin polarized |
| 198.4.20 | 261111 | Ga1Pd1 | A |
| 198.4.22 | 260982 | Co1Si1 | magnetic skyrmion |

Continued on next column

**TABLE XX– continued**

| TQC | ICSD | Stoich. Formula | SubSG |
|---|---|---|---|
| 198.4.24 | 72640 | Ca1Pt1Si1 | A |
| 198.4.25 | 63555 | Ca1Pt1Si1 | A |
| 198.4.26 | 70060 | Bi1Pd1Te1 | A |
| 198.4.27 | 59191 | Ba1P1Pt1 | A |
| 198.4.28 | 73529 | Ba1Ir1P1 | A |
| 198.4.29 | 93904 | As1Pd1Se1 | A |
| 198.4.30 | 93903 | As1Pd1S1 | A |
| 198.4.31 | 93901 | As1Ni1Se1 | A |
| 198.4.32 | 93899 | As1Ni1S1 | A |
| 198.4.33 | 59192 | As1Ba1Pt1 | A |
| 198.4.34 | 104616 | Al1Au4 | A |
| 199.3.1 | 195668 | Bi2Pt3Se2 | A |
| 199.4.1 | 417634 | Bi2Pd3S2 | A |
| 199.4.2 | 48178 | Ba3Cl2Fe2O5 | C spin polarized |
| 199.4.3 | 48179 | Ba3Br2Fe2O5 | C spin polarized |
| 202.4.1 | 179 | K2N6Ni1O12Sr1 | C spin polarized |
| 202.4.2 | 1937 | K2N6Ni1O12Pb1 | C spin polarized |
| 204.1.2 | 195439 | Cu4O12V4 | C spin polarized |
| 204.1.3 | 195438 | Cu4O12V4 | C spin polarized |
| 204.1.4 | 51893 | Cu3Na1O12Ru4 | C spin polarized |
| 204.1.6 | 248615 | Cr4Cu3O12Y1 | C spin polarized |
| 204.1.7 | 248614 | Cr4Cu3O12Y1 | C spin polarized |
| 204.1.8 | 248613 | Cr4Cu3O12Y1 | C spin polarized |
| 204.1.9 | 248612 | Cr4Cu3O12Y1 | C spin polarized |
| 204.1.10 | 248611 | Cr4Cu3O12Y1 | C spin polarized |
| 204.1.11 | 248610 | Cr4Cu3O12Y1 | C spin polarized |
| 204.1.12 | 248609 | Cr4Cu3O12Y1 | C spin polarized |
| 204.1.13 | 248608 | Cr4Cu3O12Y1 | C spin polarized |
| 204.1.14 | 248607 | Cr4Cu3La1O12 | C spin polarized |
| 204.1.15 | 248606 | Cr4Cu3La1O12 | C spin polarized |
| 204.1.16 | 248605 | Cr4Cu3La1O12 | C spin polarized |
| 204.1.17 | 248604 | Cr4Cu3La1O12 | C spin polarized |
| 204.1.18 | 248603 | Cr4Cu3La1O12 | C spin polarized |
| 204.1.19 | 248601 | Cr4Cu3La1O12 | C spin polarized |
| 204.1.20 | 248600 | Cr4Cu3La1O12 | C spin polarized |
| 204.1.21 | 248599 | Cr4Cu3La1O12 | C spin polarized |
| 204.1.23 | 174551 | Ba1Ge12Pt4 | A |
| 204.1.24 | 160797 | Ba1Ge12Pt4 | A |
| 204.4.1 | 77721 | O3Re1 | A |
| 204.4.2 | 77678 | O3Re1 | A |
| 204.4.3 | 55473 | O3Re1 | A |
| 204.4.4 | 55472 | O3Re1 | A |
| 204.4.5 | 55471 | O3Re1 | A |
| 204.4.6 | 55469 | O3Re1 | A |
| 204.4.7 | 55468 | O3Re1 | A |
| 204.4.8 | 55467 | O3Re1 | A |
| 204.4.9 | 55466 | O3Re1 | A |
| 204.4.10 | 55465 | O3Re1 | A |
| 204.4.11 | 55464 | O3Re1 | A |
| 204.4.12 | 55463 | O3Re1 | A |
| 204.4.13 | 55462 | O3Re1 | A |
| 204.4.14 | 54614 | Ni6Pb1Y12 | A |
| 204.4.17 | 19022 | Mn7Na1O12 | C spin polarized |
| 204.4.18 | 189912 | Ge12Pt4U1 | C spin polarized |
| 204.4.19 | 189911 | Ge12Pt4Th1 | C spin polarized |
| 204.4.21 | 79926 | Fe4P12U1 | C spin polarized |
| 204.4.24 | 95716 | Cu3Na1O12Ru4 | C spin polarized |
| 204.4.25 | 38419 | Cu3Mn4O12Y1 | C spin polarized |
| 204.4.26 | 38418 | Cu3Mn4O12Y1 | C spin polarized |
| 204.4.29 | 51897 | Cu3La1O12Ru4 | C spin polarized |
| 204.4.30 | 251657 | Cu3Ir4La1O12 | C spin polarized |

Continued on next column

**TABLE XX– continued**

| TQC | ICSD | Stoich. Formula | SubSG |
|---|---|---|---|
| 204.4.34 | 290765 | Cu3Fe4La1O12 | C spin polarized |
| 204.4.36 | 290763 | Cu3Fe4La1O12 | C spin polarized |
| 204.4.37 | 290762 | Cu3Fe4La1O12 | C spin polarized |
| 204.4.38 | 290761 | Cu3Fe4La1O12 | C spin polarized |
| 204.4.42 | 290757 | Cu3Fe4La1O12 | C spin polarized |
| 204.4.43 | 290756 | Cu3Fe4La1O12 | C spin polarized |
| 204.4.45 | 237399 | Cu3Fe4La1O12 | C spin polarized |
| 204.4.46 | 185584 | Cu3Fe4La1O12 | C spin polarized |
| 204.4.48 | 237420 | Cu3Fe4Ho1O12 | C spin polarized |
| 204.4.52 | 92393 | Co1P3 | A |
| 204.4.53 | 174553 | Ce1Ge12Pt4 | A |
| 204.4.54 | 237495 | Ca1O12Pd3V4 | A |
| 204.4.55 | 9188 | As3Co1 | A |
| 205.2.1 | 77109 | Pd1Sb2 | A |
| 205.2.2 | 300225 | Os1Te2 | A really we have data !!!! |
| 205.2.3 | 86354 | Ni1S2 | A |
| 205.2.4 | 68169 | Ni1S2 | A |
| 205.2.5 | 68168 | Ni1S2 | A |
| 205.2.6 | 68167 | Ni1S2 | A |
| 205.3.1 | 77111 | Bi2Pt1 | A |
| 205.4.1 | 251565 | O2Rh1 | A |
| 205.4.7 | 251567 | Ir1O2 | A |
| 205.4.8 | 100567 | F2Pd1 | A |
| 205.4.9 | 100510 | Cu1S2 | C spin polarized |
| 205.4.10 | 86351 | Co1S2 | A |
| 212.4.1 | 84932 | B1Li2Pt3 | A |
| 212.4.2 | 156466 | B1Li2Pt3 | A superconductor |
| 212.4.3 | 84931 | B1Li2Pd3 | A |
| 212.4.4 | 156465 | B1Li2Pd3 | A |
| 213.4.1 | 86170 | N1Ni2W3 | A |
| 213.4.2 | 150314 | Mo3N1Pt2 | A |
| 213.4.3 | 93502 | Mo3N1Pd2 | A |
| 213.4.4 | 96416 | Mo3N1Ni2 | A |
| 213.4.5 | 50815 | Mo3N1Ni2 | A |
| 213.4.6 | 165261 | Mo3N1Ni2 | A |
| 213.4.7 | 260022 | Mg3Ru2 | A |
| 216.2.1 | 260984 | Ni5U1 | C spin polarized |
| 216.2.4 | 418013 | La4Mg1Ru1 | A |
| 216.2.6 | 81917 | Hg1S1 | A |
| 216.2.8 | 415194 | Cd1Ce1Ni4 | C spin polarized |
| 216.3.1 | 189201 | Ni5Zr1 | A |
| 216.3.2 | 250246 | Cr4Cu1In1Se8 | C spin polarized |
| 216.3.3 | 182766 | Ce4Mg1Ru1 | A |
| 216.3.5 | 182767 | Cd1Ce4Ru1 | A |
| 216.3.6 | 54465 | Au1Mn1Sn1 | A |
| 216.4.2 | 416868 | Ir3Mg3Si8 | A |
| 216.4.3 | 418746 | Ir1La4Mg1 | A |
| 216.4.4 | 162198 | In1Sb1 | A |
| 216.4.5 | 162197 | In1Sb1 | A |
| 216.4.6 | 162196 | In1Sb1 | A |
| 216.4.7 | 201283 | I1Nb1Se1 | A |
| 216.4.8 | 81213 | Ho1Ni1Sb1 | C spin polarized |
| 216.4.9 | 290038 | Hg1Se1 | A |
| 216.4.10 | 50502 | Ge1S8V4 | skyrmions C spin polarized |
| 216.4.12 | 412294 | Cu1Mg1Sb1 | A |
| 216.4.13 | 159198 | Co4Sn1U1 | C spin polarized |
| 216.4.14 | 416990 | Co1La4Mg1 | A |
| 216.4.15 | 419070 | Cd1Ru1Y4 | A |
| 216.4.16 | 419069 | Cd1La4Ru1 | A |

Continued on next column

**TABLE XX– continued**

| TQC | ICSD | Stoich. Formula | SubSG |
|---|---|---|---|
| 216.4.17 | 419067 | Cd1Co1La4 | A |
| 216.4.18 | 51139 | Au1S1Ta5 | A |
| 216.4.19 | 248684 | Au1Bi1Ca1 | A |
| 216.4.20 | 248682 | Au1Bi1Ca1 | A |
| 216.4.21 | 248680 | Au1Bi1Ca1 | A |
| 216.4.22 | 248678 | Au1Bi1Ca1 | A |
| 216.4.23 | 248676 | Au1Bi1Ca1 | A |
| 217.1.1 | 427048 | F6Ni3O6Sb4 | A |
| 217.1.2 | 427047 | F6Ni3O6Sb4 | A |
| 217.4.1 | 77887 | Pd16S7 | A |
| 217.4.2 | 410992 | Ir4Sn4Sr3 | A |
| 217.4.3 | 104605 | Ir2Zn11 | A |
| 217.4.4 | 429725 | Ge4Ir4Sr3 | A |
| 217.4.5 | 429512 | Ga4Ir4Sr3 | A |
| 217.4.6 | 200460 | F9U2 | C spin polarized |
| 217.4.7 | 240667 | Cu5Zn8 | A |
| 217.4.8 | 2092 | Cu5Zn8 | A |
| 217.4.9 | 194374 | Co2Zn11 | A |
| 217.4.10 | 95788 | Ca3Ir4Si4 | A |
| 217.4.11 | 429727 | Ca3Ge4Ir4 | A |
| 217.4.12 | 195500 | Ca3Ge4Ir4 | A |
| 220.4.2 | 427458 | O6Rb4 | C spin polarized |
| 221.1.1 | 201036 | O1Sn1Sr3 | A |
| 221.1.2 | 181127 | Ni3Sn1 | A |
| 221.1.3 | 163195 | Mg1Pd1 | A |
| 221.1.4 | 247065 | In1N1Ni3 | A |
| 221.1.5 | 247273 | H1Pd3Tl1 | A |
| 221.1.6 | 100533 | Fe1Ti1 | A |
| 221.1.7 | 262924 | Cs1I3Sn1 | A |
| 221.1.8 | 100787 | Ca3O1Sn1 | A |
| 221.1.9 | 100789 | Ca3O1Pb1 | A |
| 221.1.10 | 402087 | Ca3Ge1O1 | A |
| 221.1.11 | 80320 | Br3Cs1Ge1 | A |
| 221.1.12 | 409846 | Br3Ce3Ga1 | C spin polarized |
| 221.1.13 | 50313 | B6Sr1 | A |
| 221.1.14 | 51495 | B6Ba1 | A |
| 221.3.1 | 85207 | Ru1Si1 | A |
| 221.3.2 | 79230 | Ru1Si1 | A |
| 221.3.4 | 96119 | Mn3Sb1 | C spin polarized |
| 221.3.6 | 79241 | C1Ru3Th1 | C spin polarized |
| 221.3.7 | 81546 | B6Th1 | C spin polarized |
| 221.3.8 | 95053 | B6Ce1 | C spin polarized |
| 221.3.9 | 84561 | B6Ce1 | C spin polarized |
| 221.3.10 | 67404 | B6Ce1 | C spin polarized |
| 221.3.11 | 241004 | B6Ce1 | C spin polarized |
| 221.3.12 | 194318 | B6Ce1 | C spin polarized |
| 221.3.13 | 194317 | B6Ce1 | C spin polarized |
| 221.3.14 | 194316 | B6Ce1 | C spin polarized |
| 221.3.15 | 194315 | B6Ce1 | C spin polarized |
| 221.3.16 | 194314 | B6Ce1 | C spin polarized |
| 221.3.17 | 194313 | B6Ce1 | C spin polarized |
| 221.3.18 | 194312 | B6Ce1 | C spin polarized |
| 221.3.19 | 194311 | B6Ce1 | C spin polarized |
| 221.3.20 | 173598 | B6Ce1 | C spin polarized |
| 221.3.21 | 173597 | B6Ce1 | C spin polarized |
| 221.4.1 | 86941 | O3Ru1Sr1 | A |
| 221.4.2 | 82984 | O3Ru1Sr1 | A |
| 221.4.3 | 82983 | O3Ru1Sr1 | A |
| 221.4.4 | 82982 | O3Ru1Sr1 | A |
| 221.4.5 | 82981 | O3Ru1Sr1 | A |
| 221.4.6 | 69360 | O3Ru1Sr1 | A |

Continued on next column

**TABLE XX– continued**

| TQC | ICSD | Stoich. Formula | SubSG |
|---|---|---|---|
| 221.4.7 | 162777 | O3Ru1Sr1 | A |
| 221.4.9 | 421118 | N1Na3 | A |
| 221.4.10 | 421117 | N1Na3 | A |
| 221.4.11 | 421116 | N1Na3 | A |
| 221.4.12 | 421115 | N1Na3 | A |
| 221.4.13 | 421113 | N1Na3 | A |
| 221.4.14 | 421112 | N1Na3 | A |
| 221.4.15 | 421111 | N1Na3 | A |
| 221.4.16 | 421110 | N1Na3 | A |
| 221.4.17 | 421109 | N1Na3 | A |
| 221.4.18 | 412313 | N1Na3 | A |
| 221.4.19 | 249566 | N1Na3 | A |
| 221.4.20 | 71994 | Mo1O3Sr1 | A |
| 221.4.21 | 174208 | Mo1O3Sr1 | A |
| 221.4.22 | 174207 | Mo1O3Sr1 | A |
| 221.4.23 | 237710 | Mn3N1Zn1 | C spin polarized |
| 221.4.26 | 92335 | Fe1O3Sr1 | C spin polarized |
| 221.4.27 | 91062 | Fe1O3Sr1 | C spin polarized |
| 221.4.28 | 73166 | F3Pd1Rb1 | A |
| 221.4.29 | 73167 | F3K1Pd1 | A |
| 221.4.30 | 37255 | F3K1Ni1 | A |
| 221.4.31 | 189369 | F3K1Ni1 | A |
| 221.4.33 | 187860 | Cu1Y1 | A |
| 221.4.34 | 245834 | Cr1O3Sr1 | C spin polarized |
| 221.4.35 | 194155 | Cr1O3Pb1 | C spin polarized |
| 221.4.36 | 160196 | Cr1O3Pb1 | C spin polarized |
| 221.4.37 | 422858 | Cd1Co3N1 | A |
| 221.4.38 | 411448 | Ca3N1Tl1 | A |
| 221.4.39 | 50164 | C1Sc3Tl1 | A |
| 221.4.40 | 50163 | C1In1Sc3 | A |
| 221.4.41 | 80956 | C1In1La3 | A |
| 221.4.42 | 72269 | Ba1O3Pb1 | A |
| 221.4.43 | 51659 | Ba1O3Pb1 | A |
| 221.4.44 | 245607 | Ba1O3Pb1 | A |
| 221.4.45 | 245606 | Ba1O3Pb1 | A |
| 221.4.46 | 43799 | Ba1Mo1O3 | A |
| 221.4.47 | 262132 | Ba1Fe1O3 | C spin polarized |
| 221.4.48 | 262131 | Ba1Fe1O3 | C spin polarized |
| 221.4.49 | 72942 | Au1Ca3N1 | A |
| 223.3.2 | 186819 | Ca1O4Pd3 | A |
| 223.4.1 | 87329 | Si1V3 | A |
| 223.4.2 | 87328 | Si1V3 | A |
| 223.4.3 | 32546 | Na1O4Pt3 | A |
| 223.4.4 | 245608 | Na1O4Pd3 | A |
| 223.4.5 | 35755 | Mo3Si1 | A |
| 223.4.6 | 84665 | La1Pd3S4 | A |
| 223.4.7 | 61031 | La1Pd3S4 | A |
| 223.4.8 | 50296 | Ir1Ti3 | A |
| 223.4.10 | 53218 | Cr3Si1 | A |
| 223.4.11 | 32509 | Cr3Si1 | A |
| 224.3.1 | 281041 | Ag2O1 | A |
| 224.3.2 | 247821 | Ag2O1 | A |
| 225.1.1 | 63098 | Pb1Te1 | A |
| 225.1.2 | 153711 | Pb1Te1 | A |
| 225.1.3 | 63097 | Pb1Se1 | A |
| 225.1.4 | 63096 | Pb1Se1 | A |
| 225.1.5 | 62196 | Pb1Se1 | A |
| 225.1.6 | 62195 | Pb1Se1 | A |
| 225.1.7 | 38294 | Pb1Se1 | A |
| 225.1.8 | 62192 | Pb1S1 | A |
| 225.1.9 | 62191 | Pb1S1 | A |

Continued on next column

**TABLE XX– continued**

| TQC | ICSD | Stoich. Formula | SubSG |
|---|---|---|---|
| 225.1.10 | 62190 | Pb1S1 | A |
| 225.1.11 | 35204 | O2U1 | C spin polarized |
| 225.1.12 | 246851 | O2U1 | C spin polarized |
| 225.1.15 | 106715 | Ga2Li1Rh1 | A |
| 225.1.16 | 160154 | Fe16Si7U6 | E |
| 225.1.22 | 159874 | C1Zr1 | A |
| 225.1.23 | 424342 | C1W1 | A |
| 225.1.24 | 159873 | C1Hf1 | A |
| 225.3.2 | 95561 | Pd20Si6U3 | C spin polarized |
| 225.3.3 | 93930 | Pd20Si6U3 | C spin polarized |
| 225.3.4 | 159262 | Ni16Sc6Si7 | A |
| 225.3.5 | 162574 | Mn6Ni16Si7 | C spin polarized |
| 225.3.6 | 162573 | Mn6Ni16Si7 | A |
| 225.3.7 | 72350 | Mg3Ni20P6 | C spin polarized |
| 225.3.8 | 10438 | Ir4Sc11 | A |
| 225.3.9 | 68998 | In2Ni21P6 | A |
| 225.3.11 | 102989 | Cu2In1La1 | A |
| 225.3.12 | 174079 | Ce2Pd21Si6 | C spin polarized |
| 225.3.13 | 249342 | Ce1Mg1Zn2 | C spin polarized |
| 225.3.14 | 26476 | C1U1 | C spin polarized |
| 225.3.15 | 44494 | C1Ti1 | A |
| 225.3.16 | 159871 | C1Ti1 | A |
| 225.3.17 | 1546 | C1Ti1 | A |
| 225.3.18 | 160155 | C1Fe16Si7U6 | C spin polarized |
| 225.3.19 | 409960 | B6Ni21Zr2 | A |
| 225.3.20 | 380216 | B6In2Ni21 | A |
| 225.3.21 | 105524 | Al1Nb1Ru2 | A |
| 225.4.1 | 261057 | Ru9Sb8Zn7 | A |
| 225.4.2 | 10439 | Os4Sc11 | A |
| 225.4.4 | 9866 | Ni1O1 | C classical MOTT |
| 225.4.5 | 260172 | Ni1 | A |
| 225.4.6 | 260169 | Ni1 | magnetic |
| 225.4.7 | 159263 | Ni16Si7Ti6 | A |
| 225.4.8 | 159264 | Nb6Ni16Si7 | A |
| 225.4.9 | 262367 | N1Np1 | E |
| 225.4.10 | 982 | N1Nb1 | A very nice band structure |
| 225.4.11 | 157027 | Mo1Ni1O6Sr2 | A |
| 225.4.12 | 155733 | Mo1Ni1O6Sr2 | Double preoswskite with rare earth Ho - spin polarized |
| 225.4.13 | 409428 | Mg2Pb1 | A |
| 225.4.14 | 10052 | Li3Tl1 | A |
| 225.4.15 | 427115 | I1Na1O6Pb2 | A |
| 225.4.16 | 63670 | H1Br1 | D acid liquid wide band gap |
| 225.4.17 | 94456 | F6Ti1Zr1 | C MOTT yellow orange wide band gap |
| 225.4.18 | 59102 | Cs2Cu1F6K1 | A or C probably wide band gap |
| 225.4.19 | 409229 | Cl6Tl2W1 | B or C looks like insulator |
| 225.4.20 | 409632 | Cl6Rb2W1 | A |
| 225.4.21 | 408777 | Cl6Mo1Tl2 | A Chevrel pahses |
| 225.4.22 | 409840 | Cl6K2W1 | A |
| 225.4.23 | 409674 | Cl6Cs2W1 | B or C looks like insulator |
| 225.4.24 | 409808 | Cl6Cs2Mo1 | A cluster Chevrel |
| 225.4.25 | 83258 | Ca2O6Pd1W1 | C spinpolarized double perovskite |

Continued on next column

**TABLE XX– continued**

| TQC | ICSD | Stoich. Formula | SubSG |
|---|---|---|---|
| 225.4.26 | 151706 | C6Cs2Fe1N6Ni1 | D non magnetic but gap - localised bands check |
| 225.4.27 | 89531 | C2Na2 | D non magnetic but gap - localised bands check |
| 225.4.28 | 89536 | C2Li2 | gap 2 |
| 225.4.29 | 89532 | C2K2 | caution |
| 225.4.30 | 402439 | Br6Rb2W1 | B or C looks like insulator |
| 225.4.31 | 402441 | Br6Cs2W1 | D non magnetic but gap - localised bands check |
| 225.4.32 | 94216 | Ba2O6Re1Y1 | Spin-Disordered Double preoswskite- spin polarized |
| 225.4.33 | 94215 | Ba2O6Re1Y1 | Spin-Disordered Ba2YReO6 spin polarized calculations |
| 225.4.34 | 168216 | Ba2O6Re1Y1 | spin-Disordered Double preoswskite- spin polarized |
| 225.4.35 | 168215 | Ba2O6Re1Y1Spin-Dis | Double preoswskite- spin polarized |
| 225.4.36 | 245142 | Ba2Ni1O6U1 | C E spinpolarized double perovskite |
| 225.4.37 | 169535 | Ba2Ni1O6U1 | E Urunaium spin-Disordered Double preoswskite- spin polarized |
| 225.4.38 | 98192 | Ba2Mo1Ni1O6 | A |
| 225.4.40 | 158359 | Ba2Ho1O6Ta1 | Double preoswskite with rare earth Ho - spin polarized |
| 225.4.41 | 84651 | Ba2Ho1O6Sb1 | Double preoswskite- spin polarized Holmium |
| 225.4.42 | 150862 | Ba2Ho1O6Sb1 | Double preoswskite with rare earth Ho - spin polarized |
| 225.4.45 | 192845 | Ba2Ca1O6Os1 | C spinpolarized double perovskite |
| 225.4.46 | 192844 | Ba2Ca1O6Os1 | C spinpolarized double perovskite |
| 225.4.47 | 171989 | Ba2Ca1O6Os1 | magnetic Double preoswskite- spin polarized |
| 225.4.48 | 171988 | Ba2Ca1O6Os1 | magnetic double preoswskite - spin polarized eventually |
| 225.4.49 | 189924 | B12U1 | E spin polarized |
| 225.4.50 | 171472 | Ag2In1Li1 | A |
| 226.3.1 | 58751 | Be13U1 | E toxic and radio active |
| 226.3.2 | 58750 | Be13U1 | E toxic and radio active |
| 226.3.3 | 58749 | Be13U1 | E toxic and radio active |
| 226.3.4 | 58748 | Be13U1 | E toxic and radio active spin polarized |
| 226.3.5 | 58747 | Be13U1 | E toxic and radio active spin polarized |
| 226.3.6 | 107528 | Be13U1 | E toxic and radio active spin polarized |
| 226.4.1 | 415889 | Rb1Zn13 | B |
| 226.4.2 | 415887 | Na1Zn13 | A |
| 226.4.3 | 421842 | Li1Zn13 | A |
| 226.4.4 | 415886 | K1Zn13 | A |

Continued on next column

**TABLE XX– continued**

| TQC | ICSD | Stoich. Formula | SubSG |
|---|---|---|---|
| 226.4.5 | 55509 | Co1Cu12La1 | A |
| 226.4.6 | 415888 | Cd13Rb1 | B |
| 226.4.7 | 415885 | Cd13K1 | A |
| 226.4.8 | 415884 | Cd13Cs1 | B |
| 226.4.9 | 58742 | Be13Th1 | E toxic and radio active |
| 226.4.10 | 58740 | Be13Th1 | E |
| 226.4.12 | 170628 | Be13Ce1 | A or C Ce spin polarized |
| 226.4.13 | 154391 | Ba1Cu13 | A |
| 227.3.10 | 83269 | Mn2O7Tl2 | magnetic |
| 227.4.1 | 160600 | O7V2Y2 | A |
| 227.4.21 | 173073 | O6Os2Rb1 | A |
| 227.4.22 | 173072 | O6Os2Rb1 | A |
| 227.4.23 | 157442 | O6Os2Rb1 | A |
| 227.4.24 | 157441 | O6Os2Rb1 | A |
| 227.4.25 | 157440 | O6Os2Rb1 | A |
| 227.4.26 | 157439 | O6Os2Rb1 | A |
| 227.4.27 | 157438 | O6Os2Rb1 | A or C Spinel para magnetic |
| 227.4.28 | 157437 | O6Os2Rb1 | A |
| 227.4.29 | 157436 | O6Os2Rb1 | A |
| 227.4.30 | 157435 | O6Os2Rb1 | A |
| 227.4.31 | 99181 | O4V2Zn1 | A or C Spinel spin glass spinpolarized - frustration in spinles |
| 227.4.32 | 55442 | O4V2Zn1 | A |
| 227.4.34 | 105162 | Na1Pt2 | A |
| 227.4.35 | 54375 | Mn2Y1 | C spin polarized |
| 227.4.36 | 202516 | Mn2O7Y2 | insulating magnetic |
| 227.4.38 | 104935 | Mn2Np1 | C spin polarized |
| 227.4.39 | 94992 | Mn1S4Sc2 | A or C |
| 227.4.40 | 248565 | Mg1O4V2 | A |
| 227.4.41 | 89352 | Li1O4V2 | A or C Spinel para magnetic |
| 227.4.42 | 89351 | Li1O4V2 | A or C Spinel para magnetic |
| 227.4.43 | 89350 | Li1O4V2 | A or C Spinel para magnetic |
| 227.4.44 | 89349 | Li1O4V2 | Spinel magnetic V3 |
| 227.4.45 | 89348 | Li1O4V2 | or C Spinel para magnetic |
| 227.4.46 | 89347 | Li1O4V2 | A or C Spinel para magnetic |
| 227.4.47 | 89346 | Li1O4V2 | A or C Spinel para magnetic a strongly temperature-dependent |
| 227.4.48 | 89345 | Li1O4V2 | A or C Spinel para magnetic |
| 227.4.49 | 89344 | Li1O4V2 | A or C Spinel para magnetic |
| 227.4.50 | 418091 | Li1O4V2 | A or C Spinel para magnetic |
| 227.4.51 | 418090 | Li1O4V2 | A |
| 227.4.52 | 201818 | Li1O4V2 | A |
| 227.4.53 | 170768 | Li1O4V2 | A or C Spinel para magnetic |
| 227.4.54 | 78482 | Li1O4Ti2 | A |
| 227.4.55 | 72859 | Li1O4Ti2 | A superconductor |
| 227.4.56 | 48127 | Li1O4Ti2 | A superconductivity |
| 227.4.57 | 154982 | Li1O4Ti2 | A superconductor !! |
| 227.4.58 | 94340 | Li1Mn2O4 | A |
| 227.4.59 | 94339 | Li1Mn2O4 | C Spinel magnetic |

Continued on next column

**TABLE XX– continued**

| TQC | ICSD | Stoich. Formula | SubSG |
|---|---|---|---|
| 227.4.60 | 89985 | Li1Mn2O4 | AFM magnetic |
| 227.4.61 | 89459 | Li1Mn2O4 | C Spinel magnetic |
| 227.4.62 | 88644 | Li1Mn2O4 | A |
| 227.4.63 | 50415 | Li1Mn2O4 | C AFM magnetic |
| 227.4.64 | 261557 | Li1Mn2O4 | C Spinel magnetic AFM |
| 227.4.65 | 192369 | Li1Mn2O4 | C spin polarized |
| 227.4.66 | 155665 | Li1Mn2O4 | A magnetic spin polarized calculations |
| 227.4.67 | 155282 | Li1Mn2O4 | C spin polarized |
| 227.4.68 | 170775 | La1Mg2 | A |
| 227.4.69 | 172620 | K1O6Os2 | A but wide gap |
| 227.4.70 | 172619 | K1O6Os2 | A |
| 227.4.71 | 172618 | K1O6Os2 | A |
| 227.4.72 | 162266 | K1O6Os2 | A |
| 227.4.73 | 160698 | K1O6Os2 | A |
| 227.4.74 | 187534 | Ir2O7Y2 | A |
| 227.4.75 | 280390 | Hf1V2 | A |
| 227.4.77 | 52179 | Ge1Ni2Zn3 | A |
| 227.4.82 | 170910 | Fe2Mn1O4 | C Spinel magnetic |
| 227.4.83 | 106646 | Fe1Hf2 | A |
| 227.4.84 | 4083 | F6Hg2Ni2S1 | A |
| 227.4.85 | 166545 | F6Hg2Ni2S1 | A |
| 227.4.86 | 27329 | F6Hg2Ni2O1 | A |
| 227.4.87 | 202047 | F3Fe1 | insulting magnetic |
| 227.4.93 | 156367 | Cu2F6Hg2S1 | A |
| 227.4.94 | 156366 | Cu2F6Hg2S1 | C spin polarized |
| 227.4.95 | 190441 | Cu1S4V2 | A all non magnetic spinels work but Crossover from incoherent-m |
| 227.4.96 | 10035 | Cu1S4V2 | A Spinel para magnetic |
| 227.4.97 | 75531 | Cu1Ir2S4 | A |
| 227.4.98 | 4206 | Cs2O12U4 | C spin polarized |
| 227.4.99 | 421173 | Cs1Fe1O2 | C spin polarized |
| 227.4.100 | 174318 | Cs1Fe1O2 | C spin polarized |
| 227.4.101 | 425260 | Cs1F6Fe2 | C spin polarized |
| 227.4.102 | 31298 | Cr2Ni1O4 | C spin polarized |
| 227.4.103 | 280061 | Cr2Ni1O4 | C spin polarized |
| 227.4.104 | 97620 | Cr2Cu1Se4 | C spin polarized |
| 227.4.105 | 87477 | Cr2Cu1Se4 | C spin polarized |
| 227.4.106 | 87476 | Cr2Cu1Se4 | C spin polarized |
| 227.4.107 | 73403 | Cr2Cu1Se4 | C spin polarized |
| 227.4.108 | 291003 | Cr2Cu1S4 | C spin polarized |
| 227.4.109 | 291002 | Cr2Cu1S4 | C spin polarized |
| 227.4.110 | 291001 | Cr2Cu1S4 | C spin polarized |
| 227.4.111 | 291000 | Cr2Cu1S4 | A |
| 227.4.112 | 290999 | Cr2Cu1S4 | C spin polarized |
| 227.4.113 | 246081 | Cr2Cu1O4 | C spin polarized |
| 227.4.114 | 63165 | Co3O4 | C Spinel magnetic |
| 227.4.115 | 63164 | Co3O4 | C spin polarized |
| 227.4.116 | 36256 | Co3O4 | C ASpinel magnetic |
| 227.4.117 | 290808 | Co1Cr2S4 | C spin polarized |
| 227.4.118 | 180008 | Cd2O7Tc2 | C spin polarized |
| 227.4.121 | 156409 | Ca2O7Ru2 | A |
| 227.4.122 | 50048 | Ca2F1O6Ta2 | A |
| 227.4.123 | 55069 | Bi2Rb1 | A |
| 227.4.124 | 55068 | Bi2K1 | A |
| 227.4.125 | 237820 | Bi2Ir2O7 | A |
| 227.4.126 | 55070 | Bi2Cs1 | A |
| 227.4.127 | 108001 | Au2Na1 | A |
| 227.4.128 | 57911 | Al2Ho1 | C spin polarized |

Continued on next column

**TABLE XX– continued**

| TQC | ICSD | Stoich. Formula | SubSG |
|---|---|---|---|
| 227.4.129 | 290133 | Al2Co1O4 | C spin polarized |
| 227.4.130 | 170868 | Ag1In2Na3 | A |
| 227.4.131 | 71695 | Ag1Cr2Te4 | C spin polarized |
| 229.1.1 | 26270 | As7Re3 | A |
| 229.3.1 | 249996 | Fe1Ga12Y4 | A |
| 229.3.2 | 247295 | Ce4Rh12Si1 | C spin polarized |
| 229.4.2 | 54511 | Ru3Sn7 | A |
| 229.4.3 | 54510 | Ru3Sn7 | A |
| 229.4.4 | 24303 | Mo3Sb7 | A |
| 229.4.5 | 164762 | Mo3Sb7 | A |
| 229.4.6 | 173131 | Mo1 | A |
| 229.4.7 | 173127 | Mo1 | A |
| 229.4.8 | 428097 | La6Pd13Zn4 | A |
| 229.4.9 | 412776 | Ir3Sn7 | A |
| 229.4.10 | 409596 | In7Pt3 | A |
| 229.4.11 | 408314 | In7Pd3 | A |
| 229.4.12 | 408313 | Ga7Ni3 | A |
| 229.4.13 | 153909 | Ga12Pt1Y4 | A |
| 229.4.17 | 421178 | Cd4La6Pd13 | A |
| 229.4.18 | 195760 | Ca5Ge6Pd6 | A |
| 229.4.20 | 90326 | As6Ru7U4 | E |

The End

TABLE XXI: The computed irreducible representations for the valence bands for the compounds of SEBR and NLC in space group 2. See the list of tables.

| TQC | ICSD | NoE | SF | | | | |
|---|---|---|---|---|---|---|---|
| 2.2.1 | 202564 | 72 | S1Ta6 | | | | |
| -GM | -R | -T | -U | -V | -X | -Y | -Z |
| 3+3 2 | 2+2 2 | 2+2 2 | 2+2 2 | 3+3 2 | 3+3 2 | 2+2 2 | 2+2 2 |
| 2+2 2 | 3+3 2 | 3+3 2 | 3+3 2 | 2+2 2 | 2+2 2 | 3+3 2 | 3+3 2 |
| 2+2 2 | 3+3 2 | 3+3 2 | 3+3 2 | 2+2 2 | 2+2 2 | 3+3 2 | 3+3 2 |
| 3+3 2 | 2+2 2 | 2+2 2 | 2+2 2 | 3+3 2 | 3+3 2 | 2+2 2 | 2+2 2 |
| 3+3 2 | 2+2 2 | 2+2 2 | 3+3 2 | 3+3 2 | 2+2 2 | 2+2 2 | 3+3 2 |
| 2+2 2 | 3+3 2 | 3+3 2 | 2+2 2 | 2+2 2 | 3+3 2 | 3+3 2 | 3+3 2 |
| 3+3 2 | 3+3 2 | 3+3 2 | 3+3 2 | 3+3 2 | 2+2 2 | 2+2 2 | 2+2 2 |
| 3+3 2 | 2+2 2 | 2+2 2 | 2+2 2 | 2+2 2 | 3+3 2 | 3+3 2 | 2+2 2 |
| 2+2 2 | 2+2 2 | 3+3 2 | 2+2 2 | 3+3 2 | 3+3 2 | 3+3 2 | 3+3 2 |
| 3+3 2 | 3+3 2 | 2+2 2 | 3+3 2 | 2+2 2 | 2+2 2 | 2+2 2 | 2+2 2 |
| 2+2 2 | 2+2 2 | 3+3 2 | 3+3 2 | 3+3 2 | 2+2 2 | 3+3 2 | 3+3 2 |
| 3+3 2 | 3+3 2 | 2+2 2 | 2+2 2 | 2+2 2 | 3+3 2 | 3+3 2 | 3+3 2 |
| 2+2 2 | 2+2 2 | 3+3 2 | 3+3 2 | 2+2 2 | 2+2 2 | 2+2 2 | 3+3 2 |
| 3+3 2 | 3+3 2 | 3+3 2 | 2+2 2 | 3+3 2 | 3+3 2 | 3+3 2 | 3+3 2 |
| 3+3 2 | 2+2 2 | 3+3 2 | 2+2 2 | 2+2 2 | 2+2 2 | 3+3 2 | 3+3 2 |
| 3+3 2 | 3+3 2 | 3+3 2 | 3+3 2 | 3+3 2 | 3+3 2 | 3+3 2 | 2+2 2 |
| 2+2 2 | 3+3 2 | 3+3 2 | 3+3 2 | 2+2 2 | 2+2 2 | 3+3 2 | 3+3 2 |
| 3+3 2 | 2+2 2 | 2+2 2 | 3+3 2 | 3+3 2 | 2+2 2 | 2+2 2 | 2+2 2 |
| 3+3 2 | 3+3 2 | 3+3 2 | 2+2 2 | 3+3 2 | 3+3 2 | 3+3 2 | 2+2 2 |
| 2+2 2 | 2+2 2 | 2+2 2 | 3+3 2 | 2+2 2 | 3+3 2 | 2+2 2 | 2+2 2 |
| 3+3 2 | 3+3 2 | 3+3 2 | 2+2 2 | 2+2 2 | 2+2 2 | 3+3 2 | 3+3 2 |
| 2+2 2 | 3+3 2 | 2+2 2 | 2+2 2 | 3+3 2 | 3+3 2 | 2+2 2 | 2+2 2 |
| 2+2 2 | 2+2 2 | 3+3 2 | 3+3 2 | 2+2 2 | 2+2 2 | 2+2 2 | 3+3 2 |
| 3+3 2 | 2+2 2 | 2+2 2 | 3+3 2 | 2+2 2 | 3+3 2 | 3+3 2 | 3+3 2 |
| 3+3 2 | 3+3 2 | 3+3 2 | 2+2 2 | 3+3 2 | 2+2 2 | 3+3 2 | 3+3 2 |
| 2+2 2 | 2+2 2 | 3+3 2 | 2+2 2 | 3+3 2 | 3+3 2 | 3+3 2 | 2+2 2 |
| 3+3 2 | 3+3 2 | 3+3 2 | 3+3 2 | 3+3 2 | 2+2 2 | 3+3 2 | 2+2 2 |
| 2+2 2 | 2+2 2 | 2+2 2 | 2+2 2 | 2+2 2 | 3+3 2 | 2+2 2 | 3+3 2 |
| 3+3 2 | 2+2 2 | 2+2 2 | 3+3 2 | 3+3 2 | 2+2 2 | 2+2 2 | 2+2 2 |
| 3+3 2 | 3+3 2 | 3+3 2 | 2+2 2 | 2+2 2 | 3+3 2 | 3+3 2 | 3+3 2 |
| 2+2 2 | 2+2 2 | 3+3 2 | 2+2 2 | 2+2 2 | 3+3 2 | 3+3 2 | 2+2 2 |
| 3+3 2 | 3+3 2 | 2+2 2 | 3+3 2 | 3+3 2 | 2+2 2 | 2+2 2 | 2+2 2 |
| 2+2 2 | 2+2 2 | 2+2 2 | 2+2 2 | 2+2 2 | 3+3 2 | 2+2 2 | 2+2 2 |
| 2+2 2 | 3+3 2 | 2+2 2 | 3+3 2 | 3+3 2 | 2+2 2 | 2+2 2 | 3+3 2 |
| 2+2 2 | 3+3 2 | 2+2 2 | 2+2 2 | 3+3 2 | 2+2 2 | 2+2 2 | 3+3 2 |
| 2+2 2 | 2+2 2 | 3+3 2 | 3+3 2 | 2+2 2 | 3+3 2 | 3+3 2 | 2+2 2 |
| 2.2.2 | 41017 | 116 | P3Tc2 | | | | |
| -GM | -R | -T | -U | -V | -X | -Y | -Z |
| 3+3 2 | 2+2 2 | 3+3 2 | 3+3 2 | 3+3 2 | 2+2 2 | 2+2 2 | 2+2 2 |
| 3+3 2 | 3+3 2 | 3+3 2 | 2+2 2 | 2+2 2 | 3+3 2 | 3+3 2 | 3+3 2 |
| 2+2 2 | 3+3 2 | 2+2 2 | 3+3 2 | 3+3 2 | 3+3 2 | 3+3 2 | 3+3 2 |
| 2+2 2 | 2+2 2 | 2+2 2 | 2+2 2 | 2+2 2 | 2+2 2 | 3+3 2 | 3+3 2 |
| 3+3 2 | 3+3 2 | 2+2 2 | 2+2 2 | 2+2 2 | 3+3 2 | 3+3 2 | 2+2 2 |
| 3+3 2 | 2+2 2 | 3+3 2 | 3+3 2 | 3+3 2 | 2+2 2 | 2+2 2 | 3+3 2 |
| 2+2 2 | 3+3 2 | 3+3 2 | 2+2 2 | 3+3 2 | 3+3 2 | 2+2 2 | 2+2 2 |
| 2+2 2 | 2+2 2 | 3+3 2 | 3+3 2 | 2+2 2 | 2+2 2 | 3+3 2 | 3+3 2 |
| 3+3 2 | 2+2 2 | 2+2 2 | 2+2 2 | 3+3 2 | 2+2 2 | 2+2 2 | 3+3 2 |
| 2+2 2 | 3+3 2 | 3+3 2 | 3+3 2 | 2+2 2 | 3+3 2 | 3+3 2 | 2+2 2 |
| 3+3 2 | 3+3 2 | 2+2 2 | 3+3 2 | 2+2 2 | 3+3 2 | 2+2 2 | 2+2 2 |
| 2+2 2 | 2+2 2 | 2+2 2 | 2+2 2 | 3+3 2 | 2+2 2 | 2+2 2 | 2+2 2 |
| 3+3 2 | 3+3 2 | 3+3 2 | 2+2 2 | 2+2 2 | 3+3 2 | 2+2 2 | 2+2 2 |
| 3+3 2 | 2+2 2 | 2+2 2 | 3+3 2 | 3+3 2 | 2+2 2 | 2+2 2 | 3+3 2 |
| 2+2 2 | 3+3 2 | 3+3 2 | 3+3 2 | 3+3 2 | 2+2 2 | 3+3 2 | 2+2 2 |
| 2+2 2 | 2+2 2 | 3+3 2 | 2+2 2 | 2+2 2 | 3+3 2 | 2+2 2 | 2+2 2 |
| 3+3 2 | 2+2 2 | 3+3 2 | 3+3 2 | 3+3 2 | 3+3 2 | 3+3 2 | 2+2 2 |
| 2+2 2 | 3+3 2 | 2+2 2 | 2+2 2 | 2+2 2 | 2+2 2 | 2+2 2 | 3+3 2 |

Continued on next column

**TABLE XXI– continued**

| TQC | ICSD | NoE | SF | | | | |
|---|---|---|---|---|---|---|---|
| 3+3 2 | 2+2 2 | 2+2 2 | 3+3 2 | 3+3 2 | 3+3 2 | 3+3 2 | 2+2 2 |
| 2+2 2 | 3+3 2 | 2+2 2 | 2+2 2 | 2+2 2 | 2+2 2 | 3+3 2 | 3+3 2 |
| 2+2 2 | 3+3 2 | 2+2 2 | 2+2 2 | 3+3 2 | 2+2 2 | 3+3 2 | 3+3 2 |
| 3+3 2 | 2+2 2 | 3+3 2 | 3+3 2 | 2+2 2 | 3+3 2 | 2+2 2 | 3+3 2 |
| 2+2 2 | 3+3 2 | 3+3 2 | 3+3 2 | 2+2 2 | 2+2 2 | 2+2 2 | 2+2 2 |
| 3+3 2 | 2+2 2 | 2+2 2 | 2+2 2 | 3+3 2 | 3+3 2 | 2+2 2 | 3+3 2 |
| 3+3 2 | 2+2 2 | 3+3 2 | 2+2 2 | 2+2 2 | 2+2 2 | 3+3 2 | 3+3 2 |
| 2+2 2 | 3+3 2 | 2+2 2 | 3+3 2 | 3+3 2 | 3+3 2 | 3+3 2 | 3+3 2 |
| 2+2 2 | 2+2 2 | 2+2 2 | 3+3 2 | 2+2 2 | 2+2 2 | 3+3 2 | 2+2 2 |
| 3+3 2 | 3+3 2 | 3+3 2 | 2+2 2 | 3+3 2 | 3+3 2 | 2+2 2 | 3+3 2 |
| 2+2 2 | 2+2 2 | 3+3 2 | 2+2 2 | 3+3 2 | 3+3 2 | 2+2 2 | 3+3 2 |
| 2+2 2 | 3+3 2 | 2+2 2 | 2+2 2 | 3+3 2 | 2+2 2 | 3+3 2 | 2+2 2 |
| 3+3 2 | 3+3 2 | 2+2 2 | 3+3 2 | 2+2 2 | 2+2 2 | 3+3 2 | 2+2 2 |
| 2+2 2 | 2+2 2 | 3+3 2 | 3+3 2 | 2+2 2 | 3+3 2 | 2+2 2 | 3+3 2 |
| 3+3 2 | 3+3 2 | 3+3 2 | 2+2 2 | 3+3 2 | 3+3 2 | 2+2 2 | 2+2 2 |
| 2+2 2 | 2+2 2 | 3+3 2 | 3+3 2 | 2+2 2 | 2+2 2 | 3+3 2 | 2+2 2 |
| 3+3 2 | 3+3 2 | 2+2 2 | 2+2 2 | 3+3 2 | 2+2 2 | 2+2 2 | 3+3 2 |
| 3+3 2 | 2+2 2 | 3+3 2 | 3+3 2 | 2+2 2 | 3+3 2 | 3+3 2 | 2+2 2 |
| 2+2 2 | 2+2 2 | 2+2 2 | 2+2 2 | 3+3 2 | 3+3 2 | 3+3 2 | 3+3 2 |
| 3+3 2 | 3+3 2 | 2+2 2 | 3+3 2 | 2+2 2 | 2+2 2 | 3+3 2 | 2+2 2 |
| 2+2 2 | 2+2 2 | 3+3 2 | 2+2 2 | 2+2 2 | 3+3 2 | 2+2 2 | 3+3 2 |
| 2+2 2 | 2+2 2 | 2+2 2 | 3+3 2 | 3+3 2 | 2+2 2 | 2+2 2 | 2+2 2 |
| 3+3 2 | 3+3 2 | 3+3 2 | 2+2 2 | 2+2 2 | 2+2 2 | 3+3 2 | 2+2 2 |
| 3+3 2 | 3+3 2 | 2+2 2 | 3+3 2 | 3+3 2 | 3+3 2 | 2+2 2 | 3+3 2 |
| 2+2 2 | 3+3 2 | 3+3 2 | 2+2 2 | 3+3 2 | 3+3 2 | 3+3 2 | 2+2 2 |
| 3+3 2 | 2+2 2 | 2+2 2 | 3+3 2 | 2+2 2 | 2+2 2 | 3+3 2 | 3+3 2 |
| 3+3 2 | 3+3 2 | 2+2 2 | 2+2 2 | 3+3 2 | 3+3 2 | 3+3 2 | 3+3 2 |
| 2+2 2 | 2+2 2 | 3+3 2 | 3+3 2 | 2+2 2 | 2+2 2 | 2+2 2 | 2+2 2 |
| 3+3 2 | 2+2 2 | 2+2 2 | 2+2 2 | 2+2 2 | 2+2 2 | 3+3 2 | 3+3 2 |
| 2+2 2 | 3+3 2 | 3+3 2 | 3+3 2 | 3+3 2 | 3+3 2 | 2+2 2 | 3+3 2 |
| 3+3 2 | 2+2 2 | 2+2 2 | 2+2 2 | 2+2 2 | 3+3 2 | 3+3 2 | 2+2 2 |
| 2+2 2 | 3+3 2 | 2+2 2 | 3+3 2 | 3+3 2 | 2+2 2 | 3+3 2 | 3+3 2 |
| 3+3 2 | 2+2 2 | 2+2 2 | 2+2 2 | 3+3 2 | 2+2 2 | 2+2 2 | 2+2 2 |
| 3+3 2 | 3+3 2 | 3+3 2 | 3+3 2 | 2+2 2 | 3+3 2 | 2+2 2 | 3+3 2 |
| 3+3 2 | 2+2 2 | 3+3 2 | 3+3 2 | 3+3 2 | 3+3 2 | 2+2 2 | 3+3 2 |
| 3+3 2 | 3+3 2 | 2+2 2 | 2+2 2 | 2+2 2 | 2+2 2 | 3+3 2 | 2+2 2 |
| 2+2 2 | 3+3 2 | 3+3 2 | 2+2 2 | 2+2 2 | 3+3 2 | 2+2 2 | 3+3 2 |
| 3+3 2 | 2+2 2 | 2+2 2 | 3+3 2 | 3+3 2 | 2+2 2 | 3+3 2 | 3+3 2 |
| 3+3 2 | 3+3 2 | 2+2 2 | 2+2 2 | 3+3 2 | 2+2 2 | 3+3 2 | 2+2 2 |
| 2+2 2 | 2+2 2 | 3+3 2 | 3+3 2 | 2+2 2 | 3+3 2 | 2+2 2 | 2+2 2 |
| 2.2.3 | 66584 | 158 | O9V5 | | | | |
| -GM | -R | -T | -U | -V | -X | -Y | -Z |
| 3+3 2 | 2+2 2 | 2+2 2 | 2+2 2 | 3+3 2 | 3+3 2 | 3+3 2 | 2+2 2 |
| 3+3 2 | 3+3 2 | 3+3 2 | 2+2 2 | 2+2 2 | 2+2 2 | 2+2 2 | 3+3 2 |
| 2+2 2 | 3+3 2 | 3+3 2 | 3+3 2 | 2+2 2 | 3+3 2 | 3+3 2 | 3+3 2 |
| 2+2 2 | 2+2 2 | 2+2 2 | 2+2 2 | 3+3 2 | 2+2 2 | 2+2 2 | 2+2 2 |
| 3+3 2 | 3+3 2 | 3+3 2 | 3+3 2 | 2+2 2 | 3+3 2 | 2+2 2 | 3+3 2 |
| 2+2 2 | 2+2 2 | 2+2 2 | 3+3 2 | 3+3 2 | 2+2 2 | 3+3 2 | 2+2 2 |
| 3+3 2 | 3+3 2 | 2+2 2 | 2+2 2 | 2+2 2 | 2+2 2 | 3+3 2 | 3+3 2 |
| 3+3 2 | 2+2 2 | 3+3 2 | 3+3 2 | 3+3 2 | 2+2 2 | 2+2 2 | 2+2 2 |
| 3+3 2 | 3+3 2 | 3+3 2 | 2+2 2 | 3+3 2 | 2+2 2 | 3+3 2 | 3+3 2 |
| 2+2 2 | 2+2 2 | 2+2 2 | 3+3 2 | 2+2 2 | 3+3 2 | 2+2 2 | 3+3 2 |
| 2+2 2 | 2+2 2 | 3+3 2 | 2+2 2 | 2+2 2 | 2+2 2 | 2+2 2 | 2+2 2 |
| 3+3 2 | 3+3 2 | 2+2 2 | 3+3 2 | 3+3 2 | 3+3 2 | 3+3 2 | 3+3 2 |
| 3+3 2 | 3+3 2 | 2+2 2 | 2+2 2 | 2+2 2 | 3+3 2 | 2+2 2 | 3+3 2 |
| 2+2 2 | 2+2 2 | 3+3 2 | 2+2 2 | 3+3 2 | 2+2 2 | 3+3 2 | 2+2 2 |
| 3+3 2 | 3+3 2 | 2+2 2 | 3+3 2 | 3+3 2 | 2+2 2 | 3+3 2 | 2+2 2 |
| 2+2 2 | 2+2 2 | 2+2 2 | 3+3 2 | 3+3 2 | 3+3 2 | 2+2 2 | 3+3 2 |
| 2+2 2 | 3+3 2 | 3+3 2 | 2+2 2 | 2+2 2 | 3+3 2 | 2+2 2 | 2+2 2 |
| 2+2 2 | 2+2 2 | 3+3 2 | 3+3 2 | 2+2 2 | 3+3 2 | 3+3 2 | 2+2 2 |
| 3+3 2 | 3+3 2 | 3+3 2 | 2+2 2 | 2+2 2 | 2+2 2 | 3+3 2 | 3+3 2 |

Continued on next column

**TABLE XXI– continued**

| TQC | ICSD | NoE | SF | | | | |
|---|---|---|---|---|---|---|---|
| | 2+2 2 | 2+2 2 | 2+2 2 | 3+3 2 | 3+3 2 | 2+2 2 | 2+2 2 | 3+3 2 |
| | 2+2 2 | 3+3 2 | 2+2 2 | 2+2 2 | 3+3 2 | 2+2 2 | 2+2 2 | 2+2 2 |
| | 3+3 2 | 2+2 2 | 3+3 2 | 3+3 2 | 2+2 2 | 3+3 2 | 3+3 2 | 2+2 2 |
| | 3+3 2 | 2+2 2 | 2+2 2 | 2+2 2 | 2+2 2 | 3+3 2 | 3+3 2 | 3+3 2 |
| | 3+3 2 | 2+2 2 | 3+3 2 | 2+2 2 | 3+3 2 | 2+2 2 | 2+2 2 | 3+3 2 |
| | 2+2 2 | 3+3 2 | 3+3 2 | 3+3 2 | 2+2 2 | 3+3 2 | 2+2 2 | 3+3 2 |
| | 3+3 2 | 3+3 2 | 2+2 2 | 2+2 2 | 3+3 2 | 2+2 2 | 3+3 2 | 2+2 2 |
| | 3+3 2 | 3+3 2 | 3+3 2 | 2+2 2 | 3+3 2 | 2+2 2 | 2+2 2 | 3+3 2 |
| | 3+3 2 | 2+2 2 | 2+2 2 | 3+3 2 | 2+2 2 | 3+3 2 | 2+2 2 | 2+2 2 |
| | 2+2 2 | 2+2 2 | 3+3 2 | 3+3 2 | 3+3 2 | 2+2 2 | 3+3 2 | 3+3 2 |
| | 3+3 2 | 2+2 2 | 2+2 2 | 2+2 2 | 2+2 2 | 3+3 2 | 3+3 2 | 2+2 2 |
| | 2+2 2 | 3+3 2 | 3+3 2 | 2+2 2 | 3+3 2 | 3+3 2 | 2+2 2 | 3+3 2 |
| | 3+3 2 | 2+2 2 | 2+2 2 | 3+3 2 | 2+2 2 | 2+2 2 | 3+3 2 | 2+2 2 |
| | 3+3 2 | 3+3 2 | 3+3 2 | 2+2 2 | 3+3 2 | 2+2 2 | 2+2 2 | 3+3 2 |
| | 2+2 2 | 2+2 2 | 2+2 2 | 3+3 2 | 2+2 2 | 3+3 2 | 3+3 2 | 2+2 2 |
| | 2+2 2 | 3+3 2 | 3+3 2 | 2+2 2 | 2+2 2 | 3+3 2 | 3+3 2 | 3+3 2 |
| | 2+2 2 | 2+2 2 | 2+2 2 | 2+2 2 | 3+3 2 | 2+2 2 | 2+2 2 | 2+2 2 |
| | 3+3 2 | 3+3 2 | 2+2 2 | 3+3 2 | 3+3 2 | 3+3 2 | 3+3 2 | 3+3 2 |
| | 3+3 2 | 3+3 2 | 3+3 2 | 2+2 2 | 2+2 2 | 2+2 2 | 2+2 2 | 2+2 2 |
| | 3+3 2 | 2+2 2 | 2+2 2 | 3+3 2 | 2+2 2 | 2+2 2 | 2+2 2 | 2+2 2 |
| | 2+2 2 | 3+3 2 | 3+3 2 | 2+2 2 | 3+3 2 | 3+3 2 | 3+3 2 | 3+3 2 |
| | 3+3 2 | 3+3 2 | 2+2 2 | 2+2 2 | 3+3 2 | 2+2 2 | 2+2 2 | 3+3 2 |
| | 3+3 2 | 2+2 2 | 3+3 2 | 3+3 2 | 3+3 2 | 2+2 2 | 3+3 2 | 2+2 2 |
| | 2+2 2 | 3+3 2 | 2+2 2 | 3+3 2 | 2+2 2 | 3+3 2 | 2+2 2 | 3+3 2 |
| | 3+3 2 | 2+2 2 | 3+3 2 | 2+2 2 | 2+2 2 | 3+3 2 | 3+3 2 | 2+2 2 |
| | 2+2 2 | 3+3 2 | 3+3 2 | 3+3 2 | 2+2 2 | 2+2 2 | 3+3 2 | 3+3 2 |
| | 2+2 2 | 2+2 2 | 2+2 2 | 2+2 2 | 3+3 2 | 3+3 2 | 3+3 2 | 2+2 2 |
| | 3+3 2 | 3+3 2 | 2+2 2 | 2+2 2 | 2+2 2 | 2+2 2 | 2+2 2 | 2+2 2 |
| | 2+2 2 | 2+2 2 | 3+3 2 | 3+3 2 | 3+3 2 | 3+3 2 | 2+2 2 | 2+2 2 |
| | 3+3 2 | 3+3 2 | 2+2 2 | 3+3 2 | 3+3 2 | 3+3 2 | 3+3 2 | 3+3 2 |
| | 2+2 2 | 2+2 2 | 3+3 2 | 2+2 2 | 2+2 2 | 3+3 2 | 2+2 2 | 2+2 2 |
| | 3+3 2 | 3+3 2 | 2+2 2 | 3+3 2 | 3+3 2 | 3+3 2 | 2+2 2 | 3+3 2 |
| | 2+2 2 | 2+2 2 | 3+3 2 | 2+2 2 | 2+2 2 | 2+2 2 | 3+3 2 | 2+2 2 |
| | 2+2 2 | 3+3 2 | 3+3 2 | 3+3 2 | 2+2 2 | 2+2 2 | 2+2 2 | 3+3 2 |
| | 3+3 2 | 2+2 2 | 2+2 2 | 2+2 2 | 3+3 2 | 3+3 2 | 3+3 2 | 3+3 2 |
| | 2+2 2 | 3+3 2 | 3+3 2 | 3+3 2 | 3+3 2 | 2+2 2 | 2+2 2 | 2+2 2 |
| | 3+3 2 | 2+2 2 | 2+2 2 | 3+3 2 | 2+2 2 | 2+2 2 | 3+3 2 | 3+3 2 |
| | 3+3 2 | 3+3 2 | 2+2 2 | 2+2 2 | 2+2 2 | 3+3 2 | 3+3 2 | 2+2 2 |
| | 2+2 2 | 2+2 2 | 3+3 2 | 3+3 2 | 2+2 2 | 3+3 2 | 2+2 2 | 2+2 2 |
| | 3+3 2 | 2+2 2 | 2+2 2 | 3+3 2 | 3+3 2 | 3+3 2 | 2+2 2 | 3+3 2 |
| | 2+2 2 | 3+3 2 | 3+3 2 | 2+2 2 | 3+3 2 | 2+2 2 | 3+3 2 | 2+2 2 |
| | 2+2 2 | 3+3 2 | 2+2 2 | 2+2 2 | 2+2 2 | 2+2 2 | 3+3 2 | 2+2 2 |
| | 3+3 2 | 2+2 2 | 2+2 2 | 3+3 2 | 3+3 2 | 3+3 2 | 2+2 2 | 3+3 2 |
| | 2+2 2 | 2+2 2 | 3+3 2 | 2+2 2 | 3+3 2 | 2+2 2 | 3+3 2 | 3+3 2 |
| | 3+3 2 | 3+3 2 | 2+2 2 | 2+2 2 | 2+2 2 | 3+3 2 | 2+2 2 | 3+3 2 |
| | 2+2 2 | 3+3 2 | 3+3 2 | 3+3 2 | 2+2 2 | 3+3 2 | 2+2 2 | 2+2 2 |
| | 2+2 2 | 2+2 2 | 3+3 2 | 2+2 2 | 3+3 2 | 2+2 2 | 3+3 2 | 3+3 2 |
| | 3+3 2 | 2+2 2 | 3+3 2 | 3+3 2 | 2+2 2 | 3+3 2 | 2+2 2 | 2+2 2 |
| | 2+2 2 | 3+3 2 | 2+2 2 | 3+3 2 | 3+3 2 | 2+2 2 | 3+3 2 | 2+2 2 |
| | 2+2 2 | 3+3 2 | 2+2 2 | 3+3 2 | 2+2 2 | 3+3 2 | 2+2 2 | 3+3 2 |
| | 2+2 2 | 2+2 2 | 3+3 2 | 3+3 2 | 3+3 2 | 2+2 2 | 3+3 2 | 2+2 2 |
| | 2+2 2 | 3+3 2 | 3+3 2 | 2+2 2 | 3+3 2 | 2+2 2 | 3+3 2 | 2+2 2 |
| | 3+3 2 | 2+2 2 | 2+2 2 | 3+3 2 | 2+2 2 | 3+3 2 | 2+2 2 | 3+3 2 |
| | 2+2 2 | 2+2 2 | 3+3 2 | 2+2 2 | 3+3 2 | 2+2 2 | 2+2 2 | 3+3 2 |
| | 3+3 2 | 3+3 2 | 2+2 2 | 2+2 2 | 2+2 2 | 2+2 2 | 3+3 2 | 3+3 2 |
| | 3+3 2 | 3+3 2 | 3+3 2 | 2+2 2 | 2+2 2 | 2+2 2 | 3+3 2 | 3+3 2 |
| | 3+3 2 | 2+2 2 | 2+2 2 | 3+3 2 | 3+3 2 | 2+2 2 | 2+2 2 | 3+3 2 |
| | 3+3 2 | 2+2 2 | 2+2 2 | 2+2 2 | 2+2 2 | 3+3 2 | 3+3 2 | 2+2 2 |
| | 3+3 2 | 3+3 2 | 3+3 2 | 2+2 2 | 3+3 2 | 2+2 2 | 2+2 2 | 3+3 2 |
| | 3+3 2 | 2+2 2 | 3+3 2 | 2+2 2 | 2+2 2 | 3+3 2 | 2+2 2 | 2+2 2 |
| 2.2.4 | 80363 | 154 | O9P3Ru1 | | | | | |

Continued on next column

**TABLE XXI– continued**

| TQC | ICSD | NoE | SF | | | | |
|---|---|---|---|---|---|---|---|
| | -GM | -R | -T | -U | -V | -X | -Y | -Z |
| | 3+3 2 | 2+2 2 | 3+3 2 | 3+3 2 | 2+2 2 | 2+2 2 | 3+3 2 | 2+2 2 |
| | 2+2 2 | 3+3 2 | 3+3 2 | 2+2 2 | 3+3 2 | 3+3 2 | 2+2 2 | 3+3 2 |
| | 3+3 2 | 2+2 2 | 2+2 2 | 2+2 2 | 2+2 2 | 3+3 2 | 2+2 2 | 2+2 2 |
| | 2+2 2 | 3+3 2 | 2+2 2 | 3+3 2 | 3+3 2 | 2+2 2 | 3+3 2 | 3+3 2 |
| | 3+3 2 | 2+2 2 | 3+3 2 | 2+2 2 | 2+2 2 | 2+2 2 | 3+3 2 | 3+3 2 |
| | 2+2 2 | 3+3 2 | 2+2 2 | 3+3 2 | 3+3 2 | 3+3 2 | 2+2 2 | 2+2 2 |
| | 3+3 2 | 3+3 2 | 3+3 2 | 3+3 2 | 3+3 2 | 3+3 2 | 2+2 2 | 2+2 2 |
| | 3+3 2 | 2+2 2 | 2+2 2 | 3+3 2 | 2+2 2 | 2+2 2 | 3+3 2 | 3+3 2 |
| | 2+2 2 | 3+3 2 | 3+3 2 | 2+2 2 | 3+3 2 | 3+3 2 | 2+2 2 | 3+3 2 |
| | 2+2 2 | 2+2 2 | 3+3 2 | 2+2 2 | 2+2 2 | 2+2 2 | 3+3 2 | 2+2 2 |
| | 3+3 2 | 3+3 2 | 2+2 2 | 3+3 2 | 2+2 2 | 3+3 2 | 3+3 2 | 3+3 2 |
| | 2+2 2 | 2+2 2 | 2+2 2 | 2+2 2 | 3+3 2 | 2+2 2 | 2+2 2 | 2+2 2 |
| | 3+3 2 | 3+3 2 | 3+3 2 | 2+2 2 | 2+2 2 | 3+3 2 | 2+2 2 | 2+2 2 |
| | 3+3 2 | 3+3 2 | 3+3 2 | 3+3 2 | 3+3 2 | 3+3 2 | 3+3 2 | 3+3 2 |
| | 3+3 2 | 2+2 2 | 3+3 2 | 3+3 2 | 3+3 2 | 2+2 2 | 2+2 2 | 2+2 2 |
| | 2+2 2 | 3+3 2 | 2+2 2 | 2+2 2 | 2+2 2 | 3+3 2 | 3+3 2 | 3+3 2 |
| | 2+2 2 | 2+2 2 | 2+2 2 | 2+2 2 | 3+3 2 | 2+2 2 | 3+3 2 | 3+3 2 |
| | 2+2 2 | 2+2 2 | 2+2 2 | 3+3 2 | 2+2 2 | 2+2 2 | 2+2 2 | 2+2 2 |
| | 3+3 2 | 3+3 2 | 3+3 2 | 3+3 2 | 3+3 2 | 3+3 2 | 2+2 2 | 2+2 2 |
| | 2+2 2 | 2+2 2 | 2+2 2 | 2+2 2 | 2+2 2 | 2+2 2 | 3+3 2 | 3+3 2 |
| | 3+3 2 | 3+3 2 | 3+3 2 | 2+2 2 | 3+3 2 | 3+3 2 | 3+3 2 | 2+2 2 |
| | 2+2 2 | 2+2 2 | 2+2 2 | 3+3 2 | 2+2 2 | 2+2 2 | 2+2 2 | 3+3 2 |
| | 3+3 2 | 3+3 2 | 3+3 2 | 2+2 2 | 2+2 2 | 2+2 2 | 2+2 2 | 3+3 2 |
| | 2+2 2 | 2+2 2 | 2+2 2 | 3+3 2 | 3+3 2 | 3+3 2 | 3+3 2 | 2+2 2 |
| | 3+3 2 | 3+3 2 | 3+3 2 | 3+3 2 | 3+3 2 | 3+3 2 | 3+3 2 | 3+3 2 |
| | 3+3 2 | 3+3 2 | 3+3 2 | 2+2 2 | 2+2 2 | 3+3 2 | 2+2 2 | 2+2 2 |
| | 3+3 2 | 3+3 2 | 3+3 2 | 2+2 2 | 2+2 2 | 3+3 2 | 2+2 2 | 2+2 2 |
| | 2+2 2 | 3+3 2 | 2+2 2 | 3+3 2 | 3+3 2 | 2+2 2 | 3+3 2 | 3+3 2 |
| | 2+2 2 | 2+2 2 | 3+3 2 | 2+2 2 | 2+2 2 | 3+3 2 | 2+2 2 | 2+2 2 |
| | 3+3 2 | 2+2 2 | 2+2 2 | 3+3 2 | 3+3 2 | 3+3 2 | 3+3 2 | 3+3 2 |
| | 2+2 2 | 3+3 2 | 2+2 2 | 2+2 2 | 2+2 2 | 2+2 2 | 2+2 2 | 2+2 2 |
| | 3+3 2 | 2+2 2 | 3+3 2 | 3+3 2 | 3+3 2 | 2+2 2 | 3+3 2 | 3+3 2 |
| | 2+2 2 | 3+3 2 | 2+2 2 | 3+3 2 | 3+3 2 | 2+2 2 | 3+3 2 | 3+3 2 |
| | 3+3 2 | 2+2 2 | 3+3 2 | 2+2 2 | 2+2 2 | 3+3 2 | 2+2 2 | 2+2 2 |
| | 2+2 2 | 2+2 2 | 2+2 2 | 2+2 2 | 2+2 2 | 2+2 2 | 3+3 2 | 3+3 2 |
| | 3+3 2 | 3+3 2 | 3+3 2 | 3+3 2 | 3+3 2 | 3+3 2 | 2+2 2 | 2+2 2 |
| | 3+3 2 | 2+2 2 | 2+2 2 | 3+3 2 | 3+3 2 | 3+3 2 | 3+3 2 | 3+3 2 |
| | 2+2 2 | 3+3 2 | 3+3 2 | 2+2 2 | 2+2 2 | 2+2 2 | 2+2 2 | 2+2 2 |
| | 3+3 2 | 3+3 2 | 3+3 2 | 2+2 2 | 3+3 2 | 2+2 2 | 3+3 2 | 3+3 2 |
| | 2+2 2 | 3+3 2 | 2+2 2 | 3+3 2 | 2+2 2 | 2+2 2 | 2+2 2 | 2+2 2 |
| | 2+2 2 | 2+2 2 | 3+3 2 | 3+3 2 | 2+2 2 | 3+3 2 | 2+2 2 | 2+2 2 |
| | 3+3 2 | 2+2 2 | 2+2 2 | 2+2 2 | 3+3 2 | 3+3 2 | 3+3 2 | 3+3 2 |
| | 2+2 2 | 2+2 2 | 2+2 2 | 2+2 2 | 2+2 2 | 2+2 2 | 3+3 2 | 2+2 2 |
| | 3+3 2 | 3+3 2 | 3+3 2 | 3+3 2 | 3+3 2 | 3+3 2 | 2+2 2 | 3+3 2 |
| | 3+3 2 | 2+2 2 | 2+2 2 | 3+3 2 | 2+2 2 | 3+3 2 | 2+2 2 | 2+2 2 |
| | 2+2 2 | 3+3 2 | 3+3 2 | 2+2 2 | 3+3 2 | 2+2 2 | 3+3 2 | 3+3 2 |
| | 3+3 2 | 2+2 2 | 2+2 2 | 2+2 2 | 3+3 2 | 2+2 2 | 3+3 2 | 3+3 2 |
| | 3+3 2 | 2+2 2 | 3+3 2 | 3+3 2 | 2+2 2 | 3+3 2 | 2+2 2 | 2+2 2 |
| | 2+2 2 | 3+3 2 | 3+3 2 | 3+3 2 | 3+3 2 | 2+2 2 | 3+3 2 | 3+3 2 |
| | 2+2 2 | 3+3 2 | 2+2 2 | 2+2 2 | 2+2 2 | 2+2 2 | 2+2 2 | 2+2 2 |
| | 3+3 2 | 2+2 2 | 3+3 2 | 3+3 2 | 3+3 2 | 3+3 2 | 2+2 2 | 2+2 2 |
| | 3+3 2 | 3+3 2 | 3+3 2 | 2+2 2 | 2+2 2 | 2+2 2 | 3+3 2 | 2+2 2 |
| | 2+2 2 | 2+2 2 | 2+2 2 | 3+3 2 | 3+3 2 | 3+3 2 | 2+2 2 | 3+3 2 |
| | 3+3 2 | 2+2 2 | 3+3 2 | 2+2 2 | 2+2 2 | 3+3 2 | 3+3 2 | 3+3 2 |
| | 2+2 2 | 3+3 2 | 2+2 2 | 3+3 2 | 3+3 2 | 2+2 2 | 3+3 2 | 2+2 2 |
| | 3+3 2 | 3+3 2 | 2+2 2 | 2+2 2 | 2+2 2 | 3+3 2 | 2+2 2 | 3+3 2 |
| | 2+2 2 | 3+3 2 | 3+3 2 | 3+3 2 | 2+2 2 | 3+3 2 | 3+3 2 | 3+3 2 |
| | 3+3 2 | 2+2 2 | 2+2 2 | 2+2 2 | 3+3 2 | 2+2 2 | 2+2 2 | 2+2 2 |
| | 2+2 2 | 3+3 2 | 3+3 2 | 2+2 2 | 3+3 2 | 2+2 2 | 2+2 2 | 3+3 2 |
| | 3+3 2 | 2+2 2 | 3+3 2 | 2+2 2 | 2+2 2 | 3+3 2 | 3+3 2 | 2+2 2 |

Continued on next column

**TABLE XXI– continued**

| TQC | ICSD | NoE | SF | | | | | |
|---|---|---|---|---|---|---|---|---|
| | 2+2 2 | 3+3 2 | 2+2 2 | 3+3 2 | 2+2 2 | 3+3 2 | 2+2 2 | 3+3 2 |
| | 2+2 2 | 3+3 2 | 2+2 2 | 3+3 2 | 3+3 2 | 3+3 2 | 3+3 2 | 2+2 2 |
| | 3+3 2 | 2+2 2 | 3+3 2 | 2+2 2 | 2+2 2 | 2+2 2 | 3+3 2 | 3+3 2 |
| | 2+2 2 | 3+3 2 | 2+2 2 | 3+3 2 | 3+3 2 | 3+3 2 | 2+2 2 | 2+2 2 |
| | 2+2 2 | 2+2 2 | 2+2 2 | 2+2 2 | 2+2 2 | 2+2 2 | 3+3 2 | 3+3 2 |
| | 3+3 2 | 2+2 2 | 2+2 2 | 3+3 2 | 3+3 2 | 2+2 2 | 2+2 2 | 2+2 2 |
| | 2+2 2 | 2+2 2 | 3+3 2 | 2+2 2 | 2+2 2 | 3+3 2 | 3+3 2 | 3+3 2 |
| | 3+3 2 | 3+3 2 | 3+3 2 | 3+3 2 | 3+3 2 | 2+2 2 | 2+2 2 | 2+2 2 |
| | 2+2 2 | 2+2 2 | 2+2 2 | 3+3 2 | 2+2 2 | 2+2 2 | 3+3 2 | 3+3 2 |
| | 2+2 2 | 3+3 2 | 2+2 2 | 2+2 2 | 3+3 2 | 3+3 2 | 2+2 2 | 2+2 2 |
| | 2+2 2 | 2+2 2 | 3+3 2 | 3+3 2 | 3+3 2 | 2+2 2 | 3+3 2 | 3+3 2 |
| | 3+3 2 | 2+2 2 | 2+2 2 | 2+2 2 | 2+2 2 | 2+2 2 | 2+2 2 | 2+2 2 |
| | 3+3 2 | 3+3 2 | 3+3 2 | 2+2 2 | 2+2 2 | 3+3 2 | 2+2 2 | 2+2 2 |
| | 3+3 2 | 3+3 2 | 3+3 2 | 3+3 2 | 3+3 2 | 3+3 2 | 3+3 2 | 3+3 2 |
| | 3+3 2 | 3+3 2 | 3+3 2 | 3+3 2 | 3+3 2 | 3+3 2 | 2+2 2 | 2+2 2 |
| | 3+3 2 | 3+3 2 | 3+3 2 | 2+2 2 | 2+2 2 | 3+3 2 | 3+3 2 | 3+3 2 |
| | 3+3 2 | 3+3 2 | 3+3 2 | 3+3 2 | 3+3 2 | 3+3 2 | 2+2 2 | 2+2 2 |
| 2.2.5 | 2212 | 124 | O7V4 | | | | | |
| | -GM | -R | -T | -U | -V | -X | -Y | -Z |
| | 3+3 2 | 2+2 2 | 3+3 2 | 2+2 2 | 3+3 2 | 2+2 2 | 2+2 2 | 2+2 2 |
| | 3+3 2 | 3+3 2 | 2+2 2 | 3+3 2 | 3+3 2 | 3+3 2 | 3+3 2 | 3+3 2 |
| | 2+2 2 | 2+2 2 | 2+2 2 | 3+3 2 | 3+3 2 | 3+3 2 | 3+3 2 | 2+2 2 |
| | 2+2 2 | 3+3 2 | 3+3 2 | 2+2 2 | 2+2 2 | 2+2 2 | 2+2 2 | 3+3 2 |
| | 3+3 2 | 2+2 2 | 3+3 2 | 2+2 2 | 2+2 2 | 3+3 2 | 3+3 2 | 3+3 2 |
| | 2+2 2 | 3+3 2 | 2+2 2 | 3+3 2 | 2+2 2 | 2+2 2 | 2+2 2 | 2+2 2 |
| | 3+3 2 | 3+3 2 | 2+2 2 | 2+2 2 | 2+2 2 | 2+2 2 | 2+2 2 | 2+2 2 |
| | 2+2 2 | 3+3 2 | 3+3 2 | 3+3 2 | 3+3 2 | 3+3 2 | 3+3 2 | 3+3 2 |
| | 2+2 2 | 2+2 2 | 2+2 2 | 3+3 2 | 2+2 2 | 3+3 2 | 2+2 2 | 3+3 2 |
| | 2+2 2 | 2+2 2 | 3+3 2 | 2+2 2 | 3+3 2 | 2+2 2 | 3+3 2 | 2+2 2 |
| | 3+3 2 | 2+2 2 | 2+2 2 | 3+3 2 | 2+2 2 | 2+2 2 | 2+2 2 | 3+3 2 |
| | 3+3 2 | 3+3 2 | 3+3 2 | 2+2 2 | 2+2 2 | 3+3 2 | 3+3 2 | 2+2 2 |
| | 3+3 2 | 2+2 2 | 2+2 2 | 2+2 2 | 3+3 2 | 2+2 2 | 2+2 2 | 2+2 2 |
| | 2+2 2 | 3+3 2 | 3+3 2 | 3+3 2 | 3+3 2 | 3+3 2 | 3+3 2 | 3+3 2 |
| | 3+3 2 | 3+3 2 | 3+3 2 | 3+3 2 | 2+2 2 | 3+3 2 | 3+3 2 | 2+2 2 |
| | 2+2 2 | 2+2 2 | 2+2 2 | 2+2 2 | 3+3 2 | 2+2 2 | 3+3 2 | 3+3 2 |
| | 2+2 2 | 2+2 2 | 2+2 2 | 3+3 2 | 2+2 2 | 3+3 2 | 2+2 2 | 2+2 2 |
| | 3+3 2 | 3+3 2 | 2+2 2 | 2+2 2 | 3+3 2 | 3+3 2 | 2+2 2 | 3+3 2 |
| | 2+2 2 | 2+2 2 | 3+3 2 | 2+2 2 | 3+3 2 | 2+2 2 | 3+3 2 | 2+2 2 |
| | 3+3 2 | 3+3 2 | 3+3 2 | 2+2 2 | 2+2 2 | 2+2 2 | 2+2 2 | 3+3 2 |
| | 3+3 2 | 2+2 2 | 3+3 2 | 3+3 2 | 3+3 2 | 3+3 2 | 3+3 2 | 3+3 2 |
| | 3+3 2 | 3+3 2 | 2+2 2 | 3+3 2 | 2+2 2 | 2+2 2 | 2+2 2 | 2+2 2 |
| | 3+3 2 | 3+3 2 | 3+3 2 | 3+3 2 | 3+3 2 | 3+3 2 | 3+3 2 | 2+2 2 |
| | 2+2 2 | 3+3 2 | 2+2 2 | 2+2 2 | 2+2 2 | 2+2 2 | 2+2 2 | 3+3 2 |
| | 2+2 2 | 2+2 2 | 3+3 2 | 2+2 2 | 3+3 2 | 2+2 2 | 3+3 2 | 2+2 2 |
| | 3+3 2 | 3+3 2 | 2+2 2 | 3+3 2 | 2+2 2 | 3+3 2 | 2+2 2 | 3+3 2 |
| | 2+2 2 | 2+2 2 | 3+3 2 | 2+2 2 | 3+3 2 | 2+2 2 | 3+3 2 | 2+2 2 |
| | 3+3 2 | 2+2 2 | 2+2 2 | 3+3 2 | 2+2 2 | 3+3 2 | 2+2 2 | 3+3 2 |
| | 2+2 2 | 3+3 2 | 2+2 2 | 3+3 2 | 3+3 2 | 2+2 2 | 2+2 2 | 2+2 2 |
| | 3+3 2 | 2+2 2 | 3+3 2 | 2+2 2 | 2+2 2 | 2+2 2 | 3+3 2 | 3+3 2 |
| | 2+2 2 | 2+2 2 | 3+3 2 | 3+3 2 | 2+2 2 | 3+3 2 | 2+2 2 | 2+2 2 |
| | 3+3 2 | 3+3 2 | 2+2 2 | 2+2 2 | 2+2 2 | 3+3 2 | 2+2 2 | 3+3 2 |
| | 2+2 2 | 3+3 2 | 2+2 2 | 3+3 2 | 3+3 2 | 2+2 2 | 3+3 2 | 2+2 2 |
| | 3+3 2 | 2+2 2 | 3+3 2 | 2+2 2 | 3+3 2 | 2+2 2 | 3+3 2 | 3+3 2 |
| | 2+2 2 | 2+2 2 | 2+2 2 | 3+3 2 | 2+2 2 | 3+3 2 | 2+2 2 | 2+2 2 |
| | 2+2 2 | 3+3 2 | 3+3 2 | 2+2 2 | 3+3 2 | 2+2 2 | 2+2 2 | 3+3 2 |
| | 3+3 2 | 2+2 2 | 2+2 2 | 2+2 2 | 3+3 2 | 3+3 2 | 3+3 2 | 3+3 2 |
| | 3+3 2 | 3+3 2 | 3+3 2 | 3+3 2 | 3+3 2 | 3+3 2 | 2+2 2 | 2+2 2 |
| | 3+3 2 | 2+2 2 | 3+3 2 | 3+3 2 | 2+2 2 | 2+2 2 | 3+3 2 | 3+3 2 |
| | 2+2 2 | 2+2 2 | 2+2 2 | 2+2 2 | 2+2 2 | 3+3 2 | 2+2 2 | 2+2 2 |
| | 2+2 2 | 3+3 2 | 3+3 2 | 2+2 2 | 3+3 2 | 2+2 2 | 3+3 2 | 2+2 2 |
| | 3+3 2 | 3+3 2 | 2+2 2 | 3+3 2 | 3+3 2 | 2+2 2 | 2+2 2 | 3+3 2 |

Continued on next column

**TABLE XXI– continued**

| TQC | ICSD | NoE | SF | | | | | |
|---|---|---|---|---|---|---|---|---|
| | 2+2 2 | 2+2 2 | 2+2 2 | 3+3 2 | 2+2 2 | 3+3 2 | 3+3 2 | 2+2 2 |
| | 3+3 2 | 2+2 2 | 3+3 2 | 3+3 2 | 3+3 2 | 2+2 2 | 3+3 2 | 3+3 2 |
| | 2+2 2 | 3+3 2 | 2+2 2 | 2+2 2 | 2+2 2 | 3+3 2 | 2+2 2 | 3+3 2 |
| | 3+3 2 | 3+3 2 | 3+3 2 | 2+2 2 | 3+3 2 | 3+3 2 | 2+2 2 | 3+3 2 |
| | 2+2 2 | 2+2 2 | 3+3 2 | 2+2 2 | 3+3 2 | 3+3 2 | 2+2 2 | 2+2 2 |
| | 3+3 2 | 3+3 2 | 2+2 2 | 2+2 2 | 3+3 2 | 2+2 2 | 3+3 2 | 2+2 2 |
| | 2+2 2 | 2+2 2 | 2+2 2 | 3+3 2 | 2+2 2 | 3+3 2 | 3+3 2 | 3+3 2 |
| | 3+3 2 | 3+3 2 | 3+3 2 | 3+3 2 | 2+2 2 | 2+2 2 | 2+2 2 | 3+3 2 |
| | 2+2 2 | 2+2 2 | 2+2 2 | 2+2 2 | 3+3 2 | 2+2 2 | 3+3 2 | 2+2 2 |
| | 3+3 2 | 3+3 2 | 3+3 2 | 3+3 2 | 2+2 2 | 2+2 2 | 3+3 2 | 2+2 2 |
| | 2+2 2 | 3+3 2 | 2+2 2 | 3+3 2 | 2+2 2 | 3+3 2 | 3+3 2 | 2+2 2 |
| | 2+2 2 | 2+2 2 | 3+3 2 | 2+2 2 | 2+2 2 | 2+2 2 | 2+2 2 | 3+3 2 |
| | 2+2 2 | 3+3 2 | 2+2 2 | 3+3 2 | 3+3 2 | 3+3 2 | 2+2 2 | 2+2 2 |
| | 3+3 2 | 2+2 2 | 3+3 2 | 2+2 2 | 2+2 2 | 3+3 2 | 3+3 2 | 3+3 2 |
| | 2+2 2 | 3+3 2 | 2+2 2 | 2+2 2 | 2+2 2 | 3+3 2 | 3+3 2 | 2+2 2 |
| | 3+3 2 | 2+2 2 | 3+3 2 | 3+3 2 | 3+3 2 | 3+3 2 | 3+3 2 | 3+3 2 |
| | 3+3 2 | 3+3 2 | 2+2 2 | 3+3 2 | 3+3 2 | 2+2 2 | 2+2 2 | 2+2 2 |
| | 2+2 2 | 2+2 2 | 3+3 2 | 2+2 2 | 2+2 2 | 2+2 2 | 2+2 2 | 3+3 2 |
| | 3+3 2 | 2+2 2 | 3+3 2 | 3+3 2 | 3+3 2 | 3+3 2 | 3+3 2 | 3+3 2 |
| | 3+3 2 | 3+3 2 | 3+3 2 | 2+2 2 | 2+2 2 | 2+2 2 | 2+2 2 | 2+2 2 |
| 2.2.6 | 19012 | 124 | O7V4 | | | | | |
| | -GM | -R | -T | -U | -V | -X | -Y | -Z |
| | 3+3 2 | 3+3 2 | 3+3 2 | 2+2 2 | 3+3 2 | 2+2 2 | 2+2 2 | 3+3 2 |
| | 3+3 2 | 2+2 2 | 2+2 2 | 3+3 2 | 3+3 2 | 3+3 2 | 3+3 2 | 2+2 2 |
| | 2+2 2 | 2+2 2 | 2+2 2 | 3+3 2 | 3+3 2 | 3+3 2 | 3+3 2 | 2+2 2 |
| | 3+3 2 | 3+3 2 | 3+3 2 | 2+2 2 | 2+2 2 | 2+2 2 | 2+2 2 | 3+3 2 |
| | 2+2 2 | 2+2 2 | 3+3 2 | 3+3 2 | 2+2 2 | 2+2 2 | 3+3 2 | 3+3 2 |
| | 2+2 2 | 3+3 2 | 2+2 2 | 2+2 2 | 2+2 2 | 3+3 2 | 2+2 2 | 2+2 2 |
| | 3+3 2 | 2+2 2 | 2+2 2 | 3+3 2 | 2+2 2 | 3+3 2 | 2+2 2 | 3+3 2 |
| | 2+2 2 | 3+3 2 | 3+3 2 | 2+2 2 | 2+2 2 | 2+2 2 | 3+3 2 | 2+2 2 |
| | 2+2 2 | 3+3 2 | 3+3 2 | 3+3 2 | 3+3 2 | 3+3 2 | 2+2 2 | 3+3 2 |
| | 2+2 2 | 2+2 2 | 2+2 2 | 2+2 2 | 3+3 2 | 2+2 2 | 3+3 2 | 2+2 2 |
| | 3+3 2 | 3+3 2 | 2+2 2 | 3+3 2 | 2+2 2 | 2+2 2 | 2+2 2 | 3+3 2 |
| | 3+3 2 | 2+2 2 | 2+2 2 | 2+2 2 | 2+2 2 | 3+3 2 | 3+3 2 | 2+2 2 |
| | 3+3 2 | 2+2 2 | 3+3 2 | 2+2 2 | 3+3 2 | 3+3 2 | 2+2 2 | 3+3 2 |
| | 2+2 2 | 3+3 2 | 3+3 2 | 3+3 2 | 3+3 2 | 2+2 2 | 3+3 2 | 2+2 2 |
| | 3+3 2 | 3+3 2 | 3+3 2 | 3+3 2 | 2+2 2 | 3+3 2 | 3+3 2 | 2+2 2 |
| | 2+2 2 | 2+2 2 | 2+2 2 | 2+2 2 | 3+3 2 | 2+2 2 | 3+3 2 | 3+3 2 |
| | 2+2 2 | 2+2 2 | 2+2 2 | 3+3 2 | 2+2 2 | 3+3 2 | 2+2 2 | 2+2 2 |
| | 3+3 2 | 3+3 2 | 2+2 2 | 2+2 2 | 3+3 2 | 3+3 2 | 2+2 2 | 3+3 2 |
| | 2+2 2 | 3+3 2 | 3+3 2 | 2+2 2 | 3+3 2 | 2+2 2 | 3+3 2 | 2+2 2 |
| | 3+3 2 | 2+2 2 | 3+3 2 | 2+2 2 | 2+2 2 | 2+2 2 | 3+3 2 | 3+3 2 |
| | 3+3 2 | 2+2 2 | 3+3 2 | 3+3 2 | 3+3 2 | 3+3 2 | 2+2 2 | 2+2 2 |
| | 3+3 2 | 3+3 2 | 3+3 2 | 3+3 2 | 2+2 2 | 3+3 2 | 2+2 2 | 3+3 2 |
| | 3+3 2 | 3+3 2 | 2+2 2 | 2+2 2 | 3+3 2 | 2+2 2 | 3+3 2 | 3+3 2 |
| | 2+2 2 | 3+3 2 | 2+2 2 | 3+3 2 | 3+3 2 | 2+2 2 | 2+2 2 | 2+2 2 |
| | 2+2 2 | 2+2 2 | 3+3 2 | 2+2 2 | 2+2 2 | 3+3 2 | 3+3 2 | 2+2 2 |
| | 3+3 2 | 3+3 2 | 3+3 2 | 3+3 2 | 3+3 2 | 2+2 2 | 2+2 2 | 3+3 2 |
| | 2+2 2 | 2+2 2 | 2+2 2 | 2+2 2 | 2+2 2 | 2+2 2 | 3+3 2 | 2+2 2 |
| | 3+3 2 | 3+3 2 | 2+2 2 | 3+3 2 | 2+2 2 | 2+2 2 | 2+2 2 | 3+3 2 |
| | 2+2 2 | 2+2 2 | 2+2 2 | 3+3 2 | 3+3 2 | 2+2 2 | 2+2 2 | 3+3 2 |
| | 3+3 2 | 2+2 2 | 3+3 2 | 2+2 2 | 2+2 2 | 3+3 2 | 3+3 2 | 2+2 2 |
| | 2+2 2 | 2+2 2 | 2+2 2 | 2+2 2 | 2+2 2 | 3+3 2 | 2+2 2 | 2+2 2 |
| | 3+3 2 | 3+3 2 | 3+3 2 | 3+3 2 | 3+3 2 | 3+3 2 | 3+3 2 | 3+3 2 |
| | 2+2 2 | 3+3 2 | 2+2 2 | 3+3 2 | 2+2 2 | 2+2 2 | 2+2 2 | 3+3 2 |
| | 3+3 2 | 2+2 2 | 2+2 2 | 2+2 2 | 3+3 2 | 3+3 2 | 2+2 2 | 2+2 2 |
| | 2+2 2 | 2+2 2 | 3+3 2 | 3+3 2 | 2+2 2 | 3+3 2 | 3+3 2 | 3+3 2 |
| | 3+3 2 | 2+2 2 | 2+2 2 | 2+2 2 | 3+3 2 | 2+2 2 | 3+3 2 | 2+2 2 |
| | 2+2 2 | 3+3 2 | 3+3 2 | 2+2 2 | 3+3 2 | 2+2 2 | 2+2 2 | 3+3 2 |
| | 3+3 2 | 3+3 2 | 3+3 2 | 3+3 2 | 2+2 2 | 3+3 2 | 2+2 2 | 2+2 2 |
| | 3+3 2 | 2+2 2 | 3+3 2 | 2+2 2 | 3+3 2 | 2+2 2 | 3+3 2 | 2+2 2 |

Continued on next column

**TABLE XXI– continued**

| TQC | ICSD | NoE | SF | | | | | |
|---|---|---|---|---|---|---|---|---|
| | 2+2 2 | 2+2 2 | 2+2 2 | 3+3 2 | 2+2 2 | 2+2 2 | 2+2 2 | 3+3 2 |
| | 3+3 2 | 3+3 2 | 3+3 2 | 3+3 2 | 3+3 2 | 2+2 2 | 3+3 2 | 2+2 2 |
| | 2+2 2 | 3+3 2 | 2+2 2 | 2+2 2 | 3+3 2 | 3+3 2 | 3+3 2 | 3+3 2 |
| | 2+2 2 | 2+2 2 | 2+2 2 | 3+3 2 | 3+3 2 | 3+3 2 | 2+2 2 | 2+2 2 |
| | 3+3 2 | 2+2 2 | 3+3 2 | 2+2 2 | 2+2 2 | 3+3 2 | 2+2 2 | 3+3 2 |
| | 2+2 2 | 3+3 2 | 3+3 2 | 3+3 2 | 2+2 2 | 2+2 2 | 3+3 2 | 3+3 2 |
| | 3+3 2 | 3+3 2 | 2+2 2 | 2+2 2 | 3+3 2 | 3+3 2 | 3+3 2 | 2+2 2 |
| | 2+2 2 | 2+2 2 | 3+3 2 | 2+2 2 | 3+3 2 | 2+2 2 | 2+2 2 | 3+3 2 |
| | 3+3 2 | 3+3 2 | 2+2 2 | 2+2 2 | 3+3 2 | 3+3 2 | 2+2 2 | 3+3 2 |
| | 2+2 2 | 2+2 2 | 2+2 2 | 3+3 2 | 2+2 2 | 3+3 2 | 3+3 2 | 3+3 2 |
| | 3+3 2 | 3+3 2 | 3+3 2 | 3+3 2 | 3+3 2 | 2+2 2 | 2+2 2 | 2+2 2 |
| | 2+2 2 | 2+2 2 | 2+2 2 | 2+2 2 | 2+2 2 | 2+2 2 | 3+3 2 | 2+2 2 |
| | 3+3 2 | 3+3 2 | 3+3 2 | 3+3 2 | 2+2 2 | 3+3 2 | 3+3 2 | 2+2 2 |
| | 2+2 2 | 3+3 2 | 3+3 2 | 3+3 2 | 2+2 2 | 2+2 2 | 3+3 2 | 2+2 2 |
| | 2+2 2 | 2+2 2 | 2+2 2 | 3+3 2 | 2+2 2 | 2+2 2 | 2+2 2 | 3+3 2 |
| | 2+2 2 | 3+3 2 | 2+2 2 | 2+2 2 | 3+3 2 | 3+3 2 | 2+2 2 | 2+2 2 |
| | 3+3 2 | 2+2 2 | 3+3 2 | 2+2 2 | 2+2 2 | 3+3 2 | 3+3 2 | 3+3 2 |
| | 2+2 2 | 3+3 2 | 3+3 2 | 2+2 2 | 2+2 2 | 3+3 2 | 3+3 2 | 2+2 2 |
| | 3+3 2 | 2+2 2 | 2+2 2 | 3+3 2 | 3+3 2 | 2+2 2 | 2+2 2 | 3+3 2 |
| | 3+3 2 | 3+3 2 | 2+2 2 | 3+3 2 | 3+3 2 | 3+3 2 | 3+3 2 | 2+2 2 |
| | 2+2 2 | 2+2 2 | 3+3 2 | 2+2 2 | 2+2 2 | 2+2 2 | 2+2 2 | 3+3 2 |
| | 3+3 2 | 2+2 2 | 2+2 2 | 2+2 2 | 3+3 2 | 3+3 2 | 3+3 2 | 2+2 2 |
| | 3+3 2 | 3+3 2 | 3+3 2 | 3+3 2 | 2+2 2 | 2+2 2 | 2+2 2 | 2+2 2 |
| 2.2.7 | 61430 | 96 | O6Pb3S1 | | | | | |
| | -GM | -R | -T | -U | -V | -X | -Y | -Z |
| | 3+3 2 | 3+3 2 | 2+2 2 | 3+3 2 | 2+2 2 | 2+2 2 | 3+3 2 | 2+2 2 |
| | 2+2 2 | 2+2 2 | 3+3 2 | 2+2 2 | 3+3 2 | 3+3 2 | 2+2 2 | 3+3 2 |
| | 3+3 2 | 2+2 2 | 2+2 2 | 2+2 2 | 3+3 2 | 3+3 2 | 3+3 2 | 2+2 2 |
| | 3+3 2 | 3+3 2 | 2+2 2 | 3+3 2 | 2+2 2 | 2+2 2 | 2+2 2 | 3+3 2 |
| | 2+2 2 | 3+3 2 | 3+3 2 | 2+2 2 | 2+2 2 | 2+2 2 | 3+3 2 | 2+2 2 |
| | 2+2 2 | 2+2 2 | 3+3 2 | 3+3 2 | 3+3 2 | 3+3 2 | 2+2 2 | 3+3 2 |
| | 3+3 2 | 2+2 2 | 2+2 2 | 3+3 2 | 2+2 2 | 3+3 2 | 2+2 2 | 3+3 2 |
| | 2+2 2 | 3+3 2 | 3+3 2 | 2+2 2 | 3+3 2 | 2+2 2 | 3+3 2 | 2+2 2 |
| | 3+3 2 | 2+2 2 | 2+2 2 | 3+3 2 | 3+3 2 | 2+2 2 | 2+2 2 | 3+3 2 |
| | 2+2 2 | 3+3 2 | 3+3 2 | 2+2 2 | 2+2 2 | 3+3 2 | 3+3 2 | 2+2 2 |
| | 3+3 2 | 2+2 2 | 3+3 2 | 3+3 2 | 2+2 2 | 3+3 2 | 3+3 2 | 3+3 2 |
| | 2+2 2 | 3+3 2 | 2+2 2 | 2+2 2 | 3+3 2 | 2+2 2 | 2+2 2 | 2+2 2 |
| | 3+3 2 | 2+2 2 | 3+3 2 | 3+3 2 | 2+2 2 | 2+2 2 | 2+2 2 | 2+2 2 |
| | 2+2 2 | 3+3 2 | 2+2 2 | 2+2 2 | 3+3 2 | 3+3 2 | 3+3 2 | 3+3 2 |
| | 2+2 2 | 2+2 2 | 3+3 2 | 2+2 2 | 3+3 2 | 2+2 2 | 2+2 2 | 3+3 2 |
| | 3+3 2 | 3+3 2 | 2+2 2 | 3+3 2 | 2+2 2 | 3+3 2 | 3+3 2 | 2+2 2 |
| | 2+2 2 | 2+2 2 | 3+3 2 | 2+2 2 | 3+3 2 | 2+2 2 | 2+2 2 | 3+3 2 |
| | 3+3 2 | 3+3 2 | 2+2 2 | 3+3 2 | 2+2 2 | 3+3 2 | 3+3 2 | 2+2 2 |
| | 3+3 2 | 3+3 2 | 2+2 2 | 3+3 2 | 2+2 2 | 2+2 2 | 3+3 2 | 2+2 2 |
| | 2+2 2 | 2+2 2 | 3+3 2 | 2+2 2 | 3+3 2 | 3+3 2 | 2+2 2 | 2+2 2 |
| | 3+3 2 | 3+3 2 | 2+2 2 | 3+3 2 | 2+2 2 | 2+2 2 | 3+3 2 | 3+3 2 |
| | 2+2 2 | 2+2 2 | 3+3 2 | 2+2 2 | 3+3 2 | 3+3 2 | 2+2 2 | 3+3 2 |
| | 3+3 2 | 3+3 2 | 2+2 2 | 3+3 2 | 2+2 2 | 3+3 2 | 3+3 2 | 2+2 2 |
| | 3+3 2 | 2+2 2 | 3+3 2 | 2+2 2 | 3+3 2 | 2+2 2 | 2+2 2 | 2+2 2 |
| | 2+2 2 | 3+3 2 | 2+2 2 | 3+3 2 | 2+2 2 | 3+3 2 | 3+3 2 | 2+2 2 |
| | 3+3 2 | 2+2 2 | 2+2 2 | 2+2 2 | 3+3 2 | 2+2 2 | 2+2 2 | 3+3 2 |
| | 2+2 2 | 3+3 2 | 3+3 2 | 2+2 2 | 2+2 2 | 2+2 2 | 3+3 2 | 3+3 2 |
| | 2+2 2 | 2+2 2 | 2+2 2 | 3+3 2 | 3+3 2 | 3+3 2 | 2+2 2 | 3+3 2 |
| | 3+3 2 | 2+2 2 | 3+3 2 | 2+2 2 | 2+2 2 | 2+2 2 | 3+3 2 | 2+2 2 |
| | 3+3 2 | 3+3 2 | 2+2 2 | 3+3 2 | 3+3 2 | 3+3 2 | 2+2 2 | 3+3 2 |
| | 2+2 2 | 2+2 2 | 3+3 2 | 2+2 2 | 3+3 2 | 2+2 2 | 3+3 2 | 2+2 2 |
| | 2+2 2 | 3+3 2 | 2+2 2 | 3+3 2 | 2+2 2 | 3+3 2 | 3+3 2 | 3+3 2 |
| | 3+3 2 | 3+3 2 | 3+3 2 | 3+3 2 | 2+2 2 | 3+3 2 | 2+2 2 | 2+2 2 |
| | 2+2 2 | 2+2 2 | 3+3 2 | 2+2 2 | 2+2 2 | 2+2 2 | 2+2 2 | 3+3 2 |
| | 3+3 2 | 3+3 2 | 2+2 2 | 3+3 2 | 3+3 2 | 3+3 2 | 3+3 2 | 2+2 2 |
| | 2+2 2 | 2+2 2 | 3+3 2 | 2+2 2 | 3+3 2 | 3+3 2 | 2+2 2 | 3+3 2 |

Continued on next column

**TABLE XXI– continued**

| TQC | ICSD | NoE | SF | | | | | |
|---|---|---|---|---|---|---|---|---|
| | 3+3 2 | 2+2 2 | 2+2 2 | 3+3 2 | 3+3 2 | 2+2 2 | 3+3 2 | 3+3 2 |
| | 3+3 2 | 3+3 2 | 2+2 2 | 2+2 2 | 2+2 2 | 2+2 2 | 3+3 2 | 2+2 2 |
| | 3+3 2 | 2+2 2 | 3+3 2 | 2+2 2 | 2+2 2 | 2+2 2 | 2+2 2 | 2+2 2 |
| | 2+2 2 | 3+3 2 | 3+3 2 | 3+3 2 | 3+3 2 | 3+3 2 | 2+2 2 | 3+3 2 |
| | 3+3 2 | 2+2 2 | 2+2 2 | 3+3 2 | 3+3 2 | 2+2 2 | 2+2 2 | 3+3 2 |
| | 2+2 2 | 3+3 2 | 3+3 2 | 2+2 2 | 2+2 2 | 3+3 2 | 3+3 2 | 2+2 2 |
| | 2+2 2 | 2+2 2 | 2+2 2 | 2+2 2 | 2+2 2 | 2+2 2 | 2+2 2 | 3+3 2 |
| | 3+3 2 | 3+3 2 | 3+3 2 | 3+3 2 | 3+3 2 | 3+3 2 | 2+2 2 | 2+2 2 |
| | 2+2 2 | 2+2 2 | 2+2 2 | 3+3 2 | 3+3 2 | 3+3 2 | 3+3 2 | 3+3 2 |
| | 3+3 2 | 3+3 2 | 3+3 2 | 2+2 2 | 2+2 2 | 2+2 2 | 3+3 2 | 2+2 2 |
| | 2+2 2 | 2+2 2 | 3+3 2 | 3+3 2 | 3+3 2 | 2+2 2 | 2+2 2 | 3+3 2 |
| | 3+3 2 | 3+3 2 | 2+2 2 | 2+2 2 | 2+2 2 | 3+3 2 | 3+3 2 | 3+3 2 |
| 2.2.8 | 82204 | 96 | O4Pb1Pt2 | | | | | |
| | -GM | -R | -T | -U | -V | -X | -Y | -Z |
| | 3+3 2 | 3+3 2 | 3+3 2 | 3+3 2 | 3+3 2 | 2+2 2 | 2+2 2 | 2+2 2 |
| | 3+3 2 | 2+2 2 | 2+2 2 | 2+2 2 | 2+2 2 | 3+3 2 | 3+3 2 | 3+3 2 |
| | 2+2 2 | 2+2 2 | 2+2 2 | 2+2 2 | 3+3 2 | 3+3 2 | 3+3 2 | 2+2 2 |
| | 3+3 2 | 3+3 2 | 3+3 2 | 3+3 2 | 2+2 2 | 2+2 2 | 2+2 2 | 2+2 2 |
| | 3+3 2 | 3+3 2 | 2+2 2 | 3+3 2 | 3+3 2 | 2+2 2 | 3+3 2 | 3+3 2 |
| | 2+2 2 | 2+2 2 | 2+2 2 | 2+2 2 | 2+2 2 | 2+2 2 | 2+2 2 | 2+2 2 |
| | 2+2 2 | 3+3 2 | 3+3 2 | 2+2 2 | 3+3 2 | 3+3 2 | 2+2 2 | 3+3 2 |
| | 2+2 2 | 2+2 2 | 3+3 2 | 3+3 2 | 2+2 2 | 3+3 2 | 3+3 2 | 3+3 2 |
| | 3+3 2 | 2+2 2 | 2+2 2 | 3+3 2 | 3+3 2 | 2+2 2 | 3+3 2 | 2+2 2 |
| | 3+3 2 | 3+3 2 | 3+3 2 | 2+2 2 | 2+2 2 | 2+2 2 | 2+2 2 | 3+3 2 |
| | 3+3 2 | 2+2 2 | 2+2 2 | 3+3 2 | 3+3 2 | 3+3 2 | 2+2 2 | 2+2 2 |
| | 2+2 2 | 3+3 2 | 2+2 2 | 3+3 2 | 3+3 2 | 2+2 2 | 3+3 2 | 2+2 2 |
| | 3+3 2 | 2+2 2 | 3+3 2 | 2+2 2 | 3+3 2 | 2+2 2 | 2+2 2 | 2+2 2 |
| | 3+3 2 | 2+2 2 | 2+2 2 | 3+3 2 | 2+2 2 | 2+2 2 | 3+3 2 | 2+2 2 |
| | 3+3 2 | 3+3 2 | 3+3 2 | 2+2 2 | 3+3 2 | 3+3 2 | 2+2 2 | 3+3 2 |
| | 3+3 2 | 3+3 2 | 2+2 2 | 3+3 2 | 3+3 2 | 2+2 2 | 2+2 2 | 2+2 2 |
| | 3+3 2 | 2+2 2 | 3+3 2 | 2+2 2 | 2+2 2 | 3+3 2 | 3+3 2 | 3+3 2 |
| | 3+3 2 | 2+2 2 | 2+2 2 | 2+2 2 | 3+3 2 | 2+2 2 | 2+2 2 | 2+2 2 |
| | 3+3 2 | 3+3 2 | 2+2 2 | 3+3 2 | 2+2 2 | 3+3 2 | 3+3 2 | 3+3 2 |
| | 2+2 2 | 3+3 2 | 3+3 2 | 2+2 2 | 3+3 2 | 2+2 2 | 2+2 2 | 2+2 2 |
| | 3+3 2 | 2+2 2 | 3+3 2 | 2+2 2 | 2+2 2 | 3+3 2 | 3+3 2 | 3+3 2 |
| | 2+2 2 | 2+2 2 | 2+2 2 | 3+3 2 | 3+3 2 | 3+3 2 | 3+3 2 | 2+2 2 |
| | 2+2 2 | 3+3 2 | 3+3 2 | 2+2 2 | 2+2 2 | 2+2 2 | 3+3 2 | 2+2 2 |
| | 3+3 2 | 3+3 2 | 2+2 2 | 3+3 2 | 3+3 2 | 3+3 2 | 3+3 2 | 3+3 2 |
| | 3+3 2 | 2+2 2 | 2+2 2 | 2+2 2 | 2+2 2 | 2+2 2 | 2+2 2 | 2+2 2 |
| | 2+2 2 | 3+3 2 | 2+2 2 | 3+3 2 | 3+3 2 | 3+3 2 | 2+2 2 | 2+2 2 |
| | 2+2 2 | 2+2 2 | 3+3 2 | 2+2 2 | 2+2 2 | 2+2 2 | 3+3 2 | 3+3 2 |
| | 2+2 2 | 3+3 2 | 2+2 2 | 3+3 2 | 2+2 2 | 3+3 2 | 2+2 2 | 2+2 2 |
| | 3+3 2 | 2+2 2 | 2+2 2 | 3+3 2 | 3+3 2 | 2+2 2 | 2+2 2 | 3+3 2 |
| | 3+3 2 | 3+3 2 | 3+3 2 | 2+2 2 | 2+2 2 | 2+2 2 | 3+3 2 | 3+3 2 |
| | 2+2 2 | 3+3 2 | 2+2 2 | 3+3 2 | 3+3 2 | 3+3 2 | 3+3 2 | 3+3 2 |
| | 2+2 2 | 3+3 2 | 3+3 2 | 2+2 2 | 2+2 2 | 3+3 2 | 2+2 2 | 2+2 2 |
| | 3+3 2 | 2+2 2 | 3+3 2 | 3+3 2 | 3+3 2 | 2+2 2 | 3+3 2 | 3+3 2 |
| | 2+2 2 | 3+3 2 | 3+3 2 | 2+2 2 | 2+2 2 | 3+3 2 | 2+2 2 | 2+2 2 |
| | 3+3 2 | 2+2 2 | 2+2 2 | 2+2 2 | 3+3 2 | 2+2 2 | 3+3 2 | 3+3 2 |
| | 2+2 2 | 2+2 2 | 2+2 2 | 3+3 2 | 2+2 2 | 3+3 2 | 2+2 2 | 3+3 2 |
| | 3+3 2 | 3+3 2 | 3+3 2 | 3+3 2 | 3+3 2 | 2+2 2 | 2+2 2 | 2+2 2 |
| | 2+2 2 | 2+2 2 | 2+2 2 | 2+2 2 | 3+3 2 | 2+2 2 | 3+3 2 | 2+2 2 |
| | 3+3 2 | 2+2 2 | 2+2 2 | 3+3 2 | 3+3 2 | 2+2 2 | 2+2 2 | 2+2 2 |
| | 3+3 2 | 3+3 2 | 3+3 2 | 2+2 2 | 2+2 2 | 3+3 2 | 3+3 2 | 3+3 2 |
| | 3+3 2 | 2+2 2 | 2+2 2 | 3+3 2 | 3+3 2 | 2+2 2 | 2+2 2 | 2+2 2 |
| | 3+3 2 | 3+3 2 | 3+3 2 | 2+2 2 | 3+3 2 | 2+2 2 | 3+3 2 | 3+3 2 |
| | 3+3 2 | 2+2 2 | 2+2 2 | 3+3 2 | 2+2 2 | 3+3 2 | 2+2 2 | 2+2 2 |
| | 2+2 2 | 3+3 2 | 2+2 2 | 2+2 2 | 3+3 2 | 2+2 2 | 3+3 2 | 2+2 2 |
| | 3+3 2 | 2+2 2 | 3+3 2 | 2+2 2 | 2+2 2 | 3+3 2 | 2+2 2 | 3+3 2 |
| | 3+3 2 | 3+3 2 | 2+2 2 | 2+2 2 | 3+3 2 | 2+2 2 | 3+3 2 | 2+2 2 |
| | 3+3 2 | 3+3 2 | 2+2 2 | 3+3 2 | 2+2 2 | 3+3 2 | 3+3 2 | 3+3 2 |

Continued on next column

**TABLE XXI– continued**

| TQC | ICSD | NoE | SF | | | | | |
|---|---|---|---|---|---|---|---|---|
| | 3+3 2 | 2+2 2 | 2+2 2 | 3+3 2 | 3+3 2 | 2+2 2 | 2+2 2 | 2+2 2 |
| 2.2.9 | 59657 | 96 | O4Pb1Pt2 | | | | | |
| | -GM | -R | -T | -U | -V | -X | -Y | -Z |
| | 3+3 2 | 3+3 2 | 2+2 2 | 3+3 2 | 3+3 2 | 2+2 2 | 2+2 2 | 2+2 2 |
| | 3+3 2 | 2+2 2 | 3+3 2 | 2+2 2 | 2+2 2 | 3+3 2 | 3+3 2 | 3+3 2 |
| | 2+2 2 | 2+2 2 | 3+3 2 | 2+2 2 | 3+3 2 | 3+3 2 | 2+2 2 | 2+2 2 |
| | 3+3 2 | 3+3 2 | 2+2 2 | 3+3 2 | 2+2 2 | 2+2 2 | 3+3 2 | 2+2 2 |
| | 3+3 2 | 2+2 2 | 2+2 2 | 2+2 2 | 3+3 2 | 2+2 2 | 3+3 2 | 3+3 2 |
| | 2+2 2 | 3+3 2 | 3+3 2 | 3+3 2 | 2+2 2 | 3+3 2 | 2+2 2 | 2+2 2 |
| | 2+2 2 | 3+3 2 | 2+2 2 | 2+2 2 | 3+3 2 | 2+2 2 | 2+2 2 | 3+3 2 |
| | 2+2 2 | 2+2 2 | 3+3 2 | 3+3 2 | 2+2 2 | 3+3 2 | 3+3 2 | 3+3 2 |
| | 3+3 2 | 2+2 2 | 2+2 2 | 3+3 2 | 3+3 2 | 2+2 2 | 3+3 2 | 2+2 2 |
| | 3+3 2 | 3+3 2 | 3+3 2 | 2+2 2 | 3+3 2 | 2+2 2 | 2+2 2 | 3+3 2 |
| | 3+3 2 | 2+2 2 | 2+2 2 | 3+3 2 | 2+2 2 | 3+3 2 | 2+2 2 | 2+2 2 |
| | 2+2 2 | 3+3 2 | 3+3 2 | 3+3 2 | 3+3 2 | 2+2 2 | 3+3 2 | 2+2 2 |
| | 3+3 2 | 2+2 2 | 2+2 2 | 2+2 2 | 3+3 2 | 2+2 2 | 2+2 2 | 2+2 2 |
| | 3+3 2 | 2+2 2 | 2+2 2 | 3+3 2 | 2+2 2 | 2+2 2 | 3+3 2 | 2+2 2 |
| | 3+3 2 | 3+3 2 | 3+3 2 | 2+2 2 | 3+3 2 | 3+3 2 | 2+2 2 | 3+3 2 |
| | 3+3 2 | 3+3 2 | 2+2 2 | 3+3 2 | 3+3 2 | 2+2 2 | 2+2 2 | 2+2 2 |
| | 3+3 2 | 2+2 2 | 3+3 2 | 2+2 2 | 2+2 2 | 3+3 2 | 3+3 2 | 3+3 2 |
| | 3+3 2 | 2+2 2 | 2+2 2 | 2+2 2 | 3+3 2 | 2+2 2 | 3+3 2 | 2+2 2 |
| | 3+3 2 | 3+3 2 | 3+3 2 | 3+3 2 | 2+2 2 | 3+3 2 | 2+2 2 | 3+3 2 |
| | 2+2 2 | 3+3 2 | 2+2 2 | 2+2 2 | 3+3 2 | 2+2 2 | 2+2 2 | 2+2 2 |
| | 3+3 2 | 2+2 2 | 3+3 2 | 2+2 2 | 2+2 2 | 3+3 2 | 3+3 2 | 3+3 2 |
| | 2+2 2 | 2+2 2 | 2+2 2 | 3+3 2 | 3+3 2 | 3+3 2 | 3+3 2 | 2+2 2 |
| | 2+2 2 | 3+3 2 | 3+3 2 | 2+2 2 | 2+2 2 | 2+2 2 | 3+3 2 | 2+2 2 |
| | 3+3 2 | 2+2 2 | 2+2 2 | 3+3 2 | 3+3 2 | 3+3 2 | 3+3 2 | 3+3 2 |
| | 3+3 2 | 3+3 2 | 2+2 2 | 2+2 2 | 2+2 2 | 2+2 2 | 2+2 2 | 2+2 2 |
| | 2+2 2 | 3+3 2 | 2+2 2 | 3+3 2 | 3+3 2 | 3+3 2 | 2+2 2 | 2+2 2 |
| | 2+2 2 | 2+2 2 | 3+3 2 | 2+2 2 | 2+2 2 | 2+2 2 | 3+3 2 | 3+3 2 |
| | 2+2 2 | 2+2 2 | 2+2 2 | 3+3 2 | 3+3 2 | 3+3 2 | 2+2 2 | 2+2 2 |
| | 3+3 2 | 3+3 2 | 2+2 2 | 3+3 2 | 2+2 2 | 2+2 2 | 2+2 2 | 3+3 2 |
| | 3+3 2 | 3+3 2 | 3+3 2 | 2+2 2 | 2+2 2 | 2+2 2 | 3+3 2 | 3+3 2 |
| | 2+2 2 | 3+3 2 | 2+2 2 | 3+3 2 | 2+2 2 | 3+3 2 | 3+3 2 | 3+3 2 |
| | 2+2 2 | 3+3 2 | 3+3 2 | 2+2 2 | 3+3 2 | 3+3 2 | 2+2 2 | 2+2 2 |
| | 3+3 2 | 2+2 2 | 3+3 2 | 3+3 2 | 3+3 2 | 2+2 2 | 3+3 2 | 2+2 2 |
| | 2+2 2 | 3+3 2 | 3+3 2 | 2+2 2 | 2+2 2 | 3+3 2 | 2+2 2 | 3+3 2 |
| | 3+3 2 | 2+2 2 | 2+2 2 | 2+2 2 | 3+3 2 | 2+2 2 | 3+3 2 | 3+3 2 |
| | 3+3 2 | 2+2 2 | 2+2 2 | 3+3 2 | 2+2 2 | 3+3 2 | 2+2 2 | 2+2 2 |
| | 2+2 2 | 3+3 2 | 3+3 2 | 3+3 2 | 3+3 2 | 2+2 2 | 3+3 2 | 3+3 2 |
| | 2+2 2 | 2+2 2 | 2+2 2 | 2+2 2 | 3+3 2 | 2+2 2 | 2+2 2 | 2+2 2 |
| | 3+3 2 | 3+3 2 | 2+2 2 | 3+3 2 | 3+3 2 | 3+3 2 | 2+2 2 | 2+2 2 |
| | 3+3 2 | 2+2 2 | 3+3 2 | 2+2 2 | 2+2 2 | 2+2 2 | 3+3 2 | 3+3 2 |
| | 3+3 2 | 2+2 2 | 2+2 2 | 2+2 2 | 3+3 2 | 2+2 2 | 2+2 2 | 2+2 2 |
| | 3+3 2 | 3+3 2 | 2+2 2 | 3+3 2 | 3+3 2 | 2+2 2 | 3+3 2 | 3+3 2 |
| | 3+3 2 | 2+2 2 | 3+3 2 | 3+3 2 | 2+2 2 | 3+3 2 | 2+2 2 | 2+2 2 |
| | 2+2 2 | 3+3 2 | 2+2 2 | 2+2 2 | 3+3 2 | 2+2 2 | 3+3 2 | 2+2 2 |
| | 3+3 2 | 2+2 2 | 3+3 2 | 2+2 2 | 2+2 2 | 3+3 2 | 2+2 2 | 3+3 2 |
| | 3+3 2 | 3+3 2 | 2+2 2 | 2+2 2 | 3+3 2 | 2+2 2 | 3+3 2 | 2+2 2 |
| | 3+3 2 | 3+3 2 | 2+2 2 | 3+3 2 | 2+2 2 | 3+3 2 | 3+3 2 | 3+3 2 |
| | 3+3 2 | 2+2 2 | 2+2 2 | 3+3 2 | 3+3 2 | 2+2 2 | 2+2 2 | 2+2 2 |
| 2.2.10 | 281617 | 98 | Na1O6Si2Ti1 | | | | | |
| | -GM | -R | -T | -U | -V | -X | -Y | -Z |
| | 3+3 2 | 2+2 2 | 2+2 2 | 3+3 2 | 3+3 2 | 2+2 2 | 3+3 2 | 2+2 2 |
| | 2+2 2 | 3+3 2 | 3+3 2 | 2+2 2 | 2+2 2 | 3+3 2 | 2+2 2 | 3+3 2 |
| | 3+3 2 | 3+3 2 | 2+2 2 | 3+3 2 | 3+3 2 | 3+3 2 | 3+3 2 | 2+2 2 |
| | 2+2 2 | 2+2 2 | 3+3 2 | 2+2 2 | 2+2 2 | 2+2 2 | 2+2 2 | 3+3 2 |
| | 3+3 2 | 3+3 2 | 3+3 2 | 3+3 2 | 3+3 2 | 2+2 2 | 2+2 2 | 3+3 2 |
| | 2+2 2 | 2+2 2 | 2+2 2 | 2+2 2 | 2+2 2 | 3+3 2 | 3+3 2 | 2+2 2 |
| | 3+3 2 | 2+2 2 | 2+2 2 | 3+3 2 | 2+2 2 | 3+3 2 | 2+2 2 | 3+3 2 |
| | 2+2 2 | 3+3 2 | 3+3 2 | 2+2 2 | 3+3 2 | 2+2 2 | 3+3 2 | 2+2 2 |

Continued on next column

**TABLE XXI– continued**

| TQC | ICSD | NoE | SF | | | | | |
|---|---|---|---|---|---|---|---|---|
| | 2+2 2 | 2+2 2 | 3+3 2 | 2+2 2 | 2+2 2 | 3+3 2 | 3+3 2 | 2+2 2 |
| | 3+3 2 | 3+3 2 | 2+2 2 | 3+3 2 | 3+3 2 | 2+2 2 | 2+2 2 | 3+3 2 |
| | 3+3 2 | 3+3 2 | 2+2 2 | 3+3 2 | 2+2 2 | 3+3 2 | 3+3 2 | 2+2 2 |
| | 2+2 2 | 2+2 2 | 3+3 2 | 2+2 2 | 3+3 2 | 2+2 2 | 2+2 2 | 3+3 2 |
| | 2+2 2 | 3+3 2 | 3+3 2 | 3+3 2 | 3+3 2 | 2+2 2 | 2+2 2 | 3+3 2 |
| | 3+3 2 | 2+2 2 | 2+2 2 | 2+2 2 | 2+2 2 | 3+3 2 | 3+3 2 | 2+2 2 |
| | 3+3 2 | 2+2 2 | 2+2 2 | 2+2 2 | 2+2 2 | 3+3 2 | 2+2 2 | 3+3 2 |
| | 3+3 2 | 3+3 2 | 3+3 2 | 3+3 2 | 3+3 2 | 2+2 2 | 3+3 2 | 2+2 2 |
| | 2+2 2 | 3+3 2 | 2+2 2 | 3+3 2 | 3+3 2 | 3+3 2 | 2+2 2 | 3+3 2 |
| | 2+2 2 | 2+2 2 | 2+2 2 | 2+2 2 | 2+2 2 | 2+2 2 | 3+3 2 | 2+2 2 |
| | 2+2 2 | 2+2 2 | 3+3 2 | 2+2 2 | 2+2 2 | 3+3 2 | 2+2 2 | 3+3 2 |
| | 3+3 2 | 3+3 2 | 3+3 2 | 3+3 2 | 3+3 2 | 2+2 2 | 3+3 2 | 2+2 2 |
| | 3+3 2 | 3+3 2 | 3+3 2 | 2+2 2 | 2+2 2 | 2+2 2 | 3+3 2 | 2+2 2 |
| | 2+2 2 | 2+2 2 | 3+3 2 | 3+3 2 | 3+3 2 | 3+3 2 | 3+3 2 | 2+2 2 |
| | 3+3 2 | 2+2 2 | 2+2 2 | 3+3 2 | 3+3 2 | 3+3 2 | 2+2 2 | 3+3 2 |
| | 2+2 2 | 3+3 2 | 2+2 2 | 2+2 2 | 2+2 2 | 2+2 2 | 2+2 2 | 3+3 2 |
| | 2+2 2 | 2+2 2 | 3+3 2 | 2+2 2 | 2+2 2 | 3+3 2 | 3+3 2 | 2+2 2 |
| | 3+3 2 | 3+3 2 | 2+2 2 | 3+3 2 | 3+3 2 | 2+2 2 | 2+2 2 | 3+3 2 |
| | 2+2 2 | 3+3 2 | 2+2 2 | 3+3 2 | 3+3 2 | 3+3 2 | 3+3 2 | 2+2 2 |
| | 3+3 2 | 2+2 2 | 3+3 2 | 3+3 2 | 3+3 2 | 2+2 2 | 2+2 2 | 3+3 2 |
| | 3+3 2 | 2+2 2 | 3+3 2 | 2+2 2 | 2+2 2 | 2+2 2 | 2+2 2 | 3+3 2 |
| | 3+3 2 | 3+3 2 | 2+2 2 | 2+2 2 | 2+2 2 | 3+3 2 | 3+3 2 | 2+2 2 |
| | 2+2 2 | 3+3 2 | 2+2 2 | 3+3 2 | 3+3 2 | 3+3 2 | 2+2 2 | 3+3 2 |
| | 3+3 2 | 2+2 2 | 3+3 2 | 2+2 2 | 2+2 2 | 2+2 2 | 3+3 2 | 2+2 2 |
| | 3+3 2 | 2+2 2 | 2+2 2 | 2+2 2 | 2+2 2 | 3+3 2 | 2+2 2 | 3+3 2 |
| | 2+2 2 | 3+3 2 | 3+3 2 | 3+3 2 | 2+2 2 | 2+2 2 | 3+3 2 | 2+2 2 |
| | 3+3 2 | 3+3 2 | 2+2 2 | 2+2 2 | 3+3 2 | 3+3 2 | 2+2 2 | 3+3 2 |
| | 3+3 2 | 2+2 2 | 3+3 2 | 3+3 2 | 3+3 2 | 2+2 2 | 3+3 2 | 2+2 2 |
| | 2+2 2 | 2+2 2 | 2+2 2 | 3+3 2 | 3+3 2 | 3+3 2 | 3+3 2 | 2+2 2 |
| | 2+2 2 | 3+3 2 | 3+3 2 | 2+2 2 | 2+2 2 | 2+2 2 | 2+2 2 | 3+3 2 |
| | 3+3 2 | 2+2 2 | 2+2 2 | 2+2 2 | 2+2 2 | 3+3 2 | 2+2 2 | 2+2 2 |
| | 2+2 2 | 3+3 2 | 3+3 2 | 3+3 2 | 3+3 2 | 2+2 2 | 3+3 2 | 3+3 2 |
| | 3+3 2 | 3+3 2 | 2+2 2 | 3+3 2 | 3+3 2 | 3+3 2 | 2+2 2 | 3+3 2 |
| | 2+2 2 | 2+2 2 | 3+3 2 | 2+2 2 | 2+2 2 | 2+2 2 | 3+3 2 | 2+2 2 |
| | 2+2 2 | 2+2 2 | 3+3 2 | 3+3 2 | 3+3 2 | 2+2 2 | 2+2 2 | 3+3 2 |
| | 2+2 2 | 3+3 2 | 2+2 2 | 2+2 2 | 2+2 2 | 3+3 2 | 3+3 2 | 2+2 2 |
| | 2+2 2 | 3+3 2 | 3+3 2 | 2+2 2 | 2+2 2 | 3+3 2 | 3+3 2 | 2+2 2 |
| | 3+3 2 | 3+3 2 | 2+2 2 | 3+3 2 | 3+3 2 | 2+2 2 | 2+2 2 | 3+3 2 |
| | 3+3 2 | 2+2 2 | 2+2 2 | 2+2 2 | 2+2 2 | 2+2 2 | 2+2 2 | 3+3 2 |
| | 2+2 2 | 2+2 2 | 3+3 2 | 3+3 2 | 3+3 2 | 3+3 2 | 3+3 2 | 2+2 2 |
| | 3+3 2 | 3+3 2 | 3+3 2 | 2+2 2 | 3+3 2 | 2+2 2 | 2+2 2 | 3+3 2 |
| 2.2.11 | 281616 | 98 | Na1O6Si2Ti1 | | | | | |
| | -GM | -R | -T | -U | -V | -X | -Y | -Z |
| | 3+3 2 | 2+2 2 | 2+2 2 | 3+3 2 | 3+3 2 | 3+3 2 | 3+3 2 | 2+2 2 |
| | 2+2 2 | 3+3 2 | 3+3 2 | 2+2 2 | 2+2 2 | 2+2 2 | 2+2 2 | 3+3 2 |
| | 3+3 2 | 3+3 2 | 2+2 2 | 3+3 2 | 2+2 2 | 2+2 2 | 3+3 2 | 2+2 2 |
| | 2+2 2 | 2+2 2 | 3+3 2 | 2+2 2 | 3+3 2 | 3+3 2 | 2+2 2 | 3+3 2 |
| | 3+3 2 | 3+3 2 | 3+3 2 | 3+3 2 | 3+3 2 | 2+2 2 | 2+2 2 | 3+3 2 |
| | 2+2 2 | 2+2 2 | 2+2 2 | 2+2 2 | 2+2 2 | 3+3 2 | 3+3 2 | 2+2 2 |
| | 3+3 2 | 2+2 2 | 2+2 2 | 3+3 2 | 2+2 2 | 3+3 2 | 2+2 2 | 3+3 2 |
| | 2+2 2 | 3+3 2 | 3+3 2 | 2+2 2 | 3+3 2 | 2+2 2 | 3+3 2 | 2+2 2 |
| | 2+2 2 | 2+2 2 | 3+3 2 | 2+2 2 | 2+2 2 | 3+3 2 | 3+3 2 | 2+2 2 |
| | 3+3 2 | 3+3 2 | 2+2 2 | 3+3 2 | 3+3 2 | 2+2 2 | 2+2 2 | 2+2 2 |
| | 3+3 2 | 3+3 2 | 2+2 2 | 3+3 2 | 2+2 2 | 3+3 2 | 3+3 2 | 3+3 2 |
| | 2+2 2 | 2+2 2 | 3+3 2 | 2+2 2 | 3+3 2 | 2+2 2 | 2+2 2 | 3+3 2 |
| | 2+2 2 | 3+3 2 | 3+3 2 | 3+3 2 | 3+3 2 | 2+2 2 | 2+2 2 | 3+3 2 |
| | 3+3 2 | 2+2 2 | 2+2 2 | 2+2 2 | 2+2 2 | 3+3 2 | 3+3 2 | 2+2 2 |
| | 3+3 2 | 2+2 2 | 2+2 2 | 2+2 2 | 3+3 2 | 3+3 2 | 2+2 2 | 3+3 2 |
| | 3+3 2 | 3+3 2 | 3+3 2 | 3+3 2 | 2+2 2 | 2+2 2 | 3+3 2 | 2+2 2 |
| | 2+2 2 | 2+2 2 | 2+2 2 | 3+3 2 | 3+3 2 | 3+3 2 | 2+2 2 | 3+3 2 |
| | 2+2 2 | 3+3 2 | 2+2 2 | 2+2 2 | 2+2 2 | 2+2 2 | 3+3 2 | 2+2 2 |

Continued on next column

**TABLE XXI– continued**

| TQC | ICSD | NoE | SF | | | | | |
|---|---|---|---|---|---|---|---|---|
| | 2+2 2 | 2+2 2 | 3+3 2 | 2+2 2 | 2+2 2 | 3+3 2 | 2+2 2 | 3+3 2 |
| | 3+3 2 | 3+3 2 | 3+3 2 | 3+3 2 | 3+3 2 | 2+2 2 | 3+3 2 | 2+2 2 |
| | 3+3 2 | 3+3 2 | 3+3 2 | 2+2 2 | 2+2 2 | 2+2 2 | 3+3 2 | 2+2 2 |
| | 2+2 2 | 2+2 2 | 3+3 2 | 3+3 2 | 3+3 2 | 3+3 2 | 3+3 2 | 2+2 2 |
| | 3+3 2 | 2+2 2 | 2+2 2 | 3+3 2 | 3+3 2 | 3+3 2 | 2+2 2 | 3+3 2 |
| | 2+2 2 | 3+3 2 | 2+2 2 | 2+2 2 | 2+2 2 | 2+2 2 | 2+2 2 | 3+3 2 |
| | 2+2 2 | 2+2 2 | 3+3 2 | 2+2 2 | 2+2 2 | 3+3 2 | 3+3 2 | 2+2 2 |
| | 3+3 2 | 3+3 2 | 2+2 2 | 3+3 2 | 3+3 2 | 2+2 2 | 2+2 2 | 3+3 2 |
| | 2+2 2 | 3+3 2 | 2+2 2 | 3+3 2 | 3+3 2 | 3+3 2 | 3+3 2 | 2+2 2 |
| | 3+3 2 | 2+2 2 | 3+3 2 | 3+3 2 | 3+3 2 | 2+2 2 | 2+2 2 | 3+3 2 |
| | 3+3 2 | 2+2 2 | 3+3 2 | 2+2 2 | 2+2 2 | 2+2 2 | 2+2 2 | 3+3 2 |
| | 3+3 2 | 3+3 2 | 2+2 2 | 2+2 2 | 2+2 2 | 3+3 2 | 3+3 2 | 2+2 2 |
| | 2+2 2 | 3+3 2 | 2+2 2 | 3+3 2 | 3+3 2 | 3+3 2 | 2+2 2 | 3+3 2 |
| | 3+3 2 | 2+2 2 | 2+2 2 | 2+2 2 | 2+2 2 | 2+2 2 | 3+3 2 | 2+2 2 |
| | 3+3 2 | 2+2 2 | 3+3 2 | 2+2 2 | 2+2 2 | 3+3 2 | 2+2 2 | 3+3 2 |
| | 2+2 2 | 3+3 2 | 3+3 2 | 3+3 2 | 2+2 2 | 2+2 2 | 3+3 2 | 2+2 2 |
| | 3+3 2 | 3+3 2 | 2+2 2 | 2+2 2 | 3+3 2 | 3+3 2 | 2+2 2 | 3+3 2 |
| | 3+3 2 | 2+2 2 | 3+3 2 | 3+3 2 | 3+3 2 | 2+2 2 | 3+3 2 | 2+2 2 |
| | 2+2 2 | 2+2 2 | 2+2 2 | 3+3 2 | 3+3 2 | 3+3 2 | 3+3 2 | 2+2 2 |
| | 2+2 2 | 3+3 2 | 3+3 2 | 2+2 2 | 2+2 2 | 2+2 2 | 2+2 2 | 3+3 2 |
| | 3+3 2 | 2+2 2 | 2+2 2 | 2+2 2 | 2+2 2 | 3+3 2 | 2+2 2 | 3+3 2 |
| | 2+2 2 | 3+3 2 | 3+3 2 | 3+3 2 | 3+3 2 | 2+2 2 | 3+3 2 | 2+2 2 |
| | 3+3 2 | 3+3 2 | 2+2 2 | 3+3 2 | 3+3 2 | 3+3 2 | 2+2 2 | 3+3 2 |
| | 2+2 2 | 2+2 2 | 3+3 2 | 2+2 2 | 2+2 2 | 2+2 2 | 3+3 2 | 2+2 2 |
| | 2+2 2 | 3+3 2 | 3+3 2 | 3+3 2 | 3+3 2 | 2+2 2 | 2+2 2 | 3+3 2 |
| | 2+2 2 | 2+2 2 | 2+2 2 | 2+2 2 | 2+2 2 | 3+3 2 | 3+3 2 | 2+2 2 |
| | 2+2 2 | 3+3 2 | 3+3 2 | 2+2 2 | 2+2 2 | 3+3 2 | 3+3 2 | 3+3 2 |
| | 3+3 2 | 2+2 2 | 2+2 2 | 3+3 2 | 3+3 2 | 2+2 2 | 2+2 2 | 2+2 2 |
| | 3+3 2 | 3+3 2 | 2+2 2 | 2+2 2 | 2+2 2 | 3+3 2 | 2+2 2 | 3+3 2 |
| | 2+2 2 | 2+2 2 | 3+3 2 | 3+3 2 | 3+3 2 | 2+2 2 | 3+3 2 | 2+2 2 |
| | 3+3 2 | 3+3 2 | 3+3 2 | 2+2 2 | 3+3 2 | 2+2 2 | 2+2 2 | 3+3 2 |
| 2.2.12 | 281615 | 98 | Na1O6Si2Ti1 | | | | | |
| | -GM | -R | -T | -U | -V | -X | -Y | -Z |
| | 3+3 2 | 2+2 2 | 2+2 2 | 3+3 2 | 3+3 2 | 2+2 2 | 3+3 2 | 2+2 2 |
| | 2+2 2 | 3+3 2 | 3+3 2 | 2+2 2 | 2+2 2 | 3+3 2 | 2+2 2 | 3+3 2 |
| | 3+3 2 | 3+3 2 | 2+2 2 | 3+3 2 | 3+3 2 | 2+2 2 | 3+3 2 | 2+2 2 |
| | 2+2 2 | 2+2 2 | 3+3 2 | 2+2 2 | 2+2 2 | 3+3 2 | 2+2 2 | 3+3 2 |
| | 3+3 2 | 3+3 2 | 3+3 2 | 2+2 2 | 3+3 2 | 2+2 2 | 2+2 2 | 3+3 2 |
| | 2+2 2 | 2+2 2 | 2+2 2 | 3+3 2 | 2+2 2 | 3+3 2 | 3+3 2 | 2+2 2 |
| | 3+3 2 | 2+2 2 | 2+2 2 | 3+3 2 | 2+2 2 | 3+3 2 | 2+2 2 | 3+3 2 |
| | 2+2 2 | 3+3 2 | 3+3 2 | 2+2 2 | 3+3 2 | 2+2 2 | 3+3 2 | 2+2 2 |
| | 2+2 2 | 2+2 2 | 3+3 2 | 2+2 2 | 2+2 2 | 3+3 2 | 2+2 2 | 2+2 2 |
| | 3+3 2 | 3+3 2 | 2+2 2 | 3+3 2 | 3+3 2 | 2+2 2 | 3+3 2 | 3+3 2 |
| | 3+3 2 | 2+2 2 | 2+2 2 | 3+3 2 | 3+3 2 | 2+2 2 | 3+3 2 | 2+2 2 |
| | 2+2 2 | 3+3 2 | 3+3 2 | 2+2 2 | 2+2 2 | 3+3 2 | 2+2 2 | 3+3 2 |
| | 2+2 2 | 3+3 2 | 3+3 2 | 3+3 2 | 3+3 2 | 2+2 2 | 2+2 2 | 3+3 2 |
| | 3+3 2 | 2+2 2 | 2+2 2 | 2+2 2 | 2+2 2 | 3+3 2 | 3+3 2 | 2+2 2 |
| | 3+3 2 | 2+2 2 | 2+2 2 | 2+2 2 | 3+3 2 | 3+3 2 | 2+2 2 | 3+3 2 |
| | 3+3 2 | 3+3 2 | 3+3 2 | 3+3 2 | 2+2 2 | 2+2 2 | 3+3 2 | 2+2 2 |
| | 2+2 2 | 3+3 2 | 2+2 2 | 3+3 2 | 3+3 2 | 2+2 2 | 2+2 2 | 3+3 2 |
| | 2+2 2 | 2+2 2 | 2+2 2 | 2+2 2 | 2+2 2 | 3+3 2 | 3+3 2 | 2+2 2 |
| | 2+2 2 | 2+2 2 | 3+3 2 | 2+2 2 | 2+2 2 | 3+3 2 | 2+2 2 | 3+3 2 |
| | 3+3 2 | 3+3 2 | 3+3 2 | 3+3 2 | 3+3 2 | 2+2 2 | 3+3 2 | 2+2 2 |
| | 3+3 2 | 3+3 2 | 3+3 2 | 2+2 2 | 2+2 2 | 2+2 2 | 3+3 2 | 2+2 2 |
| | 2+2 2 | 2+2 2 | 3+3 2 | 3+3 2 | 3+3 2 | 3+3 2 | 3+3 2 | 2+2 2 |
| | 3+3 2 | 2+2 2 | 2+2 2 | 3+3 2 | 3+3 2 | 3+3 2 | 2+2 2 | 3+3 2 |
| | 2+2 2 | 3+3 2 | 2+2 2 | 2+2 2 | 2+2 2 | 2+2 2 | 2+2 2 | 3+3 2 |
| | 2+2 2 | 2+2 2 | 3+3 2 | 2+2 2 | 2+2 2 | 3+3 2 | 3+3 2 | 2+2 2 |
| | 3+3 2 | 3+3 2 | 2+2 2 | 3+3 2 | 3+3 2 | 2+2 2 | 2+2 2 | 3+3 2 |
| | 2+2 2 | 3+3 2 | 2+2 2 | 3+3 2 | 3+3 2 | 3+3 2 | 3+3 2 | 2+2 2 |
| | 3+3 2 | 2+2 2 | 3+3 2 | 3+3 2 | 3+3 2 | 2+2 2 | 2+2 2 | 3+3 2 |

Continued on next column

**TABLE XXI– continued**

| TQC | ICSD | NoE | SF | | | | | |
|---|---|---|---|---|---|---|---|---|
| | 3+3 2 | 2+2 2 | 3+3 2 | 2+2 2 | 2+2 2 | 3+3 2 | 2+2 2 | 3+3 2 |
| | 3+3 2 | 3+3 2 | 2+2 2 | 2+2 2 | 2+2 2 | 2+2 2 | 3+3 2 | 2+2 2 |
| | 2+2 2 | 3+3 2 | 2+2 2 | 3+3 2 | 3+3 2 | 3+3 2 | 2+2 2 | 3+3 2 |
| | 3+3 2 | 2+2 2 | 3+3 2 | 2+2 2 | 2+2 2 | 2+2 2 | 3+3 2 | 2+2 2 |
| | 3+3 2 | 2+2 2 | 2+2 2 | 2+2 2 | 2+2 2 | 3+3 2 | 2+2 2 | 3+3 2 |
| | 2+2 2 | 3+3 2 | 3+3 2 | 3+3 2 | 2+2 2 | 2+2 2 | 3+3 2 | 2+2 2 |
| | 3+3 2 | 3+3 2 | 2+2 2 | 2+2 2 | 3+3 2 | 3+3 2 | 2+2 2 | 3+3 2 |
| | 3+3 2 | 2+2 2 | 3+3 2 | 3+3 2 | 3+3 2 | 2+2 2 | 3+3 2 | 2+2 2 |
| | 2+2 2 | 2+2 2 | 2+2 2 | 3+3 2 | 3+3 2 | 3+3 2 | 3+3 2 | 2+2 2 |
| | 2+2 2 | 3+3 2 | 3+3 2 | 2+2 2 | 2+2 2 | 2+2 2 | 2+2 2 | 3+3 2 |
| | 3+3 2 | 2+2 2 | 2+2 2 | 2+2 2 | 2+2 2 | 2+2 2 | 2+2 2 | 3+3 2 |
| | 2+2 2 | 3+3 2 | 3+3 2 | 3+3 2 | 3+3 2 | 3+3 2 | 3+3 2 | 2+2 2 |
| | 3+3 2 | 3+3 2 | 2+2 2 | 3+3 2 | 3+3 2 | 3+3 2 | 2+2 2 | 3+3 2 |
| | 2+2 2 | 2+2 2 | 3+3 2 | 2+2 2 | 2+2 2 | 2+2 2 | 2+2 2 | 2+2 2 |
| | 2+2 2 | 3+3 2 | 3+3 2 | 3+3 2 | 3+3 2 | 2+2 2 | 3+3 2 | 3+3 2 |
| | 2+2 2 | 2+2 2 | 2+2 2 | 2+2 2 | 2+2 2 | 3+3 2 | 3+3 2 | 2+2 2 |
| | 2+2 2 | 3+3 2 | 3+3 2 | 2+2 2 | 2+2 2 | 2+2 2 | 3+3 2 | 3+3 2 |
| | 3+3 2 | 2+2 2 | 2+2 2 | 3+3 2 | 3+3 2 | 3+3 2 | 2+2 2 | 2+2 2 |
| | 3+3 2 | 3+3 2 | 2+2 2 | 2+2 2 | 2+2 2 | 3+3 2 | 2+2 2 | 3+3 2 |
| | 2+2 2 | 2+2 2 | 3+3 2 | 3+3 2 | 3+3 2 | 2+2 2 | 3+3 2 | 2+2 2 |
| | 3+3 2 | 3+3 2 | 3+3 2 | 2+2 2 | 3+3 2 | 2+2 2 | 2+2 2 | 3+3 2 |
| 2.2.13 | 281614 | 98 | Na1O6Si2Ti1 | | | | | |
| | -GM | -R | -T | -U | -V | -X | -Y | -Z |
| | 3+3 2 | 2+2 2 | 2+2 2 | 3+3 2 | 3+3 2 | 3+3 2 | 3+3 2 | 2+2 2 |
| | 2+2 2 | 3+3 2 | 3+3 2 | 2+2 2 | 2+2 2 | 2+2 2 | 2+2 2 | 3+3 2 |
| | 3+3 2 | 3+3 2 | 2+2 2 | 3+3 2 | 2+2 2 | 2+2 2 | 3+3 2 | 2+2 2 |
| | 2+2 2 | 2+2 2 | 3+3 2 | 2+2 2 | 3+3 2 | 3+3 2 | 2+2 2 | 3+3 2 |
| | 3+3 2 | 3+3 2 | 3+3 2 | 3+3 2 | 3+3 2 | 2+2 2 | 2+2 2 | 3+3 2 |
| | 2+2 2 | 2+2 2 | 2+2 2 | 2+2 2 | 2+2 2 | 3+3 2 | 3+3 2 | 2+2 2 |
| | 3+3 2 | 2+2 2 | 2+2 2 | 3+3 2 | 2+2 2 | 3+3 2 | 2+2 2 | 3+3 2 |
| | 2+2 2 | 3+3 2 | 3+3 2 | 2+2 2 | 3+3 2 | 2+2 2 | 3+3 2 | 2+2 2 |
| | 2+2 2 | 2+2 2 | 3+3 2 | 2+2 2 | 2+2 2 | 3+3 2 | 3+3 2 | 2+2 2 |
| | 3+3 2 | 3+3 2 | 2+2 2 | 3+3 2 | 3+3 2 | 2+2 2 | 2+2 2 | 3+3 2 |
| | 3+3 2 | 3+3 2 | 2+2 2 | 3+3 2 | 2+2 2 | 2+2 2 | 3+3 2 | 2+2 2 |
| | 2+2 2 | 2+2 2 | 3+3 2 | 2+2 2 | 3+3 2 | 3+3 2 | 2+2 2 | 3+3 2 |
| | 2+2 2 | 3+3 2 | 3+3 2 | 3+3 2 | 3+3 2 | 2+2 2 | 2+2 2 | 3+3 2 |
| | 3+3 2 | 2+2 2 | 2+2 2 | 2+2 2 | 2+2 2 | 3+3 2 | 3+3 2 | 2+2 2 |
| | 3+3 2 | 2+2 2 | 2+2 2 | 2+2 2 | 3+3 2 | 3+3 2 | 2+2 2 | 3+3 2 |
| | 3+3 2 | 3+3 2 | 3+3 2 | 3+3 2 | 2+2 2 | 2+2 2 | 3+3 2 | 2+2 2 |
| | 2+2 2 | 3+3 2 | 2+2 2 | 3+3 2 | 3+3 2 | 3+3 2 | 2+2 2 | 3+3 2 |
| | 2+2 2 | 2+2 2 | 2+2 2 | 2+2 2 | 2+2 2 | 2+2 2 | 3+3 2 | 2+2 2 |
| | 2+2 2 | 2+2 2 | 3+3 2 | 2+2 2 | 2+2 2 | 3+3 2 | 2+2 2 | 3+3 2 |
| | 3+3 2 | 3+3 2 | 3+3 2 | 3+3 2 | 3+3 2 | 2+2 2 | 3+3 2 | 2+2 2 |
| | 3+3 2 | 3+3 2 | 3+3 2 | 2+2 2 | 2+2 2 | 2+2 2 | 3+3 2 | 2+2 2 |
| | 2+2 2 | 2+2 2 | 3+3 2 | 3+3 2 | 3+3 2 | 3+3 2 | 3+3 2 | 2+2 2 |
| | 3+3 2 | 3+3 2 | 2+2 2 | 3+3 2 | 3+3 2 | 3+3 2 | 2+2 2 | 3+3 2 |
| | 2+2 2 | 2+2 2 | 2+2 2 | 2+2 2 | 2+2 2 | 2+2 2 | 2+2 2 | 3+3 2 |
| | 2+2 2 | 2+2 2 | 3+3 2 | 2+2 2 | 2+2 2 | 3+3 2 | 3+3 2 | 2+2 2 |
| | 3+3 2 | 3+3 2 | 2+2 2 | 3+3 2 | 3+3 2 | 2+2 2 | 2+2 2 | 3+3 2 |
| | 2+2 2 | 2+2 2 | 2+2 2 | 3+3 2 | 3+3 2 | 2+2 2 | 3+3 2 | 2+2 2 |
| | 3+3 2 | 3+3 2 | 3+3 2 | 3+3 2 | 3+3 2 | 3+3 2 | 2+2 2 | 3+3 2 |
| | 3+3 2 | 2+2 2 | 3+3 2 | 2+2 2 | 2+2 2 | 3+3 2 | 2+2 2 | 3+3 2 |
| | 3+3 2 | 3+3 2 | 2+2 2 | 2+2 2 | 2+2 2 | 2+2 2 | 3+3 2 | 2+2 2 |
| | 2+2 2 | 2+2 2 | 2+2 2 | 3+3 2 | 3+3 2 | 3+3 2 | 2+2 2 | 3+3 2 |
| | 3+3 2 | 3+3 2 | 3+3 2 | 2+2 2 | 2+2 2 | 2+2 2 | 3+3 2 | 2+2 2 |
| | 3+3 2 | 2+2 2 | 2+2 2 | 2+2 2 | 2+2 2 | 3+3 2 | 2+2 2 | 3+3 2 |
| | 2+2 2 | 3+3 2 | 3+3 2 | 3+3 2 | 2+2 2 | 2+2 2 | 3+3 2 | 2+2 2 |
| | 3+3 2 | 3+3 2 | 2+2 2 | 2+2 2 | 3+3 2 | 3+3 2 | 2+2 2 | 3+3 2 |
| | 3+3 2 | 2+2 2 | 3+3 2 | 3+3 2 | 3+3 2 | 2+2 2 | 3+3 2 | 2+2 2 |
| | 2+2 2 | 2+2 2 | 2+2 2 | 3+3 2 | 3+3 2 | 3+3 2 | 3+3 2 | 2+2 2 |
| | 2+2 2 | 3+3 2 | 3+3 2 | 2+2 2 | 2+2 2 | 2+2 2 | 2+2 2 | 3+3 2 |

Continued on next column

**TABLE XXI– continued**

| TQC | ICSD | NoE | SF | | | | |
|---|---|---|---|---|---|---|---|
| | 3+3 2 | 2+2 2 | 2+2 2 | 2+2 2 | 2+2 2 | 2+2 2 | 2+2 2 | 3+3 2 |
| | 2+2 2 | 3+3 2 | 3+3 2 | 3+3 2 | 3+3 2 | 3+3 2 | 3+3 2 | 2+2 2 |
| | 3+3 2 | 3+3 2 | 2+2 2 | 3+3 2 | 3+3 2 | 3+3 2 | 2+2 2 | 3+3 2 |
| | 2+2 2 | 2+2 2 | 3+3 2 | 2+2 2 | 2+2 2 | 2+2 2 | 2+2 2 | 2+2 2 |
| | 2+2 2 | 3+3 2 | 3+3 2 | 3+3 2 | 3+3 2 | 2+2 2 | 3+3 2 | 3+3 2 |
| | 2+2 2 | 2+2 2 | 2+2 2 | 2+2 2 | 2+2 2 | 3+3 2 | 3+3 2 | 2+2 2 |
| | 2+2 2 | 3+3 2 | 3+3 2 | 2+2 2 | 2+2 2 | 2+2 2 | 3+3 2 | 2+2 2 |
| | 3+3 2 | 2+2 2 | 2+2 2 | 3+3 2 | 3+3 2 | 3+3 2 | 2+2 2 | 3+3 2 |
| | 3+3 2 | 3+3 2 | 2+2 2 | 2+2 2 | 2+2 2 | 3+3 2 | 2+2 2 | 3+3 2 |
| | 2+2 2 | 2+2 2 | 3+3 2 | 3+3 2 | 3+3 2 | 2+2 2 | 3+3 2 | 2+2 2 |
| | 3+3 2 | 3+3 2 | 3+3 2 | 2+2 2 | 3+3 2 | 2+2 2 | 2+2 2 | 3+3 2 |
| 2.2.14 | 281613 | 98 | Na1O6Si2Ti1 | | | | | |
| | -GM | -R | -T | -U | -V | -X | -Y | -Z |
| | 3+3 2 | 2+2 2 | 2+2 2 | 3+3 2 | 3+3 2 | 2+2 2 | 3+3 2 | 2+2 2 |
| | 2+2 2 | 3+3 2 | 3+3 2 | 2+2 2 | 2+2 2 | 3+3 2 | 2+2 2 | 3+3 2 |
| | 3+3 2 | 3+3 2 | 2+2 2 | 3+3 2 | 2+2 2 | 3+3 2 | 3+3 2 | 2+2 2 |
| | 2+2 2 | 2+2 2 | 3+3 2 | 2+2 2 | 3+3 2 | 2+2 2 | 2+2 2 | 3+3 2 |
| | 3+3 2 | 3+3 2 | 3+3 2 | 3+3 2 | 3+3 2 | 2+2 2 | 2+2 2 | 3+3 2 |
| | 2+2 2 | 2+2 2 | 2+2 2 | 2+2 2 | 2+2 2 | 3+3 2 | 3+3 2 | 2+2 2 |
| | 3+3 2 | 2+2 2 | 2+2 2 | 3+3 2 | 2+2 2 | 3+3 2 | 2+2 2 | 3+3 2 |
| | 2+2 2 | 3+3 2 | 3+3 2 | 2+2 2 | 3+3 2 | 2+2 2 | 3+3 2 | 2+2 2 |
| | 2+2 2 | 3+3 2 | 3+3 2 | 2+2 2 | 2+2 2 | 3+3 2 | 2+2 2 | 2+2 2 |
| | 3+3 2 | 2+2 2 | 2+2 2 | 3+3 2 | 3+3 2 | 2+2 2 | 3+3 2 | 3+3 2 |
| | 3+3 2 | 2+2 2 | 2+2 2 | 3+3 2 | 3+3 2 | 3+3 2 | 3+3 2 | 2+2 2 |
| | 2+2 2 | 3+3 2 | 3+3 2 | 2+2 2 | 2+2 2 | 2+2 2 | 2+2 2 | 3+3 2 |
| | 2+2 2 | 3+3 2 | 3+3 2 | 3+3 2 | 3+3 2 | 2+2 2 | 2+2 2 | 3+3 2 |
| | 3+3 2 | 2+2 2 | 2+2 2 | 2+2 2 | 2+2 2 | 3+3 2 | 3+3 2 | 2+2 2 |
| | 3+3 2 | 2+2 2 | 2+2 2 | 2+2 2 | 2+2 2 | 3+3 2 | 2+2 2 | 3+3 2 |
| | 3+3 2 | 3+3 2 | 3+3 2 | 3+3 2 | 3+3 2 | 2+2 2 | 3+3 2 | 2+2 2 |
| | 2+2 2 | 3+3 2 | 2+2 2 | 3+3 2 | 3+3 2 | 2+2 2 | 2+2 2 | 3+3 2 |
| | 2+2 2 | 2+2 2 | 2+2 2 | 2+2 2 | 2+2 2 | 3+3 2 | 3+3 2 | 2+2 2 |
| | 2+2 2 | 2+2 2 | 3+3 2 | 2+2 2 | 2+2 2 | 3+3 2 | 2+2 2 | 3+3 2 |
| | 3+3 2 | 3+3 2 | 3+3 2 | 3+3 2 | 3+3 2 | 2+2 2 | 3+3 2 | 2+2 2 |
| | 3+3 2 | 3+3 2 | 3+3 2 | 2+2 2 | 2+2 2 | 2+2 2 | 3+3 2 | 2+2 2 |
| | 2+2 2 | 2+2 2 | 3+3 2 | 3+3 2 | 3+3 2 | 3+3 2 | 3+3 2 | 2+2 2 |
| | 3+3 2 | 2+2 2 | 2+2 2 | 3+3 2 | 3+3 2 | 3+3 2 | 2+2 2 | 3+3 2 |
| | 2+2 2 | 3+3 2 | 2+2 2 | 2+2 2 | 2+2 2 | 2+2 2 | 2+2 2 | 3+3 2 |
| | 2+2 2 | 2+2 2 | 3+3 2 | 2+2 2 | 2+2 2 | 2+2 2 | 3+3 2 | 2+2 2 |
| | 3+3 2 | 3+3 2 | 2+2 2 | 3+3 2 | 3+3 2 | 3+3 2 | 2+2 2 | 3+3 2 |
| | 2+2 2 | 3+3 2 | 2+2 2 | 3+3 2 | 3+3 2 | 3+3 2 | 3+3 2 | 2+2 2 |
| | 3+3 2 | 2+2 2 | 3+3 2 | 3+3 2 | 3+3 2 | 2+2 2 | 2+2 2 | 3+3 2 |
| | 3+3 2 | 2+2 2 | 3+3 2 | 2+2 2 | 2+2 2 | 2+2 2 | 2+2 2 | 3+3 2 |
| | 3+3 2 | 3+3 2 | 2+2 2 | 2+2 2 | 2+2 2 | 3+3 2 | 3+3 2 | 2+2 2 |
| | 2+2 2 | 3+3 2 | 2+2 2 | 3+3 2 | 3+3 2 | 3+3 2 | 2+2 2 | 3+3 2 |
| | 3+3 2 | 2+2 2 | 2+2 2 | 2+2 2 | 2+2 2 | 2+2 2 | 3+3 2 | 2+2 2 |
| | 3+3 2 | 2+2 2 | 3+3 2 | 2+2 2 | 2+2 2 | 3+3 2 | 2+2 2 | 3+3 2 |
| | 2+2 2 | 3+3 2 | 3+3 2 | 2+2 2 | 3+3 2 | 2+2 2 | 3+3 2 | 2+2 2 |
| | 3+3 2 | 3+3 2 | 2+2 2 | 3+3 2 | 2+2 2 | 3+3 2 | 2+2 2 | 3+3 2 |
| | 3+3 2 | 2+2 2 | 3+3 2 | 3+3 2 | 3+3 2 | 2+2 2 | 3+3 2 | 2+2 2 |
| | 2+2 2 | 2+2 2 | 2+2 2 | 3+3 2 | 3+3 2 | 3+3 2 | 3+3 2 | 2+2 2 |
| | 2+2 2 | 3+3 2 | 3+3 2 | 2+2 2 | 2+2 2 | 2+2 2 | 2+2 2 | 3+3 2 |
| | 3+3 2 | 2+2 2 | 2+2 2 | 2+2 2 | 2+2 2 | 2+2 2 | 3+3 2 | 3+3 2 |
| | 2+2 2 | 3+3 2 | 3+3 2 | 3+3 2 | 3+3 2 | 3+3 2 | 2+2 2 | 2+2 2 |
| | 3+3 2 | 3+3 2 | 2+2 2 | 3+3 2 | 3+3 2 | 3+3 2 | 2+2 2 | 3+3 2 |
| | 2+2 2 | 2+2 2 | 3+3 2 | 2+2 2 | 2+2 2 | 2+2 2 | 3+3 2 | 2+2 2 |
| | 2+2 2 | 3+3 2 | 3+3 2 | 3+3 2 | 3+3 2 | 2+2 2 | 2+2 2 | 3+3 2 |
| | 2+2 2 | 2+2 2 | 2+2 2 | 2+2 2 | 2+2 2 | 3+3 2 | 3+3 2 | 2+2 2 |
| | 2+2 2 | 3+3 2 | 3+3 2 | 2+2 2 | 2+2 2 | 2+2 2 | 3+3 2 | 3+3 2 |
| | 3+3 2 | 2+2 2 | 2+2 2 | 3+3 2 | 3+3 2 | 3+3 2 | 2+2 2 | 2+2 2 |
| | 3+3 2 | 3+3 2 | 2+2 2 | 2+2 2 | 2+2 2 | 3+3 2 | 2+2 2 | 3+3 2 |
| | 2+2 2 | 2+2 2 | 3+3 2 | 3+3 2 | 3+3 2 | 2+2 2 | 3+3 2 | 2+2 2 |

Continued on next column

**TABLE XXI– continued**

| TQC | ICSD | NoE | SF | | | | |
|---|---|---|---|---|---|---|---|
| | 3+3 2 | 3+3 2 | 3+3 2 | 2+2 2 | 3+3 2 | 2+2 2 | 2+2 2 | 3+3 2 |
| 2.2.15 | 65941 | 94 | Mo6S8Sr1 | | | | | |
| | -GM | -R | -T | -U | -V | -X | -Y | -Z |
| | 3+3 2 | 3+3 2 | 3+3 2 | 3+3 2 | 3+3 2 | 3+3 2 | 3+3 2 | 3+3 2 |
| | 2+2 2 | 2+2 2 | 2+2 2 | 2+2 2 | 2+2 2 | 2+2 2 | 2+2 2 | 2+2 2 |
| | 2+2 2 | 2+2 2 | 2+2 2 | 2+2 2 | 2+2 2 | 2+2 2 | 2+2 2 | 2+2 2 |
| | 2+2 2 | 2+2 2 | 2+2 2 | 2+2 2 | 2+2 2 | 2+2 2 | 2+2 2 | 2+2 2 |
| | 3+3 2 | 3+3 2 | 3+3 2 | 2+2 2 | 3+3 2 | 2+2 2 | 2+2 2 | 3+3 2 |
| | 2+2 2 | 2+2 2 | 2+2 2 | 3+3 2 | 2+2 2 | 3+3 2 | 3+3 2 | 2+2 2 |
| | 2+2 2 | 3+3 2 | 3+3 2 | 3+3 2 | 2+2 2 | 3+3 2 | 3+3 2 | 3+3 2 |
| | 3+3 2 | 2+2 2 | 2+2 2 | 2+2 2 | 2+2 2 | 3+3 2 | 3+3 2 | 2+2 2 |
| | 2+2 2 | 3+3 2 | 2+2 2 | 2+2 2 | 3+3 2 | 2+2 2 | 2+2 2 | 3+3 2 |
| | 3+3 2 | 2+2 2 | 3+3 2 | 3+3 2 | 3+3 2 | 2+2 2 | 2+2 2 | 2+2 2 |
| | 2+2 2 | 3+3 2 | 3+3 2 | 3+3 2 | 3+3 2 | 3+3 2 | 3+3 2 | 3+3 2 |
| | 3+3 2 | 2+2 2 | 2+2 2 | 2+2 2 | 2+2 2 | 2+2 2 | 2+2 2 | 2+2 2 |
| | 3+3 2 | 2+2 2 | 3+3 2 | 3+3 2 | 2+2 2 | 2+2 2 | 2+2 2 | 2+2 2 |
| | 3+3 2 | 2+2 2 | 3+3 2 | 3+3 2 | 3+3 2 | 3+3 2 | 3+3 2 | 3+3 2 |
| | 3+3 2 | 3+3 2 | 2+2 2 | 2+2 2 | 3+3 2 | 2+2 2 | 2+2 2 | 2+2 2 |
| | 2+2 2 | 2+2 2 | 3+3 2 | 3+3 2 | 2+2 2 | 3+3 2 | 3+3 2 | 3+3 2 |
| | 2+2 2 | 2+2 2 | 2+2 2 | 2+2 2 | 3+3 2 | 3+3 2 | 3+3 2 | 2+2 2 |
| | 2+2 2 | 3+3 2 | 2+2 2 | 3+3 2 | 3+3 2 | 2+2 2 | 2+2 2 | 3+3 2 |
| | 3+3 2 | 2+2 2 | 3+3 2 | 2+2 2 | 2+2 2 | 2+2 2 | 3+3 2 | 3+3 2 |
| | 2+2 2 | 3+3 2 | 3+3 2 | 3+3 2 | 3+3 2 | 3+3 2 | 2+2 2 | 2+2 2 |
| | 3+3 2 | 2+2 2 | 2+2 2 | 2+2 2 | 3+3 2 | 3+3 2 | 2+2 2 | 3+3 2 |
| | 2+2 2 | 3+3 2 | 3+3 2 | 3+3 2 | 2+2 2 | 2+2 2 | 3+3 2 | 2+2 2 |
| | 2+2 2 | 3+3 2 | 2+2 2 | 2+2 2 | 2+2 2 | 3+3 2 | 2+2 2 | 3+3 2 |
| | 3+3 2 | 3+3 2 | 3+3 2 | 3+3 2 | 3+3 2 | 2+2 2 | 3+3 2 | 2+2 2 |
| | 3+3 2 | 3+3 2 | 2+2 2 | 2+2 2 | 2+2 2 | 2+2 2 | 3+3 2 | 2+2 2 |
| | 3+3 2 | 3+3 2 | 2+2 2 | 2+2 2 | 2+2 2 | 3+3 2 | 2+2 2 | 3+3 2 |
| | 2+2 2 | 2+2 2 | 3+3 2 | 2+2 2 | 3+3 2 | 2+2 2 | 2+2 2 | 3+3 2 |
| | 3+3 2 | 2+2 2 | 2+2 2 | 3+3 2 | 2+2 2 | 3+3 2 | 3+3 2 | 2+2 2 |
| | 2+2 2 | 3+3 2 | 2+2 2 | 3+3 2 | 3+3 2 | 2+2 2 | 3+3 2 | 2+2 2 |
| | 2+2 2 | 2+2 2 | 3+3 2 | 2+2 2 | 3+3 2 | 3+3 2 | 2+2 2 | 3+3 2 |
| | 3+3 2 | 3+3 2 | 2+2 2 | 2+2 2 | 2+2 2 | 2+2 2 | 2+2 2 | 3+3 2 |
| | 2+2 2 | 3+3 2 | 3+3 2 | 3+3 2 | 2+2 2 | 2+2 2 | 3+3 2 | 2+2 2 |
| | 3+3 2 | 3+3 2 | 3+3 2 | 3+3 2 | 3+3 2 | 3+3 2 | 2+2 2 | 2+2 2 |
| | 2+2 2 | 2+2 2 | 2+2 2 | 2+2 2 | 2+2 2 | 3+3 2 | 3+3 2 | 3+3 2 |
| | 3+3 2 | 2+2 2 | 3+3 2 | 2+2 2 | 2+2 2 | 2+2 2 | 2+2 2 | 2+2 2 |
| | 3+3 2 | 3+3 2 | 2+2 2 | 3+3 2 | 3+3 2 | 3+3 2 | 3+3 2 | 3+3 2 |
| | 2+2 2 | 2+2 2 | 2+2 2 | 2+2 2 | 2+2 2 | 2+2 2 | 2+2 2 | 2+2 2 |
| | 3+3 2 | 2+2 2 | 3+3 2 | 3+3 2 | 3+3 2 | 3+3 2 | 3+3 2 | 3+3 2 |
| | 3+3 2 | 3+3 2 | 2+2 2 | 2+2 2 | 2+2 2 | 2+2 2 | 2+2 2 | 3+3 2 |
| | 2+2 2 | 3+3 2 | 3+3 2 | 3+3 2 | 3+3 2 | 3+3 2 | 3+3 2 | 2+2 2 |
| | 3+3 2 | 3+3 2 | 3+3 2 | 2+2 2 | 3+3 2 | 2+2 2 | 2+2 2 | 3+3 2 |
| | 2+2 2 | 2+2 2 | 2+2 2 | 3+3 2 | 2+2 2 | 3+3 2 | 3+3 2 | 2+2 2 |
| | 2+2 2 | 2+2 2 | 2+2 2 | 2+2 2 | 2+2 2 | 3+3 2 | 3+3 2 | 3+3 2 |
| | 2+2 2 | 3+3 2 | 2+2 2 | 2+2 2 | 2+2 2 | 2+2 2 | 2+2 2 | 3+3 2 |
| | 3+3 2 | 3+3 2 | 2+2 2 | 2+2 2 | 3+3 2 | 3+3 2 | 3+3 2 | 2+2 2 |
| | 3+3 2 | 2+2 2 | 3+3 2 | 3+3 2 | 2+2 2 | 2+2 2 | 2+2 2 | 2+2 2 |
| | 2+2 2 | 2+2 2 | 3+3 2 | 3+3 2 | 3+3 2 | 3+3 2 | 3+3 2 | 3+3 2 |
| 2.2.16 | 237587 | 84 | Mo3S4 | | | | | |
| | -GM | -R | -T | -U | -V | -X | -Y | -Z |
| | 3+3 2 | 2+2 2 | 3+3 2 | 3+3 2 | 3+3 2 | 2+2 2 | 2+2 2 | 2+2 2 |
| | 3+3 2 | 3+3 2 | 2+2 2 | 2+2 2 | 2+2 2 | 3+3 2 | 3+3 2 | 3+3 2 |
| | 2+2 2 | 2+2 2 | 3+3 2 | 3+3 2 | 3+3 2 | 2+2 2 | 3+3 2 | 2+2 2 |
| | 2+2 2 | 3+3 2 | 2+2 2 | 2+2 2 | 2+2 2 | 3+3 2 | 2+2 2 | 3+3 2 |
| | 3+3 2 | 2+2 2 | 3+3 2 | 2+2 2 | 2+2 2 | 2+2 2 | 3+3 2 | 3+3 2 |
| | 2+2 2 | 3+3 2 | 3+3 2 | 3+3 2 | 3+3 2 | 3+3 2 | 2+2 2 | 3+3 2 |
| | 2+2 2 | 3+3 2 | 2+2 2 | 2+2 2 | 2+2 2 | 3+3 2 | 2+2 2 | 2+2 2 |
| | 3+3 2 | 2+2 2 | 2+2 2 | 3+3 2 | 3+3 2 | 2+2 2 | 3+3 2 | 2+2 2 |
| | 2+2 2 | 2+2 2 | 3+3 2 | 3+3 2 | 3+3 2 | 2+2 2 | 2+2 2 | 3+3 2 |

Continued on next column

**TABLE XXI– continued**

| TQC | ICSD | NoE | SF | | | | | |
|---|---|---|---|---|---|---|---|---|
| | 3+3 2 | 3+3 2 | 2+2 2 | 2+2 2 | 2+2 2 | 3+3 2 | 3+3 2 | 2+2 2 |
| | 2+2 2 | 2+2 2 | 3+3 2 | 3+3 2 | 2+2 2 | 2+2 2 | 3+3 2 | 2+2 2 |
| | 3+3 2 | 3+3 2 | 3+3 2 | 2+2 2 | 3+3 2 | 3+3 2 | 2+2 2 | 3+3 2 |
| | 3+3 2 | 2+2 2 | 2+2 2 | 3+3 2 | 3+3 2 | 3+3 2 | 2+2 2 | 2+2 2 |
| | 2+2 2 | 3+3 2 | 2+2 2 | 2+2 2 | 2+2 2 | 2+2 2 | 3+3 2 | 2+2 2 |
| | 2+2 2 | 2+2 2 | 3+3 2 | 3+3 2 | 3+3 2 | 3+3 2 | 2+2 2 | 3+3 2 |
| | 3+3 2 | 3+3 2 | 3+3 2 | 2+2 2 | 2+2 2 | 2+2 2 | 3+3 2 | 3+3 2 |
| | 3+3 2 | 2+2 2 | 3+3 2 | 2+2 2 | 3+3 2 | 2+2 2 | 2+2 2 | 2+2 2 |
| | 2+2 2 | 3+3 2 | 2+2 2 | 3+3 2 | 2+2 2 | 3+3 2 | 2+2 2 | 3+3 2 |
| | 3+3 2 | 2+2 2 | 3+3 2 | 2+2 2 | 3+3 2 | 3+3 2 | 3+3 2 | 3+3 2 |
| | 2+2 2 | 3+3 2 | 2+2 2 | 3+3 2 | 2+2 2 | 2+2 2 | 3+3 2 | 2+2 2 |
| | 3+3 2 | 2+2 2 | 3+3 2 | 2+2 2 | 3+3 2 | 3+3 2 | 2+2 2 | 3+3 2 |
| | 3+3 2 | 3+3 2 | 2+2 2 | 2+2 2 | 2+2 2 | 2+2 2 | 3+3 2 | 2+2 2 |
| | 3+3 2 | 2+2 2 | 2+2 2 | 3+3 2 | 3+3 2 | 2+2 2 | 3+3 2 | 2+2 2 |
| | 2+2 2 | 3+3 2 | 2+2 2 | 3+3 2 | 2+2 2 | 3+3 2 | 3+3 2 | 2+2 2 |
| | 3+3 2 | 2+2 2 | 3+3 2 | 2+2 2 | 3+3 2 | 3+3 2 | 2+2 2 | 3+3 2 |
| | 2+2 2 | 3+3 2 | 3+3 2 | 3+3 2 | 3+3 2 | 2+2 2 | 2+2 2 | 3+3 2 |
| | 2+2 2 | 2+2 2 | 2+2 2 | 2+2 2 | 2+2 2 | 3+3 2 | 3+3 2 | 2+2 2 |
| | 3+3 2 | 3+3 2 | 2+2 2 | 3+3 2 | 2+2 2 | 2+2 2 | 2+2 2 | 2+2 2 |
| | 3+3 2 | 2+2 2 | 2+2 2 | 2+2 2 | 3+3 2 | 3+3 2 | 3+3 2 | 3+3 2 |
| | 3+3 2 | 3+3 2 | 3+3 2 | 3+3 2 | 2+2 2 | 2+2 2 | 2+2 2 | 2+2 2 |
| | 2+2 2 | 2+2 2 | 2+2 2 | 2+2 2 | 2+2 2 | 3+3 2 | 3+3 2 | 2+2 2 |
| | 2+2 2 | 3+3 2 | 2+2 2 | 3+3 2 | 3+3 2 | 2+2 2 | 3+3 2 | 3+3 2 |
| | 2+2 2 | 3+3 2 | 3+3 2 | 2+2 2 | 3+3 2 | 2+2 2 | 2+2 2 | 3+3 2 |
| | 3+3 2 | 2+2 2 | 3+3 2 | 3+3 2 | 2+2 2 | 3+3 2 | 2+2 2 | 3+3 2 |
| | 2+2 2 | 2+2 2 | 3+3 2 | 2+2 2 | 3+3 2 | 2+2 2 | 3+3 2 | 2+2 2 |
| | 3+3 2 | 3+3 2 | 2+2 2 | 3+3 2 | 2+2 2 | 3+3 2 | 3+3 2 | 2+2 2 |
| | 2+2 2 | 2+2 2 | 2+2 2 | 2+2 2 | 3+3 2 | 2+2 2 | 2+2 2 | 3+3 2 |
| | 2+2 2 | 3+3 2 | 3+3 2 | 3+3 2 | 2+2 2 | 3+3 2 | 2+2 2 | 3+3 2 |
| | 3+3 2 | 3+3 2 | 2+2 2 | 3+3 2 | 2+2 2 | 2+2 2 | 2+2 2 | 3+3 2 |
| | 2+2 2 | 2+2 2 | 2+2 2 | 2+2 2 | 3+3 2 | 3+3 2 | 2+2 2 | 3+3 2 |
| | 3+3 2 | 3+3 2 | 3+3 2 | 3+3 2 | 2+2 2 | 3+3 2 | 3+3 2 | 3+3 2 |
| | 3+3 2 | 2+2 2 | 2+2 2 | 2+2 2 | 3+3 2 | 2+2 2 | 2+2 2 | 3+3 2 |
| 2.2.17 | 61050 | 130 | Mn3Na2O7 | | | | | |
| | -GM | -R | -T | -U | -V | -X | -Y | -Z |
| | 3+3 2 | 3+3 2 | 2+2 2 | 2+2 2 | 3+3 2 | 3+3 2 | 3+3 2 | 3+3 2 |
| | 3+3 2 | 2+2 2 | 3+3 2 | 3+3 2 | 3+3 2 | 2+2 2 | 2+2 2 | 2+2 2 |
| | 3+3 2 | 3+3 2 | 2+2 2 | 2+2 2 | 3+3 2 | 3+3 2 | 3+3 2 | 3+3 2 |
| | 2+2 2 | 2+2 2 | 3+3 2 | 3+3 2 | 2+2 2 | 2+2 2 | 2+2 2 | 2+2 2 |
| | 2+2 2 | 3+3 2 | 3+3 2 | 2+2 2 | 2+2 2 | 2+2 2 | 2+2 2 | 3+3 2 |
| | 2+2 2 | 2+2 2 | 2+2 2 | 3+3 2 | 2+2 2 | 3+3 2 | 3+3 2 | 2+2 2 |
| | 2+2 2 | 2+2 2 | 3+3 2 | 3+3 2 | 2+2 2 | 2+2 2 | 2+2 2 | 2+2 2 |
| | 3+3 2 | 2+2 2 | 2+2 2 | 2+2 2 | 2+2 2 | 3+3 2 | 3+3 2 | 2+2 2 |
| | 2+2 2 | 3+3 2 | 2+2 2 | 2+2 2 | 3+3 2 | 2+2 2 | 2+2 2 | 3+3 2 |
| | 2+2 2 | 3+3 2 | 3+3 2 | 3+3 2 | 2+2 2 | 2+2 2 | 2+2 2 | 3+3 2 |
| | 2+2 2 | 3+3 2 | 3+3 2 | 3+3 2 | 2+2 2 | 3+3 2 | 3+3 2 | 3+3 2 |
| | 3+3 2 | 2+2 2 | 2+2 2 | 2+2 2 | 3+3 2 | 3+3 2 | 3+3 2 | 2+2 2 |
| | 3+3 2 | 2+2 2 | 3+3 2 | 3+3 2 | 3+3 2 | 3+3 2 | 2+2 2 | 2+2 2 |
| | 3+3 2 | 3+3 2 | 2+2 2 | 2+2 2 | 3+3 2 | 2+2 2 | 3+3 2 | 3+3 2 |
| | 3+3 2 | 2+2 2 | 3+3 2 | 3+3 2 | 2+2 2 | 2+2 2 | 2+2 2 | 2+2 2 |
| | 2+2 2 | 3+3 2 | 2+2 2 | 2+2 2 | 2+2 2 | 3+3 2 | 3+3 2 | 3+3 2 |
| | 2+2 2 | 3+3 2 | 2+2 2 | 2+2 2 | 3+3 2 | 3+3 2 | 3+3 2 | 3+3 2 |
| | 2+2 2 | 2+2 2 | 3+3 2 | 3+3 2 | 2+2 2 | 2+2 2 | 3+3 2 | 2+2 2 |
| | 3+3 2 | 3+3 2 | 2+2 2 | 2+2 2 | 3+3 2 | 3+3 2 | 2+2 2 | 3+3 2 |
| | 3+3 2 | 2+2 2 | 3+3 2 | 3+3 2 | 3+3 2 | 2+2 2 | 2+2 2 | 2+2 2 |
| | 3+3 2 | 3+3 2 | 2+2 2 | 2+2 2 | 3+3 2 | 3+3 2 | 3+3 2 | 3+3 2 |
| | 2+2 2 | 2+2 2 | 3+3 2 | 3+3 2 | 2+2 2 | 2+2 2 | 2+2 2 | 2+2 2 |
| | 2+2 2 | 2+2 2 | 3+3 2 | 3+3 2 | 3+3 2 | 2+2 2 | 3+3 2 | 2+2 2 |
| | 3+3 2 | 3+3 2 | 2+2 2 | 2+2 2 | 2+2 2 | 3+3 2 | 2+2 2 | 3+3 2 |
| | 2+2 2 | 3+3 2 | 3+3 2 | 2+2 2 | 2+2 2 | 3+3 2 | 3+3 2 | 3+3 2 |
| | 3+3 2 | 2+2 2 | 2+2 2 | 3+3 2 | 3+3 2 | 2+2 2 | 2+2 2 | 2+2 2 |

Continued on next column

**TABLE XXI– continued**

| TQC | ICSD | NoE | SF | | | | | |
|---|---|---|---|---|---|---|---|---|
| | 3+3 2 | 2+2 2 | 3+3 2 | 3+3 2 | 3+3 2 | 3+3 2 | 3+3 2 | 2+2 2 |
| | 3+3 2 | 3+3 2 | 2+2 2 | 2+2 2 | 3+3 2 | 2+2 2 | 2+2 2 | 3+3 2 |
| | 3+3 2 | 3+3 2 | 3+3 2 | 2+2 2 | 3+3 2 | 2+2 2 | 2+2 2 | 3+3 2 |
| | 2+2 2 | 2+2 2 | 2+2 2 | 3+3 2 | 2+2 2 | 3+3 2 | 3+3 2 | 2+2 2 |
| | 2+2 2 | 3+3 2 | 2+2 2 | 2+2 2 | 2+2 2 | 3+3 2 | 3+3 2 | 3+3 2 |
| | 3+3 2 | 3+3 2 | 2+2 2 | 2+2 2 | 3+3 2 | 2+2 2 | 3+3 2 | 3+3 2 |
| | 3+3 2 | 2+2 2 | 3+3 2 | 3+3 2 | 3+3 2 | 3+3 2 | 2+2 2 | 2+2 2 |
| | 2+2 2 | 2+2 2 | 3+3 2 | 3+3 2 | 2+2 2 | 2+2 2 | 2+2 2 | 2+2 2 |
| | 3+3 2 | 2+2 2 | 3+3 2 | 3+3 2 | 3+3 2 | 2+2 2 | 2+2 2 | 2+2 2 |
| | 2+2 2 | 3+3 2 | 2+2 2 | 2+2 2 | 2+2 2 | 3+3 2 | 3+3 2 | 3+3 2 |
| | 2+2 2 | 3+3 2 | 2+2 2 | 2+2 2 | 2+2 2 | 3+3 2 | 2+2 2 | 3+3 2 |
| | 2+2 2 | 2+2 2 | 3+3 2 | 3+3 2 | 2+2 2 | 2+2 2 | 3+3 2 | 2+2 2 |
| | 3+3 2 | 3+3 2 | 2+2 2 | 2+2 2 | 3+3 2 | 3+3 2 | 3+3 2 | 3+3 2 |
| | 3+3 2 | 2+2 2 | 2+2 2 | 2+2 2 | 3+3 2 | 3+3 2 | 2+2 2 | 3+3 2 |
| | 2+2 2 | 3+3 2 | 3+3 2 | 3+3 2 | 2+2 2 | 2+2 2 | 3+3 2 | 2+2 2 |
| | 3+3 2 | 2+2 2 | 3+3 2 | 3+3 2 | 3+3 2 | 2+2 2 | 2+2 2 | 2+2 2 |
| | 2+2 2 | 2+2 2 | 3+3 2 | 3+3 2 | 2+2 2 | 2+2 2 | 2+2 2 | 2+2 2 |
| | 2+2 2 | 3+3 2 | 2+2 2 | 3+3 2 | 2+2 2 | 2+2 2 | 2+2 2 | 3+3 2 |
| | 3+3 2 | 3+3 2 | 3+3 2 | 2+2 2 | 3+3 2 | 3+3 2 | 3+3 2 | 2+2 2 |
| | 3+3 2 | 2+2 2 | 2+2 2 | 2+2 2 | 2+2 2 | 3+3 2 | 3+3 2 | 3+3 2 |
| | 3+3 2 | 3+3 2 | 2+2 2 | 2+2 2 | 3+3 2 | 3+3 2 | 3+3 2 | 3+3 2 |
| | 2+2 2 | 3+3 2 | 2+2 2 | 2+2 2 | 3+3 2 | 3+3 2 | 3+3 2 | 3+3 2 |
| | 2+2 2 | 2+2 2 | 3+3 2 | 2+2 2 | 2+2 2 | 2+2 2 | 2+2 2 | 2+2 2 |
| | 3+3 2 | 3+3 2 | 2+2 2 | 3+3 2 | 2+2 2 | 3+3 2 | 3+3 2 | 3+3 2 |
| | 2+2 2 | 2+2 2 | 3+3 2 | 3+3 2 | 3+3 2 | 2+2 2 | 2+2 2 | 2+2 2 |
| | 3+3 2 | 2+2 2 | 3+3 2 | 3+3 2 | 3+3 2 | 2+2 2 | 2+2 2 | 2+2 2 |
| | 3+3 2 | 3+3 2 | 2+2 2 | 2+2 2 | 3+3 2 | 3+3 2 | 3+3 2 | 3+3 2 |
| | 2+2 2 | 2+2 2 | 3+3 2 | 3+3 2 | 2+2 2 | 2+2 2 | 2+2 2 | 2+2 2 |
| | 2+2 2 | 3+3 2 | 3+3 2 | 2+2 2 | 2+2 2 | 2+2 2 | 2+2 2 | 2+2 2 |
| | 2+2 2 | 2+2 2 | 2+2 2 | 3+3 2 | 2+2 2 | 3+3 2 | 3+3 2 | 3+3 2 |
| | 3+3 2 | 2+2 2 | 3+3 2 | 3+3 2 | 3+3 2 | 3+3 2 | 3+3 2 | 2+2 2 |
| | 3+3 2 | 3+3 2 | 2+2 2 | 2+2 2 | 3+3 2 | 2+2 2 | 2+2 2 | 3+3 2 |
| | 2+2 2 | 3+3 2 | 2+2 2 | 2+2 2 | 2+2 2 | 3+3 2 | 3+3 2 | 3+3 2 |
| | 2+2 2 | 2+2 2 | 3+3 2 | 3+3 2 | 2+2 2 | 2+2 2 | 2+2 2 | 2+2 2 |
| | 2+2 2 | 3+3 2 | 2+2 2 | 2+2 2 | 2+2 2 | 3+3 2 | 3+3 2 | 3+3 2 |
| | 3+3 2 | 2+2 2 | 3+3 2 | 3+3 2 | 3+3 2 | 2+2 2 | 2+2 2 | 2+2 2 |
| | 3+3 2 | 3+3 2 | 2+2 2 | 2+2 2 | 3+3 2 | 3+3 2 | 3+3 2 | 3+3 2 |
| | 2+2 2 | 2+2 2 | 3+3 2 | 3+3 2 | 2+2 2 | 3+3 2 | 3+3 2 | 2+2 2 |
| | 3+3 2 | 3+3 2 | 2+2 2 | 2+2 2 | 2+2 2 | 2+2 2 | 2+2 2 | 3+3 2 |
| 2.2.18 | 59328 | 130 | Mn1O7Te1V2 | | | | | |
| | -GM | -R | -T | -U | -V | -X | -Y | -Z |
| | 3+3 2 | 2+2 2 | 2+2 2 | 2+2 2 | 3+3 2 | 3+3 2 | 3+3 2 | 2+2 2 |
| | 2+2 2 | 3+3 2 | 3+3 2 | 3+3 2 | 2+2 2 | 2+2 2 | 2+2 2 | 3+3 2 |
| | 2+2 2 | 2+2 2 | 3+3 2 | 2+2 2 | 3+3 2 | 3+3 2 | 2+2 2 | 3+3 2 |
| | 3+3 2 | 3+3 2 | 2+2 2 | 3+3 2 | 2+2 2 | 2+2 2 | 3+3 2 | 2+2 2 |
| | 3+3 2 | 2+2 2 | 3+3 2 | 3+3 2 | 3+3 2 | 2+2 2 | 2+2 2 | 2+2 2 |
| | 3+3 2 | 3+3 2 | 2+2 2 | 2+2 2 | 2+2 2 | 3+3 2 | 3+3 2 | 2+2 2 |
| | 2+2 2 | 2+2 2 | 3+3 2 | 3+3 2 | 3+3 2 | 2+2 2 | 2+2 2 | 3+3 2 |
| | 2+2 2 | 3+3 2 | 2+2 2 | 2+2 2 | 2+2 2 | 3+3 2 | 3+3 2 | 3+3 2 |
| | 3+3 2 | 2+2 2 | 3+3 2 | 2+2 2 | 3+3 2 | 2+2 2 | 3+3 2 | 3+3 2 |
| | 2+2 2 | 3+3 2 | 2+2 2 | 2+2 2 | 2+2 2 | 3+3 2 | 3+3 2 | 2+2 2 |
| | 3+3 2 | 2+2 2 | 3+3 2 | 3+3 2 | 2+2 2 | 2+2 2 | 2+2 2 | 3+3 2 |
| | 2+2 2 | 3+3 2 | 2+2 2 | 2+2 2 | 3+3 2 | 3+3 2 | 2+2 2 | 2+2 2 |
| | 3+3 2 | 3+3 2 | 2+2 2 | 3+3 2 | 3+3 2 | 2+2 2 | 2+2 2 | 3+3 2 |
| | 2+2 2 | 2+2 2 | 3+3 2 | 3+3 2 | 2+2 2 | 3+3 2 | 3+3 2 | 2+2 2 |
| | 3+3 2 | 2+2 2 | 2+2 2 | 3+3 2 | 3+3 2 | 2+2 2 | 3+3 2 | 2+2 2 |
| | 2+2 2 | 3+3 2 | 3+3 2 | 2+2 2 | 2+2 2 | 3+3 2 | 2+2 2 | 3+3 2 |
| | 2+2 2 | 3+3 2 | 3+3 2 | 3+3 2 | 2+2 2 | 2+2 2 | 2+2 2 | 3+3 2 |
| | 3+3 2 | 2+2 2 | 2+2 2 | 2+2 2 | 3+3 2 | 3+3 2 | 3+3 2 | 2+2 2 |
| | 2+2 2 | 3+3 2 | 2+2 2 | 3+3 2 | 2+2 2 | 2+2 2 | 3+3 2 | 3+3 2 |
| | 3+3 2 | 2+2 2 | 3+3 2 | 2+2 2 | 3+3 2 | 3+3 2 | 2+2 2 | 2+2 2 |

Continued on next column

**TABLE XXI– continued**

| TQC | ICSD | NoE | SF | | | | | |
|---|---|---|---|---|---|---|---|---|
| | 3+3 2 | 3+3 2 | 2+2 2 | 2+2 2 | 2+2 2 | 3+3 2 | 2+2 2 | 2+2 2 |
| | 2+2 2 | 2+2 2 | 3+3 2 | 2+2 2 | 2+2 2 | 3+3 2 | 3+3 2 | 3+3 2 |
| | 3+3 2 | 3+3 2 | 3+3 2 | 3+3 2 | 3+3 2 | 2+2 2 | 2+2 2 | 2+2 2 |
| | 2+2 2 | 2+2 2 | 2+2 2 | 3+3 2 | 3+3 2 | 2+2 2 | 3+3 2 | 3+3 2 |
| | 3+3 2 | 2+2 2 | 3+3 2 | 2+2 2 | 3+3 2 | 3+3 2 | 3+3 2 | 2+2 2 |
| | 3+3 2 | 2+2 2 | 2+2 2 | 3+3 2 | 2+2 2 | 2+2 2 | 2+2 2 | 3+3 2 |
| | 2+2 2 | 3+3 2 | 3+3 2 | 2+2 2 | 3+3 2 | 3+3 2 | 2+2 2 | 2+2 2 |
| | 2+2 2 | 2+2 2 | 3+3 2 | 3+3 2 | 3+3 2 | 2+2 2 | 3+3 2 | 3+3 2 |
| | 3+3 2 | 3+3 2 | 2+2 2 | 2+2 2 | 2+2 2 | 2+2 2 | 2+2 2 | 3+3 2 |
| | 3+3 2 | 2+2 2 | 3+3 2 | 3+3 2 | 3+3 2 | 3+3 2 | 3+3 2 | 2+2 2 |
| | 2+2 2 | 3+3 2 | 2+2 2 | 3+3 2 | 2+2 2 | 2+2 2 | 3+3 2 | 2+2 2 |
| | 3+3 2 | 2+2 2 | 2+2 2 | 3+3 2 | 2+2 2 | 2+2 2 | 2+2 2 | 3+3 2 |
| | 2+2 2 | 3+3 2 | 3+3 2 | 2+2 2 | 3+3 2 | 3+3 2 | 3+3 2 | 2+2 2 |
| | 3+3 2 | 2+2 2 | 2+2 2 | 3+3 2 | 3+3 2 | 2+2 2 | 2+2 2 | 3+3 2 |
| | 2+2 2 | 3+3 2 | 3+3 2 | 2+2 2 | 2+2 2 | 3+3 2 | 3+3 2 | 3+3 2 |
| | 3+3 2 | 3+3 2 | 2+2 2 | 2+2 2 | 2+2 2 | 3+3 2 | 2+2 2 | 2+2 2 |
| | 3+3 2 | 3+3 2 | 3+3 2 | 2+2 2 | 2+2 2 | 2+2 2 | 3+3 2 | 3+3 2 |
| | 2+2 2 | 2+2 2 | 2+2 2 | 3+3 2 | 3+3 2 | 3+3 2 | 2+2 2 | 2+2 2 |
| | 2+2 2 | 2+2 2 | 3+3 2 | 2+2 2 | 2+2 2 | 2+2 2 | 2+2 2 | 2+2 2 |
| | 3+3 2 | 2+2 2 | 2+2 2 | 2+2 2 | 3+3 2 | 2+2 2 | 3+3 2 | 2+2 2 |
| | 3+3 2 | 3+3 2 | 3+3 2 | 3+3 2 | 3+3 2 | 3+3 2 | 3+3 2 | 3+3 2 |
| | 2+2 2 | 3+3 2 | 2+2 2 | 2+2 2 | 2+2 2 | 3+3 2 | 2+2 2 | 3+3 2 |
| | 3+3 2 | 2+2 2 | 2+2 2 | 3+3 2 | 3+3 2 | 3+3 2 | 3+3 2 | 2+2 2 |
| | 2+2 2 | 3+3 2 | 3+3 2 | 3+3 2 | 2+2 2 | 2+2 2 | 2+2 2 | 3+3 2 |
| | 2+2 2 | 2+2 2 | 3+3 2 | 2+2 2 | 3+3 2 | 2+2 2 | 3+3 2 | 2+2 2 |
| | 3+3 2 | 2+2 2 | 2+2 2 | 3+3 2 | 2+2 2 | 3+3 2 | 2+2 2 | 2+2 2 |
| | 2+2 2 | 2+2 2 | 3+3 2 | 2+2 2 | 3+3 2 | 3+3 2 | 2+2 2 | 3+3 2 |
| | 2+2 2 | 3+3 2 | 2+2 2 | 3+3 2 | 3+3 2 | 2+2 2 | 3+3 2 | 2+2 2 |
| | 3+3 2 | 3+3 2 | 3+3 2 | 3+3 2 | 2+2 2 | 2+2 2 | 2+2 2 | 3+3 2 |
| | 2+2 2 | 3+3 2 | 2+2 2 | 2+2 2 | 3+3 2 | 3+3 2 | 3+3 2 | 3+3 2 |
| | 3+3 2 | 2+2 2 | 2+2 2 | 2+2 2 | 2+2 2 | 2+2 2 | 2+2 2 | 2+2 2 |
| | 3+3 2 | 2+2 2 | 2+2 2 | 2+2 2 | 2+2 2 | 3+3 2 | 3+3 2 | 3+3 2 |
| | 2+2 2 | 3+3 2 | 3+3 2 | 2+2 2 | 3+3 2 | 3+3 2 | 3+3 2 | 3+3 2 |
| | 2+2 2 | 2+2 2 | 3+3 2 | 3+3 2 | 2+2 2 | 2+2 2 | 2+2 2 | 2+2 2 |
| | 3+3 2 | 3+3 2 | 2+2 2 | 3+3 2 | 3+3 2 | 3+3 2 | 3+3 2 | 2+2 2 |
| | 3+3 2 | 3+3 2 | 2+2 2 | 3+3 2 | 2+2 2 | 2+2 2 | 3+3 2 | 3+3 2 |
| | 2+2 2 | 2+2 2 | 3+3 2 | 2+2 2 | 3+3 2 | 3+3 2 | 2+2 2 | 2+2 2 |
| | 2+2 2 | 3+3 2 | 3+3 2 | 3+3 2 | 2+2 2 | 2+2 2 | 2+2 2 | 3+3 2 |
| | 2+2 2 | 2+2 2 | 3+3 2 | 3+3 2 | 2+2 2 | 2+2 2 | 2+2 2 | 3+3 2 |
| | 3+3 2 | 3+3 2 | 2+2 2 | 2+2 2 | 3+3 2 | 3+3 2 | 2+2 2 | 2+2 2 |
| | 2+2 2 | 3+3 2 | 2+2 2 | 2+2 2 | 2+2 2 | 2+2 2 | 3+3 2 | 2+2 2 |
| | 3+3 2 | 2+2 2 | 3+3 2 | 3+3 2 | 2+2 2 | 3+3 2 | 2+2 2 | 3+3 2 |
| | 2+2 2 | 3+3 2 | 2+2 2 | 2+2 2 | 3+3 2 | 2+2 2 | 3+3 2 | 2+2 2 |
| | 3+3 2 | 2+2 2 | 3+3 2 | 3+3 2 | 3+3 2 | 2+2 2 | 3+3 2 | 3+3 2 |
| | 3+3 2 | 3+3 2 | 3+3 2 | 3+3 2 | 2+2 2 | 3+3 2 | 2+2 2 | 3+3 2 |
| 2.2.19 | 431529 | 152 | Lu1Na1P2S6 | | | | | |
| | -GM | -R | -T | -U | -V | -X | -Y | -Z |
| | 2+2 2 | 2+2 2 | 2+2 2 | 2+2 2 | 2+2 2 | 2+2 2 | 2+2 2 | 2+2 2 |
| | 3+3 2 | 3+3 2 | 3+3 2 | 3+3 2 | 3+3 2 | 3+3 2 | 3+3 2 | 3+3 2 |
| | 2+2 2 | 2+2 2 | 2+2 2 | 2+2 2 | 2+2 2 | 2+2 2 | 2+2 2 | 2+2 2 |
| | 3+3 2 | 3+3 2 | 3+3 2 | 3+3 2 | 3+3 2 | 3+3 2 | 3+3 2 | 3+3 2 |
| | 3+3 2 | 2+2 2 | 2+2 2 | 2+2 2 | 3+3 2 | 3+3 2 | 3+3 2 | 2+2 2 |
| | 2+2 2 | 3+3 2 | 3+3 2 | 3+3 2 | 2+2 2 | 2+2 2 | 2+2 2 | 3+3 2 |
| | 3+3 2 | 2+2 2 | 2+2 2 | 2+2 2 | 3+3 2 | 3+3 2 | 3+3 2 | 2+2 2 |
| | 2+2 2 | 3+3 2 | 3+3 2 | 3+3 2 | 2+2 2 | 2+2 2 | 2+2 2 | 3+3 2 |
| | 3+3 2 | 2+2 2 | 2+2 2 | 3+3 2 | 2+2 2 | 3+3 2 | 2+2 2 | 3+3 2 |
| | 3+3 2 | 3+3 2 | 2+2 2 | 3+3 2 | 2+2 2 | 2+2 2 | 3+3 2 | 2+2 2 |
| | 2+2 2 | 3+3 2 | 3+3 2 | 2+2 2 | 3+3 2 | 2+2 2 | 3+3 2 | 2+2 2 |
| | 2+2 2 | 2+2 2 | 3+3 2 | 2+2 2 | 3+3 2 | 3+3 2 | 2+2 2 | 3+3 2 |
| | 2+2 2 | 3+3 2 | 3+3 2 | 2+2 2 | 3+3 2 | 2+2 2 | 3+3 2 | 2+2 2 |
| | 2+2 2 | 3+3 2 | 3+3 2 | 2+2 2 | 3+3 2 | 2+2 2 | 3+3 2 | 2+2 2 |
| | 3+3 2 | 2+2 2 | 3+3 2 | 3+3 2 | 3+3 2 | 3+3 2 | 2+2 2 | 3+3 2 |
| | 2+2 2 | 2+2 2 | 2+2 2 | 2+2 2 | 2+2 2 | 3+3 2 | 2+2 2 | 3+3 2 |
| | 3+3 2 | 2+2 2 | 2+2 2 | 3+3 2 | 2+2 2 | 3+3 2 | 2+2 2 | 3+3 2 |
| | 2+2 2 | 3+3 2 | 2+2 2 | 3+3 2 | 2+2 2 | 2+2 2 | 3+3 2 | 2+2 2 |
| | 3+3 2 | 2+2 2 | 3+3 2 | 2+2 2 | 3+3 2 | 3+3 2 | 2+2 2 | 3+3 2 |
| | 3+3 2 | 3+3 2 | 2+2 2 | 3+3 2 | 2+2 2 | 2+2 2 | 3+3 2 | 2+2 2 |
| | 3+3 2 | 2+2 2 | 2+2 2 | 3+3 2 | 2+2 2 | 3+3 2 | 2+2 2 | 3+3 2 |
| | 3+3 2 | 3+3 2 | 2+2 2 | 3+3 2 | 2+2 2 | 2+2 2 | 3+3 2 | 2+2 2 |
| | 2+2 2 | 2+2 2 | 3+3 2 | 2+2 2 | 2+2 2 | 3+3 2 | 3+3 2 | 2+2 2 |
| | 3+3 2 | 3+3 2 | 2+2 2 | 3+3 2 | 3+3 2 | 2+2 2 | 2+2 2 | 3+3 2 |
| | 2+2 2 | 3+3 2 | 3+3 2 | 2+2 2 | 3+3 2 | 2+2 2 | 3+3 2 | 2+2 2 |
| | 2+2 2 | 2+2 2 | 3+3 2 | 2+2 2 | 3+3 2 | 3+3 2 | 2+2 2 | 3+3 2 |
| | 3+3 2 | 2+2 2 | 2+2 2 | 3+3 2 | 2+2 2 | 2+2 2 | 2+2 2 | 3+3 2 |
| | 2+2 2 | 3+3 2 | 3+3 2 | 2+2 2 | 2+2 2 | 3+3 2 | 3+3 2 | 2+2 2 |
| | 2+2 2 | 3+3 2 | 2+2 2 | 2+2 2 | 3+3 2 | 2+2 2 | 2+2 2 | 2+2 2 |
| | 3+3 2 | 2+2 2 | 3+3 2 | 3+3 2 | 3+3 2 | 3+3 2 | 3+3 2 | 2+2 2 |
| | 3+3 2 | 2+2 2 | 3+3 2 | 2+2 2 | 3+3 2 | 2+2 2 | 2+2 2 | 3+3 2 |
| | 2+2 2 | 3+3 2 | 2+2 2 | 3+3 2 | 2+2 2 | 3+3 2 | 3+3 2 | 3+3 2 |
| | 3+3 2 | 2+2 2 | 2+2 2 | 3+3 2 | 2+2 2 | 2+2 2 | 2+2 2 | 2+2 2 |
| | 2+2 2 | 3+3 2 | 3+3 2 | 2+2 2 | 3+3 2 | 3+3 2 | 3+3 2 | 3+3 2 |
| | 3+3 2 | 3+3 2 | 2+2 2 | 2+2 2 | 2+2 2 | 2+2 2 | 2+2 2 | 2+2 2 |
| | 2+2 2 | 2+2 2 | 3+3 2 | 3+3 2 | 3+3 2 | 3+3 2 | 3+3 2 | 2+2 2 |
| | 3+3 2 | 2+2 2 | 3+3 2 | 2+2 2 | 2+2 2 | 2+2 2 | 3+3 2 | 3+3 2 |
| | 2+2 2 | 3+3 2 | 2+2 2 | 3+3 2 | 3+3 2 | 3+3 2 | 2+2 2 | 3+3 2 |
| | 2+2 2 | 2+2 2 | 3+3 2 | 2+2 2 | 3+3 2 | 2+2 2 | 2+2 2 | 3+3 2 |
| | 3+3 2 | 2+2 2 | 2+2 2 | 3+3 2 | 2+2 2 | 3+3 2 | 2+2 2 | 2+2 2 |
| | 3+3 2 | 3+3 2 | 2+2 2 | 3+3 2 | 2+2 2 | 3+3 2 | 3+3 2 | 2+2 2 |
| | 3+3 2 | 3+3 2 | 2+2 2 | 3+3 2 | 2+2 2 | 2+2 2 | 3+3 2 | 3+3 2 |
| | 2+2 2 | 3+3 2 | 3+3 2 | 2+2 2 | 3+3 2 | 2+2 2 | 2+2 2 | 3+3 2 |
| | 3+3 2 | 2+2 2 | 2+2 2 | 3+3 2 | 2+2 2 | 3+3 2 | 3+3 2 | 2+2 2 |
| | 2+2 2 | 3+3 2 | 3+3 2 | 2+2 2 | 3+3 2 | 2+2 2 | 2+2 2 | 3+3 2 |
| | 3+3 2 | 2+2 2 | 2+2 2 | 3+3 2 | 2+2 2 | 3+3 2 | 3+3 2 | 2+2 2 |
| | 2+2 2 | 3+3 2 | 3+3 2 | 2+2 2 | 3+3 2 | 2+2 2 | 2+2 2 | 2+2 2 |
| | 3+3 2 | 2+2 2 | 2+2 2 | 3+3 2 | 3+3 2 | 3+3 2 | 3+3 2 | 3+3 2 |
| | 2+2 2 | 2+2 2 | 3+3 2 | 2+2 2 | 2+2 2 | 2+2 2 | 2+2 2 | 2+2 2 |
| | 2+2 2 | 3+3 2 | 2+2 2 | 3+3 2 | 2+2 2 | 2+2 2 | 3+3 2 | 3+3 2 |
| | 3+3 2 | 3+3 2 | 3+3 2 | 2+2 2 | 3+3 2 | 3+3 2 | 3+3 2 | 3+3 2 |
| | 3+3 2 | 2+2 2 | 2+2 2 | 3+3 2 | 2+2 2 | 3+3 2 | 2+2 2 | 2+2 2 |
| | 2+2 2 | 3+3 2 | 3+3 2 | 2+2 2 | 3+3 2 | 2+2 2 | 3+3 2 | 3+3 2 |
| | 3+3 2 | 3+3 2 | 2+2 2 | 3+3 2 | 3+3 2 | 3+3 2 | 2+2 2 | 3+3 2 |
| | 2+2 2 | 2+2 2 | 3+3 2 | 2+2 2 | 3+3 2 | 3+3 2 | 3+3 2 | 2+2 2 |
| | 3+3 2 | 2+2 2 | 2+2 2 | 2+2 2 | 2+2 2 | 2+2 2 | 3+3 2 | 2+2 2 |
| | 2+2 2 | 3+3 2 | 3+3 2 | 3+3 2 | 3+3 2 | 3+3 2 | 2+2 2 | 2+2 2 |
| | 3+3 2 | 2+2 2 | 3+3 2 | 3+3 2 | 3+3 2 | 2+2 2 | 2+2 2 | 3+3 2 |
| | 3+3 2 | 2+2 2 | 2+2 2 | 2+2 2 | 2+2 2 | 3+3 2 | 3+3 2 | 3+3 2 |
| | 2+2 2 | 3+3 2 | 3+3 2 | 2+2 2 | 3+3 2 | 3+3 2 | 3+3 2 | 2+2 2 |
| | 2+2 2 | 2+2 2 | 2+2 2 | 3+3 2 | 2+2 2 | 2+2 2 | 2+2 2 | 2+2 2 |
| | 2+2 2 | 3+3 2 | 3+3 2 | 2+2 2 | 2+2 2 | 2+2 2 | 2+2 2 | 2+2 2 |
| | 3+3 2 | 3+3 2 | 3+3 2 | 2+2 2 | 3+3 2 | 3+3 2 | 3+3 2 | 3+3 2 |
| | 3+3 2 | 2+2 2 | 3+3 2 | 3+3 2 | 2+2 2 | 2+2 2 | 2+2 2 | 3+3 2 |
| | 3+3 2 | 2+2 2 | 2+2 2 | 2+2 2 | 3+3 2 | 3+3 2 | 3+3 2 | 3+3 2 |
| | 2+2 2 | 3+3 2 | 2+2 2 | 2+2 2 | 2+2 2 | 2+2 2 | 2+2 2 | 2+2 2 |
| | 2+2 2 | 3+3 2 | 3+3 2 | 3+3 2 | 3+3 2 | 3+3 2 | 3+3 2 | 2+2 2 |
| | 3+3 2 | 2+2 2 | 2+2 2 | 3+3 2 | 2+2 2 | 3+3 2 | 2+2 2 | 3+3 2 |
| | 2+2 2 | 2+2 2 | 3+3 2 | 3+3 2 | 3+3 2 | 2+2 2 | 3+3 2 | 2+2 2 |
| | 2+2 2 | 3+3 2 | 2+2 2 | 2+2 2 | 2+2 2 | 2+2 2 | 2+2 2 | 3+3 2 |
| | 3+3 2 | 2+2 2 | 3+3 2 | 3+3 2 | 2+2 2 | 3+3 2 | 3+3 2 | 3+3 2 |
| | 2+2 2 | 3+3 2 | 2+2 2 | 3+3 2 | 3+3 2 | 3+3 2 | 2+2 2 | 2+2 2 |
| | 3+3 2 | 2+2 2 | 2+2 2 | 2+2 2 | 2+2 2 | 2+2 2 | 3+3 2 | 3+3 2 |
| | 3+3 2 | 3+3 2 | 2+2 2 | 3+3 2 | 2+2 2 | 2+2 2 | 2+2 2 | 2+2 2 |
| | 2+2 2 | 2+2 2 | 2+2 2 | 3+3 2 | 3+3 2 | 3+3 2 | 3+3 2 | 3+3 2 |

Continued on next column

**TABLE XXI– continued**

| TQC | ICSD | NoE | SF | | | | | |
|---|---|---|---|---|---|---|---|---|
| | 3+3 2 | 3+3 2 | 3+3 2 | 2+2 2 | 3+3 2 | 2+2 2 | 3+3 2 | 2+2 2 |
| 2.2.20 | 419880 | 122 | K1O6Os2 | | | | | |
| | -GM | -R | -T | -U | -V | -X | -Y | -Z |
| | 3+3 2 | 2+2 2 | 2+2 2 | 3+3 2 | 2+2 2 | 3+3 2 | 2+2 2 | 3+3 2 |
| | 2+2 2 | 3+3 2 | 3+3 2 | 2+2 2 | 3+3 2 | 2+2 2 | 3+3 2 | 2+2 2 |
| | 3+3 2 | 2+2 2 | 3+3 2 | 2+2 2 | 2+2 2 | 2+2 2 | 3+3 2 | 2+2 2 |
| | 2+2 2 | 3+3 2 | 2+2 2 | 3+3 2 | 2+2 2 | 3+3 2 | 2+2 2 | 3+3 2 |
| | 2+2 2 | 3+3 2 | 3+3 2 | 2+2 2 | 3+3 2 | 3+3 2 | 3+3 2 | 3+3 2 |
| | 3+3 2 | 2+2 2 | 2+2 2 | 3+3 2 | 3+3 2 | 2+2 2 | 2+2 2 | 2+2 2 |
| | 3+3 2 | 2+2 2 | 2+2 2 | 3+3 2 | 3+3 2 | 2+2 2 | 3+3 2 | 3+3 2 |
| | 2+2 2 | 3+3 2 | 3+3 2 | 2+2 2 | 2+2 2 | 3+3 2 | 2+2 2 | 2+2 2 |
| | 2+2 2 | 2+2 2 | 3+3 2 | 2+2 2 | 3+3 2 | 2+2 2 | 3+3 2 | 3+3 2 |
| | 3+3 2 | 3+3 2 | 2+2 2 | 3+3 2 | 2+2 2 | 3+3 2 | 3+3 2 | 2+2 2 |
| | 3+3 2 | 2+2 2 | 2+2 2 | 2+2 2 | 3+3 2 | 2+2 2 | 2+2 2 | 3+3 2 |
| | 2+2 2 | 3+3 2 | 3+3 2 | 3+3 2 | 2+2 2 | 3+3 2 | 2+2 2 | 2+2 2 |
| | 3+3 2 | 2+2 2 | 2+2 2 | 3+3 2 | 3+3 2 | 2+2 2 | 3+3 2 | 2+2 2 |
| | 2+2 2 | 3+3 2 | 3+3 2 | 2+2 2 | 2+2 2 | 3+3 2 | 2+2 2 | 3+3 2 |
| | 3+3 2 | 2+2 2 | 2+2 2 | 3+3 2 | 2+2 2 | 3+3 2 | 2+2 2 | 3+3 2 |
| | 2+2 2 | 3+3 2 | 3+3 2 | 2+2 2 | 3+3 2 | 2+2 2 | 3+3 2 | 2+2 2 |
| | 3+3 2 | 2+2 2 | 2+2 2 | 3+3 2 | 2+2 2 | 3+3 2 | 2+2 2 | 3+3 2 |
| | 2+2 2 | 3+3 2 | 3+3 2 | 2+2 2 | 3+3 2 | 2+2 2 | 3+3 2 | 2+2 2 |
| | 3+3 2 | 2+2 2 | 2+2 2 | 3+3 2 | 2+2 2 | 3+3 2 | 2+2 2 | 3+3 2 |
| | 2+2 2 | 3+3 2 | 3+3 2 | 2+2 2 | 3+3 2 | 2+2 2 | 3+3 2 | 2+2 2 |
| | 2+2 2 | 2+2 2 | 3+3 2 | 3+3 2 | 3+3 2 | 2+2 2 | 3+3 2 | 3+3 2 |
| | 3+3 2 | 3+3 2 | 2+2 2 | 2+2 2 | 2+2 2 | 3+3 2 | 2+2 2 | 3+3 2 |
| | 3+3 2 | 2+2 2 | 3+3 2 | 2+2 2 | 2+2 2 | 2+2 2 | 2+2 2 | 2+2 2 |
| | 2+2 2 | 3+3 2 | 2+2 2 | 3+3 2 | 3+3 2 | 3+3 2 | 3+3 2 | 2+2 2 |
| | 2+2 2 | 2+2 2 | 2+2 2 | 2+2 2 | 3+3 2 | 2+2 2 | 3+3 2 | 2+2 2 |
| | 3+3 2 | 3+3 2 | 3+3 2 | 3+3 2 | 2+2 2 | 3+3 2 | 2+2 2 | 3+3 2 |
| | 3+3 2 | 2+2 2 | 3+3 2 | 3+3 2 | 3+3 2 | 2+2 2 | 3+3 2 | 2+2 2 |
| | 2+2 2 | 2+2 2 | 2+2 2 | 3+3 2 | 2+2 2 | 3+3 2 | 2+2 2 | 3+3 2 |
| | 3+3 2 | 3+3 2 | 2+2 2 | 2+2 2 | 3+3 2 | 2+2 2 | 3+3 2 | 2+2 2 |
| | 2+2 2 | 3+3 2 | 3+3 2 | 3+3 2 | 2+2 2 | 3+3 2 | 2+2 2 | 3+3 2 |
| | 3+3 2 | 2+2 2 | 3+3 2 | 2+2 2 | 2+2 2 | 2+2 2 | 3+3 2 | 2+2 2 |
| | 2+2 2 | 3+3 2 | 2+2 2 | 2+2 2 | 3+3 2 | 3+3 2 | 2+2 2 | 3+3 2 |
| | 3+3 2 | 2+2 2 | 2+2 2 | 2+2 2 | 2+2 2 | 2+2 2 | 2+2 2 | 3+3 2 |
| | 3+3 2 | 3+3 2 | 3+3 2 | 3+3 2 | 3+3 2 | 3+3 2 | 3+3 2 | 3+3 2 |
| | 2+2 2 | 3+3 2 | 3+3 2 | 2+2 2 | 2+2 2 | 2+2 2 | 2+2 2 | 2+2 2 |
| | 2+2 2 | 2+2 2 | 2+2 2 | 2+2 2 | 2+2 2 | 3+3 2 | 3+3 2 | 3+3 2 |
| | 3+3 2 | 2+2 2 | 3+3 2 | 3+3 2 | 3+3 2 | 2+2 2 | 3+3 2 | 2+2 2 |
| | 2+2 2 | 3+3 2 | 2+2 2 | 2+2 2 | 3+3 2 | 3+3 2 | 3+3 2 | 2+2 2 |
| | 3+3 2 | 2+2 2 | 3+3 2 | 3+3 2 | 2+2 2 | 2+2 2 | 2+2 2 | 3+3 2 |
| | 3+3 2 | 3+3 2 | 2+2 2 | 3+3 2 | 3+3 2 | 3+3 2 | 2+2 2 | 2+2 2 |
| | 2+2 2 | 2+2 2 | 3+3 2 | 2+2 2 | 2+2 2 | 3+3 2 | 3+3 2 | 2+2 2 |
| | 2+2 2 | 3+3 2 | 2+2 2 | 2+2 2 | 3+3 2 | 2+2 2 | 3+3 2 | 3+3 2 |
| | 3+3 2 | 3+3 2 | 3+3 2 | 3+3 2 | 3+3 2 | 2+2 2 | 2+2 2 | 3+3 2 |
| | 3+3 2 | 2+2 2 | 3+3 2 | 2+2 2 | 2+2 2 | 3+3 2 | 2+2 2 | 2+2 2 |
| | 2+2 2 | 2+2 2 | 2+2 2 | 3+3 2 | 2+2 2 | 3+3 2 | 3+3 2 | 3+3 2 |
| | 3+3 2 | 3+3 2 | 3+3 2 | 3+3 2 | 2+2 2 | 2+2 2 | 2+2 2 | 2+2 2 |
| | 2+2 2 | 3+3 2 | 2+2 2 | 2+2 2 | 3+3 2 | 3+3 2 | 3+3 2 | 2+2 2 |
| | 3+3 2 | 2+2 2 | 2+2 2 | 3+3 2 | 2+2 2 | 2+2 2 | 2+2 2 | 3+3 2 |
| | 2+2 2 | 3+3 2 | 3+3 2 | 3+3 2 | 3+3 2 | 3+3 2 | 3+3 2 | 2+2 2 |
| | 3+3 2 | 2+2 2 | 3+3 2 | 2+2 2 | 3+3 2 | 2+2 2 | 2+2 2 | 3+3 2 |
| | 2+2 2 | 3+3 2 | 2+2 2 | 2+2 2 | 2+2 2 | 2+2 2 | 2+2 2 | 2+2 2 |
| | 3+3 2 | 2+2 2 | 2+2 2 | 2+2 2 | 3+3 2 | 3+3 2 | 3+3 2 | 2+2 2 |
| | 2+2 2 | 2+2 2 | 2+2 2 | 3+3 2 | 2+2 2 | 2+2 2 | 3+3 2 | 3+3 2 |
| | 2+2 2 | 3+3 2 | 3+3 2 | 3+3 2 | 3+3 2 | 3+3 2 | 2+2 2 | 3+3 2 |
| | 2+2 2 | 3+3 2 | 2+2 2 | 3+3 2 | 2+2 2 | 2+2 2 | 3+3 2 | 2+2 2 |
| | 3+3 2 | 2+2 2 | 3+3 2 | 2+2 2 | 3+3 2 | 3+3 2 | 2+2 2 | 3+3 2 |
| | 3+3 2 | 3+3 2 | 2+2 2 | 2+2 2 | 2+2 2 | 3+3 2 | 2+2 2 | 2+2 2 |
| | 2+2 2 | 2+2 2 | 3+3 2 | 3+3 2 | 3+3 2 | 2+2 2 | 3+3 2 | 2+2 2 |

Continued on next column

**TABLE XXI– continued**

| TQC | ICSD | NoE | SF | | | | | |
|---|---|---|---|---|---|---|---|---|
| | 3+3 2 | 2+2 2 | 2+2 2 | 3+3 2 | 3+3 2 | 3+3 2 | 2+2 2 | 3+3 2 |
| | 2+2 2 | 3+3 2 | 3+3 2 | 2+2 2 | 2+2 2 | 2+2 2 | 3+3 2 | 3+3 2 |
| | 3+3 2 | 3+3 2 | 3+3 2 | 2+2 2 | 3+3 2 | 3+3 2 | 2+2 2 | 2+2 2 |
| 2.2.21 | 39887 | 98 | H2Cu1O5Se1 | | | | | |
| | -GM | -R | -T | -U | -V | -X | -Y | -Z |
| | 3+3 2 | 2+2 2 | 2+2 2 | 3+3 2 | 2+2 2 | 2+2 2 | 2+2 2 | 3+3 2 |
| | 2+2 2 | 3+3 2 | 3+3 2 | 2+2 2 | 3+3 2 | 3+3 2 | 3+3 2 | 2+2 2 |
| | 3+3 2 | 2+2 2 | 2+2 2 | 2+2 2 | 3+3 2 | 3+3 2 | 3+3 2 | 2+2 2 |
| | 2+2 2 | 3+3 2 | 3+3 2 | 3+3 2 | 2+2 2 | 2+2 2 | 2+2 2 | 3+3 2 |
| | 3+3 2 | 3+3 2 | 2+2 2 | 3+3 2 | 2+2 2 | 2+2 2 | 3+3 2 | 2+2 2 |
| | 2+2 2 | 2+2 2 | 3+3 2 | 2+2 2 | 3+3 2 | 3+3 2 | 2+2 2 | 3+3 2 |
| | 3+3 2 | 2+2 2 | 2+2 2 | 2+2 2 | 3+3 2 | 3+3 2 | 2+2 2 | 3+3 2 |
| | 2+2 2 | 3+3 2 | 3+3 2 | 3+3 2 | 2+2 2 | 2+2 2 | 3+3 2 | 2+2 2 |
| | 2+2 2 | 3+3 2 | 2+2 2 | 3+3 2 | 2+2 2 | 3+3 2 | 2+2 2 | 2+2 2 |
| | 3+3 2 | 2+2 2 | 3+3 2 | 2+2 2 | 3+3 2 | 2+2 2 | 3+3 2 | 3+3 2 |
| | 3+3 2 | 3+3 2 | 3+3 2 | 2+2 2 | 2+2 2 | 2+2 2 | 3+3 2 | 3+3 2 |
| | 3+3 2 | 2+2 2 | 2+2 2 | 3+3 2 | 3+3 2 | 3+3 2 | 2+2 2 | 2+2 2 |
| | 2+2 2 | 3+3 2 | 2+2 2 | 3+3 2 | 2+2 2 | 3+3 2 | 3+3 2 | 2+2 2 |
| | 2+2 2 | 2+2 2 | 3+3 2 | 2+2 2 | 3+3 2 | 2+2 2 | 2+2 2 | 3+3 2 |
| | 3+3 2 | 3+3 2 | 2+2 2 | 3+3 2 | 2+2 2 | 2+2 2 | 2+2 2 | 3+3 2 |
| | 3+3 2 | 2+2 2 | 2+2 2 | 3+3 2 | 2+2 2 | 3+3 2 | 3+3 2 | 2+2 2 |
| | 2+2 2 | 2+2 2 | 3+3 2 | 2+2 2 | 3+3 2 | 2+2 2 | 2+2 2 | 2+2 2 |
| | 2+2 2 | 3+3 2 | 2+2 2 | 2+2 2 | 3+3 2 | 3+3 2 | 3+3 2 | 3+3 2 |
| | 3+3 2 | 3+3 2 | 3+3 2 | 2+2 2 | 2+2 2 | 3+3 2 | 3+3 2 | 2+2 2 |
| | 3+3 2 | 2+2 2 | 2+2 2 | 3+3 2 | 2+2 2 | 3+3 2 | 2+2 2 | 3+3 2 |
| | 2+2 2 | 2+2 2 | 3+3 2 | 3+3 2 | 3+3 2 | 2+2 2 | 2+2 2 | 3+3 2 |
| | 2+2 2 | 3+3 2 | 3+3 2 | 2+2 2 | 3+3 2 | 2+2 2 | 3+3 2 | 2+2 2 |
| | 3+3 2 | 3+3 2 | 2+2 2 | 3+3 2 | 2+2 2 | 3+3 2 | 3+3 2 | 3+3 2 |
| | 3+3 2 | 2+2 2 | 2+2 2 | 3+3 2 | 2+2 2 | 2+2 2 | 2+2 2 | 2+2 2 |
| | 2+2 2 | 2+2 2 | 3+3 2 | 2+2 2 | 3+3 2 | 2+2 2 | 3+3 2 | 2+2 2 |
| | 3+3 2 | 3+3 2 | 2+2 2 | 3+3 2 | 2+2 2 | 3+3 2 | 3+3 2 | 2+2 2 |
| | 2+2 2 | 3+3 2 | 2+2 2 | 2+2 2 | 3+3 2 | 2+2 2 | 2+2 2 | 3+3 2 |
| | 3+3 2 | 2+2 2 | 3+3 2 | 2+2 2 | 2+2 2 | 3+3 2 | 2+2 2 | 3+3 2 |
| | 3+3 2 | 3+3 2 | 2+2 2 | 3+3 2 | 3+3 2 | 2+2 2 | 3+3 2 | 2+2 2 |
| | 2+2 2 | 2+2 2 | 3+3 2 | 3+3 2 | 3+3 2 | 3+3 2 | 2+2 2 | 3+3 2 |
| | 3+3 2 | 2+2 2 | 2+2 2 | 2+2 2 | 2+2 2 | 2+2 2 | 3+3 2 | 3+3 2 |
| | 2+2 2 | 3+3 2 | 3+3 2 | 3+3 2 | 2+2 2 | 3+3 2 | 2+2 2 | 2+2 2 |
| | 3+3 2 | 2+2 2 | 3+3 2 | 3+3 2 | 2+2 2 | 3+3 2 | 3+3 2 | 2+2 2 |
| | 2+2 2 | 3+3 2 | 3+3 2 | 2+2 2 | 3+3 2 | 2+2 2 | 2+2 2 | 3+3 2 |
| | 3+3 2 | 3+3 2 | 2+2 2 | 2+2 2 | 3+3 2 | 2+2 2 | 3+3 2 | 2+2 2 |
| | 2+2 2 | 2+2 2 | 2+2 2 | 3+3 2 | 2+2 2 | 3+3 2 | 2+2 2 | 3+3 2 |
| | 2+2 2 | 2+2 2 | 2+2 2 | 2+2 2 | 2+2 2 | 2+2 2 | 3+3 2 | 2+2 2 |
| | 3+3 2 | 3+3 2 | 3+3 2 | 2+2 2 | 3+3 2 | 3+3 2 | 2+2 2 | 3+3 2 |
| | 2+2 2 | 2+2 2 | 3+3 2 | 3+3 2 | 3+3 2 | 3+3 2 | 3+3 2 | 2+2 2 |
| | 3+3 2 | 3+3 2 | 2+2 2 | 3+3 2 | 2+2 2 | 2+2 2 | 2+2 2 | 3+3 2 |
| | 3+3 2 | 3+3 2 | 2+2 2 | 3+3 2 | 3+3 2 | 3+3 2 | 3+3 2 | 2+2 2 |
| | 3+3 2 | 2+2 2 | 3+3 2 | 3+3 2 | 2+2 2 | 2+2 2 | 2+2 2 | 3+3 2 |
| | 2+2 2 | 2+2 2 | 2+2 2 | 3+3 2 | 2+2 2 | 2+2 2 | 3+3 2 | 2+2 2 |
| | 3+3 2 | 3+3 2 | 2+2 2 | 2+2 2 | 2+2 2 | 3+3 2 | 2+2 2 | 3+3 2 |
| | 3+3 2 | 3+3 2 | 2+2 2 | 3+3 2 | 2+2 2 | 2+2 2 | 3+3 2 | 2+2 2 |
| | 3+3 2 | 2+2 2 | 2+2 2 | 3+3 2 | 2+2 2 | 3+3 2 | 2+2 2 | 3+3 2 |
| | 3+3 2 | 3+3 2 | 2+2 2 | 3+3 2 | 2+2 2 | 2+2 2 | 3+3 2 | 2+2 2 |
| | 3+3 2 | 2+2 2 | 2+2 2 | 3+3 2 | 2+2 2 | 3+3 2 | 2+2 2 | 3+3 2 |
| | 3+3 2 | 3+3 2 | 2+2 2 | 3+3 2 | 2+2 2 | 2+2 2 | 2+2 2 | 3+3 2 |
| 2.2.22 | 166101 | 98 | H2Cu1O5S1 | | | | | |
| | -GM | -R | -T | -U | -V | -X | -Y | -Z |
| | 3+3 2 | 2+2 2 | 3+3 2 | 2+2 2 | 2+2 2 | 2+2 2 | 3+3 2 | 3+3 2 |
| | 2+2 2 | 3+3 2 | 2+2 2 | 3+3 2 | 3+3 2 | 3+3 2 | 2+2 2 | 2+2 2 |
| | 3+3 2 | 2+2 2 | 3+3 2 | 3+3 2 | 2+2 2 | 3+3 2 | 2+2 2 | 2+2 2 |
| | 2+2 2 | 3+3 2 | 2+2 2 | 2+2 2 | 3+3 2 | 2+2 2 | 3+3 2 | 3+3 2 |
| | 2+2 2 | 3+3 2 | 2+2 2 | 3+3 2 | 3+3 2 | 3+3 2 | 2+2 2 | 2+2 2 |

Continued on next column

**TABLE XXI– continued**

| TQC | ICSD | NoE | SF | | | | | |
|---|---|---|---|---|---|---|---|---|
| | 3+3 2 | 2+2 2 | 3+3 2 | 2+2 2 | 2+2 2 | 2+2 2 | 3+3 2 | 3+3 2 |
| | 3+3 2 | 2+2 2 | 2+2 2 | 3+3 2 | 3+3 2 | 2+2 2 | 3+3 2 | 2+2 2 |
| | 2+2 2 | 3+3 2 | 3+3 2 | 2+2 2 | 2+2 2 | 3+3 2 | 2+2 2 | 3+3 2 |
| | 3+3 2 | 2+2 2 | 3+3 2 | 2+2 2 | 2+2 2 | 2+2 2 | 3+3 2 | 3+3 2 |
| | 2+2 2 | 3+3 2 | 2+2 2 | 3+3 2 | 3+3 2 | 3+3 2 | 2+2 2 | 2+2 2 |
| | 3+3 2 | 3+3 2 | 3+3 2 | 2+2 2 | 3+3 2 | 2+2 2 | 3+3 2 | 3+3 2 |
| | 2+2 2 | 2+2 2 | 2+2 2 | 3+3 2 | 2+2 2 | 3+3 2 | 2+2 2 | 2+2 2 |
| | 3+3 2 | 3+3 2 | 3+3 2 | 3+3 2 | 2+2 2 | 3+3 2 | 2+2 2 | 2+2 2 |
| | 2+2 2 | 2+2 2 | 2+2 2 | 2+2 2 | 3+3 2 | 2+2 2 | 3+3 2 | 3+3 2 |
| | 3+3 2 | 3+3 2 | 3+3 2 | 3+3 2 | 3+3 2 | 3+3 2 | 2+2 2 | 3+3 2 |
| | 2+2 2 | 2+2 2 | 2+2 2 | 2+2 2 | 2+2 2 | 2+2 2 | 3+3 2 | 2+2 2 |
| | 3+3 2 | 2+2 2 | 3+3 2 | 3+3 2 | 3+3 2 | 2+2 2 | 3+3 2 | 3+3 2 |
| | 2+2 2 | 3+3 2 | 3+3 2 | 2+2 2 | 2+2 2 | 3+3 2 | 2+2 2 | 2+2 2 |
| | 3+3 2 | 2+2 2 | 2+2 2 | 2+2 2 | 3+3 2 | 2+2 2 | 2+2 2 | 3+3 2 |
| | 2+2 2 | 3+3 2 | 2+2 2 | 3+3 2 | 2+2 2 | 3+3 2 | 3+3 2 | 2+2 2 |
| | 3+3 2 | 2+2 2 | 3+3 2 | 3+3 2 | 3+3 2 | 3+3 2 | 3+3 2 | 2+2 2 |
| | 2+2 2 | 3+3 2 | 2+2 2 | 2+2 2 | 2+2 2 | 2+2 2 | 2+2 2 | 3+3 2 |
| | 3+3 2 | 3+3 2 | 3+3 2 | 3+3 2 | 3+3 2 | 2+2 2 | 2+2 2 | 2+2 2 |
| | 3+3 2 | 2+2 2 | 3+3 2 | 3+3 2 | 2+2 2 | 3+3 2 | 3+3 2 | 3+3 2 |
| | 2+2 2 | 2+2 2 | 2+2 2 | 2+2 2 | 3+3 2 | 2+2 2 | 2+2 2 | 3+3 2 |
| | 2+2 2 | 3+3 2 | 2+2 2 | 2+2 2 | 3+3 2 | 3+3 2 | 3+3 2 | 2+2 2 |
| | 3+3 2 | 2+2 2 | 3+3 2 | 3+3 2 | 2+2 2 | 2+2 2 | 2+2 2 | 2+2 2 |
| | 3+3 2 | 3+3 2 | 3+3 2 | 3+3 2 | 2+2 2 | 3+3 2 | 3+3 2 | 3+3 2 |
| | 2+2 2 | 3+3 2 | 2+2 2 | 2+2 2 | 3+3 2 | 2+2 2 | 3+3 2 | 2+2 2 |
| | 3+3 2 | 3+3 2 | 2+2 2 | 2+2 2 | 3+3 2 | 3+3 2 | 3+3 2 | 3+3 2 |
| | 3+3 2 | 2+2 2 | 3+3 2 | 3+3 2 | 3+3 2 | 3+3 2 | 2+2 2 | 2+2 2 |
| | 2+2 2 | 2+2 2 | 2+2 2 | 3+3 2 | 2+2 2 | 2+2 2 | 3+3 2 | 3+3 2 |
| | 2+2 2 | 2+2 2 | 3+3 2 | 2+2 2 | 2+2 2 | 3+3 2 | 2+2 2 | 2+2 2 |
| | 2+2 2 | 3+3 2 | 2+2 2 | 2+2 2 | 2+2 2 | 2+2 2 | 2+2 2 | 3+3 2 |
| | 3+3 2 | 2+2 2 | 3+3 2 | 3+3 2 | 3+3 2 | 3+3 2 | 3+3 2 | 3+3 2 |
| | 3+3 2 | 3+3 2 | 3+3 2 | 3+3 2 | 3+3 2 | 2+2 2 | 2+2 2 | 2+2 2 |
| | 2+2 2 | 3+3 2 | 3+3 2 | 3+3 2 | 3+3 2 | 3+3 2 | 2+2 2 | 3+3 2 |
| | 3+3 2 | 2+2 2 | 2+2 2 | 2+2 2 | 2+2 2 | 2+2 2 | 3+3 2 | 2+2 2 |
| | 3+3 2 | 3+3 2 | 3+3 2 | 3+3 2 | 2+2 2 | 3+3 2 | 3+3 2 | 3+3 2 |
| | 3+3 2 | 2+2 2 | 2+2 2 | 2+2 2 | 3+3 2 | 2+2 2 | 2+2 2 | 2+2 2 |
| | 2+2 2 | 3+3 2 | 3+3 2 | 3+3 2 | 3+3 2 | 3+3 2 | 3+3 2 | 3+3 2 |
| | 3+3 2 | 3+3 2 | 2+2 2 | 2+2 2 | 2+2 2 | 3+3 2 | 2+2 2 | 3+3 2 |
| | 2+2 2 | 2+2 2 | 3+3 2 | 3+3 2 | 3+3 2 | 2+2 2 | 3+3 2 | 2+2 2 |
| | 3+3 2 | 2+2 2 | 3+3 2 | 3+3 2 | 3+3 2 | 2+2 2 | 2+2 2 | 2+2 2 |
| | 3+3 2 | 2+2 2 | 3+3 2 | 3+3 2 | 3+3 2 | 2+2 2 | 3+3 2 | 2+2 2 |
| | 3+3 2 | 3+3 2 | 3+3 2 | 3+3 2 | 3+3 2 | 3+3 2 | 2+2 2 | 3+3 2 |
| | 3+3 2 | 3+3 2 | 3+3 2 | 3+3 2 | 3+3 2 | 3+3 2 | 3+3 2 | 3+3 2 |
| | 3+3 2 | 2+2 2 | 3+3 2 | 3+3 2 | 3+3 2 | 3+3 2 | 2+2 2 | 2+2 2 |
| | 3+3 2 | 3+3 2 | 3+3 2 | 3+3 2 | 3+3 2 | 2+2 2 | 2+2 2 | 2+2 2 |
| 2.2.23 | 167608 | 90 | H1Fe1Li1O5P1 | | | | | |
| | -GM | -R | -T | -U | -V | -X | -Y | -Z |
| | 3+3 2 | 3+3 2 | 3+3 2 | 2+2 2 | 2+2 2 | 2+2 2 | 2+2 2 | 3+3 2 |
| | 2+2 2 | 2+2 2 | 2+2 2 | 3+3 2 | 3+3 2 | 3+3 2 | 3+3 2 | 2+2 2 |
| | 3+3 2 | 3+3 2 | 2+2 2 | 3+3 2 | 3+3 2 | 3+3 2 | 3+3 2 | 2+2 2 |
| | 2+2 2 | 2+2 2 | 3+3 2 | 2+2 2 | 2+2 2 | 2+2 2 | 2+2 2 | 3+3 2 |
| | 3+3 2 | 3+3 2 | 2+2 2 | 3+3 2 | 2+2 2 | 3+3 2 | 2+2 2 | 2+2 2 |
| | 2+2 2 | 2+2 2 | 3+3 2 | 2+2 2 | 3+3 2 | 2+2 2 | 3+3 2 | 3+3 2 |
| | 3+3 2 | 2+2 2 | 3+3 2 | 2+2 2 | 3+3 2 | 3+3 2 | 3+3 2 | 3+3 2 |
| | 2+2 2 | 3+3 2 | 3+3 2 | 3+3 2 | 2+2 2 | 2+2 2 | 2+2 2 | 2+2 2 |
| | 2+2 2 | 2+2 2 | 2+2 2 | 2+2 2 | 2+2 2 | 2+2 2 | 3+3 2 | 3+3 2 |
| | 3+3 2 | 3+3 2 | 2+2 2 | 3+3 2 | 3+3 2 | 3+3 2 | 2+2 2 | 2+2 2 |
| | 3+3 2 | 3+3 2 | 3+3 2 | 2+2 2 | 3+3 2 | 3+3 2 | 3+3 2 | 3+3 2 |
| | 2+2 2 | 2+2 2 | 2+2 2 | 3+3 2 | 2+2 2 | 2+2 2 | 2+2 2 | 2+2 2 |
| | 3+3 2 | 3+3 2 | 2+2 2 | 3+3 2 | 3+3 2 | 3+3 2 | 3+3 2 | 2+2 2 |
| | 3+3 2 | 2+2 2 | 2+2 2 | 2+2 2 | 3+3 2 | 3+3 2 | 3+3 2 | 3+3 2 |
| | 3+3 2 | 2+2 2 | 3+3 2 | 3+3 2 | 2+2 2 | 2+2 2 | 3+3 2 | 3+3 2 |

Continued on next column

**TABLE XXI– continued**

| TQC | ICSD | NoE | SF | | | | | |
|---|---|---|---|---|---|---|---|---|
| | 2+2 2 | 3+3 2 | 3+3 2 | 2+2 2 | 2+2 2 | 3+3 2 | 2+2 2 | 2+2 2 |
| | 2+2 2 | 2+2 2 | 2+2 2 | 3+3 2 | 3+3 2 | 2+2 2 | 2+2 2 | 3+3 2 |
| | 3+3 2 | 3+3 2 | 3+3 2 | 2+2 2 | 2+2 2 | 3+3 2 | 3+3 2 | 2+2 2 |
| | 2+2 2 | 2+2 2 | 3+3 2 | 2+2 2 | 3+3 2 | 2+2 2 | 2+2 2 | 3+3 2 |
| | 2+2 2 | 3+3 2 | 2+2 2 | 3+3 2 | 2+2 2 | 2+2 2 | 2+2 2 | 2+2 2 |
| | 3+3 2 | 2+2 2 | 3+3 2 | 3+3 2 | 3+3 2 | 3+3 2 | 3+3 2 | 2+2 2 |
| | 3+3 2 | 3+3 2 | 2+2 2 | 3+3 2 | 3+3 2 | 3+3 2 | 3+3 2 | 3+3 2 |
| | 3+3 2 | 2+2 2 | 2+2 2 | 2+2 2 | 3+3 2 | 3+3 2 | 2+2 2 | 3+3 2 |
| | 2+2 2 | 3+3 2 | 3+3 2 | 2+2 2 | 2+2 2 | 2+2 2 | 3+3 2 | 2+2 2 |
| | 2+2 2 | 3+3 2 | 3+3 2 | 2+2 2 | 2+2 2 | 2+2 2 | 2+2 2 | 3+3 2 |
| | 3+3 2 | 2+2 2 | 2+2 2 | 3+3 2 | 2+2 2 | 3+3 2 | 3+3 2 | 2+2 2 |
| | 3+3 2 | 2+2 2 | 3+3 2 | 2+2 2 | 3+3 2 | 2+2 2 | 2+2 2 | 3+3 2 |
| | 2+2 2 | 3+3 2 | 2+2 2 | 2+2 2 | 3+3 2 | 2+2 2 | 3+3 2 | 2+2 2 |
| | 3+3 2 | 3+3 2 | 3+3 2 | 3+3 2 | 2+2 2 | 3+3 2 | 2+2 2 | 2+2 2 |
| | 2+2 2 | 2+2 2 | 3+3 2 | 3+3 2 | 3+3 2 | 2+2 2 | 2+2 2 | 3+3 2 |
| | 3+3 2 | 3+3 2 | 2+2 2 | 3+3 2 | 2+2 2 | 3+3 2 | 3+3 2 | 3+3 2 |
| | 2+2 2 | 2+2 2 | 2+2 2 | 2+2 2 | 3+3 2 | 3+3 2 | 2+2 2 | 2+2 2 |
| | 3+3 2 | 3+3 2 | 2+2 2 | 3+3 2 | 2+2 2 | 3+3 2 | 3+3 2 | 3+3 2 |
| | 2+2 2 | 2+2 2 | 3+3 2 | 2+2 2 | 3+3 2 | 2+2 2 | 3+3 2 | 2+2 2 |
| | 2+2 2 | 3+3 2 | 2+2 2 | 3+3 2 | 2+2 2 | 3+3 2 | 2+2 2 | 2+2 2 |
| | 2+2 2 | 2+2 2 | 2+2 2 | 2+2 2 | 3+3 2 | 2+2 2 | 3+3 2 | 3+3 2 |
| | 3+3 2 | 3+3 2 | 3+3 2 | 3+3 2 | 3+3 2 | 2+2 2 | 3+3 2 | 3+3 2 |
| | 3+3 2 | 2+2 2 | 3+3 2 | 2+2 2 | 2+2 2 | 2+2 2 | 2+2 2 | 2+2 2 |
| | 2+2 2 | 2+2 2 | 2+2 2 | 3+3 2 | 2+2 2 | 3+3 2 | 2+2 2 | 3+3 2 |
| | 2+2 2 | 3+3 2 | 3+3 2 | 2+2 2 | 2+2 2 | 2+2 2 | 2+2 2 | 2+2 2 |
| | 3+3 2 | 2+2 2 | 3+3 2 | 2+2 2 | 3+3 2 | 3+3 2 | 3+3 2 | 2+2 2 |
| | 3+3 2 | 3+3 2 | 2+2 2 | 3+3 2 | 3+3 2 | 3+3 2 | 3+3 2 | 3+3 2 |
| | 3+3 2 | 3+3 2 | 3+3 2 | 2+2 2 | 3+3 2 | 3+3 2 | 3+3 2 | 3+3 2 |
| | 3+3 2 | 2+2 2 | 2+2 2 | 3+3 2 | 3+3 2 | 3+3 2 | 3+3 2 | 2+2 2 |
| | 3+3 2 | 3+3 2 | 3+3 2 | 2+2 2 | 3+3 2 | 3+3 2 | 3+3 2 | 2+2 2 |
| 2.2.24 | 62742 | 154 | H1Cu2K3O14P4 | | | | | |
| | -GM | -R | -T | -U | -V | -X | -Y | -Z |
| | 3+3 2 | 2+2 2 | 3+3 2 | 3+3 2 | 2+2 2 | 3+3 2 | 3+3 2 | 3+3 2 |
| | 2+2 2 | 3+3 2 | 2+2 2 | 2+2 2 | 3+3 2 | 2+2 2 | 2+2 2 | 2+2 2 |
| | 3+3 2 | 2+2 2 | 2+2 2 | 2+2 2 | 3+3 2 | 3+3 2 | 3+3 2 | 2+2 2 |
| | 3+3 2 | 2+2 2 | 3+3 2 | 3+3 2 | 3+3 2 | 2+2 2 | 2+2 2 | 2+2 2 |
| | 2+2 2 | 3+3 2 | 2+2 2 | 2+2 2 | 2+2 2 | 3+3 2 | 3+3 2 | 3+3 2 |
| | 3+3 2 | 2+2 2 | 3+3 2 | 3+3 2 | 3+3 2 | 2+2 2 | 2+2 2 | 2+2 2 |
| | 2+2 2 | 3+3 2 | 2+2 2 | 2+2 2 | 2+2 2 | 3+3 2 | 3+3 2 | 3+3 2 |
| | 3+3 2 | 2+2 2 | 3+3 2 | 3+3 2 | 3+3 2 | 2+2 2 | 2+2 2 | 2+2 2 |
| | 2+2 2 | 3+3 2 | 2+2 2 | 2+2 2 | 2+2 2 | 3+3 2 | 3+3 2 | 3+3 2 |
| | 3+3 2 | 2+2 2 | 3+3 2 | 3+3 2 | 3+3 2 | 2+2 2 | 2+2 2 | 2+2 2 |
| | 2+2 2 | 3+3 2 | 3+3 2 | 3+3 2 | 2+2 2 | 2+2 2 | 3+3 2 | 2+2 2 |
| | 3+3 2 | 2+2 2 | 2+2 2 | 2+2 2 | 3+3 2 | 3+3 2 | 2+2 2 | 3+3 2 |
| | 2+2 2 | 3+3 2 | 2+2 2 | 3+3 2 | 2+2 2 | 3+3 2 | 3+3 2 | 3+3 2 |
| | 2+2 2 | 2+2 2 | 3+3 2 | 2+2 2 | 3+3 2 | 2+2 2 | 3+3 2 | 2+2 2 |
| | 3+3 2 | 3+3 2 | 2+2 2 | 2+2 2 | 2+2 2 | 3+3 2 | 2+2 2 | 3+3 2 |
| | 2+2 2 | 2+2 2 | 3+3 2 | 3+3 2 | 2+2 2 | 3+3 2 | 2+2 2 | 2+2 2 |
| | 3+3 2 | 3+3 2 | 2+2 2 | 2+2 2 | 3+3 2 | 2+2 2 | 3+3 2 | 3+3 2 |
| | 3+3 2 | 2+2 2 | 3+3 2 | 3+3 2 | 2+2 2 | 3+3 2 | 3+3 2 | 3+3 2 |
| | 2+2 2 | 3+3 2 | 2+2 2 | 2+2 2 | 3+3 2 | 2+2 2 | 2+2 2 | 2+2 2 |
| | 3+3 2 | 2+2 2 | 3+3 2 | 3+3 2 | 2+2 2 | 3+3 2 | 3+3 2 | 3+3 2 |
| | 2+2 2 | 3+3 2 | 3+3 2 | 3+3 2 | 2+2 2 | 2+2 2 | 2+2 2 | 3+3 2 |
| | 3+3 2 | 3+3 2 | 2+2 2 | 2+2 2 | 3+3 2 | 2+2 2 | 2+2 2 | 3+3 2 |
| | 2+2 2 | 3+3 2 | 3+3 2 | 2+2 2 | 2+2 2 | 2+2 2 | 3+3 2 | 2+2 2 |
| | 2+2 2 | 2+2 2 | 2+2 2 | 3+3 2 | 3+3 2 | 3+3 2 | 2+2 2 | 2+2 2 |
| | 2+2 2 | 3+3 2 | 3+3 2 | 3+3 2 | 2+2 2 | 2+2 2 | 2+2 2 | 3+3 2 |
| | 2+2 2 | 3+3 2 | 3+3 2 | 3+3 2 | 2+2 2 | 2+2 2 | 2+2 2 | 3+3 2 |
| | 3+3 2 | 2+2 2 | 3+3 2 | 3+3 2 | 3+3 2 | 2+2 2 | 2+2 2 | 2+2 2 |
| | 2+2 2 | 3+3 2 | 2+2 2 | 2+2 2 | 2+2 2 | 3+3 2 | 3+3 2 | 3+3 2 |
| | 3+3 2 | 2+2 2 | 3+3 2 | 3+3 2 | 3+3 2 | 2+2 2 | 2+2 2 | 2+2 2 |

Continued on next column

**TABLE XXI– continued**

| TQC | ICSD | NoE | SF | | | | | |
|---|---|---|---|---|---|---|---|---|
| | 2+2 2 | 3+3 2 | 2+2 2 | 2+2 2 | 2+2 2 | 3+3 2 | 3+3 2 | 3+3 2 |
| | 3+3 2 | 2+2 2 | 3+3 2 | 3+3 2 | 3+3 2 | 2+2 2 | 2+2 2 | 2+2 2 |
| | 2+2 2 | 3+3 2 | 2+2 2 | 2+2 2 | 2+2 2 | 3+3 2 | 3+3 2 | 3+3 2 |
| | 3+3 2 | 2+2 2 | 3+3 2 | 3+3 2 | 3+3 2 | 2+2 2 | 2+2 2 | 2+2 2 |
| | 2+2 2 | 3+3 2 | 2+2 2 | 2+2 2 | 2+2 2 | 3+3 2 | 3+3 2 | 3+3 2 |
| | 3+3 2 | 2+2 2 | 3+3 2 | 3+3 2 | 3+3 2 | 2+2 2 | 2+2 2 | 2+2 2 |
| | 2+2 2 | 3+3 2 | 2+2 2 | 2+2 2 | 2+2 2 | 3+3 2 | 3+3 2 | 3+3 2 |
| | 3+3 2 | 2+2 2 | 3+3 2 | 3+3 2 | 3+3 2 | 2+2 2 | 2+2 2 | 2+2 2 |
| | 2+2 2 | 3+3 2 | 2+2 2 | 2+2 2 | 2+2 2 | 3+3 2 | 3+3 2 | 3+3 2 |
| | 3+3 2 | 2+2 2 | 3+3 2 | 3+3 2 | 3+3 2 | 2+2 2 | 2+2 2 | 2+2 2 |
| | 2+2 2 | 3+3 2 | 2+2 2 | 2+2 2 | 2+2 2 | 3+3 2 | 2+2 2 | 3+3 2 |
| | 3+3 2 | 2+2 2 | 2+2 2 | 3+3 2 | 3+3 2 | 2+2 2 | 3+3 2 | 2+2 2 |
| | 2+2 2 | 3+3 2 | 3+3 2 | 2+2 2 | 2+2 2 | 3+3 2 | 3+3 2 | 3+3 2 |
| | 2+2 2 | 3+3 2 | 3+3 2 | 2+2 2 | 3+3 2 | 2+2 2 | 3+3 2 | 2+2 2 |
| | 3+3 2 | 2+2 2 | 2+2 2 | 3+3 2 | 2+2 2 | 3+3 2 | 2+2 2 | 3+3 2 |
| | 3+3 2 | 2+2 2 | 3+3 2 | 3+3 2 | 3+3 2 | 2+2 2 | 2+2 2 | 2+2 2 |
| | 3+3 2 | 3+3 2 | 2+2 2 | 2+2 2 | 2+2 2 | 3+3 2 | 3+3 2 | 3+3 2 |
| | 2+2 2 | 3+3 2 | 3+3 2 | 3+3 2 | 3+3 2 | 2+2 2 | 2+2 2 | 2+2 2 |
| | 3+3 2 | 2+2 2 | 2+2 2 | 2+2 2 | 2+2 2 | 3+3 2 | 3+3 2 | 3+3 2 |
| | 2+2 2 | 3+3 2 | 3+3 2 | 2+2 2 | 2+2 2 | 2+2 2 | 2+2 2 | 2+2 2 |
| | 2+2 2 | 2+2 2 | 2+2 2 | 3+3 2 | 3+3 2 | 3+3 2 | 3+3 2 | 3+3 2 |
| | 3+3 2 | 2+2 2 | 3+3 2 | 2+2 2 | 3+3 2 | 3+3 2 | 2+2 2 | 2+2 2 |
| | 2+2 2 | 3+3 2 | 2+2 2 | 2+2 2 | 2+2 2 | 2+2 2 | 3+3 2 | 3+3 2 |
| | 3+3 2 | 3+3 2 | 3+3 2 | 3+3 2 | 3+3 2 | 3+3 2 | 2+2 2 | 2+2 2 |
| | 3+3 2 | 2+2 2 | 2+2 2 | 3+3 2 | 2+2 2 | 2+2 2 | 3+3 2 | 3+3 2 |
| | 2+2 2 | 2+2 2 | 2+2 2 | 3+3 2 | 3+3 2 | 2+2 2 | 2+2 2 | 2+2 2 |
| | 2+2 2 | 3+3 2 | 3+3 2 | 2+2 2 | 2+2 2 | 3+3 2 | 3+3 2 | 3+3 2 |
| | 3+3 2 | 2+2 2 | 2+2 2 | 3+3 2 | 2+2 2 | 2+2 2 | 3+3 2 | 3+3 2 |
| | 2+2 2 | 3+3 2 | 3+3 2 | 2+2 2 | 3+3 2 | 3+3 2 | 2+2 2 | 2+2 2 |
| | 2+2 2 | 2+2 2 | 2+2 2 | 2+2 2 | 3+3 2 | 3+3 2 | 3+3 2 | 2+2 2 |
| | 3+3 2 | 3+3 2 | 3+3 2 | 3+3 2 | 2+2 2 | 2+2 2 | 2+2 2 | 3+3 2 |
| | 3+3 2 | 2+2 2 | 3+3 2 | 3+3 2 | 3+3 2 | 2+2 2 | 3+3 2 | 2+2 2 |
| | 2+2 2 | 3+3 2 | 2+2 2 | 2+2 2 | 2+2 2 | 3+3 2 | 2+2 2 | 3+3 2 |
| | 2+2 2 | 3+3 2 | 3+3 2 | 3+3 2 | 2+2 2 | 3+3 2 | 2+2 2 | 2+2 2 |
| | 3+3 2 | 2+2 2 | 2+2 2 | 2+2 2 | 3+3 2 | 2+2 2 | 3+3 2 | 3+3 2 |
| | 3+3 2 | 2+2 2 | 3+3 2 | 3+3 2 | 3+3 2 | 3+3 2 | 2+2 2 | 2+2 2 |
| | 2+2 2 | 3+3 2 | 2+2 2 | 2+2 2 | 2+2 2 | 2+2 2 | 3+3 2 | 3+3 2 |
| | 3+3 2 | 2+2 2 | 3+3 2 | 3+3 2 | 3+3 2 | 3+3 2 | 2+2 2 | 2+2 2 |
| | 2+2 2 | 3+3 2 | 2+2 2 | 2+2 2 | 2+2 2 | 2+2 2 | 3+3 2 | 3+3 2 |
| | 3+3 2 | 2+2 2 | 3+3 2 | 3+3 2 | 3+3 2 | 3+3 2 | 2+2 2 | 2+2 2 |
| | 2+2 2 | 3+3 2 | 2+2 2 | 2+2 2 | 2+2 2 | 2+2 2 | 3+3 2 | 3+3 2 |
| | 2+2 2 | 2+2 2 | 3+3 2 | 2+2 2 | 3+3 2 | 2+2 2 | 3+3 2 | 2+2 2 |
| | 3+3 2 | 3+3 2 | 2+2 2 | 3+3 2 | 2+2 2 | 3+3 2 | 2+2 2 | 3+3 2 |
| | 3+3 2 | 2+2 2 | 3+3 2 | 3+3 2 | 3+3 2 | 2+2 2 | 2+2 2 | 2+2 2 |
| | 2+2 2 | 3+3 2 | 2+2 2 | 2+2 2 | 2+2 2 | 3+3 2 | 3+3 2 | 3+3 2 |
| | 3+3 2 | 3+3 2 | 3+3 2 | 3+3 2 | 3+3 2 | 2+2 2 | 2+2 2 | 2+2 2 |
| | 2+2 2 | 2+2 2 | 2+2 2 | 2+2 2 | 2+2 2 | 3+3 2 | 3+3 2 | 3+3 2 |
| | 2+2 2 | 2+2 2 | 3+3 2 | 3+3 2 | 3+3 2 | 3+3 2 | 3+3 2 | 2+2 2 |
| 2.2.25 | 416823 | 114 | H1Co1O5P1Zn1 | | | | | |
| | -GM | -R | -T | -U | -V | -X | -Y | -Z |
| | 3+3 2 | 2+2 2 | 3+3 2 | 3+3 2 | 3+3 2 | 2+2 2 | 2+2 2 | 2+2 2 |
| | 2+2 2 | 3+3 2 | 2+2 2 | 2+2 2 | 2+2 2 | 3+3 2 | 3+3 2 | 3+3 2 |
| | 3+3 2 | 3+3 2 | 2+2 2 | 3+3 2 | 2+2 2 | 2+2 2 | 3+3 2 | 3+3 2 |
| | 2+2 2 | 2+2 2 | 3+3 2 | 2+2 2 | 3+3 2 | 3+3 2 | 2+2 2 | 2+2 2 |
| | 3+3 2 | 3+3 2 | 2+2 2 | 3+3 2 | 2+2 2 | 2+2 2 | 3+3 2 | 2+2 2 |
| | 2+2 2 | 2+2 2 | 3+3 2 | 2+2 2 | 3+3 2 | 3+3 2 | 2+2 2 | 3+3 2 |
| | 2+2 2 | 2+2 2 | 3+3 2 | 3+3 2 | 2+2 2 | 3+3 2 | 3+3 2 | 2+2 2 |
| | 3+3 2 | 3+3 2 | 2+2 2 | 2+2 2 | 3+3 2 | 2+2 2 | 2+2 2 | 3+3 2 |
| | 3+3 2 | 2+2 2 | 2+2 2 | 3+3 2 | 2+2 2 | 3+3 2 | 2+2 2 | 3+3 2 |
| | 2+2 2 | 3+3 2 | 3+3 2 | 2+2 2 | 3+3 2 | 2+2 2 | 3+3 2 | 2+2 2 |
| | 3+3 2 | 3+3 2 | 3+3 2 | 2+2 2 | 2+2 2 | 3+3 2 | 3+3 2 | 2+2 2 |

Continued on next column

**TABLE XXI– continued**

| TQC | ICSD | NoE | SF | | | | | |
|---|---|---|---|---|---|---|---|---|
| | 2+2 2 | 2+2 2 | 2+2 2 | 3+3 2 | 3+3 2 | 2+2 2 | 2+2 2 | 3+3 2 |
| | 3+3 2 | 3+3 2 | 2+2 2 | 3+3 2 | 2+2 2 | 2+2 2 | 3+3 2 | 2+2 2 |
| | 2+2 2 | 2+2 2 | 3+3 2 | 2+2 2 | 3+3 2 | 3+3 2 | 2+2 2 | 3+3 2 |
| | 3+3 2 | 3+3 2 | 2+2 2 | 3+3 2 | 2+2 2 | 2+2 2 | 3+3 2 | 3+3 2 |
| | 2+2 2 | 2+2 2 | 3+3 2 | 2+2 2 | 3+3 2 | 3+3 2 | 2+2 2 | 2+2 2 |
| | 3+3 2 | 2+2 2 | 3+3 2 | 3+3 2 | 3+3 2 | 2+2 2 | 3+3 2 | 2+2 2 |
| | 2+2 2 | 3+3 2 | 2+2 2 | 2+2 2 | 2+2 2 | 3+3 2 | 2+2 2 | 3+3 2 |
| | 3+3 2 | 3+3 2 | 2+2 2 | 3+3 2 | 2+2 2 | 3+3 2 | 3+3 2 | 2+2 2 |
| | 2+2 2 | 2+2 2 | 3+3 2 | 2+2 2 | 3+3 2 | 2+2 2 | 2+2 2 | 3+3 2 |
| | 3+3 2 | 3+3 2 | 3+3 2 | 3+3 2 | 2+2 2 | 2+2 2 | 3+3 2 | 2+2 2 |
| | 2+2 2 | 2+2 2 | 2+2 2 | 3+3 2 | 3+3 2 | 3+3 2 | 2+2 2 | 3+3 2 |
| | 3+3 2 | 3+3 2 | 2+2 2 | 2+2 2 | 2+2 2 | 3+3 2 | 3+3 2 | 3+3 2 |
| | 2+2 2 | 2+2 2 | 3+3 2 | 2+2 2 | 3+3 2 | 2+2 2 | 2+2 2 | 3+3 2 |
| | 3+3 2 | 2+2 2 | 3+3 2 | 2+2 2 | 3+3 2 | 3+3 2 | 3+3 2 | 2+2 2 |
| | 2+2 2 | 3+3 2 | 2+2 2 | 3+3 2 | 2+2 2 | 2+2 2 | 2+2 2 | 2+2 2 |
| | 3+3 2 | 3+3 2 | 2+2 2 | 2+2 2 | 2+2 2 | 3+3 2 | 3+3 2 | 3+3 2 |
| | 2+2 2 | 2+2 2 | 3+3 2 | 3+3 2 | 3+3 2 | 2+2 2 | 2+2 2 | 2+2 2 |
| | 3+3 2 | 3+3 2 | 2+2 2 | 3+3 2 | 2+2 2 | 2+2 2 | 3+3 2 | 2+2 2 |
| | 2+2 2 | 2+2 2 | 3+3 2 | 2+2 2 | 3+3 2 | 3+3 2 | 2+2 2 | 3+3 2 |
| | 3+3 2 | 2+2 2 | 3+3 2 | 3+3 2 | 2+2 2 | 3+3 2 | 2+2 2 | 2+2 2 |
| | 3+3 2 | 3+3 2 | 2+2 2 | 2+2 2 | 3+3 2 | 2+2 2 | 3+3 2 | 3+3 2 |
| | 2+2 2 | 2+2 2 | 3+3 2 | 3+3 2 | 3+3 2 | 2+2 2 | 3+3 2 | 3+3 2 |
| | 3+3 2 | 3+3 2 | 2+2 2 | 2+2 2 | 2+2 2 | 3+3 2 | 2+2 2 | 2+2 2 |
| | 2+2 2 | 2+2 2 | 3+3 2 | 3+3 2 | 2+2 2 | 2+2 2 | 2+2 2 | 3+3 2 |
| | 3+3 2 | 3+3 2 | 3+3 2 | 2+2 2 | 3+3 2 | 3+3 2 | 3+3 2 | 2+2 2 |
| | 2+2 2 | 2+2 2 | 2+2 2 | 3+3 2 | 3+3 2 | 3+3 2 | 2+2 2 | 3+3 2 |
| | 2+2 2 | 3+3 2 | 2+2 2 | 2+2 2 | 2+2 2 | 2+2 2 | 3+3 2 | 2+2 2 |
| | 2+2 2 | 2+2 2 | 2+2 2 | 3+3 2 | 3+3 2 | 3+3 2 | 2+2 2 | 3+3 2 |
| | 3+3 2 | 3+3 2 | 3+3 2 | 2+2 2 | 2+2 2 | 3+3 2 | 3+3 2 | 3+3 2 |
| | 3+3 2 | 3+3 2 | 2+2 2 | 3+3 2 | 2+2 2 | 2+2 2 | 3+3 2 | 2+2 2 |
| | 2+2 2 | 2+2 2 | 3+3 2 | 2+2 2 | 3+3 2 | 2+2 2 | 2+2 2 | 2+2 2 |
| | 3+3 2 | 2+2 2 | 2+2 2 | 3+3 2 | 2+2 2 | 3+3 2 | 3+3 2 | 3+3 2 |
| | 3+3 2 | 3+3 2 | 3+3 2 | 3+3 2 | 3+3 2 | 3+3 2 | 2+2 2 | 3+3 2 |
| | 2+2 2 | 3+3 2 | 2+2 2 | 2+2 2 | 3+3 2 | 2+2 2 | 2+2 2 | 2+2 2 |
| | 3+3 2 | 2+2 2 | 3+3 2 | 3+3 2 | 2+2 2 | 3+3 2 | 3+3 2 | 3+3 2 |
| | 2+2 2 | 2+2 2 | 2+2 2 | 2+2 2 | 2+2 2 | 2+2 2 | 2+2 2 | 2+2 2 |
| | 3+3 2 | 3+3 2 | 3+3 2 | 3+3 2 | 3+3 2 | 3+3 2 | 3+3 2 | 3+3 2 |
| | 2+2 2 | 2+2 2 | 2+2 2 | 2+2 2 | 2+2 2 | 2+2 2 | 3+3 2 | 2+2 2 |
| | 2+2 2 | 3+3 2 | 3+3 2 | 2+2 2 | 3+3 2 | 2+2 2 | 2+2 2 | 2+2 2 |
| | 3+3 2 | 2+2 2 | 2+2 2 | 3+3 2 | 2+2 2 | 3+3 2 | 2+2 2 | 3+3 2 |
| | 3+3 2 | 2+2 2 | 2+2 2 | 3+3 2 | 2+2 2 | 3+3 2 | 2+2 2 | 3+3 2 |
| | 3+3 2 | 3+3 2 | 3+3 2 | 3+3 2 | 3+3 2 | 3+3 2 | 3+3 2 | 3+3 2 |
| | 3+3 2 | 3+3 2 | 2+2 2 | 3+3 2 | 3+3 2 | 3+3 2 | 2+2 2 | 3+3 2 |
| | 3+3 2 | 2+2 2 | 3+3 2 | 3+3 2 | 2+2 2 | 3+3 2 | 3+3 2 | 3+3 2 |
| | 3+3 2 | 3+3 2 | 3+3 2 | 3+3 2 | 3+3 2 | 3+3 2 | 3+3 2 | 3+3 2 |
| | 3+3 2 | 2+2 2 | 3+3 2 | 3+3 2 | 2+2 2 | 3+3 2 | 3+3 2 | 3+3 2 |
| 2.2.26 | 60937 | 154 | H1As4Cu3Na5O16 | | | | | |
| | -GM | -R | -T | -U | -V | -X | -Y | -Z |
| | 3+3 2 | 2+2 2 | 3+3 2 | 2+2 2 | 2+2 2 | 2+2 2 | 3+3 2 | 3+3 2 |
| | 2+2 2 | 3+3 2 | 2+2 2 | 3+3 2 | 3+3 2 | 3+3 2 | 2+2 2 | 2+2 2 |
| | 3+3 2 | 2+2 2 | 3+3 2 | 3+3 2 | 3+3 2 | 2+2 2 | 2+2 2 | 2+2 2 |
| | 2+2 2 | 3+3 2 | 2+2 2 | 2+2 2 | 2+2 2 | 3+3 2 | 3+3 2 | 3+3 2 |
| | 2+2 2 | 3+3 2 | 2+2 2 | 3+3 2 | 3+3 2 | 3+3 2 | 2+2 2 | 2+2 2 |
| | 3+3 2 | 2+2 2 | 3+3 2 | 2+2 2 | 2+2 2 | 2+2 2 | 3+3 2 | 3+3 2 |
| | 2+2 2 | 3+3 2 | 2+2 2 | 3+3 2 | 3+3 2 | 3+3 2 | 3+3 2 | 3+3 2 |
| | 3+3 2 | 2+2 2 | 3+3 2 | 2+2 2 | 2+2 2 | 2+2 2 | 2+2 2 | 2+2 2 |
| | 2+2 2 | 3+3 2 | 2+2 2 | 2+2 2 | 2+2 2 | 3+3 2 | 3+3 2 | 3+3 2 |
| | 3+3 2 | 2+2 2 | 3+3 2 | 3+3 2 | 3+3 2 | 2+2 2 | 2+2 2 | 2+2 2 |
| | 3+3 2 | 2+2 2 | 2+2 2 | 2+2 2 | 2+2 2 | 2+2 2 | 2+2 2 | 3+3 2 |
| | 2+2 2 | 3+3 2 | 3+3 2 | 3+3 2 | 3+3 2 | 3+3 2 | 3+3 2 | 2+2 2 |
| | 2+2 2 | 2+2 2 | 2+2 2 | 3+3 2 | 3+3 2 | 2+2 2 | 3+3 2 | 3+3 2 |

Continued on next column

**TABLE XXI– continued**

| TQC | ICSD | NoE | SF | | | | | |
|---|---|---|---|---|---|---|---|---|
| | 3+3 2 | 3+3 2 | 3+3 2 | 2+2 2 | 2+2 2 | 3+3 2 | 2+2 2 | 2+2 2 |
| | 3+3 2 | 2+2 2 | 2+2 2 | 2+2 2 | 3+3 2 | 3+3 2 | 3+3 2 | 2+2 2 |
| | 2+2 2 | 3+3 2 | 3+3 2 | 3+3 2 | 2+2 2 | 2+2 2 | 2+2 2 | 3+3 2 |
| | 3+3 2 | 3+3 2 | 3+3 2 | 3+3 2 | 3+3 2 | 3+3 2 | 3+3 2 | 3+3 2 |
| | 2+2 2 | 2+2 2 | 2+2 2 | 2+2 2 | 2+2 2 | 2+2 2 | 2+2 2 | 2+2 2 |
| | 3+3 2 | 3+3 2 | 3+3 2 | 2+2 2 | 2+2 2 | 3+3 2 | 2+2 2 | 2+2 2 |
| | 2+2 2 | 2+2 2 | 2+2 2 | 3+3 2 | 3+3 2 | 2+2 2 | 3+3 2 | 3+3 2 |
| | 3+3 2 | 3+3 2 | 3+3 2 | 3+3 2 | 3+3 2 | 3+3 2 | 3+3 2 | 3+3 2 |
| | 3+3 2 | 2+2 2 | 3+3 2 | 2+2 2 | 2+2 2 | 2+2 2 | 2+2 2 | 2+2 2 |
| | 2+2 2 | 3+3 2 | 2+2 2 | 3+3 2 | 3+3 2 | 2+2 2 | 3+3 2 | 3+3 2 |
| | 2+2 2 | 2+2 2 | 2+2 2 | 2+2 2 | 2+2 2 | 3+3 2 | 2+2 2 | 2+2 2 |
| | 3+3 2 | 3+3 2 | 3+3 2 | 3+3 2 | 3+3 2 | 3+3 2 | 3+3 2 | 3+3 2 |
| | 2+2 2 | 2+2 2 | 2+2 2 | 2+2 2 | 2+2 2 | 2+2 2 | 2+2 2 | 2+2 2 |
| | 3+3 2 | 2+2 2 | 3+3 2 | 2+2 2 | 2+2 2 | 3+3 2 | 3+3 2 | 3+3 2 |
| | 2+2 2 | 3+3 2 | 2+2 2 | 3+3 2 | 3+3 2 | 2+2 2 | 2+2 2 | 2+2 2 |
| | 3+3 2 | 2+2 2 | 3+3 2 | 3+3 2 | 2+2 2 | 3+3 2 | 3+3 2 | 3+3 2 |
| | 2+2 2 | 3+3 2 | 2+2 2 | 2+2 2 | 3+3 2 | 2+2 2 | 2+2 2 | 2+2 2 |
| | 3+3 2 | 2+2 2 | 3+3 2 | 3+3 2 | 3+3 2 | 2+2 2 | 2+2 2 | 2+2 2 |
| | 3+3 2 | 3+3 2 | 3+3 2 | 3+3 2 | 3+3 2 | 3+3 2 | 3+3 2 | 3+3 2 |
| | 2+2 2 | 2+2 2 | 2+2 2 | 2+2 2 | 2+2 2 | 2+2 2 | 2+2 2 | 3+3 2 |
| | 3+3 2 | 3+3 2 | 2+2 2 | 2+2 2 | 3+3 2 | 3+3 2 | 3+3 2 | 2+2 2 |
| | 2+2 2 | 3+3 2 | 3+3 2 | 3+3 2 | 2+2 2 | 3+3 2 | 3+3 2 | 3+3 2 |
| | 2+2 2 | 2+2 2 | 2+2 2 | 2+2 2 | 2+2 2 | 2+2 2 | 2+2 2 | 2+2 2 |
| | 3+3 2 | 3+3 2 | 3+3 2 | 3+3 2 | 3+3 2 | 3+3 2 | 3+3 2 | 3+3 2 |
| | 2+2 2 | 2+2 2 | 2+2 2 | 3+3 2 | 3+3 2 | 2+2 2 | 3+3 2 | 2+2 2 |
| | 3+3 2 | 3+3 2 | 2+2 2 | 2+2 2 | 2+2 2 | 3+3 2 | 2+2 2 | 3+3 2 |
| | 2+2 2 | 3+3 2 | 3+3 2 | 2+2 2 | 2+2 2 | 3+3 2 | 3+3 2 | 3+3 2 |
| | 3+3 2 | 3+3 2 | 3+3 2 | 2+2 2 | 2+2 2 | 3+3 2 | 2+2 2 | 2+2 2 |
| | 3+3 2 | 2+2 2 | 3+3 2 | 3+3 2 | 2+2 2 | 2+2 2 | 3+3 2 | 3+3 2 |
| | 3+3 2 | 2+2 2 | 3+3 2 | 3+3 2 | 3+3 2 | 2+2 2 | 2+2 2 | 2+2 2 |
| | 2+2 2 | 3+3 2 | 2+2 2 | 2+2 2 | 3+3 2 | 2+2 2 | 3+3 2 | 3+3 2 |
| | 3+3 2 | 2+2 2 | 2+2 2 | 3+3 2 | 2+2 2 | 3+3 2 | 2+2 2 | 2+2 2 |
| | 2+2 2 | 3+3 2 | 3+3 2 | 2+2 2 | 3+3 2 | 2+2 2 | 2+2 2 | 3+3 2 |
| | 2+2 2 | 2+2 2 | 2+2 2 | 3+3 2 | 2+2 2 | 3+3 2 | 3+3 2 | 2+2 2 |
| | 3+3 2 | 2+2 2 | 2+2 2 | 2+2 2 | 3+3 2 | 3+3 2 | 2+2 2 | 2+2 2 |
| | 2+2 2 | 3+3 2 | 2+2 2 | 3+3 2 | 3+3 2 | 2+2 2 | 2+2 2 | 3+3 2 |
| | 3+3 2 | 3+3 2 | 3+3 2 | 2+2 2 | 3+3 2 | 3+3 2 | 3+3 2 | 3+3 2 |
| | 2+2 2 | 3+3 2 | 3+3 2 | 3+3 2 | 2+2 2 | 2+2 2 | 3+3 2 | 2+2 2 |
| | 3+3 2 | 2+2 2 | 2+2 2 | 2+2 2 | 2+2 2 | 3+3 2 | 2+2 2 | 2+2 2 |
| | 2+2 2 | 2+2 2 | 2+2 2 | 3+3 2 | 3+3 2 | 3+3 2 | 2+2 2 | 3+3 2 |
| | 3+3 2 | 2+2 2 | 3+3 2 | 3+3 2 | 3+3 2 | 2+2 2 | 3+3 2 | 2+2 2 |
| | 2+2 2 | 3+3 2 | 3+3 2 | 2+2 2 | 2+2 2 | 2+2 2 | 2+2 2 | 3+3 2 |
| | 3+3 2 | 3+3 2 | 2+2 2 | 2+2 2 | 3+3 2 | 3+3 2 | 3+3 2 | 2+2 2 |
| | 3+3 2 | 2+2 2 | 3+3 2 | 3+3 2 | 2+2 2 | 2+2 2 | 3+3 2 | 3+3 2 |
| | 2+2 2 | 2+2 2 | 2+2 2 | 2+2 2 | 3+3 2 | 2+2 2 | 2+2 2 | 2+2 2 |
| | 2+2 2 | 2+2 2 | 2+2 2 | 3+3 2 | 2+2 2 | 3+3 2 | 3+3 2 | 2+2 2 |
| | 3+3 2 | 3+3 2 | 3+3 2 | 2+2 2 | 3+3 2 | 2+2 2 | 3+3 2 | 3+3 2 |
| | 3+3 2 | 3+3 2 | 3+3 2 | 3+3 2 | 2+2 2 | 3+3 2 | 2+2 2 | 3+3 2 |
| | 2+2 2 | 2+2 2 | 3+3 2 | 3+3 2 | 3+3 2 | 3+3 2 | 3+3 2 | 3+3 2 |
| | 3+3 2 | 3+3 2 | 2+2 2 | 2+2 2 | 3+3 2 | 3+3 2 | 2+2 2 | 2+2 2 |
| | 2+2 2 | 2+2 2 | 3+3 2 | 3+3 2 | 2+2 2 | 2+2 2 | 3+3 2 | 3+3 2 |
| | 3+3 2 | 3+3 2 | 2+2 2 | 2+2 2 | 2+2 2 | 2+2 2 | 3+3 2 | 3+3 2 |
| | 2+2 2 | 3+3 2 | 3+3 2 | 3+3 2 | 3+3 2 | 3+3 2 | 2+2 2 | 2+2 2 |
| | 3+3 2 | 2+2 2 | 2+2 2 | 3+3 2 | 3+3 2 | 2+2 2 | 3+3 2 | 3+3 2 |
| | 3+3 2 | 3+3 2 | 3+3 2 | 3+3 2 | 3+3 2 | 3+3 2 | 3+3 2 | 2+2 2 |
| | 2+2 2 | 3+3 2 | 3+3 2 | 2+2 2 | 2+2 2 | 3+3 2 | 3+3 2 | 3+3 2 |
| | 3+3 2 | 2+2 2 | 3+3 2 | 3+3 2 | 3+3 2 | 3+3 2 | 2+2 2 | 3+3 2 |
| | 3+3 2 | 3+3 2 | 3+3 2 | 2+2 2 | 3+3 2 | 3+3 2 | 2+2 2 | 2+2 2 |
| | 2+2 2 | 3+3 2 | 2+2 2 | 3+3 2 | 2+2 2 | 2+2 2 | 3+3 2 | 3+3 2 |
| | 3+3 2 | 2+2 2 | 2+2 2 | 2+2 2 | 3+3 2 | 3+3 2 | 3+3 2 | 2+2 2 |
| | 2+2 2 | 3+3 2 | 3+3 2 | 3+3 2 | 2+2 2 | 2+2 2 | 2+2 2 | 3+3 2 |

Continued on next column

**TABLE XXI– continued**

| TQC | ICSD | NoE | SF | | | | | |
|---|---|---|---|---|---|---|---|---|
| | 3+3 2 | 2+2 2 | 2+2 2 | 2+2 2 | 3+3 2 | 3+3 2 | 3+3 2 | 3+3 2 |
| | 2+2 2 | 3+3 2 | 3+3 2 | 3+3 2 | 3+3 2 | 2+2 2 | 2+2 2 | 2+2 2 |
| | 3+3 2 | 3+3 2 | 2+2 2 | 3+3 2 | 2+2 2 | 3+3 2 | 3+3 2 | 3+3 2 |
| 2.2.27 | 8063 | 118 | H1As1O5Zn2 | | | | | |
| | -GM | -R | -T | -U | -V | -X | -Y | -Z |
| | 3+3 2 | 3+3 2 | 3+3 2 | 2+2 2 | 2+2 2 | 3+3 2 | 2+2 2 | 2+2 2 |
| | 2+2 2 | 2+2 2 | 2+2 2 | 3+3 2 | 3+3 2 | 2+2 2 | 3+3 2 | 3+3 2 |
| | 3+3 2 | 2+2 2 | 2+2 2 | 3+3 2 | 3+3 2 | 2+2 2 | 2+2 2 | 3+3 2 |
| | 2+2 2 | 2+2 2 | 3+3 2 | 3+3 2 | 2+2 2 | 3+3 2 | 3+3 2 | 2+2 2 |
| | 3+3 2 | 3+3 2 | 3+3 2 | 2+2 2 | 3+3 2 | 2+2 2 | 2+2 2 | 2+2 2 |
| | 2+2 2 | 3+3 2 | 2+2 2 | 2+2 2 | 2+2 2 | 3+3 2 | 3+3 2 | 3+3 2 |
| | 3+3 2 | 3+3 2 | 3+3 2 | 2+2 2 | 2+2 2 | 2+2 2 | 2+2 2 | 3+3 2 |
| | 2+2 2 | 2+2 2 | 2+2 2 | 3+3 2 | 3+3 2 | 3+3 2 | 3+3 2 | 2+2 2 |
| | 3+3 2 | 3+3 2 | 2+2 2 | 3+3 2 | 2+2 2 | 2+2 2 | 3+3 2 | 2+2 2 |
| | 2+2 2 | 2+2 2 | 3+3 2 | 2+2 2 | 3+3 2 | 3+3 2 | 2+2 2 | 3+3 2 |
| | 3+3 2 | 2+2 2 | 2+2 2 | 3+3 2 | 3+3 2 | 3+3 2 | 2+2 2 | 3+3 2 |
| | 2+2 2 | 3+3 2 | 3+3 2 | 2+2 2 | 2+2 2 | 2+2 2 | 3+3 2 | 2+2 2 |
| | 3+3 2 | 2+2 2 | 3+3 2 | 3+3 2 | 2+2 2 | 3+3 2 | 2+2 2 | 3+3 2 |
| | 2+2 2 | 3+3 2 | 2+2 2 | 3+3 2 | 3+3 2 | 2+2 2 | 2+2 2 | 2+2 2 |
| | 3+3 2 | 2+2 2 | 3+3 2 | 2+2 2 | 3+3 2 | 2+2 2 | 2+2 2 | 3+3 2 |
| | 2+2 2 | 3+3 2 | 2+2 2 | 2+2 2 | 2+2 2 | 3+3 2 | 3+3 2 | 2+2 2 |
| | 3+3 2 | 2+2 2 | 3+3 2 | 3+3 2 | 2+2 2 | 3+3 2 | 3+3 2 | 2+2 2 |
| | 3+3 2 | 3+3 2 | 2+2 2 | 2+2 2 | 3+3 2 | 2+2 2 | 3+3 2 | 3+3 2 |
| | 2+2 2 | 2+2 2 | 2+2 2 | 3+3 2 | 2+2 2 | 2+2 2 | 3+3 2 | 2+2 2 |
| | 2+2 2 | 3+3 2 | 3+3 2 | 3+3 2 | 2+2 2 | 3+3 2 | 2+2 2 | 3+3 2 |
| | 3+3 2 | 3+3 2 | 2+2 2 | 2+2 2 | 3+3 2 | 2+2 2 | 2+2 2 | 2+2 2 |
| | 3+3 2 | 2+2 2 | 3+3 2 | 3+3 2 | 3+3 2 | 3+3 2 | 2+2 2 | 3+3 2 |
| | 2+2 2 | 2+2 2 | 2+2 2 | 2+2 2 | 2+2 2 | 2+2 2 | 3+3 2 | 2+2 2 |
| | 3+3 2 | 3+3 2 | 3+3 2 | 3+3 2 | 3+3 2 | 3+3 2 | 3+3 2 | 3+3 2 |
| | 2+2 2 | 2+2 2 | 2+2 2 | 2+2 2 | 2+2 2 | 2+2 2 | 2+2 2 | 2+2 2 |
| | 3+3 2 | 3+3 2 | 3+3 2 | 3+3 2 | 3+3 2 | 3+3 2 | 3+3 2 | 3+3 2 |
| | 2+2 2 | 2+2 2 | 2+2 2 | 2+2 2 | 2+2 2 | 2+2 2 | 2+2 2 | 2+2 2 |
| | 2+2 2 | 3+3 2 | 3+3 2 | 2+2 2 | 3+3 2 | 3+3 2 | 3+3 2 | 3+3 2 |
| | 3+3 2 | 2+2 2 | 2+2 2 | 3+3 2 | 2+2 2 | 2+2 2 | 2+2 2 | 2+2 2 |
| | 3+3 2 | 3+3 2 | 3+3 2 | 2+2 2 | 3+3 2 | 3+3 2 | 3+3 2 | 3+3 2 |
| | 2+2 2 | 2+2 2 | 2+2 2 | 3+3 2 | 3+3 2 | 3+3 2 | 2+2 2 | 2+2 2 |
| | 3+3 2 | 3+3 2 | 3+3 2 | 2+2 2 | 2+2 2 | 2+2 2 | 3+3 2 | 2+2 2 |
| | 2+2 2 | 2+2 2 | 3+3 2 | 3+3 2 | 3+3 2 | 2+2 2 | 2+2 2 | 3+3 2 |
| | 2+2 2 | 3+3 2 | 2+2 2 | 2+2 2 | 2+2 2 | 3+3 2 | 3+3 2 | 3+3 2 |
| | 2+2 2 | 2+2 2 | 2+2 2 | 3+3 2 | 2+2 2 | 2+2 2 | 3+3 2 | 2+2 2 |
| | 3+3 2 | 3+3 2 | 3+3 2 | 2+2 2 | 3+3 2 | 3+3 2 | 2+2 2 | 3+3 2 |
| | 2+2 2 | 3+3 2 | 3+3 2 | 3+3 2 | 2+2 2 | 3+3 2 | 2+2 2 | 2+2 2 |
| | 3+3 2 | 2+2 2 | 2+2 2 | 2+2 2 | 2+2 2 | 2+2 2 | 3+3 2 | 3+3 2 |
| | 2+2 2 | 2+2 2 | 3+3 2 | 2+2 2 | 3+3 2 | 2+2 2 | 3+3 2 | 3+3 2 |
| | 3+3 2 | 3+3 2 | 2+2 2 | 3+3 2 | 3+3 2 | 3+3 2 | 2+2 2 | 2+2 2 |
| | 3+3 2 | 2+2 2 | 2+2 2 | 3+3 2 | 3+3 2 | 2+2 2 | 3+3 2 | 3+3 2 |
| | 2+2 2 | 3+3 2 | 3+3 2 | 2+2 2 | 2+2 2 | 3+3 2 | 3+3 2 | 3+3 2 |
| | 3+3 2 | 3+3 2 | 2+2 2 | 2+2 2 | 2+2 2 | 3+3 2 | 2+2 2 | 2+2 2 |
| | 2+2 2 | 2+2 2 | 3+3 2 | 3+3 2 | 3+3 2 | 2+2 2 | 2+2 2 | 2+2 2 |
| | 3+3 2 | 3+3 2 | 3+3 2 | 2+2 2 | 3+3 2 | 3+3 2 | 3+3 2 | 2+2 2 |
| | 2+2 2 | 3+3 2 | 2+2 2 | 3+3 2 | 2+2 2 | 3+3 2 | 2+2 2 | 2+2 2 |
| | 3+3 2 | 2+2 2 | 3+3 2 | 2+2 2 | 2+2 2 | 2+2 2 | 2+2 2 | 2+2 2 |
| | 2+2 2 | 2+2 2 | 2+2 2 | 2+2 2 | 3+3 2 | 2+2 2 | 3+3 2 | 3+3 2 |
| | 3+3 2 | 3+3 2 | 3+3 2 | 3+3 2 | 3+3 2 | 2+2 2 | 3+3 2 | 3+3 2 |
| | 3+3 2 | 2+2 2 | 2+2 2 | 3+3 2 | 2+2 2 | 3+3 2 | 2+2 2 | 2+2 2 |
| | 2+2 2 | 3+3 2 | 3+3 2 | 3+3 2 | 2+2 2 | 3+3 2 | 3+3 2 | 3+3 2 |
| | 2+2 2 | 2+2 2 | 2+2 2 | 2+2 2 | 3+3 2 | 2+2 2 | 2+2 2 | 2+2 2 |
| | 3+3 2 | 2+2 2 | 3+3 2 | 2+2 2 | 3+3 2 | 2+2 2 | 3+3 2 | 3+3 2 |
| | 2+2 2 | 3+3 2 | 2+2 2 | 3+3 2 | 2+2 2 | 3+3 2 | 2+2 2 | 2+2 2 |
| | 3+3 2 | 3+3 2 | 2+2 2 | 3+3 2 | 2+2 2 | 2+2 2 | 3+3 2 | 3+3 2 |
| | 2+2 2 | 2+2 2 | 3+3 2 | 2+2 2 | 3+3 2 | 3+3 2 | 2+2 2 | 3+3 2 |

Continued on next column

**TABLE XXI– continued**

| TQC | ICSD | NoE | SF | | | | |
|---|---|---|---|---|---|---|---|
| | 3+3 2 | 2+2 2 | 2+2 2 | 3+3 2 | 2+2 2 | 2+2 2 | 3+3 2 | 3+3 2 |
| | 2+2 2 | 3+3 2 | 3+3 2 | 2+2 2 | 2+2 2 | 3+3 2 | 2+2 2 | 2+2 2 |
| | 3+3 2 | 3+3 2 | 3+3 2 | 2+2 2 | 3+3 2 | 2+2 2 | 3+3 2 | 3+3 2 |
| 2.2.28 | 37436 | 118 | H1As1O5Zn2 | | | | | |
| | -GM | -R | -T | -U | -V | -X | -Y | -Z |
| | 3+3 2 | 3+3 2 | 3+3 2 | 2+2 2 | 2+2 2 | 3+3 2 | 2+2 2 | 2+2 2 |
| | 2+2 2 | 2+2 2 | 2+2 2 | 3+3 2 | 3+3 2 | 2+2 2 | 3+3 2 | 3+3 2 |
| | 3+3 2 | 2+2 2 | 2+2 2 | 3+3 2 | 3+3 2 | 2+2 2 | 3+3 2 | 3+3 2 |
| | 2+2 2 | 2+2 2 | 3+3 2 | 3+3 2 | 2+2 2 | 3+3 2 | 2+2 2 | 2+2 2 |
| | 3+3 2 | 3+3 2 | 3+3 2 | 2+2 2 | 2+2 2 | 2+2 2 | 2+2 2 | 2+2 2 |
| | 2+2 2 | 3+3 2 | 2+2 2 | 2+2 2 | 3+3 2 | 3+3 2 | 3+3 2 | 3+3 2 |
| | 3+3 2 | 3+3 2 | 3+3 2 | 2+2 2 | 2+2 2 | 2+2 2 | 2+2 2 | 3+3 2 |
| | 2+2 2 | 2+2 2 | 2+2 2 | 3+3 2 | 3+3 2 | 3+3 2 | 3+3 2 | 2+2 2 |
| | 3+3 2 | 3+3 2 | 2+2 2 | 3+3 2 | 2+2 2 | 2+2 2 | 3+3 2 | 2+2 2 |
| | 2+2 2 | 2+2 2 | 3+3 2 | 2+2 2 | 3+3 2 | 3+3 2 | 2+2 2 | 3+3 2 |
| | 3+3 2 | 2+2 2 | 2+2 2 | 3+3 2 | 3+3 2 | 3+3 2 | 2+2 2 | 3+3 2 |
| | 2+2 2 | 3+3 2 | 3+3 2 | 2+2 2 | 2+2 2 | 2+2 2 | 3+3 2 | 2+2 2 |
| | 3+3 2 | 2+2 2 | 3+3 2 | 3+3 2 | 2+2 2 | 3+3 2 | 2+2 2 | 3+3 2 |
| | 2+2 2 | 3+3 2 | 2+2 2 | 3+3 2 | 3+3 2 | 2+2 2 | 2+2 2 | 2+2 2 |
| | 3+3 2 | 2+2 2 | 3+3 2 | 2+2 2 | 3+3 2 | 2+2 2 | 3+3 2 | 3+3 2 |
| | 2+2 2 | 3+3 2 | 2+2 2 | 2+2 2 | 2+2 2 | 3+3 2 | 2+2 2 | 2+2 2 |
| | 3+3 2 | 2+2 2 | 3+3 2 | 3+3 2 | 2+2 2 | 3+3 2 | 3+3 2 | 2+2 2 |
| | 3+3 2 | 3+3 2 | 2+2 2 | 2+2 2 | 3+3 2 | 2+2 2 | 3+3 2 | 3+3 2 |
| | 2+2 2 | 3+3 2 | 2+2 2 | 3+3 2 | 2+2 2 | 2+2 2 | 3+3 2 | 2+2 2 |
| | 3+3 2 | 2+2 2 | 3+3 2 | 3+3 2 | 2+2 2 | 3+3 2 | 2+2 2 | 3+3 2 |
| | 2+2 2 | 3+3 2 | 2+2 2 | 2+2 2 | 3+3 2 | 2+2 2 | 2+2 2 | 2+2 2 |
| | 3+3 2 | 2+2 2 | 3+3 2 | 3+3 2 | 3+3 2 | 3+3 2 | 2+2 2 | 3+3 2 |
| | 2+2 2 | 2+2 2 | 2+2 2 | 2+2 2 | 2+2 2 | 2+2 2 | 3+3 2 | 2+2 2 |
| | 3+3 2 | 3+3 2 | 3+3 2 | 3+3 2 | 3+3 2 | 3+3 2 | 3+3 2 | 3+3 2 |
| | 2+2 2 | 2+2 2 | 2+2 2 | 2+2 2 | 2+2 2 | 2+2 2 | 2+2 2 | 2+2 2 |
| | 3+3 2 | 3+3 2 | 3+3 2 | 3+3 2 | 3+3 2 | 3+3 2 | 3+3 2 | 3+3 2 |
| | 2+2 2 | 2+2 2 | 2+2 2 | 2+2 2 | 2+2 2 | 2+2 2 | 2+2 2 | 2+2 2 |
| | 2+2 2 | 3+3 2 | 3+3 2 | 2+2 2 | 3+3 2 | 3+3 2 | 3+3 2 | 3+3 2 |
| | 3+3 2 | 2+2 2 | 2+2 2 | 3+3 2 | 2+2 2 | 2+2 2 | 2+2 2 | 2+2 2 |
| | 3+3 2 | 3+3 2 | 3+3 2 | 2+2 2 | 3+3 2 | 3+3 2 | 3+3 2 | 3+3 2 |
| | 2+2 2 | 2+2 2 | 2+2 2 | 3+3 2 | 3+3 2 | 3+3 2 | 2+2 2 | 2+2 2 |
| | 3+3 2 | 3+3 2 | 3+3 2 | 2+2 2 | 2+2 2 | 2+2 2 | 3+3 2 | 2+2 2 |
| | 2+2 2 | 2+2 2 | 3+3 2 | 3+3 2 | 3+3 2 | 2+2 2 | 2+2 2 | 3+3 2 |
| | 2+2 2 | 3+3 2 | 2+2 2 | 2+2 2 | 2+2 2 | 3+3 2 | 3+3 2 | 3+3 2 |
| | 2+2 2 | 2+2 2 | 2+2 2 | 3+3 2 | 2+2 2 | 2+2 2 | 3+3 2 | 2+2 2 |
| | 3+3 2 | 3+3 2 | 3+3 2 | 2+2 2 | 3+3 2 | 3+3 2 | 2+2 2 | 3+3 2 |
| | 2+2 2 | 3+3 2 | 2+2 2 | 3+3 2 | 2+2 2 | 3+3 2 | 2+2 2 | 2+2 2 |
| | 3+3 2 | 2+2 2 | 3+3 2 | 2+2 2 | 3+3 2 | 2+2 2 | 3+3 2 | 3+3 2 |
| | 2+2 2 | 2+2 2 | 3+3 2 | 2+2 2 | 2+2 2 | 2+2 2 | 3+3 2 | 3+3 2 |
| | 3+3 2 | 3+3 2 | 2+2 2 | 3+3 2 | 3+3 2 | 3+3 2 | 2+2 2 | 2+2 2 |
| | 3+3 2 | 2+2 2 | 2+2 2 | 3+3 2 | 3+3 2 | 2+2 2 | 3+3 2 | 3+3 2 |
| | 2+2 2 | 3+3 2 | 3+3 2 | 2+2 2 | 2+2 2 | 3+3 2 | 3+3 2 | 3+3 2 |
| | 3+3 2 | 3+3 2 | 2+2 2 | 2+2 2 | 2+2 2 | 3+3 2 | 2+2 2 | 2+2 2 |
| | 2+2 2 | 2+2 2 | 3+3 2 | 3+3 2 | 3+3 2 | 3+3 2 | 2+2 2 | 2+2 2 |
| | 3+3 2 | 3+3 2 | 2+2 2 | 2+2 2 | 3+3 2 | 2+2 2 | 3+3 2 | 2+2 2 |
| | 2+2 2 | 3+3 2 | 3+3 2 | 3+3 2 | 2+2 2 | 3+3 2 | 2+2 2 | 2+2 2 |
| | 3+3 2 | 2+2 2 | 3+3 2 | 2+2 2 | 2+2 2 | 2+2 2 | 2+2 2 | 2+2 2 |
| | 2+2 2 | 2+2 2 | 2+2 2 | 2+2 2 | 3+3 2 | 2+2 2 | 3+3 2 | 3+3 2 |
| | 3+3 2 | 3+3 2 | 3+3 2 | 3+3 2 | 3+3 2 | 2+2 2 | 3+3 2 | 3+3 2 |
| | 3+3 2 | 2+2 2 | 2+2 2 | 3+3 2 | 2+2 2 | 3+3 2 | 2+2 2 | 2+2 2 |
| | 2+2 2 | 3+3 2 | 3+3 2 | 3+3 2 | 2+2 2 | 3+3 2 | 3+3 2 | 3+3 2 |
| | 2+2 2 | 2+2 2 | 2+2 2 | 2+2 2 | 3+3 2 | 2+2 2 | 2+2 2 | 2+2 2 |
| | 3+3 2 | 2+2 2 | 3+3 2 | 2+2 2 | 3+3 2 | 2+2 2 | 3+3 2 | 3+3 2 |
| | 2+2 2 | 3+3 2 | 2+2 2 | 3+3 2 | 2+2 2 | 3+3 2 | 2+2 2 | 2+2 2 |
| | 3+3 2 | 3+3 2 | 3+3 2 | 3+3 2 | 2+2 2 | 2+2 2 | 3+3 2 | 3+3 2 |
| | 2+2 2 | 2+2 2 | 2+2 2 | 2+2 2 | 3+3 2 | 3+3 2 | 2+2 2 | 3+3 2 |

Continued on next column

**TABLE XXI– continued**

| TQC | ICSD | NoE | SF | | | | |
|---|---|---|---|---|---|---|---|
| | 3+3 2 | 2+2 2 | 2+2 2 | 3+3 2 | 2+2 2 | 2+2 2 | 3+3 2 | 3+3 2 |
| | 2+2 2 | 3+3 2 | 3+3 2 | 2+2 2 | 2+2 2 | 3+3 2 | 2+2 2 | 2+2 2 |
| | 3+3 2 | 3+3 2 | 3+3 2 | 2+2 2 | 3+3 2 | 2+2 2 | 3+3 2 | 3+3 2 |
| 2.2.29 | 34341 | 118 | H1As1O5Zn2 | | | | | |
| | -GM | -R | -T | -U | -V | -X | -Y | -Z |
| | 3+3 2 | 3+3 2 | 3+3 2 | 2+2 2 | 2+2 2 | 3+3 2 | 2+2 2 | 2+2 2 |
| | 2+2 2 | 2+2 2 | 2+2 2 | 3+3 2 | 3+3 2 | 2+2 2 | 3+3 2 | 3+3 2 |
| | 3+3 2 | 2+2 2 | 2+2 2 | 3+3 2 | 3+3 2 | 2+2 2 | 2+2 2 | 3+3 2 |
| | 2+2 2 | 2+2 2 | 3+3 2 | 3+3 2 | 2+2 2 | 3+3 2 | 3+3 2 | 2+2 2 |
| | 3+3 2 | 3+3 2 | 3+3 2 | 2+2 2 | 2+2 2 | 2+2 2 | 2+2 2 | 2+2 2 |
| | 2+2 2 | 3+3 2 | 2+2 2 | 2+2 2 | 3+3 2 | 3+3 2 | 3+3 2 | 3+3 2 |
| | 3+3 2 | 3+3 2 | 3+3 2 | 2+2 2 | 2+2 2 | 2+2 2 | 2+2 2 | 2+2 2 |
| | 2+2 2 | 2+2 2 | 2+2 2 | 3+3 2 | 3+3 2 | 3+3 2 | 3+3 2 | 3+3 2 |
| | 3+3 2 | 3+3 2 | 2+2 2 | 3+3 2 | 2+2 2 | 2+2 2 | 3+3 2 | 2+2 2 |
| | 2+2 2 | 2+2 2 | 3+3 2 | 2+2 2 | 3+3 2 | 3+3 2 | 2+2 2 | 3+3 2 |
| | 3+3 2 | 2+2 2 | 2+2 2 | 3+3 2 | 3+3 2 | 3+3 2 | 2+2 2 | 3+3 2 |
| | 2+2 2 | 3+3 2 | 3+3 2 | 2+2 2 | 2+2 2 | 2+2 2 | 3+3 2 | 2+2 2 |
| | 3+3 2 | 2+2 2 | 3+3 2 | 3+3 2 | 2+2 2 | 3+3 2 | 2+2 2 | 3+3 2 |
| | 2+2 2 | 3+3 2 | 2+2 2 | 3+3 2 | 3+3 2 | 2+2 2 | 2+2 2 | 2+2 2 |
| | 3+3 2 | 2+2 2 | 3+3 2 | 2+2 2 | 3+3 2 | 2+2 2 | 2+2 2 | 3+3 2 |
| | 2+2 2 | 3+3 2 | 2+2 2 | 2+2 2 | 2+2 2 | 3+3 2 | 3+3 2 | 2+2 2 |
| | 3+3 2 | 2+2 2 | 3+3 2 | 3+3 2 | 2+2 2 | 3+3 2 | 3+3 2 | 2+2 2 |
| | 3+3 2 | 3+3 2 | 2+2 2 | 2+2 2 | 3+3 2 | 2+2 2 | 3+3 2 | 3+3 2 |
| | 2+2 2 | 3+3 2 | 2+2 2 | 3+3 2 | 2+2 2 | 2+2 2 | 3+3 2 | 2+2 2 |
| | 2+2 2 | 2+2 2 | 3+3 2 | 3+3 2 | 2+2 2 | 3+3 2 | 2+2 2 | 3+3 2 |
| | 3+3 2 | 3+3 2 | 2+2 2 | 2+2 2 | 3+3 2 | 2+2 2 | 2+2 2 | 2+2 2 |
| | 3+3 2 | 2+2 2 | 3+3 2 | 3+3 2 | 3+3 2 | 3+3 2 | 2+2 2 | 3+3 2 |
| | 2+2 2 | 2+2 2 | 2+2 2 | 2+2 2 | 2+2 2 | 2+2 2 | 3+3 2 | 2+2 2 |
| | 3+3 2 | 3+3 2 | 3+3 2 | 3+3 2 | 3+3 2 | 3+3 2 | 3+3 2 | 3+3 2 |
| | 2+2 2 | 2+2 2 | 2+2 2 | 2+2 2 | 2+2 2 | 2+2 2 | 2+2 2 | 2+2 2 |
| | 3+3 2 | 3+3 2 | 3+3 2 | 3+3 2 | 3+3 2 | 3+3 2 | 3+3 2 | 3+3 2 |
| | 2+2 2 | 2+2 2 | 2+2 2 | 2+2 2 | 2+2 2 | 2+2 2 | 2+2 2 | 2+2 2 |
| | 2+2 2 | 3+3 2 | 3+3 2 | 2+2 2 | 3+3 2 | 3+3 2 | 3+3 2 | 3+3 2 |
| | 3+3 2 | 2+2 2 | 2+2 2 | 3+3 2 | 2+2 2 | 2+2 2 | 2+2 2 | 2+2 2 |
| | 3+3 2 | 3+3 2 | 3+3 2 | 2+2 2 | 3+3 2 | 3+3 2 | 3+3 2 | 3+3 2 |
| | 2+2 2 | 2+2 2 | 2+2 2 | 3+3 2 | 3+3 2 | 3+3 2 | 2+2 2 | 2+2 2 |
| | 3+3 2 | 3+3 2 | 3+3 2 | 2+2 2 | 2+2 2 | 2+2 2 | 3+3 2 | 2+2 2 |
| | 2+2 2 | 2+2 2 | 3+3 2 | 3+3 2 | 3+3 2 | 2+2 2 | 2+2 2 | 3+3 2 |
| | 2+2 2 | 3+3 2 | 2+2 2 | 2+2 2 | 2+2 2 | 3+3 2 | 3+3 2 | 3+3 2 |
| | 3+3 2 | 2+2 2 | 2+2 2 | 3+3 2 | 2+2 2 | 2+2 2 | 2+2 2 | 2+2 2 |
| | 2+2 2 | 3+3 2 | 3+3 2 | 2+2 2 | 3+3 2 | 3+3 2 | 3+3 2 | 3+3 2 |
| | 2+2 2 | 3+3 2 | 3+3 2 | 3+3 2 | 2+2 2 | 3+3 2 | 2+2 2 | 2+2 2 |
| | 3+3 2 | 2+2 2 | 2+2 2 | 2+2 2 | 3+3 2 | 2+2 2 | 3+3 2 | 3+3 2 |
| | 2+2 2 | 2+2 2 | 3+3 2 | 2+2 2 | 2+2 2 | 2+2 2 | 3+3 2 | 3+3 2 |
| | 3+3 2 | 3+3 2 | 2+2 2 | 3+3 2 | 3+3 2 | 3+3 2 | 2+2 2 | 2+2 2 |
| | 3+3 2 | 2+2 2 | 2+2 2 | 3+3 2 | 3+3 2 | 2+2 2 | 3+3 2 | 3+3 2 |
| | 2+2 2 | 3+3 2 | 3+3 2 | 2+2 2 | 2+2 2 | 3+3 2 | 3+3 2 | 3+3 2 |
| | 3+3 2 | 3+3 2 | 3+3 2 | 2+2 2 | 2+2 2 | 3+3 2 | 2+2 2 | 2+2 2 |
| | 2+2 2 | 2+2 2 | 2+2 2 | 3+3 2 | 3+3 2 | 2+2 2 | 2+2 2 | 2+2 2 |
| | 3+3 2 | 3+3 2 | 3+3 2 | 2+2 2 | 3+3 2 | 3+3 2 | 3+3 2 | 2+2 2 |
| | 2+2 2 | 3+3 2 | 2+2 2 | 3+3 2 | 2+2 2 | 3+3 2 | 2+2 2 | 2+2 2 |
| | 3+3 2 | 2+2 2 | 3+3 2 | 2+2 2 | 2+2 2 | 2+2 2 | 2+2 2 | 2+2 2 |
| | 2+2 2 | 2+2 2 | 2+2 2 | 2+2 2 | 3+3 2 | 2+2 2 | 3+3 2 | 3+3 2 |
| | 3+3 2 | 3+3 2 | 3+3 2 | 3+3 2 | 3+3 2 | 2+2 2 | 3+3 2 | 3+3 2 |
| | 3+3 2 | 2+2 2 | 2+2 2 | 3+3 2 | 2+2 2 | 3+3 2 | 2+2 2 | 2+2 2 |
| | 2+2 2 | 3+3 2 | 3+3 2 | 3+3 2 | 2+2 2 | 3+3 2 | 3+3 2 | 3+3 2 |
| | 2+2 2 | 2+2 2 | 2+2 2 | 2+2 2 | 3+3 2 | 2+2 2 | 2+2 2 | 2+2 2 |
| | 3+3 2 | 2+2 2 | 3+3 2 | 2+2 2 | 3+3 2 | 2+2 2 | 3+3 2 | 3+3 2 |
| | 2+2 2 | 3+3 2 | 2+2 2 | 3+3 2 | 2+2 2 | 3+3 2 | 2+2 2 | 2+2 2 |
| | 3+3 2 | 3+3 2 | 2+2 2 | 2+2 2 | 2+2 2 | 2+2 2 | 3+3 2 | 3+3 2 |
| | 2+2 2 | 2+2 2 | 3+3 2 | 3+3 2 | 3+3 2 | 3+3 2 | 2+2 2 | 3+3 2 |

Continued on next column

**TABLE XXI– continued**

| TQC | ICSD | NoE | SF | | | | | |
|---|---|---|---|---|---|---|---|---|
| | 3+3 2 | 2+2 2 | 2+2 2 | 3+3 2 | 2+2 2 | 2+2 2 | 3+3 2 | 3+3 2 |
| | 2+2 2 | 3+3 2 | 3+3 2 | 2+2 2 | 2+2 2 | 3+3 2 | 2+2 2 | 2+2 2 |
| | 3+3 2 | 3+3 2 | 3+3 2 | 2+2 2 | 3+3 2 | 2+2 2 | 3+3 2 | 3+3 2 |
| 2.2.30 | 8065 | 174 | H1As1Ba1Cr2O10 | | | | | |
| | -GM | -R | -T | -U | -V | -X | -Y | -Z |
| | 3+3 2 | 2+2 2 | 2+2 2 | 3+3 2 | 2+2 2 | 3+3 2 | 2+2 2 | 3+3 2 |
| | 2+2 2 | 3+3 2 | 3+3 2 | 2+2 2 | 3+3 2 | 2+2 2 | 3+3 2 | 2+2 2 |
| | 3+3 2 | 3+3 2 | 3+3 2 | 3+3 2 | 3+3 2 | 3+3 2 | 3+3 2 | 3+3 2 |
| | 2+2 2 | 2+2 2 | 2+2 2 | 2+2 2 | 2+2 2 | 2+2 2 | 2+2 2 | 2+2 2 |
| | 3+3 2 | 3+3 2 | 3+3 2 | 2+2 2 | 2+2 2 | 3+3 2 | 2+2 2 | 2+2 2 |
| | 2+2 2 | 2+2 2 | 2+2 2 | 3+3 2 | 3+3 2 | 2+2 2 | 3+3 2 | 3+3 2 |
| | 3+3 2 | 3+3 2 | 3+3 2 | 3+3 2 | 3+3 2 | 3+3 2 | 3+3 2 | 3+3 2 |
| | 2+2 2 | 2+2 2 | 2+2 2 | 2+2 2 | 2+2 2 | 2+2 2 | 2+2 2 | 2+2 2 |
| | 3+3 2 | 3+3 2 | 3+3 2 | 3+3 2 | 3+3 2 | 3+3 2 | 3+3 2 | 3+3 2 |
| | 2+2 2 | 2+2 2 | 2+2 2 | 2+2 2 | 2+2 2 | 2+2 2 | 2+2 2 | 2+2 2 |
| | 3+3 2 | 3+3 2 | 2+2 2 | 3+3 2 | 2+2 2 | 2+2 2 | 3+3 2 | 2+2 2 |
| | 2+2 2 | 2+2 2 | 3+3 2 | 2+2 2 | 3+3 2 | 3+3 2 | 2+2 2 | 3+3 2 |
| | 3+3 2 | 2+2 2 | 3+3 2 | 2+2 2 | 3+3 2 | 3+3 2 | 3+3 2 | 3+3 2 |
| | 2+2 2 | 3+3 2 | 2+2 2 | 3+3 2 | 2+2 2 | 2+2 2 | 2+2 2 | 2+2 2 |
| | 3+3 2 | 2+2 2 | 2+2 2 | 3+3 2 | 2+2 2 | 2+2 2 | 3+3 2 | 2+2 2 |
| | 2+2 2 | 3+3 2 | 3+3 2 | 2+2 2 | 3+3 2 | 3+3 2 | 2+2 2 | 3+3 2 |
| | 3+3 2 | 3+3 2 | 2+2 2 | 3+3 2 | 2+2 2 | 3+3 2 | 3+3 2 | 2+2 2 |
| | 2+2 2 | 2+2 2 | 3+3 2 | 2+2 2 | 3+3 2 | 2+2 2 | 2+2 2 | 3+3 2 |
| | 3+3 2 | 2+2 2 | 3+3 2 | 2+2 2 | 3+3 2 | 3+3 2 | 2+2 2 | 3+3 2 |
| | 2+2 2 | 3+3 2 | 2+2 2 | 3+3 2 | 2+2 2 | 2+2 2 | 3+3 2 | 2+2 2 |
| | 3+3 2 | 3+3 2 | 2+2 2 | 2+2 2 | 3+3 2 | 2+2 2 | 2+2 2 | 3+3 2 |
| | 2+2 2 | 2+2 2 | 3+3 2 | 3+3 2 | 2+2 2 | 3+3 2 | 3+3 2 | 2+2 2 |
| | 3+3 2 | 2+2 2 | 2+2 2 | 3+3 2 | 2+2 2 | 3+3 2 | 2+2 2 | 3+3 2 |
| | 2+2 2 | 3+3 2 | 3+3 2 | 2+2 2 | 3+3 2 | 2+2 2 | 3+3 2 | 2+2 2 |
| | 3+3 2 | 2+2 2 | 2+2 2 | 3+3 2 | 2+2 2 | 3+3 2 | 2+2 2 | 3+3 2 |
| | 2+2 2 | 3+3 2 | 3+3 2 | 2+2 2 | 3+3 2 | 2+2 2 | 3+3 2 | 2+2 2 |
| | 3+3 2 | 2+2 2 | 2+2 2 | 3+3 2 | 2+2 2 | 3+3 2 | 2+2 2 | 3+3 2 |
| | 2+2 2 | 3+3 2 | 3+3 2 | 2+2 2 | 3+3 2 | 2+2 2 | 3+3 2 | 2+2 2 |
| | 3+3 2 | 3+3 2 | 3+3 2 | 2+2 2 | 3+3 2 | 2+2 2 | 3+3 2 | 2+2 2 |
| | 2+2 2 | 2+2 2 | 2+2 2 | 3+3 2 | 2+2 2 | 3+3 2 | 2+2 2 | 3+3 2 |
| | 2+2 2 | 2+2 2 | 2+2 2 | 3+3 2 | 3+3 2 | 2+2 2 | 3+3 2 | 3+3 2 |
| | 3+3 2 | 3+3 2 | 3+3 2 | 2+2 2 | 2+2 2 | 3+3 2 | 2+2 2 | 2+2 2 |
| | 2+2 2 | 2+2 2 | 2+2 2 | 3+3 2 | 2+2 2 | 2+2 2 | 2+2 2 | 2+2 2 |
| | 3+3 2 | 3+3 2 | 3+3 2 | 2+2 2 | 3+3 2 | 3+3 2 | 3+3 2 | 3+3 2 |
| | 3+3 2 | 3+3 2 | 3+3 2 | 3+3 2 | 2+2 2 | 3+3 2 | 3+3 2 | 3+3 2 |
| | 2+2 2 | 2+2 2 | 2+2 2 | 2+2 2 | 3+3 2 | 2+2 2 | 2+2 2 | 2+2 2 |
| | 3+3 2 | 3+3 2 | 2+2 2 | 3+3 2 | 2+2 2 | 2+2 2 | 3+3 2 | 2+2 2 |
| | 3+3 2 | 2+2 2 | 3+3 2 | 2+2 2 | 3+3 2 | 3+3 2 | 3+3 2 | 3+3 2 |
| | 2+2 2 | 2+2 2 | 2+2 2 | 2+2 2 | 2+2 2 | 2+2 2 | 2+2 2 | 2+2 2 |
| | 3+3 2 | 2+2 2 | 3+3 2 | 2+2 2 | 3+3 2 | 3+3 2 | 2+2 2 | 3+3 2 |
| | 2+2 2 | 3+3 2 | 2+2 2 | 3+3 2 | 3+3 2 | 3+3 2 | 3+3 2 | 2+2 2 |
| | 3+3 2 | 3+3 2 | 2+2 2 | 2+2 2 | 3+3 2 | 3+3 2 | 2+2 2 | 3+3 2 |
| | 2+2 2 | 2+2 2 | 3+3 2 | 3+3 2 | 2+2 2 | 2+2 2 | 2+2 2 | 2+2 2 |
| | 2+2 2 | 3+3 2 | 3+3 2 | 3+3 2 | 2+2 2 | 2+2 2 | 3+3 2 | 3+3 2 |
| | 3+3 2 | 3+3 2 | 3+3 2 | 2+2 2 | 3+3 2 | 2+2 2 | 2+2 2 | 2+2 2 |
| | 2+2 2 | 2+2 2 | 2+2 2 | 3+3 2 | 2+2 2 | 3+3 2 | 2+2 2 | 3+3 2 |
| | 3+3 2 | 3+3 2 | 3+3 2 | 3+3 2 | 2+2 2 | 3+3 2 | 3+3 2 | 2+2 2 |
| | 2+2 2 | 3+3 2 | 2+2 2 | 2+2 2 | 3+3 2 | 2+2 2 | 3+3 2 | 3+3 2 |
| | 2+2 2 | 2+2 2 | 2+2 2 | 3+3 2 | 3+3 2 | 3+3 2 | 2+2 2 | 3+3 2 |
| | 3+3 2 | 2+2 2 | 3+3 2 | 2+2 2 | 2+2 2 | 2+2 2 | 3+3 2 | 2+2 2 |
| | 3+3 2 | 2+2 2 | 2+2 2 | 2+2 2 | 3+3 2 | 3+3 2 | 2+2 2 | 2+2 2 |
| | 2+2 2 | 3+3 2 | 2+2 2 | 3+3 2 | 2+2 2 | 2+2 2 | 3+3 2 | 3+3 2 |
| | 3+3 2 | 2+2 2 | 3+3 2 | 3+3 2 | 2+2 2 | 2+2 2 | 3+3 2 | 3+3 2 |
| | 2+2 2 | 3+3 2 | 3+3 2 | 2+2 2 | 3+3 2 | 2+2 2 | 2+2 2 | 2+2 2 |
| | 3+3 2 | 3+3 2 | 2+2 2 | 3+3 2 | 2+2 2 | 3+3 2 | 3+3 2 | 3+3 2 |
| | 2+2 2 | 2+2 2 | 3+3 2 | 2+2 2 | 3+3 2 | 3+3 2 | 2+2 2 | 2+2 2 |

Continued on next column

| TQC | ICSD | NoE | SF | | | | | |
|---|---|---|---|---|---|---|---|---|
| | 3+3 2 | 3+3 2 | 3+3 2 | 2+2 2 | 2+2 2 | 3+3 2 | 3+3 2 | 3+3 2 |
| | 2+2 2 | 2+2 2 | 2+2 2 | 3+3 2 | 3+3 2 | 2+2 2 | 2+2 2 | 2+2 2 |
| | 3+3 2 | 3+3 2 | 3+3 2 | 3+3 2 | 2+2 2 | 2+2 2 | 2+2 2 | 3+3 2 |
| | 2+2 2 | 2+2 2 | 2+2 2 | 2+2 2 | 3+3 2 | 3+3 2 | 2+2 2 | 2+2 2 |
| | 3+3 2 | 2+2 2 | 3+3 2 | 3+3 2 | 2+2 2 | 2+2 2 | 3+3 2 | 2+2 2 |
| | 3+3 2 | 3+3 2 | 3+3 2 | 2+2 2 | 3+3 2 | 3+3 2 | 2+2 2 | 3+3 2 |
| | 2+2 2 | 3+3 2 | 2+2 2 | 3+3 2 | 3+3 2 | 2+2 2 | 3+3 2 | 2+2 2 |
| | 3+3 2 | 2+2 2 | 2+2 2 | 2+2 2 | 3+3 2 | 3+3 2 | 3+3 2 | 3+3 2 |
| | 2+2 2 | 2+2 2 | 3+3 2 | 2+2 2 | 2+2 2 | 3+3 2 | 3+3 2 | 3+3 2 |
| | 3+3 2 | 3+3 2 | 2+2 2 | 3+3 2 | 2+2 2 | 2+2 2 | 2+2 2 | 2+2 2 |
| | 2+2 2 | 3+3 2 | 3+3 2 | 3+3 2 | 3+3 2 | 3+3 2 | 2+2 2 | 2+2 2 |
| | 3+3 2 | 3+3 2 | 3+3 2 | 2+2 2 | 2+2 2 | 2+2 2 | 3+3 2 | 3+3 2 |
| | 2+2 2 | 2+2 2 | 2+2 2 | 3+3 2 | 2+2 2 | 3+3 2 | 3+3 2 | 3+3 2 |
| | 3+3 2 | 2+2 2 | 2+2 2 | 2+2 2 | 3+3 2 | 2+2 2 | 2+2 2 | 3+3 2 |
| | 2+2 2 | 3+3 2 | 3+3 2 | 2+2 2 | 2+2 2 | 3+3 2 | 3+3 2 | 2+2 2 |
| | 2+2 2 | 2+2 2 | 3+3 2 | 2+2 2 | 3+3 2 | 3+3 2 | 3+3 2 | 2+2 2 |
| | 3+3 2 | 3+3 2 | 2+2 2 | 3+3 2 | 2+2 2 | 2+2 2 | 2+2 2 | 3+3 2 |
| | 2+2 2 | 2+2 2 | 2+2 2 | 2+2 2 | 3+3 2 | 2+2 2 | 2+2 2 | 2+2 2 |
| | 3+3 2 | 3+3 2 | 2+2 2 | 3+3 2 | 2+2 2 | 3+3 2 | 3+3 2 | 2+2 2 |
| | 2+2 2 | 3+3 2 | 3+3 2 | 3+3 2 | 3+3 2 | 2+2 2 | 2+2 2 | 3+3 2 |
| | 3+3 2 | 2+2 2 | 3+3 2 | 2+2 2 | 2+2 2 | 3+3 2 | 3+3 2 | 2+2 2 |
| | 2+2 2 | 3+3 2 | 2+2 2 | 3+3 2 | 3+3 2 | 2+2 2 | 2+2 2 | 3+3 2 |
| | 3+3 2 | 2+2 2 | 3+3 2 | 3+3 2 | 2+2 2 | 3+3 2 | 3+3 2 | 3+3 2 |
| | 2+2 2 | 2+2 2 | 2+2 2 | 2+2 2 | 3+3 2 | 2+2 2 | 2+2 2 | 2+2 2 |
| | 3+3 2 | 3+3 2 | 3+3 2 | 2+2 2 | 3+3 2 | 3+3 2 | 3+3 2 | 3+3 2 |
| | 3+3 2 | 2+2 2 | 2+2 2 | 3+3 2 | 2+2 2 | 2+2 2 | 2+2 2 | 2+2 2 |
| | 2+2 2 | 3+3 2 | 3+3 2 | 3+3 2 | 3+3 2 | 3+3 2 | 3+3 2 | 2+2 2 |
| | 3+3 2 | 2+2 2 | 2+2 2 | 2+2 2 | 2+2 2 | 2+2 2 | 2+2 2 | 3+3 2 |
| | 2+2 2 | 3+3 2 | 3+3 2 | 3+3 2 | 3+3 2 | 3+3 2 | 3+3 2 | 3+3 2 |
| | 2+2 2 | 3+3 2 | 2+2 2 | 3+3 2 | 2+2 2 | 2+2 2 | 3+3 2 | 2+2 2 |
| | 3+3 2 | 2+2 2 | 2+2 2 | 2+2 2 | 3+3 2 | 3+3 2 | 2+2 2 | 3+3 2 |
| 2.2.31 | 73197 | 78 | H18Al2Cl6N6 | | | | | |
| | -GM | -R | -T | -U | -V | -X | -Y | -Z |
| | 3+3 2 | 3+3 2 | 3+3 2 | 3+3 2 | 3+3 2 | 3+3 2 | 3+3 2 | 3+3 2 |
| | 3+3 2 | 2+2 2 | 3+3 2 | 3+3 2 | 3+3 2 | 2+2 2 | 2+2 2 | 2+2 2 |
| | 2+2 2 | 2+2 2 | 2+2 2 | 2+2 2 | 2+2 2 | 2+2 2 | 2+2 2 | 2+2 2 |
| | 2+2 2 | 3+3 2 | 2+2 2 | 2+2 2 | 2+2 2 | 3+3 2 | 3+3 2 | 3+3 2 |
| | 2+2 2 | 3+3 2 | 2+2 2 | 2+2 2 | 2+2 2 | 3+3 2 | 3+3 2 | 3+3 2 |
| | 3+3 2 | 2+2 2 | 3+3 2 | 3+3 2 | 3+3 2 | 2+2 2 | 2+2 2 | 2+2 2 |
| | 3+3 2 | 3+3 2 | 3+3 2 | 3+3 2 | 3+3 2 | 3+3 2 | 3+3 2 | 3+3 2 |
| | 2+2 2 | 2+2 2 | 2+2 2 | 2+2 2 | 2+2 2 | 2+2 2 | 2+2 2 | 2+2 2 |
| | 2+2 2 | 2+2 2 | 2+2 2 | 2+2 2 | 2+2 2 | 2+2 2 | 2+2 2 | 2+2 2 |
| | 2+2 2 | 3+3 2 | 2+2 2 | 2+2 2 | 2+2 2 | 3+3 2 | 3+3 2 | 3+3 2 |
| | 3+3 2 | 2+2 2 | 3+3 2 | 3+3 2 | 3+3 2 | 2+2 2 | 2+2 2 | 2+2 2 |
| | 3+3 2 | 3+3 2 | 3+3 2 | 3+3 2 | 3+3 2 | 3+3 2 | 3+3 2 | 3+3 2 |
| | 3+3 2 | 2+2 2 | 3+3 2 | 3+3 2 | 3+3 2 | 2+2 2 | 2+2 2 | 2+2 2 |
| | 3+3 2 | 3+3 2 | 3+3 2 | 3+3 2 | 3+3 2 | 3+3 2 | 3+3 2 | 3+3 2 |
| | 2+2 2 | 3+3 2 | 2+2 2 | 2+2 2 | 2+2 2 | 2+2 2 | 2+2 2 | 2+2 2 |
| | 2+2 2 | 2+2 2 | 2+2 2 | 2+2 2 | 2+2 2 | 3+3 2 | 3+3 2 | 3+3 2 |
| | 2+2 2 | 3+3 2 | 2+2 2 | 2+2 2 | 2+2 2 | 3+3 2 | 3+3 2 | 3+3 2 |
| | 3+3 2 | 2+2 2 | 3+3 2 | 3+3 2 | 3+3 2 | 2+2 2 | 3+3 2 | 3+3 2 |
| | 3+3 2 | 3+3 2 | 3+3 2 | 2+2 2 | 2+2 2 | 2+2 2 | 2+2 2 | 2+2 2 |
| | 3+3 2 | 2+2 2 | 2+2 2 | 3+3 2 | 3+3 2 | 3+3 2 | 2+2 2 | 2+2 2 |
| | 2+2 2 | 2+2 2 | 3+3 2 | 2+2 2 | 2+2 2 | 3+3 2 | 3+3 2 | 2+2 2 |
| | 2+2 2 | 2+2 2 | 3+3 2 | 3+3 2 | 3+3 2 | 2+2 2 | 2+2 2 | 3+3 2 |
| | 3+3 2 | 3+3 2 | 2+2 2 | 2+2 2 | 2+2 2 | 2+2 2 | 2+2 2 | 3+3 2 |
| | 2+2 2 | 3+3 2 | 2+2 2 | 3+3 2 | 3+3 2 | 3+3 2 | 3+3 2 | 2+2 2 |
| | 3+3 2 | 3+3 2 | 3+3 2 | 3+3 2 | 3+3 2 | 2+2 2 | 3+3 2 | 3+3 2 |
| | 2+2 2 | 2+2 2 | 2+2 2 | 3+3 2 | 3+3 2 | 3+3 2 | 2+2 2 | 2+2 2 |
| | 3+3 2 | 2+2 2 | 3+3 2 | 2+2 2 | 2+2 2 | 2+2 2 | 2+2 2 | 2+2 2 |
| | 2+2 2 | 3+3 2 | 2+2 2 | 3+3 2 | 3+3 2 | 3+3 2 | 3+3 2 | 2+2 2 |

Continued on next column

**TABLE XXI– continued**

| TQC | ICSD | NoE | SF | | | | |
|---|---|---|---|---|---|---|---|
| | 3+3 2 | 2+2 2 | 2+2 2 | 2+2 2 | 2+2 2 | 3+3 2 | 2+2 2 | 3+3 2 |
| | 2+2 2 | 3+3 2 | 3+3 2 | 3+3 2 | 2+2 2 | 2+2 2 | 3+3 2 | 3+3 2 |
| | 2+2 2 | 2+2 2 | 3+3 2 | 2+2 2 | 2+2 2 | 3+3 2 | 2+2 2 | 2+2 2 |
| | 3+3 2 | 3+3 2 | 2+2 2 | 2+2 2 | 3+3 2 | 3+3 2 | 3+3 2 | 3+3 2 |
| | 2+2 2 | 3+3 2 | 2+2 2 | 2+2 2 | 2+2 2 | 2+2 2 | 3+3 2 | 3+3 2 |
| | 3+3 2 | 2+2 2 | 2+2 2 | 2+2 2 | 2+2 2 | 3+3 2 | 3+3 2 | 2+2 2 |
| | 3+3 2 | 3+3 2 | 3+3 2 | 3+3 2 | 3+3 2 | 2+2 2 | 2+2 2 | 2+2 2 |
| | 2+2 2 | 2+2 2 | 2+2 2 | 3+3 2 | 3+3 2 | 2+2 2 | 2+2 2 | 3+3 2 |
| | 3+3 2 | 3+3 2 | 3+3 2 | 3+3 2 | 2+2 2 | 2+2 2 | 2+2 2 | 3+3 2 |
| | 2+2 2 | 2+2 2 | 2+2 2 | 2+2 2 | 3+3 2 | 3+3 2 | 3+3 2 | 2+2 2 |
| | 3+3 2 | 3+3 2 | 3+3 2 | 2+2 2 | 2+2 2 | 3+3 2 | 2+2 2 | 3+3 2 |
| 2.2.32 | 417328 | 118 | H12Cs4O6P2S6 | | | | | |
| | -GM | -R | -T | -U | -V | -X | -Y | -Z |
| | 3+3 2 | 2+2 2 | 2+2 2 | 2+2 2 | 3+3 2 | 3+3 2 | 3+3 2 | 2+2 2 |
| | 2+2 2 | 3+3 2 | 3+3 2 | 3+3 2 | 2+2 2 | 2+2 2 | 2+2 2 | 3+3 2 |
| | 3+3 2 | 2+2 2 | 3+3 2 | 2+2 2 | 2+2 2 | 2+2 2 | 3+3 2 | 3+3 2 |
| | 2+2 2 | 3+3 2 | 2+2 2 | 3+3 2 | 3+3 2 | 3+3 2 | 2+2 2 | 2+2 2 |
| | 3+3 2 | 2+2 2 | 2+2 2 | 3+3 2 | 2+2 2 | 3+3 2 | 2+2 2 | 3+3 2 |
| | 2+2 2 | 3+3 2 | 3+3 2 | 2+2 2 | 3+3 2 | 2+2 2 | 3+3 2 | 2+2 2 |
| | 2+2 2 | 3+3 2 | 3+3 2 | 2+2 2 | 3+3 2 | 2+2 2 | 3+3 2 | 2+2 2 |
| | 3+3 2 | 2+2 2 | 2+2 2 | 3+3 2 | 2+2 2 | 3+3 2 | 2+2 2 | 3+3 2 |
| | 3+3 2 | 2+2 2 | 2+2 2 | 3+3 2 | 2+2 2 | 3+3 2 | 2+2 2 | 3+3 2 |
| | 2+2 2 | 3+3 2 | 3+3 2 | 2+2 2 | 3+3 2 | 2+2 2 | 3+3 2 | 2+2 2 |
| | 3+3 2 | 3+3 2 | 2+2 2 | 3+3 2 | 2+2 2 | 2+2 2 | 3+3 2 | 2+2 2 |
| | 2+2 2 | 2+2 2 | 3+3 2 | 2+2 2 | 3+3 2 | 3+3 2 | 2+2 2 | 3+3 2 |
| | 2+2 2 | 2+2 2 | 3+3 2 | 2+2 2 | 3+3 2 | 3+3 2 | 2+2 2 | 3+3 2 |
| | 2+2 2 | 2+2 2 | 3+3 2 | 2+2 2 | 3+3 2 | 3+3 2 | 2+2 2 | 3+3 2 |
| | 3+3 2 | 3+3 2 | 2+2 2 | 3+3 2 | 2+2 2 | 2+2 2 | 3+3 2 | 2+2 2 |
| | 3+3 2 | 3+3 2 | 2+2 2 | 3+3 2 | 2+2 2 | 2+2 2 | 3+3 2 | 2+2 2 |
| | 3+3 2 | 3+3 2 | 2+2 2 | 3+3 2 | 2+2 2 | 2+2 2 | 3+3 2 | 2+2 2 |
| | 3+3 2 | 2+2 2 | 3+3 2 | 2+2 2 | 3+3 2 | 3+3 2 | 2+2 2 | 2+2 2 |
| | 2+2 2 | 3+3 2 | 2+2 2 | 3+3 2 | 2+2 2 | 2+2 2 | 3+3 2 | 3+3 2 |
| | 3+3 2 | 2+2 2 | 3+3 2 | 2+2 2 | 3+3 2 | 3+3 2 | 3+3 2 | 3+3 2 |
| | 2+2 2 | 3+3 2 | 2+2 2 | 3+3 2 | 2+2 2 | 2+2 2 | 2+2 2 | 2+2 2 |
| | 2+2 2 | 2+2 2 | 3+3 2 | 2+2 2 | 3+3 2 | 3+3 2 | 2+2 2 | 3+3 2 |
| | 3+3 2 | 2+2 2 | 2+2 2 | 2+2 2 | 3+3 2 | 3+3 2 | 3+3 2 | 2+2 2 |
| | 2+2 2 | 3+3 2 | 3+3 2 | 3+3 2 | 2+2 2 | 2+2 2 | 2+2 2 | 2+2 2 |
| | 3+3 2 | 2+2 2 | 2+2 2 | 2+2 2 | 2+2 2 | 3+3 2 | 3+3 2 | 3+3 2 |
| | 2+2 2 | 3+3 2 | 3+3 2 | 3+3 2 | 3+3 2 | 2+2 2 | 2+2 2 | 3+3 2 |
| | 3+3 2 | 3+3 2 | 3+3 2 | 3+3 2 | 2+2 2 | 2+2 2 | 3+3 2 | 3+3 2 |
| | 2+2 2 | 2+2 2 | 2+2 2 | 2+2 2 | 3+3 2 | 3+3 2 | 2+2 2 | 2+2 2 |
| | 3+3 2 | 3+3 2 | 3+3 2 | 3+3 2 | 2+2 2 | 2+2 2 | 3+3 2 | 3+3 2 |
| | 2+2 2 | 2+2 2 | 2+2 2 | 2+2 2 | 3+3 2 | 3+3 2 | 2+2 2 | 2+2 2 |
| | 3+3 2 | 3+3 2 | 2+2 2 | 3+3 2 | 2+2 2 | 2+2 2 | 3+3 2 | 2+2 2 |
| | 2+2 2 | 2+2 2 | 3+3 2 | 2+2 2 | 3+3 2 | 3+3 2 | 2+2 2 | 3+3 2 |
| | 2+2 2 | 2+2 2 | 3+3 2 | 2+2 2 | 3+3 2 | 3+3 2 | 2+2 2 | 3+3 2 |
| | 3+3 2 | 3+3 2 | 2+2 2 | 3+3 2 | 2+2 2 | 2+2 2 | 3+3 2 | 2+2 2 |
| | 3+3 2 | 3+3 2 | 2+2 2 | 3+3 2 | 2+2 2 | 2+2 2 | 3+3 2 | 2+2 2 |
| | 3+3 2 | 3+3 2 | 3+3 2 | 3+3 2 | 2+2 2 | 2+2 2 | 2+2 2 | 3+3 2 |
| | 2+2 2 | 2+2 2 | 2+2 2 | 2+2 2 | 3+3 2 | 3+3 2 | 3+3 2 | 2+2 2 |
| | 3+3 2 | 3+3 2 | 2+2 2 | 3+3 2 | 2+2 2 | 2+2 2 | 3+3 2 | 2+2 2 |
| | 2+2 2 | 2+2 2 | 3+3 2 | 2+2 2 | 3+3 2 | 3+3 2 | 2+2 2 | 3+3 2 |
| | 2+2 2 | 2+2 2 | 3+3 2 | 2+2 2 | 3+3 2 | 3+3 2 | 2+2 2 | 3+3 2 |
| | 2+2 2 | 2+2 2 | 2+2 2 | 2+2 2 | 3+3 2 | 3+3 2 | 2+2 2 | 3+3 2 |
| | 3+3 2 | 3+3 2 | 3+3 2 | 3+3 2 | 2+2 2 | 2+2 2 | 3+3 2 | 2+2 2 |
| | 2+2 2 | 3+3 2 | 2+2 2 | 2+2 2 | 3+3 2 | 2+2 2 | 3+3 2 | 3+3 2 |
| | 3+3 2 | 2+2 2 | 3+3 2 | 2+2 2 | 3+3 2 | 3+3 2 | 2+2 2 | 2+2 2 |
| | 2+2 2 | 2+2 2 | 3+3 2 | 3+3 2 | 2+2 2 | 3+3 2 | 2+2 2 | 2+2 2 |
| | 3+3 2 | 3+3 2 | 2+2 2 | 2+2 2 | 2+2 2 | 2+2 2 | 3+3 2 | 3+3 2 |
| | 2+2 2 | 2+2 2 | 2+2 2 | 3+3 2 | 3+3 2 | 3+3 2 | 2+2 2 | 3+3 2 |
| | 2+2 2 | 3+3 2 | 3+3 2 | 3+3 2 | 3+3 2 | 2+2 2 | 3+3 2 | 2+2 2 |

Continued on next column

**TABLE XXI– continued**

| TQC | ICSD | NoE | SF | | | | |
|---|---|---|---|---|---|---|---|
| | 3+3 2 | 2+2 2 | 3+3 2 | 2+2 2 | 2+2 2 | 3+3 2 | 2+2 2 | 3+3 2 |
| | 3+3 2 | 2+2 2 | 3+3 2 | 2+2 2 | 3+3 2 | 3+3 2 | 2+2 2 | 3+3 2 |
| | 2+2 2 | 3+3 2 | 2+2 2 | 3+3 2 | 2+2 2 | 2+2 2 | 3+3 2 | 2+2 2 |
| | 3+3 2 | 3+3 2 | 3+3 2 | 2+2 2 | 3+3 2 | 2+2 2 | 3+3 2 | 2+2 2 |
| | 3+3 2 | 3+3 2 | 2+2 2 | 3+3 2 | 2+2 2 | 2+2 2 | 2+2 2 | 2+2 2 |
| | 2+2 2 | 3+3 2 | 3+3 2 | 3+3 2 | 3+3 2 | 3+3 2 | 3+3 2 | 3+3 2 |
| | 2+2 2 | 2+2 2 | 2+2 2 | 2+2 2 | 2+2 2 | 2+2 2 | 2+2 2 | 3+3 2 |
| | 3+3 2 | 3+3 2 | 2+2 2 | 2+2 2 | 3+3 2 | 2+2 2 | 3+3 2 | 2+2 2 |
| | 2+2 2 | 2+2 2 | 2+2 2 | 3+3 2 | 2+2 2 | 3+3 2 | 2+2 2 | 3+3 2 |
| | 3+3 2 | 2+2 2 | 3+3 2 | 3+3 2 | 2+2 2 | 3+3 2 | 2+2 2 | 2+2 2 |
| | 2+2 2 | 2+2 2 | 2+2 2 | 3+3 2 | 2+2 2 | 3+3 2 | 3+3 2 | 2+2 2 |
| 2.2.33 | 35417 | 130 | H12Cl8Hg3Mg1O6 | | | | | |
| | -GM | -R | -T | -U | -V | -X | -Y | -Z |
| | 3+3 2 | 3+3 2 | 3+3 2 | 2+2 2 | 2+2 2 | 3+3 2 | 2+2 2 | 2+2 2 |
| | 2+2 2 | 2+2 2 | 2+2 2 | 3+3 2 | 3+3 2 | 2+2 2 | 3+3 2 | 3+3 2 |
| | 2+2 2 | 2+2 2 | 2+2 2 | 3+3 2 | 3+3 2 | 2+2 2 | 3+3 2 | 3+3 2 |
| | 2+2 2 | 2+2 2 | 2+2 2 | 3+3 2 | 3+3 2 | 2+2 2 | 3+3 2 | 3+3 2 |
| | 3+3 2 | 3+3 2 | 3+3 2 | 2+2 2 | 2+2 2 | 3+3 2 | 2+2 2 | 2+2 2 |
| | 3+3 2 | 3+3 2 | 3+3 2 | 2+2 2 | 2+2 2 | 3+3 2 | 2+2 2 | 2+2 2 |
| | 3+3 2 | 3+3 2 | 3+3 2 | 2+2 2 | 3+3 2 | 2+2 2 | 3+3 2 | 3+3 2 |
| | 3+3 2 | 2+2 2 | 3+3 2 | 3+3 2 | 2+2 2 | 3+3 2 | 3+3 2 | 3+3 2 |
| | 2+2 2 | 3+3 2 | 2+2 2 | 3+3 2 | 3+3 2 | 3+3 2 | 2+2 2 | 2+2 2 |
| | 2+2 2 | 3+3 2 | 2+2 2 | 3+3 2 | 2+2 2 | 2+2 2 | 2+2 2 | 2+2 2 |
| | 3+3 2 | 2+2 2 | 2+2 2 | 2+2 2 | 2+2 2 | 2+2 2 | 3+3 2 | 2+2 2 |
| | 2+2 2 | 2+2 2 | 3+3 2 | 2+2 2 | 3+3 2 | 3+3 2 | 2+2 2 | 3+3 2 |
| | 3+3 2 | 3+3 2 | 2+2 2 | 2+2 2 | 3+3 2 | 2+2 2 | 2+2 2 | 3+3 2 |
| | 2+2 2 | 2+2 2 | 3+3 2 | 3+3 2 | 2+2 2 | 3+3 2 | 3+3 2 | 2+2 2 |
| | 3+3 2 | 3+3 2 | 3+3 2 | 3+3 2 | 3+3 2 | 3+3 2 | 3+3 2 | 3+3 2 |
| | 3+3 2 | 2+2 2 | 3+3 2 | 2+2 2 | 2+2 2 | 2+2 2 | 3+3 2 | 3+3 2 |
| | 2+2 2 | 3+3 2 | 2+2 2 | 3+3 2 | 3+3 2 | 3+3 2 | 2+2 2 | 2+2 2 |
| | 3+3 2 | 3+3 2 | 3+3 2 | 2+2 2 | 3+3 2 | 2+2 2 | 3+3 2 | 3+3 2 |
| | 3+3 2 | 3+3 2 | 2+2 2 | 3+3 2 | 3+3 2 | 3+3 2 | 2+2 2 | 3+3 2 |
| | 2+2 2 | 2+2 2 | 3+3 2 | 3+3 2 | 2+2 2 | 3+3 2 | 3+3 2 | 2+2 2 |
| | 3+3 2 | 3+3 2 | 3+3 2 | 3+3 2 | 3+3 2 | 3+3 2 | 3+3 2 | 3+3 2 |
| | 3+3 2 | 2+2 2 | 3+3 2 | 3+3 2 | 2+2 2 | 2+2 2 | 3+3 2 | 3+3 2 |
| | 2+2 2 | 3+3 2 | 2+2 2 | 2+2 2 | 3+3 2 | 3+3 2 | 2+2 2 | 2+2 2 |
| | 3+3 2 | 3+3 2 | 3+3 2 | 2+2 2 | 3+3 2 | 2+2 2 | 3+3 2 | 3+3 2 |
| | 3+3 2 | 2+2 2 | 2+2 2 | 3+3 2 | 2+2 2 | 3+3 2 | 2+2 2 | 3+3 2 |
| | 2+2 2 | 3+3 2 | 3+3 2 | 3+3 2 | 3+3 2 | 3+3 2 | 3+3 2 | 2+2 2 |
| | 3+3 2 | 3+3 2 | 2+2 2 | 2+2 2 | 3+3 2 | 2+2 2 | 2+2 2 | 3+3 2 |
| | 2+2 2 | 2+2 2 | 3+3 2 | 3+3 2 | 2+2 2 | 3+3 2 | 3+3 2 | 2+2 2 |
| | 3+3 2 | 3+3 2 | 3+3 2 | 3+3 2 | 3+3 2 | 3+3 2 | 3+3 2 | 3+3 2 |
| | 2+2 2 | 2+2 2 | 3+3 2 | 3+3 2 | 2+2 2 | 3+3 2 | 3+3 2 | 2+2 2 |
| | 3+3 2 | 3+3 2 | 2+2 2 | 2+2 2 | 3+3 2 | 2+2 2 | 2+2 2 | 3+3 2 |
| | 3+3 2 | 3+3 2 | 3+3 2 | 3+3 2 | 3+3 2 | 3+3 2 | 3+3 2 | 3+3 2 |
| | 2+2 2 | 2+2 2 | 2+2 2 | 2+2 2 | 2+2 2 | 2+2 2 | 2+2 2 | 2+2 2 |
| | 2+2 2 | 3+3 2 | 2+2 2 | 2+2 2 | 2+2 2 | 2+2 2 | 3+3 2 | 2+2 2 |
| | 3+3 2 | 2+2 2 | 3+3 2 | 3+3 2 | 2+2 2 | 2+2 2 | 2+2 2 | 2+2 2 |
| | 3+3 2 | 3+3 2 | 3+3 2 | 2+2 2 | 2+2 2 | 3+3 2 | 2+2 2 | 3+3 2 |
| | 3+3 2 | 2+2 2 | 2+2 2 | 2+2 2 | 3+3 2 | 3+3 2 | 2+2 2 | 2+2 2 |
| | 2+2 2 | 2+2 2 | 3+3 2 | 2+2 2 | 3+3 2 | 2+2 2 | 3+3 2 | 3+3 2 |
| | 3+3 2 | 3+3 2 | 2+2 2 | 3+3 2 | 2+2 2 | 3+3 2 | 2+2 2 | 2+2 2 |
| | 2+2 2 | 3+3 2 | 2+2 2 | 3+3 2 | 2+2 2 | 2+2 2 | 3+3 2 | 3+3 2 |
| | 2+2 2 | 2+2 2 | 3+3 2 | 3+3 2 | 3+3 2 | 3+3 2 | 2+2 2 | 3+3 2 |
| | 3+3 2 | 2+2 2 | 2+2 2 | 2+2 2 | 3+3 2 | 2+2 2 | 3+3 2 | 3+3 2 |
| | 2+2 2 | 3+3 2 | 2+2 2 | 2+2 2 | 2+2 2 | 2+2 2 | 2+2 2 | 2+2 2 |
| | 2+2 2 | 2+2 2 | 3+3 2 | 3+3 2 | 3+3 2 | 3+3 2 | 3+3 2 | 3+3 2 |
| | 3+3 2 | 3+3 2 | 3+3 2 | 2+2 2 | 2+2 2 | 2+2 2 | 2+2 2 | 2+2 2 |
| | 3+3 2 | 2+2 2 | 2+2 2 | 2+2 2 | 3+3 2 | 3+3 2 | 3+3 2 | 2+2 2 |
| | 2+2 2 | 2+2 2 | 3+3 2 | 3+3 2 | 3+3 2 | 2+2 2 | 3+3 2 | 2+2 2 |
| | 2+2 2 | 3+3 2 | 2+2 2 | 2+2 2 | 2+2 2 | 2+2 2 | 2+2 2 | 3+3 2 |

Continued on next column

**TABLE XXI– continued**

| TQC | ICSD | NoE | SF | | | | | |
|---|---|---|---|---|---|---|---|---|
| | 3+3 2 | 2+2 2 | 2+2 2 | 3+3 2 | 3+3 2 | 3+3 2 | 2+2 2 | 2+2 2 |
| | 2+2 2 | 3+3 2 | 2+2 2 | 2+2 2 | 2+2 2 | 2+2 2 | 3+3 2 | 3+3 2 |
| | 2+2 2 | 2+2 2 | 3+3 2 | 3+3 2 | 2+2 2 | 3+3 2 | 2+2 2 | 2+2 2 |
| | 3+3 2 | 2+2 2 | 2+2 2 | 3+3 2 | 3+3 2 | 2+2 2 | 3+3 2 | 3+3 2 |
| | 2+2 2 | 2+2 2 | 2+2 2 | 2+2 2 | 3+3 2 | 3+3 2 | 3+3 2 | 3+3 2 |
| | 3+3 2 | 3+3 2 | 3+3 2 | 3+3 2 | 3+3 2 | 2+2 2 | 2+2 2 | 2+2 2 |
| | 2+2 2 | 3+3 2 | 3+3 2 | 3+3 2 | 2+2 2 | 3+3 2 | 2+2 2 | 2+2 2 |
| | 3+3 2 | 3+3 2 | 2+2 2 | 2+2 2 | 2+2 2 | 3+3 2 | 2+2 2 | 3+3 2 |
| | 2+2 2 | 3+3 2 | 3+3 2 | 3+3 2 | 3+3 2 | 2+2 2 | 3+3 2 | 2+2 2 |
| | 3+3 2 | 2+2 2 | 3+3 2 | 2+2 2 | 2+2 2 | 3+3 2 | 3+3 2 | 3+3 2 |
| | 2+2 2 | 3+3 2 | 3+3 2 | 3+3 2 | 2+2 2 | 2+2 2 | 3+3 2 | 2+2 2 |
| | 3+3 2 | 3+3 2 | 3+3 2 | 2+2 2 | 3+3 2 | 3+3 2 | 2+2 2 | 3+3 2 |
| | 3+3 2 | 3+3 2 | 2+2 2 | 2+2 2 | 2+2 2 | 3+3 2 | 2+2 2 | 2+2 2 |
| | 3+3 2 | 2+2 2 | 3+3 2 | 3+3 2 | 2+2 2 | 2+2 2 | 3+3 2 | 3+3 2 |
| | 2+2 2 | 2+2 2 | 2+2 2 | 3+3 2 | 3+3 2 | 3+3 2 | 3+3 2 | 2+2 2 |
| | 3+3 2 | 2+2 2 | 2+2 2 | 2+2 2 | 3+3 2 | 2+2 2 | 2+2 2 | 3+3 2 |
| | 2+2 2 | 3+3 2 | 2+2 2 | 3+3 2 | 3+3 2 | 3+3 2 | 3+3 2 | 3+3 2 |
| 2.2.34 | 194210 | 68 | H12Cl4N4Pt2 | | | | | |
| | -GM | -R | -T | -U | -V | -X | -Y | -Z |
| | 3+3 2 | 3+3 2 | 3+3 2 | 3+3 2 | 3+3 2 | 3+3 2 | 3+3 2 | 3+3 2 |
| | 2+2 2 | 2+2 2 | 2+2 2 | 2+2 2 | 2+2 2 | 2+2 2 | 2+2 2 | 2+2 2 |
| | 3+3 2 | 3+3 2 | 3+3 2 | 3+3 2 | 3+3 2 | 3+3 2 | 3+3 2 | 3+3 2 |
| | 2+2 2 | 2+2 2 | 2+2 2 | 2+2 2 | 2+2 2 | 2+2 2 | 2+2 2 | 2+2 2 |
| | 3+3 2 | 2+2 2 | 3+3 2 | 3+3 2 | 3+3 2 | 2+2 2 | 2+2 2 | 2+2 2 |
| | 3+3 2 | 2+2 2 | 3+3 2 | 3+3 2 | 3+3 2 | 2+2 2 | 2+2 2 | 2+2 2 |
| | 2+2 2 | 3+3 2 | 2+2 2 | 2+2 2 | 2+2 2 | 3+3 2 | 3+3 2 | 3+3 2 |
| | 2+2 2 | 3+3 2 | 2+2 2 | 2+2 2 | 2+2 2 | 3+3 2 | 3+3 2 | 3+3 2 |
| | 3+3 2 | 3+3 2 | 3+3 2 | 3+3 2 | 3+3 2 | 3+3 2 | 3+3 2 | 3+3 2 |
| | 3+3 2 | 3+3 2 | 3+3 2 | 3+3 2 | 3+3 2 | 2+2 2 | 2+2 2 | 3+3 2 |
| | 3+3 2 | 2+2 2 | 3+3 2 | 3+3 2 | 3+3 2 | 3+3 2 | 3+3 2 | 2+2 2 |
| | 3+3 2 | 2+2 2 | 3+3 2 | 3+3 2 | 3+3 2 | 2+2 2 | 2+2 2 | 2+2 2 |
| | 3+3 2 | 2+2 2 | 3+3 2 | 3+3 2 | 3+3 2 | 2+2 2 | 2+2 2 | 2+2 2 |
| | 3+3 2 | 3+3 2 | 3+3 2 | 3+3 2 | 3+3 2 | 3+3 2 | 3+3 2 | 3+3 2 |
| | 3+3 2 | 2+2 2 | 3+3 2 | 3+3 2 | 3+3 2 | 2+2 2 | 2+2 2 | 2+2 2 |
| | 3+3 2 | 3+3 2 | 3+3 2 | 3+3 2 | 3+3 2 | 3+3 2 | 3+3 2 | 3+3 2 |
| | 3+3 2 | 2+2 2 | 2+2 2 | 3+3 2 | 3+3 2 | 2+2 2 | 2+2 2 | 2+2 2 |
| | 2+2 2 | 2+2 2 | 3+3 2 | 2+2 2 | 2+2 2 | 2+2 2 | 2+2 2 | 2+2 2 |
| | 3+3 2 | 3+3 2 | 2+2 2 | 2+2 2 | 2+2 2 | 2+2 2 | 3+3 2 | 3+3 2 |
| | 2+2 2 | 2+2 2 | 3+3 2 | 3+3 2 | 3+3 2 | 3+3 2 | 2+2 2 | 2+2 2 |
| | 2+2 2 | 2+2 2 | 2+2 2 | 2+2 2 | 2+2 2 | 2+2 2 | 2+2 2 | 2+2 2 |
| | 2+2 2 | 2+2 2 | 2+2 2 | 2+2 2 | 2+2 2 | 3+3 2 | 3+3 2 | 3+3 2 |
| | 3+3 2 | 3+3 2 | 2+2 2 | 3+3 2 | 3+3 2 | 2+2 2 | 2+2 2 | 2+2 2 |
| | 2+2 2 | 3+3 2 | 3+3 2 | 2+2 2 | 2+2 2 | 3+3 2 | 3+3 2 | 2+2 2 |
| | 2+2 2 | 2+2 2 | 2+2 2 | 2+2 2 | 2+2 2 | 2+2 2 | 2+2 2 | 3+3 2 |
| | 2+2 2 | 3+3 2 | 3+3 2 | 3+3 2 | 3+3 2 | 3+3 2 | 3+3 2 | 2+2 2 |
| | 3+3 2 | 3+3 2 | 2+2 2 | 2+2 2 | 2+2 2 | 2+2 2 | 2+2 2 | 3+3 2 |
| | 2+2 2 | 2+2 2 | 2+2 2 | 2+2 2 | 2+2 2 | 3+3 2 | 3+3 2 | 2+2 2 |
| | 3+3 2 | 2+2 2 | 3+3 2 | 2+2 2 | 2+2 2 | 2+2 2 | 2+2 2 | 3+3 2 |
| | 3+3 2 | 3+3 2 | 2+2 2 | 3+3 2 | 3+3 2 | 3+3 2 | 3+3 2 | 3+3 2 |
| | 2+2 2 | 3+3 2 | 3+3 2 | 3+3 2 | 3+3 2 | 3+3 2 | 3+3 2 | 3+3 2 |
| | 2+2 2 | 3+3 2 | 3+3 2 | 3+3 2 | 3+3 2 | 3+3 2 | 3+3 2 | 3+3 2 |
| | 3+3 2 | 3+3 2 | 2+2 2 | 2+2 2 | 2+2 2 | 3+3 2 | 3+3 2 | 3+3 2 |
| | 3+3 2 | 3+3 2 | 2+2 2 | 2+2 2 | 2+2 2 | 3+3 2 | 3+3 2 | 3+3 2 |
| 2.2.35 | 240179 | 84 | Ga5Rh3 | | | | | |
| | -GM | -R | -T | -U | -V | -X | -Y | -Z |
| | 3+3 2 | 2+2 2 | 3+3 2 | 3+3 2 | 2+2 2 | 3+3 2 | 2+2 2 | 3+3 2 |
| | 3+3 2 | 3+3 2 | 2+2 2 | 2+2 2 | 3+3 2 | 2+2 2 | 3+3 2 | 2+2 2 |
| | 2+2 2 | 3+3 2 | 3+3 2 | 3+3 2 | 3+3 2 | 3+3 2 | 2+2 2 | 2+2 2 |
| | 3+3 2 | 2+2 2 | 2+2 2 | 2+2 2 | 2+2 2 | 2+2 2 | 3+3 2 | 3+3 2 |
| | 2+2 2 | 2+2 2 | 3+3 2 | 2+2 2 | 2+2 2 | 3+3 2 | 3+3 2 | 2+2 2 |
| | 2+2 2 | 3+3 2 | 2+2 2 | 3+3 2 | 3+3 2 | 2+2 2 | 2+2 2 | 2+2 2 |

Continued on next column

**TABLE XXI– continued**

| TQC | ICSD | NoE | SF | | | | | |
|---|---|---|---|---|---|---|---|---|
| | 3+3 2 | 2+2 2 | 2+2 2 | 2+2 2 | 3+3 2 | 2+2 2 | 2+2 2 | 3+3 2 |
| | 2+2 2 | 3+3 2 | 3+3 2 | 3+3 2 | 3+3 2 | 3+3 2 | 3+3 2 | 2+2 2 |
| | 3+3 2 | 3+3 2 | 3+3 2 | 2+2 2 | 2+2 2 | 3+3 2 | 2+2 2 | 3+3 2 |
| | 3+3 2 | 2+2 2 | 2+2 2 | 3+3 2 | 2+2 2 | 2+2 2 | 2+2 2 | 3+3 2 |
| | 2+2 2 | 2+2 2 | 2+2 2 | 2+2 2 | 2+2 2 | 3+3 2 | 3+3 2 | 3+3 2 |
| | 2+2 2 | 3+3 2 | 3+3 2 | 3+3 2 | 3+3 2 | 2+2 2 | 3+3 2 | 2+2 2 |
| | 3+3 2 | 3+3 2 | 2+2 2 | 3+3 2 | 3+3 2 | 3+3 2 | 3+3 2 | 3+3 2 |
| | 2+2 2 | 3+3 2 | 3+3 2 | 2+2 2 | 2+2 2 | 2+2 2 | 3+3 2 | 2+2 2 |
| | 3+3 2 | 2+2 2 | 2+2 2 | 2+2 2 | 3+3 2 | 2+2 2 | 2+2 2 | 3+3 2 |
| | 2+2 2 | 3+3 2 | 3+3 2 | 3+3 2 | 3+3 2 | 3+3 2 | 2+2 2 | 2+2 2 |
| | 3+3 2 | 2+2 2 | 2+2 2 | 3+3 2 | 2+2 2 | 3+3 2 | 2+2 2 | 3+3 2 |
| | 2+2 2 | 2+2 2 | 3+3 2 | 2+2 2 | 2+2 2 | 3+3 2 | 2+2 2 | 3+3 2 |
| | 3+3 2 | 3+3 2 | 3+3 2 | 2+2 2 | 3+3 2 | 2+2 2 | 3+3 2 | 2+2 2 |
| | 3+3 2 | 2+2 2 | 2+2 2 | 3+3 2 | 3+3 2 | 2+2 2 | 2+2 2 | 2+2 2 |
| | 3+3 2 | 3+3 2 | 2+2 2 | 2+2 2 | 3+3 2 | 3+3 2 | 3+3 2 | 3+3 2 |
| | 2+2 2 | 2+2 2 | 3+3 2 | 3+3 2 | 2+2 2 | 3+3 2 | 2+2 2 | 3+3 2 |
| | 2+2 2 | 2+2 2 | 3+3 2 | 3+3 2 | 3+3 2 | 2+2 2 | 2+2 2 | 2+2 2 |
| | 2+2 2 | 3+3 2 | 3+3 2 | 2+2 2 | 2+2 2 | 3+3 2 | 3+3 2 | 3+3 2 |
| | 3+3 2 | 2+2 2 | 2+2 2 | 2+2 2 | 2+2 2 | 2+2 2 | 2+2 2 | 2+2 2 |
| | 3+3 2 | 3+3 2 | 2+2 2 | 3+3 2 | 3+3 2 | 3+3 2 | 3+3 2 | 2+2 2 |
| | 3+3 2 | 2+2 2 | 3+3 2 | 3+3 2 | 3+3 2 | 2+2 2 | 3+3 2 | 2+2 2 |
| | 3+3 2 | 3+3 2 | 2+2 2 | 2+2 2 | 2+2 2 | 3+3 2 | 3+3 2 | 3+3 2 |
| | 2+2 2 | 2+2 2 | 3+3 2 | 2+2 2 | 2+2 2 | 2+2 2 | 2+2 2 | 2+2 2 |
| | 3+3 2 | 3+3 2 | 2+2 2 | 3+3 2 | 2+2 2 | 3+3 2 | 3+3 2 | 3+3 2 |
| | 2+2 2 | 3+3 2 | 3+3 2 | 3+3 2 | 3+3 2 | 2+2 2 | 2+2 2 | 3+3 2 |
| | 2+2 2 | 2+2 2 | 2+2 2 | 2+2 2 | 2+2 2 | 2+2 2 | 3+3 2 | 2+2 2 |
| | 2+2 2 | 3+3 2 | 2+2 2 | 2+2 2 | 2+2 2 | 3+3 2 | 3+3 2 | 3+3 2 |
| | 3+3 2 | 2+2 2 | 3+3 2 | 3+3 2 | 3+3 2 | 2+2 2 | 2+2 2 | 2+2 2 |
| | 3+3 2 | 3+3 2 | 3+3 2 | 3+3 2 | 2+2 2 | 2+2 2 | 3+3 2 | 3+3 2 |
| | 2+2 2 | 2+2 2 | 2+2 2 | 2+2 2 | 3+3 2 | 3+3 2 | 2+2 2 | 2+2 2 |
| | 2+2 2 | 3+3 2 | 2+2 2 | 3+3 2 | 3+3 2 | 2+2 2 | 3+3 2 | 3+3 2 |
| | 3+3 2 | 2+2 2 | 3+3 2 | 2+2 2 | 2+2 2 | 3+3 2 | 3+3 2 | 3+3 2 |
| | 2+2 2 | 3+3 2 | 2+2 2 | 2+2 2 | 2+2 2 | 3+3 2 | 2+2 2 | 2+2 2 |
| | 3+3 2 | 2+2 2 | 3+3 2 | 3+3 2 | 3+3 2 | 2+2 2 | 3+3 2 | 3+3 2 |
| | 3+3 2 | 3+3 2 | 3+3 2 | 3+3 2 | 2+2 2 | 2+2 2 | 2+2 2 | 2+2 2 |
| | 2+2 2 | 2+2 2 | 2+2 2 | 2+2 2 | 3+3 2 | 2+2 2 | 2+2 2 | 2+2 2 |
| 2.2.36 | 79719 | 130 | Fe2O14P4Sr1 | | | | | |
| | -GM | -R | -T | -U | -V | -X | -Y | -Z |
| | 3+3 2 | 2+2 2 | 3+3 2 | 2+2 2 | 2+2 2 | 2+2 2 | 3+3 2 | 3+3 2 |
| | 3+3 2 | 2+2 2 | 2+2 2 | 3+3 2 | 3+3 2 | 2+2 2 | 2+2 2 | 2+2 2 |
| | 2+2 2 | 3+3 2 | 3+3 2 | 2+2 2 | 2+2 2 | 3+3 2 | 3+3 2 | 3+3 2 |
| | 2+2 2 | 3+3 2 | 2+2 2 | 2+2 2 | 3+3 2 | 3+3 2 | 3+3 2 | 3+3 2 |
| | 3+3 2 | 2+2 2 | 3+3 2 | 3+3 2 | 2+2 2 | 2+2 2 | 2+2 2 | 2+2 2 |
| | 3+3 2 | 3+3 2 | 2+2 2 | 3+3 2 | 3+3 2 | 2+2 2 | 2+2 2 | 2+2 2 |
| | 2+2 2 | 2+2 2 | 3+3 2 | 2+2 2 | 2+2 2 | 3+3 2 | 3+3 2 | 3+3 2 |
| | 3+3 2 | 2+2 2 | 2+2 2 | 3+3 2 | 2+2 2 | 3+3 2 | 2+2 2 | 3+3 2 |
| | 2+2 2 | 3+3 2 | 2+2 2 | 3+3 2 | 2+2 2 | 2+2 2 | 3+3 2 | 2+2 2 |
| | 2+2 2 | 3+3 2 | 3+3 2 | 2+2 2 | 3+3 2 | 2+2 2 | 3+3 2 | 2+2 2 |
| | 3+3 2 | 2+2 2 | 2+2 2 | 2+2 2 | 3+3 2 | 3+3 2 | 2+2 2 | 3+3 2 |
| | 2+2 2 | 3+3 2 | 3+3 2 | 2+2 2 | 3+3 2 | 3+3 2 | 2+2 2 | 2+2 2 |
| | 2+2 2 | 2+2 2 | 3+3 2 | 3+3 2 | 2+2 2 | 2+2 2 | 3+3 2 | 3+3 2 |
| | 3+3 2 | 3+3 2 | 2+2 2 | 2+2 2 | 3+3 2 | 2+2 2 | 3+3 2 | 2+2 2 |
| | 3+3 2 | 2+2 2 | 3+3 2 | 3+3 2 | 2+2 2 | 3+3 2 | 2+2 2 | 3+3 2 |
| | 2+2 2 | 3+3 2 | 2+2 2 | 3+3 2 | 3+3 2 | 3+3 2 | 2+2 2 | 2+2 2 |
| | 2+2 2 | 3+3 2 | 2+2 2 | 3+3 2 | 3+3 2 | 3+3 2 | 2+2 2 | 2+2 2 |
| | 2+2 2 | 3+3 2 | 2+2 2 | 3+3 2 | 3+3 2 | 3+3 2 | 2+2 2 | 2+2 2 |
| | 3+3 2 | 2+2 2 | 2+2 2 | 3+3 2 | 2+2 2 | 2+2 2 | 3+3 2 | 2+2 2 |
| | 2+2 2 | 3+3 2 | 3+3 2 | 2+2 2 | 3+3 2 | 3+3 2 | 2+2 2 | 3+3 2 |
| | 3+3 2 | 2+2 2 | 2+2 2 | 3+3 2 | 2+2 2 | 3+3 2 | 2+2 2 | 3+3 2 |
| | 2+2 2 | 3+3 2 | 3+3 2 | 2+2 2 | 3+3 2 | 2+2 2 | 3+3 2 | 2+2 2 |
| | 3+3 2 | 3+3 2 | 2+2 2 | 2+2 2 | 2+2 2 | 2+2 2 | 3+3 2 | 3+3 2 |

Continued on next column

**TABLE XXI– continued**

| TQC | ICSD | NoE | SF | | | | | |
|---|---|---|---|---|---|---|---|---|
| | 2+2 2 | 2+2 2 | 3+3 2 | 3+3 2 | 3+3 2 | 3+3 2 | 2+2 2 | 2+2 2 |
| | 3+3 2 | 2+2 2 | 2+2 2 | 3+3 2 | 2+2 2 | 3+3 2 | 2+2 2 | 2+2 2 |
| | 3+3 2 | 3+3 2 | 3+3 2 | 3+3 2 | 2+2 2 | 2+2 2 | 3+3 2 | 3+3 2 |
| | 2+2 2 | 2+2 2 | 2+2 2 | 3+3 2 | 3+3 2 | 2+2 2 | 2+2 2 | 3+3 2 |
| | 2+2 2 | 3+3 2 | 3+3 2 | 2+2 2 | 2+2 2 | 3+3 2 | 3+3 2 | 2+2 2 |
| | 3+3 2 | 3+3 2 | 2+2 2 | 2+2 2 | 2+2 2 | 2+2 2 | 3+3 2 | 2+2 2 |
| | 3+3 2 | 2+2 2 | 2+2 2 | 3+3 2 | 3+3 2 | 2+2 2 | 3+3 2 | 3+3 2 |
| | 2+2 2 | 2+2 2 | 3+3 2 | 2+2 2 | 3+3 2 | 3+3 2 | 2+2 2 | 3+3 2 |
| | 2+2 2 | 3+3 2 | 3+3 2 | 2+2 2 | 3+3 2 | 3+3 2 | 2+2 2 | 2+2 2 |
| | 3+3 2 | 2+2 2 | 2+2 2 | 3+3 2 | 2+2 2 | 3+3 2 | 2+2 2 | 3+3 2 |
| | 2+2 2 | 3+3 2 | 3+3 2 | 2+2 2 | 3+3 2 | 2+2 2 | 3+3 2 | 2+2 2 |
| | 3+3 2 | 2+2 2 | 2+2 2 | 2+2 2 | 2+2 2 | 3+3 2 | 2+2 2 | 3+3 2 |
| | 3+3 2 | 3+3 2 | 2+2 2 | 3+3 2 | 2+2 2 | 2+2 2 | 3+3 2 | 2+2 2 |
| | 3+3 2 | 3+3 2 | 2+2 2 | 2+2 2 | 3+3 2 | 2+2 2 | 2+2 2 | 3+3 2 |
| | 3+3 2 | 2+2 2 | 3+3 2 | 3+3 2 | 3+3 2 | 3+3 2 | 3+3 2 | 2+2 2 |
| | 2+2 2 | 3+3 2 | 3+3 2 | 3+3 2 | 2+2 2 | 2+2 2 | 2+2 2 | 2+2 2 |
| | 3+3 2 | 2+2 2 | 3+3 2 | 3+3 2 | 3+3 2 | 3+3 2 | 3+3 2 | 3+3 2 |
| | 2+2 2 | 2+2 2 | 2+2 2 | 2+2 2 | 2+2 2 | 2+2 2 | 2+2 2 | 3+3 2 |
| | 3+3 2 | 3+3 2 | 2+2 2 | 2+2 2 | 2+2 2 | 2+2 2 | 3+3 2 | 3+3 2 |
| | 2+2 2 | 2+2 2 | 3+3 2 | 2+2 2 | 3+3 2 | 3+3 2 | 2+2 2 | 2+2 2 |
| | 3+3 2 | 3+3 2 | 2+2 2 | 3+3 2 | 3+3 2 | 3+3 2 | 3+3 2 | 3+3 2 |
| | 2+2 2 | 2+2 2 | 2+2 2 | 2+2 2 | 2+2 2 | 2+2 2 | 2+2 2 | 2+2 2 |
| | 3+3 2 | 3+3 2 | 3+3 2 | 3+3 2 | 2+2 2 | 3+3 2 | 3+3 2 | 3+3 2 |
| | 2+2 2 | 3+3 2 | 2+2 2 | 3+3 2 | 3+3 2 | 3+3 2 | 3+3 2 | 2+2 2 |
| | 2+2 2 | 2+2 2 | 2+2 2 | 2+2 2 | 3+3 2 | 2+2 2 | 2+2 2 | 2+2 2 |
| | 3+3 2 | 2+2 2 | 3+3 2 | 3+3 2 | 2+2 2 | 3+3 2 | 3+3 2 | 2+2 2 |
| | 2+2 2 | 2+2 2 | 3+3 2 | 3+3 2 | 2+2 2 | 2+2 2 | 2+2 2 | 3+3 2 |
| | 2+2 2 | 3+3 2 | 3+3 2 | 2+2 2 | 2+2 2 | 2+2 2 | 3+3 2 | 3+3 2 |
| | 2+2 2 | 3+3 2 | 2+2 2 | 3+3 2 | 3+3 2 | 3+3 2 | 2+2 2 | 2+2 2 |
| | 3+3 2 | 2+2 2 | 3+3 2 | 2+2 2 | 3+3 2 | 2+2 2 | 2+2 2 | 3+3 2 |
| | 3+3 2 | 3+3 2 | 2+2 2 | 3+3 2 | 3+3 2 | 3+3 2 | 3+3 2 | 2+2 2 |
| | 2+2 2 | 2+2 2 | 3+3 2 | 2+2 2 | 2+2 2 | 2+2 2 | 3+3 2 | 3+3 2 |
| | 2+2 2 | 3+3 2 | 3+3 2 | 3+3 2 | 2+2 2 | 3+3 2 | 2+2 2 | 2+2 2 |
| | 3+3 2 | 2+2 2 | 2+2 2 | 3+3 2 | 2+2 2 | 3+3 2 | 3+3 2 | 3+3 2 |
| | 2+2 2 | 3+3 2 | 3+3 2 | 2+2 2 | 3+3 2 | 2+2 2 | 2+2 2 | 2+2 2 |
| | 2+2 2 | 3+3 2 | 3+3 2 | 2+2 2 | 3+3 2 | 2+2 2 | 3+3 2 | 2+2 2 |
| | 3+3 2 | 2+2 2 | 2+2 2 | 2+2 2 | 3+3 2 | 3+3 2 | 2+2 2 | 3+3 2 |
| | 3+3 2 | 2+2 2 | 2+2 2 | 3+3 2 | 2+2 2 | 3+3 2 | 3+3 2 | 3+3 2 |
| | 3+3 2 | 3+3 2 | 2+2 2 | 3+3 2 | 2+2 2 | 2+2 2 | 2+2 2 | 2+2 2 |
| | 3+3 2 | 3+3 2 | 2+2 2 | 3+3 2 | 2+2 2 | 2+2 2 | 2+2 2 | 3+3 2 |
| | 3+3 2 | 2+2 2 | 2+2 2 | 3+3 2 | 2+2 2 | 3+3 2 | 3+3 2 | 2+2 2 |
| | 3+3 2 | 2+2 2 | 2+2 2 | 3+3 2 | 2+2 2 | 2+2 2 | 2+2 2 | 3+3 2 |
| 2.2.37 | 72390 | 190 | Fe1N2Sr2 | | | | | |
| | -GM | -R | -T | -U | -V | -X | -Y | -Z |
| | 3+3 2 | 2+2 2 | 3+3 2 | 3+3 2 | 3+3 2 | 2+2 2 | 2+2 2 | 2+2 2 |
| | 2+2 2 | 3+3 2 | 2+2 2 | 2+2 2 | 2+2 2 | 3+3 2 | 3+3 2 | 3+3 2 |
| | 2+2 2 | 3+3 2 | 2+2 2 | 2+2 2 | 2+2 2 | 3+3 2 | 3+3 2 | 3+3 2 |
| | 3+3 2 | 2+2 2 | 3+3 2 | 3+3 2 | 3+3 2 | 2+2 2 | 2+2 2 | 2+2 2 |
| | 2+2 2 | 3+3 2 | 2+2 2 | 2+2 2 | 2+2 2 | 3+3 2 | 3+3 2 | 3+3 2 |
| | 3+3 2 | 2+2 2 | 3+3 2 | 3+3 2 | 3+3 2 | 2+2 2 | 2+2 2 | 2+2 2 |
| | 3+3 2 | 2+2 2 | 2+2 2 | 2+2 2 | 3+3 2 | 3+3 2 | 3+3 2 | 2+2 2 |
| | 2+2 2 | 3+3 2 | 3+3 2 | 3+3 2 | 2+2 2 | 2+2 2 | 2+2 2 | 3+3 2 |
| | 3+3 2 | 2+2 2 | 2+2 2 | 3+3 2 | 2+2 2 | 3+3 2 | 2+2 2 | 3+3 2 |
| | 2+2 2 | 3+3 2 | 3+3 2 | 2+2 2 | 3+3 2 | 2+2 2 | 3+3 2 | 2+2 2 |
| | 2+2 2 | 2+2 2 | 3+3 2 | 3+3 2 | 2+2 2 | 3+3 2 | 3+3 2 | 2+2 2 |
| | 3+3 2 | 3+3 2 | 2+2 2 | 2+2 2 | 3+3 2 | 2+2 2 | 2+2 2 | 3+3 2 |
| | 3+3 2 | 3+3 2 | 2+2 2 | 2+2 2 | 3+3 2 | 2+2 2 | 2+2 2 | 3+3 2 |
| | 2+2 2 | 2+2 2 | 3+3 2 | 3+3 2 | 2+2 2 | 3+3 2 | 3+3 2 | 2+2 2 |
| | 2+2 2 | 2+2 2 | 3+3 2 | 3+3 2 | 3+3 2 | 2+2 2 | 3+3 2 | 2+2 2 |
| | 3+3 2 | 3+3 2 | 2+2 2 | 2+2 2 | 2+2 2 | 3+3 2 | 2+2 2 | 3+3 2 |
| | 3+3 2 | 2+2 2 | 2+2 2 | 2+2 2 | 3+3 2 | 3+3 2 | 3+3 2 | 2+2 2 |

Continued on next column

| TQC | ICSD | NoE | SF | | | | | |
|---|---|---|---|---|---|---|---|---|
| | 2+2 2 | 3+3 2 | 3+3 2 | 3+3 2 | 2+2 2 | 2+2 2 | 2+2 2 | 3+3 2 |
| | 3+3 2 | 2+2 2 | 2+2 2 | 3+3 2 | 2+2 2 | 3+3 2 | 2+2 2 | 3+3 2 |
| | 2+2 2 | 3+3 2 | 3+3 2 | 2+2 2 | 3+3 2 | 2+2 2 | 3+3 2 | 2+2 2 |
| | 2+2 2 | 3+3 2 | 3+3 2 | 2+2 2 | 2+2 2 | 3+3 2 | 2+2 2 | 3+3 2 |
| | 3+3 2 | 2+2 2 | 2+2 2 | 3+3 2 | 3+3 2 | 2+2 2 | 3+3 2 | 2+2 2 |
| | 3+3 2 | 2+2 2 | 3+3 2 | 3+3 2 | 3+3 2 | 2+2 2 | 2+2 2 | 3+3 2 |
| | 2+2 2 | 3+3 2 | 2+2 2 | 2+2 2 | 2+2 2 | 3+3 2 | 3+3 2 | 2+2 2 |
| | 2+2 2 | 2+2 2 | 2+2 2 | 2+2 2 | 3+3 2 | 3+3 2 | 3+3 2 | 2+2 2 |
| | 3+3 2 | 3+3 2 | 3+3 2 | 3+3 2 | 2+2 2 | 2+2 2 | 2+2 2 | 3+3 2 |
| | 3+3 2 | 2+2 2 | 2+2 2 | 2+2 2 | 2+2 2 | 3+3 2 | 3+3 2 | 3+3 2 |
| | 2+2 2 | 3+3 2 | 3+3 2 | 3+3 2 | 2+2 2 | 2+2 2 | 3+3 2 | 2+2 2 |
| | 3+3 2 | 3+3 2 | 2+2 2 | 2+2 2 | 3+3 2 | 3+3 2 | 2+2 2 | 2+2 2 |
| | 2+2 2 | 2+2 2 | 3+3 2 | 3+3 2 | 3+3 2 | 2+2 2 | 2+2 2 | 3+3 2 |
| | 3+3 2 | 3+3 2 | 2+2 2 | 2+2 2 | 3+3 2 | 3+3 2 | 2+2 2 | 2+2 2 |
| | 2+2 2 | 2+2 2 | 3+3 2 | 3+3 2 | 2+2 2 | 2+2 2 | 3+3 2 | 3+3 2 |
| | 2+2 2 | 3+3 2 | 2+2 2 | 3+3 2 | 3+3 2 | 3+3 2 | 3+3 2 | 3+3 2 |
| | 3+3 2 | 2+2 2 | 3+3 2 | 2+2 2 | 2+2 2 | 2+2 2 | 2+2 2 | 2+2 2 |
| | 3+3 2 | 2+2 2 | 2+2 2 | 2+2 2 | 3+3 2 | 3+3 2 | 3+3 2 | 2+2 2 |
| | 2+2 2 | 3+3 2 | 3+3 2 | 3+3 2 | 2+2 2 | 2+2 2 | 2+2 2 | 3+3 2 |
| | 3+3 2 | 2+2 2 | 2+2 2 | 3+3 2 | 3+3 2 | 3+3 2 | 2+2 2 | 3+3 2 |
| | 2+2 2 | 3+3 2 | 3+3 2 | 2+2 2 | 2+2 2 | 2+2 2 | 3+3 2 | 2+2 2 |
| | 3+3 2 | 2+2 2 | 2+2 2 | 3+3 2 | 2+2 2 | 3+3 2 | 2+2 2 | 3+3 2 |
| | 2+2 2 | 3+3 2 | 3+3 2 | 2+2 2 | 3+3 2 | 2+2 2 | 3+3 2 | 2+2 2 |
| | 3+3 2 | 2+2 2 | 3+3 2 | 2+2 2 | 2+2 2 | 2+2 2 | 3+3 2 | 3+3 2 |
| | 3+3 2 | 3+3 2 | 3+3 2 | 2+2 2 | 2+2 2 | 3+3 2 | 2+2 2 | 2+2 2 |
| | 2+2 2 | 3+3 2 | 2+2 2 | 3+3 2 | 3+3 2 | 3+3 2 | 2+2 2 | 2+2 2 |
| | 2+2 2 | 2+2 2 | 2+2 2 | 3+3 2 | 3+3 2 | 2+2 2 | 3+3 2 | 3+3 2 |
| | 3+3 2 | 3+3 2 | 2+2 2 | 3+3 2 | 2+2 2 | 2+2 2 | 3+3 2 | 2+2 2 |
| | 2+2 2 | 2+2 2 | 2+2 2 | 3+3 2 | 3+3 2 | 3+3 2 | 2+2 2 | 2+2 2 |
| | 3+3 2 | 3+3 2 | 3+3 2 | 2+2 2 | 2+2 2 | 2+2 2 | 3+3 2 | 3+3 2 |
| | 2+2 2 | 2+2 2 | 3+3 2 | 2+2 2 | 3+3 2 | 3+3 2 | 2+2 2 | 3+3 2 |
| | 2+2 2 | 3+3 2 | 2+2 2 | 3+3 2 | 2+2 2 | 2+2 2 | 3+3 2 | 3+3 2 |
| | 3+3 2 | 2+2 2 | 3+3 2 | 2+2 2 | 3+3 2 | 3+3 2 | 2+2 2 | 2+2 2 |
| | 3+3 2 | 2+2 2 | 3+3 2 | 2+2 2 | 2+2 2 | 2+2 2 | 2+2 2 | 3+3 2 |
| | 2+2 2 | 3+3 2 | 2+2 2 | 3+3 2 | 2+2 2 | 3+3 2 | 3+3 2 | 2+2 2 |
| | 3+3 2 | 3+3 2 | 3+3 2 | 3+3 2 | 3+3 2 | 3+3 2 | 2+2 2 | 2+2 2 |
| | 2+2 2 | 3+3 2 | 2+2 2 | 2+2 2 | 3+3 2 | 3+3 2 | 3+3 2 | 2+2 2 |
| | 2+2 2 | 2+2 2 | 2+2 2 | 3+3 2 | 3+3 2 | 2+2 2 | 2+2 2 | 3+3 2 |
| | 2+2 2 | 3+3 2 | 2+2 2 | 3+3 2 | 3+3 2 | 2+2 2 | 3+3 2 | 2+2 2 |
| | 2+2 2 | 2+2 2 | 2+2 2 | 3+3 2 | 3+3 2 | 3+3 2 | 2+2 2 | 3+3 2 |
| | 3+3 2 | 2+2 2 | 2+2 2 | 3+3 2 | 2+2 2 | 2+2 2 | 3+3 2 | 3+3 2 |
| | 2+2 2 | 3+3 2 | 3+3 2 | 2+2 2 | 3+3 2 | 2+2 2 | 3+3 2 | 2+2 2 |
| | 3+3 2 | 2+2 2 | 2+2 2 | 3+3 2 | 2+2 2 | 2+2 2 | 3+3 2 | 3+3 2 |
| | 3+3 2 | 2+2 2 | 3+3 2 | 2+2 2 | 2+2 2 | 2+2 2 | 3+3 2 | 3+3 2 |
| | 3+3 2 | 3+3 2 | 3+3 2 | 2+2 2 | 2+2 2 | 3+3 2 | 2+2 2 | 2+2 2 |
| | 3+3 2 | 3+3 2 | 3+3 2 | 2+2 2 | 2+2 2 | 3+3 2 | 2+2 2 | 2+2 2 |
| | 3+3 2 | 3+3 2 | 2+2 2 | 2+2 2 | 2+2 2 | 2+2 2 | 3+3 2 | 2+2 2 |
| | 3+3 2 | 2+2 2 | 3+3 2 | 3+3 2 | 3+3 2 | 2+2 2 | 2+2 2 | 3+3 2 |
| | 2+2 2 | 3+3 2 | 2+2 2 | 3+3 2 | 2+2 2 | 3+3 2 | 3+3 2 | 2+2 2 |
| | 2+2 2 | 2+2 2 | 2+2 2 | 2+2 2 | 2+2 2 | 3+3 2 | 3+3 2 | 2+2 2 |
| | 3+3 2 | 3+3 2 | 3+3 2 | 3+3 2 | 3+3 2 | 2+2 2 | 2+2 2 | 3+3 2 |
| | 2+2 2 | 3+3 2 | 2+2 2 | 3+3 2 | 2+2 2 | 2+2 2 | 3+3 2 | 3+3 2 |
| | 3+3 2 | 2+2 2 | 2+2 2 | 2+2 2 | 3+3 2 | 2+2 2 | 3+3 2 | 2+2 2 |
| | 2+2 2 | 2+2 2 | 3+3 2 | 3+3 2 | 2+2 2 | 3+3 2 | 2+2 2 | 2+2 2 |
| | 3+3 2 | 3+3 2 | 3+3 2 | 2+2 2 | 3+3 2 | 3+3 2 | 3+3 2 | 2+2 2 |
| | 2+2 2 | 3+3 2 | 3+3 2 | 3+3 2 | 2+2 2 | 2+2 2 | 3+3 2 | 3+3 2 |
| | 3+3 2 | 2+2 2 | 2+2 2 | 2+2 2 | 3+3 2 | 3+3 2 | 2+2 2 | 3+3 2 |
| | 3+3 2 | 3+3 2 | 3+3 2 | 2+2 2 | 3+3 2 | 2+2 2 | 2+2 2 | 2+2 2 |
| | 2+2 2 | 2+2 2 | 3+3 2 | 2+2 2 | 2+2 2 | 3+3 2 | 2+2 2 | 3+3 2 |
| | 3+3 2 | 2+2 2 | 2+2 2 | 3+3 2 | 2+2 2 | 3+3 2 | 2+2 2 | 3+3 2 |
| | 2+2 2 | 3+3 2 | 3+3 2 | 2+2 2 | 3+3 2 | 2+2 2 | 2+2 2 | 2+2 2 |

Continued on next column

**TABLE XXI– continued**

| TQC | ICSD | NoE | SF | | | | | |
|---|---|---|---|---|---|---|---|---|
| | 2+2 2 | 2+2 2 | 2+2 2 | 2+2 2 | 3+3 2 | 3+3 2 | 3+3 2 | 3+3 2 |
| | 2+2 2 | 2+2 2 | 3+3 2 | 3+3 2 | 3+3 2 | 3+3 2 | 2+2 2 | 3+3 2 |
| | 3+3 2 | 3+3 2 | 3+3 2 | 2+2 2 | 2+2 2 | 2+2 2 | 3+3 2 | 2+2 2 |
| | 3+3 2 | 2+2 2 | 3+3 2 | 3+3 2 | 2+2 2 | 3+3 2 | 2+2 2 | 3+3 2 |
| | 2+2 2 | 3+3 2 | 2+2 2 | 2+2 2 | 3+3 2 | 2+2 2 | 3+3 2 | 2+2 2 |
| | 3+3 2 | 3+3 2 | 2+2 2 | 3+3 2 | 2+2 2 | 3+3 2 | 2+2 2 | 2+2 2 |
| | 3+3 2 | 2+2 2 | 3+3 2 | 2+2 2 | 3+3 2 | 2+2 2 | 2+2 2 | 3+3 2 |
| | 2+2 2 | 2+2 2 | 3+3 2 | 3+3 2 | 2+2 2 | 3+3 2 | 3+3 2 | 2+2 2 |
| | 3+3 2 | 3+3 2 | 2+2 2 | 3+3 2 | 3+3 2 | 2+2 2 | 2+2 2 | 3+3 2 |
| | 2+2 2 | 3+3 2 | 2+2 2 | 2+2 2 | 2+2 2 | 3+3 2 | 3+3 2 | 2+2 2 |
| | 2+2 2 | 2+2 2 | 2+2 2 | 3+3 2 | 3+3 2 | 2+2 2 | 3+3 2 | 3+3 2 |
| | 2+2 2 | 3+3 2 | 2+2 2 | 3+3 2 | 2+2 2 | 2+2 2 | 3+3 2 | 2+2 2 |
| | 3+3 2 | 3+3 2 | 2+2 2 | 3+3 2 | 2+2 2 | 2+2 2 | 3+3 2 | 2+2 2 |
| | 3+3 2 | 3+3 2 | 2+2 2 | 3+3 2 | 3+3 2 | 3+3 2 | 3+3 2 | 2+2 2 |
| | 3+3 2 | 3+3 2 | 3+3 2 | 2+2 2 | 2+2 2 | 2+2 2 | 2+2 2 | 3+3 2 |
| | 3+3 2 | 2+2 2 | 2+2 2 | 3+3 2 | 3+3 2 | 2+2 2 | 3+3 2 | 2+2 2 |
| | 3+3 2 | 3+3 2 | 3+3 2 | 2+2 2 | 2+2 2 | 3+3 2 | 2+2 2 | 3+3 2 |
| 2.2.38 | 171431 | 128 | Cu4Mo6Se8 | | | | | |
| | -GM | -R | -T | -U | -V | -X | -Y | -Z |
| | 3+3 2 | 2+2 2 | 2+2 2 | 3+3 2 | 3+3 2 | 2+2 2 | 3+3 2 | 2+2 2 |
| | 2+2 2 | 3+3 2 | 3+3 2 | 2+2 2 | 3+3 2 | 3+3 2 | 2+2 2 | 3+3 2 |
| | 2+2 2 | 3+3 2 | 3+3 2 | 3+3 2 | 2+2 2 | 3+3 2 | 2+2 2 | 3+3 2 |
| | 3+3 2 | 2+2 2 | 2+2 2 | 2+2 2 | 2+2 2 | 2+2 2 | 3+3 2 | 2+2 2 |
| | 2+2 2 | 3+3 2 | 3+3 2 | 3+3 2 | 3+3 2 | 3+3 2 | 2+2 2 | 3+3 2 |
| | 3+3 2 | 2+2 2 | 2+2 2 | 2+2 2 | 3+3 2 | 2+2 2 | 3+3 2 | 3+3 2 |
| | 2+2 2 | 3+3 2 | 3+3 2 | 2+2 2 | 2+2 2 | 3+3 2 | 2+2 2 | 2+2 2 |
| | 3+3 2 | 2+2 2 | 2+2 2 | 3+3 2 | 2+2 2 | 2+2 2 | 3+3 2 | 2+2 2 |
| | 3+3 2 | 2+2 2 | 2+2 2 | 3+3 2 | 2+2 2 | 2+2 2 | 3+3 2 | 2+2 2 |
| | 3+3 2 | 3+3 2 | 3+3 2 | 2+2 2 | 3+3 2 | 3+3 2 | 2+2 2 | 3+3 2 |
| | 2+2 2 | 3+3 2 | 2+2 2 | 2+2 2 | 3+3 2 | 2+2 2 | 3+3 2 | 2+2 2 |
| | 2+2 2 | 3+3 2 | 3+3 2 | 3+3 2 | 2+2 2 | 3+3 2 | 2+2 2 | 3+3 2 |
| | 3+3 2 | 2+2 2 | 2+2 2 | 3+3 2 | 2+2 2 | 3+3 2 | 2+2 2 | 2+2 2 |
| | 2+2 2 | 3+3 2 | 3+3 2 | 2+2 2 | 3+3 2 | 2+2 2 | 3+3 2 | 3+3 2 |
| | 2+2 2 | 2+2 2 | 2+2 2 | 3+3 2 | 2+2 2 | 2+2 2 | 3+3 2 | 3+3 2 |
| | 3+3 2 | 3+3 2 | 3+3 2 | 3+3 2 | 3+3 2 | 3+3 2 | 2+2 2 | 2+2 2 |
| | 3+3 2 | 2+2 2 | 2+2 2 | 2+2 2 | 2+2 2 | 3+3 2 | 2+2 2 | 3+3 2 |
| | 2+2 2 | 3+3 2 | 3+3 2 | 2+2 2 | 3+3 2 | 2+2 2 | 3+3 2 | 3+3 2 |
| | 2+2 2 | 2+2 2 | 3+3 2 | 3+3 2 | 2+2 2 | 2+2 2 | 3+3 2 | 2+2 2 |
| | 3+3 2 | 2+2 2 | 2+2 2 | 2+2 2 | 3+3 2 | 2+2 2 | 3+3 2 | 2+2 2 |
| | 3+3 2 | 2+2 2 | 3+3 2 | 2+2 2 | 2+2 2 | 3+3 2 | 2+2 2 | 3+3 2 |
| | 2+2 2 | 3+3 2 | 2+2 2 | 3+3 2 | 2+2 2 | 2+2 2 | 3+3 2 | 2+2 2 |
| | 3+3 2 | 2+2 2 | 2+2 2 | 3+3 2 | 2+2 2 | 3+3 2 | 2+2 2 | 2+2 2 |
| | 2+2 2 | 3+3 2 | 3+3 2 | 2+2 2 | 3+3 2 | 2+2 2 | 3+3 2 | 3+3 2 |
| | 3+3 2 | 2+2 2 | 2+2 2 | 3+3 2 | 3+3 2 | 2+2 2 | 2+2 2 | 2+2 2 |
| | 3+3 2 | 2+2 2 | 2+2 2 | 2+2 2 | 2+2 2 | 3+3 2 | 3+3 2 | 3+3 2 |
| | 2+2 2 | 3+3 2 | 3+3 2 | 3+3 2 | 3+3 2 | 3+3 2 | 2+2 2 | 2+2 2 |
| | 3+3 2 | 2+2 2 | 3+3 2 | 2+2 2 | 3+3 2 | 3+3 2 | 2+2 2 | 3+3 2 |
| | 3+3 2 | 3+3 2 | 3+3 2 | 2+2 2 | 3+3 2 | 2+2 2 | 3+3 2 | 2+2 2 |
| | 3+3 2 | 3+3 2 | 2+2 2 | 3+3 2 | 2+2 2 | 2+2 2 | 3+3 2 | 3+3 2 |
| | 2+2 2 | 3+3 2 | 3+3 2 | 3+3 2 | 3+3 2 | 3+3 2 | 2+2 2 | 2+2 2 |
| | 2+2 2 | 2+2 2 | 2+2 2 | 2+2 2 | 2+2 2 | 2+2 2 | 3+3 2 | 3+3 2 |
| | 2+2 2 | 3+3 2 | 2+2 2 | 3+3 2 | 2+2 2 | 3+3 2 | 2+2 2 | 2+2 2 |
| | 3+3 2 | 2+2 2 | 3+3 2 | 2+2 2 | 3+3 2 | 2+2 2 | 3+3 2 | 3+3 2 |
| | 2+2 2 | 3+3 2 | 3+3 2 | 3+3 2 | 2+2 2 | 3+3 2 | 2+2 2 | 3+3 2 |
| | 3+3 2 | 2+2 2 | 2+2 2 | 2+2 2 | 3+3 2 | 2+2 2 | 3+3 2 | 2+2 2 |
| | 3+3 2 | 3+3 2 | 3+3 2 | 2+2 2 | 2+2 2 | 3+3 2 | 2+2 2 | 2+2 2 |
| | 3+3 2 | 2+2 2 | 2+2 2 | 3+3 2 | 3+3 2 | 3+3 2 | 3+3 2 | 3+3 2 |
| | 2+2 2 | 3+3 2 | 2+2 2 | 3+3 2 | 3+3 2 | 2+2 2 | 2+2 2 | 3+3 2 |
| | 2+2 2 | 2+2 2 | 3+3 2 | 2+2 2 | 2+2 2 | 2+2 2 | 2+2 2 | 3+3 2 |
| | 3+3 2 | 3+3 2 | 2+2 2 | 2+2 2 | 3+3 2 | 3+3 2 | 3+3 2 | 2+2 2 |
| | 2+2 2 | 3+3 2 | 3+3 2 | 3+3 2 | 2+2 2 | 3+3 2 | 2+2 2 | 2+2 2 |
| | 2+2 2 | 2+2 2 | 2+2 2 | 3+3 2 | 3+3 2 | 3+3 2 | 2+2 2 | 2+2 2 |
| | 2+2 2 | 2+2 2 | 2+2 2 | 3+3 2 | 2+2 2 | 2+2 2 | 3+3 2 | 3+3 2 |
| | 3+3 2 | 3+3 2 | 3+3 2 | 2+2 2 | 3+3 2 | 2+2 2 | 2+2 2 | 2+2 2 |
| | 3+3 2 | 2+2 2 | 2+2 2 | 2+2 2 | 2+2 2 | 3+3 2 | 3+3 2 | 3+3 2 |
| | 3+3 2 | 2+2 2 | 3+3 2 | 3+3 2 | 3+3 2 | 2+2 2 | 2+2 2 | 2+2 2 |
| | 2+2 2 | 3+3 2 | 3+3 2 | 2+2 2 | 3+3 2 | 3+3 2 | 3+3 2 | 3+3 2 |
| | 3+3 2 | 2+2 2 | 3+3 2 | 3+3 2 | 3+3 2 | 2+2 2 | 2+2 2 | 3+3 2 |
| | 2+2 2 | 3+3 2 | 3+3 2 | 2+2 2 | 2+2 2 | 3+3 2 | 3+3 2 | 2+2 2 |
| | 3+3 2 | 2+2 2 | 2+2 2 | 2+2 2 | 2+2 2 | 2+2 2 | 2+2 2 | 3+3 2 |
| | 2+2 2 | 2+2 2 | 2+2 2 | 3+3 2 | 3+3 2 | 3+3 2 | 3+3 2 | 2+2 2 |
| | 2+2 2 | 3+3 2 | 3+3 2 | 2+2 2 | 2+2 2 | 2+2 2 | 2+2 2 | 2+2 2 |
| | 3+3 2 | 3+3 2 | 2+2 2 | 3+3 2 | 3+3 2 | 3+3 2 | 3+3 2 | 3+3 2 |
| | 3+3 2 | 2+2 2 | 2+2 2 | 2+2 2 | 2+2 2 | 3+3 2 | 2+2 2 | 2+2 2 |
| | 2+2 2 | 2+2 2 | 3+3 2 | 3+3 2 | 3+3 2 | 2+2 2 | 3+3 2 | 2+2 2 |
| | 2+2 2 | 3+3 2 | 2+2 2 | 2+2 2 | 3+3 2 | 3+3 2 | 2+2 2 | 3+3 2 |
| | 3+3 2 | 2+2 2 | 3+3 2 | 3+3 2 | 2+2 2 | 3+3 2 | 3+3 2 | 3+3 2 |
| | 2+2 2 | 3+3 2 | 3+3 2 | 2+2 2 | 3+3 2 | 2+2 2 | 2+2 2 | 2+2 2 |
| | 3+3 2 | 2+2 2 | 3+3 2 | 2+2 2 | 2+2 2 | 3+3 2 | 2+2 2 | 3+3 2 |
| | 2+2 2 | 2+2 2 | 2+2 2 | 3+3 2 | 3+3 2 | 2+2 2 | 3+3 2 | 3+3 2 |
| | 3+3 2 | 3+3 2 | 3+3 2 | 2+2 2 | 2+2 2 | 3+3 2 | 3+3 2 | 3+3 2 |
| | 2+2 2 | 2+2 2 | 2+2 2 | 3+3 2 | 2+2 2 | 2+2 2 | 2+2 2 | 2+2 2 |
| | 2+2 2 | 3+3 2 | 2+2 2 | 3+3 2 | 2+2 2 | 2+2 2 | 3+3 2 | 2+2 2 |
| 2.2.39 | 190444 | 198 | Cu3Mo2O9 | | | | | |
| | -GM | -R | -T | -U | -V | -X | -Y | -Z |
| | 3+3 2 | 3+3 2 | 3+3 2 | 2+2 2 | 2+2 2 | 3+3 2 | 2+2 2 | 2+2 2 |
| | 2+2 2 | 2+2 2 | 2+2 2 | 3+3 2 | 3+3 2 | 2+2 2 | 3+3 2 | 3+3 2 |
| | 3+3 2 | 3+3 2 | 3+3 2 | 2+2 2 | 2+2 2 | 3+3 2 | 2+2 2 | 2+2 2 |
| | 2+2 2 | 2+2 2 | 2+2 2 | 3+3 2 | 3+3 2 | 2+2 2 | 3+3 2 | 3+3 2 |
| | 3+3 2 | 2+2 2 | 2+2 2 | 3+3 2 | 3+3 2 | 2+2 2 | 3+3 2 | 2+2 2 |
| | 3+3 2 | 3+3 2 | 3+3 2 | 2+2 2 | 2+2 2 | 3+3 2 | 3+3 2 | 3+3 2 |
| | 2+2 2 | 2+2 2 | 2+2 2 | 2+2 2 | 3+3 2 | 2+2 2 | 2+2 2 | 2+2 2 |
| | 3+3 2 | 3+3 2 | 3+3 2 | 2+2 2 | 2+2 2 | 3+3 2 | 2+2 2 | 3+3 2 |
| | 2+2 2 | 2+2 2 | 2+2 2 | 3+3 2 | 2+2 2 | 2+2 2 | 3+3 2 | 3+3 2 |
| | 3+3 2 | 3+3 2 | 3+3 2 | 3+3 2 | 3+3 2 | 3+3 2 | 2+2 2 | 2+2 2 |
| | 2+2 2 | 3+3 2 | 2+2 2 | 3+3 2 | 3+3 2 | 2+2 2 | 2+2 2 | 2+2 2 |
| | 2+2 2 | 2+2 2 | 3+3 2 | 2+2 2 | 2+2 2 | 3+3 2 | 3+3 2 | 3+3 2 |
| | 3+3 2 | 3+3 2 | 2+2 2 | 2+2 2 | 2+2 2 | 2+2 2 | 3+3 2 | 2+2 2 |
| | 2+2 2 | 2+2 2 | 2+2 2 | 3+3 2 | 3+3 2 | 3+3 2 | 2+2 2 | 3+3 2 |
| | 2+2 2 | 3+3 2 | 3+3 2 | 3+3 2 | 3+3 2 | 3+3 2 | 2+2 2 | 2+2 2 |
| | 3+3 2 | 2+2 2 | 3+3 2 | 2+2 2 | 2+2 2 | 2+2 2 | 3+3 2 | 3+3 2 |
| | 3+3 2 | 3+3 2 | 3+3 2 | 3+3 2 | 3+3 2 | 3+3 2 | 3+3 2 | 3+3 2 |
| | 2+2 2 | 2+2 2 | 2+2 2 | 2+2 2 | 2+2 2 | 2+2 2 | 2+2 2 | 2+2 2 |
| | 3+3 2 | 3+3 2 | 3+3 2 | 2+2 2 | 2+2 2 | 3+3 2 | 2+2 2 | 2+2 2 |
| | 2+2 2 | 2+2 2 | 2+2 2 | 3+3 2 | 3+3 2 | 2+2 2 | 3+3 2 | 3+3 2 |
| | 3+3 2 | 3+3 2 | 3+3 2 | 2+2 2 | 3+3 2 | 2+2 2 | 3+3 2 | 3+3 2 |
| | 2+2 2 | 2+2 2 | 2+2 2 | 3+3 2 | 2+2 2 | 2+2 2 | 2+2 2 | 3+3 2 |
| | 2+2 2 | 3+3 2 | 3+3 2 | 3+3 2 | 3+3 2 | 3+3 2 | 2+2 2 | 2+2 2 |
| | 3+3 2 | 2+2 2 | 2+2 2 | 2+2 2 | 2+2 2 | 3+3 2 | 3+3 2 | 2+2 2 |
| | 2+2 2 | 3+3 2 | 3+3 2 | 2+2 2 | 2+2 2 | 2+2 2 | 2+2 2 | 3+3 2 |
| | 3+3 2 | 2+2 2 | 2+2 2 | 3+3 2 | 3+3 2 | 3+3 2 | 3+3 2 | 2+2 2 |
| | 3+3 2 | 2+2 2 | 2+2 2 | 2+2 2 | 3+3 2 | 2+2 2 | 2+2 2 | 3+3 2 |
| | 2+2 2 | 3+3 2 | 3+3 2 | 3+3 2 | 2+2 2 | 3+3 2 | 3+3 2 | 2+2 2 |
| | 3+3 2 | 3+3 2 | 2+2 2 | 3+3 2 | 3+3 2 | 2+2 2 | 3+3 2 | 2+2 2 |
| | 2+2 2 | 2+2 2 | 3+3 2 | 2+2 2 | 2+2 2 | 3+3 2 | 2+2 2 | 3+3 2 |
| | 3+3 2 | 2+2 2 | 2+2 2 | 2+2 2 | 3+3 2 | 2+2 2 | 2+2 2 | 2+2 2 |
| | 3+3 2 | 3+3 2 | 3+3 2 | 2+2 2 | 3+3 2 | 3+3 2 | 3+3 2 | 3+3 2 |
| | 3+3 2 | 2+2 2 | 2+2 2 | 3+3 2 | 2+2 2 | 2+2 2 | 2+2 2 | 3+3 2 |
| | 2+2 2 | 3+3 2 | 3+3 2 | 2+2 2 | 2+2 2 | 3+3 2 | 3+3 2 | 2+2 2 |
| | 3+3 2 | 2+2 2 | 2+2 2 | 3+3 2 | 3+3 2 | 2+2 2 | 2+2 2 | 3+3 2 |
| | 2+2 2 | 3+3 2 | 3+3 2 | 3+3 2 | 2+2 2 | 3+3 2 | 3+3 2 | 2+2 2 |
| | 3+3 2 | 3+3 2 | 2+2 2 | 2+2 2 | 2+2 2 | 3+3 2 | 3+3 2 | 2+2 2 |

Continued on next column

Continued on next column

**TABLE XXI– continued**

| TQC | ICSD | NoE | SF | | | | | |
|---|---|---|---|---|---|---|---|---|
| | 3+3 2 | 2+2 2 | 3+3 2 | 2+2 2 | 3+3 2 | 2+2 2 | 2+2 2 | 3+3 2 |
| | 2+2 2 | 3+3 2 | 3+3 2 | 3+3 2 | 3+3 2 | 2+2 2 | 3+3 2 | 3+3 2 |
| | 3+3 2 | 2+2 2 | 2+2 2 | 2+2 2 | 2+2 2 | 3+3 2 | 2+2 2 | 2+2 2 |
| | 2+2 2 | 2+2 2 | 3+3 2 | 2+2 2 | 2+2 2 | 3+3 2 | 2+2 2 | 2+2 2 |
| | 2+2 2 | 3+3 2 | 3+3 2 | 3+3 2 | 3+3 2 | 2+2 2 | 3+3 2 | 3+3 2 |
| | 2+2 2 | 2+2 2 | 2+2 2 | 3+3 2 | 2+2 2 | 3+3 2 | 2+2 2 | 2+2 2 |
| | 3+3 2 | 2+2 2 | 2+2 2 | 3+3 2 | 3+3 2 | 2+2 2 | 3+3 2 | 3+3 2 |
| | 2+2 2 | 3+3 2 | 3+3 2 | 2+2 2 | 2+2 2 | 3+3 2 | 2+2 2 | 2+2 2 |
| | 3+3 2 | 3+3 2 | 2+2 2 | 3+3 2 | 3+3 2 | 3+3 2 | 3+3 2 | 3+3 2 |
| | 2+2 2 | 3+3 2 | 3+3 2 | 3+3 2 | 2+2 2 | 2+2 2 | 2+2 2 | 2+2 2 |
| | 2+2 2 | 2+2 2 | 3+3 2 | 2+2 2 | 2+2 2 | 2+2 2 | 3+3 2 | 3+3 2 |
| | 2+2 2 | 3+3 2 | 2+2 2 | 3+3 2 | 3+3 2 | 3+3 2 | 3+3 2 | 2+2 2 |
| | 3+3 2 | 2+2 2 | 2+2 2 | 2+2 2 | 3+3 2 | 3+3 2 | 2+2 2 | 3+3 2 |
| | 3+3 2 | 3+3 2 | 2+2 2 | 2+2 2 | 2+2 2 | 2+2 2 | 2+2 2 | 2+2 2 |
| | 2+2 2 | 2+2 2 | 3+3 2 | 3+3 2 | 2+2 2 | 3+3 2 | 3+3 2 | 2+2 2 |
| | 2+2 2 | 3+3 2 | 2+2 2 | 3+3 2 | 3+3 2 | 2+2 2 | 2+2 2 | 3+3 2 |
| | 3+3 2 | 2+2 2 | 3+3 2 | 2+2 2 | 3+3 2 | 2+2 2 | 3+3 2 | 3+3 2 |
| | 3+3 2 | 2+2 2 | 2+2 2 | 3+3 2 | 2+2 2 | 3+3 2 | 2+2 2 | 3+3 2 |
| | 2+2 2 | 3+3 2 | 3+3 2 | 2+2 2 | 3+3 2 | 3+3 2 | 3+3 2 | 2+2 2 |
| | 3+3 2 | 2+2 2 | 2+2 2 | 2+2 2 | 3+3 2 | 2+2 2 | 2+2 2 | 2+2 2 |
| | 2+2 2 | 3+3 2 | 2+2 2 | 3+3 2 | 2+2 2 | 2+2 2 | 3+3 2 | 2+2 2 |
| | 3+3 2 | 3+3 2 | 3+3 2 | 3+3 2 | 2+2 2 | 2+2 2 | 3+3 2 | 3+3 2 |
| | 2+2 2 | 2+2 2 | 3+3 2 | 2+2 2 | 3+3 2 | 3+3 2 | 2+2 2 | 3+3 2 |
| | 2+2 2 | 3+3 2 | 2+2 2 | 3+3 2 | 2+2 2 | 3+3 2 | 2+2 2 | 2+2 2 |
| | 3+3 2 | 2+2 2 | 3+3 2 | 2+2 2 | 3+3 2 | 2+2 2 | 3+3 2 | 3+3 2 |
| | 2+2 2 | 3+3 2 | 3+3 2 | 3+3 2 | 3+3 2 | 3+3 2 | 2+2 2 | 2+2 2 |
| | 3+3 2 | 2+2 2 | 2+2 2 | 3+3 2 | 2+2 2 | 2+2 2 | 2+2 2 | 3+3 2 |
| | 3+3 2 | 3+3 2 | 2+2 2 | 2+2 2 | 3+3 2 | 2+2 2 | 3+3 2 | 3+3 2 |
| | 3+3 2 | 2+2 2 | 3+3 2 | 2+2 2 | 2+2 2 | 3+3 2 | 3+3 2 | 2+2 2 |
| | 2+2 2 | 3+3 2 | 2+2 2 | 3+3 2 | 3+3 2 | 3+3 2 | 2+2 2 | 2+2 2 |
| | 2+2 2 | 3+3 2 | 2+2 2 | 2+2 2 | 2+2 2 | 2+2 2 | 3+3 2 | 3+3 2 |
| | 3+3 2 | 2+2 2 | 3+3 2 | 3+3 2 | 2+2 2 | 3+3 2 | 2+2 2 | 2+2 2 |
| | 2+2 2 | 2+2 2 | 3+3 2 | 3+3 2 | 3+3 2 | 3+3 2 | 3+3 2 | 3+3 2 |
| | 2+2 2 | 3+3 2 | 2+2 2 | 2+2 2 | 3+3 2 | 2+2 2 | 2+2 2 | 2+2 2 |
| | 3+3 2 | 2+2 2 | 3+3 2 | 2+2 2 | 2+2 2 | 2+2 2 | 2+2 2 | 3+3 2 |
| | 3+3 2 | 3+3 2 | 3+3 2 | 2+2 2 | 3+3 2 | 3+3 2 | 3+3 2 | 2+2 2 |
| | 2+2 2 | 2+2 2 | 2+2 2 | 2+2 2 | 2+2 2 | 2+2 2 | 3+3 2 | 3+3 2 |
| | 2+2 2 | 2+2 2 | 3+3 2 | 3+3 2 | 2+2 2 | 3+3 2 | 2+2 2 | 2+2 2 |
| | 3+3 2 | 3+3 2 | 2+2 2 | 3+3 2 | 3+3 2 | 2+2 2 | 3+3 2 | 3+3 2 |
| | 3+3 2 | 2+2 2 | 2+2 2 | 2+2 2 | 2+2 2 | 3+3 2 | 2+2 2 | 2+2 2 |
| | 2+2 2 | 2+2 2 | 3+3 2 | 3+3 2 | 3+3 2 | 2+2 2 | 3+3 2 | 3+3 2 |
| | 3+3 2 | 3+3 2 | 3+3 2 | 2+2 2 | 2+2 2 | 3+3 2 | 2+2 2 | 2+2 2 |
| | 2+2 2 | 2+2 2 | 2+2 2 | 3+3 2 | 3+3 2 | 2+2 2 | 3+3 2 | 3+3 2 |
| | 3+3 2 | 3+3 2 | 3+3 2 | 2+2 2 | 3+3 2 | 2+2 2 | 2+2 2 | 3+3 2 |
| | 2+2 2 | 3+3 2 | 2+2 2 | 3+3 2 | 2+2 2 | 3+3 2 | 3+3 2 | 2+2 2 |
| | 3+3 2 | 3+3 2 | 3+3 2 | 2+2 2 | 2+2 2 | 2+2 2 | 2+2 2 | 2+2 2 |
| | 2+2 2 | 2+2 2 | 3+3 2 | 3+3 2 | 3+3 2 | 2+2 2 | 2+2 2 | 3+3 2 |
| | 3+3 2 | 2+2 2 | 2+2 2 | 2+2 2 | 3+3 2 | 3+3 2 | 3+3 2 | 2+2 2 |
| | 3+3 2 | 3+3 2 | 3+3 2 | 3+3 2 | 2+2 2 | 3+3 2 | 3+3 2 | 3+3 2 |
| | 2+2 2 | 3+3 2 | 2+2 2 | 2+2 2 | 2+2 2 | 2+2 2 | 2+2 2 | 2+2 2 |
| | 3+3 2 | 2+2 2 | 2+2 2 | 3+3 2 | 3+3 2 | 3+3 2 | 3+3 2 | 3+3 2 |
| | 2+2 2 | 2+2 2 | 3+3 2 | 3+3 2 | 3+3 2 | 2+2 2 | 3+3 2 | 2+2 2 |
| | 2+2 2 | 3+3 2 | 2+2 2 | 2+2 2 | 2+2 2 | 3+3 2 | 2+2 2 | 3+3 2 |
| | 3+3 2 | 3+3 2 | 3+3 2 | 2+2 2 | 2+2 2 | 3+3 2 | 3+3 2 | 2+2 2 |
| | 2+2 2 | 2+2 2 | 2+2 2 | 3+3 2 | 3+3 2 | 2+2 2 | 2+2 2 | 2+2 2 |
| | 2+2 2 | 3+3 2 | 3+3 2 | 2+2 2 | 3+3 2 | 2+2 2 | 3+3 2 | 3+3 2 |
| | 3+3 2 | 2+2 2 | 2+2 2 | 3+3 2 | 2+2 2 | 2+2 2 | 2+2 2 | 3+3 2 |
| | 2+2 2 | 3+3 2 | 2+2 2 | 2+2 2 | 3+3 2 | 3+3 2 | 3+3 2 | 2+2 2 |
| | 3+3 2 | 2+2 2 | 3+3 2 | 3+3 2 | 2+2 2 | 3+3 2 | 2+2 2 | 3+3 2 |
| | 2+2 2 | 2+2 2 | 2+2 2 | 3+3 2 | 3+3 2 | 2+2 2 | 3+3 2 | 3+3 2 |
| | 3+3 2 | 3+3 2 | 3+3 2 | 2+2 2 | 2+2 2 | 3+3 2 | 2+2 2 | 2+2 2 |
| | 3+3 2 | 2+2 2 | 2+2 2 | 3+3 2 | 3+3 2 | 3+3 2 | 3+3 2 | 3+3 2 |
| 2.2.40 | 84558 | 82 | Cu1O4W1 | | | | | |
| | -GM | -R | -T | -U | -V | -X | -Y | -Z |
| | 3+3 2 | 3+3 2 | 3+3 2 | 3+3 2 | 3+3 2 | 3+3 2 | 3+3 2 | 3+3 2 |
| | 2+2 2 | 2+2 2 | 2+2 2 | 3+3 2 | 2+2 2 | 2+2 2 | 2+2 2 | 2+2 2 |
| | 3+3 2 | 2+2 2 | 3+3 2 | 2+2 2 | 3+3 2 | 2+2 2 | 2+2 2 | 2+2 2 |
| | 2+2 2 | 3+3 2 | 2+2 2 | 2+2 2 | 2+2 2 | 3+3 2 | 3+3 2 | 3+3 2 |
| | 3+3 2 | 2+2 2 | 3+3 2 | 2+2 2 | 3+3 2 | 3+3 2 | 2+2 2 | 2+2 2 |
| | 2+2 2 | 3+3 2 | 2+2 2 | 3+3 2 | 2+2 2 | 2+2 2 | 3+3 2 | 3+3 2 |
| | 3+3 2 | 2+2 2 | 2+2 2 | 2+2 2 | 3+3 2 | 2+2 2 | 2+2 2 | 3+3 2 |
| | 2+2 2 | 3+3 2 | 3+3 2 | 3+3 2 | 2+2 2 | 3+3 2 | 3+3 2 | 2+2 2 |
| | 3+3 2 | 3+3 2 | 2+2 2 | 3+3 2 | 3+3 2 | 3+3 2 | 3+3 2 | 3+3 2 |
| | 2+2 2 | 2+2 2 | 3+3 2 | 2+2 2 | 2+2 2 | 2+2 2 | 2+2 2 | 2+2 2 |
| | 3+3 2 | 2+2 2 | 3+3 2 | 3+3 2 | 3+3 2 | 2+2 2 | 2+2 2 | 3+3 2 |
| | 3+3 2 | 3+3 2 | 2+2 2 | 2+2 2 | 2+2 2 | 3+3 2 | 3+3 2 | 2+2 2 |
| | 3+3 2 | 3+3 2 | 3+3 2 | 3+3 2 | 3+3 2 | 3+3 2 | 2+2 2 | 3+3 2 |
| | 2+2 2 | 2+2 2 | 2+2 2 | 3+3 2 | 2+2 2 | 2+2 2 | 3+3 2 | 2+2 2 |
| | 2+2 2 | 2+2 2 | 3+3 2 | 2+2 2 | 3+3 2 | 3+3 2 | 3+3 2 | 3+3 2 |
| | 3+3 2 | 3+3 2 | 2+2 2 | 2+2 2 | 3+3 2 | 2+2 2 | 2+2 2 | 2+2 2 |
| | 2+2 2 | 2+2 2 | 3+3 2 | 3+3 2 | 2+2 2 | 3+3 2 | 2+2 2 | 3+3 2 |
| | 3+3 2 | 3+3 2 | 2+2 2 | 2+2 2 | 2+2 2 | 2+2 2 | 3+3 2 | 2+2 2 |
| | 2+2 2 | 2+2 2 | 3+3 2 | 3+3 2 | 2+2 2 | 3+3 2 | 2+2 2 | 2+2 2 |
| | 2+2 2 | 3+3 2 | 2+2 2 | 2+2 2 | 3+3 2 | 3+3 2 | 3+3 2 | 3+3 2 |
| | 3+3 2 | 2+2 2 | 2+2 2 | 3+3 2 | 3+3 2 | 2+2 2 | 2+2 2 | 3+3 2 |
| | 3+3 2 | 3+3 2 | 3+3 2 | 2+2 2 | 2+2 2 | 3+3 2 | 3+3 2 | 2+2 2 |
| | 3+3 2 | 2+2 2 | 2+2 2 | 3+3 2 | 2+2 2 | 2+2 2 | 2+2 2 | 3+3 2 |
| | 2+2 2 | 3+3 2 | 3+3 2 | 2+2 2 | 3+3 2 | 2+2 2 | 3+3 2 | 2+2 2 |
| | 3+3 2 | 3+3 2 | 2+2 2 | 3+3 2 | 2+2 2 | 2+2 2 | 2+2 2 | 3+3 2 |
| | 2+2 2 | 2+2 2 | 3+3 2 | 3+3 2 | 3+3 2 | 3+3 2 | 3+3 2 | 2+2 2 |
| | 2+2 2 | 3+3 2 | 2+2 2 | 2+2 2 | 2+2 2 | 3+3 2 | 2+2 2 | 3+3 2 |
| | 2+2 2 | 2+2 2 | 3+3 2 | 2+2 2 | 3+3 2 | 2+2 2 | 3+3 2 | 2+2 2 |
| | 3+3 2 | 2+2 2 | 2+2 2 | 2+2 2 | 3+3 2 | 2+2 2 | 3+3 2 | 2+2 2 |
| | 2+2 2 | 3+3 2 | 2+2 2 | 3+3 2 | 2+2 2 | 3+3 2 | 2+2 2 | 3+3 2 |
| | 3+3 2 | 3+3 2 | 3+3 2 | 2+2 2 | 3+3 2 | 2+2 2 | 2+2 2 | 2+2 2 |
| | 2+2 2 | 2+2 2 | 3+3 2 | 3+3 2 | 2+2 2 | 3+3 2 | 3+3 2 | 3+3 2 |
| | 3+3 2 | 2+2 2 | 2+2 2 | 2+2 2 | 2+2 2 | 2+2 2 | 2+2 2 | 2+2 2 |
| | 2+2 2 | 3+3 2 | 3+3 2 | 3+3 2 | 3+3 2 | 3+3 2 | 3+3 2 | 3+3 2 |
| | 3+3 2 | 2+2 2 | 2+2 2 | 2+2 2 | 3+3 2 | 2+2 2 | 3+3 2 | 3+3 2 |
| | 2+2 2 | 3+3 2 | 3+3 2 | 3+3 2 | 2+2 2 | 3+3 2 | 2+2 2 | 2+2 2 |
| | 3+3 2 | 3+3 2 | 3+3 2 | 2+2 2 | 2+2 2 | 2+2 2 | 2+2 2 | 2+2 2 |
| | 2+2 2 | 2+2 2 | 2+2 2 | 3+3 2 | 3+3 2 | 3+3 2 | 3+3 2 | 3+3 2 |
| | 3+3 2 | 3+3 2 | 2+2 2 | 3+3 2 | 2+2 2 | 3+3 2 | 2+2 2 | 3+3 2 |
| | 3+3 2 | 2+2 2 | 3+3 2 | 2+2 2 | 3+3 2 | 2+2 2 | 3+3 2 | 2+2 2 |
| | 2+2 2 | 3+3 2 | 2+2 2 | 3+3 2 | 3+3 2 | 3+3 2 | 3+3 2 | 3+3 2 |
| 2.2.41 | 69759 | 82 | Cu1O4W1 | | | | | |
| | -GM | -R | -T | -U | -V | -X | -Y | -Z |
| | 3+3 2 | 2+2 2 | 2+2 2 | 3+3 2 | 2+2 2 | 3+3 2 | 2+2 2 | 3+3 2 |
| | 2+2 2 | 3+3 2 | 3+3 2 | 3+3 2 | 3+3 2 | 2+2 2 | 3+3 2 | 2+2 2 |
| | 3+3 2 | 3+3 2 | 2+2 2 | 2+2 2 | 2+2 2 | 2+2 2 | 3+3 2 | 2+2 2 |
| | 2+2 2 | 2+2 2 | 3+3 2 | 2+2 2 | 3+3 2 | 3+3 2 | 2+2 2 | 3+3 2 |
| | 3+3 2 | 3+3 2 | 2+2 2 | 2+2 2 | 2+2 2 | 3+3 2 | 3+3 2 | 2+2 2 |
| | 2+2 2 | 2+2 2 | 3+3 2 | 3+3 2 | 3+3 2 | 2+2 2 | 2+2 2 | 3+3 2 |
| | 3+3 2 | 3+3 2 | 3+3 2 | 2+2 2 | 2+2 2 | 2+2 2 | 3+3 2 | 3+3 2 |
| | 2+2 2 | 2+2 2 | 2+2 2 | 3+3 2 | 3+3 2 | 3+3 2 | 2+2 2 | 2+2 2 |
| | 3+3 2 | 2+2 2 | 3+3 2 | 3+3 2 | 2+2 2 | 3+3 2 | 2+2 2 | 3+3 2 |
| | 2+2 2 | 3+3 2 | 2+2 2 | 2+2 2 | 3+3 2 | 2+2 2 | 3+3 2 | 2+2 2 |
| | 3+3 2 | 3+3 2 | 2+2 2 | 3+3 2 | 2+2 2 | 2+2 2 | 3+3 2 | 3+3 2 |
| | 3+3 2 | 2+2 2 | 3+3 2 | 2+2 2 | 3+3 2 | 3+3 2 | 2+2 2 | 2+2 2 |
| | 3+3 2 | 2+2 2 | 2+2 2 | 3+3 2 | 2+2 2 | 3+3 2 | 3+3 2 | 3+3 2 |
| | 2+2 2 | 3+3 2 | 3+3 2 | 3+3 2 | 3+3 2 | 2+2 2 | 2+2 2 | 2+2 2 |
| | 2+2 2 | 3+3 2 | 2+2 2 | 2+2 2 | 2+2 2 | 3+3 2 | 2+2 2 | 3+3 2 |

Continued on next column

**TABLE XXI– continued**

| TQC | ICSD | NoE | SF | | | | | |
|---|---|---|---|---|---|---|---|---|
| | 3+3 2 | 2+2 2 | 3+3 2 | 2+2 2 | 2+2 2 | 2+2 2 | 3+3 2 | 2+2 2 |
| | 2+2 2 | 3+3 2 | 2+2 2 | 3+3 2 | 3+3 2 | 3+3 2 | 3+3 2 | 3+3 2 |
| | 3+3 2 | 2+2 2 | 3+3 2 | 2+2 2 | 3+3 2 | 2+2 2 | 2+2 2 | 2+2 2 |
| | 2+2 2 | 3+3 2 | 2+2 2 | 3+3 2 | 3+3 2 | 3+3 2 | 3+3 2 | 2+2 2 |
| | 2+2 2 | 2+2 2 | 3+3 2 | 2+2 2 | 2+2 2 | 3+3 2 | 2+2 2 | 3+3 2 |
| | 3+3 2 | 3+3 2 | 3+3 2 | 3+3 2 | 2+2 2 | 2+2 2 | 3+3 2 | 3+3 2 |
| | 3+3 2 | 2+2 2 | 2+2 2 | 2+2 2 | 3+3 2 | 3+3 2 | 2+2 2 | 2+2 2 |
| | 3+3 2 | 3+3 2 | 3+3 2 | 3+3 2 | 3+3 2 | 2+2 2 | 3+3 2 | 3+3 2 |
| | 2+2 2 | 2+2 2 | 2+2 2 | 2+2 2 | 2+2 2 | 2+2 2 | 2+2 2 | 2+2 2 |
| | 3+3 2 | 2+2 2 | 2+2 2 | 3+3 2 | 3+3 2 | 2+2 2 | 3+3 2 | 3+3 2 |
| | 2+2 2 | 3+3 2 | 3+3 2 | 3+3 2 | 2+2 2 | 3+3 2 | 2+2 2 | 2+2 2 |
| | 2+2 2 | 2+2 2 | 3+3 2 | 2+2 2 | 3+3 2 | 3+3 2 | 3+3 2 | 3+3 2 |
| | 2+2 2 | 3+3 2 | 2+2 2 | 2+2 2 | 2+2 2 | 2+2 2 | 2+2 2 | 2+2 2 |
| | 3+3 2 | 3+3 2 | 3+3 2 | 2+2 2 | 2+2 2 | 2+2 2 | 2+2 2 | 3+3 2 |
| | 2+2 2 | 2+2 2 | 3+3 2 | 3+3 2 | 3+3 2 | 3+3 2 | 3+3 2 | 2+2 2 |
| | 3+3 2 | 2+2 2 | 2+2 2 | 2+2 2 | 2+2 2 | 2+2 2 | 3+3 2 | 2+2 2 |
| | 2+2 2 | 3+3 2 | 2+2 2 | 3+3 2 | 3+3 2 | 3+3 2 | 2+2 2 | 3+3 2 |
| | 3+3 2 | 3+3 2 | 3+3 2 | 2+2 2 | 3+3 2 | 2+2 2 | 3+3 2 | 2+2 2 |
| | 2+2 2 | 2+2 2 | 2+2 2 | 3+3 2 | 2+2 2 | 3+3 2 | 2+2 2 | 3+3 2 |
| | 3+3 2 | 3+3 2 | 3+3 2 | 2+2 2 | 3+3 2 | 2+2 2 | 2+2 2 | 3+3 2 |
| | 2+2 2 | 2+2 2 | 2+2 2 | 3+3 2 | 2+2 2 | 3+3 2 | 3+3 2 | 2+2 2 |
| | 3+3 2 | 2+2 2 | 2+2 2 | 2+2 2 | 3+3 2 | 2+2 2 | 3+3 2 | 2+2 2 |
| | 2+2 2 | 3+3 2 | 3+3 2 | 3+3 2 | 2+2 2 | 3+3 2 | 2+2 2 | 3+3 2 |
| | 3+3 2 | 2+2 2 | 3+3 2 | 3+3 2 | 2+2 2 | 3+3 2 | 3+3 2 | 3+3 2 |
| | 3+3 2 | 3+3 2 | 2+2 2 | 2+2 2 | 3+3 2 | 2+2 2 | 2+2 2 | 2+2 2 |
| | 2+2 2 | 2+2 2 | 3+3 2 | 3+3 2 | 2+2 2 | 3+3 2 | 2+2 2 | 3+3 2 |
| 2.2.42 | 4189 | 82 | Cu1O4W1 | | | | | |
| | -GM | -R | -T | -U | -V | -X | -Y | -Z |
| | 3+3 2 | 2+2 2 | 2+2 2 | 3+3 2 | 2+2 2 | 3+3 2 | 2+2 2 | 3+3 2 |
| | 2+2 2 | 3+3 2 | 3+3 2 | 3+3 2 | 3+3 2 | 2+2 2 | 3+3 2 | 2+2 2 |
| | 3+3 2 | 3+3 2 | 2+2 2 | 2+2 2 | 2+2 2 | 2+2 2 | 3+3 2 | 2+2 2 |
| | 2+2 2 | 2+2 2 | 3+3 2 | 2+2 2 | 3+3 2 | 3+3 2 | 2+2 2 | 3+3 2 |
| | 3+3 2 | 3+3 2 | 2+2 2 | 3+3 2 | 2+2 2 | 2+2 2 | 3+3 2 | 2+2 2 |
| | 2+2 2 | 2+2 2 | 3+3 2 | 2+2 2 | 3+3 2 | 3+3 2 | 2+2 2 | 3+3 2 |
| | 3+3 2 | 3+3 2 | 3+3 2 | 2+2 2 | 3+3 2 | 2+2 2 | 3+3 2 | 3+3 2 |
| | 2+2 2 | 2+2 2 | 2+2 2 | 3+3 2 | 2+2 2 | 3+3 2 | 2+2 2 | 2+2 2 |
| | 3+3 2 | 2+2 2 | 3+3 2 | 3+3 2 | 2+2 2 | 3+3 2 | 2+2 2 | 3+3 2 |
| | 2+2 2 | 3+3 2 | 2+2 2 | 2+2 2 | 3+3 2 | 2+2 2 | 3+3 2 | 2+2 2 |
| | 3+3 2 | 3+3 2 | 2+2 2 | 3+3 2 | 2+2 2 | 2+2 2 | 2+2 2 | 2+2 2 |
| | 3+3 2 | 2+2 2 | 3+3 2 | 2+2 2 | 3+3 2 | 3+3 2 | 3+3 2 | 3+3 2 |
| | 3+3 2 | 2+2 2 | 2+2 2 | 3+3 2 | 2+2 2 | 3+3 2 | 2+2 2 | 2+2 2 |
| | 2+2 2 | 3+3 2 | 3+3 2 | 3+3 2 | 3+3 2 | 2+2 2 | 3+3 2 | 3+3 2 |
| | 2+2 2 | 3+3 2 | 2+2 2 | 2+2 2 | 2+2 2 | 3+3 2 | 2+2 2 | 3+3 2 |
| | 3+3 2 | 2+2 2 | 3+3 2 | 2+2 2 | 2+2 2 | 2+2 2 | 3+3 2 | 2+2 2 |
| | 2+2 2 | 3+3 2 | 2+2 2 | 3+3 2 | 3+3 2 | 3+3 2 | 3+3 2 | 3+3 2 |
| | 3+3 2 | 2+2 2 | 3+3 2 | 2+2 2 | 3+3 2 | 2+2 2 | 2+2 2 | 2+2 2 |
| | 2+2 2 | 3+3 2 | 2+2 2 | 3+3 2 | 3+3 2 | 3+3 2 | 3+3 2 | 2+2 2 |
| | 2+2 2 | 2+2 2 | 3+3 2 | 3+3 2 | 2+2 2 | 2+2 2 | 2+2 2 | 3+3 2 |
| | 3+3 2 | 3+3 2 | 2+2 2 | 2+2 2 | 3+3 2 | 3+3 2 | 3+3 2 | 3+3 2 |
| | 3+3 2 | 2+2 2 | 3+3 2 | 2+2 2 | 2+2 2 | 3+3 2 | 3+3 2 | 2+2 2 |
| | 2+2 2 | 3+3 2 | 3+3 2 | 3+3 2 | 3+3 2 | 2+2 2 | 2+2 2 | 3+3 2 |
| | 3+3 2 | 2+2 2 | 2+2 2 | 2+2 2 | 2+2 2 | 2+2 2 | 2+2 2 | 2+2 2 |
| | 2+2 2 | 2+2 2 | 3+3 2 | 3+3 2 | 3+3 2 | 2+2 2 | 3+3 2 | 3+3 2 |
| | 3+3 2 | 3+3 2 | 2+2 2 | 3+3 2 | 2+2 2 | 3+3 2 | 2+2 2 | 2+2 2 |
| | 2+2 2 | 2+2 2 | 3+3 2 | 2+2 2 | 3+3 2 | 3+3 2 | 3+3 2 | 3+3 2 |
| | 2+2 2 | 3+3 2 | 2+2 2 | 2+2 2 | 2+2 2 | 2+2 2 | 2+2 2 | 2+2 2 |
| | 3+3 2 | 2+2 2 | 3+3 2 | 2+2 2 | 2+2 2 | 3+3 2 | 2+2 2 | 2+2 2 |
| | 2+2 2 | 3+3 2 | 3+3 2 | 3+3 2 | 3+3 2 | 2+2 2 | 3+3 2 | 3+3 2 |
| | 3+3 2 | 2+2 2 | 2+2 2 | 2+2 2 | 2+2 2 | 2+2 2 | 3+3 2 | 2+2 2 |
| | 2+2 2 | 3+3 2 | 2+2 2 | 3+3 2 | 3+3 2 | 3+3 2 | 2+2 2 | 3+3 2 |
| | 3+3 2 | 3+3 2 | 2+2 2 | 3+3 2 | 3+3 2 | 3+3 2 | 2+2 2 | 3+3 2 |

Continued on next column

**TABLE XXI– continued**

| TQC | ICSD | NoE | SF | | | | | |
|---|---|---|---|---|---|---|---|---|
| | 2+2 2 | 2+2 2 | 3+3 2 | 2+2 2 | 2+2 2 | 2+2 2 | 3+3 2 | 2+2 2 |
| | 3+3 2 | 3+3 2 | 3+3 2 | 2+2 2 | 2+2 2 | 2+2 2 | 2+2 2 | 3+3 2 |
| | 2+2 2 | 2+2 2 | 2+2 2 | 3+3 2 | 3+3 2 | 3+3 2 | 3+3 2 | 2+2 2 |
| | 3+3 2 | 2+2 2 | 2+2 2 | 2+2 2 | 3+3 2 | 2+2 2 | 3+3 2 | 2+2 2 |
| | 2+2 2 | 3+3 2 | 3+3 2 | 3+3 2 | 2+2 2 | 3+3 2 | 2+2 2 | 3+3 2 |
| | 3+3 2 | 3+3 2 | 3+3 2 | 3+3 2 | 3+3 2 | 3+3 2 | 3+3 2 | 3+3 2 |
| | 3+3 2 | 2+2 2 | 2+2 2 | 2+2 2 | 2+2 2 | 2+2 2 | 2+2 2 | 2+2 2 |
| | 2+2 2 | 2+2 2 | 3+3 2 | 2+2 2 | 2+2 2 | 3+3 2 | 2+2 2 | 3+3 2 |
| 2.2.43 | 24339 | 82 | Cu1O4W1 | | | | | |
| | -GM | -R | -T | -U | -V | -X | -Y | -Z |
| | 3+3 2 | 2+2 2 | 2+2 2 | 3+3 2 | 2+2 2 | 3+3 2 | 2+2 2 | 3+3 2 |
| | 3+3 2 | 3+3 2 | 3+3 2 | 3+3 2 | 2+2 2 | 2+2 2 | 3+3 2 | 2+2 2 |
| | 2+2 2 | 3+3 2 | 2+2 2 | 2+2 2 | 3+3 2 | 2+2 2 | 3+3 2 | 2+2 2 |
| | 2+2 2 | 2+2 2 | 3+3 2 | 2+2 2 | 3+3 2 | 3+3 2 | 2+2 2 | 3+3 2 |
| | 3+3 2 | 2+2 2 | 3+3 2 | 2+2 2 | 3+3 2 | 3+3 2 | 2+2 2 | 2+2 2 |
| | 2+2 2 | 3+3 2 | 2+2 2 | 3+3 2 | 2+2 2 | 2+2 2 | 3+3 2 | 3+3 2 |
| | 3+3 2 | 3+3 2 | 3+3 2 | 2+2 2 | 2+2 2 | 2+2 2 | 3+3 2 | 3+3 2 |
| | 2+2 2 | 2+2 2 | 2+2 2 | 3+3 2 | 3+3 2 | 3+3 2 | 2+2 2 | 2+2 2 |
| | 3+3 2 | 2+2 2 | 2+2 2 | 3+3 2 | 2+2 2 | 3+3 2 | 2+2 2 | 3+3 2 |
| | 2+2 2 | 3+3 2 | 3+3 2 | 3+3 2 | 2+2 2 | 2+2 2 | 3+3 2 | 2+2 2 |
| | 3+3 2 | 2+2 2 | 2+2 2 | 2+2 2 | 3+3 2 | 3+3 2 | 2+2 2 | 3+3 2 |
| | 3+3 2 | 3+3 2 | 3+3 2 | 3+3 2 | 3+3 2 | 2+2 2 | 3+3 2 | 2+2 2 |
| | 3+3 2 | 2+2 2 | 2+2 2 | 2+2 2 | 2+2 2 | 3+3 2 | 2+2 2 | 3+3 2 |
| | 2+2 2 | 3+3 2 | 3+3 2 | 3+3 2 | 3+3 2 | 2+2 2 | 3+3 2 | 2+2 2 |
| | 3+3 2 | 3+3 2 | 2+2 2 | 2+2 2 | 2+2 2 | 3+3 2 | 2+2 2 | 3+3 2 |
| | 2+2 2 | 2+2 2 | 3+3 2 | 3+3 2 | 3+3 2 | 2+2 2 | 3+3 2 | 2+2 2 |
| | 2+2 2 | 3+3 2 | 2+2 2 | 2+2 2 | 2+2 2 | 3+3 2 | 3+3 2 | 3+3 2 |
| | 3+3 2 | 2+2 2 | 3+3 2 | 3+3 2 | 3+3 2 | 2+2 2 | 2+2 2 | 2+2 2 |
| | 2+2 2 | 2+2 2 | 2+2 2 | 2+2 2 | 3+3 2 | 3+3 2 | 3+3 2 | 3+3 2 |
| | 3+3 2 | 3+3 2 | 3+3 2 | 2+2 2 | 2+2 2 | 3+3 2 | 2+2 2 | 2+2 2 |
| | 2+2 2 | 3+3 2 | 3+3 2 | 3+3 2 | 3+3 2 | 2+2 2 | 3+3 2 | 2+2 2 |
| | 3+3 2 | 2+2 2 | 2+2 2 | 2+2 2 | 2+2 2 | 2+2 2 | 2+2 2 | 3+3 2 |
| | 2+2 2 | 3+3 2 | 3+3 2 | 3+3 2 | 3+3 2 | 3+3 2 | 3+3 2 | 3+3 2 |
| | 3+3 2 | 2+2 2 | 2+2 2 | 2+2 2 | 2+2 2 | 2+2 2 | 2+2 2 | 2+2 2 |
| | 2+2 2 | 2+2 2 | 3+3 2 | 3+3 2 | 3+3 2 | 2+2 2 | 3+3 2 | 3+3 2 |
| | 3+3 2 | 3+3 2 | 2+2 2 | 3+3 2 | 2+2 2 | 3+3 2 | 2+2 2 | 2+2 2 |
| | 2+2 2 | 2+2 2 | 3+3 2 | 2+2 2 | 2+2 2 | 2+2 2 | 3+3 2 | 3+3 2 |
| | 2+2 2 | 3+3 2 | 2+2 2 | 2+2 2 | 3+3 2 | 3+3 2 | 2+2 2 | 2+2 2 |
| | 3+3 2 | 3+3 2 | 3+3 2 | 2+2 2 | 2+2 2 | 2+2 2 | 2+2 2 | 3+3 2 |
| | 2+2 2 | 2+2 2 | 3+3 2 | 3+3 2 | 2+2 2 | 3+3 2 | 3+3 2 | 2+2 2 |
| | 3+3 2 | 2+2 2 | 2+2 2 | 2+2 2 | 3+3 2 | 2+2 2 | 3+3 2 | 2+2 2 |
| | 3+3 2 | 3+3 2 | 2+2 2 | 3+3 2 | 3+3 2 | 3+3 2 | 2+2 2 | 3+3 2 |
| | 2+2 2 | 3+3 2 | 2+2 2 | 2+2 2 | 3+3 2 | 3+3 2 | 2+2 2 | 2+2 2 |
| | 2+2 2 | 2+2 2 | 3+3 2 | 3+3 2 | 2+2 2 | 2+2 2 | 3+3 2 | 3+3 2 |
| | 3+3 2 | 3+3 2 | 3+3 2 | 2+2 2 | 3+3 2 | 2+2 2 | 2+2 2 | 2+2 2 |
| | 2+2 2 | 2+2 2 | 2+2 2 | 3+3 2 | 2+2 2 | 3+3 2 | 3+3 2 | 3+3 2 |
| | 3+3 2 | 2+2 2 | 2+2 2 | 2+2 2 | 3+3 2 | 3+3 2 | 3+3 2 | 2+2 2 |
| | 2+2 2 | 3+3 2 | 3+3 2 | 3+3 2 | 2+2 2 | 2+2 2 | 2+2 2 | 3+3 2 |
| | 3+3 2 | 3+3 2 | 2+2 2 | 3+3 2 | 2+2 2 | 3+3 2 | 3+3 2 | 3+3 2 |
| | 3+3 2 | 2+2 2 | 3+3 2 | 2+2 2 | 3+3 2 | 2+2 2 | 2+2 2 | 2+2 2 |
| | 2+2 2 | 2+2 2 | 3+3 2 | 3+3 2 | 2+2 2 | 3+3 2 | 2+2 2 | 3+3 2 |
| 2.2.44 | 182753 | 82 | Cu1O4W1 | | | | | |
| | -GM | -R | -T | -U | -V | -X | -Y | -Z |
| | 3+3 2 | 2+2 2 | 2+2 2 | 3+3 2 | 2+2 2 | 3+3 2 | 2+2 2 | 3+3 2 |
| | 2+2 2 | 3+3 2 | 3+3 2 | 3+3 2 | 3+3 2 | 2+2 2 | 3+3 2 | 2+2 2 |
| | 3+3 2 | 3+3 2 | 2+2 2 | 2+2 2 | 2+2 2 | 2+2 2 | 3+3 2 | 3+3 2 |
| | 2+2 2 | 2+2 2 | 3+3 2 | 2+2 2 | 3+3 2 | 3+3 2 | 2+2 2 | 2+2 2 |
| | 3+3 2 | 3+3 2 | 2+2 2 | 2+2 2 | 2+2 2 | 3+3 2 | 3+3 2 | 2+2 2 |
| | 2+2 2 | 2+2 2 | 3+3 2 | 3+3 2 | 3+3 2 | 2+2 2 | 2+2 2 | 3+3 2 |
| | 3+3 2 | 2+2 2 | 3+3 2 | 2+2 2 | 2+2 2 | 2+2 2 | 3+3 2 | 3+3 2 |
| | 2+2 2 | 3+3 2 | 2+2 2 | 3+3 2 | 3+3 2 | 3+3 2 | 2+2 2 | 2+2 2 |

Continued on next column

**TABLE XXI– continued**

| TQC | ICSD | NoE | SF | | | | | |
|---|---|---|---|---|---|---|---|---|
| | 3+3 2 | 2+2 2 | 3+3 2 | 3+3 2 | 2+2 2 | 3+3 2 | 2+2 2 | 3+3 2 |
| | 2+2 2 | 3+3 2 | 2+2 2 | 2+2 2 | 3+3 2 | 2+2 2 | 3+3 2 | 2+2 2 |
| | 3+3 2 | 3+3 2 | 2+2 2 | 3+3 2 | 2+2 2 | 2+2 2 | 3+3 2 | 3+3 2 |
| | 3+3 2 | 2+2 2 | 3+3 2 | 2+2 2 | 3+3 2 | 3+3 2 | 2+2 2 | 2+2 2 |
| | 3+3 2 | 2+2 2 | 2+2 2 | 3+3 2 | 2+2 2 | 3+3 2 | 2+2 2 | 3+3 2 |
| | 2+2 2 | 3+3 2 | 3+3 2 | 3+3 2 | 3+3 2 | 2+2 2 | 3+3 2 | 2+2 2 |
| | 2+2 2 | 3+3 2 | 2+2 2 | 2+2 2 | 2+2 2 | 3+3 2 | 2+2 2 | 3+3 2 |
| | 3+3 2 | 2+2 2 | 3+3 2 | 2+2 2 | 2+2 2 | 2+2 2 | 3+3 2 | 2+2 2 |
| | 2+2 2 | 3+3 2 | 2+2 2 | 3+3 2 | 3+3 2 | 3+3 2 | 3+3 2 | 3+3 2 |
| | 3+3 2 | 2+2 2 | 3+3 2 | 2+2 2 | 3+3 2 | 2+2 2 | 2+2 2 | 2+2 2 |
| | 2+2 2 | 3+3 2 | 2+2 2 | 3+3 2 | 3+3 2 | 3+3 2 | 2+2 2 | 2+2 2 |
| | 2+2 2 | 2+2 2 | 3+3 2 | 2+2 2 | 2+2 2 | 3+3 2 | 3+3 2 | 3+3 2 |
| | 3+3 2 | 2+2 2 | 2+2 2 | 3+3 2 | 2+2 2 | 2+2 2 | 3+3 2 | 3+3 2 |
| | 3+3 2 | 3+3 2 | 3+3 2 | 2+2 2 | 3+3 2 | 3+3 2 | 2+2 2 | 2+2 2 |
| | 3+3 2 | 3+3 2 | 3+3 2 | 3+3 2 | 3+3 2 | 2+2 2 | 3+3 2 | 3+3 2 |
| | 2+2 2 | 2+2 2 | 2+2 2 | 2+2 2 | 2+2 2 | 2+2 2 | 2+2 2 | 2+2 2 |
| | 3+3 2 | 2+2 2 | 2+2 2 | 3+3 2 | 3+3 2 | 2+2 2 | 3+3 2 | 3+3 2 |
| | 2+2 2 | 3+3 2 | 3+3 2 | 3+3 2 | 2+2 2 | 3+3 2 | 2+2 2 | 2+2 2 |
| | 2+2 2 | 2+2 2 | 3+3 2 | 2+2 2 | 2+2 2 | 3+3 2 | 3+3 2 | 3+3 2 |
| | 2+2 2 | 3+3 2 | 2+2 2 | 2+2 2 | 3+3 2 | 2+2 2 | 2+2 2 | 2+2 2 |
| | 3+3 2 | 3+3 2 | 3+3 2 | 2+2 2 | 2+2 2 | 2+2 2 | 2+2 2 | 2+2 2 |
| | 2+2 2 | 2+2 2 | 3+3 2 | 3+3 2 | 3+3 2 | 3+3 2 | 3+3 2 | 3+3 2 |
| | 3+3 2 | 2+2 2 | 2+2 2 | 2+2 2 | 2+2 2 | 2+2 2 | 3+3 2 | 2+2 2 |
| | 2+2 2 | 3+3 2 | 2+2 2 | 3+3 2 | 3+3 2 | 3+3 2 | 2+2 2 | 3+3 2 |
| | 3+3 2 | 3+3 2 | 3+3 2 | 2+2 2 | 3+3 2 | 2+2 2 | 3+3 2 | 2+2 2 |
| | 2+2 2 | 2+2 2 | 2+2 2 | 3+3 2 | 2+2 2 | 3+3 2 | 2+2 2 | 3+3 2 |
| | 3+3 2 | 3+3 2 | 3+3 2 | 2+2 2 | 3+3 2 | 2+2 2 | 2+2 2 | 3+3 2 |
| | 2+2 2 | 2+2 2 | 2+2 2 | 3+3 2 | 2+2 2 | 3+3 2 | 3+3 2 | 2+2 2 |
| | 3+3 2 | 2+2 2 | 2+2 2 | 2+2 2 | 3+3 2 | 2+2 2 | 3+3 2 | 2+2 2 |
| | 2+2 2 | 3+3 2 | 3+3 2 | 3+3 2 | 2+2 2 | 3+3 2 | 2+2 2 | 3+3 2 |
| | 3+3 2 | 2+2 2 | 3+3 2 | 2+2 2 | 3+3 2 | 2+2 2 | 3+3 2 | 3+3 2 |
| | 3+3 2 | 3+3 2 | 2+2 2 | 3+3 2 | 2+2 2 | 3+3 2 | 2+2 2 | 2+2 2 |
| | 2+2 2 | 2+2 2 | 3+3 2 | 3+3 2 | 2+2 2 | 3+3 2 | 2+2 2 | 3+3 2 |
| 2.2.45 | 182752 | 82 | Cu1O4W1 | | | | | |
| | -GM | -R | -T | -U | -V | -X | -Y | -Z |
| | 3+3 2 | 2+2 2 | 2+2 2 | 3+3 2 | 2+2 2 | 3+3 2 | 2+2 2 | 3+3 2 |
| | 3+3 2 | 3+3 2 | 3+3 2 | 3+3 2 | 3+3 2 | 2+2 2 | 3+3 2 | 2+2 2 |
| | 2+2 2 | 3+3 2 | 2+2 2 | 2+2 2 | 2+2 2 | 2+2 2 | 3+3 2 | 2+2 2 |
| | 2+2 2 | 2+2 2 | 3+3 2 | 2+2 2 | 3+3 2 | 3+3 2 | 2+2 2 | 3+3 2 |
| | 3+3 2 | 3+3 2 | 2+2 2 | 2+2 2 | 2+2 2 | 3+3 2 | 3+3 2 | 2+2 2 |
| | 2+2 2 | 2+2 2 | 3+3 2 | 3+3 2 | 3+3 2 | 2+2 2 | 2+2 2 | 3+3 2 |
| | 3+3 2 | 2+2 2 | 3+3 2 | 2+2 2 | 2+2 2 | 2+2 2 | 3+3 2 | 3+3 2 |
| | 2+2 2 | 3+3 2 | 2+2 2 | 3+3 2 | 3+3 2 | 3+3 2 | 2+2 2 | 2+2 2 |
| | 3+3 2 | 2+2 2 | 3+3 2 | 3+3 2 | 2+2 2 | 3+3 2 | 2+2 2 | 3+3 2 |
| | 2+2 2 | 3+3 2 | 2+2 2 | 2+2 2 | 3+3 2 | 2+2 2 | 3+3 2 | 2+2 2 |
| | 3+3 2 | 3+3 2 | 2+2 2 | 3+3 2 | 2+2 2 | 2+2 2 | 3+3 2 | 3+3 2 |
| | 3+3 2 | 2+2 2 | 3+3 2 | 2+2 2 | 3+3 2 | 3+3 2 | 2+2 2 | 2+2 2 |
| | 3+3 2 | 2+2 2 | 2+2 2 | 3+3 2 | 2+2 2 | 3+3 2 | 3+3 2 | 3+3 2 |
| | 2+2 2 | 3+3 2 | 3+3 2 | 3+3 2 | 3+3 2 | 2+2 2 | 2+2 2 | 2+2 2 |
| | 2+2 2 | 3+3 2 | 2+2 2 | 2+2 2 | 2+2 2 | 3+3 2 | 2+2 2 | 3+3 2 |
| | 3+3 2 | 2+2 2 | 3+3 2 | 2+2 2 | 2+2 2 | 2+2 2 | 3+3 2 | 2+2 2 |
| | 2+2 2 | 3+3 2 | 2+2 2 | 3+3 2 | 3+3 2 | 3+3 2 | 3+3 2 | 3+3 2 |
| | 3+3 2 | 2+2 2 | 3+3 2 | 2+2 2 | 3+3 2 | 2+2 2 | 2+2 2 | 2+2 2 |
| | 2+2 2 | 3+3 2 | 2+2 2 | 3+3 2 | 3+3 2 | 3+3 2 | 3+3 2 | 2+2 2 |
| | 2+2 2 | 2+2 2 | 3+3 2 | 2+2 2 | 2+2 2 | 3+3 2 | 2+2 2 | 3+3 2 |
| | 3+3 2 | 3+3 2 | 3+3 2 | 3+3 2 | 2+2 2 | 2+2 2 | 3+3 2 | 3+3 2 |
| | 3+3 2 | 2+2 2 | 2+2 2 | 2+2 2 | 3+3 2 | 3+3 2 | 2+2 2 | 2+2 2 |
| | 3+3 2 | 3+3 2 | 3+3 2 | 3+3 2 | 3+3 2 | 2+2 2 | 3+3 2 | 3+3 2 |
| | 2+2 2 | 2+2 2 | 2+2 2 | 2+2 2 | 2+2 2 | 2+2 2 | 2+2 2 | 2+2 2 |
| | 3+3 2 | 2+2 2 | 2+2 2 | 3+3 2 | 3+3 2 | 2+2 2 | 3+3 2 | 3+3 2 |
| | 2+2 2 | 3+3 2 | 3+3 2 | 3+3 2 | 2+2 2 | 3+3 2 | 2+2 2 | 2+2 2 |
| | 2+2 2 | 2+2 2 | 3+3 2 | 2+2 2 | 3+3 2 | 3+3 2 | 3+3 2 | 3+3 2 |
| | 2+2 2 | 3+3 2 | 2+2 2 | 2+2 2 | 2+2 2 | 2+2 2 | 2+2 2 | 2+2 2 |
| | 3+3 2 | 3+3 2 | 3+3 2 | 2+2 2 | 2+2 2 | 2+2 2 | 2+2 2 | 3+3 2 |
| | 2+2 2 | 2+2 2 | 3+3 2 | 3+3 2 | 3+3 2 | 3+3 2 | 3+3 2 | 2+2 2 |
| | 3+3 2 | 2+2 2 | 2+2 2 | 2+2 2 | 2+2 2 | 2+2 2 | 3+3 2 | 2+2 2 |
| | 2+2 2 | 3+3 2 | 2+2 2 | 3+3 2 | 3+3 2 | 3+3 2 | 2+2 2 | 3+3 2 |
| | 3+3 2 | 3+3 2 | 3+3 2 | 2+2 2 | 3+3 2 | 2+2 2 | 3+3 2 | 2+2 2 |
| | 2+2 2 | 2+2 2 | 2+2 2 | 3+3 2 | 2+2 2 | 3+3 2 | 2+2 2 | 3+3 2 |
| | 3+3 2 | 3+3 2 | 3+3 2 | 2+2 2 | 3+3 2 | 2+2 2 | 2+2 2 | 2+2 2 |
| | 2+2 2 | 2+2 2 | 2+2 2 | 3+3 2 | 2+2 2 | 3+3 2 | 3+3 2 | 3+3 2 |
| | 3+3 2 | 2+2 2 | 2+2 2 | 2+2 2 | 3+3 2 | 3+3 2 | 3+3 2 | 2+2 2 |
| | 2+2 2 | 3+3 2 | 3+3 2 | 3+3 2 | 2+2 2 | 2+2 2 | 2+2 2 | 3+3 2 |
| | 3+3 2 | 2+2 2 | 3+3 2 | 3+3 2 | 2+2 2 | 3+3 2 | 3+3 2 | 3+3 2 |
| | 3+3 2 | 3+3 2 | 2+2 2 | 2+2 2 | 3+3 2 | 2+2 2 | 2+2 2 | 2+2 2 |
| | 2+2 2 | 2+2 2 | 3+3 2 | 3+3 2 | 2+2 2 | 3+3 2 | 2+2 2 | 3+3 2 |
| 2.2.46 | 182750 | 82 | Cu1O4W1 | | | | | |
| | -GM | -R | -T | -U | -V | -X | -Y | -Z |
| | 3+3 2 | 2+2 2 | 2+2 2 | 3+3 2 | 2+2 2 | 3+3 2 | 2+2 2 | 3+3 2 |
| | 3+3 2 | 3+3 2 | 3+3 2 | 3+3 2 | 3+3 2 | 2+2 2 | 3+3 2 | 2+2 2 |
| | 2+2 2 | 3+3 2 | 2+2 2 | 2+2 2 | 2+2 2 | 2+2 2 | 3+3 2 | 3+3 2 |
| | 2+2 2 | 2+2 2 | 3+3 2 | 2+2 2 | 3+3 2 | 3+3 2 | 2+2 2 | 2+2 2 |
| | 3+3 2 | 3+3 2 | 2+2 2 | 2+2 2 | 2+2 2 | 3+3 2 | 3+3 2 | 2+2 2 |
| | 2+2 2 | 2+2 2 | 3+3 2 | 3+3 2 | 3+3 2 | 2+2 2 | 2+2 2 | 3+3 2 |
| | 3+3 2 | 2+2 2 | 3+3 2 | 2+2 2 | 2+2 2 | 2+2 2 | 3+3 2 | 3+3 2 |
| | 2+2 2 | 3+3 2 | 2+2 2 | 3+3 2 | 3+3 2 | 3+3 2 | 2+2 2 | 2+2 2 |
| | 3+3 2 | 2+2 2 | 3+3 2 | 3+3 2 | 2+2 2 | 3+3 2 | 2+2 2 | 3+3 2 |
| | 2+2 2 | 3+3 2 | 2+2 2 | 2+2 2 | 3+3 2 | 2+2 2 | 3+3 2 | 2+2 2 |
| | 3+3 2 | 2+2 2 | 2+2 2 | 3+3 2 | 2+2 2 | 3+3 2 | 2+2 2 | 2+2 2 |
| | 3+3 2 | 3+3 2 | 3+3 2 | 2+2 2 | 3+3 2 | 2+2 2 | 3+3 2 | 3+3 2 |
| | 3+3 2 | 2+2 2 | 2+2 2 | 3+3 2 | 2+2 2 | 2+2 2 | 2+2 2 | 3+3 2 |
| | 2+2 2 | 3+3 2 | 3+3 2 | 3+3 2 | 3+3 2 | 3+3 2 | 3+3 2 | 2+2 2 |
| | 2+2 2 | 3+3 2 | 2+2 2 | 2+2 2 | 2+2 2 | 3+3 2 | 2+2 2 | 3+3 2 |
| | 3+3 2 | 2+2 2 | 3+3 2 | 2+2 2 | 2+2 2 | 2+2 2 | 3+3 2 | 2+2 2 |
| | 2+2 2 | 3+3 2 | 2+2 2 | 3+3 2 | 3+3 2 | 3+3 2 | 3+3 2 | 3+3 2 |
| | 3+3 2 | 2+2 2 | 3+3 2 | 2+2 2 | 3+3 2 | 2+2 2 | 2+2 2 | 2+2 2 |
| | 2+2 2 | 3+3 2 | 2+2 2 | 3+3 2 | 3+3 2 | 3+3 2 | 3+3 2 | 2+2 2 |
| | 2+2 2 | 2+2 2 | 3+3 2 | 2+2 2 | 2+2 2 | 3+3 2 | 2+2 2 | 3+3 2 |
| | 3+3 2 | 2+2 2 | 2+2 2 | 3+3 2 | 2+2 2 | 2+2 2 | 3+3 2 | 2+2 2 |
| | 3+3 2 | 3+3 2 | 3+3 2 | 2+2 2 | 3+3 2 | 3+3 2 | 2+2 2 | 3+3 2 |
| | 3+3 2 | 3+3 2 | 3+3 2 | 3+3 2 | 3+3 2 | 2+2 2 | 3+3 2 | 3+3 2 |
| | 2+2 2 | 2+2 2 | 2+2 2 | 2+2 2 | 2+2 2 | 2+2 2 | 2+2 2 | 2+2 2 |
| | 3+3 2 | 2+2 2 | 2+2 2 | 3+3 2 | 3+3 2 | 3+3 2 | 3+3 2 | 3+3 2 |
| | 2+2 2 | 3+3 2 | 3+3 2 | 3+3 2 | 2+2 2 | 2+2 2 | 3+3 2 | 2+2 2 |
| | 2+2 2 | 2+2 2 | 3+3 2 | 2+2 2 | 2+2 2 | 3+3 2 | 2+2 2 | 3+3 2 |
| | 2+2 2 | 3+3 2 | 2+2 2 | 2+2 2 | 3+3 2 | 2+2 2 | 2+2 2 | 2+2 2 |
| | 3+3 2 | 3+3 2 | 3+3 2 | 2+2 2 | 2+2 2 | 2+2 2 | 2+2 2 | 2+2 2 |
| | 2+2 2 | 2+2 2 | 3+3 2 | 3+3 2 | 3+3 2 | 3+3 2 | 3+3 2 | 3+3 2 |
| | 3+3 2 | 2+2 2 | 2+2 2 | 2+2 2 | 2+2 2 | 2+2 2 | 3+3 2 | 2+2 2 |
| | 2+2 2 | 3+3 2 | 2+2 2 | 3+3 2 | 3+3 2 | 3+3 2 | 2+2 2 | 3+3 2 |
| | 3+3 2 | 3+3 2 | 3+3 2 | 2+2 2 | 3+3 2 | 3+3 2 | 3+3 2 | 2+2 2 |
| | 2+2 2 | 2+2 2 | 2+2 2 | 3+3 2 | 2+2 2 | 2+2 2 | 2+2 2 | 3+3 2 |
| | 3+3 2 | 3+3 2 | 3+3 2 | 2+2 2 | 3+3 2 | 2+2 2 | 2+2 2 | 3+3 2 |
| | 3+3 2 | 2+2 2 | 2+2 2 | 3+3 2 | 2+2 2 | 3+3 2 | 3+3 2 | 2+2 2 |
| | 2+2 2 | 2+2 2 | 2+2 2 | 2+2 2 | 3+3 2 | 2+2 2 | 2+2 2 | 2+2 2 |
| | 2+2 2 | 3+3 2 | 3+3 2 | 3+3 2 | 2+2 2 | 3+3 2 | 3+3 2 | 3+3 2 |
| | 3+3 2 | 2+2 2 | 2+2 2 | 3+3 2 | 3+3 2 | 3+3 2 | 3+3 2 | 3+3 2 |
| | 3+3 2 | 3+3 2 | 3+3 2 | 2+2 2 | 2+2 2 | 2+2 2 | 2+2 2 | 2+2 2 |
| | 2+2 2 | 2+2 2 | 3+3 2 | 3+3 2 | 2+2 2 | 3+3 2 | 2+2 2 | 3+3 2 |
| 2.2.47 | 16009 | 82 | Cu1O4W1 | | | | | |
| | -GM | -R | -T | -U | -V | -X | -Y | -Z |
| | 3+3 2 | 2+2 2 | 2+2 2 | 3+3 2 | 2+2 2 | 3+3 2 | 2+2 2 | 3+3 2 |

Continued on next column

Continued on next column

**TABLE XXI– continued**

| TQC | ICSD | NoE | SF | | | | | |
|---|---|---|---|---|---|---|---|---|
| | 2+2 2 | 3+3 2 | 3+3 2 | 3+3 2 | 3+3 2 | 2+2 2 | 3+3 2 | 2+2 2 |
| | 3+3 2 | 3+3 2 | 2+2 2 | 2+2 2 | 2+2 2 | 2+2 2 | 3+3 2 | 3+3 2 |
| | 2+2 2 | 2+2 2 | 3+3 2 | 2+2 2 | 3+3 2 | 3+3 2 | 2+2 2 | 2+2 2 |
| | 3+3 2 | 3+3 2 | 2+2 2 | 2+2 2 | 2+2 2 | 3+3 2 | 3+3 2 | 2+2 2 |
| | 2+2 2 | 2+2 2 | 3+3 2 | 3+3 2 | 3+3 2 | 2+2 2 | 2+2 2 | 3+3 2 |
| | 3+3 2 | 3+3 2 | 3+3 2 | 2+2 2 | 2+2 2 | 2+2 2 | 3+3 2 | 3+3 2 |
| | 2+2 2 | 2+2 2 | 2+2 2 | 3+3 2 | 3+3 2 | 3+3 2 | 2+2 2 | 2+2 2 |
| | 3+3 2 | 2+2 2 | 3+3 2 | 3+3 2 | 2+2 2 | 3+3 2 | 2+2 2 | 3+3 2 |
| | 2+2 2 | 3+3 2 | 2+2 2 | 2+2 2 | 3+3 2 | 2+2 2 | 3+3 2 | 2+2 2 |
| | 3+3 2 | 3+3 2 | 2+2 2 | 3+3 2 | 2+2 2 | 2+2 2 | 3+3 2 | 3+3 2 |
| | 3+3 2 | 2+2 2 | 3+3 2 | 2+2 2 | 3+3 2 | 3+3 2 | 2+2 2 | 2+2 2 |
| | 3+3 2 | 2+2 2 | 2+2 2 | 3+3 2 | 2+2 2 | 3+3 2 | 3+3 2 | 3+3 2 |
| | 2+2 2 | 3+3 2 | 3+3 2 | 3+3 2 | 3+3 2 | 2+2 2 | 2+2 2 | 2+2 2 |
| | 2+2 2 | 3+3 2 | 2+2 2 | 2+2 2 | 2+2 2 | 3+3 2 | 2+2 2 | 3+3 2 |
| | 3+3 2 | 2+2 2 | 3+3 2 | 2+2 2 | 2+2 2 | 2+2 2 | 3+3 2 | 2+2 2 |
| | 2+2 2 | 3+3 2 | 2+2 2 | 3+3 2 | 3+3 2 | 3+3 2 | 3+3 2 | 3+3 2 |
| | 3+3 2 | 2+2 2 | 3+3 2 | 2+2 2 | 3+3 2 | 2+2 2 | 2+2 2 | 2+2 2 |
| | 2+2 2 | 3+3 2 | 2+2 2 | 3+3 2 | 3+3 2 | 3+3 2 | 3+3 2 | 2+2 2 |
| | 2+2 2 | 2+2 2 | 3+3 2 | 2+2 2 | 2+2 2 | 3+3 2 | 2+2 2 | 3+3 2 |
| | 3+3 2 | 3+3 2 | 3+3 2 | 3+3 2 | 2+2 2 | 2+2 2 | 3+3 2 | 3+3 2 |
| | 3+3 2 | 2+2 2 | 2+2 2 | 2+2 2 | 3+3 2 | 3+3 2 | 2+2 2 | 2+2 2 |
| | 3+3 2 | 3+3 2 | 3+3 2 | 3+3 2 | 3+3 2 | 2+2 2 | 3+3 2 | 3+3 2 |
| | 2+2 2 | 2+2 2 | 2+2 2 | 2+2 2 | 2+2 2 | 2+2 2 | 2+2 2 | 2+2 2 |
| | 3+3 2 | 2+2 2 | 3+3 2 | 3+3 2 | 3+3 2 | 2+2 2 | 3+3 2 | 3+3 2 |
| | 2+2 2 | 3+3 2 | 2+2 2 | 3+3 2 | 2+2 2 | 3+3 2 | 2+2 2 | 2+2 2 |
| | 2+2 2 | 2+2 2 | 3+3 2 | 2+2 2 | 3+3 2 | 3+3 2 | 3+3 2 | 3+3 2 |
| | 2+2 2 | 3+3 2 | 2+2 2 | 2+2 2 | 2+2 2 | 2+2 2 | 2+2 2 | 2+2 2 |
| | 3+3 2 | 3+3 2 | 3+3 2 | 2+2 2 | 2+2 2 | 2+2 2 | 2+2 2 | 3+3 2 |
| | 2+2 2 | 2+2 2 | 3+3 2 | 3+3 2 | 3+3 2 | 3+3 2 | 3+3 2 | 2+2 2 |
| | 3+3 2 | 2+2 2 | 2+2 2 | 2+2 2 | 2+2 2 | 2+2 2 | 3+3 2 | 2+2 2 |
| | 2+2 2 | 3+3 2 | 2+2 2 | 3+3 2 | 3+3 2 | 3+3 2 | 2+2 2 | 3+3 2 |
| | 3+3 2 | 3+3 2 | 3+3 2 | 2+2 2 | 3+3 2 | 2+2 2 | 3+3 2 | 2+2 2 |
| | 2+2 2 | 2+2 2 | 2+2 2 | 3+3 2 | 2+2 2 | 3+3 2 | 2+2 2 | 3+3 2 |
| | 3+3 2 | 3+3 2 | 3+3 2 | 2+2 2 | 3+3 2 | 2+2 2 | 2+2 2 | 2+2 2 |
| | 2+2 2 | 2+2 2 | 2+2 2 | 3+3 2 | 2+2 2 | 3+3 2 | 3+3 2 | 3+3 2 |
| | 3+3 2 | 2+2 2 | 2+2 2 | 2+2 2 | 3+3 2 | 2+2 2 | 3+3 2 | 2+2 2 |
| | 2+2 2 | 3+3 2 | 3+3 2 | 3+3 2 | 2+2 2 | 3+3 2 | 2+2 2 | 3+3 2 |
| | 3+3 2 | 2+2 2 | 3+3 2 | 3+3 2 | 3+3 2 | 3+3 2 | 3+3 2 | 3+3 2 |
| | 3+3 2 | 3+3 2 | 2+2 2 | 2+2 2 | 2+2 2 | 2+2 2 | 2+2 2 | 2+2 2 |
| | 2+2 2 | 2+2 2 | 3+3 2 | 3+3 2 | 2+2 2 | 3+3 2 | 2+2 2 | 3+3 2 |
| 2.2.48 | 39439 | 82 | Cu1Mo1O4 | | | | | |
| | -GM | -R | -T | -U | -V | -X | -Y | -Z |
| | 3+3 2 | 2+2 2 | 2+2 2 | 2+2 2 | 2+2 2 | 3+3 2 | 2+2 2 | 2+2 2 |
| | 3+3 2 | 3+3 2 | 3+3 2 | 2+2 2 | 2+2 2 | 2+2 2 | 3+3 2 | 3+3 2 |
| | 2+2 2 | 2+2 2 | 3+3 2 | 3+3 2 | 3+3 2 | 3+3 2 | 3+3 2 | 2+2 2 |
| | 2+2 2 | 3+3 2 | 3+3 2 | 3+3 2 | 3+3 2 | 2+2 2 | 2+2 2 | 3+3 2 |
| | 3+3 2 | 2+2 2 | 2+2 2 | 2+2 2 | 2+2 2 | 3+3 2 | 3+3 2 | 3+3 2 |
| | 2+2 2 | 3+3 2 | 2+2 2 | 3+3 2 | 3+3 2 | 2+2 2 | 2+2 2 | 2+2 2 |
| | 3+3 2 | 3+3 2 | 2+2 2 | 3+3 2 | 2+2 2 | 2+2 2 | 3+3 2 | 2+2 2 |
| | 2+2 2 | 2+2 2 | 3+3 2 | 2+2 2 | 3+3 2 | 3+3 2 | 2+2 2 | 3+3 2 |
| | 3+3 2 | 3+3 2 | 2+2 2 | 2+2 2 | 2+2 2 | 3+3 2 | 2+2 2 | 2+2 2 |
| | 2+2 2 | 2+2 2 | 3+3 2 | 3+3 2 | 3+3 2 | 2+2 2 | 3+3 2 | 3+3 2 |
| | 3+3 2 | 2+2 2 | 3+3 2 | 2+2 2 | 2+2 2 | 3+3 2 | 3+3 2 | 2+2 2 |
| | 3+3 2 | 3+3 2 | 2+2 2 | 3+3 2 | 3+3 2 | 2+2 2 | 2+2 2 | 3+3 2 |
| | 3+3 2 | 3+3 2 | 3+3 2 | 2+2 2 | 2+2 2 | 3+3 2 | 2+2 2 | 2+2 2 |
| | 2+2 2 | 2+2 2 | 3+3 2 | 2+2 2 | 3+3 2 | 2+2 2 | 3+3 2 | 2+2 2 |
| | 2+2 2 | 2+2 2 | 2+2 2 | 3+3 2 | 2+2 2 | 3+3 2 | 2+2 2 | 3+3 2 |
| | 3+3 2 | 3+3 2 | 2+2 2 | 3+3 2 | 2+2 2 | 2+2 2 | 3+3 2 | 3+3 2 |
| | 2+2 2 | 3+3 2 | 3+3 2 | 2+2 2 | 3+3 2 | 3+3 2 | 3+3 2 | 2+2 2 |
| | 3+3 2 | 2+2 2 | 3+3 2 | 3+3 2 | 3+3 2 | 2+2 2 | 2+2 2 | 3+3 2 |
| | 2+2 2 | 3+3 2 | 2+2 2 | 2+2 2 | 2+2 2 | 3+3 2 | 2+2 2 | 3+3 2 |

Continued on next column

**TABLE XXI– continued**

| TQC | ICSD | NoE | SF | | | | | |
|---|---|---|---|---|---|---|---|---|
| | 3+3 2 | 2+2 2 | 2+2 2 | 3+3 2 | 3+3 2 | 2+2 2 | 3+3 2 | 2+2 2 |
| | 2+2 2 | 3+3 2 | 2+2 2 | 2+2 2 | 3+3 2 | 3+3 2 | 3+3 2 | 2+2 2 |
| | 3+3 2 | 2+2 2 | 3+3 2 | 2+2 2 | 2+2 2 | 2+2 2 | 2+2 2 | 3+3 2 |
| | 3+3 2 | 2+2 2 | 2+2 2 | 3+3 2 | 3+3 2 | 2+2 2 | 3+3 2 | 2+2 2 |
| | 2+2 2 | 3+3 2 | 3+3 2 | 3+3 2 | 2+2 2 | 2+2 2 | 3+3 2 | 3+3 2 |
| | 3+3 2 | 3+3 2 | 3+3 2 | 2+2 2 | 3+3 2 | 3+3 2 | 2+2 2 | 2+2 2 |
| | 2+2 2 | 2+2 2 | 2+2 2 | 2+2 2 | 2+2 2 | 3+3 2 | 2+2 2 | 3+3 2 |
| | 2+2 2 | 3+3 2 | 2+2 2 | 3+3 2 | 2+2 2 | 3+3 2 | 3+3 2 | 2+2 2 |
| | 2+2 2 | 2+2 2 | 3+3 2 | 3+3 2 | 3+3 2 | 2+2 2 | 2+2 2 | 3+3 2 |
| | 3+3 2 | 2+2 2 | 2+2 2 | 3+3 2 | 2+2 2 | 3+3 2 | 2+2 2 | 3+3 2 |
| | 2+2 2 | 3+3 2 | 2+2 2 | 2+2 2 | 3+3 2 | 2+2 2 | 3+3 2 | 2+2 2 |
| | 3+3 2 | 3+3 2 | 3+3 2 | 3+3 2 | 2+2 2 | 2+2 2 | 3+3 2 | 3+3 2 |
| | 2+2 2 | 2+2 2 | 3+3 2 | 2+2 2 | 3+3 2 | 3+3 2 | 2+2 2 | 2+2 2 |
| | 3+3 2 | 2+2 2 | 2+2 2 | 3+3 2 | 3+3 2 | 2+2 2 | 2+2 2 | 3+3 2 |
| | 2+2 2 | 3+3 2 | 3+3 2 | 2+2 2 | 2+2 2 | 3+3 2 | 3+3 2 | 2+2 2 |
| | 3+3 2 | 2+2 2 | 2+2 2 | 3+3 2 | 2+2 2 | 2+2 2 | 2+2 2 | 3+3 2 |
| | 2+2 2 | 3+3 2 | 3+3 2 | 2+2 2 | 3+3 2 | 3+3 2 | 3+3 2 | 2+2 2 |
| | 3+3 2 | 2+2 2 | 2+2 2 | 3+3 2 | 3+3 2 | 2+2 2 | 3+3 2 | 3+3 2 |
| | 2+2 2 | 3+3 2 | 3+3 2 | 2+2 2 | 2+2 2 | 3+3 2 | 2+2 2 | 2+2 2 |
| | 3+3 2 | 3+3 2 | 3+3 2 | 3+3 2 | 2+2 2 | 2+2 2 | 3+3 2 | 3+3 2 |
| | 3+3 2 | 2+2 2 | 2+2 2 | 2+2 2 | 3+3 2 | 3+3 2 | 2+2 2 | 2+2 2 |
| | 2+2 2 | 3+3 2 | 2+2 2 | 3+3 2 | 2+2 2 | 3+3 2 | 2+2 2 | 2+2 2 |
| 2.2.49 | 62367 | 174 | Cu1I3Na1O9 | | | | | |
| | -GM | -R | -T | -U | -V | -X | -Y | -Z |
| | 3+3 2 | 3+3 2 | 3+3 2 | 3+3 2 | 3+3 2 | 2+2 2 | 2+2 2 | 2+2 2 |
| | 2+2 2 | 2+2 2 | 2+2 2 | 2+2 2 | 2+2 2 | 3+3 2 | 3+3 2 | 3+3 2 |
| | 2+2 2 | 2+2 2 | 2+2 2 | 3+3 2 | 3+3 2 | 2+2 2 | 3+3 2 | 3+3 2 |
| | 3+3 2 | 3+3 2 | 3+3 2 | 2+2 2 | 2+2 2 | 3+3 2 | 2+2 2 | 2+2 2 |
| | 2+2 2 | 2+2 2 | 2+2 2 | 3+3 2 | 2+2 2 | 3+3 2 | 2+2 2 | 3+3 2 |
| | 3+3 2 | 3+3 2 | 3+3 2 | 2+2 2 | 3+3 2 | 2+2 2 | 3+3 2 | 2+2 2 |
| | 3+3 2 | 3+3 2 | 3+3 2 | 3+3 2 | 3+3 2 | 3+3 2 | 3+3 2 | 2+2 2 |
| | 3+3 2 | 2+2 2 | 2+2 2 | 2+2 2 | 2+2 2 | 2+2 2 | 2+2 2 | 3+3 2 |
| | 2+2 2 | 2+2 2 | 3+3 2 | 3+3 2 | 3+3 2 | 2+2 2 | 2+2 2 | 2+2 2 |
| | 2+2 2 | 3+3 2 | 2+2 2 | 2+2 2 | 2+2 2 | 3+3 2 | 3+3 2 | 3+3 2 |
| | 2+2 2 | 2+2 2 | 2+2 2 | 3+3 2 | 3+3 2 | 2+2 2 | 2+2 2 | 3+3 2 |
| | 3+3 2 | 3+3 2 | 3+3 2 | 2+2 2 | 2+2 2 | 3+3 2 | 3+3 2 | 2+2 2 |
| | 2+2 2 | 2+2 2 | 2+2 2 | 2+2 2 | 3+3 2 | 3+3 2 | 2+2 2 | 3+3 2 |
| | 3+3 2 | 3+3 2 | 3+3 2 | 3+3 2 | 2+2 2 | 2+2 2 | 3+3 2 | 2+2 2 |
| | 3+3 2 | 2+2 2 | 2+2 2 | 3+3 2 | 2+2 2 | 3+3 2 | 3+3 2 | 3+3 2 |
| | 2+2 2 | 3+3 2 | 3+3 2 | 2+2 2 | 3+3 2 | 2+2 2 | 2+2 2 | 2+2 2 |
| | 2+2 2 | 3+3 2 | 3+3 2 | 2+2 2 | 3+3 2 | 2+2 2 | 3+3 2 | 2+2 2 |
| | 3+3 2 | 2+2 2 | 2+2 2 | 3+3 2 | 2+2 2 | 3+3 2 | 2+2 2 | 3+3 2 |
| | 3+3 2 | 2+2 2 | 3+3 2 | 3+3 2 | 3+3 2 | 3+3 2 | 2+2 2 | 2+2 2 |
| | 2+2 2 | 3+3 2 | 2+2 2 | 2+2 2 | 2+2 2 | 2+2 2 | 3+3 2 | 3+3 2 |
| | 3+3 2 | 3+3 2 | 2+2 2 | 2+2 2 | 3+3 2 | 2+2 2 | 3+3 2 | 3+3 2 |
| | 2+2 2 | 2+2 2 | 3+3 2 | 3+3 2 | 2+2 2 | 3+3 2 | 2+2 2 | 2+2 2 |
| | 2+2 2 | 2+2 2 | 2+2 2 | 2+2 2 | 2+2 2 | 3+3 2 | 3+3 2 | 3+3 2 |
| | 3+3 2 | 3+3 2 | 3+3 2 | 3+3 2 | 3+3 2 | 2+2 2 | 2+2 2 | 2+2 2 |
| | 3+3 2 | 3+3 2 | 2+2 2 | 2+2 2 | 3+3 2 | 3+3 2 | 3+3 2 | 3+3 2 |
| | 2+2 2 | 2+2 2 | 3+3 2 | 3+3 2 | 2+2 2 | 3+3 2 | 3+3 2 | 2+2 2 |
| | 3+3 2 | 3+3 2 | 3+3 2 | 3+3 2 | 3+3 2 | 2+2 2 | 2+2 2 | 3+3 2 |
| | 2+2 2 | 2+2 2 | 2+2 2 | 2+2 2 | 2+2 2 | 3+3 2 | 3+3 2 | 2+2 2 |
| | 3+3 2 | 2+2 2 | 2+2 2 | 3+3 2 | 3+3 2 | 2+2 2 | 2+2 2 | 2+2 2 |
| | 3+3 2 | 3+3 2 | 3+3 2 | 2+2 2 | 2+2 2 | 3+3 2 | 2+2 2 | 3+3 2 |
| | 2+2 2 | 2+2 2 | 2+2 2 | 3+3 2 | 3+3 2 | 2+2 2 | 3+3 2 | 2+2 2 |
| | 2+2 2 | 3+3 2 | 3+3 2 | 2+2 2 | 2+2 2 | 2+2 2 | 2+2 2 | 3+3 2 |
| | 3+3 2 | 2+2 2 | 2+2 2 | 3+3 2 | 3+3 2 | 3+3 2 | 3+3 2 | 2+2 2 |
| | 2+2 2 | 3+3 2 | 3+3 2 | 3+3 2 | 2+2 2 | 2+2 2 | 2+2 2 | 3+3 2 |
| | 3+3 2 | 2+2 2 | 2+2 2 | 2+2 2 | 3+3 2 | 3+3 2 | 3+3 2 | 2+2 2 |
| | 3+3 2 | 3+3 2 | 3+3 2 | 2+2 2 | 3+3 2 | 2+2 2 | 2+2 2 | 3+3 2 |
| | 2+2 2 | 3+3 2 | 2+2 2 | 3+3 2 | 2+2 2 | 3+3 2 | 3+3 2 | 3+3 2 |

Continued on next column

**TABLE XXI– continued**

| TQC | ICSD | NoE | SF | | | | | |
|---|---|---|---|---|---|---|---|---|
| | 3+3 2 | 2+2 2 | 3+3 2 | 2+2 2 | 2+2 2 | 2+2 2 | 2+2 2 | 2+2 2 |
| | 2+2 2 | 3+3 2 | 3+3 2 | 3+3 2 | 3+3 2 | 3+3 2 | 2+2 2 | 3+3 2 |
| | 3+3 2 | 2+2 2 | 2+2 2 | 2+2 2 | 2+2 2 | 3+3 2 | 3+3 2 | 2+2 2 |
| | 2+2 2 | 3+3 2 | 2+2 2 | 3+3 2 | 3+3 2 | 2+2 2 | 3+3 2 | 3+3 2 |
| | 2+2 2 | 2+2 2 | 3+3 2 | 2+2 2 | 2+2 2 | 2+2 2 | 2+2 2 | 2+2 2 |
| | 3+3 2 | 2+2 2 | 3+3 2 | 2+2 2 | 3+3 2 | 3+3 2 | 3+3 2 | 2+2 2 |
| | 3+3 2 | 3+3 2 | 3+3 2 | 3+3 2 | 3+3 2 | 3+3 2 | 3+3 2 | 3+3 2 |
| | 3+3 2 | 2+2 2 | 2+2 2 | 2+2 2 | 2+2 2 | 3+3 2 | 2+2 2 | 3+3 2 |
| | 2+2 2 | 3+3 2 | 2+2 2 | 3+3 2 | 3+3 2 | 2+2 2 | 3+3 2 | 2+2 2 |
| | 3+3 2 | 2+2 2 | 2+2 2 | 2+2 2 | 2+2 2 | 3+3 2 | 2+2 2 | 2+2 2 |
| | 3+3 2 | 3+3 2 | 3+3 2 | 3+3 2 | 3+3 2 | 2+2 2 | 3+3 2 | 3+3 2 |
| | 3+3 2 | 2+2 2 | 3+3 2 | 2+2 2 | 2+2 2 | 3+3 2 | 2+2 2 | 3+3 2 |
| | 2+2 2 | 3+3 2 | 2+2 2 | 3+3 2 | 3+3 2 | 2+2 2 | 3+3 2 | 3+3 2 |
| | 2+2 2 | 2+2 2 | 3+3 2 | 2+2 2 | 3+3 2 | 2+2 2 | 2+2 2 | 2+2 2 |
| | 3+3 2 | 3+3 2 | 2+2 2 | 2+2 2 | 2+2 2 | 3+3 2 | 3+3 2 | 2+2 2 |
| | 2+2 2 | 2+2 2 | 3+3 2 | 3+3 2 | 3+3 2 | 2+2 2 | 2+2 2 | 2+2 2 |
| | 2+2 2 | 2+2 2 | 2+2 2 | 3+3 2 | 3+3 2 | 2+2 2 | 3+3 2 | 3+3 2 |
| | 3+3 2 | 3+3 2 | 3+3 2 | 3+3 2 | 3+3 2 | 3+3 2 | 2+2 2 | 2+2 2 |
| | 2+2 2 | 3+3 2 | 2+2 2 | 3+3 2 | 2+2 2 | 2+2 2 | 3+3 2 | 3+3 2 |
| | 3+3 2 | 3+3 2 | 3+3 2 | 2+2 2 | 2+2 2 | 3+3 2 | 3+3 2 | 2+2 2 |
| | 3+3 2 | 2+2 2 | 2+2 2 | 3+3 2 | 2+2 2 | 3+3 2 | 2+2 2 | 3+3 2 |
| | 2+2 2 | 2+2 2 | 3+3 2 | 2+2 2 | 3+3 2 | 2+2 2 | 3+3 2 | 3+3 2 |
| | 3+3 2 | 3+3 2 | 3+3 2 | 2+2 2 | 3+3 2 | 3+3 2 | 2+2 2 | 2+2 2 |
| | 2+2 2 | 2+2 2 | 2+2 2 | 2+2 2 | 2+2 2 | 2+2 2 | 2+2 2 | 2+2 2 |
| | 3+3 2 | 3+3 2 | 3+3 2 | 3+3 2 | 3+3 2 | 3+3 2 | 3+3 2 | 3+3 2 |
| | 2+2 2 | 2+2 2 | 2+2 2 | 3+3 2 | 2+2 2 | 2+2 2 | 3+3 2 | 2+2 2 |
| | 3+3 2 | 2+2 2 | 2+2 2 | 2+2 2 | 3+3 2 | 3+3 2 | 2+2 2 | 3+3 2 |
| | 3+3 2 | 3+3 2 | 3+3 2 | 3+3 2 | 2+2 2 | 3+3 2 | 2+2 2 | 2+2 2 |
| | 2+2 2 | 2+2 2 | 2+2 2 | 2+2 2 | 2+2 2 | 2+2 2 | 3+3 2 | 3+3 2 |
| | 3+3 2 | 3+3 2 | 2+2 2 | 3+3 2 | 3+3 2 | 3+3 2 | 2+2 2 | 2+2 2 |
| | 2+2 2 | 2+2 2 | 3+3 2 | 2+2 2 | 2+2 2 | 3+3 2 | 3+3 2 | 3+3 2 |
| | 3+3 2 | 3+3 2 | 3+3 2 | 3+3 2 | 3+3 2 | 2+2 2 | 2+2 2 | 3+3 2 |
| | 2+2 2 | 3+3 2 | 2+2 2 | 3+3 2 | 2+2 2 | 2+2 2 | 3+3 2 | 2+2 2 |
| | 2+2 2 | 3+3 2 | 3+3 2 | 2+2 2 | 2+2 2 | 3+3 2 | 3+3 2 | 3+3 2 |
| | 2+2 2 | 3+3 2 | 2+2 2 | 3+3 2 | 3+3 2 | 2+2 2 | 2+2 2 | 2+2 2 |
| | 3+3 2 | 2+2 2 | 3+3 2 | 2+2 2 | 3+3 2 | 2+2 2 | 2+2 2 | 3+3 2 |
| | 2+2 2 | 3+3 2 | 3+3 2 | 2+2 2 | 2+2 2 | 3+3 2 | 3+3 2 | 3+3 2 |
| | 2+2 2 | 2+2 2 | 2+2 2 | 3+3 2 | 3+3 2 | 3+3 2 | 3+3 2 | 2+2 2 |
| | 2+2 2 | 2+2 2 | 2+2 2 | 3+3 2 | 2+2 2 | 2+2 2 | 2+2 2 | 3+3 2 |
| | 3+3 2 | 3+3 2 | 3+3 2 | 2+2 2 | 2+2 2 | 2+2 2 | 2+2 2 | 2+2 2 |
| | 3+3 2 | 2+2 2 | 2+2 2 | 2+2 2 | 3+3 2 | 3+3 2 | 3+3 2 | 2+2 2 |
| | 3+3 2 | 2+2 2 | 3+3 2 | 3+3 2 | 2+2 2 | 3+3 2 | 3+3 2 | 3+3 2 |
| | 2+2 2 | 3+3 2 | 3+3 2 | 2+2 2 | 3+3 2 | 2+2 2 | 2+2 2 | 2+2 2 |
| | 3+3 2 | 2+2 2 | 2+2 2 | 2+2 2 | 3+3 2 | 3+3 2 | 3+3 2 | 3+3 2 |
| | 3+3 2 | 3+3 2 | 2+2 2 | 3+3 2 | 3+3 2 | 3+3 2 | 3+3 2 | 2+2 2 |
| | 3+3 2 | 3+3 2 | 3+3 2 | 3+3 2 | 3+3 2 | 3+3 2 | 3+3 2 | 3+3 2 |
| | 3+3 2 | 2+2 2 | 2+2 2 | 2+2 2 | 3+3 2 | 3+3 2 | 3+3 2 | 2+2 2 |
| | 3+3 2 | 3+3 2 | 3+3 2 | 3+3 2 | 3+3 2 | 3+3 2 | 3+3 2 | 3+3 2 |
| | 3+3 2 | 2+2 2 | 2+2 2 | 2+2 2 | 3+3 2 | 3+3 2 | 3+3 2 | 2+2 2 |
| | 3+3 2 | 2+2 2 | 2+2 2 | 3+3 2 | 3+3 2 | 3+3 2 | 3+3 2 | 2+2 2 |
| 2.2.50 | 75006 | 178 | Cu1Gd1O8W2 | | | | | |
| | -GM | -R | -T | -U | -V | -X | -Y | -Z |
| | 3+3 2 | 3+3 2 | 3+3 2 | 2+2 2 | 2+2 2 | 3+3 2 | 2+2 2 | 2+2 2 |
| | 2+2 2 | 2+2 2 | 2+2 2 | 3+3 2 | 3+3 2 | 2+2 2 | 3+3 2 | 3+3 2 |
| | 3+3 2 | 3+3 2 | 3+3 2 | 2+2 2 | 2+2 2 | 3+3 2 | 2+2 2 | 2+2 2 |
| | 2+2 2 | 2+2 2 | 2+2 2 | 3+3 2 | 3+3 2 | 2+2 2 | 3+3 2 | 3+3 2 |
| | 3+3 2 | 2+2 2 | 3+3 2 | 2+2 2 | 2+2 2 | 2+2 2 | 3+3 2 | 3+3 2 |
| | 2+2 2 | 3+3 2 | 2+2 2 | 3+3 2 | 3+3 2 | 3+3 2 | 2+2 2 | 2+2 2 |
| | 3+3 2 | 3+3 2 | 3+3 2 | 2+2 2 | 2+2 2 | 3+3 2 | 2+2 2 | 2+2 2 |
| | 3+3 2 | 2+2 2 | 2+2 2 | 3+3 2 | 3+3 2 | 2+2 2 | 3+3 2 | 3+3 2 |
| | 2+2 2 | 2+2 2 | 3+3 2 | 2+2 2 | 2+2 2 | 2+2 2 | 3+3 2 | 3+3 2 |

Continued on next column

**TABLE XXI– continued**

| TQC | ICSD | NoE | SF | | | | | |
|---|---|---|---|---|---|---|---|---|
| | 2+2 2 | 3+3 2 | 2+2 2 | 3+3 2 | 3+3 2 | 3+3 2 | 2+2 2 | 2+2 2 |
| | 2+2 2 | 2+2 2 | 2+2 2 | 2+2 2 | 2+2 2 | 2+2 2 | 2+2 2 | 2+2 2 |
| | 3+3 2 | 3+3 2 | 3+3 2 | 3+3 2 | 3+3 2 | 3+3 2 | 3+3 2 | 3+3 2 |
| | 3+3 2 | 3+3 2 | 3+3 2 | 2+2 2 | 3+3 2 | 3+3 2 | 2+2 2 | 2+2 2 |
| | 2+2 2 | 2+2 2 | 2+2 2 | 3+3 2 | 2+2 2 | 2+2 2 | 3+3 2 | 3+3 2 |
| | 2+2 2 | 3+3 2 | 2+2 2 | 3+3 2 | 3+3 2 | 3+3 2 | 3+3 2 | 2+2 2 |
| | 3+3 2 | 2+2 2 | 3+3 2 | 2+2 2 | 2+2 2 | 2+2 2 | 2+2 2 | 3+3 2 |
| | 2+2 2 | 3+3 2 | 2+2 2 | 3+3 2 | 2+2 2 | 3+3 2 | 3+3 2 | 3+3 2 |
| | 3+3 2 | 2+2 2 | 3+3 2 | 2+2 2 | 3+3 2 | 2+2 2 | 2+2 2 | 2+2 2 |
| | 2+2 2 | 3+3 2 | 3+3 2 | 3+3 2 | 3+3 2 | 3+3 2 | 2+2 2 | 3+3 2 |
| | 3+3 2 | 2+2 2 | 2+2 2 | 2+2 2 | 2+2 2 | 2+2 2 | 3+3 2 | 2+2 2 |
| | 2+2 2 | 2+2 2 | 2+2 2 | 3+3 2 | 3+3 2 | 2+2 2 | 3+3 2 | 3+3 2 |
| | 3+3 2 | 3+3 2 | 3+3 2 | 2+2 2 | 2+2 2 | 3+3 2 | 2+2 2 | 2+2 2 |
| | 3+3 2 | 3+3 2 | 2+2 2 | 2+2 2 | 3+3 2 | 2+2 2 | 2+2 2 | 2+2 2 |
| | 2+2 2 | 2+2 2 | 3+3 2 | 3+3 2 | 2+2 2 | 3+3 2 | 3+3 2 | 3+3 2 |
| | 2+2 2 | 2+2 2 | 3+3 2 | 3+3 2 | 2+2 2 | 3+3 2 | 3+3 2 | 3+3 2 |
| | 3+3 2 | 3+3 2 | 2+2 2 | 2+2 2 | 3+3 2 | 2+2 2 | 2+2 2 | 2+2 2 |
| | 3+3 2 | 2+2 2 | 3+3 2 | 2+2 2 | 2+2 2 | 2+2 2 | 3+3 2 | 2+2 2 |
| | 2+2 2 | 3+3 2 | 2+2 2 | 3+3 2 | 2+2 2 | 3+3 2 | 2+2 2 | 3+3 2 |
| | 3+3 2 | 2+2 2 | 3+3 2 | 2+2 2 | 3+3 2 | 2+2 2 | 3+3 2 | 3+3 2 |
| | 3+3 2 | 2+2 2 | 2+2 2 | 2+2 2 | 3+3 2 | 3+3 2 | 2+2 2 | 2+2 2 |
| | 2+2 2 | 3+3 2 | 3+3 2 | 3+3 2 | 2+2 2 | 2+2 2 | 3+3 2 | 3+3 2 |
| | 2+2 2 | 3+3 2 | 2+2 2 | 3+3 2 | 3+3 2 | 3+3 2 | 2+2 2 | 2+2 2 |
| | 3+3 2 | 2+2 2 | 3+3 2 | 3+3 2 | 2+2 2 | 2+2 2 | 3+3 2 | 3+3 2 |
| | 3+3 2 | 3+3 2 | 2+2 2 | 2+2 2 | 3+3 2 | 3+3 2 | 3+3 2 | 2+2 2 |
| | 2+2 2 | 2+2 2 | 3+3 2 | 2+2 2 | 3+3 2 | 2+2 2 | 2+2 2 | 3+3 2 |
| | 2+2 2 | 3+3 2 | 2+2 2 | 3+3 2 | 2+2 2 | 3+3 2 | 2+2 2 | 2+2 2 |
| | 3+3 2 | 2+2 2 | 3+3 2 | 3+3 2 | 2+2 2 | 3+3 2 | 3+3 2 | 3+3 2 |
| | 3+3 2 | 3+3 2 | 2+2 2 | 2+2 2 | 3+3 2 | 2+2 2 | 2+2 2 | 2+2 2 |
| | 2+2 2 | 2+2 2 | 3+3 2 | 2+2 2 | 3+3 2 | 2+2 2 | 2+2 2 | 3+3 2 |
| | 3+3 2 | 3+3 2 | 2+2 2 | 2+2 2 | 2+2 2 | 3+3 2 | 3+3 2 | 2+2 2 |
| | 2+2 2 | 3+3 2 | 2+2 2 | 3+3 2 | 2+2 2 | 2+2 2 | 3+3 2 | 3+3 2 |
| | 3+3 2 | 2+2 2 | 3+3 2 | 2+2 2 | 3+3 2 | 3+3 2 | 2+2 2 | 2+2 2 |
| | 2+2 2 | 2+2 2 | 2+2 2 | 3+3 2 | 2+2 2 | 3+3 2 | 2+2 2 | 2+2 2 |
| | 2+2 2 | 3+3 2 | 3+3 2 | 3+3 2 | 3+3 2 | 2+2 2 | 2+2 2 | 3+3 2 |
| | 3+3 2 | 2+2 2 | 2+2 2 | 3+3 2 | 2+2 2 | 3+3 2 | 3+3 2 | 3+3 2 |
| | 3+3 2 | 3+3 2 | 3+3 2 | 2+2 2 | 3+3 2 | 2+2 2 | 2+2 2 | 2+2 2 |
| | 2+2 2 | 3+3 2 | 2+2 2 | 3+3 2 | 3+3 2 | 3+3 2 | 3+3 2 | 3+3 2 |
| | 2+2 2 | 2+2 2 | 3+3 2 | 2+2 2 | 2+2 2 | 2+2 2 | 3+3 2 | 2+2 2 |
| | 3+3 2 | 3+3 2 | 3+3 2 | 2+2 2 | 2+2 2 | 2+2 2 | 2+2 2 | 3+3 2 |
| | 2+2 2 | 2+2 2 | 2+2 2 | 3+3 2 | 2+2 2 | 2+2 2 | 3+3 2 | 2+2 2 |
| | 3+3 2 | 2+2 2 | 3+3 2 | 2+2 2 | 3+3 2 | 3+3 2 | 3+3 2 | 3+3 2 |
| | 3+3 2 | 3+3 2 | 3+3 2 | 2+2 2 | 2+2 2 | 2+2 2 | 2+2 2 | 3+3 2 |
| | 2+2 2 | 2+2 2 | 2+2 2 | 3+3 2 | 3+3 2 | 3+3 2 | 2+2 2 | 2+2 2 |
| | 2+2 2 | 3+3 2 | 2+2 2 | 3+3 2 | 3+3 2 | 3+3 2 | 3+3 2 | 2+2 2 |
| | 3+3 2 | 2+2 2 | 3+3 2 | 2+2 2 | 2+2 2 | 3+3 2 | 2+2 2 | 3+3 2 |
| | 3+3 2 | 3+3 2 | 2+2 2 | 3+3 2 | 3+3 2 | 2+2 2 | 3+3 2 | 2+2 2 |
| | 2+2 2 | 2+2 2 | 3+3 2 | 2+2 2 | 2+2 2 | 2+2 2 | 3+3 2 | 3+3 2 |
| | 2+2 2 | 3+3 2 | 2+2 2 | 3+3 2 | 2+2 2 | 3+3 2 | 2+2 2 | 2+2 2 |
| | 2+2 2 | 2+2 2 | 3+3 2 | 2+2 2 | 3+3 2 | 2+2 2 | 3+3 2 | 2+2 2 |
| | 3+3 2 | 3+3 2 | 2+2 2 | 3+3 2 | 3+3 2 | 3+3 2 | 2+2 2 | 3+3 2 |
| | 2+2 2 | 2+2 2 | 3+3 2 | 2+2 2 | 2+2 2 | 3+3 2 | 3+3 2 | 2+2 2 |
| | 3+3 2 | 3+3 2 | 2+2 2 | 2+2 2 | 3+3 2 | 2+2 2 | 2+2 2 | 2+2 2 |
| | 3+3 2 | 3+3 2 | 3+3 2 | 3+3 2 | 2+2 2 | 3+3 2 | 2+2 2 | 3+3 2 |
| | 2+2 2 | 2+2 2 | 2+2 2 | 3+3 2 | 3+3 2 | 2+2 2 | 3+3 2 | 3+3 2 |
| | 3+3 2 | 2+2 2 | 3+3 2 | 3+3 2 | 3+3 2 | 3+3 2 | 3+3 2 | 2+2 2 |
| | 2+2 2 | 3+3 2 | 2+2 2 | 2+2 2 | 2+2 2 | 2+2 2 | 2+2 2 | 2+2 2 |
| | 2+2 2 | 3+3 2 | 3+3 2 | 3+3 2 | 2+2 2 | 3+3 2 | 2+2 2 | 3+3 2 |
| | 3+3 2 | 2+2 2 | 2+2 2 | 2+2 2 | 3+3 2 | 2+2 2 | 2+2 2 | 2+2 2 |
| | 2+2 2 | 3+3 2 | 3+3 2 | 3+3 2 | 2+2 2 | 2+2 2 | 3+3 2 | 3+3 2 |
| | 3+3 2 | 2+2 2 | 2+2 2 | 2+2 2 | 3+3 2 | 3+3 2 | 3+3 2 | 3+3 2 |

Continued on next column

**TABLE XXI– continued**

| TQC | ICSD | NoE | SF | | | | | |
|---|---|---|---|---|---|---|---|---|
| | 2+2 2 | 3+3 2 | 3+3 2 | 3+3 2 | 2+2 2 | 2+2 2 | 3+3 2 | 2+2 2 |
| | 3+3 2 | 2+2 2 | 2+2 2 | 2+2 2 | 3+3 2 | 3+3 2 | 2+2 2 | 3+3 2 |
| | 3+3 2 | 2+2 2 | 3+3 2 | 2+2 2 | 2+2 2 | 2+2 2 | 2+2 2 | 3+3 2 |
| | 2+2 2 | 3+3 2 | 2+2 2 | 3+3 2 | 3+3 2 | 3+3 2 | 3+3 2 | 2+2 2 |
| | 3+3 2 | 2+2 2 | 2+2 2 | 3+3 2 | 2+2 2 | 3+3 2 | 2+2 2 | 3+3 2 |
| | 2+2 2 | 3+3 2 | 3+3 2 | 2+2 2 | 3+3 2 | 2+2 2 | 3+3 2 | 2+2 2 |
| | 2+2 2 | 2+2 2 | 3+3 2 | 3+3 2 | 2+2 2 | 3+3 2 | 2+2 2 | 3+3 2 |
| | 3+3 2 | 3+3 2 | 2+2 2 | 2+2 2 | 3+3 2 | 2+2 2 | 3+3 2 | 2+2 2 |
| | 3+3 2 | 3+3 2 | 3+3 2 | 2+2 2 | 2+2 2 | 3+3 2 | 2+2 2 | 2+2 2 |
| | 2+2 2 | 2+2 2 | 2+2 2 | 3+3 2 | 3+3 2 | 2+2 2 | 3+3 2 | 3+3 2 |
| | 3+3 2 | 3+3 2 | 2+2 2 | 2+2 2 | 2+2 2 | 3+3 2 | 2+2 2 | 3+3 2 |
| | 2+2 2 | 2+2 2 | 3+3 2 | 3+3 2 | 3+3 2 | 2+2 2 | 3+3 2 | 2+2 2 |
| | 3+3 2 | 2+2 2 | 2+2 2 | 2+2 2 | 3+3 2 | 3+3 2 | 3+3 2 | 2+2 2 |
| | 2+2 2 | 3+3 2 | 3+3 2 | 3+3 2 | 2+2 2 | 2+2 2 | 2+2 2 | 3+3 2 |
| | 3+3 2 | 2+2 2 | 3+3 2 | 2+2 2 | 3+3 2 | 2+2 2 | 3+3 2 | 3+3 2 |
| | 2+2 2 | 3+3 2 | 2+2 2 | 3+3 2 | 2+2 2 | 2+2 2 | 2+2 2 | 2+2 2 |
| | 3+3 2 | 2+2 2 | 3+3 2 | 2+2 2 | 2+2 2 | 3+3 2 | 3+3 2 | 2+2 2 |
| | 2+2 2 | 3+3 2 | 2+2 2 | 3+3 2 | 3+3 2 | 3+3 2 | 2+2 2 | 3+3 2 |
| | 2+2 2 | 3+3 2 | 3+3 2 | 2+2 2 | 2+2 2 | 3+3 2 | 2+2 2 | 2+2 2 |
| 2.2.51 | 64660 | 126 | Cu1F6Pt1 | | | | | |
| | -GM | -R | -T | -U | -V | -X | -Y | -Z |
| | 3+3 2 | 3+3 2 | 3+3 2 | 3+3 2 | 2+2 2 | 2+2 2 | 3+3 2 | 2+2 2 |
| | 2+2 2 | 2+2 2 | 2+2 2 | 2+2 2 | 3+3 2 | 3+3 2 | 2+2 2 | 3+3 2 |
| | 3+3 2 | 3+3 2 | 3+3 2 | 3+3 2 | 3+3 2 | 3+3 2 | 2+2 2 | 2+2 2 |
| | 2+2 2 | 2+2 2 | 2+2 2 | 2+2 2 | 3+3 2 | 2+2 2 | 3+3 2 | 3+3 2 |
| | 3+3 2 | 3+3 2 | 2+2 2 | 3+3 2 | 2+2 2 | 3+3 2 | 2+2 2 | 2+2 2 |
| | 2+2 2 | 2+2 2 | 3+3 2 | 2+2 2 | 2+2 2 | 2+2 2 | 3+3 2 | 3+3 2 |
| | 3+3 2 | 2+2 2 | 3+3 2 | 2+2 2 | 2+2 2 | 3+3 2 | 3+3 2 | 3+3 2 |
| | 2+2 2 | 3+3 2 | 2+2 2 | 3+3 2 | 3+3 2 | 2+2 2 | 2+2 2 | 2+2 2 |
| | 3+3 2 | 2+2 2 | 2+2 2 | 2+2 2 | 3+3 2 | 3+3 2 | 3+3 2 | 2+2 2 |
| | 2+2 2 | 3+3 2 | 2+2 2 | 3+3 2 | 2+2 2 | 2+2 2 | 2+2 2 | 3+3 2 |
| | 3+3 2 | 2+2 2 | 3+3 2 | 3+3 2 | 2+2 2 | 3+3 2 | 2+2 2 | 3+3 2 |
| | 2+2 2 | 3+3 2 | 3+3 2 | 2+2 2 | 3+3 2 | 2+2 2 | 3+3 2 | 2+2 2 |
| | 3+3 2 | 2+2 2 | 3+3 2 | 3+3 2 | 3+3 2 | 3+3 2 | 2+2 2 | 2+2 2 |
| | 2+2 2 | 3+3 2 | 3+3 2 | 3+3 2 | 3+3 2 | 2+2 2 | 3+3 2 | 3+3 2 |
| | 3+3 2 | 2+2 2 | 2+2 2 | 2+2 2 | 2+2 2 | 3+3 2 | 3+3 2 | 2+2 2 |
| | 2+2 2 | 3+3 2 | 2+2 2 | 2+2 2 | 2+2 2 | 2+2 2 | 2+2 2 | 3+3 2 |
| | 3+3 2 | 3+3 2 | 3+3 2 | 3+3 2 | 3+3 2 | 2+2 2 | 3+3 2 | 2+2 2 |
| | 2+2 2 | 2+2 2 | 2+2 2 | 2+2 2 | 2+2 2 | 3+3 2 | 2+2 2 | 3+3 2 |
| | 3+3 2 | 3+3 2 | 3+3 2 | 3+3 2 | 3+3 2 | 2+2 2 | 3+3 2 | 3+3 2 |
| | 3+3 2 | 2+2 2 | 3+3 2 | 3+3 2 | 3+3 2 | 3+3 2 | 2+2 2 | 2+2 2 |
| | 3+3 2 | 2+2 2 | 2+2 2 | 2+2 2 | 2+2 2 | 3+3 2 | 2+2 2 | 2+2 2 |
| | 2+2 2 | 3+3 2 | 2+2 2 | 2+2 2 | 2+2 2 | 2+2 2 | 3+3 2 | 3+3 2 |
| | 3+3 2 | 2+2 2 | 3+3 2 | 3+3 2 | 3+3 2 | 3+3 2 | 2+2 2 | 2+2 2 |
| | 2+2 2 | 3+3 2 | 3+3 2 | 3+3 2 | 3+3 2 | 2+2 2 | 3+3 2 | 3+3 2 |
| | 2+2 2 | 3+3 2 | 2+2 2 | 2+2 2 | 2+2 2 | 2+2 2 | 3+3 2 | 2+2 2 |
| | 2+2 2 | 2+2 2 | 2+2 2 | 2+2 2 | 3+3 2 | 3+3 2 | 2+2 2 | 3+3 2 |
| | 3+3 2 | 3+3 2 | 3+3 2 | 3+3 2 | 2+2 2 | 2+2 2 | 2+2 2 | 2+2 2 |
| | 3+3 2 | 2+2 2 | 3+3 2 | 3+3 2 | 3+3 2 | 3+3 2 | 3+3 2 | 3+3 2 |
| | 2+2 2 | 3+3 2 | 2+2 2 | 2+2 2 | 2+2 2 | 3+3 2 | 2+2 2 | 3+3 2 |
| | 2+2 2 | 2+2 2 | 2+2 2 | 2+2 2 | 2+2 2 | 2+2 2 | 3+3 2 | 2+2 2 |
| | 3+3 2 | 2+2 2 | 3+3 2 | 3+3 2 | 3+3 2 | 3+3 2 | 2+2 2 | 2+2 2 |
| | 3+3 2 | 3+3 2 | 3+3 2 | 2+2 2 | 3+3 2 | 2+2 2 | 3+3 2 | 3+3 2 |
| | 3+3 2 | 3+3 2 | 2+2 2 | 3+3 2 | 2+2 2 | 2+2 2 | 2+2 2 | 2+2 2 |
| | 2+2 2 | 2+2 2 | 3+3 2 | 3+3 2 | 2+2 2 | 3+3 2 | 3+3 2 | 3+3 2 |
| | 3+3 2 | 3+3 2 | 2+2 2 | 2+2 2 | 3+3 2 | 2+2 2 | 3+3 2 | 3+3 2 |
| | 3+3 2 | 2+2 2 | 2+2 2 | 2+2 2 | 2+2 2 | 3+3 2 | 2+2 2 | 2+2 2 |
| | 2+2 2 | 2+2 2 | 3+3 2 | 3+3 2 | 3+3 2 | 3+3 2 | 3+3 2 | 2+2 2 |
| | 2+2 2 | 3+3 2 | 3+3 2 | 3+3 2 | 3+3 2 | 2+2 2 | 2+2 2 | 3+3 2 |
| | 2+2 2 | 2+2 2 | 2+2 2 | 2+2 2 | 2+2 2 | 2+2 2 | 2+2 2 | 3+3 2 |
| | 3+3 2 | 3+3 2 | 3+3 2 | 2+2 2 | 3+3 2 | 3+3 2 | 3+3 2 | 2+2 2 |

Continued on next column

**TABLE XXI– continued**

| TQC | ICSD | NoE | SF | | | | | |
|---|---|---|---|---|---|---|---|---|
| | 3+3 2 | 2+2 2 | 3+3 2 | 3+3 2 | 2+2 2 | 2+2 2 | 3+3 2 | 2+2 2 |
| | 2+2 2 | 3+3 2 | 2+2 2 | 3+3 2 | 3+3 2 | 3+3 2 | 3+3 2 | 3+3 2 |
| | 3+3 2 | 2+2 2 | 2+2 2 | 2+2 2 | 2+2 2 | 3+3 2 | 2+2 2 | 3+3 2 |
| | 3+3 2 | 2+2 2 | 3+3 2 | 3+3 2 | 3+3 2 | 2+2 2 | 2+2 2 | 2+2 2 |
| | 2+2 2 | 3+3 2 | 2+2 2 | 2+2 2 | 2+2 2 | 3+3 2 | 3+3 2 | 2+2 2 |
| | 2+2 2 | 3+3 2 | 2+2 2 | 3+3 2 | 2+2 2 | 2+2 2 | 2+2 2 | 3+3 2 |
| | 2+2 2 | 2+2 2 | 2+2 2 | 2+2 2 | 3+3 2 | 3+3 2 | 3+3 2 | 2+2 2 |
| | 2+2 2 | 3+3 2 | 3+3 2 | 2+2 2 | 2+2 2 | 2+2 2 | 2+2 2 | 3+3 2 |
| | 3+3 2 | 2+2 2 | 3+3 2 | 3+3 2 | 3+3 2 | 2+2 2 | 3+3 2 | 2+2 2 |
| | 3+3 2 | 3+3 2 | 3+3 2 | 3+3 2 | 3+3 2 | 3+3 2 | 2+2 2 | 3+3 2 |
| | 3+3 2 | 2+2 2 | 2+2 2 | 2+2 2 | 2+2 2 | 3+3 2 | 3+3 2 | 3+3 2 |
| | 2+2 2 | 3+3 2 | 3+3 2 | 3+3 2 | 3+3 2 | 2+2 2 | 2+2 2 | 2+2 2 |
| | 3+3 2 | 3+3 2 | 3+3 2 | 3+3 2 | 3+3 2 | 2+2 2 | 2+2 2 | 3+3 2 |
| | 2+2 2 | 2+2 2 | 2+2 2 | 2+2 2 | 2+2 2 | 3+3 2 | 3+3 2 | 2+2 2 |
| | 3+3 2 | 3+3 2 | 3+3 2 | 3+3 2 | 3+3 2 | 3+3 2 | 2+2 2 | 2+2 2 |
| | 3+3 2 | 2+2 2 | 3+3 2 | 3+3 2 | 3+3 2 | 2+2 2 | 3+3 2 | 3+3 2 |
| | 3+3 2 | 2+2 2 | 3+3 2 | 3+3 2 | 3+3 2 | 2+2 2 | 3+3 2 | 2+2 2 |
| | 3+3 2 | 3+3 2 | 3+3 2 | 3+3 2 | 3+3 2 | 3+3 2 | 2+2 2 | 3+3 2 |
| | 3+3 2 | 2+2 2 | 3+3 2 | 2+2 2 | 3+3 2 | 3+3 2 | 3+3 2 | 2+2 2 |
| | 2+2 2 | 3+3 2 | 2+2 2 | 3+3 2 | 2+2 2 | 2+2 2 | 2+2 2 | 3+3 2 |
| | 3+3 2 | 3+3 2 | 3+3 2 | 3+3 2 | 3+3 2 | 2+2 2 | 2+2 2 | 2+2 2 |
| | 3+3 2 | 2+2 2 | 3+3 2 | 3+3 2 | 3+3 2 | 3+3 2 | 3+3 2 | 3+3 2 |
| | 3+3 2 | 2+2 2 | 3+3 2 | 3+3 2 | 3+3 2 | 2+2 2 | 3+3 2 | 2+2 2 |
| 2.2.52 | 69655 | 132 | Cu1F3Na1 | | | | | |
| | -GM | -R | -T | -U | -V | -X | -Y | -Z |
| | 3+3 2 | 3+3 2 | 2+2 2 | 2+2 2 | 2+2 2 | 3+3 2 | 3+3 2 | 2+2 2 |
| | 3+3 2 | 2+2 2 | 3+3 2 | 3+3 2 | 2+2 2 | 2+2 2 | 2+2 2 | 3+3 2 |
| | 2+2 2 | 2+2 2 | 3+3 2 | 2+2 2 | 3+3 2 | 3+3 2 | 3+3 2 | 3+3 2 |
| | 2+2 2 | 3+3 2 | 2+2 2 | 3+3 2 | 3+3 2 | 2+2 2 | 2+2 2 | 2+2 2 |
| | 3+3 2 | 2+2 2 | 2+2 2 | 2+2 2 | 2+2 2 | 2+2 2 | 2+2 2 | 2+2 2 |
| | 2+2 2 | 3+3 2 | 3+3 2 | 3+3 2 | 3+3 2 | 3+3 2 | 3+3 2 | 3+3 2 |
| | 3+3 2 | 3+3 2 | 3+3 2 | 3+3 2 | 2+2 2 | 2+2 2 | 2+2 2 | 2+2 2 |
| | 2+2 2 | 2+2 2 | 2+2 2 | 2+2 2 | 3+3 2 | 2+2 2 | 2+2 2 | 3+3 2 |
| | 3+3 2 | 3+3 2 | 2+2 2 | 3+3 2 | 3+3 2 | 3+3 2 | 3+3 2 | 2+2 2 |
| | 2+2 2 | 2+2 2 | 3+3 2 | 2+2 2 | 3+3 2 | 3+3 2 | 3+3 2 | 3+3 2 |
| | 3+3 2 | 2+2 2 | 3+3 2 | 3+3 2 | 2+2 2 | 3+3 2 | 2+2 2 | 3+3 2 |
| | 2+2 2 | 3+3 2 | 2+2 2 | 2+2 2 | 2+2 2 | 2+2 2 | 3+3 2 | 2+2 2 |
| | 3+3 2 | 3+3 2 | 3+3 2 | 2+2 2 | 2+2 2 | 3+3 2 | 2+2 2 | 2+2 2 |
| | 3+3 2 | 2+2 2 | 2+2 2 | 3+3 2 | 2+2 2 | 2+2 2 | 3+3 2 | 3+3 2 |
| | 3+3 2 | 2+2 2 | 2+2 2 | 3+3 2 | 2+2 2 | 2+2 2 | 2+2 2 | 3+3 2 |
| | 3+3 2 | 3+3 2 | 3+3 2 | 2+2 2 | 2+2 2 | 3+3 2 | 3+3 2 | 2+2 2 |
| | 3+3 2 | 2+2 2 | 3+3 2 | 3+3 2 | 2+2 2 | 2+2 2 | 3+3 2 | 3+3 2 |
| | 2+2 2 | 3+3 2 | 2+2 2 | 2+2 2 | 2+2 2 | 3+3 2 | 2+2 2 | 2+2 2 |
| | 3+3 2 | 2+2 2 | 3+3 2 | 3+3 2 | 2+2 2 | 2+2 2 | 2+2 2 | 3+3 2 |
| | 3+3 2 | 3+3 2 | 2+2 2 | 2+2 2 | 3+3 2 | 3+3 2 | 3+3 2 | 2+2 2 |
| | 3+3 2 | 3+3 2 | 2+2 2 | 2+2 2 | 2+2 2 | 3+3 2 | 3+3 2 | 2+2 2 |
| | 2+2 2 | 2+2 2 | 3+3 2 | 3+3 2 | 3+3 2 | 2+2 2 | 2+2 2 | 3+3 2 |
| | 2+2 2 | 2+2 2 | 3+3 2 | 2+2 2 | 3+3 2 | 3+3 2 | 3+3 2 | 2+2 2 |
| | 3+3 2 | 3+3 2 | 2+2 2 | 3+3 2 | 3+3 2 | 3+3 2 | 2+2 2 | 2+2 2 |
| | 2+2 2 | 3+3 2 | 3+3 2 | 3+3 2 | 3+3 2 | 2+2 2 | 2+2 2 | 3+3 2 |
| | 2+2 2 | 2+2 2 | 2+2 2 | 2+2 2 | 2+2 2 | 2+2 2 | 3+3 2 | 3+3 2 |
| | 3+3 2 | 3+3 2 | 3+3 2 | 3+3 2 | 3+3 2 | 2+2 2 | 3+3 2 | 3+3 2 |
| | 2+2 2 | 2+2 2 | 2+2 2 | 2+2 2 | 2+2 2 | 3+3 2 | 2+2 2 | 2+2 2 |
| | 2+2 2 | 3+3 2 | 2+2 2 | 3+3 2 | 3+3 2 | 3+3 2 | 3+3 2 | 2+2 2 |
| | 3+3 2 | 2+2 2 | 2+2 2 | 2+2 2 | 3+3 2 | 2+2 2 | 2+2 2 | 3+3 2 |
| | 2+2 2 | 2+2 2 | 3+3 2 | 2+2 2 | 2+2 2 | 2+2 2 | 2+2 2 | 3+3 2 |
| | 3+3 2 | 3+3 2 | 3+3 2 | 3+3 2 | 3+3 2 | 3+3 2 | 3+3 2 | 2+2 2 |
| | 3+3 2 | 3+3 2 | 2+2 2 | 2+2 2 | 2+2 2 | 3+3 2 | 2+2 2 | 2+2 2 |
| | 3+3 2 | 2+2 2 | 3+3 2 | 3+3 2 | 3+3 2 | 2+2 2 | 3+3 2 | 3+3 2 |
| | 2+2 2 | 2+2 2 | 3+3 2 | 2+2 2 | 2+2 2 | 3+3 2 | 2+2 2 | 3+3 2 |
| | 2+2 2 | 3+3 2 | 2+2 2 | 3+3 2 | 3+3 2 | 2+2 2 | 3+3 2 | 2+2 2 |

Continued on next column

**TABLE XXI– continued**

| TQC | ICSD | NoE | SF | | | | | |
|---|---|---|---|---|---|---|---|---|
| | 2+2 2 | 2+2 2 | 2+2 2 | 2+2 2 | 2+2 2 | 2+2 2 | 2+2 2 | 3+3 2 |
| | 2+2 2 | 3+3 2 | 3+3 2 | 2+2 2 | 3+3 2 | 3+3 2 | 3+3 2 | 2+2 2 |
| | 3+3 2 | 3+3 2 | 2+2 2 | 3+3 2 | 2+2 2 | 3+3 2 | 2+2 2 | 2+2 2 |
| | 3+3 2 | 2+2 2 | 3+3 2 | 3+3 2 | 3+3 2 | 2+2 2 | 3+3 2 | 3+3 2 |
| | 2+2 2 | 2+2 2 | 3+3 2 | 3+3 2 | 2+2 2 | 3+3 2 | 3+3 2 | 2+2 2 |
| | 2+2 2 | 3+3 2 | 2+2 2 | 3+3 2 | 3+3 2 | 2+2 2 | 2+2 2 | 3+3 2 |
| | 3+3 2 | 2+2 2 | 2+2 2 | 2+2 2 | 2+2 2 | 2+2 2 | 2+2 2 | 2+2 2 |
| | 2+2 2 | 3+3 2 | 3+3 2 | 2+2 2 | 3+3 2 | 3+3 2 | 3+3 2 | 3+3 2 |
| | 2+2 2 | 2+2 2 | 3+3 2 | 2+2 2 | 3+3 2 | 2+2 2 | 2+2 2 | 3+3 2 |
| | 3+3 2 | 3+3 2 | 2+2 2 | 3+3 2 | 2+2 2 | 2+2 2 | 3+3 2 | 2+2 2 |
| | 2+2 2 | 3+3 2 | 3+3 2 | 3+3 2 | 3+3 2 | 3+3 2 | 2+2 2 | 3+3 2 |
| | 2+2 2 | 2+2 2 | 2+2 2 | 2+2 2 | 3+3 2 | 3+3 2 | 3+3 2 | 2+2 2 |
| | 3+3 2 | 3+3 2 | 2+2 2 | 2+2 2 | 2+2 2 | 3+3 2 | 2+2 2 | 2+2 2 |
| | 3+3 2 | 2+2 2 | 3+3 2 | 2+2 2 | 2+2 2 | 2+2 2 | 3+3 2 | 3+3 2 |
| | 3+3 2 | 3+3 2 | 2+2 2 | 3+3 2 | 2+2 2 | 2+2 2 | 2+2 2 | 2+2 2 |
| | 3+3 2 | 2+2 2 | 3+3 2 | 3+3 2 | 2+2 2 | 3+3 2 | 3+3 2 | 3+3 2 |
| | 3+3 2 | 3+3 2 | 3+3 2 | 3+3 2 | 2+2 2 | 2+2 2 | 2+2 2 | 2+2 2 |
| | 3+3 2 | 2+2 2 | 2+2 2 | 2+2 2 | 2+2 2 | 3+3 2 | 3+3 2 | 3+3 2 |
| | 3+3 2 | 2+2 2 | 3+3 2 | 3+3 2 | 2+2 2 | 3+3 2 | 2+2 2 | 3+3 2 |
| | 3+3 2 | 3+3 2 | 2+2 2 | 2+2 2 | 2+2 2 | 2+2 2 | 3+3 2 | 2+2 2 |
| | 3+3 2 | 2+2 2 | 3+3 2 | 2+2 2 | 2+2 2 | 3+3 2 | 2+2 2 | 2+2 2 |
| | 3+3 2 | 3+3 2 | 2+2 2 | 3+3 2 | 2+2 2 | 2+2 2 | 3+3 2 | 3+3 2 |
| | 3+3 2 | 3+3 2 | 2+2 2 | 2+2 2 | 2+2 2 | 2+2 2 | 3+3 2 | 2+2 2 |
| | 3+3 2 | 2+2 2 | 3+3 2 | 3+3 2 | 2+2 2 | 3+3 2 | 2+2 2 | 3+3 2 |
| | 3+3 2 | 3+3 2 | 2+2 2 | 2+2 2 | 2+2 2 | 3+3 2 | 2+2 2 | 3+3 2 |
| | 3+3 2 | 2+2 2 | 2+2 2 | 2+2 2 | 2+2 2 | 2+2 2 | 3+3 2 | 2+2 2 |
| | 3+3 2 | 2+2 2 | 3+3 2 | 3+3 2 | 2+2 2 | 2+2 2 | 2+2 2 | 2+2 2 |
| | 3+3 2 | 3+3 2 | 3+3 2 | 3+3 2 | 2+2 2 | 3+3 2 | 3+3 2 | 3+3 2 |
| | 3+3 2 | 2+2 2 | 3+3 2 | 3+3 2 | 2+2 2 | 2+2 2 | 2+2 2 | 2+2 2 |
| | 3+3 2 | 3+3 2 | 3+3 2 | 3+3 2 | 2+2 2 | 3+3 2 | 3+3 2 | 3+3 2 |
| 2.2.53 | 200533 | 98 | Cu1F1Nb1O3 | | | | | |
| | -GM | -R | -T | -U | -V | -X | -Y | -Z |
| | 3+3 2 | 2+2 2 | 3+3 2 | 2+2 2 | 2+2 2 | 2+2 2 | 3+3 2 | 3+3 2 |
| | 2+2 2 | 3+3 2 | 2+2 2 | 3+3 2 | 3+3 2 | 3+3 2 | 2+2 2 | 2+2 2 |
| | 3+3 2 | 3+3 2 | 3+3 2 | 3+3 2 | 3+3 2 | 3+3 2 | 3+3 2 | 3+3 2 |
| | 2+2 2 | 2+2 2 | 2+2 2 | 2+2 2 | 2+2 2 | 2+2 2 | 2+2 2 | 2+2 2 |
| | 3+3 2 | 3+3 2 | 3+3 2 | 3+3 2 | 3+3 2 | 3+3 2 | 3+3 2 | 3+3 2 |
| | 2+2 2 | 2+2 2 | 2+2 2 | 2+2 2 | 2+2 2 | 2+2 2 | 2+2 2 | 2+2 2 |
| | 3+3 2 | 2+2 2 | 3+3 2 | 2+2 2 | 2+2 2 | 2+2 2 | 3+3 2 | 3+3 2 |
| | 2+2 2 | 3+3 2 | 2+2 2 | 3+3 2 | 3+3 2 | 3+3 2 | 2+2 2 | 2+2 2 |
| | 3+3 2 | 2+2 2 | 2+2 2 | 2+2 2 | 3+3 2 | 3+3 2 | 3+3 2 | 3+3 2 |
| | 2+2 2 | 3+3 2 | 3+3 2 | 3+3 2 | 2+2 2 | 2+2 2 | 2+2 2 | 2+2 2 |
| | 3+3 2 | 3+3 2 | 3+3 2 | 2+2 2 | 3+3 2 | 2+2 2 | 3+3 2 | 3+3 2 |
| | 2+2 2 | 2+2 2 | 2+2 2 | 3+3 2 | 2+2 2 | 3+3 2 | 2+2 2 | 2+2 2 |
| | 3+3 2 | 2+2 2 | 2+2 2 | 2+2 2 | 3+3 2 | 2+2 2 | 2+2 2 | 3+3 2 |
| | 2+2 2 | 3+3 2 | 3+3 2 | 3+3 2 | 2+2 2 | 3+3 2 | 3+3 2 | 2+2 2 |
| | 3+3 2 | 2+2 2 | 3+3 2 | 3+3 2 | 3+3 2 | 2+2 2 | 2+2 2 | 2+2 2 |
| | 2+2 2 | 3+3 2 | 2+2 2 | 2+2 2 | 2+2 2 | 3+3 2 | 3+3 2 | 3+3 2 |
| | 3+3 2 | 3+3 2 | 3+3 2 | 3+3 2 | 3+3 2 | 2+2 2 | 3+3 2 | 3+3 2 |
| | 3+3 2 | 2+2 2 | 2+2 2 | 2+2 2 | 2+2 2 | 3+3 2 | 2+2 2 | 2+2 2 |
| | 2+2 2 | 2+2 2 | 3+3 2 | 3+3 2 | 3+3 2 | 3+3 2 | 3+3 2 | 3+3 2 |
| | 3+3 2 | 3+3 2 | 2+2 2 | 2+2 2 | 2+2 2 | 2+2 2 | 2+2 2 | 2+2 2 |
| | 2+2 2 | 2+2 2 | 3+3 2 | 2+2 2 | 2+2 2 | 2+2 2 | 3+3 2 | 3+3 2 |
| | 2+2 2 | 3+3 2 | 2+2 2 | 3+3 2 | 3+3 2 | 3+3 2 | 2+2 2 | 2+2 2 |
| | 3+3 2 | 2+2 2 | 3+3 2 | 2+2 2 | 2+2 2 | 3+3 2 | 2+2 2 | 2+2 2 |
| | 2+2 2 | 3+3 2 | 2+2 2 | 3+3 2 | 3+3 2 | 2+2 2 | 3+3 2 | 3+3 2 |
| | 3+3 2 | 3+3 2 | 2+2 2 | 2+2 2 | 3+3 2 | 2+2 2 | 3+3 2 | 3+3 2 |
| | 3+3 2 | 2+2 2 | 3+3 2 | 3+3 2 | 3+3 2 | 3+3 2 | 2+2 2 | 3+3 2 |
| | 3+3 2 | 3+3 2 | 3+3 2 | 2+2 2 | 2+2 2 | 2+2 2 | 3+3 2 | 3+3 2 |
| | 2+2 2 | 2+2 2 | 2+2 2 | 3+3 2 | 3+3 2 | 3+3 2 | 3+3 2 | 2+2 2 |
| | 3+3 2 | 3+3 2 | 3+3 2 | 3+3 2 | 2+2 2 | 3+3 2 | 2+2 2 | 3+3 2 |

Continued on next column

**TABLE XXI– continued**

| TQC | ICSD | NoE | SF | | | | | |
|---|---|---|---|---|---|---|---|---|
| | 2+2 2 | 2+2 2 | 2+2 2 | 2+2 2 | 2+2 2 | 2+2 2 | 2+2 2 | 2+2 2 |
| | 2+2 2 | 3+3 2 | 3+3 2 | 2+2 2 | 3+3 2 | 2+2 2 | 3+3 2 | 2+2 2 |
| | 2+2 2 | 2+2 2 | 3+3 2 | 2+2 2 | 2+2 2 | 3+3 2 | 2+2 2 | 3+3 2 |
| | 3+3 2 | 3+3 2 | 2+2 2 | 3+3 2 | 3+3 2 | 3+3 2 | 2+2 2 | 2+2 2 |
| | 3+3 2 | 2+2 2 | 2+2 2 | 3+3 2 | 2+2 2 | 2+2 2 | 3+3 2 | 2+2 2 |
| | 2+2 2 | 2+2 2 | 2+2 2 | 2+2 2 | 2+2 2 | 2+2 2 | 2+2 2 | 2+2 2 |
| | 2+2 2 | 3+3 2 | 3+3 2 | 3+3 2 | 3+3 2 | 3+3 2 | 3+3 2 | 3+3 2 |
| | 3+3 2 | 3+3 2 | 3+3 2 | 3+3 2 | 3+3 2 | 2+2 2 | 2+2 2 | 3+3 2 |
| | 2+2 2 | 2+2 2 | 2+2 2 | 2+2 2 | 2+2 2 | 3+3 2 | 3+3 2 | 3+3 2 |
| | 3+3 2 | 2+2 2 | 3+3 2 | 3+3 2 | 2+2 2 | 3+3 2 | 2+2 2 | 2+2 2 |
| | 2+2 2 | 3+3 2 | 2+2 2 | 2+2 2 | 3+3 2 | 2+2 2 | 3+3 2 | 2+2 2 |
| | 3+3 2 | 2+2 2 | 2+2 2 | 2+2 2 | 3+3 2 | 2+2 2 | 2+2 2 | 2+2 2 |
| | 2+2 2 | 3+3 2 | 3+3 2 | 3+3 2 | 2+2 2 | 3+3 2 | 2+2 2 | 3+3 2 |
| | 3+3 2 | 2+2 2 | 3+3 2 | 3+3 2 | 3+3 2 | 2+2 2 | 3+3 2 | 2+2 2 |
| | 2+2 2 | 3+3 2 | 2+2 2 | 2+2 2 | 2+2 2 | 3+3 2 | 3+3 2 | 3+3 2 |
| | 3+3 2 | 2+2 2 | 3+3 2 | 3+3 2 | 2+2 2 | 3+3 2 | 2+2 2 | 3+3 2 |
| | 3+3 2 | 3+3 2 | 2+2 2 | 2+2 2 | 3+3 2 | 2+2 2 | 3+3 2 | 2+2 2 |
| | 2+2 2 | 2+2 2 | 2+2 2 | 3+3 2 | 3+3 2 | 2+2 2 | 2+2 2 | 2+2 2 |
| | 2+2 2 | 3+3 2 | 3+3 2 | 2+2 2 | 2+2 2 | 3+3 2 | 3+3 2 | 3+3 2 |
| | 2+2 2 | 2+2 2 | 2+2 2 | 3+3 2 | 2+2 2 | 3+3 2 | 3+3 2 | 2+2 2 |
| 2.2.54 | 62000 | 84 | Cs1Hg1 | | | | | |
| | -GM | -R | -T | -U | -V | -X | -Y | -Z |
| | 3+3 2 | 2+2 2 | 2+2 2 | 2+2 2 | 3+3 2 | 2+2 2 | 2+2 2 | 2+2 2 |
| | 2+2 2 | 3+3 2 | 3+3 2 | 3+3 2 | 2+2 2 | 3+3 2 | 3+3 2 | 3+3 2 |
| | 2+2 2 | 3+3 2 | 3+3 2 | 2+2 2 | 2+2 2 | 2+2 2 | 2+2 2 | 3+3 2 |
| | 3+3 2 | 2+2 2 | 2+2 2 | 3+3 2 | 3+3 2 | 3+3 2 | 3+3 2 | 2+2 2 |
| | 3+3 2 | 3+3 2 | 3+3 2 | 2+2 2 | 3+3 2 | 2+2 2 | 2+2 2 | 3+3 2 |
| | 2+2 2 | 2+2 2 | 2+2 2 | 3+3 2 | 2+2 2 | 3+3 2 | 3+3 2 | 2+2 2 |
| | 3+3 2 | 2+2 2 | 2+2 2 | 2+2 2 | 3+3 2 | 2+2 2 | 2+2 2 | 2+2 2 |
| | 2+2 2 | 3+3 2 | 3+3 2 | 3+3 2 | 2+2 2 | 3+3 2 | 3+3 2 | 3+3 2 |
| | 3+3 2 | 2+2 2 | 3+3 2 | 2+2 2 | 3+3 2 | 2+2 2 | 2+2 2 | 2+2 2 |
| | 2+2 2 | 3+3 2 | 2+2 2 | 3+3 2 | 2+2 2 | 3+3 2 | 3+3 2 | 3+3 2 |
| | 3+3 2 | 3+3 2 | 3+3 2 | 2+2 2 | 3+3 2 | 2+2 2 | 2+2 2 | 2+2 2 |
| | 2+2 2 | 2+2 2 | 2+2 2 | 3+3 2 | 2+2 2 | 3+3 2 | 3+3 2 | 3+3 2 |
| | 3+3 2 | 3+3 2 | 3+3 2 | 2+2 2 | 2+2 2 | 2+2 2 | 2+2 2 | 3+3 2 |
| | 3+3 2 | 2+2 2 | 2+2 2 | 3+3 2 | 3+3 2 | 3+3 2 | 3+3 2 | 2+2 2 |
| | 2+2 2 | 3+3 2 | 3+3 2 | 2+2 2 | 2+2 2 | 2+2 2 | 2+2 2 | 2+2 2 |
| | 2+2 2 | 2+2 2 | 2+2 2 | 3+3 2 | 3+3 2 | 3+3 2 | 3+3 2 | 3+3 2 |
| | 3+3 2 | 3+3 2 | 3+3 2 | 3+3 2 | 3+3 2 | 3+3 2 | 3+3 2 | 3+3 2 |
| | 2+2 2 | 2+2 2 | 2+2 2 | 2+2 2 | 2+2 2 | 2+2 2 | 2+2 2 | 2+2 2 |
| | 3+3 2 | 2+2 2 | 2+2 2 | 2+2 2 | 2+2 2 | 2+2 2 | 2+2 2 | 2+2 2 |
| | 2+2 2 | 3+3 2 | 3+3 2 | 3+3 2 | 3+3 2 | 3+3 2 | 3+3 2 | 3+3 2 |
| | 3+3 2 | 3+3 2 | 3+3 2 | 3+3 2 | 3+3 2 | 3+3 2 | 3+3 2 | 3+3 2 |
| | 3+3 2 | 3+3 2 | 2+2 2 | 3+3 2 | 3+3 2 | 3+3 2 | 3+3 2 | 3+3 2 |
| | 2+2 2 | 2+2 2 | 3+3 2 | 2+2 2 | 2+2 2 | 2+2 2 | 2+2 2 | 2+2 2 |
| | 2+2 2 | 2+2 2 | 2+2 2 | 2+2 2 | 2+2 2 | 2+2 2 | 2+2 2 | 2+2 2 |
| | 3+3 2 | 3+3 2 | 3+3 2 | 3+3 2 | 3+3 2 | 3+3 2 | 3+3 2 | 3+3 2 |
| | 2+2 2 | 2+2 2 | 2+2 2 | 2+2 2 | 2+2 2 | 2+2 2 | 2+2 2 | 2+2 2 |
| | 3+3 2 | 3+3 2 | 3+3 2 | 3+3 2 | 3+3 2 | 3+3 2 | 3+3 2 | 3+3 2 |
| | 2+2 2 | 2+2 2 | 2+2 2 | 2+2 2 | 2+2 2 | 2+2 2 | 2+2 2 | 2+2 2 |
| | 3+3 2 | 3+3 2 | 3+3 2 | 3+3 2 | 3+3 2 | 3+3 2 | 3+3 2 | 3+3 2 |
| | 2+2 2 | 2+2 2 | 2+2 2 | 2+2 2 | 2+2 2 | 2+2 2 | 2+2 2 | 2+2 2 |
| | 2+2 2 | 2+2 2 | 2+2 2 | 2+2 2 | 2+2 2 | 2+2 2 | 2+2 2 | 2+2 2 |
| | 3+3 2 | 3+3 2 | 3+3 2 | 3+3 2 | 3+3 2 | 3+3 2 | 3+3 2 | 3+3 2 |
| | 2+2 2 | 2+2 2 | 2+2 2 | 2+2 2 | 2+2 2 | 2+2 2 | 2+2 2 | 2+2 2 |
| | 2+2 2 | 2+2 2 | 2+2 2 | 2+2 2 | 2+2 2 | 2+2 2 | 2+2 2 | 2+2 2 |
| | 3+3 2 | 3+3 2 | 3+3 2 | 3+3 2 | 3+3 2 | 3+3 2 | 3+3 2 | 3+3 2 |
| | 3+3 2 | 3+3 2 | 3+3 2 | 3+3 2 | 3+3 2 | 3+3 2 | 3+3 2 | 3+3 2 |
| | 3+3 2 | 3+3 2 | 3+3 2 | 3+3 2 | 3+3 2 | 3+3 2 | 3+3 2 | 3+3 2 |
| | 2+2 2 | 2+2 2 | 2+2 2 | 2+2 2 | 2+2 2 | 2+2 2 | 2+2 2 | 2+2 2 |
| | 2+2 2 | 2+2 2 | 2+2 2 | 2+2 2 | 2+2 2 | 2+2 2 | 2+2 2 | 2+2 2 |

Continued on next column

**TABLE XXI– continued**

| TQC | ICSD | NoE | SF | | | | | |
|---|---|---|---|---|---|---|---|---|
| | 3+3 2 | 3+3 2 | 3+3 2 | 3+3 2 | 3+3 2 | 3+3 2 | 3+3 2 | 3+3 2 |
| | 3+3 2 | 2+2 2 | 3+3 2 | 2+2 2 | 2+2 2 | 2+2 2 | 3+3 2 | 3+3 2 |
| | 3+3 2 | 3+3 2 | 2+2 2 | 3+3 2 | 3+3 2 | 3+3 2 | 2+2 2 | 3+3 2 |
| 2.2.55 | 150974 | 84 | Cs1Hg1 | | | | | |
| | -GM | -R | -T | -U | -V | -X | -Y | -Z |
| | 3+3 2 | 3+3 2 | 2+2 2 | 2+2 2 | 3+3 2 | 3+3 2 | 3+3 2 | 3+3 2 |
| | 2+2 2 | 2+2 2 | 3+3 2 | 3+3 2 | 2+2 2 | 2+2 2 | 2+2 2 | 2+2 2 |
| | 2+2 2 | 2+2 2 | 3+3 2 | 2+2 2 | 2+2 2 | 3+3 2 | 3+3 2 | 2+2 2 |
| | 3+3 2 | 3+3 2 | 2+2 2 | 3+3 2 | 3+3 2 | 2+2 2 | 2+2 2 | 3+3 2 |
| | 3+3 2 | 2+2 2 | 3+3 2 | 2+2 2 | 3+3 2 | 3+3 2 | 3+3 2 | 2+2 2 |
| | 2+2 2 | 3+3 2 | 2+2 2 | 3+3 2 | 2+2 2 | 2+2 2 | 2+2 2 | 3+3 2 |
| | 3+3 2 | 3+3 2 | 2+2 2 | 2+2 2 | 3+3 2 | 3+3 2 | 3+3 2 | 3+3 2 |
| | 2+2 2 | 2+2 2 | 3+3 2 | 3+3 2 | 2+2 2 | 2+2 2 | 2+2 2 | 2+2 2 |
| | 3+3 2 | 3+3 2 | 3+3 2 | 2+2 2 | 3+3 2 | 3+3 2 | 3+3 2 | 3+3 2 |
| | 2+2 2 | 2+2 2 | 2+2 2 | 3+3 2 | 2+2 2 | 2+2 2 | 2+2 2 | 2+2 2 |
| | 3+3 2 | 2+2 2 | 3+3 2 | 2+2 2 | 3+3 2 | 3+3 2 | 3+3 2 | 3+3 2 |
| | 2+2 2 | 3+3 2 | 2+2 2 | 3+3 2 | 2+2 2 | 2+2 2 | 2+2 2 | 2+2 2 |
| | 3+3 2 | 2+2 2 | 3+3 2 | 2+2 2 | 2+2 2 | 3+3 2 | 3+3 2 | 2+2 2 |
| | 3+3 2 | 3+3 2 | 2+2 2 | 3+3 2 | 3+3 2 | 2+2 2 | 2+2 2 | 3+3 2 |
| | 2+2 2 | 2+2 2 | 3+3 2 | 2+2 2 | 2+2 2 | 3+3 2 | 3+3 2 | 3+3 2 |
| | 2+2 2 | 3+3 2 | 2+2 2 | 3+3 2 | 3+3 2 | 2+2 2 | 2+2 2 | 2+2 2 |
| | 3+3 2 | 2+2 2 | 3+3 2 | 3+3 2 | 3+3 2 | 2+2 2 | 2+2 2 | 2+2 2 |
| | 2+2 2 | 3+3 2 | 2+2 2 | 2+2 2 | 2+2 2 | 3+3 2 | 3+3 2 | 3+3 2 |
| | 3+3 2 | 3+3 2 | 2+2 2 | 2+2 2 | 2+2 2 | 2+2 2 | 3+3 2 | 3+3 2 |
| | 2+2 2 | 2+2 2 | 3+3 2 | 3+3 2 | 3+3 2 | 3+3 2 | 2+2 2 | 2+2 2 |
| | 3+3 2 | 2+2 2 | 3+3 2 | 3+3 2 | 3+3 2 | 2+2 2 | 2+2 2 | 2+2 2 |
| | 3+3 2 | 2+2 2 | 2+2 2 | 3+3 2 | 3+3 2 | 2+2 2 | 2+2 2 | 2+2 2 |
| | 2+2 2 | 3+3 2 | 3+3 2 | 2+2 2 | 2+2 2 | 3+3 2 | 3+3 2 | 3+3 2 |
| | 2+2 2 | 3+3 2 | 2+2 2 | 2+2 2 | 2+2 2 | 3+3 2 | 3+3 2 | 3+3 2 |
| | 3+3 2 | 2+2 2 | 3+3 2 | 3+3 2 | 3+3 2 | 2+2 2 | 2+2 2 | 2+2 2 |
| | 2+2 2 | 3+3 2 | 2+2 2 | 2+2 2 | 2+2 2 | 3+3 2 | 3+3 2 | 3+3 2 |
| | 3+3 2 | 2+2 2 | 3+3 2 | 3+3 2 | 3+3 2 | 2+2 2 | 2+2 2 | 2+2 2 |
| | 2+2 2 | 3+3 2 | 2+2 2 | 2+2 2 | 2+2 2 | 3+3 2 | 3+3 2 | 3+3 2 |
| | 3+3 2 | 2+2 2 | 3+3 2 | 3+3 2 | 3+3 2 | 2+2 2 | 2+2 2 | 2+2 2 |
| | 2+2 2 | 3+3 2 | 2+2 2 | 2+2 2 | 2+2 2 | 3+3 2 | 3+3 2 | 3+3 2 |
| | 2+2 2 | 3+3 2 | 2+2 2 | 2+2 2 | 2+2 2 | 3+3 2 | 3+3 2 | 3+3 2 |
| | 3+3 2 | 2+2 2 | 3+3 2 | 3+3 2 | 3+3 2 | 2+2 2 | 2+2 2 | 2+2 2 |
| | 2+2 2 | 3+3 2 | 2+2 2 | 2+2 2 | 2+2 2 | 3+3 2 | 3+3 2 | 3+3 2 |
| | 2+2 2 | 3+3 2 | 2+2 2 | 2+2 2 | 2+2 2 | 3+3 2 | 3+3 2 | 3+3 2 |
| | 3+3 2 | 2+2 2 | 3+3 2 | 3+3 2 | 3+3 2 | 2+2 2 | 2+2 2 | 2+2 2 |
| | 3+3 2 | 2+2 2 | 3+3 2 | 3+3 2 | 3+3 2 | 2+2 2 | 2+2 2 | 2+2 2 |
| | 3+3 2 | 2+2 2 | 3+3 2 | 3+3 2 | 3+3 2 | 2+2 2 | 2+2 2 | 2+2 2 |
| | 2+2 2 | 3+3 2 | 2+2 2 | 2+2 2 | 2+2 2 | 3+3 2 | 3+3 2 | 3+3 2 |
| | 2+2 2 | 3+3 2 | 2+2 2 | 2+2 2 | 2+2 2 | 3+3 2 | 3+3 2 | 3+3 2 |
| | 3+3 2 | 2+2 2 | 3+3 2 | 3+3 2 | 3+3 2 | 2+2 2 | 2+2 2 | 2+2 2 |
| | 3+3 2 | 3+3 2 | 3+3 2 | 2+2 2 | 2+2 2 | 3+3 2 | 2+2 2 | 2+2 2 |
| | 3+3 2 | 2+2 2 | 2+2 2 | 3+3 2 | 3+3 2 | 2+2 2 | 3+3 2 | 2+2 2 |
| 2.2.56 | 62676 | 68 | Cr1Na4O4 | | | | | |
| | -GM | -R | -T | -U | -V | -X | -Y | -Z |
| | 3+3 2 | 3+3 2 | 2+2 2 | 2+2 2 | 2+2 2 | 2+2 2 | 3+3 2 | 2+2 2 |
| | 2+2 2 | 2+2 2 | 3+3 2 | 3+3 2 | 3+3 2 | 3+3 2 | 2+2 2 | 3+3 2 |
| | 3+3 2 | 3+3 2 | 3+3 2 | 3+3 2 | 3+3 2 | 3+3 2 | 2+2 2 | 3+3 2 |
| | 3+3 2 | 2+2 2 | 2+2 2 | 2+2 2 | 2+2 2 | 2+2 2 | 3+3 2 | 2+2 2 |
| | 2+2 2 | 3+3 2 | 2+2 2 | 2+2 2 | 3+3 2 | 3+3 2 | 2+2 2 | 3+3 2 |
| | 2+2 2 | 2+2 2 | 3+3 2 | 3+3 2 | 2+2 2 | 2+2 2 | 3+3 2 | 2+2 2 |
| | 3+3 2 | 3+3 2 | 2+2 2 | 2+2 2 | 2+2 2 | 2+2 2 | 2+2 2 | 2+2 2 |
| | 2+2 2 | 2+2 2 | 3+3 2 | 3+3 2 | 3+3 2 | 3+3 2 | 3+3 2 | 3+3 2 |
| | 3+3 2 | 3+3 2 | 2+2 2 | 3+3 2 | 3+3 2 | 2+2 2 | 2+2 2 | 3+3 2 |
| | 2+2 2 | 2+2 2 | 3+3 2 | 2+2 2 | 2+2 2 | 2+2 2 | 3+3 2 | 2+2 2 |
| | 3+3 2 | 3+3 2 | 3+3 2 | 3+3 2 | 3+3 2 | 3+3 2 | 2+2 2 | 2+2 2 |
| | 3+3 2 | 2+2 2 | 2+2 2 | 2+2 2 | 2+2 2 | 3+3 2 | 3+3 2 | 3+3 2 |

Continued on next column

**TABLE XXI– continued**

| TQC | ICSD | NoE | SF | | | | | |
|---|---|---|---|---|---|---|---|---|
| | 2+2 2 | 2+2 2 | 2+2 2 | 2+2 2 | 3+3 2 | 3+3 2 | 2+2 2 | 3+3 2 |
| | 3+3 2 | 3+3 2 | 3+3 2 | 2+2 2 | 2+2 2 | 2+2 2 | 3+3 2 | 2+2 2 |
| | 2+2 2 | 2+2 2 | 2+2 2 | 3+3 2 | 3+3 2 | 3+3 2 | 3+3 2 | 2+2 2 |
| | 3+3 2 | 3+3 2 | 2+2 2 | 2+2 2 | 2+2 2 | 2+2 2 | 2+2 2 | 3+3 2 |
| | 3+3 2 | 2+2 2 | 3+3 2 | 3+3 2 | 3+3 2 | 3+3 2 | 3+3 2 | 2+2 2 |
| | 2+2 2 | 3+3 2 | 3+3 2 | 3+3 2 | 2+2 2 | 2+2 2 | 2+2 2 | 3+3 2 |
| | 2+2 2 | 2+2 2 | 3+3 2 | 3+3 2 | 3+3 2 | 3+3 2 | 3+3 2 | 2+2 2 |
| | 2+2 2 | 3+3 2 | 2+2 2 | 2+2 2 | 2+2 2 | 2+2 2 | 2+2 2 | 3+3 2 |
| | 3+3 2 | 3+3 2 | 3+3 2 | 3+3 2 | 3+3 2 | 2+2 2 | 3+3 2 | 2+2 2 |
| | 3+3 2 | 2+2 2 | 2+2 2 | 2+2 2 | 2+2 2 | 2+2 2 | 2+2 2 | 3+3 2 |
| | 3+3 2 | 2+2 2 | 2+2 2 | 3+3 2 | 2+2 2 | 3+3 2 | 3+3 2 | 3+3 2 |
| | 2+2 2 | 3+3 2 | 3+3 2 | 2+2 2 | 2+2 2 | 3+3 2 | 3+3 2 | 2+2 2 |
| | 2+2 2 | 3+3 2 | 3+3 2 | 2+2 2 | 3+3 2 | 3+3 2 | 2+2 2 | 3+3 2 |
| | 2+2 2 | 2+2 2 | 2+2 2 | 3+3 2 | 3+3 2 | 2+2 2 | 2+2 2 | 2+2 2 |
| | 3+3 2 | 3+3 2 | 2+2 2 | 3+3 2 | 2+2 2 | 3+3 2 | 2+2 2 | 3+3 2 |
| | 3+3 2 | 2+2 2 | 3+3 2 | 2+2 2 | 3+3 2 | 2+2 2 | 3+3 2 | 2+2 2 |
| | 2+2 2 | 3+3 2 | 2+2 2 | 3+3 2 | 2+2 2 | 2+2 2 | 2+2 2 | 3+3 2 |
| | 2+2 2 | 2+2 2 | 3+3 2 | 2+2 2 | 3+3 2 | 3+3 2 | 3+3 2 | 2+2 2 |
| | 2+2 2 | 2+2 2 | 2+2 2 | 2+2 2 | 3+3 2 | 3+3 2 | 2+2 2 | 2+2 2 |
| | 3+3 2 | 3+3 2 | 3+3 2 | 3+3 2 | 2+2 2 | 2+2 2 | 3+3 2 | 3+3 2 |
| | 2+2 2 | 2+2 2 | 3+3 2 | 2+2 2 | 3+3 2 | 3+3 2 | 2+2 2 | 3+3 2 |
| | 3+3 2 | 3+3 2 | 2+2 2 | 3+3 2 | 2+2 2 | 2+2 2 | 2+2 2 | 2+2 2 |
| 2.2.57 | 73191 | 174 | Co2O4Rb5 | | | | | |
| | -GM | -R | -T | -U | -V | -X | -Y | -Z |
| | 3+3 2 | 3+3 2 | 3+3 2 | 3+3 2 | 3+3 2 | 3+3 2 | 3+3 2 | 3+3 2 |
| | 2+2 2 | 2+2 2 | 2+2 2 | 2+2 2 | 2+2 2 | 2+2 2 | 2+2 2 | 2+2 2 |
| | 3+3 2 | 2+2 2 | 3+3 2 | 3+3 2 | 3+3 2 | 2+2 2 | 2+2 2 | 2+2 2 |
| | 2+2 2 | 3+3 2 | 2+2 2 | 2+2 2 | 2+2 2 | 3+3 2 | 3+3 2 | 3+3 2 |
| | 3+3 2 | 3+3 2 | 2+2 2 | 3+3 2 | 2+2 2 | 2+2 2 | 3+3 2 | 2+2 2 |
| | 2+2 2 | 2+2 2 | 3+3 2 | 2+2 2 | 3+3 2 | 3+3 2 | 2+2 2 | 3+3 2 |
| | 2+2 2 | 2+2 2 | 2+2 2 | 3+3 2 | 2+2 2 | 3+3 2 | 2+2 2 | 3+3 2 |
| | 3+3 2 | 3+3 2 | 3+3 2 | 2+2 2 | 3+3 2 | 2+2 2 | 3+3 2 | 2+2 2 |
| | 2+2 2 | 3+3 2 | 3+3 2 | 2+2 2 | 3+3 2 | 2+2 2 | 3+3 2 | 2+2 2 |
| | 3+3 2 | 2+2 2 | 2+2 2 | 3+3 2 | 2+2 2 | 3+3 2 | 2+2 2 | 3+3 2 |
| | 3+3 2 | 3+3 2 | 2+2 2 | 2+2 2 | 3+3 2 | 2+2 2 | 2+2 2 | 3+3 2 |
| | 2+2 2 | 2+2 2 | 3+3 2 | 3+3 2 | 2+2 2 | 3+3 2 | 3+3 2 | 2+2 2 |
| | 2+2 2 | 2+2 2 | 3+3 2 | 3+3 2 | 2+2 2 | 2+2 2 | 2+2 2 | 3+3 2 |
| | 3+3 2 | 3+3 2 | 2+2 2 | 2+2 2 | 3+3 2 | 3+3 2 | 3+3 2 | 2+2 2 |
| | 2+2 2 | 2+2 2 | 3+3 2 | 3+3 2 | 2+2 2 | 3+3 2 | 3+3 2 | 2+2 2 |
| | 3+3 2 | 3+3 2 | 2+2 2 | 2+2 2 | 3+3 2 | 3+3 2 | 2+2 2 | 2+2 2 |
| | 2+2 2 | 2+2 2 | 2+2 2 | 2+2 2 | 3+3 2 | 2+2 2 | 3+3 2 | 3+3 2 |
| | 3+3 2 | 3+3 2 | 3+3 2 | 3+3 2 | 2+2 2 | 2+2 2 | 2+2 2 | 3+3 2 |
| | 3+3 2 | 3+3 2 | 3+3 2 | 3+3 2 | 3+3 2 | 3+3 2 | 3+3 2 | 3+3 2 |
| | 3+3 2 | 2+2 2 | 3+3 2 | 2+2 2 | 2+2 2 | 2+2 2 | 2+2 2 | 2+2 2 |
| | 2+2 2 | 2+2 2 | 2+2 2 | 3+3 2 | 3+3 2 | 2+2 2 | 3+3 2 | 2+2 2 |
| | 2+2 2 | 3+3 2 | 2+2 2 | 2+2 2 | 2+2 2 | 3+3 2 | 2+2 2 | 3+3 2 |
| | 3+3 2 | 3+3 2 | 3+3 2 | 3+3 2 | 2+2 2 | 2+2 2 | 2+2 2 | 2+2 2 |
| | 2+2 2 | 2+2 2 | 2+2 2 | 2+2 2 | 3+3 2 | 3+3 2 | 3+3 2 | 3+3 2 |
| | 3+3 2 | 3+3 2 | 3+3 2 | 2+2 2 | 3+3 2 | 2+2 2 | 3+3 2 | 2+2 2 |
| | 2+2 2 | 2+2 2 | 2+2 2 | 3+3 2 | 2+2 2 | 3+3 2 | 2+2 2 | 3+3 2 |
| | 3+3 2 | 2+2 2 | 2+2 2 | 3+3 2 | 2+2 2 | 3+3 2 | 2+2 2 | 2+2 2 |
| | 2+2 2 | 3+3 2 | 3+3 2 | 2+2 2 | 3+3 2 | 2+2 2 | 3+3 2 | 3+3 2 |
| | 3+3 2 | 3+3 2 | 3+3 2 | 3+3 2 | 3+3 2 | 2+2 2 | 3+3 2 | 3+3 2 |
| | 3+3 2 | 2+2 2 | 2+2 2 | 2+2 2 | 2+2 2 | 3+3 2 | 2+2 2 | 2+2 2 |
| | 2+2 2 | 2+2 2 | 3+3 2 | 2+2 2 | 2+2 2 | 2+2 2 | 3+3 2 | 3+3 2 |
| | 3+3 2 | 3+3 2 | 2+2 2 | 3+3 2 | 3+3 2 | 3+3 2 | 2+2 2 | 2+2 2 |
| | 2+2 2 | 2+2 2 | 3+3 2 | 3+3 2 | 2+2 2 | 2+2 2 | 3+3 2 | 2+2 2 |
| | 3+3 2 | 3+3 2 | 3+3 2 | 2+2 2 | 3+3 2 | 3+3 2 | 2+2 2 | 3+3 2 |
| | 2+2 2 | 3+3 2 | 2+2 2 | 3+3 2 | 2+2 2 | 2+2 2 | 3+3 2 | 2+2 2 |
| | 2+2 2 | 2+2 2 | 2+2 2 | 2+2 2 | 3+3 2 | 2+2 2 | 2+2 2 | 3+3 2 |
| | 3+3 2 | 2+2 2 | 3+3 2 | 3+3 2 | 2+2 2 | 3+3 2 | 3+3 2 | 2+2 2 |

Continued on next column

**TABLE XXI– continued**

| TQC | ICSD | NoE | SF | | | | |
|---|---|---|---|---|---|---|---|
| | 2+2 2 | 3+3 2 | 2+2 2 | 2+2 2 | 3+3 2 | 2+2 2 | 2+2 2 | 3+3 2 |
| | 3+3 2 | 2+2 2 | 2+2 2 | 3+3 2 | 3+3 2 | 3+3 2 | 3+3 2 | 2+2 2 |
| | 2+2 2 | 3+3 2 | 3+3 2 | 2+2 2 | 2+2 2 | 3+3 2 | 2+2 2 | 3+3 2 |
| | 2+2 2 | 3+3 2 | 3+3 2 | 3+3 2 | 3+3 2 | 2+2 2 | 3+3 2 | 3+3 2 |
| | 3+3 2 | 2+2 2 | 2+2 2 | 2+2 2 | 2+2 2 | 3+3 2 | 2+2 2 | 2+2 2 |
| | 3+3 2 | 2+2 2 | 2+2 2 | 3+3 2 | 2+2 2 | 3+3 2 | 3+3 2 | 2+2 2 |
| | 2+2 2 | 3+3 2 | 3+3 2 | 2+2 2 | 3+3 2 | 2+2 2 | 2+2 2 | 3+3 2 |
| | 3+3 2 | 3+3 2 | 3+3 2 | 3+3 2 | 3+3 2 | 2+2 2 | 3+3 2 | 2+2 2 |
| | 2+2 2 | 2+2 2 | 2+2 2 | 2+2 2 | 2+2 2 | 3+3 2 | 2+2 2 | 3+3 2 |
| | 3+3 2 | 2+2 2 | 3+3 2 | 3+3 2 | 2+2 2 | 2+2 2 | 3+3 2 | 3+3 2 |
| | 2+2 2 | 3+3 2 | 2+2 2 | 2+2 2 | 3+3 2 | 3+3 2 | 2+2 2 | 2+2 2 |
| | 3+3 2 | 3+3 2 | 2+2 2 | 2+2 2 | 3+3 2 | 2+2 2 | 2+2 2 | 3+3 2 |
| | 2+2 2 | 2+2 2 | 3+3 2 | 3+3 2 | 2+2 2 | 3+3 2 | 2+2 2 | 2+2 2 |
| | 2+2 2 | 3+3 2 | 2+2 2 | 2+2 2 | 3+3 2 | 3+3 2 | 3+3 2 | 3+3 2 |
| | 3+3 2 | 2+2 2 | 3+3 2 | 3+3 2 | 2+2 2 | 2+2 2 | 3+3 2 | 2+2 2 |
| | 2+2 2 | 2+2 2 | 3+3 2 | 3+3 2 | 2+2 2 | 3+3 2 | 3+3 2 | 2+2 2 |
| | 3+3 2 | 3+3 2 | 2+2 2 | 2+2 2 | 3+3 2 | 2+2 2 | 2+2 2 | 3+3 2 |
| | 2+2 2 | 2+2 2 | 3+3 2 | 3+3 2 | 2+2 2 | 3+3 2 | 3+3 2 | 2+2 2 |
| | 3+3 2 | 2+2 2 | 3+3 2 | 2+2 2 | 2+2 2 | 3+3 2 | 3+3 2 | 2+2 2 |
| | 2+2 2 | 2+2 2 | 2+2 2 | 3+3 2 | 3+3 2 | 3+3 2 | 2+2 2 | 2+2 2 |
| | 2+2 2 | 3+3 2 | 3+3 2 | 2+2 2 | 2+2 2 | 2+2 2 | 3+3 2 | 3+3 2 |
| | 2+2 2 | 3+3 2 | 2+2 2 | 3+3 2 | 3+3 2 | 3+3 2 | 2+2 2 | 3+3 2 |
| | 3+3 2 | 2+2 2 | 3+3 2 | 3+3 2 | 2+2 2 | 2+2 2 | 3+3 2 | 2+2 2 |
| | 3+3 2 | 3+3 2 | 2+2 2 | 2+2 2 | 3+3 2 | 2+2 2 | 2+2 2 | 3+3 2 |
| | 3+3 2 | 3+3 2 | 2+2 2 | 2+2 2 | 3+3 2 | 2+2 2 | 2+2 2 | 3+3 2 |
| | 3+3 2 | 3+3 2 | 2+2 2 | 2+2 2 | 3+3 2 | 2+2 2 | 2+2 2 | 3+3 2 |
| | 2+2 2 | 3+3 2 | 3+3 2 | 3+3 2 | 3+3 2 | 2+2 2 | 3+3 2 | 3+3 2 |
| | 2+2 2 | 2+2 2 | 3+3 2 | 2+2 2 | 2+2 2 | 3+3 2 | 2+2 2 | 2+2 2 |
| | 3+3 2 | 2+2 2 | 2+2 2 | 2+2 2 | 3+3 2 | 3+3 2 | 2+2 2 | 2+2 2 |
| | 3+3 2 | 3+3 2 | 2+2 2 | 3+3 2 | 2+2 2 | 2+2 2 | 3+3 2 | 3+3 2 |
| | 2+2 2 | 3+3 2 | 3+3 2 | 3+3 2 | 2+2 2 | 2+2 2 | 2+2 2 | 2+2 2 |
| | 2+2 2 | 2+2 2 | 2+2 2 | 2+2 2 | 3+3 2 | 3+3 2 | 3+3 2 | 3+3 2 |
| | 3+3 2 | 2+2 2 | 3+3 2 | 2+2 2 | 2+2 2 | 3+3 2 | 2+2 2 | 3+3 2 |
| | 3+3 2 | 3+3 2 | 2+2 2 | 3+3 2 | 3+3 2 | 2+2 2 | 3+3 2 | 2+2 2 |
| | 2+2 2 | 2+2 2 | 3+3 2 | 3+3 2 | 2+2 2 | 3+3 2 | 3+3 2 | 2+2 2 |
| | 3+3 2 | 2+2 2 | 2+2 2 | 2+2 2 | 2+2 2 | 3+3 2 | 2+2 2 | 2+2 2 |
| | 2+2 2 | 3+3 2 | 3+3 2 | 2+2 2 | 3+3 2 | 2+2 2 | 3+3 2 | 3+3 2 |
| | 3+3 2 | 3+3 2 | 2+2 2 | 3+3 2 | 3+3 2 | 2+2 2 | 2+2 2 | 3+3 2 |
| | 3+3 2 | 3+3 2 | 2+2 2 | 2+2 2 | 3+3 2 | 2+2 2 | 2+2 2 | 3+3 2 |
| | 2+2 2 | 2+2 2 | 3+3 2 | 3+3 2 | 2+2 2 | 3+3 2 | 3+3 2 | 2+2 2 |
| | 2+2 2 | 2+2 2 | 3+3 2 | 3+3 2 | 2+2 2 | 3+3 2 | 3+3 2 | 2+2 2 |
| | 3+3 2 | 3+3 2 | 2+2 2 | 2+2 2 | 3+3 2 | 3+3 2 | 2+2 2 | 2+2 2 |
| | 2+2 2 | 2+2 2 | 3+3 2 | 3+3 2 | 2+2 2 | 2+2 2 | 3+3 2 | 3+3 2 |
| | 3+3 2 | 3+3 2 | 2+2 2 | 2+2 2 | 2+2 2 | 3+3 2 | 2+2 2 | 2+2 2 |
| | 2+2 2 | 2+2 2 | 3+3 2 | 3+3 2 | 3+3 2 | 2+2 2 | 3+3 2 | 3+3 2 |
| | 3+3 2 | 3+3 2 | 2+2 2 | 2+2 2 | 3+3 2 | 2+2 2 | 2+2 2 | 3+3 2 |
| | 2+2 2 | 2+2 2 | 2+2 2 | 3+3 2 | 2+2 2 | 3+3 2 | 2+2 2 | 3+3 2 |
| | 3+3 2 | 3+3 2 | 3+3 2 | 2+2 2 | 3+3 2 | 2+2 2 | 3+3 2 | 2+2 2 |
| | 3+3 2 | 2+2 2 | 2+2 2 | 2+2 2 | 3+3 2 | 3+3 2 | 2+2 2 | 2+2 2 |
| | 3+3 2 | 3+3 2 | 3+3 2 | 3+3 2 | 2+2 2 | 2+2 2 | 3+3 2 | 3+3 2 |
| 2.2.58 | 73190 | 174 | Co2O4Rb5 | | | | |
| | -GM | -R | -T | -U | -V | -X | -Y | -Z |
| | 3+3 2 | 3+3 2 | 3+3 2 | 3+3 2 | 3+3 2 | 3+3 2 | 3+3 2 | 3+3 2 |
| | 2+2 2 | 2+2 2 | 2+2 2 | 2+2 2 | 2+2 2 | 2+2 2 | 2+2 2 | 2+2 2 |
| | 3+3 2 | 2+2 2 | 3+3 2 | 3+3 2 | 3+3 2 | 2+2 2 | 2+2 2 | 2+2 2 |
| | 2+2 2 | 3+3 2 | 2+2 2 | 2+2 2 | 2+2 2 | 3+3 2 | 3+3 2 | 3+3 2 |
| | 3+3 2 | 3+3 2 | 2+2 2 | 3+3 2 | 2+2 2 | 2+2 2 | 3+3 2 | 2+2 2 |
| | 2+2 2 | 2+2 2 | 3+3 2 | 2+2 2 | 3+3 2 | 3+3 2 | 2+2 2 | 3+3 2 |
| | 2+2 2 | 2+2 2 | 2+2 2 | 3+3 2 | 2+2 2 | 3+3 2 | 2+2 2 | 3+3 2 |
| | 3+3 2 | 3+3 2 | 3+3 2 | 2+2 2 | 3+3 2 | 2+2 2 | 3+3 2 | 2+2 2 |
| | 2+2 2 | 3+3 2 | 3+3 2 | 2+2 2 | 3+3 2 | 2+2 2 | 3+3 2 | 2+2 2 |

Continued on next column

**TABLE XXI– continued**

| TQC | ICSD | NoE | SF | | | | |
|---|---|---|---|---|---|---|---|
| | 3+3 2 | 2+2 2 | 2+2 2 | 3+3 2 | 2+2 2 | 3+3 2 | 2+2 2 | 3+3 2 |
| | 3+3 2 | 3+3 2 | 2+2 2 | 2+2 2 | 3+3 2 | 2+2 2 | 2+2 2 | 3+3 2 |
| | 2+2 2 | 2+2 2 | 3+3 2 | 3+3 2 | 2+2 2 | 3+3 2 | 3+3 2 | 2+2 2 |
| | 2+2 2 | 2+2 2 | 2+2 2 | 3+3 2 | 2+2 2 | 2+2 2 | 2+2 2 | 3+3 2 |
| | 3+3 2 | 3+3 2 | 3+3 2 | 2+2 2 | 3+3 2 | 3+3 2 | 3+3 2 | 2+2 2 |
| | 3+3 2 | 2+2 2 | 3+3 2 | 2+2 2 | 2+2 2 | 3+3 2 | 2+2 2 | 2+2 2 |
| | 2+2 2 | 2+2 2 | 2+2 2 | 3+3 2 | 3+3 2 | 3+3 2 | 3+3 2 | 2+2 2 |
| | 2+2 2 | 3+3 2 | 3+3 2 | 3+3 2 | 2+2 2 | 2+2 2 | 3+3 2 | 3+3 2 |
| | 3+3 2 | 3+3 2 | 2+2 2 | 2+2 2 | 3+3 2 | 2+2 2 | 2+2 2 | 3+3 2 |
| | 3+3 2 | 3+3 2 | 3+3 2 | 3+3 2 | 3+3 2 | 3+3 2 | 3+3 2 | 3+3 2 |
| | 3+3 2 | 2+2 2 | 3+3 2 | 2+2 2 | 2+2 2 | 2+2 2 | 2+2 2 | 2+2 2 |
| | 2+2 2 | 2+2 2 | 2+2 2 | 3+3 2 | 3+3 2 | 2+2 2 | 3+3 2 | 2+2 2 |
| | 2+2 2 | 3+3 2 | 2+2 2 | 2+2 2 | 2+2 2 | 3+3 2 | 2+2 2 | 3+3 2 |
| | 3+3 2 | 3+3 2 | 3+3 2 | 3+3 2 | 2+2 2 | 2+2 2 | 2+2 2 | 2+2 2 |
| | 2+2 2 | 2+2 2 | 2+2 2 | 2+2 2 | 3+3 2 | 3+3 2 | 3+3 2 | 3+3 2 |
| | 3+3 2 | 3+3 2 | 3+3 2 | 2+2 2 | 3+3 2 | 2+2 2 | 3+3 2 | 2+2 2 |
| | 3+3 2 | 2+2 2 | 2+2 2 | 3+3 2 | 2+2 2 | 3+3 2 | 2+2 2 | 3+3 2 |
| | 2+2 2 | 2+2 2 | 2+2 2 | 3+3 2 | 2+2 2 | 3+3 2 | 2+2 2 | 2+2 2 |
| | 2+2 2 | 3+3 2 | 3+3 2 | 2+2 2 | 3+3 2 | 2+2 2 | 3+3 2 | 3+3 2 |
| | 3+3 2 | 3+3 2 | 3+3 2 | 3+3 2 | 3+3 2 | 3+3 2 | 3+3 2 | 3+3 2 |
| | 3+3 2 | 2+2 2 | 2+2 2 | 2+2 2 | 2+2 2 | 2+2 2 | 2+2 2 | 2+2 2 |
| | 2+2 2 | 2+2 2 | 3+3 2 | 2+2 2 | 2+2 2 | 2+2 2 | 3+3 2 | 3+3 2 |
| | 3+3 2 | 3+3 2 | 2+2 2 | 3+3 2 | 3+3 2 | 3+3 2 | 2+2 2 | 2+2 2 |
| | 2+2 2 | 2+2 2 | 3+3 2 | 3+3 2 | 2+2 2 | 2+2 2 | 3+3 2 | 2+2 2 |
| | 3+3 2 | 3+3 2 | 2+2 2 | 2+2 2 | 3+3 2 | 3+3 2 | 2+2 2 | 3+3 2 |
| | 2+2 2 | 3+3 2 | 3+3 2 | 3+3 2 | 2+2 2 | 2+2 2 | 3+3 2 | 2+2 2 |
| | 2+2 2 | 2+2 2 | 2+2 2 | 2+2 2 | 3+3 2 | 2+2 2 | 2+2 2 | 3+3 2 |
| | 3+3 2 | 2+2 2 | 3+3 2 | 3+3 2 | 2+2 2 | 3+3 2 | 3+3 2 | 2+2 2 |
| | 2+2 2 | 3+3 2 | 2+2 2 | 2+2 2 | 3+3 2 | 2+2 2 | 2+2 2 | 3+3 2 |
| | 3+3 2 | 2+2 2 | 2+2 2 | 3+3 2 | 3+3 2 | 3+3 2 | 3+3 2 | 2+2 2 |
| | 2+2 2 | 3+3 2 | 3+3 2 | 2+2 2 | 2+2 2 | 3+3 2 | 2+2 2 | 3+3 2 |
| | 2+2 2 | 3+3 2 | 3+3 2 | 3+3 2 | 3+3 2 | 2+2 2 | 3+3 2 | 3+3 2 |
| | 3+3 2 | 2+2 2 | 2+2 2 | 3+3 2 | 2+2 2 | 3+3 2 | 2+2 2 | 2+2 2 |
| | 3+3 2 | 2+2 2 | 2+2 2 | 2+2 2 | 2+2 2 | 3+3 2 | 3+3 2 | 2+2 2 |
| | 2+2 2 | 3+3 2 | 3+3 2 | 2+2 2 | 3+3 2 | 2+2 2 | 2+2 2 | 3+3 2 |
| | 3+3 2 | 3+3 2 | 3+3 2 | 3+3 2 | 3+3 2 | 2+2 2 | 3+3 2 | 2+2 2 |
| | 2+2 2 | 2+2 2 | 2+2 2 | 2+2 2 | 2+2 2 | 3+3 2 | 2+2 2 | 3+3 2 |
| | 3+3 2 | 2+2 2 | 3+3 2 | 3+3 2 | 2+2 2 | 2+2 2 | 3+3 2 | 2+2 2 |
| | 2+2 2 | 3+3 2 | 2+2 2 | 2+2 2 | 3+3 2 | 3+3 2 | 2+2 2 | 3+3 2 |
| | 3+3 2 | 3+3 2 | 2+2 2 | 2+2 2 | 3+3 2 | 2+2 2 | 2+2 2 | 3+3 2 |
| | 2+2 2 | 2+2 2 | 3+3 2 | 3+3 2 | 2+2 2 | 3+3 2 | 3+3 2 | 2+2 2 |
| | 2+2 2 | 2+2 2 | 2+2 2 | 2+2 2 | 3+3 2 | 3+3 2 | 2+2 2 | 3+3 2 |
| | 3+3 2 | 3+3 2 | 3+3 2 | 3+3 2 | 2+2 2 | 2+2 2 | 3+3 2 | 2+2 2 |
| | 2+2 2 | 2+2 2 | 3+3 2 | 3+3 2 | 2+2 2 | 3+3 2 | 3+3 2 | 2+2 2 |
| | 3+3 2 | 3+3 2 | 2+2 2 | 2+2 2 | 3+3 2 | 2+2 2 | 2+2 2 | 3+3 2 |
| | 2+2 2 | 2+2 2 | 3+3 2 | 3+3 2 | 2+2 2 | 3+3 2 | 3+3 2 | 2+2 2 |
| | 3+3 2 | 2+2 2 | 3+3 2 | 2+2 2 | 2+2 2 | 3+3 2 | 3+3 2 | 2+2 2 |
| | 2+2 2 | 2+2 2 | 2+2 2 | 3+3 2 | 3+3 2 | 3+3 2 | 2+2 2 | 2+2 2 |
| | 2+2 2 | 3+3 2 | 3+3 2 | 2+2 2 | 2+2 2 | 2+2 2 | 3+3 2 | 3+3 2 |
| | 2+2 2 | 3+3 2 | 2+2 2 | 3+3 2 | 3+3 2 | 3+3 2 | 2+2 2 | 3+3 2 |
| | 3+3 2 | 2+2 2 | 3+3 2 | 3+3 2 | 2+2 2 | 2+2 2 | 3+3 2 | 2+2 2 |
| | 3+3 2 | 3+3 2 | 2+2 2 | 2+2 2 | 3+3 2 | 2+2 2 | 2+2 2 | 3+3 2 |
| | 3+3 2 | 3+3 2 | 2+2 2 | 2+2 2 | 3+3 2 | 2+2 2 | 2+2 2 | 3+3 2 |
| | 3+3 2 | 3+3 2 | 2+2 2 | 2+2 2 | 3+3 2 | 2+2 2 | 2+2 2 | 3+3 2 |
| | 2+2 2 | 3+3 2 | 3+3 2 | 3+3 2 | 3+3 2 | 2+2 2 | 3+3 2 | 3+3 2 |
| | 2+2 2 | 2+2 2 | 3+3 2 | 2+2 2 | 2+2 2 | 3+3 2 | 2+2 2 | 2+2 2 |
| | 3+3 2 | 3+3 2 | 2+2 2 | 2+2 2 | 3+3 2 | 3+3 2 | 2+2 2 | 2+2 2 |
| | 3+3 2 | 2+2 2 | 2+2 2 | 3+3 2 | 2+2 2 | 2+2 2 | 3+3 2 | 3+3 2 |
| | 2+2 2 | 3+3 2 | 3+3 2 | 3+3 2 | 2+2 2 | 2+2 2 | 2+2 2 | 2+2 2 |
| | 2+2 2 | 2+2 2 | 2+2 2 | 2+2 2 | 3+3 2 | 3+3 2 | 3+3 2 | 3+3 2 |
| | 3+3 2 | 2+2 2 | 3+3 2 | 2+2 2 | 2+2 2 | 3+3 2 | 2+2 2 | 3+3 2 |

Continued on next column

**TABLE XXI– continued**

| TQC | ICSD | NoE | SF | | | | | |
|---|---|---|---|---|---|---|---|---|
| | 3+3 2 | 3+3 2 | 2+2 2 | 3+3 2 | 3+3 2 | 2+2 2 | 3+3 2 | 2+2 2 |
| | 2+2 2 | 2+2 2 | 3+3 2 | 3+3 2 | 2+2 2 | 3+3 2 | 3+3 2 | 2+2 2 |
| | 3+3 2 | 2+2 2 | 2+2 2 | 2+2 2 | 2+2 2 | 3+3 2 | 2+2 2 | 2+2 2 |
| | 3+3 2 | 3+3 2 | 3+3 2 | 2+2 2 | 3+3 2 | 2+2 2 | 3+3 2 | 3+3 2 |
| | 2+2 2 | 3+3 2 | 2+2 2 | 3+3 2 | 3+3 2 | 2+2 2 | 2+2 2 | 3+3 2 |
| | 3+3 2 | 3+3 2 | 2+2 2 | 2+2 2 | 3+3 2 | 2+2 2 | 2+2 2 | 3+3 2 |
| | 2+2 2 | 2+2 2 | 3+3 2 | 3+3 2 | 2+2 2 | 3+3 2 | 3+3 2 | 2+2 2 |
| | 2+2 2 | 2+2 2 | 3+3 2 | 3+3 2 | 2+2 2 | 3+3 2 | 3+3 2 | 2+2 2 |
| | 3+3 2 | 3+3 2 | 2+2 2 | 2+2 2 | 3+3 2 | 3+3 2 | 2+2 2 | 2+2 2 |
| | 2+2 2 | 2+2 2 | 3+3 2 | 3+3 2 | 2+2 2 | 2+2 2 | 3+3 2 | 3+3 2 |
| | 3+3 2 | 3+3 2 | 2+2 2 | 2+2 2 | 2+2 2 | 3+3 2 | 2+2 2 | 2+2 2 |
| | 2+2 2 | 2+2 2 | 3+3 2 | 3+3 2 | 3+3 2 | 2+2 2 | 3+3 2 | 3+3 2 |
| | 3+3 2 | 3+3 2 | 2+2 2 | 2+2 2 | 3+3 2 | 2+2 2 | 2+2 2 | 3+3 2 |
| | 2+2 2 | 2+2 2 | 2+2 2 | 3+3 2 | 2+2 2 | 3+3 2 | 2+2 2 | 3+3 2 |
| | 3+3 2 | 3+3 2 | 3+3 2 | 2+2 2 | 3+3 2 | 2+2 2 | 3+3 2 | 2+2 2 |
| | 3+3 2 | 2+2 2 | 2+2 2 | 2+2 2 | 3+3 2 | 3+3 2 | 2+2 2 | 2+2 2 |
| | 3+3 2 | 3+3 2 | 3+3 2 | 3+3 2 | 2+2 2 | 2+2 2 | 3+3 2 | 3+3 2 |
| 2.2.59 | 167202 | 94 | Co1F1Li1O4S1 | | | | | |
| | -GM | -R | -T | -U | -V | -X | -Y | -Z |
| | 3+3 2 | 2+2 2 | 2+2 2 | 3+3 2 | 2+2 2 | 2+2 2 | 2+2 2 | 2+2 2 |
| | 2+2 2 | 3+3 2 | 3+3 2 | 2+2 2 | 3+3 2 | 3+3 2 | 3+3 2 | 3+3 2 |
| | 3+3 2 | 3+3 2 | 3+3 2 | 3+3 2 | 3+3 2 | 3+3 2 | 3+3 2 | 3+3 2 |
| | 2+2 2 | 2+2 2 | 2+2 2 | 2+2 2 | 2+2 2 | 2+2 2 | 2+2 2 | 2+2 2 |
| | 3+3 2 | 2+2 2 | 2+2 2 | 3+3 2 | 2+2 2 | 3+3 2 | 2+2 2 | 2+2 2 |
| | 2+2 2 | 3+3 2 | 3+3 2 | 2+2 2 | 3+3 2 | 2+2 2 | 3+3 2 | 3+3 2 |
| | 2+2 2 | 3+3 2 | 3+3 2 | 2+2 2 | 3+3 2 | 3+3 2 | 3+3 2 | 2+2 2 |
| | 3+3 2 | 2+2 2 | 2+2 2 | 3+3 2 | 2+2 2 | 2+2 2 | 2+2 2 | 3+3 2 |
| | 3+3 2 | 3+3 2 | 2+2 2 | 2+2 2 | 3+3 2 | 3+3 2 | 3+3 2 | 3+3 2 |
| | 2+2 2 | 2+2 2 | 3+3 2 | 3+3 2 | 2+2 2 | 2+2 2 | 2+2 2 | 2+2 2 |
| | 3+3 2 | 2+2 2 | 2+2 2 | 3+3 2 | 2+2 2 | 3+3 2 | 3+3 2 | 2+2 2 |
| | 2+2 2 | 3+3 2 | 3+3 2 | 2+2 2 | 3+3 2 | 2+2 2 | 2+2 2 | 3+3 2 |
| | 3+3 2 | 2+2 2 | 3+3 2 | 2+2 2 | 3+3 2 | 3+3 2 | 3+3 2 | 3+3 2 |
| | 2+2 2 | 3+3 2 | 2+2 2 | 3+3 2 | 3+3 2 | 3+3 2 | 2+2 2 | 2+2 2 |
| | 3+3 2 | 2+2 2 | 3+3 2 | 2+2 2 | 2+2 2 | 2+2 2 | 3+3 2 | 3+3 2 |
| | 3+3 2 | 3+3 2 | 2+2 2 | 3+3 2 | 2+2 2 | 2+2 2 | 2+2 2 | 2+2 2 |
| | 2+2 2 | 3+3 2 | 3+3 2 | 3+3 2 | 3+3 2 | 2+2 2 | 3+3 2 | 2+2 2 |
| | 2+2 2 | 2+2 2 | 2+2 2 | 2+2 2 | 2+2 2 | 3+3 2 | 2+2 2 | 3+3 2 |
| | 3+3 2 | 3+3 2 | 2+2 2 | 2+2 2 | 3+3 2 | 3+3 2 | 3+3 2 | 3+3 2 |
| | 3+3 2 | 2+2 2 | 3+3 2 | 3+3 2 | 2+2 2 | 3+3 2 | 3+3 2 | 2+2 2 |
| | 2+2 2 | 3+3 2 | 3+3 2 | 3+3 2 | 3+3 2 | 3+3 2 | 3+3 2 | 3+3 2 |
| | 3+3 2 | 2+2 2 | 2+2 2 | 2+2 2 | 2+2 2 | 2+2 2 | 2+2 2 | 2+2 2 |
| | 2+2 2 | 2+2 2 | 3+3 2 | 3+3 2 | 3+3 2 | 2+2 2 | 2+2 2 | 3+3 2 |
| | 3+3 2 | 3+3 2 | 2+2 2 | 3+3 2 | 2+2 2 | 2+2 2 | 3+3 2 | 2+2 2 |
| | 3+3 2 | 2+2 2 | 2+2 2 | 2+2 2 | 3+3 2 | 3+3 2 | 2+2 2 | 2+2 2 |
| | 2+2 2 | 3+3 2 | 3+3 2 | 3+3 2 | 3+3 2 | 2+2 2 | 2+2 2 | 3+3 2 |
| | 3+3 2 | 3+3 2 | 2+2 2 | 2+2 2 | 2+2 2 | 3+3 2 | 3+3 2 | 3+3 2 |
| | 2+2 2 | 2+2 2 | 3+3 2 | 2+2 2 | 2+2 2 | 2+2 2 | 2+2 2 | 3+3 2 |
| | 2+2 2 | 3+3 2 | 3+3 2 | 3+3 2 | 3+3 2 | 3+3 2 | 3+3 2 | 2+2 2 |
| | 3+3 2 | 2+2 2 | 2+2 2 | 2+2 2 | 3+3 2 | 2+2 2 | 2+2 2 | 2+2 2 |
| | 2+2 2 | 3+3 2 | 2+2 2 | 3+3 2 | 2+2 2 | 3+3 2 | 3+3 2 | 3+3 2 |
| | 2+2 2 | 2+2 2 | 3+3 2 | 2+2 2 | 2+2 2 | 2+2 2 | 2+2 2 | 2+2 2 |
| | 3+3 2 | 2+2 2 | 2+2 2 | 2+2 2 | 3+3 2 | 3+3 2 | 3+3 2 | 2+2 2 |
| | 2+2 2 | 3+3 2 | 3+3 2 | 3+3 2 | 2+2 2 | 2+2 2 | 2+2 2 | 3+3 2 |
| | 3+3 2 | 3+3 2 | 3+3 2 | 2+2 2 | 3+3 2 | 3+3 2 | 3+3 2 | 3+3 2 |
| | 2+2 2 | 2+2 2 | 2+2 2 | 3+3 2 | 3+3 2 | 2+2 2 | 3+3 2 | 2+2 2 |
| | 3+3 2 | 3+3 2 | 2+2 2 | 2+2 2 | 2+2 2 | 2+2 2 | 2+2 2 | 2+2 2 |
| | 2+2 2 | 2+2 2 | 3+3 2 | 3+3 2 | 2+2 2 | 3+3 2 | 2+2 2 | 3+3 2 |
| | 3+3 2 | 3+3 2 | 2+2 2 | 2+2 2 | 2+2 2 | 3+3 2 | 3+3 2 | 2+2 2 |
| | 2+2 2 | 2+2 2 | 3+3 2 | 3+3 2 | 3+3 2 | 2+2 2 | 2+2 2 | 3+3 2 |
| | 3+3 2 | 2+2 2 | 2+2 2 | 2+2 2 | 3+3 2 | 3+3 2 | 3+3 2 | 2+2 2 |
| | 3+3 2 | 3+3 2 | 3+3 2 | 3+3 2 | 3+3 2 | 3+3 2 | 3+3 2 | 3+3 2 |

Continued on next column

**TABLE XXI– continued**

| TQC | ICSD | NoE | SF | | | | | |
|---|---|---|---|---|---|---|---|---|
| | 3+3 2 | 2+2 2 | 2+2 2 | 2+2 2 | 3+3 2 | 3+3 2 | 3+3 2 | 2+2 2 |
| | 3+3 2 | 3+3 2 | 3+3 2 | 3+3 2 | 3+3 2 | 3+3 2 | 3+3 2 | 3+3 2 |
| | 3+3 2 | 2+2 2 | 2+2 2 | 2+2 2 | 3+3 2 | 3+3 2 | 3+3 2 | 2+2 2 |
| | 3+3 2 | 3+3 2 | 3+3 2 | 3+3 2 | 3+3 2 | 3+3 2 | 3+3 2 | 3+3 2 |
| | 3+3 2 | 3+3 2 | 3+3 2 | 2+2 2 | 3+3 2 | 3+3 2 | 3+3 2 | 2+2 2 |
| 2.2.60 | 410945 | 150 | Cl9Te1W1 | | | | | |
| | -GM | -R | -T | -U | -V | -X | -Y | -Z |
| | 3+3 2 | 3+3 2 | 3+3 2 | 3+3 2 | 2+2 2 | 3+3 2 | 3+3 2 | 3+3 2 |
| | 2+2 2 | 2+2 2 | 2+2 2 | 2+2 2 | 3+3 2 | 2+2 2 | 2+2 2 | 2+2 2 |
| | 3+3 2 | 3+3 2 | 3+3 2 | 3+3 2 | 2+2 2 | 2+2 2 | 3+3 2 | 2+2 2 |
| | 2+2 2 | 2+2 2 | 2+2 2 | 2+2 2 | 3+3 2 | 3+3 2 | 2+2 2 | 3+3 2 |
| | 3+3 2 | 3+3 2 | 2+2 2 | 2+2 2 | 2+2 2 | 3+3 2 | 2+2 2 | 3+3 2 |
| | 2+2 2 | 2+2 2 | 2+2 2 | 3+3 2 | 3+3 2 | 2+2 2 | 3+3 2 | 2+2 2 |
| | 3+3 2 | 3+3 2 | 3+3 2 | 3+3 2 | 2+2 2 | 2+2 2 | 3+3 2 | 3+3 2 |
| | 2+2 2 | 2+2 2 | 3+3 2 | 2+2 2 | 3+3 2 | 3+3 2 | 2+2 2 | 2+2 2 |
| | 2+2 2 | 2+2 2 | 3+3 2 | 2+2 2 | 3+3 2 | 3+3 2 | 2+2 2 | 3+3 2 |
| | 3+3 2 | 3+3 2 | 3+3 2 | 3+3 2 | 2+2 2 | 2+2 2 | 2+2 2 | 3+3 2 |
| | 2+2 2 | 2+2 2 | 2+2 2 | 2+2 2 | 3+3 2 | 3+3 2 | 3+3 2 | 2+2 2 |
| | 3+3 2 | 3+3 2 | 2+2 2 | 3+3 2 | 2+2 2 | 2+2 2 | 3+3 2 | 2+2 2 |
| | 3+3 2 | 2+2 2 | 2+2 2 | 3+3 2 | 2+2 2 | 3+3 2 | 3+3 2 | 3+3 2 |
| | 2+2 2 | 3+3 2 | 3+3 2 | 2+2 2 | 3+3 2 | 2+2 2 | 2+2 2 | 2+2 2 |
| | 3+3 2 | 3+3 2 | 2+2 2 | 3+3 2 | 2+2 2 | 2+2 2 | 3+3 2 | 2+2 2 |
| | 2+2 2 | 2+2 2 | 3+3 2 | 2+2 2 | 3+3 2 | 3+3 2 | 2+2 2 | 3+3 2 |
| | 3+3 2 | 3+3 2 | 2+2 2 | 2+2 2 | 3+3 2 | 2+2 2 | 2+2 2 | 3+3 2 |
| | 2+2 2 | 2+2 2 | 3+3 2 | 3+3 2 | 2+2 2 | 3+3 2 | 3+3 2 | 2+2 2 |
| | 3+3 2 | 3+3 2 | 3+3 2 | 3+3 2 | 3+3 2 | 3+3 2 | 3+3 2 | 3+3 2 |
| | 2+2 2 | 2+2 2 | 2+2 2 | 2+2 2 | 2+2 2 | 2+2 2 | 2+2 2 | 2+2 2 |
| | 2+2 2 | 2+2 2 | 2+2 2 | 2+2 2 | 2+2 2 | 2+2 2 | 2+2 2 | 2+2 2 |
| | 3+3 2 | 3+3 2 | 3+3 2 | 3+3 2 | 3+3 2 | 3+3 2 | 3+3 2 | 3+3 2 |
| | 3+3 2 | 2+2 2 | 2+2 2 | 2+2 2 | 2+2 2 | 2+2 2 | 2+2 2 | 2+2 2 |
| | 2+2 2 | 3+3 2 | 3+3 2 | 3+3 2 | 3+3 2 | 3+3 2 | 3+3 2 | 3+3 2 |
| | 3+3 2 | 2+2 2 | 2+2 2 | 2+2 2 | 2+2 2 | 3+3 2 | 2+2 2 | 3+3 2 |
| | 2+2 2 | 3+3 2 | 3+3 2 | 3+3 2 | 3+3 2 | 2+2 2 | 3+3 2 | 2+2 2 |
| | 2+2 2 | 3+3 2 | 3+3 2 | 3+3 2 | 3+3 2 | 2+2 2 | 3+3 2 | 2+2 2 |
| | 3+3 2 | 2+2 2 | 2+2 2 | 2+2 2 | 2+2 2 | 3+3 2 | 2+2 2 | 3+3 2 |
| | 3+3 2 | 2+2 2 | 2+2 2 | 2+2 2 | 3+3 2 | 3+3 2 | 3+3 2 | 2+2 2 |
| | 2+2 2 | 3+3 2 | 3+3 2 | 3+3 2 | 2+2 2 | 2+2 2 | 2+2 2 | 3+3 2 |
| | 3+3 2 | 2+2 2 | 2+2 2 | 2+2 2 | 3+3 2 | 3+3 2 | 3+3 2 | 2+2 2 |
| | 2+2 2 | 3+3 2 | 3+3 2 | 3+3 2 | 2+2 2 | 2+2 2 | 2+2 2 | 3+3 2 |
| | 3+3 2 | 3+3 2 | 2+2 2 | 3+3 2 | 3+3 2 | 3+3 2 | 2+2 2 | 2+2 2 |
| | 2+2 2 | 2+2 2 | 3+3 2 | 2+2 2 | 2+2 2 | 2+2 2 | 3+3 2 | 3+3 2 |
| | 3+3 2 | 2+2 2 | 2+2 2 | 3+3 2 | 2+2 2 | 3+3 2 | 2+2 2 | 3+3 2 |
| | 2+2 2 | 3+3 2 | 3+3 2 | 2+2 2 | 3+3 2 | 2+2 2 | 3+3 2 | 2+2 2 |
| | 3+3 2 | 3+3 2 | 2+2 2 | 3+3 2 | 3+3 2 | 2+2 2 | 2+2 2 | 3+3 2 |
| | 2+2 2 | 2+2 2 | 3+3 2 | 2+2 2 | 2+2 2 | 3+3 2 | 3+3 2 | 2+2 2 |
| | 3+3 2 | 3+3 2 | 2+2 2 | 3+3 2 | 2+2 2 | 2+2 2 | 3+3 2 | 2+2 2 |
| | 3+3 2 | 3+3 2 | 3+3 2 | 2+2 2 | 2+2 2 | 3+3 2 | 2+2 2 | 3+3 2 |
| | 2+2 2 | 2+2 2 | 2+2 2 | 3+3 2 | 3+3 2 | 2+2 2 | 3+3 2 | 3+3 2 |
| | 2+2 2 | 2+2 2 | 3+3 2 | 2+2 2 | 3+3 2 | 3+3 2 | 2+2 2 | 2+2 2 |
| | 3+3 2 | 2+2 2 | 2+2 2 | 3+3 2 | 2+2 2 | 2+2 2 | 3+3 2 | 2+2 2 |
| | 2+2 2 | 3+3 2 | 3+3 2 | 2+2 2 | 3+3 2 | 3+3 2 | 2+2 2 | 3+3 2 |
| | 3+3 2 | 2+2 2 | 3+3 2 | 3+3 2 | 3+3 2 | 3+3 2 | 3+3 2 | 3+3 2 |
| | 2+2 2 | 3+3 2 | 3+3 2 | 2+2 2 | 2+2 2 | 3+3 2 | 3+3 2 | 2+2 2 |
| | 3+3 2 | 3+3 2 | 2+2 2 | 3+3 2 | 3+3 2 | 2+2 2 | 2+2 2 | 3+3 2 |
| | 2+2 2 | 2+2 2 | 2+2 2 | 2+2 2 | 2+2 2 | 2+2 2 | 2+2 2 | 2+2 2 |
| | 3+3 2 | 3+3 2 | 2+2 2 | 2+2 2 | 3+3 2 | 3+3 2 | 2+2 2 | 3+3 2 |
| | 3+3 2 | 3+3 2 | 3+3 2 | 2+2 2 | 2+2 2 | 2+2 2 | 3+3 2 | 3+3 2 |
| | 2+2 2 | 2+2 2 | 3+3 2 | 3+3 2 | 3+3 2 | 3+3 2 | 2+2 2 | 2+2 2 |
| | 3+3 2 | 3+3 2 | 2+2 2 | 2+2 2 | 2+2 2 | 2+2 2 | 3+3 2 | 2+2 2 |
| | 2+2 2 | 2+2 2 | 3+3 2 | 3+3 2 | 3+3 2 | 2+2 2 | 2+2 2 | 3+3 2 |
| | 2+2 2 | 2+2 2 | 2+2 2 | 2+2 2 | 2+2 2 | 3+3 2 | 3+3 2 | 2+2 2 |

Continued on next column

**TABLE XXI– continued**

| TQC | ICSD | NoE | SF | | | | |
|---|---|---|---|---|---|---|---|
| | 3+3 2 | 2+2 2 | 3+3 2 | 3+3 2 | 2+2 2 | 2+2 2 | 2+2 2 | 3+3 2 |
| | 3+3 2 | 3+3 2 | 2+2 2 | 3+3 2 | 3+3 2 | 3+3 2 | 2+2 2 | 3+3 2 |
| | 2+2 2 | 2+2 2 | 3+3 2 | 2+2 2 | 2+2 2 | 3+3 2 | 2+2 2 | 2+2 2 |
| | 2+2 2 | 3+3 2 | 2+2 2 | 3+3 2 | 3+3 2 | 2+2 2 | 3+3 2 | 2+2 2 |
| | 3+3 2 | 2+2 2 | 2+2 2 | 3+3 2 | 3+3 2 | 3+3 2 | 3+3 2 | 3+3 2 |
| | 2+2 2 | 2+2 2 | 3+3 2 | 2+2 2 | 2+2 2 | 2+2 2 | 2+2 2 | 3+3 2 |
| | 3+3 2 | 3+3 2 | 3+3 2 | 3+3 2 | 3+3 2 | 3+3 2 | 3+3 2 | 2+2 2 |
| | 2+2 2 | 2+2 2 | 2+2 2 | 3+3 2 | 2+2 2 | 2+2 2 | 2+2 2 | 2+2 2 |
| | 3+3 2 | 3+3 2 | 3+3 2 | 2+2 2 | 3+3 2 | 3+3 2 | 3+3 2 | 3+3 2 |
| | 3+3 2 | 3+3 2 | 2+2 2 | 2+2 2 | 2+2 2 | 2+2 2 | 3+3 2 | 2+2 2 |
| | 2+2 2 | 3+3 2 | 3+3 2 | 3+3 2 | 3+3 2 | 2+2 2 | 2+2 2 | 2+2 2 |
| | 3+3 2 | 2+2 2 | 2+2 2 | 2+2 2 | 2+2 2 | 3+3 2 | 3+3 2 | 3+3 2 |
| | 2+2 2 | 2+2 2 | 2+2 2 | 3+3 2 | 3+3 2 | 2+2 2 | 3+3 2 | 2+2 2 |
| | 2+2 2 | 3+3 2 | 3+3 2 | 2+2 2 | 2+2 2 | 3+3 2 | 2+2 2 | 3+3 2 |
| | 2+2 2 | 2+2 2 | 2+2 2 | 3+3 2 | 2+2 2 | 3+3 2 | 3+3 2 | 3+3 2 |
| | 3+3 2 | 3+3 2 | 3+3 2 | 2+2 2 | 3+3 2 | 2+2 2 | 2+2 2 | 2+2 2 |
| | 2+2 2 | 2+2 2 | 2+2 2 | 2+2 2 | 2+2 2 | 3+3 2 | 2+2 2 | 3+3 2 |
| | 3+3 2 | 3+3 2 | 3+3 2 | 3+3 2 | 3+3 2 | 2+2 2 | 3+3 2 | 2+2 2 |
| | 3+3 2 | 3+3 2 | 3+3 2 | 2+2 2 | 3+3 2 | 2+2 2 | 3+3 2 | 2+2 2 |
| | 2+2 2 | 2+2 2 | 2+2 2 | 3+3 2 | 2+2 2 | 3+3 2 | 2+2 2 | 3+3 2 |
| | 3+3 2 | 2+2 2 | 2+2 2 | 2+2 2 | 3+3 2 | 3+3 2 | 2+2 2 | 3+3 2 |
| 2.2.61 | 410101 | 134 | Cl7Mo1O1Se1 | | | | | |
| | -GM | -R | -T | -U | -V | -X | -Y | -Z |
| | 3+3 2 | 2+2 2 | 3+3 2 | 2+2 2 | 2+2 2 | 3+3 2 | 2+2 2 | 2+2 2 |
| | 2+2 2 | 3+3 2 | 2+2 2 | 3+3 2 | 3+3 2 | 2+2 2 | 3+3 2 | 3+3 2 |
| | 3+3 2 | 2+2 2 | 3+3 2 | 2+2 2 | 2+2 2 | 2+2 2 | 2+2 2 | 3+3 2 |
| | 2+2 2 | 3+3 2 | 2+2 2 | 3+3 2 | 3+3 2 | 3+3 2 | 3+3 2 | 2+2 2 |
| | 2+2 2 | 3+3 2 | 2+2 2 | 3+3 2 | 3+3 2 | 3+3 2 | 2+2 2 | 2+2 2 |
| | 3+3 2 | 2+2 2 | 3+3 2 | 2+2 2 | 2+2 2 | 2+2 2 | 3+3 2 | 3+3 2 |
| | 3+3 2 | 3+3 2 | 2+2 2 | 3+3 2 | 3+3 2 | 3+3 2 | 2+2 2 | 2+2 2 |
| | 2+2 2 | 2+2 2 | 3+3 2 | 2+2 2 | 2+2 2 | 2+2 2 | 3+3 2 | 3+3 2 |
| | 3+3 2 | 2+2 2 | 3+3 2 | 3+3 2 | 3+3 2 | 2+2 2 | 2+2 2 | 2+2 2 |
| | 2+2 2 | 3+3 2 | 2+2 2 | 2+2 2 | 2+2 2 | 3+3 2 | 3+3 2 | 3+3 2 |
| | 2+2 2 | 3+3 2 | 2+2 2 | 2+2 2 | 2+2 2 | 3+3 2 | 3+3 2 | 3+3 2 |
| | 3+3 2 | 2+2 2 | 3+3 2 | 3+3 2 | 3+3 2 | 2+2 2 | 2+2 2 | 2+2 2 |
| | 2+2 2 | 3+3 2 | 2+2 2 | 2+2 2 | 2+2 2 | 3+3 2 | 3+3 2 | 3+3 2 |
| | 3+3 2 | 2+2 2 | 3+3 2 | 3+3 2 | 3+3 2 | 2+2 2 | 2+2 2 | 2+2 2 |
| | 3+3 2 | 3+3 2 | 3+3 2 | 3+3 2 | 3+3 2 | 3+3 2 | 3+3 2 | 3+3 2 |
| | 2+2 2 | 2+2 2 | 2+2 2 | 2+2 2 | 2+2 2 | 2+2 2 | 2+2 2 | 2+2 2 |
| | 3+3 2 | 2+2 2 | 2+2 2 | 2+2 2 | 2+2 2 | 2+2 2 | 3+3 2 | 3+3 2 |
| | 2+2 2 | 3+3 2 | 3+3 2 | 3+3 2 | 3+3 2 | 3+3 2 | 2+2 2 | 2+2 2 |
| | 2+2 2 | 3+3 2 | 2+2 2 | 3+3 2 | 3+3 2 | 3+3 2 | 2+2 2 | 2+2 2 |
| | 3+3 2 | 2+2 2 | 3+3 2 | 2+2 2 | 2+2 2 | 2+2 2 | 3+3 2 | 3+3 2 |
| | 2+2 2 | 2+2 2 | 3+3 2 | 3+3 2 | 2+2 2 | 3+3 2 | 3+3 2 | 2+2 2 |
| | 3+3 2 | 3+3 2 | 2+2 2 | 2+2 2 | 3+3 2 | 2+2 2 | 2+2 2 | 3+3 2 |
| | 2+2 2 | 2+2 2 | 3+3 2 | 2+2 2 | 3+3 2 | 2+2 2 | 2+2 2 | 3+3 2 |
| | 3+3 2 | 3+3 2 | 2+2 2 | 3+3 2 | 2+2 2 | 3+3 2 | 3+3 2 | 2+2 2 |
| | 3+3 2 | 2+2 2 | 3+3 2 | 3+3 2 | 3+3 2 | 2+2 2 | 2+2 2 | 2+2 2 |
| | 2+2 2 | 3+3 2 | 3+3 2 | 2+2 2 | 2+2 2 | 3+3 2 | 2+2 2 | 2+2 2 |
| | 3+3 2 | 2+2 2 | 2+2 2 | 3+3 2 | 3+3 2 | 2+2 2 | 3+3 2 | 3+3 2 |
| | 2+2 2 | 3+3 2 | 2+2 2 | 2+2 2 | 2+2 2 | 3+3 2 | 3+3 2 | 3+3 2 |
| | 3+3 2 | 3+3 2 | 3+3 2 | 2+2 2 | 2+2 2 | 3+3 2 | 2+2 2 | 2+2 2 |
| | 2+2 2 | 2+2 2 | 2+2 2 | 3+3 2 | 3+3 2 | 2+2 2 | 3+3 2 | 3+3 2 |
| | 3+3 2 | 3+3 2 | 2+2 2 | 3+3 2 | 3+3 2 | 3+3 2 | 3+3 2 | 3+3 2 |
| | 2+2 2 | 2+2 2 | 3+3 2 | 2+2 2 | 2+2 2 | 2+2 2 | 2+2 2 | 2+2 2 |
| | 3+3 2 | 3+3 2 | 2+2 2 | 2+2 2 | 2+2 2 | 3+3 2 | 3+3 2 | 2+2 2 |
| | 2+2 2 | 2+2 2 | 3+3 2 | 3+3 2 | 3+3 2 | 2+2 2 | 2+2 2 | 3+3 2 |
| | 3+3 2 | 3+3 2 | 2+2 2 | 2+2 2 | 2+2 2 | 3+3 2 | 3+3 2 | 3+3 2 |
| | 2+2 2 | 3+3 2 | 2+2 2 | 3+3 2 | 3+3 2 | 2+2 2 | 2+2 2 | 2+2 2 |
| | 3+3 2 | 2+2 2 | 3+3 2 | 2+2 2 | 2+2 2 | 2+2 2 | 3+3 2 | 3+3 2 |
| | 2+2 2 | 3+3 2 | 3+3 2 | 3+3 2 | 3+3 2 | 3+3 2 | 2+2 2 | 2+2 2 |

Continued on next column

**TABLE XXI– continued**

| TQC | ICSD | NoE | SF | | | | |
|---|---|---|---|---|---|---|---|
| | 3+3 2 | 2+2 2 | 2+2 2 | 2+2 2 | 3+3 2 | 2+2 2 | 2+2 2 | 3+3 2 |
| | 3+3 2 | 3+3 2 | 3+3 2 | 3+3 2 | 2+2 2 | 3+3 2 | 2+2 2 | 2+2 2 |
| | 2+2 2 | 2+2 2 | 3+3 2 | 3+3 2 | 3+3 2 | 2+2 2 | 3+3 2 | 2+2 2 |
| | 2+2 2 | 2+2 2 | 2+2 2 | 2+2 2 | 3+3 2 | 2+2 2 | 3+3 2 | 3+3 2 |
| | 3+3 2 | 2+2 2 | 3+3 2 | 3+3 2 | 2+2 2 | 3+3 2 | 2+2 2 | 2+2 2 |
| | 3+3 2 | 2+2 2 | 2+2 2 | 2+2 2 | 2+2 2 | 3+3 2 | 3+3 2 | 3+3 2 |
| | 2+2 2 | 3+3 2 | 3+3 2 | 3+3 2 | 3+3 2 | 2+2 2 | 2+2 2 | 2+2 2 |
| | 2+2 2 | 2+2 2 | 2+2 2 | 2+2 2 | 2+2 2 | 2+2 2 | 3+3 2 | 3+3 2 |
| | 3+3 2 | 3+3 2 | 3+3 2 | 2+2 2 | 3+3 2 | 3+3 2 | 2+2 2 | 3+3 2 |
| | 3+3 2 | 2+2 2 | 3+3 2 | 3+3 2 | 2+2 2 | 2+2 2 | 2+2 2 | 2+2 2 |
| | 2+2 2 | 3+3 2 | 2+2 2 | 2+2 2 | 3+3 2 | 3+3 2 | 3+3 2 | 2+2 2 |
| | 3+3 2 | 2+2 2 | 2+2 2 | 3+3 2 | 3+3 2 | 2+2 2 | 2+2 2 | 3+3 2 |
| | 3+3 2 | 3+3 2 | 3+3 2 | 2+2 2 | 2+2 2 | 3+3 2 | 3+3 2 | 2+2 2 |
| | 2+2 2 | 3+3 2 | 3+3 2 | 3+3 2 | 2+2 2 | 2+2 2 | 3+3 2 | 2+2 2 |
| | 2+2 2 | 2+2 2 | 2+2 2 | 3+3 2 | 3+3 2 | 3+3 2 | 2+2 2 | 3+3 2 |
| | 2+2 2 | 3+3 2 | 2+2 2 | 2+2 2 | 2+2 2 | 2+2 2 | 3+3 2 | 3+3 2 |
| | 3+3 2 | 2+2 2 | 3+3 2 | 3+3 2 | 3+3 2 | 3+3 2 | 2+2 2 | 3+3 2 |
| | 3+3 2 | 3+3 2 | 2+2 2 | 2+2 2 | 2+2 2 | 3+3 2 | 3+3 2 | 2+2 2 |
| | 2+2 2 | 2+2 2 | 2+2 2 | 3+3 2 | 3+3 2 | 2+2 2 | 3+3 2 | 2+2 2 |
| | 2+2 2 | 3+3 2 | 3+3 2 | 2+2 2 | 2+2 2 | 3+3 2 | 2+2 2 | 3+3 2 |
| | 3+3 2 | 3+3 2 | 2+2 2 | 3+3 2 | 2+2 2 | 2+2 2 | 3+3 2 | 3+3 2 |
| | 2+2 2 | 3+3 2 | 3+3 2 | 2+2 2 | 3+3 2 | 3+3 2 | 3+3 2 | 2+2 2 |
| | 3+3 2 | 2+2 2 | 2+2 2 | 2+2 2 | 2+2 2 | 3+3 2 | 2+2 2 | 3+3 2 |
| | 2+2 2 | 2+2 2 | 3+3 2 | 3+3 2 | 3+3 2 | 2+2 2 | 2+2 2 | 2+2 2 |
| | 3+3 2 | 2+2 2 | 3+3 2 | 3+3 2 | 3+3 2 | 3+3 2 | 3+3 2 | 3+3 2 |
| | 2+2 2 | 3+3 2 | 2+2 2 | 2+2 2 | 2+2 2 | 2+2 2 | 2+2 2 | 2+2 2 |
| | 2+2 2 | 3+3 2 | 3+3 2 | 2+2 2 | 2+2 2 | 3+3 2 | 3+3 2 | 3+3 2 |
| | 3+3 2 | 2+2 2 | 2+2 2 | 3+3 2 | 3+3 2 | 2+2 2 | 2+2 2 | 2+2 2 |
| | 3+3 2 | 3+3 2 | 2+2 2 | 3+3 2 | 2+2 2 | 3+3 2 | 3+3 2 | 2+2 2 |
| 2.2.62 | 84620 | 82 | Cl5Mo1 | | | | | |
| | -GM | -R | -T | -U | -V | -X | -Y | -Z |
| | 3+3 2 | 2+2 2 | 3+3 2 | 2+2 2 | 2+2 2 | 2+2 2 | 3+3 2 | 3+3 2 |
| | 2+2 2 | 3+3 2 | 2+2 2 | 3+3 2 | 3+3 2 | 3+3 2 | 2+2 2 | 2+2 2 |
| | 3+3 2 | 3+3 2 | 3+3 2 | 2+2 2 | 2+2 2 | 3+3 2 | 2+2 2 | 2+2 2 |
| | 2+2 2 | 2+2 2 | 2+2 2 | 3+3 2 | 3+3 2 | 2+2 2 | 3+3 2 | 3+3 2 |
| | 3+3 2 | 3+3 2 | 2+2 2 | 3+3 2 | 2+2 2 | 2+2 2 | 3+3 2 | 2+2 2 |
| | 2+2 2 | 2+2 2 | 3+3 2 | 2+2 2 | 3+3 2 | 3+3 2 | 2+2 2 | 3+3 2 |
| | 2+2 2 | 2+2 2 | 2+2 2 | 3+3 2 | 3+3 2 | 2+2 2 | 3+3 2 | 3+3 2 |
| | 3+3 2 | 3+3 2 | 3+3 2 | 2+2 2 | 2+2 2 | 3+3 2 | 2+2 2 | 2+2 2 |
| | 3+3 2 | 2+2 2 | 2+2 2 | 3+3 2 | 2+2 2 | 3+3 2 | 3+3 2 | 3+3 2 |
| | 2+2 2 | 3+3 2 | 3+3 2 | 2+2 2 | 3+3 2 | 2+2 2 | 2+2 2 | 2+2 2 |
| | 3+3 2 | 3+3 2 | 3+3 2 | 2+2 2 | 2+2 2 | 2+2 2 | 2+2 2 | 2+2 2 |
| | 2+2 2 | 2+2 2 | 2+2 2 | 3+3 2 | 3+3 2 | 3+3 2 | 3+3 2 | 3+3 2 |
| | 3+3 2 | 2+2 2 | 3+3 2 | 2+2 2 | 2+2 2 | 3+3 2 | 2+2 2 | 2+2 2 |
| | 2+2 2 | 3+3 2 | 2+2 2 | 3+3 2 | 3+3 2 | 2+2 2 | 3+3 2 | 3+3 2 |
| | 3+3 2 | 3+3 2 | 3+3 2 | 2+2 2 | 3+3 2 | 2+2 2 | 2+2 2 | 2+2 2 |
| | 2+2 2 | 2+2 2 | 2+2 2 | 3+3 2 | 2+2 2 | 3+3 2 | 3+3 2 | 3+3 2 |
| | 3+3 2 | 3+3 2 | 2+2 2 | 3+3 2 | 2+2 2 | 3+3 2 | 3+3 2 | 3+3 2 |
| | 2+2 2 | 2+2 2 | 3+3 2 | 2+2 2 | 3+3 2 | 2+2 2 | 2+2 2 | 2+2 2 |
| | 2+2 2 | 3+3 2 | 2+2 2 | 2+2 2 | 2+2 2 | 2+2 2 | 2+2 2 | 3+3 2 |
| | 3+3 2 | 2+2 2 | 3+3 2 | 3+3 2 | 3+3 2 | 3+3 2 | 3+3 2 | 2+2 2 |
| | 3+3 2 | 2+2 2 | 2+2 2 | 3+3 2 | 3+3 2 | 2+2 2 | 3+3 2 | 3+3 2 |
| | 2+2 2 | 3+3 2 | 3+3 2 | 2+2 2 | 2+2 2 | 3+3 2 | 2+2 2 | 2+2 2 |
| | 3+3 2 | 3+3 2 | 3+3 2 | 2+2 2 | 2+2 2 | 3+3 2 | 2+2 2 | 3+3 2 |
| | 2+2 2 | 3+3 2 | 2+2 2 | 3+3 2 | 2+2 2 | 3+3 2 | 3+3 2 | 2+2 2 |
| | 3+3 2 | 2+2 2 | 3+3 2 | 2+2 2 | 3+3 2 | 2+2 2 | 2+2 2 | 2+2 2 |
| | 3+3 2 | 2+2 2 | 3+3 2 | 2+2 2 | 3+3 2 | 3+3 2 | 2+2 2 | 2+2 2 |
| | 2+2 2 | 3+3 2 | 2+2 2 | 3+3 2 | 2+2 2 | 2+2 2 | 3+3 2 | 3+3 2 |
| | 2+2 2 | 2+2 2 | 2+2 2 | 3+3 2 | 3+3 2 | 2+2 2 | 3+3 2 | 3+3 2 |
| | 3+3 2 | 2+2 2 | 3+3 2 | 3+3 2 | 3+3 2 | 3+3 2 | 2+2 2 | 2+2 2 |
| | 3+3 2 | 3+3 2 | 2+2 2 | 2+2 2 | 2+2 2 | 2+2 2 | 3+3 2 | 3+3 2 |

Continued on next column

**TABLE XXI– continued**

| TQC | ICSD | NoE | SF | | | | |
|---|---|---|---|---|---|---|---|
| | 2+2 2 | 3+3 2 | 3+3 2 | 3+3 2 | 2+2 2 | 3+3 2 | 3+3 2 | 2+2 2 |
| | 3+3 2 | 2+2 2 | 2+2 2 | 2+2 2 | 2+2 2 | 2+2 2 | 2+2 2 | 3+3 2 |
| | 2+2 2 | 3+3 2 | 3+3 2 | 3+3 2 | 3+3 2 | 3+3 2 | 3+3 2 | 3+3 2 |
| | 3+3 2 | 2+2 2 | 2+2 2 | 2+2 2 | 3+3 2 | 3+3 2 | 2+2 2 | 2+2 2 |
| | 3+3 2 | 2+2 2 | 2+2 2 | 3+3 2 | 3+3 2 | 2+2 2 | 2+2 2 | 3+3 2 |
| | 2+2 2 | 3+3 2 | 3+3 2 | 2+2 2 | 2+2 2 | 3+3 2 | 3+3 2 | 2+2 2 |
| | 2+2 2 | 2+2 2 | 2+2 2 | 2+2 2 | 2+2 2 | 2+2 2 | 2+2 2 | 3+3 2 |
| | 2+2 2 | 3+3 2 | 3+3 2 | 3+3 2 | 3+3 2 | 3+3 2 | 3+3 2 | 2+2 2 |
| | 3+3 2 | 2+2 2 | 2+2 2 | 3+3 2 | 2+2 2 | 2+2 2 | 3+3 2 | 3+3 2 |
| | 2+2 2 | 3+3 2 | 3+3 2 | 2+2 2 | 3+3 2 | 2+2 2 | 2+2 2 | 2+2 2 |
| | 3+3 2 | 2+2 2 | 2+2 2 | 3+3 2 | 3+3 2 | 2+2 2 | 3+3 2 | 2+2 2 |
| 2.2.63 | 431558 | 156 | Cl3Cu3O3Te1 | | | | |
| | -GM | -R | -T | -U | -V | -X | -Y | -Z |
| | 3+3 2 | 2+2 2 | 2+2 2 | 3+3 2 | 2+2 2 | 3+3 2 | 2+2 2 | 3+3 2 |
| | 2+2 2 | 3+3 2 | 3+3 2 | 2+2 2 | 3+3 2 | 2+2 2 | 3+3 2 | 2+2 2 |
| | 3+3 2 | 3+3 2 | 3+3 2 | 2+2 2 | 3+3 2 | 2+2 2 | 3+3 2 | 3+3 2 |
| | 2+2 2 | 3+3 2 | 2+2 2 | 3+3 2 | 3+3 2 | 3+3 2 | 2+2 2 | 2+2 2 |
| | 2+2 2 | 2+2 2 | 3+3 2 | 2+2 2 | 2+2 2 | 2+2 2 | 3+3 2 | 2+2 2 |
| | 3+3 2 | 2+2 2 | 2+2 2 | 3+3 2 | 2+2 2 | 3+3 2 | 2+2 2 | 3+3 2 |
| | 3+3 2 | 2+2 2 | 2+2 2 | 2+2 2 | 3+3 2 | 3+3 2 | 3+3 2 | 2+2 2 |
| | 2+2 2 | 3+3 2 | 3+3 2 | 3+3 2 | 2+2 2 | 2+2 2 | 2+2 2 | 3+3 2 |
| | 3+3 2 | 3+3 2 | 3+3 2 | 2+2 2 | 2+2 2 | 3+3 2 | 2+2 2 | 2+2 2 |
| | 2+2 2 | 2+2 2 | 2+2 2 | 3+3 2 | 3+3 2 | 2+2 2 | 3+3 2 | 3+3 2 |
| | 3+3 2 | 2+2 2 | 3+3 2 | 2+2 2 | 2+2 2 | 2+2 2 | 3+3 2 | 3+3 2 |
| | 2+2 2 | 3+3 2 | 2+2 2 | 3+3 2 | 3+3 2 | 3+3 2 | 2+2 2 | 2+2 2 |
| | 3+3 2 | 3+3 2 | 2+2 2 | 3+3 2 | 3+3 2 | 3+3 2 | 2+2 2 | 3+3 2 |
| | 2+2 2 | 2+2 2 | 3+3 2 | 2+2 2 | 2+2 2 | 2+2 2 | 3+3 2 | 2+2 2 |
| | 3+3 2 | 3+3 2 | 2+2 2 | 3+3 2 | 3+3 2 | 3+3 2 | 2+2 2 | 3+3 2 |
| | 2+2 2 | 2+2 2 | 3+3 2 | 2+2 2 | 2+2 2 | 2+2 2 | 3+3 2 | 2+2 2 |
| | 3+3 2 | 3+3 2 | 3+3 2 | 2+2 2 | 3+3 2 | 2+2 2 | 3+3 2 | 3+3 2 |
| | 2+2 2 | 2+2 2 | 2+2 2 | 3+3 2 | 2+2 2 | 3+3 2 | 2+2 2 | 2+2 2 |
| | 3+3 2 | 2+2 2 | 2+2 2 | 2+2 2 | 2+2 2 | 2+2 2 | 3+3 2 | 3+3 2 |
| | 2+2 2 | 3+3 2 | 3+3 2 | 3+3 2 | 3+3 2 | 3+3 2 | 2+2 2 | 2+2 2 |
| | 2+2 2 | 3+3 2 | 3+3 2 | 2+2 2 | 3+3 2 | 2+2 2 | 3+3 2 | 2+2 2 |
| | 3+3 2 | 2+2 2 | 2+2 2 | 3+3 2 | 3+3 2 | 3+3 2 | 2+2 2 | 3+3 2 |
| | 3+3 2 | 3+3 2 | 2+2 2 | 3+3 2 | 2+2 2 | 3+3 2 | 3+3 2 | 3+3 2 |
| | 2+2 2 | 2+2 2 | 3+3 2 | 2+2 2 | 2+2 2 | 2+2 2 | 2+2 2 | 2+2 2 |
| | 3+3 2 | 2+2 2 | 3+3 2 | 3+3 2 | 3+3 2 | 3+3 2 | 2+2 2 | 2+2 2 |
| | 2+2 2 | 3+3 2 | 2+2 2 | 3+3 2 | 3+3 2 | 2+2 2 | 3+3 2 | 3+3 2 |
| | 3+3 2 | 3+3 2 | 2+2 2 | 2+2 2 | 2+2 2 | 2+2 2 | 3+3 2 | 3+3 2 |
| | 3+3 2 | 2+2 2 | 3+3 2 | 2+2 2 | 3+3 2 | 3+3 2 | 2+2 2 | 3+3 2 |
| | 3+3 2 | 3+3 2 | 2+2 2 | 2+2 2 | 2+2 2 | 2+2 2 | 3+3 2 | 2+2 2 |
| | 2+2 2 | 3+3 2 | 3+3 2 | 2+2 2 | 3+3 2 | 3+3 2 | 2+2 2 | 2+2 2 |
| | 2+2 2 | 2+2 2 | 2+2 2 | 3+3 2 | 2+2 2 | 2+2 2 | 3+3 2 | 3+3 2 |
| | 3+3 2 | 2+2 2 | 3+3 2 | 3+3 2 | 2+2 2 | 3+3 2 | 2+2 2 | 2+2 2 |
| | 2+2 2 | 2+2 2 | 2+2 2 | 2+2 2 | 3+3 2 | 2+2 2 | 3+3 2 | 3+3 2 |
| | 2+2 2 | 3+3 2 | 3+3 2 | 3+3 2 | 2+2 2 | 3+3 2 | 2+2 2 | 2+2 2 |
| | 3+3 2 | 2+2 2 | 3+3 2 | 2+2 2 | 2+2 2 | 3+3 2 | 2+2 2 | 2+2 2 |
| | 3+3 2 | 3+3 2 | 2+2 2 | 3+3 2 | 3+3 2 | 2+2 2 | 2+2 2 | 3+3 2 |
| | 2+2 2 | 3+3 2 | 2+2 2 | 3+3 2 | 3+3 2 | 2+2 2 | 3+3 2 | 2+2 2 |
| | 2+2 2 | 3+3 2 | 3+3 2 | 2+2 2 | 2+2 2 | 3+3 2 | 3+3 2 | 3+3 2 |
| | 3+3 2 | 2+2 2 | 2+2 2 | 3+3 2 | 2+2 2 | 2+2 2 | 3+3 2 | 2+2 2 |
| | 2+2 2 | 2+2 2 | 3+3 2 | 2+2 2 | 3+3 2 | 3+3 2 | 2+2 2 | 3+3 2 |
| | 3+3 2 | 2+2 2 | 2+2 2 | 2+2 2 | 2+2 2 | 2+2 2 | 2+2 2 | 2+2 2 |
| | 2+2 2 | 3+3 2 | 3+3 2 | 2+2 2 | 3+3 2 | 3+3 2 | 3+3 2 | 2+2 2 |
| | 3+3 2 | 3+3 2 | 3+3 2 | 3+3 2 | 2+2 2 | 3+3 2 | 3+3 2 | 3+3 2 |
| | 2+2 2 | 2+2 2 | 3+3 2 | 3+3 2 | 3+3 2 | 2+2 2 | 2+2 2 | 3+3 2 |
| | 3+3 2 | 3+3 2 | 2+2 2 | 2+2 2 | 2+2 2 | 2+2 2 | 3+3 2 | 2+2 2 |
| | 2+2 2 | 2+2 2 | 2+2 2 | 3+3 2 | 2+2 2 | 3+3 2 | 2+2 2 | 3+3 2 |
| | 3+3 2 | 3+3 2 | 3+3 2 | 3+3 2 | 3+3 2 | 2+2 2 | 3+3 2 | 3+3 2 |
| | 2+2 2 | 2+2 2 | 3+3 2 | 2+2 2 | 2+2 2 | 3+3 2 | 2+2 2 | 2+2 2 |

Continued on next column

**TABLE XXI– continued**

| TQC | ICSD | NoE | SF | | | | |
|---|---|---|---|---|---|---|---|
| | 3+3 2 | 2+2 2 | 2+2 2 | 2+2 2 | 3+3 2 | 2+2 2 | 2+2 2 | 3+3 2 |
| | 2+2 2 | 3+3 2 | 2+2 2 | 3+3 2 | 3+3 2 | 3+3 2 | 3+3 2 | 2+2 2 |
| | 3+3 2 | 3+3 2 | 2+2 2 | 2+2 2 | 3+3 2 | 3+3 2 | 2+2 2 | 2+2 2 |
| | 2+2 2 | 2+2 2 | 3+3 2 | 3+3 2 | 2+2 2 | 2+2 2 | 3+3 2 | 3+3 2 |
| | 3+3 2 | 2+2 2 | 3+3 2 | 2+2 2 | 2+2 2 | 3+3 2 | 2+2 2 | 2+2 2 |
| | 2+2 2 | 3+3 2 | 2+2 2 | 3+3 2 | 2+2 2 | 2+2 2 | 3+3 2 | 3+3 2 |
| | 3+3 2 | 3+3 2 | 3+3 2 | 2+2 2 | 3+3 2 | 3+3 2 | 3+3 2 | 2+2 2 |
| | 2+2 2 | 2+2 2 | 2+2 2 | 3+3 2 | 3+3 2 | 2+2 2 | 2+2 2 | 3+3 2 |
| | 3+3 2 | 2+2 2 | 2+2 2 | 2+2 2 | 2+2 2 | 3+3 2 | 2+2 2 | 3+3 2 |
| | 2+2 2 | 3+3 2 | 3+3 2 | 3+3 2 | 3+3 2 | 2+2 2 | 3+3 2 | 2+2 2 |
| | 3+3 2 | 2+2 2 | 2+2 2 | 3+3 2 | 2+2 2 | 3+3 2 | 2+2 2 | 3+3 2 |
| | 2+2 2 | 3+3 2 | 3+3 2 | 2+2 2 | 3+3 2 | 3+3 2 | 3+3 2 | 2+2 2 |
| | 3+3 2 | 2+2 2 | 2+2 2 | 3+3 2 | 2+2 2 | 2+2 2 | 2+2 2 | 3+3 2 |
| | 3+3 2 | 3+3 2 | 3+3 2 | 2+2 2 | 3+3 2 | 2+2 2 | 3+3 2 | 3+3 2 |
| | 2+2 2 | 2+2 2 | 3+3 2 | 2+2 2 | 2+2 2 | 3+3 2 | 2+2 2 | 2+2 2 |
| | 3+3 2 | 3+3 2 | 2+2 2 | 3+3 2 | 3+3 2 | 2+2 2 | 3+3 2 | 2+2 2 |
| | 2+2 2 | 2+2 2 | 2+2 2 | 3+3 2 | 2+2 2 | 3+3 2 | 2+2 2 | 3+3 2 |
| | 3+3 2 | 3+3 2 | 3+3 2 | 2+2 2 | 3+3 2 | 2+2 2 | 3+3 2 | 2+2 2 |
| | 2+2 2 | 2+2 2 | 2+2 2 | 2+2 2 | 3+3 2 | 3+3 2 | 3+3 2 | 2+2 2 |
| | 2+2 2 | 2+2 2 | 3+3 2 | 2+2 2 | 2+2 2 | 3+3 2 | 2+2 2 | 3+3 2 |
| | 3+3 2 | 3+3 2 | 2+2 2 | 3+3 2 | 3+3 2 | 2+2 2 | 3+3 2 | 2+2 2 |
| | 2+2 2 | 2+2 2 | 3+3 2 | 3+3 2 | 2+2 2 | 2+2 2 | 3+3 2 | 3+3 2 |
| | 3+3 2 | 3+3 2 | 2+2 2 | 2+2 2 | 3+3 2 | 3+3 2 | 2+2 2 | 2+2 2 |
| | 2+2 2 | 3+3 2 | 3+3 2 | 3+3 2 | 2+2 2 | 2+2 2 | 3+3 2 | 3+3 2 |
| | 3+3 2 | 2+2 2 | 2+2 2 | 2+2 2 | 3+3 2 | 3+3 2 | 2+2 2 | 2+2 2 |
| | 2+2 2 | 2+2 2 | 3+3 2 | 2+2 2 | 2+2 2 | 2+2 2 | 3+3 2 | 2+2 2 |
| | 3+3 2 | 3+3 2 | 3+3 2 | 3+3 2 | 3+3 2 | 2+2 2 | 2+2 2 | 2+2 2 |
| | 2+2 2 | 2+2 2 | 2+2 2 | 3+3 2 | 2+2 2 | 3+3 2 | 3+3 2 | 3+3 2 |
| | 2+2 2 | 3+3 2 | 2+2 2 | 2+2 2 | 2+2 2 | 2+2 2 | 2+2 2 | 3+3 2 |
| | 3+3 2 | 2+2 2 | 3+3 2 | 3+3 2 | 2+2 2 | 3+3 2 | 2+2 2 | 2+2 2 |
| 2.2.64 | 92165 | 190 | Cl2Cu3O6Se2 | | | | |
| | -GM | -R | -T | -U | -V | -X | -Y | -Z |
| | 3+3 2 | 2+2 2 | 2+2 2 | 3+3 2 | 3+3 2 | 3+3 2 | 2+2 2 | 3+3 2 |
| | 2+2 2 | 3+3 2 | 3+3 2 | 2+2 2 | 2+2 2 | 2+2 2 | 3+3 2 | 2+2 2 |
| | 3+3 2 | 2+2 2 | 3+3 2 | 3+3 2 | 3+3 2 | 2+2 2 | 3+3 2 | 2+2 2 |
| | 2+2 2 | 3+3 2 | 2+2 2 | 2+2 2 | 2+2 2 | 3+3 2 | 2+2 2 | 3+3 2 |
| | 3+3 2 | 2+2 2 | 3+3 2 | 2+2 2 | 2+2 2 | 2+2 2 | 2+2 2 | 2+2 2 |
| | 2+2 2 | 3+3 2 | 2+2 2 | 3+3 2 | 3+3 2 | 3+3 2 | 3+3 2 | 3+3 2 |
| | 3+3 2 | 2+2 2 | 3+3 2 | 3+3 2 | 3+3 2 | 2+2 2 | 2+2 2 | 2+2 2 |
| | 3+3 2 | 2+2 2 | 2+2 2 | 2+2 2 | 2+2 2 | 3+3 2 | 3+3 2 | 3+3 2 |
| | 2+2 2 | 3+3 2 | 3+3 2 | 3+3 2 | 2+2 2 | 2+2 2 | 3+3 2 | 3+3 2 |
| | 2+2 2 | 3+3 2 | 2+2 2 | 2+2 2 | 3+3 2 | 3+3 2 | 2+2 2 | 2+2 2 |
| | 3+3 2 | 2+2 2 | 3+3 2 | 3+3 2 | 3+3 2 | 2+2 2 | 2+2 2 | 2+2 2 |
| | 2+2 2 | 3+3 2 | 2+2 2 | 2+2 2 | 2+2 2 | 3+3 2 | 3+3 2 | 3+3 2 |
| | 3+3 2 | 3+3 2 | 2+2 2 | 2+2 2 | 2+2 2 | 3+3 2 | 3+3 2 | 2+2 2 |
| | 2+2 2 | 3+3 2 | 3+3 2 | 3+3 2 | 2+2 2 | 2+2 2 | 2+2 2 | 3+3 2 |
| | 3+3 2 | 2+2 2 | 3+3 2 | 2+2 2 | 3+3 2 | 3+3 2 | 2+2 2 | 2+2 2 |
| | 2+2 2 | 2+2 2 | 2+2 2 | 3+3 2 | 3+3 2 | 2+2 2 | 3+3 2 | 3+3 2 |
| | 3+3 2 | 2+2 2 | 2+2 2 | 3+3 2 | 2+2 2 | 2+2 2 | 2+2 2 | 3+3 2 |
| | 2+2 2 | 3+3 2 | 3+3 2 | 2+2 2 | 3+3 2 | 3+3 2 | 3+3 2 | 2+2 2 |
| | 3+3 2 | 2+2 2 | 3+3 2 | 3+3 2 | 3+3 2 | 2+2 2 | 2+2 2 | 2+2 2 |
| | 2+2 2 | 3+3 2 | 2+2 2 | 2+2 2 | 2+2 2 | 3+3 2 | 3+3 2 | 3+3 2 |
| | 3+3 2 | 2+2 2 | 3+3 2 | 3+3 2 | 3+3 2 | 2+2 2 | 2+2 2 | 2+2 2 |
| | 2+2 2 | 3+3 2 | 2+2 2 | 2+2 2 | 2+2 2 | 3+3 2 | 3+3 2 | 3+3 2 |
| | 3+3 2 | 2+2 2 | 3+3 2 | 3+3 2 | 3+3 2 | 2+2 2 | 2+2 2 | 2+2 2 |
| | 3+3 2 | 3+3 2 | 2+2 2 | 2+2 2 | 2+2 2 | 3+3 2 | 3+3 2 | 3+3 2 |
| | 2+2 2 | 2+2 2 | 3+3 2 | 3+3 2 | 3+3 2 | 2+2 2 | 2+2 2 | 3+3 2 |
| | 3+3 2 | 2+2 2 | 2+2 2 | 3+3 2 | 2+2 2 | 3+3 2 | 3+3 2 | 2+2 2 |
| | 2+2 2 | 3+3 2 | 3+3 2 | 2+2 2 | 3+3 2 | 2+2 2 | 2+2 2 | 3+3 2 |
| | 3+3 2 | 2+2 2 | 2+2 2 | 2+2 2 | 2+2 2 | 3+3 2 | 3+3 2 | 3+3 2 |
| | 2+2 2 | 3+3 2 | 3+3 2 | 2+2 2 | 2+2 2 | 2+2 2 | 2+2 2 | 2+2 2 |

Continued on next column

**TABLE XXI– continued**

| TQC | ICSD | NoE | SF | | | | | |
|---|---|---|---|---|---|---|---|---|
| | 2+2 2 | 3+3 2 | 2+2 2 | 3+3 2 | 3+3 2 | 3+3 2 | 3+3 2 | 2+2 2 |
| | 3+3 2 | 2+2 2 | 3+3 2 | 3+3 2 | 3+3 2 | 2+2 2 | 2+2 2 | 2+2 2 |
| | 2+2 2 | 3+3 2 | 2+2 2 | 2+2 2 | 2+2 2 | 3+3 2 | 3+3 2 | 3+3 2 |
| | 2+2 2 | 3+3 2 | 3+3 2 | 2+2 2 | 3+3 2 | 3+3 2 | 3+3 2 | 2+2 2 |
| | 2+2 2 | 2+2 2 | 2+2 2 | 3+3 2 | 3+3 2 | 2+2 2 | 2+2 2 | 3+3 2 |
| | 3+3 2 | 2+2 2 | 3+3 2 | 3+3 2 | 2+2 2 | 2+2 2 | 3+3 2 | 2+2 2 |
| | 3+3 2 | 3+3 2 | 2+2 2 | 2+2 2 | 2+2 2 | 3+3 2 | 2+2 2 | 3+3 2 |
| | 3+3 2 | 3+3 2 | 2+2 2 | 3+3 2 | 3+3 2 | 2+2 2 | 2+2 2 | 3+3 2 |
| | 2+2 2 | 2+2 2 | 3+3 2 | 3+3 2 | 3+3 2 | 2+2 2 | 3+3 2 | 2+2 2 |
| | 3+3 2 | 2+2 2 | 3+3 2 | 2+2 2 | 2+2 2 | 3+3 2 | 2+2 2 | 2+2 2 |
| | 2+2 2 | 3+3 2 | 2+2 2 | 2+2 2 | 2+2 2 | 3+3 2 | 3+3 2 | 3+3 2 |
| | 3+3 2 | 3+3 2 | 3+3 2 | 3+3 2 | 2+2 2 | 3+3 2 | 3+3 2 | 3+3 2 |
| | 3+3 2 | 2+2 2 | 2+2 2 | 2+2 2 | 3+3 2 | 2+2 2 | 2+2 2 | 2+2 2 |
| | 2+2 2 | 3+3 2 | 2+2 2 | 2+2 2 | 3+3 2 | 2+2 2 | 3+3 2 | 3+3 2 |
| | 3+3 2 | 2+2 2 | 3+3 2 | 3+3 2 | 2+2 2 | 3+3 2 | 2+2 2 | 2+2 2 |
| | 2+2 2 | 2+2 2 | 3+3 2 | 3+3 2 | 2+2 2 | 3+3 2 | 2+2 2 | 3+3 2 |
| | 3+3 2 | 3+3 2 | 2+2 2 | 2+2 2 | 3+3 2 | 2+2 2 | 3+3 2 | 2+2 2 |
| | 2+2 2 | 2+2 2 | 2+2 2 | 3+3 2 | 2+2 2 | 2+2 2 | 3+3 2 | 3+3 2 |
| | 3+3 2 | 3+3 2 | 3+3 2 | 2+2 2 | 2+2 2 | 3+3 2 | 2+2 2 | 2+2 2 |
| | 2+2 2 | 2+2 2 | 2+2 2 | 2+2 2 | 3+3 2 | 2+2 2 | 2+2 2 | 2+2 2 |
| | 3+3 2 | 3+3 2 | 3+3 2 | 3+3 2 | 3+3 2 | 3+3 2 | 3+3 2 | 3+3 2 |
| | 2+2 2 | 2+2 2 | 3+3 2 | 3+3 2 | 2+2 2 | 2+2 2 | 3+3 2 | 2+2 2 |
| | 3+3 2 | 3+3 2 | 2+2 2 | 2+2 2 | 2+2 2 | 3+3 2 | 2+2 2 | 3+3 2 |
| | 2+2 2 | 2+2 2 | 2+2 2 | 2+2 2 | 3+3 2 | 3+3 2 | 2+2 2 | 3+3 2 |
| | 2+2 2 | 3+3 2 | 3+3 2 | 3+3 2 | 3+3 2 | 2+2 2 | 2+2 2 | 2+2 2 |
| | 3+3 2 | 3+3 2 | 2+2 2 | 2+2 2 | 2+2 2 | 2+2 2 | 3+3 2 | 3+3 2 |
| | 3+3 2 | 2+2 2 | 3+3 2 | 3+3 2 | 2+2 2 | 3+3 2 | 3+3 2 | 2+2 2 |
| | 2+2 2 | 2+2 2 | 3+3 2 | 2+2 2 | 3+3 2 | 3+3 2 | 3+3 2 | 2+2 2 |
| | 3+3 2 | 3+3 2 | 2+2 2 | 3+3 2 | 3+3 2 | 2+2 2 | 2+2 2 | 3+3 2 |
| | 2+2 2 | 2+2 2 | 2+2 2 | 2+2 2 | 2+2 2 | 3+3 2 | 2+2 2 | 3+3 2 |
| | 2+2 2 | 3+3 2 | 3+3 2 | 2+2 2 | 3+3 2 | 2+2 2 | 3+3 2 | 2+2 2 |
| | 3+3 2 | 3+3 2 | 3+3 2 | 3+3 2 | 2+2 2 | 3+3 2 | 2+2 2 | 3+3 2 |
| | 3+3 2 | 2+2 2 | 2+2 2 | 2+2 2 | 2+2 2 | 3+3 2 | 2+2 2 | 2+2 2 |
| | 2+2 2 | 3+3 2 | 2+2 2 | 3+3 2 | 3+3 2 | 2+2 2 | 3+3 2 | 3+3 2 |
| | 3+3 2 | 2+2 2 | 3+3 2 | 3+3 2 | 3+3 2 | 2+2 2 | 3+3 2 | 2+2 2 |
| | 2+2 2 | 2+2 2 | 3+3 2 | 2+2 2 | 2+2 2 | 2+2 2 | 2+2 2 | 3+3 2 |
| | 2+2 2 | 3+3 2 | 2+2 2 | 3+3 2 | 3+3 2 | 3+3 2 | 3+3 2 | 2+2 2 |
| | 3+3 2 | 3+3 2 | 2+2 2 | 3+3 2 | 2+2 2 | 2+2 2 | 2+2 2 | 2+2 2 |
| | 2+2 2 | 2+2 2 | 3+3 2 | 2+2 2 | 3+3 2 | 3+3 2 | 3+3 2 | 3+3 2 |
| | 3+3 2 | 3+3 2 | 2+2 2 | 2+2 2 | 2+2 2 | 3+3 2 | 2+2 2 | 3+3 2 |
| | 2+2 2 | 2+2 2 | 3+3 2 | 3+3 2 | 3+3 2 | 2+2 2 | 3+3 2 | 2+2 2 |
| | 3+3 2 | 3+3 2 | 2+2 2 | 2+2 2 | 3+3 2 | 3+3 2 | 2+2 2 | 2+2 2 |
| | 2+2 2 | 2+2 2 | 3+3 2 | 3+3 2 | 2+2 2 | 2+2 2 | 3+3 2 | 3+3 2 |
| | 3+3 2 | 2+2 2 | 3+3 2 | 3+3 2 | 3+3 2 | 3+3 2 | 2+2 2 | 3+3 2 |
| | 2+2 2 | 3+3 2 | 2+2 2 | 2+2 2 | 2+2 2 | 2+2 2 | 3+3 2 | 2+2 2 |
| | 3+3 2 | 3+3 2 | 2+2 2 | 3+3 2 | 2+2 2 | 2+2 2 | 3+3 2 | 2+2 2 |
| | 2+2 2 | 2+2 2 | 3+3 2 | 2+2 2 | 3+3 2 | 3+3 2 | 2+2 2 | 3+3 2 |
| | 3+3 2 | 3+3 2 | 3+3 2 | 2+2 2 | 2+2 2 | 3+3 2 | 2+2 2 | 3+3 2 |
| | 3+3 2 | 2+2 2 | 2+2 2 | 3+3 2 | 3+3 2 | 2+2 2 | 3+3 2 | 2+2 2 |
| | 2+2 2 | 3+3 2 | 2+2 2 | 2+2 2 | 2+2 2 | 3+3 2 | 2+2 2 | 2+2 2 |
| | 2+2 2 | 2+2 2 | 3+3 2 | 3+3 2 | 3+3 2 | 2+2 2 | 3+3 2 | 3+3 2 |
| | 3+3 2 | 3+3 2 | 2+2 2 | 2+2 2 | 2+2 2 | 3+3 2 | 2+2 2 | 3+3 2 |
| | 2+2 2 | 2+2 2 | 3+3 2 | 3+3 2 | 3+3 2 | 2+2 2 | 2+2 2 | 2+2 2 |
| | 3+3 2 | 2+2 2 | 2+2 2 | 2+2 2 | 2+2 2 | 3+3 2 | 3+3 2 | 2+2 2 |
| | 2+2 2 | 3+3 2 | 2+2 2 | 3+3 2 | 3+3 2 | 2+2 2 | 3+3 2 | 3+3 2 |
| | 3+3 2 | 2+2 2 | 3+3 2 | 3+3 2 | 2+2 2 | 3+3 2 | 2+2 2 | 2+2 2 |
| | 2+2 2 | 3+3 2 | 3+3 2 | 2+2 2 | 3+3 2 | 2+2 2 | 3+3 2 | 3+3 2 |
| | 3+3 2 | 2+2 2 | 2+2 2 | 2+2 2 | 2+2 2 | 2+2 2 | 2+2 2 | 2+2 2 |
| | 2+2 2 | 3+3 2 | 3+3 2 | 3+3 2 | 3+3 2 | 3+3 2 | 3+3 2 | 3+3 2 |
| | 2+2 2 | 3+3 2 | 2+2 2 | 2+2 2 | 3+3 2 | 3+3 2 | 3+3 2 | 2+2 2 |
| | 3+3 2 | 2+2 2 | 3+3 2 | 3+3 2 | 2+2 2 | 2+2 2 | 2+2 2 | 2+2 2 |

Continued on next column

**TABLE XXI– continued**

| TQC | ICSD | NoE | SF | | | | | |
|---|---|---|---|---|---|---|---|---|
| | 2+2 2 | 3+3 2 | 2+2 2 | 2+2 2 | 3+3 2 | 2+2 2 | 3+3 2 | 3+3 2 |
| | 3+3 2 | 2+2 2 | 3+3 2 | 3+3 2 | 2+2 2 | 3+3 2 | 2+2 2 | 3+3 2 |
| | 2+2 2 | 3+3 2 | 3+3 2 | 2+2 2 | 3+3 2 | 2+2 2 | 3+3 2 | 2+2 2 |
| | 3+3 2 | 3+3 2 | 2+2 2 | 2+2 2 | 2+2 2 | 3+3 2 | 2+2 2 | 3+3 2 |
| | 2+2 2 | 2+2 2 | 2+2 2 | 3+3 2 | 2+2 2 | 2+2 2 | 3+3 2 | 3+3 2 |
| 2.2.65 | 427665 | 120 | Cl1O4Tm1W1 | | | | | |
| | -GM | -R | -T | -U | -V | -X | -Y | -Z |
| | 3+3 2 | 2+2 2 | 3+3 2 | 2+2 2 | 2+2 2 | 2+2 2 | 3+3 2 | 3+3 2 |
| | 2+2 2 | 3+3 2 | 2+2 2 | 3+3 2 | 3+3 2 | 3+3 2 | 2+2 2 | 2+2 2 |
| | 3+3 2 | 2+2 2 | 3+3 2 | 2+2 2 | 2+2 2 | 2+2 2 | 3+3 2 | 3+3 2 |
| | 2+2 2 | 3+3 2 | 2+2 2 | 3+3 2 | 3+3 2 | 3+3 2 | 2+2 2 | 2+2 2 |
| | 3+3 2 | 2+2 2 | 3+3 2 | 2+2 2 | 2+2 2 | 2+2 2 | 3+3 2 | 3+3 2 |
| | 2+2 2 | 3+3 2 | 2+2 2 | 3+3 2 | 3+3 2 | 3+3 2 | 2+2 2 | 2+2 2 |
| | 3+3 2 | 3+3 2 | 3+3 2 | 3+3 2 | 3+3 2 | 3+3 2 | 3+3 2 | 3+3 2 |
| | 2+2 2 | 2+2 2 | 2+2 2 | 2+2 2 | 2+2 2 | 2+2 2 | 2+2 2 | 2+2 2 |
| | 3+3 2 | 2+2 2 | 2+2 2 | 2+2 2 | 2+2 2 | 2+2 2 | 2+2 2 | 3+3 2 |
| | 2+2 2 | 3+3 2 | 3+3 2 | 3+3 2 | 3+3 2 | 3+3 2 | 3+3 2 | 2+2 2 |
| | 2+2 2 | 2+2 2 | 3+3 2 | 2+2 2 | 2+2 2 | 2+2 2 | 3+3 2 | 3+3 2 |
| | 3+3 2 | 3+3 2 | 2+2 2 | 3+3 2 | 3+3 2 | 3+3 2 | 2+2 2 | 2+2 2 |
| | 3+3 2 | 2+2 2 | 2+2 2 | 3+3 2 | 3+3 2 | 3+3 2 | 3+3 2 | 2+2 2 |
| | 2+2 2 | 3+3 2 | 3+3 2 | 2+2 2 | 2+2 2 | 2+2 2 | 2+2 2 | 3+3 2 |
| | 2+2 2 | 3+3 2 | 2+2 2 | 3+3 2 | 2+2 2 | 3+3 2 | 2+2 2 | 2+2 2 |
| | 3+3 2 | 2+2 2 | 3+3 2 | 2+2 2 | 3+3 2 | 2+2 2 | 3+3 2 | 3+3 2 |
| | 3+3 2 | 3+3 2 | 3+3 2 | 3+3 2 | 3+3 2 | 3+3 2 | 3+3 2 | 3+3 2 |
| | 2+2 2 | 2+2 2 | 2+2 2 | 2+2 2 | 2+2 2 | 2+2 2 | 2+2 2 | 2+2 2 |
| | 3+3 2 | 3+3 2 | 2+2 2 | 2+2 2 | 3+3 2 | 2+2 2 | 3+3 2 | 2+2 2 |
| | 2+2 2 | 2+2 2 | 3+3 2 | 3+3 2 | 2+2 2 | 2+2 2 | 2+2 2 | 3+3 2 |
| | 3+3 2 | 2+2 2 | 3+3 2 | 2+2 2 | 3+3 2 | 3+3 2 | 2+2 2 | 3+3 2 |
| | 3+3 2 | 3+3 2 | 2+2 2 | 3+3 2 | 2+2 2 | 2+2 2 | 3+3 2 | 2+2 2 |
| | 2+2 2 | 2+2 2 | 3+3 2 | 3+3 2 | 2+2 2 | 3+3 2 | 2+2 2 | 2+2 2 |
| | 3+3 2 | 3+3 2 | 3+3 2 | 2+2 2 | 2+2 2 | 3+3 2 | 3+3 2 | 3+3 2 |
| | 2+2 2 | 2+2 2 | 2+2 2 | 3+3 2 | 2+2 2 | 2+2 2 | 3+3 2 | 2+2 2 |
| | 3+3 2 | 3+3 2 | 2+2 2 | 2+2 2 | 3+3 2 | 3+3 2 | 2+2 2 | 2+2 2 |
| | 2+2 2 | 2+2 2 | 3+3 2 | 3+3 2 | 3+3 2 | 3+3 2 | 3+3 2 | 3+3 2 |
| | 2+2 2 | 3+3 2 | 3+3 2 | 2+2 2 | 2+2 2 | 2+2 2 | 3+3 2 | 3+3 2 |
| | 3+3 2 | 2+2 2 | 2+2 2 | 2+2 2 | 3+3 2 | 2+2 2 | 2+2 2 | 2+2 2 |
| | 2+2 2 | 3+3 2 | 2+2 2 | 3+3 2 | 3+3 2 | 3+3 2 | 2+2 2 | 3+3 2 |
| | 3+3 2 | 3+3 2 | 3+3 2 | 2+2 2 | 3+3 2 | 2+2 2 | 2+2 2 | 3+3 2 |
| | 3+3 2 | 2+2 2 | 2+2 2 | 3+3 2 | 2+2 2 | 3+3 2 | 3+3 2 | 3+3 2 |
| | 2+2 2 | 3+3 2 | 2+2 2 | 2+2 2 | 3+3 2 | 3+3 2 | 2+2 2 | 2+2 2 |
| | 3+3 2 | 2+2 2 | 2+2 2 | 3+3 2 | 2+2 2 | 2+2 2 | 3+3 2 | 3+3 2 |
| | 2+2 2 | 3+3 2 | 3+3 2 | 2+2 2 | 3+3 2 | 3+3 2 | 3+3 2 | 2+2 2 |
| | 2+2 2 | 2+2 2 | 3+3 2 | 3+3 2 | 2+2 2 | 2+2 2 | 2+2 2 | 2+2 2 |
| | 3+3 2 | 2+2 2 | 2+2 2 | 2+2 2 | 3+3 2 | 3+3 2 | 3+3 2 | 3+3 2 |
| | 2+2 2 | 3+3 2 | 2+2 2 | 3+3 2 | 2+2 2 | 2+2 2 | 3+3 2 | 3+3 2 |
| | 3+3 2 | 3+3 2 | 3+3 2 | 2+2 2 | 3+3 2 | 3+3 2 | 2+2 2 | 2+2 2 |
| | 2+2 2 | 2+2 2 | 3+3 2 | 3+3 2 | 3+3 2 | 2+2 2 | 2+2 2 | 3+3 2 |
| | 3+3 2 | 3+3 2 | 2+2 2 | 2+2 2 | 2+2 2 | 3+3 2 | 3+3 2 | 2+2 2 |
| | 2+2 2 | 3+3 2 | 2+2 2 | 3+3 2 | 3+3 2 | 2+2 2 | 2+2 2 | 3+3 2 |
| | 3+3 2 | 2+2 2 | 3+3 2 | 3+3 2 | 2+2 2 | 3+3 2 | 3+3 2 | 2+2 2 |
| | 2+2 2 | 2+2 2 | 3+3 2 | 2+2 2 | 3+3 2 | 3+3 2 | 2+2 2 | 3+3 2 |
| | 3+3 2 | 3+3 2 | 2+2 2 | 2+2 2 | 2+2 2 | 2+2 2 | 3+3 2 | 2+2 2 |
| | 2+2 2 | 2+2 2 | 3+3 2 | 3+3 2 | 2+2 2 | 2+2 2 | 2+2 2 | 3+3 2 |
| | 3+3 2 | 2+2 2 | 2+2 2 | 3+3 2 | 3+3 2 | 3+3 2 | 3+3 2 | 2+2 2 |
| | 2+2 2 | 3+3 2 | 3+3 2 | 2+2 2 | 2+2 2 | 2+2 2 | 2+2 2 | 2+2 2 |
| | 3+3 2 | 3+3 2 | 2+2 2 | 2+2 2 | 3+3 2 | 3+3 2 | 3+3 2 | 3+3 2 |
| | 2+2 2 | 2+2 2 | 3+3 2 | 3+3 2 | 2+2 2 | 2+2 2 | 2+2 2 | 3+3 2 |
| | 3+3 2 | 3+3 2 | 2+2 2 | 2+2 2 | 2+2 2 | 3+3 2 | 3+3 2 | 2+2 2 |
| | 2+2 2 | 2+2 2 | 3+3 2 | 3+3 2 | 3+3 2 | 2+2 2 | 2+2 2 | 2+2 2 |
| | 3+3 2 | 2+2 2 | 3+3 2 | 3+3 2 | 3+3 2 | 3+3 2 | 3+3 2 | 3+3 2 |
| | 2+2 2 | 3+3 2 | 2+2 2 | 2+2 2 | 2+2 2 | 2+2 2 | 2+2 2 | 2+2 2 |

Continued on next column

**TABLE XXI– continued**

| TQC | ICSD | NoE | SF | | | | | |
|---|---|---|---|---|---|---|---|---|
| | 3+3 2 | 2+2 2 | 3+3 2 | 3+3 2 | 2+2 2 | 3+3 2 | 2+2 2 | 3+3 2 |
| | 3+3 2 | 3+3 2 | 2+2 2 | 2+2 2 | 3+3 2 | 2+2 2 | 3+3 2 | 3+3 2 |
| | 2+2 2 | 2+2 2 | 2+2 2 | 3+3 2 | 2+2 2 | 3+3 2 | 3+3 2 | 2+2 2 |
| | 2+2 2 | 3+3 2 | 3+3 2 | 2+2 2 | 3+3 2 | 2+2 2 | 3+3 2 | 3+3 2 |
| | 3+3 2 | 2+2 2 | 3+3 2 | 2+2 2 | 3+3 2 | 3+3 2 | 2+2 2 | 2+2 2 |
| | 3+3 2 | 3+3 2 | 2+2 2 | 3+3 2 | 3+3 2 | 2+2 2 | 2+2 2 | 3+3 2 |
| 2.2.66 | 50441 | 148 | Cl18Mo2P2 | | | | | |
| | -GM | -R | -T | -U | -V | -X | -Y | -Z |
| | 3+3 2 | 3+3 2 | 2+2 2 | 2+2 2 | 2+2 2 | 2+2 2 | 3+3 2 | 3+3 2 |
| | 2+2 2 | 2+2 2 | 3+3 2 | 3+3 2 | 3+3 2 | 3+3 2 | 2+2 2 | 2+2 2 |
| | 3+3 2 | 3+3 2 | 2+2 2 | 2+2 2 | 2+2 2 | 2+2 2 | 2+2 2 | 3+3 2 |
| | 3+3 2 | 2+2 2 | 3+3 2 | 3+3 2 | 3+3 2 | 2+2 2 | 3+3 2 | 2+2 2 |
| | 2+2 2 | 3+3 2 | 2+2 2 | 3+3 2 | 3+3 2 | 3+3 2 | 3+3 2 | 2+2 2 |
| | 2+2 2 | 2+2 2 | 3+3 2 | 2+2 2 | 2+2 2 | 3+3 2 | 2+2 2 | 3+3 2 |
| | 2+2 2 | 2+2 2 | 3+3 2 | 2+2 2 | 3+3 2 | 3+3 2 | 2+2 2 | 3+3 2 |
| | 3+3 2 | 3+3 2 | 2+2 2 | 3+3 2 | 2+2 2 | 2+2 2 | 3+3 2 | 2+2 2 |
| | 3+3 2 | 3+3 2 | 3+3 2 | 2+2 2 | 2+2 2 | 3+3 2 | 2+2 2 | 2+2 2 |
| | 2+2 2 | 2+2 2 | 2+2 2 | 3+3 2 | 3+3 2 | 2+2 2 | 3+3 2 | 3+3 2 |
| | 2+2 2 | 2+2 2 | 2+2 2 | 3+3 2 | 3+3 2 | 2+2 2 | 3+3 2 | 3+3 2 |
| | 3+3 2 | 3+3 2 | 3+3 2 | 2+2 2 | 2+2 2 | 3+3 2 | 2+2 2 | 2+2 2 |
| | 3+3 2 | 2+2 2 | 3+3 2 | 2+2 2 | 2+2 2 | 3+3 2 | 2+2 2 | 2+2 2 |
| | 2+2 2 | 3+3 2 | 2+2 2 | 3+3 2 | 3+3 2 | 2+2 2 | 3+3 2 | 3+3 2 |
| | 2+2 2 | 2+2 2 | 2+2 2 | 3+3 2 | 3+3 2 | 2+2 2 | 3+3 2 | 3+3 2 |
| | 3+3 2 | 3+3 2 | 3+3 2 | 2+2 2 | 3+3 2 | 3+3 2 | 3+3 2 | 2+2 2 |
| | 3+3 2 | 2+2 2 | 2+2 2 | 2+2 2 | 2+2 2 | 3+3 2 | 2+2 2 | 3+3 2 |
| | 2+2 2 | 3+3 2 | 3+3 2 | 3+3 2 | 2+2 2 | 2+2 2 | 2+2 2 | 2+2 2 |
| | 3+3 2 | 3+3 2 | 2+2 2 | 2+2 2 | 2+2 2 | 2+2 2 | 3+3 2 | 3+3 2 |
| | 2+2 2 | 2+2 2 | 3+3 2 | 3+3 2 | 3+3 2 | 3+3 2 | 2+2 2 | 2+2 2 |
| | 3+3 2 | 3+3 2 | 2+2 2 | 2+2 2 | 3+3 2 | 3+3 2 | 2+2 2 | 3+3 2 |
| | 2+2 2 | 2+2 2 | 3+3 2 | 3+3 2 | 3+3 2 | 3+3 2 | 2+2 2 | 3+3 2 |
| | 2+2 2 | 3+3 2 | 2+2 2 | 2+2 2 | 2+2 2 | 2+2 2 | 3+3 2 | 2+2 2 |
| | 2+2 2 | 2+2 2 | 3+3 2 | 2+2 2 | 2+2 2 | 2+2 2 | 2+2 2 | 2+2 2 |
| | 3+3 2 | 2+2 2 | 2+2 2 | 3+3 2 | 3+3 2 | 3+3 2 | 3+3 2 | 3+3 2 |
| | 3+3 2 | 3+3 2 | 3+3 2 | 3+3 2 | 2+2 2 | 2+2 2 | 3+3 2 | 2+2 2 |
| | 3+3 2 | 3+3 2 | 3+3 2 | 3+3 2 | 2+2 2 | 2+2 2 | 3+3 2 | 3+3 2 |
| | 3+3 2 | 2+2 2 | 2+2 2 | 2+2 2 | 3+3 2 | 2+2 2 | 2+2 2 | 2+2 2 |
| | 2+2 2 | 3+3 2 | 2+2 2 | 2+2 2 | 2+2 2 | 3+3 2 | 3+3 2 | 3+3 2 |
| | 3+3 2 | 2+2 2 | 2+2 2 | 3+3 2 | 2+2 2 | 2+2 2 | 2+2 2 | 2+2 2 |
| | 3+3 2 | 3+3 2 | 3+3 2 | 2+2 2 | 3+3 2 | 3+3 2 | 2+2 2 | 3+3 2 |
| | 2+2 2 | 2+2 2 | 3+3 2 | 3+3 2 | 2+2 2 | 2+2 2 | 3+3 2 | 2+2 2 |
| | 2+2 2 | 3+3 2 | 3+3 2 | 2+2 2 | 3+3 2 | 3+3 2 | 2+2 2 | 2+2 2 |
| | 3+3 2 | 2+2 2 | 2+2 2 | 3+3 2 | 2+2 2 | 2+2 2 | 3+3 2 | 3+3 2 |
| | 2+2 2 | 3+3 2 | 2+2 2 | 3+3 2 | 3+3 2 | 3+3 2 | 3+3 2 | 2+2 2 |
| | 2+2 2 | 2+2 2 | 3+3 2 | 2+2 2 | 3+3 2 | 2+2 2 | 2+2 2 | 3+3 2 |
| | 3+3 2 | 2+2 2 | 2+2 2 | 3+3 2 | 3+3 2 | 3+3 2 | 3+3 2 | 2+2 2 |
| | 2+2 2 | 3+3 2 | 3+3 2 | 2+2 2 | 2+2 2 | 3+3 2 | 2+2 2 | 3+3 2 |
| | 3+3 2 | 3+3 2 | 2+2 2 | 2+2 2 | 3+3 2 | 3+3 2 | 3+3 2 | 2+2 2 |
| | 3+3 2 | 2+2 2 | 3+3 2 | 2+2 2 | 2+2 2 | 3+3 2 | 2+2 2 | 2+2 2 |
| | 2+2 2 | 3+3 2 | 3+3 2 | 3+3 2 | 3+3 2 | 2+2 2 | 2+2 2 | 3+3 2 |
| | 2+2 2 | 2+2 2 | 2+2 2 | 3+3 2 | 2+2 2 | 2+2 2 | 3+3 2 | 3+3 2 |
| | 2+2 2 | 2+2 2 | 2+2 2 | 3+3 2 | 3+3 2 | 3+3 2 | 3+3 2 | 3+3 2 |
| | 3+3 2 | 3+3 2 | 3+3 2 | 2+2 2 | 2+2 2 | 2+2 2 | 2+2 2 | 2+2 2 |
| | 3+3 2 | 2+2 2 | 2+2 2 | 3+3 2 | 3+3 2 | 3+3 2 | 3+3 2 | 3+3 2 |
| | 2+2 2 | 3+3 2 | 3+3 2 | 2+2 2 | 2+2 2 | 2+2 2 | 2+2 2 | 2+2 2 |
| | 3+3 2 | 3+3 2 | 2+2 2 | 2+2 2 | 3+3 2 | 3+3 2 | 2+2 2 | 3+3 2 |
| | 2+2 2 | 2+2 2 | 3+3 2 | 3+3 2 | 2+2 2 | 2+2 2 | 3+3 2 | 2+2 2 |
| | 2+2 2 | 3+3 2 | 2+2 2 | 2+2 2 | 2+2 2 | 2+2 2 | 3+3 2 | 3+3 2 |
| | 3+3 2 | 2+2 2 | 3+3 2 | 3+3 2 | 3+3 2 | 3+3 2 | 2+2 2 | 2+2 2 |
| | 2+2 2 | 3+3 2 | 3+3 2 | 2+2 2 | 2+2 2 | 3+3 2 | 2+2 2 | 2+2 2 |
| | 3+3 2 | 3+3 2 | 2+2 2 | 2+2 2 | 3+3 2 | 2+2 2 | 3+3 2 | 3+3 2 |
| | 3+3 2 | 2+2 2 | 2+2 2 | 3+3 2 | 2+2 2 | 3+3 2 | 2+2 2 | 2+2 2 |

Continued on next column

**TABLE XXI– continued**

| TQC | ICSD | NoE | SF | | | | | |
|---|---|---|---|---|---|---|---|---|
| | 2+2 2 | 3+3 2 | 3+3 2 | 2+2 2 | 3+3 2 | 2+2 2 | 3+3 2 | 3+3 2 |
| | 3+3 2 | 2+2 2 | 3+3 2 | 3+3 2 | 3+3 2 | 3+3 2 | 2+2 2 | 2+2 2 |
| | 2+2 2 | 2+2 2 | 2+2 2 | 3+3 2 | 2+2 2 | 2+2 2 | 3+3 2 | 3+3 2 |
| | 3+3 2 | 3+3 2 | 3+3 2 | 2+2 2 | 2+2 2 | 3+3 2 | 2+2 2 | 2+2 2 |
| | 3+3 2 | 2+2 2 | 2+2 2 | 2+2 2 | 3+3 2 | 3+3 2 | 3+3 2 | 2+2 2 |
| | 2+2 2 | 3+3 2 | 3+3 2 | 3+3 2 | 2+2 2 | 2+2 2 | 2+2 2 | 3+3 2 |
| | 2+2 2 | 2+2 2 | 2+2 2 | 3+3 2 | 3+3 2 | 2+2 2 | 3+3 2 | 3+3 2 |
| | 3+3 2 | 3+3 2 | 3+3 2 | 3+3 2 | 2+2 2 | 3+3 2 | 2+2 2 | 2+2 2 |
| | 3+3 2 | 2+2 2 | 3+3 2 | 2+2 2 | 3+3 2 | 2+2 2 | 2+2 2 | 2+2 2 |
| | 3+3 2 | 3+3 2 | 2+2 2 | 2+2 2 | 2+2 2 | 3+3 2 | 3+3 2 | 3+3 2 |
| | 2+2 2 | 2+2 2 | 3+3 2 | 2+2 2 | 2+2 2 | 3+3 2 | 2+2 2 | 2+2 2 |
| | 2+2 2 | 3+3 2 | 2+2 2 | 3+3 2 | 3+3 2 | 2+2 2 | 3+3 2 | 3+3 2 |
| | 2+2 2 | 2+2 2 | 2+2 2 | 3+3 2 | 3+3 2 | 2+2 2 | 3+3 2 | 3+3 2 |
| | 3+3 2 | 3+3 2 | 3+3 2 | 2+2 2 | 2+2 2 | 3+3 2 | 3+3 2 | 3+3 2 |
| | 2+2 2 | 2+2 2 | 2+2 2 | 3+3 2 | 3+3 2 | 2+2 2 | 2+2 2 | 2+2 2 |
| | 2+2 2 | 2+2 2 | 2+2 2 | 3+3 2 | 3+3 2 | 2+2 2 | 3+3 2 | 3+3 2 |
| | 2+2 2 | 3+3 2 | 2+2 2 | 2+2 2 | 3+3 2 | 3+3 2 | 2+2 2 | 3+3 2 |
| | 3+3 2 | 2+2 2 | 3+3 2 | 3+3 2 | 2+2 2 | 2+2 2 | 3+3 2 | 2+2 2 |
| | 3+3 2 | 3+3 2 | 3+3 2 | 2+2 2 | 2+2 2 | 3+3 2 | 2+2 2 | 2+2 2 |
| | 3+3 2 | 2+2 2 | 3+3 2 | 3+3 2 | 3+3 2 | 2+2 2 | 2+2 2 | 2+2 2 |
| | 2+2 2 | 2+2 2 | 2+2 2 | 3+3 2 | 3+3 2 | 3+3 2 | 3+3 2 | 3+3 2 |
| 2.2.67 | 410649 | 148 | Cl18Mo2P2 | | | | | |
| | -GM | -R | -T | -U | -V | -X | -Y | -Z |
| | 3+3 2 | 2+2 2 | 2+2 2 | 2+2 2 | 3+3 2 | 3+3 2 | 3+3 2 | 2+2 2 |
| | 2+2 2 | 3+3 2 | 3+3 2 | 3+3 2 | 2+2 2 | 2+2 2 | 2+2 2 | 3+3 2 |
| | 3+3 2 | 2+2 2 | 2+2 2 | 3+3 2 | 2+2 2 | 3+3 2 | 2+2 2 | 3+3 2 |
| | 3+3 2 | 3+3 2 | 3+3 2 | 2+2 2 | 3+3 2 | 3+3 2 | 3+3 2 | 2+2 2 |
| | 2+2 2 | 3+3 2 | 2+2 2 | 3+3 2 | 3+3 2 | 2+2 2 | 3+3 2 | 3+3 2 |
| | 2+2 2 | 2+2 2 | 3+3 2 | 2+2 2 | 2+2 2 | 2+2 2 | 2+2 2 | 2+2 2 |
| | 2+2 2 | 2+2 2 | 2+2 2 | 3+3 2 | 2+2 2 | 2+2 2 | 2+2 2 | 2+2 2 |
| | 3+3 2 | 3+3 2 | 3+3 2 | 2+2 2 | 3+3 2 | 3+3 2 | 3+3 2 | 3+3 2 |
| | 3+3 2 | 2+2 2 | 3+3 2 | 3+3 2 | 3+3 2 | 2+2 2 | 2+2 2 | 2+2 2 |
| | 2+2 2 | 3+3 2 | 2+2 2 | 2+2 2 | 2+2 2 | 3+3 2 | 3+3 2 | 3+3 2 |
| | 2+2 2 | 3+3 2 | 2+2 2 | 2+2 2 | 2+2 2 | 3+3 2 | 3+3 2 | 3+3 2 |
| | 3+3 2 | 2+2 2 | 3+3 2 | 3+3 2 | 3+3 2 | 2+2 2 | 2+2 2 | 2+2 2 |
| | 3+3 2 | 2+2 2 | 3+3 2 | 3+3 2 | 3+3 2 | 2+2 2 | 2+2 2 | 2+2 2 |
| | 2+2 2 | 3+3 2 | 2+2 2 | 2+2 2 | 2+2 2 | 3+3 2 | 3+3 2 | 3+3 2 |
| | 2+2 2 | 3+3 2 | 2+2 2 | 2+2 2 | 2+2 2 | 3+3 2 | 3+3 2 | 3+3 2 |
| | 3+3 2 | 2+2 2 | 3+3 2 | 3+3 2 | 2+2 2 | 2+2 2 | 3+3 2 | 2+2 2 |
| | 2+2 2 | 3+3 2 | 2+2 2 | 2+2 2 | 3+3 2 | 3+3 2 | 2+2 2 | 3+3 2 |
| | 3+3 2 | 2+2 2 | 3+3 2 | 3+3 2 | 3+3 2 | 2+2 2 | 2+2 2 | 2+2 2 |
| | 3+3 2 | 2+2 2 | 2+2 2 | 3+3 2 | 3+3 2 | 3+3 2 | 3+3 2 | 3+3 2 |
| | 2+2 2 | 3+3 2 | 3+3 2 | 2+2 2 | 2+2 2 | 2+2 2 | 2+2 2 | 2+2 2 |
| | 2+2 2 | 2+2 2 | 3+3 2 | 3+3 2 | 2+2 2 | 2+2 2 | 2+2 2 | 3+3 2 |
| | 3+3 2 | 3+3 2 | 2+2 2 | 2+2 2 | 2+2 2 | 2+2 2 | 2+2 2 | 3+3 2 |
| | 2+2 2 | 2+2 2 | 2+2 2 | 3+3 2 | 3+3 2 | 3+3 2 | 3+3 2 | 2+2 2 |
| | 2+2 2 | 3+3 2 | 3+3 2 | 2+2 2 | 2+2 2 | 3+3 2 | 2+2 2 | 2+2 2 |
| | 3+3 2 | 3+3 2 | 2+2 2 | 3+3 2 | 3+3 2 | 2+2 2 | 3+3 2 | 2+2 2 |
| | 3+3 2 | 2+2 2 | 3+3 2 | 2+2 2 | 3+3 2 | 3+3 2 | 3+3 2 | 3+3 2 |
| | 3+3 2 | 2+2 2 | 3+3 2 | 2+2 2 | 3+3 2 | 3+3 2 | 3+3 2 | 3+3 2 |
| | 3+3 2 | 3+3 2 | 2+2 2 | 3+3 2 | 2+2 2 | 3+3 2 | 2+2 2 | 2+2 2 |
| | 2+2 2 | 2+2 2 | 2+2 2 | 3+3 2 | 3+3 2 | 2+2 2 | 3+3 2 | 3+3 2 |
| | 3+3 2 | 3+3 2 | 2+2 2 | 2+2 2 | 2+2 2 | 2+2 2 | 2+2 2 | 2+2 2 |
| | 3+3 2 | 2+2 2 | 3+3 2 | 2+2 2 | 3+3 2 | 3+3 2 | 2+2 2 | 3+3 2 |
| | 2+2 2 | 3+3 2 | 3+3 2 | 3+3 2 | 3+3 2 | 3+3 2 | 3+3 2 | 2+2 2 |
| | 2+2 2 | 2+2 2 | 3+3 2 | 3+3 2 | 2+2 2 | 2+2 2 | 2+2 2 | 3+3 2 |
| | 3+3 2 | 3+3 2 | 2+2 2 | 2+2 2 | 3+3 2 | 3+3 2 | 3+3 2 | 2+2 2 |
| | 2+2 2 | 2+2 2 | 2+2 2 | 2+2 2 | 2+2 2 | 2+2 2 | 3+3 2 | 2+2 2 |
| | 2+2 2 | 3+3 2 | 3+3 2 | 3+3 2 | 2+2 2 | 3+3 2 | 2+2 2 | 3+3 2 |
| | 3+3 2 | 3+3 2 | 3+3 2 | 2+2 2 | 2+2 2 | 2+2 2 | 3+3 2 | 2+2 2 |
| | 2+2 2 | 2+2 2 | 2+2 2 | 3+3 2 | 3+3 2 | 2+2 2 | 2+2 2 | 3+3 2 |

Continued on next column

**TABLE XXI– continued**

| TQC | ICSD | NoE | SF | | | | | |
|---|---|---|---|---|---|---|---|---|
| | 3+3 2 | 2+2 2 | 3+3 2 | 3+3 2 | 3+3 2 | 2+2 2 | 2+2 2 | 2+2 2 |
| | 3+3 2 | 3+3 2 | 2+2 2 | 3+3 2 | 2+2 2 | 2+2 2 | 3+3 2 | 2+2 2 |
| | 2+2 2 | 2+2 2 | 3+3 2 | 2+2 2 | 2+2 2 | 3+3 2 | 3+3 2 | 3+3 2 |
| | 2+2 2 | 3+3 2 | 2+2 2 | 2+2 2 | 3+3 2 | 3+3 2 | 2+2 2 | 3+3 2 |
| | 2+2 2 | 3+3 2 | 2+2 2 | 2+2 2 | 2+2 2 | 3+3 2 | 3+3 2 | 3+3 2 |
| | 3+3 2 | 2+2 2 | 3+3 2 | 3+3 2 | 3+3 2 | 2+2 2 | 2+2 2 | 2+2 2 |
| | 3+3 2 | 2+2 2 | 2+2 2 | 2+2 2 | 2+2 2 | 2+2 2 | 2+2 2 | 3+3 2 |
| | 2+2 2 | 3+3 2 | 3+3 2 | 3+3 2 | 3+3 2 | 3+3 2 | 3+3 2 | 2+2 2 |
| | 3+3 2 | 2+2 2 | 2+2 2 | 3+3 2 | 2+2 2 | 2+2 2 | 2+2 2 | 3+3 2 |
| | 2+2 2 | 3+3 2 | 3+3 2 | 2+2 2 | 3+3 2 | 3+3 2 | 3+3 2 | 2+2 2 |
| | 2+2 2 | 2+2 2 | 2+2 2 | 3+3 2 | 3+3 2 | 2+2 2 | 3+3 2 | 3+3 2 |
| | 3+3 2 | 3+3 2 | 3+3 2 | 2+2 2 | 2+2 2 | 3+3 2 | 2+2 2 | 2+2 2 |
| | 3+3 2 | 2+2 2 | 3+3 2 | 3+3 2 | 3+3 2 | 2+2 2 | 2+2 2 | 2+2 2 |
| | 2+2 2 | 2+2 2 | 2+2 2 | 3+3 2 | 2+2 2 | 3+3 2 | 3+3 2 | 3+3 2 |
| | 3+3 2 | 3+3 2 | 2+2 2 | 2+2 2 | 3+3 2 | 2+2 2 | 2+2 2 | 2+2 2 |
| | 2+2 2 | 2+2 2 | 3+3 2 | 3+3 2 | 2+2 2 | 3+3 2 | 3+3 2 | 3+3 2 |
| | 3+3 2 | 3+3 2 | 3+3 2 | 2+2 2 | 2+2 2 | 2+2 2 | 2+2 2 | 2+2 2 |
| | 2+2 2 | 3+3 2 | 2+2 2 | 2+2 2 | 3+3 2 | 3+3 2 | 3+3 2 | 3+3 2 |
| | 3+3 2 | 2+2 2 | 3+3 2 | 3+3 2 | 3+3 2 | 2+2 2 | 2+2 2 | 2+2 2 |
| | 3+3 2 | 3+3 2 | 2+2 2 | 3+3 2 | 2+2 2 | 2+2 2 | 3+3 2 | 2+2 2 |
| | 2+2 2 | 2+2 2 | 3+3 2 | 2+2 2 | 3+3 2 | 3+3 2 | 2+2 2 | 3+3 2 |
| | 2+2 2 | 3+3 2 | 2+2 2 | 2+2 2 | 2+2 2 | 3+3 2 | 3+3 2 | 3+3 2 |
| | 3+3 2 | 2+2 2 | 3+3 2 | 2+2 2 | 3+3 2 | 2+2 2 | 2+2 2 | 2+2 2 |
| | 3+3 2 | 3+3 2 | 3+3 2 | 3+3 2 | 2+2 2 | 3+3 2 | 2+2 2 | 2+2 2 |
| | 3+3 2 | 2+2 2 | 2+2 2 | 3+3 2 | 3+3 2 | 2+2 2 | 3+3 2 | 3+3 2 |
| | 2+2 2 | 3+3 2 | 3+3 2 | 2+2 2 | 3+3 2 | 2+2 2 | 2+2 2 | 2+2 2 |
| | 2+2 2 | 2+2 2 | 2+2 2 | 3+3 2 | 2+2 2 | 3+3 2 | 3+3 2 | 3+3 2 |
| | 2+2 2 | 3+3 2 | 2+2 2 | 2+2 2 | 2+2 2 | 3+3 2 | 3+3 2 | 3+3 2 |
| | 3+3 2 | 2+2 2 | 3+3 2 | 3+3 2 | 3+3 2 | 2+2 2 | 3+3 2 | 3+3 2 |
| | 2+2 2 | 3+3 2 | 2+2 2 | 2+2 2 | 2+2 2 | 3+3 2 | 2+2 2 | 2+2 2 |
| | 2+2 2 | 3+3 2 | 2+2 2 | 2+2 2 | 2+2 2 | 3+3 2 | 3+3 2 | 3+3 2 |
| | 2+2 2 | 2+2 2 | 2+2 2 | 3+3 2 | 2+2 2 | 3+3 2 | 3+3 2 | 3+3 2 |
| | 3+3 2 | 3+3 2 | 3+3 2 | 2+2 2 | 3+3 2 | 2+2 2 | 2+2 2 | 2+2 2 |
| | 3+3 2 | 2+2 2 | 3+3 2 | 3+3 2 | 3+3 2 | 2+2 2 | 2+2 2 | 2+2 2 |
| | 3+3 2 | 3+3 2 | 3+3 2 | 2+2 2 | 2+2 2 | 3+3 2 | 2+2 2 | 2+2 2 |
| | 2+2 2 | 3+3 2 | 2+2 2 | 2+2 2 | 2+2 2 | 2+2 2 | 3+3 2 | 3+3 2 |
| 2.2.68 | 408774 | 152 | Cl14Mo4O4Tl2 | | | | | |
| | -GM | -R | -T | -U | -V | -X | -Y | -Z |
| | 3+3 2 | 2+2 2 | 2+2 2 | 2+2 2 | 3+3 2 | 3+3 2 | 3+3 2 | 2+2 2 |
| | 2+2 2 | 3+3 2 | 3+3 2 | 3+3 2 | 2+2 2 | 2+2 2 | 2+2 2 | 3+3 2 |
| | 3+3 2 | 2+2 2 | 3+3 2 | 3+3 2 | 2+2 2 | 2+2 2 | 2+2 2 | 2+2 2 |
| | 2+2 2 | 3+3 2 | 2+2 2 | 2+2 2 | 3+3 2 | 3+3 2 | 3+3 2 | 3+3 2 |
| | 3+3 2 | 2+2 2 | 3+3 2 | 3+3 2 | 3+3 2 | 2+2 2 | 2+2 2 | 2+2 2 |
| | 2+2 2 | 3+3 2 | 2+2 2 | 2+2 2 | 2+2 2 | 3+3 2 | 3+3 2 | 3+3 2 |
| | 3+3 2 | 2+2 2 | 3+3 2 | 3+3 2 | 3+3 2 | 2+2 2 | 2+2 2 | 2+2 2 |
| | 3+3 2 | 3+3 2 | 2+2 2 | 2+2 2 | 3+3 2 | 3+3 2 | 2+2 2 | 3+3 2 |
| | 2+2 2 | 2+2 2 | 3+3 2 | 3+3 2 | 2+2 2 | 2+2 2 | 3+3 2 | 2+2 2 |
| | 2+2 2 | 3+3 2 | 2+2 2 | 2+2 2 | 2+2 2 | 3+3 2 | 3+3 2 | 3+3 2 |
| | 3+3 2 | 2+2 2 | 3+3 2 | 3+3 2 | 2+2 2 | 3+3 2 | 3+3 2 | 3+3 2 |
| | 2+2 2 | 3+3 2 | 2+2 2 | 2+2 2 | 3+3 2 | 2+2 2 | 2+2 2 | 2+2 2 |
| | 3+3 2 | 3+3 2 | 2+2 2 | 3+3 2 | 3+3 2 | 3+3 2 | 2+2 2 | 3+3 2 |
| | 2+2 2 | 2+2 2 | 3+3 2 | 2+2 2 | 2+2 2 | 2+2 2 | 3+3 2 | 2+2 2 |
| | 3+3 2 | 3+3 2 | 2+2 2 | 2+2 2 | 2+2 2 | 3+3 2 | 2+2 2 | 3+3 2 |
| | 2+2 2 | 2+2 2 | 3+3 2 | 3+3 2 | 3+3 2 | 2+2 2 | 3+3 2 | 2+2 2 |
| | 3+3 2 | 3+3 2 | 2+2 2 | 3+3 2 | 3+3 2 | 3+3 2 | 2+2 2 | 3+3 2 |
| | 2+2 2 | 2+2 2 | 3+3 2 | 2+2 2 | 2+2 2 | 2+2 2 | 3+3 2 | 2+2 2 |
| | 3+3 2 | 2+2 2 | 2+2 2 | 3+3 2 | 3+3 2 | 3+3 2 | 3+3 2 | 2+2 2 |
| | 2+2 2 | 3+3 2 | 3+3 2 | 2+2 2 | 2+2 2 | 2+2 2 | 2+2 2 | 2+2 2 |
| | 3+3 2 | 2+2 2 | 2+2 2 | 3+3 2 | 3+3 2 | 2+2 2 | 2+2 2 | 3+3 2 |
| | 2+2 2 | 3+3 2 | 3+3 2 | 2+2 2 | 2+2 2 | 3+3 2 | 3+3 2 | 2+2 2 |
| | 3+3 2 | 3+3 2 | 3+3 2 | 3+3 2 | 3+3 2 | 2+2 2 | 2+2 2 | 3+3 2 |

Continued on next column

**TABLE XXI– continued**

| TQC | ICSD | NoE | SF | | | | | |
|---|---|---|---|---|---|---|---|---|
| | 3+3 2 | 2+2 2 | 3+3 2 | 3+3 2 | 2+2 2 | 2+2 2 | 2+2 2 | 3+3 2 |
| | 2+2 2 | 2+2 2 | 2+2 2 | 2+2 2 | 3+3 2 | 3+3 2 | 3+3 2 | 2+2 2 |
| | 3+3 2 | 2+2 2 | 2+2 2 | 2+2 2 | 2+2 2 | 3+3 2 | 3+3 2 | 3+3 2 |
| | 2+2 2 | 3+3 2 | 2+2 2 | 3+3 2 | 3+3 2 | 2+2 2 | 2+2 2 | 2+2 2 |
| | 3+3 2 | 3+3 2 | 3+3 2 | 2+2 2 | 2+2 2 | 3+3 2 | 3+3 2 | 3+3 2 |
| | 2+2 2 | 3+3 2 | 3+3 2 | 2+2 2 | 2+2 2 | 2+2 2 | 2+2 2 | 2+2 2 |
| | 3+3 2 | 2+2 2 | 2+2 2 | 3+3 2 | 3+3 2 | 2+2 2 | 3+3 2 | 3+3 2 |
| | 2+2 2 | 2+2 2 | 2+2 2 | 2+2 2 | 2+2 2 | 3+3 2 | 3+3 2 | 3+3 2 |
| | 3+3 2 | 3+3 2 | 3+3 2 | 2+2 2 | 3+3 2 | 3+3 2 | 2+2 2 | 2+2 2 |
| | 2+2 2 | 2+2 2 | 2+2 2 | 3+3 2 | 3+3 2 | 2+2 2 | 3+3 2 | 2+2 2 |
| | 2+2 2 | 3+3 2 | 3+3 2 | 3+3 2 | 2+2 2 | 3+3 2 | 2+2 2 | 3+3 2 |
| | 3+3 2 | 3+3 2 | 2+2 2 | 2+2 2 | 2+2 2 | 2+2 2 | 2+2 2 | 3+3 2 |
| | 2+2 2 | 2+2 2 | 3+3 2 | 3+3 2 | 2+2 2 | 3+3 2 | 3+3 2 | 2+2 2 |
| | 2+2 2 | 3+3 2 | 3+3 2 | 2+2 2 | 3+3 2 | 2+2 2 | 2+2 2 | 2+2 2 |
| | 3+3 2 | 2+2 2 | 2+2 2 | 3+3 2 | 3+3 2 | 3+3 2 | 3+3 2 | 3+3 2 |
| | 3+3 2 | 3+3 2 | 3+3 2 | 3+3 2 | 3+3 2 | 2+2 2 | 2+2 2 | 2+2 2 |
| | 2+2 2 | 2+2 2 | 2+2 2 | 2+2 2 | 2+2 2 | 3+3 2 | 2+2 2 | 3+3 2 |
| | 3+3 2 | 2+2 2 | 3+3 2 | 3+3 2 | 3+3 2 | 2+2 2 | 3+3 2 | 3+3 2 |
| | 2+2 2 | 3+3 2 | 2+2 2 | 2+2 2 | 2+2 2 | 3+3 2 | 3+3 2 | 2+2 2 |
| | 2+2 2 | 2+2 2 | 3+3 2 | 3+3 2 | 3+3 2 | 3+3 2 | 2+2 2 | 3+3 2 |
| | 3+3 2 | 3+3 2 | 2+2 2 | 3+3 2 | 2+2 2 | 2+2 2 | 3+3 2 | 3+3 2 |
| | 2+2 2 | 3+3 2 | 3+3 2 | 2+2 2 | 3+3 2 | 3+3 2 | 3+3 2 | 2+2 2 |
| | 3+3 2 | 2+2 2 | 2+2 2 | 2+2 2 | 2+2 2 | 2+2 2 | 2+2 2 | 2+2 2 |
| | 2+2 2 | 2+2 2 | 3+3 2 | 2+2 2 | 3+3 2 | 3+3 2 | 2+2 2 | 2+2 2 |
| | 3+3 2 | 2+2 2 | 2+2 2 | 3+3 2 | 2+2 2 | 3+3 2 | 3+3 2 | 3+3 2 |
| | 2+2 2 | 3+3 2 | 3+3 2 | 3+3 2 | 3+3 2 | 2+2 2 | 3+3 2 | 2+2 2 |
| | 3+3 2 | 3+3 2 | 2+2 2 | 3+3 2 | 2+2 2 | 2+2 2 | 2+2 2 | 3+3 2 |
| | 3+3 2 | 2+2 2 | 2+2 2 | 2+2 2 | 3+3 2 | 2+2 2 | 2+2 2 | 2+2 2 |
| | 3+3 2 | 3+3 2 | 3+3 2 | 3+3 2 | 2+2 2 | 2+2 2 | 2+2 2 | 3+3 2 |
| | 2+2 2 | 2+2 2 | 3+3 2 | 2+2 2 | 3+3 2 | 3+3 2 | 3+3 2 | 2+2 2 |
| | 3+3 2 | 3+3 2 | 3+3 2 | 2+2 2 | 2+2 2 | 3+3 2 | 3+3 2 | 3+3 2 |
| | 2+2 2 | 2+2 2 | 2+2 2 | 2+2 2 | 2+2 2 | 3+3 2 | 2+2 2 | 2+2 2 |
| | 3+3 2 | 2+2 2 | 3+3 2 | 3+3 2 | 3+3 2 | 2+2 2 | 3+3 2 | 3+3 2 |
| | 2+2 2 | 3+3 2 | 2+2 2 | 2+2 2 | 3+3 2 | 3+3 2 | 2+2 2 | 3+3 2 |
| | 3+3 2 | 3+3 2 | 3+3 2 | 3+3 2 | 2+2 2 | 2+2 2 | 3+3 2 | 2+2 2 |
| | 3+3 2 | 2+2 2 | 2+2 2 | 3+3 2 | 3+3 2 | 2+2 2 | 3+3 2 | 2+2 2 |
| | 2+2 2 | 3+3 2 | 2+2 2 | 2+2 2 | 2+2 2 | 3+3 2 | 3+3 2 | 3+3 2 |
| | 2+2 2 | 3+3 2 | 3+3 2 | 2+2 2 | 3+3 2 | 2+2 2 | 2+2 2 | 3+3 2 |
| | 3+3 2 | 2+2 2 | 3+3 2 | 3+3 2 | 3+3 2 | 2+2 2 | 2+2 2 | 2+2 2 |
| | 2+2 2 | 3+3 2 | 2+2 2 | 2+2 2 | 2+2 2 | 3+3 2 | 3+3 2 | 2+2 2 |
| | 3+3 2 | 3+3 2 | 2+2 2 | 3+3 2 | 2+2 2 | 2+2 2 | 2+2 2 | 2+2 2 |
| | 2+2 2 | 2+2 2 | 2+2 2 | 2+2 2 | 3+3 2 | 3+3 2 | 2+2 2 | 3+3 2 |
| | 2+2 2 | 3+3 2 | 3+3 2 | 3+3 2 | 2+2 2 | 3+3 2 | 3+3 2 | 3+3 2 |
| | 2+2 2 | 2+2 2 | 2+2 2 | 2+2 2 | 2+2 2 | 2+2 2 | 2+2 2 | 2+2 2 |
| | 3+3 2 | 2+2 2 | 2+2 2 | 3+3 2 | 3+3 2 | 3+3 2 | 2+2 2 | 3+3 2 |
| | 3+3 2 | 2+2 2 | 3+3 2 | 2+2 2 | 2+2 2 | 3+3 2 | 2+2 2 | 2+2 2 |
| | 2+2 2 | 3+3 2 | 3+3 2 | 3+3 2 | 3+3 2 | 3+3 2 | 3+3 2 | 3+3 2 |
| | 3+3 2 | 2+2 2 | 2+2 2 | 3+3 2 | 3+3 2 | 2+2 2 | 3+3 2 | 3+3 2 |
| | 2+2 2 | 3+3 2 | 3+3 2 | 2+2 2 | 2+2 2 | 2+2 2 | 2+2 2 | 2+2 2 |
| | 3+3 2 | 2+2 2 | 2+2 2 | 3+3 2 | 2+2 2 | 3+3 2 | 3+3 2 | 2+2 2 |
| | 2+2 2 | 3+3 2 | 3+3 2 | 2+2 2 | 3+3 2 | 2+2 2 | 3+3 2 | 3+3 2 |
| | 3+3 2 | 2+2 2 | 3+3 2 | 3+3 2 | 2+2 2 | 3+3 2 | 3+3 2 | 3+3 2 |
| | 2+2 2 | 3+3 2 | 2+2 2 | 2+2 2 | 3+3 2 | 2+2 2 | 2+2 2 | 3+3 2 |
| 2.2.69 | 74479 | 34 | Ca1P3 | | | | | |
| | -GM | -R | -T | -U | -V | -X | -Y | -Z |
| | 3+3 2 | 3+3 2 | 3+3 2 | 3+3 2 | 3+3 2 | 3+3 2 | 3+3 2 | 3+3 2 |
| | 3+3 2 | 2+2 2 | 2+2 2 | 2+2 2 | 2+2 2 | 2+2 2 | 2+2 2 | 2+2 2 |
| | 2+2 2 | 2+2 2 | 3+3 2 | 2+2 2 | 3+3 2 | 3+3 2 | 3+3 2 | 2+2 2 |
| | 3+3 2 | 3+3 2 | 2+2 2 | 2+2 2 | 2+2 2 | 2+2 2 | 3+3 2 | 3+3 2 |
| | 2+2 2 | 2+2 2 | 3+3 2 | 3+3 2 | 3+3 2 | 3+3 2 | 2+2 2 | 3+3 2 |
| | 2+2 2 | 3+3 2 | 2+2 2 | 3+3 2 | 2+2 2 | 2+2 2 | 2+2 2 | 2+2 2 |

Continued on next column

**TABLE XXI– continued**

| TQC | ICSD | NoE | SF | | | | | |
|---|---|---|---|---|---|---|---|---|
| | 3+3 2 | 3+3 2 | 2+2 2 | 3+3 2 | 2+2 2 | 3+3 2 | 3+3 2 | 3+3 2 |
| | 2+2 2 | 2+2 2 | 3+3 2 | 2+2 2 | 3+3 2 | 2+2 2 | 2+2 2 | 2+2 2 |
| | 2+2 2 | 2+2 2 | 3+3 2 | 2+2 2 | 3+3 2 | 3+3 2 | 2+2 2 | 3+3 2 |
| | 3+3 2 | 3+3 2 | 2+2 2 | 3+3 2 | 2+2 2 | 2+2 2 | 3+3 2 | 2+2 2 |
| | 3+3 2 | 3+3 2 | 3+3 2 | 2+2 2 | 3+3 2 | 2+2 2 | 2+2 2 | 3+3 2 |
| | 2+2 2 | 2+2 2 | 2+2 2 | 3+3 2 | 2+2 2 | 3+3 2 | 3+3 2 | 2+2 2 |
| | 2+2 2 | 3+3 2 | 2+2 2 | 2+2 2 | 2+2 2 | 2+2 2 | 3+3 2 | 2+2 2 |
| | 3+3 2 | 3+3 2 | 3+3 2 | 3+3 2 | 3+3 2 | 3+3 2 | 2+2 2 | 3+3 2 |
| | 3+3 2 | 2+2 2 | 3+3 2 | 3+3 2 | 3+3 2 | 3+3 2 | 3+3 2 | 3+3 2 |
| | 3+3 2 | 2+2 2 | 2+2 2 | 2+2 2 | 2+2 2 | 2+2 2 | 3+3 2 | 3+3 2 |
| | 3+3 2 | 2+2 2 | 3+3 2 | 2+2 2 | 3+3 2 | 3+3 2 | 2+2 2 | 2+2 2 |
| 2.2.70 | 42698 | 100 | C4H16Cl2N8Se4Te1 | | | | | |
| | -GM | -R | -T | -U | -V | -X | -Y | -Z |
| | 3+3 2 | 2+2 2 | 3+3 2 | 3+3 2 | 3+3 2 | 2+2 2 | 2+2 2 | 2+2 2 |
| | 2+2 2 | 3+3 2 | 2+2 2 | 2+2 2 | 2+2 2 | 3+3 2 | 3+3 2 | 3+3 2 |
| | 3+3 2 | 3+3 2 | 2+2 2 | 2+2 2 | 2+2 2 | 2+2 2 | 2+2 2 | 3+3 2 |
| | 2+2 2 | 2+2 2 | 3+3 2 | 3+3 2 | 3+3 2 | 3+3 2 | 3+3 2 | 2+2 2 |
| | 3+3 2 | 2+2 2 | 3+3 2 | 3+3 2 | 3+3 2 | 2+2 2 | 2+2 2 | 2+2 2 |
| | 2+2 2 | 3+3 2 | 2+2 2 | 2+2 2 | 2+2 2 | 3+3 2 | 3+3 2 | 3+3 2 |
| | 3+3 2 | 2+2 2 | 2+2 2 | 2+2 2 | 3+3 2 | 3+3 2 | 3+3 2 | 2+2 2 |
| | 2+2 2 | 3+3 2 | 3+3 2 | 3+3 2 | 2+2 2 | 2+2 2 | 2+2 2 | 3+3 2 |
| | 3+3 2 | 3+3 2 | 3+3 2 | 3+3 2 | 3+3 2 | 3+3 2 | 3+3 2 | 3+3 2 |
| | 3+3 2 | 2+2 2 | 2+2 2 | 3+3 2 | 2+2 2 | 2+2 2 | 2+2 2 | 2+2 2 |
| | 2+2 2 | 3+3 2 | 2+2 2 | 2+2 2 | 2+2 2 | 3+3 2 | 3+3 2 | 3+3 2 |
| | 2+2 2 | 2+2 2 | 3+3 2 | 2+2 2 | 3+3 2 | 2+2 2 | 2+2 2 | 2+2 2 |
| | 3+3 2 | 2+2 2 | 2+2 2 | 2+2 2 | 2+2 2 | 2+2 2 | 3+3 2 | 2+2 2 |
| | 2+2 2 | 3+3 2 | 3+3 2 | 3+3 2 | 3+3 2 | 3+3 2 | 2+2 2 | 3+3 2 |
| | 3+3 2 | 3+3 2 | 3+3 2 | 3+3 2 | 3+3 2 | 3+3 2 | 3+3 2 | 3+3 2 |
| | 3+3 2 | 2+2 2 | 2+2 2 | 3+3 2 | 2+2 2 | 3+3 2 | 2+2 2 | 3+3 2 |
| | 2+2 2 | 2+2 2 | 2+2 2 | 2+2 2 | 2+2 2 | 2+2 2 | 2+2 2 | 2+2 2 |
| | 2+2 2 | 3+3 2 | 3+3 2 | 2+2 2 | 3+3 2 | 2+2 2 | 3+3 2 | 2+2 2 |
| | 3+3 2 | 3+3 2 | 3+3 2 | 3+3 2 | 3+3 2 | 3+3 2 | 3+3 2 | 3+3 2 |
| | 2+2 2 | 2+2 2 | 2+2 2 | 2+2 2 | 2+2 2 | 3+3 2 | 2+2 2 | 2+2 2 |
| | 3+3 2 | 2+2 2 | 3+3 2 | 3+3 2 | 3+3 2 | 2+2 2 | 3+3 2 | 2+2 2 |
| | 3+3 2 | 3+3 2 | 2+2 2 | 2+2 2 | 2+2 2 | 2+2 2 | 2+2 2 | 3+3 2 |
| | 2+2 2 | 3+3 2 | 3+3 2 | 3+3 2 | 3+3 2 | 2+2 2 | 2+2 2 | 3+3 2 |
| | 2+2 2 | 2+2 2 | 2+2 2 | 2+2 2 | 3+3 2 | 3+3 2 | 3+3 2 | 2+2 2 |
| | 3+3 2 | 3+3 2 | 3+3 2 | 3+3 2 | 2+2 2 | 3+3 2 | 3+3 2 | 3+3 2 |
| | 2+2 2 | 2+2 2 | 2+2 2 | 2+2 2 | 2+2 2 | 2+2 2 | 2+2 2 | 2+2 2 |
| | 3+3 2 | 3+3 2 | 3+3 2 | 3+3 2 | 3+3 2 | 3+3 2 | 3+3 2 | 2+2 2 |
| | 2+2 2 | 2+2 2 | 3+3 2 | 3+3 2 | 2+2 2 | 2+2 2 | 2+2 2 | 3+3 2 |
| | 3+3 2 | 3+3 2 | 2+2 2 | 2+2 2 | 3+3 2 | 3+3 2 | 3+3 2 | 2+2 2 |
| | 2+2 2 | 2+2 2 | 2+2 2 | 2+2 2 | 2+2 2 | 2+2 2 | 2+2 2 | 3+3 2 |
| | 3+3 2 | 3+3 2 | 3+3 2 | 3+3 2 | 3+3 2 | 3+3 2 | 3+3 2 | 3+3 2 |
| | 2+2 2 | 2+2 2 | 2+2 2 | 2+2 2 | 2+2 2 | 2+2 2 | 2+2 2 | 2+2 2 |
| | 2+2 2 | 2+2 2 | 2+2 2 | 2+2 2 | 2+2 2 | 2+2 2 | 2+2 2 | 2+2 2 |
| | 3+3 2 | 3+3 2 | 2+2 2 | 2+2 2 | 3+3 2 | 3+3 2 | 2+2 2 | 3+3 2 |
| | 3+3 2 | 2+2 2 | 3+3 2 | 3+3 2 | 2+2 2 | 3+3 2 | 3+3 2 | 2+2 2 |
| | 2+2 2 | 3+3 2 | 2+2 2 | 3+3 2 | 3+3 2 | 2+2 2 | 3+3 2 | 3+3 2 |
| | 2+2 2 | 3+3 2 | 3+3 2 | 2+2 2 | 2+2 2 | 2+2 2 | 2+2 2 | 2+2 2 |
| | 3+3 2 | 3+3 2 | 2+2 2 | 3+3 2 | 3+3 2 | 2+2 2 | 3+3 2 | 3+3 2 |
| | 3+3 2 | 2+2 2 | 3+3 2 | 2+2 2 | 2+2 2 | 3+3 2 | 2+2 2 | 2+2 2 |
| | 2+2 2 | 2+2 2 | 2+2 2 | 3+3 2 | 3+3 2 | 2+2 2 | 3+3 2 | 2+2 2 |
| | 2+2 2 | 3+3 2 | 3+3 2 | 2+2 2 | 2+2 2 | 3+3 2 | 3+3 2 | 3+3 2 |
| | 3+3 2 | 2+2 2 | 3+3 2 | 3+3 2 | 2+2 2 | 3+3 2 | 2+2 2 | 2+2 2 |
| | 2+2 2 | 2+2 2 | 2+2 2 | 2+2 2 | 3+3 2 | 2+2 2 | 2+2 2 | 3+3 2 |
| | 3+3 2 | 3+3 2 | 2+2 2 | 3+3 2 | 3+3 2 | 3+3 2 | 3+3 2 | 3+3 2 |
| | 2+2 2 | 2+2 2 | 3+3 2 | 2+2 2 | 2+2 2 | 2+2 2 | 2+2 2 | 2+2 2 |
| | 3+3 2 | 3+3 2 | 3+3 2 | 3+3 2 | 3+3 2 | 2+2 2 | 3+3 2 | 3+3 2 |
| | 3+3 2 | 2+2 2 | 2+2 2 | 2+2 2 | 2+2 2 | 3+3 2 | 3+3 2 | 2+2 2 |
| | 2+2 2 | 3+3 2 | 3+3 2 | 3+3 2 | 3+3 2 | 3+3 2 | 2+2 2 | 3+3 2 |
| | 3+3 2 | 3+3 2 | 2+2 2 | 3+3 2 | 2+2 2 | 3+3 2 | 3+3 2 | 2+2 2 |
| | 2+2 2 | 2+2 2 | 2+2 2 | 2+2 2 | 3+3 2 | 2+2 2 | 2+2 2 | 3+3 2 |
| 2.2.71 | 170721 | 54 | C3H6As1N1 | | | | | |
| | -GM | -R | -T | -U | -V | -X | -Y | -Z |
| | 3+3 2 | 2+2 2 | 2+2 2 | 2+2 2 | 2+2 2 | 3+3 2 | 3+3 2 | 2+2 2 |
| | 2+2 2 | 3+3 2 | 3+3 2 | 3+3 2 | 3+3 2 | 2+2 2 | 2+2 2 | 3+3 2 |
| | 3+3 2 | 3+3 2 | 2+2 2 | 3+3 2 | 2+2 2 | 2+2 2 | 2+2 2 | 2+2 2 |
| | 2+2 2 | 2+2 2 | 3+3 2 | 2+2 2 | 3+3 2 | 3+3 2 | 3+3 2 | 3+3 2 |
| | 3+3 2 | 3+3 2 | 3+3 2 | 2+2 2 | 3+3 2 | 2+2 2 | 3+3 2 | 2+2 2 |
| | 2+2 2 | 2+2 2 | 2+2 2 | 3+3 2 | 2+2 2 | 3+3 2 | 2+2 2 | 3+3 2 |
| | 3+3 2 | 2+2 2 | 2+2 2 | 2+2 2 | 3+3 2 | 3+3 2 | 3+3 2 | 3+3 2 |
| | 2+2 2 | 3+3 2 | 3+3 2 | 3+3 2 | 2+2 2 | 2+2 2 | 2+2 2 | 2+2 2 |
| | 3+3 2 | 2+2 2 | 3+3 2 | 2+2 2 | 2+2 2 | 2+2 2 | 2+2 2 | 3+3 2 |
| | 2+2 2 | 3+3 2 | 2+2 2 | 3+3 2 | 3+3 2 | 3+3 2 | 3+3 2 | 2+2 2 |
| | 3+3 2 | 2+2 2 | 3+3 2 | 2+2 2 | 2+2 2 | 3+3 2 | 3+3 2 | 2+2 2 |
| | 2+2 2 | 3+3 2 | 2+2 2 | 3+3 2 | 3+3 2 | 2+2 2 | 2+2 2 | 3+3 2 |
| | 2+2 2 | 3+3 2 | 2+2 2 | 2+2 2 | 3+3 2 | 2+2 2 | 2+2 2 | 2+2 2 |
| | 2+2 2 | 2+2 2 | 3+3 2 | 3+3 2 | 2+2 2 | 3+3 2 | 3+3 2 | 2+2 2 |
| | 3+3 2 | 3+3 2 | 3+3 2 | 3+3 2 | 3+3 2 | 3+3 2 | 3+3 2 | 3+3 2 |
| | 3+3 2 | 2+2 2 | 2+2 2 | 2+2 2 | 2+2 2 | 2+2 2 | 2+2 2 | 3+3 2 |
| | 3+3 2 | 2+2 2 | 2+2 2 | 3+3 2 | 2+2 2 | 3+3 2 | 2+2 2 | 3+3 2 |
| | 2+2 2 | 3+3 2 | 3+3 2 | 2+2 2 | 3+3 2 | 2+2 2 | 3+3 2 | 2+2 2 |
| | 3+3 2 | 3+3 2 | 2+2 2 | 2+2 2 | 3+3 2 | 3+3 2 | 3+3 2 | 2+2 2 |
| | 2+2 2 | 2+2 2 | 3+3 2 | 3+3 2 | 2+2 2 | 2+2 2 | 2+2 2 | 3+3 2 |
| | 3+3 2 | 3+3 2 | 2+2 2 | 3+3 2 | 3+3 2 | 3+3 2 | 3+3 2 | 2+2 2 |
| | 2+2 2 | 2+2 2 | 3+3 2 | 2+2 2 | 2+2 2 | 2+2 2 | 2+2 2 | 3+3 2 |
| | 3+3 2 | 3+3 2 | 3+3 2 | 3+3 2 | 2+2 2 | 2+2 2 | 2+2 2 | 2+2 2 |
| | 3+3 2 | 2+2 2 | 2+2 2 | 2+2 2 | 3+3 2 | 3+3 2 | 3+3 2 | 3+3 2 |
| | 2+2 2 | 3+3 2 | 3+3 2 | 3+3 2 | 2+2 2 | 2+2 2 | 2+2 2 | 3+3 2 |
| | 2+2 2 | 2+2 2 | 2+2 2 | 2+2 2 | 3+3 2 | 3+3 2 | 3+3 2 | 2+2 2 |
| | 2+2 2 | 2+2 2 | 3+3 2 | 2+2 2 | 3+3 2 | 2+2 2 | 3+3 2 | 3+3 2 |
| 2.2.72 | 62435 | 120 | C2H8Cl2N4Se2Te1 | | | | | |
| | -GM | -R | -T | -U | -V | -X | -Y | -Z |
| | 3+3 2 | 3+3 2 | 2+2 2 | 2+2 2 | 2+2 2 | 2+2 2 | 2+2 2 | 2+2 2 |
| | 2+2 2 | 2+2 2 | 3+3 2 | 3+3 2 | 3+3 2 | 3+3 2 | 3+3 2 | 3+3 2 |
| | 3+3 2 | 2+2 2 | 3+3 2 | 2+2 2 | 2+2 2 | 3+3 2 | 2+2 2 | 2+2 2 |
| | 2+2 2 | 3+3 2 | 2+2 2 | 3+3 2 | 3+3 2 | 2+2 2 | 3+3 2 | 3+3 2 |
| | 3+3 2 | 2+2 2 | 2+2 2 | 3+3 2 | 2+2 2 | 3+3 2 | 2+2 2 | 3+3 2 |
| | 2+2 2 | 3+3 2 | 3+3 2 | 2+2 2 | 3+3 2 | 2+2 2 | 3+3 2 | 2+2 2 |
| | 3+3 2 | 3+3 2 | 3+3 2 | 2+2 2 | 2+2 2 | 3+3 2 | 2+2 2 | 2+2 2 |
| | 2+2 2 | 2+2 2 | 2+2 2 | 3+3 2 | 3+3 2 | 2+2 2 | 3+3 2 | 3+3 2 |
| | 3+3 2 | 2+2 2 | 2+2 2 | 2+2 2 | 3+3 2 | 2+2 2 | 3+3 2 | 3+3 2 |
| | 2+2 2 | 3+3 2 | 3+3 2 | 3+3 2 | 2+2 2 | 3+3 2 | 2+2 2 | 2+2 2 |
| | 3+3 2 | 2+2 2 | 2+2 2 | 2+2 2 | 3+3 2 | 3+3 2 | 3+3 2 | 2+2 2 |
| | 2+2 2 | 3+3 2 | 3+3 2 | 3+3 2 | 2+2 2 | 2+2 2 | 2+2 2 | 3+3 2 |
| | 3+3 2 | 2+2 2 | 2+2 2 | 2+2 2 | 3+3 2 | 2+2 2 | 3+3 2 | 3+3 2 |
| | 2+2 2 | 3+3 2 | 3+3 2 | 3+3 2 | 2+2 2 | 3+3 2 | 2+2 2 | 2+2 2 |
| | 2+2 2 | 3+3 2 | 3+3 2 | 3+3 2 | 2+2 2 | 2+2 2 | 2+2 2 | 3+3 2 |
| | 3+3 2 | 2+2 2 | 2+2 2 | 2+2 2 | 3+3 2 | 3+3 2 | 3+3 2 | 2+2 2 |
| | 3+3 2 | 3+3 2 | 3+3 2 | 3+3 2 | 3+3 2 | 3+3 2 | 3+3 2 | 3+3 2 |
| | 2+2 2 | 2+2 2 | 2+2 2 | 2+2 2 | 2+2 2 | 2+2 2 | 2+2 2 | 2+2 2 |
| | 2+2 2 | 2+2 2 | 2+2 2 | 2+2 2 | 3+3 2 | 2+2 2 | 3+3 2 | 2+2 2 |
| | 3+3 2 | 3+3 2 | 3+3 2 | 3+3 2 | 2+2 2 | 3+3 2 | 2+2 2 | 3+3 2 |
| | 3+3 2 | 3+3 2 | 3+3 2 | 2+2 2 | 2+2 2 | 3+3 2 | 2+2 2 | 2+2 2 |
| | 2+2 2 | 2+2 2 | 2+2 2 | 3+3 2 | 3+3 2 | 2+2 2 | 3+3 2 | 3+3 2 |
| | 2+2 2 | 2+2 2 | 2+2 2 | 2+2 2 | 2+2 2 | 2+2 2 | 2+2 2 | 2+2 2 |
| | 3+3 2 | 3+3 2 | 3+3 2 | 3+3 2 | 3+3 2 | 3+3 2 | 3+3 2 | 3+3 2 |
| | 3+3 2 | 2+2 2 | 3+3 2 | 3+3 2 | 2+2 2 | 2+2 2 | 2+2 2 | 2+2 2 |
| | 2+2 2 | 3+3 2 | 2+2 2 | 2+2 2 | 3+3 2 | 3+3 2 | 3+3 2 | 3+3 2 |
| | 3+3 2 | 2+2 2 | 3+3 2 | 3+3 2 | 2+2 2 | 3+3 2 | 2+2 2 | 2+2 2 |
| | 2+2 2 | 3+3 2 | 2+2 2 | 2+2 2 | 3+3 2 | 2+2 2 | 3+3 2 | 3+3 2 |

Continued on next column

Continued on next column

**TABLE XXI– continued**

| TQC | ICSD | NoE | SF | | | | | |
|---|---|---|---|---|---|---|---|---|
| | 3+3 2 | 3+3 2 | 3+3 2 | 2+2 2 | 2+2 2 | 3+3 2 | 2+2 2 | 2+2 2 |
| | 2+2 2 | 2+2 2 | 2+2 2 | 3+3 2 | 3+3 2 | 2+2 2 | 3+3 2 | 3+3 2 |
| | 3+3 2 | 3+3 2 | 3+3 2 | 2+2 2 | 2+2 2 | 3+3 2 | 2+2 2 | 2+2 2 |
| | 2+2 2 | 2+2 2 | 2+2 2 | 3+3 2 | 3+3 2 | 2+2 2 | 3+3 2 | 3+3 2 |
| | 3+3 2 | 3+3 2 | 2+2 2 | 2+2 2 | 3+3 2 | 2+2 2 | 3+3 2 | 3+3 2 |
| | 2+2 2 | 2+2 2 | 3+3 2 | 3+3 2 | 2+2 2 | 3+3 2 | 2+2 2 | 2+2 2 |
| | 3+3 2 | 2+2 2 | 3+3 2 | 2+2 2 | 3+3 2 | 2+2 2 | 3+3 2 | 3+3 2 |
| | 2+2 2 | 3+3 2 | 2+2 2 | 3+3 2 | 2+2 2 | 3+3 2 | 3+3 2 | 3+3 2 |
| | 3+3 2 | 2+2 2 | 3+3 2 | 2+2 2 | 2+2 2 | 3+3 2 | 2+2 2 | 2+2 2 |
| | 2+2 2 | 3+3 2 | 2+2 2 | 3+3 2 | 3+3 2 | 2+2 2 | 2+2 2 | 2+2 2 |
| | 3+3 2 | 3+3 2 | 3+3 2 | 3+3 2 | 3+3 2 | 3+3 2 | 3+3 2 | 3+3 2 |
| | 2+2 2 | 3+3 2 | 2+2 2 | 2+2 2 | 2+2 2 | 3+3 2 | 3+3 2 | 2+2 2 |
| | 3+3 2 | 2+2 2 | 2+2 2 | 2+2 2 | 2+2 2 | 2+2 2 | 2+2 2 | 2+2 2 |
| | 2+2 2 | 2+2 2 | 3+3 2 | 3+3 2 | 3+3 2 | 2+2 2 | 3+3 2 | 3+3 2 |
| | 3+3 2 | 2+2 2 | 3+3 2 | 3+3 2 | 3+3 2 | 2+2 2 | 2+2 2 | 2+2 2 |
| | 2+2 2 | 3+3 2 | 2+2 2 | 2+2 2 | 2+2 2 | 3+3 2 | 2+2 2 | 3+3 2 |
| | 3+3 2 | 3+3 2 | 3+3 2 | 2+2 2 | 3+3 2 | 2+2 2 | 3+3 2 | 2+2 2 |
| | 3+3 2 | 2+2 2 | 2+2 2 | 3+3 2 | 2+2 2 | 3+3 2 | 2+2 2 | 3+3 2 |
| | 2+2 2 | 3+3 2 | 3+3 2 | 3+3 2 | 3+3 2 | 2+2 2 | 2+2 2 | 3+3 2 |
| | 2+2 2 | 2+2 2 | 2+2 2 | 2+2 2 | 2+2 2 | 2+2 2 | 3+3 2 | 3+3 2 |
| | 3+3 2 | 3+3 2 | 2+2 2 | 3+3 2 | 3+3 2 | 2+2 2 | 2+2 2 | 2+2 2 |
| | 3+3 2 | 3+3 2 | 3+3 2 | 2+2 2 | 2+2 2 | 3+3 2 | 3+3 2 | 2+2 2 |
| | 2+2 2 | 2+2 2 | 3+3 2 | 3+3 2 | 3+3 2 | 3+3 2 | 3+3 2 | 2+2 2 |
| | 2+2 2 | 2+2 2 | 2+2 2 | 2+2 2 | 3+3 2 | 3+3 2 | 2+2 2 | 3+3 2 |
| | 3+3 2 | 3+3 2 | 3+3 2 | 3+3 2 | 2+2 2 | 3+3 2 | 2+2 2 | 2+2 2 |
| | 2+2 2 | 2+2 2 | 2+2 2 | 2+2 2 | 2+2 2 | 2+2 2 | 2+2 2 | 3+3 2 |
| | 2+2 2 | 3+3 2 | 2+2 2 | 3+3 2 | 3+3 2 | 3+3 2 | 3+3 2 | 2+2 2 |
| | 3+3 2 | 2+2 2 | 3+3 2 | 2+2 2 | 2+2 2 | 2+2 2 | 3+3 2 | 3+3 2 |
| | 2+2 2 | 3+3 2 | 2+2 2 | 2+2 2 | 2+2 2 | 3+3 2 | 2+2 2 | 2+2 2 |
| | 3+3 2 | 2+2 2 | 3+3 2 | 2+2 2 | 3+3 2 | 2+2 2 | 2+2 2 | 3+3 2 |
| | 3+3 2 | 3+3 2 | 2+2 2 | 3+3 2 | 2+2 2 | 3+3 2 | 3+3 2 | 3+3 2 |
| | 2+2 2 | 3+3 2 | 2+2 2 | 2+2 2 | 3+3 2 | 3+3 2 | 3+3 2 | 2+2 2 |
| 2.2.73 | 62436 | 120 | C2H8Br2N4Se2Te1 | | | | | |
| | -GM | -R | -T | -U | -V | -X | -Y | -Z |
| | 3+3 2 | 3+3 2 | 3+3 2 | 2+2 2 | 2+2 2 | 3+3 2 | 2+2 2 | 2+2 2 |
| | 2+2 2 | 2+2 2 | 2+2 2 | 3+3 2 | 3+3 2 | 2+2 2 | 3+3 2 | 3+3 2 |
| | 2+2 2 | 2+2 2 | 2+2 2 | 2+2 2 | 2+2 2 | 2+2 2 | 2+2 2 | 2+2 2 |
| | 3+3 2 | 3+3 2 | 3+3 2 | 3+3 2 | 3+3 2 | 3+3 2 | 3+3 2 | 3+3 2 |
| | 3+3 2 | 3+3 2 | 3+3 2 | 2+2 2 | 2+2 2 | 3+3 2 | 2+2 2 | 2+2 2 |
| | 2+2 2 | 2+2 2 | 2+2 2 | 3+3 2 | 3+3 2 | 2+2 2 | 3+3 2 | 3+3 2 |
| | 3+3 2 | 2+2 2 | 2+2 2 | 3+3 2 | 2+2 2 | 3+3 2 | 2+2 2 | 2+2 2 |
| | 2+2 2 | 3+3 2 | 3+3 2 | 2+2 2 | 3+3 2 | 2+2 2 | 3+3 2 | 3+3 2 |
| | 3+3 2 | 2+2 2 | 2+2 2 | 2+2 2 | 3+3 2 | 2+2 2 | 3+3 2 | 3+3 2 |
| | 2+2 2 | 3+3 2 | 3+3 2 | 3+3 2 | 2+2 2 | 3+3 2 | 2+2 2 | 2+2 2 |
| | 3+3 2 | 2+2 2 | 2+2 2 | 2+2 2 | 3+3 2 | 2+2 2 | 2+2 2 | 2+2 2 |
| | 2+2 2 | 3+3 2 | 3+3 2 | 2+2 2 | 2+2 2 | 3+3 2 | 3+3 2 | 3+3 2 |
| | 3+3 2 | 2+2 2 | 2+2 2 | 3+3 2 | 3+3 2 | 3+3 2 | 3+3 2 | 2+2 2 |
| | 2+2 2 | 3+3 2 | 3+3 2 | 3+3 2 | 2+2 2 | 2+2 2 | 2+2 2 | 3+3 2 |
| | 2+2 2 | 3+3 2 | 3+3 2 | 3+3 2 | 2+2 2 | 2+2 2 | 2+2 2 | 3+3 2 |
| | 3+3 2 | 2+2 2 | 2+2 2 | 2+2 2 | 3+3 2 | 3+3 2 | 3+3 2 | 2+2 2 |
| | 3+3 2 | 3+3 2 | 3+3 2 | 3+3 2 | 3+3 2 | 3+3 2 | 3+3 2 | 3+3 2 |
| | 2+2 2 | 2+2 2 | 2+2 2 | 2+2 2 | 2+2 2 | 2+2 2 | 2+2 2 | 2+2 2 |
| | 2+2 2 | 2+2 2 | 2+2 2 | 2+2 2 | 3+3 2 | 2+2 2 | 3+3 2 | 2+2 2 |
| | 3+3 2 | 3+3 2 | 3+3 2 | 3+3 2 | 2+2 2 | 3+3 2 | 2+2 2 | 3+3 2 |
| | 3+3 2 | 3+3 2 | 3+3 2 | 2+2 2 | 2+2 2 | 3+3 2 | 2+2 2 | 2+2 2 |
| | 2+2 2 | 2+2 2 | 2+2 2 | 3+3 2 | 3+3 2 | 2+2 2 | 3+3 2 | 3+3 2 |
| | 2+2 2 | 2+2 2 | 2+2 2 | 2+2 2 | 2+2 2 | 2+2 2 | 2+2 2 | 2+2 2 |
| | 3+3 2 | 3+3 2 | 3+3 2 | 3+3 2 | 3+3 2 | 3+3 2 | 3+3 2 | 3+3 2 |
| | 3+3 2 | 3+3 2 | 3+3 2 | 3+3 2 | 3+3 2 | 3+3 2 | 2+2 2 | 2+2 2 |
| | 2+2 2 | 2+2 2 | 2+2 2 | 2+2 2 | 2+2 2 | 2+2 2 | 3+3 2 | 3+3 2 |
| | 3+3 2 | 2+2 2 | 3+3 2 | 3+3 2 | 2+2 2 | 3+3 2 | 2+2 2 | 2+2 2 |

Continued on next column

**TABLE XXI– continued**

| TQC | ICSD | NoE | SF | | | | | |
|---|---|---|---|---|---|---|---|---|
| | 2+2 2 | 3+3 2 | 2+2 2 | 2+2 2 | 3+3 2 | 2+2 2 | 3+3 2 | 3+3 2 |
| | 3+3 2 | 3+3 2 | 3+3 2 | 2+2 2 | 2+2 2 | 3+3 2 | 2+2 2 | 2+2 2 |
| | 2+2 2 | 2+2 2 | 2+2 2 | 3+3 2 | 3+3 2 | 2+2 2 | 3+3 2 | 3+3 2 |
| | 3+3 2 | 3+3 2 | 3+3 2 | 2+2 2 | 2+2 2 | 3+3 2 | 2+2 2 | 3+3 2 |
| | 2+2 2 | 3+3 2 | 2+2 2 | 3+3 2 | 3+3 2 | 2+2 2 | 3+3 2 | 2+2 2 |
| | 3+3 2 | 2+2 2 | 2+2 2 | 2+2 2 | 3+3 2 | 2+2 2 | 3+3 2 | 3+3 2 |
| | 2+2 2 | 2+2 2 | 3+3 2 | 3+3 2 | 2+2 2 | 3+3 2 | 2+2 2 | 2+2 2 |
| | 3+3 2 | 2+2 2 | 3+3 2 | 2+2 2 | 2+2 2 | 2+2 2 | 3+3 2 | 3+3 2 |
| | 2+2 2 | 3+3 2 | 2+2 2 | 3+3 2 | 3+3 2 | 3+3 2 | 2+2 2 | 2+2 2 |
| | 3+3 2 | 2+2 2 | 3+3 2 | 3+3 2 | 3+3 2 | 3+3 2 | 3+3 2 | 3+3 2 |
| | 2+2 2 | 3+3 2 | 2+2 2 | 2+2 2 | 2+2 2 | 2+2 2 | 2+2 2 | 2+2 2 |
| | 3+3 2 | 3+3 2 | 2+2 2 | 3+3 2 | 3+3 2 | 3+3 2 | 3+3 2 | 3+3 2 |
| | 2+2 2 | 3+3 2 | 3+3 2 | 2+2 2 | 2+2 2 | 3+3 2 | 3+3 2 | 2+2 2 |
| | 3+3 2 | 2+2 2 | 2+2 2 | 3+3 2 | 2+2 2 | 2+2 2 | 2+2 2 | 3+3 2 |
| | 3+3 2 | 2+2 2 | 3+3 2 | 2+2 2 | 3+3 2 | 2+2 2 | 2+2 2 | 2+2 2 |
| | 2+2 2 | 2+2 2 | 3+3 2 | 2+2 2 | 3+3 2 | 2+2 2 | 3+3 2 | 2+2 2 |
| | 2+2 2 | 3+3 2 | 2+2 2 | 3+3 2 | 2+2 2 | 3+3 2 | 2+2 2 | 3+3 2 |
| | 3+3 2 | 2+2 2 | 3+3 2 | 2+2 2 | 3+3 2 | 2+2 2 | 3+3 2 | 3+3 2 |
| | 3+3 2 | 3+3 2 | 3+3 2 | 3+3 2 | 2+2 2 | 3+3 2 | 2+2 2 | 2+2 2 |
| | 2+2 2 | 2+2 2 | 2+2 2 | 3+3 2 | 3+3 2 | 2+2 2 | 2+2 2 | 3+3 2 |
| | 2+2 2 | 3+3 2 | 2+2 2 | 2+2 2 | 2+2 2 | 2+2 2 | 3+3 2 | 2+2 2 |
| | 3+3 2 | 3+3 2 | 2+2 2 | 3+3 2 | 3+3 2 | 3+3 2 | 2+2 2 | 3+3 2 |
| | 2+2 2 | 2+2 2 | 3+3 2 | 2+2 2 | 2+2 2 | 2+2 2 | 3+3 2 | 2+2 2 |
| | 3+3 2 | 3+3 2 | 3+3 2 | 3+3 2 | 3+3 2 | 3+3 2 | 2+2 2 | 2+2 2 |
| | 2+2 2 | 2+2 2 | 2+2 2 | 2+2 2 | 2+2 2 | 3+3 2 | 3+3 2 | 3+3 2 |
| | 3+3 2 | 3+3 2 | 3+3 2 | 3+3 2 | 3+3 2 | 3+3 2 | 2+2 2 | 2+2 2 |
| | 2+2 2 | 2+2 2 | 2+2 2 | 2+2 2 | 2+2 2 | 2+2 2 | 2+2 2 | 3+3 2 |
| | 2+2 2 | 3+3 2 | 2+2 2 | 3+3 2 | 3+3 2 | 3+3 2 | 3+3 2 | 2+2 2 |
| | 3+3 2 | 3+3 2 | 3+3 2 | 2+2 2 | 2+2 2 | 2+2 2 | 3+3 2 | 3+3 2 |
| | 3+3 2 | 2+2 2 | 2+2 2 | 2+2 2 | 3+3 2 | 3+3 2 | 2+2 2 | 2+2 2 |
| | 2+2 2 | 2+2 2 | 3+3 2 | 2+2 2 | 2+2 2 | 2+2 2 | 2+2 2 | 3+3 2 |
| | 3+3 2 | 3+3 2 | 2+2 2 | 3+3 2 | 2+2 2 | 3+3 2 | 3+3 2 | 2+2 2 |
| | 2+2 2 | 3+3 2 | 3+3 2 | 3+3 2 | 3+3 2 | 3+3 2 | 3+3 2 | 3+3 2 |
| 2.2.74 | 186554 | 94 | C2H3Na1O5 | | | | | |
| | -GM | -R | -T | -U | -V | -X | -Y | -Z |
| | 3+3 2 | 3+3 2 | 2+2 2 | 3+3 2 | 2+2 2 | 2+2 2 | 3+3 2 | 2+2 2 |
| | 2+2 2 | 2+2 2 | 3+3 2 | 2+2 2 | 3+3 2 | 3+3 2 | 2+2 2 | 3+3 2 |
| | 2+2 2 | 2+2 2 | 3+3 2 | 2+2 2 | 3+3 2 | 3+3 2 | 2+2 2 | 3+3 2 |
| | 3+3 2 | 3+3 2 | 2+2 2 | 3+3 2 | 2+2 2 | 2+2 2 | 3+3 2 | 2+2 2 |
| | 3+3 2 | 2+2 2 | 3+3 2 | 2+2 2 | 2+2 2 | 3+3 2 | 3+3 2 | 3+3 2 |
| | 2+2 2 | 3+3 2 | 2+2 2 | 3+3 2 | 3+3 2 | 2+2 2 | 2+2 2 | 2+2 2 |
| | 2+2 2 | 2+2 2 | 3+3 2 | 3+3 2 | 3+3 2 | 2+2 2 | 2+2 2 | 3+3 2 |
| | 3+3 2 | 3+3 2 | 2+2 2 | 2+2 2 | 2+2 2 | 3+3 2 | 3+3 2 | 2+2 2 |
| | 3+3 2 | 2+2 2 | 2+2 2 | 3+3 2 | 2+2 2 | 3+3 2 | 2+2 2 | 3+3 2 |
| | 2+2 2 | 3+3 2 | 3+3 2 | 2+2 2 | 3+3 2 | 2+2 2 | 3+3 2 | 2+2 2 |
| | 2+2 2 | 3+3 2 | 3+3 2 | 2+2 2 | 3+3 2 | 2+2 2 | 2+2 2 | 3+3 2 |
| | 3+3 2 | 2+2 2 | 2+2 2 | 3+3 2 | 2+2 2 | 3+3 2 | 3+3 2 | 2+2 2 |
| | 2+2 2 | 2+2 2 | 3+3 2 | 2+2 2 | 3+3 2 | 3+3 2 | 3+3 2 | 2+2 2 |
| | 3+3 2 | 3+3 2 | 2+2 2 | 3+3 2 | 2+2 2 | 2+2 2 | 2+2 2 | 3+3 2 |
| | 3+3 2 | 2+2 2 | 2+2 2 | 3+3 2 | 2+2 2 | 3+3 2 | 2+2 2 | 3+3 2 |
| | 2+2 2 | 3+3 2 | 3+3 2 | 2+2 2 | 3+3 2 | 2+2 2 | 3+3 2 | 2+2 2 |
| | 3+3 2 | 3+3 2 | 2+2 2 | 3+3 2 | 2+2 2 | 2+2 2 | 3+3 2 | 2+2 2 |
| | 2+2 2 | 2+2 2 | 3+3 2 | 2+2 2 | 3+3 2 | 3+3 2 | 2+2 2 | 3+3 2 |
| | 3+3 2 | 3+3 2 | 2+2 2 | 3+3 2 | 2+2 2 | 2+2 2 | 3+3 2 | 2+2 2 |
| | 2+2 2 | 2+2 2 | 3+3 2 | 2+2 2 | 3+3 2 | 3+3 2 | 2+2 2 | 3+3 2 |
| | 3+3 2 | 2+2 2 | 3+3 2 | 3+3 2 | 2+2 2 | 2+2 2 | 2+2 2 | 2+2 2 |
| | 2+2 2 | 3+3 2 | 2+2 2 | 2+2 2 | 3+3 2 | 3+3 2 | 3+3 2 | 3+3 2 |
| | 2+2 2 | 3+3 2 | 3+3 2 | 2+2 2 | 2+2 2 | 2+2 2 | 3+3 2 | 2+2 2 |
| | 3+3 2 | 2+2 2 | 2+2 2 | 3+3 2 | 3+3 2 | 3+3 2 | 2+2 2 | 3+3 2 |
| | 3+3 2 | 2+2 2 | 3+3 2 | 3+3 2 | 3+3 2 | 2+2 2 | 2+2 2 | 3+3 2 |
| | 2+2 2 | 3+3 2 | 2+2 2 | 2+2 2 | 2+2 2 | 3+3 2 | 3+3 2 | 2+2 2 |

Continued on next column

**TABLE XXI– continued**

| TQC | ICSD | NoE | SF | | | | |
|---|---|---|---|---|---|---|---|
| | 3+3 2 | 2+2 2 | 3+3 2 | 3+3 2 | 3+3 2 | 2+2 2 | 2+2 2 | 2+2 2 |
| | 2+2 2 | 3+3 2 | 2+2 2 | 2+2 2 | 2+2 2 | 3+3 2 | 3+3 2 | 3+3 2 |
| | 3+3 2 | 2+2 2 | 3+3 2 | 2+2 2 | 3+3 2 | 2+2 2 | 3+3 2 | 3+3 2 |
| | 2+2 2 | 3+3 2 | 2+2 2 | 3+3 2 | 2+2 2 | 3+3 2 | 2+2 2 | 2+2 2 |
| | 2+2 2 | 3+3 2 | 2+2 2 | 3+3 2 | 2+2 2 | 2+2 2 | 3+3 2 | 3+3 2 |
| | 3+3 2 | 2+2 2 | 3+3 2 | 2+2 2 | 3+3 2 | 3+3 2 | 2+2 2 | 2+2 2 |
| | 3+3 2 | 3+3 2 | 3+3 2 | 2+2 2 | 3+3 2 | 3+3 2 | 2+2 2 | 3+3 2 |
| | 2+2 2 | 2+2 2 | 2+2 2 | 3+3 2 | 2+2 2 | 2+2 2 | 3+3 2 | 2+2 2 |
| | 3+3 2 | 3+3 2 | 2+2 2 | 2+2 2 | 2+2 2 | 2+2 2 | 3+3 2 | 3+3 2 |
| | 2+2 2 | 2+2 2 | 3+3 2 | 3+3 2 | 3+3 2 | 3+3 2 | 2+2 2 | 2+2 2 |
| | 2+2 2 | 2+2 2 | 3+3 2 | 2+2 2 | 2+2 2 | 3+3 2 | 3+3 2 | 3+3 2 |
| | 3+3 2 | 3+3 2 | 2+2 2 | 3+3 2 | 3+3 2 | 2+2 2 | 2+2 2 | 2+2 2 |
| | 3+3 2 | 2+2 2 | 3+3 2 | 2+2 2 | 2+2 2 | 2+2 2 | 3+3 2 | 3+3 2 |
| | 2+2 2 | 3+3 2 | 2+2 2 | 3+3 2 | 3+3 2 | 3+3 2 | 2+2 2 | 2+2 2 |
| | 3+3 2 | 3+3 2 | 3+3 2 | 2+2 2 | 3+3 2 | 2+2 2 | 3+3 2 | 2+2 2 |
| | 2+2 2 | 2+2 2 | 2+2 2 | 3+3 2 | 2+2 2 | 3+3 2 | 2+2 2 | 3+3 2 |
| | 3+3 2 | 2+2 2 | 2+2 2 | 3+3 2 | 2+2 2 | 3+3 2 | 2+2 2 | 2+2 2 |
| | 2+2 2 | 3+3 2 | 3+3 2 | 2+2 2 | 3+3 2 | 2+2 2 | 3+3 2 | 3+3 2 |
| | 3+3 2 | 3+3 2 | 2+2 2 | 3+3 2 | 2+2 2 | 3+3 2 | 2+2 2 | 2+2 2 |
| | 2+2 2 | 2+2 2 | 3+3 2 | 2+2 2 | 3+3 2 | 2+2 2 | 3+3 2 | 3+3 2 |
| | 2+2 2 | 2+2 2 | 2+2 2 | 2+2 2 | 3+3 2 | 3+3 2 | 2+2 2 | 2+2 2 |
| 2.2.75 | 196028 | 58 | C1Mn1O3 | | | | |
| | -GM | -R | -T | -U | -V | -X | -Y | -Z |
| | 3+3 2 | 2+2 2 | 3+3 2 | 2+2 2 | 2+2 2 | 2+2 2 | 3+3 2 | 3+3 2 |
| | 2+2 2 | 3+3 2 | 2+2 2 | 3+3 2 | 3+3 2 | 3+3 2 | 2+2 2 | 2+2 2 |
| | 3+3 2 | 3+3 2 | 2+2 2 | 2+2 2 | 3+3 2 | 2+2 2 | 2+2 2 | 3+3 2 |
| | 2+2 2 | 2+2 2 | 3+3 2 | 3+3 2 | 2+2 2 | 3+3 2 | 3+3 2 | 2+2 2 |
| | 3+3 2 | 2+2 2 | 3+3 2 | 3+3 2 | 3+3 2 | 2+2 2 | 2+2 2 | 2+2 2 |
| | 2+2 2 | 3+3 2 | 2+2 2 | 2+2 2 | 2+2 2 | 3+3 2 | 3+3 2 | 3+3 2 |
| | 3+3 2 | 2+2 2 | 2+2 2 | 3+3 2 | 2+2 2 | 2+2 2 | 2+2 2 | 2+2 2 |
| | 2+2 2 | 3+3 2 | 2+2 2 | 2+2 2 | 3+3 2 | 3+3 2 | 3+3 2 | 3+3 2 |
| | 3+3 2 | 3+3 2 | 3+3 2 | 2+2 2 | 3+3 2 | 2+2 2 | 2+2 2 | 3+3 2 |
| | 3+3 2 | 2+2 2 | 3+3 2 | 3+3 2 | 2+2 2 | 3+3 2 | 3+3 2 | 3+3 2 |
| | 2+2 2 | 3+3 2 | 3+3 2 | 2+2 2 | 2+2 2 | 3+3 2 | 2+2 2 | 2+2 2 |
| | 3+3 2 | 2+2 2 | 2+2 2 | 3+3 2 | 3+3 2 | 2+2 2 | 3+3 2 | 3+3 2 |
| | 2+2 2 | 2+2 2 | 3+3 2 | 2+2 2 | 2+2 2 | 3+3 2 | 3+3 2 | 2+2 2 |
| | 2+2 2 | 3+3 2 | 2+2 2 | 3+3 2 | 3+3 2 | 2+2 2 | 2+2 2 | 2+2 2 |
| | 3+3 2 | 3+3 2 | 3+3 2 | 3+3 2 | 3+3 2 | 2+2 2 | 3+3 2 | 2+2 2 |
| | 3+3 2 | 2+2 2 | 2+2 2 | 2+2 2 | 2+2 2 | 2+2 2 | 2+2 2 | 3+3 2 |
| | 3+3 2 | 3+3 2 | 2+2 2 | 3+3 2 | 3+3 2 | 3+3 2 | 2+2 2 | 3+3 2 |
| | 2+2 2 | 2+2 2 | 3+3 2 | 2+2 2 | 2+2 2 | 3+3 2 | 3+3 2 | 3+3 2 |
| | 2+2 2 | 3+3 2 | 2+2 2 | 3+3 2 | 2+2 2 | 2+2 2 | 2+2 2 | 2+2 2 |
| | 3+3 2 | 2+2 2 | 3+3 2 | 2+2 2 | 3+3 2 | 3+3 2 | 3+3 2 | 2+2 2 |
| | 2+2 2 | 2+2 2 | 3+3 2 | 2+2 2 | 2+2 2 | 2+2 2 | 2+2 2 | 2+2 2 |
| | 3+3 2 | 3+3 2 | 2+2 2 | 3+3 2 | 3+3 2 | 3+3 2 | 3+3 2 | 3+3 2 |
| | 2+2 2 | 2+2 2 | 3+3 2 | 2+2 2 | 3+3 2 | 3+3 2 | 2+2 2 | 2+2 2 |
| | 2+2 2 | 3+3 2 | 2+2 2 | 3+3 2 | 2+2 2 | 2+2 2 | 3+3 2 | 3+3 2 |
| | 2+2 2 | 3+3 2 | 3+3 2 | 2+2 2 | 3+3 2 | 3+3 2 | 3+3 2 | 2+2 2 |
| | 3+3 2 | 2+2 2 | 2+2 2 | 3+3 2 | 2+2 2 | 2+2 2 | 2+2 2 | 3+3 2 |
| | 3+3 2 | 2+2 2 | 2+2 2 | 3+3 2 | 3+3 2 | 3+3 2 | 3+3 2 | 2+2 2 |
| | 2+2 2 | 3+3 2 | 3+3 2 | 2+2 2 | 2+2 2 | 2+2 2 | 3+3 2 | 3+3 2 |
| | 2+2 2 | 2+2 2 | 2+2 2 | 3+3 2 | 2+2 2 | 3+3 2 | 2+2 2 | 3+3 2 |
| 2.2.76 | 196027 | 58 | C1Mn1O3 | | | | |
| | -GM | -R | -T | -U | -V | -X | -Y | -Z |
| | 3+3 2 | 2+2 2 | 3+3 2 | 2+2 2 | 2+2 2 | 2+2 2 | 3+3 2 | 3+3 2 |
| | 2+2 2 | 3+3 2 | 2+2 2 | 3+3 2 | 3+3 2 | 3+3 2 | 2+2 2 | 2+2 2 |
| | 3+3 2 | 2+2 2 | 3+3 2 | 3+3 2 | 3+3 2 | 2+2 2 | 2+2 2 | 2+2 2 |
| | 2+2 2 | 3+3 2 | 2+2 2 | 2+2 2 | 2+2 2 | 3+3 2 | 3+3 2 | 3+3 2 |
| | 3+3 2 | 3+3 2 | 2+2 2 | 3+3 2 | 2+2 2 | 3+3 2 | 2+2 2 | 3+3 2 |
| | 2+2 2 | 2+2 2 | 3+3 2 | 2+2 2 | 3+3 2 | 2+2 2 | 3+3 2 | 2+2 2 |
| | 3+3 2 | 2+2 2 | 2+2 2 | 2+2 2 | 2+2 2 | 2+2 2 | 3+3 2 | 2+2 2 |

Continued on next column

**TABLE XXI– continued**

| TQC | ICSD | NoE | SF | | | | |
|---|---|---|---|---|---|---|---|
| | 2+2 2 | 3+3 2 | 3+3 2 | 3+3 2 | 3+3 2 | 3+3 2 | 2+2 2 | 3+3 2 |
| | 3+3 2 | 3+3 2 | 3+3 2 | 3+3 2 | 3+3 2 | 3+3 2 | 3+3 2 | 3+3 2 |
| | 3+3 2 | 2+2 2 | 2+2 2 | 2+2 2 | 2+2 2 | 2+2 2 | 2+2 2 | 3+3 2 |
| | 2+2 2 | 3+3 2 | 2+2 2 | 2+2 2 | 3+3 2 | 3+3 2 | 2+2 2 | 2+2 2 |
| | 2+2 2 | 2+2 2 | 3+3 2 | 3+3 2 | 2+2 2 | 2+2 2 | 3+3 2 | 2+2 2 |
| | 3+3 2 | 2+2 2 | 3+3 2 | 3+3 2 | 2+2 2 | 2+2 2 | 3+3 2 | 3+3 2 |
| | 2+2 2 | 3+3 2 | 2+2 2 | 2+2 2 | 3+3 2 | 3+3 2 | 2+2 2 | 2+2 2 |
| | 3+3 2 | 2+2 2 | 3+3 2 | 3+3 2 | 2+2 2 | 2+2 2 | 2+2 2 | 3+3 2 |
| | 2+2 2 | 3+3 2 | 2+2 2 | 2+2 2 | 3+3 2 | 3+3 2 | 3+3 2 | 2+2 2 |
| | 3+3 2 | 2+2 2 | 3+3 2 | 3+3 2 | 3+3 2 | 2+2 2 | 3+3 2 | 2+2 2 |
| | 3+3 2 | 3+3 2 | 2+2 2 | 3+3 2 | 2+2 2 | 3+3 2 | 2+2 2 | 3+3 2 |
| | 2+2 2 | 2+2 2 | 2+2 2 | 2+2 2 | 3+3 2 | 2+2 2 | 3+3 2 | 3+3 2 |
| | 3+3 2 | 3+3 2 | 3+3 2 | 2+2 2 | 2+2 2 | 3+3 2 | 2+2 2 | 2+2 2 |
| | 2+2 2 | 3+3 2 | 2+2 2 | 2+2 2 | 2+2 2 | 2+2 2 | 2+2 2 | 3+3 2 |
| | 2+2 2 | 2+2 2 | 2+2 2 | 3+3 2 | 3+3 2 | 3+3 2 | 2+2 2 | 2+2 2 |
| | 3+3 2 | 2+2 2 | 3+3 2 | 2+2 2 | 2+2 2 | 3+3 2 | 3+3 2 | 2+2 2 |
| | 2+2 2 | 3+3 2 | 3+3 2 | 3+3 2 | 3+3 2 | 2+2 2 | 3+3 2 | 3+3 2 |
| | 3+3 2 | 3+3 2 | 3+3 2 | 2+2 2 | 3+3 2 | 3+3 2 | 3+3 2 | 2+2 2 |
| | 2+2 2 | 2+2 2 | 2+2 2 | 2+2 2 | 2+2 2 | 3+3 2 | 2+2 2 | 3+3 2 |
| | 3+3 2 | 2+2 2 | 2+2 2 | 3+3 2 | 2+2 2 | 2+2 2 | 3+3 2 | 2+2 2 |
| | 2+2 2 | 3+3 2 | 3+3 2 | 3+3 2 | 3+3 2 | 3+3 2 | 2+2 2 | 3+3 2 |
| | 2+2 2 | 3+3 2 | 3+3 2 | 2+2 2 | 2+2 2 | 2+2 2 | 2+2 2 | 3+3 2 |
| 2.2.77 | 196026 | 58 | C1Mn1O3 | | | | |
| | -GM | -R | -T | -U | -V | -X | -Y | -Z |
| | 3+3 2 | 2+2 2 | 3+3 2 | 2+2 2 | 2+2 2 | 2+2 2 | 3+3 2 | 3+3 2 |
| | 2+2 2 | 3+3 2 | 2+2 2 | 3+3 2 | 3+3 2 | 3+3 2 | 2+2 2 | 2+2 2 |
| | 3+3 2 | 3+3 2 | 2+2 2 | 2+2 2 | 2+2 2 | 2+2 2 | 2+2 2 | 3+3 2 |
| | 2+2 2 | 2+2 2 | 3+3 2 | 3+3 2 | 3+3 2 | 3+3 2 | 3+3 2 | 2+2 2 |
| | 3+3 2 | 2+2 2 | 3+3 2 | 2+2 2 | 3+3 2 | 2+2 2 | 3+3 2 | 3+3 2 |
| | 2+2 2 | 3+3 2 | 2+2 2 | 3+3 2 | 2+2 2 | 3+3 2 | 2+2 2 | 2+2 2 |
| | 3+3 2 | 2+2 2 | 2+2 2 | 3+3 2 | 2+2 2 | 3+3 2 | 2+2 2 | 2+2 2 |
| | 2+2 2 | 3+3 2 | 2+2 2 | 2+2 2 | 3+3 2 | 2+2 2 | 3+3 2 | 3+3 2 |
| | 2+2 2 | 2+2 2 | 3+3 2 | 2+2 2 | 2+2 2 | 2+2 2 | 2+2 2 | 3+3 2 |
| | 3+3 2 | 3+3 2 | 2+2 2 | 3+3 2 | 3+3 2 | 3+3 2 | 2+2 2 | 2+2 2 |
| | 3+3 2 | 3+3 2 | 3+3 2 | 2+2 2 | 3+3 2 | 3+3 2 | 3+3 2 | 3+3 2 |
| | 2+2 2 | 2+2 2 | 3+3 2 | 3+3 2 | 2+2 2 | 2+2 2 | 3+3 2 | 2+2 2 |
| | 3+3 2 | 2+2 2 | 3+3 2 | 3+3 2 | 2+2 2 | 3+3 2 | 3+3 2 | 3+3 2 |
| | 2+2 2 | 3+3 2 | 2+2 2 | 2+2 2 | 3+3 2 | 2+2 2 | 2+2 2 | 2+2 2 |
| | 3+3 2 | 3+3 2 | 3+3 2 | 3+3 2 | 3+3 2 | 2+2 2 | 3+3 2 | 2+2 2 |
| | 3+3 2 | 3+3 2 | 2+2 2 | 3+3 2 | 3+3 2 | 3+3 2 | 2+2 2 | 3+3 2 |
| | 3+3 2 | 2+2 2 | 2+2 2 | 2+2 2 | 2+2 2 | 2+2 2 | 3+3 2 | 3+3 2 |
| | 2+2 2 | 2+2 2 | 3+3 2 | 2+2 2 | 2+2 2 | 2+2 2 | 2+2 2 | 2+2 2 |
| | 3+3 2 | 3+3 2 | 3+3 2 | 3+3 2 | 2+2 2 | 3+3 2 | 3+3 2 | 3+3 2 |
| | 2+2 2 | 2+2 2 | 2+2 2 | 2+2 2 | 3+3 2 | 3+3 2 | 2+2 2 | 2+2 2 |
| | 2+2 2 | 2+2 2 | 2+2 2 | 2+2 2 | 2+2 2 | 2+2 2 | 2+2 2 | 3+3 2 |
| | 2+2 2 | 3+3 2 | 3+3 2 | 3+3 2 | 3+3 2 | 3+3 2 | 3+3 2 | 2+2 2 |
| | 3+3 2 | 2+2 2 | 3+3 2 | 2+2 2 | 3+3 2 | 3+3 2 | 3+3 2 | 3+3 2 |
| | 2+2 2 | 3+3 2 | 2+2 2 | 3+3 2 | 2+2 2 | 2+2 2 | 2+2 2 | 2+2 2 |
| | 2+2 2 | 3+3 2 | 3+3 2 | 3+3 2 | 2+2 2 | 3+3 2 | 3+3 2 | 2+2 2 |
| | 3+3 2 | 2+2 2 | 2+2 2 | 2+2 2 | 3+3 2 | 3+3 2 | 2+2 2 | 3+3 2 |
| | 3+3 2 | 2+2 2 | 2+2 2 | 2+2 2 | 2+2 2 | 2+2 2 | 3+3 2 | 2+2 2 |
| | 2+2 2 | 3+3 2 | 3+3 2 | 3+3 2 | 3+3 2 | 2+2 2 | 2+2 2 | 3+3 2 |
| | 2+2 2 | 3+3 2 | 3+3 2 | 2+2 2 | 2+2 2 | 3+3 2 | 3+3 2 | 2+2 2 |
| 2.2.78 | 196025 | 58 | C1Mn1O3 | | | | |
| | -GM | -R | -T | -U | -V | -X | -Y | -Z |
| | 3+3 2 | 2+2 2 | 3+3 2 | 2+2 2 | 2+2 2 | 2+2 2 | 3+3 2 | 3+3 2 |
| | 2+2 2 | 3+3 2 | 2+2 2 | 3+3 2 | 3+3 2 | 3+3 2 | 2+2 2 | 2+2 2 |
| | 3+3 2 | 3+3 2 | 2+2 2 | 2+2 2 | 2+2 2 | 2+2 2 | 2+2 2 | 3+3 2 |
| | 2+2 2 | 2+2 2 | 3+3 2 | 3+3 2 | 3+3 2 | 3+3 2 | 3+3 2 | 2+2 2 |
| | 3+3 2 | 2+2 2 | 3+3 2 | 2+2 2 | 2+2 2 | 2+2 2 | 3+3 2 | 3+3 2 |
| | 2+2 2 | 3+3 2 | 2+2 2 | 3+3 2 | 3+3 2 | 3+3 2 | 2+2 2 | 2+2 2 |

Continued on next column

**TABLE XXI– continued**

| TQC | ICSD | NoE | SF | | | | | |
|---|---|---|---|---|---|---|---|---|
| | 3+3 2 | 3+3 2 | 2+2 2 | 3+3 2 | 2+2 2 | 3+3 2 | 2+2 2 | 2+2 2 |
| | 2+2 2 | 2+2 2 | 2+2 2 | 2+2 2 | 2+2 2 | 2+2 2 | 3+3 2 | 3+3 2 |
| | 2+2 2 | 2+2 2 | 3+3 2 | 2+2 2 | 3+3 2 | 2+2 2 | 2+2 2 | 3+3 2 |
| | 3+3 2 | 3+3 2 | 2+2 2 | 2+2 2 | 3+3 2 | 3+3 2 | 2+2 2 | 2+2 2 |
| | 3+3 2 | 2+2 2 | 3+3 2 | 3+3 2 | 2+2 2 | 3+3 2 | 3+3 2 | 2+2 2 |
| | 3+3 2 | 3+3 2 | 3+3 2 | 3+3 2 | 3+3 2 | 2+2 2 | 3+3 2 | 3+3 2 |
| | 2+2 2 | 2+2 2 | 3+3 2 | 3+3 2 | 2+2 2 | 3+3 2 | 3+3 2 | 3+3 2 |
| | 2+2 2 | 3+3 2 | 2+2 2 | 2+2 2 | 3+3 2 | 2+2 2 | 2+2 2 | 2+2 2 |
| | 3+3 2 | 3+3 2 | 3+3 2 | 3+3 2 | 3+3 2 | 3+3 2 | 3+3 2 | 3+3 2 |
| | 3+3 2 | 3+3 2 | 2+2 2 | 2+2 2 | 3+3 2 | 2+2 2 | 2+2 2 | 2+2 2 |
| | 3+3 2 | 2+2 2 | 2+2 2 | 3+3 2 | 2+2 2 | 2+2 2 | 3+3 2 | 3+3 2 |
| | 2+2 2 | 2+2 2 | 3+3 2 | 2+2 2 | 2+2 2 | 2+2 2 | 2+2 2 | 2+2 2 |
| | 3+3 2 | 3+3 2 | 3+3 2 | 2+2 2 | 3+3 2 | 3+3 2 | 2+2 2 | 2+2 2 |
| | 2+2 2 | 2+2 2 | 2+2 2 | 3+3 2 | 2+2 2 | 2+2 2 | 3+3 2 | 3+3 2 |
| | 2+2 2 | 2+2 2 | 2+2 2 | 2+2 2 | 2+2 2 | 3+3 2 | 2+2 2 | 3+3 2 |
| | 2+2 2 | 3+3 2 | 3+3 2 | 3+3 2 | 3+3 2 | 3+3 2 | 3+3 2 | 2+2 2 |
| | 3+3 2 | 2+2 2 | 3+3 2 | 2+2 2 | 3+3 2 | 3+3 2 | 3+3 2 | 3+3 2 |
| | 2+2 2 | 3+3 2 | 2+2 2 | 3+3 2 | 2+2 2 | 2+2 2 | 2+2 2 | 2+2 2 |
| | 3+3 2 | 3+3 2 | 3+3 2 | 3+3 2 | 2+2 2 | 3+3 2 | 3+3 2 | 2+2 2 |
| | 2+2 2 | 2+2 2 | 2+2 2 | 2+2 2 | 3+3 2 | 2+2 2 | 2+2 2 | 3+3 2 |
| | 3+3 2 | 2+2 2 | 2+2 2 | 2+2 2 | 3+3 2 | 3+3 2 | 3+3 2 | 2+2 2 |
| | 2+2 2 | 3+3 2 | 3+3 2 | 3+3 2 | 2+2 2 | 3+3 2 | 2+2 2 | 2+2 2 |
| | 2+2 2 | 3+3 2 | 3+3 2 | 2+2 2 | 2+2 2 | 2+2 2 | 3+3 2 | 3+3 2 |
| 2.2.79 | 196024 | 58 | C1Mn1O3 | | | | | |
| | -GM | -R | -T | -U | -V | -X | -Y | -Z |
| | 3+3 2 | 2+2 2 | 3+3 2 | 2+2 2 | 2+2 2 | 2+2 2 | 3+3 2 | 3+3 2 |
| | 2+2 2 | 3+3 2 | 2+2 2 | 3+3 2 | 3+3 2 | 3+3 2 | 2+2 2 | 2+2 2 |
| | 3+3 2 | 3+3 2 | 2+2 2 | 3+3 2 | 2+2 2 | 2+2 2 | 2+2 2 | 2+2 2 |
| | 2+2 2 | 2+2 2 | 3+3 2 | 2+2 2 | 3+3 2 | 3+3 2 | 3+3 2 | 3+3 2 |
| | 3+3 2 | 2+2 2 | 3+3 2 | 2+2 2 | 3+3 2 | 3+3 2 | 3+3 2 | 3+3 2 |
| | 2+2 2 | 3+3 2 | 2+2 2 | 3+3 2 | 2+2 2 | 2+2 2 | 2+2 2 | 2+2 2 |
| | 3+3 2 | 3+3 2 | 2+2 2 | 3+3 2 | 2+2 2 | 3+3 2 | 2+2 2 | 3+3 2 |
| | 2+2 2 | 2+2 2 | 2+2 2 | 2+2 2 | 2+2 2 | 2+2 2 | 3+3 2 | 2+2 2 |
| | 3+3 2 | 2+2 2 | 3+3 2 | 2+2 2 | 3+3 2 | 2+2 2 | 2+2 2 | 3+3 2 |
| | 2+2 2 | 2+2 2 | 3+3 2 | 2+2 2 | 3+3 2 | 3+3 2 | 2+2 2 | 2+2 2 |
| | 3+3 2 | 3+3 2 | 2+2 2 | 3+3 2 | 2+2 2 | 3+3 2 | 3+3 2 | 3+3 2 |
| | 3+3 2 | 3+3 2 | 3+3 2 | 3+3 2 | 3+3 2 | 2+2 2 | 3+3 2 | 2+2 2 |
| | 2+2 2 | 2+2 2 | 3+3 2 | 2+2 2 | 3+3 2 | 3+3 2 | 3+3 2 | 3+3 2 |
| | 2+2 2 | 3+3 2 | 2+2 2 | 3+3 2 | 2+2 2 | 2+2 2 | 2+2 2 | 2+2 2 |
| | 3+3 2 | 3+3 2 | 3+3 2 | 3+3 2 | 3+3 2 | 3+3 2 | 3+3 2 | 2+2 2 |
| | 3+3 2 | 3+3 2 | 2+2 2 | 3+3 2 | 3+3 2 | 2+2 2 | 2+2 2 | 3+3 2 |
| | 3+3 2 | 2+2 2 | 3+3 2 | 2+2 2 | 2+2 2 | 2+2 2 | 3+3 2 | 3+3 2 |
| | 2+2 2 | 3+3 2 | 3+3 2 | 2+2 2 | 2+2 2 | 2+2 2 | 2+2 2 | 2+2 2 |
| | 2+2 2 | 2+2 2 | 2+2 2 | 3+3 2 | 2+2 2 | 3+3 2 | 3+3 2 | 2+2 2 |
| | 3+3 2 | 2+2 2 | 2+2 2 | 2+2 2 | 3+3 2 | 3+3 2 | 2+2 2 | 3+3 2 |
| | 2+2 2 | 2+2 2 | 2+2 2 | 2+2 2 | 2+2 2 | 2+2 2 | 2+2 2 | 3+3 2 |
| | 2+2 2 | 3+3 2 | 3+3 2 | 3+3 2 | 3+3 2 | 3+3 2 | 3+3 2 | 2+2 2 |
| | 3+3 2 | 2+2 2 | 3+3 2 | 2+2 2 | 3+3 2 | 3+3 2 | 2+2 2 | 2+2 2 |
| | 2+2 2 | 3+3 2 | 2+2 2 | 3+3 2 | 2+2 2 | 2+2 2 | 3+3 2 | 3+3 2 |
| | 2+2 2 | 3+3 2 | 3+3 2 | 3+3 2 | 3+3 2 | 3+3 2 | 3+3 2 | 2+2 2 |
| | 3+3 2 | 2+2 2 | 2+2 2 | 2+2 2 | 2+2 2 | 2+2 2 | 2+2 2 | 3+3 2 |
| | 3+3 2 | 2+2 2 | 2+2 2 | 3+3 2 | 2+2 2 | 3+3 2 | 3+3 2 | 2+2 2 |
| | 2+2 2 | 3+3 2 | 3+3 2 | 2+2 2 | 3+3 2 | 2+2 2 | 3+3 2 | 3+3 2 |
| | 2+2 2 | 3+3 2 | 3+3 2 | 3+3 2 | 2+2 2 | 3+3 2 | 2+2 2 | 2+2 2 |
| 2.2.80 | 423462 | 82 | Br5W1 | | | | | |
| | -GM | -R | -T | -U | -V | -X | -Y | -Z |
| | 3+3 2 | 2+2 2 | 2+2 2 | 3+3 2 | 2+2 2 | 3+3 2 | 2+2 2 | 3+3 2 |
| | 2+2 2 | 3+3 2 | 3+3 2 | 2+2 2 | 3+3 2 | 2+2 2 | 3+3 2 | 2+2 2 |
| | 3+3 2 | 3+3 2 | 3+3 2 | 2+2 2 | 3+3 2 | 2+2 2 | 3+3 2 | 2+2 2 |
| | 2+2 2 | 2+2 2 | 2+2 2 | 3+3 2 | 2+2 2 | 3+3 2 | 2+2 2 | 3+3 2 |
| | 3+3 2 | 3+3 2 | 3+3 2 | 2+2 2 | 2+2 2 | 2+2 2 | 3+3 2 | 2+2 2 |
| | 2+2 2 | 3+3 2 | 2+2 2 | 3+3 2 | 3+3 2 | 3+3 2 | 2+2 2 | 3+3 2 |
| | 3+3 2 | 2+2 2 | 2+2 2 | 3+3 2 | 3+3 2 | 3+3 2 | 3+3 2 | 3+3 2 |
| | 2+2 2 | 2+2 2 | 3+3 2 | 2+2 2 | 2+2 2 | 2+2 2 | 2+2 2 | 2+2 2 |
| | 2+2 2 | 3+3 2 | 3+3 2 | 2+2 2 | 3+3 2 | 2+2 2 | 3+3 2 | 2+2 2 |
| | 3+3 2 | 2+2 2 | 2+2 2 | 3+3 2 | 2+2 2 | 3+3 2 | 2+2 2 | 3+3 2 |
| | 3+3 2 | 3+3 2 | 3+3 2 | 2+2 2 | 3+3 2 | 2+2 2 | 3+3 2 | 3+3 2 |
| | 2+2 2 | 2+2 2 | 2+2 2 | 3+3 2 | 2+2 2 | 3+3 2 | 2+2 2 | 2+2 2 |
| | 2+2 2 | 2+2 2 | 3+3 2 | 3+3 2 | 2+2 2 | 3+3 2 | 2+2 2 | 2+2 2 |
| | 3+3 2 | 2+2 2 | 2+2 2 | 2+2 2 | 3+3 2 | 2+2 2 | 3+3 2 | 3+3 2 |
| | 3+3 2 | 3+3 2 | 3+3 2 | 3+3 2 | 2+2 2 | 2+2 2 | 3+3 2 | 3+3 2 |
| | 3+3 2 | 3+3 2 | 2+2 2 | 2+2 2 | 3+3 2 | 3+3 2 | 2+2 2 | 2+2 2 |
| | 2+2 2 | 2+2 2 | 2+2 2 | 3+3 2 | 2+2 2 | 2+2 2 | 2+2 2 | 3+3 2 |
| | 2+2 2 | 2+2 2 | 3+3 2 | 2+2 2 | 3+3 2 | 3+3 2 | 3+3 2 | 2+2 2 |
| | 2+2 2 | 3+3 2 | 2+2 2 | 3+3 2 | 2+2 2 | 3+3 2 | 2+2 2 | 3+3 2 |
| | 3+3 2 | 3+3 2 | 3+3 2 | 3+3 2 | 2+2 2 | 2+2 2 | 2+2 2 | 2+2 2 |
| | 3+3 2 | 2+2 2 | 2+2 2 | 2+2 2 | 3+3 2 | 3+3 2 | 3+3 2 | 2+2 2 |
| | 2+2 2 | 3+3 2 | 3+3 2 | 2+2 2 | 2+2 2 | 3+3 2 | 3+3 2 | 3+3 2 |
| | 3+3 2 | 2+2 2 | 2+2 2 | 3+3 2 | 3+3 2 | 2+2 2 | 3+3 2 | 2+2 2 |
| | 2+2 2 | 3+3 2 | 2+2 2 | 2+2 2 | 3+3 2 | 2+2 2 | 2+2 2 | 3+3 2 |
| | 3+3 2 | 2+2 2 | 3+3 2 | 3+3 2 | 2+2 2 | 3+3 2 | 2+2 2 | 3+3 2 |
| | 3+3 2 | 3+3 2 | 2+2 2 | 3+3 2 | 3+3 2 | 3+3 2 | 3+3 2 | 3+3 2 |
| | 2+2 2 | 2+2 2 | 3+3 2 | 2+2 2 | 2+2 2 | 2+2 2 | 2+2 2 | 2+2 2 |
| | 2+2 2 | 3+3 2 | 3+3 2 | 2+2 2 | 3+3 2 | 2+2 2 | 3+3 2 | 3+3 2 |
| | 3+3 2 | 3+3 2 | 3+3 2 | 2+2 2 | 2+2 2 | 3+3 2 | 2+2 2 | 2+2 2 |
| | 3+3 2 | 2+2 2 | 2+2 2 | 3+3 2 | 3+3 2 | 2+2 2 | 3+3 2 | 2+2 2 |
| | 3+3 2 | 3+3 2 | 2+2 2 | 2+2 2 | 2+2 2 | 2+2 2 | 3+3 2 | 3+3 2 |
| | 2+2 2 | 2+2 2 | 2+2 2 | 3+3 2 | 3+3 2 | 3+3 2 | 2+2 2 | 2+2 2 |
| | 2+2 2 | 3+3 2 | 3+3 2 | 3+3 2 | 2+2 2 | 3+3 2 | 3+3 2 | 2+2 2 |
| | 2+2 2 | 2+2 2 | 3+3 2 | 2+2 2 | 3+3 2 | 2+2 2 | 2+2 2 | 3+3 2 |
| | 3+3 2 | 2+2 2 | 3+3 2 | 3+3 2 | 2+2 2 | 3+3 2 | 2+2 2 | 2+2 2 |
| | 2+2 2 | 3+3 2 | 2+2 2 | 2+2 2 | 3+3 2 | 3+3 2 | 3+3 2 | 3+3 2 |
| | 3+3 2 | 3+3 2 | 3+3 2 | 3+3 2 | 2+2 2 | 2+2 2 | 3+3 2 | 2+2 2 |
| | 3+3 2 | 2+2 2 | 2+2 2 | 2+2 2 | 3+3 2 | 3+3 2 | 2+2 2 | 3+3 2 |
| | 2+2 2 | 3+3 2 | 3+3 2 | 3+3 2 | 2+2 2 | 2+2 2 | 3+3 2 | 3+3 2 |
| | 2+2 2 | 2+2 2 | 2+2 2 | 2+2 2 | 3+3 2 | 2+2 2 | 2+2 2 | 2+2 2 |
| | 3+3 2 | 3+3 2 | 3+3 2 | 2+2 2 | 3+3 2 | 2+2 2 | 2+2 2 | 3+3 2 |
| 2.2.81 | 39784 | 50 | Br3Ti1 | | | | | |
| | -GM | -R | -T | -U | -V | -X | -Y | -Z |
| | 3+3 2 | 3+3 2 | 3+3 2 | 2+2 2 | 3+3 2 | 2+2 2 | 3+3 2 | 3+3 2 |
| | 2+2 2 | 2+2 2 | 2+2 2 | 3+3 2 | 2+2 2 | 3+3 2 | 2+2 2 | 2+2 2 |
| | 3+3 2 | 2+2 2 | 3+3 2 | 3+3 2 | 3+3 2 | 3+3 2 | 3+3 2 | 3+3 2 |
| | 2+2 2 | 3+3 2 | 2+2 2 | 2+2 2 | 2+2 2 | 2+2 2 | 2+2 2 | 2+2 2 |
| | 2+2 2 | 2+2 2 | 3+3 2 | 2+2 2 | 3+3 2 | 2+2 2 | 2+2 2 | 3+3 2 |
| | 3+3 2 | 3+3 2 | 2+2 2 | 3+3 2 | 2+2 2 | 3+3 2 | 3+3 2 | 2+2 2 |
| | 3+3 2 | 2+2 2 | 2+2 2 | 3+3 2 | 2+2 2 | 3+3 2 | 2+2 2 | 2+2 2 |
| | 2+2 2 | 3+3 2 | 3+3 2 | 2+2 2 | 3+3 2 | 2+2 2 | 3+3 2 | 3+3 2 |
| | 3+3 2 | 3+3 2 | 3+3 2 | 2+2 2 | 3+3 2 | 2+2 2 | 3+3 2 | 2+2 2 |
| | 2+2 2 | 2+2 2 | 2+2 2 | 2+2 2 | 3+3 2 | 3+3 2 | 2+2 2 | 3+3 2 |
| | 3+3 2 | 3+3 2 | 3+3 2 | 3+3 2 | 2+2 2 | 2+2 2 | 3+3 2 | 2+2 2 |
| | 2+2 2 | 3+3 2 | 2+2 2 | 3+3 2 | 2+2 2 | 2+2 2 | 2+2 2 | 3+3 2 |
| | 3+3 2 | 2+2 2 | 3+3 2 | 2+2 2 | 3+3 2 | 3+3 2 | 3+3 2 | 3+3 2 |
| | 3+3 2 | 2+2 2 | 2+2 2 | 3+3 2 | 2+2 2 | 3+3 2 | 2+2 2 | 3+3 2 |
| | 3+3 2 | 2+2 2 | 3+3 2 | 2+2 2 | 3+3 2 | 3+3 2 | 2+2 2 | 3+3 2 |
| | 3+3 2 | 3+3 2 | 2+2 2 | 3+3 2 | 2+2 2 | 2+2 2 | 3+3 2 | 3+3 2 |
| | 2+2 2 | 2+2 2 | 3+3 2 | 2+2 2 | 2+2 2 | 3+3 2 | 2+2 2 | 2+2 2 |
| | 2+2 2 | 3+3 2 | 2+2 2 | 3+3 2 | 3+3 2 | 2+2 2 | 3+3 2 | 2+2 2 |
| | 2+2 2 | 3+3 2 | 3+3 2 | 2+2 2 | 3+3 2 | 2+2 2 | 3+3 2 | 2+2 2 |
| | 3+3 2 | 2+2 2 | 2+2 2 | 3+3 2 | 2+2 2 | 3+3 2 | 2+2 2 | 2+2 2 |
| | 2+2 2 | 3+3 2 | 3+3 2 | 2+2 2 | 3+3 2 | 2+2 2 | 3+3 2 | 3+3 2 |
| | 3+3 2 | 2+2 2 | 2+2 2 | 3+3 2 | 3+3 2 | 3+3 2 | 2+2 2 | 3+3 2 |
| | 2+2 2 | 3+3 2 | 3+3 2 | 2+2 2 | 2+2 2 | 2+2 2 | 3+3 2 | 2+2 2 |

Continued on next column

**TABLE XXI– continued**

| TQC | ICSD | NoE | SF | | | | | |
|---|---|---|---|---|---|---|---|---|
| | 2+2 2 | 2+2 2 | 2+2 2 | 3+3 2 | 2+2 2 | 3+3 2 | 2+2 2 | 2+2 2 |
| | 3+3 2 | 2+2 2 | 2+2 2 | 2+2 2 | 3+3 2 | 3+3 2 | 2+2 2 | 3+3 2 |
| 2.2.82 | 51779 | 110 | Bi3O7Sb1 | | | | | |
| | -GM | -R | -T | -U | -V | -X | -Y | -Z |
| | 3+3 2 | 2+2 2 | 2+2 2 | 2+2 2 | 3+3 2 | 3+3 2 | 3+3 2 | 2+2 2 |
| | 3+3 2 | 3+3 2 | 3+3 2 | 3+3 2 | 3+3 2 | 2+2 2 | 2+2 2 | 3+3 2 |
| | 3+3 2 | 2+2 2 | 3+3 2 | 3+3 2 | 3+3 2 | 3+3 2 | 3+3 2 | 2+2 2 |
| | 2+2 2 | 3+3 2 | 2+2 2 | 2+2 2 | 2+2 2 | 2+2 2 | 2+2 2 | 3+3 2 |
| | 2+2 2 | 3+3 2 | 2+2 2 | 3+3 2 | 2+2 2 | 2+2 2 | 3+3 2 | 2+2 2 |
| | 3+3 2 | 2+2 2 | 3+3 2 | 2+2 2 | 2+2 2 | 3+3 2 | 3+3 2 | 3+3 2 |
| | 3+3 2 | 2+2 2 | 3+3 2 | 3+3 2 | 3+3 2 | 2+2 2 | 2+2 2 | 2+2 2 |
| | 2+2 2 | 3+3 2 | 2+2 2 | 2+2 2 | 2+2 2 | 3+3 2 | 2+2 2 | 3+3 2 |
| | 2+2 2 | 2+2 2 | 3+3 2 | 3+3 2 | 3+3 2 | 2+2 2 | 3+3 2 | 3+3 2 |
| | 2+2 2 | 3+3 2 | 2+2 2 | 2+2 2 | 2+2 2 | 3+3 2 | 2+2 2 | 2+2 2 |
| | 2+2 2 | 2+2 2 | 3+3 2 | 2+2 2 | 2+2 2 | 2+2 2 | 2+2 2 | 2+2 2 |
| | 2+2 2 | 3+3 2 | 2+2 2 | 3+3 2 | 2+2 2 | 2+2 2 | 2+2 2 | 3+3 2 |
| | 2+2 2 | 2+2 2 | 2+2 2 | 2+2 2 | 3+3 2 | 3+3 2 | 2+2 2 | 2+2 2 |
| | 3+3 2 | 3+3 2 | 3+3 2 | 3+3 2 | 2+2 2 | 2+2 2 | 3+3 2 | 3+3 2 |
| | 3+3 2 | 2+2 2 | 3+3 2 | 3+3 2 | 3+3 2 | 3+3 2 | 3+3 2 | 2+2 2 |
| | 2+2 2 | 3+3 2 | 2+2 2 | 2+2 2 | 2+2 2 | 2+2 2 | 3+3 2 | 3+3 2 |
| | 3+3 2 | 2+2 2 | 2+2 2 | 3+3 2 | 2+2 2 | 3+3 2 | 2+2 2 | 3+3 2 |
| | 3+3 2 | 3+3 2 | 3+3 2 | 2+2 2 | 3+3 2 | 2+2 2 | 2+2 2 | 2+2 2 |
| | 2+2 2 | 3+3 2 | 2+2 2 | 2+2 2 | 2+2 2 | 2+2 2 | 2+2 2 | 3+3 2 |
| | 2+2 2 | 3+3 2 | 3+3 2 | 3+3 2 | 3+3 2 | 3+3 2 | 3+3 2 | 2+2 2 |
| | 3+3 2 | 2+2 2 | 2+2 2 | 2+2 2 | 2+2 2 | 2+2 2 | 3+3 2 | 2+2 2 |
| | 3+3 2 | 2+2 2 | 3+3 2 | 3+3 2 | 3+3 2 | 3+3 2 | 3+3 2 | 3+3 2 |
| | 3+3 2 | 3+3 2 | 2+2 2 | 2+2 2 | 2+2 2 | 2+2 2 | 2+2 2 | 2+2 2 |
| | 2+2 2 | 2+2 2 | 3+3 2 | 3+3 2 | 3+3 2 | 3+3 2 | 2+2 2 | 3+3 2 |
| | 2+2 2 | 2+2 2 | 3+3 2 | 2+2 2 | 3+3 2 | 2+2 2 | 3+3 2 | 3+3 2 |
| | 3+3 2 | 3+3 2 | 2+2 2 | 3+3 2 | 2+2 2 | 3+3 2 | 2+2 2 | 2+2 2 |
| | 3+3 2 | 2+2 2 | 2+2 2 | 3+3 2 | 3+3 2 | 3+3 2 | 3+3 2 | 3+3 2 |
| | 3+3 2 | 3+3 2 | 3+3 2 | 2+2 2 | 2+2 2 | 2+2 2 | 2+2 2 | 2+2 2 |
| | 2+2 2 | 2+2 2 | 3+3 2 | 3+3 2 | 3+3 2 | 3+3 2 | 3+3 2 | 3+3 2 |
| | 3+3 2 | 3+3 2 | 2+2 2 | 2+2 2 | 3+3 2 | 2+2 2 | 2+2 2 | 2+2 2 |
| | 2+2 2 | 2+2 2 | 2+2 2 | 3+3 2 | 2+2 2 | 3+3 2 | 3+3 2 | 2+2 2 |
| | 2+2 2 | 3+3 2 | 3+3 2 | 2+2 2 | 3+3 2 | 2+2 2 | 2+2 2 | 3+3 2 |
| | 3+3 2 | 3+3 2 | 3+3 2 | 2+2 2 | 3+3 2 | 3+3 2 | 3+3 2 | 3+3 2 |
| | 2+2 2 | 2+2 2 | 2+2 2 | 3+3 2 | 2+2 2 | 3+3 2 | 2+2 2 | 2+2 2 |
| | 2+2 2 | 3+3 2 | 3+3 2 | 2+2 2 | 2+2 2 | 2+2 2 | 3+3 2 | 3+3 2 |
| | 3+3 2 | 2+2 2 | 2+2 2 | 3+3 2 | 3+3 2 | 3+3 2 | 3+3 2 | 2+2 2 |
| | 2+2 2 | 3+3 2 | 3+3 2 | 3+3 2 | 3+3 2 | 2+2 2 | 2+2 2 | 3+3 2 |
| | 2+2 2 | 2+2 2 | 2+2 2 | 2+2 2 | 2+2 2 | 3+3 2 | 2+2 2 | 2+2 2 |
| | 3+3 2 | 2+2 2 | 3+3 2 | 3+3 2 | 3+3 2 | 2+2 2 | 3+3 2 | 2+2 2 |
| | 2+2 2 | 3+3 2 | 2+2 2 | 2+2 2 | 2+2 2 | 3+3 2 | 2+2 2 | 3+3 2 |
| | 3+3 2 | 3+3 2 | 2+2 2 | 3+3 2 | 2+2 2 | 2+2 2 | 3+3 2 | 3+3 2 |
| | 2+2 2 | 2+2 2 | 3+3 2 | 2+2 2 | 2+2 2 | 3+3 2 | 2+2 2 | 2+2 2 |
| | 3+3 2 | 3+3 2 | 3+3 2 | 2+2 2 | 3+3 2 | 2+2 2 | 3+3 2 | 2+2 2 |
| | 2+2 2 | 2+2 2 | 2+2 2 | 3+3 2 | 3+3 2 | 2+2 2 | 2+2 2 | 3+3 2 |
| | 2+2 2 | 3+3 2 | 3+3 2 | 3+3 2 | 2+2 2 | 3+3 2 | 3+3 2 | 2+2 2 |
| | 3+3 2 | 2+2 2 | 2+2 2 | 2+2 2 | 3+3 2 | 2+2 2 | 2+2 2 | 3+3 2 |
| | 3+3 2 | 2+2 2 | 2+2 2 | 3+3 2 | 2+2 2 | 2+2 2 | 3+3 2 | 2+2 2 |
| | 3+3 2 | 3+3 2 | 3+3 2 | 2+2 2 | 3+3 2 | 3+3 2 | 2+2 2 | 3+3 2 |
| | 2+2 2 | 2+2 2 | 3+3 2 | 3+3 2 | 2+2 2 | 2+2 2 | 3+3 2 | 2+2 2 |
| | 2+2 2 | 3+3 2 | 2+2 2 | 2+2 2 | 2+2 2 | 3+3 2 | 2+2 2 | 3+3 2 |
| | 3+3 2 | 2+2 2 | 2+2 2 | 3+3 2 | 3+3 2 | 2+2 2 | 3+3 2 | 2+2 2 |
| | 2+2 2 | 3+3 2 | 3+3 2 | 2+2 2 | 2+2 2 | 3+3 2 | 2+2 2 | 3+3 2 |
| | 3+3 2 | 2+2 2 | 3+3 2 | 3+3 2 | 2+2 2 | 2+2 2 | 3+3 2 | 2+2 2 |
| | 3+3 2 | 2+2 2 | 2+2 2 | 2+2 2 | 3+3 2 | 3+3 2 | 3+3 2 | 3+3 2 |
| | 2+2 2 | 3+3 2 | 2+2 2 | 3+3 2 | 3+3 2 | 2+2 2 | 2+2 2 | 2+2 2 |
| 2.2.83 | 97768 | 132 | Bi1Ni1O3 | | | | | |
| | -GM | -R | -T | -U | -V | -X | -Y | -Z |

Continued on next column

**TABLE XXI– continued**

| TQC | ICSD | NoE | SF | | | | | |
|---|---|---|---|---|---|---|---|---|
| | 3+3 2 | 2+2 2 | 3+3 2 | 3+3 2 | 3+3 2 | 2+2 2 | 2+2 2 | 2+2 2 |
| | 2+2 2 | 3+3 2 | 2+2 2 | 2+2 2 | 2+2 2 | 3+3 2 | 3+3 2 | 3+3 2 |
| | 3+3 2 | 2+2 2 | 3+3 2 | 3+3 2 | 2+2 2 | 3+3 2 | 3+3 2 | 2+2 2 |
| | 3+3 2 | 3+3 2 | 2+2 2 | 2+2 2 | 3+3 2 | 2+2 2 | 2+2 2 | 3+3 2 |
| | 2+2 2 | 3+3 2 | 3+3 2 | 3+3 2 | 3+3 2 | 2+2 2 | 2+2 2 | 3+3 2 |
| | 2+2 2 | 3+3 2 | 2+2 2 | 2+2 2 | 2+2 2 | 3+3 2 | 3+3 2 | 2+2 2 |
| | 3+3 2 | 2+2 2 | 3+3 2 | 3+3 2 | 3+3 2 | 3+3 2 | 3+3 2 | 3+3 2 |
| | 3+3 2 | 2+2 2 | 2+2 2 | 2+2 2 | 3+3 2 | 2+2 2 | 2+2 2 | 2+2 2 |
| | 2+2 2 | 2+2 2 | 3+3 2 | 3+3 2 | 2+2 2 | 2+2 2 | 2+2 2 | 3+3 2 |
| | 3+3 2 | 3+3 2 | 3+3 2 | 2+2 2 | 2+2 2 | 3+3 2 | 3+3 2 | 2+2 2 |
| | 2+2 2 | 2+2 2 | 2+2 2 | 2+2 2 | 3+3 2 | 3+3 2 | 3+3 2 | 3+3 2 |
| | 2+2 2 | 3+3 2 | 2+2 2 | 3+3 2 | 2+2 2 | 2+2 2 | 2+2 2 | 2+2 2 |
| | 2+2 2 | 3+3 2 | 2+2 2 | 2+2 2 | 2+2 2 | 3+3 2 | 3+3 2 | 2+2 2 |
| | 3+3 2 | 2+2 2 | 2+2 2 | 2+2 2 | 2+2 2 | 2+2 2 | 2+2 2 | 3+3 2 |
| | 2+2 2 | 3+3 2 | 3+3 2 | 3+3 2 | 3+3 2 | 3+3 2 | 3+3 2 | 3+3 2 |
| | 3+3 2 | 2+2 2 | 3+3 2 | 3+3 2 | 3+3 2 | 2+2 2 | 2+2 2 | 2+2 2 |
| | 3+3 2 | 2+2 2 | 3+3 2 | 3+3 2 | 2+2 2 | 2+2 2 | 3+3 2 | 2+2 2 |
| | 3+3 2 | 3+3 2 | 2+2 2 | 2+2 2 | 2+2 2 | 3+3 2 | 2+2 2 | 3+3 2 |
| | 2+2 2 | 2+2 2 | 2+2 2 | 3+3 2 | 3+3 2 | 3+3 2 | 3+3 2 | 2+2 2 |
| | 3+3 2 | 3+3 2 | 3+3 2 | 2+2 2 | 2+2 2 | 2+2 2 | 3+3 2 | 3+3 2 |
| | 3+3 2 | 2+2 2 | 3+3 2 | 2+2 2 | 2+2 2 | 3+3 2 | 2+2 2 | 2+2 2 |
| | 3+3 2 | 3+3 2 | 2+2 2 | 3+3 2 | 3+3 2 | 2+2 2 | 2+2 2 | 2+2 2 |
| | 2+2 2 | 2+2 2 | 3+3 2 | 3+3 2 | 3+3 2 | 3+3 2 | 2+2 2 | 3+3 2 |
| | 2+2 2 | 3+3 2 | 2+2 2 | 2+2 2 | 2+2 2 | 2+2 2 | 3+3 2 | 3+3 2 |
| | 3+3 2 | 2+2 2 | 3+3 2 | 2+2 2 | 3+3 2 | 3+3 2 | 2+2 2 | 3+3 2 |
| | 2+2 2 | 3+3 2 | 2+2 2 | 3+3 2 | 2+2 2 | 2+2 2 | 3+3 2 | 2+2 2 |
| | 3+3 2 | 3+3 2 | 2+2 2 | 3+3 2 | 3+3 2 | 2+2 2 | 3+3 2 | 3+3 2 |
| | 2+2 2 | 2+2 2 | 3+3 2 | 2+2 2 | 2+2 2 | 2+2 2 | 2+2 2 | 2+2 2 |
| | 3+3 2 | 3+3 2 | 3+3 2 | 3+3 2 | 2+2 2 | 3+3 2 | 3+3 2 | 3+3 2 |
| | 3+3 2 | 2+2 2 | 2+2 2 | 2+2 2 | 3+3 2 | 3+3 2 | 2+2 2 | 2+2 2 |
| | 3+3 2 | 3+3 2 | 3+3 2 | 3+3 2 | 2+2 2 | 3+3 2 | 3+3 2 | 2+2 2 |
| | 2+2 2 | 2+2 2 | 3+3 2 | 3+3 2 | 3+3 2 | 2+2 2 | 2+2 2 | 3+3 2 |
| | 3+3 2 | 2+2 2 | 2+2 2 | 2+2 2 | 2+2 2 | 2+2 2 | 3+3 2 | 2+2 2 |
| | 2+2 2 | 3+3 2 | 2+2 2 | 3+3 2 | 3+3 2 | 3+3 2 | 2+2 2 | 3+3 2 |
| | 3+3 2 | 2+2 2 | 3+3 2 | 2+2 2 | 2+2 2 | 3+3 2 | 3+3 2 | 3+3 2 |
| | 2+2 2 | 3+3 2 | 2+2 2 | 3+3 2 | 3+3 2 | 2+2 2 | 2+2 2 | 2+2 2 |
| | 2+2 2 | 2+2 2 | 3+3 2 | 2+2 2 | 3+3 2 | 3+3 2 | 2+2 2 | 3+3 2 |
| | 3+3 2 | 3+3 2 | 3+3 2 | 2+2 2 | 2+2 2 | 2+2 2 | 3+3 2 | 2+2 2 |
| | 2+2 2 | 3+3 2 | 2+2 2 | 3+3 2 | 3+3 2 | 3+3 2 | 3+3 2 | 3+3 2 |
| | 3+3 2 | 2+2 2 | 2+2 2 | 2+2 2 | 3+3 2 | 2+2 2 | 2+2 2 | 2+2 2 |
| | 3+3 2 | 2+2 2 | 2+2 2 | 3+3 2 | 2+2 2 | 3+3 2 | 2+2 2 | 3+3 2 |
| | 2+2 2 | 3+3 2 | 3+3 2 | 2+2 2 | 2+2 2 | 2+2 2 | 2+2 2 | 2+2 2 |
| | 3+3 2 | 3+3 2 | 3+3 2 | 3+3 2 | 3+3 2 | 2+2 2 | 3+3 2 | 3+3 2 |
| | 2+2 2 | 3+3 2 | 2+2 2 | 2+2 2 | 2+2 2 | 3+3 2 | 3+3 2 | 2+2 2 |
| | 2+2 2 | 2+2 2 | 2+2 2 | 2+2 2 | 2+2 2 | 2+2 2 | 2+2 2 | 2+2 2 |
| | 3+3 2 | 2+2 2 | 3+3 2 | 3+3 2 | 3+3 2 | 3+3 2 | 3+3 2 | 3+3 2 |
| | 2+2 2 | 3+3 2 | 3+3 2 | 2+2 2 | 2+2 2 | 2+2 2 | 2+2 2 | 3+3 2 |
| | 2+2 2 | 2+2 2 | 2+2 2 | 3+3 2 | 3+3 2 | 3+3 2 | 3+3 2 | 2+2 2 |
| | 2+2 2 | 2+2 2 | 2+2 2 | 3+3 2 | 3+3 2 | 2+2 2 | 3+3 2 | 3+3 2 |
| | 3+3 2 | 3+3 2 | 3+3 2 | 2+2 2 | 2+2 2 | 3+3 2 | 2+2 2 | 2+2 2 |
| | 3+3 2 | 3+3 2 | 2+2 2 | 3+3 2 | 2+2 2 | 3+3 2 | 3+3 2 | 3+3 2 |
| | 3+3 2 | 3+3 2 | 3+3 2 | 2+2 2 | 2+2 2 | 2+2 2 | 2+2 2 | 2+2 2 |
| | 3+3 2 | 2+2 2 | 3+3 2 | 3+3 2 | 2+2 2 | 3+3 2 | 2+2 2 | 2+2 2 |
| | 3+3 2 | 2+2 2 | 2+2 2 | 3+3 2 | 3+3 2 | 2+2 2 | 3+3 2 | 3+3 2 |
| | 3+3 2 | 3+3 2 | 2+2 2 | 3+3 2 | 2+2 2 | 2+2 2 | 2+2 2 | 2+2 2 |
| | 3+3 2 | 2+2 2 | 3+3 2 | 2+2 2 | 2+2 2 | 3+3 2 | 3+3 2 | 3+3 2 |
| | 3+3 2 | 3+3 2 | 2+2 2 | 3+3 2 | 2+2 2 | 2+2 2 | 3+3 2 | 2+2 2 |
| | 2+2 2 | 2+2 2 | 2+2 2 | 2+2 2 | 2+2 2 | 3+3 2 | 2+2 2 | 3+3 2 |
| | 3+3 2 | 3+3 2 | 3+3 2 | 2+2 2 | 2+2 2 | 2+2 2 | 2+2 2 | 2+2 2 |
| | 3+3 2 | 2+2 2 | 3+3 2 | 2+2 2 | 2+2 2 | 3+3 2 | 3+3 2 | 3+3 2 |
| | 3+3 2 | 3+3 2 | 2+2 2 | 3+3 2 | 2+2 2 | 3+3 2 | 3+3 2 | 2+2 2 |

Continued on next column

**TABLE XXI– continued**

| TQC | ICSD | NoE | SF | | | | | |
|---|---|---|---|---|---|---|---|---|
| | 3+3 2 | 2+2 2 | 3+3 2 | 2+2 2 | 2+2 2 | 2+2 2 | 2+2 2 | 3+3 2 |
| | 3+3 2 | 2+2 2 | 3+3 2 | 2+2 2 | 2+2 2 | 3+3 2 | 2+2 2 | 3+3 2 |
| | 3+3 2 | 3+3 2 | 2+2 2 | 3+3 2 | 2+2 2 | 2+2 2 | 3+3 2 | 2+2 2 |
| | 3+3 2 | 3+3 2 | 3+3 2 | 2+2 2 | 2+2 2 | 2+2 2 | 3+3 2 | 3+3 2 |
| | 3+3 2 | 2+2 2 | 2+2 2 | 2+2 2 | 2+2 2 | 3+3 2 | 2+2 2 | 2+2 2 |
| 2.2.84 | 85490 | 94 | Ba1Mo6S8 | | | | | |
| | -GM | -R | -T | -U | -V | -X | -Y | -Z |
| | 3+3 2 | 3+3 2 | 3+3 2 | 3+3 2 | 3+3 2 | 3+3 2 | 3+3 2 | 3+3 2 |
| | 2+2 2 | 2+2 2 | 2+2 2 | 2+2 2 | 2+2 2 | 2+2 2 | 2+2 2 | 2+2 2 |
| | 3+3 2 | 2+2 2 | 2+2 2 | 2+2 2 | 3+3 2 | 2+2 2 | 2+2 2 | 2+2 2 |
| | 2+2 2 | 2+2 2 | 3+3 2 | 3+3 2 | 2+2 2 | 3+3 2 | 3+3 2 | 3+3 2 |
| | 2+2 2 | 2+2 2 | 2+2 2 | 2+2 2 | 2+2 2 | 2+2 2 | 2+2 2 | 2+2 2 |
| | 2+2 2 | 3+3 2 | 3+3 2 | 3+3 2 | 3+3 2 | 3+3 2 | 3+3 2 | 3+3 2 |
| | 3+3 2 | 3+3 2 | 2+2 2 | 2+2 2 | 2+2 2 | 3+3 2 | 3+3 2 | 3+3 2 |
| | 3+3 2 | 3+3 2 | 3+3 2 | 3+3 2 | 3+3 2 | 3+3 2 | 2+2 2 | 2+2 2 |
| | 2+2 2 | 3+3 2 | 3+3 2 | 3+3 2 | 3+3 2 | 2+2 2 | 3+3 2 | 3+3 2 |
| | 3+3 2 | 2+2 2 | 2+2 2 | 2+2 2 | 2+2 2 | 2+2 2 | 2+2 2 | 2+2 2 |
| | 2+2 2 | 2+2 2 | 2+2 2 | 2+2 2 | 2+2 2 | 2+2 2 | 2+2 2 | 2+2 2 |
| | 2+2 2 | 2+2 2 | 2+2 2 | 2+2 2 | 2+2 2 | 2+2 2 | 2+2 2 | 2+2 2 |
| | 3+3 2 | 2+2 2 | 3+3 2 | 3+3 2 | 3+3 2 | 2+2 2 | 2+2 2 | 2+2 2 |
| | 3+3 2 | 2+2 2 | 3+3 2 | 3+3 2 | 2+2 2 | 3+3 2 | 3+3 2 | 3+3 2 |
| | 3+3 2 | 3+3 2 | 2+2 2 | 2+2 2 | 3+3 2 | 2+2 2 | 2+2 2 | 2+2 2 |
| | 2+2 2 | 2+2 2 | 3+3 2 | 3+3 2 | 3+3 2 | 3+3 2 | 3+3 2 | 3+3 2 |
| | 2+2 2 | 2+2 2 | 2+2 2 | 2+2 2 | 2+2 2 | 3+3 2 | 3+3 2 | 3+3 2 |
| | 2+2 2 | 3+3 2 | 3+3 2 | 3+3 2 | 3+3 2 | 2+2 2 | 2+2 2 | 2+2 2 |
| | 3+3 2 | 3+3 2 | 2+2 2 | 2+2 2 | 2+2 2 | 3+3 2 | 3+3 2 | 3+3 2 |
| | 2+2 2 | 2+2 2 | 3+3 2 | 3+3 2 | 3+3 2 | 2+2 2 | 2+2 2 | 2+2 2 |
| | 3+3 2 | 3+3 2 | 3+3 2 | 3+3 2 | 3+3 2 | 2+2 2 | 2+2 2 | 3+3 2 |
| | 2+2 2 | 2+2 2 | 2+2 2 | 2+2 2 | 2+2 2 | 3+3 2 | 3+3 2 | 2+2 2 |
| | 3+3 2 | 3+3 2 | 2+2 2 | 2+2 2 | 2+2 2 | 3+3 2 | 3+3 2 | 3+3 2 |
| | 2+2 2 | 3+3 2 | 3+3 2 | 2+2 2 | 3+3 2 | 2+2 2 | 2+2 2 | 2+2 2 |
| | 3+3 2 | 3+3 2 | 2+2 2 | 3+3 2 | 2+2 2 | 2+2 2 | 2+2 2 | 2+2 2 |
| | 3+3 2 | 2+2 2 | 3+3 2 | 3+3 2 | 3+3 2 | 3+3 2 | 3+3 2 | 3+3 2 |
| | 2+2 2 | 3+3 2 | 2+2 2 | 2+2 2 | 2+2 2 | 3+3 2 | 3+3 2 | 3+3 2 |
| | 3+3 2 | 2+2 2 | 2+2 2 | 2+2 2 | 2+2 2 | 2+2 2 | 2+2 2 | 2+2 2 |
| | 2+2 2 | 3+3 2 | 3+3 2 | 3+3 2 | 3+3 2 | 2+2 2 | 3+3 2 | 2+2 2 |
| | 2+2 2 | 2+2 2 | 2+2 2 | 2+2 2 | 2+2 2 | 3+3 2 | 2+2 2 | 3+3 2 |
| | 3+3 2 | 3+3 2 | 2+2 2 | 3+3 2 | 2+2 2 | 3+3 2 | 3+3 2 | 3+3 2 |
| | 2+2 2 | 3+3 2 | 3+3 2 | 2+2 2 | 3+3 2 | 2+2 2 | 2+2 2 | 2+2 2 |
| | 3+3 2 | 3+3 2 | 3+3 2 | 2+2 2 | 3+3 2 | 3+3 2 | 2+2 2 | 2+2 2 |
| | 2+2 2 | 2+2 2 | 2+2 2 | 3+3 2 | 2+2 2 | 2+2 2 | 3+3 2 | 3+3 2 |
| | 3+3 2 | 2+2 2 | 2+2 2 | 2+2 2 | 2+2 2 | 2+2 2 | 2+2 2 | 2+2 2 |
| | 3+3 2 | 3+3 2 | 3+3 2 | 3+3 2 | 3+3 2 | 3+3 2 | 3+3 2 | 3+3 2 |
| | 2+2 2 | 2+2 2 | 2+2 2 | 2+2 2 | 2+2 2 | 2+2 2 | 2+2 2 | 2+2 2 |
| | 3+3 2 | 2+2 2 | 3+3 2 | 3+3 2 | 3+3 2 | 3+3 2 | 3+3 2 | 3+3 2 |
| | 3+3 2 | 3+3 2 | 2+2 2 | 2+2 2 | 2+2 2 | 2+2 2 | 2+2 2 | 3+3 2 |
| | 2+2 2 | 3+3 2 | 2+2 2 | 3+3 2 | 3+3 2 | 3+3 2 | 3+3 2 | 2+2 2 |
| | 3+3 2 | 3+3 2 | 3+3 2 | 2+2 2 | 2+2 2 | 2+2 2 | 3+3 2 | 2+2 2 |
| | 2+2 2 | 2+2 2 | 3+3 2 | 3+3 2 | 3+3 2 | 3+3 2 | 2+2 2 | 3+3 2 |
| | 2+2 2 | 2+2 2 | 2+2 2 | 2+2 2 | 2+2 2 | 3+3 2 | 3+3 2 | 3+3 2 |
| | 2+2 2 | 3+3 2 | 2+2 2 | 2+2 2 | 2+2 2 | 2+2 2 | 3+3 2 | 3+3 2 |
| | 3+3 2 | 3+3 2 | 3+3 2 | 3+3 2 | 3+3 2 | 3+3 2 | 2+2 2 | 2+2 2 |
| | 3+3 2 | 2+2 2 | 2+2 2 | 2+2 2 | 2+2 2 | 2+2 2 | 2+2 2 | 2+2 2 |
| | 2+2 2 | 2+2 2 | 3+3 2 | 3+3 2 | 3+3 2 | 3+3 2 | 3+3 2 | 3+3 2 |
| 2.2.85 | 65775 | 94 | Ba1Mo6S8 | | | | | |
| | -GM | -R | -T | -U | -V | -X | -Y | -Z |
| | 3+3 2 | 3+3 2 | 3+3 2 | 3+3 2 | 3+3 2 | 3+3 2 | 3+3 2 | 3+3 2 |
| | 2+2 2 | 2+2 2 | 2+2 2 | 2+2 2 | 2+2 2 | 2+2 2 | 2+2 2 | 2+2 2 |
| | 3+3 2 | 2+2 2 | 2+2 2 | 2+2 2 | 3+3 2 | 2+2 2 | 2+2 2 | 2+2 2 |
| | 2+2 2 | 2+2 2 | 3+3 2 | 3+3 2 | 2+2 2 | 3+3 2 | 3+3 2 | 3+3 2 |
| | 2+2 2 | 2+2 2 | 2+2 2 | 2+2 2 | 2+2 2 | 2+2 2 | 2+2 2 | 2+2 2 |

Continued on next column

**TABLE XXI– continued**

| TQC | ICSD | NoE | SF | | | | | |
|---|---|---|---|---|---|---|---|---|
| | 2+2 2 | 3+3 2 | 3+3 2 | 3+3 2 | 3+3 2 | 3+3 2 | 3+3 2 | 3+3 2 |
| | 3+3 2 | 3+3 2 | 2+2 2 | 2+2 2 | 2+2 2 | 3+3 2 | 3+3 2 | 3+3 2 |
| | 3+3 2 | 3+3 2 | 3+3 2 | 3+3 2 | 3+3 2 | 3+3 2 | 2+2 2 | 3+3 2 |
| | 2+2 2 | 3+3 2 | 3+3 2 | 2+2 2 | 3+3 2 | 2+2 2 | 3+3 2 | 2+2 2 |
| | 3+3 2 | 2+2 2 | 2+2 2 | 3+3 2 | 2+2 2 | 2+2 2 | 2+2 2 | 2+2 2 |
| | 2+2 2 | 2+2 2 | 2+2 2 | 2+2 2 | 2+2 2 | 2+2 2 | 2+2 2 | 2+2 2 |
| | 2+2 2 | 2+2 2 | 2+2 2 | 2+2 2 | 2+2 2 | 2+2 2 | 2+2 2 | 2+2 2 |
| | 3+3 2 | 2+2 2 | 3+3 2 | 3+3 2 | 3+3 2 | 2+2 2 | 2+2 2 | 2+2 2 |
| | 3+3 2 | 2+2 2 | 3+3 2 | 3+3 2 | 2+2 2 | 3+3 2 | 3+3 2 | 3+3 2 |
| | 3+3 2 | 3+3 2 | 2+2 2 | 2+2 2 | 3+3 2 | 2+2 2 | 2+2 2 | 2+2 2 |
| | 2+2 2 | 2+2 2 | 3+3 2 | 3+3 2 | 3+3 2 | 3+3 2 | 3+3 2 | 3+3 2 |
| | 2+2 2 | 2+2 2 | 2+2 2 | 2+2 2 | 2+2 2 | 3+3 2 | 3+3 2 | 3+3 2 |
| | 2+2 2 | 3+3 2 | 3+3 2 | 3+3 2 | 3+3 2 | 2+2 2 | 2+2 2 | 2+2 2 |
| | 3+3 2 | 3+3 2 | 2+2 2 | 2+2 2 | 2+2 2 | 3+3 2 | 3+3 2 | 3+3 2 |
| | 2+2 2 | 2+2 2 | 3+3 2 | 3+3 2 | 3+3 2 | 2+2 2 | 2+2 2 | 2+2 2 |
| | 3+3 2 | 2+2 2 | 3+3 2 | 3+3 2 | 3+3 2 | 2+2 2 | 2+2 2 | 3+3 2 |
| | 2+2 2 | 3+3 2 | 2+2 2 | 2+2 2 | 2+2 2 | 3+3 2 | 3+3 2 | 2+2 2 |
| | 3+3 2 | 3+3 2 | 2+2 2 | 2+2 2 | 2+2 2 | 3+3 2 | 3+3 2 | 3+3 2 |
| | 2+2 2 | 3+3 2 | 3+3 2 | 3+3 2 | 3+3 2 | 2+2 2 | 2+2 2 | 2+2 2 |
| | 3+3 2 | 3+3 2 | 2+2 2 | 2+2 2 | 2+2 2 | 2+2 2 | 2+2 2 | 2+2 2 |
| | 3+3 2 | 2+2 2 | 3+3 2 | 2+2 2 | 2+2 2 | 3+3 2 | 3+3 2 | 3+3 2 |
| | 3+3 2 | 3+3 2 | 2+2 2 | 3+3 2 | 3+3 2 | 3+3 2 | 3+3 2 | 3+3 2 |
| | 2+2 2 | 3+3 2 | 2+2 2 | 2+2 2 | 2+2 2 | 2+2 2 | 2+2 2 | 2+2 2 |
| | 2+2 2 | 2+2 2 | 3+3 2 | 3+3 2 | 3+3 2 | 2+2 2 | 3+3 2 | 2+2 2 |
| | 2+2 2 | 2+2 2 | 2+2 2 | 2+2 2 | 2+2 2 | 3+3 2 | 2+2 2 | 3+3 2 |
| | 3+3 2 | 3+3 2 | 2+2 2 | 2+2 2 | 2+2 2 | 3+3 2 | 3+3 2 | 3+3 2 |
| | 2+2 2 | 3+3 2 | 3+3 2 | 3+3 2 | 3+3 2 | 2+2 2 | 2+2 2 | 2+2 2 |
| | 3+3 2 | 3+3 2 | 2+2 2 | 2+2 2 | 3+3 2 | 3+3 2 | 2+2 2 | 2+2 2 |
| | 2+2 2 | 2+2 2 | 3+3 2 | 3+3 2 | 2+2 2 | 2+2 2 | 3+3 2 | 3+3 2 |
| | 3+3 2 | 2+2 2 | 2+2 2 | 2+2 2 | 2+2 2 | 3+3 2 | 3+3 2 | 2+2 2 |
| | 3+3 2 | 2+2 2 | 3+3 2 | 3+3 2 | 3+3 2 | 2+2 2 | 2+2 2 | 3+3 2 |
| | 2+2 2 | 3+3 2 | 2+2 2 | 2+2 2 | 2+2 2 | 2+2 2 | 2+2 2 | 2+2 2 |
| | 3+3 2 | 2+2 2 | 3+3 2 | 3+3 2 | 3+3 2 | 3+3 2 | 3+3 2 | 3+3 2 |
| | 3+3 2 | 3+3 2 | 2+2 2 | 2+2 2 | 2+2 2 | 2+2 2 | 2+2 2 | 2+2 2 |
| | 3+3 2 | 3+3 2 | 3+3 2 | 3+3 2 | 3+3 2 | 3+3 2 | 3+3 2 | 3+3 2 |
| | 2+2 2 | 3+3 2 | 2+2 2 | 2+2 2 | 2+2 2 | 2+2 2 | 3+3 2 | 2+2 2 |
| | 2+2 2 | 2+2 2 | 3+3 2 | 3+3 2 | 3+3 2 | 3+3 2 | 2+2 2 | 3+3 2 |
| | 2+2 2 | 3+3 2 | 2+2 2 | 2+2 2 | 2+2 2 | 3+3 2 | 3+3 2 | 3+3 2 |
| | 2+2 2 | 2+2 2 | 2+2 2 | 2+2 2 | 2+2 2 | 3+3 2 | 3+3 2 | 3+3 2 |
| | 3+3 2 | 3+3 2 | 2+2 2 | 2+2 2 | 2+2 2 | 2+2 2 | 2+2 2 | 2+2 2 |
| | 3+3 2 | 2+2 2 | 3+3 2 | 3+3 2 | 3+3 2 | 2+2 2 | 2+2 2 | 2+2 2 |
| | 2+2 2 | 2+2 2 | 3+3 2 | 3+3 2 | 3+3 2 | 3+3 2 | 3+3 2 | 3+3 2 |
| 2.2.86 | 237763 | 84 | Ba1Fe2O8P2 | | | | | |
| | -GM | -R | -T | -U | -V | -X | -Y | -Z |
| | 3+3 2 | 3+3 2 | 3+3 2 | 3+3 2 | 3+3 2 | 3+3 2 | 3+3 2 | 3+3 2 |
| | 3+3 2 | 3+3 2 | 3+3 2 | 2+2 2 | 2+2 2 | 3+3 2 | 2+2 2 | 2+2 2 |
| | 2+2 2 | 2+2 2 | 2+2 2 | 3+3 2 | 3+3 2 | 2+2 2 | 3+3 2 | 3+3 2 |
| | 2+2 2 | 2+2 2 | 3+3 2 | 2+2 2 | 3+3 2 | 3+3 2 | 2+2 2 | 3+3 2 |
| | 3+3 2 | 3+3 2 | 2+2 2 | 3+3 2 | 2+2 2 | 2+2 2 | 3+3 2 | 2+2 2 |
| | 3+3 2 | 3+3 2 | 2+2 2 | 2+2 2 | 3+3 2 | 2+2 2 | 2+2 2 | 3+3 2 |
| | 3+3 2 | 2+2 2 | 3+3 2 | 3+3 2 | 3+3 2 | 2+2 2 | 2+2 2 | 2+2 2 |
| | 2+2 2 | 3+3 2 | 2+2 2 | 2+2 2 | 2+2 2 | 3+3 2 | 3+3 2 | 3+3 2 |
| | 2+2 2 | 2+2 2 | 3+3 2 | 3+3 2 | 2+2 2 | 3+3 2 | 3+3 2 | 2+2 2 |
| | 2+2 2 | 2+2 2 | 2+2 2 | 2+2 2 | 2+2 2 | 2+2 2 | 2+2 2 | 2+2 2 |
| | 3+3 2 | 2+2 2 | 2+2 2 | 2+2 2 | 2+2 2 | 2+2 2 | 2+2 2 | 2+2 2 |
| | 2+2 2 | 3+3 2 | 2+2 2 | 3+3 2 | 2+2 2 | 2+2 2 | 3+3 2 | 2+2 2 |
| | 2+2 2 | 2+2 2 | 3+3 2 | 2+2 2 | 2+2 2 | 2+2 2 | 2+2 2 | 2+2 2 |
| | 2+2 2 | 2+2 2 | 2+2 2 | 2+2 2 | 3+3 2 | 3+3 2 | 2+2 2 | 3+3 2 |
| | 2+2 2 | 2+2 2 | 3+3 2 | 2+2 2 | 3+3 2 | 3+3 2 | 2+2 2 | 3+3 2 |
| | 3+3 2 | 3+3 2 | 2+2 2 | 3+3 2 | 2+2 2 | 3+3 2 | 3+3 2 | 2+2 2 |
| | 3+3 2 | 3+3 2 | 3+3 2 | 2+2 2 | 2+2 2 | 2+2 2 | 2+2 2 | 3+3 2 |

Continued on next column

**TABLE XXI– continued**

| TQC | ICSD | NoE | SF | | | | |
|---|---|---|---|---|---|---|---|
| | 2+2 2 | 2+2 2 | 3+3 2 | 3+3 2 | 3+3 2 | 2+2 2 | 3+3 2 | 2+2 2 |
| | 2+2 2 | 2+2 2 | 2+2 2 | 3+3 2 | 2+2 2 | 2+2 2 | 2+2 2 | 2+2 2 |
| | 3+3 2 | 3+3 2 | 2+2 2 | 2+2 2 | 3+3 2 | 3+3 2 | 3+3 2 | 3+3 2 |
| | 3+3 2 | 3+3 2 | 3+3 2 | 3+3 2 | 2+2 2 | 3+3 2 | 3+3 2 | 3+3 2 |
| | 2+2 2 | 3+3 2 | 2+2 2 | 2+2 2 | 3+3 2 | 3+3 2 | 3+3 2 | 2+2 2 |
| | 3+3 2 | 2+2 2 | 2+2 2 | 2+2 2 | 3+3 2 | 2+2 2 | 3+3 2 | 2+2 2 |
| | 2+2 2 | 2+2 2 | 3+3 2 | 3+3 2 | 2+2 2 | 2+2 2 | 2+2 2 | 3+3 2 |
| | 3+3 2 | 3+3 2 | 3+3 2 | 3+3 2 | 2+2 2 | 3+3 2 | 2+2 2 | 2+2 2 |
| | 2+2 2 | 2+2 2 | 2+2 2 | 2+2 2 | 3+3 2 | 2+2 2 | 2+2 2 | 3+3 2 |
| | 3+3 2 | 2+2 2 | 3+3 2 | 3+3 2 | 2+2 2 | 2+2 2 | 3+3 2 | 2+2 2 |
| | 3+3 2 | 3+3 2 | 2+2 2 | 2+2 2 | 3+3 2 | 3+3 2 | 2+2 2 | 3+3 2 |
| | 2+2 2 | 3+3 2 | 3+3 2 | 3+3 2 | 3+3 2 | 3+3 2 | 2+2 2 | 3+3 2 |
| | 2+2 2 | 2+2 2 | 2+2 2 | 2+2 2 | 3+3 2 | 2+2 2 | 3+3 2 | 3+3 2 |
| | 3+3 2 | 3+3 2 | 3+3 2 | 2+2 2 | 2+2 2 | 3+3 2 | 2+2 2 | 2+2 2 |
| | 3+3 2 | 2+2 2 | 3+3 2 | 3+3 2 | 3+3 2 | 2+2 2 | 3+3 2 | 3+3 2 |
| | 3+3 2 | 3+3 2 | 2+2 2 | 3+3 2 | 2+2 2 | 2+2 2 | 2+2 2 | 2+2 2 |
| | 2+2 2 | 2+2 2 | 2+2 2 | 2+2 2 | 3+3 2 | 3+3 2 | 2+2 2 | 3+3 2 |
| | 2+2 2 | 3+3 2 | 2+2 2 | 2+2 2 | 2+2 2 | 2+2 2 | 3+3 2 | 2+2 2 |
| | 3+3 2 | 3+3 2 | 2+2 2 | 2+2 2 | 2+2 2 | 3+3 2 | 3+3 2 | 2+2 2 |
| | 2+2 2 | 2+2 2 | 2+2 2 | 3+3 2 | 3+3 2 | 2+2 2 | 3+3 2 | 2+2 2 |
| | 3+3 2 | 3+3 2 | 3+3 2 | 3+3 2 | 2+2 2 | 3+3 2 | 3+3 2 | 2+2 2 |
| | 3+3 2 | 2+2 2 | 3+3 2 | 2+2 2 | 2+2 2 | 3+3 2 | 2+2 2 | 2+2 2 |
| | 2+2 2 | 3+3 2 | 3+3 2 | 3+3 2 | 3+3 2 | 3+3 2 | 2+2 2 | 3+3 2 |
| | 2+2 2 | 2+2 2 | 2+2 2 | 2+2 2 | 2+2 2 | 2+2 2 | 3+3 2 | 3+3 2 |
| | 3+3 2 | 2+2 2 | 3+3 2 | 3+3 2 | 3+3 2 | 2+2 2 | 2+2 2 | 2+2 2 |
| 2.2.87 | 192060 | 146 | Ba1Cu1O7P2 | | | | |
| | -GM | -R | -T | -U | -V | -X | -Y | -Z |
| | 3+3 2 | 3+3 2 | 3+3 2 | 3+3 2 | 3+3 2 | 3+3 2 | 3+3 2 | 3+3 2 |
| | 2+2 2 | 2+2 2 | 2+2 2 | 2+2 2 | 2+2 2 | 2+2 2 | 2+2 2 | 2+2 2 |
| | 3+3 2 | 3+3 2 | 2+2 2 | 3+3 2 | 2+2 2 | 2+2 2 | 3+3 2 | 2+2 2 |
| | 2+2 2 | 2+2 2 | 3+3 2 | 2+2 2 | 3+3 2 | 3+3 2 | 2+2 2 | 3+3 2 |
| | 3+3 2 | 3+3 2 | 2+2 2 | 2+2 2 | 2+2 2 | 3+3 2 | 3+3 2 | 2+2 2 |
| | 2+2 2 | 2+2 2 | 3+3 2 | 3+3 2 | 3+3 2 | 2+2 2 | 2+2 2 | 3+3 2 |
| | 3+3 2 | 3+3 2 | 3+3 2 | 2+2 2 | 2+2 2 | 3+3 2 | 2+2 2 | 2+2 2 |
| | 2+2 2 | 2+2 2 | 2+2 2 | 3+3 2 | 3+3 2 | 2+2 2 | 3+3 2 | 3+3 2 |
| | 3+3 2 | 3+3 2 | 2+2 2 | 2+2 2 | 3+3 2 | 2+2 2 | 2+2 2 | 3+3 2 |
| | 2+2 2 | 2+2 2 | 3+3 2 | 3+3 2 | 2+2 2 | 3+3 2 | 3+3 2 | 2+2 2 |
| | 3+3 2 | 2+2 2 | 2+2 2 | 2+2 2 | 2+2 2 | 2+2 2 | 2+2 2 | 3+3 2 |
| | 2+2 2 | 3+3 2 | 3+3 2 | 3+3 2 | 3+3 2 | 3+3 2 | 3+3 2 | 2+2 2 |
| | 2+2 2 | 2+2 2 | 3+3 2 | 2+2 2 | 3+3 2 | 3+3 2 | 2+2 2 | 3+3 2 |
| | 3+3 2 | 3+3 2 | 2+2 2 | 3+3 2 | 2+2 2 | 2+2 2 | 3+3 2 | 2+2 2 |
| | 2+2 2 | 2+2 2 | 3+3 2 | 2+2 2 | 2+2 2 | 2+2 2 | 2+2 2 | 2+2 2 |
| | 3+3 2 | 3+3 2 | 2+2 2 | 3+3 2 | 3+3 2 | 3+3 2 | 3+3 2 | 3+3 2 |
| | 3+3 2 | 3+3 2 | 3+3 2 | 3+3 2 | 3+3 2 | 3+3 2 | 3+3 2 | 3+3 2 |
| | 2+2 2 | 2+2 2 | 2+2 2 | 2+2 2 | 2+2 2 | 2+2 2 | 2+2 2 | 2+2 2 |
| | 3+3 2 | 3+3 2 | 3+3 2 | 3+3 2 | 3+3 2 | 3+3 2 | 2+2 2 | 3+3 2 |
| | 2+2 2 | 2+2 2 | 2+2 2 | 2+2 2 | 2+2 2 | 2+2 2 | 3+3 2 | 2+2 2 |
| | 3+3 2 | 3+3 2 | 3+3 2 | 3+3 2 | 3+3 2 | 3+3 2 | 3+3 2 | 3+3 2 |
| | 2+2 2 | 2+2 2 | 2+2 2 | 2+2 2 | 2+2 2 | 2+2 2 | 2+2 2 | 2+2 2 |
| | 3+3 2 | 3+3 2 | 3+3 2 | 3+3 2 | 2+2 2 | 3+3 2 | 3+3 2 | 3+3 2 |
| | 2+2 2 | 2+2 2 | 2+2 2 | 2+2 2 | 3+3 2 | 2+2 2 | 2+2 2 | 2+2 2 |
| | 3+3 2 | 3+3 2 | 3+3 2 | 2+2 2 | 2+2 2 | 3+3 2 | 2+2 2 | 2+2 2 |
| | 2+2 2 | 2+2 2 | 2+2 2 | 3+3 2 | 3+3 2 | 2+2 2 | 3+3 2 | 3+3 2 |
| | 3+3 2 | 2+2 2 | 2+2 2 | 3+3 2 | 2+2 2 | 2+2 2 | 3+3 2 | 3+3 2 |
| | 2+2 2 | 3+3 2 | 3+3 2 | 2+2 2 | 3+3 2 | 3+3 2 | 2+2 2 | 2+2 2 |
| | 3+3 2 | 3+3 2 | 2+2 2 | 2+2 2 | 2+2 2 | 3+3 2 | 2+2 2 | 2+2 2 |
| | 3+3 2 | 3+3 2 | 3+3 2 | 3+3 2 | 2+2 2 | 2+2 2 | 3+3 2 | 3+3 2 |
| | 2+2 2 | 2+2 2 | 3+3 2 | 2+2 2 | 3+3 2 | 2+2 2 | 2+2 2 | 2+2 2 |
| | 2+2 2 | 2+2 2 | 2+2 2 | 3+3 2 | 3+3 2 | 3+3 2 | 3+3 2 | 3+3 2 |
| | 3+3 2 | 2+2 2 | 3+3 2 | 3+3 2 | 3+3 2 | 2+2 2 | 2+2 2 | 2+2 2 |
| | 2+2 2 | 3+3 2 | 2+2 2 | 2+2 2 | 2+2 2 | 3+3 2 | 3+3 2 | 3+3 2 |

Continued on next column

**TABLE XXI– continued**

| TQC | ICSD | NoE | SF | | | | |
|---|---|---|---|---|---|---|---|
| | 3+3 2 | 2+2 2 | 3+3 2 | 3+3 2 | 3+3 2 | 3+3 2 | 2+2 2 | 2+2 2 |
| | 2+2 2 | 3+3 2 | 2+2 2 | 2+2 2 | 2+2 2 | 2+2 2 | 3+3 2 | 3+3 2 |
| | 2+2 2 | 2+2 2 | 2+2 2 | 3+3 2 | 2+2 2 | 2+2 2 | 2+2 2 | 2+2 2 |
| | 3+3 2 | 3+3 2 | 3+3 2 | 2+2 2 | 3+3 2 | 3+3 2 | 3+3 2 | 3+3 2 |
| | 3+3 2 | 2+2 2 | 2+2 2 | 2+2 2 | 3+3 2 | 2+2 2 | 2+2 2 | 2+2 2 |
| | 2+2 2 | 3+3 2 | 3+3 2 | 3+3 2 | 2+2 2 | 3+3 2 | 3+3 2 | 3+3 2 |
| | 3+3 2 | 3+3 2 | 2+2 2 | 2+2 2 | 2+2 2 | 3+3 2 | 2+2 2 | 3+3 2 |
| | 2+2 2 | 2+2 2 | 3+3 2 | 3+3 2 | 3+3 2 | 2+2 2 | 3+3 2 | 2+2 2 |
| | 3+3 2 | 3+3 2 | 2+2 2 | 3+3 2 | 2+2 2 | 2+2 2 | 2+2 2 | 2+2 2 |
| | 2+2 2 | 2+2 2 | 3+3 2 | 2+2 2 | 3+3 2 | 3+3 2 | 3+3 2 | 3+3 2 |
| | 3+3 2 | 3+3 2 | 3+3 2 | 2+2 2 | 2+2 2 | 3+3 2 | 2+2 2 | 2+2 2 |
| | 2+2 2 | 2+2 2 | 2+2 2 | 3+3 2 | 3+3 2 | 2+2 2 | 3+3 2 | 3+3 2 |
| | 3+3 2 | 3+3 2 | 2+2 2 | 2+2 2 | 2+2 2 | 3+3 2 | 2+2 2 | 2+2 2 |
| | 3+3 2 | 3+3 2 | 2+2 2 | 3+3 2 | 2+2 2 | 2+2 2 | 3+3 2 | 3+3 2 |
| | 2+2 2 | 2+2 2 | 3+3 2 | 3+3 2 | 3+3 2 | 2+2 2 | 3+3 2 | 2+2 2 |
| | 3+3 2 | 2+2 2 | 3+3 2 | 2+2 2 | 3+3 2 | 3+3 2 | 2+2 2 | 3+3 2 |
| | 2+2 2 | 3+3 2 | 2+2 2 | 2+2 2 | 2+2 2 | 2+2 2 | 2+2 2 | 2+2 2 |
| | 2+2 2 | 2+2 2 | 3+3 2 | 3+3 2 | 3+3 2 | 3+3 2 | 3+3 2 | 3+3 2 |
| | 3+3 2 | 3+3 2 | 3+3 2 | 2+2 2 | 3+3 2 | 2+2 2 | 3+3 2 | 2+2 2 |
| | 3+3 2 | 2+2 2 | 2+2 2 | 3+3 2 | 2+2 2 | 3+3 2 | 2+2 2 | 3+3 2 |
| | 2+2 2 | 2+2 2 | 3+3 2 | 2+2 2 | 3+3 2 | 2+2 2 | 2+2 2 | 2+2 2 |
| | 2+2 2 | 3+3 2 | 2+2 2 | 3+3 2 | 2+2 2 | 3+3 2 | 2+2 2 | 3+3 2 |
| | 3+3 2 | 2+2 2 | 3+3 2 | 3+3 2 | 2+2 2 | 2+2 2 | 3+3 2 | 3+3 2 |
| | 2+2 2 | 3+3 2 | 2+2 2 | 2+2 2 | 3+3 2 | 3+3 2 | 3+3 2 | 2+2 2 |
| | 3+3 2 | 2+2 2 | 3+3 2 | 2+2 2 | 2+2 2 | 2+2 2 | 3+3 2 | 3+3 2 |
| | 2+2 2 | 3+3 2 | 2+2 2 | 3+3 2 | 3+3 2 | 3+3 2 | 2+2 2 | 2+2 2 |
| | 3+3 2 | 3+3 2 | 2+2 2 | 2+2 2 | 3+3 2 | 3+3 2 | 3+3 2 | 2+2 2 |
| | 2+2 2 | 2+2 2 | 3+3 2 | 3+3 2 | 2+2 2 | 2+2 2 | 2+2 2 | 3+3 2 |
| | 3+3 2 | 2+2 2 | 2+2 2 | 2+2 2 | 3+3 2 | 2+2 2 | 3+3 2 | 2+2 2 |
| | 2+2 2 | 3+3 2 | 3+3 2 | 3+3 2 | 2+2 2 | 3+3 2 | 2+2 2 | 3+3 2 |
| | 2+2 2 | 2+2 2 | 2+2 2 | 2+2 2 | 3+3 2 | 2+2 2 | 3+3 2 | 3+3 2 |
| | 3+3 2 | 3+3 2 | 3+3 2 | 3+3 2 | 2+2 2 | 3+3 2 | 2+2 2 | 2+2 2 |
| | 2+2 2 | 3+3 2 | 3+3 2 | 3+3 2 | 3+3 2 | 3+3 2 | 2+2 2 | 2+2 2 |
| | 3+3 2 | 2+2 2 | 3+3 2 | 2+2 2 | 2+2 2 | 2+2 2 | 2+2 2 | 3+3 2 |
| | 2+2 2 | 2+2 2 | 2+2 2 | 3+3 2 | 3+3 2 | 2+2 2 | 3+3 2 | 2+2 2 |
| | 3+3 2 | 3+3 2 | 2+2 2 | 2+2 2 | 2+2 2 | 3+3 2 | 3+3 2 | 3+3 2 |
| | 2+2 2 | 3+3 2 | 2+2 2 | 2+2 2 | 2+2 2 | 3+3 2 | 3+3 2 | 3+3 2 |
| | 3+3 2 | 2+2 2 | 3+3 2 | 3+3 2 | 3+3 2 | 2+2 2 | 2+2 2 | 2+2 2 |
| | 2+2 2 | 2+2 2 | 2+2 2 | 3+3 2 | 2+2 2 | 2+2 2 | 2+2 2 | 3+3 2 |
| 2.2.88 | 96562 | 108 | B2Co2O5 | | | | |
| | -GM | -R | -T | -U | -V | -X | -Y | -Z |
| | 3+3 2 | 3+3 2 | 2+2 2 | 3+3 2 | 2+2 2 | 2+2 2 | 3+3 2 | 2+2 2 |
| | 2+2 2 | 2+2 2 | 3+3 2 | 2+2 2 | 3+3 2 | 3+3 2 | 2+2 2 | 3+3 2 |
| | 3+3 2 | 2+2 2 | 3+3 2 | 3+3 2 | 2+2 2 | 3+3 2 | 2+2 2 | 2+2 2 |
| | 2+2 2 | 3+3 2 | 2+2 2 | 2+2 2 | 3+3 2 | 2+2 2 | 3+3 2 | 3+3 2 |
| | 3+3 2 | 3+3 2 | 2+2 2 | 2+2 2 | 3+3 2 | 2+2 2 | 2+2 2 | 3+3 2 |
| | 2+2 2 | 2+2 2 | 3+3 2 | 3+3 2 | 2+2 2 | 3+3 2 | 3+3 2 | 2+2 2 |
| | 2+2 2 | 2+2 2 | 3+3 2 | 3+3 2 | 2+2 2 | 3+3 2 | 3+3 2 | 3+3 2 |
| | 3+3 2 | 3+3 2 | 2+2 2 | 2+2 2 | 3+3 2 | 2+2 2 | 2+2 2 | 2+2 2 |
| | 3+3 2 | 2+2 2 | 3+3 2 | 3+3 2 | 2+2 2 | 3+3 2 | 2+2 2 | 2+2 2 |
| | 2+2 2 | 3+3 2 | 2+2 2 | 2+2 2 | 3+3 2 | 2+2 2 | 3+3 2 | 3+3 2 |
| | 3+3 2 | 3+3 2 | 3+3 2 | 3+3 2 | 2+2 2 | 3+3 2 | 3+3 2 | 2+2 2 |
| | 2+2 2 | 2+2 2 | 2+2 2 | 2+2 2 | 3+3 2 | 2+2 2 | 2+2 2 | 3+3 2 |
| | 3+3 2 | 2+2 2 | 2+2 2 | 2+2 2 | 3+3 2 | 2+2 2 | 2+2 2 | 3+3 2 |
| | 2+2 2 | 3+3 2 | 3+3 2 | 3+3 2 | 2+2 2 | 3+3 2 | 3+3 2 | 2+2 2 |
| | 2+2 2 | 2+2 2 | 2+2 2 | 3+3 2 | 2+2 2 | 3+3 2 | 2+2 2 | 3+3 2 |
| | 3+3 2 | 3+3 2 | 3+3 2 | 2+2 2 | 3+3 2 | 2+2 2 | 3+3 2 | 2+2 2 |
| | 3+3 2 | 2+2 2 | 2+2 2 | 3+3 2 | 2+2 2 | 2+2 2 | 3+3 2 | 3+3 2 |
| | 2+2 2 | 3+3 2 | 3+3 2 | 2+2 2 | 3+3 2 | 3+3 2 | 2+2 2 | 2+2 2 |
| | 2+2 2 | 2+2 2 | 3+3 2 | 3+3 2 | 2+2 2 | 3+3 2 | 2+2 2 | 3+3 2 |
| | 3+3 2 | 3+3 2 | 2+2 2 | 2+2 2 | 3+3 2 | 2+2 2 | 3+3 2 | 2+2 2 |

Continued on next column

**TABLE XXI– continued**

| TQC | ICSD | NoE | SF | | | | |
|---|---|---|---|---|---|---|---|
| | 3+3 2 | 2+2 2 | 3+3 2 | 3+3 2 | 3+3 2 | 2+2 2 | 3+3 2 | 3+3 2 |
| | 3+3 2 | 3+3 2 | 2+2 2 | 2+2 2 | 2+2 2 | 3+3 2 | 2+2 2 | 2+2 2 |
| | 2+2 2 | 3+3 2 | 3+3 2 | 2+2 2 | 2+2 2 | 2+2 2 | 2+2 2 | 2+2 2 |
| | 2+2 2 | 2+2 2 | 2+2 2 | 3+3 2 | 3+3 2 | 3+3 2 | 3+3 2 | 3+3 2 |
| | 3+3 2 | 3+3 2 | 2+2 2 | 2+2 2 | 3+3 2 | 2+2 2 | 2+2 2 | 3+3 2 |
| | 2+2 2 | 2+2 2 | 3+3 2 | 3+3 2 | 2+2 2 | 3+3 2 | 3+3 2 | 2+2 2 |
| | 3+3 2 | 3+3 2 | 2+2 2 | 3+3 2 | 2+2 2 | 3+3 2 | 2+2 2 | 3+3 2 |
| | 3+3 2 | 2+2 2 | 3+3 2 | 2+2 2 | 3+3 2 | 2+2 2 | 3+3 2 | 2+2 2 |
| | 2+2 2 | 3+3 2 | 2+2 2 | 3+3 2 | 2+2 2 | 2+2 2 | 2+2 2 | 2+2 2 |
| | 3+3 2 | 2+2 2 | 3+3 2 | 2+2 2 | 3+3 2 | 3+3 2 | 3+3 2 | 3+3 2 |
| | 2+2 2 | 3+3 2 | 2+2 2 | 3+3 2 | 2+2 2 | 3+3 2 | 3+3 2 | 3+3 2 |
| | 3+3 2 | 2+2 2 | 3+3 2 | 2+2 2 | 3+3 2 | 2+2 2 | 2+2 2 | 2+2 2 |
| | 2+2 2 | 3+3 2 | 2+2 2 | 3+3 2 | 3+3 2 | 2+2 2 | 2+2 2 | 3+3 2 |
| | 3+3 2 | 2+2 2 | 3+3 2 | 2+2 2 | 2+2 2 | 3+3 2 | 3+3 2 | 2+2 2 |
| | 2+2 2 | 2+2 2 | 2+2 2 | 2+2 2 | 3+3 2 | 2+2 2 | 2+2 2 | 3+3 2 |
| | 2+2 2 | 3+3 2 | 3+3 2 | 3+3 2 | 2+2 2 | 3+3 2 | 3+3 2 | 2+2 2 |
| | 3+3 2 | 3+3 2 | 2+2 2 | 3+3 2 | 3+3 2 | 2+2 2 | 2+2 2 | 3+3 2 |
| | 2+2 2 | 2+2 2 | 3+3 2 | 2+2 2 | 2+2 2 | 3+3 2 | 2+2 2 | 2+2 2 |
| | 3+3 2 | 2+2 2 | 3+3 2 | 2+2 2 | 3+3 2 | 3+3 2 | 3+3 2 | 2+2 2 |
| | 2+2 2 | 3+3 2 | 2+2 2 | 3+3 2 | 2+2 2 | 2+2 2 | 3+3 2 | 3+3 2 |
| | 2+2 2 | 2+2 2 | 3+3 2 | 2+2 2 | 2+2 2 | 2+2 2 | 3+3 2 | 2+2 2 |
| | 3+3 2 | 3+3 2 | 2+2 2 | 3+3 2 | 3+3 2 | 3+3 2 | 2+2 2 | 3+3 2 |
| | 3+3 2 | 2+2 2 | 2+2 2 | 3+3 2 | 2+2 2 | 3+3 2 | 3+3 2 | 3+3 2 |
| | 2+2 2 | 3+3 2 | 3+3 2 | 2+2 2 | 3+3 2 | 2+2 2 | 2+2 2 | 2+2 2 |
| | 3+3 2 | 3+3 2 | 2+2 2 | 2+2 2 | 3+3 2 | 3+3 2 | 2+2 2 | 3+3 2 |
| | 3+3 2 | 2+2 2 | 3+3 2 | 3+3 2 | 2+2 2 | 2+2 2 | 3+3 2 | 3+3 2 |
| | 2+2 2 | 3+3 2 | 2+2 2 | 2+2 2 | 2+2 2 | 2+2 2 | 2+2 2 | 2+2 2 |
| | 2+2 2 | 2+2 2 | 3+3 2 | 3+3 2 | 3+3 2 | 3+3 2 | 3+3 2 | 2+2 2 |
| | 3+3 2 | 3+3 2 | 2+2 2 | 3+3 2 | 3+3 2 | 2+2 2 | 2+2 2 | 3+3 2 |
| | 2+2 2 | 2+2 2 | 3+3 2 | 2+2 2 | 2+2 2 | 3+3 2 | 3+3 2 | 2+2 2 |
| | 2+2 2 | 2+2 2 | 2+2 2 | 3+3 2 | 3+3 2 | 2+2 2 | 3+3 2 | 3+3 2 |
| | 3+3 2 | 3+3 2 | 3+3 2 | 2+2 2 | 2+2 2 | 3+3 2 | 2+2 2 | 2+2 2 |
| | 3+3 2 | 3+3 2 | 3+3 2 | 3+3 2 | 2+2 2 | 2+2 2 | 3+3 2 | 3+3 2 |
| | 2+2 2 | 2+2 2 | 2+2 2 | 3+3 2 | 3+3 2 | 3+3 2 | 2+2 2 | 2+2 2 |
| 2.2.89 | 24284 | 108 | B2Co2O5 | | | | |
| | -GM | -R | -T | -U | -V | -X | -Y | -Z |
| | 3+3 2 | 3+3 2 | 2+2 2 | 3+3 2 | 2+2 2 | 2+2 2 | 2+2 2 | 3+3 2 |
| | 2+2 2 | 2+2 2 | 3+3 2 | 2+2 2 | 3+3 2 | 3+3 2 | 3+3 2 | 2+2 2 |
| | 2+2 2 | 2+2 2 | 3+3 2 | 3+3 2 | 2+2 2 | 3+3 2 | 3+3 2 | 2+2 2 |
| | 3+3 2 | 3+3 2 | 2+2 2 | 2+2 2 | 3+3 2 | 2+2 2 | 2+2 2 | 3+3 2 |
| | 3+3 2 | 3+3 2 | 2+2 2 | 2+2 2 | 3+3 2 | 2+2 2 | 2+2 2 | 3+3 2 |
| | 2+2 2 | 2+2 2 | 3+3 2 | 3+3 2 | 2+2 2 | 3+3 2 | 3+3 2 | 2+2 2 |
| | 2+2 2 | 2+2 2 | 2+2 2 | 3+3 2 | 2+2 2 | 3+3 2 | 3+3 2 | 2+2 2 |
| | 3+3 2 | 3+3 2 | 3+3 2 | 2+2 2 | 3+3 2 | 2+2 2 | 2+2 2 | 3+3 2 |
| | 3+3 2 | 2+2 2 | 3+3 2 | 3+3 2 | 3+3 2 | 3+3 2 | 2+2 2 | 2+2 2 |
| | 2+2 2 | 3+3 2 | 2+2 2 | 2+2 2 | 2+2 2 | 2+2 2 | 3+3 2 | 3+3 2 |
| | 3+3 2 | 3+3 2 | 3+3 2 | 3+3 2 | 3+3 2 | 3+3 2 | 3+3 2 | 2+2 2 |
| | 2+2 2 | 2+2 2 | 2+2 2 | 2+2 2 | 2+2 2 | 2+2 2 | 2+2 2 | 3+3 2 |
| | 3+3 2 | 2+2 2 | 2+2 2 | 2+2 2 | 2+2 2 | 2+2 2 | 2+2 2 | 3+3 2 |
| | 2+2 2 | 3+3 2 | 3+3 2 | 3+3 2 | 3+3 2 | 3+3 2 | 3+3 2 | 2+2 2 |
| | 2+2 2 | 2+2 2 | 2+2 2 | 3+3 2 | 2+2 2 | 3+3 2 | 3+3 2 | 2+2 2 |
| | 3+3 2 | 3+3 2 | 3+3 2 | 2+2 2 | 3+3 2 | 2+2 2 | 2+2 2 | 3+3 2 |
| | 2+2 2 | 2+2 2 | 2+2 2 | 3+3 2 | 2+2 2 | 2+2 2 | 3+3 2 | 2+2 2 |
| | 3+3 2 | 3+3 2 | 3+3 2 | 2+2 2 | 3+3 2 | 3+3 2 | 2+2 2 | 3+3 2 |
| | 2+2 2 | 2+2 2 | 3+3 2 | 3+3 2 | 2+2 2 | 3+3 2 | 3+3 2 | 3+3 2 |
| | 3+3 2 | 3+3 2 | 2+2 2 | 2+2 2 | 3+3 2 | 2+2 2 | 2+2 2 | 2+2 2 |
| | 3+3 2 | 2+2 2 | 3+3 2 | 2+2 2 | 3+3 2 | 2+2 2 | 2+2 2 | 3+3 2 |
| | 2+2 2 | 3+3 2 | 2+2 2 | 3+3 2 | 2+2 2 | 3+3 2 | 3+3 2 | 2+2 2 |
| | 3+3 2 | 3+3 2 | 3+3 2 | 3+3 2 | 2+2 2 | 2+2 2 | 3+3 2 | 2+2 2 |
| | 2+2 2 | 2+2 2 | 2+2 2 | 2+2 2 | 3+3 2 | 3+3 2 | 2+2 2 | 3+3 2 |
| | 3+3 2 | 3+3 2 | 2+2 2 | 2+2 2 | 3+3 2 | 2+2 2 | 2+2 2 | 2+2 2 |

Continued on next column

**TABLE XXI– continued**

| TQC | ICSD | NoE | SF | | | | |
|---|---|---|---|---|---|---|---|
| | 2+2 2 | 3+3 2 | 3+3 2 | 3+3 2 | 2+2 2 | 3+3 2 | 3+3 2 | 3+3 2 |
| | 3+3 2 | 2+2 2 | 2+2 2 | 3+3 2 | 2+2 2 | 3+3 2 | 3+3 2 | 2+2 2 |
| | 3+3 2 | 2+2 2 | 3+3 2 | 2+2 2 | 3+3 2 | 2+2 2 | 2+2 2 | 3+3 2 |
| | 2+2 2 | 3+3 2 | 2+2 2 | 3+3 2 | 2+2 2 | 3+3 2 | 3+3 2 | 3+3 2 |
| | 3+3 2 | 2+2 2 | 3+3 2 | 2+2 2 | 3+3 2 | 3+3 2 | 2+2 2 | 2+2 2 |
| | 2+2 2 | 3+3 2 | 2+2 2 | 2+2 2 | 2+2 2 | 2+2 2 | 2+2 2 | 3+3 2 |
| | 3+3 2 | 2+2 2 | 3+3 2 | 3+3 2 | 3+3 2 | 2+2 2 | 3+3 2 | 2+2 2 |
| | 2+2 2 | 3+3 2 | 2+2 2 | 3+3 2 | 2+2 2 | 2+2 2 | 2+2 2 | 3+3 2 |
| | 3+3 2 | 2+2 2 | 3+3 2 | 2+2 2 | 3+3 2 | 3+3 2 | 2+2 2 | 2+2 2 |
| | 2+2 2 | 2+2 2 | 2+2 2 | 2+2 2 | 3+3 2 | 2+2 2 | 3+3 2 | 2+2 2 |
| | 2+2 2 | 3+3 2 | 2+2 2 | 3+3 2 | 2+2 2 | 2+2 2 | 3+3 2 | 3+3 2 |
| | 3+3 2 | 2+2 2 | 3+3 2 | 2+2 2 | 2+2 2 | 3+3 2 | 2+2 2 | 2+2 2 |
| | 2+2 2 | 3+3 2 | 3+3 2 | 3+3 2 | 3+3 2 | 3+3 2 | 2+2 2 | 3+3 2 |
| | 3+3 2 | 3+3 2 | 3+3 2 | 2+2 2 | 2+2 2 | 3+3 2 | 3+3 2 | 2+2 2 |
| | 2+2 2 | 2+2 2 | 2+2 2 | 3+3 2 | 3+3 2 | 2+2 2 | 3+3 2 | 3+3 2 |
| | 2+2 2 | 2+2 2 | 3+3 2 | 3+3 2 | 2+2 2 | 3+3 2 | 3+3 2 | 2+2 2 |
| | 3+3 2 | 3+3 2 | 2+2 2 | 2+2 2 | 3+3 2 | 2+2 2 | 2+2 2 | 3+3 2 |
| | 3+3 2 | 3+3 2 | 2+2 2 | 2+2 2 | 2+2 2 | 2+2 2 | 2+2 2 | 3+3 2 |
| | 2+2 2 | 2+2 2 | 2+2 2 | 3+3 2 | 3+3 2 | 3+3 2 | 3+3 2 | 2+2 2 |
| | 3+3 2 | 3+3 2 | 3+3 2 | 2+2 2 | 2+2 2 | 3+3 2 | 2+2 2 | 3+3 2 |
| | 3+3 2 | 2+2 2 | 3+3 2 | 3+3 2 | 3+3 2 | 2+2 2 | 3+3 2 | 3+3 2 |
| | 2+2 2 | 3+3 2 | 2+2 2 | 2+2 2 | 3+3 2 | 2+2 2 | 2+2 2 | 2+2 2 |
| | 3+3 2 | 2+2 2 | 3+3 2 | 3+3 2 | 2+2 2 | 3+3 2 | 2+2 2 | 2+2 2 |
| | 2+2 2 | 3+3 2 | 2+2 2 | 2+2 2 | 3+3 2 | 2+2 2 | 3+3 2 | 3+3 2 |
| | 2+2 2 | 2+2 2 | 3+3 2 | 3+3 2 | 2+2 2 | 3+3 2 | 3+3 2 | 2+2 2 |
| | 2+2 2 | 2+2 2 | 2+2 2 | 3+3 2 | 3+3 2 | 2+2 2 | 3+3 2 | 3+3 2 |
| | 3+3 2 | 3+3 2 | 3+3 2 | 2+2 2 | 2+2 2 | 3+3 2 | 2+2 2 | 2+2 2 |
| | 2+2 2 | 3+3 2 | 2+2 2 | 2+2 2 | 3+3 2 | 3+3 2 | 3+3 2 | 3+3 2 |
| | 3+3 2 | 3+3 2 | 3+3 2 | 3+3 2 | 2+2 2 | 2+2 2 | 2+2 2 | 2+2 2 |
| 2.2.90 | 66662 | 116 | As3Tc2 | | | | |
| | -GM | -R | -T | -U | -V | -X | -Y | -Z |
| | 3+3 2 | 2+2 2 | 3+3 2 | 3+3 2 | 3+3 2 | 2+2 2 | 2+2 2 | 2+2 2 |
| | 3+3 2 | 3+3 2 | 3+3 2 | 2+2 2 | 2+2 2 | 3+3 2 | 3+3 2 | 3+3 2 |
| | 2+2 2 | 3+3 2 | 2+2 2 | 3+3 2 | 3+3 2 | 3+3 2 | 3+3 2 | 3+3 2 |
| | 2+2 2 | 2+2 2 | 2+2 2 | 2+2 2 | 2+2 2 | 2+2 2 | 2+2 2 | 3+3 2 |
| | 3+3 2 | 3+3 2 | 2+2 2 | 2+2 2 | 2+2 2 | 3+3 2 | 3+3 2 | 2+2 2 |
| | 2+2 2 | 2+2 2 | 3+3 2 | 3+3 2 | 3+3 2 | 2+2 2 | 3+3 2 | 3+3 2 |
| | 2+2 2 | 3+3 2 | 3+3 2 | 2+2 2 | 3+3 2 | 3+3 2 | 2+2 2 | 2+2 2 |
| | 3+3 2 | 2+2 2 | 2+2 2 | 2+2 2 | 2+2 2 | 2+2 2 | 3+3 2 | 3+3 2 |
| | 2+2 2 | 2+2 2 | 3+3 2 | 3+3 2 | 3+3 2 | 3+3 2 | 2+2 2 | 3+3 2 |
| | 3+3 2 | 3+3 2 | 2+2 2 | 3+3 2 | 2+2 2 | 2+2 2 | 3+3 2 | 2+2 2 |
| | 3+3 2 | 2+2 2 | 3+3 2 | 3+3 2 | 2+2 2 | 3+3 2 | 2+2 2 | 2+2 2 |
| | 2+2 2 | 3+3 2 | 2+2 2 | 2+2 2 | 3+3 2 | 2+2 2 | 2+2 2 | 2+2 2 |
| | 3+3 2 | 3+3 2 | 3+3 2 | 3+3 2 | 2+2 2 | 2+2 2 | 2+2 2 | 3+3 2 |
| | 3+3 2 | 2+2 2 | 2+2 2 | 2+2 2 | 3+3 2 | 3+3 2 | 2+2 2 | 2+2 2 |
| | 2+2 2 | 3+3 2 | 3+3 2 | 3+3 2 | 3+3 2 | 2+2 2 | 3+3 2 | 2+2 2 |
| | 2+2 2 | 2+2 2 | 3+3 2 | 2+2 2 | 2+2 2 | 3+3 2 | 3+3 2 | 3+3 2 |
| | 3+3 2 | 2+2 2 | 3+3 2 | 3+3 2 | 2+2 2 | 3+3 2 | 2+2 2 | 2+2 2 |
| | 3+3 2 | 3+3 2 | 2+2 2 | 2+2 2 | 3+3 2 | 2+2 2 | 2+2 2 | 2+2 2 |
| | 2+2 2 | 2+2 2 | 2+2 2 | 3+3 2 | 2+2 2 | 3+3 2 | 3+3 2 | 2+2 2 |
| | 2+2 2 | 3+3 2 | 2+2 2 | 2+2 2 | 3+3 2 | 2+2 2 | 3+3 2 | 3+3 2 |
| | 2+2 2 | 3+3 2 | 2+2 2 | 2+2 2 | 3+3 2 | 2+2 2 | 3+3 2 | 3+3 2 |
| | 3+3 2 | 2+2 2 | 3+3 2 | 3+3 2 | 2+2 2 | 3+3 2 | 2+2 2 | 3+3 2 |
| | 3+3 2 | 3+3 2 | 3+3 2 | 3+3 2 | 2+2 2 | 2+2 2 | 2+2 2 | 2+2 2 |
| | 2+2 2 | 2+2 2 | 3+3 2 | 2+2 2 | 2+2 2 | 3+3 2 | 2+2 2 | 3+3 2 |
| | 3+3 2 | 2+2 2 | 2+2 2 | 2+2 2 | 3+3 2 | 2+2 2 | 3+3 2 | 3+3 2 |
| | 2+2 2 | 3+3 2 | 2+2 2 | 3+3 2 | 3+3 2 | 3+3 2 | 3+3 2 | 3+3 2 |
| | 2+2 2 | 2+2 2 | 2+2 2 | 3+3 2 | 2+2 2 | 3+3 2 | 3+3 2 | 3+3 2 |
| | 3+3 2 | 3+3 2 | 3+3 2 | 2+2 2 | 3+3 2 | 2+2 2 | 2+2 2 | 2+2 2 |
| | 2+2 2 | 2+2 2 | 3+3 2 | 2+2 2 | 3+3 2 | 3+3 2 | 3+3 2 | 3+3 2 |
| | 2+2 2 | 3+3 2 | 2+2 2 | 2+2 2 | 2+2 2 | 2+2 2 | 2+2 2 | 2+2 2 |

Continued on next column

**TABLE XXI– continued**

| TQC | ICSD | NoE | SF | | | | |
|---|---|---|---|---|---|---|---|
| | 3+3 2 | 3+3 2 | 3+3 2 | 3+3 2 | 3+3 2 | 2+2 2 | 3+3 2 | 2+2 2 |
| | 2+2 2 | 2+2 2 | 2+2 2 | 3+3 2 | 2+2 2 | 3+3 2 | 2+2 2 | 3+3 2 |
| | 3+3 2 | 3+3 2 | 3+3 2 | 3+3 2 | 3+3 2 | 3+3 2 | 2+2 2 | 2+2 2 |
| | 2+2 2 | 2+2 2 | 3+3 2 | 2+2 2 | 2+2 2 | 2+2 2 | 3+3 2 | 2+2 2 |
| | 3+3 2 | 3+3 2 | 2+2 2 | 2+2 2 | 3+3 2 | 3+3 2 | 3+3 2 | 3+3 2 |
| | 3+3 2 | 2+2 2 | 3+3 2 | 3+3 2 | 2+2 2 | 2+2 2 | 2+2 2 | 3+3 2 |
| | 2+2 2 | 3+3 2 | 2+2 2 | 2+2 2 | 3+3 2 | 3+3 2 | 3+3 2 | 2+2 2 |
| | 2+2 2 | 2+2 2 | 2+2 2 | 2+2 2 | 2+2 2 | 2+2 2 | 2+2 2 | 2+2 2 |
| | 3+3 2 | 2+2 2 | 2+2 2 | 3+3 2 | 2+2 2 | 2+2 2 | 3+3 2 | 2+2 2 |
| | 2+2 2 | 3+3 2 | 3+3 2 | 2+2 2 | 3+3 2 | 3+3 2 | 2+2 2 | 3+3 2 |
| | 3+3 2 | 2+2 2 | 3+3 2 | 3+3 2 | 3+3 2 | 2+2 2 | 3+3 2 | 3+3 2 |
| | 2+2 2 | 3+3 2 | 2+2 2 | 3+3 2 | 2+2 2 | 3+3 2 | 3+3 2 | 2+2 2 |
| | 3+3 2 | 2+2 2 | 2+2 2 | 3+3 2 | 3+3 2 | 2+2 2 | 2+2 2 | 3+3 2 |
| | 3+3 2 | 3+3 2 | 3+3 2 | 2+2 2 | 2+2 2 | 3+3 2 | 3+3 2 | 2+2 2 |
| | 3+3 2 | 3+3 2 | 2+2 2 | 2+2 2 | 3+3 2 | 2+2 2 | 3+3 2 | 3+3 2 |
| | 2+2 2 | 2+2 2 | 3+3 2 | 3+3 2 | 2+2 2 | 3+3 2 | 2+2 2 | 2+2 2 |
| | 3+3 2 | 3+3 2 | 2+2 2 | 3+3 2 | 2+2 2 | 2+2 2 | 3+3 2 | 3+3 2 |
| | 2+2 2 | 2+2 2 | 3+3 2 | 2+2 2 | 3+3 2 | 3+3 2 | 3+3 2 | 3+3 2 |
| | 3+3 2 | 2+2 2 | 2+2 2 | 2+2 2 | 3+3 2 | 3+3 2 | 2+2 2 | 2+2 2 |
| | 2+2 2 | 3+3 2 | 2+2 2 | 3+3 2 | 2+2 2 | 2+2 2 | 3+3 2 | 3+3 2 |
| | 3+3 2 | 3+3 2 | 3+3 2 | 2+2 2 | 3+3 2 | 3+3 2 | 2+2 2 | 2+2 2 |
| | 3+3 2 | 2+2 2 | 2+2 2 | 3+3 2 | 2+2 2 | 2+2 2 | 2+2 2 | 3+3 2 |
| | 3+3 2 | 2+2 2 | 2+2 2 | 2+2 2 | 3+3 2 | 3+3 2 | 2+2 2 | 3+3 2 |
| | 3+3 2 | 3+3 2 | 3+3 2 | 3+3 2 | 2+2 2 | 2+2 2 | 2+2 2 | 3+3 2 |
| | 2+2 2 | 3+3 2 | 2+2 2 | 2+2 2 | 2+2 2 | 2+2 2 | 3+3 2 | 2+2 2 |
| | 3+3 2 | 2+2 2 | 3+3 2 | 3+3 2 | 3+3 2 | 3+3 2 | 3+3 2 | 3+3 2 |
| | 2+2 2 | 2+2 2 | 2+2 2 | 2+2 2 | 3+3 2 | 3+3 2 | 3+3 2 | 2+2 2 |
| | 3+3 2 | 3+3 2 | 3+3 2 | 3+3 2 | 2+2 2 | 2+2 2 | 2+2 2 | 2+2 2 |
| 2.2.91 | 86400 | 50 | As3Ca1 | | | | |
| | -GM | -R | -T | -U | -V | -X | -Y | -Z |
| | 3+3 2 | 2+2 2 | 3+3 2 | 2+2 2 | 2+2 2 | 2+2 2 | 2+2 2 | 2+2 2 |
| | 2+2 2 | 3+3 2 | 2+2 2 | 3+3 2 | 3+3 2 | 3+3 2 | 3+3 2 | 3+3 2 |
| | 3+3 2 | 3+3 2 | 3+3 2 | 3+3 2 | 2+2 2 | 2+2 2 | 2+2 2 | 2+2 2 |
| | 2+2 2 | 2+2 2 | 2+2 2 | 2+2 2 | 3+3 2 | 3+3 2 | 3+3 2 | 3+3 2 |
| | 3+3 2 | 2+2 2 | 3+3 2 | 3+3 2 | 2+2 2 | 2+2 2 | 2+2 2 | 2+2 2 |
| | 2+2 2 | 3+3 2 | 2+2 2 | 2+2 2 | 3+3 2 | 3+3 2 | 3+3 2 | 3+3 2 |
| | 3+3 2 | 3+3 2 | 2+2 2 | 2+2 2 | 2+2 2 | 2+2 2 | 2+2 2 | 2+2 2 |
| | 2+2 2 | 2+2 2 | 3+3 2 | 3+3 2 | 3+3 2 | 3+3 2 | 3+3 2 | 3+3 2 |
| | 3+3 2 | 2+2 2 | 3+3 2 | 3+3 2 | 3+3 2 | 3+3 2 | 3+3 2 | 3+3 2 |
| | 3+3 2 | 3+3 2 | 2+2 2 | 2+2 2 | 2+2 2 | 2+2 2 | 2+2 2 | 2+2 2 |
| | 2+2 2 | 2+2 2 | 2+2 2 | 3+3 2 | 3+3 2 | 3+3 2 | 3+3 2 | 3+3 2 |
| | 3+3 2 | 3+3 2 | 2+2 2 | 2+2 2 | 2+2 2 | 3+3 2 | 2+2 2 | 2+2 2 |
| | 2+2 2 | 2+2 2 | 3+3 2 | 3+3 2 | 3+3 2 | 2+2 2 | 3+3 2 | 3+3 2 |
| | 2+2 2 | 3+3 2 | 3+3 2 | 2+2 2 | 2+2 2 | 2+2 2 | 2+2 2 | 2+2 2 |
| | 3+3 2 | 3+3 2 | 3+3 2 | 2+2 2 | 2+2 2 | 2+2 2 | 3+3 2 | 3+3 2 |
| | 2+2 2 | 2+2 2 | 2+2 2 | 3+3 2 | 3+3 2 | 2+2 2 | 2+2 2 | 2+2 2 |
| | 2+2 2 | 2+2 2 | 2+2 2 | 3+3 2 | 3+3 2 | 3+3 2 | 2+2 2 | 2+2 2 |
| | 3+3 2 | 3+3 2 | 3+3 2 | 2+2 2 | 2+2 2 | 3+3 2 | 3+3 2 | 3+3 2 |
| | 3+3 2 | 3+3 2 | 2+2 2 | 3+3 2 | 3+3 2 | 3+3 2 | 2+2 2 | 3+3 2 |
| | 2+2 2 | 2+2 2 | 3+3 2 | 2+2 2 | 2+2 2 | 2+2 2 | 3+3 2 | 2+2 2 |
| | 3+3 2 | 3+3 2 | 2+2 2 | 2+2 2 | 2+2 2 | 3+3 2 | 2+2 2 | 2+2 2 |
| | 3+3 2 | 3+3 2 | 3+3 2 | 3+3 2 | 3+3 2 | 2+2 2 | 3+3 2 | 3+3 2 |
| | 2+2 2 | 2+2 2 | 3+3 2 | 3+3 2 | 3+3 2 | 3+3 2 | 3+3 2 | 3+3 2 |
| | 3+3 2 | 2+2 2 | 2+2 2 | 2+2 2 | 3+3 2 | 3+3 2 | 2+2 2 | 3+3 2 |
| | 3+3 2 | 2+2 2 | 2+2 2 | 3+3 2 | 2+2 2 | 2+2 2 | 3+3 2 | 2+2 2 |
| 2.2.92 | 193 | 34 | As3Ca1 | | | | |
| | -GM | -R | -T | -U | -V | -X | -Y | -Z |
| | 3+3 2 | 3+3 2 | 3+3 2 | 3+3 2 | 3+3 2 | 3+3 2 | 3+3 2 | 3+3 2 |
| | 3+3 2 | 2+2 2 | 2+2 2 | 2+2 2 | 2+2 2 | 2+2 2 | 2+2 2 | 2+2 2 |
| | 2+2 2 | 2+2 2 | 2+2 2 | 3+3 2 | 3+3 2 | 3+3 2 | 3+3 2 | 3+3 2 |
| | 3+3 2 | 3+3 2 | 3+3 2 | 2+2 2 | 2+2 2 | 3+3 2 | 2+2 2 | 2+2 2 |

Continued on next column

**TABLE XXI– continued**

| TQC | ICSD | NoE | SF | | | | |
|---|---|---|---|---|---|---|---|
| | 2+2 2 | 2+2 2 | 2+2 2 | 3+3 2 | 3+3 2 | 2+2 2 | 3+3 2 | 3+3 2 |
| | 2+2 2 | 3+3 2 | 3+3 2 | 2+2 2 | 2+2 2 | 2+2 2 | 2+2 2 | 2+2 2 |
| | 3+3 2 | 3+3 2 | 3+3 2 | 2+2 2 | 2+2 2 | 2+2 2 | 2+2 2 | 3+3 2 |
| | 2+2 2 | 2+2 2 | 2+2 2 | 3+3 2 | 3+3 2 | 3+3 2 | 3+3 2 | 2+2 2 |
| | 2+2 2 | 2+2 2 | 2+2 2 | 3+3 2 | 3+3 2 | 2+2 2 | 3+3 2 | 3+3 2 |
| | 3+3 2 | 3+3 2 | 3+3 2 | 2+2 2 | 2+2 2 | 3+3 2 | 2+2 2 | 2+2 2 |
| | 3+3 2 | 3+3 2 | 2+2 2 | 3+3 2 | 3+3 2 | 2+2 2 | 3+3 2 | 3+3 2 |
| | 2+2 2 | 2+2 2 | 3+3 2 | 2+2 2 | 2+2 2 | 3+3 2 | 2+2 2 | 2+2 2 |
| | 2+2 2 | 3+3 2 | 2+2 2 | 2+2 2 | 2+2 2 | 3+3 2 | 2+2 2 | 2+2 2 |
| | 3+3 2 | 2+2 2 | 3+3 2 | 3+3 2 | 3+3 2 | 2+2 2 | 3+3 2 | 3+3 2 |
| | 3+3 2 | 3+3 2 | 3+3 2 | 3+3 2 | 3+3 2 | 3+3 2 | 3+3 2 | 3+3 2 |
| | 3+3 2 | 2+2 2 | 2+2 2 | 2+2 2 | 3+3 2 | 3+3 2 | 2+2 2 | 3+3 2 |
| | 3+3 2 | 2+2 2 | 2+2 2 | 3+3 2 | 2+2 2 | 2+2 2 | 3+3 2 | 2+2 2 |
| 2.2.93 | 404850 | 216 | As2Cu4O9 | | | | |
| | -GM | -R | -T | -U | -V | -X | -Y | -Z |
| | 3+3 2 | 3+3 2 | 3+3 2 | 2+2 2 | 3+3 2 | 2+2 2 | 2+2 2 | 3+3 2 |
| | 2+2 2 | 3+3 2 | 2+2 2 | 3+3 2 | 2+2 2 | 3+3 2 | 3+3 2 | 2+2 2 |
| | 3+3 2 | 2+2 2 | 2+2 2 | 2+2 2 | 3+3 2 | 2+2 2 | 2+2 2 | 3+3 2 |
| | 2+2 2 | 2+2 2 | 3+3 2 | 3+3 2 | 2+2 2 | 3+3 2 | 3+3 2 | 2+2 2 |
| | 2+2 2 | 3+3 2 | 3+3 2 | 3+3 2 | 2+2 2 | 3+3 2 | 3+3 2 | 3+3 2 |
| | 3+3 2 | 2+2 2 | 2+2 2 | 2+2 2 | 3+3 2 | 2+2 2 | 3+3 2 | 2+2 2 |
| | 2+2 2 | 2+2 2 | 3+3 2 | 3+3 2 | 2+2 2 | 3+3 2 | 2+2 2 | 3+3 2 |
| | 3+3 2 | 3+3 2 | 2+2 2 | 3+3 2 | 3+3 2 | 3+3 2 | 2+2 2 | 2+2 2 |
| | 3+3 2 | 3+3 2 | 2+2 2 | 2+2 2 | 3+3 2 | 2+2 2 | 2+2 2 | 3+3 2 |
| | 2+2 2 | 2+2 2 | 3+3 2 | 2+2 2 | 2+2 2 | 2+2 2 | 3+3 2 | 2+2 2 |
| | 3+3 2 | 3+3 2 | 2+2 2 | 3+3 2 | 2+2 2 | 3+3 2 | 3+3 2 | 2+2 2 |
| | 2+2 2 | 2+2 2 | 3+3 2 | 2+2 2 | 3+3 2 | 2+2 2 | 3+3 2 | 3+3 2 |
| | 2+2 2 | 2+2 2 | 2+2 2 | 3+3 2 | 3+3 2 | 3+3 2 | 2+2 2 | 2+2 2 |
| | 3+3 2 | 3+3 2 | 3+3 2 | 2+2 2 | 2+2 2 | 2+2 2 | 2+2 2 | 3+3 2 |
| | 2+2 2 | 2+2 2 | 3+3 2 | 2+2 2 | 3+3 2 | 2+2 2 | 3+3 2 | 2+2 2 |
| | 3+3 2 | 3+3 2 | 2+2 2 | 3+3 2 | 2+2 2 | 3+3 2 | 2+2 2 | 3+3 2 |
| | 3+3 2 | 2+2 2 | 3+3 2 | 3+3 2 | 2+2 2 | 3+3 2 | 2+2 2 | 2+2 2 |
| | 2+2 2 | 3+3 2 | 2+2 2 | 2+2 2 | 3+3 2 | 2+2 2 | 3+3 2 | 3+3 2 |
| | 3+3 2 | 3+3 2 | 3+3 2 | 3+3 2 | 3+3 2 | 3+3 2 | 3+3 2 | 3+3 2 |
| | 3+3 2 | 2+2 2 | 2+2 2 | 2+2 2 | 2+2 2 | 2+2 2 | 2+2 2 | 2+2 2 |
| | 2+2 2 | 3+3 2 | 3+3 2 | 2+2 2 | 3+3 2 | 2+2 2 | 2+2 2 | 3+3 2 |
| | 2+2 2 | 2+2 2 | 2+2 2 | 3+3 2 | 2+2 2 | 3+3 2 | 3+3 2 | 2+2 2 |
| | 3+3 2 | 3+3 2 | 2+2 2 | 3+3 2 | 3+3 2 | 3+3 2 | 3+3 2 | 3+3 2 |
| | 3+3 2 | 2+2 2 | 3+3 2 | 2+2 2 | 2+2 2 | 2+2 2 | 2+2 2 | 3+3 2 |
| | 2+2 2 | 3+3 2 | 2+2 2 | 3+3 2 | 2+2 2 | 2+2 2 | 2+2 2 | 2+2 2 |
| | 2+2 2 | 3+3 2 | 3+3 2 | 2+2 2 | 3+3 2 | 3+3 2 | 3+3 2 | 2+2 2 |
| | 3+3 2 | 2+2 2 | 2+2 2 | 3+3 2 | 3+3 2 | 2+2 2 | 2+2 2 | 3+3 2 |
| | 2+2 2 | 2+2 2 | 3+3 2 | 3+3 2 | 2+2 2 | 3+3 2 | 3+3 2 | 2+2 2 |
| | 3+3 2 | 3+3 2 | 2+2 2 | 2+2 2 | 2+2 2 | 2+2 2 | 3+3 2 | 3+3 2 |
| | 2+2 2 | 3+3 2 | 3+3 2 | 2+2 2 | 3+3 2 | 3+3 2 | 2+2 2 | 3+3 2 |
| | 2+2 2 | 2+2 2 | 2+2 2 | 3+3 2 | 3+3 2 | 2+2 2 | 2+2 2 | 2+2 2 |
| | 3+3 2 | 2+2 2 | 3+3 2 | 2+2 2 | 3+3 2 | 3+3 2 | 3+3 2 | 2+2 2 |
| | 3+3 2 | 3+3 2 | 2+2 2 | 3+3 2 | 2+2 2 | 3+3 2 | 2+2 2 | 2+2 2 |
| | 2+2 2 | 2+2 2 | 3+3 2 | 2+2 2 | 2+2 2 | 2+2 2 | 3+3 2 | 3+3 2 |
| | 2+2 2 | 3+3 2 | 3+3 2 | 2+2 2 | 2+2 2 | 2+2 2 | 3+3 2 | 2+2 2 |
| | 3+3 2 | 2+2 2 | 3+3 2 | 3+3 2 | 3+3 2 | 3+3 2 | 2+2 2 | 2+2 2 |
| | 2+2 2 | 2+2 2 | 2+2 2 | 3+3 2 | 2+2 2 | 3+3 2 | 3+3 2 | 3+3 2 |
| | 3+3 2 | 3+3 2 | 2+2 2 | 2+2 2 | 3+3 2 | 2+2 2 | 2+2 2 | 3+3 2 |
| | 2+2 2 | 3+3 2 | 3+3 2 | 3+3 2 | 2+2 2 | 3+3 2 | 2+2 2 | 2+2 2 |
| | 3+3 2 | 2+2 2 | 3+3 2 | 2+2 2 | 3+3 2 | 2+2 2 | 3+3 2 | 3+3 2 |
| | 2+2 2 | 2+2 2 | 2+2 2 | 2+2 2 | 2+2 2 | 3+3 2 | 3+3 2 | 2+2 2 |
| | 3+3 2 | 3+3 2 | 2+2 2 | 3+3 2 | 3+3 2 | 2+2 2 | 2+2 2 | 2+2 2 |
| | 2+2 2 | 2+2 2 | 3+3 2 | 2+2 2 | 3+3 2 | 2+2 2 | 2+2 2 | 3+3 2 |
| | 3+3 2 | 3+3 2 | 2+2 2 | 3+3 2 | 3+3 2 | 2+2 2 | 3+3 2 | 3+3 2 |
| | 3+3 2 | 2+2 2 | 2+2 2 | 3+3 2 | 2+2 2 | 3+3 2 | 3+3 2 | 3+3 2 |
| | 2+2 2 | 3+3 2 | 3+3 2 | 2+2 2 | 2+2 2 | 3+3 2 | 3+3 2 | 3+3 2 |

Continued on next column

**TABLE XXI– continued**

| TQC | ICSD | NoE | SF | | | | | |
|---|---|---|---|---|---|---|---|---|
| | 2+2 2 | 3+3 2 | 3+3 2 | 3+3 2 | 3+3 2 | 3+3 2 | 2+2 2 | 2+2 2 |
| | 3+3 2 | 2+2 2 | 3+3 2 | 2+2 2 | 2+2 2 | 2+2 2 | 3+3 2 | 2+2 2 |
| | 3+3 2 | 3+3 2 | 2+2 2 | 2+2 2 | 2+2 2 | 3+3 2 | 2+2 2 | 3+3 2 |
| | 3+3 2 | 2+2 2 | 2+2 2 | 3+3 2 | 3+3 2 | 3+3 2 | 2+2 2 | 2+2 2 |
| | 3+3 2 | 3+3 2 | 3+3 2 | 2+2 2 | 2+2 2 | 2+2 2 | 3+3 2 | 3+3 2 |
| | 2+2 2 | 3+3 2 | 2+2 2 | 3+3 2 | 3+3 2 | 2+2 2 | 2+2 2 | 2+2 2 |
| | 2+2 2 | 2+2 2 | 3+3 2 | 3+3 2 | 3+3 2 | 2+2 2 | 2+2 2 | 2+2 2 |
| | 3+3 2 | 2+2 2 | 2+2 2 | 2+2 2 | 2+2 2 | 3+3 2 | 3+3 2 | 3+3 2 |
| | 2+2 2 | 3+3 2 | 3+3 2 | 2+2 2 | 3+3 2 | 3+3 2 | 2+2 2 | 2+2 2 |
| | 3+3 2 | 2+2 2 | 2+2 2 | 3+3 2 | 2+2 2 | 2+2 2 | 3+3 2 | 3+3 2 |
| | 2+2 2 | 3+3 2 | 2+2 2 | 2+2 2 | 2+2 2 | 2+2 2 | 2+2 2 | 3+3 2 |
| | 2+2 2 | 2+2 2 | 3+3 2 | 3+3 2 | 3+3 2 | 3+3 2 | 3+3 2 | 2+2 2 |
| | 3+3 2 | 3+3 2 | 2+2 2 | 3+3 2 | 2+2 2 | 2+2 2 | 2+2 2 | 2+2 2 |
| | 2+2 2 | 2+2 2 | 3+3 2 | 2+2 2 | 3+3 2 | 3+3 2 | 3+3 2 | 3+3 2 |
| | 3+3 2 | 2+2 2 | 2+2 2 | 3+3 2 | 3+3 2 | 2+2 2 | 3+3 2 | 3+3 2 |
| | 2+2 2 | 3+3 2 | 3+3 2 | 2+2 2 | 2+2 2 | 3+3 2 | 2+2 2 | 2+2 2 |
| | 3+3 2 | 2+2 2 | 3+3 2 | 3+3 2 | 2+2 2 | 3+3 2 | 3+3 2 | 2+2 2 |
| | 2+2 2 | 3+3 2 | 2+2 2 | 2+2 2 | 3+3 2 | 2+2 2 | 2+2 2 | 3+3 2 |
| | 2+2 2 | 2+2 2 | 3+3 2 | 2+2 2 | 2+2 2 | 3+3 2 | 3+3 2 | 2+2 2 |
| | 3+3 2 | 2+2 2 | 2+2 2 | 3+3 2 | 3+3 2 | 3+3 2 | 3+3 2 | 2+2 2 |
| | 3+3 2 | 3+3 2 | 3+3 2 | 3+3 2 | 2+2 2 | 2+2 2 | 2+2 2 | 3+3 2 |
| | 2+2 2 | 3+3 2 | 2+2 2 | 2+2 2 | 3+3 2 | 2+2 2 | 3+3 2 | 2+2 2 |
| | 2+2 2 | 2+2 2 | 3+3 2 | 3+3 2 | 2+2 2 | 3+3 2 | 2+2 2 | 3+3 2 |
| | 3+3 2 | 3+3 2 | 2+2 2 | 2+2 2 | 3+3 2 | 2+2 2 | 3+3 2 | 3+3 2 |
| | 2+2 2 | 3+3 2 | 3+3 2 | 3+3 2 | 3+3 2 | 2+2 2 | 2+2 2 | 2+2 2 |
| | 3+3 2 | 2+2 2 | 2+2 2 | 2+2 2 | 2+2 2 | 3+3 2 | 2+2 2 | 3+3 2 |
| | 2+2 2 | 2+2 2 | 3+3 2 | 3+3 2 | 3+3 2 | 2+2 2 | 3+3 2 | 3+3 2 |
| | 3+3 2 | 3+3 2 | 2+2 2 | 2+2 2 | 2+2 2 | 3+3 2 | 2+2 2 | 2+2 2 |
| | 3+3 2 | 3+3 2 | 2+2 2 | 3+3 2 | 3+3 2 | 3+3 2 | 3+3 2 | 3+3 2 |
| | 3+3 2 | 3+3 2 | 3+3 2 | 2+2 2 | 2+2 2 | 2+2 2 | 2+2 2 | 2+2 2 |
| | 2+2 2 | 2+2 2 | 3+3 2 | 3+3 2 | 3+3 2 | 2+2 2 | 3+3 2 | 3+3 2 |
| | 2+2 2 | 2+2 2 | 2+2 2 | 2+2 2 | 2+2 2 | 3+3 2 | 2+2 2 | 2+2 2 |
| | 3+3 2 | 3+3 2 | 3+3 2 | 3+3 2 | 3+3 2 | 3+3 2 | 2+2 2 | 3+3 2 |
| | 3+3 2 | 2+2 2 | 2+2 2 | 2+2 2 | 2+2 2 | 2+2 2 | 3+3 2 | 2+2 2 |
| | 2+2 2 | 3+3 2 | 2+2 2 | 3+3 2 | 2+2 2 | 2+2 2 | 3+3 2 | 3+3 2 |
| | 3+3 2 | 2+2 2 | 3+3 2 | 2+2 2 | 3+3 2 | 2+2 2 | 2+2 2 | 2+2 2 |
| | 2+2 2 | 3+3 2 | 2+2 2 | 3+3 2 | 3+3 2 | 3+3 2 | 3+3 2 | 2+2 2 |
| | 3+3 2 | 2+2 2 | 3+3 2 | 2+2 2 | 2+2 2 | 3+3 2 | 3+3 2 | 3+3 2 |
| | 2+2 2 | 3+3 2 | 2+2 2 | 3+3 2 | 2+2 2 | 2+2 2 | 2+2 2 | 2+2 2 |
| | 3+3 2 | 3+3 2 | 2+2 2 | 2+2 2 | 3+3 2 | 3+3 2 | 2+2 2 | 3+3 2 |
| | 2+2 2 | 2+2 2 | 3+3 2 | 3+3 2 | 3+3 2 | 3+3 2 | 3+3 2 | 2+2 2 |
| | 2+2 2 | 3+3 2 | 3+3 2 | 2+2 2 | 2+2 2 | 2+2 2 | 2+2 2 | 2+2 2 |
| | 3+3 2 | 2+2 2 | 2+2 2 | 3+3 2 | 2+2 2 | 2+2 2 | 2+2 2 | 3+3 2 |
| | 2+2 2 | 2+2 2 | 2+2 2 | 2+2 2 | 2+2 2 | 3+3 2 | 3+3 2 | 3+3 2 |
| | 3+3 2 | 3+3 2 | 3+3 2 | 3+3 2 | 3+3 2 | 2+2 2 | 2+2 2 | 3+3 2 |
| | 2+2 2 | 2+2 2 | 3+3 2 | 3+3 2 | 3+3 2 | 3+3 2 | 3+3 2 | 2+2 2 |
| | 3+3 2 | 3+3 2 | 2+2 2 | 2+2 2 | 2+2 2 | 3+3 2 | 3+3 2 | 3+3 2 |
| | 2+2 2 | 2+2 2 | 3+3 2 | 2+2 2 | 2+2 2 | 2+2 2 | 2+2 2 | 2+2 2 |
| | 3+3 2 | 3+3 2 | 2+2 2 | 3+3 2 | 3+3 2 | 2+2 2 | 2+2 2 | 2+2 2 |
| | 2+2 2 | 2+2 2 | 3+3 2 | 2+2 2 | 3+3 2 | 3+3 2 | 3+3 2 | 3+3 2 |
| | 3+3 2 | 3+3 2 | 2+2 2 | 3+3 2 | 2+2 2 | 3+3 2 | 3+3 2 | 3+3 2 |
| | 2+2 2 | 2+2 2 | 3+3 2 | 2+2 2 | 3+3 2 | 2+2 2 | 2+2 2 | 2+2 2 |
| | 2+2 2 | 2+2 2 | 2+2 2 | 2+2 2 | 2+2 2 | 3+3 2 | 2+2 2 | 3+3 2 |
| | 3+3 2 | 3+3 2 | 3+3 2 | 3+3 2 | 3+3 2 | 2+2 2 | 3+3 2 | 2+2 2 |
| | 2+2 2 | 2+2 2 | 2+2 2 | 3+3 2 | 3+3 2 | 3+3 2 | 2+2 2 | 3+3 2 |
| | 3+3 2 | 3+3 2 | 3+3 2 | 2+2 2 | 2+2 2 | 2+2 2 | 3+3 2 | 2+2 2 |
| | 3+3 2 | 3+3 2 | 2+2 2 | 3+3 2 | 3+3 2 | 2+2 2 | 2+2 2 | 3+3 2 |
| | 2+2 2 | 2+2 2 | 3+3 2 | 2+2 2 | 2+2 2 | 3+3 2 | 3+3 2 | 2+2 2 |
| | 2+2 2 | 2+2 2 | 2+2 2 | 2+2 2 | 2+2 2 | 3+3 2 | 3+3 2 | 2+2 2 |
| | 3+3 2 | 3+3 2 | 3+3 2 | 3+3 2 | 3+3 2 | 2+2 2 | 2+2 2 | 3+3 2 |
| | 2+2 2 | 2+2 2 | 3+3 2 | 3+3 2 | 2+2 2 | 3+3 2 | 3+3 2 | 3+3 2 |

Continued on next column

**TABLE XXI– continued**

| TQC | ICSD | NoE | SF | | | | | |
|---|---|---|---|---|---|---|---|---|
| | 2+2 2 | 2+2 2 | 2+2 2 | 2+2 2 | 3+3 2 | 2+2 2 | 2+2 2 | 2+2 2 |
| 2.2.94 | 31787 | 150 | As1F6N2S3 | | | | | |
| | -GM | -R | -T | -U | -V | -X | -Y | -Z |
| | 3+3 2 | 3+3 2 | 3+3 2 | 3+3 2 | 3+3 2 | 3+3 2 | 3+3 2 | 3+3 2 |
| | 2+2 2 | 2+2 2 | 2+2 2 | 2+2 2 | 2+2 2 | 2+2 2 | 2+2 2 | 2+2 2 |
| | 2+2 2 | 3+3 2 | 2+2 2 | 2+2 2 | 2+2 2 | 2+2 2 | 2+2 2 | 2+2 2 |
| | 3+3 2 | 2+2 2 | 3+3 2 | 3+3 2 | 3+3 2 | 3+3 2 | 3+3 2 | 3+3 2 |
| | 3+3 2 | 3+3 2 | 2+2 2 | 2+2 2 | 3+3 2 | 2+2 2 | 2+2 2 | 3+3 2 |
| | 2+2 2 | 2+2 2 | 3+3 2 | 3+3 2 | 2+2 2 | 3+3 2 | 3+3 2 | 2+2 2 |
| | 3+3 2 | 3+3 2 | 3+3 2 | 3+3 2 | 3+3 2 | 3+3 2 | 3+3 2 | 3+3 2 |
| | 2+2 2 | 2+2 2 | 2+2 2 | 2+2 2 | 2+2 2 | 2+2 2 | 2+2 2 | 2+2 2 |
| | 3+3 2 | 3+3 2 | 2+2 2 | 2+2 2 | 3+3 2 | 2+2 2 | 2+2 2 | 3+3 2 |
| | 2+2 2 | 2+2 2 | 3+3 2 | 3+3 2 | 2+2 2 | 3+3 2 | 3+3 2 | 2+2 2 |
| | 3+3 2 | 3+3 2 | 3+3 2 | 3+3 2 | 3+3 2 | 3+3 2 | 3+3 2 | 3+3 2 |
| | 2+2 2 | 2+2 2 | 2+2 2 | 2+2 2 | 2+2 2 | 2+2 2 | 2+2 2 | 2+2 2 |
| | 3+3 2 | 2+2 2 | 3+3 2 | 3+3 2 | 3+3 2 | 2+2 2 | 2+2 2 | 2+2 2 |
| | 2+2 2 | 3+3 2 | 2+2 2 | 2+2 2 | 2+2 2 | 3+3 2 | 3+3 2 | 3+3 2 |
| | 3+3 2 | 2+2 2 | 3+3 2 | 3+3 2 | 3+3 2 | 2+2 2 | 2+2 2 | 2+2 2 |
| | 2+2 2 | 3+3 2 | 2+2 2 | 2+2 2 | 2+2 2 | 3+3 2 | 3+3 2 | 3+3 2 |
| | 3+3 2 | 2+2 2 | 3+3 2 | 3+3 2 | 2+2 2 | 2+2 2 | 2+2 2 | 2+2 2 |
| | 2+2 2 | 3+3 2 | 2+2 2 | 2+2 2 | 3+3 2 | 3+3 2 | 3+3 2 | 3+3 2 |
| | 3+3 2 | 2+2 2 | 3+3 2 | 3+3 2 | 3+3 2 | 2+2 2 | 2+2 2 | 2+2 2 |
| | 2+2 2 | 3+3 2 | 2+2 2 | 2+2 2 | 2+2 2 | 3+3 2 | 3+3 2 | 3+3 2 |
| | 3+3 2 | 3+3 2 | 2+2 2 | 2+2 2 | 3+3 2 | 3+3 2 | 2+2 2 | 3+3 2 |
| | 2+2 2 | 2+2 2 | 3+3 2 | 3+3 2 | 2+2 2 | 2+2 2 | 3+3 2 | 2+2 2 |
| | 2+2 2 | 2+2 2 | 2+2 2 | 2+2 2 | 2+2 2 | 3+3 2 | 3+3 2 | 3+3 2 |
| | 3+3 2 | 3+3 2 | 3+3 2 | 3+3 2 | 3+3 2 | 2+2 2 | 2+2 2 | 2+2 2 |
| | 3+3 2 | 3+3 2 | 3+3 2 | 2+2 2 | 3+3 2 | 2+2 2 | 2+2 2 | 3+3 2 |
| | 2+2 2 | 2+2 2 | 2+2 2 | 3+3 2 | 2+2 2 | 3+3 2 | 3+3 2 | 2+2 2 |
| | 3+3 2 | 3+3 2 | 2+2 2 | 2+2 2 | 3+3 2 | 2+2 2 | 2+2 2 | 3+3 2 |
| | 2+2 2 | 2+2 2 | 3+3 2 | 3+3 2 | 2+2 2 | 3+3 2 | 3+3 2 | 2+2 2 |
| | 3+3 2 | 2+2 2 | 3+3 2 | 2+2 2 | 2+2 2 | 2+2 2 | 2+2 2 | 2+2 2 |
| | 2+2 2 | 3+3 2 | 2+2 2 | 3+3 2 | 3+3 2 | 3+3 2 | 3+3 2 | 3+3 2 |
| | 3+3 2 | 2+2 2 | 3+3 2 | 3+3 2 | 3+3 2 | 2+2 2 | 2+2 2 | 2+2 2 |
| | 2+2 2 | 3+3 2 | 2+2 2 | 2+2 2 | 2+2 2 | 2+2 2 | 3+3 2 | 3+3 2 |
| | 3+3 2 | 2+2 2 | 3+3 2 | 3+3 2 | 2+2 2 | 3+3 2 | 2+2 2 | 2+2 2 |
| | 2+2 2 | 3+3 2 | 2+2 2 | 2+2 2 | 3+3 2 | 3+3 2 | 3+3 2 | 3+3 2 |
| | 3+3 2 | 2+2 2 | 3+3 2 | 3+3 2 | 2+2 2 | 3+3 2 | 3+3 2 | 3+3 2 |
| | 2+2 2 | 3+3 2 | 2+2 2 | 2+2 2 | 3+3 2 | 2+2 2 | 2+2 2 | 2+2 2 |
| | 3+3 2 | 2+2 2 | 2+2 2 | 2+2 2 | 2+2 2 | 3+3 2 | 3+3 2 | 2+2 2 |
| | 2+2 2 | 3+3 2 | 3+3 2 | 3+3 2 | 3+3 2 | 2+2 2 | 2+2 2 | 3+3 2 |
| | 3+3 2 | 3+3 2 | 2+2 2 | 3+3 2 | 3+3 2 | 3+3 2 | 3+3 2 | 3+3 2 |
| | 2+2 2 | 2+2 2 | 3+3 2 | 2+2 2 | 2+2 2 | 2+2 2 | 2+2 2 | 2+2 2 |
| | 3+3 2 | 3+3 2 | 3+3 2 | 2+2 2 | 3+3 2 | 3+3 2 | 2+2 2 | 3+3 2 |
| | 2+2 2 | 2+2 2 | 2+2 2 | 3+3 2 | 2+2 2 | 2+2 2 | 3+3 2 | 2+2 2 |
| | 2+2 2 | 3+3 2 | 2+2 2 | 2+2 2 | 3+3 2 | 2+2 2 | 2+2 2 | 3+3 2 |
| | 3+3 2 | 2+2 2 | 2+2 2 | 3+3 2 | 2+2 2 | 3+3 2 | 3+3 2 | 2+2 2 |
| | 3+3 2 | 3+3 2 | 3+3 2 | 2+2 2 | 3+3 2 | 3+3 2 | 2+2 2 | 2+2 2 |
| | 2+2 2 | 2+2 2 | 3+3 2 | 3+3 2 | 2+2 2 | 2+2 2 | 3+3 2 | 3+3 2 |
| | 3+3 2 | 3+3 2 | 2+2 2 | 3+3 2 | 2+2 2 | 3+3 2 | 3+3 2 | 3+3 2 |
| | 2+2 2 | 2+2 2 | 3+3 2 | 2+2 2 | 3+3 2 | 2+2 2 | 2+2 2 | 2+2 2 |
| | 3+3 2 | 3+3 2 | 3+3 2 | 3+3 2 | 2+2 2 | 3+3 2 | 3+3 2 | 2+2 2 |
| | 2+2 2 | 2+2 2 | 2+2 2 | 2+2 2 | 3+3 2 | 2+2 2 | 2+2 2 | 3+3 2 |
| | 3+3 2 | 2+2 2 | 2+2 2 | 2+2 2 | 3+3 2 | 3+3 2 | 2+2 2 | 3+3 2 |
| | 2+2 2 | 3+3 2 | 3+3 2 | 3+3 2 | 2+2 2 | 2+2 2 | 3+3 2 | 2+2 2 |
| | 3+3 2 | 2+2 2 | 3+3 2 | 3+3 2 | 3+3 2 | 3+3 2 | 3+3 2 | 2+2 2 |
| | 2+2 2 | 3+3 2 | 2+2 2 | 2+2 2 | 2+2 2 | 2+2 2 | 2+2 2 | 3+3 2 |
| | 3+3 2 | 2+2 2 | 3+3 2 | 2+2 2 | 3+3 2 | 2+2 2 | 3+3 2 | 3+3 2 |
| | 3+3 2 | 3+3 2 | 2+2 2 | 3+3 2 | 2+2 2 | 3+3 2 | 2+2 2 | 2+2 2 |
| | 2+2 2 | 2+2 2 | 3+3 2 | 2+2 2 | 2+2 2 | 3+3 2 | 3+3 2 | 2+2 2 |
| | 2+2 2 | 3+3 2 | 2+2 2 | 3+3 2 | 3+3 2 | 2+2 2 | 3+3 2 | 3+3 2 |

Continued on next column

**TABLE XXI– continued**

| TQC | ICSD | NoE | SF | | | | |
|---|---|---|---|---|---|---|---|
| | 3+3 2 | 2+2 2 | 3+3 2 | 3+3 2 | 3+3 2 | 3+3 2 | 2+2 2 | 3+3 2 |
| | 3+3 2 | 3+3 2 | 2+2 2 | 2+2 2 | 2+2 2 | 2+2 2 | 2+2 2 | 2+2 2 |
| | 2+2 2 | 2+2 2 | 3+3 2 | 3+3 2 | 3+3 2 | 3+3 2 | 3+3 2 | 3+3 2 |
| | 2+2 2 | 3+3 2 | 2+2 2 | 2+2 2 | 2+2 2 | 2+2 2 | 2+2 2 | 2+2 2 |
| | 3+3 2 | 2+2 2 | 3+3 2 | 3+3 2 | 3+3 2 | 3+3 2 | 3+3 2 | 3+3 2 |
| | 2+2 2 | 3+3 2 | 2+2 2 | 2+2 2 | 2+2 2 | 2+2 2 | 2+2 2 | 2+2 2 |
| | 2+2 2 | 2+2 2 | 3+3 2 | 3+3 2 | 3+3 2 | 2+2 2 | 2+2 2 | 2+2 2 |
| | 3+3 2 | 2+2 2 | 2+2 2 | 2+2 2 | 2+2 2 | 3+3 2 | 3+3 2 | 2+2 2 |
| | 3+3 2 | 3+3 2 | 3+3 2 | 3+3 2 | 2+2 2 | 2+2 2 | 3+3 2 | 3+3 2 |
| | 2+2 2 | 3+3 2 | 2+2 2 | 2+2 2 | 3+3 2 | 2+2 2 | 2+2 2 | 3+3 2 |
| | 3+3 2 | 3+3 2 | 3+3 2 | 3+3 2 | 3+3 2 | 3+3 2 | 2+2 2 | 2+2 2 |
| | 2+2 2 | 2+2 2 | 2+2 2 | 2+2 2 | 2+2 2 | 3+3 2 | 3+3 2 | 3+3 2 |
| | 3+3 2 | 3+3 2 | 3+3 2 | 3+3 2 | 3+3 2 | 3+3 2 | 3+3 2 | 3+3 2 |
| | 2+2 2 | 2+2 2 | 2+2 2 | 2+2 2 | 2+2 2 | 2+2 2 | 2+2 2 | 2+2 2 |
| | 3+3 2 | 3+3 2 | 2+2 2 | 3+3 2 | 2+2 2 | 3+3 2 | 3+3 2 | 3+3 2 |
| | 2+2 2 | 2+2 2 | 3+3 2 | 2+2 2 | 3+3 2 | 2+2 2 | 2+2 2 | 2+2 2 |
| | 3+3 2 | 2+2 2 | 3+3 2 | 3+3 2 | 2+2 2 | 2+2 2 | 3+3 2 | 2+2 2 |
| 2.2.95 | 92916 | 100 | As1Bi1Ni1O5 | | | | |
| | -GM | -R | -T | -U | -V | -X | -Y | -Z |
| | 3+3 2 | 2+2 2 | 2+2 2 | 3+3 2 | 3+3 2 | 3+3 2 | 3+3 2 | 2+2 2 |
| | 2+2 2 | 3+3 2 | 3+3 2 | 2+2 2 | 2+2 2 | 2+2 2 | 2+2 2 | 3+3 2 |
| | 3+3 2 | 3+3 2 | 2+2 2 | 2+2 2 | 3+3 2 | 2+2 2 | 2+2 2 | 3+3 2 |
| | 2+2 2 | 2+2 2 | 3+3 2 | 3+3 2 | 2+2 2 | 3+3 2 | 3+3 2 | 2+2 2 |
| | 2+2 2 | 2+2 2 | 2+2 2 | 2+2 2 | 2+2 2 | 2+2 2 | 2+2 2 | 2+2 2 |
| | 3+3 2 | 3+3 2 | 3+3 2 | 3+3 2 | 3+3 2 | 3+3 2 | 3+3 2 | 3+3 2 |
| | 3+3 2 | 3+3 2 | 3+3 2 | 3+3 2 | 3+3 2 | 2+2 2 | 2+2 2 | 3+3 2 |
| | 2+2 2 | 3+3 2 | 2+2 2 | 2+2 2 | 2+2 2 | 3+3 2 | 2+2 2 | 3+3 2 |
| | 3+3 2 | 2+2 2 | 2+2 2 | 2+2 2 | 3+3 2 | 2+2 2 | 3+3 2 | 2+2 2 |
| | 2+2 2 | 2+2 2 | 3+3 2 | 3+3 2 | 2+2 2 | 3+3 2 | 3+3 2 | 2+2 2 |
| | 2+2 2 | 2+2 2 | 3+3 2 | 3+3 2 | 2+2 2 | 3+3 2 | 3+3 2 | 2+2 2 |
| | 3+3 2 | 3+3 2 | 2+2 2 | 2+2 2 | 3+3 2 | 2+2 2 | 2+2 2 | 3+3 2 |
| | 2+2 2 | 2+2 2 | 3+3 2 | 3+3 2 | 2+2 2 | 2+2 2 | 2+2 2 | 2+2 2 |
| | 3+3 2 | 3+3 2 | 2+2 2 | 2+2 2 | 3+3 2 | 3+3 2 | 3+3 2 | 3+3 2 |
| | 2+2 2 | 2+2 2 | 2+2 2 | 2+2 2 | 3+3 2 | 3+3 2 | 3+3 2 | 3+3 2 |
| | 3+3 2 | 3+3 2 | 3+3 2 | 2+2 2 | 2+2 2 | 2+2 2 | 2+2 2 | 2+2 2 |
| | 2+2 2 | 2+2 2 | 2+2 2 | 3+3 2 | 2+2 2 | 2+2 2 | 2+2 2 | 3+3 2 |
| | 3+3 2 | 3+3 2 | 3+3 2 | 3+3 2 | 3+3 2 | 3+3 2 | 3+3 2 | 2+2 2 |
| | 3+3 2 | 3+3 2 | 3+3 2 | 3+3 2 | 3+3 2 | 2+2 2 | 3+3 2 | 3+3 2 |
| | 2+2 2 | 2+2 2 | 2+2 2 | 2+2 2 | 2+2 2 | 3+3 2 | 2+2 2 | 2+2 2 |
| | 3+3 2 | 2+2 2 | 2+2 2 | 2+2 2 | 3+3 2 | 2+2 2 | 2+2 2 | 3+3 2 |
| | 3+3 2 | 3+3 2 | 3+3 2 | 3+3 2 | 2+2 2 | 3+3 2 | 3+3 2 | 3+3 2 |
| | 2+2 2 | 3+3 2 | 3+3 2 | 2+2 2 | 3+3 2 | 2+2 2 | 2+2 2 | 2+2 2 |
| | 3+3 2 | 3+3 2 | 2+2 2 | 3+3 2 | 2+2 2 | 3+3 2 | 3+3 2 | 2+2 2 |
| | 2+2 2 | 3+3 2 | 2+2 2 | 3+3 2 | 3+3 2 | 2+2 2 | 2+2 2 | 3+3 2 |
| | 3+3 2 | 2+2 2 | 3+3 2 | 2+2 2 | 2+2 2 | 3+3 2 | 3+3 2 | 2+2 2 |
| | 2+2 2 | 2+2 2 | 3+3 2 | 3+3 2 | 2+2 2 | 2+2 2 | 2+2 2 | 2+2 2 |
| | 3+3 2 | 2+2 2 | 2+2 2 | 3+3 2 | 3+3 2 | 3+3 2 | 3+3 2 | 3+3 2 |
| | 3+3 2 | 2+2 2 | 2+2 2 | 2+2 2 | 2+2 2 | 3+3 2 | 2+2 2 | 2+2 2 |
| | 2+2 2 | 3+3 2 | 3+3 2 | 3+3 2 | 3+3 2 | 2+2 2 | 3+3 2 | 3+3 2 |
| | 3+3 2 | 2+2 2 | 2+2 2 | 2+2 2 | 3+3 2 | 3+3 2 | 2+2 2 | 3+3 2 |
| | 2+2 2 | 3+3 2 | 3+3 2 | 2+2 2 | 2+2 2 | 2+2 2 | 3+3 2 | 2+2 2 |
| | 3+3 2 | 3+3 2 | 2+2 2 | 3+3 2 | 2+2 2 | 3+3 2 | 2+2 2 | 3+3 2 |
| | 2+2 2 | 2+2 2 | 3+3 2 | 2+2 2 | 3+3 2 | 2+2 2 | 2+2 2 | 2+2 2 |
| | 2+2 2 | 3+3 2 | 2+2 2 | 2+2 2 | 2+2 2 | 3+3 2 | 3+3 2 | 3+3 2 |
| | 3+3 2 | 2+2 2 | 3+3 2 | 3+3 2 | 3+3 2 | 2+2 2 | 3+3 2 | 2+2 2 |
| | 2+2 2 | 2+2 2 | 3+3 2 | 3+3 2 | 2+2 2 | 2+2 2 | 2+2 2 | 3+3 2 |
| | 2+2 2 | 3+3 2 | 2+2 2 | 2+2 2 | 2+2 2 | 3+3 2 | 2+2 2 | 2+2 2 |
| | 3+3 2 | 2+2 2 | 3+3 2 | 3+3 2 | 3+3 2 | 3+3 2 | 3+3 2 | 3+3 2 |
| | 3+3 2 | 3+3 2 | 2+2 2 | 2+2 2 | 3+3 2 | 2+2 2 | 2+2 2 | 3+3 2 |
| | 2+2 2 | 2+2 2 | 2+2 2 | 2+2 2 | 2+2 2 | 2+2 2 | 3+3 2 | 2+2 2 |
| | 2+2 2 | 3+3 2 | 3+3 2 | 3+3 2 | 3+3 2 | 3+3 2 | 3+3 2 | 2+2 2 |
| | 3+3 2 | 2+2 2 | 2+2 2 | 2+2 2 | 3+3 2 | 3+3 2 | 3+3 2 | 3+3 2 |
| | 2+2 2 | 3+3 2 | 3+3 2 | 2+2 2 | 3+3 2 | 2+2 2 | 3+3 2 | 2+2 2 |
| | 3+3 2 | 3+3 2 | 3+3 2 | 3+3 2 | 2+2 2 | 2+2 2 | 2+2 2 | 2+2 2 |
| | 2+2 2 | 2+2 2 | 2+2 2 | 3+3 2 | 2+2 2 | 3+3 2 | 2+2 2 | 3+3 2 |
| | 3+3 2 | 3+3 2 | 2+2 2 | 2+2 2 | 3+3 2 | 2+2 2 | 3+3 2 | 2+2 2 |
| | 2+2 2 | 2+2 2 | 3+3 2 | 3+3 2 | 2+2 2 | 3+3 2 | 2+2 2 | 3+3 2 |
| | 2+2 2 | 3+3 2 | 3+3 2 | 3+3 2 | 3+3 2 | 2+2 2 | 3+3 2 | 3+3 2 |
| | 3+3 2 | 3+3 2 | 2+2 2 | 2+2 2 | 2+2 2 | 2+2 2 | 2+2 2 | 2+2 2 |
| 2.2.96 | 428633 | 98 | As1Bi1Co1O5 | | | | |
| | -GM | -R | -T | -U | -V | -X | -Y | -Z |
| | 3+3 2 | 3+3 2 | 2+2 2 | 2+2 2 | 2+2 2 | 2+2 2 | 3+3 2 | 2+2 2 |
| | 2+2 2 | 2+2 2 | 3+3 2 | 3+3 2 | 3+3 2 | 3+3 2 | 2+2 2 | 3+3 2 |
| | 3+3 2 | 2+2 2 | 2+2 2 | 3+3 2 | 2+2 2 | 3+3 2 | 2+2 2 | 3+3 2 |
| | 2+2 2 | 3+3 2 | 3+3 2 | 2+2 2 | 3+3 2 | 2+2 2 | 3+3 2 | 2+2 2 |
| | 2+2 2 | 3+3 2 | 2+2 2 | 3+3 2 | 3+3 2 | 3+3 2 | 2+2 2 | 2+2 2 |
| | 3+3 2 | 2+2 2 | 3+3 2 | 2+2 2 | 2+2 2 | 2+2 2 | 3+3 2 | 3+3 2 |
| | 3+3 2 | 2+2 2 | 3+3 2 | 2+2 2 | 2+2 2 | 3+3 2 | 2+2 2 | 3+3 2 |
| | 2+2 2 | 3+3 2 | 2+2 2 | 3+3 2 | 3+3 2 | 2+2 2 | 2+2 2 | 3+3 2 |
| | 3+3 2 | 2+2 2 | 2+2 2 | 3+3 2 | 2+2 2 | 3+3 2 | 3+3 2 | 2+2 2 |
| | 2+2 2 | 3+3 2 | 3+3 2 | 2+2 2 | 3+3 2 | 2+2 2 | 3+3 2 | 2+2 2 |
| | 2+2 2 | 3+3 2 | 3+3 2 | 2+2 2 | 3+3 2 | 2+2 2 | 3+3 2 | 2+2 2 |
| | 3+3 2 | 2+2 2 | 2+2 2 | 3+3 2 | 2+2 2 | 3+3 2 | 2+2 2 | 3+3 2 |
| | 2+2 2 | 3+3 2 | 3+3 2 | 2+2 2 | 3+3 2 | 3+3 2 | 2+2 2 | 2+2 2 |
| | 3+3 2 | 2+2 2 | 2+2 2 | 3+3 2 | 2+2 2 | 2+2 2 | 3+3 2 | 3+3 2 |
| | 2+2 2 | 3+3 2 | 2+2 2 | 3+3 2 | 2+2 2 | 2+2 2 | 3+3 2 | 2+2 2 |
| | 3+3 2 | 2+2 2 | 3+3 2 | 3+3 2 | 3+3 2 | 3+3 2 | 2+2 2 | 3+3 2 |
| | 2+2 2 | 3+3 2 | 2+2 2 | 2+2 2 | 2+2 2 | 3+3 2 | 3+3 2 | 3+3 2 |
| | 3+3 2 | 2+2 2 | 3+3 2 | 2+2 2 | 3+3 2 | 2+2 2 | 2+2 2 | 2+2 2 |
| | 3+3 2 | 2+2 2 | 3+3 2 | 2+2 2 | 2+2 2 | 3+3 2 | 3+3 2 | 3+3 2 |
| | 2+2 2 | 3+3 2 | 2+2 2 | 3+3 2 | 3+3 2 | 2+2 2 | 2+2 2 | 2+2 2 |
| | 3+3 2 | 3+3 2 | 2+2 2 | 3+3 2 | 2+2 2 | 2+2 2 | 2+2 2 | 3+3 2 |
| | 2+2 2 | 2+2 2 | 3+3 2 | 2+2 2 | 2+2 2 | 3+3 2 | 3+3 2 | 3+3 2 |
| | 3+3 2 | 2+2 2 | 3+3 2 | 3+3 2 | 3+3 2 | 3+3 2 | 3+3 2 | 2+2 2 |
| | 3+3 2 | 2+2 2 | 2+2 2 | 2+2 2 | 3+3 2 | 3+3 2 | 2+2 2 | 3+3 2 |
| | 2+2 2 | 2+2 2 | 3+3 2 | 2+2 2 | 2+2 2 | 2+2 2 | 2+2 2 | 2+2 2 |
| | 3+3 2 | 3+3 2 | 2+2 2 | 3+3 2 | 3+3 2 | 2+2 2 | 3+3 2 | 2+2 2 |
| | 2+2 2 | 3+3 2 | 3+3 2 | 2+2 2 | 3+3 2 | 3+3 2 | 2+2 2 | 2+2 2 |
| | 3+3 2 | 3+3 2 | 2+2 2 | 2+2 2 | 2+2 2 | 2+2 2 | 3+3 2 | 3+3 2 |
| | 3+3 2 | 3+3 2 | 2+2 2 | 3+3 2 | 2+2 2 | 2+2 2 | 2+2 2 | 2+2 2 |
| | 2+2 2 | 2+2 2 | 3+3 2 | 2+2 2 | 3+3 2 | 3+3 2 | 2+2 2 | 3+3 2 |
| | 3+3 2 | 3+3 2 | 2+2 2 | 3+3 2 | 2+2 2 | 2+2 2 | 3+3 2 | 3+3 2 |
| | 2+2 2 | 2+2 2 | 3+3 2 | 3+3 2 | 3+3 2 | 3+3 2 | 3+3 2 | 2+2 2 |
| | 3+3 2 | 2+2 2 | 2+2 2 | 2+2 2 | 3+3 2 | 2+2 2 | 2+2 2 | 3+3 2 |
| | 2+2 2 | 3+3 2 | 3+3 2 | 3+3 2 | 2+2 2 | 3+3 2 | 2+2 2 | 2+2 2 |
| | 2+2 2 | 2+2 2 | 2+2 2 | 3+3 2 | 3+3 2 | 2+2 2 | 3+3 2 | 3+3 2 |
| | 3+3 2 | 3+3 2 | 3+3 2 | 2+2 2 | 2+2 2 | 3+3 2 | 3+3 2 | 2+2 2 |
| | 2+2 2 | 2+2 2 | 3+3 2 | 2+2 2 | 3+3 2 | 3+3 2 | 2+2 2 | 3+3 2 |
| | 3+3 2 | 3+3 2 | 2+2 2 | 3+3 2 | 3+3 2 | 2+2 2 | 2+2 2 | 2+2 2 |
| | 2+2 2 | 3+3 2 | 3+3 2 | 2+2 2 | 2+2 2 | 3+3 2 | 3+3 2 | 3+3 2 |
| | 3+3 2 | 2+2 2 | 2+2 2 | 3+3 2 | 2+2 2 | 2+2 2 | 2+2 2 | 2+2 2 |
| | 2+2 2 | 3+3 2 | 2+2 2 | 3+3 2 | 3+3 2 | 3+3 2 | 3+3 2 | 3+3 2 |
| | 2+2 2 | 2+2 2 | 3+3 2 | 2+2 2 | 2+2 2 | 2+2 2 | 3+3 2 | 2+2 2 |
| | 3+3 2 | 3+3 2 | 2+2 2 | 3+3 2 | 2+2 2 | 2+2 2 | 3+3 2 | 3+3 2 |
| | 2+2 2 | 2+2 2 | 3+3 2 | 3+3 2 | 2+2 2 | 3+3 2 | 3+3 2 | 2+2 2 |
| | 3+3 2 | 2+2 2 | 2+2 2 | 2+2 2 | 3+3 2 | 3+3 2 | 2+2 2 | 2+2 2 |
| | 2+2 2 | 3+3 2 | 3+3 2 | 2+2 2 | 3+3 2 | 2+2 2 | 2+2 2 | 3+3 2 |
| | 3+3 2 | 2+2 2 | 2+2 2 | 3+3 2 | 2+2 2 | 3+3 2 | 3+3 2 | 2+2 2 |
| | 2+2 2 | 3+3 2 | 3+3 2 | 2+2 2 | 3+3 2 | 2+2 2 | 2+2 2 | 3+3 2 |
| | 2+2 2 | 2+2 2 | 2+2 2 | 2+2 2 | 2+2 2 | 3+3 2 | 3+3 2 | 3+3 2 |
| 2.2.97 | 422342 | 84 | Al2Fe3Si3 | | | | |
| | -GM | -R | -T | -U | -V | -X | -Y | -Z |

Continued on next column

**TABLE XXI– continued**

| TQC | ICSD | NoE | SF | | | | | |
|---|---|---|---|---|---|---|---|---|
| | 3+3 2 | 2+2 2 | 2+2 2 | 3+3 2 | 3+3 2 | 2+2 2 | 3+3 2 | 2+2 2 |
| | 3+3 2 | 3+3 2 | 3+3 2 | 2+2 2 | 2+2 2 | 3+3 2 | 2+2 2 | 3+3 2 |
| | 3+3 2 | 2+2 2 | 3+3 2 | 3+3 2 | 3+3 2 | 2+2 2 | 3+3 2 | 2+2 2 |
| | 2+2 2 | 3+3 2 | 2+2 2 | 2+2 2 | 2+2 2 | 3+3 2 | 2+2 2 | 3+3 2 |
| | 2+2 2 | 2+2 2 | 2+2 2 | 3+3 2 | 3+3 2 | 2+2 2 | 3+3 2 | 2+2 2 |
| | 2+2 2 | 3+3 2 | 2+2 2 | 2+2 2 | 2+2 2 | 3+3 2 | 2+2 2 | 3+3 2 |
| | 3+3 2 | 2+2 2 | 3+3 2 | 2+2 2 | 2+2 2 | 3+3 2 | 3+3 2 | 2+2 2 |
| | 2+2 2 | 3+3 2 | 3+3 2 | 3+3 2 | 3+3 2 | 2+2 2 | 2+2 2 | 3+3 2 |
| | 3+3 2 | 3+3 2 | 3+3 2 | 3+3 2 | 3+3 2 | 2+2 2 | 2+2 2 | 3+3 2 |
| | 2+2 2 | 2+2 2 | 2+2 2 | 2+2 2 | 2+2 2 | 3+3 2 | 3+3 2 | 2+2 2 |
| | 3+3 2 | 3+3 2 | 3+3 2 | 3+3 2 | 2+2 2 | 2+2 2 | 3+3 2 | 2+2 2 |
| | 2+2 2 | 2+2 2 | 2+2 2 | 2+2 2 | 3+3 2 | 3+3 2 | 2+2 2 | 3+3 2 |
| | 2+2 2 | 3+3 2 | 3+3 2 | 2+2 2 | 2+2 2 | 2+2 2 | 2+2 2 | 3+3 2 |
| | 3+3 2 | 2+2 2 | 2+2 2 | 3+3 2 | 3+3 2 | 3+3 2 | 3+3 2 | 2+2 2 |
| | 3+3 2 | 3+3 2 | 3+3 2 | 3+3 2 | 3+3 2 | 3+3 2 | 3+3 2 | 3+3 2 |
| | 2+2 2 | 3+3 2 | 2+2 2 | 3+3 2 | 2+2 2 | 2+2 2 | 2+2 2 | 2+2 2 |
| | 3+3 2 | 2+2 2 | 3+3 2 | 2+2 2 | 3+3 2 | 2+2 2 | 3+3 2 | 3+3 2 |
| | 3+3 2 | 3+3 2 | 2+2 2 | 2+2 2 | 2+2 2 | 2+2 2 | 2+2 2 | 2+2 2 |
| | 2+2 2 | 2+2 2 | 2+2 2 | 3+3 2 | 3+3 2 | 3+3 2 | 2+2 2 | 3+3 2 |
| | 3+3 2 | 2+2 2 | 3+3 2 | 2+2 2 | 3+3 2 | 2+2 2 | 3+3 2 | 2+2 2 |
| | 3+3 2 | 2+2 2 | 3+3 2 | 2+2 2 | 3+3 2 | 3+3 2 | 3+3 2 | 3+3 2 |
| | 2+2 2 | 3+3 2 | 2+2 2 | 3+3 2 | 2+2 2 | 3+3 2 | 3+3 2 | 2+2 2 |
| | 3+3 2 | 3+3 2 | 2+2 2 | 3+3 2 | 3+3 2 | 2+2 2 | 2+2 2 | 3+3 2 |
| | 3+3 2 | 2+2 2 | 3+3 2 | 2+2 2 | 2+2 2 | 3+3 2 | 2+2 2 | 2+2 2 |
| | 3+3 2 | 2+2 2 | 2+2 2 | 3+3 2 | 3+3 2 | 2+2 2 | 3+3 2 | 3+3 2 |
| | 2+2 2 | 3+3 2 | 3+3 2 | 2+2 2 | 2+2 2 | 3+3 2 | 3+3 2 | 3+3 2 |
| | 3+3 2 | 3+3 2 | 2+2 2 | 3+3 2 | 2+2 2 | 3+3 2 | 2+2 2 | 2+2 2 |
| | 2+2 2 | 2+2 2 | 2+2 2 | 2+2 2 | 2+2 2 | 2+2 2 | 3+3 2 | 2+2 2 |
| | 2+2 2 | 2+2 2 | 3+3 2 | 2+2 2 | 3+3 2 | 2+2 2 | 2+2 2 | 2+2 2 |
| | 3+3 2 | 3+3 2 | 3+3 2 | 3+3 2 | 2+2 2 | 2+2 2 | 2+2 2 | 3+3 2 |
| | 2+2 2 | 2+2 2 | 3+3 2 | 2+2 2 | 3+3 2 | 3+3 2 | 3+3 2 | 3+3 2 |
| | 2+2 2 | 3+3 2 | 2+2 2 | 3+3 2 | 3+3 2 | 3+3 2 | 2+2 2 | 2+2 2 |
| | 2+2 2 | 3+3 2 | 3+3 2 | 2+2 2 | 2+2 2 | 2+2 2 | 3+3 2 | 3+3 2 |
| | 3+3 2 | 2+2 2 | 2+2 2 | 3+3 2 | 2+2 2 | 2+2 2 | 2+2 2 | 2+2 2 |
| | 2+2 2 | 2+2 2 | 3+3 2 | 2+2 2 | 2+2 2 | 3+3 2 | 3+3 2 | 3+3 2 |
| | 2+2 2 | 3+3 2 | 2+2 2 | 3+3 2 | 3+3 2 | 2+2 2 | 2+2 2 | 2+2 2 |
| | 3+3 2 | 3+3 2 | 3+3 2 | 2+2 2 | 2+2 2 | 3+3 2 | 2+2 2 | 2+2 2 |
| | 3+3 2 | 2+2 2 | 2+2 2 | 3+3 2 | 3+3 2 | 3+3 2 | 3+3 2 | 3+3 2 |
| | 2+2 2 | 2+2 2 | 2+2 2 | 3+3 2 | 3+3 2 | 2+2 2 | 2+2 2 | 3+3 2 |
| | 3+3 2 | 3+3 2 | 3+3 2 | 2+2 2 | 2+2 2 | 3+3 2 | 3+3 2 | 2+2 2 |
| | 2+2 2 | 3+3 2 | 2+2 2 | 2+2 2 | 2+2 2 | 3+3 2 | 2+2 2 | 2+2 2 |
| | 3+3 2 | 2+2 2 | 3+3 2 | 3+3 2 | 2+2 2 | 3+3 2 | 3+3 2 | 3+3 2 |
| 2.2.98 | 95832 | 228 | Ag2F5 | | | | | |
| | -GM | -R | -T | -U | -V | -X | -Y | -Z |
| | 3+3 2 | 2+2 2 | 2+2 2 | 2+2 2 | 2+2 2 | 2+2 2 | 2+2 2 | 3+3 2 |
| | 3+3 2 | 3+3 2 | 3+3 2 | 3+3 2 | 3+3 2 | 3+3 2 | 3+3 2 | 2+2 2 |
| | 2+2 2 | 2+2 2 | 3+3 2 | 2+2 2 | 2+2 2 | 2+2 2 | 2+2 2 | 2+2 2 |
| | 3+3 2 | 2+2 2 | 3+3 2 | 3+3 2 | 3+3 2 | 2+2 2 | 3+3 2 | 3+3 2 |
| | 2+2 2 | 3+3 2 | 2+2 2 | 3+3 2 | 2+2 2 | 2+2 2 | 2+2 2 | 2+2 2 |
| | 3+3 2 | 3+3 2 | 2+2 2 | 2+2 2 | 3+3 2 | 3+3 2 | 3+3 2 | 3+3 2 |
| | 2+2 2 | 2+2 2 | 2+2 2 | 3+3 2 | 3+3 2 | 3+3 2 | 3+3 2 | 2+2 2 |
| | 3+3 2 | 3+3 2 | 3+3 2 | 3+3 2 | 2+2 2 | 2+2 2 | 2+2 2 | 3+3 2 |
| | 3+3 2 | 2+2 2 | 3+3 2 | 2+2 2 | 3+3 2 | 2+2 2 | 2+2 2 | 2+2 2 |
| | 2+2 2 | 3+3 2 | 2+2 2 | 3+3 2 | 3+3 2 | 3+3 2 | 2+2 2 | 2+2 2 |
| | 3+3 2 | 2+2 2 | 3+3 2 | 2+2 2 | 2+2 2 | 2+2 2 | 3+3 2 | 3+3 2 |
| | 2+2 2 | 3+3 2 | 2+2 2 | 2+2 2 | 2+2 2 | 3+3 2 | 2+2 2 | 3+3 2 |
| | 2+2 2 | 3+3 2 | 3+3 2 | 3+3 2 | 3+3 2 | 3+3 2 | 3+3 2 | 2+2 2 |
| | 3+3 2 | 2+2 2 | 3+3 2 | 2+2 2 | 2+2 2 | 3+3 2 | 3+3 2 | 3+3 2 |
| | 2+2 2 | 2+2 2 | 2+2 2 | 3+3 2 | 3+3 2 | 2+2 2 | 3+3 2 | 2+2 2 |
| | 2+2 2 | 3+3 2 | 2+2 2 | 2+2 2 | 2+2 2 | 3+3 2 | 2+2 2 | 3+3 2 |
| | 3+3 2 | 2+2 2 | 3+3 2 | 2+2 2 | 2+2 2 | 2+2 2 | 3+3 2 | 3+3 2 |

Continued on next column

**TABLE XXI– continued**

| TQC | ICSD | NoE | SF | | | | | |
|---|---|---|---|---|---|---|---|---|
| | 3+3 2 | 2+2 2 | 2+2 2 | 2+2 2 | 3+3 2 | 3+3 2 | 3+3 2 | 2+2 2 |
| | 2+2 2 | 3+3 2 | 3+3 2 | 3+3 2 | 2+2 2 | 2+2 2 | 2+2 2 | 3+3 2 |
| | 2+2 2 | 3+3 2 | 2+2 2 | 3+3 2 | 3+3 2 | 3+3 2 | 2+2 2 | 2+2 2 |
| | 3+3 2 | 2+2 2 | 3+3 2 | 3+3 2 | 3+3 2 | 2+2 2 | 2+2 2 | 2+2 2 |
| | 3+3 2 | 3+3 2 | 2+2 2 | 2+2 2 | 3+3 2 | 2+2 2 | 2+2 2 | 3+3 2 |
| | 3+3 2 | 2+2 2 | 3+3 2 | 2+2 2 | 2+2 2 | 2+2 2 | 3+3 2 | 2+2 2 |
| | 2+2 2 | 2+2 2 | 3+3 2 | 3+3 2 | 2+2 2 | 3+3 2 | 3+3 2 | 3+3 2 |
| | 3+3 2 | 3+3 2 | 2+2 2 | 3+3 2 | 2+2 2 | 2+2 2 | 3+3 2 | 2+2 2 |
| | 3+3 2 | 2+2 2 | 3+3 2 | 2+2 2 | 3+3 2 | 2+2 2 | 2+2 2 | 3+3 2 |
| | 3+3 2 | 3+3 2 | 2+2 2 | 3+3 2 | 3+3 2 | 2+2 2 | 2+2 2 | 2+2 2 |
| | 3+3 2 | 2+2 2 | 3+3 2 | 2+2 2 | 3+3 2 | 3+3 2 | 2+2 2 | 3+3 2 |
| | 2+2 2 | 2+2 2 | 3+3 2 | 3+3 2 | 2+2 2 | 2+2 2 | 3+3 2 | 3+3 2 |
| | 2+2 2 | 3+3 2 | 2+2 2 | 2+2 2 | 2+2 2 | 2+2 2 | 3+3 2 | 2+2 2 |
| | 3+3 2 | 2+2 2 | 3+3 2 | 3+3 2 | 3+3 2 | 3+3 2 | 3+3 2 | 2+2 2 |
| | 3+3 2 | 2+2 2 | 3+3 2 | 2+2 2 | 2+2 2 | 2+2 2 | 2+2 2 | 3+3 2 |
| | 3+3 2 | 3+3 2 | 3+3 2 | 2+2 2 | 2+2 2 | 3+3 2 | 2+2 2 | 2+2 2 |
| | 2+2 2 | 2+2 2 | 2+2 2 | 3+3 2 | 3+3 2 | 2+2 2 | 3+3 2 | 3+3 2 |
| | 3+3 2 | 2+2 2 | 3+3 2 | 2+2 2 | 2+2 2 | 2+2 2 | 3+3 2 | 3+3 2 |
| | 3+3 2 | 3+3 2 | 2+2 2 | 3+3 2 | 3+3 2 | 2+2 2 | 2+2 2 | 3+3 2 |
| | 3+3 2 | 3+3 2 | 3+3 2 | 2+2 2 | 2+2 2 | 3+3 2 | 3+3 2 | 2+2 2 |
| | 2+2 2 | 3+3 2 | 2+2 2 | 3+3 2 | 3+3 2 | 2+2 2 | 2+2 2 | 2+2 2 |
| | 2+2 2 | 2+2 2 | 2+2 2 | 2+2 2 | 2+2 2 | 3+3 2 | 2+2 2 | 3+3 2 |
| | 3+3 2 | 2+2 2 | 3+3 2 | 3+3 2 | 3+3 2 | 2+2 2 | 3+3 2 | 2+2 2 |
| | 2+2 2 | 3+3 2 | 3+3 2 | 2+2 2 | 2+2 2 | 2+2 2 | 3+3 2 | 2+2 2 |
| | 3+3 2 | 2+2 2 | 2+2 2 | 3+3 2 | 3+3 2 | 3+3 2 | 2+2 2 | 3+3 2 |
| | 3+3 2 | 3+3 2 | 2+2 2 | 2+2 2 | 2+2 2 | 2+2 2 | 3+3 2 | 2+2 2 |
| | 3+3 2 | 2+2 2 | 3+3 2 | 3+3 2 | 3+3 2 | 2+2 2 | 2+2 2 | 3+3 2 |
| | 2+2 2 | 3+3 2 | 2+2 2 | 2+2 2 | 2+2 2 | 3+3 2 | 2+2 2 | 2+2 2 |
| | 2+2 2 | 2+2 2 | 3+3 2 | 3+3 2 | 3+3 2 | 3+3 2 | 3+3 2 | 3+3 2 |
| | 3+3 2 | 3+3 2 | 2+2 2 | 2+2 2 | 2+2 2 | 2+2 2 | 3+3 2 | 3+3 2 |
| | 2+2 2 | 3+3 2 | 3+3 2 | 2+2 2 | 2+2 2 | 3+3 2 | 2+2 2 | 2+2 2 |
| | 3+3 2 | 2+2 2 | 2+2 2 | 3+3 2 | 3+3 2 | 2+2 2 | 3+3 2 | 3+3 2 |
| | 2+2 2 | 2+2 2 | 2+2 2 | 2+2 2 | 3+3 2 | 3+3 2 | 2+2 2 | 2+2 2 |
| | 3+3 2 | 2+2 2 | 2+2 2 | 3+3 2 | 2+2 2 | 3+3 2 | 2+2 2 | 3+3 2 |
| | 2+2 2 | 3+3 2 | 3+3 2 | 3+3 2 | 3+3 2 | 2+2 2 | 3+3 2 | 2+2 2 |
| | 3+3 2 | 2+2 2 | 3+3 2 | 2+2 2 | 2+2 2 | 2+2 2 | 3+3 2 | 3+3 2 |
| | 2+2 2 | 3+3 2 | 2+2 2 | 3+3 2 | 3+3 2 | 3+3 2 | 2+2 2 | 2+2 2 |
| | 3+3 2 | 2+2 2 | 3+3 2 | 2+2 2 | 2+2 2 | 2+2 2 | 3+3 2 | 3+3 2 |
| | 2+2 2 | 3+3 2 | 3+3 2 | 3+3 2 | 3+3 2 | 3+3 2 | 2+2 2 | 2+2 2 |
| | 2+2 2 | 3+3 2 | 2+2 2 | 2+2 2 | 3+3 2 | 2+2 2 | 3+3 2 | 3+3 2 |
| | 3+3 2 | 2+2 2 | 3+3 2 | 3+3 2 | 2+2 2 | 3+3 2 | 2+2 2 | 2+2 2 |
| | 2+2 2 | 3+3 2 | 2+2 2 | 2+2 2 | 3+3 2 | 3+3 2 | 3+3 2 | 3+3 2 |
| | 3+3 2 | 3+3 2 | 2+2 2 | 2+2 2 | 2+2 2 | 3+3 2 | 2+2 2 | 2+2 2 |
| | 2+2 2 | 3+3 2 | 3+3 2 | 3+3 2 | 2+2 2 | 2+2 2 | 3+3 2 | 2+2 2 |
| | 3+3 2 | 2+2 2 | 2+2 2 | 3+3 2 | 3+3 2 | 3+3 2 | 2+2 2 | 3+3 2 |
| | 3+3 2 | 3+3 2 | 2+2 2 | 3+3 2 | 2+2 2 | 2+2 2 | 2+2 2 | 3+3 2 |
| | 2+2 2 | 3+3 2 | 2+2 2 | 2+2 2 | 3+3 2 | 3+3 2 | 3+3 2 | 2+2 2 |
| | 2+2 2 | 2+2 2 | 3+3 2 | 3+3 2 | 3+3 2 | 2+2 2 | 2+2 2 | 3+3 2 |
| | 3+3 2 | 3+3 2 | 2+2 2 | 3+3 2 | 2+2 2 | 2+2 2 | 2+2 2 | 2+2 2 |
| | 2+2 2 | 2+2 2 | 3+3 2 | 2+2 2 | 3+3 2 | 3+3 2 | 3+3 2 | 3+3 2 |
| | 3+3 2 | 3+3 2 | 3+3 2 | 2+2 2 | 2+2 2 | 2+2 2 | 3+3 2 | 3+3 2 |
| | 2+2 2 | 3+3 2 | 2+2 2 | 2+2 2 | 3+3 2 | 3+3 2 | 3+3 2 | 2+2 2 |
| | 3+3 2 | 2+2 2 | 2+2 2 | 3+3 2 | 3+3 2 | 3+3 2 | 3+3 2 | 2+2 2 |
| | 2+2 2 | 2+2 2 | 3+3 2 | 3+3 2 | 2+2 2 | 2+2 2 | 2+2 2 | 3+3 2 |
| | 2+2 2 | 3+3 2 | 2+2 2 | 2+2 2 | 2+2 2 | 3+3 2 | 2+2 2 | 2+2 2 |
| | 3+3 2 | 2+2 2 | 3+3 2 | 2+2 2 | 2+2 2 | 2+2 2 | 3+3 2 | 3+3 2 |
| | 2+2 2 | 3+3 2 | 3+3 2 | 3+3 2 | 3+3 2 | 3+3 2 | 2+2 2 | 2+2 2 |
| | 3+3 2 | 2+2 2 | 2+2 2 | 2+2 2 | 2+2 2 | 2+2 2 | 2+2 2 | 3+3 2 |
| | 2+2 2 | 3+3 2 | 3+3 2 | 3+3 2 | 3+3 2 | 2+2 2 | 3+3 2 | 2+2 2 |
| | 3+3 2 | 2+2 2 | 2+2 2 | 2+2 2 | 2+2 2 | 3+3 2 | 3+3 2 | 3+3 2 |
| | 3+3 2 | 3+3 2 | 3+3 2 | 3+3 2 | 3+3 2 | 3+3 2 | 3+3 2 | 2+2 2 |

Continued on next column

**TABLE XXI– continued**

| TQC | ICSD | NoE | SF | | | | | |
|---|---|---|---|---|---|---|---|---|
| | 2+2 2 | 2+2 2 | 2+2 2 | 2+2 2 | 2+2 2 | 2+2 2 | 2+2 2 | 3+3 2 |
| | 2+2 2 | 2+2 2 | 2+2 2 | 3+3 2 | 3+3 2 | 3+3 2 | 3+3 2 | 2+2 2 |
| | 3+3 2 | 3+3 2 | 2+2 2 | 2+2 2 | 2+2 2 | 2+2 2 | 2+2 2 | 2+2 2 |
| | 2+2 2 | 3+3 2 | 3+3 2 | 3+3 2 | 3+3 2 | 3+3 2 | 2+2 2 | 3+3 2 |
| | 3+3 2 | 3+3 2 | 2+2 2 | 3+3 2 | 2+2 2 | 2+2 2 | 3+3 2 | 3+3 2 |
| | 3+3 2 | 2+2 2 | 3+3 2 | 2+2 2 | 3+3 2 | 3+3 2 | 2+2 2 | 2+2 2 |
| | 2+2 2 | 3+3 2 | 2+2 2 | 3+3 2 | 2+2 2 | 2+2 2 | 3+3 2 | 2+2 2 |
| | 2+2 2 | 3+3 2 | 3+3 2 | 2+2 2 | 3+3 2 | 3+3 2 | 2+2 2 | 3+3 2 |
| | 3+3 2 | 2+2 2 | 2+2 2 | 3+3 2 | 2+2 2 | 2+2 2 | 3+3 2 | 2+2 2 |
| | 2+2 2 | 3+3 2 | 2+2 2 | 2+2 2 | 3+3 2 | 2+2 2 | 2+2 2 | 3+3 2 |
| | 3+3 2 | 2+2 2 | 3+3 2 | 3+3 2 | 2+2 2 | 3+3 2 | 3+3 2 | 2+2 2 |
| | 2+2 2 | 3+3 2 | 2+2 2 | 2+2 2 | 3+3 2 | 2+2 2 | 2+2 2 | 3+3 2 |
| | 3+3 2 | 2+2 2 | 3+3 2 | 3+3 2 | 2+2 2 | 2+2 2 | 2+2 2 | 2+2 2 |
| | 3+3 2 | 3+3 2 | 2+2 2 | 2+2 2 | 2+2 2 | 3+3 2 | 3+3 2 | 3+3 2 |
| | 3+3 2 | 2+2 2 | 2+2 2 | 2+2 2 | 3+3 2 | 2+2 2 | 2+2 2 | 2+2 2 |
| | 2+2 2 | 3+3 2 | 3+3 2 | 3+3 2 | 3+3 2 | 2+2 2 | 2+2 2 | 3+3 2 |
| | 3+3 2 | 2+2 2 | 2+2 2 | 2+2 2 | 2+2 2 | 3+3 2 | 3+3 2 | 3+3 2 |
| | 2+2 2 | 3+3 2 | 3+3 2 | 3+3 2 | 3+3 2 | 2+2 2 | 3+3 2 | 2+2 2 |
| | 3+3 2 | 2+2 2 | 3+3 2 | 2+2 2 | 3+3 2 | 3+3 2 | 2+2 2 | 3+3 2 |
| | 3+3 2 | 2+2 2 | 2+2 2 | 3+3 2 | 2+2 2 | 2+2 2 | 2+2 2 | 2+2 2 |
| | 2+2 2 | 3+3 2 | 3+3 2 | 3+3 2 | 3+3 2 | 3+3 2 | 3+3 2 | 2+2 2 |
| | 3+3 2 | 2+2 2 | 3+3 2 | 2+2 2 | 2+2 2 | 2+2 2 | 3+3 2 | 3+3 2 |
| | 2+2 2 | 2+2 2 | 2+2 2 | 2+2 2 | 3+3 2 | 2+2 2 | 3+3 2 | 2+2 2 |
| | 3+3 2 | 3+3 2 | 3+3 2 | 3+3 2 | 2+2 2 | 2+2 2 | 2+2 2 | 3+3 2 |
| | 3+3 2 | 3+3 2 | 3+3 2 | 3+3 2 | 3+3 2 | 3+3 2 | 2+2 2 | 2+2 2 |
| | 2+2 2 | 2+2 2 | 3+3 2 | 2+2 2 | 2+2 2 | 3+3 2 | 3+3 2 | 3+3 2 |
| | 3+3 2 | 2+2 2 | 2+2 2 | 3+3 2 | 3+3 2 | 2+2 2 | 3+3 2 | 2+2 2 |
| | 3+3 2 | 2+2 2 | 3+3 2 | 2+2 2 | 2+2 2 | 2+2 2 | 2+2 2 | 3+3 2 |
| | 3+3 2 | 3+3 2 | 2+2 2 | 3+3 2 | 3+3 2 | 2+2 2 | 2+2 2 | 3+3 2 |
| | 3+3 2 | 3+3 2 | 3+3 2 | 2+2 2 | 3+3 2 | 2+2 2 | 3+3 2 | 2+2 2 |
| | 3+3 2 | 2+2 2 | 2+2 2 | 2+2 2 | 2+2 2 | 2+2 2 | 2+2 2 | 3+3 2 |
| | 3+3 2 | 3+3 2 | 2+2 2 | 3+3 2 | 2+2 2 | 2+2 2 | 3+3 2 | 2+2 2 |
| | 2+2 2 | 2+2 2 | 2+2 2 | 3+3 2 | 2+2 2 | 3+3 2 | 3+3 2 | 2+2 2 |
| | 3+3 2 | 3+3 2 | 3+3 2 | 2+2 2 | 3+3 2 | 2+2 2 | 2+2 2 | 3+3 2 |
| | 3+3 2 | 3+3 2 | 2+2 2 | 3+3 2 | 3+3 2 | 2+2 2 | 3+3 2 | 2+2 2 |
| | 3+3 2 | 2+2 2 | 3+3 2 | 2+2 2 | 2+2 2 | 3+3 2 | 2+2 2 | 3+3 2 |
| 2.2.99 | 421341 | 82 | Ag1O4S1 | | | | | |
| | -GM | -R | -T | -U | -V | -X | -Y | -Z |
| | 3+3 2 | 2+2 2 | 2+2 2 | 2+2 2 | 2+2 2 | 3+3 2 | 2+2 2 | 3+3 2 |
| | 2+2 2 | 3+3 2 | 3+3 2 | 3+3 2 | 3+3 2 | 2+2 2 | 3+3 2 | 2+2 2 |
| | 3+3 2 | 2+2 2 | 3+3 2 | 3+3 2 | 2+2 2 | 3+3 2 | 3+3 2 | 3+3 2 |
| | 2+2 2 | 3+3 2 | 2+2 2 | 2+2 2 | 3+3 2 | 2+2 2 | 2+2 2 | 2+2 2 |
| | 3+3 2 | 3+3 2 | 3+3 2 | 2+2 2 | 3+3 2 | 2+2 2 | 3+3 2 | 3+3 2 |
| | 2+2 2 | 2+2 2 | 2+2 2 | 3+3 2 | 2+2 2 | 3+3 2 | 2+2 2 | 2+2 2 |
| | 3+3 2 | 2+2 2 | 2+2 2 | 2+2 2 | 3+3 2 | 3+3 2 | 3+3 2 | 2+2 2 |
| | 2+2 2 | 3+3 2 | 3+3 2 | 3+3 2 | 2+2 2 | 2+2 2 | 2+2 2 | 3+3 2 |
| | 3+3 2 | 2+2 2 | 2+2 2 | 2+2 2 | 3+3 2 | 3+3 2 | 3+3 2 | 2+2 2 |
| | 2+2 2 | 3+3 2 | 3+3 2 | 3+3 2 | 2+2 2 | 2+2 2 | 2+2 2 | 3+3 2 |
| | 3+3 2 | 2+2 2 | 3+3 2 | 2+2 2 | 3+3 2 | 3+3 2 | 3+3 2 | 2+2 2 |
| | 3+3 2 | 3+3 2 | 3+3 2 | 3+3 2 | 3+3 2 | 2+2 2 | 3+3 2 | 3+3 2 |
| | 2+2 2 | 2+2 2 | 2+2 2 | 2+2 2 | 3+3 2 | 3+3 2 | 3+3 2 | 2+2 2 |
| | 3+3 2 | 2+2 2 | 2+2 2 | 3+3 2 | 2+2 2 | 2+2 2 | 2+2 2 | 3+3 2 |
| | 2+2 2 | 3+3 2 | 3+3 2 | 3+3 2 | 2+2 2 | 2+2 2 | 2+2 2 | 2+2 2 |
| | 2+2 2 | 3+3 2 | 2+2 2 | 2+2 2 | 2+2 2 | 3+3 2 | 2+2 2 | 3+3 2 |
| | 3+3 2 | 3+3 2 | 3+3 2 | 3+3 2 | 3+3 2 | 3+3 2 | 3+3 2 | 3+3 2 |
| | 3+3 2 | 2+2 2 | 3+3 2 | 3+3 2 | 3+3 2 | 2+2 2 | 2+2 2 | 2+2 2 |
| | 3+3 2 | 3+3 2 | 3+3 2 | 3+3 2 | 3+3 2 | 2+2 2 | 2+2 2 | 3+3 2 |
| | 3+3 2 | 2+2 2 | 2+2 2 | 3+3 2 | 2+2 2 | 3+3 2 | 3+3 2 | 2+2 2 |
| | 2+2 2 | 2+2 2 | 2+2 2 | 2+2 2 | 3+3 2 | 3+3 2 | 3+3 2 | 3+3 2 |
| | 3+3 2 | 2+2 2 | 3+3 2 | 2+2 2 | 2+2 2 | 2+2 2 | 2+2 2 | 2+2 2 |
| | 2+2 2 | 3+3 2 | 2+2 2 | 2+2 2 | 2+2 2 | 3+3 2 | 3+3 2 | 3+3 2 |

Continued on next column

**TABLE XXI– continued**

| TQC | ICSD | NoE | SF | | | | | |
|---|---|---|---|---|---|---|---|---|
| | 3+3 2 | 2+2 2 | 3+3 2 | 3+3 2 | 2+2 2 | 3+3 2 | 2+2 2 | 2+2 2 |
| | 2+2 2 | 3+3 2 | 2+2 2 | 2+2 2 | 3+3 2 | 2+2 2 | 3+3 2 | 2+2 2 |
| | 3+3 2 | 3+3 2 | 3+3 2 | 3+3 2 | 3+3 2 | 2+2 2 | 2+2 2 | 3+3 2 |
| | 2+2 2 | 3+3 2 | 2+2 2 | 2+2 2 | 3+3 2 | 3+3 2 | 3+3 2 | 2+2 2 |
| | 3+3 2 | 2+2 2 | 3+3 2 | 3+3 2 | 3+3 2 | 2+2 2 | 3+3 2 | 3+3 2 |
| | 3+3 2 | 2+2 2 | 3+3 2 | 3+3 2 | 3+3 2 | 3+3 2 | 2+2 2 | 2+2 2 |
| | 3+3 2 | 3+3 2 | 3+3 2 | 3+3 2 | 3+3 2 | 2+2 2 | 2+2 2 | 3+3 2 |
| | 2+2 2 | 3+3 2 | 3+3 2 | 3+3 2 | 2+2 2 | 3+3 2 | 3+3 2 | 3+3 2 |
| | 3+3 2 | 2+2 2 | 3+3 2 | 3+3 2 | 3+3 2 | 2+2 2 | 2+2 2 | 2+2 2 |
| | 3+3 2 | 3+3 2 | 3+3 2 | 3+3 2 | 3+3 2 | 3+3 2 | 3+3 2 | 2+2 2 |
| | 2+2 2 | 3+3 2 | 3+3 2 | 3+3 2 | 3+3 2 | 2+2 2 | 3+3 2 | 3+3 2 |
| | 3+3 2 | 2+2 2 | 2+2 2 | 2+2 2 | 2+2 2 | 3+3 2 | 2+2 2 | 3+3 2 |
| | 2+2 2 | 2+2 2 | 3+3 2 | 3+3 2 | 3+3 2 | 2+2 2 | 2+2 2 | 2+2 2 |
| | 3+3 2 | 3+3 2 | 3+3 2 | 2+2 2 | 3+3 2 | 2+2 2 | 3+3 2 | 3+3 2 |
| | 3+3 2 | 2+2 2 | 2+2 2 | 3+3 2 | 2+2 2 | 3+3 2 | 2+2 2 | 2+2 2 |
| | 2+2 2 | 3+3 2 | 2+2 2 | 2+2 2 | 2+2 2 | 2+2 2 | 3+3 2 | 3+3 2 |
| | 3+3 2 | 2+2 2 | 3+3 2 | 3+3 2 | 3+3 2 | 3+3 2 | 2+2 2 | 2+2 2 |
| | 3+3 2 | 2+2 2 | 3+3 2 | 3+3 2 | 3+3 2 | 3+3 2 | 3+3 2 | 3+3 2 |
| 2.2.100 | 51507 | 126 | Ag1F6Pd1 | | | | | |
| | -GM | -R | -T | -U | -V | -X | -Y | -Z |
| | 3+3 2 | 2+2 2 | 3+3 2 | 3+3 2 | 2+2 2 | 2+2 2 | 3+3 2 | 3+3 2 |
| | 2+2 2 | 3+3 2 | 2+2 2 | 2+2 2 | 3+3 2 | 3+3 2 | 2+2 2 | 2+2 2 |
| | 3+3 2 | 2+2 2 | 3+3 2 | 3+3 2 | 3+3 2 | 3+3 2 | 3+3 2 | 3+3 2 |
| | 3+3 2 | 3+3 2 | 3+3 2 | 2+2 2 | 3+3 2 | 2+2 2 | 2+2 2 | 2+2 2 |
| | 2+2 2 | 2+2 2 | 2+2 2 | 3+3 2 | 2+2 2 | 2+2 2 | 2+2 2 | 2+2 2 |
| | 2+2 2 | 3+3 2 | 2+2 2 | 2+2 2 | 2+2 2 | 3+3 2 | 3+3 2 | 3+3 2 |
| | 3+3 2 | 3+3 2 | 3+3 2 | 3+3 2 | 2+2 2 | 2+2 2 | 3+3 2 | 2+2 2 |
| | 2+2 2 | 2+2 2 | 2+2 2 | 2+2 2 | 3+3 2 | 3+3 2 | 2+2 2 | 3+3 2 |
| | 3+3 2 | 3+3 2 | 2+2 2 | 2+2 2 | 3+3 2 | 2+2 2 | 2+2 2 | 3+3 2 |
| | 3+3 2 | 2+2 2 | 3+3 2 | 2+2 2 | 2+2 2 | 3+3 2 | 3+3 2 | 2+2 2 |
| | 2+2 2 | 3+3 2 | 3+3 2 | 3+3 2 | 2+2 2 | 3+3 2 | 2+2 2 | 2+2 2 |
| | 2+2 2 | 2+2 2 | 2+2 2 | 3+3 2 | 3+3 2 | 2+2 2 | 3+3 2 | 3+3 2 |
| | 3+3 2 | 3+3 2 | 3+3 2 | 3+3 2 | 3+3 2 | 3+3 2 | 2+2 2 | 3+3 2 |
| | 3+3 2 | 2+2 2 | 3+3 2 | 3+3 2 | 3+3 2 | 2+2 2 | 3+3 2 | 2+2 2 |
| | 2+2 2 | 3+3 2 | 2+2 2 | 2+2 2 | 2+2 2 | 3+3 2 | 2+2 2 | 3+3 2 |
| | 2+2 2 | 2+2 2 | 2+2 2 | 2+2 2 | 2+2 2 | 2+2 2 | 3+3 2 | 2+2 2 |
| | 3+3 2 | 3+3 2 | 3+3 2 | 3+3 2 | 3+3 2 | 3+3 2 | 3+3 2 | 2+2 2 |
| | 3+3 2 | 2+2 2 | 3+3 2 | 3+3 2 | 3+3 2 | 2+2 2 | 2+2 2 | 3+3 2 |
| | 3+3 2 | 3+3 2 | 2+2 2 | 2+2 2 | 2+2 2 | 2+2 2 | 3+3 2 | 3+3 2 |
| | 3+3 2 | 2+2 2 | 3+3 2 | 2+2 2 | 2+2 2 | 3+3 2 | 2+2 2 | 2+2 2 |
| | 2+2 2 | 3+3 2 | 2+2 2 | 3+3 2 | 3+3 2 | 2+2 2 | 3+3 2 | 2+2 2 |
| | 3+3 2 | 2+2 2 | 3+3 2 | 3+3 2 | 3+3 2 | 3+3 2 | 2+2 2 | 3+3 2 |
| | 2+2 2 | 3+3 2 | 2+2 2 | 2+2 2 | 2+2 2 | 2+2 2 | 3+3 2 | 2+2 2 |
| | 2+2 2 | 2+2 2 | 3+3 2 | 3+3 2 | 3+3 2 | 3+3 2 | 2+2 2 | 3+3 2 |
| | 2+2 2 | 3+3 2 | 2+2 2 | 2+2 2 | 2+2 2 | 3+3 2 | 2+2 2 | 3+3 2 |
| | 3+3 2 | 2+2 2 | 3+3 2 | 3+3 2 | 3+3 2 | 2+2 2 | 3+3 2 | 2+2 2 |
| | 3+3 2 | 2+2 2 | 3+3 2 | 3+3 2 | 3+3 2 | 2+2 2 | 3+3 2 | 3+3 2 |
| | 2+2 2 | 3+3 2 | 2+2 2 | 2+2 2 | 2+2 2 | 3+3 2 | 2+2 2 | 2+2 2 |
| | 2+2 2 | 2+2 2 | 3+3 2 | 2+2 2 | 3+3 2 | 3+3 2 | 2+2 2 | 2+2 2 |
| | 2+2 2 | 3+3 2 | 2+2 2 | 3+3 2 | 2+2 2 | 2+2 2 | 3+3 2 | 3+3 2 |
| | 3+3 2 | 2+2 2 | 2+2 2 | 2+2 2 | 2+2 2 | 3+3 2 | 2+2 2 | 3+3 2 |
| | 3+3 2 | 3+3 2 | 3+3 2 | 3+3 2 | 3+3 2 | 2+2 2 | 3+3 2 | 2+2 2 |
| | 3+3 2 | 2+2 2 | 3+3 2 | 3+3 2 | 3+3 2 | 2+2 2 | 3+3 2 | 2+2 2 |
| | 3+3 2 | 3+3 2 | 3+3 2 | 3+3 2 | 3+3 2 | 3+3 2 | 2+2 2 | 3+3 2 |
| | 2+2 2 | 2+2 2 | 2+2 2 | 2+2 2 | 2+2 2 | 2+2 2 | 3+3 2 | 3+3 2 |
| | 3+3 2 | 3+3 2 | 3+3 2 | 3+3 2 | 3+3 2 | 3+3 2 | 2+2 2 | 2+2 2 |
| | 2+2 2 | 3+3 2 | 2+2 2 | 2+2 2 | 2+2 2 | 2+2 2 | 2+2 2 | 2+2 2 |
| | 3+3 2 | 2+2 2 | 2+2 2 | 2+2 2 | 2+2 2 | 3+3 2 | 3+3 2 | 3+3 2 |
| | 2+2 2 | 3+3 2 | 3+3 2 | 3+3 2 | 3+3 2 | 3+3 2 | 3+3 2 | 2+2 2 |
| | 3+3 2 | 2+2 2 | 2+2 2 | 2+2 2 | 2+2 2 | 2+2 2 | 2+2 2 | 3+3 2 |
| | 2+2 2 | 3+3 2 | 3+3 2 | 3+3 2 | 3+3 2 | 2+2 2 | 3+3 2 | 3+3 2 |

Continued on next column

**TABLE XXI– continued**

| TQC | ICSD | NoE | SF | | | | |
|---|---|---|---|---|---|---|---|
| | 3+3 2 | 3+3 2 | 2+2 2 | 2+2 2 | 2+2 2 | 3+3 2 | 2+2 2 | 2+2 2 |
| | 3+3 2 | 2+2 2 | 3+3 2 | 3+3 2 | 2+2 2 | 2+2 2 | 2+2 2 | 3+3 2 |
| | 2+2 2 | 2+2 2 | 2+2 2 | 3+3 2 | 3+3 2 | 3+3 2 | 3+3 2 | 2+2 2 |
| | 3+3 2 | 3+3 2 | 3+3 2 | 2+2 2 | 3+3 2 | 2+2 2 | 2+2 2 | 2+2 2 |
| | 2+2 2 | 2+2 2 | 2+2 2 | 2+2 2 | 2+2 2 | 3+3 2 | 3+3 2 | 3+3 2 |
| | 2+2 2 | 3+3 2 | 3+3 2 | 2+2 2 | 3+3 2 | 2+2 2 | 2+2 2 | 3+3 2 |
| | 3+3 2 | 2+2 2 | 2+2 2 | 3+3 2 | 2+2 2 | 3+3 2 | 3+3 2 | 2+2 2 |
| | 2+2 2 | 2+2 2 | 3+3 2 | 2+2 2 | 3+3 2 | 3+3 2 | 2+2 2 | 3+3 2 |
| | 3+3 2 | 3+3 2 | 2+2 2 | 3+3 2 | 2+2 2 | 2+2 2 | 3+3 2 | 2+2 2 |
| | 2+2 2 | 2+2 2 | 3+3 2 | 3+3 2 | 3+3 2 | 3+3 2 | 2+2 2 | 2+2 2 |
| | 3+3 2 | 3+3 2 | 3+3 2 | 3+3 2 | 3+3 2 | 2+2 2 | 3+3 2 | 3+3 2 |
| | 3+3 2 | 2+2 2 | 3+3 2 | 3+3 2 | 3+3 2 | 3+3 2 | 3+3 2 | 2+2 2 |
| | 3+3 2 | 3+3 2 | 3+3 2 | 3+3 2 | 3+3 2 | 2+2 2 | 2+2 2 | 3+3 2 |
| | 3+3 2 | 3+3 2 | 2+2 2 | 2+2 2 | 2+2 2 | 2+2 2 | 2+2 2 | 3+3 2 |
| | 3+3 2 | 2+2 2 | 3+3 2 | 3+3 2 | 3+3 2 | 3+3 2 | 3+3 2 | 2+2 2 |
| | 2+2 2 | 2+2 2 | 3+3 2 | 3+3 2 | 3+3 2 | 3+3 2 | 2+2 2 | 2+2 2 |
| | 2+2 2 | 3+3 2 | 2+2 2 | 2+2 2 | 2+2 2 | 2+2 2 | 3+3 2 | 3+3 2 |
| | 2+2 2 | 3+3 2 | 2+2 2 | 3+3 2 | 3+3 2 | 2+2 2 | 2+2 2 | 3+3 2 |
| | 3+3 2 | 2+2 2 | 3+3 2 | 2+2 2 | 2+2 2 | 3+3 2 | 3+3 2 | 2+2 2 |
| | 3+3 2 | 2+2 2 | 3+3 2 | 3+3 2 | 3+3 2 | 3+3 2 | 2+2 2 | 2+2 2 |
| | 3+3 2 | 3+3 2 | 3+3 2 | 3+3 2 | 3+3 2 | 2+2 2 | 3+3 2 | 3+3 2 |
| | 3+3 2 | 3+3 2 | 3+3 2 | 3+3 2 | 3+3 2 | 2+2 2 | 3+3 2 | 3+3 2 |
| 2.2.101 | 174407 | 172 | Ag1Cu1F3 | | | | |
| | -GM | -R | -T | -U | -V | -X | -Y | -Z |
| | 3+3 2 | 3+3 2 | 3+3 2 | 3+3 2 | 2+2 2 | 2+2 2 | 2+2 2 | 2+2 2 |
| | 3+3 2 | 2+2 2 | 2+2 2 | 2+2 2 | 2+2 2 | 3+3 2 | 3+3 2 | 3+3 2 |
| | 2+2 2 | 2+2 2 | 2+2 2 | 2+2 2 | 3+3 2 | 2+2 2 | 2+2 2 | 3+3 2 |
| | 2+2 2 | 3+3 2 | 3+3 2 | 3+3 2 | 3+3 2 | 3+3 2 | 3+3 2 | 2+2 2 |
| | 3+3 2 | 3+3 2 | 2+2 2 | 2+2 2 | 2+2 2 | 3+3 2 | 3+3 2 | 2+2 2 |
| | 2+2 2 | 2+2 2 | 3+3 2 | 3+3 2 | 2+2 2 | 2+2 2 | 2+2 2 | 3+3 2 |
| | 3+3 2 | 2+2 2 | 3+3 2 | 3+3 2 | 3+3 2 | 3+3 2 | 3+3 2 | 2+2 2 |
| | 2+2 2 | 3+3 2 | 2+2 2 | 2+2 2 | 3+3 2 | 2+2 2 | 2+2 2 | 3+3 2 |
| | 2+2 2 | 2+2 2 | 2+2 2 | 3+3 2 | 3+3 2 | 3+3 2 | 3+3 2 | 3+3 2 |
| | 3+3 2 | 3+3 2 | 3+3 2 | 2+2 2 | 3+3 2 | 2+2 2 | 2+2 2 | 2+2 2 |
| | 2+2 2 | 3+3 2 | 3+3 2 | 3+3 2 | 2+2 2 | 3+3 2 | 3+3 2 | 2+2 2 |
| | 3+3 2 | 2+2 2 | 2+2 2 | 2+2 2 | 2+2 2 | 2+2 2 | 2+2 2 | 3+3 2 |
| | 3+3 2 | 3+3 2 | 2+2 2 | 2+2 2 | 2+2 2 | 2+2 2 | 2+2 2 | 2+2 2 |
| | 3+3 2 | 2+2 2 | 3+3 2 | 3+3 2 | 2+2 2 | 3+3 2 | 3+3 2 | 3+3 2 |
| | 3+3 2 | 3+3 2 | 3+3 2 | 2+2 2 | 2+2 2 | 2+2 2 | 3+3 2 | 2+2 2 |
| | 3+3 2 | 2+2 2 | 2+2 2 | 3+3 2 | 2+2 2 | 3+3 2 | 2+2 2 | 3+3 2 |
| | 3+3 2 | 3+3 2 | 2+2 2 | 3+3 2 | 2+2 2 | 2+2 2 | 3+3 2 | 2+2 2 |
| | 3+3 2 | 2+2 2 | 3+3 2 | 2+2 2 | 2+2 2 | 3+3 2 | 3+3 2 | 3+3 2 |
| | 2+2 2 | 2+2 2 | 3+3 2 | 2+2 2 | 2+2 2 | 3+3 2 | 2+2 2 | 2+2 2 |
| | 3+3 2 | 3+3 2 | 2+2 2 | 3+3 2 | 3+3 2 | 2+2 2 | 2+2 2 | 3+3 2 |
| | 2+2 2 | 2+2 2 | 2+2 2 | 2+2 2 | 2+2 2 | 3+3 2 | 3+3 2 | 2+2 2 |
| | 3+3 2 | 3+3 2 | 3+3 2 | 3+3 2 | 3+3 2 | 2+2 2 | 2+2 2 | 3+3 2 |
| | 3+3 2 | 2+2 2 | 3+3 2 | 3+3 2 | 3+3 2 | 2+2 2 | 2+2 2 | 3+3 2 |
| | 2+2 2 | 3+3 2 | 2+2 2 | 2+2 2 | 3+3 2 | 2+2 2 | 3+3 2 | 2+2 2 |
| | 2+2 2 | 3+3 2 | 3+3 2 | 3+3 2 | 3+3 2 | 3+3 2 | 2+2 2 | 3+3 2 |
| | 2+2 2 | 2+2 2 | 2+2 2 | 2+2 2 | 2+2 2 | 3+3 2 | 3+3 2 | 2+2 2 |
| | 3+3 2 | 3+3 2 | 3+3 2 | 2+2 2 | 3+3 2 | 2+2 2 | 2+2 2 | 2+2 2 |
| | 2+2 2 | 2+2 2 | 2+2 2 | 3+3 2 | 2+2 2 | 2+2 2 | 3+3 2 | 3+3 2 |
| | 3+3 2 | 3+3 2 | 2+2 2 | 2+2 2 | 3+3 2 | 3+3 2 | 2+2 2 | 2+2 2 |
| | 2+2 2 | 2+2 2 | 3+3 2 | 2+2 2 | 3+3 2 | 3+3 2 | 3+3 2 | 3+3 2 |
| | 3+3 2 | 2+2 2 | 2+2 2 | 3+3 2 | 2+2 2 | 3+3 2 | 3+3 2 | 2+2 2 |
| | 2+2 2 | 3+3 2 | 3+3 2 | 3+3 2 | 2+2 2 | 2+2 2 | 2+2 2 | 3+3 2 |
| | 3+3 2 | 3+3 2 | 2+2 2 | 3+3 2 | 3+3 2 | 3+3 2 | 3+3 2 | 3+3 2 |
| | 2+2 2 | 2+2 2 | 3+3 2 | 2+2 2 | 3+3 2 | 2+2 2 | 2+2 2 | 2+2 2 |
| | 3+3 2 | 3+3 2 | 2+2 2 | 2+2 2 | 2+2 2 | 2+2 2 | 2+2 2 | 2+2 2 |
| | 2+2 2 | 2+2 2 | 3+3 2 | 3+3 2 | 3+3 2 | 3+3 2 | 3+3 2 | 3+3 2 |
| | 2+2 2 | 2+2 2 | 3+3 2 | 3+3 2 | 2+2 2 | 3+3 2 | 3+3 2 | 2+2 2 |

Continued on next column

**TABLE XXI– continued**

| TQC | ICSD | NoE | SF | | | | |
|---|---|---|---|---|---|---|---|
| | 2+2 2 | 3+3 2 | 2+2 2 | 2+2 2 | 3+3 2 | 2+2 2 | 2+2 2 | 3+3 2 |
| | 2+2 2 | 2+2 2 | 2+2 2 | 3+3 2 | 3+3 2 | 3+3 2 | 2+2 2 | 2+2 2 |
| | 3+3 2 | 3+3 2 | 3+3 2 | 2+2 2 | 2+2 2 | 2+2 2 | 3+3 2 | 3+3 2 |
| | 3+3 2 | 2+2 2 | 2+2 2 | 2+2 2 | 2+2 2 | 2+2 2 | 2+2 2 | 3+3 2 |
| | 2+2 2 | 3+3 2 | 3+3 2 | 3+3 2 | 3+3 2 | 3+3 2 | 3+3 2 | 2+2 2 |
| | 2+2 2 | 3+3 2 | 2+2 2 | 2+2 2 | 3+3 2 | 3+3 2 | 3+3 2 | 3+3 2 |
| | 2+2 2 | 2+2 2 | 3+3 2 | 3+3 2 | 3+3 2 | 2+2 2 | 2+2 2 | 2+2 2 |
| | 3+3 2 | 2+2 2 | 2+2 2 | 2+2 2 | 3+3 2 | 3+3 2 | 3+3 2 | 3+3 2 |
| | 2+2 2 | 3+3 2 | 3+3 2 | 3+3 2 | 2+2 2 | 2+2 2 | 2+2 2 | 2+2 2 |
| | 2+2 2 | 3+3 2 | 3+3 2 | 2+2 2 | 3+3 2 | 3+3 2 | 3+3 2 | 2+2 2 |
| | 3+3 2 | 2+2 2 | 2+2 2 | 3+3 2 | 2+2 2 | 2+2 2 | 2+2 2 | 3+3 2 |
| | 3+3 2 | 3+3 2 | 2+2 2 | 2+2 2 | 2+2 2 | 2+2 2 | 2+2 2 | 2+2 2 |
| | 3+3 2 | 2+2 2 | 3+3 2 | 3+3 2 | 2+2 2 | 3+3 2 | 3+3 2 | 3+3 2 |
| | 3+3 2 | 2+2 2 | 2+2 2 | 2+2 2 | 2+2 2 | 2+2 2 | 2+2 2 | 3+3 2 |
| | 3+3 2 | 3+3 2 | 3+3 2 | 3+3 2 | 2+2 2 | 3+3 2 | 3+3 2 | 2+2 2 |
| | 3+3 2 | 3+3 2 | 2+2 2 | 2+2 2 | 2+2 2 | 2+2 2 | 3+3 2 | 3+3 2 |
| | 2+2 2 | 2+2 2 | 3+3 2 | 3+3 2 | 2+2 2 | 3+3 2 | 2+2 2 | 2+2 2 |
| | 3+3 2 | 3+3 2 | 3+3 2 | 2+2 2 | 3+3 2 | 2+2 2 | 3+3 2 | 3+3 2 |
| | 2+2 2 | 2+2 2 | 2+2 2 | 3+3 2 | 2+2 2 | 3+3 2 | 2+2 2 | 2+2 2 |
| | 3+3 2 | 3+3 2 | 3+3 2 | 2+2 2 | 3+3 2 | 2+2 2 | 3+3 2 | 2+2 2 |
| | 3+3 2 | 2+2 2 | 2+2 2 | 3+3 2 | 2+2 2 | 3+3 2 | 2+2 2 | 3+3 2 |
| | 2+2 2 | 3+3 2 | 2+2 2 | 3+3 2 | 2+2 2 | 2+2 2 | 2+2 2 | 2+2 2 |
| | 3+3 2 | 2+2 2 | 3+3 2 | 2+2 2 | 2+2 2 | 3+3 2 | 3+3 2 | 3+3 2 |
| | 3+3 2 | 2+2 2 | 3+3 2 | 2+2 2 | 3+3 2 | 2+2 2 | 3+3 2 | 2+2 2 |
| | 3+3 2 | 3+3 2 | 2+2 2 | 3+3 2 | 3+3 2 | 3+3 2 | 2+2 2 | 3+3 2 |
| | 2+2 2 | 2+2 2 | 2+2 2 | 2+2 2 | 2+2 2 | 2+2 2 | 3+3 2 | 2+2 2 |
| | 3+3 2 | 3+3 2 | 3+3 2 | 3+3 2 | 2+2 2 | 3+3 2 | 2+2 2 | 3+3 2 |
| | 3+3 2 | 3+3 2 | 3+3 2 | 2+2 2 | 2+2 2 | 3+3 2 | 3+3 2 | 2+2 2 |
| | 3+3 2 | 2+2 2 | 2+2 2 | 3+3 2 | 2+2 2 | 2+2 2 | 2+2 2 | 3+3 2 |
| | 3+3 2 | 2+2 2 | 3+3 2 | 3+3 2 | 2+2 2 | 2+2 2 | 3+3 2 | 2+2 2 |
| | 2+2 2 | 3+3 2 | 2+2 2 | 2+2 2 | 2+2 2 | 3+3 2 | 2+2 2 | 3+3 2 |
| | 3+3 2 | 3+3 2 | 2+2 2 | 3+3 2 | 2+2 2 | 3+3 2 | 3+3 2 | 3+3 2 |
| | 3+3 2 | 2+2 2 | 3+3 2 | 2+2 2 | 2+2 2 | 2+2 2 | 2+2 2 | 2+2 2 |
| | 3+3 2 | 2+2 2 | 3+3 2 | 2+2 2 | 3+3 2 | 2+2 2 | 2+2 2 | 2+2 2 |
| | 2+2 2 | 3+3 2 | 2+2 2 | 3+3 2 | 2+2 2 | 3+3 2 | 3+3 2 | 3+3 2 |
| | 3+3 2 | 2+2 2 | 3+3 2 | 3+3 2 | 2+2 2 | 2+2 2 | 2+2 2 | 2+2 2 |
| | 2+2 2 | 3+3 2 | 2+2 2 | 2+2 2 | 3+3 2 | 3+3 2 | 3+3 2 | 3+3 2 |
| | 3+3 2 | 3+3 2 | 2+2 2 | 3+3 2 | 2+2 2 | 3+3 2 | 3+3 2 | 2+2 2 |
| | 3+3 2 | 2+2 2 | 3+3 2 | 2+2 2 | 3+3 2 | 2+2 2 | 2+2 2 | 3+3 2 |
| | 2+2 2 | 2+2 2 | 3+3 2 | 2+2 2 | 2+2 2 | 2+2 2 | 2+2 2 | 2+2 2 |
| | 2+2 2 | 3+3 2 | 2+2 2 | 3+3 2 | 3+3 2 | 3+3 2 | 3+3 2 | 3+3 2 |
| | 3+3 2 | 3+3 2 | 2+2 2 | 2+2 2 | 3+3 2 | 2+2 2 | 2+2 2 | 3+3 2 |
| | 3+3 2 | 2+2 2 | 3+3 2 | 3+3 2 | 2+2 2 | 3+3 2 | 3+3 2 | 2+2 2 |
| | 3+3 2 | 2+2 2 | 2+2 2 | 2+2 2 | 2+2 2 | 2+2 2 | 2+2 2 | 3+3 2 |
| | 2+2 2 | 3+3 2 | 3+3 2 | 3+3 2 | 3+3 2 | 3+3 2 | 3+3 2 | 2+2 2 |
| | 3+3 2 | 3+3 2 | 2+2 2 | 2+2 2 | 2+2 2 | 3+3 2 | 2+2 2 | 2+2 2 |
| | 3+3 2 | 2+2 2 | 3+3 2 | 3+3 2 | 2+2 2 | 2+2 2 | 3+3 2 | 3+3 2 |
| | 3+3 2 | 3+3 2 | 3+3 2 | 3+3 2 | 2+2 2 | 3+3 2 | 2+2 2 | 2+2 2 |
| | 3+3 2 | 2+2 2 | 2+2 2 | 3+3 2 | 2+2 2 | 2+2 2 | 3+3 2 | 3+3 2 |

The End

TABLE XXII: The computed irreducible representations for the valence bands for the compounds of SEBR and NLC in space group 10. See the list of tables.

| TQC | ICSD | NoE | SF | | | | |
|---|---|---|---|---|---|---|---|
| 10.2.1 | 417035 | 96 | Co2Gd4Mg3 | | | | |
| | -A | -B | -C | -D | -E | -GM | -Y | -Z |
| | 3+4 2 | 3+4 2 | 5+6 2 | 5+6 2 | 5+6 2 | 3+4 2 | 3+4 2 | 5+6 2 |
| | 5+6 2 | 5+6 2 | 3+4 2 | 3+4 2 | 3+4 2 | 5+6 2 | 5+6 2 | 3+4 2 |

Continued on next column

**TABLE XXII– continued**

| TQC | ICSD | NoE | SF | | | | | |
|---|---|---|---|---|---|---|---|---|
| | 3+4 2 | 5+6 2 | 3+4 2 | 3+4 2 | 5+6 2 | 3+4 2 | 5+6 2 | 5+6 2 |
| | 5+6 2 | 3+4 2 | 5+6 2 | 5+6 2 | 3+4 2 | 5+6 2 | 3+4 2 | 3+4 2 |
| | 3+4 2 | 5+6 2 | 3+4 2 | 3+4 2 | 5+6 2 | 3+4 2 | 5+6 2 | 5+6 2 |
| | 5+6 2 | 3+4 2 | 5+6 2 | 5+6 2 | 3+4 2 | 5+6 2 | 3+4 2 | 3+4 2 |
| | 3+4 2 | 5+6 2 | 3+4 2 | 3+4 2 | 5+6 2 | 3+4 2 | 5+6 2 | 5+6 2 |
| | 5+6 2 | 3+4 2 | 5+6 2 | 5+6 2 | 3+4 2 | 5+6 2 | 3+4 2 | 3+4 2 |
| | 3+4 2 | 5+6 2 | 3+4 2 | 3+4 2 | 5+6 2 | 3+4 2 | 5+6 2 | 5+6 2 |
| | 5+6 2 | 3+4 2 | 5+6 2 | 5+6 2 | 3+4 2 | 5+6 2 | 3+4 2 | 3+4 2 |
| | 3+4 2 | 3+4 2 | 3+4 2 | 5+6 2 | 5+6 2 | 3+4 2 | 3+4 2 | 5+6 2 |
| | 5+6 2 | 5+6 2 | 5+6 2 | 3+4 2 | 3+4 2 | 5+6 2 | 5+6 2 | 3+4 2 |
| | 3+4 2 | 5+6 2 | 3+4 2 | 3+4 2 | 5+6 2 | 3+4 2 | 3+4 2 | 5+6 2 |
| | 5+6 2 | 3+4 2 | 5+6 2 | 5+6 2 | 3+4 2 | 5+6 2 | 5+6 2 | 3+4 2 |
| | 3+4 2 | 5+6 2 | 3+4 2 | 5+6 2 | 5+6 2 | 3+4 2 | 5+6 2 | 5+6 2 |
| | 5+6 2 | 3+4 2 | 5+6 2 | 3+4 2 | 3+4 2 | 5+6 2 | 3+4 2 | 3+4 2 |
| | 3+4 2 | 3+4 2 | 3+4 2 | 3+4 2 | 3+4 2 | 3+4 2 | 3+4 2 | 3+4 2 |
| | 5+6 2 | 5+6 2 | 5+6 2 | 5+6 2 | 5+6 2 | 3+4 2 | 5+6 2 | 5+6 2 |
| | 5+6 2 | 5+6 2 | 3+4 2 | 3+4 2 | 5+6 2 | 5+6 2 | 3+4 2 | 3+4 2 |
| | 3+4 2 | 3+4 2 | 5+6 2 | 5+6 2 | 5+6 2 | 3+4 2 | 5+6 2 | 3+4 2 |
| | 3+4 2 | 3+4 2 | 5+6 2 | 3+4 2 | 3+4 2 | 5+6 2 | 3+4 2 | 5+6 2 |
| | 5+6 2 | 3+4 2 | 3+4 2 | 5+6 2 | 3+4 2 | 3+4 2 | 5+6 2 | 5+6 2 |
| | 5+6 2 | 5+6 2 | 5+6 2 | 5+6 2 | 5+6 2 | 5+6 2 | 5+6 2 | 5+6 2 |
| | 3+4 2 | 5+6 2 | 3+4 2 | 5+6 2 | 3+4 2 | 3+4 2 | 5+6 2 | 3+4 2 |
| | 5+6 2 | 5+6 2 | 3+4 2 | 3+4 2 | 5+6 2 | 5+6 2 | 3+4 2 | 3+4 2 |
| | 5+6 2 | 3+4 2 | 5+6 2 | 3+4 2 | 5+6 2 | 5+6 2 | 3+4 2 | 5+6 2 |
| | 3+4 2 | 3+4 2 | 5+6 2 | 5+6 2 | 3+4 2 | 5+6 2 | 3+4 2 | 5+6 2 |
| | 3+4 2 | 5+6 2 | 3+4 2 | 3+4 2 | 3+4 2 | 3+4 2 | 5+6 2 | 3+4 2 |
| | 5+6 2 | 5+6 2 | 5+6 2 | 5+6 2 | 3+4 2 | 5+6 2 | 5+6 2 | 3+4 2 |
| | 5+6 2 | 3+4 2 | 3+4 2 | 5+6 2 | 3+4 2 | 3+4 2 | 3+4 2 | 5+6 2 |
| | 3+4 2 | 3+4 2 | 3+4 2 | 3+4 2 | 5+6 2 | 5+6 2 | 5+6 2 | 3+4 2 |
| | 5+6 2 | 5+6 2 | 5+6 2 | 5+6 2 | 5+6 2 | 3+4 2 | 3+4 2 | 5+6 2 |
| | 3+4 2 | 5+6 2 | 5+6 2 | 5+6 2 | 3+4 2 | 3+4 2 | 5+6 2 | 3+4 2 |
| | 3+4 2 | 3+4 2 | 3+4 2 | 3+4 2 | 3+4 2 | 5+6 2 | 3+4 2 | 5+6 2 |
| | 3+4 2 | 5+6 2 | 5+6 2 | 3+4 2 | 5+6 2 | 3+4 2 | 3+4 2 | 3+4 2 |
| | 5+6 2 | 3+4 2 | 3+4 2 | 3+4 2 | 5+6 2 | 3+4 2 | 5+6 2 | 3+4 2 |
| | 5+6 2 | 3+4 2 | 5+6 2 | 5+6 2 | 3+4 2 | 5+6 2 | 3+4 2 | 5+6 2 |
| | 5+6 2 | 5+6 2 | 3+4 2 | 5+6 2 | 5+6 2 | 5+6 2 | 5+6 2 | 5+6 2 |
| | 3+4 2 | 3+4 2 | 5+6 2 | 3+4 2 | 3+4 2 | 3+4 2 | 3+4 2 | 5+6 2 |
| | 3+4 2 | 5+6 2 | 3+4 2 | 5+6 2 | 5+6 2 | 5+6 2 | 5+6 2 | 3+4 2 |
| | 5+6 2 | 3+4 2 | 5+6 2 | 3+4 2 | 3+4 2 | 3+4 2 | 3+4 2 | 3+4 2 |
| | 3+4 2 | 5+6 2 | 3+4 2 | 5+6 2 | 5+6 2 | 5+6 2 | 5+6 2 | 5+6 2 |
| | 5+6 2 | 3+4 2 | 5+6 2 | 5+6 2 | 3+4 2 | 3+4 2 | 3+4 2 | 3+4 2 |
| | 5+6 2 | 5+6 2 | 3+4 2 | 3+4 2 | 3+4 2 | 5+6 2 | 5+6 2 | 5+6 2 |
| | 3+4 2 | 5+6 2 | 5+6 2 | 3+4 2 | 5+6 2 | 3+4 2 | 5+6 2 | 5+6 2 |
| | 3+4 2 | 3+4 2 | 3+4 2 | 5+6 2 | 5+6 2 | 5+6 2 | 3+4 2 | 3+4 2 |
| | 5+6 2 | 5+6 2 | 5+6 2 | 3+4 2 | 5+6 2 | 3+4 2 | 3+4 2 | 3+4 2 |
| | 3+4 2 | 3+4 2 | 3+4 2 | 3+4 2 | 3+4 2 | 3+4 2 | 5+6 2 | 5+6 2 |

The End

TABLE XXIII: The computed irreducible representations for the valence bands for the compounds of SEBR and NLC in space group 11. See the list of tables.

| TQC | ICSD | NoE | SF | | | | | |
|---|---|---|---|---|---|---|---|---|
| 11.2.1 | 42076 | 44 | Te3Zr1 | | | | | |
| | -A | -B | -C | -D | -E | -GM | -Y | -Z |
| | 3+4 2 | 3+4 2 | 2+2 4 | 2+2 4 | 2+2 4 | 3+4 2 | 5+6 2 | 2+2 4 |
| | 5+6 2 | 5+6 2 | 2+2 4 | 2+2 4 | 2+2 4 | 5+6 2 | 3+4 2 | 2+2 4 |
| | 5+6 2 | 5+6 2 | 2+2 4 | 2+2 4 | 2+2 4 | 3+4 2 | 3+4 2 | 2+2 4 |
| | 3+4 2 | 3+4 2 | 2+2 4 | 2+2 4 | 2+2 4 | 5+6 2 | 5+6 2 | 2+2 4 |
| | 5+6 2 | 3+4 2 | 2+2 4 | 2+2 4 | 2+2 4 | 3+4 2 | 3+4 2 | 2+2 4 |
| | 3+4 2 | 5+6 2 | 2+2 4 | 2+2 4 | 2+2 4 | 5+6 2 | 5+6 2 | 2+2 4 |
| | 5+6 2 | 5+6 2 | 2+2 4 | 2+2 4 | 2+2 4 | 3+4 2 | 3+4 2 | 2+2 4 |
| | 3+4 2 | 3+4 2 | 2+2 4 | 2+2 4 | 2+2 4 | 5+6 2 | 5+6 2 | 2+2 4 |
| | 5+6 2 | 3+4 2 | 2+2 4 | 2+2 4 | 2+2 4 | 3+4 2 | 3+4 2 | 2+2 4 |
| | 3+4 2 | 5+6 2 | 2+2 4 | 2+2 4 | 2+2 4 | 5+6 2 | 5+6 2 | 2+2 4 |
| | 3+4 2 | 5+6 2 | 2+2 4 | 2+2 4 | 2+2 4 | 3+4 2 | 5+6 2 | 2+2 4 |
| | 3+4 2 | 5+6 2 | | | | 3+4 2 | 3+4 2 | |
| | 5+6 2 | 3+4 2 | | | | 3+4 2 | 3+4 2 | |
| | 5+6 2 | 5+6 2 | | | | 5+6 2 | 5+6 2 | |
| | 3+4 2 | 3+4 2 | | | | 3+4 2 | 3+4 2 | |
| | 5+6 2 | 5+6 2 | | | | 5+6 2 | 5+6 2 | |
| | 5+6 2 | 3+4 2 | | | | 3+4 2 | 3+4 2 | |
| | 5+6 2 | 3+4 2 | | | | 5+6 2 | 5+6 2 | |
| | 3+4 2 | 5+6 2 | | | | 5+6 2 | 5+6 2 | |
| | 3+4 2 | 3+4 2 | | | | 3+4 2 | 3+4 2 | |
| | 3+4 2 | 3+4 2 | | | | 5+6 2 | 5+6 2 | |
| | 5+6 2 | 5+6 2 | | | | 5+6 2 | 5+6 2 | |
| 11.2.2 | 424108 | 84 | Sn2Sr1Zn2 | | | | | |
| | -A | -B | -C | -D | -E | -GM | -Y | -Z |
| | 3+4 2 | 3+4 2 | 2+2 4 | 2+2 4 | 2+2 4 | 3+4 2 | 3+4 2 | 2+2 4 |
| | 5+6 2 | 5+6 2 | 2+2 4 | 2+2 4 | 2+2 4 | 5+6 2 | 5+6 2 | 2+2 4 |
| | 3+4 2 | 3+4 2 | 2+2 4 | 2+2 4 | 2+2 4 | 3+4 2 | 3+4 2 | 2+2 4 |
| | 5+6 2 | 5+6 2 | 2+2 4 | 2+2 4 | 2+2 4 | 5+6 2 | 5+6 2 | 2+2 4 |
| | 3+4 2 | 3+4 2 | 2+2 4 | 2+2 4 | 2+2 4 | 3+4 2 | 3+4 2 | 2+2 4 |
| | 5+6 2 | 5+6 2 | 2+2 4 | 2+2 4 | 2+2 4 | 5+6 2 | 5+6 2 | 2+2 4 |
| | 3+4 2 | 3+4 2 | 2+2 4 | 2+2 4 | 2+2 4 | 3+4 2 | 3+4 2 | 2+2 4 |
| | 5+6 2 | 5+6 2 | 2+2 4 | 2+2 4 | 2+2 4 | 5+6 2 | 5+6 2 | 2+2 4 |
| | 5+6 2 | 5+6 2 | 2+2 4 | 2+2 4 | 2+2 4 | 3+4 2 | 3+4 2 | 2+2 4 |
| | 3+4 2 | 3+4 2 | 2+2 4 | 2+2 4 | 2+2 4 | 3+4 2 | 5+6 2 | 2+2 4 |
| | 5+6 2 | 5+6 2 | 2+2 4 | 2+2 4 | 2+2 4 | 5+6 2 | 3+4 2 | 2+2 4 |
| | 3+4 2 | 3+4 2 | 2+2 4 | 2+2 4 | 2+2 4 | 5+6 2 | 5+6 2 | 2+2 4 |
| | 3+4 2 | 5+6 2 | 2+2 4 | 2+2 4 | 2+2 4 | 3+4 2 | 3+4 2 | 2+2 4 |
| | 5+6 2 | 3+4 2 | 2+2 4 | 2+2 4 | 2+2 4 | 5+6 2 | 5+6 2 | 2+2 4 |
| | 5+6 2 | 5+6 2 | 2+2 4 | 2+2 4 | 2+2 4 | 3+4 2 | 3+4 2 | 2+2 4 |
| | 3+4 2 | 3+4 2 | 2+2 4 | 2+2 4 | 2+2 4 | 3+4 2 | 5+6 2 | 2+2 4 |
| | 3+4 2 | 5+6 2 | 2+2 4 | 2+2 4 | 2+2 4 | 5+6 2 | 3+4 2 | 2+2 4 |
| | 5+6 2 | 3+4 2 | 2+2 4 | 2+2 4 | 2+2 4 | 3+4 2 | 5+6 2 | 2+2 4 |
| | 3+4 2 | 5+6 2 | 2+2 4 | 2+2 4 | 2+2 4 | 5+6 2 | 3+4 2 | 2+2 4 |
| | 5+6 2 | 3+4 2 | 2+2 4 | 2+2 4 | 2+2 4 | 3+4 2 | 5+6 2 | 2+2 4 |
| | 3+4 2 | 5+6 2 | 2+2 4 | 2+2 4 | 2+2 4 | 5+6 2 | 3+4 2 | 2+2 4 |
| | 5+6 2 | 3+4 2 | | | | 3+4 2 | 5+6 2 | |
| | 3+4 2 | 5+6 2 | | | | 5+6 2 | 3+4 2 | |
| | 5+6 2 | 5+6 2 | | | | 3+4 2 | 5+6 2 | |
| | 5+6 2 | 3+4 2 | | | | 5+6 2 | 3+4 2 | |
| | 3+4 2 | 3+4 2 | | | | 3+4 2 | 5+6 2 | |
| | 5+6 2 | 3+4 2 | | | | 3+4 2 | 5+6 2 | |
| | 3+4 2 | 5+6 2 | | | | 5+6 2 | 3+4 2 | |
| | 5+6 2 | 3+4 2 | | | | 5+6 2 | 5+6 2 | |
| | 3+4 2 | 5+6 2 | | | | 3+4 2 | 3+4 2 | |
| | 3+4 2 | 5+6 2 | | | | 5+6 2 | 3+4 2 | |
| | 5+6 2 | 3+4 2 | | | | 5+6 2 | 5+6 2 | |
| | 5+6 2 | 3+4 2 | | | | 3+4 2 | 5+6 2 | |
| | 3+4 2 | 5+6 2 | | | | 5+6 2 | 3+4 2 | |
| | 5+6 2 | 3+4 2 | | | | 3+4 2 | 5+6 2 | |
| | 3+4 2 | 5+6 2 | | | | 3+4 2 | 3+4 2 | |
| | 5+6 2 | 5+6 2 | | | | 3+4 2 | 3+4 2 | |
| | 3+4 2 | 3+4 2 | | | | 3+4 2 | 5+6 2 | |
| | 3+4 2 | 3+4 2 | | | | 5+6 2 | 3+4 2 | |
| | 5+6 2 | 5+6 2 | | | | 5+6 2 | 5+6 2 | |
| | 3+4 2 | 5+6 2 | | | | 3+4 2 | 3+4 2 | |
| | 5+6 2 | 3+4 2 | | | | 5+6 2 | 5+6 2 | |

Continued on next column

TABLE XXIII– continued

Continued on next column

**TABLE XXIII– continued**

| TQC | ICSD | | NoE | | SF | | | |
|---|---|---|---|---|---|---|---|---|
| 11.2.3 | 42982 | | 56 | | Se3Ta2 | | | |
| | -A | -B | -C | -D | -E | -GM | -Y | -Z |
| | 5+6 2 | 3+4 2 | 2+2 4 | 2+2 4 | 2+2 4 | 3+4 2 | 5+6 2 | 2+2 4 |
| | 3+4 2 | 5+6 2 | 2+2 4 | 2+2 4 | 2+2 4 | 5+6 2 | 3+4 2 | 2+2 4 |
| | 5+6 2 | 5+6 2 | 2+2 4 | 2+2 4 | 2+2 4 | 3+4 2 | 5+6 2 | 2+2 4 |
| | 3+4 2 | 3+4 2 | 2+2 4 | 2+2 4 | 2+2 4 | 5+6 2 | 3+4 2 | 2+2 4 |
| | 5+6 2 | 5+6 2 | 2+2 4 | 2+2 4 | 2+2 4 | 3+4 2 | 3+4 2 | 2+2 4 |
| | 3+4 2 | 3+4 2 | 2+2 4 | 2+2 4 | 2+2 4 | 5+6 2 | 5+6 2 | 2+2 4 |
| | 5+6 2 | 3+4 2 | 2+2 4 | 2+2 4 | 2+2 4 | 3+4 2 | 3+4 2 | 2+2 4 |
| | 3+4 2 | 5+6 2 | 2+2 4 | 2+2 4 | 2+2 4 | 5+6 2 | 5+6 2 | 2+2 4 |
| | 5+6 2 | 5+6 2 | 2+2 4 | 2+2 4 | 2+2 4 | 3+4 2 | 5+6 2 | 2+2 4 |
| | 3+4 2 | 3+4 2 | 2+2 4 | 2+2 4 | 2+2 4 | 5+6 2 | 3+4 2 | 2+2 4 |
| | 5+6 2 | 3+4 2 | 2+2 4 | 2+2 4 | 2+2 4 | 3+4 2 | 5+6 2 | 2+2 4 |
| | 5+6 2 | 5+6 2 | 2+2 4 | 2+2 4 | 2+2 4 | 5+6 2 | 3+4 2 | 2+2 4 |
| | 3+4 2 | 3+4 2 | 2+2 4 | 2+2 4 | 2+2 4 | 3+4 2 | 5+6 2 | 2+2 4 |
| | 3+4 2 | 5+6 2 | 2+2 4 | 2+2 4 | 2+2 4 | 5+6 2 | 3+4 2 | 2+2 4 |
| | 5+6 2 | 5+6 2 | | | | 5+6 2 | 3+4 2 | |
| | 5+6 2 | 3+4 2 | | | | 3+4 2 | 5+6 2 | |
| | 3+4 2 | 3+4 2 | | | | 3+4 2 | 5+6 2 | |
| | 5+6 2 | 5+6 2 | | | | 5+6 2 | 3+4 2 | |
| | 3+4 2 | 5+6 2 | | | | 3+4 2 | 5+6 2 | |
| | 3+4 2 | 3+4 2 | | | | 3+4 2 | 5+6 2 | |
| | 3+4 2 | 5+6 2 | | | | 5+6 2 | 3+4 2 | |
| | 5+6 2 | 3+4 2 | | | | 3+4 2 | 3+4 2 | |
| | 5+6 2 | 3+4 2 | | | | 5+6 2 | 5+6 2 | |
| | 5+6 2 | 5+6 2 | | | | 5+6 2 | 5+6 2 | |
| | 3+4 2 | 3+4 2 | | | | 3+4 2 | 3+4 2 | |
| | 5+6 2 | 5+6 2 | | | | 5+6 2 | 3+4 2 | |
| | 3+4 2 | 5+6 2 | | | | 5+6 2 | 3+4 2 | |
| | 3+4 2 | 3+4 2 | | | | 3+4 2 | 3+4 2 | |
| 11.2.4 | 43213 | | 92 | | Se3Ta1 | | | |
| | -A | -B | -C | -D | -E | -GM | -Y | -Z |
| | 3+4 2 | 3+4 2 | 2+2 4 | 2+2 4 | 2+2 4 | 3+4 2 | 3+4 2 | 2+2 4 |
| | 5+6 2 | 5+6 2 | 2+2 4 | 2+2 4 | 2+2 4 | 5+6 2 | 5+6 2 | 2+2 4 |
| | 3+4 2 | 5+6 2 | 2+2 4 | 2+2 4 | 2+2 4 | 3+4 2 | 5+6 2 | 2+2 4 |
| | 5+6 2 | 3+4 2 | 2+2 4 | 2+2 4 | 2+2 4 | 5+6 2 | 3+4 2 | 2+2 4 |
| | 3+4 2 | 3+4 2 | 2+2 4 | 2+2 4 | 2+2 4 | 3+4 2 | 3+4 2 | 2+2 4 |
| | 3+4 2 | 5+6 2 | 2+2 4 | 2+2 4 | 2+2 4 | 3+4 2 | 5+6 2 | 2+2 4 |
| | 5+6 2 | 3+4 2 | 2+2 4 | 2+2 4 | 2+2 4 | 5+6 2 | 3+4 2 | 2+2 4 |
| | 5+6 2 | 5+6 2 | 2+2 4 | 2+2 4 | 2+2 4 | 5+6 2 | 5+6 2 | 2+2 4 |
| | 3+4 2 | 5+6 2 | 2+2 4 | 2+2 4 | 2+2 4 | 3+4 2 | 3+4 2 | 2+2 4 |
| | 5+6 2 | 3+4 2 | 2+2 4 | 2+2 4 | 2+2 4 | 5+6 2 | 5+6 2 | 2+2 4 |
| | 3+4 2 | 5+6 2 | 2+2 4 | 2+2 4 | 2+2 4 | 5+6 2 | 5+6 2 | 2+2 4 |
| | 5+6 2 | 3+4 2 | 2+2 4 | 2+2 4 | 2+2 4 | 3+4 2 | 3+4 2 | 2+2 4 |
| | 5+6 2 | 3+4 2 | 2+2 4 | 2+2 4 | 2+2 4 | 3+4 2 | 3+4 2 | 2+2 4 |
| | 3+4 2 | 5+6 2 | 2+2 4 | 2+2 4 | 2+2 4 | 5+6 2 | 5+6 2 | 2+2 4 |
| | 3+4 2 | 5+6 2 | 2+2 4 | 2+2 4 | 2+2 4 | 3+4 2 | 5+6 2 | 2+2 4 |
| | 5+6 2 | 3+4 2 | 2+2 4 | 2+2 4 | 2+2 4 | 5+6 2 | 3+4 2 | 2+2 4 |
| | 5+6 2 | 3+4 2 | 2+2 4 | 2+2 4 | 2+2 4 | 5+6 2 | 3+4 2 | 2+2 4 |
| | 3+4 2 | 5+6 2 | 2+2 4 | 2+2 4 | 2+2 4 | 3+4 2 | 5+6 2 | 2+2 4 |
| | 3+4 2 | 3+4 2 | 2+2 4 | 2+2 4 | 2+2 4 | 3+4 2 | 5+6 2 | 2+2 4 |
| | 5+6 2 | 5+6 2 | 2+2 4 | 2+2 4 | 2+2 4 | 3+4 2 | 3+4 2 | 2+2 4 |
| | 3+4 2 | 5+6 2 | 2+2 4 | 2+2 4 | 2+2 4 | 5+6 2 | 5+6 2 | 2+2 4 |
| | 3+4 2 | 3+4 2 | 2+2 4 | 2+2 4 | 2+2 4 | 3+4 2 | 3+4 2 | 2+2 4 |
| | 3+4 2 | 3+4 2 | 2+2 4 | 2+2 4 | 2+2 4 | 5+6 2 | 5+6 2 | 2+2 4 |
| | 5+6 2 | 3+4 2 | | | | 3+4 2 | 3+4 2 | |
| | 5+6 2 | 5+6 2 | | | | 5+6 2 | 3+4 2 | |
| | 3+4 2 | 5+6 2 | | | | 3+4 2 | 5+6 2 | |
| | 5+6 2 | 3+4 2 | | | | 5+6 2 | 3+4 2 | |
| | 3+4 2 | 3+4 2 | | | | 3+4 2 | 3+4 2 | |
| | 5+6 2 | 5+6 2 | | | | 3+4 2 | 5+6 2 | |

Continued on next column

**TABLE XXIII– continued**

| TQC | ICSD | | NoE | | SF | | | |
|---|---|---|---|---|---|---|---|---|
| | 3+4 2 | 5+6 2 | | | | 3+4 2 | 5+6 2 | |
| | 3+4 2 | 5+6 2 | | | | 5+6 2 | 5+6 2 | |
| | 5+6 2 | 5+6 2 | | | | 3+4 2 | 3+4 2 | |
| | 5+6 2 | 3+4 2 | | | | 5+6 2 | 3+4 2 | |
| | 3+4 2 | 3+4 2 | | | | 5+6 2 | 5+6 2 | |
| | 5+6 2 | 5+6 2 | | | | 3+4 2 | 5+6 2 | |
| | 3+4 2 | 3+4 2 | | | | 5+6 2 | 3+4 2 | |
| | 3+4 2 | 5+6 2 | | | | 3+4 2 | 3+4 2 | |
| | 5+6 2 | 3+4 2 | | | | 5+6 2 | 3+4 2 | |
| | 5+6 2 | 3+4 2 | | | | 3+4 2 | 5+6 2 | |
| | 3+4 2 | 5+6 2 | | | | 5+6 2 | 3+4 2 | |
| | 5+6 2 | 3+4 2 | | | | 5+6 2 | 5+6 2 | |
| | 5+6 2 | 5+6 2 | | | | 3+4 2 | 5+6 2 | |
| | 3+4 2 | 5+6 2 | | | | 5+6 2 | 5+6 2 | |
| | 5+6 2 | 3+4 2 | | | | 3+4 2 | 3+4 2 | |
| | 3+4 2 | 5+6 2 | | | | 5+6 2 | 5+6 2 | |
| | 5+6 2 | 5+6 2 | | | | 5+6 2 | 5+6 2 | |
| 11.2.5 | 52307 | | 40 | | Sb2Sr1 | | | |
| | -A | -B | -C | -D | -E | -GM | -Y | -Z |
| | 3+4 2 | 5+6 2 | 2+2 4 | 2+2 4 | 2+2 4 | 3+4 2 | 5+6 2 | 2+2 4 |
| | 5+6 2 | 3+4 2 | 2+2 4 | 2+2 4 | 2+2 4 | 5+6 2 | 3+4 2 | 2+2 4 |
| | 3+4 2 | 5+6 2 | 2+2 4 | 2+2 4 | 2+2 4 | 3+4 2 | 5+6 2 | 2+2 4 |
| | 5+6 2 | 3+4 2 | 2+2 4 | 2+2 4 | 2+2 4 | 5+6 2 | 3+4 2 | 2+2 4 |
| | 3+4 2 | 5+6 2 | 2+2 4 | 2+2 4 | 2+2 4 | 3+4 2 | 5+6 2 | 2+2 4 |
| | 5+6 2 | 3+4 2 | 2+2 4 | 2+2 4 | 2+2 4 | 5+6 2 | 3+4 2 | 2+2 4 |
| | 3+4 2 | 5+6 2 | 2+2 4 | 2+2 4 | 2+2 4 | 3+4 2 | 5+6 2 | 2+2 4 |
| | 5+6 2 | 3+4 2 | 2+2 4 | 2+2 4 | 2+2 4 | 5+6 2 | 3+4 2 | 2+2 4 |
| | 5+6 2 | 3+4 2 | 2+2 4 | 2+2 4 | 2+2 4 | 3+4 2 | 3+4 2 | 2+2 4 |
| | 3+4 2 | 5+6 2 | 2+2 4 | 2+2 4 | 2+2 4 | 3+4 2 | 3+4 2 | 2+2 4 |
| | 5+6 2 | 3+4 2 | | | | 5+6 2 | 5+6 2 | |
| | 3+4 2 | 5+6 2 | | | | 5+6 2 | 5+6 2 | |
| | 3+4 2 | 5+6 2 | | | | 3+4 2 | 5+6 2 | |
| | 5+6 2 | 3+4 2 | | | | 5+6 2 | 3+4 2 | |
| | 5+6 2 | 3+4 2 | | | | 3+4 2 | 3+4 2 | |
| | 3+4 2 | 3+4 2 | | | | 5+6 2 | 5+6 2 | |
| | 5+6 2 | 5+6 2 | | | | 3+4 2 | 3+4 2 | |
| | 3+4 2 | 5+6 2 | | | | 3+4 2 | 3+4 2 | |
| | 5+6 2 | 3+4 2 | | | | 3+4 2 | 3+4 2 | |
| | 3+4 2 | 5+6 2 | | | | 5+6 2 | 5+6 2 | |
| 11.2.6 | 171417 | | 64 | | S3U1 | | | |
| | -A | -B | -C | -D | -E | -GM | -Y | -Z |
| | 3+4 2 | 3+4 2 | 2+2 4 | 2+2 4 | 2+2 4 | 3+4 2 | 3+4 2 | 2+2 4 |
| | 5+6 2 | 5+6 2 | 2+2 4 | 2+2 4 | 2+2 4 | 5+6 2 | 5+6 2 | 2+2 4 |
| | 3+4 2 | 3+4 2 | 2+2 4 | 2+2 4 | 2+2 4 | 3+4 2 | 3+4 2 | 2+2 4 |
| | 5+6 2 | 5+6 2 | 2+2 4 | 2+2 4 | 2+2 4 | 5+6 2 | 5+6 2 | 2+2 4 |
| | 3+4 2 | 3+4 2 | 2+2 4 | 2+2 4 | 2+2 4 | 3+4 2 | 5+6 2 | 2+2 4 |
| | 5+6 2 | 5+6 2 | 2+2 4 | 2+2 4 | 2+2 4 | 5+6 2 | 3+4 2 | 2+2 4 |
| | 5+6 2 | 3+4 2 | 2+2 4 | 2+2 4 | 2+2 4 | 3+4 2 | 5+6 2 | 2+2 4 |
| | 3+4 2 | 5+6 2 | 2+2 4 | 2+2 4 | 2+2 4 | 5+6 2 | 3+4 2 | 2+2 4 |
| | 3+4 2 | 3+4 2 | 2+2 4 | 2+2 4 | 2+2 4 | 3+4 2 | 5+6 2 | 2+2 4 |
| | 5+6 2 | 5+6 2 | 2+2 4 | 2+2 4 | 2+2 4 | 5+6 2 | 3+4 2 | 2+2 4 |
| | 3+4 2 | 5+6 2 | 2+2 4 | 2+2 4 | 2+2 4 | 3+4 2 | 3+4 2 | 2+2 4 |
| | 5+6 2 | 3+4 2 | 2+2 4 | 2+2 4 | 2+2 4 | 5+6 2 | 5+6 2 | 2+2 4 |
| | 5+6 2 | 5+6 2 | 2+2 4 | 2+2 4 | 2+2 4 | 3+4 2 | 3+4 2 | 2+2 4 |
| | 3+4 2 | 3+4 2 | 2+2 4 | 2+2 4 | 2+2 4 | 5+6 2 | 5+6 2 | 2+2 4 |
| | 3+4 2 | 5+6 2 | 2+2 4 | 2+2 4 | 2+2 4 | 3+4 2 | 5+6 2 | 2+2 4 |
| | 5+6 2 | 3+4 2 | 2+2 4 | 2+2 4 | 2+2 4 | 5+6 2 | 3+4 2 | 2+2 4 |
| | 5+6 2 | 5+6 2 | | | | 3+4 2 | 5+6 2 | |
| | 3+4 2 | 5+6 2 | | | | 5+6 2 | 3+4 2 | |
| | 3+4 2 | 3+4 2 | | | | 3+4 2 | 5+6 2 | |
| | 5+6 2 | 3+4 2 | | | | 3+4 2 | 3+4 2 | |

Continued on next column

**TABLE XXIII– continued**

| TQC | ICSD | | NoE | | SF | | | |
|---|---|---|---|---|---|---|---|---|
| | 3+4 2 | 3+4 2 | | | | 5+6 2 | 3+4 2 | |
| | 5+6 2 | 3+4 2 | | | | 3+4 2 | 5+6 2 | |
| | 3+4 2 | 3+4 2 | | | | 3+4 2 | 5+6 2 | |
| | 5+6 2 | 5+6 2 | | | | 5+6 2 | 3+4 2 | |
| | 3+4 2 | 5+6 2 | | | | 5+6 2 | 5+6 2 | |
| | 5+6 2 | 5+6 2 | | | | 5+6 2 | 3+4 2 | |
| | 5+6 2 | 5+6 2 | | | | 3+4 2 | 3+4 2 | |
| | 5+6 2 | 3+4 2 | | | | 5+6 2 | 3+4 2 | |
| | 3+4 2 | 3+4 2 | | | | 3+4 2 | 5+6 2 | |
| | 3+4 2 | 5+6 2 | | | | 5+6 2 | 5+6 2 | |
| | 5+6 2 | 3+4 2 | | | | 3+4 2 | 5+6 2 | |
| | 3+4 2 | 3+4 2 | | | | 3+4 2 | 3+4 2 | |
| 11.2.7 | 430 | | 256 | | S3Tm2 | | | |
| | -A | -B | -C | -D | -E | -GM | -Y | -Z |
| | 3+4 2 | 3+4 2 | 2+2 4 | 2+2 4 | 2+2 4 | 3+4 2 | 3+4 2 | 2+2 4 |
| | 5+6 2 | 5+6 2 | 2+2 4 | 2+2 4 | 2+2 4 | 5+6 2 | 5+6 2 | 2+2 4 |
| | 3+4 2 | 3+4 2 | 2+2 4 | 2+2 4 | 2+2 4 | 3+4 2 | 3+4 2 | 2+2 4 |
| | 5+6 2 | 5+6 2 | 2+2 4 | 2+2 4 | 2+2 4 | 5+6 2 | 5+6 2 | 2+2 4 |
| | 3+4 2 | 3+4 2 | 2+2 4 | 2+2 4 | 2+2 4 | 3+4 2 | 3+4 2 | 2+2 4 |
| | 5+6 2 | 5+6 2 | 2+2 4 | 2+2 4 | 2+2 4 | 5+6 2 | 5+6 2 | 2+2 4 |
| | 3+4 2 | 3+4 2 | 2+2 4 | 2+2 4 | 2+2 4 | 3+4 2 | 3+4 2 | 2+2 4 |
| | 5+6 2 | 5+6 2 | 2+2 4 | 2+2 4 | 2+2 4 | 5+6 2 | 5+6 2 | 2+2 4 |
| | 5+6 2 | 5+6 2 | 2+2 4 | 2+2 4 | 2+2 4 | 3+4 2 | 3+4 2 | 2+2 4 |
| | 3+4 2 | 3+4 2 | 2+2 4 | 2+2 4 | 2+2 4 | 5+6 2 | 5+6 2 | 2+2 4 |
| | 5+6 2 | 3+4 2 | 2+2 4 | 2+2 4 | 2+2 4 | 3+4 2 | 5+6 2 | 2+2 4 |
| | 3+4 2 | 5+6 2 | 2+2 4 | 2+2 4 | 2+2 4 | 5+6 2 | 3+4 2 | 2+2 4 |
| | 3+4 2 | 3+4 2 | 2+2 4 | 2+2 4 | 2+2 4 | 3+4 2 | 3+4 2 | 2+2 4 |
| | 5+6 2 | 3+4 2 | 2+2 4 | 2+2 4 | 2+2 4 | 3+4 2 | 5+6 2 | 2+2 4 |
| | 3+4 2 | 5+6 2 | 2+2 4 | 2+2 4 | 2+2 4 | 5+6 2 | 3+4 2 | 2+2 4 |
| | 5+6 2 | 5+6 2 | 2+2 4 | 2+2 4 | 2+2 4 | 5+6 2 | 5+6 2 | 2+2 4 |
| | 5+6 2 | 5+6 2 | 2+2 4 | 2+2 4 | 2+2 4 | 3+4 2 | 3+4 2 | 2+2 4 |
| | 3+4 2 | 3+4 2 | 2+2 4 | 2+2 4 | 2+2 4 | 5+6 2 | 5+6 2 | 2+2 4 |
| | 5+6 2 | 5+6 2 | 2+2 4 | 2+2 4 | 2+2 4 | 3+4 2 | 3+4 2 | 2+2 4 |
| | 3+4 2 | 3+4 2 | 2+2 4 | 2+2 4 | 2+2 4 | 5+6 2 | 5+6 2 | 2+2 4 |
| | 5+6 2 | 3+4 2 | 2+2 4 | 2+2 4 | 2+2 4 | 3+4 2 | 5+6 2 | 2+2 4 |
| | 3+4 2 | 3+4 2 | 2+2 4 | 2+2 4 | 2+2 4 | 3+4 2 | 3+4 2 | 2+2 4 |
| | 3+4 2 | 5+6 2 | 2+2 4 | 2+2 4 | 2+2 4 | 5+6 2 | 3+4 2 | 2+2 4 |
| | 5+6 2 | 5+6 2 | 2+2 4 | 2+2 4 | 2+2 4 | 3+4 2 | 5+6 2 | 2+2 4 |
| | 3+4 2 | 5+6 2 | 2+2 4 | 2+2 4 | 2+2 4 | 5+6 2 | 5+6 2 | 2+2 4 |
| | 5+6 2 | 3+4 2 | 2+2 4 | 2+2 4 | 2+2 4 | 5+6 2 | 3+4 2 | 2+2 4 |
| | 3+4 2 | 3+4 2 | 2+2 4 | 2+2 4 | 2+2 4 | 3+4 2 | 5+6 2 | 2+2 4 |
| | 3+4 2 | 5+6 2 | 2+2 4 | 2+2 4 | 2+2 4 | 3+4 2 | 3+4 2 | 2+2 4 |
| | 5+6 2 | 3+4 2 | 2+2 4 | 2+2 4 | 2+2 4 | 5+6 2 | 3+4 2 | 2+2 4 |
| | 5+6 2 | 3+4 2 | 2+2 4 | 2+2 4 | 2+2 4 | 3+4 2 | 5+6 2 | 2+2 4 |
| | 3+4 2 | 5+6 2 | 2+2 4 | 2+2 4 | 2+2 4 | 5+6 2 | 5+6 2 | 2+2 4 |
| | 5+6 2 | 5+6 2 | 2+2 4 | 2+2 4 | 2+2 4 | 5+6 2 | 3+4 2 | 2+2 4 |
| | 5+6 2 | 5+6 2 | 2+2 4 | 2+2 4 | 2+2 4 | 3+4 2 | 3+4 2 | 2+2 4 |
| | 3+4 2 | 3+4 2 | 2+2 4 | 2+2 4 | 2+2 4 | 5+6 2 | 5+6 2 | 2+2 4 |
| | 3+4 2 | 3+4 2 | 2+2 4 | 2+2 4 | 2+2 4 | 3+4 2 | 3+4 2 | 2+2 4 |
| | 5+6 2 | 5+6 2 | 2+2 4 | 2+2 4 | 2+2 4 | 3+4 2 | 5+6 2 | 2+2 4 |
| | 5+6 2 | 5+6 2 | 2+2 4 | 2+2 4 | 2+2 4 | 3+4 2 | 5+6 2 | 2+2 4 |
| | 3+4 2 | 3+4 2 | 2+2 4 | 2+2 4 | 2+2 4 | 5+6 2 | 5+6 2 | 2+2 4 |
| | 5+6 2 | 5+6 2 | 2+2 4 | 2+2 4 | 2+2 4 | 5+6 2 | 3+4 2 | 2+2 4 |
| | 3+4 2 | 5+6 2 | 2+2 4 | 2+2 4 | 2+2 4 | 5+6 2 | 5+6 2 | 2+2 4 |
| | 5+6 2 | 3+4 2 | 2+2 4 | 2+2 4 | 2+2 4 | 3+4 2 | 3+4 2 | 2+2 4 |
| | 3+4 2 | 5+6 2 | 2+2 4 | 2+2 4 | 2+2 4 | 5+6 2 | 3+4 2 | 2+2 4 |
| | 5+6 2 | 3+4 2 | 2+2 4 | 2+2 4 | 2+2 4 | 3+4 2 | 3+4 2 | 2+2 4 |
| | 3+4 2 | 3+4 2 | 2+2 4 | 2+2 4 | 2+2 4 | 5+6 2 | 5+6 2 | 2+2 4 |
| | 3+4 2 | 3+4 2 | 2+2 4 | 2+2 4 | 2+2 4 | 5+6 2 | 3+4 2 | 2+2 4 |
| | 5+6 2 | 5+6 2 | 2+2 4 | 2+2 4 | 2+2 4 | 3+4 2 | 5+6 2 | 2+2 4 |
| | 5+6 2 | 5+6 2 | 2+2 4 | 2+2 4 | 2+2 4 | 3+4 2 | 3+4 2 | 2+2 4 |
| | 3+4 2 | 3+4 2 | 2+2 4 | 2+2 4 | 2+2 4 | 5+6 2 | 5+6 2 | 2+2 4 |
| | 3+4 2 | 5+6 2 | 2+2 4 | 2+2 4 | 2+2 4 | 3+4 2 | 3+4 2 | 2+2 4 |
| | 5+6 2 | 3+4 2 | 2+2 4 | 2+2 4 | 2+2 4 | 5+6 2 | 5+6 2 | 2+2 4 |
| | 5+6 2 | 5+6 2 | 2+2 4 | 2+2 4 | 2+2 4 | 3+4 2 | 5+6 2 | 2+2 4 |
| | 5+6 2 | 3+4 2 | 2+2 4 | 2+2 4 | 2+2 4 | 5+6 2 | 3+4 2 | 2+2 4 |
| | 3+4 2 | 5+6 2 | 2+2 4 | 2+2 4 | 2+2 4 | 3+4 2 | 5+6 2 | 2+2 4 |
| | 3+4 2 | 5+6 2 | 2+2 4 | 2+2 4 | 2+2 4 | 3+4 2 | 5+6 2 | 2+2 4 |
| | 3+4 2 | 3+4 2 | 2+2 4 | 2+2 4 | 2+2 4 | 5+6 2 | 3+4 2 | 2+2 4 |
| | 5+6 2 | 5+6 2 | 2+2 4 | 2+2 4 | 2+2 4 | 3+4 2 | 3+4 2 | 2+2 4 |
| | 5+6 2 | 5+6 2 | 2+2 4 | 2+2 4 | 2+2 4 | 3+4 2 | 3+4 2 | 2+2 4 |
| | 3+4 2 | 3+4 2 | 2+2 4 | 2+2 4 | 2+2 4 | 5+6 2 | 5+6 2 | 2+2 4 |
| | 3+4 2 | 3+4 2 | 2+2 4 | 2+2 4 | 2+2 4 | 5+6 2 | 5+6 2 | 2+2 4 |
| | 5+6 2 | 5+6 2 | 2+2 4 | 2+2 4 | 2+2 4 | 5+6 2 | 3+4 2 | 2+2 4 |
| | 5+6 2 | 3+4 2 | 2+2 4 | 2+2 4 | 2+2 4 | 3+4 2 | 3+4 2 | 2+2 4 |
| | 3+4 2 | 5+6 2 | 2+2 4 | 2+2 4 | 2+2 4 | 3+4 2 | 5+6 2 | 2+2 4 |
| | 3+4 2 | 3+4 2 | 2+2 4 | 2+2 4 | 2+2 4 | 3+4 2 | 5+6 2 | 2+2 4 |
| | 5+6 2 | 5+6 2 | 2+2 4 | 2+2 4 | 2+2 4 | 3+4 2 | 3+4 2 | 2+2 4 |
| | 3+4 2 | 3+4 2 | | | | 5+6 2 | 5+6 2 | |
| | 5+6 2 | 3+4 2 | | | | 3+4 2 | 3+4 2 | |
| | 3+4 2 | 5+6 2 | | | | 3+4 2 | 5+6 2 | |
| | 5+6 2 | 3+4 2 | | | | 3+4 2 | 3+4 2 | |
| | 5+6 2 | 3+4 2 | | | | 5+6 2 | 3+4 2 | |
| | 5+6 2 | 5+6 2 | | | | 5+6 2 | 5+6 2 | |
| | 3+4 2 | 5+6 2 | | | | 5+6 2 | 5+6 2 | |
| | 5+6 2 | 5+6 2 | | | | 3+4 2 | 3+4 2 | |
| | 3+4 2 | 3+4 2 | | | | 5+6 2 | 3+4 2 | |
| | 3+4 2 | 5+6 2 | | | | 3+4 2 | 5+6 2 | |
| | 5+6 2 | 3+4 2 | | | | 5+6 2 | 3+4 2 | |
| | 3+4 2 | 3+4 2 | | | | 5+6 2 | 5+6 2 | |
| | 3+4 2 | 5+6 2 | | | | 5+6 2 | 3+4 2 | |
| | 5+6 2 | 3+4 2 | | | | 3+4 2 | 5+6 2 | |
| | 5+6 2 | 5+6 2 | | | | 5+6 2 | 3+4 2 | |
| | 5+6 2 | 3+4 2 | | | | 3+4 2 | 5+6 2 | |
| | 3+4 2 | 5+6 2 | | | | 3+4 2 | 5+6 2 | |
| | 3+4 2 | 3+4 2 | | | | 3+4 2 | 3+4 2 | |
| | 5+6 2 | 5+6 2 | | | | 3+4 2 | 3+4 2 | |
| | 3+4 2 | 3+4 2 | | | | 5+6 2 | 5+6 2 | |
| | 5+6 2 | 5+6 2 | | | | 5+6 2 | 3+4 2 | |
| | 3+4 2 | 3+4 2 | | | | 5+6 2 | 5+6 2 | |
| | 5+6 2 | 5+6 2 | | | | 3+4 2 | 3+4 2 | |
| | 3+4 2 | 3+4 2 | | | | 3+4 2 | 5+6 2 | |
| | 5+6 2 | 5+6 2 | | | | 5+6 2 | 5+6 2 | |
| | 3+4 2 | 3+4 2 | | | | 3+4 2 | 3+4 2 | |
| | 5+6 2 | 5+6 2 | | | | 5+6 2 | 5+6 2 | |
| | 3+4 2 | 5+6 2 | | | | 3+4 2 | 3+4 2 | |
| | 5+6 2 | 3+4 2 | | | | 5+6 2 | 3+4 2 | |
| | 3+4 2 | 3+4 2 | | | | 3+4 2 | 5+6 2 | |
| | 3+4 2 | 5+6 2 | | | | 5+6 2 | 3+4 2 | |
| | 5+6 2 | 3+4 2 | | | | 3+4 2 | 5+6 2 | |
| | 3+4 2 | 5+6 2 | | | | 5+6 2 | 3+4 2 | |
| | 5+6 2 | 3+4 2 | | | | 3+4 2 | 3+4 2 | |
| | 3+4 2 | 5+6 2 | | | | 5+6 2 | 5+6 2 | |
| | 5+6 2 | 3+4 2 | | | | 3+4 2 | 3+4 2 | |
| | 3+4 2 | 5+6 2 | | | | 5+6 2 | 5+6 2 | |
| | 5+6 2 | 3+4 2 | | | | 5+6 2 | 5+6 2 | |
| | 5+6 2 | 5+6 2 | | | | 5+6 2 | 3+4 2 | |
| | 3+4 2 | 3+4 2 | | | | 5+6 2 | 5+6 2 | |
| | 5+6 2 | 3+4 2 | | | | 3+4 2 | 3+4 2 | |
| | 3+4 2 | 5+6 2 | | | | 3+4 2 | 5+6 2 | |
| | 5+6 2 | 3+4 2 | | | | 3+4 2 | 3+4 2 | |
| | 3+4 2 | 5+6 2 | | | | 5+6 2 | 5+6 2 | |

Continued on next column

Continued on next column

**TABLE XXIII– continued**

| TQC | ICSD | | NoE | | SF | | | |
|---|---|---|---|---|---|---|---|---|
| | 5+6 2 | 3+4 2 | | | | 3+4 2 | 3+4 2 | |
| | 3+4 2 | 5+6 2 | | | | 5+6 2 | 5+6 2 | |
| | 3+4 2 | 3+4 2 | | | | 3+4 2 | 3+4 2 | |
| | 5+6 2 | 5+6 2 | | | | 3+4 2 | 5+6 2 | |
| | 3+4 2 | 3+4 2 | | | | 5+6 2 | 5+6 2 | |
| | 5+6 2 | 5+6 2 | | | | 5+6 2 | 3+4 2 | |
| | 3+4 2 | 3+4 2 | | | | 3+4 2 | 3+4 2 | |
| | 5+6 2 | 5+6 2 | | | | 5+6 2 | 5+6 2 | |
| | 3+4 2 | 3+4 2 | | | | 3+4 2 | 3+4 2 | |
| | 5+6 2 | 5+6 2 | | | | 5+6 2 | 5+6 2 | |
| | 3+4 2 | 5+6 2 | | | | 5+6 2 | 5+6 2 | |
| | 3+4 2 | 3+4 2 | | | | 3+4 2 | 3+4 2 | |
| | 5+6 2 | 5+6 2 | | | | 3+4 2 | 3+4 2 | |
| | 5+6 2 | 3+4 2 | | | | 5+6 2 | 5+6 2 | |
| | 3+4 2 | 3+4 2 | | | | 3+4 2 | 3+4 2 | |
| | 5+6 2 | 5+6 2 | | | | 3+4 2 | 5+6 2 | |
| | 3+4 2 | 3+4 2 | | | | 3+4 2 | 3+4 2 | |
| | 5+6 2 | 5+6 2 | | | | 5+6 2 | 5+6 2 | |
| | 3+4 2 | 3+4 2 | | | | 5+6 2 | 3+4 2 | |
| | 3+4 2 | 5+6 2 | | | | 3+4 2 | 3+4 2 | |
| 11.2.8 | 260980 | | 384 | | S3Tm2 | | | |
| | -A | -B | -C | -D | -E | -GM | -Y | -Z |
| | 5+6 2 | 3+4 2 | 2+2 4 | 2+2 4 | 2+2 4 | 3+4 2 | 5+6 2 | 2+2 4 |
| | 3+4 2 | 5+6 2 | 2+2 4 | 2+2 4 | 2+2 4 | 5+6 2 | 3+4 2 | 2+2 4 |
| | 3+4 2 | 3+4 2 | 2+2 4 | 2+2 4 | 2+2 4 | 3+4 2 | 3+4 2 | 2+2 4 |
| | 5+6 2 | 5+6 2 | 2+2 4 | 2+2 4 | 2+2 4 | 5+6 2 | 5+6 2 | 2+2 4 |
| | 3+4 2 | 3+4 2 | 2+2 4 | 2+2 4 | 2+2 4 | 3+4 2 | 3+4 2 | 2+2 4 |
| | 5+6 2 | 5+6 2 | 2+2 4 | 2+2 4 | 2+2 4 | 5+6 2 | 5+6 2 | 2+2 4 |
| | 3+4 2 | 3+4 2 | 2+2 4 | 2+2 4 | 2+2 4 | 3+4 2 | 3+4 2 | 2+2 4 |
| | 5+6 2 | 5+6 2 | 2+2 4 | 2+2 4 | 2+2 4 | 5+6 2 | 5+6 2 | 2+2 4 |
| | 3+4 2 | 3+4 2 | 2+2 4 | 2+2 4 | 2+2 4 | 3+4 2 | 3+4 2 | 2+2 4 |
| | 5+6 2 | 5+6 2 | 2+2 4 | 2+2 4 | 2+2 4 | 5+6 2 | 5+6 2 | 2+2 4 |
| | 3+4 2 | 3+4 2 | 2+2 4 | 2+2 4 | 2+2 4 | 3+4 2 | 3+4 2 | 2+2 4 |
| | 5+6 2 | 5+6 2 | 2+2 4 | 2+2 4 | 2+2 4 | 5+6 2 | 5+6 2 | 2+2 4 |
| | 5+6 2 | 3+4 2 | 2+2 4 | 2+2 4 | 2+2 4 | 3+4 2 | 5+6 2 | 2+2 4 |
| | 3+4 2 | 5+6 2 | 2+2 4 | 2+2 4 | 2+2 4 | 5+6 2 | 3+4 2 | 2+2 4 |
| | 5+6 2 | 3+4 2 | 2+2 4 | 2+2 4 | 2+2 4 | 3+4 2 | 3+4 2 | 2+2 4 |
| | 3+4 2 | 5+6 2 | 2+2 4 | 2+2 4 | 2+2 4 | 5+6 2 | 5+6 2 | 2+2 4 |
| | 3+4 2 | 3+4 2 | 2+2 4 | 2+2 4 | 2+2 4 | 3+4 2 | 3+4 2 | 2+2 4 |
| | 5+6 2 | 5+6 2 | 2+2 4 | 2+2 4 | 2+2 4 | 5+6 2 | 5+6 2 | 2+2 4 |
| | 3+4 2 | 3+4 2 | 2+2 4 | 2+2 4 | 2+2 4 | 3+4 2 | 3+4 2 | 2+2 4 |
| | 5+6 2 | 5+6 2 | 2+2 4 | 2+2 4 | 2+2 4 | 5+6 2 | 5+6 2 | 2+2 4 |
| | 3+4 2 | 5+6 2 | 2+2 4 | 2+2 4 | 2+2 4 | 3+4 2 | 5+6 2 | 2+2 4 |
| | 5+6 2 | 3+4 2 | 2+2 4 | 2+2 4 | 2+2 4 | 5+6 2 | 3+4 2 | 2+2 4 |
| | 3+4 2 | 3+4 2 | 2+2 4 | 2+2 4 | 2+2 4 | 3+4 2 | 3+4 2 | 2+2 4 |
| | 5+6 2 | 5+6 2 | 2+2 4 | 2+2 4 | 2+2 4 | 5+6 2 | 5+6 2 | 2+2 4 |
| | 5+6 2 | 3+4 2 | 2+2 4 | 2+2 4 | 2+2 4 | 3+4 2 | 5+6 2 | 2+2 4 |
| | 3+4 2 | 5+6 2 | 2+2 4 | 2+2 4 | 2+2 4 | 5+6 2 | 3+4 2 | 2+2 4 |
| | 5+6 2 | 5+6 2 | 2+2 4 | 2+2 4 | 2+2 4 | 3+4 2 | 3+4 2 | 2+2 4 |
| | 3+4 2 | 3+4 2 | 2+2 4 | 2+2 4 | 2+2 4 | 5+6 2 | 5+6 2 | 2+2 4 |
| | 5+6 2 | 3+4 2 | 2+2 4 | 2+2 4 | 2+2 4 | 3+4 2 | 5+6 2 | 2+2 4 |
| | 3+4 2 | 5+6 2 | 2+2 4 | 2+2 4 | 2+2 4 | 5+6 2 | 3+4 2 | 2+2 4 |
| | 5+6 2 | 3+4 2 | 2+2 4 | 2+2 4 | 2+2 4 | 3+4 2 | 3+4 2 | 2+2 4 |
| | 3+4 2 | 5+6 2 | 2+2 4 | 2+2 4 | 2+2 4 | 3+4 2 | 5+6 2 | 2+2 4 |
| | 5+6 2 | 3+4 2 | 2+2 4 | 2+2 4 | 2+2 4 | 5+6 2 | 5+6 2 | 2+2 4 |
| | 3+4 2 | 5+6 2 | 2+2 4 | 2+2 4 | 2+2 4 | 3+4 2 | 3+4 2 | 2+2 4 |
| | 3+4 2 | 3+4 2 | 2+2 4 | 2+2 4 | 2+2 4 | 5+6 2 | 5+6 2 | 2+2 4 |
| | 5+6 2 | 5+6 2 | 2+2 4 | 2+2 4 | 2+2 4 | 5+6 2 | 3+4 2 | 2+2 4 |
| | 3+4 2 | 3+4 2 | 2+2 4 | 2+2 4 | 2+2 4 | 3+4 2 | 3+4 2 | 2+2 4 |
| | 5+6 2 | 5+6 2 | 2+2 4 | 2+2 4 | 2+2 4 | 3+4 2 | 5+6 2 | 2+2 4 |
| | 3+4 2 | 5+6 2 | 2+2 4 | 2+2 4 | 2+2 4 | 5+6 2 | 3+4 2 | 2+2 4 |

Continued on next column

**TABLE XXIII– continued**

| TQC | ICSD | | NoE | | SF | | | |
|---|---|---|---|---|---|---|---|---|
| | 5+6 2 | 3+4 2 | 2+2 4 | 2+2 4 | 2+2 4 | 5+6 2 | 5+6 2 | 2+2 4 |
| | 3+4 2 | 5+6 2 | 2+2 4 | 2+2 4 | 2+2 4 | 3+4 2 | 3+4 2 | 2+2 4 |
| | 5+6 2 | 3+4 2 | 2+2 4 | 2+2 4 | 2+2 4 | 3+4 2 | 5+6 2 | 2+2 4 |
| | 3+4 2 | 5+6 2 | 2+2 4 | 2+2 4 | 2+2 4 | 5+6 2 | 3+4 2 | 2+2 4 |
| | 5+6 2 | 3+4 2 | 2+2 4 | 2+2 4 | 2+2 4 | 5+6 2 | 5+6 2 | 2+2 4 |
| | 3+4 2 | 5+6 2 | 2+2 4 | 2+2 4 | 2+2 4 | 3+4 2 | 5+6 2 | 2+2 4 |
| | 5+6 2 | 3+4 2 | 2+2 4 | 2+2 4 | 2+2 4 | 5+6 2 | 3+4 2 | 2+2 4 |
| | 3+4 2 | 3+4 2 | 2+2 4 | 2+2 4 | 2+2 4 | 3+4 2 | 3+4 2 | 2+2 4 |
| | 5+6 2 | 5+6 2 | 2+2 4 | 2+2 4 | 2+2 4 | 5+6 2 | 5+6 2 | 2+2 4 |
| | 3+4 2 | 5+6 2 | 2+2 4 | 2+2 4 | 2+2 4 | 3+4 2 | 5+6 2 | 2+2 4 |
| | 5+6 2 | 3+4 2 | 2+2 4 | 2+2 4 | 2+2 4 | 3+4 2 | 3+4 2 | 2+2 4 |
| | 3+4 2 | 5+6 2 | 2+2 4 | 2+2 4 | 2+2 4 | 5+6 2 | 5+6 2 | 2+2 4 |
| | 5+6 2 | 3+4 2 | 2+2 4 | 2+2 4 | 2+2 4 | 3+4 2 | 3+4 2 | 2+2 4 |
| | 5+6 2 | 3+4 2 | 2+2 4 | 2+2 4 | 2+2 4 | 3+4 2 | 5+6 2 | 2+2 4 |
| | 3+4 2 | 5+6 2 | 2+2 4 | 2+2 4 | 2+2 4 | 5+6 2 | 3+4 2 | 2+2 4 |
| | 3+4 2 | 3+4 2 | 2+2 4 | 2+2 4 | 2+2 4 | 5+6 2 | 3+4 2 | 2+2 4 |
| | 5+6 2 | 5+6 2 | 2+2 4 | 2+2 4 | 2+2 4 | 3+4 2 | 3+4 2 | 2+2 4 |
| | 3+4 2 | 3+4 2 | 2+2 4 | 2+2 4 | 2+2 4 | 5+6 2 | 5+6 2 | 2+2 4 |
| | 5+6 2 | 5+6 2 | 2+2 4 | 2+2 4 | 2+2 4 | 3+4 2 | 5+6 2 | 2+2 4 |
| | 3+4 2 | 5+6 2 | 2+2 4 | 2+2 4 | 2+2 4 | 5+6 2 | 3+4 2 | 2+2 4 |
| | 5+6 2 | 3+4 2 | 2+2 4 | 2+2 4 | 2+2 4 | 5+6 2 | 5+6 2 | 2+2 4 |
| | 3+4 2 | 5+6 2 | 2+2 4 | 2+2 4 | 2+2 4 | 3+4 2 | 3+4 2 | 2+2 4 |
| | 5+6 2 | 3+4 2 | 2+2 4 | 2+2 4 | 2+2 4 | 5+6 2 | 3+4 2 | 2+2 4 |
| | 5+6 2 | 5+6 2 | 2+2 4 | 2+2 4 | 2+2 4 | 3+4 2 | 5+6 2 | 2+2 4 |
| | 3+4 2 | 3+4 2 | 2+2 4 | 2+2 4 | 2+2 4 | 5+6 2 | 5+6 2 | 2+2 4 |
| | 3+4 2 | 3+4 2 | 2+2 4 | 2+2 4 | 2+2 4 | 3+4 2 | 3+4 2 | 2+2 4 |
| | 5+6 2 | 5+6 2 | 2+2 4 | 2+2 4 | 2+2 4 | 5+6 2 | 5+6 2 | 2+2 4 |
| | 5+6 2 | 5+6 2 | 2+2 4 | 2+2 4 | 2+2 4 | 3+4 2 | 5+6 2 | 2+2 4 |
| | 3+4 2 | 3+4 2 | 2+2 4 | 2+2 4 | 2+2 4 | 5+6 2 | 3+4 2 | 2+2 4 |
| | 3+4 2 | 3+4 2 | 2+2 4 | 2+2 4 | 2+2 4 | 5+6 2 | 5+6 2 | 2+2 4 |
| | 5+6 2 | 5+6 2 | 2+2 4 | 2+2 4 | 2+2 4 | 5+6 2 | 3+4 2 | 2+2 4 |
| | 3+4 2 | 5+6 2 | 2+2 4 | 2+2 4 | 2+2 4 | 3+4 2 | 5+6 2 | 2+2 4 |
| | 5+6 2 | 3+4 2 | 2+2 4 | 2+2 4 | 2+2 4 | 3+4 2 | 3+4 2 | 2+2 4 |
| | 3+4 2 | 5+6 2 | 2+2 4 | 2+2 4 | 2+2 4 | 3+4 2 | 3+4 2 | 2+2 4 |
| | 5+6 2 | 3+4 2 | 2+2 4 | 2+2 4 | 2+2 4 | 5+6 2 | 5+6 2 | 2+2 4 |
| | 3+4 2 | 5+6 2 | 2+2 4 | 2+2 4 | 2+2 4 | 5+6 2 | 3+4 2 | 2+2 4 |
| | 5+6 2 | 3+4 2 | 2+2 4 | 2+2 4 | 2+2 4 | 3+4 2 | 3+4 2 | 2+2 4 |
| | 3+4 2 | 3+4 2 | 2+2 4 | 2+2 4 | 2+2 4 | 3+4 2 | 5+6 2 | 2+2 4 |
| | 5+6 2 | 5+6 2 | 2+2 4 | 2+2 4 | 2+2 4 | 3+4 2 | 3+4 2 | 2+2 4 |
| | 5+6 2 | 3+4 2 | 2+2 4 | 2+2 4 | 2+2 4 | 5+6 2 | 3+4 2 | 2+2 4 |
| | 3+4 2 | 5+6 2 | 2+2 4 | 2+2 4 | 2+2 4 | 3+4 2 | 5+6 2 | 2+2 4 |
| | 5+6 2 | 3+4 2 | 2+2 4 | 2+2 4 | 2+2 4 | 5+6 2 | 5+6 2 | 2+2 4 |
| | 3+4 2 | 5+6 2 | 2+2 4 | 2+2 4 | 2+2 4 | 3+4 2 | 5+6 2 | 2+2 4 |
| | 3+4 2 | 5+6 2 | 2+2 4 | 2+2 4 | 2+2 4 | 5+6 2 | 3+4 2 | 2+2 4 |
| | 3+4 2 | 5+6 2 | 2+2 4 | 2+2 4 | 2+2 4 | 3+4 2 | 5+6 2 | 2+2 4 |
| | 5+6 2 | 3+4 2 | 2+2 4 | 2+2 4 | 2+2 4 | 3+4 2 | 3+4 2 | 2+2 4 |
| | 3+4 2 | 5+6 2 | 2+2 4 | 2+2 4 | 2+2 4 | 3+4 2 | 3+4 2 | 2+2 4 |
| | 5+6 2 | 3+4 2 | 2+2 4 | 2+2 4 | 2+2 4 | 5+6 2 | 5+6 2 | 2+2 4 |
| | 5+6 2 | 5+6 2 | 2+2 4 | 2+2 4 | 2+2 4 | 3+4 2 | 5+6 2 | 2+2 4 |
| | 3+4 2 | 3+4 2 | 2+2 4 | 2+2 4 | 2+2 4 | 5+6 2 | 3+4 2 | 2+2 4 |
| | 5+6 2 | 5+6 2 | 2+2 4 | 2+2 4 | 2+2 4 | 5+6 2 | 5+6 2 | 2+2 4 |
| | 3+4 2 | 3+4 2 | 2+2 4 | 2+2 4 | 2+2 4 | 3+4 2 | 3+4 2 | 2+2 4 |
| | 5+6 2 | 3+4 2 | 2+2 4 | 2+2 4 | 2+2 4 | 5+6 2 | 5+6 2 | 2+2 4 |
| | 3+4 2 | 5+6 2 | 2+2 4 | 2+2 4 | 2+2 4 | 5+6 2 | 5+6 2 | 2+2 4 |
| | 5+6 2 | 5+6 2 | 2+2 4 | 2+2 4 | 2+2 4 | 3+4 2 | 3+4 2 | 2+2 4 |
| | 5+6 2 | 3+4 2 | 2+2 4 | 2+2 4 | 2+2 4 | 3+4 2 | 3+4 2 | 2+2 4 |
| | 5+6 2 | 3+4 2 | 2+2 4 | 2+2 4 | 2+2 4 | 3+4 2 | 5+6 2 | 2+2 4 |
| | 3+4 2 | 5+6 2 | | | | 5+6 2 | 3+4 2 | |
| | 3+4 2 | 3+4 2 | | | | 3+4 2 | 3+4 2 | |
| | 5+6 2 | 3+4 2 | | | | 3+4 2 | 5+6 2 | |
| | 3+4 2 | 5+6 2 | | | | 5+6 2 | 5+6 2 | |

Continued on next column

**TABLE XXIII– continued**

| TQC | ICSD | NoE | SF |
|---|---|---|---|
| | 5+6 2 3+4 2 | | 3+4 2 3+4 2 |
| | 3+4 2 3+4 2 | | 5+6 2 3+4 2 |
| | 5+6 2 5+6 2 | | 5+6 2 5+6 2 |
| | 3+4 2 3+4 2 | | 3+4 2 5+6 2 |
| | 3+4 2 5+6 2 | | 3+4 2 5+6 2 |
| | 5+6 2 5+6 2 | | 5+6 2 3+4 2 |
| | 3+4 2 5+6 2 | | 3+4 2 3+4 2 |
| | 3+4 2 3+4 2 | | 5+6 2 5+6 2 |
| | 5+6 2 3+4 2 | | 5+6 2 5+6 2 |
| | 3+4 2 5+6 2 | | 3+4 2 3+4 2 |
| | 5+6 2 5+6 2 | | 5+6 2 3+4 2 |
| | 5+6 2 3+4 2 | | 5+6 2 5+6 2 |
| | 5+6 2 3+4 2 | | 3+4 2 5+6 2 |
| | 3+4 2 5+6 2 | | 5+6 2 3+4 2 |
| | 5+6 2 3+4 2 | | 3+4 2 3+4 2 |
| | 3+4 2 5+6 2 | | 5+6 2 5+6 2 |
| | 5+6 2 3+4 2 | | 3+4 2 3+4 2 |
| | 3+4 2 5+6 2 | | 5+6 2 5+6 2 |
| | 5+6 2 3+4 2 | | 3+4 2 3+4 2 |
| | 5+6 2 5+6 2 | | 5+6 2 5+6 2 |
| | 3+4 2 3+4 2 | | 3+4 2 5+6 2 |
| | 3+4 2 5+6 2 | | 3+4 2 3+4 2 |
| | 3+4 2 3+4 2 | | 5+6 2 3+4 2 |
| | 5+6 2 5+6 2 | | 3+4 2 5+6 2 |
| | 3+4 2 3+4 2 | | 5+6 2 5+6 2 |
| | 5+6 2 5+6 2 | | 5+6 2 3+4 2 |
| | 3+4 2 3+4 2 | | 3+4 2 3+4 2 |
| | 5+6 2 5+6 2 | | 5+6 2 5+6 2 |
| | 3+4 2 3+4 2 | | 3+4 2 3+4 2 |
| | 5+6 2 5+6 2 | | 3+4 2 5+6 2 |
| | 5+6 2 3+4 2 | | 5+6 2 3+4 2 |
| | 3+4 2 5+6 2 | | 3+4 2 5+6 2 |
| | 3+4 2 3+4 2 | | 3+4 2 5+6 2 |
| | 5+6 2 5+6 2 | | 5+6 2 3+4 2 |
| | 5+6 2 3+4 2 | | 3+4 2 5+6 2 |
| | 3+4 2 5+6 2 | | 3+4 2 3+4 2 |
| | 5+6 2 3+4 2 | | 5+6 2 3+4 2 |
| | 3+4 2 5+6 2 | | 3+4 2 5+6 2 |
| | 3+4 2 3+4 2 | | 5+6 2 5+6 2 |
| | 5+6 2 3+4 2 | | 5+6 2 5+6 2 |
| | 5+6 2 5+6 2 | | 3+4 2 3+4 2 |
| | 3+4 2 5+6 2 | | 5+6 2 5+6 2 |
| | 5+6 2 3+4 2 | | 5+6 2 3+4 2 |
| | 3+4 2 5+6 2 | | 3+4 2 3+4 2 |
| | 5+6 2 3+4 2 | | 5+6 2 5+6 2 |
| | 3+4 2 5+6 2 | | 3+4 2 3+4 2 |
| | 3+4 2 3+4 2 | | 5+6 2 5+6 2 |
| | 5+6 2 5+6 2 | | 3+4 2 3+4 2 |
| | 3+4 2 3+4 2 | | 5+6 2 5+6 2 |
| | 5+6 2 5+6 2 | | 5+6 2 3+4 2 |
| | 5+6 2 5+6 2 | | 3+4 2 5+6 2 |
| | 3+4 2 3+4 2 | | 3+4 2 5+6 2 |
| | 3+4 2 5+6 2 | | 5+6 2 3+4 2 |
| | 3+4 2 3+4 2 | | 5+6 2 3+4 2 |
| | 5+6 2 5+6 2 | | 5+6 2 5+6 2 |
| | 5+6 2 3+4 2 | | 5+6 2 3+4 2 |
| | 5+6 2 5+6 2 | | 3+4 2 3+4 2 |
| | 3+4 2 3+4 2 | | 3+4 2 5+6 2 |
| | 5+6 2 3+4 2 | | 5+6 2 5+6 2 |
| | 3+4 2 5+6 2 | | 5+6 2 3+4 2 |
| | 5+6 2 5+6 2 | | 3+4 2 5+6 2 |

Continued on next column

**TABLE XXIII– continued**

| TQC | ICSD | NoE | SF |
|---|---|---|---|
| | 3+4 2 3+4 2 | | 3+4 2 3+4 2 |
| | 5+6 2 3+4 2 | | 5+6 2 5+6 2 |
| | 3+4 2 3+4 2 | | 3+4 2 3+4 2 |
| | 5+6 2 5+6 2 | | 5+6 2 5+6 2 |
| | 5+6 2 5+6 2 | | 3+4 2 3+4 2 |
| | 3+4 2 3+4 2 | | 5+6 2 3+4 2 |
| | 3+4 2 5+6 2 | | 5+6 2 5+6 2 |
| | 5+6 2 3+4 2 | | 3+4 2 5+6 2 |
| | 3+4 2 5+6 2 | | 3+4 2 3+4 2 |
| | 5+6 2 5+6 2 | | 3+4 2 3+4 2 |
| | 3+4 2 3+4 2 | | 5+6 2 5+6 2 |
| | 5+6 2 3+4 2 | | 5+6 2 3+4 2 |
| | 3+4 2 5+6 2 | | 3+4 2 5+6 2 |
| | 3+4 2 3+4 2 | | 3+4 2 3+4 2 |
| | 5+6 2 5+6 2 | | 5+6 2 5+6 2 |
| | 3+4 2 3+4 2 | | 3+4 2 5+6 2 |
| | 5+6 2 3+4 2 | | 3+4 2 5+6 2 |
| | 3+4 2 5+6 2 | | 5+6 2 3+4 2 |
| | 5+6 2 5+6 2 | | 3+4 2 3+4 2 |
| | 5+6 2 3+4 2 | | 3+4 2 5+6 2 |
| | 3+4 2 5+6 2 | | 5+6 2 5+6 2 |
| | 5+6 2 3+4 2 | | 5+6 2 3+4 2 |
| | 3+4 2 3+4 2 | | 3+4 2 3+4 2 |
| | 5+6 2 5+6 2 | | 3+4 2 5+6 2 |
| | 3+4 2 3+4 2 | | 5+6 2 5+6 2 |
| | 5+6 2 5+6 2 | | 3+4 2 3+4 2 |
| | 3+4 2 5+6 2 | | 5+6 2 5+6 2 |
| | 5+6 2 3+4 2 | | 5+6 2 3+4 2 |
| | 3+4 2 3+4 2 | | 3+4 2 3+4 2 |
| | 5+6 2 5+6 2 | | 5+6 2 5+6 2 |
| | 3+4 2 5+6 2 | | 3+4 2 3+4 2 |
| 11.2.9 | 52066 | 112 | Rh4Si5 |

| TQC | -A | -B | -C | -D | -E | -GM | -Y | -Z |
|---|---|---|---|---|---|---|---|---|
| 11.2.9 | 3+4 2 | 3+4 2 | 2+2 4 | 2+2 4 | 2+2 4 | 3+4 2 | 3+4 2 | 2+2 4 |
| | 5+6 2 | 5+6 2 | 2+2 4 | 2+2 4 | 2+2 4 | 5+6 2 | 5+6 2 | 2+2 4 |
| | 3+4 2 | 3+4 2 | 2+2 4 | 2+2 4 | 2+2 4 | 3+4 2 | 3+4 2 | 2+2 4 |
| | 3+4 2 | 5+6 2 | 2+2 4 | 2+2 4 | 2+2 4 | 3+4 2 | 5+6 2 | 2+2 4 |
| | 5+6 2 | 5+6 2 | 2+2 4 | 2+2 4 | 2+2 4 | 5+6 2 | 5+6 2 | 2+2 4 |
| | 5+6 2 | 3+4 2 | 2+2 4 | 2+2 4 | 2+2 4 | 3+4 2 | 3+4 2 | 2+2 4 |
| | 3+4 2 | 5+6 2 | 2+2 4 | 2+2 4 | 2+2 4 | 5+6 2 | 5+6 2 | 2+2 4 |
| | 5+6 2 | 3+4 2 | 2+2 4 | 2+2 4 | 2+2 4 | 5+6 2 | 3+4 2 | 2+2 4 |
| | 3+4 2 | 3+4 2 | 2+2 4 | 2+2 4 | 2+2 4 | 3+4 2 | 3+4 2 | 2+2 4 |
| | 5+6 2 | 5+6 2 | 2+2 4 | 2+2 4 | 2+2 4 | 5+6 2 | 5+6 2 | 2+2 4 |
| | 3+4 2 | 5+6 2 | 2+2 4 | 2+2 4 | 2+2 4 | 5+6 2 | 5+6 2 | 2+2 4 |
| | 5+6 2 | 3+4 2 | 2+2 4 | 2+2 4 | 2+2 4 | 3+4 2 | 5+6 2 | 2+2 4 |
| | 3+4 2 | 3+4 2 | 2+2 4 | 2+2 4 | 2+2 4 | 3+4 2 | 3+4 2 | 2+2 4 |
| | 3+4 2 | 5+6 2 | 2+2 4 | 2+2 4 | 2+2 4 | 5+6 2 | 3+4 2 | 2+2 4 |
| | 5+6 2 | 3+4 2 | 2+2 4 | 2+2 4 | 2+2 4 | 5+6 2 | 3+4 2 | 2+2 4 |
| | 5+6 2 | 3+4 2 | 2+2 4 | 2+2 4 | 2+2 4 | 3+4 2 | 5+6 2 | 2+2 4 |
| | 3+4 2 | 5+6 2 | 2+2 4 | 2+2 4 | 2+2 4 | 3+4 2 | 3+4 2 | 2+2 4 |
| | 3+4 2 | 3+4 2 | 2+2 4 | 2+2 4 | 2+2 4 | 5+6 2 | 3+4 2 | 2+2 4 |
| | 5+6 2 | 3+4 2 | 2+2 4 | 2+2 4 | 2+2 4 | 3+4 2 | 5+6 2 | 2+2 4 |
| | 3+4 2 | 5+6 2 | 2+2 4 | 2+2 4 | 2+2 4 | 5+6 2 | 5+6 2 | 2+2 4 |
| | 3+4 2 | 3+4 2 | 2+2 4 | 2+2 4 | 2+2 4 | 5+6 2 | 3+4 2 | 2+2 4 |
| | 5+6 2 | 5+6 2 | 2+2 4 | 2+2 4 | 2+2 4 | 3+4 2 | 3+4 2 | 2+2 4 |
| | 5+6 2 | 5+6 2 | 2+2 4 | 2+2 4 | 2+2 4 | 5+6 2 | 5+6 2 | 2+2 4 |
| | 3+4 2 | 5+6 2 | 2+2 4 | 2+2 4 | 2+2 4 | 3+4 2 | 3+4 2 | 2+2 4 |
| | 5+6 2 | 3+4 2 | 2+2 4 | 2+2 4 | 2+2 4 | 3+4 2 | 5+6 2 | 2+2 4 |
| | 3+4 2 | 3+4 2 | 2+2 4 | 2+2 4 | 2+2 4 | 3+4 2 | 5+6 2 | 2+2 4 |
| | 3+4 2 | 5+6 2 | 2+2 4 | 2+2 4 | 2+2 4 | 3+4 2 | 5+6 2 | 2+2 4 |
| | 5+6 2 | 5+6 2 | 2+2 4 | 2+2 4 | 2+2 4 | 5+6 2 | 3+4 2 | 2+2 4 |

Continued on next column

**TABLE XXIII– continued**

| TQC | ICSD | NoE | SF | | | | | |
|---|---|---|---|---|---|---|---|---|
| | 3+4 2 | 5+6 2 | | | | 3+4 2 | 5+6 2 | |
| | 5+6 2 | 3+4 2 | | | | 5+6 2 | 5+6 2 | |
| | 5+6 2 | 5+6 2 | | | | 5+6 2 | 3+4 2 | |
| | 3+4 2 | 3+4 2 | | | | 3+4 2 | 3+4 2 | |
| | 3+4 2 | 5+6 2 | | | | 3+4 2 | 5+6 2 | |
| | 3+4 2 | 3+4 2 | | | | 3+4 2 | 5+6 2 | |
| | 5+6 2 | 5+6 2 | | | | 5+6 2 | 3+4 2 | |
| | 5+6 2 | 3+4 2 | | | | 5+6 2 | 3+4 2 | |
| | 3+4 2 | 5+6 2 | | | | 5+6 2 | 3+4 2 | |
| | 5+6 2 | 3+4 2 | | | | 3+4 2 | 5+6 2 | |
| | 5+6 2 | 5+6 2 | | | | 3+4 2 | 3+4 2 | |
| | 3+4 2 | 3+4 2 | | | | 5+6 2 | 5+6 2 | |
| | 5+6 2 | 5+6 2 | | | | 5+6 2 | 3+4 2 | |
| | 3+4 2 | 3+4 2 | | | | 3+4 2 | 5+6 2 | |
| | 5+6 2 | 3+4 2 | | | | 3+4 2 | 5+6 2 | |
| | 3+4 2 | 5+6 2 | | | | 5+6 2 | 3+4 2 | |
| | 5+6 2 | 5+6 2 | | | | 5+6 2 | 5+6 2 | |
| | 5+6 2 | 3+4 2 | | | | 3+4 2 | 3+4 2 | |
| | 3+4 2 | 5+6 2 | | | | 5+6 2 | 5+6 2 | |
| | 5+6 2 | 3+4 2 | | | | 3+4 2 | 3+4 2 | |
| | 5+6 2 | 3+4 2 | | | | 5+6 2 | 5+6 2 | |
| | 3+4 2 | 5+6 2 | | | | 3+4 2 | 3+4 2 | |
| | 5+6 2 | 5+6 2 | | | | 5+6 2 | 5+6 2 | |
| | 3+4 2 | 3+4 2 | | | | 3+4 2 | 5+6 2 | |
| | 5+6 2 | 3+4 2 | | | | 5+6 2 | 3+4 2 | |
| | 5+6 2 | 5+6 2 | | | | 3+4 2 | 3+4 2 | |
| | 5+6 2 | 3+4 2 | | | | 3+4 2 | 3+4 2 | |
| | 3+4 2 | 5+6 2 | | | | 3+4 2 | 5+6 2 | |
| 11.2.10 | 24355 | 112 | Rh4Si5 | | | | | |
| | -A | -B | -C | -D | -E | -GM | -Y | -Z |
| | 5+6 2 | 3+4 2 | 2+2 4 | 2+2 4 | 2+2 4 | 3+4 2 | 5+6 2 | 2+2 4 |
| | 3+4 2 | 5+6 2 | 2+2 4 | 2+2 4 | 2+2 4 | 5+6 2 | 3+4 2 | 2+2 4 |
| | 5+6 2 | 3+4 2 | 2+2 4 | 2+2 4 | 2+2 4 | 3+4 2 | 5+6 2 | 2+2 4 |
| | 5+6 2 | 5+6 2 | 2+2 4 | 2+2 4 | 2+2 4 | 3+4 2 | 3+4 2 | 2+2 4 |
| | 3+4 2 | 5+6 2 | 2+2 4 | 2+2 4 | 2+2 4 | 5+6 2 | 3+4 2 | 2+2 4 |
| | 3+4 2 | 3+4 2 | 2+2 4 | 2+2 4 | 2+2 4 | 3+4 2 | 5+6 2 | 2+2 4 |
| | 5+6 2 | 5+6 2 | 2+2 4 | 2+2 4 | 2+2 4 | 5+6 2 | 3+4 2 | 2+2 4 |
| | 3+4 2 | 3+4 2 | 2+2 4 | 2+2 4 | 2+2 4 | 5+6 2 | 5+6 2 | 2+2 4 |
| | 5+6 2 | 3+4 2 | 2+2 4 | 2+2 4 | 2+2 4 | 3+4 2 | 5+6 2 | 2+2 4 |
| | 3+4 2 | 5+6 2 | 2+2 4 | 2+2 4 | 2+2 4 | 5+6 2 | 3+4 2 | 2+2 4 |
| | 5+6 2 | 5+6 2 | 2+2 4 | 2+2 4 | 2+2 4 | 5+6 2 | 3+4 2 | 2+2 4 |
| | 3+4 2 | 3+4 2 | 2+2 4 | 2+2 4 | 2+2 4 | 3+4 2 | 3+4 2 | 2+2 4 |
| | 5+6 2 | 3+4 2 | 2+2 4 | 2+2 4 | 2+2 4 | 3+4 2 | 5+6 2 | 2+2 4 |
| | 5+6 2 | 5+6 2 | 2+2 4 | 2+2 4 | 2+2 4 | 5+6 2 | 5+6 2 | 2+2 4 |
| | 3+4 2 | 3+4 2 | 2+2 4 | 2+2 4 | 2+2 4 | 5+6 2 | 5+6 2 | 2+2 4 |
| | 3+4 2 | 3+4 2 | 2+2 4 | 2+2 4 | 2+2 4 | 3+4 2 | 3+4 2 | 2+2 4 |
| | 5+6 2 | 5+6 2 | 2+2 4 | 2+2 4 | 2+2 4 | 3+4 2 | 5+6 2 | 2+2 4 |
| | 5+6 2 | 3+4 2 | 2+2 4 | 2+2 4 | 2+2 4 | 5+6 2 | 5+6 2 | 2+2 4 |
| | 3+4 2 | 3+4 2 | 2+2 4 | 2+2 4 | 2+2 4 | 3+4 2 | 3+4 2 | 2+2 4 |
| | 5+6 2 | 5+6 2 | 2+2 4 | 2+2 4 | 2+2 4 | 5+6 2 | 3+4 2 | 2+2 4 |
| | 5+6 2 | 3+4 2 | 2+2 4 | 2+2 4 | 2+2 4 | 5+6 2 | 5+6 2 | 2+2 4 |
| | 3+4 2 | 5+6 2 | 2+2 4 | 2+2 4 | 2+2 4 | 3+4 2 | 5+6 2 | 2+2 4 |
| | 3+4 2 | 5+6 2 | 2+2 4 | 2+2 4 | 2+2 4 | 5+6 2 | 3+4 2 | 2+2 4 |
| | 5+6 2 | 5+6 2 | 2+2 4 | 2+2 4 | 2+2 4 | 3+4 2 | 5+6 2 | 2+2 4 |
| | 3+4 2 | 3+4 2 | 2+2 4 | 2+2 4 | 2+2 4 | 3+4 2 | 3+4 2 | 2+2 4 |
| | 5+6 2 | 3+4 2 | 2+2 4 | 2+2 4 | 2+2 4 | 3+4 2 | 3+4 2 | 2+2 4 |
| | 5+6 2 | 5+6 2 | 2+2 4 | 2+2 4 | 2+2 4 | 3+4 2 | 3+4 2 | 2+2 4 |
| | 3+4 2 | 5+6 2 | 2+2 4 | 2+2 4 | 2+2 4 | 5+6 2 | 5+6 2 | 2+2 4 |
| | 3+4 2 | 5+6 2 | | | | 3+4 2 | 3+4 2 | |
| | 5+6 2 | 3+4 2 | | | | 5+6 2 | 3+4 2 | |
| | 3+4 2 | 5+6 2 | | | | 5+6 2 | 5+6 2 | |
| | 5+6 2 | 3+4 2 | | | | 3+4 2 | 5+6 2 | |
| | 5+6 2 | 5+6 2 | | | | 3+4 2 | 3+4 2 | |
| | 5+6 2 | 3+4 2 | | | | 3+4 2 | 3+4 2 | |
| | 3+4 2 | 5+6 2 | | | | 5+6 2 | 5+6 2 | |
| | 3+4 2 | 3+4 2 | | | | 5+6 2 | 5+6 2 | |
| | 5+6 2 | 5+6 2 | | | | 5+6 2 | 5+6 2 | |
| | 3+4 2 | 3+4 2 | | | | 3+4 2 | 3+4 2 | |
| | 3+4 2 | 5+6 2 | | | | 3+4 2 | 5+6 2 | |
| | 5+6 2 | 3+4 2 | | | | 5+6 2 | 3+4 2 | |
| | 3+4 2 | 5+6 2 | | | | 5+6 2 | 5+6 2 | |
| | 5+6 2 | 3+4 2 | | | | 3+4 2 | 3+4 2 | |
| | 3+4 2 | 3+4 2 | | | | 3+4 2 | 3+4 2 | |
| | 5+6 2 | 5+6 2 | | | | 5+6 2 | 5+6 2 | |
| | 3+4 2 | 5+6 2 | | | | 5+6 2 | 3+4 2 | |
| | 3+4 2 | 3+4 2 | | | | 3+4 2 | 5+6 2 | |
| | 5+6 2 | 5+6 2 | | | | 5+6 2 | 3+4 2 | |
| | 3+4 2 | 3+4 2 | | | | 3+4 2 | 5+6 2 | |
| | 3+4 2 | 3+4 2 | | | | 5+6 2 | 3+4 2 | |
| | 5+6 2 | 5+6 2 | | | | 3+4 2 | 5+6 2 | |
| | 3+4 2 | 5+6 2 | | | | 5+6 2 | 3+4 2 | |
| | 5+6 2 | 3+4 2 | | | | 3+4 2 | 3+4 2 | |
| | 3+4 2 | 3+4 2 | | | | 5+6 2 | 5+6 2 | |
| | 3+4 2 | 5+6 2 | | | | 3+4 2 | 5+6 2 | |
| | 3+4 2 | 3+4 2 | | | | 3+4 2 | 5+6 2 | |
| | 5+6 2 | 5+6 2 | | | | 3+4 2 | 3+4 2 | |
| 11.2.11 | 31692 | 256 | Pu1 | | | | | |
| | -A | -B | -C | -D | -E | -GM | -Y | -Z |
| | 3+4 2 | 5+6 2 | 2+2 4 | 2+2 4 | 2+2 4 | 3+4 2 | 5+6 2 | 2+2 4 |
| | 5+6 2 | 3+4 2 | 2+2 4 | 2+2 4 | 2+2 4 | 3+4 2 | 3+4 2 | 2+2 4 |
| | 3+4 2 | 5+6 2 | 2+2 4 | 2+2 4 | 2+2 4 | 5+6 2 | 5+6 2 | 2+2 4 |
| | 5+6 2 | 3+4 2 | 2+2 4 | 2+2 4 | 2+2 4 | 5+6 2 | 5+6 2 | 2+2 4 |
| | 3+4 2 | 5+6 2 | 2+2 4 | 2+2 4 | 2+2 4 | 5+6 2 | 3+4 2 | 2+2 4 |
| | 5+6 2 | 5+6 2 | 2+2 4 | 2+2 4 | 2+2 4 | 3+4 2 | 3+4 2 | 2+2 4 |
| | 3+4 2 | 3+4 2 | 2+2 4 | 2+2 4 | 2+2 4 | 5+6 2 | 5+6 2 | 2+2 4 |
| | 5+6 2 | 3+4 2 | 2+2 4 | 2+2 4 | 2+2 4 | 5+6 2 | 3+4 2 | 2+2 4 |
| | 5+6 2 | 5+6 2 | 2+2 4 | 2+2 4 | 2+2 4 | 3+4 2 | 3+4 2 | 2+2 4 |
| | 3+4 2 | 3+4 2 | 2+2 4 | 2+2 4 | 2+2 4 | 3+4 2 | 5+6 2 | 2+2 4 |
| | 3+4 2 | 3+4 2 | 2+2 4 | 2+2 4 | 2+2 4 | 5+6 2 | 3+4 2 | 2+2 4 |
| | 5+6 2 | 3+4 2 | 2+2 4 | 2+2 4 | 2+2 4 | 3+4 2 | 3+4 2 | 2+2 4 |
| | 5+6 2 | 5+6 2 | 2+2 4 | 2+2 4 | 2+2 4 | 5+6 2 | 5+6 2 | 2+2 4 |
| | 5+6 2 | 5+6 2 | 2+2 4 | 2+2 4 | 2+2 4 | 3+4 2 | 5+6 2 | 2+2 4 |
| | 3+4 2 | 3+4 2 | 2+2 4 | 2+2 4 | 2+2 4 | 3+4 2 | 5+6 2 | 2+2 4 |
| | 3+4 2 | 5+6 2 | 2+2 4 | 2+2 4 | 2+2 4 | 5+6 2 | 3+4 2 | 2+2 4 |
| | 5+6 2 | 3+4 2 | 2+2 4 | 2+2 4 | 2+2 4 | 3+4 2 | 5+6 2 | 2+2 4 |
| | 3+4 2 | 5+6 2 | 2+2 4 | 2+2 4 | 2+2 4 | 5+6 2 | 3+4 2 | 2+2 4 |
| | 5+6 2 | 3+4 2 | 2+2 4 | 2+2 4 | 2+2 4 | 5+6 2 | 3+4 2 | 2+2 4 |
| | 3+4 2 | 5+6 2 | 2+2 4 | 2+2 4 | 2+2 4 | 3+4 2 | 5+6 2 | 2+2 4 |
| | 3+4 2 | 3+4 2 | 2+2 4 | 2+2 4 | 2+2 4 | 5+6 2 | 3+4 2 | 2+2 4 |
| | 5+6 2 | 5+6 2 | 2+2 4 | 2+2 4 | 2+2 4 | 3+4 2 | 5+6 2 | 2+2 4 |
| | 5+6 2 | 5+6 2 | 2+2 4 | 2+2 4 | 2+2 4 | 5+6 2 | 5+6 2 | 2+2 4 |
| | 3+4 2 | 3+4 2 | 2+2 4 | 2+2 4 | 2+2 4 | 3+4 2 | 3+4 2 | 2+2 4 |
| | 3+4 2 | 5+6 2 | 2+2 4 | 2+2 4 | 2+2 4 | 5+6 2 | 5+6 2 | 2+2 4 |
| | 5+6 2 | 3+4 2 | 2+2 4 | 2+2 4 | 2+2 4 | 3+4 2 | 3+4 2 | 2+2 4 |
| | 3+4 2 | 5+6 2 | 2+2 4 | 2+2 4 | 2+2 4 | 3+4 2 | 5+6 2 | 2+2 4 |
| | 5+6 2 | 3+4 2 | 2+2 4 | 2+2 4 | 2+2 4 | 5+6 2 | 3+4 2 | 2+2 4 |
| | 3+4 2 | 5+6 2 | 2+2 4 | 2+2 4 | 2+2 4 | 3+4 2 | 5+6 2 | 2+2 4 |
| | 5+6 2 | 3+4 2 | 2+2 4 | 2+2 4 | 2+2 4 | 3+4 2 | 3+4 2 | 2+2 4 |
| | 3+4 2 | 5+6 2 | 2+2 4 | 2+2 4 | 2+2 4 | 5+6 2 | 3+4 2 | 2+2 4 |
| | 5+6 2 | 3+4 2 | 2+2 4 | 2+2 4 | 2+2 4 | 5+6 2 | 5+6 2 | 2+2 4 |
| | 3+4 2 | 5+6 2 | 2+2 4 | 2+2 4 | 2+2 4 | 5+6 2 | 3+4 2 | 2+2 4 |
| | 5+6 2 | 3+4 2 | 2+2 4 | 2+2 4 | 2+2 4 | 3+4 2 | 5+6 2 | 2+2 4 |

Continued on next column

**TABLE XXIII– continued**

| TQC | ICSD | | NoE | | SF | | | |
|---|---|---|---|---|---|---|---|---|
| | 3+4 2 | 5+6 2 | 2+2 4 | 2+2 4 | 2+2 4 | 3+4 2 | 5+6 2 | 2+2 4 |
| | 5+6 2 | 5+6 2 | 2+2 4 | 2+2 4 | 2+2 4 | 5+6 2 | 3+4 2 | 2+2 4 |
| | 3+4 2 | 3+4 2 | 2+2 4 | 2+2 4 | 2+2 4 | 5+6 2 | 5+6 2 | 2+2 4 |
| | 5+6 2 | 3+4 2 | 2+2 4 | 2+2 4 | 2+2 4 | 3+4 2 | 3+4 2 | 2+2 4 |
| | 5+6 2 | 5+6 2 | 2+2 4 | 2+2 4 | 2+2 4 | 5+6 2 | 5+6 2 | 2+2 4 |
| | 3+4 2 | 3+4 2 | 2+2 4 | 2+2 4 | 2+2 4 | 3+4 2 | 3+4 2 | 2+2 4 |
| | 5+6 2 | 5+6 2 | 2+2 4 | 2+2 4 | 2+2 4 | 3+4 2 | 5+6 2 | 2+2 4 |
| | 3+4 2 | 3+4 2 | 2+2 4 | 2+2 4 | 2+2 4 | 5+6 2 | 3+4 2 | 2+2 4 |
| | 5+6 2 | 3+4 2 | 2+2 4 | 2+2 4 | 2+2 4 | 3+4 2 | 5+6 2 | 2+2 4 |
| | 3+4 2 | 5+6 2 | 2+2 4 | 2+2 4 | 2+2 4 | 5+6 2 | 3+4 2 | 2+2 4 |
| | 3+4 2 | 5+6 2 | 2+2 4 | 2+2 4 | 2+2 4 | 3+4 2 | 3+4 2 | 2+2 4 |
| | 5+6 2 | 3+4 2 | 2+2 4 | 2+2 4 | 2+2 4 | 5+6 2 | 5+6 2 | 2+2 4 |
| | 3+4 2 | 3+4 2 | 2+2 4 | 2+2 4 | 2+2 4 | 3+4 2 | 3+4 2 | 2+2 4 |
| | 5+6 2 | 3+4 2 | 2+2 4 | 2+2 4 | 2+2 4 | 5+6 2 | 5+6 2 | 2+2 4 |
| | 5+6 2 | 5+6 2 | 2+2 4 | 2+2 4 | 2+2 4 | 3+4 2 | 3+4 2 | 2+2 4 |
| | 3+4 2 | 3+4 2 | 2+2 4 | 2+2 4 | 2+2 4 | 3+4 2 | 3+4 2 | 2+2 4 |
| | 3+4 2 | 5+6 2 | 2+2 4 | 2+2 4 | 2+2 4 | 5+6 2 | 5+6 2 | 2+2 4 |
| | 5+6 2 | 5+6 2 | 2+2 4 | 2+2 4 | 2+2 4 | 5+6 2 | 5+6 2 | 2+2 4 |
| | 3+4 2 | 3+4 2 | 2+2 4 | 2+2 4 | 2+2 4 | 3+4 2 | 3+4 2 | 2+2 4 |
| | 5+6 2 | 5+6 2 | 2+2 4 | 2+2 4 | 2+2 4 | 5+6 2 | 5+6 2 | 2+2 4 |
| | 5+6 2 | 3+4 2 | 2+2 4 | 2+2 4 | 2+2 4 | 3+4 2 | 3+4 2 | 2+2 4 |
| | 3+4 2 | 5+6 2 | 2+2 4 | 2+2 4 | 2+2 4 | 5+6 2 | 5+6 2 | 2+2 4 |
| | 3+4 2 | 3+4 2 | 2+2 4 | 2+2 4 | 2+2 4 | 5+6 2 | 5+6 2 | 2+2 4 |
| | 5+6 2 | 5+6 2 | 2+2 4 | 2+2 4 | 2+2 4 | 3+4 2 | 3+4 2 | 2+2 4 |
| | 5+6 2 | 5+6 2 | 2+2 4 | 2+2 4 | 2+2 4 | 3+4 2 | 5+6 2 | 2+2 4 |
| | 3+4 2 | 3+4 2 | 2+2 4 | 2+2 4 | 2+2 4 | 5+6 2 | 3+4 2 | 2+2 4 |
| | 5+6 2 | 5+6 2 | 2+2 4 | 2+2 4 | 2+2 4 | 3+4 2 | 5+6 2 | 2+2 4 |
| | 3+4 2 | 3+4 2 | 2+2 4 | 2+2 4 | 2+2 4 | 5+6 2 | 3+4 2 | 2+2 4 |
| | 5+6 2 | 5+6 2 | 2+2 4 | 2+2 4 | 2+2 4 | 3+4 2 | 3+4 2 | 2+2 4 |
| | 3+4 2 | 3+4 2 | 2+2 4 | 2+2 4 | 2+2 4 | 5+6 2 | 5+6 2 | 2+2 4 |
| | 3+4 2 | 3+4 2 | | | | 3+4 2 | 5+6 2 | |
| | 5+6 2 | 5+6 2 | | | | 5+6 2 | 3+4 2 | |
| | 3+4 2 | 5+6 2 | | | | 3+4 2 | 5+6 2 | |
| | 5+6 2 | 5+6 2 | | | | 5+6 2 | 3+4 2 | |
| | 3+4 2 | 5+6 2 | | | | 5+6 2 | 5+6 2 | |
| | 5+6 2 | 3+4 2 | | | | 3+4 2 | 3+4 2 | |
| | 3+4 2 | 5+6 2 | | | | 3+4 2 | 3+4 2 | |
| | 5+6 2 | 3+4 2 | | | | 5+6 2 | 5+6 2 | |
| | 3+4 2 | 3+4 2 | | | | 3+4 2 | 5+6 2 | |
| | 3+4 2 | 3+4 2 | | | | 5+6 2 | 5+6 2 | |
| | 5+6 2 | 5+6 2 | | | | 3+4 2 | 3+4 2 | |
| | 5+6 2 | 5+6 2 | | | | 5+6 2 | 5+6 2 | |
| | 3+4 2 | 5+6 2 | | | | 5+6 2 | 3+4 2 | |
| | 5+6 2 | 3+4 2 | | | | 3+4 2 | 3+4 2 | |
| | 3+4 2 | 5+6 2 | | | | 3+4 2 | 5+6 2 | |
| | 5+6 2 | 3+4 2 | | | | 3+4 2 | 5+6 2 | |
| | 5+6 2 | 3+4 2 | | | | 3+4 2 | 3+4 2 | |
| | 3+4 2 | 3+4 2 | | | | 5+6 2 | 3+4 2 | |
| | 3+4 2 | 5+6 2 | | | | 5+6 2 | 5+6 2 | |
| | 5+6 2 | 5+6 2 | | | | 5+6 2 | 3+4 2 | |
| | 5+6 2 | 3+4 2 | | | | 3+4 2 | 3+4 2 | |
| | 3+4 2 | 5+6 2 | | | | 5+6 2 | 5+6 2 | |
| | 5+6 2 | 3+4 2 | | | | 5+6 2 | 3+4 2 | |
| | 3+4 2 | 5+6 2 | | | | 3+4 2 | 5+6 2 | |
| | 3+4 2 | 3+4 2 | | | | 3+4 2 | 3+4 2 | |
| | 5+6 2 | 5+6 2 | | | | 5+6 2 | 5+6 2 | |
| | 3+4 2 | 3+4 2 | | | | 5+6 2 | 3+4 2 | |
| | 5+6 2 | 5+6 2 | | | | 3+4 2 | 5+6 2 | |
| | 5+6 2 | 3+4 2 | | | | 5+6 2 | 3+4 2 | |
| | 3+4 2 | 3+4 2 | | | | 5+6 2 | 3+4 2 | |
| | 5+6 2 | 5+6 2 | | | | 3+4 2 | 5+6 2 | |

Continued on next column

**TABLE XXIII– continued**

| TQC | ICSD | | NoE | | SF | | | |
|---|---|---|---|---|---|---|---|---|
| | 3+4 2 | 5+6 2 | | | | 5+6 2 | 5+6 2 | |
| | 5+6 2 | 3+4 2 | | | | 3+4 2 | 3+4 2 | |
| | 3+4 2 | 5+6 2 | | | | 5+6 2 | 3+4 2 | |
| | 5+6 2 | 3+4 2 | | | | 3+4 2 | 5+6 2 | |
| | 3+4 2 | 3+4 2 | | | | 3+4 2 | 5+6 2 | |
| | 5+6 2 | 5+6 2 | | | | 5+6 2 | 3+4 2 | |
| | 3+4 2 | 5+6 2 | | | | 5+6 2 | 5+6 2 | |
| | 5+6 2 | 3+4 2 | | | | 3+4 2 | 3+4 2 | |
| | 3+4 2 | 5+6 2 | | | | 5+6 2 | 5+6 2 | |
| | 3+4 2 | 3+4 2 | | | | 3+4 2 | 5+6 2 | |
| | 5+6 2 | 3+4 2 | | | | 5+6 2 | 5+6 2 | |
| | 3+4 2 | 5+6 2 | | | | 3+4 2 | 3+4 2 | |
| | 5+6 2 | 5+6 2 | | | | 5+6 2 | 3+4 2 | |
| | 3+4 2 | 3+4 2 | | | | 3+4 2 | 5+6 2 | |
| | 5+6 2 | 5+6 2 | | | | 5+6 2 | 3+4 2 | |
| | 3+4 2 | 5+6 2 | | | | 3+4 2 | 5+6 2 | |
| | 5+6 2 | 3+4 2 | | | | 5+6 2 | 3+4 2 | |
| | 3+4 2 | 3+4 2 | | | | 5+6 2 | 5+6 2 | |
| | 5+6 2 | 5+6 2 | | | | 3+4 2 | 3+4 2 | |
| | 5+6 2 | 5+6 2 | | | | 5+6 2 | 5+6 2 | |
| | 3+4 2 | 3+4 2 | | | | 3+4 2 | 3+4 2 | |
| | 3+4 2 | 5+6 2 | | | | 3+4 2 | 5+6 2 | |
| | 5+6 2 | 3+4 2 | | | | 3+4 2 | 3+4 2 | |
| | 3+4 2 | 5+6 2 | | | | 5+6 2 | 3+4 2 | |
| | 5+6 2 | 3+4 2 | | | | 3+4 2 | 5+6 2 | |
| | 3+4 2 | 3+4 2 | | | | 5+6 2 | 5+6 2 | |
| | 5+6 2 | 5+6 2 | | | | 3+4 2 | 3+4 2 | |
| | 3+4 2 | 3+4 2 | | | | 5+6 2 | 5+6 2 | |
| | 5+6 2 | 5+6 2 | | | | 3+4 2 | 3+4 2 | |
| | 3+4 2 | 3+4 2 | | | | 3+4 2 | 5+6 2 | |
| | 5+6 2 | 5+6 2 | | | | 5+6 2 | 3+4 2 | |
| | 5+6 2 | 3+4 2 | | | | 3+4 2 | 5+6 2 | |
| | 3+4 2 | 5+6 2 | | | | 3+4 2 | 3+4 2 | |
| 11.2.12 | 43283 | | 160 | | Pt6Si5 | | | |
| | -A | -B | -C | -D | -E | -GM | -Y | -Z |
| | 5+6 2 | 3+4 2 | 2+2 4 | 2+2 4 | 2+2 4 | 3+4 2 | 5+6 2 | 2+2 4 |
| | 3+4 2 | 5+6 2 | 2+2 4 | 2+2 4 | 2+2 4 | 3+4 2 | 5+6 2 | 2+2 4 |
| | 5+6 2 | 3+4 2 | 2+2 4 | 2+2 4 | 2+2 4 | 5+6 2 | 3+4 2 | 2+2 4 |
| | 3+4 2 | 5+6 2 | 2+2 4 | 2+2 4 | 2+2 4 | 5+6 2 | 3+4 2 | 2+2 4 |
| | 5+6 2 | 3+4 2 | 2+2 4 | 2+2 4 | 2+2 4 | 3+4 2 | 3+4 2 | 2+2 4 |
| | 3+4 2 | 3+4 2 | 2+2 4 | 2+2 4 | 2+2 4 | 5+6 2 | 5+6 2 | 2+2 4 |
| | 5+6 2 | 5+6 2 | 2+2 4 | 2+2 4 | 2+2 4 | 3+4 2 | 3+4 2 | 2+2 4 |
| | 3+4 2 | 5+6 2 | 2+2 4 | 2+2 4 | 2+2 4 | 3+4 2 | 5+6 2 | 2+2 4 |
| | 5+6 2 | 3+4 2 | 2+2 4 | 2+2 4 | 2+2 4 | 5+6 2 | 5+6 2 | 2+2 4 |
| | 3+4 2 | 5+6 2 | 2+2 4 | 2+2 4 | 2+2 4 | 5+6 2 | 3+4 2 | 2+2 4 |
| | 5+6 2 | 5+6 2 | 2+2 4 | 2+2 4 | 2+2 4 | 3+4 2 | 3+4 2 | 2+2 4 |
| | 3+4 2 | 3+4 2 | 2+2 4 | 2+2 4 | 2+2 4 | 5+6 2 | 5+6 2 | 2+2 4 |
| | 5+6 2 | 5+6 2 | 2+2 4 | 2+2 4 | 2+2 4 | 5+6 2 | 3+4 2 | 2+2 4 |
| | 5+6 2 | 3+4 2 | 2+2 4 | 2+2 4 | 2+2 4 | 3+4 2 | 5+6 2 | 2+2 4 |
| | 3+4 2 | 3+4 2 | 2+2 4 | 2+2 4 | 2+2 4 | 5+6 2 | 3+4 2 | 2+2 4 |
| | 3+4 2 | 3+4 2 | 2+2 4 | 2+2 4 | 2+2 4 | 3+4 2 | 5+6 2 | 2+2 4 |
| | 5+6 2 | 5+6 2 | 2+2 4 | 2+2 4 | 2+2 4 | 5+6 2 | 3+4 2 | 2+2 4 |
| | 5+6 2 | 3+4 2 | 2+2 4 | 2+2 4 | 2+2 4 | 5+6 2 | 5+6 2 | 2+2 4 |
| | 5+6 2 | 5+6 2 | 2+2 4 | 2+2 4 | 2+2 4 | 3+4 2 | 3+4 2 | 2+2 4 |
| | 3+4 2 | 3+4 2 | 2+2 4 | 2+2 4 | 2+2 4 | 5+6 2 | 5+6 2 | 2+2 4 |
| | 3+4 2 | 3+4 2 | 2+2 4 | 2+2 4 | 2+2 4 | 3+4 2 | 5+6 2 | 2+2 4 |
| | 5+6 2 | 5+6 2 | 2+2 4 | 2+2 4 | 2+2 4 | 3+4 2 | 3+4 2 | 2+2 4 |
| | 3+4 2 | 3+4 2 | 2+2 4 | 2+2 4 | 2+2 4 | 3+4 2 | 5+6 2 | 2+2 4 |
| | 3+4 2 | 5+6 2 | 2+2 4 | 2+2 4 | 2+2 4 | 3+4 2 | 3+4 2 | 2+2 4 |
| | 3+4 2 | 5+6 2 | 2+2 4 | 2+2 4 | 2+2 4 | 5+6 2 | 5+6 2 | 2+2 4 |
| | 5+6 2 | 5+6 2 | 2+2 4 | 2+2 4 | 2+2 4 | 5+6 2 | 3+4 2 | 2+2 4 |

Continued on next column

**TABLE XXIII– continued**

| TQC | ICSD | | NoE | SF | | | | |
|---|---|---|---|---|---|---|---|---|
| | 5+6 2 | 3+4 2 | 2+2 4 | 2+2 4 | 2+2 4 | 3+4 2 | 5+6 2 | 2+2 4 |
| | 3+4 2 | 5+6 2 | 2+2 4 | 2+2 4 | 2+2 4 | 5+6 2 | 3+4 2 | 2+2 4 |
| | 3+4 2 | 3+4 2 | 2+2 4 | 2+2 4 | 2+2 4 | 3+4 2 | 5+6 2 | 2+2 4 |
| | 3+4 2 | 5+6 2 | 2+2 4 | 2+2 4 | 2+2 4 | 3+4 2 | 5+6 2 | 2+2 4 |
| | 5+6 2 | 3+4 2 | 2+2 4 | 2+2 4 | 2+2 4 | 5+6 2 | 3+4 2 | 2+2 4 |
| | 5+6 2 | 5+6 2 | 2+2 4 | 2+2 4 | 2+2 4 | 3+4 2 | 5+6 2 | 2+2 4 |
| | 5+6 2 | 5+6 2 | 2+2 4 | 2+2 4 | 2+2 4 | 5+6 2 | 5+6 2 | 2+2 4 |
| | 3+4 2 | 3+4 2 | 2+2 4 | 2+2 4 | 2+2 4 | 5+6 2 | 3+4 2 | 2+2 4 |
| | 3+4 2 | 5+6 2 | 2+2 4 | 2+2 4 | 2+2 4 | 3+4 2 | 5+6 2 | 2+2 4 |
| | 5+6 2 | 3+4 2 | 2+2 4 | 2+2 4 | 2+2 4 | 5+6 2 | 3+4 2 | 2+2 4 |
| | 5+6 2 | 5+6 2 | 2+2 4 | 2+2 4 | 2+2 4 | 3+4 2 | 5+6 2 | 2+2 4 |
| | 3+4 2 | 3+4 2 | 2+2 4 | 2+2 4 | 2+2 4 | 3+4 2 | 3+4 2 | 2+2 4 |
| | 3+4 2 | 5+6 2 | 2+2 4 | 2+2 4 | 2+2 4 | 5+6 2 | 3+4 2 | 2+2 4 |
| | 5+6 2 | 5+6 2 | 2+2 4 | 2+2 4 | 2+2 4 | 5+6 2 | 5+6 2 | 2+2 4 |
| | 3+4 2 | 3+4 2 | | | | 3+4 2 | 5+6 2 | |
| | 5+6 2 | 3+4 2 | | | | 5+6 2 | 3+4 2 | |
| | 3+4 2 | 3+4 2 | | | | 3+4 2 | 3+4 2 | |
| | 3+4 2 | 5+6 2 | | | | 5+6 2 | 5+6 2 | |
| | 5+6 2 | 5+6 2 | | | | 3+4 2 | 5+6 2 | |
| | 5+6 2 | 3+4 2 | | | | 3+4 2 | 3+4 2 | |
| | 3+4 2 | 5+6 2 | | | | 5+6 2 | 5+6 2 | |
| | 5+6 2 | 3+4 2 | | | | 3+4 2 | 5+6 2 | |
| | 3+4 2 | 5+6 2 | | | | 3+4 2 | 3+4 2 | |
| | 5+6 2 | 3+4 2 | | | | 5+6 2 | 3+4 2 | |
| | 3+4 2 | 3+4 2 | | | | 3+4 2 | 3+4 2 | |
| | 5+6 2 | 3+4 2 | | | | 5+6 2 | 5+6 2 | |
| | 3+4 2 | 5+6 2 | | | | 3+4 2 | 3+4 2 | |
| | 5+6 2 | 5+6 2 | | | | 5+6 2 | 3+4 2 | |
| | 3+4 2 | 3+4 2 | | | | 5+6 2 | 5+6 2 | |
| | 5+6 2 | 5+6 2 | | | | 3+4 2 | 3+4 2 | |
| | 3+4 2 | 3+4 2 | | | | 5+6 2 | 5+6 2 | |
| | 5+6 2 | 5+6 2 | | | | 3+4 2 | 3+4 2 | |
| | 3+4 2 | 5+6 2 | | | | 5+6 2 | 5+6 2 | |
| | 3+4 2 | 3+4 2 | | | | 3+4 2 | 3+4 2 | |
| | 5+6 2 | 5+6 2 | | | | 5+6 2 | 5+6 2 | |
| | 3+4 2 | 3+4 2 | | | | 3+4 2 | 3+4 2 | |
| | 5+6 2 | 5+6 2 | | | | 5+6 2 | 3+4 2 | |
| | 3+4 2 | 5+6 2 | | | | 3+4 2 | 5+6 2 | |
| | 5+6 2 | 5+6 2 | | | | 3+4 2 | 3+4 2 | |
| | 5+6 2 | 3+4 2 | | | | 5+6 2 | 5+6 2 | |
| | 3+4 2 | 3+4 2 | | | | 3+4 2 | 3+4 2 | |
| | 5+6 2 | 5+6 2 | | | | 5+6 2 | 5+6 2 | |
| | 3+4 2 | 3+4 2 | | | | 5+6 2 | 3+4 2 | |
| | 3+4 2 | 3+4 2 | | | | 3+4 2 | 5+6 2 | |
| | 5+6 2 | 3+4 2 | | | | 5+6 2 | 3+4 2 | |
| | 5+6 2 | 5+6 2 | | | | 3+4 2 | 5+6 2 | |
| | 3+4 2 | 5+6 2 | | | | 5+6 2 | 3+4 2 | |
| | 5+6 2 | 3+4 2 | | | | 3+4 2 | 5+6 2 | |
| | 3+4 2 | 5+6 2 | | | | 5+6 2 | 3+4 2 | |
| | 5+6 2 | 3+4 2 | | | | 5+6 2 | 5+6 2 | |
| | 3+4 2 | 3+4 2 | | | | 3+4 2 | 3+4 2 | |
| | 5+6 2 | 5+6 2 | | | | 3+4 2 | 3+4 2 | |
| | 5+6 2 | 3+4 2 | | | | 3+4 2 | 5+6 2 | |
| | 3+4 2 | 5+6 2 | | | | 3+4 2 | 5+6 2 | |
| 11.2.13 | 105804 | | 140 | Pt3Ta1 | | | | |
| | -A | -B | -C | -D | -E | -GM | -Y | -Z |
| | 3+4 2 | 5+6 2 | 2+2 4 | 2+2 4 | 2+2 4 | 3+4 2 | 3+4 2 | 2+2 4 |
| | 5+6 2 | 3+4 2 | 2+2 4 | 2+2 4 | 2+2 4 | 3+4 2 | 5+6 2 | 2+2 4 |
| | 3+4 2 | 3+4 2 | 2+2 4 | 2+2 4 | 2+2 4 | 5+6 2 | 5+6 2 | 2+2 4 |
| | 5+6 2 | 5+6 2 | 2+2 4 | 2+2 4 | 2+2 4 | 5+6 2 | 5+6 2 | 2+2 4 |
| | 3+4 2 | 3+4 2 | 2+2 4 | 2+2 4 | 2+2 4 | 5+6 2 | 3+4 2 | 2+2 4 |

Continued on next column

**TABLE XXIII– continued**

| TQC | ICSD | | NoE | SF | | | | |
|---|---|---|---|---|---|---|---|---|
| | 3+4 2 | 5+6 2 | 2+2 4 | 2+2 4 | 2+2 4 | 3+4 2 | 5+6 2 | 2+2 4 |
| | 5+6 2 | 3+4 2 | 2+2 4 | 2+2 4 | 2+2 4 | 3+4 2 | 3+4 2 | 2+2 4 |
| | 5+6 2 | 5+6 2 | 2+2 4 | 2+2 4 | 2+2 4 | 5+6 2 | 3+4 2 | 2+2 4 |
| | 5+6 2 | 5+6 2 | 2+2 4 | 2+2 4 | 2+2 4 | 3+4 2 | 5+6 2 | 2+2 4 |
| | 5+6 2 | 3+4 2 | 2+2 4 | 2+2 4 | 2+2 4 | 5+6 2 | 3+4 2 | 2+2 4 |
| | 3+4 2 | 5+6 2 | 2+2 4 | 2+2 4 | 2+2 4 | 3+4 2 | 5+6 2 | 2+2 4 |
| | 3+4 2 | 3+4 2 | 2+2 4 | 2+2 4 | 2+2 4 | 3+4 2 | 3+4 2 | 2+2 4 |
| | 5+6 2 | 5+6 2 | 2+2 4 | 2+2 4 | 2+2 4 | 5+6 2 | 3+4 2 | 2+2 4 |
| | 3+4 2 | 5+6 2 | 2+2 4 | 2+2 4 | 2+2 4 | 3+4 2 | 5+6 2 | 2+2 4 |
| | 5+6 2 | 3+4 2 | 2+2 4 | 2+2 4 | 2+2 4 | 5+6 2 | 3+4 2 | 2+2 4 |
| | 5+6 2 | 3+4 2 | 2+2 4 | 2+2 4 | 2+2 4 | 3+4 2 | 3+4 2 | 2+2 4 |
| | 3+4 2 | 5+6 2 | 2+2 4 | 2+2 4 | 2+2 4 | 5+6 2 | 5+6 2 | 2+2 4 |
| | 5+6 2 | 3+4 2 | 2+2 4 | 2+2 4 | 2+2 4 | 3+4 2 | 3+4 2 | 2+2 4 |
| | 3+4 2 | 5+6 2 | 2+2 4 | 2+2 4 | 2+2 4 | 5+6 2 | 5+6 2 | 2+2 4 |
| | 5+6 2 | 3+4 2 | 2+2 4 | 2+2 4 | 2+2 4 | 3+4 2 | 5+6 2 | 2+2 4 |
| | 5+6 2 | 5+6 2 | 2+2 4 | 2+2 4 | 2+2 4 | 5+6 2 | 5+6 2 | 2+2 4 |
| | 3+4 2 | 3+4 2 | 2+2 4 | 2+2 4 | 2+2 4 | 5+6 2 | 3+4 2 | 2+2 4 |
| | 5+6 2 | 5+6 2 | 2+2 4 | 2+2 4 | 2+2 4 | 3+4 2 | 3+4 2 | 2+2 4 |
| | 3+4 2 | 3+4 2 | 2+2 4 | 2+2 4 | 2+2 4 | 3+4 2 | 5+6 2 | 2+2 4 |
| | 3+4 2 | 3+4 2 | 2+2 4 | 2+2 4 | 2+2 4 | 5+6 2 | 5+6 2 | 2+2 4 |
| | 5+6 2 | 5+6 2 | 2+2 4 | 2+2 4 | 2+2 4 | 5+6 2 | 3+4 2 | 2+2 4 |
| | 5+6 2 | 3+4 2 | 2+2 4 | 2+2 4 | 2+2 4 | 3+4 2 | 5+6 2 | 2+2 4 |
| | 3+4 2 | 5+6 2 | 2+2 4 | 2+2 4 | 2+2 4 | 5+6 2 | 3+4 2 | 2+2 4 |
| | 3+4 2 | 3+4 2 | 2+2 4 | 2+2 4 | 2+2 4 | 3+4 2 | 3+4 2 | 2+2 4 |
| | 3+4 2 | 5+6 2 | 2+2 4 | 2+2 4 | 2+2 4 | 5+6 2 | 3+4 2 | 2+2 4 |
| | 5+6 2 | 5+6 2 | 2+2 4 | 2+2 4 | 2+2 4 | 3+4 2 | 5+6 2 | 2+2 4 |
| | 5+6 2 | 3+4 2 | 2+2 4 | 2+2 4 | 2+2 4 | 5+6 2 | 5+6 2 | 2+2 4 |
| | 3+4 2 | 5+6 2 | 2+2 4 | 2+2 4 | 2+2 4 | 5+6 2 | 3+4 2 | 2+2 4 |
| | 3+4 2 | 3+4 2 | 2+2 4 | 2+2 4 | 2+2 4 | 3+4 2 | 5+6 2 | 2+2 4 |
| | 5+6 2 | 5+6 2 | 2+2 4 | 2+2 4 | 2+2 4 | 3+4 2 | 3+4 2 | 2+2 4 |
| | 3+4 2 | 3+4 2 | | | | 3+4 2 | 5+6 2 | |
| | 3+4 2 | 3+4 2 | | | | 5+6 2 | 5+6 2 | |
| | 3+4 2 | 5+6 2 | | | | 5+6 2 | 3+4 2 | |
| | 5+6 2 | 3+4 2 | | | | 3+4 2 | 5+6 2 | |
| | 3+4 2 | 5+6 2 | | | | 3+4 2 | 3+4 2 | |
| | 5+6 2 | 3+4 2 | | | | 5+6 2 | 5+6 2 | |
| | 5+6 2 | 3+4 2 | | | | 5+6 2 | 3+4 2 | |
| | 3+4 2 | 5+6 2 | | | | 5+6 2 | 5+6 2 | |
| | 5+6 2 | 5+6 2 | | | | 3+4 2 | 3+4 2 | |
| | 3+4 2 | 3+4 2 | | | | 3+4 2 | 5+6 2 | |
| | 3+4 2 | 5+6 2 | | | | 5+6 2 | 3+4 2 | |
| | 3+4 2 | 5+6 2 | | | | 3+4 2 | 5+6 2 | |
| | 5+6 2 | 3+4 2 | | | | 3+4 2 | 3+4 2 | |
| | 5+6 2 | 5+6 2 | | | | 5+6 2 | 5+6 2 | |
| | 3+4 2 | 3+4 2 | | | | 5+6 2 | 5+6 2 | |
| | 5+6 2 | 5+6 2 | | | | 3+4 2 | 3+4 2 | |
| | 3+4 2 | 3+4 2 | | | | 3+4 2 | 5+6 2 | |
| | 5+6 2 | 5+6 2 | | | | 5+6 2 | 3+4 2 | |
| | 5+6 2 | 3+4 2 | | | | 5+6 2 | 3+4 2 | |
| | 3+4 2 | 5+6 2 | | | | 3+4 2 | 3+4 2 | |
| | 3+4 2 | 3+4 2 | | | | 5+6 2 | 5+6 2 | |
| | 5+6 2 | 3+4 2 | | | | 3+4 2 | 5+6 2 | |
| | 3+4 2 | 5+6 2 | | | | 5+6 2 | 3+4 2 | |
| | 5+6 2 | 3+4 2 | | | | 3+4 2 | 5+6 2 | |
| | 5+6 2 | 5+6 2 | | | | 3+4 2 | 3+4 2 | |
| | 3+4 2 | 3+4 2 | | | | 3+4 2 | 5+6 2 | |
| | 3+4 2 | 5+6 2 | | | | 5+6 2 | 3+4 2 | |
| | 3+4 2 | 3+4 2 | | | | 3+4 2 | 5+6 2 | |
| | 5+6 2 | 5+6 2 | | | | 5+6 2 | 3+4 2 | |
| | 5+6 2 | 3+4 2 | | | | 3+4 2 | 3+4 2 | |
| | 5+6 2 | 5+6 2 | | | | 5+6 2 | 5+6 2 | |

Continued on next column

**TABLE XXIII– continued**

| TQC | ICSD | NoE | SF | | | | | |
|---|---|---|---|---|---|---|---|---|
| | 3+4 2 | 5+6 2 | | | | 3+4 2 | 5+6 2 | |
| | 3+4 2 | 3+4 2 | | | | 5+6 2 | 3+4 2 | |
| | 5+6 2 | 5+6 2 | | | | 5+6 2 | 3+4 2 | |
| | 5+6 2 | 3+4 2 | | | | 3+4 2 | 3+4 2 | |
| 11.2.14 | 99271 | 84 | Ni3Te2 | | | | | |
| | -A | -B | -C | -D | -E | -GM | -Y | -Z |
| | 5+6 2 | 3+4 2 | 2+2 4 | 2+2 4 | 2+2 4 | 3+4 2 | 5+6 2 | 2+2 4 |
| | 3+4 2 | 5+6 2 | 2+2 4 | 2+2 4 | 2+2 4 | 3+4 2 | 3+4 2 | 2+2 4 |
| | 3+4 2 | 5+6 2 | 2+2 4 | 2+2 4 | 2+2 4 | 5+6 2 | 5+6 2 | 2+2 4 |
| | 5+6 2 | 3+4 2 | 2+2 4 | 2+2 4 | 2+2 4 | 5+6 2 | 3+4 2 | 2+2 4 |
| | 3+4 2 | 3+4 2 | 2+2 4 | 2+2 4 | 2+2 4 | 3+4 2 | 3+4 2 | 2+2 4 |
| | 5+6 2 | 5+6 2 | 2+2 4 | 2+2 4 | 2+2 4 | 5+6 2 | 5+6 2 | 2+2 4 |
| | 5+6 2 | 5+6 2 | 2+2 4 | 2+2 4 | 2+2 4 | 3+4 2 | 3+4 2 | 2+2 4 |
| | 3+4 2 | 3+4 2 | 2+2 4 | 2+2 4 | 2+2 4 | 3+4 2 | 5+6 2 | 2+2 4 |
| | 3+4 2 | 3+4 2 | 2+2 4 | 2+2 4 | 2+2 4 | 5+6 2 | 5+6 2 | 2+2 4 |
| | 5+6 2 | 5+6 2 | 2+2 4 | 2+2 4 | 2+2 4 | 5+6 2 | 5+6 2 | 2+2 4 |
| | 3+4 2 | 3+4 2 | 2+2 4 | 2+2 4 | 2+2 4 | 3+4 2 | 5+6 2 | 2+2 4 |
| | 5+6 2 | 5+6 2 | 2+2 4 | 2+2 4 | 2+2 4 | 5+6 2 | 3+4 2 | 2+2 4 |
| | 5+6 2 | 5+6 2 | 2+2 4 | 2+2 4 | 2+2 4 | 5+6 2 | 5+6 2 | 2+2 4 |
| | 3+4 2 | 3+4 2 | 2+2 4 | 2+2 4 | 2+2 4 | 5+6 2 | 3+4 2 | 2+2 4 |
| | 5+6 2 | 5+6 2 | 2+2 4 | 2+2 4 | 2+2 4 | 3+4 2 | 3+4 2 | 2+2 4 |
| | 3+4 2 | 3+4 2 | 2+2 4 | 2+2 4 | 2+2 4 | 5+6 2 | 3+4 2 | 2+2 4 |
| | 5+6 2 | 5+6 2 | 2+2 4 | 2+2 4 | 2+2 4 | 3+4 2 | 5+6 2 | 2+2 4 |
| | 5+6 2 | 3+4 2 | 2+2 4 | 2+2 4 | 2+2 4 | 3+4 2 | 3+4 2 | 2+2 4 |
| | 3+4 2 | 5+6 2 | 2+2 4 | 2+2 4 | 2+2 4 | 3+4 2 | 5+6 2 | 2+2 4 |
| | 3+4 2 | 3+4 2 | 2+2 4 | 2+2 4 | 2+2 4 | 5+6 2 | 3+4 2 | 2+2 4 |
| | 5+6 2 | 3+4 2 | 2+2 4 | 2+2 4 | 2+2 4 | 3+4 2 | 5+6 2 | 2+2 4 |
| | 3+4 2 | 5+6 2 | | | | 3+4 2 | 3+4 2 | |
| | 5+6 2 | 3+4 2 | | | | 5+6 2 | 3+4 2 | |
| | 3+4 2 | 5+6 2 | | | | 3+4 2 | 5+6 2 | |
| | 3+4 2 | 3+4 2 | | | | 5+6 2 | 3+4 2 | |
| | 5+6 2 | 3+4 2 | | | | 3+4 2 | 5+6 2 | |
| | 3+4 2 | 3+4 2 | | | | 5+6 2 | 5+6 2 | |
| | 3+4 2 | 5+6 2 | | | | 3+4 2 | 3+4 2 | |
| | 5+6 2 | 3+4 2 | | | | 5+6 2 | 3+4 2 | |
| | 3+4 2 | 5+6 2 | | | | 3+4 2 | 5+6 2 | |
| | 5+6 2 | 5+6 2 | | | | 3+4 2 | 3+4 2 | |
| | 5+6 2 | 3+4 2 | | | | 5+6 2 | 5+6 2 | |
| | 5+6 2 | 5+6 2 | | | | 3+4 2 | 3+4 2 | |
| | 3+4 2 | 3+4 2 | | | | 5+6 2 | 5+6 2 | |
| | 5+6 2 | 5+6 2 | | | | 3+4 2 | 5+6 2 | |
| | 3+4 2 | 5+6 2 | | | | 5+6 2 | 3+4 2 | |
| | 5+6 2 | 3+4 2 | | | | 3+4 2 | 3+4 2 | |
| | 3+4 2 | 5+6 2 | | | | 3+4 2 | 3+4 2 | |
| | 5+6 2 | 3+4 2 | | | | 5+6 2 | 5+6 2 | |
| | 3+4 2 | 5+6 2 | | | | 3+4 2 | 3+4 2 | |
| | 3+4 2 | 3+4 2 | | | | 5+6 2 | 5+6 2 | |
| | 5+6 2 | 5+6 2 | | | | 5+6 2 | 5+6 2 | |
| 11.2.15 | 54304 | 140 | Ni3Ta1 | | | | | |
| | -A | -B | -C | -D | -E | -GM | -Y | -Z |
| | 3+4 2 | 5+6 2 | 2+2 4 | 2+2 4 | 2+2 4 | 3+4 2 | 5+6 2 | 2+2 4 |
| | 5+6 2 | 3+4 2 | 2+2 4 | 2+2 4 | 2+2 4 | 3+4 2 | 3+4 2 | 2+2 4 |
| | 3+4 2 | 3+4 2 | 2+2 4 | 2+2 4 | 2+2 4 | 5+6 2 | 3+4 2 | 2+2 4 |
| | 5+6 2 | 5+6 2 | 2+2 4 | 2+2 4 | 2+2 4 | 5+6 2 | 5+6 2 | 2+2 4 |
| | 3+4 2 | 5+6 2 | 2+2 4 | 2+2 4 | 2+2 4 | 5+6 2 | 5+6 2 | 2+2 4 |
| | 5+6 2 | 3+4 2 | 2+2 4 | 2+2 4 | 2+2 4 | 3+4 2 | 3+4 2 | 2+2 4 |
| | 5+6 2 | 3+4 2 | 2+2 4 | 2+2 4 | 2+2 4 | 5+6 2 | 5+6 2 | 2+2 4 |
| | 3+4 2 | 5+6 2 | 2+2 4 | 2+2 4 | 2+2 4 | 3+4 2 | 3+4 2 | 2+2 4 |
| | 5+6 2 | 3+4 2 | 2+2 4 | 2+2 4 | 2+2 4 | 3+4 2 | 3+4 2 | 2+2 4 |
| | 5+6 2 | 5+6 2 | 2+2 4 | 2+2 4 | 2+2 4 | 5+6 2 | 5+6 2 | 2+2 4 |
| | 3+4 2 | 5+6 2 | 2+2 4 | 2+2 4 | 2+2 4 | 3+4 2 | 3+4 2 | 2+2 4 |
| | 3+4 2 | 5+6 2 | 2+2 4 | 2+2 4 | 2+2 4 | 5+6 2 | 5+6 2 | 2+2 4 |
| | 3+4 2 | 3+4 2 | 2+2 4 | 2+2 4 | 2+2 4 | 3+4 2 | 3+4 2 | 2+2 4 |
| | 5+6 2 | 3+4 2 | 2+2 4 | 2+2 4 | 2+2 4 | 3+4 2 | 5+6 2 | 2+2 4 |
| | 3+4 2 | 3+4 2 | 2+2 4 | 2+2 4 | 2+2 4 | 5+6 2 | 5+6 2 | 2+2 4 |
| | 5+6 2 | 5+6 2 | 2+2 4 | 2+2 4 | 2+2 4 | 3+4 2 | 5+6 2 | 2+2 4 |
| | 3+4 2 | 3+4 2 | 2+2 4 | 2+2 4 | 2+2 4 | 5+6 2 | 3+4 2 | 2+2 4 |
| | 3+4 2 | 5+6 2 | 2+2 4 | 2+2 4 | 2+2 4 | 3+4 2 | 3+4 2 | 2+2 4 |
| | 3+4 2 | 5+6 2 | 2+2 4 | 2+2 4 | 2+2 4 | 3+4 2 | 3+4 2 | 2+2 4 |
| | 3+4 2 | 3+4 2 | 2+2 4 | 2+2 4 | 2+2 4 | 5+6 2 | 5+6 2 | 2+2 4 |
| | 3+4 2 | 5+6 2 | 2+2 4 | 2+2 4 | 2+2 4 | 3+4 2 | 3+4 2 | 2+2 4 |
| | 5+6 2 | 3+4 2 | 2+2 4 | 2+2 4 | 2+2 4 | 5+6 2 | 3+4 2 | 2+2 4 |
| | 5+6 2 | 3+4 2 | 2+2 4 | 2+2 4 | 2+2 4 | 5+6 2 | 5+6 2 | 2+2 4 |
| | 5+6 2 | 5+6 2 | 2+2 4 | 2+2 4 | 2+2 4 | 5+6 2 | 5+6 2 | 2+2 4 |
| | 5+6 2 | 5+6 2 | 2+2 4 | 2+2 4 | 2+2 4 | 3+4 2 | 3+4 2 | 2+2 4 |
| | 3+4 2 | 3+4 2 | 2+2 4 | 2+2 4 | 2+2 4 | 3+4 2 | 5+6 2 | 2+2 4 |
| | 5+6 2 | 5+6 2 | 2+2 4 | 2+2 4 | 2+2 4 | 5+6 2 | 5+6 2 | 2+2 4 |
| | 3+4 2 | 3+4 2 | 2+2 4 | 2+2 4 | 2+2 4 | 5+6 2 | 5+6 2 | 2+2 4 |
| | 3+4 2 | 5+6 2 | 2+2 4 | 2+2 4 | 2+2 4 | 3+4 2 | 3+4 2 | 2+2 4 |
| | 5+6 2 | 3+4 2 | 2+2 4 | 2+2 4 | 2+2 4 | 5+6 2 | 3+4 2 | 2+2 4 |
| | 5+6 2 | 3+4 2 | 2+2 4 | 2+2 4 | 2+2 4 | 3+4 2 | 3+4 2 | 2+2 4 |
| | 3+4 2 | 5+6 2 | 2+2 4 | 2+2 4 | 2+2 4 | 5+6 2 | 5+6 2 | 2+2 4 |
| | 5+6 2 | 3+4 2 | 2+2 4 | 2+2 4 | 2+2 4 | 3+4 2 | 3+4 2 | 2+2 4 |
| | 5+6 2 | 5+6 2 | 2+2 4 | 2+2 4 | 2+2 4 | 5+6 2 | 5+6 2 | 2+2 4 |
| | 3+4 2 | 3+4 2 | 2+2 4 | 2+2 4 | 2+2 4 | 3+4 2 | 3+4 2 | 2+2 4 |
| | 5+6 2 | 5+6 2 | | | | 5+6 2 | 5+6 2 | |
| | 5+6 2 | 5+6 2 | | | | 3+4 2 | 5+6 2 | |
| | 3+4 2 | 3+4 2 | | | | 5+6 2 | 3+4 2 | |
| | 5+6 2 | 5+6 2 | | | | 3+4 2 | 5+6 2 | |
| | 5+6 2 | 3+4 2 | | | | 3+4 2 | 3+4 2 | |
| | 3+4 2 | 5+6 2 | | | | 5+6 2 | 5+6 2 | |
| | 5+6 2 | 3+4 2 | | | | 5+6 2 | 3+4 2 | |
| | 3+4 2 | 5+6 2 | | | | 5+6 2 | 5+6 2 | |
| | 5+6 2 | 3+4 2 | | | | 3+4 2 | 3+4 2 | |
| | 3+4 2 | 5+6 2 | | | | 5+6 2 | 5+6 2 | |
| | 5+6 2 | 3+4 2 | | | | 3+4 2 | 5+6 2 | |
| | 3+4 2 | 5+6 2 | | | | 3+4 2 | 3+4 2 | |
| | 5+6 2 | 3+4 2 | | | | 5+6 2 | 5+6 2 | |
| | 5+6 2 | 3+4 2 | | | | 3+4 2 | 3+4 2 | |
| | 3+4 2 | 5+6 2 | | | | 5+6 2 | 5+6 2 | |
| | 3+4 2 | 3+4 2 | | | | 3+4 2 | 3+4 2 | |
| | 5+6 2 | 3+4 2 | | | | 3+4 2 | 3+4 2 | |
| | 5+6 2 | 5+6 2 | | | | 5+6 2 | 5+6 2 | |
| | 3+4 2 | 5+6 2 | | | | 5+6 2 | 3+4 2 | |
| | 3+4 2 | 5+6 2 | | | | 5+6 2 | 3+4 2 | |
| | 5+6 2 | 3+4 2 | | | | 3+4 2 | 5+6 2 | |
| | 3+4 2 | 5+6 2 | | | | 3+4 2 | 5+6 2 | |
| | 5+6 2 | 5+6 2 | | | | 5+6 2 | 3+4 2 | |
| | 3+4 2 | 3+4 2 | | | | 3+4 2 | 3+4 2 | |
| | 3+4 2 | 3+4 2 | | | | 3+4 2 | 5+6 2 | |
| | 5+6 2 | 5+6 2 | | | | 3+4 2 | 3+4 2 | |
| | 5+6 2 | 3+4 2 | | | | 5+6 2 | 5+6 2 | |
| | 3+4 2 | 5+6 2 | | | | 3+4 2 | 3+4 2 | |
| | 5+6 2 | 3+4 2 | | | | 5+6 2 | 5+6 2 | |
| | 3+4 2 | 3+4 2 | | | | 3+4 2 | 5+6 2 | |
| | 5+6 2 | 5+6 2 | | | | 5+6 2 | 5+6 2 | |
| | 3+4 2 | 5+6 2 | | | | 5+6 2 | 3+4 2 | |
| | 3+4 2 | 3+4 2 | | | | 3+4 2 | 3+4 2 | |
| | 5+6 2 | 3+4 2 | | | | 5+6 2 | 3+4 2 | |
| | 3+4 2 | 3+4 2 | | | | 3+4 2 | 5+6 2 | |
| 11.2.16 | 245081 | 28 | Ni1Ti1 | | | | | |
| | -A | -B | -C | -D | -E | -GM | -Y | -Z |

Continued on next column

**TABLE XXIII– continued**

| TQC | ICSD | NoE | SF | | | | |
|---|---|---|---|---|---|---|---|
| | 5+6 2 | 5+6 2 | 2+2 4 | 2+2 4 | 2+2 4 | 3+4 2 | 3+4 2 | 2+2 4 |
| | 3+4 2 | 3+4 2 | 2+2 4 | 2+2 4 | 2+2 4 | 5+6 2 | 5+6 2 | 2+2 4 |
| | 3+4 2 | 5+6 2 | 2+2 4 | 2+2 4 | 2+2 4 | 5+6 2 | 5+6 2 | 2+2 4 |
| | 3+4 2 | 3+4 2 | 2+2 4 | 2+2 4 | 2+2 4 | 3+4 2 | 5+6 2 | 2+2 4 |
| | 5+6 2 | 5+6 2 | 2+2 4 | 2+2 4 | 2+2 4 | 3+4 2 | 5+6 2 | 2+2 4 |
| | 3+4 2 | 3+4 2 | 2+2 4 | 2+2 4 | 2+2 4 | 5+6 2 | 3+4 2 | 2+2 4 |
| | 5+6 2 | 3+4 2 | 2+2 4 | 2+2 4 | 2+2 4 | 5+6 2 | 3+4 2 | 2+2 4 |
| | 3+4 2 | 3+4 2 | | | | 3+4 2 | 5+6 2 | |
| | 5+6 2 | 3+4 2 | | | | 3+4 2 | 3+4 2 | |
| | 5+6 2 | 5+6 2 | | | | 3+4 2 | 3+4 2 | |
| | 3+4 2 | 5+6 2 | | | | 5+6 2 | 3+4 2 | |
| | 5+6 2 | 5+6 2 | | | | 5+6 2 | 3+4 2 | |
| | 5+6 2 | 5+6 2 | | | | 3+4 2 | 5+6 2 | |
| | 5+6 2 | 5+6 2 | | | | 5+6 2 | 3+4 2 | |
| 11.2.17 | 240195 | 28 | Ni1Ti1 | | | | | |
| | -A | -B | -C | -D | -E | -GM | -Y | -Z |
| | 5+6 2 | 5+6 2 | 2+2 4 | 2+2 4 | 2+2 4 | 3+4 2 | 3+4 2 | 2+2 4 |
| | 3+4 2 | 3+4 2 | 2+2 4 | 2+2 4 | 2+2 4 | 5+6 2 | 5+6 2 | 2+2 4 |
| | 5+6 2 | 3+4 2 | 2+2 4 | 2+2 4 | 2+2 4 | 3+4 2 | 3+4 2 | 2+2 4 |
| | 3+4 2 | 5+6 2 | 2+2 4 | 2+2 4 | 2+2 4 | 5+6 2 | 5+6 2 | 2+2 4 |
| | 5+6 2 | 5+6 2 | 2+2 4 | 2+2 4 | 2+2 4 | 3+4 2 | 3+4 2 | 2+2 4 |
| | 5+6 2 | 3+4 2 | 2+2 4 | 2+2 4 | 2+2 4 | 5+6 2 | 5+6 2 | 2+2 4 |
| | 5+6 2 | 5+6 2 | 2+2 4 | 2+2 4 | 2+2 4 | 3+4 2 | 5+6 2 | 2+2 4 |
| | 3+4 2 | 5+6 2 | | | | 5+6 2 | 3+4 2 | |
| | 3+4 2 | 3+4 2 | | | | 3+4 2 | 3+4 2 | |
| | 3+4 2 | 3+4 2 | | | | 3+4 2 | 5+6 2 | |
| | 5+6 2 | 3+4 2 | | | | 5+6 2 | 3+4 2 | |
| | 3+4 2 | 5+6 2 | | | | 5+6 2 | 3+4 2 | |
| | 3+4 2 | 3+4 2 | | | | 3+4 2 | 5+6 2 | |
| | 3+4 2 | 3+4 2 | | | | 5+6 2 | 3+4 2 | |
| 11.2.18 | 166372 | 28 | Ni1Ti1 | | | | | |
| | -A | -B | -C | -D | -E | -GM | -Y | -Z |
| | 3+4 2 | 3+4 2 | 2+2 4 | 2+2 4 | 2+2 4 | 3+4 2 | 3+4 2 | 2+2 4 |
| | 5+6 2 | 5+6 2 | 2+2 4 | 2+2 4 | 2+2 4 | 5+6 2 | 5+6 2 | 2+2 4 |
| | 5+6 2 | 3+4 2 | 2+2 4 | 2+2 4 | 2+2 4 | 3+4 2 | 5+6 2 | 2+2 4 |
| | 5+6 2 | 3+4 2 | 2+2 4 | 2+2 4 | 2+2 4 | 5+6 2 | 5+6 2 | 2+2 4 |
| | 3+4 2 | 5+6 2 | 2+2 4 | 2+2 4 | 2+2 4 | 3+4 2 | 5+6 2 | 2+2 4 |
| | 5+6 2 | 5+6 2 | 2+2 4 | 2+2 4 | 2+2 4 | 5+6 2 | 3+4 2 | 2+2 4 |
| | 3+4 2 | 5+6 2 | 2+2 4 | 2+2 4 | 2+2 4 | 3+4 2 | 3+4 2 | 2+2 4 |
| | 3+4 2 | 5+6 2 | | | | 5+6 2 | 3+4 2 | |
| | 3+4 2 | 3+4 2 | | | | 3+4 2 | 3+4 2 | |
| | 5+6 2 | 5+6 2 | | | | 5+6 2 | 5+6 2 | |
| | 5+6 2 | 3+4 2 | | | | 3+4 2 | 3+4 2 | |
| | 3+4 2 | 3+4 2 | | | | 5+6 2 | 3+4 2 | |
| | 3+4 2 | 3+4 2 | | | | 3+4 2 | 5+6 2 | |
| | 3+4 2 | 3+4 2 | | | | 5+6 2 | 3+4 2 | |
| 11.2.19 | 166367 | 28 | Ni1Ti1 | | | | | |
| | -A | -B | -C | -D | -E | -GM | -Y | -Z |
| | 3+4 2 | 3+4 2 | 2+2 4 | 2+2 4 | 2+2 4 | 3+4 2 | 3+4 2 | 2+2 4 |
| | 5+6 2 | 5+6 2 | 2+2 4 | 2+2 4 | 2+2 4 | 5+6 2 | 5+6 2 | 2+2 4 |
| | 5+6 2 | 3+4 2 | 2+2 4 | 2+2 4 | 2+2 4 | 3+4 2 | 5+6 2 | 2+2 4 |
| | 5+6 2 | 3+4 2 | 2+2 4 | 2+2 4 | 2+2 4 | 5+6 2 | 5+6 2 | 2+2 4 |
| | 3+4 2 | 5+6 2 | 2+2 4 | 2+2 4 | 2+2 4 | 3+4 2 | 5+6 2 | 2+2 4 |
| | 5+6 2 | 5+6 2 | 2+2 4 | 2+2 4 | 2+2 4 | 3+4 2 | 3+4 2 | 2+2 4 |
| | 3+4 2 | 5+6 2 | 2+2 4 | 2+2 4 | 2+2 4 | 5+6 2 | 3+4 2 | 2+2 4 |
| | 3+4 2 | 5+6 2 | | | | 5+6 2 | 3+4 2 | |
| | 3+4 2 | 3+4 2 | | | | 3+4 2 | 5+6 2 | |
| | 5+6 2 | 5+6 2 | | | | 5+6 2 | 3+4 2 | |
| | 5+6 2 | 3+4 2 | | | | 3+4 2 | 3+4 2 | |
| | 3+4 2 | 3+4 2 | | | | 5+6 2 | 3+4 2 | |
| | 3+4 2 | 3+4 2 | | | | 3+4 2 | 5+6 2 | |

Continued on next column

**TABLE XXIII– continued**

| TQC | ICSD | NoE | SF | | | | | |
|---|---|---|---|---|---|---|---|---|
| | 3+4 2 | 3+4 2 | | | | 5+6 2 | 3+4 2 | |
| 11.2.20 | 164155 | 28 | Ni1Ti1 | | | | | |
| | -A | -B | -C | -D | -E | -GM | -Y | -Z |
| | 5+6 2 | 5+6 2 | 2+2 4 | 2+2 4 | 2+2 4 | 3+4 2 | 3+4 2 | 2+2 4 |
| | 3+4 2 | 3+4 2 | 2+2 4 | 2+2 4 | 2+2 4 | 5+6 2 | 5+6 2 | 2+2 4 |
| | 3+4 2 | 5+6 2 | 2+2 4 | 2+2 4 | 2+2 4 | 5+6 2 | 5+6 2 | 2+2 4 |
| | 3+4 2 | 3+4 2 | 2+2 4 | 2+2 4 | 2+2 4 | 3+4 2 | 5+6 2 | 2+2 4 |
| | 5+6 2 | 5+6 2 | 2+2 4 | 2+2 4 | 2+2 4 | 3+4 2 | 5+6 2 | 2+2 4 |
| | 3+4 2 | 3+4 2 | 2+2 4 | 2+2 4 | 2+2 4 | 5+6 2 | 3+4 2 | 2+2 4 |
| | 5+6 2 | 3+4 2 | 2+2 4 | 2+2 4 | 2+2 4 | 5+6 2 | 3+4 2 | 2+2 4 |
| | 3+4 2 | 3+4 2 | | | | 3+4 2 | 3+4 2 | |
| | 5+6 2 | 3+4 2 | | | | 3+4 2 | 5+6 2 | |
| | 5+6 2 | 5+6 2 | | | | 3+4 2 | 3+4 2 | |
| | 3+4 2 | 5+6 2 | | | | 5+6 2 | 3+4 2 | |
| | 5+6 2 | 5+6 2 | | | | 5+6 2 | 3+4 2 | |
| | 5+6 2 | 5+6 2 | | | | 3+4 2 | 5+6 2 | |
| | 5+6 2 | 5+6 2 | | | | 5+6 2 | 3+4 2 | |
| 11.2.21 | 164154 | 28 | Ni1Ti1 | | | | | |
| | -A | -B | -C | -D | -E | -GM | -Y | -Z |
| | 3+4 2 | 3+4 2 | 2+2 4 | 2+2 4 | 2+2 4 | 3+4 2 | 3+4 2 | 2+2 4 |
| | 5+6 2 | 5+6 2 | 2+2 4 | 2+2 4 | 2+2 4 | 5+6 2 | 5+6 2 | 2+2 4 |
| | 5+6 2 | 3+4 2 | 2+2 4 | 2+2 4 | 2+2 4 | 5+6 2 | 5+6 2 | 2+2 4 |
| | 3+4 2 | 5+6 2 | 2+2 4 | 2+2 4 | 2+2 4 | 3+4 2 | 5+6 2 | 2+2 4 |
| | 5+6 2 | 3+4 2 | 2+2 4 | 2+2 4 | 2+2 4 | 3+4 2 | 5+6 2 | 2+2 4 |
| | 5+6 2 | 5+6 2 | 2+2 4 | 2+2 4 | 2+2 4 | 5+6 2 | 3+4 2 | 2+2 4 |
| | 3+4 2 | 5+6 2 | 2+2 4 | 2+2 4 | 2+2 4 | 5+6 2 | 3+4 2 | 2+2 4 |
| | 5+6 2 | 5+6 2 | | | | 3+4 2 | 3+4 2 | |
| | 3+4 2 | 5+6 2 | | | | 3+4 2 | 5+6 2 | |
| | 3+4 2 | 3+4 2 | | | | 3+4 2 | 3+4 2 | |
| | 5+6 2 | 3+4 2 | | | | 5+6 2 | 3+4 2 | |
| | 3+4 2 | 3+4 2 | | | | 5+6 2 | 3+4 2 | |
| | 3+4 2 | 3+4 2 | | | | 3+4 2 | 5+6 2 | |
| | 3+4 2 | 3+4 2 | | | | 5+6 2 | 3+4 2 | |
| 11.2.22 | 164153 | 28 | Ni1Ti1 | | | | | |
| | -A | -B | -C | -D | -E | -GM | -Y | -Z |
| | 3+4 2 | 3+4 2 | 2+2 4 | 2+2 4 | 2+2 4 | 3+4 2 | 3+4 2 | 2+2 4 |
| | 5+6 2 | 5+6 2 | 2+2 4 | 2+2 4 | 2+2 4 | 5+6 2 | 5+6 2 | 2+2 4 |
| | 5+6 2 | 3+4 2 | 2+2 4 | 2+2 4 | 2+2 4 | 5+6 2 | 5+6 2 | 2+2 4 |
| | 5+6 2 | 5+6 2 | 2+2 4 | 2+2 4 | 2+2 4 | 3+4 2 | 5+6 2 | 2+2 4 |
| | 3+4 2 | 3+4 2 | 2+2 4 | 2+2 4 | 2+2 4 | 3+4 2 | 5+6 2 | 2+2 4 |
| | 5+6 2 | 5+6 2 | 2+2 4 | 2+2 4 | 2+2 4 | 5+6 2 | 3+4 2 | 2+2 4 |
| | 3+4 2 | 5+6 2 | 2+2 4 | 2+2 4 | 2+2 4 | 3+4 2 | 3+4 2 | 2+2 4 |
| | 5+6 2 | 5+6 2 | | | | 5+6 2 | 5+6 2 | |
| | 3+4 2 | 5+6 2 | | | | 3+4 2 | 3+4 2 | |
| | 3+4 2 | 3+4 2 | | | | 3+4 2 | 3+4 2 | |
| | 5+6 2 | 3+4 2 | | | | 5+6 2 | 3+4 2 | |
| | 3+4 2 | 3+4 2 | | | | 5+6 2 | 3+4 2 | |
| | 3+4 2 | 3+4 2 | | | | 3+4 2 | 5+6 2 | |
| | 3+4 2 | 3+4 2 | | | | 5+6 2 | 3+4 2 | |
| 11.2.23 | 164152 | 28 | Ni1Ti1 | | | | | |
| | -A | -B | -C | -D | -E | -GM | -Y | -Z |
| | 3+4 2 | 3+4 2 | 2+2 4 | 2+2 4 | 2+2 4 | 3+4 2 | 3+4 2 | 2+2 4 |
| | 5+6 2 | 5+6 2 | 2+2 4 | 2+2 4 | 2+2 4 | 5+6 2 | 5+6 2 | 2+2 4 |
| | 5+6 2 | 3+4 2 | 2+2 4 | 2+2 4 | 2+2 4 | 5+6 2 | 5+6 2 | 2+2 4 |
| | 5+6 2 | 5+6 2 | 2+2 4 | 2+2 4 | 2+2 4 | 3+4 2 | 5+6 2 | 2+2 4 |
| | 3+4 2 | 3+4 2 | 2+2 4 | 2+2 4 | 2+2 4 | 3+4 2 | 5+6 2 | 2+2 4 |
| | 5+6 2 | 5+6 2 | 2+2 4 | 2+2 4 | 2+2 4 | 5+6 2 | 3+4 2 | 2+2 4 |
| | 3+4 2 | 5+6 2 | 2+2 4 | 2+2 4 | 2+2 4 | 3+4 2 | 3+4 2 | 2+2 4 |
| | 5+6 2 | 5+6 2 | | | | 5+6 2 | 5+6 2 | |
| | 3+4 2 | 5+6 2 | | | | 3+4 2 | 3+4 2 | |
| | 3+4 2 | 3+4 2 | | | | 3+4 2 | 3+4 2 | |

Continued on next column

**TABLE XXIII– continued**

| TQC | ICSD | NoE | SF | | | | | |
|---|---|---|---|---|---|---|---|---|
| | 5+6 2 | 3+4 2 | | | | 5+6 2 | 3+4 2 | |
| | 3+4 2 | 3+4 2 | | | | 5+6 2 | 3+4 2 | |
| | 3+4 2 | 3+4 2 | | | | 3+4 2 | 5+6 2 | |
| | 3+4 2 | 3+4 2 | | | | 5+6 2 | 3+4 2 | |
| 11.2.24 | 164151 | 28 | Ni1Ti1 | | | | | |
| | -A | -B | -C | -D | -E | -GM | -Y | -Z |
| | 5+6 2 | 3+4 2 | 2+2 4 | 2+2 4 | 2+2 4 | 3+4 2 | 5+6 2 | 2+2 4 |
| | 3+4 2 | 5+6 2 | 2+2 4 | 2+2 4 | 2+2 4 | 5+6 2 | 3+4 2 | 2+2 4 |
| | 3+4 2 | 3+4 2 | 2+2 4 | 2+2 4 | 2+2 4 | 5+6 2 | 3+4 2 | 2+2 4 |
| | 3+4 2 | 5+6 2 | 2+2 4 | 2+2 4 | 2+2 4 | 3+4 2 | 3+4 2 | 2+2 4 |
| | 5+6 2 | 3+4 2 | 2+2 4 | 2+2 4 | 2+2 4 | 3+4 2 | 3+4 2 | 2+2 4 |
| | 3+4 2 | 5+6 2 | 2+2 4 | 2+2 4 | 2+2 4 | 5+6 2 | 5+6 2 | 2+2 4 |
| | 5+6 2 | 5+6 2 | 2+2 4 | 2+2 4 | 2+2 4 | 3+4 2 | 5+6 2 | 2+2 4 |
| | 3+4 2 | 5+6 2 | | | | 5+6 2 | 3+4 2 | |
| | 5+6 2 | 5+6 2 | | | | 3+4 2 | 5+6 2 | |
| | 5+6 2 | 3+4 2 | | | | 3+4 2 | 5+6 2 | |
| | 3+4 2 | 3+4 2 | | | | 5+6 2 | 5+6 2 | |
| | 5+6 2 | 3+4 2 | | | | 5+6 2 | 5+6 2 | |
| | 5+6 2 | 3+4 2 | | | | 3+4 2 | 3+4 2 | |
| | 5+6 2 | 3+4 2 | | | | 5+6 2 | 5+6 2 | |
| 11.2.25 | 157983 | 28 | Ni1Ti1 | | | | | |
| | -A | -B | -C | -D | -E | -GM | -Y | -Z |
| | 3+4 2 | 3+4 2 | 2+2 4 | 2+2 4 | 2+2 4 | 3+4 2 | 3+4 2 | 2+2 4 |
| | 5+6 2 | 5+6 2 | 2+2 4 | 2+2 4 | 2+2 4 | 5+6 2 | 5+6 2 | 2+2 4 |
| | 5+6 2 | 3+4 2 | 2+2 4 | 2+2 4 | 2+2 4 | 5+6 2 | 5+6 2 | 2+2 4 |
| | 5+6 2 | 5+6 2 | 2+2 4 | 2+2 4 | 2+2 4 | 3+4 2 | 5+6 2 | 2+2 4 |
| | 3+4 2 | 3+4 2 | 2+2 4 | 2+2 4 | 2+2 4 | 3+4 2 | 5+6 2 | 2+2 4 |
| | 5+6 2 | 5+6 2 | 2+2 4 | 2+2 4 | 2+2 4 | 5+6 2 | 3+4 2 | 2+2 4 |
| | 3+4 2 | 5+6 2 | 2+2 4 | 2+2 4 | 2+2 4 | 3+4 2 | 3+4 2 | 2+2 4 |
| | 5+6 2 | 5+6 2 | | | | 5+6 2 | 5+6 2 | |
| | 3+4 2 | 5+6 2 | | | | 3+4 2 | 3+4 2 | |
| | 3+4 2 | 3+4 2 | | | | 3+4 2 | 3+4 2 | |
| | 5+6 2 | 3+4 2 | | | | 5+6 2 | 3+4 2 | |
| | 3+4 2 | 3+4 2 | | | | 5+6 2 | 3+4 2 | |
| | 3+4 2 | 3+4 2 | | | | 3+4 2 | 5+6 2 | |
| | 3+4 2 | 3+4 2 | | | | 5+6 2 | 3+4 2 | |
| 11.2.26 | 105415 | 28 | Ni1Ti1 | | | | | |
| | -A | -B | -C | -D | -E | -GM | -Y | -Z |
| | 5+6 2 | 5+6 2 | 2+2 4 | 2+2 4 | 2+2 4 | 3+4 2 | 3+4 2 | 2+2 4 |
| | 3+4 2 | 3+4 2 | 2+2 4 | 2+2 4 | 2+2 4 | 5+6 2 | 5+6 2 | 2+2 4 |
| | 3+4 2 | 5+6 2 | 2+2 4 | 2+2 4 | 2+2 4 | 5+6 2 | 5+6 2 | 2+2 4 |
| | 3+4 2 | 3+4 2 | 2+2 4 | 2+2 4 | 2+2 4 | 3+4 2 | 5+6 2 | 2+2 4 |
| | 5+6 2 | 5+6 2 | 2+2 4 | 2+2 4 | 2+2 4 | 3+4 2 | 5+6 2 | 2+2 4 |
| | 3+4 2 | 3+4 2 | 2+2 4 | 2+2 4 | 2+2 4 | 5+6 2 | 3+4 2 | 2+2 4 |
| | 5+6 2 | 3+4 2 | 2+2 4 | 2+2 4 | 2+2 4 | 5+6 2 | 3+4 2 | 2+2 4 |
| | 3+4 2 | 3+4 2 | | | | 3+4 2 | 5+6 2 | |
| | 5+6 2 | 3+4 2 | | | | 3+4 2 | 3+4 2 | |
| | 5+6 2 | 5+6 2 | | | | 3+4 2 | 3+4 2 | |
| | 3+4 2 | 5+6 2 | | | | 5+6 2 | 3+4 2 | |
| | 5+6 2 | 5+6 2 | | | | 5+6 2 | 3+4 2 | |
| | 5+6 2 | 5+6 2 | | | | 3+4 2 | 5+6 2 | |
| | 5+6 2 | 5+6 2 | | | | 5+6 2 | 3+4 2 | |
| 11.2.27 | 105414 | 28 | Ni1Ti1 | | | | | |
| | -A | -B | -C | -D | -E | -GM | -Y | -Z |
| | 5+6 2 | 3+4 2 | 2+2 4 | 2+2 4 | 2+2 4 | 3+4 2 | 5+6 2 | 2+2 4 |
| | 3+4 2 | 5+6 2 | 2+2 4 | 2+2 4 | 2+2 4 | 5+6 2 | 3+4 2 | 2+2 4 |
| | 3+4 2 | 3+4 2 | 2+2 4 | 2+2 4 | 2+2 4 | 5+6 2 | 3+4 2 | 2+2 4 |
| | 5+6 2 | 5+6 2 | 2+2 4 | 2+2 4 | 2+2 4 | 3+4 2 | 3+4 2 | 2+2 4 |
| | 3+4 2 | 3+4 2 | 2+2 4 | 2+2 4 | 2+2 4 | 3+4 2 | 3+4 2 | 2+2 4 |
| | 3+4 2 | 5+6 2 | 2+2 4 | 2+2 4 | 2+2 4 | 5+6 2 | 5+6 2 | 2+2 4 |
| | 5+6 2 | 5+6 2 | 2+2 4 | 2+2 4 | 2+2 4 | 5+6 2 | 5+6 2 | 2+2 4 |

Continued on next column

**TABLE XXIII– continued**

| TQC | ICSD | NoE | SF | | | | | |
|---|---|---|---|---|---|---|---|---|
| | 3+4 2 | 5+6 2 | | | | 3+4 2 | 5+6 2 | |
| | 5+6 2 | 3+4 2 | | | | 3+4 2 | 5+6 2 | |
| | 5+6 2 | 5+6 2 | | | | 3+4 2 | 3+4 2 | |
| | 3+4 2 | 3+4 2 | | | | 5+6 2 | 5+6 2 | |
| | 5+6 2 | 3+4 2 | | | | 5+6 2 | 5+6 2 | |
| | 5+6 2 | 3+4 2 | | | | 3+4 2 | 3+4 2 | |
| | 5+6 2 | 5+6 2 | | | | 3+4 2 | 5+6 2 | |
| 11.2.28 | 66316 | 60 | Nd1Ru1Si2 | | | | | |
| | -A | -B | -C | -D | -E | -GM | -Y | -Z |
| | 5+6 2 | 3+4 2 | 2+2 4 | 2+2 4 | 2+2 4 | 3+4 2 | 5+6 2 | 2+2 4 |
| | 3+4 2 | 5+6 2 | 2+2 4 | 2+2 4 | 2+2 4 | 5+6 2 | 3+4 2 | 2+2 4 |
| | 5+6 2 | 3+4 2 | 2+2 4 | 2+2 4 | 2+2 4 | 3+4 2 | 5+6 2 | 2+2 4 |
| | 3+4 2 | 5+6 2 | 2+2 4 | 2+2 4 | 2+2 4 | 5+6 2 | 3+4 2 | 2+2 4 |
| | 5+6 2 | 3+4 2 | 2+2 4 | 2+2 4 | 2+2 4 | 3+4 2 | 5+6 2 | 2+2 4 |
| | 3+4 2 | 5+6 2 | 2+2 4 | 2+2 4 | 2+2 4 | 5+6 2 | 3+4 2 | 2+2 4 |
| | 5+6 2 | 3+4 2 | 2+2 4 | 2+2 4 | 2+2 4 | 3+4 2 | 5+6 2 | 2+2 4 |
| | 3+4 2 | 5+6 2 | 2+2 4 | 2+2 4 | 2+2 4 | 5+6 2 | 3+4 2 | 2+2 4 |
| | 3+4 2 | 5+6 2 | 2+2 4 | 2+2 4 | 2+2 4 | 3+4 2 | 5+6 2 | 2+2 4 |
| | 3+4 2 | 3+4 2 | 2+2 4 | 2+2 4 | 2+2 4 | 3+4 2 | 3+4 2 | 2+2 4 |
| | 5+6 2 | 3+4 2 | 2+2 4 | 2+2 4 | 2+2 4 | 5+6 2 | 3+4 2 | 2+2 4 |
| | 5+6 2 | 5+6 2 | 2+2 4 | 2+2 4 | 2+2 4 | 5+6 2 | 5+6 2 | 2+2 4 |
| | 3+4 2 | 3+4 2 | 2+2 4 | 2+2 4 | 2+2 4 | 3+4 2 | 5+6 2 | 2+2 4 |
| | 5+6 2 | 5+6 2 | 2+2 4 | 2+2 4 | 2+2 4 | 5+6 2 | 5+6 2 | 2+2 4 |
| | 3+4 2 | 5+6 2 | 2+2 4 | 2+2 4 | 2+2 4 | 3+4 2 | 5+6 2 | 2+2 4 |
| | 5+6 2 | 3+4 2 | | | | 3+4 2 | 3+4 2 | |
| | 3+4 2 | 5+6 2 | | | | 3+4 2 | 3+4 2 | |
| | 3+4 2 | 3+4 2 | | | | 5+6 2 | 5+6 2 | |
| | 3+4 2 | 5+6 2 | | | | 5+6 2 | 3+4 2 | |
| | 5+6 2 | 3+4 2 | | | | 3+4 2 | 3+4 2 | |
| | 5+6 2 | 3+4 2 | | | | 5+6 2 | 3+4 2 | |
| | 3+4 2 | 5+6 2 | | | | 3+4 2 | 3+4 2 | |
| | 5+6 2 | 5+6 2 | | | | 5+6 2 | 3+4 2 | |
| | 5+6 2 | 5+6 2 | | | | 3+4 2 | 5+6 2 | |
| | 5+6 2 | 3+4 2 | | | | 5+6 2 | 5+6 2 | |
| | 3+4 2 | 5+6 2 | | | | 3+4 2 | 3+4 2 | |
| | 3+4 2 | 3+4 2 | | | | 5+6 2 | 3+4 2 | |
| | 5+6 2 | 5+6 2 | | | | 3+4 2 | 5+6 2 | |
| | 3+4 2 | 3+4 2 | | | | 3+4 2 | 3+4 2 | |
| | 5+6 2 | 3+4 2 | | | | 5+6 2 | 5+6 2 | |
| 11.2.29 | 42981 | 88 | Nb2Se3 | | | | | |
| | -A | -B | -C | -D | -E | -GM | -Y | -Z |
| | 5+6 2 | 5+6 2 | 2+2 4 | 2+2 4 | 2+2 4 | 3+4 2 | 3+4 2 | 2+2 4 |
| | 5+6 2 | 3+4 2 | 2+2 4 | 2+2 4 | 2+2 4 | 3+4 2 | 5+6 2 | 2+2 4 |
| | 3+4 2 | 3+4 2 | 2+2 4 | 2+2 4 | 2+2 4 | 5+6 2 | 5+6 2 | 2+2 4 |
| | 3+4 2 | 5+6 2 | 2+2 4 | 2+2 4 | 2+2 4 | 5+6 2 | 3+4 2 | 2+2 4 |
| | 5+6 2 | 5+6 2 | 2+2 4 | 2+2 4 | 2+2 4 | 3+4 2 | 3+4 2 | 2+2 4 |
| | 5+6 2 | 3+4 2 | 2+2 4 | 2+2 4 | 2+2 4 | 3+4 2 | 5+6 2 | 2+2 4 |
| | 3+4 2 | 3+4 2 | 2+2 4 | 2+2 4 | 2+2 4 | 5+6 2 | 5+6 2 | 2+2 4 |
| | 3+4 2 | 5+6 2 | 2+2 4 | 2+2 4 | 2+2 4 | 5+6 2 | 3+4 2 | 2+2 4 |
| | 5+6 2 | 5+6 2 | 2+2 4 | 2+2 4 | 2+2 4 | 3+4 2 | 3+4 2 | 2+2 4 |
| | 5+6 2 | 3+4 2 | 2+2 4 | 2+2 4 | 2+2 4 | 3+4 2 | 5+6 2 | 2+2 4 |
| | 3+4 2 | 3+4 2 | 2+2 4 | 2+2 4 | 2+2 4 | 5+6 2 | 5+6 2 | 2+2 4 |
| | 5+6 2 | 5+6 2 | 2+2 4 | 2+2 4 | 2+2 4 | 3+4 2 | 3+4 2 | 2+2 4 |
| | 3+4 2 | 3+4 2 | 2+2 4 | 2+2 4 | 2+2 4 | 5+6 2 | 5+6 2 | 2+2 4 |
| | 5+6 2 | 5+6 2 | 2+2 4 | 2+2 4 | 2+2 4 | 3+4 2 | 3+4 2 | 2+2 4 |
| | 3+4 2 | 3+4 2 | 2+2 4 | 2+2 4 | 2+2 4 | 5+6 2 | 5+6 2 | 2+2 4 |
| | 3+4 2 | 5+6 2 | 2+2 4 | 2+2 4 | 2+2 4 | 5+6 2 | 3+4 2 | 2+2 4 |
| | 3+4 2 | 3+4 2 | 2+2 4 | 2+2 4 | 2+2 4 | 3+4 2 | 3+4 2 | 2+2 4 |
| | 5+6 2 | 5+6 2 | 2+2 4 | 2+2 4 | 2+2 4 | 5+6 2 | 5+6 2 | 2+2 4 |
| | 3+4 2 | 5+6 2 | 2+2 4 | 2+2 4 | 2+2 4 | 3+4 2 | 3+4 2 | 2+2 4 |
| | 5+6 2 | 3+4 2 | 2+2 4 | 2+2 4 | 2+2 4 | 5+6 2 | 5+6 2 | 2+2 4 |

Continued on next column

**TABLE XXIII– continued**

| TQC | ICSD | NoE | SF | | | | | |
|---|---|---|---|---|---|---|---|---|
| | 3+4 2 | 5+6 2 | 2+2 4 | 2+2 4 | 2+2 4 | 3+4 2 | 5+6 2 | 2+2 4 |
| | 5+6 2 | 3+4 2 | 2+2 4 | 2+2 4 | 2+2 4 | 5+6 2 | 3+4 2 | 2+2 4 |
| | 3+4 2 | 3+4 2 | | | | 3+4 2 | 5+6 2 | |
| | 3+4 2 | 5+6 2 | | | | 5+6 2 | 3+4 2 | |
| | 5+6 2 | 5+6 2 | | | | 5+6 2 | 3+4 2 | |
| | 3+4 2 | 3+4 2 | | | | 3+4 2 | 5+6 2 | |
| | 5+6 2 | 3+4 2 | | | | 3+4 2 | 3+4 2 | |
| | 3+4 2 | 5+6 2 | | | | 5+6 2 | 5+6 2 | |
| | 5+6 2 | 3+4 2 | | | | 3+4 2 | 5+6 2 | |
| | 3+4 2 | 5+6 2 | | | | 5+6 2 | 5+6 2 | |
| | 5+6 2 | 3+4 2 | | | | 3+4 2 | 3+4 2 | |
| | 3+4 2 | 5+6 2 | | | | 5+6 2 | 3+4 2 | |
| | 5+6 2 | 3+4 2 | | | | 3+4 2 | 3+4 2 | |
| | 3+4 2 | 5+6 2 | | | | 5+6 2 | 5+6 2 | |
| | 5+6 2 | 5+6 2 | | | | 3+4 2 | 3+4 2 | |
| | 5+6 2 | 3+4 2 | | | | 3+4 2 | 3+4 2 | |
| | 5+6 2 | 3+4 2 | | | | 5+6 2 | 5+6 2 | |
| | 3+4 2 | 5+6 2 | | | | 3+4 2 | 5+6 2 | |
| | 3+4 2 | 5+6 2 | | | | 5+6 2 | 3+4 2 | |
| | 3+4 2 | 3+4 2 | | | | 5+6 2 | 3+4 2 | |
| | 5+6 2 | 3+4 2 | | | | 3+4 2 | 5+6 2 | |
| | 5+6 2 | 5+6 2 | | | | 5+6 2 | 5+6 2 | |
| | 3+4 2 | 5+6 2 | | | | 5+6 2 | 5+6 2 | |
| | 5+6 2 | 3+4 2 | | | | 3+4 2 | 5+6 2 | |
| 11.2.30 | 54283 | 172 | Nb1Pt3 | | | | | |
| | -A | -B | -C | -D | -E | -GM | -Y | -Z |
| | 3+4 2 | 3+4 2 | 2+2 4 | 2+2 4 | 2+2 4 | 3+4 2 | 3+4 2 | 2+2 4 |
| | 5+6 2 | 5+6 2 | 2+2 4 | 2+2 4 | 2+2 4 | 5+6 2 | 5+6 2 | 2+2 4 |
| | 3+4 2 | 5+6 2 | 2+2 4 | 2+2 4 | 2+2 4 | 3+4 2 | 3+4 2 | 2+2 4 |
| | 5+6 2 | 3+4 2 | 2+2 4 | 2+2 4 | 2+2 4 | 5+6 2 | 5+6 2 | 2+2 4 |
| | 3+4 2 | 3+4 2 | 2+2 4 | 2+2 4 | 2+2 4 | 3+4 2 | 3+4 2 | 2+2 4 |
| | 5+6 2 | 5+6 2 | 2+2 4 | 2+2 4 | 2+2 4 | 5+6 2 | 5+6 2 | 2+2 4 |
| | 3+4 2 | 5+6 2 | 2+2 4 | 2+2 4 | 2+2 4 | 3+4 2 | 5+6 2 | 2+2 4 |
| | 5+6 2 | 3+4 2 | 2+2 4 | 2+2 4 | 2+2 4 | 5+6 2 | 3+4 2 | 2+2 4 |
| | 3+4 2 | 3+4 2 | 2+2 4 | 2+2 4 | 2+2 4 | 3+4 2 | 3+4 2 | 2+2 4 |
| | 5+6 2 | 5+6 2 | 2+2 4 | 2+2 4 | 2+2 4 | 5+6 2 | 5+6 2 | 2+2 4 |
| | 3+4 2 | 3+4 2 | 2+2 4 | 2+2 4 | 2+2 4 | 3+4 2 | 3+4 2 | 2+2 4 |
| | 5+6 2 | 5+6 2 | 2+2 4 | 2+2 4 | 2+2 4 | 5+6 2 | 5+6 2 | 2+2 4 |
| | 3+4 2 | 5+6 2 | 2+2 4 | 2+2 4 | 2+2 4 | 3+4 2 | 5+6 2 | 2+2 4 |
| | 5+6 2 | 3+4 2 | 2+2 4 | 2+2 4 | 2+2 4 | 5+6 2 | 3+4 2 | 2+2 4 |
| | 3+4 2 | 5+6 2 | 2+2 4 | 2+2 4 | 2+2 4 | 3+4 2 | 5+6 2 | 2+2 4 |
| | 5+6 2 | 3+4 2 | 2+2 4 | 2+2 4 | 2+2 4 | 5+6 2 | 3+4 2 | 2+2 4 |
| | 5+6 2 | 5+6 2 | 2+2 4 | 2+2 4 | 2+2 4 | 3+4 2 | 3+4 2 | 2+2 4 |
| | 3+4 2 | 3+4 2 | 2+2 4 | 2+2 4 | 2+2 4 | 5+6 2 | 5+6 2 | 2+2 4 |
| | 5+6 2 | 5+6 2 | 2+2 4 | 2+2 4 | 2+2 4 | 3+4 2 | 5+6 2 | 2+2 4 |
| | 3+4 2 | 3+4 2 | 2+2 4 | 2+2 4 | 2+2 4 | 5+6 2 | 5+6 2 | 2+2 4 |
| | 5+6 2 | 3+4 2 | 2+2 4 | 2+2 4 | 2+2 4 | 5+6 2 | 3+4 2 | 2+2 4 |
| | 3+4 2 | 5+6 2 | 2+2 4 | 2+2 4 | 2+2 4 | 3+4 2 | 3+4 2 | 2+2 4 |
| | 5+6 2 | 5+6 2 | 2+2 4 | 2+2 4 | 2+2 4 | 3+4 2 | 5+6 2 | 2+2 4 |
| | 3+4 2 | 3+4 2 | 2+2 4 | 2+2 4 | 2+2 4 | 5+6 2 | 5+6 2 | 2+2 4 |
| | 3+4 2 | 3+4 2 | 2+2 4 | 2+2 4 | 2+2 4 | 3+4 2 | 3+4 2 | 2+2 4 |
| | 3+4 2 | 5+6 2 | 2+2 4 | 2+2 4 | 2+2 4 | 5+6 2 | 5+6 2 | 2+2 4 |
| | 5+6 2 | 3+4 2 | 2+2 4 | 2+2 4 | 2+2 4 | 3+4 2 | 3+4 2 | 2+2 4 |
| | 5+6 2 | 5+6 2 | 2+2 4 | 2+2 4 | 2+2 4 | 5+6 2 | 3+4 2 | 2+2 4 |
| | 3+4 2 | 3+4 2 | 2+2 4 | 2+2 4 | 2+2 4 | 3+4 2 | 3+4 2 | 2+2 4 |
| | 3+4 2 | 5+6 2 | 2+2 4 | 2+2 4 | 2+2 4 | 5+6 2 | 5+6 2 | 2+2 4 |
| | 5+6 2 | 5+6 2 | 2+2 4 | 2+2 4 | 2+2 4 | 3+4 2 | 3+4 2 | 2+2 4 |
| | 3+4 2 | 3+4 2 | 2+2 4 | 2+2 4 | 2+2 4 | 3+4 2 | 3+4 2 | 2+2 4 |
| | 3+4 2 | 3+4 2 | 2+2 4 | 2+2 4 | 2+2 4 | 5+6 2 | 5+6 2 | 2+2 4 |
| | 5+6 2 | 5+6 2 | 2+2 4 | 2+2 4 | 2+2 4 | 3+4 2 | 3+4 2 | 2+2 4 |
| | 3+4 2 | 3+4 2 | 2+2 4 | 2+2 4 | 2+2 4 | 5+6 2 | 5+6 2 | 2+2 4 |
| | 3+4 2 | 5+6 2 | 2+2 4 | 2+2 4 | 2+2 4 | 3+4 2 | 5+6 2 | 2+2 4 |
| | 5+6 2 | 5+6 2 | 2+2 4 | 2+2 4 | 2+2 4 | 5+6 2 | 5+6 2 | 2+2 4 |
| | 5+6 2 | 3+4 2 | 2+2 4 | 2+2 4 | 2+2 4 | 3+4 2 | 3+4 2 | 2+2 4 |
| | 3+4 2 | 5+6 2 | 2+2 4 | 2+2 4 | 2+2 4 | 5+6 2 | 5+6 2 | 2+2 4 |
| | 5+6 2 | 3+4 2 | 2+2 4 | 2+2 4 | 2+2 4 | 3+4 2 | 3+4 2 | 2+2 4 |
| | 5+6 2 | 5+6 2 | 2+2 4 | 2+2 4 | 2+2 4 | 5+6 2 | 5+6 2 | 2+2 4 |
| | 3+4 2 | 3+4 2 | 2+2 4 | 2+2 4 | 2+2 4 | 5+6 2 | 3+4 2 | 2+2 4 |
| | 5+6 2 | 5+6 2 | 2+2 4 | 2+2 4 | 2+2 4 | 3+4 2 | 5+6 2 | 2+2 4 |
| | 3+4 2 | 3+4 2 | | | | 5+6 2 | 3+4 2 | |
| | 5+6 2 | 5+6 2 | | | | 3+4 2 | 3+4 2 | |
| | 3+4 2 | 3+4 2 | | | | 5+6 2 | 5+6 2 | |
| | 5+6 2 | 3+4 2 | | | | 3+4 2 | 3+4 2 | |
| | 3+4 2 | 5+6 2 | | | | 5+6 2 | 5+6 2 | |
| | 5+6 2 | 3+4 2 | | | | 3+4 2 | 3+4 2 | |
| | 5+6 2 | 5+6 2 | | | | 5+6 2 | 3+4 2 | |
| | 5+6 2 | 3+4 2 | | | | 3+4 2 | 5+6 2 | |
| | 3+4 2 | 5+6 2 | | | | 3+4 2 | 5+6 2 | |
| | 5+6 2 | 5+6 2 | | | | 5+6 2 | 3+4 2 | |
| | 5+6 2 | 3+4 2 | | | | 3+4 2 | 5+6 2 | |
| | 3+4 2 | 5+6 2 | | | | 5+6 2 | 5+6 2 | |
| | 5+6 2 | 3+4 2 | | | | 5+6 2 | 3+4 2 | |
| | 3+4 2 | 5+6 2 | | | | 3+4 2 | 5+6 2 | |
| | 3+4 2 | 3+4 2 | | | | 5+6 2 | 3+4 2 | |
| | 5+6 2 | 5+6 2 | | | | 5+6 2 | 5+6 2 | |
| | 3+4 2 | 3+4 2 | | | | 3+4 2 | 3+4 2 | |
| | 5+6 2 | 5+6 2 | | | | 3+4 2 | 5+6 2 | |
| | 5+6 2 | 3+4 2 | | | | 5+6 2 | 3+4 2 | |
| | 5+6 2 | 3+4 2 | | | | 3+4 2 | 5+6 2 | |
| | 3+4 2 | 5+6 2 | | | | 3+4 2 | 3+4 2 | |
| | 3+4 2 | 3+4 2 | | | | 5+6 2 | 5+6 2 | |
| | 5+6 2 | 5+6 2 | | | | 5+6 2 | 5+6 2 | |
| | 5+6 2 | 3+4 2 | | | | 3+4 2 | 3+4 2 | |
| | 3+4 2 | 5+6 2 | | | | 5+6 2 | 5+6 2 | |
| | 3+4 2 | 3+4 2 | | | | 3+4 2 | 3+4 2 | |
| | 3+4 2 | 5+6 2 | | | | 5+6 2 | 3+4 2 | |
| | 5+6 2 | 5+6 2 | | | | 3+4 2 | 3+4 2 | |
| | 5+6 2 | 3+4 2 | | | | 5+6 2 | 5+6 2 | |
| | 3+4 2 | 5+6 2 | | | | 3+4 2 | 5+6 2 | |
| | 5+6 2 | 3+4 2 | | | | 3+4 2 | 3+4 2 | |
| | 3+4 2 | 5+6 2 | | | | 5+6 2 | 5+6 2 | |
| | 3+4 2 | 3+4 2 | | | | 3+4 2 | 3+4 2 | |
| | 5+6 2 | 5+6 2 | | | | 3+4 2 | 3+4 2 | |
| | 5+6 2 | 3+4 2 | | | | 5+6 2 | 5+6 2 | |
| | 5+6 2 | 5+6 2 | | | | 3+4 2 | 3+4 2 | |
| | 3+4 2 | 3+4 2 | | | | 5+6 2 | 5+6 2 | |
| | 3+4 2 | 5+6 2 | | | | 3+4 2 | 3+4 2 | |
| | 5+6 2 | 3+4 2 | | | | 5+6 2 | 5+6 2 | |
| | 3+4 2 | 5+6 2 | | | | 5+6 2 | 5+6 2 | |
| | 5+6 2 | 3+4 2 | | | | 3+4 2 | 3+4 2 | |
| | 3+4 2 | 3+4 2 | | | | 5+6 2 | 3+4 2 | |
| | 3+4 2 | 5+6 2 | | | | 3+4 2 | 3+4 2 | |
| 11.2.31 | 93368 | 132 | Na1O8P1W2 | | | | | |
| | -A | -B | -C | -D | -E | -GM | -Y | -Z |
| | 5+6 2 | 5+6 2 | 2+2 4 | 2+2 4 | 2+2 4 | 3+4 2 | 3+4 2 | 2+2 4 |
| | 3+4 2 | 3+4 2 | 2+2 4 | 2+2 4 | 2+2 4 | 5+6 2 | 5+6 2 | 2+2 4 |
| | 3+4 2 | 3+4 2 | 2+2 4 | 2+2 4 | 2+2 4 | 5+6 2 | 5+6 2 | 2+2 4 |
| | 5+6 2 | 5+6 2 | 2+2 4 | 2+2 4 | 2+2 4 | 3+4 2 | 3+4 2 | 2+2 4 |
| | 5+6 2 | 5+6 2 | 2+2 4 | 2+2 4 | 2+2 4 | 3+4 2 | 3+4 2 | 2+2 4 |
| | 5+6 2 | 5+6 2 | 2+2 4 | 2+2 4 | 2+2 4 | 3+4 2 | 3+4 2 | 2+2 4 |
| | 3+4 2 | 3+4 2 | 2+2 4 | 2+2 4 | 2+2 4 | 5+6 2 | 5+6 2 | 2+2 4 |
| | 3+4 2 | 3+4 2 | 2+2 4 | 2+2 4 | 2+2 4 | 5+6 2 | 5+6 2 | 2+2 4 |

Continued on next column

**TABLE XXIII– continued**

| TQC | ICSD | NoE | SF | | | | | |
|---|---|---|---|---|---|---|---|---|
| | 5+6 2 | 3+4 2 | 2+2 4 | 2+2 4 | 2+2 4 | 3+4 2 | 5+6 2 | 2+2 4 |
| | 3+4 2 | 5+6 2 | 2+2 4 | 2+2 4 | 2+2 4 | 5+6 2 | 3+4 2 | 2+2 4 |
| | 5+6 2 | 3+4 2 | 2+2 4 | 2+2 4 | 2+2 4 | 3+4 2 | 5+6 2 | 2+2 4 |
| | 3+4 2 | 5+6 2 | 2+2 4 | 2+2 4 | 2+2 4 | 5+6 2 | 3+4 2 | 2+2 4 |
| | 5+6 2 | 3+4 2 | 2+2 4 | 2+2 4 | 2+2 4 | 3+4 2 | 5+6 2 | 2+2 4 |
| | 3+4 2 | 5+6 2 | 2+2 4 | 2+2 4 | 2+2 4 | 5+6 2 | 3+4 2 | 2+2 4 |
| | 5+6 2 | 3+4 2 | 2+2 4 | 2+2 4 | 2+2 4 | 3+4 2 | 5+6 2 | 2+2 4 |
| | 5+6 2 | 3+4 2 | 2+2 4 | 2+2 4 | 2+2 4 | 3+4 2 | 5+6 2 | 2+2 4 |
| | 3+4 2 | 5+6 2 | 2+2 4 | 2+2 4 | 2+2 4 | 3+4 2 | 3+4 2 | 2+2 4 |
| | 5+6 2 | 3+4 2 | 2+2 4 | 2+2 4 | 2+2 4 | 5+6 2 | 5+6 2 | 2+2 4 |
| | 3+4 2 | 3+4 2 | 2+2 4 | 2+2 4 | 2+2 4 | 5+6 2 | 5+6 2 | 2+2 4 |
| | 3+4 2 | 5+6 2 | 2+2 4 | 2+2 4 | 2+2 4 | 5+6 2 | 5+6 2 | 2+2 4 |
| | 5+6 2 | 3+4 2 | 2+2 4 | 2+2 4 | 2+2 4 | 3+4 2 | 3+4 2 | 2+2 4 |
| | 5+6 2 | 5+6 2 | 2+2 4 | 2+2 4 | 2+2 4 | 3+4 2 | 3+4 2 | 2+2 4 |
| | 5+6 2 | 3+4 2 | 2+2 4 | 2+2 4 | 2+2 4 | 3+4 2 | 3+4 2 | 2+2 4 |
| | 3+4 2 | 5+6 2 | 2+2 4 | 2+2 4 | 2+2 4 | 5+6 2 | 5+6 2 | 2+2 4 |
| | 3+4 2 | 5+6 2 | 2+2 4 | 2+2 4 | 2+2 4 | 5+6 2 | 3+4 2 | 2+2 4 |
| | 3+4 2 | 5+6 2 | 2+2 4 | 2+2 4 | 2+2 4 | 5+6 2 | 3+4 2 | 2+2 4 |
| | 5+6 2 | 5+6 2 | 2+2 4 | 2+2 4 | 2+2 4 | 3+4 2 | 3+4 2 | 2+2 4 |
| | 3+4 2 | 3+4 2 | 2+2 4 | 2+2 4 | 2+2 4 | 5+6 2 | 5+6 2 | 2+2 4 |
| | 3+4 2 | 3+4 2 | 2+2 4 | 2+2 4 | 2+2 4 | 3+4 2 | 3+4 2 | 2+2 4 |
| | 5+6 2 | 5+6 2 | 2+2 4 | 2+2 4 | 2+2 4 | 3+4 2 | 5+6 2 | 2+2 4 |
| | 5+6 2 | 3+4 2 | 2+2 4 | 2+2 4 | 2+2 4 | 5+6 2 | 3+4 2 | 2+2 4 |
| | 3+4 2 | 5+6 2 | 2+2 4 | 2+2 4 | 2+2 4 | 3+4 2 | 5+6 2 | 2+2 4 |
| | 5+6 2 | 5+6 2 | 2+2 4 | 2+2 4 | 2+2 4 | 3+4 2 | 3+4 2 | 2+2 4 |
| | 3+4 2 | 3+4 2 | | | | 5+6 2 | 5+6 2 | |
| | 5+6 2 | 5+6 2 | | | | 5+6 2 | 3+4 2 | |
| | 5+6 2 | 5+6 2 | | | | 3+4 2 | 5+6 2 | |
| | 3+4 2 | 3+4 2 | | | | 5+6 2 | 3+4 2 | |
| | 3+4 2 | 3+4 2 | | | | 5+6 2 | 3+4 2 | |
| | 5+6 2 | 5+6 2 | | | | 3+4 2 | 5+6 2 | |
| | 5+6 2 | 3+4 2 | | | | 3+4 2 | 3+4 2 | |
| | 3+4 2 | 3+4 2 | | | | 5+6 2 | 5+6 2 | |
| | 3+4 2 | 5+6 2 | | | | 5+6 2 | 5+6 2 | |
| | 3+4 2 | 3+4 2 | | | | 3+4 2 | 3+4 2 | |
| | 5+6 2 | 3+4 2 | | | | 5+6 2 | 5+6 2 | |
| | 3+4 2 | 5+6 2 | | | | 3+4 2 | 3+4 2 | |
| | 5+6 2 | 3+4 2 | | | | 5+6 2 | 5+6 2 | |
| | 3+4 2 | 5+6 2 | | | | 3+4 2 | 5+6 2 | |
| | 5+6 2 | 5+6 2 | | | | 5+6 2 | 3+4 2 | |
| | 3+4 2 | 5+6 2 | | | | 5+6 2 | 5+6 2 | |
| | 5+6 2 | 5+6 2 | | | | 5+6 2 | 5+6 2 | |
| | 5+6 2 | 5+6 2 | | | | 5+6 2 | 3+4 2 | |
| | 3+4 2 | 3+4 2 | | | | 3+4 2 | 3+4 2 | |
| | 3+4 2 | 5+6 2 | | | | 5+6 2 | 3+4 2 | |
| | 5+6 2 | 5+6 2 | | | | 3+4 2 | 3+4 2 | |
| | 3+4 2 | 3+4 2 | | | | 5+6 2 | 3+4 2 | |
| | 3+4 2 | 3+4 2 | | | | 3+4 2 | 5+6 2 | |
| | 5+6 2 | 5+6 2 | | | | 5+6 2 | 5+6 2 | |
| | 3+4 2 | 5+6 2 | | | | 5+6 2 | 3+4 2 | |
| | 5+6 2 | 3+4 2 | | | | 3+4 2 | 5+6 2 | |
| | 3+4 2 | 5+6 2 | | | | 5+6 2 | 3+4 2 | |
| | 3+4 2 | 3+4 2 | | | | 3+4 2 | 5+6 2 | |
| | 5+6 2 | 5+6 2 | | | | 5+6 2 | 3+4 2 | |
| | 3+4 2 | 3+4 2 | | | | 3+4 2 | 3+4 2 | |
| | 3+4 2 | 5+6 2 | | | | 5+6 2 | 3+4 2 | |
| | 5+6 2 | 3+4 2 | | | | 3+4 2 | 5+6 2 | |
| | 3+4 2 | 3+4 2 | | | | 3+4 2 | 3+4 2 | |
| 11.2.32 | 73453 | 60 | Mo2S3 | | | | | |
| | -A | -B | -C | -D | -E | -GM | -Y | -Z |
| | 3+4 2 | 5+6 2 | 2+2 4 | 2+2 4 | 2+2 4 | 3+4 2 | 3+4 2 | 2+2 4 |

Continued on next column

**TABLE XXIII– continued**

| TQC | ICSD | NoE | SF | | | | | |
|---|---|---|---|---|---|---|---|---|
| | 5+6 2 | 3+4 2 | 2+2 4 | 2+2 4 | 2+2 4 | 5+6 2 | 5+6 2 | 2+2 4 |
| | 5+6 2 | 3+4 2 | 2+2 4 | 2+2 4 | 2+2 4 | 3+4 2 | 3+4 2 | 2+2 4 |
| | 3+4 2 | 5+6 2 | 2+2 4 | 2+2 4 | 2+2 4 | 5+6 2 | 5+6 2 | 2+2 4 |
| | 5+6 2 | 3+4 2 | 2+2 4 | 2+2 4 | 2+2 4 | 3+4 2 | 5+6 2 | 2+2 4 |
| | 3+4 2 | 5+6 2 | 2+2 4 | 2+2 4 | 2+2 4 | 5+6 2 | 3+4 2 | 2+2 4 |
| | 5+6 2 | 3+4 2 | 2+2 4 | 2+2 4 | 2+2 4 | 5+6 2 | 5+6 2 | 2+2 4 |
| | 5+6 2 | 5+6 2 | 2+2 4 | 2+2 4 | 2+2 4 | 3+4 2 | 3+4 2 | 2+2 4 |
| | 3+4 2 | 3+4 2 | 2+2 4 | 2+2 4 | 2+2 4 | 5+6 2 | 3+4 2 | 2+2 4 |
| | 5+6 2 | 5+6 2 | 2+2 4 | 2+2 4 | 2+2 4 | 3+4 2 | 5+6 2 | 2+2 4 |
| | 5+6 2 | 5+6 2 | 2+2 4 | 2+2 4 | 2+2 4 | 5+6 2 | 3+4 2 | 2+2 4 |
| | 3+4 2 | 3+4 2 | 2+2 4 | 2+2 4 | 2+2 4 | 3+4 2 | 5+6 2 | 2+2 4 |
| | 3+4 2 | 5+6 2 | 2+2 4 | 2+2 4 | 2+2 4 | 3+4 2 | 5+6 2 | 2+2 4 |
| | 5+6 2 | 3+4 2 | 2+2 4 | 2+2 4 | 2+2 4 | 5+6 2 | 3+4 2 | 2+2 4 |
| | 5+6 2 | 5+6 2 | 2+2 4 | 2+2 4 | 2+2 4 | 3+4 2 | 5+6 2 | 2+2 4 |
| | 3+4 2 | 3+4 2 | | | | 5+6 2 | 3+4 2 | |
| | 3+4 2 | 5+6 2 | | | | 3+4 2 | 5+6 2 | |
| | 5+6 2 | 3+4 2 | | | | 5+6 2 | 3+4 2 | |
| | 3+4 2 | 5+6 2 | | | | 3+4 2 | 3+4 2 | |
| | 3+4 2 | 5+6 2 | | | | 3+4 2 | 3+4 2 | |
| | 3+4 2 | 3+4 2 | | | | 5+6 2 | 5+6 2 | |
| | 5+6 2 | 3+4 2 | | | | 5+6 2 | 5+6 2 | |
| | 5+6 2 | 3+4 2 | | | | 3+4 2 | 3+4 2 | |
| | 5+6 2 | 5+6 2 | | | | 5+6 2 | 3+4 2 | |
| | 3+4 2 | 5+6 2 | | | | 3+4 2 | 5+6 2 | |
| | 5+6 2 | 3+4 2 | | | | 5+6 2 | 5+6 2 | |
| | 3+4 2 | 3+4 2 | | | | 5+6 2 | 5+6 2 | |
| | 3+4 2 | 5+6 2 | | | | 3+4 2 | 5+6 2 | |
| | 3+4 2 | 5+6 2 | | | | 3+4 2 | 5+6 2 | |
| | 3+4 2 | 3+4 2 | | | | 3+4 2 | 3+4 2 | |
| 11.2.33 | 62486 | 60 | Mo2S3 | | | | | |
| | -A | -B | -C | -D | -E | -GM | -Y | -Z |
| | 5+6 2 | 3+4 2 | 2+2 4 | 2+2 4 | 2+2 4 | 3+4 2 | 3+4 2 | 2+2 4 |
| | 3+4 2 | 5+6 2 | 2+2 4 | 2+2 4 | 2+2 4 | 5+6 2 | 5+6 2 | 2+2 4 |
| | 3+4 2 | 5+6 2 | 2+2 4 | 2+2 4 | 2+2 4 | 3+4 2 | 3+4 2 | 2+2 4 |
| | 5+6 2 | 3+4 2 | 2+2 4 | 2+2 4 | 2+2 4 | 5+6 2 | 5+6 2 | 2+2 4 |
| | 3+4 2 | 5+6 2 | 2+2 4 | 2+2 4 | 2+2 4 | 3+4 2 | 5+6 2 | 2+2 4 |
| | 5+6 2 | 3+4 2 | 2+2 4 | 2+2 4 | 2+2 4 | 5+6 2 | 3+4 2 | 2+2 4 |
| | 3+4 2 | 5+6 2 | 2+2 4 | 2+2 4 | 2+2 4 | 5+6 2 | 5+6 2 | 2+2 4 |
| | 3+4 2 | 3+4 2 | 2+2 4 | 2+2 4 | 2+2 4 | 3+4 2 | 3+4 2 | 2+2 4 |
| | 5+6 2 | 5+6 2 | 2+2 4 | 2+2 4 | 2+2 4 | 5+6 2 | 3+4 2 | 2+2 4 |
| | 3+4 2 | 3+4 2 | 2+2 4 | 2+2 4 | 2+2 4 | 3+4 2 | 5+6 2 | 2+2 4 |
| | 3+4 2 | 3+4 2 | 2+2 4 | 2+2 4 | 2+2 4 | 5+6 2 | 3+4 2 | 2+2 4 |
| | 5+6 2 | 5+6 2 | 2+2 4 | 2+2 4 | 2+2 4 | 3+4 2 | 5+6 2 | 2+2 4 |
| | 5+6 2 | 3+4 2 | 2+2 4 | 2+2 4 | 2+2 4 | 3+4 2 | 5+6 2 | 2+2 4 |
| | 3+4 2 | 5+6 2 | 2+2 4 | 2+2 4 | 2+2 4 | 5+6 2 | 3+4 2 | 2+2 4 |
| | 3+4 2 | 3+4 2 | 2+2 4 | 2+2 4 | 2+2 4 | 3+4 2 | 5+6 2 | 2+2 4 |
| | 5+6 2 | 5+6 2 | | | | 5+6 2 | 3+4 2 | |
| | 5+6 2 | 3+4 2 | | | | 3+4 2 | 5+6 2 | |
| | 3+4 2 | 5+6 2 | | | | 5+6 2 | 3+4 2 | |
| | 5+6 2 | 3+4 2 | | | | 3+4 2 | 3+4 2 | |
| | 5+6 2 | 3+4 2 | | | | 3+4 2 | 5+6 2 | |
| | 5+6 2 | 5+6 2 | | | | 5+6 2 | 3+4 2 | |
| | 3+4 2 | 5+6 2 | | | | 5+6 2 | 5+6 2 | |
| | 3+4 2 | 5+6 2 | | | | 3+4 2 | 3+4 2 | |
| | 3+4 2 | 3+4 2 | | | | 5+6 2 | 3+4 2 | |
| | 5+6 2 | 3+4 2 | | | | 3+4 2 | 5+6 2 | |
| | 3+4 2 | 5+6 2 | | | | 5+6 2 | 5+6 2 | |
| | 5+6 2 | 5+6 2 | | | | 5+6 2 | 5+6 2 | |
| | 5+6 2 | 3+4 2 | | | | 3+4 2 | 5+6 2 | |
| | 5+6 2 | 3+4 2 | | | | 3+4 2 | 3+4 2 | |
| | 5+6 2 | 5+6 2 | | | | 3+4 2 | 5+6 2 | |

Continued on next column

**TABLE XXIII– continued**

| TQC | ICSD | NoE | SF | | | | | |
|---|---|---|---|---|---|---|---|---|
| 11.2.34 | 16261 | 60 | Mo2.06S3 | | | | | |
| | -A | -B | -C | -D | -E | -GM | -Y | -Z |
| | 3+4 2 | 5+6 2 | 2+2 4 | 2+2 4 | 2+2 4 | 3+4 2 | 5+6 2 | 2+2 4 |
| | 5+6 2 | 3+4 2 | 2+2 4 | 2+2 4 | 2+2 4 | 5+6 2 | 3+4 2 | 2+2 4 |
| | 5+6 2 | 3+4 2 | 2+2 4 | 2+2 4 | 2+2 4 | 3+4 2 | 3+4 2 | 2+2 4 |
| | 3+4 2 | 5+6 2 | 2+2 4 | 2+2 4 | 2+2 4 | 5+6 2 | 5+6 2 | 2+2 4 |
| | 5+6 2 | 3+4 2 | 2+2 4 | 2+2 4 | 2+2 4 | 3+4 2 | 5+6 2 | 2+2 4 |
| | 3+4 2 | 5+6 2 | 2+2 4 | 2+2 4 | 2+2 4 | 5+6 2 | 3+4 2 | 2+2 4 |
| | 5+6 2 | 5+6 2 | 2+2 4 | 2+2 4 | 2+2 4 | 5+6 2 | 5+6 2 | 2+2 4 |
| | 5+6 2 | 3+4 2 | 2+2 4 | 2+2 4 | 2+2 4 | 3+4 2 | 3+4 2 | 2+2 4 |
| | 3+4 2 | 3+4 2 | 2+2 4 | 2+2 4 | 2+2 4 | 5+6 2 | 3+4 2 | 2+2 4 |
| | 5+6 2 | 5+6 2 | 2+2 4 | 2+2 4 | 2+2 4 | 3+4 2 | 5+6 2 | 2+2 4 |
| | 5+6 2 | 5+6 2 | 2+2 4 | 2+2 4 | 2+2 4 | 5+6 2 | 3+4 2 | 2+2 4 |
| | 3+4 2 | 3+4 2 | 2+2 4 | 2+2 4 | 2+2 4 | 3+4 2 | 5+6 2 | 2+2 4 |
| | 3+4 2 | 3+4 2 | 2+2 4 | 2+2 4 | 2+2 4 | 3+4 2 | 5+6 2 | 2+2 4 |
| | 5+6 2 | 5+6 2 | 2+2 4 | 2+2 4 | 2+2 4 | 5+6 2 | 3+4 2 | 2+2 4 |
| | 5+6 2 | 5+6 2 | 2+2 4 | 2+2 4 | 2+2 4 | 3+4 2 | 5+6 2 | 2+2 4 |
| | 3+4 2 | 3+4 2 | | | | 5+6 2 | 5+6 2 | |
| | 3+4 2 | 5+6 2 | | | | 3+4 2 | 3+4 2 | |
| | 5+6 2 | 3+4 2 | | | | 5+6 2 | 3+4 2 | |
| | 3+4 2 | 5+6 2 | | | | 3+4 2 | 3+4 2 | |
| | 3+4 2 | 3+4 2 | | | | 3+4 2 | 3+4 2 | |
| | 3+4 2 | 5+6 2 | | | | 5+6 2 | 5+6 2 | |
| | 5+6 2 | 3+4 2 | | | | 5+6 2 | 5+6 2 | |
| | 5+6 2 | 3+4 2 | | | | 3+4 2 | 3+4 2 | |
| | 5+6 2 | 5+6 2 | | | | 5+6 2 | 5+6 2 | |
| | 3+4 2 | 5+6 2 | | | | 3+4 2 | 3+4 2 | |
| | 5+6 2 | 3+4 2 | | | | 5+6 2 | 5+6 2 | |
| | 3+4 2 | 3+4 2 | | | | 5+6 2 | 5+6 2 | |
| | 3+4 2 | 5+6 2 | | | | 3+4 2 | 3+4 2 | |
| | 3+4 2 | 5+6 2 | | | | 3+4 2 | 5+6 2 | |
| | 3+4 2 | 3+4 2 | | | | 3+4 2 | 5+6 2 | |
| 11.2.35 | 14349 | 72 | Mo1Te2 | | | | | |
| | -A | -B | -C | -D | -E | -GM | -Y | -Z |
| | 3+4 2 | 3+4 2 | 2+2 4 | 2+2 4 | 2+2 4 | 3+4 2 | 3+4 2 | 2+2 4 |
| | 3+4 2 | 5+6 2 | 2+2 4 | 2+2 4 | 2+2 4 | 3+4 2 | 5+6 2 | 2+2 4 |
| | 5+6 2 | 5+6 2 | 2+2 4 | 2+2 4 | 2+2 4 | 5+6 2 | 3+4 2 | 2+2 4 |
| | 5+6 2 | 3+4 2 | 2+2 4 | 2+2 4 | 2+2 4 | 5+6 2 | 5+6 2 | 2+2 4 |
| | 3+4 2 | 3+4 2 | 2+2 4 | 2+2 4 | 2+2 4 | 3+4 2 | 5+6 2 | 2+2 4 |
| | 3+4 2 | 5+6 2 | 2+2 4 | 2+2 4 | 2+2 4 | 3+4 2 | 3+4 2 | 2+2 4 |
| | 5+6 2 | 5+6 2 | 2+2 4 | 2+2 4 | 2+2 4 | 5+6 2 | 5+6 2 | 2+2 4 |
| | 5+6 2 | 3+4 2 | 2+2 4 | 2+2 4 | 2+2 4 | 5+6 2 | 3+4 2 | 2+2 4 |
| | 5+6 2 | 5+6 2 | 2+2 4 | 2+2 4 | 2+2 4 | 5+6 2 | 5+6 2 | 2+2 4 |
| | 3+4 2 | 3+4 2 | 2+2 4 | 2+2 4 | 2+2 4 | 3+4 2 | 3+4 2 | 2+2 4 |
| | 3+4 2 | 3+4 2 | 2+2 4 | 2+2 4 | 2+2 4 | 5+6 2 | 3+4 2 | 2+2 4 |
| | 5+6 2 | 5+6 2 | 2+2 4 | 2+2 4 | 2+2 4 | 3+4 2 | 5+6 2 | 2+2 4 |
| | 3+4 2 | 3+4 2 | 2+2 4 | 2+2 4 | 2+2 4 | 3+4 2 | 3+4 2 | 2+2 4 |
| | 3+4 2 | 5+6 2 | 2+2 4 | 2+2 4 | 2+2 4 | 3+4 2 | 5+6 2 | 2+2 4 |
| | 3+4 2 | 5+6 2 | 2+2 4 | 2+2 4 | 2+2 4 | 5+6 2 | 5+6 2 | 2+2 4 |
| | 3+4 2 | 3+4 2 | 2+2 4 | 2+2 4 | 2+2 4 | 5+6 2 | 3+4 2 | 2+2 4 |
| | 5+6 2 | 3+4 2 | 2+2 4 | 2+2 4 | 2+2 4 | 3+4 2 | 3+4 2 | 2+2 4 |
| | 5+6 2 | 5+6 2 | 2+2 4 | 2+2 4 | 2+2 4 | 5+6 2 | 5+6 2 | 2+2 4 |
| | 3+4 2 | 5+6 2 | | | | 5+6 2 | 3+4 2 | |
| | 3+4 2 | 3+4 2 | | | | 3+4 2 | 5+6 2 | |
| | 5+6 2 | 3+4 2 | | | | 5+6 2 | 5+6 2 | |
| | 5+6 2 | 5+6 2 | | | | 3+4 2 | 3+4 2 | |
| | 3+4 2 | 5+6 2 | | | | 3+4 2 | 3+4 2 | |
| | 3+4 2 | 3+4 2 | | | | 5+6 2 | 5+6 2 | |
| | 5+6 2 | 3+4 2 | | | | 3+4 2 | 5+6 2 | |
| | 5+6 2 | 5+6 2 | | | | 3+4 2 | 3+4 2 | |
| | 5+6 2 | 5+6 2 | | | | 5+6 2 | 5+6 2 | |
| | 5+6 2 | 3+4 2 | | | | 3+4 2 | 3+4 2 | |
| | 3+4 2 | 5+6 2 | | | | 5+6 2 | 5+6 2 | |
| | 3+4 2 | 3+4 2 | | | | 3+4 2 | 3+4 2 | |
| | 5+6 2 | 3+4 2 | | | | 3+4 2 | 5+6 2 | |
| | 5+6 2 | 5+6 2 | | | | 3+4 2 | 3+4 2 | |
| | 3+4 2 | 3+4 2 | | | | 5+6 2 | 3+4 2 | |
| | 3+4 2 | 5+6 2 | | | | 3+4 2 | 5+6 2 | |
| | 5+6 2 | 5+6 2 | | | | 5+6 2 | 3+4 2 | |
| | 5+6 2 | 3+4 2 | | | | 3+4 2 | 5+6 2 | |
| 11.2.36 | 59708 | 56 | Li4S4Si1 | | | | | |
| | -A | -B | -C | -D | -E | -GM | -Y | -Z |
| | 5+6 2 | 3+4 2 | 2+2 4 | 2+2 4 | 2+2 4 | 3+4 2 | 5+6 2 | 2+2 4 |
| | 3+4 2 | 5+6 2 | 2+2 4 | 2+2 4 | 2+2 4 | 5+6 2 | 3+4 2 | 2+2 4 |
| | 3+4 2 | 5+6 2 | 2+2 4 | 2+2 4 | 2+2 4 | 3+4 2 | 3+4 2 | 2+2 4 |
| | 5+6 2 | 5+6 2 | 2+2 4 | 2+2 4 | 2+2 4 | 5+6 2 | 5+6 2 | 2+2 4 |
| | 5+6 2 | 5+6 2 | 2+2 4 | 2+2 4 | 2+2 4 | 5+6 2 | 3+4 2 | 2+2 4 |
| | 3+4 2 | 3+4 2 | 2+2 4 | 2+2 4 | 2+2 4 | 5+6 2 | 5+6 2 | 2+2 4 |
| | 3+4 2 | 3+4 2 | 2+2 4 | 2+2 4 | 2+2 4 | 3+4 2 | 5+6 2 | 2+2 4 |
| | 5+6 2 | 3+4 2 | 2+2 4 | 2+2 4 | 2+2 4 | 3+4 2 | 3+4 2 | 2+2 4 |
| | 3+4 2 | 3+4 2 | 2+2 4 | 2+2 4 | 2+2 4 | 3+4 2 | 3+4 2 | 2+2 4 |
| | 5+6 2 | 5+6 2 | 2+2 4 | 2+2 4 | 2+2 4 | 5+6 2 | 5+6 2 | 2+2 4 |
| | 5+6 2 | 3+4 2 | 2+2 4 | 2+2 4 | 2+2 4 | 3+4 2 | 5+6 2 | 2+2 4 |
| | 5+6 2 | 5+6 2 | 2+2 4 | 2+2 4 | 2+2 4 | 5+6 2 | 5+6 2 | 2+2 4 |
| | 3+4 2 | 3+4 2 | 2+2 4 | 2+2 4 | 2+2 4 | 5+6 2 | 3+4 2 | 2+2 4 |
| | 5+6 2 | 5+6 2 | 2+2 4 | 2+2 4 | 2+2 4 | 3+4 2 | 3+4 2 | 2+2 4 |
| | 3+4 2 | 5+6 2 | | | | 3+4 2 | 5+6 2 | |
| | 3+4 2 | 3+4 2 | | | | 5+6 2 | 3+4 2 | |
| | 3+4 2 | 5+6 2 | | | | 3+4 2 | 3+4 2 | |
| | 3+4 2 | 3+4 2 | | | | 5+6 2 | 3+4 2 | |
| | 5+6 2 | 5+6 2 | | | | 5+6 2 | 5+6 2 | |
| | 3+4 2 | 5+6 2 | | | | 3+4 2 | 3+4 2 | |
| | 5+6 2 | 3+4 2 | | | | 3+4 2 | 5+6 2 | |
| | 5+6 2 | 3+4 2 | | | | 3+4 2 | 5+6 2 | |
| | 3+4 2 | 5+6 2 | | | | 5+6 2 | 5+6 2 | |
| | 5+6 2 | 5+6 2 | | | | 3+4 2 | 5+6 2 | |
| | 5+6 2 | 3+4 2 | | | | 5+6 2 | 3+4 2 | |
| | 3+4 2 | 3+4 2 | | | | 3+4 2 | 3+4 2 | |
| | 5+6 2 | 5+6 2 | | | | 5+6 2 | 5+6 2 | |
| | 5+6 2 | 5+6 2 | | | | 5+6 2 | 3+4 2 | |
| 11.2.37 | 68366 | 92 | In2S7Sn3 | | | | | |
| | -A | -B | -C | -D | -E | -GM | -Y | -Z |
| | 5+6 2 | 5+6 2 | 2+2 4 | 2+2 4 | 2+2 4 | 3+4 2 | 3+4 2 | 2+2 4 |
| | 3+4 2 | 3+4 2 | 2+2 4 | 2+2 4 | 2+2 4 | 3+4 2 | 5+6 2 | 2+2 4 |
| | 5+6 2 | 5+6 2 | 2+2 4 | 2+2 4 | 2+2 4 | 5+6 2 | 3+4 2 | 2+2 4 |
| | 3+4 2 | 3+4 2 | 2+2 4 | 2+2 4 | 2+2 4 | 3+4 2 | 5+6 2 | 2+2 4 |
| | 3+4 2 | 3+4 2 | 2+2 4 | 2+2 4 | 2+2 4 | 3+4 2 | 5+6 2 | 2+2 4 |
| | 5+6 2 | 5+6 2 | 2+2 4 | 2+2 4 | 2+2 4 | 5+6 2 | 3+4 2 | 2+2 4 |
| | 3+4 2 | 3+4 2 | 2+2 4 | 2+2 4 | 2+2 4 | 5+6 2 | 5+6 2 | 2+2 4 |
| | 5+6 2 | 3+4 2 | 2+2 4 | 2+2 4 | 2+2 4 | 5+6 2 | 5+6 2 | 2+2 4 |
| | 3+4 2 | 5+6 2 | 2+2 4 | 2+2 4 | 2+2 4 | 3+4 2 | 3+4 2 | 2+2 4 |
| | 5+6 2 | 5+6 2 | 2+2 4 | 2+2 4 | 2+2 4 | 3+4 2 | 3+4 2 | 2+2 4 |
| | 5+6 2 | 5+6 2 | 2+2 4 | 2+2 4 | 2+2 4 | 3+4 2 | 3+4 2 | 2+2 4 |
| | 3+4 2 | 3+4 2 | 2+2 4 | 2+2 4 | 2+2 4 | 5+6 2 | 5+6 2 | 2+2 4 |
| | 5+6 2 | 5+6 2 | 2+2 4 | 2+2 4 | 2+2 4 | 5+6 2 | 3+4 2 | 2+2 4 |
| | 3+4 2 | 3+4 2 | 2+2 4 | 2+2 4 | 2+2 4 | 5+6 2 | 5+6 2 | 2+2 4 |
| | 5+6 2 | 5+6 2 | 2+2 4 | 2+2 4 | 2+2 4 | 3+4 2 | 3+4 2 | 2+2 4 |
| | 3+4 2 | 3+4 2 | 2+2 4 | 2+2 4 | 2+2 4 | 5+6 2 | 5+6 2 | 2+2 4 |
| | 5+6 2 | 5+6 2 | 2+2 4 | 2+2 4 | 2+2 4 | 3+4 2 | 5+6 2 | 2+2 4 |
| | 3+4 2 | 3+4 2 | 2+2 4 | 2+2 4 | 2+2 4 | 5+6 2 | 3+4 2 | 2+2 4 |
| | 3+4 2 | 5+6 2 | 2+2 4 | 2+2 4 | 2+2 4 | 3+4 2 | 3+4 2 | 2+2 4 |
| | 5+6 2 | 3+4 2 | 2+2 4 | 2+2 4 | 2+2 4 | 5+6 2 | 5+6 2 | 2+2 4 |

Continued on next column

Continued on next column

**TABLE XXIII– continued**

| TQC | ICSD | | NoE | SF | | | | |
|---|---|---|---|---|---|---|---|---|
| | 5+6 2 | 3+4 2 | 2+2 4 | 2+2 4 | 2+2 4 | 3+4 2 | 3+4 2 | 2+2 4 |
| | 3+4 2 | 5+6 2 | 2+2 4 | 2+2 4 | 2+2 4 | 3+4 2 | 5+6 2 | 2+2 4 |
| | 3+4 2 | 3+4 2 | 2+2 4 | 2+2 4 | 2+2 4 | 5+6 2 | 3+4 2 | 2+2 4 |
| | 5+6 2 | 5+6 2 | | | | 5+6 2 | 5+6 2 | |
| | 3+4 2 | 3+4 2 | | | | 3+4 2 | 5+6 2 | |
| | 5+6 2 | 5+6 2 | | | | 3+4 2 | 3+4 2 | |
| | 3+4 2 | 5+6 2 | | | | 5+6 2 | 5+6 2 | |
| | 5+6 2 | 3+4 2 | | | | 3+4 2 | 3+4 2 | |
| | 5+6 2 | 5+6 2 | | | | 5+6 2 | 5+6 2 | |
| | 3+4 2 | 3+4 2 | | | | 3+4 2 | 3+4 2 | |
| | 5+6 2 | 3+4 2 | | | | 3+4 2 | 3+4 2 | |
| | 3+4 2 | 5+6 2 | | | | 5+6 2 | 5+6 2 | |
| | 5+6 2 | 5+6 2 | | | | 3+4 2 | 5+6 2 | |
| | 3+4 2 | 5+6 2 | | | | 5+6 2 | 3+4 2 | |
| | 5+6 2 | 3+4 2 | | | | 5+6 2 | 3+4 2 | |
| | 3+4 2 | 5+6 2 | | | | 3+4 2 | 5+6 2 | |
| | 3+4 2 | 3+4 2 | | | | 5+6 2 | 3+4 2 | |
| | 3+4 2 | 3+4 2 | | | | 3+4 2 | 5+6 2 | |
| | 5+6 2 | 3+4 2 | | | | 3+4 2 | 5+6 2 | |
| | 5+6 2 | 5+6 2 | | | | 3+4 2 | 3+4 2 | |
| | 3+4 2 | 3+4 2 | | | | 5+6 2 | 5+6 2 | |
| | 5+6 2 | 5+6 2 | | | | 5+6 2 | 3+4 2 | |
| | 3+4 2 | 3+4 2 | | | | 3+4 2 | 5+6 2 | |
| | 3+4 2 | 5+6 2 | | | | 5+6 2 | 5+6 2 | |
| | 5+6 2 | 3+4 2 | | | | 3+4 2 | 3+4 2 | |
| | 5+6 2 | 5+6 2 | | | | 5+6 2 | 5+6 2 | |
| 11.2.38 | 26525 | | 124 | H8Ba1O7Pt1 | | | | |
| | -A | -B | -C | -D | -E | -GM | -Y | -Z |
| | 3+4 2 | 3+4 2 | 2+2 4 | 2+2 4 | 2+2 4 | 3+4 2 | 3+4 2 | 2+2 4 |
| | 5+6 2 | 5+6 2 | 2+2 4 | 2+2 4 | 2+2 4 | 5+6 2 | 5+6 2 | 2+2 4 |
| | 5+6 2 | 5+6 2 | 2+2 4 | 2+2 4 | 2+2 4 | 3+4 2 | 3+4 2 | 2+2 4 |
| | 3+4 2 | 3+4 2 | 2+2 4 | 2+2 4 | 2+2 4 | 5+6 2 | 5+6 2 | 2+2 4 |
| | 3+4 2 | 5+6 2 | 2+2 4 | 2+2 4 | 2+2 4 | 5+6 2 | 5+6 2 | 2+2 4 |
| | 5+6 2 | 3+4 2 | 2+2 4 | 2+2 4 | 2+2 4 | 3+4 2 | 3+4 2 | 2+2 4 |
| | 3+4 2 | 3+4 2 | 2+2 4 | 2+2 4 | 2+2 4 | 3+4 2 | 5+6 2 | 2+2 4 |
| | 5+6 2 | 5+6 2 | 2+2 4 | 2+2 4 | 2+2 4 | 5+6 2 | 3+4 2 | 2+2 4 |
| | 3+4 2 | 3+4 2 | 2+2 4 | 2+2 4 | 2+2 4 | 3+4 2 | 3+4 2 | 2+2 4 |
| | 5+6 2 | 5+6 2 | 2+2 4 | 2+2 4 | 2+2 4 | 5+6 2 | 5+6 2 | 2+2 4 |
| | 5+6 2 | 3+4 2 | 2+2 4 | 2+2 4 | 2+2 4 | 3+4 2 | 5+6 2 | 2+2 4 |
| | 3+4 2 | 5+6 2 | 2+2 4 | 2+2 4 | 2+2 4 | 3+4 2 | 3+4 2 | 2+2 4 |
| | 5+6 2 | 5+6 2 | 2+2 4 | 2+2 4 | 2+2 4 | 5+6 2 | 5+6 2 | 2+2 4 |
| | 3+4 2 | 3+4 2 | 2+2 4 | 2+2 4 | 2+2 4 | 5+6 2 | 3+4 2 | 2+2 4 |
| | 3+4 2 | 5+6 2 | 2+2 4 | 2+2 4 | 2+2 4 | 3+4 2 | 5+6 2 | 2+2 4 |
| | 5+6 2 | 3+4 2 | 2+2 4 | 2+2 4 | 2+2 4 | 5+6 2 | 3+4 2 | 2+2 4 |
| | 3+4 2 | 3+4 2 | 2+2 4 | 2+2 4 | 2+2 4 | 3+4 2 | 3+4 2 | 2+2 4 |
| | 5+6 2 | 5+6 2 | 2+2 4 | 2+2 4 | 2+2 4 | 5+6 2 | 5+6 2 | 2+2 4 |
| | 3+4 2 | 3+4 2 | 2+2 4 | 2+2 4 | 2+2 4 | 3+4 2 | 3+4 2 | 2+2 4 |
| | 3+4 2 | 3+4 2 | 2+2 4 | 2+2 4 | 2+2 4 | 3+4 2 | 3+4 2 | 2+2 4 |
| | 5+6 2 | 5+6 2 | 2+2 4 | 2+2 4 | 2+2 4 | 5+6 2 | 5+6 2 | 2+2 4 |
| | 5+6 2 | 5+6 2 | 2+2 4 | 2+2 4 | 2+2 4 | 5+6 2 | 5+6 2 | 2+2 4 |
| | 5+6 2 | 5+6 2 | 2+2 4 | 2+2 4 | 2+2 4 | 5+6 2 | 5+6 2 | 2+2 4 |
| | 3+4 2 | 5+6 2 | 2+2 4 | 2+2 4 | 2+2 4 | 5+6 2 | 3+4 2 | 2+2 4 |
| | 3+4 2 | 3+4 2 | 2+2 4 | 2+2 4 | 2+2 4 | 3+4 2 | 5+6 2 | 2+2 4 |
| | 5+6 2 | 3+4 2 | 2+2 4 | 2+2 4 | 2+2 4 | 3+4 2 | 3+4 2 | 2+2 4 |
| | 5+6 2 | 5+6 2 | 2+2 4 | 2+2 4 | 2+2 4 | 5+6 2 | 5+6 2 | 2+2 4 |
| | 3+4 2 | 5+6 2 | 2+2 4 | 2+2 4 | 2+2 4 | 5+6 2 | 3+4 2 | 2+2 4 |
| | 5+6 2 | 3+4 2 | 2+2 4 | 2+2 4 | 2+2 4 | 3+4 2 | 5+6 2 | 2+2 4 |
| | 3+4 2 | 3+4 2 | 2+2 4 | 2+2 4 | 2+2 4 | 3+4 2 | 3+4 2 | 2+2 4 |
| | 5+6 2 | 5+6 2 | 2+2 4 | 2+2 4 | 2+2 4 | 5+6 2 | 3+4 2 | 2+2 4 |
| | 3+4 2 | 3+4 2 | | | | 3+4 2 | 5+6 2 | |
| | 5+6 2 | 3+4 2 | | | | 3+4 2 | 3+4 2 | |

Continued on next column

**TABLE XXIII– continued**

| TQC | ICSD | | NoE | SF | | | | |
|---|---|---|---|---|---|---|---|---|
| | 3+4 2 | 5+6 2 | | | | 5+6 2 | 5+6 2 | |
| | 3+4 2 | 3+4 2 | | | | 5+6 2 | 5+6 2 | |
| | 5+6 2 | 5+6 2 | | | | 3+4 2 | 3+4 2 | |
| | 3+4 2 | 3+4 2 | | | | 3+4 2 | 5+6 2 | |
| | 5+6 2 | 5+6 2 | | | | 5+6 2 | 3+4 2 | |
| | 3+4 2 | 3+4 2 | | | | 5+6 2 | 5+6 2 | |
| | 5+6 2 | 5+6 2 | | | | 3+4 2 | 3+4 2 | |
| | 3+4 2 | 3+4 2 | | | | 3+4 2 | 3+4 2 | |
| | 5+6 2 | 3+4 2 | | | | 3+4 2 | 5+6 2 | |
| | 5+6 2 | 5+6 2 | | | | 3+4 2 | 5+6 2 | |
| | 5+6 2 | 3+4 2 | | | | 5+6 2 | 5+6 2 | |
| | 3+4 2 | 5+6 2 | | | | 3+4 2 | 3+4 2 | |
| | 5+6 2 | 3+4 2 | | | | 5+6 2 | 3+4 2 | |
| | 3+4 2 | 5+6 2 | | | | 5+6 2 | 5+6 2 | |
| | 3+4 2 | 5+6 2 | | | | 3+4 2 | 3+4 2 | |
| | 5+6 2 | 3+4 2 | | | | 5+6 2 | 5+6 2 | |
| | 5+6 2 | 5+6 2 | | | | 3+4 2 | 3+4 2 | |
| | 3+4 2 | 3+4 2 | | | | 3+4 2 | 3+4 2 | |
| | 3+4 2 | 5+6 2 | | | | 5+6 2 | 5+6 2 | |
| | 5+6 2 | 5+6 2 | | | | 3+4 2 | 5+6 2 | |
| | 3+4 2 | 5+6 2 | | | | 5+6 2 | 3+4 2 | |
| | 3+4 2 | 3+4 2 | | | | 5+6 2 | 3+4 2 | |
| | 5+6 2 | 3+4 2 | | | | 5+6 2 | 5+6 2 | |
| | 3+4 2 | 5+6 2 | | | | 3+4 2 | 5+6 2 | |
| | 5+6 2 | 3+4 2 | | | | 5+6 2 | 3+4 2 | |
| | 3+4 2 | 5+6 2 | | | | 5+6 2 | 5+6 2 | |
| | 5+6 2 | 3+4 2 | | | | 3+4 2 | 5+6 2 | |
| | 3+4 2 | 5+6 2 | | | | 5+6 2 | 3+4 2 | |
| | 5+6 2 | 3+4 2 | | | | 3+4 2 | 5+6 2 | |
| 11.2.39 | 437 | | 72 | H6Br3N1O1Sn1 | | | | |
| | -A | -B | -C | -D | -E | -GM | -Y | -Z |
| | 3+4 2 | 5+6 2 | 2+2 4 | 2+2 4 | 2+2 4 | 3+4 2 | 5+6 2 | 2+2 4 |
| | 5+6 2 | 3+4 2 | 2+2 4 | 2+2 4 | 2+2 4 | 5+6 2 | 3+4 2 | 2+2 4 |
| | 5+6 2 | 3+4 2 | 2+2 4 | 2+2 4 | 2+2 4 | 3+4 2 | 5+6 2 | 2+2 4 |
| | 3+4 2 | 5+6 2 | 2+2 4 | 2+2 4 | 2+2 4 | 5+6 2 | 3+4 2 | 2+2 4 |
| | 3+4 2 | 5+6 2 | 2+2 4 | 2+2 4 | 2+2 4 | 5+6 2 | 3+4 2 | 2+2 4 |
| | 5+6 2 | 3+4 2 | 2+2 4 | 2+2 4 | 2+2 4 | 3+4 2 | 5+6 2 | 2+2 4 |
| | 5+6 2 | 3+4 2 | 2+2 4 | 2+2 4 | 2+2 4 | 3+4 2 | 3+4 2 | 2+2 4 |
| | 3+4 2 | 5+6 2 | 2+2 4 | 2+2 4 | 2+2 4 | 5+6 2 | 5+6 2 | 2+2 4 |
| | 3+4 2 | 3+4 2 | 2+2 4 | 2+2 4 | 2+2 4 | 5+6 2 | 5+6 2 | 2+2 4 |
| | 5+6 2 | 5+6 2 | 2+2 4 | 2+2 4 | 2+2 4 | 3+4 2 | 3+4 2 | 2+2 4 |
| | 5+6 2 | 3+4 2 | 2+2 4 | 2+2 4 | 2+2 4 | 3+4 2 | 5+6 2 | 2+2 4 |
| | 3+4 2 | 5+6 2 | 2+2 4 | 2+2 4 | 2+2 4 | 5+6 2 | 3+4 2 | 2+2 4 |
| | 5+6 2 | 3+4 2 | 2+2 4 | 2+2 4 | 2+2 4 | 3+4 2 | 5+6 2 | 2+2 4 |
| | 5+6 2 | 3+4 2 | 2+2 4 | 2+2 4 | 2+2 4 | 3+4 2 | 3+4 2 | 2+2 4 |
| | 3+4 2 | 5+6 2 | 2+2 4 | 2+2 4 | 2+2 4 | 5+6 2 | 5+6 2 | 2+2 4 |
| | 3+4 2 | 5+6 2 | 2+2 4 | 2+2 4 | 2+2 4 | 5+6 2 | 3+4 2 | 2+2 4 |
| | 3+4 2 | 5+6 2 | 2+2 4 | 2+2 4 | 2+2 4 | 3+4 2 | 3+4 2 | 2+2 4 |
| | 5+6 2 | 3+4 2 | 2+2 4 | 2+2 4 | 2+2 4 | 3+4 2 | 5+6 2 | 2+2 4 |
| | 3+4 2 | 5+6 2 | | | | 5+6 2 | 3+4 2 | |
| | 3+4 2 | 3+4 2 | | | | 5+6 2 | 5+6 2 | |
| | 5+6 2 | 5+6 2 | | | | 3+4 2 | 3+4 2 | |
| | 5+6 2 | 3+4 2 | | | | 5+6 2 | 5+6 2 | |
| | 5+6 2 | 3+4 2 | | | | 3+4 2 | 5+6 2 | |
| | 5+6 2 | 5+6 2 | | | | 5+6 2 | 3+4 2 | |
| | 3+4 2 | 3+4 2 | | | | 3+4 2 | 5+6 2 | |
| | 3+4 2 | 5+6 2 | | | | 5+6 2 | 3+4 2 | |
| | 3+4 2 | 3+4 2 | | | | 5+6 2 | 3+4 2 | |
| | 5+6 2 | 5+6 2 | | | | 3+4 2 | 5+6 2 | |
| | 3+4 2 | 3+4 2 | | | | 3+4 2 | 3+4 2 | |
| | 5+6 2 | 5+6 2 | | | | 5+6 2 | 5+6 2 | |

Continued on next column

**TABLE XXIII– continued**

| TQC | ICSD | NoE | SF | | | | | |
|---|---|---|---|---|---|---|---|---|
| | 5+6 2 | 5+6 2 | | | | 5+6 2 | 3+4 2 | |
| | 3+4 2 | 3+4 2 | | | | 3+4 2 | 3+4 2 | |
| | 3+4 2 | 5+6 2 | | | | 5+6 2 | 5+6 2 | |
| | 5+6 2 | 3+4 2 | | | | 3+4 2 | 5+6 2 | |
| | 3+4 2 | 5+6 2 | | | | 3+4 2 | 3+4 2 | |
| | 3+4 2 | 3+4 2 | | | | 5+6 2 | 5+6 2 | |
| 11.2.40 | 24075 | 204 | H5In1O7S1 | | | | | |
| | -A | -B | -C | -D | -E | -GM | -Y | -Z |
| | 5+6 2 | 5+6 2 | 2+2 4 | 2+2 4 | 2+2 4 | 5+6 2 | 5+6 2 | 2+2 4 |
| | 3+4 2 | 3+4 2 | 2+2 4 | 2+2 4 | 2+2 4 | 3+4 2 | 3+4 2 | 2+2 4 |
| | 5+6 2 | 3+4 2 | 2+2 4 | 2+2 4 | 2+2 4 | 5+6 2 | 3+4 2 | 2+2 4 |
| | 3+4 2 | 5+6 2 | 2+2 4 | 2+2 4 | 2+2 4 | 3+4 2 | 5+6 2 | 2+2 4 |
| | 5+6 2 | 5+6 2 | 2+2 4 | 2+2 4 | 2+2 4 | 5+6 2 | 5+6 2 | 2+2 4 |
| | 3+4 2 | 3+4 2 | 2+2 4 | 2+2 4 | 2+2 4 | 3+4 2 | 3+4 2 | 2+2 4 |
| | 5+6 2 | 5+6 2 | 2+2 4 | 2+2 4 | 2+2 4 | 5+6 2 | 5+6 2 | 2+2 4 |
| | 3+4 2 | 3+4 2 | 2+2 4 | 2+2 4 | 2+2 4 | 3+4 2 | 3+4 2 | 2+2 4 |
| | 5+6 2 | 3+4 2 | 2+2 4 | 2+2 4 | 2+2 4 | 5+6 2 | 3+4 2 | 2+2 4 |
| | 3+4 2 | 5+6 2 | 2+2 4 | 2+2 4 | 2+2 4 | 3+4 2 | 5+6 2 | 2+2 4 |
| | 5+6 2 | 3+4 2 | 2+2 4 | 2+2 4 | 2+2 4 | 5+6 2 | 3+4 2 | 2+2 4 |
| | 3+4 2 | 5+6 2 | 2+2 4 | 2+2 4 | 2+2 4 | 3+4 2 | 5+6 2 | 2+2 4 |
| | 5+6 2 | 3+4 2 | 2+2 4 | 2+2 4 | 2+2 4 | 3+4 2 | 3+4 2 | 2+2 4 |
| | 3+4 2 | 5+6 2 | 2+2 4 | 2+2 4 | 2+2 4 | 5+6 2 | 5+6 2 | 2+2 4 |
| | 3+4 2 | 3+4 2 | 2+2 4 | 2+2 4 | 2+2 4 | 3+4 2 | 3+4 2 | 2+2 4 |
| | 5+6 2 | 5+6 2 | 2+2 4 | 2+2 4 | 2+2 4 | 5+6 2 | 5+6 2 | 2+2 4 |
| | 3+4 2 | 3+4 2 | 2+2 4 | 2+2 4 | 2+2 4 | 3+4 2 | 5+6 2 | 2+2 4 |
| | 3+4 2 | 5+6 2 | 2+2 4 | 2+2 4 | 2+2 4 | 3+4 2 | 3+4 2 | 2+2 4 |
| | 5+6 2 | 5+6 2 | 2+2 4 | 2+2 4 | 2+2 4 | 5+6 2 | 5+6 2 | 2+2 4 |
| | 5+6 2 | 3+4 2 | 2+2 4 | 2+2 4 | 2+2 4 | 5+6 2 | 3+4 2 | 2+2 4 |
| | 5+6 2 | 5+6 2 | 2+2 4 | 2+2 4 | 2+2 4 | 5+6 2 | 5+6 2 | 2+2 4 |
| | 5+6 2 | 5+6 2 | 2+2 4 | 2+2 4 | 2+2 4 | 5+6 2 | 5+6 2 | 2+2 4 |
| | 5+6 2 | 3+4 2 | 2+2 4 | 2+2 4 | 2+2 4 | 5+6 2 | 3+4 2 | 2+2 4 |
| | 5+6 2 | 3+4 2 | 2+2 4 | 2+2 4 | 2+2 4 | 5+6 2 | 3+4 2 | 2+2 4 |
| | 3+4 2 | 3+4 2 | 2+2 4 | 2+2 4 | 2+2 4 | 3+4 2 | 3+4 2 | 2+2 4 |
| | 3+4 2 | 5+6 2 | 2+2 4 | 2+2 4 | 2+2 4 | 3+4 2 | 5+6 2 | 2+2 4 |
| | 3+4 2 | 3+4 2 | 2+2 4 | 2+2 4 | 2+2 4 | 3+4 2 | 3+4 2 | 2+2 4 |
| | 3+4 2 | 5+6 2 | 2+2 4 | 2+2 4 | 2+2 4 | 3+4 2 | 5+6 2 | 2+2 4 |
| | 5+6 2 | 5+6 2 | 2+2 4 | 2+2 4 | 2+2 4 | 5+6 2 | 5+6 2 | 2+2 4 |
| | 3+4 2 | 3+4 2 | 2+2 4 | 2+2 4 | 2+2 4 | 5+6 2 | 3+4 2 | 2+2 4 |
| | 5+6 2 | 3+4 2 | 2+2 4 | 2+2 4 | 2+2 4 | 3+4 2 | 3+4 2 | 2+2 4 |
| | 3+4 2 | 5+6 2 | 2+2 4 | 2+2 4 | 2+2 4 | 3+4 2 | 5+6 2 | 2+2 4 |
| | 3+4 2 | 3+4 2 | 2+2 4 | 2+2 4 | 2+2 4 | 3+4 2 | 3+4 2 | 2+2 4 |
| | 5+6 2 | 5+6 2 | 2+2 4 | 2+2 4 | 2+2 4 | 3+4 2 | 5+6 2 | 2+2 4 |
| | 3+4 2 | 5+6 2 | 2+2 4 | 2+2 4 | 2+2 4 | 5+6 2 | 5+6 2 | 2+2 4 |
| | 5+6 2 | 3+4 2 | 2+2 4 | 2+2 4 | 2+2 4 | 5+6 2 | 3+4 2 | 2+2 4 |
| | 5+6 2 | 5+6 2 | 2+2 4 | 2+2 4 | 2+2 4 | 3+4 2 | 5+6 2 | 2+2 4 |
| | 3+4 2 | 3+4 2 | 2+2 4 | 2+2 4 | 2+2 4 | 5+6 2 | 3+4 2 | 2+2 4 |
| | 5+6 2 | 3+4 2 | 2+2 4 | 2+2 4 | 2+2 4 | 5+6 2 | 3+4 2 | 2+2 4 |
| | 3+4 2 | 5+6 2 | 2+2 4 | 2+2 4 | 2+2 4 | 3+4 2 | 5+6 2 | 2+2 4 |
| | 5+6 2 | 3+4 2 | 2+2 4 | 2+2 4 | 2+2 4 | 3+4 2 | 5+6 2 | 2+2 4 |
| | 3+4 2 | 5+6 2 | 2+2 4 | 2+2 4 | 2+2 4 | 5+6 2 | 3+4 2 | 2+2 4 |
| | 5+6 2 | 3+4 2 | 2+2 4 | 2+2 4 | 2+2 4 | 5+6 2 | 3+4 2 | 2+2 4 |
| | 3+4 2 | 5+6 2 | 2+2 4 | 2+2 4 | 2+2 4 | 3+4 2 | 5+6 2 | 2+2 4 |
| | 3+4 2 | 3+4 2 | 2+2 4 | 2+2 4 | 2+2 4 | 3+4 2 | 5+6 2 | 2+2 4 |
| | 3+4 2 | 5+6 2 | 2+2 4 | 2+2 4 | 2+2 4 | 3+4 2 | 3+4 2 | 2+2 4 |
| | 3+4 2 | 5+6 2 | 2+2 4 | 2+2 4 | 2+2 4 | 5+6 2 | 3+4 2 | 2+2 4 |
| | 5+6 2 | 5+6 2 | 2+2 4 | 2+2 4 | 2+2 4 | 3+4 2 | 5+6 2 | 2+2 4 |
| | 3+4 2 | 3+4 2 | 2+2 4 | 2+2 4 | 2+2 4 | 5+6 2 | 5+6 2 | 2+2 4 |
| | 5+6 2 | 5+6 2 | 2+2 4 | 2+2 4 | 2+2 4 | 3+4 2 | 3+4 2 | 2+2 4 |
| | 5+6 2 | 3+4 2 | 2+2 4 | 2+2 4 | 2+2 4 | 5+6 2 | 5+6 2 | 2+2 4 |
| | 5+6 2 | 3+4 2 | | | | 5+6 2 | 3+4 2 | |
| | 5+6 2 | 3+4 2 | | | | 5+6 2 | 3+4 2 | |

Continued on next column

**TABLE XXIII– continued**

| TQC | ICSD | NoE | SF | | | | | |
|---|---|---|---|---|---|---|---|---|
| | 3+4 2 | 5+6 2 | | | | 3+4 2 | 5+6 2 | |
| | 5+6 2 | 3+4 2 | | | | 5+6 2 | 3+4 2 | |
| | 3+4 2 | 5+6 2 | | | | 3+4 2 | 5+6 2 | |
| | 5+6 2 | 5+6 2 | | | | 5+6 2 | 3+4 2 | |
| | 3+4 2 | 3+4 2 | | | | 5+6 2 | 3+4 2 | |
| | 5+6 2 | 3+4 2 | | | | 3+4 2 | 5+6 2 | |
| | 3+4 2 | 5+6 2 | | | | 3+4 2 | 5+6 2 | |
| | 5+6 2 | 5+6 2 | | | | 5+6 2 | 3+4 2 | |
| | 3+4 2 | 3+4 2 | | | | 3+4 2 | 5+6 2 | |
| | 5+6 2 | 3+4 2 | | | | 5+6 2 | 3+4 2 | |
| | 3+4 2 | 3+4 2 | | | | 3+4 2 | 5+6 2 | |
| | 5+6 2 | 5+6 2 | | | | 5+6 2 | 5+6 2 | |
| | 3+4 2 | 5+6 2 | | | | 3+4 2 | 3+4 2 | |
| | 5+6 2 | 3+4 2 | | | | 5+6 2 | 3+4 2 | |
| | 3+4 2 | 5+6 2 | | | | 3+4 2 | 5+6 2 | |
| | 5+6 2 | 5+6 2 | | | | 5+6 2 | 5+6 2 | |
| | 5+6 2 | 3+4 2 | | | | 5+6 2 | 3+4 2 | |
| | 5+6 2 | 3+4 2 | | | | 5+6 2 | 5+6 2 | |
| | 3+4 2 | 5+6 2 | | | | 3+4 2 | 5+6 2 | |
| | 5+6 2 | 3+4 2 | | | | 5+6 2 | 3+4 2 | |
| | 3+4 2 | 5+6 2 | | | | 3+4 2 | 3+4 2 | |
| | 3+4 2 | 5+6 2 | | | | 3+4 2 | 3+4 2 | |
| | 5+6 2 | 3+4 2 | | | | 5+6 2 | 5+6 2 | |
| | 3+4 2 | 3+4 2 | | | | 3+4 2 | 5+6 2 | |
| | 5+6 2 | 3+4 2 | | | | 5+6 2 | 3+4 2 | |
| | 5+6 2 | 5+6 2 | | | | 3+4 2 | 3+4 2 | |
| | 3+4 2 | 5+6 2 | | | | 5+6 2 | 5+6 2 | |
| | 3+4 2 | 3+4 2 | | | | 5+6 2 | 3+4 2 | |
| | 5+6 2 | 5+6 2 | | | | 3+4 2 | 3+4 2 | |
| | 5+6 2 | 3+4 2 | | | | 3+4 2 | 5+6 2 | |
| | 3+4 2 | 3+4 2 | | | | 3+4 2 | 5+6 2 | |
| | 5+6 2 | 5+6 2 | | | | 5+6 2 | 3+4 2 | |
| | 3+4 2 | 5+6 2 | | | | 5+6 2 | 5+6 2 | |
| | 5+6 2 | 3+4 2 | | | | 3+4 2 | 3+4 2 | |
| | 5+6 2 | 3+4 2 | | | | 5+6 2 | 3+4 2 | |
| | 5+6 2 | 5+6 2 | | | | 3+4 2 | 5+6 2 | |
| | 5+6 2 | 3+4 2 | | | | 3+4 2 | 3+4 2 | |
| | 3+4 2 | 5+6 2 | | | | 5+6 2 | 5+6 2 | |
| | 3+4 2 | 3+4 2 | | | | 5+6 2 | 5+6 2 | |
| | 5+6 2 | 5+6 2 | | | | 5+6 2 | 5+6 2 | |
| | 5+6 2 | 3+4 2 | | | | 3+4 2 | 5+6 2 | |
| | 3+4 2 | 5+6 2 | | | | 5+6 2 | 3+4 2 | |
| | 5+6 2 | 5+6 2 | | | | 3+4 2 | 3+4 2 | |
| | 3+4 2 | 3+4 2 | | | | 3+4 2 | 5+6 2 | |
| | 5+6 2 | 5+6 2 | | | | 5+6 2 | 5+6 2 | |
| | 3+4 2 | 5+6 2 | | | | 5+6 2 | 3+4 2 | |
| | 5+6 2 | 3+4 2 | | | | 5+6 2 | 3+4 2 | |
| | 3+4 2 | 3+4 2 | | | | 5+6 2 | 3+4 2 | |
| | 5+6 2 | 3+4 2 | | | | 3+4 2 | 5+6 2 | |
| 11.2.41 | 15199 | 112 | H5Fe1O7S1 | | | | | |
| | -A | -B | -C | -D | -E | -GM | -Y | -Z |
| | 3+4 2 | 5+6 2 | 2+2 4 | 2+2 4 | 2+2 4 | 3+4 2 | 5+6 2 | 2+2 4 |
| | 5+6 2 | 3+4 2 | 2+2 4 | 2+2 4 | 2+2 4 | 5+6 2 | 3+4 2 | 2+2 4 |
| | 3+4 2 | 3+4 2 | 2+2 4 | 2+2 4 | 2+2 4 | 3+4 2 | 3+4 2 | 2+2 4 |
| | 3+4 2 | 3+4 2 | 2+2 4 | 2+2 4 | 2+2 4 | 3+4 2 | 3+4 2 | 2+2 4 |
| | 5+6 2 | 5+6 2 | 2+2 4 | 2+2 4 | 2+2 4 | 5+6 2 | 5+6 2 | 2+2 4 |
| | 5+6 2 | 5+6 2 | 2+2 4 | 2+2 4 | 2+2 4 | 5+6 2 | 5+6 2 | 2+2 4 |
| | 5+6 2 | 3+4 2 | 2+2 4 | 2+2 4 | 2+2 4 | 5+6 2 | 3+4 2 | 2+2 4 |
| | 3+4 2 | 5+6 2 | 2+2 4 | 2+2 4 | 2+2 4 | 3+4 2 | 5+6 2 | 2+2 4 |
| | 3+4 2 | 3+4 2 | 2+2 4 | 2+2 4 | 2+2 4 | 3+4 2 | 3+4 2 | 2+2 4 |
| | 5+6 2 | 5+6 2 | 2+2 4 | 2+2 4 | 2+2 4 | 5+6 2 | 5+6 2 | 2+2 4 |

Continued on next column

**TABLE XXIII– continued**

| TQC | ICSD | NoE | SF | | | | | |
|---|---|---|---|---|---|---|---|---|
| | 3+4 2 | 3+4 2 | 2+2 4 | 2+2 4 | 2+2 4 | 3+4 2 | 5+6 2 | 2+2 4 |
| | 5+6 2 | 3+4 2 | 2+2 4 | 2+2 4 | 2+2 4 | 3+4 2 | 3+4 2 | 2+2 4 |
| | 3+4 2 | 5+6 2 | 2+2 4 | 2+2 4 | 2+2 4 | 5+6 2 | 3+4 2 | 2+2 4 |
| | 5+6 2 | 5+6 2 | 2+2 4 | 2+2 4 | 2+2 4 | 5+6 2 | 5+6 2 | 2+2 4 |
| | 3+4 2 | 3+4 2 | 2+2 4 | 2+2 4 | 2+2 4 | 3+4 2 | 3+4 2 | 2+2 4 |
| | 5+6 2 | 5+6 2 | 2+2 4 | 2+2 4 | 2+2 4 | 5+6 2 | 5+6 2 | 2+2 4 |
| | 3+4 2 | 3+4 2 | 2+2 4 | 2+2 4 | 2+2 4 | 3+4 2 | 3+4 2 | 2+2 4 |
| | 5+6 2 | 5+6 2 | 2+2 4 | 2+2 4 | 2+2 4 | 5+6 2 | 5+6 2 | 2+2 4 |
| | 3+4 2 | 3+4 2 | 2+2 4 | 2+2 4 | 2+2 4 | 3+4 2 | 3+4 2 | 2+2 4 |
| | 5+6 2 | 5+6 2 | 2+2 4 | 2+2 4 | 2+2 4 | 5+6 2 | 5+6 2 | 2+2 4 |
| | 3+4 2 | 3+4 2 | 2+2 4 | 2+2 4 | 2+2 4 | 3+4 2 | 3+4 2 | 2+2 4 |
| | 5+6 2 | 5+6 2 | 2+2 4 | 2+2 4 | 2+2 4 | 5+6 2 | 5+6 2 | 2+2 4 |
| | 3+4 2 | 3+4 2 | 2+2 4 | 2+2 4 | 2+2 4 | 3+4 2 | 3+4 2 | 2+2 4 |
| | 3+4 2 | 3+4 2 | 2+2 4 | 2+2 4 | 2+2 4 | 3+4 2 | 3+4 2 | 2+2 4 |
| | 5+6 2 | 3+4 2 | 2+2 4 | 2+2 4 | 2+2 4 | 3+4 2 | 3+4 2 | 2+2 4 |
| | 3+4 2 | 5+6 2 | 2+2 4 | 2+2 4 | 2+2 4 | 5+6 2 | 5+6 2 | 2+2 4 |
| | 3+4 2 | 5+6 2 | 2+2 4 | 2+2 4 | 2+2 4 | 5+6 2 | 5+6 2 | 2+2 4 |
| | 5+6 2 | 3+4 2 | 2+2 4 | 2+2 4 | 2+2 4 | 3+4 2 | 3+4 2 | 2+2 4 |
| | 3+4 2 | 5+6 2 | | | | 3+4 2 | 5+6 2 | |
| | 5+6 2 | 3+4 2 | | | | 5+6 2 | 3+4 2 | |
| | 5+6 2 | 3+4 2 | | | | 3+4 2 | 3+4 2 | |
| | 3+4 2 | 5+6 2 | | | | 5+6 2 | 5+6 2 | |
| | 3+4 2 | 3+4 2 | | | | 5+6 2 | 3+4 2 | |
| | 5+6 2 | 5+6 2 | | | | 3+4 2 | 5+6 2 | |
| | 3+4 2 | 3+4 2 | | | | 3+4 2 | 3+4 2 | |
| | 3+4 2 | 3+4 2 | | | | 3+4 2 | 3+4 2 | |
| | 3+4 2 | 5+6 2 | | | | 3+4 2 | 5+6 2 | |
| | 5+6 2 | 3+4 2 | | | | 5+6 2 | 3+4 2 | |
| | 3+4 2 | 3+4 2 | | | | 3+4 2 | 3+4 2 | |
| | 5+6 2 | 3+4 2 | | | | 3+4 2 | 5+6 2 | |
| | 3+4 2 | 5+6 2 | | | | 5+6 2 | 3+4 2 | |
| | 3+4 2 | 3+4 2 | | | | 3+4 2 | 3+4 2 | |
| | 5+6 2 | 3+4 2 | | | | 3+4 2 | 3+4 2 | |
| | 3+4 2 | 5+6 2 | | | | 5+6 2 | 5+6 2 | |
| | 5+6 2 | 5+6 2 | | | | 5+6 2 | 5+6 2 | |
| | 5+6 2 | 5+6 2 | | | | 5+6 2 | 5+6 2 | |
| | 5+6 2 | 5+6 2 | | | | 5+6 2 | 5+6 2 | |
| | 5+6 2 | 3+4 2 | | | | 5+6 2 | 5+6 2 | |
| | 5+6 2 | 5+6 2 | | | | 3+4 2 | 3+4 2 | |
| | 3+4 2 | 3+4 2 | | | | 3+4 2 | 5+6 2 | |
| | 5+6 2 | 5+6 2 | | | | 5+6 2 | 3+4 2 | |
| | 3+4 2 | 5+6 2 | | | | 5+6 2 | 5+6 2 | |
| | 3+4 2 | 5+6 2 | | | | 3+4 2 | 5+6 2 | |
| | 5+6 2 | 3+4 2 | | | | 3+4 2 | 3+4 2 | |
| | 3+4 2 | 3+4 2 | | | | 5+6 2 | 3+4 2 | |
| | 3+4 2 | 5+6 2 | | | | 5+6 2 | 5+6 2 | |
| 11.2.42 | 262728 | 192 | H4Co2O10P2Pb1 | | | | | |
| | -A | -B | -C | -D | -E | -GM | -Y | -Z |
| | 5+6 2 | 3+4 2 | 2+2 4 | 2+2 4 | 2+2 4 | 3+4 2 | 5+6 2 | 2+2 4 |
| | 3+4 2 | 5+6 2 | 2+2 4 | 2+2 4 | 2+2 4 | 5+6 2 | 3+4 2 | 2+2 4 |
| | 5+6 2 | 5+6 2 | 2+2 4 | 2+2 4 | 2+2 4 | 3+4 2 | 3+4 2 | 2+2 4 |
| | 3+4 2 | 3+4 2 | 2+2 4 | 2+2 4 | 2+2 4 | 5+6 2 | 5+6 2 | 2+2 4 |
| | 3+4 2 | 3+4 2 | 2+2 4 | 2+2 4 | 2+2 4 | 3+4 2 | 3+4 2 | 2+2 4 |
| | 5+6 2 | 5+6 2 | 2+2 4 | 2+2 4 | 2+2 4 | 5+6 2 | 5+6 2 | 2+2 4 |
| | 3+4 2 | 3+4 2 | 2+2 4 | 2+2 4 | 2+2 4 | 5+6 2 | 5+6 2 | 2+2 4 |
| | 5+6 2 | 5+6 2 | 2+2 4 | 2+2 4 | 2+2 4 | 3+4 2 | 3+4 2 | 2+2 4 |
| | 3+4 2 | 5+6 2 | 2+2 4 | 2+2 4 | 2+2 4 | 3+4 2 | 5+6 2 | 2+2 4 |
| | 5+6 2 | 3+4 2 | 2+2 4 | 2+2 4 | 2+2 4 | 5+6 2 | 3+4 2 | 2+2 4 |
| | 5+6 2 | 5+6 2 | 2+2 4 | 2+2 4 | 2+2 4 | 3+4 2 | 5+6 2 | 2+2 4 |
| | 3+4 2 | 3+4 2 | 2+2 4 | 2+2 4 | 2+2 4 | 3+4 2 | 5+6 2 | 2+2 4 |
| | 5+6 2 | 3+4 2 | 2+2 4 | 2+2 4 | 2+2 4 | 5+6 2 | 3+4 2 | 2+2 4 |
| | 3+4 2 | 5+6 2 | 2+2 4 | 2+2 4 | 2+2 4 | 5+6 2 | 3+4 2 | 2+2 4 |
| | 5+6 2 | 3+4 2 | 2+2 4 | 2+2 4 | 2+2 4 | 5+6 2 | 3+4 2 | 2+2 4 |
| | 3+4 2 | 5+6 2 | 2+2 4 | 2+2 4 | 2+2 4 | 3+4 2 | 5+6 2 | 2+2 4 |
| | 3+4 2 | 5+6 2 | 2+2 4 | 2+2 4 | 2+2 4 | 3+4 2 | 5+6 2 | 2+2 4 |
| | 5+6 2 | 3+4 2 | 2+2 4 | 2+2 4 | 2+2 4 | 5+6 2 | 3+4 2 | 2+2 4 |
| | 3+4 2 | 3+4 2 | 2+2 4 | 2+2 4 | 2+2 4 | 3+4 2 | 3+4 2 | 2+2 4 |
| | 5+6 2 | 5+6 2 | 2+2 4 | 2+2 4 | 2+2 4 | 5+6 2 | 5+6 2 | 2+2 4 |
| | 5+6 2 | 3+4 2 | 2+2 4 | 2+2 4 | 2+2 4 | 3+4 2 | 5+6 2 | 2+2 4 |
| | 3+4 2 | 5+6 2 | 2+2 4 | 2+2 4 | 2+2 4 | 5+6 2 | 3+4 2 | 2+2 4 |
| | 3+4 2 | 3+4 2 | 2+2 4 | 2+2 4 | 2+2 4 | 3+4 2 | 5+6 2 | 2+2 4 |
| | 5+6 2 | 5+6 2 | 2+2 4 | 2+2 4 | 2+2 4 | 5+6 2 | 3+4 2 | 2+2 4 |
| | 3+4 2 | 3+4 2 | 2+2 4 | 2+2 4 | 2+2 4 | 3+4 2 | 3+4 2 | 2+2 4 |
| | 5+6 2 | 5+6 2 | 2+2 4 | 2+2 4 | 2+2 4 | 3+4 2 | 5+6 2 | 2+2 4 |
| | 5+6 2 | 3+4 2 | 2+2 4 | 2+2 4 | 2+2 4 | 5+6 2 | 5+6 2 | 2+2 4 |
| | 5+6 2 | 5+6 2 | 2+2 4 | 2+2 4 | 2+2 4 | 3+4 2 | 3+4 2 | 2+2 4 |
| | 3+4 2 | 5+6 2 | 2+2 4 | 2+2 4 | 2+2 4 | 5+6 2 | 5+6 2 | 2+2 4 |
| | 3+4 2 | 3+4 2 | 2+2 4 | 2+2 4 | 2+2 4 | 5+6 2 | 5+6 2 | 2+2 4 |
| | 5+6 2 | 5+6 2 | 2+2 4 | 2+2 4 | 2+2 4 | 5+6 2 | 3+4 2 | 2+2 4 |
| | 3+4 2 | 3+4 2 | 2+2 4 | 2+2 4 | 2+2 4 | 3+4 2 | 5+6 2 | 2+2 4 |
| | 5+6 2 | 5+6 2 | 2+2 4 | 2+2 4 | 2+2 4 | 3+4 2 | 5+6 2 | 2+2 4 |
| | 3+4 2 | 3+4 2 | 2+2 4 | 2+2 4 | 2+2 4 | 5+6 2 | 3+4 2 | 2+2 4 |
| | 5+6 2 | 3+4 2 | 2+2 4 | 2+2 4 | 2+2 4 | 3+4 2 | 5+6 2 | 2+2 4 |
| | 3+4 2 | 5+6 2 | 2+2 4 | 2+2 4 | 2+2 4 | 3+4 2 | 3+4 2 | 2+2 4 |
| | 3+4 2 | 5+6 2 | 2+2 4 | 2+2 4 | 2+2 4 | 5+6 2 | 3+4 2 | 2+2 4 |
| | 5+6 2 | 3+4 2 | 2+2 4 | 2+2 4 | 2+2 4 | 3+4 2 | 3+4 2 | 2+2 4 |
| | 5+6 2 | 5+6 2 | 2+2 4 | 2+2 4 | 2+2 4 | 3+4 2 | 5+6 2 | 2+2 4 |
| | 3+4 2 | 3+4 2 | 2+2 4 | 2+2 4 | 2+2 4 | 5+6 2 | 5+6 2 | 2+2 4 |
| | 5+6 2 | 5+6 2 | 2+2 4 | 2+2 4 | 2+2 4 | 5+6 2 | 5+6 2 | 2+2 4 |
| | 3+4 2 | 3+4 2 | 2+2 4 | 2+2 4 | 2+2 4 | 5+6 2 | 3+4 2 | 2+2 4 |
| | 5+6 2 | 3+4 2 | 2+2 4 | 2+2 4 | 2+2 4 | 3+4 2 | 3+4 2 | 2+2 4 |
| | 5+6 2 | 5+6 2 | 2+2 4 | 2+2 4 | 2+2 4 | 5+6 2 | 5+6 2 | 2+2 4 |
| | 3+4 2 | 5+6 2 | 2+2 4 | 2+2 4 | 2+2 4 | 5+6 2 | 3+4 2 | 2+2 4 |
| | 3+4 2 | 3+4 2 | 2+2 4 | 2+2 4 | 2+2 4 | 3+4 2 | 3+4 2 | 2+2 4 |
| | 5+6 2 | 5+6 2 | 2+2 4 | 2+2 4 | 2+2 4 | 3+4 2 | 3+4 2 | 2+2 4 |
| | 3+4 2 | 3+4 2 | 2+2 4 | 2+2 4 | 2+2 4 | 5+6 2 | 5+6 2 | 2+2 4 |
| | 5+6 2 | 3+4 2 | | | | 3+4 2 | 5+6 2 | |
| | 3+4 2 | 5+6 2 | | | | 3+4 2 | 3+4 2 | |
| | 5+6 2 | 5+6 2 | | | | 3+4 2 | 3+4 2 | |
| | 3+4 2 | 5+6 2 | | | | 5+6 2 | 5+6 2 | |
| | 5+6 2 | 3+4 2 | | | | 3+4 2 | 5+6 2 | |
| | 3+4 2 | 3+4 2 | | | | 5+6 2 | 5+6 2 | |
| | 5+6 2 | 3+4 2 | | | | 3+4 2 | 3+4 2 | |
| | 5+6 2 | 5+6 2 | | | | 3+4 2 | 3+4 2 | |
| | 3+4 2 | 5+6 2 | | | | 5+6 2 | 5+6 2 | |
| | 3+4 2 | 3+4 2 | | | | 3+4 2 | 5+6 2 | |
| | 5+6 2 | 3+4 2 | | | | 5+6 2 | 3+4 2 | |
| | 3+4 2 | 5+6 2 | | | | 3+4 2 | 3+4 2 | |
| | 3+4 2 | 3+4 2 | | | | 5+6 2 | 5+6 2 | |
| | 3+4 2 | 3+4 2 | | | | 3+4 2 | 3+4 2 | |
| | 3+4 2 | 5+6 2 | | | | 5+6 2 | 3+4 2 | |
| | 5+6 2 | 5+6 2 | | | | 3+4 2 | 5+6 2 | |
| | 3+4 2 | 5+6 2 | | | | 5+6 2 | 5+6 2 | |
| | 5+6 2 | 3+4 2 | | | | 5+6 2 | 5+6 2 | |
| | 5+6 2 | 3+4 2 | | | | 3+4 2 | 5+6 2 | |
| | 3+4 2 | 5+6 2 | | | | 3+4 2 | 5+6 2 | |
| | 5+6 2 | 3+4 2 | | | | 5+6 2 | 3+4 2 | |
| | 5+6 2 | 5+6 2 | | | | 3+4 2 | 3+4 2 | |
| | 3+4 2 | 3+4 2 | | | | 5+6 2 | 5+6 2 | |
| | 5+6 2 | 3+4 2 | | | | 5+6 2 | 5+6 2 | |
| | 5+6 2 | 5+6 2 | | | | 3+4 2 | 3+4 2 | |
| | 3+4 2 | 5+6 2 | | | | 3+4 2 | 3+4 2 | |

Continued on next column

Continued on next column

**TABLE XXIII– continued**

| TQC | ICSD | NoE | SF | | | | | |
|---|---|---|---|---|---|---|---|---|
| | 5+6 2 | 3+4 2 | | | | 5+6 2 | 3+4 2 | |
| | 3+4 2 | 5+6 2 | | | | 3+4 2 | 5+6 2 | |
| | 3+4 2 | 3+4 2 | | | | 5+6 2 | 3+4 2 | |
| | 5+6 2 | 5+6 2 | | | | 5+6 2 | 5+6 2 | |
| | 3+4 2 | 5+6 2 | | | | 5+6 2 | 3+4 2 | |
| | 5+6 2 | 5+6 2 | | | | 3+4 2 | 5+6 2 | |
| | 5+6 2 | 3+4 2 | | | | 5+6 2 | 3+4 2 | |
| | 3+4 2 | 3+4 2 | | | | 5+6 2 | 3+4 2 | |
| | 5+6 2 | 3+4 2 | | | | 3+4 2 | 5+6 2 | |
| | 5+6 2 | 3+4 2 | | | | 3+4 2 | 5+6 2 | |
| | 5+6 2 | 3+4 2 | | | | 3+4 2 | 5+6 2 | |
| | 5+6 2 | 3+4 2 | | | | 3+4 2 | 5+6 2 | |
| | 3+4 2 | 5+6 2 | | | | 3+4 2 | 5+6 2 | |
| | 5+6 2 | 3+4 2 | | | | 3+4 2 | 5+6 2 | |
| | 3+4 2 | 5+6 2 | | | | 3+4 2 | 5+6 2 | |
| | 3+4 2 | 5+6 2 | | | | 3+4 2 | 5+6 2 | |
| | 5+6 2 | 3+4 2 | | | | 3+4 2 | 5+6 2 | |
| | 3+4 2 | 5+6 2 | | | | 3+4 2 | 5+6 2 | |
| | 5+6 2 | 3+4 2 | | | | 3+4 2 | 5+6 2 | |
| | 3+4 2 | 5+6 2 | | | | 3+4 2 | 5+6 2 | |
| | 5+6 2 | 3+4 2 | | | | 3+4 2 | 5+6 2 | |
| | 5+6 2 | 3+4 2 | | | | 3+4 2 | 5+6 2 | |
| 11.2.43 | 60659 | 60 | H4Ba1O3 | | | | | |
| | -A | -B | -C | -D | -E | -GM | -Y | -Z |
| | 3+4 2 | 5+6 2 | 2+2 4 | 2+2 4 | 2+2 4 | 3+4 2 | 5+6 2 | 2+2 4 |
| | 5+6 2 | 3+4 2 | 2+2 4 | 2+2 4 | 2+2 4 | 5+6 2 | 3+4 2 | 2+2 4 |
| | 5+6 2 | 5+6 2 | 2+2 4 | 2+2 4 | 2+2 4 | 3+4 2 | 3+4 2 | 2+2 4 |
| | 3+4 2 | 3+4 2 | 2+2 4 | 2+2 4 | 2+2 4 | 5+6 2 | 5+6 2 | 2+2 4 |
| | 3+4 2 | 5+6 2 | 2+2 4 | 2+2 4 | 2+2 4 | 3+4 2 | 5+6 2 | 2+2 4 |
| | 5+6 2 | 3+4 2 | 2+2 4 | 2+2 4 | 2+2 4 | 5+6 2 | 3+4 2 | 2+2 4 |
| | 3+4 2 | 3+4 2 | 2+2 4 | 2+2 4 | 2+2 4 | 3+4 2 | 3+4 2 | 2+2 4 |
| | 5+6 2 | 5+6 2 | 2+2 4 | 2+2 4 | 2+2 4 | 5+6 2 | 5+6 2 | 2+2 4 |
| | 3+4 2 | 5+6 2 | 2+2 4 | 2+2 4 | 2+2 4 | 3+4 2 | 5+6 2 | 2+2 4 |
| | 5+6 2 | 3+4 2 | 2+2 4 | 2+2 4 | 2+2 4 | 5+6 2 | 3+4 2 | 2+2 4 |
| | 3+4 2 | 5+6 2 | 2+2 4 | 2+2 4 | 2+2 4 | 3+4 2 | 5+6 2 | 2+2 4 |
| | 5+6 2 | 3+4 2 | 2+2 4 | 2+2 4 | 2+2 4 | 3+4 2 | 5+6 2 | 2+2 4 |
| | 3+4 2 | 5+6 2 | 2+2 4 | 2+2 4 | 2+2 4 | 5+6 2 | 3+4 2 | 2+2 4 |
| | 5+6 2 | 3+4 2 | 2+2 4 | 2+2 4 | 2+2 4 | 5+6 2 | 3+4 2 | 2+2 4 |
| | 5+6 2 | 3+4 2 | 2+2 4 | 2+2 4 | 2+2 4 | 3+4 2 | 5+6 2 | 2+2 4 |
| | 3+4 2 | 5+6 2 | | | | 5+6 2 | 3+4 2 | |
| | 5+6 2 | 5+6 2 | | | | 5+6 2 | 3+4 2 | |
| | 3+4 2 | 3+4 2 | | | | 3+4 2 | 5+6 2 | |
| | 5+6 2 | 3+4 2 | | | | 3+4 2 | 3+4 2 | |
| | 3+4 2 | 5+6 2 | | | | 5+6 2 | 5+6 2 | |
| | 5+6 2 | 3+4 2 | | | | 3+4 2 | 5+6 2 | |
| | 3+4 2 | 5+6 2 | | | | 5+6 2 | 3+4 2 | |
| | 3+4 2 | 5+6 2 | | | | 5+6 2 | 3+4 2 | |
| | 5+6 2 | 5+6 2 | | | | 3+4 2 | 5+6 2 | |
| | 3+4 2 | 3+4 2 | | | | 3+4 2 | 5+6 2 | |
| | 3+4 2 | 3+4 2 | | | | 5+6 2 | 3+4 2 | |
| | 5+6 2 | 3+4 2 | | | | 3+4 2 | 3+4 2 | |
| | 3+4 2 | 5+6 2 | | | | 3+4 2 | 5+6 2 | |
| | 5+6 2 | 3+4 2 | | | | 3+4 2 | 3+4 2 | |
| | 3+4 2 | 3+4 2 | | | | 5+6 2 | 5+6 2 | |
| 11.2.44 | 151476 | 544 | H4Ag2K4O2S9Sn3 | | | | | |
| | -A | -B | -C | -D | -E | -GM | -Y | -Z |
| | 3+4 2 | 3+4 2 | 2+2 4 | 2+2 4 | 2+2 4 | 3+4 2 | 3+4 2 | 2+2 4 |
| | 5+6 2 | 5+6 2 | 2+2 4 | 2+2 4 | 2+2 4 | 5+6 2 | 5+6 2 | 2+2 4 |
| | 3+4 2 | 3+4 2 | 2+2 4 | 2+2 4 | 2+2 4 | 3+4 2 | 3+4 2 | 2+2 4 |
| | 3+4 2 | 3+4 2 | 2+2 4 | 2+2 4 | 2+2 4 | 3+4 2 | 3+4 2 | 2+2 4 |
| | 5+6 2 | 5+6 2 | 2+2 4 | 2+2 4 | 2+2 4 | 5+6 2 | 5+6 2 | 2+2 4 |

Continued on next column

**TABLE XXIII– continued**

| TQC | ICSD | NoE | SF | | | | | |
|---|---|---|---|---|---|---|---|---|
| | 5+6 2 | 5+6 2 | 2+2 4 | 2+2 4 | 2+2 4 | 5+6 2 | 5+6 2 | 2+2 4 |
| | 3+4 2 | 3+4 2 | 2+2 4 | 2+2 4 | 2+2 4 | 3+4 2 | 3+4 2 | 2+2 4 |
| | 3+4 2 | 3+4 2 | 2+2 4 | 2+2 4 | 2+2 4 | 3+4 2 | 3+4 2 | 2+2 4 |
| | 5+6 2 | 5+6 2 | 2+2 4 | 2+2 4 | 2+2 4 | 5+6 2 | 5+6 2 | 2+2 4 |
| | 5+6 2 | 5+6 2 | 2+2 4 | 2+2 4 | 2+2 4 | 5+6 2 | 5+6 2 | 2+2 4 |
| | 3+4 2 | 3+4 2 | 2+2 4 | 2+2 4 | 2+2 4 | 3+4 2 | 3+4 2 | 2+2 4 |
| | 5+6 2 | 5+6 2 | 2+2 4 | 2+2 4 | 2+2 4 | 5+6 2 | 5+6 2 | 2+2 4 |
| | 3+4 2 | 3+4 2 | 2+2 4 | 2+2 4 | 2+2 4 | 3+4 2 | 3+4 2 | 2+2 4 |
| | 5+6 2 | 5+6 2 | 2+2 4 | 2+2 4 | 2+2 4 | 5+6 2 | 5+6 2 | 2+2 4 |
| | 3+4 2 | 3+4 2 | 2+2 4 | 2+2 4 | 2+2 4 | 3+4 2 | 3+4 2 | 2+2 4 |
| | 5+6 2 | 5+6 2 | 2+2 4 | 2+2 4 | 2+2 4 | 5+6 2 | 5+6 2 | 2+2 4 |
| | 3+4 2 | 3+4 2 | 2+2 4 | 2+2 4 | 2+2 4 | 3+4 2 | 3+4 2 | 2+2 4 |
| | 5+6 2 | 5+6 2 | 2+2 4 | 2+2 4 | 2+2 4 | 5+6 2 | 5+6 2 | 2+2 4 |
| | 3+4 2 | 3+4 2 | 2+2 4 | 2+2 4 | 2+2 4 | 3+4 2 | 3+4 2 | 2+2 4 |
| | 5+6 2 | 5+6 2 | 2+2 4 | 2+2 4 | 2+2 4 | 5+6 2 | 5+6 2 | 2+2 4 |
| | 3+4 2 | 3+4 2 | 2+2 4 | 2+2 4 | 2+2 4 | 3+4 2 | 3+4 2 | 2+2 4 |
| | 5+6 2 | 5+6 2 | 2+2 4 | 2+2 4 | 2+2 4 | 5+6 2 | 5+6 2 | 2+2 4 |
| | 3+4 2 | 3+4 2 | 2+2 4 | 2+2 4 | 2+2 4 | 3+4 2 | 3+4 2 | 2+2 4 |
| | 5+6 2 | 5+6 2 | 2+2 4 | 2+2 4 | 2+2 4 | 5+6 2 | 5+6 2 | 2+2 4 |
| | 5+6 2 | 3+4 2 | 2+2 4 | 2+2 4 | 2+2 4 | 5+6 2 | 3+4 2 | 2+2 4 |
| | 5+6 2 | 5+6 2 | 2+2 4 | 2+2 4 | 2+2 4 | 3+4 2 | 5+6 2 | 2+2 4 |
| | 3+4 2 | 3+4 2 | 2+2 4 | 2+2 4 | 2+2 4 | 5+6 2 | 3+4 2 | 2+2 4 |
| | 3+4 2 | 5+6 2 | 2+2 4 | 2+2 4 | 2+2 4 | 3+4 2 | 5+6 2 | 2+2 4 |
| | 5+6 2 | 3+4 2 | 2+2 4 | 2+2 4 | 2+2 4 | 5+6 2 | 3+4 2 | 2+2 4 |
| | 3+4 2 | 5+6 2 | 2+2 4 | 2+2 4 | 2+2 4 | 3+4 2 | 5+6 2 | 2+2 4 |
| | 5+6 2 | 3+4 2 | 2+2 4 | 2+2 4 | 2+2 4 | 5+6 2 | 3+4 2 | 2+2 4 |
| | 3+4 2 | 5+6 2 | 2+2 4 | 2+2 4 | 2+2 4 | 3+4 2 | 5+6 2 | 2+2 4 |
| | 3+4 2 | 3+4 2 | 2+2 4 | 2+2 4 | 2+2 4 | 5+6 2 | 5+6 2 | 2+2 4 |
| | 3+4 2 | 5+6 2 | 2+2 4 | 2+2 4 | 2+2 4 | 3+4 2 | 3+4 2 | 2+2 4 |
| | 5+6 2 | 5+6 2 | 2+2 4 | 2+2 4 | 2+2 4 | 5+6 2 | 3+4 2 | 2+2 4 |
| | 5+6 2 | 3+4 2 | 2+2 4 | 2+2 4 | 2+2 4 | 3+4 2 | 5+6 2 | 2+2 4 |
| | 3+4 2 | 5+6 2 | 2+2 4 | 2+2 4 | 2+2 4 | 3+4 2 | 5+6 2 | 2+2 4 |
| | 5+6 2 | 3+4 2 | 2+2 4 | 2+2 4 | 2+2 4 | 5+6 2 | 3+4 2 | 2+2 4 |
| | 5+6 2 | 5+6 2 | 2+2 4 | 2+2 4 | 2+2 4 | 3+4 2 | 5+6 2 | 2+2 4 |
| | 3+4 2 | 3+4 2 | 2+2 4 | 2+2 4 | 2+2 4 | 3+4 2 | 3+4 2 | 2+2 4 |
| | 3+4 2 | 5+6 2 | 2+2 4 | 2+2 4 | 2+2 4 | 5+6 2 | 5+6 2 | 2+2 4 |
| | 5+6 2 | 3+4 2 | 2+2 4 | 2+2 4 | 2+2 4 | 5+6 2 | 3+4 2 | 2+2 4 |
| | 3+4 2 | 5+6 2 | 2+2 4 | 2+2 4 | 2+2 4 | 3+4 2 | 3+4 2 | 2+2 4 |
| | 5+6 2 | 3+4 2 | 2+2 4 | 2+2 4 | 2+2 4 | 5+6 2 | 5+6 2 | 2+2 4 |
| | 5+6 2 | 5+6 2 | 2+2 4 | 2+2 4 | 2+2 4 | 3+4 2 | 3+4 2 | 2+2 4 |
| | 3+4 2 | 3+4 2 | 2+2 4 | 2+2 4 | 2+2 4 | 5+6 2 | 5+6 2 | 2+2 4 |
| | 3+4 2 | 3+4 2 | 2+2 4 | 2+2 4 | 2+2 4 | 3+4 2 | 3+4 2 | 2+2 4 |
| | 5+6 2 | 5+6 2 | 2+2 4 | 2+2 4 | 2+2 4 | 5+6 2 | 5+6 2 | 2+2 4 |
| | 5+6 2 | 3+4 2 | 2+2 4 | 2+2 4 | 2+2 4 | 3+4 2 | 5+6 2 | 2+2 4 |
| | 3+4 2 | 5+6 2 | 2+2 4 | 2+2 4 | 2+2 4 | 3+4 2 | 5+6 2 | 2+2 4 |
| | 5+6 2 | 3+4 2 | 2+2 4 | 2+2 4 | 2+2 4 | 5+6 2 | 3+4 2 | 2+2 4 |
| | 3+4 2 | 5+6 2 | 2+2 4 | 2+2 4 | 2+2 4 | 5+6 2 | 3+4 2 | 2+2 4 |
| | 5+6 2 | 5+6 2 | 2+2 4 | 2+2 4 | 2+2 4 | 3+4 2 | 5+6 2 | 2+2 4 |
| | 3+4 2 | 3+4 2 | 2+2 4 | 2+2 4 | 2+2 4 | 3+4 2 | 3+4 2 | 2+2 4 |
| | 3+4 2 | 3+4 2 | 2+2 4 | 2+2 4 | 2+2 4 | 3+4 2 | 3+4 2 | 2+2 4 |
| | 5+6 2 | 5+6 2 | 2+2 4 | 2+2 4 | 2+2 4 | 5+6 2 | 3+4 2 | 2+2 4 |
| | 3+4 2 | 3+4 2 | 2+2 4 | 2+2 4 | 2+2 4 | 3+4 2 | 5+6 2 | 2+2 4 |
| | 5+6 2 | 5+6 2 | 2+2 4 | 2+2 4 | 2+2 4 | 5+6 2 | 5+6 2 | 2+2 4 |
| | 3+4 2 | 3+4 2 | 2+2 4 | 2+2 4 | 2+2 4 | 3+4 2 | 3+4 2 | 2+2 4 |
| | 5+6 2 | 5+6 2 | 2+2 4 | 2+2 4 | 2+2 4 | 5+6 2 | 5+6 2 | 2+2 4 |
| | 3+4 2 | 3+4 2 | 2+2 4 | 2+2 4 | 2+2 4 | 5+6 2 | 3+4 2 | 2+2 4 |
| | 5+6 2 | 5+6 2 | 2+2 4 | 2+2 4 | 2+2 4 | 5+6 2 | 5+6 2 | 2+2 4 |
| | 5+6 2 | 5+6 2 | 2+2 4 | 2+2 4 | 2+2 4 | 3+4 2 | 3+4 2 | 2+2 4 |
| | 3+4 2 | 3+4 2 | 2+2 4 | 2+2 4 | 2+2 4 | 3+4 2 | 5+6 2 | 2+2 4 |
| | 5+6 2 | 5+6 2 | 2+2 4 | 2+2 4 | 2+2 4 | 5+6 2 | 5+6 2 | 2+2 4 |
| | 3+4 2 | 5+6 2 | 2+2 4 | 2+2 4 | 2+2 4 | 3+4 2 | 3+4 2 | 2+2 4 |

Continued on next column

**TABLE XXIII– continued**

| TQC | ICSD | NoE | SF |
|---|---|---|---|
| | 3+4 2 3+4 2 | 2+2 4 2+2 4 2+2 4 | 5+6 2 5+6 2 2+2 4 |
| | 5+6 2 3+4 2 | 2+2 4 2+2 4 2+2 4 | 5+6 2 5+6 2 2+2 4 |
| | 3+4 2 5+6 2 | 2+2 4 2+2 4 2+2 4 | 3+4 2 3+4 2 2+2 4 |
| | 5+6 2 3+4 2 | 2+2 4 2+2 4 2+2 4 | 5+6 2 3+4 2 2+2 4 |
| | 3+4 2 3+4 2 | 2+2 4 2+2 4 2+2 4 | 5+6 2 3+4 2 2+2 4 |
| | 5+6 2 5+6 2 | 2+2 4 2+2 4 2+2 4 | 3+4 2 5+6 2 2+2 4 |
| | 5+6 2 3+4 2 | 2+2 4 2+2 4 2+2 4 | 3+4 2 5+6 2 2+2 4 |
| | 3+4 2 3+4 2 | 2+2 4 2+2 4 2+2 4 | 3+4 2 3+4 2 2+2 4 |
| | 5+6 2 5+6 2 | 2+2 4 2+2 4 2+2 4 | 5+6 2 5+6 2 2+2 4 |
| | 3+4 2 5+6 2 | 2+2 4 2+2 4 2+2 4 | 5+6 2 3+4 2 2+2 4 |
| | 5+6 2 3+4 2 | 2+2 4 2+2 4 2+2 4 | 5+6 2 3+4 2 2+2 4 |
| | 3+4 2 5+6 2 | 2+2 4 2+2 4 2+2 4 | 3+4 2 5+6 2 2+2 4 |
| | 5+6 2 3+4 2 | 2+2 4 2+2 4 2+2 4 | 5+6 2 3+4 2 2+2 4 |
| | 5+6 2 3+4 2 | 2+2 4 2+2 4 2+2 4 | 5+6 2 3+4 2 2+2 4 |
| | 3+4 2 5+6 2 | 2+2 4 2+2 4 2+2 4 | 3+4 2 5+6 2 2+2 4 |
| | 3+4 2 5+6 2 | 2+2 4 2+2 4 2+2 4 | 3+4 2 5+6 2 2+2 4 |
| | 3+4 2 5+6 2 | 2+2 4 2+2 4 2+2 4 | 3+4 2 5+6 2 2+2 4 |
| | 5+6 2 3+4 2 | 2+2 4 2+2 4 2+2 4 | 3+4 2 3+4 2 2+2 4 |
| | 3+4 2 5+6 2 | 2+2 4 2+2 4 2+2 4 | 5+6 2 5+6 2 2+2 4 |
| | 5+6 2 5+6 2 | 2+2 4 2+2 4 2+2 4 | 5+6 2 3+4 2 2+2 4 |
| | 3+4 2 3+4 2 | 2+2 4 2+2 4 2+2 4 | 3+4 2 5+6 2 2+2 4 |
| | 5+6 2 3+4 2 | 2+2 4 2+2 4 2+2 4 | 5+6 2 3+4 2 2+2 4 |
| | 3+4 2 3+4 2 | 2+2 4 2+2 4 2+2 4 | 5+6 2 3+4 2 2+2 4 |
| | 5+6 2 5+6 2 | 2+2 4 2+2 4 2+2 4 | 3+4 2 3+4 2 2+2 4 |
| | 3+4 2 5+6 2 | 2+2 4 2+2 4 2+2 4 | 5+6 2 5+6 2 2+2 4 |
| | 5+6 2 3+4 2 | 2+2 4 2+2 4 2+2 4 | 3+4 2 5+6 2 2+2 4 |
| | 3+4 2 3+4 2 | 2+2 4 2+2 4 2+2 4 | 5+6 2 3+4 2 2+2 4 |
| | 5+6 2 5+6 2 | 2+2 4 2+2 4 2+2 4 | 3+4 2 5+6 2 2+2 4 |
| | 3+4 2 3+4 2 | 2+2 4 2+2 4 2+2 4 | 3+4 2 3+4 2 2+2 4 |
| | 3+4 2 3+4 2 | 2+2 4 2+2 4 2+2 4 | 3+4 2 5+6 2 2+2 4 |
| | 5+6 2 3+4 2 | 2+2 4 2+2 4 2+2 4 | 5+6 2 3+4 2 2+2 4 |
| | 5+6 2 5+6 2 | 2+2 4 2+2 4 2+2 4 | 5+6 2 5+6 2 2+2 4 |
| | 5+6 2 5+6 2 | 2+2 4 2+2 4 2+2 4 | 5+6 2 3+4 2 2+2 4 |
| | 3+4 2 5+6 2 | 2+2 4 2+2 4 2+2 4 | 3+4 2 5+6 2 2+2 4 |
| | 3+4 2 3+4 2 | 2+2 4 2+2 4 2+2 4 | 5+6 2 3+4 2 2+2 4 |
| | 5+6 2 5+6 2 | 2+2 4 2+2 4 2+2 4 | 3+4 2 5+6 2 2+2 4 |
| | 3+4 2 3+4 2 | 2+2 4 2+2 4 2+2 4 | 3+4 2 3+4 2 2+2 4 |
| | 5+6 2 5+6 2 | 2+2 4 2+2 4 2+2 4 | 5+6 2 5+6 2 2+2 4 |
| | 3+4 2 3+4 2 | 2+2 4 2+2 4 2+2 4 | 5+6 2 3+4 2 2+2 4 |
| | 3+4 2 3+4 2 | 2+2 4 2+2 4 2+2 4 | 5+6 2 3+4 2 2+2 4 |
| | 5+6 2 5+6 2 | 2+2 4 2+2 4 2+2 4 | 3+4 2 5+6 2 2+2 4 |
| | 5+6 2 5+6 2 | 2+2 4 2+2 4 2+2 4 | 3+4 2 5+6 2 2+2 4 |
| | 5+6 2 3+4 2 | 2+2 4 2+2 4 2+2 4 | 3+4 2 5+6 2 2+2 4 |
| | 5+6 2 5+6 2 | 2+2 4 2+2 4 2+2 4 | 3+4 2 5+6 2 2+2 4 |
| | 3+4 2 3+4 2 | 2+2 4 2+2 4 2+2 4 | 5+6 2 3+4 2 2+2 4 |
| | 3+4 2 5+6 2 | 2+2 4 2+2 4 2+2 4 | 5+6 2 3+4 2 2+2 4 |
| | 5+6 2 5+6 2 | 2+2 4 2+2 4 2+2 4 | 3+4 2 3+4 2 2+2 4 |
| | 5+6 2 3+4 2 | 2+2 4 2+2 4 2+2 4 | 3+4 2 3+4 2 2+2 4 |
| | 3+4 2 5+6 2 | 2+2 4 2+2 4 2+2 4 | 5+6 2 5+6 2 2+2 4 |
| | 3+4 2 3+4 2 | 2+2 4 2+2 4 2+2 4 | 5+6 2 5+6 2 2+2 4 |
| | 3+4 2 3+4 2 | 2+2 4 2+2 4 2+2 4 | 5+6 2 5+6 2 2+2 4 |
| | 3+4 2 5+6 2 | 2+2 4 2+2 4 2+2 4 | 3+4 2 5+6 2 2+2 4 |
| | 5+6 2 5+6 2 | 2+2 4 2+2 4 2+2 4 | 5+6 2 3+4 2 2+2 4 |
| | 5+6 2 3+4 2 | 2+2 4 2+2 4 2+2 4 | 3+4 2 3+4 2 2+2 4 |
| | 5+6 2 3+4 2 | 2+2 4 2+2 4 2+2 4 | 3+4 2 5+6 2 2+2 4 |
| | 3+4 2 5+6 2 | 2+2 4 2+2 4 2+2 4 | 5+6 2 3+4 2 2+2 4 |
| | 5+6 2 3+4 2 | 2+2 4 2+2 4 2+2 4 | 3+4 2 5+6 2 2+2 4 |
| | 3+4 2 5+6 2 | 2+2 4 2+2 4 2+2 4 | 5+6 2 3+4 2 2+2 4 |
| | 5+6 2 3+4 2 | 2+2 4 2+2 4 2+2 4 | 3+4 2 5+6 2 2+2 4 |
| | 3+4 2 5+6 2 | 2+2 4 2+2 4 2+2 4 | 5+6 2 3+4 2 2+2 4 |
| | 5+6 2 3+4 2 | 2+2 4 2+2 4 2+2 4 | 3+4 2 5+6 2 2+2 4 |

Continued on next column

**TABLE XXIII– continued**

| TQC | ICSD | NoE | SF |
|---|---|---|---|
| | 3+4 2 5+6 2 | 2+2 4 2+2 4 2+2 4 | 5+6 2 3+4 2 2+2 4 |
| | 3+4 2 3+4 2 | 2+2 4 2+2 4 2+2 4 | 3+4 2 3+4 2 2+2 4 |
| | 5+6 2 5+6 2 | 2+2 4 2+2 4 2+2 4 | 5+6 2 5+6 2 2+2 4 |
| | 3+4 2 3+4 2 | 2+2 4 2+2 4 2+2 4 | 3+4 2 3+4 2 2+2 4 |
| | 5+6 2 5+6 2 | 2+2 4 2+2 4 2+2 4 | 5+6 2 5+6 2 2+2 4 |
| | 3+4 2 3+4 2 | 2+2 4 2+2 4 2+2 4 | 3+4 2 5+6 2 2+2 4 |
| | 5+6 2 5+6 2 | 2+2 4 2+2 4 2+2 4 | 5+6 2 3+4 2 2+2 4 |
| | 3+4 2 3+4 2 | 2+2 4 2+2 4 2+2 4 | 5+6 2 3+4 2 2+2 4 |
| | 5+6 2 5+6 2 | 2+2 4 2+2 4 2+2 4 | 3+4 2 5+6 2 2+2 4 |
| | 3+4 2 3+4 2 | | 3+4 2 5+6 2 |
| | 5+6 2 5+6 2 | | 5+6 2 3+4 2 |
| | 5+6 2 3+4 2 | | 3+4 2 5+6 2 |
| | 3+4 2 5+6 2 | | 5+6 2 3+4 2 |
| | 5+6 2 3+4 2 | | 3+4 2 5+6 2 |
| | 3+4 2 3+4 2 | | 5+6 2 3+4 2 |
| | 5+6 2 5+6 2 | | 3+4 2 5+6 2 |
| | 3+4 2 5+6 2 | | 5+6 2 3+4 2 |
| | 5+6 2 3+4 2 | | 3+4 2 5+6 2 |
| | 5+6 2 5+6 2 | | 5+6 2 3+4 2 |
| | 3+4 2 3+4 2 | | 3+4 2 3+4 2 |
| | 3+4 2 5+6 2 | | 5+6 2 5+6 2 |
| | 5+6 2 5+6 2 | | 3+4 2 3+4 2 |
| | 5+6 2 3+4 2 | | 3+4 2 5+6 2 |
| | 3+4 2 3+4 2 | | 5+6 2 3+4 2 |
| | 3+4 2 5+6 2 | | 5+6 2 5+6 2 |
| | 5+6 2 5+6 2 | | 3+4 2 3+4 2 |
| | 3+4 2 3+4 2 | | 5+6 2 5+6 2 |
| | 5+6 2 5+6 2 | | 3+4 2 3+4 2 |
| | 3+4 2 3+4 2 | | 5+6 2 5+6 2 |
| | 5+6 2 5+6 2 | | 3+4 2 3+4 2 |
| | 3+4 2 3+4 2 | | 3+4 2 5+6 2 |
| | 5+6 2 5+6 2 | | 5+6 2 3+4 2 |
| | 3+4 2 3+4 2 | | 5+6 2 5+6 2 |
| | 5+6 2 5+6 2 | | 3+4 2 3+4 2 |
| | 3+4 2 3+4 2 | | 3+4 2 5+6 2 |
| | 5+6 2 5+6 2 | | 5+6 2 5+6 2 |
| | 3+4 2 3+4 2 | | 5+6 2 3+4 2 |
| | 5+6 2 5+6 2 | | 3+4 2 3+4 2 |
| | 3+4 2 3+4 2 | | 5+6 2 5+6 2 |
| | 5+6 2 3+4 2 | | 3+4 2 3+4 2 |
| | 3+4 2 5+6 2 | | 5+6 2 5+6 2 |
| | 5+6 2 5+6 2 | | 3+4 2 3+4 2 |
| | 5+6 2 3+4 2 | | 5+6 2 5+6 2 |
| | 3+4 2 3+4 2 | | 3+4 2 3+4 2 |
| | 3+4 2 3+4 2 | | 5+6 2 5+6 2 |
| | 5+6 2 5+6 2 | | 3+4 2 5+6 2 |
| | 5+6 2 5+6 2 | | 3+4 2 3+4 2 |
| | 3+4 2 3+4 2 | | 5+6 2 5+6 2 |
| | 3+4 2 5+6 2 | | 3+4 2 3+4 2 |
| | 5+6 2 3+4 2 | | 5+6 2 3+4 2 |
| | 5+6 2 3+4 2 | | 5+6 2 5+6 2 |
| | 3+4 2 5+6 2 | | 3+4 2 3+4 2 |
| | 3+4 2 5+6 2 | | 5+6 2 5+6 2 |
| | 5+6 2 5+6 2 | | 3+4 2 3+4 2 |
| | 3+4 2 3+4 2 | | 5+6 2 5+6 2 |
| | 3+4 2 3+4 2 | | 5+6 2 3+4 2 |
| | 5+6 2 5+6 2 | | 3+4 2 5+6 2 |
| | 3+4 2 3+4 2 | | 5+6 2 5+6 2 |
| | 3+4 2 5+6 2 | | 3+4 2 3+4 2 |
| | 5+6 2 3+4 2 | | 5+6 2 3+4 2 |
| | 3+4 2 5+6 2 | | 5+6 2 5+6 2 |

Continued on next column

**TABLE XXIII– continued**

| TQC | ICSD | | NoE | | SF | | | |
|---|---|---|---|---|---|---|---|---|
| | 5+6 2 | 3+4 2 | | | | 3+4 2 | 3+4 2 | |
| | 3+4 2 | 5+6 2 | | | | 5+6 2 | 3+4 2 | |
| | 3+4 2 | 3+4 2 | | | | 3+4 2 | 5+6 2 | |
| | 3+4 2 | 5+6 2 | | | | 3+4 2 | 5+6 2 | |
| | 5+6 2 | 3+4 2 | | | | 5+6 2 | 5+6 2 | |
| | 5+6 2 | 5+6 2 | | | | 5+6 2 | 5+6 2 | |
| | 5+6 2 | 3+4 2 | | | | 3+4 2 | 3+4 2 | |
| | 5+6 2 | 5+6 2 | | | | 3+4 2 | 3+4 2 | |
| | 5+6 2 | 5+6 2 | | | | 3+4 2 | 3+4 2 | |
| | 3+4 2 | 3+4 2 | | | | 5+6 2 | 3+4 2 | |
| | 3+4 2 | 5+6 2 | | | | 3+4 2 | 5+6 2 | |
| | 5+6 2 | 3+4 2 | | | | 5+6 2 | 5+6 2 | |
| | 5+6 2 | 3+4 2 | | | | 3+4 2 | 3+4 2 | |
| | 3+4 2 | 5+6 2 | | | | 3+4 2 | 5+6 2 | |
| | 5+6 2 | 3+4 2 | | | | 5+6 2 | 3+4 2 | |
| | 3+4 2 | 5+6 2 | | | | 5+6 2 | 5+6 2 | |
| | 5+6 2 | 3+4 2 | | | | 3+4 2 | 5+6 2 | |
| | 5+6 2 | 5+6 2 | | | | 3+4 2 | 5+6 2 | |
| | 3+4 2 | 3+4 2 | | | | 5+6 2 | 3+4 2 | |
| | 3+4 2 | 5+6 2 | | | | 3+4 2 | 3+4 2 | |
| | 5+6 2 | 3+4 2 | | | | 3+4 2 | 5+6 2 | |
| | 3+4 2 | 5+6 2 | | | | 5+6 2 | 5+6 2 | |
| | 5+6 2 | 3+4 2 | | | | 3+4 2 | 3+4 2 | |
| | 5+6 2 | 5+6 2 | | | | 5+6 2 | 3+4 2 | |
| | 3+4 2 | 5+6 2 | | | | 3+4 2 | 3+4 2 | |
| | 5+6 2 | 3+4 2 | | | | 5+6 2 | 5+6 2 | |
| | 5+6 2 | 3+4 2 | | | | 3+4 2 | 5+6 2 | |
| | 3+4 2 | 5+6 2 | | | | 3+4 2 | 5+6 2 | |
| | 3+4 2 | 3+4 2 | | | | 3+4 2 | 3+4 2 | |
| | 5+6 2 | 5+6 2 | | | | 5+6 2 | 5+6 2 | |
| | 3+4 2 | 5+6 2 | | | | 5+6 2 | 3+4 2 | |
| | 5+6 2 | 3+4 2 | | | | 3+4 2 | 3+4 2 | |
| | 3+4 2 | 5+6 2 | | | | 5+6 2 | 5+6 2 | |
| | 3+4 2 | 5+6 2 | | | | 5+6 2 | 3+4 2 | |
| | 5+6 2 | 3+4 2 | | | | 3+4 2 | 5+6 2 | |
| | 3+4 2 | 5+6 2 | | | | 5+6 2 | 3+4 2 | |
| | 5+6 2 | 3+4 2 | | | | 5+6 2 | 3+4 2 | |
| | 5+6 2 | 5+6 2 | | | | 3+4 2 | 5+6 2 | |
| | 3+4 2 | 3+4 2 | | | | 5+6 2 | 3+4 2 | |
| | 5+6 2 | 5+6 2 | | | | 3+4 2 | 5+6 2 | |
| | 5+6 2 | 5+6 2 | | | | 3+4 2 | 5+6 2 | |
| | 3+4 2 | 3+4 2 | | | | 3+4 2 | 3+4 2 | |
| | 3+4 2 | 3+4 2 | | | | 5+6 2 | 3+4 2 | |
| | 5+6 2 | 5+6 2 | | | | 5+6 2 | 3+4 2 | |
| | 3+4 2 | 3+4 2 | | | | 5+6 2 | 5+6 2 | |
| | 3+4 2 | 3+4 2 | | | | 3+4 2 | 3+4 2 | |
| | 5+6 2 | 5+6 2 | | | | 3+4 2 | 3+4 2 | |
| | 3+4 2 | 3+4 2 | | | | 5+6 2 | 3+4 2 | |
| | 5+6 2 | 5+6 2 | | | | 3+4 2 | 5+6 2 | |
| | 3+4 2 | 3+4 2 | | | | 5+6 2 | 5+6 2 | |
| | 3+4 2 | 3+4 2 | | | | 3+4 2 | 5+6 2 | |
| | 5+6 2 | 5+6 2 | | | | 5+6 2 | 5+6 2 | |
| | 3+4 2 | 3+4 2 | | | | 5+6 2 | 3+4 2 | |
| | 5+6 2 | 5+6 2 | | | | 5+6 2 | 3+4 2 | |
| | 5+6 2 | 5+6 2 | | | | 3+4 2 | 5+6 2 | |
| | 3+4 2 | 3+4 2 | | | | 3+4 2 | 5+6 2 | |
| | 3+4 2 | 5+6 2 | | | | 5+6 2 | 3+4 2 | |
| | 5+6 2 | 3+4 2 | | | | 5+6 2 | 5+6 2 | |
| | 3+4 2 | 3+4 2 | | | | 3+4 2 | 3+4 2 | |
| | 5+6 2 | 5+6 2 | | | | 5+6 2 | 5+6 2 | |
| | 3+4 2 | 3+4 2 | | | | 5+6 2 | 5+6 2 | |
| | 5+6 2 | 5+6 2 | | | | 3+4 2 | 3+4 2 | |
| | 3+4 2 | 3+4 2 | | | | 5+6 2 | 3+4 2 | |
| | 5+6 2 | 5+6 2 | | | | 5+6 2 | 5+6 2 | |
| | 5+6 2 | 3+4 2 | | | | 3+4 2 | 3+4 2 | |
| | 3+4 2 | 5+6 2 | | | | 3+4 2 | 5+6 2 | |
| | 5+6 2 | 3+4 2 | | | | 5+6 2 | 5+6 2 | |
| | 5+6 2 | 5+6 2 | | | | 5+6 2 | 3+4 2 | |
| | 3+4 2 | 3+4 2 | | | | 3+4 2 | 5+6 2 | |
| | 3+4 2 | 3+4 2 | | | | 3+4 2 | 3+4 2 | |
| | 3+4 2 | 5+6 2 | | | | 3+4 2 | 5+6 2 | |
| | 5+6 2 | 3+4 2 | | | | 5+6 2 | 3+4 2 | |
| | 3+4 2 | 5+6 2 | | | | 3+4 2 | 5+6 2 | |
| | 5+6 2 | 3+4 2 | | | | 5+6 2 | 5+6 2 | |
| | 3+4 2 | 5+6 2 | | | | 3+4 2 | 3+4 2 | |
| | 5+6 2 | 5+6 2 | | | | 5+6 2 | 3+4 2 | |
| | 5+6 2 | 3+4 2 | | | | 3+4 2 | 5+6 2 | |
| | 3+4 2 | 5+6 2 | | | | 3+4 2 | 3+4 2 | |
| | 5+6 2 | 3+4 2 | | | | 5+6 2 | 3+4 2 | |
| | 3+4 2 | 3+4 2 | | | | 3+4 2 | 5+6 2 | |
| | 3+4 2 | 5+6 2 | | | | 5+6 2 | 3+4 2 | |
| | 5+6 2 | 3+4 2 | | | | 3+4 2 | 5+6 2 | |
| | 5+6 2 | 3+4 2 | | | | 5+6 2 | 3+4 2 | |
| | 3+4 2 | 5+6 2 | | | | 5+6 2 | 5+6 2 | |
| 11.2.45 | 62901 | | 224 | H3Dy3Ge1O8 | | | | |
| | -A | -B | -C | -D | -E | -GM | -Y | -Z |
| | 3+4 2 | 3+4 2 | 2+2 4 | 2+2 4 | 2+2 4 | 3+4 2 | 5+6 2 | 2+2 4 |
| | 5+6 2 | 5+6 2 | 2+2 4 | 2+2 4 | 2+2 4 | 5+6 2 | 3+4 2 | 2+2 4 |
| | 3+4 2 | 3+4 2 | 2+2 4 | 2+2 4 | 2+2 4 | 3+4 2 | 5+6 2 | 2+2 4 |
| | 5+6 2 | 3+4 2 | 2+2 4 | 2+2 4 | 2+2 4 | 3+4 2 | 3+4 2 | 2+2 4 |
| | 3+4 2 | 5+6 2 | 2+2 4 | 2+2 4 | 2+2 4 | 5+6 2 | 5+6 2 | 2+2 4 |
| | 5+6 2 | 5+6 2 | 2+2 4 | 2+2 4 | 2+2 4 | 5+6 2 | 3+4 2 | 2+2 4 |
| | 5+6 2 | 3+4 2 | 2+2 4 | 2+2 4 | 2+2 4 | 3+4 2 | 5+6 2 | 2+2 4 |
| | 3+4 2 | 5+6 2 | 2+2 4 | 2+2 4 | 2+2 4 | 5+6 2 | 3+4 2 | 2+2 4 |
| | 3+4 2 | 3+4 2 | 2+2 4 | 2+2 4 | 2+2 4 | 3+4 2 | 5+6 2 | 2+2 4 |
| | 5+6 2 | 3+4 2 | 2+2 4 | 2+2 4 | 2+2 4 | 3+4 2 | 3+4 2 | 2+2 4 |
| | 3+4 2 | 5+6 2 | 2+2 4 | 2+2 4 | 2+2 4 | 5+6 2 | 5+6 2 | 2+2 4 |
| | 5+6 2 | 5+6 2 | 2+2 4 | 2+2 4 | 2+2 4 | 5+6 2 | 3+4 2 | 2+2 4 |
| | 5+6 2 | 5+6 2 | 2+2 4 | 2+2 4 | 2+2 4 | 3+4 2 | 5+6 2 | 2+2 4 |
| | 3+4 2 | 3+4 2 | 2+2 4 | 2+2 4 | 2+2 4 | 5+6 2 | 3+4 2 | 2+2 4 |
| | 5+6 2 | 3+4 2 | 2+2 4 | 2+2 4 | 2+2 4 | 3+4 2 | 5+6 2 | 2+2 4 |
| | 3+4 2 | 5+6 2 | 2+2 4 | 2+2 4 | 2+2 4 | 5+6 2 | 3+4 2 | 2+2 4 |
| | 3+4 2 | 3+4 2 | 2+2 4 | 2+2 4 | 2+2 4 | 3+4 2 | 5+6 2 | 2+2 4 |
| | 5+6 2 | 5+6 2 | 2+2 4 | 2+2 4 | 2+2 4 | 5+6 2 | 3+4 2 | 2+2 4 |
| | 3+4 2 | 3+4 2 | 2+2 4 | 2+2 4 | 2+2 4 | 3+4 2 | 3+4 2 | 2+2 4 |
| | 5+6 2 | 5+6 2 | 2+2 4 | 2+2 4 | 2+2 4 | 3+4 2 | 5+6 2 | 2+2 4 |
| | 3+4 2 | 3+4 2 | 2+2 4 | 2+2 4 | 2+2 4 | 5+6 2 | 5+6 2 | 2+2 4 |
| | 5+6 2 | 5+6 2 | 2+2 4 | 2+2 4 | 2+2 4 | 3+4 2 | 3+4 2 | 2+2 4 |
| | 3+4 2 | 3+4 2 | 2+2 4 | 2+2 4 | 2+2 4 | 5+6 2 | 5+6 2 | 2+2 4 |
| | 5+6 2 | 5+6 2 | 2+2 4 | 2+2 4 | 2+2 4 | 5+6 2 | 3+4 2 | 2+2 4 |
| | 5+6 2 | 3+4 2 | 2+2 4 | 2+2 4 | 2+2 4 | 3+4 2 | 3+4 2 | 2+2 4 |
| | 3+4 2 | 5+6 2 | 2+2 4 | 2+2 4 | 2+2 4 | 5+6 2 | 5+6 2 | 2+2 4 |
| | 3+4 2 | 5+6 2 | 2+2 4 | 2+2 4 | 2+2 4 | 5+6 2 | 5+6 2 | 2+2 4 |
| | 5+6 2 | 3+4 2 | 2+2 4 | 2+2 4 | 2+2 4 | 3+4 2 | 3+4 2 | 2+2 4 |
| | 3+4 2 | 3+4 2 | 2+2 4 | 2+2 4 | 2+2 4 | 3+4 2 | 5+6 2 | 2+2 4 |
| | 5+6 2 | 5+6 2 | 2+2 4 | 2+2 4 | 2+2 4 | 5+6 2 | 3+4 2 | 2+2 4 |
| | 5+6 2 | 5+6 2 | 2+2 4 | 2+2 4 | 2+2 4 | 5+6 2 | 3+4 2 | 2+2 4 |
| | 3+4 2 | 5+6 2 | 2+2 4 | 2+2 4 | 2+2 4 | 5+6 2 | 5+6 2 | 2+2 4 |
| | 3+4 2 | 3+4 2 | 2+2 4 | 2+2 4 | 2+2 4 | 3+4 2 | 3+4 2 | 2+2 4 |
| | 5+6 2 | 3+4 2 | 2+2 4 | 2+2 4 | 2+2 4 | 3+4 2 | 5+6 2 | 2+2 4 |
| | 5+6 2 | 3+4 2 | 2+2 4 | 2+2 4 | 2+2 4 | 3+4 2 | 3+4 2 | 2+2 4 |
| | 3+4 2 | 5+6 2 | 2+2 4 | 2+2 4 | 2+2 4 | 5+6 2 | 5+6 2 | 2+2 4 |

Continued on next column

**TABLE XXIII– continued**

| TQC | ICSD | NoE | SF | | | | | |
|---|---|---|---|---|---|---|---|---|
| | 5+6 2 | 5+6 2 | 2+2 4 | 2+2 4 | 2+2 4 | 5+6 2 | 3+4 2 | 2+2 4 |
| | 3+4 2 | 3+4 2 | 2+2 4 | 2+2 4 | 2+2 4 | 3+4 2 | 3+4 2 | 2+2 4 |
| | 5+6 2 | 5+6 2 | 2+2 4 | 2+2 4 | 2+2 4 | 5+6 2 | 5+6 2 | 2+2 4 |
| | 3+4 2 | 3+4 2 | 2+2 4 | 2+2 4 | 2+2 4 | 3+4 2 | 5+6 2 | 2+2 4 |
| | 5+6 2 | 3+4 2 | 2+2 4 | 2+2 4 | 2+2 4 | 3+4 2 | 5+6 2 | 2+2 4 |
| | 3+4 2 | 5+6 2 | 2+2 4 | 2+2 4 | 2+2 4 | 5+6 2 | 3+4 2 | 2+2 4 |
| | 5+6 2 | 5+6 2 | 2+2 4 | 2+2 4 | 2+2 4 | 3+4 2 | 5+6 2 | 2+2 4 |
| | 3+4 2 | 3+4 2 | 2+2 4 | 2+2 4 | 2+2 4 | 5+6 2 | 3+4 2 | 2+2 4 |
| | 5+6 2 | 3+4 2 | 2+2 4 | 2+2 4 | 2+2 4 | 3+4 2 | 5+6 2 | 2+2 4 |
| | 3+4 2 | 5+6 2 | 2+2 4 | 2+2 4 | 2+2 4 | 5+6 2 | 3+4 2 | 2+2 4 |
| | 3+4 2 | 3+4 2 | 2+2 4 | 2+2 4 | 2+2 4 | 3+4 2 | 3+4 2 | 2+2 4 |
| | 5+6 2 | 5+6 2 | 2+2 4 | 2+2 4 | 2+2 4 | 5+6 2 | 5+6 2 | 2+2 4 |
| | 3+4 2 | 3+4 2 | 2+2 4 | 2+2 4 | 2+2 4 | 5+6 2 | 3+4 2 | 2+2 4 |
| | 5+6 2 | 5+6 2 | 2+2 4 | 2+2 4 | 2+2 4 | 3+4 2 | 5+6 2 | 2+2 4 |
| | 3+4 2 | 3+4 2 | 2+2 4 | 2+2 4 | 2+2 4 | 5+6 2 | 3+4 2 | 2+2 4 |
| | 5+6 2 | 5+6 2 | 2+2 4 | 2+2 4 | 2+2 4 | 3+4 2 | 5+6 2 | 2+2 4 |
| | 3+4 2 | 3+4 2 | 2+2 4 | 2+2 4 | 2+2 4 | 3+4 2 | 3+4 2 | 2+2 4 |
| | 5+6 2 | 5+6 2 | 2+2 4 | 2+2 4 | 2+2 4 | 5+6 2 | 5+6 2 | 2+2 4 |
| | 5+6 2 | 5+6 2 | 2+2 4 | 2+2 4 | 2+2 4 | 3+4 2 | 3+4 2 | 2+2 4 |
| | 3+4 2 | 5+6 2 | 2+2 4 | 2+2 4 | 2+2 4 | 5+6 2 | 5+6 2 | 2+2 4 |
| | 5+6 2 | 3+4 2 | | | | 3+4 2 | 3+4 2 | |
| | 3+4 2 | 3+4 2 | | | | 3+4 2 | 5+6 2 | |
| | 3+4 2 | 3+4 2 | | | | 5+6 2 | 5+6 2 | |
| | 5+6 2 | 3+4 2 | | | | 3+4 2 | 3+4 2 | |
| | 3+4 2 | 5+6 2 | | | | 3+4 2 | 5+6 2 | |
| | 5+6 2 | 3+4 2 | | | | 5+6 2 | 5+6 2 | |
| | 3+4 2 | 5+6 2 | | | | 5+6 2 | 3+4 2 | |
| | 5+6 2 | 3+4 2 | | | | 3+4 2 | 3+4 2 | |
| | 5+6 2 | 5+6 2 | | | | 3+4 2 | 3+4 2 | |
| | 3+4 2 | 5+6 2 | | | | 5+6 2 | 5+6 2 | |
| | 3+4 2 | 3+4 2 | | | | 3+4 2 | 5+6 2 | |
| | 3+4 2 | 5+6 2 | | | | 3+4 2 | 5+6 2 | |
| | 5+6 2 | 3+4 2 | | | | 5+6 2 | 3+4 2 | |
| | 5+6 2 | 3+4 2 | | | | 5+6 2 | 3+4 2 | |
| | 3+4 2 | 5+6 2 | | | | 3+4 2 | 3+4 2 | |
| | 5+6 2 | 3+4 2 | | | | 5+6 2 | 5+6 2 | |
| | 3+4 2 | 5+6 2 | | | | 3+4 2 | 3+4 2 | |
| | 5+6 2 | 5+6 2 | | | | 5+6 2 | 3+4 2 | |
| | 3+4 2 | 3+4 2 | | | | 3+4 2 | 5+6 2 | |
| | 5+6 2 | 5+6 2 | | | | 5+6 2 | 3+4 2 | |
| | 3+4 2 | 3+4 2 | | | | 5+6 2 | 5+6 2 | |
| | 5+6 2 | 5+6 2 | | | | 3+4 2 | 5+6 2 | |
| | 3+4 2 | 5+6 2 | | | | 5+6 2 | 5+6 2 | |
| | 5+6 2 | 5+6 2 | | | | 3+4 2 | 3+4 2 | |
| | 3+4 2 | 3+4 2 | | | | 5+6 2 | 3+4 2 | |
| | 5+6 2 | 3+4 2 | | | | 3+4 2 | 5+6 2 | |
| | 3+4 2 | 5+6 2 | | | | 5+6 2 | 5+6 2 | |
| | 5+6 2 | 5+6 2 | | | | 3+4 2 | 3+4 2 | |
| | 3+4 2 | 3+4 2 | | | | 5+6 2 | 5+6 2 | |
| | 5+6 2 | 3+4 2 | | | | 3+4 2 | 5+6 2 | |
| | 3+4 2 | 5+6 2 | | | | 3+4 2 | 3+4 2 | |
| | 3+4 2 | 3+4 2 | | | | 5+6 2 | 5+6 2 | |
| | 5+6 2 | 3+4 2 | | | | 3+4 2 | 3+4 2 | |
| | 5+6 2 | 5+6 2 | | | | 5+6 2 | 3+4 2 | |
| | 3+4 2 | 3+4 2 | | | | 3+4 2 | 5+6 2 | |
| | 5+6 2 | 5+6 2 | | | | 5+6 2 | 3+4 2 | |
| | 3+4 2 | 3+4 2 | | | | 5+6 2 | 5+6 2 | |
| | 5+6 2 | 5+6 2 | | | | 3+4 2 | 3+4 2 | |
| | 3+4 2 | 3+4 2 | | | | 5+6 2 | 5+6 2 | |
| | 5+6 2 | 5+6 2 | | | | 3+4 2 | 3+4 2 | |
| | 5+6 2 | 5+6 2 | | | | 5+6 2 | 5+6 2 | |

Continued on next column

**TABLE XXIII– continued**

| TQC | ICSD | NoE | SF | | | | | |
|---|---|---|---|---|---|---|---|---|
| | 3+4 2 | 3+4 2 | | | | 3+4 2 | 5+6 2 | |
| | 3+4 2 | 5+6 2 | | | | 3+4 2 | 3+4 2 | |
| | 5+6 2 | 5+6 2 | | | | 5+6 2 | 3+4 2 | |
| | 5+6 2 | 5+6 2 | | | | 3+4 2 | 5+6 2 | |
| | 3+4 2 | 3+4 2 | | | | 5+6 2 | 3+4 2 | |
| | 3+4 2 | 3+4 2 | | | | 5+6 2 | 5+6 2 | |
| | 5+6 2 | 3+4 2 | | | | 3+4 2 | 3+4 2 | |
| | 3+4 2 | 3+4 2 | | | | 5+6 2 | 5+6 2 | |
| | 5+6 2 | 5+6 2 | | | | 5+6 2 | 3+4 2 | |
| | 5+6 2 | 3+4 2 | | | | 3+4 2 | 3+4 2 | |
| | 3+4 2 | 3+4 2 | | | | 3+4 2 | 5+6 2 | |
| | 5+6 2 | 5+6 2 | | | | 3+4 2 | 5+6 2 | |
| | 3+4 2 | 3+4 2 | | | | 5+6 2 | 3+4 2 | |
| | 3+4 2 | 5+6 2 | | | | 3+4 2 | 5+6 2 | |
| | 3+4 2 | 5+6 2 | | | | 5+6 2 | 3+4 2 | |
| 11.2.46 | 70101 | 100 | H3Cl1Cu2O3 | | | | | |
| | -A | -B | -C | -D | -E | -GM | -Y | -Z |
| | 5+6 2 | 5+6 2 | 2+2 4 | 2+2 4 | 2+2 4 | 3+4 2 | 3+4 2 | 2+2 4 |
| | 3+4 2 | 5+6 2 | 2+2 4 | 2+2 4 | 2+2 4 | 3+4 2 | 5+6 2 | 2+2 4 |
| | 5+6 2 | 3+4 2 | 2+2 4 | 2+2 4 | 2+2 4 | 5+6 2 | 3+4 2 | 2+2 4 |
| | 3+4 2 | 3+4 2 | 2+2 4 | 2+2 4 | 2+2 4 | 5+6 2 | 5+6 2 | 2+2 4 |
| | 3+4 2 | 5+6 2 | 2+2 4 | 2+2 4 | 2+2 4 | 3+4 2 | 5+6 2 | 2+2 4 |
| | 5+6 2 | 3+4 2 | 2+2 4 | 2+2 4 | 2+2 4 | 5+6 2 | 3+4 2 | 2+2 4 |
| | 5+6 2 | 3+4 2 | 2+2 4 | 2+2 4 | 2+2 4 | 3+4 2 | 5+6 2 | 2+2 4 |
| | 3+4 2 | 5+6 2 | 2+2 4 | 2+2 4 | 2+2 4 | 5+6 2 | 3+4 2 | 2+2 4 |
| | 5+6 2 | 5+6 2 | 2+2 4 | 2+2 4 | 2+2 4 | 3+4 2 | 3+4 2 | 2+2 4 |
| | 3+4 2 | 5+6 2 | 2+2 4 | 2+2 4 | 2+2 4 | 3+4 2 | 5+6 2 | 2+2 4 |
| | 3+4 2 | 3+4 2 | 2+2 4 | 2+2 4 | 2+2 4 | 5+6 2 | 5+6 2 | 2+2 4 |
| | 5+6 2 | 3+4 2 | 2+2 4 | 2+2 4 | 2+2 4 | 5+6 2 | 3+4 2 | 2+2 4 |
| | 3+4 2 | 5+6 2 | 2+2 4 | 2+2 4 | 2+2 4 | 3+4 2 | 5+6 2 | 2+2 4 |
| | 5+6 2 | 3+4 2 | 2+2 4 | 2+2 4 | 2+2 4 | 5+6 2 | 3+4 2 | 2+2 4 |
| | 3+4 2 | 5+6 2 | 2+2 4 | 2+2 4 | 2+2 4 | 3+4 2 | 5+6 2 | 2+2 4 |
| | 3+4 2 | 3+4 2 | 2+2 4 | 2+2 4 | 2+2 4 | 5+6 2 | 5+6 2 | 2+2 4 |
| | 5+6 2 | 5+6 2 | 2+2 4 | 2+2 4 | 2+2 4 | 3+4 2 | 3+4 2 | 2+2 4 |
| | 5+6 2 | 5+6 2 | 2+2 4 | 2+2 4 | 2+2 4 | 3+4 2 | 3+4 2 | 2+2 4 |
| | 5+6 2 | 5+6 2 | 2+2 4 | 2+2 4 | 2+2 4 | 3+4 2 | 5+6 2 | 2+2 4 |
| | 3+4 2 | 5+6 2 | 2+2 4 | 2+2 4 | 2+2 4 | 3+4 2 | 3+4 2 | 2+2 4 |
| | 3+4 2 | 3+4 2 | 2+2 4 | 2+2 4 | 2+2 4 | 5+6 2 | 3+4 2 | 2+2 4 |
| | 5+6 2 | 5+6 2 | 2+2 4 | 2+2 4 | 2+2 4 | 3+4 2 | 5+6 2 | 2+2 4 |
| | 3+4 2 | 3+4 2 | 2+2 4 | 2+2 4 | 2+2 4 | 5+6 2 | 5+6 2 | 2+2 4 |
| | 3+4 2 | 3+4 2 | 2+2 4 | 2+2 4 | 2+2 4 | 5+6 2 | 5+6 2 | 2+2 4 |
| | 5+6 2 | 3+4 2 | 2+2 4 | 2+2 4 | 2+2 4 | 5+6 2 | 3+4 2 | 2+2 4 |
| | 5+6 2 | 3+4 2 | | | | 5+6 2 | 3+4 2 | |
| | 3+4 2 | 5+6 2 | | | | 3+4 2 | 5+6 2 | |
| | 3+4 2 | 5+6 2 | | | | 3+4 2 | 5+6 2 | |
| | 3+4 2 | 3+4 2 | | | | 3+4 2 | 5+6 2 | |
| | 5+6 2 | 5+6 2 | | | | 3+4 2 | 5+6 2 | |
| | 3+4 2 | 5+6 2 | | | | 5+6 2 | 3+4 2 | |
| | 3+4 2 | 5+6 2 | | | | 3+4 2 | 5+6 2 | |
| | 3+4 2 | 5+6 2 | | | | 3+4 2 | 3+4 2 | |
| | 5+6 2 | 5+6 2 | | | | 5+6 2 | 5+6 2 | |
| | 3+4 2 | 3+4 2 | | | | 3+4 2 | 5+6 2 | |
| | 3+4 2 | 5+6 2 | | | | 3+4 2 | 3+4 2 | |
| | 5+6 2 | 5+6 2 | | | | 5+6 2 | 5+6 2 | |
| | 3+4 2 | 3+4 2 | | | | 3+4 2 | 3+4 2 | |
| | 3+4 2 | 5+6 2 | | | | 3+4 2 | 5+6 2 | |
| | 5+6 2 | 5+6 2 | | | | 3+4 2 | 5+6 2 | |
| | 3+4 2 | 3+4 2 | | | | 5+6 2 | 5+6 2 | |
| | 3+4 2 | 3+4 2 | | | | 3+4 2 | 5+6 2 | |
| | 5+6 2 | 5+6 2 | | | | 5+6 2 | 3+4 2 | |
| | 5+6 2 | 3+4 2 | | | | 5+6 2 | 3+4 2 | |

Continued on next column

**TABLE XXIII– continued**

| TQC | ICSD | | NoE | | SF | | | |
|---|---|---|---|---|---|---|---|---|
| | 5+6 2 | 5+6 2 | | | | 3+4 2 | 5+6 2 | |
| | 3+4 2 | 5+6 2 | | | | 3+4 2 | 5+6 2 | |
| | 3+4 2 | 3+4 2 | | | | 5+6 2 | 3+4 2 | |
| | 3+4 2 | 5+6 2 | | | | 3+4 2 | 5+6 2 | |
| | 3+4 2 | 3+4 2 | | | | 5+6 2 | 5+6 2 | |
| | 3+4 2 | 5+6 2 | | | | 3+4 2 | 5+6 2 | |
| 11.2.47 | 153707 | | 100 | | H3Cl1Cu2O3 | | | |
| | -A | -B | -C | -D | -E | -GM | -Y | -Z |
| | 3+4 2 | 3+4 2 | 2+2 4 | 2+2 4 | 2+2 4 | 3+4 2 | 3+4 2 | 2+2 4 |
| | 5+6 2 | 3+4 2 | 2+2 4 | 2+2 4 | 2+2 4 | 3+4 2 | 5+6 2 | 2+2 4 |
| | 3+4 2 | 5+6 2 | 2+2 4 | 2+2 4 | 2+2 4 | 5+6 2 | 3+4 2 | 2+2 4 |
| | 5+6 2 | 5+6 2 | 2+2 4 | 2+2 4 | 2+2 4 | 5+6 2 | 5+6 2 | 2+2 4 |
| | 5+6 2 | 3+4 2 | 2+2 4 | 2+2 4 | 2+2 4 | 3+4 2 | 5+6 2 | 2+2 4 |
| | 3+4 2 | 5+6 2 | 2+2 4 | 2+2 4 | 2+2 4 | 5+6 2 | 3+4 2 | 2+2 4 |
| | 3+4 2 | 5+6 2 | 2+2 4 | 2+2 4 | 2+2 4 | 3+4 2 | 5+6 2 | 2+2 4 |
| | 5+6 2 | 3+4 2 | 2+2 4 | 2+2 4 | 2+2 4 | 5+6 2 | 3+4 2 | 2+2 4 |
| | 5+6 2 | 5+6 2 | 2+2 4 | 2+2 4 | 2+2 4 | 5+6 2 | 5+6 2 | 2+2 4 |
| | 3+4 2 | 3+4 2 | 2+2 4 | 2+2 4 | 2+2 4 | 3+4 2 | 3+4 2 | 2+2 4 |
| | 5+6 2 | 3+4 2 | 2+2 4 | 2+2 4 | 2+2 4 | 3+4 2 | 5+6 2 | 2+2 4 |
| | 5+6 2 | 3+4 2 | 2+2 4 | 2+2 4 | 2+2 4 | 3+4 2 | 5+6 2 | 2+2 4 |
| | 3+4 2 | 5+6 2 | 2+2 4 | 2+2 4 | 2+2 4 | 5+6 2 | 3+4 2 | 2+2 4 |
| | 3+4 2 | 5+6 2 | 2+2 4 | 2+2 4 | 2+2 4 | 3+4 2 | 3+4 2 | 2+2 4 |
| | 3+4 2 | 3+4 2 | 2+2 4 | 2+2 4 | 2+2 4 | 5+6 2 | 5+6 2 | 2+2 4 |
| | 5+6 2 | 3+4 2 | 2+2 4 | 2+2 4 | 2+2 4 | 3+4 2 | 5+6 2 | 2+2 4 |
| | 5+6 2 | 5+6 2 | 2+2 4 | 2+2 4 | 2+2 4 | 3+4 2 | 3+4 2 | 2+2 4 |
| | 5+6 2 | 3+4 2 | 2+2 4 | 2+2 4 | 2+2 4 | 5+6 2 | 5+6 2 | 2+2 4 |
| | 3+4 2 | 3+4 2 | 2+2 4 | 2+2 4 | 2+2 4 | 3+4 2 | 3+4 2 | 2+2 4 |
| | 3+4 2 | 3+4 2 | 2+2 4 | 2+2 4 | 2+2 4 | 3+4 2 | 3+4 2 | 2+2 4 |
| | 5+6 2 | 5+6 2 | 2+2 4 | 2+2 4 | 2+2 4 | 5+6 2 | 5+6 2 | 2+2 4 |
| | 5+6 2 | 3+4 2 | 2+2 4 | 2+2 4 | 2+2 4 | 5+6 2 | 3+4 2 | 2+2 4 |
| | 5+6 2 | 3+4 2 | 2+2 4 | 2+2 4 | 2+2 4 | 3+4 2 | 5+6 2 | 2+2 4 |
| | 3+4 2 | 5+6 2 | 2+2 4 | 2+2 4 | 2+2 4 | 3+4 2 | 5+6 2 | 2+2 4 |
| | 5+6 2 | 3+4 2 | 2+2 4 | 2+2 4 | 2+2 4 | 3+4 2 | 5+6 2 | 2+2 4 |
| | 5+6 2 | 5+6 2 | | | | 3+4 2 | 3+4 2 | |
| | 3+4 2 | 5+6 2 | | | | 5+6 2 | 5+6 2 | |
| | 3+4 2 | 3+4 2 | | | | 5+6 2 | 3+4 2 | |
| | 3+4 2 | 5+6 2 | | | | 3+4 2 | 5+6 2 | |
| | 5+6 2 | 3+4 2 | | | | 5+6 2 | 5+6 2 | |
| | 5+6 2 | 5+6 2 | | | | 5+6 2 | 3+4 2 | |
| | 5+6 2 | 3+4 2 | | | | 3+4 2 | 3+4 2 | |
| | 5+6 2 | 5+6 2 | | | | 5+6 2 | 3+4 2 | |
| | 3+4 2 | 3+4 2 | | | | 3+4 2 | 5+6 2 | |
| | 3+4 2 | 5+6 2 | | | | 3+4 2 | 5+6 2 | |
| | 5+6 2 | 3+4 2 | | | | 3+4 2 | 3+4 2 | |
| | 5+6 2 | 3+4 2 | | | | 5+6 2 | 5+6 2 | |
| | 3+4 2 | 3+4 2 | | | | 3+4 2 | 5+6 2 | |
| | 5+6 2 | 5+6 2 | | | | 3+4 2 | 5+6 2 | |
| | 3+4 2 | 5+6 2 | | | | 5+6 2 | 5+6 2 | |
| | 5+6 2 | 3+4 2 | | | | 3+4 2 | 5+6 2 | |
| | 5+6 2 | 3+4 2 | | | | 3+4 2 | 3+4 2 | |
| | 3+4 2 | 3+4 2 | | | | 5+6 2 | 3+4 2 | |
| | 3+4 2 | 5+6 2 | | | | 5+6 2 | 5+6 2 | |
| | 5+6 2 | 5+6 2 | | | | 5+6 2 | 5+6 2 | |
| | 5+6 2 | 3+4 2 | | | | 3+4 2 | 5+6 2 | |
| | 5+6 2 | 3+4 2 | | | | 3+4 2 | 3+4 2 | |
| | 5+6 2 | 3+4 2 | | | | 3+4 2 | 5+6 2 | |
| | 5+6 2 | 5+6 2 | | | | 3+4 2 | 5+6 2 | |
| | 5+6 2 | 3+4 2 | | | | 5+6 2 | 5+6 2 | |
| 11.2.48 | 166218 | | 76 | | H2O4P1Rb1 | | | |
| | -A | -B | -C | -D | -E | -GM | -Y | -Z |
| | 3+4 2 | 5+6 2 | 2+2 4 | 2+2 4 | 2+2 4 | 3+4 2 | 3+4 2 | 2+2 4 |

Continued on next column

**TABLE XXIII– continued**

| TQC | ICSD | | NoE | | SF | | | |
|---|---|---|---|---|---|---|---|---|
| | 5+6 2 | 3+4 2 | 2+2 4 | 2+2 4 | 2+2 4 | 5+6 2 | 5+6 2 | 2+2 4 |
| | 5+6 2 | 3+4 2 | 2+2 4 | 2+2 4 | 2+2 4 | 3+4 2 | 5+6 2 | 2+2 4 |
| | 3+4 2 | 5+6 2 | 2+2 4 | 2+2 4 | 2+2 4 | 5+6 2 | 3+4 2 | 2+2 4 |
| | 5+6 2 | 3+4 2 | 2+2 4 | 2+2 4 | 2+2 4 | 3+4 2 | 5+6 2 | 2+2 4 |
| | 5+6 2 | 3+4 2 | 2+2 4 | 2+2 4 | 2+2 4 | 3+4 2 | 5+6 2 | 2+2 4 |
| | 3+4 2 | 5+6 2 | 2+2 4 | 2+2 4 | 2+2 4 | 5+6 2 | 3+4 2 | 2+2 4 |
| | 5+6 2 | 5+6 2 | 2+2 4 | 2+2 4 | 2+2 4 | 5+6 2 | 3+4 2 | 2+2 4 |
| | 3+4 2 | 5+6 2 | 2+2 4 | 2+2 4 | 2+2 4 | 3+4 2 | 5+6 2 | 2+2 4 |
| | 3+4 2 | 3+4 2 | 2+2 4 | 2+2 4 | 2+2 4 | 5+6 2 | 3+4 2 | 2+2 4 |
| | 3+4 2 | 3+4 2 | 2+2 4 | 2+2 4 | 2+2 4 | 3+4 2 | 3+4 2 | 2+2 4 |
| | 5+6 2 | 5+6 2 | 2+2 4 | 2+2 4 | 2+2 4 | 5+6 2 | 5+6 2 | 2+2 4 |
| | 3+4 2 | 3+4 2 | 2+2 4 | 2+2 4 | 2+2 4 | 3+4 2 | 3+4 2 | 2+2 4 |
| | 5+6 2 | 5+6 2 | 2+2 4 | 2+2 4 | 2+2 4 | 5+6 2 | 5+6 2 | 2+2 4 |
| | 3+4 2 | 5+6 2 | 2+2 4 | 2+2 4 | 2+2 4 | 3+4 2 | 5+6 2 | 2+2 4 |
| | 5+6 2 | 3+4 2 | 2+2 4 | 2+2 4 | 2+2 4 | 5+6 2 | 3+4 2 | 2+2 4 |
| | 5+6 2 | 3+4 2 | 2+2 4 | 2+2 4 | 2+2 4 | 3+4 2 | 5+6 2 | 2+2 4 |
| | 3+4 2 | 5+6 2 | 2+2 4 | 2+2 4 | 2+2 4 | 5+6 2 | 3+4 2 | 2+2 4 |
| | 3+4 2 | 3+4 2 | 2+2 4 | 2+2 4 | 2+2 4 | 3+4 2 | 3+4 2 | 2+2 4 |
| | 5+6 2 | 5+6 2 | | | | 5+6 2 | 5+6 2 | |
| | 5+6 2 | 3+4 2 | | | | 3+4 2 | 5+6 2 | |
| | 3+4 2 | 5+6 2 | | | | 3+4 2 | 3+4 2 | |
| | 3+4 2 | 3+4 2 | | | | 5+6 2 | 5+6 2 | |
| | 5+6 2 | 5+6 2 | | | | 5+6 2 | 3+4 2 | |
| | 3+4 2 | 5+6 2 | | | | 5+6 2 | 5+6 2 | |
| | 5+6 2 | 3+4 2 | | | | 3+4 2 | 3+4 2 | |
| | 5+6 2 | 5+6 2 | | | | 3+4 2 | 3+4 2 | |
| | 5+6 2 | 3+4 2 | | | | 3+4 2 | 3+4 2 | |
| | 3+4 2 | 3+4 2 | | | | 3+4 2 | 5+6 2 | |
| | 5+6 2 | 5+6 2 | | | | 5+6 2 | 5+6 2 | |
| | 3+4 2 | 3+4 2 | | | | 5+6 2 | 5+6 2 | |
| | 5+6 2 | 5+6 2 | | | | 5+6 2 | 5+6 2 | |
| | 3+4 2 | 3+4 2 | | | | 3+4 2 | 3+4 2 | |
| | 3+4 2 | 5+6 2 | | | | 5+6 2 | 3+4 2 | |
| | 5+6 2 | 3+4 2 | | | | 3+4 2 | 5+6 2 | |
| | 5+6 2 | 5+6 2 | | | | 5+6 2 | 5+6 2 | |
| | 3+4 2 | 3+4 2 | | | | 3+4 2 | 3+4 2 | |
| | 3+4 2 | 5+6 2 | | | | 3+4 2 | 3+4 2 | |
| 11.2.49 | 27406 | | 28 | | H2N1Rb1 | | | |
| | -A | -B | -C | -D | -E | -GM | -Y | -Z |
| | 5+6 2 | 5+6 2 | 2+2 4 | 2+2 4 | 2+2 4 | 3+4 2 | 3+4 2 | 2+2 4 |
| | 3+4 2 | 3+4 2 | 2+2 4 | 2+2 4 | 2+2 4 | 5+6 2 | 5+6 2 | 2+2 4 |
| | 3+4 2 | 3+4 2 | 2+2 4 | 2+2 4 | 2+2 4 | 3+4 2 | 5+6 2 | 2+2 4 |
| | 5+6 2 | 5+6 2 | 2+2 4 | 2+2 4 | 2+2 4 | 5+6 2 | 3+4 2 | 2+2 4 |
| | 5+6 2 | 5+6 2 | 2+2 4 | 2+2 4 | 2+2 4 | 3+4 2 | 3+4 2 | 2+2 4 |
| | 3+4 2 | 3+4 2 | 2+2 4 | 2+2 4 | 2+2 4 | 5+6 2 | 5+6 2 | 2+2 4 |
| | 5+6 2 | 5+6 2 | 2+2 4 | 2+2 4 | 2+2 4 | 3+4 2 | 3+4 2 | 2+2 4 |
| | 5+6 2 | 5+6 2 | | | | 3+4 2 | 5+6 2 | |
| | 3+4 2 | 3+4 2 | | | | 5+6 2 | 3+4 2 | |
| | 3+4 2 | 3+4 2 | | | | 5+6 2 | 5+6 2 | |
| | 5+6 2 | 3+4 2 | | | | 3+4 2 | 3+4 2 | |
| | 3+4 2 | 5+6 2 | | | | 3+4 2 | 5+6 2 | |
| | 3+4 2 | 3+4 2 | | | | 3+4 2 | 5+6 2 | |
| | 3+4 2 | 3+4 2 | | | | 5+6 2 | 5+6 2 | |
| 11.2.50 | 169093 | | 76 | | H2K1O4P1 | | | |
| | -A | -B | -C | -D | -E | -GM | -Y | -Z |
| | 5+6 2 | 3+4 2 | 2+2 4 | 2+2 4 | 2+2 4 | 3+4 2 | 5+6 2 | 2+2 4 |
| | 3+4 2 | 5+6 2 | 2+2 4 | 2+2 4 | 2+2 4 | 5+6 2 | 3+4 2 | 2+2 4 |
| | 3+4 2 | 3+4 2 | 2+2 4 | 2+2 4 | 2+2 4 | 3+4 2 | 3+4 2 | 2+2 4 |
| | 5+6 2 | 5+6 2 | 2+2 4 | 2+2 4 | 2+2 4 | 5+6 2 | 5+6 2 | 2+2 4 |
| | 3+4 2 | 3+4 2 | 2+2 4 | 2+2 4 | 2+2 4 | 3+4 2 | 3+4 2 | 2+2 4 |
| | 5+6 2 | 5+6 2 | 2+2 4 | 2+2 4 | 2+2 4 | 5+6 2 | 5+6 2 | 2+2 4 |

Continued on next column

**TABLE XXIII– continued**

| TQC | ICSD | NoE | SF | | | | | |
|---|---|---|---|---|---|---|---|---|
| | 3+4 2 | 3+4 2 | 2+2 4 | 2+2 4 | 2+2 4 | 3+4 2 | 3+4 2 | 2+2 4 |
| | 5+6 2 | 5+6 2 | 2+2 4 | 2+2 4 | 2+2 4 | 5+6 2 | 5+6 2 | 2+2 4 |
| | 3+4 2 | 3+4 2 | 2+2 4 | 2+2 4 | 2+2 4 | 3+4 2 | 3+4 2 | 2+2 4 |
| | 5+6 2 | 5+6 2 | 2+2 4 | 2+2 4 | 2+2 4 | 5+6 2 | 5+6 2 | 2+2 4 |
| | 5+6 2 | 3+4 2 | 2+2 4 | 2+2 4 | 2+2 4 | 3+4 2 | 5+6 2 | 2+2 4 |
| | 3+4 2 | 5+6 2 | 2+2 4 | 2+2 4 | 2+2 4 | 5+6 2 | 3+4 2 | 2+2 4 |
| | 5+6 2 | 3+4 2 | 2+2 4 | 2+2 4 | 2+2 4 | 3+4 2 | 5+6 2 | 2+2 4 |
| | 3+4 2 | 5+6 2 | 2+2 4 | 2+2 4 | 2+2 4 | 5+6 2 | 3+4 2 | 2+2 4 |
| | 5+6 2 | 3+4 2 | 2+2 4 | 2+2 4 | 2+2 4 | 3+4 2 | 5+6 2 | 2+2 4 |
| | 3+4 2 | 5+6 2 | 2+2 4 | 2+2 4 | 2+2 4 | 5+6 2 | 3+4 2 | 2+2 4 |
| | 3+4 2 | 3+4 2 | 2+2 4 | 2+2 4 | 2+2 4 | 3+4 2 | 3+4 2 | 2+2 4 |
| | 5+6 2 | 5+6 2 | 2+2 4 | 2+2 4 | 2+2 4 | 5+6 2 | 5+6 2 | 2+2 4 |
| | 3+4 2 | 3+4 2 | 2+2 4 | 2+2 4 | 2+2 4 | 3+4 2 | 3+4 2 | 2+2 4 |
| | 5+6 2 | 3+4 2 | | | | 3+4 2 | 5+6 2 | |
| | 5+6 2 | 5+6 2 | | | | 5+6 2 | 5+6 2 | |
| | 3+4 2 | 5+6 2 | | | | 5+6 2 | 3+4 2 | |
| | 3+4 2 | 5+6 2 | | | | 5+6 2 | 5+6 2 | |
| | 5+6 2 | 3+4 2 | | | | 3+4 2 | 3+4 2 | |
| | 5+6 2 | 5+6 2 | | | | 5+6 2 | 5+6 2 | |
| | 3+4 2 | 3+4 2 | | | | 3+4 2 | 5+6 2 | |
| | 5+6 2 | 5+6 2 | | | | 5+6 2 | 3+4 2 | |
| | 3+4 2 | 3+4 2 | | | | 3+4 2 | 3+4 2 | |
| | 3+4 2 | 3+4 2 | | | | 3+4 2 | 3+4 2 | |
| | 5+6 2 | 3+4 2 | | | | 3+4 2 | 5+6 2 | |
| | 3+4 2 | 5+6 2 | | | | 5+6 2 | 3+4 2 | |
| | 3+4 2 | 5+6 2 | | | | 5+6 2 | 3+4 2 | |
| | 5+6 2 | 3+4 2 | | | | 3+4 2 | 5+6 2 | |
| | 5+6 2 | 5+6 2 | | | | 5+6 2 | 5+6 2 | |
| | 3+4 2 | 3+4 2 | | | | 3+4 2 | 3+4 2 | |
| | 3+4 2 | 5+6 2 | | | | 5+6 2 | 3+4 2 | |
| | 5+6 2 | 3+4 2 | | | | 3+4 2 | 5+6 2 | |
| | 5+6 2 | 3+4 2 | | | | 3+4 2 | 5+6 2 | |
| 11.2.51 | 6163 | 284 | H2Ga11Na1O18 | | | | | |
| | -A | -B | -C | -D | -E | -GM | -Y | -Z |
| | 5+6 2 | 5+6 2 | 2+2 4 | 2+2 4 | 2+2 4 | 3+4 2 | 3+4 2 | 2+2 4 |
| | 3+4 2 | 3+4 2 | 2+2 4 | 2+2 4 | 2+2 4 | 3+4 2 | 5+6 2 | 2+2 4 |
| | 5+6 2 | 5+6 2 | 2+2 4 | 2+2 4 | 2+2 4 | 3+4 2 | 5+6 2 | 2+2 4 |
| | 5+6 2 | 5+6 2 | 2+2 4 | 2+2 4 | 2+2 4 | 5+6 2 | 3+4 2 | 2+2 4 |
| | 3+4 2 | 3+4 2 | 2+2 4 | 2+2 4 | 2+2 4 | 5+6 2 | 3+4 2 | 2+2 4 |
| | 3+4 2 | 3+4 2 | 2+2 4 | 2+2 4 | 2+2 4 | 3+4 2 | 5+6 2 | 2+2 4 |
| | 5+6 2 | 3+4 2 | 2+2 4 | 2+2 4 | 2+2 4 | 5+6 2 | 5+6 2 | 2+2 4 |
| | 3+4 2 | 5+6 2 | 2+2 4 | 2+2 4 | 2+2 4 | 3+4 2 | 3+4 2 | 2+2 4 |
| | 5+6 2 | 3+4 2 | 2+2 4 | 2+2 4 | 2+2 4 | 5+6 2 | 5+6 2 | 2+2 4 |
| | 5+6 2 | 5+6 2 | 2+2 4 | 2+2 4 | 2+2 4 | 3+4 2 | 3+4 2 | 2+2 4 |
| | 3+4 2 | 3+4 2 | 2+2 4 | 2+2 4 | 2+2 4 | 5+6 2 | 5+6 2 | 2+2 4 |
| | 3+4 2 | 5+6 2 | 2+2 4 | 2+2 4 | 2+2 4 | 3+4 2 | 3+4 2 | 2+2 4 |
| | 3+4 2 | 5+6 2 | 2+2 4 | 2+2 4 | 2+2 4 | 5+6 2 | 3+4 2 | 2+2 4 |
| | 5+6 2 | 3+4 2 | 2+2 4 | 2+2 4 | 2+2 4 | 3+4 2 | 5+6 2 | 2+2 4 |
| | 5+6 2 | 5+6 2 | 2+2 4 | 2+2 4 | 2+2 4 | 5+6 2 | 3+4 2 | 2+2 4 |
| | 3+4 2 | 3+4 2 | 2+2 4 | 2+2 4 | 2+2 4 | 5+6 2 | 5+6 2 | 2+2 4 |
| | 3+4 2 | 5+6 2 | 2+2 4 | 2+2 4 | 2+2 4 | 5+6 2 | 5+6 2 | 2+2 4 |
| | 5+6 2 | 3+4 2 | 2+2 4 | 2+2 4 | 2+2 4 | 5+6 2 | 3+4 2 | 2+2 4 |
| | 5+6 2 | 3+4 2 | 2+2 4 | 2+2 4 | 2+2 4 | 3+4 2 | 3+4 2 | 2+2 4 |
| | 3+4 2 | 5+6 2 | 2+2 4 | 2+2 4 | 2+2 4 | 3+4 2 | 5+6 2 | 2+2 4 |
| | 3+4 2 | 5+6 2 | 2+2 4 | 2+2 4 | 2+2 4 | 3+4 2 | 3+4 2 | 2+2 4 |
| | 5+6 2 | 3+4 2 | 2+2 4 | 2+2 4 | 2+2 4 | 5+6 2 | 3+4 2 | 2+2 4 |
| | 5+6 2 | 3+4 2 | 2+2 4 | 2+2 4 | 2+2 4 | 3+4 2 | 5+6 2 | 2+2 4 |
| | 5+6 2 | 5+6 2 | 2+2 4 | 2+2 4 | 2+2 4 | 5+6 2 | 5+6 2 | 2+2 4 |
| | 3+4 2 | 3+4 2 | 2+2 4 | 2+2 4 | 2+2 4 | 3+4 2 | 3+4 2 | 2+2 4 |
| | 3+4 2 | 3+4 2 | 2+2 4 | 2+2 4 | 2+2 4 | 5+6 2 | 5+6 2 | 2+2 4 |
| | 5+6 2 | 5+6 2 | 2+2 4 | 2+2 4 | 2+2 4 | 5+6 2 | 5+6 2 | 2+2 4 |

Continued on next column

**TABLE XXIII– continued**

| TQC | ICSD | NoE | SF | | | | | |
|---|---|---|---|---|---|---|---|---|
| | 3+4 2 | 5+6 2 | 2+2 4 | 2+2 4 | 2+2 4 | 3+4 2 | 3+4 2 | 2+2 4 |
| | 3+4 2 | 3+4 2 | 2+2 4 | 2+2 4 | 2+2 4 | 3+4 2 | 3+4 2 | 2+2 4 |
| | 5+6 2 | 3+4 2 | 2+2 4 | 2+2 4 | 2+2 4 | 5+6 2 | 5+6 2 | 2+2 4 |
| | 3+4 2 | 5+6 2 | 2+2 4 | 2+2 4 | 2+2 4 | 3+4 2 | 5+6 2 | 2+2 4 |
| | 5+6 2 | 5+6 2 | 2+2 4 | 2+2 4 | 2+2 4 | 5+6 2 | 3+4 2 | 2+2 4 |
| | 3+4 2 | 5+6 2 | 2+2 4 | 2+2 4 | 2+2 4 | 3+4 2 | 3+4 2 | 2+2 4 |
| | 5+6 2 | 3+4 2 | 2+2 4 | 2+2 4 | 2+2 4 | 5+6 2 | 5+6 2 | 2+2 4 |
| | 5+6 2 | 3+4 2 | 2+2 4 | 2+2 4 | 2+2 4 | 3+4 2 | 5+6 2 | 2+2 4 |
| | 3+4 2 | 5+6 2 | 2+2 4 | 2+2 4 | 2+2 4 | 5+6 2 | 3+4 2 | 2+2 4 |
| | 5+6 2 | 3+4 2 | 2+2 4 | 2+2 4 | 2+2 4 | 5+6 2 | 5+6 2 | 2+2 4 |
| | 3+4 2 | 5+6 2 | 2+2 4 | 2+2 4 | 2+2 4 | 3+4 2 | 3+4 2 | 2+2 4 |
| | 5+6 2 | 5+6 2 | 2+2 4 | 2+2 4 | 2+2 4 | 3+4 2 | 3+4 2 | 2+2 4 |
| | 3+4 2 | 3+4 2 | 2+2 4 | 2+2 4 | 2+2 4 | 3+4 2 | 5+6 2 | 2+2 4 |
| | 3+4 2 | 5+6 2 | 2+2 4 | 2+2 4 | 2+2 4 | 5+6 2 | 3+4 2 | 2+2 4 |
| | 5+6 2 | 5+6 2 | 2+2 4 | 2+2 4 | 2+2 4 | 5+6 2 | 3+4 2 | 2+2 4 |
| | 3+4 2 | 3+4 2 | 2+2 4 | 2+2 4 | 2+2 4 | 3+4 2 | 5+6 2 | 2+2 4 |
| | 5+6 2 | 3+4 2 | 2+2 4 | 2+2 4 | 2+2 4 | 3+4 2 | 5+6 2 | 2+2 4 |
| | 5+6 2 | 5+6 2 | 2+2 4 | 2+2 4 | 2+2 4 | 5+6 2 | 5+6 2 | 2+2 4 |
| | 3+4 2 | 5+6 2 | 2+2 4 | 2+2 4 | 2+2 4 | 3+4 2 | 3+4 2 | 2+2 4 |
| | 5+6 2 | 3+4 2 | 2+2 4 | 2+2 4 | 2+2 4 | 5+6 2 | 5+6 2 | 2+2 4 |
| | 3+4 2 | 3+4 2 | 2+2 4 | 2+2 4 | 2+2 4 | 3+4 2 | 5+6 2 | 2+2 4 |
| | 5+6 2 | 3+4 2 | 2+2 4 | 2+2 4 | 2+2 4 | 5+6 2 | 3+4 2 | 2+2 4 |
| | 3+4 2 | 5+6 2 | 2+2 4 | 2+2 4 | 2+2 4 | 3+4 2 | 3+4 2 | 2+2 4 |
| | 5+6 2 | 3+4 2 | 2+2 4 | 2+2 4 | 2+2 4 | 5+6 2 | 3+4 2 | 2+2 4 |
| | 3+4 2 | 5+6 2 | 2+2 4 | 2+2 4 | 2+2 4 | 5+6 2 | 5+6 2 | 2+2 4 |
| | 3+4 2 | 3+4 2 | 2+2 4 | 2+2 4 | 2+2 4 | 5+6 2 | 3+4 2 | 2+2 4 |
| | 5+6 2 | 3+4 2 | 2+2 4 | 2+2 4 | 2+2 4 | 3+4 2 | 5+6 2 | 2+2 4 |
| | 3+4 2 | 5+6 2 | 2+2 4 | 2+2 4 | 2+2 4 | 5+6 2 | 5+6 2 | 2+2 4 |
| | 5+6 2 | 3+4 2 | 2+2 4 | 2+2 4 | 2+2 4 | 5+6 2 | 3+4 2 | 2+2 4 |
| | 5+6 2 | 5+6 2 | 2+2 4 | 2+2 4 | 2+2 4 | 5+6 2 | 3+4 2 | 2+2 4 |
| | 5+6 2 | 5+6 2 | 2+2 4 | 2+2 4 | 2+2 4 | 3+4 2 | 5+6 2 | 2+2 4 |
| | 3+4 2 | 3+4 2 | 2+2 4 | 2+2 4 | 2+2 4 | 3+4 2 | 3+4 2 | 2+2 4 |
| | 3+4 2 | 5+6 2 | 2+2 4 | 2+2 4 | 2+2 4 | 3+4 2 | 3+4 2 | 2+2 4 |
| | 3+4 2 | 3+4 2 | 2+2 4 | 2+2 4 | 2+2 4 | 5+6 2 | 5+6 2 | 2+2 4 |
| | 5+6 2 | 5+6 2 | 2+2 4 | 2+2 4 | 2+2 4 | 3+4 2 | 3+4 2 | 2+2 4 |
| | 5+6 2 | 3+4 2 | 2+2 4 | 2+2 4 | 2+2 4 | 5+6 2 | 5+6 2 | 2+2 4 |
| | 5+6 2 | 5+6 2 | 2+2 4 | 2+2 4 | 2+2 4 | 3+4 2 | 3+4 2 | 2+2 4 |
| | 3+4 2 | 3+4 2 | 2+2 4 | 2+2 4 | 2+2 4 | 5+6 2 | 5+6 2 | 2+2 4 |
| | 3+4 2 | 3+4 2 | 2+2 4 | 2+2 4 | 2+2 4 | 3+4 2 | 3+4 2 | 2+2 4 |
| | 5+6 2 | 5+6 2 | 2+2 4 | 2+2 4 | 2+2 4 | 5+6 2 | 5+6 2 | 2+2 4 |
| | 3+4 2 | 5+6 2 | 2+2 4 | 2+2 4 | 2+2 4 | 3+4 2 | 3+4 2 | 2+2 4 |
| | 5+6 2 | 3+4 2 | 2+2 4 | 2+2 4 | 2+2 4 | 3+4 2 | 5+6 2 | 2+2 4 |
| | 3+4 2 | 5+6 2 | 2+2 4 | 2+2 4 | 2+2 4 | 3+4 2 | 5+6 2 | 2+2 4 |
| | 5+6 2 | 5+6 2 | 2+2 4 | 2+2 4 | 2+2 4 | 5+6 2 | 3+4 2 | 2+2 4 |
| | 3+4 2 | 3+4 2 | | | | 5+6 2 | 5+6 2 | |
| | 5+6 2 | 5+6 2 | | | | 3+4 2 | 3+4 2 | |
| | 3+4 2 | 3+4 2 | | | | 5+6 2 | 5+6 2 | |
| | 5+6 2 | 3+4 2 | | | | 3+4 2 | 3+4 2 | |
| | 3+4 2 | 5+6 2 | | | | 3+4 2 | 5+6 2 | |
| | 5+6 2 | 5+6 2 | | | | 5+6 2 | 5+6 2 | |
| | 5+6 2 | 3+4 2 | | | | 3+4 2 | 3+4 2 | |
| | 3+4 2 | 5+6 2 | | | | 5+6 2 | 5+6 2 | |
| | 5+6 2 | 3+4 2 | | | | 3+4 2 | 5+6 2 | |
| | 3+4 2 | 3+4 2 | | | | 5+6 2 | 3+4 2 | |
| | 5+6 2 | 5+6 2 | | | | 3+4 2 | 3+4 2 | |
| | 5+6 2 | 5+6 2 | | | | 3+4 2 | 3+4 2 | |
| | 3+4 2 | 5+6 2 | | | | 5+6 2 | 5+6 2 | |
| | 3+4 2 | 3+4 2 | | | | 3+4 2 | 5+6 2 | |
| | 3+4 2 | 3+4 2 | | | | 5+6 2 | 3+4 2 | |
| | 5+6 2 | 5+6 2 | | | | 3+4 2 | 5+6 2 | |
| | 3+4 2 | 3+4 2 | | | | 5+6 2 | 5+6 2 | |

Continued on next column

**TABLE XXIII– continued**

| TQC | ICSD | NoE | SF | | | | | |
|---|---|---|---|---|---|---|---|---|
| | 5+6 2 | 5+6 2 | | | | 3+4 2 | 3+4 2 | |
| | 5+6 2 | 5+6 2 | | | | 3+4 2 | 3+4 2 | |
| | 3+4 2 | 3+4 2 | | | | 5+6 2 | 5+6 2 | |
| | 5+6 2 | 5+6 2 | | | | 5+6 2 | 5+6 2 | |
| | 3+4 2 | 5+6 2 | | | | 5+6 2 | 3+4 2 | |
| | 3+4 2 | 3+4 2 | | | | 5+6 2 | 3+4 2 | |
| | 5+6 2 | 3+4 2 | | | | 3+4 2 | 5+6 2 | |
| | 5+6 2 | 5+6 2 | | | | 3+4 2 | 3+4 2 | |
| | 5+6 2 | 3+4 2 | | | | 3+4 2 | 5+6 2 | |
| | 3+4 2 | 5+6 2 | | | | 3+4 2 | 3+4 2 | |
| | 3+4 2 | 5+6 2 | | | | 5+6 2 | 3+4 2 | |
| | 3+4 2 | 3+4 2 | | | | 3+4 2 | 5+6 2 | |
| | 5+6 2 | 3+4 2 | | | | 5+6 2 | 5+6 2 | |
| | 3+4 2 | 5+6 2 | | | | 5+6 2 | 5+6 2 | |
| | 3+4 2 | 3+4 2 | | | | 5+6 2 | 3+4 2 | |
| | 5+6 2 | 5+6 2 | | | | 3+4 2 | 3+4 2 | |
| | 5+6 2 | 3+4 2 | | | | 5+6 2 | 5+6 2 | |
| | 5+6 2 | 5+6 2 | | | | 5+6 2 | 3+4 2 | |
| | 3+4 2 | 3+4 2 | | | | 3+4 2 | 3+4 2 | |
| | 3+4 2 | 5+6 2 | | | | 3+4 2 | 5+6 2 | |
| | 3+4 2 | 3+4 2 | | | | 5+6 2 | 5+6 2 | |
| | 5+6 2 | 3+4 2 | | | | 3+4 2 | 3+4 2 | |
| | 5+6 2 | 3+4 2 | | | | 5+6 2 | 5+6 2 | |
| | 3+4 2 | 5+6 2 | | | | 3+4 2 | 3+4 2 | |
| | 5+6 2 | 3+4 2 | | | | 3+4 2 | 3+4 2 | |
| | 3+4 2 | 5+6 2 | | | | 5+6 2 | 5+6 2 | |
| | 3+4 2 | 5+6 2 | | | | 5+6 2 | 5+6 2 | |
| | 5+6 2 | 3+4 2 | | | | 5+6 2 | 5+6 2 | |
| | 3+4 2 | 3+4 2 | | | | 5+6 2 | 3+4 2 | |
| | 3+4 2 | 5+6 2 | | | | 3+4 2 | 3+4 2 | |
| | 5+6 2 | 3+4 2 | | | | 3+4 2 | 5+6 2 | |
| | 3+4 2 | 5+6 2 | | | | 3+4 2 | 5+6 2 | |
| | 5+6 2 | 5+6 2 | | | | 3+4 2 | 3+4 2 | |
| | 3+4 2 | 3+4 2 | | | | 5+6 2 | 5+6 2 | |
| | 5+6 2 | 5+6 2 | | | | 5+6 2 | 5+6 2 | |
| | 5+6 2 | 5+6 2 | | | | 3+4 2 | 5+6 2 | |
| | 3+4 2 | 3+4 2 | | | | 5+6 2 | 3+4 2 | |
| | 5+6 2 | 3+4 2 | | | | 5+6 2 | 3+4 2 | |
| | 5+6 2 | 5+6 2 | | | | 3+4 2 | 3+4 2 | |
| | 3+4 2 | 3+4 2 | | | | 5+6 2 | 5+6 2 | |
| | 3+4 2 | 3+4 2 | | | | 5+6 2 | 3+4 2 | |
| | 5+6 2 | 5+6 2 | | | | 3+4 2 | 3+4 2 | |
| | 5+6 2 | 3+4 2 | | | | 3+4 2 | 5+6 2 | |
| | 3+4 2 | 5+6 2 | | | | 3+4 2 | 3+4 2 | |
| | 5+6 2 | 3+4 2 | | | | 5+6 2 | 3+4 2 | |
| | 3+4 2 | 5+6 2 | | | | 5+6 2 | 5+6 2 | |
| | 3+4 2 | 5+6 2 | | | | 3+4 2 | 3+4 2 | |
| | 5+6 2 | 3+4 2 | | | | 5+6 2 | 3+4 2 | |
| | 3+4 2 | 3+4 2 | | | | 3+4 2 | 5+6 2 | |
| | 5+6 2 | 3+4 2 | | | | 3+4 2 | 5+6 2 | |
| | 5+6 2 | 5+6 2 | | | | 5+6 2 | 5+6 2 | |
| | 3+4 2 | 5+6 2 | | | | 5+6 2 | 3+4 2 | |
| | 3+4 2 | 5+6 2 | | | | 5+6 2 | 3+4 2 | |
| | 3+4 2 | 5+6 2 | | | | 3+4 2 | 5+6 2 | |
| 11.2.52 | 36300 | 76 | H2Cs1O4P1 | | | | | |
| | -A | -B | -C | -D | -E | -GM | -Y | -Z |
| | 5+6 2 | 3+4 2 | 2+2 4 | 2+2 4 | 2+2 4 | 3+4 2 | 5+6 2 | 2+2 4 |
| | 3+4 2 | 5+6 2 | 2+2 4 | 2+2 4 | 2+2 4 | 5+6 2 | 3+4 2 | 2+2 4 |
| | 5+6 2 | 3+4 2 | 2+2 4 | 2+2 4 | 2+2 4 | 5+6 2 | 3+4 2 | 2+2 4 |
| | 3+4 2 | 5+6 2 | 2+2 4 | 2+2 4 | 2+2 4 | 3+4 2 | 5+6 2 | 2+2 4 |
| | 5+6 2 | 3+4 2 | 2+2 4 | 2+2 4 | 2+2 4 | 3+4 2 | 5+6 2 | 2+2 4 |

Continued on next column

**TABLE XXIII– continued**

| TQC | ICSD | NoE | SF | | | | | |
|---|---|---|---|---|---|---|---|---|
| | 3+4 2 | 5+6 2 | 2+2 4 | 2+2 4 | 2+2 4 | 5+6 2 | 3+4 2 | 2+2 4 |
| | 5+6 2 | 3+4 2 | 2+2 4 | 2+2 4 | 2+2 4 | 3+4 2 | 5+6 2 | 2+2 4 |
| | 3+4 2 | 5+6 2 | 2+2 4 | 2+2 4 | 2+2 4 | 5+6 2 | 3+4 2 | 2+2 4 |
| | 5+6 2 | 3+4 2 | 2+2 4 | 2+2 4 | 2+2 4 | 3+4 2 | 5+6 2 | 2+2 4 |
| | 3+4 2 | 5+6 2 | 2+2 4 | 2+2 4 | 2+2 4 | 5+6 2 | 3+4 2 | 2+2 4 |
| | 5+6 2 | 3+4 2 | 2+2 4 | 2+2 4 | 2+2 4 | 3+4 2 | 5+6 2 | 2+2 4 |
| | 3+4 2 | 5+6 2 | 2+2 4 | 2+2 4 | 2+2 4 | 5+6 2 | 3+4 2 | 2+2 4 |
| | 5+6 2 | 5+6 2 | 2+2 4 | 2+2 4 | 2+2 4 | 3+4 2 | 5+6 2 | 2+2 4 |
| | 3+4 2 | 3+4 2 | 2+2 4 | 2+2 4 | 2+2 4 | 5+6 2 | 3+4 2 | 2+2 4 |
| | 5+6 2 | 5+6 2 | 2+2 4 | 2+2 4 | 2+2 4 | 3+4 2 | 3+4 2 | 2+2 4 |
| | 3+4 2 | 3+4 2 | 2+2 4 | 2+2 4 | 2+2 4 | 5+6 2 | 3+4 2 | 2+2 4 |
| | 5+6 2 | 5+6 2 | 2+2 4 | 2+2 4 | 2+2 4 | 3+4 2 | 5+6 2 | 2+2 4 |
| | 3+4 2 | 3+4 2 | 2+2 4 | 2+2 4 | 2+2 4 | 5+6 2 | 5+6 2 | 2+2 4 |
| | 3+4 2 | 3+4 2 | 2+2 4 | 2+2 4 | 2+2 4 | 3+4 2 | 5+6 2 | 2+2 4 |
| | 5+6 2 | 3+4 2 | | | | 3+4 2 | 5+6 2 | |
| | 5+6 2 | 5+6 2 | | | | 5+6 2 | 3+4 2 | |
| | 3+4 2 | 5+6 2 | | | | 3+4 2 | 3+4 2 | |
| | 5+6 2 | 5+6 2 | | | | 5+6 2 | 3+4 2 | |
| | 3+4 2 | 3+4 2 | | | | 5+6 2 | 5+6 2 | |
| | 5+6 2 | 3+4 2 | | | | 5+6 2 | 3+4 2 | |
| | 3+4 2 | 5+6 2 | | | | 3+4 2 | 5+6 2 | |
| | 5+6 2 | 3+4 2 | | | | 3+4 2 | 5+6 2 | |
| | 3+4 2 | 5+6 2 | | | | 3+4 2 | 5+6 2 | |
| | 3+4 2 | 3+4 2 | | | | 5+6 2 | 3+4 2 | |
| | 5+6 2 | 5+6 2 | | | | 3+4 2 | 5+6 2 | |
| | 5+6 2 | 3+4 2 | | | | 5+6 2 | 3+4 2 | |
| | 5+6 2 | 5+6 2 | | | | 5+6 2 | 5+6 2 | |
| | 3+4 2 | 3+4 2 | | | | 3+4 2 | 3+4 2 | |
| | 3+4 2 | 5+6 2 | | | | 5+6 2 | 3+4 2 | |
| | 5+6 2 | 3+4 2 | | | | 3+4 2 | 5+6 2 | |
| | 5+6 2 | 5+6 2 | | | | 5+6 2 | 5+6 2 | |
| | 3+4 2 | 3+4 2 | | | | 3+4 2 | 3+4 2 | |
| | 3+4 2 | 3+4 2 | | | | 3+4 2 | 3+4 2 | |
| 11.2.53 | 20449 | 60 | H2Cl1O2Y1 | | | | | |
| | -A | -B | -C | -D | -E | -GM | -Y | -Z |
| | 3+4 2 | 3+4 2 | 2+2 4 | 2+2 4 | 2+2 4 | 3+4 2 | 3+4 2 | 2+2 4 |
| | 5+6 2 | 5+6 2 | 2+2 4 | 2+2 4 | 2+2 4 | 5+6 2 | 5+6 2 | 2+2 4 |
| | 3+4 2 | 3+4 2 | 2+2 4 | 2+2 4 | 2+2 4 | 3+4 2 | 3+4 2 | 2+2 4 |
| | 5+6 2 | 5+6 2 | 2+2 4 | 2+2 4 | 2+2 4 | 5+6 2 | 5+6 2 | 2+2 4 |
| | 3+4 2 | 3+4 2 | 2+2 4 | 2+2 4 | 2+2 4 | 3+4 2 | 3+4 2 | 2+2 4 |
| | 5+6 2 | 5+6 2 | 2+2 4 | 2+2 4 | 2+2 4 | 5+6 2 | 5+6 2 | 2+2 4 |
| | 5+6 2 | 3+4 2 | 2+2 4 | 2+2 4 | 2+2 4 | 3+4 2 | 5+6 2 | 2+2 4 |
| | 3+4 2 | 5+6 2 | 2+2 4 | 2+2 4 | 2+2 4 | 5+6 2 | 3+4 2 | 2+2 4 |
| | 3+4 2 | 3+4 2 | 2+2 4 | 2+2 4 | 2+2 4 | 3+4 2 | 3+4 2 | 2+2 4 |
| | 5+6 2 | 5+6 2 | 2+2 4 | 2+2 4 | 2+2 4 | 5+6 2 | 5+6 2 | 2+2 4 |
| | 5+6 2 | 5+6 2 | 2+2 4 | 2+2 4 | 2+2 4 | 5+6 2 | 5+6 2 | 2+2 4 |
| | 3+4 2 | 3+4 2 | 2+2 4 | 2+2 4 | 2+2 4 | 3+4 2 | 3+4 2 | 2+2 4 |
| | 3+4 2 | 5+6 2 | 2+2 4 | 2+2 4 | 2+2 4 | 3+4 2 | 5+6 2 | 2+2 4 |
| | 5+6 2 | 3+4 2 | 2+2 4 | 2+2 4 | 2+2 4 | 5+6 2 | 3+4 2 | 2+2 4 |
| | 3+4 2 | 5+6 2 | 2+2 4 | 2+2 4 | 2+2 4 | 3+4 2 | 5+6 2 | 2+2 4 |
| | 5+6 2 | 3+4 2 | | | | 5+6 2 | 3+4 2 | |
| | 5+6 2 | 5+6 2 | | | | 3+4 2 | 3+4 2 | |
| | 3+4 2 | 3+4 2 | | | | 3+4 2 | 5+6 2 | |
| | 3+4 2 | 3+4 2 | | | | 3+4 2 | 3+4 2 | |
| | 5+6 2 | 3+4 2 | | | | 5+6 2 | 5+6 2 | |
| | 3+4 2 | 5+6 2 | | | | 3+4 2 | 3+4 2 | |
| | 5+6 2 | 3+4 2 | | | | 5+6 2 | 5+6 2 | |
| | 3+4 2 | 3+4 2 | | | | 5+6 2 | 5+6 2 | |
| | 5+6 2 | 5+6 2 | | | | 3+4 2 | 5+6 2 | |
| | 5+6 2 | 5+6 2 | | | | 3+4 2 | 3+4 2 | |
| | 3+4 2 | 5+6 2 | | | | 3+4 2 | 3+4 2 | |

Continued on next column

**TABLE XXIII– continued**

| TQC | ICSD | | NoE | | SF | | | |
|---|---|---|---|---|---|---|---|---|
| | 5+6 2 | 3+4 2 | | | | 5+6 2 | 5+6 2 | |
| | 3+4 2 | 3+4 2 | | | | 3+4 2 | 3+4 2 | |
| | 3+4 2 | 5+6 2 | | | | 5+6 2 | 3+4 2 | |
| | 5+6 2 | 3+4 2 | | | | 5+6 2 | 5+6 2 | |
| 11.2.54 | 260836 | | 88 | | H2Cl1O2Tm1 | | | |
| | -A | -B | -C | -D | -E | -GM | -Y | -Z |
| | 3+4 2 | 5+6 2 | 2+2 4 | 2+2 4 | 2+2 4 | 3+4 2 | 3+4 2 | 2+2 4 |
| | 5+6 2 | 3+4 2 | 2+2 4 | 2+2 4 | 2+2 4 | 5+6 2 | 5+6 2 | 2+2 4 |
| | 3+4 2 | 5+6 2 | 2+2 4 | 2+2 4 | 2+2 4 | 3+4 2 | 3+4 2 | 2+2 4 |
| | 5+6 2 | 3+4 2 | 2+2 4 | 2+2 4 | 2+2 4 | 5+6 2 | 5+6 2 | 2+2 4 |
| | 3+4 2 | 5+6 2 | 2+2 4 | 2+2 4 | 2+2 4 | 3+4 2 | 5+6 2 | 2+2 4 |
| | 5+6 2 | 3+4 2 | 2+2 4 | 2+2 4 | 2+2 4 | 5+6 2 | 3+4 2 | 2+2 4 |
| | 5+6 2 | 5+6 2 | 2+2 4 | 2+2 4 | 2+2 4 | 3+4 2 | 3+4 2 | 2+2 4 |
| | 3+4 2 | 3+4 2 | 2+2 4 | 2+2 4 | 2+2 4 | 5+6 2 | 5+6 2 | 2+2 4 |
| | 3+4 2 | 5+6 2 | 2+2 4 | 2+2 4 | 2+2 4 | 3+4 2 | 5+6 2 | 2+2 4 |
| | 5+6 2 | 3+4 2 | 2+2 4 | 2+2 4 | 2+2 4 | 5+6 2 | 3+4 2 | 2+2 4 |
| | 5+6 2 | 3+4 2 | 2+2 4 | 2+2 4 | 2+2 4 | 5+6 2 | 3+4 2 | 2+2 4 |
| | 3+4 2 | 5+6 2 | 2+2 4 | 2+2 4 | 2+2 4 | 3+4 2 | 5+6 2 | 2+2 4 |
| | 5+6 2 | 3+4 2 | 2+2 4 | 2+2 4 | 2+2 4 | 3+4 2 | 5+6 2 | 2+2 4 |
| | 3+4 2 | 5+6 2 | 2+2 4 | 2+2 4 | 2+2 4 | 5+6 2 | 3+4 2 | 2+2 4 |
| | 5+6 2 | 5+6 2 | 2+2 4 | 2+2 4 | 2+2 4 | 3+4 2 | 3+4 2 | 2+2 4 |
| | 3+4 2 | 5+6 2 | 2+2 4 | 2+2 4 | 2+2 4 | 3+4 2 | 5+6 2 | 2+2 4 |
| | 3+4 2 | 3+4 2 | 2+2 4 | 2+2 4 | 2+2 4 | 5+6 2 | 3+4 2 | 2+2 4 |
| | 5+6 2 | 3+4 2 | 2+2 4 | 2+2 4 | 2+2 4 | 5+6 2 | 5+6 2 | 2+2 4 |
| | 3+4 2 | 3+4 2 | 2+2 4 | 2+2 4 | 2+2 4 | 5+6 2 | 5+6 2 | 2+2 4 |
| | 5+6 2 | 5+6 2 | 2+2 4 | 2+2 4 | 2+2 4 | 3+4 2 | 5+6 2 | 2+2 4 |
| | 3+4 2 | 5+6 2 | 2+2 4 | 2+2 4 | 2+2 4 | 3+4 2 | 3+4 2 | 2+2 4 |
| | 5+6 2 | 5+6 2 | 2+2 4 | 2+2 4 | 2+2 4 | 3+4 2 | 3+4 2 | 2+2 4 |
| | 3+4 2 | 5+6 2 | | | | 3+4 2 | 5+6 2 | |
| | 5+6 2 | 3+4 2 | | | | 5+6 2 | 3+4 2 | |
| | 5+6 2 | 3+4 2 | | | | 3+4 2 | 3+4 2 | |
| | 3+4 2 | 5+6 2 | | | | 5+6 2 | 3+4 2 | |
| | 5+6 2 | 3+4 2 | | | | 3+4 2 | 5+6 2 | |
| | 3+4 2 | 3+4 2 | | | | 5+6 2 | 3+4 2 | |
| | 3+4 2 | 3+4 2 | | | | 3+4 2 | 5+6 2 | |
| | 5+6 2 | 5+6 2 | | | | 5+6 2 | 5+6 2 | |
| | 3+4 2 | 5+6 2 | | | | 3+4 2 | 3+4 2 | |
| | 5+6 2 | 3+4 2 | | | | 5+6 2 | 5+6 2 | |
| | 3+4 2 | 5+6 2 | | | | 3+4 2 | 3+4 2 | |
| | 5+6 2 | 3+4 2 | | | | 5+6 2 | 5+6 2 | |
| | 5+6 2 | 3+4 2 | | | | 3+4 2 | 3+4 2 | |
| | 3+4 2 | 5+6 2 | | | | 5+6 2 | 5+6 2 | |
| | 3+4 2 | 3+4 2 | | | | 5+6 2 | 5+6 2 | |
| | 5+6 2 | 5+6 2 | | | | 5+6 2 | 3+4 2 | |
| | 5+6 2 | 5+6 2 | | | | 3+4 2 | 3+4 2 | |
| | 3+4 2 | 3+4 2 | | | | 3+4 2 | 5+6 2 | |
| | 3+4 2 | 5+6 2 | | | | 5+6 2 | 5+6 2 | |
| | 5+6 2 | 5+6 2 | | | | 3+4 2 | 3+4 2 | |
| | 5+6 2 | 3+4 2 | | | | 5+6 2 | 5+6 2 | |
| | 3+4 2 | 5+6 2 | | | | 3+4 2 | 3+4 2 | |
| 11.2.55 | 260832 | | 80 | | H2Cl1O2Tb1 | | | |
| | -A | -B | -C | -D | -E | -GM | -Y | -Z |
| | 3+4 2 | 3+4 2 | 2+2 4 | 2+2 4 | 2+2 4 | 3+4 2 | 3+4 2 | 2+2 4 |
| | 5+6 2 | 5+6 2 | 2+2 4 | 2+2 4 | 2+2 4 | 5+6 2 | 5+6 2 | 2+2 4 |
| | 3+4 2 | 3+4 2 | 2+2 4 | 2+2 4 | 2+2 4 | 3+4 2 | 3+4 2 | 2+2 4 |
| | 5+6 2 | 5+6 2 | 2+2 4 | 2+2 4 | 2+2 4 | 5+6 2 | 5+6 2 | 2+2 4 |
| | 3+4 2 | 3+4 2 | 2+2 4 | 2+2 4 | 2+2 4 | 3+4 2 | 3+4 2 | 2+2 4 |
| | 5+6 2 | 5+6 2 | 2+2 4 | 2+2 4 | 2+2 4 | 5+6 2 | 5+6 2 | 2+2 4 |
| | 5+6 2 | 3+4 2 | 2+2 4 | 2+2 4 | 2+2 4 | 3+4 2 | 5+6 2 | 2+2 4 |
| | 3+4 2 | 5+6 2 | 2+2 4 | 2+2 4 | 2+2 4 | 5+6 2 | 3+4 2 | 2+2 4 |
| | 3+4 2 | 3+4 2 | 2+2 4 | 2+2 4 | 2+2 4 | 3+4 2 | 3+4 2 | 2+2 4 |

Continued on next column

**TABLE XXIII– continued**

| TQC | ICSD | | NoE | | SF | | | |
|---|---|---|---|---|---|---|---|---|
| | 5+6 2 | 5+6 2 | 2+2 4 | 2+2 4 | 2+2 4 | 5+6 2 | 5+6 2 | 2+2 4 |
| | 5+6 2 | 5+6 2 | 2+2 4 | 2+2 4 | 2+2 4 | 5+6 2 | 5+6 2 | 2+2 4 |
| | 3+4 2 | 3+4 2 | 2+2 4 | 2+2 4 | 2+2 4 | 3+4 2 | 3+4 2 | 2+2 4 |
| | 3+4 2 | 5+6 2 | 2+2 4 | 2+2 4 | 2+2 4 | 3+4 2 | 5+6 2 | 2+2 4 |
| | 5+6 2 | 3+4 2 | 2+2 4 | 2+2 4 | 2+2 4 | 5+6 2 | 3+4 2 | 2+2 4 |
| | 5+6 2 | 3+4 2 | 2+2 4 | 2+2 4 | 2+2 4 | 3+4 2 | 5+6 2 | 2+2 4 |
| | 3+4 2 | 3+4 2 | 2+2 4 | 2+2 4 | 2+2 4 | 3+4 2 | 3+4 2 | 2+2 4 |
| | 3+4 2 | 5+6 2 | 2+2 4 | 2+2 4 | 2+2 4 | 5+6 2 | 3+4 2 | 2+2 4 |
| | 5+6 2 | 5+6 2 | 2+2 4 | 2+2 4 | 2+2 4 | 5+6 2 | 5+6 2 | 2+2 4 |
| | 3+4 2 | 5+6 2 | 2+2 4 | 2+2 4 | 2+2 4 | 5+6 2 | 3+4 2 | 2+2 4 |
| | 3+4 2 | 3+4 2 | 2+2 4 | 2+2 4 | 2+2 4 | 3+4 2 | 3+4 2 | 2+2 4 |
| | 5+6 2 | 3+4 2 | | | | 3+4 2 | 5+6 2 | |
| | 5+6 2 | 5+6 2 | | | | 3+4 2 | 5+6 2 | |
| | 5+6 2 | 3+4 2 | | | | 5+6 2 | 3+4 2 | |
| | 3+4 2 | 5+6 2 | | | | 3+4 2 | 5+6 2 | |
| | 5+6 2 | 3+4 2 | | | | 3+4 2 | 5+6 2 | |
| | 3+4 2 | 3+4 2 | | | | 3+4 2 | 5+6 2 | |
| | 5+6 2 | 5+6 2 | | | | 5+6 2 | 5+6 2 | |
| | 3+4 2 | 5+6 2 | | | | 5+6 2 | 3+4 2 | |
| | 3+4 2 | 5+6 2 | | | | 3+4 2 | 3+4 2 | |
| | 3+4 2 | 3+4 2 | | | | 5+6 2 | 5+6 2 | |
| | 5+6 2 | 5+6 2 | | | | 5+6 2 | 3+4 2 | |
| | 5+6 2 | 3+4 2 | | | | 5+6 2 | 3+4 2 | |
| | 5+6 2 | 3+4 2 | | | | 3+4 2 | 5+6 2 | |
| | 3+4 2 | 5+6 2 | | | | 3+4 2 | 3+4 2 | |
| | 3+4 2 | 3+4 2 | | | | 5+6 2 | 3+4 2 | |
| | 5+6 2 | 5+6 2 | | | | 5+6 2 | 5+6 2 | |
| | 5+6 2 | 3+4 2 | | | | 3+4 2 | 5+6 2 | |
| | 3+4 2 | 5+6 2 | | | | 5+6 2 | 3+4 2 | |
| | 5+6 2 | 3+4 2 | | | | 3+4 2 | 5+6 2 | |
| | 3+4 2 | 5+6 2 | | | | 3+4 2 | 3+4 2 | |
| 11.2.56 | 260834 | | 84 | | H2Cl1Ho1O2 | | | |
| | -A | -B | -C | -D | -E | -GM | -Y | -Z |
| | 3+4 2 | 5+6 2 | 2+2 4 | 2+2 4 | 2+2 4 | 3+4 2 | 3+4 2 | 2+2 4 |
| | 5+6 2 | 3+4 2 | 2+2 4 | 2+2 4 | 2+2 4 | 5+6 2 | 5+6 2 | 2+2 4 |
| | 3+4 2 | 5+6 2 | 2+2 4 | 2+2 4 | 2+2 4 | 3+4 2 | 5+6 2 | 2+2 4 |
| | 5+6 2 | 3+4 2 | 2+2 4 | 2+2 4 | 2+2 4 | 5+6 2 | 3+4 2 | 2+2 4 |
| | 3+4 2 | 5+6 2 | 2+2 4 | 2+2 4 | 2+2 4 | 3+4 2 | 5+6 2 | 2+2 4 |
| | 5+6 2 | 3+4 2 | 2+2 4 | 2+2 4 | 2+2 4 | 5+6 2 | 3+4 2 | 2+2 4 |
| | 5+6 2 | 5+6 2 | 2+2 4 | 2+2 4 | 2+2 4 | 3+4 2 | 3+4 2 | 2+2 4 |
| | 3+4 2 | 3+4 2 | 2+2 4 | 2+2 4 | 2+2 4 | 5+6 2 | 5+6 2 | 2+2 4 |
| | 3+4 2 | 5+6 2 | 2+2 4 | 2+2 4 | 2+2 4 | 3+4 2 | 5+6 2 | 2+2 4 |
| | 5+6 2 | 3+4 2 | 2+2 4 | 2+2 4 | 2+2 4 | 5+6 2 | 3+4 2 | 2+2 4 |
| | 5+6 2 | 3+4 2 | 2+2 4 | 2+2 4 | 2+2 4 | 5+6 2 | 3+4 2 | 2+2 4 |
| | 3+4 2 | 5+6 2 | 2+2 4 | 2+2 4 | 2+2 4 | 3+4 2 | 5+6 2 | 2+2 4 |
| | 5+6 2 | 3+4 2 | 2+2 4 | 2+2 4 | 2+2 4 | 3+4 2 | 3+4 2 | 2+2 4 |
| | 3+4 2 | 5+6 2 | 2+2 4 | 2+2 4 | 2+2 4 | 5+6 2 | 5+6 2 | 2+2 4 |
| | 5+6 2 | 5+6 2 | 2+2 4 | 2+2 4 | 2+2 4 | 3+4 2 | 3+4 2 | 2+2 4 |
| | 3+4 2 | 5+6 2 | 2+2 4 | 2+2 4 | 2+2 4 | 3+4 2 | 5+6 2 | 2+2 4 |
| | 3+4 2 | 3+4 2 | 2+2 4 | 2+2 4 | 2+2 4 | 5+6 2 | 5+6 2 | 2+2 4 |
| | 5+6 2 | 3+4 2 | 2+2 4 | 2+2 4 | 2+2 4 | 5+6 2 | 3+4 2 | 2+2 4 |
| | 3+4 2 | 3+4 2 | 2+2 4 | 2+2 4 | 2+2 4 | 5+6 2 | 5+6 2 | 2+2 4 |
| | 5+6 2 | 5+6 2 | 2+2 4 | 2+2 4 | 2+2 4 | 3+4 2 | 5+6 2 | 2+2 4 |
| | 3+4 2 | 5+6 2 | 2+2 4 | 2+2 4 | 2+2 4 | 3+4 2 | 3+4 2 | 2+2 4 |
| | 5+6 2 | 5+6 2 | | | | 3+4 2 | 3+4 2 | |
| | 5+6 2 | 3+4 2 | | | | 3+4 2 | 5+6 2 | |
| | 3+4 2 | 5+6 2 | | | | 5+6 2 | 3+4 2 | |
| | 5+6 2 | 3+4 2 | | | | 3+4 2 | 3+4 2 | |
| | 3+4 2 | 5+6 2 | | | | 3+4 2 | 3+4 2 | |
| | 5+6 2 | 3+4 2 | | | | 5+6 2 | 5+6 2 | |
| | 3+4 2 | 3+4 2 | | | | 5+6 2 | 3+4 2 | |

Continued on next column

**TABLE XXIII– continued**

| TQC | ICSD | | NoE | | SF | | | |
|---|---|---|---|---|---|---|---|---|
| | 3+4 2 | 3+4 2 | | | | 3+4 2 | 5+6 2 | |
| | 3+4 2 | 5+6 2 | | | | 5+6 2 | 5+6 2 | |
| | 5+6 2 | 3+4 2 | | | | 5+6 2 | 3+4 2 | |
| | 5+6 2 | 5+6 2 | | | | 3+4 2 | 5+6 2 | |
| | 5+6 2 | 5+6 2 | | | | 3+4 2 | 3+4 2 | |
| | 3+4 2 | 3+4 2 | | | | 5+6 2 | 5+6 2 | |
| | 3+4 2 | 3+4 2 | | | | 5+6 2 | 5+6 2 | |
| | 5+6 2 | 5+6 2 | | | | 3+4 2 | 3+4 2 | |
| | 5+6 2 | 5+6 2 | | | | 5+6 2 | 3+4 2 | |
| | 3+4 2 | 3+4 2 | | | | 5+6 2 | 5+6 2 | |
| | 5+6 2 | 5+6 2 | | | | 3+4 2 | 3+4 2 | |
| | 3+4 2 | 3+4 2 | | | | 3+4 2 | 5+6 2 | |
| | 3+4 2 | 5+6 2 | | | | 5+6 2 | 5+6 2 | |
| | 5+6 2 | 5+6 2 | | | | 3+4 2 | 3+4 2 | |
| 11.2.57 | 6164 | | 64 | | H1O2Tb1 | | | |
| | -A | -B | -C | -D | -E | -GM | -Y | -Z |
| | 3+4 2 | 3+4 2 | 2+2 4 | 2+2 4 | 2+2 4 | 3+4 2 | 5+6 2 | 2+2 4 |
| | 5+6 2 | 5+6 2 | 2+2 4 | 2+2 4 | 2+2 4 | 5+6 2 | 3+4 2 | 2+2 4 |
| | 3+4 2 | 5+6 2 | 2+2 4 | 2+2 4 | 2+2 4 | 3+4 2 | 5+6 2 | 2+2 4 |
| | 5+6 2 | 3+4 2 | 2+2 4 | 2+2 4 | 2+2 4 | 5+6 2 | 3+4 2 | 2+2 4 |
| | 5+6 2 | 5+6 2 | 2+2 4 | 2+2 4 | 2+2 4 | 3+4 2 | 3+4 2 | 2+2 4 |
| | 3+4 2 | 3+4 2 | 2+2 4 | 2+2 4 | 2+2 4 | 5+6 2 | 5+6 2 | 2+2 4 |
| | 3+4 2 | 3+4 2 | 2+2 4 | 2+2 4 | 2+2 4 | 3+4 2 | 5+6 2 | 2+2 4 |
| | 5+6 2 | 3+4 2 | 2+2 4 | 2+2 4 | 2+2 4 | 3+4 2 | 3+4 2 | 2+2 4 |
| | 5+6 2 | 5+6 2 | 2+2 4 | 2+2 4 | 2+2 4 | 5+6 2 | 5+6 2 | 2+2 4 |
| | 3+4 2 | 5+6 2 | 2+2 4 | 2+2 4 | 2+2 4 | 5+6 2 | 3+4 2 | 2+2 4 |
| | 3+4 2 | 5+6 2 | 2+2 4 | 2+2 4 | 2+2 4 | 5+6 2 | 5+6 2 | 2+2 4 |
| | 5+6 2 | 3+4 2 | 2+2 4 | 2+2 4 | 2+2 4 | 3+4 2 | 3+4 2 | 2+2 4 |
| | 3+4 2 | 5+6 2 | 2+2 4 | 2+2 4 | 2+2 4 | 3+4 2 | 5+6 2 | 2+2 4 |
| | 5+6 2 | 3+4 2 | 2+2 4 | 2+2 4 | 2+2 4 | 5+6 2 | 3+4 2 | 2+2 4 |
| | 5+6 2 | 5+6 2 | 2+2 4 | 2+2 4 | 2+2 4 | 5+6 2 | 3+4 2 | 2+2 4 |
| | 3+4 2 | 5+6 2 | 2+2 4 | 2+2 4 | 2+2 4 | 3+4 2 | 5+6 2 | 2+2 4 |
| | 3+4 2 | 3+4 2 | | | | 3+4 2 | 5+6 2 | |
| | 5+6 2 | 3+4 2 | | | | 3+4 2 | 5+6 2 | |
| | 5+6 2 | 5+6 2 | | | | 3+4 2 | 3+4 2 | |
| | 3+4 2 | 3+4 2 | | | | 5+6 2 | 3+4 2 | |
| | 3+4 2 | 3+4 2 | | | | 3+4 2 | 3+4 2 | |
| | 3+4 2 | 3+4 2 | | | | 5+6 2 | 5+6 2 | |
| | 5+6 2 | 5+6 2 | | | | 5+6 2 | 3+4 2 | |
| | 5+6 2 | 5+6 2 | | | | 5+6 2 | 5+6 2 | |
| | 3+4 2 | 3+4 2 | | | | 3+4 2 | 5+6 2 | |
| | 5+6 2 | 5+6 2 | | | | 3+4 2 | 3+4 2 | |
| | 3+4 2 | 3+4 2 | | | | 5+6 2 | 3+4 2 | |
| | 5+6 2 | 3+4 2 | | | | 3+4 2 | 5+6 2 | |
| | 3+4 2 | 5+6 2 | | | | 5+6 2 | 3+4 2 | |
| | 5+6 2 | 5+6 2 | | | | 5+6 2 | 5+6 2 | |
| | 3+4 2 | 3+4 2 | | | | 3+4 2 | 5+6 2 | |
| | 5+6 2 | 5+6 2 | | | | 3+4 2 | 3+4 2 | |
| 11.2.58 | 162390 | | 52 | | H1Nd1O2 | | | |
| | -A | -B | -C | -D | -E | -GM | -Y | -Z |
| | 5+6 2 | 5+6 2 | 2+2 4 | 2+2 4 | 2+2 4 | 3+4 2 | 5+6 2 | 2+2 4 |
| | 3+4 2 | 3+4 2 | 2+2 4 | 2+2 4 | 2+2 4 | 5+6 2 | 3+4 2 | 2+2 4 |
| | 3+4 2 | 5+6 2 | 2+2 4 | 2+2 4 | 2+2 4 | 3+4 2 | 5+6 2 | 2+2 4 |
| | 5+6 2 | 3+4 2 | 2+2 4 | 2+2 4 | 2+2 4 | 5+6 2 | 3+4 2 | 2+2 4 |
| | 5+6 2 | 5+6 2 | 2+2 4 | 2+2 4 | 2+2 4 | 3+4 2 | 5+6 2 | 2+2 4 |
| | 3+4 2 | 3+4 2 | 2+2 4 | 2+2 4 | 2+2 4 | 5+6 2 | 3+4 2 | 2+2 4 |
| | 5+6 2 | 3+4 2 | 2+2 4 | 2+2 4 | 2+2 4 | 3+4 2 | 5+6 2 | 2+2 4 |
| | 3+4 2 | 5+6 2 | 2+2 4 | 2+2 4 | 2+2 4 | 5+6 2 | 3+4 2 | 2+2 4 |
| | 3+4 2 | 3+4 2 | 2+2 4 | 2+2 4 | 2+2 4 | 3+4 2 | 3+4 2 | 2+2 4 |
| | 5+6 2 | 5+6 2 | 2+2 4 | 2+2 4 | 2+2 4 | 5+6 2 | 5+6 2 | 2+2 4 |
| | 3+4 2 | 3+4 2 | 2+2 4 | 2+2 4 | 2+2 4 | 5+6 2 | 3+4 2 | 2+2 4 |
| | 5+6 2 | 5+6 2 | 2+2 4 | 2+2 4 | 2+2 4 | 3+4 2 | 5+6 2 | 2+2 4 |
| | 3+4 2 | 3+4 2 | 2+2 4 | 2+2 4 | 2+2 4 | 5+6 2 | 5+6 2 | 2+2 4 |
| | 5+6 2 | 5+6 2 | | | | 3+4 2 | 3+4 2 | |
| | 3+4 2 | 5+6 2 | | | | 3+4 2 | 5+6 2 | |
| | 5+6 2 | 5+6 2 | | | | 3+4 2 | 3+4 2 | |
| | 3+4 2 | 3+4 2 | | | | 5+6 2 | 3+4 2 | |
| | 5+6 2 | 3+4 2 | | | | 5+6 2 | 5+6 2 | |
| | 5+6 2 | 3+4 2 | | | | 3+4 2 | 5+6 2 | |
| | 3+4 2 | 5+6 2 | | | | 3+4 2 | 3+4 2 | |
| | 3+4 2 | 3+4 2 | | | | 3+4 2 | 3+4 2 | |
| | 5+6 2 | 3+4 2 | | | | 5+6 2 | 5+6 2 | |
| | 3+4 2 | 5+6 2 | | | | 5+6 2 | 5+6 2 | |
| | 5+6 2 | 5+6 2 | | | | 5+6 2 | 3+4 2 | |
| | 5+6 2 | 5+6 2 | | | | 3+4 2 | 5+6 2 | |
| | 3+4 2 | 3+4 2 | | | | 3+4 2 | 3+4 2 | |
| 11.2.59 | 424942 | | 80 | | H1F4Mn1Na2O3P1 | | | |
| | -A | -B | -C | -D | -E | -GM | -Y | -Z |
| | 3+4 2 | 3+4 2 | 2+2 4 | 2+2 4 | 2+2 4 | 3+4 2 | 3+4 2 | 2+2 4 |
| | 5+6 2 | 5+6 2 | 2+2 4 | 2+2 4 | 2+2 4 | 5+6 2 | 5+6 2 | 2+2 4 |
| | 5+6 2 | 5+6 2 | 2+2 4 | 2+2 4 | 2+2 4 | 3+4 2 | 3+4 2 | 2+2 4 |
| | 3+4 2 | 3+4 2 | 2+2 4 | 2+2 4 | 2+2 4 | 5+6 2 | 5+6 2 | 2+2 4 |
| | 5+6 2 | 5+6 2 | 2+2 4 | 2+2 4 | 2+2 4 | 3+4 2 | 3+4 2 | 2+2 4 |
| | 3+4 2 | 3+4 2 | 2+2 4 | 2+2 4 | 2+2 4 | 5+6 2 | 5+6 2 | 2+2 4 |
| | 3+4 2 | 3+4 2 | 2+2 4 | 2+2 4 | 2+2 4 | 3+4 2 | 3+4 2 | 2+2 4 |
| | 5+6 2 | 5+6 2 | 2+2 4 | 2+2 4 | 2+2 4 | 5+6 2 | 5+6 2 | 2+2 4 |
| | 5+6 2 | 5+6 2 | 2+2 4 | 2+2 4 | 2+2 4 | 5+6 2 | 5+6 2 | 2+2 4 |
| | 3+4 2 | 3+4 2 | 2+2 4 | 2+2 4 | 2+2 4 | 3+4 2 | 3+4 2 | 2+2 4 |
| | 5+6 2 | 5+6 2 | 2+2 4 | 2+2 4 | 2+2 4 | 3+4 2 | 3+4 2 | 2+2 4 |
| | 5+6 2 | 5+6 2 | 2+2 4 | 2+2 4 | 2+2 4 | 3+4 2 | 3+4 2 | 2+2 4 |
| | 3+4 2 | 3+4 2 | 2+2 4 | 2+2 4 | 2+2 4 | 5+6 2 | 5+6 2 | 2+2 4 |
| | 3+4 2 | 3+4 2 | 2+2 4 | 2+2 4 | 2+2 4 | 5+6 2 | 5+6 2 | 2+2 4 |
| | 3+4 2 | 3+4 2 | 2+2 4 | 2+2 4 | 2+2 4 | 3+4 2 | 3+4 2 | 2+2 4 |
| | 5+6 2 | 5+6 2 | 2+2 4 | 2+2 4 | 2+2 4 | 5+6 2 | 5+6 2 | 2+2 4 |
| | 5+6 2 | 5+6 2 | 2+2 4 | 2+2 4 | 2+2 4 | 3+4 2 | 3+4 2 | 2+2 4 |
| | 3+4 2 | 3+4 2 | 2+2 4 | 2+2 4 | 2+2 4 | 3+4 2 | 5+6 2 | 2+2 4 |
| | 5+6 2 | 5+6 2 | 2+2 4 | 2+2 4 | 2+2 4 | 5+6 2 | 3+4 2 | 2+2 4 |
| | 3+4 2 | 3+4 2 | 2+2 4 | 2+2 4 | 2+2 4 | 5+6 2 | 5+6 2 | 2+2 4 |
| | 5+6 2 | 5+6 2 | | | | 3+4 2 | 3+4 2 | |
| | 3+4 2 | 3+4 2 | | | | 5+6 2 | 5+6 2 | |
| | 5+6 2 | 5+6 2 | | | | 3+4 2 | 3+4 2 | |
| | 3+4 2 | 3+4 2 | | | | 5+6 2 | 5+6 2 | |
| | 3+4 2 | 3+4 2 | | | | 3+4 2 | 3+4 2 | |
| | 5+6 2 | 5+6 2 | | | | 5+6 2 | 5+6 2 | |
| | 5+6 2 | 5+6 2 | | | | 3+4 2 | 5+6 2 | |
| | 3+4 2 | 3+4 2 | | | | 5+6 2 | 3+4 2 | |
| | 5+6 2 | 5+6 2 | | | | 3+4 2 | 3+4 2 | |
| | 3+4 2 | 5+6 2 | | | | 3+4 2 | 5+6 2 | |
| | 5+6 2 | 3+4 2 | | | | 5+6 2 | 3+4 2 | |
| | 5+6 2 | 3+4 2 | | | | 3+4 2 | 3+4 2 | |
| | 5+6 2 | 5+6 2 | | | | 3+4 2 | 5+6 2 | |
| | 3+4 2 | 5+6 2 | | | | 3+4 2 | 3+4 2 | |
| | 5+6 2 | 5+6 2 | | | | 3+4 2 | 3+4 2 | |
| | 5+6 2 | 5+6 2 | | | | 3+4 2 | 5+6 2 | |
| | 5+6 2 | 5+6 2 | | | | 3+4 2 | 3+4 2 | |
| | 5+6 2 | 5+6 2 | | | | 3+4 2 | 3+4 2 | |
| | 5+6 2 | 5+6 2 | | | | 3+4 2 | 3+4 2 | |
| | 3+4 2 | 5+6 2 | | | | 5+6 2 | 3+4 2 | |
| 11.2.60 | 201682 | | 116 | | H14Cl2N5O2Tc1 | | | |
| | -A | -B | -C | -D | -E | -GM | -Y | -Z |
| | 3+4 2 | 3+4 2 | 2+2 4 | 2+2 4 | 2+2 4 | 3+4 2 | 3+4 2 | 2+2 4 |
| | 5+6 2 | 5+6 2 | 2+2 4 | 2+2 4 | 2+2 4 | 5+6 2 | 5+6 2 | 2+2 4 |

Continued on next column

**TABLE XXIII– continued**

| TQC | ICSD | NoE | SF | | | | | |
|---|---|---|---|---|---|---|---|---|
| | 3+4 2 | 5+6 2 | 2+2 4 | 2+2 4 | 2+2 4 | 3+4 2 | 5+6 2 | 2+2 4 |
| | 5+6 2 | 3+4 2 | 2+2 4 | 2+2 4 | 2+2 4 | 5+6 2 | 3+4 2 | 2+2 4 |
| | 5+6 2 | 3+4 2 | 2+2 4 | 2+2 4 | 2+2 4 | 3+4 2 | 5+6 2 | 2+2 4 |
| | 3+4 2 | 5+6 2 | 2+2 4 | 2+2 4 | 2+2 4 | 3+4 2 | 3+4 2 | 2+2 4 |
| | 3+4 2 | 5+6 2 | 2+2 4 | 2+2 4 | 2+2 4 | 5+6 2 | 5+6 2 | 2+2 4 |
| | 5+6 2 | 3+4 2 | 2+2 4 | 2+2 4 | 2+2 4 | 5+6 2 | 3+4 2 | 2+2 4 |
| | 5+6 2 | 5+6 2 | 2+2 4 | 2+2 4 | 2+2 4 | 5+6 2 | 5+6 2 | 2+2 4 |
| | 3+4 2 | 3+4 2 | 2+2 4 | 2+2 4 | 2+2 4 | 3+4 2 | 3+4 2 | 2+2 4 |
| | 3+4 2 | 3+4 2 | 2+2 4 | 2+2 4 | 2+2 4 | 5+6 2 | 3+4 2 | 2+2 4 |
| | 5+6 2 | 5+6 2 | 2+2 4 | 2+2 4 | 2+2 4 | 3+4 2 | 5+6 2 | 2+2 4 |
| | 3+4 2 | 3+4 2 | 2+2 4 | 2+2 4 | 2+2 4 | 3+4 2 | 3+4 2 | 2+2 4 |
| | 5+6 2 | 5+6 2 | 2+2 4 | 2+2 4 | 2+2 4 | 5+6 2 | 5+6 2 | 2+2 4 |
| | 3+4 2 | 3+4 2 | 2+2 4 | 2+2 4 | 2+2 4 | 5+6 2 | 3+4 2 | 2+2 4 |
| | 5+6 2 | 5+6 2 | 2+2 4 | 2+2 4 | 2+2 4 | 3+4 2 | 5+6 2 | 2+2 4 |
| | 3+4 2 | 3+4 2 | 2+2 4 | 2+2 4 | 2+2 4 | 3+4 2 | 3+4 2 | 2+2 4 |
| | 5+6 2 | 5+6 2 | 2+2 4 | 2+2 4 | 2+2 4 | 5+6 2 | 5+6 2 | 2+2 4 |
| | 3+4 2 | 3+4 2 | 2+2 4 | 2+2 4 | 2+2 4 | 3+4 2 | 3+4 2 | 2+2 4 |
| | 5+6 2 | 5+6 2 | 2+2 4 | 2+2 4 | 2+2 4 | 5+6 2 | 5+6 2 | 2+2 4 |
| | 3+4 2 | 3+4 2 | 2+2 4 | 2+2 4 | 2+2 4 | 3+4 2 | 3+4 2 | 2+2 4 |
| | 5+6 2 | 5+6 2 | 2+2 4 | 2+2 4 | 2+2 4 | 5+6 2 | 5+6 2 | 2+2 4 |
| | 3+4 2 | 3+4 2 | 2+2 4 | 2+2 4 | 2+2 4 | 3+4 2 | 3+4 2 | 2+2 4 |
| | 5+6 2 | 5+6 2 | 2+2 4 | 2+2 4 | 2+2 4 | 5+6 2 | 5+6 2 | 2+2 4 |
| | 3+4 2 | 3+4 2 | 2+2 4 | 2+2 4 | 2+2 4 | 5+6 2 | 5+6 2 | 2+2 4 |
| | 5+6 2 | 5+6 2 | 2+2 4 | 2+2 4 | 2+2 4 | 3+4 2 | 3+4 2 | 2+2 4 |
| | 5+6 2 | 5+6 2 | 2+2 4 | 2+2 4 | 2+2 4 | 3+4 2 | 5+6 2 | 2+2 4 |
| | 3+4 2 | 5+6 2 | 2+2 4 | 2+2 4 | 2+2 4 | 5+6 2 | 3+4 2 | 2+2 4 |
| | 5+6 2 | 3+4 2 | 2+2 4 | 2+2 4 | 2+2 4 | 3+4 2 | 3+4 2 | 2+2 4 |
| | 3+4 2 | 3+4 2 | | | | 3+4 2 | 5+6 2 | |
| | 5+6 2 | 5+6 2 | | | | 5+6 2 | 5+6 2 | |
| | 3+4 2 | 3+4 2 | | | | 5+6 2 | 3+4 2 | |
| | 5+6 2 | 5+6 2 | | | | 3+4 2 | 5+6 2 | |
| | 3+4 2 | 3+4 2 | | | | 5+6 2 | 3+4 2 | |
| | 3+4 2 | 3+4 2 | | | | 5+6 2 | 5+6 2 | |
| | 5+6 2 | 5+6 2 | | | | 3+4 2 | 3+4 2 | |
| | 5+6 2 | 3+4 2 | | | | 5+6 2 | 5+6 2 | |
| | 5+6 2 | 5+6 2 | | | | 3+4 2 | 3+4 2 | |
| | 3+4 2 | 5+6 2 | | | | 3+4 2 | 3+4 2 | |
| | 3+4 2 | 3+4 2 | | | | 5+6 2 | 5+6 2 | |
| | 5+6 2 | 5+6 2 | | | | 3+4 2 | 5+6 2 | |
| | 3+4 2 | 3+4 2 | | | | 3+4 2 | 5+6 2 | |
| | 5+6 2 | 5+6 2 | | | | 5+6 2 | 3+4 2 | |
| | 3+4 2 | 3+4 2 | | | | 3+4 2 | 5+6 2 | |
| | 5+6 2 | 3+4 2 | | | | 3+4 2 | 3+4 2 | |
| | 3+4 2 | 5+6 2 | | | | 5+6 2 | 3+4 2 | |
| | 5+6 2 | 3+4 2 | | | | 5+6 2 | 5+6 2 | |
| | 3+4 2 | 5+6 2 | | | | 3+4 2 | 3+4 2 | |
| | 3+4 2 | 5+6 2 | | | | 5+6 2 | 5+6 2 | |
| | 5+6 2 | 3+4 2 | | | | 3+4 2 | 3+4 2 | |
| | 3+4 2 | 5+6 2 | | | | 5+6 2 | 3+4 2 | |
| | 5+6 2 | 3+4 2 | | | | 5+6 2 | 5+6 2 | |
| | 3+4 2 | 3+4 2 | | | | 5+6 2 | 3+4 2 | |
| | 3+4 2 | 5+6 2 | | | | 5+6 2 | 5+6 2 | |
| | 5+6 2 | 3+4 2 | | | | 3+4 2 | 3+4 2 | |
| | 3+4 2 | 3+4 2 | | | | 3+4 2 | 3+4 2 | |
| | 5+6 2 | 3+4 2 | | | | 3+4 2 | 5+6 2 | |
| | 3+4 2 | 5+6 2 | | | | 5+6 2 | 5+6 2 | |
| 11.2.61 | 54147 | 380 | H14Cd1Cu4O18S2 | | | | | |
| | -A | -B | -C | -D | -E | -GM | -Y | -Z |
| | 3+4 2 | 3+4 2 | 2+2 4 | 2+2 4 | 2+2 4 | 3+4 2 | 3+4 2 | 2+2 4 |
| | 5+6 2 | 5+6 2 | 2+2 4 | 2+2 4 | 2+2 4 | 5+6 2 | 5+6 2 | 2+2 4 |
| | 3+4 2 | 3+4 2 | 2+2 4 | 2+2 4 | 2+2 4 | 3+4 2 | 3+4 2 | 2+2 4 |

Continued on next column

**TABLE XXIII– continued**

| TQC | ICSD | NoE | SF | | | | | |
|---|---|---|---|---|---|---|---|---|
| | 5+6 2 | 5+6 2 | 2+2 4 | 2+2 4 | 2+2 4 | 5+6 2 | 5+6 2 | 2+2 4 |
| | 3+4 2 | 3+4 2 | 2+2 4 | 2+2 4 | 2+2 4 | 3+4 2 | 3+4 2 | 2+2 4 |
| | 5+6 2 | 5+6 2 | 2+2 4 | 2+2 4 | 2+2 4 | 5+6 2 | 5+6 2 | 2+2 4 |
| | 3+4 2 | 3+4 2 | 2+2 4 | 2+2 4 | 2+2 4 | 3+4 2 | 3+4 2 | 2+2 4 |
| | 5+6 2 | 5+6 2 | 2+2 4 | 2+2 4 | 2+2 4 | 5+6 2 | 5+6 2 | 2+2 4 |
| | 5+6 2 | 5+6 2 | 2+2 4 | 2+2 4 | 2+2 4 | 3+4 2 | 3+4 2 | 2+2 4 |
| | 5+6 2 | 5+6 2 | 2+2 4 | 2+2 4 | 2+2 4 | 3+4 2 | 3+4 2 | 2+2 4 |
| | 3+4 2 | 3+4 2 | 2+2 4 | 2+2 4 | 2+2 4 | 5+6 2 | 5+6 2 | 2+2 4 |
| | 3+4 2 | 3+4 2 | 2+2 4 | 2+2 4 | 2+2 4 | 5+6 2 | 5+6 2 | 2+2 4 |
| | 3+4 2 | 3+4 2 | 2+2 4 | 2+2 4 | 2+2 4 | 3+4 2 | 3+4 2 | 2+2 4 |
| | 3+4 2 | 3+4 2 | 2+2 4 | 2+2 4 | 2+2 4 | 3+4 2 | 3+4 2 | 2+2 4 |
| | 5+6 2 | 5+6 2 | 2+2 4 | 2+2 4 | 2+2 4 | 3+4 2 | 5+6 2 | 2+2 4 |
| | 5+6 2 | 3+4 2 | 2+2 4 | 2+2 4 | 2+2 4 | 5+6 2 | 5+6 2 | 2+2 4 |
| | 3+4 2 | 5+6 2 | 2+2 4 | 2+2 4 | 2+2 4 | 3+4 2 | 3+4 2 | 2+2 4 |
| | 5+6 2 | 3+4 2 | 2+2 4 | 2+2 4 | 2+2 4 | 5+6 2 | 5+6 2 | 2+2 4 |
| | 3+4 2 | 5+6 2 | 2+2 4 | 2+2 4 | 2+2 4 | 5+6 2 | 3+4 2 | 2+2 4 |
| | 5+6 2 | 5+6 2 | 2+2 4 | 2+2 4 | 2+2 4 | 5+6 2 | 5+6 2 | 2+2 4 |
| | 3+4 2 | 5+6 2 | 2+2 4 | 2+2 4 | 2+2 4 | 3+4 2 | 3+4 2 | 2+2 4 |
| | 3+4 2 | 5+6 2 | 2+2 4 | 2+2 4 | 2+2 4 | 3+4 2 | 3+4 2 | 2+2 4 |
| | 5+6 2 | 3+4 2 | 2+2 4 | 2+2 4 | 2+2 4 | 5+6 2 | 5+6 2 | 2+2 4 |
| | 5+6 2 | 3+4 2 | 2+2 4 | 2+2 4 | 2+2 4 | 5+6 2 | 5+6 2 | 2+2 4 |
| | 3+4 2 | 3+4 2 | 2+2 4 | 2+2 4 | 2+2 4 | 3+4 2 | 5+6 2 | 2+2 4 |
| | 5+6 2 | 5+6 2 | 2+2 4 | 2+2 4 | 2+2 4 | 5+6 2 | 3+4 2 | 2+2 4 |
| | 3+4 2 | 3+4 2 | 2+2 4 | 2+2 4 | 2+2 4 | 3+4 2 | 5+6 2 | 2+2 4 |
| | 5+6 2 | 5+6 2 | 2+2 4 | 2+2 4 | 2+2 4 | 5+6 2 | 3+4 2 | 2+2 4 |
| | 5+6 2 | 3+4 2 | 2+2 4 | 2+2 4 | 2+2 4 | 3+4 2 | 5+6 2 | 2+2 4 |
| | 3+4 2 | 3+4 2 | 2+2 4 | 2+2 4 | 2+2 4 | 3+4 2 | 5+6 2 | 2+2 4 |
| | 5+6 2 | 5+6 2 | 2+2 4 | 2+2 4 | 2+2 4 | 5+6 2 | 3+4 2 | 2+2 4 |
| | 3+4 2 | 5+6 2 | 2+2 4 | 2+2 4 | 2+2 4 | 5+6 2 | 3+4 2 | 2+2 4 |
| | 3+4 2 | 3+4 2 | 2+2 4 | 2+2 4 | 2+2 4 | 3+4 2 | 3+4 2 | 2+2 4 |
| | 5+6 2 | 5+6 2 | 2+2 4 | 2+2 4 | 2+2 4 | 5+6 2 | 5+6 2 | 2+2 4 |
| | 3+4 2 | 3+4 2 | 2+2 4 | 2+2 4 | 2+2 4 | 3+4 2 | 3+4 2 | 2+2 4 |
| | 5+6 2 | 5+6 2 | 2+2 4 | 2+2 4 | 2+2 4 | 5+6 2 | 5+6 2 | 2+2 4 |
| | 3+4 2 | 3+4 2 | 2+2 4 | 2+2 4 | 2+2 4 | 3+4 2 | 3+4 2 | 2+2 4 |
| | 5+6 2 | 5+6 2 | 2+2 4 | 2+2 4 | 2+2 4 | 5+6 2 | 5+6 2 | 2+2 4 |
| | 3+4 2 | 5+6 2 | 2+2 4 | 2+2 4 | 2+2 4 | 3+4 2 | 3+4 2 | 2+2 4 |
| | 5+6 2 | 3+4 2 | 2+2 4 | 2+2 4 | 2+2 4 | 5+6 2 | 5+6 2 | 2+2 4 |
| | 3+4 2 | 5+6 2 | 2+2 4 | 2+2 4 | 2+2 4 | 3+4 2 | 3+4 2 | 2+2 4 |
| | 5+6 2 | 3+4 2 | 2+2 4 | 2+2 4 | 2+2 4 | 3+4 2 | 5+6 2 | 2+2 4 |
| | 3+4 2 | 5+6 2 | 2+2 4 | 2+2 4 | 2+2 4 | 3+4 2 | 3+4 2 | 2+2 4 |
| | 5+6 2 | 3+4 2 | 2+2 4 | 2+2 4 | 2+2 4 | 5+6 2 | 5+6 2 | 2+2 4 |
| | 5+6 2 | 5+6 2 | 2+2 4 | 2+2 4 | 2+2 4 | 3+4 2 | 3+4 2 | 2+2 4 |
| | 5+6 2 | 5+6 2 | 2+2 4 | 2+2 4 | 2+2 4 | 5+6 2 | 3+4 2 | 2+2 4 |
| | 3+4 2 | 3+4 2 | 2+2 4 | 2+2 4 | 2+2 4 | 5+6 2 | 5+6 2 | 2+2 4 |
| | 3+4 2 | 3+4 2 | 2+2 4 | 2+2 4 | 2+2 4 | 3+4 2 | 5+6 2 | 2+2 4 |
| | 3+4 2 | 3+4 2 | 2+2 4 | 2+2 4 | 2+2 4 | 5+6 2 | 3+4 2 | 2+2 4 |
| | 3+4 2 | 5+6 2 | 2+2 4 | 2+2 4 | 2+2 4 | 3+4 2 | 5+6 2 | 2+2 4 |
| | 5+6 2 | 5+6 2 | 2+2 4 | 2+2 4 | 2+2 4 | 3+4 2 | 3+4 2 | 2+2 4 |
| | 5+6 2 | 3+4 2 | 2+2 4 | 2+2 4 | 2+2 4 | 3+4 2 | 5+6 2 | 2+2 4 |
| | 3+4 2 | 3+4 2 | 2+2 4 | 2+2 4 | 2+2 4 | 5+6 2 | 3+4 2 | 2+2 4 |
| | 5+6 2 | 5+6 2 | 2+2 4 | 2+2 4 | 2+2 4 | 5+6 2 | 3+4 2 | 2+2 4 |
| | 3+4 2 | 3+4 2 | 2+2 4 | 2+2 4 | 2+2 4 | 5+6 2 | 3+4 2 | 2+2 4 |
| | 3+4 2 | 3+4 2 | 2+2 4 | 2+2 4 | 2+2 4 | 3+4 2 | 5+6 2 | 2+2 4 |
| | 5+6 2 | 5+6 2 | 2+2 4 | 2+2 4 | 2+2 4 | 3+4 2 | 5+6 2 | 2+2 4 |
| | 5+6 2 | 5+6 2 | 2+2 4 | 2+2 4 | 2+2 4 | 3+4 2 | 5+6 2 | 2+2 4 |
| | 5+6 2 | 5+6 2 | 2+2 4 | 2+2 4 | 2+2 4 | 5+6 2 | 3+4 2 | 2+2 4 |
| | 5+6 2 | 3+4 2 | 2+2 4 | 2+2 4 | 2+2 4 | 5+6 2 | 5+6 2 | 2+2 4 |
| | 3+4 2 | 5+6 2 | 2+2 4 | 2+2 4 | 2+2 4 | 5+6 2 | 3+4 2 | 2+2 4 |
| | 3+4 2 | 3+4 2 | 2+2 4 | 2+2 4 | 2+2 4 | 3+4 2 | 5+6 2 | 2+2 4 |
| | 3+4 2 | 3+4 2 | 2+2 4 | 2+2 4 | 2+2 4 | 5+6 2 | 3+4 2 | 2+2 4 |
| | 5+6 2 | 5+6 2 | 2+2 4 | 2+2 4 | 2+2 4 | 3+4 2 | 5+6 2 | 2+2 4 |

Continued on next column

**TABLE XXIII– continued**

| TQC | ICSD | | NoE | | SF | | | |
|---|---|---|---|---|---|---|---|---|
| | 3+4 2 | 5+6 2 | 2+2 4 | 2+2 4 | 2+2 4 | 5+6 2 | 5+6 2 | 2+2 4 |
| | 5+6 2 | 3+4 2 | 2+2 4 | 2+2 4 | 2+2 4 | 5+6 2 | 3+4 2 | 2+2 4 |
| | 3+4 2 | 5+6 2 | 2+2 4 | 2+2 4 | 2+2 4 | 3+4 2 | 5+6 2 | 2+2 4 |
| | 5+6 2 | 3+4 2 | 2+2 4 | 2+2 4 | 2+2 4 | 5+6 2 | 3+4 2 | 2+2 4 |
| | 3+4 2 | 5+6 2 | 2+2 4 | 2+2 4 | 2+2 4 | 3+4 2 | 5+6 2 | 2+2 4 |
| | 5+6 2 | 3+4 2 | 2+2 4 | 2+2 4 | 2+2 4 | 5+6 2 | 3+4 2 | 2+2 4 |
| | 3+4 2 | 5+6 2 | 2+2 4 | 2+2 4 | 2+2 4 | 3+4 2 | 5+6 2 | 2+2 4 |
| | 5+6 2 | 3+4 2 | 2+2 4 | 2+2 4 | 2+2 4 | 5+6 2 | 3+4 2 | 2+2 4 |
| | 3+4 2 | 5+6 2 | 2+2 4 | 2+2 4 | 2+2 4 | 3+4 2 | 5+6 2 | 2+2 4 |
| | 5+6 2 | 5+6 2 | 2+2 4 | 2+2 4 | 2+2 4 | 5+6 2 | 3+4 2 | 2+2 4 |
| | 3+4 2 | 3+4 2 | 2+2 4 | 2+2 4 | 2+2 4 | 3+4 2 | 5+6 2 | 2+2 4 |
| | 5+6 2 | 3+4 2 | 2+2 4 | 2+2 4 | 2+2 4 | 5+6 2 | 3+4 2 | 2+2 4 |
| | 3+4 2 | 5+6 2 | 2+2 4 | 2+2 4 | 2+2 4 | 3+4 2 | 3+4 2 | 2+2 4 |
| | 5+6 2 | 3+4 2 | 2+2 4 | 2+2 4 | 2+2 4 | 5+6 2 | 5+6 2 | 2+2 4 |
| | 5+6 2 | 5+6 2 | 2+2 4 | 2+2 4 | 2+2 4 | 3+4 2 | 5+6 2 | 2+2 4 |
| | 3+4 2 | 3+4 2 | 2+2 4 | 2+2 4 | 2+2 4 | 3+4 2 | 5+6 2 | 2+2 4 |
| | 3+4 2 | 5+6 2 | 2+2 4 | 2+2 4 | 2+2 4 | 5+6 2 | 3+4 2 | 2+2 4 |
| | 5+6 2 | 3+4 2 | 2+2 4 | 2+2 4 | 2+2 4 | 3+4 2 | 5+6 2 | 2+2 4 |
| | 5+6 2 | 5+6 2 | 2+2 4 | 2+2 4 | 2+2 4 | 5+6 2 | 3+4 2 | 2+2 4 |
| | 3+4 2 | 3+4 2 | 2+2 4 | 2+2 4 | 2+2 4 | 3+4 2 | 5+6 2 | 2+2 4 |
| | 5+6 2 | 3+4 2 | 2+2 4 | 2+2 4 | 2+2 4 | 5+6 2 | 3+4 2 | 2+2 4 |
| | 3+4 2 | 3+4 2 | 2+2 4 | 2+2 4 | 2+2 4 | 3+4 2 | 5+6 2 | 2+2 4 |
| | 5+6 2 | 5+6 2 | 2+2 4 | 2+2 4 | 2+2 4 | 3+4 2 | 5+6 2 | 2+2 4 |
| | 5+6 2 | 3+4 2 | 2+2 4 | 2+2 4 | 2+2 4 | 5+6 2 | 5+6 2 | 2+2 4 |
| | 3+4 2 | 5+6 2 | 2+2 4 | 2+2 4 | 2+2 4 | 3+4 2 | 3+4 2 | 2+2 4 |
| | 3+4 2 | 5+6 2 | 2+2 4 | 2+2 4 | 2+2 4 | 3+4 2 | 5+6 2 | 2+2 4 |
| | 3+4 2 | 5+6 2 | 2+2 4 | 2+2 4 | 2+2 4 | 5+6 2 | 3+4 2 | 2+2 4 |
| | 3+4 2 | 5+6 2 | 2+2 4 | 2+2 4 | 2+2 4 | 5+6 2 | 3+4 2 | 2+2 4 |
| | 5+6 2 | 3+4 2 | 2+2 4 | 2+2 4 | 2+2 4 | 3+4 2 | 5+6 2 | 2+2 4 |
| | 5+6 2 | 3+4 2 | 2+2 4 | 2+2 4 | 2+2 4 | 3+4 2 | 5+6 2 | 2+2 4 |
| | 3+4 2 | 5+6 2 | 2+2 4 | 2+2 4 | 2+2 4 | 3+4 2 | 3+4 2 | 2+2 4 |
| | 3+4 2 | 5+6 2 | | | | 3+4 2 | 3+4 2 | |
| | 5+6 2 | 3+4 2 | | | | 3+4 2 | 5+6 2 | |
| | 5+6 2 | 3+4 2 | | | | 3+4 2 | 5+6 2 | |
| | 3+4 2 | 3+4 2 | | | | 5+6 2 | 3+4 2 | |
| | 3+4 2 | 3+4 2 | | | | 3+4 2 | 3+4 2 | |
| | 5+6 2 | 5+6 2 | | | | 5+6 2 | 3+4 2 | |
| | 5+6 2 | 5+6 2 | | | | 5+6 2 | 3+4 2 | |
| | 3+4 2 | 3+4 2 | | | | 5+6 2 | 5+6 2 | |
| | 5+6 2 | 3+4 2 | | | | 5+6 2 | 5+6 2 | |
| | 3+4 2 | 5+6 2 | | | | 3+4 2 | 3+4 2 | |
| | 5+6 2 | 5+6 2 | | | | 5+6 2 | 3+4 2 | |
| | 3+4 2 | 5+6 2 | | | | 3+4 2 | 5+6 2 | |
| | 5+6 2 | 5+6 2 | | | | 3+4 2 | 5+6 2 | |
| | 5+6 2 | 3+4 2 | | | | 5+6 2 | 3+4 2 | |
| | 3+4 2 | 3+4 2 | | | | 5+6 2 | 3+4 2 | |
| | 5+6 2 | 3+4 2 | | | | 3+4 2 | 3+4 2 | |
| | 3+4 2 | 3+4 2 | | | | 5+6 2 | 5+6 2 | |
| | 5+6 2 | 5+6 2 | | | | 5+6 2 | 5+6 2 | |
| | 3+4 2 | 5+6 2 | | | | 5+6 2 | 5+6 2 | |
| | 5+6 2 | 5+6 2 | | | | 5+6 2 | 3+4 2 | |
| | 3+4 2 | 3+4 2 | | | | 5+6 2 | 5+6 2 | |
| | 5+6 2 | 3+4 2 | | | | 3+4 2 | 3+4 2 | |
| | 5+6 2 | 3+4 2 | | | | 3+4 2 | 3+4 2 | |
| | 3+4 2 | 5+6 2 | | | | 3+4 2 | 5+6 2 | |
| | 3+4 2 | 5+6 2 | | | | 3+4 2 | 5+6 2 | |
| | 5+6 2 | 5+6 2 | | | | 3+4 2 | 5+6 2 | |
| | 3+4 2 | 3+4 2 | | | | 5+6 2 | 5+6 2 | |
| | 3+4 2 | 3+4 2 | | | | 3+4 2 | 3+4 2 | |
| | 3+4 2 | 3+4 2 | | | | 5+6 2 | 3+4 2 | |
| | 5+6 2 | 5+6 2 | | | | 3+4 2 | 5+6 2 | |

Continued on next column

**TABLE XXIII– continued**

| TQC | ICSD | | NoE | | SF | | | |
|---|---|---|---|---|---|---|---|---|
| | 5+6 2 | 5+6 2 | | | | 5+6 2 | 3+4 2 | |
| | 3+4 2 | 3+4 2 | | | | 3+4 2 | 3+4 2 | |
| | 5+6 2 | 5+6 2 | | | | 5+6 2 | 3+4 2 | |
| | 5+6 2 | 3+4 2 | | | | 5+6 2 | 5+6 2 | |
| | 3+4 2 | 5+6 2 | | | | 5+6 2 | 5+6 2 | |
| | 3+4 2 | 5+6 2 | | | | 5+6 2 | 3+4 2 | |
| | 5+6 2 | 3+4 2 | | | | 3+4 2 | 5+6 2 | |
| | 3+4 2 | 5+6 2 | | | | 3+4 2 | 3+4 2 | |
| | 5+6 2 | 3+4 2 | | | | 5+6 2 | 5+6 2 | |
| | 3+4 2 | 3+4 2 | | | | 3+4 2 | 3+4 2 | |
| | 5+6 2 | 5+6 2 | | | | 5+6 2 | 5+6 2 | |
| | 3+4 2 | 5+6 2 | | | | 5+6 2 | 3+4 2 | |
| | 5+6 2 | 3+4 2 | | | | 5+6 2 | 3+4 2 | |
| | 3+4 2 | 3+4 2 | | | | 3+4 2 | 5+6 2 | |
| | 5+6 2 | 5+6 2 | | | | 3+4 2 | 3+4 2 | |
| | 3+4 2 | 5+6 2 | | | | 5+6 2 | 5+6 2 | |
| | 5+6 2 | 3+4 2 | | | | 5+6 2 | 3+4 2 | |
| | 5+6 2 | 3+4 2 | | | | 3+4 2 | 5+6 2 | |
| | 5+6 2 | 3+4 2 | | | | 3+4 2 | 5+6 2 | |
| | 3+4 2 | 5+6 2 | | | | 3+4 2 | 5+6 2 | |
| | 3+4 2 | 5+6 2 | | | | 3+4 2 | 5+6 2 | |
| | 5+6 2 | 3+4 2 | | | | 3+4 2 | 3+4 2 | |
| | 5+6 2 | 3+4 2 | | | | 3+4 2 | 3+4 2 | |
| | 3+4 2 | 5+6 2 | | | | 3+4 2 | 5+6 2 | |
| | 3+4 2 | 5+6 2 | | | | 3+4 2 | 5+6 2 | |
| | 5+6 2 | 5+6 2 | | | | 5+6 2 | 5+6 2 | |
| | 5+6 2 | 5+6 2 | | | | 5+6 2 | 5+6 2 | |
| | 3+4 2 | 3+4 2 | | | | 3+4 2 | 5+6 2 | |
| | 3+4 2 | 3+4 2 | | | | 3+4 2 | 5+6 2 | |
| | 5+6 2 | 5+6 2 | | | | 3+4 2 | 5+6 2 | |
| | 5+6 2 | 3+4 2 | | | | 3+4 2 | 5+6 2 | |
| | 3+4 2 | 5+6 2 | | | | 3+4 2 | 5+6 2 | |
| | 3+4 2 | 3+4 2 | | | | 3+4 2 | 5+6 2 | |
| | 5+6 2 | 3+4 2 | | | | 5+6 2 | 3+4 2 | |
| | 5+6 2 | 3+4 2 | | | | 5+6 2 | 5+6 2 | |
| | 3+4 2 | 5+6 2 | | | | 3+4 2 | 5+6 2 | |
| | 3+4 2 | 5+6 2 | | | | 3+4 2 | 3+4 2 | |
| | 3+4 2 | 3+4 2 | | | | 3+4 2 | 5+6 2 | |
| | 5+6 2 | 3+4 2 | | | | 3+4 2 | 5+6 2 | |
| | 5+6 2 | 5+6 2 | | | | 5+6 2 | 3+4 2 | |
| | 3+4 2 | 5+6 2 | | | | 5+6 2 | 3+4 2 | |
| | 5+6 2 | 3+4 2 | | | | 3+4 2 | 5+6 2 | |
| | 5+6 2 | 3+4 2 | | | | 3+4 2 | 5+6 2 | |
| | 3+4 2 | 5+6 2 | | | | 3+4 2 | 5+6 2 | |
| | 3+4 2 | 5+6 2 | | | | 3+4 2 | 5+6 2 | |
| | 3+4 2 | 3+4 2 | | | | 5+6 2 | 3+4 2 | |
| | 5+6 2 | 5+6 2 | | | | 5+6 2 | 3+4 2 | |
| | 3+4 2 | 3+4 2 | | | | 5+6 2 | 3+4 2 | |
| | 5+6 2 | 5+6 2 | | | | 5+6 2 | 3+4 2 | |
| | 5+6 2 | 3+4 2 | | | | 3+4 2 | 5+6 2 | |
| | 3+4 2 | 5+6 2 | | | | 3+4 2 | 3+4 2 | |
| | 5+6 2 | 3+4 2 | | | | 5+6 2 | 5+6 2 | |
| | 3+4 2 | 5+6 2 | | | | 3+4 2 | 3+4 2 | |
| | 3+4 2 | 5+6 2 | | | | 5+6 2 | 5+6 2 | |
| | 3+4 2 | 5+6 2 | | | | 3+4 2 | 5+6 2 | |
| | 5+6 2 | 3+4 2 | | | | 5+6 2 | 5+6 2 | |
| | 5+6 2 | 3+4 2 | | | | 3+4 2 | 3+4 2 | |
| | 3+4 2 | 3+4 2 | | | | 5+6 2 | 5+6 2 | |
| | 3+4 2 | 3+4 2 | | | | 3+4 2 | 3+4 2 | |
| | 5+6 2 | 5+6 2 | | | | 3+4 2 | 5+6 2 | |
| | 5+6 2 | 5+6 2 | | | | 3+4 2 | 5+6 2 | |

Continued on next column

**TABLE XXIII– continued**

| TQC | ICSD | | NoE | | SF | | | |
|---|---|---|---|---|---|---|---|---|
| | 5+6 2 | 5+6 2 | | | | 5+6 2 | 5+6 2 | |
| | 5+6 2 | 5+6 2 | | | | 5+6 2 | 5+6 2 | |
| | 3+4 2 | 3+4 2 | | | | 3+4 2 | 5+6 2 | |
| | 3+4 2 | 3+4 2 | | | | 3+4 2 | 5+6 2 | |
| 11.2.62 | 156004 | | 228 | H12Cl8Hg3N2O2 | | | | |
| | -A | -B | -C | -D | -E | -GM | -Y | -Z |
| | 3+4 2 | 5+6 2 | 2+2 4 | 2+2 4 | 2+2 4 | 3+4 2 | 5+6 2 | 2+2 4 |
| | 5+6 2 | 3+4 2 | 2+2 4 | 2+2 4 | 2+2 4 | 5+6 2 | 3+4 2 | 2+2 4 |
| | 3+4 2 | 3+4 2 | 2+2 4 | 2+2 4 | 2+2 4 | 3+4 2 | 3+4 2 | 2+2 4 |
| | 5+6 2 | 5+6 2 | 2+2 4 | 2+2 4 | 2+2 4 | 5+6 2 | 5+6 2 | 2+2 4 |
| | 3+4 2 | 5+6 2 | 2+2 4 | 2+2 4 | 2+2 4 | 3+4 2 | 5+6 2 | 2+2 4 |
| | 5+6 2 | 3+4 2 | 2+2 4 | 2+2 4 | 2+2 4 | 5+6 2 | 3+4 2 | 2+2 4 |
| | 5+6 2 | 5+6 2 | 2+2 4 | 2+2 4 | 2+2 4 | 5+6 2 | 5+6 2 | 2+2 4 |
| | 3+4 2 | 3+4 2 | 2+2 4 | 2+2 4 | 2+2 4 | 3+4 2 | 3+4 2 | 2+2 4 |
| | 3+4 2 | 3+4 2 | 2+2 4 | 2+2 4 | 2+2 4 | 5+6 2 | 3+4 2 | 2+2 4 |
| | 5+6 2 | 5+6 2 | 2+2 4 | 2+2 4 | 2+2 4 | 3+4 2 | 5+6 2 | 2+2 4 |
| | 5+6 2 | 3+4 2 | 2+2 4 | 2+2 4 | 2+2 4 | 3+4 2 | 3+4 2 | 2+2 4 |
| | 3+4 2 | 3+4 2 | 2+2 4 | 2+2 4 | 2+2 4 | 5+6 2 | 3+4 2 | 2+2 4 |
| | 3+4 2 | 5+6 2 | 2+2 4 | 2+2 4 | 2+2 4 | 3+4 2 | 5+6 2 | 2+2 4 |
| | 5+6 2 | 5+6 2 | 2+2 4 | 2+2 4 | 2+2 4 | 3+4 2 | 5+6 2 | 2+2 4 |
| | 3+4 2 | 3+4 2 | 2+2 4 | 2+2 4 | 2+2 4 | 5+6 2 | 3+4 2 | 2+2 4 |
| | 5+6 2 | 5+6 2 | 2+2 4 | 2+2 4 | 2+2 4 | 5+6 2 | 5+6 2 | 2+2 4 |
| | 3+4 2 | 5+6 2 | 2+2 4 | 2+2 4 | 2+2 4 | 5+6 2 | 3+4 2 | 2+2 4 |
| | 5+6 2 | 3+4 2 | 2+2 4 | 2+2 4 | 2+2 4 | 3+4 2 | 5+6 2 | 2+2 4 |
| | 5+6 2 | 5+6 2 | 2+2 4 | 2+2 4 | 2+2 4 | 5+6 2 | 5+6 2 | 2+2 4 |
| | 3+4 2 | 3+4 2 | 2+2 4 | 2+2 4 | 2+2 4 | 3+4 2 | 3+4 2 | 2+2 4 |
| | 5+6 2 | 5+6 2 | 2+2 4 | 2+2 4 | 2+2 4 | 3+4 2 | 3+4 2 | 2+2 4 |
| | 3+4 2 | 3+4 2 | 2+2 4 | 2+2 4 | 2+2 4 | 3+4 2 | 3+4 2 | 2+2 4 |
| | 5+6 2 | 5+6 2 | 2+2 4 | 2+2 4 | 2+2 4 | 5+6 2 | 5+6 2 | 2+2 4 |
| | 3+4 2 | 3+4 2 | 2+2 4 | 2+2 4 | 2+2 4 | 5+6 2 | 5+6 2 | 2+2 4 |
| | 3+4 2 | 5+6 2 | 2+2 4 | 2+2 4 | 2+2 4 | 3+4 2 | 5+6 2 | 2+2 4 |
| | 5+6 2 | 3+4 2 | 2+2 4 | 2+2 4 | 2+2 4 | 5+6 2 | 3+4 2 | 2+2 4 |
| | 3+4 2 | 5+6 2 | 2+2 4 | 2+2 4 | 2+2 4 | 3+4 2 | 5+6 2 | 2+2 4 |
| | 5+6 2 | 3+4 2 | 2+2 4 | 2+2 4 | 2+2 4 | 5+6 2 | 3+4 2 | 2+2 4 |
| | 3+4 2 | 5+6 2 | 2+2 4 | 2+2 4 | 2+2 4 | 3+4 2 | 5+6 2 | 2+2 4 |
| | 3+4 2 | 3+4 2 | 2+2 4 | 2+2 4 | 2+2 4 | 3+4 2 | 3+4 2 | 2+2 4 |
| | 5+6 2 | 3+4 2 | 2+2 4 | 2+2 4 | 2+2 4 | 5+6 2 | 3+4 2 | 2+2 4 |
| | 5+6 2 | 5+6 2 | 2+2 4 | 2+2 4 | 2+2 4 | 5+6 2 | 5+6 2 | 2+2 4 |
| | 3+4 2 | 3+4 2 | 2+2 4 | 2+2 4 | 2+2 4 | 3+4 2 | 3+4 2 | 2+2 4 |
| | 5+6 2 | 5+6 2 | 2+2 4 | 2+2 4 | 2+2 4 | 5+6 2 | 5+6 2 | 2+2 4 |
| | 3+4 2 | 3+4 2 | 2+2 4 | 2+2 4 | 2+2 4 | 3+4 2 | 3+4 2 | 2+2 4 |
| | 5+6 2 | 5+6 2 | 2+2 4 | 2+2 4 | 2+2 4 | 5+6 2 | 5+6 2 | 2+2 4 |
| | 3+4 2 | 5+6 2 | 2+2 4 | 2+2 4 | 2+2 4 | 3+4 2 | 5+6 2 | 2+2 4 |
| | 5+6 2 | 3+4 2 | 2+2 4 | 2+2 4 | 2+2 4 | 5+6 2 | 3+4 2 | 2+2 4 |
| | 3+4 2 | 5+6 2 | 2+2 4 | 2+2 4 | 2+2 4 | 3+4 2 | 5+6 2 | 2+2 4 |
| | 5+6 2 | 3+4 2 | 2+2 4 | 2+2 4 | 2+2 4 | 5+6 2 | 3+4 2 | 2+2 4 |
| | 3+4 2 | 5+6 2 | 2+2 4 | 2+2 4 | 2+2 4 | 3+4 2 | 5+6 2 | 2+2 4 |
| | 3+4 2 | 3+4 2 | 2+2 4 | 2+2 4 | 2+2 4 | 3+4 2 | 3+4 2 | 2+2 4 |
| | 5+6 2 | 5+6 2 | 2+2 4 | 2+2 4 | 2+2 4 | 5+6 2 | 5+6 2 | 2+2 4 |
| | 5+6 2 | 3+4 2 | 2+2 4 | 2+2 4 | 2+2 4 | 5+6 2 | 3+4 2 | 2+2 4 |
| | 5+6 2 | 3+4 2 | 2+2 4 | 2+2 4 | 2+2 4 | 5+6 2 | 3+4 2 | 2+2 4 |
| | 3+4 2 | 5+6 2 | 2+2 4 | 2+2 4 | 2+2 4 | 3+4 2 | 5+6 2 | 2+2 4 |
| | 3+4 2 | 3+4 2 | 2+2 4 | 2+2 4 | 2+2 4 | 5+6 2 | 3+4 2 | 2+2 4 |
| | 5+6 2 | 5+6 2 | 2+2 4 | 2+2 4 | 2+2 4 | 3+4 2 | 5+6 2 | 2+2 4 |
| | 5+6 2 | 3+4 2 | 2+2 4 | 2+2 4 | 2+2 4 | 5+6 2 | 3+4 2 | 2+2 4 |
| | 3+4 2 | 5+6 2 | 2+2 4 | 2+2 4 | 2+2 4 | 3+4 2 | 5+6 2 | 2+2 4 |
| | 3+4 2 | 3+4 2 | 2+2 4 | 2+2 4 | 2+2 4 | 3+4 2 | 3+4 2 | 2+2 4 |
| | 5+6 2 | 5+6 2 | 2+2 4 | 2+2 4 | 2+2 4 | 5+6 2 | 5+6 2 | 2+2 4 |
| | 3+4 2 | 3+4 2 | 2+2 4 | 2+2 4 | 2+2 4 | 3+4 2 | 3+4 2 | 2+2 4 |
| | 5+6 2 | 5+6 2 | 2+2 4 | 2+2 4 | 2+2 4 | 5+6 2 | 5+6 2 | 2+2 4 |
| | 5+6 2 | 3+4 2 | 2+2 4 | 2+2 4 | 2+2 4 | 5+6 2 | 3+4 2 | 2+2 4 |

Continued on next column

**TABLE XXIII– continued**

| TQC | ICSD | | NoE | | SF | | | |
|---|---|---|---|---|---|---|---|---|
| | 3+4 2 | 5+6 2 | 2+2 4 | 2+2 4 | 2+2 4 | 3+4 2 | 5+6 2 | 2+2 4 |
| | 5+6 2 | 3+4 2 | 2+2 4 | 2+2 4 | 2+2 4 | 5+6 2 | 3+4 2 | 2+2 4 |
| | 3+4 2 | 5+6 2 | | | | 3+4 2 | 5+6 2 | |
| | 3+4 2 | 5+6 2 | | | | 3+4 2 | 5+6 2 | |
| | 5+6 2 | 3+4 2 | | | | 5+6 2 | 3+4 2 | |
| | 3+4 2 | 5+6 2 | | | | 3+4 2 | 5+6 2 | |
| | 5+6 2 | 3+4 2 | | | | 5+6 2 | 3+4 2 | |
| | 3+4 2 | 3+4 2 | | | | 3+4 2 | 3+4 2 | |
| | 5+6 2 | 5+6 2 | | | | 3+4 2 | 5+6 2 | |
| | 3+4 2 | 5+6 2 | | | | 5+6 2 | 5+6 2 | |
| | 5+6 2 | 5+6 2 | | | | 3+4 2 | 3+4 2 | |
| | 3+4 2 | 3+4 2 | | | | 5+6 2 | 5+6 2 | |
| | 5+6 2 | 3+4 2 | | | | 5+6 2 | 3+4 2 | |
| | 3+4 2 | 5+6 2 | | | | 3+4 2 | 5+6 2 | |
| | 3+4 2 | 3+4 2 | | | | 3+4 2 | 3+4 2 | |
| | 3+4 2 | 3+4 2 | | | | 3+4 2 | 3+4 2 | |
| | 5+6 2 | 5+6 2 | | | | 5+6 2 | 5+6 2 | |
| | 3+4 2 | 5+6 2 | | | | 5+6 2 | 3+4 2 | |
| | 5+6 2 | 3+4 2 | | | | 5+6 2 | 5+6 2 | |
| | 5+6 2 | 5+6 2 | | | | 3+4 2 | 3+4 2 | |
| | 3+4 2 | 3+4 2 | | | | 3+4 2 | 5+6 2 | |
| | 5+6 2 | 3+4 2 | | | | 5+6 2 | 3+4 2 | |
| | 5+6 2 | 3+4 2 | | | | 5+6 2 | 5+6 2 | |
| | 3+4 2 | 5+6 2 | | | | 3+4 2 | 3+4 2 | |
| | 5+6 2 | 3+4 2 | | | | 5+6 2 | 3+4 2 | |
| | 3+4 2 | 5+6 2 | | | | 3+4 2 | 3+4 2 | |
| | 5+6 2 | 3+4 2 | | | | 5+6 2 | 5+6 2 | |
| | 3+4 2 | 5+6 2 | | | | 3+4 2 | 5+6 2 | |
| | 5+6 2 | 3+4 2 | | | | 5+6 2 | 3+4 2 | |
| | 3+4 2 | 5+6 2 | | | | 5+6 2 | 5+6 2 | |
| | 5+6 2 | 5+6 2 | | | | 3+4 2 | 5+6 2 | |
| | 5+6 2 | 5+6 2 | | | | 5+6 2 | 3+4 2 | |
| | 3+4 2 | 3+4 2 | | | | 3+4 2 | 5+6 2 | |
| | 5+6 2 | 3+4 2 | | | | 3+4 2 | 5+6 2 | |
| | 3+4 2 | 5+6 2 | | | | 5+6 2 | 3+4 2 | |
| | 5+6 2 | 3+4 2 | | | | 3+4 2 | 5+6 2 | |
| | 3+4 2 | 5+6 2 | | | | 5+6 2 | 3+4 2 | |
| | 3+4 2 | 3+4 2 | | | | 3+4 2 | 5+6 2 | |
| | 5+6 2 | 5+6 2 | | | | 5+6 2 | 3+4 2 | |
| | 5+6 2 | 5+6 2 | | | | 5+6 2 | 5+6 2 | |
| | 3+4 2 | 3+4 2 | | | | 3+4 2 | 3+4 2 | |
| | 3+4 2 | 3+4 2 | | | | 3+4 2 | 3+4 2 | |
| | 5+6 2 | 5+6 2 | | | | 5+6 2 | 5+6 2 | |
| | 3+4 2 | 3+4 2 | | | | 5+6 2 | 3+4 2 | |
| | 5+6 2 | 5+6 2 | | | | 3+4 2 | 5+6 2 | |
| | 3+4 2 | 5+6 2 | | | | 3+4 2 | 5+6 2 | |
| | 5+6 2 | 5+6 2 | | | | 5+6 2 | 3+4 2 | |
| | 5+6 2 | 3+4 2 | | | | 3+4 2 | 3+4 2 | |
| | 5+6 2 | 5+6 2 | | | | 5+6 2 | 5+6 2 | |
| | 3+4 2 | 3+4 2 | | | | 3+4 2 | 3+4 2 | |
| | 3+4 2 | 3+4 2 | | | | 5+6 2 | 5+6 2 | |
| | 3+4 2 | 5+6 2 | | | | 3+4 2 | 3+4 2 | |
| | 3+4 2 | 3+4 2 | | | | 3+4 2 | 5+6 2 | |
| | 5+6 2 | 5+6 2 | | | | 5+6 2 | 3+4 2 | |
| | 3+4 2 | 3+4 2 | | | | 3+4 2 | 5+6 2 | |
| | 5+6 2 | 5+6 2 | | | | 5+6 2 | 3+4 2 | |
| | 5+6 2 | 5+6 2 | | | | 3+4 2 | 5+6 2 | |
| | 5+6 2 | 3+4 2 | | | | 3+4 2 | 3+4 2 | |
| | 3+4 2 | 3+4 2 | | | | 5+6 2 | 5+6 2 | |
| 11.2.63 | 63219 | | 188 | H10Mo3N2O11 | | | | |
| | -A | -B | -C | -D | -E | -GM | -Y | -Z |

Continued on next column

TABLE XXIII– continued

| TQC | ICSD | NoE | SF | | | | | |
|---|---|---|---|---|---|---|---|---|
| | 3+4 2 | 3+4 2 | 2+2 4 | 2+2 4 | 2+2 4 | 3+4 2 | 3+4 2 | 2+2 4 |
| | 5+6 2 | 5+6 2 | 2+2 4 | 2+2 4 | 2+2 4 | 5+6 2 | 5+6 2 | 2+2 4 |
| | 3+4 2 | 3+4 2 | 2+2 4 | 2+2 4 | 2+2 4 | 3+4 2 | 3+4 2 | 2+2 4 |
| | 5+6 2 | 5+6 2 | 2+2 4 | 2+2 4 | 2+2 4 | 5+6 2 | 5+6 2 | 2+2 4 |
| | 3+4 2 | 3+4 2 | 2+2 4 | 2+2 4 | 2+2 4 | 3+4 2 | 3+4 2 | 2+2 4 |
| | 5+6 2 | 5+6 2 | 2+2 4 | 2+2 4 | 2+2 4 | 5+6 2 | 5+6 2 | 2+2 4 |
| | 5+6 2 | 5+6 2 | 2+2 4 | 2+2 4 | 2+2 4 | 3+4 2 | 5+6 2 | 2+2 4 |
| | 3+4 2 | 3+4 2 | 2+2 4 | 2+2 4 | 2+2 4 | 5+6 2 | 3+4 2 | 2+2 4 |
| | 5+6 2 | 5+6 2 | 2+2 4 | 2+2 4 | 2+2 4 | 5+6 2 | 5+6 2 | 2+2 4 |
| | 3+4 2 | 3+4 2 | 2+2 4 | 2+2 4 | 2+2 4 | 3+4 2 | 3+4 2 | 2+2 4 |
| | 3+4 2 | 3+4 2 | 2+2 4 | 2+2 4 | 2+2 4 | 5+6 2 | 5+6 2 | 2+2 4 |
| | 5+6 2 | 5+6 2 | 2+2 4 | 2+2 4 | 2+2 4 | 3+4 2 | 3+4 2 | 2+2 4 |
| | 3+4 2 | 3+4 2 | 2+2 4 | 2+2 4 | 2+2 4 | 5+6 2 | 5+6 2 | 2+2 4 |
| | 5+6 2 | 5+6 2 | 2+2 4 | 2+2 4 | 2+2 4 | 3+4 2 | 3+4 2 | 2+2 4 |
| | 3+4 2 | 3+4 2 | 2+2 4 | 2+2 4 | 2+2 4 | 3+4 2 | 3+4 2 | 2+2 4 |
| | 5+6 2 | 5+6 2 | 2+2 4 | 2+2 4 | 2+2 4 | 5+6 2 | 5+6 2 | 2+2 4 |
| | 5+6 2 | 5+6 2 | 2+2 4 | 2+2 4 | 2+2 4 | 3+4 2 | 3+4 2 | 2+2 4 |
| | 3+4 2 | 3+4 2 | 2+2 4 | 2+2 4 | 2+2 4 | 5+6 2 | 5+6 2 | 2+2 4 |
| | 3+4 2 | 3+4 2 | 2+2 4 | 2+2 4 | 2+2 4 | 3+4 2 | 3+4 2 | 2+2 4 |
| | 5+6 2 | 5+6 2 | 2+2 4 | 2+2 4 | 2+2 4 | 5+6 2 | 5+6 2 | 2+2 4 |
| | 3+4 2 | 3+4 2 | 2+2 4 | 2+2 4 | 2+2 4 | 3+4 2 | 3+4 2 | 2+2 4 |
| | 5+6 2 | 5+6 2 | 2+2 4 | 2+2 4 | 2+2 4 | 5+6 2 | 5+6 2 | 2+2 4 |
| | 3+4 2 | 3+4 2 | 2+2 4 | 2+2 4 | 2+2 4 | 3+4 2 | 3+4 2 | 2+2 4 |
| | 5+6 2 | 5+6 2 | 2+2 4 | 2+2 4 | 2+2 4 | 5+6 2 | 5+6 2 | 2+2 4 |
| | 3+4 2 | 3+4 2 | 2+2 4 | 2+2 4 | 2+2 4 | 3+4 2 | 3+4 2 | 2+2 4 |
| | 5+6 2 | 5+6 2 | 2+2 4 | 2+2 4 | 2+2 4 | 5+6 2 | 5+6 2 | 2+2 4 |
| | 3+4 2 | 3+4 2 | 2+2 4 | 2+2 4 | 2+2 4 | 3+4 2 | 3+4 2 | 2+2 4 |
| | 5+6 2 | 5+6 2 | 2+2 4 | 2+2 4 | 2+2 4 | 5+6 2 | 5+6 2 | 2+2 4 |
| | 3+4 2 | 3+4 2 | 2+2 4 | 2+2 4 | 2+2 4 | 3+4 2 | 3+4 2 | 2+2 4 |
| | 5+6 2 | 5+6 2 | 2+2 4 | 2+2 4 | 2+2 4 | 5+6 2 | 5+6 2 | 2+2 4 |
| | 3+4 2 | 3+4 2 | 2+2 4 | 2+2 4 | 2+2 4 | 3+4 2 | 3+4 2 | 2+2 4 |
| | 5+6 2 | 5+6 2 | 2+2 4 | 2+2 4 | 2+2 4 | 5+6 2 | 5+6 2 | 2+2 4 |
| | 3+4 2 | 3+4 2 | 2+2 4 | 2+2 4 | 2+2 4 | 3+4 2 | 3+4 2 | 2+2 4 |
| | 3+4 2 | 3+4 2 | 2+2 4 | 2+2 4 | 2+2 4 | 3+4 2 | 3+4 2 | 2+2 4 |
| | 5+6 2 | 5+6 2 | 2+2 4 | 2+2 4 | 2+2 4 | 5+6 2 | 5+6 2 | 2+2 4 |
| | 5+6 2 | 5+6 2 | 2+2 4 | 2+2 4 | 2+2 4 | 5+6 2 | 5+6 2 | 2+2 4 |
| | 3+4 2 | 3+4 2 | 2+2 4 | 2+2 4 | 2+2 4 | 3+4 2 | 3+4 2 | 2+2 4 |
| | 5+6 2 | 5+6 2 | 2+2 4 | 2+2 4 | 2+2 4 | 5+6 2 | 5+6 2 | 2+2 4 |
| | 3+4 2 | 3+4 2 | 2+2 4 | 2+2 4 | 2+2 4 | 3+4 2 | 3+4 2 | 2+2 4 |
| | 5+6 2 | 5+6 2 | 2+2 4 | 2+2 4 | 2+2 4 | 5+6 2 | 5+6 2 | 2+2 4 |
| | 3+4 2 | 3+4 2 | 2+2 4 | 2+2 4 | 2+2 4 | 3+4 2 | 3+4 2 | 2+2 4 |
| | 5+6 2 | 5+6 2 | 2+2 4 | 2+2 4 | 2+2 4 | 5+6 2 | 5+6 2 | 2+2 4 |
| | 5+6 2 | 5+6 2 | 2+2 4 | 2+2 4 | 2+2 4 | 5+6 2 | 5+6 2 | 2+2 4 |
| | 3+4 2 | 3+4 2 | 2+2 4 | 2+2 4 | 2+2 4 | 3+4 2 | 3+4 2 | 2+2 4 |
| | 3+4 2 | 3+4 2 | 2+2 4 | 2+2 4 | 2+2 4 | 5+6 2 | 5+6 2 | 2+2 4 |
| | 5+6 2 | 5+6 2 | 2+2 4 | 2+2 4 | 2+2 4 | 3+4 2 | 3+4 2 | 2+2 4 |
| | 3+4 2 | 3+4 2 | 2+2 4 | 2+2 4 | 2+2 4 | 3+4 2 | 5+6 2 | 2+2 4 |
| | 5+6 2 | 5+6 2 | | | | 5+6 2 | 3+4 2 | |
| | 3+4 2 | 3+4 2 | | | | 3+4 2 | 3+4 2 | |
| | 5+6 2 | 5+6 2 | | | | 5+6 2 | 5+6 2 | |
| | 3+4 2 | 3+4 2 | | | | 3+4 2 | 3+4 2 | |
| | 5+6 2 | 5+6 2 | | | | 5+6 2 | 5+6 2 | |
| | 5+6 2 | 5+6 2 | | | | 5+6 2 | 5+6 2 | |
| | 5+6 2 | 5+6 2 | | | | 3+4 2 | 3+4 2 | |
| | 3+4 2 | 3+4 2 | | | | 5+6 2 | 5+6 2 | |
| | 3+4 2 | 3+4 2 | | | | 3+4 2 | 3+4 2 | |
| | 5+6 2 | 5+6 2 | | | | 5+6 2 | 5+6 2 | |
| | 3+4 2 | 3+4 2 | | | | 3+4 2 | 3+4 2 | |
| | 5+6 2 | 5+6 2 | | | | 5+6 2 | 5+6 2 | |
| | 3+4 2 | 3+4 2 | | | | 3+4 2 | 3+4 2 | |
| | 5+6 2 | 5+6 2 | | | | 5+6 2 | 5+6 2 | |
| | 3+4 2 | 3+4 2 | | | | 3+4 2 | 3+4 2 | |
| | 3+4 2 | 3+4 2 | | | | 3+4 2 | 3+4 2 | |
| | 5+6 2 | 5+6 2 | | | | 5+6 2 | 5+6 2 | |
| | 3+4 2 | 5+6 2 | | | | 3+4 2 | 3+4 2 | |
| | 5+6 2 | 3+4 2 | | | | 5+6 2 | 5+6 2 | |
| | 3+4 2 | 3+4 2 | | | | 3+4 2 | 3+4 2 | |
| | 5+6 2 | 5+6 2 | | | | 5+6 2 | 5+6 2 | |
| | 5+6 2 | 5+6 2 | | | | 3+4 2 | 3+4 2 | |
| | 5+6 2 | 5+6 2 | | | | 5+6 2 | 3+4 2 | |
| | 3+4 2 | 3+4 2 | | | | 5+6 2 | 5+6 2 | |
| | 5+6 2 | 3+4 2 | | | | 3+4 2 | 5+6 2 | |
| | 3+4 2 | 5+6 2 | | | | 5+6 2 | 5+6 2 | |
| | 3+4 2 | 3+4 2 | | | | 3+4 2 | 3+4 2 | |
| | 5+6 2 | 5+6 2 | | | | 3+4 2 | 5+6 2 | |
| | 5+6 2 | 3+4 2 | | | | 5+6 2 | 3+4 2 | |
| | 3+4 2 | 5+6 2 | | | | 3+4 2 | 5+6 2 | |
| | 3+4 2 | 3+4 2 | | | | 5+6 2 | 3+4 2 | |
| | 3+4 2 | 3+4 2 | | | | 3+4 2 | 3+4 2 | |
| | 5+6 2 | 5+6 2 | | | | 5+6 2 | 5+6 2 | |
| | 3+4 2 | 3+4 2 | | | | 3+4 2 | 3+4 2 | |
| | 3+4 2 | 5+6 2 | | | | 3+4 2 | 5+6 2 | |
| | 5+6 2 | 3+4 2 | | | | 5+6 2 | 3+4 2 | |
| | 5+6 2 | 3+4 2 | | | | 3+4 2 | 3+4 2 | |
| | 3+4 2 | 5+6 2 | | | | 5+6 2 | 5+6 2 | |
| | 5+6 2 | 5+6 2 | | | | 5+6 2 | 5+6 2 | |
| | 5+6 2 | 3+4 2 | | | | 3+4 2 | 3+4 2 | |
| | 3+4 2 | 5+6 2 | | | | 5+6 2 | 5+6 2 | |
| | 5+6 2 | 5+6 2 | | | | 3+4 2 | 3+4 2 | |
| | 3+4 2 | 3+4 2 | | | | 5+6 2 | 5+6 2 | |
| | 3+4 2 | 3+4 2 | | | | 3+4 2 | 5+6 2 | |
| | 3+4 2 | 3+4 2 | | | | 5+6 2 | 3+4 2 | |
| | 5+6 2 | 5+6 2 | | | | 3+4 2 | 3+4 2 | |
| | 3+4 2 | 3+4 2 | | | | 5+6 2 | 5+6 2 | |
| 11.2.64 | 418027 | 108 | Ge1Hf1Pd1 | | | | | |
| | -A | -B | -C | -D | -E | -GM | -Y | -Z |
| | 3+4 2 | 3+4 2 | 2+2 4 | 2+2 4 | 2+2 4 | 3+4 2 | 3+4 2 | 2+2 4 |
| | 5+6 2 | 5+6 2 | 2+2 4 | 2+2 4 | 2+2 4 | 5+6 2 | 5+6 2 | 2+2 4 |
| | 5+6 2 | 3+4 2 | 2+2 4 | 2+2 4 | 2+2 4 | 3+4 2 | 3+4 2 | 2+2 4 |
| | 3+4 2 | 5+6 2 | 2+2 4 | 2+2 4 | 2+2 4 | 5+6 2 | 5+6 2 | 2+2 4 |
| | 5+6 2 | 5+6 2 | 2+2 4 | 2+2 4 | 2+2 4 | 5+6 2 | 3+4 2 | 2+2 4 |
| | 3+4 2 | 3+4 2 | 2+2 4 | 2+2 4 | 2+2 4 | 3+4 2 | 5+6 2 | 2+2 4 |
| | 5+6 2 | 5+6 2 | 2+2 4 | 2+2 4 | 2+2 4 | 3+4 2 | 5+6 2 | 2+2 4 |
| | 3+4 2 | 3+4 2 | 2+2 4 | 2+2 4 | 2+2 4 | 5+6 2 | 3+4 2 | 2+2 4 |
| | 5+6 2 | 3+4 2 | 2+2 4 | 2+2 4 | 2+2 4 | 5+6 2 | 3+4 2 | 2+2 4 |
| | 3+4 2 | 5+6 2 | 2+2 4 | 2+2 4 | 2+2 4 | 3+4 2 | 5+6 2 | 2+2 4 |
| | 3+4 2 | 5+6 2 | 2+2 4 | 2+2 4 | 2+2 4 | 5+6 2 | 3+4 2 | 2+2 4 |
| | 5+6 2 | 3+4 2 | 2+2 4 | 2+2 4 | 2+2 4 | 3+4 2 | 5+6 2 | 2+2 4 |
| | 5+6 2 | 5+6 2 | 2+2 4 | 2+2 4 | 2+2 4 | 5+6 2 | 5+6 2 | 2+2 4 |
| | 3+4 2 | 3+4 2 | 2+2 4 | 2+2 4 | 2+2 4 | 3+4 2 | 5+6 2 | 2+2 4 |
| | 5+6 2 | 5+6 2 | 2+2 4 | 2+2 4 | 2+2 4 | 5+6 2 | 5+6 2 | 2+2 4 |
| | 3+4 2 | 3+4 2 | 2+2 4 | 2+2 4 | 2+2 4 | 3+4 2 | 5+6 2 | 2+2 4 |
| | 5+6 2 | 5+6 2 | 2+2 4 | 2+2 4 | 2+2 4 | 5+6 2 | 5+6 2 | 2+2 4 |
| | 3+4 2 | 3+4 2 | 2+2 4 | 2+2 4 | 2+2 4 | 5+6 2 | 3+4 2 | 2+2 4 |
| | 3+4 2 | 3+4 2 | 2+2 4 | 2+2 4 | 2+2 4 | 3+4 2 | 3+4 2 | 2+2 4 |
| | 5+6 2 | 5+6 2 | 2+2 4 | 2+2 4 | 2+2 4 | 5+6 2 | 3+4 2 | 2+2 4 |
| | 3+4 2 | 5+6 2 | 2+2 4 | 2+2 4 | 2+2 4 | 3+4 2 | 3+4 2 | 2+2 4 |
| | 3+4 2 | 3+4 2 | 2+2 4 | 2+2 4 | 2+2 4 | 3+4 2 | 3+4 2 | 2+2 4 |
| | 5+6 2 | 3+4 2 | 2+2 4 | 2+2 4 | 2+2 4 | 3+4 2 | 5+6 2 | 2+2 4 |
| | 5+6 2 | 5+6 2 | 2+2 4 | 2+2 4 | 2+2 4 | 5+6 2 | 3+4 2 | 2+2 4 |
| | 3+4 2 | 3+4 2 | 2+2 4 | 2+2 4 | 2+2 4 | 3+4 2 | 3+4 2 | 2+2 4 |
| | 5+6 2 | 5+6 2 | 2+2 4 | 2+2 4 | 2+2 4 | 5+6 2 | 5+6 2 | 2+2 4 |

Continued on next column

**TABLE XXIII– continued**

| TQC | ICSD | NoE | SF | | | | | |
|---|---|---|---|---|---|---|---|---|
| | 5+6 2 | 3+4 2 | 2+2 4 | 2+2 4 | 2+2 4 | 3+4 2 | 5+6 2 | 2+2 4 |
| | 3+4 2 | 5+6 2 | | | | 5+6 2 | 3+4 2 | |
| | 5+6 2 | 3+4 2 | | | | 5+6 2 | 5+6 2 | |
| | 5+6 2 | 5+6 2 | | | | 3+4 2 | 3+4 2 | |
| | 3+4 2 | 5+6 2 | | | | 5+6 2 | 3+4 2 | |
| | 3+4 2 | 3+4 2 | | | | 3+4 2 | 5+6 2 | |
| | 3+4 2 | 5+6 2 | | | | 3+4 2 | 3+4 2 | |
| | 5+6 2 | 3+4 2 | | | | 5+6 2 | 3+4 2 | |
| | 3+4 2 | 5+6 2 | | | | 3+4 2 | 5+6 2 | |
| | 5+6 2 | 3+4 2 | | | | 5+6 2 | 3+4 2 | |
| | 3+4 2 | 5+6 2 | | | | 3+4 2 | 3+4 2 | |
| | 5+6 2 | 3+4 2 | | | | 5+6 2 | 5+6 2 | |
| | 5+6 2 | 5+6 2 | | | | 5+6 2 | 5+6 2 | |
| | 3+4 2 | 3+4 2 | | | | 3+4 2 | 5+6 2 | |
| | 3+4 2 | 5+6 2 | | | | 3+4 2 | 3+4 2 | |
| | 5+6 2 | 3+4 2 | | | | 5+6 2 | 5+6 2 | |
| | 5+6 2 | 5+6 2 | | | | 3+4 2 | 3+4 2 | |
| | 3+4 2 | 3+4 2 | | | | 3+4 2 | 3+4 2 | |
| | 3+4 2 | 3+4 2 | | | | 5+6 2 | 5+6 2 | |
| | 5+6 2 | 5+6 2 | | | | 3+4 2 | 3+4 2 | |
| | 5+6 2 | 5+6 2 | | | | 5+6 2 | 3+4 2 | |
| | 3+4 2 | 3+4 2 | | | | 3+4 2 | 5+6 2 | |
| | 3+4 2 | 3+4 2 | | | | 3+4 2 | 3+4 2 | |
| | 5+6 2 | 5+6 2 | | | | 5+6 2 | 5+6 2 | |
| | 3+4 2 | 5+6 2 | | | | 5+6 2 | 3+4 2 | |
| | 5+6 2 | 3+4 2 | | | | 5+6 2 | 5+6 2 | |
| | 5+6 2 | 5+6 2 | | | | 3+4 2 | 3+4 2 | |
| | 3+4 2 | 3+4 2 | | | | 3+4 2 | 3+4 2 | |
| 11.2.65 | 180364 | 200 | Ga5Pt1 | | | | | |
| | -A | -B | -C | -D | -E | -GM | -Y | -Z |
| | 3+4 2 | 3+4 2 | 2+2 4 | 2+2 4 | 2+2 4 | 3+4 2 | 5+6 2 | 2+2 4 |
| | 5+6 2 | 5+6 2 | 2+2 4 | 2+2 4 | 2+2 4 | 3+4 2 | 3+4 2 | 2+2 4 |
| | 5+6 2 | 5+6 2 | 2+2 4 | 2+2 4 | 2+2 4 | 5+6 2 | 3+4 2 | 2+2 4 |
| | 3+4 2 | 3+4 2 | 2+2 4 | 2+2 4 | 2+2 4 | 5+6 2 | 3+4 2 | 2+2 4 |
| | 5+6 2 | 5+6 2 | 2+2 4 | 2+2 4 | 2+2 4 | 5+6 2 | 5+6 2 | 2+2 4 |
| | 3+4 2 | 3+4 2 | 2+2 4 | 2+2 4 | 2+2 4 | 3+4 2 | 5+6 2 | 2+2 4 |
| | 3+4 2 | 3+4 2 | 2+2 4 | 2+2 4 | 2+2 4 | 5+6 2 | 3+4 2 | 2+2 4 |
| | 5+6 2 | 5+6 2 | 2+2 4 | 2+2 4 | 2+2 4 | 3+4 2 | 5+6 2 | 2+2 4 |
| | 5+6 2 | 3+4 2 | 2+2 4 | 2+2 4 | 2+2 4 | 3+4 2 | 3+4 2 | 2+2 4 |
| | 5+6 2 | 5+6 2 | 2+2 4 | 2+2 4 | 2+2 4 | 5+6 2 | 5+6 2 | 2+2 4 |
| | 3+4 2 | 3+4 2 | 2+2 4 | 2+2 4 | 2+2 4 | 3+4 2 | 3+4 2 | 2+2 4 |
| | 3+4 2 | 5+6 2 | 2+2 4 | 2+2 4 | 2+2 4 | 3+4 2 | 5+6 2 | 2+2 4 |
| | 3+4 2 | 5+6 2 | 2+2 4 | 2+2 4 | 2+2 4 | 3+4 2 | 3+4 2 | 2+2 4 |
| | 5+6 2 | 5+6 2 | 2+2 4 | 2+2 4 | 2+2 4 | 5+6 2 | 5+6 2 | 2+2 4 |
| | 3+4 2 | 3+4 2 | 2+2 4 | 2+2 4 | 2+2 4 | 3+4 2 | 5+6 2 | 2+2 4 |
| | 5+6 2 | 3+4 2 | 2+2 4 | 2+2 4 | 2+2 4 | 5+6 2 | 3+4 2 | 2+2 4 |
| | 3+4 2 | 3+4 2 | 2+2 4 | 2+2 4 | 2+2 4 | 3+4 2 | 5+6 2 | 2+2 4 |
| | 3+4 2 | 5+6 2 | 2+2 4 | 2+2 4 | 2+2 4 | 3+4 2 | 5+6 2 | 2+2 4 |
| | 5+6 2 | 5+6 2 | 2+2 4 | 2+2 4 | 2+2 4 | 3+4 2 | 5+6 2 | 2+2 4 |
| | 5+6 2 | 3+4 2 | 2+2 4 | 2+2 4 | 2+2 4 | 5+6 2 | 3+4 2 | 2+2 4 |
| | 3+4 2 | 3+4 2 | 2+2 4 | 2+2 4 | 2+2 4 | 5+6 2 | 5+6 2 | 2+2 4 |
| | 5+6 2 | 5+6 2 | 2+2 4 | 2+2 4 | 2+2 4 | 3+4 2 | 3+4 2 | 2+2 4 |
| | 5+6 2 | 5+6 2 | 2+2 4 | 2+2 4 | 2+2 4 | 5+6 2 | 5+6 2 | 2+2 4 |
| | 5+6 2 | 3+4 2 | 2+2 4 | 2+2 4 | 2+2 4 | 3+4 2 | 3+4 2 | 2+2 4 |
| | 3+4 2 | 3+4 2 | 2+2 4 | 2+2 4 | 2+2 4 | 5+6 2 | 3+4 2 | 2+2 4 |
| | 3+4 2 | 5+6 2 | 2+2 4 | 2+2 4 | 2+2 4 | 5+6 2 | 5+6 2 | 2+2 4 |
| | 3+4 2 | 3+4 2 | 2+2 4 | 2+2 4 | 2+2 4 | 5+6 2 | 3+4 2 | 2+2 4 |
| | 5+6 2 | 5+6 2 | 2+2 4 | 2+2 4 | 2+2 4 | 5+6 2 | 3+4 2 | 2+2 4 |
| | 3+4 2 | 3+4 2 | 2+2 4 | 2+2 4 | 2+2 4 | 3+4 2 | 5+6 2 | 2+2 4 |
| | 5+6 2 | 5+6 2 | 2+2 4 | 2+2 4 | 2+2 4 | 3+4 2 | 3+4 2 | 2+2 4 |
| | 3+4 2 | 3+4 2 | 2+2 4 | 2+2 4 | 2+2 4 | 3+4 2 | 5+6 2 | 2+2 4 |

Continued on next column

**TABLE XXIII– continued**

| TQC | ICSD | NoE | SF | | | | | |
|---|---|---|---|---|---|---|---|---|
| | 5+6 2 | 5+6 2 | 2+2 4 | 2+2 4 | 2+2 4 | 5+6 2 | 3+4 2 | 2+2 4 |
| | 3+4 2 | 5+6 2 | 2+2 4 | 2+2 4 | 2+2 4 | 3+4 2 | 3+4 2 | 2+2 4 |
| | 3+4 2 | 3+4 2 | 2+2 4 | 2+2 4 | 2+2 4 | 5+6 2 | 5+6 2 | 2+2 4 |
| | 5+6 2 | 5+6 2 | 2+2 4 | 2+2 4 | 2+2 4 | 5+6 2 | 3+4 2 | 2+2 4 |
| | 5+6 2 | 3+4 2 | 2+2 4 | 2+2 4 | 2+2 4 | 3+4 2 | 5+6 2 | 2+2 4 |
| | 5+6 2 | 5+6 2 | 2+2 4 | 2+2 4 | 2+2 4 | 5+6 2 | 3+4 2 | 2+2 4 |
| | 3+4 2 | 5+6 2 | 2+2 4 | 2+2 4 | 2+2 4 | 3+4 2 | 3+4 2 | 2+2 4 |
| | 3+4 2 | 3+4 2 | 2+2 4 | 2+2 4 | 2+2 4 | 5+6 2 | 5+6 2 | 2+2 4 |
| | 5+6 2 | 5+6 2 | 2+2 4 | 2+2 4 | 2+2 4 | 3+4 2 | 3+4 2 | 2+2 4 |
| | 5+6 2 | 3+4 2 | 2+2 4 | 2+2 4 | 2+2 4 | 5+6 2 | 5+6 2 | 2+2 4 |
| | 3+4 2 | 3+4 2 | 2+2 4 | 2+2 4 | 2+2 4 | 3+4 2 | 5+6 2 | 2+2 4 |
| | 5+6 2 | 5+6 2 | 2+2 4 | 2+2 4 | 2+2 4 | 5+6 2 | 3+4 2 | 2+2 4 |
| | 3+4 2 | 3+4 2 | 2+2 4 | 2+2 4 | 2+2 4 | 5+6 2 | 3+4 2 | 2+2 4 |
| | 5+6 2 | 5+6 2 | 2+2 4 | 2+2 4 | 2+2 4 | 3+4 2 | 5+6 2 | 2+2 4 |
| | 3+4 2 | 3+4 2 | 2+2 4 | 2+2 4 | 2+2 4 | 3+4 2 | 3+4 2 | 2+2 4 |
| | 5+6 2 | 3+4 2 | 2+2 4 | 2+2 4 | 2+2 4 | 5+6 2 | 3+4 2 | 2+2 4 |
| | 3+4 2 | 5+6 2 | 2+2 4 | 2+2 4 | 2+2 4 | 3+4 2 | 5+6 2 | 2+2 4 |
| | 5+6 2 | 3+4 2 | 2+2 4 | 2+2 4 | 2+2 4 | 5+6 2 | 5+6 2 | 2+2 4 |
| | 3+4 2 | 5+6 2 | 2+2 4 | 2+2 4 | 2+2 4 | 3+4 2 | 3+4 2 | 2+2 4 |
| | 5+6 2 | 3+4 2 | | | | 5+6 2 | 5+6 2 | |
| | 3+4 2 | 5+6 2 | | | | 5+6 2 | 3+4 2 | |
| | 5+6 2 | 3+4 2 | | | | 3+4 2 | 3+4 2 | |
| | 3+4 2 | 5+6 2 | | | | 5+6 2 | 5+6 2 | |
| | 3+4 2 | 3+4 2 | | | | 3+4 2 | 5+6 2 | |
| | 5+6 2 | 5+6 2 | | | | 5+6 2 | 5+6 2 | |
| | 5+6 2 | 3+4 2 | | | | 3+4 2 | 3+4 2 | |
| | 3+4 2 | 5+6 2 | | | | 5+6 2 | 3+4 2 | |
| | 3+4 2 | 3+4 2 | | | | 3+4 2 | 5+6 2 | |
| | 5+6 2 | 5+6 2 | | | | 5+6 2 | 3+4 2 | |
| | 3+4 2 | 3+4 2 | | | | 3+4 2 | 5+6 2 | |
| | 5+6 2 | 5+6 2 | | | | 5+6 2 | 5+6 2 | |
| | 5+6 2 | 5+6 2 | | | | 3+4 2 | 3+4 2 | |
| | 3+4 2 | 3+4 2 | | | | 3+4 2 | 5+6 2 | |
| | 3+4 2 | 5+6 2 | | | | 5+6 2 | 3+4 2 | |
| | 5+6 2 | 3+4 2 | | | | 3+4 2 | 3+4 2 | |
| | 3+4 2 | 3+4 2 | | | | 5+6 2 | 5+6 2 | |
| | 5+6 2 | 5+6 2 | | | | 3+4 2 | 3+4 2 | |
| | 5+6 2 | 3+4 2 | | | | 5+6 2 | 3+4 2 | |
| | 3+4 2 | 5+6 2 | | | | 3+4 2 | 5+6 2 | |
| | 3+4 2 | 3+4 2 | | | | 5+6 2 | 5+6 2 | |
| | 5+6 2 | 5+6 2 | | | | 3+4 2 | 3+4 2 | |
| | 5+6 2 | 3+4 2 | | | | 5+6 2 | 5+6 2 | |
| | 3+4 2 | 5+6 2 | | | | 3+4 2 | 3+4 2 | |
| | 3+4 2 | 3+4 2 | | | | 5+6 2 | 5+6 2 | |
| | 5+6 2 | 5+6 2 | | | | 5+6 2 | 3+4 2 | |
| | 5+6 2 | 3+4 2 | | | | 3+4 2 | 5+6 2 | |
| | 3+4 2 | 5+6 2 | | | | 5+6 2 | 3+4 2 | |
| | 5+6 2 | 5+6 2 | | | | 3+4 2 | 3+4 2 | |
| | 3+4 2 | 3+4 2 | | | | 5+6 2 | 5+6 2 | |
| | 3+4 2 | 5+6 2 | | | | 3+4 2 | 5+6 2 | |
| | 3+4 2 | 3+4 2 | | | | 3+4 2 | 5+6 2 | |
| | 5+6 2 | 5+6 2 | | | | 5+6 2 | 3+4 2 | |
| | 5+6 2 | 3+4 2 | | | | 3+4 2 | 5+6 2 | |
| | 3+4 2 | 3+4 2 | | | | 5+6 2 | 3+4 2 | |
| | 3+4 2 | 5+6 2 | | | | 3+4 2 | 5+6 2 | |
| | 5+6 2 | 5+6 2 | | | | 3+4 2 | 5+6 2 | |
| | 5+6 2 | 3+4 2 | | | | 3+4 2 | 3+4 2 | |
| | 5+6 2 | 5+6 2 | | | | 3+4 2 | 5+6 2 | |
| | 3+4 2 | 3+4 2 | | | | 5+6 2 | 3+4 2 | |
| | 5+6 2 | 3+4 2 | | | | 5+6 2 | 5+6 2 | |
| | 3+4 2 | 5+6 2 | | | | 3+4 2 | 5+6 2 | |

Continued on next column

**TABLE XXIII– continued**

| TQC | ICSD | NoE | SF | | | | | |
|---|---|---|---|---|---|---|---|---|
| | 5+6 2 | 3+4 2 | | | | 3+4 2 | 5+6 2 | |
| | 3+4 2 | 5+6 2 | | | | 5+6 2 | 3+4 2 | |
| | 5+6 2 | 5+6 2 | | | | 3+4 2 | 5+6 2 | |
| | 3+4 2 | 3+4 2 | | | | 5+6 2 | 3+4 2 | |
| | 3+4 2 | 3+4 2 | | | | 3+4 2 | 5+6 2 | |
| | 5+6 2 | 5+6 2 | | | | 5+6 2 | 3+4 2 | |
| | 3+4 2 | 3+4 2 | | | | 3+4 2 | 5+6 2 | |
| | 5+6 2 | 5+6 2 | | | | 5+6 2 | 3+4 2 | |
| 11.2.66 | 89343 | 148 | F3K1Mn1 | | | | | |
| | -A | -B | -C | -D | -E | -GM | -Y | -Z |
| | 3+4 2 | 5+6 2 | 2+2 4 | 2+2 4 | 2+2 4 | 3+4 2 | 5+6 2 | 2+2 4 |
| | 5+6 2 | 3+4 2 | 2+2 4 | 2+2 4 | 2+2 4 | 5+6 2 | 3+4 2 | 2+2 4 |
| | 3+4 2 | 3+4 2 | 2+2 4 | 2+2 4 | 2+2 4 | 3+4 2 | 3+4 2 | 2+2 4 |
| | 5+6 2 | 5+6 2 | 2+2 4 | 2+2 4 | 2+2 4 | 5+6 2 | 5+6 2 | 2+2 4 |
| | 5+6 2 | 3+4 2 | 2+2 4 | 2+2 4 | 2+2 4 | 3+4 2 | 3+4 2 | 2+2 4 |
| | 3+4 2 | 3+4 2 | 2+2 4 | 2+2 4 | 2+2 4 | 3+4 2 | 5+6 2 | 2+2 4 |
| | 3+4 2 | 5+6 2 | 2+2 4 | 2+2 4 | 2+2 4 | 5+6 2 | 3+4 2 | 2+2 4 |
| | 5+6 2 | 5+6 2 | 2+2 4 | 2+2 4 | 2+2 4 | 5+6 2 | 5+6 2 | 2+2 4 |
| | 5+6 2 | 5+6 2 | 2+2 4 | 2+2 4 | 2+2 4 | 5+6 2 | 5+6 2 | 2+2 4 |
| | 5+6 2 | 5+6 2 | 2+2 4 | 2+2 4 | 2+2 4 | 3+4 2 | 3+4 2 | 2+2 4 |
| | 3+4 2 | 3+4 2 | 2+2 4 | 2+2 4 | 2+2 4 | 5+6 2 | 5+6 2 | 2+2 4 |
| | 5+6 2 | 5+6 2 | 2+2 4 | 2+2 4 | 2+2 4 | 5+6 2 | 5+6 2 | 2+2 4 |
| | 3+4 2 | 3+4 2 | 2+2 4 | 2+2 4 | 2+2 4 | 3+4 2 | 3+4 2 | 2+2 4 |
| | 3+4 2 | 3+4 2 | 2+2 4 | 2+2 4 | 2+2 4 | 3+4 2 | 3+4 2 | 2+2 4 |
| | 3+4 2 | 3+4 2 | 2+2 4 | 2+2 4 | 2+2 4 | 3+4 2 | 3+4 2 | 2+2 4 |
| | 5+6 2 | 5+6 2 | 2+2 4 | 2+2 4 | 2+2 4 | 5+6 2 | 5+6 2 | 2+2 4 |
| | 3+4 2 | 5+6 2 | 2+2 4 | 2+2 4 | 2+2 4 | 3+4 2 | 5+6 2 | 2+2 4 |
| | 5+6 2 | 3+4 2 | 2+2 4 | 2+2 4 | 2+2 4 | 5+6 2 | 3+4 2 | 2+2 4 |
| | 3+4 2 | 5+6 2 | 2+2 4 | 2+2 4 | 2+2 4 | 3+4 2 | 5+6 2 | 2+2 4 |
| | 5+6 2 | 3+4 2 | 2+2 4 | 2+2 4 | 2+2 4 | 5+6 2 | 3+4 2 | 2+2 4 |
| | 3+4 2 | 3+4 2 | 2+2 4 | 2+2 4 | 2+2 4 | 3+4 2 | 5+6 2 | 2+2 4 |
| | 5+6 2 | 5+6 2 | 2+2 4 | 2+2 4 | 2+2 4 | 5+6 2 | 3+4 2 | 2+2 4 |
| | 3+4 2 | 3+4 2 | 2+2 4 | 2+2 4 | 2+2 4 | 3+4 2 | 3+4 2 | 2+2 4 |
| | 5+6 2 | 5+6 2 | 2+2 4 | 2+2 4 | 2+2 4 | 5+6 2 | 5+6 2 | 2+2 4 |
| | 3+4 2 | 3+4 2 | 2+2 4 | 2+2 4 | 2+2 4 | 3+4 2 | 3+4 2 | 2+2 4 |
| | 5+6 2 | 5+6 2 | 2+2 4 | 2+2 4 | 2+2 4 | 5+6 2 | 5+6 2 | 2+2 4 |
| | 3+4 2 | 3+4 2 | 2+2 4 | 2+2 4 | 2+2 4 | 3+4 2 | 3+4 2 | 2+2 4 |
| | 5+6 2 | 5+6 2 | 2+2 4 | 2+2 4 | 2+2 4 | 5+6 2 | 5+6 2 | 2+2 4 |
| | 5+6 2 | 3+4 2 | 2+2 4 | 2+2 4 | 2+2 4 | 3+4 2 | 3+4 2 | 2+2 4 |
| | 5+6 2 | 3+4 2 | 2+2 4 | 2+2 4 | 2+2 4 | 3+4 2 | 5+6 2 | 2+2 4 |
| | 5+6 2 | 5+6 2 | 2+2 4 | 2+2 4 | 2+2 4 | 5+6 2 | 3+4 2 | 2+2 4 |
| | 5+6 2 | 5+6 2 | 2+2 4 | 2+2 4 | 2+2 4 | 3+4 2 | 5+6 2 | 2+2 4 |
| | 3+4 2 | 5+6 2 | 2+2 4 | 2+2 4 | 2+2 4 | 3+4 2 | 5+6 2 | 2+2 4 |
| | 3+4 2 | 3+4 2 | 2+2 4 | 2+2 4 | 2+2 4 | 5+6 2 | 5+6 2 | 2+2 4 |
| | 5+6 2 | 5+6 2 | 2+2 4 | 2+2 4 | 2+2 4 | 5+6 2 | 5+6 2 | 2+2 4 |
| | 5+6 2 | 3+4 2 | 2+2 4 | 2+2 4 | 2+2 4 | 3+4 2 | 3+4 2 | 2+2 4 |
| | 3+4 2 | 5+6 2 | 2+2 4 | 2+2 4 | 2+2 4 | 3+4 2 | 5+6 2 | 2+2 4 |
| | 5+6 2 | 3+4 2 | | | | 3+4 2 | 3+4 2 | |
| | 5+6 2 | 3+4 2 | | | | 5+6 2 | 3+4 2 | |
| | 3+4 2 | 3+4 2 | | | | 5+6 2 | 5+6 2 | |
| | 3+4 2 | 3+4 2 | | | | 3+4 2 | 3+4 2 | |
| | 3+4 2 | 5+6 2 | | | | 3+4 2 | 3+4 2 | |
| | 3+4 2 | 5+6 2 | | | | 3+4 2 | 5+6 2 | |
| | 3+4 2 | 3+4 2 | | | | 5+6 2 | 3+4 2 | |
| | 5+6 2 | 3+4 2 | | | | 5+6 2 | 5+6 2 | |
| | 5+6 2 | 5+6 2 | | | | 3+4 2 | 3+4 2 | |
| | 3+4 2 | 3+4 2 | | | | 5+6 2 | 5+6 2 | |
| | 3+4 2 | 5+6 2 | | | | 3+4 2 | 3+4 2 | |
| | 5+6 2 | 3+4 2 | | | | 3+4 2 | 5+6 2 | |
| | 5+6 2 | 5+6 2 | | | | 5+6 2 | 3+4 2 | |
| | 5+6 2 | 5+6 2 | | | | 5+6 2 | 5+6 2 | |

Continued on next column

**TABLE XXIII– continued**

| TQC | ICSD | NoE | SF | | | | | |
|---|---|---|---|---|---|---|---|---|
| | 3+4 2 | 3+4 2 | | | | 3+4 2 | 3+4 2 | |
| | 5+6 2 | 5+6 2 | | | | 5+6 2 | 3+4 2 | |
| | 3+4 2 | 5+6 2 | | | | 5+6 2 | 5+6 2 | |
| | 5+6 2 | 5+6 2 | | | | 5+6 2 | 3+4 2 | |
| | 3+4 2 | 5+6 2 | | | | 5+6 2 | 3+4 2 | |
| | 5+6 2 | 5+6 2 | | | | 5+6 2 | 5+6 2 | |
| | 3+4 2 | 5+6 2 | | | | 3+4 2 | 3+4 2 | |
| | 3+4 2 | 3+4 2 | | | | 5+6 2 | 5+6 2 | |
| | 3+4 2 | 3+4 2 | | | | 3+4 2 | 3+4 2 | |
| | 5+6 2 | 3+4 2 | | | | 3+4 2 | 5+6 2 | |
| | 3+4 2 | 3+4 2 | | | | 5+6 2 | 5+6 2 | |
| | 3+4 2 | 3+4 2 | | | | 5+6 2 | 3+4 2 | |
| | 5+6 2 | 5+6 2 | | | | 3+4 2 | 5+6 2 | |
| | 5+6 2 | 3+4 2 | | | | 3+4 2 | 5+6 2 | |
| | 5+6 2 | 5+6 2 | | | | 3+4 2 | 3+4 2 | |
| | 5+6 2 | 5+6 2 | | | | 3+4 2 | 3+4 2 | |
| | 5+6 2 | 3+4 2 | | | | 3+4 2 | 5+6 2 | |
| | 5+6 2 | 3+4 2 | | | | 3+4 2 | 3+4 2 | |
| | 5+6 2 | 5+6 2 | | | | 3+4 2 | 5+6 2 | |
| | 5+6 2 | 3+4 2 | | | | 3+4 2 | 3+4 2 | |
| | 5+6 2 | 5+6 2 | | | | 3+4 2 | 3+4 2 | |
| | 5+6 2 | 3+4 2 | | | | 3+4 2 | 5+6 2 | |
| | 5+6 2 | 5+6 2 | | | | 3+4 2 | 5+6 2 | |
| 11.2.67 | 94536 | 188 | Eu1K3O8P2 | | | | | |
| | -A | -B | -C | -D | -E | -GM | -Y | -Z |
| | 5+6 2 | 5+6 2 | 2+2 4 | 2+2 4 | 2+2 4 | 3+4 2 | 3+4 2 | 2+2 4 |
| | 3+4 2 | 3+4 2 | 2+2 4 | 2+2 4 | 2+2 4 | 5+6 2 | 5+6 2 | 2+2 4 |
| | 3+4 2 | 3+4 2 | 2+2 4 | 2+2 4 | 2+2 4 | 3+4 2 | 3+4 2 | 2+2 4 |
| | 5+6 2 | 5+6 2 | 2+2 4 | 2+2 4 | 2+2 4 | 5+6 2 | 5+6 2 | 2+2 4 |
| | 5+6 2 | 3+4 2 | 2+2 4 | 2+2 4 | 2+2 4 | 3+4 2 | 5+6 2 | 2+2 4 |
| | 3+4 2 | 5+6 2 | 2+2 4 | 2+2 4 | 2+2 4 | 5+6 2 | 3+4 2 | 2+2 4 |
| | 3+4 2 | 5+6 2 | 2+2 4 | 2+2 4 | 2+2 4 | 3+4 2 | 5+6 2 | 2+2 4 |
| | 5+6 2 | 3+4 2 | 2+2 4 | 2+2 4 | 2+2 4 | 5+6 2 | 3+4 2 | 2+2 4 |
| | 5+6 2 | 3+4 2 | 2+2 4 | 2+2 4 | 2+2 4 | 3+4 2 | 5+6 2 | 2+2 4 |
| | 3+4 2 | 5+6 2 | 2+2 4 | 2+2 4 | 2+2 4 | 5+6 2 | 3+4 2 | 2+2 4 |
| | 3+4 2 | 5+6 2 | 2+2 4 | 2+2 4 | 2+2 4 | 3+4 2 | 5+6 2 | 2+2 4 |
| | 5+6 2 | 3+4 2 | 2+2 4 | 2+2 4 | 2+2 4 | 5+6 2 | 3+4 2 | 2+2 4 |
| | 3+4 2 | 5+6 2 | 2+2 4 | 2+2 4 | 2+2 4 | 3+4 2 | 3+4 2 | 2+2 4 |
| | 5+6 2 | 3+4 2 | 2+2 4 | 2+2 4 | 2+2 4 | 5+6 2 | 5+6 2 | 2+2 4 |
| | 3+4 2 | 3+4 2 | 2+2 4 | 2+2 4 | 2+2 4 | 3+4 2 | 3+4 2 | 2+2 4 |
| | 3+4 2 | 5+6 2 | 2+2 4 | 2+2 4 | 2+2 4 | 5+6 2 | 5+6 2 | 2+2 4 |
| | 5+6 2 | 3+4 2 | 2+2 4 | 2+2 4 | 2+2 4 | 3+4 2 | 5+6 2 | 2+2 4 |
| | 5+6 2 | 5+6 2 | 2+2 4 | 2+2 4 | 2+2 4 | 5+6 2 | 3+4 2 | 2+2 4 |
| | 3+4 2 | 5+6 2 | 2+2 4 | 2+2 4 | 2+2 4 | 3+4 2 | 5+6 2 | 2+2 4 |
| | 5+6 2 | 3+4 2 | 2+2 4 | 2+2 4 | 2+2 4 | 5+6 2 | 3+4 2 | 2+2 4 |
| | 3+4 2 | 5+6 2 | 2+2 4 | 2+2 4 | 2+2 4 | 3+4 2 | 3+4 2 | 2+2 4 |
| | 5+6 2 | 3+4 2 | 2+2 4 | 2+2 4 | 2+2 4 | 5+6 2 | 5+6 2 | 2+2 4 |
| | 3+4 2 | 5+6 2 | 2+2 4 | 2+2 4 | 2+2 4 | 3+4 2 | 5+6 2 | 2+2 4 |
| | 5+6 2 | 3+4 2 | 2+2 4 | 2+2 4 | 2+2 4 | 5+6 2 | 3+4 2 | 2+2 4 |
| | 3+4 2 | 3+4 2 | 2+2 4 | 2+2 4 | 2+2 4 | 3+4 2 | 3+4 2 | 2+2 4 |
| | 5+6 2 | 5+6 2 | 2+2 4 | 2+2 4 | 2+2 4 | 5+6 2 | 5+6 2 | 2+2 4 |
| | 5+6 2 | 5+6 2 | 2+2 4 | 2+2 4 | 2+2 4 | 3+4 2 | 3+4 2 | 2+2 4 |
| | 3+4 2 | 3+4 2 | 2+2 4 | 2+2 4 | 2+2 4 | 5+6 2 | 5+6 2 | 2+2 4 |
| | 3+4 2 | 3+4 2 | 2+2 4 | 2+2 4 | 2+2 4 | 3+4 2 | 3+4 2 | 2+2 4 |
| | 5+6 2 | 5+6 2 | 2+2 4 | 2+2 4 | 2+2 4 | 5+6 2 | 5+6 2 | 2+2 4 |
| | 5+6 2 | 3+4 2 | 2+2 4 | 2+2 4 | 2+2 4 | 3+4 2 | 3+4 2 | 2+2 4 |
| | 3+4 2 | 5+6 2 | 2+2 4 | 2+2 4 | 2+2 4 | 5+6 2 | 5+6 2 | 2+2 4 |
| | 5+6 2 | 5+6 2 | 2+2 4 | 2+2 4 | 2+2 4 | 3+4 2 | 3+4 2 | 2+2 4 |
| | 5+6 2 | 5+6 2 | 2+2 4 | 2+2 4 | 2+2 4 | 5+6 2 | 5+6 2 | 2+2 4 |
| | 3+4 2 | 3+4 2 | 2+2 4 | 2+2 4 | 2+2 4 | 3+4 2 | 5+6 2 | 2+2 4 |
| | 5+6 2 | 5+6 2 | 2+2 4 | 2+2 4 | 2+2 4 | 5+6 2 | 3+4 2 | 2+2 4 |

Continued on next column

**TABLE XXIII– continued**

| TQC | ICSD | NoE | SF | | | | | |
|---|---|---|---|---|---|---|---|---|
| | 3+4 2 | 3+4 2 | 2+2 4 | 2+2 4 | 2+2 4 | 3+4 2 | 3+4 2 | 2+2 4 |
| | 3+4 2 | 3+4 2 | 2+2 4 | 2+2 4 | 2+2 4 | 5+6 2 | 5+6 2 | 2+2 4 |
| | 5+6 2 | 5+6 2 | 2+2 4 | 2+2 4 | 2+2 4 | 3+4 2 | 5+6 2 | 2+2 4 |
| | 3+4 2 | 3+4 2 | 2+2 4 | 2+2 4 | 2+2 4 | 5+6 2 | 3+4 2 | 2+2 4 |
| | 5+6 2 | 5+6 2 | 2+2 4 | 2+2 4 | 2+2 4 | 5+6 2 | 5+6 2 | 2+2 4 |
| | 3+4 2 | 3+4 2 | 2+2 4 | 2+2 4 | 2+2 4 | 3+4 2 | 3+4 2 | 2+2 4 |
| | 5+6 2 | 3+4 2 | 2+2 4 | 2+2 4 | 2+2 4 | 3+4 2 | 3+4 2 | 2+2 4 |
| | 3+4 2 | 5+6 2 | 2+2 4 | 2+2 4 | 2+2 4 | 5+6 2 | 5+6 2 | 2+2 4 |
| | 5+6 2 | 3+4 2 | 2+2 4 | 2+2 4 | 2+2 4 | 5+6 2 | 3+4 2 | 2+2 4 |
| | 3+4 2 | 5+6 2 | 2+2 4 | 2+2 4 | 2+2 4 | 3+4 2 | 5+6 2 | 2+2 4 |
| | 5+6 2 | 3+4 2 | 2+2 4 | 2+2 4 | 2+2 4 | 5+6 2 | 3+4 2 | 2+2 4 |
| | 3+4 2 | 5+6 2 | | | | 3+4 2 | 5+6 2 | |
| | 3+4 2 | 5+6 2 | | | | 3+4 2 | 5+6 2 | |
| | 5+6 2 | 3+4 2 | | | | 5+6 2 | 3+4 2 | |
| | 5+6 2 | 3+4 2 | | | | 5+6 2 | 3+4 2 | |
| | 3+4 2 | 5+6 2 | | | | 3+4 2 | 5+6 2 | |
| | 3+4 2 | 3+4 2 | | | | 5+6 2 | 3+4 2 | |
| | 3+4 2 | 5+6 2 | | | | 3+4 2 | 5+6 2 | |
| | 5+6 2 | 5+6 2 | | | | 3+4 2 | 5+6 2 | |
| | 3+4 2 | 3+4 2 | | | | 5+6 2 | 3+4 2 | |
| | 5+6 2 | 5+6 2 | | | | 3+4 2 | 5+6 2 | |
| | 3+4 2 | 5+6 2 | | | | 5+6 2 | 3+4 2 | |
| | 5+6 2 | 3+4 2 | | | | 3+4 2 | 5+6 2 | |
| | 3+4 2 | 3+4 2 | | | | 5+6 2 | 3+4 2 | |
| | 5+6 2 | 5+6 2 | | | | 3+4 2 | 3+4 2 | |
| | 5+6 2 | 3+4 2 | | | | 5+6 2 | 5+6 2 | |
| | 3+4 2 | 3+4 2 | | | | 3+4 2 | 3+4 2 | |
| | 5+6 2 | 5+6 2 | | | | 5+6 2 | 5+6 2 | |
| | 5+6 2 | 3+4 2 | | | | 3+4 2 | 5+6 2 | |
| | 5+6 2 | 3+4 2 | | | | 3+4 2 | 3+4 2 | |
| | 3+4 2 | 5+6 2 | | | | 5+6 2 | 5+6 2 | |
| | 3+4 2 | 3+4 2 | | | | 5+6 2 | 5+6 2 | |
| | 5+6 2 | 5+6 2 | | | | 3+4 2 | 3+4 2 | |
| | 3+4 2 | 5+6 2 | | | | 3+4 2 | 3+4 2 | |
| | 3+4 2 | 5+6 2 | | | | 5+6 2 | 5+6 2 | |
| | 5+6 2 | 3+4 2 | | | | 3+4 2 | 3+4 2 | |
| | 3+4 2 | 3+4 2 | | | | 5+6 2 | 5+6 2 | |
| | 5+6 2 | 5+6 2 | | | | 5+6 2 | 3+4 2 | |
| | 5+6 2 | 3+4 2 | | | | 3+4 2 | 5+6 2 | |
| | 3+4 2 | 5+6 2 | | | | 3+4 2 | 3+4 2 | |
| | 3+4 2 | 3+4 2 | | | | 3+4 2 | 5+6 2 | |
| | 5+6 2 | 5+6 2 | | | | 3+4 2 | 5+6 2 | |
| | 3+4 2 | 5+6 2 | | | | 5+6 2 | 3+4 2 | |
| | 3+4 2 | 3+4 2 | | | | 5+6 2 | 3+4 2 | |
| | 5+6 2 | 5+6 2 | | | | 3+4 2 | 5+6 2 | |
| | 5+6 2 | 3+4 2 | | | | 5+6 2 | 5+6 2 | |
| | 3+4 2 | 5+6 2 | | | | 5+6 2 | 3+4 2 | |
| | 5+6 2 | 3+4 2 | | | | 5+6 2 | 3+4 2 | |
| | 3+4 2 | 5+6 2 | | | | 5+6 2 | 3+4 2 | |
| | 3+4 2 | 5+6 2 | | | | 5+6 2 | 3+4 2 | |
| | 3+4 2 | 3+4 2 | | | | 3+4 2 | 5+6 2 | |
| | 5+6 2 | 3+4 2 | | | | 3+4 2 | 5+6 2 | |
| | 5+6 2 | 5+6 2 | | | | 3+4 2 | 3+4 2 | |
| | 3+4 2 | 3+4 2 | | | | 5+6 2 | 3+4 2 | |
| | 5+6 2 | 3+4 2 | | | | 5+6 2 | 5+6 2 | |
| | 5+6 2 | 5+6 2 | | | | 3+4 2 | 3+4 2 | |
| | 3+4 2 | 5+6 2 | | | | 3+4 2 | 5+6 2 | |
| | 3+4 2 | 3+4 2 | | | | 5+6 2 | 5+6 2 | |
| 11.2.68 | 190487 | 76 | D2O4P1Rb1 | | | | | |
| | -A | -B | -C | -D | -E | -GM | -Y | -Z |
| | 3+4 2 | 3+4 2 | 2+2 4 | 2+2 4 | 2+2 4 | 3+4 2 | 3+4 2 | 2+2 4 |

Continued on next column

**TABLE XXIII– continued**

| TQC | ICSD | NoE | SF | | | | | |
|---|---|---|---|---|---|---|---|---|
| | 5+6 2 | 5+6 2 | 2+2 4 | 2+2 4 | 2+2 4 | 5+6 2 | 5+6 2 | 2+2 4 |
| | 5+6 2 | 3+4 2 | 2+2 4 | 2+2 4 | 2+2 4 | 3+4 2 | 5+6 2 | 2+2 4 |
| | 3+4 2 | 5+6 2 | 2+2 4 | 2+2 4 | 2+2 4 | 5+6 2 | 3+4 2 | 2+2 4 |
| | 5+6 2 | 3+4 2 | 2+2 4 | 2+2 4 | 2+2 4 | 3+4 2 | 5+6 2 | 2+2 4 |
| | 5+6 2 | 3+4 2 | 2+2 4 | 2+2 4 | 2+2 4 | 3+4 2 | 5+6 2 | 2+2 4 |
| | 3+4 2 | 5+6 2 | 2+2 4 | 2+2 4 | 2+2 4 | 5+6 2 | 3+4 2 | 2+2 4 |
| | 5+6 2 | 5+6 2 | 2+2 4 | 2+2 4 | 2+2 4 | 5+6 2 | 5+6 2 | 2+2 4 |
| | 3+4 2 | 5+6 2 | 2+2 4 | 2+2 4 | 2+2 4 | 3+4 2 | 3+4 2 | 2+2 4 |
| | 3+4 2 | 3+4 2 | 2+2 4 | 2+2 4 | 2+2 4 | 5+6 2 | 3+4 2 | 2+2 4 |
| | 3+4 2 | 3+4 2 | 2+2 4 | 2+2 4 | 2+2 4 | 3+4 2 | 3+4 2 | 2+2 4 |
| | 5+6 2 | 5+6 2 | 2+2 4 | 2+2 4 | 2+2 4 | 5+6 2 | 5+6 2 | 2+2 4 |
| | 3+4 2 | 3+4 2 | 2+2 4 | 2+2 4 | 2+2 4 | 3+4 2 | 3+4 2 | 2+2 4 |
| | 5+6 2 | 5+6 2 | 2+2 4 | 2+2 4 | 2+2 4 | 5+6 2 | 5+6 2 | 2+2 4 |
| | 3+4 2 | 5+6 2 | 2+2 4 | 2+2 4 | 2+2 4 | 3+4 2 | 3+4 2 | 2+2 4 |
| | 5+6 2 | 3+4 2 | 2+2 4 | 2+2 4 | 2+2 4 | 5+6 2 | 5+6 2 | 2+2 4 |
| | 5+6 2 | 3+4 2 | 2+2 4 | 2+2 4 | 2+2 4 | 3+4 2 | 5+6 2 | 2+2 4 |
| | 3+4 2 | 5+6 2 | 2+2 4 | 2+2 4 | 2+2 4 | 5+6 2 | 3+4 2 | 2+2 4 |
| | 5+6 2 | 3+4 2 | 2+2 4 | 2+2 4 | 2+2 4 | 3+4 2 | 5+6 2 | 2+2 4 |
| | 3+4 2 | 5+6 2 | | | | 3+4 2 | 3+4 2 | |
| | 5+6 2 | 3+4 2 | | | | 5+6 2 | 5+6 2 | |
| | 3+4 2 | 5+6 2 | | | | 3+4 2 | 3+4 2 | |
| | 3+4 2 | 3+4 2 | | | | 5+6 2 | 3+4 2 | |
| | 5+6 2 | 5+6 2 | | | | 5+6 2 | 5+6 2 | |
| | 5+6 2 | 5+6 2 | | | | 5+6 2 | 5+6 2 | |
| | 3+4 2 | 3+4 2 | | | | 3+4 2 | 3+4 2 | |
| | 5+6 2 | 3+4 2 | | | | 3+4 2 | 5+6 2 | |
| | 5+6 2 | 5+6 2 | | | | 3+4 2 | 3+4 2 | |
| | 3+4 2 | 3+4 2 | | | | 3+4 2 | 3+4 2 | |
| | 3+4 2 | 5+6 2 | | | | 5+6 2 | 5+6 2 | |
| | 5+6 2 | 3+4 2 | | | | 5+6 2 | 5+6 2 | |
| | 5+6 2 | 5+6 2 | | | | 5+6 2 | 5+6 2 | |
| | 3+4 2 | 3+4 2 | | | | 3+4 2 | 3+4 2 | |
| | 3+4 2 | 5+6 2 | | | | 5+6 2 | 3+4 2 | |
| | 5+6 2 | 3+4 2 | | | | 3+4 2 | 5+6 2 | |
| | 5+6 2 | 5+6 2 | | | | 5+6 2 | 5+6 2 | |
| | 3+4 2 | 3+4 2 | | | | 3+4 2 | 3+4 2 | |
| | 3+4 2 | 3+4 2 | | | | 3+4 2 | 3+4 2 | |
| 11.2.69 | 67816 | 296 | Cu4K1Mg1O13V3 | | | | | |
| | -A | -B | -C | -D | -E | -GM | -Y | -Z |
| | 3+4 2 | 3+4 2 | 2+2 4 | 2+2 4 | 2+2 4 | 3+4 2 | 3+4 2 | 2+2 4 |
| | 5+6 2 | 5+6 2 | 2+2 4 | 2+2 4 | 2+2 4 | 5+6 2 | 5+6 2 | 2+2 4 |
| | 3+4 2 | 5+6 2 | 2+2 4 | 2+2 4 | 2+2 4 | 3+4 2 | 5+6 2 | 2+2 4 |
| | 5+6 2 | 3+4 2 | 2+2 4 | 2+2 4 | 2+2 4 | 5+6 2 | 3+4 2 | 2+2 4 |
| | 5+6 2 | 3+4 2 | 2+2 4 | 2+2 4 | 2+2 4 | 5+6 2 | 3+4 2 | 2+2 4 |
| | 3+4 2 | 5+6 2 | 2+2 4 | 2+2 4 | 2+2 4 | 3+4 2 | 5+6 2 | 2+2 4 |
| | 3+4 2 | 5+6 2 | 2+2 4 | 2+2 4 | 2+2 4 | 3+4 2 | 5+6 2 | 2+2 4 |
| | 5+6 2 | 3+4 2 | 2+2 4 | 2+2 4 | 2+2 4 | 5+6 2 | 3+4 2 | 2+2 4 |
| | 3+4 2 | 5+6 2 | 2+2 4 | 2+2 4 | 2+2 4 | 3+4 2 | 5+6 2 | 2+2 4 |
| | 5+6 2 | 3+4 2 | 2+2 4 | 2+2 4 | 2+2 4 | 5+6 2 | 3+4 2 | 2+2 4 |
| | 3+4 2 | 3+4 2 | 2+2 4 | 2+2 4 | 2+2 4 | 5+6 2 | 5+6 2 | 2+2 4 |
| | 5+6 2 | 3+4 2 | 2+2 4 | 2+2 4 | 2+2 4 | 3+4 2 | 3+4 2 | 2+2 4 |
| | 3+4 2 | 5+6 2 | 2+2 4 | 2+2 4 | 2+2 4 | 5+6 2 | 5+6 2 | 2+2 4 |
| | 5+6 2 | 5+6 2 | 2+2 4 | 2+2 4 | 2+2 4 | 3+4 2 | 3+4 2 | 2+2 4 |
| | 5+6 2 | 3+4 2 | 2+2 4 | 2+2 4 | 2+2 4 | 5+6 2 | 3+4 2 | 2+2 4 |
| | 3+4 2 | 5+6 2 | 2+2 4 | 2+2 4 | 2+2 4 | 3+4 2 | 3+4 2 | 2+2 4 |
| | 5+6 2 | 3+4 2 | 2+2 4 | 2+2 4 | 2+2 4 | 5+6 2 | 5+6 2 | 2+2 4 |
| | 3+4 2 | 5+6 2 | 2+2 4 | 2+2 4 | 2+2 4 | 3+4 2 | 5+6 2 | 2+2 4 |
| | 3+4 2 | 5+6 2 | 2+2 4 | 2+2 4 | 2+2 4 | 3+4 2 | 5+6 2 | 2+2 4 |
| | 5+6 2 | 3+4 2 | 2+2 4 | 2+2 4 | 2+2 4 | 5+6 2 | 3+4 2 | 2+2 4 |
| | 5+6 2 | 3+4 2 | 2+2 4 | 2+2 4 | 2+2 4 | 3+4 2 | 5+6 2 | 2+2 4 |
| | 3+4 2 | 5+6 2 | 2+2 4 | 2+2 4 | 2+2 4 | 5+6 2 | 3+4 2 | 2+2 4 |

Continued on next column

**TABLE XXIII– continued**

| TQC | ICSD | NoE | SF |
|---|---|---|---|
| | 5+6 2 3+4 2 | 2+2 4 2+2 4 2+2 4 | 3+4 2 5+6 2 2+2 4 |
| | 3+4 2 5+6 2 | 2+2 4 2+2 4 2+2 4 | 5+6 2 3+4 2 2+2 4 |
| | 3+4 2 5+6 2 | 2+2 4 2+2 4 2+2 4 | 5+6 2 3+4 2 2+2 4 |
| | 3+4 2 3+4 2 | 2+2 4 2+2 4 2+2 4 | 3+4 2 3+4 2 2+2 4 |
| | 5+6 2 5+6 2 | 2+2 4 2+2 4 2+2 4 | 5+6 2 5+6 2 2+2 4 |
| | 5+6 2 3+4 2 | 2+2 4 2+2 4 2+2 4 | 3+4 2 5+6 2 2+2 4 |
| | 3+4 2 3+4 2 | 2+2 4 2+2 4 2+2 4 | 3+4 2 3+4 2 2+2 4 |
| | 5+6 2 5+6 2 | 2+2 4 2+2 4 2+2 4 | 5+6 2 5+6 2 2+2 4 |
| | 3+4 2 3+4 2 | 2+2 4 2+2 4 2+2 4 | 3+4 2 3+4 2 2+2 4 |
| | 3+4 2 3+4 2 | 2+2 4 2+2 4 2+2 4 | 3+4 2 3+4 2 2+2 4 |
| | 5+6 2 5+6 2 | 2+2 4 2+2 4 2+2 4 | 5+6 2 5+6 2 2+2 4 |
| | 5+6 2 5+6 2 | 2+2 4 2+2 4 2+2 4 | 5+6 2 5+6 2 2+2 4 |
| | 3+4 2 5+6 2 | 2+2 4 2+2 4 2+2 4 | 3+4 2 5+6 2 2+2 4 |
| | 5+6 2 3+4 2 | 2+2 4 2+2 4 2+2 4 | 5+6 2 3+4 2 2+2 4 |
| | 5+6 2 5+6 2 | 2+2 4 2+2 4 2+2 4 | 3+4 2 3+4 2 2+2 4 |
| | 3+4 2 3+4 2 | 2+2 4 2+2 4 2+2 4 | 5+6 2 5+6 2 2+2 4 |
| | 3+4 2 5+6 2 | 2+2 4 2+2 4 2+2 4 | 3+4 2 5+6 2 2+2 4 |
| | 5+6 2 3+4 2 | 2+2 4 2+2 4 2+2 4 | 5+6 2 3+4 2 2+2 4 |
| | 3+4 2 5+6 2 | 2+2 4 2+2 4 2+2 4 | 3+4 2 5+6 2 2+2 4 |
| | 5+6 2 3+4 2 | 2+2 4 2+2 4 2+2 4 | 5+6 2 3+4 2 2+2 4 |
| | 3+4 2 3+4 2 | 2+2 4 2+2 4 2+2 4 | 5+6 2 3+4 2 2+2 4 |
| | 5+6 2 5+6 2 | 2+2 4 2+2 4 2+2 4 | 3+4 2 5+6 2 2+2 4 |
| | 3+4 2 5+6 2 | 2+2 4 2+2 4 2+2 4 | 3+4 2 5+6 2 2+2 4 |
| | 3+4 2 5+6 2 | 2+2 4 2+2 4 2+2 4 | 3+4 2 5+6 2 2+2 4 |
| | 5+6 2 3+4 2 | 2+2 4 2+2 4 2+2 4 | 5+6 2 3+4 2 2+2 4 |
| | 3+4 2 5+6 2 | 2+2 4 2+2 4 2+2 4 | 3+4 2 5+6 2 2+2 4 |
| | 5+6 2 5+6 2 | 2+2 4 2+2 4 2+2 4 | 3+4 2 3+4 2 2+2 4 |
| | 5+6 2 3+4 2 | 2+2 4 2+2 4 2+2 4 | 5+6 2 3+4 2 2+2 4 |
| | 5+6 2 3+4 2 | 2+2 4 2+2 4 2+2 4 | 3+4 2 3+4 2 2+2 4 |
| | 3+4 2 5+6 2 | 2+2 4 2+2 4 2+2 4 | 5+6 2 5+6 2 2+2 4 |
| | 3+4 2 3+4 2 | 2+2 4 2+2 4 2+2 4 | 5+6 2 5+6 2 2+2 4 |
| | 3+4 2 3+4 2 | 2+2 4 2+2 4 2+2 4 | 5+6 2 3+4 2 2+2 4 |
| | 5+6 2 5+6 2 | 2+2 4 2+2 4 2+2 4 | 5+6 2 5+6 2 2+2 4 |
| | 3+4 2 3+4 2 | 2+2 4 2+2 4 2+2 4 | 3+4 2 5+6 2 2+2 4 |
| | 3+4 2 5+6 2 | 2+2 4 2+2 4 2+2 4 | 3+4 2 5+6 2 2+2 4 |
| | 5+6 2 3+4 2 | 2+2 4 2+2 4 2+2 4 | 5+6 2 3+4 2 2+2 4 |
| | 5+6 2 5+6 2 | 2+2 4 2+2 4 2+2 4 | 3+4 2 3+4 2 2+2 4 |
| | 3+4 2 5+6 2 | 2+2 4 2+2 4 2+2 4 | 3+4 2 3+4 2 2+2 4 |
| | 5+6 2 3+4 2 | 2+2 4 2+2 4 2+2 4 | 5+6 2 5+6 2 2+2 4 |
| | 5+6 2 5+6 2 | 2+2 4 2+2 4 2+2 4 | 3+4 2 3+4 2 2+2 4 |
| | 3+4 2 3+4 2 | 2+2 4 2+2 4 2+2 4 | 5+6 2 5+6 2 2+2 4 |
| | 5+6 2 3+4 2 | 2+2 4 2+2 4 2+2 4 | 5+6 2 3+4 2 2+2 4 |
| | 5+6 2 3+4 2 | 2+2 4 2+2 4 2+2 4 | 3+4 2 3+4 2 2+2 4 |
| | 3+4 2 3+4 2 | 2+2 4 2+2 4 2+2 4 | 5+6 2 5+6 2 2+2 4 |
| | 3+4 2 5+6 2 | 2+2 4 2+2 4 2+2 4 | 5+6 2 3+4 2 2+2 4 |
| | 3+4 2 5+6 2 | 2+2 4 2+2 4 2+2 4 | 3+4 2 5+6 2 2+2 4 |
| | 5+6 2 3+4 2 | 2+2 4 2+2 4 2+2 4 | 3+4 2 3+4 2 2+2 4 |
| | 5+6 2 5+6 2 | 2+2 4 2+2 4 2+2 4 | 5+6 2 3+4 2 2+2 4 |
| | 5+6 2 3+4 2 | 2+2 4 2+2 4 2+2 4 | 5+6 2 5+6 2 2+2 4 |
| | 3+4 2 5+6 2 | 2+2 4 2+2 4 2+2 4 | 5+6 2 5+6 2 2+2 4 |
| | 5+6 2 3+4 2 | 2+2 4 2+2 4 2+2 4 | 3+4 2 5+6 2 2+2 4 |
| | 3+4 2 5+6 2 | 2+2 4 2+2 4 2+2 4 | 3+4 2 3+4 2 2+2 4 |
| | 5+6 2 5+6 2 | | 5+6 2 3+4 2 |
| | 3+4 2 3+4 2 | | 3+4 2 5+6 2 |
| | 3+4 2 3+4 2 | | 5+6 2 5+6 2 |
| | 5+6 2 5+6 2 | | 3+4 2 3+4 2 |
| | 3+4 2 5+6 2 | | 5+6 2 5+6 2 |
| | 5+6 2 3+4 2 | | 3+4 2 3+4 2 |
| | 3+4 2 5+6 2 | | 3+4 2 3+4 2 |
| | 5+6 2 3+4 2 | | 5+6 2 5+6 2 |
| | 3+4 2 3+4 2 | | 3+4 2 5+6 2 |

Continued on next column

| TQC | ICSD | NoE | SF |
|---|---|---|---|
| | 5+6 2 5+6 2 | | 3+4 2 3+4 2 |
| | 5+6 2 5+6 2 | | 5+6 2 5+6 2 |
| | 5+6 2 3+4 2 | | 5+6 2 3+4 2 |
| | 3+4 2 3+4 2 | | 3+4 2 5+6 2 |
| | 3+4 2 5+6 2 | | 5+6 2 3+4 2 |
| | 5+6 2 3+4 2 | | 5+6 2 5+6 2 |
| | 5+6 2 5+6 2 | | 3+4 2 5+6 2 |
| | 3+4 2 3+4 2 | | 3+4 2 3+4 2 |
| | 5+6 2 3+4 2 | | 5+6 2 3+4 2 |
| | 3+4 2 5+6 2 | | 3+4 2 3+4 2 |
| | 5+6 2 5+6 2 | | 3+4 2 3+4 2 |
| | 3+4 2 3+4 2 | | 5+6 2 5+6 2 |
| | 5+6 2 5+6 2 | | 5+6 2 5+6 2 |
| | 3+4 2 3+4 2 | | 5+6 2 3+4 2 |
| | 5+6 2 3+4 2 | | 3+4 2 5+6 2 |
| | 3+4 2 5+6 2 | | 3+4 2 5+6 2 |
| | 3+4 2 5+6 2 | | 3+4 2 3+4 2 |
| | 5+6 2 3+4 2 | | 5+6 2 3+4 2 |
| | 5+6 2 5+6 2 | | 3+4 2 5+6 2 |
| | 3+4 2 5+6 2 | | 5+6 2 3+4 2 |
| | 5+6 2 3+4 2 | | 3+4 2 3+4 2 |
| | 3+4 2 3+4 2 | | 5+6 2 5+6 2 |
| | 3+4 2 5+6 2 | | 5+6 2 5+6 2 |
| | 5+6 2 3+4 2 | | 5+6 2 5+6 2 |
| | 3+4 2 5+6 2 | | 3+4 2 3+4 2 |
| | 5+6 2 5+6 2 | | 3+4 2 5+6 2 |
| | 3+4 2 3+4 2 | | 5+6 2 3+4 2 |
| | 5+6 2 3+4 2 | | 3+4 2 5+6 2 |
| | 5+6 2 3+4 2 | | 3+4 2 5+6 2 |
| | 3+4 2 5+6 2 | | 5+6 2 3+4 2 |
| | 5+6 2 3+4 2 | | 5+6 2 3+4 2 |
| | 3+4 2 5+6 2 | | 5+6 2 5+6 2 |
| | 3+4 2 5+6 2 | | 3+4 2 3+4 2 |
| | 5+6 2 3+4 2 | | 3+4 2 5+6 2 |
| | 5+6 2 5+6 2 | | 5+6 2 3+4 2 |
| | 3+4 2 5+6 2 | | 3+4 2 3+4 2 |
| | 5+6 2 3+4 2 | | 3+4 2 3+4 2 |
| | 3+4 2 5+6 2 | | 5+6 2 5+6 2 |
| | 5+6 2 3+4 2 | | 5+6 2 5+6 2 |
| | 3+4 2 3+4 2 | | 3+4 2 3+4 2 |
| | 5+6 2 5+6 2 | | 5+6 2 5+6 2 |
| | 3+4 2 5+6 2 | | 5+6 2 3+4 2 |
| | 3+4 2 3+4 2 | | 3+4 2 5+6 2 |
| | 5+6 2 5+6 2 | | 3+4 2 3+4 2 |
| | 3+4 2 3+4 2 | | 5+6 2 5+6 2 |
| | 5+6 2 5+6 2 | | 5+6 2 3+4 2 |
| | 3+4 2 3+4 2 | | 3+4 2 5+6 2 |
| | 3+4 2 5+6 2 | | 3+4 2 3+4 2 |
| | 5+6 2 3+4 2 | | 3+4 2 5+6 2 |
| | 3+4 2 5+6 2 | | 5+6 2 3+4 2 |
| | 5+6 2 3+4 2 | | 5+6 2 5+6 2 |
| | 5+6 2 3+4 2 | | 5+6 2 3+4 2 |
| | 3+4 2 3+4 2 | | 3+4 2 5+6 2 |
| | 5+6 2 5+6 2 | | 5+6 2 5+6 2 |
| | 3+4 2 3+4 2 | | 5+6 2 3+4 2 |
| | 3+4 2 3+4 2 | | 5+6 2 5+6 2 |
| | 3+4 2 5+6 2 | | 3+4 2 5+6 2 |
| | 5+6 2 3+4 2 | | 5+6 2 3+4 2 |
| | 5+6 2 3+4 2 | | 3+4 2 3+4 2 |
| | 3+4 2 5+6 2 | | 5+6 2 5+6 2 |
| | 5+6 2 5+6 2 | | 3+4 2 3+4 2 |

Continued on next column

**TABLE XXIII– continued**

| TQC | ICSD | NoE | SF | | | | | |
|---|---|---|---|---|---|---|---|---|
| | 3+4 2 | 5+6 2 | | | | 3+4 2 | 5+6 2 | |
| | 5+6 2 | 5+6 2 | | | | 3+4 2 | 3+4 2 | |
| | 5+6 2 | 5+6 2 | | | | 3+4 2 | 3+4 2 | |
| | 5+6 2 | 3+4 2 | | | | 5+6 2 | 3+4 2 | |
| 11.2.70 | 414338 | 72 | Cu1Nb1Te2 | | | | | |
| | -A | -B | -C | -D | -E | -GM | -Y | -Z |
| | 3+4 2 | 3+4 2 | 2+2 4 | 2+2 4 | 2+2 4 | 3+4 2 | 3+4 2 | 2+2 4 |
| | 5+6 2 | 5+6 2 | 2+2 4 | 2+2 4 | 2+2 4 | 5+6 2 | 5+6 2 | 2+2 4 |
| | 3+4 2 | 3+4 2 | 2+2 4 | 2+2 4 | 2+2 4 | 3+4 2 | 3+4 2 | 2+2 4 |
| | 5+6 2 | 5+6 2 | 2+2 4 | 2+2 4 | 2+2 4 | 5+6 2 | 5+6 2 | 2+2 4 |
| | 3+4 2 | 3+4 2 | 2+2 4 | 2+2 4 | 2+2 4 | 3+4 2 | 3+4 2 | 2+2 4 |
| | 5+6 2 | 5+6 2 | 2+2 4 | 2+2 4 | 2+2 4 | 5+6 2 | 5+6 2 | 2+2 4 |
| | 3+4 2 | 3+4 2 | 2+2 4 | 2+2 4 | 2+2 4 | 3+4 2 | 3+4 2 | 2+2 4 |
| | 5+6 2 | 5+6 2 | 2+2 4 | 2+2 4 | 2+2 4 | 5+6 2 | 5+6 2 | 2+2 4 |
| | 5+6 2 | 3+4 2 | 2+2 4 | 2+2 4 | 2+2 4 | 3+4 2 | 5+6 2 | 2+2 4 |
| | 3+4 2 | 5+6 2 | 2+2 4 | 2+2 4 | 2+2 4 | 5+6 2 | 3+4 2 | 2+2 4 |
| | 5+6 2 | 5+6 2 | 2+2 4 | 2+2 4 | 2+2 4 | 3+4 2 | 3+4 2 | 2+2 4 |
| | 3+4 2 | 3+4 2 | 2+2 4 | 2+2 4 | 2+2 4 | 5+6 2 | 5+6 2 | 2+2 4 |
| | 3+4 2 | 5+6 2 | 2+2 4 | 2+2 4 | 2+2 4 | 3+4 2 | 5+6 2 | 2+2 4 |
| | 5+6 2 | 3+4 2 | 2+2 4 | 2+2 4 | 2+2 4 | 5+6 2 | 3+4 2 | 2+2 4 |
| | 5+6 2 | 3+4 2 | 2+2 4 | 2+2 4 | 2+2 4 | 3+4 2 | 5+6 2 | 2+2 4 |
| | 3+4 2 | 5+6 2 | 2+2 4 | 2+2 4 | 2+2 4 | 5+6 2 | 3+4 2 | 2+2 4 |
| | 5+6 2 | 5+6 2 | 2+2 4 | 2+2 4 | 2+2 4 | 5+6 2 | 5+6 2 | 2+2 4 |
| | 5+6 2 | 3+4 2 | 2+2 4 | 2+2 4 | 2+2 4 | 3+4 2 | 3+4 2 | 2+2 4 |
| | 5+6 2 | 3+4 2 | | | | 5+6 2 | 3+4 2 | |
| | 3+4 2 | 5+6 2 | | | | 3+4 2 | 5+6 2 | |
| | 3+4 2 | 5+6 2 | | | | 3+4 2 | 3+4 2 | |
| | 3+4 2 | 3+4 2 | | | | 3+4 2 | 5+6 2 | |
| | 5+6 2 | 5+6 2 | | | | 5+6 2 | 5+6 2 | |
| | 3+4 2 | 3+4 2 | | | | 5+6 2 | 5+6 2 | |
| | 5+6 2 | 5+6 2 | | | | 5+6 2 | 3+4 2 | |
| | 3+4 2 | 3+4 2 | | | | 3+4 2 | 3+4 2 | |
| | 5+6 2 | 5+6 2 | | | | 3+4 2 | 5+6 2 | |
| | 3+4 2 | 5+6 2 | | | | 3+4 2 | 3+4 2 | |
| | 5+6 2 | 3+4 2 | | | | 5+6 2 | 5+6 2 | |
| | 3+4 2 | 3+4 2 | | | | 3+4 2 | 3+4 2 | |
| | 5+6 2 | 5+6 2 | | | | 3+4 2 | 5+6 2 | |
| | 3+4 2 | 3+4 2 | | | | 5+6 2 | 3+4 2 | |
| | 5+6 2 | 3+4 2 | | | | 5+6 2 | 5+6 2 | |
| | 5+6 2 | 5+6 2 | | | | 3+4 2 | 5+6 2 | |
| | 3+4 2 | 5+6 2 | | | | 5+6 2 | 3+4 2 | |
| | 3+4 2 | 3+4 2 | | | | 3+4 2 | 3+4 2 | |
| 11.2.71 | 69990 | 292 | Cu1La2O6Sn1 | | | | | |
| | -A | -B | -C | -D | -E | -GM | -Y | -Z |
| | 3+4 2 | 3+4 2 | 2+2 4 | 2+2 4 | 2+2 4 | 3+4 2 | 3+4 2 | 2+2 4 |
| | 5+6 2 | 5+6 2 | 2+2 4 | 2+2 4 | 2+2 4 | 5+6 2 | 5+6 2 | 2+2 4 |
| | 5+6 2 | 5+6 2 | 2+2 4 | 2+2 4 | 2+2 4 | 5+6 2 | 5+6 2 | 2+2 4 |
| | 3+4 2 | 3+4 2 | 2+2 4 | 2+2 4 | 2+2 4 | 3+4 2 | 3+4 2 | 2+2 4 |
| | 3+4 2 | 3+4 2 | 2+2 4 | 2+2 4 | 2+2 4 | 3+4 2 | 3+4 2 | 2+2 4 |
| | 5+6 2 | 5+6 2 | 2+2 4 | 2+2 4 | 2+2 4 | 5+6 2 | 5+6 2 | 2+2 4 |
| | 3+4 2 | 3+4 2 | 2+2 4 | 2+2 4 | 2+2 4 | 3+4 2 | 3+4 2 | 2+2 4 |
| | 5+6 2 | 5+6 2 | 2+2 4 | 2+2 4 | 2+2 4 | 5+6 2 | 5+6 2 | 2+2 4 |
| | 5+6 2 | 3+4 2 | 2+2 4 | 2+2 4 | 2+2 4 | 3+4 2 | 5+6 2 | 2+2 4 |
| | 3+4 2 | 5+6 2 | 2+2 4 | 2+2 4 | 2+2 4 | 5+6 2 | 3+4 2 | 2+2 4 |
| | 3+4 2 | 5+6 2 | 2+2 4 | 2+2 4 | 2+2 4 | 3+4 2 | 5+6 2 | 2+2 4 |
| | 5+6 2 | 3+4 2 | 2+2 4 | 2+2 4 | 2+2 4 | 3+4 2 | 3+4 2 | 2+2 4 |
| | 5+6 2 | 3+4 2 | 2+2 4 | 2+2 4 | 2+2 4 | 3+4 2 | 5+6 2 | 2+2 4 |
| | 3+4 2 | 5+6 2 | 2+2 4 | 2+2 4 | 2+2 4 | 5+6 2 | 3+4 2 | 2+2 4 |
| | 3+4 2 | 5+6 2 | 2+2 4 | 2+2 4 | 2+2 4 | 5+6 2 | 3+4 2 | 2+2 4 |
| | 3+4 2 | 5+6 2 | 2+2 4 | 2+2 4 | 2+2 4 | 3+4 2 | 5+6 2 | 2+2 4 |
| | 5+6 2 | 5+6 2 | 2+2 4 | 2+2 4 | 2+2 4 | 5+6 2 | 3+4 2 | 2+2 4 |
| | 5+6 2 | 3+4 2 | 2+2 4 | 2+2 4 | 2+2 4 | 5+6 2 | 3+4 2 | 2+2 4 |
| | 5+6 2 | 3+4 2 | 2+2 4 | 2+2 4 | 2+2 4 | 5+6 2 | 5+6 2 | 2+2 4 |
| | 3+4 2 | 5+6 2 | 2+2 4 | 2+2 4 | 2+2 4 | 3+4 2 | 5+6 2 | 2+2 4 |
| | 5+6 2 | 3+4 2 | 2+2 4 | 2+2 4 | 2+2 4 | 5+6 2 | 3+4 2 | 2+2 4 |
| | 5+6 2 | 3+4 2 | 2+2 4 | 2+2 4 | 2+2 4 | 3+4 2 | 3+4 2 | 2+2 4 |
| | 3+4 2 | 3+4 2 | 2+2 4 | 2+2 4 | 2+2 4 | 5+6 2 | 5+6 2 | 2+2 4 |
| | 5+6 2 | 3+4 2 | 2+2 4 | 2+2 4 | 2+2 4 | 5+6 2 | 5+6 2 | 2+2 4 |
| | 3+4 2 | 3+4 2 | 2+2 4 | 2+2 4 | 2+2 4 | 5+6 2 | 3+4 2 | 2+2 4 |
| | 3+4 2 | 3+4 2 | 2+2 4 | 2+2 4 | 2+2 4 | 3+4 2 | 5+6 2 | 2+2 4 |
| | 5+6 2 | 5+6 2 | 2+2 4 | 2+2 4 | 2+2 4 | 5+6 2 | 5+6 2 | 2+2 4 |
| | 3+4 2 | 3+4 2 | 2+2 4 | 2+2 4 | 2+2 4 | 3+4 2 | 3+4 2 | 2+2 4 |
| | 3+4 2 | 5+6 2 | 2+2 4 | 2+2 4 | 2+2 4 | 5+6 2 | 5+6 2 | 2+2 4 |
| | 3+4 2 | 5+6 2 | 2+2 4 | 2+2 4 | 2+2 4 | 3+4 2 | 5+6 2 | 2+2 4 |
| | 5+6 2 | 5+6 2 | 2+2 4 | 2+2 4 | 2+2 4 | 3+4 2 | 3+4 2 | 2+2 4 |
| | 5+6 2 | 5+6 2 | 2+2 4 | 2+2 4 | 2+2 4 | 3+4 2 | 3+4 2 | 2+2 4 |
| | 3+4 2 | 3+4 2 | 2+2 4 | 2+2 4 | 2+2 4 | 5+6 2 | 3+4 2 | 2+2 4 |
| | 5+6 2 | 5+6 2 | 2+2 4 | 2+2 4 | 2+2 4 | 3+4 2 | 5+6 2 | 2+2 4 |
| | 3+4 2 | 3+4 2 | 2+2 4 | 2+2 4 | 2+2 4 | 3+4 2 | 5+6 2 | 2+2 4 |
| | 5+6 2 | 5+6 2 | 2+2 4 | 2+2 4 | 2+2 4 | 5+6 2 | 3+4 2 | 2+2 4 |
| | 3+4 2 | 3+4 2 | 2+2 4 | 2+2 4 | 2+2 4 | 3+4 2 | 5+6 2 | 2+2 4 |
| | 5+6 2 | 5+6 2 | 2+2 4 | 2+2 4 | 2+2 4 | 5+6 2 | 3+4 2 | 2+2 4 |
| | 5+6 2 | 3+4 2 | 2+2 4 | 2+2 4 | 2+2 4 | 3+4 2 | 3+4 2 | 2+2 4 |
| | 3+4 2 | 5+6 2 | 2+2 4 | 2+2 4 | 2+2 4 | 5+6 2 | 5+6 2 | 2+2 4 |
| | 5+6 2 | 3+4 2 | 2+2 4 | 2+2 4 | 2+2 4 | 3+4 2 | 5+6 2 | 2+2 4 |
| | 3+4 2 | 3+4 2 | 2+2 4 | 2+2 4 | 2+2 4 | 3+4 2 | 3+4 2 | 2+2 4 |
| | 5+6 2 | 5+6 2 | 2+2 4 | 2+2 4 | 2+2 4 | 5+6 2 | 3+4 2 | 2+2 4 |
| | 3+4 2 | 5+6 2 | 2+2 4 | 2+2 4 | 2+2 4 | 3+4 2 | 5+6 2 | 2+2 4 |
| | 3+4 2 | 3+4 2 | 2+2 4 | 2+2 4 | 2+2 4 | 5+6 2 | 5+6 2 | 2+2 4 |
| | 5+6 2 | 5+6 2 | 2+2 4 | 2+2 4 | 2+2 4 | 3+4 2 | 3+4 2 | 2+2 4 |
| | 3+4 2 | 3+4 2 | 2+2 4 | 2+2 4 | 2+2 4 | 5+6 2 | 3+4 2 | 2+2 4 |
| | 3+4 2 | 3+4 2 | 2+2 4 | 2+2 4 | 2+2 4 | 3+4 2 | 5+6 2 | 2+2 4 |
| | 5+6 2 | 5+6 2 | 2+2 4 | 2+2 4 | 2+2 4 | 5+6 2 | 3+4 2 | 2+2 4 |
| | 5+6 2 | 5+6 2 | 2+2 4 | 2+2 4 | 2+2 4 | 5+6 2 | 5+6 2 | 2+2 4 |
| | 5+6 2 | 5+6 2 | 2+2 4 | 2+2 4 | 2+2 4 | 5+6 2 | 5+6 2 | 2+2 4 |
| | 3+4 2 | 3+4 2 | 2+2 4 | 2+2 4 | 2+2 4 | 5+6 2 | 3+4 2 | 2+2 4 |
| | 5+6 2 | 5+6 2 | 2+2 4 | 2+2 4 | 2+2 4 | 3+4 2 | 5+6 2 | 2+2 4 |
| | 3+4 2 | 5+6 2 | 2+2 4 | 2+2 4 | 2+2 4 | 3+4 2 | 3+4 2 | 2+2 4 |
| | 5+6 2 | 3+4 2 | 2+2 4 | 2+2 4 | 2+2 4 | 5+6 2 | 3+4 2 | 2+2 4 |
| | 3+4 2 | 3+4 2 | 2+2 4 | 2+2 4 | 2+2 4 | 3+4 2 | 5+6 2 | 2+2 4 |
| | 5+6 2 | 3+4 2 | 2+2 4 | 2+2 4 | 2+2 4 | 3+4 2 | 5+6 2 | 2+2 4 |
| | 5+6 2 | 3+4 2 | 2+2 4 | 2+2 4 | 2+2 4 | 3+4 2 | 5+6 2 | 2+2 4 |
| | 5+6 2 | 3+4 2 | 2+2 4 | 2+2 4 | 2+2 4 | 3+4 2 | 5+6 2 | 2+2 4 |
| | 5+6 2 | 3+4 2 | 2+2 4 | 2+2 4 | 2+2 4 | 3+4 2 | 5+6 2 | 2+2 4 |
| | 5+6 2 | 3+4 2 | 2+2 4 | 2+2 4 | 2+2 4 | 3+4 2 | 5+6 2 | 2+2 4 |
| | 5+6 2 | 3+4 2 | 2+2 4 | 2+2 4 | 2+2 4 | 3+4 2 | 5+6 2 | 2+2 4 |
| | 5+6 2 | 3+4 2 | 2+2 4 | 2+2 4 | 2+2 4 | 3+4 2 | 5+6 2 | 2+2 4 |
| | 5+6 2 | 3+4 2 | 2+2 4 | 2+2 4 | 2+2 4 | 3+4 2 | 5+6 2 | 2+2 4 |
| | 5+6 2 | 3+4 2 | 2+2 4 | 2+2 4 | 2+2 4 | 3+4 2 | 5+6 2 | 2+2 4 |
| | 5+6 2 | 3+4 2 | 2+2 4 | 2+2 4 | 2+2 4 | 3+4 2 | 5+6 2 | 2+2 4 |
| | 3+4 2 | 3+4 2 | 2+2 4 | 2+2 4 | 2+2 4 | 3+4 2 | 3+4 2 | 2+2 4 |
| | 3+4 2 | 3+4 2 | 2+2 4 | 2+2 4 | 2+2 4 | 3+4 2 | 3+4 2 | 2+2 4 |
| | 3+4 2 | 5+6 2 | 2+2 4 | 2+2 4 | 2+2 4 | 3+4 2 | 3+4 2 | 2+2 4 |
| | 3+4 2 | 3+4 2 | 2+2 4 | 2+2 4 | 2+2 4 | 3+4 2 | 3+4 2 | 2+2 4 |
| | 3+4 2 | 3+4 2 | 2+2 4 | 2+2 4 | 2+2 4 | 3+4 2 | 5+6 2 | 2+2 4 |
| | 3+4 2 | 3+4 2 | 2+2 4 | 2+2 4 | 2+2 4 | 3+4 2 | 3+4 2 | 2+2 4 |
| | 3+4 2 | 3+4 2 | 2+2 4 | 2+2 4 | 2+2 4 | 3+4 2 | 3+4 2 | 2+2 4 |
| | 3+4 2 | 3+4 2 | | | | 3+4 2 | 3+4 2 | |
| | 3+4 2 | 3+4 2 | | | | 3+4 2 | 3+4 2 | |
| | 5+6 2 | 3+4 2 | | | | 3+4 2 | 3+4 2 | |
| | 3+4 2 | 3+4 2 | | | | 3+4 2 | 3+4 2 | |
| | 5+6 2 | 3+4 2 | | | | 3+4 2 | 3+4 2 | |

Continued on next column

**TABLE XXIII– continued**

| TQC | ICSD | NoE | SF | |
|---|---|---|---|---|
| | 3+4 2 5+6 2 | | | 3+4 2 5+6 2 |
| | 3+4 2 5+6 2 | | | 3+4 2 5+6 2 |
| | 5+6 2 5+6 2 | | | 5+6 2 3+4 2 |
| | 3+4 2 3+4 2 | | | 5+6 2 3+4 2 |
| | 3+4 2 3+4 2 | | | 5+6 2 3+4 2 |
| | 5+6 2 5+6 2 | | | 5+6 2 3+4 2 |
| | 5+6 2 3+4 2 | | | 5+6 2 5+6 2 |
| | 3+4 2 3+4 2 | | | 3+4 2 5+6 2 |
| | 3+4 2 5+6 2 | | | 5+6 2 3+4 2 |
| | 5+6 2 3+4 2 | | | 5+6 2 3+4 2 |
| | 3+4 2 3+4 2 | | | 3+4 2 3+4 2 |
| | 5+6 2 3+4 2 | | | 3+4 2 5+6 2 |
| | 5+6 2 5+6 2 | | | 5+6 2 5+6 2 |
| | 3+4 2 5+6 2 | | | 5+6 2 5+6 2 |
| | 5+6 2 5+6 2 | | | 5+6 2 5+6 2 |
| | 5+6 2 3+4 2 | | | 5+6 2 3+4 2 |
| | 3+4 2 3+4 2 | | | 3+4 2 5+6 2 |
| | 3+4 2 3+4 2 | | | 5+6 2 3+4 2 |
| | 5+6 2 5+6 2 | | | 3+4 2 5+6 2 |
| | 3+4 2 5+6 2 | | | 3+4 2 3+4 2 |
| | 5+6 2 3+4 2 | | | 5+6 2 3+4 2 |
| | 5+6 2 5+6 2 | | | 5+6 2 5+6 2 |
| | 3+4 2 3+4 2 | | | 5+6 2 5+6 2 |
| | 3+4 2 5+6 2 | | | 5+6 2 3+4 2 |
| | 3+4 2 5+6 2 | | | 3+4 2 3+4 2 |
| | 3+4 2 3+4 2 | | | 5+6 2 5+6 2 |
| | 5+6 2 5+6 2 | | | 3+4 2 5+6 2 |
| | 3+4 2 5+6 2 | | | 3+4 2 3+4 2 |
| | 5+6 2 3+4 2 | | | 5+6 2 5+6 2 |
| | 5+6 2 3+4 2 | | | 5+6 2 3+4 2 |
| | 3+4 2 5+6 2 | | | 5+6 2 5+6 2 |
| | 5+6 2 3+4 2 | | | 5+6 2 3+4 2 |
| | 3+4 2 3+4 2 | | | 3+4 2 3+4 2 |
| | 3+4 2 5+6 2 | | | 3+4 2 3+4 2 |
| | 5+6 2 5+6 2 | | | 3+4 2 5+6 2 |
| | 5+6 2 3+4 2 | | | 5+6 2 5+6 2 |
| | 5+6 2 5+6 2 | | | 3+4 2 3+4 2 |
| | 5+6 2 3+4 2 | | | 3+4 2 5+6 2 |
| | 3+4 2 5+6 2 | | | 3+4 2 3+4 2 |
| | 5+6 2 3+4 2 | | | 5+6 2 5+6 2 |
| | 3+4 2 3+4 2 | | | 5+6 2 5+6 2 |
| | 5+6 2 5+6 2 | | | 5+6 2 3+4 2 |
| | 3+4 2 3+4 2 | | | 5+6 2 5+6 2 |
| | 5+6 2 5+6 2 | | | 3+4 2 3+4 2 |
| | 3+4 2 5+6 2 | | | 3+4 2 5+6 2 |
| | 5+6 2 5+6 2 | | | 3+4 2 5+6 2 |
| | 5+6 2 3+4 2 | | | 5+6 2 3+4 2 |
| | 3+4 2 3+4 2 | | | 3+4 2 5+6 2 |
| | 3+4 2 3+4 2 | | | 5+6 2 3+4 2 |
| | 5+6 2 3+4 2 | | | 5+6 2 5+6 2 |
| | 3+4 2 5+6 2 | | | 3+4 2 3+4 2 |
| | 5+6 2 3+4 2 | | | 5+6 2 5+6 2 |
| | 3+4 2 5+6 2 | | | 3+4 2 5+6 2 |
| | 3+4 2 3+4 2 | | | 5+6 2 3+4 2 |
| | 3+4 2 5+6 2 | | | 3+4 2 3+4 2 |
| | 5+6 2 5+6 2 | | | 5+6 2 5+6 2 |
| | 5+6 2 3+4 2 | | | 3+4 2 5+6 2 |
| | 3+4 2 5+6 2 | | | 3+4 2 3+4 2 |
| | 3+4 2 5+6 2 | | | 5+6 2 3+4 2 |
| | 5+6 2 3+4 2 | | | 3+4 2 3+4 2 |
| | 3+4 2 3+4 2 | | | 3+4 2 5+6 2 |

Continued on next column

**TABLE XXIII– continued**

| TQC | ICSD | | NoE | | SF | | | |
|---|---|---|---|---|---|---|---|---|
| | 5+6 2 | 3+4 2 | | | | 5+6 2 | 5+6 2 | |
| | 5+6 2 | 5+6 2 | | | | 3+4 2 | 3+4 2 | |
| | 3+4 2 | 3+4 2 | | | | 3+4 2 | 5+6 2 | |
| | 5+6 2 | 3+4 2 | | | | 3+4 2 | 3+4 2 | |
| | 3+4 2 | 5+6 2 | | | | 5+6 2 | 3+4 2 | |
| | 5+6 2 | 5+6 2 | | | | 3+4 2 | 5+6 2 | |
| | 3+4 2 | 5+6 2 | | | | 5+6 2 | 5+6 2 | |
| 11.2.72 | 108896 | | 148 | | Co1O3Sr1 | | | |
| | -A | -B | -C | -D | -E | -GM | -Y | -Z |
| | 3+4 2 | 5+6 2 | 2+2 4 | 2+2 4 | 2+2 4 | 3+4 2 | 5+6 2 | 2+2 4 |
| | 5+6 2 | 3+4 2 | 2+2 4 | 2+2 4 | 2+2 4 | 5+6 2 | 3+4 2 | 2+2 4 |
| | 5+6 2 | 3+4 2 | 2+2 4 | 2+2 4 | 2+2 4 | 5+6 2 | 3+4 2 | 2+2 4 |
| | 3+4 2 | 5+6 2 | 2+2 4 | 2+2 4 | 2+2 4 | 3+4 2 | 5+6 2 | 2+2 4 |
| | 5+6 2 | 3+4 2 | 2+2 4 | 2+2 4 | 2+2 4 | 3+4 2 | 5+6 2 | 2+2 4 |
| | 3+4 2 | 5+6 2 | 2+2 4 | 2+2 4 | 2+2 4 | 5+6 2 | 3+4 2 | 2+2 4 |
| | 3+4 2 | 5+6 2 | 2+2 4 | 2+2 4 | 2+2 4 | 5+6 2 | 3+4 2 | 2+2 4 |
| | 5+6 2 | 3+4 2 | 2+2 4 | 2+2 4 | 2+2 4 | 3+4 2 | 3+4 2 | 2+2 4 |
| | 3+4 2 | 3+4 2 | 2+2 4 | 2+2 4 | 2+2 4 | 3+4 2 | 5+6 2 | 2+2 4 |
| | 5+6 2 | 3+4 2 | 2+2 4 | 2+2 4 | 2+2 4 | 3+4 2 | 5+6 2 | 2+2 4 |
| | 3+4 2 | 5+6 2 | 2+2 4 | 2+2 4 | 2+2 4 | 5+6 2 | 3+4 2 | 2+2 4 |
| | 5+6 2 | 5+6 2 | 2+2 4 | 2+2 4 | 2+2 4 | 5+6 2 | 5+6 2 | 2+2 4 |
| | 3+4 2 | 5+6 2 | 2+2 4 | 2+2 4 | 2+2 4 | 3+4 2 | 5+6 2 | 2+2 4 |
| | 5+6 2 | 3+4 2 | 2+2 4 | 2+2 4 | 2+2 4 | 5+6 2 | 3+4 2 | 2+2 4 |
| | 5+6 2 | 5+6 2 | 2+2 4 | 2+2 4 | 2+2 4 | 3+4 2 | 3+4 2 | 2+2 4 |
| | 3+4 2 | 3+4 2 | 2+2 4 | 2+2 4 | 2+2 4 | 5+6 2 | 5+6 2 | 2+2 4 |
| | 5+6 2 | 5+6 2 | 2+2 4 | 2+2 4 | 2+2 4 | 3+4 2 | 3+4 2 | 2+2 4 |
| | 3+4 2 | 3+4 2 | 2+2 4 | 2+2 4 | 2+2 4 | 5+6 2 | 5+6 2 | 2+2 4 |
| | 3+4 2 | 5+6 2 | 2+2 4 | 2+2 4 | 2+2 4 | 3+4 2 | 5+6 2 | 2+2 4 |
| | 5+6 2 | 3+4 2 | 2+2 4 | 2+2 4 | 2+2 4 | 5+6 2 | 3+4 2 | 2+2 4 |
| | 5+6 2 | 3+4 2 | 2+2 4 | 2+2 4 | 2+2 4 | 5+6 2 | 5+6 2 | 2+2 4 |
| | 3+4 2 | 5+6 2 | 2+2 4 | 2+2 4 | 2+2 4 | 3+4 2 | 3+4 2 | 2+2 4 |
| | 3+4 2 | 5+6 2 | 2+2 4 | 2+2 4 | 2+2 4 | 3+4 2 | 5+6 2 | 2+2 4 |
| | 5+6 2 | 3+4 2 | 2+2 4 | 2+2 4 | 2+2 4 | 3+4 2 | 3+4 2 | 2+2 4 |
| | 5+6 2 | 5+6 2 | 2+2 4 | 2+2 4 | 2+2 4 | 5+6 2 | 3+4 2 | 2+2 4 |
| | 3+4 2 | 3+4 2 | 2+2 4 | 2+2 4 | 2+2 4 | 3+4 2 | 5+6 2 | 2+2 4 |
| | 3+4 2 | 5+6 2 | 2+2 4 | 2+2 4 | 2+2 4 | 5+6 2 | 5+6 2 | 2+2 4 |
| | 5+6 2 | 3+4 2 | 2+2 4 | 2+2 4 | 2+2 4 | 5+6 2 | 3+4 2 | 2+2 4 |
| | 5+6 2 | 3+4 2 | 2+2 4 | 2+2 4 | 2+2 4 | 3+4 2 | 5+6 2 | 2+2 4 |
| | 5+6 2 | 3+4 2 | 2+2 4 | 2+2 4 | 2+2 4 | 3+4 2 | 5+6 2 | 2+2 4 |
| | 5+6 2 | 3+4 2 | 2+2 4 | 2+2 4 | 2+2 4 | 3+4 2 | 5+6 2 | 2+2 4 |
| | 3+4 2 | 5+6 2 | 2+2 4 | 2+2 4 | 2+2 4 | 5+6 2 | 3+4 2 | 2+2 4 |
| | 5+6 2 | 5+6 2 | 2+2 4 | 2+2 4 | 2+2 4 | 5+6 2 | 3+4 2 | 2+2 4 |
| | 3+4 2 | 3+4 2 | 2+2 4 | 2+2 4 | 2+2 4 | 5+6 2 | 5+6 2 | 2+2 4 |
| | 3+4 2 | 5+6 2 | 2+2 4 | 2+2 4 | 2+2 4 | 3+4 2 | 3+4 2 | 2+2 4 |
| | 3+4 2 | 3+4 2 | 2+2 4 | 2+2 4 | 2+2 4 | 3+4 2 | 3+4 2 | 2+2 4 |
| | 5+6 2 | 3+4 2 | 2+2 4 | 2+2 4 | 2+2 4 | 3+4 2 | 5+6 2 | 2+2 4 |
| | 3+4 2 | 3+4 2 | | | | 3+4 2 | 3+4 2 | |
| | 5+6 2 | 3+4 2 | | | | 3+4 2 | 5+6 2 | |
| | 5+6 2 | 5+6 2 | | | | 5+6 2 | 5+6 2 | |
| | 5+6 2 | 3+4 2 | | | | 3+4 2 | 5+6 2 | |
| | 5+6 2 | 3+4 2 | | | | 3+4 2 | 5+6 2 | |
| | 3+4 2 | 5+6 2 | | | | 5+6 2 | 3+4 2 | |
| | 5+6 2 | 5+6 2 | | | | 5+6 2 | 5+6 2 | |
| | 3+4 2 | 3+4 2 | | | | 3+4 2 | 3+4 2 | |
| | 3+4 2 | 3+4 2 | | | | 3+4 2 | 3+4 2 | |
| | 3+4 2 | 5+6 2 | | | | 5+6 2 | 3+4 2 | |
| | 3+4 2 | 5+6 2 | | | | 5+6 2 | 3+4 2 | |
| | 3+4 2 | 5+6 2 | | | | 5+6 2 | 3+4 2 | |
| | 5+6 2 | 5+6 2 | | | | 5+6 2 | 5+6 2 | |
| | 3+4 2 | 5+6 2 | | | | 5+6 2 | 3+4 2 | |
| | 5+6 2 | 5+6 2 | | | | 5+6 2 | 5+6 2 | |

Continued on next column

**TABLE XXIII– continued**

| TQC | ICSD | NoE | SF | | | | | |
|---|---|---|---|---|---|---|---|---|
| | 5+6 2 | 3+4 2 | | | | 3+4 2 | 5+6 2 | |
| | 5+6 2 | 3+4 2 | | | | 3+4 2 | 5+6 2 | |
| | 5+6 2 | 3+4 2 | | | | 3+4 2 | 3+4 2 | |
| | 3+4 2 | 5+6 2 | | | | 3+4 2 | 5+6 2 | |
| | 5+6 2 | 3+4 2 | | | | 5+6 2 | 3+4 2 | |
| | 3+4 2 | 5+6 2 | | | | 3+4 2 | 5+6 2 | |
| | 5+6 2 | 5+6 2 | | | | 3+4 2 | 5+6 2 | |
| | 3+4 2 | 3+4 2 | | | | 3+4 2 | 3+4 2 | |
| | 3+4 2 | 3+4 2 | | | | 3+4 2 | 5+6 2 | |
| | 5+6 2 | 3+4 2 | | | | 5+6 2 | 3+4 2 | |
| | 5+6 2 | 3+4 2 | | | | 3+4 2 | 5+6 2 | |
| | 5+6 2 | 3+4 2 | | | | 5+6 2 | 5+6 2 | |
| | 5+6 2 | 5+6 2 | | | | 5+6 2 | 5+6 2 | |
| | 5+6 2 | 5+6 2 | | | | 3+4 2 | 5+6 2 | |
| | 3+4 2 | 3+4 2 | | | | 5+6 2 | 3+4 2 | |
| | 5+6 2 | 5+6 2 | | | | 3+4 2 | 3+4 2 | |
| | 3+4 2 | 3+4 2 | | | | 5+6 2 | 5+6 2 | |
| | 5+6 2 | 5+6 2 | | | | 3+4 2 | 3+4 2 | |
| | 3+4 2 | 3+4 2 | | | | 3+4 2 | 5+6 2 | |
| | 3+4 2 | 3+4 2 | | | | 3+4 2 | 3+4 2 | |
| | 5+6 2 | 5+6 2 | | | | 5+6 2 | 5+6 2 | |
| | 5+6 2 | 5+6 2 | | | | 5+6 2 | 5+6 2 | |
| 11.2.73 | 411140 | 132 | Cl7Cs1Ti2 | | | | | |
| | -A | -B | -C | -D | -E | -GM | -Y | -Z |
| | 3+4 2 | 3+4 2 | 2+2 4 | 2+2 4 | 2+2 4 | 3+4 2 | 3+4 2 | 2+2 4 |
| | 5+6 2 | 5+6 2 | 2+2 4 | 2+2 4 | 2+2 4 | 5+6 2 | 5+6 2 | 2+2 4 |
| | 3+4 2 | 3+4 2 | 2+2 4 | 2+2 4 | 2+2 4 | 3+4 2 | 3+4 2 | 2+2 4 |
| | 5+6 2 | 5+6 2 | 2+2 4 | 2+2 4 | 2+2 4 | 3+4 2 | 5+6 2 | 2+2 4 |
| | 3+4 2 | 3+4 2 | 2+2 4 | 2+2 4 | 2+2 4 | 5+6 2 | 3+4 2 | 2+2 4 |
| | 5+6 2 | 5+6 2 | 2+2 4 | 2+2 4 | 2+2 4 | 5+6 2 | 5+6 2 | 2+2 4 |
| | 5+6 2 | 5+6 2 | 2+2 4 | 2+2 4 | 2+2 4 | 5+6 2 | 5+6 2 | 2+2 4 |
| | 3+4 2 | 3+4 2 | 2+2 4 | 2+2 4 | 2+2 4 | 3+4 2 | 3+4 2 | 2+2 4 |
| | 3+4 2 | 5+6 2 | 2+2 4 | 2+2 4 | 2+2 4 | 3+4 2 | 3+4 2 | 2+2 4 |
| | 5+6 2 | 3+4 2 | 2+2 4 | 2+2 4 | 2+2 4 | 5+6 2 | 5+6 2 | 2+2 4 |
| | 5+6 2 | 5+6 2 | 2+2 4 | 2+2 4 | 2+2 4 | 5+6 2 | 5+6 2 | 2+2 4 |
| | 3+4 2 | 3+4 2 | 2+2 4 | 2+2 4 | 2+2 4 | 3+4 2 | 3+4 2 | 2+2 4 |
| | 5+6 2 | 5+6 2 | 2+2 4 | 2+2 4 | 2+2 4 | 3+4 2 | 5+6 2 | 2+2 4 |
| | 3+4 2 | 5+6 2 | 2+2 4 | 2+2 4 | 2+2 4 | 3+4 2 | 3+4 2 | 2+2 4 |
| | 3+4 2 | 3+4 2 | 2+2 4 | 2+2 4 | 2+2 4 | 5+6 2 | 5+6 2 | 2+2 4 |
| | 5+6 2 | 3+4 2 | 2+2 4 | 2+2 4 | 2+2 4 | 5+6 2 | 3+4 2 | 2+2 4 |
| | 3+4 2 | 3+4 2 | 2+2 4 | 2+2 4 | 2+2 4 | 5+6 2 | 3+4 2 | 2+2 4 |
| | 5+6 2 | 5+6 2 | 2+2 4 | 2+2 4 | 2+2 4 | 3+4 2 | 5+6 2 | 2+2 4 |
| | 5+6 2 | 3+4 2 | 2+2 4 | 2+2 4 | 2+2 4 | 3+4 2 | 3+4 2 | 2+2 4 |
| | 3+4 2 | 5+6 2 | 2+2 4 | 2+2 4 | 2+2 4 | 5+6 2 | 5+6 2 | 2+2 4 |
| | 3+4 2 | 3+4 2 | 2+2 4 | 2+2 4 | 2+2 4 | 5+6 2 | 3+4 2 | 2+2 4 |
| | 5+6 2 | 5+6 2 | 2+2 4 | 2+2 4 | 2+2 4 | 3+4 2 | 5+6 2 | 2+2 4 |
| | 5+6 2 | 3+4 2 | 2+2 4 | 2+2 4 | 2+2 4 | 3+4 2 | 5+6 2 | 2+2 4 |
| | 5+6 2 | 5+6 2 | 2+2 4 | 2+2 4 | 2+2 4 | 5+6 2 | 5+6 2 | 2+2 4 |
| | 3+4 2 | 5+6 2 | 2+2 4 | 2+2 4 | 2+2 4 | 5+6 2 | 3+4 2 | 2+2 4 |
| | 3+4 2 | 3+4 2 | 2+2 4 | 2+2 4 | 2+2 4 | 3+4 2 | 3+4 2 | 2+2 4 |
| | 3+4 2 | 3+4 2 | 2+2 4 | 2+2 4 | 2+2 4 | 3+4 2 | 5+6 2 | 2+2 4 |
| | 5+6 2 | 5+6 2 | 2+2 4 | 2+2 4 | 2+2 4 | 3+4 2 | 3+4 2 | 2+2 4 |
| | 5+6 2 | 3+4 2 | 2+2 4 | 2+2 4 | 2+2 4 | 3+4 2 | 3+4 2 | 2+2 4 |
| | 3+4 2 | 5+6 2 | 2+2 4 | 2+2 4 | 2+2 4 | 5+6 2 | 3+4 2 | 2+2 4 |
| | 5+6 2 | 3+4 2 | 2+2 4 | 2+2 4 | 2+2 4 | 5+6 2 | 5+6 2 | 2+2 4 |
| | 3+4 2 | 3+4 2 | 2+2 4 | 2+2 4 | 2+2 4 | 3+4 2 | 5+6 2 | 2+2 4 |
| | 3+4 2 | 3+4 2 | 2+2 4 | 2+2 4 | 2+2 4 | 5+6 2 | 5+6 2 | 2+2 4 |
| | 3+4 2 | 5+6 2 | | | | 3+4 2 | 3+4 2 | |
| | 5+6 2 | 5+6 2 | | | | 5+6 2 | 3+4 2 | |
| | 3+4 2 | 5+6 2 | | | | 3+4 2 | 3+4 2 | |
| | 5+6 2 | 5+6 2 | | | | 5+6 2 | 5+6 2 | |

Continued on next column

**TABLE XXIII– continued**

| TQC | ICSD | NoE | SF | | | | | |
|---|---|---|---|---|---|---|---|---|
| | 3+4 2 | 3+4 2 | | | | 3+4 2 | 5+6 2 | |
| | 3+4 2 | 3+4 2 | | | | 5+6 2 | 3+4 2 | |
| | 5+6 2 | 3+4 2 | | | | 3+4 2 | 5+6 2 | |
| | 3+4 2 | 5+6 2 | | | | 3+4 2 | 3+4 2 | |
| | 5+6 2 | 3+4 2 | | | | 5+6 2 | 3+4 2 | |
| | 5+6 2 | 5+6 2 | | | | 5+6 2 | 5+6 2 | |
| | 5+6 2 | 5+6 2 | | | | 5+6 2 | 5+6 2 | |
| | 3+4 2 | 5+6 2 | | | | 3+4 2 | 5+6 2 | |
| | 5+6 2 | 3+4 2 | | | | 5+6 2 | 3+4 2 | |
| | 3+4 2 | 3+4 2 | | | | 3+4 2 | 3+4 2 | |
| | 5+6 2 | 5+6 2 | | | | 5+6 2 | 5+6 2 | |
| | 3+4 2 | 5+6 2 | | | | 3+4 2 | 5+6 2 | |
| | 3+4 2 | 3+4 2 | | | | 3+4 2 | 3+4 2 | |
| | 5+6 2 | 3+4 2 | | | | 5+6 2 | 3+4 2 | |
| | 5+6 2 | 5+6 2 | | | | 3+4 2 | 5+6 2 | |
| | 3+4 2 | 3+4 2 | | | | 3+4 2 | 3+4 2 | |
| | 5+6 2 | 3+4 2 | | | | 3+4 2 | 3+4 2 | |
| | 3+4 2 | 5+6 2 | | | | 5+6 2 | 5+6 2 | |
| | 5+6 2 | 5+6 2 | | | | 3+4 2 | 3+4 2 | |
| | 3+4 2 | 5+6 2 | | | | 5+6 2 | 3+4 2 | |
| | 5+6 2 | 3+4 2 | | | | 5+6 2 | 5+6 2 | |
| | 5+6 2 | 5+6 2 | | | | 3+4 2 | 5+6 2 | |
| | 3+4 2 | 5+6 2 | | | | 5+6 2 | 5+6 2 | |
| | 5+6 2 | 3+4 2 | | | | 5+6 2 | 3+4 2 | |
| | 5+6 2 | 3+4 2 | | | | 3+4 2 | 5+6 2 | |
| | 3+4 2 | 5+6 2 | | | | 5+6 2 | 5+6 2 | |
| | 3+4 2 | 3+4 2 | | | | 5+6 2 | 3+4 2 | |
| | 3+4 2 | 3+4 2 | | | | 3+4 2 | 3+4 2 | |
| | 3+4 2 | 3+4 2 | | | | 5+6 2 | 3+4 2 | |
| 11.2.74 | 415798 | 112 | Cl2Co2O3Te1 | | | | | |
| | -A | -B | -C | -D | -E | -GM | -Y | -Z |
| | 5+6 2 | 3+4 2 | 2+2 4 | 2+2 4 | 2+2 4 | 3+4 2 | 5+6 2 | 2+2 4 |
| | 3+4 2 | 5+6 2 | 2+2 4 | 2+2 4 | 2+2 4 | 5+6 2 | 3+4 2 | 2+2 4 |
| | 3+4 2 | 3+4 2 | 2+2 4 | 2+2 4 | 2+2 4 | 3+4 2 | 3+4 2 | 2+2 4 |
| | 5+6 2 | 5+6 2 | 2+2 4 | 2+2 4 | 2+2 4 | 5+6 2 | 5+6 2 | 2+2 4 |
| | 3+4 2 | 5+6 2 | 2+2 4 | 2+2 4 | 2+2 4 | 5+6 2 | 3+4 2 | 2+2 4 |
| | 5+6 2 | 3+4 2 | 2+2 4 | 2+2 4 | 2+2 4 | 3+4 2 | 5+6 2 | 2+2 4 |
| | 3+4 2 | 3+4 2 | 2+2 4 | 2+2 4 | 2+2 4 | 3+4 2 | 3+4 2 | 2+2 4 |
| | 5+6 2 | 5+6 2 | 2+2 4 | 2+2 4 | 2+2 4 | 5+6 2 | 5+6 2 | 2+2 4 |
| | 5+6 2 | 5+6 2 | 2+2 4 | 2+2 4 | 2+2 4 | 3+4 2 | 3+4 2 | 2+2 4 |
| | 3+4 2 | 3+4 2 | 2+2 4 | 2+2 4 | 2+2 4 | 5+6 2 | 5+6 2 | 2+2 4 |
| | 5+6 2 | 3+4 2 | 2+2 4 | 2+2 4 | 2+2 4 | 3+4 2 | 5+6 2 | 2+2 4 |
| | 3+4 2 | 5+6 2 | 2+2 4 | 2+2 4 | 2+2 4 | 5+6 2 | 3+4 2 | 2+2 4 |
| | 3+4 2 | 5+6 2 | 2+2 4 | 2+2 4 | 2+2 4 | 5+6 2 | 3+4 2 | 2+2 4 |
| | 3+4 2 | 3+4 2 | 2+2 4 | 2+2 4 | 2+2 4 | 3+4 2 | 3+4 2 | 2+2 4 |
| | 5+6 2 | 5+6 2 | 2+2 4 | 2+2 4 | 2+2 4 | 5+6 2 | 5+6 2 | 2+2 4 |
| | 3+4 2 | 3+4 2 | 2+2 4 | 2+2 4 | 2+2 4 | 3+4 2 | 5+6 2 | 2+2 4 |
| | 5+6 2 | 3+4 2 | 2+2 4 | 2+2 4 | 2+2 4 | 3+4 2 | 3+4 2 | 2+2 4 |
| | 5+6 2 | 5+6 2 | 2+2 4 | 2+2 4 | 2+2 4 | 5+6 2 | 5+6 2 | 2+2 4 |
| | 3+4 2 | 3+4 2 | 2+2 4 | 2+2 4 | 2+2 4 | 3+4 2 | 5+6 2 | 2+2 4 |
| | 5+6 2 | 3+4 2 | 2+2 4 | 2+2 4 | 2+2 4 | 3+4 2 | 5+6 2 | 2+2 4 |
| | 5+6 2 | 5+6 2 | 2+2 4 | 2+2 4 | 2+2 4 | 5+6 2 | 3+4 2 | 2+2 4 |
| | 3+4 2 | 3+4 2 | 2+2 4 | 2+2 4 | 2+2 4 | 3+4 2 | 3+4 2 | 2+2 4 |
| | 3+4 2 | 5+6 2 | 2+2 4 | 2+2 4 | 2+2 4 | 5+6 2 | 3+4 2 | 2+2 4 |
| | 3+4 2 | 5+6 2 | 2+2 4 | 2+2 4 | 2+2 4 | 3+4 2 | 3+4 2 | 2+2 4 |
| | 5+6 2 | 3+4 2 | 2+2 4 | 2+2 4 | 2+2 4 | 5+6 2 | 5+6 2 | 2+2 4 |
| | 3+4 2 | 3+4 2 | 2+2 4 | 2+2 4 | 2+2 4 | 3+4 2 | 3+4 2 | 2+2 4 |
| | 3+4 2 | 3+4 2 | 2+2 4 | 2+2 4 | 2+2 4 | 3+4 2 | 5+6 2 | 2+2 4 |
| | 3+4 2 | 3+4 2 | 2+2 4 | 2+2 4 | 2+2 4 | 5+6 2 | 3+4 2 | 2+2 4 |
| | 5+6 2 | 5+6 2 | | | | 3+4 2 | 3+4 2 | |
| | 5+6 2 | 5+6 2 | | | | 5+6 2 | 5+6 2 | |

Continued on next column

**TABLE XXIII– continued**

| TQC | ICSD | NoE | SF | | | | | |
|---|---|---|---|---|---|---|---|---|
| | 5+6 2 | 3+4 2 | | | | 3+4 2 | 5+6 2 | |
| | 3+4 2 | 5+6 2 | | | | 3+4 2 | 5+6 2 | |
| | 5+6 2 | 5+6 2 | | | | 5+6 2 | 5+6 2 | |
| | 5+6 2 | 3+4 2 | | | | 5+6 2 | 3+4 2 | |
| | 3+4 2 | 5+6 2 | | | | 3+4 2 | 3+4 2 | |
| | 5+6 2 | 3+4 2 | | | | 5+6 2 | 5+6 2 | |
| | 5+6 2 | 5+6 2 | | | | 5+6 2 | 5+6 2 | |
| | 3+4 2 | 5+6 2 | | | | 3+4 2 | 3+4 2 | |
| | 5+6 2 | 5+6 2 | | | | 3+4 2 | 3+4 2 | |
| | 3+4 2 | 3+4 2 | | | | 5+6 2 | 3+4 2 | |
| | 5+6 2 | 5+6 2 | | | | 5+6 2 | 5+6 2 | |
| | 3+4 2 | 3+4 2 | | | | 5+6 2 | 5+6 2 | |
| | 3+4 2 | 3+4 2 | | | | 3+4 2 | 3+4 2 | |
| | 3+4 2 | 3+4 2 | | | | 3+4 2 | 3+4 2 | |
| | 3+4 2 | 3+4 2 | | | | 3+4 2 | 3+4 2 | |
| | 3+4 2 | 3+4 2 | | | | 3+4 2 | 3+4 2 | |
| | 3+4 2 | 5+6 2 | | | | 5+6 2 | 3+4 2 | |
| | 3+4 2 | 3+4 2 | | | | 3+4 2 | 3+4 2 | |
| | 3+4 2 | 5+6 2 | | | | 3+4 2 | 5+6 2 | |
| | 5+6 2 | 3+4 2 | | | | 5+6 2 | 3+4 2 | |
| | 3+4 2 | 3+4 2 | | | | 3+4 2 | 5+6 2 | |
| | 5+6 2 | 5+6 2 | | | | 5+6 2 | 3+4 2 | |
| | 5+6 2 | 3+4 2 | | | | 3+4 2 | 5+6 2 | |
| | 3+4 2 | 3+4 2 | | | | 3+4 2 | 3+4 2 | |
| | 3+4 2 | 5+6 2 | | | | 3+4 2 | 3+4 2 | |
| | 3+4 2 | 3+4 2 | | | | 5+6 2 | 5+6 2 | |
| 11.2.75 | 65346 | 92 | Cl1Fe1Li1Mo1O4 | | | | | |
| | -A | -B | -C | -D | -E | -GM | -Y | -Z |
| | 5+6 2 | 5+6 2 | 2+2 4 | 2+2 4 | 2+2 4 | 3+4 2 | 3+4 2 | 2+2 4 |
| | 3+4 2 | 3+4 2 | 2+2 4 | 2+2 4 | 2+2 4 | 5+6 2 | 5+6 2 | 2+2 4 |
| | 5+6 2 | 5+6 2 | 2+2 4 | 2+2 4 | 2+2 4 | 5+6 2 | 5+6 2 | 2+2 4 |
| | 3+4 2 | 3+4 2 | 2+2 4 | 2+2 4 | 2+2 4 | 3+4 2 | 3+4 2 | 2+2 4 |
| | 5+6 2 | 5+6 2 | 2+2 4 | 2+2 4 | 2+2 4 | 5+6 2 | 5+6 2 | 2+2 4 |
| | 3+4 2 | 3+4 2 | 2+2 4 | 2+2 4 | 2+2 4 | 3+4 2 | 3+4 2 | 2+2 4 |
| | 3+4 2 | 3+4 2 | 2+2 4 | 2+2 4 | 2+2 4 | 5+6 2 | 5+6 2 | 2+2 4 |
| | 5+6 2 | 5+6 2 | 2+2 4 | 2+2 4 | 2+2 4 | 3+4 2 | 3+4 2 | 2+2 4 |
| | 5+6 2 | 3+4 2 | 2+2 4 | 2+2 4 | 2+2 4 | 5+6 2 | 3+4 2 | 2+2 4 |
| | 3+4 2 | 5+6 2 | 2+2 4 | 2+2 4 | 2+2 4 | 3+4 2 | 5+6 2 | 2+2 4 |
| | 5+6 2 | 5+6 2 | 2+2 4 | 2+2 4 | 2+2 4 | 3+4 2 | 3+4 2 | 2+2 4 |
| | 3+4 2 | 3+4 2 | 2+2 4 | 2+2 4 | 2+2 4 | 3+4 2 | 3+4 2 | 2+2 4 |
| | 5+6 2 | 5+6 2 | 2+2 4 | 2+2 4 | 2+2 4 | 5+6 2 | 5+6 2 | 2+2 4 |
| | 3+4 2 | 3+4 2 | 2+2 4 | 2+2 4 | 2+2 4 | 3+4 2 | 3+4 2 | 2+2 4 |
| | 3+4 2 | 3+4 2 | 2+2 4 | 2+2 4 | 2+2 4 | 5+6 2 | 5+6 2 | 2+2 4 |
| | 5+6 2 | 5+6 2 | 2+2 4 | 2+2 4 | 2+2 4 | 5+6 2 | 5+6 2 | 2+2 4 |
| | 5+6 2 | 3+4 2 | 2+2 4 | 2+2 4 | 2+2 4 | 3+4 2 | 3+4 2 | 2+2 4 |
| | 3+4 2 | 5+6 2 | 2+2 4 | 2+2 4 | 2+2 4 | 5+6 2 | 3+4 2 | 2+2 4 |
| | 5+6 2 | 5+6 2 | 2+2 4 | 2+2 4 | 2+2 4 | 3+4 2 | 5+6 2 | 2+2 4 |
| | 3+4 2 | 3+4 2 | 2+2 4 | 2+2 4 | 2+2 4 | 3+4 2 | 5+6 2 | 2+2 4 |
| | 5+6 2 | 5+6 2 | 2+2 4 | 2+2 4 | 2+2 4 | 5+6 2 | 3+4 2 | 2+2 4 |
| | 3+4 2 | 3+4 2 | 2+2 4 | 2+2 4 | 2+2 4 | 5+6 2 | 5+6 2 | 2+2 4 |
| | 3+4 2 | 5+6 2 | 2+2 4 | 2+2 4 | 2+2 4 | 5+6 2 | 3+4 2 | 2+2 4 |
| | 5+6 2 | 3+4 2 | | | | 3+4 2 | 5+6 2 | |
| | 3+4 2 | 3+4 2 | | | | 5+6 2 | 5+6 2 | |
| | 5+6 2 | 5+6 2 | | | | 3+4 2 | 3+4 2 | |
| | 5+6 2 | 3+4 2 | | | | 3+4 2 | 3+4 2 | |
| | 3+4 2 | 5+6 2 | | | | 5+6 2 | 5+6 2 | |
| | 5+6 2 | 3+4 2 | | | | 5+6 2 | 3+4 2 | |
| | 3+4 2 | 3+4 2 | | | | 5+6 2 | 5+6 2 | |
| | 5+6 2 | 5+6 2 | | | | 5+6 2 | 5+6 2 | |
| | 3+4 2 | 5+6 2 | | | | 3+4 2 | 5+6 2 | |
| | 5+6 2 | 5+6 2 | | | | 5+6 2 | 5+6 2 | |
| | 3+4 2 | 3+4 2 | | | | 3+4 2 | 3+4 2 | |
| | 3+4 2 | 3+4 2 | | | | 3+4 2 | 3+4 2 | |
| | 3+4 2 | 5+6 2 | | | | 5+6 2 | 3+4 2 | |
| | 5+6 2 | 3+4 2 | | | | 5+6 2 | 3+4 2 | |
| | 5+6 2 | 3+4 2 | | | | 3+4 2 | 3+4 2 | |
| | 5+6 2 | 5+6 2 | | | | 3+4 2 | 5+6 2 | |
| | 3+4 2 | 5+6 2 | | | | 3+4 2 | 5+6 2 | |
| | 5+6 2 | 3+4 2 | | | | 3+4 2 | 3+4 2 | |
| | 3+4 2 | 5+6 2 | | | | 3+4 2 | 5+6 2 | |
| | 3+4 2 | 5+6 2 | | | | 5+6 2 | 5+6 2 | |
| | 5+6 2 | 3+4 2 | | | | 5+6 2 | 3+4 2 | |
| | 3+4 2 | 3+4 2 | | | | 3+4 2 | 3+4 2 | |
| | 5+6 2 | 5+6 2 | | | | 3+4 2 | 5+6 2 | |
| 11.2.76 | 54553 | 152 | Ce1In2Pt2 | | | | | |
| | -A | -B | -C | -D | -E | -GM | -Y | -Z |
| | 3+4 2 | 3+4 2 | 2+2 4 | 2+2 4 | 2+2 4 | 3+4 2 | 3+4 2 | 2+2 4 |
| | 5+6 2 | 5+6 2 | 2+2 4 | 2+2 4 | 2+2 4 | 5+6 2 | 5+6 2 | 2+2 4 |
| | 3+4 2 | 3+4 2 | 2+2 4 | 2+2 4 | 2+2 4 | 3+4 2 | 3+4 2 | 2+2 4 |
| | 5+6 2 | 5+6 2 | 2+2 4 | 2+2 4 | 2+2 4 | 5+6 2 | 5+6 2 | 2+2 4 |
| | 3+4 2 | 3+4 2 | 2+2 4 | 2+2 4 | 2+2 4 | 3+4 2 | 5+6 2 | 2+2 4 |
| | 5+6 2 | 5+6 2 | 2+2 4 | 2+2 4 | 2+2 4 | 5+6 2 | 3+4 2 | 2+2 4 |
| | 5+6 2 | 3+4 2 | 2+2 4 | 2+2 4 | 2+2 4 | 3+4 2 | 3+4 2 | 2+2 4 |
| | 3+4 2 | 5+6 2 | 2+2 4 | 2+2 4 | 2+2 4 | 5+6 2 | 5+6 2 | 2+2 4 |
| | 3+4 2 | 3+4 2 | 2+2 4 | 2+2 4 | 2+2 4 | 3+4 2 | 5+6 2 | 2+2 4 |
| | 5+6 2 | 5+6 2 | 2+2 4 | 2+2 4 | 2+2 4 | 5+6 2 | 3+4 2 | 2+2 4 |
| | 5+6 2 | 5+6 2 | 2+2 4 | 2+2 4 | 2+2 4 | 3+4 2 | 3+4 2 | 2+2 4 |
| | 3+4 2 | 3+4 2 | 2+2 4 | 2+2 4 | 2+2 4 | 5+6 2 | 5+6 2 | 2+2 4 |
| | 3+4 2 | 3+4 2 | 2+2 4 | 2+2 4 | 2+2 4 | 3+4 2 | 3+4 2 | 2+2 4 |
| | 5+6 2 | 5+6 2 | 2+2 4 | 2+2 4 | 2+2 4 | 5+6 2 | 5+6 2 | 2+2 4 |
| | 3+4 2 | 3+4 2 | 2+2 4 | 2+2 4 | 2+2 4 | 3+4 2 | 5+6 2 | 2+2 4 |
| | 5+6 2 | 5+6 2 | 2+2 4 | 2+2 4 | 2+2 4 | 5+6 2 | 3+4 2 | 2+2 4 |
| | 3+4 2 | 3+4 2 | 2+2 4 | 2+2 4 | 2+2 4 | 3+4 2 | 5+6 2 | 2+2 4 |
| | 5+6 2 | 5+6 2 | 2+2 4 | 2+2 4 | 2+2 4 | 5+6 2 | 3+4 2 | 2+2 4 |
| | 5+6 2 | 5+6 2 | 2+2 4 | 2+2 4 | 2+2 4 | 3+4 2 | 3+4 2 | 2+2 4 |
| | 3+4 2 | 3+4 2 | 2+2 4 | 2+2 4 | 2+2 4 | 3+4 2 | 5+6 2 | 2+2 4 |
| | 5+6 2 | 5+6 2 | 2+2 4 | 2+2 4 | 2+2 4 | 3+4 2 | 5+6 2 | 2+2 4 |
| | 3+4 2 | 3+4 2 | 2+2 4 | 2+2 4 | 2+2 4 | 5+6 2 | 3+4 2 | 2+2 4 |
| | 5+6 2 | 5+6 2 | 2+2 4 | 2+2 4 | 2+2 4 | 5+6 2 | 3+4 2 | 2+2 4 |
| | 3+4 2 | 3+4 2 | 2+2 4 | 2+2 4 | 2+2 4 | 5+6 2 | 5+6 2 | 2+2 4 |
| | 5+6 2 | 3+4 2 | 2+2 4 | 2+2 4 | 2+2 4 | 5+6 2 | 3+4 2 | 2+2 4 |
| | 3+4 2 | 3+4 2 | 2+2 4 | 2+2 4 | 2+2 4 | 3+4 2 | 5+6 2 | 2+2 4 |
| | 5+6 2 | 5+6 2 | 2+2 4 | 2+2 4 | 2+2 4 | 5+6 2 | 5+6 2 | 2+2 4 |
| | 3+4 2 | 5+6 2 | 2+2 4 | 2+2 4 | 2+2 4 | 3+4 2 | 3+4 2 | 2+2 4 |
| | 5+6 2 | 3+4 2 | 2+2 4 | 2+2 4 | 2+2 4 | 3+4 2 | 5+6 2 | 2+2 4 |
| | 3+4 2 | 5+6 2 | 2+2 4 | 2+2 4 | 2+2 4 | 5+6 2 | 3+4 2 | 2+2 4 |
| | 5+6 2 | 5+6 2 | 2+2 4 | 2+2 4 | 2+2 4 | 3+4 2 | 3+4 2 | 2+2 4 |
| | 3+4 2 | 3+4 2 | 2+2 4 | 2+2 4 | 2+2 4 | 3+4 2 | 5+6 2 | 2+2 4 |
| | 5+6 2 | 3+4 2 | 2+2 4 | 2+2 4 | 2+2 4 | 5+6 2 | 3+4 2 | 2+2 4 |
| | 3+4 2 | 5+6 2 | 2+2 4 | 2+2 4 | 2+2 4 | 3+4 2 | 5+6 2 | 2+2 4 |
| | 5+6 2 | 5+6 2 | 2+2 4 | 2+2 4 | 2+2 4 | 5+6 2 | 5+6 2 | 2+2 4 |
| | 3+4 2 | 5+6 2 | 2+2 4 | 2+2 4 | 2+2 4 | 3+4 2 | 3+4 2 | 2+2 4 |
| | 3+4 2 | 3+4 2 | 2+2 4 | 2+2 4 | 2+2 4 | 5+6 2 | 5+6 2 | 2+2 4 |
| | 5+6 2 | 3+4 2 | 2+2 4 | 2+2 4 | 2+2 4 | 3+4 2 | 3+4 2 | 2+2 4 |
| | 3+4 2 | 3+4 2 | | | | 5+6 2 | 5+6 2 | |
| | 5+6 2 | 5+6 2 | | | | 3+4 2 | 5+6 2 | |
| | 5+6 2 | 5+6 2 | | | | 5+6 2 | 3+4 2 | |
| | 3+4 2 | 5+6 2 | | | | 3+4 2 | 5+6 2 | |
| | 5+6 2 | 3+4 2 | | | | 5+6 2 | 3+4 2 | |
| | 3+4 2 | 3+4 2 | | | | 3+4 2 | 3+4 2 | |
| | 5+6 2 | 5+6 2 | | | | 3+4 2 | 3+4 2 | |
| | 3+4 2 | 5+6 2 | | | | 5+6 2 | 5+6 2 | |

Continued on next column

Continued on next column

**TABLE XXIII– continued**

| TQC | ICSD | NoE | SF | | | | | |
|---|---|---|---|---|---|---|---|---|
| | 5+6 2 | 3+4 2 | | | | 5+6 2 | 3+4 2 | |
| | 3+4 2 | 3+4 2 | | | | 3+4 2 | 5+6 2 | |
| | 3+4 2 | 5+6 2 | | | | 3+4 2 | 3+4 2 | |
| | 5+6 2 | 3+4 2 | | | | 5+6 2 | 5+6 2 | |
| | 5+6 2 | 5+6 2 | | | | 5+6 2 | 3+4 2 | |
| | 3+4 2 | 3+4 2 | | | | 3+4 2 | 5+6 2 | |
| | 5+6 2 | 5+6 2 | | | | 3+4 2 | 5+6 2 | |
| | 5+6 2 | 5+6 2 | | | | 5+6 2 | 3+4 2 | |
| | 3+4 2 | 3+4 2 | | | | 5+6 2 | 5+6 2 | |
| | 3+4 2 | 3+4 2 | | | | 3+4 2 | 3+4 2 | |
| | 5+6 2 | 5+6 2 | | | | 5+6 2 | 3+4 2 | |
| | 3+4 2 | 5+6 2 | | | | 3+4 2 | 5+6 2 | |
| | 5+6 2 | 3+4 2 | | | | 5+6 2 | 5+6 2 | |
| | 3+4 2 | 5+6 2 | | | | 3+4 2 | 3+4 2 | |
| | 3+4 2 | 3+4 2 | | | | 5+6 2 | 5+6 2 | |
| | 5+6 2 | 5+6 2 | | | | 3+4 2 | 3+4 2 | |
| | 5+6 2 | 3+4 2 | | | | 5+6 2 | 3+4 2 | |
| | 3+4 2 | 5+6 2 | | | | 3+4 2 | 5+6 2 | |
| | 5+6 2 | 3+4 2 | | | | 5+6 2 | 5+6 2 | |
| | 3+4 2 | 3+4 2 | | | | 3+4 2 | 3+4 2 | |
| | 5+6 2 | 3+4 2 | | | | 5+6 2 | 5+6 2 | |
| | 3+4 2 | 3+4 2 | | | | 3+4 2 | 5+6 2 | |
| | 3+4 2 | 5+6 2 | | | | 5+6 2 | 3+4 2 | |
| | 5+6 2 | 5+6 2 | | | | 3+4 2 | 3+4 2 | |
| | 3+4 2 | 3+4 2 | | | | 3+4 2 | 5+6 2 | |
| | 5+6 2 | 5+6 2 | | | | 3+4 2 | 3+4 2 | |
| | 5+6 2 | 3+4 2 | | | | 5+6 2 | 3+4 2 | |
| | 3+4 2 | 3+4 2 | | | | 3+4 2 | 5+6 2 | |
| | 5+6 2 | 5+6 2 | | | | 5+6 2 | 3+4 2 | |
| | 3+4 2 | 5+6 2 | | | | 5+6 2 | 5+6 2 | |
| 11.2.77 | 862 | 24 | Ca1Sb2 | | | | | |
| | -A | -B | -C | -D | -E | -GM | -Y | -Z |
| | 3+4 2 | 3+4 2 | 2+2 4 | 2+2 4 | 2+2 4 | 3+4 2 | 5+6 2 | 2+2 4 |
| | 3+4 2 | 5+6 2 | 2+2 4 | 2+2 4 | 2+2 4 | 3+4 2 | 3+4 2 | 2+2 4 |
| | 5+6 2 | 3+4 2 | 2+2 4 | 2+2 4 | 2+2 4 | 5+6 2 | 5+6 2 | 2+2 4 |
| | 5+6 2 | 5+6 2 | 2+2 4 | 2+2 4 | 2+2 4 | 5+6 2 | 3+4 2 | 2+2 4 |
| | 5+6 2 | 5+6 2 | 2+2 4 | 2+2 4 | 2+2 4 | 3+4 2 | 3+4 2 | 2+2 4 |
| | 3+4 2 | 3+4 2 | 2+2 4 | 2+2 4 | 2+2 4 | 5+6 2 | 5+6 2 | 2+2 4 |
| | 5+6 2 | 3+4 2 | | | | 3+4 2 | 5+6 2 | |
| | 3+4 2 | 3+4 2 | | | | 5+6 2 | 3+4 2 | |
| | 3+4 2 | 5+6 2 | | | | 3+4 2 | 5+6 2 | |
| | 3+4 2 | 5+6 2 | | | | 3+4 2 | 3+4 2 | |
| | 3+4 2 | 3+4 2 | | | | 3+4 2 | 5+6 2 | |
| | 5+6 2 | 5+6 2 | | | | 3+4 2 | 3+4 2 | |
| 11.2.78 | 62428 | 300 | C4H2As1F17N2 | | | | | |
| | -A | -B | -C | -D | -E | -GM | -Y | -Z |
| | 3+4 2 | 3+4 2 | 2+2 4 | 2+2 4 | 2+2 4 | 3+4 2 | 3+4 2 | 2+2 4 |
| | 5+6 2 | 5+6 2 | 2+2 4 | 2+2 4 | 2+2 4 | 5+6 2 | 5+6 2 | 2+2 4 |
| | 3+4 2 | 3+4 2 | 2+2 4 | 2+2 4 | 2+2 4 | 3+4 2 | 3+4 2 | 2+2 4 |
| | 5+6 2 | 5+6 2 | 2+2 4 | 2+2 4 | 2+2 4 | 5+6 2 | 5+6 2 | 2+2 4 |
| | 3+4 2 | 3+4 2 | 2+2 4 | 2+2 4 | 2+2 4 | 3+4 2 | 3+4 2 | 2+2 4 |
| | 5+6 2 | 5+6 2 | 2+2 4 | 2+2 4 | 2+2 4 | 5+6 2 | 5+6 2 | 2+2 4 |
| | 3+4 2 | 3+4 2 | 2+2 4 | 2+2 4 | 2+2 4 | 3+4 2 | 3+4 2 | 2+2 4 |
| | 5+6 2 | 5+6 2 | 2+2 4 | 2+2 4 | 2+2 4 | 5+6 2 | 5+6 2 | 2+2 4 |
| | 5+6 2 | 5+6 2 | 2+2 4 | 2+2 4 | 2+2 4 | 5+6 2 | 5+6 2 | 2+2 4 |
| | 3+4 2 | 3+4 2 | 2+2 4 | 2+2 4 | 2+2 4 | 3+4 2 | 3+4 2 | 2+2 4 |
| | 3+4 2 | 3+4 2 | 2+2 4 | 2+2 4 | 2+2 4 | 3+4 2 | 3+4 2 | 2+2 4 |
| | 5+6 2 | 5+6 2 | 2+2 4 | 2+2 4 | 2+2 4 | 5+6 2 | 5+6 2 | 2+2 4 |
| | 3+4 2 | 3+4 2 | 2+2 4 | 2+2 4 | 2+2 4 | 5+6 2 | 5+6 2 | 2+2 4 |
| | 5+6 2 | 5+6 2 | 2+2 4 | 2+2 4 | 2+2 4 | 3+4 2 | 3+4 2 | 2+2 4 |
| | 3+4 2 | 5+6 2 | 2+2 4 | 2+2 4 | 2+2 4 | 3+4 2 | 3+4 2 | 2+2 4 |

Continued on next column

**TABLE XXIII– continued**

| TQC | ICSD | NoE | SF | | | | | |
|---|---|---|---|---|---|---|---|---|
| | 5+6 2 | 3+4 2 | 2+2 4 | 2+2 4 | 2+2 4 | 5+6 2 | 5+6 2 | 2+2 4 |
| | 3+4 2 | 3+4 2 | 2+2 4 | 2+2 4 | 2+2 4 | 3+4 2 | 3+4 2 | 2+2 4 |
| | 5+6 2 | 5+6 2 | 2+2 4 | 2+2 4 | 2+2 4 | 5+6 2 | 5+6 2 | 2+2 4 |
| | 3+4 2 | 3+4 2 | 2+2 4 | 2+2 4 | 2+2 4 | 3+4 2 | 3+4 2 | 2+2 4 |
| | 5+6 2 | 5+6 2 | 2+2 4 | 2+2 4 | 2+2 4 | 5+6 2 | 5+6 2 | 2+2 4 |
| | 3+4 2 | 5+6 2 | 2+2 4 | 2+2 4 | 2+2 4 | 3+4 2 | 3+4 2 | 2+2 4 |
| | 5+6 2 | 3+4 2 | 2+2 4 | 2+2 4 | 2+2 4 | 5+6 2 | 5+6 2 | 2+2 4 |
| | 3+4 2 | 3+4 2 | 2+2 4 | 2+2 4 | 2+2 4 | 3+4 2 | 3+4 2 | 2+2 4 |
| | 5+6 2 | 5+6 2 | 2+2 4 | 2+2 4 | 2+2 4 | 5+6 2 | 5+6 2 | 2+2 4 |
| | 3+4 2 | 3+4 2 | 2+2 4 | 2+2 4 | 2+2 4 | 3+4 2 | 3+4 2 | 2+2 4 |
| | 5+6 2 | 5+6 2 | 2+2 4 | 2+2 4 | 2+2 4 | 5+6 2 | 5+6 2 | 2+2 4 |
| | 3+4 2 | 3+4 2 | 2+2 4 | 2+2 4 | 2+2 4 | 3+4 2 | 3+4 2 | 2+2 4 |
| | 5+6 2 | 5+6 2 | 2+2 4 | 2+2 4 | 2+2 4 | 5+6 2 | 5+6 2 | 2+2 4 |
| | 3+4 2 | 3+4 2 | 2+2 4 | 2+2 4 | 2+2 4 | 3+4 2 | 3+4 2 | 2+2 4 |
| | 5+6 2 | 5+6 2 | 2+2 4 | 2+2 4 | 2+2 4 | 5+6 2 | 5+6 2 | 2+2 4 |
| | 3+4 2 | 3+4 2 | 2+2 4 | 2+2 4 | 2+2 4 | 3+4 2 | 3+4 2 | 2+2 4 |
| | 5+6 2 | 5+6 2 | 2+2 4 | 2+2 4 | 2+2 4 | 5+6 2 | 5+6 2 | 2+2 4 |
| | 3+4 2 | 3+4 2 | 2+2 4 | 2+2 4 | 2+2 4 | 3+4 2 | 3+4 2 | 2+2 4 |
| | 5+6 2 | 5+6 2 | 2+2 4 | 2+2 4 | 2+2 4 | 5+6 2 | 5+6 2 | 2+2 4 |
| | 3+4 2 | 3+4 2 | 2+2 4 | 2+2 4 | 2+2 4 | 3+4 2 | 3+4 2 | 2+2 4 |
| | 5+6 2 | 5+6 2 | 2+2 4 | 2+2 4 | 2+2 4 | 5+6 2 | 5+6 2 | 2+2 4 |
| | 3+4 2 | 3+4 2 | 2+2 4 | 2+2 4 | 2+2 4 | 3+4 2 | 3+4 2 | 2+2 4 |
| | 5+6 2 | 5+6 2 | 2+2 4 | 2+2 4 | 2+2 4 | 5+6 2 | 5+6 2 | 2+2 4 |
| | 3+4 2 | 3+4 2 | 2+2 4 | 2+2 4 | 2+2 4 | 3+4 2 | 3+4 2 | 2+2 4 |
| | 5+6 2 | 5+6 2 | 2+2 4 | 2+2 4 | 2+2 4 | 5+6 2 | 5+6 2 | 2+2 4 |
| | 5+6 2 | 3+4 2 | 2+2 4 | 2+2 4 | 2+2 4 | 3+4 2 | 3+4 2 | 2+2 4 |
| | 3+4 2 | 5+6 2 | 2+2 4 | 2+2 4 | 2+2 4 | 5+6 2 | 5+6 2 | 2+2 4 |
| | 3+4 2 | 5+6 2 | 2+2 4 | 2+2 4 | 2+2 4 | 5+6 2 | 3+4 2 | 2+2 4 |
| | 5+6 2 | 3+4 2 | 2+2 4 | 2+2 4 | 2+2 4 | 3+4 2 | 5+6 2 | 2+2 4 |
| | 5+6 2 | 3+4 2 | 2+2 4 | 2+2 4 | 2+2 4 | 3+4 2 | 5+6 2 | 2+2 4 |
| | 3+4 2 | 5+6 2 | 2+2 4 | 2+2 4 | 2+2 4 | 5+6 2 | 3+4 2 | 2+2 4 |
| | 5+6 2 | 3+4 2 | 2+2 4 | 2+2 4 | 2+2 4 | 3+4 2 | 5+6 2 | 2+2 4 |
| | 3+4 2 | 5+6 2 | 2+2 4 | 2+2 4 | 2+2 4 | 5+6 2 | 3+4 2 | 2+2 4 |
| | 5+6 2 | 3+4 2 | 2+2 4 | 2+2 4 | 2+2 4 | 3+4 2 | 3+4 2 | 2+2 4 |
| | 3+4 2 | 5+6 2 | 2+2 4 | 2+2 4 | 2+2 4 | 5+6 2 | 5+6 2 | 2+2 4 |
| | 3+4 2 | 5+6 2 | 2+2 4 | 2+2 4 | 2+2 4 | 5+6 2 | 3+4 2 | 2+2 4 |
| | 5+6 2 | 3+4 2 | 2+2 4 | 2+2 4 | 2+2 4 | 3+4 2 | 5+6 2 | 2+2 4 |
| | 5+6 2 | 5+6 2 | 2+2 4 | 2+2 4 | 2+2 4 | 5+6 2 | 5+6 2 | 2+2 4 |
| | 3+4 2 | 3+4 2 | 2+2 4 | 2+2 4 | 2+2 4 | 3+4 2 | 3+4 2 | 2+2 4 |
| | 5+6 2 | 5+6 2 | 2+2 4 | 2+2 4 | 2+2 4 | 5+6 2 | 5+6 2 | 2+2 4 |
| | 3+4 2 | 3+4 2 | 2+2 4 | 2+2 4 | 2+2 4 | 3+4 2 | 3+4 2 | 2+2 4 |
| | 5+6 2 | 5+6 2 | 2+2 4 | 2+2 4 | 2+2 4 | 5+6 2 | 5+6 2 | 2+2 4 |
| | 3+4 2 | 3+4 2 | 2+2 4 | 2+2 4 | 2+2 4 | 5+6 2 | 3+4 2 | 2+2 4 |
| | 5+6 2 | 5+6 2 | 2+2 4 | 2+2 4 | 2+2 4 | 3+4 2 | 5+6 2 | 2+2 4 |
| | 3+4 2 | 3+4 2 | 2+2 4 | 2+2 4 | 2+2 4 | 3+4 2 | 3+4 2 | 2+2 4 |
| | 3+4 2 | 5+6 2 | 2+2 4 | 2+2 4 | 2+2 4 | 5+6 2 | 3+4 2 | 2+2 4 |
| | 5+6 2 | 3+4 2 | 2+2 4 | 2+2 4 | 2+2 4 | 3+4 2 | 5+6 2 | 2+2 4 |
| | 3+4 2 | 3+4 2 | 2+2 4 | 2+2 4 | 2+2 4 | 5+6 2 | 5+6 2 | 2+2 4 |
| | 5+6 2 | 5+6 2 | 2+2 4 | 2+2 4 | 2+2 4 | 3+4 2 | 3+4 2 | 2+2 4 |
| | 3+4 2 | 5+6 2 | 2+2 4 | 2+2 4 | 2+2 4 | 5+6 2 | 5+6 2 | 2+2 4 |
| | 5+6 2 | 3+4 2 | 2+2 4 | 2+2 4 | 2+2 4 | 3+4 2 | 3+4 2 | 2+2 4 |
| | 5+6 2 | 3+4 2 | 2+2 4 | 2+2 4 | 2+2 4 | 3+4 2 | 5+6 2 | 2+2 4 |
| | 3+4 2 | 5+6 2 | 2+2 4 | 2+2 4 | 2+2 4 | 5+6 2 | 3+4 2 | 2+2 4 |
| | 5+6 2 | 5+6 2 | 2+2 4 | 2+2 4 | 2+2 4 | 3+4 2 | 3+4 2 | 2+2 4 |
| | 3+4 2 | 3+4 2 | 2+2 4 | 2+2 4 | 2+2 4 | 5+6 2 | 5+6 2 | 2+2 4 |
| | 5+6 2 | 5+6 2 | 2+2 4 | 2+2 4 | 2+2 4 | 3+4 2 | 3+4 2 | 2+2 4 |
| | 3+4 2 | 3+4 2 | 2+2 4 | 2+2 4 | 2+2 4 | 5+6 2 | 5+6 2 | 2+2 4 |
| | 5+6 2 | 3+4 2 | 2+2 4 | 2+2 4 | 2+2 4 | 3+4 2 | 3+4 2 | 2+2 4 |
| | 3+4 2 | 5+6 2 | 2+2 4 | 2+2 4 | 2+2 4 | 5+6 2 | 5+6 2 | 2+2 4 |
| | 3+4 2 | 5+6 2 | 2+2 4 | 2+2 4 | 2+2 4 | 3+4 2 | 3+4 2 | 2+2 4 |
| | 5+6 2 | 3+4 2 | | | | 5+6 2 | 5+6 2 | |

Continued on next column

**TABLE XXIII– continued**

| TQC | ICSD | NoE | SF | | | | | |
|---|---|---|---|---|---|---|---|---|
| | 3+4 2 | 3+4 2 | | | | 3+4 2 | 3+4 2 | |
| | 3+4 2 | 3+4 2 | | | | 3+4 2 | 3+4 2 | |
| | 5+6 2 | 5+6 2 | | | | 5+6 2 | 5+6 2 | |
| | 5+6 2 | 5+6 2 | | | | 5+6 2 | 3+4 2 | |
| | 3+4 2 | 5+6 2 | | | | 3+4 2 | 5+6 2 | |
| | 5+6 2 | 3+4 2 | | | | 5+6 2 | 5+6 2 | |
| | 5+6 2 | 5+6 2 | | | | 3+4 2 | 3+4 2 | |
| | 3+4 2 | 3+4 2 | | | | 5+6 2 | 5+6 2 | |
| | 5+6 2 | 3+4 2 | | | | 3+4 2 | 5+6 2 | |
| | 3+4 2 | 5+6 2 | | | | 5+6 2 | 3+4 2 | |
| | 5+6 2 | 3+4 2 | | | | 3+4 2 | 3+4 2 | |
| | 3+4 2 | 5+6 2 | | | | 5+6 2 | 5+6 2 | |
| | 3+4 2 | 3+4 2 | | | | 3+4 2 | 3+4 2 | |
| | 5+6 2 | 5+6 2 | | | | 5+6 2 | 5+6 2 | |
| | 3+4 2 | 5+6 2 | | | | 3+4 2 | 3+4 2 | |
| | 5+6 2 | 3+4 2 | | | | 5+6 2 | 5+6 2 | |
| | 3+4 2 | 3+4 2 | | | | 3+4 2 | 3+4 2 | |
| | 5+6 2 | 5+6 2 | | | | 5+6 2 | 5+6 2 | |
| | 3+4 2 | 5+6 2 | | | | 3+4 2 | 3+4 2 | |
| | 3+4 2 | 3+4 2 | | | | 5+6 2 | 5+6 2 | |
| | 5+6 2 | 3+4 2 | | | | 3+4 2 | 3+4 2 | |
| | 5+6 2 | 3+4 2 | | | | 5+6 2 | 5+6 2 | |
| | 3+4 2 | 5+6 2 | | | | 3+4 2 | 3+4 2 | |
| | 5+6 2 | 5+6 2 | | | | 5+6 2 | 5+6 2 | |
| | 3+4 2 | 3+4 2 | | | | 3+4 2 | 3+4 2 | |
| | 3+4 2 | 5+6 2 | | | | 5+6 2 | 5+6 2 | |
| | 5+6 2 | 5+6 2 | | | | 3+4 2 | 5+6 2 | |
| | 5+6 2 | 3+4 2 | | | | 5+6 2 | 3+4 2 | |
| | 3+4 2 | 5+6 2 | | | | 5+6 2 | 5+6 2 | |
| | 5+6 2 | 3+4 2 | | | | 3+4 2 | 5+6 2 | |
| | 5+6 2 | 5+6 2 | | | | 5+6 2 | 3+4 2 | |
| | 5+6 2 | 3+4 2 | | | | 5+6 2 | 3+4 2 | |
| | 3+4 2 | 3+4 2 | | | | 3+4 2 | 3+4 2 | |
| | 3+4 2 | 5+6 2 | | | | 3+4 2 | 5+6 2 | |
| | 5+6 2 | 5+6 2 | | | | 5+6 2 | 5+6 2 | |
| | 3+4 2 | 3+4 2 | | | | 3+4 2 | 3+4 2 | |
| | 5+6 2 | 3+4 2 | | | | 3+4 2 | 5+6 2 | |
| | 3+4 2 | 5+6 2 | | | | 5+6 2 | 5+6 2 | |
| | 3+4 2 | 5+6 2 | | | | 5+6 2 | 3+4 2 | |
| | 5+6 2 | 3+4 2 | | | | 3+4 2 | 3+4 2 | |
| | 3+4 2 | 3+4 2 | | | | 5+6 2 | 5+6 2 | |
| | 3+4 2 | 5+6 2 | | | | 5+6 2 | 5+6 2 | |
| | 5+6 2 | 3+4 2 | | | | 3+4 2 | 3+4 2 | |
| | 5+6 2 | 5+6 2 | | | | 3+4 2 | 3+4 2 | |
| | 5+6 2 | 5+6 2 | | | | 3+4 2 | 3+4 2 | |
| | 3+4 2 | 3+4 2 | | | | 5+6 2 | 5+6 2 | |
| | 3+4 2 | 3+4 2 | | | | 3+4 2 | 3+4 2 | |
| | 5+6 2 | 5+6 2 | | | | 5+6 2 | 5+6 2 | |
| | 5+6 2 | 3+4 2 | | | | 5+6 2 | 3+4 2 | |
| | 3+4 2 | 5+6 2 | | | | 3+4 2 | 5+6 2 | |
| | 5+6 2 | 3+4 2 | | | | 5+6 2 | 3+4 2 | |
| | 3+4 2 | 5+6 2 | | | | 3+4 2 | 5+6 2 | |
| | 3+4 2 | 5+6 2 | | | | 5+6 2 | 5+6 2 | |
| | 5+6 2 | 3+4 2 | | | | 3+4 2 | 3+4 2 | |
| | 5+6 2 | 3+4 2 | | | | 3+4 2 | 3+4 2 | |
| | 3+4 2 | 5+6 2 | | | | 5+6 2 | 5+6 2 | |
| | 3+4 2 | 5+6 2 | | | | 3+4 2 | 3+4 2 | |
| | 5+6 2 | 3+4 2 | | | | 5+6 2 | 5+6 2 | |
| | 5+6 2 | 5+6 2 | | | | 3+4 2 | 3+4 2 | |
| | 3+4 2 | 3+4 2 | | | | 5+6 2 | 5+6 2 | |
| | 3+4 2 | 5+6 2 | | | | 3+4 2 | 5+6 2 | |
| | 5+6 2 | 3+4 2 | | | | 5+6 2 | 3+4 2 | |
| | 5+6 2 | 5+6 2 | | | | 3+4 2 | 3+4 2 | |
| | 3+4 2 | 3+4 2 | | | | 5+6 2 | 5+6 2 | |
| | 5+6 2 | 3+4 2 | | | | 5+6 2 | 3+4 2 | |
| | 3+4 2 | 5+6 2 | | | | 3+4 2 | 5+6 2 | |
| | 5+6 2 | 3+4 2 | | | | 5+6 2 | 3+4 2 | |
| | 3+4 2 | 5+6 2 | | | | 3+4 2 | 5+6 2 | |
| | 3+4 2 | 5+6 2 | | | | 3+4 2 | 5+6 2 | |
| | 5+6 2 | 3+4 2 | | | | 5+6 2 | 3+4 2 | |
| | 3+4 2 | 3+4 2 | | | | 3+4 2 | 5+6 2 | |
| | 5+6 2 | 5+6 2 | | | | 5+6 2 | 3+4 2 | |
| | 5+6 2 | 3+4 2 | | | | 5+6 2 | 3+4 2 | |
| | 3+4 2 | 3+4 2 | | | | 5+6 2 | 5+6 2 | |
| 11.2.79 | 110059 | 84 | C4H12Cd1Cl3N1 | | | | | |
| | -A | -B | -C | -D | -E | -GM | -Y | -Z |
| | 3+4 2 | 3+4 2 | 2+2 4 | 2+2 4 | 2+2 4 | 3+4 2 | 3+4 2 | 2+2 4 |
| | 5+6 2 | 5+6 2 | 2+2 4 | 2+2 4 | 2+2 4 | 5+6 2 | 5+6 2 | 2+2 4 |
| | 3+4 2 | 3+4 2 | 2+2 4 | 2+2 4 | 2+2 4 | 3+4 2 | 5+6 2 | 2+2 4 |
| | 5+6 2 | 5+6 2 | 2+2 4 | 2+2 4 | 2+2 4 | 5+6 2 | 3+4 2 | 2+2 4 |
| | 3+4 2 | 3+4 2 | 2+2 4 | 2+2 4 | 2+2 4 | 3+4 2 | 3+4 2 | 2+2 4 |
| | 5+6 2 | 5+6 2 | 2+2 4 | 2+2 4 | 2+2 4 | 5+6 2 | 5+6 2 | 2+2 4 |
| | 3+4 2 | 3+4 2 | 2+2 4 | 2+2 4 | 2+2 4 | 3+4 2 | 3+4 2 | 2+2 4 |
| | 5+6 2 | 5+6 2 | 2+2 4 | 2+2 4 | 2+2 4 | 5+6 2 | 5+6 2 | 2+2 4 |
| | 3+4 2 | 3+4 2 | 2+2 4 | 2+2 4 | 2+2 4 | 3+4 2 | 3+4 2 | 2+2 4 |
| | 3+4 2 | 3+4 2 | 2+2 4 | 2+2 4 | 2+2 4 | 3+4 2 | 3+4 2 | 2+2 4 |
| | 3+4 2 | 3+4 2 | 2+2 4 | 2+2 4 | 2+2 4 | 3+4 2 | 3+4 2 | 2+2 4 |
| | 3+4 2 | 3+4 2 | 2+2 4 | 2+2 4 | 2+2 4 | 3+4 2 | 3+4 2 | 2+2 4 |
| | 3+4 2 | 3+4 2 | 2+2 4 | 2+2 4 | 2+2 4 | 3+4 2 | 3+4 2 | 2+2 4 |
| | 3+4 2 | 3+4 2 | 2+2 4 | 2+2 4 | 2+2 4 | 3+4 2 | 3+4 2 | 2+2 4 |
| | 3+4 2 | 3+4 2 | 2+2 4 | 2+2 4 | 2+2 4 | 3+4 2 | 3+4 2 | 2+2 4 |
| | 3+4 2 | 3+4 2 | 2+2 4 | 2+2 4 | 2+2 4 | 3+4 2 | 3+4 2 | 2+2 4 |
| | 3+4 2 | 3+4 2 | 2+2 4 | 2+2 4 | 2+2 4 | 3+4 2 | 3+4 2 | 2+2 4 |
| | 3+4 2 | 3+4 2 | 2+2 4 | 2+2 4 | 2+2 4 | 3+4 2 | 3+4 2 | 2+2 4 |
| | 3+4 2 | 5+6 2 | 2+2 4 | 2+2 4 | 2+2 4 | 3+4 2 | 5+6 2 | 2+2 4 |
| | 5+6 2 | 3+4 2 | 2+2 4 | 2+2 4 | 2+2 4 | 5+6 2 | 3+4 2 | 2+2 4 |
| | 3+4 2 | 3+4 2 | 2+2 4 | 2+2 4 | 2+2 4 | 3+4 2 | 3+4 2 | 2+2 4 |
| | 3+4 2 | 3+4 2 | | | | 3+4 2 | 3+4 2 | |
| | 5+6 2 | 5+6 2 | | | | 5+6 2 | 5+6 2 | |
| | 5+6 2 | 5+6 2 | | | | 5+6 2 | 5+6 2 | |
| | 5+6 2 | 5+6 2 | | | | 5+6 2 | 5+6 2 | |
| | 5+6 2 | 5+6 2 | | | | 5+6 2 | 5+6 2 | |
| | 5+6 2 | 5+6 2 | | | | 5+6 2 | 5+6 2 | |
| | 3+4 2 | 3+4 2 | | | | 3+4 2 | 3+4 2 | |
| | 3+4 2 | 3+4 2 | | | | 3+4 2 | 3+4 2 | |
| | 5+6 2 | 5+6 2 | | | | 5+6 2 | 5+6 2 | |
| | 5+6 2 | 5+6 2 | | | | 5+6 2 | 5+6 2 | |
| | 5+6 2 | 5+6 2 | | | | 5+6 2 | 5+6 2 | |
| | 5+6 2 | 5+6 2 | | | | 5+6 2 | 5+6 2 | |
| | 3+4 2 | 3+4 2 | | | | 3+4 2 | 3+4 2 | |
| | 3+4 2 | 3+4 2 | | | | 3+4 2 | 3+4 2 | |
| | 3+4 2 | 5+6 2 | | | | 3+4 2 | 5+6 2 | |
| | 5+6 2 | 3+4 2 | | | | 5+6 2 | 3+4 2 | |
| | 3+4 2 | 3+4 2 | | | | 5+6 2 | 3+4 2 | |
| | 3+4 2 | 5+6 2 | | | | 3+4 2 | 5+6 2 | |
| | 5+6 2 | 5+6 2 | | | | 3+4 2 | 3+4 2 | |
| | 3+4 2 | 3+4 2 | | | | 3+4 2 | 3+4 2 | |
| | 3+4 2 | 3+4 2 | | | | 5+6 2 | 5+6 2 | |
| 11.2.80 | 110260 | 48 | C3H10Cl1N1 | | | | | |
| | -A | -B | -C | -D | -E | -GM | -Y | -Z |
| | 3+4 2 | 3+4 2 | 2+2 4 | 2+2 4 | 2+2 4 | 3+4 2 | 3+4 2 | 2+2 4 |
| | 5+6 2 | 5+6 2 | 2+2 4 | 2+2 4 | 2+2 4 | 5+6 2 | 5+6 2 | 2+2 4 |

Continued on next column

**TABLE XXIII– continued**

| TQC | ICSD | NoE | SF | | | | | |
|---|---|---|---|---|---|---|---|---|
| | 5+6 2 | 3+4 2 | 2+2 4 | 2+2 4 | 2+2 4 | 3+4 2 | 5+6 2 | 2+2 4 |
| | 3+4 2 | 5+6 2 | 2+2 4 | 2+2 4 | 2+2 4 | 5+6 2 | 3+4 2 | 2+2 4 |
| | 3+4 2 | 3+4 2 | 2+2 4 | 2+2 4 | 2+2 4 | 3+4 2 | 3+4 2 | 2+2 4 |
| | 5+6 2 | 5+6 2 | 2+2 4 | 2+2 4 | 2+2 4 | 5+6 2 | 5+6 2 | 2+2 4 |
| | 3+4 2 | 3+4 2 | 2+2 4 | 2+2 4 | 2+2 4 | 3+4 2 | 3+4 2 | 2+2 4 |
| | 5+6 2 | 5+6 2 | 2+2 4 | 2+2 4 | 2+2 4 | 5+6 2 | 5+6 2 | 2+2 4 |
| | 3+4 2 | 5+6 2 | 2+2 4 | 2+2 4 | 2+2 4 | 3+4 2 | 5+6 2 | 2+2 4 |
| | 5+6 2 | 3+4 2 | 2+2 4 | 2+2 4 | 2+2 4 | 5+6 2 | 3+4 2 | 2+2 4 |
| | 3+4 2 | 5+6 2 | 2+2 4 | 2+2 4 | 2+2 4 | 3+4 2 | 3+4 2 | 2+2 4 |
| | 5+6 2 | 3+4 2 | 2+2 4 | 2+2 4 | 2+2 4 | 5+6 2 | 5+6 2 | 2+2 4 |
| | 5+6 2 | 3+4 2 | | | | 5+6 2 | 3+4 2 | |
| | 5+6 2 | 5+6 2 | | | | 3+4 2 | 3+4 2 | |
| | 3+4 2 | 3+4 2 | | | | 5+6 2 | 5+6 2 | |
| | 3+4 2 | 5+6 2 | | | | 3+4 2 | 5+6 2 | |
| | 3+4 2 | 5+6 2 | | | | 3+4 2 | 5+6 2 | |
| | 5+6 2 | 3+4 2 | | | | 3+4 2 | 3+4 2 | |
| | 3+4 2 | 5+6 2 | | | | 5+6 2 | 5+6 2 | |
| | 5+6 2 | 3+4 2 | | | | 5+6 2 | 3+4 2 | |
| | 3+4 2 | 5+6 2 | | | | 3+4 2 | 5+6 2 | |
| | 5+6 2 | 3+4 2 | | | | 5+6 2 | 3+4 2 | |
| | 5+6 2 | 3+4 2 | | | | 3+4 2 | 5+6 2 | |
| | 3+4 2 | 3+4 2 | | | | 3+4 2 | 3+4 2 | |
| 11.2.81 | 20741 | 188 | C2H8N4O7Se1U1 | | | | | |
| | -A | -B | -C | -D | -E | -GM | -Y | -Z |
| | 3+4 2 | 3+4 2 | 2+2 4 | 2+2 4 | 2+2 4 | 3+4 2 | 3+4 2 | 2+2 4 |
| | 5+6 2 | 5+6 2 | 2+2 4 | 2+2 4 | 2+2 4 | 5+6 2 | 5+6 2 | 2+2 4 |
| | 3+4 2 | 3+4 2 | 2+2 4 | 2+2 4 | 2+2 4 | 3+4 2 | 3+4 2 | 2+2 4 |
| | 5+6 2 | 5+6 2 | 2+2 4 | 2+2 4 | 2+2 4 | 5+6 2 | 5+6 2 | 2+2 4 |
| | 3+4 2 | 3+4 2 | 2+2 4 | 2+2 4 | 2+2 4 | 3+4 2 | 3+4 2 | 2+2 4 |
| | 5+6 2 | 5+6 2 | 2+2 4 | 2+2 4 | 2+2 4 | 5+6 2 | 5+6 2 | 2+2 4 |
| | 3+4 2 | 5+6 2 | 2+2 4 | 2+2 4 | 2+2 4 | 3+4 2 | 3+4 2 | 2+2 4 |
| | 5+6 2 | 3+4 2 | 2+2 4 | 2+2 4 | 2+2 4 | 5+6 2 | 5+6 2 | 2+2 4 |
| | 3+4 2 | 5+6 2 | 2+2 4 | 2+2 4 | 2+2 4 | 3+4 2 | 3+4 2 | 2+2 4 |
| | 5+6 2 | 3+4 2 | 2+2 4 | 2+2 4 | 2+2 4 | 5+6 2 | 5+6 2 | 2+2 4 |
| | 3+4 2 | 3+4 2 | 2+2 4 | 2+2 4 | 2+2 4 | 3+4 2 | 3+4 2 | 2+2 4 |
| | 5+6 2 | 5+6 2 | 2+2 4 | 2+2 4 | 2+2 4 | 5+6 2 | 5+6 2 | 2+2 4 |
| | 3+4 2 | 3+4 2 | 2+2 4 | 2+2 4 | 2+2 4 | 3+4 2 | 3+4 2 | 2+2 4 |
| | 5+6 2 | 5+6 2 | 2+2 4 | 2+2 4 | 2+2 4 | 5+6 2 | 5+6 2 | 2+2 4 |
| | 3+4 2 | 5+6 2 | 2+2 4 | 2+2 4 | 2+2 4 | 3+4 2 | 3+4 2 | 2+2 4 |
| | 5+6 2 | 3+4 2 | 2+2 4 | 2+2 4 | 2+2 4 | 5+6 2 | 5+6 2 | 2+2 4 |
| | 3+4 2 | 5+6 2 | 2+2 4 | 2+2 4 | 2+2 4 | 3+4 2 | 3+4 2 | 2+2 4 |
| | 5+6 2 | 3+4 2 | 2+2 4 | 2+2 4 | 2+2 4 | 5+6 2 | 5+6 2 | 2+2 4 |
| | 3+4 2 | 3+4 2 | 2+2 4 | 2+2 4 | 2+2 4 | 3+4 2 | 3+4 2 | 2+2 4 |
| | 5+6 2 | 5+6 2 | 2+2 4 | 2+2 4 | 2+2 4 | 5+6 2 | 5+6 2 | 2+2 4 |
| | 3+4 2 | 5+6 2 | 2+2 4 | 2+2 4 | 2+2 4 | 3+4 2 | 3+4 2 | 2+2 4 |
| | 5+6 2 | 3+4 2 | 2+2 4 | 2+2 4 | 2+2 4 | 5+6 2 | 5+6 2 | 2+2 4 |
| | 3+4 2 | 5+6 2 | 2+2 4 | 2+2 4 | 2+2 4 | 3+4 2 | 5+6 2 | 2+2 4 |
| | 5+6 2 | 3+4 2 | 2+2 4 | 2+2 4 | 2+2 4 | 5+6 2 | 3+4 2 | 2+2 4 |
| | 3+4 2 | 5+6 2 | 2+2 4 | 2+2 4 | 2+2 4 | 3+4 2 | 5+6 2 | 2+2 4 |
| | 5+6 2 | 3+4 2 | 2+2 4 | 2+2 4 | 2+2 4 | 5+6 2 | 3+4 2 | 2+2 4 |
| | 5+6 2 | 3+4 2 | 2+2 4 | 2+2 4 | 2+2 4 | 5+6 2 | 3+4 2 | 2+2 4 |
| | 3+4 2 | 3+4 2 | 2+2 4 | 2+2 4 | 2+2 4 | 5+6 2 | 5+6 2 | 2+2 4 |
| | 5+6 2 | 5+6 2 | 2+2 4 | 2+2 4 | 2+2 4 | 3+4 2 | 3+4 2 | 2+2 4 |
| | 3+4 2 | 5+6 2 | 2+2 4 | 2+2 4 | 2+2 4 | 3+4 2 | 5+6 2 | 2+2 4 |
| | 3+4 2 | 3+4 2 | 2+2 4 | 2+2 4 | 2+2 4 | 3+4 2 | 3+4 2 | 2+2 4 |
| | 5+6 2 | 5+6 2 | 2+2 4 | 2+2 4 | 2+2 4 | 5+6 2 | 5+6 2 | 2+2 4 |
| | 5+6 2 | 5+6 2 | 2+2 4 | 2+2 4 | 2+2 4 | 3+4 2 | 3+4 2 | 2+2 4 |
| | 3+4 2 | 3+4 2 | 2+2 4 | 2+2 4 | 2+2 4 | 5+6 2 | 5+6 2 | 2+2 4 |
| | 5+6 2 | 5+6 2 | 2+2 4 | 2+2 4 | 2+2 4 | 3+4 2 | 5+6 2 | 2+2 4 |
| | 3+4 2 | 3+4 2 | 2+2 4 | 2+2 4 | 2+2 4 | 5+6 2 | 3+4 2 | 2+2 4 |
| | 5+6 2 | 5+6 2 | 2+2 4 | 2+2 4 | 2+2 4 | 3+4 2 | 3+4 2 | 2+2 4 |
| | 3+4 2 | 3+4 2 | 2+2 4 | 2+2 4 | 2+2 4 | 5+6 2 | 5+6 2 | 2+2 4 |
| | 3+4 2 | 3+4 2 | 2+2 4 | 2+2 4 | 2+2 4 | 5+6 2 | 5+6 2 | 2+2 4 |
| | 5+6 2 | 5+6 2 | 2+2 4 | 2+2 4 | 2+2 4 | 3+4 2 | 3+4 2 | 2+2 4 |
| | 5+6 2 | 5+6 2 | 2+2 4 | 2+2 4 | 2+2 4 | 3+4 2 | 3+4 2 | 2+2 4 |
| | 3+4 2 | 3+4 2 | 2+2 4 | 2+2 4 | 2+2 4 | 5+6 2 | 5+6 2 | 2+2 4 |
| | 3+4 2 | 5+6 2 | 2+2 4 | 2+2 4 | 2+2 4 | 3+4 2 | 5+6 2 | 2+2 4 |
| | 5+6 2 | 3+4 2 | 2+2 4 | 2+2 4 | 2+2 4 | 5+6 2 | 3+4 2 | 2+2 4 |
| | 3+4 2 | 3+4 2 | 2+2 4 | 2+2 4 | 2+2 4 | 3+4 2 | 3+4 2 | 2+2 4 |
| | 5+6 2 | 5+6 2 | 2+2 4 | 2+2 4 | 2+2 4 | 5+6 2 | 5+6 2 | 2+2 4 |
| | 3+4 2 | 5+6 2 | 2+2 4 | 2+2 4 | 2+2 4 | 3+4 2 | 5+6 2 | 2+2 4 |
| | 5+6 2 | 3+4 2 | | | | 5+6 2 | 3+4 2 | |
| | 3+4 2 | 5+6 2 | | | | 3+4 2 | 5+6 2 | |
| | 5+6 2 | 3+4 2 | | | | 5+6 2 | 3+4 2 | |
| | 5+6 2 | 5+6 2 | | | | 3+4 2 | 3+4 2 | |
| | 5+6 2 | 5+6 2 | | | | 3+4 2 | 3+4 2 | |
| | 3+4 2 | 3+4 2 | | | | 5+6 2 | 5+6 2 | |
| | 3+4 2 | 3+4 2 | | | | 5+6 2 | 5+6 2 | |
| | 3+4 2 | 5+6 2 | | | | 3+4 2 | 5+6 2 | |
| | 3+4 2 | 3+4 2 | | | | 5+6 2 | 5+6 2 | |
| | 5+6 2 | 5+6 2 | | | | 3+4 2 | 3+4 2 | |
| | 5+6 2 | 3+4 2 | | | | 5+6 2 | 3+4 2 | |
| | 3+4 2 | 3+4 2 | | | | 5+6 2 | 5+6 2 | |
| | 5+6 2 | 5+6 2 | | | | 3+4 2 | 3+4 2 | |
| | 5+6 2 | 3+4 2 | | | | 3+4 2 | 3+4 2 | |
| | 5+6 2 | 5+6 2 | | | | 5+6 2 | 3+4 2 | |
| | 3+4 2 | 3+4 2 | | | | 3+4 2 | 5+6 2 | |
| | 3+4 2 | 5+6 2 | | | | 5+6 2 | 5+6 2 | |
| | 3+4 2 | 3+4 2 | | | | 5+6 2 | 5+6 2 | |
| | 5+6 2 | 5+6 2 | | | | 3+4 2 | 3+4 2 | |
| | 3+4 2 | 5+6 2 | | | | 3+4 2 | 3+4 2 | |
| | 5+6 2 | 3+4 2 | | | | 5+6 2 | 5+6 2 | |
| | 3+4 2 | 3+4 2 | | | | 3+4 2 | 5+6 2 | |
| | 5+6 2 | 5+6 2 | | | | 5+6 2 | 3+4 2 | |
| | 5+6 2 | 5+6 2 | | | | 3+4 2 | 5+6 2 | |
| | 3+4 2 | 5+6 2 | | | | 3+4 2 | 3+4 2 | |
| | 5+6 2 | 3+4 2 | | | | 5+6 2 | 3+4 2 | |
| | 3+4 2 | 3+4 2 | | | | 5+6 2 | 5+6 2 | |
| | 5+6 2 | 5+6 2 | | | | 3+4 2 | 3+4 2 | |
| | 3+4 2 | 3+4 2 | | | | 5+6 2 | 5+6 2 | |
| | 3+4 2 | 5+6 2 | | | | 3+4 2 | 5+6 2 | |
| | 5+6 2 | 3+4 2 | | | | 5+6 2 | 3+4 2 | |
| | 5+6 2 | 3+4 2 | | | | 3+4 2 | 3+4 2 | |
| | 3+4 2 | 5+6 2 | | | | 5+6 2 | 5+6 2 | |
| | 5+6 2 | 5+6 2 | | | | 3+4 2 | 5+6 2 | |
| | 3+4 2 | 3+4 2 | | | | 5+6 2 | 3+4 2 | |
| | 3+4 2 | 3+4 2 | | | | 3+4 2 | 5+6 2 | |
| | 5+6 2 | 5+6 2 | | | | 5+6 2 | 3+4 2 | |
| | 3+4 2 | 3+4 2 | | | | 5+6 2 | 5+6 2 | |
| | 5+6 2 | 5+6 2 | | | | 3+4 2 | 3+4 2 | |
| | 5+6 2 | 3+4 2 | | | | 5+6 2 | 5+6 2 | |
| | 3+4 2 | 5+6 2 | | | | 3+4 2 | 3+4 2 | |
| | 5+6 2 | 5+6 2 | | | | 5+6 2 | 3+4 2 | |
| | 3+4 2 | 3+4 2 | | | | 3+4 2 | 5+6 2 | |
| | 5+6 2 | 5+6 2 | | | | 5+6 2 | 3+4 2 | |
| | 3+4 2 | 3+4 2 | | | | 3+4 2 | 5+6 2 | |
| | 3+4 2 | 5+6 2 | | | | 5+6 2 | 3+4 2 | |
| | 5+6 2 | 3+4 2 | | | | 5+6 2 | 5+6 2 | |
| 11.2.82 | 416300 | 80 | Bi2Pd2Sr1 | | | | | |
| | -A | -B | -C | -D | -E | -GM | -Y | -Z |
| | 3+4 2 | 3+4 2 | 2+2 4 | 2+2 4 | 2+2 4 | 3+4 2 | 3+4 2 | 2+2 4 |
| | 5+6 2 | 5+6 2 | 2+2 4 | 2+2 4 | 2+2 4 | 5+6 2 | 5+6 2 | 2+2 4 |

Continued on next column

Continued on next column

**TABLE XXIII– continued**

| TQC | ICSD | NoE | SF | | | | | |
|---|---|---|---|---|---|---|---|---|
| | 3+4 2 | 3+4 2 | 2+2 4 | 2+2 4 | 2+2 4 | 3+4 2 | 3+4 2 | 2+2 4 |
| | 5+6 2 | 5+6 2 | 2+2 4 | 2+2 4 | 2+2 4 | 5+6 2 | 5+6 2 | 2+2 4 |
| | 3+4 2 | 3+4 2 | 2+2 4 | 2+2 4 | 2+2 4 | 3+4 2 | 3+4 2 | 2+2 4 |
| | 5+6 2 | 5+6 2 | 2+2 4 | 2+2 4 | 2+2 4 | 5+6 2 | 5+6 2 | 2+2 4 |
| | 5+6 2 | 3+4 2 | 2+2 4 | 2+2 4 | 2+2 4 | 3+4 2 | 3+4 2 | 2+2 4 |
| | 3+4 2 | 5+6 2 | 2+2 4 | 2+2 4 | 2+2 4 | 5+6 2 | 5+6 2 | 2+2 4 |
| | 3+4 2 | 5+6 2 | 2+2 4 | 2+2 4 | 2+2 4 | 3+4 2 | 3+4 2 | 2+2 4 |
| | 5+6 2 | 3+4 2 | 2+2 4 | 2+2 4 | 2+2 4 | 3+4 2 | 5+6 2 | 2+2 4 |
| | 3+4 2 | 5+6 2 | 2+2 4 | 2+2 4 | 2+2 4 | 5+6 2 | 3+4 2 | 2+2 4 |
| | 5+6 2 | 3+4 2 | 2+2 4 | 2+2 4 | 2+2 4 | 5+6 2 | 5+6 2 | 2+2 4 |
| | 5+6 2 | 3+4 2 | 2+2 4 | 2+2 4 | 2+2 4 | 5+6 2 | 5+6 2 | 2+2 4 |
| | 3+4 2 | 5+6 2 | 2+2 4 | 2+2 4 | 2+2 4 | 3+4 2 | 3+4 2 | 2+2 4 |
| | 3+4 2 | 5+6 2 | 2+2 4 | 2+2 4 | 2+2 4 | 3+4 2 | 3+4 2 | 2+2 4 |
| | 3+4 2 | 3+4 2 | 2+2 4 | 2+2 4 | 2+2 4 | 5+6 2 | 5+6 2 | 2+2 4 |
| | 5+6 2 | 5+6 2 | 2+2 4 | 2+2 4 | 2+2 4 | 3+4 2 | 3+4 2 | 2+2 4 |
| | 3+4 2 | 5+6 2 | 2+2 4 | 2+2 4 | 2+2 4 | 5+6 2 | 3+4 2 | 2+2 4 |
| | 5+6 2 | 5+6 2 | 2+2 4 | 2+2 4 | 2+2 4 | 5+6 2 | 5+6 2 | 2+2 4 |
| | 5+6 2 | 3+4 2 | 2+2 4 | 2+2 4 | 2+2 4 | 3+4 2 | 3+4 2 | 2+2 4 |
| | 3+4 2 | 5+6 2 | | | | 3+4 2 | 5+6 2 | |
| | 5+6 2 | 3+4 2 | | | | 3+4 2 | 5+6 2 | |
| | 5+6 2 | 3+4 2 | | | | 5+6 2 | 5+6 2 | |
| | 3+4 2 | 3+4 2 | | | | 5+6 2 | 3+4 2 | |
| | 5+6 2 | 5+6 2 | | | | 3+4 2 | 5+6 2 | |
| | 3+4 2 | 3+4 2 | | | | 3+4 2 | 3+4 2 | |
| | 3+4 2 | 5+6 2 | | | | 3+4 2 | 3+4 2 | |
| | 5+6 2 | 3+4 2 | | | | 5+6 2 | 5+6 2 | |
| | 3+4 2 | 3+4 2 | | | | 5+6 2 | 3+4 2 | |
| | 5+6 2 | 5+6 2 | | | | 3+4 2 | 3+4 2 | |
| | 3+4 2 | 3+4 2 | | | | 5+6 2 | 5+6 2 | |
| | 5+6 2 | 5+6 2 | | | | 5+6 2 | 5+6 2 | |
| | 3+4 2 | 3+4 2 | | | | 5+6 2 | 3+4 2 | |
| | 5+6 2 | 3+4 2 | | | | 3+4 2 | 5+6 2 | |
| | 5+6 2 | 5+6 2 | | | | 3+4 2 | 5+6 2 | |
| | 3+4 2 | 5+6 2 | | | | 3+4 2 | 3+4 2 | |
| | 5+6 2 | 3+4 2 | | | | 3+4 2 | 5+6 2 | |
| | 3+4 2 | 3+4 2 | | | | 3+4 2 | 3+4 2 | |
| | 5+6 2 | 5+6 2 | | | | 5+6 2 | 3+4 2 | |
| | 3+4 2 | 5+6 2 | | | | 3+4 2 | 5+6 2 | |
| 11.2.83 | 42679 | 20 | Bi1 | | | | | |
| | -A | -B | -C | -D | -E | -GM | -Y | -Z |
| | 3+4 2 | 5+6 2 | 2+2 4 | 2+2 4 | 2+2 4 | 3+4 2 | 5+6 2 | 2+2 4 |
| | 5+6 2 | 3+4 2 | 2+2 4 | 2+2 4 | 2+2 4 | 5+6 2 | 3+4 2 | 2+2 4 |
| | 3+4 2 | 3+4 2 | 2+2 4 | 2+2 4 | 2+2 4 | 3+4 2 | 5+6 2 | 2+2 4 |
| | 5+6 2 | 5+6 2 | 2+2 4 | 2+2 4 | 2+2 4 | 5+6 2 | 3+4 2 | 2+2 4 |
| | 5+6 2 | 5+6 2 | 2+2 4 | 2+2 4 | 2+2 4 | 3+4 2 | 3+4 2 | 2+2 4 |
| | 3+4 2 | 3+4 2 | | | | 3+4 2 | 5+6 2 | |
| | 3+4 2 | 5+6 2 | | | | 5+6 2 | 3+4 2 | |
| | 5+6 2 | 5+6 2 | | | | 3+4 2 | 5+6 2 | |
| | 3+4 2 | 3+4 2 | | | | 5+6 2 | 3+4 2 | |
| | 3+4 2 | 3+4 2 | | | | 5+6 2 | 5+6 2 | |
| 11.2.84 | 81819 | 96 | Ba3Ge2N2 | | | | | |
| | -A | -B | -C | -D | -E | -GM | -Y | -Z |
| | 5+6 2 | 3+4 2 | 2+2 4 | 2+2 4 | 2+2 4 | 3+4 2 | 5+6 2 | 2+2 4 |
| | 3+4 2 | 5+6 2 | 2+2 4 | 2+2 4 | 2+2 4 | 5+6 2 | 3+4 2 | 2+2 4 |
| | 5+6 2 | 5+6 2 | 2+2 4 | 2+2 4 | 2+2 4 | 3+4 2 | 3+4 2 | 2+2 4 |
| | 3+4 2 | 3+4 2 | 2+2 4 | 2+2 4 | 2+2 4 | 3+4 2 | 3+4 2 | 2+2 4 |
| | 3+4 2 | 3+4 2 | 2+2 4 | 2+2 4 | 2+2 4 | 5+6 2 | 5+6 2 | 2+2 4 |
| | 5+6 2 | 5+6 2 | 2+2 4 | 2+2 4 | 2+2 4 | 5+6 2 | 5+6 2 | 2+2 4 |
| | 5+6 2 | 5+6 2 | 2+2 4 | 2+2 4 | 2+2 4 | 5+6 2 | 5+6 2 | 2+2 4 |
| | 3+4 2 | 3+4 2 | 2+2 4 | 2+2 4 | 2+2 4 | 3+4 2 | 3+4 2 | 2+2 4 |
| | 3+4 2 | 3+4 2 | 2+2 4 | 2+2 4 | 2+2 4 | 3+4 2 | 3+4 2 | 2+2 4 |

Continued on next column

**TABLE XXIII– continued**

| TQC | ICSD | NoE | SF | | | | | |
|---|---|---|---|---|---|---|---|---|
| | 5+6 2 | 5+6 2 | 2+2 4 | 2+2 4 | 2+2 4 | 5+6 2 | 5+6 2 | 2+2 4 |
| | 5+6 2 | 3+4 2 | 2+2 4 | 2+2 4 | 2+2 4 | 3+4 2 | 3+4 2 | 2+2 4 |
| | 3+4 2 | 5+6 2 | 2+2 4 | 2+2 4 | 2+2 4 | 5+6 2 | 5+6 2 | 2+2 4 |
| | 3+4 2 | 3+4 2 | 2+2 4 | 2+2 4 | 2+2 4 | 3+4 2 | 3+4 2 | 2+2 4 |
| | 5+6 2 | 5+6 2 | 2+2 4 | 2+2 4 | 2+2 4 | 5+6 2 | 5+6 2 | 2+2 4 |
| | 5+6 2 | 5+6 2 | 2+2 4 | 2+2 4 | 2+2 4 | 3+4 2 | 5+6 2 | 2+2 4 |
| | 3+4 2 | 3+4 2 | 2+2 4 | 2+2 4 | 2+2 4 | 5+6 2 | 3+4 2 | 2+2 4 |
| | 5+6 2 | 3+4 2 | 2+2 4 | 2+2 4 | 2+2 4 | 3+4 2 | 3+4 2 | 2+2 4 |
| | 3+4 2 | 5+6 2 | 2+2 4 | 2+2 4 | 2+2 4 | 5+6 2 | 5+6 2 | 2+2 4 |
| | 5+6 2 | 3+4 2 | 2+2 4 | 2+2 4 | 2+2 4 | 3+4 2 | 3+4 2 | 2+2 4 |
| | 3+4 2 | 5+6 2 | 2+2 4 | 2+2 4 | 2+2 4 | 3+4 2 | 3+4 2 | 2+2 4 |
| | 5+6 2 | 5+6 2 | 2+2 4 | 2+2 4 | 2+2 4 | 5+6 2 | 5+6 2 | 2+2 4 |
| | 3+4 2 | 3+4 2 | 2+2 4 | 2+2 4 | 2+2 4 | 3+4 2 | 3+4 2 | 2+2 4 |
| | 3+4 2 | 5+6 2 | 2+2 4 | 2+2 4 | 2+2 4 | 3+4 2 | 5+6 2 | 2+2 4 |
| | 5+6 2 | 3+4 2 | 2+2 4 | 2+2 4 | 2+2 4 | 5+6 2 | 3+4 2 | 2+2 4 |
| | 3+4 2 | 3+4 2 | | | | 5+6 2 | 5+6 2 | |
| | 5+6 2 | 5+6 2 | | | | 5+6 2 | 5+6 2 | |
| | 5+6 2 | 5+6 2 | | | | 5+6 2 | 5+6 2 | |
| | 3+4 2 | 3+4 2 | | | | 3+4 2 | 3+4 2 | |
| | 3+4 2 | 5+6 2 | | | | 3+4 2 | 5+6 2 | |
| | 3+4 2 | 3+4 2 | | | | 5+6 2 | 5+6 2 | |
| | 5+6 2 | 5+6 2 | | | | 3+4 2 | 3+4 2 | |
| | 5+6 2 | 3+4 2 | | | | 5+6 2 | 3+4 2 | |
| | 5+6 2 | 3+4 2 | | | | 5+6 2 | 5+6 2 | |
| | 3+4 2 | 5+6 2 | | | | 3+4 2 | 5+6 2 | |
| | 5+6 2 | 3+4 2 | | | | 5+6 2 | 3+4 2 | |
| | 3+4 2 | 5+6 2 | | | | 3+4 2 | 3+4 2 | |
| | 5+6 2 | 5+6 2 | | | | 5+6 2 | 5+6 2 | |
| | 3+4 2 | 3+4 2 | | | | 3+4 2 | 3+4 2 | |
| | 5+6 2 | 3+4 2 | | | | 5+6 2 | 3+4 2 | |
| | 3+4 2 | 3+4 2 | | | | 3+4 2 | 3+4 2 | |
| | 5+6 2 | 5+6 2 | | | | 3+4 2 | 5+6 2 | |
| | 3+4 2 | 3+4 2 | | | | 5+6 2 | 5+6 2 | |
| | 5+6 2 | 5+6 2 | | | | 3+4 2 | 3+4 2 | |
| | 5+6 2 | 3+4 2 | | | | 5+6 2 | 5+6 2 | |
| | 3+4 2 | 5+6 2 | | | | 3+4 2 | 3+4 2 | |
| | 3+4 2 | 5+6 2 | | | | 5+6 2 | 5+6 2 | |
| | 3+4 2 | 5+6 2 | | | | 3+4 2 | 5+6 2 | |
| | 3+4 2 | 5+6 2 | | | | 5+6 2 | 3+4 2 | |
| 11.2.85 | 409517 | 40 | Ba1Sb2 | | | | | |
| | -A | -B | -C | -D | -E | -GM | -Y | -Z |
| | 5+6 2 | 5+6 2 | 2+2 4 | 2+2 4 | 2+2 4 | 3+4 2 | 3+4 2 | 2+2 4 |
| | 3+4 2 | 3+4 2 | 2+2 4 | 2+2 4 | 2+2 4 | 5+6 2 | 5+6 2 | 2+2 4 |
| | 5+6 2 | 5+6 2 | 2+2 4 | 2+2 4 | 2+2 4 | 3+4 2 | 3+4 2 | 2+2 4 |
| | 3+4 2 | 3+4 2 | 2+2 4 | 2+2 4 | 2+2 4 | 5+6 2 | 5+6 2 | 2+2 4 |
| | 3+4 2 | 5+6 2 | 2+2 4 | 2+2 4 | 2+2 4 | 3+4 2 | 5+6 2 | 2+2 4 |
| | 5+6 2 | 3+4 2 | 2+2 4 | 2+2 4 | 2+2 4 | 5+6 2 | 3+4 2 | 2+2 4 |
| | 5+6 2 | 5+6 2 | 2+2 4 | 2+2 4 | 2+2 4 | 3+4 2 | 3+4 2 | 2+2 4 |
| | 3+4 2 | 3+4 2 | 2+2 4 | 2+2 4 | 2+2 4 | 5+6 2 | 5+6 2 | 2+2 4 |
| | 3+4 2 | 5+6 2 | 2+2 4 | 2+2 4 | 2+2 4 | 3+4 2 | 5+6 2 | 2+2 4 |
| | 5+6 2 | 3+4 2 | 2+2 4 | 2+2 4 | 2+2 4 | 3+4 2 | 5+6 2 | 2+2 4 |
| | 3+4 2 | 3+4 2 | | | | 5+6 2 | 3+4 2 | |
| | 5+6 2 | 5+6 2 | | | | 5+6 2 | 3+4 2 | |
| | 5+6 2 | 5+6 2 | | | | 3+4 2 | 3+4 2 | |
| | 3+4 2 | 3+4 2 | | | | 5+6 2 | 5+6 2 | |
| | 3+4 2 | 3+4 2 | | | | 3+4 2 | 3+4 2 | |
| | 5+6 2 | 5+6 2 | | | | 5+6 2 | 5+6 2 | |
| | 3+4 2 | 3+4 2 | | | | 3+4 2 | 5+6 2 | |
| | 3+4 2 | 3+4 2 | | | | 3+4 2 | 5+6 2 | |
| | 5+6 2 | 5+6 2 | | | | 3+4 2 | 5+6 2 | |
| | 5+6 2 | 5+6 2 | | | | 5+6 2 | 3+4 2 | |

Continued on next column

**TABLE XXIII– continued**

| TQC | ICSD | NoE | SF | | | | | |
|---|---|---|---|---|---|---|---|---|
| 11.2.86 | 194985 | 440 | Ba1Nd2O10Ti3 | | | | | |
| | -A | -B | -C | -D | -E | -GM | -Y | -Z |
| | 3+4 2 | 3+4 2 | 2+2 4 | 2+2 4 | 2+2 4 | 3+4 2 | 3+4 2 | 2+2 4 |
| | 5+6 2 | 5+6 2 | 2+2 4 | 2+2 4 | 2+2 4 | 5+6 2 | 5+6 2 | 2+2 4 |
| | 3+4 2 | 3+4 2 | 2+2 4 | 2+2 4 | 2+2 4 | 3+4 2 | 3+4 2 | 2+2 4 |
| | 5+6 2 | 5+6 2 | 2+2 4 | 2+2 4 | 2+2 4 | 5+6 2 | 5+6 2 | 2+2 4 |
| | 3+4 2 | 3+4 2 | 2+2 4 | 2+2 4 | 2+2 4 | 3+4 2 | 3+4 2 | 2+2 4 |
| | 5+6 2 | 5+6 2 | 2+2 4 | 2+2 4 | 2+2 4 | 5+6 2 | 5+6 2 | 2+2 4 |
| | 3+4 2 | 3+4 2 | 2+2 4 | 2+2 4 | 2+2 4 | 3+4 2 | 3+4 2 | 2+2 4 |
| | 5+6 2 | 5+6 2 | 2+2 4 | 2+2 4 | 2+2 4 | 5+6 2 | 5+6 2 | 2+2 4 |
| | 5+6 2 | 5+6 2 | 2+2 4 | 2+2 4 | 2+2 4 | 3+4 2 | 3+4 2 | 2+2 4 |
| | 3+4 2 | 3+4 2 | 2+2 4 | 2+2 4 | 2+2 4 | 5+6 2 | 5+6 2 | 2+2 4 |
| | 3+4 2 | 5+6 2 | 2+2 4 | 2+2 4 | 2+2 4 | 3+4 2 | 5+6 2 | 2+2 4 |
| | 5+6 2 | 3+4 2 | 2+2 4 | 2+2 4 | 2+2 4 | 5+6 2 | 3+4 2 | 2+2 4 |
| | 3+4 2 | 3+4 2 | 2+2 4 | 2+2 4 | 2+2 4 | 3+4 2 | 3+4 2 | 2+2 4 |
| | 5+6 2 | 5+6 2 | 2+2 4 | 2+2 4 | 2+2 4 | 5+6 2 | 5+6 2 | 2+2 4 |
| | 3+4 2 | 3+4 2 | 2+2 4 | 2+2 4 | 2+2 4 | 3+4 2 | 3+4 2 | 2+2 4 |
| | 5+6 2 | 5+6 2 | 2+2 4 | 2+2 4 | 2+2 4 | 5+6 2 | 5+6 2 | 2+2 4 |
| | 3+4 2 | 3+4 2 | 2+2 4 | 2+2 4 | 2+2 4 | 3+4 2 | 3+4 2 | 2+2 4 |
| | 5+6 2 | 5+6 2 | 2+2 4 | 2+2 4 | 2+2 4 | 5+6 2 | 5+6 2 | 2+2 4 |
| | 3+4 2 | 3+4 2 | 2+2 4 | 2+2 4 | 2+2 4 | 3+4 2 | 3+4 2 | 2+2 4 |
| | 5+6 2 | 5+6 2 | 2+2 4 | 2+2 4 | 2+2 4 | 5+6 2 | 5+6 2 | 2+2 4 |
| | 5+6 2 | 3+4 2 | 2+2 4 | 2+2 4 | 2+2 4 | 3+4 2 | 5+6 2 | 2+2 4 |
| | 3+4 2 | 3+4 2 | 2+2 4 | 2+2 4 | 2+2 4 | 3+4 2 | 3+4 2 | 2+2 4 |
| | 3+4 2 | 5+6 2 | 2+2 4 | 2+2 4 | 2+2 4 | 5+6 2 | 3+4 2 | 2+2 4 |
| | 5+6 2 | 5+6 2 | 2+2 4 | 2+2 4 | 2+2 4 | 5+6 2 | 5+6 2 | 2+2 4 |
| | 3+4 2 | 3+4 2 | 2+2 4 | 2+2 4 | 2+2 4 | 3+4 2 | 3+4 2 | 2+2 4 |
| | 5+6 2 | 5+6 2 | 2+2 4 | 2+2 4 | 2+2 4 | 5+6 2 | 5+6 2 | 2+2 4 |
| | 3+4 2 | 3+4 2 | 2+2 4 | 2+2 4 | 2+2 4 | 3+4 2 | 3+4 2 | 2+2 4 |
| | 5+6 2 | 5+6 2 | 2+2 4 | 2+2 4 | 2+2 4 | 5+6 2 | 5+6 2 | 2+2 4 |
| | 3+4 2 | 3+4 2 | 2+2 4 | 2+2 4 | 2+2 4 | 3+4 2 | 3+4 2 | 2+2 4 |
| | 5+6 2 | 5+6 2 | 2+2 4 | 2+2 4 | 2+2 4 | 5+6 2 | 5+6 2 | 2+2 4 |
| | 5+6 2 | 3+4 2 | 2+2 4 | 2+2 4 | 2+2 4 | 3+4 2 | 5+6 2 | 2+2 4 |
| | 3+4 2 | 3+4 2 | 2+2 4 | 2+2 4 | 2+2 4 | 3+4 2 | 3+4 2 | 2+2 4 |
| | 3+4 2 | 5+6 2 | 2+2 4 | 2+2 4 | 2+2 4 | 5+6 2 | 3+4 2 | 2+2 4 |
| | 3+4 2 | 3+4 2 | 2+2 4 | 2+2 4 | 2+2 4 | 3+4 2 | 3+4 2 | 2+2 4 |
| | 5+6 2 | 3+4 2 | 2+2 4 | 2+2 4 | 2+2 4 | 3+4 2 | 5+6 2 | 2+2 4 |
| | 3+4 2 | 5+6 2 | 2+2 4 | 2+2 4 | 2+2 4 | 5+6 2 | 3+4 2 | 2+2 4 |
| | 5+6 2 | 5+6 2 | 2+2 4 | 2+2 4 | 2+2 4 | 5+6 2 | 5+6 2 | 2+2 4 |
| | 3+4 2 | 3+4 2 | 2+2 4 | 2+2 4 | 2+2 4 | 3+4 2 | 3+4 2 | 2+2 4 |
| | 5+6 2 | 5+6 2 | 2+2 4 | 2+2 4 | 2+2 4 | 3+4 2 | 5+6 2 | 2+2 4 |
| | 5+6 2 | 5+6 2 | 2+2 4 | 2+2 4 | 2+2 4 | 5+6 2 | 5+6 2 | 2+2 4 |
| | 3+4 2 | 5+6 2 | 2+2 4 | 2+2 4 | 2+2 4 | 5+6 2 | 3+4 2 | 2+2 4 |
| | 5+6 2 | 3+4 2 | 2+2 4 | 2+2 4 | 2+2 4 | 3+4 2 | 5+6 2 | 2+2 4 |
| | 5+6 2 | 3+4 2 | 2+2 4 | 2+2 4 | 2+2 4 | 5+6 2 | 3+4 2 | 2+2 4 |
| | 3+4 2 | 5+6 2 | 2+2 4 | 2+2 4 | 2+2 4 | 3+4 2 | 5+6 2 | 2+2 4 |
| | 5+6 2 | 3+4 2 | 2+2 4 | 2+2 4 | 2+2 4 | 5+6 2 | 3+4 2 | 2+2 4 |
| | 3+4 2 | 5+6 2 | 2+2 4 | 2+2 4 | 2+2 4 | 3+4 2 | 5+6 2 | 2+2 4 |
| | 5+6 2 | 5+6 2 | 2+2 4 | 2+2 4 | 2+2 4 | 5+6 2 | 3+4 2 | 2+2 4 |
| | 3+4 2 | 3+4 2 | 2+2 4 | 2+2 4 | 2+2 4 | 5+6 2 | 5+6 2 | 2+2 4 |
| | 5+6 2 | 3+4 2 | 2+2 4 | 2+2 4 | 2+2 4 | 3+4 2 | 3+4 2 | 2+2 4 |
| | 3+4 2 | 5+6 2 | 2+2 4 | 2+2 4 | 2+2 4 | 3+4 2 | 3+4 2 | 2+2 4 |
| | 5+6 2 | 5+6 2 | 2+2 4 | 2+2 4 | 2+2 4 | 5+6 2 | 5+6 2 | 2+2 4 |
| | 3+4 2 | 3+4 2 | 2+2 4 | 2+2 4 | 2+2 4 | 3+4 2 | 5+6 2 | 2+2 4 |
| | 5+6 2 | 5+6 2 | 2+2 4 | 2+2 4 | 2+2 4 | 5+6 2 | 3+4 2 | 2+2 4 |
| | 3+4 2 | 3+4 2 | 2+2 4 | 2+2 4 | 2+2 4 | 5+6 2 | 5+6 2 | 2+2 4 |
| | 3+4 2 | 3+4 2 | 2+2 4 | 2+2 4 | 2+2 4 | 3+4 2 | 3+4 2 | 2+2 4 |
| | 5+6 2 | 5+6 2 | 2+2 4 | 2+2 4 | 2+2 4 | 5+6 2 | 5+6 2 | 2+2 4 |
| | 3+4 2 | 5+6 2 | 2+2 4 | 2+2 4 | 2+2 4 | 3+4 2 | 5+6 2 | 2+2 4 |
| | 5+6 2 | 3+4 2 | 2+2 4 | 2+2 4 | 2+2 4 | 3+4 2 | 3+4 2 | 2+2 4 |
| | 3+4 2 | 5+6 2 | 2+2 4 | 2+2 4 | 2+2 4 | 5+6 2 | 5+6 2 | 2+2 4 |
| | 5+6 2 | 3+4 2 | 2+2 4 | 2+2 4 | 2+2 4 | 5+6 2 | 3+4 2 | 2+2 4 |
| | 3+4 2 | 5+6 2 | 2+2 4 | 2+2 4 | 2+2 4 | 3+4 2 | 5+6 2 | 2+2 4 |
| | 5+6 2 | 3+4 2 | 2+2 4 | 2+2 4 | 2+2 4 | 3+4 2 | 3+4 2 | 2+2 4 |
| | 3+4 2 | 5+6 2 | 2+2 4 | 2+2 4 | 2+2 4 | 5+6 2 | 3+4 2 | 2+2 4 |
| | 5+6 2 | 3+4 2 | 2+2 4 | 2+2 4 | 2+2 4 | 5+6 2 | 5+6 2 | 2+2 4 |
| | 3+4 2 | 3+4 2 | 2+2 4 | 2+2 4 | 2+2 4 | 3+4 2 | 3+4 2 | 2+2 4 |
| | 5+6 2 | 5+6 2 | 2+2 4 | 2+2 4 | 2+2 4 | 5+6 2 | 5+6 2 | 2+2 4 |
| | 3+4 2 | 3+4 2 | 2+2 4 | 2+2 4 | 2+2 4 | 3+4 2 | 3+4 2 | 2+2 4 |
| | 5+6 2 | 5+6 2 | 2+2 4 | 2+2 4 | 2+2 4 | 5+6 2 | 5+6 2 | 2+2 4 |
| | 3+4 2 | 3+4 2 | 2+2 4 | 2+2 4 | 2+2 4 | 3+4 2 | 3+4 2 | 2+2 4 |
| | 5+6 2 | 5+6 2 | 2+2 4 | 2+2 4 | 2+2 4 | 5+6 2 | 5+6 2 | 2+2 4 |
| | 3+4 2 | 3+4 2 | 2+2 4 | 2+2 4 | 2+2 4 | 3+4 2 | 3+4 2 | 2+2 4 |
| | 5+6 2 | 5+6 2 | 2+2 4 | 2+2 4 | 2+2 4 | 5+6 2 | 5+6 2 | 2+2 4 |
| | 3+4 2 | 3+4 2 | 2+2 4 | 2+2 4 | 2+2 4 | 3+4 2 | 3+4 2 | 2+2 4 |
| | 5+6 2 | 5+6 2 | 2+2 4 | 2+2 4 | 2+2 4 | 5+6 2 | 5+6 2 | 2+2 4 |
| | 3+4 2 | 3+4 2 | 2+2 4 | 2+2 4 | 2+2 4 | 3+4 2 | 3+4 2 | 2+2 4 |
| | 5+6 2 | 5+6 2 | 2+2 4 | 2+2 4 | 2+2 4 | 5+6 2 | 5+6 2 | 2+2 4 |
| | 3+4 2 | 5+6 2 | 2+2 4 | 2+2 4 | 2+2 4 | 3+4 2 | 5+6 2 | 2+2 4 |
| | 5+6 2 | 3+4 2 | 2+2 4 | 2+2 4 | 2+2 4 | 5+6 2 | 3+4 2 | 2+2 4 |
| | 5+6 2 | 5+6 2 | 2+2 4 | 2+2 4 | 2+2 4 | 3+4 2 | 3+4 2 | 2+2 4 |
| | 3+4 2 | 3+4 2 | 2+2 4 | 2+2 4 | 2+2 4 | 5+6 2 | 5+6 2 | 2+2 4 |
| | 5+6 2 | 5+6 2 | 2+2 4 | 2+2 4 | 2+2 4 | 3+4 2 | 3+4 2 | 2+2 4 |
| | 3+4 2 | 3+4 2 | 2+2 4 | 2+2 4 | 2+2 4 | 5+6 2 | 5+6 2 | 2+2 4 |
| | 3+4 2 | 5+6 2 | 2+2 4 | 2+2 4 | 2+2 4 | 3+4 2 | 3+4 2 | 2+2 4 |
| | 5+6 2 | 3+4 2 | 2+2 4 | 2+2 4 | 2+2 4 | 5+6 2 | 5+6 2 | 2+2 4 |
| | 5+6 2 | 5+6 2 | 2+2 4 | 2+2 4 | 2+2 4 | 3+4 2 | 3+4 2 | 2+2 4 |
| | 3+4 2 | 3+4 2 | 2+2 4 | 2+2 4 | 2+2 4 | 5+6 2 | 5+6 2 | 2+2 4 |
| | 3+4 2 | 3+4 2 | 2+2 4 | 2+2 4 | 2+2 4 | 5+6 2 | 5+6 2 | 2+2 4 |
| | 5+6 2 | 5+6 2 | 2+2 4 | 2+2 4 | 2+2 4 | 3+4 2 | 3+4 2 | 2+2 4 |
| | 3+4 2 | 3+4 2 | 2+2 4 | 2+2 4 | 2+2 4 | 3+4 2 | 5+6 2 | 2+2 4 |
| | 5+6 2 | 3+4 2 | 2+2 4 | 2+2 4 | 2+2 4 | 3+4 2 | 3+4 2 | 2+2 4 |
| | 5+6 2 | 5+6 2 | 2+2 4 | 2+2 4 | 2+2 4 | 5+6 2 | 5+6 2 | 2+2 4 |
| | 3+4 2 | 3+4 2 | 2+2 4 | 2+2 4 | 2+2 4 | 3+4 2 | 3+4 2 | 2+2 4 |
| | 3+4 2 | 3+4 2 | 2+2 4 | 2+2 4 | 2+2 4 | 3+4 2 | 3+4 2 | 2+2 4 |
| | 5+6 2 | 5+6 2 | 2+2 4 | 2+2 4 | 2+2 4 | 5+6 2 | 5+6 2 | 2+2 4 |
| | 3+4 2 | 3+4 2 | 2+2 4 | 2+2 4 | 2+2 4 | 3+4 2 | 3+4 2 | 2+2 4 |
| | 3+4 2 | 5+6 2 | 2+2 4 | 2+2 4 | 2+2 4 | 5+6 2 | 3+4 2 | 2+2 4 |
| | 5+6 2 | 3+4 2 | 2+2 4 | 2+2 4 | 2+2 4 | 3+4 2 | 5+6 2 | 2+2 4 |
| | 3+4 2 | 3+4 2 | 2+2 4 | 2+2 4 | 2+2 4 | 3+4 2 | 3+4 2 | 2+2 4 |
| | 3+4 2 | 3+4 2 | 2+2 4 | 2+2 4 | 2+2 4 | 3+4 2 | 3+4 2 | 2+2 4 |
| | 5+6 2 | 5+6 2 | 2+2 4 | 2+2 4 | 2+2 4 | 5+6 2 | 5+6 2 | 2+2 4 |
| | 3+4 2 | 3+4 2 | 2+2 4 | 2+2 4 | 2+2 4 | 3+4 2 | 3+4 2 | 2+2 4 |
| | 5+6 2 | 5+6 2 | 2+2 4 | 2+2 4 | 2+2 4 | 5+6 2 | 5+6 2 | 2+2 4 |
| | 3+4 2 | 3+4 2 | 2+2 4 | 2+2 4 | 2+2 4 | 3+4 2 | 3+4 2 | 2+2 4 |
| | 5+6 2 | 3+4 2 | 2+2 4 | 2+2 4 | 2+2 4 | 3+4 2 | 5+6 2 | 2+2 4 |
| | 3+4 2 | 3+4 2 | 2+2 4 | 2+2 4 | 2+2 4 | 5+6 2 | 3+4 2 | 2+2 4 |
| | 5+6 2 | 3+4 2 | 2+2 4 | 2+2 4 | 2+2 4 | 3+4 2 | 5+6 2 | 2+2 4 |
| | 3+4 2 | 5+6 2 | 2+2 4 | 2+2 4 | 2+2 4 | 3+4 2 | 3+4 2 | 2+2 4 |
| | 5+6 2 | 5+6 2 | 2+2 4 | 2+2 4 | 2+2 4 | 5+6 2 | 5+6 2 | 2+2 4 |
| | 3+4 2 | 5+6 2 | 2+2 4 | 2+2 4 | 2+2 4 | 5+6 2 | 3+4 2 | 2+2 4 |
| | 3+4 2 | 3+4 2 | 2+2 4 | 2+2 4 | 2+2 4 | 3+4 2 | 3+4 2 | 2+2 4 |
| | 5+6 2 | 5+6 2 | | | | 5+6 2 | 5+6 2 | |
| | 3+4 2 | 3+4 2 | | | | 3+4 2 | 3+4 2 | |
| | 5+6 2 | 5+6 2 | | | | 5+6 2 | 5+6 2 | |
| | 5+6 2 | 3+4 2 | | | | 5+6 2 | 5+6 2 | |
| | 5+6 2 | 5+6 2 | | | | 3+4 2 | 5+6 2 | |
| | 5+6 2 | 3+4 2 | | | | 3+4 2 | 3+4 2 | |
| | 3+4 2 | 5+6 2 | | | | 3+4 2 | 5+6 2 | |
| | 3+4 2 | 3+4 2 | | | | 5+6 2 | 3+4 2 | |
| | 5+6 2 | 5+6 2 | | | | 5+6 2 | 5+6 2 | |
| | 3+4 2 | 3+4 2 | | | | 5+6 2 | 5+6 2 | |

Continued on next column

Continued on next column

**TABLE XXIII– continued**

| TQC | ICSD | NoE | SF |
|---|---|---|---|
| | 5+6 2 5+6 2 | | 3+4 2 3+4 2 |
| | 5+6 2 3+4 2 | | 3+4 2 5+6 2 |
| | 3+4 2 3+4 2 | | 3+4 2 3+4 2 |
| | 3+4 2 5+6 2 | | 5+6 2 3+4 2 |
| | 5+6 2 3+4 2 | | 5+6 2 5+6 2 |
| | 5+6 2 5+6 2 | | 3+4 2 5+6 2 |
| | 3+4 2 5+6 2 | | 5+6 2 3+4 2 |
| | 5+6 2 3+4 2 | | 3+4 2 5+6 2 |
| | 3+4 2 3+4 2 | | 3+4 2 3+4 2 |
| | 3+4 2 5+6 2 | | 5+6 2 3+4 2 |
| | 5+6 2 3+4 2 | | 3+4 2 3+4 2 |
| | 3+4 2 5+6 2 | | 5+6 2 5+6 2 |
| | 5+6 2 5+6 2 | | 5+6 2 5+6 2 |
| | 3+4 2 3+4 2 | | 3+4 2 3+4 2 |
| | 5+6 2 5+6 2 | | 5+6 2 5+6 2 |
| | 3+4 2 3+4 2 | | 3+4 2 3+4 2 |
| | 5+6 2 5+6 2 | | 5+6 2 5+6 2 |
| | 3+4 2 3+4 2 | | 3+4 2 5+6 2 |
| | 5+6 2 3+4 2 | | 5+6 2 5+6 2 |
| | 5+6 2 3+4 2 | | 3+4 2 3+4 2 |
| | 3+4 2 5+6 2 | | 3+4 2 3+4 2 |
| | 3+4 2 5+6 2 | | 5+6 2 3+4 2 |
| | 5+6 2 3+4 2 | | 3+4 2 5+6 2 |
| | 3+4 2 5+6 2 | | 3+4 2 3+4 2 |
| | 5+6 2 3+4 2 | | 5+6 2 5+6 2 |
| | 3+4 2 5+6 2 | | 5+6 2 3+4 2 |
| | 3+4 2 5+6 2 | | 5+6 2 5+6 2 |
| | 5+6 2 5+6 2 | | 5+6 2 3+4 2 |
| | 5+6 2 5+6 2 | | 3+4 2 3+4 2 |
| | 3+4 2 3+4 2 | | 3+4 2 5+6 2 |
| | 5+6 2 5+6 2 | | 5+6 2 5+6 2 |
| | 5+6 2 5+6 2 | | 3+4 2 3+4 2 |
| | 3+4 2 3+4 2 | | 5+6 2 3+4 2 |
| | 3+4 2 5+6 2 | | 3+4 2 5+6 2 |
| | 3+4 2 3+4 2 | | 5+6 2 3+4 2 |
| | 5+6 2 5+6 2 | | 3+4 2 5+6 2 |
| | 5+6 2 3+4 2 | | 5+6 2 5+6 2 |
| | 5+6 2 3+4 2 | | 5+6 2 5+6 2 |
| | 5+6 2 5+6 2 | | 3+4 2 3+4 2 |
| | 3+4 2 3+4 2 | | 5+6 2 5+6 2 |
| | 3+4 2 5+6 2 | | 3+4 2 3+4 2 |
| | 5+6 2 3+4 2 | | 5+6 2 3+4 2 |
| | 3+4 2 5+6 2 | | 5+6 2 3+4 2 |
| | 3+4 2 5+6 2 | | 3+4 2 5+6 2 |
| | 5+6 2 3+4 2 | | 5+6 2 3+4 2 |
| | 5+6 2 3+4 2 | | 3+4 2 5+6 2 |
| | 3+4 2 5+6 2 | | 5+6 2 5+6 2 |
| | 5+6 2 5+6 2 | | 3+4 2 3+4 2 |
| | 3+4 2 3+4 2 | | 3+4 2 5+6 2 |
| | 5+6 2 5+6 2 | | 5+6 2 3+4 2 |
| | 3+4 2 5+6 2 | | 3+4 2 5+6 2 |
| | 5+6 2 3+4 2 | | 5+6 2 5+6 2 |
| | 5+6 2 5+6 2 | | 3+4 2 3+4 2 |
| | 3+4 2 3+4 2 | | 5+6 2 5+6 2 |
| | 5+6 2 5+6 2 | | 3+4 2 5+6 2 |
| | 3+4 2 5+6 2 | | 5+6 2 3+4 2 |
| | 5+6 2 5+6 2 | | 5+6 2 3+4 2 |
| | 3+4 2 5+6 2 | | 5+6 2 5+6 2 |
| | 5+6 2 3+4 2 | | 3+4 2 5+6 2 |
| | 5+6 2 5+6 2 | | 3+4 2 3+4 2 |
| | 3+4 2 3+4 2 | | 5+6 2 5+6 2 |

Continued on next column

**TABLE XXIII– continued**

| TQC | ICSD | NoE | SF |
|---|---|---|---|
| | 5+6 2 5+6 2 | | 5+6 2 3+4 2 |
| | 3+4 2 3+4 2 | | 3+4 2 3+4 2 |
| | 5+6 2 3+4 2 | | 5+6 2 3+4 2 |
| | 3+4 2 5+6 2 | | 3+4 2 5+6 2 |
| | 3+4 2 3+4 2 | | 3+4 2 3+4 2 |
| | 3+4 2 3+4 2 | | 5+6 2 5+6 2 |
| | 5+6 2 5+6 2 | | 3+4 2 3+4 2 |
| | 3+4 2 3+4 2 | | 3+4 2 5+6 2 |
| | 3+4 2 5+6 2 | | 5+6 2 5+6 2 |
| | 5+6 2 3+4 2 | | 5+6 2 3+4 2 |
| | 3+4 2 3+4 2 | | 3+4 2 3+4 2 |
| | 5+6 2 5+6 2 | | 5+6 2 5+6 2 |
| | 5+6 2 5+6 2 | | 3+4 2 5+6 2 |
| | 3+4 2 5+6 2 | | 3+4 2 5+6 2 |
| | 5+6 2 3+4 2 | | 5+6 2 3+4 2 |
| | 3+4 2 3+4 2 | | 5+6 2 3+4 2 |
| | 5+6 2 5+6 2 | | 5+6 2 5+6 2 |
| | 3+4 2 5+6 2 | | 3+4 2 5+6 2 |
| | 5+6 2 3+4 2 | | 5+6 2 3+4 2 |
| | 3+4 2 5+6 2 | | 5+6 2 3+4 2 |
| | 3+4 2 3+4 2 | | 3+4 2 5+6 2 |
| | 5+6 2 5+6 2 | | 5+6 2 3+4 2 |
| | 3+4 2 3+4 2 | | 3+4 2 5+6 2 |
| | 5+6 2 3+4 2 | | 5+6 2 3+4 2 |
| | 3+4 2 5+6 2 | | 5+6 2 5+6 2 |
| | 5+6 2 3+4 2 | | 3+4 2 5+6 2 |
| | 5+6 2 5+6 2 | | 5+6 2 3+4 2 |
| | 5+6 2 3+4 2 | | 3+4 2 5+6 2 |
| | 3+4 2 5+6 2 | | 5+6 2 3+4 2 |
| | 3+4 2 3+4 2 | | 3+4 2 3+4 2 |
| | 5+6 2 5+6 2 | | 5+6 2 5+6 2 |
| | 5+6 2 3+4 2 | | 3+4 2 5+6 2 |
| | 3+4 2 5+6 2 | | 5+6 2 3+4 2 |
| | 5+6 2 3+4 2 | | 3+4 2 5+6 2 |
| | 3+4 2 5+6 2 | | 5+6 2 3+4 2 |
| | 5+6 2 3+4 2 | | 3+4 2 5+6 2 |
| | 3+4 2 3+4 2 | | 3+4 2 3+4 2 |
| | 3+4 2 5+6 2 | | 5+6 2 3+4 2 |
| | 5+6 2 3+4 2 | | 3+4 2 5+6 2 |

| TQC | ICSD | NoE | SF | | | | | |
|---|---|---|---|---|---|---|---|---|
| 11.2.87 | 429621 | 432 | Ba1Ho4O17Si5 | | | | | |
| | -A | -B | -C | -D | -E | -GM | -Y | -Z |
| | 3+4 2 | 3+4 2 | 2+2 4 | 2+2 4 | 2+2 4 | 3+4 2 | 3+4 2 | 2+2 4 |
| | 5+6 2 | 5+6 2 | 2+2 4 | 2+2 4 | 2+2 4 | 5+6 2 | 5+6 2 | 2+2 4 |
| | 3+4 2 | 3+4 2 | 2+2 4 | 2+2 4 | 2+2 4 | 3+4 2 | 3+4 2 | 2+2 4 |
| | 5+6 2 | 5+6 2 | 2+2 4 | 2+2 4 | 2+2 4 | 5+6 2 | 5+6 2 | 2+2 4 |
| | 5+6 2 | 3+4 2 | 2+2 4 | 2+2 4 | 2+2 4 | 3+4 2 | 5+6 2 | 2+2 4 |
| | 3+4 2 | 3+4 2 | 2+2 4 | 2+2 4 | 2+2 4 | 3+4 2 | 3+4 2 | 2+2 4 |
| | 5+6 2 | 5+6 2 | 2+2 4 | 2+2 4 | 2+2 4 | 5+6 2 | 5+6 2 | 2+2 4 |
| | 3+4 2 | 5+6 2 | 2+2 4 | 2+2 4 | 2+2 4 | 5+6 2 | 3+4 2 | 2+2 4 |
| | 3+4 2 | 3+4 2 | 2+2 4 | 2+2 4 | 2+2 4 | 3+4 2 | 3+4 2 | 2+2 4 |
| | 5+6 2 | 5+6 2 | 2+2 4 | 2+2 4 | 2+2 4 | 5+6 2 | 5+6 2 | 2+2 4 |
| | 3+4 2 | 3+4 2 | 2+2 4 | 2+2 4 | 2+2 4 | 3+4 2 | 3+4 2 | 2+2 4 |
| | 5+6 2 | 5+6 2 | 2+2 4 | 2+2 4 | 2+2 4 | 5+6 2 | 5+6 2 | 2+2 4 |
| | 3+4 2 | 3+4 2 | 2+2 4 | 2+2 4 | 2+2 4 | 3+4 2 | 3+4 2 | 2+2 4 |
| | 5+6 2 | 3+4 2 | 2+2 4 | 2+2 4 | 2+2 4 | 3+4 2 | 5+6 2 | 2+2 4 |
| | 5+6 2 | 5+6 2 | 2+2 4 | 2+2 4 | 2+2 4 | 5+6 2 | 5+6 2 | 2+2 4 |
| | 3+4 2 | 3+4 2 | 2+2 4 | 2+2 4 | 2+2 4 | 3+4 2 | 3+4 2 | 2+2 4 |
| | 5+6 2 | 5+6 2 | 2+2 4 | 2+2 4 | 2+2 4 | 5+6 2 | 5+6 2 | 2+2 4 |
| | 3+4 2 | 5+6 2 | 2+2 4 | 2+2 4 | 2+2 4 | 5+6 2 | 3+4 2 | 2+2 4 |
| | 5+6 2 | 3+4 2 | 2+2 4 | 2+2 4 | 2+2 4 | 3+4 2 | 5+6 2 | 2+2 4 |
| | 3+4 2 | 5+6 2 | 2+2 4 | 2+2 4 | 2+2 4 | 5+6 2 | 3+4 2 | 2+2 4 |

Continued on next column

**TABLE XXIII– continued**

| TQC | ICSD | | NoE | | SF | | | | | | | | | | | |
|---|---|---|---|---|---|---|---|---|---|---|---|---|---|---|---|---|
| | 5+6 | 2 | 3+4 | 2 | 2+2 | 4 | 2+2 | 4 | 2+2 | 4 | 3+4 | 2 | 5+6 | 2 | 2+2 | 4 |
| | 3+4 | 2 | 3+4 | 2 | 2+2 | 4 | 2+2 | 4 | 2+2 | 4 | 3+4 | 2 | 3+4 | 2 | 2+2 | 4 |
| | 3+4 | 2 | 5+6 | 2 | 2+2 | 4 | 2+2 | 4 | 2+2 | 4 | 5+6 | 2 | 3+4 | 2 | 2+2 | 4 |
| | 5+6 | 2 | 5+6 | 2 | 2+2 | 4 | 2+2 | 4 | 2+2 | 4 | 5+6 | 2 | 5+6 | 2 | 2+2 | 4 |
| | 3+4 | 2 | 3+4 | 2 | 2+2 | 4 | 2+2 | 4 | 2+2 | 4 | 3+4 | 2 | 3+4 | 2 | 2+2 | 4 |
| | 5+6 | 2 | 5+6 | 2 | 2+2 | 4 | 2+2 | 4 | 2+2 | 4 | 3+4 | 2 | 5+6 | 2 | 2+2 | 4 |
| | 5+6 | 2 | 3+4 | 2 | 2+2 | 4 | 2+2 | 4 | 2+2 | 4 | 5+6 | 2 | 5+6 | 2 | 2+2 | 4 |
| | 3+4 | 2 | 5+6 | 2 | 2+2 | 4 | 2+2 | 4 | 2+2 | 4 | 5+6 | 2 | 3+4 | 2 | 2+2 | 4 |
| | 5+6 | 2 | 5+6 | 2 | 2+2 | 4 | 2+2 | 4 | 2+2 | 4 | 3+4 | 2 | 5+6 | 2 | 2+2 | 4 |
| | 3+4 | 2 | 3+4 | 2 | 2+2 | 4 | 2+2 | 4 | 2+2 | 4 | 5+6 | 2 | 3+4 | 2 | 2+2 | 4 |
| | 3+4 | 2 | 3+4 | 2 | 2+2 | 4 | 2+2 | 4 | 2+2 | 4 | 3+4 | 2 | 5+6 | 2 | 2+2 | 4 |
| | 5+6 | 2 | 3+4 | 2 | 2+2 | 4 | 2+2 | 4 | 2+2 | 4 | 3+4 | 2 | 3+4 | 2 | 2+2 | 4 |
| | 3+4 | 2 | 5+6 | 2 | 2+2 | 4 | 2+2 | 4 | 2+2 | 4 | 5+6 | 2 | 3+4 | 2 | 2+2 | 4 |
| | 5+6 | 2 | 3+4 | 2 | 2+2 | 4 | 2+2 | 4 | 2+2 | 4 | 3+4 | 2 | 5+6 | 2 | 2+2 | 4 |
| | 3+4 | 2 | 5+6 | 2 | 2+2 | 4 | 2+2 | 4 | 2+2 | 4 | 5+6 | 2 | 3+4 | 2 | 2+2 | 4 |
| | 5+6 | 2 | 5+6 | 2 | 2+2 | 4 | 2+2 | 4 | 2+2 | 4 | 5+6 | 2 | 5+6 | 2 | 2+2 | 4 |
| | 5+6 | 2 | 5+6 | 2 | 2+2 | 4 | 2+2 | 4 | 2+2 | 4 | 3+4 | 2 | 3+4 | 2 | 2+2 | 4 |
| | 5+6 | 2 | 5+6 | 2 | 2+2 | 4 | 2+2 | 4 | 2+2 | 4 | 3+4 | 2 | 3+4 | 2 | 2+2 | 4 |
| | 5+6 | 2 | 3+4 | 2 | 2+2 | 4 | 2+2 | 4 | 2+2 | 4 | 3+4 | 2 | 5+6 | 2 | 2+2 | 4 |
| | 3+4 | 2 | 5+6 | 2 | 2+2 | 4 | 2+2 | 4 | 2+2 | 4 | 5+6 | 2 | 3+4 | 2 | 2+2 | 4 |
| | 3+4 | 2 | 3+4 | 2 | 2+2 | 4 | 2+2 | 4 | 2+2 | 4 | 5+6 | 2 | 5+6 | 2 | 2+2 | 4 |
| | 5+6 | 2 | 5+6 | 2 | 2+2 | 4 | 2+2 | 4 | 2+2 | 4 | 3+4 | 2 | 3+4 | 2 | 2+2 | 4 |
| | 3+4 | 2 | 3+4 | 2 | 2+2 | 4 | 2+2 | 4 | 2+2 | 4 | 5+6 | 2 | 5+6 | 2 | 2+2 | 4 |
| | 5+6 | 2 | 3+4 | 2 | 2+2 | 4 | 2+2 | 4 | 2+2 | 4 | 3+4 | 2 | 5+6 | 2 | 2+2 | 4 |
| | 3+4 | 2 | 3+4 | 2 | 2+2 | 4 | 2+2 | 4 | 2+2 | 4 | 5+6 | 2 | 5+6 | 2 | 2+2 | 4 |
| | 5+6 | 2 | 5+6 | 2 | 2+2 | 4 | 2+2 | 4 | 2+2 | 4 | 3+4 | 2 | 3+4 | 2 | 2+2 | 4 |
| | 3+4 | 2 | 3+4 | 2 | 2+2 | 4 | 2+2 | 4 | 2+2 | 4 | 5+6 | 2 | 3+4 | 2 | 2+2 | 4 |
| | 5+6 | 2 | 5+6 | 2 | 2+2 | 4 | 2+2 | 4 | 2+2 | 4 | 5+6 | 2 | 5+6 | 2 | 2+2 | 4 |
| | 3+4 | 2 | 5+6 | 2 | 2+2 | 4 | 2+2 | 4 | 2+2 | 4 | 3+4 | 2 | 3+4 | 2 | 2+2 | 4 |
| | 3+4 | 2 | 3+4 | 2 | 2+2 | 4 | 2+2 | 4 | 2+2 | 4 | 3+4 | 2 | 3+4 | 2 | 2+2 | 4 |
| | 3+4 | 2 | 5+6 | 2 | 2+2 | 4 | 2+2 | 4 | 2+2 | 4 | 5+6 | 2 | 5+6 | 2 | 2+2 | 4 |
| | 5+6 | 2 | 5+6 | 2 | 2+2 | 4 | 2+2 | 4 | 2+2 | 4 | 3+4 | 2 | 3+4 | 2 | 2+2 | 4 |
| | 3+4 | 2 | 3+4 | 2 | 2+2 | 4 | 2+2 | 4 | 2+2 | 4 | 5+6 | 2 | 3+4 | 2 | 2+2 | 4 |
| | 5+6 | 2 | 5+6 | 2 | 2+2 | 4 | 2+2 | 4 | 2+2 | 4 | 5+6 | 2 | 3+4 | 2 | 2+2 | 4 |
| | 5+6 | 2 | 5+6 | 2 | 2+2 | 4 | 2+2 | 4 | 2+2 | 4 | 3+4 | 2 | 5+6 | 2 | 2+2 | 4 |
| | 5+6 | 2 | 3+4 | 2 | 2+2 | 4 | 2+2 | 4 | 2+2 | 4 | 3+4 | 2 | 5+6 | 2 | 2+2 | 4 |
| | 3+4 | 2 | 5+6 | 2 | 2+2 | 4 | 2+2 | 4 | 2+2 | 4 | 3+4 | 2 | 5+6 | 2 | 2+2 | 4 |
| | 5+6 | 2 | 3+4 | 2 | 2+2 | 4 | 2+2 | 4 | 2+2 | 4 | 5+6 | 2 | 3+4 | 2 | 2+2 | 4 |
| | 3+4 | 2 | 5+6 | 2 | 2+2 | 4 | 2+2 | 4 | 2+2 | 4 | 5+6 | 2 | 5+6 | 2 | 2+2 | 4 |
| | 5+6 | 2 | 3+4 | 2 | 2+2 | 4 | 2+2 | 4 | 2+2 | 4 | 3+4 | 2 | 3+4 | 2 | 2+2 | 4 |
| | 3+4 | 2 | 5+6 | 2 | 2+2 | 4 | 2+2 | 4 | 2+2 | 4 | 5+6 | 2 | 5+6 | 2 | 2+2 | 4 |
| | 3+4 | 2 | 5+6 | 2 | 2+2 | 4 | 2+2 | 4 | 2+2 | 4 | 5+6 | 2 | 3+4 | 2 | 2+2 | 4 |
| | 5+6 | 2 | 3+4 | 2 | 2+2 | 4 | 2+2 | 4 | 2+2 | 4 | 3+4 | 2 | 3+4 | 2 | 2+2 | 4 |
| | 5+6 | 2 | 5+6 | 2 | 2+2 | 4 | 2+2 | 4 | 2+2 | 4 | 3+4 | 2 | 5+6 | 2 | 2+2 | 4 |
| | 3+4 | 2 | 3+4 | 2 | 2+2 | 4 | 2+2 | 4 | 2+2 | 4 | 5+6 | 2 | 5+6 | 2 | 2+2 | 4 |
| | 5+6 | 2 | 5+6 | 2 | 2+2 | 4 | 2+2 | 4 | 2+2 | 4 | 5+6 | 2 | 3+4 | 2 | 2+2 | 4 |
| | 5+6 | 2 | 3+4 | 2 | 2+2 | 4 | 2+2 | 4 | 2+2 | 4 | 3+4 | 2 | 5+6 | 2 | 2+2 | 4 |
| | 3+4 | 2 | 3+4 | 2 | 2+2 | 4 | 2+2 | 4 | 2+2 | 4 | 3+4 | 2 | 3+4 | 2 | 2+2 | 4 |
| | 3+4 | 2 | 3+4 | 2 | 2+2 | 4 | 2+2 | 4 | 2+2 | 4 | 3+4 | 2 | 5+6 | 2 | 2+2 | 4 |
| | 5+6 | 2 | 5+6 | 2 | 2+2 | 4 | 2+2 | 4 | 2+2 | 4 | 5+6 | 2 | 3+4 | 2 | 2+2 | 4 |
| | 3+4 | 2 | 3+4 | 2 | 2+2 | 4 | 2+2 | 4 | 2+2 | 4 | 5+6 | 2 | 5+6 | 2 | 2+2 | 4 |
| | 5+6 | 2 | 5+6 | 2 | 2+2 | 4 | 2+2 | 4 | 2+2 | 4 | 3+4 | 2 | 3+4 | 2 | 2+2 | 4 |
| | 3+4 | 2 | 3+4 | 2 | 2+2 | 4 | 2+2 | 4 | 2+2 | 4 | 5+6 | 2 | 5+6 | 2 | 2+2 | 4 |
| | 5+6 | 2 | 5+6 | 2 | 2+2 | 4 | 2+2 | 4 | 2+2 | 4 | 3+4 | 2 | 3+4 | 2 | 2+2 | 4 |
| | 5+6 | 2 | 3+4 | 2 | 2+2 | 4 | 2+2 | 4 | 2+2 | 4 | 3+4 | 2 | 5+6 | 2 | 2+2 | 4 |
| | 3+4 | 2 | 5+6 | 2 | 2+2 | 4 | 2+2 | 4 | 2+2 | 4 | 5+6 | 2 | 3+4 | 2 | 2+2 | 4 |
| | 3+4 | 2 | 3+4 | 2 | 2+2 | 4 | 2+2 | 4 | 2+2 | 4 | 5+6 | 2 | 5+6 | 2 | 2+2 | 4 |
| | 3+4 | 2 | 3+4 | 2 | 2+2 | 4 | 2+2 | 4 | 2+2 | 4 | 5+6 | 2 | 5+6 | 2 | 2+2 | 4 |
| | 5+6 | 2 | 3+4 | 2 | 2+2 | 4 | 2+2 | 4 | 2+2 | 4 | 3+4 | 2 | 5+6 | 2 | 2+2 | 4 |
| | 3+4 | 2 | 5+6 | 2 | 2+2 | 4 | 2+2 | 4 | 2+2 | 4 | 5+6 | 2 | 3+4 | 2 | 2+2 | 4 |
| | 5+6 | 2 | 3+4 | 2 | 2+2 | 4 | 2+2 | 4 | 2+2 | 4 | 3+4 | 2 | 3+4 | 2 | 2+2 | 4 |
| | 3+4 | 2 | 5+6 | 2 | 2+2 | 4 | 2+2 | 4 | 2+2 | 4 | 5+6 | 2 | 5+6 | 2 | 2+2 | 4 |
| | 5+6 | 2 | 5+6 | 2 | 2+2 | 4 | 2+2 | 4 | 2+2 | 4 | 3+4 | 2 | 3+4 | 2 | 2+2 | 4 |
| | 5+6 | 2 | 5+6 | 2 | 2+2 | 4 | 2+2 | 4 | 2+2 | 4 | 3+4 | 2 | 3+4 | 2 | 2+2 | 4 |
| | 3+4 | 2 | 3+4 | 2 | 2+2 | 4 | 2+2 | 4 | 2+2 | 4 | 5+6 | 2 | 5+6 | 2 | 2+2 | 4 |
| | 3+4 | 2 | 3+4 | 2 | 2+2 | 4 | 2+2 | 4 | 2+2 | 4 | 3+4 | 2 | 3+4 | 2 | 2+2 | 4 |
| | 5+6 | 2 | 5+6 | 2 | 2+2 | 4 | 2+2 | 4 | 2+2 | 4 | 5+6 | 2 | 5+6 | 2 | 2+2 | 4 |
| | 5+6 | 2 | 5+6 | 2 | 2+2 | 4 | 2+2 | 4 | 2+2 | 4 | 3+4 | 2 | 5+6 | 2 | 2+2 | 4 |
| | 3+4 | 2 | 3+4 | 2 | 2+2 | 4 | 2+2 | 4 | 2+2 | 4 | 5+6 | 2 | 3+4 | 2 | 2+2 | 4 |
| | 3+4 | 2 | 3+4 | 2 | 2+2 | 4 | 2+2 | 4 | 2+2 | 4 | 5+6 | 2 | 5+6 | 2 | 2+2 | 4 |
| | 3+4 | 2 | 3+4 | 2 | 2+2 | 4 | 2+2 | 4 | 2+2 | 4 | 3+4 | 2 | 3+4 | 2 | 2+2 | 4 |
| | 5+6 | 2 | 5+6 | 2 | 2+2 | 4 | 2+2 | 4 | 2+2 | 4 | 5+6 | 2 | 5+6 | 2 | 2+2 | 4 |
| | 3+4 | 2 | 3+4 | 2 | 2+2 | 4 | 2+2 | 4 | 2+2 | 4 | 5+6 | 2 | 3+4 | 2 | 2+2 | 4 |
| | 3+4 | 2 | 3+4 | 2 | 2+2 | 4 | 2+2 | 4 | 2+2 | 4 | 5+6 | 2 | 5+6 | 2 | 2+2 | 4 |
| | 3+4 | 2 | 3+4 | 2 | 2+2 | 4 | 2+2 | 4 | 2+2 | 4 | 3+4 | 2 | 5+6 | 2 | 2+2 | 4 |
| | 5+6 | 2 | 5+6 | 2 | 2+2 | 4 | 2+2 | 4 | 2+2 | 4 | 3+4 | 2 | 5+6 | 2 | 2+2 | 4 |
| | 5+6 | 2 | 3+4 | 2 | 2+2 | 4 | 2+2 | 4 | 2+2 | 4 | 3+4 | 2 | 3+4 | 2 | 2+2 | 4 |
| | 3+4 | 2 | 3+4 | 2 | 2+2 | 4 | 2+2 | 4 | 2+2 | 4 | 5+6 | 2 | 5+6 | 2 | 2+2 | 4 |
| | 3+4 | 2 | 5+6 | 2 | 2+2 | 4 | 2+2 | 4 | 2+2 | 4 | 3+4 | 2 | 3+4 | 2 | 2+2 | 4 |
| | 5+6 | 2 | 3+4 | 2 | 2+2 | 4 | 2+2 | 4 | 2+2 | 4 | 5+6 | 2 | 3+4 | 2 | 2+2 | 4 |
| | 5+6 | 2 | 5+6 | 2 | 2+2 | 4 | 2+2 | 4 | 2+2 | 4 | 5+6 | 2 | 3+4 | 2 | 2+2 | 4 |
| | 3+4 | 2 | 5+6 | 2 | 2+2 | 4 | 2+2 | 4 | 2+2 | 4 | 3+4 | 2 | 5+6 | 2 | 2+2 | 4 |
| | 3+4 | 2 | 3+4 | 2 | 2+2 | 4 | 2+2 | 4 | 2+2 | 4 | 5+6 | 2 | 3+4 | 2 | 2+2 | 4 |
| | 5+6 | 2 | 5+6 | 2 | 2+2 | 4 | 2+2 | 4 | 2+2 | 4 | 5+6 | 2 | 5+6 | 2 | 2+2 | 4 |
| | 5+6 | 2 | 5+6 | 2 | 2+2 | 4 | 2+2 | 4 | 2+2 | 4 | 3+4 | 2 | 5+6 | 2 | 2+2 | 4 |
| | 3+4 | 2 | 5+6 | 2 | 2+2 | 4 | 2+2 | 4 | 2+2 | 4 | 3+4 | 2 | 5+6 | 2 | 2+2 | 4 |
| | 5+6 | 2 | 3+4 | 2 | 2+2 | 4 | 2+2 | 4 | 2+2 | 4 | 5+6 | 2 | 3+4 | 2 | 2+2 | 4 |
| | 3+4 | 2 | 3+4 | 2 | 2+2 | 4 | 2+2 | 4 | 2+2 | 4 | 3+4 | 2 | 3+4 | 2 | 2+2 | 4 |
| | 5+6 | 2 | 5+6 | 2 | | | | | | | 3+4 | 2 | 5+6 | 2 | | |
| | 3+4 | 2 | 5+6 | 2 | | | | | | | 5+6 | 2 | 5+6 | 2 | | |
| | 5+6 | 2 | 3+4 | 2 | | | | | | | 5+6 | 2 | 3+4 | 2 | | |
| | 5+6 | 2 | 5+6 | 2 | | | | | | | 3+4 | 2 | 3+4 | 2 | | |
| | 3+4 | 2 | 3+4 | 2 | | | | | | | 3+4 | 2 | 3+4 | 2 | | |
| | 3+4 | 2 | 5+6 | 2 | | | | | | | 5+6 | 2 | 5+6 | 2 | | |
| | 3+4 | 2 | 3+4 | 2 | | | | | | | 5+6 | 2 | 3+4 | 2 | | |
| | 5+6 | 2 | 3+4 | 2 | | | | | | | 5+6 | 2 | 5+6 | 2 | | |
| | 5+6 | 2 | 5+6 | 2 | | | | | | | 3+4 | 2 | 3+4 | 2 | | |
| | 5+6 | 2 | 3+4 | 2 | | | | | | | 3+4 | 2 | 3+4 | 2 | | |
| | 5+6 | 2 | 3+4 | 2 | | | | | | | 3+4 | 2 | 5+6 | 2 | | |
| | 3+4 | 2 | 5+6 | 2 | | | | | | | 5+6 | 2 | 5+6 | 2 | | |
| | 3+4 | 2 | 3+4 | 2 | | | | | | | 5+6 | 2 | 5+6 | 2 | | |
| | 5+6 | 2 | 5+6 | 2 | | | | | | | 5+6 | 2 | 5+6 | 2 | | |
| | 5+6 | 2 | 3+4 | 2 | | | | | | | 3+4 | 2 | 3+4 | 2 | | |
| | 3+4 | 2 | 5+6 | 2 | | | | | | | 5+6 | 2 | 3+4 | 2 | | |
| | 3+4 | 2 | 5+6 | 2 | | | | | | | 3+4 | 2 | 5+6 | 2 | | |
| | 3+4 | 2 | 3+4 | 2 | | | | | | | 5+6 | 2 | 3+4 | 2 | | |
| | 3+4 | 2 | 5+6 | 2 | | | | | | | 3+4 | 2 | 5+6 | 2 | | |
| | 5+6 | 2 | 5+6 | 2 | | | | | | | 3+4 | 2 | 5+6 | 2 | | |
| | 5+6 | 2 | 5+6 | 2 | | | | | | | 5+6 | 2 | 3+4 | 2 | | |
| | 5+6 | 2 | 5+6 | 2 | | | | | | | 3+4 | 2 | 5+6 | 2 | | |
| | 3+4 | 2 | 3+4 | 2 | | | | | | | 5+6 | 2 | 3+4 | 2 | | |
| | 3+4 | 2 | 3+4 | 2 | | | | | | | 5+6 | 2 | 5+6 | 2 | | |
| | 5+6 | 2 | 5+6 | 2 | | | | | | | 3+4 | 2 | 5+6 | 2 | | |
| | 3+4 | 2 | 3+4 | 2 | | | | | | | 3+4 | 2 | 3+4 | 2 | | |
| | 5+6 | 2 | 3+4 | 2 | | | | | | | 3+4 | 2 | 5+6 | 2 | | |
| | 3+4 | 2 | 3+4 | 2 | | | | | | | 3+4 | 2 | 3+4 | 2 | | |
| | 3+4 | 2 | 5+6 | 2 | | | | | | | 5+6 | 2 | 5+6 | 2 | | |
| | 5+6 | 2 | 3+4 | 2 | | | | | | | 5+6 | 2 | 5+6 | 2 | | |
| | 3+4 | 2 | 5+6 | 2 | | | | | | | 5+6 | 2 | 3+4 | 2 | | |
| | 5+6 | 2 | 3+4 | 2 | | | | | | | 3+4 | 2 | 5+6 | 2 | | |
| | 3+4 | 2 | 3+4 | 2 | | | | | | | 3+4 | 2 | 3+4 | 2 | | |
| | 5+6 | 2 | 3+4 | 2 | | | | | | | 5+6 | 2 | 5+6 | 2 | | |

Continued on next column

Continued on next column

**TABLE XXIII– continued**

| TQC | ICSD | NoE | SF | | | | | |
|---|---|---|---|---|---|---|---|---|
| | 3+4 2 | 5+6 2 | | | | 5+6 2 | 3+4 2 | |
| | 3+4 2 | 5+6 2 | | | | 3+4 2 | 3+4 2 | |
| | 5+6 2 | 5+6 2 | | | | 5+6 2 | 3+4 2 | |
| | 3+4 2 | 3+4 2 | | | | 3+4 2 | 5+6 2 | |
| | 5+6 2 | 5+6 2 | | | | 5+6 2 | 5+6 2 | |
| | 3+4 2 | 3+4 2 | | | | 5+6 2 | 5+6 2 | |
| | 3+4 2 | 3+4 2 | | | | 5+6 2 | 3+4 2 | |
| | 5+6 2 | 3+4 2 | | | | 3+4 2 | 3+4 2 | |
| | 3+4 2 | 5+6 2 | | | | 5+6 2 | 5+6 2 | |
| | 3+4 2 | 3+4 2 | | | | 5+6 2 | 3+4 2 | |
| | 5+6 2 | 5+6 2 | | | | 3+4 2 | 5+6 2 | |
| | 5+6 2 | 5+6 2 | | | | 3+4 2 | 5+6 2 | |
| | 5+6 2 | 5+6 2 | | | | 5+6 2 | 3+4 2 | |
| | 3+4 2 | 3+4 2 | | | | 3+4 2 | 3+4 2 | |
| | 5+6 2 | 5+6 2 | | | | 5+6 2 | 5+6 2 | |
| | 3+4 2 | 3+4 2 | | | | 3+4 2 | 3+4 2 | |
| | 5+6 2 | 5+6 2 | | | | 3+4 2 | 3+4 2 | |
| | 5+6 2 | 3+4 2 | | | | 5+6 2 | 5+6 2 | |
| | 5+6 2 | 5+6 2 | | | | 3+4 2 | 3+4 2 | |
| | 5+6 2 | 5+6 2 | | | | 3+4 2 | 3+4 2 | |
| | 3+4 2 | 3+4 2 | | | | 5+6 2 | 5+6 2 | |
| | 3+4 2 | 3+4 2 | | | | 3+4 2 | 5+6 2 | |
| | 5+6 2 | 5+6 2 | | | | 5+6 2 | 3+4 2 | |
| | 3+4 2 | 5+6 2 | | | | 5+6 2 | 3+4 2 | |
| | 5+6 2 | 3+4 2 | | | | 5+6 2 | 5+6 2 | |
| | 3+4 2 | 3+4 2 | | | | 3+4 2 | 5+6 2 | |
| | 3+4 2 | 5+6 2 | | | | 5+6 2 | 3+4 2 | |
| | 3+4 2 | 3+4 2 | | | | 3+4 2 | 5+6 2 | |
| | 5+6 2 | 5+6 2 | | | | 3+4 2 | 5+6 2 | |
| | 5+6 2 | 3+4 2 | | | | 3+4 2 | 5+6 2 | |
| | 3+4 2 | 3+4 2 | | | | 5+6 2 | 3+4 2 | |
| | 5+6 2 | 3+4 2 | | | | 5+6 2 | 3+4 2 | |
| | 3+4 2 | 5+6 2 | | | | 5+6 2 | 5+6 2 | |
| | 3+4 2 | 3+4 2 | | | | 5+6 2 | 5+6 2 | |
| | 5+6 2 | 5+6 2 | | | | 3+4 2 | 3+4 2 | |
| | 3+4 2 | 5+6 2 | | | | 3+4 2 | 5+6 2 | |
| | 5+6 2 | 3+4 2 | | | | 5+6 2 | 3+4 2 | |
| | 3+4 2 | 3+4 2 | | | | 5+6 2 | 5+6 2 | |
| | 5+6 2 | 3+4 2 | | | | 3+4 2 | 5+6 2 | |
| | 5+6 2 | 5+6 2 | | | | 3+4 2 | 5+6 2 | |
| | 3+4 2 | 3+4 2 | | | | 5+6 2 | 3+4 2 | |
| | 5+6 2 | 5+6 2 | | | | 5+6 2 | 3+4 2 | |
| | 3+4 2 | 5+6 2 | | | | 3+4 2 | 3+4 2 | |
| | 3+4 2 | 3+4 2 | | | | 5+6 2 | 5+6 2 | |
| | 5+6 2 | 5+6 2 | | | | 3+4 2 | 3+4 2 | |
| | 3+4 2 | 3+4 2 | | | | 5+6 2 | 5+6 2 | |
| | 5+6 2 | 3+4 2 | | | | 3+4 2 | 3+4 2 | |
| | 5+6 2 | 5+6 2 | | | | 3+4 2 | 5+6 2 | |
| | 3+4 2 | 3+4 2 | | | | 3+4 2 | 3+4 2 | |
| | 3+4 2 | 3+4 2 | | | | 5+6 2 | 5+6 2 | |
| | 5+6 2 | 5+6 2 | | | | 3+4 2 | 3+4 2 | |
| | 3+4 2 | 5+6 2 | | | | 5+6 2 | 5+6 2 | |
| | 5+6 2 | 3+4 2 | | | | 3+4 2 | 3+4 2 | |
| | 3+4 2 | 3+4 2 | | | | 5+6 2 | 5+6 2 | |
| | 5+6 2 | 5+6 2 | | | | 3+4 2 | 3+4 2 | |
| | 3+4 2 | 5+6 2 | | | | 5+6 2 | 5+6 2 | |
| | 5+6 2 | 5+6 2 | | | | 5+6 2 | 3+4 2 | |
| | 3+4 2 | 3+4 2 | | | | 5+6 2 | 5+6 2 | |
| | 5+6 2 | 5+6 2 | | | | 3+4 2 | 3+4 2 | |
| | 3+4 2 | 3+4 2 | | | | 5+6 2 | 5+6 2 | |
| | 5+6 2 | 5+6 2 | | | | 3+4 2 | 3+4 2 | |
| | 3+4 2 | 3+4 2 | | | | 5+6 2 | 5+6 2 | |
| | 5+6 2 | 5+6 2 | | | | 3+4 2 | 3+4 2 | |
| | 3+4 2 | 3+4 2 | | | | 5+6 2 | 5+6 2 | |
| | 5+6 2 | 5+6 2 | | | | 3+4 2 | 3+4 2 | |
| | 3+4 2 | 3+4 2 | | | | 5+6 2 | 5+6 2 | |
| | 5+6 2 | 3+4 2 | | | | 3+4 2 | 5+6 2 | |
| | 5+6 2 | 3+4 2 | | | | 3+4 2 | 3+4 2 | |
| | 3+4 2 | 5+6 2 | | | | 5+6 2 | 3+4 2 | |
| | 3+4 2 | 5+6 2 | | | | 5+6 2 | 5+6 2 | |
| | 5+6 2 | 3+4 2 | | | | 3+4 2 | 5+6 2 | |
| | 5+6 2 | 3+4 2 | | | | 3+4 2 | 5+6 2 | |
| | 3+4 2 | 5+6 2 | | | | 3+4 2 | 3+4 2 | |
| | 5+6 2 | 3+4 2 | | | | 5+6 2 | 3+4 2 | |
| 11.2.88 | 167615 | 252 | Ba1Er2O10Si3 | | | | | |
| | -A | -B | -C | -D | -E | -GM | -Y | -Z |
| | 5+6 2 | 5+6 2 | 2+2 4 | 2+2 4 | 2+2 4 | 3+4 2 | 3+4 2 | 2+2 4 |
| | 3+4 2 | 3+4 2 | 2+2 4 | 2+2 4 | 2+2 4 | 5+6 2 | 5+6 2 | 2+2 4 |
| | 5+6 2 | 5+6 2 | 2+2 4 | 2+2 4 | 2+2 4 | 3+4 2 | 3+4 2 | 2+2 4 |
| | 3+4 2 | 3+4 2 | 2+2 4 | 2+2 4 | 2+2 4 | 5+6 2 | 5+6 2 | 2+2 4 |
| | 3+4 2 | 3+4 2 | 2+2 4 | 2+2 4 | 2+2 4 | 3+4 2 | 3+4 2 | 2+2 4 |
| | 5+6 2 | 5+6 2 | 2+2 4 | 2+2 4 | 2+2 4 | 5+6 2 | 5+6 2 | 2+2 4 |
| | 5+6 2 | 5+6 2 | 2+2 4 | 2+2 4 | 2+2 4 | 3+4 2 | 3+4 2 | 2+2 4 |
| | 5+6 2 | 3+4 2 | 2+2 4 | 2+2 4 | 2+2 4 | 3+4 2 | 3+4 2 | 2+2 4 |
| | 3+4 2 | 5+6 2 | 2+2 4 | 2+2 4 | 2+2 4 | 5+6 2 | 5+6 2 | 2+2 4 |
| | 3+4 2 | 3+4 2 | 2+2 4 | 2+2 4 | 2+2 4 | 5+6 2 | 5+6 2 | 2+2 4 |
| | 5+6 2 | 5+6 2 | 2+2 4 | 2+2 4 | 2+2 4 | 3+4 2 | 3+4 2 | 2+2 4 |
| | 3+4 2 | 3+4 2 | 2+2 4 | 2+2 4 | 2+2 4 | 5+6 2 | 5+6 2 | 2+2 4 |
| | 5+6 2 | 5+6 2 | 2+2 4 | 2+2 4 | 2+2 4 | 3+4 2 | 3+4 2 | 2+2 4 |
| | 3+4 2 | 3+4 2 | 2+2 4 | 2+2 4 | 2+2 4 | 3+4 2 | 5+6 2 | 2+2 4 |
| | 5+6 2 | 5+6 2 | 2+2 4 | 2+2 4 | 2+2 4 | 5+6 2 | 3+4 2 | 2+2 4 |
| | 3+4 2 | 3+4 2 | 2+2 4 | 2+2 4 | 2+2 4 | 5+6 2 | 5+6 2 | 2+2 4 |
| | 5+6 2 | 5+6 2 | 2+2 4 | 2+2 4 | 2+2 4 | 3+4 2 | 3+4 2 | 2+2 4 |
| | 3+4 2 | 3+4 2 | 2+2 4 | 2+2 4 | 2+2 4 | 5+6 2 | 5+6 2 | 2+2 4 |
| | 5+6 2 | 5+6 2 | 2+2 4 | 2+2 4 | 2+2 4 | 3+4 2 | 3+4 2 | 2+2 4 |
| | 3+4 2 | 3+4 2 | 2+2 4 | 2+2 4 | 2+2 4 | 5+6 2 | 5+6 2 | 2+2 4 |
| | 3+4 2 | 5+6 2 | 2+2 4 | 2+2 4 | 2+2 4 | 3+4 2 | 5+6 2 | 2+2 4 |
| | 5+6 2 | 3+4 2 | 2+2 4 | 2+2 4 | 2+2 4 | 5+6 2 | 3+4 2 | 2+2 4 |
| | 3+4 2 | 3+4 2 | 2+2 4 | 2+2 4 | 2+2 4 | 5+6 2 | 5+6 2 | 2+2 4 |
| | 5+6 2 | 5+6 2 | 2+2 4 | 2+2 4 | 2+2 4 | 3+4 2 | 3+4 2 | 2+2 4 |
| | 5+6 2 | 5+6 2 | 2+2 4 | 2+2 4 | 2+2 4 | 3+4 2 | 3+4 2 | 2+2 4 |
| | 3+4 2 | 3+4 2 | 2+2 4 | 2+2 4 | 2+2 4 | 3+4 2 | 3+4 2 | 2+2 4 |
| | 5+6 2 | 5+6 2 | 2+2 4 | 2+2 4 | 2+2 4 | 5+6 2 | 5+6 2 | 2+2 4 |
| | 5+6 2 | 3+4 2 | 2+2 4 | 2+2 4 | 2+2 4 | 5+6 2 | 5+6 2 | 2+2 4 |
| | 3+4 2 | 5+6 2 | 2+2 4 | 2+2 4 | 2+2 4 | 3+4 2 | 5+6 2 | 2+2 4 |
| | 3+4 2 | 3+4 2 | 2+2 4 | 2+2 4 | 2+2 4 | 5+6 2 | 3+4 2 | 2+2 4 |
| | 5+6 2 | 3+4 2 | 2+2 4 | 2+2 4 | 2+2 4 | 5+6 2 | 5+6 2 | 2+2 4 |
| | 3+4 2 | 5+6 2 | 2+2 4 | 2+2 4 | 2+2 4 | 3+4 2 | 3+4 2 | 2+2 4 |
| | 3+4 2 | 3+4 2 | 2+2 4 | 2+2 4 | 2+2 4 | 5+6 2 | 5+6 2 | 2+2 4 |
| | 5+6 2 | 5+6 2 | 2+2 4 | 2+2 4 | 2+2 4 | 3+4 2 | 3+4 2 | 2+2 4 |
| | 5+6 2 | 3+4 2 | 2+2 4 | 2+2 4 | 2+2 4 | 5+6 2 | 5+6 2 | 2+2 4 |
| | 3+4 2 | 5+6 2 | 2+2 4 | 2+2 4 | 2+2 4 | 3+4 2 | 3+4 2 | 2+2 4 |
| | 5+6 2 | 3+4 2 | 2+2 4 | 2+2 4 | 2+2 4 | 5+6 2 | 5+6 2 | 2+2 4 |
| | 3+4 2 | 5+6 2 | 2+2 4 | 2+2 4 | 2+2 4 | 3+4 2 | 3+4 2 | 2+2 4 |
| | 3+4 2 | 3+4 2 | 2+2 4 | 2+2 4 | 2+2 4 | 3+4 2 | 3+4 2 | 2+2 4 |
| | 5+6 2 | 5+6 2 | 2+2 4 | 2+2 4 | 2+2 4 | 5+6 2 | 5+6 2 | 2+2 4 |
| | 3+4 2 | 5+6 2 | 2+2 4 | 2+2 4 | 2+2 4 | 5+6 2 | 3+4 2 | 2+2 4 |
| | 5+6 2 | 3+4 2 | 2+2 4 | 2+2 4 | 2+2 4 | 3+4 2 | 5+6 2 | 2+2 4 |
| | 3+4 2 | 3+4 2 | 2+2 4 | 2+2 4 | 2+2 4 | 3+4 2 | 3+4 2 | 2+2 4 |
| | 5+6 2 | 5+6 2 | 2+2 4 | 2+2 4 | 2+2 4 | 5+6 2 | 5+6 2 | 2+2 4 |
| | 3+4 2 | 3+4 2 | 2+2 4 | 2+2 4 | 2+2 4 | 3+4 2 | 5+6 2 | 2+2 4 |
| | 5+6 2 | 5+6 2 | 2+2 4 | 2+2 4 | 2+2 4 | 5+6 2 | 3+4 2 | 2+2 4 |

Continued on next column

**TABLE XXIII– continued**

| TQC | ICSD | NoE | SF | | | | |
|---|---|---|---|---|---|---|---|
| | 3+4 2 | 5+6 2 | 2+2 4 | 2+2 4 | 2+2 4 | 3+4 2 | 5+6 2 | 2+2 4 |
| | 5+6 2 | 3+4 2 | 2+2 4 | 2+2 4 | 2+2 4 | 5+6 2 | 3+4 2 | 2+2 4 |
| | 5+6 2 | 5+6 2 | 2+2 4 | 2+2 4 | 2+2 4 | 3+4 2 | 3+4 2 | 2+2 4 |
| | 3+4 2 | 3+4 2 | 2+2 4 | 2+2 4 | 2+2 4 | 5+6 2 | 5+6 2 | 2+2 4 |
| | 3+4 2 | 3+4 2 | 2+2 4 | 2+2 4 | 2+2 4 | 3+4 2 | 3+4 2 | 2+2 4 |
| | 5+6 2 | 5+6 2 | 2+2 4 | 2+2 4 | 2+2 4 | 5+6 2 | 5+6 2 | 2+2 4 |
| | 5+6 2 | 3+4 2 | 2+2 4 | 2+2 4 | 2+2 4 | 3+4 2 | 3+4 2 | 2+2 4 |
| | 3+4 2 | 5+6 2 | 2+2 4 | 2+2 4 | 2+2 4 | 5+6 2 | 5+6 2 | 2+2 4 |
| | 3+4 2 | 3+4 2 | 2+2 4 | 2+2 4 | 2+2 4 | 5+6 2 | 5+6 2 | 2+2 4 |
| | 3+4 2 | 5+6 2 | 2+2 4 | 2+2 4 | 2+2 4 | 3+4 2 | 3+4 2 | 2+2 4 |
| | 5+6 2 | 5+6 2 | 2+2 4 | 2+2 4 | 2+2 4 | 3+4 2 | 5+6 2 | 2+2 4 |
| | 3+4 2 | 3+4 2 | 2+2 4 | 2+2 4 | 2+2 4 | 5+6 2 | 5+6 2 | 2+2 4 |
| | 5+6 2 | 5+6 2 | 2+2 4 | 2+2 4 | 2+2 4 | 3+4 2 | 3+4 2 | 2+2 4 |
| | 5+6 2 | 3+4 2 | 2+2 4 | 2+2 4 | 2+2 4 | 5+6 2 | 3+4 2 | 2+2 4 |
| | 3+4 2 | 3+4 2 | 2+2 4 | 2+2 4 | 2+2 4 | 3+4 2 | 5+6 2 | 2+2 4 |
| | 3+4 2 | 5+6 2 | 2+2 4 | 2+2 4 | 2+2 4 | 5+6 2 | 5+6 2 | 2+2 4 |
| | 5+6 2 | 3+4 2 | 2+2 4 | 2+2 4 | 2+2 4 | 3+4 2 | 3+4 2 | 2+2 4 |
| | 5+6 2 | 5+6 2 | | | | 5+6 2 | 3+4 2 | |
| | 5+6 2 | 3+4 2 | | | | 3+4 2 | 3+4 2 | |
| | 3+4 2 | 5+6 2 | | | | 5+6 2 | 5+6 2 | |
| | 5+6 2 | 5+6 2 | | | | 5+6 2 | 3+4 2 | |
| | 5+6 2 | 3+4 2 | | | | 3+4 2 | 3+4 2 | |
| | 3+4 2 | 5+6 2 | | | | 3+4 2 | 5+6 2 | |
| | 5+6 2 | 5+6 2 | | | | 5+6 2 | 3+4 2 | |
| | 3+4 2 | 3+4 2 | | | | 3+4 2 | 5+6 2 | |
| | 3+4 2 | 3+4 2 | | | | 5+6 2 | 5+6 2 | |
| | 3+4 2 | 5+6 2 | | | | 3+4 2 | 3+4 2 | |
| | 5+6 2 | 5+6 2 | | | | 3+4 2 | 5+6 2 | |
| | 5+6 2 | 3+4 2 | | | | 5+6 2 | 3+4 2 | |
| | 3+4 2 | 3+4 2 | | | | 3+4 2 | 5+6 2 | |
| | 5+6 2 | 3+4 2 | | | | 5+6 2 | 5+6 2 | |
| | 5+6 2 | 3+4 2 | | | | 5+6 2 | 3+4 2 | |
| | 3+4 2 | 5+6 2 | | | | 3+4 2 | 3+4 2 | |
| | 5+6 2 | 5+6 2 | | | | 3+4 2 | 5+6 2 | |
| | 3+4 2 | 3+4 2 | | | | 3+4 2 | 3+4 2 | |
| | 3+4 2 | 3+4 2 | | | | 5+6 2 | 5+6 2 | |
| | 5+6 2 | 5+6 2 | | | | 5+6 2 | 5+6 2 | |
| | 5+6 2 | 3+4 2 | | | | 3+4 2 | 3+4 2 | |
| | 3+4 2 | 5+6 2 | | | | 5+6 2 | 5+6 2 | |
| | 3+4 2 | 5+6 2 | | | | 3+4 2 | 3+4 2 | |
| | 5+6 2 | 3+4 2 | | | | 5+6 2 | 5+6 2 | |
| | 3+4 2 | 3+4 2 | | | | 3+4 2 | 3+4 2 | |
| | 3+4 2 | 5+6 2 | | | | 5+6 2 | 3+4 2 | |
| | 5+6 2 | 3+4 2 | | | | 5+6 2 | 5+6 2 | |
| | 5+6 2 | 5+6 2 | | | | 3+4 2 | 5+6 2 | |
| | 3+4 2 | 5+6 2 | | | | 5+6 2 | 3+4 2 | |
| | 3+4 2 | 5+6 2 | | | | 3+4 2 | 3+4 2 | |
| | 3+4 2 | 3+4 2 | | | | 3+4 2 | 5+6 2 | |
| | 5+6 2 | 3+4 2 | | | | 5+6 2 | 3+4 2 | |
| | 5+6 2 | 5+6 2 | | | | 5+6 2 | 5+6 2 | |
| | 5+6 2 | 3+4 2 | | | | 3+4 2 | 5+6 2 | |
| | 3+4 2 | 5+6 2 | | | | 5+6 2 | 3+4 2 | |
| | 5+6 2 | 3+4 2 | | | | 5+6 2 | 5+6 2 | |
| | 5+6 2 | 5+6 2 | | | | 3+4 2 | 3+4 2 | |
| | 5+6 2 | 3+4 2 | | | | 3+4 2 | 5+6 2 | |
| | 3+4 2 | 3+4 2 | | | | 3+4 2 | 3+4 2 | |
| | 3+4 2 | 5+6 2 | | | | 5+6 2 | 3+4 2 | |
| | 5+6 2 | 5+6 2 | | | | 5+6 2 | 5+6 2 | |
| | 3+4 2 | 3+4 2 | | | | 3+4 2 | 5+6 2 | |
| | 5+6 2 | 5+6 2 | | | | 5+6 2 | 3+4 2 | |
| | 3+4 2 | 3+4 2 | | | | 3+4 2 | 5+6 2 | |
| | 5+6 2 | 5+6 2 | | | | 5+6 2 | 3+4 2 | |
| | 5+6 2 | 5+6 2 | | | | 3+4 2 | 3+4 2 | |
| | 3+4 2 | 3+4 2 | | | | 5+6 2 | 5+6 2 | |
| | 3+4 2 | 3+4 2 | | | | 3+4 2 | 3+4 2 | |
| | 3+4 2 | 5+6 2 | | | | 5+6 2 | 3+4 2 | |
| | 5+6 2 | 3+4 2 | | | | 5+6 2 | 5+6 2 | |
| | 3+4 2 | 5+6 2 | | | | 3+4 2 | 5+6 2 | |
| | 5+6 2 | 5+6 2 | | | | 5+6 2 | 3+4 2 | |
| | 3+4 2 | 3+4 2 | | | | 3+4 2 | 5+6 2 | |
| | 3+4 2 | 3+4 2 | | | | 5+6 2 | 5+6 2 | |
| | 5+6 2 | 3+4 2 | | | | 5+6 2 | 3+4 2 | |
| | 3+4 2 | 5+6 2 | | | | 3+4 2 | 5+6 2 | |
| | 5+6 2 | 5+6 2 | | | | 3+4 2 | 3+4 2 | |
| | 5+6 2 | 5+6 2 | | | | 3+4 2 | 3+4 2 | |
| | 5+6 2 | 5+6 2 | | | | 3+4 2 | 3+4 2 | |
| | 3+4 2 | 3+4 2 | | | | 5+6 2 | 5+6 2 | |
| | 3+4 2 | 3+4 2 | | | | 5+6 2 | 5+6 2 | |
| | 3+4 2 | 5+6 2 | | | | 3+4 2 | 3+4 2 | |
| | 5+6 2 | 5+6 2 | | | | 5+6 2 | 5+6 2 | |
| 11.2.89 | 416299 | 80 | Ba1Bi2Pd2 | | | | | |
| | -A | -B | -C | -D | -E | -GM | -Y | -Z |
| | 3+4 2 | 3+4 2 | 2+2 4 | 2+2 4 | 2+2 4 | 3+4 2 | 3+4 2 | 2+2 4 |
| | 5+6 2 | 5+6 2 | 2+2 4 | 2+2 4 | 2+2 4 | 5+6 2 | 5+6 2 | 2+2 4 |
| | 3+4 2 | 3+4 2 | 2+2 4 | 2+2 4 | 2+2 4 | 3+4 2 | 3+4 2 | 2+2 4 |
| | 5+6 2 | 5+6 2 | 2+2 4 | 2+2 4 | 2+2 4 | 5+6 2 | 5+6 2 | 2+2 4 |
| | 5+6 2 | 3+4 2 | 2+2 4 | 2+2 4 | 2+2 4 | 3+4 2 | 3+4 2 | 2+2 4 |
| | 3+4 2 | 5+6 2 | 2+2 4 | 2+2 4 | 2+2 4 | 5+6 2 | 5+6 2 | 2+2 4 |
| | 3+4 2 | 3+4 2 | 2+2 4 | 2+2 4 | 2+2 4 | 3+4 2 | 3+4 2 | 2+2 4 |
| | 5+6 2 | 5+6 2 | 2+2 4 | 2+2 4 | 2+2 4 | 5+6 2 | 5+6 2 | 2+2 4 |
| | 5+6 2 | 3+4 2 | 2+2 4 | 2+2 4 | 2+2 4 | 3+4 2 | 3+4 2 | 2+2 4 |
| | 3+4 2 | 5+6 2 | 2+2 4 | 2+2 4 | 2+2 4 | 3+4 2 | 5+6 2 | 2+2 4 |
| | 5+6 2 | 3+4 2 | 2+2 4 | 2+2 4 | 2+2 4 | 5+6 2 | 3+4 2 | 2+2 4 |
| | 3+4 2 | 5+6 2 | 2+2 4 | 2+2 4 | 2+2 4 | 5+6 2 | 5+6 2 | 2+2 4 |
| | 3+4 2 | 3+4 2 | 2+2 4 | 2+2 4 | 2+2 4 | 3+4 2 | 3+4 2 | 2+2 4 |
| | 5+6 2 | 5+6 2 | 2+2 4 | 2+2 4 | 2+2 4 | 5+6 2 | 5+6 2 | 2+2 4 |
| | 5+6 2 | 3+4 2 | 2+2 4 | 2+2 4 | 2+2 4 | 3+4 2 | 3+4 2 | 2+2 4 |
| | 3+4 2 | 5+6 2 | 2+2 4 | 2+2 4 | 2+2 4 | 3+4 2 | 5+6 2 | 2+2 4 |
| | 5+6 2 | 3+4 2 | 2+2 4 | 2+2 4 | 2+2 4 | 5+6 2 | 3+4 2 | 2+2 4 |
| | 5+6 2 | 3+4 2 | 2+2 4 | 2+2 4 | 2+2 4 | 5+6 2 | 5+6 2 | 2+2 4 |
| | 3+4 2 | 3+4 2 | 2+2 4 | 2+2 4 | 2+2 4 | 3+4 2 | 3+4 2 | 2+2 4 |
| | 3+4 2 | 5+6 2 | 2+2 4 | 2+2 4 | 2+2 4 | 5+6 2 | 3+4 2 | 2+2 4 |
| | 5+6 2 | 3+4 2 | | | | 3+4 2 | 5+6 2 | |
| | 3+4 2 | 5+6 2 | | | | 3+4 2 | 5+6 2 | |
| | 3+4 2 | 5+6 2 | | | | 5+6 2 | 5+6 2 | |
| | 5+6 2 | 5+6 2 | | | | 5+6 2 | 3+4 2 | |
| | 3+4 2 | 3+4 2 | | | | 3+4 2 | 5+6 2 | |
| | 5+6 2 | 5+6 2 | | | | 3+4 2 | 3+4 2 | |
| | 3+4 2 | 3+4 2 | | | | 3+4 2 | 3+4 2 | |
| | 5+6 2 | 5+6 2 | | | | 5+6 2 | 5+6 2 | |
| | 5+6 2 | 3+4 2 | | | | 5+6 2 | 3+4 2 | |
| | 3+4 2 | 5+6 2 | | | | 3+4 2 | 3+4 2 | |
| | 5+6 2 | 5+6 2 | | | | 5+6 2 | 5+6 2 | |
| | 3+4 2 | 3+4 2 | | | | 5+6 2 | 5+6 2 | |
| | 5+6 2 | 5+6 2 | | | | 5+6 2 | 3+4 2 | |
| | 3+4 2 | 5+6 2 | | | | 3+4 2 | 5+6 2 | |
| | 3+4 2 | 3+4 2 | | | | 3+4 2 | 5+6 2 | |
| | 5+6 2 | 3+4 2 | | | | 3+4 2 | 3+4 2 | |
| | 3+4 2 | 5+6 2 | | | | 3+4 2 | 5+6 2 | |
| | 5+6 2 | 5+6 2 | | | | 3+4 2 | 3+4 2 | |
| | 5+6 2 | 3+4 2 | | | | 5+6 2 | 5+6 2 | |

Continued on next column

**TABLE XXIII– continued**

| TQC | ICSD | NoE | SF | | | | | |
|---|---|---|---|---|---|---|---|---|
| | 3+4 2 | 3+4 2 | | | | 3+4 2 | 3+4 2 | |
| 11.2.90 | 413519 | 120 | Ba1.5Sb3 | | | | | |
| | -A | -B | -C | -D | -E | -GM | -Y | -Z |
| | 3+4 2 | 3+4 2 | 2+2 4 | 2+2 4 | 2+2 4 | 3+4 2 | 3+4 2 | 2+2 4 |
| | 5+6 2 | 5+6 2 | 2+2 4 | 2+2 4 | 2+2 4 | 5+6 2 | 5+6 2 | 2+2 4 |
| | 3+4 2 | 3+4 2 | 2+2 4 | 2+2 4 | 2+2 4 | 3+4 2 | 3+4 2 | 2+2 4 |
| | 5+6 2 | 5+6 2 | 2+2 4 | 2+2 4 | 2+2 4 | 5+6 2 | 5+6 2 | 2+2 4 |
| | 5+6 2 | 3+4 2 | 2+2 4 | 2+2 4 | 2+2 4 | 3+4 2 | 5+6 2 | 2+2 4 |
| | 3+4 2 | 5+6 2 | 2+2 4 | 2+2 4 | 2+2 4 | 5+6 2 | 3+4 2 | 2+2 4 |
| | 3+4 2 | 3+4 2 | 2+2 4 | 2+2 4 | 2+2 4 | 3+4 2 | 5+6 2 | 2+2 4 |
| | 5+6 2 | 5+6 2 | 2+2 4 | 2+2 4 | 2+2 4 | 5+6 2 | 3+4 2 | 2+2 4 |
| | 5+6 2 | 5+6 2 | 2+2 4 | 2+2 4 | 2+2 4 | 3+4 2 | 3+4 2 | 2+2 4 |
| | 3+4 2 | 3+4 2 | 2+2 4 | 2+2 4 | 2+2 4 | 5+6 2 | 5+6 2 | 2+2 4 |
| | 5+6 2 | 3+4 2 | 2+2 4 | 2+2 4 | 2+2 4 | 3+4 2 | 5+6 2 | 2+2 4 |
| | 3+4 2 | 5+6 2 | 2+2 4 | 2+2 4 | 2+2 4 | 5+6 2 | 3+4 2 | 2+2 4 |
| | 5+6 2 | 5+6 2 | 2+2 4 | 2+2 4 | 2+2 4 | 5+6 2 | 3+4 2 | 2+2 4 |
| | 3+4 2 | 3+4 2 | 2+2 4 | 2+2 4 | 2+2 4 | 3+4 2 | 5+6 2 | 2+2 4 |
| | 3+4 2 | 5+6 2 | 2+2 4 | 2+2 4 | 2+2 4 | 3+4 2 | 3+4 2 | 2+2 4 |
| | 5+6 2 | 3+4 2 | 2+2 4 | 2+2 4 | 2+2 4 | 5+6 2 | 5+6 2 | 2+2 4 |
| | 5+6 2 | 3+4 2 | 2+2 4 | 2+2 4 | 2+2 4 | 3+4 2 | 5+6 2 | 2+2 4 |
| | 3+4 2 | 5+6 2 | 2+2 4 | 2+2 4 | 2+2 4 | 5+6 2 | 3+4 2 | 2+2 4 |
| | 3+4 2 | 3+4 2 | 2+2 4 | 2+2 4 | 2+2 4 | 5+6 2 | 5+6 2 | 2+2 4 |
| | 5+6 2 | 5+6 2 | 2+2 4 | 2+2 4 | 2+2 4 | 3+4 2 | 3+4 2 | 2+2 4 |
| | 3+4 2 | 3+4 2 | 2+2 4 | 2+2 4 | 2+2 4 | 3+4 2 | 5+6 2 | 2+2 4 |
| | 5+6 2 | 3+4 2 | 2+2 4 | 2+2 4 | 2+2 4 | 3+4 2 | 3+4 2 | 2+2 4 |
| | 5+6 2 | 5+6 2 | 2+2 4 | 2+2 4 | 2+2 4 | 5+6 2 | 5+6 2 | 2+2 4 |
| | 3+4 2 | 5+6 2 | 2+2 4 | 2+2 4 | 2+2 4 | 5+6 2 | 3+4 2 | 2+2 4 |
| | 5+6 2 | 3+4 2 | 2+2 4 | 2+2 4 | 2+2 4 | 3+4 2 | 5+6 2 | 2+2 4 |
| | 3+4 2 | 5+6 2 | 2+2 4 | 2+2 4 | 2+2 4 | 5+6 2 | 3+4 2 | 2+2 4 |
| | 5+6 2 | 3+4 2 | 2+2 4 | 2+2 4 | 2+2 4 | 3+4 2 | 3+4 2 | 2+2 4 |
| | 3+4 2 | 5+6 2 | 2+2 4 | 2+2 4 | 2+2 4 | 5+6 2 | 3+4 2 | 2+2 4 |
| | 5+6 2 | 3+4 2 | 2+2 4 | 2+2 4 | 2+2 4 | 3+4 2 | 5+6 2 | 2+2 4 |
| | 3+4 2 | 5+6 2 | 2+2 4 | 2+2 4 | 2+2 4 | 3+4 2 | 3+4 2 | 2+2 4 |
| | 5+6 2 | 5+6 2 | | | | 5+6 2 | 5+6 2 | |
| | 3+4 2 | 3+4 2 | | | | 3+4 2 | 3+4 2 | |
| | 3+4 2 | 5+6 2 | | | | 5+6 2 | 5+6 2 | |
| | 5+6 2 | 3+4 2 | | | | 5+6 2 | 5+6 2 | |
| | 3+4 2 | 3+4 2 | | | | 3+4 2 | 3+4 2 | |
| | 5+6 2 | 5+6 2 | | | | 5+6 2 | 5+6 2 | |
| | 5+6 2 | 3+4 2 | | | | 3+4 2 | 5+6 2 | |
| | 3+4 2 | 5+6 2 | | | | 5+6 2 | 3+4 2 | |
| | 5+6 2 | 3+4 2 | | | | 3+4 2 | 5+6 2 | |
| | 3+4 2 | 5+6 2 | | | | 5+6 2 | 3+4 2 | |
| | 3+4 2 | 5+6 2 | | | | 3+4 2 | 3+4 2 | |
| | 5+6 2 | 3+4 2 | | | | 5+6 2 | 5+6 2 | |
| | 3+4 2 | 5+6 2 | | | | 3+4 2 | 5+6 2 | |
| | 5+6 2 | 3+4 2 | | | | 5+6 2 | 3+4 2 | |
| | 3+4 2 | 5+6 2 | | | | 3+4 2 | 5+6 2 | |
| | 5+6 2 | 3+4 2 | | | | 5+6 2 | 3+4 2 | |
| | 3+4 2 | 5+6 2 | | | | 3+4 2 | 5+6 2 | |
| | 5+6 2 | 3+4 2 | | | | 5+6 2 | 3+4 2 | |
| | 3+4 2 | 3+4 2 | | | | 3+4 2 | 5+6 2 | |
| | 5+6 2 | 5+6 2 | | | | 3+4 2 | 3+4 2 | |
| | 3+4 2 | 5+6 2 | | | | 5+6 2 | 3+4 2 | |
| | 5+6 2 | 3+4 2 | | | | 3+4 2 | 3+4 2 | |
| | 3+4 2 | 5+6 2 | | | | 5+6 2 | 5+6 2 | |
| | 5+6 2 | 5+6 2 | | | | 3+4 2 | 3+4 2 | |
| | 5+6 2 | 3+4 2 | | | | 5+6 2 | 5+6 2 | |
| | 3+4 2 | 5+6 2 | | | | 3+4 2 | 3+4 2 | |
| | 3+4 2 | 3+4 2 | | | | 3+4 2 | 3+4 2 | |
| | 3+4 2 | 3+4 2 | | | | 5+6 2 | 5+6 2 | |

Continued on next column

**TABLE XXIII– continued**

| TQC | ICSD | NoE | SF | | | | | |
|---|---|---|---|---|---|---|---|---|
| | 5+6 2 | 5+6 2 | | | | 3+4 2 | 5+6 2 | |
| | 5+6 2 | 3+4 2 | | | | 5+6 2 | 3+4 2 | |
| 11.2.91 | 261042 | 116 | Au2In4Yb1 | | | | | |
| | -A | -B | -C | -D | -E | -GM | -Y | -Z |
| | 3+4 2 | 3+4 2 | 2+2 4 | 2+2 4 | 2+2 4 | 3+4 2 | 3+4 2 | 2+2 4 |
| | 5+6 2 | 5+6 2 | 2+2 4 | 2+2 4 | 2+2 4 | 5+6 2 | 5+6 2 | 2+2 4 |
| | 3+4 2 | 3+4 2 | 2+2 4 | 2+2 4 | 2+2 4 | 3+4 2 | 3+4 2 | 2+2 4 |
| | 5+6 2 | 5+6 2 | 2+2 4 | 2+2 4 | 2+2 4 | 5+6 2 | 5+6 2 | 2+2 4 |
| | 5+6 2 | 3+4 2 | 2+2 4 | 2+2 4 | 2+2 4 | 3+4 2 | 3+4 2 | 2+2 4 |
| | 3+4 2 | 5+6 2 | 2+2 4 | 2+2 4 | 2+2 4 | 5+6 2 | 5+6 2 | 2+2 4 |
| | 3+4 2 | 3+4 2 | 2+2 4 | 2+2 4 | 2+2 4 | 3+4 2 | 3+4 2 | 2+2 4 |
| | 5+6 2 | 5+6 2 | 2+2 4 | 2+2 4 | 2+2 4 | 5+6 2 | 5+6 2 | 2+2 4 |
| | 3+4 2 | 3+4 2 | 2+2 4 | 2+2 4 | 2+2 4 | 3+4 2 | 5+6 2 | 2+2 4 |
| | 5+6 2 | 5+6 2 | 2+2 4 | 2+2 4 | 2+2 4 | 3+4 2 | 3+4 2 | 2+2 4 |
| | 5+6 2 | 5+6 2 | 2+2 4 | 2+2 4 | 2+2 4 | 5+6 2 | 3+4 2 | 2+2 4 |
| | 3+4 2 | 3+4 2 | 2+2 4 | 2+2 4 | 2+2 4 | 5+6 2 | 5+6 2 | 2+2 4 |
| | 3+4 2 | 5+6 2 | 2+2 4 | 2+2 4 | 2+2 4 | 3+4 2 | 5+6 2 | 2+2 4 |
| | 5+6 2 | 3+4 2 | 2+2 4 | 2+2 4 | 2+2 4 | 3+4 2 | 3+4 2 | 2+2 4 |
| | 5+6 2 | 3+4 2 | 2+2 4 | 2+2 4 | 2+2 4 | 5+6 2 | 3+4 2 | 2+2 4 |
| | 3+4 2 | 3+4 2 | 2+2 4 | 2+2 4 | 2+2 4 | 3+4 2 | 5+6 2 | 2+2 4 |
| | 3+4 2 | 5+6 2 | 2+2 4 | 2+2 4 | 2+2 4 | 5+6 2 | 3+4 2 | 2+2 4 |
| | 5+6 2 | 5+6 2 | 2+2 4 | 2+2 4 | 2+2 4 | 5+6 2 | 5+6 2 | 2+2 4 |
| | 5+6 2 | 5+6 2 | 2+2 4 | 2+2 4 | 2+2 4 | 3+4 2 | 3+4 2 | 2+2 4 |
| | 3+4 2 | 3+4 2 | 2+2 4 | 2+2 4 | 2+2 4 | 3+4 2 | 5+6 2 | 2+2 4 |
| | 3+4 2 | 5+6 2 | 2+2 4 | 2+2 4 | 2+2 4 | 5+6 2 | 3+4 2 | 2+2 4 |
| | 5+6 2 | 3+4 2 | 2+2 4 | 2+2 4 | 2+2 4 | 3+4 2 | 5+6 2 | 2+2 4 |
| | 5+6 2 | 3+4 2 | 2+2 4 | 2+2 4 | 2+2 4 | 5+6 2 | 3+4 2 | 2+2 4 |
| | 3+4 2 | 3+4 2 | 2+2 4 | 2+2 4 | 2+2 4 | 3+4 2 | 5+6 2 | 2+2 4 |
| | 5+6 2 | 5+6 2 | 2+2 4 | 2+2 4 | 2+2 4 | 5+6 2 | 3+4 2 | 2+2 4 |
| | 3+4 2 | 5+6 2 | 2+2 4 | 2+2 4 | 2+2 4 | 5+6 2 | 5+6 2 | 2+2 4 |
| | 3+4 2 | 5+6 2 | 2+2 4 | 2+2 4 | 2+2 4 | 3+4 2 | 3+4 2 | 2+2 4 |
| | 5+6 2 | 3+4 2 | 2+2 4 | 2+2 4 | 2+2 4 | 3+4 2 | 5+6 2 | 2+2 4 |
| | 3+4 2 | 3+4 2 | 2+2 4 | 2+2 4 | 2+2 4 | 3+4 2 | 5+6 2 | 2+2 4 |
| | 5+6 2 | 5+6 2 | | | | 5+6 2 | 3+4 2 | |
| | 5+6 2 | 5+6 2 | | | | 3+4 2 | 5+6 2 | |
| | 3+4 2 | 3+4 2 | | | | 5+6 2 | 3+4 2 | |
| | 3+4 2 | 3+4 2 | | | | 5+6 2 | 3+4 2 | |
| | 5+6 2 | 5+6 2 | | | | 3+4 2 | 5+6 2 | |
| | 5+6 2 | 3+4 2 | | | | 5+6 2 | 3+4 2 | |
| | 3+4 2 | 5+6 2 | | | | 3+4 2 | 5+6 2 | |
| | 3+4 2 | 5+6 2 | | | | 5+6 2 | 5+6 2 | |
| | 5+6 2 | 3+4 2 | | | | 3+4 2 | 3+4 2 | |
| | 3+4 2 | 3+4 2 | | | | 5+6 2 | 5+6 2 | |
| | 5+6 2 | 5+6 2 | | | | 3+4 2 | 3+4 2 | |
| | 3+4 2 | 5+6 2 | | | | 5+6 2 | 5+6 2 | |
| | 5+6 2 | 3+4 2 | | | | 3+4 2 | 3+4 2 | |
| | 3+4 2 | 3+4 2 | | | | 5+6 2 | 3+4 2 | |
| | 5+6 2 | 5+6 2 | | | | 3+4 2 | 5+6 2 | |
| | 3+4 2 | 3+4 2 | | | | 3+4 2 | 5+6 2 | |
| | 5+6 2 | 5+6 2 | | | | 3+4 2 | 5+6 2 | |
| | 3+4 2 | 3+4 2 | | | | 5+6 2 | 3+4 2 | |
| | 5+6 2 | 5+6 2 | | | | 5+6 2 | 3+4 2 | |
| | 5+6 2 | 5+6 2 | | | | 3+4 2 | 3+4 2 | |
| | 3+4 2 | 3+4 2 | | | | 5+6 2 | 3+4 2 | |
| | 5+6 2 | 3+4 2 | | | | 5+6 2 | 5+6 2 | |
| | 3+4 2 | 5+6 2 | | | | 3+4 2 | 5+6 2 | |
| | 3+4 2 | 5+6 2 | | | | 3+4 2 | 3+4 2 | |
| | 5+6 2 | 3+4 2 | | | | 3+4 2 | 5+6 2 | |
| | 3+4 2 | 3+4 2 | | | | 3+4 2 | 3+4 2 | |
| | 5+6 2 | 5+6 2 | | | | 5+6 2 | 5+6 2 | |
| | 5+6 2 | 3+4 2 | | | | 5+6 2 | 3+4 2 | |

Continued on next column

**TABLE XXIII– continued**

| TQC | ICSD | NoE | SF | | | | | |
|---|---|---|---|---|---|---|---|---|
| | 5+6 2 | 3+4 2 | | | | 5+6 2 | 3+4 2 | |
| 11.2.92 | 410884 | 152 | Au2Eu2Sn5 | | | | | |
| | -A | -B | -C | -D | -E | -GM | -Y | -Z |
| | 3+4 2 | 3+4 2 | 2+2 4 | 2+2 4 | 2+2 4 | 3+4 2 | 3+4 2 | 2+2 4 |
| | 5+6 2 | 5+6 2 | 2+2 4 | 2+2 4 | 2+2 4 | 5+6 2 | 5+6 2 | 2+2 4 |
| | 3+4 2 | 3+4 2 | 2+2 4 | 2+2 4 | 2+2 4 | 3+4 2 | 3+4 2 | 2+2 4 |
| | 5+6 2 | 5+6 2 | 2+2 4 | 2+2 4 | 2+2 4 | 5+6 2 | 5+6 2 | 2+2 4 |
| | 3+4 2 | 3+4 2 | 2+2 4 | 2+2 4 | 2+2 4 | 3+4 2 | 3+4 2 | 2+2 4 |
| | 5+6 2 | 5+6 2 | 2+2 4 | 2+2 4 | 2+2 4 | 5+6 2 | 5+6 2 | 2+2 4 |
| | 3+4 2 | 3+4 2 | 2+2 4 | 2+2 4 | 2+2 4 | 3+4 2 | 3+4 2 | 2+2 4 |
| | 5+6 2 | 5+6 2 | 2+2 4 | 2+2 4 | 2+2 4 | 5+6 2 | 5+6 2 | 2+2 4 |
| | 3+4 2 | 3+4 2 | 2+2 4 | 2+2 4 | 2+2 4 | 3+4 2 | 3+4 2 | 2+2 4 |
| | 5+6 2 | 5+6 2 | 2+2 4 | 2+2 4 | 2+2 4 | 5+6 2 | 5+6 2 | 2+2 4 |
| | 5+6 2 | 3+4 2 | 2+2 4 | 2+2 4 | 2+2 4 | 3+4 2 | 3+4 2 | 2+2 4 |
| | 3+4 2 | 5+6 2 | 2+2 4 | 2+2 4 | 2+2 4 | 5+6 2 | 5+6 2 | 2+2 4 |
| | 3+4 2 | 3+4 2 | 2+2 4 | 2+2 4 | 2+2 4 | 3+4 2 | 5+6 2 | 2+2 4 |
| | 5+6 2 | 5+6 2 | 2+2 4 | 2+2 4 | 2+2 4 | 5+6 2 | 3+4 2 | 2+2 4 |
| | 3+4 2 | 3+4 2 | 2+2 4 | 2+2 4 | 2+2 4 | 3+4 2 | 5+6 2 | 2+2 4 |
| | 5+6 2 | 5+6 2 | 2+2 4 | 2+2 4 | 2+2 4 | 5+6 2 | 3+4 2 | 2+2 4 |
| | 5+6 2 | 3+4 2 | 2+2 4 | 2+2 4 | 2+2 4 | 3+4 2 | 3+4 2 | 2+2 4 |
| | 3+4 2 | 5+6 2 | 2+2 4 | 2+2 4 | 2+2 4 | 3+4 2 | 5+6 2 | 2+2 4 |
| | 5+6 2 | 5+6 2 | 2+2 4 | 2+2 4 | 2+2 4 | 5+6 2 | 3+4 2 | 2+2 4 |
| | 3+4 2 | 3+4 2 | 2+2 4 | 2+2 4 | 2+2 4 | 3+4 2 | 5+6 2 | 2+2 4 |
| | 5+6 2 | 5+6 2 | 2+2 4 | 2+2 4 | 2+2 4 | 5+6 2 | 5+6 2 | 2+2 4 |
| | 3+4 2 | 3+4 2 | 2+2 4 | 2+2 4 | 2+2 4 | 3+4 2 | 3+4 2 | 2+2 4 |
| | 3+4 2 | 5+6 2 | 2+2 4 | 2+2 4 | 2+2 4 | 5+6 2 | 5+6 2 | 2+2 4 |
| | 5+6 2 | 3+4 2 | 2+2 4 | 2+2 4 | 2+2 4 | 5+6 2 | 3+4 2 | 2+2 4 |
| | 5+6 2 | 5+6 2 | 2+2 4 | 2+2 4 | 2+2 4 | 3+4 2 | 5+6 2 | 2+2 4 |
| | 3+4 2 | 3+4 2 | 2+2 4 | 2+2 4 | 2+2 4 | 5+6 2 | 3+4 2 | 2+2 4 |
| | 5+6 2 | 5+6 2 | 2+2 4 | 2+2 4 | 2+2 4 | 3+4 2 | 5+6 2 | 2+2 4 |
| | 3+4 2 | 3+4 2 | 2+2 4 | 2+2 4 | 2+2 4 | 3+4 2 | 3+4 2 | 2+2 4 |
| | 3+4 2 | 3+4 2 | 2+2 4 | 2+2 4 | 2+2 4 | 5+6 2 | 5+6 2 | 2+2 4 |
| | 5+6 2 | 5+6 2 | 2+2 4 | 2+2 4 | 2+2 4 | 3+4 2 | 3+4 2 | 2+2 4 |
| | 5+6 2 | 5+6 2 | 2+2 4 | 2+2 4 | 2+2 4 | 5+6 2 | 3+4 2 | 2+2 4 |
| | 3+4 2 | 3+4 2 | 2+2 4 | 2+2 4 | 2+2 4 | 3+4 2 | 5+6 2 | 2+2 4 |
| | 3+4 2 | 5+6 2 | 2+2 4 | 2+2 4 | 2+2 4 | 5+6 2 | 3+4 2 | 2+2 4 |
| | 5+6 2 | 3+4 2 | 2+2 4 | 2+2 4 | 2+2 4 | 3+4 2 | 5+6 2 | 2+2 4 |
| | 5+6 2 | 3+4 2 | 2+2 4 | 2+2 4 | 2+2 4 | 5+6 2 | 5+6 2 | 2+2 4 |
| | 3+4 2 | 5+6 2 | 2+2 4 | 2+2 4 | 2+2 4 | 3+4 2 | 3+4 2 | 2+2 4 |
| | 3+4 2 | 5+6 2 | 2+2 4 | 2+2 4 | 2+2 4 | 5+6 2 | 5+6 2 | 2+2 4 |
| | 5+6 2 | 3+4 2 | 2+2 4 | 2+2 4 | 2+2 4 | 3+4 2 | 3+4 2 | 2+2 4 |
| | 5+6 2 | 3+4 2 | | | | 5+6 2 | 5+6 2 | |
| | 3+4 2 | 5+6 2 | | | | 3+4 2 | 3+4 2 | |
| | 5+6 2 | 3+4 2 | | | | 5+6 2 | 3+4 2 | |
| | 3+4 2 | 5+6 2 | | | | 3+4 2 | 5+6 2 | |
| | 5+6 2 | 5+6 2 | | | | 5+6 2 | 3+4 2 | |
| | 3+4 2 | 3+4 2 | | | | 3+4 2 | 5+6 2 | |
| | 3+4 2 | 5+6 2 | | | | 5+6 2 | 3+4 2 | |
| | 5+6 2 | 3+4 2 | | | | 3+4 2 | 5+6 2 | |
| | 5+6 2 | 3+4 2 | | | | 5+6 2 | 5+6 2 | |
| | 3+4 2 | 5+6 2 | | | | 5+6 2 | 3+4 2 | |
| | 3+4 2 | 5+6 2 | | | | 5+6 2 | 5+6 2 | |
| | 5+6 2 | 3+4 2 | | | | 3+4 2 | 3+4 2 | |
| | 3+4 2 | 3+4 2 | | | | 3+4 2 | 3+4 2 | |
| | 3+4 2 | 5+6 2 | | | | 3+4 2 | 5+6 2 | |
| | 5+6 2 | 5+6 2 | | | | 3+4 2 | 3+4 2 | |
| | 5+6 2 | 3+4 2 | | | | 3+4 2 | 3+4 2 | |
| | 3+4 2 | 5+6 2 | | | | 5+6 2 | 5+6 2 | |
| | 5+6 2 | 3+4 2 | | | | 5+6 2 | 5+6 2 | |
| | 3+4 2 | 5+6 2 | | | | 3+4 2 | 3+4 2 | |
| | 5+6 2 | 3+4 2 | | | | 5+6 2 | 3+4 2 | |

Continued on next column

| TQC | ICSD | NoE | SF | | | | | |
|---|---|---|---|---|---|---|---|---|
| | 3+4 2 | 3+4 2 | | | | 3+4 2 | 5+6 2 | |
| | 5+6 2 | 5+6 2 | | | | 3+4 2 | 3+4 2 | |
| | 5+6 2 | 5+6 2 | | | | 5+6 2 | 5+6 2 | |
| | 3+4 2 | 3+4 2 | | | | 5+6 2 | 5+6 2 | |
| | 5+6 2 | 5+6 2 | | | | 3+4 2 | 5+6 2 | |
| | 3+4 2 | 3+4 2 | | | | 3+4 2 | 3+4 2 | |
| | 5+6 2 | 3+4 2 | | | | 5+6 2 | 5+6 2 | |
| | 3+4 2 | 5+6 2 | | | | 5+6 2 | 3+4 2 | |
| | 5+6 2 | 3+4 2 | | | | 3+4 2 | 5+6 2 | |
| | 3+4 2 | 5+6 2 | | | | 3+4 2 | 3+4 2 | |
| | 3+4 2 | 3+4 2 | | | | 5+6 2 | 3+4 2 | |
| | 5+6 2 | 5+6 2 | | | | 3+4 2 | 3+4 2 | |
| | 3+4 2 | 5+6 2 | | | | 5+6 2 | 5+6 2 | |
| | 5+6 2 | 3+4 2 | | | | 3+4 2 | 3+4 2 | |
| | 3+4 2 | 5+6 2 | | | | 5+6 2 | 5+6 2 | |
| | 5+6 2 | 3+4 2 | | | | 5+6 2 | 5+6 2 | |
| | 5+6 2 | 5+6 2 | | | | 5+6 2 | 5+6 2 | |
| | 3+4 2 | 3+4 2 | | | | 3+4 2 | 3+4 2 | |
| 11.2.93 | 416298 | 84 | Au2Ba1Sb2 | | | | | |
| | -A | -B | -C | -D | -E | -GM | -Y | -Z |
| | 3+4 2 | 3+4 2 | 2+2 4 | 2+2 4 | 2+2 4 | 3+4 2 | 3+4 2 | 2+2 4 |
| | 5+6 2 | 5+6 2 | 2+2 4 | 2+2 4 | 2+2 4 | 5+6 2 | 5+6 2 | 2+2 4 |
| | 3+4 2 | 5+6 2 | 2+2 4 | 2+2 4 | 2+2 4 | 3+4 2 | 3+4 2 | 2+2 4 |
| | 5+6 2 | 3+4 2 | 2+2 4 | 2+2 4 | 2+2 4 | 5+6 2 | 5+6 2 | 2+2 4 |
| | 5+6 2 | 5+6 2 | 2+2 4 | 2+2 4 | 2+2 4 | 3+4 2 | 3+4 2 | 2+2 4 |
| | 3+4 2 | 3+4 2 | 2+2 4 | 2+2 4 | 2+2 4 | 5+6 2 | 5+6 2 | 2+2 4 |
| | 3+4 2 | 5+6 2 | 2+2 4 | 2+2 4 | 2+2 4 | 3+4 2 | 3+4 2 | 2+2 4 |
| | 5+6 2 | 3+4 2 | 2+2 4 | 2+2 4 | 2+2 4 | 5+6 2 | 5+6 2 | 2+2 4 |
| | 3+4 2 | 5+6 2 | 2+2 4 | 2+2 4 | 2+2 4 | 3+4 2 | 5+6 2 | 2+2 4 |
| | 5+6 2 | 3+4 2 | 2+2 4 | 2+2 4 | 2+2 4 | 3+4 2 | 3+4 2 | 2+2 4 |
| | 5+6 2 | 5+6 2 | 2+2 4 | 2+2 4 | 2+2 4 | 5+6 2 | 5+6 2 | 2+2 4 |
| | 3+4 2 | 3+4 2 | 2+2 4 | 2+2 4 | 2+2 4 | 5+6 2 | 3+4 2 | 2+2 4 |
| | 3+4 2 | 3+4 2 | 2+2 4 | 2+2 4 | 2+2 4 | 3+4 2 | 3+4 2 | 2+2 4 |
| | 5+6 2 | 5+6 2 | 2+2 4 | 2+2 4 | 2+2 4 | 5+6 2 | 5+6 2 | 2+2 4 |
| | 5+6 2 | 5+6 2 | 2+2 4 | 2+2 4 | 2+2 4 | 3+4 2 | 5+6 2 | 2+2 4 |
| | 3+4 2 | 3+4 2 | 2+2 4 | 2+2 4 | 2+2 4 | 5+6 2 | 3+4 2 | 2+2 4 |
| | 5+6 2 | 5+6 2 | 2+2 4 | 2+2 4 | 2+2 4 | 3+4 2 | 3+4 2 | 2+2 4 |
| | 3+4 2 | 3+4 2 | 2+2 4 | 2+2 4 | 2+2 4 | 5+6 2 | 5+6 2 | 2+2 4 |
| | 3+4 2 | 5+6 2 | 2+2 4 | 2+2 4 | 2+2 4 | 3+4 2 | 5+6 2 | 2+2 4 |
| | 5+6 2 | 5+6 2 | 2+2 4 | 2+2 4 | 2+2 4 | 3+4 2 | 3+4 2 | 2+2 4 |
| | 5+6 2 | 3+4 2 | 2+2 4 | 2+2 4 | 2+2 4 | 5+6 2 | 3+4 2 | 2+2 4 |
| | 3+4 2 | 5+6 2 | | | | 5+6 2 | 5+6 2 | |
| | 3+4 2 | 3+4 2 | | | | 3+4 2 | 3+4 2 | |
| | 5+6 2 | 3+4 2 | | | | 5+6 2 | 5+6 2 | |
| | 5+6 2 | 5+6 2 | | | | 3+4 2 | 5+6 2 | |
| | 3+4 2 | 3+4 2 | | | | 3+4 2 | 3+4 2 | |
| | 3+4 2 | 5+6 2 | | | | 5+6 2 | 5+6 2 | |
| | 5+6 2 | 3+4 2 | | | | 3+4 2 | 3+4 2 | |
| | 3+4 2 | 5+6 2 | | | | 5+6 2 | 5+6 2 | |
| | 5+6 2 | 3+4 2 | | | | 3+4 2 | 3+4 2 | |
| | 5+6 2 | 3+4 2 | | | | 5+6 2 | 5+6 2 | |
| | 3+4 2 | 5+6 2 | | | | 5+6 2 | 3+4 2 | |
| | 5+6 2 | 3+4 2 | | | | 3+4 2 | 5+6 2 | |
| | 3+4 2 | 5+6 2 | | | | 5+6 2 | 3+4 2 | |
| | 3+4 2 | 3+4 2 | | | | 3+4 2 | 3+4 2 | |
| | 5+6 2 | 5+6 2 | | | | 3+4 2 | 5+6 2 | |
| | 3+4 2 | 3+4 2 | | | | 3+4 2 | 3+4 2 | |
| | 5+6 2 | 5+6 2 | | | | 3+4 2 | 5+6 2 | |
| | 5+6 2 | 3+4 2 | | | | 5+6 2 | 5+6 2 | |
| | 3+4 2 | 5+6 2 | | | | 5+6 2 | 3+4 2 | |
| | 3+4 2 | 5+6 2 | | | | 3+4 2 | 3+4 2 | |

Continued on next column

**TABLE XXIII– continued**

| TQC | ICSD | NoE | SF | | | | | |
|---|---|---|---|---|---|---|---|---|
| | 5+6 2 | 3+4 2 | | | | 5+6 2 | 5+6 2 | |
| 11.2.94 | 902 | 108 | As1Cl1Co2O4 | | | | | |
| | -A | -B | -C | -D | -E | -GM | -Y | -Z |
| | 5+6 2 | 5+6 2 | 2+2 4 | 2+2 4 | 2+2 4 | 3+4 2 | 3+4 2 | 2+2 4 |
| | 3+4 2 | 3+4 2 | 2+2 4 | 2+2 4 | 2+2 4 | 5+6 2 | 5+6 2 | 2+2 4 |
| | 3+4 2 | 3+4 2 | 2+2 4 | 2+2 4 | 2+2 4 | 5+6 2 | 5+6 2 | 2+2 4 |
| | 5+6 2 | 5+6 2 | 2+2 4 | 2+2 4 | 2+2 4 | 3+4 2 | 3+4 2 | 2+2 4 |
| | 3+4 2 | 5+6 2 | 2+2 4 | 2+2 4 | 2+2 4 | 3+4 2 | 3+4 2 | 2+2 4 |
| | 5+6 2 | 3+4 2 | 2+2 4 | 2+2 4 | 2+2 4 | 5+6 2 | 5+6 2 | 2+2 4 |
| | 5+6 2 | 5+6 2 | 2+2 4 | 2+2 4 | 2+2 4 | 3+4 2 | 3+4 2 | 2+2 4 |
| | 3+4 2 | 3+4 2 | 2+2 4 | 2+2 4 | 2+2 4 | 5+6 2 | 5+6 2 | 2+2 4 |
| | 3+4 2 | 5+6 2 | 2+2 4 | 2+2 4 | 2+2 4 | 3+4 2 | 5+6 2 | 2+2 4 |
| | 5+6 2 | 3+4 2 | 2+2 4 | 2+2 4 | 2+2 4 | 5+6 2 | 3+4 2 | 2+2 4 |
| | 5+6 2 | 5+6 2 | 2+2 4 | 2+2 4 | 2+2 4 | 3+4 2 | 3+4 2 | 2+2 4 |
| | 3+4 2 | 3+4 2 | 2+2 4 | 2+2 4 | 2+2 4 | 5+6 2 | 5+6 2 | 2+2 4 |
| | 3+4 2 | 3+4 2 | 2+2 4 | 2+2 4 | 2+2 4 | 5+6 2 | 5+6 2 | 2+2 4 |
| | 3+4 2 | 5+6 2 | 2+2 4 | 2+2 4 | 2+2 4 | 3+4 2 | 5+6 2 | 2+2 4 |
| | 5+6 2 | 3+4 2 | 2+2 4 | 2+2 4 | 2+2 4 | 5+6 2 | 5+6 2 | 2+2 4 |
| | 5+6 2 | 5+6 2 | 2+2 4 | 2+2 4 | 2+2 4 | 3+4 2 | 3+4 2 | 2+2 4 |
| | 3+4 2 | 5+6 2 | 2+2 4 | 2+2 4 | 2+2 4 | 3+4 2 | 3+4 2 | 2+2 4 |
| | 5+6 2 | 3+4 2 | 2+2 4 | 2+2 4 | 2+2 4 | 5+6 2 | 3+4 2 | 2+2 4 |
| | 3+4 2 | 5+6 2 | 2+2 4 | 2+2 4 | 2+2 4 | 3+4 2 | 3+4 2 | 2+2 4 |
| | 5+6 2 | 5+6 2 | 2+2 4 | 2+2 4 | 2+2 4 | 3+4 2 | 3+4 2 | 2+2 4 |
| | 5+6 2 | 3+4 2 | 2+2 4 | 2+2 4 | 2+2 4 | 5+6 2 | 5+6 2 | 2+2 4 |
| | 3+4 2 | 5+6 2 | 2+2 4 | 2+2 4 | 2+2 4 | 3+4 2 | 5+6 2 | 2+2 4 |
| | 5+6 2 | 3+4 2 | 2+2 4 | 2+2 4 | 2+2 4 | 5+6 2 | 3+4 2 | 2+2 4 |
| | 3+4 2 | 5+6 2 | 2+2 4 | 2+2 4 | 2+2 4 | 3+4 2 | 5+6 2 | 2+2 4 |
| | 3+4 2 | 3+4 2 | 2+2 4 | 2+2 4 | 2+2 4 | 3+4 2 | 5+6 2 | 2+2 4 |
| | 3+4 2 | 5+6 2 | 2+2 4 | 2+2 4 | 2+2 4 | 5+6 2 | 5+6 2 | 2+2 4 |
| | 5+6 2 | 5+6 2 | 2+2 4 | 2+2 4 | 2+2 4 | 3+4 2 | 3+4 2 | 2+2 4 |
| | 3+4 2 | 3+4 2 | | | | 5+6 2 | 3+4 2 | |
| | 5+6 2 | 3+4 2 | | | | 3+4 2 | 5+6 2 | |
| | 3+4 2 | 5+6 2 | | | | 5+6 2 | 5+6 2 | |
| | 5+6 2 | 3+4 2 | | | | 3+4 2 | 3+4 2 | |
| | 5+6 2 | 3+4 2 | | | | 3+4 2 | 3+4 2 | |
| | 3+4 2 | 5+6 2 | | | | 5+6 2 | 3+4 2 | |
| | 5+6 2 | 3+4 2 | | | | 5+6 2 | 5+6 2 | |
| | 3+4 2 | 5+6 2 | | | | 5+6 2 | 5+6 2 | |
| | 3+4 2 | 3+4 2 | | | | 5+6 2 | 3+4 2 | |
| | 3+4 2 | 5+6 2 | | | | 3+4 2 | 5+6 2 | |
| | 5+6 2 | 3+4 2 | | | | 5+6 2 | 5+6 2 | |
| | 5+6 2 | 5+6 2 | | | | 5+6 2 | 3+4 2 | |
| | 5+6 2 | 3+4 2 | | | | 3+4 2 | 3+4 2 | |
| | 3+4 2 | 5+6 2 | | | | 3+4 2 | 5+6 2 | |
| | 3+4 2 | 5+6 2 | | | | 3+4 2 | 5+6 2 | |
| | 3+4 2 | 5+6 2 | | | | 3+4 2 | 5+6 2 | |
| | 3+4 2 | 3+4 2 | | | | 3+4 2 | 5+6 2 | |
| | 3+4 2 | 5+6 2 | | | | 5+6 2 | 5+6 2 | |
| | 3+4 2 | 5+6 2 | | | | 3+4 2 | 5+6 2 | |
| | 3+4 2 | 5+6 2 | | | | 3+4 2 | 5+6 2 | |
| | 5+6 2 | 3+4 2 | | | | 5+6 2 | 5+6 2 | |
| | 5+6 2 | 5+6 2 | | | | 3+4 2 | 3+4 2 | |
| | 3+4 2 | 5+6 2 | | | | 3+4 2 | 3+4 2 | |
| | 3+4 2 | 3+4 2 | | | | 5+6 2 | 5+6 2 | |
| | 5+6 2 | 5+6 2 | | | | 3+4 2 | 3+4 2 | |
| | 3+4 2 | 5+6 2 | | | | 5+6 2 | 5+6 2 | |
| | 3+4 2 | 3+4 2 | | | | 3+4 2 | 3+4 2 | |
| 11.2.95 | 43456 | 68 | Am1 | | | | | |
| | -A | -B | -C | -D | -E | -GM | -Y | -Z |
| | 3+4 2 | 3+4 2 | 2+2 4 | 2+2 4 | 2+2 4 | 3+4 2 | 3+4 2 | 2+2 4 |
| | 3+4 2 | 3+4 2 | 2+2 4 | 2+2 4 | 2+2 4 | 3+4 2 | 3+4 2 | 2+2 4 |
| | 5+6 2 | 5+6 2 | 2+2 4 | 2+2 4 | 2+2 4 | 5+6 2 | 5+6 2 | 2+2 4 |
| | 3+4 2 | 3+4 2 | 2+2 4 | 2+2 4 | 2+2 4 | 3+4 2 | 3+4 2 | 2+2 4 |
| | 5+6 2 | 5+6 2 | 2+2 4 | 2+2 4 | 2+2 4 | 5+6 2 | 5+6 2 | 2+2 4 |
| | 5+6 2 | 5+6 2 | 2+2 4 | 2+2 4 | 2+2 4 | 5+6 2 | 5+6 2 | 2+2 4 |
| | 3+4 2 | 3+4 2 | 2+2 4 | 2+2 4 | 2+2 4 | 3+4 2 | 3+4 2 | 2+2 4 |
| | 5+6 2 | 5+6 2 | 2+2 4 | 2+2 4 | 2+2 4 | 5+6 2 | 5+6 2 | 2+2 4 |
| | 5+6 2 | 5+6 2 | 2+2 4 | 2+2 4 | 2+2 4 | 5+6 2 | 5+6 2 | 2+2 4 |
| | 5+6 2 | 5+6 2 | 2+2 4 | 2+2 4 | 2+2 4 | 5+6 2 | 5+6 2 | 2+2 4 |
| | 3+4 2 | 3+4 2 | 2+2 4 | 2+2 4 | 2+2 4 | 3+4 2 | 3+4 2 | 2+2 4 |
| | 5+6 2 | 5+6 2 | 2+2 4 | 2+2 4 | 2+2 4 | 5+6 2 | 5+6 2 | 2+2 4 |
| | 5+6 2 | 5+6 2 | 2+2 4 | 2+2 4 | 2+2 4 | 5+6 2 | 5+6 2 | 2+2 4 |
| | 3+4 2 | 3+4 2 | 2+2 4 | 2+2 4 | 2+2 4 | 3+4 2 | 3+4 2 | 2+2 4 |
| | 5+6 2 | 5+6 2 | 2+2 4 | 2+2 4 | 2+2 4 | 5+6 2 | 5+6 2 | 2+2 4 |
| | 5+6 2 | 5+6 2 | 2+2 4 | 2+2 4 | 2+2 4 | 5+6 2 | 5+6 2 | 2+2 4 |
| | 3+4 2 | 3+4 2 | 2+2 4 | 2+2 4 | 2+2 4 | 3+4 2 | 3+4 2 | 2+2 4 |
| | 3+4 2 | 3+4 2 | | | | 3+4 2 | 3+4 2 | |
| | 3+4 2 | 3+4 2 | | | | 5+6 2 | 3+4 2 | |
| | 5+6 2 | 5+6 2 | | | | 3+4 2 | 5+6 2 | |
| | 3+4 2 | 3+4 2 | | | | 5+6 2 | 5+6 2 | |
| | 5+6 2 | 5+6 2 | | | | 5+6 2 | 3+4 2 | |
| | 3+4 2 | 5+6 2 | | | | 3+4 2 | 5+6 2 | |
| | 5+6 2 | 5+6 2 | | | | 5+6 2 | 5+6 2 | |
| | 3+4 2 | 3+4 2 | | | | 5+6 2 | 3+4 2 | |
| | 5+6 2 | 5+6 2 | | | | 5+6 2 | 5+6 2 | |
| | 5+6 2 | 5+6 2 | | | | 3+4 2 | 5+6 2 | |
| | 5+6 2 | 5+6 2 | | | | 5+6 2 | 5+6 2 | |
| | 5+6 2 | 3+4 2 | | | | 3+4 2 | 3+4 2 | |
| | 3+4 2 | 5+6 2 | | | | 5+6 2 | 5+6 2 | |
| | 5+6 2 | 5+6 2 | | | | 5+6 2 | 5+6 2 | |
| | 5+6 2 | 3+4 2 | | | | 5+6 2 | 3+4 2 | |
| | 5+6 2 | 5+6 2 | | | | 3+4 2 | 5+6 2 | |
| | 3+4 2 | 5+6 2 | | | | 3+4 2 | 5+6 2 | |
| 11.2.96 | 104626 | 156 | Al1Ce3 | | | | | |
| | -A | -B | -C | -D | -E | -GM | -Y | -Z |
| | 3+4 2 | 5+6 2 | 2+2 4 | 2+2 4 | 2+2 4 | 3+4 2 | 3+4 2 | 2+2 4 |
| | 3+4 2 | 3+4 2 | 2+2 4 | 2+2 4 | 2+2 4 | 3+4 2 | 5+6 2 | 2+2 4 |
| | 5+6 2 | 3+4 2 | 2+2 4 | 2+2 4 | 2+2 4 | 5+6 2 | 3+4 2 | 2+2 4 |
| | 5+6 2 | 5+6 2 | 2+2 4 | 2+2 4 | 2+2 4 | 5+6 2 | 5+6 2 | 2+2 4 |
| | 5+6 2 | 3+4 2 | 2+2 4 | 2+2 4 | 2+2 4 | 5+6 2 | 3+4 2 | 2+2 4 |
| | 3+4 2 | 5+6 2 | 2+2 4 | 2+2 4 | 2+2 4 | 5+6 2 | 5+6 2 | 2+2 4 |
| | 5+6 2 | 5+6 2 | 2+2 4 | 2+2 4 | 2+2 4 | 3+4 2 | 5+6 2 | 2+2 4 |
| | 3+4 2 | 3+4 2 | 2+2 4 | 2+2 4 | 2+2 4 | 3+4 2 | 3+4 2 | 2+2 4 |
| | 5+6 2 | 5+6 2 | 2+2 4 | 2+2 4 | 2+2 4 | 5+6 2 | 5+6 2 | 2+2 4 |
| | 3+4 2 | 3+4 2 | 2+2 4 | 2+2 4 | 2+2 4 | 3+4 2 | 3+4 2 | 2+2 4 |
| | 5+6 2 | 3+4 2 | 2+2 4 | 2+2 4 | 2+2 4 | 5+6 2 | 3+4 2 | 2+2 4 |
| | 3+4 2 | 5+6 2 | 2+2 4 | 2+2 4 | 2+2 4 | 3+4 2 | 5+6 2 | 2+2 4 |
| | 3+4 2 | 5+6 2 | 2+2 4 | 2+2 4 | 2+2 4 | 5+6 2 | 5+6 2 | 2+2 4 |
| | 5+6 2 | 3+4 2 | 2+2 4 | 2+2 4 | 2+2 4 | 3+4 2 | 3+4 2 | 2+2 4 |
| | 3+4 2 | 5+6 2 | 2+2 4 | 2+2 4 | 2+2 4 | 5+6 2 | 3+4 2 | 2+2 4 |
| | 5+6 2 | 3+4 2 | 2+2 4 | 2+2 4 | 2+2 4 | 3+4 2 | 5+6 2 | 2+2 4 |
| | 5+6 2 | 5+6 2 | 2+2 4 | 2+2 4 | 2+2 4 | 3+4 2 | 3+4 2 | 2+2 4 |
| | 3+4 2 | 3+4 2 | 2+2 4 | 2+2 4 | 2+2 4 | 5+6 2 | 5+6 2 | 2+2 4 |
| | 3+4 2 | 3+4 2 | 2+2 4 | 2+2 4 | 2+2 4 | 5+6 2 | 5+6 2 | 2+2 4 |
| | 5+6 2 | 5+6 2 | 2+2 4 | 2+2 4 | 2+2 4 | 3+4 2 | 3+4 2 | 2+2 4 |
| | 3+4 2 | 3+4 2 | 2+2 4 | 2+2 4 | 2+2 4 | 3+4 2 | 3+4 2 | 2+2 4 |
| | 5+6 2 | 5+6 2 | 2+2 4 | 2+2 4 | 2+2 4 | 3+4 2 | 5+6 2 | 2+2 4 |
| | 3+4 2 | 5+6 2 | 2+2 4 | 2+2 4 | 2+2 4 | 5+6 2 | 3+4 2 | 2+2 4 |
| | 5+6 2 | 3+4 2 | 2+2 4 | 2+2 4 | 2+2 4 | 5+6 2 | 5+6 2 | 2+2 4 |
| | 3+4 2 | 5+6 2 | 2+2 4 | 2+2 4 | 2+2 4 | 3+4 2 | 5+6 2 | 2+2 4 |
| | 3+4 2 | 3+4 2 | 2+2 4 | 2+2 4 | 2+2 4 | 3+4 2 | 3+4 2 | 2+2 4 |
| | 5+6 2 | 5+6 2 | 2+2 4 | 2+2 4 | 2+2 4 | 5+6 2 | 3+4 2 | 2+2 4 |

Continued on next column

**TABLE XXIII– continued**

| TQC | ICSD | NoE | SF | | | | | |
|---|---|---|---|---|---|---|---|---|
| | 5+6 2 | 3+4 2 | 2+2 4 | 2+2 4 | 2+2 4 | 5+6 2 | 5+6 2 | 2+2 4 |
| | 5+6 2 | 5+6 2 | 2+2 4 | 2+2 4 | 2+2 4 | 5+6 2 | 5+6 2 | 2+2 4 |
| | 3+4 2 | 3+4 2 | 2+2 4 | 2+2 4 | 2+2 4 | 3+4 2 | 3+4 2 | 2+2 4 |
| | 5+6 2 | 3+4 2 | 2+2 4 | 2+2 4 | 2+2 4 | 5+6 2 | 3+4 2 | 2+2 4 |
| | 3+4 2 | 3+4 2 | 2+2 4 | 2+2 4 | 2+2 4 | 3+4 2 | 5+6 2 | 2+2 4 |
| | 5+6 2 | 5+6 2 | 2+2 4 | 2+2 4 | 2+2 4 | 5+6 2 | 3+4 2 | 2+2 4 |
| | 3+4 2 | 5+6 2 | 2+2 4 | 2+2 4 | 2+2 4 | 5+6 2 | 5+6 2 | 2+2 4 |
| | 5+6 2 | 3+4 2 | 2+2 4 | 2+2 4 | 2+2 4 | 3+4 2 | 5+6 2 | 2+2 4 |
| | 3+4 2 | 5+6 2 | 2+2 4 | 2+2 4 | 2+2 4 | 5+6 2 | 3+4 2 | 2+2 4 |
| | 5+6 2 | 5+6 2 | 2+2 4 | 2+2 4 | 2+2 4 | 3+4 2 | 5+6 2 | 2+2 4 |
| | 3+4 2 | 3+4 2 | 2+2 4 | 2+2 4 | 2+2 4 | 3+4 2 | 3+4 2 | 2+2 4 |
| | 5+6 2 | 5+6 2 | 2+2 4 | 2+2 4 | 2+2 4 | 5+6 2 | 5+6 2 | 2+2 4 |
| | 5+6 2 | 3+4 2 | | | | 3+4 2 | 3+4 2 | |
| | 3+4 2 | 5+6 2 | | | | 5+6 2 | 3+4 2 | |
| | 3+4 2 | 3+4 2 | | | | 3+4 2 | 5+6 2 | |
| | 5+6 2 | 5+6 2 | | | | 3+4 2 | 3+4 2 | |
| | 3+4 2 | 3+4 2 | | | | 3+4 2 | 5+6 2 | |
| | 3+4 2 | 3+4 2 | | | | 5+6 2 | 3+4 2 | |
| | 3+4 2 | 5+6 2 | | | | 3+4 2 | 5+6 2 | |
| | 5+6 2 | 5+6 2 | | | | 5+6 2 | 5+6 2 | |
| | 5+6 2 | 3+4 2 | | | | 5+6 2 | 3+4 2 | |
| | 5+6 2 | 3+4 2 | | | | 3+4 2 | 3+4 2 | |
| | 3+4 2 | 5+6 2 | | | | 5+6 2 | 5+6 2 | |
| | 5+6 2 | 5+6 2 | | | | 5+6 2 | 5+6 2 | |
| | 3+4 2 | 3+4 2 | | | | 3+4 2 | 3+4 2 | |
| | 3+4 2 | 3+4 2 | | | | 3+4 2 | 5+6 2 | |
| | 3+4 2 | 5+6 2 | | | | 3+4 2 | 3+4 2 | |
| | 5+6 2 | 5+6 2 | | | | 5+6 2 | 3+4 2 | |
| | 3+4 2 | 3+4 2 | | | | 3+4 2 | 5+6 2 | |
| | 5+6 2 | 5+6 2 | | | | 3+4 2 | 5+6 2 | |
| | 5+6 2 | 3+4 2 | | | | 5+6 2 | 3+4 2 | |
| | 5+6 2 | 5+6 2 | | | | 3+4 2 | 5+6 2 | |
| | 3+4 2 | 3+4 2 | | | | 5+6 2 | 3+4 2 | |
| | 5+6 2 | 5+6 2 | | | | 5+6 2 | 3+4 2 | |
| | 3+4 2 | 3+4 2 | | | | 3+4 2 | 5+6 2 | |
| | 5+6 2 | 3+4 2 | | | | 5+6 2 | 5+6 2 | |
| | 3+4 2 | 3+4 2 | | | | 3+4 2 | 3+4 2 | |
| | 5+6 2 | 5+6 2 | | | | 5+6 2 | 5+6 2 | |
| | 3+4 2 | 5+6 2 | | | | 5+6 2 | 5+6 2 | |
| | 5+6 2 | 5+6 2 | | | | 3+4 2 | 3+4 2 | |
| | 3+4 2 | 3+4 2 | | | | 3+4 2 | 3+4 2 | |
| | 3+4 2 | 3+4 2 | | | | 3+4 2 | 5+6 2 | |
| | 5+6 2 | 3+4 2 | | | | 5+6 2 | 3+4 2 | |
| | 3+4 2 | 5+6 2 | | | | 5+6 2 | 3+4 2 | |
| | 5+6 2 | 5+6 2 | | | | 3+4 2 | 5+6 2 | |
| | 3+4 2 | 5+6 2 | | | | 3+4 2 | 5+6 2 | |
| | 3+4 2 | 3+4 2 | | | | 3+4 2 | 5+6 2 | |
| | 5+6 2 | 5+6 2 | | | | 3+4 2 | 3+4 2 | |
| | 3+4 2 | 3+4 2 | | | | 5+6 2 | 5+6 2 | |
| | 3+4 2 | 5+6 2 | | | | 3+4 2 | 3+4 2 | |
| | 5+6 2 | 5+6 2 | | | | 5+6 2 | 5+6 2 | |
| 11.2.97 | 57494 | 56 | Al1Au1 | | | | | |
| | -A | -B | -C | -D | -E | -GM | -Y | -Z |
| | 3+4 2 | 5+6 2 | 2+2 4 | 2+2 4 | 2+2 4 | 3+4 2 | 3+4 2 | 2+2 4 |
| | 5+6 2 | 3+4 2 | 2+2 4 | 2+2 4 | 2+2 4 | 5+6 2 | 5+6 2 | 2+2 4 |
| | 3+4 2 | 3+4 2 | 2+2 4 | 2+2 4 | 2+2 4 | 3+4 2 | 5+6 2 | 2+2 4 |
| | 5+6 2 | 5+6 2 | 2+2 4 | 2+2 4 | 2+2 4 | 5+6 2 | 3+4 2 | 2+2 4 |
| | 5+6 2 | 3+4 2 | 2+2 4 | 2+2 4 | 2+2 4 | 3+4 2 | 5+6 2 | 2+2 4 |
| | 5+6 2 | 5+6 2 | 2+2 4 | 2+2 4 | 2+2 4 | 5+6 2 | 3+4 2 | 2+2 4 |
| | 3+4 2 | 5+6 2 | 2+2 4 | 2+2 4 | 2+2 4 | 3+4 2 | 5+6 2 | 2+2 4 |
| | 3+4 2 | 3+4 2 | 2+2 4 | 2+2 4 | 2+2 4 | 5+6 2 | 3+4 2 | 2+2 4 |

Continued on next column

**TABLE XXIII– continued**

| TQC | ICSD | NoE | SF | | | | | |
|---|---|---|---|---|---|---|---|---|
| | 5+6 2 | 5+6 2 | 2+2 4 | 2+2 4 | 2+2 4 | 3+4 2 | 5+6 2 | 2+2 4 |
| | 5+6 2 | 3+4 2 | 2+2 4 | 2+2 4 | 2+2 4 | 5+6 2 | 3+4 2 | 2+2 4 |
| | 3+4 2 | 5+6 2 | 2+2 4 | 2+2 4 | 2+2 4 | 3+4 2 | 5+6 2 | 2+2 4 |
| | 3+4 2 | 3+4 2 | 2+2 4 | 2+2 4 | 2+2 4 | 5+6 2 | 3+4 2 | 2+2 4 |
| | 5+6 2 | 5+6 2 | 2+2 4 | 2+2 4 | 2+2 4 | 3+4 2 | 5+6 2 | 2+2 4 |
| | 5+6 2 | 3+4 2 | 2+2 4 | 2+2 4 | 2+2 4 | 5+6 2 | 3+4 2 | 2+2 4 |
| | 3+4 2 | 3+4 2 | | | | 3+4 2 | 3+4 2 | |
| | 5+6 2 | 5+6 2 | | | | 3+4 2 | 5+6 2 | |
| | 3+4 2 | 3+4 2 | | | | 3+4 2 | 5+6 2 | |
| | 5+6 2 | 5+6 2 | | | | 5+6 2 | 3+4 2 | |
| | 3+4 2 | 3+4 2 | | | | 5+6 2 | 5+6 2 | |
| | 3+4 2 | 5+6 2 | | | | 3+4 2 | 3+4 2 | |
| | 3+4 2 | 3+4 2 | | | | 3+4 2 | 3+4 2 | |
| | 3+4 2 | 5+6 2 | | | | 5+6 2 | 5+6 2 | |
| | 5+6 2 | 5+6 2 | | | | 5+6 2 | 5+6 2 | |
| | 5+6 2 | 3+4 2 | | | | 5+6 2 | 3+4 2 | |
| | 3+4 2 | 3+4 2 | | | | 3+4 2 | 3+4 2 | |
| | 5+6 2 | 5+6 2 | | | | 3+4 2 | 5+6 2 | |
| | 3+4 2 | 3+4 2 | | | | 5+6 2 | 5+6 2 | |
| | 5+6 2 | 5+6 2 | | | | 3+4 2 | 3+4 2 | |
| 11.2.98 | 104649 | 56 | Al1Au1 | | | | | |
| | -A | -B | -C | -D | -E | -GM | -Y | -Z |
| | 5+6 2 | 3+4 2 | 2+2 4 | 2+2 4 | 2+2 4 | 3+4 2 | 5+6 2 | 2+2 4 |
| | 3+4 2 | 5+6 2 | 2+2 4 | 2+2 4 | 2+2 4 | 3+4 2 | 3+4 2 | 2+2 4 |
| | 3+4 2 | 5+6 2 | 2+2 4 | 2+2 4 | 2+2 4 | 5+6 2 | 3+4 2 | 2+2 4 |
| | 5+6 2 | 3+4 2 | 2+2 4 | 2+2 4 | 2+2 4 | 5+6 2 | 5+6 2 | 2+2 4 |
| | 5+6 2 | 5+6 2 | 2+2 4 | 2+2 4 | 2+2 4 | 5+6 2 | 3+4 2 | 2+2 4 |
| | 5+6 2 | 3+4 2 | 2+2 4 | 2+2 4 | 2+2 4 | 3+4 2 | 5+6 2 | 2+2 4 |
| | 3+4 2 | 3+4 2 | 2+2 4 | 2+2 4 | 2+2 4 | 3+4 2 | 5+6 2 | 2+2 4 |
| | 3+4 2 | 5+6 2 | 2+2 4 | 2+2 4 | 2+2 4 | 5+6 2 | 3+4 2 | 2+2 4 |
| | 5+6 2 | 3+4 2 | 2+2 4 | 2+2 4 | 2+2 4 | 3+4 2 | 3+4 2 | 2+2 4 |
| | 5+6 2 | 5+6 2 | 2+2 4 | 2+2 4 | 2+2 4 | 5+6 2 | 5+6 2 | 2+2 4 |
| | 3+4 2 | 3+4 2 | 2+2 4 | 2+2 4 | 2+2 4 | 3+4 2 | 3+4 2 | 2+2 4 |
| | 3+4 2 | 5+6 2 | 2+2 4 | 2+2 4 | 2+2 4 | 5+6 2 | 5+6 2 | 2+2 4 |
| | 5+6 2 | 3+4 2 | 2+2 4 | 2+2 4 | 2+2 4 | 3+4 2 | 3+4 2 | 2+2 4 |
| | 5+6 2 | 5+6 2 | 2+2 4 | 2+2 4 | 2+2 4 | 5+6 2 | 5+6 2 | 2+2 4 |
| | 3+4 2 | 5+6 2 | | | | 3+4 2 | 5+6 2 | |
| | 5+6 2 | 3+4 2 | | | | 3+4 2 | 3+4 2 | |
| | 3+4 2 | 3+4 2 | | | | 3+4 2 | 3+4 2 | |
| | 5+6 2 | 5+6 2 | | | | 5+6 2 | 5+6 2 | |
| | 3+4 2 | 5+6 2 | | | | 5+6 2 | 3+4 2 | |
| | 3+4 2 | 3+4 2 | | | | 3+4 2 | 5+6 2 | |
| | 5+6 2 | 5+6 2 | | | | 5+6 2 | 5+6 2 | |
| | 3+4 2 | 3+4 2 | | | | 3+4 2 | 3+4 2 | |
| | 3+4 2 | 5+6 2 | | | | 5+6 2 | 3+4 2 | |
| | 5+6 2 | 3+4 2 | | | | 5+6 2 | 5+6 2 | |
| | 3+4 2 | 3+4 2 | | | | 3+4 2 | 5+6 2 | |
| | 5+6 2 | 5+6 2 | | | | 3+4 2 | 3+4 2 | |
| | 3+4 2 | 5+6 2 | | | | 5+6 2 | 3+4 2 | |
| | 5+6 2 | 3+4 2 | | | | 3+4 2 | 5+6 2 | |

The End

TABLE XXIV: The computed irreducible representations for the valence bands for the compounds of SEBR and NLC in space group 12. See the list of tables.

| TQC | ICSD | NoE | SF | | |
|---|---|---|---|---|---|
| 12.1.1 | 187638 | 22 | Y1 | | |
| | -A | -GM | -M | -Y | -L | -V |
| | 3+4 2 | 3+4 2 | 3+4 2 | 3+4 2 | 2+2 2 | 2+2 2 |

Continued on next column

**TABLE XXIV– continued**

| TQC | ICSD | NoE | SF | | | |
|---|---|---|---|---|---|---|
| | 5+6 2 | 5+6 2 | 5+6 2 | 5+6 2 | 3+3 2 | 3+3 2 |
| | 3+4 2 | 3+4 2 | 3+4 2 | 3+4 2 | 3+3 2 | 2+2 2 |
| | 5+6 2 | 5+6 2 | 5+6 2 | 5+6 2 | 2+2 2 | 3+3 2 |
| | 5+6 2 | 5+6 2 | 5+6 2 | 5+6 2 | 3+3 2 | 3+3 2 |
| | 3+4 2 | 3+4 2 | 3+4 2 | 3+4 2 | 2+2 2 | 2+2 2 |
| | 3+4 2 | 3+4 2 | 3+4 2 | 3+4 2 | 2+2 2 | 2+2 2 |
| | 5+6 2 | 5+6 2 | 5+6 2 | 5+6 2 | 3+3 2 | 3+3 2 |
| | 3+4 2 | 3+4 2 | 3+4 2 | 3+4 2 | 2+2 2 | 2+2 2 |
| | 5+6 2 | 5+6 2 | 5+6 2 | 5+6 2 | 3+3 2 | 3+3 2 |
| | 3+4 2 | 3+4 2 | 5+6 2 | 5+6 2 | 3+3 2 | 3+3 2 |
| 12.1.2 | 43412 | 36 | Te4Ti3 | | | |
| | -A | -GM | -M | -Y | -L | -V |
| | 3+4 2 | 3+4 2 | 3+4 2 | 5+6 2 | 2+2 2 | 2+2 2 |
| | 5+6 2 | 5+6 2 | 5+6 2 | 3+4 2 | 3+3 2 | 3+3 2 |
| | 5+6 2 | 3+4 2 | 3+4 2 | 5+6 2 | 2+2 2 | 3+3 2 |
| | 3+4 2 | 5+6 2 | 5+6 2 | 3+4 2 | 3+3 2 | 2+2 2 |
| | 3+4 2 | 3+4 2 | 3+4 2 | 3+4 2 | 3+3 2 | 3+3 2 |
| | 5+6 2 | 5+6 2 | 5+6 2 | 5+6 2 | 2+2 2 | 2+2 2 |
| | 3+4 2 | 3+4 2 | 3+4 2 | 3+4 2 | 3+3 2 | 3+3 2 |
| | 5+6 2 | 3+4 2 | 5+6 2 | 5+6 2 | 2+2 2 | 2+2 2 |
| | 3+4 2 | 3+4 2 | 3+4 2 | 5+6 2 | 3+3 2 | 3+3 2 |
| | 5+6 2 | 5+6 2 | 5+6 2 | 3+4 2 | 2+2 2 | 2+2 2 |
| | 3+4 2 | 5+6 2 | 3+4 2 | 3+4 2 | 3+3 2 | 3+3 2 |
| | 5+6 2 | 3+4 2 | 5+6 2 | 5+6 2 | 2+2 2 | 2+2 2 |
| | 3+4 2 | 3+4 2 | 3+4 2 | 5+6 2 | 2+2 2 | 3+3 2 |
| | 5+6 2 | 3+4 2 | 3+4 2 | 5+6 2 | 2+2 2 | 2+2 2 |
| | 5+6 2 | 5+6 2 | 5+6 2 | 3+4 2 | 3+3 2 | 3+3 2 |
| | 3+4 2 | 5+6 2 | 3+4 2 | 3+4 2 | 3+3 2 | 2+2 2 |
| | 3+4 2 | 3+4 2 | 5+6 2 | 3+4 2 | 3+3 2 | 3+3 2 |
| | 5+6 2 | 3+4 2 | 5+6 2 | 5+6 2 | 2+2 2 | 3+3 2 |
| 12.1.3 | 41047 | 56 | Ta2Te3 | | | |
| | -A | -GM | -M | -Y | -L | -V |
| | 3+4 2 | 3+4 2 | 3+4 2 | 3+4 2 | 3+3 2 | 2+2 2 |
| | 5+6 2 | 5+6 2 | 5+6 2 | 5+6 2 | 2+2 2 | 3+3 2 |
| | 3+4 2 | 3+4 2 | 3+4 2 | 3+4 2 | 3+3 2 | 3+3 2 |
| | 5+6 2 | 5+6 2 | 5+6 2 | 5+6 2 | 2+2 2 | 2+2 2 |
| | 3+4 2 | 3+4 2 | 3+4 2 | 3+4 2 | 3+3 2 | 3+3 2 |
| | 5+6 2 | 5+6 2 | 5+6 2 | 5+6 2 | 2+2 2 | 2+2 2 |
| | 5+6 2 | 3+4 2 | 5+6 2 | 5+6 2 | 2+2 2 | 2+2 2 |
| | 3+4 2 | 5+6 2 | 3+4 2 | 3+4 2 | 3+3 2 | 3+3 2 |
| | 5+6 2 | 5+6 2 | 5+6 2 | 5+6 2 | 2+2 2 | 2+2 2 |
| | 3+4 2 | 3+4 2 | 3+4 2 | 3+4 2 | 3+3 2 | 3+3 2 |
| | 3+4 2 | 3+4 2 | 3+4 2 | 3+4 2 | 3+3 2 | 3+3 2 |
| | 3+4 2 | 5+6 2 | 5+6 2 | 3+4 2 | 2+2 2 | 2+2 2 |
| | 5+6 2 | 3+4 2 | 3+4 2 | 5+6 2 | 3+3 2 | 3+3 2 |
| | 5+6 2 | 5+6 2 | 3+4 2 | 5+6 2 | 3+3 2 | 2+2 2 |
| | 3+4 2 | 3+4 2 | 5+6 2 | 3+4 2 | 2+2 2 | 3+3 2 |
| | 3+4 2 | 5+6 2 | 5+6 2 | 5+6 2 | 3+3 2 | 2+2 2 |
| | 5+6 2 | 3+4 2 | 5+6 2 | 3+4 2 | 2+2 2 | 3+3 2 |
| | 5+6 2 | 5+6 2 | 3+4 2 | 5+6 2 | 2+2 2 | 3+3 2 |
| | 5+6 2 | 3+4 2 | 3+4 2 | 5+6 2 | 3+3 2 | 2+2 2 |
| | 3+4 2 | 5+6 2 | 5+6 2 | 3+4 2 | 3+3 2 | 3+3 2 |
| | 3+4 2 | 5+6 2 | 5+6 2 | 5+6 2 | 2+2 2 | 2+2 2 |
| | 3+4 2 | 3+4 2 | 3+4 2 | 3+4 2 | 3+3 2 | 2+2 2 |
| | 5+6 2 | 3+4 2 | 5+6 2 | 3+4 2 | 2+2 2 | 3+3 2 |
| | 3+4 2 | 5+6 2 | 3+4 2 | 3+4 2 | 2+2 2 | 2+2 2 |
| | 5+6 2 | 3+4 2 | 5+6 2 | 5+6 2 | 2+2 2 | 3+3 2 |
| | 5+6 2 | 5+6 2 | 3+4 2 | 3+4 2 | 3+3 2 | 2+2 2 |
| | 3+4 2 | 3+4 2 | 5+6 2 | 5+6 2 | 2+2 2 | 3+3 2 |
| | 3+4 2 | 5+6 2 | 3+4 2 | 3+4 2 | 3+3 2 | 2+2 2 |
| 12.1.4 | 37122 | 76 | Se8Tl1V5 | | | |

Continued on next column

**TABLE XXIV– continued**

| TQC | ICSD | NoE | SF | | | |
|---|---|---|---|---|---|---|
| | -A | -GM | -M | -Y | -L | -V |
| | 5+6 2 | 3+4 2 | 5+6 2 | 5+6 2 | 2+2 2 | 3+3 2 |
| | 3+4 2 | 5+6 2 | 3+4 2 | 3+4 2 | 3+3 2 | 2+2 2 |
| | 5+6 2 | 3+4 2 | 5+6 2 | 5+6 2 | 2+2 2 | 3+3 2 |
| | 3+4 2 | 5+6 2 | 3+4 2 | 3+4 2 | 3+3 2 | 2+2 2 |
| | 5+6 2 | 3+4 2 | 5+6 2 | 5+6 2 | 3+3 2 | 2+2 2 |
| | 3+4 2 | 5+6 2 | 3+4 2 | 3+4 2 | 2+2 2 | 3+3 2 |
| | 3+4 2 | 3+4 2 | 3+4 2 | 3+4 2 | 3+3 2 | 2+2 2 |
| | 5+6 2 | 5+6 2 | 5+6 2 | 5+6 2 | 2+2 2 | 3+3 2 |
| | 3+4 2 | 3+4 2 | 3+4 2 | 3+4 2 | 3+3 2 | 2+2 2 |
| | 5+6 2 | 3+4 2 | 5+6 2 | 3+4 2 | 3+3 2 | 3+3 2 |
| | 5+6 2 | 3+4 2 | 3+4 2 | 5+6 2 | 2+2 2 | 3+3 2 |
| | 5+6 2 | 5+6 2 | 5+6 2 | 3+4 2 | 3+3 2 | 2+2 2 |
| | 3+4 2 | 5+6 2 | 5+6 2 | 5+6 2 | 2+2 2 | 2+2 2 |
| | 3+4 2 | 3+4 2 | 3+4 2 | 5+6 2 | 2+2 2 | 3+3 2 |
| | 5+6 2 | 5+6 2 | 3+4 2 | 5+6 2 | 3+3 2 | 2+2 2 |
| | 3+4 2 | 3+4 2 | 5+6 2 | 3+4 2 | 2+2 2 | 3+3 2 |
| | 3+4 2 | 5+6 2 | 3+4 2 | 3+4 2 | 2+2 2 | 2+2 2 |
| | 5+6 2 | 5+6 2 | 5+6 2 | 5+6 2 | 3+3 2 | 3+3 2 |
| | 5+6 2 | 3+4 2 | 3+4 2 | 3+4 2 | 2+2 2 | 2+2 2 |
| | 5+6 2 | 3+4 2 | 3+4 2 | 5+6 2 | 3+3 2 | 3+3 2 |
| | 3+4 2 | 5+6 2 | 5+6 2 | 3+4 2 | 3+3 2 | 2+2 2 |
| | 5+6 2 | 3+4 2 | 5+6 2 | 5+6 2 | 2+2 2 | 3+3 2 |
| | 3+4 2 | 3+4 2 | 5+6 2 | 3+4 2 | 3+3 2 | 3+3 2 |
| | 3+4 2 | 3+4 2 | 3+4 2 | 3+4 2 | 2+2 2 | 2+2 2 |
| | 5+6 2 | 5+6 2 | 5+6 2 | 5+6 2 | 3+3 2 | 2+2 2 |
| | 5+6 2 | 3+4 2 | 5+6 2 | 5+6 2 | 3+3 2 | 3+3 2 |
| | 3+4 2 | 3+4 2 | 3+4 2 | 3+4 2 | 2+2 2 | 2+2 2 |
| | 5+6 2 | 5+6 2 | 5+6 2 | 5+6 2 | 3+3 2 | 2+2 2 |
| | 3+4 2 | 5+6 2 | 3+4 2 | 3+4 2 | 2+2 2 | 3+3 2 |
| | 3+4 2 | 3+4 2 | 3+4 2 | 5+6 2 | 3+3 2 | 2+2 2 |
| | 5+6 2 | 5+6 2 | 3+4 2 | 5+6 2 | 3+3 2 | 3+3 2 |
| | 5+6 2 | 5+6 2 | 5+6 2 | 3+4 2 | 2+2 2 | 3+3 2 |
| | 3+4 2 | 3+4 2 | 5+6 2 | 3+4 2 | 2+2 2 | 3+3 2 |
| | 3+4 2 | 5+6 2 | 3+4 2 | 3+4 2 | 3+3 2 | 2+2 2 |
| | 5+6 2 | 5+6 2 | 3+4 2 | 5+6 2 | 2+2 2 | 2+2 2 |
| | 3+4 2 | 3+4 2 | 5+6 2 | 3+4 2 | 3+3 2 | 2+2 2 |
| | 5+6 2 | 3+4 2 | 5+6 2 | 5+6 2 | 3+3 2 | 3+3 2 |
| | 5+6 2 | 3+4 2 | 3+4 2 | 3+4 2 | 2+2 2 | 3+3 2 |
| 12.1.5 | 601347 | 36 | Se4Ti3 | | | |
| | -A | -GM | -M | -Y | -L | -V |
| | 5+6 2 | 3+4 2 | 3+4 2 | 5+6 2 | 2+2 2 | 3+3 2 |
| | 3+4 2 | 5+6 2 | 5+6 2 | 3+4 2 | 2+2 2 | 2+2 2 |
| | 5+6 2 | 3+4 2 | 3+4 2 | 5+6 2 | 3+3 2 | 3+3 2 |
| | 3+4 2 | 5+6 2 | 5+6 2 | 3+4 2 | 3+3 2 | 2+2 2 |
| | 3+4 2 | 3+4 2 | 3+4 2 | 3+4 2 | 3+3 2 | 3+3 2 |
| | 5+6 2 | 5+6 2 | 5+6 2 | 5+6 2 | 2+2 2 | 2+2 2 |
| | 5+6 2 | 3+4 2 | 3+4 2 | 3+4 2 | 3+3 2 | 2+2 2 |
| | 5+6 2 | 3+4 2 | 5+6 2 | 5+6 2 | 2+2 2 | 3+3 2 |
| | 3+4 2 | 5+6 2 | 3+4 2 | 5+6 2 | 3+3 2 | 3+3 2 |
| | 3+4 2 | 3+4 2 | 3+4 2 | 3+4 2 | 2+2 2 | 3+3 2 |
| | 3+4 2 | 5+6 2 | 5+6 2 | 3+4 2 | 3+3 2 | 2+2 2 |
| | 5+6 2 | 3+4 2 | 3+4 2 | 5+6 2 | 2+2 2 | 3+3 2 |
| | 3+4 2 | 3+4 2 | 3+4 2 | 5+6 2 | 2+2 2 | 2+2 2 |
| | 3+4 2 | 5+6 2 | 5+6 2 | 3+4 2 | 2+2 2 | 3+3 2 |
| | 5+6 2 | 3+4 2 | 5+6 2 | 5+6 2 | 3+3 2 | 2+2 2 |
| | 5+6 2 | 5+6 2 | 5+6 2 | 3+4 2 | 3+3 2 | 2+2 2 |
| | 3+4 2 | 5+6 2 | 3+4 2 | 3+4 2 | 3+3 2 | 3+3 2 |
| | 5+6 2 | 5+6 2 | 3+4 2 | 3+4 2 | 2+2 2 | 2+2 2 |
| 12.1.6 | 48176 | 78 | Rh1Si2Y2 | | | |
| | -A | -GM | -M | -Y | -L | -V |

Continued on next column

**TABLE XXIV– continued**

| TQC | ICSD | | NoE | | SF | |
|---|---|---|---|---|---|---|
| | 3+4 2 | 3+4 2 | 3+4 2 | 3+4 2 | 3+3 2 | 3+3 2 |
| | 5+6 2 | 5+6 2 | 5+6 2 | 5+6 2 | 2+2 2 | 2+2 2 |
| | 5+6 2 | 3+4 2 | 5+6 2 | 3+4 2 | 2+2 2 | 3+3 2 |
| | 3+4 2 | 5+6 2 | 3+4 2 | 5+6 2 | 3+3 2 | 2+2 2 |
| | 3+4 2 | 3+4 2 | 3+4 2 | 3+4 2 | 3+3 2 | 3+3 2 |
| | 5+6 2 | 5+6 2 | 5+6 2 | 5+6 2 | 2+2 2 | 2+2 2 |
| | 5+6 2 | 3+4 2 | 5+6 2 | 3+4 2 | 2+2 2 | 3+3 2 |
| | 3+4 2 | 5+6 2 | 3+4 2 | 5+6 2 | 3+3 2 | 2+2 2 |
| | 3+4 2 | 3+4 2 | 3+4 2 | 3+4 2 | 3+3 2 | 3+3 2 |
| | 5+6 2 | 3+4 2 | 5+6 2 | 5+6 2 | 2+2 2 | 2+2 2 |
| | 3+4 2 | 5+6 2 | 5+6 2 | 3+4 2 | 3+3 2 | 3+3 2 |
| | 5+6 2 | 5+6 2 | 3+4 2 | 5+6 2 | 2+2 2 | 2+2 2 |
| | 5+6 2 | 3+4 2 | 5+6 2 | 3+4 2 | 2+2 2 | 3+3 2 |
| | 3+4 2 | 5+6 2 | 3+4 2 | 5+6 2 | 3+3 2 | 2+2 2 |
| | 3+4 2 | 3+4 2 | 3+4 2 | 3+4 2 | 2+2 2 | 3+3 2 |
| | 5+6 2 | 5+6 2 | 5+6 2 | 5+6 2 | 3+3 2 | 2+2 2 |
| | 5+6 2 | 3+4 2 | 5+6 2 | 3+4 2 | 3+3 2 | 2+2 2 |
| | 3+4 2 | 3+4 2 | 3+4 2 | 3+4 2 | 2+2 2 | 2+2 2 |
| | 3+4 2 | 5+6 2 | 3+4 2 | 5+6 2 | 2+2 2 | 3+3 2 |
| | 5+6 2 | 5+6 2 | 5+6 2 | 5+6 2 | 3+3 2 | 3+3 2 |
| | 5+6 2 | 3+4 2 | 5+6 2 | 3+4 2 | 2+2 2 | 3+3 2 |
| | 3+4 2 | 5+6 2 | 3+4 2 | 5+6 2 | 3+3 2 | 2+2 2 |
| | 3+4 2 | 5+6 2 | 5+6 2 | 5+6 2 | 3+3 2 | 3+3 2 |
| | 5+6 2 | 3+4 2 | 3+4 2 | 3+4 2 | 2+2 2 | 2+2 2 |
| | 3+4 2 | 5+6 2 | 5+6 2 | 3+4 2 | 3+3 2 | 2+2 2 |
| | 5+6 2 | 3+4 2 | 5+6 2 | 5+6 2 | 2+2 2 | 3+3 2 |
| | 5+6 2 | 3+4 2 | 3+4 2 | 3+4 2 | 2+2 2 | 3+3 2 |
| | 5+6 2 | 5+6 2 | 3+4 2 | 5+6 2 | 3+3 2 | 2+2 2 |
| | 3+4 2 | 3+4 2 | 3+4 2 | 5+6 2 | 2+2 2 | 2+2 2 |
| | 3+4 2 | 5+6 2 | 5+6 2 | 3+4 2 | 3+3 2 | 3+3 2 |
| | 5+6 2 | 3+4 2 | 3+4 2 | 5+6 2 | 3+3 2 | 2+2 2 |
| | 3+4 2 | 5+6 2 | 5+6 2 | 3+4 2 | 3+3 2 | 3+3 2 |
| | 5+6 2 | 5+6 2 | 5+6 2 | 3+4 2 | 2+2 2 | 2+2 2 |
| | 5+6 2 | 3+4 2 | 3+4 2 | 5+6 2 | 2+2 2 | 2+2 2 |
| | 3+4 2 | 3+4 2 | 3+4 2 | 5+6 2 | 3+3 2 | 3+3 2 |
| | 3+4 2 | 3+4 2 | 3+4 2 | 3+4 2 | 2+2 2 | 2+2 2 |
| | 5+6 2 | 3+4 2 | 5+6 2 | 5+6 2 | 2+2 2 | 3+3 2 |
| | 5+6 2 | 5+6 2 | 3+4 2 | 3+4 2 | 3+3 2 | 2+2 2 |
| | 3+4 2 | 5+6 2 | 3+4 2 | 3+4 2 | 3+3 2 | 3+3 2 |
| 12.1.7 | 42077 | | 30 | | P2V1 | |
| | -A | -GM | -M | -Y | -L | -V |
| | 3+4 2 | 3+4 2 | 3+4 2 | 3+4 2 | 3+3 2 | 3+3 2 |
| | 5+6 2 | 3+4 2 | 5+6 2 | 5+6 2 | 3+3 2 | 2+2 2 |
| | 3+4 2 | 5+6 2 | 3+4 2 | 5+6 2 | 2+2 2 | 3+3 2 |
| | 5+6 2 | 5+6 2 | 5+6 2 | 3+4 2 | 2+2 2 | 2+2 2 |
| | 3+4 2 | 3+4 2 | 5+6 2 | 5+6 2 | 3+3 2 | 2+2 2 |
| | 5+6 2 | 3+4 2 | 3+4 2 | 3+4 2 | 2+2 2 | 3+3 2 |
| | 5+6 2 | 5+6 2 | 3+4 2 | 5+6 2 | 3+3 2 | 2+2 2 |
| | 3+4 2 | 3+4 2 | 5+6 2 | 3+4 2 | 2+2 2 | 3+3 2 |
| | 3+4 2 | 5+6 2 | 3+4 2 | 5+6 2 | 3+3 2 | 2+2 2 |
| | 3+4 2 | 3+4 2 | 5+6 2 | 5+6 2 | 2+2 2 | 3+3 2 |
| | 5+6 2 | 5+6 2 | 5+6 2 | 3+4 2 | 3+3 2 | 2+2 2 |
| | 5+6 2 | 3+4 2 | 3+4 2 | 5+6 2 | 3+3 2 | 3+3 2 |
| | 5+6 2 | 3+4 2 | 3+4 2 | 3+4 2 | 2+2 2 | 2+2 2 |
| | 3+4 2 | 3+4 2 | 5+6 2 | 3+4 2 | 2+2 2 | 3+3 2 |
| | 3+4 2 | 5+6 2 | 5+6 2 | 3+4 2 | 3+3 2 | 3+3 2 |
| 12.1.8 | 56694 | | 50 | | O5Ti5 | |
| | -A | -GM | -M | -Y | -L | -V |
| | 5+6 2 | 3+4 2 | 3+4 2 | 5+6 2 | 2+2 2 | 3+3 2 |
| | 3+4 2 | 3+4 2 | 5+6 2 | 3+4 2 | 2+2 2 | 2+2 2 |
| | 5+6 2 | 5+6 2 | 5+6 2 | 3+4 2 | 3+3 2 | 3+3 2 |

Continued on next column

**TABLE XXIV– continued**

| TQC | ICSD | | NoE | | SF | |
|---|---|---|---|---|---|---|
| | 3+4 2 | 5+6 2 | 3+4 2 | 5+6 2 | 3+3 2 | 2+2 2 |
| | 3+4 2 | 3+4 2 | 3+4 2 | 3+4 2 | 2+2 2 | 2+2 2 |
| | 5+6 2 | 3+4 2 | 5+6 2 | 3+4 2 | 2+2 2 | 3+3 2 |
| | 5+6 2 | 5+6 2 | 5+6 2 | 5+6 2 | 3+3 2 | 3+3 2 |
| | 3+4 2 | 5+6 2 | 3+4 2 | 5+6 2 | 2+2 2 | 2+2 2 |
| | 5+6 2 | 3+4 2 | 3+4 2 | 3+4 2 | 3+3 2 | 2+2 2 |
| | 5+6 2 | 3+4 2 | 5+6 2 | 5+6 2 | 2+2 2 | 3+3 2 |
| | 3+4 2 | 5+6 2 | 5+6 2 | 5+6 2 | 3+3 2 | 3+3 2 |
| | 3+4 2 | 3+4 2 | 3+4 2 | 3+4 2 | 3+3 2 | 3+3 2 |
| | 5+6 2 | 5+6 2 | 5+6 2 | 3+4 2 | 3+3 2 | 2+2 2 |
| | 5+6 2 | 3+4 2 | 3+4 2 | 5+6 2 | 2+2 2 | 2+2 2 |
| | 3+4 2 | 5+6 2 | 5+6 2 | 5+6 2 | 3+3 2 | 3+3 2 |
| | 5+6 2 | 5+6 2 | 3+4 2 | 5+6 2 | 2+2 2 | 2+2 2 |
| | 5+6 2 | 3+4 2 | 5+6 2 | 5+6 2 | 2+2 2 | 3+3 2 |
| | 3+4 2 | 5+6 2 | 3+4 2 | 3+4 2 | 3+3 2 | 2+2 2 |
| | 3+4 2 | 5+6 2 | 5+6 2 | 3+4 2 | 3+3 2 | 3+3 2 |
| | 5+6 2 | 5+6 2 | 5+6 2 | 5+6 2 | 3+3 2 | 3+3 2 |
| | 5+6 2 | 3+4 2 | 5+6 2 | 3+4 2 | 3+3 2 | 2+2 2 |
| | 3+4 2 | 3+4 2 | 5+6 2 | 3+4 2 | 2+2 2 | 3+3 2 |
| | 5+6 2 | 5+6 2 | 5+6 2 | 3+4 2 | 2+2 2 | 2+2 2 |
| | 3+4 2 | 3+4 2 | 3+4 2 | 5+6 2 | 2+2 2 | 3+3 2 |
| | 5+6 2 | 3+4 2 | 3+4 2 | 3+4 2 | 3+3 2 | 3+3 2 |
| 12.1.9 | 34417 | | 68 | | O2V1 | |
| | -A | -GM | -M | -Y | -L | -V |
| | 5+6 2 | 3+4 2 | 3+4 2 | 5+6 2 | 3+3 2 | 3+3 2 |
| | 3+4 2 | 5+6 2 | 3+4 2 | 3+4 2 | 2+2 2 | 2+2 2 |
| | 5+6 2 | 5+6 2 | 5+6 2 | 5+6 2 | 2+2 2 | 3+3 2 |
| | 3+4 2 | 3+4 2 | 5+6 2 | 3+4 2 | 3+3 2 | 2+2 2 |
| | 5+6 2 | 3+4 2 | 5+6 2 | 3+4 2 | 3+3 2 | 2+2 2 |
| | 3+4 2 | 5+6 2 | 3+4 2 | 5+6 2 | 2+2 2 | 3+3 2 |
| | 3+4 2 | 3+4 2 | 5+6 2 | 3+4 2 | 3+3 2 | 3+3 2 |
| | 5+6 2 | 5+6 2 | 3+4 2 | 5+6 2 | 2+2 2 | 2+2 2 |
| | 5+6 2 | 3+4 2 | 3+4 2 | 5+6 2 | 3+3 2 | 3+3 2 |
| | 5+6 2 | 3+4 2 | 3+4 2 | 3+4 2 | 2+2 2 | 2+2 2 |
| | 3+4 2 | 5+6 2 | 5+6 2 | 5+6 2 | 3+3 2 | 3+3 2 |
| | 3+4 2 | 3+4 2 | 5+6 2 | 3+4 2 | 2+2 2 | 2+2 2 |
| | 5+6 2 | 5+6 2 | 5+6 2 | 5+6 2 | 3+3 2 | 3+3 2 |
| | 3+4 2 | 5+6 2 | 5+6 2 | 5+6 2 | 2+2 2 | 2+2 2 |
| | 3+4 2 | 5+6 2 | 3+4 2 | 3+4 2 | 3+3 2 | 3+3 2 |
| | 5+6 2 | 3+4 2 | 3+4 2 | 3+4 2 | 2+2 2 | 2+2 2 |
| | 3+4 2 | 3+4 2 | 3+4 2 | 5+6 2 | 2+2 2 | 3+3 2 |
| | 5+6 2 | 3+4 2 | 3+4 2 | 5+6 2 | 3+3 2 | 2+2 2 |
| | 5+6 2 | 3+4 2 | 5+6 2 | 5+6 2 | 2+2 2 | 2+2 2 |
| | 5+6 2 | 3+4 2 | 3+4 2 | 3+4 2 | 3+3 2 | 3+3 2 |
| | 3+4 2 | 5+6 2 | 3+4 2 | 3+4 2 | 2+2 2 | 3+3 2 |
| | 3+4 2 | 5+6 2 | 5+6 2 | 3+4 2 | 3+3 2 | 2+2 2 |
| | 3+4 2 | 3+4 2 | 3+4 2 | 3+4 2 | 2+2 2 | 3+3 2 |
| | 5+6 2 | 5+6 2 | 5+6 2 | 5+6 2 | 3+3 2 | 3+3 2 |
| | 5+6 2 | 5+6 2 | 5+6 2 | 3+4 2 | 3+3 2 | 2+2 2 |
| | 3+4 2 | 5+6 2 | 3+4 2 | 5+6 2 | 3+3 2 | 2+2 2 |
| | 5+6 2 | 3+4 2 | 5+6 2 | 3+4 2 | 2+2 2 | 3+3 2 |
| | 3+4 2 | 3+4 2 | 5+6 2 | 5+6 2 | 2+2 2 | 2+2 2 |
| | 5+6 2 | 5+6 2 | 5+6 2 | 3+4 2 | 3+3 2 | 3+3 2 |
| | 3+4 2 | 5+6 2 | 5+6 2 | 5+6 2 | 2+2 2 | 2+2 2 |
| | 5+6 2 | 5+6 2 | 3+4 2 | 3+4 2 | 2+2 2 | 3+3 2 |
| | 3+4 2 | 3+4 2 | 3+4 2 | 5+6 2 | 3+3 2 | 2+2 2 |
| | 5+6 2 | 3+4 2 | 3+4 2 | 5+6 2 | 2+2 2 | 2+2 2 |
| | 3+4 2 | 3+4 2 | 3+4 2 | 3+4 2 | 2+2 2 | 2+2 2 |
| 12.1.10 | 34416 | | 68 | | O2V1 | |
| | -A | -GM | -M | -Y | -L | -V |
| | 5+6 2 | 3+4 2 | 3+4 2 | 5+6 2 | 3+3 2 | 3+3 2 |

Continued on next column

**TABLE XXIV– continued**

| TQC | ICSD | NoE | SF | | | |
|---|---|---|---|---|---|---|
| | 3+4 2 | 5+6 2 | 3+4 2 | 3+4 2 | 2+2 2 | 2+2 2 |
| | 5+6 2 | 5+6 2 | 5+6 2 | 5+6 2 | 3+3 2 | 3+3 2 |
| | 3+4 2 | 3+4 2 | 5+6 2 | 3+4 2 | 2+2 2 | 2+2 2 |
| | 5+6 2 | 3+4 2 | 5+6 2 | 3+4 2 | 3+3 2 | 3+3 2 |
| | 3+4 2 | 3+4 2 | 5+6 2 | 5+6 2 | 2+2 2 | 2+2 2 |
| | 5+6 2 | 5+6 2 | 3+4 2 | 3+4 2 | 3+3 2 | 3+3 2 |
| | 3+4 2 | 5+6 2 | 3+4 2 | 5+6 2 | 2+2 2 | 2+2 2 |
| | 5+6 2 | 3+4 2 | 3+4 2 | 3+4 2 | 3+3 2 | 3+3 2 |
| | 5+6 2 | 3+4 2 | 3+4 2 | 5+6 2 | 2+2 2 | 2+2 2 |
| | 3+4 2 | 5+6 2 | 5+6 2 | 5+6 2 | 3+3 2 | 3+3 2 |
| | 3+4 2 | 3+4 2 | 5+6 2 | 3+4 2 | 2+2 2 | 3+3 2 |
| | 5+6 2 | 5+6 2 | 5+6 2 | 5+6 2 | 3+3 2 | 2+2 2 |
| | 3+4 2 | 5+6 2 | 5+6 2 | 5+6 2 | 2+2 2 | 2+2 2 |
| | 3+4 2 | 5+6 2 | 3+4 2 | 3+4 2 | 3+3 2 | 3+3 2 |
| | 5+6 2 | 3+4 2 | 3+4 2 | 3+4 2 | 2+2 2 | 2+2 2 |
| | 3+4 2 | 3+4 2 | 3+4 2 | 5+6 2 | 2+2 2 | 3+3 2 |
| | 5+6 2 | 3+4 2 | 3+4 2 | 5+6 2 | 3+3 2 | 2+2 2 |
| | 5+6 2 | 3+4 2 | 5+6 2 | 5+6 2 | 2+2 2 | 2+2 2 |
| | 5+6 2 | 5+6 2 | 3+4 2 | 3+4 2 | 3+3 2 | 3+3 2 |
| | 3+4 2 | 3+4 2 | 3+4 2 | 3+4 2 | 2+2 2 | 3+3 2 |
| | 3+4 2 | 5+6 2 | 5+6 2 | 3+4 2 | 3+3 2 | 2+2 2 |
| | 3+4 2 | 3+4 2 | 3+4 2 | 3+4 2 | 2+2 2 | 3+3 2 |
| | 5+6 2 | 5+6 2 | 3+4 2 | 5+6 2 | 3+3 2 | 3+3 2 |
| | 5+6 2 | 5+6 2 | 5+6 2 | 3+4 2 | 3+3 2 | 2+2 2 |
| | 3+4 2 | 5+6 2 | 5+6 2 | 5+6 2 | 3+3 2 | 2+2 2 |
| | 3+4 2 | 3+4 2 | 5+6 2 | 5+6 2 | 2+2 2 | 3+3 2 |
| | 5+6 2 | 3+4 2 | 5+6 2 | 3+4 2 | 2+2 2 | 2+2 2 |
| | 5+6 2 | 5+6 2 | 5+6 2 | 3+4 2 | 3+3 2 | 3+3 2 |
| | 3+4 2 | 5+6 2 | 5+6 2 | 5+6 2 | 2+2 2 | 2+2 2 |
| | 5+6 2 | 5+6 2 | 3+4 2 | 3+4 2 | 2+2 2 | 3+3 2 |
| | 3+4 2 | 3+4 2 | 3+4 2 | 5+6 2 | 3+3 2 | 2+2 2 |
| | 5+6 2 | 3+4 2 | 3+4 2 | 5+6 2 | 2+2 2 | 2+2 2 |
| | 3+4 2 | 3+4 2 | 3+4 2 | 3+4 2 | 2+2 2 | 2+2 2 |
| 12.1.11 | 50409 | 108 | O13V6 | | | |
| | -A | -GM | -M | -Y | -L | -V |
| | 5+6 2 | 3+4 2 | 3+4 2 | 5+6 2 | 3+3 2 | 2+2 2 |
| | 3+4 2 | 3+4 2 | 5+6 2 | 3+4 2 | 2+2 2 | 3+3 2 |
| | 3+4 2 | 5+6 2 | 5+6 2 | 5+6 2 | 2+2 2 | 2+2 2 |
| | 5+6 2 | 3+4 2 | 3+4 2 | 3+4 2 | 3+3 2 | 3+3 2 |
| | 5+6 2 | 5+6 2 | 5+6 2 | 3+4 2 | 2+2 2 | 2+2 2 |
| | 3+4 2 | 5+6 2 | 3+4 2 | 5+6 2 | 3+3 2 | 3+3 2 |
| | 3+4 2 | 3+4 2 | 5+6 2 | 3+4 2 | 2+2 2 | 2+2 2 |
| | 5+6 2 | 5+6 2 | 3+4 2 | 3+4 2 | 3+3 2 | 3+3 2 |
| | 3+4 2 | 3+4 2 | 3+4 2 | 5+6 2 | 3+3 2 | 2+2 2 |
| | 3+4 2 | 5+6 2 | 5+6 2 | 3+4 2 | 2+2 2 | 3+3 2 |
| | 5+6 2 | 3+4 2 | 3+4 2 | 5+6 2 | 2+2 2 | 3+3 2 |
| | 5+6 2 | 5+6 2 | 5+6 2 | 5+6 2 | 3+3 2 | 2+2 2 |
| | 3+4 2 | 3+4 2 | 3+4 2 | 3+4 2 | 2+2 2 | 2+2 2 |
| | 5+6 2 | 3+4 2 | 3+4 2 | 5+6 2 | 3+3 2 | 2+2 2 |
| | 3+4 2 | 5+6 2 | 3+4 2 | 5+6 2 | 2+2 2 | 3+3 2 |
| | 5+6 2 | 3+4 2 | 5+6 2 | 5+6 2 | 3+3 2 | 3+3 2 |
| | 5+6 2 | 3+4 2 | 5+6 2 | 3+4 2 | 2+2 2 | 3+3 2 |
| | 3+4 2 | 5+6 2 | 5+6 2 | 3+4 2 | 2+2 2 | 2+2 2 |
| | 5+6 2 | 5+6 2 | 3+4 2 | 5+6 2 | 3+3 2 | 2+2 2 |
| | 5+6 2 | 5+6 2 | 5+6 2 | 3+4 2 | 3+3 2 | 2+2 2 |
| | 3+4 2 | 5+6 2 | 3+4 2 | 3+4 2 | 2+2 2 | 3+3 2 |
| | 3+4 2 | 3+4 2 | 5+6 2 | 3+4 2 | 3+3 2 | 3+3 2 |
| | 5+6 2 | 3+4 2 | 5+6 2 | 5+6 2 | 3+3 2 | 2+2 2 |
| | 3+4 2 | 3+4 2 | 5+6 2 | 5+6 2 | 2+2 2 | 3+3 2 |
| | 3+4 2 | 5+6 2 | 3+4 2 | 5+6 2 | 3+3 2 | 2+2 2 |
| | 5+6 2 | 3+4 2 | 5+6 2 | 3+4 2 | 3+3 2 | 3+3 2 |

Continued on next column

**TABLE XXIV– continued**

| TQC | ICSD | NoE | SF | | | |
|---|---|---|---|---|---|---|
| | 3+4 2 | 5+6 2 | 3+4 2 | 5+6 2 | 2+2 2 | 2+2 2 |
| | 5+6 2 | 3+4 2 | 5+6 2 | 5+6 2 | 2+2 2 | 3+3 2 |
| | 3+4 2 | 3+4 2 | 5+6 2 | 3+4 2 | 3+3 2 | 3+3 2 |
| | 5+6 2 | 3+4 2 | 3+4 2 | 5+6 2 | 2+2 2 | 2+2 2 |
| | 3+4 2 | 5+6 2 | 3+4 2 | 5+6 2 | 3+3 2 | 3+3 2 |
| | 5+6 2 | 5+6 2 | 3+4 2 | 3+4 2 | 2+2 2 | 2+2 2 |
| | 5+6 2 | 5+6 2 | 5+6 2 | 5+6 2 | 3+3 2 | 3+3 2 |
| | 3+4 2 | 3+4 2 | 3+4 2 | 5+6 2 | 2+2 2 | 2+2 2 |
| | 5+6 2 | 5+6 2 | 5+6 2 | 3+4 2 | 2+2 2 | 3+3 2 |
| | 3+4 2 | 3+4 2 | 5+6 2 | 5+6 2 | 3+3 2 | 2+2 2 |
| | 5+6 2 | 5+6 2 | 3+4 2 | 5+6 2 | 3+3 2 | 3+3 2 |
| | 3+4 2 | 5+6 2 | 3+4 2 | 3+4 2 | 3+3 2 | 3+3 2 |
| | 5+6 2 | 3+4 2 | 5+6 2 | 5+6 2 | 2+2 2 | 2+2 2 |
| | 5+6 2 | 3+4 2 | 5+6 2 | 3+4 2 | 3+3 2 | 2+2 2 |
| | 5+6 2 | 5+6 2 | 5+6 2 | 5+6 2 | 2+2 2 | 2+2 2 |
| | 3+4 2 | 3+4 2 | 3+4 2 | 3+4 2 | 2+2 2 | 3+3 2 |
| | 3+4 2 | 3+4 2 | 3+4 2 | 3+4 2 | 3+3 2 | 2+2 2 |
| | 5+6 2 | 5+6 2 | 3+4 2 | 3+4 2 | 3+3 2 | 3+3 2 |
| | 5+6 2 | 5+6 2 | 5+6 2 | 5+6 2 | 2+2 2 | 2+2 2 |
| | 3+4 2 | 5+6 2 | 3+4 2 | 3+4 2 | 3+3 2 | 3+3 2 |
| | 5+6 2 | 5+6 2 | 5+6 2 | 3+4 2 | 2+2 2 | 2+2 2 |
| | 3+4 2 | 5+6 2 | 3+4 2 | 3+4 2 | 3+3 2 | 3+3 2 |
| | 5+6 2 | 3+4 2 | 5+6 2 | 5+6 2 | 3+3 2 | 3+3 2 |
| | 3+4 2 | 5+6 2 | 5+6 2 | 5+6 2 | 2+2 2 | 3+3 2 |
| | 3+4 2 | 5+6 2 | 5+6 2 | 3+4 2 | 3+3 2 | 2+2 2 |
| | 5+6 2 | 3+4 2 | 3+4 2 | 5+6 2 | 2+2 2 | 3+3 2 |
| | 5+6 2 | 5+6 2 | 3+4 2 | 3+4 2 | 2+2 2 | 2+2 2 |
| | 3+4 2 | 3+4 2 | 5+6 2 | 5+6 2 | 2+2 2 | 3+3 2 |
| 12.1.12 | 15028 | 108 | O13V6 | | | |
| | -A | -GM | -M | -Y | -L | -V |
| | 5+6 2 | 3+4 2 | 3+4 2 | 5+6 2 | 3+3 2 | 2+2 2 |
| | 3+4 2 | 3+4 2 | 5+6 2 | 3+4 2 | 2+2 2 | 3+3 2 |
| | 3+4 2 | 5+6 2 | 5+6 2 | 5+6 2 | 2+2 2 | 2+2 2 |
| | 5+6 2 | 3+4 2 | 3+4 2 | 3+4 2 | 3+3 2 | 3+3 2 |
| | 5+6 2 | 5+6 2 | 5+6 2 | 3+4 2 | 2+2 2 | 2+2 2 |
| | 3+4 2 | 5+6 2 | 5+6 2 | 5+6 2 | 3+3 2 | 3+3 2 |
| | 3+4 2 | 3+4 2 | 3+4 2 | 3+4 2 | 2+2 2 | 2+2 2 |
| | 5+6 2 | 5+6 2 | 3+4 2 | 3+4 2 | 3+3 2 | 3+3 2 |
| | 3+4 2 | 3+4 2 | 5+6 2 | 5+6 2 | 3+3 2 | 2+2 2 |
| | 3+4 2 | 5+6 2 | 3+4 2 | 3+4 2 | 2+2 2 | 3+3 2 |
| | 5+6 2 | 3+4 2 | 3+4 2 | 5+6 2 | 2+2 2 | 3+3 2 |
| | 5+6 2 | 5+6 2 | 5+6 2 | 5+6 2 | 3+3 2 | 2+2 2 |
| | 3+4 2 | 3+4 2 | 3+4 2 | 3+4 2 | 2+2 2 | 2+2 2 |
| | 5+6 2 | 3+4 2 | 3+4 2 | 5+6 2 | 3+3 2 | 2+2 2 |
| | 3+4 2 | 5+6 2 | 3+4 2 | 5+6 2 | 2+2 2 | 3+3 2 |
| | 5+6 2 | 3+4 2 | 5+6 2 | 3+4 2 | 3+3 2 | 3+3 2 |
| | 5+6 2 | 3+4 2 | 5+6 2 | 5+6 2 | 2+2 2 | 3+3 2 |
| | 3+4 2 | 5+6 2 | 5+6 2 | 3+4 2 | 2+2 2 | 2+2 2 |
| | 5+6 2 | 5+6 2 | 3+4 2 | 5+6 2 | 3+3 2 | 2+2 2 |
| | 5+6 2 | 5+6 2 | 5+6 2 | 3+4 2 | 3+3 2 | 2+2 2 |
| | 3+4 2 | 5+6 2 | 3+4 2 | 3+4 2 | 2+2 2 | 3+3 2 |
| | 3+4 2 | 3+4 2 | 5+6 2 | 3+4 2 | 3+3 2 | 3+3 2 |
| | 5+6 2 | 3+4 2 | 5+6 2 | 5+6 2 | 3+3 2 | 2+2 2 |
| | 3+4 2 | 3+4 2 | 5+6 2 | 5+6 2 | 2+2 2 | 3+3 2 |
| | 3+4 2 | 5+6 2 | 3+4 2 | 5+6 2 | 3+3 2 | 3+3 2 |
| | 5+6 2 | 3+4 2 | 5+6 2 | 3+4 2 | 3+3 2 | 2+2 2 |
| | 3+4 2 | 5+6 2 | 3+4 2 | 5+6 2 | 2+2 2 | 2+2 2 |
| | 5+6 2 | 3+4 2 | 5+6 2 | 5+6 2 | 2+2 2 | 3+3 2 |
| | 3+4 2 | 3+4 2 | 5+6 2 | 3+4 2 | 3+3 2 | 3+3 2 |
| | 5+6 2 | 3+4 2 | 3+4 2 | 5+6 2 | 2+2 2 | 2+2 2 |
| | 3+4 2 | 5+6 2 | 3+4 2 | 5+6 2 | 3+3 2 | 3+3 2 |

Continued on next column

**TABLE XXIV– continued**

| TQC | ICSD | | NoE | SF | | |
|---|---|---|---|---|---|---|
| | 5+6 2 | 5+6 2 | 3+4 2 | 3+4 2 | 2+2 2 | 2+2 2 |
| | 5+6 2 | 5+6 2 | 5+6 2 | 5+6 2 | 3+3 2 | 3+3 2 |
| | 3+4 2 | 3+4 2 | 3+4 2 | 5+6 2 | 2+2 2 | 2+2 2 |
| | 5+6 2 | 5+6 2 | 5+6 2 | 3+4 2 | 2+2 2 | 3+3 2 |
| | 3+4 2 | 3+4 2 | 5+6 2 | 5+6 2 | 3+3 2 | 2+2 2 |
| | 5+6 2 | 5+6 2 | 3+4 2 | 5+6 2 | 3+3 2 | 3+3 2 |
| | 3+4 2 | 5+6 2 | 3+4 2 | 3+4 2 | 3+3 2 | 3+3 2 |
| | 5+6 2 | 3+4 2 | 5+6 2 | 5+6 2 | 2+2 2 | 2+2 2 |
| | 5+6 2 | 3+4 2 | 5+6 2 | 3+4 2 | 2+2 2 | 2+2 2 |
| | 5+6 2 | 5+6 2 | 3+4 2 | 5+6 2 | 3+3 2 | 2+2 2 |
| | 3+4 2 | 3+4 2 | 5+6 2 | 3+4 2 | 2+2 2 | 3+3 2 |
| | 3+4 2 | 5+6 2 | 3+4 2 | 3+4 2 | 3+3 2 | 2+2 2 |
| | 5+6 2 | 3+4 2 | 3+4 2 | 3+4 2 | 3+3 2 | 3+3 2 |
| | 5+6 2 | 5+6 2 | 5+6 2 | 5+6 2 | 2+2 2 | 2+2 2 |
| | 3+4 2 | 5+6 2 | 3+4 2 | 3+4 2 | 3+3 2 | 3+3 2 |
| | 5+6 2 | 5+6 2 | 5+6 2 | 3+4 2 | 2+2 2 | 2+2 2 |
| | 3+4 2 | 5+6 2 | 3+4 2 | 3+4 2 | 3+3 2 | 3+3 2 |
| | 5+6 2 | 3+4 2 | 5+6 2 | 5+6 2 | 3+3 2 | 3+3 2 |
| | 3+4 2 | 5+6 2 | 5+6 2 | 5+6 2 | 2+2 2 | 3+3 2 |
| | 3+4 2 | 5+6 2 | 5+6 2 | 3+4 2 | 3+3 2 | 2+2 2 |
| | 5+6 2 | 3+4 2 | 3+4 2 | 5+6 2 | 2+2 2 | 3+3 2 |
| | 5+6 2 | 5+6 2 | 3+4 2 | 3+4 2 | 2+2 2 | 2+2 2 |
| | 3+4 2 | 3+4 2 | 5+6 2 | 5+6 2 | 2+2 2 | 3+3 2 |
| 12.1.13 | 646547 | | 42 | Ni1Se4Ti2 | | |
| | -A | -GM | -M | -Y | -L | -V |
| | 3+4 2 | 3+4 2 | 5+6 2 | 5+6 2 | 2+2 2 | 3+3 2 |
| | 5+6 2 | 5+6 2 | 3+4 2 | 3+4 2 | 3+3 2 | 2+2 2 |
| | 3+4 2 | 3+4 2 | 5+6 2 | 5+6 2 | 2+2 2 | 2+2 2 |
| | 5+6 2 | 5+6 2 | 3+4 2 | 3+4 2 | 3+3 2 | 3+3 2 |
| | 3+4 2 | 3+4 2 | 3+4 2 | 3+4 2 | 3+3 2 | 3+3 2 |
| | 5+6 2 | 5+6 2 | 5+6 2 | 5+6 2 | 2+2 2 | 2+2 2 |
| | 3+4 2 | 3+4 2 | 5+6 2 | 3+4 2 | 3+3 2 | 2+2 2 |
| | 5+6 2 | 3+4 2 | 5+6 2 | 5+6 2 | 2+2 2 | 3+3 2 |
| | 3+4 2 | 5+6 2 | 3+4 2 | 5+6 2 | 3+3 2 | 3+3 2 |
| | 3+4 2 | 3+4 2 | 3+4 2 | 3+4 2 | 2+2 2 | 3+3 2 |
| | 5+6 2 | 5+6 2 | 3+4 2 | 3+4 2 | 3+3 2 | 2+2 2 |
| | 3+4 2 | 3+4 2 | 5+6 2 | 5+6 2 | 2+2 2 | 2+2 2 |
| | 5+6 2 | 3+4 2 | 3+4 2 | 5+6 2 | 2+2 2 | 3+3 2 |
| | 3+4 2 | 3+4 2 | 3+4 2 | 5+6 2 | 3+3 2 | 3+3 2 |
| | 3+4 2 | 3+4 2 | 5+6 2 | 3+4 2 | 2+2 2 | 2+2 2 |
| | 5+6 2 | 3+4 2 | 3+4 2 | 3+4 2 | 3+3 2 | 2+2 2 |
| | 5+6 2 | 5+6 2 | 5+6 2 | 3+4 2 | 3+3 2 | 3+3 2 |
| | 3+4 2 | 5+6 2 | 3+4 2 | 3+4 2 | 3+3 2 | 3+3 2 |
| | 3+4 2 | 3+4 2 | 3+4 2 | 3+4 2 | 3+3 2 | 3+3 2 |
| | 3+4 2 | 5+6 2 | 5+6 2 | 3+4 2 | 3+3 2 | 3+3 2 |
| | 3+4 2 | 3+4 2 | 3+4 2 | 5+6 2 | 2+2 2 | 3+3 2 |
| 12.1.14 | 646544 | | 42 | Ni1Se4Ti2 | | |
| | -A | -GM | -M | -Y | -L | -V |
| | 3+4 2 | 3+4 2 | 5+6 2 | 5+6 2 | 2+2 2 | 3+3 2 |
| | 5+6 2 | 5+6 2 | 3+4 2 | 3+4 2 | 3+3 2 | 2+2 2 |
| | 3+4 2 | 3+4 2 | 5+6 2 | 5+6 2 | 2+2 2 | 2+2 2 |
| | 5+6 2 | 5+6 2 | 3+4 2 | 3+4 2 | 3+3 2 | 3+3 2 |
| | 3+4 2 | 3+4 2 | 3+4 2 | 3+4 2 | 3+3 2 | 3+3 2 |
| | 5+6 2 | 5+6 2 | 5+6 2 | 5+6 2 | 2+2 2 | 2+2 2 |
| | 3+4 2 | 3+4 2 | 5+6 2 | 3+4 2 | 3+3 2 | 2+2 2 |
| | 5+6 2 | 3+4 2 | 5+6 2 | 5+6 2 | 2+2 2 | 3+3 2 |
| | 3+4 2 | 5+6 2 | 3+4 2 | 5+6 2 | 3+3 2 | 3+3 2 |
| | 3+4 2 | 3+4 2 | 3+4 2 | 3+4 2 | 2+2 2 | 3+3 2 |
| | 5+6 2 | 5+6 2 | 3+4 2 | 3+4 2 | 3+3 2 | 2+2 2 |
| | 3+4 2 | 3+4 2 | 5+6 2 | 5+6 2 | 2+2 2 | 2+2 2 |
| | 3+4 2 | 3+4 2 | 3+4 2 | 5+6 2 | 2+2 2 | 3+3 2 |

Continued on next column

**TABLE XXIV– continued**

| TQC | ICSD | | NoE | SF | | |
|---|---|---|---|---|---|---|
| | 5+6 2 | 3+4 2 | 3+4 2 | 5+6 2 | 3+3 2 | 3+3 2 |
| | 3+4 2 | 3+4 2 | 5+6 2 | 3+4 2 | 2+2 2 | 2+2 2 |
| | 5+6 2 | 3+4 2 | 3+4 2 | 3+4 2 | 3+3 2 | 2+2 2 |
| | 5+6 2 | 5+6 2 | 5+6 2 | 3+4 2 | 3+3 2 | 3+3 2 |
| | 3+4 2 | 3+4 2 | 3+4 2 | 3+4 2 | 3+3 2 | 3+3 2 |
| | 3+4 2 | 5+6 2 | 3+4 2 | 3+4 2 | 3+3 2 | 3+3 2 |
| | 3+4 2 | 5+6 2 | 5+6 2 | 3+4 2 | 3+3 2 | 3+3 2 |
| | 3+4 2 | 3+4 2 | 3+4 2 | 5+6 2 | 2+2 2 | 3+3 2 |
| 12.1.15 | 646533 | | 42 | Ni1Se4Si2 | | |
| | -A | -GM | -M | -Y | -L | -V |
| | 5+6 2 | 3+4 2 | 3+4 2 | 5+6 2 | 2+2 2 | 3+3 2 |
| | 3+4 2 | 5+6 2 | 5+6 2 | 3+4 2 | 3+3 2 | 2+2 2 |
| | 3+4 2 | 3+4 2 | 5+6 2 | 5+6 2 | 2+2 2 | 2+2 2 |
| | 5+6 2 | 5+6 2 | 3+4 2 | 3+4 2 | 3+3 2 | 3+3 2 |
| | 5+6 2 | 5+6 2 | 3+4 2 | 3+4 2 | 2+2 2 | 3+3 2 |
| | 3+4 2 | 3+4 2 | 5+6 2 | 5+6 2 | 3+3 2 | 2+2 2 |
| | 3+4 2 | 3+4 2 | 3+4 2 | 5+6 2 | 3+3 2 | 3+3 2 |
| | 5+6 2 | 3+4 2 | 5+6 2 | 3+4 2 | 2+2 2 | 2+2 2 |
| | 5+6 2 | 3+4 2 | 3+4 2 | 5+6 2 | 2+2 2 | 2+2 2 |
| | 3+4 2 | 5+6 2 | 5+6 2 | 5+6 2 | 3+3 2 | 3+3 2 |
| | 5+6 2 | 3+4 2 | 3+4 2 | 3+4 2 | 2+2 2 | 3+3 2 |
| | 3+4 2 | 5+6 2 | 3+4 2 | 3+4 2 | 3+3 2 | 2+2 2 |
| | 3+4 2 | 3+4 2 | 5+6 2 | 3+4 2 | 3+3 2 | 3+3 2 |
| | 3+4 2 | 5+6 2 | 3+4 2 | 3+4 2 | 2+2 2 | 2+2 2 |
| | 3+4 2 | 3+4 2 | 5+6 2 | 5+6 2 | 3+3 2 | 3+3 2 |
| | 3+4 2 | 5+6 2 | 5+6 2 | 3+4 2 | 2+2 2 | 2+2 2 |
| | 5+6 2 | 3+4 2 | 3+4 2 | 5+6 2 | 3+3 2 | 3+3 2 |
| | 3+4 2 | 3+4 2 | 3+4 2 | 3+4 2 | 3+3 2 | 3+3 2 |
| | 5+6 2 | 5+6 2 | 3+4 2 | 3+4 2 | 3+3 2 | 3+3 2 |
| | 3+4 2 | 5+6 2 | 5+6 2 | 3+4 2 | 3+3 2 | 3+3 2 |
| | 3+4 2 | 3+4 2 | 3+4 2 | 3+4 2 | 2+2 2 | 3+3 2 |
| 12.1.16 | 48004 | | 62 | Ni1Sc3Si3 | | |
| | -A | -GM | -M | -Y | -L | -V |
| | 5+6 2 | 3+4 2 | 3+4 2 | 3+4 2 | 3+3 2 | 3+3 2 |
| | 3+4 2 | 3+4 2 | 5+6 2 | 3+4 2 | 2+2 2 | 3+3 2 |
| | 5+6 2 | 5+6 2 | 3+4 2 | 5+6 2 | 2+2 2 | 2+2 2 |
| | 3+4 2 | 3+4 2 | 5+6 2 | 5+6 2 | 3+3 2 | 2+2 2 |
| | 3+4 2 | 5+6 2 | 3+4 2 | 3+4 2 | 3+3 2 | 3+3 2 |
| | 5+6 2 | 5+6 2 | 5+6 2 | 5+6 2 | 2+2 2 | 2+2 2 |
| | 5+6 2 | 3+4 2 | 5+6 2 | 5+6 2 | 3+3 2 | 2+2 2 |
| | 3+4 2 | 5+6 2 | 3+4 2 | 3+4 2 | 2+2 2 | 3+3 2 |
| | 5+6 2 | 5+6 2 | 5+6 2 | 3+4 2 | 2+2 2 | 2+2 2 |
| | 3+4 2 | 3+4 2 | 3+4 2 | 5+6 2 | 3+3 2 | 3+3 2 |
| | 3+4 2 | 5+6 2 | 3+4 2 | 5+6 2 | 2+2 2 | 2+2 2 |
| | 5+6 2 | 3+4 2 | 5+6 2 | 3+4 2 | 2+2 2 | 3+3 2 |
| | 3+4 2 | 3+4 2 | 5+6 2 | 3+4 2 | 3+3 2 | 2+2 2 |
| | 5+6 2 | 3+4 2 | 3+4 2 | 3+4 2 | 3+3 2 | 3+3 2 |
| | 3+4 2 | 3+4 2 | 5+6 2 | 5+6 2 | 3+3 2 | 3+3 2 |
| | 5+6 2 | 5+6 2 | 3+4 2 | 5+6 2 | 2+2 2 | 2+2 2 |
| | 5+6 2 | 5+6 2 | 3+4 2 | 5+6 2 | 3+3 2 | 2+2 2 |
| | 3+4 2 | 3+4 2 | 3+4 2 | 3+4 2 | 2+2 2 | 3+3 2 |
| | 3+4 2 | 5+6 2 | 5+6 2 | 3+4 2 | 3+3 2 | 3+3 2 |
| | 5+6 2 | 5+6 2 | 5+6 2 | 5+6 2 | 2+2 2 | 2+2 2 |
| | 5+6 2 | 3+4 2 | 5+6 2 | 5+6 2 | 2+2 2 | 3+3 2 |
| | 3+4 2 | 5+6 2 | 3+4 2 | 5+6 2 | 3+3 2 | 3+3 2 |
| | 5+6 2 | 3+4 2 | 3+4 2 | 3+4 2 | 2+2 2 | 2+2 2 |
| | 3+4 2 | 3+4 2 | 3+4 2 | 3+4 2 | 3+3 2 | 3+3 2 |
| | 5+6 2 | 5+6 2 | 3+4 2 | 5+6 2 | 2+2 2 | 2+2 2 |
| | 5+6 2 | 3+4 2 | 3+4 2 | 3+4 2 | 3+3 2 | 3+3 2 |
| | 3+4 2 | 5+6 2 | 5+6 2 | 5+6 2 | 3+3 2 | 2+2 2 |
| | 3+4 2 | 3+4 2 | 5+6 2 | 5+6 2 | 2+2 2 | 3+3 2 |

Continued on next column

**TABLE XXIV– continued**

| TQC | ICSD | | NoE | SF | | |
|---|---|---|---|---|---|---|
| | 5+6 2 | 5+6 2 | 5+6 2 | 3+4 2 | 3+3 2 | 2+2 2 |
| | 5+6 2 | 3+4 2 | 3+4 2 | 3+4 2 | 2+2 2 | 3+3 2 |
| | 5+6 2 | 5+6 2 | 5+6 2 | 3+4 2 | 3+3 2 | 2+2 2 |
| 12.1.17 | 646389 | | 74 | Ni1S8Ti4 | | |
| | -A | -GM | -M | -Y | -L | -V |
| | 3+4 2 | 3+4 2 | 5+6 2 | 5+6 2 | 3+3 2 | 3+3 2 |
| | 5+6 2 | 5+6 2 | 3+4 2 | 3+4 2 | 2+2 2 | 2+2 2 |
| | 5+6 2 | 5+6 2 | 5+6 2 | 3+4 2 | 2+2 2 | 2+2 2 |
| | 3+4 2 | 3+4 2 | 5+6 2 | 5+6 2 | 3+3 2 | 3+3 2 |
| | 5+6 2 | 3+4 2 | 3+4 2 | 5+6 2 | 2+2 2 | 3+3 2 |
| | 3+4 2 | 5+6 2 | 3+4 2 | 3+4 2 | 3+3 2 | 2+2 2 |
| | 5+6 2 | 5+6 2 | 5+6 2 | 3+4 2 | 3+3 2 | 2+2 2 |
| | 3+4 2 | 3+4 2 | 3+4 2 | 5+6 2 | 2+2 2 | 3+3 2 |
| | 3+4 2 | 3+4 2 | 3+4 2 | 3+4 2 | 3+3 2 | 3+3 2 |
| | 5+6 2 | 5+6 2 | 5+6 2 | 3+4 2 | 2+2 2 | 2+2 2 |
| | 5+6 2 | 5+6 2 | 3+4 2 | 3+4 2 | 3+3 2 | 3+3 2 |
| | 3+4 2 | 3+4 2 | 3+4 2 | 5+6 2 | 2+2 2 | 2+2 2 |
| | 3+4 2 | 3+4 2 | 5+6 2 | 3+4 2 | 2+2 2 | 3+3 2 |
| | 5+6 2 | 3+4 2 | 5+6 2 | 5+6 2 | 3+3 2 | 2+2 2 |
| | 3+4 2 | 5+6 2 | 5+6 2 | 5+6 2 | 3+3 2 | 3+3 2 |
| | 3+4 2 | 3+4 2 | 3+4 2 | 5+6 2 | 2+2 2 | 2+2 2 |
| | 5+6 2 | 3+4 2 | 5+6 2 | 5+6 2 | 2+2 2 | 3+3 2 |
| | 3+4 2 | 5+6 2 | 3+4 2 | 3+4 2 | 3+3 2 | 2+2 2 |
| | 5+6 2 | 5+6 2 | 3+4 2 | 5+6 2 | 3+3 2 | 3+3 2 |
| | 3+4 2 | 5+6 2 | 5+6 2 | 3+4 2 | 3+3 2 | 2+2 2 |
| | 3+4 2 | 3+4 2 | 5+6 2 | 5+6 2 | 3+3 2 | 3+3 2 |
| | 3+4 2 | 3+4 2 | 3+4 2 | 3+4 2 | 2+2 2 | 2+2 2 |
| | 5+6 2 | 5+6 2 | 5+6 2 | 5+6 2 | 2+2 2 | 3+3 2 |
| | 5+6 2 | 5+6 2 | 5+6 2 | 5+6 2 | 2+2 2 | 2+2 2 |
| | 3+4 2 | 3+4 2 | 3+4 2 | 3+4 2 | 2+2 2 | 3+3 2 |
| | 5+6 2 | 3+4 2 | 5+6 2 | 3+4 2 | 3+3 2 | 3+3 2 |
| | 5+6 2 | 3+4 2 | 3+4 2 | 3+4 2 | 3+3 2 | 2+2 2 |
| | 5+6 2 | 5+6 2 | 3+4 2 | 5+6 2 | 3+3 2 | 2+2 2 |
| | 3+4 2 | 3+4 2 | 5+6 2 | 3+4 2 | 2+2 2 | 3+3 2 |
| | 3+4 2 | 3+4 2 | 5+6 2 | 3+4 2 | 3+3 2 | 2+2 2 |
| | 5+6 2 | 3+4 2 | 3+4 2 | 5+6 2 | 3+3 2 | 3+3 2 |
| | 3+4 2 | 3+4 2 | 3+4 2 | 3+4 2 | 2+2 2 | 2+2 2 |
| | 5+6 2 | 5+6 2 | 3+4 2 | 5+6 2 | 3+3 2 | 3+3 2 |
| | 3+4 2 | 5+6 2 | 3+4 2 | 3+4 2 | 2+2 2 | 3+3 2 |
| | 3+4 2 | 5+6 2 | 3+4 2 | 3+4 2 | 3+3 2 | 3+3 2 |
| | 3+4 2 | 3+4 2 | 5+6 2 | 3+4 2 | 3+3 2 | 3+3 2 |
| | 3+4 2 | 5+6 2 | 3+4 2 | 3+4 2 | 2+2 2 | 3+3 2 |
| 12.1.18 | 646388 | | 42 | Ni1S4Ti2 | | |
| | -A | -GM | -M | -Y | -L | -V |
| | 3+4 2 | 3+4 2 | 3+4 2 | 5+6 2 | 2+2 2 | 3+3 2 |
| | 5+6 2 | 5+6 2 | 5+6 2 | 3+4 2 | 3+3 2 | 2+2 2 |
| | 3+4 2 | 5+6 2 | 5+6 2 | 3+4 2 | 2+2 2 | 2+2 2 |
| | 5+6 2 | 3+4 2 | 3+4 2 | 5+6 2 | 3+3 2 | 3+3 2 |
| | 3+4 2 | 3+4 2 | 3+4 2 | 3+4 2 | 3+3 2 | 3+3 2 |
| | 5+6 2 | 5+6 2 | 5+6 2 | 5+6 2 | 2+2 2 | 2+2 2 |
| | 3+4 2 | 3+4 2 | 5+6 2 | 3+4 2 | 3+3 2 | 3+3 2 |
| | 5+6 2 | 3+4 2 | 3+4 2 | 5+6 2 | 2+2 2 | 2+2 2 |
| | 3+4 2 | 3+4 2 | 5+6 2 | 3+4 2 | 3+3 2 | 3+3 2 |
| | 3+4 2 | 5+6 2 | 3+4 2 | 5+6 2 | 2+2 2 | 3+3 2 |
| | 5+6 2 | 5+6 2 | 3+4 2 | 3+4 2 | 3+3 2 | 2+2 2 |
| | 3+4 2 | 3+4 2 | 5+6 2 | 5+6 2 | 2+2 2 | 3+3 2 |
| | 5+6 2 | 3+4 2 | 3+4 2 | 5+6 2 | 2+2 2 | 2+2 2 |
| | 3+4 2 | 3+4 2 | 3+4 2 | 5+6 2 | 3+3 2 | 2+2 2 |
| | 5+6 2 | 3+4 2 | 5+6 2 | 3+4 2 | 2+2 2 | 3+3 2 |
| | 5+6 2 | 3+4 2 | 3+4 2 | 3+4 2 | 3+3 2 | 2+2 2 |
| | 3+4 2 | 5+6 2 | 5+6 2 | 3+4 2 | 3+3 2 | 3+3 2 |

Continued on next column

**TABLE XXIV– continued**

| TQC | ICSD | | NoE | SF | | |
|---|---|---|---|---|---|---|
| | 3+4 2 | 5+6 2 | 3+4 2 | 3+4 2 | 3+3 2 | 3+3 2 |
| | 3+4 2 | 5+6 2 | 3+4 2 | 3+4 2 | 3+3 2 | 3+3 2 |
| | 3+4 2 | 3+4 2 | 3+4 2 | 3+4 2 | 3+3 2 | 3+3 2 |
| | 3+4 2 | 3+4 2 | 5+6 2 | 5+6 2 | 2+2 2 | 3+3 2 |
| 12.1.19 | 646321 | | 52 | Ni1Rh2Se4 | | |
| | -A | -GM | -M | -Y | -L | -V |
| | 3+4 2 | 3+4 2 | 3+4 2 | 5+6 2 | 2+2 2 | 3+3 2 |
| | 5+6 2 | 5+6 2 | 5+6 2 | 3+4 2 | 2+2 2 | 2+2 2 |
| | 3+4 2 | 3+4 2 | 5+6 2 | 3+4 2 | 3+3 2 | 2+2 2 |
| | 5+6 2 | 5+6 2 | 3+4 2 | 5+6 2 | 3+3 2 | 3+3 2 |
| | 3+4 2 | 3+4 2 | 3+4 2 | 3+4 2 | 3+3 2 | 3+3 2 |
| | 5+6 2 | 5+6 2 | 5+6 2 | 5+6 2 | 2+2 2 | 2+2 2 |
| | 3+4 2 | 3+4 2 | 3+4 2 | 3+4 2 | 3+3 2 | 3+3 2 |
| | 5+6 2 | 3+4 2 | 5+6 2 | 5+6 2 | 2+2 2 | 2+2 2 |
| | 3+4 2 | 3+4 2 | 5+6 2 | 3+4 2 | 3+3 2 | 3+3 2 |
| | 5+6 2 | 5+6 2 | 3+4 2 | 5+6 2 | 3+3 2 | 2+2 2 |
| | 3+4 2 | 5+6 2 | 5+6 2 | 5+6 2 | 2+2 2 | 3+3 2 |
| | 5+6 2 | 3+4 2 | 3+4 2 | 3+4 2 | 2+2 2 | 2+2 2 |
| | 3+4 2 | 3+4 2 | 3+4 2 | 5+6 2 | 2+2 2 | 2+2 2 |
| | 3+4 2 | 3+4 2 | 3+4 2 | 5+6 2 | 3+3 2 | 3+3 2 |
| | 5+6 2 | 5+6 2 | 5+6 2 | 3+4 2 | 2+2 2 | 2+2 2 |
| | 3+4 2 | 3+4 2 | 3+4 2 | 3+4 2 | 3+3 2 | 3+3 2 |
| | 5+6 2 | 3+4 2 | 5+6 2 | 5+6 2 | 3+3 2 | 3+3 2 |
| | 3+4 2 | 5+6 2 | 3+4 2 | 3+4 2 | 2+2 2 | 2+2 2 |
| | 5+6 2 | 3+4 2 | 3+4 2 | 5+6 2 | 3+3 2 | 3+3 2 |
| | 3+4 2 | 5+6 2 | 3+4 2 | 3+4 2 | 2+2 2 | 3+3 2 |
| | 3+4 2 | 3+4 2 | 3+4 2 | 3+4 2 | 3+3 2 | 2+2 2 |
| | 5+6 2 | 5+6 2 | 5+6 2 | 3+4 2 | 3+3 2 | 3+3 2 |
| | 5+6 2 | 3+4 2 | 5+6 2 | 3+4 2 | 3+3 2 | 3+3 2 |
| | 3+4 2 | 3+4 2 | 3+4 2 | 5+6 2 | 2+2 2 | 3+3 2 |
| | 3+4 2 | 5+6 2 | 3+4 2 | 3+4 2 | 3+3 2 | 3+3 2 |
| | 3+4 2 | 3+4 2 | 3+4 2 | 3+4 2 | 3+3 2 | 2+2 2 |
| 12.1.20 | 646318 | | 52 | Ni1Rh2Se4 | | |
| | -A | -GM | -M | -Y | -L | -V |
| | 3+4 2 | 3+4 2 | 3+4 2 | 5+6 2 | 2+2 2 | 3+3 2 |
| | 5+6 2 | 5+6 2 | 5+6 2 | 3+4 2 | 2+2 2 | 2+2 2 |
| | 3+4 2 | 3+4 2 | 5+6 2 | 3+4 2 | 3+3 2 | 2+2 2 |
| | 5+6 2 | 5+6 2 | 3+4 2 | 5+6 2 | 3+3 2 | 3+3 2 |
| | 3+4 2 | 3+4 2 | 3+4 2 | 3+4 2 | 3+3 2 | 3+3 2 |
| | 5+6 2 | 5+6 2 | 5+6 2 | 5+6 2 | 2+2 2 | 2+2 2 |
| | 3+4 2 | 3+4 2 | 3+4 2 | 3+4 2 | 3+3 2 | 3+3 2 |
| | 5+6 2 | 3+4 2 | 5+6 2 | 5+6 2 | 2+2 2 | 2+2 2 |
| | 3+4 2 | 3+4 2 | 5+6 2 | 3+4 2 | 3+3 2 | 3+3 2 |
| | 5+6 2 | 5+6 2 | 3+4 2 | 5+6 2 | 3+3 2 | 2+2 2 |
| | 3+4 2 | 5+6 2 | 5+6 2 | 5+6 2 | 2+2 2 | 3+3 2 |
| | 5+6 2 | 3+4 2 | 3+4 2 | 3+4 2 | 2+2 2 | 2+2 2 |
| | 3+4 2 | 3+4 2 | 3+4 2 | 5+6 2 | 2+2 2 | 2+2 2 |
| | 3+4 2 | 3+4 2 | 3+4 2 | 5+6 2 | 3+3 2 | 3+3 2 |
| | 5+6 2 | 5+6 2 | 5+6 2 | 3+4 2 | 2+2 2 | 2+2 2 |
| | 3+4 2 | 3+4 2 | 3+4 2 | 3+4 2 | 3+3 2 | 3+3 2 |
| | 5+6 2 | 3+4 2 | 5+6 2 | 5+6 2 | 3+3 2 | 3+3 2 |
| | 3+4 2 | 5+6 2 | 3+4 2 | 3+4 2 | 2+2 2 | 2+2 2 |
| | 5+6 2 | 3+4 2 | 3+4 2 | 5+6 2 | 3+3 2 | 3+3 2 |
| | 3+4 2 | 5+6 2 | 3+4 2 | 3+4 2 | 2+2 2 | 3+3 2 |
| | 3+4 2 | 3+4 2 | 3+4 2 | 3+4 2 | 3+3 2 | 2+2 2 |
| | 5+6 2 | 5+6 2 | 5+6 2 | 3+4 2 | 3+3 2 | 3+3 2 |
| | 5+6 2 | 3+4 2 | 5+6 2 | 3+4 2 | 3+3 2 | 3+3 2 |
| | 3+4 2 | 3+4 2 | 3+4 2 | 5+6 2 | 2+2 2 | 3+3 2 |
| | 3+4 2 | 5+6 2 | 3+4 2 | 3+4 2 | 3+3 2 | 3+3 2 |
| | 3+4 2 | 3+4 2 | 3+4 2 | 3+4 2 | 3+3 2 | 2+2 2 |
| 12.1.21 | 15214 | | 52 | Ni1Rh2Se4 | | |

Continued on next column

**TABLE XXIV– continued**

| TQC | ICSD | | NoE | | SF | |
|---|---|---|---|---|---|---|
| | -A | -GM | -M | -Y | -L | -V |
| | 3+4 2 | 3+4 2 | 3+4 2 | 5+6 2 | 2+2 2 | 3+3 2 |
| | 5+6 2 | 5+6 2 | 5+6 2 | 3+4 2 | 2+2 2 | 2+2 2 |
| | 3+4 2 | 3+4 2 | 5+6 2 | 3+4 2 | 3+3 2 | 2+2 2 |
| | 5+6 2 | 5+6 2 | 3+4 2 | 5+6 2 | 3+3 2 | 3+3 2 |
| | 3+4 2 | 3+4 2 | 3+4 2 | 3+4 2 | 3+3 2 | 3+3 2 |
| | 5+6 2 | 5+6 2 | 5+6 2 | 5+6 2 | 2+2 2 | 2+2 2 |
| | 3+4 2 | 3+4 2 | 3+4 2 | 3+4 2 | 3+3 2 | 3+3 2 |
| | 5+6 2 | 3+4 2 | 5+6 2 | 5+6 2 | 2+2 2 | 2+2 2 |
| | 3+4 2 | 3+4 2 | 5+6 2 | 3+4 2 | 3+3 2 | 3+3 2 |
| | 5+6 2 | 5+6 2 | 3+4 2 | 5+6 2 | 3+3 2 | 2+2 2 |
| | 3+4 2 | 5+6 2 | 5+6 2 | 5+6 2 | 2+2 2 | 3+3 2 |
| | 5+6 2 | 3+4 2 | 3+4 2 | 3+4 2 | 2+2 2 | 2+2 2 |
| | 3+4 2 | 3+4 2 | 3+4 2 | 5+6 2 | 2+2 2 | 2+2 2 |
| | 3+4 2 | 3+4 2 | 3+4 2 | 5+6 2 | 3+3 2 | 3+3 2 |
| | 5+6 2 | 5+6 2 | 5+6 2 | 3+4 2 | 2+2 2 | 2+2 2 |
| | 3+4 2 | 3+4 2 | 3+4 2 | 3+4 2 | 3+3 2 | 3+3 2 |
| | 5+6 2 | 3+4 2 | 5+6 2 | 5+6 2 | 3+3 2 | 3+3 2 |
| | 3+4 2 | 5+6 2 | 3+4 2 | 3+4 2 | 2+2 2 | 2+2 2 |
| | 5+6 2 | 3+4 2 | 3+4 2 | 5+6 2 | 3+3 2 | 3+3 2 |
| | 3+4 2 | 5+6 2 | 3+4 2 | 3+4 2 | 2+2 2 | 3+3 2 |
| | 3+4 2 | 3+4 2 | 3+4 2 | 3+4 2 | 3+3 2 | 2+2 2 |
| | 5+6 2 | 5+6 2 | 5+6 2 | 3+4 2 | 3+3 2 | 3+3 2 |
| | 5+6 2 | 3+4 2 | 5+6 2 | 3+4 2 | 3+3 2 | 3+3 2 |
| | 3+4 2 | 3+4 2 | 3+4 2 | 5+6 2 | 2+2 2 | 3+3 2 |
| | 3+4 2 | 5+6 2 | 3+4 2 | 3+4 2 | 3+3 2 | 3+3 2 |
| | 3+4 2 | 3+4 2 | 3+4 2 | 3+4 2 | 3+3 2 | 2+2 2 |
| 12.1.22 | 41238 | | 222 | | Nb7P4 | |
| | -A | -GM | -M | -Y | -L | -V |
| | 5+6 2 | 3+4 2 | 3+4 2 | 5+6 2 | 2+2 2 | 2+2 2 |
| | 3+4 2 | 5+6 2 | 5+6 2 | 3+4 2 | 3+3 2 | 3+3 2 |
| | 3+4 2 | 3+4 2 | 5+6 2 | 5+6 2 | 2+2 2 | 2+2 2 |
| | 5+6 2 | 5+6 2 | 3+4 2 | 3+4 2 | 3+3 2 | 3+3 2 |
| | 3+4 2 | 3+4 2 | 3+4 2 | 3+4 2 | 2+2 2 | 2+2 2 |
| | 5+6 2 | 5+6 2 | 5+6 2 | 5+6 2 | 3+3 2 | 3+3 2 |
| | 5+6 2 | 3+4 2 | 5+6 2 | 3+4 2 | 2+2 2 | 2+2 2 |
| | 3+4 2 | 5+6 2 | 3+4 2 | 3+4 2 | 3+3 2 | 3+3 2 |
| | 5+6 2 | 3+4 2 | 5+6 2 | 5+6 2 | 3+3 2 | 2+2 2 |
| | 3+4 2 | 5+6 2 | 3+4 2 | 5+6 2 | 2+2 2 | 2+2 2 |
| | 3+4 2 | 3+4 2 | 3+4 2 | 3+4 2 | 3+3 2 | 3+3 2 |
| | 5+6 2 | 3+4 2 | 5+6 2 | 3+4 2 | 3+3 2 | 3+3 2 |
| | 3+4 2 | 3+4 2 | 3+4 2 | 3+4 2 | 2+2 2 | 2+2 2 |
| | 5+6 2 | 5+6 2 | 5+6 2 | 5+6 2 | 3+3 2 | 3+3 2 |
| | 5+6 2 | 3+4 2 | 3+4 2 | 5+6 2 | 2+2 2 | 3+3 2 |
| | 3+4 2 | 5+6 2 | 5+6 2 | 5+6 2 | 3+3 2 | 2+2 2 |
| | 3+4 2 | 3+4 2 | 5+6 2 | 3+4 2 | 2+2 2 | 2+2 2 |
| | 5+6 2 | 5+6 2 | 3+4 2 | 5+6 2 | 3+3 2 | 3+3 2 |
| | 5+6 2 | 5+6 2 | 5+6 2 | 3+4 2 | 2+2 2 | 2+2 2 |
| | 3+4 2 | 3+4 2 | 3+4 2 | 3+4 2 | 3+3 2 | 3+3 2 |
| | 3+4 2 | 5+6 2 | 3+4 2 | 5+6 2 | 2+2 2 | 3+3 2 |
| | 3+4 2 | 3+4 2 | 5+6 2 | 5+6 2 | 3+3 2 | 2+2 2 |
| | 5+6 2 | 5+6 2 | 3+4 2 | 3+4 2 | 2+2 2 | 3+3 2 |
| | 5+6 2 | 3+4 2 | 5+6 2 | 3+4 2 | 2+2 2 | 3+3 2 |
| | 5+6 2 | 5+6 2 | 5+6 2 | 5+6 2 | 2+2 2 | 2+2 2 |
| | 3+4 2 | 5+6 2 | 3+4 2 | 5+6 2 | 3+3 2 | 2+2 2 |
| | 5+6 2 | 3+4 2 | 3+4 2 | 5+6 2 | 3+3 2 | 3+3 2 |
| | 3+4 2 | 5+6 2 | 5+6 2 | 3+4 2 | 2+2 2 | 2+2 2 |
| | 3+4 2 | 5+6 2 | 3+4 2 | 5+6 2 | 2+2 2 | 3+3 2 |
| | 5+6 2 | 3+4 2 | 5+6 2 | 5+6 2 | 2+2 2 | 2+2 2 |
| | 5+6 2 | 3+4 2 | 5+6 2 | 3+4 2 | 3+3 2 | 2+2 2 |
| | 3+4 2 | 5+6 2 | 3+4 2 | 5+6 2 | 3+3 2 | 3+3 2 |

Continued on next column

**TABLE XXIV– continued**

| TQC | ICSD | | NoE | | SF | |
|---|---|---|---|---|---|---|
| | 5+6 2 | 3+4 2 | 3+4 2 | 5+6 2 | 2+2 2 | 2+2 2 |
| | 3+4 2 | 5+6 2 | 5+6 2 | 3+4 2 | 3+3 2 | 3+3 2 |
| | 3+4 2 | 3+4 2 | 3+4 2 | 5+6 2 | 2+2 2 | 3+3 2 |
| | 5+6 2 | 5+6 2 | 5+6 2 | 5+6 2 | 3+3 2 | 2+2 2 |
| | 3+4 2 | 5+6 2 | 3+4 2 | 3+4 2 | 2+2 2 | 2+2 2 |
| | 5+6 2 | 5+6 2 | 5+6 2 | 3+4 2 | 2+2 2 | 3+3 2 |
| | 3+4 2 | 3+4 2 | 5+6 2 | 5+6 2 | 3+3 2 | 3+3 2 |
| | 5+6 2 | 5+6 2 | 3+4 2 | 3+4 2 | 2+2 2 | 2+2 2 |
| | 3+4 2 | 3+4 2 | 5+6 2 | 5+6 2 | 3+3 2 | 3+3 2 |
| | 3+4 2 | 3+4 2 | 3+4 2 | 3+4 2 | 2+2 2 | 2+2 2 |
| | 5+6 2 | 5+6 2 | 5+6 2 | 5+6 2 | 3+3 2 | 3+3 2 |
| | 3+4 2 | 3+4 2 | 3+4 2 | 3+4 2 | 2+2 2 | 3+3 2 |
| | 5+6 2 | 5+6 2 | 3+4 2 | 3+4 2 | 3+3 2 | 2+2 2 |
| | 3+4 2 | 3+4 2 | 5+6 2 | 5+6 2 | 2+2 2 | 2+2 2 |
| | 5+6 2 | 5+6 2 | 3+4 2 | 3+4 2 | 3+3 2 | 3+3 2 |
| | 5+6 2 | 3+4 2 | 5+6 2 | 5+6 2 | 3+3 2 | 2+2 2 |
| | 3+4 2 | 5+6 2 | 3+4 2 | 3+4 2 | 2+2 2 | 2+2 2 |
| | 5+6 2 | 5+6 2 | 5+6 2 | 5+6 2 | 2+2 2 | 2+2 2 |
| | 5+6 2 | 5+6 2 | 5+6 2 | 5+6 2 | 2+2 2 | 3+3 2 |
| | 3+4 2 | 3+4 2 | 3+4 2 | 5+6 2 | 2+2 2 | 3+3 2 |
| | 5+6 2 | 5+6 2 | 5+6 2 | 3+4 2 | 3+3 2 | 2+2 2 |
| | 3+4 2 | 3+4 2 | 3+4 2 | 3+4 2 | 2+2 2 | 3+3 2 |
| | 3+4 2 | 5+6 2 | 3+4 2 | 5+6 2 | 3+3 2 | 2+2 2 |
| | 5+6 2 | 5+6 2 | 5+6 2 | 5+6 2 | 2+2 2 | 3+3 2 |
| | 5+6 2 | 3+4 2 | 5+6 2 | 3+4 2 | 3+3 2 | 2+2 2 |
| | 3+4 2 | 5+6 2 | 3+4 2 | 5+6 2 | 3+3 2 | 3+3 2 |
| | 5+6 2 | 3+4 2 | 5+6 2 | 3+4 2 | 2+2 2 | 2+2 2 |
| | 3+4 2 | 5+6 2 | 3+4 2 | 5+6 2 | 2+2 2 | 3+3 2 |
| | 3+4 2 | 5+6 2 | 3+4 2 | 3+4 2 | 3+3 2 | 2+2 2 |
| | 5+6 2 | 3+4 2 | 5+6 2 | 5+6 2 | 3+3 2 | 3+3 2 |
| | 5+6 2 | 5+6 2 | 5+6 2 | 5+6 2 | 2+2 2 | 3+3 2 |
| | 3+4 2 | 3+4 2 | 3+4 2 | 3+4 2 | 2+2 2 | 2+2 2 |
| | 3+4 2 | 3+4 2 | 3+4 2 | 3+4 2 | 2+2 2 | 3+3 2 |
| | 5+6 2 | 3+4 2 | 5+6 2 | 3+4 2 | 3+3 2 | 2+2 2 |
| | 5+6 2 | 3+4 2 | 5+6 2 | 3+4 2 | 2+2 2 | 2+2 2 |
| | 3+4 2 | 5+6 2 | 3+4 2 | 5+6 2 | 3+3 2 | 3+3 2 |
| | 3+4 2 | 5+6 2 | 5+6 2 | 5+6 2 | 3+3 2 | 2+2 2 |
| | 5+6 2 | 3+4 2 | 3+4 2 | 3+4 2 | 3+3 2 | 3+3 2 |
| | 5+6 2 | 3+4 2 | 3+4 2 | 3+4 2 | 2+2 2 | 2+2 2 |
| | 3+4 2 | 5+6 2 | 5+6 2 | 5+6 2 | 3+3 2 | 3+3 2 |
| | 5+6 2 | 5+6 2 | 5+6 2 | 5+6 2 | 3+3 2 | 2+2 2 |
| | 3+4 2 | 3+4 2 | 3+4 2 | 5+6 2 | 3+3 2 | 3+3 2 |
| | 5+6 2 | 5+6 2 | 3+4 2 | 3+4 2 | 3+3 2 | 3+3 2 |
| | 3+4 2 | 5+6 2 | 3+4 2 | 5+6 2 | 2+2 2 | 2+2 2 |
| | 3+4 2 | 5+6 2 | 5+6 2 | 5+6 2 | 2+2 2 | 2+2 2 |
| | 5+6 2 | 3+4 2 | 5+6 2 | 3+4 2 | 2+2 2 | 3+3 2 |
| | 5+6 2 | 5+6 2 | 5+6 2 | 3+4 2 | 3+3 2 | 2+2 2 |
| | 3+4 2 | 3+4 2 | 3+4 2 | 5+6 2 | 2+2 2 | 3+3 2 |
| | 5+6 2 | 3+4 2 | 3+4 2 | 3+4 2 | 3+3 2 | 2+2 2 |
| | 5+6 2 | 5+6 2 | 5+6 2 | 3+4 2 | 2+2 2 | 3+3 2 |
| | 3+4 2 | 5+6 2 | 3+4 2 | 5+6 2 | 3+3 2 | 2+2 2 |
| | 3+4 2 | 3+4 2 | 5+6 2 | 3+4 2 | 2+2 2 | 3+3 2 |
| | 5+6 2 | 3+4 2 | 5+6 2 | 3+4 2 | 2+2 2 | 3+3 2 |
| | 3+4 2 | 3+4 2 | 5+6 2 | 5+6 2 | 3+3 2 | 2+2 2 |
| | 3+4 2 | 3+4 2 | 3+4 2 | 3+4 2 | 2+2 2 | 3+3 2 |
| | 5+6 2 | 3+4 2 | 3+4 2 | 3+4 2 | 2+2 2 | 2+2 2 |
| | 3+4 2 | 5+6 2 | 5+6 2 | 3+4 2 | 3+3 2 | 3+3 2 |
| | 3+4 2 | 5+6 2 | 3+4 2 | 5+6 2 | 3+3 2 | 2+2 2 |
| | 5+6 2 | 3+4 2 | 5+6 2 | 3+4 2 | 3+3 2 | 2+2 2 |
| | 5+6 2 | 5+6 2 | 3+4 2 | 5+6 2 | 2+2 2 | 3+3 2 |
| | 5+6 2 | 3+4 2 | 3+4 2 | 3+4 2 | 2+2 2 | 2+2 2 |

Continued on next column

**TABLE XXIV– continued**

| TQC | ICSD | NoE | SF | | | |
|---|---|---|---|---|---|---|
| | 5+6 2 | 5+6 2 | 3+4 2 | 5+6 2 | 3+3 2 | 3+3 2 |
| | 3+4 2 | 5+6 2 | 5+6 2 | 5+6 2 | 3+3 2 | 3+3 2 |
| | 3+4 2 | 5+6 2 | 3+4 2 | 3+4 2 | 3+3 2 | 2+2 2 |
| | 5+6 2 | 3+4 2 | 5+6 2 | 5+6 2 | 2+2 2 | 2+2 2 |
| | 3+4 2 | 3+4 2 | 5+6 2 | 5+6 2 | 3+3 2 | 3+3 2 |
| | 3+4 2 | 5+6 2 | 3+4 2 | 5+6 2 | 2+2 2 | 3+3 2 |
| | 5+6 2 | 3+4 2 | 5+6 2 | 3+4 2 | 2+2 2 | 2+2 2 |
| | 5+6 2 | 3+4 2 | 3+4 2 | 5+6 2 | 3+3 2 | 2+2 2 |
| | 3+4 2 | 3+4 2 | 5+6 2 | 3+4 2 | 2+2 2 | 3+3 2 |
| | 5+6 2 | 5+6 2 | 5+6 2 | 3+4 2 | 3+3 2 | 2+2 2 |
| | 3+4 2 | 3+4 2 | 3+4 2 | 5+6 2 | 2+2 2 | 2+2 2 |
| | 5+6 2 | 5+6 2 | 3+4 2 | 3+4 2 | 3+3 2 | 3+3 2 |
| | 3+4 2 | 3+4 2 | 5+6 2 | 5+6 2 | 2+2 2 | 3+3 2 |
| | 5+6 2 | 3+4 2 | 5+6 2 | 3+4 2 | 3+3 2 | 2+2 2 |
| | 3+4 2 | 5+6 2 | 3+4 2 | 5+6 2 | 2+2 2 | 3+3 2 |
| | 3+4 2 | 5+6 2 | 3+4 2 | 3+4 2 | 2+2 2 | 3+3 2 |
| | 3+4 2 | 5+6 2 | 5+6 2 | 3+4 2 | 3+3 2 | 3+3 2 |
| | 5+6 2 | 3+4 2 | 3+4 2 | 5+6 2 | 2+2 2 | 2+2 2 |
| 12.1.23 | 26183 | 128 | Nb2Se1 | | | |
| | -A | -GM | -M | -Y | -L | -V |
| | 3+4 2 | 3+4 2 | 5+6 2 | 5+6 2 | 2+2 2 | 2+2 2 |
| | 5+6 2 | 5+6 2 | 3+4 2 | 3+4 2 | 3+3 2 | 3+3 2 |
| | 5+6 2 | 3+4 2 | 5+6 2 | 3+4 2 | 3+3 2 | 2+2 2 |
| | 3+4 2 | 5+6 2 | 3+4 2 | 5+6 2 | 2+2 2 | 3+3 2 |
| | 5+6 2 | 3+4 2 | 3+4 2 | 5+6 2 | 2+2 2 | 3+3 2 |
| | 3+4 2 | 5+6 2 | 5+6 2 | 3+4 2 | 3+3 2 | 2+2 2 |
| | 5+6 2 | 3+4 2 | 3+4 2 | 5+6 2 | 2+2 2 | 2+2 2 |
| | 3+4 2 | 5+6 2 | 5+6 2 | 3+4 2 | 3+3 2 | 3+3 2 |
| | 3+4 2 | 3+4 2 | 5+6 2 | 5+6 2 | 2+2 2 | 2+2 2 |
| | 5+6 2 | 5+6 2 | 3+4 2 | 3+4 2 | 3+3 2 | 3+3 2 |
| | 3+4 2 | 5+6 2 | 3+4 2 | 5+6 2 | 3+3 2 | 2+2 2 |
| | 5+6 2 | 3+4 2 | 5+6 2 | 3+4 2 | 2+2 2 | 3+3 2 |
| | 5+6 2 | 3+4 2 | 5+6 2 | 3+4 2 | 2+2 2 | 3+3 2 |
| | 3+4 2 | 5+6 2 | 3+4 2 | 5+6 2 | 3+3 2 | 2+2 2 |
| | 5+6 2 | 3+4 2 | 3+4 2 | 5+6 2 | 2+2 2 | 3+3 2 |
| | 3+4 2 | 5+6 2 | 5+6 2 | 3+4 2 | 3+3 2 | 2+2 2 |
| | 3+4 2 | 3+4 2 | 5+6 2 | 5+6 2 | 2+2 2 | 2+2 2 |
| | 5+6 2 | 3+4 2 | 3+4 2 | 5+6 2 | 3+3 2 | 3+3 2 |
| | 3+4 2 | 5+6 2 | 5+6 2 | 3+4 2 | 2+2 2 | 2+2 2 |
| | 5+6 2 | 5+6 2 | 3+4 2 | 3+4 2 | 3+3 2 | 3+3 2 |
| | 3+4 2 | 5+6 2 | 3+4 2 | 5+6 2 | 2+2 2 | 2+2 2 |
| | 5+6 2 | 3+4 2 | 5+6 2 | 3+4 2 | 3+3 2 | 3+3 2 |
| | 5+6 2 | 3+4 2 | 3+4 2 | 5+6 2 | 3+3 2 | 2+2 2 |
| | 3+4 2 | 5+6 2 | 5+6 2 | 3+4 2 | 2+2 2 | 3+3 2 |
| | 5+6 2 | 3+4 2 | 3+4 2 | 5+6 2 | 2+2 2 | 3+3 2 |
| | 3+4 2 | 5+6 2 | 5+6 2 | 3+4 2 | 3+3 2 | 2+2 2 |
| | 5+6 2 | 3+4 2 | 5+6 2 | 3+4 2 | 2+2 2 | 3+3 2 |
| | 5+6 2 | 3+4 2 | 3+4 2 | 5+6 2 | 3+3 2 | 2+2 2 |
| | 3+4 2 | 5+6 2 | 3+4 2 | 5+6 2 | 2+2 2 | 3+3 2 |
| | 3+4 2 | 3+4 2 | 5+6 2 | 3+4 2 | 3+3 2 | 2+2 2 |
| | 5+6 2 | 5+6 2 | 3+4 2 | 5+6 2 | 2+2 2 | 3+3 2 |
| | 3+4 2 | 5+6 2 | 5+6 2 | 3+4 2 | 3+3 2 | 2+2 2 |
| | 3+4 2 | 3+4 2 | 5+6 2 | 5+6 2 | 3+3 2 | 3+3 2 |
| | 5+6 2 | 5+6 2 | 3+4 2 | 3+4 2 | 2+2 2 | 2+2 2 |
| | 5+6 2 | 5+6 2 | 5+6 2 | 5+6 2 | 2+2 2 | 2+2 2 |
| | 3+4 2 | 3+4 2 | 3+4 2 | 3+4 2 | 3+3 2 | 3+3 2 |
| | 3+4 2 | 3+4 2 | 3+4 2 | 3+4 2 | 3+3 2 | 3+3 2 |
| | 5+6 2 | 5+6 2 | 5+6 2 | 5+6 2 | 2+2 2 | 2+2 2 |
| | 5+6 2 | 3+4 2 | 3+4 2 | 3+4 2 | 3+3 2 | 3+3 2 |
| | 3+4 2 | 5+6 2 | 5+6 2 | 5+6 2 | 2+2 2 | 2+2 2 |
| | 5+6 2 | 5+6 2 | 3+4 2 | 3+4 2 | 3+3 2 | 3+3 2 |

Continued on next column

**TABLE XXIV– continued**

| TQC | ICSD | NoE | SF | | | |
|---|---|---|---|---|---|---|
| | 3+4 2 | 3+4 2 | 5+6 2 | 5+6 2 | 2+2 2 | 2+2 2 |
| | 5+6 2 | 5+6 2 | 5+6 2 | 5+6 2 | 2+2 2 | 2+2 2 |
| | 3+4 2 | 5+6 2 | 3+4 2 | 3+4 2 | 3+3 2 | 3+3 2 |
| | 5+6 2 | 3+4 2 | 3+4 2 | 3+4 2 | 2+2 2 | 3+3 2 |
| | 3+4 2 | 3+4 2 | 5+6 2 | 3+4 2 | 3+3 2 | 2+2 2 |
| | 3+4 2 | 3+4 2 | 5+6 2 | 5+6 2 | 2+2 2 | 2+2 2 |
| | 5+6 2 | 5+6 2 | 3+4 2 | 5+6 2 | 3+3 2 | 3+3 2 |
| | 5+6 2 | 3+4 2 | 3+4 2 | 5+6 2 | 3+3 2 | 3+3 2 |
| | 5+6 2 | 3+4 2 | 5+6 2 | 3+4 2 | 2+2 2 | 2+2 2 |
| | 3+4 2 | 5+6 2 | 5+6 2 | 3+4 2 | 3+3 2 | 2+2 2 |
| | 3+4 2 | 5+6 2 | 3+4 2 | 3+4 2 | 3+3 2 | 2+2 2 |
| | 3+4 2 | 5+6 2 | 5+6 2 | 5+6 2 | 2+2 2 | 3+3 2 |
| | 5+6 2 | 3+4 2 | 3+4 2 | 5+6 2 | 3+3 2 | 3+3 2 |
| | 5+6 2 | 3+4 2 | 3+4 2 | 3+4 2 | 2+2 2 | 3+3 2 |
| | 3+4 2 | 5+6 2 | 5+6 2 | 3+4 2 | 2+2 2 | 2+2 2 |
| | 3+4 2 | 5+6 2 | 5+6 2 | 5+6 2 | 2+2 2 | 2+2 2 |
| | 5+6 2 | 3+4 2 | 5+6 2 | 5+6 2 | 3+3 2 | 2+2 2 |
| | 3+4 2 | 3+4 2 | 5+6 2 | 3+4 2 | 2+2 2 | 3+3 2 |
| | 5+6 2 | 5+6 2 | 3+4 2 | 5+6 2 | 3+3 2 | 2+2 2 |
| | 3+4 2 | 3+4 2 | 3+4 2 | 3+4 2 | 3+3 2 | 3+3 2 |
| | 5+6 2 | 5+6 2 | 5+6 2 | 5+6 2 | 3+3 2 | 2+2 2 |
| | 3+4 2 | 3+4 2 | 3+4 2 | 5+6 2 | 2+2 2 | 3+3 2 |
| | 5+6 2 | 3+4 2 | 5+6 2 | 3+4 2 | 2+2 2 | 3+3 2 |
| 12.1.24 | 77752 | 46 | Nb1Sb2 | | | |
| | -A | -GM | -M | -Y | -L | -V |
| | 3+4 2 | 3+4 2 | 3+4 2 | 3+4 2 | 3+3 2 | 3+3 2 |
| | 5+6 2 | 5+6 2 | 5+6 2 | 5+6 2 | 2+2 2 | 2+2 2 |
| | 3+4 2 | 3+4 2 | 3+4 2 | 3+4 2 | 3+3 2 | 3+3 2 |
| | 5+6 2 | 5+6 2 | 5+6 2 | 5+6 2 | 2+2 2 | 2+2 2 |
| | 3+4 2 | 3+4 2 | 3+4 2 | 3+4 2 | 2+2 2 | 2+2 2 |
| | 5+6 2 | 5+6 2 | 5+6 2 | 5+6 2 | 3+3 2 | 3+3 2 |
| | 3+4 2 | 3+4 2 | 3+4 2 | 3+4 2 | 3+3 2 | 3+3 2 |
| | 5+6 2 | 5+6 2 | 5+6 2 | 5+6 2 | 2+2 2 | 2+2 2 |
| | 5+6 2 | 3+4 2 | 3+4 2 | 5+6 2 | 2+2 2 | 3+3 2 |
| | 3+4 2 | 3+4 2 | 5+6 2 | 3+4 2 | 2+2 2 | 2+2 2 |
| | 5+6 2 | 5+6 2 | 5+6 2 | 5+6 2 | 3+3 2 | 3+3 2 |
| | 3+4 2 | 5+6 2 | 3+4 2 | 3+4 2 | 3+3 2 | 2+2 2 |
| | 5+6 2 | 3+4 2 | 5+6 2 | 5+6 2 | 2+2 2 | 2+2 2 |
| | 3+4 2 | 5+6 2 | 3+4 2 | 3+4 2 | 3+3 2 | 3+3 2 |
| | 3+4 2 | 3+4 2 | 5+6 2 | 5+6 2 | 3+3 2 | 2+2 2 |
| | 5+6 2 | 3+4 2 | 3+4 2 | 3+4 2 | 2+2 2 | 3+3 2 |
| | 5+6 2 | 5+6 2 | 3+4 2 | 5+6 2 | 2+2 2 | 2+2 2 |
| | 3+4 2 | 5+6 2 | 5+6 2 | 5+6 2 | 3+3 2 | 2+2 2 |
| | 5+6 2 | 3+4 2 | 3+4 2 | 3+4 2 | 2+2 2 | 3+3 2 |
| | 3+4 2 | 3+4 2 | 5+6 2 | 3+4 2 | 2+2 2 | 3+3 2 |
| | 3+4 2 | 3+4 2 | 3+4 2 | 5+6 2 | 3+3 2 | 2+2 2 |
| | 5+6 2 | 3+4 2 | 5+6 2 | 3+4 2 | 3+3 2 | 3+3 2 |
| | 5+6 2 | 5+6 2 | 3+4 2 | 3+4 2 | 2+2 2 | 3+3 2 |
| 12.1.25 | 645350 | 46 | Nb1Sb2 | | | |
| | -A | -GM | -M | -Y | -L | -V |
| | 3+4 2 | 3+4 2 | 3+4 2 | 3+4 2 | 3+3 2 | 3+3 2 |
| | 5+6 2 | 5+6 2 | 5+6 2 | 5+6 2 | 2+2 2 | 2+2 2 |
| | 3+4 2 | 3+4 2 | 3+4 2 | 3+4 2 | 3+3 2 | 3+3 2 |
| | 5+6 2 | 5+6 2 | 5+6 2 | 5+6 2 | 2+2 2 | 2+2 2 |
| | 3+4 2 | 3+4 2 | 3+4 2 | 3+4 2 | 2+2 2 | 2+2 2 |
| | 5+6 2 | 5+6 2 | 5+6 2 | 5+6 2 | 3+3 2 | 3+3 2 |
| | 3+4 2 | 3+4 2 | 3+4 2 | 3+4 2 | 3+3 2 | 3+3 2 |
| | 5+6 2 | 5+6 2 | 5+6 2 | 5+6 2 | 2+2 2 | 2+2 2 |
| | 5+6 2 | 3+4 2 | 3+4 2 | 5+6 2 | 2+2 2 | 3+3 2 |
| | 3+4 2 | 3+4 2 | 5+6 2 | 3+4 2 | 2+2 2 | 2+2 2 |
| | 5+6 2 | 5+6 2 | 5+6 2 | 5+6 2 | 3+3 2 | 3+3 2 |

Continued on next column

**TABLE XXIV– continued**

| TQC | ICSD | | NoE | SF | | |
|---|---|---|---|---|---|---|
| | 3+4 2 | 5+6 2 | 3+4 2 | 3+4 2 | 3+3 2 | 2+2 2 |
| | 5+6 2 | 3+4 2 | 5+6 2 | 5+6 2 | 2+2 2 | 2+2 2 |
| | 3+4 2 | 5+6 2 | 3+4 2 | 3+4 2 | 3+3 2 | 3+3 2 |
| | 3+4 2 | 3+4 2 | 5+6 2 | 5+6 2 | 3+3 2 | 2+2 2 |
| | 5+6 2 | 3+4 2 | 3+4 2 | 3+4 2 | 2+2 2 | 3+3 2 |
| | 5+6 2 | 5+6 2 | 3+4 2 | 5+6 2 | 2+2 2 | 2+2 2 |
| | 3+4 2 | 5+6 2 | 5+6 2 | 5+6 2 | 3+3 2 | 2+2 2 |
| | 5+6 2 | 3+4 2 | 3+4 2 | 3+4 2 | 2+2 2 | 3+3 2 |
| | 3+4 2 | 3+4 2 | 5+6 2 | 3+4 2 | 2+2 2 | 3+3 2 |
| | 3+4 2 | 3+4 2 | 5+6 2 | 5+6 2 | 3+3 2 | 2+2 2 |
| | 5+6 2 | 3+4 2 | 3+4 2 | 3+4 2 | 3+3 2 | 3+3 2 |
| | 5+6 2 | 5+6 2 | 3+4 2 | 3+4 2 | 2+2 2 | 3+3 2 |
| 12.1.26 | 57430 | | 46 | Nb1Sb2 | | |
| | -A | -GM | -M | -Y | -L | -V |
| | 3+4 2 | 3+4 2 | 3+4 2 | 3+4 2 | 3+3 2 | 3+3 2 |
| | 5+6 2 | 5+6 2 | 5+6 2 | 5+6 2 | 2+2 2 | 2+2 2 |
| | 3+4 2 | 3+4 2 | 3+4 2 | 3+4 2 | 3+3 2 | 3+3 2 |
| | 5+6 2 | 5+6 2 | 5+6 2 | 5+6 2 | 2+2 2 | 2+2 2 |
| | 3+4 2 | 3+4 2 | 3+4 2 | 3+4 2 | 2+2 2 | 2+2 2 |
| | 5+6 2 | 5+6 2 | 5+6 2 | 5+6 2 | 3+3 2 | 3+3 2 |
| | 3+4 2 | 3+4 2 | 3+4 2 | 3+4 2 | 3+3 2 | 3+3 2 |
| | 5+6 2 | 5+6 2 | 5+6 2 | 5+6 2 | 2+2 2 | 2+2 2 |
| | 5+6 2 | 3+4 2 | 3+4 2 | 5+6 2 | 2+2 2 | 3+3 2 |
| | 3+4 2 | 3+4 2 | 5+6 2 | 3+4 2 | 2+2 2 | 2+2 2 |
| | 5+6 2 | 5+6 2 | 5+6 2 | 5+6 2 | 3+3 2 | 3+3 2 |
| | 3+4 2 | 5+6 2 | 3+4 2 | 3+4 2 | 3+3 2 | 2+2 2 |
| | 5+6 2 | 3+4 2 | 5+6 2 | 5+6 2 | 2+2 2 | 2+2 2 |
| | 3+4 2 | 5+6 2 | 3+4 2 | 3+4 2 | 3+3 2 | 3+3 2 |
| | 3+4 2 | 3+4 2 | 5+6 2 | 5+6 2 | 3+3 2 | 2+2 2 |
| | 5+6 2 | 3+4 2 | 3+4 2 | 3+4 2 | 2+2 2 | 3+3 2 |
| | 5+6 2 | 5+6 2 | 3+4 2 | 5+6 2 | 2+2 2 | 2+2 2 |
| | 3+4 2 | 5+6 2 | 5+6 2 | 5+6 2 | 3+3 2 | 2+2 2 |
| | 5+6 2 | 3+4 2 | 3+4 2 | 3+4 2 | 2+2 2 | 3+3 2 |
| | 3+4 2 | 3+4 2 | 5+6 2 | 3+4 2 | 2+2 2 | 3+3 2 |
| | 3+4 2 | 3+4 2 | 3+4 2 | 5+6 2 | 3+3 2 | 2+2 2 |
| | 5+6 2 | 3+4 2 | 5+6 2 | 3+4 2 | 3+3 2 | 3+3 2 |
| | 5+6 2 | 5+6 2 | 3+4 2 | 3+4 2 | 2+2 2 | 3+3 2 |
| 12.1.27 | 190180 | | 46 | Nb1Sb2 | | |
| | -A | -GM | -M | -Y | -L | -V |
| | 5+6 2 | 3+4 2 | 5+6 2 | 3+4 2 | 2+2 2 | 3+3 2 |
| | 3+4 2 | 5+6 2 | 3+4 2 | 5+6 2 | 3+3 2 | 2+2 2 |
| | 5+6 2 | 3+4 2 | 5+6 2 | 3+4 2 | 2+2 2 | 3+3 2 |
| | 3+4 2 | 5+6 2 | 3+4 2 | 5+6 2 | 3+3 2 | 2+2 2 |
| | 5+6 2 | 3+4 2 | 5+6 2 | 3+4 2 | 2+2 2 | 2+2 2 |
| | 3+4 2 | 5+6 2 | 3+4 2 | 5+6 2 | 3+3 2 | 3+3 2 |
| | 5+6 2 | 3+4 2 | 5+6 2 | 3+4 2 | 2+2 2 | 3+3 2 |
| | 3+4 2 | 5+6 2 | 3+4 2 | 5+6 2 | 3+3 2 | 2+2 2 |
| | 3+4 2 | 3+4 2 | 5+6 2 | 5+6 2 | 3+3 2 | 3+3 2 |
| | 5+6 2 | 3+4 2 | 3+4 2 | 3+4 2 | 3+3 2 | 2+2 2 |
| | 3+4 2 | 5+6 2 | 3+4 2 | 5+6 2 | 2+2 2 | 3+3 2 |
| | 5+6 2 | 5+6 2 | 5+6 2 | 3+4 2 | 2+2 2 | 2+2 2 |
| | 3+4 2 | 3+4 2 | 3+4 2 | 5+6 2 | 3+3 2 | 2+2 2 |
| | 5+6 2 | 5+6 2 | 5+6 2 | 3+4 2 | 2+2 2 | 3+3 2 |
| | 5+6 2 | 3+4 2 | 3+4 2 | 5+6 2 | 2+2 2 | 2+2 2 |
| | 3+4 2 | 3+4 2 | 5+6 2 | 3+4 2 | 3+3 2 | 3+3 2 |
| | 3+4 2 | 5+6 2 | 5+6 2 | 5+6 2 | 3+3 2 | 2+2 2 |
| | 5+6 2 | 5+6 2 | 3+4 2 | 5+6 2 | 2+2 2 | 2+2 2 |
| | 3+4 2 | 3+4 2 | 5+6 2 | 3+4 2 | 3+3 2 | 3+3 2 |
| | 5+6 2 | 3+4 2 | 3+4 2 | 3+4 2 | 3+3 2 | 3+3 2 |
| | 5+6 2 | 3+4 2 | 5+6 2 | 5+6 2 | 2+2 2 | 2+2 2 |
| | 3+4 2 | 3+4 2 | 3+4 2 | 3+4 2 | 2+2 2 | 3+3 2 |

Continued on next column

**TABLE XXIV– continued**

| TQC | ICSD | | NoE | SF | | |
|---|---|---|---|---|---|---|
| | 3+4 2 | 5+6 2 | 5+6 2 | 3+4 2 | 3+3 2 | 3+3 2 |
| 12.1.28 | 18144 | | 46 | Nb1Sb2 | | |
| | -A | -GM | -M | -Y | -L | -V |
| | 5+6 2 | 3+4 2 | 5+6 2 | 3+4 2 | 2+2 2 | 3+3 2 |
| | 3+4 2 | 5+6 2 | 3+4 2 | 5+6 2 | 3+3 2 | 2+2 2 |
| | 5+6 2 | 3+4 2 | 5+6 2 | 3+4 2 | 2+2 2 | 3+3 2 |
| | 3+4 2 | 5+6 2 | 3+4 2 | 5+6 2 | 3+3 2 | 2+2 2 |
| | 5+6 2 | 3+4 2 | 5+6 2 | 3+4 2 | 3+3 2 | 2+2 2 |
| | 3+4 2 | 5+6 2 | 3+4 2 | 5+6 2 | 2+2 2 | 3+3 2 |
| | 5+6 2 | 3+4 2 | 5+6 2 | 3+4 2 | 2+2 2 | 3+3 2 |
| | 3+4 2 | 5+6 2 | 3+4 2 | 5+6 2 | 3+3 2 | 2+2 2 |
| | 3+4 2 | 3+4 2 | 5+6 2 | 5+6 2 | 3+3 2 | 3+3 2 |
| | 5+6 2 | 3+4 2 | 3+4 2 | 3+4 2 | 3+3 2 | 2+2 2 |
| | 3+4 2 | 5+6 2 | 3+4 2 | 5+6 2 | 2+2 2 | 3+3 2 |
| | 5+6 2 | 5+6 2 | 5+6 2 | 3+4 2 | 2+2 2 | 2+2 2 |
| | 3+4 2 | 3+4 2 | 3+4 2 | 5+6 2 | 3+3 2 | 3+3 2 |
| | 5+6 2 | 5+6 2 | 5+6 2 | 3+4 2 | 2+2 2 | 2+2 2 |
| | 5+6 2 | 3+4 2 | 3+4 2 | 5+6 2 | 2+2 2 | 2+2 2 |
| | 3+4 2 | 3+4 2 | 5+6 2 | 3+4 2 | 3+3 2 | 3+3 2 |
| | 3+4 2 | 5+6 2 | 5+6 2 | 5+6 2 | 3+3 2 | 2+2 2 |
| | 5+6 2 | 3+4 2 | 3+4 2 | 5+6 2 | 2+2 2 | 2+2 2 |
| | 3+4 2 | 5+6 2 | 5+6 2 | 3+4 2 | 3+3 2 | 3+3 2 |
| | 5+6 2 | 3+4 2 | 3+4 2 | 3+4 2 | 3+3 2 | 3+3 2 |
| | 5+6 2 | 3+4 2 | 5+6 2 | 5+6 2 | 2+2 2 | 2+2 2 |
| | 3+4 2 | 3+4 2 | 3+4 2 | 3+4 2 | 2+2 2 | 3+3 2 |
| | 3+4 2 | 5+6 2 | 5+6 2 | 3+4 2 | 3+3 2 | 3+3 2 |
| 12.1.29 | 645169 | | 46 | Nb1P2 | | |
| | -A | -GM | -M | -Y | -L | -V |
| | 3+4 2 | 3+4 2 | 3+4 2 | 3+4 2 | 3+3 2 | 3+3 2 |
| | 5+6 2 | 5+6 2 | 5+6 2 | 5+6 2 | 2+2 2 | 2+2 2 |
| | 3+4 2 | 3+4 2 | 3+4 2 | 3+4 2 | 3+3 2 | 3+3 2 |
| | 5+6 2 | 5+6 2 | 5+6 2 | 5+6 2 | 2+2 2 | 2+2 2 |
| | 3+4 2 | 3+4 2 | 3+4 2 | 3+4 2 | 3+3 2 | 3+3 2 |
| | 5+6 2 | 5+6 2 | 5+6 2 | 5+6 2 | 2+2 2 | 2+2 2 |
| | 3+4 2 | 3+4 2 | 3+4 2 | 3+4 2 | 3+3 2 | 3+3 2 |
| | 5+6 2 | 5+6 2 | 5+6 2 | 5+6 2 | 2+2 2 | 2+2 2 |
| | 5+6 2 | 3+4 2 | 3+4 2 | 3+4 2 | 2+2 2 | 3+3 2 |
| | 3+4 2 | 3+4 2 | 5+6 2 | 5+6 2 | 2+2 2 | 2+2 2 |
| | 5+6 2 | 5+6 2 | 5+6 2 | 5+6 2 | 3+3 2 | 3+3 2 |
| | 3+4 2 | 5+6 2 | 3+4 2 | 3+4 2 | 3+3 2 | 2+2 2 |
| | 5+6 2 | 3+4 2 | 5+6 2 | 5+6 2 | 2+2 2 | 2+2 2 |
| | 3+4 2 | 5+6 2 | 3+4 2 | 5+6 2 | 3+3 2 | 3+3 2 |
| | 3+4 2 | 3+4 2 | 5+6 2 | 3+4 2 | 2+2 2 | 2+2 2 |
| | 5+6 2 | 3+4 2 | 5+6 2 | 3+4 2 | 3+3 2 | 3+3 2 |
| | 5+6 2 | 5+6 2 | 3+4 2 | 5+6 2 | 2+2 2 | 2+2 2 |
| | 5+6 2 | 5+6 2 | 3+4 2 | 5+6 2 | 3+3 2 | 2+2 2 |
| | 3+4 2 | 3+4 2 | 3+4 2 | 3+4 2 | 2+2 2 | 3+3 2 |
| | 3+4 2 | 3+4 2 | 5+6 2 | 3+4 2 | 2+2 2 | 3+3 2 |
| | 3+4 2 | 3+4 2 | 5+6 2 | 5+6 2 | 3+3 2 | 2+2 2 |
| | 5+6 2 | 3+4 2 | 3+4 2 | 3+4 2 | 3+3 2 | 3+3 2 |
| | 5+6 2 | 5+6 2 | 3+4 2 | 3+4 2 | 2+2 2 | 3+3 2 |
| 12.1.30 | 29383 | | 16 | N3Na1 | | |
| | -A | -GM | -M | -Y | -L | -V |
| | 5+6 2 | 3+4 2 | 5+6 2 | 3+4 2 | 2+2 2 | 3+3 2 |
| | 3+4 2 | 5+6 2 | 3+4 2 | 5+6 2 | 3+3 2 | 2+2 2 |
| | 5+6 2 | 3+4 2 | 5+6 2 | 3+4 2 | 3+3 2 | 2+2 2 |
| | 3+4 2 | 5+6 2 | 3+4 2 | 5+6 2 | 2+2 2 | 3+3 2 |
| | 5+6 2 | 3+4 2 | 3+4 2 | 5+6 2 | 2+2 2 | 3+3 2 |
| | 3+4 2 | 5+6 2 | 5+6 2 | 3+4 2 | 3+3 2 | 2+2 2 |
| | 3+4 2 | 5+6 2 | 3+4 2 | 3+4 2 | 2+2 2 | 3+3 2 |
| | 5+6 2 | 3+4 2 | 5+6 2 | 5+6 2 | 3+3 2 | 3+3 2 |

Continued on next column

**TABLE XXIV– continued**

| TQC | ICSD | NoE | SF | | | |
|---|---|---|---|---|---|---|
| 12.1.31 | 29382 | 16 | N3Na1 | | | |
| | -A | -GM | -M | -Y | -L | -V |
| | 5+6 2 | 3+4 2 | 5+6 2 | 3+4 2 | 2+2 2 | 3+3 2 |
| | 3+4 2 | 5+6 2 | 3+4 2 | 5+6 2 | 3+3 2 | 2+2 2 |
| | 5+6 2 | 3+4 2 | 5+6 2 | 3+4 2 | 3+3 2 | 2+2 2 |
| | 3+4 2 | 5+6 2 | 3+4 2 | 5+6 2 | 2+2 2 | 3+3 2 |
| | 3+4 2 | 3+4 2 | 3+4 2 | 5+6 2 | 2+2 2 | 3+3 2 |
| | 3+4 2 | 5+6 2 | 5+6 2 | 3+4 2 | 3+3 2 | 2+2 2 |
| | 5+6 2 | 5+6 2 | 3+4 2 | 3+4 2 | 2+2 2 | 3+3 2 |
| | 5+6 2 | 3+4 2 | 5+6 2 | 5+6 2 | 3+3 2 | 3+3 2 |
| 12.1.32 | 29381 | 16 | N3Na1 | | | |
| | -A | -GM | -M | -Y | -L | -V |
| | 5+6 2 | 3+4 2 | 5+6 2 | 3+4 2 | 2+2 2 | 3+3 2 |
| | 3+4 2 | 5+6 2 | 3+4 2 | 5+6 2 | 3+3 2 | 2+2 2 |
| | 5+6 2 | 3+4 2 | 5+6 2 | 3+4 2 | 3+3 2 | 2+2 2 |
| | 3+4 2 | 5+6 2 | 3+4 2 | 5+6 2 | 2+2 2 | 3+3 2 |
| | 5+6 2 | 3+4 2 | 3+4 2 | 5+6 2 | 2+2 2 | 3+3 2 |
| | 3+4 2 | 5+6 2 | 5+6 2 | 3+4 2 | 3+3 2 | 2+2 2 |
| | 3+4 2 | 5+6 2 | 3+4 2 | 3+4 2 | 2+2 2 | 3+3 2 |
| | 5+6 2 | 3+4 2 | 5+6 2 | 5+6 2 | 3+3 2 | 3+3 2 |
| 12.1.33 | 29380 | 16 | N3Na1 | | | |
| | -A | -GM | -M | -Y | -L | -V |
| | 5+6 2 | 3+4 2 | 5+6 2 | 3+4 2 | 2+2 2 | 3+3 2 |
| | 3+4 2 | 5+6 2 | 3+4 2 | 5+6 2 | 3+3 2 | 2+2 2 |
| | 5+6 2 | 3+4 2 | 5+6 2 | 3+4 2 | 3+3 2 | 2+2 2 |
| | 3+4 2 | 5+6 2 | 3+4 2 | 5+6 2 | 2+2 2 | 3+3 2 |
| | 5+6 2 | 3+4 2 | 3+4 2 | 5+6 2 | 2+2 2 | 3+3 2 |
| | 3+4 2 | 5+6 2 | 5+6 2 | 3+4 2 | 3+3 2 | 2+2 2 |
| | 3+4 2 | 5+6 2 | 3+4 2 | 3+4 2 | 2+2 2 | 3+3 2 |
| | 5+6 2 | 3+4 2 | 5+6 2 | 5+6 2 | 3+3 2 | 3+3 2 |
| 12.1.34 | 29379 | 16 | N3Na1 | | | |
| | -A | -GM | -M | -Y | -L | -V |
| | 5+6 2 | 3+4 2 | 5+6 2 | 3+4 2 | 2+2 2 | 3+3 2 |
| | 3+4 2 | 5+6 2 | 3+4 2 | 5+6 2 | 3+3 2 | 2+2 2 |
| | 5+6 2 | 3+4 2 | 5+6 2 | 3+4 2 | 3+3 2 | 2+2 2 |
| | 3+4 2 | 5+6 2 | 3+4 2 | 5+6 2 | 2+2 2 | 3+3 2 |
| | 5+6 2 | 3+4 2 | 3+4 2 | 5+6 2 | 2+2 2 | 3+3 2 |
| | 3+4 2 | 5+6 2 | 5+6 2 | 3+4 2 | 3+3 2 | 2+2 2 |
| | 3+4 2 | 5+6 2 | 3+4 2 | 3+4 2 | 2+2 2 | 3+3 2 |
| | 5+6 2 | 3+4 2 | 5+6 2 | 5+6 2 | 3+3 2 | 3+3 2 |
| 12.1.35 | 29378 | 16 | N3Na1 | | | |
| | -A | -GM | -M | -Y | -L | -V |
| | 5+6 2 | 3+4 2 | 5+6 2 | 3+4 2 | 2+2 2 | 3+3 2 |
| | 3+4 2 | 5+6 2 | 3+4 2 | 5+6 2 | 3+3 2 | 2+2 2 |
| | 5+6 2 | 3+4 2 | 5+6 2 | 3+4 2 | 3+3 2 | 2+2 2 |
| | 3+4 2 | 5+6 2 | 3+4 2 | 5+6 2 | 2+2 2 | 3+3 2 |
| | 5+6 2 | 3+4 2 | 3+4 2 | 5+6 2 | 2+2 2 | 3+3 2 |
| | 3+4 2 | 5+6 2 | 5+6 2 | 3+4 2 | 3+3 2 | 2+2 2 |
| | 3+4 2 | 5+6 2 | 3+4 2 | 3+4 2 | 2+2 2 | 3+3 2 |
| | 5+6 2 | 3+4 2 | 5+6 2 | 5+6 2 | 3+3 2 | 3+3 2 |
| 12.1.36 | 29377 | 16 | N3Na1 | | | |
| | -A | -GM | -M | -Y | -L | -V |
| | 5+6 2 | 3+4 2 | 5+6 2 | 3+4 2 | 2+2 2 | 3+3 2 |
| | 3+4 2 | 5+6 2 | 3+4 2 | 5+6 2 | 3+3 2 | 2+2 2 |
| | 5+6 2 | 3+4 2 | 5+6 2 | 3+4 2 | 3+3 2 | 2+2 2 |
| | 3+4 2 | 5+6 2 | 3+4 2 | 5+6 2 | 2+2 2 | 3+3 2 |
| | 5+6 2 | 3+4 2 | 3+4 2 | 5+6 2 | 2+2 2 | 3+3 2 |
| | 3+4 2 | 5+6 2 | 5+6 2 | 3+4 2 | 3+3 2 | 2+2 2 |
| | 3+4 2 | 5+6 2 | 3+4 2 | 3+4 2 | 2+2 2 | 3+3 2 |
| | 5+6 2 | 3+4 2 | 5+6 2 | 5+6 2 | 3+3 2 | 3+3 2 |

Continued on next column

**TABLE XXIV– continued**

| TQC | ICSD | NoE | SF | | | |
|---|---|---|---|---|---|---|
| 12.1.37 | 29376 | 16 | N3Na1 | | | |
| | -A | -GM | -M | -Y | -L | -V |
| | 5+6 2 | 3+4 2 | 5+6 2 | 3+4 2 | 2+2 2 | 3+3 2 |
| | 3+4 2 | 5+6 2 | 3+4 2 | 5+6 2 | 3+3 2 | 2+2 2 |
| | 5+6 2 | 3+4 2 | 5+6 2 | 3+4 2 | 3+3 2 | 2+2 2 |
| | 3+4 2 | 5+6 2 | 3+4 2 | 5+6 2 | 2+2 2 | 3+3 2 |
| | 3+4 2 | 3+4 2 | 3+4 2 | 5+6 2 | 2+2 2 | 3+3 2 |
| | 5+6 2 | 5+6 2 | 5+6 2 | 3+4 2 | 3+3 2 | 2+2 2 |
| | 3+4 2 | 5+6 2 | 3+4 2 | 3+4 2 | 2+2 2 | 3+3 2 |
| | 5+6 2 | 3+4 2 | 5+6 2 | 5+6 2 | 3+3 2 | 3+3 2 |
| 12.1.38 | 29375 | 16 | N3Na1 | | | |
| | -A | -GM | -M | -Y | -L | -V |
| | 5+6 2 | 3+4 2 | 5+6 2 | 3+4 2 | 2+2 2 | 3+3 2 |
| | 3+4 2 | 5+6 2 | 3+4 2 | 5+6 2 | 3+3 2 | 2+2 2 |
| | 5+6 2 | 3+4 2 | 5+6 2 | 3+4 2 | 3+3 2 | 2+2 2 |
| | 3+4 2 | 5+6 2 | 3+4 2 | 5+6 2 | 2+2 2 | 3+3 2 |
| | 3+4 2 | 3+4 2 | 3+4 2 | 5+6 2 | 2+2 2 | 3+3 2 |
| | 5+6 2 | 5+6 2 | 5+6 2 | 3+4 2 | 3+3 2 | 2+2 2 |
| | 3+4 2 | 5+6 2 | 3+4 2 | 3+4 2 | 2+2 2 | 3+3 2 |
| | 5+6 2 | 3+4 2 | 5+6 2 | 5+6 2 | 3+3 2 | 3+3 2 |
| 12.1.39 | 29374 | 16 | N3Na1 | | | |
| | -A | -GM | -M | -Y | -L | -V |
| | 5+6 2 | 3+4 2 | 5+6 2 | 3+4 2 | 2+2 2 | 3+3 2 |
| | 3+4 2 | 5+6 2 | 3+4 2 | 5+6 2 | 3+3 2 | 2+2 2 |
| | 5+6 2 | 3+4 2 | 5+6 2 | 3+4 2 | 3+3 2 | 2+2 2 |
| | 3+4 2 | 5+6 2 | 3+4 2 | 5+6 2 | 2+2 2 | 3+3 2 |
| | 3+4 2 | 3+4 2 | 3+4 2 | 5+6 2 | 2+2 2 | 3+3 2 |
| | 5+6 2 | 5+6 2 | 5+6 2 | 3+4 2 | 3+3 2 | 2+2 2 |
| | 3+4 2 | 5+6 2 | 3+4 2 | 3+4 2 | 2+2 2 | 3+3 2 |
| | 5+6 2 | 3+4 2 | 5+6 2 | 5+6 2 | 3+3 2 | 3+3 2 |
| 12.1.40 | 29373 | 16 | N3Na1 | | | |
| | -A | -GM | -M | -Y | -L | -V |
| | 5+6 2 | 3+4 2 | 5+6 2 | 3+4 2 | 2+2 2 | 3+3 2 |
| | 3+4 2 | 5+6 2 | 3+4 2 | 5+6 2 | 3+3 2 | 2+2 2 |
| | 5+6 2 | 3+4 2 | 5+6 2 | 3+4 2 | 3+3 2 | 2+2 2 |
| | 3+4 2 | 5+6 2 | 3+4 2 | 5+6 2 | 2+2 2 | 3+3 2 |
| | 3+4 2 | 3+4 2 | 3+4 2 | 5+6 2 | 2+2 2 | 3+3 2 |
| | 5+6 2 | 5+6 2 | 5+6 2 | 3+4 2 | 3+3 2 | 2+2 2 |
| | 3+4 2 | 5+6 2 | 3+4 2 | 3+4 2 | 2+2 2 | 3+3 2 |
| | 5+6 2 | 3+4 2 | 5+6 2 | 5+6 2 | 3+3 2 | 3+3 2 |
| 12.1.41 | 29372 | 16 | N3Na1 | | | |
| | -A | -GM | -M | -Y | -L | -V |
| | 5+6 2 | 3+4 2 | 5+6 2 | 3+4 2 | 2+2 2 | 3+3 2 |
| | 3+4 2 | 5+6 2 | 3+4 2 | 5+6 2 | 3+3 2 | 2+2 2 |
| | 5+6 2 | 3+4 2 | 5+6 2 | 3+4 2 | 3+3 2 | 2+2 2 |
| | 3+4 2 | 5+6 2 | 3+4 2 | 5+6 2 | 2+2 2 | 3+3 2 |
| | 5+6 2 | 3+4 2 | 3+4 2 | 5+6 2 | 2+2 2 | 3+3 2 |
| | 3+4 2 | 5+6 2 | 5+6 2 | 3+4 2 | 3+3 2 | 2+2 2 |
| | 3+4 2 | 5+6 2 | 3+4 2 | 3+4 2 | 2+2 2 | 3+3 2 |
| | 5+6 2 | 3+4 2 | 5+6 2 | 5+6 2 | 3+3 2 | 3+3 2 |
| 12.1.42 | 29371 | 16 | N3Na1 | | | |
| | -A | -GM | -M | -Y | -L | -V |
| | 5+6 2 | 3+4 2 | 5+6 2 | 3+4 2 | 2+2 2 | 3+3 2 |
| | 3+4 2 | 5+6 2 | 3+4 2 | 5+6 2 | 3+3 2 | 2+2 2 |
| | 5+6 2 | 3+4 2 | 5+6 2 | 3+4 2 | 3+3 2 | 2+2 2 |
| | 3+4 2 | 5+6 2 | 3+4 2 | 5+6 2 | 2+2 2 | 3+3 2 |
| | 3+4 2 | 3+4 2 | 3+4 2 | 5+6 2 | 2+2 2 | 3+3 2 |
| | 5+6 2 | 5+6 2 | 5+6 2 | 3+4 2 | 3+3 2 | 2+2 2 |
| | 3+4 2 | 5+6 2 | 3+4 2 | 3+4 2 | 2+2 2 | 3+3 2 |
| | 5+6 2 | 3+4 2 | 5+6 2 | 5+6 2 | 3+3 2 | 3+3 2 |

Continued on next column

**TABLE XXIV– continued**

| TQC | ICSD | | NoE | | SF | |
|---|---|---|---|---|---|---|
| 12.1.43 | 29370 | | 16 | | N3Na1 | |
| | -A | -GM | -M | -Y | -L | -V |
| | 5+6 2 | 3+4 2 | 5+6 2 | 3+4 2 | 2+2 2 | 3+3 2 |
| | 3+4 2 | 5+6 2 | 3+4 2 | 5+6 2 | 3+3 2 | 2+2 2 |
| | 5+6 2 | 3+4 2 | 5+6 2 | 3+4 2 | 3+3 2 | 2+2 2 |
| | 3+4 2 | 5+6 2 | 3+4 2 | 5+6 2 | 2+2 2 | 3+3 2 |
| | 3+4 2 | 3+4 2 | 3+4 2 | 5+6 2 | 2+2 2 | 3+3 2 |
| | 5+6 2 | 5+6 2 | 5+6 2 | 3+4 2 | 3+3 2 | 2+2 2 |
| | 3+4 2 | 5+6 2 | 3+4 2 | 3+4 2 | 2+2 2 | 3+3 2 |
| | 5+6 2 | 3+4 2 | 5+6 2 | 5+6 2 | 3+3 2 | 3+3 2 |
| 12.1.44 | 151588 | | 122 | | Mn7Na1O12 | |
| | -A | -GM | -M | -Y | -L | -V |
| | 5+6 2 | 3+4 2 | 5+6 2 | 5+6 2 | 2+2 2 | 2+2 2 |
| | 3+4 2 | 3+4 2 | 3+4 2 | 5+6 2 | 3+3 2 | 3+3 2 |
| | 5+6 2 | 3+4 2 | 3+4 2 | 5+6 2 | 3+3 2 | 3+3 2 |
| | 5+6 2 | 3+4 2 | 5+6 2 | 3+4 2 | 2+2 2 | 2+2 2 |
| | 3+4 2 | 3+4 2 | 5+6 2 | 3+4 2 | 2+2 2 | 2+2 2 |
| | 3+4 2 | 3+4 2 | 3+4 2 | 3+4 2 | 3+3 2 | 3+3 2 |
| | 5+6 2 | 5+6 2 | 5+6 2 | 3+4 2 | 2+2 2 | 2+2 2 |
| | 3+4 2 | 5+6 2 | 3+4 2 | 3+4 2 | 3+3 2 | 3+3 2 |
| | 5+6 2 | 5+6 2 | 5+6 2 | 3+4 2 | 2+2 2 | 2+2 2 |
| | 3+4 2 | 5+6 2 | 3+4 2 | 5+6 2 | 3+3 2 | 3+3 2 |
| | 5+6 2 | 5+6 2 | 5+6 2 | 5+6 2 | 2+2 2 | 3+3 2 |
| | 3+4 2 | 5+6 2 | 3+4 2 | 5+6 2 | 3+3 2 | 2+2 2 |
| | 5+6 2 | 3+4 2 | 5+6 2 | 5+6 2 | 2+2 2 | 2+2 2 |
| | 5+6 2 | 3+4 2 | 3+4 2 | 5+6 2 | 2+2 2 | 3+3 2 |
| | 5+6 2 | 3+4 2 | 5+6 2 | 5+6 2 | 3+3 2 | 2+2 2 |
| | 3+4 2 | 3+4 2 | 3+4 2 | 5+6 2 | 3+3 2 | 3+3 2 |
| | 3+4 2 | 3+4 2 | 5+6 2 | 5+6 2 | 2+2 2 | 2+2 2 |
| | 5+6 2 | 3+4 2 | 3+4 2 | 5+6 2 | 3+3 2 | 3+3 2 |
| | 3+4 2 | 3+4 2 | 5+6 2 | 5+6 2 | 2+2 2 | 2+2 2 |
| | 5+6 2 | 3+4 2 | 3+4 2 | 3+4 2 | 2+2 2 | 2+2 2 |
| | 3+4 2 | 3+4 2 | 5+6 2 | 3+4 2 | 3+3 2 | 3+3 2 |
| | 3+4 2 | 5+6 2 | 3+4 2 | 3+4 2 | 3+3 2 | 3+3 2 |
| | 5+6 2 | 5+6 2 | 3+4 2 | 3+4 2 | 2+2 2 | 3+3 2 |
| | 5+6 2 | 5+6 2 | 3+4 2 | 5+6 2 | 3+3 2 | 2+2 2 |
| | 5+6 2 | 3+4 2 | 5+6 2 | 3+4 2 | 2+2 2 | 3+3 2 |
| | 3+4 2 | 3+4 2 | 5+6 2 | 3+4 2 | 3+3 2 | 2+2 2 |
| | 3+4 2 | 3+4 2 | 3+4 2 | 5+6 2 | 2+2 2 | 2+2 2 |
| | 5+6 2 | 3+4 2 | 3+4 2 | 3+4 2 | 3+3 2 | 3+3 2 |
| | 5+6 2 | 5+6 2 | 5+6 2 | 3+4 2 | 3+3 2 | 2+2 2 |
| | 5+6 2 | 5+6 2 | 5+6 2 | 3+4 2 | 2+2 2 | 2+2 2 |
| | 3+4 2 | 3+4 2 | 5+6 2 | 3+4 2 | 2+2 2 | 3+3 2 |
| | 3+4 2 | 5+6 2 | 5+6 2 | 3+4 2 | 2+2 2 | 2+2 2 |
| | 5+6 2 | 5+6 2 | 3+4 2 | 3+4 2 | 3+3 2 | 2+2 2 |
| | 3+4 2 | 3+4 2 | 5+6 2 | 5+6 2 | 2+2 2 | 3+3 2 |
| | 3+4 2 | 5+6 2 | 3+4 2 | 5+6 2 | 3+3 2 | 3+3 2 |
| | 3+4 2 | 5+6 2 | 3+4 2 | 5+6 2 | 3+3 2 | 2+2 2 |
| | 5+6 2 | 5+6 2 | 5+6 2 | 3+4 2 | 2+2 2 | 3+3 2 |
| | 5+6 2 | 5+6 2 | 5+6 2 | 3+4 2 | 2+2 2 | 2+2 2 |
| | 3+4 2 | 3+4 2 | 3+4 2 | 5+6 2 | 3+3 2 | 3+3 2 |
| | 3+4 2 | 3+4 2 | 3+4 2 | 3+4 2 | 3+3 2 | 3+3 2 |
| | 5+6 2 | 5+6 2 | 5+6 2 | 5+6 2 | 2+2 2 | 2+2 2 |
| | 5+6 2 | 3+4 2 | 3+4 2 | 5+6 2 | 2+2 2 | 2+2 2 |
| | 3+4 2 | 5+6 2 | 5+6 2 | 5+6 2 | 3+3 2 | 3+3 2 |
| | 3+4 2 | 5+6 2 | 3+4 2 | 5+6 2 | 3+3 2 | 3+3 2 |
| | 5+6 2 | 5+6 2 | 5+6 2 | 5+6 2 | 2+2 2 | 2+2 2 |
| | 3+4 2 | 5+6 2 | 3+4 2 | 3+4 2 | 3+3 2 | 3+3 2 |
| | 3+4 2 | 5+6 2 | 5+6 2 | 3+4 2 | 2+2 2 | 3+3 2 |
| | 5+6 2 | 5+6 2 | 3+4 2 | 3+4 2 | 3+3 2 | 2+2 2 |
| | 3+4 2 | 3+4 2 | 3+4 2 | 5+6 2 | 2+2 2 | 2+2 2 |

Continued on next column

**TABLE XXIV– continued**

| TQC | ICSD | | NoE | | SF | |
|---|---|---|---|---|---|---|
| | 5+6 2 | 3+4 2 | 5+6 2 | 5+6 2 | 3+3 2 | 3+3 2 |
| | 5+6 2 | 3+4 2 | 5+6 2 | 5+6 2 | 2+2 2 | 2+2 2 |
| | 3+4 2 | 3+4 2 | 3+4 2 | 5+6 2 | 2+2 2 | 3+3 2 |
| | 5+6 2 | 3+4 2 | 5+6 2 | 5+6 2 | 3+3 2 | 2+2 2 |
| | 5+6 2 | 3+4 2 | 5+6 2 | 5+6 2 | 2+2 2 | 2+2 2 |
| | 3+4 2 | 3+4 2 | 3+4 2 | 3+4 2 | 3+3 2 | 3+3 2 |
| | 3+4 2 | 3+4 2 | 3+4 2 | 3+4 2 | 2+2 2 | 3+3 2 |
| | 5+6 2 | 3+4 2 | 5+6 2 | 3+4 2 | 3+3 2 | 2+2 2 |
| | 5+6 2 | 3+4 2 | 5+6 2 | 3+4 2 | 2+2 2 | 3+3 2 |
| | 3+4 2 | 3+4 2 | 3+4 2 | 3+4 2 | 3+3 2 | 2+2 2 |
| | 5+6 2 | 3+4 2 | 5+6 2 | 3+4 2 | 2+2 2 | 2+2 2 |
| | 5+6 2 | 3+4 2 | 5+6 2 | 3+4 2 | 2+2 2 | 2+2 2 |
| 12.1.45 | 78328 | | 148 | | Mn1Mo1O4 | |
| | -A | -GM | -M | -Y | -L | -V |
| | 5+6 2 | 3+4 2 | 5+6 2 | 3+4 2 | 2+2 2 | 3+3 2 |
| | 3+4 2 | 5+6 2 | 3+4 2 | 5+6 2 | 3+3 2 | 2+2 2 |
| | 3+4 2 | 3+4 2 | 3+4 2 | 3+4 2 | 3+3 2 | 2+2 2 |
| | 5+6 2 | 5+6 2 | 5+6 2 | 3+4 2 | 2+2 2 | 3+3 2 |
| | 3+4 2 | 3+4 2 | 5+6 2 | 5+6 2 | 3+3 2 | 3+3 2 |
| | 3+4 2 | 5+6 2 | 3+4 2 | 5+6 2 | 2+2 2 | 2+2 2 |
| | 5+6 2 | 3+4 2 | 5+6 2 | 3+4 2 | 2+2 2 | 3+3 2 |
| | 3+4 2 | 5+6 2 | 3+4 2 | 5+6 2 | 3+3 2 | 2+2 2 |
| | 5+6 2 | 5+6 2 | 3+4 2 | 5+6 2 | 2+2 2 | 3+3 2 |
| | 5+6 2 | 3+4 2 | 5+6 2 | 5+6 2 | 3+3 2 | 2+2 2 |
| | 3+4 2 | 5+6 2 | 3+4 2 | 3+4 2 | 3+3 2 | 2+2 2 |
| | 5+6 2 | 3+4 2 | 5+6 2 | 3+4 2 | 2+2 2 | 2+2 2 |
| | 3+4 2 | 5+6 2 | 5+6 2 | 5+6 2 | 3+3 2 | 3+3 2 |
| | 5+6 2 | 5+6 2 | 3+4 2 | 5+6 2 | 3+3 2 | 3+3 2 |
| | 5+6 2 | 3+4 2 | 5+6 2 | 3+4 2 | 2+2 2 | 3+3 2 |
| | 3+4 2 | 3+4 2 | 3+4 2 | 3+4 2 | 2+2 2 | 2+2 2 |
| | 5+6 2 | 3+4 2 | 3+4 2 | 3+4 2 | 2+2 2 | 3+3 2 |
| | 5+6 2 | 5+6 2 | 5+6 2 | 5+6 2 | 3+3 2 | 2+2 2 |
| | 3+4 2 | 3+4 2 | 3+4 2 | 5+6 2 | 3+3 2 | 3+3 2 |
| | 3+4 2 | 3+4 2 | 5+6 2 | 3+4 2 | 2+2 2 | 3+3 2 |
| | 3+4 2 | 5+6 2 | 5+6 2 | 3+4 2 | 2+2 2 | 2+2 2 |
| | 5+6 2 | 5+6 2 | 3+4 2 | 3+4 2 | 3+3 2 | 2+2 2 |
| | 3+4 2 | 3+4 2 | 3+4 2 | 5+6 2 | 2+2 2 | 3+3 2 |
| | 5+6 2 | 5+6 2 | 5+6 2 | 5+6 2 | 3+3 2 | 2+2 2 |
| | 3+4 2 | 3+4 2 | 5+6 2 | 3+4 2 | 2+2 2 | 2+2 2 |
| | 5+6 2 | 3+4 2 | 3+4 2 | 3+4 2 | 3+3 2 | 2+2 2 |
| | 5+6 2 | 3+4 2 | 5+6 2 | 5+6 2 | 2+2 2 | 3+3 2 |
| | 3+4 2 | 3+4 2 | 5+6 2 | 3+4 2 | 3+3 2 | 2+2 2 |
| | 5+6 2 | 5+6 2 | 3+4 2 | 5+6 2 | 2+2 2 | 3+3 2 |
| | 3+4 2 | 5+6 2 | 3+4 2 | 3+4 2 | 3+3 2 | 3+3 2 |
| | 5+6 2 | 5+6 2 | 5+6 2 | 5+6 2 | 2+2 2 | 3+3 2 |
| | 3+4 2 | 3+4 2 | 3+4 2 | 5+6 2 | 3+3 2 | 2+2 2 |
| | 5+6 2 | 3+4 2 | 3+4 2 | 5+6 2 | 2+2 2 | 3+3 2 |
| | 5+6 2 | 3+4 2 | 5+6 2 | 3+4 2 | 3+3 2 | 2+2 2 |
| | 3+4 2 | 5+6 2 | 5+6 2 | 5+6 2 | 3+3 2 | 2+2 2 |
| | 5+6 2 | 5+6 2 | 3+4 2 | 3+4 2 | 2+2 2 | 3+3 2 |
| | 3+4 2 | 5+6 2 | 5+6 2 | 3+4 2 | 2+2 2 | 3+3 2 |
| | 3+4 2 | 5+6 2 | 3+4 2 | 5+6 2 | 3+3 2 | 2+2 2 |
| | 3+4 2 | 3+4 2 | 3+4 2 | 3+4 2 | 3+3 2 | 2+2 2 |
| | 5+6 2 | 3+4 2 | 5+6 2 | 5+6 2 | 2+2 2 | 3+3 2 |
| | 5+6 2 | 5+6 2 | 3+4 2 | 3+4 2 | 3+3 2 | 3+3 2 |
| | 3+4 2 | 5+6 2 | 3+4 2 | 3+4 2 | 2+2 2 | 2+2 2 |
| | 3+4 2 | 3+4 2 | 5+6 2 | 5+6 2 | 3+3 2 | 2+2 2 |
| | 3+4 2 | 3+4 2 | 5+6 2 | 5+6 2 | 2+2 2 | 3+3 2 |
| | 5+6 2 | 5+6 2 | 5+6 2 | 5+6 2 | 3+3 2 | 3+3 2 |
| | 5+6 2 | 5+6 2 | 3+4 2 | 3+4 2 | 2+2 2 | 2+2 2 |
| | 3+4 2 | 3+4 2 | 5+6 2 | 3+4 2 | 3+3 2 | 2+2 2 |

Continued on next column

**TABLE XXIV– continued**

| TQC | ICSD | | NoE | | SF | |
|---|---|---|---|---|---|---|
| | 5+6 2 | 5+6 2 | 3+4 2 | 5+6 2 | 2+2 2 | 3+3 2 |
| | 3+4 2 | 5+6 2 | 3+4 2 | 3+4 2 | 2+2 2 | 3+3 2 |
| | 5+6 2 | 5+6 2 | 5+6 2 | 5+6 2 | 3+3 2 | 3+3 2 |
| | 3+4 2 | 5+6 2 | 5+6 2 | 5+6 2 | 3+3 2 | 3+3 2 |
| | 5+6 2 | 3+4 2 | 3+4 2 | 5+6 2 | 3+3 2 | 2+2 2 |
| | 3+4 2 | 3+4 2 | 3+4 2 | 5+6 2 | 2+2 2 | 2+2 2 |
| | 5+6 2 | 5+6 2 | 5+6 2 | 3+4 2 | 3+3 2 | 3+3 2 |
| | 3+4 2 | 3+4 2 | 3+4 2 | 5+6 2 | 2+2 2 | 2+2 2 |
| | 3+4 2 | 5+6 2 | 3+4 2 | 3+4 2 | 2+2 2 | 2+2 2 |
| | 3+4 2 | 3+4 2 | 5+6 2 | 3+4 2 | 3+3 2 | 3+3 2 |
| | 5+6 2 | 5+6 2 | 5+6 2 | 3+4 2 | 2+2 2 | 3+3 2 |
| | 5+6 2 | 3+4 2 | 5+6 2 | 5+6 2 | 3+3 2 | 2+2 2 |
| | 5+6 2 | 5+6 2 | 3+4 2 | 5+6 2 | 2+2 2 | 2+2 2 |
| | 3+4 2 | 5+6 2 | 3+4 2 | 5+6 2 | 3+3 2 | 3+3 2 |
| | 5+6 2 | 3+4 2 | 5+6 2 | 3+4 2 | 2+2 2 | 2+2 2 |
| | 3+4 2 | 3+4 2 | 3+4 2 | 3+4 2 | 2+2 2 | 3+3 2 |
| | 5+6 2 | 3+4 2 | 5+6 2 | 3+4 2 | 3+3 2 | 2+2 2 |
| | 5+6 2 | 3+4 2 | 3+4 2 | 5+6 2 | 2+2 2 | 2+2 2 |
| | 3+4 2 | 5+6 2 | 5+6 2 | 3+4 2 | 2+2 2 | 3+3 2 |
| | 5+6 2 | 3+4 2 | 5+6 2 | 3+4 2 | 3+3 2 | 2+2 2 |
| | 3+4 2 | 3+4 2 | 3+4 2 | 3+4 2 | 3+3 2 | 3+3 2 |
| | 5+6 2 | 5+6 2 | 5+6 2 | 5+6 2 | 3+3 2 | 2+2 2 |
| | 3+4 2 | 3+4 2 | 3+4 2 | 3+4 2 | 2+2 2 | 2+2 2 |
| | 3+4 2 | 5+6 2 | 3+4 2 | 5+6 2 | 3+3 2 | 3+3 2 |
| | 5+6 2 | 3+4 2 | 5+6 2 | 3+4 2 | 2+2 2 | 2+2 2 |
| | 3+4 2 | 5+6 2 | 3+4 2 | 3+4 2 | 2+2 2 | 3+3 2 |
| | 5+6 2 | 3+4 2 | 5+6 2 | 5+6 2 | 3+3 2 | 3+3 2 |
| 12.1.46 | 239971 | | 54 | | Li2Mn1O3 | |
| | -A | -GM | -M | -Y | -L | -V |
| | 3+4 2 | 3+4 2 | 5+6 2 | 5+6 2 | 3+3 2 | 3+3 2 |
| | 5+6 2 | 5+6 2 | 3+4 2 | 3+4 2 | 2+2 2 | 2+2 2 |
| | 5+6 2 | 5+6 2 | 5+6 2 | 5+6 2 | 3+3 2 | 2+2 2 |
| | 3+4 2 | 3+4 2 | 3+4 2 | 3+4 2 | 2+2 2 | 3+3 2 |
| | 3+4 2 | 3+4 2 | 3+4 2 | 3+4 2 | 3+3 2 | 3+3 2 |
| | 5+6 2 | 5+6 2 | 5+6 2 | 5+6 2 | 2+2 2 | 2+2 2 |
| | 5+6 2 | 3+4 2 | 3+4 2 | 3+4 2 | 2+2 2 | 2+2 2 |
| | 5+6 2 | 5+6 2 | 5+6 2 | 5+6 2 | 3+3 2 | 3+3 2 |
| | 3+4 2 | 5+6 2 | 3+4 2 | 5+6 2 | 3+3 2 | 2+2 2 |
| | 3+4 2 | 3+4 2 | 5+6 2 | 5+6 2 | 3+3 2 | 3+3 2 |
| | 5+6 2 | 5+6 2 | 3+4 2 | 3+4 2 | 2+2 2 | 3+3 2 |
| | 3+4 2 | 3+4 2 | 5+6 2 | 5+6 2 | 3+3 2 | 2+2 2 |
| | 3+4 2 | 3+4 2 | 5+6 2 | 3+4 2 | 2+2 2 | 3+3 2 |
| | 3+4 2 | 3+4 2 | 3+4 2 | 3+4 2 | 2+2 2 | 2+2 2 |
| | 5+6 2 | 3+4 2 | 3+4 2 | 3+4 2 | 3+3 2 | 2+2 2 |
| | 5+6 2 | 5+6 2 | 5+6 2 | 5+6 2 | 2+2 2 | 3+3 2 |
| | 3+4 2 | 5+6 2 | 3+4 2 | 3+4 2 | 3+3 2 | 3+3 2 |
| | 5+6 2 | 3+4 2 | 5+6 2 | 5+6 2 | 2+2 2 | 2+2 2 |
| | 3+4 2 | 5+6 2 | 5+6 2 | 5+6 2 | 3+3 2 | 3+3 2 |
| | 3+4 2 | 3+4 2 | 3+4 2 | 3+4 2 | 3+3 2 | 2+2 2 |
| | 5+6 2 | 5+6 2 | 5+6 2 | 5+6 2 | 2+2 2 | 3+3 2 |
| | 3+4 2 | 5+6 2 | 3+4 2 | 5+6 2 | 3+3 2 | 2+2 2 |
| | 5+6 2 | 3+4 2 | 3+4 2 | 3+4 2 | 2+2 2 | 3+3 2 |
| | 5+6 2 | 5+6 2 | 5+6 2 | 3+4 2 | 2+2 2 | 2+2 2 |
| | 5+6 2 | 5+6 2 | 5+6 2 | 5+6 2 | 3+3 2 | 2+2 2 |
| | 3+4 2 | 3+4 2 | 3+4 2 | 3+4 2 | 2+2 2 | 3+3 2 |
| | 3+4 2 | 3+4 2 | 5+6 2 | 5+6 2 | 2+2 2 | 2+2 2 |
| 12.1.47 | 165326 | | 38 | | Li1Mo1O2 | |
| | -A | -GM | -M | -Y | -L | -V |
| | 3+4 2 | 3+4 2 | 3+4 2 | 5+6 2 | 2+2 2 | 2+2 2 |
| | 5+6 2 | 5+6 2 | 5+6 2 | 3+4 2 | 3+3 2 | 3+3 2 |
| | 5+6 2 | 5+6 2 | 3+4 2 | 3+4 2 | 2+2 2 | 3+3 2 |

Continued on next column

**TABLE XXIV– continued**

| TQC | ICSD | | NoE | | SF | |
|---|---|---|---|---|---|---|
| | 3+4 2 | 3+4 2 | 5+6 2 | 5+6 2 | 3+3 2 | 2+2 2 |
| | 5+6 2 | 5+6 2 | 3+4 2 | 3+4 2 | 3+3 2 | 2+2 2 |
| | 3+4 2 | 3+4 2 | 5+6 2 | 5+6 2 | 2+2 2 | 3+3 2 |
| | 5+6 2 | 5+6 2 | 3+4 2 | 5+6 2 | 2+2 2 | 3+3 2 |
| | 5+6 2 | 3+4 2 | 5+6 2 | 3+4 2 | 2+2 2 | 2+2 2 |
| | 3+4 2 | 5+6 2 | 3+4 2 | 3+4 2 | 3+3 2 | 2+2 2 |
| | 5+6 2 | 3+4 2 | 3+4 2 | 5+6 2 | 2+2 2 | 3+3 2 |
| | 3+4 2 | 3+4 2 | 5+6 2 | 5+6 2 | 3+3 2 | 2+2 2 |
| | 3+4 2 | 3+4 2 | 5+6 2 | 3+4 2 | 3+3 2 | 3+3 2 |
| | 5+6 2 | 5+6 2 | 3+4 2 | 5+6 2 | 2+2 2 | 3+3 2 |
| | 5+6 2 | 3+4 2 | 3+4 2 | 5+6 2 | 3+3 2 | 2+2 2 |
| | 3+4 2 | 5+6 2 | 5+6 2 | 3+4 2 | 2+2 2 | 2+2 2 |
| | 3+4 2 | 5+6 2 | 5+6 2 | 3+4 2 | 3+3 2 | 3+3 2 |
| | 5+6 2 | 5+6 2 | 3+4 2 | 3+4 2 | 3+3 2 | 3+3 2 |
| | 3+4 2 | 3+4 2 | 5+6 2 | 5+6 2 | 2+2 2 | 2+2 2 |
| | 3+4 2 | 3+4 2 | 5+6 2 | 5+6 2 | 2+2 2 | 3+3 2 |
| 12.1.48 | 49021 | | 58 | | K1Li1Mn1O2 | |
| | -A | -GM | -M | -Y | -L | -V |
| | 5+6 2 | 3+4 2 | 5+6 2 | 3+4 2 | 2+2 2 | 3+3 2 |
| | 3+4 2 | 5+6 2 | 3+4 2 | 5+6 2 | 3+3 2 | 2+2 2 |
| | 3+4 2 | 3+4 2 | 3+4 2 | 3+4 2 | 3+3 2 | 3+3 2 |
| | 5+6 2 | 5+6 2 | 5+6 2 | 5+6 2 | 2+2 2 | 2+2 2 |
| | 5+6 2 | 5+6 2 | 5+6 2 | 5+6 2 | 3+3 2 | 2+2 2 |
| | 3+4 2 | 3+4 2 | 3+4 2 | 3+4 2 | 2+2 2 | 3+3 2 |
| | 5+6 2 | 3+4 2 | 5+6 2 | 3+4 2 | 2+2 2 | 3+3 2 |
| | 5+6 2 | 5+6 2 | 3+4 2 | 5+6 2 | 3+3 2 | 2+2 2 |
| | 3+4 2 | 3+4 2 | 5+6 2 | 3+4 2 | 2+2 2 | 3+3 2 |
| | 3+4 2 | 5+6 2 | 3+4 2 | 5+6 2 | 3+3 2 | 2+2 2 |
| | 5+6 2 | 3+4 2 | 3+4 2 | 5+6 2 | 2+2 2 | 3+3 2 |
| | 3+4 2 | 5+6 2 | 5+6 2 | 3+4 2 | 3+3 2 | 2+2 2 |
| | 3+4 2 | 3+4 2 | 5+6 2 | 5+6 2 | 2+2 2 | 3+3 2 |
| | 5+6 2 | 5+6 2 | 3+4 2 | 3+4 2 | 3+3 2 | 2+2 2 |
| | 3+4 2 | 3+4 2 | 5+6 2 | 3+4 2 | 2+2 2 | 2+2 2 |
| | 5+6 2 | 3+4 2 | 3+4 2 | 5+6 2 | 3+3 2 | 3+3 2 |
| | 5+6 2 | 5+6 2 | 5+6 2 | 5+6 2 | 3+3 2 | 2+2 2 |
| | 3+4 2 | 5+6 2 | 3+4 2 | 3+4 2 | 2+2 2 | 3+3 2 |
| | 3+4 2 | 3+4 2 | 3+4 2 | 3+4 2 | 2+2 2 | 2+2 2 |
| | 5+6 2 | 5+6 2 | 5+6 2 | 5+6 2 | 3+3 2 | 3+3 2 |
| | 3+4 2 | 3+4 2 | 5+6 2 | 5+6 2 | 2+2 2 | 2+2 2 |
| | 3+4 2 | 3+4 2 | 3+4 2 | 5+6 2 | 3+3 2 | 2+2 2 |
| | 5+6 2 | 5+6 2 | 5+6 2 | 3+4 2 | 3+3 2 | 3+3 2 |
| | 5+6 2 | 5+6 2 | 3+4 2 | 3+4 2 | 2+2 2 | 3+3 2 |
| | 3+4 2 | 3+4 2 | 3+4 2 | 3+4 2 | 3+3 2 | 3+3 2 |
| | 5+6 2 | 5+6 2 | 5+6 2 | 5+6 2 | 2+2 2 | 2+2 2 |
| | 3+4 2 | 3+4 2 | 3+4 2 | 3+4 2 | 3+3 2 | 3+3 2 |
| | 3+4 2 | 3+4 2 | 3+4 2 | 3+4 2 | 2+2 2 | 2+2 2 |
| | 5+6 2 | 3+4 2 | 5+6 2 | 3+4 2 | 3+3 2 | 3+3 2 |
| 12.1.49 | 240135 | | 44 | | In4Sr1 | |
| | -A | -GM | -M | -Y | -L | -V |
| | 3+4 2 | 3+4 2 | 3+4 2 | 3+4 2 | 2+2 2 | 2+2 2 |
| | 5+6 2 | 5+6 2 | 5+6 2 | 5+6 2 | 3+3 2 | 3+3 2 |
| | 5+6 2 | 3+4 2 | 5+6 2 | 3+4 2 | 2+2 2 | 3+3 2 |
| | 3+4 2 | 5+6 2 | 3+4 2 | 5+6 2 | 3+3 2 | 2+2 2 |
| | 5+6 2 | 3+4 2 | 5+6 2 | 3+4 2 | 2+2 2 | 3+3 2 |
| | 3+4 2 | 5+6 2 | 3+4 2 | 5+6 2 | 3+3 2 | 2+2 2 |
| | 5+6 2 | 3+4 2 | 5+6 2 | 3+4 2 | 2+2 2 | 3+3 2 |
| | 3+4 2 | 5+6 2 | 3+4 2 | 5+6 2 | 3+3 2 | 2+2 2 |
| | 3+4 2 | 3+4 2 | 3+4 2 | 3+4 2 | 3+3 2 | 2+2 2 |
| | 5+6 2 | 3+4 2 | 5+6 2 | 5+6 2 | 2+2 2 | 3+3 2 |
| | 3+4 2 | 5+6 2 | 3+4 2 | 5+6 2 | 3+3 2 | 2+2 2 |
| | 5+6 2 | 5+6 2 | 5+6 2 | 3+4 2 | 2+2 2 | 3+3 2 |

Continued on next column

**TABLE XXIV– continued**

| TQC | ICSD | | NoE | | SF | |
|---|---|---|---|---|---|---|
| | 5+6 2 | 3+4 2 | 5+6 2 | 3+4 2 | 3+3 2 | 2+2 2 |
| | 3+4 2 | 5+6 2 | 3+4 2 | 5+6 2 | 2+2 2 | 3+3 2 |
| | 5+6 2 | 3+4 2 | 5+6 2 | 3+4 2 | 2+2 2 | 2+2 2 |
| | 3+4 2 | 5+6 2 | 5+6 2 | 5+6 2 | 3+3 2 | 3+3 2 |
| | 5+6 2 | 3+4 2 | 3+4 2 | 5+6 2 | 3+3 2 | 3+3 2 |
| | 3+4 2 | 5+6 2 | 3+4 2 | 5+6 2 | 2+2 2 | 2+2 2 |
| | 5+6 2 | 3+4 2 | 3+4 2 | 3+4 2 | 2+2 2 | 2+2 2 |
| | 3+4 2 | 3+4 2 | 5+6 2 | 3+4 2 | 3+3 2 | 2+2 2 |
| | 3+4 2 | 3+4 2 | 3+4 2 | 3+4 2 | 3+3 2 | 3+3 2 |
| | 5+6 2 | 3+4 2 | 5+6 2 | 5+6 2 | 2+2 2 | 3+3 2 |
| 12.1.50 | 240133 | | 44 | | In4Sr1 | |
| | -A | -GM | -M | -Y | -L | -V |
| | 3+4 2 | 3+4 2 | 3+4 2 | 3+4 2 | 2+2 2 | 2+2 2 |
| | 5+6 2 | 5+6 2 | 5+6 2 | 5+6 2 | 3+3 2 | 3+3 2 |
| | 5+6 2 | 3+4 2 | 5+6 2 | 3+4 2 | 2+2 2 | 3+3 2 |
| | 3+4 2 | 5+6 2 | 3+4 2 | 5+6 2 | 3+3 2 | 2+2 2 |
| | 5+6 2 | 3+4 2 | 5+6 2 | 3+4 2 | 2+2 2 | 3+3 2 |
| | 3+4 2 | 5+6 2 | 3+4 2 | 5+6 2 | 3+3 2 | 2+2 2 |
| | 5+6 2 | 3+4 2 | 5+6 2 | 3+4 2 | 2+2 2 | 3+3 2 |
| | 3+4 2 | 5+6 2 | 3+4 2 | 5+6 2 | 3+3 2 | 2+2 2 |
| | 3+4 2 | 3+4 2 | 3+4 2 | 3+4 2 | 3+3 2 | 2+2 2 |
| | 5+6 2 | 3+4 2 | 5+6 2 | 5+6 2 | 2+2 2 | 3+3 2 |
| | 3+4 2 | 5+6 2 | 3+4 2 | 5+6 2 | 3+3 2 | 2+2 2 |
| | 5+6 2 | 5+6 2 | 5+6 2 | 3+4 2 | 2+2 2 | 3+3 2 |
| | 5+6 2 | 3+4 2 | 5+6 2 | 3+4 2 | 3+3 2 | 2+2 2 |
| | 3+4 2 | 5+6 2 | 3+4 2 | 5+6 2 | 2+2 2 | 3+3 2 |
| | 5+6 2 | 3+4 2 | 5+6 2 | 3+4 2 | 2+2 2 | 2+2 2 |
| | 3+4 2 | 5+6 2 | 5+6 2 | 5+6 2 | 3+3 2 | 3+3 2 |
| | 3+4 2 | 3+4 2 | 3+4 2 | 5+6 2 | 3+3 2 | 3+3 2 |
| | 5+6 2 | 5+6 2 | 3+4 2 | 5+6 2 | 2+2 2 | 2+2 2 |
| | 5+6 2 | 3+4 2 | 3+4 2 | 3+4 2 | 2+2 2 | 2+2 2 |
| | 3+4 2 | 3+4 2 | 5+6 2 | 3+4 2 | 3+3 2 | 2+2 2 |
| | 3+4 2 | 3+4 2 | 3+4 2 | 5+6 2 | 3+3 2 | 3+3 2 |
| | 5+6 2 | 3+4 2 | 5+6 2 | 3+4 2 | 2+2 2 | 3+3 2 |
| 12.1.51 | 407248 | | 192 | | I3La5Si5 | |
| | -A | -GM | -M | -Y | -L | -V |
| | 3+4 2 | 3+4 2 | 3+4 2 | 3+4 2 | 2+2 2 | 2+2 2 |
| | 5+6 2 | 5+6 2 | 5+6 2 | 5+6 2 | 3+3 2 | 3+3 2 |
| | 3+4 2 | 3+4 2 | 3+4 2 | 3+4 2 | 2+2 2 | 2+2 2 |
| | 5+6 2 | 5+6 2 | 5+6 2 | 5+6 2 | 3+3 2 | 3+3 2 |
| | 3+4 2 | 3+4 2 | 5+6 2 | 5+6 2 | 2+2 2 | 2+2 2 |
| | 5+6 2 | 5+6 2 | 3+4 2 | 3+4 2 | 3+3 2 | 3+3 2 |
| | 5+6 2 | 3+4 2 | 5+6 2 | 3+4 2 | 2+2 2 | 3+3 2 |
| | 3+4 2 | 5+6 2 | 3+4 2 | 5+6 2 | 3+3 2 | 2+2 2 |
| | 3+4 2 | 3+4 2 | 3+4 2 | 3+4 2 | 2+2 2 | 2+2 2 |
| | 5+6 2 | 5+6 2 | 5+6 2 | 5+6 2 | 3+3 2 | 3+3 2 |
| | 3+4 2 | 5+6 2 | 3+4 2 | 5+6 2 | 2+2 2 | 2+2 2 |
| | 5+6 2 | 3+4 2 | 5+6 2 | 3+4 2 | 3+3 2 | 3+3 2 |
| | 5+6 2 | 3+4 2 | 3+4 2 | 5+6 2 | 2+2 2 | 2+2 2 |
| | 3+4 2 | 5+6 2 | 5+6 2 | 3+4 2 | 3+3 2 | 3+3 2 |
| | 3+4 2 | 3+4 2 | 5+6 2 | 5+6 2 | 2+2 2 | 2+2 2 |
| | 5+6 2 | 5+6 2 | 3+4 2 | 3+4 2 | 3+3 2 | 3+3 2 |
| | 5+6 2 | 3+4 2 | 5+6 2 | 3+4 2 | 2+2 2 | 3+3 2 |
| | 3+4 2 | 5+6 2 | 3+4 2 | 5+6 2 | 3+3 2 | 2+2 2 |
| | 3+4 2 | 3+4 2 | 5+6 2 | 5+6 2 | 2+2 2 | 2+2 2 |
| | 5+6 2 | 5+6 2 | 3+4 2 | 3+4 2 | 3+3 2 | 3+3 2 |
| | 3+4 2 | 5+6 2 | 3+4 2 | 5+6 2 | 2+2 2 | 3+3 2 |
| | 5+6 2 | 3+4 2 | 5+6 2 | 3+4 2 | 3+3 2 | 2+2 2 |
| | 5+6 2 | 3+4 2 | 3+4 2 | 5+6 2 | 2+2 2 | 3+3 2 |
| | 3+4 2 | 5+6 2 | 5+6 2 | 3+4 2 | 3+3 2 | 2+2 2 |
| | 5+6 2 | 3+4 2 | 5+6 2 | 3+4 2 | 3+3 2 | 2+2 2 |

Continued on next column

**TABLE XXIV– continued**

| TQC | ICSD | | NoE | | SF | |
|---|---|---|---|---|---|---|
| | 3+4 2 | 5+6 2 | 3+4 2 | 5+6 2 | 2+2 2 | 3+3 2 |
| | 5+6 2 | 3+4 2 | 3+4 2 | 5+6 2 | 3+3 2 | 2+2 2 |
| | 3+4 2 | 5+6 2 | 5+6 2 | 3+4 2 | 2+2 2 | 3+3 2 |
| | 3+4 2 | 3+4 2 | 5+6 2 | 5+6 2 | 2+2 2 | 3+3 2 |
| | 5+6 2 | 5+6 2 | 3+4 2 | 3+4 2 | 3+3 2 | 2+2 2 |
| | 3+4 2 | 3+4 2 | 5+6 2 | 5+6 2 | 3+3 2 | 2+2 2 |
| | 5+6 2 | 5+6 2 | 3+4 2 | 3+4 2 | 2+2 2 | 3+3 2 |
| | 5+6 2 | 3+4 2 | 5+6 2 | 3+4 2 | 3+3 2 | 2+2 2 |
| | 3+4 2 | 5+6 2 | 3+4 2 | 5+6 2 | 2+2 2 | 3+3 2 |
| | 5+6 2 | 3+4 2 | 5+6 2 | 3+4 2 | 2+2 2 | 3+3 2 |
| | 3+4 2 | 5+6 2 | 3+4 2 | 5+6 2 | 3+3 2 | 2+2 2 |
| | 5+6 2 | 3+4 2 | 5+6 2 | 3+4 2 | 2+2 2 | 3+3 2 |
| | 3+4 2 | 5+6 2 | 3+4 2 | 5+6 2 | 3+3 2 | 2+2 2 |
| | 5+6 2 | 3+4 2 | 5+6 2 | 3+4 2 | 2+2 2 | 3+3 2 |
| | 3+4 2 | 5+6 2 | 3+4 2 | 5+6 2 | 3+3 2 | 2+2 2 |
| | 5+6 2 | 3+4 2 | 3+4 2 | 5+6 2 | 2+2 2 | 3+3 2 |
| | 3+4 2 | 5+6 2 | 5+6 2 | 3+4 2 | 3+3 2 | 2+2 2 |
| | 3+4 2 | 3+4 2 | 3+4 2 | 3+4 2 | 2+2 2 | 2+2 2 |
| | 5+6 2 | 5+6 2 | 5+6 2 | 5+6 2 | 3+3 2 | 3+3 2 |
| | 3+4 2 | 3+4 2 | 3+4 2 | 3+4 2 | 3+3 2 | 3+3 2 |
| | 5+6 2 | 5+6 2 | 5+6 2 | 5+6 2 | 2+2 2 | 2+2 2 |
| | 3+4 2 | 3+4 2 | 5+6 2 | 5+6 2 | 3+3 2 | 2+2 2 |
| | 5+6 2 | 3+4 2 | 5+6 2 | 3+4 2 | 2+2 2 | 3+3 2 |
| | 3+4 2 | 5+6 2 | 3+4 2 | 5+6 2 | 3+3 2 | 2+2 2 |
| | 5+6 2 | 3+4 2 | 5+6 2 | 3+4 2 | 2+2 2 | 3+3 2 |
| | 3+4 2 | 5+6 2 | 3+4 2 | 5+6 2 | 3+3 2 | 2+2 2 |
| | 5+6 2 | 5+6 2 | 3+4 2 | 3+4 2 | 2+2 2 | 3+3 2 |
| | 3+4 2 | 3+4 2 | 5+6 2 | 5+6 2 | 3+3 2 | 2+2 2 |
| | 5+6 2 | 3+4 2 | 5+6 2 | 3+4 2 | 2+2 2 | 3+3 2 |
| | 5+6 2 | 5+6 2 | 3+4 2 | 5+6 2 | 3+3 2 | 2+2 2 |
| | 3+4 2 | 3+4 2 | 3+4 2 | 3+4 2 | 2+2 2 | 3+3 2 |
| | 3+4 2 | 5+6 2 | 5+6 2 | 5+6 2 | 3+3 2 | 2+2 2 |
| | 5+6 2 | 3+4 2 | 3+4 2 | 3+4 2 | 2+2 2 | 3+3 2 |
| | 3+4 2 | 5+6 2 | 5+6 2 | 5+6 2 | 3+3 2 | 2+2 2 |
| | 5+6 2 | 3+4 2 | 3+4 2 | 3+4 2 | 2+2 2 | 3+3 2 |
| | 3+4 2 | 5+6 2 | 5+6 2 | 5+6 2 | 3+3 2 | 2+2 2 |
| | 5+6 2 | 3+4 2 | 3+4 2 | 3+4 2 | 2+2 2 | 3+3 2 |
| | 5+6 2 | 5+6 2 | 5+6 2 | 5+6 2 | 2+2 2 | 2+2 2 |
| | 3+4 2 | 5+6 2 | 3+4 2 | 3+4 2 | 3+3 2 | 3+3 2 |
| | 3+4 2 | 3+4 2 | 3+4 2 | 3+4 2 | 2+2 2 | 2+2 2 |
| | 5+6 2 | 5+6 2 | 5+6 2 | 5+6 2 | 3+3 2 | 3+3 2 |
| | 5+6 2 | 3+4 2 | 3+4 2 | 5+6 2 | 3+3 2 | 3+3 2 |
| | 3+4 2 | 3+4 2 | 5+6 2 | 5+6 2 | 2+2 2 | 2+2 2 |
| | 5+6 2 | 5+6 2 | 3+4 2 | 3+4 2 | 3+3 2 | 3+3 2 |
| | 3+4 2 | 3+4 2 | 5+6 2 | 5+6 2 | 2+2 2 | 2+2 2 |
| | 3+4 2 | 3+4 2 | 3+4 2 | 3+4 2 | 3+3 2 | 3+3 2 |
| | 5+6 2 | 5+6 2 | 5+6 2 | 5+6 2 | 2+2 2 | 2+2 2 |
| | 3+4 2 | 3+4 2 | 5+6 2 | 3+4 2 | 3+3 2 | 3+3 2 |
| | 5+6 2 | 5+6 2 | 3+4 2 | 5+6 2 | 2+2 2 | 2+2 2 |
| | 3+4 2 | 5+6 2 | 5+6 2 | 3+4 2 | 3+3 2 | 3+3 2 |
| | 3+4 2 | 3+4 2 | 3+4 2 | 5+6 2 | 2+2 2 | 2+2 2 |
| | 5+6 2 | 3+4 2 | 5+6 2 | 3+4 2 | 2+2 2 | 3+3 2 |
| | 3+4 2 | 5+6 2 | 5+6 2 | 5+6 2 | 3+3 2 | 2+2 2 |
| | 5+6 2 | 5+6 2 | 3+4 2 | 3+4 2 | 3+3 2 | 2+2 2 |
| | 3+4 2 | 3+4 2 | 5+6 2 | 5+6 2 | 2+2 2 | 3+3 2 |
| | 3+4 2 | 5+6 2 | 3+4 2 | 3+4 2 | 2+2 2 | 2+2 2 |
| | 5+6 2 | 5+6 2 | 3+4 2 | 5+6 2 | 3+3 2 | 3+3 2 |
| | 5+6 2 | 3+4 2 | 5+6 2 | 3+4 2 | 3+3 2 | 2+2 2 |
| | 3+4 2 | 5+6 2 | 3+4 2 | 5+6 2 | 2+2 2 | 3+3 2 |
| | 5+6 2 | 3+4 2 | 3+4 2 | 5+6 2 | 2+2 2 | 2+2 2 |
| | 3+4 2 | 5+6 2 | 5+6 2 | 3+4 2 | 3+3 2 | 3+3 2 |

Continued on next column

**TABLE XXIV– continued**

| TQC | ICSD | NoE | SF | | | |
|---|---|---|---|---|---|---|
| | 5+6 2 | 3+4 2 | 5+6 2 | 3+4 2 | 3+3 2 | 2+2 2 |
| | 5+6 2 | 3+4 2 | 3+4 2 | 3+4 2 | 2+2 2 | 3+3 2 |
| | 5+6 2 | 5+6 2 | 5+6 2 | 3+4 2 | 3+3 2 | 2+2 2 |
| | 3+4 2 | 5+6 2 | 3+4 2 | 5+6 2 | 2+2 2 | 3+3 2 |
| | 3+4 2 | 3+4 2 | 3+4 2 | 5+6 2 | 3+3 2 | 2+2 2 |
| | 3+4 2 | 5+6 2 | 5+6 2 | 3+4 2 | 2+2 2 | 3+3 2 |
| | 5+6 2 | 3+4 2 | 5+6 2 | 3+4 2 | 3+3 2 | 2+2 2 |
| | 5+6 2 | 3+4 2 | 5+6 2 | 5+6 2 | 2+2 2 | 3+3 2 |
| | 5+6 2 | 5+6 2 | 3+4 2 | 3+4 2 | 3+3 2 | 2+2 2 |
| | 3+4 2 | 3+4 2 | 5+6 2 | 5+6 2 | 2+2 2 | 3+3 2 |
| 12.1.52 | 407247 | 162 | I3La4Si4 | | | |
| | -A | -GM | -M | -Y | -L | -V |
| | 3+4 2 | 3+4 2 | 3+4 2 | 3+4 2 | 2+2 2 | 2+2 2 |
| | 5+6 2 | 5+6 2 | 5+6 2 | 5+6 2 | 3+3 2 | 3+3 2 |
| | 5+6 2 | 3+4 2 | 5+6 2 | 3+4 2 | 2+2 2 | 3+3 2 |
| | 3+4 2 | 5+6 2 | 3+4 2 | 5+6 2 | 3+3 2 | 2+2 2 |
| | 5+6 2 | 3+4 2 | 5+6 2 | 3+4 2 | 2+2 2 | 2+2 2 |
| | 3+4 2 | 5+6 2 | 3+4 2 | 5+6 2 | 3+3 2 | 3+3 2 |
| | 3+4 2 | 3+4 2 | 3+4 2 | 3+4 2 | 2+2 2 | 2+2 2 |
| | 5+6 2 | 5+6 2 | 5+6 2 | 5+6 2 | 3+3 2 | 3+3 2 |
| | 3+4 2 | 3+4 2 | 5+6 2 | 5+6 2 | 2+2 2 | 2+2 2 |
| | 5+6 2 | 5+6 2 | 3+4 2 | 3+4 2 | 3+3 2 | 3+3 2 |
| | 5+6 2 | 3+4 2 | 5+6 2 | 3+4 2 | 2+2 2 | 3+3 2 |
| | 3+4 2 | 5+6 2 | 3+4 2 | 5+6 2 | 3+3 2 | 2+2 2 |
| | 5+6 2 | 3+4 2 | 5+6 2 | 3+4 2 | 2+2 2 | 3+3 2 |
| | 3+4 2 | 5+6 2 | 3+4 2 | 5+6 2 | 3+3 2 | 2+2 2 |
| | 3+4 2 | 3+4 2 | 3+4 2 | 3+4 2 | 2+2 2 | 2+2 2 |
| | 5+6 2 | 5+6 2 | 5+6 2 | 5+6 2 | 3+3 2 | 3+3 2 |
| | 3+4 2 | 3+4 2 | 5+6 2 | 5+6 2 | 2+2 2 | 2+2 2 |
| | 5+6 2 | 5+6 2 | 3+4 2 | 3+4 2 | 3+3 2 | 3+3 2 |
| | 3+4 2 | 3+4 2 | 5+6 2 | 5+6 2 | 3+3 2 | 3+3 2 |
| | 5+6 2 | 5+6 2 | 3+4 2 | 3+4 2 | 2+2 2 | 2+2 2 |
| | 5+6 2 | 3+4 2 | 5+6 2 | 3+4 2 | 2+2 2 | 2+2 2 |
| | 3+4 2 | 5+6 2 | 3+4 2 | 5+6 2 | 3+3 2 | 3+3 2 |
| | 5+6 2 | 3+4 2 | 5+6 2 | 3+4 2 | 2+2 2 | 3+3 2 |
| | 3+4 2 | 5+6 2 | 3+4 2 | 5+6 2 | 3+3 2 | 2+2 2 |
| | 5+6 2 | 3+4 2 | 5+6 2 | 3+4 2 | 2+2 2 | 3+3 2 |
| | 3+4 2 | 5+6 2 | 3+4 2 | 5+6 2 | 3+3 2 | 2+2 2 |
| | 5+6 2 | 3+4 2 | 5+6 2 | 3+4 2 | 2+2 2 | 3+3 2 |
| | 3+4 2 | 5+6 2 | 3+4 2 | 5+6 2 | 3+3 2 | 2+2 2 |
| | 3+4 2 | 3+4 2 | 3+4 2 | 3+4 2 | 3+3 2 | 3+3 2 |
| | 5+6 2 | 5+6 2 | 5+6 2 | 5+6 2 | 2+2 2 | 2+2 2 |
| | 3+4 2 | 3+4 2 | 3+4 2 | 3+4 2 | 2+2 2 | 2+2 2 |
| | 5+6 2 | 5+6 2 | 5+6 2 | 5+6 2 | 3+3 2 | 3+3 2 |
| | 3+4 2 | 3+4 2 | 5+6 2 | 5+6 2 | 2+2 2 | 2+2 2 |
| | 5+6 2 | 5+6 2 | 3+4 2 | 3+4 2 | 3+3 2 | 3+3 2 |
| | 3+4 2 | 3+4 2 | 5+6 2 | 5+6 2 | 3+3 2 | 2+2 2 |
| | 5+6 2 | 5+6 2 | 3+4 2 | 3+4 2 | 2+2 2 | 3+3 2 |
| | 5+6 2 | 3+4 2 | 3+4 2 | 5+6 2 | 2+2 2 | 2+2 2 |
| | 3+4 2 | 5+6 2 | 5+6 2 | 3+4 2 | 3+3 2 | 3+3 2 |
| | 5+6 2 | 3+4 2 | 5+6 2 | 3+4 2 | 3+3 2 | 3+3 2 |
| | 3+4 2 | 5+6 2 | 3+4 2 | 5+6 2 | 3+3 2 | 2+2 2 |
| | 3+4 2 | 3+4 2 | 3+4 2 | 3+4 2 | 2+2 2 | 2+2 2 |
| | 5+6 2 | 5+6 2 | 5+6 2 | 5+6 2 | 3+3 2 | 3+3 2 |
| | 3+4 2 | 3+4 2 | 3+4 2 | 3+4 2 | 2+2 2 | 2+2 2 |
| | 5+6 2 | 5+6 2 | 5+6 2 | 5+6 2 | 2+2 2 | 3+3 2 |
| | 3+4 2 | 3+4 2 | 3+4 2 | 3+4 2 | 3+3 2 | 3+3 2 |
| | 5+6 2 | 3+4 2 | 5+6 2 | 5+6 2 | 2+2 2 | 2+2 2 |
| | 5+6 2 | 5+6 2 | 5+6 2 | 3+4 2 | 3+3 2 | 3+3 2 |
| | 3+4 2 | 5+6 2 | 3+4 2 | 5+6 2 | 2+2 2 | 2+2 2 |
| | 5+6 2 | 3+4 2 | 5+6 2 | 3+4 2 | 3+3 2 | 3+3 2 |

Continued on next column

**TABLE XXIV– continued**

| TQC | ICSD | NoE | SF | | | |
|---|---|---|---|---|---|---|
| | 3+4 2 | 3+4 2 | 3+4 2 | 5+6 2 | 2+2 2 | 2+2 2 |
| | 5+6 2 | 5+6 2 | 5+6 2 | 3+4 2 | 3+3 2 | 2+2 2 |
| | 3+4 2 | 3+4 2 | 3+4 2 | 5+6 2 | 2+2 2 | 3+3 2 |
| | 3+4 2 | 5+6 2 | 5+6 2 | 5+6 2 | 3+3 2 | 2+2 2 |
| | 5+6 2 | 5+6 2 | 3+4 2 | 5+6 2 | 2+2 2 | 3+3 2 |
| | 3+4 2 | 3+4 2 | 3+4 2 | 3+4 2 | 3+3 2 | 2+2 2 |
| | 3+4 2 | 3+4 2 | 5+6 2 | 5+6 2 | 2+2 2 | 3+3 2 |
| | 5+6 2 | 3+4 2 | 3+4 2 | 3+4 2 | 3+3 2 | 2+2 2 |
| | 3+4 2 | 5+6 2 | 5+6 2 | 5+6 2 | 2+2 2 | 2+2 2 |
| | 5+6 2 | 5+6 2 | 5+6 2 | 3+4 2 | 3+3 2 | 3+3 2 |
| | 5+6 2 | 3+4 2 | 5+6 2 | 5+6 2 | 3+3 2 | 3+3 2 |
| | 3+4 2 | 3+4 2 | 3+4 2 | 5+6 2 | 2+2 2 | 2+2 2 |
| | 5+6 2 | 5+6 2 | 3+4 2 | 3+4 2 | 3+3 2 | 3+3 2 |
| | 5+6 2 | 5+6 2 | 5+6 2 | 3+4 2 | 2+2 2 | 2+2 2 |
| | 3+4 2 | 3+4 2 | 3+4 2 | 5+6 2 | 2+2 2 | 3+3 2 |
| | 5+6 2 | 3+4 2 | 3+4 2 | 3+4 2 | 3+3 2 | 2+2 2 |
| | 3+4 2 | 5+6 2 | 5+6 2 | 5+6 2 | 3+3 2 | 2+2 2 |
| | 3+4 2 | 5+6 2 | 5+6 2 | 3+4 2 | 2+2 2 | 3+3 2 |
| | 5+6 2 | 3+4 2 | 3+4 2 | 5+6 2 | 2+2 2 | 2+2 2 |
| | 5+6 2 | 5+6 2 | 5+6 2 | 3+4 2 | 3+3 2 | 3+3 2 |
| | 3+4 2 | 5+6 2 | 3+4 2 | 3+4 2 | 2+2 2 | 3+3 2 |
| | 5+6 2 | 5+6 2 | 3+4 2 | 5+6 2 | 3+3 2 | 3+3 2 |
| | 3+4 2 | 3+4 2 | 5+6 2 | 3+4 2 | 2+2 2 | 2+2 2 |
| | 5+6 2 | 3+4 2 | 5+6 2 | 3+4 2 | 3+3 2 | 3+3 2 |
| | 3+4 2 | 3+4 2 | 3+4 2 | 3+4 2 | 2+2 2 | 2+2 2 |
| | 5+6 2 | 3+4 2 | 5+6 2 | 3+4 2 | 3+3 2 | 2+2 2 |
| | 3+4 2 | 5+6 2 | 3+4 2 | 5+6 2 | 2+2 2 | 3+3 2 |
| | 3+4 2 | 5+6 2 | 3+4 2 | 5+6 2 | 3+3 2 | 3+3 2 |
| | 5+6 2 | 5+6 2 | 5+6 2 | 5+6 2 | 3+3 2 | 2+2 2 |
| | 5+6 2 | 3+4 2 | 5+6 2 | 5+6 2 | 2+2 2 | 3+3 2 |
| | 5+6 2 | 5+6 2 | 5+6 2 | 3+4 2 | 2+2 2 | 3+3 2 |
| | 3+4 2 | 3+4 2 | 3+4 2 | 3+4 2 | 3+3 2 | 2+2 2 |
| 12.1.53 | 83965 | 56 | Hf2Te1 | | | |
| | -A | -GM | -M | -Y | -L | -V |
| | 5+6 2 | 3+4 2 | 5+6 2 | 3+4 2 | 3+3 2 | 2+2 2 |
| | 3+4 2 | 5+6 2 | 3+4 2 | 5+6 2 | 2+2 2 | 3+3 2 |
| | 3+4 2 | 5+6 2 | 3+4 2 | 5+6 2 | 2+2 2 | 3+3 2 |
| | 5+6 2 | 3+4 2 | 5+6 2 | 3+4 2 | 3+3 2 | 2+2 2 |
| | 5+6 2 | 3+4 2 | 5+6 2 | 3+4 2 | 3+3 2 | 2+2 2 |
| | 3+4 2 | 3+4 2 | 5+6 2 | 5+6 2 | 2+2 2 | 3+3 2 |
| | 3+4 2 | 5+6 2 | 3+4 2 | 3+4 2 | 2+2 2 | 3+3 2 |
| | 5+6 2 | 5+6 2 | 3+4 2 | 5+6 2 | 3+3 2 | 2+2 2 |
| | 3+4 2 | 5+6 2 | 3+4 2 | 5+6 2 | 3+3 2 | 3+3 2 |
| | 5+6 2 | 3+4 2 | 5+6 2 | 3+4 2 | 2+2 2 | 2+2 2 |
| | 3+4 2 | 3+4 2 | 3+4 2 | 5+6 2 | 3+3 2 | 2+2 2 |
| | 5+6 2 | 5+6 2 | 5+6 2 | 3+4 2 | 2+2 2 | 3+3 2 |
| | 3+4 2 | 3+4 2 | 3+4 2 | 5+6 2 | 3+3 2 | 3+3 2 |
| | 5+6 2 | 5+6 2 | 5+6 2 | 3+4 2 | 2+2 2 | 2+2 2 |
| | 3+4 2 | 3+4 2 | 3+4 2 | 3+4 2 | 3+3 2 | 2+2 2 |
| | 5+6 2 | 5+6 2 | 5+6 2 | 5+6 2 | 2+2 2 | 3+3 2 |
| | 5+6 2 | 3+4 2 | 5+6 2 | 3+4 2 | 3+3 2 | 3+3 2 |
| | 3+4 2 | 5+6 2 | 3+4 2 | 3+4 2 | 2+2 2 | 2+2 2 |
| | 5+6 2 | 3+4 2 | 5+6 2 | 5+6 2 | 2+2 2 | 3+3 2 |
| | 3+4 2 | 3+4 2 | 3+4 2 | 3+4 2 | 3+3 2 | 2+2 2 |
| | 3+4 2 | 5+6 2 | 3+4 2 | 5+6 2 | 3+3 2 | 2+2 2 |
| | 5+6 2 | 5+6 2 | 5+6 2 | 5+6 2 | 2+2 2 | 3+3 2 |
| | 3+4 2 | 5+6 2 | 5+6 2 | 3+4 2 | 2+2 2 | 2+2 2 |
| | 5+6 2 | 3+4 2 | 3+4 2 | 3+4 2 | 3+3 2 | 2+2 2 |
| | 3+4 2 | 3+4 2 | 3+4 2 | 5+6 2 | 2+2 2 | 3+3 2 |
| | 5+6 2 | 5+6 2 | 3+4 2 | 5+6 2 | 2+2 2 | 3+3 2 |
| | 5+6 2 | 5+6 2 | 5+6 2 | 3+4 2 | 3+3 2 | 2+2 2 |

Continued on next column

**TABLE XXIV– continued**

| TQC | ICSD | | NoE | | SF | |
|---|---|---|---|---|---|---|
| | 5+6 2 | 3+4 2 | 3+4 2 | 5+6 2 | 3+3 2 | 2+2 2 |
| 12.1.54 | 20755 | | 72 | | H4Cl2Ni1O2 | |
| | -A | -GM | -M | -Y | -L | -V |
| | 3+4 2 | 3+4 2 | 3+4 2 | 3+4 2 | 2+2 2 | 2+2 2 |
| | 5+6 2 | 5+6 2 | 5+6 2 | 5+6 2 | 3+3 2 | 3+3 2 |
| | 5+6 2 | 5+6 2 | 5+6 2 | 5+6 2 | 3+3 2 | 3+3 2 |
| | 3+4 2 | 3+4 2 | 3+4 2 | 3+4 2 | 2+2 2 | 2+2 2 |
| | 3+4 2 | 3+4 2 | 3+4 2 | 3+4 2 | 3+3 2 | 2+2 2 |
| | 5+6 2 | 3+4 2 | 5+6 2 | 5+6 2 | 2+2 2 | 3+3 2 |
| | 3+4 2 | 5+6 2 | 3+4 2 | 3+4 2 | 3+3 2 | 3+3 2 |
| | 5+6 2 | 5+6 2 | 5+6 2 | 5+6 2 | 2+2 2 | 2+2 2 |
| | 5+6 2 | 5+6 2 | 5+6 2 | 5+6 2 | 3+3 2 | 2+2 2 |
| | 3+4 2 | 3+4 2 | 3+4 2 | 3+4 2 | 2+2 2 | 3+3 2 |
| | 3+4 2 | 3+4 2 | 3+4 2 | 3+4 2 | 2+2 2 | 2+2 2 |
| | 3+4 2 | 3+4 2 | 5+6 2 | 3+4 2 | 3+3 2 | 3+3 2 |
| | 5+6 2 | 3+4 2 | 3+4 2 | 3+4 2 | 2+2 2 | 2+2 2 |
| | 3+4 2 | 5+6 2 | 3+4 2 | 5+6 2 | 3+3 2 | 3+3 2 |
| | 5+6 2 | 3+4 2 | 5+6 2 | 3+4 2 | 2+2 2 | 2+2 2 |
| | 5+6 2 | 5+6 2 | 5+6 2 | 5+6 2 | 3+3 2 | 3+3 2 |
| | 3+4 2 | 3+4 2 | 3+4 2 | 5+6 2 | 3+3 2 | 3+3 2 |
| | 3+4 2 | 5+6 2 | 3+4 2 | 3+4 2 | 2+2 2 | 2+2 2 |
| | 5+6 2 | 5+6 2 | 5+6 2 | 5+6 2 | 2+2 2 | 2+2 2 |
| | 5+6 2 | 5+6 2 | 5+6 2 | 3+4 2 | 3+3 2 | 3+3 2 |
| | 3+4 2 | 3+4 2 | 3+4 2 | 5+6 2 | 3+3 2 | 2+2 2 |
| | 5+6 2 | 5+6 2 | 5+6 2 | 3+4 2 | 2+2 2 | 3+3 2 |
| | 3+4 2 | 3+4 2 | 3+4 2 | 5+6 2 | 2+2 2 | 2+2 2 |
| | 5+6 2 | 5+6 2 | 5+6 2 | 5+6 2 | 3+3 2 | 3+3 2 |
| | 5+6 2 | 5+6 2 | 3+4 2 | 5+6 2 | 2+2 2 | 2+2 2 |
| | 3+4 2 | 3+4 2 | 5+6 2 | 3+4 2 | 3+3 2 | 3+3 2 |
| | 3+4 2 | 3+4 2 | 3+4 2 | 3+4 2 | 2+2 2 | 3+3 2 |
| | 5+6 2 | 5+6 2 | 5+6 2 | 5+6 2 | 3+3 2 | 2+2 2 |
| | 3+4 2 | 3+4 2 | 3+4 2 | 5+6 2 | 3+3 2 | 3+3 2 |
| | 5+6 2 | 3+4 2 | 5+6 2 | 3+4 2 | 2+2 2 | 2+2 2 |
| | 5+6 2 | 3+4 2 | 3+4 2 | 3+4 2 | 2+2 2 | 2+2 2 |
| | 3+4 2 | 5+6 2 | 3+4 2 | 5+6 2 | 3+3 2 | 3+3 2 |
| | 3+4 2 | 5+6 2 | 5+6 2 | 5+6 2 | 2+2 2 | 2+2 2 |
| | 5+6 2 | 5+6 2 | 5+6 2 | 5+6 2 | 3+3 2 | 3+3 2 |
| | 5+6 2 | 5+6 2 | 3+4 2 | 3+4 2 | 2+2 2 | 3+3 2 |
| | 5+6 2 | 3+4 2 | 5+6 2 | 5+6 2 | 3+3 2 | 2+2 2 |
| 12.1.55 | 23361 | | 50 | | H3N1Na2S3 | |
| | -A | -GM | -M | -Y | -L | -V |
| | 5+6 2 | 3+4 2 | 3+4 2 | 5+6 2 | 2+2 2 | 3+3 2 |
| | 3+4 2 | 5+6 2 | 5+6 2 | 3+4 2 | 3+3 2 | 2+2 2 |
| | 5+6 2 | 3+4 2 | 3+4 2 | 5+6 2 | 3+3 2 | 2+2 2 |
| | 3+4 2 | 5+6 2 | 5+6 2 | 3+4 2 | 2+2 2 | 3+3 2 |
| | 3+4 2 | 3+4 2 | 5+6 2 | 5+6 2 | 2+2 2 | 2+2 2 |
| | 5+6 2 | 5+6 2 | 3+4 2 | 3+4 2 | 3+3 2 | 3+3 2 |
| | 3+4 2 | 5+6 2 | 5+6 2 | 3+4 2 | 3+3 2 | 3+3 2 |
| | 5+6 2 | 3+4 2 | 3+4 2 | 5+6 2 | 2+2 2 | 2+2 2 |
| | 5+6 2 | 3+4 2 | 3+4 2 | 5+6 2 | 2+2 2 | 3+3 2 |
| | 5+6 2 | 3+4 2 | 3+4 2 | 5+6 2 | 3+3 2 | 2+2 2 |
| | 3+4 2 | 5+6 2 | 5+6 2 | 3+4 2 | 3+3 2 | 2+2 2 |
| | 3+4 2 | 5+6 2 | 5+6 2 | 3+4 2 | 2+2 2 | 3+3 2 |
| | 5+6 2 | 3+4 2 | 5+6 2 | 3+4 2 | 2+2 2 | 3+3 2 |
| | 3+4 2 | 5+6 2 | 3+4 2 | 5+6 2 | 3+3 2 | 2+2 2 |
| | 3+4 2 | 5+6 2 | 5+6 2 | 3+4 2 | 3+3 2 | 2+2 2 |
| | 3+4 2 | 5+6 2 | 3+4 2 | 5+6 2 | 2+2 2 | 3+3 2 |
| | 5+6 2 | 3+4 2 | 3+4 2 | 5+6 2 | 3+3 2 | 2+2 2 |
| | 5+6 2 | 3+4 2 | 5+6 2 | 3+4 2 | 3+3 2 | 3+3 2 |
| | 5+6 2 | 3+4 2 | 3+4 2 | 5+6 2 | 2+2 2 | 2+2 2 |
| | 3+4 2 | 5+6 2 | 5+6 2 | 3+4 2 | 2+2 2 | 3+3 2 |

Continued on next column

**TABLE XXIV– continued**

| TQC | ICSD | | NoE | | SF | |
|---|---|---|---|---|---|---|
| | 5+6 2 | 3+4 2 | 3+4 2 | 3+4 2 | 3+3 2 | 2+2 2 |
| | 3+4 2 | 5+6 2 | 5+6 2 | 5+6 2 | 2+2 2 | 3+3 2 |
| | 3+4 2 | 3+4 2 | 5+6 2 | 5+6 2 | 3+3 2 | 3+3 2 |
| | 3+4 2 | 3+4 2 | 3+4 2 | 3+4 2 | 2+2 2 | 2+2 2 |
| | 5+6 2 | 5+6 2 | 5+6 2 | 5+6 2 | 2+2 2 | 2+2 2 |
| 12.1.56 | 62646 | | 106 | | H3Cu2F2K1O8P2 | |
| | -A | -GM | -M | -Y | -L | -V |
| | 3+4 2 | 3+4 2 | 3+4 2 | 3+4 2 | 3+3 2 | 3+3 2 |
| | 3+4 2 | 3+4 2 | 5+6 2 | 5+6 2 | 2+2 2 | 2+2 2 |
| | 5+6 2 | 5+6 2 | 3+4 2 | 3+4 2 | 3+3 2 | 3+3 2 |
| | 5+6 2 | 3+4 2 | 5+6 2 | 3+4 2 | 2+2 2 | 3+3 2 |
| | 3+4 2 | 5+6 2 | 3+4 2 | 5+6 2 | 3+3 2 | 2+2 2 |
| | 5+6 2 | 3+4 2 | 5+6 2 | 3+4 2 | 2+2 2 | 3+3 2 |
| | 3+4 2 | 5+6 2 | 3+4 2 | 5+6 2 | 3+3 2 | 2+2 2 |
| | 5+6 2 | 3+4 2 | 3+4 2 | 5+6 2 | 2+2 2 | 3+3 2 |
| | 3+4 2 | 5+6 2 | 5+6 2 | 3+4 2 | 3+3 2 | 2+2 2 |
| | 3+4 2 | 5+6 2 | 3+4 2 | 5+6 2 | 2+2 2 | 3+3 2 |
| | 5+6 2 | 3+4 2 | 5+6 2 | 3+4 2 | 3+3 2 | 2+2 2 |
| | 5+6 2 | 5+6 2 | 5+6 2 | 5+6 2 | 2+2 2 | 2+2 2 |
| | 5+6 2 | 5+6 2 | 5+6 2 | 5+6 2 | 2+2 2 | 2+2 2 |
| | 5+6 2 | 5+6 2 | 5+6 2 | 5+6 2 | 2+2 2 | 2+2 2 |
| | 5+6 2 | 3+4 2 | 3+4 2 | 3+4 2 | 2+2 2 | 2+2 2 |
| | 3+4 2 | 5+6 2 | 5+6 2 | 5+6 2 | 3+3 2 | 3+3 2 |
| | 5+6 2 | 3+4 2 | 5+6 2 | 3+4 2 | 2+2 2 | 3+3 2 |
| | 3+4 2 | 5+6 2 | 3+4 2 | 5+6 2 | 3+3 2 | 2+2 2 |
| | 5+6 2 | 3+4 2 | 3+4 2 | 5+6 2 | 3+3 2 | 3+3 2 |
| | 5+6 2 | 3+4 2 | 5+6 2 | 5+6 2 | 2+2 2 | 2+2 2 |
| | 3+4 2 | 3+4 2 | 3+4 2 | 3+4 2 | 2+2 2 | 3+3 2 |
| | 3+4 2 | 5+6 2 | 3+4 2 | 3+4 2 | 2+2 2 | 2+2 2 |
| | 3+4 2 | 5+6 2 | 5+6 2 | 3+4 2 | 3+3 2 | 2+2 2 |
| | 5+6 2 | 5+6 2 | 5+6 2 | 5+6 2 | 3+3 2 | 3+3 2 |
| | 5+6 2 | 3+4 2 | 5+6 2 | 3+4 2 | 2+2 2 | 3+3 2 |
| | 5+6 2 | 3+4 2 | 3+4 2 | 5+6 2 | 3+3 2 | 3+3 2 |
| | 3+4 2 | 5+6 2 | 3+4 2 | 5+6 2 | 3+3 2 | 2+2 2 |
| | 5+6 2 | 3+4 2 | 3+4 2 | 5+6 2 | 2+2 2 | 2+2 2 |
| | 5+6 2 | 3+4 2 | 5+6 2 | 3+4 2 | 3+3 2 | 2+2 2 |
| | 5+6 2 | 5+6 2 | 3+4 2 | 5+6 2 | 2+2 2 | 3+3 2 |
| | 3+4 2 | 3+4 2 | 5+6 2 | 3+4 2 | 3+3 2 | 2+2 2 |
| | 3+4 2 | 5+6 2 | 5+6 2 | 3+4 2 | 2+2 2 | 3+3 2 |
| | 3+4 2 | 5+6 2 | 3+4 2 | 5+6 2 | 2+2 2 | 3+3 2 |
| | 5+6 2 | 3+4 2 | 5+6 2 | 3+4 2 | 3+3 2 | 2+2 2 |
| | 5+6 2 | 3+4 2 | 3+4 2 | 5+6 2 | 2+2 2 | 3+3 2 |
| | 3+4 2 | 5+6 2 | 5+6 2 | 3+4 2 | 3+3 2 | 2+2 2 |
| | 3+4 2 | 3+4 2 | 3+4 2 | 5+6 2 | 2+2 2 | 3+3 2 |
| | 5+6 2 | 5+6 2 | 5+6 2 | 3+4 2 | 3+3 2 | 2+2 2 |
| | 3+4 2 | 5+6 2 | 3+4 2 | 5+6 2 | 3+3 2 | 3+3 2 |
| | 5+6 2 | 3+4 2 | 3+4 2 | 5+6 2 | 2+2 2 | 2+2 2 |
| | 3+4 2 | 5+6 2 | 3+4 2 | 3+4 2 | 2+2 2 | 3+3 2 |
| | 5+6 2 | 3+4 2 | 5+6 2 | 5+6 2 | 3+3 2 | 2+2 2 |
| | 3+4 2 | 5+6 2 | 5+6 2 | 3+4 2 | 3+3 2 | 3+3 2 |
| | 5+6 2 | 3+4 2 | 3+4 2 | 5+6 2 | 2+2 2 | 2+2 2 |
| | 5+6 2 | 3+4 2 | 5+6 2 | 3+4 2 | 3+3 2 | 3+3 2 |
| | 5+6 2 | 3+4 2 | 3+4 2 | 5+6 2 | 2+2 2 | 2+2 2 |
| | 5+6 2 | 3+4 2 | 3+4 2 | 5+6 2 | 2+2 2 | 3+3 2 |
| | 3+4 2 | 5+6 2 | 3+4 2 | 5+6 2 | 3+3 2 | 2+2 2 |
| | 5+6 2 | 3+4 2 | 3+4 2 | 5+6 2 | 3+3 2 | 2+2 2 |
| | 5+6 2 | 3+4 2 | 3+4 2 | 5+6 2 | 2+2 2 | 3+3 2 |
| | 5+6 2 | 3+4 2 | 3+4 2 | 5+6 2 | 2+2 2 | 3+3 2 |
| | 5+6 2 | 3+4 2 | 3+4 2 | 5+6 2 | 3+3 2 | 2+2 2 |
| | 5+6 2 | 3+4 2 | 3+4 2 | 5+6 2 | 2+2 2 | 3+3 2 |
| 12.1.57 | 422475 | | 94 | | H3Co2Mo2Na1O10 | |

Continued on next column

**TABLE XXIV– continued**

| TQC | ICSD | NoE | SF | | | |
|---|---|---|---|---|---|---|
| | -A | -GM | -M | -Y | -L | -V |
| | 3+4 2 | 3+4 2 | 3+4 2 | 3+4 2 | 3+3 2 | 3+3 2 |
| | 5+6 2 | 5+6 2 | 5+6 2 | 5+6 2 | 2+2 2 | 2+2 2 |
| | 3+4 2 | 3+4 2 | 3+4 2 | 3+4 2 | 2+2 2 | 2+2 2 |
| | 5+6 2 | 5+6 2 | 5+6 2 | 5+6 2 | 3+3 2 | 3+3 2 |
| | 5+6 2 | 5+6 2 | 5+6 2 | 5+6 2 | 3+3 2 | 3+3 2 |
| | 3+4 2 | 3+4 2 | 3+4 2 | 3+4 2 | 2+2 2 | 2+2 2 |
| | 3+4 2 | 3+4 2 | 5+6 2 | 5+6 2 | 2+2 2 | 3+3 2 |
| | 5+6 2 | 5+6 2 | 3+4 2 | 3+4 2 | 3+3 2 | 2+2 2 |
| | 5+6 2 | 3+4 2 | 3+4 2 | 5+6 2 | 2+2 2 | 3+3 2 |
| | 3+4 2 | 5+6 2 | 5+6 2 | 3+4 2 | 3+3 2 | 2+2 2 |
| | 3+4 2 | 3+4 2 | 3+4 2 | 3+4 2 | 3+3 2 | 3+3 2 |
| | 5+6 2 | 5+6 2 | 5+6 2 | 5+6 2 | 2+2 2 | 2+2 2 |
| | 3+4 2 | 3+4 2 | 5+6 2 | 5+6 2 | 3+3 2 | 3+3 2 |
| | 3+4 2 | 3+4 2 | 3+4 2 | 3+4 2 | 2+2 2 | 2+2 2 |
| | 5+6 2 | 5+6 2 | 5+6 2 | 5+6 2 | 3+3 2 | 3+3 2 |
| | 5+6 2 | 5+6 2 | 3+4 2 | 3+4 2 | 2+2 2 | 2+2 2 |
| | 3+4 2 | 3+4 2 | 3+4 2 | 5+6 2 | 3+3 2 | 3+3 2 |
| | 3+4 2 | 3+4 2 | 5+6 2 | 3+4 2 | 2+2 2 | 2+2 2 |
| | 5+6 2 | 3+4 2 | 3+4 2 | 5+6 2 | 2+2 2 | 3+3 2 |
| | 3+4 2 | 3+4 2 | 5+6 2 | 3+4 2 | 3+3 2 | 2+2 2 |
| | 3+4 2 | 3+4 2 | 3+4 2 | 5+6 2 | 2+2 2 | 3+3 2 |
| | 5+6 2 | 5+6 2 | 3+4 2 | 3+4 2 | 3+3 2 | 2+2 2 |
| | 3+4 2 | 5+6 2 | 5+6 2 | 5+6 2 | 3+3 2 | 2+2 2 |
| | 5+6 2 | 5+6 2 | 5+6 2 | 3+4 2 | 2+2 2 | 3+3 2 |
| | 5+6 2 | 3+4 2 | 5+6 2 | 5+6 2 | 3+3 2 | 3+3 2 |
| | 3+4 2 | 5+6 2 | 3+4 2 | 3+4 2 | 2+2 2 | 2+2 2 |
| | 3+4 2 | 3+4 2 | 5+6 2 | 5+6 2 | 2+2 2 | 2+2 2 |
| | 5+6 2 | 5+6 2 | 5+6 2 | 5+6 2 | 3+3 2 | 3+3 2 |
| | 5+6 2 | 5+6 2 | 5+6 2 | 3+4 2 | 3+3 2 | 3+3 2 |
| | 3+4 2 | 3+4 2 | 3+4 2 | 3+4 2 | 2+2 2 | 2+2 2 |
| | 5+6 2 | 5+6 2 | 3+4 2 | 5+6 2 | 3+3 2 | 2+2 2 |
| | 5+6 2 | 3+4 2 | 3+4 2 | 3+4 2 | 2+2 2 | 3+3 2 |
| | 3+4 2 | 5+6 2 | 5+6 2 | 3+4 2 | 2+2 2 | 2+2 2 |
| | 3+4 2 | 3+4 2 | 5+6 2 | 5+6 2 | 3+3 2 | 3+3 2 |
| | 5+6 2 | 3+4 2 | 3+4 2 | 5+6 2 | 3+3 2 | 2+2 2 |
| | 5+6 2 | 5+6 2 | 3+4 2 | 5+6 2 | 2+2 2 | 3+3 2 |
| | 3+4 2 | 3+4 2 | 3+4 2 | 3+4 2 | 3+3 2 | 2+2 2 |
| | 5+6 2 | 5+6 2 | 5+6 2 | 5+6 2 | 2+2 2 | 3+3 2 |
| | 5+6 2 | 5+6 2 | 3+4 2 | 3+4 2 | 2+2 2 | 2+2 2 |
| | 3+4 2 | 5+6 2 | 5+6 2 | 3+4 2 | 3+3 2 | 3+3 2 |
| | 3+4 2 | 3+4 2 | 5+6 2 | 5+6 2 | 3+3 2 | 3+3 2 |
| | 3+4 2 | 3+4 2 | 5+6 2 | 5+6 2 | 2+2 2 | 2+2 2 |
| | 3+4 2 | 3+4 2 | 5+6 2 | 5+6 2 | 3+3 2 | 3+3 2 |
| | 3+4 2 | 3+4 2 | 5+6 2 | 5+6 2 | 2+2 2 | 2+2 2 |
| | 3+4 2 | 3+4 2 | 5+6 2 | 5+6 2 | 2+2 2 | 2+2 2 |
| | 3+4 2 | 3+4 2 | 5+6 2 | 5+6 2 | 3+3 2 | 3+3 2 |
| | 3+4 2 | 3+4 2 | 5+6 2 | 5+6 2 | 2+2 2 | 2+2 2 |
| 12.1.58 | 24759 | 126 | H2I2O12Sb8 | | | |
| | -A | -GM | -M | -Y | -L | -V |
| | 3+4 2 | 3+4 2 | 5+6 2 | 5+6 2 | 2+2 2 | 2+2 2 |
| | 5+6 2 | 5+6 2 | 3+4 2 | 3+4 2 | 3+3 2 | 3+3 2 |
| | 5+6 2 | 5+6 2 | 5+6 2 | 5+6 2 | 2+2 2 | 2+2 2 |
| | 3+4 2 | 3+4 2 | 3+4 2 | 3+4 2 | 3+3 2 | 3+3 2 |
| | 3+4 2 | 3+4 2 | 3+4 2 | 3+4 2 | 2+2 2 | 2+2 2 |
| | 5+6 2 | 5+6 2 | 5+6 2 | 5+6 2 | 3+3 2 | 3+3 2 |
| | 5+6 2 | 5+6 2 | 5+6 2 | 5+6 2 | 3+3 2 | 3+3 2 |
| | 3+4 2 | 3+4 2 | 3+4 2 | 3+4 2 | 2+2 2 | 2+2 2 |
| | 5+6 2 | 3+4 2 | 3+4 2 | 3+4 2 | 3+3 2 | 3+3 2 |
| | 3+4 2 | 5+6 2 | 3+4 2 | 3+4 2 | 2+2 2 | 2+2 2 |
| | 3+4 2 | 3+4 2 | 5+6 2 | 5+6 2 | 3+3 2 | 3+3 2 |

Continued on next column

**TABLE XXIV– continued**

| TQC | ICSD | NoE | SF | | | |
|---|---|---|---|---|---|---|
| | 5+6 2 | 5+6 2 | 5+6 2 | 5+6 2 | 2+2 2 | 2+2 2 |
| | 5+6 2 | 3+4 2 | 5+6 2 | 3+4 2 | 2+2 2 | 3+3 2 |
| | 3+4 2 | 5+6 2 | 3+4 2 | 5+6 2 | 3+3 2 | 2+2 2 |
| | 3+4 2 | 3+4 2 | 3+4 2 | 3+4 2 | 3+3 2 | 3+3 2 |
| | 5+6 2 | 5+6 2 | 5+6 2 | 5+6 2 | 2+2 2 | 2+2 2 |
| | 5+6 2 | 5+6 2 | 5+6 2 | 5+6 2 | 3+3 2 | 3+3 2 |
| | 3+4 2 | 3+4 2 | 3+4 2 | 3+4 2 | 2+2 2 | 2+2 2 |
| | 3+4 2 | 3+4 2 | 3+4 2 | 3+4 2 | 3+3 2 | 3+3 2 |
| | 5+6 2 | 5+6 2 | 5+6 2 | 5+6 2 | 2+2 2 | 2+2 2 |
| | 5+6 2 | 5+6 2 | 5+6 2 | 5+6 2 | 3+3 2 | 3+3 2 |
| | 3+4 2 | 3+4 2 | 3+4 2 | 3+4 2 | 2+2 2 | 2+2 2 |
| | 5+6 2 | 5+6 2 | 5+6 2 | 5+6 2 | 3+3 2 | 3+3 2 |
| | 5+6 2 | 5+6 2 | 3+4 2 | 3+4 2 | 2+2 2 | 2+2 2 |
| | 3+4 2 | 3+4 2 | 3+4 2 | 5+6 2 | 3+3 2 | 3+3 2 |
| | 3+4 2 | 3+4 2 | 5+6 2 | 3+4 2 | 2+2 2 | 2+2 2 |
| | 3+4 2 | 3+4 2 | 3+4 2 | 3+4 2 | 3+3 2 | 3+3 2 |
| | 5+6 2 | 3+4 2 | 5+6 2 | 5+6 2 | 2+2 2 | 3+3 2 |
| | 3+4 2 | 5+6 2 | 5+6 2 | 5+6 2 | 3+3 2 | 2+2 2 |
| | 5+6 2 | 3+4 2 | 3+4 2 | 3+4 2 | 2+2 2 | 2+2 2 |
| | 5+6 2 | 5+6 2 | 5+6 2 | 5+6 2 | 3+3 2 | 3+3 2 |
| | 3+4 2 | 3+4 2 | 3+4 2 | 5+6 2 | 2+2 2 | 2+2 2 |
| | 3+4 2 | 5+6 2 | 5+6 2 | 3+4 2 | 2+2 2 | 2+2 2 |
| | 5+6 2 | 3+4 2 | 3+4 2 | 3+4 2 | 3+3 2 | 2+2 2 |
| | 3+4 2 | 5+6 2 | 3+4 2 | 3+4 2 | 3+3 2 | 3+3 2 |
| | 3+4 2 | 3+4 2 | 5+6 2 | 5+6 2 | 2+2 2 | 3+3 2 |
| | 5+6 2 | 5+6 2 | 3+4 2 | 3+4 2 | 2+2 2 | 2+2 2 |
| | 5+6 2 | 5+6 2 | 5+6 2 | 5+6 2 | 3+3 2 | 2+2 2 |
| | 3+4 2 | 3+4 2 | 5+6 2 | 5+6 2 | 2+2 2 | 3+3 2 |
| | 5+6 2 | 5+6 2 | 3+4 2 | 3+4 2 | 3+3 2 | 3+3 2 |
| | 3+4 2 | 3+4 2 | 3+4 2 | 3+4 2 | 3+3 2 | 3+3 2 |
| | 5+6 2 | 3+4 2 | 5+6 2 | 5+6 2 | 2+2 2 | 2+2 2 |
| | 3+4 2 | 5+6 2 | 3+4 2 | 3+4 2 | 2+2 2 | 2+2 2 |
| | 3+4 2 | 3+4 2 | 5+6 2 | 5+6 2 | 3+3 2 | 2+2 2 |
| | 5+6 2 | 5+6 2 | 5+6 2 | 5+6 2 | 2+2 2 | 3+3 2 |
| | 5+6 2 | 3+4 2 | 5+6 2 | 3+4 2 | 3+3 2 | 2+2 2 |
| | 3+4 2 | 3+4 2 | 3+4 2 | 3+4 2 | 2+2 2 | 3+3 2 |
| | 5+6 2 | 5+6 2 | 3+4 2 | 5+6 2 | 3+3 2 | 3+3 2 |
| | 3+4 2 | 5+6 2 | 5+6 2 | 5+6 2 | 2+2 2 | 2+2 2 |
| | 5+6 2 | 5+6 2 | 3+4 2 | 3+4 2 | 3+3 2 | 3+3 2 |
| | 3+4 2 | 3+4 2 | 3+4 2 | 3+4 2 | 2+2 2 | 2+2 2 |
| | 3+4 2 | 5+6 2 | 5+6 2 | 5+6 2 | 3+3 2 | 3+3 2 |
| | 5+6 2 | 3+4 2 | 5+6 2 | 3+4 2 | 2+2 2 | 3+3 2 |
| | 5+6 2 | 3+4 2 | 3+4 2 | 5+6 2 | 3+3 2 | 2+2 2 |
| | 3+4 2 | 5+6 2 | 5+6 2 | 3+4 2 | 3+3 2 | 3+3 2 |
| | 5+6 2 | 5+6 2 | 3+4 2 | 5+6 2 | 2+2 2 | 3+3 2 |
| | 5+6 2 | 3+4 2 | 5+6 2 | 3+4 2 | 3+3 2 | 2+2 2 |
| | 3+4 2 | 5+6 2 | 3+4 2 | 5+6 2 | 2+2 2 | 3+3 2 |
| | 5+6 2 | 3+4 2 | 3+4 2 | 5+6 2 | 2+2 2 | 2+2 2 |
| | 3+4 2 | 5+6 2 | 5+6 2 | 3+4 2 | 3+3 2 | 3+3 2 |
| | 3+4 2 | 5+6 2 | 5+6 2 | 3+4 2 | 3+3 2 | 2+2 2 |
| | 5+6 2 | 3+4 2 | 3+4 2 | 5+6 2 | 2+2 2 | 3+3 2 |
| | 3+4 2 | 5+6 2 | 5+6 2 | 3+4 2 | 2+2 2 | 2+2 2 |
| 12.1.59 | 24760 | 126 | H2Br2O12Sb8 | | | |
| | -A | -GM | -M | -Y | -L | -V |
| | 3+4 2 | 3+4 2 | 5+6 2 | 5+6 2 | 2+2 2 | 2+2 2 |
| | 5+6 2 | 5+6 2 | 3+4 2 | 3+4 2 | 3+3 2 | 3+3 2 |
| | 5+6 2 | 5+6 2 | 5+6 2 | 5+6 2 | 2+2 2 | 2+2 2 |
| | 3+4 2 | 3+4 2 | 3+4 2 | 3+4 2 | 3+3 2 | 3+3 2 |
| | 3+4 2 | 3+4 2 | 3+4 2 | 3+4 2 | 2+2 2 | 2+2 2 |
| | 5+6 2 | 5+6 2 | 5+6 2 | 5+6 2 | 3+3 2 | 3+3 2 |
| | 5+6 2 | 5+6 2 | 5+6 2 | 5+6 2 | 3+3 2 | 3+3 2 |

Continued on next column

**TABLE XXIV– continued**

| TQC | ICSD | NoE | SF | | | |
|---|---|---|---|---|---|---|
| | 3+4 2 | 3+4 2 | 3+4 2 | 3+4 2 | 2+2 2 | 2+2 2 |
| | 3+4 2 | 3+4 2 | 3+4 2 | 3+4 2 | 3+3 2 | 3+3 2 |
| | 5+6 2 | 5+6 2 | 3+4 2 | 5+6 2 | 2+2 2 | 2+2 2 |
| | 5+6 2 | 5+6 2 | 5+6 2 | 3+4 2 | 3+3 2 | 3+3 2 |
| | 3+4 2 | 3+4 2 | 5+6 2 | 5+6 2 | 2+2 2 | 2+2 2 |
| | 5+6 2 | 3+4 2 | 5+6 2 | 3+4 2 | 2+2 2 | 3+3 2 |
| | 3+4 2 | 5+6 2 | 3+4 2 | 5+6 2 | 3+3 2 | 2+2 2 |
| | 3+4 2 | 3+4 2 | 3+4 2 | 3+4 2 | 3+3 2 | 3+3 2 |
| | 5+6 2 | 5+6 2 | 5+6 2 | 5+6 2 | 2+2 2 | 2+2 2 |
| | 5+6 2 | 5+6 2 | 5+6 2 | 5+6 2 | 3+3 2 | 3+3 2 |
| | 3+4 2 | 3+4 2 | 3+4 2 | 3+4 2 | 2+2 2 | 2+2 2 |
| | 3+4 2 | 3+4 2 | 3+4 2 | 3+4 2 | 3+3 2 | 3+3 2 |
| | 5+6 2 | 5+6 2 | 5+6 2 | 5+6 2 | 2+2 2 | 2+2 2 |
| | 5+6 2 | 5+6 2 | 5+6 2 | 5+6 2 | 3+3 2 | 2+2 2 |
| | 5+6 2 | 5+6 2 | 3+4 2 | 3+4 2 | 2+2 2 | 3+3 2 |
| | 3+4 2 | 3+4 2 | 3+4 2 | 5+6 2 | 3+3 2 | 3+3 2 |
| | 3+4 2 | 3+4 2 | 5+6 2 | 3+4 2 | 2+2 2 | 2+2 2 |
| | 5+6 2 | 5+6 2 | 3+4 2 | 5+6 2 | 3+3 2 | 3+3 2 |
| | 3+4 2 | 3+4 2 | 5+6 2 | 3+4 2 | 2+2 2 | 2+2 2 |
| | 3+4 2 | 3+4 2 | 3+4 2 | 5+6 2 | 3+3 2 | 3+3 2 |
| | 5+6 2 | 3+4 2 | 5+6 2 | 3+4 2 | 2+2 2 | 3+3 2 |
| | 3+4 2 | 5+6 2 | 3+4 2 | 5+6 2 | 3+3 2 | 2+2 2 |
| | 5+6 2 | 3+4 2 | 5+6 2 | 3+4 2 | 2+2 2 | 2+2 2 |
| | 3+4 2 | 5+6 2 | 5+6 2 | 5+6 2 | 3+3 2 | 3+3 2 |
| | 3+4 2 | 3+4 2 | 5+6 2 | 5+6 2 | 2+2 2 | 2+2 2 |
| | 5+6 2 | 5+6 2 | 3+4 2 | 3+4 2 | 2+2 2 | 2+2 2 |
| | 5+6 2 | 3+4 2 | 3+4 2 | 3+4 2 | 3+3 2 | 2+2 2 |
| | 3+4 2 | 3+4 2 | 3+4 2 | 3+4 2 | 3+3 2 | 3+3 2 |
| | 5+6 2 | 5+6 2 | 5+6 2 | 5+6 2 | 2+2 2 | 3+3 2 |
| | 5+6 2 | 5+6 2 | 3+4 2 | 3+4 2 | 2+2 2 | 2+2 2 |
| | 3+4 2 | 5+6 2 | 5+6 2 | 5+6 2 | 2+2 2 | 2+2 2 |
| | 5+6 2 | 3+4 2 | 5+6 2 | 5+6 2 | 3+3 2 | 3+3 2 |
| | 3+4 2 | 5+6 2 | 3+4 2 | 3+4 2 | 3+3 2 | 3+3 2 |
| | 3+4 2 | 3+4 2 | 3+4 2 | 5+6 2 | 3+3 2 | 3+3 2 |
| | 5+6 2 | 3+4 2 | 5+6 2 | 3+4 2 | 2+2 2 | 2+2 2 |
| | 3+4 2 | 3+4 2 | 5+6 2 | 3+4 2 | 2+2 2 | 2+2 2 |
| | 3+4 2 | 5+6 2 | 3+4 2 | 5+6 2 | 3+3 2 | 2+2 2 |
| | 5+6 2 | 5+6 2 | 5+6 2 | 5+6 2 | 3+3 2 | 3+3 2 |
| | 5+6 2 | 5+6 2 | 3+4 2 | 3+4 2 | 2+2 2 | 2+2 2 |
| | 3+4 2 | 3+4 2 | 5+6 2 | 5+6 2 | 2+2 2 | 3+3 2 |
| | 5+6 2 | 5+6 2 | 3+4 2 | 3+4 2 | 3+3 2 | 3+3 2 |
| | 3+4 2 | 3+4 2 | 5+6 2 | 5+6 2 | 3+3 2 | 2+2 2 |
| | 5+6 2 | 5+6 2 | 3+4 2 | 3+4 2 | 2+2 2 | 3+3 2 |
| | 3+4 2 | 3+4 2 | 5+6 2 | 3+4 2 | 2+2 2 | 2+2 2 |
| | 5+6 2 | 5+6 2 | 3+4 2 | 5+6 2 | 3+3 2 | 3+3 2 |
| | 3+4 2 | 3+4 2 | 3+4 2 | 3+4 2 | 2+2 2 | 3+3 2 |
| | 5+6 2 | 5+6 2 | 5+6 2 | 5+6 2 | 3+3 2 | 3+3 2 |
| | 3+4 2 | 3+4 2 | 5+6 2 | 5+6 2 | 3+3 2 | 2+2 2 |
| | 5+6 2 | 5+6 2 | 3+4 2 | 3+4 2 | 2+2 2 | 2+2 2 |
| | 3+4 2 | 3+4 2 | 5+6 2 | 5+6 2 | 3+3 2 | 3+3 2 |
| | 5+6 2 | 5+6 2 | 3+4 2 | 3+4 2 | 2+2 2 | 2+2 2 |
| | 3+4 2 | 3+4 2 | 3+4 2 | 3+4 2 | 3+3 2 | 3+3 2 |
| | 5+6 2 | 5+6 2 | 5+6 2 | 5+6 2 | 2+2 2 | 3+3 2 |
| | 5+6 2 | 3+4 2 | 5+6 2 | 3+4 2 | 3+3 2 | 2+2 2 |
| | 3+4 2 | 5+6 2 | 5+6 2 | 3+4 2 | 2+2 2 | 3+3 2 |
| | 3+4 2 | 5+6 2 | 3+4 2 | 5+6 2 | 2+2 2 | 2+2 2 |
| 12.1.60 | 16023 | 166 | H12N2O16Zn5 | | | |
| | -A | -GM | -M | -Y | -L | -V |
| | 5+6 2 | 3+4 2 | 3+4 2 | 5+6 2 | 2+2 2 | 2+2 2 |
| | 3+4 2 | 5+6 2 | 5+6 2 | 3+4 2 | 3+3 2 | 3+3 2 |
| | 3+4 2 | 5+6 2 | 5+6 2 | 3+4 2 | 2+2 2 | 2+2 2 |

Continued on next column

**TABLE XXIV– continued**

| TQC | ICSD | NoE | SF | | | |
|---|---|---|---|---|---|---|
| | 5+6 2 | 3+4 2 | 3+4 2 | 5+6 2 | 3+3 2 | 3+3 2 |
| | 3+4 2 | 5+6 2 | 5+6 2 | 3+4 2 | 2+2 2 | 2+2 2 |
| | 5+6 2 | 3+4 2 | 3+4 2 | 5+6 2 | 3+3 2 | 3+3 2 |
| | 5+6 2 | 3+4 2 | 5+6 2 | 3+4 2 | 3+3 2 | 2+2 2 |
| | 3+4 2 | 3+4 2 | 3+4 2 | 3+4 2 | 2+2 2 | 3+3 2 |
| | 5+6 2 | 5+6 2 | 5+6 2 | 5+6 2 | 2+2 2 | 3+3 2 |
| | 3+4 2 | 5+6 2 | 3+4 2 | 5+6 2 | 3+3 2 | 2+2 2 |
| | 5+6 2 | 5+6 2 | 5+6 2 | 5+6 2 | 2+2 2 | 2+2 2 |
| | 3+4 2 | 3+4 2 | 3+4 2 | 3+4 2 | 3+3 2 | 2+2 2 |
| | 5+6 2 | 5+6 2 | 5+6 2 | 5+6 2 | 2+2 2 | 3+3 2 |
| | 3+4 2 | 3+4 2 | 3+4 2 | 3+4 2 | 3+3 2 | 3+3 2 |
| | 3+4 2 | 3+4 2 | 5+6 2 | 5+6 2 | 3+3 2 | 3+3 2 |
| | 5+6 2 | 5+6 2 | 3+4 2 | 3+4 2 | 2+2 2 | 2+2 2 |
| | 5+6 2 | 3+4 2 | 3+4 2 | 5+6 2 | 2+2 2 | 2+2 2 |
| | 3+4 2 | 5+6 2 | 5+6 2 | 3+4 2 | 3+3 2 | 3+3 2 |
| | 5+6 2 | 3+4 2 | 3+4 2 | 5+6 2 | 2+2 2 | 2+2 2 |
| | 3+4 2 | 5+6 2 | 5+6 2 | 3+4 2 | 3+3 2 | 3+3 2 |
| | 5+6 2 | 3+4 2 | 3+4 2 | 5+6 2 | 2+2 2 | 2+2 2 |
| | 3+4 2 | 5+6 2 | 5+6 2 | 3+4 2 | 3+3 2 | 3+3 2 |
| | 5+6 2 | 3+4 2 | 3+4 2 | 5+6 2 | 2+2 2 | 2+2 2 |
| | 3+4 2 | 5+6 2 | 5+6 2 | 3+4 2 | 3+3 2 | 3+3 2 |
| | 5+6 2 | 3+4 2 | 5+6 2 | 3+4 2 | 2+2 2 | 2+2 2 |
| | 3+4 2 | 3+4 2 | 3+4 2 | 3+4 2 | 2+2 2 | 2+2 2 |
| | 3+4 2 | 3+4 2 | 3+4 2 | 3+4 2 | 2+2 2 | 3+3 2 |
| | 3+4 2 | 5+6 2 | 3+4 2 | 5+6 2 | 3+3 2 | 2+2 2 |
| | 5+6 2 | 5+6 2 | 5+6 2 | 5+6 2 | 3+3 2 | 2+2 2 |
| | 3+4 2 | 3+4 2 | 3+4 2 | 3+4 2 | 2+2 2 | 3+3 2 |
| | 5+6 2 | 3+4 2 | 5+6 2 | 3+4 2 | 2+2 2 | 3+3 2 |
| | 3+4 2 | 5+6 2 | 3+4 2 | 5+6 2 | 3+3 2 | 3+3 2 |
| | 3+4 2 | 5+6 2 | 3+4 2 | 5+6 2 | 2+2 2 | 2+2 2 |
| | 5+6 2 | 3+4 2 | 5+6 2 | 3+4 2 | 2+2 2 | 2+2 2 |
| | 3+4 2 | 3+4 2 | 3+4 2 | 3+4 2 | 3+3 2 | 2+2 2 |
| | 5+6 2 | 5+6 2 | 5+6 2 | 5+6 2 | 2+2 2 | 3+3 2 |
| | 5+6 2 | 3+4 2 | 5+6 2 | 3+4 2 | 3+3 2 | 2+2 2 |
| | 3+4 2 | 3+4 2 | 3+4 2 | 3+4 2 | 2+2 2 | 2+2 2 |
| | 3+4 2 | 5+6 2 | 3+4 2 | 5+6 2 | 3+3 2 | 3+3 2 |
| | 5+6 2 | 5+6 2 | 5+6 2 | 3+4 2 | 2+2 2 | 3+3 2 |
| | 3+4 2 | 3+4 2 | 3+4 2 | 5+6 2 | 3+3 2 | 2+2 2 |
| | 5+6 2 | 3+4 2 | 5+6 2 | 3+4 2 | 3+3 2 | 2+2 2 |
| | 3+4 2 | 3+4 2 | 3+4 2 | 3+4 2 | 2+2 2 | 3+3 2 |
| | 3+4 2 | 5+6 2 | 3+4 2 | 5+6 2 | 2+2 2 | 2+2 2 |
| | 5+6 2 | 3+4 2 | 5+6 2 | 3+4 2 | 3+3 2 | 3+3 2 |
| | 3+4 2 | 5+6 2 | 3+4 2 | 5+6 2 | 2+2 2 | 2+2 2 |
| | 5+6 2 | 3+4 2 | 5+6 2 | 3+4 2 | 3+3 2 | 2+2 2 |
| | 3+4 2 | 5+6 2 | 3+4 2 | 5+6 2 | 2+2 2 | 2+2 2 |
| | 3+4 2 | 3+4 2 | 3+4 2 | 3+4 2 | 2+2 2 | 3+3 2 |
| | 3+4 2 | 3+4 2 | 3+4 2 | 3+4 2 | 3+3 2 | 3+3 2 |
| | 3+4 2 | 3+4 2 | 3+4 2 | 3+4 2 | 3+3 2 | 2+2 2 |
| | 5+6 2 | 5+6 2 | 5+6 2 | 5+6 2 | 2+2 2 | 2+2 2 |
| | 5+6 2 | 5+6 2 | 5+6 2 | 5+6 2 | 2+2 2 | 3+3 2 |
| | 3+4 2 | 3+4 2 | 3+4 2 | 3+4 2 | 2+2 2 | 3+3 2 |
| | 5+6 2 | 5+6 2 | 5+6 2 | 3+4 2 | 3+3 2 | 2+2 2 |
| | 3+4 2 | 3+4 2 | 5+6 2 | 5+6 2 | 2+2 2 | 2+2 2 |
| | 5+6 2 | 3+4 2 | 3+4 2 | 3+4 2 | 3+3 2 | 3+3 2 |
| | 5+6 2 | 5+6 2 | 5+6 2 | 5+6 2 | 3+3 2 | 3+3 2 |
| | 3+4 2 | 3+4 2 | 3+4 2 | 3+4 2 | 2+2 2 | 3+3 2 |
| | 3+4 2 | 5+6 2 | 5+6 2 | 5+6 2 | 3+3 2 | 2+2 2 |
| | 5+6 2 | 5+6 2 | 5+6 2 | 3+4 2 | 2+2 2 | 3+3 2 |
| | 5+6 2 | 5+6 2 | 3+4 2 | 5+6 2 | 3+3 2 | 2+2 2 |
| | 3+4 2 | 3+4 2 | 5+6 2 | 3+4 2 | 3+3 2 | 3+3 2 |
| | 3+4 2 | 5+6 2 | 5+6 2 | 3+4 2 | 2+2 2 | 2+2 2 |

Continued on next column

**TABLE XXIV– continued**

| TQC | ICSD | NoE | SF | | | |
|---|---|---|---|---|---|---|
| | 5+6 2 | 5+6 2 | 5+6 2 | 5+6 2 | 3+3 2 | 2+2 2 |
| | 5+6 2 | 3+4 2 | 3+4 2 | 3+4 2 | 2+2 2 | 2+2 2 |
| | 3+4 2 | 3+4 2 | 3+4 2 | 5+6 2 | 3+3 2 | 3+3 2 |
| | 5+6 2 | 3+4 2 | 5+6 2 | 5+6 2 | 2+2 2 | 2+2 2 |
| | 5+6 2 | 5+6 2 | 3+4 2 | 5+6 2 | 2+2 2 | 3+3 2 |
| | 5+6 2 | 5+6 2 | 5+6 2 | 5+6 2 | 3+3 2 | 3+3 2 |
| | 3+4 2 | 3+4 2 | 3+4 2 | 3+4 2 | 3+3 2 | 2+2 2 |
| | 5+6 2 | 5+6 2 | 3+4 2 | 5+6 2 | 3+3 2 | 2+2 2 |
| | 3+4 2 | 5+6 2 | 5+6 2 | 5+6 2 | 2+2 2 | 3+3 2 |
| | 5+6 2 | 3+4 2 | 5+6 2 | 5+6 2 | 2+2 2 | 3+3 2 |
| | 3+4 2 | 5+6 2 | 3+4 2 | 3+4 2 | 2+2 2 | 2+2 2 |
| | 5+6 2 | 3+4 2 | 3+4 2 | 3+4 2 | 3+3 2 | 2+2 2 |
| | 3+4 2 | 3+4 2 | 5+6 2 | 3+4 2 | 3+3 2 | 3+3 2 |
| | 5+6 2 | 5+6 2 | 5+6 2 | 5+6 2 | 2+2 2 | 3+3 2 |
| | 3+4 2 | 3+4 2 | 3+4 2 | 5+6 2 | 2+2 2 | 3+3 2 |
| | 3+4 2 | 5+6 2 | 5+6 2 | 5+6 2 | 3+3 2 | 2+2 2 |
| | 5+6 2 | 3+4 2 | 3+4 2 | 3+4 2 | 2+2 2 | 3+3 2 |
| | 3+4 2 | 5+6 2 | 3+4 2 | 5+6 2 | 2+2 2 | 3+3 2 |
| | 5+6 2 | 3+4 2 | 3+4 2 | 3+4 2 | 3+3 2 | 2+2 2 |
| 12.1.61 | 262220 | 82 | Ge6La4Mg7 | | | |
| | -A | -GM | -M | -Y | -L | -V |
| | 3+4 2 | 3+4 2 | 5+6 2 | 5+6 2 | 2+2 2 | 2+2 2 |
| | 5+6 2 | 5+6 2 | 3+4 2 | 3+4 2 | 3+3 2 | 3+3 2 |
| | 5+6 2 | 3+4 2 | 3+4 2 | 5+6 2 | 2+2 2 | 2+2 2 |
| | 3+4 2 | 5+6 2 | 5+6 2 | 3+4 2 | 3+3 2 | 3+3 2 |
| | 3+4 2 | 3+4 2 | 5+6 2 | 5+6 2 | 2+2 2 | 2+2 2 |
| | 5+6 2 | 5+6 2 | 3+4 2 | 3+4 2 | 3+3 2 | 3+3 2 |
| | 5+6 2 | 3+4 2 | 3+4 2 | 5+6 2 | 2+2 2 | 2+2 2 |
| | 3+4 2 | 5+6 2 | 5+6 2 | 3+4 2 | 3+3 2 | 3+3 2 |
| | 3+4 2 | 3+4 2 | 3+4 2 | 5+6 2 | 3+3 2 | 2+2 2 |
| | 5+6 2 | 5+6 2 | 5+6 2 | 3+4 2 | 2+2 2 | 3+3 2 |
| | 3+4 2 | 3+4 2 | 5+6 2 | 5+6 2 | 2+2 2 | 2+2 2 |
| | 5+6 2 | 3+4 2 | 3+4 2 | 3+4 2 | 3+3 2 | 3+3 2 |
| | 5+6 2 | 5+6 2 | 3+4 2 | 5+6 2 | 2+2 2 | 3+3 2 |
| | 3+4 2 | 5+6 2 | 5+6 2 | 3+4 2 | 3+3 2 | 2+2 2 |
| | 5+6 2 | 3+4 2 | 3+4 2 | 5+6 2 | 2+2 2 | 2+2 2 |
| | 3+4 2 | 5+6 2 | 5+6 2 | 3+4 2 | 3+3 2 | 3+3 2 |
| | 5+6 2 | 3+4 2 | 5+6 2 | 3+4 2 | 2+2 2 | 3+3 2 |
| | 3+4 2 | 5+6 2 | 3+4 2 | 5+6 2 | 3+3 2 | 3+3 2 |
| | 5+6 2 | 3+4 2 | 5+6 2 | 3+4 2 | 3+3 2 | 2+2 2 |
| | 3+4 2 | 5+6 2 | 3+4 2 | 5+6 2 | 2+2 2 | 3+3 2 |
| | 5+6 2 | 3+4 2 | 5+6 2 | 3+4 2 | 2+2 2 | 2+2 2 |
| | 3+4 2 | 5+6 2 | 3+4 2 | 5+6 2 | 3+3 2 | 2+2 2 |
| | 5+6 2 | 3+4 2 | 5+6 2 | 3+4 2 | 2+2 2 | 2+2 2 |
| | 3+4 2 | 5+6 2 | 3+4 2 | 5+6 2 | 3+3 2 | 3+3 2 |
| | 3+4 2 | 3+4 2 | 5+6 2 | 3+4 2 | 3+3 2 | 2+2 2 |
| | 5+6 2 | 5+6 2 | 3+4 2 | 5+6 2 | 2+2 2 | 2+2 2 |
| | 3+4 2 | 3+4 2 | 5+6 2 | 3+4 2 | 3+3 2 | 3+3 2 |
| | 5+6 2 | 5+6 2 | 3+4 2 | 5+6 2 | 2+2 2 | 3+3 2 |
| | 3+4 2 | 5+6 2 | 5+6 2 | 3+4 2 | 3+3 2 | 2+2 2 |
| | 3+4 2 | 3+4 2 | 3+4 2 | 5+6 2 | 2+2 2 | 3+3 2 |
| | 5+6 2 | 5+6 2 | 3+4 2 | 5+6 2 | 2+2 2 | 3+3 2 |
| | 5+6 2 | 5+6 2 | 5+6 2 | 3+4 2 | 3+3 2 | 3+3 2 |
| | 3+4 2 | 3+4 2 | 5+6 2 | 3+4 2 | 2+2 2 | 3+3 2 |
| | 5+6 2 | 3+4 2 | 3+4 2 | 3+4 2 | 3+3 2 | 2+2 2 |
| | 5+6 2 | 3+4 2 | 3+4 2 | 5+6 2 | 3+3 2 | 2+2 2 |
| | 3+4 2 | 3+4 2 | 5+6 2 | 5+6 2 | 2+2 2 | 3+3 2 |
| | 3+4 2 | 3+4 2 | 5+6 2 | 5+6 2 | 3+3 2 | 2+2 2 |
| | 3+4 2 | 5+6 2 | 3+4 2 | 5+6 2 | 2+2 2 | 2+2 2 |
| | 3+4 2 | 5+6 2 | 5+6 2 | 3+4 2 | 3+3 2 | 3+3 2 |
| | 5+6 2 | 5+6 2 | 3+4 2 | 3+4 2 | 2+2 2 | 2+2 2 |

Continued on next column

**TABLE XXIV– continued**

| TQC | ICSD | NoE | SF | | | |
|---|---|---|---|---|---|---|
| | 5+6 2 | 5+6 2 | 5+6 2 | 3+4 2 | 3+3 2 | 3+3 2 |
| 12.1.62 | 637671 | 30 | Ge2Re1 | | | |
| | -A | -GM | -M | -Y | -L | -V |
| | 3+4 2 | 3+4 2 | 5+6 2 | 5+6 2 | 3+3 2 | 3+3 2 |
| | 5+6 2 | 3+4 2 | 3+4 2 | 5+6 2 | 3+3 2 | 2+2 2 |
| | 3+4 2 | 5+6 2 | 3+4 2 | 3+4 2 | 2+2 2 | 3+3 2 |
| | 5+6 2 | 5+6 2 | 5+6 2 | 3+4 2 | 2+2 2 | 2+2 2 |
| | 3+4 2 | 3+4 2 | 3+4 2 | 5+6 2 | 3+3 2 | 2+2 2 |
| | 5+6 2 | 5+6 2 | 5+6 2 | 5+6 2 | 2+2 2 | 3+3 2 |
| | 5+6 2 | 3+4 2 | 5+6 2 | 3+4 2 | 3+3 2 | 3+3 2 |
| | 3+4 2 | 3+4 2 | 3+4 2 | 3+4 2 | 2+2 2 | 2+2 2 |
| | 3+4 2 | 5+6 2 | 5+6 2 | 5+6 2 | 3+3 2 | 2+2 2 |
| | 5+6 2 | 3+4 2 | 3+4 2 | 3+4 2 | 3+3 2 | 2+2 2 |
| | 5+6 2 | 5+6 2 | 3+4 2 | 3+4 2 | 2+2 2 | 2+2 2 |
| | 3+4 2 | 3+4 2 | 5+6 2 | 5+6 2 | 2+2 2 | 3+3 2 |
| | 5+6 2 | 3+4 2 | 5+6 2 | 3+4 2 | 3+3 2 | 3+3 2 |
| | 3+4 2 | 5+6 2 | 3+4 2 | 3+4 2 | 2+2 2 | 3+3 2 |
| | 3+4 2 | 3+4 2 | 5+6 2 | 5+6 2 | 3+3 2 | 3+3 2 |
| 12.1.63 | 183378 | 82 | Gd1O6Ru1Sr2 | | | |
| | -A | -GM | -M | -Y | -L | -V |
| | 3+4 2 | 3+4 2 | 3+4 2 | 3+4 2 | 3+3 2 | 3+3 2 |
| | 3+4 2 | 3+4 2 | 5+6 2 | 5+6 2 | 2+2 2 | 2+2 2 |
| | 5+6 2 | 5+6 2 | 3+4 2 | 3+4 2 | 3+3 2 | 3+3 2 |
| | 5+6 2 | 5+6 2 | 5+6 2 | 5+6 2 | 2+2 2 | 2+2 2 |
| | 5+6 2 | 5+6 2 | 5+6 2 | 5+6 2 | 2+2 2 | 2+2 2 |
| | 5+6 2 | 5+6 2 | 5+6 2 | 5+6 2 | 2+2 2 | 2+2 2 |
| | 3+4 2 | 3+4 2 | 3+4 2 | 3+4 2 | 3+3 2 | 3+3 2 |
| | 3+4 2 | 3+4 2 | 3+4 2 | 3+4 2 | 3+3 2 | 3+3 2 |
| | 5+6 2 | 3+4 2 | 5+6 2 | 3+4 2 | 3+3 2 | 2+2 2 |
| | 3+4 2 | 3+4 2 | 3+4 2 | 5+6 2 | 2+2 2 | 3+3 2 |
| | 5+6 2 | 5+6 2 | 5+6 2 | 3+4 2 | 3+3 2 | 2+2 2 |
| | 3+4 2 | 5+6 2 | 3+4 2 | 5+6 2 | 2+2 2 | 3+3 2 |
| | 5+6 2 | 3+4 2 | 5+6 2 | 5+6 2 | 2+2 2 | 2+2 2 |
| | 5+6 2 | 3+4 2 | 5+6 2 | 3+4 2 | 3+3 2 | 2+2 2 |
| | 3+4 2 | 5+6 2 | 3+4 2 | 5+6 2 | 2+2 2 | 3+3 2 |
| | 5+6 2 | 5+6 2 | 5+6 2 | 3+4 2 | 3+3 2 | 2+2 2 |
| | 3+4 2 | 5+6 2 | 5+6 2 | 5+6 2 | 2+2 2 | 3+3 2 |
| | 5+6 2 | 5+6 2 | 3+4 2 | 5+6 2 | 2+2 2 | 2+2 2 |
| | 5+6 2 | 3+4 2 | 5+6 2 | 3+4 2 | 3+3 2 | 2+2 2 |
| | 5+6 2 | 3+4 2 | 5+6 2 | 3+4 2 | 3+3 2 | 2+2 2 |
| | 5+6 2 | 3+4 2 | 5+6 2 | 3+4 2 | 3+3 2 | 2+2 2 |
| | 5+6 2 | 3+4 2 | 5+6 2 | 3+4 2 | 3+3 2 | 2+2 2 |
| | 5+6 2 | 3+4 2 | 5+6 2 | 3+4 2 | 3+3 2 | 2+2 2 |
| | 3+4 2 | 3+4 2 | 3+4 2 | 3+4 2 | 3+3 2 | 3+3 2 |
| | 3+4 2 | 5+6 2 | 3+4 2 | 5+6 2 | 2+2 2 | 3+3 2 |
| | 5+6 2 | 5+6 2 | 5+6 2 | 5+6 2 | 2+2 2 | 2+2 2 |
| | 3+4 2 | 5+6 2 | 3+4 2 | 5+6 2 | 2+2 2 | 3+3 2 |
| | 3+4 2 | 5+6 2 | 3+4 2 | 5+6 2 | 2+2 2 | 3+3 2 |
| | 3+4 2 | 5+6 2 | 3+4 2 | 3+4 2 | 3+3 2 | 3+3 2 |
| | 3+4 2 | 5+6 2 | 3+4 2 | 3+4 2 | 2+2 2 | 3+3 2 |
| | 3+4 2 | 5+6 2 | 3+4 2 | 5+6 2 | 3+3 2 | 3+3 2 |
| | 3+4 2 | 5+6 2 | 3+4 2 | 5+6 2 | 2+2 2 | 3+3 2 |
| | 3+4 2 | 5+6 2 | 5+6 2 | 5+6 2 | 2+2 2 | 2+2 2 |
| | 5+6 2 | 3+4 2 | 3+4 2 | 5+6 2 | 2+2 2 | 2+2 2 |
| | 5+6 2 | 3+4 2 | 5+6 2 | 5+6 2 | 2+2 2 | 3+3 2 |
| | 5+6 2 | 3+4 2 | 5+6 2 | 3+4 2 | 3+3 2 | 2+2 2 |
| | 5+6 2 | 3+4 2 | 5+6 2 | 3+4 2 | 3+3 2 | 2+2 2 |
| | 5+6 2 | 3+4 2 | 5+6 2 | 5+6 2 | 2+2 2 | 2+2 2 |
| | 5+6 2 | 5+6 2 | 5+6 2 | 5+6 2 | 2+2 2 | 2+2 2 |
| | 5+6 2 | 3+4 2 | 5+6 2 | 5+6 2 | 2+2 2 | 2+2 2 |
| | 5+6 2 | 5+6 2 | 5+6 2 | 5+6 2 | 2+2 2 | 2+2 2 |

Continued on next column

**TABLE XXIV– continued**

| TQC | ICSD | NoE | SF | | | |
|---|---|---|---|---|---|---|
| 12.1.64 | 430563 | 82 | Fe5O7 | | | |
| | -A | -GM | -M | -Y | -L | -V |
| | 3+4 2 | 3+4 2 | 5+6 2 | 5+6 2 | 2+2 2 | 3+3 2 |
| | 5+6 2 | 3+4 2 | 3+4 2 | 3+4 2 | 3+3 2 | 2+2 2 |
| | 5+6 2 | 5+6 2 | 3+4 2 | 5+6 2 | 3+3 2 | 3+3 2 |
| | 3+4 2 | 5+6 2 | 5+6 2 | 3+4 2 | 2+2 2 | 2+2 2 |
| | 3+4 2 | 3+4 2 | 3+4 2 | 5+6 2 | 3+3 2 | 2+2 2 |
| | 5+6 2 | 3+4 2 | 5+6 2 | 3+4 2 | 3+3 2 | 2+2 2 |
| | 5+6 2 | 5+6 2 | 5+6 2 | 3+4 2 | 2+2 2 | 3+3 2 |
| | 3+4 2 | 3+4 2 | 3+4 2 | 3+4 2 | 2+2 2 | 3+3 2 |
| | 5+6 2 | 5+6 2 | 3+4 2 | 5+6 2 | 3+3 2 | 3+3 2 |
| | 3+4 2 | 5+6 2 | 5+6 2 | 5+6 2 | 3+3 2 | 2+2 2 |
| | 3+4 2 | 3+4 2 | 5+6 2 | 3+4 2 | 2+2 2 | 3+3 2 |
| | 5+6 2 | 3+4 2 | 3+4 2 | 5+6 2 | 2+2 2 | 3+3 2 |
| | 3+4 2 | 5+6 2 | 3+4 2 | 5+6 2 | 3+3 2 | 2+2 2 |
| | 5+6 2 | 3+4 2 | 5+6 2 | 3+4 2 | 2+2 2 | 2+2 2 |
| | 3+4 2 | 5+6 2 | 3+4 2 | 5+6 2 | 3+3 2 | 2+2 2 |
| | 3+4 2 | 3+4 2 | 3+4 2 | 5+6 2 | 2+2 2 | 3+3 2 |
| | 3+4 2 | 5+6 2 | 3+4 2 | 3+4 2 | 3+3 2 | 3+3 2 |
| | 5+6 2 | 3+4 2 | 5+6 2 | 3+4 2 | 2+2 2 | 2+2 2 |
| | 5+6 2 | 3+4 2 | 3+4 2 | 5+6 2 | 3+3 2 | 3+3 2 |
| | 5+6 2 | 5+6 2 | 5+6 2 | 5+6 2 | 2+2 2 | 3+3 2 |
| | 3+4 2 | 5+6 2 | 3+4 2 | 5+6 2 | 3+3 2 | 3+3 2 |
| | 3+4 2 | 3+4 2 | 5+6 2 | 3+4 2 | 2+2 2 | 2+2 2 |
| | 5+6 2 | 3+4 2 | 5+6 2 | 3+4 2 | 3+3 2 | 3+3 2 |
| | 3+4 2 | 5+6 2 | 3+4 2 | 5+6 2 | 2+2 2 | 3+3 2 |
| | 3+4 2 | 5+6 2 | 5+6 2 | 3+4 2 | 2+2 2 | 2+2 2 |
| | 5+6 2 | 5+6 2 | 3+4 2 | 5+6 2 | 3+3 2 | 2+2 2 |
| | 5+6 2 | 5+6 2 | 5+6 2 | 3+4 2 | 2+2 2 | 2+2 2 |
| | 3+4 2 | 5+6 2 | 3+4 2 | 5+6 2 | 2+2 2 | 3+3 2 |
| | 5+6 2 | 3+4 2 | 5+6 2 | 3+4 2 | 3+3 2 | 2+2 2 |
| | 3+4 2 | 3+4 2 | 3+4 2 | 3+4 2 | 3+3 2 | 3+3 2 |
| | 5+6 2 | 5+6 2 | 5+6 2 | 5+6 2 | 2+2 2 | 2+2 2 |
| | 3+4 2 | 3+4 2 | 3+4 2 | 5+6 2 | 3+3 2 | 3+3 2 |
| | 5+6 2 | 3+4 2 | 3+4 2 | 3+4 2 | 3+3 2 | 3+3 2 |
| | 5+6 2 | 3+4 2 | 5+6 2 | 5+6 2 | 3+3 2 | 2+2 2 |
| | 3+4 2 | 5+6 2 | 3+4 2 | 3+4 2 | 2+2 2 | 3+3 2 |
| | 3+4 2 | 3+4 2 | 3+4 2 | 3+4 2 | 2+2 2 | 3+3 2 |
| | 5+6 2 | 3+4 2 | 3+4 2 | 3+4 2 | 3+3 2 | 2+2 2 |
| | 3+4 2 | 5+6 2 | 5+6 2 | 3+4 2 | 2+2 2 | 3+3 2 |
| | 3+4 2 | 5+6 2 | 5+6 2 | 3+4 2 | 3+3 2 | 2+2 2 |
| | 5+6 2 | 3+4 2 | 3+4 2 | 3+4 2 | 3+3 2 | 3+3 2 |
| | 3+4 2 | 5+6 2 | 3+4 2 | 5+6 2 | 3+3 2 | 2+2 2 |
| 12.1.65 | 633477 | 48 | Fe3Se4 | | | |
| | -A | -GM | -M | -Y | -L | -V |
| | 3+4 2 | 3+4 2 | 3+4 2 | 5+6 2 | 2+2 2 | 3+3 2 |
| | 5+6 2 | 3+4 2 | 5+6 2 | 3+4 2 | 2+2 2 | 2+2 2 |
| | 3+4 2 | 5+6 2 | 5+6 2 | 5+6 2 | 3+3 2 | 3+3 2 |
| | 5+6 2 | 5+6 2 | 3+4 2 | 3+4 2 | 3+3 2 | 2+2 2 |
| | 3+4 2 | 3+4 2 | 3+4 2 | 3+4 2 | 3+3 2 | 3+3 2 |
| | 5+6 2 | 5+6 2 | 5+6 2 | 5+6 2 | 2+2 2 | 2+2 2 |
| | 3+4 2 | 3+4 2 | 5+6 2 | 5+6 2 | 3+3 2 | 3+3 2 |
| | 5+6 2 | 3+4 2 | 3+4 2 | 3+4 2 | 2+2 2 | 2+2 2 |
| | 3+4 2 | 3+4 2 | 5+6 2 | 5+6 2 | 2+2 2 | 3+3 2 |
| | 3+4 2 | 5+6 2 | 3+4 2 | 3+4 2 | 3+3 2 | 3+3 2 |
| | 5+6 2 | 5+6 2 | 3+4 2 | 3+4 2 | 3+3 2 | 2+2 2 |
| | 3+4 2 | 3+4 2 | 5+6 2 | 5+6 2 | 2+2 2 | 2+2 2 |
| | 5+6 2 | 3+4 2 | 3+4 2 | 5+6 2 | 2+2 2 | 3+3 2 |
| | 3+4 2 | 3+4 2 | 3+4 2 | 5+6 2 | 2+2 2 | 3+3 2 |
| | 5+6 2 | 5+6 2 | 5+6 2 | 3+4 2 | 3+3 2 | 2+2 2 |
| | 5+6 2 | 5+6 2 | 3+4 2 | 3+4 2 | 3+3 2 | 2+2 2 |

Continued on next column

**TABLE XXIV– continued**

| TQC | ICSD | NoE | SF | | | |
|---|---|---|---|---|---|---|
| | 3+4 2 | 3+4 2 | 5+6 2 | 3+4 2 | 3+3 2 | 3+3 2 |
| | 3+4 2 | 3+4 2 | 5+6 2 | 5+6 2 | 2+2 2 | 2+2 2 |
| | 5+6 2 | 3+4 2 | 3+4 2 | 3+4 2 | 3+3 2 | 3+3 2 |
| | 5+6 2 | 3+4 2 | 3+4 2 | 5+6 2 | 3+3 2 | 3+3 2 |
| | 3+4 2 | 5+6 2 | 3+4 2 | 3+4 2 | 3+3 2 | 3+3 2 |
| | 5+6 2 | 5+6 2 | 3+4 2 | 3+4 2 | 2+2 2 | 3+3 2 |
| | 3+4 2 | 3+4 2 | 3+4 2 | 5+6 2 | 3+3 2 | 2+2 2 |
| | 3+4 2 | 5+6 2 | 5+6 2 | 3+4 2 | 3+3 2 | 3+3 2 |
| 12.1.66 | 24560 | 48 | Fe3Se4 | | | |
| | -A | -GM | -M | -Y | -L | -V |
| | 3+4 2 | 3+4 2 | 3+4 2 | 5+6 2 | 3+3 2 | 2+2 2 |
| | 5+6 2 | 5+6 2 | 5+6 2 | 3+4 2 | 2+2 2 | 3+3 2 |
| | 3+4 2 | 3+4 2 | 5+6 2 | 3+4 2 | 3+3 2 | 2+2 2 |
| | 5+6 2 | 5+6 2 | 3+4 2 | 5+6 2 | 2+2 2 | 3+3 2 |
| | 3+4 2 | 3+4 2 | 3+4 2 | 3+4 2 | 2+2 2 | 2+2 2 |
| | 5+6 2 | 5+6 2 | 5+6 2 | 5+6 2 | 3+3 2 | 3+3 2 |
| | 3+4 2 | 3+4 2 | 5+6 2 | 3+4 2 | 2+2 2 | 2+2 2 |
| | 5+6 2 | 3+4 2 | 3+4 2 | 5+6 2 | 3+3 2 | 3+3 2 |
| | 3+4 2 | 5+6 2 | 5+6 2 | 3+4 2 | 3+3 2 | 2+2 2 |
| | 3+4 2 | 3+4 2 | 3+4 2 | 3+4 2 | 2+2 2 | 3+3 2 |
| | 5+6 2 | 5+6 2 | 5+6 2 | 5+6 2 | 3+3 2 | 2+2 2 |
| | 5+6 2 | 3+4 2 | 3+4 2 | 5+6 2 | 2+2 2 | 3+3 2 |
| | 3+4 2 | 3+4 2 | 3+4 2 | 5+6 2 | 3+3 2 | 2+2 2 |
| | 3+4 2 | 3+4 2 | 3+4 2 | 5+6 2 | 2+2 2 | 3+3 2 |
| | 5+6 2 | 5+6 2 | 5+6 2 | 3+4 2 | 3+3 2 | 2+2 2 |
| | 3+4 2 | 3+4 2 | 3+4 2 | 3+4 2 | 2+2 2 | 3+3 2 |
| | 5+6 2 | 5+6 2 | 5+6 2 | 3+4 2 | 2+2 2 | 2+2 2 |
| | 3+4 2 | 3+4 2 | 5+6 2 | 5+6 2 | 3+3 2 | 3+3 2 |
| | 5+6 2 | 5+6 2 | 3+4 2 | 5+6 2 | 2+2 2 | 2+2 2 |
| | 5+6 2 | 3+4 2 | 3+4 2 | 3+4 2 | 2+2 2 | 2+2 2 |
| | 3+4 2 | 3+4 2 | 3+4 2 | 3+4 2 | 2+2 2 | 2+2 2 |
| | 5+6 2 | 5+6 2 | 3+4 2 | 3+4 2 | 3+3 2 | 3+3 2 |
| | 3+4 2 | 3+4 2 | 3+4 2 | 5+6 2 | 2+2 2 | 2+2 2 |
| | 3+4 2 | 5+6 2 | 5+6 2 | 3+4 2 | 2+2 2 | 2+2 2 |
| 12.1.67 | 150568 | 48 | Fe3Se4 | | | |
| | -A | -GM | -M | -Y | -L | -V |
| | 3+4 2 | 3+4 2 | 5+6 2 | 5+6 2 | 3+3 2 | 2+2 2 |
| | 5+6 2 | 3+4 2 | 3+4 2 | 3+4 2 | 3+3 2 | 3+3 2 |
| | 3+4 2 | 5+6 2 | 5+6 2 | 5+6 2 | 2+2 2 | 2+2 2 |
| | 5+6 2 | 5+6 2 | 3+4 2 | 3+4 2 | 2+2 2 | 3+3 2 |
| | 3+4 2 | 3+4 2 | 3+4 2 | 3+4 2 | 2+2 2 | 2+2 2 |
| | 5+6 2 | 5+6 2 | 5+6 2 | 5+6 2 | 3+3 2 | 3+3 2 |
| | 3+4 2 | 3+4 2 | 5+6 2 | 5+6 2 | 2+2 2 | 3+3 2 |
| | 5+6 2 | 3+4 2 | 5+6 2 | 3+4 2 | 3+3 2 | 2+2 2 |
| | 3+4 2 | 3+4 2 | 3+4 2 | 5+6 2 | 2+2 2 | 2+2 2 |
| | 3+4 2 | 5+6 2 | 3+4 2 | 3+4 2 | 3+3 2 | 3+3 2 |
| | 5+6 2 | 5+6 2 | 3+4 2 | 3+4 2 | 2+2 2 | 2+2 2 |
| | 3+4 2 | 3+4 2 | 5+6 2 | 5+6 2 | 3+3 2 | 3+3 2 |
| | 5+6 2 | 3+4 2 | 3+4 2 | 5+6 2 | 3+3 2 | 2+2 2 |
| | 3+4 2 | 3+4 2 | 3+4 2 | 5+6 2 | 3+3 2 | 2+2 2 |
| | 5+6 2 | 5+6 2 | 5+6 2 | 3+4 2 | 2+2 2 | 3+3 2 |
| | 5+6 2 | 5+6 2 | 3+4 2 | 3+4 2 | 2+2 2 | 3+3 2 |
| | 3+4 2 | 3+4 2 | 5+6 2 | 3+4 2 | 2+2 2 | 2+2 2 |
| | 3+4 2 | 3+4 2 | 5+6 2 | 5+6 2 | 3+3 2 | 3+3 2 |
| | 5+6 2 | 3+4 2 | 3+4 2 | 3+4 2 | 2+2 2 | 2+2 2 |
| | 5+6 2 | 3+4 2 | 3+4 2 | 5+6 2 | 2+2 2 | 2+2 2 |
| | 3+4 2 | 5+6 2 | 3+4 2 | 3+4 2 | 2+2 2 | 2+2 2 |
| | 5+6 2 | 5+6 2 | 3+4 2 | 3+4 2 | 3+3 2 | 3+3 2 |
| | 3+4 2 | 3+4 2 | 3+4 2 | 5+6 2 | 2+2 2 | 2+2 2 |
| | 3+4 2 | 5+6 2 | 5+6 2 | 3+4 2 | 3+3 2 | 2+2 2 |
| 12.1.68 | 240050 | 164 | Fe1Zn13 | | | |

Continued on next column

**TABLE XXIV– continued**

| TQC | ICSD -A | -GM | NoE -M | -Y | SF -L | -V |
|---|---|---|---|---|---|---|
| | 3+4 2 | 3+4 2 | 5+6 2 | 5+6 2 | 2+2 2 | 3+3 2 |
| | 5+6 2 | 5+6 2 | 3+4 2 | 3+4 2 | 3+3 2 | 2+2 2 |
| | 3+4 2 | 5+6 2 | 5+6 2 | 3+4 2 | 2+2 2 | 3+3 2 |
| | 5+6 2 | 3+4 2 | 3+4 2 | 5+6 2 | 3+3 2 | 3+3 2 |
| | 5+6 2 | 3+4 2 | 5+6 2 | 5+6 2 | 2+2 2 | 2+2 2 |
| | 3+4 2 | 5+6 2 | 3+4 2 | 3+4 2 | 3+3 2 | 2+2 2 |
| | 3+4 2 | 3+4 2 | 3+4 2 | 3+4 2 | 3+3 2 | 3+3 2 |
| | 5+6 2 | 3+4 2 | 3+4 2 | 5+6 2 | 3+3 2 | 2+2 2 |
| | 3+4 2 | 5+6 2 | 5+6 2 | 3+4 2 | 2+2 2 | 3+3 2 |
| | 5+6 2 | 3+4 2 | 3+4 2 | 5+6 2 | 2+2 2 | 3+3 2 |
| | 5+6 2 | 5+6 2 | 5+6 2 | 3+4 2 | 3+3 2 | 2+2 2 |
| | 3+4 2 | 5+6 2 | 3+4 2 | 5+6 2 | 2+2 2 | 3+3 2 |
| | 5+6 2 | 3+4 2 | 5+6 2 | 3+4 2 | 3+3 2 | 2+2 2 |
| | 5+6 2 | 3+4 2 | 3+4 2 | 5+6 2 | 3+3 2 | 2+2 2 |
| | 3+4 2 | 5+6 2 | 3+4 2 | 3+4 2 | 2+2 2 | 3+3 2 |
| | 3+4 2 | 3+4 2 | 5+6 2 | 3+4 2 | 2+2 2 | 3+3 2 |
| | 5+6 2 | 3+4 2 | 3+4 2 | 5+6 2 | 3+3 2 | 3+3 2 |
| | 3+4 2 | 3+4 2 | 5+6 2 | 3+4 2 | 3+3 2 | 2+2 2 |
| | 5+6 2 | 5+6 2 | 3+4 2 | 3+4 2 | 2+2 2 | 3+3 2 |
| | 3+4 2 | 3+4 2 | 3+4 2 | 5+6 2 | 3+3 2 | 2+2 2 |
| | 5+6 2 | 5+6 2 | 5+6 2 | 3+4 2 | 2+2 2 | 3+3 2 |
| | 3+4 2 | 3+4 2 | 5+6 2 | 5+6 2 | 3+3 2 | 2+2 2 |
| | 5+6 2 | 5+6 2 | 5+6 2 | 3+4 2 | 2+2 2 | 3+3 2 |
| | 3+4 2 | 3+4 2 | 3+4 2 | 3+4 2 | 3+3 2 | 2+2 2 |
| | 3+4 2 | 3+4 2 | 3+4 2 | 5+6 2 | 2+2 2 | 3+3 2 |
| | 5+6 2 | 5+6 2 | 5+6 2 | 5+6 2 | 3+3 2 | 3+3 2 |
| | 3+4 2 | 5+6 2 | 5+6 2 | 5+6 2 | 2+2 2 | 2+2 2 |
| | 3+4 2 | 3+4 2 | 3+4 2 | 3+4 2 | 3+3 2 | 2+2 2 |
| | 5+6 2 | 3+4 2 | 3+4 2 | 5+6 2 | 3+3 2 | 3+3 2 |
| | 3+4 2 | 5+6 2 | 3+4 2 | 3+4 2 | 2+2 2 | 2+2 2 |
| | 5+6 2 | 3+4 2 | 5+6 2 | 3+4 2 | 3+3 2 | 3+3 2 |
| | 3+4 2 | 5+6 2 | 5+6 2 | 5+6 2 | 2+2 2 | 3+3 2 |
| | 3+4 2 | 3+4 2 | 3+4 2 | 3+4 2 | 3+3 2 | 2+2 2 |
| | 3+4 2 | 5+6 2 | 5+6 2 | 5+6 2 | 2+2 2 | 3+3 2 |
| | 3+4 2 | 3+4 2 | 5+6 2 | 3+4 2 | 3+3 2 | 2+2 2 |
| | 5+6 2 | 5+6 2 | 3+4 2 | 5+6 2 | 2+2 2 | 3+3 2 |
| | 3+4 2 | 3+4 2 | 5+6 2 | 3+4 2 | 3+3 2 | 2+2 2 |
| | 5+6 2 | 5+6 2 | 3+4 2 | 5+6 2 | 2+2 2 | 2+2 2 |
| | 5+6 2 | 3+4 2 | 5+6 2 | 3+4 2 | 3+3 2 | 3+3 2 |
| | 3+4 2 | 5+6 2 | 3+4 2 | 5+6 2 | 2+2 2 | 2+2 2 |
| | 3+4 2 | 5+6 2 | 5+6 2 | 3+4 2 | 3+3 2 | 2+2 2 |
| | 5+6 2 | 5+6 2 | 3+4 2 | 5+6 2 | 2+2 2 | 3+3 2 |
| | 3+4 2 | 3+4 2 | 5+6 2 | 5+6 2 | 3+3 2 | 2+2 2 |
| | 5+6 2 | 5+6 2 | 3+4 2 | 3+4 2 | 2+2 2 | 3+3 2 |
| | 3+4 2 | 3+4 2 | 5+6 2 | 5+6 2 | 3+3 2 | 2+2 2 |
| | 5+6 2 | 5+6 2 | 5+6 2 | 3+4 2 | 2+2 2 | 3+3 2 |
| | 3+4 2 | 5+6 2 | 3+4 2 | 3+4 2 | 3+3 2 | 2+2 2 |
| | 5+6 2 | 3+4 2 | 3+4 2 | 5+6 2 | 2+2 2 | 3+3 2 |
| | 3+4 2 | 5+6 2 | 5+6 2 | 3+4 2 | 3+3 2 | 2+2 2 |
| | 5+6 2 | 3+4 2 | 3+4 2 | 5+6 2 | 2+2 2 | 3+3 2 |
| | 3+4 2 | 3+4 2 | 5+6 2 | 5+6 2 | 2+2 2 | 2+2 2 |
| | 5+6 2 | 3+4 2 | 5+6 2 | 3+4 2 | 3+3 2 | 3+3 2 |
| | 3+4 2 | 5+6 2 | 3+4 2 | 3+4 2 | 3+3 2 | 2+2 2 |
| | 3+4 2 | 5+6 2 | 3+4 2 | 5+6 2 | 3+3 2 | 2+2 2 |
| | 5+6 2 | 3+4 2 | 3+4 2 | 5+6 2 | 2+2 2 | 3+3 2 |
| | 5+6 2 | 5+6 2 | 5+6 2 | 3+4 2 | 2+2 2 | 2+2 2 |
| | 3+4 2 | 3+4 2 | 3+4 2 | 3+4 2 | 3+3 2 | 3+3 2 |
| | 3+4 2 | 5+6 2 | 3+4 2 | 3+4 2 | 2+2 2 | 3+3 2 |
| | 5+6 2 | 3+4 2 | 5+6 2 | 5+6 2 | 3+3 2 | 2+2 2 |
| | 5+6 2 | 5+6 2 | 3+4 2 | 3+4 2 | 3+3 2 | 3+3 2 |

Continued on next column

**TABLE XXIV– continued**

| TQC | ICSD | NoE | SF | | | |
|---|---|---|---|---|---|---|
| | 3+4 2 | 5+6 2 | 5+6 2 | 3+4 2 | 2+2 2 | 2+2 2 |
| | 3+4 2 | 3+4 2 | 3+4 2 | 5+6 2 | 3+3 2 | 3+3 2 |
| | 5+6 2 | 3+4 2 | 5+6 2 | 3+4 2 | 2+2 2 | 2+2 2 |
| | 5+6 2 | 3+4 2 | 3+4 2 | 5+6 2 | 2+2 2 | 3+3 2 |
| | 3+4 2 | 3+4 2 | 5+6 2 | 5+6 2 | 3+3 2 | 3+3 2 |
| | 5+6 2 | 3+4 2 | 3+4 2 | 3+4 2 | 3+3 2 | 3+3 2 |
| | 3+4 2 | 3+4 2 | 3+4 2 | 3+4 2 | 3+3 2 | 3+3 2 |
| | 5+6 2 | 5+6 2 | 5+6 2 | 5+6 2 | 2+2 2 | 2+2 2 |
| | 5+6 2 | 5+6 2 | 3+4 2 | 3+4 2 | 3+3 2 | 3+3 2 |
| | 3+4 2 | 3+4 2 | 5+6 2 | 3+4 2 | 2+2 2 | 3+3 2 |
| | 3+4 2 | 3+4 2 | 3+4 2 | 5+6 2 | 3+3 2 | 2+2 2 |
| | 5+6 2 | 5+6 2 | 3+4 2 | 3+4 2 | 2+2 2 | 3+3 2 |
| | 5+6 2 | 3+4 2 | 5+6 2 | 5+6 2 | 3+3 2 | 2+2 2 |
| | 3+4 2 | 5+6 2 | 5+6 2 | 3+4 2 | 2+2 2 | 3+3 2 |
| | 3+4 2 | 3+4 2 | 3+4 2 | 5+6 2 | 3+3 2 | 2+2 2 |
| | 3+4 2 | 3+4 2 | 5+6 2 | 3+4 2 | 2+2 2 | 2+2 2 |
| | 5+6 2 | 3+4 2 | 3+4 2 | 3+4 2 | 3+3 2 | 3+3 2 |
| | 3+4 2 | 5+6 2 | 5+6 2 | 5+6 2 | 2+2 2 | 3+3 2 |
| | 5+6 2 | 3+4 2 | 5+6 2 | 5+6 2 | 2+2 2 | 3+3 2 |
| | 5+6 2 | 5+6 2 | 5+6 2 | 5+6 2 | 2+2 2 | 2+2 2 |
| | 5+6 2 | 3+4 2 | 5+6 2 | 3+4 2 | 3+3 2 | 2+2 2 |
| | 5+6 2 | 5+6 2 | 5+6 2 | 3+4 2 | 2+2 2 | 3+3 2 |
| 12.1.69 | 163222 | 164 | Fe1Zn13 | | | |
| | -A | -GM | -M | -Y | -L | -V |
| | 3+4 2 | 3+4 2 | 5+6 2 | 5+6 2 | 2+2 2 | 3+3 2 |
| | 5+6 2 | 5+6 2 | 3+4 2 | 3+4 2 | 3+3 2 | 2+2 2 |
| | 3+4 2 | 3+4 2 | 5+6 2 | 3+4 2 | 2+2 2 | 3+3 2 |
| | 5+6 2 | 5+6 2 | 3+4 2 | 5+6 2 | 3+3 2 | 3+3 2 |
| | 5+6 2 | 3+4 2 | 5+6 2 | 5+6 2 | 2+2 2 | 2+2 2 |
| | 3+4 2 | 5+6 2 | 3+4 2 | 3+4 2 | 3+3 2 | 2+2 2 |
| | 3+4 2 | 3+4 2 | 3+4 2 | 3+4 2 | 3+3 2 | 3+3 2 |
| | 5+6 2 | 3+4 2 | 3+4 2 | 5+6 2 | 3+3 2 | 2+2 2 |
| | 3+4 2 | 5+6 2 | 5+6 2 | 3+4 2 | 2+2 2 | 3+3 2 |
| | 5+6 2 | 3+4 2 | 3+4 2 | 5+6 2 | 2+2 2 | 3+3 2 |
| | 5+6 2 | 5+6 2 | 5+6 2 | 3+4 2 | 3+3 2 | 2+2 2 |
| | 3+4 2 | 5+6 2 | 5+6 2 | 3+4 2 | 3+3 2 | 3+3 2 |
| | 5+6 2 | 3+4 2 | 3+4 2 | 5+6 2 | 2+2 2 | 2+2 2 |
| | 5+6 2 | 3+4 2 | 3+4 2 | 5+6 2 | 3+3 2 | 3+3 2 |
| | 3+4 2 | 5+6 2 | 3+4 2 | 3+4 2 | 2+2 2 | 2+2 2 |
| | 3+4 2 | 3+4 2 | 3+4 2 | 3+4 2 | 2+2 2 | 3+3 2 |
| | 5+6 2 | 3+4 2 | 5+6 2 | 5+6 2 | 3+3 2 | 3+3 2 |
| | 3+4 2 | 3+4 2 | 5+6 2 | 3+4 2 | 3+3 2 | 2+2 2 |
| | 5+6 2 | 5+6 2 | 3+4 2 | 3+4 2 | 2+2 2 | 3+3 2 |
| | 3+4 2 | 3+4 2 | 3+4 2 | 5+6 2 | 3+3 2 | 2+2 2 |
| | 3+4 2 | 3+4 2 | 5+6 2 | 3+4 2 | 2+2 2 | 3+3 2 |
| | 5+6 2 | 5+6 2 | 5+6 2 | 5+6 2 | 3+3 2 | 3+3 2 |
| | 3+4 2 | 5+6 2 | 3+4 2 | 3+4 2 | 2+2 2 | 2+2 2 |
| | 5+6 2 | 3+4 2 | 5+6 2 | 3+4 2 | 3+3 2 | 2+2 2 |
| | 3+4 2 | 3+4 2 | 3+4 2 | 5+6 2 | 2+2 2 | 3+3 2 |
| | 5+6 2 | 5+6 2 | 5+6 2 | 5+6 2 | 3+3 2 | 3+3 2 |
| | 3+4 2 | 5+6 2 | 3+4 2 | 3+4 2 | 2+2 2 | 2+2 2 |
| | 3+4 2 | 3+4 2 | 5+6 2 | 5+6 2 | 3+3 2 | 2+2 2 |
| | 5+6 2 | 3+4 2 | 3+4 2 | 5+6 2 | 3+3 2 | 3+3 2 |
| | 3+4 2 | 5+6 2 | 5+6 2 | 3+4 2 | 2+2 2 | 2+2 2 |
| | 5+6 2 | 3+4 2 | 3+4 2 | 3+4 2 | 3+3 2 | 3+3 2 |
| | 3+4 2 | 5+6 2 | 5+6 2 | 5+6 2 | 2+2 2 | 3+3 2 |
| | 3+4 2 | 3+4 2 | 5+6 2 | 3+4 2 | 3+3 2 | 2+2 2 |
| | 3+4 2 | 3+4 2 | 3+4 2 | 5+6 2 | 2+2 2 | 3+3 2 |
| | 3+4 2 | 5+6 2 | 5+6 2 | 5+6 2 | 3+3 2 | 2+2 2 |
| | 5+6 2 | 5+6 2 | 3+4 2 | 3+4 2 | 2+2 2 | 3+3 2 |
| | 3+4 2 | 3+4 2 | 5+6 2 | 3+4 2 | 3+3 2 | 2+2 2 |

Continued on next column

**TABLE XXIV– continued**

| TQC | ICSD NoE SF |
|---|---|
| | 5+6 2 5+6 2 3+4 2 5+6 2 2+2 2 2+2 2 |
| | 3+4 2 3+4 2 5+6 2 3+4 2 3+3 2 3+3 2 |
| | 5+6 2 5+6 2 3+4 2 5+6 2 2+2 2 2+2 2 |
| | 3+4 2 5+6 2 5+6 2 3+4 2 3+3 2 2+2 2 |
| | 5+6 2 5+6 2 3+4 2 5+6 2 2+2 2 3+3 2 |
| | 3+4 2 3+4 2 5+6 2 3+4 2 3+3 2 2+2 2 |
| | 5+6 2 5+6 2 3+4 2 5+6 2 2+2 2 3+3 2 |
| | 3+4 2 3+4 2 5+6 2 5+6 2 3+3 2 2+2 2 |
| | 5+6 2 5+6 2 5+6 2 3+4 2 2+2 2 3+3 2 |
| | 3+4 2 5+6 2 3+4 2 3+4 2 3+3 2 2+2 2 |
| | 5+6 2 3+4 2 3+4 2 5+6 2 2+2 2 3+3 2 |
| | 3+4 2 3+4 2 5+6 2 3+4 2 3+3 2 2+2 2 |
| | 5+6 2 5+6 2 5+6 2 5+6 2 2+2 2 3+3 2 |
| | 3+4 2 3+4 2 3+4 2 5+6 2 2+2 2 2+2 2 |
| | 5+6 2 3+4 2 5+6 2 3+4 2 3+3 2 3+3 2 |
| | 3+4 2 5+6 2 3+4 2 3+4 2 3+3 2 2+2 2 |
| | 3+4 2 5+6 2 3+4 2 5+6 2 3+3 2 2+2 2 |
| | 5+6 2 3+4 2 3+4 2 5+6 2 2+2 2 3+3 2 |
| | 5+6 2 5+6 2 5+6 2 3+4 2 2+2 2 2+2 2 |
| | 3+4 2 3+4 2 3+4 2 3+4 2 3+3 2 3+3 2 |
| | 3+4 2 5+6 2 3+4 2 3+4 2 2+2 2 3+3 2 |
| | 5+6 2 3+4 2 5+6 2 5+6 2 3+3 2 2+2 2 |
| | 5+6 2 5+6 2 3+4 2 3+4 2 3+3 2 3+3 2 |
| | 3+4 2 3+4 2 5+6 2 3+4 2 2+2 2 2+2 2 |
| | 5+6 2 5+6 2 3+4 2 5+6 2 3+3 2 3+3 2 |
| | 3+4 2 3+4 2 5+6 2 3+4 2 2+2 2 2+2 2 |
| | 5+6 2 3+4 2 3+4 2 5+6 2 2+2 2 3+3 2 |
| | 3+4 2 3+4 2 5+6 2 5+6 2 3+3 2 3+3 2 |
| | 5+6 2 3+4 2 3+4 2 3+4 2 3+3 2 3+3 2 |
| | 3+4 2 3+4 2 3+4 2 3+4 2 3+3 2 3+3 2 |
| | 5+6 2 5+6 2 5+6 2 5+6 2 2+2 2 2+2 2 |
| | 5+6 2 5+6 2 3+4 2 3+4 2 3+3 2 3+3 2 |
| | 3+4 2 3+4 2 5+6 2 3+4 2 2+2 2 3+3 2 |
| | 3+4 2 3+4 2 3+4 2 5+6 2 3+3 2 2+2 2 |
| | 5+6 2 5+6 2 3+4 2 3+4 2 2+2 2 3+3 2 |
| | 5+6 2 3+4 2 5+6 2 5+6 2 3+3 2 2+2 2 |
| | 3+4 2 5+6 2 3+4 2 3+4 2 2+2 2 3+3 2 |
| | 3+4 2 3+4 2 5+6 2 5+6 2 3+3 2 2+2 2 |
| | 3+4 2 3+4 2 5+6 2 3+4 2 2+2 2 2+2 2 |
| | 5+6 2 3+4 2 3+4 2 3+4 2 3+3 2 3+3 2 |
| | 3+4 2 5+6 2 5+6 2 5+6 2 2+2 2 3+3 2 |
| | 5+6 2 3+4 2 5+6 2 5+6 2 2+2 2 3+3 2 |
| | 5+6 2 5+6 2 5+6 2 5+6 2 2+2 2 2+2 2 |
| | 5+6 2 3+4 2 5+6 2 3+4 2 3+3 2 2+2 2 |
| | 5+6 2 5+6 2 5+6 2 3+4 2 2+2 2 3+3 2 |
| 12.1.70 | 633326 72 Fe1S8Ti4 |
| | -A -GM -M -Y -L -V |
| | 3+4 2 3+4 2 5+6 2 5+6 2 3+3 2 3+3 2 |
| | 5+6 2 5+6 2 3+4 2 3+4 2 2+2 2 2+2 2 |
| | 5+6 2 5+6 2 5+6 2 3+4 2 2+2 2 2+2 2 |
| | 3+4 2 3+4 2 5+6 2 5+6 2 3+3 2 3+3 2 |
| | 3+4 2 3+4 2 3+4 2 5+6 2 2+2 2 3+3 2 |
| | 5+6 2 5+6 2 3+4 2 3+4 2 3+3 2 2+2 2 |
| | 5+6 2 5+6 2 5+6 2 3+4 2 3+3 2 2+2 2 |
| | 3+4 2 3+4 2 3+4 2 5+6 2 2+2 2 3+3 2 |
| | 3+4 2 3+4 2 3+4 2 3+4 2 3+3 2 3+3 2 |
| | 5+6 2 5+6 2 5+6 2 3+4 2 3+3 2 2+2 2 |
| | 5+6 2 5+6 2 3+4 2 3+4 2 2+2 2 3+3 2 |
| | 3+4 2 3+4 2 3+4 2 5+6 2 2+2 2 3+3 2 |
| | 3+4 2 3+4 2 5+6 2 3+4 2 2+2 2 2+2 2 |
| | 5+6 2 3+4 2 5+6 2 5+6 2 3+3 2 2+2 2 |

Continued on next column

**TABLE XXIV– continued**

| TQC | ICSD NoE SF |
|---|---|
| | 3+4 2 5+6 2 5+6 2 5+6 2 3+3 2 3+3 2 |
| | 3+4 2 3+4 2 5+6 2 5+6 2 2+2 2 2+2 2 |
| | 5+6 2 3+4 2 3+4 2 5+6 2 2+2 2 3+3 2 |
| | 3+4 2 5+6 2 3+4 2 3+4 2 3+3 2 2+2 2 |
| | 5+6 2 5+6 2 3+4 2 5+6 2 3+3 2 3+3 2 |
| | 3+4 2 5+6 2 5+6 2 3+4 2 3+3 2 2+2 2 |
| | 3+4 2 3+4 2 5+6 2 5+6 2 3+3 2 3+3 2 |
| | 3+4 2 3+4 2 3+4 2 3+4 2 2+2 2 2+2 2 |
| | 5+6 2 5+6 2 5+6 2 5+6 2 2+2 2 3+3 2 |
| | 5+6 2 5+6 2 5+6 2 5+6 2 2+2 2 2+2 2 |
| | 3+4 2 3+4 2 3+4 2 3+4 2 2+2 2 3+3 2 |
| | 5+6 2 3+4 2 5+6 2 3+4 2 3+3 2 3+3 2 |
| | 5+6 2 5+6 2 3+4 2 3+4 2 3+3 2 2+2 2 |
| | 5+6 2 3+4 2 3+4 2 5+6 2 3+3 2 2+2 2 |
| | 3+4 2 3+4 2 5+6 2 3+4 2 2+2 2 3+3 2 |
| | 3+4 2 3+4 2 5+6 2 3+4 2 3+3 2 2+2 2 |
| | 5+6 2 5+6 2 3+4 2 5+6 2 2+2 2 3+3 2 |
| | 5+6 2 3+4 2 3+4 2 5+6 2 2+2 2 2+2 2 |
| | 3+4 2 5+6 2 3+4 2 3+4 2 3+3 2 3+3 2 |
| | 3+4 2 3+4 2 3+4 2 3+4 2 3+3 2 3+3 2 |
| | 3+4 2 5+6 2 3+4 2 3+4 2 3+3 2 3+3 2 |
| | 3+4 2 5+6 2 5+6 2 3+4 2 2+2 2 3+3 2 |
| 12.1.71 | 53534 72 Fe1S8Ti4 |
| | -A -GM -M -Y -L -V |
| | 3+4 2 3+4 2 5+6 2 5+6 2 3+3 2 3+3 2 |
| | 5+6 2 5+6 2 3+4 2 3+4 2 2+2 2 2+2 2 |
| | 5+6 2 5+6 2 5+6 2 3+4 2 2+2 2 2+2 2 |
| | 3+4 2 3+4 2 5+6 2 5+6 2 3+3 2 3+3 2 |
| | 5+6 2 3+4 2 3+4 2 5+6 2 3+3 2 2+2 2 |
| | 3+4 2 5+6 2 3+4 2 3+4 2 2+2 2 3+3 2 |
| | 5+6 2 5+6 2 5+6 2 3+4 2 2+2 2 3+3 2 |
| | 3+4 2 3+4 2 3+4 2 5+6 2 3+3 2 2+2 2 |
| | 3+4 2 3+4 2 3+4 2 3+4 2 3+3 2 3+3 2 |
| | 5+6 2 5+6 2 5+6 2 3+4 2 2+2 2 3+3 2 |
| | 5+6 2 5+6 2 3+4 2 3+4 2 3+3 2 2+2 2 |
| | 3+4 2 3+4 2 3+4 2 5+6 2 3+3 2 2+2 2 |
| | 3+4 2 3+4 2 5+6 2 3+4 2 2+2 2 2+2 2 |
| | 5+6 2 3+4 2 5+6 2 5+6 2 2+2 2 3+3 2 |
| | 3+4 2 5+6 2 5+6 2 5+6 2 3+3 2 3+3 2 |
| | 3+4 2 3+4 2 5+6 2 5+6 2 2+2 2 2+2 2 |
| | 5+6 2 3+4 2 3+4 2 5+6 2 3+3 2 2+2 2 |
| | 3+4 2 5+6 2 3+4 2 3+4 2 2+2 2 3+3 2 |
| | 5+6 2 5+6 2 3+4 2 5+6 2 3+3 2 3+3 2 |
| | 3+4 2 5+6 2 5+6 2 3+4 2 2+2 2 3+3 2 |
| | 3+4 2 3+4 2 5+6 2 5+6 2 3+3 2 3+3 2 |
| | 3+4 2 3+4 2 3+4 2 3+4 2 2+2 2 2+2 2 |
| | 5+6 2 5+6 2 5+6 2 5+6 2 3+3 2 2+2 2 |
| | 5+6 2 5+6 2 5+6 2 5+6 2 2+2 2 2+2 2 |
| | 3+4 2 3+4 2 3+4 2 3+4 2 3+3 2 2+2 2 |
| | 5+6 2 3+4 2 5+6 2 3+4 2 3+3 2 3+3 2 |
| | 5+6 2 5+6 2 3+4 2 3+4 2 2+2 2 3+3 2 |
| | 5+6 2 3+4 2 3+4 2 5+6 2 2+2 2 3+3 2 |
| | 3+4 2 3+4 2 5+6 2 3+4 2 3+3 2 2+2 2 |
| | 5+6 2 3+4 2 5+6 2 3+4 2 2+2 2 3+3 2 |
| | 3+4 2 5+6 2 3+4 2 5+6 2 3+3 2 2+2 2 |
| | 5+6 2 3+4 2 3+4 2 5+6 2 2+2 2 2+2 2 |
| | 3+4 2 5+6 2 3+4 2 3+4 2 3+3 2 3+3 2 |
| | 3+4 2 3+4 2 3+4 2 3+4 2 3+3 2 3+3 2 |
| | 3+4 2 5+6 2 3+4 2 3+4 2 3+3 2 3+3 2 |
| | 3+4 2 5+6 2 5+6 2 3+4 2 3+3 2 2+2 2 |
| 12.1.72 | 42563 40 Fe1S4Ti2 |

Continued on next column

**TABLE XXIV– continued**

| TQC | ICSD | | NoE | SF | | |
|---|---|---|---|---|---|---|
| | -A | -GM | -M | -Y | -L | -V |
| | 3+4 2 | 3+4 2 | 3+4 2 | 5+6 2 | 2+2 2 | 2+2 2 |
| | 5+6 2 | 5+6 2 | 5+6 2 | 3+4 2 | 3+3 2 | 3+3 2 |
| | 5+6 2 | 5+6 2 | 3+4 2 | 5+6 2 | 2+2 2 | 3+3 2 |
| | 3+4 2 | 3+4 2 | 5+6 2 | 3+4 2 | 3+3 2 | 2+2 2 |
| | 3+4 2 | 3+4 2 | 3+4 2 | 3+4 2 | 3+3 2 | 3+3 2 |
| | 5+6 2 | 5+6 2 | 5+6 2 | 5+6 2 | 2+2 2 | 2+2 2 |
| | 3+4 2 | 3+4 2 | 3+4 2 | 3+4 2 | 3+3 2 | 2+2 2 |
| | 5+6 2 | 3+4 2 | 5+6 2 | 5+6 2 | 2+2 2 | 3+3 2 |
| | 3+4 2 | 3+4 2 | 3+4 2 | 3+4 2 | 3+3 2 | 3+3 2 |
| | 3+4 2 | 5+6 2 | 3+4 2 | 5+6 2 | 2+2 2 | 3+3 2 |
| | 5+6 2 | 5+6 2 | 5+6 2 | 3+4 2 | 3+3 2 | 2+2 2 |
| | 5+6 2 | 3+4 2 | 5+6 2 | 5+6 2 | 2+2 2 | 2+2 2 |
| | 3+4 2 | 3+4 2 | 3+4 2 | 5+6 2 | 2+2 2 | 3+3 2 |
| | 5+6 2 | 3+4 2 | 3+4 2 | 5+6 2 | 3+3 2 | 2+2 2 |
| | 3+4 2 | 5+6 2 | 5+6 2 | 3+4 2 | 3+3 2 | 3+3 2 |
| | 5+6 2 | 5+6 2 | 3+4 2 | 3+4 2 | 2+2 2 | 2+2 2 |
| | 3+4 2 | 5+6 2 | 5+6 2 | 3+4 2 | 3+3 2 | 3+3 2 |
| | 3+4 2 | 3+4 2 | 3+4 2 | 3+4 2 | 3+3 2 | 3+3 2 |
| | 3+4 2 | 3+4 2 | 3+4 2 | 3+4 2 | 2+2 2 | 3+3 2 |
| | 3+4 2 | 5+6 2 | 5+6 2 | 3+4 2 | 3+3 2 | 3+3 2 |
| 12.1.73 | 103658 | | 160 | Fe17Th2 | | |
| | -A | -GM | -M | -Y | -L | -V |
| | 5+6 2 | 3+4 2 | 5+6 2 | 3+4 2 | 2+2 2 | 3+3 2 |
| | 3+4 2 | 5+6 2 | 3+4 2 | 5+6 2 | 3+3 2 | 2+2 2 |
| | 5+6 2 | 3+4 2 | 5+6 2 | 3+4 2 | 2+2 2 | 3+3 2 |
| | 3+4 2 | 5+6 2 | 3+4 2 | 5+6 2 | 3+3 2 | 2+2 2 |
| | 5+6 2 | 3+4 2 | 3+4 2 | 3+4 2 | 3+3 2 | 3+3 2 |
| | 3+4 2 | 5+6 2 | 5+6 2 | 5+6 2 | 2+2 2 | 2+2 2 |
| | 5+6 2 | 3+4 2 | 5+6 2 | 3+4 2 | 2+2 2 | 3+3 2 |
| | 3+4 2 | 5+6 2 | 3+4 2 | 5+6 2 | 3+3 2 | 2+2 2 |
| | 5+6 2 | 3+4 2 | 5+6 2 | 3+4 2 | 3+3 2 | 2+2 2 |
| | 3+4 2 | 5+6 2 | 3+4 2 | 5+6 2 | 2+2 2 | 3+3 2 |
| | 3+4 2 | 5+6 2 | 5+6 2 | 5+6 2 | 2+2 2 | 3+3 2 |
| | 5+6 2 | 3+4 2 | 5+6 2 | 3+4 2 | 3+3 2 | 2+2 2 |
| | 3+4 2 | 5+6 2 | 3+4 2 | 5+6 2 | 2+2 2 | 2+2 2 |
| | 3+4 2 | 3+4 2 | 3+4 2 | 3+4 2 | 3+3 2 | 2+2 2 |
| | 5+6 2 | 3+4 2 | 5+6 2 | 5+6 2 | 3+3 2 | 3+3 2 |
| | 5+6 2 | 3+4 2 | 5+6 2 | 3+4 2 | 3+3 2 | 2+2 2 |
| | 3+4 2 | 5+6 2 | 3+4 2 | 3+4 2 | 2+2 2 | 3+3 2 |
| | 5+6 2 | 3+4 2 | 5+6 2 | 5+6 2 | 2+2 2 | 3+3 2 |
| | 5+6 2 | 3+4 2 | 3+4 2 | 3+4 2 | 3+3 2 | 2+2 2 |
| | 3+4 2 | 3+4 2 | 5+6 2 | 3+4 2 | 3+3 2 | 2+2 2 |
| | 3+4 2 | 3+4 2 | 3+4 2 | 3+4 2 | 2+2 2 | 3+3 2 |
| | 5+6 2 | 5+6 2 | 5+6 2 | 5+6 2 | 3+3 2 | 2+2 2 |
| | 5+6 2 | 3+4 2 | 5+6 2 | 5+6 2 | 2+2 2 | 3+3 2 |
| | 5+6 2 | 5+6 2 | 5+6 2 | 3+4 2 | 2+2 2 | 3+3 2 |
| | 3+4 2 | 3+4 2 | 3+4 2 | 3+4 2 | 3+3 2 | 2+2 2 |
| | 3+4 2 | 5+6 2 | 5+6 2 | 5+6 2 | 3+3 2 | 2+2 2 |
| | 5+6 2 | 5+6 2 | 3+4 2 | 5+6 2 | 3+3 2 | 2+2 2 |
| | 5+6 2 | 3+4 2 | 5+6 2 | 3+4 2 | 2+2 2 | 2+2 2 |
| | 3+4 2 | 3+4 2 | 5+6 2 | 5+6 2 | 3+3 2 | 3+3 2 |
| | 5+6 2 | 5+6 2 | 3+4 2 | 3+4 2 | 3+3 2 | 2+2 2 |
| | 3+4 2 | 3+4 2 | 5+6 2 | 5+6 2 | 2+2 2 | 3+3 2 |
| | 5+6 2 | 3+4 2 | 5+6 2 | 3+4 2 | 2+2 2 | 3+3 2 |
| | 3+4 2 | 5+6 2 | 3+4 2 | 5+6 2 | 3+3 2 | 3+3 2 |
| | 5+6 2 | 3+4 2 | 5+6 2 | 5+6 2 | 2+2 2 | 2+2 2 |
| | 5+6 2 | 5+6 2 | 3+4 2 | 5+6 2 | 2+2 2 | 3+3 2 |
| | 3+4 2 | 5+6 2 | 5+6 2 | 3+4 2 | 3+3 2 | 2+2 2 |
| | 3+4 2 | 3+4 2 | 3+4 2 | 5+6 2 | 2+2 2 | 2+2 2 |
| | 3+4 2 | 3+4 2 | 3+4 2 | 5+6 2 | 3+3 2 | 2+2 2 |
| | 5+6 2 | 3+4 2 | 3+4 2 | 5+6 2 | 2+2 2 | 3+3 2 |
| | 3+4 2 | 5+6 2 | 5+6 2 | 3+4 2 | 3+3 2 | 2+2 2 |
| | 3+4 2 | 5+6 2 | 5+6 2 | 5+6 2 | 2+2 2 | 3+3 2 |
| | 3+4 2 | 3+4 2 | 5+6 2 | 5+6 2 | 3+3 2 | 2+2 2 |
| | 5+6 2 | 3+4 2 | 3+4 2 | 3+4 2 | 3+3 2 | 3+3 2 |
| | 3+4 2 | 5+6 2 | 5+6 2 | 3+4 2 | 2+2 2 | 3+3 2 |
| | 3+4 2 | 3+4 2 | 5+6 2 | 3+4 2 | 3+3 2 | 2+2 2 |
| | 5+6 2 | 5+6 2 | 3+4 2 | 5+6 2 | 2+2 2 | 3+3 2 |
| | 3+4 2 | 3+4 2 | 3+4 2 | 5+6 2 | 3+3 2 | 3+3 2 |
| | 3+4 2 | 5+6 2 | 3+4 2 | 3+4 2 | 2+2 2 | 2+2 2 |
| | 5+6 2 | 5+6 2 | 5+6 2 | 5+6 2 | 2+2 2 | 3+3 2 |
| | 5+6 2 | 3+4 2 | 5+6 2 | 3+4 2 | 3+3 2 | 3+3 2 |
| | 5+6 2 | 5+6 2 | 3+4 2 | 5+6 2 | 3+3 2 | 2+2 2 |
| | 5+6 2 | 5+6 2 | 5+6 2 | 3+4 2 | 2+2 2 | 2+2 2 |
| | 3+4 2 | 3+4 2 | 5+6 2 | 3+4 2 | 3+3 2 | 2+2 2 |
| | 3+4 2 | 3+4 2 | 3+4 2 | 5+6 2 | 2+2 2 | 3+3 2 |
| | 3+4 2 | 3+4 2 | 3+4 2 | 3+4 2 | 2+2 2 | 3+3 2 |
| | 5+6 2 | 5+6 2 | 3+4 2 | 5+6 2 | 3+3 2 | 2+2 2 |
| | 5+6 2 | 3+4 2 | 5+6 2 | 3+4 2 | 2+2 2 | 3+3 2 |
| | 3+4 2 | 5+6 2 | 5+6 2 | 5+6 2 | 3+3 2 | 2+2 2 |
| | 5+6 2 | 3+4 2 | 3+4 2 | 5+6 2 | 2+2 2 | 3+3 2 |
| | 5+6 2 | 5+6 2 | 5+6 2 | 5+6 2 | 3+3 2 | 2+2 2 |
| | 5+6 2 | 3+4 2 | 5+6 2 | 5+6 2 | 2+2 2 | 3+3 2 |
| | 5+6 2 | 3+4 2 | 5+6 2 | 3+4 2 | 3+3 2 | 2+2 2 |
| | 3+4 2 | 3+4 2 | 3+4 2 | 3+4 2 | 3+3 2 | 2+2 2 |
| | 3+4 2 | 5+6 2 | 3+4 2 | 5+6 2 | 2+2 2 | 3+3 2 |
| | 5+6 2 | 5+6 2 | 3+4 2 | 3+4 2 | 2+2 2 | 2+2 2 |
| | 3+4 2 | 5+6 2 | 3+4 2 | 5+6 2 | 3+3 2 | 3+3 2 |
| | 3+4 2 | 3+4 2 | 5+6 2 | 3+4 2 | 2+2 2 | 2+2 2 |
| | 5+6 2 | 3+4 2 | 5+6 2 | 3+4 2 | 3+3 2 | 2+2 2 |
| | 3+4 2 | 5+6 2 | 5+6 2 | 3+4 2 | 3+3 2 | 3+3 2 |
| | 5+6 2 | 3+4 2 | 5+6 2 | 5+6 2 | 2+2 2 | 3+3 2 |
| | 3+4 2 | 5+6 2 | 3+4 2 | 5+6 2 | 2+2 2 | 2+2 2 |
| | 3+4 2 | 3+4 2 | 3+4 2 | 3+4 2 | 3+3 2 | 3+3 2 |
| | 3+4 2 | 5+6 2 | 5+6 2 | 5+6 2 | 2+2 2 | 3+3 2 |
| | 5+6 2 | 3+4 2 | 5+6 2 | 3+4 2 | 3+3 2 | 3+3 2 |
| | 3+4 2 | 5+6 2 | 3+4 2 | 5+6 2 | 3+3 2 | 2+2 2 |
| | 5+6 2 | 3+4 2 | 3+4 2 | 3+4 2 | 2+2 2 | 2+2 2 |
| | 3+4 2 | 3+4 2 | 3+4 2 | 5+6 2 | 3+3 2 | 2+2 2 |
| | 5+6 2 | 5+6 2 | 5+6 2 | 3+4 2 | 2+2 2 | 3+3 2 |
| | 3+4 2 | 3+4 2 | 5+6 2 | 5+6 2 | 2+2 2 | 2+2 2 |
| | 5+6 2 | 3+4 2 | 5+6 2 | 5+6 2 | 3+3 2 | 3+3 2 |
| 12.1.74 | 633332 | | 40 | Fe0.5S2Ti1 | | |
| | -A | -GM | -M | -Y | -L | -V |
| | 3+4 2 | 3+4 2 | 3+4 2 | 5+6 2 | 2+2 2 | 2+2 2 |
| | 5+6 2 | 5+6 2 | 5+6 2 | 3+4 2 | 3+3 2 | 3+3 2 |
| | 3+4 2 | 5+6 2 | 5+6 2 | 5+6 2 | 2+2 2 | 3+3 2 |
| | 5+6 2 | 3+4 2 | 3+4 2 | 3+4 2 | 3+3 2 | 2+2 2 |
| | 3+4 2 | 3+4 2 | 3+4 2 | 3+4 2 | 3+3 2 | 3+3 2 |
| | 5+6 2 | 5+6 2 | 5+6 2 | 5+6 2 | 2+2 2 | 2+2 2 |
| | 3+4 2 | 3+4 2 | 3+4 2 | 3+4 2 | 3+3 2 | 2+2 2 |
| | 5+6 2 | 3+4 2 | 5+6 2 | 5+6 2 | 2+2 2 | 3+3 2 |
| | 3+4 2 | 3+4 2 | 3+4 2 | 3+4 2 | 3+3 2 | 3+3 2 |
| | 5+6 2 | 5+6 2 | 3+4 2 | 5+6 2 | 2+2 2 | 3+3 2 |
| | 3+4 2 | 5+6 2 | 5+6 2 | 3+4 2 | 3+3 2 | 2+2 2 |
| | 5+6 2 | 3+4 2 | 5+6 2 | 5+6 2 | 2+2 2 | 2+2 2 |
| | 3+4 2 | 3+4 2 | 3+4 2 | 5+6 2 | 2+2 2 | 3+3 2 |
| | 3+4 2 | 3+4 2 | 5+6 2 | 5+6 2 | 3+3 2 | 2+2 2 |
| | 5+6 2 | 5+6 2 | 3+4 2 | 3+4 2 | 3+3 2 | 3+3 2 |
| | 3+4 2 | 5+6 2 | 5+6 2 | 3+4 2 | 2+2 2 | 2+2 2 |
| | 5+6 2 | 5+6 2 | 3+4 2 | 3+4 2 | 3+3 2 | 3+3 2 |

Continued on next column

**TABLE XXIV– continued**

| TQC | ICSD | | NoE | | SF | |
|---|---|---|---|---|---|---|
| | 3+4 2 | 3+4 2 | 3+4 2 | 3+4 2 | 3+3 2 | 3+3 2 |
| | 3+4 2 | 3+4 2 | 3+4 2 | 3+4 2 | 2+2 2 | 3+3 2 |
| | 5+6 2 | 5+6 2 | 3+4 2 | 3+4 2 | 3+3 2 | 3+3 2 |
| 12.1.75 | 192506 | | 122 | | F5Fe1K2 | |
| | -A | -GM | -M | -Y | -L | -V |
| | 3+4 2 | 3+4 2 | 3+4 2 | 3+4 2 | 2+2 2 | 2+2 2 |
| | 5+6 2 | 5+6 2 | 5+6 2 | 5+6 2 | 3+3 2 | 3+3 2 |
| | 3+4 2 | 3+4 2 | 5+6 2 | 5+6 2 | 2+2 2 | 3+3 2 |
| | 5+6 2 | 5+6 2 | 3+4 2 | 3+4 2 | 3+3 2 | 2+2 2 |
| | 5+6 2 | 3+4 2 | 3+4 2 | 5+6 2 | 3+3 2 | 3+3 2 |
| | 3+4 2 | 5+6 2 | 5+6 2 | 3+4 2 | 2+2 2 | 2+2 2 |
| | 5+6 2 | 3+4 2 | 3+4 2 | 5+6 2 | 2+2 2 | 2+2 2 |
| | 5+6 2 | 3+4 2 | 3+4 2 | 5+6 2 | 3+3 2 | 3+3 2 |
| | 3+4 2 | 5+6 2 | 5+6 2 | 5+6 2 | 2+2 2 | 2+2 2 |
| | 5+6 2 | 3+4 2 | 3+4 2 | 3+4 2 | 3+3 2 | 3+3 2 |
| | 3+4 2 | 5+6 2 | 5+6 2 | 3+4 2 | 2+2 2 | 2+2 2 |
| | 5+6 2 | 3+4 2 | 3+4 2 | 3+4 2 | 3+3 2 | 3+3 2 |
| | 3+4 2 | 5+6 2 | 5+6 2 | 5+6 2 | 2+2 2 | 3+3 2 |
| | 5+6 2 | 3+4 2 | 3+4 2 | 5+6 2 | 3+3 2 | 2+2 2 |
| | 3+4 2 | 3+4 2 | 3+4 2 | 3+4 2 | 2+2 2 | 2+2 2 |
| | 5+6 2 | 5+6 2 | 5+6 2 | 5+6 2 | 3+3 2 | 3+3 2 |
| | 3+4 2 | 3+4 2 | 3+4 2 | 3+4 2 | 2+2 2 | 2+2 2 |
| | 3+4 2 | 5+6 2 | 5+6 2 | 5+6 2 | 2+2 2 | 3+3 2 |
| | 5+6 2 | 3+4 2 | 5+6 2 | 3+4 2 | 3+3 2 | 2+2 2 |
| | 5+6 2 | 5+6 2 | 3+4 2 | 5+6 2 | 3+3 2 | 3+3 2 |
| | 3+4 2 | 3+4 2 | 3+4 2 | 3+4 2 | 3+3 2 | 2+2 2 |
| | 5+6 2 | 5+6 2 | 5+6 2 | 5+6 2 | 2+2 2 | 3+3 2 |
| | 3+4 2 | 3+4 2 | 5+6 2 | 5+6 2 | 2+2 2 | 3+3 2 |
| | 5+6 2 | 3+4 2 | 5+6 2 | 3+4 2 | 3+3 2 | 2+2 2 |
| | 3+4 2 | 5+6 2 | 3+4 2 | 5+6 2 | 2+2 2 | 3+3 2 |
| | 5+6 2 | 5+6 2 | 3+4 2 | 3+4 2 | 3+3 2 | 2+2 2 |
| | 3+4 2 | 5+6 2 | 5+6 2 | 3+4 2 | 3+3 2 | 3+3 2 |
| | 3+4 2 | 5+6 2 | 5+6 2 | 3+4 2 | 2+2 2 | 2+2 2 |
| | 5+6 2 | 3+4 2 | 3+4 2 | 5+6 2 | 3+3 2 | 3+3 2 |
| | 3+4 2 | 5+6 2 | 5+6 2 | 3+4 2 | 2+2 2 | 2+2 2 |
| | 5+6 2 | 3+4 2 | 5+6 2 | 3+4 2 | 2+2 2 | 3+3 2 |
| | 5+6 2 | 5+6 2 | 3+4 2 | 5+6 2 | 3+3 2 | 2+2 2 |
| | 3+4 2 | 5+6 2 | 3+4 2 | 5+6 2 | 2+2 2 | 2+2 2 |
| | 3+4 2 | 3+4 2 | 5+6 2 | 3+4 2 | 3+3 2 | 3+3 2 |
| | 3+4 2 | 5+6 2 | 5+6 2 | 3+4 2 | 3+3 2 | 3+3 2 |
| | 3+4 2 | 5+6 2 | 5+6 2 | 3+4 2 | 2+2 2 | 2+2 2 |
| | 3+4 2 | 5+6 2 | 5+6 2 | 3+4 2 | 3+3 2 | 2+2 2 |
| | 3+4 2 | 3+4 2 | 3+4 2 | 5+6 2 | 2+2 2 | 3+3 2 |
| | 5+6 2 | 3+4 2 | 3+4 2 | 3+4 2 | 3+3 2 | 3+3 2 |
| | 5+6 2 | 5+6 2 | 5+6 2 | 5+6 2 | 2+2 2 | 2+2 2 |
| | 5+6 2 | 3+4 2 | 3+4 2 | 5+6 2 | 2+2 2 | 3+3 2 |
| | 5+6 2 | 3+4 2 | 3+4 2 | 5+6 2 | 3+3 2 | 2+2 2 |
| | 3+4 2 | 5+6 2 | 5+6 2 | 3+4 2 | 3+3 2 | 3+3 2 |
| | 3+4 2 | 3+4 2 | 5+6 2 | 5+6 2 | 2+2 2 | 2+2 2 |
| | 5+6 2 | 5+6 2 | 3+4 2 | 3+4 2 | 2+2 2 | 3+3 2 |
| | 3+4 2 | 3+4 2 | 5+6 2 | 3+4 2 | 3+3 2 | 2+2 2 |
| | 3+4 2 | 5+6 2 | 3+4 2 | 5+6 2 | 2+2 2 | 2+2 2 |
| | 5+6 2 | 5+6 2 | 5+6 2 | 3+4 2 | 3+3 2 | 3+3 2 |
| | 5+6 2 | 3+4 2 | 3+4 2 | 5+6 2 | 3+3 2 | 3+3 2 |
| | 5+6 2 | 5+6 2 | 5+6 2 | 3+4 2 | 2+2 2 | 2+2 2 |
| | 5+6 2 | 3+4 2 | 5+6 2 | 5+6 2 | 2+2 2 | 3+3 2 |
| | 3+4 2 | 3+4 2 | 3+4 2 | 3+4 2 | 3+3 2 | 2+2 2 |
| | 3+4 2 | 5+6 2 | 3+4 2 | 5+6 2 | 2+2 2 | 2+2 2 |
| | 3+4 2 | 5+6 2 | 5+6 2 | 3+4 2 | 3+3 2 | 3+3 2 |
| | 3+4 2 | 5+6 2 | 5+6 2 | 3+4 2 | 2+2 2 | 2+2 2 |
| | 3+4 2 | 5+6 2 | 5+6 2 | 3+4 2 | 3+3 2 | 3+3 2 |
| | 5+6 2 | 3+4 2 | 3+4 2 | 5+6 2 | 2+2 2 | 2+2 2 |
| | 5+6 2 | 3+4 2 | 3+4 2 | 5+6 2 | 3+3 2 | 3+3 2 |
| | 5+6 2 | 3+4 2 | 3+4 2 | 5+6 2 | 2+2 2 | 2+2 2 |
| | 3+4 2 | 5+6 2 | 5+6 2 | 3+4 2 | 3+3 2 | 3+3 2 |
| | 3+4 2 | 5+6 2 | 5+6 2 | 3+4 2 | 3+3 2 | 3+3 2 |
| 12.1.76 | 106627 | | 58 | | Eu1In4 | |
| | -A | -GM | -M | -Y | -L | -V |
| | 3+4 2 | 3+4 2 | 3+4 2 | 3+4 2 | 2+2 2 | 2+2 2 |
| | 5+6 2 | 5+6 2 | 5+6 2 | 5+6 2 | 3+3 2 | 3+3 2 |
| | 5+6 2 | 3+4 2 | 5+6 2 | 3+4 2 | 2+2 2 | 3+3 2 |
| | 3+4 2 | 5+6 2 | 3+4 2 | 5+6 2 | 3+3 2 | 2+2 2 |
| | 5+6 2 | 3+4 2 | 5+6 2 | 3+4 2 | 2+2 2 | 3+3 2 |
| | 3+4 2 | 5+6 2 | 3+4 2 | 5+6 2 | 3+3 2 | 2+2 2 |
| | 5+6 2 | 3+4 2 | 5+6 2 | 3+4 2 | 2+2 2 | 3+3 2 |
| | 3+4 2 | 5+6 2 | 3+4 2 | 5+6 2 | 3+3 2 | 2+2 2 |
| | 3+4 2 | 3+4 2 | 3+4 2 | 3+4 2 | 3+3 2 | 2+2 2 |
| | 5+6 2 | 3+4 2 | 5+6 2 | 5+6 2 | 2+2 2 | 3+3 2 |
| | 3+4 2 | 5+6 2 | 5+6 2 | 5+6 2 | 3+3 2 | 2+2 2 |
| | 5+6 2 | 5+6 2 | 3+4 2 | 3+4 2 | 2+2 2 | 3+3 2 |
| | 5+6 2 | 3+4 2 | 5+6 2 | 3+4 2 | 3+3 2 | 2+2 2 |
| | 3+4 2 | 5+6 2 | 3+4 2 | 5+6 2 | 2+2 2 | 3+3 2 |
| | 5+6 2 | 3+4 2 | 5+6 2 | 3+4 2 | 2+2 2 | 2+2 2 |
| | 3+4 2 | 5+6 2 | 5+6 2 | 5+6 2 | 3+3 2 | 3+3 2 |
| | 5+6 2 | 3+4 2 | 3+4 2 | 5+6 2 | 3+3 2 | 3+3 2 |
| | 3+4 2 | 5+6 2 | 3+4 2 | 5+6 2 | 2+2 2 | 2+2 2 |
| | 5+6 2 | 3+4 2 | 3+4 2 | 3+4 2 | 2+2 2 | 2+2 2 |
| | 3+4 2 | 3+4 2 | 5+6 2 | 3+4 2 | 3+3 2 | 2+2 2 |
| | 3+4 2 | 3+4 2 | 5+6 2 | 5+6 2 | 3+3 2 | 3+3 2 |
| | 5+6 2 | 3+4 2 | 3+4 2 | 3+4 2 | 2+2 2 | 2+2 2 |
| | 5+6 2 | 5+6 2 | 5+6 2 | 5+6 2 | 3+3 2 | 3+3 2 |
| | 3+4 2 | 5+6 2 | 3+4 2 | 3+4 2 | 2+2 2 | 2+2 2 |
| | 5+6 2 | 5+6 2 | 3+4 2 | 5+6 2 | 3+3 2 | 3+3 2 |
| | 3+4 2 | 3+4 2 | 5+6 2 | 3+4 2 | 2+2 2 | 2+2 2 |
| | 3+4 2 | 3+4 2 | 5+6 2 | 5+6 2 | 2+2 2 | 3+3 2 |
| | 5+6 2 | 5+6 2 | 3+4 2 | 3+4 2 | 3+3 2 | 2+2 2 |
| | 3+4 2 | 3+4 2 | 5+6 2 | 3+4 2 | 3+3 2 | 3+3 2 |
| 12.1.77 | 151123 | | 94 | | Er2O7Si2 | |
| | -A | -GM | -M | -Y | -L | -V |
| | 3+4 2 | 3+4 2 | 5+6 2 | 5+6 2 | 2+2 2 | 2+2 2 |
| | 5+6 2 | 5+6 2 | 3+4 2 | 3+4 2 | 3+3 2 | 3+3 2 |
| | 3+4 2 | 3+4 2 | 5+6 2 | 5+6 2 | 2+2 2 | 2+2 2 |
| | 5+6 2 | 5+6 2 | 3+4 2 | 3+4 2 | 3+3 2 | 3+3 2 |
| | 3+4 2 | 3+4 2 | 3+4 2 | 3+4 2 | 2+2 2 | 2+2 2 |
| | 5+6 2 | 5+6 2 | 5+6 2 | 5+6 2 | 3+3 2 | 3+3 2 |
| | 3+4 2 | 3+4 2 | 5+6 2 | 5+6 2 | 2+2 2 | 2+2 2 |
| | 5+6 2 | 5+6 2 | 3+4 2 | 3+4 2 | 3+3 2 | 3+3 2 |
| | 5+6 2 | 3+4 2 | 5+6 2 | 3+4 2 | 2+2 2 | 3+3 2 |
| | 3+4 2 | 5+6 2 | 3+4 2 | 5+6 2 | 3+3 2 | 2+2 2 |
| | 3+4 2 | 3+4 2 | 5+6 2 | 3+4 2 | 2+2 2 | 3+3 2 |
| | 5+6 2 | 5+6 2 | 5+6 2 | 5+6 2 | 3+3 2 | 2+2 2 |
| | 5+6 2 | 3+4 2 | 3+4 2 | 3+4 2 | 2+2 2 | 3+3 2 |
| | 5+6 2 | 5+6 2 | 3+4 2 | 5+6 2 | 3+3 2 | 3+3 2 |
| | 3+4 2 | 3+4 2 | 5+6 2 | 3+4 2 | 2+2 2 | 2+2 2 |
| | 3+4 2 | 5+6 2 | 3+4 2 | 5+6 2 | 3+3 2 | 2+2 2 |
| | 5+6 2 | 3+4 2 | 5+6 2 | 3+4 2 | 2+2 2 | 3+3 2 |
| | 3+4 2 | 5+6 2 | 3+4 2 | 5+6 2 | 3+3 2 | 2+2 2 |
| | 3+4 2 | 3+4 2 | 5+6 2 | 5+6 2 | 2+2 2 | 2+2 2 |
| | 3+4 2 | 5+6 2 | 3+4 2 | 3+4 2 | 3+3 2 | 3+3 2 |
| | 5+6 2 | 3+4 2 | 3+4 2 | 5+6 2 | 3+3 2 | 2+2 2 |
| | 5+6 2 | 5+6 2 | 5+6 2 | 3+4 2 | 2+2 2 | 3+3 2 |
| | 5+6 2 | 3+4 2 | 5+6 2 | 3+4 2 | 2+2 2 | 2+2 2 |

Continued on next column

**TABLE XXIV– continued**

| TQC | ICSD | NoE | SF | | | |
|---|---|---|---|---|---|---|
| | 3+4 2 | 5+6 2 | 5+6 2 | 5+6 2 | 3+3 2 | 3+3 2 |
| | 5+6 2 | 5+6 2 | 3+4 2 | 5+6 2 | 2+2 2 | 2+2 2 |
| | 3+4 2 | 3+4 2 | 3+4 2 | 3+4 2 | 3+3 2 | 3+3 2 |
| | 3+4 2 | 3+4 2 | 3+4 2 | 5+6 2 | 3+3 2 | 2+2 2 |
| | 3+4 2 | 3+4 2 | 3+4 2 | 3+4 2 | 2+2 2 | 2+2 2 |
| | 5+6 2 | 5+6 2 | 5+6 2 | 5+6 2 | 3+3 2 | 3+3 2 |
| | 5+6 2 | 5+6 2 | 5+6 2 | 3+4 2 | 3+3 2 | 2+2 2 |
| | 3+4 2 | 5+6 2 | 3+4 2 | 5+6 2 | 2+2 2 | 3+3 2 |
| | 5+6 2 | 3+4 2 | 5+6 2 | 3+4 2 | 2+2 2 | 3+3 2 |
| | 5+6 2 | 5+6 2 | 3+4 2 | 5+6 2 | 3+3 2 | 2+2 2 |
| | 3+4 2 | 3+4 2 | 5+6 2 | 5+6 2 | 2+2 2 | 2+2 2 |
| | 3+4 2 | 5+6 2 | 3+4 2 | 3+4 2 | 3+3 2 | 3+3 2 |
| | 3+4 2 | 5+6 2 | 3+4 2 | 5+6 2 | 3+3 2 | 2+2 2 |
| | 5+6 2 | 3+4 2 | 3+4 2 | 3+4 2 | 2+2 2 | 3+3 2 |
| | 3+4 2 | 5+6 2 | 5+6 2 | 5+6 2 | 2+2 2 | 2+2 2 |
| | 5+6 2 | 3+4 2 | 5+6 2 | 3+4 2 | 3+3 2 | 3+3 2 |
| | 3+4 2 | 3+4 2 | 3+4 2 | 5+6 2 | 3+3 2 | 2+2 2 |
| | 5+6 2 | 5+6 2 | 5+6 2 | 3+4 2 | 2+2 2 | 3+3 2 |
| | 3+4 2 | 5+6 2 | 3+4 2 | 5+6 2 | 3+3 2 | 2+2 2 |
| | 3+4 2 | 3+4 2 | 5+6 2 | 3+4 2 | 2+2 2 | 2+2 2 |
| | 5+6 2 | 3+4 2 | 3+4 2 | 5+6 2 | 2+2 2 | 3+3 2 |
| | 3+4 2 | 5+6 2 | 5+6 2 | 3+4 2 | 3+3 2 | 2+2 2 |
| | 3+4 2 | 3+4 2 | 3+4 2 | 5+6 2 | 3+3 2 | 3+3 2 |
| | 5+6 2 | 5+6 2 | 5+6 2 | 5+6 2 | 2+2 2 | 3+3 2 |
| 12.1.78 | 419435 | 186 | Er2O3 | | | |
| | -A | -GM | -M | -Y | -L | -V |
| | 5+6 2 | 3+4 2 | 3+4 2 | 5+6 2 | 2+2 2 | 3+3 2 |
| | 3+4 2 | 5+6 2 | 5+6 2 | 3+4 2 | 3+3 2 | 2+2 2 |
| | 3+4 2 | 3+4 2 | 5+6 2 | 5+6 2 | 2+2 2 | 2+2 2 |
| | 5+6 2 | 5+6 2 | 3+4 2 | 3+4 2 | 3+3 2 | 3+3 2 |
| | 3+4 2 | 3+4 2 | 3+4 2 | 3+4 2 | 2+2 2 | 2+2 2 |
| | 5+6 2 | 5+6 2 | 5+6 2 | 5+6 2 | 3+3 2 | 3+3 2 |
| | 3+4 2 | 3+4 2 | 5+6 2 | 5+6 2 | 3+3 2 | 3+3 2 |
| | 5+6 2 | 5+6 2 | 3+4 2 | 3+4 2 | 2+2 2 | 2+2 2 |
| | 5+6 2 | 3+4 2 | 5+6 2 | 5+6 2 | 2+2 2 | 3+3 2 |
| | 3+4 2 | 5+6 2 | 3+4 2 | 3+4 2 | 3+3 2 | 2+2 2 |
| | 3+4 2 | 3+4 2 | 3+4 2 | 5+6 2 | 2+2 2 | 2+2 2 |
| | 5+6 2 | 5+6 2 | 5+6 2 | 3+4 2 | 3+3 2 | 3+3 2 |
| | 3+4 2 | 3+4 2 | 5+6 2 | 5+6 2 | 3+3 2 | 3+3 2 |
| | 5+6 2 | 5+6 2 | 3+4 2 | 3+4 2 | 2+2 2 | 2+2 2 |
| | 5+6 2 | 3+4 2 | 5+6 2 | 5+6 2 | 3+3 2 | 3+3 2 |
| | 3+4 2 | 5+6 2 | 3+4 2 | 3+4 2 | 2+2 2 | 2+2 2 |
| | 5+6 2 | 3+4 2 | 5+6 2 | 5+6 2 | 2+2 2 | 2+2 2 |
| | 3+4 2 | 5+6 2 | 3+4 2 | 3+4 2 | 3+3 2 | 3+3 2 |
| | 5+6 2 | 3+4 2 | 3+4 2 | 5+6 2 | 2+2 2 | 2+2 2 |
| | 3+4 2 | 3+4 2 | 5+6 2 | 3+4 2 | 3+3 2 | 3+3 2 |
| | 3+4 2 | 5+6 2 | 5+6 2 | 5+6 2 | 2+2 2 | 2+2 2 |
| | 5+6 2 | 3+4 2 | 3+4 2 | 3+4 2 | 3+3 2 | 3+3 2 |
| | 3+4 2 | 5+6 2 | 3+4 2 | 5+6 2 | 2+2 2 | 3+3 2 |
| | 5+6 2 | 5+6 2 | 5+6 2 | 3+4 2 | 3+3 2 | 2+2 2 |
| | 3+4 2 | 3+4 2 | 3+4 2 | 3+4 2 | 3+3 2 | 3+3 2 |
| | 3+4 2 | 5+6 2 | 3+4 2 | 5+6 2 | 3+3 2 | 3+3 2 |
| | 5+6 2 | 3+4 2 | 5+6 2 | 3+4 2 | 2+2 2 | 2+2 2 |
| | 5+6 2 | 5+6 2 | 5+6 2 | 3+4 2 | 2+2 2 | 2+2 2 |
| | 3+4 2 | 5+6 2 | 5+6 2 | 5+6 2 | 2+2 2 | 3+3 2 |
| | 5+6 2 | 3+4 2 | 3+4 2 | 5+6 2 | 3+3 2 | 2+2 2 |
| | 3+4 2 | 5+6 2 | 3+4 2 | 3+4 2 | 3+3 2 | 2+2 2 |
| | 5+6 2 | 3+4 2 | 5+6 2 | 5+6 2 | 2+2 2 | 3+3 2 |
| | 3+4 2 | 3+4 2 | 3+4 2 | 3+4 2 | 3+3 2 | 3+3 2 |
| | 5+6 2 | 5+6 2 | 5+6 2 | 3+4 2 | 2+2 2 | 2+2 2 |
| | 5+6 2 | 3+4 2 | 3+4 2 | 5+6 2 | 3+3 2 | 3+3 2 |

Continued on next column

| TQC | ICSD | NoE | SF | | | |
|---|---|---|---|---|---|---|
| | 3+4 2 | 3+4 2 | 5+6 2 | 3+4 2 | 3+3 2 | 2+2 2 |
| | 3+4 2 | 5+6 2 | 5+6 2 | 5+6 2 | 2+2 2 | 3+3 2 |
| | 3+4 2 | 5+6 2 | 5+6 2 | 3+4 2 | 3+3 2 | 2+2 2 |
| | 3+4 2 | 3+4 2 | 5+6 2 | 5+6 2 | 2+2 2 | 3+3 2 |
| | 5+6 2 | 3+4 2 | 3+4 2 | 5+6 2 | 3+3 2 | 3+3 2 |
| | 5+6 2 | 5+6 2 | 3+4 2 | 3+4 2 | 2+2 2 | 2+2 2 |
| | 3+4 2 | 3+4 2 | 5+6 2 | 5+6 2 | 2+2 2 | 2+2 2 |
| | 5+6 2 | 5+6 2 | 3+4 2 | 3+4 2 | 3+3 2 | 3+3 2 |
| | 5+6 2 | 5+6 2 | 5+6 2 | 5+6 2 | 2+2 2 | 3+3 2 |
| | 3+4 2 | 5+6 2 | 3+4 2 | 5+6 2 | 2+2 2 | 2+2 2 |
| | 5+6 2 | 3+4 2 | 3+4 2 | 5+6 2 | 2+2 2 | 2+2 2 |
| | 5+6 2 | 3+4 2 | 5+6 2 | 3+4 2 | 3+3 2 | 3+3 2 |
| | 3+4 2 | 5+6 2 | 3+4 2 | 3+4 2 | 2+2 2 | 3+3 2 |
| | 3+4 2 | 3+4 2 | 5+6 2 | 5+6 2 | 3+3 2 | 2+2 2 |
| | 5+6 2 | 3+4 2 | 5+6 2 | 5+6 2 | 2+2 2 | 2+2 2 |
| | 5+6 2 | 5+6 2 | 3+4 2 | 5+6 2 | 2+2 2 | 3+3 2 |
| | 5+6 2 | 3+4 2 | 3+4 2 | 3+4 2 | 3+3 2 | 2+2 2 |
| | 3+4 2 | 3+4 2 | 5+6 2 | 5+6 2 | 2+2 2 | 2+2 2 |
| | 5+6 2 | 5+6 2 | 3+4 2 | 3+4 2 | 2+2 2 | 3+3 2 |
| | 3+4 2 | 5+6 2 | 5+6 2 | 5+6 2 | 3+3 2 | 2+2 2 |
| | 5+6 2 | 5+6 2 | 3+4 2 | 3+4 2 | 3+3 2 | 3+3 2 |
| | 3+4 2 | 5+6 2 | 5+6 2 | 5+6 2 | 3+3 2 | 2+2 2 |
| | 5+6 2 | 3+4 2 | 3+4 2 | 3+4 2 | 2+2 2 | 2+2 2 |
| | 3+4 2 | 5+6 2 | 5+6 2 | 5+6 2 | 2+2 2 | 3+3 2 |
| | 5+6 2 | 5+6 2 | 5+6 2 | 3+4 2 | 3+3 2 | 2+2 2 |
| | 3+4 2 | 3+4 2 | 3+4 2 | 3+4 2 | 2+2 2 | 3+3 2 |
| | 5+6 2 | 5+6 2 | 5+6 2 | 5+6 2 | 3+3 2 | 2+2 2 |
| | 3+4 2 | 3+4 2 | 3+4 2 | 3+4 2 | 2+2 2 | 3+3 2 |
| | 3+4 2 | 3+4 2 | 5+6 2 | 5+6 2 | 3+3 2 | 2+2 2 |
| | 5+6 2 | 5+6 2 | 5+6 2 | 3+4 2 | 3+3 2 | 3+3 2 |
| | 5+6 2 | 3+4 2 | 3+4 2 | 5+6 2 | 2+2 2 | 2+2 2 |
| | 3+4 2 | 3+4 2 | 3+4 2 | 5+6 2 | 3+3 2 | 3+3 2 |
| | 5+6 2 | 5+6 2 | 5+6 2 | 3+4 2 | 2+2 2 | 2+2 2 |
| | 3+4 2 | 3+4 2 | 3+4 2 | 5+6 2 | 3+3 2 | 3+3 2 |
| | 5+6 2 | 3+4 2 | 3+4 2 | 3+4 2 | 2+2 2 | 3+3 2 |
| | 3+4 2 | 5+6 2 | 3+4 2 | 5+6 2 | 3+3 2 | 2+2 2 |
| | 5+6 2 | 3+4 2 | 5+6 2 | 3+4 2 | 2+2 2 | 3+3 2 |
| | 3+4 2 | 5+6 2 | 3+4 2 | 3+4 2 | 3+3 2 | 2+2 2 |
| | 3+4 2 | 5+6 2 | 5+6 2 | 5+6 2 | 2+2 2 | 2+2 2 |
| | 5+6 2 | 5+6 2 | 5+6 2 | 3+4 2 | 2+2 2 | 3+3 2 |
| | 3+4 2 | 3+4 2 | 5+6 2 | 5+6 2 | 3+3 2 | 2+2 2 |
| | 5+6 2 | 5+6 2 | 3+4 2 | 3+4 2 | 2+2 2 | 3+3 2 |
| | 5+6 2 | 5+6 2 | 5+6 2 | 5+6 2 | 3+3 2 | 2+2 2 |
| | 3+4 2 | 3+4 2 | 5+6 2 | 3+4 2 | 3+3 2 | 3+3 2 |
| | 5+6 2 | 5+6 2 | 3+4 2 | 5+6 2 | 2+2 2 | 2+2 2 |
| | 3+4 2 | 3+4 2 | 3+4 2 | 3+4 2 | 3+3 2 | 3+3 2 |
| | 5+6 2 | 3+4 2 | 5+6 2 | 5+6 2 | 2+2 2 | 2+2 2 |
| | 3+4 2 | 3+4 2 | 3+4 2 | 5+6 2 | 3+3 2 | 3+3 2 |
| | 5+6 2 | 5+6 2 | 5+6 2 | 3+4 2 | 2+2 2 | 2+2 2 |
| | 3+4 2 | 3+4 2 | 3+4 2 | 5+6 2 | 3+3 2 | 3+3 2 |
| | 5+6 2 | 5+6 2 | 5+6 2 | 3+4 2 | 3+3 2 | 2+2 2 |
| | 3+4 2 | 3+4 2 | 3+4 2 | 3+4 2 | 2+2 2 | 3+3 2 |
| | 5+6 2 | 3+4 2 | 3+4 2 | 5+6 2 | 3+3 2 | 2+2 2 |
| | 3+4 2 | 5+6 2 | 5+6 2 | 3+4 2 | 2+2 2 | 3+3 2 |
| | 3+4 2 | 3+4 2 | 3+4 2 | 5+6 2 | 2+2 2 | 3+3 2 |
| | 5+6 2 | 3+4 2 | 3+4 2 | 5+6 2 | 3+3 2 | 2+2 2 |
| | 3+4 2 | 5+6 2 | 5+6 2 | 3+4 2 | 3+3 2 | 3+3 2 |
| | 5+6 2 | 3+4 2 | 3+4 2 | 5+6 2 | 2+2 2 | 2+2 2 |
| 12.1.79 | 180224 | 48 | Dy1Si1 | | | |
| | -A | -GM | -M | -Y | -L | -V |
| | 5+6 2 | 3+4 2 | 3+4 2 | 5+6 2 | 2+2 2 | 3+3 2 |

Continued on next column

**TABLE XXIV– continued**

| TQC | ICSD | | NoE | | SF | |
|---|---|---|---|---|---|---|
| | 3+4 2 | 5+6 2 | 5+6 2 | 3+4 2 | 3+3 2 | 2+2 2 |
| | 5+6 2 | 3+4 2 | 3+4 2 | 5+6 2 | 2+2 2 | 3+3 2 |
| | 3+4 2 | 5+6 2 | 5+6 2 | 3+4 2 | 3+3 2 | 2+2 2 |
| | 5+6 2 | 3+4 2 | 3+4 2 | 5+6 2 | 2+2 2 | 3+3 2 |
| | 3+4 2 | 3+4 2 | 5+6 2 | 3+4 2 | 3+3 2 | 2+2 2 |
| | 5+6 2 | 5+6 2 | 3+4 2 | 5+6 2 | 2+2 2 | 3+3 2 |
| | 3+4 2 | 5+6 2 | 5+6 2 | 3+4 2 | 3+3 2 | 2+2 2 |
| | 3+4 2 | 3+4 2 | 3+4 2 | 3+4 2 | 2+2 2 | 2+2 2 |
| | 5+6 2 | 5+6 2 | 5+6 2 | 5+6 2 | 3+3 2 | 3+3 2 |
| | 3+4 2 | 3+4 2 | 5+6 2 | 3+4 2 | 3+3 2 | 3+3 2 |
| | 5+6 2 | 5+6 2 | 3+4 2 | 5+6 2 | 2+2 2 | 2+2 2 |
| | 5+6 2 | 3+4 2 | 5+6 2 | 3+4 2 | 3+3 2 | 2+2 2 |
| | 3+4 2 | 5+6 2 | 3+4 2 | 5+6 2 | 2+2 2 | 2+2 2 |
| | 3+4 2 | 5+6 2 | 3+4 2 | 3+4 2 | 3+3 2 | 3+3 2 |
| | 5+6 2 | 3+4 2 | 5+6 2 | 5+6 2 | 2+2 2 | 2+2 2 |
| | 3+4 2 | 3+4 2 | 3+4 2 | 3+4 2 | 2+2 2 | 3+3 2 |
| | 5+6 2 | 5+6 2 | 5+6 2 | 5+6 2 | 3+3 2 | 2+2 2 |
| | 3+4 2 | 5+6 2 | 5+6 2 | 3+4 2 | 3+3 2 | 3+3 2 |
| | 5+6 2 | 3+4 2 | 3+4 2 | 3+4 2 | 2+2 2 | 2+2 2 |
| | 3+4 2 | 5+6 2 | 5+6 2 | 5+6 2 | 2+2 2 | 3+3 2 |
| | 5+6 2 | 3+4 2 | 3+4 2 | 3+4 2 | 3+3 2 | 2+2 2 |
| | 3+4 2 | 5+6 2 | 5+6 2 | 5+6 2 | 2+2 2 | 3+3 2 |
| | 5+6 2 | 5+6 2 | 3+4 2 | 3+4 2 | 3+3 2 | 2+2 2 |
| 12.1.80 | 180223 | | 48 | | Dy1Si1 | |
| | -A | -GM | -M | -Y | -L | -V |
| | 5+6 2 | 3+4 2 | 3+4 2 | 5+6 2 | 2+2 2 | 3+3 2 |
| | 3+4 2 | 5+6 2 | 5+6 2 | 3+4 2 | 3+3 2 | 2+2 2 |
| | 5+6 2 | 3+4 2 | 3+4 2 | 5+6 2 | 2+2 2 | 3+3 2 |
| | 3+4 2 | 5+6 2 | 5+6 2 | 3+4 2 | 3+3 2 | 2+2 2 |
| | 5+6 2 | 3+4 2 | 3+4 2 | 5+6 2 | 2+2 2 | 3+3 2 |
| | 3+4 2 | 3+4 2 | 5+6 2 | 3+4 2 | 3+3 2 | 2+2 2 |
| | 5+6 2 | 5+6 2 | 3+4 2 | 5+6 2 | 2+2 2 | 3+3 2 |
| | 3+4 2 | 5+6 2 | 5+6 2 | 3+4 2 | 3+3 2 | 2+2 2 |
| | 3+4 2 | 3+4 2 | 3+4 2 | 3+4 2 | 2+2 2 | 2+2 2 |
| | 5+6 2 | 5+6 2 | 5+6 2 | 5+6 2 | 3+3 2 | 3+3 2 |
| | 3+4 2 | 3+4 2 | 5+6 2 | 3+4 2 | 3+3 2 | 3+3 2 |
| | 5+6 2 | 5+6 2 | 3+4 2 | 5+6 2 | 2+2 2 | 2+2 2 |
| | 5+6 2 | 3+4 2 | 5+6 2 | 3+4 2 | 3+3 2 | 2+2 2 |
| | 3+4 2 | 5+6 2 | 3+4 2 | 5+6 2 | 2+2 2 | 2+2 2 |
| | 3+4 2 | 5+6 2 | 3+4 2 | 3+4 2 | 3+3 2 | 3+3 2 |
| | 5+6 2 | 3+4 2 | 5+6 2 | 5+6 2 | 2+2 2 | 2+2 2 |
| | 3+4 2 | 3+4 2 | 3+4 2 | 3+4 2 | 2+2 2 | 3+3 2 |
| | 5+6 2 | 5+6 2 | 5+6 2 | 5+6 2 | 3+3 2 | 2+2 2 |
| | 3+4 2 | 5+6 2 | 5+6 2 | 3+4 2 | 3+3 2 | 3+3 2 |
| | 5+6 2 | 3+4 2 | 3+4 2 | 3+4 2 | 2+2 2 | 2+2 2 |
| | 3+4 2 | 5+6 2 | 5+6 2 | 5+6 2 | 2+2 2 | 3+3 2 |
| | 5+6 2 | 3+4 2 | 3+4 2 | 3+4 2 | 3+3 2 | 2+2 2 |
| | 3+4 2 | 5+6 2 | 5+6 2 | 5+6 2 | 3+3 2 | 3+3 2 |
| | 5+6 2 | 5+6 2 | 3+4 2 | 3+4 2 | 2+2 2 | 2+2 2 |
| 12.1.81 | 180222 | | 48 | | Dy1Si1 | |
| | -A | -GM | -M | -Y | -L | -V |
| | 5+6 2 | 3+4 2 | 3+4 2 | 5+6 2 | 2+2 2 | 3+3 2 |
| | 3+4 2 | 5+6 2 | 5+6 2 | 3+4 2 | 3+3 2 | 2+2 2 |
| | 5+6 2 | 3+4 2 | 3+4 2 | 5+6 2 | 2+2 2 | 3+3 2 |
| | 3+4 2 | 5+6 2 | 5+6 2 | 3+4 2 | 3+3 2 | 2+2 2 |
| | 5+6 2 | 3+4 2 | 3+4 2 | 5+6 2 | 2+2 2 | 3+3 2 |
| | 3+4 2 | 3+4 2 | 5+6 2 | 3+4 2 | 3+3 2 | 2+2 2 |
| | 5+6 2 | 5+6 2 | 3+4 2 | 5+6 2 | 2+2 2 | 3+3 2 |
| | 3+4 2 | 5+6 2 | 5+6 2 | 3+4 2 | 3+3 2 | 2+2 2 |
| | 3+4 2 | 3+4 2 | 3+4 2 | 3+4 2 | 2+2 2 | 2+2 2 |
| | 5+6 2 | 5+6 2 | 5+6 2 | 5+6 2 | 3+3 2 | 3+3 2 |

Continued on next column

**TABLE XXIV– continued**

| TQC | ICSD | | NoE | | SF | |
|---|---|---|---|---|---|---|
| | 3+4 2 | 3+4 2 | 5+6 2 | 3+4 2 | 3+3 2 | 3+3 2 |
| | 5+6 2 | 5+6 2 | 3+4 2 | 5+6 2 | 2+2 2 | 2+2 2 |
| | 5+6 2 | 3+4 2 | 5+6 2 | 3+4 2 | 3+3 2 | 2+2 2 |
| | 3+4 2 | 5+6 2 | 3+4 2 | 5+6 2 | 2+2 2 | 2+2 2 |
| | 3+4 2 | 5+6 2 | 3+4 2 | 3+4 2 | 3+3 2 | 3+3 2 |
| | 5+6 2 | 3+4 2 | 5+6 2 | 5+6 2 | 2+2 2 | 2+2 2 |
| | 3+4 2 | 3+4 2 | 3+4 2 | 3+4 2 | 2+2 2 | 3+3 2 |
| | 5+6 2 | 5+6 2 | 5+6 2 | 5+6 2 | 3+3 2 | 2+2 2 |
| | 3+4 2 | 3+4 2 | 5+6 2 | 3+4 2 | 3+3 2 | 3+3 2 |
| | 5+6 2 | 5+6 2 | 3+4 2 | 3+4 2 | 2+2 2 | 2+2 2 |
| | 3+4 2 | 5+6 2 | 5+6 2 | 5+6 2 | 2+2 2 | 3+3 2 |
| | 5+6 2 | 3+4 2 | 3+4 2 | 3+4 2 | 3+3 2 | 2+2 2 |
| | 3+4 2 | 5+6 2 | 5+6 2 | 5+6 2 | 2+2 2 | 3+3 2 |
| | 5+6 2 | 5+6 2 | 3+4 2 | 3+4 2 | 3+3 2 | 2+2 2 |
| 12.1.82 | 180221 | | 48 | | Dy1Si1 | |
| | -A | -GM | -M | -Y | -L | -V |
| | 5+6 2 | 3+4 2 | 3+4 2 | 5+6 2 | 2+2 2 | 3+3 2 |
| | 3+4 2 | 5+6 2 | 5+6 2 | 3+4 2 | 3+3 2 | 2+2 2 |
| | 5+6 2 | 3+4 2 | 3+4 2 | 5+6 2 | 2+2 2 | 3+3 2 |
| | 3+4 2 | 5+6 2 | 5+6 2 | 3+4 2 | 3+3 2 | 2+2 2 |
| | 5+6 2 | 3+4 2 | 3+4 2 | 5+6 2 | 2+2 2 | 3+3 2 |
| | 3+4 2 | 3+4 2 | 5+6 2 | 3+4 2 | 3+3 2 | 2+2 2 |
| | 5+6 2 | 5+6 2 | 3+4 2 | 5+6 2 | 2+2 2 | 3+3 2 |
| | 3+4 2 | 5+6 2 | 5+6 2 | 3+4 2 | 3+3 2 | 2+2 2 |
| | 3+4 2 | 3+4 2 | 3+4 2 | 3+4 2 | 2+2 2 | 2+2 2 |
| | 5+6 2 | 5+6 2 | 5+6 2 | 5+6 2 | 3+3 2 | 3+3 2 |
| | 3+4 2 | 3+4 2 | 5+6 2 | 3+4 2 | 3+3 2 | 3+3 2 |
| | 5+6 2 | 5+6 2 | 3+4 2 | 5+6 2 | 2+2 2 | 2+2 2 |
| | 5+6 2 | 3+4 2 | 5+6 2 | 3+4 2 | 3+3 2 | 2+2 2 |
| | 3+4 2 | 5+6 2 | 3+4 2 | 5+6 2 | 2+2 2 | 2+2 2 |
| | 3+4 2 | 5+6 2 | 3+4 2 | 3+4 2 | 3+3 2 | 3+3 2 |
| | 5+6 2 | 3+4 2 | 5+6 2 | 5+6 2 | 2+2 2 | 2+2 2 |
| | 3+4 2 | 3+4 2 | 3+4 2 | 3+4 2 | 2+2 2 | 3+3 2 |
| | 5+6 2 | 5+6 2 | 5+6 2 | 5+6 2 | 3+3 2 | 2+2 2 |
| | 3+4 2 | 3+4 2 | 5+6 2 | 3+4 2 | 3+3 2 | 3+3 2 |
| | 5+6 2 | 5+6 2 | 3+4 2 | 3+4 2 | 2+2 2 | 2+2 2 |
| | 3+4 2 | 5+6 2 | 5+6 2 | 5+6 2 | 2+2 2 | 3+3 2 |
| | 5+6 2 | 3+4 2 | 3+4 2 | 3+4 2 | 3+3 2 | 2+2 2 |
| | 3+4 2 | 5+6 2 | 5+6 2 | 5+6 2 | 2+2 2 | 3+3 2 |
| | 5+6 2 | 5+6 2 | 3+4 2 | 3+4 2 | 3+3 2 | 2+2 2 |
| 12.1.83 | 418227 | | 120 | | Dy1P2Pt8 | |
| | -A | -GM | -M | -Y | -L | -V |
| | 5+6 2 | 3+4 2 | 5+6 2 | 3+4 2 | 3+3 2 | 2+2 2 |
| | 3+4 2 | 5+6 2 | 3+4 2 | 5+6 2 | 2+2 2 | 3+3 2 |
| | 3+4 2 | 5+6 2 | 3+4 2 | 5+6 2 | 2+2 2 | 3+3 2 |
| | 3+4 2 | 5+6 2 | 3+4 2 | 5+6 2 | 2+2 2 | 3+3 2 |
| | 3+4 2 | 3+4 2 | 3+4 2 | 3+4 2 | 2+2 2 | 3+3 2 |
| | 5+6 2 | 5+6 2 | 5+6 2 | 5+6 2 | 3+3 2 | 2+2 2 |
| | 3+4 2 | 5+6 2 | 5+6 2 | 3+4 2 | 2+2 2 | 3+3 2 |
| | 5+6 2 | 5+6 2 | 3+4 2 | 5+6 2 | 3+3 2 | 2+2 2 |
| | 5+6 2 | 3+4 2 | 5+6 2 | 3+4 2 | 3+3 2 | 2+2 2 |
| | 5+6 2 | 3+4 2 | 3+4 2 | 5+6 2 | 2+2 2 | 3+3 2 |
| | 3+4 2 | 5+6 2 | 5+6 2 | 3+4 2 | 2+2 2 | 2+2 2 |
| | 3+4 2 | 3+4 2 | 5+6 2 | 5+6 2 | 3+3 2 | 3+3 2 |
| | 5+6 2 | 5+6 2 | 3+4 2 | 3+4 2 | 3+3 2 | 3+3 2 |
| | 3+4 2 | 3+4 2 | 3+4 2 | 5+6 2 | 2+2 2 | 2+2 2 |
| | 5+6 2 | 3+4 2 | 5+6 2 | 5+6 2 | 3+3 2 | 2+2 2 |
| | 3+4 2 | 5+6 2 | 3+4 2 | 5+6 2 | 2+2 2 | 3+3 2 |
| | 5+6 2 | 5+6 2 | 5+6 2 | 3+4 2 | 3+3 2 | 2+2 2 |
| | 3+4 2 | 3+4 2 | 3+4 2 | 3+4 2 | 2+2 2 | 2+2 2 |
| | 5+6 2 | 3+4 2 | 5+6 2 | 5+6 2 | 3+3 2 | 3+3 2 |

Continued on next column

**TABLE XXIV– continued**

| TQC | ICSD | NoE | SF | | | |
|---|---|---|---|---|---|---|
| | 3+4 2 | 3+4 2 | 5+6 2 | 3+4 2 | 2+2 2 | 3+3 2 |
| | 5+6 2 | 5+6 2 | 3+4 2 | 3+4 2 | 3+3 2 | 2+2 2 |
| | 3+4 2 | 3+4 2 | 3+4 2 | 3+4 2 | 2+2 2 | 2+2 2 |
| | 5+6 2 | 5+6 2 | 3+4 2 | 5+6 2 | 3+3 2 | 3+3 2 |
| | 3+4 2 | 3+4 2 | 5+6 2 | 5+6 2 | 2+2 2 | 3+3 2 |
| | 3+4 2 | 5+6 2 | 3+4 2 | 3+4 2 | 3+3 2 | 2+2 2 |
| | 5+6 2 | 5+6 2 | 3+4 2 | 5+6 2 | 2+2 2 | 3+3 2 |
| | 5+6 2 | 3+4 2 | 5+6 2 | 3+4 2 | 3+3 2 | 2+2 2 |
| | 3+4 2 | 3+4 2 | 5+6 2 | 5+6 2 | 2+2 2 | 3+3 2 |
| | 5+6 2 | 5+6 2 | 3+4 2 | 3+4 2 | 3+3 2 | 2+2 2 |
| | 3+4 2 | 5+6 2 | 5+6 2 | 3+4 2 | 2+2 2 | 3+3 2 |
| | 5+6 2 | 3+4 2 | 5+6 2 | 5+6 2 | 3+3 2 | 2+2 2 |
| | 3+4 2 | 5+6 2 | 5+6 2 | 3+4 2 | 2+2 2 | 3+3 2 |
| | 5+6 2 | 3+4 2 | 3+4 2 | 5+6 2 | 3+3 2 | 3+3 2 |
| | 3+4 2 | 5+6 2 | 3+4 2 | 3+4 2 | 2+2 2 | 2+2 2 |
| | 3+4 2 | 3+4 2 | 5+6 2 | 5+6 2 | 2+2 2 | 2+2 2 |
| | 3+4 2 | 3+4 2 | 3+4 2 | 5+6 2 | 3+3 2 | 3+3 2 |
| | 5+6 2 | 5+6 2 | 5+6 2 | 5+6 2 | 3+3 2 | 2+2 2 |
| | 3+4 2 | 3+4 2 | 3+4 2 | 3+4 2 | 2+2 2 | 3+3 2 |
| | 5+6 2 | 3+4 2 | 3+4 2 | 5+6 2 | 3+3 2 | 3+3 2 |
| | 5+6 2 | 5+6 2 | 5+6 2 | 3+4 2 | 3+3 2 | 2+2 2 |
| | 3+4 2 | 3+4 2 | 3+4 2 | 5+6 2 | 2+2 2 | 2+2 2 |
| | 5+6 2 | 5+6 2 | 5+6 2 | 3+4 2 | 3+3 2 | 3+3 2 |
| | 3+4 2 | 3+4 2 | 5+6 2 | 5+6 2 | 2+2 2 | 3+3 2 |
| | 5+6 2 | 3+4 2 | 3+4 2 | 3+4 2 | 2+2 2 | 2+2 2 |
| | 5+6 2 | 5+6 2 | 3+4 2 | 5+6 2 | 2+2 2 | 2+2 2 |
| | 3+4 2 | 5+6 2 | 5+6 2 | 3+4 2 | 2+2 2 | 3+3 2 |
| | 5+6 2 | 3+4 2 | 3+4 2 | 5+6 2 | 3+3 2 | 2+2 2 |
| | 3+4 2 | 5+6 2 | 5+6 2 | 3+4 2 | 2+2 2 | 3+3 2 |
| | 5+6 2 | 3+4 2 | 3+4 2 | 5+6 2 | 3+3 2 | 2+2 2 |
| | 3+4 2 | 5+6 2 | 3+4 2 | 3+4 2 | 2+2 2 | 3+3 2 |
| | 3+4 2 | 3+4 2 | 3+4 2 | 5+6 2 | 3+3 2 | 3+3 2 |
| | 5+6 2 | 5+6 2 | 5+6 2 | 5+6 2 | 2+2 2 | 3+3 2 |
| | 3+4 2 | 3+4 2 | 3+4 2 | 5+6 2 | 2+2 2 | 3+3 2 |
| | 3+4 2 | 5+6 2 | 3+4 2 | 3+4 2 | 3+3 2 | 2+2 2 |
| | 3+4 2 | 5+6 2 | 3+4 2 | 5+6 2 | 2+2 2 | 3+3 2 |
| | 5+6 2 | 5+6 2 | 5+6 2 | 5+6 2 | 3+3 2 | 2+2 2 |
| | 3+4 2 | 3+4 2 | 5+6 2 | 3+4 2 | 2+2 2 | 3+3 2 |
| | 5+6 2 | 5+6 2 | 3+4 2 | 5+6 2 | 2+2 2 | 3+3 2 |
| | 3+4 2 | 5+6 2 | 3+4 2 | 5+6 2 | 2+2 2 | 2+2 2 |
| | 3+4 2 | 3+4 2 | 3+4 2 | 5+6 2 | 3+3 2 | 3+3 2 |
| 12.1.84 | 421466 | 138 | Cu4Na1O12Te2Tl3 | | | |
| | -A | -GM | -M | -Y | -L | -V |
| | 3+4 2 | 3+4 2 | 5+6 2 | 5+6 2 | 2+2 2 | 3+3 2 |
| | 5+6 2 | 5+6 2 | 3+4 2 | 3+4 2 | 3+3 2 | 2+2 2 |
| | 5+6 2 | 3+4 2 | 3+4 2 | 5+6 2 | 3+3 2 | 2+2 2 |
| | 3+4 2 | 3+4 2 | 5+6 2 | 5+6 2 | 2+2 2 | 3+3 2 |
| | 5+6 2 | 5+6 2 | 3+4 2 | 3+4 2 | 2+2 2 | 3+3 2 |
| | 3+4 2 | 5+6 2 | 5+6 2 | 3+4 2 | 3+3 2 | 2+2 2 |
| | 5+6 2 | 3+4 2 | 3+4 2 | 3+4 2 | 2+2 2 | 3+3 2 |
| | 3+4 2 | 5+6 2 | 5+6 2 | 5+6 2 | 3+3 2 | 2+2 2 |
| | 5+6 2 | 3+4 2 | 3+4 2 | 5+6 2 | 2+2 2 | 2+2 2 |
| | 3+4 2 | 5+6 2 | 5+6 2 | 3+4 2 | 3+3 2 | 3+3 2 |
| | 5+6 2 | 3+4 2 | 3+4 2 | 5+6 2 | 3+3 2 | 3+3 2 |
| | 3+4 2 | 5+6 2 | 5+6 2 | 3+4 2 | 2+2 2 | 2+2 2 |
| | 5+6 2 | 3+4 2 | 5+6 2 | 5+6 2 | 2+2 2 | 3+3 2 |
| | 3+4 2 | 5+6 2 | 3+4 2 | 3+4 2 | 3+3 2 | 2+2 2 |
| | 3+4 2 | 3+4 2 | 3+4 2 | 3+4 2 | 2+2 2 | 3+3 2 |
| | 5+6 2 | 5+6 2 | 5+6 2 | 5+6 2 | 3+3 2 | 2+2 2 |
| | 3+4 2 | 5+6 2 | 3+4 2 | 3+4 2 | 2+2 2 | 2+2 2 |
| | 5+6 2 | 3+4 2 | 5+6 2 | 5+6 2 | 3+3 2 | 3+3 2 |

Continued on next column

**TABLE XXIV– continued**

| TQC | ICSD | NoE | SF | | | |
|---|---|---|---|---|---|---|
| | 3+4 2 | 3+4 2 | 5+6 2 | 5+6 2 | 3+3 2 | 2+2 2 |
| | 5+6 2 | 3+4 2 | 3+4 2 | 3+4 2 | 2+2 2 | 2+2 2 |
| | 3+4 2 | 5+6 2 | 3+4 2 | 3+4 2 | 3+3 2 | 3+3 2 |
| | 5+6 2 | 5+6 2 | 5+6 2 | 3+4 2 | 2+2 2 | 3+3 2 |
| | 3+4 2 | 5+6 2 | 3+4 2 | 5+6 2 | 2+2 2 | 2+2 2 |
| | 5+6 2 | 3+4 2 | 5+6 2 | 5+6 2 | 3+3 2 | 3+3 2 |
| | 3+4 2 | 3+4 2 | 5+6 2 | 5+6 2 | 2+2 2 | 3+3 2 |
| | 5+6 2 | 3+4 2 | 3+4 2 | 3+4 2 | 3+3 2 | 2+2 2 |
| | 3+4 2 | 5+6 2 | 3+4 2 | 5+6 2 | 2+2 2 | 2+2 2 |
| | 5+6 2 | 3+4 2 | 3+4 2 | 5+6 2 | 2+2 2 | 2+2 2 |
| | 5+6 2 | 3+4 2 | 3+4 2 | 3+4 2 | 3+3 2 | 3+3 2 |
| | 3+4 2 | 5+6 2 | 5+6 2 | 3+4 2 | 3+3 2 | 3+3 2 |
| | 3+4 2 | 3+4 2 | 5+6 2 | 5+6 2 | 2+2 2 | 3+3 2 |
| | 5+6 2 | 5+6 2 | 3+4 2 | 5+6 2 | 3+3 2 | 2+2 2 |
| | 5+6 2 | 5+6 2 | 5+6 2 | 3+4 2 | 3+3 2 | 2+2 2 |
| | 5+6 2 | 3+4 2 | 3+4 2 | 3+4 2 | 2+2 2 | 2+2 2 |
| | 5+6 2 | 3+4 2 | 5+6 2 | 5+6 2 | 2+2 2 | 3+3 2 |
| | 3+4 2 | 5+6 2 | 3+4 2 | 5+6 2 | 3+3 2 | 3+3 2 |
| | 3+4 2 | 3+4 2 | 3+4 2 | 3+4 2 | 2+2 2 | 3+3 2 |
| | 5+6 2 | 3+4 2 | 5+6 2 | 3+4 2 | 3+3 2 | 2+2 2 |
| | 3+4 2 | 5+6 2 | 5+6 2 | 5+6 2 | 3+3 2 | 2+2 2 |
| | 5+6 2 | 5+6 2 | 5+6 2 | 3+4 2 | 2+2 2 | 3+3 2 |
| | 3+4 2 | 5+6 2 | 3+4 2 | 3+4 2 | 2+2 2 | 3+3 2 |
| | 3+4 2 | 3+4 2 | 5+6 2 | 3+4 2 | 3+3 2 | 2+2 2 |
| | 5+6 2 | 5+6 2 | 3+4 2 | 5+6 2 | 2+2 2 | 2+2 2 |
| | 3+4 2 | 5+6 2 | 5+6 2 | 3+4 2 | 3+3 2 | 3+3 2 |
| | 3+4 2 | 3+4 2 | 5+6 2 | 5+6 2 | 2+2 2 | 3+3 2 |
| | 3+4 2 | 3+4 2 | 3+4 2 | 5+6 2 | 3+3 2 | 2+2 2 |
| | 3+4 2 | 3+4 2 | 3+4 2 | 5+6 2 | 2+2 2 | 3+3 2 |
| | 5+6 2 | 5+6 2 | 5+6 2 | 3+4 2 | 3+3 2 | 2+2 2 |
| | 5+6 2 | 5+6 2 | 5+6 2 | 5+6 2 | 2+2 2 | 2+2 2 |
| | 5+6 2 | 3+4 2 | 3+4 2 | 3+4 2 | 2+2 2 | 3+3 2 |
| | 3+4 2 | 5+6 2 | 3+4 2 | 5+6 2 | 3+3 2 | 2+2 2 |
| | 5+6 2 | 3+4 2 | 3+4 2 | 3+4 2 | 3+3 2 | 3+3 2 |
| | 5+6 2 | 5+6 2 | 5+6 2 | 3+4 2 | 2+2 2 | 2+2 2 |
| | 3+4 2 | 3+4 2 | 5+6 2 | 5+6 2 | 3+3 2 | 3+3 2 |
| | 5+6 2 | 5+6 2 | 3+4 2 | 3+4 2 | 2+2 2 | 2+2 2 |
| | 3+4 2 | 3+4 2 | 5+6 2 | 5+6 2 | 2+2 2 | 2+2 2 |
| | 3+4 2 | 3+4 2 | 3+4 2 | 5+6 2 | 3+3 2 | 3+3 2 |
| | 3+4 2 | 5+6 2 | 5+6 2 | 5+6 2 | 3+3 2 | 3+3 2 |
| | 5+6 2 | 5+6 2 | 5+6 2 | 3+4 2 | 2+2 2 | 2+2 2 |
| | 5+6 2 | 3+4 2 | 3+4 2 | 3+4 2 | 3+3 2 | 3+3 2 |
| | 3+4 2 | 5+6 2 | 3+4 2 | 5+6 2 | 2+2 2 | 2+2 2 |
| | 5+6 2 | 3+4 2 | 3+4 2 | 3+4 2 | 3+3 2 | 3+3 2 |
| | 3+4 2 | 5+6 2 | 5+6 2 | 3+4 2 | 2+2 2 | 2+2 2 |
| | 5+6 2 | 3+4 2 | 5+6 2 | 5+6 2 | 3+3 2 | 3+3 2 |
| | 3+4 2 | 3+4 2 | 3+4 2 | 3+4 2 | 2+2 2 | 2+2 2 |
| | 5+6 2 | 5+6 2 | 5+6 2 | 5+6 2 | 3+3 2 | 3+3 2 |
| | 3+4 2 | 5+6 2 | 3+4 2 | 3+4 2 | 2+2 2 | 2+2 2 |
| | 3+4 2 | 5+6 2 | 5+6 2 | 3+4 2 | 3+3 2 | 3+3 2 |
| | 5+6 2 | 5+6 2 | 3+4 2 | 3+4 2 | 2+2 2 | 2+2 2 |
| 12.1.85 | 71624 | 106 | Cs2Ir1Na2O4 | | | |
| | -A | -GM | -M | -Y | -L | -V |
| | 3+4 2 | 3+4 2 | 5+6 2 | 5+6 2 | 2+2 2 | 2+2 2 |
| | 5+6 2 | 5+6 2 | 3+4 2 | 3+4 2 | 3+3 2 | 3+3 2 |
| | 5+6 2 | 5+6 2 | 5+6 2 | 5+6 2 | 3+3 2 | 3+3 2 |
| | 3+4 2 | 3+4 2 | 3+4 2 | 3+4 2 | 2+2 2 | 2+2 2 |
| | 5+6 2 | 3+4 2 | 5+6 2 | 3+4 2 | 3+3 2 | 2+2 2 |
| | 3+4 2 | 3+4 2 | 3+4 2 | 3+4 2 | 3+3 2 | 3+3 2 |
| | 5+6 2 | 3+4 2 | 5+6 2 | 3+4 2 | 3+3 2 | 2+2 2 |
| | 3+4 2 | 3+4 2 | 3+4 2 | 3+4 2 | 3+3 2 | 3+3 2 |

Continued on next column

**TABLE XXIV– continued**

| TQC | ICSD | NoE | SF | | | |
|---|---|---|---|---|---|---|
| | 5+6 2 | 5+6 2 | 3+4 2 | 5+6 2 | 2+2 2 | 2+2 2 |
| | 3+4 2 | 5+6 2 | 5+6 2 | 5+6 2 | 2+2 2 | 3+3 2 |
| | 3+4 2 | 5+6 2 | 3+4 2 | 5+6 2 | 2+2 2 | 3+3 2 |
| | 5+6 2 | 5+6 2 | 5+6 2 | 5+6 2 | 2+2 2 | 2+2 2 |
| | 5+6 2 | 5+6 2 | 3+4 2 | 3+4 2 | 3+3 2 | 3+3 2 |
| | 3+4 2 | 3+4 2 | 5+6 2 | 5+6 2 | 2+2 2 | 2+2 2 |
| | 3+4 2 | 3+4 2 | 5+6 2 | 5+6 2 | 2+2 2 | 2+2 2 |
| | 5+6 2 | 5+6 2 | 3+4 2 | 3+4 2 | 3+3 2 | 3+3 2 |
| | 3+4 2 | 3+4 2 | 3+4 2 | 3+4 2 | 3+3 2 | 3+3 2 |
| | 5+6 2 | 5+6 2 | 5+6 2 | 5+6 2 | 2+2 2 | 2+2 2 |
| | 5+6 2 | 5+6 2 | 5+6 2 | 5+6 2 | 2+2 2 | 2+2 2 |
| | 3+4 2 | 3+4 2 | 3+4 2 | 3+4 2 | 3+3 2 | 3+3 2 |
| | 3+4 2 | 3+4 2 | 5+6 2 | 5+6 2 | 3+3 2 | 3+3 2 |
| | 3+4 2 | 3+4 2 | 3+4 2 | 3+4 2 | 2+2 2 | 2+2 2 |
| | 5+6 2 | 5+6 2 | 5+6 2 | 5+6 2 | 3+3 2 | 2+2 2 |
| | 5+6 2 | 5+6 2 | 3+4 2 | 3+4 2 | 2+2 2 | 3+3 2 |
| | 5+6 2 | 3+4 2 | 3+4 2 | 3+4 2 | 3+3 2 | 2+2 2 |
| | 3+4 2 | 3+4 2 | 5+6 2 | 3+4 2 | 3+3 2 | 3+3 2 |
| | 5+6 2 | 3+4 2 | 5+6 2 | 3+4 2 | 3+3 2 | 2+2 2 |
| | 5+6 2 | 3+4 2 | 3+4 2 | 3+4 2 | 3+3 2 | 3+3 2 |
| | 3+4 2 | 3+4 2 | 5+6 2 | 3+4 2 | 3+3 2 | 3+3 2 |
| | 3+4 2 | 3+4 2 | 3+4 2 | 3+4 2 | 3+3 2 | 2+2 2 |
| | 5+6 2 | 3+4 2 | 5+6 2 | 3+4 2 | 3+3 2 | 2+2 2 |
| | 5+6 2 | 3+4 2 | 5+6 2 | 3+4 2 | 3+3 2 | 3+3 2 |
| | 3+4 2 | 3+4 2 | 3+4 2 | 3+4 2 | 3+3 2 | 2+2 2 |
| | 3+4 2 | 3+4 2 | 3+4 2 | 3+4 2 | 3+3 2 | 3+3 2 |
| | 3+4 2 | 5+6 2 | 3+4 2 | 5+6 2 | 2+2 2 | 3+3 2 |
| | 5+6 2 | 5+6 2 | 5+6 2 | 5+6 2 | 2+2 2 | 2+2 2 |
| | 3+4 2 | 5+6 2 | 3+4 2 | 5+6 2 | 2+2 2 | 2+2 2 |
| | 5+6 2 | 5+6 2 | 3+4 2 | 5+6 2 | 2+2 2 | 2+2 2 |
| | 3+4 2 | 5+6 2 | 5+6 2 | 5+6 2 | 2+2 2 | 3+3 2 |
| | 5+6 2 | 5+6 2 | 5+6 2 | 5+6 2 | 2+2 2 | 3+3 2 |
| | 5+6 2 | 5+6 2 | 3+4 2 | 5+6 2 | 2+2 2 | 3+3 2 |
| | 3+4 2 | 5+6 2 | 5+6 2 | 5+6 2 | 2+2 2 | 3+3 2 |
| | 3+4 2 | 3+4 2 | 3+4 2 | 5+6 2 | 2+2 2 | 2+2 2 |
| | 3+4 2 | 5+6 2 | 3+4 2 | 3+4 2 | 2+2 2 | 2+2 2 |
| | 5+6 2 | 3+4 2 | 5+6 2 | 3+4 2 | 2+2 2 | 3+3 2 |
| | 5+6 2 | 5+6 2 | 5+6 2 | 5+6 2 | 3+3 2 | 2+2 2 |
| | 5+6 2 | 5+6 2 | 3+4 2 | 5+6 2 | 3+3 2 | 3+3 2 |
| | 3+4 2 | 5+6 2 | 5+6 2 | 5+6 2 | 2+2 2 | 2+2 2 |
| | 3+4 2 | 3+4 2 | 3+4 2 | 3+4 2 | 3+3 2 | 2+2 2 |
| | 5+6 2 | 3+4 2 | 5+6 2 | 3+4 2 | 3+3 2 | 3+3 2 |
| | 5+6 2 | 3+4 2 | 5+6 2 | 3+4 2 | 3+3 2 | 3+3 2 |
| | 5+6 2 | 3+4 2 | 5+6 2 | 3+4 2 | 3+3 2 | 3+3 2 |
| | 3+4 2 | 3+4 2 | 3+4 2 | 3+4 2 | 3+3 2 | 2+2 2 |
| 12.1.86 | 26977 | 90 | Cr7Se8 | | | |
| | -A | -GM | -M | -Y | -L | -V |
| | 5+6 2 | 3+4 2 | 5+6 2 | 3+4 2 | 2+2 2 | 2+2 2 |
| | 3+4 2 | 5+6 2 | 3+4 2 | 5+6 2 | 3+3 2 | 3+3 2 |
| | 5+6 2 | 3+4 2 | 5+6 2 | 5+6 2 | 2+2 2 | 3+3 2 |
| | 3+4 2 | 5+6 2 | 5+6 2 | 3+4 2 | 3+3 2 | 2+2 2 |
| | 3+4 2 | 3+4 2 | 5+6 2 | 3+4 2 | 3+3 2 | 2+2 2 |
| | 3+4 2 | 3+4 2 | 3+4 2 | 5+6 2 | 2+2 2 | 3+3 2 |
| | 5+6 2 | 5+6 2 | 3+4 2 | 5+6 2 | 2+2 2 | 2+2 2 |
| | 5+6 2 | 5+6 2 | 3+4 2 | 3+4 2 | 3+3 2 | 3+3 2 |
| | 5+6 2 | 3+4 2 | 3+4 2 | 5+6 2 | 2+2 2 | 2+2 2 |
| | 3+4 2 | 3+4 2 | 3+4 2 | 3+4 2 | 2+2 2 | 2+2 2 |
| | 5+6 2 | 5+6 2 | 5+6 2 | 5+6 2 | 3+3 2 | 3+3 2 |
| | 3+4 2 | 3+4 2 | 3+4 2 | 3+4 2 | 3+3 2 | 3+3 2 |
| | 5+6 2 | 5+6 2 | 5+6 2 | 5+6 2 | 2+2 2 | 2+2 2 |
| | 3+4 2 | 3+4 2 | 5+6 2 | 3+4 2 | 3+3 2 | 3+3 2 |

Continued on next column

| TQC | ICSD | NoE | SF | | | |
|---|---|---|---|---|---|---|
| | 5+6 2 | 5+6 2 | 5+6 2 | 5+6 2 | 2+2 2 | 2+2 2 |
| | 3+4 2 | 3+4 2 | 3+4 2 | 3+4 2 | 3+3 2 | 3+3 2 |
| | 3+4 2 | 3+4 2 | 5+6 2 | 5+6 2 | 2+2 2 | 2+2 2 |
| | 5+6 2 | 5+6 2 | 5+6 2 | 3+4 2 | 3+3 2 | 3+3 2 |
| | 5+6 2 | 3+4 2 | 3+4 2 | 5+6 2 | 2+2 2 | 2+2 2 |
| | 3+4 2 | 3+4 2 | 3+4 2 | 3+4 2 | 3+3 2 | 3+3 2 |
| | 5+6 2 | 5+6 2 | 3+4 2 | 5+6 2 | 2+2 2 | 3+3 2 |
| | 5+6 2 | 3+4 2 | 5+6 2 | 3+4 2 | 3+3 2 | 2+2 2 |
| | 3+4 2 | 5+6 2 | 3+4 2 | 5+6 2 | 3+3 2 | 3+3 2 |
| | 3+4 2 | 3+4 2 | 5+6 2 | 3+4 2 | 2+2 2 | 2+2 2 |
| | 5+6 2 | 5+6 2 | 3+4 2 | 5+6 2 | 3+3 2 | 2+2 2 |
| | 3+4 2 | 5+6 2 | 3+4 2 | 3+4 2 | 2+2 2 | 3+3 2 |
| | 5+6 2 | 3+4 2 | 5+6 2 | 3+4 2 | 2+2 2 | 2+2 2 |
| | 3+4 2 | 5+6 2 | 5+6 2 | 5+6 2 | 2+2 2 | 3+3 2 |
| | 5+6 2 | 3+4 2 | 5+6 2 | 5+6 2 | 3+3 2 | 2+2 2 |
| | 3+4 2 | 3+4 2 | 5+6 2 | 5+6 2 | 2+2 2 | 3+3 2 |
| | 5+6 2 | 5+6 2 | 3+4 2 | 3+4 2 | 3+3 2 | 2+2 2 |
| | 3+4 2 | 3+4 2 | 3+4 2 | 3+4 2 | 3+3 2 | 3+3 2 |
| | 5+6 2 | 5+6 2 | 3+4 2 | 5+6 2 | 2+2 2 | 2+2 2 |
| | 3+4 2 | 5+6 2 | 3+4 2 | 5+6 2 | 3+3 2 | 3+3 2 |
| | 5+6 2 | 5+6 2 | 5+6 2 | 3+4 2 | 2+2 2 | 2+2 2 |
| | 5+6 2 | 3+4 2 | 3+4 2 | 5+6 2 | 2+2 2 | 2+2 2 |
| | 3+4 2 | 5+6 2 | 5+6 2 | 3+4 2 | 3+3 2 | 2+2 2 |
| | 3+4 2 | 3+4 2 | 3+4 2 | 3+4 2 | 2+2 2 | 3+3 2 |
| | 5+6 2 | 3+4 2 | 5+6 2 | 5+6 2 | 3+3 2 | 2+2 2 |
| | 3+4 2 | 3+4 2 | 3+4 2 | 3+4 2 | 2+2 2 | 3+3 2 |
| | 5+6 2 | 3+4 2 | 3+4 2 | 5+6 2 | 2+2 2 | 2+2 2 |
| | 5+6 2 | 3+4 2 | 5+6 2 | 5+6 2 | 3+3 2 | 3+3 2 |
| | 3+4 2 | 3+4 2 | 3+4 2 | 3+4 2 | 2+2 2 | 3+3 2 |
| | 5+6 2 | 3+4 2 | 3+4 2 | 5+6 2 | 2+2 2 | 2+2 2 |
| | 3+4 2 | 5+6 2 | 3+4 2 | 5+6 2 | 3+3 2 | 3+3 2 |
| 12.1.87 | 626888 | 42 | Cr3Te4 | | | |
| | -A | -GM | -M | -Y | -L | -V |
| | 3+4 2 | 3+4 2 | 3+4 2 | 5+6 2 | 2+2 2 | 2+2 2 |
| | 5+6 2 | 5+6 2 | 5+6 2 | 3+4 2 | 3+3 2 | 3+3 2 |
| | 3+4 2 | 3+4 2 | 5+6 2 | 5+6 2 | 2+2 2 | 3+3 2 |
| | 5+6 2 | 5+6 2 | 3+4 2 | 3+4 2 | 3+3 2 | 2+2 2 |
| | 3+4 2 | 3+4 2 | 3+4 2 | 3+4 2 | 3+3 2 | 3+3 2 |
| | 5+6 2 | 5+6 2 | 5+6 2 | 5+6 2 | 2+2 2 | 2+2 2 |
| | 3+4 2 | 3+4 2 | 5+6 2 | 5+6 2 | 3+3 2 | 3+3 2 |
| | 5+6 2 | 3+4 2 | 3+4 2 | 3+4 2 | 2+2 2 | 2+2 2 |
| | 3+4 2 | 3+4 2 | 5+6 2 | 5+6 2 | 2+2 2 | 3+3 2 |
| | 5+6 2 | 5+6 2 | 3+4 2 | 3+4 2 | 3+3 2 | 2+2 2 |
| | 3+4 2 | 5+6 2 | 5+6 2 | 3+4 2 | 3+3 2 | 3+3 2 |
| | 5+6 2 | 3+4 2 | 3+4 2 | 5+6 2 | 2+2 2 | 2+2 2 |
| | 3+4 2 | 3+4 2 | 3+4 2 | 5+6 2 | 2+2 2 | 3+3 2 |
| | 5+6 2 | 3+4 2 | 3+4 2 | 5+6 2 | 3+3 2 | 2+2 2 |
| | 3+4 2 | 5+6 2 | 5+6 2 | 3+4 2 | 2+2 2 | 3+3 2 |
| | 5+6 2 | 5+6 2 | 3+4 2 | 3+4 2 | 3+3 2 | 2+2 2 |
| | 3+4 2 | 3+4 2 | 5+6 2 | 3+4 2 | 3+3 2 | 3+3 2 |
| | 3+4 2 | 3+4 2 | 3+4 2 | 5+6 2 | 2+2 2 | 3+3 2 |
| | 3+4 2 | 3+4 2 | 5+6 2 | 5+6 2 | 3+3 2 | 2+2 2 |
| | 5+6 2 | 3+4 2 | 3+4 2 | 3+4 2 | 3+3 2 | 3+3 2 |
| | 3+4 2 | 5+6 2 | 3+4 2 | 3+4 2 | 2+2 2 | 3+3 2 |
| 12.1.88 | 626884 | 42 | Cr3Te4 | | | |
| | -A | -GM | -M | -Y | -L | -V |
| | 3+4 2 | 3+4 2 | 3+4 2 | 5+6 2 | 2+2 2 | 2+2 2 |
| | 5+6 2 | 3+4 2 | 5+6 2 | 3+4 2 | 3+3 2 | 3+3 2 |
| | 5+6 2 | 5+6 2 | 5+6 2 | 5+6 2 | 2+2 2 | 3+3 2 |
| | 3+4 2 | 5+6 2 | 3+4 2 | 3+4 2 | 3+3 2 | 2+2 2 |
| | 3+4 2 | 3+4 2 | 3+4 2 | 3+4 2 | 3+3 2 | 3+3 2 |

Continued on next column

**TABLE XXIV– continued**

| TQC | ICSD | | NoE | | SF | |
|---|---|---|---|---|---|---|
| | 5+6 2 | 5+6 2 | 5+6 2 | 5+6 2 | 2+2 2 | 2+2 2 |
| | 3+4 2 | 3+4 2 | 3+4 2 | 3+4 2 | 3+3 2 | 3+3 2 |
| | 5+6 2 | 3+4 2 | 5+6 2 | 5+6 2 | 2+2 2 | 2+2 2 |
| | 5+6 2 | 3+4 2 | 5+6 2 | 5+6 2 | 2+2 2 | 3+3 2 |
| | 3+4 2 | 5+6 2 | 3+4 2 | 3+4 2 | 3+3 2 | 2+2 2 |
| | 3+4 2 | 3+4 2 | 3+4 2 | 3+4 2 | 3+3 2 | 3+3 2 |
| | 5+6 2 | 5+6 2 | 5+6 2 | 5+6 2 | 2+2 2 | 2+2 2 |
| | 3+4 2 | 3+4 2 | 3+4 2 | 5+6 2 | 2+2 2 | 3+3 2 |
| | 5+6 2 | 3+4 2 | 3+4 2 | 5+6 2 | 3+3 2 | 2+2 2 |
| | 3+4 2 | 5+6 2 | 5+6 2 | 3+4 2 | 2+2 2 | 3+3 2 |
| | 5+6 2 | 5+6 2 | 3+4 2 | 3+4 2 | 3+3 2 | 2+2 2 |
| | 3+4 2 | 3+4 2 | 5+6 2 | 3+4 2 | 3+3 2 | 3+3 2 |
| | 3+4 2 | 3+4 2 | 5+6 2 | 5+6 2 | 2+2 2 | 3+3 2 |
| | 3+4 2 | 3+4 2 | 3+4 2 | 5+6 2 | 3+3 2 | 2+2 2 |
| | 5+6 2 | 3+4 2 | 3+4 2 | 3+4 2 | 3+3 2 | 3+3 2 |
| | 3+4 2 | 5+6 2 | 3+4 2 | 3+4 2 | 2+2 2 | 3+3 2 |
| 12.1.89 | 626874 | | 42 | | Cr3Te4 | |
| | -A | -GM | -M | -Y | -L | -V |
| | 3+4 2 | 3+4 2 | 3+4 2 | 5+6 2 | 2+2 2 | 2+2 2 |
| | 5+6 2 | 3+4 2 | 5+6 2 | 3+4 2 | 3+3 2 | 3+3 2 |
| | 3+4 2 | 5+6 2 | 5+6 2 | 5+6 2 | 2+2 2 | 3+3 2 |
| | 5+6 2 | 5+6 2 | 3+4 2 | 3+4 2 | 3+3 2 | 2+2 2 |
| | 3+4 2 | 3+4 2 | 3+4 2 | 3+4 2 | 3+3 2 | 3+3 2 |
| | 5+6 2 | 5+6 2 | 5+6 2 | 5+6 2 | 2+2 2 | 2+2 2 |
| | 3+4 2 | 3+4 2 | 3+4 2 | 5+6 2 | 3+3 2 | 3+3 2 |
| | 5+6 2 | 3+4 2 | 5+6 2 | 3+4 2 | 2+2 2 | 2+2 2 |
| | 3+4 2 | 3+4 2 | 5+6 2 | 5+6 2 | 2+2 2 | 3+3 2 |
| | 5+6 2 | 5+6 2 | 3+4 2 | 3+4 2 | 3+3 2 | 2+2 2 |
| | 3+4 2 | 5+6 2 | 5+6 2 | 3+4 2 | 3+3 2 | 3+3 2 |
| | 5+6 2 | 3+4 2 | 3+4 2 | 5+6 2 | 2+2 2 | 2+2 2 |
| | 3+4 2 | 3+4 2 | 3+4 2 | 5+6 2 | 2+2 2 | 3+3 2 |
| | 5+6 2 | 3+4 2 | 3+4 2 | 5+6 2 | 3+3 2 | 2+2 2 |
| | 3+4 2 | 5+6 2 | 5+6 2 | 3+4 2 | 2+2 2 | 3+3 2 |
| | 5+6 2 | 5+6 2 | 3+4 2 | 3+4 2 | 3+3 2 | 2+2 2 |
| | 3+4 2 | 3+4 2 | 5+6 2 | 3+4 2 | 3+3 2 | 3+3 2 |
| | 3+4 2 | 3+4 2 | 3+4 2 | 5+6 2 | 2+2 2 | 3+3 2 |
| | 3+4 2 | 3+4 2 | 5+6 2 | 5+6 2 | 3+3 2 | 2+2 2 |
| | 5+6 2 | 3+4 2 | 3+4 2 | 3+4 2 | 3+3 2 | 3+3 2 |
| | 3+4 2 | 5+6 2 | 3+4 2 | 3+4 2 | 2+2 2 | 3+3 2 |
| 12.1.90 | 626716 | | 42 | | Cr3Se4 | |
| | -A | -GM | -M | -Y | -L | -V |
| | 3+4 2 | 3+4 2 | 5+6 2 | 5+6 2 | 2+2 2 | 3+3 2 |
| | 5+6 2 | 5+6 2 | 3+4 2 | 3+4 2 | 3+3 2 | 2+2 2 |
| | 3+4 2 | 3+4 2 | 5+6 2 | 5+6 2 | 2+2 2 | 3+3 2 |
| | 5+6 2 | 5+6 2 | 3+4 2 | 3+4 2 | 3+3 2 | 2+2 2 |
| | 3+4 2 | 3+4 2 | 3+4 2 | 3+4 2 | 3+3 2 | 3+3 2 |
| | 5+6 2 | 5+6 2 | 5+6 2 | 5+6 2 | 2+2 2 | 2+2 2 |
| | 3+4 2 | 3+4 2 | 5+6 2 | 5+6 2 | 3+3 2 | 3+3 2 |
| | 5+6 2 | 3+4 2 | 3+4 2 | 3+4 2 | 2+2 2 | 2+2 2 |
| | 3+4 2 | 3+4 2 | 5+6 2 | 3+4 2 | 3+3 2 | 3+3 2 |
| | 3+4 2 | 5+6 2 | 3+4 2 | 5+6 2 | 2+2 2 | 3+3 2 |
| | 5+6 2 | 5+6 2 | 3+4 2 | 3+4 2 | 3+3 2 | 2+2 2 |
| | 3+4 2 | 3+4 2 | 5+6 2 | 5+6 2 | 2+2 2 | 2+2 2 |
| | 5+6 2 | 3+4 2 | 3+4 2 | 5+6 2 | 2+2 2 | 3+3 2 |
| | 3+4 2 | 3+4 2 | 3+4 2 | 5+6 2 | 3+3 2 | 2+2 2 |
| | 5+6 2 | 5+6 2 | 5+6 2 | 3+4 2 | 2+2 2 | 3+3 2 |
| | 5+6 2 | 5+6 2 | 5+6 2 | 3+4 2 | 3+3 2 | 2+2 2 |
| | 3+4 2 | 3+4 2 | 3+4 2 | 3+4 2 | 3+3 2 | 3+3 2 |
| | 3+4 2 | 5+6 2 | 5+6 2 | 5+6 2 | 2+2 2 | 2+2 2 |
| | 3+4 2 | 3+4 2 | 3+4 2 | 5+6 2 | 3+3 2 | 3+3 2 |
| | 5+6 2 | 3+4 2 | 3+4 2 | 3+4 2 | 3+3 2 | 3+3 2 |
| | 5+6 2 | 5+6 2 | 3+4 2 | 3+4 2 | 2+2 2 | 3+3 2 |
| 12.1.91 | 626712 | | 42 | | Cr3Se4 | |
| | -A | -GM | -M | -Y | -L | -V |
| | 3+4 2 | 3+4 2 | 5+6 2 | 5+6 2 | 2+2 2 | 3+3 2 |
| | 5+6 2 | 5+6 2 | 3+4 2 | 3+4 2 | 3+3 2 | 2+2 2 |
| | 3+4 2 | 3+4 2 | 5+6 2 | 5+6 2 | 2+2 2 | 3+3 2 |
| | 5+6 2 | 5+6 2 | 3+4 2 | 3+4 2 | 3+3 2 | 2+2 2 |
| | 3+4 2 | 3+4 2 | 3+4 2 | 3+4 2 | 3+3 2 | 3+3 2 |
| | 5+6 2 | 5+6 2 | 5+6 2 | 5+6 2 | 2+2 2 | 2+2 2 |
| | 3+4 2 | 3+4 2 | 5+6 2 | 5+6 2 | 3+3 2 | 3+3 2 |
| | 5+6 2 | 3+4 2 | 3+4 2 | 3+4 2 | 2+2 2 | 2+2 2 |
| | 3+4 2 | 3+4 2 | 5+6 2 | 3+4 2 | 3+3 2 | 3+3 2 |
| | 5+6 2 | 5+6 2 | 3+4 2 | 5+6 2 | 2+2 2 | 3+3 2 |
| | 3+4 2 | 5+6 2 | 3+4 2 | 3+4 2 | 3+3 2 | 2+2 2 |
| | 3+4 2 | 3+4 2 | 5+6 2 | 5+6 2 | 2+2 2 | 2+2 2 |
| | 5+6 2 | 3+4 2 | 3+4 2 | 5+6 2 | 2+2 2 | 3+3 2 |
| | 3+4 2 | 3+4 2 | 3+4 2 | 5+6 2 | 3+3 2 | 2+2 2 |
| | 5+6 2 | 5+6 2 | 5+6 2 | 3+4 2 | 2+2 2 | 3+3 2 |
| | 5+6 2 | 5+6 2 | 5+6 2 | 3+4 2 | 3+3 2 | 2+2 2 |
| | 3+4 2 | 3+4 2 | 3+4 2 | 3+4 2 | 3+3 2 | 3+3 2 |
| | 3+4 2 | 5+6 2 | 5+6 2 | 5+6 2 | 2+2 2 | 2+2 2 |
| | 3+4 2 | 3+4 2 | 3+4 2 | 5+6 2 | 3+3 2 | 3+3 2 |
| | 5+6 2 | 3+4 2 | 3+4 2 | 3+4 2 | 3+3 2 | 3+3 2 |
| | 5+6 2 | 5+6 2 | 3+4 2 | 3+4 2 | 2+2 2 | 3+3 2 |
| 12.1.92 | 26976 | | 42 | | Cr3Se4 | |
| | -A | -GM | -M | -Y | -L | -V |
| | 5+6 2 | 3+4 2 | 5+6 2 | 5+6 2 | 2+2 2 | 2+2 2 |
| | 3+4 2 | 3+4 2 | 3+4 2 | 3+4 2 | 2+2 2 | 3+3 2 |
| | 5+6 2 | 5+6 2 | 3+4 2 | 5+6 2 | 3+3 2 | 2+2 2 |
| | 3+4 2 | 5+6 2 | 5+6 2 | 3+4 2 | 3+3 2 | 3+3 2 |
| | 5+6 2 | 3+4 2 | 5+6 2 | 3+4 2 | 3+3 2 | 2+2 2 |
| | 3+4 2 | 5+6 2 | 3+4 2 | 5+6 2 | 2+2 2 | 3+3 2 |
| | 5+6 2 | 3+4 2 | 5+6 2 | 5+6 2 | 3+3 2 | 2+2 2 |
| | 3+4 2 | 3+4 2 | 3+4 2 | 3+4 2 | 2+2 2 | 3+3 2 |
| | 5+6 2 | 3+4 2 | 3+4 2 | 5+6 2 | 3+3 2 | 2+2 2 |
| | 3+4 2 | 5+6 2 | 5+6 2 | 3+4 2 | 2+2 2 | 3+3 2 |
| | 5+6 2 | 5+6 2 | 5+6 2 | 3+4 2 | 3+3 2 | 2+2 2 |
| | 3+4 2 | 3+4 2 | 3+4 2 | 5+6 2 | 2+2 2 | 3+3 2 |
| | 5+6 2 | 3+4 2 | 5+6 2 | 5+6 2 | 2+2 2 | 2+2 2 |
| | 5+6 2 | 3+4 2 | 5+6 2 | 5+6 2 | 3+3 2 | 3+3 2 |
| | 3+4 2 | 5+6 2 | 3+4 2 | 3+4 2 | 2+2 2 | 2+2 2 |
| | 3+4 2 | 5+6 2 | 3+4 2 | 3+4 2 | 3+3 2 | 3+3 2 |
| | 5+6 2 | 5+6 2 | 5+6 2 | 3+4 2 | 3+3 2 | 2+2 2 |
| | 5+6 2 | 3+4 2 | 3+4 2 | 5+6 2 | 2+2 2 | 2+2 2 |
| | 5+6 2 | 3+4 2 | 5+6 2 | 5+6 2 | 3+3 2 | 3+3 2 |
| | 3+4 2 | 5+6 2 | 5+6 2 | 3+4 2 | 3+3 2 | 2+2 2 |
| | 3+4 2 | 3+4 2 | 5+6 2 | 3+4 2 | 2+2 2 | 2+2 2 |
| 12.1.93 | 81886 | | 42 | | Cr3S4 | |
| | -A | -GM | -M | -Y | -L | -V |
| | 3+4 2 | 3+4 2 | 3+4 2 | 5+6 2 | 2+2 2 | 3+3 2 |
| | 5+6 2 | 5+6 2 | 5+6 2 | 3+4 2 | 3+3 2 | 2+2 2 |
| | 3+4 2 | 3+4 2 | 5+6 2 | 5+6 2 | 2+2 2 | 3+3 2 |
| | 5+6 2 | 5+6 2 | 3+4 2 | 3+4 2 | 3+3 2 | 2+2 2 |
| | 3+4 2 | 3+4 2 | 3+4 2 | 3+4 2 | 3+3 2 | 3+3 2 |
| | 5+6 2 | 5+6 2 | 5+6 2 | 5+6 2 | 2+2 2 | 2+2 2 |
| | 3+4 2 | 3+4 2 | 5+6 2 | 5+6 2 | 3+3 2 | 3+3 2 |
| | 5+6 2 | 3+4 2 | 3+4 2 | 3+4 2 | 2+2 2 | 2+2 2 |
| | 3+4 2 | 3+4 2 | 5+6 2 | 3+4 2 | 3+3 2 | 3+3 2 |
| | 3+4 2 | 5+6 2 | 3+4 2 | 5+6 2 | 2+2 2 | 3+3 2 |
| | 5+6 2 | 5+6 2 | 3+4 2 | 3+4 2 | 3+3 2 | 2+2 2 |
| | 5+6 2 | 3+4 2 | 5+6 2 | 5+6 2 | 2+2 2 | 2+2 2 |

Continued on next column

Continued on next column

**TABLE XXIV– continued**

| TQC | ICSD | | NoE | SF | | |
|---|---|---|---|---|---|---|
| | 3+4 2 | 3+4 2 | 3+4 2 | 5+6 2 | 2+2 2 | 3+3 2 |
| | 3+4 2 | 3+4 2 | 3+4 2 | 5+6 2 | 3+3 2 | 2+2 2 |
| | 5+6 2 | 5+6 2 | 5+6 2 | 3+4 2 | 2+2 2 | 3+3 2 |
| | 5+6 2 | 5+6 2 | 5+6 2 | 3+4 2 | 3+3 2 | 2+2 2 |
| | 3+4 2 | 5+6 2 | 3+4 2 | 3+4 2 | 3+3 2 | 3+3 2 |
| | 3+4 2 | 5+6 2 | 5+6 2 | 5+6 2 | 2+2 2 | 3+3 2 |
| | 3+4 2 | 3+4 2 | 3+4 2 | 5+6 2 | 3+3 2 | 2+2 2 |
| | 5+6 2 | 3+4 2 | 3+4 2 | 3+4 2 | 3+3 2 | 3+3 2 |
| | 5+6 2 | 3+4 2 | 3+4 2 | 3+4 2 | 2+2 2 | 3+3 2 |
| 12.1.94 | 626598 | | 42 | Cr3S4 | | |
| | -A | -GM | -M | -Y | -L | -V |
| | 3+4 2 | 3+4 2 | 3+4 2 | 5+6 2 | 2+2 2 | 2+2 2 |
| | 5+6 2 | 5+6 2 | 5+6 2 | 3+4 2 | 3+3 2 | 3+3 2 |
| | 3+4 2 | 3+4 2 | 5+6 2 | 5+6 2 | 2+2 2 | 3+3 2 |
| | 5+6 2 | 5+6 2 | 3+4 2 | 3+4 2 | 3+3 2 | 2+2 2 |
| | 3+4 2 | 3+4 2 | 3+4 2 | 3+4 2 | 3+3 2 | 3+3 2 |
| | 3+4 2 | 5+6 2 | 5+6 2 | 5+6 2 | 2+2 2 | 2+2 2 |
| | 5+6 2 | 3+4 2 | 5+6 2 | 3+4 2 | 3+3 2 | 3+3 2 |
| | 5+6 2 | 3+4 2 | 3+4 2 | 5+6 2 | 2+2 2 | 2+2 2 |
| | 3+4 2 | 3+4 2 | 5+6 2 | 3+4 2 | 3+3 2 | 3+3 2 |
| | 5+6 2 | 5+6 2 | 3+4 2 | 3+4 2 | 2+2 2 | 2+2 2 |
| | 3+4 2 | 5+6 2 | 5+6 2 | 5+6 2 | 3+3 2 | 3+3 2 |
| | 5+6 2 | 3+4 2 | 3+4 2 | 5+6 2 | 2+2 2 | 3+3 2 |
| | 3+4 2 | 3+4 2 | 3+4 2 | 5+6 2 | 2+2 2 | 2+2 2 |
| | 5+6 2 | 3+4 2 | 3+4 2 | 5+6 2 | 3+3 2 | 2+2 2 |
| | 3+4 2 | 5+6 2 | 5+6 2 | 3+4 2 | 2+2 2 | 3+3 2 |
| | 5+6 2 | 5+6 2 | 5+6 2 | 3+4 2 | 3+3 2 | 2+2 2 |
| | 3+4 2 | 5+6 2 | 3+4 2 | 3+4 2 | 3+3 2 | 3+3 2 |
| | 3+4 2 | 5+6 2 | 5+6 2 | 5+6 2 | 2+2 2 | 3+3 2 |
| | 3+4 2 | 3+4 2 | 3+4 2 | 5+6 2 | 3+3 2 | 2+2 2 |
| | 5+6 2 | 3+4 2 | 3+4 2 | 3+4 2 | 3+3 2 | 3+3 2 |
| | 5+6 2 | 3+4 2 | 3+4 2 | 3+4 2 | 2+2 2 | 2+2 2 |
| 12.1.95 | 192944 | | 42 | Cr3S4 | | |
| | -A | -GM | -M | -Y | -L | -V |
| | 3+4 2 | 3+4 2 | 3+4 2 | 5+6 2 | 2+2 2 | 3+3 2 |
| | 5+6 2 | 5+6 2 | 5+6 2 | 3+4 2 | 3+3 2 | 2+2 2 |
| | 3+4 2 | 3+4 2 | 5+6 2 | 5+6 2 | 2+2 2 | 3+3 2 |
| | 5+6 2 | 5+6 2 | 3+4 2 | 3+4 2 | 3+3 2 | 2+2 2 |
| | 3+4 2 | 3+4 2 | 3+4 2 | 3+4 2 | 3+3 2 | 3+3 2 |
| | 5+6 2 | 5+6 2 | 5+6 2 | 5+6 2 | 2+2 2 | 2+2 2 |
| | 3+4 2 | 3+4 2 | 5+6 2 | 5+6 2 | 3+3 2 | 3+3 2 |
| | 5+6 2 | 3+4 2 | 3+4 2 | 3+4 2 | 2+2 2 | 2+2 2 |
| | 3+4 2 | 3+4 2 | 5+6 2 | 5+6 2 | 3+3 2 | 3+3 2 |
| | 3+4 2 | 5+6 2 | 3+4 2 | 3+4 2 | 2+2 2 | 3+3 2 |
| | 5+6 2 | 5+6 2 | 3+4 2 | 3+4 2 | 3+3 2 | 2+2 2 |
| | 5+6 2 | 3+4 2 | 5+6 2 | 5+6 2 | 2+2 2 | 2+2 2 |
| | 3+4 2 | 3+4 2 | 3+4 2 | 5+6 2 | 2+2 2 | 3+3 2 |
| | 3+4 2 | 3+4 2 | 3+4 2 | 5+6 2 | 3+3 2 | 2+2 2 |
| | 5+6 2 | 5+6 2 | 5+6 2 | 3+4 2 | 2+2 2 | 3+3 2 |
| | 5+6 2 | 5+6 2 | 5+6 2 | 3+4 2 | 3+3 2 | 2+2 2 |
| | 3+4 2 | 5+6 2 | 3+4 2 | 3+4 2 | 3+3 2 | 3+3 2 |
| | 3+4 2 | 5+6 2 | 5+6 2 | 5+6 2 | 2+2 2 | 3+3 2 |
| | 3+4 2 | 3+4 2 | 3+4 2 | 5+6 2 | 3+3 2 | 2+2 2 |
| | 5+6 2 | 3+4 2 | 3+4 2 | 3+4 2 | 3+3 2 | 3+3 2 |
| | 5+6 2 | 3+4 2 | 3+4 2 | 3+4 2 | 2+2 2 | 3+3 2 |
| 12.1.96 | 16722 | | 42 | Cr3S4 | | |
| | -A | -GM | -M | -Y | -L | -V |
| | 3+4 2 | 3+4 2 | 3+4 2 | 5+6 2 | 2+2 2 | 2+2 2 |
| | 5+6 2 | 5+6 2 | 5+6 2 | 3+4 2 | 3+3 2 | 3+3 2 |
| | 3+4 2 | 3+4 2 | 5+6 2 | 5+6 2 | 2+2 2 | 3+3 2 |
| | 5+6 2 | 5+6 2 | 3+4 2 | 3+4 2 | 3+3 2 | 2+2 2 |

Continued on next column

**TABLE XXIV– continued**

| TQC | ICSD | | NoE | SF | | |
|---|---|---|---|---|---|---|
| | 3+4 2 | 3+4 2 | 3+4 2 | 3+4 2 | 3+3 2 | 3+3 2 |
| | 3+4 2 | 5+6 2 | 5+6 2 | 5+6 2 | 2+2 2 | 2+2 2 |
| | 5+6 2 | 3+4 2 | 5+6 2 | 3+4 2 | 3+3 2 | 3+3 2 |
| | 5+6 2 | 3+4 2 | 3+4 2 | 5+6 2 | 2+2 2 | 2+2 2 |
| | 3+4 2 | 3+4 2 | 5+6 2 | 3+4 2 | 3+3 2 | 3+3 2 |
| | 5+6 2 | 5+6 2 | 3+4 2 | 3+4 2 | 2+2 2 | 2+2 2 |
| | 3+4 2 | 5+6 2 | 5+6 2 | 5+6 2 | 3+3 2 | 3+3 2 |
| | 5+6 2 | 3+4 2 | 3+4 2 | 5+6 2 | 2+2 2 | 3+3 2 |
| | 3+4 2 | 3+4 2 | 3+4 2 | 5+6 2 | 2+2 2 | 2+2 2 |
| | 5+6 2 | 3+4 2 | 3+4 2 | 5+6 2 | 3+3 2 | 2+2 2 |
| | 3+4 2 | 5+6 2 | 5+6 2 | 3+4 2 | 2+2 2 | 3+3 2 |
| | 5+6 2 | 5+6 2 | 5+6 2 | 3+4 2 | 3+3 2 | 2+2 2 |
| | 3+4 2 | 5+6 2 | 3+4 2 | 3+4 2 | 3+3 2 | 3+3 2 |
| | 3+4 2 | 5+6 2 | 5+6 2 | 5+6 2 | 2+2 2 | 3+3 2 |
| | 3+4 2 | 3+4 2 | 3+4 2 | 5+6 2 | 3+3 2 | 2+2 2 |
| | 5+6 2 | 3+4 2 | 3+4 2 | 3+4 2 | 3+3 2 | 3+3 2 |
| | 5+6 2 | 3+4 2 | 3+4 2 | 3+4 2 | 2+2 2 | 2+2 2 |
| 12.1.97 | 74296 | | 70 | Cr2S8Tl1V3 | | |
| | -A | -GM | -M | -Y | -L | -V |
| | 3+4 2 | 3+4 2 | 3+4 2 | 3+4 2 | 3+3 2 | 3+3 2 |
| | 5+6 2 | 3+4 2 | 5+6 2 | 5+6 2 | 3+3 2 | 2+2 2 |
| | 3+4 2 | 3+4 2 | 3+4 2 | 3+4 2 | 2+2 2 | 2+2 2 |
| | 5+6 2 | 5+6 2 | 5+6 2 | 5+6 2 | 3+3 2 | 3+3 2 |
| | 5+6 2 | 5+6 2 | 5+6 2 | 3+4 2 | 2+2 2 | 2+2 2 |
| | 3+4 2 | 3+4 2 | 3+4 2 | 3+4 2 | 3+3 2 | 3+3 2 |
| | 5+6 2 | 5+6 2 | 5+6 2 | 5+6 2 | 2+2 2 | 2+2 2 |
| | 5+6 2 | 5+6 2 | 3+4 2 | 3+4 2 | 2+2 2 | 3+3 2 |
| | 3+4 2 | 3+4 2 | 5+6 2 | 5+6 2 | 3+3 2 | 3+3 2 |
| | 3+4 2 | 5+6 2 | 3+4 2 | 5+6 2 | 3+3 2 | 2+2 2 |
| | 5+6 2 | 3+4 2 | 5+6 2 | 3+4 2 | 2+2 2 | 2+2 2 |
| | 3+4 2 | 3+4 2 | 3+4 2 | 5+6 2 | 2+2 2 | 3+3 2 |
| | 5+6 2 | 5+6 2 | 5+6 2 | 3+4 2 | 3+3 2 | 2+2 2 |
| | 3+4 2 | 3+4 2 | 3+4 2 | 5+6 2 | 2+2 2 | 3+3 2 |
| | 3+4 2 | 3+4 2 | 5+6 2 | 5+6 2 | 3+3 2 | 2+2 2 |
| | 5+6 2 | 3+4 2 | 5+6 2 | 3+4 2 | 3+3 2 | 2+2 2 |
| | 3+4 2 | 3+4 2 | 5+6 2 | 3+4 2 | 2+2 2 | 3+3 2 |
| | 5+6 2 | 5+6 2 | 5+6 2 | 5+6 2 | 3+3 2 | 2+2 2 |
| | 5+6 2 | 3+4 2 | 3+4 2 | 5+6 2 | 2+2 2 | 3+3 2 |
| | 3+4 2 | 3+4 2 | 3+4 2 | 3+4 2 | 2+2 2 | 2+2 2 |
| | 5+6 2 | 5+6 2 | 3+4 2 | 5+6 2 | 3+3 2 | 3+3 2 |
| | 5+6 2 | 3+4 2 | 5+6 2 | 3+4 2 | 3+3 2 | 2+2 2 |
| | 3+4 2 | 5+6 2 | 3+4 2 | 3+4 2 | 2+2 2 | 2+2 2 |
| | 5+6 2 | 3+4 2 | 5+6 2 | 5+6 2 | 3+3 2 | 3+3 2 |
| | 3+4 2 | 5+6 2 | 5+6 2 | 3+4 2 | 2+2 2 | 2+2 2 |
| | 5+6 2 | 5+6 2 | 3+4 2 | 3+4 2 | 3+3 2 | 2+2 2 |
| | 5+6 2 | 3+4 2 | 5+6 2 | 5+6 2 | 3+3 2 | 3+3 2 |
| | 3+4 2 | 3+4 2 | 5+6 2 | 3+4 2 | 2+2 2 | 3+3 2 |
| | 5+6 2 | 3+4 2 | 5+6 2 | 5+6 2 | 2+2 2 | 2+2 2 |
| | 3+4 2 | 3+4 2 | 3+4 2 | 3+4 2 | 2+2 2 | 2+2 2 |
| | 5+6 2 | 5+6 2 | 5+6 2 | 5+6 2 | 3+3 2 | 3+3 2 |
| | 5+6 2 | 5+6 2 | 3+4 2 | 3+4 2 | 2+2 2 | 2+2 2 |
| | 5+6 2 | 5+6 2 | 5+6 2 | 5+6 2 | 3+3 2 | 3+3 2 |
| | 3+4 2 | 5+6 2 | 3+4 2 | 5+6 2 | 3+3 2 | 2+2 2 |
| | 3+4 2 | 5+6 2 | 5+6 2 | 3+4 2 | 2+2 2 | 3+3 2 |
| 12.1.98 | 626640 | | 40 | Cr2S4Ti1 | | |
| | -A | -GM | -M | -Y | -L | -V |
| | 3+4 2 | 3+4 2 | 3+4 2 | 5+6 2 | 3+3 2 | 3+3 2 |
| | 5+6 2 | 5+6 2 | 5+6 2 | 3+4 2 | 2+2 2 | 2+2 2 |
| | 3+4 2 | 3+4 2 | 5+6 2 | 5+6 2 | 3+3 2 | 3+3 2 |
| | 5+6 2 | 5+6 2 | 3+4 2 | 3+4 2 | 2+2 2 | 2+2 2 |
| | 3+4 2 | 3+4 2 | 3+4 2 | 3+4 2 | 3+3 2 | 3+3 2 |

Continued on next column

**TABLE XXIV– continued**

| TQC | ICSD | NoE | SF | | | |
|---|---|---|---|---|---|---|
| | 3+4 2 | 5+6 2 | 5+6 2 | 5+6 2 | 2+2 2 | 2+2 2 |
| | 5+6 2 | 3+4 2 | 3+4 2 | 3+4 2 | 3+3 2 | 3+3 2 |
| | 5+6 2 | 3+4 2 | 5+6 2 | 5+6 2 | 2+2 2 | 2+2 2 |
| | 3+4 2 | 3+4 2 | 5+6 2 | 3+4 2 | 3+3 2 | 3+3 2 |
| | 5+6 2 | 5+6 2 | 3+4 2 | 3+4 2 | 2+2 2 | 3+3 2 |
| | 3+4 2 | 5+6 2 | 3+4 2 | 5+6 2 | 3+3 2 | 2+2 2 |
| | 5+6 2 | 3+4 2 | 5+6 2 | 5+6 2 | 2+2 2 | 3+3 2 |
| | 3+4 2 | 3+4 2 | 3+4 2 | 5+6 2 | 2+2 2 | 2+2 2 |
| | 5+6 2 | 5+6 2 | 3+4 2 | 5+6 2 | 3+3 2 | 2+2 2 |
| | 3+4 2 | 3+4 2 | 5+6 2 | 3+4 2 | 2+2 2 | 3+3 2 |
| | 5+6 2 | 5+6 2 | 5+6 2 | 3+4 2 | 3+3 2 | 2+2 2 |
| | 3+4 2 | 5+6 2 | 3+4 2 | 3+4 2 | 3+3 2 | 3+3 2 |
| | 5+6 2 | 5+6 2 | 5+6 2 | 5+6 2 | 2+2 2 | 2+2 2 |
| | 3+4 2 | 3+4 2 | 3+4 2 | 5+6 2 | 3+3 2 | 3+3 2 |
| | 3+4 2 | 3+4 2 | 3+4 2 | 5+6 2 | 2+2 2 | 3+3 2 |
| 12.1.99 | 625968 | 44 | Cr2Fe1Se4 | | | |
| | -A | -GM | -M | -Y | -L | -V |
| | 3+4 2 | 3+4 2 | 3+4 2 | 5+6 2 | 2+2 2 | 3+3 2 |
| | 5+6 2 | 5+6 2 | 5+6 2 | 3+4 2 | 2+2 2 | 2+2 2 |
| | 3+4 2 | 3+4 2 | 5+6 2 | 5+6 2 | 3+3 2 | 3+3 2 |
| | 5+6 2 | 5+6 2 | 3+4 2 | 3+4 2 | 3+3 2 | 2+2 2 |
| | 3+4 2 | 3+4 2 | 3+4 2 | 3+4 2 | 3+3 2 | 3+3 2 |
| | 5+6 2 | 5+6 2 | 5+6 2 | 5+6 2 | 2+2 2 | 2+2 2 |
| | 3+4 2 | 3+4 2 | 5+6 2 | 5+6 2 | 3+3 2 | 3+3 2 |
| | 5+6 2 | 3+4 2 | 3+4 2 | 3+4 2 | 2+2 2 | 2+2 2 |
| | 3+4 2 | 3+4 2 | 5+6 2 | 3+4 2 | 3+3 2 | 3+3 2 |
| | 5+6 2 | 5+6 2 | 3+4 2 | 5+6 2 | 2+2 2 | 2+2 2 |
| | 3+4 2 | 5+6 2 | 5+6 2 | 3+4 2 | 3+3 2 | 3+3 2 |
| | 5+6 2 | 3+4 2 | 3+4 2 | 5+6 2 | 2+2 2 | 2+2 2 |
| | 3+4 2 | 3+4 2 | 3+4 2 | 5+6 2 | 2+2 2 | 3+3 2 |
| | 3+4 2 | 3+4 2 | 3+4 2 | 5+6 2 | 3+3 2 | 2+2 2 |
| | 5+6 2 | 5+6 2 | 5+6 2 | 3+4 2 | 2+2 2 | 3+3 2 |
| | 5+6 2 | 5+6 2 | 3+4 2 | 3+4 2 | 3+3 2 | 2+2 2 |
| | 3+4 2 | 3+4 2 | 5+6 2 | 3+4 2 | 3+3 2 | 3+3 2 |
| | 3+4 2 | 3+4 2 | 5+6 2 | 3+4 2 | 2+2 2 | 3+3 2 |
| | 3+4 2 | 3+4 2 | 3+4 2 | 5+6 2 | 3+3 2 | 3+3 2 |
| | 5+6 2 | 5+6 2 | 3+4 2 | 5+6 2 | 3+3 2 | 2+2 2 |
| | 3+4 2 | 3+4 2 | 3+4 2 | 3+4 2 | 3+3 2 | 3+3 2 |
| | 5+6 2 | 5+6 2 | 3+4 2 | 3+4 2 | 2+2 2 | 3+3 2 |
| 12.1.100 | 155121 | 44 | Cr2Fe1Se4 | | | |
| | -A | -GM | -M | -Y | -L | -V |
| | 3+4 2 | 3+4 2 | 3+4 2 | 5+6 2 | 2+2 2 | 3+3 2 |
| | 5+6 2 | 5+6 2 | 5+6 2 | 3+4 2 | 2+2 2 | 2+2 2 |
| | 3+4 2 | 3+4 2 | 5+6 2 | 5+6 2 | 3+3 2 | 3+3 2 |
| | 5+6 2 | 5+6 2 | 3+4 2 | 3+4 2 | 3+3 2 | 2+2 2 |
| | 3+4 2 | 3+4 2 | 3+4 2 | 3+4 2 | 3+3 2 | 3+3 2 |
| | 5+6 2 | 5+6 2 | 5+6 2 | 5+6 2 | 2+2 2 | 2+2 2 |
| | 3+4 2 | 3+4 2 | 5+6 2 | 5+6 2 | 3+3 2 | 3+3 2 |
| | 5+6 2 | 3+4 2 | 3+4 2 | 3+4 2 | 2+2 2 | 2+2 2 |
| | 3+4 2 | 3+4 2 | 5+6 2 | 3+4 2 | 3+3 2 | 3+3 2 |
| | 5+6 2 | 5+6 2 | 3+4 2 | 5+6 2 | 2+2 2 | 2+2 2 |
| | 3+4 2 | 5+6 2 | 5+6 2 | 3+4 2 | 3+3 2 | 3+3 2 |
| | 5+6 2 | 3+4 2 | 3+4 2 | 5+6 2 | 2+2 2 | 2+2 2 |
| | 3+4 2 | 3+4 2 | 3+4 2 | 5+6 2 | 2+2 2 | 3+3 2 |
| | 3+4 2 | 3+4 2 | 3+4 2 | 5+6 2 | 3+3 2 | 2+2 2 |
| | 5+6 2 | 5+6 2 | 5+6 2 | 3+4 2 | 2+2 2 | 3+3 2 |
| | 5+6 2 | 5+6 2 | 3+4 2 | 3+4 2 | 3+3 2 | 2+2 2 |
| | 3+4 2 | 3+4 2 | 5+6 2 | 3+4 2 | 3+3 2 | 3+3 2 |
| | 3+4 2 | 3+4 2 | 5+6 2 | 3+4 2 | 2+2 2 | 3+3 2 |
| | 3+4 2 | 3+4 2 | 3+4 2 | 5+6 2 | 3+3 2 | 3+3 2 |
| | 5+6 2 | 5+6 2 | 3+4 2 | 5+6 2 | 3+3 2 | 2+2 2 |

Continued on next column

**TABLE XXIV– continued**

| TQC | ICSD | NoE | SF | | | |
|---|---|---|---|---|---|---|
| | 3+4 2 | 3+4 2 | 3+4 2 | 3+4 2 | 3+3 2 | 3+3 2 |
| | 5+6 2 | 5+6 2 | 3+4 2 | 3+4 2 | 2+2 2 | 3+3 2 |
| 12.1.101 | 625931 | 44 | Cr2Fe1S4 | | | |
| | -A | -GM | -M | -Y | -L | -V |
| | 3+4 2 | 3+4 2 | 3+4 2 | 5+6 2 | 2+2 2 | 3+3 2 |
| | 5+6 2 | 5+6 2 | 5+6 2 | 3+4 2 | 2+2 2 | 2+2 2 |
| | 3+4 2 | 3+4 2 | 5+6 2 | 5+6 2 | 3+3 2 | 3+3 2 |
| | 5+6 2 | 5+6 2 | 3+4 2 | 3+4 2 | 3+3 2 | 2+2 2 |
| | 3+4 2 | 3+4 2 | 3+4 2 | 3+4 2 | 3+3 2 | 3+3 2 |
| | 5+6 2 | 5+6 2 | 5+6 2 | 5+6 2 | 2+2 2 | 2+2 2 |
| | 3+4 2 | 3+4 2 | 5+6 2 | 3+4 2 | 3+3 2 | 3+3 2 |
| | 5+6 2 | 3+4 2 | 3+4 2 | 5+6 2 | 2+2 2 | 2+2 2 |
| | 3+4 2 | 5+6 2 | 5+6 2 | 5+6 2 | 3+3 2 | 3+3 2 |
| | 3+4 2 | 3+4 2 | 3+4 2 | 3+4 2 | 2+2 2 | 3+3 2 |
| | 5+6 2 | 5+6 2 | 5+6 2 | 3+4 2 | 3+3 2 | 2+2 2 |
| | 3+4 2 | 3+4 2 | 3+4 2 | 5+6 2 | 2+2 2 | 2+2 2 |
| | 5+6 2 | 3+4 2 | 3+4 2 | 5+6 2 | 2+2 2 | 3+3 2 |
| | 3+4 2 | 3+4 2 | 3+4 2 | 5+6 2 | 3+3 2 | 3+3 2 |
| | 5+6 2 | 5+6 2 | 5+6 2 | 3+4 2 | 2+2 2 | 2+2 2 |
| | 5+6 2 | 5+6 2 | 5+6 2 | 3+4 2 | 3+3 2 | 2+2 2 |
| | 3+4 2 | 5+6 2 | 3+4 2 | 3+4 2 | 3+3 2 | 3+3 2 |
| | 3+4 2 | 5+6 2 | 5+6 2 | 3+4 2 | 2+2 2 | 3+3 2 |
| | 3+4 2 | 3+4 2 | 3+4 2 | 5+6 2 | 3+3 2 | 2+2 2 |
| | 5+6 2 | 3+4 2 | 3+4 2 | 3+4 2 | 3+3 2 | 3+3 2 |
| | 3+4 2 | 3+4 2 | 3+4 2 | 3+4 2 | 3+3 2 | 3+3 2 |
| | 5+6 2 | 3+4 2 | 3+4 2 | 5+6 2 | 3+3 2 | 3+3 2 |
| 12.1.102 | 251718 | 18 | Cr1Se2 | | | |
| | -A | -GM | -M | -Y | -L | -V |
| | 3+4 2 | 3+4 2 | 5+6 2 | 5+6 2 | 2+2 2 | 2+2 2 |
| | 5+6 2 | 5+6 2 | 3+4 2 | 3+4 2 | 3+3 2 | 3+3 2 |
| | 3+4 2 | 3+4 2 | 3+4 2 | 3+4 2 | 3+3 2 | 3+3 2 |
| | 3+4 2 | 3+4 2 | 5+6 2 | 5+6 2 | 2+2 2 | 2+2 2 |
| | 3+4 2 | 3+4 2 | 3+4 2 | 3+4 2 | 3+3 2 | 3+3 2 |
| | 5+6 2 | 5+6 2 | 3+4 2 | 3+4 2 | 3+3 2 | 3+3 2 |
| | 3+4 2 | 3+4 2 | 5+6 2 | 5+6 2 | 2+2 2 | 2+2 2 |
| | 3+4 2 | 3+4 2 | 5+6 2 | 5+6 2 | 2+2 2 | 2+2 2 |
| | 5+6 2 | 5+6 2 | 3+4 2 | 3+4 2 | 3+3 2 | 3+3 2 |
| 12.1.103 | 75420 | 18 | Cr1S2 | | | |
| | -A | -GM | -M | -Y | -L | -V |
| | 3+4 2 | 3+4 2 | 3+4 2 | 3+4 2 | 2+2 2 | 2+2 2 |
| | 5+6 2 | 5+6 2 | 5+6 2 | 5+6 2 | 3+3 2 | 3+3 2 |
| | 3+4 2 | 3+4 2 | 3+4 2 | 3+4 2 | 3+3 2 | 3+3 2 |
| | 3+4 2 | 3+4 2 | 5+6 2 | 5+6 2 | 2+2 2 | 2+2 2 |
| | 3+4 2 | 3+4 2 | 3+4 2 | 3+4 2 | 3+3 2 | 3+3 2 |
| | 5+6 2 | 5+6 2 | 3+4 2 | 3+4 2 | 3+3 2 | 3+3 2 |
| | 3+4 2 | 3+4 2 | 5+6 2 | 5+6 2 | 2+2 2 | 2+2 2 |
| | 3+4 2 | 3+4 2 | 5+6 2 | 5+6 2 | 2+2 2 | 2+2 2 |
| | 5+6 2 | 5+6 2 | 3+4 2 | 3+4 2 | 3+3 2 | 3+3 2 |
| 12.1.104 | 626579 | 48 | Cr1Rh2Te4 | | | |
| | -A | -GM | -M | -Y | -L | -V |
| | 5+6 2 | 3+4 2 | 3+4 2 | 5+6 2 | 2+2 2 | 3+3 2 |
| | 3+4 2 | 5+6 2 | 5+6 2 | 3+4 2 | 3+3 2 | 2+2 2 |
| | 3+4 2 | 3+4 2 | 5+6 2 | 5+6 2 | 2+2 2 | 2+2 2 |
| | 5+6 2 | 5+6 2 | 3+4 2 | 3+4 2 | 3+3 2 | 3+3 2 |
| | 3+4 2 | 3+4 2 | 3+4 2 | 3+4 2 | 2+2 2 | 3+3 2 |
| | 5+6 2 | 5+6 2 | 5+6 2 | 5+6 2 | 3+3 2 | 3+3 2 |
| | 3+4 2 | 3+4 2 | 5+6 2 | 3+4 2 | 3+3 2 | 2+2 2 |
| | 5+6 2 | 3+4 2 | 3+4 2 | 5+6 2 | 2+2 2 | 2+2 2 |
| | 3+4 2 | 5+6 2 | 5+6 2 | 3+4 2 | 3+3 2 | 3+3 2 |
| | 5+6 2 | 3+4 2 | 3+4 2 | 5+6 2 | 2+2 2 | 3+3 2 |
| | 3+4 2 | 5+6 2 | 5+6 2 | 5+6 2 | 2+2 2 | 2+2 2 |

Continued on next column

**TABLE XXIV– continued**

| TQC | ICSD | | NoE | | SF | |
|---|---|---|---|---|---|---|
| | 3+4 2 | 3+4 2 | 3+4 2 | 5+6 2 | 3+3 2 | 2+2 2 |
| | 5+6 2 | 3+4 2 | 3+4 2 | 3+4 2 | 2+2 2 | 2+2 2 |
| | 3+4 2 | 5+6 2 | 5+6 2 | 3+4 2 | 3+3 2 | 3+3 2 |
| | 5+6 2 | 3+4 2 | 3+4 2 | 5+6 2 | 2+2 2 | 3+3 2 |
| | 3+4 2 | 5+6 2 | 5+6 2 | 3+4 2 | 3+3 2 | 2+2 2 |
| | 5+6 2 | 3+4 2 | 3+4 2 | 5+6 2 | 2+2 2 | 3+3 2 |
| | 5+6 2 | 5+6 2 | 3+4 2 | 3+4 2 | 3+3 2 | 2+2 2 |
| | 5+6 2 | 3+4 2 | 3+4 2 | 5+6 2 | 2+2 2 | 3+3 2 |
| | 3+4 2 | 3+4 2 | 5+6 2 | 3+4 2 | 3+3 2 | 2+2 2 |
| | 5+6 2 | 3+4 2 | 3+4 2 | 3+4 2 | 3+3 2 | 3+3 2 |
| | 3+4 2 | 5+6 2 | 3+4 2 | 5+6 2 | 3+3 2 | 2+2 2 |
| | 3+4 2 | 3+4 2 | 5+6 2 | 3+4 2 | 2+2 2 | 3+3 2 |
| | 3+4 2 | 5+6 2 | 3+4 2 | 3+4 2 | 3+3 2 | 3+3 2 |
| 12.1.105 | 626577 | | 48 | | Cr1Rh2Se4 | |
| | -A | -GM | -M | -Y | -L | -V |
| | 5+6 2 | 3+4 2 | 3+4 2 | 5+6 2 | 2+2 2 | 3+3 2 |
| | 3+4 2 | 5+6 2 | 5+6 2 | 3+4 2 | 3+3 2 | 2+2 2 |
| | 3+4 2 | 3+4 2 | 5+6 2 | 5+6 2 | 2+2 2 | 2+2 2 |
| | 5+6 2 | 5+6 2 | 3+4 2 | 3+4 2 | 3+3 2 | 3+3 2 |
| | 3+4 2 | 3+4 2 | 5+6 2 | 3+4 2 | 3+3 2 | 3+3 2 |
| | 5+6 2 | 5+6 2 | 3+4 2 | 5+6 2 | 2+2 2 | 2+2 2 |
| | 3+4 2 | 3+4 2 | 5+6 2 | 3+4 2 | 3+3 2 | 3+3 2 |
| | 3+4 2 | 5+6 2 | 5+6 2 | 3+4 2 | 2+2 2 | 2+2 2 |
| | 5+6 2 | 3+4 2 | 3+4 2 | 5+6 2 | 3+3 2 | 3+3 2 |
| | 5+6 2 | 5+6 2 | 3+4 2 | 5+6 2 | 2+2 2 | 3+3 2 |
| | 3+4 2 | 3+4 2 | 5+6 2 | 5+6 2 | 3+3 2 | 2+2 2 |
| | 3+4 2 | 3+4 2 | 3+4 2 | 5+6 2 | 2+2 2 | 2+2 2 |
| | 5+6 2 | 3+4 2 | 3+4 2 | 3+4 2 | 2+2 2 | 2+2 2 |
| | 3+4 2 | 5+6 2 | 3+4 2 | 3+4 2 | 3+3 2 | 3+3 2 |
| | 5+6 2 | 3+4 2 | 5+6 2 | 5+6 2 | 2+2 2 | 3+3 2 |
| | 3+4 2 | 5+6 2 | 5+6 2 | 3+4 2 | 3+3 2 | 2+2 2 |
| | 5+6 2 | 3+4 2 | 3+4 2 | 5+6 2 | 2+2 2 | 3+3 2 |
| | 5+6 2 | 3+4 2 | 3+4 2 | 3+4 2 | 3+3 2 | 2+2 2 |
| | 5+6 2 | 5+6 2 | 3+4 2 | 5+6 2 | 3+3 2 | 3+3 2 |
| | 5+6 2 | 3+4 2 | 5+6 2 | 3+4 2 | 2+2 2 | 2+2 2 |
| | 3+4 2 | 3+4 2 | 3+4 2 | 5+6 2 | 3+3 2 | 3+3 2 |
| | 3+4 2 | 5+6 2 | 3+4 2 | 3+4 2 | 3+3 2 | 3+3 2 |
| | 3+4 2 | 5+6 2 | 3+4 2 | 3+4 2 | 3+3 2 | 2+2 2 |
| | 3+4 2 | 3+4 2 | 5+6 2 | 3+4 2 | 2+2 2 | 3+3 2 |
| 12.1.106 | 626541 | | 56 | | Cr1Pr2Se4 | |
| | -A | -GM | -M | -Y | -L | -V |
| | 5+6 2 | 3+4 2 | 3+4 2 | 5+6 2 | 2+2 2 | 2+2 2 |
| | 3+4 2 | 5+6 2 | 5+6 2 | 3+4 2 | 3+3 2 | 3+3 2 |
| | 3+4 2 | 3+4 2 | 5+6 2 | 5+6 2 | 2+2 2 | 3+3 2 |
| | 5+6 2 | 5+6 2 | 3+4 2 | 3+4 2 | 3+3 2 | 2+2 2 |
| | 5+6 2 | 3+4 2 | 3+4 2 | 5+6 2 | 3+3 2 | 3+3 2 |
| | 3+4 2 | 3+4 2 | 5+6 2 | 5+6 2 | 2+2 2 | 2+2 2 |
| | 5+6 2 | 5+6 2 | 3+4 2 | 3+4 2 | 3+3 2 | 3+3 2 |
| | 3+4 2 | 3+4 2 | 5+6 2 | 3+4 2 | 2+2 2 | 2+2 2 |
| | 5+6 2 | 5+6 2 | 3+4 2 | 3+4 2 | 2+2 2 | 3+3 2 |
| | 3+4 2 | 5+6 2 | 5+6 2 | 5+6 2 | 3+3 2 | 2+2 2 |
| | 5+6 2 | 3+4 2 | 3+4 2 | 5+6 2 | 2+2 2 | 3+3 2 |
| | 3+4 2 | 5+6 2 | 5+6 2 | 3+4 2 | 3+3 2 | 2+2 2 |
| | 3+4 2 | 3+4 2 | 5+6 2 | 3+4 2 | 3+3 2 | 3+3 2 |
| | 3+4 2 | 5+6 2 | 3+4 2 | 3+4 2 | 2+2 2 | 2+2 2 |
| | 3+4 2 | 3+4 2 | 5+6 2 | 5+6 2 | 3+3 2 | 2+2 2 |
| | 5+6 2 | 5+6 2 | 3+4 2 | 5+6 2 | 2+2 2 | 3+3 2 |
| | 5+6 2 | 3+4 2 | 5+6 2 | 3+4 2 | 3+3 2 | 3+3 2 |
| | 3+4 2 | 3+4 2 | 3+4 2 | 3+4 2 | 2+2 2 | 2+2 2 |
| | 5+6 2 | 5+6 2 | 5+6 2 | 5+6 2 | 3+3 2 | 3+3 2 |
| | 3+4 2 | 3+4 2 | 3+4 2 | 3+4 2 | 2+2 2 | 2+2 2 |

Continued on next column

**TABLE XXIV– continued**

| TQC | ICSD | | NoE | | SF | |
|---|---|---|---|---|---|---|
| | 3+4 2 | 5+6 2 | 5+6 2 | 5+6 2 | 2+2 2 | 3+3 2 |
| | 5+6 2 | 3+4 2 | 3+4 2 | 3+4 2 | 3+3 2 | 3+3 2 |
| | 3+4 2 | 5+6 2 | 3+4 2 | 3+4 2 | 3+3 2 | 2+2 2 |
| | 5+6 2 | 3+4 2 | 5+6 2 | 5+6 2 | 2+2 2 | 2+2 2 |
| | 5+6 2 | 5+6 2 | 3+4 2 | 5+6 2 | 3+3 2 | 3+3 2 |
| | 3+4 2 | 5+6 2 | 5+6 2 | 3+4 2 | 3+3 2 | 3+3 2 |
| | 3+4 2 | 3+4 2 | 3+4 2 | 3+4 2 | 3+3 2 | 2+2 2 |
| | 3+4 2 | 3+4 2 | 3+4 2 | 3+4 2 | 2+2 2 | 3+3 2 |
| 12.1.107 | 821 | | 70 | | Cr1P1S4 | |
| | -A | -GM | -M | -Y | -L | -V |
| | 3+4 2 | 3+4 2 | 5+6 2 | 5+6 2 | 2+2 2 | 2+2 2 |
| | 5+6 2 | 5+6 2 | 3+4 2 | 3+4 2 | 3+3 2 | 3+3 2 |
| | 5+6 2 | 3+4 2 | 3+4 2 | 3+4 2 | 2+2 2 | 3+3 2 |
| | 3+4 2 | 5+6 2 | 3+4 2 | 5+6 2 | 3+3 2 | 2+2 2 |
| | 5+6 2 | 3+4 2 | 5+6 2 | 3+4 2 | 3+3 2 | 3+3 2 |
| | 3+4 2 | 5+6 2 | 5+6 2 | 5+6 2 | 2+2 2 | 2+2 2 |
| | 5+6 2 | 5+6 2 | 3+4 2 | 3+4 2 | 3+3 2 | 3+3 2 |
| | 3+4 2 | 3+4 2 | 5+6 2 | 5+6 2 | 2+2 2 | 2+2 2 |
| | 3+4 2 | 3+4 2 | 3+4 2 | 5+6 2 | 2+2 2 | 2+2 2 |
| | 5+6 2 | 5+6 2 | 5+6 2 | 3+4 2 | 3+3 2 | 3+3 2 |
| | 3+4 2 | 3+4 2 | 5+6 2 | 3+4 2 | 3+3 2 | 3+3 2 |
| | 3+4 2 | 3+4 2 | 3+4 2 | 5+6 2 | 2+2 2 | 2+2 2 |
| | 5+6 2 | 5+6 2 | 5+6 2 | 5+6 2 | 2+2 2 | 3+3 2 |
| | 3+4 2 | 5+6 2 | 5+6 2 | 5+6 2 | 3+3 2 | 2+2 2 |
| | 5+6 2 | 3+4 2 | 3+4 2 | 3+4 2 | 2+2 2 | 2+2 2 |
| | 5+6 2 | 5+6 2 | 3+4 2 | 3+4 2 | 3+3 2 | 3+3 2 |
| | 5+6 2 | 5+6 2 | 3+4 2 | 3+4 2 | 3+3 2 | 2+2 2 |
| | 3+4 2 | 3+4 2 | 5+6 2 | 5+6 2 | 2+2 2 | 3+3 2 |
| | 3+4 2 | 3+4 2 | 3+4 2 | 3+4 2 | 3+3 2 | 3+3 2 |
| | 5+6 2 | 5+6 2 | 3+4 2 | 5+6 2 | 2+2 2 | 2+2 2 |
| | 3+4 2 | 3+4 2 | 5+6 2 | 3+4 2 | 2+2 2 | 3+3 2 |
| | 3+4 2 | 3+4 2 | 3+4 2 | 3+4 2 | 3+3 2 | 2+2 2 |
| | 5+6 2 | 5+6 2 | 5+6 2 | 5+6 2 | 3+3 2 | 2+2 2 |
| | 5+6 2 | 5+6 2 | 3+4 2 | 3+4 2 | 2+2 2 | 3+3 2 |
| | 5+6 2 | 3+4 2 | 5+6 2 | 5+6 2 | 2+2 2 | 2+2 2 |
| | 3+4 2 | 3+4 2 | 3+4 2 | 3+4 2 | 3+3 2 | 3+3 2 |
| | 3+4 2 | 3+4 2 | 5+6 2 | 5+6 2 | 3+3 2 | 2+2 2 |
| | 3+4 2 | 5+6 2 | 3+4 2 | 5+6 2 | 2+2 2 | 3+3 2 |
| | 5+6 2 | 5+6 2 | 5+6 2 | 5+6 2 | 2+2 2 | 2+2 2 |
| | 5+6 2 | 5+6 2 | 5+6 2 | 3+4 2 | 3+3 2 | 3+3 2 |
| | 5+6 2 | 3+4 2 | 3+4 2 | 3+4 2 | 2+2 2 | 3+3 2 |
| | 3+4 2 | 5+6 2 | 5+6 2 | 5+6 2 | 3+3 2 | 2+2 2 |
| | 3+4 2 | 3+4 2 | 3+4 2 | 3+4 2 | 3+3 2 | 2+2 2 |
| | 5+6 2 | 5+6 2 | 5+6 2 | 5+6 2 | 2+2 2 | 3+3 2 |
| | 3+4 2 | 3+4 2 | 5+6 2 | 5+6 2 | 3+3 2 | 3+3 2 |
| 12.1.108 | 56835 | | 70 | | Cr1P1S4 | |
| | -A | -GM | -M | -Y | -L | -V |
| | 3+4 2 | 3+4 2 | 5+6 2 | 5+6 2 | 2+2 2 | 2+2 2 |
| | 5+6 2 | 5+6 2 | 3+4 2 | 3+4 2 | 3+3 2 | 3+3 2 |
| | 5+6 2 | 3+4 2 | 3+4 2 | 3+4 2 | 2+2 2 | 3+3 2 |
| | 3+4 2 | 5+6 2 | 3+4 2 | 5+6 2 | 3+3 2 | 2+2 2 |
| | 5+6 2 | 3+4 2 | 5+6 2 | 3+4 2 | 3+3 2 | 3+3 2 |
| | 3+4 2 | 5+6 2 | 5+6 2 | 5+6 2 | 2+2 2 | 2+2 2 |
| | 5+6 2 | 5+6 2 | 3+4 2 | 3+4 2 | 3+3 2 | 3+3 2 |
| | 3+4 2 | 3+4 2 | 5+6 2 | 5+6 2 | 2+2 2 | 2+2 2 |
| | 3+4 2 | 3+4 2 | 3+4 2 | 5+6 2 | 2+2 2 | 2+2 2 |
| | 5+6 2 | 5+6 2 | 5+6 2 | 3+4 2 | 3+3 2 | 3+3 2 |
| | 3+4 2 | 3+4 2 | 5+6 2 | 3+4 2 | 3+3 2 | 3+3 2 |
| | 3+4 2 | 3+4 2 | 3+4 2 | 5+6 2 | 2+2 2 | 2+2 2 |
| | 5+6 2 | 5+6 2 | 5+6 2 | 5+6 2 | 2+2 2 | 3+3 2 |
| | 3+4 2 | 5+6 2 | 5+6 2 | 5+6 2 | 3+3 2 | 2+2 2 |

Continued on next column

**TABLE XXIV– continued**

| TQC | ICSD | NoE | SF | | | |
|---|---|---|---|---|---|---|
| | 5+6 2 | 3+4 2 | 3+4 2 | 3+4 2 | 2+2 2 | 2+2 2 |
| | 5+6 2 | 5+6 2 | 3+4 2 | 3+4 2 | 3+3 2 | 3+3 2 |
| | 5+6 2 | 5+6 2 | 3+4 2 | 3+4 2 | 3+3 2 | 2+2 2 |
| | 3+4 2 | 3+4 2 | 5+6 2 | 5+6 2 | 2+2 2 | 3+3 2 |
| | 3+4 2 | 3+4 2 | 3+4 2 | 3+4 2 | 3+3 2 | 3+3 2 |
| | 5+6 2 | 5+6 2 | 3+4 2 | 5+6 2 | 2+2 2 | 2+2 2 |
| | 3+4 2 | 3+4 2 | 5+6 2 | 3+4 2 | 2+2 2 | 3+3 2 |
| | 3+4 2 | 3+4 2 | 3+4 2 | 3+4 2 | 3+3 2 | 2+2 2 |
| | 5+6 2 | 5+6 2 | 5+6 2 | 5+6 2 | 3+3 2 | 2+2 2 |
| | 5+6 2 | 5+6 2 | 3+4 2 | 3+4 2 | 2+2 2 | 3+3 2 |
| | 5+6 2 | 3+4 2 | 5+6 2 | 5+6 2 | 2+2 2 | 2+2 2 |
| | 3+4 2 | 3+4 2 | 3+4 2 | 3+4 2 | 3+3 2 | 3+3 2 |
| | 3+4 2 | 3+4 2 | 5+6 2 | 5+6 2 | 3+3 2 | 2+2 2 |
| | 3+4 2 | 5+6 2 | 3+4 2 | 5+6 2 | 2+2 2 | 3+3 2 |
| | 5+6 2 | 5+6 2 | 5+6 2 | 5+6 2 | 2+2 2 | 2+2 2 |
| | 5+6 2 | 5+6 2 | 5+6 2 | 3+4 2 | 3+3 2 | 3+3 2 |
| | 5+6 2 | 3+4 2 | 3+4 2 | 3+4 2 | 2+2 2 | 3+3 2 |
| | 3+4 2 | 5+6 2 | 5+6 2 | 5+6 2 | 3+3 2 | 2+2 2 |
| | 3+4 2 | 3+4 2 | 3+4 2 | 3+4 2 | 3+3 2 | 2+2 2 |
| | 5+6 2 | 5+6 2 | 5+6 2 | 5+6 2 | 2+2 2 | 3+3 2 |
| | 3+4 2 | 3+4 2 | 5+6 2 | 5+6 2 | 3+3 2 | 3+3 2 |
| 12.1.109 | 173875 | | 74 | | Cr1O4Re1 | |
| | -A | -GM | -M | -Y | -L | -V |
| | 3+4 2 | 3+4 2 | 3+4 2 | 3+4 2 | 2+2 2 | 3+3 2 |
| | 5+6 2 | 5+6 2 | 3+4 2 | 5+6 2 | 3+3 2 | 2+2 2 |
| | 5+6 2 | 5+6 2 | 5+6 2 | 3+4 2 | 3+3 2 | 2+2 2 |
| | 3+4 2 | 3+4 2 | 5+6 2 | 5+6 2 | 2+2 2 | 3+3 2 |
| | 5+6 2 | 5+6 2 | 5+6 2 | 5+6 2 | 2+2 2 | 2+2 2 |
| | 3+4 2 | 3+4 2 | 3+4 2 | 3+4 2 | 3+3 2 | 3+3 2 |
| | 3+4 2 | 3+4 2 | 5+6 2 | 3+4 2 | 2+2 2 | 2+2 2 |
| | 5+6 2 | 5+6 2 | 3+4 2 | 5+6 2 | 3+3 2 | 3+3 2 |
| | 3+4 2 | 3+4 2 | 3+4 2 | 3+4 2 | 2+2 2 | 2+2 2 |
| | 5+6 2 | 3+4 2 | 3+4 2 | 5+6 2 | 3+3 2 | 3+3 2 |
| | 5+6 2 | 5+6 2 | 5+6 2 | 5+6 2 | 2+2 2 | 2+2 2 |
| | 5+6 2 | 5+6 2 | 5+6 2 | 5+6 2 | 3+3 2 | 3+3 2 |
| | 3+4 2 | 3+4 2 | 5+6 2 | 3+4 2 | 2+2 2 | 3+3 2 |
| | 3+4 2 | 5+6 2 | 5+6 2 | 5+6 2 | 2+2 2 | 2+2 2 |
| | 5+6 2 | 5+6 2 | 3+4 2 | 3+4 2 | 3+3 2 | 2+2 2 |
| | 5+6 2 | 3+4 2 | 3+4 2 | 5+6 2 | 3+3 2 | 2+2 2 |
| | 3+4 2 | 3+4 2 | 3+4 2 | 3+4 2 | 2+2 2 | 3+3 2 |
| | 3+4 2 | 3+4 2 | 3+4 2 | 3+4 2 | 3+3 2 | 3+3 2 |
| | 3+4 2 | 3+4 2 | 3+4 2 | 3+4 2 | 3+3 2 | 3+3 2 |
| | 3+4 2 | 5+6 2 | 3+4 2 | 5+6 2 | 2+2 2 | 2+2 2 |
| | 5+6 2 | 5+6 2 | 5+6 2 | 3+4 2 | 2+2 2 | 2+2 2 |
| | 5+6 2 | 5+6 2 | 5+6 2 | 5+6 2 | 3+3 2 | 3+3 2 |
| | 3+4 2 | 3+4 2 | 5+6 2 | 3+4 2 | 2+2 2 | 2+2 2 |
| | 5+6 2 | 3+4 2 | 3+4 2 | 5+6 2 | 3+3 2 | 2+2 2 |
| | 5+6 2 | 5+6 2 | 3+4 2 | 5+6 2 | 3+3 2 | 3+3 2 |
| | 3+4 2 | 5+6 2 | 5+6 2 | 3+4 2 | 3+3 2 | 3+3 2 |
| | 3+4 2 | 3+4 2 | 5+6 2 | 3+4 2 | 2+2 2 | 2+2 2 |
| | 5+6 2 | 3+4 2 | 5+6 2 | 5+6 2 | 2+2 2 | 2+2 2 |
| | 5+6 2 | 5+6 2 | 5+6 2 | 5+6 2 | 2+2 2 | 3+3 2 |
| | 3+4 2 | 5+6 2 | 5+6 2 | 3+4 2 | 3+3 2 | 3+3 2 |
| | 5+6 2 | 3+4 2 | 3+4 2 | 3+4 2 | 3+3 2 | 3+3 2 |
| | 3+4 2 | 5+6 2 | 3+4 2 | 5+6 2 | 2+2 2 | 2+2 2 |
| | 3+4 2 | 3+4 2 | 3+4 2 | 3+4 2 | 3+3 2 | 3+3 2 |
| | 5+6 2 | 5+6 2 | 5+6 2 | 5+6 2 | 2+2 2 | 2+2 2 |
| | 5+6 2 | 3+4 2 | 5+6 2 | 3+4 2 | 2+2 2 | 3+3 2 |
| | 5+6 2 | 5+6 2 | 3+4 2 | 5+6 2 | 3+3 2 | 2+2 2 |
| | 3+4 2 | 3+4 2 | 3+4 2 | 5+6 2 | 3+3 2 | 3+3 2 |
| 12.1.110 | 624485 | | 52 | | Co2Ni1Se4 | |

Continued on next column

**TABLE XXIV– continued**

| TQC | ICSD | NoE | SF | | | |
|---|---|---|---|---|---|---|
| | -A | -GM | -M | -Y | -L | -V |
| | 3+4 2 | 3+4 2 | 5+6 2 | 5+6 2 | 2+2 2 | 3+3 2 |
| | 5+6 2 | 3+4 2 | 3+4 2 | 3+4 2 | 2+2 2 | 2+2 2 |
| | 3+4 2 | 5+6 2 | 5+6 2 | 5+6 2 | 3+3 2 | 3+3 2 |
| | 5+6 2 | 5+6 2 | 3+4 2 | 3+4 2 | 3+3 2 | 2+2 2 |
| | 3+4 2 | 3+4 2 | 3+4 2 | 3+4 2 | 3+3 2 | 3+3 2 |
| | 5+6 2 | 5+6 2 | 5+6 2 | 5+6 2 | 2+2 2 | 2+2 2 |
| | 3+4 2 | 3+4 2 | 3+4 2 | 5+6 2 | 3+3 2 | 3+3 2 |
| | 5+6 2 | 3+4 2 | 5+6 2 | 3+4 2 | 2+2 2 | 2+2 2 |
| | 3+4 2 | 3+4 2 | 5+6 2 | 5+6 2 | 3+3 2 | 3+3 2 |
| | 3+4 2 | 5+6 2 | 3+4 2 | 3+4 2 | 2+2 2 | 2+2 2 |
| | 5+6 2 | 3+4 2 | 5+6 2 | 3+4 2 | 3+3 2 | 3+3 2 |
| | 5+6 2 | 5+6 2 | 3+4 2 | 5+6 2 | 2+2 2 | 2+2 2 |
| | 3+4 2 | 3+4 2 | 3+4 2 | 5+6 2 | 2+2 2 | 3+3 2 |
| | 3+4 2 | 3+4 2 | 3+4 2 | 5+6 2 | 2+2 2 | 3+3 2 |
| | 5+6 2 | 3+4 2 | 5+6 2 | 3+4 2 | 3+3 2 | 2+2 2 |
| | 3+4 2 | 3+4 2 | 3+4 2 | 3+4 2 | 3+3 2 | 2+2 2 |
| | 5+6 2 | 5+6 2 | 3+4 2 | 3+4 2 | 3+3 2 | 3+3 2 |
| | 3+4 2 | 3+4 2 | 3+4 2 | 3+4 2 | 3+3 2 | 3+3 2 |
| | 3+4 2 | 5+6 2 | 5+6 2 | 3+4 2 | 2+2 2 | 3+3 2 |
| | 3+4 2 | 3+4 2 | 3+4 2 | 5+6 2 | 3+3 2 | 2+2 2 |
| | 5+6 2 | 5+6 2 | 3+4 2 | 5+6 2 | 3+3 2 | 3+3 2 |
| | 5+6 2 | 5+6 2 | 5+6 2 | 3+4 2 | 2+2 2 | 3+3 2 |
| | 5+6 2 | 3+4 2 | 5+6 2 | 3+4 2 | 3+3 2 | 2+2 2 |
| | 3+4 2 | 3+4 2 | 3+4 2 | 5+6 2 | 2+2 2 | 3+3 2 |
| | 3+4 2 | 5+6 2 | 3+4 2 | 3+4 2 | 3+3 2 | 3+3 2 |
| | 3+4 2 | 3+4 2 | 3+4 2 | 3+4 2 | 3+3 2 | 2+2 2 |
| 12.1.111 | 66006 | | 70 | | Co2Na7O6Rb1 | |
| | -A | -GM | -M | -Y | -L | -V |
| | 3+4 2 | 3+4 2 | 3+4 2 | 3+4 2 | 3+3 2 | 3+3 2 |
| | 3+4 2 | 3+4 2 | 5+6 2 | 5+6 2 | 2+2 2 | 2+2 2 |
| | 5+6 2 | 5+6 2 | 3+4 2 | 3+4 2 | 3+3 2 | 3+3 2 |
| | 3+4 2 | 3+4 2 | 5+6 2 | 5+6 2 | 3+3 2 | 3+3 2 |
| | 5+6 2 | 5+6 2 | 3+4 2 | 3+4 2 | 2+2 2 | 2+2 2 |
| | 5+6 2 | 3+4 2 | 3+4 2 | 5+6 2 | 3+3 2 | 2+2 2 |
| | 3+4 2 | 5+6 2 | 5+6 2 | 3+4 2 | 2+2 2 | 3+3 2 |
| | 5+6 2 | 5+6 2 | 5+6 2 | 5+6 2 | 2+2 2 | 2+2 2 |
| | 5+6 2 | 5+6 2 | 5+6 2 | 5+6 2 | 2+2 2 | 2+2 2 |
| | 5+6 2 | 5+6 2 | 5+6 2 | 5+6 2 | 2+2 2 | 2+2 2 |
| | 3+4 2 | 3+4 2 | 5+6 2 | 5+6 2 | 2+2 2 | 2+2 2 |
| | 5+6 2 | 5+6 2 | 3+4 2 | 3+4 2 | 3+3 2 | 3+3 2 |
| | 5+6 2 | 3+4 2 | 5+6 2 | 5+6 2 | 3+3 2 | 2+2 2 |
| | 3+4 2 | 3+4 2 | 3+4 2 | 3+4 2 | 2+2 2 | 3+3 2 |
| | 5+6 2 | 5+6 2 | 3+4 2 | 5+6 2 | 3+3 2 | 2+2 2 |
| | 3+4 2 | 5+6 2 | 5+6 2 | 3+4 2 | 2+2 2 | 3+3 2 |
| | 3+4 2 | 3+4 2 | 3+4 2 | 3+4 2 | 3+3 2 | 3+3 2 |
| | 5+6 2 | 5+6 2 | 5+6 2 | 5+6 2 | 2+2 2 | 2+2 2 |
| | 3+4 2 | 3+4 2 | 5+6 2 | 3+4 2 | 3+3 2 | 3+3 2 |
| | 5+6 2 | 3+4 2 | 3+4 2 | 5+6 2 | 2+2 2 | 3+3 2 |
| | 3+4 2 | 5+6 2 | 3+4 2 | 3+4 2 | 2+2 2 | 2+2 2 |
| | 5+6 2 | 5+6 2 | 5+6 2 | 5+6 2 | 3+3 2 | 2+2 2 |
| | 3+4 2 | 3+4 2 | 5+6 2 | 5+6 2 | 2+2 2 | 2+2 2 |
| | 5+6 2 | 3+4 2 | 3+4 2 | 3+4 2 | 3+3 2 | 3+3 2 |
| | 5+6 2 | 5+6 2 | 3+4 2 | 5+6 2 | 3+3 2 | 2+2 2 |
| | 3+4 2 | 5+6 2 | 5+6 2 | 3+4 2 | 2+2 2 | 3+3 2 |
| | 3+4 2 | 5+6 2 | 5+6 2 | 3+4 2 | 3+3 2 | 3+3 2 |
| | 5+6 2 | 3+4 2 | 3+4 2 | 5+6 2 | 2+2 2 | 2+2 2 |
| | 3+4 2 | 3+4 2 | 3+4 2 | 3+4 2 | 3+3 2 | 3+3 2 |
| | 5+6 2 | 5+6 2 | 5+6 2 | 5+6 2 | 2+2 2 | 2+2 2 |
| | 5+6 2 | 3+4 2 | 3+4 2 | 5+6 2 | 3+3 2 | 2+2 2 |
| | 3+4 2 | 3+4 2 | 5+6 2 | 3+4 2 | 2+2 2 | 3+3 2 |

Continued on next column

**TABLE XXIV– continued**

| TQC | ICSD | NoE | SF | | | |
|---|---|---|---|---|---|---|
| | 3+4 2 | 5+6 2 | 5+6 2 | 3+4 2 | 2+2 2 | 2+2 2 |
| | 5+6 2 | 3+4 2 | 3+4 2 | 5+6 2 | 3+3 2 | 3+3 2 |
| | 3+4 2 | 5+6 2 | 5+6 2 | 5+6 2 | 3+3 2 | 3+3 2 |
| 12.1.112 | 2371 | 50 | Co1Si2Zr2 | | | |
| | -A | -GM | -M | -Y | -L | -V |
| | 5+6 2 | 3+4 2 | 5+6 2 | 3+4 2 | 3+3 2 | 2+2 2 |
| | 3+4 2 | 3+4 2 | 3+4 2 | 3+4 2 | 2+2 2 | 2+2 2 |
| | 3+4 2 | 5+6 2 | 3+4 2 | 5+6 2 | 2+2 2 | 3+3 2 |
| | 5+6 2 | 5+6 2 | 5+6 2 | 5+6 2 | 3+3 2 | 3+3 2 |
| | 5+6 2 | 3+4 2 | 3+4 2 | 3+4 2 | 2+2 2 | 3+3 2 |
| | 3+4 2 | 5+6 2 | 5+6 2 | 5+6 2 | 3+3 2 | 2+2 2 |
| | 5+6 2 | 5+6 2 | 5+6 2 | 5+6 2 | 3+3 2 | 3+3 2 |
| | 3+4 2 | 5+6 2 | 3+4 2 | 5+6 2 | 2+2 2 | 2+2 2 |
| | 5+6 2 | 3+4 2 | 5+6 2 | 3+4 2 | 3+3 2 | 2+2 2 |
| | 3+4 2 | 5+6 2 | 5+6 2 | 3+4 2 | 2+2 2 | 2+2 2 |
| | 3+4 2 | 3+4 2 | 3+4 2 | 3+4 2 | 2+2 2 | 3+3 2 |
| | 5+6 2 | 3+4 2 | 3+4 2 | 5+6 2 | 3+3 2 | 3+3 2 |
| | 5+6 2 | 5+6 2 | 5+6 2 | 3+4 2 | 3+3 2 | 2+2 2 |
| | 3+4 2 | 3+4 2 | 3+4 2 | 5+6 2 | 2+2 2 | 3+3 2 |
| | 5+6 2 | 3+4 2 | 5+6 2 | 5+6 2 | 2+2 2 | 2+2 2 |
| | 5+6 2 | 3+4 2 | 3+4 2 | 3+4 2 | 3+3 2 | 2+2 2 |
| | 3+4 2 | 3+4 2 | 3+4 2 | 3+4 2 | 2+2 2 | 3+3 2 |
| | 3+4 2 | 5+6 2 | 3+4 2 | 3+4 2 | 3+3 2 | 2+2 2 |
| | 5+6 2 | 5+6 2 | 5+6 2 | 5+6 2 | 2+2 2 | 3+3 2 |
| | 5+6 2 | 3+4 2 | 5+6 2 | 5+6 2 | 3+3 2 | 2+2 2 |
| | 3+4 2 | 5+6 2 | 3+4 2 | 5+6 2 | 2+2 2 | 2+2 2 |
| | 5+6 2 | 3+4 2 | 3+4 2 | 3+4 2 | 3+3 2 | 3+3 2 |
| | 3+4 2 | 5+6 2 | 3+4 2 | 5+6 2 | 3+3 2 | 3+3 2 |
| | 5+6 2 | 3+4 2 | 5+6 2 | 3+4 2 | 2+2 2 | 2+2 2 |
| | 3+4 2 | 5+6 2 | 5+6 2 | 5+6 2 | 2+2 2 | 3+3 2 |
| 12.1.113 | 624952 | 46 | Co1Sc2Si2 | | | |
| | -A | -GM | -M | -Y | -L | -V |
| | 5+6 2 | 3+4 2 | 3+4 2 | 3+4 2 | 3+3 2 | 2+2 2 |
| | 3+4 2 | 3+4 2 | 5+6 2 | 3+4 2 | 2+2 2 | 2+2 2 |
| | 3+4 2 | 5+6 2 | 3+4 2 | 5+6 2 | 2+2 2 | 3+3 2 |
| | 5+6 2 | 5+6 2 | 5+6 2 | 5+6 2 | 3+3 2 | 3+3 2 |
| | 5+6 2 | 3+4 2 | 3+4 2 | 3+4 2 | 2+2 2 | 3+3 2 |
| | 3+4 2 | 5+6 2 | 5+6 2 | 5+6 2 | 3+3 2 | 2+2 2 |
| | 3+4 2 | 5+6 2 | 5+6 2 | 3+4 2 | 3+3 2 | 3+3 2 |
| | 5+6 2 | 5+6 2 | 3+4 2 | 5+6 2 | 2+2 2 | 2+2 2 |
| | 3+4 2 | 3+4 2 | 5+6 2 | 5+6 2 | 3+3 2 | 2+2 2 |
| | 5+6 2 | 3+4 2 | 5+6 2 | 3+4 2 | 2+2 2 | 3+3 2 |
| | 5+6 2 | 3+4 2 | 3+4 2 | 3+4 2 | 2+2 2 | 2+2 2 |
| | 3+4 2 | 5+6 2 | 3+4 2 | 5+6 2 | 3+3 2 | 3+3 2 |
| | 3+4 2 | 5+6 2 | 3+4 2 | 3+4 2 | 2+2 2 | 2+2 2 |
| | 5+6 2 | 3+4 2 | 5+6 2 | 5+6 2 | 3+3 2 | 3+3 2 |
| | 3+4 2 | 3+4 2 | 3+4 2 | 5+6 2 | 3+3 2 | 2+2 2 |
| | 5+6 2 | 5+6 2 | 5+6 2 | 3+4 2 | 2+2 2 | 2+2 2 |
| | 5+6 2 | 3+4 2 | 3+4 2 | 5+6 2 | 3+3 2 | 3+3 2 |
| | 3+4 2 | 3+4 2 | 3+4 2 | 3+4 2 | 2+2 2 | 3+3 2 |
| | 5+6 2 | 5+6 2 | 3+4 2 | 5+6 2 | 3+3 2 | 2+2 2 |
| | 5+6 2 | 3+4 2 | 5+6 2 | 3+4 2 | 2+2 2 | 2+2 2 |
| | 5+6 2 | 3+4 2 | 5+6 2 | 5+6 2 | 2+2 2 | 3+3 2 |
| | 3+4 2 | 5+6 2 | 3+4 2 | 3+4 2 | 3+3 2 | 2+2 2 |
| | 3+4 2 | 5+6 2 | 3+4 2 | 3+4 2 | 2+2 2 | 3+3 2 |
| 12.1.114 | 623423 | 52 | Co1Ge1 | | | |
| | -A | -GM | -M | -Y | -L | -V |
| | 3+4 2 | 3+4 2 | 5+6 2 | 3+4 2 | 2+2 2 | 2+2 2 |
| | 5+6 2 | 3+4 2 | 3+4 2 | 5+6 2 | 3+3 2 | 3+3 2 |
| | 3+4 2 | 5+6 2 | 5+6 2 | 3+4 2 | 3+3 2 | 2+2 2 |
| | 5+6 2 | 5+6 2 | 3+4 2 | 5+6 2 | 2+2 2 | 3+3 2 |

Continued on next column

**TABLE XXIV– continued**

| TQC | ICSD | NoE | SF | | | |
|---|---|---|---|---|---|---|
| | 5+6 2 | 3+4 2 | 3+4 2 | 3+4 2 | 2+2 2 | 3+3 2 |
| | 3+4 2 | 3+4 2 | 5+6 2 | 3+4 2 | 3+3 2 | 3+3 2 |
| | 3+4 2 | 3+4 2 | 3+4 2 | 3+4 2 | 3+3 2 | 2+2 2 |
| | 3+4 2 | 5+6 2 | 5+6 2 | 5+6 2 | 2+2 2 | 3+3 2 |
| | 5+6 2 | 3+4 2 | 5+6 2 | 3+4 2 | 2+2 2 | 3+3 2 |
| | 3+4 2 | 5+6 2 | 3+4 2 | 3+4 2 | 3+3 2 | 2+2 2 |
| | 3+4 2 | 3+4 2 | 5+6 2 | 5+6 2 | 2+2 2 | 3+3 2 |
| | 5+6 2 | 3+4 2 | 5+6 2 | 3+4 2 | 3+3 2 | 2+2 2 |
| | 3+4 2 | 5+6 2 | 5+6 2 | 3+4 2 | 3+3 2 | 2+2 2 |
| | 5+6 2 | 3+4 2 | 3+4 2 | 5+6 2 | 3+3 2 | 3+3 2 |
| | 3+4 2 | 3+4 2 | 3+4 2 | 5+6 2 | 2+2 2 | 2+2 2 |
| | 5+6 2 | 3+4 2 | 3+4 2 | 5+6 2 | 3+3 2 | 3+3 2 |
| | 5+6 2 | 3+4 2 | 5+6 2 | 3+4 2 | 2+2 2 | 3+3 2 |
| | 3+4 2 | 5+6 2 | 3+4 2 | 5+6 2 | 2+2 2 | 2+2 2 |
| | 5+6 2 | 5+6 2 | 5+6 2 | 3+4 2 | 3+3 2 | 3+3 2 |
| | 5+6 2 | 3+4 2 | 3+4 2 | 3+4 2 | 2+2 2 | 2+2 2 |
| | 5+6 2 | 5+6 2 | 3+4 2 | 3+4 2 | 3+3 2 | 3+3 2 |
| | 3+4 2 | 3+4 2 | 5+6 2 | 5+6 2 | 2+2 2 | 3+3 2 |
| | 5+6 2 | 3+4 2 | 5+6 2 | 3+4 2 | 2+2 2 | 3+3 2 |
| | 3+4 2 | 3+4 2 | 3+4 2 | 3+4 2 | 3+3 2 | 2+2 2 |
| | 5+6 2 | 3+4 2 | 3+4 2 | 3+4 2 | 3+3 2 | 3+3 2 |
| | 3+4 2 | 3+4 2 | 5+6 2 | 5+6 2 | 2+2 2 | 3+3 2 |
| 12.1.115 | 43677 | 52 | Co1Ge1 | | | |
| | -A | -GM | -M | -Y | -L | -V |
| | 3+4 2 | 3+4 2 | 5+6 2 | 3+4 2 | 2+2 2 | 2+2 2 |
| | 5+6 2 | 3+4 2 | 3+4 2 | 5+6 2 | 3+3 2 | 3+3 2 |
| | 3+4 2 | 5+6 2 | 5+6 2 | 3+4 2 | 3+3 2 | 2+2 2 |
| | 5+6 2 | 5+6 2 | 3+4 2 | 5+6 2 | 2+2 2 | 3+3 2 |
| | 5+6 2 | 3+4 2 | 3+4 2 | 3+4 2 | 2+2 2 | 3+3 2 |
| | 3+4 2 | 3+4 2 | 5+6 2 | 3+4 2 | 3+3 2 | 3+3 2 |
| | 3+4 2 | 3+4 2 | 3+4 2 | 3+4 2 | 3+3 2 | 2+2 2 |
| | 3+4 2 | 5+6 2 | 5+6 2 | 5+6 2 | 2+2 2 | 3+3 2 |
| | 5+6 2 | 3+4 2 | 5+6 2 | 3+4 2 | 2+2 2 | 3+3 2 |
| | 3+4 2 | 5+6 2 | 3+4 2 | 3+4 2 | 3+3 2 | 2+2 2 |
| | 3+4 2 | 3+4 2 | 5+6 2 | 5+6 2 | 2+2 2 | 3+3 2 |
| | 5+6 2 | 3+4 2 | 5+6 2 | 3+4 2 | 3+3 2 | 2+2 2 |
| | 3+4 2 | 5+6 2 | 5+6 2 | 3+4 2 | 3+3 2 | 2+2 2 |
| | 5+6 2 | 3+4 2 | 3+4 2 | 5+6 2 | 3+3 2 | 3+3 2 |
| | 3+4 2 | 3+4 2 | 3+4 2 | 5+6 2 | 2+2 2 | 2+2 2 |
| | 5+6 2 | 3+4 2 | 3+4 2 | 5+6 2 | 3+3 2 | 3+3 2 |
| | 5+6 2 | 3+4 2 | 5+6 2 | 3+4 2 | 2+2 2 | 3+3 2 |
| | 3+4 2 | 5+6 2 | 3+4 2 | 5+6 2 | 2+2 2 | 2+2 2 |
| | 5+6 2 | 5+6 2 | 5+6 2 | 3+4 2 | 3+3 2 | 3+3 2 |
| | 5+6 2 | 3+4 2 | 3+4 2 | 3+4 2 | 2+2 2 | 2+2 2 |
| | 5+6 2 | 5+6 2 | 3+4 2 | 3+4 2 | 3+3 2 | 3+3 2 |
| | 3+4 2 | 3+4 2 | 5+6 2 | 5+6 2 | 2+2 2 | 3+3 2 |
| | 5+6 2 | 3+4 2 | 5+6 2 | 3+4 2 | 2+2 2 | 3+3 2 |
| | 3+4 2 | 3+4 2 | 3+4 2 | 3+4 2 | 3+3 2 | 2+2 2 |
| | 5+6 2 | 3+4 2 | 3+4 2 | 3+4 2 | 3+3 2 | 3+3 2 |
| | 3+4 2 | 3+4 2 | 5+6 2 | 5+6 2 | 2+2 2 | 3+3 2 |
| 12.1.116 | 184254 | 52 | Co1Ge1 | | | |
| | -A | -GM | -M | -Y | -L | -V |
| | 3+4 2 | 3+4 2 | 5+6 2 | 3+4 2 | 2+2 2 | 2+2 2 |
| | 5+6 2 | 3+4 2 | 3+4 2 | 5+6 2 | 3+3 2 | 3+3 2 |
| | 3+4 2 | 5+6 2 | 5+6 2 | 3+4 2 | 3+3 2 | 2+2 2 |
| | 5+6 2 | 5+6 2 | 3+4 2 | 5+6 2 | 2+2 2 | 3+3 2 |
| | 5+6 2 | 3+4 2 | 3+4 2 | 3+4 2 | 2+2 2 | 3+3 2 |
| | 3+4 2 | 3+4 2 | 5+6 2 | 3+4 2 | 3+3 2 | 3+3 2 |
| | 3+4 2 | 3+4 2 | 3+4 2 | 3+4 2 | 3+3 2 | 2+2 2 |
| | 3+4 2 | 5+6 2 | 5+6 2 | 5+6 2 | 2+2 2 | 3+3 2 |
| | 5+6 2 | 3+4 2 | 5+6 2 | 3+4 2 | 2+2 2 | 3+3 2 |

Continued on next column

**TABLE XXIV– continued**

| TQC | ICSD | NoE | SF | | | |
|---|---|---|---|---|---|---|
| | 3+4 2 | 5+6 2 | 5+6 2 | 3+4 2 | 3+3 2 | 2+2 2 |
| | 3+4 2 | 3+4 2 | 3+4 2 | 5+6 2 | 2+2 2 | 2+2 2 |
| | 5+6 2 | 3+4 2 | 5+6 2 | 3+4 2 | 3+3 2 | 3+3 2 |
| | 3+4 2 | 5+6 2 | 5+6 2 | 5+6 2 | 3+3 2 | 2+2 2 |
| | 5+6 2 | 3+4 2 | 3+4 2 | 3+4 2 | 3+3 2 | 3+3 2 |
| | 3+4 2 | 3+4 2 | 3+4 2 | 5+6 2 | 2+2 2 | 2+2 2 |
| | 5+6 2 | 3+4 2 | 3+4 2 | 5+6 2 | 3+3 2 | 3+3 2 |
| | 5+6 2 | 3+4 2 | 5+6 2 | 3+4 2 | 2+2 2 | 3+3 2 |
| | 3+4 2 | 5+6 2 | 3+4 2 | 5+6 2 | 2+2 2 | 2+2 2 |
| | 5+6 2 | 5+6 2 | 5+6 2 | 3+4 2 | 3+3 2 | 3+3 2 |
| | 5+6 2 | 3+4 2 | 3+4 2 | 3+4 2 | 2+2 2 | 2+2 2 |
| | 5+6 2 | 5+6 2 | 5+6 2 | 3+4 2 | 2+2 2 | 3+3 2 |
| | 3+4 2 | 3+4 2 | 3+4 2 | 5+6 2 | 3+3 2 | 3+3 2 |
| | 5+6 2 | 3+4 2 | 5+6 2 | 3+4 2 | 2+2 2 | 3+3 2 |
| | 3+4 2 | 3+4 2 | 3+4 2 | 3+4 2 | 3+3 2 | 2+2 2 |
| | 5+6 2 | 3+4 2 | 3+4 2 | 3+4 2 | 3+3 2 | 3+3 2 |
| | 3+4 2 | 3+4 2 | 5+6 2 | 5+6 2 | 2+2 2 | 3+3 2 |
| 12.1.117 | 183325 | 106 | Co1Gd2Si2 | | | |
| | -A | -GM | -M | -Y | -L | -V |
| | 3+4 2 | 3+4 2 | 5+6 2 | 5+6 2 | 3+3 2 | 3+3 2 |
| | 5+6 2 | 5+6 2 | 3+4 2 | 3+4 2 | 2+2 2 | 2+2 2 |
| | 5+6 2 | 3+4 2 | 5+6 2 | 3+4 2 | 2+2 2 | 3+3 2 |
| | 3+4 2 | 5+6 2 | 3+4 2 | 5+6 2 | 3+3 2 | 2+2 2 |
| | 3+4 2 | 3+4 2 | 5+6 2 | 5+6 2 | 3+3 2 | 3+3 2 |
| | 5+6 2 | 5+6 2 | 3+4 2 | 3+4 2 | 2+2 2 | 2+2 2 |
| | 5+6 2 | 3+4 2 | 5+6 2 | 3+4 2 | 2+2 2 | 3+3 2 |
| | 3+4 2 | 5+6 2 | 3+4 2 | 5+6 2 | 3+3 2 | 2+2 2 |
| | 3+4 2 | 3+4 2 | 3+4 2 | 3+4 2 | 3+3 2 | 3+3 2 |
| | 5+6 2 | 5+6 2 | 5+6 2 | 5+6 2 | 2+2 2 | 2+2 2 |
| | 3+4 2 | 3+4 2 | 5+6 2 | 5+6 2 | 2+2 2 | 3+3 2 |
| | 5+6 2 | 3+4 2 | 5+6 2 | 3+4 2 | 3+3 2 | 2+2 2 |
| | 3+4 2 | 5+6 2 | 3+4 2 | 3+4 2 | 2+2 2 | 3+3 2 |
| | 5+6 2 | 3+4 2 | 3+4 2 | 5+6 2 | 3+3 2 | 2+2 2 |
| | 3+4 2 | 5+6 2 | 5+6 2 | 5+6 2 | 2+2 2 | 3+3 2 |
| | 5+6 2 | 5+6 2 | 3+4 2 | 3+4 2 | 3+3 2 | 2+2 2 |
| | 5+6 2 | 3+4 2 | 5+6 2 | 3+4 2 | 3+3 2 | 2+2 2 |
| | 3+4 2 | 3+4 2 | 3+4 2 | 5+6 2 | 2+2 2 | 2+2 2 |
| | 3+4 2 | 5+6 2 | 3+4 2 | 3+4 2 | 2+2 2 | 3+3 2 |
| | 5+6 2 | 5+6 2 | 5+6 2 | 5+6 2 | 3+3 2 | 3+3 2 |
| | 5+6 2 | 3+4 2 | 3+4 2 | 3+4 2 | 2+2 2 | 3+3 2 |
| | 3+4 2 | 5+6 2 | 5+6 2 | 5+6 2 | 3+3 2 | 2+2 2 |
| | 5+6 2 | 5+6 2 | 5+6 2 | 5+6 2 | 3+3 2 | 3+3 2 |
| | 3+4 2 | 3+4 2 | 3+4 2 | 5+6 2 | 2+2 2 | 2+2 2 |
| | 5+6 2 | 3+4 2 | 3+4 2 | 3+4 2 | 3+3 2 | 2+2 2 |
| | 3+4 2 | 5+6 2 | 5+6 2 | 3+4 2 | 2+2 2 | 2+2 2 |
| | 5+6 2 | 3+4 2 | 5+6 2 | 3+4 2 | 2+2 2 | 3+3 2 |
| | 3+4 2 | 5+6 2 | 3+4 2 | 3+4 2 | 3+3 2 | 2+2 2 |
| | 5+6 2 | 5+6 2 | 3+4 2 | 5+6 2 | 3+3 2 | 3+3 2 |
| | 3+4 2 | 3+4 2 | 3+4 2 | 5+6 2 | 2+2 2 | 2+2 2 |
| | 5+6 2 | 3+4 2 | 5+6 2 | 3+4 2 | 3+3 2 | 3+3 2 |
| | 5+6 2 | 3+4 2 | 5+6 2 | 5+6 2 | 2+2 2 | 2+2 2 |
| | 3+4 2 | 3+4 2 | 3+4 2 | 5+6 2 | 2+2 2 | 2+2 2 |
| | 5+6 2 | 5+6 2 | 3+4 2 | 5+6 2 | 3+3 2 | 3+3 2 |
| | 3+4 2 | 5+6 2 | 5+6 2 | 3+4 2 | 2+2 2 | 3+3 2 |
| | 5+6 2 | 5+6 2 | 3+4 2 | 3+4 2 | 3+3 2 | 2+2 2 |
| | 3+4 2 | 3+4 2 | 5+6 2 | 3+4 2 | 2+2 2 | 3+3 2 |
| | 3+4 2 | 5+6 2 | 3+4 2 | 5+6 2 | 3+3 2 | 2+2 2 |
| | 5+6 2 | 5+6 2 | 5+6 2 | 5+6 2 | 2+2 2 | 3+3 2 |
| | 3+4 2 | 3+4 2 | 3+4 2 | 3+4 2 | 3+3 2 | 2+2 2 |
| | 5+6 2 | 5+6 2 | 5+6 2 | 5+6 2 | 2+2 2 | 3+3 2 |
| | 5+6 2 | 3+4 2 | 5+6 2 | 3+4 2 | 3+3 2 | 2+2 2 |

Continued on next column

**TABLE XXIV– continued**

| TQC | ICSD | NoE | SF | | | |
|---|---|---|---|---|---|---|
| | 3+4 2 | 3+4 2 | 3+4 2 | 3+4 2 | 3+3 2 | 3+3 2 |
| | 3+4 2 | 5+6 2 | 5+6 2 | 5+6 2 | 2+2 2 | 2+2 2 |
| | 5+6 2 | 3+4 2 | 3+4 2 | 3+4 2 | 2+2 2 | 3+3 2 |
| | 3+4 2 | 5+6 2 | 5+6 2 | 5+6 2 | 3+3 2 | 2+2 2 |
| | 5+6 2 | 3+4 2 | 3+4 2 | 3+4 2 | 2+2 2 | 3+3 2 |
| | 3+4 2 | 3+4 2 | 5+6 2 | 5+6 2 | 3+3 2 | 2+2 2 |
| | 5+6 2 | 5+6 2 | 3+4 2 | 3+4 2 | 2+2 2 | 3+3 2 |
| | 5+6 2 | 3+4 2 | 3+4 2 | 5+6 2 | 3+3 2 | 2+2 2 |
| | 3+4 2 | 5+6 2 | 5+6 2 | 5+6 2 | 3+3 2 | 3+3 2 |
| | 3+4 2 | 5+6 2 | 3+4 2 | 3+4 2 | 2+2 2 | 2+2 2 |
| | 5+6 2 | 3+4 2 | 5+6 2 | 5+6 2 | 3+3 2 | 3+3 2 |
| 12.1.118 | 184277 | 106 | Co1Gd2Ge2 | | | |
| | -A | -GM | -M | -Y | -L | -V |
| | 3+4 2 | 3+4 2 | 3+4 2 | 3+4 2 | 3+3 2 | 3+3 2 |
| | 5+6 2 | 5+6 2 | 5+6 2 | 5+6 2 | 2+2 2 | 2+2 2 |
| | 5+6 2 | 3+4 2 | 5+6 2 | 3+4 2 | 2+2 2 | 3+3 2 |
| | 3+4 2 | 5+6 2 | 3+4 2 | 5+6 2 | 3+3 2 | 2+2 2 |
| | 3+4 2 | 3+4 2 | 3+4 2 | 3+4 2 | 3+3 2 | 3+3 2 |
| | 5+6 2 | 5+6 2 | 5+6 2 | 5+6 2 | 2+2 2 | 2+2 2 |
| | 5+6 2 | 3+4 2 | 5+6 2 | 3+4 2 | 2+2 2 | 3+3 2 |
| | 3+4 2 | 5+6 2 | 3+4 2 | 5+6 2 | 3+3 2 | 2+2 2 |
| | 3+4 2 | 3+4 2 | 3+4 2 | 3+4 2 | 3+3 2 | 3+3 2 |
| | 5+6 2 | 5+6 2 | 5+6 2 | 5+6 2 | 2+2 2 | 2+2 2 |
| | 3+4 2 | 3+4 2 | 5+6 2 | 3+4 2 | 2+2 2 | 3+3 2 |
| | 5+6 2 | 3+4 2 | 3+4 2 | 5+6 2 | 3+3 2 | 2+2 2 |
| | 5+6 2 | 5+6 2 | 5+6 2 | 3+4 2 | 3+3 2 | 3+3 2 |
| | 3+4 2 | 3+4 2 | 3+4 2 | 5+6 2 | 2+2 2 | 2+2 2 |
| | 3+4 2 | 5+6 2 | 3+4 2 | 5+6 2 | 2+2 2 | 3+3 2 |
| | 5+6 2 | 5+6 2 | 5+6 2 | 3+4 2 | 3+3 2 | 2+2 2 |
| | 5+6 2 | 3+4 2 | 3+4 2 | 5+6 2 | 2+2 2 | 2+2 2 |
| | 3+4 2 | 3+4 2 | 5+6 2 | 3+4 2 | 3+3 2 | 2+2 2 |
| | 3+4 2 | 5+6 2 | 3+4 2 | 3+4 2 | 2+2 2 | 3+3 2 |
| | 5+6 2 | 5+6 2 | 5+6 2 | 5+6 2 | 3+3 2 | 3+3 2 |
| | 3+4 2 | 3+4 2 | 5+6 2 | 3+4 2 | 2+2 2 | 2+2 2 |
| | 5+6 2 | 5+6 2 | 3+4 2 | 5+6 2 | 3+3 2 | 3+3 2 |
| | 5+6 2 | 3+4 2 | 5+6 2 | 5+6 2 | 3+3 2 | 3+3 2 |
| | 5+6 2 | 3+4 2 | 3+4 2 | 5+6 2 | 2+2 2 | 2+2 2 |
| | 5+6 2 | 5+6 2 | 3+4 2 | 3+4 2 | 2+2 2 | 2+2 2 |
| | 3+4 2 | 3+4 2 | 5+6 2 | 3+4 2 | 3+3 2 | 2+2 2 |
| | 3+4 2 | 5+6 2 | 3+4 2 | 3+4 2 | 2+2 2 | 3+3 2 |
| | 3+4 2 | 5+6 2 | 5+6 2 | 5+6 2 | 3+3 2 | 3+3 2 |
| | 5+6 2 | 3+4 2 | 3+4 2 | 3+4 2 | 3+3 2 | 2+2 2 |
| | 3+4 2 | 5+6 2 | 3+4 2 | 5+6 2 | 2+2 2 | 2+2 2 |
| | 3+4 2 | 3+4 2 | 5+6 2 | 5+6 2 | 2+2 2 | 3+3 2 |
| | 5+6 2 | 3+4 2 | 5+6 2 | 3+4 2 | 3+3 2 | 2+2 2 |
| | 5+6 2 | 5+6 2 | 3+4 2 | 5+6 2 | 2+2 2 | 3+3 2 |
| | 5+6 2 | 3+4 2 | 3+4 2 | 5+6 2 | 3+3 2 | 3+3 2 |
| | 3+4 2 | 5+6 2 | 5+6 2 | 3+4 2 | 3+3 2 | 2+2 2 |
| | 3+4 2 | 3+4 2 | 3+4 2 | 3+4 2 | 2+2 2 | 2+2 2 |
| | 5+6 2 | 5+6 2 | 5+6 2 | 5+6 2 | 2+2 2 | 3+3 2 |
| | 3+4 2 | 5+6 2 | 3+4 2 | 3+4 2 | 3+3 2 | 2+2 2 |
| | 5+6 2 | 3+4 2 | 5+6 2 | 5+6 2 | 2+2 2 | 3+3 2 |
| | 3+4 2 | 5+6 2 | 5+6 2 | 3+4 2 | 3+3 2 | 2+2 2 |
| | 5+6 2 | 3+4 2 | 3+4 2 | 5+6 2 | 3+3 2 | 3+3 2 |
| | 5+6 2 | 5+6 2 | 5+6 2 | 3+4 2 | 2+2 2 | 2+2 2 |
| | 3+4 2 | 3+4 2 | 3+4 2 | 3+4 2 | 2+2 2 | 3+3 2 |
| | 3+4 2 | 5+6 2 | 5+6 2 | 5+6 2 | 3+3 2 | 2+2 2 |
| | 5+6 2 | 5+6 2 | 3+4 2 | 3+4 2 | 2+2 2 | 3+3 2 |
| | 3+4 2 | 3+4 2 | 5+6 2 | 5+6 2 | 3+3 2 | 3+3 2 |
| | 3+4 2 | 3+4 2 | 3+4 2 | 5+6 2 | 2+2 2 | 2+2 2 |
| | 5+6 2 | 3+4 2 | 3+4 2 | 3+4 2 | 3+3 2 | 3+3 2 |

Continued on next column

**TABLE XXIV– continued**

| TQC | ICSD | NoE | SF | | | |
|---|---|---|---|---|---|---|
| | 5+6 2 | 5+6 2 | 5+6 2 | 5+6 2 | 2+2 2 | 2+2 2 |
| | 5+6 2 | 3+4 2 | 5+6 2 | 3+4 2 | 3+3 2 | 2+2 2 |
| | 3+4 2 | 5+6 2 | 3+4 2 | 5+6 2 | 2+2 2 | 3+3 2 |
| | 3+4 2 | 5+6 2 | 3+4 2 | 3+4 2 | 3+3 2 | 2+2 2 |
| | 5+6 2 | 3+4 2 | 5+6 2 | 5+6 2 | 3+3 2 | 3+3 2 |
| 12.1.119 | 183326 | 106 | Co1Gd2Ge2 | | | |
| | -A | -GM | -M | -Y | -L | -V |
| | 3+4 2 | 3+4 2 | 3+4 2 | 3+4 2 | 3+3 2 | 3+3 2 |
| | 5+6 2 | 5+6 2 | 5+6 2 | 5+6 2 | 2+2 2 | 2+2 2 |
| | 5+6 2 | 3+4 2 | 5+6 2 | 3+4 2 | 2+2 2 | 3+3 2 |
| | 3+4 2 | 5+6 2 | 3+4 2 | 5+6 2 | 3+3 2 | 2+2 2 |
| | 3+4 2 | 3+4 2 | 3+4 2 | 3+4 2 | 3+3 2 | 3+3 2 |
| | 5+6 2 | 5+6 2 | 5+6 2 | 5+6 2 | 2+2 2 | 2+2 2 |
| | 5+6 2 | 3+4 2 | 5+6 2 | 3+4 2 | 2+2 2 | 3+3 2 |
| | 3+4 2 | 5+6 2 | 3+4 2 | 5+6 2 | 3+3 2 | 2+2 2 |
| | 3+4 2 | 3+4 2 | 3+4 2 | 3+4 2 | 3+3 2 | 3+3 2 |
| | 5+6 2 | 5+6 2 | 5+6 2 | 5+6 2 | 2+2 2 | 2+2 2 |
| | 3+4 2 | 3+4 2 | 5+6 2 | 3+4 2 | 2+2 2 | 3+3 2 |
| | 5+6 2 | 3+4 2 | 3+4 2 | 5+6 2 | 3+3 2 | 2+2 2 |
| | 5+6 2 | 5+6 2 | 5+6 2 | 3+4 2 | 3+3 2 | 3+3 2 |
| | 3+4 2 | 3+4 2 | 3+4 2 | 5+6 2 | 2+2 2 | 2+2 2 |
| | 3+4 2 | 5+6 2 | 3+4 2 | 3+4 2 | 2+2 2 | 3+3 2 |
| | 5+6 2 | 5+6 2 | 5+6 2 | 5+6 2 | 3+3 2 | 2+2 2 |
| | 5+6 2 | 3+4 2 | 3+4 2 | 5+6 2 | 3+3 2 | 2+2 2 |
| | 3+4 2 | 3+4 2 | 5+6 2 | 3+4 2 | 2+2 2 | 2+2 2 |
| | 3+4 2 | 5+6 2 | 3+4 2 | 3+4 2 | 2+2 2 | 3+3 2 |
| | 5+6 2 | 5+6 2 | 5+6 2 | 5+6 2 | 3+3 2 | 3+3 2 |
| | 3+4 2 | 3+4 2 | 5+6 2 | 3+4 2 | 2+2 2 | 2+2 2 |
| | 5+6 2 | 5+6 2 | 3+4 2 | 5+6 2 | 3+3 2 | 3+3 2 |
| | 5+6 2 | 3+4 2 | 5+6 2 | 5+6 2 | 3+3 2 | 3+3 2 |
| | 5+6 2 | 3+4 2 | 3+4 2 | 5+6 2 | 2+2 2 | 2+2 2 |
| | 5+6 2 | 5+6 2 | 3+4 2 | 3+4 2 | 2+2 2 | 2+2 2 |
| | 3+4 2 | 3+4 2 | 5+6 2 | 3+4 2 | 3+3 2 | 2+2 2 |
| | 3+4 2 | 5+6 2 | 3+4 2 | 3+4 2 | 2+2 2 | 3+3 2 |
| | 3+4 2 | 5+6 2 | 5+6 2 | 5+6 2 | 3+3 2 | 3+3 2 |
| | 5+6 2 | 3+4 2 | 3+4 2 | 3+4 2 | 3+3 2 | 2+2 2 |
| | 3+4 2 | 5+6 2 | 3+4 2 | 5+6 2 | 2+2 2 | 2+2 2 |
| | 3+4 2 | 3+4 2 | 5+6 2 | 5+6 2 | 2+2 2 | 3+3 2 |
| | 5+6 2 | 3+4 2 | 5+6 2 | 3+4 2 | 3+3 2 | 2+2 2 |
| | 5+6 2 | 5+6 2 | 3+4 2 | 5+6 2 | 2+2 2 | 3+3 2 |
| | 5+6 2 | 3+4 2 | 3+4 2 | 5+6 2 | 3+3 2 | 2+2 2 |
| | 3+4 2 | 5+6 2 | 5+6 2 | 3+4 2 | 3+3 2 | 3+3 2 |
| | 3+4 2 | 3+4 2 | 3+4 2 | 3+4 2 | 2+2 2 | 2+2 2 |
| | 5+6 2 | 5+6 2 | 5+6 2 | 5+6 2 | 2+2 2 | 3+3 2 |
| | 3+4 2 | 5+6 2 | 3+4 2 | 3+4 2 | 3+3 2 | 2+2 2 |
| | 5+6 2 | 3+4 2 | 5+6 2 | 5+6 2 | 2+2 2 | 3+3 2 |
| | 3+4 2 | 5+6 2 | 5+6 2 | 3+4 2 | 3+3 2 | 2+2 2 |
| | 5+6 2 | 3+4 2 | 3+4 2 | 5+6 2 | 3+3 2 | 3+3 2 |
| | 5+6 2 | 5+6 2 | 5+6 2 | 3+4 2 | 2+2 2 | 2+2 2 |
| | 3+4 2 | 3+4 2 | 3+4 2 | 3+4 2 | 2+2 2 | 3+3 2 |
| | 3+4 2 | 5+6 2 | 5+6 2 | 5+6 2 | 3+3 2 | 2+2 2 |
| | 5+6 2 | 5+6 2 | 3+4 2 | 3+4 2 | 2+2 2 | 3+3 2 |
| | 3+4 2 | 3+4 2 | 5+6 2 | 5+6 2 | 3+3 2 | 3+3 2 |
| | 3+4 2 | 3+4 2 | 3+4 2 | 5+6 2 | 2+2 2 | 2+2 2 |
| | 5+6 2 | 3+4 2 | 3+4 2 | 3+4 2 | 3+3 2 | 3+3 2 |
| | 5+6 2 | 5+6 2 | 5+6 2 | 5+6 2 | 2+2 2 | 2+2 2 |
| | 5+6 2 | 3+4 2 | 5+6 2 | 3+4 2 | 3+3 2 | 2+2 2 |
| | 3+4 2 | 5+6 2 | 3+4 2 | 5+6 2 | 2+2 2 | 3+3 2 |
| | 3+4 2 | 5+6 2 | 3+4 2 | 3+4 2 | 3+3 2 | 2+2 2 |
| | 5+6 2 | 3+4 2 | 5+6 2 | 5+6 2 | 3+3 2 | 3+3 2 |
| 12.1.120 | 427661 | 110 | Cl1Gd1O4W1 | | | |

Continued on next column

**TABLE XXIV– continued**

| TQC | ICSD | NoE | SF | | | |
|---|---|---|---|---|---|---|
| | -A | -GM | -M | -Y | -L | -V |
| | 3+4 2 | 3+4 2 | 5+6 2 | 5+6 2 | 2+2 2 | 2+2 2 |
| | 5+6 2 | 5+6 2 | 3+4 2 | 3+4 2 | 3+3 2 | 3+3 2 |
| | 3+4 2 | 3+4 2 | 5+6 2 | 5+6 2 | 3+3 2 | 3+3 2 |
| | 5+6 2 | 5+6 2 | 3+4 2 | 3+4 2 | 2+2 2 | 2+2 2 |
| | 3+4 2 | 3+4 2 | 5+6 2 | 5+6 2 | 2+2 2 | 2+2 2 |
| | 5+6 2 | 5+6 2 | 3+4 2 | 3+4 2 | 3+3 2 | 3+3 2 |
| | 3+4 2 | 3+4 2 | 3+4 2 | 3+4 2 | 3+3 2 | 2+2 2 |
| | 5+6 2 | 5+6 2 | 5+6 2 | 5+6 2 | 2+2 2 | 3+3 2 |
| | 3+4 2 | 3+4 2 | 5+6 2 | 5+6 2 | 2+2 2 | 3+3 2 |
| | 5+6 2 | 5+6 2 | 3+4 2 | 3+4 2 | 3+3 2 | 2+2 2 |
| | 3+4 2 | 5+6 2 | 3+4 2 | 5+6 2 | 2+2 2 | 3+3 2 |
| | 5+6 2 | 3+4 2 | 5+6 2 | 3+4 2 | 3+3 2 | 2+2 2 |
| | 5+6 2 | 3+4 2 | 3+4 2 | 3+4 2 | 3+3 2 | 3+3 2 |
| | 3+4 2 | 5+6 2 | 5+6 2 | 5+6 2 | 2+2 2 | 2+2 2 |
| | 5+6 2 | 5+6 2 | 3+4 2 | 3+4 2 | 3+3 2 | 2+2 2 |
| | 3+4 2 | 3+4 2 | 5+6 2 | 5+6 2 | 2+2 2 | 3+3 2 |
| | 3+4 2 | 3+4 2 | 3+4 2 | 3+4 2 | 3+3 2 | 3+3 2 |
| | 5+6 2 | 5+6 2 | 5+6 2 | 5+6 2 | 2+2 2 | 2+2 2 |
| | 5+6 2 | 3+4 2 | 3+4 2 | 3+4 2 | 2+2 2 | 2+2 2 |
| | 3+4 2 | 5+6 2 | 5+6 2 | 5+6 2 | 3+3 2 | 3+3 2 |
| | 3+4 2 | 3+4 2 | 5+6 2 | 3+4 2 | 2+2 2 | 2+2 2 |
| | 5+6 2 | 3+4 2 | 3+4 2 | 5+6 2 | 3+3 2 | 2+2 2 |
| | 3+4 2 | 5+6 2 | 5+6 2 | 5+6 2 | 3+3 2 | 3+3 2 |
| | 5+6 2 | 5+6 2 | 3+4 2 | 5+6 2 | 2+2 2 | 2+2 2 |
| | 3+4 2 | 3+4 2 | 3+4 2 | 3+4 2 | 3+3 2 | 3+3 2 |
| | 5+6 2 | 3+4 2 | 5+6 2 | 5+6 2 | 2+2 2 | 3+3 2 |
| | 5+6 2 | 5+6 2 | 5+6 2 | 3+4 2 | 3+3 2 | 3+3 2 |
| | 3+4 2 | 5+6 2 | 5+6 2 | 5+6 2 | 2+2 2 | 2+2 2 |
| | 3+4 2 | 3+4 2 | 3+4 2 | 3+4 2 | 3+3 2 | 3+3 2 |
| | 5+6 2 | 5+6 2 | 3+4 2 | 3+4 2 | 2+2 2 | 2+2 2 |
| | 3+4 2 | 3+4 2 | 3+4 2 | 3+4 2 | 2+2 2 | 2+2 2 |
| | 3+4 2 | 3+4 2 | 5+6 2 | 5+6 2 | 3+3 2 | 3+3 2 |
| | 5+6 2 | 5+6 2 | 5+6 2 | 5+6 2 | 2+2 2 | 2+2 2 |
| | 3+4 2 | 5+6 2 | 3+4 2 | 3+4 2 | 3+3 2 | 3+3 2 |
| | 5+6 2 | 3+4 2 | 3+4 2 | 3+4 2 | 2+2 2 | 3+3 2 |
| | 5+6 2 | 5+6 2 | 5+6 2 | 5+6 2 | 3+3 2 | 2+2 2 |
| | 3+4 2 | 3+4 2 | 3+4 2 | 3+4 2 | 2+2 2 | 3+3 2 |
| | 3+4 2 | 5+6 2 | 5+6 2 | 5+6 2 | 3+3 2 | 2+2 2 |
| | 3+4 2 | 3+4 2 | 3+4 2 | 3+4 2 | 2+2 2 | 3+3 2 |
| | 5+6 2 | 5+6 2 | 5+6 2 | 3+4 2 | 2+2 2 | 2+2 2 |
| | 5+6 2 | 3+4 2 | 3+4 2 | 3+4 2 | 3+3 2 | 3+3 2 |
| | 3+4 2 | 3+4 2 | 3+4 2 | 5+6 2 | 3+3 2 | 2+2 2 |
| | 5+6 2 | 5+6 2 | 5+6 2 | 5+6 2 | 3+3 2 | 3+3 2 |
| | 3+4 2 | 5+6 2 | 3+4 2 | 5+6 2 | 2+2 2 | 3+3 2 |
| | 5+6 2 | 3+4 2 | 5+6 2 | 5+6 2 | 2+2 2 | 2+2 2 |
| | 5+6 2 | 5+6 2 | 5+6 2 | 3+4 2 | 3+3 2 | 2+2 2 |
| | 3+4 2 | 3+4 2 | 5+6 2 | 3+4 2 | 3+3 2 | 3+3 2 |
| | 5+6 2 | 5+6 2 | 3+4 2 | 5+6 2 | 2+2 2 | 2+2 2 |
| | 3+4 2 | 3+4 2 | 5+6 2 | 5+6 2 | 2+2 2 | 2+2 2 |
| | 5+6 2 | 3+4 2 | 3+4 2 | 3+4 2 | 3+3 2 | 3+3 2 |
| | 3+4 2 | 5+6 2 | 5+6 2 | 3+4 2 | 3+3 2 | 3+3 2 |
| | 3+4 2 | 5+6 2 | 3+4 2 | 5+6 2 | 2+2 2 | 2+2 2 |
| | 5+6 2 | 3+4 2 | 5+6 2 | 3+4 2 | 3+3 2 | 3+3 2 |
| | 5+6 2 | 5+6 2 | 3+4 2 | 5+6 2 | 2+2 2 | 2+2 2 |
| | 3+4 2 | 3+4 2 | 5+6 2 | 5+6 2 | 3+3 2 | 3+3 2 |
| 12.1.121 | 35292 | 110 | Cl1Gd1O4W1 | | | |
| | -A | -GM | -M | -Y | -L | -V |
| | 3+4 2 | 3+4 2 | 5+6 2 | 5+6 2 | 2+2 2 | 2+2 2 |
| | 5+6 2 | 5+6 2 | 3+4 2 | 3+4 2 | 3+3 2 | 3+3 2 |
| | 3+4 2 | 3+4 2 | 5+6 2 | 5+6 2 | 3+3 2 | 3+3 2 |

Continued on next column

**TABLE XXIV– continued**

| TQC | ICSD | NoE | SF | | | |
|---|---|---|---|---|---|---|
| | 5+6 2 | 5+6 2 | 3+4 2 | 3+4 2 | 2+2 2 | 2+2 2 |
| | 3+4 2 | 3+4 2 | 5+6 2 | 5+6 2 | 2+2 2 | 2+2 2 |
| | 5+6 2 | 5+6 2 | 3+4 2 | 3+4 2 | 3+3 2 | 3+3 2 |
| | 3+4 2 | 3+4 2 | 5+6 2 | 3+4 2 | 3+3 2 | 2+2 2 |
| | 5+6 2 | 5+6 2 | 3+4 2 | 5+6 2 | 2+2 2 | 3+3 2 |
| | 3+4 2 | 3+4 2 | 5+6 2 | 5+6 2 | 2+2 2 | 3+3 2 |
| | 5+6 2 | 5+6 2 | 3+4 2 | 3+4 2 | 3+3 2 | 2+2 2 |
| | 3+4 2 | 5+6 2 | 3+4 2 | 5+6 2 | 2+2 2 | 3+3 2 |
| | 5+6 2 | 3+4 2 | 5+6 2 | 3+4 2 | 3+3 2 | 2+2 2 |
| | 5+6 2 | 3+4 2 | 3+4 2 | 3+4 2 | 3+3 2 | 3+3 2 |
| | 3+4 2 | 5+6 2 | 5+6 2 | 5+6 2 | 2+2 2 | 2+2 2 |
| | 5+6 2 | 5+6 2 | 3+4 2 | 3+4 2 | 3+3 2 | 2+2 2 |
| | 3+4 2 | 3+4 2 | 5+6 2 | 5+6 2 | 2+2 2 | 3+3 2 |
| | 3+4 2 | 3+4 2 | 3+4 2 | 3+4 2 | 3+3 2 | 3+3 2 |
| | 5+6 2 | 5+6 2 | 5+6 2 | 5+6 2 | 2+2 2 | 2+2 2 |
| | 5+6 2 | 3+4 2 | 3+4 2 | 3+4 2 | 2+2 2 | 2+2 2 |
| | 3+4 2 | 5+6 2 | 5+6 2 | 5+6 2 | 3+3 2 | 3+3 2 |
| | 3+4 2 | 3+4 2 | 5+6 2 | 5+6 2 | 2+2 2 | 2+2 2 |
| | 3+4 2 | 3+4 2 | 3+4 2 | 3+4 2 | 3+3 2 | 3+3 2 |
| | 5+6 2 | 5+6 2 | 5+6 2 | 5+6 2 | 3+3 2 | 2+2 2 |
| | 5+6 2 | 5+6 2 | 3+4 2 | 5+6 2 | 2+2 2 | 3+3 2 |
| | 3+4 2 | 3+4 2 | 5+6 2 | 3+4 2 | 3+3 2 | 2+2 2 |
| | 5+6 2 | 3+4 2 | 3+4 2 | 5+6 2 | 2+2 2 | 3+3 2 |
| | 5+6 2 | 5+6 2 | 5+6 2 | 3+4 2 | 3+3 2 | 3+3 2 |
| | 3+4 2 | 3+4 2 | 5+6 2 | 5+6 2 | 2+2 2 | 2+2 2 |
| | 3+4 2 | 5+6 2 | 3+4 2 | 3+4 2 | 3+3 2 | 3+3 2 |
| | 5+6 2 | 5+6 2 | 3+4 2 | 3+4 2 | 2+2 2 | 2+2 2 |
| | 3+4 2 | 3+4 2 | 3+4 2 | 3+4 2 | 2+2 2 | 2+2 2 |
| | 3+4 2 | 3+4 2 | 5+6 2 | 5+6 2 | 3+3 2 | 3+3 2 |
| | 5+6 2 | 5+6 2 | 5+6 2 | 3+4 2 | 2+2 2 | 2+2 2 |
| | 3+4 2 | 3+4 2 | 3+4 2 | 5+6 2 | 3+3 2 | 3+3 2 |
| | 5+6 2 | 5+6 2 | 3+4 2 | 3+4 2 | 2+2 2 | 2+2 2 |
| | 5+6 2 | 5+6 2 | 5+6 2 | 5+6 2 | 3+3 2 | 3+3 2 |
| | 3+4 2 | 3+4 2 | 3+4 2 | 3+4 2 | 2+2 2 | 3+3 2 |
| | 3+4 2 | 5+6 2 | 5+6 2 | 5+6 2 | 3+3 2 | 2+2 2 |
| | 3+4 2 | 5+6 2 | 5+6 2 | 3+4 2 | 2+2 2 | 3+3 2 |
| | 5+6 2 | 3+4 2 | 3+4 2 | 3+4 2 | 2+2 2 | 2+2 2 |
| | 5+6 2 | 3+4 2 | 3+4 2 | 5+6 2 | 3+3 2 | 3+3 2 |
| | 5+6 2 | 3+4 2 | 3+4 2 | 3+4 2 | 3+3 2 | 2+2 2 |
| | 3+4 2 | 5+6 2 | 5+6 2 | 5+6 2 | 3+3 2 | 3+3 2 |
| | 3+4 2 | 5+6 2 | 3+4 2 | 5+6 2 | 2+2 2 | 3+3 2 |
| | 5+6 2 | 3+4 2 | 5+6 2 | 5+6 2 | 2+2 2 | 2+2 2 |
| | 5+6 2 | 5+6 2 | 5+6 2 | 3+4 2 | 3+3 2 | 2+2 2 |
| | 3+4 2 | 3+4 2 | 5+6 2 | 3+4 2 | 3+3 2 | 3+3 2 |
| | 5+6 2 | 5+6 2 | 3+4 2 | 5+6 2 | 2+2 2 | 2+2 2 |
| | 3+4 2 | 3+4 2 | 5+6 2 | 5+6 2 | 3+3 2 | 2+2 2 |
| | 5+6 2 | 5+6 2 | 3+4 2 | 3+4 2 | 2+2 2 | 3+3 2 |
| | 3+4 2 | 3+4 2 | 5+6 2 | 3+4 2 | 3+3 2 | 3+3 2 |
| | 3+4 2 | 5+6 2 | 3+4 2 | 5+6 2 | 2+2 2 | 2+2 2 |
| | 5+6 2 | 3+4 2 | 5+6 2 | 3+4 2 | 3+3 2 | 3+3 2 |
| | 5+6 2 | 5+6 2 | 3+4 2 | 5+6 2 | 2+2 2 | 2+2 2 |
| | 3+4 2 | 3+4 2 | 5+6 2 | 5+6 2 | 3+3 2 | 3+3 2 |
| 12.1.122 | 420055 | 110 | Cl1Gd1Mo1O4 | | | |
| | -A | -GM | -M | -Y | -L | -V |
| | 3+4 2 | 3+4 2 | 5+6 2 | 5+6 2 | 2+2 2 | 2+2 2 |
| | 5+6 2 | 5+6 2 | 3+4 2 | 3+4 2 | 3+3 2 | 3+3 2 |
| | 3+4 2 | 3+4 2 | 5+6 2 | 5+6 2 | 3+3 2 | 3+3 2 |
| | 5+6 2 | 5+6 2 | 3+4 2 | 3+4 2 | 2+2 2 | 2+2 2 |
| | 3+4 2 | 3+4 2 | 5+6 2 | 5+6 2 | 2+2 2 | 2+2 2 |
| | 5+6 2 | 5+6 2 | 3+4 2 | 3+4 2 | 3+3 2 | 3+3 2 |
| | 3+4 2 | 3+4 2 | 3+4 2 | 3+4 2 | 3+3 2 | 3+3 2 |

Continued on next column

**TABLE XXIV– continued**

| TQC | ICSD | NoE | SF | | | |
|---|---|---|---|---|---|---|
| | 5+6 2 | 5+6 2 | 5+6 2 | 5+6 2 | 2+2 2 | 2+2 2 |
| | 3+4 2 | 3+4 2 | 5+6 2 | 5+6 2 | 2+2 2 | 2+2 2 |
| | 5+6 2 | 5+6 2 | 3+4 2 | 3+4 2 | 3+3 2 | 3+3 2 |
| | 3+4 2 | 5+6 2 | 3+4 2 | 5+6 2 | 2+2 2 | 3+3 2 |
| | 5+6 2 | 3+4 2 | 5+6 2 | 3+4 2 | 3+3 2 | 2+2 2 |
| | 5+6 2 | 3+4 2 | 3+4 2 | 3+4 2 | 3+3 2 | 2+2 2 |
| | 3+4 2 | 5+6 2 | 5+6 2 | 5+6 2 | 2+2 2 | 3+3 2 |
| | 5+6 2 | 5+6 2 | 3+4 2 | 3+4 2 | 3+3 2 | 2+2 2 |
| | 3+4 2 | 3+4 2 | 5+6 2 | 5+6 2 | 2+2 2 | 3+3 2 |
| | 3+4 2 | 3+4 2 | 3+4 2 | 3+4 2 | 3+3 2 | 3+3 2 |
| | 5+6 2 | 5+6 2 | 5+6 2 | 5+6 2 | 2+2 2 | 2+2 2 |
| | 5+6 2 | 3+4 2 | 3+4 2 | 3+4 2 | 2+2 2 | 2+2 2 |
| | 3+4 2 | 3+4 2 | 5+6 2 | 3+4 2 | 3+3 2 | 3+3 2 |
| | 3+4 2 | 5+6 2 | 5+6 2 | 5+6 2 | 2+2 2 | 2+2 2 |
| | 5+6 2 | 3+4 2 | 5+6 2 | 5+6 2 | 3+3 2 | 2+2 2 |
| | 3+4 2 | 5+6 2 | 3+4 2 | 5+6 2 | 3+3 2 | 3+3 2 |
| | 5+6 2 | 5+6 2 | 3+4 2 | 5+6 2 | 2+2 2 | 3+3 2 |
| | 5+6 2 | 3+4 2 | 3+4 2 | 5+6 2 | 3+3 2 | 2+2 2 |
| | 3+4 2 | 5+6 2 | 5+6 2 | 3+4 2 | 2+2 2 | 3+3 2 |
| | 5+6 2 | 5+6 2 | 5+6 2 | 3+4 2 | 3+3 2 | 3+3 2 |
| | 3+4 2 | 3+4 2 | 3+4 2 | 3+4 2 | 2+2 2 | 2+2 2 |
| | 3+4 2 | 3+4 2 | 5+6 2 | 5+6 2 | 3+3 2 | 2+2 2 |
| | 5+6 2 | 3+4 2 | 3+4 2 | 3+4 2 | 2+2 2 | 3+3 2 |
| | 3+4 2 | 5+6 2 | 5+6 2 | 3+4 2 | 2+2 2 | 2+2 2 |
| | 3+4 2 | 3+4 2 | 3+4 2 | 5+6 2 | 3+3 2 | 3+3 2 |
| | 5+6 2 | 5+6 2 | 5+6 2 | 3+4 2 | 2+2 2 | 2+2 2 |
| | 3+4 2 | 5+6 2 | 3+4 2 | 5+6 2 | 3+3 2 | 3+3 2 |
| | 5+6 2 | 3+4 2 | 3+4 2 | 3+4 2 | 2+2 2 | 2+2 2 |
| | 5+6 2 | 5+6 2 | 5+6 2 | 5+6 2 | 3+3 2 | 3+3 2 |
| | 3+4 2 | 3+4 2 | 3+4 2 | 3+4 2 | 2+2 2 | 3+3 2 |
| | 3+4 2 | 5+6 2 | 5+6 2 | 5+6 2 | 3+3 2 | 3+3 2 |
| | 3+4 2 | 5+6 2 | 3+4 2 | 3+4 2 | 2+2 2 | 2+2 2 |
| | 5+6 2 | 3+4 2 | 5+6 2 | 3+4 2 | 2+2 2 | 2+2 2 |
| | 5+6 2 | 3+4 2 | 3+4 2 | 3+4 2 | 3+3 2 | 3+3 2 |
| | 3+4 2 | 3+4 2 | 3+4 2 | 5+6 2 | 3+3 2 | 2+2 2 |
| | 5+6 2 | 5+6 2 | 5+6 2 | 5+6 2 | 3+3 2 | 3+3 2 |
| | 3+4 2 | 5+6 2 | 3+4 2 | 5+6 2 | 2+2 2 | 3+3 2 |
| | 5+6 2 | 3+4 2 | 5+6 2 | 5+6 2 | 2+2 2 | 2+2 2 |
| | 5+6 2 | 5+6 2 | 5+6 2 | 3+4 2 | 3+3 2 | 2+2 2 |
| | 3+4 2 | 3+4 2 | 5+6 2 | 3+4 2 | 3+3 2 | 3+3 2 |
| | 5+6 2 | 5+6 2 | 3+4 2 | 5+6 2 | 2+2 2 | 2+2 2 |
| | 3+4 2 | 3+4 2 | 5+6 2 | 5+6 2 | 3+3 2 | 2+2 2 |
| | 5+6 2 | 3+4 2 | 3+4 2 | 3+4 2 | 2+2 2 | 3+3 2 |
| | 3+4 2 | 5+6 2 | 5+6 2 | 3+4 2 | 3+3 2 | 3+3 2 |
| | 3+4 2 | 5+6 2 | 3+4 2 | 5+6 2 | 2+2 2 | 2+2 2 |
| | 5+6 2 | 3+4 2 | 5+6 2 | 3+4 2 | 3+3 2 | 3+3 2 |
| | 5+6 2 | 5+6 2 | 3+4 2 | 5+6 2 | 2+2 2 | 2+2 2 |
| | 3+4 2 | 3+4 2 | 5+6 2 | 5+6 2 | 3+3 2 | 3+3 2 |
| 12.1.123 | 73096 | 98 | Cl1Dy1O3Ti1 | | | |
| | -A | -GM | -M | -Y | -L | -V |
| | 3+4 2 | 3+4 2 | 3+4 2 | 3+4 2 | 2+2 2 | 2+2 2 |
| | 5+6 2 | 5+6 2 | 5+6 2 | 5+6 2 | 3+3 2 | 3+3 2 |
| | 3+4 2 | 3+4 2 | 5+6 2 | 5+6 2 | 3+3 2 | 2+2 2 |
| | 5+6 2 | 5+6 2 | 3+4 2 | 3+4 2 | 2+2 2 | 3+3 2 |
| | 5+6 2 | 3+4 2 | 5+6 2 | 5+6 2 | 3+3 2 | 2+2 2 |
| | 3+4 2 | 5+6 2 | 3+4 2 | 3+4 2 | 2+2 2 | 3+3 2 |
| | 5+6 2 | 3+4 2 | 5+6 2 | 5+6 2 | 3+3 2 | 2+2 2 |
| | 3+4 2 | 5+6 2 | 3+4 2 | 3+4 2 | 2+2 2 | 3+3 2 |
| | 5+6 2 | 3+4 2 | 3+4 2 | 5+6 2 | 3+3 2 | 2+2 2 |
| | 3+4 2 | 5+6 2 | 5+6 2 | 3+4 2 | 2+2 2 | 3+3 2 |
| | 5+6 2 | 3+4 2 | 5+6 2 | 3+4 2 | 3+3 2 | 2+2 2 |

Continued on next column

**TABLE XXIV– continued**

| TQC | ICSD | | NoE | | SF | |
|---|---|---|---|---|---|---|
| | 3+4 2 | 5+6 2 | 3+4 2 | 5+6 2 | 2+2 2 | 3+3 2 |
| | 3+4 2 | 3+4 2 | 5+6 2 | 3+4 2 | 2+2 2 | 3+3 2 |
| | 5+6 2 | 5+6 2 | 3+4 2 | 5+6 2 | 3+3 2 | 2+2 2 |
| | 3+4 2 | 3+4 2 | 5+6 2 | 5+6 2 | 3+3 2 | 3+3 2 |
| | 5+6 2 | 5+6 2 | 3+4 2 | 3+4 2 | 2+2 2 | 2+2 2 |
| | 3+4 2 | 3+4 2 | 3+4 2 | 3+4 2 | 3+3 2 | 2+2 2 |
| | 5+6 2 | 3+4 2 | 5+6 2 | 5+6 2 | 2+2 2 | 3+3 2 |
| | 5+6 2 | 5+6 2 | 3+4 2 | 5+6 2 | 2+2 2 | 3+3 2 |
| | 5+6 2 | 3+4 2 | 3+4 2 | 5+6 2 | 3+3 2 | 2+2 2 |
| | 3+4 2 | 3+4 2 | 5+6 2 | 3+4 2 | 2+2 2 | 2+2 2 |
| | 5+6 2 | 5+6 2 | 5+6 2 | 3+4 2 | 3+3 2 | 3+3 2 |
| | 3+4 2 | 3+4 2 | 5+6 2 | 5+6 2 | 3+3 2 | 2+2 2 |
| | 3+4 2 | 5+6 2 | 5+6 2 | 3+4 2 | 2+2 2 | 3+3 2 |
| | 5+6 2 | 5+6 2 | 3+4 2 | 3+4 2 | 2+2 2 | 3+3 2 |
| | 5+6 2 | 3+4 2 | 3+4 2 | 3+4 2 | 3+3 2 | 3+3 2 |
| | 3+4 2 | 3+4 2 | 5+6 2 | 5+6 2 | 3+3 2 | 2+2 2 |
| | 3+4 2 | 5+6 2 | 3+4 2 | 5+6 2 | 2+2 2 | 3+3 2 |
| | 3+4 2 | 3+4 2 | 3+4 2 | 5+6 2 | 3+3 2 | 2+2 2 |
| | 5+6 2 | 5+6 2 | 5+6 2 | 5+6 2 | 3+3 2 | 2+2 2 |
| | 3+4 2 | 5+6 2 | 3+4 2 | 3+4 2 | 2+2 2 | 2+2 2 |
| | 5+6 2 | 3+4 2 | 5+6 2 | 5+6 2 | 3+3 2 | 3+3 2 |
| | 3+4 2 | 3+4 2 | 5+6 2 | 3+4 2 | 2+2 2 | 2+2 2 |
| | 5+6 2 | 3+4 2 | 3+4 2 | 5+6 2 | 3+3 2 | 3+3 2 |
| | 5+6 2 | 5+6 2 | 5+6 2 | 5+6 2 | 2+2 2 | 3+3 2 |
| | 3+4 2 | 5+6 2 | 3+4 2 | 3+4 2 | 2+2 2 | 2+2 2 |
| | 5+6 2 | 3+4 2 | 3+4 2 | 3+4 2 | 3+3 2 | 3+3 2 |
| | 3+4 2 | 5+6 2 | 5+6 2 | 5+6 2 | 3+3 2 | 2+2 2 |
| | 3+4 2 | 5+6 2 | 3+4 2 | 3+4 2 | 2+2 2 | 2+2 2 |
| | 5+6 2 | 5+6 2 | 5+6 2 | 3+4 2 | 2+2 2 | 3+3 2 |
| | 5+6 2 | 3+4 2 | 3+4 2 | 3+4 2 | 3+3 2 | 2+2 2 |
| | 3+4 2 | 5+6 2 | 5+6 2 | 5+6 2 | 2+2 2 | 3+3 2 |
| | 3+4 2 | 3+4 2 | 3+4 2 | 5+6 2 | 3+3 2 | 2+2 2 |
| | 5+6 2 | 5+6 2 | 5+6 2 | 3+4 2 | 2+2 2 | 3+3 2 |
| | 3+4 2 | 3+4 2 | 3+4 2 | 5+6 2 | 3+3 2 | 2+2 2 |
| | 5+6 2 | 5+6 2 | 5+6 2 | 3+4 2 | 2+2 2 | 3+3 2 |
| | 3+4 2 | 3+4 2 | 3+4 2 | 3+4 2 | 3+3 2 | 3+3 2 |
| | 5+6 2 | 5+6 2 | 5+6 2 | 5+6 2 | 2+2 2 | 2+2 2 |
| | 3+4 2 | 3+4 2 | 5+6 2 | 5+6 2 | 3+3 2 | 2+2 2 |
| 12.1.124 | 424719 | | 86 | | Ce4Ge6Mg7 | |
| | -A | -GM | -M | -Y | -L | -V |
| | 3+4 2 | 3+4 2 | 5+6 2 | 5+6 2 | 2+2 2 | 2+2 2 |
| | 5+6 2 | 5+6 2 | 3+4 2 | 3+4 2 | 3+3 2 | 3+3 2 |
| | 5+6 2 | 3+4 2 | 3+4 2 | 5+6 2 | 2+2 2 | 2+2 2 |
| | 3+4 2 | 5+6 2 | 5+6 2 | 3+4 2 | 3+3 2 | 3+3 2 |
| | 3+4 2 | 3+4 2 | 5+6 2 | 5+6 2 | 2+2 2 | 2+2 2 |
| | 5+6 2 | 5+6 2 | 3+4 2 | 3+4 2 | 3+3 2 | 3+3 2 |
| | 5+6 2 | 3+4 2 | 3+4 2 | 5+6 2 | 2+2 2 | 2+2 2 |
| | 3+4 2 | 5+6 2 | 5+6 2 | 3+4 2 | 3+3 2 | 3+3 2 |
| | 3+4 2 | 3+4 2 | 3+4 2 | 5+6 2 | 3+3 2 | 2+2 2 |
| | 5+6 2 | 5+6 2 | 5+6 2 | 3+4 2 | 2+2 2 | 3+3 2 |
| | 3+4 2 | 3+4 2 | 5+6 2 | 5+6 2 | 2+2 2 | 2+2 2 |
| | 5+6 2 | 3+4 2 | 3+4 2 | 3+4 2 | 3+3 2 | 3+3 2 |
| | 5+6 2 | 5+6 2 | 3+4 2 | 5+6 2 | 2+2 2 | 3+3 2 |
| | 3+4 2 | 5+6 2 | 5+6 2 | 3+4 2 | 3+3 2 | 2+2 2 |
| | 5+6 2 | 3+4 2 | 3+4 2 | 5+6 2 | 2+2 2 | 2+2 2 |
| | 3+4 2 | 5+6 2 | 5+6 2 | 3+4 2 | 3+3 2 | 3+3 2 |
| | 5+6 2 | 3+4 2 | 5+6 2 | 3+4 2 | 2+2 2 | 3+3 2 |
| | 3+4 2 | 5+6 2 | 3+4 2 | 5+6 2 | 3+3 2 | 3+3 2 |
| | 5+6 2 | 3+4 2 | 5+6 2 | 3+4 2 | 2+2 2 | 2+2 2 |
| | 3+4 2 | 5+6 2 | 3+4 2 | 5+6 2 | 3+3 2 | 3+3 2 |
| | 5+6 2 | 3+4 2 | 5+6 2 | 3+4 2 | 2+2 2 | 2+2 2 |

Continued on next column

**TABLE XXIV– continued**

| TQC | ICSD | | NoE | | SF | |
|---|---|---|---|---|---|---|
| | 3+4 2 | 5+6 2 | 3+4 2 | 5+6 2 | 3+3 2 | 2+2 2 |
| | 5+6 2 | 3+4 2 | 5+6 2 | 3+4 2 | 2+2 2 | 2+2 2 |
| | 3+4 2 | 5+6 2 | 3+4 2 | 3+4 2 | 3+3 2 | 3+3 2 |
| | 3+4 2 | 3+4 2 | 5+6 2 | 5+6 2 | 3+3 2 | 2+2 2 |
| | 5+6 2 | 5+6 2 | 3+4 2 | 5+6 2 | 2+2 2 | 2+2 2 |
| | 3+4 2 | 3+4 2 | 5+6 2 | 3+4 2 | 3+3 2 | 3+3 2 |
| | 5+6 2 | 5+6 2 | 3+4 2 | 5+6 2 | 2+2 2 | 3+3 2 |
| | 3+4 2 | 3+4 2 | 5+6 2 | 3+4 2 | 3+3 2 | 2+2 2 |
| | 3+4 2 | 5+6 2 | 3+4 2 | 5+6 2 | 2+2 2 | 3+3 2 |
| | 5+6 2 | 5+6 2 | 3+4 2 | 5+6 2 | 3+3 2 | 3+3 2 |
| | 5+6 2 | 5+6 2 | 5+6 2 | 3+4 2 | 2+2 2 | 3+3 2 |
| | 5+6 2 | 3+4 2 | 5+6 2 | 3+4 2 | 2+2 2 | 2+2 2 |
| | 3+4 2 | 3+4 2 | 3+4 2 | 3+4 2 | 3+3 2 | 3+3 2 |
| | 5+6 2 | 3+4 2 | 3+4 2 | 5+6 2 | 3+3 2 | 2+2 2 |
| | 3+4 2 | 3+4 2 | 5+6 2 | 5+6 2 | 2+2 2 | 3+3 2 |
| | 3+4 2 | 3+4 2 | 5+6 2 | 5+6 2 | 3+3 2 | 2+2 2 |
| | 3+4 2 | 5+6 2 | 3+4 2 | 5+6 2 | 2+2 2 | 2+2 2 |
| | 5+6 2 | 5+6 2 | 5+6 2 | 3+4 2 | 2+2 2 | 3+3 2 |
| | 3+4 2 | 5+6 2 | 3+4 2 | 3+4 2 | 3+3 2 | 2+2 2 |
| | 5+6 2 | 5+6 2 | 5+6 2 | 3+4 2 | 3+3 2 | 2+2 2 |
| | 5+6 2 | 5+6 2 | 3+4 2 | 5+6 2 | 2+2 2 | 3+3 2 |
| | 5+6 2 | 3+4 2 | 3+4 2 | 3+4 2 | 2+2 2 | 3+3 2 |
| 12.1.125 | 100827 | | 126 | | Ca2Fe7O11 | |
| | -A | -GM | -M | -Y | -L | -V |
| | 3+4 2 | 3+4 2 | 3+4 2 | 3+4 2 | 2+2 2 | 2+2 2 |
| | 5+6 2 | 3+4 2 | 5+6 2 | 5+6 2 | 3+3 2 | 3+3 2 |
| | 5+6 2 | 5+6 2 | 3+4 2 | 5+6 2 | 2+2 2 | 2+2 2 |
| | 3+4 2 | 5+6 2 | 5+6 2 | 5+6 2 | 2+2 2 | 2+2 2 |
| | 3+4 2 | 3+4 2 | 5+6 2 | 3+4 2 | 3+3 2 | 3+3 2 |
| | 3+4 2 | 3+4 2 | 3+4 2 | 3+4 2 | 2+2 2 | 2+2 2 |
| | 5+6 2 | 5+6 2 | 3+4 2 | 3+4 2 | 3+3 2 | 3+3 2 |
| | 3+4 2 | 5+6 2 | 5+6 2 | 3+4 2 | 2+2 2 | 3+3 2 |
| | 5+6 2 | 3+4 2 | 3+4 2 | 5+6 2 | 3+3 2 | 2+2 2 |
| | 3+4 2 | 5+6 2 | 3+4 2 | 5+6 2 | 2+2 2 | 3+3 2 |
| | 5+6 2 | 3+4 2 | 5+6 2 | 3+4 2 | 3+3 2 | 2+2 2 |
| | 5+6 2 | 5+6 2 | 5+6 2 | 5+6 2 | 3+3 2 | 3+3 2 |
| | 5+6 2 | 3+4 2 | 5+6 2 | 3+4 2 | 2+2 2 | 3+3 2 |
| | 3+4 2 | 3+4 2 | 3+4 2 | 3+4 2 | 3+3 2 | 3+3 2 |
| | 5+6 2 | 5+6 2 | 5+6 2 | 5+6 2 | 2+2 2 | 2+2 2 |
| | 3+4 2 | 5+6 2 | 5+6 2 | 3+4 2 | 3+3 2 | 2+2 2 |
| | 5+6 2 | 3+4 2 | 3+4 2 | 5+6 2 | 3+3 2 | 3+3 2 |
| | 3+4 2 | 3+4 2 | 5+6 2 | 5+6 2 | 2+2 2 | 2+2 2 |
| | 3+4 2 | 5+6 2 | 3+4 2 | 3+4 2 | 3+3 2 | 2+2 2 |
| | 5+6 2 | 5+6 2 | 3+4 2 | 5+6 2 | 2+2 2 | 3+3 2 |
| | 5+6 2 | 3+4 2 | 3+4 2 | 5+6 2 | 3+3 2 | 3+3 2 |
| | 3+4 2 | 3+4 2 | 5+6 2 | 3+4 2 | 2+2 2 | 2+2 2 |
| | 5+6 2 | 5+6 2 | 5+6 2 | 5+6 2 | 2+2 2 | 2+2 2 |
| | 5+6 2 | 5+6 2 | 3+4 2 | 5+6 2 | 3+3 2 | 3+3 2 |
| | 5+6 2 | 3+4 2 | 5+6 2 | 3+4 2 | 2+2 2 | 2+2 2 |
| | 3+4 2 | 3+4 2 | 5+6 2 | 3+4 2 | 3+3 2 | 3+3 2 |
| | 5+6 2 | 5+6 2 | 3+4 2 | 3+4 2 | 2+2 2 | 3+3 2 |
| | 3+4 2 | 5+6 2 | 5+6 2 | 5+6 2 | 3+3 2 | 3+3 2 |
| | 5+6 2 | 5+6 2 | 3+4 2 | 5+6 2 | 2+2 2 | 2+2 2 |
| | 3+4 2 | 3+4 2 | 5+6 2 | 5+6 2 | 2+2 2 | 3+3 2 |
| | 3+4 2 | 3+4 2 | 3+4 2 | 3+4 2 | 3+3 2 | 3+3 2 |
| | 3+4 2 | 3+4 2 | 5+6 2 | 5+6 2 | 3+3 2 | 2+2 2 |
| | 5+6 2 | 3+4 2 | 3+4 2 | 3+4 2 | 2+2 2 | 3+3 2 |
| | 5+6 2 | 5+6 2 | 5+6 2 | 5+6 2 | 2+2 2 | 2+2 2 |
| | 3+4 2 | 3+4 2 | 5+6 2 | 3+4 2 | 3+3 2 | 3+3 2 |
| | 5+6 2 | 3+4 2 | 3+4 2 | 5+6 2 | 2+2 2 | 2+2 2 |
| | 5+6 2 | 5+6 2 | 3+4 2 | 3+4 2 | 3+3 2 | 2+2 2 |

Continued on next column

**TABLE XXIV– continued**

| TQC | ICSD | | NoE | | SF | |
|---|---|---|---|---|---|---|
| | 3+4 2 | 5+6 2 | 5+6 2 | 5+6 2 | 3+3 2 | 3+3 2 |
| | 5+6 2 | 3+4 2 | 3+4 2 | 3+4 2 | 3+3 2 | 3+3 2 |
| | 3+4 2 | 5+6 2 | 5+6 2 | 5+6 2 | 2+2 2 | 2+2 2 |
| | 5+6 2 | 5+6 2 | 3+4 2 | 3+4 2 | 3+3 2 | 3+3 2 |
| | 5+6 2 | 5+6 2 | 5+6 2 | 5+6 2 | 3+3 2 | 3+3 2 |
| | 3+4 2 | 5+6 2 | 5+6 2 | 3+4 2 | 2+2 2 | 2+2 2 |
| | 3+4 2 | 5+6 2 | 3+4 2 | 5+6 2 | 3+3 2 | 2+2 2 |
| | 5+6 2 | 3+4 2 | 3+4 2 | 3+4 2 | 2+2 2 | 2+2 2 |
| | 3+4 2 | 3+4 2 | 5+6 2 | 3+4 2 | 2+2 2 | 3+3 2 |
| | 5+6 2 | 5+6 2 | 5+6 2 | 5+6 2 | 3+3 2 | 3+3 2 |
| | 3+4 2 | 5+6 2 | 3+4 2 | 5+6 2 | 2+2 2 | 2+2 2 |
| | 3+4 2 | 3+4 2 | 3+4 2 | 3+4 2 | 3+3 2 | 3+3 2 |
| | 5+6 2 | 3+4 2 | 5+6 2 | 5+6 2 | 3+3 2 | 2+2 2 |
| | 3+4 2 | 3+4 2 | 5+6 2 | 3+4 2 | 2+2 2 | 3+3 2 |
| | 5+6 2 | 5+6 2 | 3+4 2 | 5+6 2 | 2+2 2 | 2+2 2 |
| | 3+4 2 | 3+4 2 | 5+6 2 | 3+4 2 | 3+3 2 | 3+3 2 |
| | 5+6 2 | 3+4 2 | 5+6 2 | 5+6 2 | 2+2 2 | 3+3 2 |
| | 5+6 2 | 5+6 2 | 5+6 2 | 3+4 2 | 3+3 2 | 2+2 2 |
| | 3+4 2 | 3+4 2 | 3+4 2 | 3+4 2 | 2+2 2 | 3+3 2 |
| | 5+6 2 | 5+6 2 | 5+6 2 | 5+6 2 | 2+2 2 | 3+3 2 |
| | 5+6 2 | 3+4 2 | 3+4 2 | 3+4 2 | 3+3 2 | 2+2 2 |
| | 3+4 2 | 5+6 2 | 3+4 2 | 3+4 2 | 2+2 2 | 2+2 2 |
| | 5+6 2 | 3+4 2 | 5+6 2 | 5+6 2 | 3+3 2 | 3+3 2 |
| | 5+6 2 | 3+4 2 | 5+6 2 | 3+4 2 | 2+2 2 | 3+3 2 |
| | 3+4 2 | 3+4 2 | 3+4 2 | 5+6 2 | 3+3 2 | 2+2 2 |
| | 3+4 2 | 5+6 2 | 5+6 2 | 3+4 2 | 2+2 2 | 3+3 2 |
| 12.1.126 | 20762 | | 34 | | C6H5Cl2Tl1 | |
| | -A | -GM | -M | -Y | -L | -V |
| | 3+4 2 | 3+4 2 | 3+4 2 | 3+4 2 | 3+3 2 | 3+3 2 |
| | 5+6 2 | 5+6 2 | 5+6 2 | 5+6 2 | 2+2 2 | 2+2 2 |
| | 3+4 2 | 3+4 2 | 3+4 2 | 3+4 2 | 2+2 2 | 2+2 2 |
| | 5+6 2 | 5+6 2 | 5+6 2 | 5+6 2 | 3+3 2 | 3+3 2 |
| | 3+4 2 | 3+4 2 | 3+4 2 | 3+4 2 | 3+3 2 | 3+3 2 |
| | 5+6 2 | 5+6 2 | 5+6 2 | 5+6 2 | 2+2 2 | 2+2 2 |
| | 5+6 2 | 5+6 2 | 5+6 2 | 5+6 2 | 3+3 2 | 3+3 2 |
| | 3+4 2 | 3+4 2 | 3+4 2 | 3+4 2 | 2+2 2 | 2+2 2 |
| | 3+4 2 | 3+4 2 | 3+4 2 | 3+4 2 | 2+2 2 | 3+3 2 |
| | 5+6 2 | 5+6 2 | 5+6 2 | 5+6 2 | 3+3 2 | 2+2 2 |
| | 3+4 2 | 3+4 2 | 3+4 2 | 3+4 2 | 3+3 2 | 3+3 2 |
| | 3+4 2 | 3+4 2 | 3+4 2 | 3+4 2 | 2+2 2 | 2+2 2 |
| | 5+6 2 | 5+6 2 | 5+6 2 | 5+6 2 | 2+2 2 | 2+2 2 |
| | 5+6 2 | 5+6 2 | 5+6 2 | 5+6 2 | 3+3 2 | 2+2 2 |
| | 5+6 2 | 3+4 2 | 5+6 2 | 3+4 2 | 2+2 2 | 3+3 2 |
| | 3+4 2 | 3+4 2 | 3+4 2 | 3+4 2 | 3+3 2 | 3+3 2 |
| | 3+4 2 | 5+6 2 | 3+4 2 | 5+6 2 | 2+2 2 | 3+3 2 |
| 12.1.127 | 63503 | | 98 | | C5Nb6 | |
| | -A | -GM | -M | -Y | -L | -V |
| | 3+4 2 | 3+4 2 | 3+4 2 | 3+4 2 | 2+2 2 | 3+3 2 |
| | 5+6 2 | 5+6 2 | 5+6 2 | 5+6 2 | 3+3 2 | 2+2 2 |
| | 3+4 2 | 3+4 2 | 3+4 2 | 5+6 2 | 3+3 2 | 2+2 2 |
| | 5+6 2 | 5+6 2 | 5+6 2 | 5+6 2 | 2+2 2 | 3+3 2 |
| | 5+6 2 | 5+6 2 | 3+4 2 | 3+4 2 | 2+2 2 | 2+2 2 |
| | 3+4 2 | 3+4 2 | 5+6 2 | 3+4 2 | 3+3 2 | 3+3 2 |
| | 3+4 2 | 3+4 2 | 3+4 2 | 3+4 2 | 2+2 2 | 2+2 2 |
| | 5+6 2 | 5+6 2 | 5+6 2 | 5+6 2 | 3+3 2 | 3+3 2 |
| | 5+6 2 | 3+4 2 | 3+4 2 | 5+6 2 | 2+2 2 | 2+2 2 |
| | 3+4 2 | 5+6 2 | 5+6 2 | 3+4 2 | 3+3 2 | 3+3 2 |
| | 3+4 2 | 3+4 2 | 5+6 2 | 5+6 2 | 3+3 2 | 3+3 2 |
| | 5+6 2 | 5+6 2 | 3+4 2 | 3+4 2 | 2+2 2 | 2+2 2 |
| | 3+4 2 | 3+4 2 | 3+4 2 | 3+4 2 | 2+2 2 | 3+3 2 |
| | 5+6 2 | 5+6 2 | 5+6 2 | 5+6 2 | 3+3 2 | 2+2 2 |

Continued on next column

**TABLE XXIV– continued**

| TQC | ICSD | | NoE | | SF | |
|---|---|---|---|---|---|---|
| | 5+6 2 | 3+4 2 | 5+6 2 | 3+4 2 | 2+2 2 | 2+2 2 |
| | 3+4 2 | 5+6 2 | 3+4 2 | 5+6 2 | 3+3 2 | 3+3 2 |
| | 5+6 2 | 3+4 2 | 5+6 2 | 5+6 2 | 3+3 2 | 3+3 2 |
| | 5+6 2 | 5+6 2 | 3+4 2 | 3+4 2 | 2+2 2 | 2+2 2 |
| | 3+4 2 | 5+6 2 | 3+4 2 | 5+6 2 | 2+2 2 | 2+2 2 |
| | 3+4 2 | 3+4 2 | 5+6 2 | 3+4 2 | 2+2 2 | 3+3 2 |
| | 5+6 2 | 3+4 2 | 3+4 2 | 5+6 2 | 3+3 2 | 2+2 2 |
| | 3+4 2 | 3+4 2 | 3+4 2 | 3+4 2 | 3+3 2 | 3+3 2 |
| | 3+4 2 | 5+6 2 | 5+6 2 | 5+6 2 | 2+2 2 | 3+3 2 |
| | 5+6 2 | 5+6 2 | 5+6 2 | 3+4 2 | 3+3 2 | 2+2 2 |
| | 5+6 2 | 3+4 2 | 3+4 2 | 5+6 2 | 3+3 2 | 2+2 2 |
| | 3+4 2 | 3+4 2 | 5+6 2 | 3+4 2 | 3+3 2 | 3+3 2 |
| | 5+6 2 | 5+6 2 | 5+6 2 | 3+4 2 | 2+2 2 | 2+2 2 |
| | 3+4 2 | 5+6 2 | 5+6 2 | 5+6 2 | 2+2 2 | 3+3 2 |
| | 5+6 2 | 3+4 2 | 3+4 2 | 3+4 2 | 3+3 2 | 2+2 2 |
| | 3+4 2 | 5+6 2 | 3+4 2 | 5+6 2 | 2+2 2 | 3+3 2 |
| | 3+4 2 | 3+4 2 | 3+4 2 | 3+4 2 | 3+3 2 | 3+3 2 |
| | 5+6 2 | 5+6 2 | 5+6 2 | 3+4 2 | 3+3 2 | 2+2 2 |
| | 5+6 2 | 3+4 2 | 3+4 2 | 5+6 2 | 2+2 2 | 3+3 2 |
| | 3+4 2 | 5+6 2 | 5+6 2 | 5+6 2 | 2+2 2 | 2+2 2 |
| | 3+4 2 | 5+6 2 | 5+6 2 | 5+6 2 | 2+2 2 | 3+3 2 |
| | 5+6 2 | 5+6 2 | 3+4 2 | 5+6 2 | 2+2 2 | 2+2 2 |
| | 5+6 2 | 3+4 2 | 3+4 2 | 5+6 2 | 3+3 2 | 3+3 2 |
| | 3+4 2 | 3+4 2 | 5+6 2 | 3+4 2 | 3+3 2 | 2+2 2 |
| | 3+4 2 | 5+6 2 | 3+4 2 | 5+6 2 | 2+2 2 | 3+3 2 |
| | 3+4 2 | 3+4 2 | 5+6 2 | 3+4 2 | 3+3 2 | 3+3 2 |
| | 5+6 2 | 3+4 2 | 5+6 2 | 3+4 2 | 2+2 2 | 2+2 2 |
| | 5+6 2 | 3+4 2 | 3+4 2 | 3+4 2 | 3+3 2 | 2+2 2 |
| | 3+4 2 | 5+6 2 | 3+4 2 | 3+4 2 | 2+2 2 | 3+3 2 |
| | 3+4 2 | 5+6 2 | 3+4 2 | 5+6 2 | 3+3 2 | 3+3 2 |
| | 3+4 2 | 5+6 2 | 3+4 2 | 5+6 2 | 3+3 2 | 3+3 2 |
| | 5+6 2 | 3+4 2 | 5+6 2 | 3+4 2 | 2+2 2 | 2+2 2 |
| | 5+6 2 | 3+4 2 | 5+6 2 | 5+6 2 | 2+2 2 | 2+2 2 |
| | 5+6 2 | 3+4 2 | 3+4 2 | 3+4 2 | 3+3 2 | 2+2 2 |
| | 3+4 2 | 3+4 2 | 5+6 2 | 5+6 2 | 3+3 2 | 3+3 2 |
| 12.1.128 | 39297 | | 98 | | C5Nb6 | |
| | -A | -GM | -M | -Y | -L | -V |
| | 5+6 2 | 3+4 2 | 5+6 2 | 3+4 2 | 2+2 2 | 2+2 2 |
| | 3+4 2 | 5+6 2 | 3+4 2 | 5+6 2 | 3+3 2 | 3+3 2 |
| | 5+6 2 | 3+4 2 | 3+4 2 | 5+6 2 | 3+3 2 | 3+3 2 |
| | 3+4 2 | 5+6 2 | 5+6 2 | 5+6 2 | 2+2 2 | 2+2 2 |
| | 3+4 2 | 3+4 2 | 5+6 2 | 3+4 2 | 2+2 2 | 3+3 2 |
| | 5+6 2 | 5+6 2 | 3+4 2 | 3+4 2 | 3+3 2 | 2+2 2 |
| | 3+4 2 | 3+4 2 | 5+6 2 | 3+4 2 | 2+2 2 | 3+3 2 |
| | 5+6 2 | 5+6 2 | 3+4 2 | 5+6 2 | 3+3 2 | 2+2 2 |
| | 3+4 2 | 3+4 2 | 5+6 2 | 5+6 2 | 2+2 2 | 2+2 2 |
| | 5+6 2 | 5+6 2 | 3+4 2 | 3+4 2 | 3+3 2 | 3+3 2 |
| | 5+6 2 | 3+4 2 | 3+4 2 | 5+6 2 | 3+3 2 | 2+2 2 |
| | 3+4 2 | 5+6 2 | 5+6 2 | 3+4 2 | 2+2 2 | 3+3 2 |
| | 3+4 2 | 3+4 2 | 5+6 2 | 3+4 2 | 2+2 2 | 2+2 2 |
| | 5+6 2 | 5+6 2 | 3+4 2 | 5+6 2 | 3+3 2 | 3+3 2 |
| | 5+6 2 | 3+4 2 | 5+6 2 | 3+4 2 | 2+2 2 | 3+3 2 |
| | 3+4 2 | 5+6 2 | 3+4 2 | 5+6 2 | 3+3 2 | 2+2 2 |
| | 3+4 2 | 3+4 2 | 5+6 2 | 5+6 2 | 3+3 2 | 2+2 2 |
| | 5+6 2 | 3+4 2 | 3+4 2 | 3+4 2 | 2+2 2 | 3+3 2 |
| | 3+4 2 | 5+6 2 | 3+4 2 | 5+6 2 | 2+2 2 | 3+3 2 |
| | 5+6 2 | 3+4 2 | 5+6 2 | 3+4 2 | 2+2 2 | 2+2 2 |
| | 5+6 2 | 5+6 2 | 3+4 2 | 5+6 2 | 3+3 2 | 3+3 2 |
| | 3+4 2 | 3+4 2 | 5+6 2 | 5+6 2 | 3+3 2 | 2+2 2 |
| | 5+6 2 | 5+6 2 | 3+4 2 | 3+4 2 | 2+2 2 | 2+2 2 |
| | 3+4 2 | 5+6 2 | 5+6 2 | 3+4 2 | 3+3 2 | 3+3 2 |

Continued on next column

**TABLE XXIV– continued**

| TQC | ICSD | | NoE | | SF | |
|---|---|---|---|---|---|---|
| | 5+6 2 | 3+4 2 | 5+6 2 | 3+4 2 | 3+3 2 | 2+2 2 |
| | 3+4 2 | 3+4 2 | 3+4 2 | 5+6 2 | 2+2 2 | 3+3 2 |
| | 5+6 2 | 5+6 2 | 5+6 2 | 3+4 2 | 3+3 2 | 2+2 2 |
| | 3+4 2 | 5+6 2 | 5+6 2 | 5+6 2 | 2+2 2 | 2+2 2 |
| | 5+6 2 | 3+4 2 | 3+4 2 | 3+4 2 | 3+3 2 | 3+3 2 |
| | 3+4 2 | 5+6 2 | 3+4 2 | 5+6 2 | 2+2 2 | 3+3 2 |
| | 3+4 2 | 3+4 2 | 3+4 2 | 3+4 2 | 3+3 2 | 3+3 2 |
| | 3+4 2 | 3+4 2 | 3+4 2 | 3+4 2 | 3+3 2 | 3+3 2 |
| | 5+6 2 | 5+6 2 | 5+6 2 | 5+6 2 | 2+2 2 | 2+2 2 |
| | 5+6 2 | 5+6 2 | 5+6 2 | 5+6 2 | 2+2 2 | 2+2 2 |
| | 5+6 2 | 5+6 2 | 5+6 2 | 5+6 2 | 2+2 2 | 2+2 2 |
| | 3+4 2 | 5+6 2 | 3+4 2 | 5+6 2 | 2+2 2 | 3+3 2 |
| | 3+4 2 | 3+4 2 | 3+4 2 | 3+4 2 | 3+3 2 | 3+3 2 |
| | 5+6 2 | 3+4 2 | 5+6 2 | 3+4 2 | 3+3 2 | 2+2 2 |
| | 3+4 2 | 5+6 2 | 3+4 2 | 5+6 2 | 2+2 2 | 3+3 2 |
| | 3+4 2 | 3+4 2 | 5+6 2 | 3+4 2 | 3+3 2 | 2+2 2 |
| | 5+6 2 | 3+4 2 | 3+4 2 | 3+4 2 | 3+3 2 | 3+3 2 |
| | 3+4 2 | 5+6 2 | 3+4 2 | 3+4 2 | 2+2 2 | 2+2 2 |
| | 5+6 2 | 5+6 2 | 5+6 2 | 5+6 2 | 3+3 2 | 3+3 2 |
| | 3+4 2 | 5+6 2 | 3+4 2 | 5+6 2 | 2+2 2 | 3+3 2 |
| | 3+4 2 | 3+4 2 | 5+6 2 | 5+6 2 | 2+2 2 | 3+3 2 |
| | 5+6 2 | 3+4 2 | 3+4 2 | 3+4 2 | 3+3 2 | 2+2 2 |
| | 5+6 2 | 3+4 2 | 5+6 2 | 5+6 2 | 2+2 2 | 2+2 2 |
| | 5+6 2 | 3+4 2 | 5+6 2 | 5+6 2 | 2+2 2 | 2+2 2 |
| | 5+6 2 | 5+6 2 | 3+4 2 | 3+4 2 | 3+3 2 | 3+3 2 |
| 12.1.129 | 20695 | | 98 | | C5Nb6 | |
| | -A | -GM | -M | -Y | -L | -V |
| | 3+4 2 | 3+4 2 | 3+4 2 | 3+4 2 | 2+2 2 | 2+2 2 |
| | 5+6 2 | 5+6 2 | 5+6 2 | 5+6 2 | 3+3 2 | 3+3 2 |
| | 3+4 2 | 3+4 2 | 3+4 2 | 5+6 2 | 3+3 2 | 2+2 2 |
| | 5+6 2 | 5+6 2 | 5+6 2 | 5+6 2 | 2+2 2 | 3+3 2 |
| | 3+4 2 | 3+4 2 | 3+4 2 | 3+4 2 | 2+2 2 | 2+2 2 |
| | 5+6 2 | 5+6 2 | 5+6 2 | 3+4 2 | 3+3 2 | 3+3 2 |
| | 3+4 2 | 3+4 2 | 3+4 2 | 3+4 2 | 2+2 2 | 3+3 2 |
| | 5+6 2 | 5+6 2 | 5+6 2 | 5+6 2 | 3+3 2 | 2+2 2 |
| | 3+4 2 | 3+4 2 | 5+6 2 | 5+6 2 | 2+2 2 | 2+2 2 |
| | 5+6 2 | 5+6 2 | 3+4 2 | 3+4 2 | 3+3 2 | 3+3 2 |
| | 5+6 2 | 3+4 2 | 5+6 2 | 5+6 2 | 3+3 2 | 2+2 2 |
| | 3+4 2 | 5+6 2 | 3+4 2 | 3+4 2 | 2+2 2 | 3+3 2 |
| | 3+4 2 | 3+4 2 | 3+4 2 | 3+4 2 | 2+2 2 | 3+3 2 |
| | 5+6 2 | 5+6 2 | 5+6 2 | 5+6 2 | 3+3 2 | 2+2 2 |
| | 3+4 2 | 3+4 2 | 5+6 2 | 3+4 2 | 2+2 2 | 3+3 2 |
| | 5+6 2 | 5+6 2 | 3+4 2 | 5+6 2 | 3+3 2 | 2+2 2 |
| | 3+4 2 | 3+4 2 | 5+6 2 | 5+6 2 | 3+3 2 | 2+2 2 |
| | 5+6 2 | 5+6 2 | 3+4 2 | 3+4 2 | 2+2 2 | 3+3 2 |
| | 5+6 2 | 3+4 2 | 5+6 2 | 5+6 2 | 2+2 2 | 2+2 2 |
| | 3+4 2 | 5+6 2 | 3+4 2 | 3+4 2 | 2+2 2 | 3+3 2 |
| | 5+6 2 | 3+4 2 | 3+4 2 | 5+6 2 | 3+3 2 | 3+3 2 |
| | 3+4 2 | 3+4 2 | 5+6 2 | 5+6 2 | 3+3 2 | 2+2 2 |
| | 3+4 2 | 5+6 2 | 3+4 2 | 3+4 2 | 2+2 2 | 3+3 2 |
| | 5+6 2 | 5+6 2 | 5+6 2 | 3+4 2 | 3+3 2 | 2+2 2 |
| | 5+6 2 | 3+4 2 | 3+4 2 | 3+4 2 | 3+3 2 | 2+2 2 |
| | 3+4 2 | 3+4 2 | 5+6 2 | 5+6 2 | 2+2 2 | 3+3 2 |
| | 3+4 2 | 5+6 2 | 5+6 2 | 3+4 2 | 3+3 2 | 2+2 2 |
| | 5+6 2 | 5+6 2 | 3+4 2 | 5+6 2 | 2+2 2 | 3+3 2 |
| | 5+6 2 | 3+4 2 | 5+6 2 | 3+4 2 | 3+3 2 | 2+2 2 |
| | 3+4 2 | 5+6 2 | 3+4 2 | 5+6 2 | 2+2 2 | 3+3 2 |
| | 3+4 2 | 3+4 2 | 3+4 2 | 3+4 2 | 3+3 2 | 3+3 2 |
| | 5+6 2 | 3+4 2 | 3+4 2 | 5+6 2 | 3+3 2 | 3+3 2 |
| | 5+6 2 | 5+6 2 | 5+6 2 | 3+4 2 | 2+2 2 | 2+2 2 |
| | 3+4 2 | 5+6 2 | 5+6 2 | 5+6 2 | 2+2 2 | 2+2 2 |

Continued on next column

**TABLE XXIV– continued**

| TQC | ICSD | | NoE | | SF | |
|---|---|---|---|---|---|---|
| | 5+6 2 | 5+6 2 | 5+6 2 | 5+6 2 | 2+2 2 | 2+2 2 |
| | 3+4 2 | 5+6 2 | 3+4 2 | 5+6 2 | 2+2 2 | 3+3 2 |
| | 3+4 2 | 3+4 2 | 3+4 2 | 3+4 2 | 3+3 2 | 3+3 2 |
| | 5+6 2 | 3+4 2 | 5+6 2 | 3+4 2 | 3+3 2 | 2+2 2 |
| | 3+4 2 | 5+6 2 | 5+6 2 | 3+4 2 | 3+3 2 | 3+3 2 |
| | 3+4 2 | 3+4 2 | 3+4 2 | 5+6 2 | 2+2 2 | 2+2 2 |
| | 5+6 2 | 3+4 2 | 3+4 2 | 3+4 2 | 3+3 2 | 3+3 2 |
| | 5+6 2 | 5+6 2 | 3+4 2 | 5+6 2 | 2+2 2 | 2+2 2 |
| | 3+4 2 | 5+6 2 | 5+6 2 | 3+4 2 | 3+3 2 | 3+3 2 |
| | 3+4 2 | 3+4 2 | 3+4 2 | 5+6 2 | 2+2 2 | 3+3 2 |
| | 3+4 2 | 5+6 2 | 5+6 2 | 5+6 2 | 2+2 2 | 3+3 2 |
| | 5+6 2 | 3+4 2 | 3+4 2 | 3+4 2 | 3+3 2 | 2+2 2 |
| | 5+6 2 | 3+4 2 | 5+6 2 | 5+6 2 | 2+2 2 | 2+2 2 |
| | 5+6 2 | 3+4 2 | 5+6 2 | 5+6 2 | 2+2 2 | 2+2 2 |
| | 5+6 2 | 5+6 2 | 3+4 2 | 3+4 2 | 3+3 2 | 3+3 2 |
| 12.1.130 | 167353 | | 98 | | C5Nb6 | |
| | -A | -GM | -M | -Y | -L | -V |
| | 3+4 2 | 3+4 2 | 3+4 2 | 3+4 2 | 2+2 2 | 3+3 2 |
| | 5+6 2 | 5+6 2 | 5+6 2 | 5+6 2 | 3+3 2 | 2+2 2 |
| | 3+4 2 | 3+4 2 | 3+4 2 | 5+6 2 | 3+3 2 | 2+2 2 |
| | 5+6 2 | 5+6 2 | 5+6 2 | 5+6 2 | 2+2 2 | 3+3 2 |
| | 5+6 2 | 5+6 2 | 3+4 2 | 3+4 2 | 2+2 2 | 2+2 2 |
| | 3+4 2 | 3+4 2 | 5+6 2 | 3+4 2 | 3+3 2 | 3+3 2 |
| | 3+4 2 | 3+4 2 | 3+4 2 | 3+4 2 | 2+2 2 | 2+2 2 |
| | 5+6 2 | 5+6 2 | 5+6 2 | 5+6 2 | 3+3 2 | 3+3 2 |
| | 5+6 2 | 3+4 2 | 3+4 2 | 5+6 2 | 2+2 2 | 2+2 2 |
| | 3+4 2 | 5+6 2 | 5+6 2 | 3+4 2 | 3+3 2 | 3+3 2 |
| | 3+4 2 | 3+4 2 | 5+6 2 | 5+6 2 | 3+3 2 | 3+3 2 |
| | 5+6 2 | 5+6 2 | 3+4 2 | 3+4 2 | 2+2 2 | 2+2 2 |
| | 3+4 2 | 3+4 2 | 3+4 2 | 3+4 2 | 2+2 2 | 3+3 2 |
| | 5+6 2 | 5+6 2 | 5+6 2 | 5+6 2 | 3+3 2 | 2+2 2 |
| | 5+6 2 | 3+4 2 | 5+6 2 | 3+4 2 | 2+2 2 | 2+2 2 |
| | 3+4 2 | 5+6 2 | 3+4 2 | 5+6 2 | 3+3 2 | 3+3 2 |
| | 5+6 2 | 3+4 2 | 5+6 2 | 5+6 2 | 3+3 2 | 3+3 2 |
| | 5+6 2 | 5+6 2 | 3+4 2 | 3+4 2 | 2+2 2 | 2+2 2 |
| | 3+4 2 | 5+6 2 | 3+4 2 | 5+6 2 | 2+2 2 | 2+2 2 |
| | 3+4 2 | 3+4 2 | 5+6 2 | 3+4 2 | 2+2 2 | 3+3 2 |
| | 5+6 2 | 3+4 2 | 3+4 2 | 5+6 2 | 3+3 2 | 2+2 2 |
| | 3+4 2 | 3+4 2 | 3+4 2 | 3+4 2 | 3+3 2 | 3+3 2 |
| | 3+4 2 | 5+6 2 | 5+6 2 | 5+6 2 | 2+2 2 | 3+3 2 |
| | 5+6 2 | 5+6 2 | 5+6 2 | 3+4 2 | 3+3 2 | 2+2 2 |
| | 5+6 2 | 3+4 2 | 3+4 2 | 5+6 2 | 3+3 2 | 2+2 2 |
| | 3+4 2 | 3+4 2 | 5+6 2 | 3+4 2 | 3+3 2 | 3+3 2 |
| | 5+6 2 | 5+6 2 | 5+6 2 | 3+4 2 | 2+2 2 | 2+2 2 |
| | 3+4 2 | 5+6 2 | 5+6 2 | 5+6 2 | 2+2 2 | 3+3 2 |
| | 5+6 2 | 3+4 2 | 3+4 2 | 3+4 2 | 3+3 2 | 2+2 2 |
| | 3+4 2 | 5+6 2 | 3+4 2 | 5+6 2 | 2+2 2 | 3+3 2 |
| | 3+4 2 | 3+4 2 | 3+4 2 | 3+4 2 | 3+3 2 | 3+3 2 |
| | 5+6 2 | 5+6 2 | 5+6 2 | 3+4 2 | 3+3 2 | 2+2 2 |
| | 5+6 2 | 3+4 2 | 3+4 2 | 5+6 2 | 2+2 2 | 3+3 2 |
| | 3+4 2 | 5+6 2 | 5+6 2 | 5+6 2 | 2+2 2 | 2+2 2 |
| | 3+4 2 | 5+6 2 | 5+6 2 | 5+6 2 | 2+2 2 | 3+3 2 |
| | 5+6 2 | 5+6 2 | 3+4 2 | 5+6 2 | 2+2 2 | 2+2 2 |
| | 5+6 2 | 3+4 2 | 3+4 2 | 5+6 2 | 3+3 2 | 3+3 2 |
| | 3+4 2 | 3+4 2 | 5+6 2 | 3+4 2 | 3+3 2 | 2+2 2 |
| | 3+4 2 | 5+6 2 | 3+4 2 | 5+6 2 | 2+2 2 | 3+3 2 |
| | 3+4 2 | 3+4 2 | 5+6 2 | 3+4 2 | 3+3 2 | 3+3 2 |
| | 5+6 2 | 3+4 2 | 5+6 2 | 3+4 2 | 2+2 2 | 2+2 2 |
| | 5+6 2 | 3+4 2 | 3+4 2 | 3+4 2 | 3+3 2 | 2+2 2 |
| | 3+4 2 | 5+6 2 | 3+4 2 | 3+4 2 | 2+2 2 | 3+3 2 |
| | 3+4 2 | 5+6 2 | 3+4 2 | 5+6 2 | 3+3 2 | 3+3 2 |

Continued on next column

**TABLE XXIV– continued**

| TQC | ICSD | NoE | SF | | | |
|---|---|---|---|---|---|---|
| | 3+4 2 | 5+6 2 | 3+4 2 | 5+6 2 | 3+3 2 | 3+3 2 |
| | 5+6 2 | 3+4 2 | 5+6 2 | 3+4 2 | 2+2 2 | 2+2 2 |
| | 5+6 2 | 3+4 2 | 5+6 2 | 5+6 2 | 2+2 2 | 2+2 2 |
| | 5+6 2 | 3+4 2 | 3+4 2 | 3+4 2 | 3+3 2 | 2+2 2 |
| | 3+4 2 | 3+4 2 | 5+6 2 | 5+6 2 | 3+3 2 | 3+3 2 |
| 12.1.131 | 619030 | 52 | C3U2W2 | | | |
| | -A | -GM | -M | -Y | -L | -V |
| | 3+4 2 | 3+4 2 | 5+6 2 | 5+6 2 | 3+3 2 | 2+2 2 |
| | 5+6 2 | 5+6 2 | 3+4 2 | 3+4 2 | 2+2 2 | 3+3 2 |
| | 3+4 2 | 3+4 2 | 5+6 2 | 5+6 2 | 3+3 2 | 2+2 2 |
| | 5+6 2 | 5+6 2 | 3+4 2 | 3+4 2 | 2+2 2 | 3+3 2 |
| | 5+6 2 | 3+4 2 | 5+6 2 | 3+4 2 | 2+2 2 | 3+3 2 |
| | 3+4 2 | 5+6 2 | 3+4 2 | 5+6 2 | 3+3 2 | 2+2 2 |
| | 3+4 2 | 3+4 2 | 5+6 2 | 5+6 2 | 3+3 2 | 2+2 2 |
| | 5+6 2 | 5+6 2 | 3+4 2 | 3+4 2 | 2+2 2 | 3+3 2 |
| | 5+6 2 | 3+4 2 | 3+4 2 | 5+6 2 | 3+3 2 | 2+2 2 |
| | 3+4 2 | 5+6 2 | 5+6 2 | 3+4 2 | 2+2 2 | 3+3 2 |
| | 5+6 2 | 3+4 2 | 5+6 2 | 3+4 2 | 2+2 2 | 3+3 2 |
| | 3+4 2 | 5+6 2 | 5+6 2 | 3+4 2 | 3+3 2 | 2+2 2 |
| | 5+6 2 | 5+6 2 | 3+4 2 | 5+6 2 | 2+2 2 | 3+3 2 |
| | 3+4 2 | 3+4 2 | 3+4 2 | 5+6 2 | 3+3 2 | 2+2 2 |
| | 3+4 2 | 5+6 2 | 3+4 2 | 3+4 2 | 3+3 2 | 3+3 2 |
| | 5+6 2 | 3+4 2 | 5+6 2 | 5+6 2 | 2+2 2 | 2+2 2 |
| | 3+4 2 | 5+6 2 | 5+6 2 | 3+4 2 | 3+3 2 | 3+3 2 |
| | 5+6 2 | 3+4 2 | 3+4 2 | 5+6 2 | 2+2 2 | 2+2 2 |
| | 3+4 2 | 5+6 2 | 3+4 2 | 5+6 2 | 3+3 2 | 2+2 2 |
| | 3+4 2 | 5+6 2 | 3+4 2 | 5+6 2 | 3+3 2 | 2+2 2 |
| | 5+6 2 | 3+4 2 | 3+4 2 | 3+4 2 | 2+2 2 | 3+3 2 |
| | 3+4 2 | 3+4 2 | 5+6 2 | 3+4 2 | 2+2 2 | 3+3 2 |
| | 5+6 2 | 5+6 2 | 3+4 2 | 5+6 2 | 3+3 2 | 2+2 2 |
| | 5+6 2 | 3+4 2 | 3+4 2 | 5+6 2 | 3+3 2 | 3+3 2 |
| | 3+4 2 | 5+6 2 | 5+6 2 | 3+4 2 | 2+2 2 | 2+2 2 |
| | 5+6 2 | 3+4 2 | 5+6 2 | 5+6 2 | 3+3 2 | 3+3 2 |
| 12.1.132 | 618707 | 54 | C3Re2U2 | | | |
| | -A | -GM | -M | -Y | -L | -V |
| | 3+4 2 | 3+4 2 | 5+6 2 | 5+6 2 | 3+3 2 | 2+2 2 |
| | 5+6 2 | 5+6 2 | 3+4 2 | 3+4 2 | 2+2 2 | 3+3 2 |
| | 3+4 2 | 3+4 2 | 5+6 2 | 5+6 2 | 3+3 2 | 2+2 2 |
| | 5+6 2 | 5+6 2 | 3+4 2 | 3+4 2 | 2+2 2 | 3+3 2 |
| | 5+6 2 | 3+4 2 | 5+6 2 | 3+4 2 | 2+2 2 | 3+3 2 |
| | 3+4 2 | 5+6 2 | 3+4 2 | 5+6 2 | 3+3 2 | 2+2 2 |
| | 3+4 2 | 3+4 2 | 5+6 2 | 5+6 2 | 3+3 2 | 2+2 2 |
| | 5+6 2 | 5+6 2 | 3+4 2 | 3+4 2 | 2+2 2 | 3+3 2 |
| | 5+6 2 | 3+4 2 | 3+4 2 | 5+6 2 | 3+3 2 | 2+2 2 |
| | 3+4 2 | 5+6 2 | 5+6 2 | 3+4 2 | 2+2 2 | 3+3 2 |
| | 5+6 2 | 3+4 2 | 5+6 2 | 3+4 2 | 2+2 2 | 3+3 2 |
| | 3+4 2 | 5+6 2 | 5+6 2 | 3+4 2 | 3+3 2 | 2+2 2 |
| | 5+6 2 | 5+6 2 | 3+4 2 | 5+6 2 | 2+2 2 | 3+3 2 |
| | 3+4 2 | 5+6 2 | 3+4 2 | 5+6 2 | 3+3 2 | 2+2 2 |
| | 5+6 2 | 3+4 2 | 3+4 2 | 3+4 2 | 2+2 2 | 3+3 2 |
| | 3+4 2 | 3+4 2 | 5+6 2 | 5+6 2 | 3+3 2 | 2+2 2 |
| | 3+4 2 | 5+6 2 | 5+6 2 | 5+6 2 | 3+3 2 | 3+3 2 |
| | 5+6 2 | 3+4 2 | 3+4 2 | 3+4 2 | 2+2 2 | 2+2 2 |
| | 3+4 2 | 5+6 2 | 3+4 2 | 5+6 2 | 3+3 2 | 2+2 2 |
| | 3+4 2 | 5+6 2 | 3+4 2 | 5+6 2 | 3+3 2 | 2+2 2 |
| | 5+6 2 | 3+4 2 | 3+4 2 | 3+4 2 | 2+2 2 | 3+3 2 |
| | 5+6 2 | 3+4 2 | 5+6 2 | 5+6 2 | 2+2 2 | 3+3 2 |
| | 5+6 2 | 3+4 2 | 3+4 2 | 3+4 2 | 3+3 2 | 2+2 2 |
| | 3+4 2 | 5+6 2 | 3+4 2 | 5+6 2 | 3+3 2 | 3+3 2 |
| | 5+6 2 | 5+6 2 | 5+6 2 | 3+4 2 | 2+2 2 | 2+2 2 |
| | 3+4 2 | 3+4 2 | 5+6 2 | 5+6 2 | 3+3 2 | 3+3 2 |
| | 3+4 2 | 3+4 2 | 3+4 2 | 3+4 2 | 2+2 2 | 2+2 2 |
| 12.1.133 | 618357 | 52 | C3Mo2U2 | | | |
| | -A | -GM | -M | -Y | -L | -V |
| | 3+4 2 | 3+4 2 | 5+6 2 | 5+6 2 | 3+3 2 | 2+2 2 |
| | 5+6 2 | 5+6 2 | 3+4 2 | 3+4 2 | 2+2 2 | 3+3 2 |
| | 3+4 2 | 3+4 2 | 5+6 2 | 5+6 2 | 3+3 2 | 2+2 2 |
| | 5+6 2 | 5+6 2 | 3+4 2 | 3+4 2 | 2+2 2 | 3+3 2 |
| | 3+4 2 | 3+4 2 | 5+6 2 | 3+4 2 | 2+2 2 | 3+3 2 |
| | 5+6 2 | 5+6 2 | 3+4 2 | 5+6 2 | 3+3 2 | 2+2 2 |
| | 3+4 2 | 3+4 2 | 5+6 2 | 5+6 2 | 3+3 2 | 2+2 2 |
| | 5+6 2 | 5+6 2 | 3+4 2 | 3+4 2 | 2+2 2 | 3+3 2 |
| | 5+6 2 | 3+4 2 | 5+6 2 | 3+4 2 | 3+3 2 | 3+3 2 |
| | 3+4 2 | 5+6 2 | 3+4 2 | 5+6 2 | 2+2 2 | 2+2 2 |
| | 5+6 2 | 3+4 2 | 5+6 2 | 3+4 2 | 2+2 2 | 3+3 2 |
| | 3+4 2 | 5+6 2 | 3+4 2 | 5+6 2 | 3+3 2 | 2+2 2 |
| | 5+6 2 | 5+6 2 | 5+6 2 | 3+4 2 | 2+2 2 | 2+2 2 |
| | 3+4 2 | 5+6 2 | 3+4 2 | 5+6 2 | 3+3 2 | 3+3 2 |
| | 3+4 2 | 3+4 2 | 3+4 2 | 3+4 2 | 3+3 2 | 3+3 2 |
| | 5+6 2 | 3+4 2 | 5+6 2 | 5+6 2 | 2+2 2 | 2+2 2 |
| | 3+4 2 | 5+6 2 | 3+4 2 | 3+4 2 | 3+3 2 | 3+3 2 |
| | 5+6 2 | 3+4 2 | 5+6 2 | 5+6 2 | 2+2 2 | 2+2 2 |
| | 3+4 2 | 5+6 2 | 3+4 2 | 5+6 2 | 3+3 2 | 2+2 2 |
| | 3+4 2 | 5+6 2 | 3+4 2 | 5+6 2 | 3+3 2 | 2+2 2 |
| | 5+6 2 | 3+4 2 | 3+4 2 | 3+4 2 | 2+2 2 | 3+3 2 |
| | 5+6 2 | 3+4 2 | 5+6 2 | 5+6 2 | 2+2 2 | 3+3 2 |
| | 3+4 2 | 5+6 2 | 3+4 2 | 3+4 2 | 3+3 2 | 2+2 2 |
| | 5+6 2 | 3+4 2 | 3+4 2 | 5+6 2 | 3+3 2 | 3+3 2 |
| | 5+6 2 | 5+6 2 | 5+6 2 | 3+4 2 | 2+2 2 | 2+2 2 |
| | 3+4 2 | 3+4 2 | 5+6 2 | 5+6 2 | 3+3 2 | 3+3 2 |
| 12.1.134 | 618223 | 74 | C3Lu2Mo2 | | | |
| | -A | -GM | -M | -Y | -L | -V |
| | 3+4 2 | 3+4 2 | 5+6 2 | 5+6 2 | 2+2 2 | 2+2 2 |
| | 5+6 2 | 5+6 2 | 3+4 2 | 3+4 2 | 3+3 2 | 3+3 2 |
| | 3+4 2 | 3+4 2 | 5+6 2 | 5+6 2 | 3+3 2 | 2+2 2 |
| | 5+6 2 | 5+6 2 | 3+4 2 | 3+4 2 | 2+2 2 | 3+3 2 |
| | 3+4 2 | 3+4 2 | 5+6 2 | 3+4 2 | 2+2 2 | 2+2 2 |
| | 5+6 2 | 5+6 2 | 3+4 2 | 5+6 2 | 3+3 2 | 3+3 2 |
| | 3+4 2 | 3+4 2 | 5+6 2 | 5+6 2 | 3+3 2 | 2+2 2 |
| | 5+6 2 | 5+6 2 | 3+4 2 | 3+4 2 | 2+2 2 | 3+3 2 |
| | 5+6 2 | 3+4 2 | 3+4 2 | 5+6 2 | 3+3 2 | 3+3 2 |
| | 3+4 2 | 5+6 2 | 5+6 2 | 3+4 2 | 2+2 2 | 2+2 2 |
| | 5+6 2 | 3+4 2 | 5+6 2 | 3+4 2 | 2+2 2 | 3+3 2 |
| | 3+4 2 | 5+6 2 | 5+6 2 | 5+6 2 | 3+3 2 | 2+2 2 |
| | 5+6 2 | 3+4 2 | 3+4 2 | 3+4 2 | 2+2 2 | 2+2 2 |
| | 3+4 2 | 5+6 2 | 3+4 2 | 3+4 2 | 3+3 2 | 3+3 2 |
| | 5+6 2 | 5+6 2 | 5+6 2 | 5+6 2 | 2+2 2 | 2+2 2 |
| | 3+4 2 | 3+4 2 | 3+4 2 | 5+6 2 | 3+3 2 | 3+3 2 |
| | 5+6 2 | 5+6 2 | 5+6 2 | 3+4 2 | 2+2 2 | 2+2 2 |
| | 3+4 2 | 3+4 2 | 5+6 2 | 5+6 2 | 3+3 2 | 3+3 2 |
| | 3+4 2 | 3+4 2 | 3+4 2 | 3+4 2 | 2+2 2 | 2+2 2 |
| | 5+6 2 | 5+6 2 | 3+4 2 | 5+6 2 | 3+3 2 | 3+3 2 |
| | 3+4 2 | 5+6 2 | 5+6 2 | 3+4 2 | 3+3 2 | 3+3 2 |
| | 5+6 2 | 3+4 2 | 5+6 2 | 3+4 2 | 3+3 2 | 2+2 2 |
| | 3+4 2 | 5+6 2 | 3+4 2 | 5+6 2 | 2+2 2 | 3+3 2 |
| | 5+6 2 | 5+6 2 | 3+4 2 | 5+6 2 | 2+2 2 | 2+2 2 |
| | 3+4 2 | 3+4 2 | 3+4 2 | 3+4 2 | 3+3 2 | 2+2 2 |
| | 3+4 2 | 5+6 2 | 5+6 2 | 5+6 2 | 3+3 2 | 3+3 2 |
| | 5+6 2 | 3+4 2 | 3+4 2 | 3+4 2 | 2+2 2 | 2+2 2 |
| | 5+6 2 | 3+4 2 | 5+6 2 | 5+6 2 | 3+3 2 | 3+3 2 |
| | 3+4 2 | 5+6 2 | 5+6 2 | 3+4 2 | 2+2 2 | 2+2 2 |
| | 5+6 2 | 3+4 2 | 3+4 2 | 5+6 2 | 3+3 2 | 3+3 2 |

Continued on next column

**TABLE XXIV– continued**

| TQC | ICSD | NoE | SF | | | |
|---|---|---|---|---|---|---|
| | 3+4 2 | 3+4 2 | 5+6 2 | 3+4 2 | 2+2 2 | 2+2 2 |
| | 5+6 2 | 5+6 2 | 3+4 2 | 5+6 2 | 2+2 2 | 3+3 2 |
| | 3+4 2 | 5+6 2 | 3+4 2 | 5+6 2 | 3+3 2 | 2+2 2 |
| | 3+4 2 | 3+4 2 | 3+4 2 | 5+6 2 | 3+3 2 | 2+2 2 |
| | 5+6 2 | 5+6 2 | 3+4 2 | 3+4 2 | 2+2 2 | 3+3 2 |
| | 3+4 2 | 3+4 2 | 5+6 2 | 3+4 2 | 3+3 2 | 3+3 2 |
| | 5+6 2 | 3+4 2 | 3+4 2 | 5+6 2 | 2+2 2 | 2+2 2 |
| 12.1.135 | 88513 | 66 | C3Ho2Mo2 | | | |
| | -A | -GM | -M | -Y | -L | -V |
| | 3+4 2 | 3+4 2 | 5+6 2 | 5+6 2 | 2+2 2 | 2+2 2 |
| | 5+6 2 | 5+6 2 | 3+4 2 | 3+4 2 | 3+3 2 | 3+3 2 |
| | 3+4 2 | 3+4 2 | 5+6 2 | 5+6 2 | 2+2 2 | 2+2 2 |
| | 5+6 2 | 5+6 2 | 3+4 2 | 3+4 2 | 3+3 2 | 3+3 2 |
| | 3+4 2 | 3+4 2 | 5+6 2 | 3+4 2 | 3+3 2 | 2+2 2 |
| | 5+6 2 | 5+6 2 | 3+4 2 | 5+6 2 | 2+2 2 | 3+3 2 |
| | 3+4 2 | 3+4 2 | 5+6 2 | 5+6 2 | 2+2 2 | 2+2 2 |
| | 5+6 2 | 5+6 2 | 3+4 2 | 3+4 2 | 3+3 2 | 3+3 2 |
| | 5+6 2 | 3+4 2 | 3+4 2 | 5+6 2 | 3+3 2 | 3+3 2 |
| | 3+4 2 | 5+6 2 | 5+6 2 | 3+4 2 | 2+2 2 | 2+2 2 |
| | 3+4 2 | 3+4 2 | 3+4 2 | 3+4 2 | 3+3 2 | 3+3 2 |
| | 3+4 2 | 5+6 2 | 5+6 2 | 5+6 2 | 2+2 2 | 2+2 2 |
| | 5+6 2 | 3+4 2 | 5+6 2 | 3+4 2 | 3+3 2 | 2+2 2 |
| | 5+6 2 | 5+6 2 | 3+4 2 | 5+6 2 | 2+2 2 | 3+3 2 |
| | 5+6 2 | 5+6 2 | 3+4 2 | 5+6 2 | 2+2 2 | 2+2 2 |
| | 3+4 2 | 3+4 2 | 5+6 2 | 3+4 2 | 3+3 2 | 3+3 2 |
| | 3+4 2 | 3+4 2 | 5+6 2 | 3+4 2 | 2+2 2 | 2+2 2 |
| | 5+6 2 | 5+6 2 | 3+4 2 | 5+6 2 | 3+3 2 | 3+3 2 |
| | 5+6 2 | 5+6 2 | 5+6 2 | 5+6 2 | 2+2 2 | 2+2 2 |
| | 5+6 2 | 5+6 2 | 5+6 2 | 3+4 2 | 3+3 2 | 2+2 2 |
| | 5+6 2 | 5+6 2 | 3+4 2 | 5+6 2 | 2+2 2 | 3+3 2 |
| | 3+4 2 | 3+4 2 | 5+6 2 | 3+4 2 | 2+2 2 | 3+3 2 |
| | 3+4 2 | 3+4 2 | 3+4 2 | 5+6 2 | 3+3 2 | 2+2 2 |
| | 3+4 2 | 3+4 2 | 3+4 2 | 3+4 2 | 2+2 2 | 2+2 2 |
| | 5+6 2 | 5+6 2 | 5+6 2 | 5+6 2 | 3+3 2 | 3+3 2 |
| | 5+6 2 | 3+4 2 | 5+6 2 | 3+4 2 | 2+2 2 | 2+2 2 |
| | 5+6 2 | 5+6 2 | 3+4 2 | 5+6 2 | 3+3 2 | 3+3 2 |
| | 3+4 2 | 3+4 2 | 5+6 2 | 3+4 2 | 2+2 2 | 3+3 2 |
| | 5+6 2 | 5+6 2 | 3+4 2 | 5+6 2 | 3+3 2 | 2+2 2 |
| | 3+4 2 | 3+4 2 | 3+4 2 | 5+6 2 | 2+2 2 | 3+3 2 |
| | 3+4 2 | 3+4 2 | 3+4 2 | 3+4 2 | 3+3 2 | 2+2 2 |
| | 5+6 2 | 5+6 2 | 5+6 2 | 5+6 2 | 2+2 2 | 3+3 2 |
| | 5+6 2 | 3+4 2 | 5+6 2 | 3+4 2 | 3+3 2 | 2+2 2 |
| 12.1.136 | 88512 | 66 | C3Ho2Mo2 | | | |
| | -A | -GM | -M | -Y | -L | -V |
| | 3+4 2 | 3+4 2 | 5+6 2 | 5+6 2 | 2+2 2 | 2+2 2 |
| | 5+6 2 | 5+6 2 | 3+4 2 | 3+4 2 | 3+3 2 | 3+3 2 |
| | 3+4 2 | 3+4 2 | 5+6 2 | 5+6 2 | 2+2 2 | 2+2 2 |
| | 5+6 2 | 5+6 2 | 3+4 2 | 3+4 2 | 3+3 2 | 3+3 2 |
| | 3+4 2 | 3+4 2 | 5+6 2 | 3+4 2 | 3+3 2 | 2+2 2 |
| | 5+6 2 | 5+6 2 | 3+4 2 | 5+6 2 | 2+2 2 | 3+3 2 |
| | 3+4 2 | 3+4 2 | 5+6 2 | 5+6 2 | 2+2 2 | 2+2 2 |
| | 5+6 2 | 5+6 2 | 3+4 2 | 3+4 2 | 3+3 2 | 3+3 2 |
| | 5+6 2 | 3+4 2 | 3+4 2 | 5+6 2 | 3+3 2 | 3+3 2 |
| | 3+4 2 | 5+6 2 | 5+6 2 | 3+4 2 | 2+2 2 | 2+2 2 |
| | 3+4 2 | 3+4 2 | 3+4 2 | 3+4 2 | 3+3 2 | 3+3 2 |
| | 3+4 2 | 5+6 2 | 5+6 2 | 5+6 2 | 2+2 2 | 2+2 2 |
| | 5+6 2 | 3+4 2 | 5+6 2 | 3+4 2 | 3+3 2 | 2+2 2 |
| | 5+6 2 | 5+6 2 | 3+4 2 | 5+6 2 | 2+2 2 | 3+3 2 |
| | 5+6 2 | 3+4 2 | 3+4 2 | 5+6 2 | 2+2 2 | 2+2 2 |
| | 3+4 2 | 5+6 2 | 5+6 2 | 3+4 2 | 3+3 2 | 3+3 2 |
| | 3+4 2 | 3+4 2 | 5+6 2 | 3+4 2 | 2+2 2 | 2+2 2 |
| | 5+6 2 | 5+6 2 | 3+4 2 | 5+6 2 | 3+3 2 | 3+3 2 |
| | 5+6 2 | 5+6 2 | 5+6 2 | 5+6 2 | 2+2 2 | 2+2 2 |
| | 5+6 2 | 5+6 2 | 5+6 2 | 3+4 2 | 3+3 2 | 2+2 2 |
| | 5+6 2 | 5+6 2 | 3+4 2 | 5+6 2 | 2+2 2 | 3+3 2 |
| | 3+4 2 | 3+4 2 | 5+6 2 | 3+4 2 | 2+2 2 | 3+3 2 |
| | 3+4 2 | 3+4 2 | 3+4 2 | 5+6 2 | 3+3 2 | 2+2 2 |
| | 3+4 2 | 3+4 2 | 5+6 2 | 3+4 2 | 2+2 2 | 2+2 2 |
| | 5+6 2 | 5+6 2 | 3+4 2 | 5+6 2 | 3+3 2 | 3+3 2 |
| | 5+6 2 | 3+4 2 | 5+6 2 | 3+4 2 | 2+2 2 | 2+2 2 |
| | 5+6 2 | 5+6 2 | 3+4 2 | 5+6 2 | 3+3 2 | 3+3 2 |
| | 3+4 2 | 3+4 2 | 5+6 2 | 3+4 2 | 2+2 2 | 3+3 2 |
| | 5+6 2 | 5+6 2 | 3+4 2 | 5+6 2 | 3+3 2 | 2+2 2 |
| | 3+4 2 | 3+4 2 | 3+4 2 | 5+6 2 | 2+2 2 | 3+3 2 |
| | 3+4 2 | 3+4 2 | 3+4 2 | 3+4 2 | 3+3 2 | 2+2 2 |
| | 5+6 2 | 5+6 2 | 5+6 2 | 5+6 2 | 2+2 2 | 3+3 2 |
| | 5+6 2 | 3+4 2 | 5+6 2 | 3+4 2 | 3+3 2 | 2+2 2 |
| 12.1.137 | 88511 | 66 | C3Ho2Mo2 | | | |
| | -A | -GM | -M | -Y | -L | -V |
| | 3+4 2 | 3+4 2 | 5+6 2 | 5+6 2 | 2+2 2 | 2+2 2 |
| | 5+6 2 | 5+6 2 | 3+4 2 | 3+4 2 | 3+3 2 | 3+3 2 |
| | 3+4 2 | 3+4 2 | 5+6 2 | 5+6 2 | 2+2 2 | 2+2 2 |
| | 5+6 2 | 5+6 2 | 3+4 2 | 3+4 2 | 3+3 2 | 3+3 2 |
| | 3+4 2 | 3+4 2 | 5+6 2 | 3+4 2 | 3+3 2 | 2+2 2 |
| | 5+6 2 | 5+6 2 | 3+4 2 | 5+6 2 | 2+2 2 | 3+3 2 |
| | 3+4 2 | 3+4 2 | 5+6 2 | 5+6 2 | 2+2 2 | 2+2 2 |
| | 5+6 2 | 5+6 2 | 3+4 2 | 3+4 2 | 3+3 2 | 3+3 2 |
| | 5+6 2 | 3+4 2 | 3+4 2 | 5+6 2 | 3+3 2 | 3+3 2 |
| | 3+4 2 | 5+6 2 | 5+6 2 | 3+4 2 | 2+2 2 | 2+2 2 |
| | 3+4 2 | 3+4 2 | 3+4 2 | 3+4 2 | 3+3 2 | 3+3 2 |
| | 3+4 2 | 5+6 2 | 5+6 2 | 5+6 2 | 2+2 2 | 2+2 2 |
| | 5+6 2 | 3+4 2 | 5+6 2 | 3+4 2 | 3+3 2 | 2+2 2 |
| | 5+6 2 | 5+6 2 | 3+4 2 | 5+6 2 | 2+2 2 | 3+3 2 |
| | 5+6 2 | 3+4 2 | 3+4 2 | 5+6 2 | 2+2 2 | 2+2 2 |
| | 3+4 2 | 5+6 2 | 5+6 2 | 3+4 2 | 3+3 2 | 3+3 2 |
| | 3+4 2 | 3+4 2 | 5+6 2 | 3+4 2 | 2+2 2 | 2+2 2 |
| | 5+6 2 | 5+6 2 | 3+4 2 | 5+6 2 | 3+3 2 | 3+3 2 |
| | 5+6 2 | 5+6 2 | 5+6 2 | 5+6 2 | 2+2 2 | 2+2 2 |
| | 5+6 2 | 5+6 2 | 5+6 2 | 3+4 2 | 3+3 2 | 2+2 2 |
| | 5+6 2 | 5+6 2 | 3+4 2 | 5+6 2 | 2+2 2 | 3+3 2 |
| | 3+4 2 | 3+4 2 | 5+6 2 | 3+4 2 | 2+2 2 | 3+3 2 |
| | 3+4 2 | 3+4 2 | 3+4 2 | 5+6 2 | 3+3 2 | 2+2 2 |
| | 3+4 2 | 3+4 2 | 5+6 2 | 3+4 2 | 2+2 2 | 2+2 2 |
| | 5+6 2 | 5+6 2 | 3+4 2 | 5+6 2 | 3+3 2 | 3+3 2 |
| | 5+6 2 | 3+4 2 | 5+6 2 | 3+4 2 | 2+2 2 | 2+2 2 |
| | 5+6 2 | 5+6 2 | 3+4 2 | 5+6 2 | 3+3 2 | 3+3 2 |
| | 3+4 2 | 3+4 2 | 5+6 2 | 3+4 2 | 2+2 2 | 3+3 2 |
| | 5+6 2 | 5+6 2 | 3+4 2 | 5+6 2 | 3+3 2 | 2+2 2 |
| | 3+4 2 | 3+4 2 | 3+4 2 | 5+6 2 | 2+2 2 | 3+3 2 |
| | 3+4 2 | 3+4 2 | 3+4 2 | 3+4 2 | 3+3 2 | 2+2 2 |
| | 5+6 2 | 5+6 2 | 5+6 2 | 5+6 2 | 2+2 2 | 3+3 2 |
| | 5+6 2 | 3+4 2 | 5+6 2 | 3+4 2 | 3+3 2 | 2+2 2 |
| 12.1.138 | 602482 | 72 | C3Cr2Yb2 | | | |
| | -A | -GM | -M | -Y | -L | -V |
| | 3+4 2 | 3+4 2 | 5+6 2 | 3+4 2 | 2+2 2 | 2+2 2 |
| | 5+6 2 | 5+6 2 | 3+4 2 | 5+6 2 | 3+3 2 | 3+3 2 |
| | 3+4 2 | 3+4 2 | 5+6 2 | 3+4 2 | 2+2 2 | 2+2 2 |
| | 5+6 2 | 5+6 2 | 3+4 2 | 5+6 2 | 3+3 2 | 3+3 2 |
| | 3+4 2 | 3+4 2 | 5+6 2 | 3+4 2 | 2+2 2 | 2+2 2 |
| | 5+6 2 | 5+6 2 | 3+4 2 | 5+6 2 | 3+3 2 | 3+3 2 |
| | 3+4 2 | 3+4 2 | 5+6 2 | 5+6 2 | 2+2 2 | 2+2 2 |
| | 5+6 2 | 5+6 2 | 3+4 2 | 3+4 2 | 3+3 2 | 3+3 2 |

Continued on next column

**TABLE XXIV– continued**

| TQC | ICSD | NoE | SF | | | |
|---|---|---|---|---|---|---|
| | 5+6 2 | 3+4 2 | 3+4 2 | 5+6 2 | 2+2 2 | 3+3 2 |
| | 3+4 2 | 5+6 2 | 5+6 2 | 3+4 2 | 3+3 2 | 2+2 2 |
| | 5+6 2 | 3+4 2 | 5+6 2 | 3+4 2 | 2+2 2 | 3+3 2 |
| | 3+4 2 | 5+6 2 | 3+4 2 | 5+6 2 | 3+3 2 | 2+2 2 |
| | 5+6 2 | 3+4 2 | 5+6 2 | 3+4 2 | 2+2 2 | 2+2 2 |
| | 3+4 2 | 5+6 2 | 3+4 2 | 5+6 2 | 3+3 2 | 3+3 2 |
| | 3+4 2 | 5+6 2 | 3+4 2 | 5+6 2 | 3+3 2 | 2+2 2 |
| | 3+4 2 | 3+4 2 | 5+6 2 | 3+4 2 | 3+3 2 | 3+3 2 |
| | 5+6 2 | 5+6 2 | 5+6 2 | 3+4 2 | 2+2 2 | 3+3 2 |
| | 5+6 2 | 3+4 2 | 3+4 2 | 5+6 2 | 3+3 2 | 2+2 2 |
| | 3+4 2 | 5+6 2 | 3+4 2 | 5+6 2 | 2+2 2 | 2+2 2 |
| | 3+4 2 | 5+6 2 | 3+4 2 | 5+6 2 | 3+3 2 | 2+2 2 |
| | 5+6 2 | 3+4 2 | 5+6 2 | 3+4 2 | 2+2 2 | 3+3 2 |
| | 3+4 2 | 3+4 2 | 5+6 2 | 3+4 2 | 3+3 2 | 2+2 2 |
| | 5+6 2 | 5+6 2 | 5+6 2 | 5+6 2 | 2+2 2 | 3+3 2 |
| | 5+6 2 | 3+4 2 | 3+4 2 | 3+4 2 | 3+3 2 | 3+3 2 |
| | 3+4 2 | 5+6 2 | 3+4 2 | 5+6 2 | 2+2 2 | 2+2 2 |
| | 3+4 2 | 5+6 2 | 3+4 2 | 5+6 2 | 3+3 2 | 2+2 2 |
| | 5+6 2 | 3+4 2 | 3+4 2 | 3+4 2 | 2+2 2 | 3+3 2 |
| | 5+6 2 | 3+4 2 | 5+6 2 | 5+6 2 | 3+3 2 | 3+3 2 |
| | 3+4 2 | 3+4 2 | 3+4 2 | 3+4 2 | 2+2 2 | 2+2 2 |
| | 5+6 2 | 5+6 2 | 5+6 2 | 5+6 2 | 3+3 2 | 3+3 2 |
| | 3+4 2 | 3+4 2 | 5+6 2 | 3+4 2 | 2+2 2 | 2+2 2 |
| | 5+6 2 | 5+6 2 | 5+6 2 | 5+6 2 | 3+3 2 | 3+3 2 |
| | 3+4 2 | 3+4 2 | 3+4 2 | 3+4 2 | 2+2 2 | 2+2 2 |
| | 5+6 2 | 3+4 2 | 3+4 2 | 5+6 2 | 3+3 2 | 3+3 2 |
| | 3+4 2 | 5+6 2 | 5+6 2 | 3+4 2 | 2+2 2 | 2+2 2 |
| | 5+6 2 | 5+6 2 | 3+4 2 | 5+6 2 | 3+3 2 | 3+3 2 |
| 12.1.139 | 617537 | 62 | C3Cr2Tb2 | | | |
| | -A | -GM | -M | -Y | -L | -V |
| | 3+4 2 | 3+4 2 | 5+6 2 | 3+4 2 | 2+2 2 | 2+2 2 |
| | 5+6 2 | 5+6 2 | 3+4 2 | 5+6 2 | 3+3 2 | 3+3 2 |
| | 3+4 2 | 3+4 2 | 5+6 2 | 3+4 2 | 2+2 2 | 2+2 2 |
| | 5+6 2 | 5+6 2 | 3+4 2 | 5+6 2 | 3+3 2 | 3+3 2 |
| | 3+4 2 | 3+4 2 | 5+6 2 | 3+4 2 | 2+2 2 | 2+2 2 |
| | 5+6 2 | 5+6 2 | 3+4 2 | 5+6 2 | 3+3 2 | 3+3 2 |
| | 3+4 2 | 3+4 2 | 5+6 2 | 5+6 2 | 2+2 2 | 2+2 2 |
| | 5+6 2 | 5+6 2 | 3+4 2 | 3+4 2 | 3+3 2 | 3+3 2 |
| | 5+6 2 | 3+4 2 | 3+4 2 | 5+6 2 | 2+2 2 | 3+3 2 |
| | 3+4 2 | 5+6 2 | 5+6 2 | 3+4 2 | 3+3 2 | 2+2 2 |
| | 5+6 2 | 3+4 2 | 5+6 2 | 3+4 2 | 2+2 2 | 3+3 2 |
| | 3+4 2 | 5+6 2 | 3+4 2 | 5+6 2 | 3+3 2 | 2+2 2 |
| | 5+6 2 | 3+4 2 | 5+6 2 | 3+4 2 | 2+2 2 | 2+2 2 |
| | 3+4 2 | 5+6 2 | 3+4 2 | 5+6 2 | 3+3 2 | 3+3 2 |
| | 3+4 2 | 5+6 2 | 3+4 2 | 5+6 2 | 3+3 2 | 2+2 2 |
| | 3+4 2 | 3+4 2 | 5+6 2 | 3+4 2 | 3+3 2 | 3+3 2 |
| | 5+6 2 | 5+6 2 | 3+4 2 | 3+4 2 | 3+3 2 | 3+3 2 |
| | 5+6 2 | 3+4 2 | 5+6 2 | 5+6 2 | 2+2 2 | 2+2 2 |
| | 3+4 2 | 5+6 2 | 3+4 2 | 5+6 2 | 2+2 2 | 2+2 2 |
| | 3+4 2 | 5+6 2 | 3+4 2 | 5+6 2 | 3+3 2 | 2+2 2 |
| | 5+6 2 | 3+4 2 | 3+4 2 | 3+4 2 | 2+2 2 | 3+3 2 |
| | 5+6 2 | 3+4 2 | 3+4 2 | 5+6 2 | 2+2 2 | 3+3 2 |
| | 3+4 2 | 5+6 2 | 5+6 2 | 3+4 2 | 3+3 2 | 2+2 2 |
| | 5+6 2 | 3+4 2 | 5+6 2 | 5+6 2 | 3+3 2 | 3+3 2 |
| | 3+4 2 | 5+6 2 | 5+6 2 | 3+4 2 | 2+2 2 | 2+2 2 |
| | 3+4 2 | 3+4 2 | 3+4 2 | 5+6 2 | 2+2 2 | 2+2 2 |
| | 5+6 2 | 5+6 2 | 3+4 2 | 3+4 2 | 3+3 2 | 3+3 2 |
| | 5+6 2 | 3+4 2 | 5+6 2 | 5+6 2 | 3+3 2 | 3+3 2 |
| | 5+6 2 | 3+4 2 | 3+4 2 | 3+4 2 | 2+2 2 | 2+2 2 |
| | 3+4 2 | 5+6 2 | 5+6 2 | 5+6 2 | 3+3 2 | 3+3 2 |
| | 5+6 2 | 3+4 2 | 5+6 2 | 3+4 2 | 2+2 2 | 3+3 2 |
| 12.1.140 | 617514 | 74 | C3Cr2Lu2 | | | |
| | -A | -GM | -M | -Y | -L | -V |
| | 3+4 2 | 3+4 2 | 5+6 2 | 3+4 2 | 2+2 2 | 2+2 2 |
| | 5+6 2 | 5+6 2 | 3+4 2 | 5+6 2 | 3+3 2 | 3+3 2 |
| | 3+4 2 | 3+4 2 | 5+6 2 | 3+4 2 | 2+2 2 | 2+2 2 |
| | 5+6 2 | 5+6 2 | 3+4 2 | 5+6 2 | 3+3 2 | 3+3 2 |
| | 3+4 2 | 3+4 2 | 5+6 2 | 3+4 2 | 2+2 2 | 2+2 2 |
| | 5+6 2 | 5+6 2 | 3+4 2 | 5+6 2 | 3+3 2 | 3+3 2 |
| | 3+4 2 | 3+4 2 | 5+6 2 | 5+6 2 | 3+3 2 | 2+2 2 |
| | 5+6 2 | 5+6 2 | 3+4 2 | 3+4 2 | 2+2 2 | 3+3 2 |
| | 5+6 2 | 3+4 2 | 3+4 2 | 5+6 2 | 2+2 2 | 3+3 2 |
| | 3+4 2 | 5+6 2 | 5+6 2 | 3+4 2 | 3+3 2 | 2+2 2 |
| | 5+6 2 | 3+4 2 | 5+6 2 | 3+4 2 | 2+2 2 | 3+3 2 |
| | 3+4 2 | 5+6 2 | 3+4 2 | 5+6 2 | 3+3 2 | 2+2 2 |
| | 3+4 2 | 5+6 2 | 5+6 2 | 3+4 2 | 2+2 2 | 2+2 2 |
| | 5+6 2 | 3+4 2 | 3+4 2 | 5+6 2 | 3+3 2 | 3+3 2 |
| | 5+6 2 | 5+6 2 | 3+4 2 | 3+4 2 | 2+2 2 | 3+3 2 |
| | 3+4 2 | 3+4 2 | 5+6 2 | 5+6 2 | 3+3 2 | 2+2 2 |
| | 5+6 2 | 3+4 2 | 5+6 2 | 3+4 2 | 2+2 2 | 3+3 2 |
| | 3+4 2 | 5+6 2 | 3+4 2 | 5+6 2 | 3+3 2 | 2+2 2 |
| | 3+4 2 | 3+4 2 | 5+6 2 | 3+4 2 | 2+2 2 | 2+2 2 |
| | 5+6 2 | 5+6 2 | 3+4 2 | 5+6 2 | 3+3 2 | 3+3 2 |
| | 3+4 2 | 5+6 2 | 3+4 2 | 3+4 2 | 3+3 2 | 2+2 2 |
| | 5+6 2 | 5+6 2 | 5+6 2 | 5+6 2 | 2+2 2 | 3+3 2 |
| | 3+4 2 | 3+4 2 | 3+4 2 | 5+6 2 | 3+3 2 | 2+2 2 |
| | 5+6 2 | 3+4 2 | 5+6 2 | 3+4 2 | 2+2 2 | 3+3 2 |
| | 3+4 2 | 5+6 2 | 3+4 2 | 5+6 2 | 3+3 2 | 2+2 2 |
| | 3+4 2 | 3+4 2 | 5+6 2 | 3+4 2 | 2+2 2 | 3+3 2 |
| | 5+6 2 | 3+4 2 | 5+6 2 | 3+4 2 | 3+3 2 | 3+3 2 |
| | 5+6 2 | 5+6 2 | 3+4 2 | 5+6 2 | 2+2 2 | 2+2 2 |
| | 3+4 2 | 5+6 2 | 5+6 2 | 5+6 2 | 3+3 2 | 2+2 2 |
| | 3+4 2 | 3+4 2 | 3+4 2 | 3+4 2 | 3+3 2 | 3+3 2 |
| | 5+6 2 | 5+6 2 | 5+6 2 | 3+4 2 | 2+2 2 | 3+3 2 |
| | 5+6 2 | 3+4 2 | 3+4 2 | 5+6 2 | 3+3 2 | 2+2 2 |
| | 3+4 2 | 5+6 2 | 3+4 2 | 5+6 2 | 2+2 2 | 2+2 2 |
| | 3+4 2 | 3+4 2 | 3+4 2 | 5+6 2 | 3+3 2 | 2+2 2 |
| | 5+6 2 | 5+6 2 | 5+6 2 | 3+4 2 | 2+2 2 | 3+3 2 |
| | 5+6 2 | 3+4 2 | 3+4 2 | 5+6 2 | 3+3 2 | 3+3 2 |
| | 3+4 2 | 3+4 2 | 3+4 2 | 3+4 2 | 2+2 2 | 2+2 2 |
| 12.1.141 | 617492 | 64 | C3Cr2Dy2 | | | |
| | -A | -GM | -M | -Y | -L | -V |
| | 3+4 2 | 3+4 2 | 5+6 2 | 3+4 2 | 2+2 2 | 2+2 2 |
| | 5+6 2 | 5+6 2 | 3+4 2 | 5+6 2 | 3+3 2 | 3+3 2 |
| | 3+4 2 | 3+4 2 | 5+6 2 | 3+4 2 | 2+2 2 | 2+2 2 |
| | 5+6 2 | 5+6 2 | 3+4 2 | 5+6 2 | 3+3 2 | 3+3 2 |
| | 3+4 2 | 3+4 2 | 5+6 2 | 3+4 2 | 2+2 2 | 2+2 2 |
| | 5+6 2 | 5+6 2 | 3+4 2 | 5+6 2 | 3+3 2 | 3+3 2 |
| | 3+4 2 | 3+4 2 | 5+6 2 | 5+6 2 | 2+2 2 | 2+2 2 |
| | 5+6 2 | 5+6 2 | 3+4 2 | 3+4 2 | 3+3 2 | 3+3 2 |
| | 5+6 2 | 3+4 2 | 3+4 2 | 5+6 2 | 2+2 2 | 3+3 2 |
| | 3+4 2 | 5+6 2 | 5+6 2 | 3+4 2 | 3+3 2 | 2+2 2 |
| | 5+6 2 | 3+4 2 | 5+6 2 | 3+4 2 | 2+2 2 | 3+3 2 |
| | 3+4 2 | 5+6 2 | 3+4 2 | 5+6 2 | 3+3 2 | 2+2 2 |
| | 5+6 2 | 3+4 2 | 5+6 2 | 3+4 2 | 2+2 2 | 2+2 2 |
| | 3+4 2 | 5+6 2 | 3+4 2 | 5+6 2 | 3+3 2 | 3+3 2 |
| | 3+4 2 | 5+6 2 | 3+4 2 | 5+6 2 | 3+3 2 | 2+2 2 |
| | 3+4 2 | 3+4 2 | 5+6 2 | 3+4 2 | 3+3 2 | 3+3 2 |
| | 5+6 2 | 5+6 2 | 3+4 2 | 3+4 2 | 3+3 2 | 3+3 2 |
| | 5+6 2 | 3+4 2 | 5+6 2 | 5+6 2 | 2+2 2 | 2+2 2 |
| | 3+4 2 | 5+6 2 | 3+4 2 | 5+6 2 | 2+2 2 | 2+2 2 |
| | 3+4 2 | 5+6 2 | 3+4 2 | 5+6 2 | 3+3 2 | 2+2 2 |

Continued on next column

Continued on next column

**TABLE XXIV– continued**

| TQC | ICSD | | NoE | | SF | |
|---|---|---|---|---|---|---|
| | 5+6 2 | 3+4 2 | 3+4 2 | 3+4 2 | 2+2 2 | 3+3 2 |
| | 3+4 2 | 3+4 2 | 5+6 2 | 3+4 2 | 2+2 2 | 3+3 2 |
| | 5+6 2 | 5+6 2 | 5+6 2 | 5+6 2 | 3+3 2 | 2+2 2 |
| | 5+6 2 | 3+4 2 | 3+4 2 | 5+6 2 | 3+3 2 | 3+3 2 |
| | 3+4 2 | 5+6 2 | 5+6 2 | 3+4 2 | 2+2 2 | 2+2 2 |
| | 3+4 2 | 3+4 2 | 3+4 2 | 5+6 2 | 2+2 2 | 2+2 2 |
| | 5+6 2 | 5+6 2 | 3+4 2 | 3+4 2 | 3+3 2 | 3+3 2 |
| | 5+6 2 | 3+4 2 | 5+6 2 | 5+6 2 | 3+3 2 | 3+3 2 |
| | 5+6 2 | 5+6 2 | 3+4 2 | 3+4 2 | 2+2 2 | 2+2 2 |
| | 3+4 2 | 3+4 2 | 5+6 2 | 5+6 2 | 3+3 2 | 3+3 2 |
| | 5+6 2 | 3+4 2 | 5+6 2 | 5+6 2 | 2+2 2 | 2+2 2 |
| | 3+4 2 | 5+6 2 | 3+4 2 | 3+4 2 | 3+3 2 | 3+3 2 |
| 12.1.142 | 43018 | | 64 | | C3Cr2Dy2 | |
| | -A | -GM | -M | -Y | -L | -V |
| | 3+4 2 | 3+4 2 | 5+6 2 | 3+4 2 | 2+2 2 | 2+2 2 |
| | 5+6 2 | 5+6 2 | 3+4 2 | 5+6 2 | 3+3 2 | 3+3 2 |
| | 3+4 2 | 3+4 2 | 5+6 2 | 3+4 2 | 2+2 2 | 2+2 2 |
| | 5+6 2 | 5+6 2 | 3+4 2 | 5+6 2 | 3+3 2 | 3+3 2 |
| | 3+4 2 | 3+4 2 | 5+6 2 | 3+4 2 | 2+2 2 | 2+2 2 |
| | 5+6 2 | 5+6 2 | 3+4 2 | 5+6 2 | 3+3 2 | 3+3 2 |
| | 3+4 2 | 3+4 2 | 5+6 2 | 5+6 2 | 2+2 2 | 2+2 2 |
| | 5+6 2 | 5+6 2 | 3+4 2 | 3+4 2 | 3+3 2 | 3+3 2 |
| | 5+6 2 | 3+4 2 | 3+4 2 | 5+6 2 | 2+2 2 | 3+3 2 |
| | 3+4 2 | 5+6 2 | 5+6 2 | 3+4 2 | 3+3 2 | 2+2 2 |
| | 5+6 2 | 3+4 2 | 5+6 2 | 3+4 2 | 2+2 2 | 3+3 2 |
| | 3+4 2 | 5+6 2 | 3+4 2 | 5+6 2 | 3+3 2 | 2+2 2 |
| | 5+6 2 | 3+4 2 | 5+6 2 | 3+4 2 | 2+2 2 | 2+2 2 |
| | 3+4 2 | 5+6 2 | 3+4 2 | 5+6 2 | 3+3 2 | 3+3 2 |
| | 3+4 2 | 5+6 2 | 3+4 2 | 5+6 2 | 3+3 2 | 2+2 2 |
| | 3+4 2 | 3+4 2 | 5+6 2 | 3+4 2 | 3+3 2 | 3+3 2 |
| | 5+6 2 | 5+6 2 | 3+4 2 | 3+4 2 | 3+3 2 | 3+3 2 |
| | 5+6 2 | 3+4 2 | 5+6 2 | 5+6 2 | 2+2 2 | 2+2 2 |
| | 3+4 2 | 5+6 2 | 3+4 2 | 5+6 2 | 2+2 2 | 2+2 2 |
| | 3+4 2 | 5+6 2 | 3+4 2 | 5+6 2 | 3+3 2 | 2+2 2 |
| | 5+6 2 | 3+4 2 | 3+4 2 | 3+4 2 | 2+2 2 | 3+3 2 |
| | 5+6 2 | 3+4 2 | 5+6 2 | 5+6 2 | 2+2 2 | 3+3 2 |
| | 3+4 2 | 5+6 2 | 5+6 2 | 3+4 2 | 3+3 2 | 2+2 2 |
| | 5+6 2 | 3+4 2 | 3+4 2 | 5+6 2 | 3+3 2 | 3+3 2 |
| | 3+4 2 | 5+6 2 | 5+6 2 | 3+4 2 | 2+2 2 | 2+2 2 |
| | 3+4 2 | 3+4 2 | 3+4 2 | 5+6 2 | 2+2 2 | 2+2 2 |
| | 5+6 2 | 5+6 2 | 3+4 2 | 3+4 2 | 3+3 2 | 3+3 2 |
| | 5+6 2 | 3+4 2 | 5+6 2 | 5+6 2 | 3+3 2 | 3+3 2 |
| | 5+6 2 | 3+4 2 | 3+4 2 | 3+4 2 | 2+2 2 | 2+2 2 |
| | 3+4 2 | 5+6 2 | 5+6 2 | 5+6 2 | 3+3 2 | 3+3 2 |
| | 5+6 2 | 3+4 2 | 5+6 2 | 5+6 2 | 2+2 2 | 3+3 2 |
| | 3+4 2 | 5+6 2 | 3+4 2 | 3+4 2 | 3+3 2 | 2+2 2 |
| 12.1.143 | 417827 | | 48 | | C3Ce2Mo2 | |
| | -A | -GM | -M | -Y | -L | -V |
| | 3+4 2 | 3+4 2 | 5+6 2 | 3+4 2 | 3+3 2 | 2+2 2 |
| | 5+6 2 | 5+6 2 | 3+4 2 | 5+6 2 | 2+2 2 | 3+3 2 |
| | 3+4 2 | 3+4 2 | 5+6 2 | 3+4 2 | 3+3 2 | 2+2 2 |
| | 5+6 2 | 5+6 2 | 3+4 2 | 5+6 2 | 2+2 2 | 3+3 2 |
| | 3+4 2 | 3+4 2 | 5+6 2 | 3+4 2 | 3+3 2 | 2+2 2 |
| | 5+6 2 | 5+6 2 | 3+4 2 | 5+6 2 | 2+2 2 | 3+3 2 |
| | 3+4 2 | 3+4 2 | 5+6 2 | 5+6 2 | 3+3 2 | 2+2 2 |
| | 5+6 2 | 5+6 2 | 3+4 2 | 3+4 2 | 2+2 2 | 3+3 2 |
| | 5+6 2 | 3+4 2 | 3+4 2 | 3+4 2 | 3+3 2 | 2+2 2 |
| | 3+4 2 | 5+6 2 | 5+6 2 | 5+6 2 | 2+2 2 | 3+3 2 |
| | 3+4 2 | 3+4 2 | 3+4 2 | 3+4 2 | 3+3 2 | 3+3 2 |
| | 5+6 2 | 5+6 2 | 5+6 2 | 5+6 2 | 2+2 2 | 2+2 2 |
| | 3+4 2 | 3+4 2 | 5+6 2 | 3+4 2 | 3+3 2 | 2+2 2 |

Continued on next column

**TABLE XXIV– continued**

| TQC | ICSD | | NoE | | SF | |
|---|---|---|---|---|---|---|
| | 5+6 2 | 5+6 2 | 3+4 2 | 5+6 2 | 2+2 2 | 3+3 2 |
| | 5+6 2 | 5+6 2 | 5+6 2 | 3+4 2 | 2+2 2 | 2+2 2 |
| | 3+4 2 | 3+4 2 | 3+4 2 | 5+6 2 | 3+3 2 | 3+3 2 |
| | 3+4 2 | 3+4 2 | 5+6 2 | 3+4 2 | 2+2 2 | 2+2 2 |
| | 5+6 2 | 5+6 2 | 3+4 2 | 5+6 2 | 3+3 2 | 3+3 2 |
| | 5+6 2 | 5+6 2 | 5+6 2 | 3+4 2 | 3+3 2 | 2+2 2 |
| | 5+6 2 | 5+6 2 | 5+6 2 | 5+6 2 | 2+2 2 | 2+2 2 |
| | 3+4 2 | 5+6 2 | 3+4 2 | 5+6 2 | 2+2 2 | 3+3 2 |
| | 5+6 2 | 3+4 2 | 5+6 2 | 5+6 2 | 2+2 2 | 3+3 2 |
| | 3+4 2 | 3+4 2 | 3+4 2 | 3+4 2 | 3+3 2 | 3+3 2 |
| | 5+6 2 | 3+4 2 | 3+4 2 | 5+6 2 | 2+2 2 | 2+2 2 |
| 12.1.144 | 618692 | | 56 | | C3.5Pu2W2 | |
| | -A | -GM | -M | -Y | -L | -V |
| | 3+4 2 | 3+4 2 | 5+6 2 | 5+6 2 | 3+3 2 | 2+2 2 |
| | 5+6 2 | 5+6 2 | 3+4 2 | 3+4 2 | 2+2 2 | 3+3 2 |
| | 3+4 2 | 3+4 2 | 5+6 2 | 5+6 2 | 3+3 2 | 2+2 2 |
| | 5+6 2 | 5+6 2 | 3+4 2 | 3+4 2 | 2+2 2 | 3+3 2 |
| | 5+6 2 | 3+4 2 | 5+6 2 | 3+4 2 | 2+2 2 | 3+3 2 |
| | 3+4 2 | 5+6 2 | 3+4 2 | 5+6 2 | 3+3 2 | 2+2 2 |
| | 3+4 2 | 3+4 2 | 5+6 2 | 5+6 2 | 3+3 2 | 2+2 2 |
| | 5+6 2 | 5+6 2 | 3+4 2 | 3+4 2 | 2+2 2 | 3+3 2 |
| | 5+6 2 | 3+4 2 | 3+4 2 | 5+6 2 | 3+3 2 | 2+2 2 |
| | 3+4 2 | 5+6 2 | 5+6 2 | 3+4 2 | 2+2 2 | 3+3 2 |
| | 5+6 2 | 3+4 2 | 5+6 2 | 3+4 2 | 2+2 2 | 3+3 2 |
| | 3+4 2 | 5+6 2 | 5+6 2 | 3+4 2 | 3+3 2 | 2+2 2 |
| | 5+6 2 | 5+6 2 | 3+4 2 | 5+6 2 | 2+2 2 | 3+3 2 |
| | 3+4 2 | 3+4 2 | 3+4 2 | 5+6 2 | 3+3 2 | 2+2 2 |
| | 3+4 2 | 5+6 2 | 3+4 2 | 3+4 2 | 3+3 2 | 3+3 2 |
| | 5+6 2 | 3+4 2 | 5+6 2 | 5+6 2 | 2+2 2 | 2+2 2 |
| | 3+4 2 | 5+6 2 | 5+6 2 | 3+4 2 | 3+3 2 | 3+3 2 |
| | 5+6 2 | 3+4 2 | 3+4 2 | 5+6 2 | 2+2 2 | 2+2 2 |
| | 3+4 2 | 5+6 2 | 3+4 2 | 5+6 2 | 3+3 2 | 2+2 2 |
| | 3+4 2 | 5+6 2 | 3+4 2 | 5+6 2 | 3+3 2 | 2+2 2 |
| | 5+6 2 | 3+4 2 | 3+4 2 | 3+4 2 | 2+2 2 | 3+3 2 |
| | 5+6 2 | 3+4 2 | 5+6 2 | 3+4 2 | 2+2 2 | 3+3 2 |
| | 3+4 2 | 5+6 2 | 3+4 2 | 5+6 2 | 3+3 2 | 2+2 2 |
| | 5+6 2 | 3+4 2 | 3+4 2 | 3+4 2 | 3+3 2 | 3+3 2 |
| | 3+4 2 | 5+6 2 | 5+6 2 | 5+6 2 | 2+2 2 | 2+2 2 |
| | 5+6 2 | 3+4 2 | 5+6 2 | 5+6 2 | 3+3 2 | 3+3 2 |
| | 3+4 2 | 3+4 2 | 3+4 2 | 3+4 2 | 2+2 2 | 2+2 2 |
| | 3+4 2 | 5+6 2 | 5+6 2 | 3+4 2 | 2+2 2 | 3+3 2 |
| 12.1.145 | 262897 | | 126 | | C1H3La1O6P2 | |
| | -A | -GM | -M | -Y | -L | -V |
| | 3+4 2 | 3+4 2 | 5+6 2 | 5+6 2 | 2+2 2 | 2+2 2 |
| | 5+6 2 | 5+6 2 | 3+4 2 | 3+4 2 | 3+3 2 | 3+3 2 |
| | 5+6 2 | 3+4 2 | 5+6 2 | 3+4 2 | 3+3 2 | 3+3 2 |
| | 3+4 2 | 5+6 2 | 3+4 2 | 5+6 2 | 2+2 2 | 2+2 2 |
| | 3+4 2 | 3+4 2 | 3+4 2 | 5+6 2 | 2+2 2 | 2+2 2 |
| | 5+6 2 | 5+6 2 | 5+6 2 | 3+4 2 | 3+3 2 | 3+3 2 |
| | 3+4 2 | 5+6 2 | 5+6 2 | 5+6 2 | 2+2 2 | 3+3 2 |
| | 5+6 2 | 3+4 2 | 3+4 2 | 3+4 2 | 3+3 2 | 2+2 2 |
| | 3+4 2 | 3+4 2 | 5+6 2 | 5+6 2 | 3+3 2 | 2+2 2 |
| | 5+6 2 | 5+6 2 | 3+4 2 | 3+4 2 | 2+2 2 | 3+3 2 |
| | 3+4 2 | 3+4 2 | 5+6 2 | 5+6 2 | 2+2 2 | 3+3 2 |
| | 5+6 2 | 5+6 2 | 3+4 2 | 3+4 2 | 3+3 2 | 2+2 2 |
| | 3+4 2 | 3+4 2 | 3+4 2 | 5+6 2 | 3+3 2 | 2+2 2 |
| | 5+6 2 | 5+6 2 | 5+6 2 | 3+4 2 | 2+2 2 | 3+3 2 |
| | 5+6 2 | 5+6 2 | 3+4 2 | 3+4 2 | 3+3 2 | 3+3 2 |
| | 3+4 2 | 3+4 2 | 5+6 2 | 5+6 2 | 2+2 2 | 2+2 2 |
| | 5+6 2 | 5+6 2 | 5+6 2 | 5+6 2 | 2+2 2 | 3+3 2 |
| | 3+4 2 | 3+4 2 | 3+4 2 | 3+4 2 | 3+3 2 | 2+2 2 |

Continued on next column

**TABLE XXIV– continued**

| TQC | ICSD | NoE | SF | | | |
|---|---|---|---|---|---|---|
| | 5+6 2 | 3+4 2 | 5+6 2 | 5+6 2 | 3+3 2 | 2+2 2 |
| | 3+4 2 | 5+6 2 | 3+4 2 | 3+4 2 | 2+2 2 | 3+3 2 |
| | 3+4 2 | 3+4 2 | 3+4 2 | 3+4 2 | 2+2 2 | 2+2 2 |
| | 5+6 2 | 5+6 2 | 5+6 2 | 5+6 2 | 3+3 2 | 3+3 2 |
| | 3+4 2 | 5+6 2 | 5+6 2 | 5+6 2 | 3+3 2 | 2+2 2 |
| | 5+6 2 | 3+4 2 | 3+4 2 | 3+4 2 | 2+2 2 | 3+3 2 |
| | 3+4 2 | 3+4 2 | 5+6 2 | 3+4 2 | 2+2 2 | 3+3 2 |
| | 5+6 2 | 5+6 2 | 3+4 2 | 5+6 2 | 3+3 2 | 2+2 2 |
| | 5+6 2 | 3+4 2 | 5+6 2 | 3+4 2 | 3+3 2 | 2+2 2 |
| | 3+4 2 | 3+4 2 | 3+4 2 | 5+6 2 | 2+2 2 | 3+3 2 |
| | 3+4 2 | 5+6 2 | 5+6 2 | 5+6 2 | 2+2 2 | 2+2 2 |
| | 5+6 2 | 5+6 2 | 3+4 2 | 3+4 2 | 3+3 2 | 3+3 2 |
| | 5+6 2 | 5+6 2 | 5+6 2 | 5+6 2 | 2+2 2 | 3+3 2 |
| | 3+4 2 | 3+4 2 | 3+4 2 | 5+6 2 | 3+3 2 | 2+2 2 |
| | 3+4 2 | 5+6 2 | 5+6 2 | 5+6 2 | 3+3 2 | 3+3 2 |
| | 5+6 2 | 3+4 2 | 3+4 2 | 3+4 2 | 2+2 2 | 2+2 2 |
| | 3+4 2 | 3+4 2 | 3+4 2 | 3+4 2 | 3+3 2 | 3+3 2 |
| | 5+6 2 | 5+6 2 | 3+4 2 | 5+6 2 | 2+2 2 | 2+2 2 |
| | 3+4 2 | 3+4 2 | 5+6 2 | 3+4 2 | 3+3 2 | 2+2 2 |
| | 5+6 2 | 3+4 2 | 5+6 2 | 3+4 2 | 2+2 2 | 3+3 2 |
| | 3+4 2 | 5+6 2 | 5+6 2 | 3+4 2 | 3+3 2 | 2+2 2 |
| | 5+6 2 | 5+6 2 | 3+4 2 | 5+6 2 | 2+2 2 | 3+3 2 |
| | 3+4 2 | 5+6 2 | 5+6 2 | 3+4 2 | 2+2 2 | 3+3 2 |
| | 5+6 2 | 3+4 2 | 3+4 2 | 5+6 2 | 3+3 2 | 2+2 2 |
| | 3+4 2 | 5+6 2 | 5+6 2 | 3+4 2 | 3+3 2 | 3+3 2 |
| | 5+6 2 | 3+4 2 | 3+4 2 | 5+6 2 | 2+2 2 | 2+2 2 |
| | 5+6 2 | 5+6 2 | 3+4 2 | 5+6 2 | 2+2 2 | 2+2 2 |
| | 5+6 2 | 3+4 2 | 5+6 2 | 5+6 2 | 2+2 2 | 3+3 2 |
| | 3+4 2 | 3+4 2 | 3+4 2 | 3+4 2 | 3+3 2 | 3+3 2 |
| | 3+4 2 | 5+6 2 | 5+6 2 | 3+4 2 | 3+3 2 | 2+2 2 |
| | 3+4 2 | 3+4 2 | 5+6 2 | 3+4 2 | 2+2 2 | 3+3 2 |
| | 5+6 2 | 5+6 2 | 5+6 2 | 5+6 2 | 3+3 2 | 2+2 2 |
| | 3+4 2 | 5+6 2 | 3+4 2 | 5+6 2 | 3+3 2 | 2+2 2 |
| | 5+6 2 | 3+4 2 | 3+4 2 | 3+4 2 | 2+2 2 | 3+3 2 |
| | 5+6 2 | 5+6 2 | 5+6 2 | 5+6 2 | 2+2 2 | 2+2 2 |
| | 3+4 2 | 3+4 2 | 3+4 2 | 3+4 2 | 3+3 2 | 3+3 2 |
| | 3+4 2 | 5+6 2 | 5+6 2 | 5+6 2 | 3+3 2 | 2+2 2 |
| | 5+6 2 | 3+4 2 | 3+4 2 | 3+4 2 | 2+2 2 | 3+3 2 |
| | 3+4 2 | 5+6 2 | 3+4 2 | 5+6 2 | 3+3 2 | 2+2 2 |
| | 3+4 2 | 3+4 2 | 3+4 2 | 3+4 2 | 2+2 2 | 3+3 2 |
| | 5+6 2 | 3+4 2 | 5+6 2 | 5+6 2 | 2+2 2 | 3+3 2 |
| | 5+6 2 | 3+4 2 | 5+6 2 | 3+4 2 | 3+3 2 | 2+2 2 |
| | 5+6 2 | 5+6 2 | 5+6 2 | 5+6 2 | 2+2 2 | 3+3 2 |
| | 3+4 2 | 5+6 2 | 3+4 2 | 3+4 2 | 3+3 2 | 2+2 2 |
| | 3+4 2 | 3+4 2 | 5+6 2 | 5+6 2 | 3+3 2 | 3+3 2 |
| 12.1.146 | 79394 | 74 | C1Fe2Si2Tm2 | | | |
| | -A | -GM | -M | -Y | -L | -V |
| | 3+4 2 | 3+4 2 | 3+4 2 | 3+4 2 | 2+2 2 | 2+2 2 |
| | 5+6 2 | 5+6 2 | 5+6 2 | 5+6 2 | 3+3 2 | 3+3 2 |
| | 3+4 2 | 3+4 2 | 3+4 2 | 3+4 2 | 2+2 2 | 2+2 2 |
| | 5+6 2 | 5+6 2 | 5+6 2 | 5+6 2 | 3+3 2 | 3+3 2 |
| | 3+4 2 | 3+4 2 | 3+4 2 | 3+4 2 | 2+2 2 | 2+2 2 |
| | 5+6 2 | 3+4 2 | 5+6 2 | 5+6 2 | 3+3 2 | 3+3 2 |
| | 5+6 2 | 5+6 2 | 3+4 2 | 5+6 2 | 2+2 2 | 2+2 2 |
| | 3+4 2 | 5+6 2 | 5+6 2 | 3+4 2 | 3+3 2 | 3+3 2 |
| | 3+4 2 | 3+4 2 | 3+4 2 | 3+4 2 | 3+3 2 | 3+3 2 |
| | 5+6 2 | 3+4 2 | 5+6 2 | 3+4 2 | 2+2 2 | 3+3 2 |
| | 3+4 2 | 5+6 2 | 3+4 2 | 5+6 2 | 3+3 2 | 2+2 2 |
| | 5+6 2 | 5+6 2 | 5+6 2 | 5+6 2 | 2+2 2 | 2+2 2 |
| | 5+6 2 | 3+4 2 | 3+4 2 | 3+4 2 | 3+3 2 | 2+2 2 |
| | 5+6 2 | 5+6 2 | 5+6 2 | 5+6 2 | 2+2 2 | 3+3 2 |

Continued on next column

**TABLE XXIV– continued**

| TQC | ICSD | NoE | SF | | | |
|---|---|---|---|---|---|---|
| | 5+6 2 | 5+6 2 | 5+6 2 | 5+6 2 | 2+2 2 | 2+2 2 |
| | 5+6 2 | 5+6 2 | 5+6 2 | 5+6 2 | 2+2 2 | 2+2 2 |
| | 3+4 2 | 5+6 2 | 3+4 2 | 3+4 2 | 3+3 2 | 3+3 2 |
| | 3+4 2 | 3+4 2 | 3+4 2 | 3+4 2 | 2+2 2 | 2+2 2 |
| | 3+4 2 | 5+6 2 | 5+6 2 | 5+6 2 | 2+2 2 | 3+3 2 |
| | 5+6 2 | 3+4 2 | 5+6 2 | 5+6 2 | 3+3 2 | 3+3 2 |
| | 3+4 2 | 3+4 2 | 5+6 2 | 5+6 2 | 3+3 2 | 3+3 2 |
| | 3+4 2 | 5+6 2 | 3+4 2 | 3+4 2 | 2+2 2 | 2+2 2 |
| | 5+6 2 | 3+4 2 | 3+4 2 | 3+4 2 | 3+3 2 | 3+3 2 |
| | 5+6 2 | 3+4 2 | 5+6 2 | 5+6 2 | 2+2 2 | 2+2 2 |
| | 3+4 2 | 5+6 2 | 3+4 2 | 3+4 2 | 3+3 2 | 2+2 2 |
| | 3+4 2 | 5+6 2 | 5+6 2 | 5+6 2 | 2+2 2 | 3+3 2 |
| | 5+6 2 | 3+4 2 | 5+6 2 | 3+4 2 | 3+3 2 | 2+2 2 |
| | 3+4 2 | 3+4 2 | 3+4 2 | 5+6 2 | 2+2 2 | 3+3 2 |
| | 3+4 2 | 3+4 2 | 5+6 2 | 3+4 2 | 3+3 2 | 3+3 2 |
| | 3+4 2 | 5+6 2 | 3+4 2 | 3+4 2 | 3+3 2 | 2+2 2 |
| | 5+6 2 | 3+4 2 | 5+6 2 | 5+6 2 | 2+2 2 | 3+3 2 |
| | 5+6 2 | 5+6 2 | 3+4 2 | 3+4 2 | 2+2 2 | 2+2 2 |
| | 5+6 2 | 3+4 2 | 5+6 2 | 5+6 2 | 3+3 2 | 2+2 2 |
| | 3+4 2 | 5+6 2 | 3+4 2 | 5+6 2 | 2+2 2 | 3+3 2 |
| | 3+4 2 | 3+4 2 | 3+4 2 | 3+4 2 | 3+3 2 | 2+2 2 |
| | 5+6 2 | 5+6 2 | 5+6 2 | 3+4 2 | 2+2 2 | 3+3 2 |
| | 3+4 2 | 3+4 2 | 3+4 2 | 5+6 2 | 3+3 2 | 3+3 2 |
| 12.1.147 | 195010 | 64 | C1Fe2Gd2Si2 | | | |
| | -A | -GM | -M | -Y | -L | -V |
| | 5+6 2 | 3+4 2 | 3+4 2 | 3+4 2 | 3+3 2 | 2+2 2 |
| | 3+4 2 | 5+6 2 | 5+6 2 | 5+6 2 | 2+2 2 | 3+3 2 |
| | 5+6 2 | 3+4 2 | 3+4 2 | 3+4 2 | 3+3 2 | 2+2 2 |
| | 3+4 2 | 5+6 2 | 5+6 2 | 5+6 2 | 2+2 2 | 3+3 2 |
| | 5+6 2 | 3+4 2 | 3+4 2 | 3+4 2 | 3+3 2 | 2+2 2 |
| | 3+4 2 | 3+4 2 | 5+6 2 | 5+6 2 | 2+2 2 | 3+3 2 |
| | 5+6 2 | 5+6 2 | 3+4 2 | 3+4 2 | 3+3 2 | 2+2 2 |
| | 3+4 2 | 5+6 2 | 5+6 2 | 5+6 2 | 2+2 2 | 3+3 2 |
| | 3+4 2 | 3+4 2 | 3+4 2 | 3+4 2 | 3+3 2 | 3+3 2 |
| | 5+6 2 | 3+4 2 | 5+6 2 | 5+6 2 | 2+2 2 | 3+3 2 |
| | 3+4 2 | 5+6 2 | 3+4 2 | 3+4 2 | 3+3 2 | 2+2 2 |
| | 5+6 2 | 5+6 2 | 5+6 2 | 5+6 2 | 2+2 2 | 2+2 2 |
| | 5+6 2 | 3+4 2 | 5+6 2 | 3+4 2 | 3+3 2 | 2+2 2 |
| | 5+6 2 | 5+6 2 | 3+4 2 | 5+6 2 | 2+2 2 | 3+3 2 |
| | 5+6 2 | 5+6 2 | 5+6 2 | 5+6 2 | 2+2 2 | 2+2 2 |
| | 3+4 2 | 3+4 2 | 5+6 2 | 5+6 2 | 2+2 2 | 2+2 2 |
| | 3+4 2 | 5+6 2 | 3+4 2 | 3+4 2 | 3+3 2 | 3+3 2 |
| | 3+4 2 | 3+4 2 | 5+6 2 | 3+4 2 | 2+2 2 | 3+3 2 |
| | 5+6 2 | 5+6 2 | 5+6 2 | 5+6 2 | 2+2 2 | 2+2 2 |
| | 5+6 2 | 3+4 2 | 3+4 2 | 5+6 2 | 3+3 2 | 3+3 2 |
| | 3+4 2 | 5+6 2 | 5+6 2 | 5+6 2 | 3+3 2 | 3+3 2 |
| | 3+4 2 | 3+4 2 | 3+4 2 | 3+4 2 | 3+3 2 | 3+3 2 |
| | 3+4 2 | 3+4 2 | 3+4 2 | 3+4 2 | 2+2 2 | 2+2 2 |
| | 5+6 2 | 5+6 2 | 5+6 2 | 5+6 2 | 3+3 2 | 3+3 2 |
| | 3+4 2 | 3+4 2 | 3+4 2 | 3+4 2 | 2+2 2 | 2+2 2 |
| | 5+6 2 | 5+6 2 | 5+6 2 | 5+6 2 | 2+2 2 | 2+2 2 |
| | 5+6 2 | 3+4 2 | 3+4 2 | 3+4 2 | 3+3 2 | 2+2 2 |
| | 3+4 2 | 5+6 2 | 5+6 2 | 5+6 2 | 2+2 2 | 3+3 2 |
| | 3+4 2 | 3+4 2 | 3+4 2 | 3+4 2 | 3+3 2 | 3+3 2 |
| | 5+6 2 | 5+6 2 | 5+6 2 | 5+6 2 | 3+3 2 | 2+2 2 |
| | 3+4 2 | 3+4 2 | 5+6 2 | 3+4 2 | 2+2 2 | 3+3 2 |
| | 5+6 2 | 5+6 2 | 3+4 2 | 5+6 2 | 2+2 2 | 2+2 2 |
| 12.1.148 | 67467 | 68 | C1Dy2Fe2Si2 | | | |
| | -A | -GM | -M | -Y | -L | -V |
| | 5+6 2 | 3+4 2 | 5+6 2 | 3+4 2 | 2+2 2 | 3+3 2 |
| | 3+4 2 | 5+6 2 | 3+4 2 | 5+6 2 | 3+3 2 | 2+2 2 |

Continued on next column

**TABLE XXIV– continued**

| TQC | ICSD | NoE | SF | | | |
|---|---|---|---|---|---|---|
| | 5+6 2 | 3+4 2 | 5+6 2 | 3+4 2 | 2+2 2 | 3+3 2 |
| | 3+4 2 | 5+6 2 | 3+4 2 | 5+6 2 | 3+3 2 | 2+2 2 |
| | 5+6 2 | 3+4 2 | 5+6 2 | 3+4 2 | 2+2 2 | 3+3 2 |
| | 3+4 2 | 3+4 2 | 3+4 2 | 5+6 2 | 3+3 2 | 2+2 2 |
| | 3+4 2 | 5+6 2 | 5+6 2 | 5+6 2 | 2+2 2 | 3+3 2 |
| | 5+6 2 | 5+6 2 | 3+4 2 | 3+4 2 | 3+3 2 | 2+2 2 |
| | 5+6 2 | 3+4 2 | 5+6 2 | 3+4 2 | 3+3 2 | 2+2 2 |
| | 3+4 2 | 3+4 2 | 3+4 2 | 3+4 2 | 2+2 2 | 2+2 2 |
| | 5+6 2 | 5+6 2 | 5+6 2 | 5+6 2 | 3+3 2 | 3+3 2 |
| | 3+4 2 | 5+6 2 | 3+4 2 | 5+6 2 | 2+2 2 | 3+3 2 |
| | 3+4 2 | 3+4 2 | 5+6 2 | 3+4 2 | 3+3 2 | 3+3 2 |
| | 3+4 2 | 5+6 2 | 3+4 2 | 5+6 2 | 2+2 2 | 2+2 2 |
| | 3+4 2 | 5+6 2 | 3+4 2 | 5+6 2 | 2+2 2 | 3+3 2 |
| | 3+4 2 | 5+6 2 | 3+4 2 | 5+6 2 | 2+2 2 | 3+3 2 |
| | 5+6 2 | 5+6 2 | 5+6 2 | 3+4 2 | 3+3 2 | 2+2 2 |
| | 5+6 2 | 3+4 2 | 5+6 2 | 3+4 2 | 2+2 2 | 3+3 2 |
| | 5+6 2 | 5+6 2 | 3+4 2 | 5+6 2 | 2+2 2 | 2+2 2 |
| | 3+4 2 | 3+4 2 | 3+4 2 | 5+6 2 | 3+3 2 | 2+2 2 |
| | 5+6 2 | 3+4 2 | 3+4 2 | 5+6 2 | 3+3 2 | 2+2 2 |
| | 5+6 2 | 3+4 2 | 5+6 2 | 3+4 2 | 2+2 2 | 3+3 2 |
| | 5+6 2 | 3+4 2 | 3+4 2 | 3+4 2 | 3+3 2 | 2+2 2 |
| | 3+4 2 | 5+6 2 | 5+6 2 | 5+6 2 | 3+3 2 | 3+3 2 |
| | 5+6 2 | 5+6 2 | 5+6 2 | 3+4 2 | 2+2 2 | 2+2 2 |
| | 3+4 2 | 3+4 2 | 3+4 2 | 5+6 2 | 2+2 2 | 3+3 2 |
| | 3+4 2 | 3+4 2 | 3+4 2 | 3+4 2 | 3+3 2 | 3+3 2 |
| | 5+6 2 | 5+6 2 | 5+6 2 | 5+6 2 | 2+2 2 | 2+2 2 |
| | 5+6 2 | 3+4 2 | 5+6 2 | 3+4 2 | 3+3 2 | 2+2 2 |
| | 5+6 2 | 5+6 2 | 3+4 2 | 3+4 2 | 3+3 2 | 3+3 2 |
| | 3+4 2 | 3+4 2 | 3+4 2 | 5+6 2 | 2+2 2 | 2+2 2 |
| | 3+4 2 | 5+6 2 | 5+6 2 | 3+4 2 | 2+2 2 | 3+3 2 |
| | 3+4 2 | 3+4 2 | 3+4 2 | 5+6 2 | 3+3 2 | 3+3 2 |
| | 5+6 2 | 5+6 2 | 5+6 2 | 5+6 2 | 2+2 2 | 2+2 2 |
| 12.1.149 | 391256 | 62 | C1Cl2Eu2N2 | | | |
| | -A | -GM | -M | -Y | -L | -V |
| | 5+6 2 | 3+4 2 | 3+4 2 | 5+6 2 | 2+2 2 | 2+2 2 |
| | 3+4 2 | 5+6 2 | 5+6 2 | 3+4 2 | 3+3 2 | 3+3 2 |
| | 5+6 2 | 3+4 2 | 3+4 2 | 5+6 2 | 2+2 2 | 3+3 2 |
| | 3+4 2 | 5+6 2 | 5+6 2 | 3+4 2 | 3+3 2 | 2+2 2 |
| | 5+6 2 | 3+4 2 | 5+6 2 | 3+4 2 | 2+2 2 | 3+3 2 |
| | 3+4 2 | 5+6 2 | 3+4 2 | 5+6 2 | 3+3 2 | 2+2 2 |
| | 3+4 2 | 3+4 2 | 3+4 2 | 5+6 2 | 2+2 2 | 2+2 2 |
| | 5+6 2 | 3+4 2 | 5+6 2 | 3+4 2 | 3+3 2 | 3+3 2 |
| | 5+6 2 | 5+6 2 | 5+6 2 | 3+4 2 | 2+2 2 | 3+3 2 |
| | 3+4 2 | 5+6 2 | 3+4 2 | 5+6 2 | 3+3 2 | 2+2 2 |
| | 3+4 2 | 3+4 2 | 5+6 2 | 5+6 2 | 2+2 2 | 2+2 2 |
| | 5+6 2 | 5+6 2 | 3+4 2 | 3+4 2 | 3+3 2 | 3+3 2 |
| | 5+6 2 | 3+4 2 | 5+6 2 | 3+4 2 | 2+2 2 | 3+3 2 |
| | 3+4 2 | 5+6 2 | 3+4 2 | 5+6 2 | 3+3 2 | 2+2 2 |
| | 3+4 2 | 5+6 2 | 3+4 2 | 5+6 2 | 3+3 2 | 2+2 2 |
| | 3+4 2 | 5+6 2 | 3+4 2 | 5+6 2 | 3+3 2 | 2+2 2 |
| | 3+4 2 | 3+4 2 | 5+6 2 | 3+4 2 | 2+2 2 | 3+3 2 |
| | 3+4 2 | 3+4 2 | 3+4 2 | 5+6 2 | 2+2 2 | 2+2 2 |
| | 5+6 2 | 3+4 2 | 5+6 2 | 5+6 2 | 3+3 2 | 2+2 2 |
| | 3+4 2 | 5+6 2 | 5+6 2 | 3+4 2 | 2+2 2 | 3+3 2 |
| | 5+6 2 | 5+6 2 | 3+4 2 | 5+6 2 | 3+3 2 | 3+3 2 |
| | 5+6 2 | 5+6 2 | 3+4 2 | 3+4 2 | 3+3 2 | 2+2 2 |
| | 5+6 2 | 3+4 2 | 5+6 2 | 3+4 2 | 2+2 2 | 3+3 2 |
| | 5+6 2 | 3+4 2 | 5+6 2 | 3+4 2 | 2+2 2 | 3+3 2 |
| | 3+4 2 | 3+4 2 | 3+4 2 | 5+6 2 | 3+3 2 | 2+2 2 |
| | 5+6 2 | 3+4 2 | 5+6 2 | 3+4 2 | 2+2 2 | 3+3 2 |
| | 3+4 2 | 5+6 2 | 3+4 2 | 5+6 2 | 3+3 2 | 2+2 2 |

Continued on next column

**TABLE XXIV– continued**

| TQC | ICSD | NoE | SF | | | |
|---|---|---|---|---|---|---|
| | 5+6 2 | 5+6 2 | 5+6 2 | 3+4 2 | 2+2 2 | 3+3 2 |
| | 3+4 2 | 3+4 2 | 3+4 2 | 5+6 2 | 3+3 2 | 2+2 2 |
| | 5+6 2 | 5+6 2 | 5+6 2 | 3+4 2 | 2+2 2 | 3+3 2 |
| | 3+4 2 | 3+4 2 | 3+4 2 | 3+4 2 | 3+3 2 | 2+2 2 |
| 12.1.150 | 603998 | 52 | C1.66Mo1U1 | | | |
| | -A | -GM | -M | -Y | -L | -V |
| | 3+4 2 | 3+4 2 | 5+6 2 | 5+6 2 | 3+3 2 | 2+2 2 |
| | 5+6 2 | 5+6 2 | 3+4 2 | 3+4 2 | 2+2 2 | 3+3 2 |
| | 3+4 2 | 3+4 2 | 5+6 2 | 3+4 2 | 3+3 2 | 2+2 2 |
| | 5+6 2 | 5+6 2 | 3+4 2 | 5+6 2 | 2+2 2 | 3+3 2 |
| | 5+6 2 | 3+4 2 | 5+6 2 | 3+4 2 | 2+2 2 | 3+3 2 |
| | 3+4 2 | 5+6 2 | 3+4 2 | 5+6 2 | 3+3 2 | 2+2 2 |
| | 3+4 2 | 3+4 2 | 5+6 2 | 5+6 2 | 3+3 2 | 2+2 2 |
| | 5+6 2 | 5+6 2 | 3+4 2 | 3+4 2 | 2+2 2 | 3+3 2 |
| | 5+6 2 | 3+4 2 | 5+6 2 | 3+4 2 | 3+3 2 | 3+3 2 |
| | 3+4 2 | 5+6 2 | 3+4 2 | 5+6 2 | 2+2 2 | 2+2 2 |
| | 5+6 2 | 3+4 2 | 5+6 2 | 3+4 2 | 2+2 2 | 3+3 2 |
| | 3+4 2 | 5+6 2 | 5+6 2 | 3+4 2 | 3+3 2 | 2+2 2 |
| | 5+6 2 | 5+6 2 | 3+4 2 | 5+6 2 | 2+2 2 | 2+2 2 |
| | 3+4 2 | 5+6 2 | 3+4 2 | 5+6 2 | 3+3 2 | 3+3 2 |
| | 3+4 2 | 3+4 2 | 3+4 2 | 3+4 2 | 3+3 2 | 3+3 2 |
| | 5+6 2 | 3+4 2 | 5+6 2 | 5+6 2 | 2+2 2 | 2+2 2 |
| | 3+4 2 | 5+6 2 | 5+6 2 | 3+4 2 | 3+3 2 | 3+3 2 |
| | 5+6 2 | 3+4 2 | 3+4 2 | 5+6 2 | 2+2 2 | 2+2 2 |
| | 3+4 2 | 5+6 2 | 3+4 2 | 5+6 2 | 3+3 2 | 2+2 2 |
| | 3+4 2 | 5+6 2 | 3+4 2 | 5+6 2 | 3+3 2 | 2+2 2 |
| | 5+6 2 | 3+4 2 | 3+4 2 | 3+4 2 | 2+2 2 | 3+3 2 |
| | 5+6 2 | 3+4 2 | 5+6 2 | 5+6 2 | 2+2 2 | 3+3 2 |
| | 3+4 2 | 3+4 2 | 3+4 2 | 3+4 2 | 3+3 2 | 2+2 2 |
| | 5+6 2 | 5+6 2 | 3+4 2 | 5+6 2 | 3+3 2 | 3+3 2 |
| | 5+6 2 | 5+6 2 | 5+6 2 | 3+4 2 | 2+2 2 | 2+2 2 |
| | 3+4 2 | 3+4 2 | 5+6 2 | 5+6 2 | 3+3 2 | 3+3 2 |
| 12.1.151 | 416907 | 156 | C10Cl4La8 | | | |
| | -A | -GM | -M | -Y | -L | -V |
| | 5+6 2 | 3+4 2 | 5+6 2 | 3+4 2 | 2+2 2 | 3+3 2 |
| | 3+4 2 | 3+4 2 | 5+6 2 | 5+6 2 | 2+2 2 | 2+2 2 |
| | 5+6 2 | 5+6 2 | 3+4 2 | 3+4 2 | 3+3 2 | 3+3 2 |
| | 3+4 2 | 5+6 2 | 3+4 2 | 5+6 2 | 3+3 2 | 2+2 2 |
| | 5+6 2 | 3+4 2 | 3+4 2 | 5+6 2 | 2+2 2 | 2+2 2 |
| | 3+4 2 | 5+6 2 | 5+6 2 | 3+4 2 | 3+3 2 | 3+3 2 |
| | 3+4 2 | 5+6 2 | 5+6 2 | 3+4 2 | 2+2 2 | 2+2 2 |
| | 5+6 2 | 3+4 2 | 3+4 2 | 5+6 2 | 3+3 2 | 3+3 2 |
| | 5+6 2 | 3+4 2 | 5+6 2 | 3+4 2 | 3+3 2 | 2+2 2 |
| | 3+4 2 | 5+6 2 | 3+4 2 | 5+6 2 | 2+2 2 | 3+3 2 |
| | 5+6 2 | 3+4 2 | 5+6 2 | 3+4 2 | 3+3 2 | 2+2 2 |
| | 3+4 2 | 5+6 2 | 3+4 2 | 5+6 2 | 2+2 2 | 3+3 2 |
| | 3+4 2 | 3+4 2 | 5+6 2 | 3+4 2 | 3+3 2 | 2+2 2 |
| | 5+6 2 | 5+6 2 | 3+4 2 | 5+6 2 | 2+2 2 | 3+3 2 |
| | 5+6 2 | 3+4 2 | 3+4 2 | 5+6 2 | 3+3 2 | 2+2 2 |
| | 3+4 2 | 5+6 2 | 5+6 2 | 3+4 2 | 2+2 2 | 3+3 2 |
| | 3+4 2 | 5+6 2 | 5+6 2 | 5+6 2 | 2+2 2 | 3+3 2 |
| | 5+6 2 | 3+4 2 | 3+4 2 | 3+4 2 | 3+3 2 | 2+2 2 |
| | 3+4 2 | 3+4 2 | 3+4 2 | 3+4 2 | 2+2 2 | 2+2 2 |
| | 5+6 2 | 5+6 2 | 5+6 2 | 5+6 2 | 3+3 2 | 3+3 2 |
| | 5+6 2 | 3+4 2 | 5+6 2 | 3+4 2 | 3+3 2 | 2+2 2 |
| | 3+4 2 | 5+6 2 | 3+4 2 | 5+6 2 | 2+2 2 | 3+3 2 |
| | 5+6 2 | 3+4 2 | 5+6 2 | 3+4 2 | 3+3 2 | 2+2 2 |
| | 3+4 2 | 5+6 2 | 5+6 2 | 5+6 2 | 2+2 2 | 3+3 2 |
| | 5+6 2 | 3+4 2 | 3+4 2 | 3+4 2 | 3+3 2 | 2+2 2 |
| | 3+4 2 | 3+4 2 | 3+4 2 | 5+6 2 | 2+2 2 | 3+3 2 |
| | 3+4 2 | 5+6 2 | 5+6 2 | 5+6 2 | 2+2 2 | 3+3 2 |

Continued on next column

**TABLE XXIV– continued**

| TQC | ICSD | NoE | SF | | | |
|---|---|---|---|---|---|---|
| | 5+6 2 | 3+4 2 | 3+4 2 | 3+4 2 | 3+3 2 | 2+2 2 |
| | 5+6 2 | 5+6 2 | 5+6 2 | 5+6 2 | 2+2 2 | 3+3 2 |
| | 3+4 2 | 3+4 2 | 3+4 2 | 3+4 2 | 3+3 2 | 2+2 2 |
| | 5+6 2 | 5+6 2 | 3+4 2 | 5+6 2 | 2+2 2 | 2+2 2 |
| | 3+4 2 | 5+6 2 | 5+6 2 | 3+4 2 | 3+3 2 | 3+3 2 |
| | 5+6 2 | 3+4 2 | 5+6 2 | 3+4 2 | 3+3 2 | 2+2 2 |
| | 3+4 2 | 3+4 2 | 3+4 2 | 5+6 2 | 2+2 2 | 3+3 2 |
| | 5+6 2 | 5+6 2 | 3+4 2 | 5+6 2 | 2+2 2 | 3+3 2 |
| | 3+4 2 | 5+6 2 | 5+6 2 | 3+4 2 | 3+3 2 | 2+2 2 |
| | 3+4 2 | 3+4 2 | 5+6 2 | 5+6 2 | 3+3 2 | 3+3 2 |
| | 5+6 2 | 5+6 2 | 3+4 2 | 3+4 2 | 2+2 2 | 2+2 2 |
| | 5+6 2 | 3+4 2 | 5+6 2 | 3+4 2 | 3+3 2 | 2+2 2 |
| | 3+4 2 | 5+6 2 | 3+4 2 | 5+6 2 | 2+2 2 | 3+3 2 |
| | 5+6 2 | 3+4 2 | 5+6 2 | 3+4 2 | 3+3 2 | 2+2 2 |
| | 3+4 2 | 5+6 2 | 3+4 2 | 5+6 2 | 2+2 2 | 3+3 2 |
| | 5+6 2 | 5+6 2 | 3+4 2 | 3+4 2 | 3+3 2 | 2+2 2 |
| | 3+4 2 | 3+4 2 | 5+6 2 | 5+6 2 | 2+2 2 | 3+3 2 |
| | 5+6 2 | 5+6 2 | 5+6 2 | 3+4 2 | 3+3 2 | 3+3 2 |
| | 3+4 2 | 3+4 2 | 3+4 2 | 5+6 2 | 2+2 2 | 2+2 2 |
| | 5+6 2 | 3+4 2 | 5+6 2 | 3+4 2 | 3+3 2 | 2+2 2 |
| | 3+4 2 | 3+4 2 | 3+4 2 | 5+6 2 | 2+2 2 | 3+3 2 |
| | 5+6 2 | 5+6 2 | 3+4 2 | 3+4 2 | 3+3 2 | 2+2 2 |
| | 3+4 2 | 5+6 2 | 5+6 2 | 5+6 2 | 2+2 2 | 3+3 2 |
| | 3+4 2 | 5+6 2 | 3+4 2 | 5+6 2 | 2+2 2 | 2+2 2 |
| | 5+6 2 | 3+4 2 | 5+6 2 | 3+4 2 | 3+3 2 | 3+3 2 |
| | 5+6 2 | 3+4 2 | 3+4 2 | 3+4 2 | 3+3 2 | 3+3 2 |
| | 3+4 2 | 3+4 2 | 5+6 2 | 5+6 2 | 3+3 2 | 2+2 2 |
| | 5+6 2 | 5+6 2 | 3+4 2 | 3+4 2 | 2+2 2 | 3+3 2 |
| | 3+4 2 | 5+6 2 | 5+6 2 | 5+6 2 | 3+3 2 | 2+2 2 |
| | 3+4 2 | 5+6 2 | 3+4 2 | 5+6 2 | 2+2 2 | 3+3 2 |
| | 5+6 2 | 3+4 2 | 3+4 2 | 3+4 2 | 2+2 2 | 2+2 2 |
| | 3+4 2 | 3+4 2 | 5+6 2 | 5+6 2 | 2+2 2 | 3+3 2 |
| | 5+6 2 | 5+6 2 | 5+6 2 | 3+4 2 | 3+3 2 | 2+2 2 |
| | 5+6 2 | 5+6 2 | 3+4 2 | 3+4 2 | 3+3 2 | 3+3 2 |
| | 3+4 2 | 3+4 2 | 5+6 2 | 5+6 2 | 2+2 2 | 2+2 2 |
| | 3+4 2 | 5+6 2 | 3+4 2 | 5+6 2 | 2+2 2 | 3+3 2 |
| | 3+4 2 | 3+4 2 | 3+4 2 | 5+6 2 | 3+3 2 | 2+2 2 |
| | 5+6 2 | 5+6 2 | 5+6 2 | 3+4 2 | 2+2 2 | 3+3 2 |
| | 3+4 2 | 3+4 2 | 3+4 2 | 5+6 2 | 3+3 2 | 3+3 2 |
| | 5+6 2 | 5+6 2 | 5+6 2 | 3+4 2 | 2+2 2 | 2+2 2 |
| | 3+4 2 | 3+4 2 | 3+4 2 | 5+6 2 | 2+2 2 | 3+3 2 |
| | 5+6 2 | 5+6 2 | 5+6 2 | 3+4 2 | 3+3 2 | 2+2 2 |
| | 3+4 2 | 3+4 2 | 3+4 2 | 5+6 2 | 2+2 2 | 3+3 2 |
| | 5+6 2 | 5+6 2 | 5+6 2 | 5+6 2 | 3+3 2 | 2+2 2 |
| | 3+4 2 | 5+6 2 | 3+4 2 | 3+4 2 | 2+2 2 | 3+3 2 |
| | 5+6 2 | 3+4 2 | 5+6 2 | 3+4 2 | 3+3 2 | 2+2 2 |
| | 5+6 2 | 5+6 2 | 5+6 2 | 3+4 2 | 3+3 2 | 2+2 2 |
| | 3+4 2 | 3+4 2 | 5+6 2 | 5+6 2 | 3+3 2 | 3+3 2 |
| | 5+6 2 | 3+4 2 | 3+4 2 | 3+4 2 | 2+2 2 | 2+2 2 |
| | 5+6 2 | 3+4 2 | 5+6 2 | 5+6 2 | 2+2 2 | 2+2 2 |
| | 3+4 2 | 3+4 2 | 5+6 2 | 3+4 2 | 3+3 2 | 3+3 2 |
| 12.1.152 | 617015 | 90 | Bi2Rh3Si2 | | | |
| | -A | -GM | -M | -Y | -L | -V |
| | 5+6 2 | 3+4 2 | 5+6 2 | 3+4 2 | 2+2 2 | 3+3 2 |
| | 3+4 2 | 5+6 2 | 3+4 2 | 5+6 2 | 3+3 2 | 2+2 2 |
| | 5+6 2 | 3+4 2 | 5+6 2 | 3+4 2 | 2+2 2 | 3+3 2 |
| | 3+4 2 | 5+6 2 | 3+4 2 | 5+6 2 | 3+3 2 | 2+2 2 |
| | 5+6 2 | 3+4 2 | 3+4 2 | 5+6 2 | 2+2 2 | 3+3 2 |
| | 5+6 2 | 5+6 2 | 3+4 2 | 3+4 2 | 3+3 2 | 2+2 2 |
| | 3+4 2 | 3+4 2 | 5+6 2 | 5+6 2 | 3+3 2 | 2+2 2 |
| | 3+4 2 | 5+6 2 | 5+6 2 | 3+4 2 | 2+2 2 | 3+3 2 |

Continued on next column

**TABLE XXIV– continued**

| TQC | ICSD | NoE | SF | | | |
|---|---|---|---|---|---|---|
| | 5+6 2 | 3+4 2 | 3+4 2 | 5+6 2 | 3+3 2 | 2+2 2 |
| | 3+4 2 | 3+4 2 | 3+4 2 | 5+6 2 | 2+2 2 | 3+3 2 |
| | 5+6 2 | 5+6 2 | 5+6 2 | 3+4 2 | 3+3 2 | 2+2 2 |
| | 5+6 2 | 3+4 2 | 3+4 2 | 5+6 2 | 2+2 2 | 3+3 2 |
| | 3+4 2 | 3+4 2 | 5+6 2 | 5+6 2 | 2+2 2 | 3+3 2 |
| | 5+6 2 | 5+6 2 | 3+4 2 | 5+6 2 | 3+3 2 | 2+2 2 |
| | 5+6 2 | 3+4 2 | 3+4 2 | 3+4 2 | 2+2 2 | 3+3 2 |
| | 5+6 2 | 3+4 2 | 3+4 2 | 5+6 2 | 3+3 2 | 2+2 2 |
| | 3+4 2 | 3+4 2 | 5+6 2 | 3+4 2 | 2+2 2 | 3+3 2 |
| | 5+6 2 | 3+4 2 | 5+6 2 | 3+4 2 | 3+3 2 | 2+2 2 |
| | 3+4 2 | 3+4 2 | 3+4 2 | 5+6 2 | 2+2 2 | 3+3 2 |
| | 5+6 2 | 5+6 2 | 3+4 2 | 3+4 2 | 3+3 2 | 2+2 2 |
| | 5+6 2 | 5+6 2 | 5+6 2 | 3+4 2 | 2+2 2 | 3+3 2 |
| | 3+4 2 | 3+4 2 | 5+6 2 | 5+6 2 | 3+3 2 | 2+2 2 |
| | 5+6 2 | 3+4 2 | 3+4 2 | 5+6 2 | 2+2 2 | 3+3 2 |
| | 5+6 2 | 5+6 2 | 3+4 2 | 5+6 2 | 3+3 2 | 2+2 2 |
| | 3+4 2 | 5+6 2 | 5+6 2 | 3+4 2 | 3+3 2 | 3+3 2 |
| | 5+6 2 | 3+4 2 | 5+6 2 | 5+6 2 | 2+2 2 | 2+2 2 |
| | 3+4 2 | 5+6 2 | 3+4 2 | 3+4 2 | 3+3 2 | 3+3 2 |
| | 3+4 2 | 3+4 2 | 5+6 2 | 3+4 2 | 2+2 2 | 3+3 2 |
| | 3+4 2 | 3+4 2 | 3+4 2 | 5+6 2 | 2+2 2 | 2+2 2 |
| | 5+6 2 | 5+6 2 | 3+4 2 | 3+4 2 | 3+3 2 | 2+2 2 |
| | 5+6 2 | 5+6 2 | 5+6 2 | 5+6 2 | 3+3 2 | 3+3 2 |
| | 5+6 2 | 3+4 2 | 3+4 2 | 3+4 2 | 2+2 2 | 2+2 2 |
| | 3+4 2 | 5+6 2 | 3+4 2 | 5+6 2 | 3+3 2 | 3+3 2 |
| | 5+6 2 | 3+4 2 | 3+4 2 | 3+4 2 | 2+2 2 | 2+2 2 |
| | 3+4 2 | 5+6 2 | 5+6 2 | 5+6 2 | 3+3 2 | 2+2 2 |
| | 5+6 2 | 3+4 2 | 3+4 2 | 3+4 2 | 3+3 2 | 3+3 2 |
| | 5+6 2 | 3+4 2 | 5+6 2 | 5+6 2 | 2+2 2 | 3+3 2 |
| | 5+6 2 | 3+4 2 | 3+4 2 | 5+6 2 | 3+3 2 | 2+2 2 |
| | 5+6 2 | 5+6 2 | 5+6 2 | 5+6 2 | 2+2 2 | 3+3 2 |
| | 3+4 2 | 5+6 2 | 3+4 2 | 5+6 2 | 3+3 2 | 2+2 2 |
| | 3+4 2 | 3+4 2 | 3+4 2 | 3+4 2 | 2+2 2 | 3+3 2 |
| | 3+4 2 | 3+4 2 | 3+4 2 | 5+6 2 | 2+2 2 | 2+2 2 |
| | 5+6 2 | 5+6 2 | 3+4 2 | 5+6 2 | 3+3 2 | 2+2 2 |
| | 3+4 2 | 3+4 2 | 5+6 2 | 5+6 2 | 3+3 2 | 3+3 2 |
| | 5+6 2 | 5+6 2 | 5+6 2 | 3+4 2 | 2+2 2 | 2+2 2 |
| 12.1.153 | 420725 | 98 | Bi2Rh3S2 | | | |
| | -A | -GM | -M | -Y | -L | -V |
| | 5+6 2 | 3+4 2 | 5+6 2 | 5+6 2 | 2+2 2 | 3+3 2 |
| | 3+4 2 | 3+4 2 | 3+4 2 | 5+6 2 | 3+3 2 | 2+2 2 |
| | 3+4 2 | 5+6 2 | 3+4 2 | 3+4 2 | 3+3 2 | 2+2 2 |
| | 5+6 2 | 5+6 2 | 5+6 2 | 3+4 2 | 2+2 2 | 3+3 2 |
| | 5+6 2 | 3+4 2 | 3+4 2 | 3+4 2 | 2+2 2 | 3+3 2 |
| | 3+4 2 | 3+4 2 | 5+6 2 | 3+4 2 | 3+3 2 | 2+2 2 |
| | 5+6 2 | 5+6 2 | 3+4 2 | 5+6 2 | 3+3 2 | 2+2 2 |
| | 3+4 2 | 5+6 2 | 5+6 2 | 5+6 2 | 2+2 2 | 3+3 2 |
| | 5+6 2 | 3+4 2 | 3+4 2 | 5+6 2 | 3+3 2 | 2+2 2 |
| | 3+4 2 | 5+6 2 | 3+4 2 | 5+6 2 | 2+2 2 | 3+3 2 |
| | 5+6 2 | 3+4 2 | 5+6 2 | 5+6 2 | 3+3 2 | 2+2 2 |
| | 5+6 2 | 5+6 2 | 3+4 2 | 3+4 2 | 2+2 2 | 2+2 2 |
| | 3+4 2 | 3+4 2 | 3+4 2 | 5+6 2 | 3+3 2 | 3+3 2 |
| | 5+6 2 | 3+4 2 | 5+6 2 | 3+4 2 | 2+2 2 | 3+3 2 |
| | 3+4 2 | 3+4 2 | 5+6 2 | 5+6 2 | 2+2 2 | 3+3 2 |
| | 5+6 2 | 3+4 2 | 3+4 2 | 3+4 2 | 3+3 2 | 2+2 2 |
| | 3+4 2 | 5+6 2 | 5+6 2 | 3+4 2 | 2+2 2 | 3+3 2 |
| | 5+6 2 | 3+4 2 | 5+6 2 | 5+6 2 | 3+3 2 | 2+2 2 |
| | 5+6 2 | 5+6 2 | 3+4 2 | 5+6 2 | 2+2 2 | 3+3 2 |
| | 5+6 2 | 3+4 2 | 3+4 2 | 3+4 2 | 2+2 2 | 2+2 2 |
| | 3+4 2 | 5+6 2 | 5+6 2 | 5+6 2 | 3+3 2 | 3+3 2 |
| | 5+6 2 | 3+4 2 | 3+4 2 | 3+4 2 | 2+2 2 | 2+2 2 |

Continued on next column

**TABLE XXIV– continued**

| TQC | ICSD | NoE | SF | | | |
|---|---|---|---|---|---|---|
| | 3+4 2 | 3+4 2 | 5+6 2 | 3+4 2 | 3+3 2 | 3+3 2 |
| | 5+6 2 | 5+6 2 | 3+4 2 | 5+6 2 | 2+2 2 | 3+3 2 |
| | 5+6 2 | 3+4 2 | 3+4 2 | 3+4 2 | 3+3 2 | 2+2 2 |
| | 3+4 2 | 5+6 2 | 3+4 2 | 5+6 2 | 2+2 2 | 2+2 2 |
| | 5+6 2 | 3+4 2 | 5+6 2 | 3+4 2 | 3+3 2 | 3+3 2 |
| | 3+4 2 | 5+6 2 | 3+4 2 | 5+6 2 | 3+3 2 | 2+2 2 |
| | 5+6 2 | 3+4 2 | 5+6 2 | 5+6 2 | 2+2 2 | 3+3 2 |
| | 3+4 2 | 5+6 2 | 3+4 2 | 3+4 2 | 3+3 2 | 2+2 2 |
| | 5+6 2 | 3+4 2 | 5+6 2 | 5+6 2 | 2+2 2 | 2+2 2 |
| | 3+4 2 | 3+4 2 | 3+4 2 | 5+6 2 | 3+3 2 | 3+3 2 |
| | 3+4 2 | 5+6 2 | 5+6 2 | 3+4 2 | 2+2 2 | 2+2 2 |
| | 5+6 2 | 3+4 2 | 3+4 2 | 3+4 2 | 3+3 2 | 3+3 2 |
| | 3+4 2 | 5+6 2 | 5+6 2 | 5+6 2 | 3+3 2 | 2+2 2 |
| | 3+4 2 | 5+6 2 | 3+4 2 | 5+6 2 | 2+2 2 | 2+2 2 |
| | 5+6 2 | 3+4 2 | 3+4 2 | 3+4 2 | 3+3 2 | 3+3 2 |
| | 5+6 2 | 3+4 2 | 3+4 2 | 3+4 2 | 2+2 2 | 2+2 2 |
| | 5+6 2 | 3+4 2 | 5+6 2 | 5+6 2 | 2+2 2 | 3+3 2 |
| | 5+6 2 | 3+4 2 | 3+4 2 | 3+4 2 | 3+3 2 | 2+2 2 |
| | 5+6 2 | 3+4 2 | 5+6 2 | 5+6 2 | 3+3 2 | 3+3 2 |
| | 3+4 2 | 5+6 2 | 3+4 2 | 5+6 2 | 2+2 2 | 3+3 2 |
| | 5+6 2 | 3+4 2 | 5+6 2 | 5+6 2 | 2+2 2 | 2+2 2 |
| | 3+4 2 | 5+6 2 | 3+4 2 | 5+6 2 | 3+3 2 | 3+3 2 |
| | 5+6 2 | 3+4 2 | 5+6 2 | 3+4 2 | 3+3 2 | 3+3 2 |
| | 3+4 2 | 5+6 2 | 3+4 2 | 3+4 2 | 2+2 2 | 2+2 2 |
| | 5+6 2 | 5+6 2 | 3+4 2 | 5+6 2 | 3+3 2 | 2+2 2 |
| | 5+6 2 | 3+4 2 | 3+4 2 | 5+6 2 | 2+2 2 | 3+3 2 |
| | 3+4 2 | 5+6 2 | 5+6 2 | 5+6 2 | 2+2 2 | 3+3 2 |
| 12.1.154 | 261457 | 104 | Bi2Pd3Se2 | | | |
| | -A | -GM | -M | -Y | -L | -V |
| | 5+6 2 | 3+4 2 | 5+6 2 | 3+4 2 | 3+3 2 | 2+2 2 |
| | 3+4 2 | 5+6 2 | 3+4 2 | 5+6 2 | 2+2 2 | 3+3 2 |
| | 3+4 2 | 5+6 2 | 5+6 2 | 5+6 2 | 3+3 2 | 3+3 2 |
| | 5+6 2 | 3+4 2 | 3+4 2 | 3+4 2 | 2+2 2 | 2+2 2 |
| | 3+4 2 | 3+4 2 | 3+4 2 | 3+4 2 | 2+2 2 | 2+2 2 |
| | 5+6 2 | 5+6 2 | 5+6 2 | 5+6 2 | 3+3 2 | 3+3 2 |
| | 3+4 2 | 3+4 2 | 3+4 2 | 3+4 2 | 3+3 2 | 3+3 2 |
| | 5+6 2 | 5+6 2 | 5+6 2 | 5+6 2 | 2+2 2 | 2+2 2 |
| | 5+6 2 | 3+4 2 | 5+6 2 | 5+6 2 | 2+2 2 | 2+2 2 |
| | 3+4 2 | 5+6 2 | 5+6 2 | 5+6 2 | 3+3 2 | 3+3 2 |
| | 3+4 2 | 3+4 2 | 5+6 2 | 3+4 2 | 2+2 2 | 2+2 2 |
| | 3+4 2 | 5+6 2 | 3+4 2 | 5+6 2 | 3+3 2 | 3+3 2 |
| | 5+6 2 | 3+4 2 | 3+4 2 | 5+6 2 | 3+3 2 | 3+3 2 |
| | 3+4 2 | 3+4 2 | 5+6 2 | 5+6 2 | 2+2 2 | 2+2 2 |
| | 3+4 2 | 3+4 2 | 5+6 2 | 3+4 2 | 3+3 2 | 2+2 2 |
| | 5+6 2 | 3+4 2 | 3+4 2 | 3+4 2 | 2+2 2 | 3+3 2 |
| | 3+4 2 | 5+6 2 | 5+6 2 | 5+6 2 | 3+3 2 | 3+3 2 |
| | 5+6 2 | 3+4 2 | 3+4 2 | 3+4 2 | 3+3 2 | 3+3 2 |
| | 3+4 2 | 3+4 2 | 5+6 2 | 5+6 2 | 2+2 2 | 2+2 2 |
| | 3+4 2 | 5+6 2 | 3+4 2 | 3+4 2 | 2+2 2 | 2+2 2 |
| | 3+4 2 | 3+4 2 | 3+4 2 | 3+4 2 | 2+2 2 | 3+3 2 |
| | 5+6 2 | 3+4 2 | 5+6 2 | 5+6 2 | 3+3 2 | 2+2 2 |
| | 3+4 2 | 5+6 2 | 3+4 2 | 5+6 2 | 2+2 2 | 2+2 2 |
| | 5+6 2 | 3+4 2 | 5+6 2 | 5+6 2 | 3+3 2 | 3+3 2 |
| | 3+4 2 | 5+6 2 | 3+4 2 | 3+4 2 | 3+3 2 | 2+2 2 |
| | 5+6 2 | 3+4 2 | 5+6 2 | 5+6 2 | 2+2 2 | 3+3 2 |
| | 3+4 2 | 5+6 2 | 5+6 2 | 3+4 2 | 3+3 2 | 3+3 2 |
| | 3+4 2 | 3+4 2 | 5+6 2 | 5+6 2 | 2+2 2 | 2+2 2 |
| | 3+4 2 | 3+4 2 | 3+4 2 | 3+4 2 | 3+3 2 | 3+3 2 |
| | 5+6 2 | 5+6 2 | 5+6 2 | 5+6 2 | 2+2 2 | 2+2 2 |
| | 5+6 2 | 5+6 2 | 3+4 2 | 3+4 2 | 2+2 2 | 2+2 2 |
| | 3+4 2 | 3+4 2 | 5+6 2 | 5+6 2 | 3+3 2 | 3+3 2 |

Continued on next column

**TABLE XXIV– continued**

| TQC | ICSD | NoE | SF | | | |
|---|---|---|---|---|---|---|
| | 5+6 2 | 5+6 2 | 3+4 2 | 5+6 2 | 2+2 2 | 2+2 2 |
| | 3+4 2 | 5+6 2 | 5+6 2 | 3+4 2 | 3+3 2 | 3+3 2 |
| | 5+6 2 | 3+4 2 | 5+6 2 | 5+6 2 | 2+2 2 | 2+2 2 |
| | 3+4 2 | 5+6 2 | 5+6 2 | 5+6 2 | 3+3 2 | 3+3 2 |
| | 3+4 2 | 3+4 2 | 5+6 2 | 3+4 2 | 2+2 2 | 2+2 2 |
| | 5+6 2 | 3+4 2 | 3+4 2 | 5+6 2 | 2+2 2 | 3+3 2 |
| | 3+4 2 | 3+4 2 | 5+6 2 | 5+6 2 | 3+3 2 | 2+2 2 |
| | 5+6 2 | 3+4 2 | 5+6 2 | 3+4 2 | 2+2 2 | 3+3 2 |
| | 3+4 2 | 5+6 2 | 3+4 2 | 5+6 2 | 3+3 2 | 2+2 2 |
| | 5+6 2 | 3+4 2 | 5+6 2 | 5+6 2 | 3+3 2 | 3+3 2 |
| | 5+6 2 | 5+6 2 | 5+6 2 | 3+4 2 | 2+2 2 | 2+2 2 |
| | 3+4 2 | 3+4 2 | 3+4 2 | 5+6 2 | 3+3 2 | 3+3 2 |
| | 3+4 2 | 3+4 2 | 3+4 2 | 3+4 2 | 2+2 2 | 2+2 2 |
| | 5+6 2 | 5+6 2 | 5+6 2 | 3+4 2 | 3+3 2 | 3+3 2 |
| | 3+4 2 | 3+4 2 | 3+4 2 | 5+6 2 | 2+2 2 | 2+2 2 |
| | 3+4 2 | 5+6 2 | 5+6 2 | 3+4 2 | 3+3 2 | 3+3 2 |
| | 5+6 2 | 3+4 2 | 5+6 2 | 3+4 2 | 3+3 2 | 3+3 2 |
| | 3+4 2 | 3+4 2 | 3+4 2 | 5+6 2 | 2+2 2 | 2+2 2 |
| | 3+4 2 | 5+6 2 | 3+4 2 | 5+6 2 | 2+2 2 | 3+3 2 |
| | 5+6 2 | 3+4 2 | 5+6 2 | 3+4 2 | 3+3 2 | 2+2 2 |
| 12.1.155 | 251720 | 22 | Bi1Se1 | | | |
| | -A | -GM | -M | -Y | -L | -V |
| | 5+6 2 | 3+4 2 | 3+4 2 | 5+6 2 | 3+3 2 | 2+2 2 |
| | 3+4 2 | 5+6 2 | 5+6 2 | 3+4 2 | 2+2 2 | 3+3 2 |
| | 3+4 2 | 5+6 2 | 5+6 2 | 3+4 2 | 2+2 2 | 3+3 2 |
| | 5+6 2 | 3+4 2 | 3+4 2 | 5+6 2 | 3+3 2 | 2+2 2 |
| | 5+6 2 | 3+4 2 | 3+4 2 | 5+6 2 | 3+3 2 | 2+2 2 |
| | 5+6 2 | 3+4 2 | 5+6 2 | 3+4 2 | 2+2 2 | 3+3 2 |
| | 5+6 2 | 3+4 2 | 5+6 2 | 3+4 2 | 3+3 2 | 2+2 2 |
| | 3+4 2 | 5+6 2 | 3+4 2 | 5+6 2 | 2+2 2 | 3+3 2 |
| | 3+4 2 | 5+6 2 | 3+4 2 | 5+6 2 | 2+2 2 | 3+3 2 |
| | 3+4 2 | 5+6 2 | 5+6 2 | 3+4 2 | 3+3 2 | 2+2 2 |
| | 3+4 2 | 5+6 2 | 5+6 2 | 3+4 2 | 2+2 2 | 3+3 2 |
| 12.1.156 | 251717 | 22 | Bi1Se1 | | | |
| | -A | -GM | -M | -Y | -L | -V |
| | 5+6 2 | 3+4 2 | 3+4 2 | 5+6 2 | 2+2 2 | 3+3 2 |
| | 3+4 2 | 5+6 2 | 5+6 2 | 3+4 2 | 3+3 2 | 2+2 2 |
| | 3+4 2 | 5+6 2 | 5+6 2 | 3+4 2 | 3+3 2 | 2+2 2 |
| | 5+6 2 | 3+4 2 | 3+4 2 | 5+6 2 | 2+2 2 | 3+3 2 |
| | 5+6 2 | 3+4 2 | 3+4 2 | 5+6 2 | 2+2 2 | 3+3 2 |
| | 5+6 2 | 3+4 2 | 5+6 2 | 3+4 2 | 3+3 2 | 2+2 2 |
| | 5+6 2 | 3+4 2 | 5+6 2 | 3+4 2 | 2+2 2 | 3+3 2 |
| | 3+4 2 | 5+6 2 | 3+4 2 | 5+6 2 | 3+3 2 | 2+2 2 |
| | 3+4 2 | 5+6 2 | 3+4 2 | 5+6 2 | 3+3 2 | 2+2 2 |
| | 3+4 2 | 5+6 2 | 5+6 2 | 3+4 2 | 2+2 2 | 3+3 2 |
| | 3+4 2 | 5+6 2 | 5+6 2 | 3+4 2 | 3+3 2 | 2+2 2 |
| 12.1.157 | 1559 | 48 | Bi1I1 | | | |
| | -A | -GM | -M | -Y | -L | -V |
| | 5+6 2 | 3+4 2 | 3+4 2 | 5+6 2 | 2+2 2 | 3+3 2 |
| | 3+4 2 | 5+6 2 | 5+6 2 | 3+4 2 | 3+3 2 | 2+2 2 |
| | 3+4 2 | 3+4 2 | 5+6 2 | 3+4 2 | 2+2 2 | 3+3 2 |
| | 5+6 2 | 5+6 2 | 3+4 2 | 5+6 2 | 3+3 2 | 2+2 2 |
| | 3+4 2 | 3+4 2 | 5+6 2 | 5+6 2 | 3+3 2 | 3+3 2 |
| | 5+6 2 | 5+6 2 | 3+4 2 | 3+4 2 | 2+2 2 | 2+2 2 |
| | 3+4 2 | 3+4 2 | 5+6 2 | 5+6 2 | 2+2 2 | 2+2 2 |
| | 5+6 2 | 5+6 2 | 3+4 2 | 3+4 2 | 3+3 2 | 3+3 2 |
| | 5+6 2 | 3+4 2 | 3+4 2 | 5+6 2 | 3+3 2 | 3+3 2 |
| | 5+6 2 | 5+6 2 | 5+6 2 | 3+4 2 | 2+2 2 | 2+2 2 |
| | 3+4 2 | 3+4 2 | 3+4 2 | 5+6 2 | 3+3 2 | 2+2 2 |
| | 3+4 2 | 5+6 2 | 5+6 2 | 3+4 2 | 2+2 2 | 3+3 2 |
| | 5+6 2 | 3+4 2 | 5+6 2 | 3+4 2 | 3+3 2 | 3+3 2 |

Continued on next column

**TABLE XXIV– continued**

| TQC | ICSD | | NoE | | SF | |
|---|---|---|---|---|---|---|
| | 3+4 2 | 5+6 2 | 3+4 2 | 5+6 2 | 2+2 2 | 2+2 2 |
| | 5+6 2 | 3+4 2 | 3+4 2 | 3+4 2 | 2+2 2 | 2+2 2 |
| | 3+4 2 | 3+4 2 | 5+6 2 | 5+6 2 | 3+3 2 | 3+3 2 |
| | 5+6 2 | 5+6 2 | 5+6 2 | 3+4 2 | 2+2 2 | 2+2 2 |
| | 3+4 2 | 5+6 2 | 3+4 2 | 5+6 2 | 3+3 2 | 3+3 2 |
| | 3+4 2 | 3+4 2 | 5+6 2 | 5+6 2 | 2+2 2 | 3+3 2 |
| | 5+6 2 | 5+6 2 | 5+6 2 | 3+4 2 | 3+3 2 | 2+2 2 |
| | 3+4 2 | 3+4 2 | 3+4 2 | 3+4 2 | 2+2 2 | 3+3 2 |
| | 5+6 2 | 5+6 2 | 3+4 2 | 5+6 2 | 3+3 2 | 2+2 2 |
| | 3+4 2 | 3+4 2 | 5+6 2 | 5+6 2 | 2+2 2 | 2+2 2 |
| | 3+4 2 | 3+4 2 | 3+4 2 | 3+4 2 | 3+3 2 | 3+3 2 |
| 12.1.158 | 1560 | | 96 | | Bi1Br1 | |
| | -A | -GM | -M | -Y | -L | -V |
| | 5+6 2 | 3+4 2 | 5+6 2 | 3+4 2 | 3+3 2 | 2+2 2 |
| | 3+4 2 | 5+6 2 | 3+4 2 | 5+6 2 | 2+2 2 | 3+3 2 |
| | 3+4 2 | 3+4 2 | 3+4 2 | 3+4 2 | 2+2 2 | 2+2 2 |
| | 5+6 2 | 5+6 2 | 5+6 2 | 5+6 2 | 3+3 2 | 3+3 2 |
| | 5+6 2 | 3+4 2 | 5+6 2 | 3+4 2 | 3+3 2 | 2+2 2 |
| | 3+4 2 | 5+6 2 | 3+4 2 | 5+6 2 | 2+2 2 | 3+3 2 |
| | 3+4 2 | 3+4 2 | 3+4 2 | 3+4 2 | 2+2 2 | 2+2 2 |
| | 5+6 2 | 5+6 2 | 5+6 2 | 5+6 2 | 3+3 2 | 3+3 2 |
| | 3+4 2 | 3+4 2 | 5+6 2 | 5+6 2 | 2+2 2 | 3+3 2 |
| | 5+6 2 | 3+4 2 | 3+4 2 | 5+6 2 | 3+3 2 | 3+3 2 |
| | 3+4 2 | 5+6 2 | 5+6 2 | 3+4 2 | 3+3 2 | 2+2 2 |
| | 5+6 2 | 5+6 2 | 3+4 2 | 3+4 2 | 2+2 2 | 2+2 2 |
| | 5+6 2 | 3+4 2 | 3+4 2 | 5+6 2 | 2+2 2 | 2+2 2 |
| | 3+4 2 | 3+4 2 | 5+6 2 | 5+6 2 | 3+3 2 | 3+3 2 |
| | 3+4 2 | 5+6 2 | 5+6 2 | 3+4 2 | 3+3 2 | 2+2 2 |
| | 5+6 2 | 5+6 2 | 3+4 2 | 3+4 2 | 2+2 2 | 3+3 2 |
| | 3+4 2 | 5+6 2 | 5+6 2 | 3+4 2 | 2+2 2 | 3+3 2 |
| | 5+6 2 | 3+4 2 | 3+4 2 | 3+4 2 | 3+3 2 | 3+3 2 |
| | 5+6 2 | 5+6 2 | 3+4 2 | 5+6 2 | 3+3 2 | 2+2 2 |
| | 3+4 2 | 3+4 2 | 5+6 2 | 5+6 2 | 2+2 2 | 2+2 2 |
| | 3+4 2 | 5+6 2 | 3+4 2 | 3+4 2 | 3+3 2 | 2+2 2 |
| | 5+6 2 | 5+6 2 | 5+6 2 | 3+4 2 | 2+2 2 | 3+3 2 |
| | 5+6 2 | 3+4 2 | 5+6 2 | 5+6 2 | 2+2 2 | 2+2 2 |
| | 3+4 2 | 3+4 2 | 3+4 2 | 5+6 2 | 3+3 2 | 3+3 2 |
| | 5+6 2 | 3+4 2 | 3+4 2 | 5+6 2 | 2+2 2 | 2+2 2 |
| | 3+4 2 | 5+6 2 | 5+6 2 | 5+6 2 | 3+3 2 | 3+3 2 |
| | 3+4 2 | 3+4 2 | 5+6 2 | 3+4 2 | 3+3 2 | 2+2 2 |
| | 5+6 2 | 3+4 2 | 3+4 2 | 3+4 2 | 2+2 2 | 3+3 2 |
| | 5+6 2 | 5+6 2 | 5+6 2 | 5+6 2 | 3+3 2 | 2+2 2 |
| | 3+4 2 | 3+4 2 | 3+4 2 | 3+4 2 | 2+2 2 | 3+3 2 |
| | 5+6 2 | 5+6 2 | 5+6 2 | 3+4 2 | 2+2 2 | 2+2 2 |
| | 3+4 2 | 5+6 2 | 3+4 2 | 5+6 2 | 3+3 2 | 3+3 2 |
| | 3+4 2 | 3+4 2 | 3+4 2 | 5+6 2 | 2+2 2 | 2+2 2 |
| | 5+6 2 | 5+6 2 | 5+6 2 | 3+4 2 | 3+3 2 | 3+3 2 |
| | 3+4 2 | 3+4 2 | 5+6 2 | 3+4 2 | 2+2 2 | 2+2 2 |
| | 5+6 2 | 5+6 2 | 3+4 2 | 5+6 2 | 3+3 2 | 3+3 2 |
| | 5+6 2 | 3+4 2 | 3+4 2 | 5+6 2 | 3+3 2 | 3+3 2 |
| | 3+4 2 | 5+6 2 | 5+6 2 | 3+4 2 | 2+2 2 | 2+2 2 |
| | 3+4 2 | 3+4 2 | 3+4 2 | 3+4 2 | 3+3 2 | 3+3 2 |
| | 5+6 2 | 5+6 2 | 5+6 2 | 5+6 2 | 2+2 2 | 2+2 2 |
| | 3+4 2 | 3+4 2 | 5+6 2 | 5+6 2 | 3+3 2 | 3+3 2 |
| | 5+6 2 | 3+4 2 | 3+4 2 | 5+6 2 | 2+2 2 | 2+2 2 |
| | 5+6 2 | 5+6 2 | 3+4 2 | 3+4 2 | 3+3 2 | 2+2 2 |
| | 3+4 2 | 5+6 2 | 5+6 2 | 3+4 2 | 2+2 2 | 3+3 2 |
| | 5+6 2 | 3+4 2 | 3+4 2 | 5+6 2 | 3+3 2 | 2+2 2 |
| | 3+4 2 | 3+4 2 | 5+6 2 | 5+6 2 | 2+2 2 | 3+3 2 |
| | 5+6 2 | 3+4 2 | 5+6 2 | 3+4 2 | 2+2 2 | 2+2 2 |
| | 3+4 2 | 3+4 2 | 3+4 2 | 3+4 2 | 3+3 2 | 3+3 2 |

Continued on next column

**TABLE XXIV– continued**

| TQC | ICSD | | NoE | | SF | |
|---|---|---|---|---|---|---|
| 12.1.159 | 65614 | | 144 | | Ba3Cu2O10Pt1Y2 | |
| | -A | -GM | -M | -Y | -L | -V |
| | 3+4 2 | 3+4 2 | 5+6 2 | 5+6 2 | 2+2 2 | 2+2 2 |
| | 5+6 2 | 5+6 2 | 3+4 2 | 3+4 2 | 3+3 2 | 3+3 2 |
| | 5+6 2 | 3+4 2 | 5+6 2 | 3+4 2 | 2+2 2 | 3+3 2 |
| | 3+4 2 | 5+6 2 | 3+4 2 | 5+6 2 | 3+3 2 | 2+2 2 |
| | 3+4 2 | 3+4 2 | 3+4 2 | 3+4 2 | 3+3 2 | 3+3 2 |
| | 3+4 2 | 3+4 2 | 5+6 2 | 5+6 2 | 3+3 2 | 2+2 2 |
| | 5+6 2 | 5+6 2 | 3+4 2 | 3+4 2 | 2+2 2 | 3+3 2 |
| | 3+4 2 | 3+4 2 | 5+6 2 | 5+6 2 | 3+3 2 | 2+2 2 |
| | 5+6 2 | 5+6 2 | 3+4 2 | 3+4 2 | 2+2 2 | 3+3 2 |
| | 3+4 2 | 3+4 2 | 5+6 2 | 5+6 2 | 3+3 2 | 2+2 2 |
| | 5+6 2 | 5+6 2 | 3+4 2 | 3+4 2 | 2+2 2 | 3+3 2 |
| | 5+6 2 | 3+4 2 | 5+6 2 | 3+4 2 | 3+3 2 | 2+2 2 |
| | 3+4 2 | 5+6 2 | 3+4 2 | 5+6 2 | 2+2 2 | 2+2 2 |
| | 5+6 2 | 3+4 2 | 5+6 2 | 3+4 2 | 3+3 2 | 3+3 2 |
| | 5+6 2 | 3+4 2 | 5+6 2 | 3+4 2 | 3+3 2 | 2+2 2 |
| | 3+4 2 | 5+6 2 | 3+4 2 | 5+6 2 | 2+2 2 | 3+3 2 |
| | 3+4 2 | 5+6 2 | 3+4 2 | 5+6 2 | 2+2 2 | 3+3 2 |
| | 5+6 2 | 3+4 2 | 5+6 2 | 3+4 2 | 3+3 2 | 2+2 2 |
| | 3+4 2 | 5+6 2 | 3+4 2 | 5+6 2 | 2+2 2 | 3+3 2 |
| | 5+6 2 | 5+6 2 | 3+4 2 | 3+4 2 | 2+2 2 | 2+2 2 |
| | 3+4 2 | 3+4 2 | 5+6 2 | 5+6 2 | 3+3 2 | 3+3 2 |
| | 5+6 2 | 3+4 2 | 5+6 2 | 5+6 2 | 2+2 2 | 3+3 2 |
| | 3+4 2 | 5+6 2 | 3+4 2 | 3+4 2 | 3+3 2 | 2+2 2 |
| | 5+6 2 | 5+6 2 | 5+6 2 | 5+6 2 | 2+2 2 | 2+2 2 |
| | 5+6 2 | 3+4 2 | 5+6 2 | 5+6 2 | 2+2 2 | 3+3 2 |
| | 5+6 2 | 3+4 2 | 3+4 2 | 3+4 2 | 3+3 2 | 2+2 2 |
| | 3+4 2 | 5+6 2 | 5+6 2 | 3+4 2 | 2+2 2 | 3+3 2 |
| | 3+4 2 | 5+6 2 | 3+4 2 | 5+6 2 | 3+3 2 | 2+2 2 |
| | 5+6 2 | 5+6 2 | 5+6 2 | 5+6 2 | 2+2 2 | 2+2 2 |
| | 5+6 2 | 5+6 2 | 5+6 2 | 5+6 2 | 2+2 2 | 2+2 2 |
| | 5+6 2 | 3+4 2 | 5+6 2 | 3+4 2 | 3+3 2 | 2+2 2 |
| | 5+6 2 | 3+4 2 | 5+6 2 | 3+4 2 | 3+3 2 | 2+2 2 |
| | 5+6 2 | 3+4 2 | 5+6 2 | 3+4 2 | 3+3 2 | 2+2 2 |
| | 5+6 2 | 3+4 2 | 5+6 2 | 3+4 2 | 3+3 2 | 2+2 2 |
| | 3+4 2 | 3+4 2 | 5+6 2 | 3+4 2 | 3+3 2 | 2+2 2 |
| | 5+6 2 | 3+4 2 | 5+6 2 | 3+4 2 | 3+3 2 | 2+2 2 |
| | 5+6 2 | 5+6 2 | 3+4 2 | 5+6 2 | 2+2 2 | 3+3 2 |
| | 3+4 2 | 3+4 2 | 3+4 2 | 5+6 2 | 2+2 2 | 3+3 2 |
| | 3+4 2 | 5+6 2 | 3+4 2 | 3+4 2 | 3+3 2 | 3+3 2 |
| | 3+4 2 | 3+4 2 | 5+6 2 | 5+6 2 | 2+2 2 | 2+2 2 |
| | 5+6 2 | 5+6 2 | 5+6 2 | 3+4 2 | 3+3 2 | 3+3 2 |
| | 3+4 2 | 5+6 2 | 3+4 2 | 5+6 2 | 2+2 2 | 2+2 2 |
| | 5+6 2 | 5+6 2 | 3+4 2 | 5+6 2 | 2+2 2 | 3+3 2 |
| | 3+4 2 | 3+4 2 | 3+4 2 | 5+6 2 | 3+3 2 | 3+3 2 |
| | 3+4 2 | 5+6 2 | 3+4 2 | 3+4 2 | 2+2 2 | 3+3 2 |
| | 5+6 2 | 5+6 2 | 5+6 2 | 5+6 2 | 3+3 2 | 2+2 2 |
| | 3+4 2 | 3+4 2 | 5+6 2 | 5+6 2 | 2+2 2 | 3+3 2 |
| | 3+4 2 | 5+6 2 | 3+4 2 | 3+4 2 | 3+3 2 | 3+3 2 |
| | 3+4 2 | 5+6 2 | 3+4 2 | 3+4 2 | 2+2 2 | 2+2 2 |
| | 5+6 2 | 3+4 2 | 5+6 2 | 5+6 2 | 2+2 2 | 3+3 2 |
| | 3+4 2 | 3+4 2 | 3+4 2 | 3+4 2 | 3+3 2 | 2+2 2 |
| | 5+6 2 | 3+4 2 | 3+4 2 | 5+6 2 | 2+2 2 | 3+3 2 |
| | 5+6 2 | 5+6 2 | 5+6 2 | 5+6 2 | 2+2 2 | 2+2 2 |
| | 3+4 2 | 5+6 2 | 5+6 2 | 3+4 2 | 2+2 2 | 2+2 2 |
| | 5+6 2 | 5+6 2 | 3+4 2 | 5+6 2 | 3+3 2 | 3+3 2 |
| | 3+4 2 | 3+4 2 | 3+4 2 | 3+4 2 | 3+3 2 | 3+3 2 |
| | 5+6 2 | 3+4 2 | 5+6 2 | 5+6 2 | 2+2 2 | 3+3 2 |
| | 3+4 2 | 5+6 2 | 3+4 2 | 3+4 2 | 3+3 2 | 2+2 2 |
| | 5+6 2 | 3+4 2 | 5+6 2 | 3+4 2 | 2+2 2 | 3+3 2 |

Continued on next column

**TABLE XXIV– continued**

| TQC | ICSD | | NoE | SF | | |
|---|---|---|---|---|---|---|
| | 3+4 2 | 3+4 2 | 5+6 2 | 5+6 2 | 3+3 2 | 2+2 2 |
| | 3+4 2 | 5+6 2 | 3+4 2 | 5+6 2 | 2+2 2 | 2+2 2 |
| | 3+4 2 | 5+6 2 | 5+6 2 | 5+6 2 | 2+2 2 | 2+2 2 |
| | 5+6 2 | 3+4 2 | 3+4 2 | 3+4 2 | 3+3 2 | 3+3 2 |
| | 5+6 2 | 5+6 2 | 3+4 2 | 5+6 2 | 3+3 2 | 3+3 2 |
| | 5+6 2 | 5+6 2 | 3+4 2 | 3+4 2 | 2+2 2 | 2+2 2 |
| | 3+4 2 | 3+4 2 | 5+6 2 | 3+4 2 | 3+3 2 | 3+3 2 |
| | 5+6 2 | 5+6 2 | 5+6 2 | 5+6 2 | 2+2 2 | 2+2 2 |
| | 3+4 2 | 3+4 2 | 3+4 2 | 5+6 2 | 2+2 2 | 3+3 2 |
| | 5+6 2 | 5+6 2 | 5+6 2 | 3+4 2 | 3+3 2 | 2+2 2 |
| | 5+6 2 | 3+4 2 | 5+6 2 | 3+4 2 | 3+3 2 | 2+2 2 |
| | 5+6 2 | 3+4 2 | 5+6 2 | 3+4 2 | 3+3 2 | 2+2 2 |
| | 3+4 2 | 3+4 2 | 5+6 2 | 5+6 2 | 2+2 2 | 3+3 2 |
| 12.1.160 | 69569 | | 166 | Ba3Cu2Er2O10Pt1 | | |
| | -A | -GM | -M | -Y | -L | -V |
| | 5+6 2 | 3+4 2 | 3+4 2 | 5+6 2 | 2+2 2 | 2+2 2 |
| | 3+4 2 | 5+6 2 | 5+6 2 | 3+4 2 | 3+3 2 | 3+3 2 |
| | 3+4 2 | 3+4 2 | 5+6 2 | 5+6 2 | 3+3 2 | 3+3 2 |
| | 5+6 2 | 5+6 2 | 3+4 2 | 3+4 2 | 2+2 2 | 2+2 2 |
| | 5+6 2 | 3+4 2 | 5+6 2 | 3+4 2 | 2+2 2 | 3+3 2 |
| | 5+6 2 | 3+4 2 | 3+4 2 | 5+6 2 | 3+3 2 | 2+2 2 |
| | 3+4 2 | 5+6 2 | 5+6 2 | 3+4 2 | 2+2 2 | 3+3 2 |
| | 5+6 2 | 3+4 2 | 3+4 2 | 5+6 2 | 2+2 2 | 2+2 2 |
| | 3+4 2 | 5+6 2 | 5+6 2 | 3+4 2 | 3+3 2 | 3+3 2 |
| | 5+6 2 | 3+4 2 | 3+4 2 | 5+6 2 | 2+2 2 | 2+2 2 |
| | 3+4 2 | 5+6 2 | 5+6 2 | 3+4 2 | 3+3 2 | 3+3 2 |
| | 3+4 2 | 3+4 2 | 3+4 2 | 3+4 2 | 2+2 2 | 2+2 2 |
| | 5+6 2 | 5+6 2 | 5+6 2 | 5+6 2 | 3+3 2 | 2+2 2 |
| | 3+4 2 | 3+4 2 | 3+4 2 | 3+4 2 | 2+2 2 | 3+3 2 |
| | 3+4 2 | 3+4 2 | 3+4 2 | 3+4 2 | 2+2 2 | 2+2 2 |
| | 5+6 2 | 5+6 2 | 5+6 2 | 5+6 2 | 3+3 2 | 3+3 2 |
| | 5+6 2 | 5+6 2 | 5+6 2 | 5+6 2 | 3+3 2 | 3+3 2 |
| | 3+4 2 | 3+4 2 | 3+4 2 | 5+6 2 | 3+3 2 | 2+2 2 |
| | 5+6 2 | 5+6 2 | 5+6 2 | 3+4 2 | 2+2 2 | 3+3 2 |
| | 3+4 2 | 5+6 2 | 5+6 2 | 3+4 2 | 3+3 2 | 2+2 2 |
| | 5+6 2 | 3+4 2 | 3+4 2 | 5+6 2 | 2+2 2 | 3+3 2 |
| | 3+4 2 | 3+4 2 | 3+4 2 | 5+6 2 | 3+3 2 | 3+3 2 |
| | 5+6 2 | 5+6 2 | 5+6 2 | 3+4 2 | 2+2 2 | 2+2 2 |
| | 3+4 2 | 5+6 2 | 3+4 2 | 5+6 2 | 3+3 2 | 2+2 2 |
| | 3+4 2 | 3+4 2 | 3+4 2 | 5+6 2 | 3+3 2 | 3+3 2 |
| | 3+4 2 | 3+4 2 | 5+6 2 | 3+4 2 | 2+2 2 | 2+2 2 |
| | 5+6 2 | 5+6 2 | 3+4 2 | 3+4 2 | 3+3 2 | 3+3 2 |
| | 5+6 2 | 5+6 2 | 5+6 2 | 5+6 2 | 2+2 2 | 2+2 2 |
| | 3+4 2 | 5+6 2 | 3+4 2 | 5+6 2 | 3+3 2 | 2+2 2 |
| | 3+4 2 | 5+6 2 | 3+4 2 | 5+6 2 | 3+3 2 | 2+2 2 |
| | 3+4 2 | 3+4 2 | 3+4 2 | 3+4 2 | 2+2 2 | 2+2 2 |
| | 3+4 2 | 3+4 2 | 3+4 2 | 3+4 2 | 2+2 2 | 2+2 2 |
| | 3+4 2 | 3+4 2 | 3+4 2 | 3+4 2 | 2+2 2 | 2+2 2 |
| | 3+4 2 | 3+4 2 | 3+4 2 | 3+4 2 | 2+2 2 | 2+2 2 |
| | 5+6 2 | 3+4 2 | 3+4 2 | 3+4 2 | 2+2 2 | 2+2 2 |
| | 3+4 2 | 3+4 2 | 5+6 2 | 3+4 2 | 2+2 2 | 2+2 2 |
| | 3+4 2 | 5+6 2 | 3+4 2 | 5+6 2 | 3+3 2 | 3+3 2 |
| | 5+6 2 | 3+4 2 | 5+6 2 | 5+6 2 | 3+3 2 | 3+3 2 |
| | 5+6 2 | 5+6 2 | 3+4 2 | 3+4 2 | 2+2 2 | 3+3 2 |
| | 3+4 2 | 3+4 2 | 5+6 2 | 5+6 2 | 2+2 2 | 2+2 2 |
| | 5+6 2 | 5+6 2 | 3+4 2 | 3+4 2 | 3+3 2 | 3+3 2 |
| | 5+6 2 | 5+6 2 | 5+6 2 | 5+6 2 | 3+3 2 | 2+2 2 |
| | 3+4 2 | 3+4 2 | 5+6 2 | 5+6 2 | 3+3 2 | 3+3 2 |
| | 5+6 2 | 5+6 2 | 5+6 2 | 3+4 2 | 2+2 2 | 3+3 2 |
| | 5+6 2 | 5+6 2 | 5+6 2 | 5+6 2 | 3+3 2 | 3+3 2 |
| | 3+4 2 | 5+6 2 | 3+4 2 | 5+6 2 | 2+2 2 | 2+2 2 |

Continued on next column

**TABLE XXIV– continued**

| TQC | ICSD | | NoE | SF | | |
|---|---|---|---|---|---|---|
| | 5+6 2 | 3+4 2 | 3+4 2 | 3+4 2 | 3+3 2 | 3+3 2 |
| | 5+6 2 | 5+6 2 | 5+6 2 | 5+6 2 | 2+2 2 | 3+3 2 |
| | 5+6 2 | 5+6 2 | 5+6 2 | 3+4 2 | 3+3 2 | 2+2 2 |
| | 3+4 2 | 3+4 2 | 3+4 2 | 3+4 2 | 3+3 2 | 3+3 2 |
| | 5+6 2 | 3+4 2 | 5+6 2 | 5+6 2 | 2+2 2 | 2+2 2 |
| | 3+4 2 | 3+4 2 | 5+6 2 | 5+6 2 | 3+3 2 | 3+3 2 |
| | 3+4 2 | 5+6 2 | 3+4 2 | 5+6 2 | 3+3 2 | 2+2 2 |
| | 3+4 2 | 5+6 2 | 3+4 2 | 3+4 2 | 3+3 2 | 2+2 2 |
| | 5+6 2 | 3+4 2 | 5+6 2 | 5+6 2 | 2+2 2 | 3+3 2 |
| | 5+6 2 | 5+6 2 | 5+6 2 | 3+4 2 | 2+2 2 | 3+3 2 |
| | 3+4 2 | 5+6 2 | 3+4 2 | 5+6 2 | 2+2 2 | 3+3 2 |
| | 3+4 2 | 3+4 2 | 3+4 2 | 3+4 2 | 3+3 2 | 2+2 2 |
| | 5+6 2 | 3+4 2 | 5+6 2 | 5+6 2 | 3+3 2 | 3+3 2 |
| | 5+6 2 | 5+6 2 | 3+4 2 | 3+4 2 | 2+2 2 | 2+2 2 |
| | 5+6 2 | 3+4 2 | 5+6 2 | 3+4 2 | 3+3 2 | 2+2 2 |
| | 3+4 2 | 5+6 2 | 3+4 2 | 5+6 2 | 2+2 2 | 2+2 2 |
| | 3+4 2 | 5+6 2 | 5+6 2 | 5+6 2 | 3+3 2 | 3+3 2 |
| | 5+6 2 | 3+4 2 | 5+6 2 | 3+4 2 | 2+2 2 | 3+3 2 |
| | 3+4 2 | 5+6 2 | 5+6 2 | 5+6 2 | 3+3 2 | 2+2 2 |
| | 5+6 2 | 3+4 2 | 5+6 2 | 3+4 2 | 3+3 2 | 3+3 2 |
| | 5+6 2 | 5+6 2 | 3+4 2 | 5+6 2 | 2+2 2 | 2+2 2 |
| | 3+4 2 | 3+4 2 | 3+4 2 | 5+6 2 | 3+3 2 | 2+2 2 |
| | 3+4 2 | 5+6 2 | 3+4 2 | 3+4 2 | 2+2 2 | 3+3 2 |
| | 3+4 2 | 3+4 2 | 3+4 2 | 3+4 2 | 2+2 2 | 2+2 2 |
| | 5+6 2 | 3+4 2 | 5+6 2 | 5+6 2 | 3+3 2 | 3+3 2 |
| | 5+6 2 | 5+6 2 | 3+4 2 | 5+6 2 | 2+2 2 | 3+3 2 |
| | 3+4 2 | 5+6 2 | 5+6 2 | 3+4 2 | 3+3 2 | 2+2 2 |
| | 3+4 2 | 3+4 2 | 3+4 2 | 5+6 2 | 2+2 2 | 2+2 2 |
| | 5+6 2 | 3+4 2 | 5+6 2 | 3+4 2 | 3+3 2 | 3+3 2 |
| | 3+4 2 | 5+6 2 | 3+4 2 | 3+4 2 | 2+2 2 | 2+2 2 |
| | 3+4 2 | 3+4 2 | 3+4 2 | 3+4 2 | 2+2 2 | 2+2 2 |
| | 5+6 2 | 3+4 2 | 5+6 2 | 5+6 2 | 3+3 2 | 2+2 2 |
| | 3+4 2 | 5+6 2 | 3+4 2 | 3+4 2 | 2+2 2 | 3+3 2 |
| | 5+6 2 | 3+4 2 | 3+4 2 | 5+6 2 | 3+3 2 | 3+3 2 |
| | 5+6 2 | 3+4 2 | 5+6 2 | 5+6 2 | 2+2 2 | 2+2 2 |
| | 3+4 2 | 5+6 2 | 3+4 2 | 5+6 2 | 3+3 2 | 3+3 2 |
| | 5+6 2 | 5+6 2 | 5+6 2 | 3+4 2 | 2+2 2 | 2+2 2 |
| 12.1.161 | 173685 | | 74 | Ba3Cd2Sb4 | | |
| | -A | -GM | -M | -Y | -L | -V |
| | 5+6 2 | 3+4 2 | 3+4 2 | 3+4 2 | 2+2 2 | 3+3 2 |
| | 3+4 2 | 5+6 2 | 5+6 2 | 5+6 2 | 3+3 2 | 2+2 2 |
| | 3+4 2 | 3+4 2 | 3+4 2 | 3+4 2 | 3+3 2 | 3+3 2 |
| | 5+6 2 | 3+4 2 | 5+6 2 | 5+6 2 | 2+2 2 | 2+2 2 |
| | 3+4 2 | 5+6 2 | 3+4 2 | 3+4 2 | 3+3 2 | 3+3 2 |
| | 5+6 2 | 5+6 2 | 5+6 2 | 5+6 2 | 2+2 2 | 2+2 2 |
| | 5+6 2 | 3+4 2 | 5+6 2 | 5+6 2 | 3+3 2 | 2+2 2 |
| | 3+4 2 | 5+6 2 | 3+4 2 | 3+4 2 | 2+2 2 | 3+3 2 |
| | 5+6 2 | 3+4 2 | 5+6 2 | 5+6 2 | 2+2 2 | 2+2 2 |
| | 3+4 2 | 5+6 2 | 3+4 2 | 3+4 2 | 3+3 2 | 3+3 2 |
| | 5+6 2 | 5+6 2 | 5+6 2 | 5+6 2 | 2+2 2 | 2+2 2 |
| | 5+6 2 | 5+6 2 | 5+6 2 | 5+6 2 | 2+2 2 | 2+2 2 |
| | 5+6 2 | 3+4 2 | 5+6 2 | 3+4 2 | 3+3 2 | 2+2 2 |
| | 5+6 2 | 5+6 2 | 3+4 2 | 3+4 2 | 3+3 2 | 3+3 2 |
| | 3+4 2 | 3+4 2 | 5+6 2 | 5+6 2 | 2+2 2 | 2+2 2 |
| | 3+4 2 | 5+6 2 | 3+4 2 | 5+6 2 | 2+2 2 | 3+3 2 |
| | 5+6 2 | 3+4 2 | 5+6 2 | 3+4 2 | 3+3 2 | 2+2 2 |
| | 3+4 2 | 3+4 2 | 5+6 2 | 5+6 2 | 2+2 2 | 3+3 2 |
| | 5+6 2 | 5+6 2 | 3+4 2 | 3+4 2 | 3+3 2 | 2+2 2 |
| | 3+4 2 | 5+6 2 | 3+4 2 | 5+6 2 | 2+2 2 | 3+3 2 |
| | 3+4 2 | 3+4 2 | 5+6 2 | 5+6 2 | 3+3 2 | 2+2 2 |
| | 5+6 2 | 3+4 2 | 5+6 2 | 3+4 2 | 2+2 2 | 3+3 2 |

Continued on next column

**TABLE XXIV– continued**

| TQC | ICSD | NoE | SF | | | |
|---|---|---|---|---|---|---|
| | 3+4 2 | 5+6 2 | 3+4 2 | 5+6 2 | 3+3 2 | 2+2 2 |
| | 5+6 2 | 3+4 2 | 5+6 2 | 3+4 2 | 3+3 2 | 3+3 2 |
| | 3+4 2 | 5+6 2 | 3+4 2 | 3+4 2 | 2+2 2 | 2+2 2 |
| | 5+6 2 | 5+6 2 | 3+4 2 | 5+6 2 | 2+2 2 | 3+3 2 |
| | 3+4 2 | 5+6 2 | 3+4 2 | 3+4 2 | 3+3 2 | 3+3 2 |
| | 5+6 2 | 3+4 2 | 5+6 2 | 5+6 2 | 2+2 2 | 2+2 2 |
| | 3+4 2 | 3+4 2 | 5+6 2 | 3+4 2 | 3+3 2 | 3+3 2 |
| | 5+6 2 | 3+4 2 | 3+4 2 | 5+6 2 | 2+2 2 | 2+2 2 |
| | 3+4 2 | 5+6 2 | 5+6 2 | 5+6 2 | 3+3 2 | 3+3 2 |
| | 3+4 2 | 3+4 2 | 5+6 2 | 5+6 2 | 2+2 2 | 2+2 2 |
| | 3+4 2 | 3+4 2 | 3+4 2 | 5+6 2 | 3+3 2 | 3+3 2 |
| | 5+6 2 | 5+6 2 | 3+4 2 | 3+4 2 | 3+3 2 | 3+3 2 |
| | 5+6 2 | 3+4 2 | 5+6 2 | 3+4 2 | 2+2 2 | 2+2 2 |
| | 5+6 2 | 5+6 2 | 3+4 2 | 3+4 2 | 2+2 2 | 2+2 2 |
| | 3+4 2 | 5+6 2 | 3+4 2 | 3+4 2 | 3+3 2 | 2+2 2 |
| 12.1.162 | 423268 | 86 | Ba2Ge4Ni5 | | | |
| | -A | -GM | -M | -Y | -L | -V |
| | 3+4 2 | 3+4 2 | 3+4 2 | 3+4 2 | 3+3 2 | 3+3 2 |
| | 5+6 2 | 5+6 2 | 5+6 2 | 5+6 2 | 2+2 2 | 2+2 2 |
| | 3+4 2 | 3+4 2 | 3+4 2 | 3+4 2 | 3+3 2 | 3+3 2 |
| | 5+6 2 | 5+6 2 | 5+6 2 | 5+6 2 | 2+2 2 | 2+2 2 |
| | 3+4 2 | 3+4 2 | 3+4 2 | 3+4 2 | 3+3 2 | 3+3 2 |
| | 5+6 2 | 5+6 2 | 5+6 2 | 5+6 2 | 2+2 2 | 2+2 2 |
| | 3+4 2 | 3+4 2 | 5+6 2 | 5+6 2 | 3+3 2 | 3+3 2 |
| | 5+6 2 | 5+6 2 | 3+4 2 | 3+4 2 | 2+2 2 | 2+2 2 |
| | 5+6 2 | 3+4 2 | 5+6 2 | 3+4 2 | 2+2 2 | 3+3 2 |
| | 3+4 2 | 5+6 2 | 3+4 2 | 5+6 2 | 3+3 2 | 2+2 2 |
| | 3+4 2 | 3+4 2 | 3+4 2 | 5+6 2 | 3+3 2 | 2+2 2 |
| | 5+6 2 | 5+6 2 | 5+6 2 | 3+4 2 | 2+2 2 | 3+3 2 |
| | 5+6 2 | 3+4 2 | 5+6 2 | 3+4 2 | 3+3 2 | 2+2 2 |
| | 3+4 2 | 3+4 2 | 3+4 2 | 3+4 2 | 2+2 2 | 3+3 2 |
| | 3+4 2 | 5+6 2 | 5+6 2 | 5+6 2 | 3+3 2 | 2+2 2 |
| | 5+6 2 | 3+4 2 | 5+6 2 | 3+4 2 | 2+2 2 | 3+3 2 |
| | 5+6 2 | 5+6 2 | 3+4 2 | 3+4 2 | 3+3 2 | 2+2 2 |
| | 3+4 2 | 3+4 2 | 5+6 2 | 5+6 2 | 2+2 2 | 3+3 2 |
| | 5+6 2 | 5+6 2 | 3+4 2 | 5+6 2 | 3+3 2 | 2+2 2 |
| | 3+4 2 | 3+4 2 | 5+6 2 | 3+4 2 | 2+2 2 | 3+3 2 |
| | 5+6 2 | 5+6 2 | 5+6 2 | 5+6 2 | 3+3 2 | 2+2 2 |
| | 5+6 2 | 3+4 2 | 3+4 2 | 3+4 2 | 3+3 2 | 2+2 2 |
| | 3+4 2 | 5+6 2 | 3+4 2 | 5+6 2 | 2+2 2 | 3+3 2 |
| | 5+6 2 | 3+4 2 | 5+6 2 | 3+4 2 | 3+3 2 | 2+2 2 |
| | 3+4 2 | 5+6 2 | 3+4 2 | 3+4 2 | 2+2 2 | 3+3 2 |
| | 5+6 2 | 3+4 2 | 5+6 2 | 5+6 2 | 2+2 2 | 3+3 2 |
| | 3+4 2 | 5+6 2 | 3+4 2 | 5+6 2 | 2+2 2 | 2+2 2 |
| | 5+6 2 | 3+4 2 | 5+6 2 | 3+4 2 | 3+3 2 | 3+3 2 |
| | 3+4 2 | 5+6 2 | 5+6 2 | 3+4 2 | 3+3 2 | 2+2 2 |
| | 5+6 2 | 3+4 2 | 3+4 2 | 5+6 2 | 3+3 2 | 2+2 2 |
| | 5+6 2 | 3+4 2 | 5+6 2 | 3+4 2 | 2+2 2 | 2+2 2 |
| | 3+4 2 | 5+6 2 | 3+4 2 | 5+6 2 | 3+3 2 | 3+3 2 |
| | 5+6 2 | 3+4 2 | 3+4 2 | 5+6 2 | 2+2 2 | 3+3 2 |
| | 5+6 2 | 3+4 2 | 5+6 2 | 3+4 2 | 3+3 2 | 2+2 2 |
| | 3+4 2 | 5+6 2 | 3+4 2 | 3+4 2 | 2+2 2 | 2+2 2 |
| | 3+4 2 | 3+4 2 | 5+6 2 | 5+6 2 | 2+2 2 | 2+2 2 |
| | 5+6 2 | 5+6 2 | 3+4 2 | 3+4 2 | 3+3 2 | 3+3 2 |
| | 5+6 2 | 3+4 2 | 5+6 2 | 5+6 2 | 3+3 2 | 2+2 2 |
| | 5+6 2 | 3+4 2 | 3+4 2 | 3+4 2 | 2+2 2 | 3+3 2 |
| | 3+4 2 | 5+6 2 | 5+6 2 | 3+4 2 | 3+3 2 | 2+2 2 |
| | 5+6 2 | 3+4 2 | 5+6 2 | 5+6 2 | 3+3 2 | 2+2 2 |
| | 3+4 2 | 5+6 2 | 3+4 2 | 3+4 2 | 2+2 2 | 3+3 2 |
| | 5+6 2 | 3+4 2 | 5+6 2 | 5+6 2 | 3+3 2 | 2+2 2 |
| 12.1.163 | 238228 | 96 | Ba1Te4U1 | | | |
| | -A | -GM | -M | -Y | -L | -V |
| | 3+4 2 | 3+4 2 | 5+6 2 | 3+4 2 | 2+2 2 | 2+2 2 |
| | 5+6 2 | 5+6 2 | 3+4 2 | 5+6 2 | 3+3 2 | 3+3 2 |
| | 3+4 2 | 3+4 2 | 3+4 2 | 3+4 2 | 3+3 2 | 3+3 2 |
| | 5+6 2 | 5+6 2 | 5+6 2 | 5+6 2 | 2+2 2 | 2+2 2 |
| | 3+4 2 | 3+4 2 | 5+6 2 | 3+4 2 | 3+3 2 | 2+2 2 |
| | 5+6 2 | 5+6 2 | 3+4 2 | 5+6 2 | 2+2 2 | 3+3 2 |
| | 5+6 2 | 3+4 2 | 5+6 2 | 3+4 2 | 3+3 2 | 2+2 2 |
| | 3+4 2 | 5+6 2 | 3+4 2 | 5+6 2 | 2+2 2 | 3+3 2 |
| | 3+4 2 | 3+4 2 | 5+6 2 | 5+6 2 | 3+3 2 | 2+2 2 |
| | 5+6 2 | 5+6 2 | 3+4 2 | 3+4 2 | 2+2 2 | 3+3 2 |
| | 5+6 2 | 3+4 2 | 3+4 2 | 3+4 2 | 2+2 2 | 3+3 2 |
| | 3+4 2 | 5+6 2 | 5+6 2 | 5+6 2 | 3+3 2 | 2+2 2 |
| | 5+6 2 | 3+4 2 | 5+6 2 | 3+4 2 | 2+2 2 | 3+3 2 |
| | 3+4 2 | 3+4 2 | 3+4 2 | 5+6 2 | 3+3 2 | 2+2 2 |
| | 5+6 2 | 5+6 2 | 3+4 2 | 3+4 2 | 2+2 2 | 3+3 2 |
| | 3+4 2 | 5+6 2 | 5+6 2 | 5+6 2 | 3+3 2 | 2+2 2 |
| | 5+6 2 | 3+4 2 | 5+6 2 | 3+4 2 | 2+2 2 | 3+3 2 |
| | 3+4 2 | 5+6 2 | 3+4 2 | 5+6 2 | 3+3 2 | 2+2 2 |
| | 5+6 2 | 3+4 2 | 5+6 2 | 3+4 2 | 2+2 2 | 3+3 2 |
| | 5+6 2 | 5+6 2 | 3+4 2 | 5+6 2 | 3+3 2 | 2+2 2 |
| | 3+4 2 | 3+4 2 | 5+6 2 | 3+4 2 | 2+2 2 | 3+3 2 |
| | 3+4 2 | 5+6 2 | 3+4 2 | 5+6 2 | 3+3 2 | 2+2 2 |
| | 5+6 2 | 5+6 2 | 5+6 2 | 5+6 2 | 2+2 2 | 3+3 2 |
| | 3+4 2 | 3+4 2 | 3+4 2 | 3+4 2 | 3+3 2 | 2+2 2 |
| | 5+6 2 | 5+6 2 | 3+4 2 | 3+4 2 | 2+2 2 | 3+3 2 |
| | 3+4 2 | 3+4 2 | 5+6 2 | 5+6 2 | 3+3 2 | 2+2 2 |
| | 5+6 2 | 3+4 2 | 5+6 2 | 3+4 2 | 2+2 2 | 2+2 2 |
| | 3+4 2 | 5+6 2 | 3+4 2 | 5+6 2 | 3+3 2 | 3+3 2 |
| | 3+4 2 | 5+6 2 | 3+4 2 | 5+6 2 | 2+2 2 | 2+2 2 |
| | 3+4 2 | 3+4 2 | 5+6 2 | 3+4 2 | 3+3 2 | 3+3 2 |
| | 5+6 2 | 5+6 2 | 5+6 2 | 5+6 2 | 2+2 2 | 3+3 2 |
| | 3+4 2 | 3+4 2 | 3+4 2 | 5+6 2 | 3+3 2 | 2+2 2 |
| | 5+6 2 | 3+4 2 | 3+4 2 | 3+4 2 | 3+3 2 | 3+3 2 |
| | 5+6 2 | 3+4 2 | 5+6 2 | 5+6 2 | 2+2 2 | 2+2 2 |
| | 3+4 2 | 5+6 2 | 5+6 2 | 3+4 2 | 3+3 2 | 3+3 2 |
| | 5+6 2 | 3+4 2 | 3+4 2 | 3+4 2 | 2+2 2 | 2+2 2 |
| | 3+4 2 | 5+6 2 | 5+6 2 | 5+6 2 | 3+3 2 | 2+2 2 |
| | 5+6 2 | 3+4 2 | 5+6 2 | 5+6 2 | 2+2 2 | 3+3 2 |
| | 3+4 2 | 5+6 2 | 3+4 2 | 3+4 2 | 3+3 2 | 3+3 2 |
| | 5+6 2 | 3+4 2 | 3+4 2 | 3+4 2 | 2+2 2 | 2+2 2 |
| | 3+4 2 | 5+6 2 | 5+6 2 | 5+6 2 | 3+3 2 | 3+3 2 |
| | 5+6 2 | 3+4 2 | 5+6 2 | 3+4 2 | 2+2 2 | 2+2 2 |
| | 5+6 2 | 5+6 2 | 3+4 2 | 5+6 2 | 3+3 2 | 2+2 2 |
| | 5+6 2 | 5+6 2 | 5+6 2 | 3+4 2 | 2+2 2 | 3+3 2 |
| | 3+4 2 | 3+4 2 | 3+4 2 | 3+4 2 | 3+3 2 | 2+2 2 |
| | 3+4 2 | 3+4 2 | 3+4 2 | 5+6 2 | 2+2 2 | 3+3 2 |
| | 3+4 2 | 5+6 2 | 5+6 2 | 5+6 2 | 2+2 2 | 3+3 2 |
| | 5+6 2 | 3+4 2 | 5+6 2 | 3+4 2 | 3+3 2 | 2+2 2 |
| 12.1.164 | 195017 | 80 | Ba1Pt2Sb2 | | | |
| | -A | -GM | -M | -Y | -L | -V |
| | 3+4 2 | 3+4 2 | 3+4 2 | 3+4 2 | 2+2 2 | 2+2 2 |
| | 5+6 2 | 5+6 2 | 5+6 2 | 5+6 2 | 3+3 2 | 3+3 2 |
| | 5+6 2 | 3+4 2 | 3+4 2 | 3+4 2 | 2+2 2 | 2+2 2 |
| | 3+4 2 | 5+6 2 | 5+6 2 | 5+6 2 | 3+3 2 | 3+3 2 |
| | 5+6 2 | 3+4 2 | 3+4 2 | 3+4 2 | 2+2 2 | 2+2 2 |
| | 3+4 2 | 5+6 2 | 5+6 2 | 5+6 2 | 3+3 2 | 3+3 2 |
| | 5+6 2 | 3+4 2 | 3+4 2 | 3+4 2 | 2+2 2 | 2+2 2 |
| | 3+4 2 | 5+6 2 | 5+6 2 | 5+6 2 | 3+3 2 | 3+3 2 |
| | 5+6 2 | 3+4 2 | 3+4 2 | 3+4 2 | 3+3 2 | 2+2 2 |
| | 3+4 2 | 3+4 2 | 3+4 2 | 5+6 2 | 2+2 2 | 3+3 2 |

Continued on next column

**TABLE XXIV– continued**

| TQC | ICSD | | NoE | | SF | |
|---|---|---|---|---|---|---|
| | 5+6 2 | 5+6 2 | 5+6 2 | 5+6 2 | 3+3 2 | 3+3 2 |
| | 3+4 2 | 5+6 2 | 5+6 2 | 3+4 2 | 2+2 2 | 2+2 2 |
| | 3+4 2 | 5+6 2 | 3+4 2 | 3+4 2 | 2+2 2 | 2+2 2 |
| | 5+6 2 | 3+4 2 | 5+6 2 | 5+6 2 | 3+3 2 | 3+3 2 |
| | 5+6 2 | 3+4 2 | 3+4 2 | 3+4 2 | 3+3 2 | 3+3 2 |
| | 3+4 2 | 3+4 2 | 5+6 2 | 5+6 2 | 2+2 2 | 2+2 2 |
| | 5+6 2 | 5+6 2 | 5+6 2 | 5+6 2 | 3+3 2 | 3+3 2 |
| | 5+6 2 | 3+4 2 | 3+4 2 | 3+4 2 | 2+2 2 | 3+3 2 |
| | 5+6 2 | 5+6 2 | 5+6 2 | 3+4 2 | 3+3 2 | 2+2 2 |
| | 3+4 2 | 3+4 2 | 3+4 2 | 5+6 2 | 2+2 2 | 2+2 2 |
| | 5+6 2 | 5+6 2 | 5+6 2 | 5+6 2 | 3+3 2 | 3+3 2 |
| | 3+4 2 | 3+4 2 | 3+4 2 | 3+4 2 | 2+2 2 | 2+2 2 |
| | 3+4 2 | 5+6 2 | 3+4 2 | 3+4 2 | 2+2 2 | 3+3 2 |
| | 3+4 2 | 5+6 2 | 5+6 2 | 5+6 2 | 3+3 2 | 2+2 2 |
| | 5+6 2 | 3+4 2 | 3+4 2 | 3+4 2 | 2+2 2 | 2+2 2 |
| | 3+4 2 | 3+4 2 | 5+6 2 | 5+6 2 | 3+3 2 | 3+3 2 |
| | 5+6 2 | 3+4 2 | 3+4 2 | 5+6 2 | 2+2 2 | 2+2 2 |
| | 3+4 2 | 5+6 2 | 5+6 2 | 3+4 2 | 3+3 2 | 3+3 2 |
| | 3+4 2 | 5+6 2 | 5+6 2 | 3+4 2 | 3+3 2 | 3+3 2 |
| | 5+6 2 | 3+4 2 | 3+4 2 | 5+6 2 | 2+2 2 | 2+2 2 |
| | 3+4 2 | 5+6 2 | 5+6 2 | 3+4 2 | 3+3 2 | 3+3 2 |
| | 5+6 2 | 3+4 2 | 3+4 2 | 5+6 2 | 2+2 2 | 2+2 2 |
| | 3+4 2 | 5+6 2 | 5+6 2 | 3+4 2 | 2+2 2 | 3+3 2 |
| | 5+6 2 | 3+4 2 | 3+4 2 | 5+6 2 | 3+3 2 | 2+2 2 |
| | 3+4 2 | 3+4 2 | 3+4 2 | 3+4 2 | 3+3 2 | 3+3 2 |
| | 5+6 2 | 3+4 2 | 5+6 2 | 5+6 2 | 2+2 2 | 2+2 2 |
| | 3+4 2 | 5+6 2 | 3+4 2 | 3+4 2 | 2+2 2 | 2+2 2 |
| | 3+4 2 | 3+4 2 | 5+6 2 | 5+6 2 | 3+3 2 | 3+3 2 |
| | 5+6 2 | 5+6 2 | 5+6 2 | 5+6 2 | 3+3 2 | 2+2 2 |
| | 5+6 2 | 3+4 2 | 3+4 2 | 3+4 2 | 2+2 2 | 3+3 2 |
| 12.1.165 | 154011 | | 142 | | Ba1Mn2O6Y1 | |
| | -A | -GM | -M | -Y | -L | -V |
| | 5+6 2 | 3+4 2 | 3+4 2 | 3+4 2 | 2+2 2 | 3+3 2 |
| | 3+4 2 | 5+6 2 | 5+6 2 | 5+6 2 | 3+3 2 | 2+2 2 |
| | 3+4 2 | 3+4 2 | 5+6 2 | 5+6 2 | 2+2 2 | 2+2 2 |
| | 5+6 2 | 5+6 2 | 3+4 2 | 3+4 2 | 3+3 2 | 3+3 2 |
| | 5+6 2 | 3+4 2 | 5+6 2 | 3+4 2 | 2+2 2 | 3+3 2 |
| | 3+4 2 | 5+6 2 | 3+4 2 | 5+6 2 | 3+3 2 | 2+2 2 |
| | 5+6 2 | 3+4 2 | 5+6 2 | 3+4 2 | 2+2 2 | 3+3 2 |
| | 3+4 2 | 5+6 2 | 3+4 2 | 5+6 2 | 3+3 2 | 2+2 2 |
| | 5+6 2 | 3+4 2 | 5+6 2 | 3+4 2 | 2+2 2 | 3+3 2 |
| | 3+4 2 | 5+6 2 | 3+4 2 | 5+6 2 | 3+3 2 | 2+2 2 |
| | 5+6 2 | 3+4 2 | 3+4 2 | 3+4 2 | 2+2 2 | 3+3 2 |
| | 3+4 2 | 5+6 2 | 5+6 2 | 5+6 2 | 3+3 2 | 2+2 2 |
| | 3+4 2 | 5+6 2 | 5+6 2 | 5+6 2 | 3+3 2 | 2+2 2 |
| | 5+6 2 | 3+4 2 | 3+4 2 | 3+4 2 | 2+2 2 | 3+3 2 |
| | 3+4 2 | 5+6 2 | 3+4 2 | 3+4 2 | 3+3 2 | 3+3 2 |
| | 3+4 2 | 5+6 2 | 5+6 2 | 5+6 2 | 2+2 2 | 2+2 2 |
| | 5+6 2 | 3+4 2 | 3+4 2 | 5+6 2 | 2+2 2 | 3+3 2 |
| | 3+4 2 | 5+6 2 | 5+6 2 | 3+4 2 | 3+3 2 | 2+2 2 |
| | 5+6 2 | 3+4 2 | 5+6 2 | 5+6 2 | 3+3 2 | 2+2 2 |
| | 5+6 2 | 3+4 2 | 3+4 2 | 3+4 2 | 2+2 2 | 3+3 2 |
| | 3+4 2 | 5+6 2 | 3+4 2 | 3+4 2 | 2+2 2 | 2+2 2 |
| | 5+6 2 | 3+4 2 | 5+6 2 | 5+6 2 | 3+3 2 | 3+3 2 |
| | 3+4 2 | 3+4 2 | 5+6 2 | 5+6 2 | 2+2 2 | 2+2 2 |
| | 5+6 2 | 5+6 2 | 3+4 2 | 3+4 2 | 3+3 2 | 3+3 2 |
| | 3+4 2 | 3+4 2 | 5+6 2 | 5+6 2 | 2+2 2 | 2+2 2 |
| | 3+4 2 | 3+4 2 | 3+4 2 | 3+4 2 | 3+3 2 | 3+3 2 |
| | 5+6 2 | 5+6 2 | 5+6 2 | 5+6 2 | 3+3 2 | 2+2 2 |
| | 5+6 2 | 5+6 2 | 3+4 2 | 3+4 2 | 2+2 2 | 3+3 2 |
| | 5+6 2 | 3+4 2 | 5+6 2 | 3+4 2 | 2+2 2 | 3+3 2 |
| | 3+4 2 | 5+6 2 | 3+4 2 | 5+6 2 | 3+3 2 | 2+2 2 |
| | 3+4 2 | 5+6 2 | 3+4 2 | 3+4 2 | 2+2 2 | 3+3 2 |
| | 5+6 2 | 5+6 2 | 5+6 2 | 5+6 2 | 3+3 2 | 2+2 2 |
| | 3+4 2 | 3+4 2 | 3+4 2 | 5+6 2 | 3+3 2 | 2+2 2 |
| | 5+6 2 | 5+6 2 | 5+6 2 | 3+4 2 | 2+2 2 | 3+3 2 |
| | 3+4 2 | 3+4 2 | 3+4 2 | 3+4 2 | 3+3 2 | 2+2 2 |
| | 5+6 2 | 3+4 2 | 5+6 2 | 5+6 2 | 2+2 2 | 3+3 2 |
| | 5+6 2 | 5+6 2 | 5+6 2 | 3+4 2 | 3+3 2 | 2+2 2 |
| | 3+4 2 | 3+4 2 | 3+4 2 | 5+6 2 | 2+2 2 | 3+3 2 |
| | 3+4 2 | 5+6 2 | 5+6 2 | 5+6 2 | 2+2 2 | 2+2 2 |
| | 5+6 2 | 3+4 2 | 3+4 2 | 3+4 2 | 3+3 2 | 3+3 2 |
| | 5+6 2 | 3+4 2 | 5+6 2 | 3+4 2 | 2+2 2 | 2+2 2 |
| | 3+4 2 | 3+4 2 | 3+4 2 | 5+6 2 | 3+3 2 | 2+2 2 |
| | 5+6 2 | 5+6 2 | 3+4 2 | 3+4 2 | 2+2 2 | 3+3 2 |
| | 5+6 2 | 5+6 2 | 5+6 2 | 5+6 2 | 3+3 2 | 3+3 2 |
| | 3+4 2 | 5+6 2 | 5+6 2 | 5+6 2 | 3+3 2 | 3+3 2 |
| | 3+4 2 | 3+4 2 | 3+4 2 | 3+4 2 | 2+2 2 | 2+2 2 |
| | 5+6 2 | 3+4 2 | 3+4 2 | 5+6 2 | 3+3 2 | 3+3 2 |
| | 5+6 2 | 3+4 2 | 5+6 2 | 3+4 2 | 2+2 2 | 2+2 2 |
| | 3+4 2 | 5+6 2 | 3+4 2 | 3+4 2 | 3+3 2 | 2+2 2 |
| | 5+6 2 | 5+6 2 | 5+6 2 | 5+6 2 | 2+2 2 | 3+3 2 |
| | 3+4 2 | 5+6 2 | 5+6 2 | 5+6 2 | 3+3 2 | 2+2 2 |
| | 5+6 2 | 3+4 2 | 5+6 2 | 3+4 2 | 2+2 2 | 3+3 2 |
| | 3+4 2 | 5+6 2 | 3+4 2 | 3+4 2 | 2+2 2 | 3+3 2 |
| | 3+4 2 | 5+6 2 | 3+4 2 | 5+6 2 | 3+3 2 | 3+3 2 |
| | 5+6 2 | 3+4 2 | 3+4 2 | 3+4 2 | 2+2 2 | 2+2 2 |
| | 5+6 2 | 3+4 2 | 5+6 2 | 5+6 2 | 2+2 2 | 2+2 2 |
| | 5+6 2 | 3+4 2 | 5+6 2 | 3+4 2 | 3+3 2 | 3+3 2 |
| | 3+4 2 | 5+6 2 | 3+4 2 | 5+6 2 | 3+3 2 | 3+3 2 |
| | 3+4 2 | 3+4 2 | 5+6 2 | 3+4 2 | 3+3 2 | 2+2 2 |
| | 5+6 2 | 3+4 2 | 3+4 2 | 3+4 2 | 2+2 2 | 2+2 2 |
| | 3+4 2 | 5+6 2 | 5+6 2 | 5+6 2 | 3+3 2 | 2+2 2 |
| | 3+4 2 | 5+6 2 | 3+4 2 | 5+6 2 | 2+2 2 | 3+3 2 |
| | 3+4 2 | 3+4 2 | 5+6 2 | 5+6 2 | 3+3 2 | 2+2 2 |
| | 5+6 2 | 5+6 2 | 3+4 2 | 3+4 2 | 2+2 2 | 3+3 2 |
| | 3+4 2 | 5+6 2 | 5+6 2 | 5+6 2 | 3+3 2 | 2+2 2 |
| | 5+6 2 | 3+4 2 | 3+4 2 | 3+4 2 | 2+2 2 | 3+3 2 |
| | 3+4 2 | 3+4 2 | 3+4 2 | 5+6 2 | 2+2 2 | 3+3 2 |
| | 3+4 2 | 3+4 2 | 5+6 2 | 3+4 2 | 3+3 2 | 2+2 2 |
| | 5+6 2 | 5+6 2 | 3+4 2 | 5+6 2 | 3+3 2 | 3+3 2 |
| | 5+6 2 | 5+6 2 | 5+6 2 | 3+4 2 | 2+2 2 | 2+2 2 |
| | 5+6 2 | 3+4 2 | 3+4 2 | 5+6 2 | 3+3 2 | 2+2 2 |
| 12.1.166 | 173848 | | 222 | | Ba1Ir1O3 | |
| | -A | -GM | -M | -Y | -L | -V |
| | 3+4 2 | 3+4 2 | 5+6 2 | 5+6 2 | 3+3 2 | 3+3 2 |
| | 5+6 2 | 5+6 2 | 3+4 2 | 3+4 2 | 2+2 2 | 2+2 2 |
| | 3+4 2 | 3+4 2 | 5+6 2 | 5+6 2 | 2+2 2 | 2+2 2 |
| | 5+6 2 | 5+6 2 | 3+4 2 | 3+4 2 | 3+3 2 | 3+3 2 |
| | 5+6 2 | 3+4 2 | 5+6 2 | 3+4 2 | 3+3 2 | 2+2 2 |
| | 3+4 2 | 5+6 2 | 3+4 2 | 5+6 2 | 2+2 2 | 3+3 2 |
| | 3+4 2 | 3+4 2 | 3+4 2 | 3+4 2 | 2+2 2 | 2+2 2 |
| | 5+6 2 | 5+6 2 | 5+6 2 | 5+6 2 | 3+3 2 | 3+3 2 |
| | 3+4 2 | 3+4 2 | 5+6 2 | 3+4 2 | 2+2 2 | 2+2 2 |
| | 5+6 2 | 5+6 2 | 3+4 2 | 5+6 2 | 3+3 2 | 3+3 2 |
| | 3+4 2 | 3+4 2 | 3+4 2 | 3+4 2 | 2+2 2 | 3+3 2 |
| | 5+6 2 | 5+6 2 | 5+6 2 | 5+6 2 | 3+3 2 | 2+2 2 |
| | 3+4 2 | 3+4 2 | 3+4 2 | 3+4 2 | 2+2 2 | 3+3 2 |
| | 5+6 2 | 3+4 2 | 5+6 2 | 3+4 2 | 2+2 2 | 2+2 2 |
| | 5+6 2 | 5+6 2 | 5+6 2 | 5+6 2 | 2+2 2 | 2+2 2 |
| | 3+4 2 | 5+6 2 | 5+6 2 | 3+4 2 | 3+3 2 | 2+2 2 |
| | 3+4 2 | 3+4 2 | 3+4 2 | 3+4 2 | 3+3 2 | 3+3 2 |

Continued on next column

**TABLE XXIV– continued**

| TQC | ICSD | NoE | SF | | | |
|---|---|---|---|---|---|---|
| | 5+6 2 | 3+4 2 | 3+4 2 | 5+6 2 | 3+3 2 | 3+3 2 |
| | 5+6 2 | 5+6 2 | 5+6 2 | 5+6 2 | 2+2 2 | 3+3 2 |
| | 3+4 2 | 5+6 2 | 3+4 2 | 3+4 2 | 3+3 2 | 2+2 2 |
| | 3+4 2 | 3+4 2 | 3+4 2 | 5+6 2 | 2+2 2 | 3+3 2 |
| | 5+6 2 | 3+4 2 | 3+4 2 | 5+6 2 | 2+2 2 | 2+2 2 |
| | 5+6 2 | 5+6 2 | 5+6 2 | 3+4 2 | 3+3 2 | 3+3 2 |
| | 3+4 2 | 5+6 2 | 5+6 2 | 5+6 2 | 3+3 2 | 2+2 2 |
| | 3+4 2 | 3+4 2 | 5+6 2 | 5+6 2 | 3+3 2 | 3+3 2 |
| | 5+6 2 | 5+6 2 | 3+4 2 | 3+4 2 | 2+2 2 | 2+2 2 |
| | 3+4 2 | 3+4 2 | 5+6 2 | 5+6 2 | 3+3 2 | 3+3 2 |
| | 5+6 2 | 3+4 2 | 3+4 2 | 3+4 2 | 2+2 2 | 2+2 2 |
| | 5+6 2 | 5+6 2 | 5+6 2 | 3+4 2 | 3+3 2 | 2+2 2 |
| | 3+4 2 | 5+6 2 | 3+4 2 | 5+6 2 | 2+2 2 | 3+3 2 |
| | 3+4 2 | 3+4 2 | 5+6 2 | 5+6 2 | 3+3 2 | 3+3 2 |
| | 5+6 2 | 5+6 2 | 3+4 2 | 3+4 2 | 2+2 2 | 2+2 2 |
| | 3+4 2 | 3+4 2 | 5+6 2 | 5+6 2 | 3+3 2 | 3+3 2 |
| | 5+6 2 | 3+4 2 | 3+4 2 | 3+4 2 | 2+2 2 | 2+2 2 |
| | 3+4 2 | 3+4 2 | 3+4 2 | 3+4 2 | 3+3 2 | 2+2 2 |
| | 5+6 2 | 5+6 2 | 5+6 2 | 5+6 2 | 2+2 2 | 3+3 2 |
| | 3+4 2 | 3+4 2 | 5+6 2 | 5+6 2 | 3+3 2 | 2+2 2 |
| | 5+6 2 | 5+6 2 | 3+4 2 | 3+4 2 | 2+2 2 | 3+3 2 |
| | 5+6 2 | 3+4 2 | 5+6 2 | 3+4 2 | 3+3 2 | 2+2 2 |
| | 5+6 2 | 5+6 2 | 3+4 2 | 5+6 2 | 2+2 2 | 3+3 2 |
| | 3+4 2 | 5+6 2 | 5+6 2 | 3+4 2 | 3+3 2 | 2+2 2 |
| | 3+4 2 | 5+6 2 | 3+4 2 | 5+6 2 | 2+2 2 | 3+3 2 |
| | 3+4 2 | 3+4 2 | 5+6 2 | 3+4 2 | 2+2 2 | 2+2 2 |
| | 5+6 2 | 3+4 2 | 3+4 2 | 5+6 2 | 2+2 2 | 3+3 2 |
| | 3+4 2 | 5+6 2 | 3+4 2 | 3+4 2 | 3+3 2 | 2+2 2 |
| | 5+6 2 | 3+4 2 | 5+6 2 | 3+4 2 | 2+2 2 | 3+3 2 |
| | 3+4 2 | 5+6 2 | 3+4 2 | 5+6 2 | 3+3 2 | 2+2 2 |
| | 5+6 2 | 5+6 2 | 5+6 2 | 5+6 2 | 3+3 2 | 3+3 2 |
| | 3+4 2 | 3+4 2 | 3+4 2 | 3+4 2 | 2+2 2 | 2+2 2 |
| | 5+6 2 | 3+4 2 | 5+6 2 | 3+4 2 | 2+2 2 | 3+3 2 |
| | 5+6 2 | 5+6 2 | 5+6 2 | 3+4 2 | 2+2 2 | 2+2 2 |
| | 3+4 2 | 3+4 2 | 3+4 2 | 5+6 2 | 3+3 2 | 3+3 2 |
| | 3+4 2 | 3+4 2 | 5+6 2 | 3+4 2 | 2+2 2 | 3+3 2 |
| | 5+6 2 | 3+4 2 | 3+4 2 | 3+4 2 | 2+2 2 | 2+2 2 |
| | 3+4 2 | 5+6 2 | 5+6 2 | 5+6 2 | 3+3 2 | 3+3 2 |
| | 5+6 2 | 5+6 2 | 3+4 2 | 3+4 2 | 2+2 2 | 2+2 2 |
| | 5+6 2 | 3+4 2 | 3+4 2 | 3+4 2 | 2+2 2 | 3+3 2 |
| | 5+6 2 | 3+4 2 | 5+6 2 | 3+4 2 | 2+2 2 | 2+2 2 |
| | 3+4 2 | 5+6 2 | 5+6 2 | 5+6 2 | 3+3 2 | 3+3 2 |
| | 3+4 2 | 3+4 2 | 3+4 2 | 3+4 2 | 2+2 2 | 2+2 2 |
| | 5+6 2 | 3+4 2 | 5+6 2 | 5+6 2 | 3+3 2 | 3+3 2 |
| | 3+4 2 | 3+4 2 | 5+6 2 | 3+4 2 | 2+2 2 | 3+3 2 |
| | 5+6 2 | 5+6 2 | 3+4 2 | 3+4 2 | 2+2 2 | 2+2 2 |
| | 3+4 2 | 3+4 2 | 3+4 2 | 5+6 2 | 3+3 2 | 3+3 2 |
| | 3+4 2 | 5+6 2 | 3+4 2 | 3+4 2 | 2+2 2 | 2+2 2 |
| | 5+6 2 | 3+4 2 | 3+4 2 | 5+6 2 | 2+2 2 | 2+2 2 |
| | 5+6 2 | 5+6 2 | 5+6 2 | 3+4 2 | 3+3 2 | 3+3 2 |
| | 3+4 2 | 3+4 2 | 3+4 2 | 5+6 2 | 3+3 2 | 2+2 2 |
| | 3+4 2 | 3+4 2 | 5+6 2 | 3+4 2 | 2+2 2 | 3+3 2 |
| | 5+6 2 | 5+6 2 | 3+4 2 | 3+4 2 | 3+3 2 | 2+2 2 |
| | 5+6 2 | 3+4 2 | 5+6 2 | 5+6 2 | 2+2 2 | 3+3 2 |
| | 3+4 2 | 5+6 2 | 5+6 2 | 5+6 2 | 3+3 2 | 2+2 2 |
| | 3+4 2 | 5+6 2 | 3+4 2 | 3+4 2 | 2+2 2 | 2+2 2 |
| | 5+6 2 | 5+6 2 | 5+6 2 | 5+6 2 | 3+3 2 | 2+2 2 |
| | 3+4 2 | 3+4 2 | 5+6 2 | 5+6 2 | 3+3 2 | 3+3 2 |
| | 5+6 2 | 5+6 2 | 3+4 2 | 3+4 2 | 2+2 2 | 3+3 2 |
| | 3+4 2 | 3+4 2 | 3+4 2 | 5+6 2 | 3+3 2 | 2+2 2 |
| | 5+6 2 | 3+4 2 | 5+6 2 | 5+6 2 | 2+2 2 | 3+3 2 |

Continued on next column

**TABLE XXIV– continued**

| TQC | ICSD | NoE | SF | | | |
|---|---|---|---|---|---|---|
| | 5+6 2 | 5+6 2 | 5+6 2 | 3+4 2 | 3+3 2 | 2+2 2 |
| | 3+4 2 | 5+6 2 | 3+4 2 | 5+6 2 | 3+3 2 | 3+3 2 |
| | 5+6 2 | 3+4 2 | 5+6 2 | 5+6 2 | 3+3 2 | 2+2 2 |
| | 3+4 2 | 3+4 2 | 3+4 2 | 3+4 2 | 2+2 2 | 3+3 2 |
| | 5+6 2 | 5+6 2 | 5+6 2 | 5+6 2 | 3+3 2 | 3+3 2 |
| | 3+4 2 | 3+4 2 | 3+4 2 | 5+6 2 | 2+2 2 | 2+2 2 |
| | 5+6 2 | 5+6 2 | 3+4 2 | 3+4 2 | 3+3 2 | 3+3 2 |
| | 3+4 2 | 5+6 2 | 5+6 2 | 3+4 2 | 2+2 2 | 2+2 2 |
| | 5+6 2 | 5+6 2 | 3+4 2 | 5+6 2 | 3+3 2 | 3+3 2 |
| | 3+4 2 | 3+4 2 | 5+6 2 | 5+6 2 | 2+2 2 | 2+2 2 |
| | 3+4 2 | 5+6 2 | 3+4 2 | 3+4 2 | 3+3 2 | 3+3 2 |
| | 5+6 2 | 3+4 2 | 5+6 2 | 3+4 2 | 3+3 2 | 3+3 2 |
| | 5+6 2 | 5+6 2 | 3+4 2 | 5+6 2 | 2+2 2 | 2+2 2 |
| | 3+4 2 | 3+4 2 | 5+6 2 | 5+6 2 | 3+3 2 | 2+2 2 |
| | 5+6 2 | 5+6 2 | 5+6 2 | 5+6 2 | 3+3 2 | 2+2 2 |
| | 3+4 2 | 5+6 2 | 3+4 2 | 3+4 2 | 2+2 2 | 3+3 2 |
| | 3+4 2 | 5+6 2 | 5+6 2 | 3+4 2 | 3+3 2 | 3+3 2 |
| | 5+6 2 | 3+4 2 | 3+4 2 | 5+6 2 | 2+2 2 | 2+2 2 |
| | 3+4 2 | 3+4 2 | 5+6 2 | 5+6 2 | 3+3 2 | 3+3 2 |
| | 3+4 2 | 5+6 2 | 3+4 2 | 5+6 2 | 2+2 2 | 2+2 2 |
| | 5+6 2 | 3+4 2 | 5+6 2 | 3+4 2 | 3+3 2 | 3+3 2 |
| | 5+6 2 | 5+6 2 | 3+4 2 | 3+4 2 | 2+2 2 | 2+2 2 |
| | 5+6 2 | 3+4 2 | 3+4 2 | 3+4 2 | 2+2 2 | 2+2 2 |
| | 3+4 2 | 5+6 2 | 5+6 2 | 5+6 2 | 2+2 2 | 3+3 2 |
| | 3+4 2 | 5+6 2 | 3+4 2 | 3+4 2 | 3+3 2 | 2+2 2 |
| | 5+6 2 | 3+4 2 | 5+6 2 | 3+4 2 | 2+2 2 | 3+3 2 |
| | 3+4 2 | 3+4 2 | 3+4 2 | 5+6 2 | 3+3 2 | 2+2 2 |
| | 5+6 2 | 3+4 2 | 5+6 2 | 3+4 2 | 2+2 2 | 3+3 2 |
| | 3+4 2 | 3+4 2 | 3+4 2 | 5+6 2 | 3+3 2 | 2+2 2 |
| | 5+6 2 | 5+6 2 | 5+6 2 | 3+4 2 | 2+2 2 | 3+3 2 |
| | 5+6 2 | 3+4 2 | 5+6 2 | 5+6 2 | 3+3 2 | 2+2 2 |
| | 3+4 2 | 5+6 2 | 3+4 2 | 3+4 2 | 2+2 2 | 3+3 2 |
| | 3+4 2 | 3+4 2 | 3+4 2 | 3+4 2 | 2+2 2 | 3+3 2 |
| 12.1.167 | 151699 | 110 | Ba1Er2F8 | | | |
| | -A | -GM | -M | -Y | -L | -V |
| | 5+6 2 | 3+4 2 | 3+4 2 | 5+6 2 | 3+3 2 | 2+2 2 |
| | 3+4 2 | 5+6 2 | 5+6 2 | 3+4 2 | 2+2 2 | 3+3 2 |
| | 5+6 2 | 3+4 2 | 3+4 2 | 5+6 2 | 3+3 2 | 2+2 2 |
| | 3+4 2 | 5+6 2 | 5+6 2 | 3+4 2 | 2+2 2 | 3+3 2 |
| | 3+4 2 | 3+4 2 | 3+4 2 | 3+4 2 | 3+3 2 | 3+3 2 |
| | 5+6 2 | 3+4 2 | 3+4 2 | 5+6 2 | 2+2 2 | 2+2 2 |
| | 3+4 2 | 5+6 2 | 5+6 2 | 3+4 2 | 3+3 2 | 3+3 2 |
| | 5+6 2 | 3+4 2 | 3+4 2 | 5+6 2 | 3+3 2 | 2+2 2 |
| | 3+4 2 | 5+6 2 | 5+6 2 | 3+4 2 | 2+2 2 | 3+3 2 |
| | 5+6 2 | 3+4 2 | 3+4 2 | 5+6 2 | 2+2 2 | 2+2 2 |
| | 3+4 2 | 3+4 2 | 5+6 2 | 3+4 2 | 3+3 2 | 3+3 2 |
| | 3+4 2 | 5+6 2 | 3+4 2 | 3+4 2 | 2+2 2 | 2+2 2 |
| | 5+6 2 | 5+6 2 | 5+6 2 | 5+6 2 | 3+3 2 | 3+3 2 |
| | 3+4 2 | 5+6 2 | 5+6 2 | 3+4 2 | 2+2 2 | 2+2 2 |
| | 5+6 2 | 3+4 2 | 3+4 2 | 5+6 2 | 3+3 2 | 3+3 2 |
| | 3+4 2 | 5+6 2 | 5+6 2 | 3+4 2 | 2+2 2 | 3+3 2 |
| | 5+6 2 | 3+4 2 | 3+4 2 | 5+6 2 | 3+3 2 | 2+2 2 |
| | 5+6 2 | 5+6 2 | 5+6 2 | 5+6 2 | 2+2 2 | 2+2 2 |
| | 5+6 2 | 5+6 2 | 5+6 2 | 5+6 2 | 2+2 2 | 2+2 2 |
| | 5+6 2 | 5+6 2 | 5+6 2 | 5+6 2 | 2+2 2 | 2+2 2 |
| | 3+4 2 | 3+4 2 | 5+6 2 | 3+4 2 | 2+2 2 | 3+3 2 |
| | 3+4 2 | 5+6 2 | 3+4 2 | 5+6 2 | 3+3 2 | 2+2 2 |
| | 5+6 2 | 3+4 2 | 3+4 2 | 5+6 2 | 3+3 2 | 3+3 2 |
| | 5+6 2 | 3+4 2 | 3+4 2 | 3+4 2 | 3+3 2 | 2+2 2 |
| | 3+4 2 | 5+6 2 | 5+6 2 | 3+4 2 | 2+2 2 | 3+3 2 |
| | 5+6 2 | 3+4 2 | 5+6 2 | 5+6 2 | 3+3 2 | 2+2 2 |

Continued on next column

**TABLE XXIV– continued**

| TQC | ICSD | | NoE | | SF | |
|---|---|---|---|---|---|---|
| | 3+4 2 | 3+4 2 | 3+4 2 | 5+6 2 | 2+2 2 | 3+3 2 |
| | 5+6 2 | 5+6 2 | 5+6 2 | 3+4 2 | 3+3 2 | 2+2 2 |
| | 5+6 2 | 3+4 2 | 3+4 2 | 3+4 2 | 2+2 2 | 3+3 2 |
| | 5+6 2 | 3+4 2 | 5+6 2 | 5+6 2 | 3+3 2 | 2+2 2 |
| | 3+4 2 | 5+6 2 | 3+4 2 | 3+4 2 | 3+3 2 | 2+2 2 |
| | 3+4 2 | 5+6 2 | 3+4 2 | 5+6 2 | 2+2 2 | 3+3 2 |
| | 3+4 2 | 5+6 2 | 5+6 2 | 3+4 2 | 2+2 2 | 2+2 2 |
| | 5+6 2 | 3+4 2 | 3+4 2 | 5+6 2 | 3+3 2 | 3+3 2 |
| | 5+6 2 | 3+4 2 | 3+4 2 | 3+4 2 | 2+2 2 | 2+2 2 |
| | 5+6 2 | 5+6 2 | 5+6 2 | 5+6 2 | 2+2 2 | 3+3 2 |
| | 5+6 2 | 3+4 2 | 3+4 2 | 3+4 2 | 3+3 2 | 3+3 2 |
| | 3+4 2 | 5+6 2 | 5+6 2 | 5+6 2 | 2+2 2 | 2+2 2 |
| | 3+4 2 | 5+6 2 | 5+6 2 | 3+4 2 | 3+3 2 | 3+3 2 |
| | 3+4 2 | 3+4 2 | 3+4 2 | 5+6 2 | 2+2 2 | 2+2 2 |
| | 5+6 2 | 3+4 2 | 3+4 2 | 3+4 2 | 3+3 2 | 2+2 2 |
| | 3+4 2 | 5+6 2 | 5+6 2 | 5+6 2 | 2+2 2 | 3+3 2 |
| | 3+4 2 | 5+6 2 | 5+6 2 | 3+4 2 | 2+2 2 | 2+2 2 |
| | 5+6 2 | 5+6 2 | 5+6 2 | 5+6 2 | 3+3 2 | 3+3 2 |
| | 5+6 2 | 3+4 2 | 3+4 2 | 3+4 2 | 3+3 2 | 2+2 2 |
| | 3+4 2 | 5+6 2 | 5+6 2 | 5+6 2 | 2+2 2 | 3+3 2 |
| | 3+4 2 | 3+4 2 | 3+4 2 | 3+4 2 | 3+3 2 | 2+2 2 |
| | 5+6 2 | 5+6 2 | 5+6 2 | 5+6 2 | 2+2 2 | 3+3 2 |
| | 3+4 2 | 3+4 2 | 3+4 2 | 3+4 2 | 3+3 2 | 2+2 2 |
| | 5+6 2 | 5+6 2 | 5+6 2 | 5+6 2 | 2+2 2 | 3+3 2 |
| | 5+6 2 | 3+4 2 | 3+4 2 | 3+4 2 | 3+3 2 | 2+2 2 |
| | 3+4 2 | 5+6 2 | 5+6 2 | 5+6 2 | 2+2 2 | 3+3 2 |
| | 5+6 2 | 3+4 2 | 5+6 2 | 3+4 2 | 3+3 2 | 2+2 2 |
| | 3+4 2 | 5+6 2 | 3+4 2 | 5+6 2 | 2+2 2 | 3+3 2 |
| | 5+6 2 | 3+4 2 | 3+4 2 | 5+6 2 | 3+3 2 | 2+2 2 |
| 12.1.168 | 61499 | | 66 | | Ba1Bi1O3 | |
| | -A | -GM | -M | -Y | -L | -V |
| | 5+6 2 | 3+4 2 | 5+6 2 | 5+6 2 | 2+2 2 | 3+3 2 |
| | 3+4 2 | 5+6 2 | 3+4 2 | 3+4 2 | 3+3 2 | 2+2 2 |
| | 5+6 2 | 3+4 2 | 5+6 2 | 3+4 2 | 3+3 2 | 2+2 2 |
| | 3+4 2 | 5+6 2 | 3+4 2 | 5+6 2 | 2+2 2 | 3+3 2 |
| | 3+4 2 | 5+6 2 | 3+4 2 | 3+4 2 | 3+3 2 | 2+2 2 |
| | 5+6 2 | 5+6 2 | 3+4 2 | 5+6 2 | 2+2 2 | 3+3 2 |
| | 3+4 2 | 3+4 2 | 5+6 2 | 5+6 2 | 2+2 2 | 3+3 2 |
| | 5+6 2 | 3+4 2 | 5+6 2 | 3+4 2 | 3+3 2 | 2+2 2 |
| | 3+4 2 | 3+4 2 | 5+6 2 | 5+6 2 | 2+2 2 | 3+3 2 |
| | 5+6 2 | 5+6 2 | 3+4 2 | 3+4 2 | 3+3 2 | 2+2 2 |
| | 3+4 2 | 3+4 2 | 5+6 2 | 5+6 2 | 2+2 2 | 3+3 2 |
| | 5+6 2 | 3+4 2 | 3+4 2 | 3+4 2 | 3+3 2 | 2+2 2 |
| | 5+6 2 | 3+4 2 | 5+6 2 | 5+6 2 | 2+2 2 | 3+3 2 |
| | 3+4 2 | 5+6 2 | 3+4 2 | 3+4 2 | 3+3 2 | 2+2 2 |
| | 5+6 2 | 5+6 2 | 5+6 2 | 3+4 2 | 3+3 2 | 3+3 2 |
| | 3+4 2 | 5+6 2 | 3+4 2 | 3+4 2 | 3+3 2 | 2+2 2 |
| | 3+4 2 | 3+4 2 | 3+4 2 | 5+6 2 | 2+2 2 | 2+2 2 |
| | 5+6 2 | 5+6 2 | 5+6 2 | 5+6 2 | 2+2 2 | 3+3 2 |
| | 3+4 2 | 5+6 2 | 3+4 2 | 5+6 2 | 2+2 2 | 2+2 2 |
| | 5+6 2 | 3+4 2 | 5+6 2 | 3+4 2 | 3+3 2 | 3+3 2 |
| | 3+4 2 | 3+4 2 | 3+4 2 | 5+6 2 | 2+2 2 | 2+2 2 |
| | 5+6 2 | 3+4 2 | 5+6 2 | 5+6 2 | 2+2 2 | 3+3 2 |
| | 5+6 2 | 5+6 2 | 5+6 2 | 5+6 2 | 3+3 2 | 3+3 2 |
| | 3+4 2 | 5+6 2 | 3+4 2 | 3+4 2 | 2+2 2 | 2+2 2 |
| | 5+6 2 | 3+4 2 | 5+6 2 | 3+4 2 | 3+3 2 | 2+2 2 |
| | 3+4 2 | 3+4 2 | 3+4 2 | 5+6 2 | 2+2 2 | 3+3 2 |
| | 5+6 2 | 5+6 2 | 5+6 2 | 5+6 2 | 3+3 2 | 3+3 2 |
| | 3+4 2 | 5+6 2 | 3+4 2 | 3+4 2 | 2+2 2 | 2+2 2 |
| | 3+4 2 | 5+6 2 | 3+4 2 | 5+6 2 | 3+3 2 | 3+3 2 |
| | 5+6 2 | 5+6 2 | 5+6 2 | 3+4 2 | 2+2 2 | 2+2 2 |

Continued on next column

**TABLE XXIV– continued**

| TQC | ICSD | | NoE | | SF | |
|---|---|---|---|---|---|---|
| | 3+4 2 | 3+4 2 | 5+6 2 | 3+4 2 | 3+3 2 | 2+2 2 |
| | 5+6 2 | 3+4 2 | 3+4 2 | 3+4 2 | 3+3 2 | 3+3 2 |
| | 5+6 2 | 3+4 2 | 5+6 2 | 3+4 2 | 3+3 2 | 3+3 2 |
| 12.1.169 | 63510 | | 90 | | B4Nd2Ni5 | |
| | -A | -GM | -M | -Y | -L | -V |
| | 5+6 2 | 3+4 2 | 3+4 2 | 5+6 2 | 3+3 2 | 2+2 2 |
| | 3+4 2 | 5+6 2 | 5+6 2 | 3+4 2 | 2+2 2 | 3+3 2 |
| | 5+6 2 | 3+4 2 | 3+4 2 | 5+6 2 | 3+3 2 | 2+2 2 |
| | 3+4 2 | 5+6 2 | 5+6 2 | 3+4 2 | 2+2 2 | 3+3 2 |
| | 5+6 2 | 3+4 2 | 3+4 2 | 5+6 2 | 3+3 2 | 2+2 2 |
| | 5+6 2 | 3+4 2 | 5+6 2 | 3+4 2 | 2+2 2 | 3+3 2 |
| | 3+4 2 | 5+6 2 | 3+4 2 | 5+6 2 | 3+3 2 | 2+2 2 |
| | 3+4 2 | 5+6 2 | 5+6 2 | 3+4 2 | 2+2 2 | 3+3 2 |
| | 3+4 2 | 3+4 2 | 3+4 2 | 3+4 2 | 2+2 2 | 2+2 2 |
| | 5+6 2 | 5+6 2 | 5+6 2 | 5+6 2 | 3+3 2 | 3+3 2 |
| | 3+4 2 | 3+4 2 | 3+4 2 | 3+4 2 | 2+2 2 | 2+2 2 |
| | 5+6 2 | 5+6 2 | 5+6 2 | 5+6 2 | 3+3 2 | 3+3 2 |
| | 3+4 2 | 3+4 2 | 5+6 2 | 5+6 2 | 3+3 2 | 3+3 2 |
| | 5+6 2 | 5+6 2 | 5+6 2 | 3+4 2 | 2+2 2 | 3+3 2 |
| | 5+6 2 | 3+4 2 | 3+4 2 | 5+6 2 | 2+2 2 | 2+2 2 |
| | 5+6 2 | 3+4 2 | 3+4 2 | 3+4 2 | 3+3 2 | 2+2 2 |
| | 3+4 2 | 5+6 2 | 3+4 2 | 3+4 2 | 2+2 2 | 3+3 2 |
| | 3+4 2 | 3+4 2 | 5+6 2 | 5+6 2 | 2+2 2 | 2+2 2 |
| | 3+4 2 | 3+4 2 | 3+4 2 | 3+4 2 | 3+3 2 | 3+3 2 |
| | 5+6 2 | 5+6 2 | 5+6 2 | 3+4 2 | 3+3 2 | 2+2 2 |
| | 3+4 2 | 5+6 2 | 5+6 2 | 3+4 2 | 2+2 2 | 3+3 2 |
| | 3+4 2 | 3+4 2 | 3+4 2 | 5+6 2 | 3+3 2 | 2+2 2 |
| | 5+6 2 | 5+6 2 | 5+6 2 | 3+4 2 | 2+2 2 | 3+3 2 |
| | 3+4 2 | 3+4 2 | 5+6 2 | 5+6 2 | 2+2 2 | 3+3 2 |
| | 5+6 2 | 3+4 2 | 3+4 2 | 3+4 2 | 3+3 2 | 2+2 2 |
| | 5+6 2 | 5+6 2 | 5+6 2 | 5+6 2 | 2+2 2 | 3+3 2 |
| | 5+6 2 | 3+4 2 | 5+6 2 | 3+4 2 | 3+3 2 | 3+3 2 |
| | 5+6 2 | 3+4 2 | 5+6 2 | 3+4 2 | 2+2 2 | 2+2 2 |
| | 3+4 2 | 5+6 2 | 3+4 2 | 5+6 2 | 3+3 2 | 3+3 2 |
| | 3+4 2 | 5+6 2 | 3+4 2 | 5+6 2 | 2+2 2 | 2+2 2 |
| | 5+6 2 | 5+6 2 | 5+6 2 | 3+4 2 | 2+2 2 | 2+2 2 |
| | 3+4 2 | 3+4 2 | 5+6 2 | 5+6 2 | 3+3 2 | 3+3 2 |
| | 5+6 2 | 5+6 2 | 3+4 2 | 3+4 2 | 3+3 2 | 3+3 2 |
| | 5+6 2 | 3+4 2 | 3+4 2 | 5+6 2 | 2+2 2 | 2+2 2 |
| | 3+4 2 | 5+6 2 | 5+6 2 | 3+4 2 | 2+2 2 | 2+2 2 |
| | 5+6 2 | 3+4 2 | 3+4 2 | 5+6 2 | 3+3 2 | 3+3 2 |
| | 3+4 2 | 3+4 2 | 5+6 2 | 3+4 2 | 2+2 2 | 2+2 2 |
| | 5+6 2 | 3+4 2 | 5+6 2 | 5+6 2 | 2+2 2 | 3+3 2 |
| | 5+6 2 | 3+4 2 | 5+6 2 | 3+4 2 | 2+2 2 | 3+3 2 |
| | 3+4 2 | 5+6 2 | 3+4 2 | 3+4 2 | 3+3 2 | 3+3 2 |
| | 5+6 2 | 3+4 2 | 5+6 2 | 5+6 2 | 2+2 2 | 3+3 2 |
| | 3+4 2 | 3+4 2 | 5+6 2 | 3+4 2 | 3+3 2 | 2+2 2 |
| | 5+6 2 | 5+6 2 | 3+4 2 | 5+6 2 | 2+2 2 | 3+3 2 |
| | 5+6 2 | 5+6 2 | 3+4 2 | 5+6 2 | 3+3 2 | 2+2 2 |
| | 3+4 2 | 3+4 2 | 5+6 2 | 3+4 2 | 3+3 2 | 3+3 2 |
| 12.1.170 | 170618 | | 84 | | B4La2Ni5 | |
| | -A | -GM | -M | -Y | -L | -V |
| | 5+6 2 | 3+4 2 | 3+4 2 | 5+6 2 | 3+3 2 | 2+2 2 |
| | 3+4 2 | 5+6 2 | 5+6 2 | 3+4 2 | 2+2 2 | 3+3 2 |
| | 5+6 2 | 3+4 2 | 3+4 2 | 5+6 2 | 3+3 2 | 2+2 2 |
| | 3+4 2 | 5+6 2 | 5+6 2 | 3+4 2 | 2+2 2 | 3+3 2 |
| | 5+6 2 | 3+4 2 | 3+4 2 | 5+6 2 | 3+3 2 | 2+2 2 |
| | 3+4 2 | 3+4 2 | 5+6 2 | 3+4 2 | 2+2 2 | 3+3 2 |
| | 5+6 2 | 5+6 2 | 3+4 2 | 5+6 2 | 3+3 2 | 2+2 2 |
| | 3+4 2 | 5+6 2 | 5+6 2 | 3+4 2 | 2+2 2 | 3+3 2 |
| | 3+4 2 | 3+4 2 | 3+4 2 | 3+4 2 | 2+2 2 | 2+2 2 |

Continued on next column

**TABLE XXIV– continued**

| TQC | ICSD | NoE | SF | | | |
|---|---|---|---|---|---|---|
| | 5+6 2 | 5+6 2 | 5+6 2 | 5+6 2 | 3+3 2 | 3+3 2 |
| | 5+6 2 | 3+4 2 | 3+4 2 | 3+4 2 | 2+2 2 | 2+2 2 |
| | 3+4 2 | 5+6 2 | 5+6 2 | 5+6 2 | 3+3 2 | 3+3 2 |
| | 3+4 2 | 3+4 2 | 5+6 2 | 5+6 2 | 3+3 2 | 3+3 2 |
| | 5+6 2 | 5+6 2 | 5+6 2 | 3+4 2 | 2+2 2 | 3+3 2 |
| | 5+6 2 | 3+4 2 | 3+4 2 | 3+4 2 | 2+2 2 | 2+2 2 |
| | 3+4 2 | 3+4 2 | 3+4 2 | 3+4 2 | 3+3 2 | 2+2 2 |
| | 5+6 2 | 5+6 2 | 3+4 2 | 5+6 2 | 2+2 2 | 3+3 2 |
| | 3+4 2 | 3+4 2 | 5+6 2 | 5+6 2 | 3+3 2 | 2+2 2 |
| | 5+6 2 | 3+4 2 | 3+4 2 | 3+4 2 | 2+2 2 | 3+3 2 |
| | 3+4 2 | 5+6 2 | 5+6 2 | 3+4 2 | 3+3 2 | 2+2 2 |
| | 5+6 2 | 5+6 2 | 5+6 2 | 3+4 2 | 2+2 2 | 3+3 2 |
| | 3+4 2 | 3+4 2 | 5+6 2 | 5+6 2 | 3+3 2 | 3+3 2 |
| | 3+4 2 | 5+6 2 | 5+6 2 | 3+4 2 | 2+2 2 | 2+2 2 |
| | 3+4 2 | 3+4 2 | 3+4 2 | 5+6 2 | 2+2 2 | 3+3 2 |
| | 5+6 2 | 3+4 2 | 5+6 2 | 3+4 2 | 3+3 2 | 2+2 2 |
| | 5+6 2 | 5+6 2 | 3+4 2 | 3+4 2 | 2+2 2 | 3+3 2 |
| | 5+6 2 | 3+4 2 | 5+6 2 | 5+6 2 | 3+3 2 | 3+3 2 |
| | 5+6 2 | 5+6 2 | 5+6 2 | 3+4 2 | 2+2 2 | 2+2 2 |
| | 3+4 2 | 3+4 2 | 3+4 2 | 5+6 2 | 3+3 2 | 3+3 2 |
| | 3+4 2 | 5+6 2 | 3+4 2 | 5+6 2 | 2+2 2 | 2+2 2 |
| | 5+6 2 | 5+6 2 | 5+6 2 | 5+6 2 | 2+2 2 | 2+2 2 |
| | 5+6 2 | 3+4 2 | 3+4 2 | 3+4 2 | 3+3 2 | 3+3 2 |
| | 3+4 2 | 5+6 2 | 5+6 2 | 3+4 2 | 2+2 2 | 2+2 2 |
| | 3+4 2 | 3+4 2 | 3+4 2 | 5+6 2 | 3+3 2 | 3+3 2 |
| | 5+6 2 | 3+4 2 | 5+6 2 | 3+4 2 | 2+2 2 | 2+2 2 |
| | 5+6 2 | 3+4 2 | 3+4 2 | 5+6 2 | 3+3 2 | 3+3 2 |
| | 3+4 2 | 5+6 2 | 5+6 2 | 3+4 2 | 2+2 2 | 2+2 2 |
| | 5+6 2 | 5+6 2 | 5+6 2 | 5+6 2 | 2+2 2 | 3+3 2 |
| | 3+4 2 | 3+4 2 | 5+6 2 | 3+4 2 | 2+2 2 | 3+3 2 |
| | 5+6 2 | 3+4 2 | 5+6 2 | 3+4 2 | 3+3 2 | 3+3 2 |
| | 5+6 2 | 3+4 2 | 5+6 2 | 3+4 2 | 2+2 2 | 3+3 2 |
| | 5+6 2 | 3+4 2 | 3+4 2 | 5+6 2 | 3+3 2 | 2+2 2 |
| 12.1.171 | 405291 | | 182 | B4Br5La4 | | |
| | -A | -GM | -M | -Y | -L | -V |
| | 3+4 2 | 3+4 2 | 3+4 2 | 3+4 2 | 2+2 2 | 2+2 2 |
| | 5+6 2 | 5+6 2 | 5+6 2 | 5+6 2 | 3+3 2 | 3+3 2 |
| | 3+4 2 | 3+4 2 | 3+4 2 | 3+4 2 | 2+2 2 | 2+2 2 |
| | 5+6 2 | 5+6 2 | 5+6 2 | 5+6 2 | 3+3 2 | 3+3 2 |
| | 3+4 2 | 3+4 2 | 3+4 2 | 3+4 2 | 2+2 2 | 2+2 2 |
| | 5+6 2 | 5+6 2 | 5+6 2 | 5+6 2 | 3+3 2 | 3+3 2 |
| | 5+6 2 | 3+4 2 | 5+6 2 | 3+4 2 | 2+2 2 | 2+2 2 |
| | 3+4 2 | 5+6 2 | 3+4 2 | 5+6 2 | 3+3 2 | 3+3 2 |
| | 3+4 2 | 3+4 2 | 3+4 2 | 3+4 2 | 2+2 2 | 2+2 2 |
| | 5+6 2 | 5+6 2 | 5+6 2 | 5+6 2 | 3+3 2 | 3+3 2 |
| | 5+6 2 | 5+6 2 | 5+6 2 | 5+6 2 | 2+2 2 | 2+2 2 |
| | 3+4 2 | 3+4 2 | 3+4 2 | 3+4 2 | 3+3 2 | 3+3 2 |
| | 3+4 2 | 3+4 2 | 3+4 2 | 3+4 2 | 2+2 2 | 2+2 2 |
| | 5+6 2 | 5+6 2 | 5+6 2 | 5+6 2 | 3+3 2 | 3+3 2 |
| | 5+6 2 | 3+4 2 | 5+6 2 | 3+4 2 | 2+2 2 | 3+3 2 |
| | 3+4 2 | 5+6 2 | 3+4 2 | 5+6 2 | 3+3 2 | 2+2 2 |
| | 3+4 2 | 3+4 2 | 5+6 2 | 5+6 2 | 3+3 2 | 3+3 2 |
| | 5+6 2 | 5+6 2 | 3+4 2 | 3+4 2 | 2+2 2 | 2+2 2 |
| | 3+4 2 | 3+4 2 | 5+6 2 | 5+6 2 | 2+2 2 | 2+2 2 |
| | 5+6 2 | 5+6 2 | 3+4 2 | 3+4 2 | 3+3 2 | 3+3 2 |
| | 3+4 2 | 3+4 2 | 5+6 2 | 3+4 2 | 2+2 2 | 3+3 2 |
| | 5+6 2 | 3+4 2 | 3+4 2 | 5+6 2 | 3+3 2 | 2+2 2 |
| | 3+4 2 | 5+6 2 | 5+6 2 | 3+4 2 | 2+2 2 | 3+3 2 |
| | 5+6 2 | 5+6 2 | 3+4 2 | 5+6 2 | 3+3 2 | 2+2 2 |
| | 5+6 2 | 3+4 2 | 5+6 2 | 3+4 2 | 2+2 2 | 3+3 2 |
| | 3+4 2 | 5+6 2 | 3+4 2 | 5+6 2 | 3+3 2 | 3+3 2 |

Continued on next column

**TABLE XXIV– continued**

| TQC | ICSD | NoE | SF | | | |
|---|---|---|---|---|---|---|
| | 5+6 2 | 3+4 2 | 5+6 2 | 3+4 2 | 2+2 2 | 2+2 2 |
| | 3+4 2 | 5+6 2 | 3+4 2 | 5+6 2 | 3+3 2 | 2+2 2 |
| | 5+6 2 | 3+4 2 | 5+6 2 | 3+4 2 | 2+2 2 | 3+3 2 |
| | 3+4 2 | 3+4 2 | 3+4 2 | 3+4 2 | 2+2 2 | 2+2 2 |
| | 5+6 2 | 5+6 2 | 5+6 2 | 5+6 2 | 3+3 2 | 2+2 2 |
| | 3+4 2 | 3+4 2 | 3+4 2 | 5+6 2 | 3+3 2 | 3+3 2 |
| | 5+6 2 | 5+6 2 | 5+6 2 | 3+4 2 | 2+2 2 | 3+3 2 |
| | 3+4 2 | 5+6 2 | 3+4 2 | 5+6 2 | 3+3 2 | 2+2 2 |
| | 5+6 2 | 3+4 2 | 5+6 2 | 3+4 2 | 2+2 2 | 3+3 2 |
| | 3+4 2 | 5+6 2 | 3+4 2 | 5+6 2 | 3+3 2 | 2+2 2 |
| | 5+6 2 | 3+4 2 | 5+6 2 | 3+4 2 | 2+2 2 | 3+3 2 |
| | 3+4 2 | 5+6 2 | 3+4 2 | 5+6 2 | 3+3 2 | 2+2 2 |
| | 5+6 2 | 3+4 2 | 5+6 2 | 3+4 2 | 2+2 2 | 3+3 2 |
| | 3+4 2 | 3+4 2 | 3+4 2 | 3+4 2 | 2+2 2 | 2+2 2 |
| | 5+6 2 | 3+4 2 | 3+4 2 | 5+6 2 | 3+3 2 | 2+2 2 |
| | 3+4 2 | 5+6 2 | 5+6 2 | 3+4 2 | 2+2 2 | 3+3 2 |
| | 3+4 2 | 5+6 2 | 3+4 2 | 5+6 2 | 3+3 2 | 2+2 2 |
| | 5+6 2 | 3+4 2 | 5+6 2 | 3+4 2 | 2+2 2 | 3+3 2 |
| | 5+6 2 | 5+6 2 | 3+4 2 | 5+6 2 | 3+3 2 | 2+2 2 |
| | 3+4 2 | 3+4 2 | 5+6 2 | 3+4 2 | 2+2 2 | 3+3 2 |
| | 3+4 2 | 5+6 2 | 3+4 2 | 5+6 2 | 2+2 2 | 2+2 2 |
| | 5+6 2 | 3+4 2 | 5+6 2 | 3+4 2 | 3+3 2 | 3+3 2 |
| | 3+4 2 | 5+6 2 | 3+4 2 | 5+6 2 | 3+3 2 | 3+3 2 |
| | 5+6 2 | 3+4 2 | 5+6 2 | 3+4 2 | 3+3 2 | 3+3 2 |
| | 3+4 2 | 5+6 2 | 5+6 2 | 5+6 2 | 2+2 2 | 2+2 2 |
| | 5+6 2 | 5+6 2 | 5+6 2 | 5+6 2 | 2+2 2 | 2+2 2 |
| | 5+6 2 | 3+4 2 | 3+4 2 | 3+4 2 | 3+3 2 | 2+2 2 |
| | 3+4 2 | 3+4 2 | 3+4 2 | 5+6 2 | 3+3 2 | 3+3 2 |
| | 5+6 2 | 5+6 2 | 3+4 2 | 3+4 2 | 3+3 2 | 2+2 2 |
| | 3+4 2 | 3+4 2 | 5+6 2 | 3+4 2 | 2+2 2 | 3+3 2 |
| | 3+4 2 | 5+6 2 | 3+4 2 | 5+6 2 | 3+3 2 | 2+2 2 |
| | 5+6 2 | 3+4 2 | 5+6 2 | 5+6 2 | 2+2 2 | 3+3 2 |
| | 3+4 2 | 3+4 2 | 5+6 2 | 5+6 2 | 2+2 2 | 2+2 2 |
| | 5+6 2 | 5+6 2 | 5+6 2 | 3+4 2 | 3+3 2 | 3+3 2 |
| | 3+4 2 | 3+4 2 | 5+6 2 | 5+6 2 | 2+2 2 | 2+2 2 |
| | 5+6 2 | 5+6 2 | 3+4 2 | 5+6 2 | 3+3 2 | 3+3 2 |
| | 3+4 2 | 3+4 2 | 3+4 2 | 3+4 2 | 3+3 2 | 2+2 2 |
| | 3+4 2 | 5+6 2 | 3+4 2 | 3+4 2 | 2+2 2 | 3+3 2 |
| | 3+4 2 | 5+6 2 | 5+6 2 | 5+6 2 | 2+2 2 | 2+2 2 |
| | 5+6 2 | 3+4 2 | 3+4 2 | 5+6 2 | 3+3 2 | 3+3 2 |
| | 5+6 2 | 5+6 2 | 3+4 2 | 3+4 2 | 3+3 2 | 2+2 2 |
| | 3+4 2 | 3+4 2 | 3+4 2 | 3+4 2 | 2+2 2 | 2+2 2 |
| | 3+4 2 | 5+6 2 | 5+6 2 | 5+6 2 | 3+3 2 | 3+3 2 |
| | 5+6 2 | 3+4 2 | 3+4 2 | 5+6 2 | 2+2 2 | 2+2 2 |
| | 3+4 2 | 5+6 2 | 5+6 2 | 5+6 2 | 3+3 2 | 3+3 2 |
| | 5+6 2 | 5+6 2 | 3+4 2 | 3+4 2 | 2+2 2 | 3+3 2 |
| | 5+6 2 | 3+4 2 | 5+6 2 | 5+6 2 | 3+3 2 | 2+2 2 |
| | 3+4 2 | 5+6 2 | 3+4 2 | 3+4 2 | 3+3 2 | 2+2 2 |
| | 5+6 2 | 5+6 2 | 5+6 2 | 5+6 2 | 2+2 2 | 3+3 2 |
| | 3+4 2 | 3+4 2 | 3+4 2 | 5+6 2 | 3+3 2 | 2+2 2 |
| | 5+6 2 | 5+6 2 | 5+6 2 | 3+4 2 | 3+3 2 | 3+3 2 |
| | 3+4 2 | 5+6 2 | 3+4 2 | 5+6 2 | 2+2 2 | 3+3 2 |
| | 5+6 2 | 5+6 2 | 5+6 2 | 3+4 2 | 3+3 2 | 2+2 2 |
| | 5+6 2 | 5+6 2 | 3+4 2 | 5+6 2 | 3+3 2 | 3+3 2 |
| | 3+4 2 | 3+4 2 | 5+6 2 | 3+4 2 | 2+2 2 | 2+2 2 |
| | 5+6 2 | 5+6 2 | 5+6 2 | 5+6 2 | 3+3 2 | 3+3 2 |
| | 5+6 2 | 3+4 2 | 5+6 2 | 3+4 2 | 2+2 2 | 2+2 2 |
| | 3+4 2 | 3+4 2 | 3+4 2 | 3+4 2 | 3+3 2 | 3+3 2 |
| | 3+4 2 | 5+6 2 | 5+6 2 | 5+6 2 | 2+2 2 | 2+2 2 |
| | 5+6 2 | 5+6 2 | 3+4 2 | 5+6 2 | 3+3 2 | 3+3 2 |
| | 5+6 2 | 3+4 2 | 3+4 2 | 3+4 2 | 2+2 2 | 3+3 2 |

Continued on next column

**TABLE XXIV– continued**

| TQC | ICSD | | NoE | | SF | |
|---|---|---|---|---|---|---|
| | 3+4 2 | 5+6 2 | 5+6 2 | 5+6 2 | 3+3 2 | 2+2 2 |
| | 3+4 2 | 3+4 2 | 3+4 2 | 3+4 2 | 2+2 2 | 2+2 2 |
| | 5+6 2 | 5+6 2 | 5+6 2 | 5+6 2 | 3+3 2 | 3+3 2 |
| | 5+6 2 | 5+6 2 | 3+4 2 | 3+4 2 | 2+2 2 | 3+3 2 |
| 12.1.172 | 615340 | | 44 | | B2Rh3Y1 | |
| | -A | -GM | -M | -Y | -L | -V |
| | 5+6 2 | 3+4 2 | 5+6 2 | 3+4 2 | 3+3 2 | 2+2 2 |
| | 3+4 2 | 5+6 2 | 3+4 2 | 5+6 2 | 2+2 2 | 3+3 2 |
| | 3+4 2 | 5+6 2 | 3+4 2 | 5+6 2 | 2+2 2 | 3+3 2 |
| | 3+4 2 | 5+6 2 | 3+4 2 | 5+6 2 | 2+2 2 | 3+3 2 |
| | 3+4 2 | 3+4 2 | 5+6 2 | 5+6 2 | 3+3 2 | 3+3 2 |
| | 3+4 2 | 3+4 2 | 3+4 2 | 3+4 2 | 2+2 2 | 2+2 2 |
| | 5+6 2 | 3+4 2 | 5+6 2 | 5+6 2 | 3+3 2 | 3+3 2 |
| | 3+4 2 | 5+6 2 | 5+6 2 | 5+6 2 | 3+3 2 | 3+3 2 |
| | 3+4 2 | 3+4 2 | 3+4 2 | 5+6 2 | 2+2 2 | 3+3 2 |
| | 3+4 2 | 3+4 2 | 5+6 2 | 3+4 2 | 3+3 2 | 2+2 2 |
| | 3+4 2 | 3+4 2 | 5+6 2 | 5+6 2 | 3+3 2 | 3+3 2 |
| | 3+4 2 | 3+4 2 | 3+4 2 | 3+4 2 | 2+2 2 | 2+2 2 |
| | 3+4 2 | 3+4 2 | 5+6 2 | 3+4 2 | 3+3 2 | 2+2 2 |
| | 5+6 2 | 3+4 2 | 5+6 2 | 3+4 2 | 3+3 2 | 2+2 2 |
| | 3+4 2 | 3+4 2 | 3+4 2 | 5+6 2 | 2+2 2 | 3+3 2 |
| | 5+6 2 | 3+4 2 | 3+4 2 | 5+6 2 | 2+2 2 | 3+3 2 |
| | 5+6 2 | 3+4 2 | 3+4 2 | 5+6 2 | 3+3 2 | 3+3 2 |
| | 3+4 2 | 3+4 2 | 5+6 2 | 3+4 2 | 2+2 2 | 2+2 2 |
| | 3+4 2 | 3+4 2 | 5+6 2 | 3+4 2 | 3+3 2 | 3+3 2 |
| | 3+4 2 | 5+6 2 | 5+6 2 | 5+6 2 | 3+3 2 | 2+2 2 |
| | 5+6 2 | 3+4 2 | 5+6 2 | 5+6 2 | 3+3 2 | 3+3 2 |
| | 5+6 2 | 5+6 2 | 5+6 2 | 5+6 2 | 3+3 2 | 3+3 2 |
| 12.1.173 | 261932 | | 96 | | B2Ni3Zn1 | |
| | -A | -GM | -M | -Y | -L | -V |
| | 3+4 2 | 3+4 2 | 3+4 2 | 3+4 2 | 3+3 2 | 3+3 2 |
| | 5+6 2 | 5+6 2 | 5+6 2 | 5+6 2 | 2+2 2 | 2+2 2 |
| | 3+4 2 | 3+4 2 | 3+4 2 | 3+4 2 | 2+2 2 | 2+2 2 |
| | 5+6 2 | 3+4 2 | 5+6 2 | 5+6 2 | 3+3 2 | 3+3 2 |
| | 3+4 2 | 5+6 2 | 5+6 2 | 3+4 2 | 3+3 2 | 2+2 2 |
| | 5+6 2 | 5+6 2 | 3+4 2 | 5+6 2 | 2+2 2 | 3+3 2 |
| | 3+4 2 | 3+4 2 | 3+4 2 | 3+4 2 | 3+3 2 | 3+3 2 |
| | 5+6 2 | 5+6 2 | 5+6 2 | 5+6 2 | 2+2 2 | 2+2 2 |
| | 3+4 2 | 5+6 2 | 5+6 2 | 5+6 2 | 3+3 2 | 2+2 2 |
| | 5+6 2 | 3+4 2 | 3+4 2 | 3+4 2 | 2+2 2 | 2+2 2 |
| | 3+4 2 | 5+6 2 | 5+6 2 | 5+6 2 | 2+2 2 | 3+3 2 |
| | 5+6 2 | 3+4 2 | 3+4 2 | 3+4 2 | 2+2 2 | 3+3 2 |
| | 5+6 2 | 3+4 2 | 3+4 2 | 5+6 2 | 3+3 2 | 3+3 2 |
| | 5+6 2 | 5+6 2 | 5+6 2 | 3+4 2 | 3+3 2 | 2+2 2 |
| | 3+4 2 | 3+4 2 | 3+4 2 | 3+4 2 | 3+3 2 | 3+3 2 |
| | 3+4 2 | 5+6 2 | 5+6 2 | 5+6 2 | 2+2 2 | 2+2 2 |
| | 5+6 2 | 3+4 2 | 3+4 2 | 3+4 2 | 2+2 2 | 2+2 2 |
| | 3+4 2 | 3+4 2 | 5+6 2 | 5+6 2 | 3+3 2 | 3+3 2 |
| | 5+6 2 | 5+6 2 | 3+4 2 | 5+6 2 | 3+3 2 | 3+3 2 |
| | 5+6 2 | 5+6 2 | 3+4 2 | 3+4 2 | 2+2 2 | 3+3 2 |
| | 3+4 2 | 5+6 2 | 3+4 2 | 3+4 2 | 3+3 2 | 2+2 2 |
| | 5+6 2 | 3+4 2 | 5+6 2 | 3+4 2 | 2+2 2 | 2+2 2 |
| | 3+4 2 | 3+4 2 | 3+4 2 | 5+6 2 | 3+3 2 | 3+3 2 |
| | 3+4 2 | 3+4 2 | 5+6 2 | 3+4 2 | 2+2 2 | 3+3 2 |
| | 5+6 2 | 5+6 2 | 5+6 2 | 5+6 2 | 3+3 2 | 2+2 2 |
| | 5+6 2 | 5+6 2 | 5+6 2 | 5+6 2 | 2+2 2 | 2+2 2 |
| | 3+4 2 | 3+4 2 | 3+4 2 | 5+6 2 | 3+3 2 | 3+3 2 |
| | 3+4 2 | 5+6 2 | 5+6 2 | 3+4 2 | 2+2 2 | 2+2 2 |
| | 5+6 2 | 3+4 2 | 5+6 2 | 3+4 2 | 3+3 2 | 3+3 2 |
| | 3+4 2 | 3+4 2 | 3+4 2 | 3+4 2 | 2+2 2 | 2+2 2 |
| | 3+4 2 | 3+4 2 | 3+4 2 | 5+6 2 | 3+3 2 | 3+3 2 |

Continued on next column

**TABLE XXIV– continued**

| TQC | ICSD | | NoE | | SF | |
|---|---|---|---|---|---|---|
| | 5+6 2 | 5+6 2 | 5+6 2 | 5+6 2 | 2+2 2 | 2+2 2 |
| | 5+6 2 | 3+4 2 | 5+6 2 | 3+4 2 | 2+2 2 | 3+3 2 |
| | 5+6 2 | 5+6 2 | 5+6 2 | 5+6 2 | 3+3 2 | 2+2 2 |
| | 3+4 2 | 5+6 2 | 3+4 2 | 5+6 2 | 2+2 2 | 2+2 2 |
| | 5+6 2 | 3+4 2 | 5+6 2 | 3+4 2 | 3+3 2 | 2+2 2 |
| | 3+4 2 | 3+4 2 | 3+4 2 | 3+4 2 | 2+2 2 | 3+3 2 |
| | 3+4 2 | 5+6 2 | 5+6 2 | 5+6 2 | 2+2 2 | 3+3 2 |
| | 3+4 2 | 5+6 2 | 3+4 2 | 3+4 2 | 3+3 2 | 2+2 2 |
| | 5+6 2 | 3+4 2 | 3+4 2 | 5+6 2 | 2+2 2 | 3+3 2 |
| | 3+4 2 | 3+4 2 | 5+6 2 | 5+6 2 | 3+3 2 | 3+3 2 |
| | 3+4 2 | 5+6 2 | 3+4 2 | 3+4 2 | 3+3 2 | 2+2 2 |
| | 3+4 2 | 3+4 2 | 3+4 2 | 5+6 2 | 2+2 2 | 2+2 2 |
| | 5+6 2 | 5+6 2 | 5+6 2 | 3+4 2 | 2+2 2 | 3+3 2 |
| | 5+6 2 | 5+6 2 | 3+4 2 | 5+6 2 | 3+3 2 | 3+3 2 |
| | 3+4 2 | 3+4 2 | 5+6 2 | 3+4 2 | 3+3 2 | 2+2 2 |
| | 5+6 2 | 5+6 2 | 5+6 2 | 3+4 2 | 2+2 2 | 2+2 2 |
| | 3+4 2 | 3+4 2 | 3+4 2 | 5+6 2 | 3+3 2 | 2+2 2 |
| 12.1.174 | 107555 | | 254 | | Au8Ga1K4 | |
| | -A | -GM | -M | -Y | -L | -V |
| | 5+6 2 | 3+4 2 | 3+4 2 | 5+6 2 | 2+2 2 | 2+2 2 |
| | 3+4 2 | 5+6 2 | 5+6 2 | 3+4 2 | 3+3 2 | 3+3 2 |
| | 3+4 2 | 3+4 2 | 5+6 2 | 5+6 2 | 2+2 2 | 2+2 2 |
| | 5+6 2 | 5+6 2 | 3+4 2 | 3+4 2 | 3+3 2 | 3+3 2 |
| | 5+6 2 | 3+4 2 | 5+6 2 | 3+4 2 | 2+2 2 | 3+3 2 |
| | 3+4 2 | 5+6 2 | 3+4 2 | 5+6 2 | 3+3 2 | 2+2 2 |
| | 3+4 2 | 3+4 2 | 3+4 2 | 3+4 2 | 2+2 2 | 2+2 2 |
| | 5+6 2 | 5+6 2 | 5+6 2 | 5+6 2 | 3+3 2 | 3+3 2 |
| | 5+6 2 | 3+4 2 | 3+4 2 | 5+6 2 | 2+2 2 | 2+2 2 |
| | 3+4 2 | 5+6 2 | 5+6 2 | 3+4 2 | 3+3 2 | 3+3 2 |
| | 5+6 2 | 3+4 2 | 3+4 2 | 5+6 2 | 2+2 2 | 2+2 2 |
| | 3+4 2 | 3+4 2 | 5+6 2 | 5+6 2 | 3+3 2 | 3+3 2 |
| | 5+6 2 | 5+6 2 | 3+4 2 | 3+4 2 | 2+2 2 | 2+2 2 |
| | 3+4 2 | 3+4 2 | 5+6 2 | 5+6 2 | 3+3 2 | 3+3 2 |
| | 5+6 2 | 5+6 2 | 3+4 2 | 3+4 2 | 2+2 2 | 2+2 2 |
| | 3+4 2 | 5+6 2 | 5+6 2 | 3+4 2 | 3+3 2 | 3+3 2 |
| | 5+6 2 | 3+4 2 | 3+4 2 | 5+6 2 | 2+2 2 | 2+2 2 |
| | 3+4 2 | 5+6 2 | 5+6 2 | 3+4 2 | 3+3 2 | 3+3 2 |
| | 3+4 2 | 3+4 2 | 5+6 2 | 5+6 2 | 2+2 2 | 2+2 2 |
| | 3+4 2 | 3+4 2 | 3+4 2 | 3+4 2 | 3+3 2 | 3+3 2 |
| | 5+6 2 | 5+6 2 | 5+6 2 | 5+6 2 | 2+2 2 | 2+2 2 |
| | 3+4 2 | 3+4 2 | 3+4 2 | 3+4 2 | 3+3 2 | 3+3 2 |
| | 5+6 2 | 5+6 2 | 5+6 2 | 5+6 2 | 2+2 2 | 2+2 2 |
| | 5+6 2 | 5+6 2 | 3+4 2 | 3+4 2 | 3+3 2 | 3+3 2 |
| | 3+4 2 | 3+4 2 | 3+4 2 | 3+4 2 | 2+2 2 | 2+2 2 |
| | 5+6 2 | 5+6 2 | 5+6 2 | 5+6 2 | 2+2 2 | 3+3 2 |
| | 5+6 2 | 3+4 2 | 5+6 2 | 3+4 2 | 3+3 2 | 2+2 2 |
| | 3+4 2 | 5+6 2 | 3+4 2 | 5+6 2 | 3+3 2 | 3+3 2 |
| | 3+4 2 | 3+4 2 | 3+4 2 | 3+4 2 | 2+2 2 | 2+2 2 |
| | 5+6 2 | 5+6 2 | 5+6 2 | 5+6 2 | 3+3 2 | 3+3 2 |
| | 3+4 2 | 3+4 2 | 3+4 2 | 3+4 2 | 2+2 2 | 2+2 2 |
| | 5+6 2 | 5+6 2 | 5+6 2 | 5+6 2 | 3+3 2 | 3+3 2 |
| | 3+4 2 | 3+4 2 | 3+4 2 | 3+4 2 | 3+3 2 | 2+2 2 |
| | 5+6 2 | 5+6 2 | 5+6 2 | 5+6 2 | 2+2 2 | 3+3 2 |
| | 5+6 2 | 3+4 2 | 5+6 2 | 3+4 2 | 3+3 2 | 3+3 2 |
| | 3+4 2 | 5+6 2 | 3+4 2 | 5+6 2 | 3+3 2 | 2+2 2 |
| | 3+4 2 | 5+6 2 | 5+6 2 | 3+4 2 | 2+2 2 | 2+2 2 |
| | 5+6 2 | 3+4 2 | 3+4 2 | 5+6 2 | 3+3 2 | 3+3 2 |
| | 3+4 2 | 3+4 2 | 3+4 2 | 3+4 2 | 2+2 2 | 3+3 2 |
| | 5+6 2 | 3+4 2 | 5+6 2 | 3+4 2 | 3+3 2 | 2+2 2 |
| | 5+6 2 | 3+4 2 | 3+4 2 | 5+6 2 | 2+2 2 | 3+3 2 |
| | 3+4 2 | 3+4 2 | 5+6 2 | 3+4 2 | 3+3 2 | 2+2 2 |

Continued on next column

**TABLE XXIV– continued**

| TQC | ICSD | NoE | SF | | | |
|---|---|---|---|---|---|---|
| | 5+6 2 | 5+6 2 | 5+6 2 | 3+4 2 | 3+3 2 | 3+3 2 |
| | 3+4 2 | 5+6 2 | 3+4 2 | 5+6 2 | 3+3 2 | 2+2 2 |
| | 5+6 2 | 3+4 2 | 3+4 2 | 5+6 2 | 2+2 2 | 2+2 2 |
| | 3+4 2 | 5+6 2 | 5+6 2 | 3+4 2 | 3+3 2 | 3+3 2 |
| | 5+6 2 | 3+4 2 | 3+4 2 | 5+6 2 | 2+2 2 | 2+2 2 |
| | 3+4 2 | 5+6 2 | 3+4 2 | 3+4 2 | 3+3 2 | 3+3 2 |
| | 3+4 2 | 3+4 2 | 5+6 2 | 3+4 2 | 2+2 2 | 3+3 2 |
| | 5+6 2 | 3+4 2 | 5+6 2 | 3+4 2 | 3+3 2 | 2+2 2 |
| | 3+4 2 | 5+6 2 | 5+6 2 | 3+4 2 | 2+2 2 | 2+2 2 |
| | 5+6 2 | 3+4 2 | 5+6 2 | 5+6 2 | 3+3 2 | 3+3 2 |
| | 5+6 2 | 3+4 2 | 3+4 2 | 3+4 2 | 3+3 2 | 3+3 2 |
| | 3+4 2 | 5+6 2 | 3+4 2 | 5+6 2 | 2+2 2 | 2+2 2 |
| | 5+6 2 | 3+4 2 | 5+6 2 | 5+6 2 | 3+3 2 | 3+3 2 |
| | 3+4 2 | 5+6 2 | 3+4 2 | 3+4 2 | 3+3 2 | 2+2 2 |
| | 5+6 2 | 5+6 2 | 3+4 2 | 5+6 2 | 2+2 2 | 3+3 2 |
| | 3+4 2 | 3+4 2 | 5+6 2 | 3+4 2 | 2+2 2 | 2+2 2 |
| | 3+4 2 | 3+4 2 | 3+4 2 | 3+4 2 | 3+3 2 | 2+2 2 |
| | 5+6 2 | 3+4 2 | 5+6 2 | 5+6 2 | 2+2 2 | 3+3 2 |
| | 3+4 2 | 5+6 2 | 3+4 2 | 3+4 2 | 3+3 2 | 2+2 2 |
| | 5+6 2 | 3+4 2 | 5+6 2 | 5+6 2 | 2+2 2 | 3+3 2 |
| | 3+4 2 | 5+6 2 | 3+4 2 | 3+4 2 | 3+3 2 | 2+2 2 |
| | 5+6 2 | 5+6 2 | 5+6 2 | 5+6 2 | 2+2 2 | 3+3 2 |
| | 5+6 2 | 3+4 2 | 3+4 2 | 5+6 2 | 3+3 2 | 2+2 2 |
| | 3+4 2 | 5+6 2 | 5+6 2 | 3+4 2 | 2+2 2 | 3+3 2 |
| | 3+4 2 | 5+6 2 | 3+4 2 | 5+6 2 | 2+2 2 | 3+3 2 |
| | 5+6 2 | 3+4 2 | 5+6 2 | 3+4 2 | 3+3 2 | 2+2 2 |
| | 3+4 2 | 3+4 2 | 3+4 2 | 3+4 2 | 3+3 2 | 3+3 2 |
| | 5+6 2 | 5+6 2 | 5+6 2 | 5+6 2 | 2+2 2 | 2+2 2 |
| | 5+6 2 | 3+4 2 | 3+4 2 | 5+6 2 | 3+3 2 | 2+2 2 |
| | 3+4 2 | 5+6 2 | 5+6 2 | 3+4 2 | 2+2 2 | 3+3 2 |
| | 3+4 2 | 3+4 2 | 3+4 2 | 3+4 2 | 3+3 2 | 2+2 2 |
| | 5+6 2 | 3+4 2 | 5+6 2 | 5+6 2 | 2+2 2 | 3+3 2 |
| | 5+6 2 | 5+6 2 | 5+6 2 | 3+4 2 | 3+3 2 | 2+2 2 |
| | 3+4 2 | 5+6 2 | 3+4 2 | 5+6 2 | 2+2 2 | 3+3 2 |
| | 3+4 2 | 3+4 2 | 5+6 2 | 3+4 2 | 3+3 2 | 3+3 2 |
| | 5+6 2 | 5+6 2 | 3+4 2 | 5+6 2 | 2+2 2 | 2+2 2 |
| | 3+4 2 | 3+4 2 | 3+4 2 | 3+4 2 | 3+3 2 | 2+2 2 |
| | 5+6 2 | 5+6 2 | 5+6 2 | 5+6 2 | 2+2 2 | 3+3 2 |
| | 5+6 2 | 3+4 2 | 3+4 2 | 3+4 2 | 3+3 2 | 3+3 2 |
| | 3+4 2 | 5+6 2 | 5+6 2 | 5+6 2 | 3+3 2 | 2+2 2 |
| | 3+4 2 | 3+4 2 | 3+4 2 | 5+6 2 | 2+2 2 | 2+2 2 |
| | 5+6 2 | 3+4 2 | 5+6 2 | 3+4 2 | 3+3 2 | 3+3 2 |
| | 5+6 2 | 5+6 2 | 3+4 2 | 5+6 2 | 2+2 2 | 2+2 2 |
| | 3+4 2 | 3+4 2 | 5+6 2 | 3+4 2 | 2+2 2 | 3+3 2 |
| | 5+6 2 | 5+6 2 | 5+6 2 | 3+4 2 | 2+2 2 | 2+2 2 |
| | 3+4 2 | 3+4 2 | 3+4 2 | 5+6 2 | 3+3 2 | 3+3 2 |
| | 5+6 2 | 5+6 2 | 5+6 2 | 3+4 2 | 2+2 2 | 2+2 2 |
| | 3+4 2 | 3+4 2 | 3+4 2 | 3+4 2 | 3+3 2 | 3+3 2 |
| | 5+6 2 | 3+4 2 | 5+6 2 | 5+6 2 | 3+3 2 | 2+2 2 |
| | 3+4 2 | 3+4 2 | 3+4 2 | 3+4 2 | 2+2 2 | 3+3 2 |
| | 3+4 2 | 5+6 2 | 5+6 2 | 5+6 2 | 2+2 2 | 2+2 2 |
| | 5+6 2 | 3+4 2 | 3+4 2 | 3+4 2 | 3+3 2 | 3+3 2 |
| | 3+4 2 | 5+6 2 | 3+4 2 | 5+6 2 | 3+3 2 | 2+2 2 |
| | 5+6 2 | 5+6 2 | 5+6 2 | 3+4 2 | 2+2 2 | 3+3 2 |
| | 5+6 2 | 3+4 2 | 5+6 2 | 5+6 2 | 3+3 2 | 2+2 2 |
| | 3+4 2 | 5+6 2 | 3+4 2 | 3+4 2 | 2+2 2 | 3+3 2 |
| | 3+4 2 | 3+4 2 | 5+6 2 | 5+6 2 | 3+3 2 | 2+2 2 |
| | 5+6 2 | 3+4 2 | 3+4 2 | 3+4 2 | 3+3 2 | 3+3 2 |
| | 3+4 2 | 5+6 2 | 5+6 2 | 3+4 2 | 2+2 2 | 2+2 2 |
| | 5+6 2 | 3+4 2 | 3+4 2 | 5+6 2 | 2+2 2 | 3+3 2 |
| | 5+6 2 | 5+6 2 | 5+6 2 | 3+4 2 | 3+3 2 | 2+2 2 |

Continued on next column

**TABLE XXIV– continued**

| TQC | ICSD | NoE | SF | | | |
|---|---|---|---|---|---|---|
| | 3+4 2 | 3+4 2 | 3+4 2 | 3+4 2 | 3+3 2 | 3+3 2 |
| | 5+6 2 | 3+4 2 | 5+6 2 | 5+6 2 | 3+3 2 | 2+2 2 |
| | 3+4 2 | 5+6 2 | 3+4 2 | 3+4 2 | 2+2 2 | 3+3 2 |
| | 5+6 2 | 5+6 2 | 5+6 2 | 5+6 2 | 3+3 2 | 3+3 2 |
| | 3+4 2 | 3+4 2 | 3+4 2 | 3+4 2 | 2+2 2 | 2+2 2 |
| | 5+6 2 | 5+6 2 | 5+6 2 | 5+6 2 | 3+3 2 | 3+3 2 |
| | 3+4 2 | 3+4 2 | 3+4 2 | 3+4 2 | 2+2 2 | 2+2 2 |
| | 5+6 2 | 3+4 2 | 5+6 2 | 3+4 2 | 3+3 2 | 3+3 2 |
| | 3+4 2 | 5+6 2 | 3+4 2 | 5+6 2 | 3+3 2 | 2+2 2 |
| | 5+6 2 | 3+4 2 | 5+6 2 | 5+6 2 | 2+2 2 | 3+3 2 |
| | 3+4 2 | 5+6 2 | 3+4 2 | 3+4 2 | 2+2 2 | 2+2 2 |
| | 5+6 2 | 5+6 2 | 3+4 2 | 3+4 2 | 3+3 2 | 3+3 2 |
| | 3+4 2 | 3+4 2 | 5+6 2 | 5+6 2 | 3+3 2 | 2+2 2 |
| | 5+6 2 | 3+4 2 | 3+4 2 | 3+4 2 | 2+2 2 | 3+3 2 |
| | 3+4 2 | 3+4 2 | 5+6 2 | 3+4 2 | 3+3 2 | 2+2 2 |
| | 3+4 2 | 5+6 2 | 3+4 2 | 5+6 2 | 2+2 2 | 2+2 2 |
| | 5+6 2 | 3+4 2 | 3+4 2 | 3+4 2 | 3+3 2 | 3+3 2 |
| | 3+4 2 | 3+4 2 | 5+6 2 | 3+4 2 | 3+3 2 | 2+2 2 |
| | 5+6 2 | 5+6 2 | 5+6 2 | 3+4 2 | 3+3 2 | 3+3 2 |
| | 3+4 2 | 5+6 2 | 5+6 2 | 5+6 2 | 2+2 2 | 3+3 2 |
| | 5+6 2 | 3+4 2 | 3+4 2 | 5+6 2 | 3+3 2 | 2+2 2 |
| | 5+6 2 | 3+4 2 | 5+6 2 | 3+4 2 | 3+3 2 | 3+3 2 |
| | 3+4 2 | 5+6 2 | 3+4 2 | 5+6 2 | 2+2 2 | 2+2 2 |
| | 3+4 2 | 3+4 2 | 5+6 2 | 5+6 2 | 2+2 2 | 3+3 2 |
| 12.1.175 | 425401 | 102 | Au1Ce2O1P2 | | | |
| | -A | -GM | -M | -Y | -L | -V |
| | 5+6 2 | 3+4 2 | 3+4 2 | 5+6 2 | 2+2 2 | 3+3 2 |
| | 3+4 2 | 5+6 2 | 5+6 2 | 3+4 2 | 3+3 2 | 2+2 2 |
| | 3+4 2 | 3+4 2 | 5+6 2 | 3+4 2 | 2+2 2 | 3+3 2 |
| | 5+6 2 | 5+6 2 | 3+4 2 | 5+6 2 | 3+3 2 | 2+2 2 |
| | 5+6 2 | 3+4 2 | 3+4 2 | 5+6 2 | 2+2 2 | 3+3 2 |
| | 3+4 2 | 5+6 2 | 5+6 2 | 3+4 2 | 3+3 2 | 2+2 2 |
| | 5+6 2 | 3+4 2 | 3+4 2 | 5+6 2 | 2+2 2 | 3+3 2 |
| | 3+4 2 | 5+6 2 | 5+6 2 | 3+4 2 | 3+3 2 | 2+2 2 |
| | 3+4 2 | 3+4 2 | 5+6 2 | 3+4 2 | 3+3 2 | 3+3 2 |
| | 5+6 2 | 5+6 2 | 3+4 2 | 5+6 2 | 2+2 2 | 2+2 2 |
| | 5+6 2 | 3+4 2 | 3+4 2 | 5+6 2 | 3+3 2 | 2+2 2 |
| | 5+6 2 | 3+4 2 | 5+6 2 | 5+6 2 | 2+2 2 | 3+3 2 |
| | 3+4 2 | 5+6 2 | 3+4 2 | 3+4 2 | 2+2 2 | 3+3 2 |
| | 3+4 2 | 5+6 2 | 5+6 2 | 3+4 2 | 3+3 2 | 2+2 2 |
| | 5+6 2 | 3+4 2 | 3+4 2 | 5+6 2 | 2+2 2 | 3+3 2 |
| | 3+4 2 | 5+6 2 | 5+6 2 | 3+4 2 | 3+3 2 | 2+2 2 |
| | 3+4 2 | 5+6 2 | 5+6 2 | 3+4 2 | 3+3 2 | 3+3 2 |
| | 5+6 2 | 3+4 2 | 3+4 2 | 5+6 2 | 2+2 2 | 2+2 2 |
| | 5+6 2 | 3+4 2 | 5+6 2 | 3+4 2 | 3+3 2 | 2+2 2 |
| | 3+4 2 | 3+4 2 | 3+4 2 | 5+6 2 | 3+3 2 | 3+3 2 |
| | 5+6 2 | 5+6 2 | 5+6 2 | 3+4 2 | 2+2 2 | 2+2 2 |
| | 3+4 2 | 5+6 2 | 3+4 2 | 5+6 2 | 2+2 2 | 3+3 2 |
| | 3+4 2 | 3+4 2 | 5+6 2 | 5+6 2 | 3+3 2 | 2+2 2 |
| | 5+6 2 | 5+6 2 | 3+4 2 | 3+4 2 | 2+2 2 | 3+3 2 |
| | 3+4 2 | 3+4 2 | 5+6 2 | 5+6 2 | 3+3 2 | 2+2 2 |
| | 5+6 2 | 5+6 2 | 3+4 2 | 3+4 2 | 2+2 2 | 3+3 2 |
| | 3+4 2 | 3+4 2 | 5+6 2 | 5+6 2 | 2+2 2 | 2+2 2 |
| | 5+6 2 | 5+6 2 | 5+6 2 | 3+4 2 | 3+3 2 | 3+3 2 |
| | 3+4 2 | 5+6 2 | 3+4 2 | 3+4 2 | 3+3 2 | 2+2 2 |
| | 5+6 2 | 3+4 2 | 3+4 2 | 5+6 2 | 2+2 2 | 3+3 2 |
| | 3+4 2 | 3+4 2 | 3+4 2 | 3+4 2 | 2+2 2 | 3+3 2 |
| | 5+6 2 | 3+4 2 | 5+6 2 | 5+6 2 | 3+3 2 | 2+2 2 |
| | 5+6 2 | 5+6 2 | 5+6 2 | 5+6 2 | 3+3 2 | 3+3 2 |
| | 5+6 2 | 3+4 2 | 3+4 2 | 5+6 2 | 2+2 2 | 2+2 2 |
| | 3+4 2 | 5+6 2 | 3+4 2 | 3+4 2 | 2+2 2 | 2+2 2 |

Continued on next column

**TABLE XXIV– continued**

| TQC | ICSD | NoE | SF | | | |
|---|---|---|---|---|---|---|
| | 5+6 2 | 3+4 2 | 3+4 2 | 3+4 2 | 3+3 2 | 3+3 2 |
| | 3+4 2 | 5+6 2 | 5+6 2 | 5+6 2 | 2+2 2 | 2+2 2 |
| | 3+4 2 | 5+6 2 | 5+6 2 | 3+4 2 | 3+3 2 | 3+3 2 |
| | 5+6 2 | 3+4 2 | 3+4 2 | 5+6 2 | 2+2 2 | 3+3 2 |
| | 3+4 2 | 5+6 2 | 5+6 2 | 3+4 2 | 3+3 2 | 2+2 2 |
| | 3+4 2 | 5+6 2 | 3+4 2 | 5+6 2 | 2+2 2 | 3+3 2 |
| | 5+6 2 | 3+4 2 | 5+6 2 | 3+4 2 | 3+3 2 | 2+2 2 |
| | 3+4 2 | 3+4 2 | 5+6 2 | 5+6 2 | 2+2 2 | 3+3 2 |
| | 3+4 2 | 3+4 2 | 5+6 2 | 5+6 2 | 3+3 2 | 2+2 2 |
| | 5+6 2 | 5+6 2 | 3+4 2 | 3+4 2 | 2+2 2 | 3+3 2 |
| | 5+6 2 | 5+6 2 | 3+4 2 | 3+4 2 | 3+3 2 | 2+2 2 |
| | 3+4 2 | 5+6 2 | 3+4 2 | 3+4 2 | 3+3 2 | 2+2 2 |
| | 5+6 2 | 3+4 2 | 5+6 2 | 5+6 2 | 3+3 2 | 3+3 2 |
| | 5+6 2 | 3+4 2 | 5+6 2 | 3+4 2 | 2+2 2 | 2+2 2 |
| | 5+6 2 | 3+4 2 | 5+6 2 | 3+4 2 | 2+2 2 | 3+3 2 |
| | 3+4 2 | 5+6 2 | 3+4 2 | 5+6 2 | 3+3 2 | 3+3 2 |
| 12.1.176 | 424760 | 74 | As4Ba3Zn2 | | | |
| | -A | -GM | -M | -Y | -L | -V |
| | 5+6 2 | 3+4 2 | 3+4 2 | 3+4 2 | 2+2 2 | 3+3 2 |
| | 3+4 2 | 5+6 2 | 5+6 2 | 5+6 2 | 3+3 2 | 2+2 2 |
| | 3+4 2 | 3+4 2 | 3+4 2 | 3+4 2 | 3+3 2 | 3+3 2 |
| | 5+6 2 | 5+6 2 | 5+6 2 | 5+6 2 | 2+2 2 | 2+2 2 |
| | 3+4 2 | 3+4 2 | 3+4 2 | 3+4 2 | 3+3 2 | 3+3 2 |
| | 5+6 2 | 5+6 2 | 5+6 2 | 5+6 2 | 2+2 2 | 2+2 2 |
| | 5+6 2 | 3+4 2 | 5+6 2 | 3+4 2 | 3+3 2 | 2+2 2 |
| | 3+4 2 | 5+6 2 | 5+6 2 | 5+6 2 | 2+2 2 | 2+2 2 |
| | 5+6 2 | 3+4 2 | 3+4 2 | 5+6 2 | 2+2 2 | 3+3 2 |
| | 3+4 2 | 5+6 2 | 3+4 2 | 3+4 2 | 2+2 2 | 2+2 2 |
| | 5+6 2 | 5+6 2 | 5+6 2 | 5+6 2 | 3+3 2 | 3+3 2 |
| | 5+6 2 | 5+6 2 | 5+6 2 | 5+6 2 | 2+2 2 | 2+2 2 |
| | 5+6 2 | 3+4 2 | 5+6 2 | 3+4 2 | 3+3 2 | 2+2 2 |
| | 3+4 2 | 3+4 2 | 5+6 2 | 5+6 2 | 3+3 2 | 3+3 2 |
| | 5+6 2 | 5+6 2 | 3+4 2 | 3+4 2 | 2+2 2 | 2+2 2 |
| | 3+4 2 | 5+6 2 | 3+4 2 | 5+6 2 | 2+2 2 | 3+3 2 |
| | 5+6 2 | 3+4 2 | 5+6 2 | 3+4 2 | 3+3 2 | 2+2 2 |
| | 3+4 2 | 3+4 2 | 3+4 2 | 5+6 2 | 2+2 2 | 3+3 2 |
| | 3+4 2 | 5+6 2 | 5+6 2 | 5+6 2 | 3+3 2 | 2+2 2 |
| | 5+6 2 | 5+6 2 | 3+4 2 | 3+4 2 | 2+2 2 | 3+3 2 |
| | 3+4 2 | 3+4 2 | 5+6 2 | 5+6 2 | 3+3 2 | 2+2 2 |
| | 5+6 2 | 3+4 2 | 3+4 2 | 3+4 2 | 2+2 2 | 3+3 2 |
| | 3+4 2 | 5+6 2 | 5+6 2 | 5+6 2 | 3+3 2 | 2+2 2 |
| | 5+6 2 | 3+4 2 | 5+6 2 | 3+4 2 | 2+2 2 | 3+3 2 |
| | 3+4 2 | 5+6 2 | 3+4 2 | 5+6 2 | 3+3 2 | 3+3 2 |
| | 5+6 2 | 5+6 2 | 3+4 2 | 3+4 2 | 2+2 2 | 2+2 2 |
| | 3+4 2 | 5+6 2 | 3+4 2 | 3+4 2 | 3+3 2 | 2+2 2 |
| | 5+6 2 | 3+4 2 | 5+6 2 | 5+6 2 | 2+2 2 | 3+3 2 |
| | 3+4 2 | 3+4 2 | 5+6 2 | 3+4 2 | 3+3 2 | 3+3 2 |
| | 5+6 2 | 3+4 2 | 3+4 2 | 5+6 2 | 2+2 2 | 2+2 2 |
| | 3+4 2 | 3+4 2 | 5+6 2 | 5+6 2 | 3+3 2 | 3+3 2 |
| | 3+4 2 | 5+6 2 | 3+4 2 | 5+6 2 | 2+2 2 | 2+2 2 |
| | 3+4 2 | 3+4 2 | 5+6 2 | 5+6 2 | 3+3 2 | 3+3 2 |
| | 5+6 2 | 5+6 2 | 3+4 2 | 3+4 2 | 3+3 2 | 3+3 2 |
| | 5+6 2 | 3+4 2 | 5+6 2 | 3+4 2 | 2+2 2 | 2+2 2 |
| | 5+6 2 | 5+6 2 | 3+4 2 | 3+4 2 | 2+2 2 | 2+2 2 |
| | 3+4 2 | 5+6 2 | 3+4 2 | 3+4 2 | 3+3 2 | 2+2 2 |
| 12.1.177 | 424761 | 74 | As4Ba3Cd2 | | | |
| | -A | -GM | -M | -Y | -L | -V |
| | 3+4 2 | 3+4 2 | 3+4 2 | 3+4 2 | 3+3 2 | 3+3 2 |
| | 5+6 2 | 5+6 2 | 5+6 2 | 5+6 2 | 2+2 2 | 2+2 2 |
| | 3+4 2 | 3+4 2 | 3+4 2 | 3+4 2 | 3+3 2 | 3+3 2 |
| | 5+6 2 | 5+6 2 | 5+6 2 | 5+6 2 | 2+2 2 | 2+2 2 |

Continued on next column

**TABLE XXIV– continued**

| TQC | ICSD | NoE | SF | | | |
|---|---|---|---|---|---|---|
| | 5+6 2 | 3+4 2 | 3+4 2 | 3+4 2 | 3+3 2 | 2+2 2 |
| | 3+4 2 | 5+6 2 | 5+6 2 | 5+6 2 | 2+2 2 | 3+3 2 |
| | 5+6 2 | 3+4 2 | 5+6 2 | 3+4 2 | 3+3 2 | 2+2 2 |
| | 5+6 2 | 5+6 2 | 5+6 2 | 5+6 2 | 2+2 2 | 2+2 2 |
| | 3+4 2 | 5+6 2 | 3+4 2 | 5+6 2 | 2+2 2 | 3+3 2 |
| | 5+6 2 | 3+4 2 | 3+4 2 | 3+4 2 | 2+2 2 | 2+2 2 |
| | 3+4 2 | 5+6 2 | 5+6 2 | 5+6 2 | 3+3 2 | 3+3 2 |
| | 5+6 2 | 5+6 2 | 5+6 2 | 5+6 2 | 2+2 2 | 2+2 2 |
| | 5+6 2 | 3+4 2 | 5+6 2 | 3+4 2 | 3+3 2 | 2+2 2 |
| | 3+4 2 | 3+4 2 | 3+4 2 | 3+4 2 | 3+3 2 | 3+3 2 |
| | 5+6 2 | 5+6 2 | 5+6 2 | 5+6 2 | 2+2 2 | 2+2 2 |
| | 3+4 2 | 5+6 2 | 3+4 2 | 5+6 2 | 2+2 2 | 3+3 2 |
| | 5+6 2 | 3+4 2 | 5+6 2 | 3+4 2 | 3+3 2 | 2+2 2 |
| | 3+4 2 | 3+4 2 | 5+6 2 | 5+6 2 | 2+2 2 | 3+3 2 |
| | 3+4 2 | 5+6 2 | 3+4 2 | 3+4 2 | 3+3 2 | 2+2 2 |
| | 5+6 2 | 5+6 2 | 3+4 2 | 5+6 2 | 2+2 2 | 3+3 2 |
| | 3+4 2 | 3+4 2 | 5+6 2 | 5+6 2 | 3+3 2 | 2+2 2 |
| | 5+6 2 | 3+4 2 | 5+6 2 | 3+4 2 | 2+2 2 | 3+3 2 |
| | 3+4 2 | 5+6 2 | 3+4 2 | 5+6 2 | 3+3 2 | 2+2 2 |
| | 5+6 2 | 3+4 2 | 5+6 2 | 3+4 2 | 2+2 2 | 3+3 2 |
| | 3+4 2 | 5+6 2 | 3+4 2 | 3+4 2 | 3+3 2 | 3+3 2 |
| | 5+6 2 | 5+6 2 | 3+4 2 | 5+6 2 | 2+2 2 | 2+2 2 |
| | 3+4 2 | 3+4 2 | 3+4 2 | 3+4 2 | 3+3 2 | 3+3 2 |
| | 5+6 2 | 5+6 2 | 5+6 2 | 5+6 2 | 2+2 2 | 2+2 2 |
| | 3+4 2 | 3+4 2 | 5+6 2 | 3+4 2 | 3+3 2 | 3+3 2 |
| | 5+6 2 | 5+6 2 | 3+4 2 | 5+6 2 | 2+2 2 | 2+2 2 |
| | 3+4 2 | 3+4 2 | 5+6 2 | 5+6 2 | 2+2 2 | 3+3 2 |
| | 3+4 2 | 3+4 2 | 5+6 2 | 5+6 2 | 3+3 2 | 2+2 2 |
| | 3+4 2 | 3+4 2 | 3+4 2 | 5+6 2 | 3+3 2 | 3+3 2 |
| | 5+6 2 | 5+6 2 | 3+4 2 | 3+4 2 | 3+3 2 | 3+3 2 |
| | 5+6 2 | 3+4 2 | 5+6 2 | 3+4 2 | 2+2 2 | 2+2 2 |
| | 5+6 2 | 5+6 2 | 3+4 2 | 3+4 2 | 2+2 2 | 2+2 2 |
| | 3+4 2 | 5+6 2 | 3+4 2 | 3+4 2 | 3+3 2 | 2+2 2 |
| 12.1.178 | 43185 | 54 | As3W2 | | | |
| | -A | -GM | -M | -Y | -L | -V |
| | 5+6 2 | 3+4 2 | 5+6 2 | 3+4 2 | 2+2 2 | 3+3 2 |
| | 3+4 2 | 3+4 2 | 3+4 2 | 5+6 2 | 3+3 2 | 3+3 2 |
| | 5+6 2 | 5+6 2 | 3+4 2 | 3+4 2 | 2+2 2 | 2+2 2 |
| | 3+4 2 | 5+6 2 | 5+6 2 | 5+6 2 | 3+3 2 | 3+3 2 |
| | 3+4 2 | 5+6 2 | 5+6 2 | 3+4 2 | 2+2 2 | 2+2 2 |
| | 5+6 2 | 3+4 2 | 3+4 2 | 5+6 2 | 3+3 2 | 2+2 2 |
| | 3+4 2 | 3+4 2 | 5+6 2 | 3+4 2 | 3+3 2 | 2+2 2 |
| | 5+6 2 | 5+6 2 | 3+4 2 | 3+4 2 | 2+2 2 | 3+3 2 |
| | 5+6 2 | 3+4 2 | 3+4 2 | 5+6 2 | 3+3 2 | 2+2 2 |
| | 3+4 2 | 5+6 2 | 5+6 2 | 5+6 2 | 2+2 2 | 3+3 2 |
| | 3+4 2 | 5+6 2 | 5+6 2 | 3+4 2 | 2+2 2 | 2+2 2 |
| | 5+6 2 | 3+4 2 | 3+4 2 | 5+6 2 | 3+3 2 | 3+3 2 |
| | 3+4 2 | 5+6 2 | 5+6 2 | 3+4 2 | 3+3 2 | 2+2 2 |
| | 3+4 2 | 3+4 2 | 3+4 2 | 3+4 2 | 2+2 2 | 3+3 2 |
| | 5+6 2 | 5+6 2 | 3+4 2 | 5+6 2 | 2+2 2 | 3+3 2 |
| | 5+6 2 | 5+6 2 | 5+6 2 | 5+6 2 | 3+3 2 | 3+3 2 |
| | 3+4 2 | 3+4 2 | 5+6 2 | 3+4 2 | 2+2 2 | 2+2 2 |
| | 5+6 2 | 3+4 2 | 3+4 2 | 3+4 2 | 3+3 2 | 3+3 2 |
| | 5+6 2 | 5+6 2 | 3+4 2 | 5+6 2 | 2+2 2 | 2+2 2 |
| | 3+4 2 | 3+4 2 | 5+6 2 | 5+6 2 | 3+3 2 | 3+3 2 |
| | 5+6 2 | 5+6 2 | 5+6 2 | 5+6 2 | 2+2 2 | 2+2 2 |
| | 5+6 2 | 3+4 2 | 3+4 2 | 3+4 2 | 3+3 2 | 2+2 2 |
| | 3+4 2 | 3+4 2 | 3+4 2 | 3+4 2 | 2+2 2 | 3+3 2 |
| | 3+4 2 | 3+4 2 | 5+6 2 | 5+6 2 | 3+3 2 | 3+3 2 |
| | 3+4 2 | 3+4 2 | 3+4 2 | 3+4 2 | 2+2 2 | 2+2 2 |
| | 5+6 2 | 3+4 2 | 5+6 2 | 5+6 2 | 3+3 2 | 3+3 2 |

Continued on next column

**TABLE XXIV– continued**

| TQC | ICSD | | NoE | | SF | |
|---|---|---|---|---|---|---|
| | 3+4 2 | 3+4 2 | 5+6 2 | 5+6 2 | 2+2 2 | 2+2 2 |
| 12.1.179 | 15049 | | 70 | | As3V4 | |
| | -A | -GM | -M | -Y | -L | -V |
| | 5+6 2 | 3+4 2 | 3+4 2 | 5+6 2 | 2+2 2 | 3+3 2 |
| | 3+4 2 | 5+6 2 | 5+6 2 | 3+4 2 | 3+3 2 | 2+2 2 |
| | 3+4 2 | 3+4 2 | 3+4 2 | 3+4 2 | 3+3 2 | 3+3 2 |
| | 5+6 2 | 5+6 2 | 5+6 2 | 5+6 2 | 2+2 2 | 2+2 2 |
| | 3+4 2 | 3+4 2 | 5+6 2 | 5+6 2 | 2+2 2 | 2+2 2 |
| | 5+6 2 | 5+6 2 | 3+4 2 | 3+4 2 | 3+3 2 | 3+3 2 |
| | 3+4 2 | 3+4 2 | 3+4 2 | 3+4 2 | 3+3 2 | 2+2 2 |
| | 5+6 2 | 3+4 2 | 5+6 2 | 3+4 2 | 2+2 2 | 3+3 2 |
| | 5+6 2 | 5+6 2 | 5+6 2 | 5+6 2 | 3+3 2 | 3+3 2 |
| | 3+4 2 | 3+4 2 | 5+6 2 | 5+6 2 | 2+2 2 | 3+3 2 |
| | 3+4 2 | 5+6 2 | 3+4 2 | 3+4 2 | 2+2 2 | 2+2 2 |
| | 5+6 2 | 5+6 2 | 3+4 2 | 5+6 2 | 3+3 2 | 2+2 2 |
| | 5+6 2 | 3+4 2 | 5+6 2 | 5+6 2 | 2+2 2 | 3+3 2 |
| | 3+4 2 | 3+4 2 | 3+4 2 | 3+4 2 | 3+3 2 | 3+3 2 |
| | 3+4 2 | 5+6 2 | 5+6 2 | 3+4 2 | 3+3 2 | 3+3 2 |
| | 3+4 2 | 5+6 2 | 3+4 2 | 5+6 2 | 2+2 2 | 2+2 2 |
| | 3+4 2 | 3+4 2 | 5+6 2 | 3+4 2 | 3+3 2 | 2+2 2 |
| | 5+6 2 | 3+4 2 | 5+6 2 | 5+6 2 | 2+2 2 | 3+3 2 |
| | 5+6 2 | 3+4 2 | 3+4 2 | 3+4 2 | 2+2 2 | 2+2 2 |
| | 5+6 2 | 3+4 2 | 5+6 2 | 5+6 2 | 3+3 2 | 2+2 2 |
| | 5+6 2 | 5+6 2 | 5+6 2 | 3+4 2 | 2+2 2 | 3+3 2 |
| | 3+4 2 | 3+4 2 | 3+4 2 | 3+4 2 | 3+3 2 | 2+2 2 |
| | 3+4 2 | 5+6 2 | 3+4 2 | 3+4 2 | 2+2 2 | 2+2 2 |
| | 5+6 2 | 5+6 2 | 5+6 2 | 3+4 2 | 3+3 2 | 3+3 2 |
| | 3+4 2 | 3+4 2 | 3+4 2 | 5+6 2 | 2+2 2 | 3+3 2 |
| | 5+6 2 | 5+6 2 | 3+4 2 | 5+6 2 | 3+3 2 | 3+3 2 |
| | 5+6 2 | 3+4 2 | 5+6 2 | 3+4 2 | 2+2 2 | 3+3 2 |
| | 3+4 2 | 5+6 2 | 3+4 2 | 5+6 2 | 3+3 2 | 2+2 2 |
| | 3+4 2 | 3+4 2 | 5+6 2 | 3+4 2 | 3+3 2 | 2+2 2 |
| | 5+6 2 | 3+4 2 | 3+4 2 | 5+6 2 | 2+2 2 | 3+3 2 |
| | 3+4 2 | 5+6 2 | 5+6 2 | 5+6 2 | 3+3 2 | 3+3 2 |
| | 5+6 2 | 3+4 2 | 3+4 2 | 5+6 2 | 2+2 2 | 2+2 2 |
| | 3+4 2 | 3+4 2 | 5+6 2 | 3+4 2 | 2+2 2 | 3+3 2 |
| | 5+6 2 | 5+6 2 | 3+4 2 | 3+4 2 | 3+3 2 | 2+2 2 |
| | 3+4 2 | 3+4 2 | 3+4 2 | 5+6 2 | 3+3 2 | 2+2 2 |
| 12.1.180 | 43184 | | 54 | | As3Mo2 | |
| | -A | -GM | -M | -Y | -L | -V |
| | 5+6 2 | 3+4 2 | 5+6 2 | 3+4 2 | 2+2 2 | 3+3 2 |
| | 3+4 2 | 3+4 2 | 3+4 2 | 5+6 2 | 3+3 2 | 3+3 2 |
| | 5+6 2 | 5+6 2 | 3+4 2 | 3+4 2 | 3+3 2 | 2+2 2 |
| | 3+4 2 | 5+6 2 | 5+6 2 | 5+6 2 | 2+2 2 | 3+3 2 |
| | 5+6 2 | 5+6 2 | 5+6 2 | 3+4 2 | 2+2 2 | 2+2 2 |
| | 3+4 2 | 3+4 2 | 3+4 2 | 5+6 2 | 3+3 2 | 2+2 2 |
| | 3+4 2 | 3+4 2 | 5+6 2 | 3+4 2 | 3+3 2 | 2+2 2 |
| | 5+6 2 | 5+6 2 | 3+4 2 | 3+4 2 | 2+2 2 | 3+3 2 |
| | 5+6 2 | 3+4 2 | 3+4 2 | 5+6 2 | 3+3 2 | 2+2 2 |
| | 3+4 2 | 5+6 2 | 5+6 2 | 3+4 2 | 2+2 2 | 3+3 2 |
| | 3+4 2 | 3+4 2 | 5+6 2 | 5+6 2 | 2+2 2 | 2+2 2 |
| | 5+6 2 | 5+6 2 | 3+4 2 | 5+6 2 | 3+3 2 | 3+3 2 |
| | 3+4 2 | 5+6 2 | 5+6 2 | 3+4 2 | 3+3 2 | 2+2 2 |
| | 3+4 2 | 3+4 2 | 3+4 2 | 3+4 2 | 2+2 2 | 3+3 2 |
| | 5+6 2 | 5+6 2 | 3+4 2 | 5+6 2 | 2+2 2 | 3+3 2 |
| | 5+6 2 | 5+6 2 | 5+6 2 | 5+6 2 | 3+3 2 | 3+3 2 |
| | 5+6 2 | 3+4 2 | 5+6 2 | 3+4 2 | 2+2 2 | 2+2 2 |
| | 3+4 2 | 3+4 2 | 3+4 2 | 3+4 2 | 3+3 2 | 3+3 2 |
| | 5+6 2 | 5+6 2 | 3+4 2 | 5+6 2 | 2+2 2 | 2+2 2 |
| | 3+4 2 | 3+4 2 | 5+6 2 | 5+6 2 | 3+3 2 | 2+2 2 |
| | 5+6 2 | 5+6 2 | 5+6 2 | 5+6 2 | 2+2 2 | 3+3 2 |

Continued on next column

**TABLE XXIV– continued**

| TQC | ICSD | | NoE | | SF | |
|---|---|---|---|---|---|---|
| | 5+6 2 | 3+4 2 | 3+4 2 | 3+4 2 | 3+3 2 | 2+2 2 |
| | 3+4 2 | 3+4 2 | 3+4 2 | 5+6 2 | 2+2 2 | 3+3 2 |
| | 3+4 2 | 3+4 2 | 5+6 2 | 3+4 2 | 3+3 2 | 2+2 2 |
| | 3+4 2 | 3+4 2 | 3+4 2 | 3+4 2 | 2+2 2 | 3+3 2 |
| | 5+6 2 | 3+4 2 | 5+6 2 | 5+6 2 | 2+2 2 | 3+3 2 |
| | 3+4 2 | 3+4 2 | 5+6 2 | 5+6 2 | 3+3 2 | 2+2 2 |
| 12.1.181 | 41015 | | 54 | | As3Mo2 | |
| | -A | -GM | -M | -Y | -L | -V |
| | 5+6 2 | 3+4 2 | 5+6 2 | 3+4 2 | 2+2 2 | 3+3 2 |
| | 3+4 2 | 3+4 2 | 3+4 2 | 3+4 2 | 3+3 2 | 3+3 2 |
| | 5+6 2 | 5+6 2 | 3+4 2 | 5+6 2 | 3+3 2 | 2+2 2 |
| | 3+4 2 | 5+6 2 | 5+6 2 | 5+6 2 | 2+2 2 | 3+3 2 |
| | 5+6 2 | 5+6 2 | 5+6 2 | 3+4 2 | 2+2 2 | 2+2 2 |
| | 3+4 2 | 3+4 2 | 3+4 2 | 5+6 2 | 3+3 2 | 2+2 2 |
| | 3+4 2 | 3+4 2 | 5+6 2 | 3+4 2 | 3+3 2 | 2+2 2 |
| | 5+6 2 | 5+6 2 | 3+4 2 | 3+4 2 | 2+2 2 | 3+3 2 |
| | 5+6 2 | 3+4 2 | 3+4 2 | 5+6 2 | 3+3 2 | 2+2 2 |
| | 3+4 2 | 5+6 2 | 5+6 2 | 3+4 2 | 2+2 2 | 3+3 2 |
| | 3+4 2 | 3+4 2 | 5+6 2 | 5+6 2 | 2+2 2 | 2+2 2 |
| | 5+6 2 | 5+6 2 | 3+4 2 | 5+6 2 | 3+3 2 | 3+3 2 |
| | 3+4 2 | 5+6 2 | 5+6 2 | 3+4 2 | 3+3 2 | 2+2 2 |
| | 3+4 2 | 3+4 2 | 3+4 2 | 3+4 2 | 2+2 2 | 3+3 2 |
| | 5+6 2 | 5+6 2 | 3+4 2 | 5+6 2 | 2+2 2 | 3+3 2 |
| | 5+6 2 | 5+6 2 | 5+6 2 | 5+6 2 | 3+3 2 | 3+3 2 |
| | 5+6 2 | 3+4 2 | 5+6 2 | 3+4 2 | 2+2 2 | 2+2 2 |
| | 3+4 2 | 3+4 2 | 3+4 2 | 3+4 2 | 3+3 2 | 3+3 2 |
| | 5+6 2 | 5+6 2 | 3+4 2 | 5+6 2 | 2+2 2 | 2+2 2 |
| | 3+4 2 | 3+4 2 | 5+6 2 | 5+6 2 | 3+3 2 | 2+2 2 |
| | 5+6 2 | 5+6 2 | 5+6 2 | 5+6 2 | 2+2 2 | 3+3 2 |
| | 5+6 2 | 3+4 2 | 3+4 2 | 3+4 2 | 3+3 2 | 2+2 2 |
| | 3+4 2 | 3+4 2 | 3+4 2 | 5+6 2 | 2+2 2 | 3+3 2 |
| | 3+4 2 | 3+4 2 | 5+6 2 | 3+4 2 | 3+3 2 | 2+2 2 |
| | 3+4 2 | 3+4 2 | 3+4 2 | 3+4 2 | 2+2 2 | 3+3 2 |
| | 5+6 2 | 3+4 2 | 5+6 2 | 5+6 2 | 3+3 2 | 3+3 2 |
| | 3+4 2 | 3+4 2 | 5+6 2 | 5+6 2 | 2+2 2 | 2+2 2 |
| 12.1.182 | 610428 | | 64 | | As3Eu1 | |
| | -A | -GM | -M | -Y | -L | -V |
| | 5+6 2 | 3+4 2 | 5+6 2 | 3+4 2 | 3+3 2 | 2+2 2 |
| | 3+4 2 | 5+6 2 | 3+4 2 | 5+6 2 | 2+2 2 | 3+3 2 |
| | 5+6 2 | 3+4 2 | 5+6 2 | 3+4 2 | 3+3 2 | 2+2 2 |
| | 3+4 2 | 5+6 2 | 3+4 2 | 5+6 2 | 2+2 2 | 3+3 2 |
| | 5+6 2 | 3+4 2 | 5+6 2 | 5+6 2 | 3+3 2 | 2+2 2 |
| | 3+4 2 | 5+6 2 | 3+4 2 | 3+4 2 | 2+2 2 | 3+3 2 |
| | 5+6 2 | 3+4 2 | 5+6 2 | 3+4 2 | 3+3 2 | 2+2 2 |
| | 3+4 2 | 5+6 2 | 3+4 2 | 5+6 2 | 2+2 2 | 3+3 2 |
| | 3+4 2 | 3+4 2 | 3+4 2 | 3+4 2 | 2+2 2 | 2+2 2 |
| | 5+6 2 | 3+4 2 | 5+6 2 | 5+6 2 | 3+3 2 | 3+3 2 |
| | 3+4 2 | 5+6 2 | 5+6 2 | 5+6 2 | 2+2 2 | 2+2 2 |
| | 3+4 2 | 3+4 2 | 5+6 2 | 5+6 2 | 3+3 2 | 3+3 2 |
| | 5+6 2 | 5+6 2 | 3+4 2 | 3+4 2 | 2+2 2 | 2+2 2 |
| | 5+6 2 | 5+6 2 | 3+4 2 | 3+4 2 | 3+3 2 | 3+3 2 |
| | 3+4 2 | 3+4 2 | 5+6 2 | 3+4 2 | 3+3 2 | 2+2 2 |
| | 5+6 2 | 5+6 2 | 3+4 2 | 5+6 2 | 2+2 2 | 3+3 2 |
| | 5+6 2 | 5+6 2 | 5+6 2 | 3+4 2 | 2+2 2 | 2+2 2 |
| | 3+4 2 | 3+4 2 | 3+4 2 | 5+6 2 | 3+3 2 | 3+3 2 |
| | 3+4 2 | 3+4 2 | 3+4 2 | 3+4 2 | 3+3 2 | 3+3 2 |
| | 5+6 2 | 3+4 2 | 5+6 2 | 5+6 2 | 2+2 2 | 2+2 2 |
| | 3+4 2 | 5+6 2 | 3+4 2 | 3+4 2 | 3+3 2 | 3+3 2 |
| | 3+4 2 | 5+6 2 | 5+6 2 | 5+6 2 | 2+2 2 | 2+2 2 |
| | 5+6 2 | 3+4 2 | 5+6 2 | 5+6 2 | 2+2 2 | 2+2 2 |
| | 5+6 2 | 3+4 2 | 3+4 2 | 3+4 2 | 2+2 2 | 2+2 2 |

Continued on next column

**TABLE XXIV– continued**

| TQC | ICSD | | NoE | | SF | |
|---|---|---|---|---|---|---|
| | 3+4 2 | 5+6 2 | 5+6 2 | 5+6 2 | 3+3 2 | 3+3 2 |
| | 3+4 2 | 3+4 2 | 3+4 2 | 3+4 2 | 3+3 2 | 2+2 2 |
| | 5+6 2 | 5+6 2 | 5+6 2 | 5+6 2 | 2+2 2 | 3+3 2 |
| | 5+6 2 | 3+4 2 | 3+4 2 | 3+4 2 | 3+3 2 | 3+3 2 |
| | 3+4 2 | 3+4 2 | 5+6 2 | 3+4 2 | 2+2 2 | 2+2 2 |
| | 5+6 2 | 5+6 2 | 3+4 2 | 5+6 2 | 3+3 2 | 3+3 2 |
| | 3+4 2 | 5+6 2 | 3+4 2 | 5+6 2 | 2+2 2 | 2+2 2 |
| | 3+4 2 | 3+4 2 | 5+6 2 | 3+4 2 | 3+3 2 | 3+3 2 |
| 12.1.183 | 611574 | | 30 | | As2V1 | |
| | -A | -GM | -M | -Y | -L | -V |
| | 5+6 2 | 3+4 2 | 5+6 2 | 3+4 2 | 2+2 2 | 3+3 2 |
| | 3+4 2 | 3+4 2 | 3+4 2 | 5+6 2 | 2+2 2 | 2+2 2 |
| | 5+6 2 | 5+6 2 | 5+6 2 | 5+6 2 | 3+3 2 | 3+3 2 |
| | 3+4 2 | 5+6 2 | 3+4 2 | 3+4 2 | 3+3 2 | 2+2 2 |
| | 5+6 2 | 3+4 2 | 3+4 2 | 5+6 2 | 2+2 2 | 2+2 2 |
| | 3+4 2 | 5+6 2 | 5+6 2 | 3+4 2 | 3+3 2 | 3+3 2 |
| | 3+4 2 | 3+4 2 | 5+6 2 | 5+6 2 | 3+3 2 | 2+2 2 |
| | 5+6 2 | 3+4 2 | 3+4 2 | 3+4 2 | 2+2 2 | 3+3 2 |
| | 5+6 2 | 5+6 2 | 3+4 2 | 5+6 2 | 2+2 2 | 2+2 2 |
| | 5+6 2 | 3+4 2 | 5+6 2 | 5+6 2 | 3+3 2 | 3+3 2 |
| | 3+4 2 | 3+4 2 | 3+4 2 | 3+4 2 | 2+2 2 | 2+2 2 |
| | 3+4 2 | 5+6 2 | 5+6 2 | 5+6 2 | 2+2 2 | 3+3 2 |
| | 5+6 2 | 3+4 2 | 5+6 2 | 3+4 2 | 3+3 2 | 2+2 2 |
| | 3+4 2 | 3+4 2 | 3+4 2 | 3+4 2 | 3+3 2 | 3+3 2 |
| | 5+6 2 | 5+6 2 | 3+4 2 | 3+4 2 | 2+2 2 | 3+3 2 |
| 12.1.184 | 44069 | | 30 | | As2Ta1 | |
| | -A | -GM | -M | -Y | -L | -V |
| | 5+6 2 | 3+4 2 | 3+4 2 | 3+4 2 | 2+2 2 | 3+3 2 |
| | 3+4 2 | 3+4 2 | 5+6 2 | 5+6 2 | 2+2 2 | 2+2 2 |
| | 5+6 2 | 5+6 2 | 5+6 2 | 5+6 2 | 3+3 2 | 3+3 2 |
| | 3+4 2 | 5+6 2 | 3+4 2 | 3+4 2 | 3+3 2 | 2+2 2 |
| | 5+6 2 | 3+4 2 | 5+6 2 | 5+6 2 | 3+3 2 | 3+3 2 |
| | 3+4 2 | 5+6 2 | 3+4 2 | 3+4 2 | 2+2 2 | 2+2 2 |
| | 3+4 2 | 3+4 2 | 5+6 2 | 5+6 2 | 3+3 2 | 2+2 2 |
| | 5+6 2 | 3+4 2 | 3+4 2 | 3+4 2 | 2+2 2 | 3+3 2 |
| | 5+6 2 | 5+6 2 | 3+4 2 | 5+6 2 | 2+2 2 | 2+2 2 |
| | 3+4 2 | 5+6 2 | 5+6 2 | 5+6 2 | 3+3 2 | 2+2 2 |
| | 5+6 2 | 3+4 2 | 3+4 2 | 3+4 2 | 2+2 2 | 3+3 2 |
| | 3+4 2 | 3+4 2 | 5+6 2 | 5+6 2 | 2+2 2 | 3+3 2 |
| | 3+4 2 | 3+4 2 | 5+6 2 | 3+4 2 | 3+3 2 | 2+2 2 |
| | 5+6 2 | 3+4 2 | 3+4 2 | 3+4 2 | 3+3 2 | 3+3 2 |
| | 5+6 2 | 5+6 2 | 3+4 2 | 3+4 2 | 2+2 2 | 3+3 2 |
| 12.1.185 | 69002 | | 72 | | As2Ni2O7 | |
| | -A | -GM | -M | -Y | -L | -V |
| | 3+4 2 | 3+4 2 | 3+4 2 | 3+4 2 | 3+3 2 | 3+3 2 |
| | 5+6 2 | 5+6 2 | 5+6 2 | 5+6 2 | 2+2 2 | 2+2 2 |
| | 3+4 2 | 3+4 2 | 3+4 2 | 3+4 2 | 3+3 2 | 3+3 2 |
| | 5+6 2 | 3+4 2 | 5+6 2 | 5+6 2 | 3+3 2 | 2+2 2 |
| | 5+6 2 | 5+6 2 | 3+4 2 | 5+6 2 | 2+2 2 | 3+3 2 |
| | 3+4 2 | 3+4 2 | 3+4 2 | 3+4 2 | 3+3 2 | 2+2 2 |
| | 3+4 2 | 5+6 2 | 5+6 2 | 3+4 2 | 2+2 2 | 3+3 2 |
| | 5+6 2 | 5+6 2 | 5+6 2 | 5+6 2 | 2+2 2 | 2+2 2 |
| | 3+4 2 | 3+4 2 | 3+4 2 | 3+4 2 | 3+3 2 | 3+3 2 |
| | 5+6 2 | 5+6 2 | 5+6 2 | 5+6 2 | 2+2 2 | 2+2 2 |
| | 5+6 2 | 3+4 2 | 3+4 2 | 5+6 2 | 3+3 2 | 2+2 2 |
| | 5+6 2 | 5+6 2 | 5+6 2 | 3+4 2 | 2+2 2 | 2+2 2 |
| | 3+4 2 | 3+4 2 | 5+6 2 | 5+6 2 | 2+2 2 | 3+3 2 |
| | 3+4 2 | 5+6 2 | 3+4 2 | 3+4 2 | 3+3 2 | 3+3 2 |
| | 3+4 2 | 5+6 2 | 3+4 2 | 3+4 2 | 3+3 2 | 2+2 2 |
| | 3+4 2 | 3+4 2 | 5+6 2 | 5+6 2 | 2+2 2 | 3+3 2 |
| | 5+6 2 | 5+6 2 | 5+6 2 | 5+6 2 | 2+2 2 | 2+2 2 |

Continued on next column

**TABLE XXIV– continued**

| TQC | ICSD | | NoE | | SF | |
|---|---|---|---|---|---|---|
| | 5+6 2 | 3+4 2 | 3+4 2 | 5+6 2 | 2+2 2 | 3+3 2 |
| | 5+6 2 | 3+4 2 | 5+6 2 | 3+4 2 | 3+3 2 | 2+2 2 |
| | 5+6 2 | 3+4 2 | 3+4 2 | 5+6 2 | 2+2 2 | 2+2 2 |
| | 3+4 2 | 5+6 2 | 5+6 2 | 3+4 2 | 3+3 2 | 3+3 2 |
| | 3+4 2 | 5+6 2 | 5+6 2 | 5+6 2 | 2+2 2 | 2+2 2 |
| | 5+6 2 | 3+4 2 | 3+4 2 | 3+4 2 | 3+3 2 | 3+3 2 |
| | 5+6 2 | 5+6 2 | 5+6 2 | 3+4 2 | 2+2 2 | 2+2 2 |
| | 3+4 2 | 5+6 2 | 3+4 2 | 5+6 2 | 3+3 2 | 3+3 2 |
| | 3+4 2 | 5+6 2 | 5+6 2 | 5+6 2 | 3+3 2 | 2+2 2 |
| | 5+6 2 | 3+4 2 | 5+6 2 | 3+4 2 | 2+2 2 | 2+2 2 |
| | 5+6 2 | 5+6 2 | 3+4 2 | 5+6 2 | 2+2 2 | 3+3 2 |
| | 5+6 2 | 3+4 2 | 5+6 2 | 3+4 2 | 2+2 2 | 3+3 2 |
| | 3+4 2 | 5+6 2 | 3+4 2 | 5+6 2 | 3+3 2 | 2+2 2 |
| | 5+6 2 | 3+4 2 | 3+4 2 | 3+4 2 | 3+3 2 | 3+3 2 |
| | 3+4 2 | 3+4 2 | 5+6 2 | 5+6 2 | 2+2 2 | 2+2 2 |
| | 3+4 2 | 5+6 2 | 5+6 2 | 3+4 2 | 3+3 2 | 3+3 2 |
| | 5+6 2 | 5+6 2 | 3+4 2 | 5+6 2 | 2+2 2 | 2+2 2 |
| | 3+4 2 | 5+6 2 | 3+4 2 | 3+4 2 | 3+3 2 | 3+3 2 |
| | 3+4 2 | 5+6 2 | 5+6 2 | 5+6 2 | 2+2 2 | 2+2 2 |
| 12.1.186 | 81218 | | 46 | | As2Nb1 | |
| | -A | -GM | -M | -Y | -L | -V |
| | 3+4 2 | 3+4 2 | 3+4 2 | 3+4 2 | 3+3 2 | 3+3 2 |
| | 5+6 2 | 5+6 2 | 5+6 2 | 5+6 2 | 2+2 2 | 2+2 2 |
| | 3+4 2 | 3+4 2 | 3+4 2 | 3+4 2 | 3+3 2 | 3+3 2 |
| | 5+6 2 | 5+6 2 | 5+6 2 | 5+6 2 | 2+2 2 | 2+2 2 |
| | 3+4 2 | 3+4 2 | 3+4 2 | 3+4 2 | 3+3 2 | 3+3 2 |
| | 5+6 2 | 5+6 2 | 5+6 2 | 5+6 2 | 2+2 2 | 2+2 2 |
| | 3+4 2 | 3+4 2 | 3+4 2 | 3+4 2 | 3+3 2 | 3+3 2 |
| | 5+6 2 | 5+6 2 | 5+6 2 | 5+6 2 | 2+2 2 | 2+2 2 |
| | 5+6 2 | 3+4 2 | 5+6 2 | 3+4 2 | 2+2 2 | 3+3 2 |
| | 3+4 2 | 3+4 2 | 3+4 2 | 5+6 2 | 2+2 2 | 2+2 2 |
| | 5+6 2 | 5+6 2 | 5+6 2 | 5+6 2 | 3+3 2 | 3+3 2 |
| | 3+4 2 | 5+6 2 | 3+4 2 | 3+4 2 | 3+3 2 | 2+2 2 |
| | 5+6 2 | 3+4 2 | 5+6 2 | 5+6 2 | 2+2 2 | 2+2 2 |
| | 3+4 2 | 5+6 2 | 3+4 2 | 3+4 2 | 3+3 2 | 3+3 2 |
| | 3+4 2 | 3+4 2 | 5+6 2 | 5+6 2 | 3+3 2 | 2+2 2 |
| | 5+6 2 | 3+4 2 | 3+4 2 | 3+4 2 | 2+2 2 | 3+3 2 |
| | 5+6 2 | 5+6 2 | 5+6 2 | 5+6 2 | 2+2 2 | 2+2 2 |
| | 5+6 2 | 5+6 2 | 3+4 2 | 5+6 2 | 3+3 2 | 2+2 2 |
| | 3+4 2 | 3+4 2 | 3+4 2 | 3+4 2 | 2+2 2 | 3+3 2 |
| | 3+4 2 | 3+4 2 | 5+6 2 | 3+4 2 | 2+2 2 | 3+3 2 |
| | 3+4 2 | 3+4 2 | 5+6 2 | 5+6 2 | 3+3 2 | 2+2 2 |
| | 5+6 2 | 3+4 2 | 3+4 2 | 3+4 2 | 3+3 2 | 3+3 2 |
| | 5+6 2 | 5+6 2 | 3+4 2 | 3+4 2 | 2+2 2 | 3+3 2 |
| 12.1.187 | 69003 | | 66 | | As2Mn2O7 | |
| | -A | -GM | -M | -Y | -L | -V |
| | 3+4 2 | 3+4 2 | 3+4 2 | 3+4 2 | 3+3 2 | 3+3 2 |
| | 5+6 2 | 5+6 2 | 5+6 2 | 5+6 2 | 2+2 2 | 2+2 2 |
| | 3+4 2 | 3+4 2 | 3+4 2 | 3+4 2 | 3+3 2 | 3+3 2 |
| | 5+6 2 | 5+6 2 | 5+6 2 | 5+6 2 | 3+3 2 | 3+3 2 |
| | 3+4 2 | 3+4 2 | 3+4 2 | 5+6 2 | 2+2 2 | 2+2 2 |
| | 5+6 2 | 3+4 2 | 3+4 2 | 3+4 2 | 3+3 2 | 2+2 2 |
| | 3+4 2 | 5+6 2 | 5+6 2 | 3+4 2 | 2+2 2 | 3+3 2 |
| | 5+6 2 | 5+6 2 | 5+6 2 | 5+6 2 | 2+2 2 | 2+2 2 |
| | 3+4 2 | 3+4 2 | 3+4 2 | 3+4 2 | 3+3 2 | 3+3 2 |
| | 5+6 2 | 5+6 2 | 5+6 2 | 5+6 2 | 2+2 2 | 2+2 2 |
| | 5+6 2 | 3+4 2 | 3+4 2 | 5+6 2 | 3+3 2 | 2+2 2 |
| | 5+6 2 | 3+4 2 | 5+6 2 | 5+6 2 | 2+2 2 | 2+2 2 |
| | 3+4 2 | 5+6 2 | 5+6 2 | 3+4 2 | 2+2 2 | 3+3 2 |
| | 3+4 2 | 5+6 2 | 3+4 2 | 3+4 2 | 3+3 2 | 3+3 2 |
| | 3+4 2 | 5+6 2 | 3+4 2 | 3+4 2 | 3+3 2 | 2+2 2 |

Continued on next column

**TABLE XXIV– continued**

| TQC | ICSD | | NoE | SF | | |
|---|---|---|---|---|---|---|
| | 5+6 2 | 3+4 2 | 5+6 2 | 5+6 2 | 2+2 2 | 3+3 2 |
| | 5+6 2 | 5+6 2 | 5+6 2 | 5+6 2 | 2+2 2 | 2+2 2 |
| | 3+4 2 | 3+4 2 | 5+6 2 | 5+6 2 | 2+2 2 | 3+3 2 |
| | 5+6 2 | 3+4 2 | 3+4 2 | 3+4 2 | 3+3 2 | 2+2 2 |
| | 5+6 2 | 3+4 2 | 5+6 2 | 3+4 2 | 2+2 2 | 2+2 2 |
| | 3+4 2 | 5+6 2 | 3+4 2 | 5+6 2 | 3+3 2 | 3+3 2 |
| | 3+4 2 | 5+6 2 | 5+6 2 | 5+6 2 | 2+2 2 | 2+2 2 |
| | 5+6 2 | 3+4 2 | 3+4 2 | 3+4 2 | 3+3 2 | 3+3 2 |
| | 5+6 2 | 5+6 2 | 5+6 2 | 3+4 2 | 2+2 2 | 2+2 2 |
| | 3+4 2 | 5+6 2 | 3+4 2 | 5+6 2 | 3+3 2 | 3+3 2 |
| | 3+4 2 | 5+6 2 | 5+6 2 | 5+6 2 | 2+2 2 | 2+2 2 |
| | 5+6 2 | 3+4 2 | 5+6 2 | 3+4 2 | 3+3 2 | 2+2 2 |
| | 5+6 2 | 5+6 2 | 3+4 2 | 5+6 2 | 2+2 2 | 3+3 2 |
| | 5+6 2 | 3+4 2 | 5+6 2 | 3+4 2 | 3+3 2 | 3+3 2 |
| | 5+6 2 | 5+6 2 | 3+4 2 | 5+6 2 | 3+3 2 | 2+2 2 |
| | 3+4 2 | 3+4 2 | 3+4 2 | 5+6 2 | 2+2 2 | 3+3 2 |
| | 3+4 2 | 3+4 2 | 5+6 2 | 3+4 2 | 2+2 2 | 2+2 2 |
| | 5+6 2 | 5+6 2 | 5+6 2 | 3+4 2 | 3+3 2 | 3+3 2 |
| 12.1.188 | 261143 | | 66 | As2Mn2O7 | | |
| | -A | -GM | -M | -Y | -L | -V |
| | 5+6 2 | 3+4 2 | 5+6 2 | 3+4 2 | 2+2 2 | 3+3 2 |
| | 3+4 2 | 5+6 2 | 3+4 2 | 5+6 2 | 3+3 2 | 2+2 2 |
| | 5+6 2 | 3+4 2 | 5+6 2 | 3+4 2 | 2+2 2 | 3+3 2 |
| | 3+4 2 | 3+4 2 | 3+4 2 | 5+6 2 | 2+2 2 | 2+2 2 |
| | 5+6 2 | 5+6 2 | 5+6 2 | 5+6 2 | 3+3 2 | 3+3 2 |
| | 3+4 2 | 3+4 2 | 5+6 2 | 3+4 2 | 2+2 2 | 2+2 2 |
| | 5+6 2 | 5+6 2 | 3+4 2 | 3+4 2 | 3+3 2 | 3+3 2 |
| | 3+4 2 | 5+6 2 | 3+4 2 | 5+6 2 | 3+3 2 | 2+2 2 |
| | 5+6 2 | 3+4 2 | 5+6 2 | 3+4 2 | 2+2 2 | 3+3 2 |
| | 3+4 2 | 5+6 2 | 3+4 2 | 5+6 2 | 3+3 2 | 2+2 2 |
| | 3+4 2 | 3+4 2 | 5+6 2 | 5+6 2 | 2+2 2 | 2+2 2 |
| | 3+4 2 | 3+4 2 | 3+4 2 | 3+4 2 | 3+3 2 | 2+2 2 |
| | 5+6 2 | 5+6 2 | 3+4 2 | 5+6 2 | 3+3 2 | 3+3 2 |
| | 5+6 2 | 5+6 2 | 5+6 2 | 3+4 2 | 2+2 2 | 3+3 2 |
| | 5+6 2 | 5+6 2 | 5+6 2 | 3+4 2 | 2+2 2 | 2+2 2 |
| | 5+6 2 | 3+4 2 | 3+4 2 | 5+6 2 | 3+3 2 | 3+3 2 |
| | 3+4 2 | 5+6 2 | 3+4 2 | 5+6 2 | 3+3 2 | 2+2 2 |
| | 3+4 2 | 3+4 2 | 3+4 2 | 5+6 2 | 3+3 2 | 3+3 2 |
| | 3+4 2 | 3+4 2 | 5+6 2 | 3+4 2 | 2+2 2 | 2+2 2 |
| | 5+6 2 | 3+4 2 | 3+4 2 | 3+4 2 | 3+3 2 | 2+2 2 |
| | 3+4 2 | 5+6 2 | 5+6 2 | 5+6 2 | 2+2 2 | 3+3 2 |
| | 5+6 2 | 5+6 2 | 3+4 2 | 3+4 2 | 3+3 2 | 2+2 2 |
| | 3+4 2 | 3+4 2 | 5+6 2 | 5+6 2 | 2+2 2 | 3+3 2 |
| | 5+6 2 | 5+6 2 | 5+6 2 | 3+4 2 | 3+3 2 | 2+2 2 |
| | 3+4 2 | 5+6 2 | 3+4 2 | 5+6 2 | 2+2 2 | 3+3 2 |
| | 5+6 2 | 5+6 2 | 3+4 2 | 5+6 2 | 2+2 2 | 2+2 2 |
| | 3+4 2 | 3+4 2 | 5+6 2 | 3+4 2 | 3+3 2 | 2+2 2 |
| | 3+4 2 | 5+6 2 | 3+4 2 | 5+6 2 | 3+3 2 | 3+3 2 |
| | 3+4 2 | 3+4 2 | 3+4 2 | 3+4 2 | 2+2 2 | 3+3 2 |
| | 3+4 2 | 5+6 2 | 5+6 2 | 5+6 2 | 3+3 2 | 2+2 2 |
| | 5+6 2 | 3+4 2 | 5+6 2 | 5+6 2 | 2+2 2 | 3+3 2 |
| | 5+6 2 | 3+4 2 | 3+4 2 | 3+4 2 | 3+3 2 | 2+2 2 |
| | 3+4 2 | 5+6 2 | 3+4 2 | 3+4 2 | 2+2 2 | 3+3 2 |
| 12.1.189 | 79441 | | 174 | As2Gd1Pd3 | | |
| | -A | -GM | -M | -Y | -L | -V |
| | 3+4 2 | 3+4 2 | 3+4 2 | 3+4 2 | 2+2 2 | 2+2 2 |
| | 5+6 2 | 5+6 2 | 5+6 2 | 5+6 2 | 3+3 2 | 3+3 2 |
| | 3+4 2 | 3+4 2 | 3+4 2 | 3+4 2 | 3+3 2 | 3+3 2 |
| | 3+4 2 | 3+4 2 | 3+4 2 | 3+4 2 | 2+2 2 | 2+2 2 |
| | 5+6 2 | 5+6 2 | 5+6 2 | 5+6 2 | 3+3 2 | 3+3 2 |
| | 5+6 2 | 5+6 2 | 5+6 2 | 5+6 2 | 2+2 2 | 2+2 2 |

Continued on next column

**TABLE XXIV– continued**

| TQC | ICSD | | NoE | SF | | |
|---|---|---|---|---|---|---|
| | 5+6 2 | 5+6 2 | 5+6 2 | 5+6 2 | 2+2 2 | 2+2 2 |
| | 3+4 2 | 3+4 2 | 3+4 2 | 3+4 2 | 3+3 2 | 3+3 2 |
| | 5+6 2 | 5+6 2 | 5+6 2 | 5+6 2 | 2+2 2 | 2+2 2 |
| | 3+4 2 | 3+4 2 | 3+4 2 | 3+4 2 | 2+2 2 | 2+2 2 |
| | 5+6 2 | 5+6 2 | 5+6 2 | 5+6 2 | 3+3 2 | 3+3 2 |
| | 5+6 2 | 5+6 2 | 5+6 2 | 5+6 2 | 2+2 2 | 2+2 2 |
| | 5+6 2 | 3+4 2 | 5+6 2 | 3+4 2 | 3+3 2 | 2+2 2 |
| | 3+4 2 | 5+6 2 | 3+4 2 | 5+6 2 | 2+2 2 | 3+3 2 |
| | 3+4 2 | 3+4 2 | 5+6 2 | 5+6 2 | 3+3 2 | 3+3 2 |
| | 5+6 2 | 5+6 2 | 3+4 2 | 3+4 2 | 2+2 2 | 2+2 2 |
| | 3+4 2 | 3+4 2 | 3+4 2 | 3+4 2 | 3+3 2 | 2+2 2 |
| | 5+6 2 | 5+6 2 | 5+6 2 | 5+6 2 | 2+2 2 | 3+3 2 |
| | 5+6 2 | 3+4 2 | 5+6 2 | 3+4 2 | 3+3 2 | 3+3 2 |
| | 3+4 2 | 3+4 2 | 3+4 2 | 5+6 2 | 2+2 2 | 2+2 2 |
| | 5+6 2 | 5+6 2 | 5+6 2 | 3+4 2 | 3+3 2 | 3+3 2 |
| | 3+4 2 | 3+4 2 | 5+6 2 | 3+4 2 | 3+3 2 | 2+2 2 |
| | 3+4 2 | 5+6 2 | 3+4 2 | 3+4 2 | 2+2 2 | 2+2 2 |
| | 3+4 2 | 5+6 2 | 5+6 2 | 5+6 2 | 2+2 2 | 2+2 2 |
| | 5+6 2 | 5+6 2 | 3+4 2 | 3+4 2 | 3+3 2 | 3+3 2 |
| | 5+6 2 | 3+4 2 | 3+4 2 | 5+6 2 | 3+3 2 | 2+2 2 |
| | 3+4 2 | 3+4 2 | 5+6 2 | 3+4 2 | 3+3 2 | 3+3 2 |
| | 5+6 2 | 5+6 2 | 3+4 2 | 5+6 2 | 2+2 2 | 3+3 2 |
| | 5+6 2 | 5+6 2 | 5+6 2 | 5+6 2 | 2+2 2 | 2+2 2 |
| | 3+4 2 | 3+4 2 | 3+4 2 | 3+4 2 | 3+3 2 | 3+3 2 |
| | 5+6 2 | 3+4 2 | 3+4 2 | 5+6 2 | 2+2 2 | 2+2 2 |
| | 3+4 2 | 5+6 2 | 5+6 2 | 3+4 2 | 3+3 2 | 2+2 2 |
| | 5+6 2 | 3+4 2 | 3+4 2 | 5+6 2 | 2+2 2 | 3+3 2 |
| | 3+4 2 | 3+4 2 | 5+6 2 | 5+6 2 | 3+3 2 | 2+2 2 |
| | 5+6 2 | 3+4 2 | 5+6 2 | 3+4 2 | 2+2 2 | 3+3 2 |
| | 5+6 2 | 5+6 2 | 3+4 2 | 3+4 2 | 3+3 2 | 3+3 2 |
| | 5+6 2 | 3+4 2 | 5+6 2 | 3+4 2 | 3+3 2 | 2+2 2 |
| | 3+4 2 | 3+4 2 | 3+4 2 | 3+4 2 | 2+2 2 | 2+2 2 |
| | 5+6 2 | 5+6 2 | 5+6 2 | 5+6 2 | 2+2 2 | 3+3 2 |
| | 3+4 2 | 3+4 2 | 5+6 2 | 3+4 2 | 3+3 2 | 2+2 2 |
| | 3+4 2 | 5+6 2 | 5+6 2 | 5+6 2 | 2+2 2 | 3+3 2 |
| | 5+6 2 | 5+6 2 | 3+4 2 | 5+6 2 | 3+3 2 | 2+2 2 |
| | 5+6 2 | 3+4 2 | 3+4 2 | 3+4 2 | 2+2 2 | 3+3 2 |
| | 3+4 2 | 5+6 2 | 5+6 2 | 3+4 2 | 3+3 2 | 3+3 2 |
| | 3+4 2 | 3+4 2 | 5+6 2 | 5+6 2 | 3+3 2 | 2+2 2 |
| | 3+4 2 | 3+4 2 | 3+4 2 | 5+6 2 | 2+2 2 | 3+3 2 |
| | 3+4 2 | 3+4 2 | 5+6 2 | 3+4 2 | 3+3 2 | 2+2 2 |
| | 5+6 2 | 5+6 2 | 3+4 2 | 5+6 2 | 2+2 2 | 2+2 2 |
| | 5+6 2 | 5+6 2 | 5+6 2 | 3+4 2 | 2+2 2 | 3+3 2 |
| | 3+4 2 | 3+4 2 | 3+4 2 | 5+6 2 | 3+3 2 | 2+2 2 |
| | 5+6 2 | 5+6 2 | 5+6 2 | 3+4 2 | 2+2 2 | 3+3 2 |
| | 5+6 2 | 3+4 2 | 5+6 2 | 3+4 2 | 3+3 2 | 2+2 2 |
| | 5+6 2 | 5+6 2 | 5+6 2 | 5+6 2 | 2+2 2 | 3+3 2 |
| | 3+4 2 | 3+4 2 | 3+4 2 | 3+4 2 | 3+3 2 | 2+2 2 |
| | 3+4 2 | 3+4 2 | 3+4 2 | 5+6 2 | 3+3 2 | 3+3 2 |
| | 5+6 2 | 3+4 2 | 5+6 2 | 3+4 2 | 2+2 2 | 2+2 2 |
| | 5+6 2 | 5+6 2 | 3+4 2 | 5+6 2 | 3+3 2 | 2+2 2 |
| | 5+6 2 | 3+4 2 | 3+4 2 | 5+6 2 | 2+2 2 | 2+2 2 |
| | 3+4 2 | 5+6 2 | 5+6 2 | 3+4 2 | 3+3 2 | 3+3 2 |
| | 5+6 2 | 3+4 2 | 3+4 2 | 3+4 2 | 3+3 2 | 2+2 2 |
| | 3+4 2 | 5+6 2 | 5+6 2 | 5+6 2 | 2+2 2 | 3+3 2 |
| | 5+6 2 | 5+6 2 | 3+4 2 | 5+6 2 | 2+2 2 | 2+2 2 |
| | 5+6 2 | 3+4 2 | 5+6 2 | 5+6 2 | 3+3 2 | 3+3 2 |
| | 3+4 2 | 5+6 2 | 3+4 2 | 3+4 2 | 3+3 2 | 2+2 2 |
| | 3+4 2 | 3+4 2 | 5+6 2 | 3+4 2 | 2+2 2 | 3+3 2 |
| | 5+6 2 | 3+4 2 | 5+6 2 | 5+6 2 | 2+2 2 | 2+2 2 |
| | 5+6 2 | 3+4 2 | 5+6 2 | 3+4 2 | 3+3 2 | 3+3 2 |

Continued on next column

**TABLE XXIV– continued**

| TQC | ICSD | NoE | SF | | | |
|---|---|---|---|---|---|---|
| | 5+6 2 | 5+6 2 | 3+4 2 | 5+6 2 | 3+3 2 | 2+2 2 |
| | 3+4 2 | 5+6 2 | 3+4 2 | 5+6 2 | 3+3 2 | 3+3 2 |
| | 3+4 2 | 5+6 2 | 5+6 2 | 3+4 2 | 2+2 2 | 3+3 2 |
| | 3+4 2 | 3+4 2 | 3+4 2 | 3+4 2 | 3+3 2 | 2+2 2 |
| | 5+6 2 | 3+4 2 | 5+6 2 | 5+6 2 | 2+2 2 | 3+3 2 |
| | 3+4 2 | 5+6 2 | 5+6 2 | 5+6 2 | 2+2 2 | 2+2 2 |
| | 5+6 2 | 3+4 2 | 3+4 2 | 3+4 2 | 3+3 2 | 3+3 2 |
| | 5+6 2 | 5+6 2 | 5+6 2 | 5+6 2 | 2+2 2 | 2+2 2 |
| | 5+6 2 | 5+6 2 | 5+6 2 | 5+6 2 | 2+2 2 | 2+2 2 |
| | 3+4 2 | 3+4 2 | 3+4 2 | 3+4 2 | 3+3 2 | 3+3 2 |
| | 5+6 2 | 5+6 2 | 5+6 2 | 5+6 2 | 2+2 2 | 2+2 2 |
| | 5+6 2 | 3+4 2 | 5+6 2 | 3+4 2 | 3+3 2 | 2+2 2 |
| | 3+4 2 | 5+6 2 | 3+4 2 | 5+6 2 | 2+2 2 | 3+3 2 |
| | 5+6 2 | 5+6 2 | 5+6 2 | 5+6 2 | 2+2 2 | 2+2 2 |
| | 3+4 2 | 3+4 2 | 3+4 2 | 5+6 2 | 3+3 2 | 2+2 2 |
| | 5+6 2 | 5+6 2 | 5+6 2 | 3+4 2 | 2+2 2 | 3+3 2 |
| | 3+4 2 | 3+4 2 | 5+6 2 | 3+4 2 | 3+3 2 | 2+2 2 |
| | 5+6 2 | 3+4 2 | 5+6 2 | 5+6 2 | 2+2 2 | 3+3 2 |
| | 5+6 2 | 5+6 2 | 3+4 2 | 5+6 2 | 2+2 2 | 2+2 2 |
| | 3+4 2 | 5+6 2 | 5+6 2 | 5+6 2 | 3+3 2 | 2+2 2 |
| 12.1.190 | 81775 | 156 | As2Ce1Pd3 | | | |
| | -A | -GM | -M | -Y | -L | -V |
| | 3+4 2 | 3+4 2 | 3+4 2 | 3+4 2 | 2+2 2 | 2+2 2 |
| | 5+6 2 | 5+6 2 | 5+6 2 | 5+6 2 | 3+3 2 | 3+3 2 |
| | 3+4 2 | 3+4 2 | 3+4 2 | 3+4 2 | 3+3 2 | 3+3 2 |
| | 5+6 2 | 3+4 2 | 5+6 2 | 5+6 2 | 2+2 2 | 2+2 2 |
| | 3+4 2 | 5+6 2 | 3+4 2 | 3+4 2 | 3+3 2 | 3+3 2 |
| | 5+6 2 | 5+6 2 | 5+6 2 | 5+6 2 | 2+2 2 | 2+2 2 |
| | 5+6 2 | 5+6 2 | 5+6 2 | 5+6 2 | 3+3 2 | 2+2 2 |
| | 3+4 2 | 3+4 2 | 3+4 2 | 3+4 2 | 2+2 2 | 3+3 2 |
| | 5+6 2 | 5+6 2 | 5+6 2 | 5+6 2 | 2+2 2 | 2+2 2 |
| | 3+4 2 | 3+4 2 | 3+4 2 | 3+4 2 | 2+2 2 | 2+2 2 |
| | 5+6 2 | 5+6 2 | 5+6 2 | 5+6 2 | 3+3 2 | 3+3 2 |
| | 5+6 2 | 5+6 2 | 5+6 2 | 5+6 2 | 2+2 2 | 2+2 2 |
| | 5+6 2 | 3+4 2 | 5+6 2 | 3+4 2 | 3+3 2 | 2+2 2 |
| | 3+4 2 | 5+6 2 | 3+4 2 | 5+6 2 | 2+2 2 | 3+3 2 |
| | 3+4 2 | 3+4 2 | 5+6 2 | 5+6 2 | 3+3 2 | 3+3 2 |
| | 5+6 2 | 5+6 2 | 3+4 2 | 3+4 2 | 2+2 2 | 2+2 2 |
| | 3+4 2 | 3+4 2 | 3+4 2 | 3+4 2 | 3+3 2 | 2+2 2 |
| | 5+6 2 | 5+6 2 | 5+6 2 | 5+6 2 | 2+2 2 | 3+3 2 |
| | 5+6 2 | 3+4 2 | 5+6 2 | 3+4 2 | 3+3 2 | 3+3 2 |
| | 3+4 2 | 3+4 2 | 3+4 2 | 5+6 2 | 2+2 2 | 2+2 2 |
| | 5+6 2 | 5+6 2 | 5+6 2 | 3+4 2 | 3+3 2 | 3+3 2 |
| | 3+4 2 | 5+6 2 | 5+6 2 | 3+4 2 | 3+3 2 | 2+2 2 |
| | 3+4 2 | 3+4 2 | 3+4 2 | 3+4 2 | 2+2 2 | 2+2 2 |
| | 5+6 2 | 5+6 2 | 5+6 2 | 5+6 2 | 2+2 2 | 2+2 2 |
| | 3+4 2 | 3+4 2 | 3+4 2 | 3+4 2 | 3+3 2 | 3+3 2 |
| | 5+6 2 | 5+6 2 | 3+4 2 | 5+6 2 | 3+3 2 | 3+3 2 |
| | 5+6 2 | 5+6 2 | 5+6 2 | 5+6 2 | 3+3 2 | 2+2 2 |
| | 3+4 2 | 3+4 2 | 3+4 2 | 3+4 2 | 2+2 2 | 3+3 2 |
| | 5+6 2 | 5+6 2 | 5+6 2 | 5+6 2 | 2+2 2 | 3+3 2 |
| | 3+4 2 | 3+4 2 | 3+4 2 | 3+4 2 | 3+3 2 | 2+2 2 |
| | 5+6 2 | 3+4 2 | 5+6 2 | 5+6 2 | 2+2 2 | 2+2 2 |
| | 3+4 2 | 3+4 2 | 3+4 2 | 3+4 2 | 3+3 2 | 2+2 2 |
| | 5+6 2 | 5+6 2 | 3+4 2 | 5+6 2 | 2+2 2 | 3+3 2 |
| | 3+4 2 | 3+4 2 | 5+6 2 | 3+4 2 | 3+3 2 | 2+2 2 |
| | 5+6 2 | 3+4 2 | 5+6 2 | 5+6 2 | 3+3 2 | 3+3 2 |
| | 5+6 2 | 3+4 2 | 3+4 2 | 3+4 2 | 2+2 2 | 3+3 2 |
| | 5+6 2 | 3+4 2 | 5+6 2 | 3+4 2 | 3+3 2 | 2+2 2 |
| | 3+4 2 | 5+6 2 | 3+4 2 | 3+4 2 | 2+2 2 | 2+2 2 |
| | 5+6 2 | 5+6 2 | 5+6 2 | 5+6 2 | 3+3 2 | 3+3 2 |
| | 3+4 2 | 3+4 2 | 5+6 2 | 3+4 2 | 2+2 2 | 3+3 2 |
| | 3+4 2 | 5+6 2 | 5+6 2 | 5+6 2 | 2+2 2 | 2+2 2 |
| | 5+6 2 | 5+6 2 | 3+4 2 | 3+4 2 | 3+3 2 | 2+2 2 |
| | 3+4 2 | 3+4 2 | 5+6 2 | 5+6 2 | 2+2 2 | 3+3 2 |
| | 5+6 2 | 5+6 2 | 3+4 2 | 3+4 2 | 3+3 2 | 3+3 2 |
| | 3+4 2 | 3+4 2 | 3+4 2 | 5+6 2 | 3+3 2 | 2+2 2 |
| | 3+4 2 | 3+4 2 | 5+6 2 | 5+6 2 | 2+2 2 | 3+3 2 |
| | 3+4 2 | 3+4 2 | 5+6 2 | 3+4 2 | 3+3 2 | 2+2 2 |
| | 5+6 2 | 5+6 2 | 3+4 2 | 5+6 2 | 2+2 2 | 2+2 2 |
| | 5+6 2 | 5+6 2 | 5+6 2 | 3+4 2 | 3+3 2 | 3+3 2 |
| | 3+4 2 | 3+4 2 | 3+4 2 | 3+4 2 | 2+2 2 | 2+2 2 |
| | 5+6 2 | 5+6 2 | 5+6 2 | 5+6 2 | 2+2 2 | 3+3 2 |
| | 5+6 2 | 3+4 2 | 5+6 2 | 5+6 2 | 3+3 2 | 2+2 2 |
| | 3+4 2 | 5+6 2 | 3+4 2 | 3+4 2 | 3+3 2 | 3+3 2 |
| | 5+6 2 | 3+4 2 | 5+6 2 | 3+4 2 | 2+2 2 | 2+2 2 |
| | 3+4 2 | 3+4 2 | 3+4 2 | 5+6 2 | 3+3 2 | 3+3 2 |
| | 5+6 2 | 3+4 2 | 5+6 2 | 3+4 2 | 2+2 2 | 2+2 2 |
| | 5+6 2 | 5+6 2 | 3+4 2 | 5+6 2 | 3+3 2 | 2+2 2 |
| | 3+4 2 | 3+4 2 | 5+6 2 | 3+4 2 | 3+3 2 | 2+2 2 |
| | 5+6 2 | 5+6 2 | 3+4 2 | 5+6 2 | 2+2 2 | 3+3 2 |
| | 5+6 2 | 3+4 2 | 3+4 2 | 3+4 2 | 3+3 2 | 2+2 2 |
| | 3+4 2 | 5+6 2 | 3+4 2 | 5+6 2 | 2+2 2 | 3+3 2 |
| | 5+6 2 | 5+6 2 | 5+6 2 | 5+6 2 | 3+3 2 | 2+2 2 |
| | 3+4 2 | 3+4 2 | 5+6 2 | 5+6 2 | 2+2 2 | 3+3 2 |
| | 5+6 2 | 3+4 2 | 3+4 2 | 3+4 2 | 3+3 2 | 2+2 2 |
| | 3+4 2 | 5+6 2 | 5+6 2 | 3+4 2 | 2+2 2 | 3+3 2 |
| | 5+6 2 | 3+4 2 | 5+6 2 | 5+6 2 | 3+3 2 | 3+3 2 |
| | 5+6 2 | 3+4 2 | 3+4 2 | 3+4 2 | 2+2 2 | 2+2 2 |
| | 3+4 2 | 5+6 2 | 5+6 2 | 5+6 2 | 3+3 2 | 2+2 2 |
| | 5+6 2 | 5+6 2 | 3+4 2 | 5+6 2 | 3+3 2 | 3+3 2 |
| | 3+4 2 | 5+6 2 | 5+6 2 | 3+4 2 | 2+2 2 | 3+3 2 |
| | 3+4 2 | 3+4 2 | 3+4 2 | 3+4 2 | 3+3 2 | 3+3 2 |
| | 5+6 2 | 3+4 2 | 5+6 2 | 5+6 2 | 2+2 2 | 2+2 2 |
| | 3+4 2 | 5+6 2 | 5+6 2 | 5+6 2 | 2+2 2 | 2+2 2 |
| | 3+4 2 | 3+4 2 | 3+4 2 | 3+4 2 | 3+3 2 | 3+3 2 |
| | 5+6 2 | 3+4 2 | 5+6 2 | 3+4 2 | 2+2 2 | 2+2 2 |
| | 5+6 2 | 3+4 2 | 3+4 2 | 5+6 2 | 3+3 2 | 3+3 2 |
| | 5+6 2 | 5+6 2 | 5+6 2 | 3+4 2 | 2+2 2 | 2+2 2 |
| | 3+4 2 | 5+6 2 | 3+4 2 | 5+6 2 | 3+3 2 | 2+2 2 |
| 12.1.191 | 75454 | 50 | As2Ba2Mn2O1 | | | |
| | -A | -GM | -M | -Y | -L | -V |
| | 5+6 2 | 3+4 2 | 3+4 2 | 3+4 2 | 2+2 2 | 3+3 2 |
| | 3+4 2 | 5+6 2 | 5+6 2 | 5+6 2 | 3+3 2 | 2+2 2 |
| | 3+4 2 | 3+4 2 | 3+4 2 | 3+4 2 | 2+2 2 | 2+2 2 |
| | 5+6 2 | 3+4 2 | 5+6 2 | 3+4 2 | 3+3 2 | 3+3 2 |
| | 3+4 2 | 5+6 2 | 3+4 2 | 5+6 2 | 2+2 2 | 2+2 2 |
| | 5+6 2 | 3+4 2 | 5+6 2 | 3+4 2 | 3+3 2 | 3+3 2 |
| | 3+4 2 | 5+6 2 | 3+4 2 | 5+6 2 | 2+2 2 | 2+2 2 |
| | 5+6 2 | 3+4 2 | 5+6 2 | 5+6 2 | 3+3 2 | 3+3 2 |
| | 3+4 2 | 5+6 2 | 3+4 2 | 3+4 2 | 2+2 2 | 2+2 2 |
| | 3+4 2 | 3+4 2 | 5+6 2 | 5+6 2 | 2+2 2 | 3+3 2 |
| | 5+6 2 | 5+6 2 | 3+4 2 | 3+4 2 | 3+3 2 | 2+2 2 |
| | 5+6 2 | 5+6 2 | 5+6 2 | 5+6 2 | 3+3 2 | 3+3 2 |
| | 5+6 2 | 5+6 2 | 5+6 2 | 5+6 2 | 3+3 2 | 3+3 2 |
| | 5+6 2 | 5+6 2 | 5+6 2 | 5+6 2 | 3+3 2 | 3+3 2 |
| | 3+4 2 | 3+4 2 | 3+4 2 | 3+4 2 | 2+2 2 | 2+2 2 |
| | 5+6 2 | 5+6 2 | 5+6 2 | 5+6 2 | 3+3 2 | 3+3 2 |
| | 5+6 2 | 5+6 2 | 5+6 2 | 3+4 2 | 3+3 2 | 2+2 2 |
| | 5+6 2 | 5+6 2 | 3+4 2 | 5+6 2 | 3+3 2 | 2+2 2 |
| | 3+4 2 | 3+4 2 | 3+4 2 | 3+4 2 | 2+2 2 | 3+3 2 |
| | 3+4 2 | 3+4 2 | 5+6 2 | 5+6 2 | 2+2 2 | 3+3 2 |

Continued on next column

**TABLE XXIV– continued**

| TQC | ICSD | NoE | SF | | | |
|---|---|---|---|---|---|---|
| | 3+4 2 | 3+4 2 | 5+6 2 | 5+6 2 | 3+3 2 | 2+2 2 |
| | 3+4 2 | 3+4 2 | 3+4 2 | 3+4 2 | 2+2 2 | 2+2 2 |
| | 3+4 2 | 3+4 2 | 5+6 2 | 3+4 2 | 2+2 2 | 3+3 2 |
| | 5+6 2 | 5+6 2 | 3+4 2 | 5+6 2 | 2+2 2 | 2+2 2 |
| | 5+6 2 | 3+4 2 | 5+6 2 | 3+4 2 | 3+3 2 | 3+3 2 |
| 12.1.192 | 68906 | 64 | As1F6N1O2 | | | |
| | -A | -GM | -M | -Y | -L | -V |
| | 5+6 2 | 3+4 2 | 5+6 2 | 3+4 2 | 3+3 2 | 2+2 2 |
| | 3+4 2 | 5+6 2 | 3+4 2 | 5+6 2 | 2+2 2 | 3+3 2 |
| | 3+4 2 | 3+4 2 | 5+6 2 | 5+6 2 | 3+3 2 | 3+3 2 |
| | 5+6 2 | 5+6 2 | 5+6 2 | 5+6 2 | 2+2 2 | 2+2 2 |
| | 5+6 2 | 5+6 2 | 3+4 2 | 3+4 2 | 2+2 2 | 2+2 2 |
| | 3+4 2 | 3+4 2 | 3+4 2 | 3+4 2 | 3+3 2 | 3+3 2 |
| | 5+6 2 | 5+6 2 | 3+4 2 | 3+4 2 | 2+2 2 | 2+2 2 |
| | 3+4 2 | 3+4 2 | 5+6 2 | 5+6 2 | 3+3 2 | 3+3 2 |
| | 5+6 2 | 3+4 2 | 5+6 2 | 3+4 2 | 3+3 2 | 2+2 2 |
| | 3+4 2 | 5+6 2 | 3+4 2 | 5+6 2 | 2+2 2 | 3+3 2 |
| | 3+4 2 | 5+6 2 | 3+4 2 | 5+6 2 | 2+2 2 | 3+3 2 |
| | 3+4 2 | 5+6 2 | 3+4 2 | 5+6 2 | 2+2 2 | 3+3 2 |
| | 3+4 2 | 3+4 2 | 3+4 2 | 3+4 2 | 3+3 2 | 3+3 2 |
| | 5+6 2 | 5+6 2 | 5+6 2 | 5+6 2 | 2+2 2 | 2+2 2 |
| | 5+6 2 | 5+6 2 | 5+6 2 | 3+4 2 | 2+2 2 | 2+2 2 |
| | 3+4 2 | 3+4 2 | 5+6 2 | 3+4 2 | 3+3 2 | 2+2 2 |
| | 5+6 2 | 3+4 2 | 3+4 2 | 5+6 2 | 3+3 2 | 2+2 2 |
| | 5+6 2 | 3+4 2 | 5+6 2 | 3+4 2 | 3+3 2 | 3+3 2 |
| | 3+4 2 | 3+4 2 | 3+4 2 | 3+4 2 | 2+2 2 | 2+2 2 |
| | 5+6 2 | 5+6 2 | 5+6 2 | 5+6 2 | 3+3 2 | 3+3 2 |
| | 5+6 2 | 5+6 2 | 5+6 2 | 5+6 2 | 2+2 2 | 2+2 2 |
| | 3+4 2 | 3+4 2 | 3+4 2 | 3+4 2 | 3+3 2 | 3+3 2 |
| | 5+6 2 | 5+6 2 | 5+6 2 | 5+6 2 | 2+2 2 | 2+2 2 |
| | 3+4 2 | 3+4 2 | 5+6 2 | 5+6 2 | 3+3 2 | 3+3 2 |
| | 3+4 2 | 3+4 2 | 3+4 2 | 3+4 2 | 2+2 2 | 2+2 2 |
| | 3+4 2 | 3+4 2 | 5+6 2 | 5+6 2 | 3+3 2 | 3+3 2 |
| | 5+6 2 | 5+6 2 | 5+6 2 | 5+6 2 | 2+2 2 | 2+2 2 |
| | 3+4 2 | 3+4 2 | 3+4 2 | 3+4 2 | 3+3 2 | 3+3 2 |
| | 5+6 2 | 5+6 2 | 3+4 2 | 3+4 2 | 2+2 2 | 2+2 2 |
| | 5+6 2 | 5+6 2 | 3+4 2 | 3+4 2 | 3+3 2 | 3+3 2 |
| | 3+4 2 | 5+6 2 | 5+6 2 | 5+6 2 | 3+3 2 | 3+3 2 |
| | 3+4 2 | 5+6 2 | 3+4 2 | 3+4 2 | 2+2 2 | 2+2 2 |
| 12.1.193 | 418388 | 186 | Al6O1Si8Sr13 | | | |
| | -A | -GM | -M | -Y | -L | -V |
| | 3+4 2 | 3+4 2 | 3+4 2 | 3+4 2 | 2+2 2 | 2+2 2 |
| | 5+6 2 | 5+6 2 | 5+6 2 | 5+6 2 | 3+3 2 | 3+3 2 |
| | 3+4 2 | 3+4 2 | 3+4 2 | 3+4 2 | 2+2 2 | 2+2 2 |
| | 5+6 2 | 5+6 2 | 5+6 2 | 5+6 2 | 3+3 2 | 3+3 2 |
| | 5+6 2 | 3+4 2 | 3+4 2 | 5+6 2 | 2+2 2 | 2+2 2 |
| | 3+4 2 | 3+4 2 | 5+6 2 | 3+4 2 | 3+3 2 | 3+3 2 |
| | 3+4 2 | 5+6 2 | 5+6 2 | 5+6 2 | 2+2 2 | 2+2 2 |
| | 5+6 2 | 5+6 2 | 3+4 2 | 3+4 2 | 3+3 2 | 3+3 2 |
| | 3+4 2 | 3+4 2 | 3+4 2 | 3+4 2 | 3+3 2 | 3+3 2 |
| | 5+6 2 | 5+6 2 | 5+6 2 | 5+6 2 | 2+2 2 | 2+2 2 |
| | 5+6 2 | 3+4 2 | 5+6 2 | 3+4 2 | 2+2 2 | 2+2 2 |
| | 3+4 2 | 5+6 2 | 3+4 2 | 5+6 2 | 3+3 2 | 3+3 2 |
| | 5+6 2 | 3+4 2 | 5+6 2 | 3+4 2 | 3+3 2 | 2+2 2 |
| | 5+6 2 | 3+4 2 | 5+6 2 | 3+4 2 | 2+2 2 | 3+3 2 |
| | 3+4 2 | 5+6 2 | 3+4 2 | 5+6 2 | 3+3 2 | 2+2 2 |
| | 5+6 2 | 5+6 2 | 3+4 2 | 3+4 2 | 2+2 2 | 3+3 2 |
| | 3+4 2 | 3+4 2 | 5+6 2 | 5+6 2 | 2+2 2 | 2+2 2 |
| | 5+6 2 | 3+4 2 | 5+6 2 | 3+4 2 | 3+3 2 | 3+3 2 |
| | 3+4 2 | 3+4 2 | 3+4 2 | 5+6 2 | 2+2 2 | 2+2 2 |
| | 5+6 2 | 3+4 2 | 5+6 2 | 5+6 2 | 3+3 2 | 3+3 2 |

Continued on next column

**TABLE XXIV– continued**

| TQC | ICSD | NoE | SF | | | |
|---|---|---|---|---|---|---|
| | 3+4 2 | 5+6 2 | 3+4 2 | 3+4 2 | 2+2 2 | 2+2 2 |
| | 5+6 2 | 5+6 2 | 5+6 2 | 3+4 2 | 3+3 2 | 3+3 2 |
| | 3+4 2 | 3+4 2 | 3+4 2 | 3+4 2 | 3+3 2 | 3+3 2 |
| | 5+6 2 | 5+6 2 | 5+6 2 | 5+6 2 | 2+2 2 | 2+2 2 |
| | 3+4 2 | 5+6 2 | 3+4 2 | 5+6 2 | 3+3 2 | 2+2 2 |
| | 3+4 2 | 5+6 2 | 3+4 2 | 5+6 2 | 2+2 2 | 3+3 2 |
| | 5+6 2 | 3+4 2 | 5+6 2 | 3+4 2 | 3+3 2 | 2+2 2 |
| | 3+4 2 | 5+6 2 | 3+4 2 | 5+6 2 | 2+2 2 | 3+3 2 |
| | 5+6 2 | 3+4 2 | 5+6 2 | 3+4 2 | 3+3 2 | 2+2 2 |
| | 3+4 2 | 5+6 2 | 3+4 2 | 5+6 2 | 2+2 2 | 3+3 2 |
| | 5+6 2 | 3+4 2 | 5+6 2 | 3+4 2 | 3+3 2 | 3+3 2 |
| | 5+6 2 | 5+6 2 | 3+4 2 | 3+4 2 | 2+2 2 | 2+2 2 |
| | 3+4 2 | 3+4 2 | 5+6 2 | 5+6 2 | 2+2 2 | 2+2 2 |
| | 3+4 2 | 3+4 2 | 3+4 2 | 5+6 2 | 3+3 2 | 3+3 2 |
| | 5+6 2 | 5+6 2 | 5+6 2 | 3+4 2 | 2+2 2 | 3+3 2 |
| | 3+4 2 | 3+4 2 | 3+4 2 | 5+6 2 | 3+3 2 | 2+2 2 |
| | 5+6 2 | 3+4 2 | 5+6 2 | 5+6 2 | 2+2 2 | 3+3 2 |
| | 5+6 2 | 5+6 2 | 3+4 2 | 3+4 2 | 3+3 2 | 2+2 2 |
| | 3+4 2 | 3+4 2 | 5+6 2 | 5+6 2 | 2+2 2 | 3+3 2 |
| | 3+4 2 | 5+6 2 | 3+4 2 | 3+4 2 | 3+3 2 | 2+2 2 |
| | 5+6 2 | 3+4 2 | 5+6 2 | 5+6 2 | 2+2 2 | 3+3 2 |
| | 3+4 2 | 5+6 2 | 3+4 2 | 3+4 2 | 3+3 2 | 2+2 2 |
| | 5+6 2 | 5+6 2 | 5+6 2 | 3+4 2 | 2+2 2 | 3+3 2 |
| | 3+4 2 | 3+4 2 | 3+4 2 | 3+4 2 | 3+3 2 | 3+3 2 |
| | 5+6 2 | 5+6 2 | 5+6 2 | 5+6 2 | 2+2 2 | 2+2 2 |
| | 3+4 2 | 3+4 2 | 3+4 2 | 3+4 2 | 3+3 2 | 3+3 2 |
| | 5+6 2 | 5+6 2 | 5+6 2 | 5+6 2 | 2+2 2 | 2+2 2 |
| | 3+4 2 | 5+6 2 | 3+4 2 | 5+6 2 | 2+2 2 | 3+3 2 |
| | 5+6 2 | 3+4 2 | 5+6 2 | 3+4 2 | 3+3 2 | 2+2 2 |
| | 3+4 2 | 5+6 2 | 3+4 2 | 5+6 2 | 3+3 2 | 2+2 2 |
| | 3+4 2 | 5+6 2 | 3+4 2 | 5+6 2 | 2+2 2 | 3+3 2 |
| | 3+4 2 | 5+6 2 | 3+4 2 | 5+6 2 | 2+2 2 | 3+3 2 |
| | 5+6 2 | 3+4 2 | 5+6 2 | 3+4 2 | 2+2 2 | 3+3 2 |
| | 3+4 2 | 3+4 2 | 3+4 2 | 3+4 2 | 2+2 2 | 2+2 2 |
| | 5+6 2 | 5+6 2 | 5+6 2 | 5+6 2 | 3+3 2 | 3+3 2 |
| | 5+6 2 | 5+6 2 | 5+6 2 | 5+6 2 | 3+3 2 | 3+3 2 |
| | 3+4 2 | 3+4 2 | 3+4 2 | 3+4 2 | 2+2 2 | 2+2 2 |
| | 3+4 2 | 3+4 2 | 3+4 2 | 3+4 2 | 2+2 2 | 2+2 2 |
| | 5+6 2 | 5+6 2 | 5+6 2 | 5+6 2 | 3+3 2 | 3+3 2 |
| | 5+6 2 | 5+6 2 | 5+6 2 | 5+6 2 | 3+3 2 | 3+3 2 |
| | 3+4 2 | 3+4 2 | 3+4 2 | 3+4 2 | 2+2 2 | 2+2 2 |
| | 3+4 2 | 3+4 2 | 3+4 2 | 3+4 2 | 2+2 2 | 2+2 2 |
| | 5+6 2 | 3+4 2 | 5+6 2 | 3+4 2 | 3+3 2 | 3+3 2 |
| | 3+4 2 | 5+6 2 | 3+4 2 | 5+6 2 | 2+2 2 | 2+2 2 |
| | 5+6 2 | 5+6 2 | 5+6 2 | 5+6 2 | 3+3 2 | 3+3 2 |
| | 3+4 2 | 5+6 2 | 3+4 2 | 5+6 2 | 3+3 2 | 2+2 2 |
| | 3+4 2 | 5+6 2 | 3+4 2 | 3+4 2 | 3+3 2 | 2+2 2 |
| | 3+4 2 | 3+4 2 | 5+6 2 | 5+6 2 | 3+3 2 | 2+2 2 |
| | 5+6 2 | 3+4 2 | 3+4 2 | 3+4 2 | 3+3 2 | 3+3 2 |
| | 3+4 2 | 5+6 2 | 3+4 2 | 5+6 2 | 2+2 2 | 2+2 2 |
| | 5+6 2 | 5+6 2 | 5+6 2 | 5+6 2 | 2+2 2 | 2+2 2 |
| | 3+4 2 | 3+4 2 | 3+4 2 | 3+4 2 | 3+3 2 | 2+2 2 |
| | 5+6 2 | 5+6 2 | 5+6 2 | 5+6 2 | 2+2 2 | 3+3 2 |
| | 3+4 2 | 5+6 2 | 3+4 2 | 5+6 2 | 2+2 2 | 3+3 2 |
| | 3+4 2 | 5+6 2 | 5+6 2 | 5+6 2 | 3+3 2 | 2+2 2 |
| | 5+6 2 | 3+4 2 | 3+4 2 | 3+4 2 | 3+3 2 | 3+3 2 |
| | 5+6 2 | 3+4 2 | 5+6 2 | 3+4 2 | 3+3 2 | 3+3 2 |
| | 3+4 2 | 5+6 2 | 3+4 2 | 5+6 2 | 2+2 2 | 2+2 2 |
| | 5+6 2 | 3+4 2 | 5+6 2 | 3+4 2 | 3+3 2 | 3+3 2 |
| | 3+4 2 | 5+6 2 | 3+4 2 | 3+4 2 | 2+2 2 | 2+2 2 |
| | 5+6 2 | 3+4 2 | 3+4 2 | 5+6 2 | 2+2 2 | 2+2 2 |

Continued on next column

**TABLE XXIV– continued**

| TQC | ICSD | | NoE | | SF | |
|---|---|---|---|---|---|---|
| | 3+4 2 | 3+4 2 | 5+6 2 | 3+4 2 | 3+3 2 | 2+2 2 |
| | 5+6 2 | 3+4 2 | 5+6 2 | 5+6 2 | 3+3 2 | 3+3 2 |
| | 3+4 2 | 5+6 2 | 3+4 2 | 5+6 2 | 2+2 2 | 2+2 2 |
| | 3+4 2 | 5+6 2 | 5+6 2 | 5+6 2 | 3+3 2 | 2+2 2 |
| | 5+6 2 | 5+6 2 | 3+4 2 | 3+4 2 | 2+2 2 | 3+3 2 |
| | 3+4 2 | 3+4 2 | 5+6 2 | 5+6 2 | 3+3 2 | 3+3 2 |
| | 5+6 2 | 5+6 2 | 3+4 2 | 3+4 2 | 3+3 2 | 3+3 2 |
| | 5+6 2 | 3+4 2 | 5+6 2 | 3+4 2 | 2+2 2 | 3+3 2 |
| | 3+4 2 | 3+4 2 | 3+4 2 | 3+4 2 | 2+2 2 | 2+2 2 |
| | 3+4 2 | 5+6 2 | 3+4 2 | 5+6 2 | 2+2 2 | 3+3 2 |
| | 5+6 2 | 5+6 2 | 5+6 2 | 3+4 2 | 3+3 2 | 2+2 2 |
| | 3+4 2 | 5+6 2 | 5+6 2 | 5+6 2 | 3+3 2 | 2+2 2 |
| 12.1.194 | 409319 | | 156 | | Al4Na4P12Sr8 | |
| | -A | -GM | -M | -Y | -L | -V |
| | 5+6 2 | 3+4 2 | 3+4 2 | 5+6 2 | 2+2 2 | 2+2 2 |
| | 5+6 2 | 3+4 2 | 3+4 2 | 5+6 2 | 3+3 2 | 3+3 2 |
| | 3+4 2 | 3+4 2 | 5+6 2 | 3+4 2 | 2+2 2 | 2+2 2 |
| | 5+6 2 | 5+6 2 | 3+4 2 | 5+6 2 | 3+3 2 | 3+3 2 |
| | 3+4 2 | 5+6 2 | 3+4 2 | 5+6 2 | 2+2 2 | 2+2 2 |
| | 5+6 2 | 3+4 2 | 5+6 2 | 3+4 2 | 3+3 2 | 3+3 2 |
| | 3+4 2 | 3+4 2 | 5+6 2 | 5+6 2 | 2+2 2 | 2+2 2 |
| | 3+4 2 | 3+4 2 | 5+6 2 | 5+6 2 | 3+3 2 | 3+3 2 |
| | 3+4 2 | 5+6 2 | 5+6 2 | 3+4 2 | 2+2 2 | 2+2 2 |
| | 3+4 2 | 5+6 2 | 5+6 2 | 3+4 2 | 3+3 2 | 3+3 2 |
| | 3+4 2 | 5+6 2 | 3+4 2 | 3+4 2 | 2+2 2 | 2+2 2 |
| | 5+6 2 | 3+4 2 | 5+6 2 | 5+6 2 | 3+3 2 | 3+3 2 |
| | 5+6 2 | 3+4 2 | 5+6 2 | 3+4 2 | 3+3 2 | 2+2 2 |
| | 3+4 2 | 5+6 2 | 3+4 2 | 5+6 2 | 2+2 2 | 3+3 2 |
| | 5+6 2 | 5+6 2 | 3+4 2 | 3+4 2 | 2+2 2 | 2+2 2 |
| | 5+6 2 | 5+6 2 | 3+4 2 | 3+4 2 | 3+3 2 | 3+3 2 |
| | 3+4 2 | 5+6 2 | 5+6 2 | 3+4 2 | 3+3 2 | 2+2 2 |
| | 3+4 2 | 5+6 2 | 5+6 2 | 3+4 2 | 2+2 2 | 3+3 2 |
| | 3+4 2 | 5+6 2 | 5+6 2 | 3+4 2 | 2+2 2 | 2+2 2 |
| | 5+6 2 | 3+4 2 | 5+6 2 | 3+4 2 | 3+3 2 | 3+3 2 |
| | 3+4 2 | 5+6 2 | 3+4 2 | 5+6 2 | 3+3 2 | 2+2 2 |
| | 5+6 2 | 3+4 2 | 3+4 2 | 5+6 2 | 2+2 2 | 3+3 2 |
| | 3+4 2 | 5+6 2 | 5+6 2 | 3+4 2 | 2+2 2 | 3+3 2 |
| | 5+6 2 | 3+4 2 | 3+4 2 | 5+6 2 | 3+3 2 | 2+2 2 |
| | 3+4 2 | 5+6 2 | 5+6 2 | 3+4 2 | 3+3 2 | 2+2 2 |
| | 5+6 2 | 3+4 2 | 5+6 2 | 3+4 2 | 2+2 2 | 3+3 2 |
| | 3+4 2 | 5+6 2 | 3+4 2 | 5+6 2 | 2+2 2 | 3+3 2 |
| | 3+4 2 | 5+6 2 | 5+6 2 | 3+4 2 | 3+3 2 | 2+2 2 |
| | 5+6 2 | 5+6 2 | 3+4 2 | 3+4 2 | 2+2 2 | 2+2 2 |
| | 5+6 2 | 5+6 2 | 3+4 2 | 3+4 2 | 3+3 2 | 3+3 2 |
| | 5+6 2 | 5+6 2 | 3+4 2 | 3+4 2 | 2+2 2 | 2+2 2 |
| | 5+6 2 | 5+6 2 | 3+4 2 | 3+4 2 | 3+3 2 | 3+3 2 |
| | 5+6 2 | 3+4 2 | 5+6 2 | 3+4 2 | 3+3 2 | 2+2 2 |
| | 3+4 2 | 3+4 2 | 3+4 2 | 5+6 2 | 3+3 2 | 3+3 2 |
| | 5+6 2 | 5+6 2 | 5+6 2 | 3+4 2 | 2+2 2 | 2+2 2 |
| | 5+6 2 | 3+4 2 | 5+6 2 | 3+4 2 | 2+2 2 | 3+3 2 |
| | 3+4 2 | 3+4 2 | 5+6 2 | 5+6 2 | 2+2 2 | 2+2 2 |
| | 5+6 2 | 5+6 2 | 3+4 2 | 3+4 2 | 3+3 2 | 3+3 2 |
| | 3+4 2 | 3+4 2 | 3+4 2 | 3+4 2 | 3+3 2 | 3+3 2 |
| | 3+4 2 | 3+4 2 | 5+6 2 | 5+6 2 | 2+2 2 | 2+2 2 |
| | 5+6 2 | 5+6 2 | 3+4 2 | 3+4 2 | 3+3 2 | 3+3 2 |
| | 5+6 2 | 5+6 2 | 5+6 2 | 5+6 2 | 2+2 2 | 2+2 2 |
| | 3+4 2 | 5+6 2 | 3+4 2 | 5+6 2 | 2+2 2 | 3+3 2 |
| | 3+4 2 | 5+6 2 | 3+4 2 | 5+6 2 | 3+3 2 | 2+2 2 |
| | 3+4 2 | 3+4 2 | 3+4 2 | 5+6 2 | 2+2 2 | 2+2 2 |
| | 5+6 2 | 3+4 2 | 5+6 2 | 3+4 2 | 3+3 2 | 3+3 2 |
| | 3+4 2 | 5+6 2 | 3+4 2 | 5+6 2 | 3+3 2 | 3+3 2 |

Continued on next column

**TABLE XXIV– continued**

| TQC | ICSD | | NoE | | SF | |
|---|---|---|---|---|---|---|
| | 5+6 2 | 5+6 2 | 5+6 2 | 3+4 2 | 2+2 2 | 2+2 2 |
| | 3+4 2 | 3+4 2 | 3+4 2 | 5+6 2 | 3+3 2 | 2+2 2 |
| | 5+6 2 | 5+6 2 | 3+4 2 | 3+4 2 | 2+2 2 | 3+3 2 |
| | 3+4 2 | 5+6 2 | 5+6 2 | 3+4 2 | 2+2 2 | 3+3 2 |
| | 5+6 2 | 3+4 2 | 3+4 2 | 5+6 2 | 3+3 2 | 2+2 2 |
| | 3+4 2 | 3+4 2 | 5+6 2 | 5+6 2 | 2+2 2 | 3+3 2 |
| | 5+6 2 | 5+6 2 | 5+6 2 | 5+6 2 | 3+3 2 | 2+2 2 |
| | 3+4 2 | 3+4 2 | 3+4 2 | 3+4 2 | 3+3 2 | 3+3 2 |
| | 5+6 2 | 3+4 2 | 3+4 2 | 3+4 2 | 2+2 2 | 3+3 2 |
| | 3+4 2 | 3+4 2 | 5+6 2 | 5+6 2 | 2+2 2 | 2+2 2 |
| | 5+6 2 | 5+6 2 | 5+6 2 | 3+4 2 | 3+3 2 | 2+2 2 |
| | 5+6 2 | 3+4 2 | 3+4 2 | 5+6 2 | 2+2 2 | 3+3 2 |
| | 3+4 2 | 5+6 2 | 5+6 2 | 5+6 2 | 3+3 2 | 2+2 2 |
| | 3+4 2 | 5+6 2 | 3+4 2 | 3+4 2 | 3+3 2 | 2+2 2 |
| | 5+6 2 | 3+4 2 | 3+4 2 | 5+6 2 | 2+2 2 | 3+3 2 |
| | 3+4 2 | 5+6 2 | 5+6 2 | 3+4 2 | 3+3 2 | 3+3 2 |
| | 3+4 2 | 3+4 2 | 5+6 2 | 5+6 2 | 2+2 2 | 3+3 2 |
| | 5+6 2 | 5+6 2 | 5+6 2 | 5+6 2 | 3+3 2 | 2+2 2 |
| | 3+4 2 | 3+4 2 | 3+4 2 | 3+4 2 | 3+3 2 | 2+2 2 |
| | 5+6 2 | 3+4 2 | 3+4 2 | 3+4 2 | 2+2 2 | 2+2 2 |
| | 3+4 2 | 5+6 2 | 5+6 2 | 5+6 2 | 3+3 2 | 3+3 2 |
| | 5+6 2 | 3+4 2 | 5+6 2 | 5+6 2 | 2+2 2 | 2+2 2 |
| | 3+4 2 | 5+6 2 | 3+4 2 | 3+4 2 | 3+3 2 | 2+2 2 |
| | 5+6 2 | 3+4 2 | 5+6 2 | 3+4 2 | 2+2 2 | 3+3 2 |
| | 3+4 2 | 3+4 2 | 5+6 2 | 5+6 2 | 3+3 2 | 3+3 2 |
| | 5+6 2 | 5+6 2 | 5+6 2 | 5+6 2 | 2+2 2 | 2+2 2 |
| | 3+4 2 | 5+6 2 | 3+4 2 | 3+4 2 | 3+3 2 | 3+3 2 |
| | 5+6 2 | 5+6 2 | 3+4 2 | 3+4 2 | 3+3 2 | 2+2 2 |
| | 5+6 2 | 3+4 2 | 3+4 2 | 3+4 2 | 2+2 2 | 2+2 2 |
| | 5+6 2 | 5+6 2 | 5+6 2 | 3+4 2 | 2+2 2 | 3+3 2 |
| | 5+6 2 | 3+4 2 | 3+4 2 | 5+6 2 | 2+2 2 | 3+3 2 |
| 12.1.195 | 27569 | | 150 | | Al4Cl51Zr12 | |
| | -A | -GM | -M | -Y | -L | -V |
| | 3+4 2 | 3+4 2 | 5+6 2 | 5+6 2 | 2+2 2 | 3+3 2 |
| | 5+6 2 | 3+4 2 | 5+6 2 | 3+4 2 | 3+3 2 | 2+2 2 |
| | 3+4 2 | 5+6 2 | 3+4 2 | 5+6 2 | 2+2 2 | 3+3 2 |
| | 5+6 2 | 3+4 2 | 5+6 2 | 3+4 2 | 3+3 2 | 2+2 2 |
| | 5+6 2 | 5+6 2 | 3+4 2 | 5+6 2 | 2+2 2 | 3+3 2 |
| | 3+4 2 | 5+6 2 | 3+4 2 | 3+4 2 | 3+3 2 | 2+2 2 |
| | 3+4 2 | 5+6 2 | 3+4 2 | 3+4 2 | 3+3 2 | 3+3 2 |
| | 5+6 2 | 3+4 2 | 5+6 2 | 3+4 2 | 2+2 2 | 2+2 2 |
| | 5+6 2 | 5+6 2 | 3+4 2 | 5+6 2 | 2+2 2 | 3+3 2 |
| | 5+6 2 | 3+4 2 | 3+4 2 | 5+6 2 | 3+3 2 | 2+2 2 |
| | 3+4 2 | 5+6 2 | 5+6 2 | 5+6 2 | 2+2 2 | 2+2 2 |
| | 3+4 2 | 3+4 2 | 5+6 2 | 3+4 2 | 3+3 2 | 3+3 2 |
| | 5+6 2 | 3+4 2 | 3+4 2 | 3+4 2 | 3+3 2 | 3+3 2 |
| | 3+4 2 | 5+6 2 | 5+6 2 | 5+6 2 | 2+2 2 | 2+2 2 |
| | 5+6 2 | 3+4 2 | 3+4 2 | 5+6 2 | 2+2 2 | 2+2 2 |
| | 3+4 2 | 5+6 2 | 5+6 2 | 3+4 2 | 3+3 2 | 3+3 2 |
| | 5+6 2 | 3+4 2 | 5+6 2 | 5+6 2 | 2+2 2 | 2+2 2 |
| | 3+4 2 | 5+6 2 | 3+4 2 | 3+4 2 | 3+3 2 | 3+3 2 |
| | 5+6 2 | 5+6 2 | 3+4 2 | 3+4 2 | 3+3 2 | 2+2 2 |
| | 3+4 2 | 3+4 2 | 3+4 2 | 5+6 2 | 2+2 2 | 3+3 2 |
| | 5+6 2 | 5+6 2 | 5+6 2 | 3+4 2 | 3+3 2 | 2+2 2 |
| | 3+4 2 | 3+4 2 | 5+6 2 | 5+6 2 | 2+2 2 | 3+3 2 |
| | 3+4 2 | 5+6 2 | 5+6 2 | 3+4 2 | 3+3 2 | 2+2 2 |
| | 3+4 2 | 3+4 2 | 5+6 2 | 5+6 2 | 2+2 2 | 3+3 2 |
| | 5+6 2 | 3+4 2 | 3+4 2 | 5+6 2 | 3+3 2 | 3+3 2 |
| | 5+6 2 | 3+4 2 | 5+6 2 | 3+4 2 | 2+2 2 | 2+2 2 |
| | 3+4 2 | 3+4 2 | 3+4 2 | 5+6 2 | 2+2 2 | 3+3 2 |
| | 5+6 2 | 5+6 2 | 5+6 2 | 3+4 2 | 3+3 2 | 2+2 2 |

Continued on next column

**TABLE XXIV– continued**

| TQC | ICSD | NoE | SF | | | |
|---|---|---|---|---|---|---|
| | 3+4 2 | 5+6 2 | 3+4 2 | 3+4 2 | 2+2 2 | 3+3 2 |
| | 3+4 2 | 3+4 2 | 5+6 2 | 5+6 2 | 3+3 2 | 3+3 2 |
| | 5+6 2 | 5+6 2 | 3+4 2 | 5+6 2 | 2+2 2 | 2+2 2 |
| | 3+4 2 | 3+4 2 | 5+6 2 | 5+6 2 | 2+2 2 | 3+3 2 |
| | 5+6 2 | 3+4 2 | 5+6 2 | 5+6 2 | 3+3 2 | 2+2 2 |
| | 3+4 2 | 3+4 2 | 5+6 2 | 3+4 2 | 2+2 2 | 3+3 2 |
| | 5+6 2 | 5+6 2 | 3+4 2 | 3+4 2 | 3+3 2 | 2+2 2 |
| | 5+6 2 | 5+6 2 | 3+4 2 | 3+4 2 | 3+3 2 | 2+2 2 |
| | 5+6 2 | 3+4 2 | 3+4 2 | 5+6 2 | 2+2 2 | 3+3 2 |
| | 3+4 2 | 3+4 2 | 3+4 2 | 5+6 2 | 3+3 2 | 3+3 2 |
| | 3+4 2 | 5+6 2 | 5+6 2 | 5+6 2 | 2+2 2 | 2+2 2 |
| | 3+4 2 | 5+6 2 | 3+4 2 | 3+4 2 | 3+3 2 | 2+2 2 |
| | 5+6 2 | 3+4 2 | 5+6 2 | 3+4 2 | 2+2 2 | 3+3 2 |
| | 5+6 2 | 5+6 2 | 5+6 2 | 3+4 2 | 3+3 2 | 2+2 2 |
| | 3+4 2 | 3+4 2 | 3+4 2 | 5+6 2 | 2+2 2 | 3+3 2 |
| | 5+6 2 | 5+6 2 | 5+6 2 | 5+6 2 | 3+3 2 | 3+3 2 |
| | 3+4 2 | 3+4 2 | 5+6 2 | 3+4 2 | 2+2 2 | 2+2 2 |
| | 5+6 2 | 5+6 2 | 3+4 2 | 3+4 2 | 3+3 2 | 3+3 2 |
| | 3+4 2 | 3+4 2 | 3+4 2 | 5+6 2 | 2+2 2 | 2+2 2 |
| | 5+6 2 | 5+6 2 | 5+6 2 | 3+4 2 | 3+3 2 | 2+2 2 |
| | 5+6 2 | 5+6 2 | 3+4 2 | 3+4 2 | 2+2 2 | 3+3 2 |
| | 3+4 2 | 5+6 2 | 3+4 2 | 5+6 2 | 2+2 2 | 2+2 2 |
| | 5+6 2 | 3+4 2 | 5+6 2 | 5+6 2 | 3+3 2 | 2+2 2 |
| | 3+4 2 | 3+4 2 | 3+4 2 | 3+4 2 | 3+3 2 | 3+3 2 |
| | 5+6 2 | 3+4 2 | 5+6 2 | 5+6 2 | 2+2 2 | 3+3 2 |
| | 3+4 2 | 5+6 2 | 3+4 2 | 5+6 2 | 3+3 2 | 3+3 2 |
| | 3+4 2 | 3+4 2 | 5+6 2 | 3+4 2 | 2+2 2 | 2+2 2 |
| | 5+6 2 | 5+6 2 | 3+4 2 | 3+4 2 | 3+3 2 | 3+3 2 |
| | 5+6 2 | 3+4 2 | 5+6 2 | 5+6 2 | 2+2 2 | 2+2 2 |
| | 3+4 2 | 5+6 2 | 3+4 2 | 3+4 2 | 3+3 2 | 3+3 2 |
| | 5+6 2 | 3+4 2 | 5+6 2 | 5+6 2 | 2+2 2 | 2+2 2 |
| | 3+4 2 | 5+6 2 | 3+4 2 | 3+4 2 | 3+3 2 | 2+2 2 |
| | 3+4 2 | 5+6 2 | 5+6 2 | 3+4 2 | 2+2 2 | 3+3 2 |
| | 5+6 2 | 3+4 2 | 3+4 2 | 5+6 2 | 3+3 2 | 2+2 2 |
| | 5+6 2 | 5+6 2 | 5+6 2 | 3+4 2 | 2+2 2 | 3+3 2 |
| | 3+4 2 | 5+6 2 | 5+6 2 | 5+6 2 | 3+3 2 | 2+2 2 |
| | 3+4 2 | 5+6 2 | 3+4 2 | 3+4 2 | 2+2 2 | 2+2 2 |
| | 5+6 2 | 3+4 2 | 3+4 2 | 5+6 2 | 3+3 2 | 3+3 2 |
| | 3+4 2 | 3+4 2 | 3+4 2 | 3+4 2 | 3+3 2 | 3+3 2 |
| | 5+6 2 | 5+6 2 | 5+6 2 | 5+6 2 | 2+2 2 | 2+2 2 |
| | 5+6 2 | 5+6 2 | 3+4 2 | 5+6 2 | 2+2 2 | 2+2 2 |
| | 3+4 2 | 3+4 2 | 5+6 2 | 3+4 2 | 3+3 2 | 3+3 2 |
| | 5+6 2 | 5+6 2 | 3+4 2 | 5+6 2 | 2+2 2 | 2+2 2 |
| | 3+4 2 | 3+4 2 | 5+6 2 | 3+4 2 | 3+3 2 | 3+3 2 |
| | 5+6 2 | 5+6 2 | 3+4 2 | 5+6 2 | 3+3 2 | 2+2 2 |
| | 3+4 2 | 3+4 2 | 5+6 2 | 3+4 2 | 2+2 2 | 3+3 2 |
| | 3+4 2 | 3+4 2 | 5+6 2 | 5+6 2 | 2+2 2 | 3+3 2 |
| 12.1.196 | 40332 | 70 | Al1Cu1 | | | |
| | -A | -GM | -M | -Y | -L | -V |
| | 3+4 2 | 3+4 2 | 3+4 2 | 3+4 2 | 2+2 2 | 3+3 2 |
| | 5+6 2 | 3+4 2 | 5+6 2 | 5+6 2 | 3+3 2 | 2+2 2 |
| | 3+4 2 | 5+6 2 | 5+6 2 | 3+4 2 | 3+3 2 | 2+2 2 |
| | 5+6 2 | 5+6 2 | 3+4 2 | 5+6 2 | 2+2 2 | 3+3 2 |
| | 5+6 2 | 3+4 2 | 5+6 2 | 3+4 2 | 2+2 2 | 2+2 2 |
| | 5+6 2 | 5+6 2 | 5+6 2 | 3+4 2 | 3+3 2 | 3+3 2 |
| | 3+4 2 | 3+4 2 | 3+4 2 | 5+6 2 | 2+2 2 | 3+3 2 |
| | 5+6 2 | 3+4 2 | 5+6 2 | 3+4 2 | 2+2 2 | 2+2 2 |
| | 3+4 2 | 3+4 2 | 5+6 2 | 5+6 2 | 3+3 2 | 3+3 2 |
| | 5+6 2 | 3+4 2 | 3+4 2 | 3+4 2 | 2+2 2 | 3+3 2 |
| | 5+6 2 | 5+6 2 | 3+4 2 | 5+6 2 | 2+2 2 | 2+2 2 |
| | 5+6 2 | 3+4 2 | 5+6 2 | 3+4 2 | 2+2 2 | 3+3 2 |

Continued on next column

**TABLE XXIV– continued**

| TQC | ICSD | NoE | SF | | | |
|---|---|---|---|---|---|---|
| | 3+4 2 | 5+6 2 | 5+6 2 | 5+6 2 | 3+3 2 | 2+2 2 |
| | 5+6 2 | 3+4 2 | 3+4 2 | 5+6 2 | 3+3 2 | 3+3 2 |
| | 3+4 2 | 3+4 2 | 3+4 2 | 3+4 2 | 2+2 2 | 2+2 2 |
| | 5+6 2 | 3+4 2 | 5+6 2 | 3+4 2 | 3+3 2 | 3+3 2 |
| | 3+4 2 | 5+6 2 | 5+6 2 | 3+4 2 | 2+2 2 | 2+2 2 |
| | 3+4 2 | 5+6 2 | 5+6 2 | 3+4 2 | 3+3 2 | 3+3 2 |
| | 5+6 2 | 3+4 2 | 3+4 2 | 3+4 2 | 2+2 2 | 3+3 2 |
| | 3+4 2 | 5+6 2 | 5+6 2 | 5+6 2 | 3+3 2 | 2+2 2 |
| | 5+6 2 | 3+4 2 | 5+6 2 | 3+4 2 | 2+2 2 | 3+3 2 |
| | 3+4 2 | 5+6 2 | 3+4 2 | 5+6 2 | 2+2 2 | 3+3 2 |
| | 3+4 2 | 5+6 2 | 3+4 2 | 3+4 2 | 3+3 2 | 3+3 2 |
| | 5+6 2 | 5+6 2 | 5+6 2 | 5+6 2 | 3+3 2 | 2+2 2 |
| | 3+4 2 | 3+4 2 | 3+4 2 | 5+6 2 | 2+2 2 | 2+2 2 |
| | 5+6 2 | 3+4 2 | 3+4 2 | 5+6 2 | 2+2 2 | 3+3 2 |
| | 5+6 2 | 3+4 2 | 5+6 2 | 3+4 2 | 3+3 2 | 3+3 2 |
| | 5+6 2 | 3+4 2 | 5+6 2 | 3+4 2 | 2+2 2 | 3+3 2 |
| | 5+6 2 | 5+6 2 | 5+6 2 | 3+4 2 | 2+2 2 | 2+2 2 |
| | 3+4 2 | 3+4 2 | 3+4 2 | 5+6 2 | 2+2 2 | 3+3 2 |
| | 5+6 2 | 5+6 2 | 5+6 2 | 5+6 2 | 3+3 2 | 2+2 2 |
| | 5+6 2 | 3+4 2 | 3+4 2 | 5+6 2 | 3+3 2 | 2+2 2 |
| | 3+4 2 | 3+4 2 | 5+6 2 | 3+4 2 | 3+3 2 | 3+3 2 |
| | 3+4 2 | 5+6 2 | 5+6 2 | 3+4 2 | 3+3 2 | 2+2 2 |
| | 5+6 2 | 5+6 2 | 3+4 2 | 3+4 2 | 2+2 2 | 3+3 2 |
| 12.1.197 | 165239 | 68 | Al14Ca13 | | | |
| | -A | -GM | -M | -Y | -L | -V |
| | 3+4 2 | 3+4 2 | 3+4 2 | 3+4 2 | 3+3 2 | 2+2 2 |
| | 5+6 2 | 3+4 2 | 5+6 2 | 5+6 2 | 2+2 2 | 3+3 2 |
| | 5+6 2 | 5+6 2 | 5+6 2 | 3+4 2 | 3+3 2 | 3+3 2 |
| | 3+4 2 | 3+4 2 | 3+4 2 | 3+4 2 | 3+3 2 | 2+2 2 |
| | 5+6 2 | 5+6 2 | 5+6 2 | 5+6 2 | 2+2 2 | 2+2 2 |
| | 3+4 2 | 5+6 2 | 3+4 2 | 5+6 2 | 2+2 2 | 3+3 2 |
| | 5+6 2 | 3+4 2 | 5+6 2 | 3+4 2 | 2+2 2 | 3+3 2 |
| | 3+4 2 | 3+4 2 | 3+4 2 | 5+6 2 | 3+3 2 | 2+2 2 |
| | 5+6 2 | 5+6 2 | 5+6 2 | 5+6 2 | 3+3 2 | 3+3 2 |
| | 3+4 2 | 5+6 2 | 3+4 2 | 3+4 2 | 2+2 2 | 2+2 2 |
| | 3+4 2 | 5+6 2 | 3+4 2 | 5+6 2 | 2+2 2 | 2+2 2 |
| | 5+6 2 | 3+4 2 | 5+6 2 | 3+4 2 | 2+2 2 | 3+3 2 |
| | 5+6 2 | 5+6 2 | 3+4 2 | 5+6 2 | 3+3 2 | 3+3 2 |
| | 3+4 2 | 3+4 2 | 5+6 2 | 3+4 2 | 3+3 2 | 2+2 2 |
| | 5+6 2 | 5+6 2 | 5+6 2 | 5+6 2 | 2+2 2 | 3+3 2 |
| | 3+4 2 | 3+4 2 | 3+4 2 | 3+4 2 | 3+3 2 | 2+2 2 |
| | 5+6 2 | 3+4 2 | 5+6 2 | 5+6 2 | 3+3 2 | 3+3 2 |
| | 3+4 2 | 5+6 2 | 3+4 2 | 3+4 2 | 2+2 2 | 3+3 2 |
| | 5+6 2 | 3+4 2 | 5+6 2 | 5+6 2 | 3+3 2 | 2+2 2 |
| | 3+4 2 | 3+4 2 | 5+6 2 | 3+4 2 | 2+2 2 | 2+2 2 |
| | 5+6 2 | 5+6 2 | 3+4 2 | 3+4 2 | 2+2 2 | 3+3 2 |
| | 3+4 2 | 5+6 2 | 5+6 2 | 5+6 2 | 3+3 2 | 2+2 2 |
| | 3+4 2 | 3+4 2 | 3+4 2 | 3+4 2 | 2+2 2 | 3+3 2 |
| | 5+6 2 | 5+6 2 | 3+4 2 | 5+6 2 | 3+3 2 | 3+3 2 |
| | 3+4 2 | 3+4 2 | 5+6 2 | 5+6 2 | 3+3 2 | 2+2 2 |
| | 5+6 2 | 5+6 2 | 3+4 2 | 3+4 2 | 2+2 2 | 2+2 2 |
| | 3+4 2 | 3+4 2 | 5+6 2 | 3+4 2 | 3+3 2 | 3+3 2 |
| | 5+6 2 | 3+4 2 | 5+6 2 | 5+6 2 | 3+3 2 | 3+3 2 |
| | 3+4 2 | 3+4 2 | 3+4 2 | 3+4 2 | 2+2 2 | 2+2 2 |
| | 3+4 2 | 3+4 2 | 5+6 2 | 5+6 2 | 2+2 2 | 2+2 2 |
| | 5+6 2 | 5+6 2 | 3+4 2 | 3+4 2 | 3+3 2 | 3+3 2 |
| | 5+6 2 | 5+6 2 | 5+6 2 | 5+6 2 | 2+2 2 | 2+2 2 |
| | 3+4 2 | 5+6 2 | 5+6 2 | 3+4 2 | 2+2 2 | 3+3 2 |
| | 5+6 2 | 3+4 2 | 3+4 2 | 5+6 2 | 3+3 2 | 2+2 2 |
| 12.1.198 | 429000 | 116 | Ag1Ba1S3Tb1 | | | |
| | -A | -GM | -M | -Y | -L | -V |

Continued on next column

**TABLE XXIV– continued**

| TQC | ICSD | NoE | SF | | | |
|---|---|---|---|---|---|---|
| | 3+4 2 | 3+4 2 | 5+6 2 | 5+6 2 | 2+2 2 | 2+2 2 |
| | 5+6 2 | 5+6 2 | 3+4 2 | 3+4 2 | 3+3 2 | 3+3 2 |
| | 5+6 2 | 3+4 2 | 3+4 2 | 3+4 2 | 2+2 2 | 3+3 2 |
| | 3+4 2 | 5+6 2 | 5+6 2 | 5+6 2 | 3+3 2 | 2+2 2 |
| | 3+4 2 | 3+4 2 | 5+6 2 | 5+6 2 | 3+3 2 | 3+3 2 |
| | 5+6 2 | 5+6 2 | 3+4 2 | 3+4 2 | 2+2 2 | 2+2 2 |
| | 3+4 2 | 3+4 2 | 5+6 2 | 5+6 2 | 3+3 2 | 3+3 2 |
| | 5+6 2 | 5+6 2 | 3+4 2 | 3+4 2 | 2+2 2 | 2+2 2 |
| | 3+4 2 | 3+4 2 | 5+6 2 | 5+6 2 | 3+3 2 | 3+3 2 |
| | 5+6 2 | 5+6 2 | 3+4 2 | 3+4 2 | 2+2 2 | 2+2 2 |
| | 5+6 2 | 3+4 2 | 5+6 2 | 3+4 2 | 3+3 2 | 2+2 2 |
| | 3+4 2 | 5+6 2 | 3+4 2 | 5+6 2 | 2+2 2 | 3+3 2 |
| | 5+6 2 | 3+4 2 | 5+6 2 | 3+4 2 | 3+3 2 | 2+2 2 |
| | 3+4 2 | 3+4 2 | 3+4 2 | 5+6 2 | 2+2 2 | 3+3 2 |
| | 3+4 2 | 5+6 2 | 3+4 2 | 3+4 2 | 3+3 2 | 3+3 2 |
| | 5+6 2 | 3+4 2 | 5+6 2 | 5+6 2 | 3+3 2 | 2+2 2 |
| | 5+6 2 | 5+6 2 | 3+4 2 | 3+4 2 | 2+2 2 | 3+3 2 |
| | 3+4 2 | 3+4 2 | 5+6 2 | 3+4 2 | 3+3 2 | 2+2 2 |
| | 3+4 2 | 5+6 2 | 3+4 2 | 5+6 2 | 2+2 2 | 3+3 2 |
| | 5+6 2 | 3+4 2 | 5+6 2 | 3+4 2 | 3+3 2 | 2+2 2 |
| | 3+4 2 | 5+6 2 | 3+4 2 | 5+6 2 | 2+2 2 | 3+3 2 |
| | 5+6 2 | 3+4 2 | 5+6 2 | 3+4 2 | 3+3 2 | 2+2 2 |
| | 5+6 2 | 5+6 2 | 5+6 2 | 5+6 2 | 2+2 2 | 2+2 2 |
| | 3+4 2 | 5+6 2 | 3+4 2 | 5+6 2 | 2+2 2 | 2+2 2 |
| | 5+6 2 | 5+6 2 | 5+6 2 | 5+6 2 | 2+2 2 | 3+3 2 |
| | 5+6 2 | 5+6 2 | 5+6 2 | 5+6 2 | 3+3 2 | 2+2 2 |
| | 3+4 2 | 3+4 2 | 3+4 2 | 3+4 2 | 2+2 2 | 3+3 2 |
| | 5+6 2 | 3+4 2 | 5+6 2 | 3+4 2 | 2+2 2 | 2+2 2 |
| | 3+4 2 | 5+6 2 | 3+4 2 | 5+6 2 | 3+3 2 | 3+3 2 |
| | 3+4 2 | 3+4 2 | 3+4 2 | 3+4 2 | 3+3 2 | 3+3 2 |
| | 5+6 2 | 3+4 2 | 5+6 2 | 5+6 2 | 2+2 2 | 3+3 2 |
| | 3+4 2 | 5+6 2 | 3+4 2 | 3+4 2 | 3+3 2 | 2+2 2 |
| | 3+4 2 | 5+6 2 | 3+4 2 | 3+4 2 | 3+3 2 | 3+3 2 |
| | 5+6 2 | 3+4 2 | 5+6 2 | 5+6 2 | 2+2 2 | 2+2 2 |
| | 3+4 2 | 3+4 2 | 3+4 2 | 3+4 2 | 2+2 2 | 2+2 2 |
| | 5+6 2 | 5+6 2 | 5+6 2 | 5+6 2 | 3+3 2 | 3+3 2 |
| | 3+4 2 | 3+4 2 | 5+6 2 | 5+6 2 | 3+3 2 | 2+2 2 |
| | 5+6 2 | 3+4 2 | 3+4 2 | 3+4 2 | 2+2 2 | 3+3 2 |
| | 3+4 2 | 5+6 2 | 3+4 2 | 3+4 2 | 2+2 2 | 3+3 2 |
| | 5+6 2 | 3+4 2 | 5+6 2 | 5+6 2 | 2+2 2 | 2+2 2 |
| | 5+6 2 | 3+4 2 | 3+4 2 | 5+6 2 | 3+3 2 | 3+3 2 |
| | 3+4 2 | 3+4 2 | 5+6 2 | 3+4 2 | 3+3 2 | 2+2 2 |
| | 3+4 2 | 5+6 2 | 5+6 2 | 5+6 2 | 2+2 2 | 2+2 2 |
| | 5+6 2 | 5+6 2 | 5+6 2 | 5+6 2 | 3+3 2 | 2+2 2 |
| | 5+6 2 | 5+6 2 | 3+4 2 | 5+6 2 | 2+2 2 | 3+3 2 |
| | 3+4 2 | 5+6 2 | 3+4 2 | 3+4 2 | 3+3 2 | 3+3 2 |
| | 3+4 2 | 5+6 2 | 3+4 2 | 3+4 2 | 2+2 2 | 2+2 2 |
| | 5+6 2 | 5+6 2 | 3+4 2 | 5+6 2 | 2+2 2 | 3+3 2 |
| | 3+4 2 | 3+4 2 | 5+6 2 | 5+6 2 | 2+2 2 | 2+2 2 |
| | 5+6 2 | 3+4 2 | 5+6 2 | 3+4 2 | 2+2 2 | 3+3 2 |
| | 3+4 2 | 5+6 2 | 3+4 2 | 5+6 2 | 3+3 2 | 2+2 2 |
| | 3+4 2 | 5+6 2 | 5+6 2 | 3+4 2 | 3+3 2 | 3+3 2 |
| | 5+6 2 | 3+4 2 | 5+6 2 | 5+6 2 | 2+2 2 | 3+3 2 |
| | 3+4 2 | 5+6 2 | 3+4 2 | 5+6 2 | 3+3 2 | 2+2 2 |
| | 5+6 2 | 5+6 2 | 3+4 2 | 3+4 2 | 2+2 2 | 2+2 2 |
| | 5+6 2 | 5+6 2 | 5+6 2 | 5+6 2 | 2+2 2 | 3+3 2 |
| | 3+4 2 | 3+4 2 | 3+4 2 | 3+4 2 | 3+3 2 | 3+3 2 |
| | 3+4 2 | 3+4 2 | 5+6 2 | 5+6 2 | 2+2 2 | 2+2 2 |
| 12.2.1 | 187496 | 56 | Sb2Te3 | | | |
| | -A | -GM | -M | -Y | -L | -V |
| | 3+4 2 | 3+4 2 | 3+4 2 | 3+4 2 | 3+3 2 | 3+3 2 |

Continued on next column

**TABLE XXIV– continued**

| TQC | ICSD | NoE | SF | | | |
|---|---|---|---|---|---|---|
| | 5+6 2 | 5+6 2 | 5+6 2 | 5+6 2 | 2+2 2 | 2+2 2 |
| | 3+4 2 | 3+4 2 | 3+4 2 | 3+4 2 | 2+2 2 | 2+2 2 |
| | 5+6 2 | 5+6 2 | 5+6 2 | 5+6 2 | 3+3 2 | 3+3 2 |
| | 5+6 2 | 5+6 2 | 5+6 2 | 5+6 2 | 2+2 2 | 2+2 2 |
| | 3+4 2 | 3+4 2 | 3+4 2 | 3+4 2 | 3+3 2 | 3+3 2 |
| | 5+6 2 | 5+6 2 | 5+6 2 | 3+4 2 | 3+3 2 | 2+2 2 |
| | 3+4 2 | 3+4 2 | 3+4 2 | 5+6 2 | 2+2 2 | 3+3 2 |
| | 3+4 2 | 3+4 2 | 3+4 2 | 3+4 2 | 3+3 2 | 3+3 2 |
| | 5+6 2 | 5+6 2 | 5+6 2 | 5+6 2 | 2+2 2 | 2+2 2 |
| | 5+6 2 | 3+4 2 | 3+4 2 | 3+4 2 | 3+3 2 | 3+3 2 |
| | 3+4 2 | 5+6 2 | 5+6 2 | 5+6 2 | 2+2 2 | 2+2 2 |
| | 5+6 2 | 3+4 2 | 5+6 2 | 5+6 2 | 3+3 2 | 3+3 2 |
| | 3+4 2 | 5+6 2 | 3+4 2 | 3+4 2 | 2+2 2 | 2+2 2 |
| | 3+4 2 | 3+4 2 | 3+4 2 | 5+6 2 | 2+2 2 | 2+2 2 |
| | 5+6 2 | 5+6 2 | 5+6 2 | 3+4 2 | 3+3 2 | 3+3 2 |
| | 5+6 2 | 3+4 2 | 3+4 2 | 3+4 2 | 3+3 2 | 2+2 2 |
| | 3+4 2 | 5+6 2 | 5+6 2 | 5+6 2 | 2+2 2 | 2+2 2 |
| | 5+6 2 | 3+4 2 | 5+6 2 | 5+6 2 | 2+2 2 | 3+3 2 |
| | 5+6 2 | 5+6 2 | 3+4 2 | 3+4 2 | 3+3 2 | 2+2 2 |
| | 3+4 2 | 3+4 2 | 5+6 2 | 3+4 2 | 2+2 2 | 3+3 2 |
| | 3+4 2 | 3+4 2 | 3+4 2 | 3+4 2 | 3+3 2 | 2+2 2 |
| | 3+4 2 | 3+4 2 | 5+6 2 | 5+6 2 | 2+2 2 | 3+3 2 |
| | 5+6 2 | 3+4 2 | 3+4 2 | 3+4 2 | 3+3 2 | 3+3 2 |
| | 3+4 2 | 5+6 2 | 3+4 2 | 3+4 2 | 2+2 2 | 2+2 2 |
| | 5+6 2 | 5+6 2 | 3+4 2 | 5+6 2 | 2+2 2 | 3+3 2 |
| | 3+4 2 | 5+6 2 | 5+6 2 | 5+6 2 | 3+3 2 | 2+2 2 |
| | 5+6 2 | 5+6 2 | 3+4 2 | 5+6 2 | 3+3 2 | 3+3 2 |
| 12.2.2 | 420133 | 44 | Ru1Sc2Si2 | | | |
| | -A | -GM | -M | -Y | -L | -V |
| | 5+6 2 | 3+4 2 | 5+6 2 | 3+4 2 | 3+3 2 | 2+2 2 |
| | 3+4 2 | 3+4 2 | 3+4 2 | 3+4 2 | 2+2 2 | 2+2 2 |
| | 3+4 2 | 5+6 2 | 3+4 2 | 5+6 2 | 2+2 2 | 3+3 2 |
| | 5+6 2 | 5+6 2 | 5+6 2 | 5+6 2 | 3+3 2 | 3+3 2 |
| | 5+6 2 | 3+4 2 | 5+6 2 | 3+4 2 | 2+2 2 | 3+3 2 |
| | 3+4 2 | 5+6 2 | 3+4 2 | 5+6 2 | 3+3 2 | 2+2 2 |
| | 3+4 2 | 5+6 2 | 5+6 2 | 5+6 2 | 3+3 2 | 3+3 2 |
| | 5+6 2 | 3+4 2 | 3+4 2 | 3+4 2 | 2+2 2 | 2+2 2 |
| | 3+4 2 | 5+6 2 | 5+6 2 | 3+4 2 | 2+2 2 | 2+2 2 |
| | 5+6 2 | 3+4 2 | 5+6 2 | 5+6 2 | 3+3 2 | 3+3 2 |
| | 5+6 2 | 3+4 2 | 3+4 2 | 3+4 2 | 2+2 2 | 3+3 2 |
| | 5+6 2 | 5+6 2 | 3+4 2 | 5+6 2 | 3+3 2 | 2+2 2 |
| | 3+4 2 | 3+4 2 | 3+4 2 | 5+6 2 | 2+2 2 | 3+3 2 |
| | 5+6 2 | 5+6 2 | 5+6 2 | 3+4 2 | 3+3 2 | 2+2 2 |
| | 3+4 2 | 3+4 2 | 3+4 2 | 5+6 2 | 3+3 2 | 2+2 2 |
| | 3+4 2 | 5+6 2 | 5+6 2 | 3+4 2 | 2+2 2 | 3+3 2 |
| | 5+6 2 | 5+6 2 | 5+6 2 | 5+6 2 | 3+3 2 | 2+2 2 |
| | 3+4 2 | 3+4 2 | 3+4 2 | 3+4 2 | 3+3 2 | 2+2 2 |
| | 5+6 2 | 3+4 2 | 3+4 2 | 5+6 2 | 2+2 2 | 3+3 2 |
| | 5+6 2 | 3+4 2 | 3+4 2 | 3+4 2 | 2+2 2 | 2+2 2 |
| | 3+4 2 | 3+4 2 | 5+6 2 | 5+6 2 | 2+2 2 | 3+3 2 |
| | 5+6 2 | 5+6 2 | 3+4 2 | 3+4 2 | 3+3 2 | 2+2 2 |
| 12.2.3 | 61353 | 124 | Pt1Se7Ta2 | | | |
| | -A | -GM | -M | -Y | -L | -V |
| | 5+6 2 | 3+4 2 | 3+4 2 | 5+6 2 | 3+3 2 | 2+2 2 |
| | 3+4 2 | 5+6 2 | 5+6 2 | 3+4 2 | 2+2 2 | 3+3 2 |
| | 3+4 2 | 3+4 2 | 5+6 2 | 5+6 2 | 3+3 2 | 3+3 2 |
| | 5+6 2 | 5+6 2 | 3+4 2 | 3+4 2 | 2+2 2 | 2+2 2 |
| | 3+4 2 | 3+4 2 | 5+6 2 | 5+6 2 | 2+2 2 | 2+2 2 |
| | 5+6 2 | 5+6 2 | 3+4 2 | 3+4 2 | 3+3 2 | 3+3 2 |
| | 3+4 2 | 3+4 2 | 5+6 2 | 5+6 2 | 3+3 2 | 3+3 2 |
| | 5+6 2 | 5+6 2 | 3+4 2 | 3+4 2 | 2+2 2 | 2+2 2 |

Continued on next column

**TABLE XXIV– continued**

| TQC | ICSD | | NoE | SF | | |
|---|---|---|---|---|---|---|
| | 5+6 2 | 3+4 2 | 3+4 2 | 5+6 2 | 2+2 2 | 3+3 2 |
| | 3+4 2 | 3+4 2 | 5+6 2 | 5+6 2 | 3+3 2 | 2+2 2 |
| | 5+6 2 | 5+6 2 | 3+4 2 | 3+4 2 | 2+2 2 | 2+2 2 |
| | 3+4 2 | 5+6 2 | 5+6 2 | 3+4 2 | 3+3 2 | 3+3 2 |
| | 5+6 2 | 3+4 2 | 5+6 2 | 3+4 2 | 3+3 2 | 2+2 2 |
| | 3+4 2 | 5+6 2 | 3+4 2 | 5+6 2 | 2+2 2 | 3+3 2 |
| | 3+4 2 | 3+4 2 | 3+4 2 | 3+4 2 | 2+2 2 | 2+2 2 |
| | 5+6 2 | 5+6 2 | 5+6 2 | 5+6 2 | 3+3 2 | 3+3 2 |
| | 3+4 2 | 5+6 2 | 3+4 2 | 3+4 2 | 3+3 2 | 3+3 2 |
| | 5+6 2 | 3+4 2 | 5+6 2 | 5+6 2 | 2+2 2 | 2+2 2 |
| | 5+6 2 | 3+4 2 | 3+4 2 | 3+4 2 | 2+2 2 | 2+2 2 |
| | 3+4 2 | 5+6 2 | 5+6 2 | 5+6 2 | 3+3 2 | 3+3 2 |
| | 5+6 2 | 3+4 2 | 3+4 2 | 3+4 2 | 3+3 2 | 2+2 2 |
| | 3+4 2 | 5+6 2 | 5+6 2 | 5+6 2 | 2+2 2 | 3+3 2 |
| | 3+4 2 | 5+6 2 | 5+6 2 | 3+4 2 | 3+3 2 | 2+2 2 |
| | 5+6 2 | 3+4 2 | 5+6 2 | 5+6 2 | 2+2 2 | 3+3 2 |
| | 3+4 2 | 3+4 2 | 3+4 2 | 3+4 2 | 2+2 2 | 2+2 2 |
| | 5+6 2 | 5+6 2 | 3+4 2 | 5+6 2 | 3+3 2 | 3+3 2 |
| | 3+4 2 | 3+4 2 | 3+4 2 | 5+6 2 | 3+3 2 | 3+3 2 |
| | 5+6 2 | 3+4 2 | 5+6 2 | 3+4 2 | 2+2 2 | 2+2 2 |
| | 5+6 2 | 5+6 2 | 5+6 2 | 5+6 2 | 3+3 2 | 2+2 2 |
| | 3+4 2 | 3+4 2 | 3+4 2 | 5+6 2 | 2+2 2 | 3+3 2 |
| | 5+6 2 | 5+6 2 | 3+4 2 | 3+4 2 | 3+3 2 | 3+3 2 |
| | 3+4 2 | 5+6 2 | 5+6 2 | 3+4 2 | 2+2 2 | 2+2 2 |
| | 3+4 2 | 3+4 2 | 3+4 2 | 5+6 2 | 2+2 2 | 3+3 2 |
| | 5+6 2 | 3+4 2 | 5+6 2 | 3+4 2 | 3+3 2 | 2+2 2 |
| | 3+4 2 | 5+6 2 | 5+6 2 | 5+6 2 | 3+3 2 | 2+2 2 |
| | 5+6 2 | 3+4 2 | 3+4 2 | 5+6 2 | 2+2 2 | 3+3 2 |
| | 3+4 2 | 5+6 2 | 3+4 2 | 5+6 2 | 3+3 2 | 2+2 2 |
| | 5+6 2 | 5+6 2 | 3+4 2 | 3+4 2 | 2+2 2 | 3+3 2 |
| | 3+4 2 | 3+4 2 | 5+6 2 | 3+4 2 | 2+2 2 | 2+2 2 |
| | 5+6 2 | 5+6 2 | 5+6 2 | 5+6 2 | 3+3 2 | 3+3 2 |
| | 5+6 2 | 3+4 2 | 3+4 2 | 3+4 2 | 3+3 2 | 2+2 2 |
| | 3+4 2 | 3+4 2 | 5+6 2 | 5+6 2 | 2+2 2 | 3+3 2 |
| | 5+6 2 | 5+6 2 | 5+6 2 | 3+4 2 | 3+3 2 | 2+2 2 |
| | 3+4 2 | 5+6 2 | 3+4 2 | 3+4 2 | 2+2 2 | 2+2 2 |
| | 5+6 2 | 3+4 2 | 5+6 2 | 5+6 2 | 2+2 2 | 3+3 2 |
| | 3+4 2 | 3+4 2 | 3+4 2 | 3+4 2 | 3+3 2 | 2+2 2 |
| | 5+6 2 | 3+4 2 | 5+6 2 | 5+6 2 | 2+2 2 | 3+3 2 |
| | 3+4 2 | 3+4 2 | 3+4 2 | 5+6 2 | 3+3 2 | 3+3 2 |
| | 5+6 2 | 5+6 2 | 3+4 2 | 5+6 2 | 2+2 2 | 2+2 2 |
| | 3+4 2 | 3+4 2 | 5+6 2 | 3+4 2 | 3+3 2 | 3+3 2 |
| | 5+6 2 | 3+4 2 | 5+6 2 | 5+6 2 | 2+2 2 | 2+2 2 |
| | 3+4 2 | 5+6 2 | 3+4 2 | 5+6 2 | 3+3 2 | 3+3 2 |
| | 3+4 2 | 5+6 2 | 5+6 2 | 3+4 2 | 3+3 2 | 2+2 2 |
| | 5+6 2 | 5+6 2 | 5+6 2 | 3+4 2 | 2+2 2 | 3+3 2 |
| | 3+4 2 | 3+4 2 | 3+4 2 | 5+6 2 | 3+3 2 | 2+2 2 |
| | 5+6 2 | 3+4 2 | 3+4 2 | 5+6 2 | 2+2 2 | 3+3 2 |
| | 5+6 2 | 5+6 2 | 5+6 2 | 3+4 2 | 3+3 2 | 2+2 2 |
| | 3+4 2 | 3+4 2 | 3+4 2 | 5+6 2 | 2+2 2 | 3+3 2 |
| | 5+6 2 | 5+6 2 | 3+4 2 | 3+4 2 | 3+3 2 | 2+2 2 |
| | 3+4 2 | 5+6 2 | 5+6 2 | 3+4 2 | 2+2 2 | 3+3 2 |
| | 5+6 2 | 5+6 2 | 3+4 2 | 3+4 2 | 3+3 2 | 2+2 2 |
| | 3+4 2 | 3+4 2 | 5+6 2 | 3+4 2 | 2+2 2 | 3+3 2 |
| 12.2.4 | 248500 | | 26 | Pr1 | | |
| | -A | -GM | -M | -Y | -L | -V |
| | 5+6 2 | 3+4 2 | 5+6 2 | 5+6 2 | 3+3 2 | 2+2 2 |
| | 3+4 2 | 5+6 2 | 3+4 2 | 3+4 2 | 2+2 2 | 3+3 2 |
| | 5+6 2 | 3+4 2 | 5+6 2 | 5+6 2 | 3+3 2 | 2+2 2 |
| | 3+4 2 | 5+6 2 | 3+4 2 | 3+4 2 | 2+2 2 | 3+3 2 |
| | 5+6 2 | 3+4 2 | 5+6 2 | 5+6 2 | 3+3 2 | 2+2 2 |

Continued on next column

**TABLE XXIV– continued**

| TQC | ICSD | | NoE | SF | | |
|---|---|---|---|---|---|---|
| | 3+4 2 | 3+4 2 | 3+4 2 | 3+4 2 | 2+2 2 | 3+3 2 |
| | 3+4 2 | 5+6 2 | 5+6 2 | 5+6 2 | 3+3 2 | 2+2 2 |
| | 5+6 2 | 5+6 2 | 3+4 2 | 3+4 2 | 2+2 2 | 3+3 2 |
| | 3+4 2 | 3+4 2 | 5+6 2 | 3+4 2 | 2+2 2 | 2+2 2 |
| | 5+6 2 | 5+6 2 | 3+4 2 | 3+4 2 | 3+3 2 | 3+3 2 |
| | 3+4 2 | 5+6 2 | 3+4 2 | 5+6 2 | 2+2 2 | 3+3 2 |
| | 5+6 2 | 3+4 2 | 5+6 2 | 3+4 2 | 2+2 2 | 2+2 2 |
| | 3+4 2 | 5+6 2 | 5+6 2 | 3+4 2 | 3+3 2 | 2+2 2 |
| 12.2.5 | 61005 | | 56 | Pd1Se6Ta2 | | |
| | -A | -GM | -M | -Y | -L | -V |
| | 5+6 2 | 3+4 2 | 3+4 2 | 5+6 2 | 3+3 2 | 3+3 2 |
| | 3+4 2 | 5+6 2 | 5+6 2 | 3+4 2 | 2+2 2 | 2+2 2 |
| | 3+4 2 | 3+4 2 | 5+6 2 | 5+6 2 | 3+3 2 | 3+3 2 |
| | 5+6 2 | 5+6 2 | 3+4 2 | 3+4 2 | 2+2 2 | 2+2 2 |
| | 5+6 2 | 3+4 2 | 3+4 2 | 5+6 2 | 3+3 2 | 2+2 2 |
| | 3+4 2 | 5+6 2 | 5+6 2 | 3+4 2 | 2+2 2 | 3+3 2 |
| | 3+4 2 | 3+4 2 | 3+4 2 | 3+4 2 | 3+3 2 | 3+3 2 |
| | 3+4 2 | 5+6 2 | 5+6 2 | 5+6 2 | 2+2 2 | 2+2 2 |
| | 5+6 2 | 3+4 2 | 3+4 2 | 3+4 2 | 3+3 2 | 3+3 2 |
| | 5+6 2 | 3+4 2 | 3+4 2 | 3+4 2 | 3+3 2 | 3+3 2 |
| | 3+4 2 | 5+6 2 | 5+6 2 | 5+6 2 | 2+2 2 | 2+2 2 |
| | 5+6 2 | 3+4 2 | 3+4 2 | 5+6 2 | 3+3 2 | 3+3 2 |
| | 3+4 2 | 3+4 2 | 3+4 2 | 3+4 2 | 2+2 2 | 3+3 2 |
| | 3+4 2 | 3+4 2 | 3+4 2 | 3+4 2 | 3+3 2 | 2+2 2 |
| | 5+6 2 | 3+4 2 | 5+6 2 | 3+4 2 | 3+3 2 | 2+2 2 |
| | 3+4 2 | 5+6 2 | 3+4 2 | 5+6 2 | 2+2 2 | 3+3 2 |
| | 3+4 2 | 3+4 2 | 5+6 2 | 3+4 2 | 2+2 2 | 2+2 2 |
| | 3+4 2 | 3+4 2 | 3+4 2 | 5+6 2 | 3+3 2 | 3+3 2 |
| | 5+6 2 | 5+6 2 | 5+6 2 | 5+6 2 | 3+3 2 | 3+3 2 |
| | 5+6 2 | 3+4 2 | 3+4 2 | 3+4 2 | 2+2 2 | 2+2 2 |
| | 3+4 2 | 3+4 2 | 3+4 2 | 5+6 2 | 3+3 2 | 3+3 2 |
| | 3+4 2 | 5+6 2 | 5+6 2 | 3+4 2 | 3+3 2 | 2+2 2 |
| | 5+6 2 | 3+4 2 | 5+6 2 | 5+6 2 | 3+3 2 | 3+3 2 |
| | 5+6 2 | 5+6 2 | 3+4 2 | 3+4 2 | 2+2 2 | 3+3 2 |
| | 3+4 2 | 5+6 2 | 5+6 2 | 3+4 2 | 2+2 2 | 2+2 2 |
| | 3+4 2 | 5+6 2 | 3+4 2 | 5+6 2 | 3+3 2 | 3+3 2 |
| | 5+6 2 | 3+4 2 | 3+4 2 | 3+4 2 | 2+2 2 | 2+2 2 |
| | 5+6 2 | 3+4 2 | 3+4 2 | 3+4 2 | 3+3 2 | 3+3 2 |
| 12.2.6 | 40281 | | 96 | P4Zr7 | | |
| | -A | -GM | -M | -Y | -L | -V |
| | 5+6 2 | 3+4 2 | 5+6 2 | 3+4 2 | 3+3 2 | 2+2 2 |
| | 3+4 2 | 3+4 2 | 3+4 2 | 5+6 2 | 3+3 2 | 3+3 2 |
| | 3+4 2 | 5+6 2 | 5+6 2 | 3+4 2 | 2+2 2 | 2+2 2 |
| | 5+6 2 | 5+6 2 | 3+4 2 | 5+6 2 | 3+3 2 | 3+3 2 |
| | 3+4 2 | 5+6 2 | 3+4 2 | 3+4 2 | 2+2 2 | 2+2 2 |
| | 5+6 2 | 5+6 2 | 5+6 2 | 5+6 2 | 3+3 2 | 3+3 2 |
| | 5+6 2 | 3+4 2 | 5+6 2 | 5+6 2 | 2+2 2 | 3+3 2 |
| | 3+4 2 | 3+4 2 | 3+4 2 | 3+4 2 | 2+2 2 | 2+2 2 |
| | 3+4 2 | 3+4 2 | 3+4 2 | 3+4 2 | 2+2 2 | 3+3 2 |
| | 5+6 2 | 3+4 2 | 5+6 2 | 3+4 2 | 3+3 2 | 2+2 2 |
| | 5+6 2 | 3+4 2 | 5+6 2 | 3+4 2 | 2+2 2 | 2+2 2 |
| | 3+4 2 | 5+6 2 | 3+4 2 | 5+6 2 | 3+3 2 | 3+3 2 |
| | 3+4 2 | 5+6 2 | 5+6 2 | 5+6 2 | 3+3 2 | 2+2 2 |
| | 5+6 2 | 3+4 2 | 3+4 2 | 3+4 2 | 3+3 2 | 3+3 2 |
| | 3+4 2 | 3+4 2 | 3+4 2 | 3+4 2 | 2+2 2 | 2+2 2 |
| | 5+6 2 | 5+6 2 | 5+6 2 | 5+6 2 | 3+3 2 | 3+3 2 |
| | 5+6 2 | 5+6 2 | 5+6 2 | 5+6 2 | 3+3 2 | 2+2 2 |
| | 3+4 2 | 3+4 2 | 3+4 2 | 5+6 2 | 2+2 2 | 3+3 2 |
| | 5+6 2 | 5+6 2 | 3+4 2 | 3+4 2 | 3+3 2 | 3+3 2 |
| | 3+4 2 | 5+6 2 | 3+4 2 | 5+6 2 | 3+3 2 | 2+2 2 |
| | 5+6 2 | 5+6 2 | 5+6 2 | 5+6 2 | 2+2 2 | 2+2 2 |

Continued on next column

**TABLE XXIV– continued**

| TQC | ICSD | NoE | SF | | | |
|---|---|---|---|---|---|---|
| | 3+4 2 | 5+6 2 | 5+6 2 | 3+4 2 | 2+2 2 | 3+3 2 |
| | 3+4 2 | 3+4 2 | 5+6 2 | 3+4 2 | 3+3 2 | 2+2 2 |
| | 5+6 2 | 3+4 2 | 3+4 2 | 5+6 2 | 2+2 2 | 3+3 2 |
| | 5+6 2 | 3+4 2 | 5+6 2 | 3+4 2 | 2+2 2 | 2+2 2 |
| | 3+4 2 | 5+6 2 | 3+4 2 | 5+6 2 | 3+3 2 | 3+3 2 |
| | 5+6 2 | 3+4 2 | 3+4 2 | 5+6 2 | 3+3 2 | 2+2 2 |
| | 5+6 2 | 5+6 2 | 5+6 2 | 3+4 2 | 2+2 2 | 3+3 2 |
| | 3+4 2 | 3+4 2 | 5+6 2 | 3+4 2 | 2+2 2 | 3+3 2 |
| | 3+4 2 | 3+4 2 | 3+4 2 | 3+4 2 | 3+3 2 | 2+2 2 |
| | 3+4 2 | 3+4 2 | 3+4 2 | 3+4 2 | 2+2 2 | 3+3 2 |
| | 5+6 2 | 3+4 2 | 5+6 2 | 3+4 2 | 2+2 2 | 2+2 2 |
| | 3+4 2 | 5+6 2 | 5+6 2 | 3+4 2 | 3+3 2 | 3+3 2 |
| | 3+4 2 | 5+6 2 | 3+4 2 | 5+6 2 | 3+3 2 | 2+2 2 |
| | 5+6 2 | 5+6 2 | 3+4 2 | 5+6 2 | 3+3 2 | 3+3 2 |
| | 5+6 2 | 3+4 2 | 3+4 2 | 3+4 2 | 2+2 2 | 2+2 2 |
| | 5+6 2 | 5+6 2 | 5+6 2 | 5+6 2 | 2+2 2 | 2+2 2 |
| | 5+6 2 | 3+4 2 | 5+6 2 | 3+4 2 | 3+3 2 | 3+3 2 |
| | 3+4 2 | 5+6 2 | 3+4 2 | 5+6 2 | 3+3 2 | 2+2 2 |
| | 3+4 2 | 5+6 2 | 5+6 2 | 5+6 2 | 2+2 2 | 3+3 2 |
| | 3+4 2 | 3+4 2 | 3+4 2 | 3+4 2 | 3+3 2 | 2+2 2 |
| | 5+6 2 | 3+4 2 | 5+6 2 | 5+6 2 | 3+3 2 | 3+3 2 |
| | 3+4 2 | 5+6 2 | 3+4 2 | 5+6 2 | 2+2 2 | 3+3 2 |
| | 5+6 2 | 3+4 2 | 5+6 2 | 3+4 2 | 2+2 2 | 2+2 2 |
| | 3+4 2 | 3+4 2 | 3+4 2 | 5+6 2 | 3+3 2 | 2+2 2 |
| | 5+6 2 | 5+6 2 | 5+6 2 | 3+4 2 | 2+2 2 | 3+3 2 |
| | 5+6 2 | 3+4 2 | 5+6 2 | 3+4 2 | 2+2 2 | 2+2 2 |
| | 5+6 2 | 3+4 2 | 3+4 2 | 3+4 2 | 3+3 2 | 2+2 2 |
| 12.2.7 | 41016 | 58 | P3Tc2 | | | |
| | -A | -GM | -M | -Y | -L | -V |
| | 5+6 2 | 3+4 2 | 5+6 2 | 3+4 2 | 2+2 2 | 3+3 2 |
| | 3+4 2 | 3+4 2 | 3+4 2 | 3+4 2 | 3+3 2 | 3+3 2 |
| | 5+6 2 | 5+6 2 | 3+4 2 | 5+6 2 | 3+3 2 | 2+2 2 |
| | 3+4 2 | 5+6 2 | 5+6 2 | 5+6 2 | 2+2 2 | 3+3 2 |
| | 3+4 2 | 3+4 2 | 3+4 2 | 3+4 2 | 3+3 2 | 2+2 2 |
| | 5+6 2 | 5+6 2 | 5+6 2 | 5+6 2 | 2+2 2 | 2+2 2 |
| | 3+4 2 | 3+4 2 | 5+6 2 | 3+4 2 | 3+3 2 | 2+2 2 |
| | 5+6 2 | 5+6 2 | 3+4 2 | 3+4 2 | 2+2 2 | 2+2 2 |
| | 5+6 2 | 5+6 2 | 3+4 2 | 5+6 2 | 3+3 2 | 3+3 2 |
| | 3+4 2 | 3+4 2 | 5+6 2 | 3+4 2 | 2+2 2 | 2+2 2 |
| | 3+4 2 | 3+4 2 | 5+6 2 | 5+6 2 | 2+2 2 | 3+3 2 |
| | 3+4 2 | 5+6 2 | 3+4 2 | 5+6 2 | 3+3 2 | 3+3 2 |
| | 3+4 2 | 5+6 2 | 5+6 2 | 3+4 2 | 2+2 2 | 2+2 2 |
| | 5+6 2 | 5+6 2 | 3+4 2 | 3+4 2 | 3+3 2 | 3+3 2 |
| | 5+6 2 | 3+4 2 | 5+6 2 | 5+6 2 | 2+2 2 | 3+3 2 |
| | 5+6 2 | 3+4 2 | 3+4 2 | 3+4 2 | 3+3 2 | 2+2 2 |
| | 5+6 2 | 5+6 2 | 5+6 2 | 5+6 2 | 2+2 2 | 3+3 2 |
| | 3+4 2 | 5+6 2 | 3+4 2 | 5+6 2 | 3+3 2 | 2+2 2 |
| | 5+6 2 | 3+4 2 | 5+6 2 | 3+4 2 | 2+2 2 | 3+3 2 |
| | 5+6 2 | 3+4 2 | 3+4 2 | 5+6 2 | 3+3 2 | 3+3 2 |
| | 3+4 2 | 3+4 2 | 3+4 2 | 3+4 2 | 2+2 2 | 2+2 2 |
| | 3+4 2 | 3+4 2 | 3+4 2 | 5+6 2 | 3+3 2 | 2+2 2 |
| | 5+6 2 | 5+6 2 | 5+6 2 | 5+6 2 | 2+2 2 | 3+3 2 |
| | 3+4 2 | 3+4 2 | 3+4 2 | 5+6 2 | 3+3 2 | 3+3 2 |
| | 5+6 2 | 3+4 2 | 5+6 2 | 3+4 2 | 2+2 2 | 2+2 2 |
| | 3+4 2 | 3+4 2 | 5+6 2 | 3+4 2 | 3+3 2 | 3+3 2 |
| | 5+6 2 | 3+4 2 | 5+6 2 | 5+6 2 | 2+2 2 | 3+3 2 |
| | 3+4 2 | 5+6 2 | 3+4 2 | 3+4 2 | 3+3 2 | 2+2 2 |
| | 3+4 2 | 3+4 2 | 3+4 2 | 5+6 2 | 3+3 2 | 2+2 2 |
| 12.2.8 | 74780 | 98 | O7Si2Yb2 | | | |
| | -A | -GM | -M | -Y | -L | -V |
| | 3+4 2 | 3+4 2 | 5+6 2 | 5+6 2 | 2+2 2 | 2+2 2 |
| | 5+6 2 | 5+6 2 | 3+4 2 | 3+4 2 | 3+3 2 | 3+3 2 |
| | 3+4 2 | 3+4 2 | 5+6 2 | 5+6 2 | 2+2 2 | 2+2 2 |
| | 5+6 2 | 5+6 2 | 3+4 2 | 3+4 2 | 3+3 2 | 3+3 2 |
| | 3+4 2 | 3+4 2 | 5+6 2 | 3+4 2 | 2+2 2 | 2+2 2 |
| | 5+6 2 | 5+6 2 | 3+4 2 | 5+6 2 | 3+3 2 | 3+3 2 |
| | 3+4 2 | 3+4 2 | 5+6 2 | 5+6 2 | 2+2 2 | 2+2 2 |
| | 5+6 2 | 5+6 2 | 3+4 2 | 3+4 2 | 3+3 2 | 3+3 2 |
| | 5+6 2 | 3+4 2 | 5+6 2 | 3+4 2 | 2+2 2 | 3+3 2 |
| | 3+4 2 | 5+6 2 | 3+4 2 | 5+6 2 | 3+3 2 | 2+2 2 |
| | 5+6 2 | 3+4 2 | 5+6 2 | 3+4 2 | 2+2 2 | 3+3 2 |
| | 5+6 2 | 5+6 2 | 5+6 2 | 5+6 2 | 3+3 2 | 3+3 2 |
| | 3+4 2 | 3+4 2 | 3+4 2 | 3+4 2 | 2+2 2 | 2+2 2 |
| | 5+6 2 | 5+6 2 | 3+4 2 | 5+6 2 | 3+3 2 | 3+3 2 |
| | 3+4 2 | 3+4 2 | 5+6 2 | 3+4 2 | 2+2 2 | 2+2 2 |
| | 3+4 2 | 5+6 2 | 3+4 2 | 5+6 2 | 3+3 2 | 2+2 2 |
| | 3+4 2 | 3+4 2 | 5+6 2 | 3+4 2 | 2+2 2 | 3+3 2 |
| | 5+6 2 | 5+6 2 | 3+4 2 | 5+6 2 | 3+3 2 | 2+2 2 |
| | 3+4 2 | 3+4 2 | 5+6 2 | 5+6 2 | 2+2 2 | 2+2 2 |
| | 3+4 2 | 5+6 2 | 3+4 2 | 5+6 2 | 3+3 2 | 2+2 2 |
| | 5+6 2 | 3+4 2 | 3+4 2 | 3+4 2 | 3+3 2 | 3+3 2 |
| | 5+6 2 | 5+6 2 | 5+6 2 | 3+4 2 | 2+2 2 | 3+3 2 |
| | 3+4 2 | 5+6 2 | 5+6 2 | 3+4 2 | 2+2 2 | 2+2 2 |
| | 5+6 2 | 3+4 2 | 3+4 2 | 5+6 2 | 3+3 2 | 3+3 2 |
| | 3+4 2 | 5+6 2 | 5+6 2 | 5+6 2 | 3+3 2 | 2+2 2 |
| | 5+6 2 | 3+4 2 | 3+4 2 | 5+6 2 | 2+2 2 | 3+3 2 |
| | 3+4 2 | 3+4 2 | 3+4 2 | 3+4 2 | 3+3 2 | 2+2 2 |
| | 3+4 2 | 5+6 2 | 3+4 2 | 3+4 2 | 3+3 2 | 2+2 2 |
| | 5+6 2 | 3+4 2 | 5+6 2 | 5+6 2 | 2+2 2 | 3+3 2 |
| | 5+6 2 | 5+6 2 | 5+6 2 | 3+4 2 | 3+3 2 | 2+2 2 |
| | 3+4 2 | 5+6 2 | 3+4 2 | 5+6 2 | 2+2 2 | 3+3 2 |
| | 5+6 2 | 3+4 2 | 3+4 2 | 5+6 2 | 3+3 2 | 2+2 2 |
| | 3+4 2 | 5+6 2 | 5+6 2 | 5+6 2 | 3+3 2 | 3+3 2 |
| | 5+6 2 | 3+4 2 | 3+4 2 | 3+4 2 | 2+2 2 | 2+2 2 |
| | 3+4 2 | 5+6 2 | 5+6 2 | 3+4 2 | 2+2 2 | 3+3 2 |
| | 3+4 2 | 3+4 2 | 3+4 2 | 5+6 2 | 3+3 2 | 2+2 2 |
| | 5+6 2 | 5+6 2 | 5+6 2 | 3+4 2 | 3+3 2 | 3+3 2 |
| | 5+6 2 | 3+4 2 | 3+4 2 | 5+6 2 | 2+2 2 | 2+2 2 |
| | 3+4 2 | 5+6 2 | 5+6 2 | 3+4 2 | 2+2 2 | 3+3 2 |
| | 5+6 2 | 3+4 2 | 3+4 2 | 5+6 2 | 3+3 2 | 2+2 2 |
| | 3+4 2 | 5+6 2 | 5+6 2 | 3+4 2 | 2+2 2 | 2+2 2 |
| | 3+4 2 | 5+6 2 | 3+4 2 | 5+6 2 | 3+3 2 | 3+3 2 |
| | 3+4 2 | 3+4 2 | 5+6 2 | 3+4 2 | 2+2 2 | 2+2 2 |
| | 5+6 2 | 3+4 2 | 3+4 2 | 5+6 2 | 2+2 2 | 3+3 2 |
| | 5+6 2 | 5+6 2 | 5+6 2 | 3+4 2 | 3+3 2 | 2+2 2 |
| | 3+4 2 | 3+4 2 | 3+4 2 | 5+6 2 | 3+3 2 | 3+3 2 |
| | 3+4 2 | 5+6 2 | 5+6 2 | 3+4 2 | 2+2 2 | 2+2 2 |
| | 5+6 2 | 3+4 2 | 3+4 2 | 5+6 2 | 3+3 2 | 3+3 2 |
| | 3+4 2 | 5+6 2 | 5+6 2 | 3+4 2 | 2+2 2 | 2+2 2 |
| 12.2.9 | 65363 | 98 | O7Si2Yb2 | | | |
| | -A | -GM | -M | -Y | -L | -V |
| | 3+4 2 | 3+4 2 | 5+6 2 | 5+6 2 | 2+2 2 | 2+2 2 |
| | 5+6 2 | 5+6 2 | 3+4 2 | 3+4 2 | 3+3 2 | 3+3 2 |
| | 3+4 2 | 3+4 2 | 5+6 2 | 5+6 2 | 2+2 2 | 2+2 2 |
| | 5+6 2 | 5+6 2 | 3+4 2 | 3+4 2 | 3+3 2 | 3+3 2 |
| | 3+4 2 | 3+4 2 | 5+6 2 | 5+6 2 | 2+2 2 | 2+2 2 |
| | 5+6 2 | 5+6 2 | 3+4 2 | 3+4 2 | 3+3 2 | 3+3 2 |
| | 3+4 2 | 3+4 2 | 5+6 2 | 5+6 2 | 2+2 2 | 2+2 2 |
| | 5+6 2 | 5+6 2 | 3+4 2 | 3+4 2 | 3+3 2 | 3+3 2 |
| | 5+6 2 | 3+4 2 | 5+6 2 | 3+4 2 | 2+2 2 | 3+3 2 |
| | 3+4 2 | 5+6 2 | 3+4 2 | 5+6 2 | 3+3 2 | 2+2 2 |
| | 5+6 2 | 3+4 2 | 5+6 2 | 3+4 2 | 2+2 2 | 3+3 2 |

Continued on next column

**TABLE XXIV– continued**

| TQC | ICSD | NoE | SF | | | |
|---|---|---|---|---|---|---|
| | 3+4 2 | 5+6 2 | 5+6 2 | 3+4 2 | 2+2 2 | 3+3 2 |
| | 5+6 2 | 3+4 2 | 3+4 2 | 5+6 2 | 3+3 2 | 2+2 2 |
| | 5+6 2 | 5+6 2 | 3+4 2 | 5+6 2 | 3+3 2 | 3+3 2 |
| | 3+4 2 | 3+4 2 | 5+6 2 | 3+4 2 | 2+2 2 | 2+2 2 |
| | 3+4 2 | 5+6 2 | 3+4 2 | 5+6 2 | 3+3 2 | 2+2 2 |
| | 3+4 2 | 3+4 2 | 5+6 2 | 3+4 2 | 2+2 2 | 3+3 2 |
| | 5+6 2 | 5+6 2 | 3+4 2 | 5+6 2 | 3+3 2 | 2+2 2 |
| | 3+4 2 | 3+4 2 | 5+6 2 | 5+6 2 | 2+2 2 | 2+2 2 |
| | 3+4 2 | 5+6 2 | 3+4 2 | 3+4 2 | 3+3 2 | 2+2 2 |
| | 5+6 2 | 3+4 2 | 3+4 2 | 5+6 2 | 3+3 2 | 3+3 2 |
| | 5+6 2 | 5+6 2 | 5+6 2 | 3+4 2 | 2+2 2 | 3+3 2 |
| | 3+4 2 | 5+6 2 | 5+6 2 | 3+4 2 | 2+2 2 | 2+2 2 |
| | 5+6 2 | 3+4 2 | 3+4 2 | 5+6 2 | 3+3 2 | 3+3 2 |
| | 5+6 2 | 5+6 2 | 5+6 2 | 5+6 2 | 3+3 2 | 2+2 2 |
| | 3+4 2 | 3+4 2 | 3+4 2 | 5+6 2 | 2+2 2 | 3+3 2 |
| | 3+4 2 | 3+4 2 | 3+4 2 | 3+4 2 | 3+3 2 | 2+2 2 |
| | 5+6 2 | 5+6 2 | 3+4 2 | 3+4 2 | 3+3 2 | 2+2 2 |
| | 3+4 2 | 3+4 2 | 5+6 2 | 5+6 2 | 2+2 2 | 3+3 2 |
| | 5+6 2 | 5+6 2 | 5+6 2 | 3+4 2 | 3+3 2 | 2+2 2 |
| | 5+6 2 | 3+4 2 | 3+4 2 | 5+6 2 | 2+2 2 | 3+3 2 |
| | 3+4 2 | 5+6 2 | 5+6 2 | 5+6 2 | 3+3 2 | 2+2 2 |
| | 3+4 2 | 5+6 2 | 3+4 2 | 5+6 2 | 2+2 2 | 3+3 2 |
| | 5+6 2 | 3+4 2 | 3+4 2 | 3+4 2 | 3+3 2 | 2+2 2 |
| | 3+4 2 | 5+6 2 | 5+6 2 | 3+4 2 | 2+2 2 | 3+3 2 |
| | 5+6 2 | 3+4 2 | 3+4 2 | 5+6 2 | 3+3 2 | 2+2 2 |
| | 3+4 2 | 3+4 2 | 5+6 2 | 3+4 2 | 2+2 2 | 3+3 2 |
| | 5+6 2 | 5+6 2 | 3+4 2 | 5+6 2 | 3+3 2 | 2+2 2 |
| | 3+4 2 | 5+6 2 | 5+6 2 | 3+4 2 | 2+2 2 | 3+3 2 |
| | 5+6 2 | 3+4 2 | 3+4 2 | 5+6 2 | 2+2 2 | 2+2 2 |
| | 3+4 2 | 5+6 2 | 5+6 2 | 3+4 2 | 3+3 2 | 3+3 2 |
| | 3+4 2 | 5+6 2 | 3+4 2 | 5+6 2 | 3+3 2 | 2+2 2 |
| | 3+4 2 | 3+4 2 | 5+6 2 | 3+4 2 | 2+2 2 | 2+2 2 |
| | 5+6 2 | 3+4 2 | 3+4 2 | 5+6 2 | 2+2 2 | 3+3 2 |
| | 5+6 2 | 5+6 2 | 5+6 2 | 3+4 2 | 3+3 2 | 2+2 2 |
| | 3+4 2 | 3+4 2 | 3+4 2 | 5+6 2 | 2+2 2 | 3+3 2 |
| | 3+4 2 | 5+6 2 | 5+6 2 | 3+4 2 | 3+3 2 | 3+3 2 |
| | 5+6 2 | 3+4 2 | 3+4 2 | 5+6 2 | 3+3 2 | 2+2 2 |
| | 3+4 2 | 5+6 2 | 5+6 2 | 3+4 2 | 2+2 2 | 2+2 2 |
| 12.2.10 | 62356 | 98 | O7Si2Yb2 | | | |
| | -A | -GM | -M | -Y | -L | -V |
| | 3+4 2 | 3+4 2 | 5+6 2 | 5+6 2 | 2+2 2 | 2+2 2 |
| | 5+6 2 | 5+6 2 | 3+4 2 | 3+4 2 | 3+3 2 | 3+3 2 |
| | 3+4 2 | 3+4 2 | 5+6 2 | 5+6 2 | 2+2 2 | 2+2 2 |
| | 5+6 2 | 5+6 2 | 3+4 2 | 3+4 2 | 3+3 2 | 3+3 2 |
| | 3+4 2 | 3+4 2 | 5+6 2 | 5+6 2 | 2+2 2 | 2+2 2 |
| | 5+6 2 | 5+6 2 | 3+4 2 | 3+4 2 | 3+3 2 | 3+3 2 |
| | 3+4 2 | 3+4 2 | 5+6 2 | 5+6 2 | 2+2 2 | 2+2 2 |
| | 5+6 2 | 5+6 2 | 3+4 2 | 3+4 2 | 3+3 2 | 3+3 2 |
| | 5+6 2 | 3+4 2 | 5+6 2 | 3+4 2 | 2+2 2 | 3+3 2 |
| | 3+4 2 | 5+6 2 | 3+4 2 | 5+6 2 | 3+3 2 | 2+2 2 |
| | 5+6 2 | 3+4 2 | 5+6 2 | 3+4 2 | 2+2 2 | 3+3 2 |
| | 3+4 2 | 5+6 2 | 5+6 2 | 3+4 2 | 2+2 2 | 3+3 2 |
| | 5+6 2 | 3+4 2 | 3+4 2 | 5+6 2 | 3+3 2 | 2+2 2 |
| | 5+6 2 | 5+6 2 | 3+4 2 | 5+6 2 | 3+3 2 | 3+3 2 |
| | 3+4 2 | 3+4 2 | 5+6 2 | 3+4 2 | 2+2 2 | 2+2 2 |
| | 3+4 2 | 5+6 2 | 3+4 2 | 5+6 2 | 3+3 2 | 2+2 2 |
| | 3+4 2 | 3+4 2 | 5+6 2 | 3+4 2 | 2+2 2 | 3+3 2 |
| | 5+6 2 | 5+6 2 | 3+4 2 | 5+6 2 | 3+3 2 | 2+2 2 |
| | 3+4 2 | 3+4 2 | 5+6 2 | 5+6 2 | 2+2 2 | 2+2 2 |
| | 3+4 2 | 5+6 2 | 3+4 2 | 3+4 2 | 3+3 2 | 2+2 2 |
| | 5+6 2 | 3+4 2 | 3+4 2 | 5+6 2 | 3+3 2 | 3+3 2 |

Continued on next column

**TABLE XXIV– continued**

| TQC | ICSD | NoE | SF | | | |
|---|---|---|---|---|---|---|
| | 5+6 2 | 5+6 2 | 5+6 2 | 3+4 2 | 2+2 2 | 3+3 2 |
| | 3+4 2 | 5+6 2 | 5+6 2 | 3+4 2 | 2+2 2 | 2+2 2 |
| | 5+6 2 | 3+4 2 | 3+4 2 | 5+6 2 | 3+3 2 | 3+3 2 |
| | 5+6 2 | 5+6 2 | 5+6 2 | 5+6 2 | 3+3 2 | 2+2 2 |
| | 3+4 2 | 3+4 2 | 3+4 2 | 5+6 2 | 2+2 2 | 3+3 2 |
| | 3+4 2 | 3+4 2 | 3+4 2 | 3+4 2 | 3+3 2 | 2+2 2 |
| | 5+6 2 | 5+6 2 | 3+4 2 | 3+4 2 | 3+3 2 | 2+2 2 |
| | 3+4 2 | 3+4 2 | 5+6 2 | 5+6 2 | 2+2 2 | 3+3 2 |
| | 5+6 2 | 5+6 2 | 5+6 2 | 3+4 2 | 3+3 2 | 2+2 2 |
| | 5+6 2 | 3+4 2 | 3+4 2 | 5+6 2 | 2+2 2 | 3+3 2 |
| | 3+4 2 | 5+6 2 | 5+6 2 | 5+6 2 | 3+3 2 | 2+2 2 |
| | 3+4 2 | 5+6 2 | 3+4 2 | 5+6 2 | 2+2 2 | 3+3 2 |
| | 5+6 2 | 3+4 2 | 3+4 2 | 3+4 2 | 3+3 2 | 2+2 2 |
| | 3+4 2 | 5+6 2 | 5+6 2 | 3+4 2 | 2+2 2 | 3+3 2 |
| | 5+6 2 | 3+4 2 | 3+4 2 | 5+6 2 | 3+3 2 | 2+2 2 |
| | 3+4 2 | 3+4 2 | 5+6 2 | 3+4 2 | 2+2 2 | 3+3 2 |
| | 5+6 2 | 5+6 2 | 3+4 2 | 5+6 2 | 3+3 2 | 2+2 2 |
| | 3+4 2 | 5+6 2 | 5+6 2 | 3+4 2 | 2+2 2 | 3+3 2 |
| | 5+6 2 | 3+4 2 | 3+4 2 | 5+6 2 | 2+2 2 | 2+2 2 |
| | 3+4 2 | 5+6 2 | 5+6 2 | 3+4 2 | 3+3 2 | 3+3 2 |
| | 3+4 2 | 5+6 2 | 3+4 2 | 5+6 2 | 3+3 2 | 2+2 2 |
| | 3+4 2 | 3+4 2 | 5+6 2 | 3+4 2 | 2+2 2 | 2+2 2 |
| | 5+6 2 | 3+4 2 | 3+4 2 | 5+6 2 | 2+2 2 | 3+3 2 |
| | 5+6 2 | 5+6 2 | 5+6 2 | 3+4 2 | 3+3 2 | 2+2 2 |
| | 3+4 2 | 3+4 2 | 3+4 2 | 5+6 2 | 2+2 2 | 3+3 2 |
| | 3+4 2 | 5+6 2 | 5+6 2 | 3+4 2 | 3+3 2 | 3+3 2 |
| | 5+6 2 | 3+4 2 | 3+4 2 | 5+6 2 | 3+3 2 | 2+2 2 |
| | 3+4 2 | 5+6 2 | 5+6 2 | 3+4 2 | 2+2 2 | 2+2 2 |
| 12.2.11 | 248719 | 96 | O7Si2Tm2 | | | |
| | -A | -GM | -M | -Y | -L | -V |
| | 5+6 2 | 3+4 2 | 3+4 2 | 5+6 2 | 3+3 2 | 2+2 2 |
| | 3+4 2 | 5+6 2 | 5+6 2 | 3+4 2 | 2+2 2 | 3+3 2 |
| | 5+6 2 | 3+4 2 | 3+4 2 | 5+6 2 | 3+3 2 | 2+2 2 |
| | 3+4 2 | 5+6 2 | 5+6 2 | 3+4 2 | 2+2 2 | 3+3 2 |
| | 5+6 2 | 3+4 2 | 3+4 2 | 5+6 2 | 3+3 2 | 2+2 2 |
| | 3+4 2 | 5+6 2 | 5+6 2 | 3+4 2 | 2+2 2 | 3+3 2 |
| | 5+6 2 | 3+4 2 | 3+4 2 | 5+6 2 | 3+3 2 | 2+2 2 |
| | 3+4 2 | 5+6 2 | 5+6 2 | 3+4 2 | 2+2 2 | 3+3 2 |
| | 3+4 2 | 3+4 2 | 3+4 2 | 3+4 2 | 3+3 2 | 3+3 2 |
| | 5+6 2 | 5+6 2 | 5+6 2 | 5+6 2 | 2+2 2 | 2+2 2 |
| | 3+4 2 | 3+4 2 | 3+4 2 | 3+4 2 | 3+3 2 | 3+3 2 |
| | 5+6 2 | 5+6 2 | 3+4 2 | 5+6 2 | 2+2 2 | 2+2 2 |
| | 3+4 2 | 3+4 2 | 5+6 2 | 3+4 2 | 3+3 2 | 3+3 2 |
| | 3+4 2 | 5+6 2 | 5+6 2 | 5+6 2 | 2+2 2 | 3+3 2 |
| | 5+6 2 | 3+4 2 | 3+4 2 | 3+4 2 | 3+3 2 | 2+2 2 |
| | 5+6 2 | 5+6 2 | 5+6 2 | 5+6 2 | 2+2 2 | 2+2 2 |
| | 5+6 2 | 3+4 2 | 3+4 2 | 3+4 2 | 3+3 2 | 3+3 2 |
| | 3+4 2 | 5+6 2 | 5+6 2 | 5+6 2 | 2+2 2 | 2+2 2 |
| | 5+6 2 | 3+4 2 | 3+4 2 | 5+6 2 | 3+3 2 | 2+2 2 |
| | 5+6 2 | 5+6 2 | 5+6 2 | 3+4 2 | 2+2 2 | 3+3 2 |
| | 3+4 2 | 3+4 2 | 5+6 2 | 5+6 2 | 2+2 2 | 2+2 2 |
| | 3+4 2 | 5+6 2 | 3+4 2 | 3+4 2 | 3+3 2 | 3+3 2 |
| | 5+6 2 | 5+6 2 | 3+4 2 | 3+4 2 | 3+3 2 | 2+2 2 |
| | 3+4 2 | 3+4 2 | 5+6 2 | 5+6 2 | 2+2 2 | 3+3 2 |
| | 3+4 2 | 5+6 2 | 3+4 2 | 5+6 2 | 2+2 2 | 2+2 2 |
| | 5+6 2 | 3+4 2 | 5+6 2 | 5+6 2 | 3+3 2 | 3+3 2 |
| | 5+6 2 | 3+4 2 | 5+6 2 | 3+4 2 | 2+2 2 | 2+2 2 |
| | 3+4 2 | 3+4 2 | 5+6 2 | 3+4 2 | 2+2 2 | 2+2 2 |
| | 5+6 2 | 5+6 2 | 3+4 2 | 5+6 2 | 3+3 2 | 3+3 2 |
| | 3+4 2 | 5+6 2 | 3+4 2 | 3+4 2 | 2+2 2 | 2+2 2 |
| | 5+6 2 | 5+6 2 | 5+6 2 | 5+6 2 | 3+3 2 | 3+3 2 |

Continued on next column

**TABLE XXIV– continued**

| TQC | ICSD | NoE | SF | | | |
|---|---|---|---|---|---|---|
| | 3+4 2 | 3+4 2 | 3+4 2 | 5+6 2 | 2+2 2 | 3+3 2 |
| | 3+4 2 | 5+6 2 | 5+6 2 | 3+4 2 | 3+3 2 | 2+2 2 |
| | 5+6 2 | 3+4 2 | 5+6 2 | 5+6 2 | 2+2 2 | 2+2 2 |
| | 5+6 2 | 5+6 2 | 3+4 2 | 3+4 2 | 3+3 2 | 3+3 2 |
| | 5+6 2 | 5+6 2 | 5+6 2 | 5+6 2 | 2+2 2 | 2+2 2 |
| | 3+4 2 | 3+4 2 | 3+4 2 | 5+6 2 | 3+3 2 | 2+2 2 |
| | 5+6 2 | 5+6 2 | 5+6 2 | 3+4 2 | 2+2 2 | 3+3 2 |
| | 3+4 2 | 3+4 2 | 3+4 2 | 3+4 2 | 3+3 2 | 3+3 2 |
| | 5+6 2 | 3+4 2 | 5+6 2 | 3+4 2 | 2+2 2 | 2+2 2 |
| | 3+4 2 | 5+6 2 | 3+4 2 | 5+6 2 | 3+3 2 | 2+2 2 |
| | 5+6 2 | 5+6 2 | 5+6 2 | 5+6 2 | 2+2 2 | 3+3 2 |
| | 5+6 2 | 3+4 2 | 3+4 2 | 3+4 2 | 3+3 2 | 2+2 2 |
| | 3+4 2 | 3+4 2 | 5+6 2 | 5+6 2 | 3+3 2 | 3+3 2 |
| | 5+6 2 | 5+6 2 | 3+4 2 | 3+4 2 | 2+2 2 | 2+2 2 |
| | 3+4 2 | 3+4 2 | 5+6 2 | 5+6 2 | 2+2 2 | 3+3 2 |
| | 5+6 2 | 5+6 2 | 3+4 2 | 3+4 2 | 3+3 2 | 2+2 2 |
| | 3+4 2 | 3+4 2 | 5+6 2 | 5+6 2 | 2+2 2 | 3+3 2 |
| 12.2.12 | 75194 | 84 | O5Ti3 | | | |
| | -A | -GM | -M | -Y | -L | -V |
| | 3+4 2 | 3+4 2 | 5+6 2 | 5+6 2 | 3+3 2 | 3+3 2 |
| | 5+6 2 | 5+6 2 | 3+4 2 | 3+4 2 | 2+2 2 | 2+2 2 |
| | 3+4 2 | 3+4 2 | 5+6 2 | 3+4 2 | 2+2 2 | 2+2 2 |
| | 5+6 2 | 5+6 2 | 3+4 2 | 5+6 2 | 3+3 2 | 3+3 2 |
| | 3+4 2 | 3+4 2 | 5+6 2 | 3+4 2 | 2+2 2 | 2+2 2 |
| | 5+6 2 | 3+4 2 | 3+4 2 | 5+6 2 | 3+3 2 | 3+3 2 |
| | 5+6 2 | 5+6 2 | 3+4 2 | 5+6 2 | 2+2 2 | 3+3 2 |
| | 3+4 2 | 5+6 2 | 5+6 2 | 3+4 2 | 3+3 2 | 2+2 2 |
| | 3+4 2 | 3+4 2 | 5+6 2 | 3+4 2 | 3+3 2 | 2+2 2 |
| | 5+6 2 | 5+6 2 | 3+4 2 | 5+6 2 | 2+2 2 | 3+3 2 |
| | 5+6 2 | 5+6 2 | 3+4 2 | 3+4 2 | 3+3 2 | 3+3 2 |
| | 3+4 2 | 3+4 2 | 5+6 2 | 5+6 2 | 2+2 2 | 2+2 2 |
| | 5+6 2 | 5+6 2 | 5+6 2 | 5+6 2 | 2+2 2 | 2+2 2 |
| | 3+4 2 | 3+4 2 | 5+6 2 | 3+4 2 | 3+3 2 | 2+2 2 |
| | 3+4 2 | 3+4 2 | 3+4 2 | 5+6 2 | 2+2 2 | 3+3 2 |
| | 5+6 2 | 3+4 2 | 3+4 2 | 5+6 2 | 3+3 2 | 3+3 2 |
| | 5+6 2 | 5+6 2 | 5+6 2 | 3+4 2 | 3+3 2 | 2+2 2 |
| | 3+4 2 | 5+6 2 | 3+4 2 | 5+6 2 | 2+2 2 | 3+3 2 |
| | 5+6 2 | 3+4 2 | 5+6 2 | 3+4 2 | 2+2 2 | 3+3 2 |
| | 3+4 2 | 5+6 2 | 3+4 2 | 3+4 2 | 3+3 2 | 2+2 2 |
| | 5+6 2 | 3+4 2 | 3+4 2 | 5+6 2 | 2+2 2 | 3+3 2 |
| | 5+6 2 | 3+4 2 | 5+6 2 | 3+4 2 | 3+3 2 | 2+2 2 |
| | 3+4 2 | 5+6 2 | 5+6 2 | 5+6 2 | 2+2 2 | 2+2 2 |
| | 3+4 2 | 3+4 2 | 3+4 2 | 3+4 2 | 2+2 2 | 2+2 2 |
| | 3+4 2 | 3+4 2 | 5+6 2 | 5+6 2 | 3+3 2 | 3+3 2 |
| | 5+6 2 | 5+6 2 | 3+4 2 | 3+4 2 | 3+3 2 | 3+3 2 |
| | 5+6 2 | 3+4 2 | 5+6 2 | 3+4 2 | 3+3 2 | 3+3 2 |
| | 3+4 2 | 5+6 2 | 5+6 2 | 5+6 2 | 2+2 2 | 2+2 2 |
| | 3+4 2 | 5+6 2 | 3+4 2 | 5+6 2 | 2+2 2 | 2+2 2 |
| | 3+4 2 | 5+6 2 | 3+4 2 | 5+6 2 | 3+3 2 | 2+2 2 |
| | 5+6 2 | 3+4 2 | 3+4 2 | 3+4 2 | 3+3 2 | 3+3 2 |
| | 3+4 2 | 3+4 2 | 5+6 2 | 3+4 2 | 2+2 2 | 3+3 2 |
| | 5+6 2 | 5+6 2 | 5+6 2 | 5+6 2 | 2+2 2 | 2+2 2 |
| | 5+6 2 | 3+4 2 | 5+6 2 | 3+4 2 | 3+3 2 | 3+3 2 |
| | 3+4 2 | 5+6 2 | 3+4 2 | 5+6 2 | 3+3 2 | 3+3 2 |
| | 3+4 2 | 5+6 2 | 5+6 2 | 3+4 2 | 2+2 2 | 2+2 2 |
| | 5+6 2 | 5+6 2 | 3+4 2 | 5+6 2 | 3+3 2 | 3+3 2 |
| | 5+6 2 | 3+4 2 | 5+6 2 | 3+4 2 | 2+2 2 | 2+2 2 |
| | 3+4 2 | 3+4 2 | 3+4 2 | 3+4 2 | 3+3 2 | 3+3 2 |
| | 5+6 2 | 5+6 2 | 3+4 2 | 5+6 2 | 2+2 2 | 2+2 2 |
| | 3+4 2 | 3+4 2 | 3+4 2 | 3+4 2 | 3+3 2 | 3+3 2 |
| | 3+4 2 | 5+6 2 | 5+6 2 | 5+6 2 | 2+2 2 | 2+2 2 |

Continued on next column

**TABLE XXIV– continued**

| TQC | ICSD | NoE | SF | | | |
|---|---|---|---|---|---|---|
| 12.2.13 | 75193 | 84 | O5Ti3 | | | |
| | -A | -GM | -M | -Y | -L | -V |
| | 5+6 2 | 3+4 2 | 3+4 2 | 5+6 2 | 2+2 2 | 3+3 2 |
| | 3+4 2 | 3+4 2 | 5+6 2 | 3+4 2 | 3+3 2 | 2+2 2 |
| | 3+4 2 | 5+6 2 | 5+6 2 | 5+6 2 | 3+3 2 | 3+3 2 |
| | 5+6 2 | 5+6 2 | 5+6 2 | 3+4 2 | 2+2 2 | 2+2 2 |
| | 5+6 2 | 3+4 2 | 3+4 2 | 3+4 2 | 3+3 2 | 3+3 2 |
| | 3+4 2 | 5+6 2 | 3+4 2 | 5+6 2 | 2+2 2 | 2+2 2 |
| | 5+6 2 | 3+4 2 | 5+6 2 | 3+4 2 | 2+2 2 | 3+3 2 |
| | 3+4 2 | 3+4 2 | 3+4 2 | 5+6 2 | 2+2 2 | 2+2 2 |
| | 5+6 2 | 5+6 2 | 5+6 2 | 3+4 2 | 3+3 2 | 3+3 2 |
| | 3+4 2 | 5+6 2 | 3+4 2 | 5+6 2 | 3+3 2 | 2+2 2 |
| | 5+6 2 | 5+6 2 | 5+6 2 | 3+4 2 | 2+2 2 | 3+3 2 |
| | 3+4 2 | 3+4 2 | 3+4 2 | 5+6 2 | 3+3 2 | 2+2 2 |
| | 3+4 2 | 3+4 2 | 3+4 2 | 5+6 2 | 2+2 2 | 3+3 2 |
| | 5+6 2 | 3+4 2 | 5+6 2 | 5+6 2 | 2+2 2 | 2+2 2 |
| | 5+6 2 | 5+6 2 | 3+4 2 | 3+4 2 | 3+3 2 | 2+2 2 |
| | 5+6 2 | 3+4 2 | 3+4 2 | 3+4 2 | 3+3 2 | 3+3 2 |
| | 3+4 2 | 5+6 2 | 5+6 2 | 5+6 2 | 3+3 2 | 2+2 2 |
| | 5+6 2 | 3+4 2 | 5+6 2 | 3+4 2 | 2+2 2 | 3+3 2 |
| | 3+4 2 | 5+6 2 | 5+6 2 | 5+6 2 | 2+2 2 | 3+3 2 |
| | 3+4 2 | 3+4 2 | 3+4 2 | 3+4 2 | 3+3 2 | 2+2 2 |
| | 3+4 2 | 3+4 2 | 5+6 2 | 5+6 2 | 2+2 2 | 3+3 2 |
| | 5+6 2 | 5+6 2 | 5+6 2 | 3+4 2 | 2+2 2 | 2+2 2 |
| | 3+4 2 | 3+4 2 | 3+4 2 | 5+6 2 | 3+3 2 | 2+2 2 |
| | 5+6 2 | 3+4 2 | 3+4 2 | 3+4 2 | 3+3 2 | 3+3 2 |
| | 5+6 2 | 3+4 2 | 3+4 2 | 5+6 2 | 2+2 2 | 3+3 2 |
| | 3+4 2 | 5+6 2 | 5+6 2 | 3+4 2 | 3+3 2 | 2+2 2 |
| | 3+4 2 | 5+6 2 | 3+4 2 | 3+4 2 | 2+2 2 | 2+2 2 |
| | 5+6 2 | 5+6 2 | 5+6 2 | 5+6 2 | 3+3 2 | 3+3 2 |
| | 5+6 2 | 5+6 2 | 5+6 2 | 5+6 2 | 2+2 2 | 3+3 2 |
| | 3+4 2 | 3+4 2 | 3+4 2 | 5+6 2 | 3+3 2 | 2+2 2 |
| | 5+6 2 | 5+6 2 | 3+4 2 | 3+4 2 | 2+2 2 | 3+3 2 |
| | 3+4 2 | 3+4 2 | 5+6 2 | 3+4 2 | 3+3 2 | 2+2 2 |
| | 3+4 2 | 5+6 2 | 5+6 2 | 5+6 2 | 2+2 2 | 3+3 2 |
| | 5+6 2 | 3+4 2 | 3+4 2 | 5+6 2 | 3+3 2 | 2+2 2 |
| | 5+6 2 | 3+4 2 | 3+4 2 | 3+4 2 | 2+2 2 | 3+3 2 |
| | 3+4 2 | 3+4 2 | 5+6 2 | 5+6 2 | 3+3 2 | 2+2 2 |
| | 5+6 2 | 5+6 2 | 3+4 2 | 5+6 2 | 2+2 2 | 3+3 2 |
| | 3+4 2 | 5+6 2 | 5+6 2 | 3+4 2 | 3+3 2 | 3+3 2 |
| | 3+4 2 | 5+6 2 | 5+6 2 | 3+4 2 | 2+2 2 | 2+2 2 |
| | 5+6 2 | 5+6 2 | 3+4 2 | 3+4 2 | 3+3 2 | 2+2 2 |
| | 5+6 2 | 3+4 2 | 3+4 2 | 5+6 2 | 3+3 2 | 3+3 2 |
| | 3+4 2 | 5+6 2 | 3+4 2 | 3+4 2 | 2+2 2 | 2+2 2 |
| 12.2.14 | 38223 | 84 | O3Rb2Te1 | | | |
| | -A | -GM | -M | -Y | -L | -V |
| | 3+4 2 | 3+4 2 | 5+6 2 | 5+6 2 | 2+2 2 | 2+2 2 |
| | 5+6 2 | 5+6 2 | 3+4 2 | 3+4 2 | 3+3 2 | 3+3 2 |
| | 3+4 2 | 3+4 2 | 3+4 2 | 3+4 2 | 3+3 2 | 3+3 2 |
| | 5+6 2 | 3+4 2 | 5+6 2 | 3+4 2 | 2+2 2 | 3+3 2 |
| | 5+6 2 | 3+4 2 | 3+4 2 | 5+6 2 | 3+3 2 | 2+2 2 |
| | 3+4 2 | 5+6 2 | 5+6 2 | 3+4 2 | 2+2 2 | 3+3 2 |
| | 5+6 2 | 3+4 2 | 3+4 2 | 5+6 2 | 3+3 2 | 2+2 2 |
| | 3+4 2 | 5+6 2 | 5+6 2 | 3+4 2 | 2+2 2 | 3+3 2 |
| | 5+6 2 | 5+6 2 | 5+6 2 | 5+6 2 | 2+2 2 | 2+2 2 |
| | 3+4 2 | 3+4 2 | 3+4 2 | 3+4 2 | 3+3 2 | 3+3 2 |
| | 5+6 2 | 5+6 2 | 5+6 2 | 5+6 2 | 2+2 2 | 2+2 2 |
| | 3+4 2 | 3+4 2 | 5+6 2 | 5+6 2 | 3+3 2 | 3+3 2 |
| | 5+6 2 | 5+6 2 | 3+4 2 | 3+4 2 | 2+2 2 | 2+2 2 |
| | 5+6 2 | 5+6 2 | 5+6 2 | 5+6 2 | 2+2 2 | 2+2 2 |
| | 3+4 2 | 5+6 2 | 5+6 2 | 5+6 2 | 3+3 2 | 2+2 2 |

Continued on next column

**TABLE XXIV– continued**

| TQC | ICSD | NoE | SF | | | |
|---|---|---|---|---|---|---|
| | 5+6 2 | 3+4 2 | 3+4 2 | 3+4 2 | 2+2 2 | 3+3 2 |
| | 3+4 2 | 3+4 2 | 3+4 2 | 3+4 2 | 3+3 2 | 2+2 2 |
| | 3+4 2 | 5+6 2 | 3+4 2 | 5+6 2 | 2+2 2 | 2+2 2 |
| | 5+6 2 | 5+6 2 | 5+6 2 | 5+6 2 | 3+3 2 | 3+3 2 |
| | 5+6 2 | 5+6 2 | 5+6 2 | 5+6 2 | 2+2 2 | 2+2 2 |
| | 3+4 2 | 5+6 2 | 3+4 2 | 5+6 2 | 3+3 2 | 2+2 2 |
| | 3+4 2 | 3+4 2 | 3+4 2 | 3+4 2 | 3+3 2 | 2+2 2 |
| | 3+4 2 | 5+6 2 | 5+6 2 | 5+6 2 | 3+3 2 | 3+3 2 |
| | 5+6 2 | 5+6 2 | 3+4 2 | 5+6 2 | 2+2 2 | 2+2 2 |
| | 3+4 2 | 3+4 2 | 5+6 2 | 3+4 2 | 3+3 2 | 3+3 2 |
| | 5+6 2 | 5+6 2 | 3+4 2 | 5+6 2 | 2+2 2 | 2+2 2 |
| | 3+4 2 | 3+4 2 | 5+6 2 | 3+4 2 | 3+3 2 | 3+3 2 |
| | 5+6 2 | 3+4 2 | 3+4 2 | 3+4 2 | 2+2 2 | 2+2 2 |
| | 5+6 2 | 5+6 2 | 5+6 2 | 5+6 2 | 3+3 2 | 2+2 2 |
| | 3+4 2 | 5+6 2 | 3+4 2 | 5+6 2 | 2+2 2 | 3+3 2 |
| | 5+6 2 | 3+4 2 | 3+4 2 | 5+6 2 | 2+2 2 | 3+3 2 |
| | 5+6 2 | 3+4 2 | 3+4 2 | 3+4 2 | 3+3 2 | 3+3 2 |
| | 3+4 2 | 3+4 2 | 5+6 2 | 5+6 2 | 2+2 2 | 2+2 2 |
| | 5+6 2 | 5+6 2 | 5+6 2 | 3+4 2 | 3+3 2 | 3+3 2 |
| | 5+6 2 | 3+4 2 | 3+4 2 | 3+4 2 | 2+2 2 | 2+2 2 |
| | 3+4 2 | 5+6 2 | 5+6 2 | 5+6 2 | 3+3 2 | 2+2 2 |
| | 3+4 2 | 5+6 2 | 3+4 2 | 5+6 2 | 3+3 2 | 3+3 2 |
| | 3+4 2 | 5+6 2 | 5+6 2 | 3+4 2 | 2+2 2 | 2+2 2 |
| | 5+6 2 | 3+4 2 | 3+4 2 | 5+6 2 | 3+3 2 | 3+3 2 |
| | 3+4 2 | 5+6 2 | 5+6 2 | 3+4 2 | 2+2 2 | 2+2 2 |
| | 5+6 2 | 3+4 2 | 5+6 2 | 3+4 2 | 3+3 2 | 2+2 2 |
| | 3+4 2 | 3+4 2 | 3+4 2 | 5+6 2 | 2+2 2 | 3+3 2 |
| 12.2.15 | 16779 | 108 | O26V12 | | | |
| | -A | -GM | -M | -Y | -L | -V |
| | 5+6 2 | 3+4 2 | 3+4 2 | 5+6 2 | 3+3 2 | 2+2 2 |
| | 3+4 2 | 5+6 2 | 5+6 2 | 3+4 2 | 2+2 2 | 3+3 2 |
| | 3+4 2 | 5+6 2 | 5+6 2 | 3+4 2 | 2+2 2 | 3+3 2 |
| | 5+6 2 | 3+4 2 | 3+4 2 | 5+6 2 | 3+3 2 | 2+2 2 |
| | 3+4 2 | 3+4 2 | 5+6 2 | 5+6 2 | 3+3 2 | 2+2 2 |
| | 5+6 2 | 5+6 2 | 5+6 2 | 3+4 2 | 2+2 2 | 3+3 2 |
| | 5+6 2 | 3+4 2 | 3+4 2 | 3+4 2 | 2+2 2 | 2+2 2 |
| | 3+4 2 | 5+6 2 | 3+4 2 | 5+6 2 | 3+3 2 | 3+3 2 |
| | 3+4 2 | 3+4 2 | 3+4 2 | 3+4 2 | 2+2 2 | 2+2 2 |
| | 5+6 2 | 5+6 2 | 5+6 2 | 5+6 2 | 3+3 2 | 3+3 2 |
| | 3+4 2 | 3+4 2 | 3+4 2 | 3+4 2 | 2+2 2 | 2+2 2 |
| | 5+6 2 | 5+6 2 | 3+4 2 | 3+4 2 | 3+3 2 | 3+3 2 |
| | 3+4 2 | 3+4 2 | 5+6 2 | 5+6 2 | 2+2 2 | 2+2 2 |
| | 5+6 2 | 3+4 2 | 5+6 2 | 5+6 2 | 2+2 2 | 3+3 2 |
| | 3+4 2 | 5+6 2 | 3+4 2 | 3+4 2 | 3+3 2 | 2+2 2 |
| | 3+4 2 | 3+4 2 | 5+6 2 | 5+6 2 | 3+3 2 | 2+2 2 |
| | 5+6 2 | 5+6 2 | 3+4 2 | 3+4 2 | 2+2 2 | 3+3 2 |
| | 5+6 2 | 3+4 2 | 5+6 2 | 5+6 2 | 3+3 2 | 3+3 2 |
| | 5+6 2 | 5+6 2 | 5+6 2 | 3+4 2 | 2+2 2 | 2+2 2 |
| | 3+4 2 | 5+6 2 | 3+4 2 | 5+6 2 | 2+2 2 | 3+3 2 |
| | 5+6 2 | 5+6 2 | 3+4 2 | 5+6 2 | 3+3 2 | 2+2 2 |
| | 5+6 2 | 3+4 2 | 5+6 2 | 3+4 2 | 2+2 2 | 3+3 2 |
| | 3+4 2 | 3+4 2 | 5+6 2 | 3+4 2 | 3+3 2 | 2+2 2 |
| | 3+4 2 | 3+4 2 | 5+6 2 | 5+6 2 | 3+3 2 | 3+3 2 |
| | 5+6 2 | 5+6 2 | 3+4 2 | 5+6 2 | 3+3 2 | 2+2 2 |
| | 3+4 2 | 3+4 2 | 5+6 2 | 5+6 2 | 2+2 2 | 3+3 2 |
| | 5+6 2 | 5+6 2 | 3+4 2 | 3+4 2 | 3+3 2 | 3+3 2 |
| | 3+4 2 | 3+4 2 | 3+4 2 | 5+6 2 | 2+2 2 | 2+2 2 |
| | 3+4 2 | 5+6 2 | 5+6 2 | 5+6 2 | 3+3 2 | 3+3 2 |
| | 5+6 2 | 3+4 2 | 3+4 2 | 3+4 2 | 2+2 2 | 2+2 2 |
| | 3+4 2 | 5+6 2 | 5+6 2 | 3+4 2 | 2+2 2 | 3+3 2 |
| | 5+6 2 | 3+4 2 | 5+6 2 | 5+6 2 | 3+3 2 | 2+2 2 |

Continued on next column

**TABLE XXIV– continued**

| TQC | ICSD | NoE | SF | | | |
|---|---|---|---|---|---|---|
| | 5+6 2 | 3+4 2 | 3+4 2 | 5+6 2 | 3+3 2 | 2+2 2 |
| | 5+6 2 | 5+6 2 | 3+4 2 | 3+4 2 | 2+2 2 | 3+3 2 |
| | 3+4 2 | 5+6 2 | 5+6 2 | 3+4 2 | 3+3 2 | 3+3 2 |
| | 5+6 2 | 5+6 2 | 3+4 2 | 5+6 2 | 3+3 2 | 2+2 2 |
| | 3+4 2 | 3+4 2 | 5+6 2 | 5+6 2 | 2+2 2 | 3+3 2 |
| | 3+4 2 | 5+6 2 | 3+4 2 | 5+6 2 | 2+2 2 | 2+2 2 |
| | 5+6 2 | 3+4 2 | 5+6 2 | 3+4 2 | 3+3 2 | 3+3 2 |
| | 5+6 2 | 3+4 2 | 5+6 2 | 3+4 2 | 3+3 2 | 2+2 2 |
| | 3+4 2 | 5+6 2 | 3+4 2 | 5+6 2 | 2+2 2 | 2+2 2 |
| | 5+6 2 | 3+4 2 | 5+6 2 | 3+4 2 | 2+2 2 | 3+3 2 |
| | 3+4 2 | 5+6 2 | 3+4 2 | 5+6 2 | 3+3 2 | 3+3 2 |
| | 3+4 2 | 5+6 2 | 3+4 2 | 3+4 2 | 2+2 2 | 2+2 2 |
| | 3+4 2 | 3+4 2 | 5+6 2 | 3+4 2 | 3+3 2 | 2+2 2 |
| | 3+4 2 | 5+6 2 | 5+6 2 | 5+6 2 | 2+2 2 | 3+3 2 |
| | 5+6 2 | 3+4 2 | 5+6 2 | 5+6 2 | 3+3 2 | 2+2 2 |
| | 5+6 2 | 5+6 2 | 3+4 2 | 3+4 2 | 2+2 2 | 3+3 2 |
| | 5+6 2 | 5+6 2 | 5+6 2 | 5+6 2 | 3+3 2 | 3+3 2 |
| | 5+6 2 | 5+6 2 | 3+4 2 | 3+4 2 | 3+3 2 | 3+3 2 |
| | 3+4 2 | 3+4 2 | 3+4 2 | 3+4 2 | 2+2 2 | 2+2 2 |
| | 5+6 2 | 5+6 2 | 5+6 2 | 5+6 2 | 3+3 2 | 3+3 2 |
| | 3+4 2 | 5+6 2 | 3+4 2 | 3+4 2 | 2+2 2 | 2+2 2 |
| | 5+6 2 | 3+4 2 | 5+6 2 | 3+4 2 | 3+3 2 | 3+3 2 |
| 12.2.16 | 10455 | 122 | O13Sr2Ti6 | | | |
| | -A | -GM | -M | -Y | -L | -V |
| | 3+4 2 | 3+4 2 | 3+4 2 | 3+4 2 | 2+2 2 | 2+2 2 |
| | 5+6 2 | 5+6 2 | 5+6 2 | 5+6 2 | 3+3 2 | 3+3 2 |
| | 5+6 2 | 3+4 2 | 3+4 2 | 5+6 2 | 2+2 2 | 3+3 2 |
| | 3+4 2 | 5+6 2 | 5+6 2 | 3+4 2 | 3+3 2 | 2+2 2 |
| | 3+4 2 | 5+6 2 | 3+4 2 | 3+4 2 | 3+3 2 | 2+2 2 |
| | 5+6 2 | 3+4 2 | 5+6 2 | 5+6 2 | 2+2 2 | 3+3 2 |
| | 5+6 2 | 3+4 2 | 5+6 2 | 3+4 2 | 3+3 2 | 3+3 2 |
| | 3+4 2 | 5+6 2 | 3+4 2 | 5+6 2 | 2+2 2 | 3+3 2 |
| | 5+6 2 | 3+4 2 | 3+4 2 | 3+4 2 | 2+2 2 | 2+2 2 |
| | 3+4 2 | 3+4 2 | 3+4 2 | 5+6 2 | 3+3 2 | 3+3 2 |
| | 3+4 2 | 5+6 2 | 5+6 2 | 3+4 2 | 3+3 2 | 2+2 2 |
| | 3+4 2 | 3+4 2 | 3+4 2 | 3+4 2 | 2+2 2 | 2+2 2 |
| | 5+6 2 | 5+6 2 | 5+6 2 | 5+6 2 | 3+3 2 | 3+3 2 |
| | 3+4 2 | 5+6 2 | 5+6 2 | 5+6 2 | 2+2 2 | 2+2 2 |
| | 5+6 2 | 3+4 2 | 3+4 2 | 3+4 2 | 3+3 2 | 3+3 2 |
| | 5+6 2 | 5+6 2 | 5+6 2 | 5+6 2 | 3+3 2 | 3+3 2 |
| | 3+4 2 | 3+4 2 | 3+4 2 | 3+4 2 | 2+2 2 | 2+2 2 |
| | 5+6 2 | 5+6 2 | 5+6 2 | 5+6 2 | 3+3 2 | 2+2 2 |
| | 3+4 2 | 3+4 2 | 3+4 2 | 3+4 2 | 2+2 2 | 3+3 2 |
| | 5+6 2 | 5+6 2 | 5+6 2 | 5+6 2 | 3+3 2 | 3+3 2 |
| | 3+4 2 | 3+4 2 | 3+4 2 | 3+4 2 | 2+2 2 | 2+2 2 |
| | 5+6 2 | 3+4 2 | 5+6 2 | 3+4 2 | 2+2 2 | 2+2 2 |
| | 5+6 2 | 5+6 2 | 3+4 2 | 5+6 2 | 3+3 2 | 3+3 2 |
| | 3+4 2 | 5+6 2 | 5+6 2 | 3+4 2 | 3+3 2 | 2+2 2 |
| | 5+6 2 | 3+4 2 | 3+4 2 | 5+6 2 | 2+2 2 | 2+2 2 |
| | 3+4 2 | 3+4 2 | 3+4 2 | 5+6 2 | 2+2 2 | 3+3 2 |
| | 5+6 2 | 5+6 2 | 5+6 2 | 5+6 2 | 3+3 2 | 3+3 2 |
| | 5+6 2 | 5+6 2 | 3+4 2 | 5+6 2 | 2+2 2 | 3+3 2 |
| | 3+4 2 | 3+4 2 | 5+6 2 | 3+4 2 | 2+2 2 | 2+2 2 |
| | 3+4 2 | 3+4 2 | 5+6 2 | 3+4 2 | 3+3 2 | 2+2 2 |
| | 3+4 2 | 5+6 2 | 3+4 2 | 5+6 2 | 3+3 2 | 3+3 2 |
| | 5+6 2 | 5+6 2 | 5+6 2 | 5+6 2 | 2+2 2 | 2+2 2 |
| | 3+4 2 | 3+4 2 | 5+6 2 | 3+4 2 | 2+2 2 | 2+2 2 |
| | 5+6 2 | 3+4 2 | 3+4 2 | 5+6 2 | 3+3 2 | 2+2 2 |
| | 5+6 2 | 5+6 2 | 5+6 2 | 3+4 2 | 2+2 2 | 3+3 2 |
| | 3+4 2 | 3+4 2 | 3+4 2 | 5+6 2 | 2+2 2 | 3+3 2 |
| | 5+6 2 | 5+6 2 | 5+6 2 | 3+4 2 | 3+3 2 | 2+2 2 |

Continued on next column

**TABLE XXIV– continued**

| TQC | ICSD | | NoE | | SF | |
|---|---|---|---|---|---|---|
| | 5+6 2 | 5+6 2 | 3+4 2 | 3+4 2 | 2+2 2 | 2+2 2 |
| | 3+4 2 | 3+4 2 | 5+6 2 | 5+6 2 | 3+3 2 | 3+3 2 |
| | 3+4 2 | 3+4 2 | 3+4 2 | 5+6 2 | 2+2 2 | 2+2 2 |
| | 5+6 2 | 5+6 2 | 5+6 2 | 3+4 2 | 2+2 2 | 3+3 2 |
| | 5+6 2 | 5+6 2 | 3+4 2 | 5+6 2 | 3+3 2 | 2+2 2 |
| | 3+4 2 | 5+6 2 | 3+4 2 | 5+6 2 | 3+3 2 | 3+3 2 |
| | 5+6 2 | 3+4 2 | 3+4 2 | 5+6 2 | 2+2 2 | 2+2 2 |
| | 5+6 2 | 5+6 2 | 5+6 2 | 3+4 2 | 3+3 2 | 3+3 2 |
| | 3+4 2 | 3+4 2 | 5+6 2 | 3+4 2 | 3+3 2 | 2+2 2 |
| | 3+4 2 | 3+4 2 | 5+6 2 | 5+6 2 | 2+2 2 | 3+3 2 |
| | 5+6 2 | 5+6 2 | 5+6 2 | 3+4 2 | 3+3 2 | 3+3 2 |
| | 3+4 2 | 3+4 2 | 3+4 2 | 3+4 2 | 2+2 2 | 2+2 2 |
| | 5+6 2 | 5+6 2 | 3+4 2 | 5+6 2 | 3+3 2 | 3+3 2 |
| | 3+4 2 | 5+6 2 | 3+4 2 | 3+4 2 | 2+2 2 | 2+2 2 |
| | 5+6 2 | 3+4 2 | 5+6 2 | 5+6 2 | 3+3 2 | 3+3 2 |
| | 3+4 2 | 5+6 2 | 5+6 2 | 3+4 2 | 2+2 2 | 2+2 2 |
| | 5+6 2 | 3+4 2 | 3+4 2 | 5+6 2 | 3+3 2 | 3+3 2 |
| | 5+6 2 | 5+6 2 | 3+4 2 | 3+4 2 | 2+2 2 | 2+2 2 |
| | 5+6 2 | 5+6 2 | 5+6 2 | 5+6 2 | 3+3 2 | 3+3 2 |
| | 3+4 2 | 3+4 2 | 3+4 2 | 3+4 2 | 2+2 2 | 2+2 2 |
| | 3+4 2 | 5+6 2 | 5+6 2 | 5+6 2 | 2+2 2 | 2+2 2 |
| | 5+6 2 | 3+4 2 | 5+6 2 | 5+6 2 | 3+3 2 | 3+3 2 |
| | 3+4 2 | 5+6 2 | 5+6 2 | 3+4 2 | 2+2 2 | 2+2 2 |
| | 3+4 2 | 3+4 2 | 3+4 2 | 5+6 2 | 3+3 2 | 3+3 2 |
| 12.2.17 | 646757 | | 46 | | Ni3Sn4 | |
| | -A | -GM | -M | -Y | -L | -V |
| | 5+6 2 | 3+4 2 | 5+6 2 | 5+6 2 | 3+3 2 | 2+2 2 |
| | 3+4 2 | 3+4 2 | 3+4 2 | 3+4 2 | 2+2 2 | 3+3 2 |
| | 3+4 2 | 5+6 2 | 5+6 2 | 3+4 2 | 3+3 2 | 2+2 2 |
| | 5+6 2 | 5+6 2 | 3+4 2 | 5+6 2 | 2+2 2 | 3+3 2 |
| | 5+6 2 | 3+4 2 | 5+6 2 | 3+4 2 | 2+2 2 | 3+3 2 |
| | 3+4 2 | 3+4 2 | 3+4 2 | 5+6 2 | 3+3 2 | 2+2 2 |
| | 3+4 2 | 5+6 2 | 5+6 2 | 3+4 2 | 3+3 2 | 3+3 2 |
| | 3+4 2 | 5+6 2 | 3+4 2 | 3+4 2 | 2+2 2 | 3+3 2 |
| | 5+6 2 | 5+6 2 | 5+6 2 | 5+6 2 | 2+2 2 | 2+2 2 |
| | 3+4 2 | 3+4 2 | 3+4 2 | 5+6 2 | 3+3 2 | 2+2 2 |
| | 3+4 2 | 3+4 2 | 3+4 2 | 3+4 2 | 3+3 2 | 3+3 2 |
| | 5+6 2 | 3+4 2 | 3+4 2 | 5+6 2 | 2+2 2 | 2+2 2 |
| | 3+4 2 | 3+4 2 | 5+6 2 | 3+4 2 | 2+2 2 | 3+3 2 |
| | 3+4 2 | 3+4 2 | 3+4 2 | 3+4 2 | 3+3 2 | 2+2 2 |
| | 5+6 2 | 5+6 2 | 3+4 2 | 5+6 2 | 3+3 2 | 3+3 2 |
| | 5+6 2 | 3+4 2 | 5+6 2 | 5+6 2 | 3+3 2 | 2+2 2 |
| | 3+4 2 | 5+6 2 | 5+6 2 | 3+4 2 | 2+2 2 | 3+3 2 |
| | 3+4 2 | 3+4 2 | 3+4 2 | 3+4 2 | 3+3 2 | 3+3 2 |
| | 3+4 2 | 5+6 2 | 3+4 2 | 3+4 2 | 3+3 2 | 2+2 2 |
| | 5+6 2 | 3+4 2 | 3+4 2 | 3+4 2 | 3+3 2 | 3+3 2 |
| | 3+4 2 | 3+4 2 | 3+4 2 | 5+6 2 | 2+2 2 | 3+3 2 |
| | 5+6 2 | 3+4 2 | 5+6 2 | 3+4 2 | 3+3 2 | 2+2 2 |
| | 5+6 2 | 3+4 2 | 3+4 2 | 5+6 2 | 3+3 2 | 3+3 2 |
| 12.2.18 | 105364 | | 46 | | Ni3Sn4 | |
| | -A | -GM | -M | -Y | -L | -V |
| | 5+6 2 | 3+4 2 | 5+6 2 | 5+6 2 | 3+3 2 | 2+2 2 |
| | 3+4 2 | 3+4 2 | 3+4 2 | 3+4 2 | 2+2 2 | 3+3 2 |
| | 3+4 2 | 5+6 2 | 5+6 2 | 5+6 2 | 3+3 2 | 2+2 2 |
| | 5+6 2 | 5+6 2 | 3+4 2 | 3+4 2 | 2+2 2 | 3+3 2 |
| | 5+6 2 | 3+4 2 | 3+4 2 | 3+4 2 | 2+2 2 | 3+3 2 |
| | 3+4 2 | 3+4 2 | 5+6 2 | 3+4 2 | 3+3 2 | 3+3 2 |
| | 3+4 2 | 5+6 2 | 5+6 2 | 3+4 2 | 3+3 2 | 2+2 2 |
| | 3+4 2 | 5+6 2 | 3+4 2 | 5+6 2 | 2+2 2 | 3+3 2 |
| | 3+4 2 | 5+6 2 | 3+4 2 | 5+6 2 | 2+2 2 | 2+2 2 |
| | 5+6 2 | 3+4 2 | 3+4 2 | 3+4 2 | 3+3 2 | 3+3 2 |

Continued on next column

**TABLE XXIV– continued**

| TQC | ICSD | | NoE | | SF | |
|---|---|---|---|---|---|---|
| | 3+4 2 | 3+4 2 | 5+6 2 | 5+6 2 | 3+3 2 | 2+2 2 |
| | 3+4 2 | 3+4 2 | 3+4 2 | 3+4 2 | 3+3 2 | 3+3 2 |
| | 5+6 2 | 3+4 2 | 3+4 2 | 5+6 2 | 2+2 2 | 2+2 2 |
| | 3+4 2 | 3+4 2 | 5+6 2 | 3+4 2 | 3+3 2 | 2+2 2 |
| | 3+4 2 | 3+4 2 | 3+4 2 | 5+6 2 | 2+2 2 | 3+3 2 |
| | 5+6 2 | 5+6 2 | 5+6 2 | 3+4 2 | 3+3 2 | 3+3 2 |
| | 5+6 2 | 3+4 2 | 5+6 2 | 5+6 2 | 3+3 2 | 2+2 2 |
| | 3+4 2 | 5+6 2 | 3+4 2 | 3+4 2 | 3+3 2 | 3+3 2 |
| | 3+4 2 | 5+6 2 | 3+4 2 | 3+4 2 | 2+2 2 | 2+2 2 |
| | 5+6 2 | 3+4 2 | 3+4 2 | 3+4 2 | 3+3 2 | 3+3 2 |
| | 3+4 2 | 3+4 2 | 3+4 2 | 5+6 2 | 2+2 2 | 3+3 2 |
| | 5+6 2 | 3+4 2 | 5+6 2 | 3+4 2 | 3+3 2 | 2+2 2 |
| | 5+6 2 | 3+4 2 | 3+4 2 | 5+6 2 | 3+3 2 | 3+3 2 |
| 12.2.19 | 105363 | | 46 | | Ni3Sn4 | |
| | -A | -GM | -M | -Y | -L | -V |
| | 5+6 2 | 3+4 2 | 5+6 2 | 5+6 2 | 3+3 2 | 2+2 2 |
| | 3+4 2 | 3+4 2 | 3+4 2 | 3+4 2 | 2+2 2 | 3+3 2 |
| | 3+4 2 | 5+6 2 | 5+6 2 | 3+4 2 | 3+3 2 | 2+2 2 |
| | 5+6 2 | 5+6 2 | 3+4 2 | 5+6 2 | 2+2 2 | 3+3 2 |
| | 5+6 2 | 3+4 2 | 5+6 2 | 3+4 2 | 2+2 2 | 3+3 2 |
| | 3+4 2 | 3+4 2 | 3+4 2 | 3+4 2 | 3+3 2 | 2+2 2 |
| | 3+4 2 | 5+6 2 | 5+6 2 | 5+6 2 | 3+3 2 | 3+3 2 |
| | 3+4 2 | 5+6 2 | 3+4 2 | 3+4 2 | 2+2 2 | 3+3 2 |
| | 5+6 2 | 5+6 2 | 3+4 2 | 5+6 2 | 2+2 2 | 2+2 2 |
| | 3+4 2 | 3+4 2 | 5+6 2 | 5+6 2 | 3+3 2 | 2+2 2 |
| | 3+4 2 | 3+4 2 | 3+4 2 | 3+4 2 | 3+3 2 | 3+3 2 |
| | 5+6 2 | 3+4 2 | 3+4 2 | 5+6 2 | 2+2 2 | 2+2 2 |
| | 3+4 2 | 3+4 2 | 5+6 2 | 3+4 2 | 3+3 2 | 3+3 2 |
| | 3+4 2 | 3+4 2 | 3+4 2 | 3+4 2 | 2+2 2 | 2+2 2 |
| | 5+6 2 | 5+6 2 | 5+6 2 | 5+6 2 | 3+3 2 | 3+3 2 |
| | 5+6 2 | 3+4 2 | 3+4 2 | 5+6 2 | 3+3 2 | 2+2 2 |
| | 3+4 2 | 5+6 2 | 5+6 2 | 3+4 2 | 2+2 2 | 3+3 2 |
| | 3+4 2 | 3+4 2 | 3+4 2 | 3+4 2 | 3+3 2 | 3+3 2 |
| | 3+4 2 | 5+6 2 | 3+4 2 | 3+4 2 | 3+3 2 | 2+2 2 |
| | 5+6 2 | 3+4 2 | 3+4 2 | 3+4 2 | 3+3 2 | 3+3 2 |
| | 3+4 2 | 3+4 2 | 3+4 2 | 5+6 2 | 2+2 2 | 3+3 2 |
| | 5+6 2 | 3+4 2 | 5+6 2 | 3+4 2 | 3+3 2 | 2+2 2 |
| | 5+6 2 | 3+4 2 | 3+4 2 | 5+6 2 | 3+3 2 | 3+3 2 |
| 12.2.20 | 30433 | | 72 | | Ni2O7P2 | |
| | -A | -GM | -M | -Y | -L | -V |
| | 3+4 2 | 3+4 2 | 3+4 2 | 3+4 2 | 3+3 2 | 3+3 2 |
| | 5+6 2 | 5+6 2 | 5+6 2 | 5+6 2 | 2+2 2 | 2+2 2 |
| | 3+4 2 | 3+4 2 | 3+4 2 | 3+4 2 | 3+3 2 | 3+3 2 |
| | 5+6 2 | 3+4 2 | 5+6 2 | 5+6 2 | 3+3 2 | 2+2 2 |
| | 3+4 2 | 3+4 2 | 3+4 2 | 3+4 2 | 2+2 2 | 2+2 2 |
| | 5+6 2 | 5+6 2 | 5+6 2 | 5+6 2 | 3+3 2 | 3+3 2 |
| | 3+4 2 | 5+6 2 | 3+4 2 | 3+4 2 | 2+2 2 | 3+3 2 |
| | 5+6 2 | 5+6 2 | 5+6 2 | 5+6 2 | 2+2 2 | 2+2 2 |
| | 5+6 2 | 3+4 2 | 5+6 2 | 3+4 2 | 3+3 2 | 3+3 2 |
| | 5+6 2 | 5+6 2 | 3+4 2 | 5+6 2 | 2+2 2 | 2+2 2 |
| | 3+4 2 | 3+4 2 | 3+4 2 | 5+6 2 | 3+3 2 | 2+2 2 |
| | 5+6 2 | 5+6 2 | 5+6 2 | 5+6 2 | 2+2 2 | 2+2 2 |
| | 3+4 2 | 5+6 2 | 5+6 2 | 3+4 2 | 2+2 2 | 3+3 2 |
| | 5+6 2 | 3+4 2 | 3+4 2 | 3+4 2 | 3+3 2 | 3+3 2 |
| | 3+4 2 | 5+6 2 | 3+4 2 | 5+6 2 | 2+2 2 | 2+2 2 |
| | 5+6 2 | 3+4 2 | 5+6 2 | 3+4 2 | 3+3 2 | 2+2 2 |
| | 3+4 2 | 5+6 2 | 5+6 2 | 5+6 2 | 2+2 2 | 3+3 2 |
| | 3+4 2 | 3+4 2 | 5+6 2 | 5+6 2 | 2+2 2 | 2+2 2 |
| | 5+6 2 | 3+4 2 | 3+4 2 | 3+4 2 | 2+2 2 | 3+3 2 |
| | 5+6 2 | 5+6 2 | 5+6 2 | 5+6 2 | 3+3 2 | 2+2 2 |
| | 3+4 2 | 3+4 2 | 5+6 2 | 3+4 2 | 3+3 2 | 3+3 2 |

Continued on next column

**TABLE XXIV– continued**

| TQC | ICSD | NoE | SF | | | |
|---|---|---|---|---|---|---|
| | 3+4 2 | 5+6 2 | 3+4 2 | 3+4 2 | 2+2 2 | 2+2 2 |
| | 5+6 2 | 3+4 2 | 3+4 2 | 3+4 2 | 3+3 2 | 3+3 2 |
| | 5+6 2 | 5+6 2 | 3+4 2 | 5+6 2 | 2+2 2 | 2+2 2 |
| | 3+4 2 | 5+6 2 | 5+6 2 | 5+6 2 | 3+3 2 | 3+3 2 |
| | 3+4 2 | 3+4 2 | 3+4 2 | 5+6 2 | 3+3 2 | 2+2 2 |
| | 5+6 2 | 5+6 2 | 5+6 2 | 5+6 2 | 2+2 2 | 2+2 2 |
| | 5+6 2 | 5+6 2 | 5+6 2 | 3+4 2 | 2+2 2 | 3+3 2 |
| | 5+6 2 | 3+4 2 | 5+6 2 | 3+4 2 | 2+2 2 | 3+3 2 |
| | 3+4 2 | 5+6 2 | 3+4 2 | 5+6 2 | 3+3 2 | 2+2 2 |
| | 5+6 2 | 3+4 2 | 3+4 2 | 3+4 2 | 3+3 2 | 3+3 2 |
| | 3+4 2 | 5+6 2 | 5+6 2 | 5+6 2 | 2+2 2 | 2+2 2 |
| | 3+4 2 | 3+4 2 | 3+4 2 | 3+4 2 | 3+3 2 | 2+2 2 |
| | 5+6 2 | 5+6 2 | 5+6 2 | 5+6 2 | 2+2 2 | 3+3 2 |
| | 3+4 2 | 5+6 2 | 3+4 2 | 3+4 2 | 3+3 2 | 3+3 2 |
| | 5+6 2 | 5+6 2 | 5+6 2 | 5+6 2 | 2+2 2 | 2+2 2 |
| 12.2.21 | 646556 | 44 | Ni1Se4V2 | | | |
| | -A | -GM | -M | -Y | -L | -V |
| | 3+4 2 | 3+4 2 | 5+6 2 | 5+6 2 | 2+2 2 | 3+3 2 |
| | 5+6 2 | 5+6 2 | 3+4 2 | 3+4 2 | 2+2 2 | 2+2 2 |
| | 3+4 2 | 3+4 2 | 5+6 2 | 5+6 2 | 3+3 2 | 3+3 2 |
| | 5+6 2 | 5+6 2 | 3+4 2 | 3+4 2 | 3+3 2 | 2+2 2 |
| | 3+4 2 | 3+4 2 | 3+4 2 | 3+4 2 | 3+3 2 | 3+3 2 |
| | 5+6 2 | 5+6 2 | 5+6 2 | 5+6 2 | 2+2 2 | 2+2 2 |
| | 3+4 2 | 3+4 2 | 5+6 2 | 3+4 2 | 3+3 2 | 2+2 2 |
| | 5+6 2 | 3+4 2 | 5+6 2 | 5+6 2 | 2+2 2 | 3+3 2 |
| | 3+4 2 | 5+6 2 | 3+4 2 | 5+6 2 | 3+3 2 | 3+3 2 |
| | 5+6 2 | 3+4 2 | 3+4 2 | 3+4 2 | 2+2 2 | 3+3 2 |
| | 3+4 2 | 5+6 2 | 3+4 2 | 3+4 2 | 3+3 2 | 2+2 2 |
| | 3+4 2 | 3+4 2 | 5+6 2 | 5+6 2 | 2+2 2 | 2+2 2 |
| | 3+4 2 | 3+4 2 | 3+4 2 | 5+6 2 | 2+2 2 | 3+3 2 |
| | 5+6 2 | 3+4 2 | 3+4 2 | 3+4 2 | 2+2 2 | 3+3 2 |
| | 3+4 2 | 3+4 2 | 3+4 2 | 3+4 2 | 3+3 2 | 2+2 2 |
| | 5+6 2 | 3+4 2 | 5+6 2 | 5+6 2 | 3+3 2 | 2+2 2 |
| | 5+6 2 | 5+6 2 | 3+4 2 | 3+4 2 | 3+3 2 | 3+3 2 |
| | 3+4 2 | 5+6 2 | 5+6 2 | 3+4 2 | 3+3 2 | 3+3 2 |
| | 3+4 2 | 3+4 2 | 3+4 2 | 3+4 2 | 3+3 2 | 3+3 2 |
| | 3+4 2 | 3+4 2 | 5+6 2 | 3+4 2 | 3+3 2 | 3+3 2 |
| | 3+4 2 | 5+6 2 | 3+4 2 | 5+6 2 | 2+2 2 | 3+3 2 |
| | 5+6 2 | 5+6 2 | 3+4 2 | 5+6 2 | 3+3 2 | 2+2 2 |
| 12.2.22 | 646555 | 44 | Ni1Se4V2 | | | |
| | -A | -GM | -M | -Y | -L | -V |
| | 3+4 2 | 3+4 2 | 5+6 2 | 5+6 2 | 2+2 2 | 3+3 2 |
| | 5+6 2 | 5+6 2 | 3+4 2 | 3+4 2 | 2+2 2 | 2+2 2 |
| | 3+4 2 | 3+4 2 | 5+6 2 | 5+6 2 | 3+3 2 | 3+3 2 |
| | 5+6 2 | 5+6 2 | 3+4 2 | 3+4 2 | 3+3 2 | 2+2 2 |
| | 3+4 2 | 3+4 2 | 3+4 2 | 3+4 2 | 3+3 2 | 3+3 2 |
| | 5+6 2 | 5+6 2 | 5+6 2 | 5+6 2 | 2+2 2 | 2+2 2 |
| | 3+4 2 | 3+4 2 | 5+6 2 | 3+4 2 | 3+3 2 | 2+2 2 |
| | 5+6 2 | 3+4 2 | 5+6 2 | 5+6 2 | 2+2 2 | 3+3 2 |
| | 3+4 2 | 5+6 2 | 3+4 2 | 5+6 2 | 3+3 2 | 3+3 2 |
| | 5+6 2 | 3+4 2 | 3+4 2 | 3+4 2 | 2+2 2 | 3+3 2 |
| | 3+4 2 | 5+6 2 | 3+4 2 | 3+4 2 | 3+3 2 | 2+2 2 |
| | 3+4 2 | 3+4 2 | 5+6 2 | 5+6 2 | 2+2 2 | 2+2 2 |
| | 3+4 2 | 3+4 2 | 3+4 2 | 5+6 2 | 2+2 2 | 3+3 2 |
| | 5+6 2 | 3+4 2 | 3+4 2 | 3+4 2 | 2+2 2 | 3+3 2 |
| | 3+4 2 | 3+4 2 | 3+4 2 | 3+4 2 | 3+3 2 | 2+2 2 |
| | 5+6 2 | 3+4 2 | 5+6 2 | 5+6 2 | 3+3 2 | 2+2 2 |
| | 5+6 2 | 5+6 2 | 3+4 2 | 3+4 2 | 3+3 2 | 3+3 2 |
| | 3+4 2 | 5+6 2 | 5+6 2 | 3+4 2 | 3+3 2 | 3+3 2 |
| | 3+4 2 | 3+4 2 | 3+4 2 | 3+4 2 | 3+3 2 | 3+3 2 |
| | 3+4 2 | 3+4 2 | 5+6 2 | 3+4 2 | 3+3 2 | 3+3 2 |

Continued on next column

**TABLE XXIV– continued**

| TQC | ICSD | NoE | SF | | | |
|---|---|---|---|---|---|---|
| | 3+4 2 | 5+6 2 | 3+4 2 | 5+6 2 | 2+2 2 | 3+3 2 |
| | 5+6 2 | 5+6 2 | 3+4 2 | 5+6 2 | 3+3 2 | 2+2 2 |
| 12.2.23 | 23970 | 44 | Ni1Se4V2 | | | |
| | -A | -GM | -M | -Y | -L | -V |
| | 3+4 2 | 3+4 2 | 5+6 2 | 5+6 2 | 2+2 2 | 3+3 2 |
| | 5+6 2 | 5+6 2 | 3+4 2 | 3+4 2 | 2+2 2 | 2+2 2 |
| | 3+4 2 | 3+4 2 | 5+6 2 | 5+6 2 | 3+3 2 | 3+3 2 |
| | 5+6 2 | 5+6 2 | 3+4 2 | 3+4 2 | 3+3 2 | 2+2 2 |
| | 3+4 2 | 3+4 2 | 3+4 2 | 3+4 2 | 3+3 2 | 3+3 2 |
| | 5+6 2 | 5+6 2 | 5+6 2 | 5+6 2 | 2+2 2 | 2+2 2 |
| | 3+4 2 | 3+4 2 | 5+6 2 | 3+4 2 | 3+3 2 | 2+2 2 |
| | 5+6 2 | 3+4 2 | 5+6 2 | 5+6 2 | 2+2 2 | 3+3 2 |
| | 3+4 2 | 5+6 2 | 3+4 2 | 5+6 2 | 3+3 2 | 3+3 2 |
| | 5+6 2 | 3+4 2 | 3+4 2 | 3+4 2 | 2+2 2 | 3+3 2 |
| | 3+4 2 | 5+6 2 | 3+4 2 | 3+4 2 | 3+3 2 | 2+2 2 |
| | 3+4 2 | 3+4 2 | 5+6 2 | 5+6 2 | 2+2 2 | 2+2 2 |
| | 3+4 2 | 3+4 2 | 3+4 2 | 5+6 2 | 2+2 2 | 3+3 2 |
| | 5+6 2 | 3+4 2 | 3+4 2 | 3+4 2 | 2+2 2 | 3+3 2 |
| | 3+4 2 | 3+4 2 | 3+4 2 | 3+4 2 | 3+3 2 | 2+2 2 |
| | 5+6 2 | 3+4 2 | 5+6 2 | 5+6 2 | 3+3 2 | 2+2 2 |
| | 5+6 2 | 5+6 2 | 3+4 2 | 3+4 2 | 3+3 2 | 3+3 2 |
| | 3+4 2 | 5+6 2 | 5+6 2 | 3+4 2 | 3+3 2 | 3+3 2 |
| | 3+4 2 | 3+4 2 | 3+4 2 | 3+4 2 | 3+3 2 | 3+3 2 |
| | 3+4 2 | 3+4 2 | 5+6 2 | 3+4 2 | 3+3 2 | 3+3 2 |
| | 3+4 2 | 5+6 2 | 3+4 2 | 5+6 2 | 2+2 2 | 3+3 2 |
| | 5+6 2 | 5+6 2 | 3+4 2 | 5+6 2 | 3+3 2 | 2+2 2 |
| 12.2.24 | 35141 | 78 | Ni1S8V4 | | | |
| | -A | -GM | -M | -Y | -L | -V |
| | 5+6 2 | 3+4 2 | 5+6 2 | 3+4 2 | 2+2 2 | 3+3 2 |
| | 3+4 2 | 3+4 2 | 3+4 2 | 5+6 2 | 3+3 2 | 2+2 2 |
| | 5+6 2 | 5+6 2 | 5+6 2 | 3+4 2 | 2+2 2 | 3+3 2 |
| | 5+6 2 | 5+6 2 | 5+6 2 | 5+6 2 | 3+3 2 | 2+2 2 |
| | 3+4 2 | 3+4 2 | 3+4 2 | 3+4 2 | 2+2 2 | 3+3 2 |
| | 3+4 2 | 5+6 2 | 5+6 2 | 5+6 2 | 3+3 2 | 3+3 2 |
| | 5+6 2 | 3+4 2 | 3+4 2 | 3+4 2 | 2+2 2 | 2+2 2 |
| | 3+4 2 | 5+6 2 | 3+4 2 | 5+6 2 | 3+3 2 | 2+2 2 |
| | 3+4 2 | 3+4 2 | 3+4 2 | 3+4 2 | 3+3 2 | 3+3 2 |
| | 5+6 2 | 5+6 2 | 3+4 2 | 5+6 2 | 2+2 2 | 2+2 2 |
| | 5+6 2 | 5+6 2 | 5+6 2 | 5+6 2 | 3+3 2 | 3+3 2 |
| | 3+4 2 | 3+4 2 | 3+4 2 | 3+4 2 | 2+2 2 | 2+2 2 |
| | 3+4 2 | 3+4 2 | 5+6 2 | 5+6 2 | 3+3 2 | 2+2 2 |
| | 5+6 2 | 3+4 2 | 5+6 2 | 3+4 2 | 2+2 2 | 3+3 2 |
| | 3+4 2 | 5+6 2 | 5+6 2 | 5+6 2 | 3+3 2 | 3+3 2 |
| | 3+4 2 | 3+4 2 | 3+4 2 | 3+4 2 | 2+2 2 | 2+2 2 |
| | 5+6 2 | 3+4 2 | 5+6 2 | 5+6 2 | 2+2 2 | 3+3 2 |
| | 5+6 2 | 5+6 2 | 3+4 2 | 3+4 2 | 3+3 2 | 3+3 2 |
| | 3+4 2 | 5+6 2 | 5+6 2 | 3+4 2 | 3+3 2 | 2+2 2 |
| | 3+4 2 | 5+6 2 | 5+6 2 | 5+6 2 | 2+2 2 | 3+3 2 |
| | 5+6 2 | 5+6 2 | 3+4 2 | 3+4 2 | 2+2 2 | 2+2 2 |
| | 3+4 2 | 3+4 2 | 5+6 2 | 3+4 2 | 3+3 2 | 3+3 2 |
| | 5+6 2 | 3+4 2 | 3+4 2 | 5+6 2 | 2+2 2 | 2+2 2 |
| | 3+4 2 | 5+6 2 | 3+4 2 | 5+6 2 | 3+3 2 | 2+2 2 |
| | 3+4 2 | 3+4 2 | 3+4 2 | 5+6 2 | 3+3 2 | 3+3 2 |
| | 5+6 2 | 3+4 2 | 5+6 2 | 5+6 2 | 3+3 2 | 3+3 2 |
| | 3+4 2 | 5+6 2 | 5+6 2 | 3+4 2 | 2+2 2 | 3+3 2 |
| | 5+6 2 | 3+4 2 | 3+4 2 | 5+6 2 | 2+2 2 | 2+2 2 |
| | 5+6 2 | 3+4 2 | 3+4 2 | 3+4 2 | 3+3 2 | 2+2 2 |
| | 3+4 2 | 3+4 2 | 5+6 2 | 3+4 2 | 2+2 2 | 2+2 2 |
| | 5+6 2 | 5+6 2 | 5+6 2 | 3+4 2 | 3+3 2 | 3+3 2 |
| | 5+6 2 | 3+4 2 | 3+4 2 | 3+4 2 | 3+3 2 | 3+3 2 |
| | 3+4 2 | 3+4 2 | 3+4 2 | 3+4 2 | 2+2 2 | 3+3 2 |

Continued on next column

**TABLE XXIV– continued**

| TQC | ICSD | NoE | SF | | | |
|---|---|---|---|---|---|---|
| | 3+4 2 | 5+6 2 | 3+4 2 | 3+4 2 | 3+3 2 | 3+3 2 |
| | 3+4 2 | 5+6 2 | 3+4 2 | 5+6 2 | 3+3 2 | 2+2 2 |
| | 5+6 2 | 3+4 2 | 5+6 2 | 3+4 2 | 3+3 2 | 3+3 2 |
| | 3+4 2 | 5+6 2 | 3+4 2 | 3+4 2 | 2+2 2 | 3+3 2 |
| | 3+4 2 | 3+4 2 | 3+4 2 | 5+6 2 | 3+3 2 | 2+2 2 |
| | 3+4 2 | 3+4 2 | 5+6 2 | 3+4 2 | 3+3 2 | 3+3 2 |
| 12.2.25 | 646403 | 44 | Ni1S4V2 | | | |
| | -A | -GM | -M | -Y | -L | -V |
| | 3+4 2 | 3+4 2 | 5+6 2 | 5+6 2 | 2+2 2 | 3+3 2 |
| | 5+6 2 | 5+6 2 | 3+4 2 | 3+4 2 | 2+2 2 | 2+2 2 |
| | 3+4 2 | 3+4 2 | 5+6 2 | 5+6 2 | 3+3 2 | 3+3 2 |
| | 5+6 2 | 5+6 2 | 3+4 2 | 3+4 2 | 3+3 2 | 2+2 2 |
| | 3+4 2 | 3+4 2 | 3+4 2 | 3+4 2 | 3+3 2 | 3+3 2 |
| | 5+6 2 | 5+6 2 | 5+6 2 | 5+6 2 | 2+2 2 | 2+2 2 |
| | 3+4 2 | 3+4 2 | 5+6 2 | 3+4 2 | 3+3 2 | 2+2 2 |
| | 5+6 2 | 3+4 2 | 5+6 2 | 5+6 2 | 2+2 2 | 3+3 2 |
| | 3+4 2 | 3+4 2 | 3+4 2 | 5+6 2 | 3+3 2 | 3+3 2 |
| | 3+4 2 | 5+6 2 | 3+4 2 | 3+4 2 | 2+2 2 | 3+3 2 |
| | 5+6 2 | 5+6 2 | 3+4 2 | 3+4 2 | 3+3 2 | 2+2 2 |
| | 3+4 2 | 3+4 2 | 5+6 2 | 5+6 2 | 2+2 2 | 2+2 2 |
| | 3+4 2 | 3+4 2 | 3+4 2 | 5+6 2 | 2+2 2 | 3+3 2 |
| | 5+6 2 | 3+4 2 | 3+4 2 | 3+4 2 | 2+2 2 | 3+3 2 |
| | 3+4 2 | 3+4 2 | 3+4 2 | 3+4 2 | 3+3 2 | 2+2 2 |
| | 5+6 2 | 3+4 2 | 5+6 2 | 5+6 2 | 3+3 2 | 2+2 2 |
| | 5+6 2 | 5+6 2 | 5+6 2 | 3+4 2 | 3+3 2 | 3+3 2 |
| | 3+4 2 | 5+6 2 | 3+4 2 | 3+4 2 | 3+3 2 | 3+3 2 |
| | 3+4 2 | 3+4 2 | 3+4 2 | 3+4 2 | 3+3 2 | 3+3 2 |
| | 3+4 2 | 5+6 2 | 5+6 2 | 3+4 2 | 3+3 2 | 3+3 2 |
| | 3+4 2 | 3+4 2 | 3+4 2 | 5+6 2 | 2+2 2 | 2+2 2 |
| | 5+6 2 | 5+6 2 | 3+4 2 | 5+6 2 | 3+3 2 | 3+3 2 |
| 12.2.26 | 35140 | 44 | Ni1S4V2 | | | |
| | -A | -GM | -M | -Y | -L | -V |
| | 3+4 2 | 3+4 2 | 5+6 2 | 5+6 2 | 2+2 2 | 3+3 2 |
| | 5+6 2 | 5+6 2 | 3+4 2 | 3+4 2 | 2+2 2 | 2+2 2 |
| | 3+4 2 | 3+4 2 | 5+6 2 | 5+6 2 | 3+3 2 | 3+3 2 |
| | 5+6 2 | 5+6 2 | 3+4 2 | 3+4 2 | 3+3 2 | 2+2 2 |
| | 3+4 2 | 3+4 2 | 3+4 2 | 3+4 2 | 3+3 2 | 3+3 2 |
| | 5+6 2 | 5+6 2 | 5+6 2 | 5+6 2 | 2+2 2 | 2+2 2 |
| | 3+4 2 | 3+4 2 | 5+6 2 | 3+4 2 | 3+3 2 | 2+2 2 |
| | 5+6 2 | 3+4 2 | 5+6 2 | 5+6 2 | 2+2 2 | 3+3 2 |
| | 3+4 2 | 5+6 2 | 3+4 2 | 3+4 2 | 3+3 2 | 3+3 2 |
| | 3+4 2 | 3+4 2 | 3+4 2 | 5+6 2 | 2+2 2 | 3+3 2 |
| | 5+6 2 | 5+6 2 | 3+4 2 | 3+4 2 | 3+3 2 | 2+2 2 |
| | 3+4 2 | 3+4 2 | 5+6 2 | 5+6 2 | 2+2 2 | 2+2 2 |
| | 3+4 2 | 3+4 2 | 3+4 2 | 5+6 2 | 2+2 2 | 3+3 2 |
| | 5+6 2 | 3+4 2 | 3+4 2 | 3+4 2 | 3+3 2 | 3+3 2 |
| | 3+4 2 | 3+4 2 | 3+4 2 | 3+4 2 | 2+2 2 | 2+2 2 |
| | 5+6 2 | 3+4 2 | 5+6 2 | 5+6 2 | 3+3 2 | 2+2 2 |
| | 5+6 2 | 5+6 2 | 5+6 2 | 3+4 2 | 3+3 2 | 3+3 2 |
| | 3+4 2 | 5+6 2 | 3+4 2 | 3+4 2 | 3+3 2 | 3+3 2 |
| | 3+4 2 | 5+6 2 | 3+4 2 | 3+4 2 | 3+3 2 | 3+3 2 |
| | 3+4 2 | 3+4 2 | 5+6 2 | 3+4 2 | 2+2 2 | 3+3 2 |
| | 3+4 2 | 3+4 2 | 3+4 2 | 5+6 2 | 3+3 2 | 2+2 2 |
| | 5+6 2 | 5+6 2 | 3+4 2 | 5+6 2 | 3+3 2 | 3+3 2 |
| 12.2.27 | 646394 | 42 | Ni1S4Ti2 | | | |
| | -A | -GM | -M | -Y | -L | -V |
| | 3+4 2 | 3+4 2 | 3+4 2 | 5+6 2 | 2+2 2 | 3+3 2 |
| | 5+6 2 | 5+6 2 | 5+6 2 | 3+4 2 | 3+3 2 | 2+2 2 |
| | 3+4 2 | 5+6 2 | 5+6 2 | 3+4 2 | 2+2 2 | 2+2 2 |
| | 5+6 2 | 3+4 2 | 3+4 2 | 5+6 2 | 3+3 2 | 3+3 2 |
| | 3+4 2 | 3+4 2 | 3+4 2 | 3+4 2 | 3+3 2 | 3+3 2 |

Continued on next column

**TABLE XXIV– continued**

| TQC | ICSD | NoE | SF | | | |
|---|---|---|---|---|---|---|
| | 5+6 2 | 5+6 2 | 5+6 2 | 5+6 2 | 2+2 2 | 2+2 2 |
| | 3+4 2 | 3+4 2 | 5+6 2 | 3+4 2 | 3+3 2 | 2+2 2 |
| | 5+6 2 | 3+4 2 | 3+4 2 | 5+6 2 | 2+2 2 | 3+3 2 |
| | 3+4 2 | 5+6 2 | 5+6 2 | 3+4 2 | 3+3 2 | 3+3 2 |
| | 3+4 2 | 3+4 2 | 3+4 2 | 5+6 2 | 2+2 2 | 3+3 2 |
| | 5+6 2 | 5+6 2 | 3+4 2 | 3+4 2 | 3+3 2 | 2+2 2 |
| | 3+4 2 | 3+4 2 | 5+6 2 | 5+6 2 | 2+2 2 | 2+2 2 |
| | 5+6 2 | 3+4 2 | 3+4 2 | 5+6 2 | 2+2 2 | 3+3 2 |
| | 3+4 2 | 3+4 2 | 3+4 2 | 5+6 2 | 3+3 2 | 2+2 2 |
| | 5+6 2 | 3+4 2 | 5+6 2 | 3+4 2 | 2+2 2 | 3+3 2 |
| | 5+6 2 | 3+4 2 | 3+4 2 | 3+4 2 | 3+3 2 | 2+2 2 |
| | 3+4 2 | 5+6 2 | 5+6 2 | 3+4 2 | 3+3 2 | 3+3 2 |
| | 3+4 2 | 5+6 2 | 3+4 2 | 3+4 2 | 3+3 2 | 3+3 2 |
| | 3+4 2 | 5+6 2 | 3+4 2 | 3+4 2 | 3+3 2 | 3+3 2 |
| | 3+4 2 | 3+4 2 | 5+6 2 | 3+4 2 | 3+3 2 | 3+3 2 |
| | 3+4 2 | 5+6 2 | 3+4 2 | 5+6 2 | 2+2 2 | 3+3 2 |
| 12.2.28 | 193869 | 42 | Ni0.5Se2Ti1 | | | |
| | -A | -GM | -M | -Y | -L | -V |
| | 3+4 2 | 3+4 2 | 3+4 2 | 5+6 2 | 2+2 2 | 2+2 2 |
| | 5+6 2 | 5+6 2 | 5+6 2 | 3+4 2 | 3+3 2 | 3+3 2 |
| | 5+6 2 | 5+6 2 | 5+6 2 | 5+6 2 | 2+2 2 | 3+3 2 |
| | 3+4 2 | 3+4 2 | 3+4 2 | 3+4 2 | 3+3 2 | 2+2 2 |
| | 3+4 2 | 3+4 2 | 3+4 2 | 3+4 2 | 3+3 2 | 3+3 2 |
| | 5+6 2 | 5+6 2 | 5+6 2 | 5+6 2 | 2+2 2 | 2+2 2 |
| | 3+4 2 | 3+4 2 | 3+4 2 | 3+4 2 | 3+3 2 | 3+3 2 |
| | 5+6 2 | 3+4 2 | 5+6 2 | 5+6 2 | 2+2 2 | 2+2 2 |
| | 5+6 2 | 3+4 2 | 3+4 2 | 3+4 2 | 3+3 2 | 3+3 2 |
| | 3+4 2 | 5+6 2 | 3+4 2 | 5+6 2 | 2+2 2 | 2+2 2 |
| | 3+4 2 | 5+6 2 | 5+6 2 | 3+4 2 | 3+3 2 | 3+3 2 |
| | 5+6 2 | 3+4 2 | 3+4 2 | 5+6 2 | 2+2 2 | 2+2 2 |
| | 3+4 2 | 3+4 2 | 5+6 2 | 5+6 2 | 2+2 2 | 3+3 2 |
| | 3+4 2 | 3+4 2 | 5+6 2 | 5+6 2 | 3+3 2 | 2+2 2 |
| | 3+4 2 | 3+4 2 | 3+4 2 | 3+4 2 | 3+3 2 | 3+3 2 |
| | 5+6 2 | 3+4 2 | 5+6 2 | 3+4 2 | 2+2 2 | 2+2 2 |
| | 3+4 2 | 5+6 2 | 3+4 2 | 3+4 2 | 3+3 2 | 3+3 2 |
| | 3+4 2 | 5+6 2 | 3+4 2 | 3+4 2 | 3+3 2 | 3+3 2 |
| | 5+6 2 | 3+4 2 | 3+4 2 | 3+4 2 | 3+3 2 | 3+3 2 |
| | 5+6 2 | 5+6 2 | 3+4 2 | 3+4 2 | 2+2 2 | 3+3 2 |
| | 3+4 2 | 5+6 2 | 3+4 2 | 5+6 2 | 3+3 2 | 3+3 2 |
| 12.2.29 | 73315 | 234 | Nb8Pt1Se20 | | | |
| | -A | -GM | -M | -Y | -L | -V |
| | 3+4 2 | 3+4 2 | 3+4 2 | 3+4 2 | 2+2 2 | 2+2 2 |
| | 5+6 2 | 5+6 2 | 5+6 2 | 5+6 2 | 3+3 2 | 3+3 2 |
| | 3+4 2 | 3+4 2 | 3+4 2 | 3+4 2 | 2+2 2 | 2+2 2 |
| | 5+6 2 | 5+6 2 | 5+6 2 | 5+6 2 | 3+3 2 | 3+3 2 |
| | 3+4 2 | 3+4 2 | 3+4 2 | 3+4 2 | 2+2 2 | 2+2 2 |
| | 5+6 2 | 5+6 2 | 5+6 2 | 5+6 2 | 3+3 2 | 3+3 2 |
| | 3+4 2 | 3+4 2 | 3+4 2 | 3+4 2 | 2+2 2 | 2+2 2 |
| | 5+6 2 | 5+6 2 | 5+6 2 | 5+6 2 | 3+3 2 | 3+3 2 |
| | 5+6 2 | 3+4 2 | 3+4 2 | 5+6 2 | 2+2 2 | 2+2 2 |
| | 3+4 2 | 5+6 2 | 5+6 2 | 3+4 2 | 3+3 2 | 3+3 2 |
| | 3+4 2 | 3+4 2 | 3+4 2 | 3+4 2 | 2+2 2 | 2+2 2 |
| | 5+6 2 | 5+6 2 | 5+6 2 | 5+6 2 | 3+3 2 | 3+3 2 |
| | 3+4 2 | 3+4 2 | 3+4 2 | 3+4 2 | 2+2 2 | 2+2 2 |
| | 5+6 2 | 5+6 2 | 5+6 2 | 5+6 2 | 3+3 2 | 3+3 2 |
| | 3+4 2 | 3+4 2 | 3+4 2 | 3+4 2 | 2+2 2 | 2+2 2 |
| | 5+6 2 | 5+6 2 | 5+6 2 | 5+6 2 | 3+3 2 | 3+3 2 |
| | 5+6 2 | 3+4 2 | 3+4 2 | 5+6 2 | 2+2 2 | 2+2 2 |
| | 3+4 2 | 5+6 2 | 5+6 2 | 3+4 2 | 3+3 2 | 3+3 2 |
| | 5+6 2 | 3+4 2 | 3+4 2 | 5+6 2 | 2+2 2 | 2+2 2 |
| | 3+4 2 | 5+6 2 | 5+6 2 | 3+4 2 | 3+3 2 | 3+3 2 |

Continued on next column

**TABLE XXIV– continued**

| TQC | ICSD NoE SF |
|---|---|
| | 3+4 2 3+4 2 3+4 2 3+4 2 2+2 2 2+2 2 |
| | 5+6 2 5+6 2 5+6 2 5+6 2 3+3 2 3+3 2 |
| | 3+4 2 3+4 2 3+4 2 3+4 2 2+2 2 2+2 2 |
| | 5+6 2 5+6 2 5+6 2 5+6 2 3+3 2 3+3 2 |
| | 3+4 2 3+4 2 3+4 2 3+4 2 2+2 2 2+2 2 |
| | 5+6 2 5+6 2 5+6 2 5+6 2 3+3 2 3+3 2 |
| | 3+4 2 3+4 2 3+4 2 3+4 2 2+2 2 2+2 2 |
| | 5+6 2 5+6 2 5+6 2 5+6 2 3+3 2 3+3 2 |
| | 3+4 2 3+4 2 3+4 2 3+4 2 2+2 2 2+2 2 |
| | 5+6 2 5+6 2 5+6 2 5+6 2 3+3 2 3+3 2 |
| | 3+4 2 3+4 2 3+4 2 3+4 2 2+2 2 2+2 2 |
| | 5+6 2 5+6 2 5+6 2 5+6 2 3+3 2 3+3 2 |
| | 5+6 2 3+4 2 3+4 2 5+6 2 3+3 2 2+2 2 |
| | 3+4 2 5+6 2 5+6 2 3+4 2 2+2 2 3+3 2 |
| | 3+4 2 3+4 2 3+4 2 5+6 2 2+2 2 2+2 2 |
| | 5+6 2 3+4 2 5+6 2 3+4 2 2+2 2 2+2 2 |
| | 3+4 2 5+6 2 3+4 2 3+4 2 3+3 2 3+3 2 |
| | 5+6 2 5+6 2 5+6 2 5+6 2 2+2 2 3+3 2 |
| | 5+6 2 3+4 2 3+4 2 5+6 2 3+3 2 2+2 2 |
| | 3+4 2 5+6 2 5+6 2 3+4 2 3+3 2 3+3 2 |
| | 5+6 2 3+4 2 3+4 2 5+6 2 2+2 2 2+2 2 |
| | 3+4 2 3+4 2 5+6 2 3+4 2 2+2 2 3+3 2 |
| | 3+4 2 5+6 2 3+4 2 5+6 2 3+3 2 2+2 2 |
| | 5+6 2 3+4 2 3+4 2 3+4 2 3+3 2 3+3 2 |
| | 5+6 2 5+6 2 5+6 2 3+4 2 2+2 2 2+2 2 |
| | 3+4 2 5+6 2 5+6 2 5+6 2 3+3 2 3+3 2 |
| | 5+6 2 3+4 2 3+4 2 3+4 2 2+2 2 2+2 2 |
| | 3+4 2 5+6 2 5+6 2 5+6 2 3+3 2 3+3 2 |
| | 3+4 2 3+4 2 5+6 2 5+6 2 2+2 2 2+2 2 |
| | 5+6 2 5+6 2 3+4 2 3+4 2 3+3 2 3+3 2 |
| | 3+4 2 5+6 2 5+6 2 3+4 2 3+3 2 2+2 2 |
| | 5+6 2 3+4 2 3+4 2 5+6 2 2+2 2 3+3 2 |
| | 3+4 2 3+4 2 3+4 2 3+4 2 2+2 2 3+3 2 |
| | 5+6 2 3+4 2 5+6 2 3+4 2 3+3 2 2+2 2 |
| | 3+4 2 5+6 2 3+4 2 5+6 2 2+2 2 2+2 2 |
| | 3+4 2 3+4 2 3+4 2 3+4 2 2+2 2 3+3 2 |
| | 5+6 2 3+4 2 5+6 2 3+4 2 3+3 2 2+2 2 |
| | 3+4 2 5+6 2 3+4 2 5+6 2 2+2 2 2+2 2 |
| | 5+6 2 3+4 2 3+4 2 5+6 2 3+3 2 3+3 2 |
| | 3+4 2 3+4 2 5+6 2 3+4 2 2+2 2 2+2 2 |
| | 3+4 2 5+6 2 3+4 2 5+6 2 2+2 2 2+2 2 |
| | 5+6 2 5+6 2 5+6 2 3+4 2 3+3 2 3+3 2 |
| | 5+6 2 3+4 2 5+6 2 5+6 2 3+3 2 2+2 2 |
| | 5+6 2 5+6 2 3+4 2 5+6 2 2+2 2 3+3 2 |
| | 3+4 2 3+4 2 3+4 2 3+4 2 2+2 2 2+2 2 |
| | 3+4 2 5+6 2 5+6 2 3+4 2 3+3 2 2+2 2 |
| | 5+6 2 3+4 2 3+4 2 3+4 2 2+2 2 3+3 2 |
| | 3+4 2 5+6 2 5+6 2 5+6 2 3+3 2 2+2 2 |
| | 5+6 2 3+4 2 3+4 2 3+4 2 2+2 2 3+3 2 |
| | 3+4 2 5+6 2 5+6 2 5+6 2 3+3 2 3+3 2 |
| | 5+6 2 3+4 2 3+4 2 5+6 2 2+2 2 2+2 2 |
| | 3+4 2 5+6 2 5+6 2 3+4 2 3+3 2 3+3 2 |
| | 3+4 2 3+4 2 3+4 2 5+6 2 2+2 2 2+2 2 |
| | 5+6 2 3+4 2 3+4 2 3+4 2 2+2 2 3+3 2 |
| | 5+6 2 3+4 2 5+6 2 3+4 2 3+3 2 2+2 2 |
| | 3+4 2 5+6 2 3+4 2 5+6 2 3+3 2 3+3 2 |
| | 3+4 2 5+6 2 5+6 2 3+4 2 2+2 2 2+2 2 |
| | 3+4 2 3+4 2 3+4 2 5+6 2 3+3 2 2+2 2 |
| | 5+6 2 3+4 2 3+4 2 3+4 2 2+2 2 3+3 2 |
| | 3+4 2 5+6 2 5+6 2 3+4 2 3+3 2 2+2 2 |
| | 5+6 2 3+4 2 3+4 2 5+6 2 2+2 2 3+3 2 |

Continued on next column

**TABLE XXIV– continued**

| TQC | ICSD NoE SF |
|---|---|
| | 5+6 2 5+6 2 5+6 2 5+6 2 3+3 2 2+2 2 |
| | 3+4 2 3+4 2 5+6 2 5+6 2 2+2 2 3+3 2 |
| | 5+6 2 5+6 2 3+4 2 3+4 2 3+3 2 2+2 2 |
| | 3+4 2 3+4 2 3+4 2 5+6 2 2+2 2 3+3 2 |
| | 5+6 2 5+6 2 5+6 2 3+4 2 3+3 2 3+3 2 |
| | 3+4 2 3+4 2 5+6 2 3+4 2 2+2 2 2+2 2 |
| | 5+6 2 5+6 2 3+4 2 3+4 2 3+3 2 3+3 2 |
| | 3+4 2 3+4 2 3+4 2 5+6 2 2+2 2 2+2 2 |
| | 5+6 2 5+6 2 5+6 2 5+6 2 3+3 2 3+3 2 |
| | 3+4 2 3+4 2 3+4 2 5+6 2 3+3 2 2+2 2 |
| | 5+6 2 3+4 2 5+6 2 3+4 2 2+2 2 3+3 2 |
| | 3+4 2 3+4 2 3+4 2 3+4 2 2+2 2 3+3 2 |
| | 5+6 2 5+6 2 3+4 2 5+6 2 2+2 2 2+2 2 |
| | 3+4 2 5+6 2 5+6 2 5+6 2 3+3 2 3+3 2 |
| | 5+6 2 3+4 2 5+6 2 3+4 2 3+3 2 2+2 2 |
| | 3+4 2 5+6 2 3+4 2 5+6 2 2+2 2 2+2 2 |
| | 5+6 2 3+4 2 5+6 2 3+4 2 3+3 2 3+3 2 |
| | 3+4 2 3+4 2 3+4 2 5+6 2 2+2 2 2+2 2 |
| | 3+4 2 5+6 2 5+6 2 3+4 2 3+3 2 2+2 2 |
| | 5+6 2 5+6 2 3+4 2 5+6 2 2+2 2 3+3 2 |
| | 5+6 2 3+4 2 5+6 2 3+4 2 3+3 2 3+3 2 |
| | 3+4 2 5+6 2 5+6 2 5+6 2 2+2 2 2+2 2 |
| | 3+4 2 5+6 2 3+4 2 3+4 2 3+3 2 3+3 2 |
| | 5+6 2 3+4 2 5+6 2 5+6 2 2+2 2 2+2 2 |
| | 3+4 2 5+6 2 5+6 2 5+6 2 3+3 2 3+3 2 |
| | 5+6 2 3+4 2 3+4 2 3+4 2 2+2 2 2+2 2 |
| | 3+4 2 5+6 2 3+4 2 3+4 2 3+3 2 3+3 2 |
| | 5+6 2 3+4 2 5+6 2 5+6 2 2+2 2 2+2 2 |
| | 5+6 2 5+6 2 5+6 2 3+4 2 2+2 2 2+2 2 |
| | 3+4 2 5+6 2 3+4 2 3+4 2 3+3 2 3+3 2 |
| | 3+4 2 3+4 2 5+6 2 3+4 2 2+2 2 2+2 2 |
| | 5+6 2 3+4 2 3+4 2 5+6 2 3+3 2 3+3 2 |
| | 3+4 2 5+6 2 5+6 2 3+4 2 2+2 2 2+2 2 |
| | 5+6 2 5+6 2 3+4 2 5+6 2 3+3 2 3+3 2 |
| | 5+6 2 3+4 2 3+4 2 5+6 2 2+2 2 2+2 2 |
| | 3+4 2 3+4 2 3+4 2 5+6 2 3+3 2 3+3 2 |
| 12.2.30 | 280016 88 Nb2Te3 |
| | -A -GM -M -Y -L -V |
| | 5+6 2 3+4 2 3+4 2 5+6 2 2+2 2 2+2 2 |
| | 3+4 2 3+4 2 5+6 2 5+6 2 3+3 2 3+3 2 |
| | 3+4 2 5+6 2 5+6 2 3+4 2 2+2 2 2+2 2 |
| | 5+6 2 5+6 2 3+4 2 3+4 2 3+3 2 3+3 2 |
| | 5+6 2 3+4 2 3+4 2 5+6 2 2+2 2 2+2 2 |
| | 3+4 2 3+4 2 5+6 2 5+6 2 3+3 2 3+3 2 |
| | 3+4 2 5+6 2 5+6 2 3+4 2 2+2 2 2+2 2 |
| | 5+6 2 5+6 2 3+4 2 3+4 2 3+3 2 3+3 2 |
| | 5+6 2 3+4 2 3+4 2 5+6 2 2+2 2 2+2 2 |
| | 3+4 2 3+4 2 5+6 2 5+6 2 3+3 2 3+3 2 |
| | 3+4 2 5+6 2 5+6 2 3+4 2 2+2 2 2+2 2 |
| | 3+4 2 3+4 2 3+4 2 5+6 2 3+3 2 3+3 2 |
| | 5+6 2 3+4 2 5+6 2 3+4 2 2+2 2 2+2 2 |
| | 5+6 2 5+6 2 3+4 2 5+6 2 3+3 2 3+3 2 |
| | 3+4 2 5+6 2 5+6 2 3+4 2 2+2 2 2+2 2 |
| | 5+6 2 5+6 2 3+4 2 3+4 2 3+3 2 3+3 2 |
| | 5+6 2 3+4 2 5+6 2 3+4 2 2+2 2 3+3 2 |
| | 3+4 2 3+4 2 3+4 2 5+6 2 3+3 2 3+3 2 |
| | 5+6 2 5+6 2 5+6 2 3+4 2 2+2 2 2+2 2 |
| | 3+4 2 5+6 2 5+6 2 5+6 2 3+3 2 3+3 2 |
| | 5+6 2 3+4 2 3+4 2 3+4 2 2+2 2 2+2 2 |
| | 3+4 2 5+6 2 3+4 2 5+6 2 3+3 2 2+2 2 |
| | 3+4 2 3+4 2 5+6 2 3+4 2 3+3 2 2+2 2 |

Continued on next column

**TABLE XXIV– continued**

| TQC | ICSD | | NoE | | SF | |
|---|---|---|---|---|---|---|
| | 5+6 2 | 5+6 2 | 3+4 2 | 3+4 2 | 2+2 2 | 3+3 2 |
| | 5+6 2 | 3+4 2 | 3+4 2 | 5+6 2 | 3+3 2 | 2+2 2 |
| | 3+4 2 | 5+6 2 | 5+6 2 | 5+6 2 | 2+2 2 | 3+3 2 |
| | 3+4 2 | 3+4 2 | 5+6 2 | 5+6 2 | 3+3 2 | 2+2 2 |
| | 5+6 2 | 5+6 2 | 3+4 2 | 3+4 2 | 2+2 2 | 3+3 2 |
| | 3+4 2 | 5+6 2 | 3+4 2 | 3+4 2 | 3+3 2 | 2+2 2 |
| | 5+6 2 | 3+4 2 | 5+6 2 | 3+4 2 | 2+2 2 | 3+3 2 |
| | 3+4 2 | 5+6 2 | 5+6 2 | 5+6 2 | 2+2 2 | 3+3 2 |
| | 5+6 2 | 3+4 2 | 3+4 2 | 3+4 2 | 3+3 2 | 3+3 2 |
| | 5+6 2 | 5+6 2 | 3+4 2 | 5+6 2 | 2+2 2 | 2+2 2 |
| | 3+4 2 | 3+4 2 | 5+6 2 | 3+4 2 | 3+3 2 | 3+3 2 |
| | 5+6 2 | 5+6 2 | 5+6 2 | 5+6 2 | 2+2 2 | 3+3 2 |
| | 5+6 2 | 3+4 2 | 3+4 2 | 3+4 2 | 3+3 2 | 2+2 2 |
| | 3+4 2 | 3+4 2 | 5+6 2 | 3+4 2 | 2+2 2 | 3+3 2 |
| | 5+6 2 | 5+6 2 | 3+4 2 | 5+6 2 | 3+3 2 | 2+2 2 |
| | 3+4 2 | 3+4 2 | 3+4 2 | 5+6 2 | 2+2 2 | 2+2 2 |
| | 3+4 2 | 5+6 2 | 3+4 2 | 5+6 2 | 2+2 2 | 2+2 2 |
| | 5+6 2 | 3+4 2 | 5+6 2 | 3+4 2 | 3+3 2 | 3+3 2 |
| | 3+4 2 | 3+4 2 | 5+6 2 | 5+6 2 | 3+3 2 | 3+3 2 |
| | 3+4 2 | 5+6 2 | 5+6 2 | 3+4 2 | 2+2 2 | 2+2 2 |
| | 5+6 2 | 3+4 2 | 3+4 2 | 5+6 2 | 3+3 2 | 2+2 2 |
| 12.2.31 | 192215 | | 132 | | Nb2Pd1S5 | |
| | -A | -GM | -M | -Y | -L | -V |
| | 3+4 2 | 3+4 2 | 3+4 2 | 3+4 2 | 2+2 2 | 2+2 2 |
| | 5+6 2 | 5+6 2 | 5+6 2 | 5+6 2 | 3+3 2 | 3+3 2 |
| | 3+4 2 | 3+4 2 | 3+4 2 | 3+4 2 | 2+2 2 | 2+2 2 |
| | 5+6 2 | 5+6 2 | 5+6 2 | 5+6 2 | 3+3 2 | 3+3 2 |
| | 5+6 2 | 3+4 2 | 3+4 2 | 5+6 2 | 2+2 2 | 2+2 2 |
| | 3+4 2 | 5+6 2 | 5+6 2 | 3+4 2 | 3+3 2 | 3+3 2 |
| | 3+4 2 | 3+4 2 | 3+4 2 | 3+4 2 | 2+2 2 | 2+2 2 |
| | 5+6 2 | 5+6 2 | 5+6 2 | 5+6 2 | 3+3 2 | 3+3 2 |
| | 5+6 2 | 3+4 2 | 3+4 2 | 3+4 2 | 2+2 2 | 2+2 2 |
| | 3+4 2 | 5+6 2 | 5+6 2 | 5+6 2 | 3+3 2 | 3+3 2 |
| | 5+6 2 | 3+4 2 | 3+4 2 | 5+6 2 | 2+2 2 | 2+2 2 |
| | 3+4 2 | 5+6 2 | 5+6 2 | 3+4 2 | 3+3 2 | 3+3 2 |
| | 3+4 2 | 3+4 2 | 3+4 2 | 3+4 2 | 2+2 2 | 2+2 2 |
| | 5+6 2 | 5+6 2 | 5+6 2 | 5+6 2 | 3+3 2 | 3+3 2 |
| | 3+4 2 | 3+4 2 | 3+4 2 | 3+4 2 | 2+2 2 | 2+2 2 |
| | 5+6 2 | 5+6 2 | 5+6 2 | 5+6 2 | 3+3 2 | 3+3 2 |
| | 5+6 2 | 3+4 2 | 3+4 2 | 5+6 2 | 2+2 2 | 3+3 2 |
| | 3+4 2 | 3+4 2 | 5+6 2 | 3+4 2 | 3+3 2 | 2+2 2 |
| | 5+6 2 | 5+6 2 | 3+4 2 | 3+4 2 | 3+3 2 | 3+3 2 |
| | 3+4 2 | 5+6 2 | 5+6 2 | 5+6 2 | 2+2 2 | 2+2 2 |
| | 3+4 2 | 3+4 2 | 3+4 2 | 5+6 2 | 3+3 2 | 3+3 2 |
| | 5+6 2 | 5+6 2 | 5+6 2 | 3+4 2 | 2+2 2 | 3+3 2 |
| | 5+6 2 | 3+4 2 | 3+4 2 | 3+4 2 | 3+3 2 | 3+3 2 |
| | 3+4 2 | 5+6 2 | 5+6 2 | 5+6 2 | 2+2 2 | 2+2 2 |
| | 3+4 2 | 3+4 2 | 3+4 2 | 3+4 2 | 3+3 2 | 2+2 2 |
| | 5+6 2 | 5+6 2 | 5+6 2 | 5+6 2 | 2+2 2 | 2+2 2 |
| | 5+6 2 | 3+4 2 | 5+6 2 | 3+4 2 | 2+2 2 | 3+3 2 |
| | 3+4 2 | 3+4 2 | 3+4 2 | 3+4 2 | 3+3 2 | 2+2 2 |
| | 5+6 2 | 5+6 2 | 5+6 2 | 5+6 2 | 2+2 2 | 3+3 2 |
| | 3+4 2 | 3+4 2 | 3+4 2 | 3+4 2 | 3+3 2 | 3+3 2 |
| | 5+6 2 | 3+4 2 | 3+4 2 | 5+6 2 | 2+2 2 | 2+2 2 |
| | 3+4 2 | 5+6 2 | 3+4 2 | 3+4 2 | 3+3 2 | 3+3 2 |
| | 5+6 2 | 3+4 2 | 5+6 2 | 3+4 2 | 2+2 2 | 3+3 2 |
| | 3+4 2 | 3+4 2 | 3+4 2 | 3+4 2 | 3+3 2 | 3+3 2 |
| | 3+4 2 | 5+6 2 | 5+6 2 | 5+6 2 | 3+3 2 | 2+2 2 |
| | 5+6 2 | 3+4 2 | 5+6 2 | 3+4 2 | 2+2 2 | 3+3 2 |
| | 5+6 2 | 3+4 2 | 3+4 2 | 5+6 2 | 2+2 2 | 2+2 2 |
| | 5+6 2 | 5+6 2 | 5+6 2 | 5+6 2 | 3+3 2 | 3+3 2 |

Continued on next column

**TABLE XXIV– continued**

| TQC | ICSD | | NoE | | SF | |
|---|---|---|---|---|---|---|
| | 3+4 2 | 3+4 2 | 3+4 2 | 3+4 2 | 3+3 2 | 2+2 2 |
| | 3+4 2 | 3+4 2 | 3+4 2 | 3+4 2 | 3+3 2 | 3+3 2 |
| | 5+6 2 | 5+6 2 | 5+6 2 | 3+4 2 | 2+2 2 | 2+2 2 |
| | 3+4 2 | 3+4 2 | 3+4 2 | 5+6 2 | 3+3 2 | 3+3 2 |
| | 3+4 2 | 3+4 2 | 5+6 2 | 3+4 2 | 3+3 2 | 2+2 2 |
| | 5+6 2 | 5+6 2 | 3+4 2 | 5+6 2 | 2+2 2 | 2+2 2 |
| | 3+4 2 | 3+4 2 | 5+6 2 | 5+6 2 | 3+3 2 | 3+3 2 |
| | 5+6 2 | 3+4 2 | 3+4 2 | 3+4 2 | 2+2 2 | 3+3 2 |
| | 3+4 2 | 3+4 2 | 3+4 2 | 3+4 2 | 2+2 2 | 2+2 2 |
| | 3+4 2 | 5+6 2 | 5+6 2 | 5+6 2 | 3+3 2 | 3+3 2 |
| | 5+6 2 | 3+4 2 | 5+6 2 | 3+4 2 | 2+2 2 | 2+2 2 |
| | 5+6 2 | 5+6 2 | 5+6 2 | 3+4 2 | 2+2 2 | 2+2 2 |
| | 5+6 2 | 5+6 2 | 5+6 2 | 5+6 2 | 3+3 2 | 3+3 2 |
| | 3+4 2 | 3+4 2 | 3+4 2 | 5+6 2 | 2+2 2 | 3+3 2 |
| | 5+6 2 | 5+6 2 | 3+4 2 | 3+4 2 | 3+3 2 | 3+3 2 |
| | 3+4 2 | 3+4 2 | 3+4 2 | 3+4 2 | 3+3 2 | 3+3 2 |
| | 5+6 2 | 5+6 2 | 5+6 2 | 3+4 2 | 2+2 2 | 2+2 2 |
| | 3+4 2 | 3+4 2 | 5+6 2 | 5+6 2 | 2+2 2 | 3+3 2 |
| | 3+4 2 | 3+4 2 | 3+4 2 | 3+4 2 | 3+3 2 | 2+2 2 |
| | 5+6 2 | 5+6 2 | 3+4 2 | 5+6 2 | 2+2 2 | 3+3 2 |
| | 5+6 2 | 5+6 2 | 5+6 2 | 3+4 2 | 2+2 2 | 2+2 2 |
| | 5+6 2 | 3+4 2 | 5+6 2 | 5+6 2 | 3+3 2 | 3+3 2 |
| | 3+4 2 | 5+6 2 | 5+6 2 | 5+6 2 | 2+2 2 | 2+2 2 |
| | 3+4 2 | 5+6 2 | 3+4 2 | 3+4 2 | 3+3 2 | 3+3 2 |
| | 3+4 2 | 3+4 2 | 5+6 2 | 3+4 2 | 3+3 2 | 3+3 2 |
| | 5+6 2 | 3+4 2 | 3+4 2 | 5+6 2 | 2+2 2 | 3+3 2 |
| | 5+6 2 | 3+4 2 | 3+4 2 | 3+4 2 | 2+2 2 | 3+3 2 |
| | 5+6 2 | 5+6 2 | 5+6 2 | 3+4 2 | 3+3 2 | 2+2 2 |
| 12.2.32 | 81059 | | 160 | | Mo1Ni1O4 | |
| | -A | -GM | -M | -Y | -L | -V |
| | 3+4 2 | 3+4 2 | 3+4 2 | 3+4 2 | 2+2 2 | 2+2 2 |
| | 5+6 2 | 5+6 2 | 5+6 2 | 5+6 2 | 3+3 2 | 3+3 2 |
| | 5+6 2 | 5+6 2 | 5+6 2 | 3+4 2 | 3+3 2 | 3+3 2 |
| | 3+4 2 | 3+4 2 | 3+4 2 | 5+6 2 | 2+2 2 | 2+2 2 |
| | 5+6 2 | 5+6 2 | 5+6 2 | 5+6 2 | 2+2 2 | 3+3 2 |
| | 3+4 2 | 3+4 2 | 3+4 2 | 5+6 2 | 3+3 2 | 2+2 2 |
| | 3+4 2 | 3+4 2 | 3+4 2 | 5+6 2 | 2+2 2 | 2+2 2 |
| | 5+6 2 | 3+4 2 | 5+6 2 | 3+4 2 | 3+3 2 | 3+3 2 |
| | 5+6 2 | 5+6 2 | 3+4 2 | 3+4 2 | 3+3 2 | 2+2 2 |
| | 3+4 2 | 5+6 2 | 5+6 2 | 3+4 2 | 2+2 2 | 3+3 2 |
| | 3+4 2 | 3+4 2 | 5+6 2 | 5+6 2 | 3+3 2 | 2+2 2 |
| | 5+6 2 | 5+6 2 | 3+4 2 | 5+6 2 | 2+2 2 | 3+3 2 |
| | 5+6 2 | 5+6 2 | 5+6 2 | 3+4 2 | 3+3 2 | 3+3 2 |
| | 3+4 2 | 3+4 2 | 3+4 2 | 3+4 2 | 2+2 2 | 2+2 2 |
| | 3+4 2 | 3+4 2 | 5+6 2 | 5+6 2 | 2+2 2 | 2+2 2 |
| | 5+6 2 | 5+6 2 | 3+4 2 | 3+4 2 | 3+3 2 | 3+3 2 |
| | 3+4 2 | 3+4 2 | 5+6 2 | 3+4 2 | 2+2 2 | 2+2 2 |
| | 5+6 2 | 3+4 2 | 3+4 2 | 5+6 2 | 3+3 2 | 3+3 2 |
| | 3+4 2 | 5+6 2 | 3+4 2 | 5+6 2 | 2+2 2 | 2+2 2 |
| | 5+6 2 | 5+6 2 | 3+4 2 | 3+4 2 | 2+2 2 | 3+3 2 |
| | 3+4 2 | 3+4 2 | 3+4 2 | 3+4 2 | 3+3 2 | 2+2 2 |
| | 5+6 2 | 5+6 2 | 5+6 2 | 5+6 2 | 3+3 2 | 3+3 2 |
| | 3+4 2 | 3+4 2 | 5+6 2 | 3+4 2 | 2+2 2 | 3+3 2 |
| | 3+4 2 | 5+6 2 | 5+6 2 | 3+4 2 | 2+2 2 | 2+2 2 |
| | 5+6 2 | 3+4 2 | 3+4 2 | 5+6 2 | 3+3 2 | 3+3 2 |
| | 5+6 2 | 3+4 2 | 3+4 2 | 3+4 2 | 3+3 2 | 2+2 2 |
| | 5+6 2 | 5+6 2 | 5+6 2 | 5+6 2 | 2+2 2 | 3+3 2 |
| | 3+4 2 | 5+6 2 | 5+6 2 | 5+6 2 | 3+3 2 | 2+2 2 |
| | 5+6 2 | 3+4 2 | 5+6 2 | 5+6 2 | 2+2 2 | 3+3 2 |
| | 3+4 2 | 5+6 2 | 3+4 2 | 3+4 2 | 3+3 2 | 2+2 2 |
| | 3+4 2 | 3+4 2 | 3+4 2 | 5+6 2 | 3+3 2 | 3+3 2 |

Continued on next column

**TABLE XXIV– continued**

| TQC | ICSD | NoE | SF | | | |
|---|---|---|---|---|---|---|
| | 5+6 2 | 5+6 2 | 5+6 2 | 3+4 2 | 3+3 2 | 2+2 2 |
| | 5+6 2 | 3+4 2 | 5+6 2 | 3+4 2 | 2+2 2 | 3+3 2 |
| | 3+4 2 | 5+6 2 | 3+4 2 | 5+6 2 | 2+2 2 | 2+2 2 |
| | 3+4 2 | 3+4 2 | 3+4 2 | 3+4 2 | 3+3 2 | 3+3 2 |
| | 3+4 2 | 3+4 2 | 5+6 2 | 3+4 2 | 2+2 2 | 2+2 2 |
| | 5+6 2 | 5+6 2 | 5+6 2 | 5+6 2 | 2+2 2 | 2+2 2 |
| | 5+6 2 | 3+4 2 | 5+6 2 | 5+6 2 | 3+3 2 | 2+2 2 |
| | 5+6 2 | 3+4 2 | 5+6 2 | 5+6 2 | 3+3 2 | 3+3 2 |
| | 5+6 2 | 5+6 2 | 3+4 2 | 3+4 2 | 2+2 2 | 3+3 2 |
| | 3+4 2 | 3+4 2 | 3+4 2 | 3+4 2 | 3+3 2 | 2+2 2 |
| | 5+6 2 | 5+6 2 | 3+4 2 | 3+4 2 | 2+2 2 | 3+3 2 |
| | 3+4 2 | 3+4 2 | 3+4 2 | 5+6 2 | 3+3 2 | 2+2 2 |
| | 3+4 2 | 5+6 2 | 5+6 2 | 5+6 2 | 2+2 2 | 3+3 2 |
| | 5+6 2 | 5+6 2 | 3+4 2 | 3+4 2 | 3+3 2 | 2+2 2 |
| | 3+4 2 | 3+4 2 | 5+6 2 | 5+6 2 | 2+2 2 | 2+2 2 |
| | 5+6 2 | 5+6 2 | 5+6 2 | 5+6 2 | 3+3 2 | 3+3 2 |
| | 3+4 2 | 3+4 2 | 3+4 2 | 5+6 2 | 2+2 2 | 2+2 2 |
| | 5+6 2 | 5+6 2 | 3+4 2 | 3+4 2 | 3+3 2 | 3+3 2 |
| | 3+4 2 | 3+4 2 | 5+6 2 | 3+4 2 | 2+2 2 | 3+3 2 |
| | 5+6 2 | 5+6 2 | 3+4 2 | 3+4 2 | 2+2 2 | 2+2 2 |
| | 3+4 2 | 3+4 2 | 5+6 2 | 5+6 2 | 3+3 2 | 2+2 2 |
| | 5+6 2 | 5+6 2 | 3+4 2 | 5+6 2 | 2+2 2 | 2+2 2 |
| | 3+4 2 | 5+6 2 | 5+6 2 | 3+4 2 | 3+3 2 | 3+3 2 |
| | 5+6 2 | 5+6 2 | 5+6 2 | 3+4 2 | 3+3 2 | 2+2 2 |
| | 3+4 2 | 3+4 2 | 5+6 2 | 5+6 2 | 2+2 2 | 3+3 2 |
| | 3+4 2 | 5+6 2 | 3+4 2 | 5+6 2 | 3+3 2 | 2+2 2 |
| | 5+6 2 | 5+6 2 | 3+4 2 | 5+6 2 | 2+2 2 | 3+3 2 |
| | 3+4 2 | 3+4 2 | 3+4 2 | 3+4 2 | 2+2 2 | 2+2 2 |
| | 5+6 2 | 5+6 2 | 3+4 2 | 5+6 2 | 3+3 2 | 3+3 2 |
| | 5+6 2 | 3+4 2 | 5+6 2 | 3+4 2 | 2+2 2 | 3+3 2 |
| | 3+4 2 | 3+4 2 | 5+6 2 | 5+6 2 | 3+3 2 | 2+2 2 |
| | 3+4 2 | 5+6 2 | 5+6 2 | 3+4 2 | 2+2 2 | 3+3 2 |
| | 5+6 2 | 3+4 2 | 3+4 2 | 3+4 2 | 3+3 2 | 3+3 2 |
| | 3+4 2 | 5+6 2 | 3+4 2 | 3+4 2 | 2+2 2 | 3+3 2 |
| | 5+6 2 | 3+4 2 | 5+6 2 | 3+4 2 | 3+3 2 | 3+3 2 |
| | 3+4 2 | 3+4 2 | 3+4 2 | 5+6 2 | 2+2 2 | 2+2 2 |
| | 5+6 2 | 3+4 2 | 5+6 2 | 3+4 2 | 2+2 2 | 3+3 2 |
| | 5+6 2 | 3+4 2 | 3+4 2 | 5+6 2 | 3+3 2 | 2+2 2 |
| | 3+4 2 | 5+6 2 | 5+6 2 | 3+4 2 | 3+3 2 | 2+2 2 |
| | 5+6 2 | 5+6 2 | 5+6 2 | 5+6 2 | 3+3 2 | 3+3 2 |
| | 3+4 2 | 5+6 2 | 3+4 2 | 3+4 2 | 2+2 2 | 3+3 2 |
| | 3+4 2 | 3+4 2 | 3+4 2 | 5+6 2 | 3+3 2 | 2+2 2 |
| | 5+6 2 | 5+6 2 | 5+6 2 | 3+4 2 | 2+2 2 | 2+2 2 |
| | 3+4 2 | 3+4 2 | 5+6 2 | 5+6 2 | 3+3 2 | 3+3 2 |
| | 5+6 2 | 5+6 2 | 3+4 2 | 5+6 2 | 2+2 2 | 2+2 2 |
| | 5+6 2 | 3+4 2 | 3+4 2 | 5+6 2 | 2+2 2 | 3+3 2 |
| | 3+4 2 | 5+6 2 | 5+6 2 | 5+6 2 | 3+3 2 | 2+2 2 |
| | 3+4 2 | 5+6 2 | 3+4 2 | 3+4 2 | 3+3 2 | 3+3 2 |
| | 3+4 2 | 3+4 2 | 5+6 2 | 3+4 2 | 2+2 2 | 2+2 2 |
| 12.2.33 | 20296 | 66 | Mn2O7P2 | | | |
| | -A | -GM | -M | -Y | -L | -V |
| | 5+6 2 | 3+4 2 | 5+6 2 | 3+4 2 | 2+2 2 | 3+3 2 |
| | 3+4 2 | 5+6 2 | 3+4 2 | 5+6 2 | 3+3 2 | 2+2 2 |
| | 5+6 2 | 3+4 2 | 5+6 2 | 3+4 2 | 2+2 2 | 3+3 2 |
| | 3+4 2 | 3+4 2 | 5+6 2 | 5+6 2 | 2+2 2 | 2+2 2 |
| | 5+6 2 | 5+6 2 | 3+4 2 | 3+4 2 | 3+3 2 | 3+3 2 |
| | 3+4 2 | 3+4 2 | 3+4 2 | 3+4 2 | 2+2 2 | 2+2 2 |
| | 5+6 2 | 5+6 2 | 5+6 2 | 5+6 2 | 3+3 2 | 3+3 2 |
| | 3+4 2 | 5+6 2 | 3+4 2 | 5+6 2 | 3+3 2 | 2+2 2 |
| | 3+4 2 | 3+4 2 | 5+6 2 | 3+4 2 | 2+2 2 | 3+3 2 |
| | 5+6 2 | 5+6 2 | 3+4 2 | 5+6 2 | 3+3 2 | 2+2 2 |

Continued on next column

**TABLE XXIV– continued**

| TQC | ICSD | NoE | SF | | | |
|---|---|---|---|---|---|---|
| | 3+4 2 | 5+6 2 | 5+6 2 | 5+6 2 | 3+3 2 | 2+2 2 |
| | 3+4 2 | 3+4 2 | 3+4 2 | 5+6 2 | 2+2 2 | 2+2 2 |
| | 5+6 2 | 3+4 2 | 3+4 2 | 3+4 2 | 3+3 2 | 3+3 2 |
| | 5+6 2 | 5+6 2 | 5+6 2 | 5+6 2 | 2+2 2 | 3+3 2 |
| | 3+4 2 | 5+6 2 | 5+6 2 | 3+4 2 | 3+3 2 | 2+2 2 |
| | 3+4 2 | 3+4 2 | 3+4 2 | 3+4 2 | 3+3 2 | 2+2 2 |
| | 5+6 2 | 5+6 2 | 3+4 2 | 5+6 2 | 2+2 2 | 3+3 2 |
| | 5+6 2 | 3+4 2 | 3+4 2 | 5+6 2 | 3+3 2 | 2+2 2 |
| | 3+4 2 | 5+6 2 | 3+4 2 | 5+6 2 | 2+2 2 | 3+3 2 |
| | 5+6 2 | 5+6 2 | 5+6 2 | 3+4 2 | 2+2 2 | 3+3 2 |
| | 3+4 2 | 3+4 2 | 3+4 2 | 3+4 2 | 3+3 2 | 2+2 2 |
| | 5+6 2 | 3+4 2 | 5+6 2 | 5+6 2 | 3+3 2 | 2+2 2 |
| | 3+4 2 | 5+6 2 | 5+6 2 | 3+4 2 | 3+3 2 | 2+2 2 |
| | 3+4 2 | 5+6 2 | 3+4 2 | 5+6 2 | 2+2 2 | 3+3 2 |
| | 5+6 2 | 3+4 2 | 5+6 2 | 3+4 2 | 2+2 2 | 2+2 2 |
| | 5+6 2 | 3+4 2 | 5+6 2 | 5+6 2 | 3+3 2 | 3+3 2 |
| | 3+4 2 | 5+6 2 | 3+4 2 | 3+4 2 | 3+3 2 | 2+2 2 |
| | 3+4 2 | 5+6 2 | 3+4 2 | 5+6 2 | 2+2 2 | 3+3 2 |
| | 3+4 2 | 5+6 2 | 5+6 2 | 3+4 2 | 3+3 2 | 3+3 2 |
| | 3+4 2 | 3+4 2 | 3+4 2 | 5+6 2 | 2+2 2 | 2+2 2 |
| | 5+6 2 | 3+4 2 | 5+6 2 | 5+6 2 | 2+2 2 | 3+3 2 |
| | 5+6 2 | 3+4 2 | 3+4 2 | 3+4 2 | 3+3 2 | 2+2 2 |
| | 5+6 2 | 5+6 2 | 3+4 2 | 3+4 2 | 2+2 2 | 3+3 2 |
| 12.2.34 | 15615 | 148 | Mn1Mo1O4 | | | |
| | -A | -GM | -M | -Y | -L | -V |
| | 5+6 2 | 3+4 2 | 5+6 2 | 3+4 2 | 2+2 2 | 3+3 2 |
| | 3+4 2 | 5+6 2 | 3+4 2 | 5+6 2 | 3+3 2 | 2+2 2 |
| | 3+4 2 | 3+4 2 | 3+4 2 | 5+6 2 | 3+3 2 | 2+2 2 |
| | 5+6 2 | 5+6 2 | 5+6 2 | 3+4 2 | 2+2 2 | 3+3 2 |
| | 5+6 2 | 3+4 2 | 5+6 2 | 3+4 2 | 3+3 2 | 2+2 2 |
| | 5+6 2 | 5+6 2 | 3+4 2 | 5+6 2 | 2+2 2 | 2+2 2 |
| | 3+4 2 | 5+6 2 | 3+4 2 | 5+6 2 | 2+2 2 | 3+3 2 |
| | 3+4 2 | 3+4 2 | 5+6 2 | 3+4 2 | 2+2 2 | 2+2 2 |
| | 5+6 2 | 3+4 2 | 3+4 2 | 5+6 2 | 3+3 2 | 3+3 2 |
| | 3+4 2 | 3+4 2 | 5+6 2 | 3+4 2 | 3+3 2 | 2+2 2 |
| | 5+6 2 | 5+6 2 | 3+4 2 | 5+6 2 | 2+2 2 | 3+3 2 |
| | 3+4 2 | 5+6 2 | 5+6 2 | 3+4 2 | 3+3 2 | 3+3 2 |
| | 5+6 2 | 5+6 2 | 3+4 2 | 5+6 2 | 3+3 2 | 3+3 2 |
| | 3+4 2 | 5+6 2 | 5+6 2 | 5+6 2 | 2+2 2 | 3+3 2 |
| | 5+6 2 | 3+4 2 | 3+4 2 | 3+4 2 | 2+2 2 | 2+2 2 |
| | 3+4 2 | 3+4 2 | 5+6 2 | 3+4 2 | 3+3 2 | 2+2 2 |
| | 3+4 2 | 3+4 2 | 3+4 2 | 3+4 2 | 3+3 2 | 3+3 2 |
| | 5+6 2 | 5+6 2 | 5+6 2 | 5+6 2 | 2+2 2 | 2+2 2 |
| | 3+4 2 | 3+4 2 | 3+4 2 | 5+6 2 | 2+2 2 | 3+3 2 |
| | 5+6 2 | 3+4 2 | 5+6 2 | 3+4 2 | 3+3 2 | 2+2 2 |
| | 3+4 2 | 5+6 2 | 3+4 2 | 5+6 2 | 3+3 2 | 3+3 2 |
| | 5+6 2 | 5+6 2 | 5+6 2 | 3+4 2 | 2+2 2 | 2+2 2 |
| | 5+6 2 | 3+4 2 | 3+4 2 | 3+4 2 | 3+3 2 | 3+3 2 |
| | 3+4 2 | 3+4 2 | 5+6 2 | 5+6 2 | 2+2 2 | 2+2 2 |
| | 3+4 2 | 5+6 2 | 3+4 2 | 3+4 2 | 2+2 2 | 2+2 2 |
| | 5+6 2 | 5+6 2 | 5+6 2 | 5+6 2 | 3+3 2 | 3+3 2 |
| | 5+6 2 | 3+4 2 | 5+6 2 | 3+4 2 | 3+3 2 | 2+2 2 |
| | 3+4 2 | 3+4 2 | 3+4 2 | 5+6 2 | 2+2 2 | 2+2 2 |
| | 5+6 2 | 5+6 2 | 3+4 2 | 5+6 2 | 3+3 2 | 3+3 2 |
| | 3+4 2 | 3+4 2 | 5+6 2 | 3+4 2 | 2+2 2 | 3+3 2 |
| | 3+4 2 | 3+4 2 | 5+6 2 | 3+4 2 | 3+3 2 | 2+2 2 |
| | 5+6 2 | 5+6 2 | 3+4 2 | 5+6 2 | 2+2 2 | 3+3 2 |
| | 3+4 2 | 3+4 2 | 3+4 2 | 5+6 2 | 3+3 2 | 3+3 2 |
| | 5+6 2 | 5+6 2 | 5+6 2 | 3+4 2 | 2+2 2 | 2+2 2 |
| | 3+4 2 | 5+6 2 | 3+4 2 | 3+4 2 | 2+2 2 | 2+2 2 |
| | 3+4 2 | 3+4 2 | 3+4 2 | 5+6 2 | 3+3 2 | 3+3 2 |

Continued on next column

**TABLE XXIV– continued**

| TQC | ICSD | | NoE | SF | | |
|---|---|---|---|---|---|---|
| | 5+6 2 | 5+6 2 | 5+6 2 | 3+4 2 | 3+3 2 | 3+3 2 |
| | 5+6 2 | 5+6 2 | 5+6 2 | 5+6 2 | 2+2 2 | 2+2 2 |
| | 5+6 2 | 3+4 2 | 5+6 2 | 3+4 2 | 2+2 2 | 2+2 2 |
| | 5+6 2 | 5+6 2 | 5+6 2 | 5+6 2 | 3+3 2 | 3+3 2 |
| | 3+4 2 | 3+4 2 | 3+4 2 | 3+4 2 | 2+2 2 | 3+3 2 |
| | 3+4 2 | 5+6 2 | 5+6 2 | 3+4 2 | 3+3 2 | 2+2 2 |
| | 5+6 2 | 3+4 2 | 3+4 2 | 5+6 2 | 2+2 2 | 3+3 2 |
| | 5+6 2 | 3+4 2 | 3+4 2 | 5+6 2 | 3+3 2 | 2+2 2 |
| | 3+4 2 | 5+6 2 | 3+4 2 | 5+6 2 | 2+2 2 | 2+2 2 |
| | 3+4 2 | 5+6 2 | 5+6 2 | 3+4 2 | 3+3 2 | 3+3 2 |
| | 5+6 2 | 3+4 2 | 5+6 2 | 5+6 2 | 2+2 2 | 2+2 2 |
| | 5+6 2 | 5+6 2 | 3+4 2 | 3+4 2 | 3+3 2 | 3+3 2 |
| | 3+4 2 | 5+6 2 | 5+6 2 | 3+4 2 | 3+3 2 | 3+3 2 |
| | 3+4 2 | 5+6 2 | 3+4 2 | 5+6 2 | 2+2 2 | 3+3 2 |
| | 5+6 2 | 5+6 2 | 3+4 2 | 5+6 2 | 2+2 2 | 2+2 2 |
| | 5+6 2 | 3+4 2 | 5+6 2 | 5+6 2 | 2+2 2 | 3+3 2 |
| | 3+4 2 | 3+4 2 | 5+6 2 | 3+4 2 | 3+3 2 | 2+2 2 |
| | 3+4 2 | 5+6 2 | 3+4 2 | 5+6 2 | 2+2 2 | 2+2 2 |
| | 5+6 2 | 3+4 2 | 3+4 2 | 5+6 2 | 3+3 2 | 3+3 2 |
| | 5+6 2 | 5+6 2 | 5+6 2 | 3+4 2 | 3+3 2 | 2+2 2 |
| | 3+4 2 | 5+6 2 | 5+6 2 | 3+4 2 | 2+2 2 | 3+3 2 |
| | 5+6 2 | 3+4 2 | 3+4 2 | 5+6 2 | 3+3 2 | 2+2 2 |
| | 3+4 2 | 3+4 2 | 3+4 2 | 3+4 2 | 2+2 2 | 3+3 2 |
| | 3+4 2 | 5+6 2 | 5+6 2 | 5+6 2 | 3+3 2 | 2+2 2 |
| | 5+6 2 | 5+6 2 | 5+6 2 | 3+4 2 | 2+2 2 | 3+3 2 |
| | 3+4 2 | 3+4 2 | 3+4 2 | 5+6 2 | 3+3 2 | 3+3 2 |
| | 5+6 2 | 3+4 2 | 3+4 2 | 3+4 2 | 3+3 2 | 2+2 2 |
| | 3+4 2 | 3+4 2 | 5+6 2 | 3+4 2 | 2+2 2 | 2+2 2 |
| | 3+4 2 | 3+4 2 | 3+4 2 | 5+6 2 | 3+3 2 | 2+2 2 |
| | 5+6 2 | 5+6 2 | 5+6 2 | 3+4 2 | 3+3 2 | 3+3 2 |
| | 3+4 2 | 3+4 2 | 5+6 2 | 3+4 2 | 2+2 2 | 2+2 2 |
| | 5+6 2 | 3+4 2 | 3+4 2 | 3+4 2 | 2+2 2 | 3+3 2 |
| | 5+6 2 | 3+4 2 | 5+6 2 | 5+6 2 | 2+2 2 | 2+2 2 |
| | 3+4 2 | 5+6 2 | 3+4 2 | 3+4 2 | 3+3 2 | 3+3 2 |
| | 3+4 2 | 5+6 2 | 5+6 2 | 5+6 2 | 2+2 2 | 2+2 2 |
| | 5+6 2 | 5+6 2 | 3+4 2 | 3+4 2 | 3+3 2 | 2+2 2 |
| | 5+6 2 | 3+4 2 | 5+6 2 | 5+6 2 | 2+2 2 | 3+3 2 |
| | 3+4 2 | 5+6 2 | 3+4 2 | 3+4 2 | 3+3 2 | 3+3 2 |
| 12.2.35 | 50001 | | 154 | Mg2Si10Sr11 | | |
| | -A | -GM | -M | -Y | -L | -V |
| | 3+4 2 | 3+4 2 | 3+4 2 | 3+4 2 | 2+2 2 | 2+2 2 |
| | 5+6 2 | 5+6 2 | 5+6 2 | 5+6 2 | 3+3 2 | 3+3 2 |
| | 3+4 2 | 3+4 2 | 3+4 2 | 3+4 2 | 2+2 2 | 2+2 2 |
| | 5+6 2 | 5+6 2 | 5+6 2 | 5+6 2 | 2+2 2 | 2+2 2 |
| | 5+6 2 | 3+4 2 | 5+6 2 | 3+4 2 | 3+3 2 | 3+3 2 |
| | 3+4 2 | 3+4 2 | 3+4 2 | 3+4 2 | 3+3 2 | 2+2 2 |
| | 3+4 2 | 5+6 2 | 3+4 2 | 5+6 2 | 2+2 2 | 3+3 2 |
| | 5+6 2 | 3+4 2 | 3+4 2 | 5+6 2 | 2+2 2 | 2+2 2 |
| | 3+4 2 | 5+6 2 | 5+6 2 | 3+4 2 | 3+3 2 | 3+3 2 |
| | 3+4 2 | 3+4 2 | 3+4 2 | 3+4 2 | 2+2 2 | 2+2 2 |
| | 5+6 2 | 5+6 2 | 5+6 2 | 5+6 2 | 3+3 2 | 3+3 2 |
| | 5+6 2 | 5+6 2 | 5+6 2 | 5+6 2 | 3+3 2 | 3+3 2 |
| | 3+4 2 | 3+4 2 | 3+4 2 | 3+4 2 | 2+2 2 | 2+2 2 |
| | 5+6 2 | 5+6 2 | 5+6 2 | 5+6 2 | 3+3 2 | 3+3 2 |
| | 5+6 2 | 3+4 2 | 5+6 2 | 3+4 2 | 3+3 2 | 3+3 2 |
| | 3+4 2 | 3+4 2 | 3+4 2 | 3+4 2 | 3+3 2 | 2+2 2 |
| | 3+4 2 | 5+6 2 | 3+4 2 | 5+6 2 | 2+2 2 | 2+2 2 |
| | 5+6 2 | 5+6 2 | 5+6 2 | 5+6 2 | 2+2 2 | 3+3 2 |
| | 5+6 2 | 3+4 2 | 3+4 2 | 5+6 2 | 2+2 2 | 2+2 2 |
| | 3+4 2 | 5+6 2 | 5+6 2 | 3+4 2 | 3+3 2 | 3+3 2 |
| | 5+6 2 | 3+4 2 | 5+6 2 | 3+4 2 | 2+2 2 | 2+2 2 |

Continued on next column

**TABLE XXIV– continued**

| TQC | ICSD | | NoE | SF | | |
|---|---|---|---|---|---|---|
| | 3+4 2 | 5+6 2 | 3+4 2 | 5+6 2 | 3+3 2 | 3+3 2 |
| | 5+6 2 | 5+6 2 | 5+6 2 | 5+6 2 | 3+3 2 | 3+3 2 |
| | 3+4 2 | 3+4 2 | 3+4 2 | 3+4 2 | 2+2 2 | 2+2 2 |
| | 5+6 2 | 5+6 2 | 5+6 2 | 5+6 2 | 3+3 2 | 3+3 2 |
| | 3+4 2 | 3+4 2 | 3+4 2 | 3+4 2 | 2+2 2 | 2+2 2 |
| | 5+6 2 | 5+6 2 | 5+6 2 | 5+6 2 | 3+3 2 | 3+3 2 |
| | 5+6 2 | 5+6 2 | 5+6 2 | 5+6 2 | 3+3 2 | 3+3 2 |
| | 5+6 2 | 3+4 2 | 5+6 2 | 3+4 2 | 3+3 2 | 2+2 2 |
| | 3+4 2 | 3+4 2 | 3+4 2 | 3+4 2 | 2+2 2 | 3+3 2 |
| | 3+4 2 | 5+6 2 | 3+4 2 | 5+6 2 | 2+2 2 | 2+2 2 |
| | 5+6 2 | 3+4 2 | 5+6 2 | 5+6 2 | 3+3 2 | 2+2 2 |
| | 3+4 2 | 5+6 2 | 3+4 2 | 3+4 2 | 2+2 2 | 2+2 2 |
| | 5+6 2 | 3+4 2 | 5+6 2 | 5+6 2 | 2+2 2 | 3+3 2 |
| | 3+4 2 | 5+6 2 | 3+4 2 | 3+4 2 | 3+3 2 | 3+3 2 |
| | 5+6 2 | 3+4 2 | 3+4 2 | 5+6 2 | 3+3 2 | 3+3 2 |
| | 5+6 2 | 5+6 2 | 5+6 2 | 5+6 2 | 2+2 2 | 2+2 2 |
| | 5+6 2 | 3+4 2 | 5+6 2 | 3+4 2 | 3+3 2 | 3+3 2 |
| | 3+4 2 | 5+6 2 | 3+4 2 | 5+6 2 | 2+2 2 | 2+2 2 |
| | 3+4 2 | 5+6 2 | 5+6 2 | 3+4 2 | 3+3 2 | 3+3 2 |
| | 3+4 2 | 5+6 2 | 3+4 2 | 5+6 2 | 2+2 2 | 3+3 2 |
| | 5+6 2 | 3+4 2 | 5+6 2 | 3+4 2 | 3+3 2 | 2+2 2 |
| | 5+6 2 | 3+4 2 | 5+6 2 | 3+4 2 | 3+3 2 | 2+2 2 |
| | 3+4 2 | 5+6 2 | 3+4 2 | 5+6 2 | 2+2 2 | 3+3 2 |
| | 5+6 2 | 3+4 2 | 5+6 2 | 3+4 2 | 2+2 2 | 3+3 2 |
| | 3+4 2 | 5+6 2 | 3+4 2 | 5+6 2 | 3+3 2 | 2+2 2 |
| | 5+6 2 | 3+4 2 | 5+6 2 | 3+4 2 | 2+2 2 | 3+3 2 |
| | 3+4 2 | 5+6 2 | 3+4 2 | 5+6 2 | 3+3 2 | 2+2 2 |
| | 3+4 2 | 3+4 2 | 5+6 2 | 5+6 2 | 2+2 2 | 2+2 2 |
| | 5+6 2 | 3+4 2 | 3+4 2 | 3+4 2 | 2+2 2 | 3+3 2 |
| | 5+6 2 | 5+6 2 | 5+6 2 | 3+4 2 | 3+3 2 | 3+3 2 |
| | 3+4 2 | 5+6 2 | 3+4 2 | 5+6 2 | 3+3 2 | 2+2 2 |
| | 5+6 2 | 3+4 2 | 3+4 2 | 3+4 2 | 2+2 2 | 3+3 2 |
| | 3+4 2 | 5+6 2 | 5+6 2 | 5+6 2 | 3+3 2 | 2+2 2 |
| | 3+4 2 | 5+6 2 | 3+4 2 | 3+4 2 | 3+3 2 | 2+2 2 |
| | 5+6 2 | 3+4 2 | 5+6 2 | 5+6 2 | 3+3 2 | 3+3 2 |
| | 3+4 2 | 3+4 2 | 5+6 2 | 3+4 2 | 2+2 2 | 2+2 2 |
| | 5+6 2 | 5+6 2 | 3+4 2 | 5+6 2 | 2+2 2 | 3+3 2 |
| | 3+4 2 | 3+4 2 | 5+6 2 | 5+6 2 | 3+3 2 | 2+2 2 |
| | 5+6 2 | 3+4 2 | 3+4 2 | 3+4 2 | 2+2 2 | 2+2 2 |
| | 3+4 2 | 5+6 2 | 3+4 2 | 5+6 2 | 3+3 2 | 3+3 2 |
| | 3+4 2 | 5+6 2 | 5+6 2 | 3+4 2 | 2+2 2 | 3+3 2 |
| | 5+6 2 | 3+4 2 | 3+4 2 | 3+4 2 | 3+3 2 | 2+2 2 |
| | 5+6 2 | 3+4 2 | 5+6 2 | 5+6 2 | 2+2 2 | 3+3 2 |
| | 3+4 2 | 5+6 2 | 3+4 2 | 5+6 2 | 2+2 2 | 2+2 2 |
| | 3+4 2 | 3+4 2 | 5+6 2 | 3+4 2 | 2+2 2 | 3+3 2 |
| | 5+6 2 | 5+6 2 | 3+4 2 | 5+6 2 | 3+3 2 | 2+2 2 |
| | 5+6 2 | 3+4 2 | 5+6 2 | 3+4 2 | 3+3 2 | 3+3 2 |
| | 3+4 2 | 3+4 2 | 5+6 2 | 5+6 2 | 2+2 2 | 2+2 2 |
| | 5+6 2 | 3+4 2 | 5+6 2 | 3+4 2 | 3+3 2 | 3+3 2 |
| | 3+4 2 | 5+6 2 | 3+4 2 | 3+4 2 | 2+2 2 | 2+2 2 |
| | 3+4 2 | 5+6 2 | 3+4 2 | 5+6 2 | 2+2 2 | 3+3 2 |
| | 3+4 2 | 5+6 2 | 3+4 2 | 3+4 2 | 3+3 2 | 2+2 2 |
| | 5+6 2 | 3+4 2 | 5+6 2 | 5+6 2 | 3+3 2 | 2+2 2 |
| | 5+6 2 | 5+6 2 | 3+4 2 | 3+4 2 | 2+2 2 | 3+3 2 |
| | 5+6 2 | 3+4 2 | 5+6 2 | 3+4 2 | 2+2 2 | 3+3 2 |
| | 3+4 2 | 3+4 2 | 5+6 2 | 3+4 2 | 3+3 2 | 3+3 2 |
| 12.2.36 | 180011 | | 66 | Li0.5O2Ti1 | | |
| | -A | -GM | -M | -Y | -L | -V |
| | 5+6 2 | 3+4 2 | 5+6 2 | 3+4 2 | 3+3 2 | 2+2 2 |
| | 3+4 2 | 5+6 2 | 3+4 2 | 5+6 2 | 2+2 2 | 3+3 2 |
| | 5+6 2 | 3+4 2 | 3+4 2 | 5+6 2 | 2+2 2 | 3+3 2 |

Continued on next column

**TABLE XXIV– continued**

| TQC | ICSD | NoE | SF |
|---|---|---|---|
| | 3+4 2 5+6 2 5+6 2 3+4 2 3+3 2 2+2 2 | | |
| | 3+4 2 5+6 2 5+6 2 3+4 2 2+2 2 3+3 2 | | |
| | 5+6 2 3+4 2 3+4 2 5+6 2 3+3 2 2+2 2 | | |
| | 3+4 2 3+4 2 5+6 2 5+6 2 2+2 2 2+2 2 | | |
| | 5+6 2 5+6 2 3+4 2 3+4 2 3+3 2 3+3 2 | | |
| | 5+6 2 3+4 2 5+6 2 3+4 2 3+3 2 2+2 2 | | |
| | 3+4 2 5+6 2 3+4 2 5+6 2 2+2 2 3+3 2 | | |
| | 3+4 2 5+6 2 3+4 2 5+6 2 3+3 2 2+2 2 | | |
| | 5+6 2 3+4 2 5+6 2 3+4 2 2+2 2 3+3 2 | | |
| | 5+6 2 3+4 2 5+6 2 3+4 2 3+3 2 2+2 2 | | |
| | 3+4 2 5+6 2 5+6 2 3+4 2 2+2 2 3+3 2 | | |
| | 5+6 2 3+4 2 3+4 2 5+6 2 3+3 2 2+2 2 | | |
| | 5+6 2 3+4 2 3+4 2 5+6 2 2+2 2 3+3 2 | | |
| | 3+4 2 5+6 2 3+4 2 5+6 2 3+3 2 2+2 2 | | |
| | 5+6 2 3+4 2 3+4 2 5+6 2 2+2 2 3+3 2 | | |
| | 3+4 2 5+6 2 5+6 2 3+4 2 3+3 2 2+2 2 | | |
| | 3+4 2 5+6 2 5+6 2 3+4 2 2+2 2 3+3 2 | | |
| | 5+6 2 3+4 2 5+6 2 5+6 2 3+3 2 2+2 2 | | |
| | 3+4 2 3+4 2 3+4 2 3+4 2 2+2 2 3+3 2 | | |
| | 5+6 2 5+6 2 3+4 2 3+4 2 2+2 2 3+3 2 | | |
| | 5+6 2 3+4 2 3+4 2 5+6 2 3+3 2 2+2 2 | | |
| | 3+4 2 5+6 2 5+6 2 5+6 2 3+3 2 2+2 2 | | |
| | 3+4 2 3+4 2 5+6 2 3+4 2 2+2 2 3+3 2 | | |
| | 5+6 2 5+6 2 5+6 2 5+6 2 3+3 2 2+2 2 | | |
| | 3+4 2 5+6 2 3+4 2 3+4 2 2+2 2 3+3 2 | | |
| | 3+4 2 3+4 2 5+6 2 5+6 2 3+3 2 2+2 2 | | |
| | 3+4 2 3+4 2 5+6 2 5+6 2 2+2 2 3+3 2 | | |
| | 5+6 2 5+6 2 3+4 2 3+4 2 3+3 2 3+3 2 | | |
| | 5+6 2 5+6 2 3+4 2 3+4 2 2+2 2 2+2 2 | | |
| | 5+6 2 3+4 2 3+4 2 3+4 2 2+2 2 2+2 2 | | |
| 12.2.37 | 246807 | 62 | La1Si5 |
| | -A -GM -M -Y -L -V | | |
| | 3+4 2 3+4 2 3+4 2 3+4 2 2+2 2 2+2 2 | | |
| | 5+6 2 5+6 2 5+6 2 5+6 2 3+3 2 3+3 2 | | |
| | 3+4 2 3+4 2 3+4 2 3+4 2 2+2 2 2+2 2 | | |
| | 5+6 2 5+6 2 5+6 2 5+6 2 3+3 2 3+3 2 | | |
| | 5+6 2 3+4 2 3+4 2 3+4 2 3+3 2 3+3 2 | | |
| | 3+4 2 5+6 2 5+6 2 5+6 2 2+2 2 2+2 2 | | |
| | 3+4 2 3+4 2 3+4 2 3+4 2 2+2 2 2+2 2 | | |
| | 5+6 2 5+6 2 5+6 2 5+6 2 3+3 2 3+3 2 | | |
| | 5+6 2 3+4 2 5+6 2 3+4 2 3+3 2 2+2 2 | | |
| | 3+4 2 5+6 2 3+4 2 5+6 2 2+2 2 3+3 2 | | |
| | 5+6 2 3+4 2 3+4 2 5+6 2 2+2 2 3+3 2 | | |
| | 3+4 2 3+4 2 5+6 2 3+4 2 3+3 2 3+3 2 | | |
| | 5+6 2 5+6 2 3+4 2 5+6 2 2+2 2 2+2 2 | | |
| | 3+4 2 3+4 2 5+6 2 3+4 2 3+3 2 3+3 2 | | |
| | 5+6 2 5+6 2 3+4 2 3+4 2 2+2 2 2+2 2 | | |
| | 3+4 2 3+4 2 5+6 2 5+6 2 3+3 2 3+3 2 | | |
| | 3+4 2 5+6 2 5+6 2 3+4 2 3+3 2 2+2 2 | | |
| | 5+6 2 5+6 2 3+4 2 5+6 2 2+2 2 2+2 2 | | |
| | 3+4 2 3+4 2 3+4 2 3+4 2 3+3 2 3+3 2 | | |
| | 5+6 2 3+4 2 5+6 2 5+6 2 2+2 2 2+2 2 | | |
| | 5+6 2 5+6 2 5+6 2 3+4 2 3+3 2 3+3 2 | | |
| | 3+4 2 3+4 2 3+4 2 5+6 2 2+2 2 2+2 2 | | |
| | 5+6 2 5+6 2 5+6 2 5+6 2 2+2 2 3+3 2 | | |
| | 3+4 2 5+6 2 3+4 2 3+4 2 3+3 2 2+2 2 | | |
| | 5+6 2 3+4 2 3+4 2 5+6 2 3+3 2 2+2 2 | | |
| | 3+4 2 3+4 2 5+6 2 5+6 2 2+2 2 3+3 2 | | |
| | 3+4 2 5+6 2 3+4 2 3+4 2 2+2 2 3+3 2 | | |
| | 5+6 2 3+4 2 5+6 2 3+4 2 3+3 2 2+2 2 | | |
| | 5+6 2 3+4 2 3+4 2 5+6 2 2+2 2 3+3 2 | | |

Continued on next column

**TABLE XXIV– continued**

| TQC | ICSD | NoE | SF |
|---|---|---|---|
| | 3+4 2 3+4 2 3+4 2 3+4 2 3+3 2 2+2 2 | | |
| | 5+6 2 5+6 2 5+6 2 5+6 2 2+2 2 3+3 2 | | |
| 12.2.38 | 428765 | 84 | Hg2Sn2Sr1 |
| | -A -GM -M -Y -L -V | | |
| | 3+4 2 3+4 2 3+4 2 3+4 2 2+2 2 2+2 2 | | |
| | 5+6 2 5+6 2 5+6 2 5+6 2 3+3 2 3+3 2 | | |
| | 5+6 2 3+4 2 5+6 2 3+4 2 2+2 2 3+3 2 | | |
| | 3+4 2 5+6 2 3+4 2 5+6 2 3+3 2 2+2 2 | | |
| | 5+6 2 3+4 2 5+6 2 3+4 2 2+2 2 2+2 2 | | |
| | 3+4 2 5+6 2 3+4 2 5+6 2 3+3 2 3+3 2 | | |
| | 5+6 2 3+4 2 5+6 2 3+4 2 2+2 2 3+3 2 | | |
| | 3+4 2 5+6 2 3+4 2 5+6 2 3+3 2 2+2 2 | | |
| | 3+4 2 3+4 2 3+4 2 3+4 2 3+3 2 3+3 2 | | |
| | 5+6 2 3+4 2 5+6 2 3+4 2 3+3 2 2+2 2 | | |
| | 3+4 2 5+6 2 3+4 2 5+6 2 2+2 2 3+3 2 | | |
| | 5+6 2 5+6 2 5+6 2 5+6 2 2+2 2 2+2 2 | | |
| | 3+4 2 3+4 2 5+6 2 3+4 2 3+3 2 2+2 2 | | |
| | 5+6 2 5+6 2 3+4 2 5+6 2 2+2 2 3+3 2 | | |
| | 3+4 2 3+4 2 3+4 2 3+4 2 3+3 2 3+3 2 | | |
| | 5+6 2 5+6 2 5+6 2 5+6 2 2+2 2 2+2 2 | | |
| | 5+6 2 3+4 2 3+4 2 3+4 2 3+3 2 2+2 2 | | |
| | 3+4 2 3+4 2 5+6 2 5+6 2 3+3 2 3+3 2 | | |
| | 3+4 2 5+6 2 5+6 2 5+6 2 2+2 2 3+3 2 | | |
| | 5+6 2 5+6 2 3+4 2 3+4 2 2+2 2 2+2 2 | | |
| | 5+6 2 5+6 2 5+6 2 5+6 2 2+2 2 2+2 2 | | |
| | 3+4 2 3+4 2 3+4 2 3+4 2 3+3 2 3+3 2 | | |
| | 5+6 2 3+4 2 3+4 2 3+4 2 3+3 2 3+3 2 | | |
| | 3+4 2 5+6 2 5+6 2 5+6 2 2+2 2 2+2 2 | | |
| | 5+6 2 3+4 2 5+6 2 5+6 2 3+3 2 3+3 2 | | |
| | 3+4 2 3+4 2 3+4 2 3+4 2 2+2 2 2+2 2 | | |
| | 3+4 2 5+6 2 5+6 2 5+6 2 2+2 2 2+2 2 | | |
| | 5+6 2 5+6 2 3+4 2 3+4 2 3+3 2 3+3 2 | | |
| | 5+6 2 3+4 2 3+4 2 5+6 2 3+3 2 3+3 2 | | |
| | 3+4 2 3+4 2 5+6 2 3+4 2 2+2 2 2+2 2 | | |
| | 3+4 2 5+6 2 5+6 2 3+4 2 3+3 2 3+3 2 | | |
| | 5+6 2 5+6 2 3+4 2 5+6 2 2+2 2 2+2 2 | | |
| | 5+6 2 3+4 2 5+6 2 3+4 2 3+3 2 2+2 2 | | |
| | 3+4 2 5+6 2 3+4 2 5+6 2 2+2 2 3+3 2 | | |
| | 5+6 2 3+4 2 5+6 2 3+4 2 2+2 2 2+2 2 | | |
| | 3+4 2 3+4 2 3+4 2 5+6 2 3+3 2 3+3 2 | | |
| | 5+6 2 5+6 2 5+6 2 5+6 2 3+3 2 3+3 2 | | |
| | 3+4 2 5+6 2 3+4 2 5+6 2 2+2 2 2+2 2 | | |
| | 3+4 2 3+4 2 3+4 2 3+4 2 2+2 2 2+2 2 | | |
| | 5+6 2 3+4 2 5+6 2 3+4 2 3+3 2 2+2 2 | | |
| | 3+4 2 3+4 2 3+4 2 3+4 2 3+3 2 3+3 2 | | |
| | 3+4 2 5+6 2 3+4 2 5+6 2 3+3 2 3+3 2 | | |
| 12.2.39 | 157861 | 36 | Hg1 |
| | -A -GM -M -Y -L -V | | |
| | 3+4 2 3+4 2 3+4 2 3+4 2 3+3 2 3+3 2 | | |
| | 5+6 2 3+4 2 3+4 2 3+4 2 3+3 2 2+2 2 | | |
| | 3+4 2 5+6 2 5+6 2 5+6 2 2+2 2 3+3 2 | | |
| | 3+4 2 3+4 2 3+4 2 3+4 2 3+3 2 3+3 2 | | |
| | 3+4 2 5+6 2 5+6 2 3+4 2 3+3 2 2+2 2 | | |
| | 5+6 2 3+4 2 3+4 2 5+6 2 2+2 2 3+3 2 | | |
| | 3+4 2 3+4 2 3+4 2 5+6 2 2+2 2 3+3 2 | | |
| | 5+6 2 3+4 2 3+4 2 3+4 2 3+3 2 2+2 2 | | |
| | 3+4 2 5+6 2 5+6 2 3+4 2 3+3 2 3+3 2 | | |
| | 3+4 2 3+4 2 3+4 2 5+6 2 2+2 2 3+3 2 | | |
| | 5+6 2 5+6 2 5+6 2 3+4 2 3+3 2 2+2 2 | | |
| | 5+6 2 3+4 2 3+4 2 3+4 2 3+3 2 3+3 2 | | |
| | 3+4 2 3+4 2 3+4 2 5+6 2 2+2 2 3+3 2 | | |

Continued on next column

**TABLE XXIV– continued**

| TQC | ICSD | | NoE | SF | | |
|---|---|---|---|---|---|---|
| | 3+4 2 | 3+4 2 | 5+6 2 | 3+4 2 | 3+3 2 | 2+2 2 |
| | 5+6 2 | 3+4 2 | 3+4 2 | 3+4 2 | 3+3 2 | 3+3 2 |
| | 3+4 2 | 5+6 2 | 5+6 2 | 5+6 2 | 3+3 2 | 2+2 2 |
| | 3+4 2 | 5+6 2 | 3+4 2 | 3+4 2 | 2+2 2 | 2+2 2 |
| | 5+6 2 | 3+4 2 | 5+6 2 | 5+6 2 | 2+2 2 | 3+3 2 |
| 12.2.40 | 405208 | | 96 | Hf7P4 | | |
| | -A | -GM | -M | -Y | -L | -V |
| | 5+6 2 | 3+4 2 | 5+6 2 | 3+4 2 | 3+3 2 | 2+2 2 |
| | 3+4 2 | 3+4 2 | 3+4 2 | 5+6 2 | 3+3 2 | 3+3 2 |
| | 3+4 2 | 5+6 2 | 3+4 2 | 3+4 2 | 3+3 2 | 2+2 2 |
| | 5+6 2 | 5+6 2 | 5+6 2 | 5+6 2 | 2+2 2 | 3+3 2 |
| | 3+4 2 | 5+6 2 | 3+4 2 | 3+4 2 | 2+2 2 | 3+3 2 |
| | 5+6 2 | 5+6 2 | 5+6 2 | 5+6 2 | 3+3 2 | 2+2 2 |
| | 5+6 2 | 3+4 2 | 5+6 2 | 5+6 2 | 2+2 2 | 3+3 2 |
| | 3+4 2 | 3+4 2 | 3+4 2 | 3+4 2 | 2+2 2 | 2+2 2 |
| | 5+6 2 | 3+4 2 | 3+4 2 | 3+4 2 | 2+2 2 | 3+3 2 |
| | 3+4 2 | 3+4 2 | 5+6 2 | 3+4 2 | 3+3 2 | 2+2 2 |
| | 5+6 2 | 3+4 2 | 5+6 2 | 5+6 2 | 2+2 2 | 2+2 2 |
| | 3+4 2 | 5+6 2 | 5+6 2 | 3+4 2 | 3+3 2 | 3+3 2 |
| | 3+4 2 | 5+6 2 | 3+4 2 | 5+6 2 | 3+3 2 | 2+2 2 |
| | 5+6 2 | 3+4 2 | 3+4 2 | 3+4 2 | 3+3 2 | 3+3 2 |
| | 3+4 2 | 5+6 2 | 3+4 2 | 3+4 2 | 2+2 2 | 2+2 2 |
| | 5+6 2 | 5+6 2 | 5+6 2 | 5+6 2 | 3+3 2 | 3+3 2 |
| | 5+6 2 | 3+4 2 | 5+6 2 | 5+6 2 | 3+3 2 | 2+2 2 |
| | 3+4 2 | 5+6 2 | 3+4 2 | 5+6 2 | 2+2 2 | 3+3 2 |
| | 5+6 2 | 3+4 2 | 3+4 2 | 5+6 2 | 3+3 2 | 3+3 2 |
| | 3+4 2 | 5+6 2 | 5+6 2 | 3+4 2 | 3+3 2 | 2+2 2 |
| | 3+4 2 | 3+4 2 | 5+6 2 | 3+4 2 | 2+2 2 | 3+3 2 |
| | 5+6 2 | 5+6 2 | 3+4 2 | 5+6 2 | 2+2 2 | 2+2 2 |
| | 5+6 2 | 3+4 2 | 3+4 2 | 5+6 2 | 2+2 2 | 2+2 2 |
| | 3+4 2 | 5+6 2 | 3+4 2 | 3+4 2 | 3+3 2 | 3+3 2 |
| | 3+4 2 | 5+6 2 | 5+6 2 | 5+6 2 | 2+2 2 | 2+2 2 |
| | 5+6 2 | 3+4 2 | 5+6 2 | 3+4 2 | 3+3 2 | 3+3 2 |
| | 5+6 2 | 3+4 2 | 3+4 2 | 5+6 2 | 2+2 2 | 2+2 2 |
| | 5+6 2 | 3+4 2 | 5+6 2 | 3+4 2 | 3+3 2 | 3+3 2 |
| | 3+4 2 | 5+6 2 | 3+4 2 | 3+4 2 | 2+2 2 | 3+3 2 |
| | 5+6 2 | 3+4 2 | 5+6 2 | 3+4 2 | 3+3 2 | 2+2 2 |
| | 3+4 2 | 3+4 2 | 5+6 2 | 3+4 2 | 2+2 2 | 3+3 2 |
| | 3+4 2 | 3+4 2 | 3+4 2 | 3+4 2 | 2+2 2 | 2+2 2 |
| | 3+4 2 | 5+6 2 | 5+6 2 | 3+4 2 | 3+3 2 | 3+3 2 |
| | 3+4 2 | 5+6 2 | 3+4 2 | 5+6 2 | 3+3 2 | 2+2 2 |
| | 5+6 2 | 5+6 2 | 3+4 2 | 5+6 2 | 3+3 2 | 2+2 2 |
| | 5+6 2 | 3+4 2 | 5+6 2 | 3+4 2 | 2+2 2 | 3+3 2 |
| | 5+6 2 | 5+6 2 | 3+4 2 | 5+6 2 | 2+2 2 | 2+2 2 |
| | 3+4 2 | 3+4 2 | 5+6 2 | 3+4 2 | 2+2 2 | 3+3 2 |
| | 5+6 2 | 5+6 2 | 3+4 2 | 5+6 2 | 3+3 2 | 2+2 2 |
| | 3+4 2 | 3+4 2 | 5+6 2 | 5+6 2 | 3+3 2 | 3+3 2 |
| | 3+4 2 | 5+6 2 | 3+4 2 | 3+4 2 | 2+2 2 | 2+2 2 |
| | 5+6 2 | 3+4 2 | 5+6 2 | 5+6 2 | 3+3 2 | 3+3 2 |
| | 5+6 2 | 5+6 2 | 3+4 2 | 5+6 2 | 3+3 2 | 3+3 2 |
| | 3+4 2 | 3+4 2 | 5+6 2 | 3+4 2 | 2+2 2 | 2+2 2 |
| | 5+6 2 | 3+4 2 | 3+4 2 | 5+6 2 | 3+3 2 | 3+3 2 |
| | 3+4 2 | 3+4 2 | 3+4 2 | 3+4 2 | 2+2 2 | 2+2 2 |
| | 3+4 2 | 5+6 2 | 5+6 2 | 3+4 2 | 2+2 2 | 3+3 2 |
| | 5+6 2 | 3+4 2 | 5+6 2 | 5+6 2 | 3+3 2 | 2+2 2 |
| 12.2.41 | 162649 | | 106 | H8Cu2Mg1O12V2 | | |
| | -A | -GM | -M | -Y | -L | -V |
| | 3+4 2 | 3+4 2 | 3+4 2 | 3+4 2 | 2+2 2 | 2+2 2 |
| | 5+6 2 | 3+4 2 | 5+6 2 | 5+6 2 | 3+3 2 | 3+3 2 |
| | 3+4 2 | 5+6 2 | 5+6 2 | 5+6 2 | 3+3 2 | 2+2 2 |
| | 5+6 2 | 5+6 2 | 3+4 2 | 3+4 2 | 2+2 2 | 3+3 2 |

Continued on next column

**TABLE XXIV– continued**

| TQC | ICSD | | NoE | SF | | |
|---|---|---|---|---|---|---|
| | 3+4 2 | 3+4 2 | 3+4 2 | 3+4 2 | 2+2 2 | 2+2 2 |
| | 5+6 2 | 5+6 2 | 5+6 2 | 3+4 2 | 3+3 2 | 3+3 2 |
| | 3+4 2 | 3+4 2 | 3+4 2 | 5+6 2 | 2+2 2 | 2+2 2 |
| | 5+6 2 | 5+6 2 | 5+6 2 | 5+6 2 | 3+3 2 | 3+3 2 |
| | 3+4 2 | 3+4 2 | 3+4 2 | 3+4 2 | 2+2 2 | 2+2 2 |
| | 5+6 2 | 3+4 2 | 5+6 2 | 3+4 2 | 3+3 2 | 2+2 2 |
| | 5+6 2 | 3+4 2 | 3+4 2 | 5+6 2 | 2+2 2 | 3+3 2 |
| | 3+4 2 | 5+6 2 | 5+6 2 | 3+4 2 | 3+3 2 | 2+2 2 |
| | 5+6 2 | 5+6 2 | 5+6 2 | 5+6 2 | 3+3 2 | 3+3 2 |
| | 5+6 2 | 5+6 2 | 5+6 2 | 5+6 2 | 3+3 2 | 3+3 2 |
| | 5+6 2 | 5+6 2 | 5+6 2 | 5+6 2 | 3+3 2 | 3+3 2 |
| | 3+4 2 | 3+4 2 | 5+6 2 | 5+6 2 | 3+3 2 | 3+3 2 |
| | 3+4 2 | 3+4 2 | 3+4 2 | 3+4 2 | 2+2 2 | 2+2 2 |
| | 3+4 2 | 5+6 2 | 5+6 2 | 5+6 2 | 2+2 2 | 2+2 2 |
| | 5+6 2 | 3+4 2 | 5+6 2 | 5+6 2 | 3+3 2 | 3+3 2 |
| | 3+4 2 | 3+4 2 | 3+4 2 | 3+4 2 | 2+2 2 | 3+3 2 |
| | 3+4 2 | 5+6 2 | 3+4 2 | 3+4 2 | 2+2 2 | 2+2 2 |
| | 5+6 2 | 3+4 2 | 3+4 2 | 5+6 2 | 3+3 2 | 2+2 2 |
| | 3+4 2 | 5+6 2 | 5+6 2 | 5+6 2 | 2+2 2 | 3+3 2 |
| | 5+6 2 | 3+4 2 | 5+6 2 | 5+6 2 | 3+3 2 | 2+2 2 |
| | 3+4 2 | 5+6 2 | 3+4 2 | 3+4 2 | 2+2 2 | 3+3 2 |
| | 5+6 2 | 5+6 2 | 3+4 2 | 5+6 2 | 2+2 2 | 2+2 2 |
| | 3+4 2 | 3+4 2 | 5+6 2 | 3+4 2 | 3+3 2 | 3+3 2 |
| | 3+4 2 | 3+4 2 | 3+4 2 | 5+6 2 | 2+2 2 | 2+2 2 |
| | 5+6 2 | 3+4 2 | 5+6 2 | 3+4 2 | 2+2 2 | 2+2 2 |
| | 3+4 2 | 5+6 2 | 5+6 2 | 5+6 2 | 3+3 2 | 3+3 2 |
| | 5+6 2 | 5+6 2 | 3+4 2 | 3+4 2 | 2+2 2 | 2+2 2 |
| | 3+4 2 | 3+4 2 | 5+6 2 | 5+6 2 | 3+3 2 | 3+3 2 |
| | 3+4 2 | 3+4 2 | 3+4 2 | 3+4 2 | 2+2 2 | 2+2 2 |
| | 3+4 2 | 3+4 2 | 5+6 2 | 3+4 2 | 2+2 2 | 3+3 2 |
| | 3+4 2 | 3+4 2 | 3+4 2 | 5+6 2 | 3+3 2 | 2+2 2 |
| | 3+4 2 | 3+4 2 | 3+4 2 | 5+6 2 | 3+3 2 | 3+3 2 |
| | 3+4 2 | 5+6 2 | 5+6 2 | 3+4 2 | 2+2 2 | 3+3 2 |
| | 3+4 2 | 3+4 2 | 5+6 2 | 3+4 2 | 3+3 2 | 2+2 2 |
| | 5+6 2 | 3+4 2 | 5+6 2 | 5+6 2 | 2+2 2 | 3+3 2 |
| | 5+6 2 | 5+6 2 | 5+6 2 | 5+6 2 | 3+3 2 | 2+2 2 |
| | 3+4 2 | 3+4 2 | 3+4 2 | 5+6 2 | 3+3 2 | 3+3 2 |
| | 3+4 2 | 5+6 2 | 5+6 2 | 5+6 2 | 2+2 2 | 2+2 2 |
| | 5+6 2 | 3+4 2 | 5+6 2 | 5+6 2 | 2+2 2 | 3+3 2 |
| | 5+6 2 | 5+6 2 | 5+6 2 | 5+6 2 | 3+3 2 | 2+2 2 |
| | 3+4 2 | 3+4 2 | 3+4 2 | 5+6 2 | 2+2 2 | 3+3 2 |
| | 5+6 2 | 5+6 2 | 5+6 2 | 3+4 2 | 3+3 2 | 3+3 2 |
| | 3+4 2 | 5+6 2 | 3+4 2 | 5+6 2 | 3+3 2 | 2+2 2 |
| | 5+6 2 | 3+4 2 | 5+6 2 | 5+6 2 | 2+2 2 | 2+2 2 |
| | 3+4 2 | 3+4 2 | 3+4 2 | 5+6 2 | 2+2 2 | 3+3 2 |
| | 3+4 2 | 3+4 2 | 5+6 2 | 5+6 2 | 3+3 2 | 3+3 2 |
| | 5+6 2 | 5+6 2 | 3+4 2 | 3+4 2 | 2+2 2 | 3+3 2 |
| | 3+4 2 | 5+6 2 | 5+6 2 | 5+6 2 | 2+2 2 | 3+3 2 |
| | 3+4 2 | 3+4 2 | 3+4 2 | 3+4 2 | 3+3 2 | 2+2 2 |
| 12.2.42 | 33293 | | 42 | H6Au1Br3N2 | | |
| | -A | -GM | -M | -Y | -L | -V |
| | 3+4 2 | 3+4 2 | 3+4 2 | 3+4 2 | 3+3 2 | 3+3 2 |
| | 5+6 2 | 5+6 2 | 5+6 2 | 5+6 2 | 2+2 2 | 2+2 2 |
| | 3+4 2 | 3+4 2 | 3+4 2 | 3+4 2 | 3+3 2 | 3+3 2 |
| | 5+6 2 | 3+4 2 | 5+6 2 | 3+4 2 | 3+3 2 | 2+2 2 |
| | 5+6 2 | 5+6 2 | 5+6 2 | 5+6 2 | 2+2 2 | 2+2 2 |
| | 3+4 2 | 3+4 2 | 3+4 2 | 3+4 2 | 3+3 2 | 3+3 2 |
| | 3+4 2 | 3+4 2 | 3+4 2 | 3+4 2 | 3+3 2 | 3+3 2 |
| | 3+4 2 | 3+4 2 | 3+4 2 | 3+4 2 | 3+3 2 | 3+3 2 |
| | 3+4 2 | 3+4 2 | 3+4 2 | 3+4 2 | 3+3 2 | 3+3 2 |
| | 3+4 2 | 3+4 2 | 3+4 2 | 3+4 2 | 3+3 2 | 3+3 2 |

Continued on next column

**TABLE XXIV– continued**

| TQC | ICSD | | NoE | SF | | |
|---|---|---|---|---|---|---|
| | 5+6 2 | 5+6 2 | 3+4 2 | 5+6 2 | 2+2 2 | 2+2 2 |
| | 3+4 2 | 3+4 2 | 5+6 2 | 3+4 2 | 3+3 2 | 3+3 2 |
| | 5+6 2 | 5+6 2 | 5+6 2 | 5+6 2 | 2+2 2 | 2+2 2 |
| | 3+4 2 | 5+6 2 | 5+6 2 | 5+6 2 | 2+2 2 | 2+2 2 |
| | 3+4 2 | 3+4 2 | 3+4 2 | 5+6 2 | 2+2 2 | 3+3 2 |
| | 5+6 2 | 5+6 2 | 3+4 2 | 3+4 2 | 3+3 2 | 3+3 2 |
| | 5+6 2 | 5+6 2 | 5+6 2 | 5+6 2 | 2+2 2 | 3+3 2 |
| | 3+4 2 | 3+4 2 | 3+4 2 | 3+4 2 | 2+2 2 | 3+3 2 |
| | 3+4 2 | 5+6 2 | 5+6 2 | 5+6 2 | 3+3 2 | 2+2 2 |
| | 3+4 2 | 5+6 2 | 3+4 2 | 5+6 2 | 2+2 2 | 2+2 2 |
| | 5+6 2 | 5+6 2 | 5+6 2 | 5+6 2 | 2+2 2 | 2+2 2 |
| 12.2.43 | 2559 | | 60 | H4F6Li2O2Sn1 | | |
| | -A | -GM | -M | -Y | -L | -V |
| | 3+4 2 | 3+4 2 | 3+4 2 | 3+4 2 | 3+3 2 | 3+3 2 |
| | 5+6 2 | 5+6 2 | 5+6 2 | 5+6 2 | 2+2 2 | 2+2 2 |
| | 5+6 2 | 5+6 2 | 5+6 2 | 5+6 2 | 2+2 2 | 2+2 2 |
| | 5+6 2 | 5+6 2 | 5+6 2 | 5+6 2 | 2+2 2 | 2+2 2 |
| | 3+4 2 | 3+4 2 | 3+4 2 | 3+4 2 | 3+3 2 | 3+3 2 |
| | 3+4 2 | 3+4 2 | 3+4 2 | 3+4 2 | 3+3 2 | 3+3 2 |
| | 3+4 2 | 5+6 2 | 5+6 2 | 3+4 2 | 3+3 2 | 2+2 2 |
| | 5+6 2 | 3+4 2 | 3+4 2 | 5+6 2 | 2+2 2 | 3+3 2 |
| | 3+4 2 | 3+4 2 | 3+4 2 | 3+4 2 | 3+3 2 | 3+3 2 |
| | 5+6 2 | 5+6 2 | 5+6 2 | 5+6 2 | 2+2 2 | 2+2 2 |
| | 5+6 2 | 5+6 2 | 5+6 2 | 5+6 2 | 2+2 2 | 2+2 2 |
| | 5+6 2 | 5+6 2 | 5+6 2 | 5+6 2 | 2+2 2 | 2+2 2 |
| | 3+4 2 | 3+4 2 | 3+4 2 | 3+4 2 | 3+3 2 | 3+3 2 |
| | 3+4 2 | 3+4 2 | 3+4 2 | 3+4 2 | 3+3 2 | 3+3 2 |
| | 5+6 2 | 3+4 2 | 5+6 2 | 5+6 2 | 3+3 2 | 2+2 2 |
| | 3+4 2 | 5+6 2 | 3+4 2 | 3+4 2 | 2+2 2 | 3+3 2 |
| | 3+4 2 | 5+6 2 | 5+6 2 | 3+4 2 | 2+2 2 | 3+3 2 |
| | 5+6 2 | 3+4 2 | 5+6 2 | 5+6 2 | 2+2 2 | 2+2 2 |
| | 5+6 2 | 5+6 2 | 3+4 2 | 5+6 2 | 3+3 2 | 2+2 2 |
| | 3+4 2 | 3+4 2 | 3+4 2 | 3+4 2 | 3+3 2 | 2+2 2 |
| | 5+6 2 | 5+6 2 | 5+6 2 | 5+6 2 | 2+2 2 | 3+3 2 |
| | 3+4 2 | 5+6 2 | 5+6 2 | 5+6 2 | 3+3 2 | 2+2 2 |
| | 5+6 2 | 5+6 2 | 3+4 2 | 3+4 2 | 2+2 2 | 3+3 2 |
| | 5+6 2 | 3+4 2 | 3+4 2 | 3+4 2 | 3+3 2 | 2+2 2 |
| | 3+4 2 | 3+4 2 | 5+6 2 | 5+6 2 | 2+2 2 | 3+3 2 |
| | 3+4 2 | 3+4 2 | 3+4 2 | 3+4 2 | 3+3 2 | 3+3 2 |
| | 3+4 2 | 3+4 2 | 5+6 2 | 3+4 2 | 3+3 2 | 2+2 2 |
| | 5+6 2 | 5+6 2 | 3+4 2 | 3+4 2 | 2+2 2 | 3+3 2 |
| | 3+4 2 | 5+6 2 | 5+6 2 | 5+6 2 | 2+2 2 | 2+2 2 |
| | 5+6 2 | 3+4 2 | 5+6 2 | 5+6 2 | 3+3 2 | 3+3 2 |
| 12.2.44 | 201563 | | 116 | H4Cl8N1O2Rb3Ru2 | | |
| | -A | -GM | -M | -Y | -L | -V |
| | 3+4 2 | 3+4 2 | 3+4 2 | 3+4 2 | 3+3 2 | 3+3 2 |
| | 5+6 2 | 5+6 2 | 5+6 2 | 5+6 2 | 2+2 2 | 2+2 2 |
| | 5+6 2 | 3+4 2 | 5+6 2 | 3+4 2 | 3+3 2 | 2+2 2 |
| | 5+6 2 | 3+4 2 | 5+6 2 | 3+4 2 | 2+2 2 | 3+3 2 |
| | 3+4 2 | 5+6 2 | 3+4 2 | 5+6 2 | 3+3 2 | 2+2 2 |
| | 5+6 2 | 3+4 2 | 5+6 2 | 3+4 2 | 2+2 2 | 3+3 2 |
| | 3+4 2 | 5+6 2 | 3+4 2 | 5+6 2 | 3+3 2 | 2+2 2 |
| | 5+6 2 | 3+4 2 | 5+6 2 | 3+4 2 | 2+2 2 | 3+3 2 |
| | 3+4 2 | 3+4 2 | 5+6 2 | 5+6 2 | 2+2 2 | 2+2 2 |
| | 3+4 2 | 5+6 2 | 3+4 2 | 5+6 2 | 3+3 2 | 3+3 2 |
| | 5+6 2 | 5+6 2 | 3+4 2 | 3+4 2 | 3+3 2 | 2+2 2 |
| | 5+6 2 | 3+4 2 | 5+6 2 | 3+4 2 | 2+2 2 | 3+3 2 |
| | 3+4 2 | 5+6 2 | 5+6 2 | 5+6 2 | 2+2 2 | 2+2 2 |
| | 5+6 2 | 3+4 2 | 3+4 2 | 3+4 2 | 3+3 2 | 3+3 2 |
| | 3+4 2 | 3+4 2 | 3+4 2 | 3+4 2 | 3+3 2 | 3+3 2 |
| | 5+6 2 | 5+6 2 | 5+6 2 | 5+6 2 | 2+2 2 | 2+2 2 |

Continued on next column

**TABLE XXIV– continued**

| TQC | ICSD | | NoE | SF | | |
|---|---|---|---|---|---|---|
| | 3+4 2 | 5+6 2 | 3+4 2 | 5+6 2 | 2+2 2 | 3+3 2 |
| | 3+4 2 | 3+4 2 | 3+4 2 | 3+4 2 | 3+3 2 | 3+3 2 |
| | 5+6 2 | 5+6 2 | 5+6 2 | 5+6 2 | 2+2 2 | 2+2 2 |
| | 3+4 2 | 3+4 2 | 3+4 2 | 5+6 2 | 3+3 2 | 3+3 2 |
| | 5+6 2 | 5+6 2 | 5+6 2 | 3+4 2 | 2+2 2 | 2+2 2 |
| | 3+4 2 | 5+6 2 | 3+4 2 | 5+6 2 | 2+2 2 | 3+3 2 |
| | 3+4 2 | 5+6 2 | 3+4 2 | 5+6 2 | 2+2 2 | 3+3 2 |
| | 3+4 2 | 5+6 2 | 3+4 2 | 5+6 2 | 3+3 2 | 2+2 2 |
| | 3+4 2 | 5+6 2 | 3+4 2 | 5+6 2 | 3+3 2 | 2+2 2 |
| | 3+4 2 | 5+6 2 | 3+4 2 | 5+6 2 | 3+3 2 | 2+2 2 |
| | 5+6 2 | 3+4 2 | 5+6 2 | 3+4 2 | 2+2 2 | 3+3 2 |
| | 5+6 2 | 3+4 2 | 5+6 2 | 5+6 2 | 3+3 2 | 2+2 2 |
| | 3+4 2 | 5+6 2 | 3+4 2 | 3+4 2 | 2+2 2 | 3+3 2 |
| | 3+4 2 | 5+6 2 | 3+4 2 | 5+6 2 | 3+3 2 | 2+2 2 |
| | 5+6 2 | 3+4 2 | 5+6 2 | 3+4 2 | 3+3 2 | 3+3 2 |
| | 3+4 2 | 5+6 2 | 3+4 2 | 5+6 2 | 2+2 2 | 2+2 2 |
| | 5+6 2 | 3+4 2 | 5+6 2 | 3+4 2 | 2+2 2 | 3+3 2 |
| | 5+6 2 | 3+4 2 | 5+6 2 | 5+6 2 | 2+2 2 | 3+3 2 |
| | 3+4 2 | 5+6 2 | 3+4 2 | 5+6 2 | 3+3 2 | 2+2 2 |
| | 3+4 2 | 5+6 2 | 3+4 2 | 3+4 2 | 3+3 2 | 2+2 2 |
| | 5+6 2 | 3+4 2 | 5+6 2 | 3+4 2 | 2+2 2 | 3+3 2 |
| | 5+6 2 | 3+4 2 | 5+6 2 | 3+4 2 | 2+2 2 | 3+3 2 |
| | 5+6 2 | 3+4 2 | 5+6 2 | 3+4 2 | 2+2 2 | 3+3 2 |
| | 3+4 2 | 5+6 2 | 3+4 2 | 5+6 2 | 3+3 2 | 2+2 2 |
| | 3+4 2 | 5+6 2 | 5+6 2 | 3+4 2 | 3+3 2 | 2+2 2 |
| | 5+6 2 | 3+4 2 | 3+4 2 | 5+6 2 | 2+2 2 | 3+3 2 |
| | 3+4 2 | 3+4 2 | 5+6 2 | 3+4 2 | 3+3 2 | 2+2 2 |
| | 5+6 2 | 5+6 2 | 3+4 2 | 5+6 2 | 2+2 2 | 3+3 2 |
| | 3+4 2 | 3+4 2 | 5+6 2 | 3+4 2 | 3+3 2 | 2+2 2 |
| | 5+6 2 | 5+6 2 | 3+4 2 | 5+6 2 | 3+3 2 | 3+3 2 |
| | 3+4 2 | 3+4 2 | 3+4 2 | 5+6 2 | 2+2 2 | 2+2 2 |
| | 5+6 2 | 5+6 2 | 3+4 2 | 5+6 2 | 3+3 2 | 2+2 2 |
| | 3+4 2 | 5+6 2 | 5+6 2 | 5+6 2 | 2+2 2 | 2+2 2 |
| | 5+6 2 | 3+4 2 | 5+6 2 | 3+4 2 | 3+3 2 | 3+3 2 |
| | 3+4 2 | 5+6 2 | 3+4 2 | 3+4 2 | 2+2 2 | 3+3 2 |
| | 3+4 2 | 5+6 2 | 3+4 2 | 5+6 2 | 3+3 2 | 2+2 2 |
| | 5+6 2 | 3+4 2 | 5+6 2 | 3+4 2 | 2+2 2 | 3+3 2 |
| | 5+6 2 | 3+4 2 | 5+6 2 | 3+4 2 | 3+3 2 | 3+3 2 |
| | 3+4 2 | 3+4 2 | 3+4 2 | 5+6 2 | 2+2 2 | 2+2 2 |
| | 3+4 2 | 5+6 2 | 3+4 2 | 3+4 2 | 3+3 2 | 3+3 2 |
| | 5+6 2 | 5+6 2 | 5+6 2 | 5+6 2 | 2+2 2 | 2+2 2 |
| | 3+4 2 | 3+4 2 | 3+4 2 | 5+6 2 | 2+2 2 | 2+2 2 |
| 12.2.45 | 16130 | | 116 | H4Cl8K3N1O2Ru2 | | |
| | -A | -GM | -M | -Y | -L | -V |
| | 5+6 2 | 3+4 2 | 5+6 2 | 3+4 2 | 2+2 2 | 3+3 2 |
| | 3+4 2 | 5+6 2 | 3+4 2 | 5+6 2 | 3+3 2 | 2+2 2 |
| | 3+4 2 | 3+4 2 | 3+4 2 | 3+4 2 | 2+2 2 | 2+2 2 |
| | 3+4 2 | 3+4 2 | 3+4 2 | 3+4 2 | 3+3 2 | 3+3 2 |
| | 5+6 2 | 5+6 2 | 5+6 2 | 5+6 2 | 2+2 2 | 2+2 2 |
| | 3+4 2 | 3+4 2 | 3+4 2 | 3+4 2 | 3+3 2 | 3+3 2 |
| | 5+6 2 | 3+4 2 | 5+6 2 | 5+6 2 | 2+2 2 | 3+3 2 |
| | 3+4 2 | 5+6 2 | 3+4 2 | 3+4 2 | 3+3 2 | 2+2 2 |
| | 5+6 2 | 5+6 2 | 3+4 2 | 5+6 2 | 3+3 2 | 2+2 2 |
| | 5+6 2 | 3+4 2 | 5+6 2 | 5+6 2 | 2+2 2 | 3+3 2 |
| | 3+4 2 | 3+4 2 | 3+4 2 | 3+4 2 | 3+3 2 | 2+2 2 |
| | 5+6 2 | 5+6 2 | 5+6 2 | 3+4 2 | 2+2 2 | 3+3 2 |
| | 3+4 2 | 5+6 2 | 3+4 2 | 5+6 2 | 2+2 2 | 2+2 2 |
| | 3+4 2 | 3+4 2 | 5+6 2 | 3+4 2 | 3+3 2 | 3+3 2 |
| | 5+6 2 | 3+4 2 | 5+6 2 | 3+4 2 | 2+2 2 | 3+3 2 |
| | 3+4 2 | 5+6 2 | 3+4 2 | 5+6 2 | 3+3 2 | 2+2 2 |
| | 5+6 2 | 5+6 2 | 5+6 2 | 5+6 2 | 3+3 2 | 3+3 2 |

Continued on next column

**TABLE XXIV– continued**

| TQC | ICSD | NoE | SF | | | |
|---|---|---|---|---|---|---|
| | 5+6 2 | 5+6 2 | 5+6 2 | 3+4 2 | 2+2 2 | 3+3 2 |
| | 3+4 2 | 3+4 2 | 3+4 2 | 5+6 2 | 3+3 2 | 2+2 2 |
| | 5+6 2 | 3+4 2 | 5+6 2 | 3+4 2 | 2+2 2 | 3+3 2 |
| | 3+4 2 | 5+6 2 | 3+4 2 | 5+6 2 | 3+3 2 | 2+2 2 |
| | 5+6 2 | 5+6 2 | 5+6 2 | 5+6 2 | 3+3 2 | 3+3 2 |
| | 5+6 2 | 5+6 2 | 5+6 2 | 5+6 2 | 3+3 2 | 3+3 2 |
| | 5+6 2 | 5+6 2 | 5+6 2 | 5+6 2 | 2+2 2 | 2+2 2 |
| | 5+6 2 | 5+6 2 | 5+6 2 | 5+6 2 | 2+2 2 | 2+2 2 |
| | 5+6 2 | 5+6 2 | 5+6 2 | 5+6 2 | 2+2 2 | 2+2 2 |
| | 3+4 2 | 3+4 2 | 3+4 2 | 3+4 2 | 3+3 2 | 3+3 2 |
| | 3+4 2 | 3+4 2 | 3+4 2 | 5+6 2 | 2+2 2 | 2+2 2 |
| | 5+6 2 | 5+6 2 | 5+6 2 | 5+6 2 | 3+3 2 | 3+3 2 |
| | 5+6 2 | 5+6 2 | 5+6 2 | 3+4 2 | 2+2 2 | 2+2 2 |
| | 3+4 2 | 3+4 2 | 3+4 2 | 3+4 2 | 2+2 2 | 3+3 2 |
| | 5+6 2 | 5+6 2 | 5+6 2 | 5+6 2 | 3+3 2 | 2+2 2 |
| | 3+4 2 | 3+4 2 | 3+4 2 | 3+4 2 | 3+3 2 | 3+3 2 |
| | 3+4 2 | 3+4 2 | 5+6 2 | 5+6 2 | 3+3 2 | 3+3 2 |
| | 5+6 2 | 5+6 2 | 3+4 2 | 5+6 2 | 2+2 2 | 2+2 2 |
| | 5+6 2 | 5+6 2 | 5+6 2 | 3+4 2 | 2+2 2 | 2+2 2 |
| | 3+4 2 | 3+4 2 | 3+4 2 | 3+4 2 | 3+3 2 | 3+3 2 |
| | 3+4 2 | 3+4 2 | 3+4 2 | 3+4 2 | 3+3 2 | 3+3 2 |
| | 3+4 2 | 3+4 2 | 3+4 2 | 3+4 2 | 3+3 2 | 3+3 2 |
| | 5+6 2 | 5+6 2 | 5+6 2 | 5+6 2 | 2+2 2 | 2+2 2 |
| | 5+6 2 | 5+6 2 | 3+4 2 | 3+4 2 | 3+3 2 | 2+2 2 |
| | 5+6 2 | 3+4 2 | 5+6 2 | 5+6 2 | 2+2 2 | 3+3 2 |
| | 3+4 2 | 3+4 2 | 3+4 2 | 5+6 2 | 3+3 2 | 3+3 2 |
| | 3+4 2 | 3+4 2 | 5+6 2 | 3+4 2 | 2+2 2 | 2+2 2 |
| | 5+6 2 | 5+6 2 | 5+6 2 | 5+6 2 | 2+2 2 | 2+2 2 |
| | 3+4 2 | 5+6 2 | 3+4 2 | 3+4 2 | 2+2 2 | 3+3 2 |
| | 5+6 2 | 5+6 2 | 5+6 2 | 5+6 2 | 3+3 2 | 2+2 2 |
| | 5+6 2 | 5+6 2 | 3+4 2 | 3+4 2 | 2+2 2 | 2+2 2 |
| | 3+4 2 | 3+4 2 | 5+6 2 | 5+6 2 | 3+3 2 | 3+3 2 |
| | 3+4 2 | 3+4 2 | 3+4 2 | 5+6 2 | 3+3 2 | 2+2 2 |
| | 5+6 2 | 5+6 2 | 5+6 2 | 3+4 2 | 2+2 2 | 3+3 2 |
| | 5+6 2 | 5+6 2 | 5+6 2 | 5+6 2 | 2+2 2 | 2+2 2 |
| | 3+4 2 | 3+4 2 | 3+4 2 | 3+4 2 | 3+3 2 | 3+3 2 |
| | 3+4 2 | 3+4 2 | 5+6 2 | 3+4 2 | 2+2 2 | 3+3 2 |
| | 5+6 2 | 3+4 2 | 3+4 2 | 5+6 2 | 3+3 2 | 2+2 2 |
| | 5+6 2 | 5+6 2 | 5+6 2 | 5+6 2 | 2+2 2 | 2+2 2 |
| | 3+4 2 | 5+6 2 | 3+4 2 | 3+4 2 | 3+3 2 | 3+3 2 |
| | 5+6 2 | 3+4 2 | 5+6 2 | 5+6 2 | 2+2 2 | 2+2 2 |
| 12.2.46 | 200012 | 84 | H4Al1Cr2K1O10 | | | |
| | -A | -GM | -M | -Y | -L | -V |
| | 5+6 2 | 3+4 2 | 5+6 2 | 3+4 2 | 2+2 2 | 3+3 2 |
| | 3+4 2 | 3+4 2 | 3+4 2 | 3+4 2 | 3+3 2 | 3+3 2 |
| | 5+6 2 | 5+6 2 | 5+6 2 | 5+6 2 | 2+2 2 | 2+2 2 |
| | 5+6 2 | 5+6 2 | 5+6 2 | 5+6 2 | 3+3 2 | 2+2 2 |
| | 3+4 2 | 3+4 2 | 3+4 2 | 3+4 2 | 2+2 2 | 3+3 2 |
| | 5+6 2 | 3+4 2 | 3+4 2 | 5+6 2 | 3+3 2 | 2+2 2 |
| | 3+4 2 | 5+6 2 | 5+6 2 | 3+4 2 | 2+2 2 | 3+3 2 |
| | 3+4 2 | 5+6 2 | 5+6 2 | 3+4 2 | 3+3 2 | 2+2 2 |
| | 5+6 2 | 3+4 2 | 3+4 2 | 5+6 2 | 2+2 2 | 3+3 2 |
| | 5+6 2 | 5+6 2 | 5+6 2 | 5+6 2 | 2+2 2 | 2+2 2 |
| | 3+4 2 | 3+4 2 | 3+4 2 | 3+4 2 | 3+3 2 | 3+3 2 |
| | 3+4 2 | 5+6 2 | 3+4 2 | 5+6 2 | 3+3 2 | 2+2 2 |
| | 3+4 2 | 5+6 2 | 3+4 2 | 5+6 2 | 3+3 2 | 2+2 2 |
| | 3+4 2 | 5+6 2 | 3+4 2 | 5+6 2 | 3+3 2 | 2+2 2 |
| | 3+4 2 | 3+4 2 | 3+4 2 | 3+4 2 | 3+3 2 | 3+3 2 |
| | 5+6 2 | 5+6 2 | 5+6 2 | 5+6 2 | 2+2 2 | 2+2 2 |
| | 5+6 2 | 5+6 2 | 5+6 2 | 5+6 2 | 2+2 2 | 2+2 2 |
| | 3+4 2 | 3+4 2 | 3+4 2 | 3+4 2 | 3+3 2 | 3+3 2 |

Continued on next column

**TABLE XXIV– continued**

| TQC | ICSD | NoE | SF | | | |
|---|---|---|---|---|---|---|
| | 3+4 2 | 3+4 2 | 5+6 2 | 3+4 2 | 3+3 2 | 3+3 2 |
| | 5+6 2 | 5+6 2 | 3+4 2 | 3+4 2 | 3+3 2 | 3+3 2 |
| | 5+6 2 | 3+4 2 | 5+6 2 | 5+6 2 | 2+2 2 | 2+2 2 |
| | 3+4 2 | 3+4 2 | 3+4 2 | 5+6 2 | 3+3 2 | 2+2 2 |
| | 5+6 2 | 5+6 2 | 5+6 2 | 3+4 2 | 2+2 2 | 2+2 2 |
| | 3+4 2 | 5+6 2 | 3+4 2 | 5+6 2 | 2+2 2 | 3+3 2 |
| | 5+6 2 | 3+4 2 | 3+4 2 | 5+6 2 | 2+2 2 | 3+3 2 |
| | 3+4 2 | 5+6 2 | 5+6 2 | 3+4 2 | 3+3 2 | 2+2 2 |
| | 3+4 2 | 3+4 2 | 3+4 2 | 3+4 2 | 2+2 2 | 2+2 2 |
| | 5+6 2 | 3+4 2 | 3+4 2 | 5+6 2 | 3+3 2 | 3+3 2 |
| | 5+6 2 | 3+4 2 | 5+6 2 | 5+6 2 | 2+2 2 | 2+2 2 |
| | 3+4 2 | 5+6 2 | 5+6 2 | 3+4 2 | 3+3 2 | 3+3 2 |
| | 3+4 2 | 5+6 2 | 3+4 2 | 3+4 2 | 2+2 2 | 2+2 2 |
| | 5+6 2 | 5+6 2 | 5+6 2 | 5+6 2 | 3+3 2 | 3+3 2 |
| | 5+6 2 | 3+4 2 | 5+6 2 | 5+6 2 | 2+2 2 | 2+2 2 |
| | 5+6 2 | 5+6 2 | 5+6 2 | 5+6 2 | 3+3 2 | 3+3 2 |
| | 3+4 2 | 3+4 2 | 3+4 2 | 3+4 2 | 3+3 2 | 3+3 2 |
| | 3+4 2 | 5+6 2 | 3+4 2 | 3+4 2 | 2+2 2 | 2+2 2 |
| | 5+6 2 | 5+6 2 | 5+6 2 | 5+6 2 | 2+2 2 | 3+3 2 |
| | 5+6 2 | 3+4 2 | 5+6 2 | 3+4 2 | 3+3 2 | 2+2 2 |
| | 3+4 2 | 5+6 2 | 3+4 2 | 3+4 2 | 3+3 2 | 2+2 2 |
| | 3+4 2 | 3+4 2 | 3+4 2 | 3+4 2 | 2+2 2 | 3+3 2 |
| | 5+6 2 | 5+6 2 | 3+4 2 | 5+6 2 | 3+3 2 | 2+2 2 |
| | 5+6 2 | 3+4 2 | 5+6 2 | 5+6 2 | 2+2 2 | 3+3 2 |
| 12.2.47 | 35156 | 26 | H3Li1O2 | | | |
| | -A | -GM | -M | -Y | -L | -V |
| | 3+4 2 | 3+4 2 | 3+4 2 | 3+4 2 | 3+3 2 | 3+3 2 |
| | 5+6 2 | 5+6 2 | 5+6 2 | 5+6 2 | 2+2 2 | 2+2 2 |
| | 5+6 2 | 5+6 2 | 5+6 2 | 5+6 2 | 2+2 2 | 2+2 2 |
| | 3+4 2 | 3+4 2 | 3+4 2 | 3+4 2 | 3+3 2 | 3+3 2 |
| | 5+6 2 | 3+4 2 | 3+4 2 | 3+4 2 | 2+2 2 | 2+2 2 |
| | 3+4 2 | 3+4 2 | 3+4 2 | 5+6 2 | 3+3 2 | 3+3 2 |
| | 3+4 2 | 5+6 2 | 5+6 2 | 3+4 2 | 3+3 2 | 2+2 2 |
| | 5+6 2 | 5+6 2 | 5+6 2 | 5+6 2 | 2+2 2 | 3+3 2 |
| | 3+4 2 | 3+4 2 | 5+6 2 | 3+4 2 | 3+3 2 | 2+2 2 |
| | 3+4 2 | 3+4 2 | 5+6 2 | 5+6 2 | 3+3 2 | 3+3 2 |
| | 5+6 2 | 5+6 2 | 3+4 2 | 5+6 2 | 2+2 2 | 3+3 2 |
| | 5+6 2 | 5+6 2 | 3+4 2 | 3+4 2 | 2+2 2 | 2+2 2 |
| | 5+6 2 | 3+4 2 | 5+6 2 | 5+6 2 | 3+3 2 | 2+2 2 |
| 12.2.48 | 30223 | 26 | H3Li1O2 | | | |
| | -A | -GM | -M | -Y | -L | -V |
| | 3+4 2 | 3+4 2 | 3+4 2 | 3+4 2 | 3+3 2 | 3+3 2 |
| | 5+6 2 | 5+6 2 | 5+6 2 | 5+6 2 | 2+2 2 | 2+2 2 |
| | 5+6 2 | 5+6 2 | 5+6 2 | 5+6 2 | 2+2 2 | 2+2 2 |
| | 3+4 2 | 3+4 2 | 3+4 2 | 3+4 2 | 3+3 2 | 3+3 2 |
| | 5+6 2 | 3+4 2 | 3+4 2 | 3+4 2 | 2+2 2 | 2+2 2 |
| | 3+4 2 | 3+4 2 | 3+4 2 | 5+6 2 | 3+3 2 | 3+3 2 |
| | 3+4 2 | 5+6 2 | 5+6 2 | 3+4 2 | 3+3 2 | 2+2 2 |
| | 5+6 2 | 5+6 2 | 5+6 2 | 5+6 2 | 2+2 2 | 3+3 2 |
| | 3+4 2 | 3+4 2 | 5+6 2 | 3+4 2 | 3+3 2 | 2+2 2 |
| | 3+4 2 | 5+6 2 | 5+6 2 | 5+6 2 | 3+3 2 | 3+3 2 |
| | 5+6 2 | 3+4 2 | 3+4 2 | 5+6 2 | 2+2 2 | 3+3 2 |
| | 5+6 2 | 5+6 2 | 3+4 2 | 3+4 2 | 2+2 2 | 2+2 2 |
| | 5+6 2 | 3+4 2 | 5+6 2 | 5+6 2 | 3+3 2 | 2+2 2 |
| 12.2.49 | 27761 | 28 | H3Li1O2 | | | |
| | -A | -GM | -M | -Y | -L | -V |
| | 5+6 2 | 3+4 2 | 5+6 2 | 3+4 2 | 3+3 2 | 2+2 2 |
| | 3+4 2 | 5+6 2 | 3+4 2 | 5+6 2 | 2+2 2 | 3+3 2 |
| | 3+4 2 | 3+4 2 | 5+6 2 | 5+6 2 | 3+3 2 | 2+2 2 |
| | 5+6 2 | 5+6 2 | 3+4 2 | 3+4 2 | 2+2 2 | 3+3 2 |
| | 3+4 2 | 3+4 2 | 3+4 2 | 3+4 2 | 2+2 2 | 2+2 2 |

Continued on next column

**TABLE XXIV– continued**

| TQC | ICSD | | NoE | | SF | |
|---|---|---|---|---|---|---|
| | 5+6 2 | 5+6 2 | 5+6 2 | 5+6 2 | 3+3 2 | 3+3 2 |
| | 5+6 2 | 3+4 2 | 3+4 2 | 3+4 2 | 3+3 2 | 2+2 2 |
| | 3+4 2 | 3+4 2 | 5+6 2 | 5+6 2 | 3+3 2 | 3+3 2 |
| | 5+6 2 | 5+6 2 | 3+4 2 | 5+6 2 | 2+2 2 | 2+2 2 |
| | 3+4 2 | 5+6 2 | 5+6 2 | 3+4 2 | 2+2 2 | 3+3 2 |
| | 3+4 2 | 3+4 2 | 5+6 2 | 3+4 2 | 3+3 2 | 2+2 2 |
| | 5+6 2 | 5+6 2 | 3+4 2 | 5+6 2 | 2+2 2 | 3+3 2 |
| | 5+6 2 | 3+4 2 | 5+6 2 | 5+6 2 | 2+2 2 | 2+2 2 |
| | 3+4 2 | 5+6 2 | 5+6 2 | 3+4 2 | 3+3 2 | 3+3 2 |
| 12.2.50 | 159510 | | 26 | | H3Li1O2 | |
| | -A | -GM | -M | -Y | -L | -V |
| | 3+4 2 | 3+4 2 | 3+4 2 | 3+4 2 | 3+3 2 | 3+3 2 |
| | 5+6 2 | 5+6 2 | 5+6 2 | 5+6 2 | 2+2 2 | 2+2 2 |
| | 5+6 2 | 5+6 2 | 5+6 2 | 5+6 2 | 2+2 2 | 2+2 2 |
| | 3+4 2 | 3+4 2 | 3+4 2 | 3+4 2 | 3+3 2 | 3+3 2 |
| | 5+6 2 | 3+4 2 | 3+4 2 | 3+4 2 | 2+2 2 | 2+2 2 |
| | 3+4 2 | 3+4 2 | 3+4 2 | 5+6 2 | 3+3 2 | 3+3 2 |
| | 3+4 2 | 5+6 2 | 5+6 2 | 3+4 2 | 3+3 2 | 2+2 2 |
| | 5+6 2 | 5+6 2 | 5+6 2 | 5+6 2 | 2+2 2 | 3+3 2 |
| | 3+4 2 | 3+4 2 | 5+6 2 | 3+4 2 | 3+3 2 | 2+2 2 |
| | 3+4 2 | 5+6 2 | 5+6 2 | 5+6 2 | 3+3 2 | 3+3 2 |
| | 5+6 2 | 3+4 2 | 3+4 2 | 5+6 2 | 2+2 2 | 3+3 2 |
| | 5+6 2 | 5+6 2 | 3+4 2 | 3+4 2 | 2+2 2 | 2+2 2 |
| | 5+6 2 | 3+4 2 | 5+6 2 | 5+6 2 | 3+3 2 | 2+2 2 |
| 12.2.51 | 24430 | | 106 | | H3Al2O7P1 | |
| | -A | -GM | -M | -Y | -L | -V |
| | 3+4 2 | 3+4 2 | 3+4 2 | 3+4 2 | 2+2 2 | 2+2 2 |
| | 5+6 2 | 5+6 2 | 5+6 2 | 5+6 2 | 3+3 2 | 3+3 2 |
| | 3+4 2 | 5+6 2 | 5+6 2 | 3+4 2 | 3+3 2 | 3+3 2 |
| | 5+6 2 | 3+4 2 | 3+4 2 | 5+6 2 | 2+2 2 | 2+2 2 |
| | 3+4 2 | 5+6 2 | 5+6 2 | 3+4 2 | 3+3 2 | 3+3 2 |
| | 5+6 2 | 3+4 2 | 3+4 2 | 5+6 2 | 2+2 2 | 2+2 2 |
| | 3+4 2 | 3+4 2 | 3+4 2 | 3+4 2 | 2+2 2 | 2+2 2 |
| | 5+6 2 | 5+6 2 | 5+6 2 | 5+6 2 | 3+3 2 | 3+3 2 |
| | 3+4 2 | 3+4 2 | 3+4 2 | 3+4 2 | 3+3 2 | 3+3 2 |
| | 5+6 2 | 5+6 2 | 5+6 2 | 5+6 2 | 2+2 2 | 2+2 2 |
| | 5+6 2 | 5+6 2 | 5+6 2 | 5+6 2 | 2+2 2 | 2+2 2 |
| | 3+4 2 | 3+4 2 | 3+4 2 | 3+4 2 | 3+3 2 | 3+3 2 |
| | 3+4 2 | 3+4 2 | 3+4 2 | 3+4 2 | 3+3 2 | 3+3 2 |
| | 5+6 2 | 5+6 2 | 5+6 2 | 5+6 2 | 2+2 2 | 2+2 2 |
| | 3+4 2 | 3+4 2 | 3+4 2 | 3+4 2 | 2+2 2 | 2+2 2 |
| | 5+6 2 | 5+6 2 | 5+6 2 | 5+6 2 | 3+3 2 | 3+3 2 |
| | 3+4 2 | 3+4 2 | 5+6 2 | 5+6 2 | 3+3 2 | 3+3 2 |
| | 5+6 2 | 5+6 2 | 5+6 2 | 3+4 2 | 2+2 2 | 2+2 2 |
| | 5+6 2 | 5+6 2 | 3+4 2 | 5+6 2 | 3+3 2 | 3+3 2 |
| | 3+4 2 | 3+4 2 | 3+4 2 | 3+4 2 | 2+2 2 | 2+2 2 |
| | 3+4 2 | 3+4 2 | 3+4 2 | 5+6 2 | 2+2 2 | 3+3 2 |
| | 5+6 2 | 5+6 2 | 5+6 2 | 3+4 2 | 3+3 2 | 2+2 2 |
| | 5+6 2 | 3+4 2 | 3+4 2 | 3+4 2 | 3+3 2 | 3+3 2 |
| | 5+6 2 | 5+6 2 | 5+6 2 | 5+6 2 | 2+2 2 | 2+2 2 |
| | 3+4 2 | 5+6 2 | 5+6 2 | 5+6 2 | 2+2 2 | 2+2 2 |
| | 3+4 2 | 3+4 2 | 3+4 2 | 3+4 2 | 3+3 2 | 2+2 2 |
| | 5+6 2 | 3+4 2 | 3+4 2 | 3+4 2 | 2+2 2 | 3+3 2 |
| | 3+4 2 | 5+6 2 | 5+6 2 | 3+4 2 | 2+2 2 | 3+3 2 |
| | 3+4 2 | 3+4 2 | 3+4 2 | 5+6 2 | 3+3 2 | 2+2 2 |
| | 3+4 2 | 5+6 2 | 5+6 2 | 5+6 2 | 3+3 2 | 3+3 2 |
| | 5+6 2 | 3+4 2 | 3+4 2 | 3+4 2 | 2+2 2 | 2+2 2 |
| | 3+4 2 | 5+6 2 | 3+4 2 | 5+6 2 | 3+3 2 | 2+2 2 |
| | 5+6 2 | 5+6 2 | 5+6 2 | 3+4 2 | 3+3 2 | 3+3 2 |
| | 3+4 2 | 3+4 2 | 5+6 2 | 5+6 2 | 2+2 2 | 2+2 2 |
| | 3+4 2 | 3+4 2 | 3+4 2 | 3+4 2 | 2+2 2 | 3+3 2 |

Continued on next column

**TABLE XXIV– continued**

| TQC | ICSD | | NoE | | SF | |
|---|---|---|---|---|---|---|
| | 5+6 2 | 3+4 2 | 5+6 2 | 3+4 2 | 3+3 2 | 3+3 2 |
| | 5+6 2 | 5+6 2 | 5+6 2 | 5+6 2 | 2+2 2 | 3+3 2 |
| | 3+4 2 | 3+4 2 | 5+6 2 | 5+6 2 | 3+3 2 | 3+3 2 |
| | 5+6 2 | 5+6 2 | 3+4 2 | 5+6 2 | 3+3 2 | 2+2 2 |
| | 5+6 2 | 3+4 2 | 3+4 2 | 3+4 2 | 2+2 2 | 2+2 2 |
| | 3+4 2 | 5+6 2 | 3+4 2 | 3+4 2 | 3+3 2 | 3+3 2 |
| | 5+6 2 | 3+4 2 | 5+6 2 | 5+6 2 | 2+2 2 | 3+3 2 |
| | 5+6 2 | 5+6 2 | 3+4 2 | 3+4 2 | 3+3 2 | 2+2 2 |
| | 3+4 2 | 5+6 2 | 5+6 2 | 5+6 2 | 2+2 2 | 3+3 2 |
| | 3+4 2 | 3+4 2 | 3+4 2 | 3+4 2 | 3+3 2 | 2+2 2 |
| | 3+4 2 | 3+4 2 | 5+6 2 | 5+6 2 | 3+3 2 | 3+3 2 |
| | 5+6 2 | 5+6 2 | 5+6 2 | 3+4 2 | 3+3 2 | 3+3 2 |
| | 5+6 2 | 3+4 2 | 3+4 2 | 5+6 2 | 2+2 2 | 2+2 2 |
| | 3+4 2 | 5+6 2 | 3+4 2 | 5+6 2 | 3+3 2 | 2+2 2 |
| | 5+6 2 | 5+6 2 | 5+6 2 | 3+4 2 | 2+2 2 | 3+3 2 |
| | 3+4 2 | 5+6 2 | 5+6 2 | 3+4 2 | 2+2 2 | 2+2 2 |
| | 3+4 2 | 3+4 2 | 3+4 2 | 5+6 2 | 3+3 2 | 2+2 2 |
| | 5+6 2 | 5+6 2 | 3+4 2 | 3+4 2 | 2+2 2 | 3+3 2 |
| 12.2.52 | 421319 | | 64 | | H2Bi1Cl3O1 | |
| | -A | -GM | -M | -Y | -L | -V |
| | 3+4 2 | 3+4 2 | 3+4 2 | 3+4 2 | 2+2 2 | 2+2 2 |
| | 5+6 2 | 5+6 2 | 5+6 2 | 5+6 2 | 3+3 2 | 3+3 2 |
| | 3+4 2 | 3+4 2 | 5+6 2 | 5+6 2 | 3+3 2 | 3+3 2 |
| | 5+6 2 | 5+6 2 | 3+4 2 | 3+4 2 | 2+2 2 | 2+2 2 |
| | 3+4 2 | 3+4 2 | 5+6 2 | 5+6 2 | 2+2 2 | 3+3 2 |
| | 3+4 2 | 3+4 2 | 5+6 2 | 3+4 2 | 3+3 2 | 2+2 2 |
| | 5+6 2 | 5+6 2 | 3+4 2 | 5+6 2 | 3+3 2 | 3+3 2 |
| | 5+6 2 | 5+6 2 | 3+4 2 | 3+4 2 | 2+2 2 | 2+2 2 |
| | 5+6 2 | 5+6 2 | 3+4 2 | 3+4 2 | 2+2 2 | 2+2 2 |
| | 3+4 2 | 3+4 2 | 5+6 2 | 5+6 2 | 3+3 2 | 3+3 2 |
| | 3+4 2 | 3+4 2 | 5+6 2 | 3+4 2 | 3+3 2 | 3+3 2 |
| | 5+6 2 | 5+6 2 | 3+4 2 | 5+6 2 | 2+2 2 | 2+2 2 |
| | 3+4 2 | 3+4 2 | 5+6 2 | 5+6 2 | 2+2 2 | 3+3 2 |
| | 5+6 2 | 5+6 2 | 5+6 2 | 3+4 2 | 3+3 2 | 2+2 2 |
| | 3+4 2 | 3+4 2 | 3+4 2 | 5+6 2 | 3+3 2 | 3+3 2 |
| | 5+6 2 | 3+4 2 | 3+4 2 | 3+4 2 | 2+2 2 | 2+2 2 |
| | 5+6 2 | 5+6 2 | 5+6 2 | 5+6 2 | 2+2 2 | 3+3 2 |
| | 3+4 2 | 3+4 2 | 3+4 2 | 3+4 2 | 3+3 2 | 2+2 2 |
| | 3+4 2 | 3+4 2 | 3+4 2 | 5+6 2 | 3+3 2 | 2+2 2 |
| | 5+6 2 | 5+6 2 | 5+6 2 | 5+6 2 | 3+3 2 | 3+3 2 |
| | 3+4 2 | 5+6 2 | 3+4 2 | 3+4 2 | 2+2 2 | 3+3 2 |
| | 5+6 2 | 3+4 2 | 5+6 2 | 3+4 2 | 2+2 2 | 2+2 2 |
| | 3+4 2 | 3+4 2 | 5+6 2 | 5+6 2 | 2+2 2 | 3+3 2 |
| | 3+4 2 | 5+6 2 | 5+6 2 | 3+4 2 | 3+3 2 | 2+2 2 |
| | 5+6 2 | 5+6 2 | 3+4 2 | 3+4 2 | 3+3 2 | 3+3 2 |
| | 5+6 2 | 5+6 2 | 5+6 2 | 5+6 2 | 2+2 2 | 2+2 2 |
| | 3+4 2 | 5+6 2 | 3+4 2 | 5+6 2 | 2+2 2 | 2+2 2 |
| | 5+6 2 | 3+4 2 | 3+4 2 | 3+4 2 | 3+3 2 | 3+3 2 |
| | 3+4 2 | 3+4 2 | 3+4 2 | 5+6 2 | 2+2 2 | 2+2 2 |
| | 5+6 2 | 3+4 2 | 5+6 2 | 3+4 2 | 2+2 2 | 3+3 2 |
| | 3+4 2 | 3+4 2 | 3+4 2 | 3+4 2 | 3+3 2 | 2+2 2 |
| | 3+4 2 | 5+6 2 | 5+6 2 | 3+4 2 | 3+3 2 | 3+3 2 |
| 12.2.53 | 26919 | | 60 | | H12Cl2Ni1O6 | |
| | -A | -GM | -M | -Y | -L | -V |
| | 3+4 2 | 3+4 2 | 3+4 2 | 3+4 2 | 3+3 2 | 3+3 2 |
| | 5+6 2 | 5+6 2 | 5+6 2 | 5+6 2 | 2+2 2 | 2+2 2 |
| | 5+6 2 | 5+6 2 | 5+6 2 | 5+6 2 | 2+2 2 | 2+2 2 |
| | 3+4 2 | 3+4 2 | 3+4 2 | 3+4 2 | 3+3 2 | 3+3 2 |
| | 3+4 2 | 5+6 2 | 5+6 2 | 3+4 2 | 3+3 2 | 2+2 2 |
| | 5+6 2 | 3+4 2 | 3+4 2 | 5+6 2 | 2+2 2 | 3+3 2 |
| | 3+4 2 | 3+4 2 | 5+6 2 | 3+4 2 | 3+3 2 | 3+3 2 |

Continued on next column

**TABLE XXIV– continued**

| TQC | ICSD | | NoE | SF | | |
|---|---|---|---|---|---|---|
| | 5+6 2 | 5+6 2 | 3+4 2 | 5+6 2 | 2+2 2 | 2+2 2 |
| | 3+4 2 | 3+4 2 | 3+4 2 | 3+4 2 | 3+3 2 | 3+3 2 |
| | 3+4 2 | 3+4 2 | 3+4 2 | 3+4 2 | 3+3 2 | 3+3 2 |
| | 3+4 2 | 3+4 2 | 3+4 2 | 3+4 2 | 3+3 2 | 3+3 2 |
| | 3+4 2 | 3+4 2 | 3+4 2 | 3+4 2 | 3+3 2 | 3+3 2 |
| | 5+6 2 | 3+4 2 | 3+4 2 | 3+4 2 | 3+3 2 | 3+3 2 |
| | 3+4 2 | 3+4 2 | 3+4 2 | 3+4 2 | 3+3 2 | 2+2 2 |
| | 3+4 2 | 5+6 2 | 5+6 2 | 5+6 2 | 2+2 2 | 3+3 2 |
| | 5+6 2 | 5+6 2 | 5+6 2 | 5+6 2 | 2+2 2 | 2+2 2 |
| | 5+6 2 | 5+6 2 | 5+6 2 | 5+6 2 | 2+2 2 | 2+2 2 |
| | 5+6 2 | 5+6 2 | 5+6 2 | 5+6 2 | 2+2 2 | 2+2 2 |
| | 3+4 2 | 3+4 2 | 5+6 2 | 5+6 2 | 2+2 2 | 2+2 2 |
| | 5+6 2 | 3+4 2 | 3+4 2 | 5+6 2 | 3+3 2 | 3+3 2 |
| | 5+6 2 | 3+4 2 | 5+6 2 | 3+4 2 | 2+2 2 | 2+2 2 |
| | 3+4 2 | 5+6 2 | 3+4 2 | 3+4 2 | 3+3 2 | 3+3 2 |
| | 3+4 2 | 5+6 2 | 5+6 2 | 5+6 2 | 2+2 2 | 3+3 2 |
| | 5+6 2 | 3+4 2 | 3+4 2 | 3+4 2 | 3+3 2 | 3+3 2 |
| | 3+4 2 | 5+6 2 | 5+6 2 | 5+6 2 | 3+3 2 | 2+2 2 |
| | 5+6 2 | 5+6 2 | 3+4 2 | 3+4 2 | 2+2 2 | 2+2 2 |
| | 3+4 2 | 3+4 2 | 3+4 2 | 3+4 2 | 3+3 2 | 3+3 2 |
| | 3+4 2 | 5+6 2 | 5+6 2 | 5+6 2 | 2+2 2 | 2+2 2 |
| | 5+6 2 | 5+6 2 | 5+6 2 | 3+4 2 | 3+3 2 | 3+3 2 |
| | 5+6 2 | 3+4 2 | 3+4 2 | 3+4 2 | 2+2 2 | 2+2 2 |
| 12.2.54 | 49911 | | 70 | H12Cl2N2O6Pt1 | | |
| | -A | -GM | -M | -Y | -L | -V |
| | 5+6 2 | 3+4 2 | 5+6 2 | 3+4 2 | 2+2 2 | 3+3 2 |
| | 3+4 2 | 5+6 2 | 3+4 2 | 5+6 2 | 3+3 2 | 2+2 2 |
| | 3+4 2 | 3+4 2 | 3+4 2 | 3+4 2 | 3+3 2 | 3+3 2 |
| | 5+6 2 | 5+6 2 | 5+6 2 | 5+6 2 | 2+2 2 | 2+2 2 |
| | 3+4 2 | 3+4 2 | 3+4 2 | 5+6 2 | 2+2 2 | 3+3 2 |
| | 5+6 2 | 5+6 2 | 5+6 2 | 3+4 2 | 3+3 2 | 2+2 2 |
| | 3+4 2 | 3+4 2 | 3+4 2 | 3+4 2 | 3+3 2 | 3+3 2 |
| | 5+6 2 | 5+6 2 | 5+6 2 | 5+6 2 | 2+2 2 | 2+2 2 |
| | 3+4 2 | 3+4 2 | 3+4 2 | 3+4 2 | 3+3 2 | 3+3 2 |
| | 5+6 2 | 5+6 2 | 5+6 2 | 5+6 2 | 2+2 2 | 2+2 2 |
| | 3+4 2 | 3+4 2 | 3+4 2 | 3+4 2 | 3+3 2 | 3+3 2 |
| | 3+4 2 | 3+4 2 | 3+4 2 | 3+4 2 | 3+3 2 | 3+3 2 |
| | 5+6 2 | 5+6 2 | 5+6 2 | 5+6 2 | 2+2 2 | 2+2 2 |
| | 3+4 2 | 3+4 2 | 3+4 2 | 3+4 2 | 3+3 2 | 3+3 2 |
| | 3+4 2 | 3+4 2 | 3+4 2 | 3+4 2 | 3+3 2 | 3+3 2 |
| | 3+4 2 | 3+4 2 | 3+4 2 | 3+4 2 | 3+3 2 | 3+3 2 |
| | 3+4 2 | 3+4 2 | 3+4 2 | 3+4 2 | 3+3 2 | 3+3 2 |
| | 5+6 2 | 5+6 2 | 5+6 2 | 3+4 2 | 2+2 2 | 2+2 2 |
| | 3+4 2 | 3+4 2 | 5+6 2 | 5+6 2 | 2+2 2 | 3+3 2 |
| | 5+6 2 | 5+6 2 | 3+4 2 | 5+6 2 | 3+3 2 | 2+2 2 |
| | 3+4 2 | 3+4 2 | 3+4 2 | 3+4 2 | 3+3 2 | 3+3 2 |
| | 3+4 2 | 3+4 2 | 3+4 2 | 3+4 2 | 3+3 2 | 3+3 2 |
| | 5+6 2 | 5+6 2 | 5+6 2 | 5+6 2 | 2+2 2 | 2+2 2 |
| | 5+6 2 | 5+6 2 | 5+6 2 | 5+6 2 | 2+2 2 | 2+2 2 |
| | 5+6 2 | 5+6 2 | 5+6 2 | 5+6 2 | 2+2 2 | 2+2 2 |
| | 5+6 2 | 5+6 2 | 5+6 2 | 5+6 2 | 2+2 2 | 2+2 2 |
| | 5+6 2 | 5+6 2 | 5+6 2 | 5+6 2 | 2+2 2 | 2+2 2 |
| | 5+6 2 | 5+6 2 | 5+6 2 | 5+6 2 | 2+2 2 | 2+2 2 |
| | 3+4 2 | 3+4 2 | 3+4 2 | 5+6 2 | 3+3 2 | 2+2 2 |
| | 3+4 2 | 5+6 2 | 3+4 2 | 3+4 2 | 3+3 2 | 3+3 2 |
| | 5+6 2 | 3+4 2 | 5+6 2 | 5+6 2 | 2+2 2 | 3+3 2 |
| | 3+4 2 | 3+4 2 | 5+6 2 | 3+4 2 | 2+2 2 | 3+3 2 |
| | 5+6 2 | 3+4 2 | 3+4 2 | 3+4 2 | 3+3 2 | 3+3 2 |
| | 3+4 2 | 5+6 2 | 3+4 2 | 5+6 2 | 3+3 2 | 2+2 2 |
| | 5+6 2 | 3+4 2 | 5+6 2 | 3+4 2 | 2+2 2 | 2+2 2 |
| 12.2.55 | 200906 | | 182 | H10F6Mn1O5Zr1 | | |

Continued on next column

**TABLE XXIV– continued**

| TQC | ICSD | | NoE | SF | | |
|---|---|---|---|---|---|---|
| | -A | -GM | -M | -Y | -L | -V |
| | 5+6 2 | 3+4 2 | 5+6 2 | 3+4 2 | 2+2 2 | 2+2 2 |
| | 3+4 2 | 5+6 2 | 3+4 2 | 5+6 2 | 3+3 2 | 3+3 2 |
| | 5+6 2 | 3+4 2 | 5+6 2 | 3+4 2 | 3+3 2 | 2+2 2 |
| | 3+4 2 | 5+6 2 | 3+4 2 | 5+6 2 | 2+2 2 | 3+3 2 |
| | 3+4 2 | 5+6 2 | 3+4 2 | 5+6 2 | 3+3 2 | 2+2 2 |
| | 5+6 2 | 3+4 2 | 5+6 2 | 3+4 2 | 2+2 2 | 3+3 2 |
| | 5+6 2 | 3+4 2 | 5+6 2 | 3+4 2 | 2+2 2 | 3+3 2 |
| | 3+4 2 | 5+6 2 | 3+4 2 | 5+6 2 | 3+3 2 | 2+2 2 |
| | 5+6 2 | 3+4 2 | 5+6 2 | 3+4 2 | 3+3 2 | 2+2 2 |
| | 3+4 2 | 5+6 2 | 3+4 2 | 5+6 2 | 2+2 2 | 3+3 2 |
| | 5+6 2 | 3+4 2 | 5+6 2 | 3+4 2 | 2+2 2 | 3+3 2 |
| | 3+4 2 | 5+6 2 | 3+4 2 | 5+6 2 | 3+3 2 | 2+2 2 |
| | 5+6 2 | 3+4 2 | 5+6 2 | 3+4 2 | 3+3 2 | 2+2 2 |
| | 3+4 2 | 5+6 2 | 3+4 2 | 5+6 2 | 2+2 2 | 3+3 2 |
| | 5+6 2 | 3+4 2 | 5+6 2 | 3+4 2 | 2+2 2 | 3+3 2 |
| | 3+4 2 | 5+6 2 | 3+4 2 | 5+6 2 | 3+3 2 | 2+2 2 |
| | 3+4 2 | 5+6 2 | 3+4 2 | 5+6 2 | 2+2 2 | 3+3 2 |
| | 5+6 2 | 3+4 2 | 5+6 2 | 3+4 2 | 3+3 2 | 2+2 2 |
| | 3+4 2 | 5+6 2 | 3+4 2 | 5+6 2 | 3+3 2 | 2+2 2 |
| | 5+6 2 | 3+4 2 | 5+6 2 | 3+4 2 | 2+2 2 | 3+3 2 |
| | 5+6 2 | 3+4 2 | 3+4 2 | 5+6 2 | 3+3 2 | 2+2 2 |
| | 3+4 2 | 5+6 2 | 5+6 2 | 3+4 2 | 2+2 2 | 3+3 2 |
| | 5+6 2 | 5+6 2 | 5+6 2 | 5+6 2 | 2+2 2 | 2+2 2 |
| | 3+4 2 | 3+4 2 | 3+4 2 | 3+4 2 | 3+3 2 | 3+3 2 |
| | 5+6 2 | 5+6 2 | 3+4 2 | 3+4 2 | 3+3 2 | 3+3 2 |
| | 3+4 2 | 3+4 2 | 5+6 2 | 5+6 2 | 2+2 2 | 2+2 2 |
| | 3+4 2 | 5+6 2 | 5+6 2 | 3+4 2 | 2+2 2 | 2+2 2 |
| | 5+6 2 | 3+4 2 | 3+4 2 | 5+6 2 | 3+3 2 | 3+3 2 |
| | 3+4 2 | 3+4 2 | 5+6 2 | 3+4 2 | 2+2 2 | 2+2 2 |
| | 5+6 2 | 5+6 2 | 3+4 2 | 5+6 2 | 3+3 2 | 3+3 2 |
| | 5+6 2 | 3+4 2 | 5+6 2 | 3+4 2 | 3+3 2 | 2+2 2 |
| | 3+4 2 | 5+6 2 | 3+4 2 | 5+6 2 | 2+2 2 | 3+3 2 |
| | 3+4 2 | 5+6 2 | 3+4 2 | 5+6 2 | 2+2 2 | 3+3 2 |
| | 5+6 2 | 3+4 2 | 5+6 2 | 3+4 2 | 2+2 2 | 3+3 2 |
| | 5+6 2 | 3+4 2 | 5+6 2 | 3+4 2 | 3+3 2 | 2+2 2 |
| | 3+4 2 | 5+6 2 | 3+4 2 | 5+6 2 | 3+3 2 | 2+2 2 |
| | 5+6 2 | 3+4 2 | 5+6 2 | 3+4 2 | 3+3 2 | 2+2 2 |
| | 5+6 2 | 3+4 2 | 5+6 2 | 3+4 2 | 2+2 2 | 3+3 2 |
| | 5+6 2 | 3+4 2 | 5+6 2 | 3+4 2 | 3+3 2 | 2+2 2 |
| | 3+4 2 | 5+6 2 | 3+4 2 | 5+6 2 | 3+3 2 | 3+3 2 |
| | 3+4 2 | 5+6 2 | 3+4 2 | 5+6 2 | 2+2 2 | 2+2 2 |
| | 3+4 2 | 5+6 2 | 5+6 2 | 3+4 2 | 2+2 2 | 3+3 2 |
| | 5+6 2 | 3+4 2 | 3+4 2 | 5+6 2 | 2+2 2 | 3+3 2 |
| | 3+4 2 | 5+6 2 | 3+4 2 | 5+6 2 | 3+3 2 | 2+2 2 |
| | 3+4 2 | 5+6 2 | 3+4 2 | 5+6 2 | 3+3 2 | 2+2 2 |
| | 5+6 2 | 3+4 2 | 5+6 2 | 3+4 2 | 2+2 2 | 3+3 2 |
| | 5+6 2 | 3+4 2 | 5+6 2 | 3+4 2 | 3+3 2 | 2+2 2 |
| | 5+6 2 | 3+4 2 | 5+6 2 | 5+6 2 | 3+3 2 | 3+3 2 |
| | 5+6 2 | 5+6 2 | 3+4 2 | 3+4 2 | 2+2 2 | 2+2 2 |
| | 3+4 2 | 3+4 2 | 3+4 2 | 3+4 2 | 2+2 2 | 3+3 2 |
| | 3+4 2 | 5+6 2 | 5+6 2 | 5+6 2 | 2+2 2 | 3+3 2 |
| | 3+4 2 | 5+6 2 | 5+6 2 | 5+6 2 | 3+3 2 | 2+2 2 |
| | 5+6 2 | 3+4 2 | 3+4 2 | 3+4 2 | 2+2 2 | 2+2 2 |
| | 5+6 2 | 3+4 2 | 5+6 2 | 3+4 2 | 3+3 2 | 3+3 2 |
| | 3+4 2 | 5+6 2 | 3+4 2 | 5+6 2 | 3+3 2 | 2+2 2 |
| | 5+6 2 | 3+4 2 | 3+4 2 | 5+6 2 | 3+3 2 | 3+3 2 |
| | 3+4 2 | 5+6 2 | 3+4 2 | 3+4 2 | 2+2 2 | 2+2 2 |
| | 3+4 2 | 5+6 2 | 5+6 2 | 3+4 2 | 3+3 2 | 2+2 2 |
| | 3+4 2 | 5+6 2 | 5+6 2 | 5+6 2 | 2+2 2 | 3+3 2 |
| | 5+6 2 | 3+4 2 | 5+6 2 | 5+6 2 | 2+2 2 | 3+3 2 |

Continued on next column

**TABLE XXIV– continued**

| TQC | ICSD | | NoE | | SF | |
|---|---|---|---|---|---|---|
| | 3+4 2 | 5+6 2 | 3+4 2 | 3+4 2 | 3+3 2 | 3+3 2 |
| | 5+6 2 | 3+4 2 | 3+4 2 | 5+6 2 | 2+2 2 | 2+2 2 |
| | 5+6 2 | 5+6 2 | 3+4 2 | 3+4 2 | 2+2 2 | 3+3 2 |
| | 3+4 2 | 3+4 2 | 5+6 2 | 5+6 2 | 2+2 2 | 2+2 2 |
| | 5+6 2 | 3+4 2 | 5+6 2 | 5+6 2 | 3+3 2 | 2+2 2 |
| | 3+4 2 | 5+6 2 | 3+4 2 | 3+4 2 | 3+3 2 | 3+3 2 |
| | 5+6 2 | 3+4 2 | 5+6 2 | 5+6 2 | 2+2 2 | 3+3 2 |
| | 5+6 2 | 3+4 2 | 5+6 2 | 3+4 2 | 3+3 2 | 2+2 2 |
| | 3+4 2 | 5+6 2 | 3+4 2 | 3+4 2 | 3+3 2 | 3+3 2 |
| | 3+4 2 | 3+4 2 | 3+4 2 | 5+6 2 | 3+3 2 | 2+2 2 |
| | 5+6 2 | 5+6 2 | 5+6 2 | 3+4 2 | 2+2 2 | 2+2 2 |
| | 3+4 2 | 3+4 2 | 3+4 2 | 3+4 2 | 3+3 2 | 3+3 2 |
| | 5+6 2 | 5+6 2 | 3+4 2 | 3+4 2 | 2+2 2 | 2+2 2 |
| | 3+4 2 | 3+4 2 | 5+6 2 | 5+6 2 | 2+2 2 | 3+3 2 |
| | 5+6 2 | 5+6 2 | 3+4 2 | 5+6 2 | 3+3 2 | 3+3 2 |
| | 5+6 2 | 3+4 2 | 5+6 2 | 5+6 2 | 2+2 2 | 2+2 2 |
| | 5+6 2 | 5+6 2 | 5+6 2 | 3+4 2 | 3+3 2 | 3+3 2 |
| | 3+4 2 | 5+6 2 | 5+6 2 | 3+4 2 | 2+2 2 | 2+2 2 |
| | 3+4 2 | 5+6 2 | 3+4 2 | 3+4 2 | 2+2 2 | 3+3 2 |
| | 3+4 2 | 3+4 2 | 3+4 2 | 5+6 2 | 2+2 2 | 2+2 2 |
| | 5+6 2 | 3+4 2 | 5+6 2 | 3+4 2 | 3+3 2 | 2+2 2 |
| | 3+4 2 | 3+4 2 | 5+6 2 | 5+6 2 | 2+2 2 | 3+3 2 |
| | 5+6 2 | 5+6 2 | 3+4 2 | 5+6 2 | 3+3 2 | 2+2 2 |
| | 3+4 2 | 3+4 2 | 3+4 2 | 5+6 2 | 3+3 2 | 3+3 2 |
| | 5+6 2 | 5+6 2 | 5+6 2 | 5+6 2 | 2+2 2 | 3+3 2 |
| | 3+4 2 | 5+6 2 | 5+6 2 | 5+6 2 | 2+2 2 | 2+2 2 |
| | 3+4 2 | 5+6 2 | 5+6 2 | 3+4 2 | 3+3 2 | 3+3 2 |
| | 3+4 2 | 5+6 2 | 3+4 2 | 3+4 2 | 3+3 2 | 2+2 2 |
| | 5+6 2 | 3+4 2 | 3+4 2 | 3+4 2 | 2+2 2 | 2+2 2 |
| | 5+6 2 | 3+4 2 | 5+6 2 | 3+4 2 | 3+3 2 | 3+3 2 |
| | 3+4 2 | 3+4 2 | 3+4 2 | 5+6 2 | 2+2 2 | 3+3 2 |
| 12.2.56 | 76350 | | 116 | | Ge2Ho2Os1 | |
| | -A | -GM | -M | -Y | -L | -V |
| | 3+4 2 | 3+4 2 | 3+4 2 | 3+4 2 | 3+3 2 | 3+3 2 |
| | 5+6 2 | 5+6 2 | 5+6 2 | 5+6 2 | 2+2 2 | 2+2 2 |
| | 5+6 2 | 3+4 2 | 5+6 2 | 3+4 2 | 2+2 2 | 3+3 2 |
| | 3+4 2 | 5+6 2 | 3+4 2 | 5+6 2 | 3+3 2 | 2+2 2 |
| | 3+4 2 | 3+4 2 | 3+4 2 | 3+4 2 | 3+3 2 | 3+3 2 |
| | 5+6 2 | 5+6 2 | 5+6 2 | 5+6 2 | 2+2 2 | 2+2 2 |
| | 5+6 2 | 3+4 2 | 5+6 2 | 3+4 2 | 2+2 2 | 3+3 2 |
| | 3+4 2 | 5+6 2 | 3+4 2 | 5+6 2 | 3+3 2 | 2+2 2 |
| | 3+4 2 | 3+4 2 | 3+4 2 | 3+4 2 | 3+3 2 | 3+3 2 |
| | 5+6 2 | 5+6 2 | 5+6 2 | 5+6 2 | 2+2 2 | 2+2 2 |
| | 3+4 2 | 3+4 2 | 5+6 2 | 3+4 2 | 2+2 2 | 3+3 2 |
| | 5+6 2 | 3+4 2 | 3+4 2 | 5+6 2 | 3+3 2 | 2+2 2 |
| | 5+6 2 | 5+6 2 | 5+6 2 | 3+4 2 | 3+3 2 | 3+3 2 |
| | 3+4 2 | 5+6 2 | 3+4 2 | 5+6 2 | 2+2 2 | 2+2 2 |
| | 3+4 2 | 3+4 2 | 5+6 2 | 3+4 2 | 2+2 2 | 3+3 2 |
| | 5+6 2 | 5+6 2 | 3+4 2 | 5+6 2 | 3+3 2 | 2+2 2 |
| | 5+6 2 | 3+4 2 | 5+6 2 | 3+4 2 | 3+3 2 | 2+2 2 |
| | 3+4 2 | 3+4 2 | 3+4 2 | 5+6 2 | 2+2 2 | 2+2 2 |
| | 3+4 2 | 5+6 2 | 3+4 2 | 3+4 2 | 2+2 2 | 3+3 2 |
| | 5+6 2 | 5+6 2 | 5+6 2 | 5+6 2 | 3+3 2 | 3+3 2 |
| | 5+6 2 | 3+4 2 | 5+6 2 | 3+4 2 | 2+2 2 | 3+3 2 |
| | 3+4 2 | 5+6 2 | 3+4 2 | 5+6 2 | 3+3 2 | 2+2 2 |
| | 3+4 2 | 5+6 2 | 5+6 2 | 5+6 2 | 3+3 2 | 3+3 2 |
| | 5+6 2 | 3+4 2 | 3+4 2 | 3+4 2 | 2+2 2 | 2+2 2 |
| | 3+4 2 | 5+6 2 | 5+6 2 | 3+4 2 | 3+3 2 | 2+2 2 |
| | 5+6 2 | 3+4 2 | 5+6 2 | 5+6 2 | 2+2 2 | 3+3 2 |
| | 5+6 2 | 3+4 2 | 3+4 2 | 3+4 2 | 3+3 2 | 2+2 2 |
| | 3+4 2 | 5+6 2 | 3+4 2 | 5+6 2 | 2+2 2 | 3+3 2 |

Continued on next column

**TABLE XXIV– continued**

| TQC | ICSD | | NoE | | SF | |
|---|---|---|---|---|---|---|
| | 5+6 2 | 3+4 2 | 3+4 2 | 5+6 2 | 3+3 2 | 2+2 2 |
| | 3+4 2 | 5+6 2 | 3+4 2 | 3+4 2 | 2+2 2 | 3+3 2 |
| | 3+4 2 | 3+4 2 | 5+6 2 | 5+6 2 | 3+3 2 | 2+2 2 |
| | 5+6 2 | 5+6 2 | 5+6 2 | 3+4 2 | 2+2 2 | 2+2 2 |
| | 5+6 2 | 3+4 2 | 3+4 2 | 5+6 2 | 3+3 2 | 3+3 2 |
| | 3+4 2 | 3+4 2 | 5+6 2 | 5+6 2 | 2+2 2 | 2+2 2 |
| | 5+6 2 | 3+4 2 | 3+4 2 | 3+4 2 | 3+3 2 | 3+3 2 |
| | 3+4 2 | 5+6 2 | 3+4 2 | 3+4 2 | 2+2 2 | 2+2 2 |
| | 3+4 2 | 5+6 2 | 5+6 2 | 5+6 2 | 2+2 2 | 3+3 2 |
| | 3+4 2 | 5+6 2 | 3+4 2 | 3+4 2 | 3+3 2 | 2+2 2 |
| | 5+6 2 | 3+4 2 | 5+6 2 | 5+6 2 | 2+2 2 | 3+3 2 |
| | 5+6 2 | 5+6 2 | 5+6 2 | 3+4 2 | 3+3 2 | 2+2 2 |
| | 5+6 2 | 3+4 2 | 3+4 2 | 5+6 2 | 2+2 2 | 3+3 2 |
| | 5+6 2 | 5+6 2 | 5+6 2 | 3+4 2 | 3+3 2 | 2+2 2 |
| | 3+4 2 | 3+4 2 | 3+4 2 | 5+6 2 | 3+3 2 | 3+3 2 |
| | 3+4 2 | 5+6 2 | 5+6 2 | 3+4 2 | 2+2 2 | 2+2 2 |
| | 5+6 2 | 3+4 2 | 3+4 2 | 5+6 2 | 3+3 2 | 3+3 2 |
| | 3+4 2 | 3+4 2 | 5+6 2 | 3+4 2 | 2+2 2 | 3+3 2 |
| | 5+6 2 | 5+6 2 | 3+4 2 | 5+6 2 | 3+3 2 | 2+2 2 |
| | 3+4 2 | 3+4 2 | 3+4 2 | 3+4 2 | 2+2 2 | 2+2 2 |
| | 5+6 2 | 3+4 2 | 5+6 2 | 5+6 2 | 2+2 2 | 3+3 2 |
| | 3+4 2 | 5+6 2 | 5+6 2 | 3+4 2 | 3+3 2 | 2+2 2 |
| | 5+6 2 | 5+6 2 | 3+4 2 | 5+6 2 | 3+3 2 | 3+3 2 |
| | 3+4 2 | 5+6 2 | 3+4 2 | 3+4 2 | 2+2 2 | 2+2 2 |
| | 5+6 2 | 3+4 2 | 5+6 2 | 5+6 2 | 2+2 2 | 3+3 2 |
| | 3+4 2 | 5+6 2 | 3+4 2 | 3+4 2 | 3+3 2 | 2+2 2 |
| | 5+6 2 | 3+4 2 | 5+6 2 | 3+4 2 | 3+3 2 | 3+3 2 |
| | 5+6 2 | 5+6 2 | 5+6 2 | 5+6 2 | 2+2 2 | 2+2 2 |
| | 3+4 2 | 3+4 2 | 3+4 2 | 5+6 2 | 2+2 2 | 3+3 2 |
| | 5+6 2 | 5+6 2 | 5+6 2 | 3+4 2 | 3+3 2 | 2+2 2 |
| 12.2.57 | 183379 | | 82 | | Gd1O6Ru1Sr2 | |
| | -A | -GM | -M | -Y | -L | -V |
| | 3+4 2 | 3+4 2 | 3+4 2 | 3+4 2 | 3+3 2 | 3+3 2 |
| | 5+6 2 | 3+4 2 | 5+6 2 | 5+6 2 | 2+2 2 | 3+3 2 |
| | 3+4 2 | 5+6 2 | 3+4 2 | 3+4 2 | 3+3 2 | 2+2 2 |
| | 5+6 2 | 5+6 2 | 5+6 2 | 5+6 2 | 2+2 2 | 2+2 2 |
| | 5+6 2 | 5+6 2 | 5+6 2 | 5+6 2 | 2+2 2 | 2+2 2 |
| | 5+6 2 | 5+6 2 | 5+6 2 | 5+6 2 | 2+2 2 | 2+2 2 |
| | 3+4 2 | 3+4 2 | 3+4 2 | 3+4 2 | 3+3 2 | 3+3 2 |
| | 5+6 2 | 3+4 2 | 5+6 2 | 3+4 2 | 3+3 2 | 2+2 2 |
| | 3+4 2 | 3+4 2 | 3+4 2 | 3+4 2 | 3+3 2 | 3+3 2 |
| | 3+4 2 | 5+6 2 | 5+6 2 | 5+6 2 | 2+2 2 | 2+2 2 |
| | 5+6 2 | 3+4 2 | 3+4 2 | 5+6 2 | 2+2 2 | 3+3 2 |
| | 3+4 2 | 5+6 2 | 3+4 2 | 3+4 2 | 3+3 2 | 2+2 2 |
| | 5+6 2 | 5+6 2 | 5+6 2 | 5+6 2 | 2+2 2 | 3+3 2 |
| | 5+6 2 | 3+4 2 | 5+6 2 | 5+6 2 | 2+2 2 | 2+2 2 |
| | 5+6 2 | 3+4 2 | 5+6 2 | 3+4 2 | 3+3 2 | 2+2 2 |
| | 3+4 2 | 5+6 2 | 3+4 2 | 3+4 2 | 3+3 2 | 3+3 2 |
| | 5+6 2 | 5+6 2 | 5+6 2 | 5+6 2 | 2+2 2 | 2+2 2 |
| | 3+4 2 | 5+6 2 | 3+4 2 | 5+6 2 | 2+2 2 | 3+3 2 |
| | 5+6 2 | 3+4 2 | 5+6 2 | 3+4 2 | 3+3 2 | 2+2 2 |
| | 5+6 2 | 3+4 2 | 5+6 2 | 3+4 2 | 3+3 2 | 2+2 2 |
| | 5+6 2 | 3+4 2 | 5+6 2 | 3+4 2 | 3+3 2 | 2+2 2 |
| | 5+6 2 | 3+4 2 | 5+6 2 | 3+4 2 | 3+3 2 | 2+2 2 |
| | 5+6 2 | 3+4 2 | 5+6 2 | 3+4 2 | 3+3 2 | 2+2 2 |
| | 3+4 2 | 3+4 2 | 3+4 2 | 3+4 2 | 3+3 2 | 3+3 2 |
| | 5+6 2 | 5+6 2 | 5+6 2 | 5+6 2 | 2+2 2 | 2+2 2 |
| | 3+4 2 | 5+6 2 | 3+4 2 | 5+6 2 | 2+2 2 | 3+3 2 |
| | 3+4 2 | 5+6 2 | 3+4 2 | 5+6 2 | 2+2 2 | 3+3 2 |
| | 3+4 2 | 5+6 2 | 3+4 2 | 5+6 2 | 2+2 2 | 3+3 2 |
| | 3+4 2 | 5+6 2 | 3+4 2 | 5+6 2 | 2+2 2 | 3+3 2 |

Continued on next column

**TABLE XXIV– continued**

| TQC | ICSD | NoE | SF | | | |
|---|---|---|---|---|---|---|
| | 3+4 2 | 5+6 2 | 3+4 2 | 3+4 2 | 3+3 2 | 3+3 2 |
| | 3+4 2 | 5+6 2 | 3+4 2 | 5+6 2 | 2+2 2 | 3+3 2 |
| | 3+4 2 | 5+6 2 | 3+4 2 | 3+4 2 | 2+2 2 | 3+3 2 |
| | 3+4 2 | 5+6 2 | 3+4 2 | 5+6 2 | 3+3 2 | 2+2 2 |
| | 5+6 2 | 3+4 2 | 5+6 2 | 5+6 2 | 2+2 2 | 2+2 2 |
| | 5+6 2 | 3+4 2 | 5+6 2 | 5+6 2 | 2+2 2 | 3+3 2 |
| | 5+6 2 | 3+4 2 | 5+6 2 | 3+4 2 | 3+3 2 | 2+2 2 |
| | 5+6 2 | 3+4 2 | 5+6 2 | 3+4 2 | 3+3 2 | 2+2 2 |
| | 5+6 2 | 5+6 2 | 5+6 2 | 5+6 2 | 2+2 2 | 2+2 2 |
| | 5+6 2 | 3+4 2 | 5+6 2 | 5+6 2 | 2+2 2 | 2+2 2 |
| | 5+6 2 | 5+6 2 | 5+6 2 | 5+6 2 | 2+2 2 | 2+2 2 |
| | 5+6 2 | 5+6 2 | 5+6 2 | 5+6 2 | 2+2 2 | 2+2 2 |
| 12.2.58 | 635673 | 36 | Ga4Yb1 | | | |
| | -A | -GM | -M | -Y | -L | -V |
| | 3+4 2 | 3+4 2 | 3+4 2 | 3+4 2 | 3+3 2 | 3+3 2 |
| | 5+6 2 | 5+6 2 | 5+6 2 | 5+6 2 | 2+2 2 | 2+2 2 |
| | 5+6 2 | 5+6 2 | 5+6 2 | 5+6 2 | 2+2 2 | 2+2 2 |
| | 5+6 2 | 5+6 2 | 5+6 2 | 5+6 2 | 2+2 2 | 2+2 2 |
| | 5+6 2 | 3+4 2 | 3+4 2 | 3+4 2 | 2+2 2 | 2+2 2 |
| | 3+4 2 | 5+6 2 | 5+6 2 | 3+4 2 | 3+3 2 | 3+3 2 |
| | 3+4 2 | 3+4 2 | 3+4 2 | 5+6 2 | 2+2 2 | 2+2 2 |
| | 5+6 2 | 5+6 2 | 5+6 2 | 5+6 2 | 3+3 2 | 3+3 2 |
| | 5+6 2 | 3+4 2 | 3+4 2 | 3+4 2 | 2+2 2 | 2+2 2 |
| | 5+6 2 | 3+4 2 | 5+6 2 | 5+6 2 | 3+3 2 | 3+3 2 |
| | 5+6 2 | 5+6 2 | 3+4 2 | 3+4 2 | 2+2 2 | 2+2 2 |
| | 5+6 2 | 5+6 2 | 5+6 2 | 5+6 2 | 2+2 2 | 2+2 2 |
| | 3+4 2 | 5+6 2 | 5+6 2 | 5+6 2 | 2+2 2 | 2+2 2 |
| | 3+4 2 | 5+6 2 | 5+6 2 | 5+6 2 | 2+2 2 | 2+2 2 |
| | 5+6 2 | 3+4 2 | 5+6 2 | 5+6 2 | 2+2 2 | 2+2 2 |
| | 5+6 2 | 5+6 2 | 5+6 2 | 5+6 2 | 2+2 2 | 2+2 2 |
| | 5+6 2 | 5+6 2 | 5+6 2 | 5+6 2 | 2+2 2 | 2+2 2 |
| | 5+6 2 | 5+6 2 | 5+6 2 | 5+6 2 | 2+2 2 | 2+2 2 |
| 12.2.59 | 165189 | 48 | Ga3Sc5 | | | |
| | -A | -GM | -M | -Y | -L | -V |
| | 3+4 2 | 3+4 2 | 3+4 2 | 5+6 2 | 3+3 2 | 3+3 2 |
| | 5+6 2 | 3+4 2 | 3+4 2 | 3+4 2 | 2+2 2 | 2+2 2 |
| | 5+6 2 | 3+4 2 | 5+6 2 | 5+6 2 | 3+3 2 | 3+3 2 |
| | 3+4 2 | 5+6 2 | 3+4 2 | 3+4 2 | 2+2 2 | 2+2 2 |
| | 3+4 2 | 5+6 2 | 5+6 2 | 5+6 2 | 3+3 2 | 3+3 2 |
| | 5+6 2 | 5+6 2 | 5+6 2 | 3+4 2 | 2+2 2 | 2+2 2 |
| | 5+6 2 | 5+6 2 | 5+6 2 | 3+4 2 | 2+2 2 | 2+2 2 |
| | 3+4 2 | 3+4 2 | 3+4 2 | 5+6 2 | 3+3 2 | 3+3 2 |
| | 3+4 2 | 3+4 2 | 5+6 2 | 3+4 2 | 2+2 2 | 2+2 2 |
| | 5+6 2 | 5+6 2 | 3+4 2 | 5+6 2 | 3+3 2 | 3+3 2 |
| | 3+4 2 | 5+6 2 | 5+6 2 | 3+4 2 | 2+2 2 | 2+2 2 |
| | 5+6 2 | 3+4 2 | 3+4 2 | 5+6 2 | 3+3 2 | 3+3 2 |
| | 5+6 2 | 3+4 2 | 3+4 2 | 5+6 2 | 3+3 2 | 3+3 2 |
| | 3+4 2 | 3+4 2 | 5+6 2 | 5+6 2 | 2+2 2 | 2+2 2 |
| | 3+4 2 | 3+4 2 | 3+4 2 | 3+4 2 | 3+3 2 | 3+3 2 |
| | 5+6 2 | 3+4 2 | 5+6 2 | 5+6 2 | 2+2 2 | 3+3 2 |
| | 5+6 2 | 3+4 2 | 5+6 2 | 5+6 2 | 2+2 2 | 2+2 2 |
| | 3+4 2 | 5+6 2 | 3+4 2 | 3+4 2 | 3+3 2 | 3+3 2 |
| | 3+4 2 | 3+4 2 | 3+4 2 | 3+4 2 | 3+3 2 | 2+2 2 |
| | 5+6 2 | 3+4 2 | 3+4 2 | 3+4 2 | 2+2 2 | 2+2 2 |
| | 5+6 2 | 5+6 2 | 5+6 2 | 5+6 2 | 2+2 2 | 2+2 2 |
| | 3+4 2 | 3+4 2 | 5+6 2 | 3+4 2 | 3+3 2 | 3+3 2 |
| | 5+6 2 | 5+6 2 | 3+4 2 | 3+4 2 | 3+3 2 | 2+2 2 |
| | 3+4 2 | 5+6 2 | 5+6 2 | 3+4 2 | 2+2 2 | 3+3 2 |
| 12.2.60 | 15044 | 48 | Fe3Se4 | | | |
| | -A | -GM | -M | -Y | -L | -V |
| | 5+6 2 | 3+4 2 | 3+4 2 | 5+6 2 | 2+2 2 | 2+2 2 |

Continued on next column

**TABLE XXIV– continued**

| TQC | ICSD | NoE | SF | | | |
|---|---|---|---|---|---|---|
| | 3+4 2 | 3+4 2 | 5+6 2 | 3+4 2 | 2+2 2 | 3+3 2 |
| | 5+6 2 | 5+6 2 | 3+4 2 | 5+6 2 | 3+3 2 | 2+2 2 |
| | 3+4 2 | 5+6 2 | 5+6 2 | 3+4 2 | 3+3 2 | 3+3 2 |
| | 5+6 2 | 3+4 2 | 5+6 2 | 3+4 2 | 3+3 2 | 2+2 2 |
| | 3+4 2 | 5+6 2 | 3+4 2 | 5+6 2 | 2+2 2 | 3+3 2 |
| | 5+6 2 | 3+4 2 | 3+4 2 | 3+4 2 | 3+3 2 | 3+3 2 |
| | 3+4 2 | 3+4 2 | 3+4 2 | 5+6 2 | 2+2 2 | 2+2 2 |
| | 5+6 2 | 3+4 2 | 5+6 2 | 5+6 2 | 2+2 2 | 2+2 2 |
| | 3+4 2 | 5+6 2 | 5+6 2 | 3+4 2 | 3+3 2 | 2+2 2 |
| | 5+6 2 | 5+6 2 | 5+6 2 | 3+4 2 | 3+3 2 | 3+3 2 |
| | 5+6 2 | 3+4 2 | 3+4 2 | 5+6 2 | 2+2 2 | 3+3 2 |
| | 3+4 2 | 3+4 2 | 5+6 2 | 5+6 2 | 2+2 2 | 2+2 2 |
| | 5+6 2 | 3+4 2 | 5+6 2 | 5+6 2 | 2+2 2 | 2+2 2 |
| | 3+4 2 | 5+6 2 | 3+4 2 | 3+4 2 | 3+3 2 | 3+3 2 |
| | 3+4 2 | 5+6 2 | 3+4 2 | 3+4 2 | 3+3 2 | 3+3 2 |
| | 5+6 2 | 3+4 2 | 5+6 2 | 3+4 2 | 3+3 2 | 2+2 2 |
| | 5+6 2 | 3+4 2 | 3+4 2 | 5+6 2 | 2+2 2 | 3+3 2 |
| | 3+4 2 | 3+4 2 | 5+6 2 | 3+4 2 | 3+3 2 | 2+2 2 |
| | 3+4 2 | 3+4 2 | 5+6 2 | 5+6 2 | 3+3 2 | 2+2 2 |
| | 5+6 2 | 5+6 2 | 5+6 2 | 3+4 2 | 3+3 2 | 2+2 2 |
| | 3+4 2 | 5+6 2 | 5+6 2 | 5+6 2 | 2+2 2 | 2+2 2 |
| | 5+6 2 | 3+4 2 | 5+6 2 | 3+4 2 | 3+3 2 | 3+3 2 |
| | 5+6 2 | 5+6 2 | 5+6 2 | 3+4 2 | 3+3 2 | 2+2 2 |
| 12.2.61 | 15043 | 48 | Fe3Se4 | | | |
| | -A | -GM | -M | -Y | -L | -V |
| | 5+6 2 | 3+4 2 | 3+4 2 | 5+6 2 | 2+2 2 | 2+2 2 |
| | 3+4 2 | 3+4 2 | 5+6 2 | 3+4 2 | 2+2 2 | 3+3 2 |
| | 5+6 2 | 5+6 2 | 3+4 2 | 5+6 2 | 3+3 2 | 2+2 2 |
| | 3+4 2 | 5+6 2 | 5+6 2 | 3+4 2 | 3+3 2 | 3+3 2 |
| | 5+6 2 | 3+4 2 | 5+6 2 | 3+4 2 | 3+3 2 | 2+2 2 |
| | 3+4 2 | 5+6 2 | 3+4 2 | 5+6 2 | 2+2 2 | 3+3 2 |
| | 5+6 2 | 3+4 2 | 3+4 2 | 5+6 2 | 3+3 2 | 2+2 2 |
| | 3+4 2 | 3+4 2 | 3+4 2 | 3+4 2 | 2+2 2 | 3+3 2 |
| | 5+6 2 | 3+4 2 | 5+6 2 | 5+6 2 | 2+2 2 | 2+2 2 |
| | 3+4 2 | 5+6 2 | 5+6 2 | 3+4 2 | 3+3 2 | 2+2 2 |
| | 5+6 2 | 5+6 2 | 5+6 2 | 3+4 2 | 3+3 2 | 3+3 2 |
| | 5+6 2 | 3+4 2 | 3+4 2 | 5+6 2 | 2+2 2 | 3+3 2 |
| | 3+4 2 | 3+4 2 | 5+6 2 | 5+6 2 | 2+2 2 | 2+2 2 |
| | 5+6 2 | 3+4 2 | 5+6 2 | 5+6 2 | 2+2 2 | 2+2 2 |
| | 3+4 2 | 5+6 2 | 3+4 2 | 3+4 2 | 3+3 2 | 3+3 2 |
| | 3+4 2 | 5+6 2 | 5+6 2 | 3+4 2 | 3+3 2 | 3+3 2 |
| | 5+6 2 | 3+4 2 | 3+4 2 | 3+4 2 | 3+3 2 | 2+2 2 |
| | 5+6 2 | 3+4 2 | 3+4 2 | 5+6 2 | 2+2 2 | 3+3 2 |
| | 3+4 2 | 3+4 2 | 5+6 2 | 3+4 2 | 3+3 2 | 2+2 2 |
| | 3+4 2 | 3+4 2 | 5+6 2 | 5+6 2 | 3+3 2 | 2+2 2 |
| | 5+6 2 | 5+6 2 | 5+6 2 | 3+4 2 | 3+3 2 | 2+2 2 |
| | 3+4 2 | 5+6 2 | 5+6 2 | 5+6 2 | 2+2 2 | 2+2 2 |
| | 5+6 2 | 3+4 2 | 5+6 2 | 3+4 2 | 3+3 2 | 3+3 2 |
| | 5+6 2 | 5+6 2 | 5+6 2 | 3+4 2 | 3+3 2 | 2+2 2 |
| 12.2.62 | 152533 | 178 | Fe1Si3Tm3 | | | |
| | -A | -GM | -M | -Y | -L | -V |
| | 3+4 2 | 3+4 2 | 3+4 2 | 3+4 2 | 2+2 2 | 2+2 2 |
| | 5+6 2 | 5+6 2 | 5+6 2 | 5+6 2 | 3+3 2 | 3+3 2 |
| | 3+4 2 | 3+4 2 | 3+4 2 | 3+4 2 | 2+2 2 | 2+2 2 |
| | 5+6 2 | 5+6 2 | 5+6 2 | 5+6 2 | 3+3 2 | 3+3 2 |
| | 5+6 2 | 3+4 2 | 5+6 2 | 3+4 2 | 3+3 2 | 2+2 2 |
| | 3+4 2 | 5+6 2 | 3+4 2 | 5+6 2 | 2+2 2 | 3+3 2 |
| | 3+4 2 | 3+4 2 | 3+4 2 | 3+4 2 | 3+3 2 | 3+3 2 |
| | 5+6 2 | 5+6 2 | 5+6 2 | 5+6 2 | 2+2 2 | 2+2 2 |
| | 3+4 2 | 3+4 2 | 5+6 2 | 5+6 2 | 2+2 2 | 2+2 2 |
| | 5+6 2 | 5+6 2 | 3+4 2 | 3+4 2 | 3+3 2 | 3+3 2 |

Continued on next column

**TABLE XXIV– continued**

| TQC | ICSD | | NoE | | SF | |
|---|---|---|---|---|---|---|
| | 5+6 2 | 3+4 2 | 5+6 2 | 3+4 2 | 3+3 2 | 2+2 2 |
| | 3+4 2 | 5+6 2 | 3+4 2 | 5+6 2 | 2+2 2 | 3+3 2 |
| | 3+4 2 | 3+4 2 | 5+6 2 | 5+6 2 | 3+3 2 | 2+2 2 |
| | 5+6 2 | 5+6 2 | 3+4 2 | 3+4 2 | 2+2 2 | 3+3 2 |
| | 3+4 2 | 3+4 2 | 3+4 2 | 3+4 2 | 2+2 2 | 2+2 2 |
| | 5+6 2 | 5+6 2 | 5+6 2 | 5+6 2 | 3+3 2 | 3+3 2 |
| | 3+4 2 | 3+4 2 | 5+6 2 | 5+6 2 | 2+2 2 | 2+2 2 |
| | 5+6 2 | 3+4 2 | 3+4 2 | 3+4 2 | 3+3 2 | 3+3 2 |
| | 3+4 2 | 5+6 2 | 5+6 2 | 3+4 2 | 3+3 2 | 2+2 2 |
| | 5+6 2 | 3+4 2 | 5+6 2 | 5+6 2 | 2+2 2 | 3+3 2 |
| | 5+6 2 | 5+6 2 | 3+4 2 | 3+4 2 | 2+2 2 | 2+2 2 |
| | 3+4 2 | 5+6 2 | 3+4 2 | 5+6 2 | 3+3 2 | 3+3 2 |
| | 5+6 2 | 3+4 2 | 5+6 2 | 3+4 2 | 3+3 2 | 2+2 2 |
| | 3+4 2 | 5+6 2 | 3+4 2 | 5+6 2 | 2+2 2 | 3+3 2 |
| | 3+4 2 | 3+4 2 | 3+4 2 | 3+4 2 | 3+3 2 | 3+3 2 |
| | 5+6 2 | 3+4 2 | 5+6 2 | 3+4 2 | 2+2 2 | 3+3 2 |
| | 5+6 2 | 5+6 2 | 5+6 2 | 5+6 2 | 2+2 2 | 2+2 2 |
| | 3+4 2 | 3+4 2 | 3+4 2 | 5+6 2 | 3+3 2 | 2+2 2 |
| | 3+4 2 | 5+6 2 | 3+4 2 | 3+4 2 | 3+3 2 | 3+3 2 |
| | 5+6 2 | 5+6 2 | 5+6 2 | 5+6 2 | 2+2 2 | 2+2 2 |
| | 5+6 2 | 3+4 2 | 3+4 2 | 5+6 2 | 3+3 2 | 2+2 2 |
| | 3+4 2 | 5+6 2 | 5+6 2 | 3+4 2 | 2+2 2 | 3+3 2 |
| | 5+6 2 | 3+4 2 | 3+4 2 | 3+4 2 | 2+2 2 | 2+2 2 |
| | 3+4 2 | 5+6 2 | 5+6 2 | 5+6 2 | 3+3 2 | 3+3 2 |
| | 3+4 2 | 5+6 2 | 5+6 2 | 5+6 2 | 2+2 2 | 2+2 2 |
| | 5+6 2 | 3+4 2 | 3+4 2 | 3+4 2 | 3+3 2 | 3+3 2 |
| | 3+4 2 | 3+4 2 | 5+6 2 | 5+6 2 | 2+2 2 | 3+3 2 |
| | 3+4 2 | 3+4 2 | 5+6 2 | 3+4 2 | 3+3 2 | 2+2 2 |
| | 5+6 2 | 5+6 2 | 3+4 2 | 3+4 2 | 3+3 2 | 3+3 2 |
| | 5+6 2 | 5+6 2 | 3+4 2 | 5+6 2 | 2+2 2 | 2+2 2 |
| | 5+6 2 | 3+4 2 | 3+4 2 | 3+4 2 | 3+3 2 | 3+3 2 |
| | 3+4 2 | 3+4 2 | 5+6 2 | 5+6 2 | 2+2 2 | 2+2 2 |
| | 3+4 2 | 5+6 2 | 3+4 2 | 3+4 2 | 2+2 2 | 3+3 2 |
| | 5+6 2 | 5+6 2 | 3+4 2 | 5+6 2 | 3+3 2 | 3+3 2 |
| | 3+4 2 | 3+4 2 | 5+6 2 | 5+6 2 | 3+3 2 | 2+2 2 |
| | 5+6 2 | 5+6 2 | 5+6 2 | 3+4 2 | 2+2 2 | 2+2 2 |
| | 3+4 2 | 3+4 2 | 3+4 2 | 5+6 2 | 2+2 2 | 3+3 2 |
| | 5+6 2 | 5+6 2 | 3+4 2 | 3+4 2 | 3+3 2 | 3+3 2 |
| | 5+6 2 | 3+4 2 | 5+6 2 | 5+6 2 | 3+3 2 | 2+2 2 |
| | 3+4 2 | 5+6 2 | 5+6 2 | 3+4 2 | 2+2 2 | 3+3 2 |
| | 3+4 2 | 3+4 2 | 3+4 2 | 5+6 2 | 3+3 2 | 2+2 2 |
| | 5+6 2 | 5+6 2 | 5+6 2 | 3+4 2 | 2+2 2 | 3+3 2 |
| | 3+4 2 | 3+4 2 | 5+6 2 | 3+4 2 | 3+3 2 | 2+2 2 |
| | 5+6 2 | 5+6 2 | 3+4 2 | 5+6 2 | 2+2 2 | 3+3 2 |
| | 3+4 2 | 5+6 2 | 3+4 2 | 3+4 2 | 3+3 2 | 2+2 2 |
| | 5+6 2 | 3+4 2 | 5+6 2 | 5+6 2 | 2+2 2 | 3+3 2 |
| | 5+6 2 | 3+4 2 | 3+4 2 | 5+6 2 | 3+3 2 | 2+2 2 |
| | 3+4 2 | 5+6 2 | 5+6 2 | 3+4 2 | 2+2 2 | 2+2 2 |
| | 3+4 2 | 3+4 2 | 3+4 2 | 5+6 2 | 3+3 2 | 3+3 2 |
| | 5+6 2 | 5+6 2 | 3+4 2 | 3+4 2 | 2+2 2 | 3+3 2 |
| | 3+4 2 | 3+4 2 | 5+6 2 | 5+6 2 | 3+3 2 | 2+2 2 |
| | 5+6 2 | 5+6 2 | 5+6 2 | 3+4 2 | 2+2 2 | 2+2 2 |
| | 5+6 2 | 3+4 2 | 3+4 2 | 3+4 2 | 3+3 2 | 3+3 2 |
| | 5+6 2 | 5+6 2 | 5+6 2 | 5+6 2 | 2+2 2 | 3+3 2 |
| | 3+4 2 | 3+4 2 | 3+4 2 | 5+6 2 | 3+3 2 | 2+2 2 |
| | 3+4 2 | 3+4 2 | 3+4 2 | 3+4 2 | 2+2 2 | 3+3 2 |
| | 5+6 2 | 5+6 2 | 5+6 2 | 5+6 2 | 3+3 2 | 2+2 2 |
| | 5+6 2 | 3+4 2 | 3+4 2 | 3+4 2 | 2+2 2 | 3+3 2 |
| | 3+4 2 | 3+4 2 | 3+4 2 | 3+4 2 | 3+3 2 | 2+2 2 |
| | 5+6 2 | 5+6 2 | 5+6 2 | 5+6 2 | 2+2 2 | 3+3 2 |
| | 3+4 2 | 3+4 2 | 5+6 2 | 3+4 2 | 2+2 2 | 2+2 2 |

Continued on next column

**TABLE XXIV– continued**

| TQC | ICSD | | NoE | | SF | |
|---|---|---|---|---|---|---|
| | 5+6 2 | 5+6 2 | 5+6 2 | 5+6 2 | 3+3 2 | 3+3 2 |
| | 3+4 2 | 5+6 2 | 3+4 2 | 5+6 2 | 2+2 2 | 2+2 2 |
| | 3+4 2 | 5+6 2 | 3+4 2 | 3+4 2 | 3+3 2 | 3+3 2 |
| | 5+6 2 | 3+4 2 | 5+6 2 | 3+4 2 | 2+2 2 | 2+2 2 |
| | 3+4 2 | 5+6 2 | 3+4 2 | 5+6 2 | 3+3 2 | 3+3 2 |
| | 5+6 2 | 3+4 2 | 5+6 2 | 3+4 2 | 2+2 2 | 2+2 2 |
| | 5+6 2 | 5+6 2 | 5+6 2 | 5+6 2 | 3+3 2 | 3+3 2 |
| | 3+4 2 | 3+4 2 | 3+4 2 | 3+4 2 | 3+3 2 | 2+2 2 |
| | 3+4 2 | 5+6 2 | 3+4 2 | 3+4 2 | 2+2 2 | 2+2 2 |
| | 5+6 2 | 3+4 2 | 5+6 2 | 5+6 2 | 3+3 2 | 3+3 2 |
| | 3+4 2 | 5+6 2 | 3+4 2 | 3+4 2 | 3+3 2 | 2+2 2 |
| | 5+6 2 | 3+4 2 | 5+6 2 | 5+6 2 | 2+2 2 | 3+3 2 |
| | 5+6 2 | 5+6 2 | 5+6 2 | 5+6 2 | 2+2 2 | 2+2 2 |
| | 3+4 2 | 3+4 2 | 3+4 2 | 3+4 2 | 3+3 2 | 3+3 2 |
| | 5+6 2 | 3+4 2 | 5+6 2 | 5+6 2 | 2+2 2 | 3+3 2 |
| | 3+4 2 | 5+6 2 | 3+4 2 | 3+4 2 | 3+3 2 | 2+2 2 |
| | 5+6 2 | 3+4 2 | 3+4 2 | 5+6 2 | 2+2 2 | 2+2 2 |
| | 3+4 2 | 5+6 2 | 5+6 2 | 3+4 2 | 3+3 2 | 3+3 2 |
| 12.2.63 | 84205 | | 44 | | Fe1Sc2Si2 | |
| | -A | -GM | -M | -Y | -L | -V |
| | 5+6 2 | 3+4 2 | 3+4 2 | 3+4 2 | 3+3 2 | 2+2 2 |
| | 3+4 2 | 3+4 2 | 5+6 2 | 3+4 2 | 2+2 2 | 2+2 2 |
| | 3+4 2 | 5+6 2 | 3+4 2 | 5+6 2 | 2+2 2 | 3+3 2 |
| | 5+6 2 | 5+6 2 | 5+6 2 | 5+6 2 | 3+3 2 | 3+3 2 |
| | 5+6 2 | 3+4 2 | 3+4 2 | 3+4 2 | 2+2 2 | 3+3 2 |
| | 3+4 2 | 5+6 2 | 5+6 2 | 5+6 2 | 3+3 2 | 2+2 2 |
| | 5+6 2 | 5+6 2 | 5+6 2 | 5+6 2 | 3+3 2 | 3+3 2 |
| | 3+4 2 | 3+4 2 | 3+4 2 | 3+4 2 | 2+2 2 | 2+2 2 |
| | 3+4 2 | 5+6 2 | 5+6 2 | 5+6 2 | 3+3 2 | 2+2 2 |
| | 5+6 2 | 3+4 2 | 5+6 2 | 3+4 2 | 2+2 2 | 3+3 2 |
| | 5+6 2 | 3+4 2 | 3+4 2 | 3+4 2 | 2+2 2 | 3+3 2 |
| | 3+4 2 | 5+6 2 | 3+4 2 | 5+6 2 | 3+3 2 | 2+2 2 |
| | 5+6 2 | 5+6 2 | 3+4 2 | 3+4 2 | 3+3 2 | 2+2 2 |
| | 3+4 2 | 3+4 2 | 5+6 2 | 5+6 2 | 2+2 2 | 3+3 2 |
| | 3+4 2 | 3+4 2 | 3+4 2 | 5+6 2 | 3+3 2 | 2+2 2 |
| | 5+6 2 | 5+6 2 | 5+6 2 | 3+4 2 | 2+2 2 | 2+2 2 |
| | 5+6 2 | 3+4 2 | 3+4 2 | 5+6 2 | 2+2 2 | 3+3 2 |
| | 3+4 2 | 3+4 2 | 3+4 2 | 3+4 2 | 3+3 2 | 2+2 2 |
| | 5+6 2 | 5+6 2 | 5+6 2 | 5+6 2 | 2+2 2 | 3+3 2 |
| | 5+6 2 | 3+4 2 | 3+4 2 | 3+4 2 | 3+3 2 | 2+2 2 |
| | 5+6 2 | 3+4 2 | 3+4 2 | 5+6 2 | 2+2 2 | 3+3 2 |
| | 3+4 2 | 5+6 2 | 5+6 2 | 3+4 2 | 3+3 2 | 2+2 2 |
| 12.2.64 | 76348 | | 44 | | Fe1Sc2Si2 | |
| | -A | -GM | -M | -Y | -L | -V |
| | 5+6 2 | 3+4 2 | 3+4 2 | 3+4 2 | 3+3 2 | 2+2 2 |
| | 3+4 2 | 3+4 2 | 5+6 2 | 3+4 2 | 2+2 2 | 2+2 2 |
| | 3+4 2 | 5+6 2 | 3+4 2 | 5+6 2 | 2+2 2 | 3+3 2 |
| | 5+6 2 | 5+6 2 | 5+6 2 | 5+6 2 | 3+3 2 | 3+3 2 |
| | 5+6 2 | 3+4 2 | 3+4 2 | 3+4 2 | 2+2 2 | 3+3 2 |
| | 3+4 2 | 5+6 2 | 5+6 2 | 5+6 2 | 3+3 2 | 2+2 2 |
| | 5+6 2 | 5+6 2 | 5+6 2 | 5+6 2 | 3+3 2 | 3+3 2 |
| | 3+4 2 | 3+4 2 | 3+4 2 | 3+4 2 | 2+2 2 | 2+2 2 |
| | 3+4 2 | 5+6 2 | 5+6 2 | 5+6 2 | 3+3 2 | 2+2 2 |
| | 5+6 2 | 3+4 2 | 5+6 2 | 3+4 2 | 2+2 2 | 3+3 2 |
| | 5+6 2 | 3+4 2 | 3+4 2 | 3+4 2 | 2+2 2 | 3+3 2 |
| | 3+4 2 | 5+6 2 | 3+4 2 | 5+6 2 | 3+3 2 | 2+2 2 |
| | 5+6 2 | 5+6 2 | 3+4 2 | 3+4 2 | 3+3 2 | 2+2 2 |
| | 3+4 2 | 3+4 2 | 5+6 2 | 5+6 2 | 2+2 2 | 3+3 2 |
| | 3+4 2 | 3+4 2 | 3+4 2 | 5+6 2 | 3+3 2 | 2+2 2 |
| | 5+6 2 | 5+6 2 | 5+6 2 | 3+4 2 | 2+2 2 | 2+2 2 |
| | 5+6 2 | 3+4 2 | 3+4 2 | 5+6 2 | 2+2 2 | 3+3 2 |

Continued on next column

**TABLE XXIV– continued**

| TQC | ICSD | | NoE | | SF | |
|---|---|---|---|---|---|---|
| | 3+4 2 | 3+4 2 | 3+4 2 | 3+4 2 | 3+3 2 | 2+2 2 |
| | 5+6 2 | 5+6 2 | 5+6 2 | 5+6 2 | 2+2 2 | 3+3 2 |
| | 5+6 2 | 3+4 2 | 3+4 2 | 3+4 2 | 3+3 2 | 2+2 2 |
| | 5+6 2 | 3+4 2 | 3+4 2 | 5+6 2 | 2+2 2 | 3+3 2 |
| | 3+4 2 | 5+6 2 | 5+6 2 | 3+4 2 | 3+3 2 | 2+2 2 |
| 12.2.65 | 633455 | | 44 | | Fe1Sc2Si2 | |
| | -A | -GM | -M | -Y | -L | -V |
| | 5+6 2 | 3+4 2 | 5+6 2 | 3+4 2 | 2+2 2 | 3+3 2 |
| | 3+4 2 | 3+4 2 | 3+4 2 | 3+4 2 | 3+3 2 | 3+3 2 |
| | 3+4 2 | 5+6 2 | 3+4 2 | 5+6 2 | 3+3 2 | 2+2 2 |
| | 5+6 2 | 5+6 2 | 5+6 2 | 5+6 2 | 2+2 2 | 2+2 2 |
| | 5+6 2 | 3+4 2 | 3+4 2 | 3+4 2 | 3+3 2 | 2+2 2 |
| | 3+4 2 | 5+6 2 | 5+6 2 | 5+6 2 | 2+2 2 | 3+3 2 |
| | 5+6 2 | 3+4 2 | 5+6 2 | 5+6 2 | 2+2 2 | 2+2 2 |
| | 3+4 2 | 5+6 2 | 3+4 2 | 3+4 2 | 3+3 2 | 3+3 2 |
| | 3+4 2 | 5+6 2 | 3+4 2 | 5+6 2 | 3+3 2 | 3+3 2 |
| | 5+6 2 | 3+4 2 | 5+6 2 | 3+4 2 | 2+2 2 | 2+2 2 |
| | 5+6 2 | 3+4 2 | 5+6 2 | 3+4 2 | 3+3 2 | 3+3 2 |
| | 3+4 2 | 5+6 2 | 3+4 2 | 5+6 2 | 2+2 2 | 2+2 2 |
| | 3+4 2 | 3+4 2 | 3+4 2 | 3+4 2 | 2+2 2 | 3+3 2 |
| | 5+6 2 | 5+6 2 | 5+6 2 | 5+6 2 | 3+3 2 | 2+2 2 |
| | 3+4 2 | 3+4 2 | 3+4 2 | 5+6 2 | 3+3 2 | 3+3 2 |
| | 5+6 2 | 5+6 2 | 5+6 2 | 5+6 2 | 2+2 2 | 3+3 2 |
| | 5+6 2 | 3+4 2 | 3+4 2 | 3+4 2 | 3+3 2 | 2+2 2 |
| | 3+4 2 | 3+4 2 | 3+4 2 | 3+4 2 | 2+2 2 | 2+2 2 |
| | 5+6 2 | 3+4 2 | 3+4 2 | 5+6 2 | 3+3 2 | 3+3 2 |
| | 5+6 2 | 5+6 2 | 5+6 2 | 3+4 2 | 2+2 2 | 3+3 2 |
| | 5+6 2 | 3+4 2 | 3+4 2 | 5+6 2 | 3+3 2 | 2+2 2 |
| | 3+4 2 | 5+6 2 | 5+6 2 | 3+4 2 | 2+2 2 | 3+3 2 |
| 12.2.66 | 20661 | | 44 | | Fe1Sc2Si2 | |
| | -A | -GM | -M | -Y | -L | -V |
| | 5+6 2 | 3+4 2 | 5+6 2 | 3+4 2 | 2+2 2 | 3+3 2 |
| | 3+4 2 | 3+4 2 | 3+4 2 | 3+4 2 | 3+3 2 | 3+3 2 |
| | 5+6 2 | 5+6 2 | 3+4 2 | 5+6 2 | 3+3 2 | 2+2 2 |
| | 3+4 2 | 5+6 2 | 5+6 2 | 5+6 2 | 2+2 2 | 2+2 2 |
| | 5+6 2 | 3+4 2 | 3+4 2 | 3+4 2 | 3+3 2 | 2+2 2 |
| | 3+4 2 | 5+6 2 | 5+6 2 | 5+6 2 | 2+2 2 | 3+3 2 |
| | 3+4 2 | 3+4 2 | 5+6 2 | 5+6 2 | 2+2 2 | 2+2 2 |
| | 5+6 2 | 5+6 2 | 3+4 2 | 3+4 2 | 3+3 2 | 3+3 2 |
| | 3+4 2 | 5+6 2 | 3+4 2 | 5+6 2 | 3+3 2 | 3+3 2 |
| | 5+6 2 | 3+4 2 | 5+6 2 | 3+4 2 | 2+2 2 | 2+2 2 |
| | 5+6 2 | 3+4 2 | 5+6 2 | 3+4 2 | 3+3 2 | 3+3 2 |
| | 3+4 2 | 5+6 2 | 3+4 2 | 5+6 2 | 2+2 2 | 2+2 2 |
| | 5+6 2 | 5+6 2 | 5+6 2 | 3+4 2 | 3+3 2 | 3+3 2 |
| | 3+4 2 | 3+4 2 | 3+4 2 | 5+6 2 | 2+2 2 | 2+2 2 |
| | 5+6 2 | 3+4 2 | 3+4 2 | 5+6 2 | 3+3 2 | 3+3 2 |
| | 3+4 2 | 3+4 2 | 5+6 2 | 5+6 2 | 2+2 2 | 3+3 2 |
| | 5+6 2 | 3+4 2 | 3+4 2 | 3+4 2 | 3+3 2 | 2+2 2 |
| | 3+4 2 | 5+6 2 | 3+4 2 | 3+4 2 | 3+3 2 | 2+2 2 |
| | 5+6 2 | 5+6 2 | 3+4 2 | 5+6 2 | 2+2 2 | 3+3 2 |
| | 5+6 2 | 3+4 2 | 5+6 2 | 3+4 2 | 2+2 2 | 3+3 2 |
| | 5+6 2 | 3+4 2 | 3+4 2 | 5+6 2 | 3+3 2 | 2+2 2 |
| | 3+4 2 | 5+6 2 | 5+6 2 | 3+4 2 | 2+2 2 | 3+3 2 |
| 12.2.67 | 633361 | | 42 | | Fe1S4V2 | |
| | -A | -GM | -M | -Y | -L | -V |
| | 5+6 2 | 3+4 2 | 5+6 2 | 5+6 2 | 3+3 2 | 2+2 2 |
| | 3+4 2 | 5+6 2 | 3+4 2 | 3+4 2 | 2+2 2 | 3+3 2 |
| | 5+6 2 | 5+6 2 | 3+4 2 | 3+4 2 | 3+3 2 | 3+3 2 |
| | 3+4 2 | 3+4 2 | 5+6 2 | 5+6 2 | 2+2 2 | 2+2 2 |
| | 5+6 2 | 3+4 2 | 5+6 2 | 3+4 2 | 2+2 2 | 3+3 2 |
| | 3+4 2 | 5+6 2 | 3+4 2 | 5+6 2 | 3+3 2 | 2+2 2 |

Continued on next column

**TABLE XXIV– continued**

| TQC | ICSD | | NoE | | SF | |
|---|---|---|---|---|---|---|
| | 5+6 2 | 3+4 2 | 3+4 2 | 3+4 2 | 2+2 2 | 3+3 2 |
| | 3+4 2 | 3+4 2 | 5+6 2 | 5+6 2 | 3+3 2 | 2+2 2 |
| | 5+6 2 | 3+4 2 | 3+4 2 | 3+4 2 | 2+2 2 | 3+3 2 |
| | 3+4 2 | 5+6 2 | 5+6 2 | 5+6 2 | 3+3 2 | 2+2 2 |
| | 5+6 2 | 5+6 2 | 3+4 2 | 3+4 2 | 2+2 2 | 3+3 2 |
| | 5+6 2 | 3+4 2 | 5+6 2 | 5+6 2 | 3+3 2 | 3+3 2 |
| | 3+4 2 | 3+4 2 | 5+6 2 | 5+6 2 | 3+3 2 | 2+2 2 |
| | 5+6 2 | 3+4 2 | 5+6 2 | 5+6 2 | 2+2 2 | 2+2 2 |
| | 3+4 2 | 5+6 2 | 3+4 2 | 3+4 2 | 3+3 2 | 3+3 2 |
| | 3+4 2 | 5+6 2 | 3+4 2 | 3+4 2 | 2+2 2 | 2+2 2 |
| | 5+6 2 | 5+6 2 | 5+6 2 | 3+4 2 | 2+2 2 | 3+3 2 |
| | 5+6 2 | 5+6 2 | 3+4 2 | 3+4 2 | 3+3 2 | 3+3 2 |
| | 5+6 2 | 3+4 2 | 5+6 2 | 3+4 2 | 2+2 2 | 3+3 2 |
| | 3+4 2 | 3+4 2 | 5+6 2 | 5+6 2 | 2+2 2 | 2+2 2 |
| | 5+6 2 | 3+4 2 | 5+6 2 | 5+6 2 | 2+2 2 | 3+3 2 |
| 12.2.68 | 603354 | | 50 | | Fe1Rh2Se4 | |
| | -A | -GM | -M | -Y | -L | -V |
| | 5+6 2 | 3+4 2 | 5+6 2 | 5+6 2 | 2+2 2 | 3+3 2 |
| | 3+4 2 | 5+6 2 | 3+4 2 | 3+4 2 | 3+3 2 | 2+2 2 |
| | 3+4 2 | 3+4 2 | 5+6 2 | 5+6 2 | 2+2 2 | 2+2 2 |
| | 5+6 2 | 5+6 2 | 3+4 2 | 3+4 2 | 3+3 2 | 3+3 2 |
| | 3+4 2 | 3+4 2 | 3+4 2 | 3+4 2 | 3+3 2 | 3+3 2 |
| | 5+6 2 | 5+6 2 | 5+6 2 | 5+6 2 | 2+2 2 | 2+2 2 |
| | 3+4 2 | 3+4 2 | 5+6 2 | 3+4 2 | 3+3 2 | 2+2 2 |
| | 5+6 2 | 3+4 2 | 5+6 2 | 3+4 2 | 2+2 2 | 3+3 2 |
| | 3+4 2 | 5+6 2 | 3+4 2 | 5+6 2 | 3+3 2 | 3+3 2 |
| | 3+4 2 | 3+4 2 | 5+6 2 | 5+6 2 | 2+2 2 | 3+3 2 |
| | 5+6 2 | 5+6 2 | 3+4 2 | 5+6 2 | 3+3 2 | 2+2 2 |
| | 5+6 2 | 3+4 2 | 3+4 2 | 5+6 2 | 2+2 2 | 2+2 2 |
| | 3+4 2 | 3+4 2 | 5+6 2 | 3+4 2 | 2+2 2 | 3+3 2 |
| | 3+4 2 | 5+6 2 | 3+4 2 | 3+4 2 | 2+2 2 | 2+2 2 |
| | 3+4 2 | 3+4 2 | 3+4 2 | 5+6 2 | 3+3 2 | 3+3 2 |
| | 5+6 2 | 5+6 2 | 5+6 2 | 3+4 2 | 3+3 2 | 2+2 2 |
| | 5+6 2 | 3+4 2 | 3+4 2 | 3+4 2 | 2+2 2 | 3+3 2 |
| | 5+6 2 | 3+4 2 | 3+4 2 | 5+6 2 | 3+3 2 | 2+2 2 |
| | 5+6 2 | 5+6 2 | 5+6 2 | 5+6 2 | 3+3 2 | 3+3 2 |
| | 5+6 2 | 3+4 2 | 3+4 2 | 3+4 2 | 2+2 2 | 2+2 2 |
| | 3+4 2 | 3+4 2 | 3+4 2 | 3+4 2 | 3+3 2 | 3+3 2 |
| | 3+4 2 | 5+6 2 | 3+4 2 | 5+6 2 | 3+3 2 | 3+3 2 |
| | 3+4 2 | 3+4 2 | 3+4 2 | 3+4 2 | 3+3 2 | 3+3 2 |
| | 3+4 2 | 5+6 2 | 5+6 2 | 3+4 2 | 2+2 2 | 2+2 2 |
| | 3+4 2 | 5+6 2 | 3+4 2 | 3+4 2 | 3+3 2 | 3+3 2 |
| 12.2.69 | 632972 | | 52 | | Fe1Ni2Se4 | |
| | -A | -GM | -M | -Y | -L | -V |
| | 3+4 2 | 3+4 2 | 3+4 2 | 5+6 2 | 2+2 2 | 3+3 2 |
| | 5+6 2 | 3+4 2 | 5+6 2 | 3+4 2 | 3+3 2 | 2+2 2 |
| | 3+4 2 | 5+6 2 | 5+6 2 | 5+6 2 | 2+2 2 | 2+2 2 |
| | 5+6 2 | 5+6 2 | 3+4 2 | 3+4 2 | 3+3 2 | 3+3 2 |
| | 3+4 2 | 3+4 2 | 3+4 2 | 3+4 2 | 2+2 2 | 3+3 2 |
| | 5+6 2 | 5+6 2 | 5+6 2 | 5+6 2 | 3+3 2 | 2+2 2 |
| | 3+4 2 | 3+4 2 | 5+6 2 | 5+6 2 | 3+3 2 | 3+3 2 |
| | 5+6 2 | 3+4 2 | 3+4 2 | 3+4 2 | 2+2 2 | 2+2 2 |
| | 3+4 2 | 3+4 2 | 5+6 2 | 5+6 2 | 2+2 2 | 3+3 2 |
| | 3+4 2 | 5+6 2 | 3+4 2 | 3+4 2 | 3+3 2 | 2+2 2 |
| | 5+6 2 | 5+6 2 | 5+6 2 | 5+6 2 | 3+3 2 | 3+3 2 |
| | 5+6 2 | 3+4 2 | 3+4 2 | 5+6 2 | 2+2 2 | 2+2 2 |
| | 3+4 2 | 3+4 2 | 3+4 2 | 3+4 2 | 2+2 2 | 3+3 2 |
| | 3+4 2 | 3+4 2 | 3+4 2 | 5+6 2 | 2+2 2 | 2+2 2 |
| | 5+6 2 | 5+6 2 | 5+6 2 | 3+4 2 | 3+3 2 | 3+3 2 |
| | 3+4 2 | 5+6 2 | 5+6 2 | 3+4 2 | 3+3 2 | 2+2 2 |
| | 5+6 2 | 3+4 2 | 3+4 2 | 3+4 2 | 3+3 2 | 3+3 2 |

Continued on next column

**TABLE XXIV– continued**

| TQC | ICSD | | NoE | SF | | |
|---|---|---|---|---|---|---|
| | 5+6 2 | 3+4 2 | 3+4 2 | 5+6 2 | 2+2 2 | 2+2 2 |
| | 5+6 2 | 5+6 2 | 3+4 2 | 5+6 2 | 2+2 2 | 3+3 2 |
| | 3+4 2 | 3+4 2 | 5+6 2 | 3+4 2 | 3+3 2 | 2+2 2 |
| | 5+6 2 | 3+4 2 | 3+4 2 | 3+4 2 | 3+3 2 | 3+3 2 |
| | 3+4 2 | 5+6 2 | 3+4 2 | 5+6 2 | 2+2 2 | 3+3 2 |
| | 3+4 2 | 3+4 2 | 5+6 2 | 3+4 2 | 3+3 2 | 2+2 2 |
| | 3+4 2 | 5+6 2 | 3+4 2 | 3+4 2 | 3+3 2 | 3+3 2 |
| | 3+4 2 | 3+4 2 | 5+6 2 | 3+4 2 | 3+3 2 | 3+3 2 |
| | 5+6 2 | 5+6 2 | 3+4 2 | 3+4 2 | 3+3 2 | 2+2 2 |
| 12.2.70 | 90460 | | 72 | Fe1Ge1Yb1 | | |
| | -A | -GM | -M | -Y | -L | -V |
| | 5+6 2 | 3+4 2 | 5+6 2 | 3+4 2 | 3+3 2 | 2+2 2 |
| | 3+4 2 | 5+6 2 | 3+4 2 | 5+6 2 | 2+2 2 | 3+3 2 |
| | 5+6 2 | 3+4 2 | 5+6 2 | 3+4 2 | 3+3 2 | 2+2 2 |
| | 3+4 2 | 5+6 2 | 3+4 2 | 5+6 2 | 2+2 2 | 3+3 2 |
| | 5+6 2 | 3+4 2 | 5+6 2 | 3+4 2 | 3+3 2 | 2+2 2 |
| | 3+4 2 | 5+6 2 | 3+4 2 | 5+6 2 | 2+2 2 | 3+3 2 |
| | 3+4 2 | 3+4 2 | 5+6 2 | 3+4 2 | 3+3 2 | 2+2 2 |
| | 5+6 2 | 5+6 2 | 3+4 2 | 5+6 2 | 2+2 2 | 3+3 2 |
| | 3+4 2 | 3+4 2 | 3+4 2 | 3+4 2 | 3+3 2 | 3+3 2 |
| | 5+6 2 | 5+6 2 | 5+6 2 | 5+6 2 | 2+2 2 | 2+2 2 |
| | 3+4 2 | 3+4 2 | 3+4 2 | 3+4 2 | 2+2 2 | 2+2 2 |
| | 5+6 2 | 5+6 2 | 5+6 2 | 5+6 2 | 3+3 2 | 3+3 2 |
| | 3+4 2 | 5+6 2 | 5+6 2 | 3+4 2 | 3+3 2 | 2+2 2 |
| | 3+4 2 | 3+4 2 | 5+6 2 | 5+6 2 | 2+2 2 | 3+3 2 |
| | 5+6 2 | 5+6 2 | 3+4 2 | 3+4 2 | 3+3 2 | 3+3 2 |
| | 5+6 2 | 3+4 2 | 3+4 2 | 5+6 2 | 2+2 2 | 3+3 2 |
| | 5+6 2 | 3+4 2 | 5+6 2 | 3+4 2 | 3+3 2 | 2+2 2 |
| | 3+4 2 | 5+6 2 | 3+4 2 | 5+6 2 | 2+2 2 | 2+2 2 |
| | 3+4 2 | 5+6 2 | 5+6 2 | 3+4 2 | 2+2 2 | 3+3 2 |
| | 5+6 2 | 3+4 2 | 3+4 2 | 5+6 2 | 3+3 2 | 2+2 2 |
| | 3+4 2 | 5+6 2 | 5+6 2 | 3+4 2 | 3+3 2 | 3+3 2 |
| | 5+6 2 | 3+4 2 | 3+4 2 | 5+6 2 | 2+2 2 | 2+2 2 |
| | 5+6 2 | 3+4 2 | 5+6 2 | 3+4 2 | 3+3 2 | 3+3 2 |
| | 3+4 2 | 3+4 2 | 3+4 2 | 5+6 2 | 2+2 2 | 2+2 2 |
| | 5+6 2 | 5+6 2 | 3+4 2 | 5+6 2 | 2+2 2 | 3+3 2 |
| | 3+4 2 | 3+4 2 | 5+6 2 | 3+4 2 | 3+3 2 | 3+3 2 |
| | 3+4 2 | 5+6 2 | 3+4 2 | 3+4 2 | 3+3 2 | 2+2 2 |
| | 5+6 2 | 3+4 2 | 5+6 2 | 5+6 2 | 2+2 2 | 3+3 2 |
| | 5+6 2 | 5+6 2 | 3+4 2 | 3+4 2 | 3+3 2 | 2+2 2 |
| | 5+6 2 | 5+6 2 | 5+6 2 | 5+6 2 | 2+2 2 | 3+3 2 |
| | 3+4 2 | 3+4 2 | 3+4 2 | 3+4 2 | 3+3 2 | 2+2 2 |
| | 3+4 2 | 5+6 2 | 3+4 2 | 5+6 2 | 2+2 2 | 2+2 2 |
| | 5+6 2 | 3+4 2 | 5+6 2 | 5+6 2 | 3+3 2 | 3+3 2 |
| | 3+4 2 | 3+4 2 | 5+6 2 | 3+4 2 | 2+2 2 | 2+2 2 |
| | 3+4 2 | 5+6 2 | 3+4 2 | 5+6 2 | 3+3 2 | 3+3 2 |
| | 5+6 2 | 3+4 2 | 5+6 2 | 3+4 2 | 2+2 2 | 3+3 2 |
| 12.2.71 | 631983 | | 48 | Fe1Ge1 | | |
| | -A | -GM | -M | -Y | -L | -V |
| | 3+4 2 | 3+4 2 | 3+4 2 | 3+4 2 | 2+2 2 | 2+2 2 |
| | 5+6 2 | 3+4 2 | 5+6 2 | 5+6 2 | 3+3 2 | 3+3 2 |
| | 3+4 2 | 5+6 2 | 5+6 2 | 3+4 2 | 2+2 2 | 2+2 2 |
| | 5+6 2 | 5+6 2 | 3+4 2 | 5+6 2 | 3+3 2 | 3+3 2 |
| | 5+6 2 | 3+4 2 | 3+4 2 | 3+4 2 | 2+2 2 | 3+3 2 |
| | 3+4 2 | 3+4 2 | 5+6 2 | 3+4 2 | 3+3 2 | 3+3 2 |
| | 3+4 2 | 3+4 2 | 3+4 2 | 3+4 2 | 3+3 2 | 2+2 2 |
| | 3+4 2 | 5+6 2 | 5+6 2 | 3+4 2 | 2+2 2 | 3+3 2 |
| | 5+6 2 | 3+4 2 | 5+6 2 | 3+4 2 | 2+2 2 | 3+3 2 |
| | 3+4 2 | 5+6 2 | 5+6 2 | 5+6 2 | 3+3 2 | 2+2 2 |
| | 3+4 2 | 3+4 2 | 5+6 2 | 5+6 2 | 2+2 2 | 2+2 2 |
| | 5+6 2 | 3+4 2 | 3+4 2 | 3+4 2 | 3+3 2 | 3+3 2 |

Continued on next column

**TABLE XXIV– continued**

| TQC | ICSD | | NoE | SF | | |
|---|---|---|---|---|---|---|
| | 3+4 2 | 3+4 2 | 5+6 2 | 3+4 2 | 3+3 2 | 2+2 2 |
| | 5+6 2 | 5+6 2 | 3+4 2 | 5+6 2 | 3+3 2 | 3+3 2 |
| | 3+4 2 | 3+4 2 | 3+4 2 | 5+6 2 | 2+2 2 | 3+3 2 |
| | 5+6 2 | 3+4 2 | 3+4 2 | 5+6 2 | 3+3 2 | 2+2 2 |
| | 5+6 2 | 3+4 2 | 5+6 2 | 5+6 2 | 2+2 2 | 3+3 2 |
| | 3+4 2 | 5+6 2 | 3+4 2 | 3+4 2 | 2+2 2 | 2+2 2 |
| | 5+6 2 | 5+6 2 | 5+6 2 | 3+4 2 | 3+3 2 | 3+3 2 |
| | 5+6 2 | 5+6 2 | 5+6 2 | 3+4 2 | 2+2 2 | 2+2 2 |
| | 5+6 2 | 3+4 2 | 3+4 2 | 3+4 2 | 2+2 2 | 3+3 2 |
| | 5+6 2 | 3+4 2 | 3+4 2 | 5+6 2 | 3+3 2 | 3+3 2 |
| | 3+4 2 | 3+4 2 | 5+6 2 | 3+4 2 | 2+2 2 | 3+3 2 |
| | 3+4 2 | 3+4 2 | 3+4 2 | 3+4 2 | 3+3 2 | 2+2 2 |
| 12.2.72 | 53458 | | 48 | Fe1Ge1 | | |
| | -A | -GM | -M | -Y | -L | -V |
| | 3+4 2 | 3+4 2 | 5+6 2 | 5+6 2 | 2+2 2 | 2+2 2 |
| | 5+6 2 | 3+4 2 | 3+4 2 | 3+4 2 | 3+3 2 | 3+3 2 |
| | 3+4 2 | 5+6 2 | 5+6 2 | 5+6 2 | 3+3 2 | 2+2 2 |
| | 5+6 2 | 5+6 2 | 3+4 2 | 3+4 2 | 2+2 2 | 3+3 2 |
| | 5+6 2 | 3+4 2 | 3+4 2 | 3+4 2 | 2+2 2 | 3+3 2 |
| | 3+4 2 | 3+4 2 | 5+6 2 | 3+4 2 | 3+3 2 | 3+3 2 |
| | 3+4 2 | 3+4 2 | 5+6 2 | 3+4 2 | 3+3 2 | 2+2 2 |
| | 3+4 2 | 5+6 2 | 5+6 2 | 5+6 2 | 2+2 2 | 3+3 2 |
| | 5+6 2 | 3+4 2 | 3+4 2 | 3+4 2 | 2+2 2 | 3+3 2 |
| | 3+4 2 | 5+6 2 | 5+6 2 | 3+4 2 | 3+3 2 | 2+2 2 |
| | 5+6 2 | 5+6 2 | 5+6 2 | 5+6 2 | 2+2 2 | 2+2 2 |
| | 3+4 2 | 3+4 2 | 3+4 2 | 5+6 2 | 3+3 2 | 3+3 2 |
| | 5+6 2 | 3+4 2 | 5+6 2 | 3+4 2 | 3+3 2 | 2+2 2 |
| | 5+6 2 | 3+4 2 | 3+4 2 | 3+4 2 | 2+2 2 | 2+2 2 |
| | 3+4 2 | 3+4 2 | 3+4 2 | 5+6 2 | 3+3 2 | 3+3 2 |
| | 3+4 2 | 3+4 2 | 3+4 2 | 5+6 2 | 2+2 2 | 3+3 2 |
| | 5+6 2 | 3+4 2 | 5+6 2 | 5+6 2 | 3+3 2 | 3+3 2 |
| | 5+6 2 | 5+6 2 | 3+4 2 | 3+4 2 | 2+2 2 | 2+2 2 |
| | 3+4 2 | 5+6 2 | 5+6 2 | 3+4 2 | 2+2 2 | 3+3 2 |
| | 5+6 2 | 5+6 2 | 3+4 2 | 3+4 2 | 3+3 2 | 2+2 2 |
| | 3+4 2 | 3+4 2 | 5+6 2 | 3+4 2 | 2+2 2 | 3+3 2 |
| | 5+6 2 | 3+4 2 | 3+4 2 | 5+6 2 | 3+3 2 | 3+3 2 |
| | 3+4 2 | 3+4 2 | 5+6 2 | 3+4 2 | 3+3 2 | 3+3 2 |
| | 5+6 2 | 3+4 2 | 3+4 2 | 3+4 2 | 2+2 2 | 2+2 2 |
| 12.2.73 | 53457 | | 48 | Fe1Ge1 | | |
| | -A | -GM | -M | -Y | -L | -V |
| | 3+4 2 | 3+4 2 | 5+6 2 | 5+6 2 | 2+2 2 | 2+2 2 |
| | 5+6 2 | 3+4 2 | 3+4 2 | 3+4 2 | 3+3 2 | 3+3 2 |
| | 3+4 2 | 5+6 2 | 5+6 2 | 5+6 2 | 3+3 2 | 2+2 2 |
| | 5+6 2 | 5+6 2 | 3+4 2 | 3+4 2 | 2+2 2 | 3+3 2 |
| | 5+6 2 | 3+4 2 | 3+4 2 | 3+4 2 | 2+2 2 | 3+3 2 |
| | 3+4 2 | 3+4 2 | 5+6 2 | 3+4 2 | 3+3 2 | 3+3 2 |
| | 3+4 2 | 3+4 2 | 5+6 2 | 3+4 2 | 3+3 2 | 2+2 2 |
| | 3+4 2 | 5+6 2 | 5+6 2 | 5+6 2 | 2+2 2 | 3+3 2 |
| | 5+6 2 | 3+4 2 | 3+4 2 | 3+4 2 | 2+2 2 | 3+3 2 |
| | 3+4 2 | 5+6 2 | 5+6 2 | 3+4 2 | 3+3 2 | 2+2 2 |
| | 5+6 2 | 5+6 2 | 5+6 2 | 5+6 2 | 2+2 2 | 2+2 2 |
| | 3+4 2 | 3+4 2 | 3+4 2 | 5+6 2 | 3+3 2 | 3+3 2 |
| | 5+6 2 | 3+4 2 | 5+6 2 | 3+4 2 | 3+3 2 | 2+2 2 |
| | 5+6 2 | 3+4 2 | 3+4 2 | 3+4 2 | 2+2 2 | 2+2 2 |
| | 3+4 2 | 3+4 2 | 3+4 2 | 5+6 2 | 3+3 2 | 3+3 2 |
| | 3+4 2 | 3+4 2 | 3+4 2 | 5+6 2 | 2+2 2 | 3+3 2 |
| | 5+6 2 | 3+4 2 | 5+6 2 | 5+6 2 | 3+3 2 | 3+3 2 |
| | 5+6 2 | 5+6 2 | 3+4 2 | 3+4 2 | 2+2 2 | 2+2 2 |
| | 3+4 2 | 5+6 2 | 5+6 2 | 3+4 2 | 2+2 2 | 3+3 2 |
| | 5+6 2 | 5+6 2 | 3+4 2 | 3+4 2 | 3+3 2 | 2+2 2 |
| | 3+4 2 | 3+4 2 | 5+6 2 | 3+4 2 | 2+2 2 | 3+3 2 |

Continued on next column

**TABLE XXIV– continued**

| TQC | ICSD | NoE | SF | | | |
|---|---|---|---|---|---|---|
| | 5+6 2 | 3+4 2 | 3+4 2 | 5+6 2 | 3+3 2 | 3+3 2 |
| | 3+4 2 | 3+4 2 | 5+6 2 | 3+4 2 | 3+3 2 | 3+3 2 |
| | 5+6 2 | 3+4 2 | 3+4 2 | 3+4 2 | 2+2 2 | 2+2 2 |
| 12.2.74 | 53456 | 48 | Fe1Ge1 | | | |
| | -A | -GM | -M | -Y | -L | -V |
| | 3+4 2 | 3+4 2 | 5+6 2 | 5+6 2 | 2+2 2 | 2+2 2 |
| | 5+6 2 | 3+4 2 | 3+4 2 | 3+4 2 | 3+3 2 | 3+3 2 |
| | 3+4 2 | 5+6 2 | 5+6 2 | 5+6 2 | 3+3 2 | 2+2 2 |
| | 5+6 2 | 5+6 2 | 3+4 2 | 3+4 2 | 2+2 2 | 3+3 2 |
| | 5+6 2 | 3+4 2 | 3+4 2 | 3+4 2 | 2+2 2 | 3+3 2 |
| | 3+4 2 | 3+4 2 | 5+6 2 | 3+4 2 | 3+3 2 | 3+3 2 |
| | 3+4 2 | 3+4 2 | 5+6 2 | 3+4 2 | 3+3 2 | 2+2 2 |
| | 3+4 2 | 5+6 2 | 5+6 2 | 5+6 2 | 2+2 2 | 3+3 2 |
| | 5+6 2 | 3+4 2 | 3+4 2 | 3+4 2 | 2+2 2 | 3+3 2 |
| | 3+4 2 | 5+6 2 | 5+6 2 | 3+4 2 | 3+3 2 | 2+2 2 |
| | 5+6 2 | 5+6 2 | 5+6 2 | 5+6 2 | 2+2 2 | 2+2 2 |
| | 3+4 2 | 3+4 2 | 3+4 2 | 5+6 2 | 3+3 2 | 3+3 2 |
| | 5+6 2 | 3+4 2 | 5+6 2 | 3+4 2 | 3+3 2 | 2+2 2 |
| | 5+6 2 | 3+4 2 | 3+4 2 | 3+4 2 | 2+2 2 | 2+2 2 |
| | 3+4 2 | 3+4 2 | 3+4 2 | 5+6 2 | 3+3 2 | 3+3 2 |
| | 3+4 2 | 3+4 2 | 3+4 2 | 5+6 2 | 2+2 2 | 3+3 2 |
| | 5+6 2 | 3+4 2 | 5+6 2 | 5+6 2 | 3+3 2 | 3+3 2 |
| | 5+6 2 | 5+6 2 | 3+4 2 | 3+4 2 | 2+2 2 | 2+2 2 |
| | 3+4 2 | 5+6 2 | 5+6 2 | 3+4 2 | 2+2 2 | 3+3 2 |
| | 5+6 2 | 5+6 2 | 3+4 2 | 3+4 2 | 3+3 2 | 2+2 2 |
| | 3+4 2 | 3+4 2 | 5+6 2 | 3+4 2 | 2+2 2 | 3+3 2 |
| | 5+6 2 | 3+4 2 | 5+6 2 | 5+6 2 | 3+3 2 | 3+3 2 |
| | 3+4 2 | 3+4 2 | 3+4 2 | 3+4 2 | 3+3 2 | 3+3 2 |
| | 5+6 2 | 3+4 2 | 3+4 2 | 3+4 2 | 2+2 2 | 2+2 2 |
| 12.2.75 | 53455 | 48 | Fe1Ge1 | | | |
| | -A | -GM | -M | -Y | -L | -V |
| | 3+4 2 | 3+4 2 | 5+6 2 | 5+6 2 | 2+2 2 | 2+2 2 |
| | 5+6 2 | 3+4 2 | 3+4 2 | 3+4 2 | 3+3 2 | 3+3 2 |
| | 3+4 2 | 5+6 2 | 5+6 2 | 5+6 2 | 3+3 2 | 2+2 2 |
| | 5+6 2 | 5+6 2 | 3+4 2 | 3+4 2 | 2+2 2 | 3+3 2 |
| | 5+6 2 | 3+4 2 | 3+4 2 | 3+4 2 | 2+2 2 | 3+3 2 |
| | 3+4 2 | 3+4 2 | 5+6 2 | 3+4 2 | 3+3 2 | 3+3 2 |
| | 3+4 2 | 3+4 2 | 5+6 2 | 3+4 2 | 3+3 2 | 2+2 2 |
| | 3+4 2 | 5+6 2 | 5+6 2 | 5+6 2 | 2+2 2 | 3+3 2 |
| | 5+6 2 | 3+4 2 | 3+4 2 | 3+4 2 | 2+2 2 | 3+3 2 |
| | 3+4 2 | 5+6 2 | 5+6 2 | 3+4 2 | 3+3 2 | 2+2 2 |
| | 5+6 2 | 5+6 2 | 5+6 2 | 5+6 2 | 2+2 2 | 2+2 2 |
| | 5+6 2 | 3+4 2 | 3+4 2 | 5+6 2 | 3+3 2 | 3+3 2 |
| | 3+4 2 | 3+4 2 | 5+6 2 | 3+4 2 | 3+3 2 | 2+2 2 |
| | 5+6 2 | 3+4 2 | 3+4 2 | 3+4 2 | 2+2 2 | 2+2 2 |
| | 3+4 2 | 3+4 2 | 3+4 2 | 5+6 2 | 3+3 2 | 3+3 2 |
| | 3+4 2 | 3+4 2 | 3+4 2 | 5+6 2 | 2+2 2 | 3+3 2 |
| | 5+6 2 | 3+4 2 | 5+6 2 | 5+6 2 | 3+3 2 | 3+3 2 |
| | 5+6 2 | 5+6 2 | 3+4 2 | 3+4 2 | 2+2 2 | 2+2 2 |
| | 3+4 2 | 5+6 2 | 5+6 2 | 3+4 2 | 2+2 2 | 3+3 2 |
| | 5+6 2 | 5+6 2 | 3+4 2 | 3+4 2 | 3+3 2 | 2+2 2 |
| | 3+4 2 | 3+4 2 | 5+6 2 | 3+4 2 | 2+2 2 | 3+3 2 |
| | 5+6 2 | 3+4 2 | 5+6 2 | 5+6 2 | 3+3 2 | 3+3 2 |
| | 3+4 2 | 3+4 2 | 3+4 2 | 3+4 2 | 3+3 2 | 3+3 2 |
| | 5+6 2 | 3+4 2 | 3+4 2 | 3+4 2 | 2+2 2 | 2+2 2 |
| 12.2.76 | 43055 | 48 | Fe1Ge1 | | | |
| | -A | -GM | -M | -Y | -L | -V |
| | 3+4 2 | 3+4 2 | 5+6 2 | 5+6 2 | 2+2 2 | 2+2 2 |
| | 5+6 2 | 3+4 2 | 3+4 2 | 3+4 2 | 3+3 2 | 3+3 2 |
| | 3+4 2 | 5+6 2 | 5+6 2 | 5+6 2 | 3+3 2 | 2+2 2 |
| | 5+6 2 | 5+6 2 | 3+4 2 | 3+4 2 | 2+2 2 | 3+3 2 |

Continued on next column

**TABLE XXIV– continued**

| TQC | ICSD | NoE | SF | | | |
|---|---|---|---|---|---|---|
| | 5+6 2 | 3+4 2 | 3+4 2 | 3+4 2 | 2+2 2 | 3+3 2 |
| | 3+4 2 | 3+4 2 | 5+6 2 | 3+4 2 | 3+3 2 | 3+3 2 |
| | 3+4 2 | 3+4 2 | 5+6 2 | 3+4 2 | 3+3 2 | 2+2 2 |
| | 3+4 2 | 5+6 2 | 5+6 2 | 5+6 2 | 2+2 2 | 3+3 2 |
| | 5+6 2 | 3+4 2 | 3+4 2 | 3+4 2 | 2+2 2 | 3+3 2 |
| | 3+4 2 | 5+6 2 | 5+6 2 | 3+4 2 | 3+3 2 | 2+2 2 |
| | 5+6 2 | 5+6 2 | 5+6 2 | 5+6 2 | 2+2 2 | 2+2 2 |
| | 5+6 2 | 3+4 2 | 3+4 2 | 5+6 2 | 3+3 2 | 3+3 2 |
| | 3+4 2 | 3+4 2 | 5+6 2 | 3+4 2 | 3+3 2 | 2+2 2 |
| | 5+6 2 | 3+4 2 | 3+4 2 | 3+4 2 | 2+2 2 | 2+2 2 |
| | 3+4 2 | 3+4 2 | 3+4 2 | 5+6 2 | 3+3 2 | 3+3 2 |
| | 3+4 2 | 3+4 2 | 3+4 2 | 5+6 2 | 2+2 2 | 3+3 2 |
| | 5+6 2 | 3+4 2 | 5+6 2 | 5+6 2 | 3+3 2 | 3+3 2 |
| | 5+6 2 | 5+6 2 | 3+4 2 | 3+4 2 | 2+2 2 | 2+2 2 |
| | 3+4 2 | 5+6 2 | 5+6 2 | 3+4 2 | 2+2 2 | 3+3 2 |
| | 5+6 2 | 5+6 2 | 3+4 2 | 3+4 2 | 3+3 2 | 2+2 2 |
| | 3+4 2 | 3+4 2 | 5+6 2 | 3+4 2 | 2+2 2 | 3+3 2 |
| | 5+6 2 | 3+4 2 | 5+6 2 | 5+6 2 | 3+3 2 | 3+3 2 |
| | 3+4 2 | 3+4 2 | 3+4 2 | 3+4 2 | 3+3 2 | 3+3 2 |
| | 5+6 2 | 3+4 2 | 3+4 2 | 3+4 2 | 2+2 2 | 2+2 2 |
| 12.2.77 | 416629 | 94 | Eu2Li1Si3 | | | |
| | -A | -GM | -M | -Y | -L | -V |
| | 3+4 2 | 3+4 2 | 3+4 2 | 3+4 2 | 2+2 2 | 2+2 2 |
| | 5+6 2 | 5+6 2 | 5+6 2 | 5+6 2 | 3+3 2 | 3+3 2 |
| | 5+6 2 | 3+4 2 | 5+6 2 | 3+4 2 | 3+3 2 | 2+2 2 |
| | 3+4 2 | 5+6 2 | 3+4 2 | 5+6 2 | 2+2 2 | 3+3 2 |
| | 3+4 2 | 3+4 2 | 3+4 2 | 5+6 2 | 2+2 2 | 2+2 2 |
| | 5+6 2 | 5+6 2 | 5+6 2 | 3+4 2 | 3+3 2 | 3+3 2 |
| | 5+6 2 | 3+4 2 | 5+6 2 | 3+4 2 | 3+3 2 | 2+2 2 |
| | 3+4 2 | 5+6 2 | 3+4 2 | 5+6 2 | 2+2 2 | 3+3 2 |
| | 3+4 2 | 3+4 2 | 5+6 2 | 5+6 2 | 2+2 2 | 2+2 2 |
| | 5+6 2 | 5+6 2 | 3+4 2 | 3+4 2 | 3+3 2 | 3+3 2 |
| | 3+4 2 | 3+4 2 | 3+4 2 | 3+4 2 | 2+2 2 | 2+2 2 |
| | 5+6 2 | 5+6 2 | 5+6 2 | 5+6 2 | 3+3 2 | 3+3 2 |
| | 5+6 2 | 3+4 2 | 5+6 2 | 3+4 2 | 3+3 2 | 2+2 2 |
| | 3+4 2 | 5+6 2 | 3+4 2 | 5+6 2 | 2+2 2 | 3+3 2 |
| | 5+6 2 | 3+4 2 | 5+6 2 | 3+4 2 | 3+3 2 | 2+2 2 |
| | 3+4 2 | 5+6 2 | 3+4 2 | 5+6 2 | 2+2 2 | 3+3 2 |
| | 5+6 2 | 3+4 2 | 5+6 2 | 3+4 2 | 2+2 2 | 3+3 2 |
| | 3+4 2 | 5+6 2 | 3+4 2 | 5+6 2 | 3+3 2 | 2+2 2 |
| | 3+4 2 | 3+4 2 | 3+4 2 | 3+4 2 | 3+3 2 | 3+3 2 |
| | 5+6 2 | 5+6 2 | 5+6 2 | 5+6 2 | 2+2 2 | 2+2 2 |
| | 5+6 2 | 3+4 2 | 5+6 2 | 3+4 2 | 2+2 2 | 3+3 2 |
| | 3+4 2 | 5+6 2 | 3+4 2 | 5+6 2 | 3+3 2 | 2+2 2 |
| | 5+6 2 | 3+4 2 | 5+6 2 | 5+6 2 | 3+3 2 | 2+2 2 |
| | 3+4 2 | 5+6 2 | 3+4 2 | 3+4 2 | 2+2 2 | 3+3 2 |
| | 3+4 2 | 5+6 2 | 5+6 2 | 3+4 2 | 2+2 2 | 2+2 2 |
| | 5+6 2 | 3+4 2 | 3+4 2 | 5+6 2 | 3+3 2 | 3+3 2 |
| | 5+6 2 | 3+4 2 | 3+4 2 | 3+4 2 | 2+2 2 | 3+3 2 |
| | 3+4 2 | 5+6 2 | 5+6 2 | 3+4 2 | 2+2 2 | 2+2 2 |
| | 5+6 2 | 3+4 2 | 3+4 2 | 5+6 2 | 3+3 2 | 3+3 2 |
| | 5+6 2 | 3+4 2 | 5+6 2 | 5+6 2 | 3+3 2 | 3+3 2 |
| | 3+4 2 | 5+6 2 | 3+4 2 | 5+6 2 | 3+3 2 | 2+2 2 |
| | 3+4 2 | 3+4 2 | 5+6 2 | 3+4 2 | 2+2 2 | 2+2 2 |
| | 5+6 2 | 5+6 2 | 3+4 2 | 5+6 2 | 2+2 2 | 3+3 2 |
| | 3+4 2 | 5+6 2 | 5+6 2 | 3+4 2 | 3+3 2 | 3+3 2 |
| | 3+4 2 | 3+4 2 | 5+6 2 | 3+4 2 | 2+2 2 | 2+2 2 |
| | 5+6 2 | 5+6 2 | 3+4 2 | 5+6 2 | 3+3 2 | 2+2 2 |
| | 3+4 2 | 3+4 2 | 3+4 2 | 3+4 2 | 3+3 2 | 3+3 2 |
| | 5+6 2 | 5+6 2 | 5+6 2 | 5+6 2 | 2+2 2 | 3+3 2 |
| | 3+4 2 | 3+4 2 | 3+4 2 | 3+4 2 | 2+2 2 | 2+2 2 |

Continued on next column

**TABLE XXIV– continued**

| TQC | ICSD | NoE | SF | | | |
|---|---|---|---|---|---|---|
| | 5+6 2 | 5+6 2 | 5+6 2 | 5+6 2 | 3+3 2 | 2+2 2 |
| | 5+6 2 | 3+4 2 | 3+4 2 | 3+4 2 | 2+2 2 | 3+3 2 |
| | 5+6 2 | 3+4 2 | 5+6 2 | 5+6 2 | 3+3 2 | 2+2 2 |
| | 3+4 2 | 5+6 2 | 3+4 2 | 3+4 2 | 2+2 2 | 3+3 2 |
| | 3+4 2 | 5+6 2 | 5+6 2 | 5+6 2 | 3+3 2 | 2+2 2 |
| | 5+6 2 | 3+4 2 | 3+4 2 | 3+4 2 | 3+3 2 | 3+3 2 |
| | 3+4 2 | 3+4 2 | 5+6 2 | 5+6 2 | 2+2 2 | 3+3 2 |
| | 3+4 2 | 5+6 2 | 3+4 2 | 5+6 2 | 2+2 2 | 2+2 2 |
| 12.2.78 | 425141 | | 166 | Dy3O3Sb1 | | |
| | -A | -GM | -M | -Y | -L | -V |
| | 5+6 2 | 3+4 2 | 3+4 2 | 5+6 2 | 2+2 2 | 3+3 2 |
| | 3+4 2 | 5+6 2 | 5+6 2 | 3+4 2 | 3+3 2 | 2+2 2 |
| | 3+4 2 | 3+4 2 | 5+6 2 | 5+6 2 | 2+2 2 | 2+2 2 |
| | 5+6 2 | 5+6 2 | 3+4 2 | 3+4 2 | 3+3 2 | 3+3 2 |
| | 3+4 2 | 5+6 2 | 3+4 2 | 3+4 2 | 2+2 2 | 2+2 2 |
| | 5+6 2 | 3+4 2 | 5+6 2 | 5+6 2 | 3+3 2 | 3+3 2 |
| | 5+6 2 | 3+4 2 | 3+4 2 | 5+6 2 | 2+2 2 | 2+2 2 |
| | 3+4 2 | 5+6 2 | 5+6 2 | 5+6 2 | 3+3 2 | 3+3 2 |
| | 5+6 2 | 3+4 2 | 5+6 2 | 3+4 2 | 2+2 2 | 3+3 2 |
| | 3+4 2 | 5+6 2 | 3+4 2 | 3+4 2 | 3+3 2 | 2+2 2 |
| | 3+4 2 | 3+4 2 | 3+4 2 | 5+6 2 | 2+2 2 | 2+2 2 |
| | 5+6 2 | 5+6 2 | 5+6 2 | 3+4 2 | 3+3 2 | 3+3 2 |
| | 5+6 2 | 3+4 2 | 3+4 2 | 5+6 2 | 3+3 2 | 2+2 2 |
| | 3+4 2 | 5+6 2 | 5+6 2 | 3+4 2 | 2+2 2 | 3+3 2 |
| | 5+6 2 | 3+4 2 | 5+6 2 | 5+6 2 | 2+2 2 | 3+3 2 |
| | 3+4 2 | 5+6 2 | 3+4 2 | 3+4 2 | 3+3 2 | 2+2 2 |
| | 3+4 2 | 3+4 2 | 3+4 2 | 5+6 2 | 3+3 2 | 3+3 2 |
| | 5+6 2 | 5+6 2 | 5+6 2 | 3+4 2 | 2+2 2 | 2+2 2 |
| | 5+6 2 | 3+4 2 | 3+4 2 | 5+6 2 | 3+3 2 | 2+2 2 |
| | 3+4 2 | 3+4 2 | 5+6 2 | 5+6 2 | 2+2 2 | 3+3 2 |
| | 3+4 2 | 5+6 2 | 5+6 2 | 3+4 2 | 2+2 2 | 3+3 2 |
| | 5+6 2 | 3+4 2 | 3+4 2 | 5+6 2 | 3+3 2 | 2+2 2 |
| | 3+4 2 | 5+6 2 | 3+4 2 | 3+4 2 | 2+2 2 | 3+3 2 |
| | 5+6 2 | 5+6 2 | 5+6 2 | 3+4 2 | 3+3 2 | 2+2 2 |
| | 3+4 2 | 3+4 2 | 5+6 2 | 3+4 2 | 2+2 2 | 3+3 2 |
| | 5+6 2 | 5+6 2 | 3+4 2 | 5+6 2 | 3+3 2 | 2+2 2 |
| | 3+4 2 | 5+6 2 | 5+6 2 | 3+4 2 | 3+3 2 | 2+2 2 |
| | 5+6 2 | 3+4 2 | 3+4 2 | 5+6 2 | 2+2 2 | 3+3 2 |
| | 5+6 2 | 5+6 2 | 3+4 2 | 3+4 2 | 2+2 2 | 3+3 2 |
| | 3+4 2 | 3+4 2 | 5+6 2 | 5+6 2 | 3+3 2 | 2+2 2 |
| | 5+6 2 | 3+4 2 | 5+6 2 | 3+4 2 | 3+3 2 | 2+2 2 |
| | 3+4 2 | 5+6 2 | 3+4 2 | 5+6 2 | 2+2 2 | 3+3 2 |
| | 3+4 2 | 5+6 2 | 3+4 2 | 3+4 2 | 2+2 2 | 2+2 2 |
| | 5+6 2 | 3+4 2 | 5+6 2 | 5+6 2 | 3+3 2 | 3+3 2 |
| | 5+6 2 | 5+6 2 | 5+6 2 | 3+4 2 | 3+3 2 | 3+3 2 |
| | 3+4 2 | 3+4 2 | 3+4 2 | 5+6 2 | 2+2 2 | 2+2 2 |
| | 5+6 2 | 3+4 2 | 3+4 2 | 5+6 2 | 2+2 2 | 3+3 2 |
| | 3+4 2 | 3+4 2 | 5+6 2 | 5+6 2 | 3+3 2 | 2+2 2 |
| | 5+6 2 | 5+6 2 | 3+4 2 | 3+4 2 | 3+3 2 | 2+2 2 |
| | 3+4 2 | 5+6 2 | 5+6 2 | 5+6 2 | 2+2 2 | 3+3 2 |
| | 5+6 2 | 3+4 2 | 3+4 2 | 3+4 2 | 2+2 2 | 2+2 2 |
| | 5+6 2 | 5+6 2 | 3+4 2 | 3+4 2 | 2+2 2 | 3+3 2 |
| | 3+4 2 | 3+4 2 | 3+4 2 | 5+6 2 | 3+3 2 | 3+3 2 |
| | 3+4 2 | 3+4 2 | 5+6 2 | 3+4 2 | 3+3 2 | 2+2 2 |
| | 3+4 2 | 3+4 2 | 3+4 2 | 5+6 2 | 3+3 2 | 3+3 2 |
| | 5+6 2 | 5+6 2 | 5+6 2 | 5+6 2 | 2+2 2 | 2+2 2 |
| | 3+4 2 | 3+4 2 | 5+6 2 | 3+4 2 | 3+3 2 | 2+2 2 |
| | 5+6 2 | 5+6 2 | 5+6 2 | 3+4 2 | 2+2 2 | 3+3 2 |
| | 5+6 2 | 5+6 2 | 3+4 2 | 5+6 2 | 2+2 2 | 3+3 2 |
| | 3+4 2 | 5+6 2 | 5+6 2 | 3+4 2 | 3+3 2 | 2+2 2 |
| | 5+6 2 | 3+4 2 | 5+6 2 | 3+4 2 | 2+2 2 | 3+3 2 |

Continued on next column

**TABLE XXIV– continued**

| TQC | ICSD | NoE | SF | | | |
|---|---|---|---|---|---|---|
| | 3+4 2 | 3+4 2 | 3+4 2 | 5+6 2 | 3+3 2 | 2+2 2 |
| | 5+6 2 | 5+6 2 | 5+6 2 | 3+4 2 | 3+3 2 | 2+2 2 |
| | 3+4 2 | 5+6 2 | 3+4 2 | 5+6 2 | 2+2 2 | 3+3 2 |
| | 5+6 2 | 3+4 2 | 3+4 2 | 5+6 2 | 3+3 2 | 2+2 2 |
| | 3+4 2 | 3+4 2 | 5+6 2 | 3+4 2 | 2+2 2 | 3+3 2 |
| | 5+6 2 | 5+6 2 | 3+4 2 | 5+6 2 | 3+3 2 | 2+2 2 |
| | 3+4 2 | 3+4 2 | 5+6 2 | 5+6 2 | 2+2 2 | 3+3 2 |
| | 5+6 2 | 5+6 2 | 3+4 2 | 3+4 2 | 3+3 2 | 2+2 2 |
| | 3+4 2 | 3+4 2 | 5+6 2 | 5+6 2 | 2+2 2 | 3+3 2 |
| | 3+4 2 | 5+6 2 | 5+6 2 | 3+4 2 | 3+3 2 | 2+2 2 |
| | 5+6 2 | 3+4 2 | 3+4 2 | 3+4 2 | 2+2 2 | 2+2 2 |
| | 3+4 2 | 5+6 2 | 3+4 2 | 5+6 2 | 3+3 2 | 3+3 2 |
| | 5+6 2 | 3+4 2 | 5+6 2 | 3+4 2 | 2+2 2 | 3+3 2 |
| | 5+6 2 | 3+4 2 | 3+4 2 | 5+6 2 | 3+3 2 | 2+2 2 |
| | 3+4 2 | 5+6 2 | 5+6 2 | 5+6 2 | 2+2 2 | 3+3 2 |
| | 3+4 2 | 5+6 2 | 5+6 2 | 3+4 2 | 3+3 2 | 2+2 2 |
| | 5+6 2 | 3+4 2 | 3+4 2 | 3+4 2 | 2+2 2 | 3+3 2 |
| | 5+6 2 | 5+6 2 | 5+6 2 | 5+6 2 | 3+3 2 | 2+2 2 |
| | 3+4 2 | 5+6 2 | 3+4 2 | 3+4 2 | 2+2 2 | 3+3 2 |
| | 5+6 2 | 3+4 2 | 5+6 2 | 3+4 2 | 3+3 2 | 2+2 2 |
| | 3+4 2 | 5+6 2 | 3+4 2 | 5+6 2 | 2+2 2 | 3+3 2 |
| | 5+6 2 | 3+4 2 | 3+4 2 | 5+6 2 | 3+3 2 | 2+2 2 |
| | 3+4 2 | 5+6 2 | 5+6 2 | 3+4 2 | 2+2 2 | 3+3 2 |
| | 5+6 2 | 3+4 2 | 3+4 2 | 5+6 2 | 3+3 2 | 2+2 2 |
| | 3+4 2 | 3+4 2 | 5+6 2 | 3+4 2 | 2+2 2 | 3+3 2 |
| | 3+4 2 | 5+6 2 | 3+4 2 | 5+6 2 | 3+3 2 | 2+2 2 |
| | 5+6 2 | 3+4 2 | 5+6 2 | 3+4 2 | 2+2 2 | 2+2 2 |
| | 5+6 2 | 5+6 2 | 3+4 2 | 5+6 2 | 3+3 2 | 3+3 2 |
| | 3+4 2 | 3+4 2 | 5+6 2 | 5+6 2 | 2+2 2 | 3+3 2 |
| | 3+4 2 | 5+6 2 | 3+4 2 | 3+4 2 | 3+3 2 | 2+2 2 |
| | 5+6 2 | 3+4 2 | 5+6 2 | 5+6 2 | 2+2 2 | 3+3 2 |
| | 3+4 2 | 5+6 2 | 3+4 2 | 3+4 2 | 2+2 2 | 3+3 2 |
| 12.2.79 | 422453 | | 40 | Cu2Sn2Sr1 | | |
| | -A | -GM | -M | -Y | -L | -V |
| | 3+4 2 | 3+4 2 | 3+4 2 | 3+4 2 | 3+3 2 | 3+3 2 |
| | 5+6 2 | 5+6 2 | 5+6 2 | 5+6 2 | 2+2 2 | 2+2 2 |
| | 5+6 2 | 5+6 2 | 5+6 2 | 5+6 2 | 2+2 2 | 2+2 2 |
| | 5+6 2 | 5+6 2 | 5+6 2 | 5+6 2 | 2+2 2 | 2+2 2 |
| | 5+6 2 | 3+4 2 | 3+4 2 | 5+6 2 | 3+3 2 | 2+2 2 |
| | 3+4 2 | 5+6 2 | 5+6 2 | 3+4 2 | 2+2 2 | 3+3 2 |
| | 3+4 2 | 3+4 2 | 5+6 2 | 3+4 2 | 3+3 2 | 3+3 2 |
| | 5+6 2 | 3+4 2 | 3+4 2 | 5+6 2 | 2+2 2 | 2+2 2 |
| | 5+6 2 | 5+6 2 | 3+4 2 | 3+4 2 | 3+3 2 | 2+2 2 |
| | 5+6 2 | 5+6 2 | 5+6 2 | 5+6 2 | 2+2 2 | 3+3 2 |
| | 3+4 2 | 3+4 2 | 3+4 2 | 3+4 2 | 3+3 2 | 2+2 2 |
| | 3+4 2 | 3+4 2 | 5+6 2 | 5+6 2 | 2+2 2 | 3+3 2 |
| | 5+6 2 | 5+6 2 | 3+4 2 | 5+6 2 | 3+3 2 | 2+2 2 |
| | 5+6 2 | 3+4 2 | 3+4 2 | 3+4 2 | 2+2 2 | 3+3 2 |
| | 3+4 2 | 5+6 2 | 5+6 2 | 5+6 2 | 3+3 2 | 3+3 2 |
| | 3+4 2 | 5+6 2 | 5+6 2 | 3+4 2 | 2+2 2 | 2+2 2 |
| | 5+6 2 | 3+4 2 | 5+6 2 | 5+6 2 | 3+3 2 | 2+2 2 |
| | 3+4 2 | 3+4 2 | 3+4 2 | 3+4 2 | 2+2 2 | 3+3 2 |
| | 3+4 2 | 3+4 2 | 3+4 2 | 5+6 2 | 3+3 2 | 3+3 2 |
| | 5+6 2 | 5+6 2 | 5+6 2 | 3+4 2 | 2+2 2 | 2+2 2 |
| 12.2.80 | 182102 | | 40 | Cu2Sn2Sr1 | | |
| | -A | -GM | -M | -Y | -L | -V |
| | 3+4 2 | 3+4 2 | 3+4 2 | 3+4 2 | 3+3 2 | 3+3 2 |
| | 5+6 2 | 5+6 2 | 5+6 2 | 5+6 2 | 2+2 2 | 2+2 2 |
| | 5+6 2 | 5+6 2 | 5+6 2 | 5+6 2 | 2+2 2 | 2+2 2 |
| | 5+6 2 | 5+6 2 | 5+6 2 | 5+6 2 | 2+2 2 | 2+2 2 |
| | 5+6 2 | 3+4 2 | 3+4 2 | 5+6 2 | 3+3 2 | 2+2 2 |

Continued on next column

**TABLE XXIV– continued**

| TQC | ICSD | NoE | SF | | | |
|---|---|---|---|---|---|---|
| | 3+4 2 | 5+6 2 | 5+6 2 | 3+4 2 | 2+2 2 | 3+3 2 |
| | 3+4 2 | 3+4 2 | 5+6 2 | 3+4 2 | 3+3 2 | 3+3 2 |
| | 5+6 2 | 3+4 2 | 3+4 2 | 5+6 2 | 2+2 2 | 2+2 2 |
| | 5+6 2 | 5+6 2 | 3+4 2 | 3+4 2 | 3+3 2 | 2+2 2 |
| | 5+6 2 | 5+6 2 | 5+6 2 | 5+6 2 | 2+2 2 | 3+3 2 |
| | 3+4 2 | 3+4 2 | 3+4 2 | 3+4 2 | 3+3 2 | 2+2 2 |
| | 3+4 2 | 3+4 2 | 5+6 2 | 5+6 2 | 2+2 2 | 3+3 2 |
| | 5+6 2 | 5+6 2 | 3+4 2 | 5+6 2 | 3+3 2 | 2+2 2 |
| | 5+6 2 | 3+4 2 | 3+4 2 | 3+4 2 | 2+2 2 | 3+3 2 |
| | 3+4 2 | 5+6 2 | 5+6 2 | 5+6 2 | 3+3 2 | 3+3 2 |
| | 3+4 2 | 5+6 2 | 5+6 2 | 3+4 2 | 2+2 2 | 2+2 2 |
| | 5+6 2 | 3+4 2 | 5+6 2 | 5+6 2 | 3+3 2 | 2+2 2 |
| | 3+4 2 | 3+4 2 | 3+4 2 | 3+4 2 | 2+2 2 | 3+3 2 |
| | 3+4 2 | 3+4 2 | 3+4 2 | 5+6 2 | 3+3 2 | 3+3 2 |
| | 5+6 2 | 5+6 2 | 5+6 2 | 3+4 2 | 2+2 2 | 2+2 2 |
| 12.2.81 | 79536 | 52 | Cu2Na2S4Zr1 | | | |
| | -A | -GM | -M | -Y | -L | -V |
| | 3+4 2 | 3+4 2 | 3+4 2 | 3+4 2 | 3+3 2 | 3+3 2 |
| | 5+6 2 | 3+4 2 | 5+6 2 | 5+6 2 | 3+3 2 | 2+2 2 |
| | 3+4 2 | 5+6 2 | 5+6 2 | 3+4 2 | 2+2 2 | 3+3 2 |
| | 5+6 2 | 5+6 2 | 3+4 2 | 5+6 2 | 2+2 2 | 2+2 2 |
| | 3+4 2 | 3+4 2 | 3+4 2 | 3+4 2 | 3+3 2 | 3+3 2 |
| | 3+4 2 | 3+4 2 | 5+6 2 | 5+6 2 | 2+2 2 | 3+3 2 |
| | 5+6 2 | 5+6 2 | 3+4 2 | 3+4 2 | 3+3 2 | 2+2 2 |
| | 5+6 2 | 3+4 2 | 3+4 2 | 3+4 2 | 2+2 2 | 2+2 2 |
| | 3+4 2 | 5+6 2 | 5+6 2 | 5+6 2 | 3+3 2 | 3+3 2 |
| | 5+6 2 | 3+4 2 | 3+4 2 | 3+4 2 | 2+2 2 | 2+2 2 |
| | 3+4 2 | 5+6 2 | 3+4 2 | 3+4 2 | 3+3 2 | 3+3 2 |
| | 5+6 2 | 5+6 2 | 3+4 2 | 5+6 2 | 3+3 2 | 2+2 2 |
| | 3+4 2 | 3+4 2 | 5+6 2 | 3+4 2 | 2+2 2 | 3+3 2 |
| | 3+4 2 | 3+4 2 | 5+6 2 | 5+6 2 | 2+2 2 | 2+2 2 |
| | 5+6 2 | 3+4 2 | 5+6 2 | 5+6 2 | 3+3 2 | 3+3 2 |
| | 3+4 2 | 5+6 2 | 5+6 2 | 5+6 2 | 2+2 2 | 2+2 2 |
| | 5+6 2 | 5+6 2 | 3+4 2 | 3+4 2 | 3+3 2 | 3+3 2 |
| | 3+4 2 | 3+4 2 | 5+6 2 | 5+6 2 | 2+2 2 | 2+2 2 |
| | 5+6 2 | 3+4 2 | 5+6 2 | 5+6 2 | 3+3 2 | 3+3 2 |
| | 3+4 2 | 5+6 2 | 3+4 2 | 3+4 2 | 3+3 2 | 2+2 2 |
| | 5+6 2 | 5+6 2 | 5+6 2 | 3+4 2 | 2+2 2 | 3+3 2 |
| | 5+6 2 | 3+4 2 | 3+4 2 | 5+6 2 | 3+3 2 | 3+3 2 |
| | 3+4 2 | 3+4 2 | 3+4 2 | 5+6 2 | 2+2 2 | 2+2 2 |
| | 3+4 2 | 5+6 2 | 5+6 2 | 3+4 2 | 3+3 2 | 2+2 2 |
| | 5+6 2 | 5+6 2 | 5+6 2 | 5+6 2 | 2+2 2 | 3+3 2 |
| | 5+6 2 | 3+4 2 | 3+4 2 | 3+4 2 | 2+2 2 | 2+2 2 |
| 12.2.82 | 426133 | 166 | Cu1Ge6Yb2 | | | |
| | -A | -GM | -M | -Y | -L | -V |
| | 5+6 2 | 3+4 2 | 3+4 2 | 3+4 2 | 2+2 2 | 2+2 2 |
| | 3+4 2 | 5+6 2 | 3+4 2 | 5+6 2 | 2+2 2 | 2+2 2 |
| | 3+4 2 | 3+4 2 | 5+6 2 | 5+6 2 | 3+3 2 | 3+3 2 |
| | 5+6 2 | 5+6 2 | 5+6 2 | 3+4 2 | 3+3 2 | 3+3 2 |
| | 3+4 2 | 3+4 2 | 3+4 2 | 5+6 2 | 2+2 2 | 2+2 2 |
| | 5+6 2 | 5+6 2 | 3+4 2 | 3+4 2 | 3+3 2 | 3+3 2 |
| | 5+6 2 | 3+4 2 | 5+6 2 | 5+6 2 | 2+2 2 | 2+2 2 |
| | 3+4 2 | 5+6 2 | 5+6 2 | 3+4 2 | 3+3 2 | 3+3 2 |
| | 3+4 2 | 3+4 2 | 3+4 2 | 5+6 2 | 2+2 2 | 2+2 2 |
| | 5+6 2 | 5+6 2 | 5+6 2 | 3+4 2 | 3+3 2 | 3+3 2 |
| | 5+6 2 | 3+4 2 | 3+4 2 | 3+4 2 | 2+2 2 | 2+2 2 |
| | 3+4 2 | 3+4 2 | 5+6 2 | 5+6 2 | 3+3 2 | 3+3 2 |
| | 5+6 2 | 5+6 2 | 3+4 2 | 3+4 2 | 2+2 2 | 2+2 2 |
| | 3+4 2 | 5+6 2 | 5+6 2 | 5+6 2 | 3+3 2 | 3+3 2 |
| | 5+6 2 | 3+4 2 | 3+4 2 | 5+6 2 | 2+2 2 | 2+2 2 |
| | 3+4 2 | 5+6 2 | 5+6 2 | 3+4 2 | 3+3 2 | 3+3 2 |

Continued on next column

**TABLE XXIV– continued**

| TQC | ICSD | NoE | SF | | | |
|---|---|---|---|---|---|---|
| | 3+4 2 | 3+4 2 | 3+4 2 | 3+4 2 | 3+3 2 | 3+3 2 |
| | 5+6 2 | 3+4 2 | 5+6 2 | 5+6 2 | 2+2 2 | 2+2 2 |
| | 5+6 2 | 5+6 2 | 5+6 2 | 3+4 2 | 3+3 2 | 2+2 2 |
| | 3+4 2 | 3+4 2 | 5+6 2 | 5+6 2 | 2+2 2 | 3+3 2 |
| | 3+4 2 | 3+4 2 | 3+4 2 | 3+4 2 | 3+3 2 | 3+3 2 |
| | 3+4 2 | 5+6 2 | 3+4 2 | 5+6 2 | 2+2 2 | 2+2 2 |
| | 5+6 2 | 3+4 2 | 5+6 2 | 3+4 2 | 3+3 2 | 3+3 2 |
| | 3+4 2 | 5+6 2 | 5+6 2 | 5+6 2 | 2+2 2 | 2+2 2 |
| | 5+6 2 | 5+6 2 | 3+4 2 | 3+4 2 | 2+2 2 | 3+3 2 |
| | 5+6 2 | 5+6 2 | 3+4 2 | 3+4 2 | 3+3 2 | 2+2 2 |
| | 3+4 2 | 3+4 2 | 5+6 2 | 5+6 2 | 3+3 2 | 3+3 2 |
| | 5+6 2 | 5+6 2 | 3+4 2 | 5+6 2 | 2+2 2 | 2+2 2 |
| | 5+6 2 | 5+6 2 | 3+4 2 | 5+6 2 | 3+3 2 | 2+2 2 |
| | 5+6 2 | 3+4 2 | 3+4 2 | 3+4 2 | 2+2 2 | 3+3 2 |
| | 3+4 2 | 5+6 2 | 5+6 2 | 5+6 2 | 3+3 2 | 2+2 2 |
| | 3+4 2 | 3+4 2 | 5+6 2 | 3+4 2 | 2+2 2 | 3+3 2 |
| | 3+4 2 | 3+4 2 | 3+4 2 | 5+6 2 | 3+3 2 | 3+3 2 |
| | 5+6 2 | 5+6 2 | 5+6 2 | 3+4 2 | 2+2 2 | 2+2 2 |
| | 5+6 2 | 5+6 2 | 5+6 2 | 5+6 2 | 3+3 2 | 3+3 2 |
| | 3+4 2 | 3+4 2 | 3+4 2 | 3+4 2 | 2+2 2 | 2+2 2 |
| | 5+6 2 | 3+4 2 | 5+6 2 | 3+4 2 | 2+2 2 | 3+3 2 |
| | 3+4 2 | 3+4 2 | 3+4 2 | 5+6 2 | 3+3 2 | 2+2 2 |
| | 3+4 2 | 5+6 2 | 5+6 2 | 5+6 2 | 2+2 2 | 3+3 2 |
| | 5+6 2 | 3+4 2 | 3+4 2 | 3+4 2 | 3+3 2 | 2+2 2 |
| | 3+4 2 | 5+6 2 | 3+4 2 | 5+6 2 | 3+3 2 | 2+2 2 |
| | 5+6 2 | 5+6 2 | 5+6 2 | 3+4 2 | 2+2 2 | 3+3 2 |
| | 3+4 2 | 3+4 2 | 5+6 2 | 3+4 2 | 2+2 2 | 2+2 2 |
| | 5+6 2 | 5+6 2 | 3+4 2 | 3+4 2 | 3+3 2 | 3+3 2 |
| | 5+6 2 | 5+6 2 | 5+6 2 | 5+6 2 | 3+3 2 | 2+2 2 |
| | 3+4 2 | 3+4 2 | 3+4 2 | 3+4 2 | 2+2 2 | 3+3 2 |
| | 5+6 2 | 3+4 2 | 5+6 2 | 5+6 2 | 3+3 2 | 3+3 2 |
| | 3+4 2 | 3+4 2 | 5+6 2 | 5+6 2 | 2+2 2 | 2+2 2 |
| | 5+6 2 | 5+6 2 | 3+4 2 | 3+4 2 | 2+2 2 | 2+2 2 |
| | 5+6 2 | 3+4 2 | 5+6 2 | 3+4 2 | 3+3 2 | 3+3 2 |
| | 3+4 2 | 5+6 2 | 3+4 2 | 5+6 2 | 3+3 2 | 2+2 2 |
| | 5+6 2 | 3+4 2 | 3+4 2 | 3+4 2 | 2+2 2 | 3+3 2 |
| | 3+4 2 | 5+6 2 | 5+6 2 | 5+6 2 | 2+2 2 | 2+2 2 |
| | 3+4 2 | 3+4 2 | 3+4 2 | 3+4 2 | 3+3 2 | 3+3 2 |
| | 3+4 2 | 3+4 2 | 5+6 2 | 5+6 2 | 2+2 2 | 2+2 2 |
| | 3+4 2 | 5+6 2 | 5+6 2 | 3+4 2 | 3+3 2 | 3+3 2 |
| | 5+6 2 | 5+6 2 | 3+4 2 | 5+6 2 | 2+2 2 | 2+2 2 |
| | 3+4 2 | 5+6 2 | 3+4 2 | 5+6 2 | 3+3 2 | 3+3 2 |
| | 5+6 2 | 3+4 2 | 3+4 2 | 3+4 2 | 2+2 2 | 2+2 2 |
| | 5+6 2 | 5+6 2 | 5+6 2 | 3+4 2 | 3+3 2 | 3+3 2 |
| | 3+4 2 | 5+6 2 | 5+6 2 | 5+6 2 | 3+3 2 | 3+3 2 |
| | 5+6 2 | 3+4 2 | 3+4 2 | 3+4 2 | 2+2 2 | 2+2 2 |
| | 5+6 2 | 3+4 2 | 5+6 2 | 5+6 2 | 2+2 2 | 2+2 2 |
| | 3+4 2 | 3+4 2 | 5+6 2 | 5+6 2 | 3+3 2 | 3+3 2 |
| | 5+6 2 | 3+4 2 | 5+6 2 | 3+4 2 | 3+3 2 | 2+2 2 |
| | 3+4 2 | 5+6 2 | 3+4 2 | 3+4 2 | 2+2 2 | 3+3 2 |
| | 3+4 2 | 3+4 2 | 3+4 2 | 5+6 2 | 2+2 2 | 3+3 2 |
| | 3+4 2 | 3+4 2 | 5+6 2 | 3+4 2 | 3+3 2 | 2+2 2 |
| | 3+4 2 | 5+6 2 | 3+4 2 | 5+6 2 | 3+3 2 | 2+2 2 |
| | 3+4 2 | 3+4 2 | 5+6 2 | 5+6 2 | 2+2 2 | 3+3 2 |
| | 5+6 2 | 5+6 2 | 3+4 2 | 3+4 2 | 2+2 2 | 2+2 2 |
| | 5+6 2 | 5+6 2 | 5+6 2 | 3+4 2 | 3+3 2 | 3+3 2 |
| | 3+4 2 | 3+4 2 | 3+4 2 | 5+6 2 | 2+2 2 | 2+2 2 |
| | 3+4 2 | 5+6 2 | 5+6 2 | 3+4 2 | 3+3 2 | 3+3 2 |
| | 5+6 2 | 5+6 2 | 3+4 2 | 5+6 2 | 2+2 2 | 2+2 2 |
| | 5+6 2 | 3+4 2 | 5+6 2 | 3+4 2 | 3+3 2 | 3+3 2 |
| | 3+4 2 | 3+4 2 | 3+4 2 | 5+6 2 | 2+2 2 | 2+2 2 |

Continued on next column

**TABLE XXIV– continued**

| TQC | ICSD | | NoE | | SF | |
|---|---|---|---|---|---|---|
| | 5+6 2 | 5+6 2 | 3+4 2 | 5+6 2 | 3+3 2 | 3+3 2 |
| | 5+6 2 | 3+4 2 | 5+6 2 | 3+4 2 | 2+2 2 | 3+3 2 |
| | 3+4 2 | 3+4 2 | 5+6 2 | 5+6 2 | 3+3 2 | 2+2 2 |
| | 5+6 2 | 5+6 2 | 3+4 2 | 3+4 2 | 3+3 2 | 2+2 2 |
| | 3+4 2 | 3+4 2 | 5+6 2 | 5+6 2 | 2+2 2 | 3+3 2 |
| | 3+4 2 | 5+6 2 | 3+4 2 | 3+4 2 | 2+2 2 | 3+3 2 |
| 12.2.83 | 380416 | | 68 | | Cs4Ge2Se6 | |
| | -A | -GM | -M | -Y | -L | -V |
| | 5+6 2 | 3+4 2 | 5+6 2 | 3+4 2 | 2+2 2 | 3+3 2 |
| | 3+4 2 | 5+6 2 | 3+4 2 | 5+6 2 | 3+3 2 | 2+2 2 |
| | 5+6 2 | 3+4 2 | 3+4 2 | 3+4 2 | 3+3 2 | 2+2 2 |
| | 3+4 2 | 5+6 2 | 5+6 2 | 5+6 2 | 2+2 2 | 3+3 2 |
| | 5+6 2 | 3+4 2 | 5+6 2 | 3+4 2 | 2+2 2 | 3+3 2 |
| | 3+4 2 | 5+6 2 | 3+4 2 | 5+6 2 | 3+3 2 | 2+2 2 |
| | 5+6 2 | 5+6 2 | 3+4 2 | 3+4 2 | 2+2 2 | 3+3 2 |
| | 3+4 2 | 3+4 2 | 5+6 2 | 5+6 2 | 3+3 2 | 2+2 2 |
| | 3+4 2 | 5+6 2 | 5+6 2 | 3+4 2 | 2+2 2 | 2+2 2 |
| | 5+6 2 | 3+4 2 | 3+4 2 | 5+6 2 | 3+3 2 | 3+3 2 |
| | 3+4 2 | 5+6 2 | 3+4 2 | 5+6 2 | 3+3 2 | 2+2 2 |
| | 5+6 2 | 3+4 2 | 5+6 2 | 3+4 2 | 2+2 2 | 3+3 2 |
| | 5+6 2 | 3+4 2 | 5+6 2 | 3+4 2 | 2+2 2 | 3+3 2 |
| | 5+6 2 | 3+4 2 | 5+6 2 | 3+4 2 | 2+2 2 | 3+3 2 |
| | 3+4 2 | 5+6 2 | 3+4 2 | 5+6 2 | 3+3 2 | 2+2 2 |
| | 5+6 2 | 3+4 2 | 5+6 2 | 3+4 2 | 2+2 2 | 3+3 2 |
| | 3+4 2 | 5+6 2 | 3+4 2 | 5+6 2 | 3+3 2 | 2+2 2 |
| | 3+4 2 | 3+4 2 | 3+4 2 | 5+6 2 | 3+3 2 | 2+2 2 |
| | 5+6 2 | 3+4 2 | 5+6 2 | 5+6 2 | 2+2 2 | 3+3 2 |
| | 3+4 2 | 5+6 2 | 3+4 2 | 3+4 2 | 3+3 2 | 2+2 2 |
| | 3+4 2 | 5+6 2 | 5+6 2 | 5+6 2 | 2+2 2 | 3+3 2 |
| | 5+6 2 | 5+6 2 | 3+4 2 | 3+4 2 | 3+3 2 | 2+2 2 |
| | 5+6 2 | 3+4 2 | 5+6 2 | 5+6 2 | 3+3 2 | 3+3 2 |
| | 3+4 2 | 3+4 2 | 3+4 2 | 3+4 2 | 2+2 2 | 2+2 2 |
| | 3+4 2 | 5+6 2 | 5+6 2 | 3+4 2 | 3+3 2 | 2+2 2 |
| | 5+6 2 | 5+6 2 | 3+4 2 | 5+6 2 | 2+2 2 | 3+3 2 |
| | 3+4 2 | 5+6 2 | 3+4 2 | 5+6 2 | 2+2 2 | 2+2 2 |
| | 5+6 2 | 3+4 2 | 5+6 2 | 3+4 2 | 3+3 2 | 3+3 2 |
| | 3+4 2 | 5+6 2 | 3+4 2 | 5+6 2 | 3+3 2 | 2+2 2 |
| | 5+6 2 | 3+4 2 | 5+6 2 | 3+4 2 | 2+2 2 | 3+3 2 |
| | 5+6 2 | 5+6 2 | 5+6 2 | 3+4 2 | 2+2 2 | 2+2 2 |
| | 3+4 2 | 3+4 2 | 3+4 2 | 5+6 2 | 3+3 2 | 3+3 2 |
| | 3+4 2 | 5+6 2 | 5+6 2 | 3+4 2 | 2+2 2 | 2+2 2 |
| | 5+6 2 | 3+4 2 | 3+4 2 | 3+4 2 | 3+3 2 | 3+3 2 |
| 12.2.84 | 626873 | | 42 | | Cr3Te4 | |
| | -A | -GM | -M | -Y | -L | -V |
| | 5+6 2 | 3+4 2 | 3+4 2 | 5+6 2 | 3+3 2 | 2+2 2 |
| | 3+4 2 | 5+6 2 | 5+6 2 | 3+4 2 | 2+2 2 | 3+3 2 |
| | 3+4 2 | 3+4 2 | 5+6 2 | 5+6 2 | 2+2 2 | 2+2 2 |
| | 5+6 2 | 5+6 2 | 3+4 2 | 3+4 2 | 3+3 2 | 3+3 2 |
| | 5+6 2 | 3+4 2 | 5+6 2 | 3+4 2 | 2+2 2 | 3+3 2 |
| | 3+4 2 | 5+6 2 | 3+4 2 | 5+6 2 | 3+3 2 | 2+2 2 |
| | 5+6 2 | 3+4 2 | 5+6 2 | 3+4 2 | 2+2 2 | 3+3 2 |
| | 5+6 2 | 3+4 2 | 3+4 2 | 5+6 2 | 3+3 2 | 2+2 2 |
| | 3+4 2 | 3+4 2 | 3+4 2 | 5+6 2 | 3+3 2 | 2+2 2 |
| | 3+4 2 | 5+6 2 | 5+6 2 | 3+4 2 | 2+2 2 | 3+3 2 |
| | 5+6 2 | 5+6 2 | 5+6 2 | 3+4 2 | 3+3 2 | 3+3 2 |
| | 5+6 2 | 3+4 2 | 3+4 2 | 5+6 2 | 2+2 2 | 2+2 2 |
| | 5+6 2 | 3+4 2 | 5+6 2 | 5+6 2 | 2+2 2 | 2+2 2 |
| | 3+4 2 | 3+4 2 | 3+4 2 | 3+4 2 | 3+3 2 | 3+3 2 |
| | 5+6 2 | 3+4 2 | 5+6 2 | 5+6 2 | 2+2 2 | 2+2 2 |
| | 5+6 2 | 5+6 2 | 3+4 2 | 3+4 2 | 3+3 2 | 3+3 2 |
| | 3+4 2 | 3+4 2 | 5+6 2 | 3+4 2 | 2+2 2 | 3+3 2 |

Continued on next column

**TABLE XXIV– continued**

| TQC | ICSD | | NoE | | SF | |
|---|---|---|---|---|---|---|
| | 3+4 2 | 5+6 2 | 3+4 2 | 5+6 2 | 3+3 2 | 2+2 2 |
| | 3+4 2 | 5+6 2 | 5+6 2 | 3+4 2 | 2+2 2 | 3+3 2 |
| | 5+6 2 | 3+4 2 | 5+6 2 | 3+4 2 | 2+2 2 | 3+3 2 |
| | 5+6 2 | 3+4 2 | 5+6 2 | 3+4 2 | 2+2 2 | 3+3 2 |
| 12.2.85 | 601348 | | 42 | | Cr3Se4 | |
| | -A | -GM | -M | -Y | -L | -V |
| | 3+4 2 | 3+4 2 | 5+6 2 | 5+6 2 | 2+2 2 | 3+3 2 |
| | 5+6 2 | 3+4 2 | 3+4 2 | 3+4 2 | 2+2 2 | 2+2 2 |
| | 3+4 2 | 5+6 2 | 5+6 2 | 5+6 2 | 3+3 2 | 3+3 2 |
| | 5+6 2 | 5+6 2 | 3+4 2 | 3+4 2 | 3+3 2 | 2+2 2 |
| | 3+4 2 | 3+4 2 | 3+4 2 | 3+4 2 | 3+3 2 | 3+3 2 |
| | 5+6 2 | 5+6 2 | 5+6 2 | 5+6 2 | 2+2 2 | 2+2 2 |
| | 3+4 2 | 3+4 2 | 5+6 2 | 3+4 2 | 3+3 2 | 2+2 2 |
| | 5+6 2 | 3+4 2 | 5+6 2 | 5+6 2 | 2+2 2 | 3+3 2 |
| | 3+4 2 | 3+4 2 | 3+4 2 | 5+6 2 | 2+2 2 | 3+3 2 |
| | 3+4 2 | 5+6 2 | 3+4 2 | 3+4 2 | 3+3 2 | 3+3 2 |
| | 5+6 2 | 5+6 2 | 3+4 2 | 3+4 2 | 3+3 2 | 2+2 2 |
| | 3+4 2 | 3+4 2 | 5+6 2 | 5+6 2 | 2+2 2 | 2+2 2 |
| | 3+4 2 | 3+4 2 | 3+4 2 | 5+6 2 | 2+2 2 | 3+3 2 |
| | 5+6 2 | 5+6 2 | 3+4 2 | 5+6 2 | 2+2 2 | 3+3 2 |
| | 5+6 2 | 3+4 2 | 5+6 2 | 3+4 2 | 3+3 2 | 2+2 2 |
| | 5+6 2 | 5+6 2 | 5+6 2 | 3+4 2 | 3+3 2 | 2+2 2 |
| | 3+4 2 | 5+6 2 | 3+4 2 | 3+4 2 | 3+3 2 | 3+3 2 |
| | 3+4 2 | 3+4 2 | 5+6 2 | 5+6 2 | 2+2 2 | 2+2 2 |
| | 5+6 2 | 3+4 2 | 3+4 2 | 3+4 2 | 3+3 2 | 3+3 2 |
| | 5+6 2 | 5+6 2 | 3+4 2 | 3+4 2 | 3+3 2 | 3+3 2 |
| | 5+6 2 | 3+4 2 | 3+4 2 | 5+6 2 | 2+2 2 | 3+3 2 |
| 12.2.86 | 73143 | | 114 | | Cr3Rb1S5 | |
| | -A | -GM | -M | -Y | -L | -V |
| | 3+4 2 | 3+4 2 | 5+6 2 | 5+6 2 | 2+2 2 | 2+2 2 |
| | 5+6 2 | 5+6 2 | 3+4 2 | 3+4 2 | 3+3 2 | 3+3 2 |
| | 5+6 2 | 3+4 2 | 3+4 2 | 3+4 2 | 3+3 2 | 3+3 2 |
| | 3+4 2 | 5+6 2 | 5+6 2 | 5+6 2 | 3+3 2 | 2+2 2 |
| | 5+6 2 | 3+4 2 | 3+4 2 | 3+4 2 | 2+2 2 | 3+3 2 |
| | 3+4 2 | 3+4 2 | 5+6 2 | 5+6 2 | 2+2 2 | 2+2 2 |
| | 5+6 2 | 5+6 2 | 3+4 2 | 3+4 2 | 3+3 2 | 3+3 2 |
| | 3+4 2 | 5+6 2 | 5+6 2 | 5+6 2 | 2+2 2 | 2+2 2 |
| | 5+6 2 | 3+4 2 | 5+6 2 | 5+6 2 | 2+2 2 | 2+2 2 |
| | 3+4 2 | 5+6 2 | 3+4 2 | 3+4 2 | 2+2 2 | 3+3 2 |
| | 5+6 2 | 3+4 2 | 5+6 2 | 5+6 2 | 3+3 2 | 3+3 2 |
| | 3+4 2 | 5+6 2 | 3+4 2 | 3+4 2 | 3+3 2 | 2+2 2 |
| | 5+6 2 | 3+4 2 | 5+6 2 | 5+6 2 | 2+2 2 | 3+3 2 |
| | 3+4 2 | 5+6 2 | 3+4 2 | 3+4 2 | 3+3 2 | 2+2 2 |
| | 3+4 2 | 3+4 2 | 5+6 2 | 5+6 2 | 2+2 2 | 3+3 2 |
| | 5+6 2 | 5+6 2 | 3+4 2 | 3+4 2 | 3+3 2 | 2+2 2 |
| | 3+4 2 | 3+4 2 | 5+6 2 | 5+6 2 | 2+2 2 | 2+2 2 |
| | 5+6 2 | 5+6 2 | 3+4 2 | 3+4 2 | 3+3 2 | 3+3 2 |
| | 3+4 2 | 3+4 2 | 3+4 2 | 3+4 2 | 3+3 2 | 2+2 2 |
| | 3+4 2 | 5+6 2 | 5+6 2 | 3+4 2 | 2+2 2 | 2+2 2 |
| | 5+6 2 | 3+4 2 | 5+6 2 | 5+6 2 | 2+2 2 | 3+3 2 |
| | 5+6 2 | 3+4 2 | 3+4 2 | 3+4 2 | 3+3 2 | 2+2 2 |
| | 5+6 2 | 5+6 2 | 3+4 2 | 5+6 2 | 3+3 2 | 3+3 2 |
| | 3+4 2 | 3+4 2 | 5+6 2 | 5+6 2 | 2+2 2 | 2+2 2 |
| | 5+6 2 | 5+6 2 | 3+4 2 | 3+4 2 | 3+3 2 | 3+3 2 |
| | 5+6 2 | 3+4 2 | 5+6 2 | 5+6 2 | 2+2 2 | 3+3 2 |
| | 3+4 2 | 3+4 2 | 3+4 2 | 3+4 2 | 3+3 2 | 2+2 2 |
| | 3+4 2 | 5+6 2 | 5+6 2 | 5+6 2 | 2+2 2 | 3+3 2 |
| | 5+6 2 | 3+4 2 | 3+4 2 | 3+4 2 | 2+2 2 | 3+3 2 |
| | 5+6 2 | 5+6 2 | 5+6 2 | 5+6 2 | 3+3 2 | 2+2 2 |
| | 3+4 2 | 5+6 2 | 3+4 2 | 3+4 2 | 3+3 2 | 2+2 2 |
| | 3+4 2 | 3+4 2 | 3+4 2 | 3+4 2 | 2+2 2 | 3+3 2 |

Continued on next column

**TABLE XXIV– continued**

| TQC | ICSD | NoE | SF | | | |
|---|---|---|---|---|---|---|
| | 3+4 2 | 3+4 2 | 5+6 2 | 3+4 2 | 2+2 2 | 2+2 2 |
| | 5+6 2 | 5+6 2 | 5+6 2 | 3+4 2 | 3+3 2 | 2+2 2 |
| | 5+6 2 | 3+4 2 | 5+6 2 | 5+6 2 | 2+2 2 | 3+3 2 |
| | 5+6 2 | 3+4 2 | 3+4 2 | 3+4 2 | 3+3 2 | 2+2 2 |
| | 3+4 2 | 3+4 2 | 3+4 2 | 5+6 2 | 3+3 2 | 3+3 2 |
| | 5+6 2 | 3+4 2 | 5+6 2 | 5+6 2 | 2+2 2 | 2+2 2 |
| | 3+4 2 | 5+6 2 | 3+4 2 | 3+4 2 | 3+3 2 | 3+3 2 |
| | 5+6 2 | 5+6 2 | 5+6 2 | 5+6 2 | 2+2 2 | 2+2 2 |
| | 3+4 2 | 5+6 2 | 3+4 2 | 5+6 2 | 2+2 2 | 3+3 2 |
| | 5+6 2 | 3+4 2 | 5+6 2 | 3+4 2 | 3+3 2 | 2+2 2 |
| | 3+4 2 | 3+4 2 | 5+6 2 | 5+6 2 | 2+2 2 | 2+2 2 |
| | 3+4 2 | 5+6 2 | 3+4 2 | 5+6 2 | 3+3 2 | 2+2 2 |
| | 5+6 2 | 5+6 2 | 3+4 2 | 3+4 2 | 2+2 2 | 3+3 2 |
| | 5+6 2 | 5+6 2 | 3+4 2 | 3+4 2 | 3+3 2 | 3+3 2 |
| | 3+4 2 | 3+4 2 | 5+6 2 | 5+6 2 | 2+2 2 | 3+3 2 |
| | 3+4 2 | 5+6 2 | 5+6 2 | 5+6 2 | 3+3 2 | 3+3 2 |
| | 5+6 2 | 5+6 2 | 3+4 2 | 3+4 2 | 2+2 2 | 2+2 2 |
| | 3+4 2 | 3+4 2 | 5+6 2 | 5+6 2 | 3+3 2 | 2+2 2 |
| | 3+4 2 | 5+6 2 | 3+4 2 | 3+4 2 | 2+2 2 | 3+3 2 |
| | 5+6 2 | 3+4 2 | 3+4 2 | 3+4 2 | 3+3 2 | 2+2 2 |
| | 5+6 2 | 3+4 2 | 5+6 2 | 3+4 2 | 2+2 2 | 2+2 2 |
| | 3+4 2 | 3+4 2 | 5+6 2 | 3+4 2 | 3+3 2 | 3+3 2 |
| | 3+4 2 | 5+6 2 | 3+4 2 | 3+4 2 | 3+3 2 | 2+2 2 |
| | 5+6 2 | 3+4 2 | 3+4 2 | 5+6 2 | 2+2 2 | 2+2 2 |
| | 3+4 2 | 3+4 2 | 5+6 2 | 5+6 2 | 3+3 2 | 2+2 2 |
| 12.2.87 | 1824 | 90 | Cr3Ga4 | | | |
| | -A | -GM | -M | -Y | -L | -V |
| | 5+6 2 | 3+4 2 | 3+4 2 | 5+6 2 | 3+3 2 | 3+3 2 |
| | 3+4 2 | 3+4 2 | 5+6 2 | 3+4 2 | 2+2 2 | 2+2 2 |
| | 5+6 2 | 5+6 2 | 3+4 2 | 5+6 2 | 2+2 2 | 2+2 2 |
| | 5+6 2 | 3+4 2 | 5+6 2 | 3+4 2 | 3+3 2 | 3+3 2 |
| | 3+4 2 | 5+6 2 | 3+4 2 | 3+4 2 | 2+2 2 | 3+3 2 |
| | 3+4 2 | 3+4 2 | 5+6 2 | 5+6 2 | 3+3 2 | 2+2 2 |
| | 3+4 2 | 5+6 2 | 5+6 2 | 5+6 2 | 2+2 2 | 3+3 2 |
| | 5+6 2 | 5+6 2 | 3+4 2 | 3+4 2 | 3+3 2 | 2+2 2 |
| | 5+6 2 | 3+4 2 | 3+4 2 | 3+4 2 | 3+3 2 | 3+3 2 |
| | 3+4 2 | 5+6 2 | 5+6 2 | 5+6 2 | 2+2 2 | 2+2 2 |
| | 5+6 2 | 5+6 2 | 3+4 2 | 5+6 2 | 3+3 2 | 3+3 2 |
| | 3+4 2 | 3+4 2 | 5+6 2 | 3+4 2 | 2+2 2 | 2+2 2 |
| | 3+4 2 | 3+4 2 | 3+4 2 | 3+4 2 | 3+3 2 | 3+3 2 |
| | 5+6 2 | 5+6 2 | 3+4 2 | 3+4 2 | 2+2 2 | 2+2 2 |
| | 3+4 2 | 3+4 2 | 5+6 2 | 5+6 2 | 2+2 2 | 3+3 2 |
| | 3+4 2 | 3+4 2 | 3+4 2 | 3+4 2 | 3+3 2 | 3+3 2 |
| | 3+4 2 | 5+6 2 | 5+6 2 | 5+6 2 | 2+2 2 | 2+2 2 |
| | 5+6 2 | 5+6 2 | 3+4 2 | 5+6 2 | 3+3 2 | 2+2 2 |
| | 5+6 2 | 3+4 2 | 5+6 2 | 3+4 2 | 3+3 2 | 3+3 2 |
| | 5+6 2 | 5+6 2 | 5+6 2 | 3+4 2 | 3+3 2 | 3+3 2 |
| | 3+4 2 | 3+4 2 | 3+4 2 | 5+6 2 | 2+2 2 | 2+2 2 |
| | 5+6 2 | 5+6 2 | 3+4 2 | 5+6 2 | 3+3 2 | 3+3 2 |
| | 5+6 2 | 5+6 2 | 3+4 2 | 3+4 2 | 2+2 2 | 2+2 2 |
| | 5+6 2 | 3+4 2 | 5+6 2 | 3+4 2 | 2+2 2 | 3+3 2 |
| | 5+6 2 | 5+6 2 | 3+4 2 | 5+6 2 | 3+3 2 | 2+2 2 |
| | 3+4 2 | 3+4 2 | 3+4 2 | 5+6 2 | 2+2 2 | 3+3 2 |
| | 3+4 2 | 3+4 2 | 5+6 2 | 3+4 2 | 3+3 2 | 3+3 2 |
| | 3+4 2 | 3+4 2 | 5+6 2 | 5+6 2 | 3+3 2 | 2+2 2 |
| | 5+6 2 | 3+4 2 | 3+4 2 | 3+4 2 | 3+3 2 | 2+2 2 |
| | 5+6 2 | 3+4 2 | 3+4 2 | 3+4 2 | 2+2 2 | 3+3 2 |
| | 3+4 2 | 5+6 2 | 5+6 2 | 5+6 2 | 2+2 2 | 2+2 2 |
| | 3+4 2 | 3+4 2 | 3+4 2 | 3+4 2 | 2+2 2 | 3+3 2 |
| | 3+4 2 | 5+6 2 | 5+6 2 | 5+6 2 | 3+3 2 | 3+3 2 |
| | 3+4 2 | 3+4 2 | 5+6 2 | 5+6 2 | 3+3 2 | 2+2 2 |

Continued on next column

| TQC | ICSD | NoE | SF | | | |
|---|---|---|---|---|---|---|
| | 5+6 2 | 5+6 2 | 3+4 2 | 3+4 2 | 2+2 2 | 3+3 2 |
| | 5+6 2 | 3+4 2 | 3+4 2 | 3+4 2 | 3+3 2 | 2+2 2 |
| | 3+4 2 | 5+6 2 | 5+6 2 | 5+6 2 | 3+3 2 | 3+3 2 |
| | 3+4 2 | 3+4 2 | 3+4 2 | 3+4 2 | 2+2 2 | 2+2 2 |
| | 5+6 2 | 5+6 2 | 5+6 2 | 5+6 2 | 2+2 2 | 3+3 2 |
| | 3+4 2 | 5+6 2 | 3+4 2 | 3+4 2 | 2+2 2 | 2+2 2 |
| | 5+6 2 | 3+4 2 | 5+6 2 | 5+6 2 | 3+3 2 | 3+3 2 |
| | 3+4 2 | 3+4 2 | 5+6 2 | 3+4 2 | 3+3 2 | 2+2 2 |
| | 5+6 2 | 5+6 2 | 5+6 2 | 3+4 2 | 2+2 2 | 2+2 2 |
| | 3+4 2 | 5+6 2 | 5+6 2 | 5+6 2 | 3+3 2 | 3+3 2 |
| | 5+6 2 | 3+4 2 | 3+4 2 | 5+6 2 | 2+2 2 | 3+3 2 |
| 12.2.88 | 626896 | 40 | Cr2Te4Ti1 | | | |
| | -A | -GM | -M | -Y | -L | -V |
| | 3+4 2 | 3+4 2 | 3+4 2 | 5+6 2 | 2+2 2 | 2+2 2 |
| | 5+6 2 | 5+6 2 | 5+6 2 | 3+4 2 | 3+3 2 | 3+3 2 |
| | 5+6 2 | 3+4 2 | 5+6 2 | 5+6 2 | 2+2 2 | 3+3 2 |
| | 3+4 2 | 5+6 2 | 3+4 2 | 3+4 2 | 3+3 2 | 2+2 2 |
| | 3+4 2 | 3+4 2 | 3+4 2 | 3+4 2 | 3+3 2 | 3+3 2 |
| | 5+6 2 | 5+6 2 | 5+6 2 | 5+6 2 | 2+2 2 | 2+2 2 |
| | 3+4 2 | 3+4 2 | 5+6 2 | 3+4 2 | 3+3 2 | 3+3 2 |
| | 5+6 2 | 3+4 2 | 3+4 2 | 5+6 2 | 2+2 2 | 2+2 2 |
| | 3+4 2 | 3+4 2 | 5+6 2 | 5+6 2 | 2+2 2 | 3+3 2 |
| | 5+6 2 | 5+6 2 | 3+4 2 | 3+4 2 | 3+3 2 | 2+2 2 |
| | 3+4 2 | 5+6 2 | 5+6 2 | 3+4 2 | 3+3 2 | 3+3 2 |
| | 5+6 2 | 3+4 2 | 3+4 2 | 5+6 2 | 2+2 2 | 2+2 2 |
| | 3+4 2 | 3+4 2 | 3+4 2 | 5+6 2 | 2+2 2 | 3+3 2 |
| | 5+6 2 | 3+4 2 | 3+4 2 | 5+6 2 | 3+3 2 | 2+2 2 |
| | 3+4 2 | 5+6 2 | 5+6 2 | 3+4 2 | 2+2 2 | 3+3 2 |
| | 5+6 2 | 5+6 2 | 3+4 2 | 3+4 2 | 3+3 2 | 2+2 2 |
| | 3+4 2 | 3+4 2 | 5+6 2 | 3+4 2 | 3+3 2 | 3+3 2 |
| | 5+6 2 | 3+4 2 | 5+6 2 | 5+6 2 | 2+2 2 | 2+2 2 |
| | 3+4 2 | 5+6 2 | 3+4 2 | 5+6 2 | 3+3 2 | 3+3 2 |
| | 3+4 2 | 5+6 2 | 3+4 2 | 3+4 2 | 2+2 2 | 3+3 2 |
| 12.2.89 | 16884 | 46 | Cr2Ni1S4 | | | |
| | -A | -GM | -M | -Y | -L | -V |
| | 5+6 2 | 3+4 2 | 5+6 2 | 5+6 2 | 2+2 2 | 3+3 2 |
| | 3+4 2 | 5+6 2 | 3+4 2 | 3+4 2 | 3+3 2 | 2+2 2 |
| | 3+4 2 | 5+6 2 | 3+4 2 | 5+6 2 | 2+2 2 | 2+2 2 |
| | 5+6 2 | 3+4 2 | 5+6 2 | 3+4 2 | 3+3 2 | 3+3 2 |
| | 5+6 2 | 3+4 2 | 5+6 2 | 3+4 2 | 3+3 2 | 2+2 2 |
| | 5+6 2 | 5+6 2 | 3+4 2 | 5+6 2 | 2+2 2 | 3+3 2 |
| | 3+4 2 | 3+4 2 | 5+6 2 | 3+4 2 | 3+3 2 | 2+2 2 |
| | 3+4 2 | 3+4 2 | 3+4 2 | 5+6 2 | 2+2 2 | 3+3 2 |
| | 5+6 2 | 3+4 2 | 3+4 2 | 3+4 2 | 3+3 2 | 2+2 2 |
| | 5+6 2 | 5+6 2 | 5+6 2 | 3+4 2 | 2+2 2 | 3+3 2 |
| | 3+4 2 | 5+6 2 | 5+6 2 | 5+6 2 | 3+3 2 | 2+2 2 |
| | 5+6 2 | 3+4 2 | 3+4 2 | 5+6 2 | 2+2 2 | 2+2 2 |
| | 3+4 2 | 3+4 2 | 5+6 2 | 5+6 2 | 2+2 2 | 3+3 2 |
| | 5+6 2 | 3+4 2 | 5+6 2 | 5+6 2 | 3+3 2 | 3+3 2 |
| | 3+4 2 | 3+4 2 | 5+6 2 | 3+4 2 | 3+3 2 | 2+2 2 |
| | 3+4 2 | 3+4 2 | 3+4 2 | 3+4 2 | 2+2 2 | 3+3 2 |
| | 5+6 2 | 5+6 2 | 5+6 2 | 3+4 2 | 3+3 2 | 2+2 2 |
| | 5+6 2 | 5+6 2 | 5+6 2 | 3+4 2 | 3+3 2 | 2+2 2 |
| | 5+6 2 | 5+6 2 | 3+4 2 | 3+4 2 | 3+3 2 | 2+2 2 |
| | 5+6 2 | 3+4 2 | 3+4 2 | 5+6 2 | 2+2 2 | 2+2 2 |
| | 3+4 2 | 5+6 2 | 5+6 2 | 5+6 2 | 3+3 2 | 2+2 2 |
| | 5+6 2 | 3+4 2 | 5+6 2 | 3+4 2 | 3+3 2 | 3+3 2 |
| | 5+6 2 | 3+4 2 | 3+4 2 | 5+6 2 | 3+3 2 | 2+2 2 |
| 12.2.90 | 625971 | 44 | Cr2Fe1Se4 | | | |
| | -A | -GM | -M | -Y | -L | -V |
| | 3+4 2 | 3+4 2 | 5+6 2 | 5+6 2 | 2+2 2 | 2+2 2 |

Continued on next column

**TABLE XXIV– continued**

| TQC | ICSD | NoE | SF | | | |
|---|---|---|---|---|---|---|
| | 5+6 2 | 5+6 2 | 3+4 2 | 3+4 2 | 3+3 2 | 3+3 2 |
| | 5+6 2 | 3+4 2 | 5+6 2 | 5+6 2 | 2+2 2 | 3+3 2 |
| | 3+4 2 | 5+6 2 | 3+4 2 | 3+4 2 | 3+3 2 | 2+2 2 |
| | 5+6 2 | 3+4 2 | 5+6 2 | 3+4 2 | 3+3 2 | 2+2 2 |
| | 3+4 2 | 5+6 2 | 3+4 2 | 5+6 2 | 2+2 2 | 2+2 2 |
| | 5+6 2 | 3+4 2 | 5+6 2 | 5+6 2 | 3+3 2 | 3+3 2 |
| | 3+4 2 | 3+4 2 | 3+4 2 | 3+4 2 | 2+2 2 | 3+3 2 |
| | 3+4 2 | 3+4 2 | 5+6 2 | 5+6 2 | 2+2 2 | 2+2 2 |
| | 5+6 2 | 5+6 2 | 3+4 2 | 3+4 2 | 3+3 2 | 3+3 2 |
| | 5+6 2 | 5+6 2 | 5+6 2 | 3+4 2 | 3+3 2 | 2+2 2 |
| | 3+4 2 | 3+4 2 | 3+4 2 | 5+6 2 | 2+2 2 | 3+3 2 |
| | 5+6 2 | 3+4 2 | 5+6 2 | 5+6 2 | 2+2 2 | 2+2 2 |
| | 5+6 2 | 5+6 2 | 5+6 2 | 3+4 2 | 3+3 2 | 3+3 2 |
| | 3+4 2 | 3+4 2 | 3+4 2 | 5+6 2 | 3+3 2 | 2+2 2 |
| | 3+4 2 | 5+6 2 | 3+4 2 | 3+4 2 | 2+2 2 | 3+3 2 |
| | 5+6 2 | 3+4 2 | 5+6 2 | 3+4 2 | 3+3 2 | 2+2 2 |
| | 5+6 2 | 5+6 2 | 5+6 2 | 3+4 2 | 2+2 2 | 2+2 2 |
| | 5+6 2 | 3+4 2 | 5+6 2 | 5+6 2 | 3+3 2 | 2+2 2 |
| | 5+6 2 | 3+4 2 | 5+6 2 | 3+4 2 | 3+3 2 | 2+2 2 |
| | 3+4 2 | 3+4 2 | 3+4 2 | 5+6 2 | 3+3 2 | 3+3 2 |
| | 5+6 2 | 5+6 2 | 5+6 2 | 3+4 2 | 2+2 2 | 3+3 2 |
| 12.2.91 | 626655 | 40 | Cr1S4V2 | | | |
| | -A | -GM | -M | -Y | -L | -V |
| | 3+4 2 | 3+4 2 | 3+4 2 | 5+6 2 | 2+2 2 | 2+2 2 |
| | 5+6 2 | 5+6 2 | 5+6 2 | 3+4 2 | 3+3 2 | 3+3 2 |
| | 5+6 2 | 3+4 2 | 5+6 2 | 3+4 2 | 2+2 2 | 3+3 2 |
| | 3+4 2 | 5+6 2 | 3+4 2 | 5+6 2 | 3+3 2 | 2+2 2 |
| | 3+4 2 | 3+4 2 | 3+4 2 | 3+4 2 | 3+3 2 | 3+3 2 |
| | 3+4 2 | 5+6 2 | 5+6 2 | 5+6 2 | 2+2 2 | 2+2 2 |
| | 5+6 2 | 3+4 2 | 5+6 2 | 3+4 2 | 3+3 2 | 3+3 2 |
| | 5+6 2 | 3+4 2 | 3+4 2 | 5+6 2 | 2+2 2 | 2+2 2 |
| | 3+4 2 | 3+4 2 | 5+6 2 | 3+4 2 | 3+3 2 | 3+3 2 |
| | 5+6 2 | 5+6 2 | 3+4 2 | 3+4 2 | 2+2 2 | 2+2 2 |
| | 3+4 2 | 5+6 2 | 3+4 2 | 5+6 2 | 3+3 2 | 3+3 2 |
| | 3+4 2 | 3+4 2 | 5+6 2 | 5+6 2 | 2+2 2 | 3+3 2 |
| | 5+6 2 | 3+4 2 | 3+4 2 | 5+6 2 | 2+2 2 | 2+2 2 |
| | 5+6 2 | 3+4 2 | 3+4 2 | 5+6 2 | 3+3 2 | 2+2 2 |
| | 3+4 2 | 5+6 2 | 5+6 2 | 3+4 2 | 2+2 2 | 3+3 2 |
| | 5+6 2 | 5+6 2 | 5+6 2 | 3+4 2 | 3+3 2 | 2+2 2 |
| | 3+4 2 | 5+6 2 | 3+4 2 | 3+4 2 | 3+3 2 | 3+3 2 |
| | 3+4 2 | 5+6 2 | 5+6 2 | 3+4 2 | 2+2 2 | 3+3 2 |
| | 3+4 2 | 3+4 2 | 3+4 2 | 5+6 2 | 3+3 2 | 3+3 2 |
| | 5+6 2 | 3+4 2 | 3+4 2 | 5+6 2 | 3+3 2 | 2+2 2 |
| 12.2.92 | 626543 | 56 | Cr1Pr2Se4 | | | |
| | -A | -GM | -M | -Y | -L | -V |
| | 5+6 2 | 3+4 2 | 3+4 2 | 5+6 2 | 2+2 2 | 2+2 2 |
| | 3+4 2 | 5+6 2 | 5+6 2 | 3+4 2 | 3+3 2 | 3+3 2 |
| | 3+4 2 | 3+4 2 | 5+6 2 | 5+6 2 | 2+2 2 | 3+3 2 |
| | 5+6 2 | 5+6 2 | 3+4 2 | 3+4 2 | 3+3 2 | 2+2 2 |
| | 5+6 2 | 3+4 2 | 3+4 2 | 5+6 2 | 3+3 2 | 3+3 2 |
| | 3+4 2 | 5+6 2 | 5+6 2 | 3+4 2 | 2+2 2 | 2+2 2 |
| | 5+6 2 | 3+4 2 | 3+4 2 | 5+6 2 | 3+3 2 | 3+3 2 |
| | 3+4 2 | 5+6 2 | 5+6 2 | 3+4 2 | 2+2 2 | 2+2 2 |
| | 3+4 2 | 3+4 2 | 3+4 2 | 3+4 2 | 2+2 2 | 3+3 2 |
| | 5+6 2 | 5+6 2 | 5+6 2 | 5+6 2 | 3+3 2 | 2+2 2 |
| | 5+6 2 | 3+4 2 | 3+4 2 | 3+4 2 | 2+2 2 | 3+3 2 |
| | 3+4 2 | 5+6 2 | 5+6 2 | 5+6 2 | 3+3 2 | 2+2 2 |
| | 3+4 2 | 3+4 2 | 5+6 2 | 3+4 2 | 3+3 2 | 3+3 2 |
| | 3+4 2 | 5+6 2 | 3+4 2 | 5+6 2 | 2+2 2 | 2+2 2 |
| | 5+6 2 | 3+4 2 | 3+4 2 | 3+4 2 | 3+3 2 | 3+3 2 |
| | 3+4 2 | 5+6 2 | 5+6 2 | 3+4 2 | 3+3 2 | 2+2 2 |

Continued on next column

**TABLE XXIV– continued**

| TQC | ICSD | NoE | SF | | | |
|---|---|---|---|---|---|---|
| | 5+6 2 | 3+4 2 | 5+6 2 | 5+6 2 | 2+2 2 | 3+3 2 |
| | 3+4 2 | 3+4 2 | 3+4 2 | 3+4 2 | 2+2 2 | 2+2 2 |
| | 5+6 2 | 5+6 2 | 5+6 2 | 5+6 2 | 3+3 2 | 3+3 2 |
| | 3+4 2 | 5+6 2 | 3+4 2 | 5+6 2 | 2+2 2 | 2+2 2 |
| | 5+6 2 | 3+4 2 | 3+4 2 | 3+4 2 | 3+3 2 | 3+3 2 |
| | 3+4 2 | 5+6 2 | 5+6 2 | 3+4 2 | 3+3 2 | 2+2 2 |
| | 3+4 2 | 3+4 2 | 3+4 2 | 3+4 2 | 2+2 2 | 3+3 2 |
| | 5+6 2 | 3+4 2 | 3+4 2 | 5+6 2 | 3+3 2 | 2+2 2 |
| | 5+6 2 | 5+6 2 | 5+6 2 | 3+4 2 | 2+2 2 | 3+3 2 |
| | 3+4 2 | 5+6 2 | 5+6 2 | 5+6 2 | 2+2 2 | 3+3 2 |
| | 3+4 2 | 3+4 2 | 3+4 2 | 3+4 2 | 3+3 2 | 2+2 2 |
| | 5+6 2 | 3+4 2 | 3+4 2 | 3+4 2 | 3+3 2 | 3+3 2 |
| 12.2.93 | 626536 | 44 | Cr1Pr2S4 | | | |
| | -A | -GM | -M | -Y | -L | -V |
| | 5+6 2 | 3+4 2 | 3+4 2 | 5+6 2 | 2+2 2 | 2+2 2 |
| | 3+4 2 | 5+6 2 | 5+6 2 | 3+4 2 | 3+3 2 | 3+3 2 |
| | 5+6 2 | 3+4 2 | 3+4 2 | 5+6 2 | 3+3 2 | 3+3 2 |
| | 3+4 2 | 5+6 2 | 5+6 2 | 3+4 2 | 2+2 2 | 2+2 2 |
| | 5+6 2 | 3+4 2 | 3+4 2 | 5+6 2 | 3+3 2 | 3+3 2 |
| | 3+4 2 | 3+4 2 | 5+6 2 | 5+6 2 | 2+2 2 | 2+2 2 |
| | 5+6 2 | 5+6 2 | 3+4 2 | 3+4 2 | 3+3 2 | 3+3 2 |
| | 3+4 2 | 5+6 2 | 5+6 2 | 3+4 2 | 2+2 2 | 2+2 2 |
| | 5+6 2 | 3+4 2 | 3+4 2 | 3+4 2 | 3+3 2 | 3+3 2 |
| | 3+4 2 | 5+6 2 | 5+6 2 | 5+6 2 | 2+2 2 | 2+2 2 |
| | 3+4 2 | 3+4 2 | 3+4 2 | 3+4 2 | 3+3 2 | 3+3 2 |
| | 5+6 2 | 3+4 2 | 5+6 2 | 5+6 2 | 2+2 2 | 2+2 2 |
| | 3+4 2 | 5+6 2 | 3+4 2 | 3+4 2 | 3+3 2 | 3+3 2 |
| | 3+4 2 | 3+4 2 | 5+6 2 | 5+6 2 | 2+2 2 | 3+3 2 |
| | 5+6 2 | 5+6 2 | 3+4 2 | 3+4 2 | 3+3 2 | 2+2 2 |
| | 3+4 2 | 5+6 2 | 5+6 2 | 3+4 2 | 2+2 2 | 2+2 2 |
| | 5+6 2 | 3+4 2 | 5+6 2 | 5+6 2 | 3+3 2 | 3+3 2 |
| | 3+4 2 | 5+6 2 | 3+4 2 | 3+4 2 | 3+3 2 | 3+3 2 |
| | 3+4 2 | 3+4 2 | 3+4 2 | 3+4 2 | 2+2 2 | 3+3 2 |
| | 5+6 2 | 3+4 2 | 5+6 2 | 5+6 2 | 3+3 2 | 2+2 2 |
| | 3+4 2 | 5+6 2 | 3+4 2 | 3+4 2 | 3+3 2 | 3+3 2 |
| | 5+6 2 | 3+4 2 | 5+6 2 | 5+6 2 | 2+2 2 | 2+2 2 |
| 12.2.94 | 16087 | 28 | Cr1O5Pb2 | | | |
| | -A | -GM | -M | -Y | -L | -V |
| | 3+4 2 | 3+4 2 | 3+4 2 | 3+4 2 | 2+2 2 | 2+2 2 |
| | 5+6 2 | 5+6 2 | 5+6 2 | 5+6 2 | 3+3 2 | 3+3 2 |
| | 5+6 2 | 3+4 2 | 5+6 2 | 3+4 2 | 3+3 2 | 2+2 2 |
| | 3+4 2 | 5+6 2 | 3+4 2 | 5+6 2 | 2+2 2 | 3+3 2 |
| | 3+4 2 | 3+4 2 | 3+4 2 | 3+4 2 | 2+2 2 | 3+3 2 |
| | 5+6 2 | 5+6 2 | 5+6 2 | 5+6 2 | 3+3 2 | 2+2 2 |
| | 5+6 2 | 3+4 2 | 5+6 2 | 5+6 2 | 3+3 2 | 3+3 2 |
| | 3+4 2 | 5+6 2 | 5+6 2 | 3+4 2 | 2+2 2 | 2+2 2 |
| | 3+4 2 | 3+4 2 | 3+4 2 | 3+4 2 | 3+3 2 | 2+2 2 |
| | 3+4 2 | 3+4 2 | 5+6 2 | 5+6 2 | 2+2 2 | 3+3 2 |
| | 5+6 2 | 3+4 2 | 3+4 2 | 3+4 2 | 2+2 2 | 2+2 2 |
| | 5+6 2 | 5+6 2 | 3+4 2 | 5+6 2 | 3+3 2 | 3+3 2 |
| | 3+4 2 | 3+4 2 | 3+4 2 | 5+6 2 | 2+2 2 | 3+3 2 |
| | 3+4 2 | 5+6 2 | 3+4 2 | 3+4 2 | 3+3 2 | 2+2 2 |
| 12.2.95 | 79865 | 140 | Cr1O4V1 | | | |
| | -A | -GM | -M | -Y | -L | -V |
| | 5+6 2 | 3+4 2 | 5+6 2 | 3+4 2 | 2+2 2 | 3+3 2 |
| | 3+4 2 | 5+6 2 | 3+4 2 | 5+6 2 | 3+3 2 | 2+2 2 |
| | 3+4 2 | 3+4 2 | 3+4 2 | 5+6 2 | 3+3 2 | 2+2 2 |
| | 5+6 2 | 5+6 2 | 5+6 2 | 3+4 2 | 2+2 2 | 3+3 2 |
| | 3+4 2 | 5+6 2 | 3+4 2 | 5+6 2 | 3+3 2 | 2+2 2 |
| | 5+6 2 | 3+4 2 | 5+6 2 | 3+4 2 | 2+2 2 | 2+2 2 |
| | 5+6 2 | 3+4 2 | 3+4 2 | 3+4 2 | 3+3 2 | 3+3 2 |

Continued on next column

**TABLE XXIV– continued**

| TQC | ICSD | | NoE | | SF | |
|---|---|---|---|---|---|---|
| | 3+4 2 | 3+4 2 | 5+6 2 | 3+4 2 | 2+2 2 | 2+2 2 |
| | 5+6 2 | 5+6 2 | 3+4 2 | 5+6 2 | 2+2 2 | 3+3 2 |
| | 3+4 2 | 5+6 2 | 5+6 2 | 5+6 2 | 3+3 2 | 3+3 2 |
| | 5+6 2 | 3+4 2 | 3+4 2 | 5+6 2 | 3+3 2 | 3+3 2 |
| | 3+4 2 | 5+6 2 | 5+6 2 | 5+6 2 | 2+2 2 | 2+2 2 |
| | 5+6 2 | 5+6 2 | 5+6 2 | 3+4 2 | 3+3 2 | 3+3 2 |
| | 3+4 2 | 3+4 2 | 3+4 2 | 5+6 2 | 2+2 2 | 2+2 2 |
| | 3+4 2 | 5+6 2 | 3+4 2 | 3+4 2 | 2+2 2 | 3+3 2 |
| | 5+6 2 | 3+4 2 | 5+6 2 | 3+4 2 | 3+3 2 | 2+2 2 |
| | 3+4 2 | 3+4 2 | 3+4 2 | 3+4 2 | 3+3 2 | 3+3 2 |
| | 5+6 2 | 5+6 2 | 5+6 2 | 5+6 2 | 2+2 2 | 2+2 2 |
| | 3+4 2 | 5+6 2 | 3+4 2 | 5+6 2 | 3+3 2 | 2+2 2 |
| | 5+6 2 | 3+4 2 | 5+6 2 | 3+4 2 | 2+2 2 | 3+3 2 |
| | 5+6 2 | 3+4 2 | 3+4 2 | 3+4 2 | 2+2 2 | 3+3 2 |
| | 3+4 2 | 3+4 2 | 5+6 2 | 3+4 2 | 3+3 2 | 2+2 2 |
| | 5+6 2 | 3+4 2 | 5+6 2 | 5+6 2 | 2+2 2 | 3+3 2 |
| | 3+4 2 | 5+6 2 | 3+4 2 | 5+6 2 | 3+3 2 | 2+2 2 |
| | 3+4 2 | 3+4 2 | 5+6 2 | 3+4 2 | 2+2 2 | 3+3 2 |
| | 5+6 2 | 5+6 2 | 3+4 2 | 5+6 2 | 3+3 2 | 2+2 2 |
| | 5+6 2 | 3+4 2 | 5+6 2 | 3+4 2 | 2+2 2 | 2+2 2 |
| | 3+4 2 | 5+6 2 | 3+4 2 | 3+4 2 | 3+3 2 | 3+3 2 |
| | 3+4 2 | 5+6 2 | 5+6 2 | 5+6 2 | 2+2 2 | 2+2 2 |
| | 5+6 2 | 3+4 2 | 3+4 2 | 5+6 2 | 3+3 2 | 2+2 2 |
| | 3+4 2 | 5+6 2 | 5+6 2 | 3+4 2 | 3+3 2 | 3+3 2 |
| | 5+6 2 | 5+6 2 | 5+6 2 | 3+4 2 | 2+2 2 | 3+3 2 |
| | 3+4 2 | 5+6 2 | 3+4 2 | 5+6 2 | 3+3 2 | 3+3 2 |
| | 3+4 2 | 3+4 2 | 3+4 2 | 5+6 2 | 2+2 2 | 2+2 2 |
| | 5+6 2 | 3+4 2 | 3+4 2 | 5+6 2 | 2+2 2 | 2+2 2 |
| | 5+6 2 | 3+4 2 | 5+6 2 | 3+4 2 | 3+3 2 | 3+3 2 |
| | 5+6 2 | 5+6 2 | 5+6 2 | 3+4 2 | 3+3 2 | 3+3 2 |
| | 3+4 2 | 3+4 2 | 3+4 2 | 5+6 2 | 2+2 2 | 2+2 2 |
| | 3+4 2 | 3+4 2 | 3+4 2 | 3+4 2 | 2+2 2 | 3+3 2 |
| | 5+6 2 | 3+4 2 | 5+6 2 | 5+6 2 | 2+2 2 | 2+2 2 |
| | 5+6 2 | 5+6 2 | 3+4 2 | 3+4 2 | 3+3 2 | 3+3 2 |
| | 5+6 2 | 5+6 2 | 5+6 2 | 5+6 2 | 3+3 2 | 2+2 2 |
| | 3+4 2 | 3+4 2 | 5+6 2 | 3+4 2 | 2+2 2 | 2+2 2 |
| | 5+6 2 | 5+6 2 | 5+6 2 | 3+4 2 | 3+3 2 | 3+3 2 |
| | 3+4 2 | 3+4 2 | 3+4 2 | 5+6 2 | 2+2 2 | 2+2 2 |
| | 3+4 2 | 5+6 2 | 3+4 2 | 5+6 2 | 2+2 2 | 3+3 2 |
| | 5+6 2 | 5+6 2 | 5+6 2 | 3+4 2 | 3+3 2 | 2+2 2 |
| | 3+4 2 | 5+6 2 | 5+6 2 | 3+4 2 | 3+3 2 | 3+3 2 |
| | 5+6 2 | 5+6 2 | 3+4 2 | 5+6 2 | 2+2 2 | 3+3 2 |
| | 5+6 2 | 5+6 2 | 3+4 2 | 5+6 2 | 3+3 2 | 3+3 2 |
| | 3+4 2 | 3+4 2 | 3+4 2 | 5+6 2 | 2+2 2 | 2+2 2 |
| | 3+4 2 | 5+6 2 | 3+4 2 | 3+4 2 | 3+3 2 | 3+3 2 |
| | 5+6 2 | 3+4 2 | 5+6 2 | 5+6 2 | 2+2 2 | 2+2 2 |
| | 5+6 2 | 5+6 2 | 5+6 2 | 5+6 2 | 2+2 2 | 2+2 2 |
| | 3+4 2 | 3+4 2 | 5+6 2 | 5+6 2 | 3+3 2 | 3+3 2 |
| | 3+4 2 | 5+6 2 | 3+4 2 | 3+4 2 | 3+3 2 | 2+2 2 |
| | 5+6 2 | 3+4 2 | 5+6 2 | 3+4 2 | 3+3 2 | 3+3 2 |
| | 5+6 2 | 5+6 2 | 3+4 2 | 5+6 2 | 2+2 2 | 3+3 2 |
| | 3+4 2 | 3+4 2 | 3+4 2 | 3+4 2 | 2+2 2 | 2+2 2 |
| | 3+4 2 | 5+6 2 | 5+6 2 | 5+6 2 | 3+3 2 | 2+2 2 |
| | 5+6 2 | 3+4 2 | 5+6 2 | 3+4 2 | 3+3 2 | 3+3 2 |
| | 3+4 2 | 5+6 2 | 3+4 2 | 5+6 2 | 2+2 2 | 3+3 2 |
| | 5+6 2 | 3+4 2 | 3+4 2 | 3+4 2 | 3+3 2 | 2+2 2 |
| | 3+4 2 | 3+4 2 | 5+6 2 | 3+4 2 | 2+2 2 | 2+2 2 |
| | 3+4 2 | 3+4 2 | 3+4 2 | 5+6 2 | 3+3 2 | 2+2 2 |
| | 5+6 2 | 5+6 2 | 5+6 2 | 3+4 2 | 2+2 2 | 3+3 2 |
| | 3+4 2 | 3+4 2 | 3+4 2 | 3+4 2 | 3+3 2 | 2+2 2 |
| | 5+6 2 | 3+4 2 | 5+6 2 | 3+4 2 | 2+2 2 | 3+3 2 |

Continued on next column

**TABLE XXIV– continued**

| TQC | ICSD | | NoE | | SF | |
|---|---|---|---|---|---|---|
| | 3+4 2 | 5+6 2 | 5+6 2 | 5+6 2 | 2+2 2 | 2+2 2 |
| | 5+6 2 | 5+6 2 | 3+4 2 | 3+4 2 | 3+3 2 | 3+3 2 |
| 12.2.96 | 202377 | | 76 | | Cr1Nd1O1S2 | |
| | -A | -GM | -M | -Y | -L | -V |
| | 3+4 2 | 3+4 2 | 5+6 2 | 5+6 2 | 2+2 2 | 2+2 2 |
| | 5+6 2 | 5+6 2 | 3+4 2 | 3+4 2 | 3+3 2 | 3+3 2 |
| | 3+4 2 | 3+4 2 | 5+6 2 | 5+6 2 | 3+3 2 | 3+3 2 |
| | 5+6 2 | 5+6 2 | 3+4 2 | 3+4 2 | 2+2 2 | 2+2 2 |
| | 3+4 2 | 3+4 2 | 5+6 2 | 5+6 2 | 3+3 2 | 3+3 2 |
| | 5+6 2 | 5+6 2 | 3+4 2 | 3+4 2 | 2+2 2 | 2+2 2 |
| | 3+4 2 | 3+4 2 | 5+6 2 | 5+6 2 | 2+2 2 | 2+2 2 |
| | 5+6 2 | 5+6 2 | 3+4 2 | 3+4 2 | 3+3 2 | 3+3 2 |
| | 5+6 2 | 5+6 2 | 3+4 2 | 3+4 2 | 3+3 2 | 3+3 2 |
| | 3+4 2 | 3+4 2 | 5+6 2 | 5+6 2 | 2+2 2 | 2+2 2 |
| | 5+6 2 | 3+4 2 | 3+4 2 | 3+4 2 | 2+2 2 | 2+2 2 |
| | 3+4 2 | 3+4 2 | 5+6 2 | 5+6 2 | 3+3 2 | 3+3 2 |
| | 5+6 2 | 5+6 2 | 3+4 2 | 3+4 2 | 2+2 2 | 3+3 2 |
| | 3+4 2 | 5+6 2 | 5+6 2 | 5+6 2 | 3+3 2 | 2+2 2 |
| | 3+4 2 | 3+4 2 | 3+4 2 | 3+4 2 | 3+3 2 | 3+3 2 |
| | 5+6 2 | 3+4 2 | 5+6 2 | 3+4 2 | 3+3 2 | 2+2 2 |
| | 3+4 2 | 5+6 2 | 3+4 2 | 5+6 2 | 2+2 2 | 2+2 2 |
| | 5+6 2 | 3+4 2 | 5+6 2 | 5+6 2 | 3+3 2 | 3+3 2 |
| | 3+4 2 | 3+4 2 | 3+4 2 | 3+4 2 | 2+2 2 | 3+3 2 |
| | 3+4 2 | 3+4 2 | 5+6 2 | 5+6 2 | 3+3 2 | 2+2 2 |
| | 5+6 2 | 5+6 2 | 3+4 2 | 3+4 2 | 3+3 2 | 3+3 2 |
| | 5+6 2 | 5+6 2 | 5+6 2 | 5+6 2 | 2+2 2 | 2+2 2 |
| | 5+6 2 | 3+4 2 | 5+6 2 | 3+4 2 | 2+2 2 | 3+3 2 |
| | 3+4 2 | 3+4 2 | 3+4 2 | 3+4 2 | 2+2 2 | 2+2 2 |
| | 5+6 2 | 5+6 2 | 5+6 2 | 3+4 2 | 3+3 2 | 2+2 2 |
| | 3+4 2 | 3+4 2 | 3+4 2 | 5+6 2 | 2+2 2 | 3+3 2 |
| | 5+6 2 | 3+4 2 | 5+6 2 | 3+4 2 | 3+3 2 | 2+2 2 |
| | 3+4 2 | 5+6 2 | 3+4 2 | 5+6 2 | 3+3 2 | 3+3 2 |
| | 5+6 2 | 5+6 2 | 5+6 2 | 5+6 2 | 2+2 2 | 2+2 2 |
| | 5+6 2 | 5+6 2 | 3+4 2 | 3+4 2 | 3+3 2 | 3+3 2 |
| | 3+4 2 | 5+6 2 | 3+4 2 | 5+6 2 | 2+2 2 | 2+2 2 |
| | 3+4 2 | 3+4 2 | 5+6 2 | 5+6 2 | 2+2 2 | 3+3 2 |
| | 3+4 2 | 5+6 2 | 5+6 2 | 3+4 2 | 3+3 2 | 2+2 2 |
| | 5+6 2 | 3+4 2 | 3+4 2 | 3+4 2 | 3+3 2 | 3+3 2 |
| | 3+4 2 | 3+4 2 | 3+4 2 | 3+4 2 | 3+3 2 | 3+3 2 |
| | 5+6 2 | 3+4 2 | 5+6 2 | 3+4 2 | 3+3 2 | 2+2 2 |
| | 3+4 2 | 3+4 2 | 5+6 2 | 5+6 2 | 3+3 2 | 3+3 2 |
| | 3+4 2 | 5+6 2 | 3+4 2 | 3+4 2 | 2+2 2 | 2+2 2 |
| 12.2.97 | 151873 | | 46 | | Cr1Fe2Se4 | |
| | -A | -GM | -M | -Y | -L | -V |
| | 5+6 2 | 3+4 2 | 5+6 2 | 5+6 2 | 3+3 2 | 3+3 2 |
| | 3+4 2 | 3+4 2 | 3+4 2 | 3+4 2 | 3+3 2 | 2+2 2 |
| | 5+6 2 | 5+6 2 | 3+4 2 | 5+6 2 | 2+2 2 | 3+3 2 |
| | 3+4 2 | 5+6 2 | 5+6 2 | 3+4 2 | 2+2 2 | 2+2 2 |
| | 5+6 2 | 3+4 2 | 5+6 2 | 3+4 2 | 2+2 2 | 3+3 2 |
| | 3+4 2 | 5+6 2 | 3+4 2 | 5+6 2 | 3+3 2 | 2+2 2 |
| | 5+6 2 | 3+4 2 | 3+4 2 | 5+6 2 | 2+2 2 | 3+3 2 |
| | 3+4 2 | 3+4 2 | 5+6 2 | 3+4 2 | 3+3 2 | 2+2 2 |
| | 5+6 2 | 3+4 2 | 3+4 2 | 5+6 2 | 3+3 2 | 3+3 2 |
| | 5+6 2 | 5+6 2 | 5+6 2 | 3+4 2 | 2+2 2 | 3+3 2 |
| | 3+4 2 | 5+6 2 | 5+6 2 | 3+4 2 | 2+2 2 | 2+2 2 |
| | 5+6 2 | 3+4 2 | 3+4 2 | 5+6 2 | 3+3 2 | 2+2 2 |
| | 3+4 2 | 3+4 2 | 5+6 2 | 5+6 2 | 3+3 2 | 3+3 2 |
| | 5+6 2 | 3+4 2 | 5+6 2 | 5+6 2 | 2+2 2 | 3+3 2 |
| | 3+4 2 | 5+6 2 | 3+4 2 | 3+4 2 | 3+3 2 | 2+2 2 |
| | 3+4 2 | 5+6 2 | 3+4 2 | 3+4 2 | 2+2 2 | 2+2 2 |
| | 5+6 2 | 3+4 2 | 5+6 2 | 3+4 2 | 2+2 2 | 3+3 2 |

Continued on next column

**TABLE XXIV– continued**

| TQC | ICSD | NoE | SF | | | |
|---|---|---|---|---|---|---|
| | 3+4 2 | 3+4 2 | 3+4 2 | 5+6 2 | 3+3 2 | 2+2 2 |
| | 5+6 2 | 5+6 2 | 5+6 2 | 5+6 2 | 2+2 2 | 3+3 2 |
| | 3+4 2 | 3+4 2 | 5+6 2 | 3+4 2 | 2+2 2 | 3+3 2 |
| | 3+4 2 | 3+4 2 | 5+6 2 | 5+6 2 | 3+3 2 | 2+2 2 |
| | 5+6 2 | 5+6 2 | 5+6 2 | 3+4 2 | 2+2 2 | 3+3 2 |
| | 5+6 2 | 5+6 2 | 5+6 2 | 3+4 2 | 2+2 2 | 3+3 2 |
| 12.2.98 | 74542 | 70 | Co2O7P2 | | | |
| | -A | -GM | -M | -Y | -L | -V |
| | 3+4 2 | 3+4 2 | 3+4 2 | 3+4 2 | 3+3 2 | 3+3 2 |
| | 5+6 2 | 5+6 2 | 5+6 2 | 5+6 2 | 2+2 2 | 2+2 2 |
| | 3+4 2 | 3+4 2 | 3+4 2 | 3+4 2 | 3+3 2 | 3+3 2 |
| | 5+6 2 | 3+4 2 | 5+6 2 | 5+6 2 | 3+3 2 | 2+2 2 |
| | 3+4 2 | 3+4 2 | 3+4 2 | 3+4 2 | 2+2 2 | 3+3 2 |
| | 3+4 2 | 5+6 2 | 3+4 2 | 5+6 2 | 3+3 2 | 2+2 2 |
| | 5+6 2 | 5+6 2 | 5+6 2 | 3+4 2 | 2+2 2 | 3+3 2 |
| | 5+6 2 | 5+6 2 | 5+6 2 | 5+6 2 | 2+2 2 | 2+2 2 |
| | 5+6 2 | 3+4 2 | 5+6 2 | 3+4 2 | 3+3 2 | 3+3 2 |
| | 3+4 2 | 5+6 2 | 3+4 2 | 5+6 2 | 2+2 2 | 2+2 2 |
| | 5+6 2 | 3+4 2 | 3+4 2 | 5+6 2 | 3+3 2 | 2+2 2 |
| | 5+6 2 | 5+6 2 | 5+6 2 | 5+6 2 | 2+2 2 | 2+2 2 |
| | 3+4 2 | 5+6 2 | 5+6 2 | 3+4 2 | 2+2 2 | 3+3 2 |
| | 3+4 2 | 3+4 2 | 3+4 2 | 3+4 2 | 3+3 2 | 3+3 2 |
| | 5+6 2 | 5+6 2 | 3+4 2 | 3+4 2 | 2+2 2 | 2+2 2 |
| | 5+6 2 | 3+4 2 | 5+6 2 | 5+6 2 | 3+3 2 | 2+2 2 |
| | 3+4 2 | 5+6 2 | 5+6 2 | 5+6 2 | 2+2 2 | 3+3 2 |
| | 3+4 2 | 3+4 2 | 5+6 2 | 5+6 2 | 2+2 2 | 2+2 2 |
| | 5+6 2 | 3+4 2 | 3+4 2 | 3+4 2 | 2+2 2 | 3+3 2 |
| | 3+4 2 | 5+6 2 | 5+6 2 | 5+6 2 | 3+3 2 | 2+2 2 |
| | 3+4 2 | 3+4 2 | 5+6 2 | 3+4 2 | 3+3 2 | 3+3 2 |
| | 5+6 2 | 5+6 2 | 3+4 2 | 3+4 2 | 2+2 2 | 2+2 2 |
| | 5+6 2 | 3+4 2 | 3+4 2 | 5+6 2 | 3+3 2 | 3+3 2 |
| | 3+4 2 | 5+6 2 | 3+4 2 | 3+4 2 | 2+2 2 | 2+2 2 |
| | 5+6 2 | 5+6 2 | 5+6 2 | 5+6 2 | 3+3 2 | 3+3 2 |
| | 3+4 2 | 3+4 2 | 5+6 2 | 5+6 2 | 3+3 2 | 2+2 2 |
| | 5+6 2 | 5+6 2 | 5+6 2 | 5+6 2 | 2+2 2 | 2+2 2 |
| | 5+6 2 | 5+6 2 | 3+4 2 | 3+4 2 | 2+2 2 | 3+3 2 |
| | 5+6 2 | 3+4 2 | 5+6 2 | 3+4 2 | 2+2 2 | 3+3 2 |
| | 5+6 2 | 5+6 2 | 3+4 2 | 5+6 2 | 3+3 2 | 2+2 2 |
| | 3+4 2 | 3+4 2 | 3+4 2 | 3+4 2 | 3+3 2 | 3+3 2 |
| | 3+4 2 | 5+6 2 | 5+6 2 | 5+6 2 | 2+2 2 | 2+2 2 |
| | 5+6 2 | 3+4 2 | 5+6 2 | 3+4 2 | 3+3 2 | 2+2 2 |
| | 3+4 2 | 5+6 2 | 3+4 2 | 5+6 2 | 2+2 2 | 3+3 2 |
| | 3+4 2 | 5+6 2 | 3+4 2 | 3+4 2 | 3+3 2 | 3+3 2 |
| 12.2.99 | 41026 | 152 | Co2Nb4Pd1Se12 | | | |
| | -A | -GM | -M | -Y | -L | -V |
| | 5+6 2 | 3+4 2 | 3+4 2 | 5+6 2 | 2+2 2 | 2+2 2 |
| | 3+4 2 | 5+6 2 | 5+6 2 | 3+4 2 | 3+3 2 | 3+3 2 |
| | 3+4 2 | 3+4 2 | 3+4 2 | 3+4 2 | 2+2 2 | 2+2 2 |
| | 5+6 2 | 5+6 2 | 5+6 2 | 5+6 2 | 3+3 2 | 3+3 2 |
| | 5+6 2 | 3+4 2 | 3+4 2 | 5+6 2 | 2+2 2 | 2+2 2 |
| | 3+4 2 | 5+6 2 | 5+6 2 | 3+4 2 | 3+3 2 | 3+3 2 |
| | 3+4 2 | 3+4 2 | 3+4 2 | 3+4 2 | 2+2 2 | 2+2 2 |
| | 5+6 2 | 5+6 2 | 5+6 2 | 5+6 2 | 3+3 2 | 3+3 2 |
| | 5+6 2 | 3+4 2 | 3+4 2 | 5+6 2 | 2+2 2 | 2+2 2 |
| | 3+4 2 | 5+6 2 | 5+6 2 | 3+4 2 | 3+3 2 | 3+3 2 |
| | 5+6 2 | 3+4 2 | 3+4 2 | 5+6 2 | 2+2 2 | 2+2 2 |
| | 3+4 2 | 5+6 2 | 5+6 2 | 3+4 2 | 3+3 2 | 3+3 2 |
| | 3+4 2 | 3+4 2 | 3+4 2 | 3+4 2 | 2+2 2 | 2+2 2 |
| | 5+6 2 | 5+6 2 | 5+6 2 | 5+6 2 | 3+3 2 | 3+3 2 |
| | 3+4 2 | 3+4 2 | 3+4 2 | 3+4 2 | 2+2 2 | 2+2 2 |
| | 5+6 2 | 5+6 2 | 5+6 2 | 5+6 2 | 3+3 2 | 3+3 2 |
| | 5+6 2 | 3+4 2 | 3+4 2 | 5+6 2 | 2+2 2 | 3+3 2 |
| | 3+4 2 | 5+6 2 | 5+6 2 | 3+4 2 | 3+3 2 | 2+2 2 |
| | 5+6 2 | 3+4 2 | 5+6 2 | 3+4 2 | 3+3 2 | 3+3 2 |
| | 3+4 2 | 3+4 2 | 3+4 2 | 5+6 2 | 3+3 2 | 2+2 2 |
| | 3+4 2 | 5+6 2 | 3+4 2 | 5+6 2 | 2+2 2 | 3+3 2 |
| | 5+6 2 | 5+6 2 | 3+4 2 | 3+4 2 | 2+2 2 | 2+2 2 |
| | 5+6 2 | 3+4 2 | 5+6 2 | 3+4 2 | 3+3 2 | 3+3 2 |
| | 3+4 2 | 5+6 2 | 5+6 2 | 5+6 2 | 2+2 2 | 3+3 2 |
| | 3+4 2 | 5+6 2 | 3+4 2 | 5+6 2 | 3+3 2 | 2+2 2 |
| | 3+4 2 | 3+4 2 | 5+6 2 | 3+4 2 | 2+2 2 | 2+2 2 |
| | 5+6 2 | 5+6 2 | 5+6 2 | 5+6 2 | 3+3 2 | 3+3 2 |
| | 5+6 2 | 3+4 2 | 3+4 2 | 3+4 2 | 2+2 2 | 2+2 2 |
| | 3+4 2 | 3+4 2 | 5+6 2 | 3+4 2 | 3+3 2 | 2+2 2 |
| | 5+6 2 | 5+6 2 | 3+4 2 | 3+4 2 | 2+2 2 | 3+3 2 |
| | 3+4 2 | 3+4 2 | 3+4 2 | 5+6 2 | 3+3 2 | 3+3 2 |
| | 3+4 2 | 3+4 2 | 5+6 2 | 3+4 2 | 2+2 2 | 2+2 2 |
| | 5+6 2 | 5+6 2 | 3+4 2 | 5+6 2 | 3+3 2 | 3+3 2 |
| | 3+4 2 | 3+4 2 | 3+4 2 | 3+4 2 | 3+3 2 | 3+3 2 |
| | 5+6 2 | 3+4 2 | 3+4 2 | 5+6 2 | 3+3 2 | 2+2 2 |
| | 5+6 2 | 5+6 2 | 5+6 2 | 3+4 2 | 2+2 2 | 3+3 2 |
| | 3+4 2 | 5+6 2 | 3+4 2 | 5+6 2 | 2+2 2 | 2+2 2 |
| | 5+6 2 | 5+6 2 | 5+6 2 | 3+4 2 | 2+2 2 | 3+3 2 |
| | 3+4 2 | 3+4 2 | 3+4 2 | 5+6 2 | 3+3 2 | 2+2 2 |
| | 3+4 2 | 3+4 2 | 5+6 2 | 5+6 2 | 3+3 2 | 3+3 2 |
| | 5+6 2 | 3+4 2 | 3+4 2 | 3+4 2 | 2+2 2 | 3+3 2 |
| | 3+4 2 | 5+6 2 | 5+6 2 | 3+4 2 | 3+3 2 | 2+2 2 |
| | 3+4 2 | 3+4 2 | 3+4 2 | 3+4 2 | 3+3 2 | 3+3 2 |
| | 5+6 2 | 5+6 2 | 5+6 2 | 3+4 2 | 2+2 2 | 3+3 2 |
| | 5+6 2 | 3+4 2 | 3+4 2 | 5+6 2 | 3+3 2 | 2+2 2 |
| | 3+4 2 | 3+4 2 | 3+4 2 | 5+6 2 | 3+3 2 | 2+2 2 |
| | 3+4 2 | 5+6 2 | 5+6 2 | 3+4 2 | 2+2 2 | 3+3 2 |
| | 5+6 2 | 3+4 2 | 5+6 2 | 5+6 2 | 2+2 2 | 3+3 2 |
| | 3+4 2 | 3+4 2 | 3+4 2 | 5+6 2 | 2+2 2 | 2+2 2 |
| | 3+4 2 | 5+6 2 | 5+6 2 | 3+4 2 | 3+3 2 | 2+2 2 |
| | 5+6 2 | 3+4 2 | 3+4 2 | 5+6 2 | 2+2 2 | 2+2 2 |
| | 3+4 2 | 5+6 2 | 3+4 2 | 3+4 2 | 3+3 2 | 3+3 2 |
| | 5+6 2 | 3+4 2 | 5+6 2 | 5+6 2 | 3+3 2 | 3+3 2 |
| | 5+6 2 | 5+6 2 | 3+4 2 | 3+4 2 | 2+2 2 | 2+2 2 |
| | 3+4 2 | 3+4 2 | 3+4 2 | 3+4 2 | 2+2 2 | 2+2 2 |
| | 5+6 2 | 3+4 2 | 5+6 2 | 5+6 2 | 3+3 2 | 3+3 2 |
| | 3+4 2 | 5+6 2 | 5+6 2 | 3+4 2 | 2+2 2 | 3+3 2 |
| | 5+6 2 | 5+6 2 | 3+4 2 | 5+6 2 | 3+3 2 | 2+2 2 |
| | 3+4 2 | 3+4 2 | 3+4 2 | 3+4 2 | 3+3 2 | 2+2 2 |
| | 5+6 2 | 5+6 2 | 5+6 2 | 5+6 2 | 2+2 2 | 3+3 2 |
| | 3+4 2 | 3+4 2 | 5+6 2 | 3+4 2 | 3+3 2 | 2+2 2 |
| | 5+6 2 | 5+6 2 | 3+4 2 | 5+6 2 | 3+3 2 | 3+3 2 |
| | 3+4 2 | 3+4 2 | 5+6 2 | 3+4 2 | 3+3 2 | 3+3 2 |
| | 3+4 2 | 5+6 2 | 3+4 2 | 3+4 2 | 2+2 2 | 2+2 2 |
| | 5+6 2 | 5+6 2 | 3+4 2 | 5+6 2 | 2+2 2 | 3+3 2 |
| | 5+6 2 | 3+4 2 | 3+4 2 | 3+4 2 | 3+3 2 | 3+3 2 |
| | 3+4 2 | 3+4 2 | 5+6 2 | 5+6 2 | 2+2 2 | 2+2 2 |
| | 3+4 2 | 5+6 2 | 5+6 2 | 5+6 2 | 3+3 2 | 2+2 2 |
| | 5+6 2 | 3+4 2 | 5+6 2 | 3+4 2 | 3+3 2 | 3+3 2 |
| | 5+6 2 | 3+4 2 | 3+4 2 | 3+4 2 | 2+2 2 | 3+3 2 |
| | 3+4 2 | 5+6 2 | 3+4 2 | 5+6 2 | 2+2 2 | 2+2 2 |
| | 5+6 2 | 5+6 2 | 5+6 2 | 5+6 2 | 3+3 2 | 3+3 2 |
| | 3+4 2 | 3+4 2 | 3+4 2 | 5+6 2 | 2+2 2 | 2+2 2 |
| | 5+6 2 | 5+6 2 | 5+6 2 | 3+4 2 | 3+3 2 | 3+3 2 |
| | 3+4 2 | 3+4 2 | 3+4 2 | 3+4 2 | 2+2 2 | 2+2 2 |
| | 5+6 2 | 5+6 2 | 5+6 2 | 3+4 2 | 3+3 2 | 3+3 2 |
| 12.2.100 | 408031 | 118 | Cl2La3Si3 | | | |

Continued on next column

Continued on next column

**TABLE XXIV– continued**

| TQC | ICSD | NoE | SF | | | |
|---|---|---|---|---|---|---|
| | -A | -GM | -M | -Y | -L | -V |
| | 3+4 2 | 3+4 2 | 5+6 2 | 5+6 2 | 2+2 2 | 2+2 2 |
| | 5+6 2 | 5+6 2 | 3+4 2 | 3+4 2 | 3+3 2 | 3+3 2 |
| | 5+6 2 | 3+4 2 | 5+6 2 | 3+4 2 | 3+3 2 | 2+2 2 |
| | 3+4 2 | 5+6 2 | 3+4 2 | 5+6 2 | 2+2 2 | 3+3 2 |
| | 3+4 2 | 3+4 2 | 3+4 2 | 3+4 2 | 2+2 2 | 2+2 2 |
| | 5+6 2 | 5+6 2 | 5+6 2 | 5+6 2 | 3+3 2 | 3+3 2 |
| | 3+4 2 | 3+4 2 | 5+6 2 | 5+6 2 | 2+2 2 | 2+2 2 |
| | 5+6 2 | 5+6 2 | 3+4 2 | 3+4 2 | 3+3 2 | 3+3 2 |
| | 5+6 2 | 3+4 2 | 3+4 2 | 5+6 2 | 3+3 2 | 2+2 2 |
| | 3+4 2 | 5+6 2 | 5+6 2 | 3+4 2 | 2+2 2 | 3+3 2 |
| | 3+4 2 | 3+4 2 | 3+4 2 | 3+4 2 | 2+2 2 | 2+2 2 |
| | 5+6 2 | 5+6 2 | 5+6 2 | 5+6 2 | 3+3 2 | 3+3 2 |
| | 3+4 2 | 3+4 2 | 5+6 2 | 5+6 2 | 2+2 2 | 2+2 2 |
| | 5+6 2 | 5+6 2 | 3+4 2 | 3+4 2 | 3+3 2 | 3+3 2 |
| | 5+6 2 | 3+4 2 | 3+4 2 | 5+6 2 | 3+3 2 | 2+2 2 |
| | 3+4 2 | 5+6 2 | 5+6 2 | 3+4 2 | 2+2 2 | 3+3 2 |
| | 5+6 2 | 3+4 2 | 3+4 2 | 5+6 2 | 2+2 2 | 3+3 2 |
| | 3+4 2 | 5+6 2 | 5+6 2 | 3+4 2 | 3+3 2 | 2+2 2 |
| | 5+6 2 | 5+6 2 | 5+6 2 | 3+4 2 | 3+3 2 | 3+3 2 |
| | 3+4 2 | 3+4 2 | 3+4 2 | 5+6 2 | 2+2 2 | 2+2 2 |
| | 5+6 2 | 5+6 2 | 3+4 2 | 3+4 2 | 3+3 2 | 2+2 2 |
| | 3+4 2 | 3+4 2 | 5+6 2 | 5+6 2 | 2+2 2 | 3+3 2 |
| | 3+4 2 | 3+4 2 | 3+4 2 | 3+4 2 | 2+2 2 | 3+3 2 |
| | 5+6 2 | 5+6 2 | 5+6 2 | 3+4 2 | 3+3 2 | 2+2 2 |
| | 3+4 2 | 3+4 2 | 3+4 2 | 5+6 2 | 2+2 2 | 2+2 2 |
| | 5+6 2 | 5+6 2 | 5+6 2 | 5+6 2 | 3+3 2 | 3+3 2 |
| | 3+4 2 | 5+6 2 | 3+4 2 | 3+4 2 | 2+2 2 | 2+2 2 |
| | 5+6 2 | 3+4 2 | 5+6 2 | 5+6 2 | 3+3 2 | 3+3 2 |
| | 3+4 2 | 3+4 2 | 3+4 2 | 3+4 2 | 2+2 2 | 3+3 2 |
| | 5+6 2 | 5+6 2 | 5+6 2 | 5+6 2 | 3+3 2 | 3+3 2 |
| | 5+6 2 | 3+4 2 | 5+6 2 | 3+4 2 | 3+3 2 | 2+2 2 |
| | 3+4 2 | 5+6 2 | 3+4 2 | 5+6 2 | 2+2 2 | 3+3 2 |
| | 3+4 2 | 3+4 2 | 3+4 2 | 3+4 2 | 2+2 2 | 2+2 2 |
| | 5+6 2 | 3+4 2 | 5+6 2 | 5+6 2 | 3+3 2 | 2+2 2 |
| | 3+4 2 | 5+6 2 | 3+4 2 | 3+4 2 | 3+3 2 | 3+3 2 |
| | 5+6 2 | 3+4 2 | 5+6 2 | 5+6 2 | 2+2 2 | 3+3 2 |
| | 3+4 2 | 5+6 2 | 3+4 2 | 5+6 2 | 3+3 2 | 2+2 2 |
| | 3+4 2 | 3+4 2 | 5+6 2 | 3+4 2 | 2+2 2 | 2+2 2 |
| | 5+6 2 | 3+4 2 | 5+6 2 | 5+6 2 | 3+3 2 | 3+3 2 |
| | 5+6 2 | 5+6 2 | 3+4 2 | 5+6 2 | 2+2 2 | 2+2 2 |
| | 3+4 2 | 3+4 2 | 5+6 2 | 3+4 2 | 2+2 2 | 3+3 2 |
| | 5+6 2 | 3+4 2 | 3+4 2 | 5+6 2 | 3+3 2 | 2+2 2 |
| | 5+6 2 | 5+6 2 | 3+4 2 | 3+4 2 | 3+3 2 | 3+3 2 |
| | 5+6 2 | 5+6 2 | 3+4 2 | 5+6 2 | 2+2 2 | 2+2 2 |
| | 3+4 2 | 5+6 2 | 5+6 2 | 3+4 2 | 2+2 2 | 3+3 2 |
| | 3+4 2 | 5+6 2 | 5+6 2 | 3+4 2 | 3+3 2 | 2+2 2 |
| | 5+6 2 | 3+4 2 | 3+4 2 | 3+4 2 | 3+3 2 | 2+2 2 |
| | 3+4 2 | 5+6 2 | 5+6 2 | 5+6 2 | 2+2 2 | 3+3 2 |
| | 5+6 2 | 3+4 2 | 5+6 2 | 3+4 2 | 3+3 2 | 2+2 2 |
| | 3+4 2 | 5+6 2 | 3+4 2 | 5+6 2 | 2+2 2 | 3+3 2 |
| | 3+4 2 | 5+6 2 | 5+6 2 | 5+6 2 | 2+2 2 | 3+3 2 |
| | 5+6 2 | 3+4 2 | 3+4 2 | 5+6 2 | 3+3 2 | 2+2 2 |
| | 5+6 2 | 5+6 2 | 5+6 2 | 3+4 2 | 3+3 2 | 2+2 2 |
| | 3+4 2 | 3+4 2 | 5+6 2 | 3+4 2 | 2+2 2 | 3+3 2 |
| | 5+6 2 | 3+4 2 | 3+4 2 | 3+4 2 | 3+3 2 | 2+2 2 |
| | 3+4 2 | 5+6 2 | 3+4 2 | 5+6 2 | 2+2 2 | 3+3 2 |
| | 5+6 2 | 5+6 2 | 3+4 2 | 3+4 2 | 3+3 2 | 3+3 2 |
| | 3+4 2 | 3+4 2 | 5+6 2 | 5+6 2 | 2+2 2 | 3+3 2 |
| | 3+4 2 | 3+4 2 | 5+6 2 | 3+4 2 | 2+2 2 | 2+2 2 |
| 12.2.101 | 420061 | 122 | Cl1Mo1O4Yb1 | | | |

Continued on next column

**TABLE XXIV– continued**

| TQC | ICSD | NoE | SF | | | |
|---|---|---|---|---|---|---|
| | -A | -GM | -M | -Y | -L | -V |
| | 3+4 2 | 3+4 2 | 3+4 2 | 3+4 2 | 2+2 2 | 2+2 2 |
| | 5+6 2 | 5+6 2 | 5+6 2 | 5+6 2 | 3+3 2 | 3+3 2 |
| | 3+4 2 | 3+4 2 | 5+6 2 | 5+6 2 | 3+3 2 | 3+3 2 |
| | 5+6 2 | 5+6 2 | 3+4 2 | 3+4 2 | 2+2 2 | 2+2 2 |
| | 3+4 2 | 3+4 2 | 5+6 2 | 5+6 2 | 3+3 2 | 3+3 2 |
| | 5+6 2 | 5+6 2 | 3+4 2 | 3+4 2 | 2+2 2 | 2+2 2 |
| | 3+4 2 | 3+4 2 | 3+4 2 | 3+4 2 | 3+3 2 | 3+3 2 |
| | 5+6 2 | 5+6 2 | 5+6 2 | 5+6 2 | 2+2 2 | 2+2 2 |
| | 3+4 2 | 3+4 2 | 5+6 2 | 5+6 2 | 2+2 2 | 2+2 2 |
| | 5+6 2 | 5+6 2 | 3+4 2 | 3+4 2 | 3+3 2 | 3+3 2 |
| | 3+4 2 | 5+6 2 | 5+6 2 | 3+4 2 | 2+2 2 | 2+2 2 |
| | 5+6 2 | 3+4 2 | 3+4 2 | 5+6 2 | 3+3 2 | 3+3 2 |
| | 5+6 2 | 5+6 2 | 3+4 2 | 3+4 2 | 2+2 2 | 3+3 2 |
| | 3+4 2 | 3+4 2 | 5+6 2 | 5+6 2 | 3+3 2 | 2+2 2 |
| | 5+6 2 | 5+6 2 | 3+4 2 | 5+6 2 | 3+3 2 | 3+3 2 |
| | 3+4 2 | 3+4 2 | 5+6 2 | 3+4 2 | 2+2 2 | 2+2 2 |
| | 3+4 2 | 3+4 2 | 3+4 2 | 3+4 2 | 3+3 2 | 3+3 2 |
| | 5+6 2 | 5+6 2 | 5+6 2 | 5+6 2 | 2+2 2 | 2+2 2 |
| | 5+6 2 | 3+4 2 | 3+4 2 | 3+4 2 | 2+2 2 | 2+2 2 |
| | 3+4 2 | 5+6 2 | 5+6 2 | 3+4 2 | 3+3 2 | 3+3 2 |
| | 3+4 2 | 3+4 2 | 5+6 2 | 5+6 2 | 2+2 2 | 2+2 2 |
| | 5+6 2 | 3+4 2 | 5+6 2 | 5+6 2 | 3+3 2 | 2+2 2 |
| | 3+4 2 | 5+6 2 | 3+4 2 | 5+6 2 | 3+3 2 | 3+3 2 |
| | 5+6 2 | 3+4 2 | 3+4 2 | 5+6 2 | 2+2 2 | 3+3 2 |
| | 5+6 2 | 5+6 2 | 5+6 2 | 5+6 2 | 3+3 2 | 2+2 2 |
| | 5+6 2 | 5+6 2 | 3+4 2 | 3+4 2 | 2+2 2 | 3+3 2 |
| | 3+4 2 | 5+6 2 | 5+6 2 | 3+4 2 | 3+3 2 | 3+3 2 |
| | 3+4 2 | 3+4 2 | 3+4 2 | 3+4 2 | 2+2 2 | 2+2 2 |
| | 3+4 2 | 3+4 2 | 5+6 2 | 5+6 2 | 3+3 2 | 2+2 2 |
| | 5+6 2 | 3+4 2 | 3+4 2 | 3+4 2 | 2+2 2 | 3+3 2 |
| | 3+4 2 | 5+6 2 | 3+4 2 | 3+4 2 | 2+2 2 | 2+2 2 |
| | 3+4 2 | 3+4 2 | 5+6 2 | 5+6 2 | 3+3 2 | 3+3 2 |
| | 3+4 2 | 5+6 2 | 5+6 2 | 3+4 2 | 2+2 2 | 3+3 2 |
| | 5+6 2 | 3+4 2 | 3+4 2 | 5+6 2 | 2+2 2 | 2+2 2 |
| | 5+6 2 | 5+6 2 | 3+4 2 | 3+4 2 | 3+3 2 | 3+3 2 |
| | 5+6 2 | 5+6 2 | 5+6 2 | 3+4 2 | 2+2 2 | 2+2 2 |
| | 3+4 2 | 3+4 2 | 3+4 2 | 5+6 2 | 3+3 2 | 3+3 2 |
| | 3+4 2 | 5+6 2 | 3+4 2 | 5+6 2 | 3+3 2 | 2+2 2 |
| | 5+6 2 | 3+4 2 | 5+6 2 | 3+4 2 | 2+2 2 | 3+3 2 |
| | 3+4 2 | 5+6 2 | 3+4 2 | 5+6 2 | 3+3 2 | 3+3 2 |
| | 3+4 2 | 3+4 2 | 5+6 2 | 3+4 2 | 2+2 2 | 2+2 2 |
| | 5+6 2 | 5+6 2 | 5+6 2 | 3+4 2 | 3+3 2 | 3+3 2 |
| | 3+4 2 | 3+4 2 | 3+4 2 | 5+6 2 | 2+2 2 | 2+2 2 |
| | 5+6 2 | 5+6 2 | 5+6 2 | 3+4 2 | 3+3 2 | 2+2 2 |
| | 5+6 2 | 5+6 2 | 3+4 2 | 5+6 2 | 2+2 2 | 3+3 2 |
| | 3+4 2 | 3+4 2 | 5+6 2 | 3+4 2 | 3+3 2 | 2+2 2 |
| | 5+6 2 | 3+4 2 | 3+4 2 | 5+6 2 | 2+2 2 | 3+3 2 |
| | 3+4 2 | 3+4 2 | 5+6 2 | 3+4 2 | 3+3 2 | 2+2 2 |
| | 3+4 2 | 5+6 2 | 3+4 2 | 5+6 2 | 2+2 2 | 3+3 2 |
| | 5+6 2 | 5+6 2 | 5+6 2 | 5+6 2 | 3+3 2 | 3+3 2 |
| | 5+6 2 | 5+6 2 | 5+6 2 | 5+6 2 | 2+2 2 | 2+2 2 |
| | 3+4 2 | 3+4 2 | 3+4 2 | 3+4 2 | 3+3 2 | 2+2 2 |
| | 5+6 2 | 3+4 2 | 3+4 2 | 3+4 2 | 3+3 2 | 3+3 2 |
| | 5+6 2 | 5+6 2 | 5+6 2 | 5+6 2 | 2+2 2 | 2+2 2 |
| | 3+4 2 | 3+4 2 | 5+6 2 | 5+6 2 | 2+2 2 | 3+3 2 |
| | 3+4 2 | 5+6 2 | 3+4 2 | 3+4 2 | 3+3 2 | 2+2 2 |
| | 5+6 2 | 3+4 2 | 5+6 2 | 5+6 2 | 2+2 2 | 3+3 2 |
| | 3+4 2 | 5+6 2 | 3+4 2 | 3+4 2 | 3+3 2 | 2+2 2 |
| | 5+6 2 | 3+4 2 | 5+6 2 | 3+4 2 | 2+2 2 | 3+3 2 |
| | 3+4 2 | 3+4 2 | 3+4 2 | 5+6 2 | 3+3 2 | 2+2 2 |

Continued on next column

**TABLE XXIV– continued**

| TQC | ICSD | | NoE | | SF | |
|---|---|---|---|---|---|---|
| | 5+6 2 | 5+6 2 | 5+6 2 | 3+4 2 | 3+3 2 | 3+3 2 |
| 12.2.102 | 420060 | | 120 | | Cl1Mo1O4Tm1 | |
| | -A | -GM | -M | -Y | -L | -V |
| | 3+4 2 | 3+4 2 | 5+6 2 | 5+6 2 | 2+2 2 | 2+2 2 |
| | 5+6 2 | 5+6 2 | 3+4 2 | 3+4 2 | 3+3 2 | 3+3 2 |
| | 3+4 2 | 3+4 2 | 5+6 2 | 5+6 2 | 3+3 2 | 3+3 2 |
| | 5+6 2 | 5+6 2 | 3+4 2 | 3+4 2 | 2+2 2 | 2+2 2 |
| | 3+4 2 | 3+4 2 | 5+6 2 | 5+6 2 | 3+3 2 | 3+3 2 |
| | 5+6 2 | 5+6 2 | 3+4 2 | 3+4 2 | 2+2 2 | 2+2 2 |
| | 3+4 2 | 3+4 2 | 3+4 2 | 3+4 2 | 3+3 2 | 3+3 2 |
| | 5+6 2 | 5+6 2 | 5+6 2 | 5+6 2 | 2+2 2 | 2+2 2 |
| | 3+4 2 | 3+4 2 | 5+6 2 | 5+6 2 | 2+2 2 | 2+2 2 |
| | 5+6 2 | 5+6 2 | 3+4 2 | 3+4 2 | 3+3 2 | 3+3 2 |
| | 3+4 2 | 5+6 2 | 5+6 2 | 3+4 2 | 2+2 2 | 2+2 2 |
| | 5+6 2 | 3+4 2 | 3+4 2 | 5+6 2 | 3+3 2 | 3+3 2 |
| | 5+6 2 | 5+6 2 | 3+4 2 | 3+4 2 | 3+3 2 | 3+3 2 |
| | 3+4 2 | 3+4 2 | 5+6 2 | 5+6 2 | 2+2 2 | 2+2 2 |
| | 5+6 2 | 3+4 2 | 3+4 2 | 5+6 2 | 3+3 2 | 2+2 2 |
| | 3+4 2 | 5+6 2 | 5+6 2 | 3+4 2 | 2+2 2 | 3+3 2 |
| | 3+4 2 | 3+4 2 | 3+4 2 | 3+4 2 | 3+3 2 | 3+3 2 |
| | 5+6 2 | 5+6 2 | 5+6 2 | 5+6 2 | 2+2 2 | 2+2 2 |
| | 5+6 2 | 3+4 2 | 3+4 2 | 3+4 2 | 2+2 2 | 2+2 2 |
| | 3+4 2 | 5+6 2 | 5+6 2 | 3+4 2 | 3+3 2 | 3+3 2 |
| | 3+4 2 | 3+4 2 | 5+6 2 | 5+6 2 | 2+2 2 | 2+2 2 |
| | 5+6 2 | 3+4 2 | 5+6 2 | 5+6 2 | 3+3 2 | 2+2 2 |
| | 3+4 2 | 5+6 2 | 3+4 2 | 5+6 2 | 3+3 2 | 3+3 2 |
| | 5+6 2 | 3+4 2 | 3+4 2 | 5+6 2 | 2+2 2 | 3+3 2 |
| | 5+6 2 | 5+6 2 | 5+6 2 | 5+6 2 | 3+3 2 | 2+2 2 |
| | 3+4 2 | 5+6 2 | 3+4 2 | 3+4 2 | 2+2 2 | 3+3 2 |
| | 5+6 2 | 5+6 2 | 5+6 2 | 3+4 2 | 3+3 2 | 3+3 2 |
| | 3+4 2 | 3+4 2 | 3+4 2 | 3+4 2 | 2+2 2 | 2+2 2 |
| | 3+4 2 | 3+4 2 | 5+6 2 | 5+6 2 | 3+3 2 | 2+2 2 |
| | 5+6 2 | 3+4 2 | 3+4 2 | 3+4 2 | 2+2 2 | 3+3 2 |
| | 3+4 2 | 5+6 2 | 3+4 2 | 3+4 2 | 2+2 2 | 2+2 2 |
| | 3+4 2 | 3+4 2 | 5+6 2 | 5+6 2 | 3+3 2 | 3+3 2 |
| | 5+6 2 | 5+6 2 | 5+6 2 | 3+4 2 | 2+2 2 | 2+2 2 |
| | 3+4 2 | 3+4 2 | 3+4 2 | 5+6 2 | 2+2 2 | 3+3 2 |
| | 5+6 2 | 5+6 2 | 3+4 2 | 3+4 2 | 3+3 2 | 3+3 2 |
| | 5+6 2 | 5+6 2 | 5+6 2 | 3+4 2 | 2+2 2 | 2+2 2 |
| | 3+4 2 | 3+4 2 | 3+4 2 | 5+6 2 | 3+3 2 | 3+3 2 |
| | 3+4 2 | 5+6 2 | 5+6 2 | 5+6 2 | 3+3 2 | 3+3 2 |
| | 3+4 2 | 5+6 2 | 3+4 2 | 3+4 2 | 2+2 2 | 2+2 2 |
| | 5+6 2 | 3+4 2 | 5+6 2 | 3+4 2 | 2+2 2 | 2+2 2 |
| | 5+6 2 | 3+4 2 | 3+4 2 | 3+4 2 | 3+3 2 | 3+3 2 |
| | 3+4 2 | 3+4 2 | 3+4 2 | 5+6 2 | 3+3 2 | 2+2 2 |
| | 5+6 2 | 5+6 2 | 5+6 2 | 5+6 2 | 3+3 2 | 3+3 2 |
| | 3+4 2 | 5+6 2 | 3+4 2 | 3+4 2 | 2+2 2 | 3+3 2 |
| | 5+6 2 | 5+6 2 | 5+6 2 | 5+6 2 | 2+2 2 | 2+2 2 |
| | 3+4 2 | 3+4 2 | 5+6 2 | 5+6 2 | 3+3 2 | 2+2 2 |
| | 5+6 2 | 3+4 2 | 5+6 2 | 3+4 2 | 2+2 2 | 3+3 2 |
| | 3+4 2 | 5+6 2 | 3+4 2 | 5+6 2 | 3+3 2 | 2+2 2 |
| | 5+6 2 | 3+4 2 | 3+4 2 | 3+4 2 | 2+2 2 | 3+3 2 |
| | 3+4 2 | 5+6 2 | 5+6 2 | 5+6 2 | 3+3 2 | 3+3 2 |
| | 5+6 2 | 3+4 2 | 5+6 2 | 5+6 2 | 2+2 2 | 2+2 2 |
| | 5+6 2 | 5+6 2 | 3+4 2 | 3+4 2 | 3+3 2 | 2+2 2 |
| | 3+4 2 | 3+4 2 | 3+4 2 | 3+4 2 | 3+3 2 | 3+3 2 |
| | 5+6 2 | 5+6 2 | 5+6 2 | 5+6 2 | 2+2 2 | 2+2 2 |
| | 3+4 2 | 3+4 2 | 5+6 2 | 5+6 2 | 3+3 2 | 3+3 2 |
| | 5+6 2 | 3+4 2 | 3+4 2 | 3+4 2 | 2+2 2 | 2+2 2 |
| | 3+4 2 | 5+6 2 | 5+6 2 | 5+6 2 | 3+3 2 | 3+3 2 |
| | 3+4 2 | 5+6 2 | 3+4 2 | 3+4 2 | 2+2 2 | 2+2 2 |

Continued on next column

**TABLE XXIV– continued**

| TQC | ICSD | | NoE | | SF | |
|---|---|---|---|---|---|---|
| | 5+6 2 | 3+4 2 | 5+6 2 | 3+4 2 | 2+2 2 | 3+3 2 |
| | 3+4 2 | 3+4 2 | 3+4 2 | 3+4 2 | 3+3 2 | 2+2 2 |
| 12.2.103 | 420059 | | 112 | | Cl1Mo1O4Tb1 | |
| | -A | -GM | -M | -Y | -L | -V |
| | 3+4 2 | 3+4 2 | 5+6 2 | 5+6 2 | 2+2 2 | 2+2 2 |
| | 5+6 2 | 5+6 2 | 3+4 2 | 3+4 2 | 3+3 2 | 3+3 2 |
| | 3+4 2 | 3+4 2 | 5+6 2 | 5+6 2 | 3+3 2 | 3+3 2 |
| | 5+6 2 | 5+6 2 | 3+4 2 | 3+4 2 | 2+2 2 | 2+2 2 |
| | 3+4 2 | 3+4 2 | 5+6 2 | 5+6 2 | 2+2 2 | 2+2 2 |
| | 5+6 2 | 5+6 2 | 3+4 2 | 3+4 2 | 3+3 2 | 3+3 2 |
| | 3+4 2 | 3+4 2 | 3+4 2 | 3+4 2 | 3+3 2 | 3+3 2 |
| | 5+6 2 | 5+6 2 | 5+6 2 | 5+6 2 | 2+2 2 | 2+2 2 |
| | 3+4 2 | 3+4 2 | 5+6 2 | 5+6 2 | 2+2 2 | 2+2 2 |
| | 5+6 2 | 5+6 2 | 3+4 2 | 3+4 2 | 3+3 2 | 3+3 2 |
| | 3+4 2 | 5+6 2 | 5+6 2 | 3+4 2 | 2+2 2 | 2+2 2 |
| | 5+6 2 | 3+4 2 | 3+4 2 | 5+6 2 | 3+3 2 | 3+3 2 |
| | 5+6 2 | 3+4 2 | 3+4 2 | 3+4 2 | 2+2 2 | 3+3 2 |
| | 3+4 2 | 5+6 2 | 5+6 2 | 5+6 2 | 3+3 2 | 2+2 2 |
| | 5+6 2 | 5+6 2 | 3+4 2 | 3+4 2 | 3+3 2 | 2+2 2 |
| | 3+4 2 | 3+4 2 | 5+6 2 | 5+6 2 | 2+2 2 | 3+3 2 |
| | 3+4 2 | 3+4 2 | 3+4 2 | 3+4 2 | 3+3 2 | 3+3 2 |
| | 5+6 2 | 5+6 2 | 5+6 2 | 5+6 2 | 2+2 2 | 2+2 2 |
| | 5+6 2 | 3+4 2 | 3+4 2 | 3+4 2 | 2+2 2 | 2+2 2 |
| | 3+4 2 | 3+4 2 | 5+6 2 | 3+4 2 | 3+3 2 | 3+3 2 |
| | 3+4 2 | 5+6 2 | 5+6 2 | 5+6 2 | 2+2 2 | 2+2 2 |
| | 5+6 2 | 3+4 2 | 5+6 2 | 5+6 2 | 3+3 2 | 2+2 2 |
| | 3+4 2 | 5+6 2 | 3+4 2 | 5+6 2 | 3+3 2 | 3+3 2 |
| | 5+6 2 | 5+6 2 | 3+4 2 | 5+6 2 | 2+2 2 | 3+3 2 |
| | 5+6 2 | 3+4 2 | 3+4 2 | 5+6 2 | 3+3 2 | 2+2 2 |
| | 3+4 2 | 5+6 2 | 5+6 2 | 3+4 2 | 2+2 2 | 3+3 2 |
| | 5+6 2 | 5+6 2 | 5+6 2 | 3+4 2 | 3+3 2 | 3+3 2 |
| | 3+4 2 | 3+4 2 | 3+4 2 | 3+4 2 | 2+2 2 | 2+2 2 |
| | 3+4 2 | 3+4 2 | 5+6 2 | 5+6 2 | 3+3 2 | 2+2 2 |
| | 5+6 2 | 3+4 2 | 3+4 2 | 3+4 2 | 2+2 2 | 3+3 2 |
| | 3+4 2 | 5+6 2 | 5+6 2 | 3+4 2 | 2+2 2 | 2+2 2 |
| | 3+4 2 | 3+4 2 | 3+4 2 | 5+6 2 | 3+3 2 | 3+3 2 |
| | 5+6 2 | 5+6 2 | 5+6 2 | 3+4 2 | 2+2 2 | 2+2 2 |
| | 3+4 2 | 3+4 2 | 3+4 2 | 5+6 2 | 3+3 2 | 3+3 2 |
| | 5+6 2 | 5+6 2 | 3+4 2 | 3+4 2 | 2+2 2 | 2+2 2 |
| | 5+6 2 | 5+6 2 | 5+6 2 | 5+6 2 | 2+2 2 | 3+3 2 |
| | 3+4 2 | 3+4 2 | 3+4 2 | 3+4 2 | 3+3 2 | 3+3 2 |
| | 3+4 2 | 5+6 2 | 5+6 2 | 5+6 2 | 3+3 2 | 3+3 2 |
| | 3+4 2 | 5+6 2 | 3+4 2 | 3+4 2 | 2+2 2 | 2+2 2 |
| | 5+6 2 | 3+4 2 | 5+6 2 | 3+4 2 | 2+2 2 | 2+2 2 |
| | 5+6 2 | 3+4 2 | 3+4 2 | 3+4 2 | 3+3 2 | 3+3 2 |
| | 3+4 2 | 3+4 2 | 3+4 2 | 5+6 2 | 3+3 2 | 2+2 2 |
| | 5+6 2 | 5+6 2 | 5+6 2 | 5+6 2 | 3+3 2 | 3+3 2 |
| | 3+4 2 | 5+6 2 | 3+4 2 | 5+6 2 | 2+2 2 | 3+3 2 |
| | 5+6 2 | 3+4 2 | 5+6 2 | 5+6 2 | 2+2 2 | 2+2 2 |
| | 5+6 2 | 5+6 2 | 5+6 2 | 3+4 2 | 3+3 2 | 2+2 2 |
| | 3+4 2 | 3+4 2 | 5+6 2 | 3+4 2 | 3+3 2 | 3+3 2 |
| | 5+6 2 | 5+6 2 | 3+4 2 | 5+6 2 | 2+2 2 | 2+2 2 |
| | 3+4 2 | 3+4 2 | 5+6 2 | 5+6 2 | 2+2 2 | 2+2 2 |
| | 5+6 2 | 5+6 2 | 3+4 2 | 3+4 2 | 3+3 2 | 3+3 2 |
| | 3+4 2 | 3+4 2 | 5+6 2 | 3+4 2 | 3+3 2 | 3+3 2 |
| | 3+4 2 | 5+6 2 | 3+4 2 | 5+6 2 | 2+2 2 | 2+2 2 |
| | 5+6 2 | 3+4 2 | 5+6 2 | 3+4 2 | 3+3 2 | 3+3 2 |
| | 5+6 2 | 5+6 2 | 3+4 2 | 5+6 2 | 2+2 2 | 2+2 2 |
| | 3+4 2 | 3+4 2 | 5+6 2 | 5+6 2 | 3+3 2 | 3+3 2 |
| | 5+6 2 | 3+4 2 | 3+4 2 | 3+4 2 | 2+2 2 | 2+2 2 |
| 12.2.104 | 427663 | | 116 | | Cl1Ho1O4W1 | |

Continued on next column

**TABLE XXIV– continued**

| TQC | ICSD | | NoE | | SF | |
|---|---|---|---|---|---|---|
| | -A | -GM | -M | -Y | -L | -V |
| | 3+4 2 | 3+4 2 | 5+6 2 | 5+6 2 | 2+2 2 | 2+2 2 |
| | 5+6 2 | 5+6 2 | 3+4 2 | 3+4 2 | 3+3 2 | 3+3 2 |
| | 3+4 2 | 3+4 2 | 5+6 2 | 5+6 2 | 3+3 2 | 3+3 2 |
| | 5+6 2 | 5+6 2 | 3+4 2 | 3+4 2 | 2+2 2 | 2+2 2 |
| | 3+4 2 | 3+4 2 | 5+6 2 | 5+6 2 | 2+2 2 | 2+2 2 |
| | 5+6 2 | 5+6 2 | 3+4 2 | 3+4 2 | 3+3 2 | 3+3 2 |
| | 3+4 2 | 3+4 2 | 3+4 2 | 3+4 2 | 3+3 2 | 3+3 2 |
| | 5+6 2 | 5+6 2 | 5+6 2 | 5+6 2 | 2+2 2 | 2+2 2 |
| | 3+4 2 | 3+4 2 | 5+6 2 | 5+6 2 | 2+2 2 | 2+2 2 |
| | 5+6 2 | 5+6 2 | 3+4 2 | 3+4 2 | 3+3 2 | 3+3 2 |
| | 3+4 2 | 5+6 2 | 5+6 2 | 3+4 2 | 2+2 2 | 2+2 2 |
| | 5+6 2 | 3+4 2 | 3+4 2 | 5+6 2 | 3+3 2 | 3+3 2 |
| | 5+6 2 | 3+4 2 | 3+4 2 | 3+4 2 | 3+3 2 | 3+3 2 |
| | 3+4 2 | 5+6 2 | 5+6 2 | 5+6 2 | 2+2 2 | 2+2 2 |
| | 5+6 2 | 3+4 2 | 3+4 2 | 5+6 2 | 3+3 2 | 2+2 2 |
| | 3+4 2 | 5+6 2 | 5+6 2 | 3+4 2 | 2+2 2 | 3+3 2 |
| | 3+4 2 | 3+4 2 | 3+4 2 | 3+4 2 | 3+3 2 | 3+3 2 |
| | 5+6 2 | 5+6 2 | 5+6 2 | 5+6 2 | 2+2 2 | 2+2 2 |
| | 5+6 2 | 3+4 2 | 3+4 2 | 3+4 2 | 2+2 2 | 2+2 2 |
| | 3+4 2 | 5+6 2 | 5+6 2 | 5+6 2 | 3+3 2 | 3+3 2 |
| | 3+4 2 | 3+4 2 | 5+6 2 | 3+4 2 | 2+2 2 | 2+2 2 |
| | 5+6 2 | 3+4 2 | 3+4 2 | 5+6 2 | 3+3 2 | 2+2 2 |
| | 3+4 2 | 5+6 2 | 5+6 2 | 5+6 2 | 3+3 2 | 3+3 2 |
| | 5+6 2 | 3+4 2 | 3+4 2 | 5+6 2 | 2+2 2 | 3+3 2 |
| | 3+4 2 | 5+6 2 | 3+4 2 | 3+4 2 | 3+3 2 | 2+2 2 |
| | 5+6 2 | 3+4 2 | 5+6 2 | 5+6 2 | 2+2 2 | 3+3 2 |
| | 5+6 2 | 5+6 2 | 5+6 2 | 3+4 2 | 3+3 2 | 3+3 2 |
| | 3+4 2 | 5+6 2 | 5+6 2 | 5+6 2 | 2+2 2 | 2+2 2 |
| | 3+4 2 | 3+4 2 | 3+4 2 | 3+4 2 | 3+3 2 | 3+3 2 |
| | 5+6 2 | 5+6 2 | 3+4 2 | 3+4 2 | 2+2 2 | 2+2 2 |
| | 3+4 2 | 3+4 2 | 3+4 2 | 3+4 2 | 2+2 2 | 2+2 2 |
| | 3+4 2 | 3+4 2 | 5+6 2 | 5+6 2 | 3+3 2 | 3+3 2 |
| | 5+6 2 | 5+6 2 | 5+6 2 | 3+4 2 | 2+2 2 | 2+2 2 |
| | 3+4 2 | 3+4 2 | 3+4 2 | 5+6 2 | 3+3 2 | 3+3 2 |
| | 5+6 2 | 5+6 2 | 3+4 2 | 3+4 2 | 2+2 2 | 3+3 2 |
| | 5+6 2 | 5+6 2 | 5+6 2 | 3+4 2 | 3+3 2 | 2+2 2 |
| | 3+4 2 | 5+6 2 | 3+4 2 | 5+6 2 | 2+2 2 | 3+3 2 |
| | 3+4 2 | 3+4 2 | 5+6 2 | 5+6 2 | 3+3 2 | 2+2 2 |
| | 3+4 2 | 5+6 2 | 3+4 2 | 3+4 2 | 2+2 2 | 3+3 2 |
| | 5+6 2 | 3+4 2 | 5+6 2 | 3+4 2 | 2+2 2 | 2+2 2 |
| | 5+6 2 | 3+4 2 | 3+4 2 | 3+4 2 | 3+3 2 | 3+3 2 |
| | 3+4 2 | 3+4 2 | 5+6 2 | 5+6 2 | 3+3 2 | 2+2 2 |
| | 3+4 2 | 5+6 2 | 3+4 2 | 5+6 2 | 3+3 2 | 3+3 2 |
| | 5+6 2 | 5+6 2 | 3+4 2 | 5+6 2 | 2+2 2 | 3+3 2 |
| | 5+6 2 | 3+4 2 | 5+6 2 | 5+6 2 | 2+2 2 | 2+2 2 |
| | 5+6 2 | 5+6 2 | 5+6 2 | 3+4 2 | 3+3 2 | 2+2 2 |
| | 3+4 2 | 3+4 2 | 5+6 2 | 5+6 2 | 3+3 2 | 3+3 2 |
| | 5+6 2 | 5+6 2 | 3+4 2 | 3+4 2 | 2+2 2 | 2+2 2 |
| | 3+4 2 | 3+4 2 | 5+6 2 | 5+6 2 | 2+2 2 | 2+2 2 |
| | 5+6 2 | 5+6 2 | 3+4 2 | 3+4 2 | 3+3 2 | 3+3 2 |
| | 3+4 2 | 3+4 2 | 3+4 2 | 3+4 2 | 2+2 2 | 3+3 2 |
| | 5+6 2 | 5+6 2 | 5+6 2 | 5+6 2 | 3+3 2 | 2+2 2 |
| | 3+4 2 | 3+4 2 | 5+6 2 | 3+4 2 | 3+3 2 | 3+3 2 |
| | 5+6 2 | 5+6 2 | 3+4 2 | 5+6 2 | 2+2 2 | 2+2 2 |
| | 3+4 2 | 3+4 2 | 5+6 2 | 5+6 2 | 3+3 2 | 3+3 2 |
| | 3+4 2 | 3+4 2 | 3+4 2 | 3+4 2 | 2+2 2 | 2+2 2 |
| | 5+6 2 | 5+6 2 | 5+6 2 | 5+6 2 | 3+3 2 | 3+3 2 |
| | 3+4 2 | 5+6 2 | 3+4 2 | 3+4 2 | 2+2 2 | 2+2 2 |
| 12.2.105 | 427664 | | 118 | | Cl1Er1O4W1 | |
| | -A | -GM | -M | -Y | -L | -V |

Continued on next column

**TABLE XXIV– continued**

| TQC | ICSD | | NoE | | SF | |
|---|---|---|---|---|---|---|
| | 3+4 2 | 3+4 2 | 5+6 2 | 5+6 2 | 2+2 2 | 2+2 2 |
| | 5+6 2 | 5+6 2 | 3+4 2 | 3+4 2 | 3+3 2 | 3+3 2 |
| | 3+4 2 | 3+4 2 | 5+6 2 | 5+6 2 | 3+3 2 | 3+3 2 |
| | 5+6 2 | 5+6 2 | 3+4 2 | 3+4 2 | 2+2 2 | 2+2 2 |
| | 3+4 2 | 3+4 2 | 5+6 2 | 5+6 2 | 2+2 2 | 2+2 2 |
| | 5+6 2 | 5+6 2 | 3+4 2 | 3+4 2 | 3+3 2 | 3+3 2 |
| | 3+4 2 | 3+4 2 | 3+4 2 | 3+4 2 | 3+3 2 | 3+3 2 |
| | 5+6 2 | 5+6 2 | 5+6 2 | 5+6 2 | 2+2 2 | 2+2 2 |
| | 3+4 2 | 3+4 2 | 5+6 2 | 5+6 2 | 2+2 2 | 2+2 2 |
| | 5+6 2 | 5+6 2 | 3+4 2 | 3+4 2 | 3+3 2 | 3+3 2 |
| | 3+4 2 | 5+6 2 | 5+6 2 | 3+4 2 | 2+2 2 | 3+3 2 |
| | 5+6 2 | 3+4 2 | 3+4 2 | 5+6 2 | 3+3 2 | 2+2 2 |
| | 5+6 2 | 3+4 2 | 3+4 2 | 3+4 2 | 2+2 2 | 3+3 2 |
| | 3+4 2 | 5+6 2 | 5+6 2 | 5+6 2 | 3+3 2 | 2+2 2 |
| | 5+6 2 | 5+6 2 | 3+4 2 | 5+6 2 | 3+3 2 | 2+2 2 |
| | 3+4 2 | 3+4 2 | 5+6 2 | 3+4 2 | 2+2 2 | 3+3 2 |
| | 3+4 2 | 3+4 2 | 3+4 2 | 3+4 2 | 3+3 2 | 3+3 2 |
| | 5+6 2 | 5+6 2 | 5+6 2 | 5+6 2 | 2+2 2 | 2+2 2 |
| | 5+6 2 | 3+4 2 | 3+4 2 | 3+4 2 | 2+2 2 | 2+2 2 |
| | 3+4 2 | 5+6 2 | 5+6 2 | 3+4 2 | 3+3 2 | 3+3 2 |
| | 3+4 2 | 3+4 2 | 5+6 2 | 5+6 2 | 2+2 2 | 2+2 2 |
| | 5+6 2 | 3+4 2 | 3+4 2 | 5+6 2 | 3+3 2 | 2+2 2 |
| | 3+4 2 | 5+6 2 | 5+6 2 | 5+6 2 | 3+3 2 | 3+3 2 |
| | 5+6 2 | 3+4 2 | 3+4 2 | 5+6 2 | 2+2 2 | 3+3 2 |
| | 5+6 2 | 5+6 2 | 3+4 2 | 5+6 2 | 3+3 2 | 2+2 2 |
| | 3+4 2 | 3+4 2 | 5+6 2 | 3+4 2 | 2+2 2 | 3+3 2 |
| | 5+6 2 | 5+6 2 | 5+6 2 | 3+4 2 | 3+3 2 | 3+3 2 |
| | 3+4 2 | 3+4 2 | 5+6 2 | 5+6 2 | 2+2 2 | 2+2 2 |
| | 3+4 2 | 5+6 2 | 3+4 2 | 3+4 2 | 3+3 2 | 2+2 2 |
| | 5+6 2 | 5+6 2 | 3+4 2 | 3+4 2 | 2+2 2 | 3+3 2 |
| | 3+4 2 | 3+4 2 | 3+4 2 | 3+4 2 | 2+2 2 | 2+2 2 |
| | 3+4 2 | 3+4 2 | 5+6 2 | 5+6 2 | 3+3 2 | 3+3 2 |
| | 5+6 2 | 5+6 2 | 5+6 2 | 3+4 2 | 2+2 2 | 2+2 2 |
| | 3+4 2 | 3+4 2 | 3+4 2 | 5+6 2 | 3+3 2 | 3+3 2 |
| | 5+6 2 | 5+6 2 | 3+4 2 | 3+4 2 | 2+2 2 | 3+3 2 |
| | 5+6 2 | 5+6 2 | 5+6 2 | 3+4 2 | 3+3 2 | 2+2 2 |
| | 3+4 2 | 5+6 2 | 3+4 2 | 5+6 2 | 2+2 2 | 3+3 2 |
| | 3+4 2 | 3+4 2 | 5+6 2 | 5+6 2 | 3+3 2 | 3+3 2 |
| | 3+4 2 | 5+6 2 | 3+4 2 | 3+4 2 | 2+2 2 | 2+2 2 |
| | 5+6 2 | 3+4 2 | 5+6 2 | 3+4 2 | 2+2 2 | 2+2 2 |
| | 5+6 2 | 3+4 2 | 3+4 2 | 3+4 2 | 3+3 2 | 3+3 2 |
| | 3+4 2 | 3+4 2 | 3+4 2 | 5+6 2 | 3+3 2 | 2+2 2 |
| | 3+4 2 | 5+6 2 | 5+6 2 | 5+6 2 | 2+2 2 | 3+3 2 |
| | 5+6 2 | 5+6 2 | 3+4 2 | 5+6 2 | 3+3 2 | 3+3 2 |
| | 5+6 2 | 5+6 2 | 5+6 2 | 3+4 2 | 2+2 2 | 2+2 2 |
| | 3+4 2 | 3+4 2 | 5+6 2 | 5+6 2 | 3+3 2 | 2+2 2 |
| | 5+6 2 | 3+4 2 | 5+6 2 | 5+6 2 | 3+3 2 | 3+3 2 |
| | 3+4 2 | 5+6 2 | 3+4 2 | 5+6 2 | 2+2 2 | 2+2 2 |
| | 5+6 2 | 3+4 2 | 3+4 2 | 3+4 2 | 2+2 2 | 3+3 2 |
| | 3+4 2 | 5+6 2 | 5+6 2 | 3+4 2 | 3+3 2 | 2+2 2 |
| | 5+6 2 | 3+4 2 | 5+6 2 | 5+6 2 | 2+2 2 | 3+3 2 |
| | 5+6 2 | 5+6 2 | 3+4 2 | 3+4 2 | 3+3 2 | 2+2 2 |
| | 3+4 2 | 3+4 2 | 5+6 2 | 3+4 2 | 3+3 2 | 3+3 2 |
| | 5+6 2 | 5+6 2 | 3+4 2 | 5+6 2 | 2+2 2 | 2+2 2 |
| | 3+4 2 | 3+4 2 | 5+6 2 | 5+6 2 | 3+3 2 | 3+3 2 |
| | 3+4 2 | 3+4 2 | 3+4 2 | 3+4 2 | 2+2 2 | 2+2 2 |
| | 5+6 2 | 5+6 2 | 5+6 2 | 5+6 2 | 3+3 2 | 3+3 2 |
| | 3+4 2 | 5+6 2 | 3+4 2 | 3+4 2 | 2+2 2 | 2+2 2 |
| | 5+6 2 | 3+4 2 | 5+6 2 | 3+4 2 | 2+2 2 | 3+3 2 |
| 12.2.106 | 420053 | | 118 | | Cl1Er1Mo1O4 | |
| | -A | -GM | -M | -Y | -L | -V |

Continued on next column

**TABLE XXIV– continued**

| TQC | ICSD | NoE | SF | | | |
|---|---|---|---|---|---|---|
| | 3+4 2 | 3+4 2 | 5+6 2 | 5+6 2 | 2+2 2 | 2+2 2 |
| | 5+6 2 | 5+6 2 | 3+4 2 | 3+4 2 | 3+3 2 | 3+3 2 |
| | 3+4 2 | 3+4 2 | 5+6 2 | 5+6 2 | 3+3 2 | 3+3 2 |
| | 5+6 2 | 5+6 2 | 3+4 2 | 3+4 2 | 2+2 2 | 2+2 2 |
| | 3+4 2 | 3+4 2 | 5+6 2 | 5+6 2 | 3+3 2 | 3+3 2 |
| | 5+6 2 | 5+6 2 | 3+4 2 | 3+4 2 | 2+2 2 | 2+2 2 |
| | 3+4 2 | 3+4 2 | 3+4 2 | 3+4 2 | 3+3 2 | 3+3 2 |
| | 5+6 2 | 5+6 2 | 5+6 2 | 5+6 2 | 2+2 2 | 2+2 2 |
| | 3+4 2 | 3+4 2 | 5+6 2 | 5+6 2 | 2+2 2 | 2+2 2 |
| | 5+6 2 | 5+6 2 | 3+4 2 | 3+4 2 | 3+3 2 | 3+3 2 |
| | 3+4 2 | 5+6 2 | 5+6 2 | 3+4 2 | 2+2 2 | 2+2 2 |
| | 5+6 2 | 3+4 2 | 3+4 2 | 5+6 2 | 3+3 2 | 3+3 2 |
| | 5+6 2 | 5+6 2 | 3+4 2 | 3+4 2 | 2+2 2 | 3+3 2 |
| | 3+4 2 | 3+4 2 | 5+6 2 | 5+6 2 | 3+3 2 | 2+2 2 |
| | 5+6 2 | 5+6 2 | 3+4 2 | 3+4 2 | 3+3 2 | 3+3 2 |
| | 3+4 2 | 3+4 2 | 5+6 2 | 5+6 2 | 2+2 2 | 2+2 2 |
| | 3+4 2 | 3+4 2 | 3+4 2 | 3+4 2 | 3+3 2 | 3+3 2 |
| | 5+6 2 | 5+6 2 | 5+6 2 | 5+6 2 | 2+2 2 | 2+2 2 |
| | 5+6 2 | 3+4 2 | 3+4 2 | 3+4 2 | 2+2 2 | 2+2 2 |
| | 3+4 2 | 5+6 2 | 5+6 2 | 3+4 2 | 3+3 2 | 3+3 2 |
| | 3+4 2 | 3+4 2 | 5+6 2 | 5+6 2 | 2+2 2 | 2+2 2 |
| | 5+6 2 | 3+4 2 | 5+6 2 | 5+6 2 | 3+3 2 | 2+2 2 |
| | 3+4 2 | 5+6 2 | 3+4 2 | 5+6 2 | 3+3 2 | 3+3 2 |
| | 5+6 2 | 5+6 2 | 3+4 2 | 5+6 2 | 2+2 2 | 3+3 2 |
| | 5+6 2 | 3+4 2 | 5+6 2 | 5+6 2 | 3+3 2 | 2+2 2 |
| | 3+4 2 | 5+6 2 | 3+4 2 | 3+4 2 | 2+2 2 | 3+3 2 |
| | 5+6 2 | 5+6 2 | 5+6 2 | 3+4 2 | 3+3 2 | 3+3 2 |
| | 3+4 2 | 3+4 2 | 3+4 2 | 3+4 2 | 2+2 2 | 2+2 2 |
| | 3+4 2 | 3+4 2 | 5+6 2 | 5+6 2 | 3+3 2 | 2+2 2 |
| | 5+6 2 | 3+4 2 | 3+4 2 | 3+4 2 | 2+2 2 | 3+3 2 |
| | 3+4 2 | 5+6 2 | 3+4 2 | 3+4 2 | 2+2 2 | 2+2 2 |
| | 3+4 2 | 3+4 2 | 5+6 2 | 5+6 2 | 3+3 2 | 3+3 2 |
| | 3+4 2 | 5+6 2 | 5+6 2 | 3+4 2 | 2+2 2 | 2+2 2 |
| | 5+6 2 | 3+4 2 | 3+4 2 | 5+6 2 | 2+2 2 | 3+3 2 |
| | 5+6 2 | 5+6 2 | 3+4 2 | 3+4 2 | 3+3 2 | 3+3 2 |
| | 5+6 2 | 5+6 2 | 5+6 2 | 3+4 2 | 2+2 2 | 2+2 2 |
| | 3+4 2 | 3+4 2 | 3+4 2 | 5+6 2 | 3+3 2 | 3+3 2 |
| | 3+4 2 | 5+6 2 | 5+6 2 | 5+6 2 | 3+3 2 | 3+3 2 |
| | 3+4 2 | 5+6 2 | 3+4 2 | 3+4 2 | 2+2 2 | 2+2 2 |
| | 5+6 2 | 3+4 2 | 5+6 2 | 3+4 2 | 2+2 2 | 2+2 2 |
| | 5+6 2 | 3+4 2 | 3+4 2 | 3+4 2 | 3+3 2 | 3+3 2 |
| | 3+4 2 | 3+4 2 | 3+4 2 | 5+6 2 | 3+3 2 | 2+2 2 |
| | 5+6 2 | 5+6 2 | 5+6 2 | 5+6 2 | 3+3 2 | 3+3 2 |
| | 3+4 2 | 5+6 2 | 3+4 2 | 5+6 2 | 2+2 2 | 3+3 2 |
| | 5+6 2 | 5+6 2 | 5+6 2 | 5+6 2 | 2+2 2 | 2+2 2 |
| | 3+4 2 | 3+4 2 | 5+6 2 | 3+4 2 | 3+3 2 | 2+2 2 |
| | 5+6 2 | 3+4 2 | 5+6 2 | 5+6 2 | 3+3 2 | 3+3 2 |
| | 3+4 2 | 5+6 2 | 3+4 2 | 5+6 2 | 2+2 2 | 2+2 2 |
| | 5+6 2 | 3+4 2 | 3+4 2 | 3+4 2 | 2+2 2 | 3+3 2 |
| | 3+4 2 | 5+6 2 | 5+6 2 | 3+4 2 | 3+3 2 | 2+2 2 |
| | 5+6 2 | 3+4 2 | 5+6 2 | 5+6 2 | 2+2 2 | 3+3 2 |
| | 5+6 2 | 5+6 2 | 3+4 2 | 3+4 2 | 3+3 2 | 2+2 2 |
| | 3+4 2 | 3+4 2 | 5+6 2 | 3+4 2 | 3+3 2 | 3+3 2 |
| | 5+6 2 | 5+6 2 | 3+4 2 | 5+6 2 | 2+2 2 | 2+2 2 |
| | 3+4 2 | 3+4 2 | 5+6 2 | 5+6 2 | 3+3 2 | 3+3 2 |
| | 5+6 2 | 3+4 2 | 3+4 2 | 3+4 2 | 2+2 2 | 2+2 2 |
| | 3+4 2 | 5+6 2 | 5+6 2 | 3+4 2 | 3+3 2 | 3+3 2 |
| | 3+4 2 | 5+6 2 | 3+4 2 | 5+6 2 | 2+2 2 | 2+2 2 |
| | 5+6 2 | 3+4 2 | 5+6 2 | 3+4 2 | 2+2 2 | 3+3 2 |
| 12.2.107 | 424718 | 124 | Ce5Ge8Mg8 | | | |
| | -A | -GM | -M | -Y | -L | -V |

Continued on next column

**TABLE XXIV– continued**

| TQC | ICSD | NoE | SF | | | |
|---|---|---|---|---|---|---|
| | 3+4 2 | 3+4 2 | 3+4 2 | 3+4 2 | 2+2 2 | 2+2 2 |
| | 5+6 2 | 3+4 2 | 5+6 2 | 3+4 2 | 2+2 2 | 2+2 2 |
| | 3+4 2 | 5+6 2 | 3+4 2 | 5+6 2 | 3+3 2 | 3+3 2 |
| | 5+6 2 | 3+4 2 | 5+6 2 | 3+4 2 | 2+2 2 | 3+3 2 |
| | 3+4 2 | 5+6 2 | 3+4 2 | 5+6 2 | 3+3 2 | 2+2 2 |
| | 5+6 2 | 5+6 2 | 5+6 2 | 5+6 2 | 3+3 2 | 3+3 2 |
| | 3+4 2 | 3+4 2 | 3+4 2 | 3+4 2 | 3+3 2 | 2+2 2 |
| | 5+6 2 | 5+6 2 | 5+6 2 | 5+6 2 | 2+2 2 | 3+3 2 |
| | 5+6 2 | 3+4 2 | 5+6 2 | 3+4 2 | 2+2 2 | 3+3 2 |
| | 3+4 2 | 5+6 2 | 3+4 2 | 5+6 2 | 3+3 2 | 2+2 2 |
| | 5+6 2 | 3+4 2 | 5+6 2 | 3+4 2 | 2+2 2 | 3+3 2 |
| | 5+6 2 | 5+6 2 | 5+6 2 | 5+6 2 | 3+3 2 | 3+3 2 |
| | 3+4 2 | 5+6 2 | 3+4 2 | 5+6 2 | 3+3 2 | 2+2 2 |
| | 5+6 2 | 3+4 2 | 5+6 2 | 3+4 2 | 3+3 2 | 3+3 2 |
| | 5+6 2 | 5+6 2 | 5+6 2 | 5+6 2 | 2+2 2 | 2+2 2 |
| | 3+4 2 | 3+4 2 | 3+4 2 | 3+4 2 | 3+3 2 | 3+3 2 |
| | 5+6 2 | 5+6 2 | 3+4 2 | 5+6 2 | 2+2 2 | 2+2 2 |
| | 3+4 2 | 5+6 2 | 5+6 2 | 5+6 2 | 3+3 2 | 3+3 2 |
| | 5+6 2 | 3+4 2 | 5+6 2 | 3+4 2 | 2+2 2 | 3+3 2 |
| | 3+4 2 | 5+6 2 | 3+4 2 | 5+6 2 | 3+3 2 | 2+2 2 |
| | 5+6 2 | 3+4 2 | 3+4 2 | 5+6 2 | 2+2 2 | 2+2 2 |
| | 3+4 2 | 5+6 2 | 5+6 2 | 3+4 2 | 3+3 2 | 3+3 2 |
| | 3+4 2 | 3+4 2 | 3+4 2 | 5+6 2 | 3+3 2 | 2+2 2 |
| | 5+6 2 | 3+4 2 | 5+6 2 | 3+4 2 | 2+2 2 | 3+3 2 |
| | 3+4 2 | 5+6 2 | 5+6 2 | 5+6 2 | 2+2 2 | 3+3 2 |
| | 5+6 2 | 3+4 2 | 3+4 2 | 3+4 2 | 3+3 2 | 2+2 2 |
| | 5+6 2 | 5+6 2 | 5+6 2 | 3+4 2 | 3+3 2 | 2+2 2 |
| | 3+4 2 | 5+6 2 | 3+4 2 | 5+6 2 | 2+2 2 | 2+2 2 |
| | 5+6 2 | 5+6 2 | 3+4 2 | 3+4 2 | 3+3 2 | 3+3 2 |
| | 3+4 2 | 3+4 2 | 5+6 2 | 5+6 2 | 2+2 2 | 3+3 2 |
| | 5+6 2 | 5+6 2 | 5+6 2 | 3+4 2 | 3+3 2 | 2+2 2 |
| | 3+4 2 | 3+4 2 | 3+4 2 | 5+6 2 | 2+2 2 | 3+3 2 |
| | 3+4 2 | 3+4 2 | 5+6 2 | 3+4 2 | 3+3 2 | 2+2 2 |
| | 5+6 2 | 5+6 2 | 3+4 2 | 5+6 2 | 2+2 2 | 3+3 2 |
| | 3+4 2 | 3+4 2 | 5+6 2 | 3+4 2 | 3+3 2 | 2+2 2 |
| | 5+6 2 | 5+6 2 | 3+4 2 | 5+6 2 | 2+2 2 | 3+3 2 |
| | 5+6 2 | 3+4 2 | 3+4 2 | 5+6 2 | 3+3 2 | 3+3 2 |
| | 5+6 2 | 3+4 2 | 5+6 2 | 3+4 2 | 2+2 2 | 2+2 2 |
| | 3+4 2 | 5+6 2 | 3+4 2 | 5+6 2 | 2+2 2 | 2+2 2 |
| | 3+4 2 | 3+4 2 | 5+6 2 | 3+4 2 | 3+3 2 | 3+3 2 |
| | 3+4 2 | 5+6 2 | 3+4 2 | 5+6 2 | 2+2 2 | 2+2 2 |
| | 5+6 2 | 3+4 2 | 3+4 2 | 5+6 2 | 3+3 2 | 2+2 2 |
| | 3+4 2 | 3+4 2 | 3+4 2 | 3+4 2 | 3+3 2 | 3+3 2 |
| | 5+6 2 | 3+4 2 | 5+6 2 | 5+6 2 | 2+2 2 | 3+3 2 |
| | 3+4 2 | 5+6 2 | 5+6 2 | 5+6 2 | 3+3 2 | 2+2 2 |
| | 3+4 2 | 5+6 2 | 5+6 2 | 3+4 2 | 2+2 2 | 2+2 2 |
| | 5+6 2 | 3+4 2 | 3+4 2 | 3+4 2 | 2+2 2 | 3+3 2 |
| | 5+6 2 | 5+6 2 | 5+6 2 | 3+4 2 | 3+3 2 | 2+2 2 |
| | 5+6 2 | 5+6 2 | 3+4 2 | 3+4 2 | 2+2 2 | 3+3 2 |
| | 3+4 2 | 3+4 2 | 3+4 2 | 5+6 2 | 2+2 2 | 3+3 2 |
| | 5+6 2 | 3+4 2 | 3+4 2 | 5+6 2 | 3+3 2 | 2+2 2 |
| | 3+4 2 | 5+6 2 | 5+6 2 | 3+4 2 | 3+3 2 | 3+3 2 |
| | 3+4 2 | 5+6 2 | 5+6 2 | 3+4 2 | 3+3 2 | 2+2 2 |
| | 5+6 2 | 3+4 2 | 5+6 2 | 3+4 2 | 2+2 2 | 3+3 2 |
| | 3+4 2 | 5+6 2 | 5+6 2 | 5+6 2 | 2+2 2 | 2+2 2 |
| | 5+6 2 | 3+4 2 | 3+4 2 | 3+4 2 | 3+3 2 | 2+2 2 |
| | 5+6 2 | 5+6 2 | 5+6 2 | 5+6 2 | 2+2 2 | 3+3 2 |
| | 3+4 2 | 5+6 2 | 3+4 2 | 3+4 2 | 3+3 2 | 2+2 2 |
| | 5+6 2 | 3+4 2 | 3+4 2 | 5+6 2 | 3+3 2 | 3+3 2 |
| | 3+4 2 | 3+4 2 | 5+6 2 | 5+6 2 | 2+2 2 | 3+3 2 |
| | 5+6 2 | 5+6 2 | 3+4 2 | 5+6 2 | 3+3 2 | 2+2 2 |

Continued on next column

**TABLE XXIV– continued**

| TQC | ICSD | NoE | SF | | | |
|---|---|---|---|---|---|---|
| | 5+6 2 | 3+4 2 | 3+4 2 | 3+4 2 | 2+2 2 | 3+3 2 |
| 12.2.108 | 417665 | 96 | Ce1Ru1Sn1 | | | |
| | -A | -GM | -M | -Y | -L | -V |
| | 3+4 2 | 3+4 2 | 5+6 2 | 5+6 2 | 3+3 2 | 3+3 2 |
| | 5+6 2 | 5+6 2 | 3+4 2 | 3+4 2 | 2+2 2 | 2+2 2 |
| | 5+6 2 | 3+4 2 | 3+4 2 | 5+6 2 | 2+2 2 | 3+3 2 |
| | 3+4 2 | 5+6 2 | 5+6 2 | 3+4 2 | 3+3 2 | 2+2 2 |
| | 3+4 2 | 3+4 2 | 5+6 2 | 5+6 2 | 3+3 2 | 3+3 2 |
| | 5+6 2 | 5+6 2 | 3+4 2 | 3+4 2 | 2+2 2 | 2+2 2 |
| | 5+6 2 | 3+4 2 | 3+4 2 | 5+6 2 | 2+2 2 | 3+3 2 |
| | 3+4 2 | 5+6 2 | 5+6 2 | 3+4 2 | 3+3 2 | 2+2 2 |
| | 3+4 2 | 3+4 2 | 3+4 2 | 5+6 2 | 3+3 2 | 3+3 2 |
| | 5+6 2 | 3+4 2 | 5+6 2 | 3+4 2 | 2+2 2 | 2+2 2 |
| | 3+4 2 | 5+6 2 | 3+4 2 | 5+6 2 | 2+2 2 | 3+3 2 |
| | 5+6 2 | 3+4 2 | 5+6 2 | 3+4 2 | 3+3 2 | 2+2 2 |
| | 5+6 2 | 5+6 2 | 5+6 2 | 5+6 2 | 3+3 2 | 3+3 2 |
| | 3+4 2 | 3+4 2 | 3+4 2 | 3+4 2 | 2+2 2 | 2+2 2 |
| | 5+6 2 | 5+6 2 | 3+4 2 | 5+6 2 | 2+2 2 | 3+3 2 |
| | 3+4 2 | 5+6 2 | 5+6 2 | 3+4 2 | 3+3 2 | 2+2 2 |
| | 5+6 2 | 3+4 2 | 5+6 2 | 3+4 2 | 3+3 2 | 2+2 2 |
| | 3+4 2 | 5+6 2 | 3+4 2 | 5+6 2 | 2+2 2 | 3+3 2 |
| | 5+6 2 | 3+4 2 | 5+6 2 | 3+4 2 | 3+3 2 | 2+2 2 |
| | 3+4 2 | 5+6 2 | 3+4 2 | 5+6 2 | 2+2 2 | 3+3 2 |
| | 5+6 2 | 5+6 2 | 3+4 2 | 3+4 2 | 3+3 2 | 2+2 2 |
| | 3+4 2 | 3+4 2 | 5+6 2 | 5+6 2 | 2+2 2 | 3+3 2 |
| | 3+4 2 | 3+4 2 | 5+6 2 | 3+4 2 | 3+3 2 | 2+2 2 |
| | 5+6 2 | 5+6 2 | 3+4 2 | 5+6 2 | 2+2 2 | 3+3 2 |
| | 3+4 2 | 3+4 2 | 3+4 2 | 5+6 2 | 2+2 2 | 3+3 2 |
| | 5+6 2 | 3+4 2 | 5+6 2 | 3+4 2 | 2+2 2 | 2+2 2 |
| | 5+6 2 | 5+6 2 | 3+4 2 | 3+4 2 | 3+3 2 | 3+3 2 |
| | 3+4 2 | 5+6 2 | 3+4 2 | 5+6 2 | 3+3 2 | 2+2 2 |
| | 5+6 2 | 3+4 2 | 5+6 2 | 3+4 2 | 2+2 2 | 3+3 2 |
| | 3+4 2 | 5+6 2 | 5+6 2 | 3+4 2 | 3+3 2 | 2+2 2 |
| | 5+6 2 | 3+4 2 | 3+4 2 | 5+6 2 | 3+3 2 | 2+2 2 |
| | 3+4 2 | 5+6 2 | 5+6 2 | 3+4 2 | 2+2 2 | 3+3 2 |
| | 5+6 2 | 5+6 2 | 3+4 2 | 5+6 2 | 3+3 2 | 2+2 2 |
| | 3+4 2 | 3+4 2 | 5+6 2 | 5+6 2 | 2+2 2 | 3+3 2 |
| | 3+4 2 | 5+6 2 | 5+6 2 | 3+4 2 | 2+2 2 | 2+2 2 |
| | 3+4 2 | 3+4 2 | 5+6 2 | 3+4 2 | 3+3 2 | 3+3 2 |
| | 5+6 2 | 5+6 2 | 3+4 2 | 3+4 2 | 2+2 2 | 2+2 2 |
| | 5+6 2 | 3+4 2 | 5+6 2 | 5+6 2 | 2+2 2 | 3+3 2 |
| | 3+4 2 | 3+4 2 | 3+4 2 | 5+6 2 | 3+3 2 | 2+2 2 |
| | 5+6 2 | 5+6 2 | 3+4 2 | 5+6 2 | 3+3 2 | 3+3 2 |
| | 5+6 2 | 3+4 2 | 5+6 2 | 3+4 2 | 2+2 2 | 2+2 2 |
| | 5+6 2 | 5+6 2 | 3+4 2 | 5+6 2 | 3+3 2 | 3+3 2 |
| | 3+4 2 | 3+4 2 | 3+4 2 | 3+4 2 | 2+2 2 | 2+2 2 |
| | 5+6 2 | 3+4 2 | 5+6 2 | 5+6 2 | 3+3 2 | 3+3 2 |
| | 3+4 2 | 5+6 2 | 3+4 2 | 3+4 2 | 2+2 2 | 2+2 2 |
| | 3+4 2 | 3+4 2 | 5+6 2 | 5+6 2 | 3+3 2 | 2+2 2 |
| | 5+6 2 | 5+6 2 | 5+6 2 | 3+4 2 | 2+2 2 | 3+3 2 |
| | 3+4 2 | 5+6 2 | 3+4 2 | 3+4 2 | 3+3 2 | 2+2 2 |
| 12.2.109 | 192243 | 144 | Ce1Ru1Sn1 | | | |
| | -A | -GM | -M | -Y | -L | -V |
| | 5+6 2 | 3+4 2 | 3+4 2 | 5+6 2 | 2+2 2 | 3+3 2 |
| | 3+4 2 | 5+6 2 | 5+6 2 | 3+4 2 | 3+3 2 | 2+2 2 |
| | 3+4 2 | 3+4 2 | 3+4 2 | 3+4 2 | 3+3 2 | 3+3 2 |
| | 5+6 2 | 5+6 2 | 5+6 2 | 5+6 2 | 2+2 2 | 2+2 2 |
| | 5+6 2 | 5+6 2 | 3+4 2 | 3+4 2 | 2+2 2 | 2+2 2 |
| | 3+4 2 | 3+4 2 | 5+6 2 | 5+6 2 | 3+3 2 | 3+3 2 |
| | 5+6 2 | 3+4 2 | 3+4 2 | 5+6 2 | 2+2 2 | 3+3 2 |
| | 3+4 2 | 5+6 2 | 5+6 2 | 3+4 2 | 3+3 2 | 2+2 2 |

Continued on next column

**TABLE XXIV– continued**

| TQC | ICSD | NoE | SF | | | |
|---|---|---|---|---|---|---|
| | 3+4 2 | 3+4 2 | 3+4 2 | 3+4 2 | 3+3 2 | 3+3 2 |
| | 5+6 2 | 5+6 2 | 5+6 2 | 5+6 2 | 2+2 2 | 2+2 2 |
| | 3+4 2 | 3+4 2 | 5+6 2 | 5+6 2 | 3+3 2 | 3+3 2 |
| | 5+6 2 | 5+6 2 | 3+4 2 | 3+4 2 | 2+2 2 | 2+2 2 |
| | 5+6 2 | 3+4 2 | 3+4 2 | 3+4 2 | 2+2 2 | 3+3 2 |
| | 3+4 2 | 3+4 2 | 5+6 2 | 5+6 2 | 3+3 2 | 2+2 2 |
| | 5+6 2 | 5+6 2 | 3+4 2 | 3+4 2 | 2+2 2 | 3+3 2 |
| | 3+4 2 | 3+4 2 | 5+6 2 | 5+6 2 | 3+3 2 | 2+2 2 |
| | 5+6 2 | 5+6 2 | 3+4 2 | 5+6 2 | 3+3 2 | 3+3 2 |
| | 5+6 2 | 3+4 2 | 5+6 2 | 3+4 2 | 2+2 2 | 2+2 2 |
| | 3+4 2 | 5+6 2 | 3+4 2 | 5+6 2 | 2+2 2 | 3+3 2 |
| | 3+4 2 | 3+4 2 | 5+6 2 | 3+4 2 | 3+3 2 | 2+2 2 |
| | 5+6 2 | 5+6 2 | 3+4 2 | 5+6 2 | 3+3 2 | 3+3 2 |
| | 3+4 2 | 5+6 2 | 5+6 2 | 3+4 2 | 2+2 2 | 2+2 2 |
| | 3+4 2 | 3+4 2 | 3+4 2 | 3+4 2 | 3+3 2 | 3+3 2 |
| | 5+6 2 | 5+6 2 | 5+6 2 | 5+6 2 | 2+2 2 | 2+2 2 |
| | 5+6 2 | 3+4 2 | 3+4 2 | 3+4 2 | 2+2 2 | 2+2 2 |
| | 3+4 2 | 5+6 2 | 5+6 2 | 5+6 2 | 3+3 2 | 3+3 2 |
| | 3+4 2 | 3+4 2 | 3+4 2 | 3+4 2 | 2+2 2 | 2+2 2 |
| | 5+6 2 | 5+6 2 | 5+6 2 | 5+6 2 | 3+3 2 | 3+3 2 |
| | 3+4 2 | 3+4 2 | 3+4 2 | 3+4 2 | 3+3 2 | 2+2 2 |
| | 5+6 2 | 5+6 2 | 5+6 2 | 5+6 2 | 2+2 2 | 3+3 2 |
| | 3+4 2 | 5+6 2 | 3+4 2 | 3+4 2 | 2+2 2 | 2+2 2 |
| | 5+6 2 | 3+4 2 | 5+6 2 | 5+6 2 | 3+3 2 | 3+3 2 |
| | 3+4 2 | 3+4 2 | 5+6 2 | 3+4 2 | 2+2 2 | 2+2 2 |
| | 5+6 2 | 3+4 2 | 3+4 2 | 5+6 2 | 3+3 2 | 3+3 2 |
| | 3+4 2 | 5+6 2 | 3+4 2 | 3+4 2 | 2+2 2 | 2+2 2 |
| | 5+6 2 | 3+4 2 | 5+6 2 | 5+6 2 | 3+3 2 | 3+3 2 |
| | 5+6 2 | 3+4 2 | 3+4 2 | 3+4 2 | 3+3 2 | 3+3 2 |
| | 3+4 2 | 5+6 2 | 5+6 2 | 5+6 2 | 3+3 2 | 2+2 2 |
| | 5+6 2 | 5+6 2 | 3+4 2 | 3+4 2 | 2+2 2 | 2+2 2 |
| | 3+4 2 | 5+6 2 | 5+6 2 | 5+6 2 | 2+2 2 | 3+3 2 |
| | 5+6 2 | 5+6 2 | 5+6 2 | 3+4 2 | 2+2 2 | 3+3 2 |
| | 5+6 2 | 3+4 2 | 3+4 2 | 5+6 2 | 2+2 2 | 2+2 2 |
| | 3+4 2 | 5+6 2 | 5+6 2 | 3+4 2 | 3+3 2 | 2+2 2 |
| | 3+4 2 | 3+4 2 | 3+4 2 | 3+4 2 | 3+3 2 | 3+3 2 |
| | 3+4 2 | 3+4 2 | 5+6 2 | 3+4 2 | 3+3 2 | 2+2 2 |
| | 5+6 2 | 5+6 2 | 3+4 2 | 5+6 2 | 2+2 2 | 3+3 2 |
| | 3+4 2 | 3+4 2 | 5+6 2 | 5+6 2 | 3+3 2 | 2+2 2 |
| | 5+6 2 | 5+6 2 | 3+4 2 | 5+6 2 | 2+2 2 | 3+3 2 |
| | 5+6 2 | 5+6 2 | 5+6 2 | 3+4 2 | 3+3 2 | 3+3 2 |
| | 3+4 2 | 3+4 2 | 3+4 2 | 5+6 2 | 2+2 2 | 2+2 2 |
| | 3+4 2 | 5+6 2 | 5+6 2 | 3+4 2 | 3+3 2 | 3+3 2 |
| | 5+6 2 | 5+6 2 | 3+4 2 | 5+6 2 | 2+2 2 | 2+2 2 |
| | 5+6 2 | 3+4 2 | 3+4 2 | 3+4 2 | 3+3 2 | 2+2 2 |
| | 3+4 2 | 3+4 2 | 5+6 2 | 5+6 2 | 2+2 2 | 3+3 2 |
| | 5+6 2 | 3+4 2 | 5+6 2 | 3+4 2 | 3+3 2 | 2+2 2 |
| | 3+4 2 | 5+6 2 | 5+6 2 | 5+6 2 | 2+2 2 | 2+2 2 |
| | 5+6 2 | 3+4 2 | 3+4 2 | 5+6 2 | 3+3 2 | 3+3 2 |
| | 3+4 2 | 5+6 2 | 3+4 2 | 3+4 2 | 2+2 2 | 3+3 2 |
| | 5+6 2 | 3+4 2 | 3+4 2 | 3+4 2 | 3+3 2 | 2+2 2 |
| | 3+4 2 | 5+6 2 | 5+6 2 | 5+6 2 | 2+2 2 | 3+3 2 |
| | 3+4 2 | 3+4 2 | 5+6 2 | 3+4 2 | 2+2 2 | 2+2 2 |
| | 5+6 2 | 5+6 2 | 3+4 2 | 5+6 2 | 3+3 2 | 2+2 2 |
| | 3+4 2 | 5+6 2 | 3+4 2 | 3+4 2 | 2+2 2 | 3+3 2 |
| | 3+4 2 | 3+4 2 | 5+6 2 | 5+6 2 | 2+2 2 | 2+2 2 |
| | 5+6 2 | 3+4 2 | 5+6 2 | 5+6 2 | 3+3 2 | 3+3 2 |
| | 5+6 2 | 3+4 2 | 5+6 2 | 3+4 2 | 3+3 2 | 2+2 2 |
| | 3+4 2 | 5+6 2 | 3+4 2 | 5+6 2 | 2+2 2 | 3+3 2 |
| | 5+6 2 | 3+4 2 | 3+4 2 | 3+4 2 | 3+3 2 | 3+3 2 |
| | 3+4 2 | 3+4 2 | 5+6 2 | 5+6 2 | 3+3 2 | 2+2 2 |

Continued on next column

**TABLE XXIV– continued**

| TQC | ICSD | | NoE | | SF | |
|---|---|---|---|---|---|---|
| | 3+4 2 | 5+6 2 | 5+6 2 | 3+4 2 | 2+2 2 | 2+2 2 |
| | 5+6 2 | 3+4 2 | 5+6 2 | 3+4 2 | 3+3 2 | 3+3 2 |
| | 3+4 2 | 5+6 2 | 3+4 2 | 5+6 2 | 2+2 2 | 2+2 2 |
| 12.2.110 | 192242 | | 96 | | Ce1Ru1Sn1 | |
| | -A | -GM | -M | -Y | -L | -V |
| | 3+4 2 | 3+4 2 | 5+6 2 | 5+6 2 | 3+3 2 | 3+3 2 |
| | 5+6 2 | 5+6 2 | 3+4 2 | 3+4 2 | 2+2 2 | 2+2 2 |
| | 5+6 2 | 3+4 2 | 3+4 2 | 5+6 2 | 2+2 2 | 3+3 2 |
| | 3+4 2 | 5+6 2 | 5+6 2 | 3+4 2 | 3+3 2 | 2+2 2 |
| | 3+4 2 | 3+4 2 | 5+6 2 | 5+6 2 | 3+3 2 | 3+3 2 |
| | 5+6 2 | 5+6 2 | 3+4 2 | 3+4 2 | 2+2 2 | 2+2 2 |
| | 5+6 2 | 3+4 2 | 3+4 2 | 5+6 2 | 2+2 2 | 3+3 2 |
| | 3+4 2 | 5+6 2 | 5+6 2 | 3+4 2 | 3+3 2 | 2+2 2 |
| | 3+4 2 | 3+4 2 | 3+4 2 | 5+6 2 | 3+3 2 | 3+3 2 |
| | 5+6 2 | 3+4 2 | 5+6 2 | 3+4 2 | 2+2 2 | 2+2 2 |
| | 3+4 2 | 5+6 2 | 3+4 2 | 5+6 2 | 2+2 2 | 3+3 2 |
| | 5+6 2 | 3+4 2 | 5+6 2 | 3+4 2 | 3+3 2 | 2+2 2 |
| | 5+6 2 | 3+4 2 | 5+6 2 | 5+6 2 | 3+3 2 | 3+3 2 |
| | 3+4 2 | 5+6 2 | 3+4 2 | 3+4 2 | 2+2 2 | 2+2 2 |
| | 5+6 2 | 5+6 2 | 3+4 2 | 5+6 2 | 2+2 2 | 3+3 2 |
| | 3+4 2 | 5+6 2 | 5+6 2 | 3+4 2 | 3+3 2 | 2+2 2 |
| | 5+6 2 | 3+4 2 | 5+6 2 | 3+4 2 | 3+3 2 | 2+2 2 |
| | 3+4 2 | 5+6 2 | 3+4 2 | 5+6 2 | 2+2 2 | 3+3 2 |
| | 5+6 2 | 3+4 2 | 5+6 2 | 3+4 2 | 3+3 2 | 2+2 2 |
| | 3+4 2 | 5+6 2 | 3+4 2 | 5+6 2 | 2+2 2 | 3+3 2 |
| | 5+6 2 | 5+6 2 | 3+4 2 | 3+4 2 | 3+3 2 | 2+2 2 |
| | 3+4 2 | 3+4 2 | 5+6 2 | 5+6 2 | 2+2 2 | 3+3 2 |
| | 3+4 2 | 3+4 2 | 5+6 2 | 3+4 2 | 3+3 2 | 3+3 2 |
| | 5+6 2 | 3+4 2 | 3+4 2 | 5+6 2 | 2+2 2 | 2+2 2 |
| | 3+4 2 | 5+6 2 | 3+4 2 | 5+6 2 | 2+2 2 | 3+3 2 |
| | 5+6 2 | 3+4 2 | 5+6 2 | 3+4 2 | 2+2 2 | 2+2 2 |
| | 5+6 2 | 5+6 2 | 3+4 2 | 3+4 2 | 3+3 2 | 3+3 2 |
| | 5+6 2 | 5+6 2 | 3+4 2 | 5+6 2 | 3+3 2 | 2+2 2 |
| | 3+4 2 | 3+4 2 | 5+6 2 | 3+4 2 | 2+2 2 | 3+3 2 |
| | 3+4 2 | 5+6 2 | 5+6 2 | 3+4 2 | 3+3 2 | 2+2 2 |
| | 5+6 2 | 3+4 2 | 3+4 2 | 5+6 2 | 3+3 2 | 2+2 2 |
| | 3+4 2 | 5+6 2 | 5+6 2 | 5+6 2 | 2+2 2 | 2+2 2 |
| | 5+6 2 | 5+6 2 | 3+4 2 | 3+4 2 | 3+3 2 | 3+3 2 |
| | 3+4 2 | 3+4 2 | 5+6 2 | 5+6 2 | 2+2 2 | 3+3 2 |
| | 3+4 2 | 5+6 2 | 5+6 2 | 3+4 2 | 2+2 2 | 2+2 2 |
| | 5+6 2 | 5+6 2 | 3+4 2 | 3+4 2 | 3+3 2 | 3+3 2 |
| | 3+4 2 | 3+4 2 | 5+6 2 | 3+4 2 | 2+2 2 | 2+2 2 |
| | 5+6 2 | 3+4 2 | 5+6 2 | 5+6 2 | 2+2 2 | 3+3 2 |
| | 3+4 2 | 3+4 2 | 3+4 2 | 5+6 2 | 3+3 2 | 2+2 2 |
| | 5+6 2 | 5+6 2 | 3+4 2 | 5+6 2 | 3+3 2 | 3+3 2 |
| | 5+6 2 | 3+4 2 | 5+6 2 | 3+4 2 | 2+2 2 | 2+2 2 |
| | 5+6 2 | 5+6 2 | 3+4 2 | 5+6 2 | 3+3 2 | 3+3 2 |
| | 3+4 2 | 3+4 2 | 3+4 2 | 3+4 2 | 3+3 2 | 2+2 2 |
| | 5+6 2 | 3+4 2 | 5+6 2 | 3+4 2 | 2+2 2 | 3+3 2 |
| | 3+4 2 | 3+4 2 | 3+4 2 | 5+6 2 | 2+2 2 | 2+2 2 |
| | 3+4 2 | 5+6 2 | 5+6 2 | 5+6 2 | 3+3 2 | 2+2 2 |
| | 5+6 2 | 5+6 2 | 5+6 2 | 3+4 2 | 2+2 2 | 3+3 2 |
| | 3+4 2 | 5+6 2 | 3+4 2 | 3+4 2 | 3+3 2 | 2+2 2 |
| 12.2.111 | 189952 | | 144 | | Ce1Ru1Sn1 | |
| | -A | -GM | -M | -Y | -L | -V |
| | 3+4 2 | 3+4 2 | 5+6 2 | 5+6 2 | 2+2 2 | 2+2 2 |
| | 5+6 2 | 5+6 2 | 3+4 2 | 3+4 2 | 3+3 2 | 3+3 2 |
| | 5+6 2 | 3+4 2 | 5+6 2 | 3+4 2 | 3+3 2 | 2+2 2 |
| | 3+4 2 | 5+6 2 | 3+4 2 | 5+6 2 | 2+2 2 | 3+3 2 |
| | 3+4 2 | 5+6 2 | 5+6 2 | 3+4 2 | 2+2 2 | 3+3 2 |
| | 5+6 2 | 3+4 2 | 3+4 2 | 5+6 2 | 3+3 2 | 2+2 2 |
| | 3+4 2 | 3+4 2 | 5+6 2 | 5+6 2 | 2+2 2 | 2+2 2 |
| | 5+6 2 | 5+6 2 | 3+4 2 | 3+4 2 | 3+3 2 | 3+3 2 |
| | 3+4 2 | 3+4 2 | 5+6 2 | 3+4 2 | 3+3 2 | 2+2 2 |
| | 5+6 2 | 5+6 2 | 3+4 2 | 5+6 2 | 2+2 2 | 3+3 2 |
| | 5+6 2 | 3+4 2 | 3+4 2 | 5+6 2 | 3+3 2 | 2+2 2 |
| | 3+4 2 | 5+6 2 | 5+6 2 | 3+4 2 | 2+2 2 | 3+3 2 |
| | 3+4 2 | 3+4 2 | 5+6 2 | 3+4 2 | 2+2 2 | 2+2 2 |
| | 3+4 2 | 3+4 2 | 3+4 2 | 5+6 2 | 3+3 2 | 3+3 2 |
| | 5+6 2 | 5+6 2 | 5+6 2 | 3+4 2 | 2+2 2 | 2+2 2 |
| | 5+6 2 | 3+4 2 | 3+4 2 | 5+6 2 | 3+3 2 | 3+3 2 |
| | 3+4 2 | 5+6 2 | 3+4 2 | 5+6 2 | 3+3 2 | 2+2 2 |
| | 5+6 2 | 3+4 2 | 5+6 2 | 3+4 2 | 2+2 2 | 3+3 2 |
| | 3+4 2 | 5+6 2 | 5+6 2 | 5+6 2 | 2+2 2 | 2+2 2 |
| | 5+6 2 | 3+4 2 | 3+4 2 | 3+4 2 | 3+3 2 | 3+3 2 |
| | 5+6 2 | 5+6 2 | 5+6 2 | 5+6 2 | 3+3 2 | 2+2 2 |
| | 3+4 2 | 5+6 2 | 3+4 2 | 3+4 2 | 2+2 2 | 3+3 2 |
| | 5+6 2 | 3+4 2 | 5+6 2 | 3+4 2 | 3+3 2 | 2+2 2 |
| | 3+4 2 | 5+6 2 | 3+4 2 | 5+6 2 | 2+2 2 | 3+3 2 |
| | 5+6 2 | 3+4 2 | 5+6 2 | 3+4 2 | 2+2 2 | 3+3 2 |
| | 3+4 2 | 5+6 2 | 3+4 2 | 5+6 2 | 3+3 2 | 2+2 2 |
| | 5+6 2 | 3+4 2 | 5+6 2 | 3+4 2 | 2+2 2 | 3+3 2 |
| | 3+4 2 | 5+6 2 | 3+4 2 | 5+6 2 | 3+3 2 | 2+2 2 |
| | 5+6 2 | 3+4 2 | 5+6 2 | 5+6 2 | 2+2 2 | 3+3 2 |
| | 3+4 2 | 5+6 2 | 3+4 2 | 3+4 2 | 3+3 2 | 2+2 2 |
| | 5+6 2 | 5+6 2 | 3+4 2 | 3+4 2 | 2+2 2 | 3+3 2 |
| | 3+4 2 | 3+4 2 | 5+6 2 | 5+6 2 | 3+3 2 | 2+2 2 |
| | 5+6 2 | 3+4 2 | 3+4 2 | 3+4 2 | 2+2 2 | 2+2 2 |
| | 3+4 2 | 3+4 2 | 5+6 2 | 5+6 2 | 3+3 2 | 3+3 2 |
| | 3+4 2 | 5+6 2 | 5+6 2 | 3+4 2 | 3+3 2 | 2+2 2 |
| | 3+4 2 | 5+6 2 | 5+6 2 | 3+4 2 | 2+2 2 | 3+3 2 |
| | 5+6 2 | 3+4 2 | 3+4 2 | 5+6 2 | 3+3 2 | 2+2 2 |
| | 5+6 2 | 3+4 2 | 3+4 2 | 5+6 2 | 2+2 2 | 3+3 2 |
| | 3+4 2 | 5+6 2 | 5+6 2 | 3+4 2 | 3+3 2 | 3+3 2 |
| | 5+6 2 | 5+6 2 | 3+4 2 | 3+4 2 | 2+2 2 | 2+2 2 |
| | 3+4 2 | 5+6 2 | 3+4 2 | 3+4 2 | 2+2 2 | 3+3 2 |
| | 5+6 2 | 3+4 2 | 5+6 2 | 5+6 2 | 3+3 2 | 2+2 2 |
| | 5+6 2 | 3+4 2 | 5+6 2 | 5+6 2 | 2+2 2 | 3+3 2 |
| | 3+4 2 | 5+6 2 | 3+4 2 | 3+4 2 | 3+3 2 | 2+2 2 |
| | 5+6 2 | 3+4 2 | 3+4 2 | 5+6 2 | 3+3 2 | 2+2 2 |
| | 3+4 2 | 5+6 2 | 5+6 2 | 3+4 2 | 2+2 2 | 3+3 2 |
| | 5+6 2 | 3+4 2 | 3+4 2 | 5+6 2 | 3+3 2 | 2+2 2 |
| | 3+4 2 | 5+6 2 | 5+6 2 | 5+6 2 | 2+2 2 | 3+3 2 |
| | 3+4 2 | 5+6 2 | 3+4 2 | 3+4 2 | 3+3 2 | 2+2 2 |
| | 3+4 2 | 5+6 2 | 5+6 2 | 5+6 2 | 2+2 2 | 3+3 2 |
| | 5+6 2 | 3+4 2 | 5+6 2 | 3+4 2 | 3+3 2 | 2+2 2 |
| | 3+4 2 | 5+6 2 | 3+4 2 | 5+6 2 | 3+3 2 | 3+3 2 |
| | 5+6 2 | 3+4 2 | 5+6 2 | 3+4 2 | 2+2 2 | 2+2 2 |
| | 5+6 2 | 3+4 2 | 3+4 2 | 3+4 2 | 3+3 2 | 3+3 2 |
| | 3+4 2 | 3+4 2 | 3+4 2 | 5+6 2 | 2+2 2 | 3+3 2 |
| | 5+6 2 | 5+6 2 | 5+6 2 | 5+6 2 | 2+2 2 | 3+3 2 |
| | 3+4 2 | 3+4 2 | 5+6 2 | 3+4 2 | 3+3 2 | 2+2 2 |
| | 5+6 2 | 5+6 2 | 3+4 2 | 5+6 2 | 2+2 2 | 2+2 2 |
| | 3+4 2 | 3+4 2 | 5+6 2 | 3+4 2 | 3+3 2 | 3+3 2 |
| | 5+6 2 | 5+6 2 | 3+4 2 | 5+6 2 | 2+2 2 | 2+2 2 |
| | 5+6 2 | 3+4 2 | 3+4 2 | 5+6 2 | 3+3 2 | 3+3 2 |
| | 3+4 2 | 5+6 2 | 5+6 2 | 3+4 2 | 2+2 2 | 2+2 2 |
| | 5+6 2 | 3+4 2 | 5+6 2 | 3+4 2 | 2+2 2 | 3+3 2 |
| | 5+6 2 | 5+6 2 | 3+4 2 | 5+6 2 | 3+3 2 | 3+3 2 |
| | 3+4 2 | 3+4 2 | 3+4 2 | 3+4 2 | 2+2 2 | 2+2 2 |
| | 3+4 2 | 3+4 2 | 3+4 2 | 5+6 2 | 2+2 2 | 3+3 2 |
| | 5+6 2 | 5+6 2 | 5+6 2 | 3+4 2 | 3+3 2 | 2+2 2 |

Continued on next column

**TABLE XXIV– continued**

| TQC | ICSD | | NoE | | SF | |
|---|---|---|---|---|---|---|
| | 5+6 2 | 3+4 2 | 5+6 2 | 5+6 2 | 3+3 2 | 3+3 2 |
| | 3+4 2 | 3+4 2 | 3+4 2 | 5+6 2 | 3+3 2 | 2+2 2 |
| | 5+6 2 | 5+6 2 | 3+4 2 | 3+4 2 | 2+2 2 | 3+3 2 |
| | 3+4 2 | 3+4 2 | 3+4 2 | 3+4 2 | 3+3 2 | 2+2 2 |
| | 5+6 2 | 5+6 2 | 5+6 2 | 5+6 2 | 2+2 2 | 3+3 2 |
| 12.2.112 | 189412 | | 144 | | Ce1Ru1Sn1 | |
| | -A | -GM | -M | -Y | -L | -V |
| | 3+4 2 | 3+4 2 | 5+6 2 | 5+6 2 | 3+3 2 | 3+3 2 |
| | 5+6 2 | 5+6 2 | 3+4 2 | 3+4 2 | 2+2 2 | 2+2 2 |
| | 5+6 2 | 3+4 2 | 5+6 2 | 3+4 2 | 2+2 2 | 3+3 2 |
| | 3+4 2 | 5+6 2 | 3+4 2 | 5+6 2 | 3+3 2 | 2+2 2 |
| | 3+4 2 | 3+4 2 | 5+6 2 | 3+4 2 | 2+2 2 | 2+2 2 |
| | 5+6 2 | 5+6 2 | 3+4 2 | 5+6 2 | 3+3 2 | 3+3 2 |
| | 3+4 2 | 3+4 2 | 5+6 2 | 5+6 2 | 3+3 2 | 3+3 2 |
| | 5+6 2 | 3+4 2 | 3+4 2 | 3+4 2 | 2+2 2 | 2+2 2 |
| | 3+4 2 | 5+6 2 | 5+6 2 | 3+4 2 | 2+2 2 | 3+3 2 |
| | 5+6 2 | 5+6 2 | 3+4 2 | 5+6 2 | 3+3 2 | 2+2 2 |
| | 5+6 2 | 3+4 2 | 3+4 2 | 5+6 2 | 2+2 2 | 3+3 2 |
| | 3+4 2 | 5+6 2 | 5+6 2 | 3+4 2 | 3+3 2 | 2+2 2 |
| | 3+4 2 | 3+4 2 | 5+6 2 | 5+6 2 | 3+3 2 | 3+3 2 |
| | 5+6 2 | 3+4 2 | 3+4 2 | 3+4 2 | 2+2 2 | 2+2 2 |
| | 3+4 2 | 5+6 2 | 5+6 2 | 3+4 2 | 2+2 2 | 3+3 2 |
| | 3+4 2 | 3+4 2 | 3+4 2 | 5+6 2 | 3+3 2 | 2+2 2 |
| | 5+6 2 | 5+6 2 | 5+6 2 | 3+4 2 | 2+2 2 | 3+3 2 |
| | 5+6 2 | 3+4 2 | 5+6 2 | 5+6 2 | 3+3 2 | 2+2 2 |
| | 3+4 2 | 5+6 2 | 3+4 2 | 5+6 2 | 3+3 2 | 3+3 2 |
| | 5+6 2 | 3+4 2 | 3+4 2 | 3+4 2 | 2+2 2 | 2+2 2 |
| | 3+4 2 | 5+6 2 | 5+6 2 | 5+6 2 | 2+2 2 | 3+3 2 |
| | 5+6 2 | 5+6 2 | 3+4 2 | 3+4 2 | 3+3 2 | 2+2 2 |
| | 5+6 2 | 3+4 2 | 5+6 2 | 3+4 2 | 2+2 2 | 3+3 2 |
| | 3+4 2 | 5+6 2 | 3+4 2 | 5+6 2 | 3+3 2 | 2+2 2 |
| | 5+6 2 | 3+4 2 | 5+6 2 | 3+4 2 | 3+3 2 | 2+2 2 |
| | 3+4 2 | 5+6 2 | 3+4 2 | 5+6 2 | 2+2 2 | 3+3 2 |
| | 5+6 2 | 3+4 2 | 3+4 2 | 3+4 2 | 3+3 2 | 2+2 2 |
| | 3+4 2 | 5+6 2 | 5+6 2 | 5+6 2 | 2+2 2 | 3+3 2 |
| | 3+4 2 | 3+4 2 | 3+4 2 | 5+6 2 | 3+3 2 | 2+2 2 |
| | 5+6 2 | 5+6 2 | 5+6 2 | 3+4 2 | 2+2 2 | 3+3 2 |
| | 5+6 2 | 3+4 2 | 3+4 2 | 3+4 2 | 3+3 2 | 2+2 2 |
| | 3+4 2 | 5+6 2 | 5+6 2 | 5+6 2 | 2+2 2 | 3+3 2 |
| | 3+4 2 | 3+4 2 | 3+4 2 | 3+4 2 | 3+3 2 | 3+3 2 |
| | 5+6 2 | 5+6 2 | 5+6 2 | 5+6 2 | 2+2 2 | 2+2 2 |
| | 3+4 2 | 3+4 2 | 5+6 2 | 3+4 2 | 3+3 2 | 2+2 2 |
| | 5+6 2 | 3+4 2 | 5+6 2 | 3+4 2 | 2+2 2 | 3+3 2 |
| | 3+4 2 | 5+6 2 | 3+4 2 | 5+6 2 | 2+2 2 | 2+2 2 |
| | 5+6 2 | 5+6 2 | 3+4 2 | 3+4 2 | 3+3 2 | 3+3 2 |
| | 3+4 2 | 3+4 2 | 5+6 2 | 5+6 2 | 3+3 2 | 2+2 2 |
| | 5+6 2 | 5+6 2 | 3+4 2 | 3+4 2 | 2+2 2 | 3+3 2 |
| | 3+4 2 | 3+4 2 | 3+4 2 | 5+6 2 | 3+3 2 | 2+2 2 |
| | 5+6 2 | 5+6 2 | 5+6 2 | 5+6 2 | 2+2 2 | 3+3 2 |
| | 5+6 2 | 3+4 2 | 3+4 2 | 3+4 2 | 2+2 2 | 2+2 2 |
| | 3+4 2 | 5+6 2 | 5+6 2 | 3+4 2 | 3+3 2 | 3+3 2 |
| | 5+6 2 | 3+4 2 | 3+4 2 | 5+6 2 | 2+2 2 | 2+2 2 |
| | 3+4 2 | 5+6 2 | 3+4 2 | 5+6 2 | 3+3 2 | 2+2 2 |
| | 5+6 2 | 5+6 2 | 5+6 2 | 3+4 2 | 2+2 2 | 3+3 2 |
| | 3+4 2 | 3+4 2 | 5+6 2 | 3+4 2 | 2+2 2 | 3+3 2 |
| | 3+4 2 | 5+6 2 | 3+4 2 | 5+6 2 | 3+3 2 | 2+2 2 |
| | 5+6 2 | 5+6 2 | 5+6 2 | 5+6 2 | 3+3 2 | 3+3 2 |
| | 3+4 2 | 3+4 2 | 3+4 2 | 3+4 2 | 2+2 2 | 2+2 2 |
| | 5+6 2 | 3+4 2 | 5+6 2 | 3+4 2 | 3+3 2 | 3+3 2 |
| | 3+4 2 | 3+4 2 | 3+4 2 | 5+6 2 | 2+2 2 | 2+2 2 |
| | 3+4 2 | 5+6 2 | 5+6 2 | 5+6 2 | 2+2 2 | 3+3 2 |

Continued on next column

**TABLE XXIV– continued**

| TQC | ICSD | | NoE | | SF | |
|---|---|---|---|---|---|---|
| | 5+6 2 | 3+4 2 | 3+4 2 | 3+4 2 | 3+3 2 | 2+2 2 |
| | 5+6 2 | 5+6 2 | 5+6 2 | 5+6 2 | 3+3 2 | 2+2 2 |
| | 3+4 2 | 5+6 2 | 3+4 2 | 5+6 2 | 2+2 2 | 3+3 2 |
| | 5+6 2 | 3+4 2 | 5+6 2 | 3+4 2 | 3+3 2 | 3+3 2 |
| | 3+4 2 | 5+6 2 | 5+6 2 | 3+4 2 | 2+2 2 | 2+2 2 |
| | 5+6 2 | 3+4 2 | 3+4 2 | 5+6 2 | 3+3 2 | 3+3 2 |
| | 5+6 2 | 3+4 2 | 5+6 2 | 3+4 2 | 2+2 2 | 2+2 2 |
| | 3+4 2 | 5+6 2 | 3+4 2 | 5+6 2 | 3+3 2 | 2+2 2 |
| | 5+6 2 | 3+4 2 | 5+6 2 | 3+4 2 | 3+3 2 | 3+3 2 |
| | 5+6 2 | 3+4 2 | 3+4 2 | 3+4 2 | 3+3 2 | 2+2 2 |
| | 3+4 2 | 5+6 2 | 5+6 2 | 5+6 2 | 2+2 2 | 3+3 2 |
| | 5+6 2 | 3+4 2 | 3+4 2 | 5+6 2 | 2+2 2 | 2+2 2 |
| | 3+4 2 | 5+6 2 | 3+4 2 | 3+4 2 | 3+3 2 | 3+3 2 |
| | 3+4 2 | 3+4 2 | 3+4 2 | 5+6 2 | 2+2 2 | 3+3 2 |
| | 5+6 2 | 3+4 2 | 5+6 2 | 5+6 2 | 3+3 2 | 2+2 2 |
| | 5+6 2 | 5+6 2 | 5+6 2 | 3+4 2 | 2+2 2 | 2+2 2 |
| | 5+6 2 | 3+4 2 | 3+4 2 | 5+6 2 | 3+3 2 | 3+3 2 |
| | 3+4 2 | 5+6 2 | 3+4 2 | 3+4 2 | 2+2 2 | 2+2 2 |
| 12.2.113 | 189411 | | 96 | | Ce1Ru1Sn1 | |
| | -A | -GM | -M | -Y | -L | -V |
| | 3+4 2 | 3+4 2 | 5+6 2 | 5+6 2 | 3+3 2 | 3+3 2 |
| | 5+6 2 | 5+6 2 | 3+4 2 | 3+4 2 | 2+2 2 | 2+2 2 |
| | 5+6 2 | 3+4 2 | 3+4 2 | 5+6 2 | 2+2 2 | 3+3 2 |
| | 3+4 2 | 5+6 2 | 5+6 2 | 3+4 2 | 3+3 2 | 2+2 2 |
| | 3+4 2 | 3+4 2 | 5+6 2 | 5+6 2 | 3+3 2 | 3+3 2 |
| | 5+6 2 | 5+6 2 | 3+4 2 | 3+4 2 | 2+2 2 | 2+2 2 |
| | 5+6 2 | 3+4 2 | 3+4 2 | 5+6 2 | 2+2 2 | 3+3 2 |
| | 3+4 2 | 5+6 2 | 5+6 2 | 3+4 2 | 3+3 2 | 2+2 2 |
| | 3+4 2 | 3+4 2 | 5+6 2 | 5+6 2 | 3+3 2 | 3+3 2 |
| | 3+4 2 | 3+4 2 | 3+4 2 | 3+4 2 | 2+2 2 | 2+2 2 |
| | 5+6 2 | 5+6 2 | 5+6 2 | 5+6 2 | 2+2 2 | 3+3 2 |
| | 5+6 2 | 3+4 2 | 3+4 2 | 3+4 2 | 3+3 2 | 2+2 2 |
| | 5+6 2 | 5+6 2 | 5+6 2 | 5+6 2 | 3+3 2 | 3+3 2 |
| | 3+4 2 | 3+4 2 | 3+4 2 | 3+4 2 | 2+2 2 | 2+2 2 |
| | 5+6 2 | 5+6 2 | 3+4 2 | 5+6 2 | 2+2 2 | 3+3 2 |
| | 3+4 2 | 5+6 2 | 5+6 2 | 3+4 2 | 3+3 2 | 2+2 2 |
| | 5+6 2 | 3+4 2 | 5+6 2 | 3+4 2 | 3+3 2 | 2+2 2 |
| | 3+4 2 | 5+6 2 | 3+4 2 | 5+6 2 | 2+2 2 | 3+3 2 |
| | 5+6 2 | 3+4 2 | 3+4 2 | 3+4 2 | 2+2 2 | 2+2 2 |
| | 3+4 2 | 5+6 2 | 5+6 2 | 5+6 2 | 3+3 2 | 3+3 2 |
| | 5+6 2 | 5+6 2 | 3+4 2 | 3+4 2 | 3+3 2 | 2+2 2 |
| | 3+4 2 | 3+4 2 | 5+6 2 | 5+6 2 | 2+2 2 | 3+3 2 |
| | 3+4 2 | 3+4 2 | 5+6 2 | 3+4 2 | 3+3 2 | 2+2 2 |
| | 5+6 2 | 3+4 2 | 3+4 2 | 5+6 2 | 2+2 2 | 3+3 2 |
| | 3+4 2 | 5+6 2 | 3+4 2 | 5+6 2 | 2+2 2 | 3+3 2 |
| | 5+6 2 | 3+4 2 | 5+6 2 | 3+4 2 | 2+2 2 | 2+2 2 |
| | 5+6 2 | 5+6 2 | 3+4 2 | 3+4 2 | 3+3 2 | 3+3 2 |
| | 5+6 2 | 3+4 2 | 3+4 2 | 5+6 2 | 3+3 2 | 2+2 2 |
| | 3+4 2 | 5+6 2 | 5+6 2 | 3+4 2 | 2+2 2 | 3+3 2 |
| | 3+4 2 | 5+6 2 | 5+6 2 | 3+4 2 | 3+3 2 | 2+2 2 |
| | 5+6 2 | 3+4 2 | 3+4 2 | 5+6 2 | 3+3 2 | 2+2 2 |
| | 3+4 2 | 5+6 2 | 5+6 2 | 3+4 2 | 2+2 2 | 3+3 2 |
| | 5+6 2 | 3+4 2 | 3+4 2 | 5+6 2 | 3+3 2 | 2+2 2 |
| | 3+4 2 | 5+6 2 | 5+6 2 | 5+6 2 | 2+2 2 | 3+3 2 |
| | 3+4 2 | 5+6 2 | 5+6 2 | 3+4 2 | 3+3 2 | 2+2 2 |
| | 3+4 2 | 3+4 2 | 5+6 2 | 3+4 2 | 2+2 2 | 3+3 2 |
| | 5+6 2 | 5+6 2 | 3+4 2 | 3+4 2 | 2+2 2 | 2+2 2 |
| | 5+6 2 | 3+4 2 | 5+6 2 | 5+6 2 | 3+3 2 | 2+2 2 |
| | 3+4 2 | 3+4 2 | 3+4 2 | 5+6 2 | 2+2 2 | 3+3 2 |
| | 5+6 2 | 5+6 2 | 3+4 2 | 5+6 2 | 3+3 2 | 3+3 2 |
| | 5+6 2 | 3+4 2 | 5+6 2 | 3+4 2 | 3+3 2 | 2+2 2 |

Continued on next column

**TABLE XXIV– continued**

| TQC | ICSD | | NoE | SF | | |
|---|---|---|---|---|---|---|
| | 5+6 2 | 5+6 2 | 3+4 2 | 5+6 2 | 2+2 2 | 3+3 2 |
| | 3+4 2 | 3+4 2 | 3+4 2 | 3+4 2 | 3+3 2 | 2+2 2 |
| | 5+6 2 | 5+6 2 | 5+6 2 | 5+6 2 | 2+2 2 | 3+3 2 |
| | 3+4 2 | 3+4 2 | 3+4 2 | 3+4 2 | 2+2 2 | 2+2 2 |
| | 3+4 2 | 5+6 2 | 5+6 2 | 5+6 2 | 3+3 2 | 2+2 2 |
| | 5+6 2 | 3+4 2 | 5+6 2 | 3+4 2 | 2+2 2 | 3+3 2 |
| | 3+4 2 | 5+6 2 | 3+4 2 | 3+4 2 | 3+3 2 | 2+2 2 |
| 12.2.114 | 2284 | | 24 | Ce1 | | |
| | -A | -GM | -M | -Y | -L | -V |
| | 5+6 2 | 3+4 2 | 5+6 2 | 5+6 2 | 3+3 2 | 2+2 2 |
| | 3+4 2 | 5+6 2 | 3+4 2 | 3+4 2 | 2+2 2 | 3+3 2 |
| | 5+6 2 | 3+4 2 | 5+6 2 | 3+4 2 | 2+2 2 | 2+2 2 |
| | 3+4 2 | 5+6 2 | 3+4 2 | 5+6 2 | 3+3 2 | 3+3 2 |
| | 5+6 2 | 3+4 2 | 5+6 2 | 5+6 2 | 3+3 2 | 2+2 2 |
| | 3+4 2 | 3+4 2 | 3+4 2 | 3+4 2 | 2+2 2 | 3+3 2 |
| | 3+4 2 | 5+6 2 | 5+6 2 | 5+6 2 | 3+3 2 | 2+2 2 |
| | 5+6 2 | 5+6 2 | 3+4 2 | 3+4 2 | 2+2 2 | 3+3 2 |
| | 5+6 2 | 3+4 2 | 3+4 2 | 3+4 2 | 2+2 2 | 2+2 2 |
| | 3+4 2 | 5+6 2 | 5+6 2 | 3+4 2 | 2+2 2 | 3+3 2 |
| | 3+4 2 | 5+6 2 | 3+4 2 | 5+6 2 | 3+3 2 | 3+3 2 |
| | 5+6 2 | 3+4 2 | 5+6 2 | 3+4 2 | 2+2 2 | 2+2 2 |
| 12.2.115 | 620729 | | 48 | Ce1Co1Ga1 | | |
| | -A | -GM | -M | -Y | -L | -V |
| | 5+6 2 | 3+4 2 | 3+4 2 | 5+6 2 | 2+2 2 | 3+3 2 |
| | 3+4 2 | 5+6 2 | 5+6 2 | 3+4 2 | 3+3 2 | 2+2 2 |
| | 5+6 2 | 3+4 2 | 3+4 2 | 5+6 2 | 2+2 2 | 3+3 2 |
| | 3+4 2 | 5+6 2 | 5+6 2 | 3+4 2 | 3+3 2 | 2+2 2 |
| | 5+6 2 | 3+4 2 | 3+4 2 | 3+4 2 | 2+2 2 | 3+3 2 |
| | 3+4 2 | 3+4 2 | 5+6 2 | 5+6 2 | 3+3 2 | 2+2 2 |
| | 5+6 2 | 5+6 2 | 3+4 2 | 5+6 2 | 2+2 2 | 3+3 2 |
| | 3+4 2 | 5+6 2 | 5+6 2 | 3+4 2 | 3+3 2 | 2+2 2 |
| | 3+4 2 | 3+4 2 | 5+6 2 | 3+4 2 | 2+2 2 | 2+2 2 |
| | 5+6 2 | 5+6 2 | 3+4 2 | 5+6 2 | 3+3 2 | 3+3 2 |
| | 3+4 2 | 3+4 2 | 5+6 2 | 3+4 2 | 2+2 2 | 2+2 2 |
| | 5+6 2 | 5+6 2 | 3+4 2 | 3+4 2 | 3+3 2 | 3+3 2 |
| | 5+6 2 | 3+4 2 | 3+4 2 | 5+6 2 | 2+2 2 | 2+2 2 |
| | 3+4 2 | 5+6 2 | 5+6 2 | 3+4 2 | 3+3 2 | 3+3 2 |
| | 5+6 2 | 5+6 2 | 5+6 2 | 5+6 2 | 3+3 2 | 3+3 2 |
| | 3+4 2 | 3+4 2 | 3+4 2 | 5+6 2 | 2+2 2 | 2+2 2 |
| | 5+6 2 | 5+6 2 | 3+4 2 | 5+6 2 | 3+3 2 | 3+3 2 |
| | 3+4 2 | 5+6 2 | 5+6 2 | 3+4 2 | 2+2 2 | 2+2 2 |
| | 5+6 2 | 3+4 2 | 3+4 2 | 5+6 2 | 3+3 2 | 2+2 2 |
| | 3+4 2 | 3+4 2 | 5+6 2 | 3+4 2 | 2+2 2 | 2+2 2 |
| | 3+4 2 | 3+4 2 | 3+4 2 | 3+4 2 | 2+2 2 | 3+3 2 |
| | 3+4 2 | 3+4 2 | 5+6 2 | 5+6 2 | 3+3 2 | 2+2 2 |
| | 5+6 2 | 3+4 2 | 5+6 2 | 3+4 2 | 2+2 2 | 3+3 2 |
| | 3+4 2 | 5+6 2 | 5+6 2 | 5+6 2 | 3+3 2 | 2+2 2 |
| 12.2.116 | 620726 | | 48 | Ce1Co1Ga1 | | |
| | -A | -GM | -M | -Y | -L | -V |
| | 5+6 2 | 3+4 2 | 3+4 2 | 5+6 2 | 2+2 2 | 3+3 2 |
| | 3+4 2 | 5+6 2 | 5+6 2 | 3+4 2 | 3+3 2 | 2+2 2 |
| | 5+6 2 | 3+4 2 | 3+4 2 | 5+6 2 | 2+2 2 | 3+3 2 |
| | 3+4 2 | 5+6 2 | 5+6 2 | 3+4 2 | 3+3 2 | 2+2 2 |
| | 5+6 2 | 3+4 2 | 3+4 2 | 3+4 2 | 2+2 2 | 3+3 2 |
| | 3+4 2 | 3+4 2 | 5+6 2 | 5+6 2 | 3+3 2 | 2+2 2 |
| | 5+6 2 | 5+6 2 | 3+4 2 | 5+6 2 | 2+2 2 | 3+3 2 |
| | 3+4 2 | 5+6 2 | 5+6 2 | 3+4 2 | 3+3 2 | 2+2 2 |
| | 3+4 2 | 3+4 2 | 5+6 2 | 3+4 2 | 2+2 2 | 2+2 2 |
| | 5+6 2 | 5+6 2 | 3+4 2 | 5+6 2 | 3+3 2 | 3+3 2 |
| | 3+4 2 | 3+4 2 | 5+6 2 | 3+4 2 | 2+2 2 | 2+2 2 |
| | 5+6 2 | 5+6 2 | 3+4 2 | 3+4 2 | 3+3 2 | 3+3 2 |

Continued on next column

**TABLE XXIV– continued**

| TQC | ICSD | | NoE | SF | | |
|---|---|---|---|---|---|---|
| | 5+6 2 | 3+4 2 | 3+4 2 | 5+6 2 | 2+2 2 | 2+2 2 |
| | 3+4 2 | 5+6 2 | 5+6 2 | 3+4 2 | 3+3 2 | 3+3 2 |
| | 5+6 2 | 5+6 2 | 5+6 2 | 5+6 2 | 3+3 2 | 3+3 2 |
| | 3+4 2 | 3+4 2 | 3+4 2 | 5+6 2 | 2+2 2 | 2+2 2 |
| | 5+6 2 | 5+6 2 | 3+4 2 | 5+6 2 | 3+3 2 | 3+3 2 |
| | 3+4 2 | 5+6 2 | 5+6 2 | 3+4 2 | 2+2 2 | 2+2 2 |
| | 5+6 2 | 3+4 2 | 3+4 2 | 5+6 2 | 3+3 2 | 2+2 2 |
| | 3+4 2 | 3+4 2 | 5+6 2 | 3+4 2 | 2+2 2 | 2+2 2 |
| | 3+4 2 | 3+4 2 | 3+4 2 | 3+4 2 | 2+2 2 | 3+3 2 |
| | 3+4 2 | 3+4 2 | 5+6 2 | 5+6 2 | 3+3 2 | 2+2 2 |
| | 5+6 2 | 3+4 2 | 5+6 2 | 3+4 2 | 2+2 2 | 3+3 2 |
| | 3+4 2 | 5+6 2 | 5+6 2 | 5+6 2 | 3+3 2 | 2+2 2 |
| 12.2.117 | 429884 | | 48 | Ce1Co1Ga1 | | |
| | -A | -GM | -M | -Y | -L | -V |
| | 5+6 2 | 3+4 2 | 3+4 2 | 5+6 2 | 2+2 2 | 3+3 2 |
| | 3+4 2 | 5+6 2 | 5+6 2 | 3+4 2 | 3+3 2 | 2+2 2 |
| | 5+6 2 | 3+4 2 | 3+4 2 | 5+6 2 | 2+2 2 | 3+3 2 |
| | 3+4 2 | 5+6 2 | 5+6 2 | 3+4 2 | 3+3 2 | 2+2 2 |
| | 5+6 2 | 3+4 2 | 3+4 2 | 3+4 2 | 2+2 2 | 3+3 2 |
| | 3+4 2 | 3+4 2 | 5+6 2 | 5+6 2 | 3+3 2 | 2+2 2 |
| | 5+6 2 | 5+6 2 | 3+4 2 | 5+6 2 | 3+3 2 | 3+3 2 |
| | 3+4 2 | 5+6 2 | 5+6 2 | 3+4 2 | 2+2 2 | 2+2 2 |
| | 3+4 2 | 3+4 2 | 5+6 2 | 3+4 2 | 2+2 2 | 2+2 2 |
| | 5+6 2 | 5+6 2 | 3+4 2 | 5+6 2 | 3+3 2 | 3+3 2 |
| | 3+4 2 | 3+4 2 | 5+6 2 | 3+4 2 | 2+2 2 | 2+2 2 |
| | 5+6 2 | 5+6 2 | 3+4 2 | 3+4 2 | 3+3 2 | 3+3 2 |
| | 5+6 2 | 5+6 2 | 3+4 2 | 3+4 2 | 2+2 2 | 2+2 2 |
| | 3+4 2 | 3+4 2 | 5+6 2 | 5+6 2 | 3+3 2 | 3+3 2 |
| | 5+6 2 | 5+6 2 | 5+6 2 | 5+6 2 | 3+3 2 | 2+2 2 |
| | 3+4 2 | 5+6 2 | 3+4 2 | 5+6 2 | 2+2 2 | 3+3 2 |
| | 5+6 2 | 3+4 2 | 3+4 2 | 5+6 2 | 3+3 2 | 2+2 2 |
| | 3+4 2 | 5+6 2 | 5+6 2 | 3+4 2 | 2+2 2 | 3+3 2 |
| | 5+6 2 | 3+4 2 | 3+4 2 | 5+6 2 | 3+3 2 | 2+2 2 |
| | 3+4 2 | 3+4 2 | 5+6 2 | 3+4 2 | 2+2 2 | 2+2 2 |
| | 3+4 2 | 3+4 2 | 3+4 2 | 3+4 2 | 2+2 2 | 3+3 2 |
| | 3+4 2 | 3+4 2 | 5+6 2 | 5+6 2 | 3+3 2 | 2+2 2 |
| | 3+4 2 | 3+4 2 | 5+6 2 | 3+4 2 | 2+2 2 | 3+3 2 |
| | 5+6 2 | 5+6 2 | 5+6 2 | 5+6 2 | 3+3 2 | 2+2 2 |
| 12.2.118 | 187960 | | 48 | Ce1Co1Ga1 | | |
| | -A | -GM | -M | -Y | -L | -V |
| | 5+6 2 | 3+4 2 | 3+4 2 | 5+6 2 | 2+2 2 | 3+3 2 |
| | 3+4 2 | 5+6 2 | 5+6 2 | 3+4 2 | 3+3 2 | 2+2 2 |
| | 5+6 2 | 3+4 2 | 3+4 2 | 5+6 2 | 2+2 2 | 3+3 2 |
| | 3+4 2 | 5+6 2 | 5+6 2 | 3+4 2 | 3+3 2 | 2+2 2 |
| | 5+6 2 | 3+4 2 | 3+4 2 | 3+4 2 | 2+2 2 | 3+3 2 |
| | 3+4 2 | 3+4 2 | 5+6 2 | 5+6 2 | 3+3 2 | 2+2 2 |
| | 5+6 2 | 5+6 2 | 3+4 2 | 5+6 2 | 3+3 2 | 3+3 2 |
| | 3+4 2 | 5+6 2 | 5+6 2 | 3+4 2 | 2+2 2 | 2+2 2 |
| | 3+4 2 | 3+4 2 | 5+6 2 | 3+4 2 | 2+2 2 | 2+2 2 |
| | 5+6 2 | 5+6 2 | 3+4 2 | 5+6 2 | 3+3 2 | 3+3 2 |
| | 3+4 2 | 3+4 2 | 5+6 2 | 3+4 2 | 2+2 2 | 2+2 2 |
| | 5+6 2 | 5+6 2 | 3+4 2 | 3+4 2 | 3+3 2 | 3+3 2 |
| | 5+6 2 | 5+6 2 | 3+4 2 | 3+4 2 | 2+2 2 | 2+2 2 |
| | 3+4 2 | 3+4 2 | 5+6 2 | 5+6 2 | 3+3 2 | 3+3 2 |
| | 5+6 2 | 5+6 2 | 5+6 2 | 5+6 2 | 3+3 2 | 2+2 2 |
| | 3+4 2 | 5+6 2 | 3+4 2 | 5+6 2 | 2+2 2 | 3+3 2 |
| | 5+6 2 | 3+4 2 | 3+4 2 | 3+4 2 | 3+3 2 | 2+2 2 |
| | 3+4 2 | 5+6 2 | 5+6 2 | 5+6 2 | 2+2 2 | 3+3 2 |
| | 5+6 2 | 3+4 2 | 3+4 2 | 5+6 2 | 3+3 2 | 2+2 2 |
| | 3+4 2 | 3+4 2 | 5+6 2 | 3+4 2 | 2+2 2 | 2+2 2 |
| | 3+4 2 | 3+4 2 | 3+4 2 | 3+4 2 | 2+2 2 | 3+3 2 |

Continued on next column

**TABLE XXIV– continued**

| TQC | ICSD | | NoE | SF | | |
|---|---|---|---|---|---|---|
| | 3+4 2 | 3+4 2 | 5+6 2 | 5+6 2 | 3+3 2 | 2+2 2 |
| | 5+6 2 | 3+4 2 | 5+6 2 | 3+4 2 | 2+2 2 | 3+3 2 |
| | 3+4 2 | 5+6 2 | 5+6 2 | 5+6 2 | 3+3 2 | 2+2 2 |
| 12.2.119 | 171243 | | 62 | Ca6Cu2Sn7 | | |
| | -A | -GM | -M | -Y | -L | -V |
| | 5+6 2 | 3+4 2 | 3+4 2 | 3+4 2 | 3+3 2 | 3+3 2 |
| | 3+4 2 | 3+4 2 | 5+6 2 | 5+6 2 | 2+2 2 | 3+3 2 |
| | 5+6 2 | 5+6 2 | 3+4 2 | 3+4 2 | 3+3 2 | 2+2 2 |
| | 3+4 2 | 5+6 2 | 5+6 2 | 5+6 2 | 2+2 2 | 2+2 2 |
| | 3+4 2 | 3+4 2 | 3+4 2 | 3+4 2 | 2+2 2 | 3+3 2 |
| | 5+6 2 | 3+4 2 | 5+6 2 | 5+6 2 | 3+3 2 | 3+3 2 |
| | 5+6 2 | 5+6 2 | 5+6 2 | 3+4 2 | 2+2 2 | 2+2 2 |
| | 3+4 2 | 5+6 2 | 5+6 2 | 3+4 2 | 3+3 2 | 2+2 2 |
| | 5+6 2 | 5+6 2 | 3+4 2 | 5+6 2 | 2+2 2 | 2+2 2 |
| | 3+4 2 | 3+4 2 | 3+4 2 | 5+6 2 | 3+3 2 | 3+3 2 |
| | 5+6 2 | 3+4 2 | 3+4 2 | 5+6 2 | 2+2 2 | 2+2 2 |
| | 3+4 2 | 5+6 2 | 5+6 2 | 5+6 2 | 3+3 2 | 2+2 2 |
| | 5+6 2 | 3+4 2 | 3+4 2 | 3+4 2 | 3+3 2 | 3+3 2 |
| | 3+4 2 | 5+6 2 | 5+6 2 | 5+6 2 | 2+2 2 | 3+3 2 |
| | 3+4 2 | 3+4 2 | 5+6 2 | 3+4 2 | 2+2 2 | 2+2 2 |
| | 5+6 2 | 5+6 2 | 3+4 2 | 5+6 2 | 3+3 2 | 3+3 2 |
| | 5+6 2 | 3+4 2 | 3+4 2 | 3+4 2 | 2+2 2 | 3+3 2 |
| | 3+4 2 | 5+6 2 | 5+6 2 | 3+4 2 | 3+3 2 | 2+2 2 |
| | 5+6 2 | 3+4 2 | 3+4 2 | 3+4 2 | 2+2 2 | 2+2 2 |
| | 3+4 2 | 5+6 2 | 5+6 2 | 5+6 2 | 2+2 2 | 3+3 2 |
| | 5+6 2 | 3+4 2 | 3+4 2 | 3+4 2 | 3+3 2 | 2+2 2 |
| | 3+4 2 | 3+4 2 | 3+4 2 | 5+6 2 | 3+3 2 | 3+3 2 |
| | 3+4 2 | 5+6 2 | 5+6 2 | 5+6 2 | 2+2 2 | 3+3 2 |
| | 5+6 2 | 3+4 2 | 3+4 2 | 5+6 2 | 3+3 2 | 2+2 2 |
| | 5+6 2 | 5+6 2 | 5+6 2 | 3+4 2 | 3+3 2 | 3+3 2 |
| | 3+4 2 | 3+4 2 | 5+6 2 | 5+6 2 | 2+2 2 | 2+2 2 |
| | 5+6 2 | 5+6 2 | 3+4 2 | 3+4 2 | 3+3 2 | 2+2 2 |
| | 3+4 2 | 5+6 2 | 3+4 2 | 3+4 2 | 3+3 2 | 3+3 2 |
| | 3+4 2 | 3+4 2 | 5+6 2 | 3+4 2 | 2+2 2 | 2+2 2 |
| | 5+6 2 | 3+4 2 | 3+4 2 | 5+6 2 | 2+2 2 | 3+3 2 |
| | 3+4 2 | 5+6 2 | 5+6 2 | 5+6 2 | 3+3 2 | 3+3 2 |
| 12.2.120 | 85505 | | 88 | Ca1K2N4O8 | | |
| | -A | -GM | -M | -Y | -L | -V |
| | 3+4 2 | 3+4 2 | 3+4 2 | 3+4 2 | 2+2 2 | 2+2 2 |
| | 5+6 2 | 5+6 2 | 5+6 2 | 5+6 2 | 3+3 2 | 3+3 2 |
| | 3+4 2 | 3+4 2 | 3+4 2 | 3+4 2 | 2+2 2 | 2+2 2 |
| | 5+6 2 | 5+6 2 | 5+6 2 | 5+6 2 | 3+3 2 | 3+3 2 |
| | 3+4 2 | 3+4 2 | 3+4 2 | 3+4 2 | 2+2 2 | 2+2 2 |
| | 5+6 2 | 5+6 2 | 5+6 2 | 5+6 2 | 3+3 2 | 3+3 2 |
| | 5+6 2 | 5+6 2 | 5+6 2 | 5+6 2 | 3+3 2 | 3+3 2 |
| | 3+4 2 | 3+4 2 | 3+4 2 | 3+4 2 | 2+2 2 | 2+2 2 |
| | 5+6 2 | 5+6 2 | 5+6 2 | 5+6 2 | 3+3 2 | 3+3 2 |
| | 3+4 2 | 3+4 2 | 3+4 2 | 3+4 2 | 2+2 2 | 2+2 2 |
| | 3+4 2 | 3+4 2 | 3+4 2 | 3+4 2 | 2+2 2 | 2+2 2 |
| | 5+6 2 | 5+6 2 | 5+6 2 | 5+6 2 | 3+3 2 | 3+3 2 |
| | 3+4 2 | 3+4 2 | 3+4 2 | 3+4 2 | 2+2 2 | 2+2 2 |
| | 5+6 2 | 3+4 2 | 5+6 2 | 5+6 2 | 3+3 2 | 3+3 2 |
| | 3+4 2 | 5+6 2 | 3+4 2 | 3+4 2 | 2+2 2 | 2+2 2 |
| | 5+6 2 | 5+6 2 | 5+6 2 | 5+6 2 | 3+3 2 | 3+3 2 |
| | 3+4 2 | 3+4 2 | 3+4 2 | 3+4 2 | 2+2 2 | 2+2 2 |
| | 5+6 2 | 5+6 2 | 5+6 2 | 5+6 2 | 3+3 2 | 3+3 2 |
| | 3+4 2 | 3+4 2 | 3+4 2 | 3+4 2 | 2+2 2 | 2+2 2 |
| | 5+6 2 | 5+6 2 | 5+6 2 | 5+6 2 | 3+3 2 | 3+3 2 |
| | 3+4 2 | 3+4 2 | 3+4 2 | 3+4 2 | 3+3 2 | 3+3 2 |
| | 5+6 2 | 5+6 2 | 5+6 2 | 5+6 2 | 2+2 2 | 2+2 2 |
| | 3+4 2 | 3+4 2 | 3+4 2 | 3+4 2 | 2+2 2 | 2+2 2 |

Continued on next column

**TABLE XXIV– continued**

| TQC | ICSD | | NoE | SF | | |
|---|---|---|---|---|---|---|
| | 5+6 2 | 5+6 2 | 5+6 2 | 5+6 2 | 3+3 2 | 3+3 2 |
| | 3+4 2 | 3+4 2 | 3+4 2 | 3+4 2 | 2+2 2 | 2+2 2 |
| | 3+4 2 | 3+4 2 | 5+6 2 | 5+6 2 | 3+3 2 | 3+3 2 |
| | 5+6 2 | 5+6 2 | 5+6 2 | 5+6 2 | 2+2 2 | 2+2 2 |
| | 5+6 2 | 5+6 2 | 3+4 2 | 3+4 2 | 3+3 2 | 3+3 2 |
| | 3+4 2 | 3+4 2 | 3+4 2 | 3+4 2 | 2+2 2 | 2+2 2 |
| | 5+6 2 | 5+6 2 | 5+6 2 | 5+6 2 | 3+3 2 | 3+3 2 |
| | 3+4 2 | 3+4 2 | 3+4 2 | 3+4 2 | 2+2 2 | 2+2 2 |
| | 5+6 2 | 5+6 2 | 5+6 2 | 5+6 2 | 3+3 2 | 3+3 2 |
| | 3+4 2 | 3+4 2 | 3+4 2 | 3+4 2 | 2+2 2 | 2+2 2 |
| | 3+4 2 | 3+4 2 | 5+6 2 | 5+6 2 | 3+3 2 | 3+3 2 |
| | 5+6 2 | 5+6 2 | 3+4 2 | 3+4 2 | 3+3 2 | 3+3 2 |
| | 3+4 2 | 3+4 2 | 3+4 2 | 3+4 2 | 2+2 2 | 2+2 2 |
| | 5+6 2 | 5+6 2 | 5+6 2 | 5+6 2 | 3+3 2 | 3+3 2 |
| | 3+4 2 | 3+4 2 | 3+4 2 | 3+4 2 | 2+2 2 | 2+2 2 |
| | 5+6 2 | 5+6 2 | 5+6 2 | 5+6 2 | 2+2 2 | 2+2 2 |
| | 5+6 2 | 5+6 2 | 5+6 2 | 5+6 2 | 3+3 2 | 3+3 2 |
| | 5+6 2 | 5+6 2 | 3+4 2 | 3+4 2 | 3+3 2 | 3+3 2 |
| | 3+4 2 | 3+4 2 | 5+6 2 | 5+6 2 | 2+2 2 | 2+2 2 |
| | 3+4 2 | 3+4 2 | 3+4 2 | 3+4 2 | 2+2 2 | 2+2 2 |
| | 5+6 2 | 3+4 2 | 3+4 2 | 3+4 2 | 2+2 2 | 2+2 2 |
| 12.2.121 | 619274 | | 14 | Ca1Ga4 | | |
| | -A | -GM | -M | -Y | -L | -V |
| | 5+6 2 | 3+4 2 | 3+4 2 | 3+4 2 | 2+2 2 | 2+2 2 |
| | 3+4 2 | 5+6 2 | 5+6 2 | 5+6 2 | 3+3 2 | 3+3 2 |
| | 3+4 2 | 3+4 2 | 3+4 2 | 3+4 2 | 2+2 2 | 2+2 2 |
| | 5+6 2 | 5+6 2 | 5+6 2 | 5+6 2 | 3+3 2 | 3+3 2 |
| | 5+6 2 | 3+4 2 | 3+4 2 | 3+4 2 | 3+3 2 | 3+3 2 |
| | 3+4 2 | 3+4 2 | 5+6 2 | 5+6 2 | 2+2 2 | 2+2 2 |
| | 3+4 2 | 3+4 2 | 3+4 2 | 3+4 2 | 2+2 2 | 2+2 2 |
| 12.2.122 | 58893 | | 14 | Ca1Ga4 | | |
| | -A | -GM | -M | -Y | -L | -V |
| | 5+6 2 | 3+4 2 | 3+4 2 | 3+4 2 | 2+2 2 | 2+2 2 |
| | 3+4 2 | 5+6 2 | 5+6 2 | 5+6 2 | 3+3 2 | 3+3 2 |
| | 3+4 2 | 3+4 2 | 3+4 2 | 3+4 2 | 2+2 2 | 2+2 2 |
| | 5+6 2 | 5+6 2 | 5+6 2 | 5+6 2 | 3+3 2 | 3+3 2 |
| | 5+6 2 | 3+4 2 | 3+4 2 | 3+4 2 | 3+3 2 | 3+3 2 |
| | 3+4 2 | 3+4 2 | 5+6 2 | 5+6 2 | 2+2 2 | 2+2 2 |
| | 3+4 2 | 3+4 2 | 3+4 2 | 3+4 2 | 2+2 2 | 2+2 2 |
| 12.2.123 | 54994 | | 14 | Ca1Ga4 | | |
| | -A | -GM | -M | -Y | -L | -V |
| | 5+6 2 | 3+4 2 | 3+4 2 | 3+4 2 | 2+2 2 | 2+2 2 |
| | 3+4 2 | 5+6 2 | 5+6 2 | 5+6 2 | 3+3 2 | 3+3 2 |
| | 3+4 2 | 3+4 2 | 3+4 2 | 3+4 2 | 2+2 2 | 2+2 2 |
| | 5+6 2 | 5+6 2 | 5+6 2 | 5+6 2 | 3+3 2 | 3+3 2 |
| | 5+6 2 | 3+4 2 | 3+4 2 | 3+4 2 | 3+3 2 | 3+3 2 |
| | 3+4 2 | 3+4 2 | 5+6 2 | 5+6 2 | 2+2 2 | 2+2 2 |
| | 3+4 2 | 3+4 2 | 3+4 2 | 3+4 2 | 2+2 2 | 2+2 2 |
| 12.2.124 | 417717 | | 32 | Ca1Cu2Sn2 | | |
| | -A | -GM | -M | -Y | -L | -V |
| | 5+6 2 | 3+4 2 | 3+4 2 | 5+6 2 | 3+3 2 | 2+2 2 |
| | 3+4 2 | 5+6 2 | 5+6 2 | 3+4 2 | 2+2 2 | 3+3 2 |
| | 3+4 2 | 3+4 2 | 5+6 2 | 3+4 2 | 2+2 2 | 3+3 2 |
| | 5+6 2 | 3+4 2 | 3+4 2 | 5+6 2 | 3+3 2 | 2+2 2 |
| | 5+6 2 | 5+6 2 | 3+4 2 | 3+4 2 | 3+3 2 | 2+2 2 |
| | 5+6 2 | 5+6 2 | 5+6 2 | 5+6 2 | 2+2 2 | 3+3 2 |
| | 3+4 2 | 3+4 2 | 3+4 2 | 3+4 2 | 3+3 2 | 2+2 2 |
| | 3+4 2 | 3+4 2 | 5+6 2 | 5+6 2 | 2+2 2 | 3+3 2 |
| | 5+6 2 | 5+6 2 | 3+4 2 | 5+6 2 | 3+3 2 | 2+2 2 |
| | 5+6 2 | 3+4 2 | 3+4 2 | 3+4 2 | 2+2 2 | 3+3 2 |
| | 3+4 2 | 5+6 2 | 5+6 2 | 5+6 2 | 3+3 2 | 3+3 2 |

Continued on next column

**TABLE XXIV– continued**

| TQC | ICSD | | NoE | SF | | |
|---|---|---|---|---|---|---|
| | 3+4 2 | 5+6 2 | 5+6 2 | 3+4 2 | 2+2 2 | 2+2 2 |
| | 5+6 2 | 3+4 2 | 5+6 2 | 5+6 2 | 3+3 2 | 2+2 2 |
| | 3+4 2 | 3+4 2 | 3+4 2 | 3+4 2 | 2+2 2 | 3+3 2 |
| | 5+6 2 | 5+6 2 | 3+4 2 | 5+6 2 | 3+3 2 | 3+3 2 |
| | 3+4 2 | 3+4 2 | 5+6 2 | 3+4 2 | 2+2 2 | 2+2 2 |
| 12.2.125 | 236391 | | 68 | C4Co1Sc3 | | |
| | -A | -GM | -M | -Y | -L | -V |
| | 5+6 2 | 3+4 2 | 5+6 2 | 3+4 2 | 2+2 2 | 3+3 2 |
| | 3+4 2 | 5+6 2 | 3+4 2 | 5+6 2 | 2+2 2 | 2+2 2 |
| | 5+6 2 | 3+4 2 | 5+6 2 | 5+6 2 | 3+3 2 | 3+3 2 |
| | 3+4 2 | 5+6 2 | 3+4 2 | 3+4 2 | 3+3 2 | 2+2 2 |
| | 5+6 2 | 5+6 2 | 5+6 2 | 5+6 2 | 3+3 2 | 2+2 2 |
| | 3+4 2 | 3+4 2 | 3+4 2 | 3+4 2 | 2+2 2 | 3+3 2 |
| | 5+6 2 | 3+4 2 | 5+6 2 | 3+4 2 | 3+3 2 | 2+2 2 |
| | 3+4 2 | 5+6 2 | 3+4 2 | 3+4 2 | 2+2 2 | 3+3 2 |
| | 3+4 2 | 3+4 2 | 5+6 2 | 5+6 2 | 3+3 2 | 2+2 2 |
| | 5+6 2 | 3+4 2 | 3+4 2 | 5+6 2 | 2+2 2 | 3+3 2 |
| | 3+4 2 | 5+6 2 | 3+4 2 | 5+6 2 | 2+2 2 | 3+3 2 |
| | 5+6 2 | 5+6 2 | 5+6 2 | 3+4 2 | 2+2 2 | 2+2 2 |
| | 3+4 2 | 3+4 2 | 3+4 2 | 3+4 2 | 3+3 2 | 3+3 2 |
| | 5+6 2 | 3+4 2 | 5+6 2 | 5+6 2 | 3+3 2 | 2+2 2 |
| | 5+6 2 | 5+6 2 | 3+4 2 | 3+4 2 | 2+2 2 | 2+2 2 |
| | 3+4 2 | 5+6 2 | 5+6 2 | 3+4 2 | 2+2 2 | 3+3 2 |
| | 3+4 2 | 3+4 2 | 3+4 2 | 5+6 2 | 3+3 2 | 3+3 2 |
| | 5+6 2 | 5+6 2 | 5+6 2 | 5+6 2 | 3+3 2 | 2+2 2 |
| | 3+4 2 | 3+4 2 | 5+6 2 | 3+4 2 | 3+3 2 | 3+3 2 |
| | 5+6 2 | 5+6 2 | 3+4 2 | 5+6 2 | 2+2 2 | 2+2 2 |
| | 5+6 2 | 5+6 2 | 5+6 2 | 5+6 2 | 3+3 2 | 2+2 2 |
| | 3+4 2 | 3+4 2 | 3+4 2 | 3+4 2 | 2+2 2 | 3+3 2 |
| | 3+4 2 | 5+6 2 | 5+6 2 | 5+6 2 | 3+3 2 | 3+3 2 |
| | 3+4 2 | 3+4 2 | 3+4 2 | 3+4 2 | 2+2 2 | 3+3 2 |
| | 5+6 2 | 5+6 2 | 3+4 2 | 5+6 2 | 3+3 2 | 2+2 2 |
| | 5+6 2 | 3+4 2 | 5+6 2 | 3+4 2 | 3+3 2 | 2+2 2 |
| | 5+6 2 | 3+4 2 | 5+6 2 | 3+4 2 | 2+2 2 | 2+2 2 |
| | 3+4 2 | 3+4 2 | 3+4 2 | 5+6 2 | 3+3 2 | 3+3 2 |
| | 3+4 2 | 3+4 2 | 3+4 2 | 3+4 2 | 2+2 2 | 3+3 2 |
| | 5+6 2 | 5+6 2 | 3+4 2 | 5+6 2 | 3+3 2 | 2+2 2 |
| | 5+6 2 | 5+6 2 | 5+6 2 | 3+4 2 | 3+3 2 | 3+3 2 |
| | 3+4 2 | 3+4 2 | 5+6 2 | 5+6 2 | 2+2 2 | 2+2 2 |
| | 3+4 2 | 5+6 2 | 5+6 2 | 3+4 2 | 3+3 2 | 2+2 2 |
| | 5+6 2 | 3+4 2 | 3+4 2 | 5+6 2 | 2+2 2 | 3+3 2 |
| 12.2.126 | 618350 | | 70 | C3Mo2Tm2 | | |
| | -A | -GM | -M | -Y | -L | -V |
| | 3+4 2 | 3+4 2 | 5+6 2 | 5+6 2 | 3+3 2 | 2+2 2 |
| | 5+6 2 | 5+6 2 | 3+4 2 | 3+4 2 | 2+2 2 | 3+3 2 |
| | 3+4 2 | 3+4 2 | 5+6 2 | 5+6 2 | 3+3 2 | 2+2 2 |
| | 5+6 2 | 5+6 2 | 3+4 2 | 3+4 2 | 2+2 2 | 3+3 2 |
| | 3+4 2 | 3+4 2 | 5+6 2 | 3+4 2 | 2+2 2 | 2+2 2 |
| | 5+6 2 | 5+6 2 | 3+4 2 | 5+6 2 | 3+3 2 | 3+3 2 |
| | 3+4 2 | 3+4 2 | 5+6 2 | 5+6 2 | 3+3 2 | 2+2 2 |
| | 5+6 2 | 5+6 2 | 3+4 2 | 3+4 2 | 2+2 2 | 3+3 2 |
| | 5+6 2 | 3+4 2 | 3+4 2 | 5+6 2 | 3+3 2 | 2+2 2 |
| | 3+4 2 | 5+6 2 | 5+6 2 | 3+4 2 | 2+2 2 | 3+3 2 |
| | 5+6 2 | 3+4 2 | 5+6 2 | 3+4 2 | 2+2 2 | 3+3 2 |
| | 3+4 2 | 5+6 2 | 5+6 2 | 5+6 2 | 3+3 2 | 2+2 2 |
| | 5+6 2 | 3+4 2 | 3+4 2 | 3+4 2 | 2+2 2 | 2+2 2 |
| | 3+4 2 | 5+6 2 | 3+4 2 | 5+6 2 | 3+3 2 | 3+3 2 |
| | 3+4 2 | 5+6 2 | 5+6 2 | 3+4 2 | 3+3 2 | 3+3 2 |
| | 5+6 2 | 3+4 2 | 5+6 2 | 3+4 2 | 3+3 2 | 2+2 2 |
| | 3+4 2 | 5+6 2 | 3+4 2 | 5+6 2 | 2+2 2 | 3+3 2 |
| | 5+6 2 | 3+4 2 | 3+4 2 | 5+6 2 | 2+2 2 | 2+2 2 |
| | 3+4 2 | 5+6 2 | 3+4 2 | 5+6 2 | 3+3 2 | 2+2 2 |
| | 3+4 2 | 5+6 2 | 3+4 2 | 5+6 2 | 3+3 2 | 2+2 2 |
| | 5+6 2 | 3+4 2 | 5+6 2 | 3+4 2 | 2+2 2 | 3+3 2 |
| | 3+4 2 | 3+4 2 | 5+6 2 | 3+4 2 | 2+2 2 | 2+2 2 |
| | 5+6 2 | 5+6 2 | 3+4 2 | 5+6 2 | 3+3 2 | 3+3 2 |
| | 5+6 2 | 3+4 2 | 5+6 2 | 3+4 2 | 3+3 2 | 2+2 2 |
| | 3+4 2 | 5+6 2 | 3+4 2 | 5+6 2 | 2+2 2 | 3+3 2 |
| | 3+4 2 | 3+4 2 | 3+4 2 | 5+6 2 | 2+2 2 | 2+2 2 |
| | 5+6 2 | 5+6 2 | 3+4 2 | 3+4 2 | 3+3 2 | 3+3 2 |
| | 5+6 2 | 3+4 2 | 3+4 2 | 5+6 2 | 3+3 2 | 3+3 2 |
| | 3+4 2 | 3+4 2 | 5+6 2 | 3+4 2 | 2+2 2 | 2+2 2 |
| | 5+6 2 | 5+6 2 | 5+6 2 | 5+6 2 | 3+3 2 | 3+3 2 |
| | 3+4 2 | 3+4 2 | 5+6 2 | 3+4 2 | 2+2 2 | 2+2 2 |
| | 3+4 2 | 5+6 2 | 5+6 2 | 5+6 2 | 3+3 2 | 3+3 2 |
| | 5+6 2 | 3+4 2 | 3+4 2 | 3+4 2 | 2+2 2 | 2+2 2 |
| | 3+4 2 | 3+4 2 | 3+4 2 | 3+4 2 | 3+3 2 | 3+3 2 |
| | 5+6 2 | 5+6 2 | 5+6 2 | 5+6 2 | 2+2 2 | 2+2 2 |
| 12.2.127 | 88516 | | 68 | C3Er2Mo2 | | |
| | -A | -GM | -M | -Y | -L | -V |
| | 3+4 2 | 3+4 2 | 5+6 2 | 5+6 2 | 2+2 2 | 2+2 2 |
| | 5+6 2 | 5+6 2 | 3+4 2 | 3+4 2 | 3+3 2 | 3+3 2 |
| | 3+4 2 | 3+4 2 | 5+6 2 | 5+6 2 | 2+2 2 | 2+2 2 |
| | 5+6 2 | 5+6 2 | 3+4 2 | 3+4 2 | 3+3 2 | 3+3 2 |
| | 3+4 2 | 3+4 2 | 5+6 2 | 3+4 2 | 3+3 2 | 2+2 2 |
| | 5+6 2 | 5+6 2 | 3+4 2 | 5+6 2 | 2+2 2 | 3+3 2 |
| | 3+4 2 | 3+4 2 | 5+6 2 | 5+6 2 | 2+2 2 | 2+2 2 |
| | 5+6 2 | 5+6 2 | 3+4 2 | 3+4 2 | 3+3 2 | 3+3 2 |
| | 5+6 2 | 3+4 2 | 3+4 2 | 5+6 2 | 3+3 2 | 3+3 2 |
| | 3+4 2 | 5+6 2 | 5+6 2 | 3+4 2 | 2+2 2 | 2+2 2 |
| | 3+4 2 | 3+4 2 | 3+4 2 | 3+4 2 | 3+3 2 | 3+3 2 |
| | 5+6 2 | 5+6 2 | 5+6 2 | 5+6 2 | 2+2 2 | 2+2 2 |
| | 3+4 2 | 3+4 2 | 5+6 2 | 3+4 2 | 3+3 2 | 2+2 2 |
| | 5+6 2 | 5+6 2 | 3+4 2 | 5+6 2 | 2+2 2 | 3+3 2 |
| | 3+4 2 | 3+4 2 | 3+4 2 | 5+6 2 | 2+2 2 | 2+2 2 |
| | 5+6 2 | 5+6 2 | 5+6 2 | 3+4 2 | 3+3 2 | 3+3 2 |
| | 3+4 2 | 3+4 2 | 5+6 2 | 3+4 2 | 2+2 2 | 2+2 2 |
| | 5+6 2 | 5+6 2 | 3+4 2 | 5+6 2 | 3+3 2 | 3+3 2 |
| | 5+6 2 | 5+6 2 | 5+6 2 | 5+6 2 | 2+2 2 | 2+2 2 |
| | 5+6 2 | 5+6 2 | 5+6 2 | 3+4 2 | 2+2 2 | 2+2 2 |
| | 3+4 2 | 5+6 2 | 3+4 2 | 5+6 2 | 3+3 2 | 3+3 2 |
| | 5+6 2 | 3+4 2 | 3+4 2 | 3+4 2 | 2+2 2 | 2+2 2 |
| | 3+4 2 | 3+4 2 | 5+6 2 | 5+6 2 | 3+3 2 | 3+3 2 |
| | 3+4 2 | 3+4 2 | 5+6 2 | 3+4 2 | 2+2 2 | 2+2 2 |
| | 5+6 2 | 5+6 2 | 3+4 2 | 5+6 2 | 3+3 2 | 3+3 2 |
| | 5+6 2 | 5+6 2 | 5+6 2 | 3+4 2 | 2+2 2 | 2+2 2 |
| | 5+6 2 | 3+4 2 | 3+4 2 | 5+6 2 | 3+3 2 | 3+3 2 |
| | 3+4 2 | 5+6 2 | 3+4 2 | 3+4 2 | 2+2 2 | 3+3 2 |
| | 5+6 2 | 3+4 2 | 5+6 2 | 5+6 2 | 3+3 2 | 2+2 2 |
| | 3+4 2 | 3+4 2 | 3+4 2 | 5+6 2 | 2+2 2 | 3+3 2 |
| | 3+4 2 | 3+4 2 | 3+4 2 | 3+4 2 | 3+3 2 | 2+2 2 |
| | 5+6 2 | 5+6 2 | 5+6 2 | 5+6 2 | 2+2 2 | 3+3 2 |
| | 5+6 2 | 3+4 2 | 5+6 2 | 3+4 2 | 3+3 2 | 2+2 2 |
| | 3+4 2 | 5+6 2 | 3+4 2 | 3+4 2 | 2+2 2 | 3+3 2 |
| 12.2.128 | 617669 | | 68 | C3Er2Mo2 | | |
| | -A | -GM | -M | -Y | -L | -V |
| | 3+4 2 | 3+4 2 | 5+6 2 | 5+6 2 | 3+3 2 | 2+2 2 |
| | 5+6 2 | 5+6 2 | 3+4 2 | 3+4 2 | 2+2 2 | 3+3 2 |
| | 3+4 2 | 3+4 2 | 5+6 2 | 5+6 2 | 3+3 2 | 2+2 2 |
| | 5+6 2 | 5+6 2 | 3+4 2 | 3+4 2 | 2+2 2 | 3+3 2 |
| | 3+4 2 | 3+4 2 | 5+6 2 | 3+4 2 | 2+2 2 | 2+2 2 |
| | 5+6 2 | 5+6 2 | 3+4 2 | 5+6 2 | 3+3 2 | 3+3 2 |

Continued on next column

**TABLE XXIV– continued**

| TQC | ICSD | NoE | SF | | | |
|---|---|---|---|---|---|---|
| | 3+4 2 | 3+4 2 | 5+6 2 | 5+6 2 | 3+3 2 | 2+2 2 |
| | 5+6 2 | 5+6 2 | 3+4 2 | 3+4 2 | 2+2 2 | 3+3 2 |
| | 5+6 2 | 3+4 2 | 3+4 2 | 5+6 2 | 3+3 2 | 2+2 2 |
| | 3+4 2 | 5+6 2 | 5+6 2 | 3+4 2 | 2+2 2 | 3+3 2 |
| | 5+6 2 | 3+4 2 | 5+6 2 | 3+4 2 | 2+2 2 | 3+3 2 |
| | 3+4 2 | 5+6 2 | 5+6 2 | 5+6 2 | 3+3 2 | 2+2 2 |
| | 5+6 2 | 3+4 2 | 3+4 2 | 3+4 2 | 2+2 2 | 2+2 2 |
| | 3+4 2 | 5+6 2 | 3+4 2 | 5+6 2 | 3+3 2 | 3+3 2 |
| | 3+4 2 | 5+6 2 | 5+6 2 | 3+4 2 | 3+3 2 | 3+3 2 |
| | 5+6 2 | 3+4 2 | 5+6 2 | 3+4 2 | 2+2 2 | 2+2 2 |
| | 3+4 2 | 5+6 2 | 3+4 2 | 5+6 2 | 3+3 2 | 3+3 2 |
| | 5+6 2 | 3+4 2 | 3+4 2 | 5+6 2 | 2+2 2 | 2+2 2 |
| | 3+4 2 | 5+6 2 | 3+4 2 | 5+6 2 | 3+3 2 | 2+2 2 |
| | 3+4 2 | 5+6 2 | 3+4 2 | 5+6 2 | 3+3 2 | 2+2 2 |
| | 5+6 2 | 3+4 2 | 5+6 2 | 3+4 2 | 2+2 2 | 3+3 2 |
| | 3+4 2 | 3+4 2 | 5+6 2 | 3+4 2 | 2+2 2 | 2+2 2 |
| | 5+6 2 | 5+6 2 | 3+4 2 | 5+6 2 | 3+3 2 | 3+3 2 |
| | 5+6 2 | 3+4 2 | 5+6 2 | 3+4 2 | 3+3 2 | 3+3 2 |
| | 3+4 2 | 5+6 2 | 3+4 2 | 5+6 2 | 2+2 2 | 2+2 2 |
| | 3+4 2 | 3+4 2 | 3+4 2 | 5+6 2 | 2+2 2 | 2+2 2 |
| | 5+6 2 | 5+6 2 | 3+4 2 | 3+4 2 | 3+3 2 | 3+3 2 |
| | 5+6 2 | 3+4 2 | 3+4 2 | 5+6 2 | 2+2 2 | 3+3 2 |
| | 3+4 2 | 5+6 2 | 5+6 2 | 3+4 2 | 3+3 2 | 2+2 2 |
| | 5+6 2 | 3+4 2 | 5+6 2 | 5+6 2 | 3+3 2 | 3+3 2 |
| | 3+4 2 | 3+4 2 | 5+6 2 | 3+4 2 | 2+2 2 | 2+2 2 |
| | 3+4 2 | 5+6 2 | 3+4 2 | 5+6 2 | 3+3 2 | 3+3 2 |
| | 5+6 2 | 3+4 2 | 5+6 2 | 3+4 2 | 2+2 2 | 2+2 2 |
| | 3+4 2 | 3+4 2 | 3+4 2 | 3+4 2 | 3+3 2 | 3+3 2 |
| 12.2.129 | 617541 | 70 | C3Cr2Tm2 | | | |
| | -A | -GM | -M | -Y | -L | -V |
| | 3+4 2 | 3+4 2 | 5+6 2 | 3+4 2 | 2+2 2 | 2+2 2 |
| | 5+6 2 | 5+6 2 | 3+4 2 | 5+6 2 | 3+3 2 | 3+3 2 |
| | 3+4 2 | 3+4 2 | 5+6 2 | 3+4 2 | 2+2 2 | 2+2 2 |
| | 5+6 2 | 5+6 2 | 3+4 2 | 5+6 2 | 3+3 2 | 3+3 2 |
| | 3+4 2 | 3+4 2 | 5+6 2 | 3+4 2 | 2+2 2 | 2+2 2 |
| | 5+6 2 | 5+6 2 | 3+4 2 | 5+6 2 | 3+3 2 | 3+3 2 |
| | 3+4 2 | 3+4 2 | 5+6 2 | 5+6 2 | 2+2 2 | 2+2 2 |
| | 5+6 2 | 5+6 2 | 3+4 2 | 3+4 2 | 3+3 2 | 3+3 2 |
| | 5+6 2 | 3+4 2 | 3+4 2 | 5+6 2 | 2+2 2 | 3+3 2 |
| | 3+4 2 | 5+6 2 | 5+6 2 | 3+4 2 | 3+3 2 | 2+2 2 |
| | 5+6 2 | 3+4 2 | 5+6 2 | 3+4 2 | 2+2 2 | 3+3 2 |
| | 3+4 2 | 5+6 2 | 3+4 2 | 5+6 2 | 3+3 2 | 2+2 2 |
| | 5+6 2 | 3+4 2 | 5+6 2 | 3+4 2 | 2+2 2 | 2+2 2 |
| | 3+4 2 | 5+6 2 | 3+4 2 | 5+6 2 | 3+3 2 | 3+3 2 |
| | 3+4 2 | 5+6 2 | 3+4 2 | 5+6 2 | 3+3 2 | 2+2 2 |
| | 3+4 2 | 3+4 2 | 5+6 2 | 3+4 2 | 3+3 2 | 3+3 2 |
| | 5+6 2 | 5+6 2 | 3+4 2 | 3+4 2 | 2+2 2 | 3+3 2 |
| | 5+6 2 | 3+4 2 | 5+6 2 | 5+6 2 | 3+3 2 | 2+2 2 |
| | 3+4 2 | 5+6 2 | 3+4 2 | 5+6 2 | 2+2 2 | 2+2 2 |
| | 3+4 2 | 5+6 2 | 3+4 2 | 5+6 2 | 3+3 2 | 2+2 2 |
| | 5+6 2 | 3+4 2 | 5+6 2 | 3+4 2 | 2+2 2 | 3+3 2 |
| | 3+4 2 | 3+4 2 | 5+6 2 | 3+4 2 | 3+3 2 | 2+2 2 |
| | 5+6 2 | 5+6 2 | 3+4 2 | 5+6 2 | 2+2 2 | 3+3 2 |
| | 5+6 2 | 3+4 2 | 5+6 2 | 3+4 2 | 3+3 2 | 3+3 2 |
| | 3+4 2 | 5+6 2 | 3+4 2 | 5+6 2 | 2+2 2 | 2+2 2 |
| | 3+4 2 | 5+6 2 | 3+4 2 | 5+6 2 | 3+3 2 | 2+2 2 |
| | 5+6 2 | 3+4 2 | 3+4 2 | 3+4 2 | 2+2 2 | 3+3 2 |
| | 5+6 2 | 3+4 2 | 5+6 2 | 5+6 2 | 3+3 2 | 3+3 2 |
| | 3+4 2 | 3+4 2 | 3+4 2 | 3+4 2 | 2+2 2 | 2+2 2 |
| | 5+6 2 | 5+6 2 | 5+6 2 | 5+6 2 | 3+3 2 | 3+3 2 |
| | 5+6 2 | 3+4 2 | 5+6 2 | 3+4 2 | 2+2 2 | 2+2 2 |

Continued on next column

**TABLE XXIV– continued**

| TQC | ICSD | NoE | SF | | | |
|---|---|---|---|---|---|---|
| | 3+4 2 | 5+6 2 | 3+4 2 | 5+6 2 | 3+3 2 | 3+3 2 |
| | 3+4 2 | 3+4 2 | 5+6 2 | 3+4 2 | 2+2 2 | 2+2 2 |
| | 5+6 2 | 5+6 2 | 3+4 2 | 5+6 2 | 3+3 2 | 3+3 2 |
| | 3+4 2 | 3+4 2 | 5+6 2 | 3+4 2 | 2+2 2 | 2+2 2 |
| 12.2.130 | 617494 | 68 | C3Cr2Er2 | | | |
| | -A | -GM | -M | -Y | -L | -V |
| | 3+4 2 | 3+4 2 | 5+6 2 | 3+4 2 | 2+2 2 | 2+2 2 |
| | 5+6 2 | 5+6 2 | 3+4 2 | 5+6 2 | 3+3 2 | 3+3 2 |
| | 3+4 2 | 3+4 2 | 5+6 2 | 3+4 2 | 2+2 2 | 2+2 2 |
| | 5+6 2 | 5+6 2 | 3+4 2 | 5+6 2 | 3+3 2 | 3+3 2 |
| | 3+4 2 | 3+4 2 | 5+6 2 | 3+4 2 | 2+2 2 | 2+2 2 |
| | 5+6 2 | 5+6 2 | 3+4 2 | 5+6 2 | 3+3 2 | 3+3 2 |
| | 3+4 2 | 3+4 2 | 5+6 2 | 5+6 2 | 2+2 2 | 2+2 2 |
| | 5+6 2 | 5+6 2 | 3+4 2 | 3+4 2 | 3+3 2 | 3+3 2 |
| | 5+6 2 | 3+4 2 | 3+4 2 | 5+6 2 | 2+2 2 | 3+3 2 |
| | 3+4 2 | 5+6 2 | 5+6 2 | 3+4 2 | 3+3 2 | 2+2 2 |
| | 5+6 2 | 3+4 2 | 5+6 2 | 3+4 2 | 2+2 2 | 3+3 2 |
| | 3+4 2 | 5+6 2 | 3+4 2 | 5+6 2 | 3+3 2 | 2+2 2 |
| | 5+6 2 | 3+4 2 | 5+6 2 | 3+4 2 | 2+2 2 | 2+2 2 |
| | 3+4 2 | 5+6 2 | 3+4 2 | 5+6 2 | 3+3 2 | 3+3 2 |
| | 3+4 2 | 5+6 2 | 3+4 2 | 5+6 2 | 3+3 2 | 2+2 2 |
| | 3+4 2 | 3+4 2 | 5+6 2 | 3+4 2 | 3+3 2 | 3+3 2 |
| | 5+6 2 | 5+6 2 | 3+4 2 | 3+4 2 | 3+3 2 | 3+3 2 |
| | 5+6 2 | 3+4 2 | 5+6 2 | 5+6 2 | 2+2 2 | 2+2 2 |
| | 3+4 2 | 5+6 2 | 3+4 2 | 5+6 2 | 2+2 2 | 2+2 2 |
| | 3+4 2 | 5+6 2 | 3+4 2 | 5+6 2 | 3+3 2 | 2+2 2 |
| | 5+6 2 | 3+4 2 | 5+6 2 | 3+4 2 | 2+2 2 | 3+3 2 |
| | 3+4 2 | 3+4 2 | 3+4 2 | 3+4 2 | 3+3 2 | 2+2 2 |
| | 5+6 2 | 5+6 2 | 5+6 2 | 5+6 2 | 2+2 2 | 3+3 2 |
| | 5+6 2 | 3+4 2 | 5+6 2 | 3+4 2 | 3+3 2 | 3+3 2 |
| | 3+4 2 | 5+6 2 | 3+4 2 | 5+6 2 | 2+2 2 | 2+2 2 |
| | 3+4 2 | 5+6 2 | 3+4 2 | 5+6 2 | 3+3 2 | 2+2 2 |
| | 5+6 2 | 3+4 2 | 3+4 2 | 3+4 2 | 2+2 2 | 3+3 2 |
| | 5+6 2 | 3+4 2 | 5+6 2 | 5+6 2 | 3+3 2 | 3+3 2 |
| | 3+4 2 | 5+6 2 | 3+4 2 | 3+4 2 | 2+2 2 | 2+2 2 |
| | 5+6 2 | 3+4 2 | 5+6 2 | 5+6 2 | 3+3 2 | 3+3 2 |
| | 5+6 2 | 3+4 2 | 5+6 2 | 5+6 2 | 2+2 2 | 2+2 2 |
| | 3+4 2 | 5+6 2 | 3+4 2 | 3+4 2 | 3+3 2 | 3+3 2 |
| | 3+4 2 | 3+4 2 | 5+6 2 | 3+4 2 | 2+2 2 | 2+2 2 |
| | 5+6 2 | 5+6 2 | 5+6 2 | 5+6 2 | 3+3 2 | 3+3 2 |
| 12.2.131 | 252737 | 20 | C2Ca1 | | | |
| | -A | -GM | -M | -Y | -L | -V |
| | 5+6 2 | 3+4 2 | 5+6 2 | 3+4 2 | 3+3 2 | 2+2 2 |
| | 3+4 2 | 5+6 2 | 3+4 2 | 5+6 2 | 2+2 2 | 3+3 2 |
| | 3+4 2 | 3+4 2 | 3+4 2 | 5+6 2 | 2+2 2 | 3+3 2 |
| | 5+6 2 | 5+6 2 | 5+6 2 | 5+6 2 | 2+2 2 | 3+3 2 |
| | 3+4 2 | 5+6 2 | 3+4 2 | 3+4 2 | 3+3 2 | 2+2 2 |
| | 5+6 2 | 3+4 2 | 5+6 2 | 5+6 2 | 3+3 2 | 2+2 2 |
| | 5+6 2 | 3+4 2 | 5+6 2 | 3+4 2 | 3+3 2 | 2+2 2 |
| | 3+4 2 | 3+4 2 | 3+4 2 | 3+4 2 | 2+2 2 | 3+3 2 |
| | 5+6 2 | 5+6 2 | 3+4 2 | 5+6 2 | 2+2 2 | 3+3 2 |
| | 3+4 2 | 5+6 2 | 5+6 2 | 5+6 2 | 3+3 2 | 2+2 2 |
| 12.2.132 | 252734 | 20 | C2Ca1 | | | |
| | -A | -GM | -M | -Y | -L | -V |
| | 5+6 2 | 3+4 2 | 5+6 2 | 3+4 2 | 2+2 2 | 3+3 2 |
| | 3+4 2 | 3+4 2 | 3+4 2 | 3+4 2 | 2+2 2 | 2+2 2 |
| | 5+6 2 | 5+6 2 | 5+6 2 | 5+6 2 | 3+3 2 | 3+3 2 |
| | 3+4 2 | 5+6 2 | 3+4 2 | 5+6 2 | 3+3 2 | 2+2 2 |
| | 3+4 2 | 3+4 2 | 3+4 2 | 5+6 2 | 3+3 2 | 2+2 2 |
| | 5+6 2 | 5+6 2 | 5+6 2 | 3+4 2 | 3+3 2 | 3+3 2 |
| | 3+4 2 | 5+6 2 | 3+4 2 | 5+6 2 | 2+2 2 | 2+2 2 |

Continued on next column

**TABLE XXIV– continued**

| TQC | ICSD | | NoE | SF | | |
|---|---|---|---|---|---|---|
| | 3+4 2 | 3+4 2 | 5+6 2 | 5+6 2 | 3+3 2 | 2+2 2 |
| | 5+6 2 | 5+6 2 | 3+4 2 | 3+4 2 | 2+2 2 | 3+3 2 |
| | 3+4 2 | 5+6 2 | 5+6 2 | 5+6 2 | 3+3 2 | 3+3 2 |
| 12.2.133 | 252716 | | 20 | C2Ca1 | | |
| | -A | -GM | -M | -Y | -L | -V |
| | 5+6 2 | 3+4 2 | 5+6 2 | 3+4 2 | 2+2 2 | 3+3 2 |
| | 5+6 2 | 3+4 2 | 5+6 2 | 3+4 2 | 3+3 2 | 2+2 2 |
| | 3+4 2 | 5+6 2 | 3+4 2 | 5+6 2 | 2+2 2 | 3+3 2 |
| | 3+4 2 | 5+6 2 | 3+4 2 | 5+6 2 | 3+3 2 | 2+2 2 |
| | 3+4 2 | 5+6 2 | 3+4 2 | 5+6 2 | 3+3 2 | 2+2 2 |
| | 5+6 2 | 3+4 2 | 5+6 2 | 3+4 2 | 2+2 2 | 3+3 2 |
| | 5+6 2 | 3+4 2 | 5+6 2 | 3+4 2 | 2+2 2 | 3+3 2 |
| | 3+4 2 | 5+6 2 | 3+4 2 | 5+6 2 | 3+3 2 | 2+2 2 |
| | 3+4 2 | 5+6 2 | 3+4 2 | 5+6 2 | 3+3 2 | 2+2 2 |
| | 3+4 2 | 3+4 2 | 5+6 2 | 3+4 2 | 2+2 2 | 3+3 2 |
| 12.2.134 | 617825 | | 50 | C1Fe2Si2Y2 | | |
| | -A | -GM | -M | -Y | -L | -V |
| | 5+6 2 | 3+4 2 | 5+6 2 | 3+4 2 | 2+2 2 | 3+3 2 |
| | 3+4 2 | 5+6 2 | 3+4 2 | 5+6 2 | 3+3 2 | 2+2 2 |
| | 5+6 2 | 3+4 2 | 5+6 2 | 3+4 2 | 2+2 2 | 3+3 2 |
| | 3+4 2 | 5+6 2 | 3+4 2 | 5+6 2 | 3+3 2 | 2+2 2 |
| | 5+6 2 | 3+4 2 | 5+6 2 | 3+4 2 | 2+2 2 | 3+3 2 |
| | 3+4 2 | 3+4 2 | 3+4 2 | 5+6 2 | 3+3 2 | 2+2 2 |
| | 3+4 2 | 5+6 2 | 5+6 2 | 5+6 2 | 2+2 2 | 3+3 2 |
| | 5+6 2 | 5+6 2 | 3+4 2 | 3+4 2 | 3+3 2 | 2+2 2 |
| | 5+6 2 | 3+4 2 | 5+6 2 | 3+4 2 | 3+3 2 | 2+2 2 |
| | 3+4 2 | 3+4 2 | 3+4 2 | 3+4 2 | 2+2 2 | 2+2 2 |
| | 5+6 2 | 5+6 2 | 5+6 2 | 5+6 2 | 3+3 2 | 3+3 2 |
| | 3+4 2 | 5+6 2 | 3+4 2 | 5+6 2 | 2+2 2 | 3+3 2 |
| | 3+4 2 | 3+4 2 | 5+6 2 | 3+4 2 | 3+3 2 | 3+3 2 |
| | 3+4 2 | 5+6 2 | 3+4 2 | 5+6 2 | 2+2 2 | 2+2 2 |
| | 3+4 2 | 5+6 2 | 3+4 2 | 5+6 2 | 2+2 2 | 3+3 2 |
| | 3+4 2 | 5+6 2 | 3+4 2 | 3+4 2 | 2+2 2 | 3+3 2 |
| | 5+6 2 | 5+6 2 | 5+6 2 | 5+6 2 | 3+3 2 | 2+2 2 |
| | 5+6 2 | 3+4 2 | 5+6 2 | 3+4 2 | 2+2 2 | 3+3 2 |
| | 5+6 2 | 5+6 2 | 3+4 2 | 5+6 2 | 2+2 2 | 2+2 2 |
| | 3+4 2 | 3+4 2 | 3+4 2 | 5+6 2 | 3+3 2 | 2+2 2 |
| | 5+6 2 | 3+4 2 | 3+4 2 | 5+6 2 | 2+2 2 | 3+3 2 |
| | 5+6 2 | 3+4 2 | 5+6 2 | 3+4 2 | 3+3 2 | 2+2 2 |
| | 5+6 2 | 3+4 2 | 5+6 2 | 3+4 2 | 3+3 2 | 2+2 2 |
| | 5+6 2 | 5+6 2 | 3+4 2 | 5+6 2 | 3+3 2 | 3+3 2 |
| | 3+4 2 | 3+4 2 | 3+4 2 | 3+4 2 | 2+2 2 | 2+2 2 |
| 12.2.135 | 617663 | | 72 | C1Er2Fe2Si2 | | |
| | -A | -GM | -M | -Y | -L | -V |
| | 5+6 2 | 3+4 2 | 5+6 2 | 3+4 2 | 2+2 2 | 3+3 2 |
| | 3+4 2 | 5+6 2 | 3+4 2 | 5+6 2 | 3+3 2 | 2+2 2 |
| | 5+6 2 | 3+4 2 | 5+6 2 | 3+4 2 | 2+2 2 | 3+3 2 |
| | 3+4 2 | 5+6 2 | 3+4 2 | 5+6 2 | 3+3 2 | 2+2 2 |
| | 5+6 2 | 3+4 2 | 5+6 2 | 3+4 2 | 2+2 2 | 3+3 2 |
| | 3+4 2 | 3+4 2 | 3+4 2 | 5+6 2 | 3+3 2 | 2+2 2 |
| | 3+4 2 | 5+6 2 | 5+6 2 | 5+6 2 | 2+2 2 | 3+3 2 |
| | 5+6 2 | 5+6 2 | 3+4 2 | 3+4 2 | 3+3 2 | 2+2 2 |
| | 5+6 2 | 3+4 2 | 5+6 2 | 3+4 2 | 3+3 2 | 2+2 2 |
| | 3+4 2 | 3+4 2 | 3+4 2 | 3+4 2 | 2+2 2 | 2+2 2 |
| | 5+6 2 | 5+6 2 | 5+6 2 | 5+6 2 | 3+3 2 | 3+3 2 |
| | 3+4 2 | 5+6 2 | 3+4 2 | 5+6 2 | 2+2 2 | 3+3 2 |
| | 3+4 2 | 3+4 2 | 5+6 2 | 3+4 2 | 3+3 2 | 3+3 2 |
| | 3+4 2 | 5+6 2 | 3+4 2 | 5+6 2 | 2+2 2 | 2+2 2 |
| | 3+4 2 | 5+6 2 | 3+4 2 | 5+6 2 | 2+2 2 | 3+3 2 |
| | 3+4 2 | 5+6 2 | 3+4 2 | 5+6 2 | 2+2 2 | 3+3 2 |
| | 5+6 2 | 5+6 2 | 5+6 2 | 3+4 2 | 3+3 2 | 2+2 2 |

Continued on next column

**TABLE XXIV– continued**

| TQC | ICSD | | NoE | SF | | |
|---|---|---|---|---|---|---|
| | 5+6 2 | 3+4 2 | 5+6 2 | 3+4 2 | 2+2 2 | 3+3 2 |
| | 5+6 2 | 5+6 2 | 3+4 2 | 5+6 2 | 2+2 2 | 2+2 2 |
| | 3+4 2 | 3+4 2 | 3+4 2 | 5+6 2 | 3+3 2 | 2+2 2 |
| | 5+6 2 | 3+4 2 | 3+4 2 | 5+6 2 | 3+3 2 | 2+2 2 |
| | 5+6 2 | 5+6 2 | 5+6 2 | 3+4 2 | 2+2 2 | 3+3 2 |
| | 3+4 2 | 3+4 2 | 5+6 2 | 3+4 2 | 3+3 2 | 2+2 2 |
| | 5+6 2 | 3+4 2 | 3+4 2 | 5+6 2 | 2+2 2 | 3+3 2 |
| | 3+4 2 | 5+6 2 | 5+6 2 | 3+4 2 | 3+3 2 | 3+3 2 |
| | 5+6 2 | 3+4 2 | 3+4 2 | 5+6 2 | 2+2 2 | 2+2 2 |
| | 3+4 2 | 5+6 2 | 3+4 2 | 3+4 2 | 3+3 2 | 3+3 2 |
| | 5+6 2 | 3+4 2 | 5+6 2 | 5+6 2 | 2+2 2 | 2+2 2 |
| | 5+6 2 | 3+4 2 | 5+6 2 | 3+4 2 | 3+3 2 | 2+2 2 |
| | 5+6 2 | 5+6 2 | 3+4 2 | 3+4 2 | 3+3 2 | 3+3 2 |
| | 3+4 2 | 3+4 2 | 3+4 2 | 5+6 2 | 2+2 2 | 2+2 2 |
| | 3+4 2 | 5+6 2 | 5+6 2 | 3+4 2 | 2+2 2 | 3+3 2 |
| | 3+4 2 | 3+4 2 | 3+4 2 | 5+6 2 | 3+3 2 | 3+3 2 |
| | 5+6 2 | 5+6 2 | 5+6 2 | 5+6 2 | 2+2 2 | 2+2 2 |
| | 5+6 2 | 3+4 2 | 5+6 2 | 3+4 2 | 3+3 2 | 3+3 2 |
| | 3+4 2 | 5+6 2 | 3+4 2 | 3+4 2 | 2+2 2 | 2+2 2 |
| 12.2.136 | 61141 | | 144 | C10Fe22Mo12 | | |
| | -A | -GM | -M | -Y | -L | -V |
| | 5+6 2 | 3+4 2 | 5+6 2 | 5+6 2 | 3+3 2 | 3+3 2 |
| | 3+4 2 | 5+6 2 | 3+4 2 | 3+4 2 | 2+2 2 | 2+2 2 |
| | 3+4 2 | 3+4 2 | 3+4 2 | 5+6 2 | 3+3 2 | 3+3 2 |
| | 5+6 2 | 3+4 2 | 3+4 2 | 3+4 2 | 2+2 2 | 2+2 2 |
| | 3+4 2 | 5+6 2 | 5+6 2 | 3+4 2 | 3+3 2 | 3+3 2 |
| | 5+6 2 | 5+6 2 | 5+6 2 | 3+4 2 | 2+2 2 | 2+2 2 |
| | 3+4 2 | 5+6 2 | 3+4 2 | 5+6 2 | 2+2 2 | 3+3 2 |
| | 5+6 2 | 3+4 2 | 5+6 2 | 5+6 2 | 3+3 2 | 2+2 2 |
| | 5+6 2 | 5+6 2 | 5+6 2 | 3+4 2 | 2+2 2 | 2+2 2 |
| | 5+6 2 | 3+4 2 | 3+4 2 | 5+6 2 | 3+3 2 | 3+3 2 |
| | 3+4 2 | 3+4 2 | 3+4 2 | 5+6 2 | 2+2 2 | 3+3 2 |
| | 3+4 2 | 5+6 2 | 3+4 2 | 3+4 2 | 3+3 2 | 2+2 2 |
| | 5+6 2 | 3+4 2 | 5+6 2 | 3+4 2 | 2+2 2 | 2+2 2 |
| | 3+4 2 | 5+6 2 | 5+6 2 | 5+6 2 | 3+3 2 | 2+2 2 |
| | 3+4 2 | 3+4 2 | 5+6 2 | 3+4 2 | 2+2 2 | 3+3 2 |
| | 5+6 2 | 5+6 2 | 3+4 2 | 3+4 2 | 3+3 2 | 2+2 2 |
| | 3+4 2 | 5+6 2 | 5+6 2 | 5+6 2 | 2+2 2 | 3+3 2 |
| | 5+6 2 | 5+6 2 | 3+4 2 | 3+4 2 | 2+2 2 | 2+2 2 |
| | 5+6 2 | 5+6 2 | 5+6 2 | 5+6 2 | 2+2 2 | 3+3 2 |
| | 3+4 2 | 3+4 2 | 3+4 2 | 5+6 2 | 3+3 2 | 2+2 2 |
| | 5+6 2 | 3+4 2 | 5+6 2 | 5+6 2 | 2+2 2 | 2+2 2 |
| | 3+4 2 | 3+4 2 | 5+6 2 | 3+4 2 | 3+3 2 | 3+3 2 |
| | 3+4 2 | 5+6 2 | 5+6 2 | 3+4 2 | 2+2 2 | 2+2 2 |
| | 5+6 2 | 5+6 2 | 5+6 2 | 3+4 2 | 3+3 2 | 3+3 2 |
| | 5+6 2 | 3+4 2 | 3+4 2 | 5+6 2 | 2+2 2 | 2+2 2 |
| | 3+4 2 | 3+4 2 | 3+4 2 | 5+6 2 | 3+3 2 | 3+3 2 |
| | 3+4 2 | 5+6 2 | 3+4 2 | 5+6 2 | 3+3 2 | 2+2 2 |
| | 5+6 2 | 3+4 2 | 3+4 2 | 5+6 2 | 2+2 2 | 3+3 2 |
| | 3+4 2 | 5+6 2 | 3+4 2 | 3+4 2 | 2+2 2 | 3+3 2 |
| | 3+4 2 | 5+6 2 | 3+4 2 | 3+4 2 | 3+3 2 | 2+2 2 |
| | 5+6 2 | 3+4 2 | 3+4 2 | 3+4 2 | 3+3 2 | 3+3 2 |
| | 5+6 2 | 3+4 2 | 5+6 2 | 3+4 2 | 2+2 2 | 3+3 2 |
| | 5+6 2 | 3+4 2 | 3+4 2 | 5+6 2 | 2+2 2 | 2+2 2 |
| | 3+4 2 | 3+4 2 | 5+6 2 | 3+4 2 | 3+3 2 | 2+2 2 |
| | 3+4 2 | 5+6 2 | 3+4 2 | 5+6 2 | 2+2 2 | 3+3 2 |
| | 5+6 2 | 5+6 2 | 3+4 2 | 5+6 2 | 2+2 2 | 2+2 2 |
| | 5+6 2 | 3+4 2 | 5+6 2 | 5+6 2 | 3+3 2 | 2+2 2 |
| | 3+4 2 | 3+4 2 | 5+6 2 | 5+6 2 | 3+3 2 | 3+3 2 |
| | 3+4 2 | 5+6 2 | 3+4 2 | 3+4 2 | 2+2 2 | 2+2 2 |
| | 5+6 2 | 3+4 2 | 3+4 2 | 3+4 2 | 2+2 2 | 2+2 2 |

Continued on next column

**TABLE XXIV– continued**

| TQC | ICSD | NoE | SF | | | |
|---|---|---|---|---|---|---|
| | 3+4 2 | 5+6 2 | 3+4 2 | 5+6 2 | 2+2 2 | 3+3 2 |
| | 3+4 2 | 5+6 2 | 5+6 2 | 3+4 2 | 3+3 2 | 2+2 2 |
| | 5+6 2 | 3+4 2 | 5+6 2 | 3+4 2 | 3+3 2 | 3+3 2 |
| | 5+6 2 | 5+6 2 | 3+4 2 | 5+6 2 | 2+2 2 | 2+2 2 |
| | 3+4 2 | 3+4 2 | 5+6 2 | 3+4 2 | 2+2 2 | 3+3 2 |
| | 3+4 2 | 3+4 2 | 5+6 2 | 3+4 2 | 3+3 2 | 3+3 2 |
| | 5+6 2 | 3+4 2 | 3+4 2 | 5+6 2 | 3+3 2 | 2+2 2 |
| | 5+6 2 | 5+6 2 | 5+6 2 | 5+6 2 | 2+2 2 | 3+3 2 |
| | 3+4 2 | 5+6 2 | 3+4 2 | 5+6 2 | 3+3 2 | 2+2 2 |
| | 5+6 2 | 3+4 2 | 3+4 2 | 3+4 2 | 3+3 2 | 3+3 2 |
| | 3+4 2 | 5+6 2 | 5+6 2 | 5+6 2 | 2+2 2 | 2+2 2 |
| | 3+4 2 | 3+4 2 | 5+6 2 | 3+4 2 | 3+3 2 | 3+3 2 |
| | 5+6 2 | 3+4 2 | 5+6 2 | 3+4 2 | 2+2 2 | 2+2 2 |
| | 3+4 2 | 5+6 2 | 3+4 2 | 5+6 2 | 2+2 2 | 3+3 2 |
| | 5+6 2 | 5+6 2 | 5+6 2 | 3+4 2 | 3+3 2 | 2+2 2 |
| | 5+6 2 | 5+6 2 | 5+6 2 | 5+6 2 | 2+2 2 | 3+3 2 |
| | 3+4 2 | 3+4 2 | 3+4 2 | 5+6 2 | 3+3 2 | 2+2 2 |
| | 3+4 2 | 3+4 2 | 5+6 2 | 3+4 2 | 3+3 2 | 2+2 2 |
| | 5+6 2 | 5+6 2 | 5+6 2 | 3+4 2 | 3+3 2 | 3+3 2 |
| | 5+6 2 | 3+4 2 | 3+4 2 | 3+4 2 | 2+2 2 | 3+3 2 |
| | 5+6 2 | 5+6 2 | 3+4 2 | 5+6 2 | 3+3 2 | 3+3 2 |
| | 5+6 2 | 3+4 2 | 5+6 2 | 3+4 2 | 2+2 2 | 2+2 2 |
| | 3+4 2 | 3+4 2 | 3+4 2 | 5+6 2 | 3+3 2 | 2+2 2 |
| | 3+4 2 | 3+4 2 | 3+4 2 | 3+4 2 | 2+2 2 | 3+3 2 |
| | 3+4 2 | 5+6 2 | 5+6 2 | 5+6 2 | 3+3 2 | 2+2 2 |
| | 5+6 2 | 3+4 2 | 5+6 2 | 5+6 2 | 2+2 2 | 3+3 2 |
| | 3+4 2 | 3+4 2 | 3+4 2 | 3+4 2 | 3+3 2 | 3+3 2 |
| | 5+6 2 | 5+6 2 | 5+6 2 | 3+4 2 | 2+2 2 | 2+2 2 |
| | 3+4 2 | 3+4 2 | 3+4 2 | 3+4 2 | 2+2 2 | 3+3 2 |
| | 5+6 2 | 3+4 2 | 5+6 2 | 5+6 2 | 3+3 2 | 2+2 2 |
| | 3+4 2 | 5+6 2 | 3+4 2 | 5+6 2 | 2+2 2 | 2+2 2 |
| | 5+6 2 | 5+6 2 | 5+6 2 | 3+4 2 | 3+3 2 | 3+3 2 |
| 12.2.137 | 65114 | 166 | C10Au2Cl8S8 | | | |
| | -A | -GM | -M | -Y | -L | -V |
| | 5+6 2 | 3+4 2 | 5+6 2 | 3+4 2 | 3+3 2 | 2+2 2 |
| | 3+4 2 | 5+6 2 | 3+4 2 | 5+6 2 | 2+2 2 | 3+3 2 |
| | 5+6 2 | 3+4 2 | 5+6 2 | 3+4 2 | 3+3 2 | 2+2 2 |
| | 3+4 2 | 5+6 2 | 3+4 2 | 5+6 2 | 2+2 2 | 3+3 2 |
| | 5+6 2 | 3+4 2 | 5+6 2 | 3+4 2 | 3+3 2 | 2+2 2 |
| | 3+4 2 | 5+6 2 | 3+4 2 | 5+6 2 | 2+2 2 | 3+3 2 |
| | 3+4 2 | 5+6 2 | 3+4 2 | 5+6 2 | 2+2 2 | 3+3 2 |
| | 5+6 2 | 3+4 2 | 5+6 2 | 3+4 2 | 3+3 2 | 2+2 2 |
| | 5+6 2 | 3+4 2 | 3+4 2 | 5+6 2 | 2+2 2 | 3+3 2 |
| | 3+4 2 | 5+6 2 | 5+6 2 | 3+4 2 | 3+3 2 | 2+2 2 |
| | 5+6 2 | 3+4 2 | 5+6 2 | 3+4 2 | 3+3 2 | 2+2 2 |
| | 3+4 2 | 3+4 2 | 3+4 2 | 3+4 2 | 3+3 2 | 3+3 2 |
| | 5+6 2 | 5+6 2 | 5+6 2 | 5+6 2 | 2+2 2 | 2+2 2 |
| | 5+6 2 | 3+4 2 | 3+4 2 | 3+4 2 | 3+3 2 | 3+3 2 |
| | 3+4 2 | 5+6 2 | 5+6 2 | 5+6 2 | 2+2 2 | 2+2 2 |
| | 3+4 2 | 3+4 2 | 3+4 2 | 3+4 2 | 3+3 2 | 3+3 2 |
| | 5+6 2 | 5+6 2 | 5+6 2 | 5+6 2 | 2+2 2 | 2+2 2 |
| | 3+4 2 | 5+6 2 | 3+4 2 | 5+6 2 | 2+2 2 | 3+3 2 |
| | 5+6 2 | 3+4 2 | 5+6 2 | 3+4 2 | 3+3 2 | 2+2 2 |
| | 3+4 2 | 5+6 2 | 3+4 2 | 5+6 2 | 2+2 2 | 3+3 2 |
| | 5+6 2 | 3+4 2 | 5+6 2 | 5+6 2 | 3+3 2 | 2+2 2 |
| | 3+4 2 | 5+6 2 | 3+4 2 | 3+4 2 | 2+2 2 | 3+3 2 |
| | 5+6 2 | 3+4 2 | 5+6 2 | 3+4 2 | 3+3 2 | 2+2 2 |
| | 3+4 2 | 5+6 2 | 3+4 2 | 5+6 2 | 2+2 2 | 3+3 2 |
| | 3+4 2 | 5+6 2 | 3+4 2 | 5+6 2 | 2+2 2 | 3+3 2 |
| | 5+6 2 | 3+4 2 | 5+6 2 | 3+4 2 | 3+3 2 | 2+2 2 |
| | 5+6 2 | 3+4 2 | 5+6 2 | 3+4 2 | 3+3 2 | 2+2 2 |

Continued on next column

**TABLE XXIV– continued**

| TQC | ICSD | NoE | SF | | | |
|---|---|---|---|---|---|---|
| | 3+4 2 | 5+6 2 | 3+4 2 | 5+6 2 | 2+2 2 | 3+3 2 |
| | 5+6 2 | 3+4 2 | 5+6 2 | 3+4 2 | 3+3 2 | 2+2 2 |
| | 3+4 2 | 3+4 2 | 3+4 2 | 3+4 2 | 3+3 2 | 2+2 2 |
| | 5+6 2 | 5+6 2 | 5+6 2 | 3+4 2 | 2+2 2 | 3+3 2 |
| | 5+6 2 | 5+6 2 | 3+4 2 | 5+6 2 | 2+2 2 | 2+2 2 |
| | 3+4 2 | 3+4 2 | 5+6 2 | 5+6 2 | 3+3 2 | 3+3 2 |
| | 3+4 2 | 5+6 2 | 3+4 2 | 5+6 2 | 2+2 2 | 3+3 2 |
| | 5+6 2 | 3+4 2 | 3+4 2 | 5+6 2 | 3+3 2 | 3+3 2 |
| | 3+4 2 | 5+6 2 | 3+4 2 | 3+4 2 | 2+2 2 | 2+2 2 |
| | 3+4 2 | 5+6 2 | 5+6 2 | 5+6 2 | 2+2 2 | 3+3 2 |
| | 5+6 2 | 3+4 2 | 5+6 2 | 3+4 2 | 3+3 2 | 2+2 2 |
| | 5+6 2 | 3+4 2 | 5+6 2 | 3+4 2 | 3+3 2 | 2+2 2 |
| | 3+4 2 | 5+6 2 | 3+4 2 | 5+6 2 | 2+2 2 | 3+3 2 |
| | 5+6 2 | 3+4 2 | 3+4 2 | 3+4 2 | 3+3 2 | 2+2 2 |
| | 3+4 2 | 5+6 2 | 5+6 2 | 5+6 2 | 2+2 2 | 3+3 2 |
| | 3+4 2 | 5+6 2 | 3+4 2 | 5+6 2 | 2+2 2 | 3+3 2 |
| | 5+6 2 | 3+4 2 | 5+6 2 | 3+4 2 | 3+3 2 | 3+3 2 |
| | 3+4 2 | 5+6 2 | 3+4 2 | 5+6 2 | 2+2 2 | 2+2 2 |
| | 5+6 2 | 5+6 2 | 5+6 2 | 5+6 2 | 3+3 2 | 3+3 2 |
| | 5+6 2 | 3+4 2 | 5+6 2 | 3+4 2 | 3+3 2 | 2+2 2 |
| | 3+4 2 | 3+4 2 | 3+4 2 | 3+4 2 | 2+2 2 | 2+2 2 |
| | 5+6 2 | 3+4 2 | 5+6 2 | 5+6 2 | 3+3 2 | 3+3 2 |
| | 5+6 2 | 5+6 2 | 5+6 2 | 3+4 2 | 3+3 2 | 2+2 2 |
| | 3+4 2 | 3+4 2 | 3+4 2 | 3+4 2 | 2+2 2 | 2+2 2 |
| | 3+4 2 | 3+4 2 | 5+6 2 | 3+4 2 | 3+3 2 | 3+3 2 |
| | 5+6 2 | 5+6 2 | 3+4 2 | 5+6 2 | 2+2 2 | 2+2 2 |
| | 3+4 2 | 3+4 2 | 5+6 2 | 3+4 2 | 2+2 2 | 2+2 2 |
| | 5+6 2 | 5+6 2 | 3+4 2 | 5+6 2 | 3+3 2 | 3+3 2 |
| | 3+4 2 | 3+4 2 | 3+4 2 | 3+4 2 | 2+2 2 | 2+2 2 |
| | 3+4 2 | 3+4 2 | 5+6 2 | 5+6 2 | 2+2 2 | 3+3 2 |
| | 5+6 2 | 3+4 2 | 5+6 2 | 3+4 2 | 3+3 2 | 2+2 2 |
| | 5+6 2 | 5+6 2 | 3+4 2 | 5+6 2 | 3+3 2 | 3+3 2 |
| | 3+4 2 | 5+6 2 | 5+6 2 | 3+4 2 | 2+2 2 | 2+2 2 |
| | 5+6 2 | 3+4 2 | 3+4 2 | 5+6 2 | 3+3 2 | 3+3 2 |
| | 3+4 2 | 5+6 2 | 5+6 2 | 3+4 2 | 2+2 2 | 3+3 2 |
| | 5+6 2 | 5+6 2 | 3+4 2 | 5+6 2 | 3+3 2 | 2+2 2 |
| | 5+6 2 | 3+4 2 | 5+6 2 | 5+6 2 | 2+2 2 | 3+3 2 |
| | 3+4 2 | 5+6 2 | 3+4 2 | 3+4 2 | 2+2 2 | 3+3 2 |
| | 3+4 2 | 5+6 2 | 3+4 2 | 5+6 2 | 3+3 2 | 2+2 2 |
| | 5+6 2 | 3+4 2 | 3+4 2 | 3+4 2 | 3+3 2 | 2+2 2 |
| | 5+6 2 | 3+4 2 | 5+6 2 | 5+6 2 | 2+2 2 | 3+3 2 |
| | 3+4 2 | 3+4 2 | 3+4 2 | 5+6 2 | 3+3 2 | 2+2 2 |
| | 5+6 2 | 3+4 2 | 3+4 2 | 3+4 2 | 2+2 2 | 3+3 2 |
| | 5+6 2 | 5+6 2 | 5+6 2 | 5+6 2 | 3+3 2 | 2+2 2 |
| | 3+4 2 | 5+6 2 | 5+6 2 | 5+6 2 | 2+2 2 | 3+3 2 |
| | 5+6 2 | 3+4 2 | 5+6 2 | 3+4 2 | 3+3 2 | 2+2 2 |
| | 3+4 2 | 5+6 2 | 3+4 2 | 3+4 2 | 2+2 2 | 3+3 2 |
| | 3+4 2 | 5+6 2 | 5+6 2 | 3+4 2 | 3+3 2 | 2+2 2 |
| | 3+4 2 | 5+6 2 | 3+4 2 | 5+6 2 | 3+3 2 | 3+3 2 |
| | 5+6 2 | 3+4 2 | 5+6 2 | 3+4 2 | 2+2 2 | 2+2 2 |
| | 3+4 2 | 5+6 2 | 3+4 2 | 5+6 2 | 3+3 2 | 2+2 2 |
| | 5+6 2 | 3+4 2 | 5+6 2 | 3+4 2 | 2+2 2 | 3+3 2 |
| | 3+4 2 | 5+6 2 | 3+4 2 | 3+4 2 | 3+3 2 | 3+3 2 |
| | 5+6 2 | 3+4 2 | 5+6 2 | 5+6 2 | 2+2 2 | 2+2 2 |
| | 5+6 2 | 3+4 2 | 5+6 2 | 5+6 2 | 2+2 2 | 2+2 2 |
| | 3+4 2 | 5+6 2 | 3+4 2 | 5+6 2 | 3+3 2 | 3+3 2 |
| 12.2.138 | 240479 | 70 | Br4Cs2I2Pd1 | | | |
| | -A | -GM | -M | -Y | -L | -V |
| | 5+6 2 | 3+4 2 | 3+4 2 | 3+4 2 | 2+2 2 | 2+2 2 |
| | 3+4 2 | 5+6 2 | 5+6 2 | 5+6 2 | 3+3 2 | 3+3 2 |
| | 3+4 2 | 3+4 2 | 3+4 2 | 3+4 2 | 3+3 2 | 3+3 2 |

Continued on next column

**TABLE XXIV– continued**

| TQC | ICSD | | NoE | | SF | |
|---|---|---|---|---|---|---|
| | 5+6 2 | 5+6 2 | 5+6 2 | 5+6 2 | 2+2 2 | 2+2 2 |
| | 5+6 2 | 5+6 2 | 5+6 2 | 5+6 2 | 3+3 2 | 3+3 2 |
| | 3+4 2 | 3+4 2 | 3+4 2 | 3+4 2 | 2+2 2 | 2+2 2 |
| | 5+6 2 | 3+4 2 | 5+6 2 | 3+4 2 | 3+3 2 | 2+2 2 |
| | 3+4 2 | 5+6 2 | 3+4 2 | 5+6 2 | 2+2 2 | 3+3 2 |
| | 5+6 2 | 3+4 2 | 3+4 2 | 3+4 2 | 2+2 2 | 3+3 2 |
| | 3+4 2 | 5+6 2 | 5+6 2 | 5+6 2 | 3+3 2 | 2+2 2 |
| | 3+4 2 | 3+4 2 | 3+4 2 | 3+4 2 | 2+2 2 | 3+3 2 |
| | 5+6 2 | 3+4 2 | 5+6 2 | 3+4 2 | 3+3 2 | 2+2 2 |
| | 3+4 2 | 5+6 2 | 3+4 2 | 5+6 2 | 2+2 2 | 2+2 2 |
| | 5+6 2 | 5+6 2 | 5+6 2 | 5+6 2 | 3+3 2 | 3+3 2 |
| | 3+4 2 | 3+4 2 | 3+4 2 | 3+4 2 | 3+3 2 | 3+3 2 |
| | 3+4 2 | 3+4 2 | 3+4 2 | 3+4 2 | 3+3 2 | 3+3 2 |
| | 5+6 2 | 3+4 2 | 5+6 2 | 3+4 2 | 3+3 2 | 2+2 2 |
| | 3+4 2 | 3+4 2 | 3+4 2 | 3+4 2 | 3+3 2 | 3+3 2 |
| | 3+4 2 | 5+6 2 | 3+4 2 | 3+4 2 | 3+3 2 | 3+3 2 |
| | 3+4 2 | 3+4 2 | 3+4 2 | 5+6 2 | 2+2 2 | 3+3 2 |
| | 5+6 2 | 3+4 2 | 3+4 2 | 3+4 2 | 3+3 2 | 3+3 2 |
| | 3+4 2 | 5+6 2 | 5+6 2 | 5+6 2 | 2+2 2 | 2+2 2 |
| | 5+6 2 | 5+6 2 | 5+6 2 | 5+6 2 | 2+2 2 | 2+2 2 |
| | 5+6 2 | 5+6 2 | 5+6 2 | 5+6 2 | 2+2 2 | 2+2 2 |
| | 5+6 2 | 5+6 2 | 5+6 2 | 5+6 2 | 3+3 2 | 2+2 2 |
| | 3+4 2 | 3+4 2 | 3+4 2 | 3+4 2 | 2+2 2 | 3+3 2 |
| | 5+6 2 | 3+4 2 | 5+6 2 | 5+6 2 | 2+2 2 | 2+2 2 |
| | 3+4 2 | 5+6 2 | 5+6 2 | 3+4 2 | 3+3 2 | 3+3 2 |
| | 3+4 2 | 3+4 2 | 3+4 2 | 5+6 2 | 3+3 2 | 3+3 2 |
| | 5+6 2 | 3+4 2 | 3+4 2 | 3+4 2 | 2+2 2 | 2+2 2 |
| | 3+4 2 | 5+6 2 | 3+4 2 | 3+4 2 | 3+3 2 | 3+3 2 |
| | 3+4 2 | 3+4 2 | 3+4 2 | 5+6 2 | 3+3 2 | 3+3 2 |
| | 3+4 2 | 5+6 2 | 5+6 2 | 3+4 2 | 2+2 2 | 2+2 2 |
| | 5+6 2 | 3+4 2 | 3+4 2 | 3+4 2 | 3+3 2 | 3+3 2 |
| | 5+6 2 | 3+4 2 | 5+6 2 | 5+6 2 | 3+3 2 | 2+2 2 |
| 12.2.139 | 172782 | | 62 | | Br1Mn1S2Sb1 | |
| | -A | -GM | -M | -Y | -L | -V |
| | 5+6 2 | 3+4 2 | 5+6 2 | 3+4 2 | 3+3 2 | 2+2 2 |
| | 3+4 2 | 5+6 2 | 3+4 2 | 5+6 2 | 2+2 2 | 3+3 2 |
| | 3+4 2 | 3+4 2 | 5+6 2 | 5+6 2 | 2+2 2 | 2+2 2 |
| | 5+6 2 | 5+6 2 | 3+4 2 | 3+4 2 | 3+3 2 | 3+3 2 |
| | 3+4 2 | 3+4 2 | 5+6 2 | 5+6 2 | 2+2 2 | 3+3 2 |
| | 5+6 2 | 5+6 2 | 3+4 2 | 3+4 2 | 3+3 2 | 2+2 2 |
| | 5+6 2 | 5+6 2 | 3+4 2 | 3+4 2 | 3+3 2 | 3+3 2 |
| | 3+4 2 | 3+4 2 | 5+6 2 | 5+6 2 | 2+2 2 | 2+2 2 |
| | 3+4 2 | 3+4 2 | 5+6 2 | 3+4 2 | 2+2 2 | 2+2 2 |
| | 5+6 2 | 3+4 2 | 3+4 2 | 5+6 2 | 3+3 2 | 3+3 2 |
| | 3+4 2 | 5+6 2 | 5+6 2 | 3+4 2 | 2+2 2 | 3+3 2 |
| | 5+6 2 | 3+4 2 | 5+6 2 | 5+6 2 | 3+3 2 | 3+3 2 |
| | 3+4 2 | 3+4 2 | 3+4 2 | 5+6 2 | 2+2 2 | 2+2 2 |
| | 5+6 2 | 5+6 2 | 5+6 2 | 5+6 2 | 3+3 2 | 2+2 2 |
| | 5+6 2 | 3+4 2 | 3+4 2 | 3+4 2 | 3+3 2 | 3+3 2 |
| | 3+4 2 | 5+6 2 | 3+4 2 | 3+4 2 | 3+3 2 | 2+2 2 |
| | 5+6 2 | 3+4 2 | 5+6 2 | 3+4 2 | 2+2 2 | 3+3 2 |
| | 3+4 2 | 3+4 2 | 3+4 2 | 5+6 2 | 2+2 2 | 2+2 2 |
| | 5+6 2 | 3+4 2 | 5+6 2 | 3+4 2 | 3+3 2 | 2+2 2 |
| | 3+4 2 | 5+6 2 | 3+4 2 | 3+4 2 | 2+2 2 | 3+3 2 |
| | 3+4 2 | 5+6 2 | 3+4 2 | 5+6 2 | 2+2 2 | 3+3 2 |
| | 3+4 2 | 3+4 2 | 3+4 2 | 3+4 2 | 3+3 2 | 2+2 2 |
| | 5+6 2 | 5+6 2 | 5+6 2 | 5+6 2 | 3+3 2 | 3+3 2 |
| | 5+6 2 | 5+6 2 | 5+6 2 | 3+4 2 | 2+2 2 | 3+3 2 |
| | 3+4 2 | 5+6 2 | 3+4 2 | 5+6 2 | 3+3 2 | 2+2 2 |
| | 5+6 2 | 5+6 2 | 5+6 2 | 5+6 2 | 2+2 2 | 2+2 2 |
| | 5+6 2 | 3+4 2 | 3+4 2 | 3+4 2 | 2+2 2 | 3+3 2 |

Continued on next column

**TABLE XXIV– continued**

| TQC | ICSD | | NoE | | SF | |
|---|---|---|---|---|---|---|
| | 3+4 2 | 3+4 2 | 5+6 2 | 3+4 2 | 3+3 2 | 3+3 2 |
| | 5+6 2 | 3+4 2 | 5+6 2 | 3+4 2 | 3+3 2 | 3+3 2 |
| | 3+4 2 | 3+4 2 | 3+4 2 | 3+4 2 | 2+2 2 | 3+3 2 |
| | 5+6 2 | 3+4 2 | 3+4 2 | 3+4 2 | 2+2 2 | 3+3 2 |
| 12.2.140 | 427411 | | 126 | | Bi4Pb7Se13 | |
| | -A | -GM | -M | -Y | -L | -V |
| | 3+4 2 | 3+4 2 | 5+6 2 | 5+6 2 | 3+3 2 | 3+3 2 |
| | 5+6 2 | 3+4 2 | 5+6 2 | 3+4 2 | 2+2 2 | 2+2 2 |
| | 3+4 2 | 5+6 2 | 3+4 2 | 5+6 2 | 3+3 2 | 3+3 2 |
| | 5+6 2 | 5+6 2 | 3+4 2 | 3+4 2 | 2+2 2 | 2+2 2 |
| | 3+4 2 | 3+4 2 | 3+4 2 | 3+4 2 | 3+3 2 | 3+3 2 |
| | 5+6 2 | 5+6 2 | 5+6 2 | 5+6 2 | 2+2 2 | 2+2 2 |
| | 5+6 2 | 3+4 2 | 3+4 2 | 3+4 2 | 3+3 2 | 3+3 2 |
| | 3+4 2 | 5+6 2 | 5+6 2 | 5+6 2 | 2+2 2 | 2+2 2 |
| | 5+6 2 | 3+4 2 | 5+6 2 | 3+4 2 | 2+2 2 | 2+2 2 |
| | 3+4 2 | 5+6 2 | 3+4 2 | 5+6 2 | 2+2 2 | 3+3 2 |
| | 5+6 2 | 3+4 2 | 3+4 2 | 5+6 2 | 3+3 2 | 3+3 2 |
| | 5+6 2 | 3+4 2 | 5+6 2 | 3+4 2 | 3+3 2 | 2+2 2 |
| | 3+4 2 | 5+6 2 | 5+6 2 | 3+4 2 | 2+2 2 | 3+3 2 |
| | 5+6 2 | 5+6 2 | 3+4 2 | 3+4 2 | 3+3 2 | 2+2 2 |
| | 3+4 2 | 3+4 2 | 3+4 2 | 5+6 2 | 2+2 2 | 2+2 2 |
| | 3+4 2 | 5+6 2 | 5+6 2 | 5+6 2 | 3+3 2 | 3+3 2 |
| | 5+6 2 | 3+4 2 | 5+6 2 | 3+4 2 | 2+2 2 | 3+3 2 |
| | 3+4 2 | 3+4 2 | 3+4 2 | 5+6 2 | 3+3 2 | 3+3 2 |
| | 5+6 2 | 5+6 2 | 3+4 2 | 3+4 2 | 2+2 2 | 2+2 2 |
| | 3+4 2 | 5+6 2 | 5+6 2 | 3+4 2 | 2+2 2 | 2+2 2 |
| | 3+4 2 | 3+4 2 | 5+6 2 | 5+6 2 | 3+3 2 | 2+2 2 |
| | 5+6 2 | 3+4 2 | 3+4 2 | 3+4 2 | 3+3 2 | 3+3 2 |
| | 3+4 2 | 5+6 2 | 3+4 2 | 5+6 2 | 2+2 2 | 2+2 2 |
| | 5+6 2 | 3+4 2 | 5+6 2 | 3+4 2 | 2+2 2 | 3+3 2 |
| | 3+4 2 | 5+6 2 | 5+6 2 | 5+6 2 | 3+3 2 | 3+3 2 |
| | 3+4 2 | 3+4 2 | 3+4 2 | 5+6 2 | 3+3 2 | 2+2 2 |
| | 5+6 2 | 3+4 2 | 5+6 2 | 5+6 2 | 2+2 2 | 2+2 2 |
| | 5+6 2 | 5+6 2 | 3+4 2 | 3+4 2 | 2+2 2 | 3+3 2 |
| | 3+4 2 | 5+6 2 | 5+6 2 | 3+4 2 | 3+3 2 | 2+2 2 |
| | 5+6 2 | 3+4 2 | 3+4 2 | 5+6 2 | 2+2 2 | 3+3 2 |
| | 3+4 2 | 3+4 2 | 5+6 2 | 3+4 2 | 3+3 2 | 2+2 2 |
| | 5+6 2 | 5+6 2 | 3+4 2 | 5+6 2 | 2+2 2 | 3+3 2 |
| | 5+6 2 | 3+4 2 | 5+6 2 | 3+4 2 | 3+3 2 | 2+2 2 |
| | 3+4 2 | 5+6 2 | 3+4 2 | 5+6 2 | 2+2 2 | 3+3 2 |
| | 5+6 2 | 5+6 2 | 3+4 2 | 3+4 2 | 3+3 2 | 3+3 2 |
| | 5+6 2 | 3+4 2 | 5+6 2 | 5+6 2 | 3+3 2 | 2+2 2 |
| | 3+4 2 | 5+6 2 | 3+4 2 | 3+4 2 | 2+2 2 | 2+2 2 |
| | 3+4 2 | 5+6 2 | 5+6 2 | 5+6 2 | 3+3 2 | 3+3 2 |
| | 5+6 2 | 3+4 2 | 3+4 2 | 3+4 2 | 3+3 2 | 2+2 2 |
| | 3+4 2 | 5+6 2 | 5+6 2 | 5+6 2 | 2+2 2 | 3+3 2 |
| | 5+6 2 | 3+4 2 | 5+6 2 | 5+6 2 | 3+3 2 | 2+2 2 |
| | 3+4 2 | 5+6 2 | 3+4 2 | 3+4 2 | 3+3 2 | 3+3 2 |
| | 5+6 2 | 3+4 2 | 5+6 2 | 3+4 2 | 2+2 2 | 3+3 2 |
| | 3+4 2 | 3+4 2 | 3+4 2 | 5+6 2 | 2+2 2 | 2+2 2 |
| | 5+6 2 | 5+6 2 | 5+6 2 | 5+6 2 | 3+3 2 | 3+3 2 |
| | 3+4 2 | 5+6 2 | 3+4 2 | 5+6 2 | 3+3 2 | 2+2 2 |
| | 5+6 2 | 3+4 2 | 3+4 2 | 3+4 2 | 2+2 2 | 2+2 2 |
| | 3+4 2 | 5+6 2 | 5+6 2 | 3+4 2 | 2+2 2 | 3+3 2 |
| | 5+6 2 | 3+4 2 | 3+4 2 | 5+6 2 | 3+3 2 | 3+3 2 |
| | 3+4 2 | 5+6 2 | 5+6 2 | 3+4 2 | 2+2 2 | 2+2 2 |
| | 5+6 2 | 3+4 2 | 3+4 2 | 5+6 2 | 3+3 2 | 3+3 2 |
| | 5+6 2 | 5+6 2 | 3+4 2 | 3+4 2 | 2+2 2 | 2+2 2 |
| | 3+4 2 | 3+4 2 | 5+6 2 | 5+6 2 | 3+3 2 | 3+3 2 |
| | 3+4 2 | 5+6 2 | 5+6 2 | 3+4 2 | 2+2 2 | 2+2 2 |
| | 3+4 2 | 3+4 2 | 3+4 2 | 5+6 2 | 3+3 2 | 2+2 2 |

Continued on next column

**TABLE XXIV– continued**

| TQC | ICSD | NoE | SF | | | |
|---|---|---|---|---|---|---|
| | 5+6 2 | 5+6 2 | 5+6 2 | 3+4 2 | 2+2 2 | 2+2 2 |
| | 3+4 2 | 5+6 2 | 3+4 2 | 5+6 2 | 3+3 2 | 3+3 2 |
| | 3+4 2 | 5+6 2 | 3+4 2 | 5+6 2 | 2+2 2 | 3+3 2 |
| | 5+6 2 | 3+4 2 | 5+6 2 | 3+4 2 | 3+3 2 | 2+2 2 |
| | 3+4 2 | 5+6 2 | 3+4 2 | 5+6 2 | 2+2 2 | 2+2 2 |
| | 3+4 2 | 3+4 2 | 5+6 2 | 3+4 2 | 3+3 2 | 3+3 2 |
| | 5+6 2 | 5+6 2 | 3+4 2 | 3+4 2 | 3+3 2 | 2+2 2 |
| | 3+4 2 | 3+4 2 | 5+6 2 | 5+6 2 | 2+2 2 | 2+2 2 |
| 12.2.141 | 58840 | 120 | Bi2Pd5 | | | |
| | -A | -GM | -M | -Y | -L | -V |
| | 3+4 2 | 3+4 2 | 3+4 2 | 3+4 2 | 2+2 2 | 2+2 2 |
| | 5+6 2 | 5+6 2 | 5+6 2 | 5+6 2 | 3+3 2 | 3+3 2 |
| | 5+6 2 | 3+4 2 | 3+4 2 | 5+6 2 | 2+2 2 | 2+2 2 |
| | 3+4 2 | 5+6 2 | 5+6 2 | 3+4 2 | 3+3 2 | 3+3 2 |
| | 5+6 2 | 3+4 2 | 5+6 2 | 3+4 2 | 3+3 2 | 2+2 2 |
| | 3+4 2 | 5+6 2 | 3+4 2 | 5+6 2 | 2+2 2 | 3+3 2 |
| | 5+6 2 | 3+4 2 | 5+6 2 | 3+4 2 | 3+3 2 | 3+3 2 |
| | 3+4 2 | 5+6 2 | 3+4 2 | 5+6 2 | 2+2 2 | 2+2 2 |
| | 3+4 2 | 3+4 2 | 3+4 2 | 3+4 2 | 2+2 2 | 3+3 2 |
| | 5+6 2 | 5+6 2 | 3+4 2 | 5+6 2 | 3+3 2 | 2+2 2 |
| | 3+4 2 | 5+6 2 | 5+6 2 | 5+6 2 | 2+2 2 | 2+2 2 |
| | 5+6 2 | 3+4 2 | 3+4 2 | 5+6 2 | 3+3 2 | 2+2 2 |
| | 3+4 2 | 5+6 2 | 3+4 2 | 3+4 2 | 2+2 2 | 3+3 2 |
| | 5+6 2 | 3+4 2 | 5+6 2 | 3+4 2 | 3+3 2 | 3+3 2 |
| | 3+4 2 | 3+4 2 | 5+6 2 | 5+6 2 | 2+2 2 | 2+2 2 |
| | 5+6 2 | 5+6 2 | 5+6 2 | 5+6 2 | 3+3 2 | 3+3 2 |
| | 5+6 2 | 5+6 2 | 3+4 2 | 3+4 2 | 2+2 2 | 2+2 2 |
| | 3+4 2 | 5+6 2 | 5+6 2 | 5+6 2 | 3+3 2 | 3+3 2 |
| | 3+4 2 | 3+4 2 | 3+4 2 | 3+4 2 | 2+2 2 | 2+2 2 |
| | 5+6 2 | 3+4 2 | 3+4 2 | 3+4 2 | 2+2 2 | 3+3 2 |
| | 5+6 2 | 5+6 2 | 5+6 2 | 5+6 2 | 3+3 2 | 2+2 2 |
| | 3+4 2 | 3+4 2 | 3+4 2 | 3+4 2 | 3+3 2 | 3+3 2 |
| | 5+6 2 | 3+4 2 | 5+6 2 | 5+6 2 | 2+2 2 | 2+2 2 |
| | 3+4 2 | 5+6 2 | 3+4 2 | 5+6 2 | 3+3 2 | 3+3 2 |
| | 5+6 2 | 3+4 2 | 3+4 2 | 3+4 2 | 2+2 2 | 2+2 2 |
| | 3+4 2 | 5+6 2 | 5+6 2 | 3+4 2 | 3+3 2 | 3+3 2 |
| | 5+6 2 | 3+4 2 | 3+4 2 | 5+6 2 | 2+2 2 | 2+2 2 |
| | 5+6 2 | 3+4 2 | 5+6 2 | 3+4 2 | 3+3 2 | 3+3 2 |
| | 3+4 2 | 5+6 2 | 5+6 2 | 5+6 2 | 3+3 2 | 3+3 2 |
| | 5+6 2 | 3+4 2 | 3+4 2 | 5+6 2 | 2+2 2 | 2+2 2 |
| | 3+4 2 | 5+6 2 | 5+6 2 | 3+4 2 | 2+2 2 | 3+3 2 |
| | 5+6 2 | 5+6 2 | 5+6 2 | 3+4 2 | 3+3 2 | 2+2 2 |
| | 3+4 2 | 3+4 2 | 5+6 2 | 3+4 2 | 2+2 2 | 2+2 2 |
| | 3+4 2 | 5+6 2 | 3+4 2 | 5+6 2 | 3+3 2 | 3+3 2 |
| | 3+4 2 | 3+4 2 | 5+6 2 | 3+4 2 | 2+2 2 | 2+2 2 |
| | 5+6 2 | 3+4 2 | 3+4 2 | 5+6 2 | 3+3 2 | 3+3 2 |
| | 5+6 2 | 5+6 2 | 5+6 2 | 3+4 2 | 2+2 2 | 2+2 2 |
| | 3+4 2 | 3+4 2 | 3+4 2 | 5+6 2 | 3+3 2 | 3+3 2 |
| | 3+4 2 | 5+6 2 | 3+4 2 | 5+6 2 | 3+3 2 | 2+2 2 |
| | 5+6 2 | 3+4 2 | 5+6 2 | 3+4 2 | 2+2 2 | 3+3 2 |
| | 5+6 2 | 5+6 2 | 5+6 2 | 5+6 2 | 3+3 2 | 3+3 2 |
| | 3+4 2 | 3+4 2 | 5+6 2 | 5+6 2 | 2+2 2 | 2+2 2 |
| | 5+6 2 | 5+6 2 | 3+4 2 | 3+4 2 | 2+2 2 | 3+3 2 |
| | 3+4 2 | 3+4 2 | 5+6 2 | 5+6 2 | 3+3 2 | 3+3 2 |
| | 5+6 2 | 5+6 2 | 3+4 2 | 3+4 2 | 2+2 2 | 2+2 2 |
| | 3+4 2 | 3+4 2 | 3+4 2 | 3+4 2 | 2+2 2 | 3+3 2 |
| | 5+6 2 | 5+6 2 | 5+6 2 | 5+6 2 | 3+3 2 | 2+2 2 |
| | 3+4 2 | 3+4 2 | 3+4 2 | 3+4 2 | 3+3 2 | 2+2 2 |
| | 5+6 2 | 3+4 2 | 5+6 2 | 3+4 2 | 3+3 2 | 3+3 2 |
| | 3+4 2 | 5+6 2 | 3+4 2 | 3+4 2 | 2+2 2 | 2+2 2 |
| | 5+6 2 | 3+4 2 | 5+6 2 | 5+6 2 | 3+3 2 | 2+2 2 |

Continued on next column

**TABLE XXIV– continued**

| TQC | ICSD | NoE | SF | | | |
|---|---|---|---|---|---|---|
| | 3+4 2 | 3+4 2 | 5+6 2 | 5+6 2 | 2+2 2 | 3+3 2 |
| | 3+4 2 | 5+6 2 | 3+4 2 | 3+4 2 | 3+3 2 | 2+2 2 |
| | 5+6 2 | 3+4 2 | 5+6 2 | 5+6 2 | 2+2 2 | 3+3 2 |
| | 5+6 2 | 5+6 2 | 3+4 2 | 3+4 2 | 3+3 2 | 2+2 2 |
| | 3+4 2 | 5+6 2 | 3+4 2 | 3+4 2 | 2+2 2 | 3+3 2 |
| | 3+4 2 | 3+4 2 | 3+4 2 | 5+6 2 | 3+3 2 | 3+3 2 |
| | 5+6 2 | 5+6 2 | 5+6 2 | 5+6 2 | 2+2 2 | 2+2 2 |
| | 5+6 2 | 3+4 2 | 5+6 2 | 3+4 2 | 3+3 2 | 3+3 2 |
| | 3+4 2 | 5+6 2 | 3+4 2 | 5+6 2 | 2+2 2 | 2+2 2 |
| 12.2.142 | 616957 | 104 | Bi2Pd3Se2 | | | |
| | -A | -GM | -M | -Y | -L | -V |
| | 5+6 2 | 3+4 2 | 3+4 2 | 3+4 2 | 2+2 2 | 3+3 2 |
| | 3+4 2 | 5+6 2 | 5+6 2 | 5+6 2 | 3+3 2 | 2+2 2 |
| | 5+6 2 | 5+6 2 | 3+4 2 | 5+6 2 | 3+3 2 | 2+2 2 |
| | 3+4 2 | 3+4 2 | 5+6 2 | 3+4 2 | 2+2 2 | 3+3 2 |
| | 5+6 2 | 3+4 2 | 5+6 2 | 3+4 2 | 2+2 2 | 3+3 2 |
| | 3+4 2 | 3+4 2 | 3+4 2 | 3+4 2 | 3+3 2 | 2+2 2 |
| | 5+6 2 | 5+6 2 | 5+6 2 | 5+6 2 | 2+2 2 | 3+3 2 |
| | 3+4 2 | 5+6 2 | 3+4 2 | 5+6 2 | 3+3 2 | 2+2 2 |
| | 5+6 2 | 3+4 2 | 3+4 2 | 5+6 2 | 3+3 2 | 2+2 2 |
| | 3+4 2 | 3+4 2 | 3+4 2 | 5+6 2 | 2+2 2 | 3+3 2 |
| | 5+6 2 | 5+6 2 | 5+6 2 | 5+6 2 | 3+3 2 | 2+2 2 |
| | 5+6 2 | 3+4 2 | 3+4 2 | 3+4 2 | 2+2 2 | 3+3 2 |
| | 5+6 2 | 5+6 2 | 3+4 2 | 5+6 2 | 3+3 2 | 2+2 2 |
| | 3+4 2 | 3+4 2 | 5+6 2 | 5+6 2 | 2+2 2 | 3+3 2 |
| | 5+6 2 | 3+4 2 | 3+4 2 | 3+4 2 | 2+2 2 | 3+3 2 |
| | 3+4 2 | 3+4 2 | 5+6 2 | 5+6 2 | 3+3 2 | 2+2 2 |
| | 5+6 2 | 5+6 2 | 3+4 2 | 3+4 2 | 2+2 2 | 3+3 2 |
| | 5+6 2 | 3+4 2 | 5+6 2 | 3+4 2 | 3+3 2 | 2+2 2 |
| | 3+4 2 | 3+4 2 | 5+6 2 | 5+6 2 | 2+2 2 | 3+3 2 |
| | 5+6 2 | 5+6 2 | 3+4 2 | 5+6 2 | 2+2 2 | 2+2 2 |
| | 5+6 2 | 3+4 2 | 3+4 2 | 3+4 2 | 3+3 2 | 3+3 2 |
| | 3+4 2 | 3+4 2 | 5+6 2 | 5+6 2 | 2+2 2 | 3+3 2 |
| | 5+6 2 | 5+6 2 | 3+4 2 | 3+4 2 | 3+3 2 | 2+2 2 |
| | 3+4 2 | 3+4 2 | 5+6 2 | 5+6 2 | 2+2 2 | 3+3 2 |
| | 5+6 2 | 3+4 2 | 3+4 2 | 3+4 2 | 3+3 2 | 2+2 2 |
| | 5+6 2 | 5+6 2 | 3+4 2 | 5+6 2 | 2+2 2 | 3+3 2 |
| | 3+4 2 | 5+6 2 | 5+6 2 | 3+4 2 | 3+3 2 | 2+2 2 |
| | 5+6 2 | 3+4 2 | 3+4 2 | 3+4 2 | 2+2 2 | 2+2 2 |
| | 3+4 2 | 3+4 2 | 5+6 2 | 3+4 2 | 3+3 2 | 3+3 2 |
| | 3+4 2 | 5+6 2 | 5+6 2 | 5+6 2 | 3+3 2 | 2+2 2 |
| | 5+6 2 | 3+4 2 | 3+4 2 | 5+6 2 | 3+3 2 | 2+2 2 |
| | 3+4 2 | 5+6 2 | 3+4 2 | 5+6 2 | 3+3 2 | 3+3 2 |
| | 5+6 2 | 5+6 2 | 3+4 2 | 5+6 2 | 2+2 2 | 3+3 2 |
| | 5+6 2 | 3+4 2 | 3+4 2 | 3+4 2 | 2+2 2 | 2+2 2 |
| | 3+4 2 | 5+6 2 | 3+4 2 | 5+6 2 | 3+3 2 | 2+2 2 |
| | 3+4 2 | 5+6 2 | 5+6 2 | 3+4 2 | 2+2 2 | 3+3 2 |
| | 5+6 2 | 3+4 2 | 3+4 2 | 5+6 2 | 3+3 2 | 3+3 2 |
| | 5+6 2 | 3+4 2 | 5+6 2 | 5+6 2 | 3+3 2 | 2+2 2 |
| | 5+6 2 | 3+4 2 | 3+4 2 | 5+6 2 | 3+3 2 | 3+3 2 |
| | 3+4 2 | 3+4 2 | 5+6 2 | 3+4 2 | 2+2 2 | 2+2 2 |
| | 5+6 2 | 5+6 2 | 3+4 2 | 5+6 2 | 2+2 2 | 2+2 2 |
| | 3+4 2 | 3+4 2 | 3+4 2 | 5+6 2 | 3+3 2 | 3+3 2 |
| | 5+6 2 | 5+6 2 | 5+6 2 | 5+6 2 | 2+2 2 | 2+2 2 |
| | 3+4 2 | 3+4 2 | 3+4 2 | 3+4 2 | 2+2 2 | 2+2 2 |
| | 5+6 2 | 5+6 2 | 3+4 2 | 3+4 2 | 3+3 2 | 3+3 2 |
| | 5+6 2 | 3+4 2 | 5+6 2 | 3+4 2 | 2+2 2 | 3+3 2 |
| | 3+4 2 | 3+4 2 | 5+6 2 | 3+4 2 | 3+3 2 | 3+3 2 |
| | 5+6 2 | 5+6 2 | 3+4 2 | 5+6 2 | 2+2 2 | 2+2 2 |
| | 3+4 2 | 3+4 2 | 5+6 2 | 3+4 2 | 2+2 2 | 3+3 2 |
| | 5+6 2 | 3+4 2 | 3+4 2 | 5+6 2 | 3+3 2 | 2+2 2 |

Continued on next column

**TABLE XXIV– continued**

| TQC | ICSD | NoE | SF | | | |
|---|---|---|---|---|---|---|
| | 5+6 2 | 5+6 2 | 5+6 2 | 5+6 2 | 3+3 2 | 3+3 2 |
| | 5+6 2 | 3+4 2 | 3+4 2 | 3+4 2 | 2+2 2 | 2+2 2 |
| 12.2.143 | 421873 | 104 | Bi2Pd3Se2 | | | |
| | -A | -GM | -M | -Y | -L | -V |
| | 5+6 2 | 3+4 2 | 5+6 2 | 3+4 2 | 3+3 2 | 2+2 2 |
| | 3+4 2 | 5+6 2 | 3+4 2 | 5+6 2 | 2+2 2 | 3+3 2 |
| | 3+4 2 | 5+6 2 | 5+6 2 | 5+6 2 | 3+3 2 | 3+3 2 |
| | 5+6 2 | 3+4 2 | 3+4 2 | 3+4 2 | 2+2 2 | 2+2 2 |
| | 3+4 2 | 3+4 2 | 3+4 2 | 3+4 2 | 3+3 2 | 3+3 2 |
| | 5+6 2 | 5+6 2 | 5+6 2 | 3+4 2 | 2+2 2 | 2+2 2 |
| | 3+4 2 | 3+4 2 | 3+4 2 | 5+6 2 | 3+3 2 | 3+3 2 |
| | 5+6 2 | 5+6 2 | 5+6 2 | 5+6 2 | 2+2 2 | 2+2 2 |
| | 3+4 2 | 3+4 2 | 5+6 2 | 5+6 2 | 2+2 2 | 2+2 2 |
| | 5+6 2 | 3+4 2 | 5+6 2 | 5+6 2 | 3+3 2 | 2+2 2 |
| | 3+4 2 | 5+6 2 | 3+4 2 | 3+4 2 | 2+2 2 | 3+3 2 |
| | 3+4 2 | 3+4 2 | 5+6 2 | 5+6 2 | 3+3 2 | 3+3 2 |
| | 5+6 2 | 5+6 2 | 3+4 2 | 5+6 2 | 3+3 2 | 3+3 2 |
| | 3+4 2 | 3+4 2 | 5+6 2 | 5+6 2 | 2+2 2 | 2+2 2 |
| | 3+4 2 | 3+4 2 | 5+6 2 | 3+4 2 | 3+3 2 | 2+2 2 |
| | 5+6 2 | 3+4 2 | 3+4 2 | 5+6 2 | 2+2 2 | 3+3 2 |
| | 3+4 2 | 3+4 2 | 5+6 2 | 3+4 2 | 3+3 2 | 3+3 2 |
| | 5+6 2 | 5+6 2 | 3+4 2 | 3+4 2 | 3+3 2 | 3+3 2 |
| | 3+4 2 | 3+4 2 | 5+6 2 | 5+6 2 | 2+2 2 | 2+2 2 |
| | 3+4 2 | 5+6 2 | 3+4 2 | 5+6 2 | 2+2 2 | 2+2 2 |
| | 3+4 2 | 3+4 2 | 5+6 2 | 3+4 2 | 3+3 2 | 3+3 2 |
| | 5+6 2 | 3+4 2 | 3+4 2 | 5+6 2 | 2+2 2 | 3+3 2 |
| | 3+4 2 | 5+6 2 | 5+6 2 | 3+4 2 | 3+3 2 | 2+2 2 |
| | 5+6 2 | 3+4 2 | 5+6 2 | 5+6 2 | 3+3 2 | 3+3 2 |
| | 3+4 2 | 3+4 2 | 3+4 2 | 3+4 2 | 2+2 2 | 2+2 2 |
| | 5+6 2 | 5+6 2 | 5+6 2 | 5+6 2 | 2+2 2 | 3+3 2 |
| | 3+4 2 | 5+6 2 | 3+4 2 | 3+4 2 | 3+3 2 | 2+2 2 |
| | 3+4 2 | 3+4 2 | 5+6 2 | 3+4 2 | 3+3 2 | 3+3 2 |
| | 5+6 2 | 3+4 2 | 3+4 2 | 3+4 2 | 2+2 2 | 2+2 2 |
| | 5+6 2 | 5+6 2 | 3+4 2 | 5+6 2 | 2+2 2 | 2+2 2 |
| | 3+4 2 | 5+6 2 | 5+6 2 | 5+6 2 | 2+2 2 | 3+3 2 |
| | 3+4 2 | 3+4 2 | 5+6 2 | 5+6 2 | 3+3 2 | 2+2 2 |
| | 5+6 2 | 5+6 2 | 5+6 2 | 3+4 2 | 2+2 2 | 3+3 2 |
| | 3+4 2 | 3+4 2 | 5+6 2 | 5+6 2 | 3+3 2 | 2+2 2 |
| | 5+6 2 | 5+6 2 | 5+6 2 | 5+6 2 | 2+2 2 | 2+2 2 |
| | 5+6 2 | 5+6 2 | 3+4 2 | 3+4 2 | 2+2 2 | 3+3 2 |
| | 3+4 2 | 3+4 2 | 5+6 2 | 5+6 2 | 3+3 2 | 3+3 2 |
| | 3+4 2 | 3+4 2 | 3+4 2 | 5+6 2 | 2+2 2 | 2+2 2 |
| | 3+4 2 | 3+4 2 | 5+6 2 | 5+6 2 | 2+2 2 | 3+3 2 |
| | 3+4 2 | 3+4 2 | 3+4 2 | 3+4 2 | 3+3 2 | 2+2 2 |
| | 5+6 2 | 5+6 2 | 5+6 2 | 5+6 2 | 3+3 2 | 2+2 2 |
| | 3+4 2 | 3+4 2 | 5+6 2 | 5+6 2 | 2+2 2 | 3+3 2 |
| | 5+6 2 | 5+6 2 | 3+4 2 | 3+4 2 | 3+3 2 | 2+2 2 |
| | 5+6 2 | 3+4 2 | 5+6 2 | 5+6 2 | 3+3 2 | 3+3 2 |
| | 3+4 2 | 5+6 2 | 5+6 2 | 3+4 2 | 2+2 2 | 2+2 2 |
| | 3+4 2 | 3+4 2 | 3+4 2 | 3+4 2 | 3+3 2 | 3+3 2 |
| | 5+6 2 | 3+4 2 | 3+4 2 | 5+6 2 | 2+2 2 | 3+3 2 |
| | 3+4 2 | 5+6 2 | 5+6 2 | 3+4 2 | 3+3 2 | 2+2 2 |
| | 5+6 2 | 3+4 2 | 3+4 2 | 3+4 2 | 3+3 2 | 3+3 2 |
| | 3+4 2 | 3+4 2 | 5+6 2 | 5+6 2 | 2+2 2 | 2+2 2 |
| | 3+4 2 | 5+6 2 | 3+4 2 | 5+6 2 | 2+2 2 | 3+3 2 |
| | 3+4 2 | 3+4 2 | 5+6 2 | 3+4 2 | 3+3 2 | 2+2 2 |
| 12.2.144 | 261459 | 104 | Bi2Pd3Se2 | | | |
| | -A | -GM | -M | -Y | -L | -V |
| | 3+4 2 | 3+4 2 | 3+4 2 | 3+4 2 | 2+2 2 | 2+2 2 |
| | 5+6 2 | 5+6 2 | 5+6 2 | 5+6 2 | 3+3 2 | 3+3 2 |
| | 5+6 2 | 5+6 2 | 3+4 2 | 5+6 2 | 2+2 2 | 3+3 2 |

Continued on next column

**TABLE XXIV– continued**

| TQC | ICSD | NoE | SF | | | |
|---|---|---|---|---|---|---|
| | 3+4 2 | 3+4 2 | 5+6 2 | 3+4 2 | 3+3 2 | 2+2 2 |
| | 5+6 2 | 3+4 2 | 5+6 2 | 3+4 2 | 2+2 2 | 3+3 2 |
| | 3+4 2 | 5+6 2 | 3+4 2 | 3+4 2 | 3+3 2 | 2+2 2 |
| | 5+6 2 | 3+4 2 | 5+6 2 | 5+6 2 | 2+2 2 | 3+3 2 |
| | 3+4 2 | 5+6 2 | 3+4 2 | 5+6 2 | 3+3 2 | 2+2 2 |
| | 5+6 2 | 3+4 2 | 3+4 2 | 5+6 2 | 3+3 2 | 2+2 2 |
| | 3+4 2 | 3+4 2 | 3+4 2 | 5+6 2 | 2+2 2 | 2+2 2 |
| | 5+6 2 | 5+6 2 | 5+6 2 | 3+4 2 | 3+3 2 | 3+3 2 |
| | 5+6 2 | 5+6 2 | 3+4 2 | 5+6 2 | 2+2 2 | 3+3 2 |
| | 3+4 2 | 3+4 2 | 5+6 2 | 5+6 2 | 2+2 2 | 3+3 2 |
| | 5+6 2 | 3+4 2 | 3+4 2 | 5+6 2 | 3+3 2 | 2+2 2 |
| | 5+6 2 | 3+4 2 | 3+4 2 | 3+4 2 | 2+2 2 | 2+2 2 |
| | 3+4 2 | 3+4 2 | 5+6 2 | 5+6 2 | 3+3 2 | 3+3 2 |
| | 5+6 2 | 3+4 2 | 3+4 2 | 3+4 2 | 2+2 2 | 3+3 2 |
| | 3+4 2 | 5+6 2 | 5+6 2 | 3+4 2 | 2+2 2 | 3+3 2 |
| | 5+6 2 | 3+4 2 | 3+4 2 | 5+6 2 | 3+3 2 | 2+2 2 |
| | 5+6 2 | 5+6 2 | 5+6 2 | 5+6 2 | 3+3 2 | 2+2 2 |
| | 5+6 2 | 3+4 2 | 3+4 2 | 3+4 2 | 2+2 2 | 3+3 2 |
| | 3+4 2 | 3+4 2 | 5+6 2 | 5+6 2 | 3+3 2 | 3+3 2 |
| | 5+6 2 | 5+6 2 | 3+4 2 | 5+6 2 | 2+2 2 | 2+2 2 |
| | 3+4 2 | 3+4 2 | 3+4 2 | 3+4 2 | 2+2 2 | 3+3 2 |
| | 5+6 2 | 3+4 2 | 5+6 2 | 3+4 2 | 3+3 2 | 2+2 2 |
| | 3+4 2 | 5+6 2 | 3+4 2 | 5+6 2 | 3+3 2 | 3+3 2 |
| | 5+6 2 | 5+6 2 | 5+6 2 | 3+4 2 | 2+2 2 | 2+2 2 |
| | 5+6 2 | 3+4 2 | 3+4 2 | 3+4 2 | 2+2 2 | 3+3 2 |
| | 3+4 2 | 3+4 2 | 5+6 2 | 3+4 2 | 3+3 2 | 2+2 2 |
| | 3+4 2 | 5+6 2 | 5+6 2 | 5+6 2 | 3+3 2 | 2+2 2 |
| | 5+6 2 | 5+6 2 | 3+4 2 | 5+6 2 | 3+3 2 | 3+3 2 |
| | 5+6 2 | 3+4 2 | 3+4 2 | 5+6 2 | 2+2 2 | 2+2 2 |
| | 3+4 2 | 5+6 2 | 3+4 2 | 3+4 2 | 3+3 2 | 3+3 2 |
| | 5+6 2 | 3+4 2 | 3+4 2 | 5+6 2 | 2+2 2 | 2+2 2 |
| | 3+4 2 | 5+6 2 | 3+4 2 | 5+6 2 | 3+3 2 | 2+2 2 |
| | 3+4 2 | 5+6 2 | 5+6 2 | 3+4 2 | 3+3 2 | 3+3 2 |
| | 5+6 2 | 3+4 2 | 3+4 2 | 5+6 2 | 2+2 2 | 3+3 2 |
| | 5+6 2 | 3+4 2 | 5+6 2 | 5+6 2 | 3+3 2 | 2+2 2 |
| | 5+6 2 | 3+4 2 | 3+4 2 | 5+6 2 | 3+3 2 | 3+3 2 |
| | 3+4 2 | 3+4 2 | 5+6 2 | 3+4 2 | 2+2 2 | 2+2 2 |
| | 5+6 2 | 5+6 2 | 3+4 2 | 5+6 2 | 2+2 2 | 2+2 2 |
| | 5+6 2 | 3+4 2 | 3+4 2 | 5+6 2 | 3+3 2 | 3+3 2 |
| | 3+4 2 | 5+6 2 | 5+6 2 | 3+4 2 | 2+2 2 | 2+2 2 |
| | 3+4 2 | 3+4 2 | 3+4 2 | 5+6 2 | 2+2 2 | 3+3 2 |
| | 5+6 2 | 5+6 2 | 3+4 2 | 3+4 2 | 3+3 2 | 2+2 2 |
| | 5+6 2 | 3+4 2 | 5+6 2 | 3+4 2 | 2+2 2 | 3+3 2 |
| | 3+4 2 | 3+4 2 | 5+6 2 | 5+6 2 | 3+3 2 | 3+3 2 |
| | 5+6 2 | 5+6 2 | 3+4 2 | 3+4 2 | 2+2 2 | 2+2 2 |
| | 3+4 2 | 3+4 2 | 5+6 2 | 3+4 2 | 2+2 2 | 3+3 2 |
| | 5+6 2 | 3+4 2 | 3+4 2 | 5+6 2 | 3+3 2 | 2+2 2 |
| | 5+6 2 | 5+6 2 | 5+6 2 | 5+6 2 | 3+3 2 | 3+3 2 |
| | 5+6 2 | 3+4 2 | 3+4 2 | 3+4 2 | 2+2 2 | 2+2 2 |
| 12.2.145 | 261458 | 104 | Bi2Pd3Se2 | | | |
| | -A | -GM | -M | -Y | -L | -V |
| | 3+4 2 | 3+4 2 | 3+4 2 | 3+4 2 | 3+3 2 | 3+3 2 |
| | 5+6 2 | 5+6 2 | 5+6 2 | 5+6 2 | 2+2 2 | 2+2 2 |
| | 5+6 2 | 5+6 2 | 3+4 2 | 5+6 2 | 3+3 2 | 2+2 2 |
| | 3+4 2 | 3+4 2 | 5+6 2 | 3+4 2 | 2+2 2 | 3+3 2 |
| | 5+6 2 | 3+4 2 | 5+6 2 | 3+4 2 | 3+3 2 | 2+2 2 |
| | 3+4 2 | 5+6 2 | 3+4 2 | 3+4 2 | 2+2 2 | 3+3 2 |
| | 5+6 2 | 3+4 2 | 5+6 2 | 5+6 2 | 3+3 2 | 2+2 2 |
| | 3+4 2 | 5+6 2 | 3+4 2 | 5+6 2 | 2+2 2 | 3+3 2 |
| | 5+6 2 | 3+4 2 | 3+4 2 | 5+6 2 | 2+2 2 | 3+3 2 |
| | 3+4 2 | 3+4 2 | 3+4 2 | 5+6 2 | 3+3 2 | 3+3 2 |

Continued on next column

**TABLE XXIV– continued**

| TQC | ICSD | NoE | SF | | | |
|---|---|---|---|---|---|---|
| | 5+6 2 | 5+6 2 | 5+6 2 | 3+4 2 | 2+2 2 | 2+2 2 |
| | 5+6 2 | 3+4 2 | 3+4 2 | 5+6 2 | 3+3 2 | 2+2 2 |
| | 3+4 2 | 5+6 2 | 5+6 2 | 5+6 2 | 3+3 2 | 2+2 2 |
| | 5+6 2 | 3+4 2 | 3+4 2 | 5+6 2 | 2+2 2 | 3+3 2 |
| | 5+6 2 | 3+4 2 | 3+4 2 | 3+4 2 | 3+3 2 | 3+3 2 |
| | 3+4 2 | 3+4 2 | 5+6 2 | 5+6 2 | 2+2 2 | 2+2 2 |
| | 5+6 2 | 3+4 2 | 3+4 2 | 3+4 2 | 3+3 2 | 2+2 2 |
| | 3+4 2 | 5+6 2 | 5+6 2 | 3+4 2 | 3+3 2 | 2+2 2 |
| | 5+6 2 | 3+4 2 | 3+4 2 | 5+6 2 | 2+2 2 | 3+3 2 |
| | 5+6 2 | 5+6 2 | 5+6 2 | 5+6 2 | 2+2 2 | 3+3 2 |
| | 5+6 2 | 3+4 2 | 3+4 2 | 3+4 2 | 3+3 2 | 2+2 2 |
| | 3+4 2 | 3+4 2 | 5+6 2 | 5+6 2 | 2+2 2 | 2+2 2 |
| | 5+6 2 | 3+4 2 | 3+4 2 | 5+6 2 | 3+3 2 | 3+3 2 |
| | 3+4 2 | 5+6 2 | 3+4 2 | 3+4 2 | 3+3 2 | 2+2 2 |
| | 5+6 2 | 3+4 2 | 5+6 2 | 3+4 2 | 2+2 2 | 3+3 2 |
| | 3+4 2 | 5+6 2 | 3+4 2 | 5+6 2 | 2+2 2 | 2+2 2 |
| | 5+6 2 | 5+6 2 | 5+6 2 | 3+4 2 | 3+3 2 | 3+3 2 |
| | 5+6 2 | 3+4 2 | 3+4 2 | 3+4 2 | 3+3 2 | 2+2 2 |
| | 3+4 2 | 3+4 2 | 5+6 2 | 3+4 2 | 2+2 2 | 3+3 2 |
| | 3+4 2 | 5+6 2 | 5+6 2 | 5+6 2 | 2+2 2 | 3+3 2 |
| | 5+6 2 | 5+6 2 | 3+4 2 | 5+6 2 | 2+2 2 | 2+2 2 |
| | 5+6 2 | 3+4 2 | 3+4 2 | 5+6 2 | 3+3 2 | 3+3 2 |
| | 3+4 2 | 5+6 2 | 3+4 2 | 3+4 2 | 2+2 2 | 2+2 2 |
| | 5+6 2 | 3+4 2 | 3+4 2 | 5+6 2 | 3+3 2 | 3+3 2 |
| | 3+4 2 | 5+6 2 | 3+4 2 | 5+6 2 | 2+2 2 | 3+3 2 |
| | 3+4 2 | 5+6 2 | 5+6 2 | 3+4 2 | 2+2 2 | 2+2 2 |
| | 5+6 2 | 3+4 2 | 3+4 2 | 5+6 2 | 3+3 2 | 2+2 2 |
| | 5+6 2 | 3+4 2 | 5+6 2 | 5+6 2 | 2+2 2 | 3+3 2 |
| | 5+6 2 | 3+4 2 | 3+4 2 | 5+6 2 | 2+2 2 | 2+2 2 |
| | 5+6 2 | 3+4 2 | 5+6 2 | 3+4 2 | 3+3 2 | 3+3 2 |
| | 3+4 2 | 5+6 2 | 3+4 2 | 5+6 2 | 3+3 2 | 3+3 2 |
| | 5+6 2 | 3+4 2 | 3+4 2 | 5+6 2 | 2+2 2 | 2+2 2 |
| | 3+4 2 | 5+6 2 | 5+6 2 | 3+4 2 | 3+3 2 | 3+3 2 |
| | 3+4 2 | 3+4 2 | 3+4 2 | 5+6 2 | 3+3 2 | 2+2 2 |
| | 5+6 2 | 5+6 2 | 3+4 2 | 3+4 2 | 2+2 2 | 3+3 2 |
| | 5+6 2 | 3+4 2 | 5+6 2 | 3+4 2 | 3+3 2 | 2+2 2 |
| | 3+4 2 | 3+4 2 | 5+6 2 | 5+6 2 | 2+2 2 | 2+2 2 |
| | 5+6 2 | 5+6 2 | 3+4 2 | 3+4 2 | 3+3 2 | 3+3 2 |
| | 3+4 2 | 3+4 2 | 5+6 2 | 3+4 2 | 3+3 2 | 2+2 2 |
| | 5+6 2 | 3+4 2 | 3+4 2 | 5+6 2 | 2+2 2 | 3+3 2 |
| | 5+6 2 | 5+6 2 | 5+6 2 | 5+6 2 | 2+2 2 | 2+2 2 |
| | 5+6 2 | 3+4 2 | 3+4 2 | 3+4 2 | 3+3 2 | 3+3 2 |
| 12.2.146 | 261456 | 104 | Bi2Pd3Se2 | | | |
| | -A | -GM | -M | -Y | -L | -V |
| | 5+6 2 | 3+4 2 | 5+6 2 | 3+4 2 | 3+3 2 | 2+2 2 |
| | 3+4 2 | 5+6 2 | 3+4 2 | 5+6 2 | 2+2 2 | 3+3 2 |
| | 3+4 2 | 5+6 2 | 5+6 2 | 5+6 2 | 3+3 2 | 3+3 2 |
| | 5+6 2 | 3+4 2 | 3+4 2 | 3+4 2 | 2+2 2 | 2+2 2 |
| | 3+4 2 | 3+4 2 | 3+4 2 | 3+4 2 | 3+3 2 | 3+3 2 |
| | 5+6 2 | 3+4 2 | 5+6 2 | 3+4 2 | 2+2 2 | 2+2 2 |
| | 3+4 2 | 5+6 2 | 3+4 2 | 5+6 2 | 3+3 2 | 3+3 2 |
| | 5+6 2 | 5+6 2 | 5+6 2 | 5+6 2 | 2+2 2 | 2+2 2 |
| | 5+6 2 | 3+4 2 | 5+6 2 | 5+6 2 | 2+2 2 | 2+2 2 |
| | 3+4 2 | 5+6 2 | 5+6 2 | 5+6 2 | 3+3 2 | 2+2 2 |
| | 3+4 2 | 3+4 2 | 3+4 2 | 3+4 2 | 2+2 2 | 3+3 2 |
| | 3+4 2 | 5+6 2 | 5+6 2 | 5+6 2 | 3+3 2 | 3+3 2 |
| | 5+6 2 | 3+4 2 | 3+4 2 | 5+6 2 | 3+3 2 | 3+3 2 |
| | 3+4 2 | 3+4 2 | 5+6 2 | 5+6 2 | 2+2 2 | 2+2 2 |
| | 3+4 2 | 3+4 2 | 5+6 2 | 3+4 2 | 3+3 2 | 2+2 2 |
| | 5+6 2 | 3+4 2 | 3+4 2 | 5+6 2 | 2+2 2 | 3+3 2 |
| | 3+4 2 | 3+4 2 | 5+6 2 | 3+4 2 | 3+3 2 | 3+3 2 |

Continued on next column

**TABLE XXIV– continued**

| TQC | ICSD | NoE | SF | | | |
|---|---|---|---|---|---|---|
| | 5+6 2 | 5+6 2 | 3+4 2 | 3+4 2 | 3+3 2 | 3+3 2 |
| | 3+4 2 | 3+4 2 | 5+6 2 | 5+6 2 | 2+2 2 | 2+2 2 |
| | 3+4 2 | 5+6 2 | 3+4 2 | 3+4 2 | 2+2 2 | 2+2 2 |
| | 5+6 2 | 3+4 2 | 3+4 2 | 5+6 2 | 3+3 2 | 3+3 2 |
| | 3+4 2 | 3+4 2 | 5+6 2 | 5+6 2 | 2+2 2 | 3+3 2 |
| | 3+4 2 | 3+4 2 | 5+6 2 | 3+4 2 | 3+3 2 | 2+2 2 |
| | 5+6 2 | 5+6 2 | 5+6 2 | 5+6 2 | 3+3 2 | 3+3 2 |
| | 3+4 2 | 3+4 2 | 3+4 2 | 3+4 2 | 2+2 2 | 2+2 2 |
| | 5+6 2 | 5+6 2 | 3+4 2 | 5+6 2 | 2+2 2 | 3+3 2 |
| | 3+4 2 | 5+6 2 | 5+6 2 | 3+4 2 | 3+3 2 | 2+2 2 |
| | 3+4 2 | 3+4 2 | 5+6 2 | 3+4 2 | 3+3 2 | 3+3 2 |
| | 5+6 2 | 3+4 2 | 3+4 2 | 3+4 2 | 2+2 2 | 2+2 2 |
| | 5+6 2 | 5+6 2 | 3+4 2 | 5+6 2 | 2+2 2 | 2+2 2 |
| | 3+4 2 | 5+6 2 | 5+6 2 | 5+6 2 | 2+2 2 | 3+3 2 |
| | 3+4 2 | 3+4 2 | 5+6 2 | 5+6 2 | 3+3 2 | 2+2 2 |
| | 5+6 2 | 5+6 2 | 5+6 2 | 3+4 2 | 2+2 2 | 3+3 2 |
| | 3+4 2 | 3+4 2 | 5+6 2 | 5+6 2 | 3+3 2 | 2+2 2 |
| | 5+6 2 | 5+6 2 | 3+4 2 | 5+6 2 | 2+2 2 | 2+2 2 |
| | 5+6 2 | 5+6 2 | 5+6 2 | 3+4 2 | 2+2 2 | 3+3 2 |
| | 3+4 2 | 3+4 2 | 5+6 2 | 5+6 2 | 3+3 2 | 3+3 2 |
| | 3+4 2 | 3+4 2 | 3+4 2 | 5+6 2 | 2+2 2 | 2+2 2 |
| | 3+4 2 | 3+4 2 | 5+6 2 | 5+6 2 | 2+2 2 | 3+3 2 |
| | 5+6 2 | 3+4 2 | 3+4 2 | 3+4 2 | 3+3 2 | 2+2 2 |
| | 3+4 2 | 5+6 2 | 5+6 2 | 5+6 2 | 3+3 2 | 2+2 2 |
| | 3+4 2 | 3+4 2 | 5+6 2 | 5+6 2 | 2+2 2 | 3+3 2 |
| | 5+6 2 | 5+6 2 | 3+4 2 | 3+4 2 | 3+3 2 | 2+2 2 |
| | 5+6 2 | 3+4 2 | 5+6 2 | 5+6 2 | 3+3 2 | 2+2 2 |
| | 3+4 2 | 5+6 2 | 5+6 2 | 3+4 2 | 2+2 2 | 3+3 2 |
| | 3+4 2 | 3+4 2 | 3+4 2 | 3+4 2 | 3+3 2 | 3+3 2 |
| | 5+6 2 | 3+4 2 | 3+4 2 | 5+6 2 | 2+2 2 | 3+3 2 |
| | 3+4 2 | 5+6 2 | 5+6 2 | 3+4 2 | 3+3 2 | 2+2 2 |
| | 5+6 2 | 3+4 2 | 3+4 2 | 3+4 2 | 3+3 2 | 3+3 2 |
| | 3+4 2 | 3+4 2 | 5+6 2 | 5+6 2 | 2+2 2 | 2+2 2 |
| | 3+4 2 | 5+6 2 | 3+4 2 | 5+6 2 | 2+2 2 | 3+3 2 |
| | 3+4 2 | 3+4 2 | 5+6 2 | 3+4 2 | 3+3 2 | 2+2 2 |
| 12.2.147 | 616954 | 104 | Bi2Pd3S2 | | | |
| | -A | -GM | -M | -Y | -L | -V |
| | 3+4 2 | 3+4 2 | 5+6 2 | 5+6 2 | 2+2 2 | 3+3 2 |
| | 5+6 2 | 3+4 2 | 3+4 2 | 3+4 2 | 3+3 2 | 2+2 2 |
| | 3+4 2 | 5+6 2 | 5+6 2 | 5+6 2 | 3+3 2 | 2+2 2 |
| | 5+6 2 | 5+6 2 | 3+4 2 | 3+4 2 | 2+2 2 | 3+3 2 |
| | 5+6 2 | 3+4 2 | 5+6 2 | 3+4 2 | 2+2 2 | 3+3 2 |
| | 3+4 2 | 3+4 2 | 3+4 2 | 3+4 2 | 3+3 2 | 2+2 2 |
| | 5+6 2 | 5+6 2 | 5+6 2 | 5+6 2 | 2+2 2 | 3+3 2 |
| | 3+4 2 | 5+6 2 | 3+4 2 | 5+6 2 | 3+3 2 | 2+2 2 |
| | 3+4 2 | 3+4 2 | 5+6 2 | 5+6 2 | 3+3 2 | 2+2 2 |
| | 3+4 2 | 3+4 2 | 3+4 2 | 5+6 2 | 2+2 2 | 3+3 2 |
| | 3+4 2 | 5+6 2 | 5+6 2 | 5+6 2 | 3+3 2 | 2+2 2 |
| | 5+6 2 | 3+4 2 | 5+6 2 | 3+4 2 | 3+3 2 | 3+3 2 |
| | 3+4 2 | 5+6 2 | 5+6 2 | 5+6 2 | 2+2 2 | 2+2 2 |
| | 5+6 2 | 3+4 2 | 3+4 2 | 5+6 2 | 2+2 2 | 3+3 2 |
| | 5+6 2 | 3+4 2 | 5+6 2 | 3+4 2 | 2+2 2 | 3+3 2 |
| | 3+4 2 | 3+4 2 | 3+4 2 | 5+6 2 | 3+3 2 | 2+2 2 |
| | 3+4 2 | 3+4 2 | 5+6 2 | 3+4 2 | 2+2 2 | 3+3 2 |
| | 5+6 2 | 5+6 2 | 5+6 2 | 3+4 2 | 3+3 2 | 2+2 2 |
| | 3+4 2 | 3+4 2 | 5+6 2 | 5+6 2 | 2+2 2 | 3+3 2 |
| | 5+6 2 | 5+6 2 | 3+4 2 | 5+6 2 | 2+2 2 | 2+2 2 |
| | 3+4 2 | 3+4 2 | 5+6 2 | 3+4 2 | 3+3 2 | 3+3 2 |
| | 5+6 2 | 3+4 2 | 3+4 2 | 5+6 2 | 2+2 2 | 3+3 2 |
| | 3+4 2 | 3+4 2 | 5+6 2 | 3+4 2 | 3+3 2 | 2+2 2 |
| | 5+6 2 | 5+6 2 | 3+4 2 | 5+6 2 | 2+2 2 | 3+3 2 |

Continued on next column

**TABLE XXIV– continued**

| TQC | ICSD | | NoE | SF | | |
|---|---|---|---|---|---|---|
| | 3+4 2 | 3+4 2 | 5+6 2 | 3+4 2 | 3+3 2 | 2+2 2 |
| | 3+4 2 | 5+6 2 | 5+6 2 | 5+6 2 | 2+2 2 | 3+3 2 |
| | 5+6 2 | 5+6 2 | 3+4 2 | 3+4 2 | 3+3 2 | 2+2 2 |
| | 3+4 2 | 3+4 2 | 3+4 2 | 3+4 2 | 2+2 2 | 2+2 2 |
| | 5+6 2 | 5+6 2 | 5+6 2 | 3+4 2 | 3+3 2 | 3+3 2 |
| | 5+6 2 | 3+4 2 | 3+4 2 | 5+6 2 | 3+3 2 | 2+2 2 |
| | 3+4 2 | 5+6 2 | 5+6 2 | 5+6 2 | 3+3 2 | 2+2 2 |
| | 3+4 2 | 3+4 2 | 3+4 2 | 5+6 2 | 3+3 2 | 3+3 2 |
| | 3+4 2 | 5+6 2 | 5+6 2 | 5+6 2 | 2+2 2 | 3+3 2 |
| | 3+4 2 | 3+4 2 | 5+6 2 | 3+4 2 | 3+3 2 | 2+2 2 |
| | 5+6 2 | 5+6 2 | 3+4 2 | 5+6 2 | 2+2 2 | 2+2 2 |
| | 3+4 2 | 5+6 2 | 3+4 2 | 5+6 2 | 2+2 2 | 3+3 2 |
| | 3+4 2 | 3+4 2 | 5+6 2 | 3+4 2 | 3+3 2 | 3+3 2 |
| | 3+4 2 | 3+4 2 | 5+6 2 | 5+6 2 | 3+3 2 | 2+2 2 |
| | 5+6 2 | 3+4 2 | 3+4 2 | 5+6 2 | 3+3 2 | 3+3 2 |
| | 5+6 2 | 3+4 2 | 5+6 2 | 3+4 2 | 2+2 2 | 2+2 2 |
| | 3+4 2 | 5+6 2 | 5+6 2 | 5+6 2 | 2+2 2 | 2+2 2 |
| | 3+4 2 | 3+4 2 | 3+4 2 | 5+6 2 | 3+3 2 | 3+3 2 |
| | 5+6 2 | 5+6 2 | 5+6 2 | 5+6 2 | 2+2 2 | 2+2 2 |
| | 3+4 2 | 3+4 2 | 3+4 2 | 3+4 2 | 2+2 2 | 2+2 2 |
| | 5+6 2 | 5+6 2 | 5+6 2 | 3+4 2 | 3+3 2 | 3+3 2 |
| | 3+4 2 | 3+4 2 | 5+6 2 | 3+4 2 | 2+2 2 | 3+3 2 |
| | 5+6 2 | 5+6 2 | 3+4 2 | 3+4 2 | 3+3 2 | 3+3 2 |
| | 3+4 2 | 3+4 2 | 5+6 2 | 5+6 2 | 2+2 2 | 2+2 2 |
| | 5+6 2 | 3+4 2 | 3+4 2 | 3+4 2 | 2+2 2 | 3+3 2 |
| | 3+4 2 | 3+4 2 | 5+6 2 | 3+4 2 | 3+3 2 | 2+2 2 |
| | 5+6 2 | 5+6 2 | 5+6 2 | 5+6 2 | 3+3 2 | 3+3 2 |
| | 3+4 2 | 3+4 2 | 5+6 2 | 5+6 2 | 2+2 2 | 2+2 2 |
| 12.2.148 | 616946 | | 40 | Bi2Pd1 | | |
| | -A | -GM | -M | -Y | -L | -V |
| | 5+6 2 | 3+4 2 | 3+4 2 | 5+6 2 | 2+2 2 | 2+2 2 |
| | 3+4 2 | 5+6 2 | 5+6 2 | 3+4 2 | 3+3 2 | 3+3 2 |
| | 5+6 2 | 5+6 2 | 3+4 2 | 3+4 2 | 2+2 2 | 2+2 2 |
| | 3+4 2 | 3+4 2 | 5+6 2 | 5+6 2 | 3+3 2 | 3+3 2 |
| | 3+4 2 | 3+4 2 | 3+4 2 | 3+4 2 | 3+3 2 | 3+3 2 |
| | 5+6 2 | 5+6 2 | 5+6 2 | 5+6 2 | 2+2 2 | 2+2 2 |
| | 5+6 2 | 3+4 2 | 3+4 2 | 3+4 2 | 2+2 2 | 3+3 2 |
| | 5+6 2 | 5+6 2 | 5+6 2 | 5+6 2 | 3+3 2 | 2+2 2 |
| | 3+4 2 | 5+6 2 | 3+4 2 | 5+6 2 | 2+2 2 | 3+3 2 |
| | 3+4 2 | 3+4 2 | 5+6 2 | 3+4 2 | 3+3 2 | 2+2 2 |
| | 3+4 2 | 3+4 2 | 5+6 2 | 5+6 2 | 2+2 2 | 3+3 2 |
| | 5+6 2 | 3+4 2 | 3+4 2 | 5+6 2 | 3+3 2 | 2+2 2 |
| | 5+6 2 | 5+6 2 | 3+4 2 | 3+4 2 | 3+3 2 | 2+2 2 |
| | 3+4 2 | 3+4 2 | 5+6 2 | 3+4 2 | 2+2 2 | 3+3 2 |
| | 3+4 2 | 3+4 2 | 3+4 2 | 5+6 2 | 2+2 2 | 2+2 2 |
| | 5+6 2 | 5+6 2 | 5+6 2 | 3+4 2 | 3+3 2 | 3+3 2 |
| | 3+4 2 | 5+6 2 | 5+6 2 | 5+6 2 | 3+3 2 | 2+2 2 |
| | 5+6 2 | 3+4 2 | 3+4 2 | 3+4 2 | 2+2 2 | 2+2 2 |
| | 3+4 2 | 3+4 2 | 5+6 2 | 5+6 2 | 3+3 2 | 3+3 2 |
| | 5+6 2 | 5+6 2 | 3+4 2 | 3+4 2 | 2+2 2 | 3+3 2 |
| 12.2.149 | 42565 | | 40 | Bi2Pd1 | | |
| | -A | -GM | -M | -Y | -L | -V |
| | 3+4 2 | 3+4 2 | 5+6 2 | 5+6 2 | 2+2 2 | 2+2 2 |
| | 5+6 2 | 5+6 2 | 3+4 2 | 3+4 2 | 3+3 2 | 3+3 2 |
| | 3+4 2 | 5+6 2 | 5+6 2 | 3+4 2 | 3+3 2 | 2+2 2 |
| | 5+6 2 | 3+4 2 | 3+4 2 | 5+6 2 | 2+2 2 | 3+3 2 |
| | 5+6 2 | 3+4 2 | 5+6 2 | 3+4 2 | 2+2 2 | 3+3 2 |
| | 3+4 2 | 5+6 2 | 3+4 2 | 5+6 2 | 3+3 2 | 2+2 2 |
| | 3+4 2 | 3+4 2 | 5+6 2 | 3+4 2 | 3+3 2 | 3+3 2 |
| | 3+4 2 | 5+6 2 | 3+4 2 | 5+6 2 | 2+2 2 | 2+2 2 |
| | 5+6 2 | 5+6 2 | 5+6 2 | 5+6 2 | 3+3 2 | 3+3 2 |

Continued on next column

| TQC | ICSD | | NoE | SF | | |
|---|---|---|---|---|---|---|
| | 5+6 2 | 3+4 2 | 3+4 2 | 3+4 2 | 2+2 2 | 2+2 2 |
| | 5+6 2 | 3+4 2 | 3+4 2 | 5+6 2 | 3+3 2 | 3+3 2 |
| | 3+4 2 | 3+4 2 | 5+6 2 | 5+6 2 | 2+2 2 | 2+2 2 |
| | 3+4 2 | 5+6 2 | 5+6 2 | 3+4 2 | 2+2 2 | 2+2 2 |
| | 5+6 2 | 3+4 2 | 3+4 2 | 3+4 2 | 3+3 2 | 3+3 2 |
| | 5+6 2 | 3+4 2 | 5+6 2 | 5+6 2 | 3+3 2 | 2+2 2 |
| | 3+4 2 | 5+6 2 | 3+4 2 | 3+4 2 | 2+2 2 | 3+3 2 |
| | 5+6 2 | 5+6 2 | 3+4 2 | 5+6 2 | 2+2 2 | 2+2 2 |
| | 3+4 2 | 3+4 2 | 5+6 2 | 3+4 2 | 3+3 2 | 2+2 2 |
| | 5+6 2 | 3+4 2 | 3+4 2 | 5+6 2 | 2+2 2 | 3+3 2 |
| | 3+4 2 | 5+6 2 | 5+6 2 | 3+4 2 | 3+3 2 | 3+3 2 |
| 12.2.150 | 1024 | | 104 | Bi2Ni3Se2 | | |
| | -A | -GM | -M | -Y | -L | -V |
| | 3+4 2 | 3+4 2 | 3+4 2 | 3+4 2 | 2+2 2 | 2+2 2 |
| | 5+6 2 | 5+6 2 | 5+6 2 | 5+6 2 | 3+3 2 | 3+3 2 |
| | 5+6 2 | 5+6 2 | 5+6 2 | 5+6 2 | 3+3 2 | 3+3 2 |
| | 3+4 2 | 3+4 2 | 3+4 2 | 3+4 2 | 2+2 2 | 2+2 2 |
| | 5+6 2 | 3+4 2 | 5+6 2 | 3+4 2 | 2+2 2 | 3+3 2 |
| | 3+4 2 | 3+4 2 | 3+4 2 | 3+4 2 | 3+3 2 | 2+2 2 |
| | 5+6 2 | 5+6 2 | 5+6 2 | 5+6 2 | 2+2 2 | 3+3 2 |
| | 3+4 2 | 5+6 2 | 3+4 2 | 5+6 2 | 3+3 2 | 2+2 2 |
| | 3+4 2 | 3+4 2 | 3+4 2 | 5+6 2 | 3+3 2 | 2+2 2 |
| | 5+6 2 | 5+6 2 | 3+4 2 | 5+6 2 | 2+2 2 | 2+2 2 |
| | 5+6 2 | 3+4 2 | 5+6 2 | 5+6 2 | 3+3 2 | 3+3 2 |
| | 3+4 2 | 5+6 2 | 3+4 2 | 3+4 2 | 2+2 2 | 3+3 2 |
| | 5+6 2 | 3+4 2 | 5+6 2 | 5+6 2 | 2+2 2 | 2+2 2 |
| | 3+4 2 | 3+4 2 | 5+6 2 | 5+6 2 | 3+3 2 | 3+3 2 |
| | 5+6 2 | 5+6 2 | 3+4 2 | 3+4 2 | 3+3 2 | 2+2 2 |
| | 3+4 2 | 5+6 2 | 5+6 2 | 3+4 2 | 2+2 2 | 3+3 2 |
| | 5+6 2 | 3+4 2 | 3+4 2 | 5+6 2 | 2+2 2 | 3+3 2 |
| | 5+6 2 | 5+6 2 | 5+6 2 | 3+4 2 | 3+3 2 | 2+2 2 |
| | 3+4 2 | 5+6 2 | 3+4 2 | 3+4 2 | 2+2 2 | 3+3 2 |
| | 5+6 2 | 3+4 2 | 3+4 2 | 5+6 2 | 3+3 2 | 2+2 2 |
| | 3+4 2 | 3+4 2 | 5+6 2 | 3+4 2 | 2+2 2 | 3+3 2 |
| | 5+6 2 | 3+4 2 | 3+4 2 | 3+4 2 | 3+3 2 | 2+2 2 |
| | 3+4 2 | 5+6 2 | 5+6 2 | 5+6 2 | 3+3 2 | 3+3 2 |
| | 5+6 2 | 3+4 2 | 5+6 2 | 3+4 2 | 2+2 2 | 2+2 2 |
| | 5+6 2 | 5+6 2 | 3+4 2 | 5+6 2 | 2+2 2 | 3+3 2 |
| | 3+4 2 | 3+4 2 | 3+4 2 | 3+4 2 | 3+3 2 | 2+2 2 |
| | 3+4 2 | 3+4 2 | 5+6 2 | 5+6 2 | 2+2 2 | 3+3 2 |
| | 5+6 2 | 3+4 2 | 3+4 2 | 3+4 2 | 3+3 2 | 2+2 2 |
| | 5+6 2 | 5+6 2 | 5+6 2 | 3+4 2 | 2+2 2 | 3+3 2 |
| | 5+6 2 | 3+4 2 | 3+4 2 | 5+6 2 | 3+3 2 | 3+3 2 |
| | 5+6 2 | 3+4 2 | 3+4 2 | 3+4 2 | 3+3 2 | 2+2 2 |
| | 3+4 2 | 3+4 2 | 3+4 2 | 5+6 2 | 2+2 2 | 2+2 2 |
| | 5+6 2 | 5+6 2 | 5+6 2 | 5+6 2 | 2+2 2 | 3+3 2 |
| | 3+4 2 | 5+6 2 | 3+4 2 | 5+6 2 | 3+3 2 | 2+2 2 |
| | 5+6 2 | 3+4 2 | 5+6 2 | 5+6 2 | 3+3 2 | 2+2 2 |
| | 3+4 2 | 5+6 2 | 3+4 2 | 5+6 2 | 3+3 2 | 3+3 2 |
| | 3+4 2 | 3+4 2 | 3+4 2 | 3+4 2 | 2+2 2 | 2+2 2 |
| | 5+6 2 | 3+4 2 | 5+6 2 | 3+4 2 | 2+2 2 | 3+3 2 |
| | 5+6 2 | 3+4 2 | 5+6 2 | 5+6 2 | 3+3 2 | 3+3 2 |
| | 5+6 2 | 3+4 2 | 3+4 2 | 5+6 2 | 3+3 2 | 2+2 2 |
| | 5+6 2 | 5+6 2 | 3+4 2 | 5+6 2 | 2+2 2 | 3+3 2 |
| | 3+4 2 | 5+6 2 | 3+4 2 | 5+6 2 | 2+2 2 | 2+2 2 |
| | 5+6 2 | 3+4 2 | 3+4 2 | 3+4 2 | 3+3 2 | 2+2 2 |
| | 3+4 2 | 3+4 2 | 3+4 2 | 5+6 2 | 2+2 2 | 3+3 2 |
| | 5+6 2 | 5+6 2 | 5+6 2 | 5+6 2 | 3+3 2 | 2+2 2 |
| | 3+4 2 | 3+4 2 | 3+4 2 | 3+4 2 | 2+2 2 | 3+3 2 |
| | 5+6 2 | 5+6 2 | 5+6 2 | 3+4 2 | 3+3 2 | 3+3 2 |
| | 3+4 2 | 3+4 2 | 3+4 2 | 5+6 2 | 2+2 2 | 2+2 2 |

Continued on next column

**TABLE XXIV– continued**

| TQC | ICSD | | NoE | | SF | |
|---|---|---|---|---|---|---|
| | 5+6 2 | 3+4 2 | 5+6 2 | 3+4 2 | 2+2 2 | 3+3 2 |
| | 5+6 2 | 3+4 2 | 3+4 2 | 5+6 2 | 3+3 2 | 2+2 2 |
| | 5+6 2 | 5+6 2 | 5+6 2 | 5+6 2 | 3+3 2 | 3+3 2 |
| | 5+6 2 | 3+4 2 | 3+4 2 | 3+4 2 | 2+2 2 | 2+2 2 |
| 12.2.151 | 195678 | | 92 | | Bi2Hg1S4 | |
| | -A | -GM | -M | -Y | -L | -V |
| | 3+4 2 | 3+4 2 | 5+6 2 | 5+6 2 | 2+2 2 | 3+3 2 |
| | 5+6 2 | 5+6 2 | 3+4 2 | 3+4 2 | 3+3 2 | 3+3 2 |
| | 3+4 2 | 3+4 2 | 5+6 2 | 5+6 2 | 3+3 2 | 2+2 2 |
| | 5+6 2 | 5+6 2 | 3+4 2 | 3+4 2 | 2+2 2 | 3+3 2 |
| | 3+4 2 | 3+4 2 | 5+6 2 | 5+6 2 | 3+3 2 | 2+2 2 |
| | 5+6 2 | 5+6 2 | 3+4 2 | 3+4 2 | 2+2 2 | 2+2 2 |
| | 5+6 2 | 3+4 2 | 3+4 2 | 3+4 2 | 3+3 2 | 3+3 2 |
| | 5+6 2 | 5+6 2 | 3+4 2 | 5+6 2 | 2+2 2 | 2+2 2 |
| | 3+4 2 | 3+4 2 | 5+6 2 | 3+4 2 | 2+2 2 | 3+3 2 |
| | 5+6 2 | 5+6 2 | 5+6 2 | 3+4 2 | 3+3 2 | 2+2 2 |
| | 3+4 2 | 3+4 2 | 3+4 2 | 5+6 2 | 3+3 2 | 2+2 2 |
| | 3+4 2 | 5+6 2 | 5+6 2 | 5+6 2 | 2+2 2 | 3+3 2 |
| | 3+4 2 | 3+4 2 | 3+4 2 | 3+4 2 | 3+3 2 | 3+3 2 |
| | 3+4 2 | 3+4 2 | 3+4 2 | 3+4 2 | 2+2 2 | 3+3 2 |
| | 5+6 2 | 3+4 2 | 5+6 2 | 3+4 2 | 3+3 2 | 3+3 2 |
| | 5+6 2 | 3+4 2 | 5+6 2 | 3+4 2 | 2+2 2 | 3+3 2 |
| | 3+4 2 | 3+4 2 | 3+4 2 | 3+4 2 | 3+3 2 | 3+3 2 |
| | 3+4 2 | 3+4 2 | 3+4 2 | 3+4 2 | 2+2 2 | 3+3 2 |
| | 5+6 2 | 3+4 2 | 5+6 2 | 3+4 2 | 3+3 2 | 3+3 2 |
| | 3+4 2 | 3+4 2 | 3+4 2 | 3+4 2 | 3+3 2 | 3+3 2 |
| | 5+6 2 | 3+4 2 | 5+6 2 | 3+4 2 | 2+2 2 | 3+3 2 |
| | 5+6 2 | 3+4 2 | 5+6 2 | 5+6 2 | 2+2 2 | 2+2 2 |
| | 3+4 2 | 3+4 2 | 5+6 2 | 3+4 2 | 2+2 2 | 3+3 2 |
| | 3+4 2 | 3+4 2 | 3+4 2 | 3+4 2 | 3+3 2 | 3+3 2 |
| | 3+4 2 | 5+6 2 | 5+6 2 | 3+4 2 | 3+3 2 | 2+2 2 |
| | 5+6 2 | 3+4 2 | 3+4 2 | 5+6 2 | 3+3 2 | 3+3 2 |
| | 5+6 2 | 5+6 2 | 5+6 2 | 5+6 2 | 2+2 2 | 2+2 2 |
| | 5+6 2 | 3+4 2 | 3+4 2 | 3+4 2 | 2+2 2 | 3+3 2 |
| | 5+6 2 | 5+6 2 | 5+6 2 | 5+6 2 | 2+2 2 | 2+2 2 |
| | 3+4 2 | 5+6 2 | 5+6 2 | 3+4 2 | 3+3 2 | 3+3 2 |
| | 3+4 2 | 3+4 2 | 3+4 2 | 5+6 2 | 3+3 2 | 3+3 2 |
| | 5+6 2 | 3+4 2 | 3+4 2 | 3+4 2 | 2+2 2 | 2+2 2 |
| | 3+4 2 | 5+6 2 | 5+6 2 | 5+6 2 | 3+3 2 | 3+3 2 |
| | 5+6 2 | 5+6 2 | 3+4 2 | 3+4 2 | 2+2 2 | 2+2 2 |
| | 5+6 2 | 5+6 2 | 5+6 2 | 5+6 2 | 2+2 2 | 2+2 2 |
| | 3+4 2 | 3+4 2 | 3+4 2 | 3+4 2 | 3+3 2 | 2+2 2 |
| | 5+6 2 | 5+6 2 | 5+6 2 | 5+6 2 | 2+2 2 | 3+3 2 |
| | 5+6 2 | 3+4 2 | 3+4 2 | 5+6 2 | 3+3 2 | 2+2 2 |
| | 3+4 2 | 5+6 2 | 5+6 2 | 3+4 2 | 2+2 2 | 3+3 2 |
| | 3+4 2 | 3+4 2 | 3+4 2 | 5+6 2 | 3+3 2 | 2+2 2 |
| | 5+6 2 | 3+4 2 | 3+4 2 | 3+4 2 | 3+3 2 | 3+3 2 |
| | 3+4 2 | 5+6 2 | 3+4 2 | 5+6 2 | 3+3 2 | 3+3 2 |
| | 5+6 2 | 5+6 2 | 5+6 2 | 3+4 2 | 2+2 2 | 2+2 2 |
| | 3+4 2 | 3+4 2 | 3+4 2 | 5+6 2 | 2+2 2 | 2+2 2 |
| | 5+6 2 | 5+6 2 | 5+6 2 | 3+4 2 | 3+3 2 | 3+3 2 |
| | 3+4 2 | 3+4 2 | 5+6 2 | 5+6 2 | 2+2 2 | 3+3 2 |
| 12.2.152 | 415138 | | 62 | | Bi1I1Mn1Se2 | |
| | -A | -GM | -M | -Y | -L | -V |
| | 3+4 2 | 3+4 2 | 5+6 2 | 5+6 2 | 3+3 2 | 3+3 2 |
| | 5+6 2 | 5+6 2 | 3+4 2 | 3+4 2 | 2+2 2 | 2+2 2 |
| | 5+6 2 | 3+4 2 | 5+6 2 | 5+6 2 | 3+3 2 | 2+2 2 |
| | 3+4 2 | 5+6 2 | 3+4 2 | 5+6 2 | 2+2 2 | 3+3 2 |
| | 3+4 2 | 3+4 2 | 5+6 2 | 3+4 2 | 2+2 2 | 3+3 2 |
| | 5+6 2 | 5+6 2 | 3+4 2 | 3+4 2 | 3+3 2 | 2+2 2 |
| | 5+6 2 | 5+6 2 | 3+4 2 | 3+4 2 | 2+2 2 | 2+2 2 |

Continued on next column

**TABLE XXIV– continued**

| TQC | ICSD | | NoE | | SF | |
|---|---|---|---|---|---|---|
| | 3+4 2 | 3+4 2 | 5+6 2 | 5+6 2 | 3+3 2 | 3+3 2 |
| | 3+4 2 | 3+4 2 | 5+6 2 | 3+4 2 | 3+3 2 | 3+3 2 |
| | 5+6 2 | 3+4 2 | 3+4 2 | 5+6 2 | 2+2 2 | 2+2 2 |
| | 3+4 2 | 5+6 2 | 5+6 2 | 3+4 2 | 3+3 2 | 2+2 2 |
| | 5+6 2 | 3+4 2 | 3+4 2 | 5+6 2 | 2+2 2 | 2+2 2 |
| | 3+4 2 | 3+4 2 | 5+6 2 | 5+6 2 | 2+2 2 | 3+3 2 |
| | 5+6 2 | 5+6 2 | 5+6 2 | 5+6 2 | 3+3 2 | 3+3 2 |
| | 3+4 2 | 3+4 2 | 3+4 2 | 3+4 2 | 2+2 2 | 2+2 2 |
| | 5+6 2 | 5+6 2 | 3+4 2 | 3+4 2 | 3+3 2 | 3+3 2 |
| | 3+4 2 | 3+4 2 | 5+6 2 | 3+4 2 | 2+2 2 | 2+2 2 |
| | 5+6 2 | 5+6 2 | 3+4 2 | 5+6 2 | 3+3 2 | 3+3 2 |
| | 3+4 2 | 3+4 2 | 5+6 2 | 3+4 2 | 2+2 2 | 3+3 2 |
| | 5+6 2 | 3+4 2 | 3+4 2 | 3+4 2 | 3+3 2 | 2+2 2 |
| | 3+4 2 | 3+4 2 | 3+4 2 | 3+4 2 | 3+3 2 | 2+2 2 |
| | 5+6 2 | 5+6 2 | 5+6 2 | 5+6 2 | 2+2 2 | 3+3 2 |
| | 3+4 2 | 5+6 2 | 3+4 2 | 3+4 2 | 3+3 2 | 2+2 2 |
| | 5+6 2 | 5+6 2 | 3+4 2 | 5+6 2 | 2+2 2 | 2+2 2 |
| | 5+6 2 | 5+6 2 | 5+6 2 | 5+6 2 | 2+2 2 | 3+3 2 |
| | 3+4 2 | 5+6 2 | 5+6 2 | 5+6 2 | 3+3 2 | 3+3 2 |
| | 5+6 2 | 3+4 2 | 5+6 2 | 3+4 2 | 3+3 2 | 2+2 2 |
| | 3+4 2 | 3+4 2 | 3+4 2 | 3+4 2 | 2+2 2 | 2+2 2 |
| | 5+6 2 | 3+4 2 | 3+4 2 | 3+4 2 | 3+3 2 | 2+2 2 |
| | 3+4 2 | 3+4 2 | 5+6 2 | 3+4 2 | 2+2 2 | 2+2 2 |
| | 5+6 2 | 3+4 2 | 3+4 2 | 3+4 2 | 2+2 2 | 2+2 2 |
| 12.2.153 | 429459 | | 154 | | Bi1Gd3O3 | |
| | -A | -GM | -M | -Y | -L | -V |
| | 5+6 2 | 3+4 2 | 3+4 2 | 5+6 2 | 2+2 2 | 2+2 2 |
| | 3+4 2 | 3+4 2 | 5+6 2 | 5+6 2 | 3+3 2 | 3+3 2 |
| | 3+4 2 | 5+6 2 | 5+6 2 | 3+4 2 | 2+2 2 | 2+2 2 |
| | 5+6 2 | 5+6 2 | 3+4 2 | 3+4 2 | 3+3 2 | 3+3 2 |
| | 3+4 2 | 5+6 2 | 5+6 2 | 3+4 2 | 2+2 2 | 2+2 2 |
| | 5+6 2 | 3+4 2 | 3+4 2 | 5+6 2 | 3+3 2 | 3+3 2 |
| | 5+6 2 | 3+4 2 | 3+4 2 | 5+6 2 | 2+2 2 | 2+2 2 |
| | 3+4 2 | 5+6 2 | 5+6 2 | 5+6 2 | 3+3 2 | 3+3 2 |
| | 5+6 2 | 3+4 2 | 5+6 2 | 3+4 2 | 2+2 2 | 3+3 2 |
| | 3+4 2 | 5+6 2 | 3+4 2 | 3+4 2 | 3+3 2 | 2+2 2 |
| | 3+4 2 | 3+4 2 | 3+4 2 | 5+6 2 | 3+3 2 | 2+2 2 |
| | 5+6 2 | 5+6 2 | 5+6 2 | 3+4 2 | 2+2 2 | 3+3 2 |
| | 5+6 2 | 3+4 2 | 3+4 2 | 5+6 2 | 3+3 2 | 2+2 2 |
| | 3+4 2 | 5+6 2 | 5+6 2 | 3+4 2 | 2+2 2 | 3+3 2 |
| | 5+6 2 | 3+4 2 | 5+6 2 | 5+6 2 | 2+2 2 | 2+2 2 |
| | 3+4 2 | 5+6 2 | 3+4 2 | 3+4 2 | 3+3 2 | 3+3 2 |
| | 3+4 2 | 3+4 2 | 3+4 2 | 5+6 2 | 3+3 2 | 3+3 2 |
| | 5+6 2 | 5+6 2 | 5+6 2 | 3+4 2 | 2+2 2 | 2+2 2 |
| | 5+6 2 | 3+4 2 | 3+4 2 | 5+6 2 | 3+3 2 | 2+2 2 |
| | 3+4 2 | 3+4 2 | 5+6 2 | 5+6 2 | 2+2 2 | 3+3 2 |
| | 5+6 2 | 5+6 2 | 5+6 2 | 3+4 2 | 2+2 2 | 3+3 2 |
| | 3+4 2 | 3+4 2 | 3+4 2 | 5+6 2 | 3+3 2 | 2+2 2 |
| | 3+4 2 | 5+6 2 | 3+4 2 | 3+4 2 | 2+2 2 | 3+3 2 |
| | 5+6 2 | 5+6 2 | 5+6 2 | 3+4 2 | 3+3 2 | 2+2 2 |
| | 3+4 2 | 3+4 2 | 5+6 2 | 3+4 2 | 2+2 2 | 3+3 2 |
| | 5+6 2 | 5+6 2 | 3+4 2 | 5+6 2 | 3+3 2 | 2+2 2 |
| | 3+4 2 | 5+6 2 | 5+6 2 | 3+4 2 | 3+3 2 | 2+2 2 |
| | 5+6 2 | 3+4 2 | 3+4 2 | 5+6 2 | 2+2 2 | 3+3 2 |
| | 5+6 2 | 5+6 2 | 5+6 2 | 3+4 2 | 2+2 2 | 3+3 2 |
| | 3+4 2 | 3+4 2 | 3+4 2 | 5+6 2 | 3+3 2 | 2+2 2 |
| | 5+6 2 | 3+4 2 | 5+6 2 | 3+4 2 | 3+3 2 | 2+2 2 |
| | 3+4 2 | 5+6 2 | 3+4 2 | 5+6 2 | 2+2 2 | 3+3 2 |
| | 3+4 2 | 5+6 2 | 3+4 2 | 3+4 2 | 2+2 2 | 2+2 2 |
| | 5+6 2 | 3+4 2 | 5+6 2 | 5+6 2 | 3+3 2 | 3+3 2 |
| | 5+6 2 | 5+6 2 | 5+6 2 | 3+4 2 | 3+3 2 | 3+3 2 |

Continued on next column

**TABLE XXIV– continued**

| TQC | ICSD | NoE | SF | | | |
|---|---|---|---|---|---|---|
| | 3+4 2 | 3+4 2 | 5+6 2 | 5+6 2 | 2+2 2 | 2+2 2 |
| | 3+4 2 | 3+4 2 | 3+4 2 | 5+6 2 | 2+2 2 | 3+3 2 |
| | 5+6 2 | 3+4 2 | 3+4 2 | 3+4 2 | 3+3 2 | 2+2 2 |
| | 5+6 2 | 5+6 2 | 3+4 2 | 5+6 2 | 3+3 2 | 2+2 2 |
| | 5+6 2 | 5+6 2 | 5+6 2 | 5+6 2 | 2+2 2 | 3+3 2 |
| | 5+6 2 | 5+6 2 | 3+4 2 | 3+4 2 | 2+2 2 | 2+2 2 |
| | 3+4 2 | 3+4 2 | 3+4 2 | 3+4 2 | 3+3 2 | 3+3 2 |
| | 3+4 2 | 3+4 2 | 3+4 2 | 5+6 2 | 2+2 2 | 3+3 2 |
| | 3+4 2 | 3+4 2 | 5+6 2 | 3+4 2 | 3+3 2 | 2+2 2 |
| | 3+4 2 | 3+4 2 | 3+4 2 | 5+6 2 | 3+3 2 | 3+3 2 |
| | 5+6 2 | 5+6 2 | 5+6 2 | 3+4 2 | 2+2 2 | 2+2 2 |
| | 3+4 2 | 3+4 2 | 5+6 2 | 5+6 2 | 3+3 2 | 2+2 2 |
| | 5+6 2 | 5+6 2 | 5+6 2 | 3+4 2 | 2+2 2 | 3+3 2 |
| | 5+6 2 | 5+6 2 | 3+4 2 | 5+6 2 | 2+2 2 | 3+3 2 |
| | 3+4 2 | 5+6 2 | 5+6 2 | 3+4 2 | 3+3 2 | 2+2 2 |
| | 5+6 2 | 3+4 2 | 5+6 2 | 3+4 2 | 3+3 2 | 2+2 2 |
| | 3+4 2 | 5+6 2 | 3+4 2 | 5+6 2 | 2+2 2 | 3+3 2 |
| | 5+6 2 | 3+4 2 | 5+6 2 | 3+4 2 | 3+3 2 | 2+2 2 |
| | 3+4 2 | 5+6 2 | 3+4 2 | 5+6 2 | 2+2 2 | 3+3 2 |
| | 3+4 2 | 3+4 2 | 3+4 2 | 5+6 2 | 3+3 2 | 2+2 2 |
| | 5+6 2 | 5+6 2 | 5+6 2 | 3+4 2 | 2+2 2 | 3+3 2 |
| | 5+6 2 | 3+4 2 | 3+4 2 | 5+6 2 | 3+3 2 | 2+2 2 |
| | 3+4 2 | 3+4 2 | 5+6 2 | 3+4 2 | 2+2 2 | 3+3 2 |
| | 3+4 2 | 5+6 2 | 3+4 2 | 5+6 2 | 3+3 2 | 2+2 2 |
| | 5+6 2 | 3+4 2 | 3+4 2 | 5+6 2 | 2+2 2 | 3+3 2 |
| | 3+4 2 | 5+6 2 | 5+6 2 | 3+4 2 | 3+3 2 | 2+2 2 |
| | 5+6 2 | 3+4 2 | 5+6 2 | 3+4 2 | 2+2 2 | 3+3 2 |
| | 5+6 2 | 5+6 2 | 3+4 2 | 5+6 2 | 3+3 2 | 2+2 2 |
| | 3+4 2 | 3+4 2 | 5+6 2 | 3+4 2 | 2+2 2 | 3+3 2 |
| | 3+4 2 | 5+6 2 | 3+4 2 | 5+6 2 | 3+3 2 | 2+2 2 |
| | 5+6 2 | 3+4 2 | 5+6 2 | 3+4 2 | 2+2 2 | 3+3 2 |
| | 3+4 2 | 5+6 2 | 5+6 2 | 5+6 2 | 3+3 2 | 2+2 2 |
| | 5+6 2 | 3+4 2 | 3+4 2 | 3+4 2 | 2+2 2 | 3+3 2 |
| | 5+6 2 | 5+6 2 | 5+6 2 | 5+6 2 | 3+3 2 | 2+2 2 |
| | 3+4 2 | 5+6 2 | 3+4 2 | 3+4 2 | 2+2 2 | 3+3 2 |
| | 5+6 2 | 3+4 2 | 5+6 2 | 3+4 2 | 3+3 2 | 2+2 2 |
| | 3+4 2 | 5+6 2 | 3+4 2 | 5+6 2 | 2+2 2 | 3+3 2 |
| | 5+6 2 | 3+4 2 | 3+4 2 | 3+4 2 | 3+3 2 | 2+2 2 |
| | 3+4 2 | 3+4 2 | 3+4 2 | 5+6 2 | 2+2 2 | 3+3 2 |
| | 5+6 2 | 5+6 2 | 5+6 2 | 5+6 2 | 3+3 2 | 2+2 2 |
| | 3+4 2 | 3+4 2 | 3+4 2 | 5+6 2 | 2+2 2 | 3+3 2 |
| | 3+4 2 | 5+6 2 | 5+6 2 | 3+4 2 | 3+3 2 | 2+2 2 |
| 12.2.154 | 164224 | 186 | Bi1Fe1O3 | | | |
| | -A | -GM | -M | -Y | -L | -V |
| | 5+6 2 | 3+4 2 | 3+4 2 | 5+6 2 | 2+2 2 | 2+2 2 |
| | 3+4 2 | 5+6 2 | 5+6 2 | 3+4 2 | 3+3 2 | 3+3 2 |
| | 5+6 2 | 3+4 2 | 3+4 2 | 5+6 2 | 3+3 2 | 2+2 2 |
| | 5+6 2 | 3+4 2 | 3+4 2 | 5+6 2 | 2+2 2 | 3+3 2 |
| | 3+4 2 | 3+4 2 | 5+6 2 | 3+4 2 | 3+3 2 | 2+2 2 |
| | 3+4 2 | 5+6 2 | 5+6 2 | 5+6 2 | 2+2 2 | 3+3 2 |
| | 5+6 2 | 5+6 2 | 5+6 2 | 3+4 2 | 3+3 2 | 2+2 2 |
| | 3+4 2 | 5+6 2 | 3+4 2 | 3+4 2 | 2+2 2 | 3+3 2 |
| | 5+6 2 | 3+4 2 | 3+4 2 | 5+6 2 | 3+3 2 | 2+2 2 |
| | 5+6 2 | 3+4 2 | 3+4 2 | 5+6 2 | 2+2 2 | 3+3 2 |
| | 3+4 2 | 3+4 2 | 5+6 2 | 3+4 2 | 3+3 2 | 3+3 2 |
| | 5+6 2 | 5+6 2 | 3+4 2 | 3+4 2 | 2+2 2 | 2+2 2 |
| | 3+4 2 | 5+6 2 | 5+6 2 | 5+6 2 | 2+2 2 | 2+2 2 |
| | 5+6 2 | 5+6 2 | 5+6 2 | 5+6 2 | 3+3 2 | 3+3 2 |
| | 3+4 2 | 3+4 2 | 3+4 2 | 3+4 2 | 2+2 2 | 2+2 2 |
| | 3+4 2 | 5+6 2 | 5+6 2 | 3+4 2 | 3+3 2 | 3+3 2 |
| | 3+4 2 | 5+6 2 | 3+4 2 | 5+6 2 | 3+3 2 | 3+3 2 |

Continued on next column

**TABLE XXIV– continued**

| TQC | ICSD | NoE | SF | | | |
|---|---|---|---|---|---|---|
| | 5+6 2 | 3+4 2 | 5+6 2 | 3+4 2 | 2+2 2 | 2+2 2 |
| | 5+6 2 | 3+4 2 | 3+4 2 | 5+6 2 | 2+2 2 | 3+3 2 |
| | 5+6 2 | 3+4 2 | 3+4 2 | 5+6 2 | 3+3 2 | 2+2 2 |
| | 5+6 2 | 3+4 2 | 3+4 2 | 5+6 2 | 2+2 2 | 2+2 2 |
| | 5+6 2 | 3+4 2 | 3+4 2 | 5+6 2 | 3+3 2 | 3+3 2 |
| | 3+4 2 | 5+6 2 | 5+6 2 | 3+4 2 | 2+2 2 | 3+3 2 |
| | 3+4 2 | 5+6 2 | 5+6 2 | 3+4 2 | 3+3 2 | 2+2 2 |
| | 5+6 2 | 3+4 2 | 3+4 2 | 5+6 2 | 2+2 2 | 2+2 2 |
| | 3+4 2 | 3+4 2 | 5+6 2 | 5+6 2 | 3+3 2 | 3+3 2 |
| | 5+6 2 | 5+6 2 | 3+4 2 | 3+4 2 | 2+2 2 | 3+3 2 |
| | 3+4 2 | 5+6 2 | 5+6 2 | 3+4 2 | 3+3 2 | 2+2 2 |
| | 5+6 2 | 3+4 2 | 3+4 2 | 5+6 2 | 2+2 2 | 3+3 2 |
| | 3+4 2 | 5+6 2 | 5+6 2 | 3+4 2 | 3+3 2 | 2+2 2 |
| | 3+4 2 | 5+6 2 | 3+4 2 | 3+4 2 | 2+2 2 | 3+3 2 |
| | 5+6 2 | 3+4 2 | 5+6 2 | 5+6 2 | 3+3 2 | 2+2 2 |
| | 3+4 2 | 5+6 2 | 5+6 2 | 3+4 2 | 3+3 2 | 3+3 2 |
| | 5+6 2 | 3+4 2 | 3+4 2 | 5+6 2 | 2+2 2 | 2+2 2 |
| | 3+4 2 | 3+4 2 | 5+6 2 | 5+6 2 | 3+3 2 | 2+2 2 |
| | 5+6 2 | 5+6 2 | 5+6 2 | 3+4 2 | 2+2 2 | 3+3 2 |
| | 3+4 2 | 3+4 2 | 3+4 2 | 3+4 2 | 3+3 2 | 2+2 2 |
| | 5+6 2 | 5+6 2 | 5+6 2 | 5+6 2 | 2+2 2 | 3+3 2 |
| | 3+4 2 | 3+4 2 | 3+4 2 | 3+4 2 | 3+3 2 | 3+3 2 |
| | 5+6 2 | 5+6 2 | 5+6 2 | 5+6 2 | 2+2 2 | 2+2 2 |
| | 3+4 2 | 3+4 2 | 3+4 2 | 5+6 2 | 3+3 2 | 2+2 2 |
| | 5+6 2 | 5+6 2 | 3+4 2 | 3+4 2 | 2+2 2 | 3+3 2 |
| | 3+4 2 | 3+4 2 | 5+6 2 | 3+4 2 | 2+2 2 | 3+3 2 |
| | 5+6 2 | 5+6 2 | 3+4 2 | 3+4 2 | 3+3 2 | 2+2 2 |
| | 3+4 2 | 3+4 2 | 5+6 2 | 5+6 2 | 3+3 2 | 2+2 2 |
| | 5+6 2 | 3+4 2 | 5+6 2 | 3+4 2 | 2+2 2 | 3+3 2 |
| | 3+4 2 | 5+6 2 | 3+4 2 | 5+6 2 | 3+3 2 | 3+3 2 |
| | 3+4 2 | 5+6 2 | 3+4 2 | 5+6 2 | 2+2 2 | 3+3 2 |
| | 5+6 2 | 3+4 2 | 5+6 2 | 5+6 2 | 3+3 2 | 2+2 2 |
| | 5+6 2 | 5+6 2 | 5+6 2 | 3+4 2 | 2+2 2 | 2+2 2 |
| | 5+6 2 | 3+4 2 | 3+4 2 | 3+4 2 | 3+3 2 | 2+2 2 |
| | 3+4 2 | 3+4 2 | 3+4 2 | 5+6 2 | 2+2 2 | 3+3 2 |
| | 5+6 2 | 3+4 2 | 3+4 2 | 3+4 2 | 3+3 2 | 2+2 2 |
| | 3+4 2 | 5+6 2 | 5+6 2 | 5+6 2 | 2+2 2 | 3+3 2 |
| | 3+4 2 | 5+6 2 | 5+6 2 | 3+4 2 | 2+2 2 | 3+3 2 |
| | 5+6 2 | 3+4 2 | 3+4 2 | 5+6 2 | 3+3 2 | 2+2 2 |
| | 3+4 2 | 5+6 2 | 3+4 2 | 5+6 2 | 2+2 2 | 2+2 2 |
| | 5+6 2 | 3+4 2 | 5+6 2 | 3+4 2 | 3+3 2 | 3+3 2 |
| | 3+4 2 | 3+4 2 | 5+6 2 | 5+6 2 | 2+2 2 | 2+2 2 |
| | 5+6 2 | 5+6 2 | 3+4 2 | 3+4 2 | 3+3 2 | 3+3 2 |
| | 5+6 2 | 5+6 2 | 3+4 2 | 5+6 2 | 2+2 2 | 3+3 2 |
| | 5+6 2 | 3+4 2 | 3+4 2 | 5+6 2 | 3+3 2 | 2+2 2 |
| | 5+6 2 | 5+6 2 | 5+6 2 | 3+4 2 | 2+2 2 | 2+2 2 |
| | 3+4 2 | 3+4 2 | 5+6 2 | 3+4 2 | 3+3 2 | 3+3 2 |
| | 5+6 2 | 5+6 2 | 5+6 2 | 5+6 2 | 2+2 2 | 2+2 2 |
| | 3+4 2 | 5+6 2 | 3+4 2 | 3+4 2 | 3+3 2 | 2+2 2 |
| | 3+4 2 | 3+4 2 | 3+4 2 | 5+6 2 | 2+2 2 | 3+3 2 |
| | 3+4 2 | 3+4 2 | 3+4 2 | 5+6 2 | 3+3 2 | 3+3 2 |
| | 5+6 2 | 3+4 2 | 5+6 2 | 5+6 2 | 3+3 2 | 2+2 2 |
| | 5+6 2 | 5+6 2 | 3+4 2 | 5+6 2 | 2+2 2 | 3+3 2 |
| | 5+6 2 | 5+6 2 | 3+4 2 | 3+4 2 | 3+3 2 | 2+2 2 |
| | 3+4 2 | 3+4 2 | 5+6 2 | 3+4 2 | 2+2 2 | 3+3 2 |
| | 3+4 2 | 5+6 2 | 5+6 2 | 3+4 2 | 3+3 2 | 2+2 2 |
| | 5+6 2 | 5+6 2 | 3+4 2 | 5+6 2 | 2+2 2 | 3+3 2 |
| | 3+4 2 | 3+4 2 | 5+6 2 | 3+4 2 | 3+3 2 | 2+2 2 |
| | 5+6 2 | 5+6 2 | 3+4 2 | 5+6 2 | 2+2 2 | 3+3 2 |
| | 5+6 2 | 3+4 2 | 3+4 2 | 5+6 2 | 3+3 2 | 2+2 2 |
| | 3+4 2 | 3+4 2 | 5+6 2 | 3+4 2 | 2+2 2 | 3+3 2 |

Continued on next column

**TABLE XXIV– continued**

| TQC | ICSD | NoE | SF | | | |
|---|---|---|---|---|---|---|
| | 5+6 2 | 5+6 2 | 3+4 2 | 5+6 2 | 3+3 2 | 3+3 2 |
| | 3+4 2 | 3+4 2 | 3+4 2 | 5+6 2 | 2+2 2 | 2+2 2 |
| | 5+6 2 | 3+4 2 | 5+6 2 | 3+4 2 | 3+3 2 | 3+3 2 |
| | 3+4 2 | 5+6 2 | 5+6 2 | 5+6 2 | 2+2 2 | 3+3 2 |
| | 5+6 2 | 3+4 2 | 3+4 2 | 3+4 2 | 3+3 2 | 2+2 2 |
| | 5+6 2 | 5+6 2 | 5+6 2 | 3+4 2 | 2+2 2 | 3+3 2 |
| | 5+6 2 | 3+4 2 | 3+4 2 | 5+6 2 | 3+3 2 | 2+2 2 |
| | 3+4 2 | 5+6 2 | 5+6 2 | 3+4 2 | 2+2 2 | 3+3 2 |
| | 5+6 2 | 3+4 2 | 5+6 2 | 3+4 2 | 3+3 2 | 2+2 2 |
| | 5+6 2 | 5+6 2 | 3+4 2 | 5+6 2 | 2+2 2 | 2+2 2 |
| | 3+4 2 | 3+4 2 | 3+4 2 | 5+6 2 | 2+2 2 | 3+3 2 |
| | 3+4 2 | 3+4 2 | 3+4 2 | 3+4 2 | 3+3 2 | 2+2 2 |
| | 3+4 2 | 5+6 2 | 5+6 2 | 5+6 2 | 2+2 2 | 3+3 2 |
| | 3+4 2 | 3+4 2 | 3+4 2 | 3+4 2 | 3+3 2 | 2+2 2 |
| | 5+6 2 | 5+6 2 | 3+4 2 | 5+6 2 | 2+2 2 | 3+3 2 |
| 12.2.155 | 164223 | 186 | Bi1Fe1O3 | | | |
| | -A | -GM | -M | -Y | -L | -V |
| | 5+6 2 | 3+4 2 | 3+4 2 | 5+6 2 | 2+2 2 | 2+2 2 |
| | 3+4 2 | 5+6 2 | 5+6 2 | 3+4 2 | 3+3 2 | 3+3 2 |
| | 5+6 2 | 3+4 2 | 3+4 2 | 5+6 2 | 3+3 2 | 2+2 2 |
| | 5+6 2 | 3+4 2 | 3+4 2 | 5+6 2 | 2+2 2 | 3+3 2 |
| | 3+4 2 | 3+4 2 | 5+6 2 | 3+4 2 | 3+3 2 | 2+2 2 |
| | 3+4 2 | 5+6 2 | 5+6 2 | 5+6 2 | 2+2 2 | 3+3 2 |
| | 5+6 2 | 5+6 2 | 5+6 2 | 3+4 2 | 3+3 2 | 2+2 2 |
| | 3+4 2 | 5+6 2 | 3+4 2 | 3+4 2 | 2+2 2 | 3+3 2 |
| | 5+6 2 | 3+4 2 | 3+4 2 | 5+6 2 | 3+3 2 | 2+2 2 |
| | 5+6 2 | 3+4 2 | 3+4 2 | 5+6 2 | 2+2 2 | 3+3 2 |
| | 3+4 2 | 3+4 2 | 5+6 2 | 3+4 2 | 3+3 2 | 3+3 2 |
| | 3+4 2 | 5+6 2 | 3+4 2 | 3+4 2 | 2+2 2 | 2+2 2 |
| | 5+6 2 | 5+6 2 | 5+6 2 | 5+6 2 | 2+2 2 | 2+2 2 |
| | 5+6 2 | 5+6 2 | 5+6 2 | 5+6 2 | 3+3 2 | 3+3 2 |
| | 3+4 2 | 3+4 2 | 5+6 2 | 3+4 2 | 2+2 2 | 2+2 2 |
| | 3+4 2 | 5+6 2 | 3+4 2 | 3+4 2 | 3+3 2 | 3+3 2 |
| | 3+4 2 | 5+6 2 | 3+4 2 | 5+6 2 | 3+3 2 | 3+3 2 |
| | 5+6 2 | 3+4 2 | 5+6 2 | 3+4 2 | 2+2 2 | 2+2 2 |
| | 5+6 2 | 3+4 2 | 3+4 2 | 5+6 2 | 2+2 2 | 3+3 2 |
| | 5+6 2 | 3+4 2 | 3+4 2 | 5+6 2 | 3+3 2 | 2+2 2 |
| | 5+6 2 | 3+4 2 | 3+4 2 | 5+6 2 | 2+2 2 | 2+2 2 |
| | 5+6 2 | 3+4 2 | 3+4 2 | 5+6 2 | 3+3 2 | 3+3 2 |
| | 3+4 2 | 5+6 2 | 5+6 2 | 3+4 2 | 2+2 2 | 3+3 2 |
| | 3+4 2 | 5+6 2 | 5+6 2 | 3+4 2 | 3+3 2 | 2+2 2 |
| | 5+6 2 | 3+4 2 | 3+4 2 | 5+6 2 | 2+2 2 | 2+2 2 |
| | 3+4 2 | 3+4 2 | 5+6 2 | 3+4 2 | 3+3 2 | 3+3 2 |
| | 5+6 2 | 5+6 2 | 3+4 2 | 5+6 2 | 2+2 2 | 3+3 2 |
| | 3+4 2 | 5+6 2 | 5+6 2 | 3+4 2 | 3+3 2 | 2+2 2 |
| | 5+6 2 | 3+4 2 | 3+4 2 | 5+6 2 | 2+2 2 | 3+3 2 |
| | 3+4 2 | 5+6 2 | 5+6 2 | 3+4 2 | 3+3 2 | 2+2 2 |
| | 5+6 2 | 5+6 2 | 3+4 2 | 3+4 2 | 2+2 2 | 3+3 2 |
| | 3+4 2 | 3+4 2 | 5+6 2 | 5+6 2 | 3+3 2 | 2+2 2 |
| | 3+4 2 | 5+6 2 | 5+6 2 | 3+4 2 | 3+3 2 | 3+3 2 |
| | 5+6 2 | 3+4 2 | 3+4 2 | 5+6 2 | 2+2 2 | 2+2 2 |
| | 3+4 2 | 3+4 2 | 5+6 2 | 3+4 2 | 3+3 2 | 2+2 2 |
| | 5+6 2 | 5+6 2 | 5+6 2 | 5+6 2 | 2+2 2 | 3+3 2 |
| | 3+4 2 | 3+4 2 | 3+4 2 | 3+4 2 | 3+3 2 | 2+2 2 |
| | 5+6 2 | 5+6 2 | 5+6 2 | 5+6 2 | 2+2 2 | 3+3 2 |
| | 3+4 2 | 3+4 2 | 3+4 2 | 3+4 2 | 3+3 2 | 3+3 2 |
| | 5+6 2 | 5+6 2 | 5+6 2 | 5+6 2 | 2+2 2 | 2+2 2 |
| | 3+4 2 | 3+4 2 | 3+4 2 | 5+6 2 | 3+3 2 | 2+2 2 |
| | 5+6 2 | 5+6 2 | 3+4 2 | 3+4 2 | 2+2 2 | 3+3 2 |
| | 3+4 2 | 5+6 2 | 5+6 2 | 3+4 2 | 2+2 2 | 3+3 2 |
| | 5+6 2 | 3+4 2 | 3+4 2 | 3+4 2 | 3+3 2 | 2+2 2 |

Continued on next column

**TABLE XXIV– continued**

| TQC | ICSD | NoE | SF | | | |
|---|---|---|---|---|---|---|
| | 3+4 2 | 3+4 2 | 5+6 2 | 5+6 2 | 3+3 2 | 2+2 2 |
| | 5+6 2 | 3+4 2 | 5+6 2 | 3+4 2 | 2+2 2 | 3+3 2 |
| | 3+4 2 | 5+6 2 | 3+4 2 | 5+6 2 | 3+3 2 | 3+3 2 |
| | 3+4 2 | 5+6 2 | 3+4 2 | 5+6 2 | 2+2 2 | 3+3 2 |
| | 5+6 2 | 3+4 2 | 5+6 2 | 5+6 2 | 3+3 2 | 2+2 2 |
| | 5+6 2 | 5+6 2 | 5+6 2 | 3+4 2 | 2+2 2 | 2+2 2 |
| | 5+6 2 | 3+4 2 | 3+4 2 | 3+4 2 | 3+3 2 | 2+2 2 |
| | 3+4 2 | 3+4 2 | 3+4 2 | 5+6 2 | 2+2 2 | 3+3 2 |
| | 5+6 2 | 3+4 2 | 3+4 2 | 3+4 2 | 3+3 2 | 2+2 2 |
| | 3+4 2 | 5+6 2 | 5+6 2 | 5+6 2 | 2+2 2 | 3+3 2 |
| | 3+4 2 | 5+6 2 | 5+6 2 | 3+4 2 | 2+2 2 | 3+3 2 |
| | 5+6 2 | 3+4 2 | 3+4 2 | 5+6 2 | 3+3 2 | 2+2 2 |
| | 3+4 2 | 5+6 2 | 3+4 2 | 5+6 2 | 2+2 2 | 2+2 2 |
| | 5+6 2 | 3+4 2 | 5+6 2 | 3+4 2 | 3+3 2 | 3+3 2 |
| | 3+4 2 | 3+4 2 | 5+6 2 | 5+6 2 | 2+2 2 | 2+2 2 |
| | 5+6 2 | 5+6 2 | 3+4 2 | 3+4 2 | 3+3 2 | 3+3 2 |
| | 5+6 2 | 5+6 2 | 3+4 2 | 5+6 2 | 2+2 2 | 3+3 2 |
| | 5+6 2 | 3+4 2 | 3+4 2 | 5+6 2 | 3+3 2 | 2+2 2 |
| | 5+6 2 | 5+6 2 | 5+6 2 | 3+4 2 | 2+2 2 | 2+2 2 |
| | 3+4 2 | 3+4 2 | 5+6 2 | 3+4 2 | 3+3 2 | 3+3 2 |
| | 5+6 2 | 5+6 2 | 5+6 2 | 5+6 2 | 2+2 2 | 2+2 2 |
| | 3+4 2 | 5+6 2 | 3+4 2 | 3+4 2 | 3+3 2 | 2+2 2 |
| | 3+4 2 | 3+4 2 | 3+4 2 | 5+6 2 | 2+2 2 | 3+3 2 |
| | 5+6 2 | 3+4 2 | 3+4 2 | 5+6 2 | 3+3 2 | 3+3 2 |
| | 3+4 2 | 3+4 2 | 5+6 2 | 5+6 2 | 3+3 2 | 2+2 2 |
| | 5+6 2 | 5+6 2 | 3+4 2 | 5+6 2 | 2+2 2 | 3+3 2 |
| | 5+6 2 | 5+6 2 | 3+4 2 | 3+4 2 | 3+3 2 | 3+3 2 |
| | 3+4 2 | 3+4 2 | 5+6 2 | 3+4 2 | 2+2 2 | 2+2 2 |
| | 3+4 2 | 5+6 2 | 5+6 2 | 3+4 2 | 3+3 2 | 2+2 2 |
| | 5+6 2 | 5+6 2 | 3+4 2 | 5+6 2 | 2+2 2 | 3+3 2 |
| | 3+4 2 | 3+4 2 | 5+6 2 | 3+4 2 | 3+3 2 | 2+2 2 |
| | 5+6 2 | 5+6 2 | 3+4 2 | 5+6 2 | 2+2 2 | 3+3 2 |
| | 5+6 2 | 3+4 2 | 3+4 2 | 5+6 2 | 3+3 2 | 2+2 2 |
| | 3+4 2 | 3+4 2 | 5+6 2 | 3+4 2 | 2+2 2 | 3+3 2 |
| | 5+6 2 | 5+6 2 | 3+4 2 | 5+6 2 | 3+3 2 | 3+3 2 |
| | 3+4 2 | 3+4 2 | 3+4 2 | 5+6 2 | 2+2 2 | 2+2 2 |
| | 5+6 2 | 3+4 2 | 5+6 2 | 3+4 2 | 3+3 2 | 3+3 2 |
| | 3+4 2 | 5+6 2 | 5+6 2 | 5+6 2 | 2+2 2 | 3+3 2 |
| | 5+6 2 | 3+4 2 | 3+4 2 | 3+4 2 | 3+3 2 | 2+2 2 |
| | 5+6 2 | 5+6 2 | 5+6 2 | 3+4 2 | 2+2 2 | 3+3 2 |
| | 5+6 2 | 3+4 2 | 3+4 2 | 5+6 2 | 3+3 2 | 2+2 2 |
| | 3+4 2 | 5+6 2 | 5+6 2 | 3+4 2 | 2+2 2 | 3+3 2 |
| | 5+6 2 | 3+4 2 | 5+6 2 | 3+4 2 | 3+3 2 | 2+2 2 |
| | 5+6 2 | 5+6 2 | 3+4 2 | 5+6 2 | 2+2 2 | 2+2 2 |
| | 3+4 2 | 3+4 2 | 3+4 2 | 5+6 2 | 2+2 2 | 3+3 2 |
| | 3+4 2 | 3+4 2 | 3+4 2 | 3+4 2 | 3+3 2 | 2+2 2 |
| | 3+4 2 | 5+6 2 | 5+6 2 | 5+6 2 | 2+2 2 | 3+3 2 |
| | 3+4 2 | 3+4 2 | 3+4 2 | 3+4 2 | 3+3 2 | 2+2 2 |
| | 5+6 2 | 5+6 2 | 3+4 2 | 5+6 2 | 2+2 2 | 3+3 2 |
| 12.2.156 | 246145 | 108 | Bi16I4 | | | |
| | -A | -GM | -M | -Y | -L | -V |
| | 5+6 2 | 3+4 2 | 3+4 2 | 5+6 2 | 2+2 2 | 2+2 2 |
| | 3+4 2 | 5+6 2 | 5+6 2 | 3+4 2 | 3+3 2 | 3+3 2 |
| | 3+4 2 | 3+4 2 | 5+6 2 | 5+6 2 | 2+2 2 | 2+2 2 |
| | 5+6 2 | 5+6 2 | 3+4 2 | 3+4 2 | 3+3 2 | 3+3 2 |
| | 5+6 2 | 3+4 2 | 3+4 2 | 3+4 2 | 2+2 2 | 2+2 2 |
| | 3+4 2 | 5+6 2 | 5+6 2 | 5+6 2 | 3+3 2 | 3+3 2 |
| | 5+6 2 | 3+4 2 | 5+6 2 | 3+4 2 | 2+2 2 | 3+3 2 |
| | 3+4 2 | 5+6 2 | 3+4 2 | 5+6 2 | 3+3 2 | 2+2 2 |
| | 3+4 2 | 3+4 2 | 5+6 2 | 3+4 2 | 2+2 2 | 2+2 2 |
| | 5+6 2 | 5+6 2 | 3+4 2 | 5+6 2 | 2+2 2 | 3+3 2 |

Continued on next column

**TABLE XXIV– continued**

| TQC | ICSD | NoE | SF | | | |
|---|---|---|---|---|---|---|
| | 5+6 2 | 3+4 2 | 5+6 2 | 3+4 2 | 3+3 2 | 3+3 2 |
| | 3+4 2 | 5+6 2 | 3+4 2 | 5+6 2 | 2+2 2 | 2+2 2 |
| | 5+6 2 | 5+6 2 | 5+6 2 | 5+6 2 | 3+3 2 | 3+3 2 |
| | 3+4 2 | 3+4 2 | 3+4 2 | 3+4 2 | 2+2 2 | 2+2 2 |
| | 5+6 2 | 3+4 2 | 5+6 2 | 3+4 2 | 3+3 2 | 2+2 2 |
| | 3+4 2 | 5+6 2 | 3+4 2 | 5+6 2 | 3+3 2 | 3+3 2 |
| | 3+4 2 | 3+4 2 | 3+4 2 | 3+4 2 | 2+2 2 | 3+3 2 |
| | 5+6 2 | 5+6 2 | 5+6 2 | 5+6 2 | 2+2 2 | 2+2 2 |
| | 5+6 2 | 3+4 2 | 5+6 2 | 3+4 2 | 3+3 2 | 2+2 2 |
| | 3+4 2 | 5+6 2 | 3+4 2 | 5+6 2 | 3+3 2 | 3+3 2 |
| | 5+6 2 | 3+4 2 | 3+4 2 | 5+6 2 | 3+3 2 | 2+2 2 |
| | 3+4 2 | 5+6 2 | 5+6 2 | 3+4 2 | 2+2 2 | 3+3 2 |
| | 3+4 2 | 3+4 2 | 3+4 2 | 5+6 2 | 2+2 2 | 3+3 2 |
| | 5+6 2 | 3+4 2 | 5+6 2 | 3+4 2 | 3+3 2 | 2+2 2 |
| | 5+6 2 | 5+6 2 | 5+6 2 | 5+6 2 | 3+3 2 | 2+2 2 |
| | 3+4 2 | 3+4 2 | 3+4 2 | 3+4 2 | 3+3 2 | 3+3 2 |
| | 3+4 2 | 5+6 2 | 3+4 2 | 3+4 2 | 2+2 2 | 3+3 2 |
| | 5+6 2 | 5+6 2 | 5+6 2 | 5+6 2 | 2+2 2 | 2+2 2 |
| | 3+4 2 | 3+4 2 | 3+4 2 | 5+6 2 | 3+3 2 | 3+3 2 |
| | 5+6 2 | 5+6 2 | 5+6 2 | 3+4 2 | 3+3 2 | 2+2 2 |
| | 5+6 2 | 3+4 2 | 3+4 2 | 5+6 2 | 2+2 2 | 3+3 2 |
| | 3+4 2 | 5+6 2 | 5+6 2 | 3+4 2 | 2+2 2 | 2+2 2 |
| | 5+6 2 | 3+4 2 | 3+4 2 | 5+6 2 | 2+2 2 | 3+3 2 |
| | 5+6 2 | 3+4 2 | 5+6 2 | 5+6 2 | 3+3 2 | 2+2 2 |
| | 3+4 2 | 5+6 2 | 5+6 2 | 3+4 2 | 2+2 2 | 3+3 2 |
| | 3+4 2 | 3+4 2 | 3+4 2 | 3+4 2 | 3+3 2 | 2+2 2 |
| | 3+4 2 | 5+6 2 | 3+4 2 | 5+6 2 | 2+2 2 | 2+2 2 |
| | 5+6 2 | 5+6 2 | 5+6 2 | 3+4 2 | 3+3 2 | 3+3 2 |
| | 3+4 2 | 3+4 2 | 3+4 2 | 5+6 2 | 3+3 2 | 2+2 2 |
| | 5+6 2 | 5+6 2 | 5+6 2 | 3+4 2 | 2+2 2 | 3+3 2 |
| | 5+6 2 | 5+6 2 | 5+6 2 | 3+4 2 | 2+2 2 | 2+2 2 |
| | 5+6 2 | 3+4 2 | 3+4 2 | 5+6 2 | 3+3 2 | 3+3 2 |
| | 3+4 2 | 5+6 2 | 3+4 2 | 3+4 2 | 2+2 2 | 2+2 2 |
| | 3+4 2 | 5+6 2 | 5+6 2 | 5+6 2 | 2+2 2 | 3+3 2 |
| | 5+6 2 | 3+4 2 | 5+6 2 | 3+4 2 | 3+3 2 | 3+3 2 |
| | 3+4 2 | 3+4 2 | 3+4 2 | 5+6 2 | 3+3 2 | 2+2 2 |
| | 5+6 2 | 5+6 2 | 5+6 2 | 3+4 2 | 2+2 2 | 2+2 2 |
| | 3+4 2 | 3+4 2 | 3+4 2 | 5+6 2 | 3+3 2 | 3+3 2 |
| | 5+6 2 | 5+6 2 | 3+4 2 | 3+4 2 | 2+2 2 | 2+2 2 |
| | 3+4 2 | 3+4 2 | 5+6 2 | 5+6 2 | 2+2 2 | 3+3 2 |
| | 3+4 2 | 3+4 2 | 5+6 2 | 3+4 2 | 3+3 2 | 2+2 2 |
| | 5+6 2 | 3+4 2 | 3+4 2 | 3+4 2 | 2+2 2 | 3+3 2 |
| | 3+4 2 | 5+6 2 | 5+6 2 | 5+6 2 | 3+3 2 | 2+2 2 |
| | 3+4 2 | 3+4 2 | 3+4 2 | 3+4 2 | 2+2 2 | 3+3 2 |
| 12.2.157 | 423458 | 106 | Ba5Cd2O2Sb4 | | | |
| | -A | -GM | -M | -Y | -L | -V |
| | 3+4 2 | 3+4 2 | 3+4 2 | 3+4 2 | 3+3 2 | 3+3 2 |
| | 5+6 2 | 3+4 2 | 3+4 2 | 5+6 2 | 2+2 2 | 2+2 2 |
| | 3+4 2 | 5+6 2 | 5+6 2 | 3+4 2 | 3+3 2 | 3+3 2 |
| | 5+6 2 | 3+4 2 | 5+6 2 | 3+4 2 | 2+2 2 | 3+3 2 |
| | 3+4 2 | 5+6 2 | 3+4 2 | 5+6 2 | 3+3 2 | 2+2 2 |
| | 5+6 2 | 5+6 2 | 5+6 2 | 3+4 2 | 2+2 2 | 2+2 2 |
| | 3+4 2 | 3+4 2 | 3+4 2 | 5+6 2 | 3+3 2 | 3+3 2 |
| | 5+6 2 | 5+6 2 | 5+6 2 | 5+6 2 | 2+2 2 | 2+2 2 |
| | 5+6 2 | 3+4 2 | 3+4 2 | 5+6 2 | 2+2 2 | 2+2 2 |
| | 3+4 2 | 3+4 2 | 5+6 2 | 3+4 2 | 3+3 2 | 3+3 2 |
| | 5+6 2 | 5+6 2 | 3+4 2 | 5+6 2 | 3+3 2 | 3+3 2 |
| | 3+4 2 | 5+6 2 | 5+6 2 | 3+4 2 | 2+2 2 | 2+2 2 |
| | 5+6 2 | 5+6 2 | 5+6 2 | 5+6 2 | 2+2 2 | 2+2 2 |
| | 5+6 2 | 3+4 2 | 3+4 2 | 5+6 2 | 2+2 2 | 2+2 2 |
| | 5+6 2 | 5+6 2 | 5+6 2 | 3+4 2 | 3+3 2 | 3+3 2 |

Continued on next column

**TABLE XXIV– continued**

| TQC | ICSD | NoE | SF | | | |
|---|---|---|---|---|---|---|
| | 3+4 2 | 3+4 2 | 3+4 2 | 5+6 2 | 2+2 2 | 2+2 2 |
| | 5+6 2 | 5+6 2 | 5+6 2 | 5+6 2 | 3+3 2 | 3+3 2 |
| | 3+4 2 | 3+4 2 | 3+4 2 | 3+4 2 | 2+2 2 | 2+2 2 |
| | 5+6 2 | 5+6 2 | 5+6 2 | 5+6 2 | 3+3 2 | 3+3 2 |
| | 3+4 2 | 3+4 2 | 5+6 2 | 3+4 2 | 2+2 2 | 2+2 2 |
| | 5+6 2 | 5+6 2 | 3+4 2 | 3+4 2 | 3+3 2 | 3+3 2 |
| | 3+4 2 | 5+6 2 | 5+6 2 | 5+6 2 | 2+2 2 | 2+2 2 |
| | 5+6 2 | 3+4 2 | 5+6 2 | 3+4 2 | 3+3 2 | 2+2 2 |
| | 5+6 2 | 5+6 2 | 3+4 2 | 3+4 2 | 2+2 2 | 3+3 2 |
| | 3+4 2 | 3+4 2 | 5+6 2 | 5+6 2 | 3+3 2 | 2+2 2 |
| | 3+4 2 | 5+6 2 | 3+4 2 | 5+6 2 | 2+2 2 | 3+3 2 |
| | 3+4 2 | 3+4 2 | 5+6 2 | 3+4 2 | 3+3 2 | 2+2 2 |
| | 5+6 2 | 3+4 2 | 3+4 2 | 5+6 2 | 2+2 2 | 3+3 2 |
| | 5+6 2 | 5+6 2 | 5+6 2 | 3+4 2 | 3+3 2 | 2+2 2 |
| | 3+4 2 | 3+4 2 | 3+4 2 | 5+6 2 | 2+2 2 | 3+3 2 |
| | 3+4 2 | 5+6 2 | 5+6 2 | 5+6 2 | 3+3 2 | 2+2 2 |
| | 3+4 2 | 3+4 2 | 5+6 2 | 3+4 2 | 2+2 2 | 3+3 2 |
| | 5+6 2 | 5+6 2 | 3+4 2 | 5+6 2 | 3+3 2 | 2+2 2 |
| | 5+6 2 | 3+4 2 | 5+6 2 | 3+4 2 | 3+3 2 | 3+3 2 |
| | 5+6 2 | 5+6 2 | 3+4 2 | 3+4 2 | 2+2 2 | 2+2 2 |
| | 3+4 2 | 5+6 2 | 3+4 2 | 5+6 2 | 2+2 2 | 3+3 2 |
| | 3+4 2 | 5+6 2 | 3+4 2 | 3+4 2 | 3+3 2 | 3+3 2 |
| | 5+6 2 | 3+4 2 | 5+6 2 | 5+6 2 | 2+2 2 | 2+2 2 |
| | 3+4 2 | 3+4 2 | 5+6 2 | 3+4 2 | 3+3 2 | 3+3 2 |
| | 5+6 2 | 3+4 2 | 3+4 2 | 3+4 2 | 3+3 2 | 2+2 2 |
| | 3+4 2 | 5+6 2 | 5+6 2 | 5+6 2 | 2+2 2 | 3+3 2 |
| | 3+4 2 | 3+4 2 | 3+4 2 | 3+4 2 | 3+3 2 | 2+2 2 |
| | 3+4 2 | 3+4 2 | 5+6 2 | 5+6 2 | 2+2 2 | 3+3 2 |
| | 3+4 2 | 5+6 2 | 3+4 2 | 5+6 2 | 2+2 2 | 2+2 2 |
| | 5+6 2 | 5+6 2 | 5+6 2 | 5+6 2 | 3+3 2 | 3+3 2 |
| | 5+6 2 | 3+4 2 | 5+6 2 | 3+4 2 | 2+2 2 | 2+2 2 |
| | 3+4 2 | 3+4 2 | 3+4 2 | 5+6 2 | 3+3 2 | 3+3 2 |
| | 5+6 2 | 5+6 2 | 3+4 2 | 3+4 2 | 2+2 2 | 3+3 2 |
| | 3+4 2 | 5+6 2 | 5+6 2 | 5+6 2 | 3+3 2 | 2+2 2 |
| | 5+6 2 | 3+4 2 | 3+4 2 | 3+4 2 | 2+2 2 | 2+2 2 |
| | 5+6 2 | 5+6 2 | 3+4 2 | 5+6 2 | 2+2 2 | 3+3 2 |
| | 5+6 2 | 5+6 2 | 5+6 2 | 3+4 2 | 3+3 2 | 2+2 2 |
| | 5+6 2 | 3+4 2 | 3+4 2 | 3+4 2 | 3+3 2 | 2+2 2 |
| 12.2.158 | 240016 | 66 | Ba3Li4Sn8 | | | |
| | -A | -GM | -M | -Y | -L | -V |
| | 5+6 2 | 3+4 2 | 3+4 2 | 5+6 2 | 2+2 2 | 2+2 2 |
| | 3+4 2 | 5+6 2 | 5+6 2 | 3+4 2 | 3+3 2 | 3+3 2 |
| | 3+4 2 | 3+4 2 | 3+4 2 | 3+4 2 | 3+3 2 | 3+3 2 |
| | 5+6 2 | 3+4 2 | 3+4 2 | 5+6 2 | 2+2 2 | 2+2 2 |
| | 3+4 2 | 5+6 2 | 5+6 2 | 3+4 2 | 3+3 2 | 3+3 2 |
| | 5+6 2 | 5+6 2 | 5+6 2 | 5+6 2 | 2+2 2 | 2+2 2 |
| | 5+6 2 | 3+4 2 | 3+4 2 | 5+6 2 | 2+2 2 | 2+2 2 |
| | 3+4 2 | 3+4 2 | 5+6 2 | 3+4 2 | 3+3 2 | 3+3 2 |
| | 5+6 2 | 5+6 2 | 3+4 2 | 5+6 2 | 2+2 2 | 2+2 2 |
| | 3+4 2 | 5+6 2 | 5+6 2 | 3+4 2 | 3+3 2 | 3+3 2 |
| | 5+6 2 | 5+6 2 | 5+6 2 | 5+6 2 | 2+2 2 | 2+2 2 |
| | 5+6 2 | 5+6 2 | 5+6 2 | 5+6 2 | 2+2 2 | 2+2 2 |
| | 3+4 2 | 3+4 2 | 3+4 2 | 3+4 2 | 2+2 2 | 2+2 2 |
| | 5+6 2 | 5+6 2 | 5+6 2 | 5+6 2 | 3+3 2 | 3+3 2 |
| | 3+4 2 | 3+4 2 | 3+4 2 | 3+4 2 | 3+3 2 | 3+3 2 |
| | 5+6 2 | 5+6 2 | 5+6 2 | 5+6 2 | 2+2 2 | 2+2 2 |
| | 3+4 2 | 3+4 2 | 5+6 2 | 3+4 2 | 2+2 2 | 2+2 2 |
| | 3+4 2 | 3+4 2 | 5+6 2 | 5+6 2 | 3+3 2 | 3+3 2 |
| | 5+6 2 | 5+6 2 | 3+4 2 | 3+4 2 | 3+3 2 | 2+2 2 |
| | 5+6 2 | 5+6 2 | 3+4 2 | 5+6 2 | 2+2 2 | 3+3 2 |
| | 3+4 2 | 3+4 2 | 3+4 2 | 3+4 2 | 3+3 2 | 3+3 2 |

Continued on next column

**TABLE XXIV– continued**

| TQC | ICSD | NoE | SF | | | |
|---|---|---|---|---|---|---|
| | 5+6 2 | 5+6 2 | 5+6 2 | 5+6 2 | 2+2 2 | 2+2 2 |
| | 3+4 2 | 3+4 2 | 3+4 2 | 3+4 2 | 3+3 2 | 3+3 2 |
| | 5+6 2 | 3+4 2 | 5+6 2 | 5+6 2 | 2+2 2 | 2+2 2 |
| | 3+4 2 | 5+6 2 | 5+6 2 | 5+6 2 | 2+2 2 | 3+3 2 |
| | 5+6 2 | 3+4 2 | 5+6 2 | 3+4 2 | 3+3 2 | 2+2 2 |
| | 3+4 2 | 5+6 2 | 3+4 2 | 3+4 2 | 3+3 2 | 3+3 2 |
| | 3+4 2 | 3+4 2 | 3+4 2 | 5+6 2 | 2+2 2 | 2+2 2 |
| | 5+6 2 | 3+4 2 | 5+6 2 | 3+4 2 | 3+3 2 | 3+3 2 |
| | 3+4 2 | 5+6 2 | 5+6 2 | 5+6 2 | 2+2 2 | 2+2 2 |
| | 5+6 2 | 5+6 2 | 3+4 2 | 3+4 2 | 3+3 2 | 3+3 2 |
| | 5+6 2 | 3+4 2 | 3+4 2 | 5+6 2 | 2+2 2 | 2+2 2 |
| | 3+4 2 | 3+4 2 | 3+4 2 | 3+4 2 | 2+2 2 | 2+2 2 |
| 12.2.159 | 62390 | 144 | Ba3Cu2O10Pt1Y2 | | | |
| | -A | -GM | -M | -Y | -L | -V |
| | 3+4 2 | 3+4 2 | 5+6 2 | 5+6 2 | 2+2 2 | 2+2 2 |
| | 5+6 2 | 5+6 2 | 3+4 2 | 3+4 2 | 3+3 2 | 3+3 2 |
| | 5+6 2 | 3+4 2 | 3+4 2 | 3+4 2 | 2+2 2 | 3+3 2 |
| | 3+4 2 | 5+6 2 | 5+6 2 | 5+6 2 | 3+3 2 | 2+2 2 |
| | 3+4 2 | 3+4 2 | 3+4 2 | 3+4 2 | 3+3 2 | 3+3 2 |
| | 3+4 2 | 3+4 2 | 5+6 2 | 5+6 2 | 3+3 2 | 2+2 2 |
| | 5+6 2 | 5+6 2 | 3+4 2 | 3+4 2 | 2+2 2 | 3+3 2 |
| | 3+4 2 | 3+4 2 | 5+6 2 | 5+6 2 | 3+3 2 | 2+2 2 |
| | 5+6 2 | 5+6 2 | 3+4 2 | 3+4 2 | 2+2 2 | 3+3 2 |
| | 3+4 2 | 3+4 2 | 5+6 2 | 5+6 2 | 3+3 2 | 2+2 2 |
| | 5+6 2 | 5+6 2 | 3+4 2 | 3+4 2 | 2+2 2 | 3+3 2 |
| | 5+6 2 | 3+4 2 | 5+6 2 | 3+4 2 | 3+3 2 | 2+2 2 |
| | 3+4 2 | 5+6 2 | 3+4 2 | 5+6 2 | 2+2 2 | 2+2 2 |
| | 5+6 2 | 3+4 2 | 5+6 2 | 3+4 2 | 3+3 2 | 3+3 2 |
| | 3+4 2 | 5+6 2 | 5+6 2 | 5+6 2 | 2+2 2 | 3+3 2 |
| | 5+6 2 | 3+4 2 | 3+4 2 | 3+4 2 | 3+3 2 | 2+2 2 |
| | 3+4 2 | 5+6 2 | 3+4 2 | 5+6 2 | 2+2 2 | 3+3 2 |
| | 5+6 2 | 3+4 2 | 5+6 2 | 5+6 2 | 2+2 2 | 2+2 2 |
| | 3+4 2 | 5+6 2 | 3+4 2 | 3+4 2 | 3+3 2 | 3+3 2 |
| | 5+6 2 | 5+6 2 | 3+4 2 | 3+4 2 | 2+2 2 | 2+2 2 |
| | 3+4 2 | 3+4 2 | 5+6 2 | 5+6 2 | 3+3 2 | 3+3 2 |
| | 5+6 2 | 3+4 2 | 5+6 2 | 5+6 2 | 2+2 2 | 3+3 2 |
| | 3+4 2 | 5+6 2 | 3+4 2 | 3+4 2 | 3+3 2 | 2+2 2 |
| | 5+6 2 | 5+6 2 | 5+6 2 | 5+6 2 | 2+2 2 | 2+2 2 |
| | 5+6 2 | 3+4 2 | 3+4 2 | 5+6 2 | 2+2 2 | 3+3 2 |
| | 5+6 2 | 3+4 2 | 5+6 2 | 3+4 2 | 3+3 2 | 2+2 2 |
| | 3+4 2 | 5+6 2 | 5+6 2 | 3+4 2 | 2+2 2 | 3+3 2 |
| | 3+4 2 | 5+6 2 | 3+4 2 | 5+6 2 | 3+3 2 | 2+2 2 |
| | 5+6 2 | 5+6 2 | 5+6 2 | 5+6 2 | 2+2 2 | 2+2 2 |
| | 5+6 2 | 5+6 2 | 5+6 2 | 5+6 2 | 2+2 2 | 2+2 2 |
| | 5+6 2 | 3+4 2 | 5+6 2 | 3+4 2 | 3+3 2 | 2+2 2 |
| | 5+6 2 | 3+4 2 | 5+6 2 | 3+4 2 | 3+3 2 | 2+2 2 |
| | 5+6 2 | 3+4 2 | 5+6 2 | 3+4 2 | 3+3 2 | 2+2 2 |
| | 5+6 2 | 3+4 2 | 5+6 2 | 3+4 2 | 3+3 2 | 2+2 2 |
| | 3+4 2 | 3+4 2 | 5+6 2 | 3+4 2 | 3+3 2 | 2+2 2 |
| | 5+6 2 | 3+4 2 | 5+6 2 | 3+4 2 | 3+3 2 | 2+2 2 |
| | 5+6 2 | 5+6 2 | 3+4 2 | 5+6 2 | 2+2 2 | 3+3 2 |
| | 3+4 2 | 3+4 2 | 3+4 2 | 5+6 2 | 2+2 2 | 3+3 2 |
| | 3+4 2 | 5+6 2 | 3+4 2 | 3+4 2 | 3+3 2 | 3+3 2 |
| | 3+4 2 | 3+4 2 | 5+6 2 | 5+6 2 | 2+2 2 | 2+2 2 |
| | 5+6 2 | 5+6 2 | 5+6 2 | 3+4 2 | 3+3 2 | 3+3 2 |
| | 3+4 2 | 5+6 2 | 3+4 2 | 5+6 2 | 2+2 2 | 2+2 2 |
| | 5+6 2 | 3+4 2 | 3+4 2 | 5+6 2 | 2+2 2 | 3+3 2 |
| | 3+4 2 | 5+6 2 | 3+4 2 | 3+4 2 | 3+3 2 | 3+3 2 |
| | 3+4 2 | 5+6 2 | 3+4 2 | 5+6 2 | 2+2 2 | 3+3 2 |
| | 5+6 2 | 5+6 2 | 5+6 2 | 5+6 2 | 3+3 2 | 3+3 2 |
| | 3+4 2 | 3+4 2 | 5+6 2 | 5+6 2 | 2+2 2 | 2+2 2 |

Continued on next column

**TABLE XXIV– continued**

| TQC | ICSD | NoE | SF | | | |
|---|---|---|---|---|---|---|
| | 3+4 2 | 5+6 2 | 3+4 2 | 3+4 2 | 3+3 2 | 3+3 2 |
| | 3+4 2 | 5+6 2 | 5+6 2 | 3+4 2 | 2+2 2 | 2+2 2 |
| | 5+6 2 | 3+4 2 | 3+4 2 | 5+6 2 | 2+2 2 | 3+3 2 |
| | 3+4 2 | 3+4 2 | 3+4 2 | 3+4 2 | 3+3 2 | 2+2 2 |
| | 5+6 2 | 3+4 2 | 3+4 2 | 5+6 2 | 2+2 2 | 3+3 2 |
| | 5+6 2 | 5+6 2 | 5+6 2 | 5+6 2 | 2+2 2 | 2+2 2 |
| | 5+6 2 | 5+6 2 | 5+6 2 | 3+4 2 | 2+2 2 | 2+2 2 |
| | 3+4 2 | 5+6 2 | 3+4 2 | 5+6 2 | 3+3 2 | 3+3 2 |
| | 3+4 2 | 3+4 2 | 3+4 2 | 5+6 2 | 3+3 2 | 3+3 2 |
| | 5+6 2 | 3+4 2 | 5+6 2 | 3+4 2 | 2+2 2 | 3+3 2 |
| | 3+4 2 | 5+6 2 | 5+6 2 | 3+4 2 | 3+3 2 | 2+2 2 |
| | 5+6 2 | 3+4 2 | 3+4 2 | 3+4 2 | 2+2 2 | 3+3 2 |
| | 3+4 2 | 5+6 2 | 5+6 2 | 5+6 2 | 3+3 2 | 2+2 2 |
| | 3+4 2 | 3+4 2 | 3+4 2 | 5+6 2 | 2+2 2 | 2+2 2 |
| | 3+4 2 | 5+6 2 | 3+4 2 | 5+6 2 | 2+2 2 | 2+2 2 |
| | 5+6 2 | 3+4 2 | 5+6 2 | 3+4 2 | 3+3 2 | 3+3 2 |
| | 5+6 2 | 5+6 2 | 3+4 2 | 3+4 2 | 3+3 2 | 3+3 2 |
| | 5+6 2 | 5+6 2 | 3+4 2 | 5+6 2 | 2+2 2 | 2+2 2 |
| | 3+4 2 | 3+4 2 | 5+6 2 | 3+4 2 | 3+3 2 | 3+3 2 |
| | 5+6 2 | 5+6 2 | 3+4 2 | 5+6 2 | 2+2 2 | 2+2 2 |
| | 3+4 2 | 3+4 2 | 5+6 2 | 5+6 2 | 3+3 2 | 2+2 2 |
| | 5+6 2 | 3+4 2 | 5+6 2 | 3+4 2 | 2+2 2 | 3+3 2 |
| | 5+6 2 | 5+6 2 | 5+6 2 | 3+4 2 | 3+3 2 | 2+2 2 |
| | 5+6 2 | 3+4 2 | 5+6 2 | 3+4 2 | 3+3 2 | 2+2 2 |
| | 3+4 2 | 3+4 2 | 3+4 2 | 5+6 2 | 2+2 2 | 3+3 2 |
| 12.2.160 | 62391 | 164 | Ba3Cu2Ho2O10Pt1 | | | |
| | -A | -GM | -M | -Y | -L | -V |
| | 3+4 2 | 3+4 2 | 5+6 2 | 5+6 2 | 2+2 2 | 2+2 2 |
| | 5+6 2 | 5+6 2 | 3+4 2 | 3+4 2 | 3+3 2 | 3+3 2 |
| | 5+6 2 | 3+4 2 | 3+4 2 | 5+6 2 | 2+2 2 | 3+3 2 |
| | 3+4 2 | 5+6 2 | 5+6 2 | 3+4 2 | 3+3 2 | 2+2 2 |
| | 3+4 2 | 3+4 2 | 3+4 2 | 3+4 2 | 3+3 2 | 3+3 2 |
| | 3+4 2 | 3+4 2 | 5+6 2 | 5+6 2 | 2+2 2 | 2+2 2 |
| | 5+6 2 | 5+6 2 | 3+4 2 | 3+4 2 | 3+3 2 | 3+3 2 |
| | 3+4 2 | 3+4 2 | 5+6 2 | 5+6 2 | 3+3 2 | 2+2 2 |
| | 5+6 2 | 5+6 2 | 3+4 2 | 3+4 2 | 2+2 2 | 3+3 2 |
| | 3+4 2 | 3+4 2 | 5+6 2 | 5+6 2 | 3+3 2 | 2+2 2 |
| | 5+6 2 | 5+6 2 | 3+4 2 | 3+4 2 | 2+2 2 | 3+3 2 |
| | 5+6 2 | 3+4 2 | 5+6 2 | 3+4 2 | 3+3 2 | 2+2 2 |
| | 3+4 2 | 5+6 2 | 3+4 2 | 5+6 2 | 2+2 2 | 2+2 2 |
| | 5+6 2 | 3+4 2 | 5+6 2 | 3+4 2 | 3+3 2 | 3+3 2 |
| | 5+6 2 | 3+4 2 | 5+6 2 | 3+4 2 | 2+2 2 | 2+2 2 |
| | 3+4 2 | 5+6 2 | 3+4 2 | 5+6 2 | 3+3 2 | 3+3 2 |
| | 3+4 2 | 5+6 2 | 3+4 2 | 5+6 2 | 2+2 2 | 3+3 2 |
| | 5+6 2 | 5+6 2 | 5+6 2 | 3+4 2 | 2+2 2 | 2+2 2 |
| | 3+4 2 | 3+4 2 | 3+4 2 | 5+6 2 | 3+3 2 | 3+3 2 |
| | 5+6 2 | 5+6 2 | 3+4 2 | 3+4 2 | 2+2 2 | 2+2 2 |
| | 3+4 2 | 3+4 2 | 5+6 2 | 5+6 2 | 3+3 2 | 3+3 2 |
| | 5+6 2 | 3+4 2 | 5+6 2 | 5+6 2 | 2+2 2 | 3+3 2 |
| | 3+4 2 | 5+6 2 | 3+4 2 | 3+4 2 | 3+3 2 | 2+2 2 |
| | 5+6 2 | 5+6 2 | 5+6 2 | 5+6 2 | 2+2 2 | 2+2 2 |
| | 5+6 2 | 3+4 2 | 3+4 2 | 5+6 2 | 2+2 2 | 3+3 2 |
| | 5+6 2 | 3+4 2 | 5+6 2 | 3+4 2 | 3+3 2 | 2+2 2 |
| | 3+4 2 | 5+6 2 | 5+6 2 | 3+4 2 | 2+2 2 | 3+3 2 |
| | 3+4 2 | 5+6 2 | 3+4 2 | 5+6 2 | 3+3 2 | 2+2 2 |
| | 5+6 2 | 5+6 2 | 5+6 2 | 5+6 2 | 2+2 2 | 2+2 2 |
| | 5+6 2 | 5+6 2 | 5+6 2 | 5+6 2 | 2+2 2 | 2+2 2 |
| | 5+6 2 | 3+4 2 | 5+6 2 | 3+4 2 | 3+3 2 | 2+2 2 |
| | 5+6 2 | 3+4 2 | 5+6 2 | 3+4 2 | 3+3 2 | 2+2 2 |
| | 5+6 2 | 3+4 2 | 5+6 2 | 3+4 2 | 3+3 2 | 2+2 2 |
| | 5+6 2 | 3+4 2 | 5+6 2 | 3+4 2 | 3+3 2 | 2+2 2 |

Continued on next column

**TABLE XXIV– continued**

| TQC | ICSD | NoE | SF | | | |
|---|---|---|---|---|---|---|
| | 3+4 2 | 3+4 2 | 5+6 2 | 3+4 2 | 3+3 2 | 2+2 2 |
| | 5+6 2 | 3+4 2 | 3+4 2 | 3+4 2 | 3+3 2 | 2+2 2 |
| | 5+6 2 | 5+6 2 | 5+6 2 | 5+6 2 | 2+2 2 | 3+3 2 |
| | 3+4 2 | 3+4 2 | 3+4 2 | 5+6 2 | 2+2 2 | 3+3 2 |
| | 3+4 2 | 5+6 2 | 3+4 2 | 3+4 2 | 3+3 2 | 3+3 2 |
| | 5+6 2 | 3+4 2 | 5+6 2 | 5+6 2 | 3+3 2 | 2+2 2 |
| | 3+4 2 | 5+6 2 | 3+4 2 | 3+4 2 | 2+2 2 | 3+3 2 |
| | 3+4 2 | 5+6 2 | 5+6 2 | 5+6 2 | 2+2 2 | 2+2 2 |
| | 5+6 2 | 3+4 2 | 3+4 2 | 5+6 2 | 2+2 2 | 3+3 2 |
| | 3+4 2 | 5+6 2 | 3+4 2 | 3+4 2 | 3+3 2 | 3+3 2 |
| | 3+4 2 | 5+6 2 | 3+4 2 | 5+6 2 | 2+2 2 | 3+3 2 |
| | 5+6 2 | 5+6 2 | 5+6 2 | 5+6 2 | 3+3 2 | 3+3 2 |
| | 3+4 2 | 3+4 2 | 5+6 2 | 5+6 2 | 2+2 2 | 2+2 2 |
| | 3+4 2 | 5+6 2 | 3+4 2 | 3+4 2 | 3+3 2 | 3+3 2 |
| | 3+4 2 | 5+6 2 | 3+4 2 | 3+4 2 | 2+2 2 | 2+2 2 |
| | 5+6 2 | 3+4 2 | 5+6 2 | 3+4 2 | 2+2 2 | 3+3 2 |
| | 3+4 2 | 3+4 2 | 3+4 2 | 5+6 2 | 3+3 2 | 2+2 2 |
| | 5+6 2 | 3+4 2 | 3+4 2 | 5+6 2 | 2+2 2 | 3+3 2 |
| | 5+6 2 | 5+6 2 | 5+6 2 | 5+6 2 | 2+2 2 | 2+2 2 |
| | 5+6 2 | 5+6 2 | 5+6 2 | 3+4 2 | 2+2 2 | 2+2 2 |
| | 3+4 2 | 5+6 2 | 3+4 2 | 5+6 2 | 3+3 2 | 3+3 2 |
| | 3+4 2 | 3+4 2 | 3+4 2 | 5+6 2 | 3+3 2 | 3+3 2 |
| | 5+6 2 | 5+6 2 | 5+6 2 | 3+4 2 | 3+3 2 | 3+3 2 |
| | 5+6 2 | 3+4 2 | 5+6 2 | 3+4 2 | 2+2 2 | 2+2 2 |
| | 3+4 2 | 3+4 2 | 3+4 2 | 5+6 2 | 2+2 2 | 3+3 2 |
| | 3+4 2 | 5+6 2 | 5+6 2 | 3+4 2 | 3+3 2 | 2+2 2 |
| | 3+4 2 | 3+4 2 | 3+4 2 | 5+6 2 | 2+2 2 | 2+2 2 |
| | 5+6 2 | 5+6 2 | 5+6 2 | 3+4 2 | 3+3 2 | 2+2 2 |
| | 5+6 2 | 5+6 2 | 3+4 2 | 5+6 2 | 2+2 2 | 3+3 2 |
| | 3+4 2 | 3+4 2 | 3+4 2 | 3+4 2 | 3+3 2 | 3+3 2 |
| | 5+6 2 | 5+6 2 | 3+4 2 | 5+6 2 | 2+2 2 | 2+2 2 |
| | 3+4 2 | 3+4 2 | 3+4 2 | 3+4 2 | 2+2 2 | 3+3 2 |
| | 3+4 2 | 5+6 2 | 5+6 2 | 5+6 2 | 3+3 2 | 2+2 2 |
| | 5+6 2 | 3+4 2 | 5+6 2 | 5+6 2 | 2+2 2 | 2+2 2 |
| | 5+6 2 | 3+4 2 | 5+6 2 | 3+4 2 | 3+3 2 | 3+3 2 |
| | 5+6 2 | 5+6 2 | 5+6 2 | 3+4 2 | 3+3 2 | 2+2 2 |
| | 3+4 2 | 3+4 2 | 3+4 2 | 5+6 2 | 2+2 2 | 3+3 2 |
| | 5+6 2 | 5+6 2 | 5+6 2 | 3+4 2 | 3+3 2 | 2+2 2 |
| | 3+4 2 | 3+4 2 | 3+4 2 | 5+6 2 | 3+3 2 | 3+3 2 |
| | 5+6 2 | 5+6 2 | 5+6 2 | 5+6 2 | 2+2 2 | 2+2 2 |
| | 3+4 2 | 3+4 2 | 5+6 2 | 3+4 2 | 3+3 2 | 2+2 2 |
| | 5+6 2 | 3+4 2 | 3+4 2 | 3+4 2 | 2+2 2 | 3+3 2 |
| | 5+6 2 | 5+6 2 | 5+6 2 | 3+4 2 | 3+3 2 | 2+2 2 |
| | 3+4 2 | 3+4 2 | 3+4 2 | 5+6 2 | 2+2 2 | 3+3 2 |
| | 5+6 2 | 5+6 2 | 5+6 2 | 3+4 2 | 3+3 2 | 2+2 2 |
| | 3+4 2 | 3+4 2 | 5+6 2 | 5+6 2 | 2+2 2 | 3+3 2 |
| | 3+4 2 | 3+4 2 | 3+4 2 | 5+6 2 | 3+3 2 | 2+2 2 |
| | 5+6 2 | 5+6 2 | 5+6 2 | 5+6 2 | 2+2 2 | 3+3 2 |
| 12.2.161 | 408938 | 84 | Ba2Mg3Si4 | | | |
| | -A | -GM | -M | -Y | -L | -V |
| | 5+6 2 | 3+4 2 | 5+6 2 | 3+4 2 | 2+2 2 | 3+3 2 |
| | 3+4 2 | 5+6 2 | 3+4 2 | 5+6 2 | 3+3 2 | 2+2 2 |
| | 3+4 2 | 3+4 2 | 5+6 2 | 5+6 2 | 3+3 2 | 3+3 2 |
| | 5+6 2 | 5+6 2 | 3+4 2 | 3+4 2 | 2+2 2 | 2+2 2 |
| | 5+6 2 | 3+4 2 | 5+6 2 | 3+4 2 | 2+2 2 | 3+3 2 |
| | 3+4 2 | 3+4 2 | 3+4 2 | 5+6 2 | 3+3 2 | 2+2 2 |
| | 3+4 2 | 5+6 2 | 5+6 2 | 5+6 2 | 3+3 2 | 3+3 2 |
| | 5+6 2 | 5+6 2 | 3+4 2 | 3+4 2 | 2+2 2 | 2+2 2 |
| | 5+6 2 | 3+4 2 | 5+6 2 | 3+4 2 | 2+2 2 | 2+2 2 |
| | 3+4 2 | 3+4 2 | 3+4 2 | 5+6 2 | 3+3 2 | 3+3 2 |
| | 3+4 2 | 5+6 2 | 3+4 2 | 3+4 2 | 2+2 2 | 2+2 2 |

Continued on next column

**TABLE XXIV– continued**

| TQC | ICSD | NoE | SF | | | |
|---|---|---|---|---|---|---|
| | 5+6 2 | 5+6 2 | 5+6 2 | 5+6 2 | 3+3 2 | 3+3 2 |
| | 3+4 2 | 3+4 2 | 5+6 2 | 3+4 2 | 2+2 2 | 3+3 2 |
| | 5+6 2 | 3+4 2 | 3+4 2 | 5+6 2 | 3+3 2 | 2+2 2 |
| | 3+4 2 | 5+6 2 | 5+6 2 | 5+6 2 | 3+3 2 | 3+3 2 |
| | 5+6 2 | 5+6 2 | 3+4 2 | 3+4 2 | 2+2 2 | 2+2 2 |
| | 3+4 2 | 3+4 2 | 3+4 2 | 3+4 2 | 2+2 2 | 2+2 2 |
| | 5+6 2 | 5+6 2 | 5+6 2 | 5+6 2 | 3+3 2 | 3+3 2 |
| | 5+6 2 | 3+4 2 | 5+6 2 | 3+4 2 | 3+3 2 | 2+2 2 |
| | 3+4 2 | 3+4 2 | 3+4 2 | 3+4 2 | 2+2 2 | 2+2 2 |
| | 5+6 2 | 5+6 2 | 5+6 2 | 5+6 2 | 3+3 2 | 3+3 2 |
| | 3+4 2 | 5+6 2 | 3+4 2 | 5+6 2 | 2+2 2 | 3+3 2 |
| | 3+4 2 | 3+4 2 | 3+4 2 | 3+4 2 | 2+2 2 | 2+2 2 |
| | 5+6 2 | 5+6 2 | 5+6 2 | 5+6 2 | 3+3 2 | 3+3 2 |
| | 3+4 2 | 3+4 2 | 5+6 2 | 5+6 2 | 3+3 2 | 3+3 2 |
| | 5+6 2 | 3+4 2 | 3+4 2 | 3+4 2 | 2+2 2 | 2+2 2 |
| | 5+6 2 | 5+6 2 | 3+4 2 | 3+4 2 | 3+3 2 | 2+2 2 |
| | 3+4 2 | 5+6 2 | 5+6 2 | 5+6 2 | 2+2 2 | 3+3 2 |
| | 5+6 2 | 3+4 2 | 5+6 2 | 3+4 2 | 2+2 2 | 2+2 2 |
| | 5+6 2 | 5+6 2 | 3+4 2 | 5+6 2 | 3+3 2 | 3+3 2 |
| | 3+4 2 | 5+6 2 | 3+4 2 | 5+6 2 | 2+2 2 | 2+2 2 |
| | 5+6 2 | 3+4 2 | 3+4 2 | 3+4 2 | 3+3 2 | 3+3 2 |
| | 3+4 2 | 5+6 2 | 3+4 2 | 5+6 2 | 3+3 2 | 2+2 2 |
| | 3+4 2 | 3+4 2 | 5+6 2 | 5+6 2 | 2+2 2 | 3+3 2 |
| | 3+4 2 | 3+4 2 | 5+6 2 | 3+4 2 | 3+3 2 | 3+3 2 |
| | 5+6 2 | 3+4 2 | 5+6 2 | 5+6 2 | 2+2 2 | 2+2 2 |
| | 3+4 2 | 5+6 2 | 3+4 2 | 3+4 2 | 3+3 2 | 3+3 2 |
| | 5+6 2 | 3+4 2 | 5+6 2 | 3+4 2 | 2+2 2 | 2+2 2 |
| | 5+6 2 | 5+6 2 | 3+4 2 | 3+4 2 | 2+2 2 | 2+2 2 |
| | 3+4 2 | 3+4 2 | 5+6 2 | 3+4 2 | 3+3 2 | 3+3 2 |
| | 5+6 2 | 5+6 2 | 3+4 2 | 5+6 2 | 2+2 2 | 2+2 2 |
| | 3+4 2 | 3+4 2 | 3+4 2 | 3+4 2 | 3+3 2 | 2+2 2 |
| 12.2.162 | 420834 | 118 | Ba2Cd2Sb3 | | | |
| | -A | -GM | -M | -Y | -L | -V |
| | 3+4 2 | 3+4 2 | 3+4 2 | 3+4 2 | 2+2 2 | 2+2 2 |
| | 5+6 2 | 5+6 2 | 5+6 2 | 5+6 2 | 3+3 2 | 3+3 2 |
| | 5+6 2 | 3+4 2 | 3+4 2 | 5+6 2 | 2+2 2 | 2+2 2 |
| | 3+4 2 | 5+6 2 | 5+6 2 | 3+4 2 | 3+3 2 | 3+3 2 |
| | 3+4 2 | 3+4 2 | 3+4 2 | 3+4 2 | 2+2 2 | 2+2 2 |
| | 5+6 2 | 5+6 2 | 5+6 2 | 5+6 2 | 3+3 2 | 3+3 2 |
| | 3+4 2 | 3+4 2 | 3+4 2 | 5+6 2 | 2+2 2 | 2+2 2 |
| | 5+6 2 | 5+6 2 | 5+6 2 | 3+4 2 | 3+3 2 | 3+3 2 |
| | 3+4 2 | 3+4 2 | 3+4 2 | 3+4 2 | 3+3 2 | 3+3 2 |
| | 5+6 2 | 5+6 2 | 5+6 2 | 5+6 2 | 2+2 2 | 2+2 2 |
| | 3+4 2 | 3+4 2 | 3+4 2 | 3+4 2 | 3+3 2 | 3+3 2 |
| | 5+6 2 | 5+6 2 | 5+6 2 | 5+6 2 | 2+2 2 | 2+2 2 |
| | 5+6 2 | 3+4 2 | 5+6 2 | 5+6 2 | 3+3 2 | 3+3 2 |
| | 3+4 2 | 3+4 2 | 3+4 2 | 5+6 2 | 2+2 2 | 2+2 2 |
| | 5+6 2 | 5+6 2 | 3+4 2 | 3+4 2 | 2+2 2 | 2+2 2 |
| | 3+4 2 | 5+6 2 | 5+6 2 | 3+4 2 | 3+3 2 | 3+3 2 |
| | 3+4 2 | 3+4 2 | 3+4 2 | 3+4 2 | 3+3 2 | 3+3 2 |
| | 5+6 2 | 5+6 2 | 5+6 2 | 5+6 2 | 2+2 2 | 2+2 2 |
| | 3+4 2 | 3+4 2 | 3+4 2 | 3+4 2 | 3+3 2 | 3+3 2 |
| | 5+6 2 | 5+6 2 | 5+6 2 | 5+6 2 | 2+2 2 | 2+2 2 |
| | 3+4 2 | 3+4 2 | 3+4 2 | 3+4 2 | 3+3 2 | 3+3 2 |
| | 5+6 2 | 5+6 2 | 3+4 2 | 5+6 2 | 2+2 2 | 2+2 2 |
| | 3+4 2 | 3+4 2 | 5+6 2 | 3+4 2 | 2+2 2 | 3+3 2 |
| | 5+6 2 | 5+6 2 | 5+6 2 | 5+6 2 | 3+3 2 | 2+2 2 |
| | 3+4 2 | 3+4 2 | 3+4 2 | 3+4 2 | 3+3 2 | 3+3 2 |
| | 5+6 2 | 3+4 2 | 5+6 2 | 3+4 2 | 3+3 2 | 2+2 2 |
| | 3+4 2 | 5+6 2 | 3+4 2 | 5+6 2 | 2+2 2 | 3+3 2 |
| | 3+4 2 | 5+6 2 | 3+4 2 | 5+6 2 | 3+3 2 | 2+2 2 |

Continued on next column

**TABLE XXIV– continued**

| TQC | ICSD | | NoE | | SF | |
|---|---|---|---|---|---|---|
| | 5+6 2 | 3+4 2 | 5+6 2 | 3+4 2 | 2+2 2 | 3+3 2 |
| | 5+6 2 | 5+6 2 | 5+6 2 | 5+6 2 | 2+2 2 | 2+2 2 |
| | 3+4 2 | 3+4 2 | 3+4 2 | 3+4 2 | 3+3 2 | 2+2 2 |
| | 5+6 2 | 5+6 2 | 5+6 2 | 5+6 2 | 2+2 2 | 3+3 2 |
| | 3+4 2 | 3+4 2 | 3+4 2 | 3+4 2 | 3+3 2 | 3+3 2 |
| | 5+6 2 | 5+6 2 | 5+6 2 | 5+6 2 | 3+3 2 | 2+2 2 |
| | 5+6 2 | 3+4 2 | 5+6 2 | 3+4 2 | 2+2 2 | 2+2 2 |
| | 3+4 2 | 3+4 2 | 3+4 2 | 3+4 2 | 3+3 2 | 3+3 2 |
| | 5+6 2 | 5+6 2 | 3+4 2 | 5+6 2 | 2+2 2 | 3+3 2 |
| | 3+4 2 | 3+4 2 | 5+6 2 | 3+4 2 | 3+3 2 | 2+2 2 |
| | 3+4 2 | 5+6 2 | 3+4 2 | 5+6 2 | 2+2 2 | 3+3 2 |
| | 3+4 2 | 5+6 2 | 3+4 2 | 5+6 2 | 2+2 2 | 2+2 2 |
| | 5+6 2 | 3+4 2 | 5+6 2 | 3+4 2 | 3+3 2 | 3+3 2 |
| | 5+6 2 | 5+6 2 | 5+6 2 | 5+6 2 | 2+2 2 | 2+2 2 |
| | 5+6 2 | 3+4 2 | 5+6 2 | 3+4 2 | 3+3 2 | 3+3 2 |
| | 3+4 2 | 5+6 2 | 3+4 2 | 5+6 2 | 2+2 2 | 2+2 2 |
| | 5+6 2 | 3+4 2 | 5+6 2 | 3+4 2 | 2+2 2 | 3+3 2 |
| | 3+4 2 | 5+6 2 | 3+4 2 | 5+6 2 | 3+3 2 | 2+2 2 |
| | 3+4 2 | 3+4 2 | 3+4 2 | 3+4 2 | 3+3 2 | 2+2 2 |
| | 5+6 2 | 5+6 2 | 5+6 2 | 5+6 2 | 2+2 2 | 3+3 2 |
| | 3+4 2 | 5+6 2 | 3+4 2 | 5+6 2 | 3+3 2 | 3+3 2 |
| | 3+4 2 | 5+6 2 | 5+6 2 | 3+4 2 | 2+2 2 | 2+2 2 |
| | 5+6 2 | 3+4 2 | 3+4 2 | 5+6 2 | 3+3 2 | 2+2 2 |
| | 5+6 2 | 3+4 2 | 5+6 2 | 3+4 2 | 2+2 2 | 2+2 2 |
| | 5+6 2 | 3+4 2 | 3+4 2 | 5+6 2 | 3+3 2 | 3+3 2 |
| | 3+4 2 | 5+6 2 | 5+6 2 | 3+4 2 | 2+2 2 | 2+2 2 |
| | 5+6 2 | 3+4 2 | 5+6 2 | 5+6 2 | 3+3 2 | 3+3 2 |
| | 3+4 2 | 3+4 2 | 3+4 2 | 3+4 2 | 2+2 2 | 2+2 2 |
| | 5+6 2 | 5+6 2 | 5+6 2 | 5+6 2 | 3+3 2 | 3+3 2 |
| | 3+4 2 | 3+4 2 | 3+4 2 | 3+4 2 | 2+2 2 | 3+3 2 |
| | 3+4 2 | 5+6 2 | 3+4 2 | 5+6 2 | 3+3 2 | 3+3 2 |
| 12.2.163 | 80901 | | 80 | | Ba2Bi1O6Tb1 | |
| | -A | -GM | -M | -Y | -L | -V |
| | 5+6 2 | 3+4 2 | 5+6 2 | 3+4 2 | 2+2 2 | 3+3 2 |
| | 3+4 2 | 3+4 2 | 3+4 2 | 5+6 2 | 3+3 2 | 2+2 2 |
| | 5+6 2 | 5+6 2 | 5+6 2 | 3+4 2 | 2+2 2 | 3+3 2 |
| | 3+4 2 | 5+6 2 | 3+4 2 | 5+6 2 | 3+3 2 | 2+2 2 |
| | 3+4 2 | 5+6 2 | 3+4 2 | 5+6 2 | 3+3 2 | 2+2 2 |
| | 3+4 2 | 5+6 2 | 3+4 2 | 5+6 2 | 3+3 2 | 2+2 2 |
| | 5+6 2 | 3+4 2 | 5+6 2 | 3+4 2 | 2+2 2 | 3+3 2 |
| | 5+6 2 | 3+4 2 | 5+6 2 | 3+4 2 | 2+2 2 | 2+2 2 |
| | 3+4 2 | 5+6 2 | 3+4 2 | 5+6 2 | 3+3 2 | 3+3 2 |
| | 5+6 2 | 3+4 2 | 5+6 2 | 3+4 2 | 2+2 2 | 3+3 2 |
| | 3+4 2 | 5+6 2 | 3+4 2 | 5+6 2 | 3+3 2 | 2+2 2 |
| | 3+4 2 | 5+6 2 | 3+4 2 | 5+6 2 | 3+3 2 | 2+2 2 |
| | 5+6 2 | 3+4 2 | 5+6 2 | 5+6 2 | 3+3 2 | 3+3 2 |
| | 3+4 2 | 5+6 2 | 3+4 2 | 3+4 2 | 2+2 2 | 2+2 2 |
| | 5+6 2 | 3+4 2 | 5+6 2 | 5+6 2 | 3+3 2 | 3+3 2 |
| | 3+4 2 | 3+4 2 | 3+4 2 | 3+4 2 | 2+2 2 | 2+2 2 |
| | 3+4 2 | 5+6 2 | 3+4 2 | 5+6 2 | 3+3 2 | 2+2 2 |
| | 5+6 2 | 5+6 2 | 5+6 2 | 3+4 2 | 2+2 2 | 3+3 2 |
| | 3+4 2 | 3+4 2 | 3+4 2 | 3+4 2 | 2+2 2 | 2+2 2 |
| | 5+6 2 | 5+6 2 | 5+6 2 | 5+6 2 | 3+3 2 | 3+3 2 |
| | 5+6 2 | 5+6 2 | 5+6 2 | 5+6 2 | 3+3 2 | 3+3 2 |
| | 5+6 2 | 5+6 2 | 5+6 2 | 5+6 2 | 2+2 2 | 3+3 2 |
| | 3+4 2 | 3+4 2 | 5+6 2 | 3+4 2 | 3+3 2 | 3+3 2 |
| | 5+6 2 | 3+4 2 | 3+4 2 | 3+4 2 | 2+2 2 | 2+2 2 |
| | 5+6 2 | 3+4 2 | 5+6 2 | 3+4 2 | 2+2 2 | 3+3 2 |
| | 5+6 2 | 3+4 2 | 5+6 2 | 3+4 2 | 2+2 2 | 3+3 2 |
| | 3+4 2 | 3+4 2 | 3+4 2 | 3+4 2 | 3+3 2 | 2+2 2 |
| | 3+4 2 | 5+6 2 | 3+4 2 | 5+6 2 | 2+2 2 | 2+2 2 |
| | 3+4 2 | 5+6 2 | 3+4 2 | 5+6 2 | 3+3 2 | 2+2 2 |
| | 5+6 2 | 5+6 2 | 5+6 2 | 5+6 2 | 3+3 2 | 3+3 2 |
| | 3+4 2 | 5+6 2 | 3+4 2 | 5+6 2 | 3+3 2 | 2+2 2 |
| | 3+4 2 | 5+6 2 | 3+4 2 | 3+4 2 | 2+2 2 | 2+2 2 |
| | 5+6 2 | 5+6 2 | 3+4 2 | 3+4 2 | 2+2 2 | 3+3 2 |
| | 3+4 2 | 3+4 2 | 5+6 2 | 5+6 2 | 3+3 2 | 2+2 2 |
| | 3+4 2 | 3+4 2 | 5+6 2 | 5+6 2 | 3+3 2 | 2+2 2 |
| | 5+6 2 | 3+4 2 | 3+4 2 | 3+4 2 | 2+2 2 | 3+3 2 |
| | 3+4 2 | 5+6 2 | 3+4 2 | 5+6 2 | 3+3 2 | 2+2 2 |
| | 3+4 2 | 5+6 2 | 3+4 2 | 5+6 2 | 3+3 2 | 2+2 2 |
| | 3+4 2 | 5+6 2 | 3+4 2 | 5+6 2 | 3+3 2 | 2+2 2 |
| | 3+4 2 | 5+6 2 | 3+4 2 | 3+4 2 | 3+3 2 | 2+2 2 |
| 12.2.164 | 261206 | | 44 | | Ba1Tl4 | |
| | -A | -GM | -M | -Y | -L | -V |
| | 5+6 2 | 3+4 2 | 5+6 2 | 3+4 2 | 2+2 2 | 3+3 2 |
| | 3+4 2 | 5+6 2 | 3+4 2 | 5+6 2 | 3+3 2 | 2+2 2 |
| | 5+6 2 | 3+4 2 | 5+6 2 | 5+6 2 | 2+2 2 | 3+3 2 |
| | 3+4 2 | 5+6 2 | 3+4 2 | 3+4 2 | 3+3 2 | 2+2 2 |
| | 5+6 2 | 3+4 2 | 5+6 2 | 5+6 2 | 3+3 2 | 2+2 2 |
| | 3+4 2 | 5+6 2 | 3+4 2 | 3+4 2 | 2+2 2 | 3+3 2 |
| | 5+6 2 | 3+4 2 | 5+6 2 | 5+6 2 | 2+2 2 | 3+3 2 |
| | 3+4 2 | 5+6 2 | 3+4 2 | 3+4 2 | 3+3 2 | 2+2 2 |
| | 3+4 2 | 3+4 2 | 3+4 2 | 3+4 2 | 3+3 2 | 2+2 2 |
| | 5+6 2 | 3+4 2 | 5+6 2 | 5+6 2 | 2+2 2 | 3+3 2 |
| | 3+4 2 | 5+6 2 | 5+6 2 | 5+6 2 | 3+3 2 | 2+2 2 |
| | 5+6 2 | 5+6 2 | 3+4 2 | 3+4 2 | 2+2 2 | 3+3 2 |
| | 5+6 2 | 3+4 2 | 5+6 2 | 3+4 2 | 3+3 2 | 2+2 2 |
| | 3+4 2 | 5+6 2 | 3+4 2 | 3+4 2 | 2+2 2 | 2+2 2 |
| | 5+6 2 | 3+4 2 | 5+6 2 | 5+6 2 | 2+2 2 | 3+3 2 |
| | 3+4 2 | 5+6 2 | 3+4 2 | 5+6 2 | 3+3 2 | 3+3 2 |
| | 3+4 2 | 3+4 2 | 5+6 2 | 5+6 2 | 3+3 2 | 3+3 2 |
| | 5+6 2 | 5+6 2 | 3+4 2 | 5+6 2 | 2+2 2 | 2+2 2 |
| | 5+6 2 | 3+4 2 | 3+4 2 | 3+4 2 | 3+3 2 | 2+2 2 |
| | 3+4 2 | 3+4 2 | 5+6 2 | 5+6 2 | 2+2 2 | 3+3 2 |
| | 3+4 2 | 3+4 2 | 3+4 2 | 3+4 2 | 2+2 2 | 2+2 2 |
| | 5+6 2 | 5+6 2 | 5+6 2 | 3+4 2 | 3+3 2 | 3+3 2 |
| 12.2.165 | 238227 | | 92 | | Ba1Te4Th1 | |
| | -A | -GM | -M | -Y | -L | -V |
| | 3+4 2 | 3+4 2 | 5+6 2 | 3+4 2 | 2+2 2 | 2+2 2 |
| | 5+6 2 | 5+6 2 | 3+4 2 | 5+6 2 | 3+3 2 | 3+3 2 |
| | 3+4 2 | 3+4 2 | 3+4 2 | 3+4 2 | 2+2 2 | 3+3 2 |
| | 5+6 2 | 5+6 2 | 5+6 2 | 5+6 2 | 3+3 2 | 2+2 2 |
| | 3+4 2 | 3+4 2 | 5+6 2 | 3+4 2 | 3+3 2 | 2+2 2 |
| | 5+6 2 | 5+6 2 | 3+4 2 | 5+6 2 | 2+2 2 | 3+3 2 |
| | 5+6 2 | 3+4 2 | 5+6 2 | 3+4 2 | 3+3 2 | 2+2 2 |
| | 3+4 2 | 5+6 2 | 3+4 2 | 5+6 2 | 2+2 2 | 3+3 2 |
| | 3+4 2 | 3+4 2 | 5+6 2 | 5+6 2 | 3+3 2 | 2+2 2 |
| | 5+6 2 | 5+6 2 | 3+4 2 | 3+4 2 | 2+2 2 | 3+3 2 |
| | 5+6 2 | 3+4 2 | 3+4 2 | 3+4 2 | 2+2 2 | 3+3 2 |
| | 3+4 2 | 5+6 2 | 5+6 2 | 5+6 2 | 3+3 2 | 2+2 2 |
| | 5+6 2 | 3+4 2 | 5+6 2 | 3+4 2 | 2+2 2 | 3+3 2 |
| | 3+4 2 | 3+4 2 | 3+4 2 | 5+6 2 | 3+3 2 | 2+2 2 |
| | 5+6 2 | 5+6 2 | 3+4 2 | 5+6 2 | 2+2 2 | 3+3 2 |
| | 3+4 2 | 5+6 2 | 5+6 2 | 3+4 2 | 3+3 2 | 2+2 2 |
| | 5+6 2 | 3+4 2 | 5+6 2 | 3+4 2 | 2+2 2 | 3+3 2 |
| | 3+4 2 | 5+6 2 | 3+4 2 | 5+6 2 | 3+3 2 | 2+2 2 |
| | 5+6 2 | 3+4 2 | 5+6 2 | 3+4 2 | 2+2 2 | 3+3 2 |
| | 5+6 2 | 5+6 2 | 3+4 2 | 5+6 2 | 3+3 2 | 2+2 2 |
| | 3+4 2 | 3+4 2 | 5+6 2 | 3+4 2 | 2+2 2 | 3+3 2 |
| | 3+4 2 | 5+6 2 | 3+4 2 | 5+6 2 | 3+3 2 | 2+2 2 |
| | 5+6 2 | 5+6 2 | 5+6 2 | 5+6 2 | 2+2 2 | 3+3 2 |

Continued on next column

Continued on next column

**TABLE XXIV– continued**

| TQC | ICSD | | NoE | | SF | |
|---|---|---|---|---|---|---|
| | 3+4 2 | 3+4 2 | 3+4 2 | 3+4 2 | 3+3 2 | 2+2 2 |
| | 5+6 2 | 3+4 2 | 3+4 2 | 3+4 2 | 2+2 2 | 3+3 2 |
| | 3+4 2 | 5+6 2 | 5+6 2 | 5+6 2 | 3+3 2 | 2+2 2 |
| | 5+6 2 | 3+4 2 | 5+6 2 | 3+4 2 | 2+2 2 | 2+2 2 |
| | 3+4 2 | 5+6 2 | 3+4 2 | 5+6 2 | 3+3 2 | 3+3 2 |
| | 3+4 2 | 5+6 2 | 3+4 2 | 5+6 2 | 2+2 2 | 2+2 2 |
| | 3+4 2 | 3+4 2 | 5+6 2 | 3+4 2 | 3+3 2 | 3+3 2 |
| | 5+6 2 | 5+6 2 | 5+6 2 | 5+6 2 | 2+2 2 | 3+3 2 |
| | 3+4 2 | 3+4 2 | 3+4 2 | 5+6 2 | 3+3 2 | 2+2 2 |
| | 5+6 2 | 3+4 2 | 3+4 2 | 3+4 2 | 2+2 2 | 3+3 2 |
| | 5+6 2 | 3+4 2 | 5+6 2 | 5+6 2 | 3+3 2 | 2+2 2 |
| | 3+4 2 | 5+6 2 | 5+6 2 | 3+4 2 | 3+3 2 | 3+3 2 |
| | 5+6 2 | 3+4 2 | 3+4 2 | 3+4 2 | 2+2 2 | 2+2 2 |
| | 3+4 2 | 5+6 2 | 5+6 2 | 5+6 2 | 3+3 2 | 2+2 2 |
| | 5+6 2 | 3+4 2 | 5+6 2 | 5+6 2 | 2+2 2 | 3+3 2 |
| | 3+4 2 | 5+6 2 | 3+4 2 | 3+4 2 | 3+3 2 | 3+3 2 |
| | 5+6 2 | 3+4 2 | 3+4 2 | 3+4 2 | 2+2 2 | 2+2 2 |
| | 3+4 2 | 5+6 2 | 5+6 2 | 5+6 2 | 3+3 2 | 2+2 2 |
| | 5+6 2 | 3+4 2 | 5+6 2 | 3+4 2 | 2+2 2 | 3+3 2 |
| | 5+6 2 | 5+6 2 | 3+4 2 | 5+6 2 | 3+3 2 | 2+2 2 |
| | 5+6 2 | 5+6 2 | 5+6 2 | 3+4 2 | 2+2 2 | 3+3 2 |
| | 3+4 2 | 3+4 2 | 3+4 2 | 3+4 2 | 3+3 2 | 2+2 2 |
| | 3+4 2 | 3+4 2 | 3+4 2 | 5+6 2 | 2+2 2 | 3+3 2 |
| 12.2.166 | 49000 | | 50 | | Ba1Sb3 | |
| | -A | -GM | -M | -Y | -L | -V |
| | 5+6 2 | 3+4 2 | 5+6 2 | 3+4 2 | 3+3 2 | 2+2 2 |
| | 3+4 2 | 5+6 2 | 3+4 2 | 5+6 2 | 2+2 2 | 3+3 2 |
| | 5+6 2 | 3+4 2 | 5+6 2 | 3+4 2 | 3+3 2 | 2+2 2 |
| | 3+4 2 | 5+6 2 | 3+4 2 | 5+6 2 | 2+2 2 | 3+3 2 |
| | 5+6 2 | 3+4 2 | 3+4 2 | 3+4 2 | 2+2 2 | 3+3 2 |
| | 3+4 2 | 3+4 2 | 5+6 2 | 5+6 2 | 3+3 2 | 2+2 2 |
| | 5+6 2 | 5+6 2 | 5+6 2 | 5+6 2 | 3+3 2 | 2+2 2 |
| | 3+4 2 | 5+6 2 | 3+4 2 | 3+4 2 | 2+2 2 | 3+3 2 |
| | 5+6 2 | 3+4 2 | 5+6 2 | 3+4 2 | 2+2 2 | 3+3 2 |
| | 3+4 2 | 3+4 2 | 3+4 2 | 5+6 2 | 3+3 2 | 2+2 2 |
| | 5+6 2 | 5+6 2 | 3+4 2 | 5+6 2 | 2+2 2 | 3+3 2 |
| | 5+6 2 | 3+4 2 | 5+6 2 | 3+4 2 | 3+3 2 | 2+2 2 |
| | 3+4 2 | 5+6 2 | 3+4 2 | 5+6 2 | 2+2 2 | 3+3 2 |
| | 3+4 2 | 5+6 2 | 5+6 2 | 3+4 2 | 3+3 2 | 2+2 2 |
| | 5+6 2 | 3+4 2 | 3+4 2 | 3+4 2 | 3+3 2 | 3+3 2 |
| | 3+4 2 | 5+6 2 | 5+6 2 | 5+6 2 | 2+2 2 | 2+2 2 |
| | 3+4 2 | 5+6 2 | 3+4 2 | 5+6 2 | 3+3 2 | 3+3 2 |
| | 5+6 2 | 3+4 2 | 5+6 2 | 3+4 2 | 2+2 2 | 2+2 2 |
| | 5+6 2 | 3+4 2 | 3+4 2 | 3+4 2 | 3+3 2 | 3+3 2 |
| | 3+4 2 | 5+6 2 | 5+6 2 | 5+6 2 | 2+2 2 | 2+2 2 |
| | 5+6 2 | 3+4 2 | 5+6 2 | 5+6 2 | 3+3 2 | 2+2 2 |
| | 5+6 2 | 5+6 2 | 5+6 2 | 3+4 2 | 2+2 2 | 3+3 2 |
| | 3+4 2 | 3+4 2 | 3+4 2 | 3+4 2 | 2+2 2 | 3+3 2 |
| | 5+6 2 | 3+4 2 | 3+4 2 | 5+6 2 | 2+2 2 | 3+3 2 |
| | 3+4 2 | 3+4 2 | 3+4 2 | 5+6 2 | 3+3 2 | 2+2 2 |
| 12.2.167 | 16634 | | 180 | | Ba1O5Ti2 | |
| | -A | -GM | -M | -Y | -L | -V |
| | 3+4 2 | 3+4 2 | 5+6 2 | 5+6 2 | 2+2 2 | 2+2 2 |
| | 5+6 2 | 5+6 2 | 3+4 2 | 3+4 2 | 3+3 2 | 3+3 2 |
| | 3+4 2 | 3+4 2 | 5+6 2 | 5+6 2 | 3+3 2 | 2+2 2 |
| | 5+6 2 | 5+6 2 | 3+4 2 | 3+4 2 | 2+2 2 | 3+3 2 |
| | 3+4 2 | 3+4 2 | 5+6 2 | 5+6 2 | 3+3 2 | 3+3 2 |
| | 3+4 2 | 3+4 2 | 3+4 2 | 3+4 2 | 3+3 2 | 2+2 2 |
| | 5+6 2 | 5+6 2 | 3+4 2 | 3+4 2 | 2+2 2 | 3+3 2 |
| | 5+6 2 | 3+4 2 | 3+4 2 | 5+6 2 | 2+2 2 | 3+3 2 |
| | 3+4 2 | 5+6 2 | 5+6 2 | 3+4 2 | 3+3 2 | 2+2 2 |
| | 5+6 2 | 5+6 2 | 5+6 2 | 5+6 2 | 3+3 2 | 3+3 2 |
| | 3+4 2 | 3+4 2 | 3+4 2 | 3+4 2 | 2+2 2 | 2+2 2 |
| | 5+6 2 | 5+6 2 | 3+4 2 | 3+4 2 | 3+3 2 | 2+2 2 |
| | 3+4 2 | 3+4 2 | 5+6 2 | 5+6 2 | 2+2 2 | 3+3 2 |
| | 3+4 2 | 3+4 2 | 5+6 2 | 3+4 2 | 3+3 2 | 2+2 2 |
| | 5+6 2 | 5+6 2 | 3+4 2 | 5+6 2 | 2+2 2 | 3+3 2 |
| | 5+6 2 | 3+4 2 | 5+6 2 | 3+4 2 | 2+2 2 | 2+2 2 |
| | 5+6 2 | 5+6 2 | 3+4 2 | 3+4 2 | 3+3 2 | 3+3 2 |
| | 3+4 2 | 3+4 2 | 5+6 2 | 5+6 2 | 3+3 2 | 2+2 2 |
| | 3+4 2 | 5+6 2 | 3+4 2 | 5+6 2 | 2+2 2 | 2+2 2 |
| | 5+6 2 | 3+4 2 | 5+6 2 | 3+4 2 | 3+3 2 | 3+3 2 |
| | 5+6 2 | 5+6 2 | 3+4 2 | 5+6 2 | 2+2 2 | 3+3 2 |
| | 3+4 2 | 3+4 2 | 5+6 2 | 3+4 2 | 3+3 2 | 2+2 2 |
| | 3+4 2 | 3+4 2 | 5+6 2 | 5+6 2 | 3+3 2 | 2+2 2 |
| | 5+6 2 | 5+6 2 | 3+4 2 | 3+4 2 | 2+2 2 | 3+3 2 |
| | 3+4 2 | 3+4 2 | 5+6 2 | 5+6 2 | 2+2 2 | 3+3 2 |
| | 5+6 2 | 5+6 2 | 3+4 2 | 3+4 2 | 3+3 2 | 2+2 2 |
| | 3+4 2 | 5+6 2 | 3+4 2 | 5+6 2 | 2+2 2 | 2+2 2 |
| | 5+6 2 | 3+4 2 | 5+6 2 | 5+6 2 | 2+2 2 | 3+3 2 |
| | 3+4 2 | 3+4 2 | 5+6 2 | 3+4 2 | 2+2 2 | 2+2 2 |
| | 5+6 2 | 5+6 2 | 3+4 2 | 3+4 2 | 3+3 2 | 2+2 2 |
| | 3+4 2 | 5+6 2 | 3+4 2 | 5+6 2 | 2+2 2 | 3+3 2 |
| | 5+6 2 | 3+4 2 | 5+6 2 | 3+4 2 | 2+2 2 | 2+2 2 |
| | 3+4 2 | 5+6 2 | 5+6 2 | 5+6 2 | 3+3 2 | 3+3 2 |
| | 5+6 2 | 5+6 2 | 3+4 2 | 5+6 2 | 2+2 2 | 2+2 2 |
| | 3+4 2 | 3+4 2 | 5+6 2 | 5+6 2 | 3+3 2 | 2+2 2 |
| | 5+6 2 | 5+6 2 | 5+6 2 | 3+4 2 | 2+2 2 | 3+3 2 |
| | 5+6 2 | 3+4 2 | 3+4 2 | 5+6 2 | 2+2 2 | 3+3 2 |
| | 3+4 2 | 5+6 2 | 3+4 2 | 5+6 2 | 3+3 2 | 2+2 2 |
| | 5+6 2 | 5+6 2 | 5+6 2 | 3+4 2 | 3+3 2 | 2+2 2 |
| | 5+6 2 | 3+4 2 | 5+6 2 | 5+6 2 | 2+2 2 | 3+3 2 |
| | 3+4 2 | 5+6 2 | 3+4 2 | 3+4 2 | 3+3 2 | 3+3 2 |
| | 5+6 2 | 3+4 2 | 5+6 2 | 5+6 2 | 2+2 2 | 2+2 2 |
| | 5+6 2 | 5+6 2 | 5+6 2 | 5+6 2 | 3+3 2 | 2+2 2 |
| | 3+4 2 | 3+4 2 | 5+6 2 | 3+4 2 | 2+2 2 | 3+3 2 |
| | 3+4 2 | 5+6 2 | 3+4 2 | 3+4 2 | 3+3 2 | 2+2 2 |
| | 5+6 2 | 5+6 2 | 3+4 2 | 5+6 2 | 3+3 2 | 3+3 2 |
| | 3+4 2 | 3+4 2 | 5+6 2 | 3+4 2 | 2+2 2 | 2+2 2 |
| | 5+6 2 | 5+6 2 | 3+4 2 | 5+6 2 | 3+3 2 | 3+3 2 |
| | 3+4 2 | 3+4 2 | 5+6 2 | 3+4 2 | 2+2 2 | 2+2 2 |
| | 5+6 2 | 5+6 2 | 3+4 2 | 5+6 2 | 2+2 2 | 3+3 2 |
| | 5+6 2 | 3+4 2 | 5+6 2 | 3+4 2 | 3+3 2 | 2+2 2 |
| | 3+4 2 | 3+4 2 | 3+4 2 | 5+6 2 | 2+2 2 | 3+3 2 |
| | 5+6 2 | 5+6 2 | 3+4 2 | 3+4 2 | 3+3 2 | 3+3 2 |
| | 3+4 2 | 3+4 2 | 5+6 2 | 3+4 2 | 2+2 2 | 2+2 2 |
| | 3+4 2 | 3+4 2 | 3+4 2 | 5+6 2 | 3+3 2 | 3+3 2 |
| | 5+6 2 | 5+6 2 | 3+4 2 | 3+4 2 | 2+2 2 | 3+3 2 |
| | 5+6 2 | 5+6 2 | 3+4 2 | 5+6 2 | 3+3 2 | 2+2 2 |
| | 5+6 2 | 5+6 2 | 5+6 2 | 3+4 2 | 3+3 2 | 2+2 2 |
| | 3+4 2 | 3+4 2 | 5+6 2 | 5+6 2 | 2+2 2 | 3+3 2 |
| | 3+4 2 | 5+6 2 | 3+4 2 | 5+6 2 | 2+2 2 | 3+3 2 |
| | 5+6 2 | 5+6 2 | 3+4 2 | 3+4 2 | 3+3 2 | 2+2 2 |
| | 5+6 2 | 3+4 2 | 5+6 2 | 5+6 2 | 3+3 2 | 2+2 2 |
| | 3+4 2 | 5+6 2 | 5+6 2 | 3+4 2 | 2+2 2 | 2+2 2 |
| | 5+6 2 | 5+6 2 | 3+4 2 | 3+4 2 | 2+2 2 | 3+3 2 |
| | 3+4 2 | 3+4 2 | 3+4 2 | 5+6 2 | 3+3 2 | 3+3 2 |
| | 3+4 2 | 3+4 2 | 5+6 2 | 5+6 2 | 2+2 2 | 2+2 2 |
| | 3+4 2 | 3+4 2 | 5+6 2 | 3+4 2 | 2+2 2 | 3+3 2 |
| | 3+4 2 | 5+6 2 | 3+4 2 | 3+4 2 | 3+3 2 | 2+2 2 |
| | 3+4 2 | 3+4 2 | 3+4 2 | 5+6 2 | 3+3 2 | 3+3 2 |
| | 5+6 2 | 5+6 2 | 3+4 2 | 5+6 2 | 2+2 2 | 2+2 2 |

Continued on next column

**TABLE XXIV– continued**

| TQC | ICSD | NoE | SF | | | |
|---|---|---|---|---|---|---|
| | 5+6 2 | 5+6 2 | 5+6 2 | 3+4 2 | 2+2 2 | 3+3 2 |
| | 5+6 2 | 5+6 2 | 5+6 2 | 5+6 2 | 3+3 2 | 2+2 2 |
| | 3+4 2 | 5+6 2 | 3+4 2 | 5+6 2 | 3+3 2 | 2+2 2 |
| | 3+4 2 | 3+4 2 | 5+6 2 | 3+4 2 | 2+2 2 | 3+3 2 |
| | 5+6 2 | 5+6 2 | 3+4 2 | 5+6 2 | 3+3 2 | 3+3 2 |
| | 5+6 2 | 3+4 2 | 5+6 2 | 5+6 2 | 2+2 2 | 3+3 2 |
| | 3+4 2 | 3+4 2 | 5+6 2 | 3+4 2 | 2+2 2 | 2+2 2 |
| | 5+6 2 | 5+6 2 | 5+6 2 | 3+4 2 | 3+3 2 | 2+2 2 |
| | 3+4 2 | 3+4 2 | 5+6 2 | 5+6 2 | 2+2 2 | 3+3 2 |
| | 3+4 2 | 5+6 2 | 3+4 2 | 5+6 2 | 3+3 2 | 3+3 2 |
| | 5+6 2 | 3+4 2 | 3+4 2 | 3+4 2 | 2+2 2 | 3+3 2 |
| | 3+4 2 | 3+4 2 | 3+4 2 | 5+6 2 | 2+2 2 | 2+2 2 |
| | 5+6 2 | 3+4 2 | 3+4 2 | 3+4 2 | 3+3 2 | 2+2 2 |
| | 3+4 2 | 5+6 2 | 5+6 2 | 5+6 2 | 3+3 2 | 3+3 2 |
| | 5+6 2 | 5+6 2 | 5+6 2 | 3+4 2 | 2+2 2 | 2+2 2 |
| | 3+4 2 | 5+6 2 | 3+4 2 | 5+6 2 | 3+3 2 | 3+3 2 |
| | 5+6 2 | 3+4 2 | 5+6 2 | 3+4 2 | 2+2 2 | 2+2 2 |
| | 3+4 2 | 5+6 2 | 3+4 2 | 5+6 2 | 2+2 2 | 3+3 2 |
| | 5+6 2 | 3+4 2 | 5+6 2 | 3+4 2 | 3+3 2 | 2+2 2 |
| | 3+4 2 | 3+4 2 | 3+4 2 | 5+6 2 | 2+2 2 | 3+3 2 |
| 12.2.168 | 83618 | 280 | Ba1Cu10P4 | | | |
| | -A | -GM | -M | -Y | -L | -V |
| | 3+4 2 | 3+4 2 | 3+4 2 | 3+4 2 | 2+2 2 | 2+2 2 |
| | 5+6 2 | 5+6 2 | 5+6 2 | 5+6 2 | 3+3 2 | 3+3 2 |
| | 3+4 2 | 3+4 2 | 5+6 2 | 5+6 2 | 2+2 2 | 2+2 2 |
| | 5+6 2 | 5+6 2 | 3+4 2 | 3+4 2 | 3+3 2 | 3+3 2 |
| | 3+4 2 | 3+4 2 | 5+6 2 | 3+4 2 | 2+2 2 | 2+2 2 |
| | 5+6 2 | 5+6 2 | 3+4 2 | 5+6 2 | 3+3 2 | 3+3 2 |
| | 3+4 2 | 3+4 2 | 5+6 2 | 3+4 2 | 2+2 2 | 3+3 2 |
| | 5+6 2 | 3+4 2 | 3+4 2 | 5+6 2 | 3+3 2 | 2+2 2 |
| | 3+4 2 | 5+6 2 | 5+6 2 | 5+6 2 | 2+2 2 | 2+2 2 |
| | 5+6 2 | 5+6 2 | 3+4 2 | 3+4 2 | 3+3 2 | 3+3 2 |
| | 3+4 2 | 5+6 2 | 5+6 2 | 5+6 2 | 2+2 2 | 2+2 2 |
| | 5+6 2 | 3+4 2 | 3+4 2 | 3+4 2 | 3+3 2 | 3+3 2 |
| | 3+4 2 | 3+4 2 | 5+6 2 | 3+4 2 | 2+2 2 | 3+3 2 |
| | 5+6 2 | 5+6 2 | 3+4 2 | 5+6 2 | 3+3 2 | 2+2 2 |
| | 5+6 2 | 3+4 2 | 5+6 2 | 3+4 2 | 2+2 2 | 3+3 2 |
| | 3+4 2 | 5+6 2 | 3+4 2 | 5+6 2 | 3+3 2 | 2+2 2 |
| | 5+6 2 | 3+4 2 | 3+4 2 | 5+6 2 | 3+3 2 | 2+2 2 |
| | 3+4 2 | 3+4 2 | 5+6 2 | 3+4 2 | 2+2 2 | 3+3 2 |
| | 3+4 2 | 5+6 2 | 3+4 2 | 3+4 2 | 2+2 2 | 3+3 2 |
| | 3+4 2 | 3+4 2 | 5+6 2 | 5+6 2 | 3+3 2 | 2+2 2 |
| | 5+6 2 | 3+4 2 | 5+6 2 | 5+6 2 | 3+3 2 | 3+3 2 |
| | 5+6 2 | 5+6 2 | 3+4 2 | 3+4 2 | 2+2 2 | 2+2 2 |
| | 3+4 2 | 5+6 2 | 5+6 2 | 5+6 2 | 2+2 2 | 3+3 2 |
| | 5+6 2 | 5+6 2 | 3+4 2 | 3+4 2 | 3+3 2 | 2+2 2 |
| | 3+4 2 | 5+6 2 | 5+6 2 | 5+6 2 | 2+2 2 | 3+3 2 |
| | 5+6 2 | 5+6 2 | 3+4 2 | 3+4 2 | 3+3 2 | 2+2 2 |
| | 5+6 2 | 3+4 2 | 3+4 2 | 5+6 2 | 3+3 2 | 2+2 2 |
| | 3+4 2 | 5+6 2 | 5+6 2 | 5+6 2 | 2+2 2 | 2+2 2 |
| | 3+4 2 | 5+6 2 | 3+4 2 | 3+4 2 | 2+2 2 | 3+3 2 |
| | 5+6 2 | 3+4 2 | 5+6 2 | 5+6 2 | 3+3 2 | 3+3 2 |
| | 3+4 2 | 3+4 2 | 3+4 2 | 3+4 2 | 2+2 2 | 2+2 2 |
| | 5+6 2 | 3+4 2 | 5+6 2 | 3+4 2 | 3+3 2 | 3+3 2 |
| | 5+6 2 | 3+4 2 | 3+4 2 | 3+4 2 | 2+2 2 | 3+3 2 |
| | 3+4 2 | 5+6 2 | 5+6 2 | 5+6 2 | 3+3 2 | 2+2 2 |
| | 5+6 2 | 5+6 2 | 3+4 2 | 3+4 2 | 2+2 2 | 2+2 2 |
| | 5+6 2 | 3+4 2 | 5+6 2 | 5+6 2 | 3+3 2 | 3+3 2 |
| | 3+4 2 | 5+6 2 | 3+4 2 | 3+4 2 | 3+3 2 | 3+3 2 |
| | 5+6 2 | 3+4 2 | 3+4 2 | 5+6 2 | 2+2 2 | 2+2 2 |
| | 3+4 2 | 3+4 2 | 5+6 2 | 5+6 2 | 3+3 2 | 2+2 2 |
| | 3+4 2 | 5+6 2 | 5+6 2 | 3+4 2 | 2+2 2 | 3+3 2 |
| | 5+6 2 | 3+4 2 | 3+4 2 | 3+4 2 | 2+2 2 | 2+2 2 |
| | 3+4 2 | 3+4 2 | 3+4 2 | 5+6 2 | 3+3 2 | 3+3 2 |
| | 5+6 2 | 5+6 2 | 5+6 2 | 3+4 2 | 3+3 2 | 3+3 2 |
| | 3+4 2 | 5+6 2 | 3+4 2 | 3+4 2 | 2+2 2 | 2+2 2 |
| | 3+4 2 | 5+6 2 | 5+6 2 | 5+6 2 | 3+3 2 | 2+2 2 |
| | 5+6 2 | 3+4 2 | 5+6 2 | 5+6 2 | 2+2 2 | 3+3 2 |
| | 3+4 2 | 3+4 2 | 5+6 2 | 3+4 2 | 3+3 2 | 2+2 2 |
| | 5+6 2 | 5+6 2 | 5+6 2 | 5+6 2 | 2+2 2 | 3+3 2 |
| | 3+4 2 | 5+6 2 | 3+4 2 | 5+6 2 | 3+3 2 | 2+2 2 |
| | 5+6 2 | 5+6 2 | 5+6 2 | 3+4 2 | 2+2 2 | 3+3 2 |
| | 3+4 2 | 3+4 2 | 3+4 2 | 3+4 2 | 2+2 2 | 2+2 2 |
| | 3+4 2 | 5+6 2 | 5+6 2 | 5+6 2 | 3+3 2 | 3+3 2 |
| | 5+6 2 | 3+4 2 | 3+4 2 | 5+6 2 | 2+2 2 | 3+3 2 |
| | 3+4 2 | 5+6 2 | 5+6 2 | 3+4 2 | 3+3 2 | 2+2 2 |
| | 5+6 2 | 3+4 2 | 5+6 2 | 5+6 2 | 2+2 2 | 2+2 2 |
| | 3+4 2 | 5+6 2 | 3+4 2 | 3+4 2 | 3+3 2 | 3+3 2 |
| | 5+6 2 | 3+4 2 | 3+4 2 | 3+4 2 | 2+2 2 | 2+2 2 |
| | 3+4 2 | 5+6 2 | 5+6 2 | 5+6 2 | 3+3 2 | 3+3 2 |
| | 5+6 2 | 3+4 2 | 3+4 2 | 5+6 2 | 3+3 2 | 2+2 2 |
| | 5+6 2 | 3+4 2 | 5+6 2 | 5+6 2 | 2+2 2 | 3+3 2 |
| | 5+6 2 | 3+4 2 | 3+4 2 | 3+4 2 | 3+3 2 | 2+2 2 |
| | 3+4 2 | 5+6 2 | 3+4 2 | 5+6 2 | 2+2 2 | 3+3 2 |
| | 3+4 2 | 5+6 2 | 5+6 2 | 3+4 2 | 3+3 2 | 2+2 2 |
| | 5+6 2 | 5+6 2 | 3+4 2 | 3+4 2 | 2+2 2 | 2+2 2 |
| | 5+6 2 | 3+4 2 | 5+6 2 | 3+4 2 | 3+3 2 | 3+3 2 |
| | 5+6 2 | 3+4 2 | 3+4 2 | 5+6 2 | 2+2 2 | 3+3 2 |
| | 3+4 2 | 3+4 2 | 3+4 2 | 3+4 2 | 3+3 2 | 2+2 2 |
| | 3+4 2 | 5+6 2 | 5+6 2 | 5+6 2 | 2+2 2 | 3+3 2 |
| | 5+6 2 | 3+4 2 | 3+4 2 | 3+4 2 | 3+3 2 | 2+2 2 |
| | 5+6 2 | 3+4 2 | 5+6 2 | 5+6 2 | 2+2 2 | 3+3 2 |
| | 3+4 2 | 5+6 2 | 5+6 2 | 3+4 2 | 3+3 2 | 2+2 2 |
| | 5+6 2 | 5+6 2 | 3+4 2 | 5+6 2 | 2+2 2 | 3+3 2 |
| | 3+4 2 | 3+4 2 | 3+4 2 | 3+4 2 | 3+3 2 | 3+3 2 |
| | 3+4 2 | 5+6 2 | 3+4 2 | 5+6 2 | 2+2 2 | 2+2 2 |
| | 5+6 2 | 3+4 2 | 5+6 2 | 3+4 2 | 3+3 2 | 3+3 2 |
| | 3+4 2 | 5+6 2 | 5+6 2 | 3+4 2 | 2+2 2 | 2+2 2 |
| | 5+6 2 | 3+4 2 | 3+4 2 | 5+6 2 | 2+2 2 | 3+3 2 |
| | 5+6 2 | 5+6 2 | 5+6 2 | 3+4 2 | 3+3 2 | 3+3 2 |
| | 3+4 2 | 5+6 2 | 3+4 2 | 3+4 2 | 2+2 2 | 2+2 2 |
| | 5+6 2 | 3+4 2 | 5+6 2 | 5+6 2 | 3+3 2 | 3+3 2 |
| | 3+4 2 | 5+6 2 | 5+6 2 | 5+6 2 | 2+2 2 | 2+2 2 |
| | 3+4 2 | 3+4 2 | 3+4 2 | 5+6 2 | 3+3 2 | 3+3 2 |
| | 5+6 2 | 3+4 2 | 5+6 2 | 5+6 2 | 2+2 2 | 2+2 2 |
| | 5+6 2 | 5+6 2 | 3+4 2 | 3+4 2 | 3+3 2 | 2+2 2 |
| | 3+4 2 | 3+4 2 | 3+4 2 | 3+4 2 | 3+3 2 | 3+3 2 |
| | 3+4 2 | 5+6 2 | 5+6 2 | 5+6 2 | 2+2 2 | 3+3 2 |
| | 5+6 2 | 5+6 2 | 3+4 2 | 3+4 2 | 3+3 2 | 2+2 2 |
| | 5+6 2 | 3+4 2 | 5+6 2 | 5+6 2 | 2+2 2 | 2+2 2 |
| | 3+4 2 | 3+4 2 | 3+4 2 | 3+4 2 | 3+3 2 | 3+3 2 |
| | 5+6 2 | 3+4 2 | 3+4 2 | 3+4 2 | 2+2 2 | 2+2 2 |
| | 3+4 2 | 5+6 2 | 3+4 2 | 5+6 2 | 2+2 2 | 3+3 2 |
| | 5+6 2 | 3+4 2 | 5+6 2 | 3+4 2 | 3+3 2 | 2+2 2 |
| | 5+6 2 | 3+4 2 | 3+4 2 | 5+6 2 | 2+2 2 | 3+3 2 |
| | 3+4 2 | 3+4 2 | 5+6 2 | 5+6 2 | 3+3 2 | 3+3 2 |
| | 5+6 2 | 5+6 2 | 3+4 2 | 3+4 2 | 3+3 2 | 2+2 2 |
| | 3+4 2 | 5+6 2 | 5+6 2 | 3+4 2 | 2+2 2 | 2+2 2 |
| | 3+4 2 | 3+4 2 | 3+4 2 | 3+4 2 | 3+3 2 | 3+3 2 |
| | 5+6 2 | 3+4 2 | 5+6 2 | 5+6 2 | 2+2 2 | 2+2 2 |
| | 3+4 2 | 5+6 2 | 5+6 2 | 3+4 2 | 3+3 2 | 3+3 2 |
| | 5+6 2 | 3+4 2 | 3+4 2 | 5+6 2 | 2+2 2 | 2+2 2 |

Continued on next column

**TABLE XXIV– continued**

| TQC | ICSD | NoE | SF | | | |
|---|---|---|---|---|---|---|
| | 3+4 2 | 5+6 2 | 5+6 2 | 5+6 2 | 3+3 2 | 3+3 2 |
| | 5+6 2 | 3+4 2 | 3+4 2 | 3+4 2 | 2+2 2 | 2+2 2 |
| | 3+4 2 | 5+6 2 | 5+6 2 | 5+6 2 | 2+2 2 | 2+2 2 |
| | 3+4 2 | 3+4 2 | 3+4 2 | 5+6 2 | 3+3 2 | 3+3 2 |
| | 5+6 2 | 3+4 2 | 5+6 2 | 5+6 2 | 2+2 2 | 2+2 2 |
| | 3+4 2 | 5+6 2 | 3+4 2 | 3+4 2 | 3+3 2 | 3+3 2 |
| | 5+6 2 | 3+4 2 | 3+4 2 | 5+6 2 | 3+3 2 | 2+2 2 |
| | 3+4 2 | 5+6 2 | 5+6 2 | 5+6 2 | 2+2 2 | 3+3 2 |
| | 5+6 2 | 3+4 2 | 5+6 2 | 3+4 2 | 2+2 2 | 2+2 2 |
| | 3+4 2 | 5+6 2 | 5+6 2 | 5+6 2 | 2+2 2 | 3+3 2 |
| | 5+6 2 | 3+4 2 | 5+6 2 | 5+6 2 | 3+3 2 | 2+2 2 |
| | 5+6 2 | 5+6 2 | 3+4 2 | 3+4 2 | 3+3 2 | 2+2 2 |
| | 3+4 2 | 5+6 2 | 3+4 2 | 3+4 2 | 2+2 2 | 3+3 2 |
| | 3+4 2 | 3+4 2 | 3+4 2 | 3+4 2 | 3+3 2 | 3+3 2 |
| | 3+4 2 | 5+6 2 | 5+6 2 | 5+6 2 | 2+2 2 | 2+2 2 |
| | 5+6 2 | 5+6 2 | 3+4 2 | 3+4 2 | 3+3 2 | 3+3 2 |
| | 3+4 2 | 3+4 2 | 5+6 2 | 5+6 2 | 2+2 2 | 2+2 2 |
| | 3+4 2 | 5+6 2 | 5+6 2 | 3+4 2 | 3+3 2 | 3+3 2 |
| | 5+6 2 | 3+4 2 | 5+6 2 | 3+4 2 | 2+2 2 | 2+2 2 |
| | 3+4 2 | 5+6 2 | 5+6 2 | 5+6 2 | 3+3 2 | 3+3 2 |
| | 5+6 2 | 5+6 2 | 3+4 2 | 3+4 2 | 2+2 2 | 3+3 2 |
| | 5+6 2 | 3+4 2 | 5+6 2 | 5+6 2 | 3+3 2 | 3+3 2 |
| | 5+6 2 | 3+4 2 | 3+4 2 | 3+4 2 | 3+3 2 | 2+2 2 |
| | 3+4 2 | 5+6 2 | 3+4 2 | 5+6 2 | 2+2 2 | 3+3 2 |
| | 3+4 2 | 3+4 2 | 3+4 2 | 3+4 2 | 2+2 2 | 2+2 2 |
| | 5+6 2 | 5+6 2 | 5+6 2 | 3+4 2 | 3+3 2 | 3+3 2 |
| | 3+4 2 | 5+6 2 | 3+4 2 | 5+6 2 | 2+2 2 | 2+2 2 |
| | 5+6 2 | 3+4 2 | 5+6 2 | 5+6 2 | 3+3 2 | 3+3 2 |
| | 3+4 2 | 5+6 2 | 5+6 2 | 3+4 2 | 2+2 2 | 2+2 2 |
| | 5+6 2 | 3+4 2 | 3+4 2 | 3+4 2 | 3+3 2 | 2+2 2 |
| | 3+4 2 | 3+4 2 | 5+6 2 | 5+6 2 | 2+2 2 | 2+2 2 |
| | 5+6 2 | 5+6 2 | 5+6 2 | 5+6 2 | 3+3 2 | 3+3 2 |
| | 3+4 2 | 5+6 2 | 5+6 2 | 5+6 2 | 2+2 2 | 2+2 2 |
| | 5+6 2 | 3+4 2 | 3+4 2 | 3+4 2 | 3+3 2 | 3+3 2 |
| | 3+4 2 | 5+6 2 | 3+4 2 | 5+6 2 | 2+2 2 | 3+3 2 |
| | 5+6 2 | 5+6 2 | 3+4 2 | 5+6 2 | 3+3 2 | 2+2 2 |
| | 3+4 2 | 5+6 2 | 5+6 2 | 3+4 2 | 2+2 2 | 3+3 2 |
| | 5+6 2 | 3+4 2 | 3+4 2 | 5+6 2 | 3+3 2 | 2+2 2 |
| | 3+4 2 | 3+4 2 | 3+4 2 | 3+4 2 | 2+2 2 | 2+2 2 |
| | 5+6 2 | 3+4 2 | 5+6 2 | 3+4 2 | 3+3 2 | 3+3 2 |
| 12.2.169 | 418886 | 242 | Ba11Cd6Sb12 | | | |
| | -A | -GM | -M | -Y | -L | -V |
| | 5+6 2 | 3+4 2 | 3+4 2 | 5+6 2 | 2+2 2 | 2+2 2 |
| | 3+4 2 | 5+6 2 | 5+6 2 | 3+4 2 | 3+3 2 | 3+3 2 |
| | 3+4 2 | 3+4 2 | 3+4 2 | 3+4 2 | 2+2 2 | 2+2 2 |
| | 5+6 2 | 5+6 2 | 5+6 2 | 5+6 2 | 3+3 2 | 3+3 2 |
| | 3+4 2 | 3+4 2 | 5+6 2 | 5+6 2 | 2+2 2 | 2+2 2 |
| | 5+6 2 | 5+6 2 | 3+4 2 | 3+4 2 | 3+3 2 | 3+3 2 |
| | 5+6 2 | 3+4 2 | 5+6 2 | 3+4 2 | 2+2 2 | 3+3 2 |
| | 3+4 2 | 5+6 2 | 3+4 2 | 5+6 2 | 3+3 2 | 2+2 2 |
| | 3+4 2 | 5+6 2 | 3+4 2 | 5+6 2 | 3+3 2 | 2+2 2 |
| | 5+6 2 | 3+4 2 | 5+6 2 | 3+4 2 | 2+2 2 | 3+3 2 |
| | 3+4 2 | 3+4 2 | 3+4 2 | 3+4 2 | 3+3 2 | 3+3 2 |
| | 3+4 2 | 3+4 2 | 5+6 2 | 5+6 2 | 2+2 2 | 2+2 2 |
| | 5+6 2 | 3+4 2 | 3+4 2 | 5+6 2 | 3+3 2 | 3+3 2 |
| | 3+4 2 | 5+6 2 | 5+6 2 | 3+4 2 | 2+2 2 | 2+2 2 |
| | 5+6 2 | 5+6 2 | 3+4 2 | 3+4 2 | 3+3 2 | 3+3 2 |
| | 3+4 2 | 3+4 2 | 5+6 2 | 5+6 2 | 2+2 2 | 2+2 2 |
| | 5+6 2 | 5+6 2 | 3+4 2 | 3+4 2 | 3+3 2 | 3+3 2 |
| | 5+6 2 | 5+6 2 | 5+6 2 | 5+6 2 | 3+3 2 | 2+2 2 |
| | 3+4 2 | 3+4 2 | 3+4 2 | 3+4 2 | 2+2 2 | 3+3 2 |
| | 5+6 2 | 3+4 2 | 5+6 2 | 3+4 2 | 2+2 2 | 3+3 2 |
| | 3+4 2 | 5+6 2 | 3+4 2 | 5+6 2 | 3+3 2 | 2+2 2 |
| | 5+6 2 | 5+6 2 | 5+6 2 | 5+6 2 | 2+2 2 | 2+2 2 |
| | 3+4 2 | 3+4 2 | 5+6 2 | 5+6 2 | 2+2 2 | 2+2 2 |
| | 5+6 2 | 3+4 2 | 5+6 2 | 5+6 2 | 3+3 2 | 3+3 2 |
| | 3+4 2 | 5+6 2 | 3+4 2 | 3+4 2 | 2+2 2 | 2+2 2 |
| | 5+6 2 | 5+6 2 | 3+4 2 | 3+4 2 | 3+3 2 | 3+3 2 |
| | 3+4 2 | 3+4 2 | 5+6 2 | 5+6 2 | 2+2 2 | 2+2 2 |
| | 5+6 2 | 5+6 2 | 5+6 2 | 3+4 2 | 3+3 2 | 3+3 2 |
| | 5+6 2 | 3+4 2 | 3+4 2 | 5+6 2 | 2+2 2 | 2+2 2 |
| | 3+4 2 | 3+4 2 | 5+6 2 | 5+6 2 | 3+3 2 | 3+3 2 |
| | 5+6 2 | 5+6 2 | 3+4 2 | 3+4 2 | 3+3 2 | 2+2 2 |
| | 3+4 2 | 3+4 2 | 5+6 2 | 5+6 2 | 2+2 2 | 2+2 2 |
| | 5+6 2 | 5+6 2 | 3+4 2 | 3+4 2 | 3+3 2 | 3+3 2 |
| | 5+6 2 | 3+4 2 | 3+4 2 | 3+4 2 | 2+2 2 | 3+3 2 |
| | 3+4 2 | 5+6 2 | 3+4 2 | 5+6 2 | 2+2 2 | 2+2 2 |
| | 3+4 2 | 5+6 2 | 5+6 2 | 3+4 2 | 3+3 2 | 3+3 2 |
| | 5+6 2 | 3+4 2 | 5+6 2 | 3+4 2 | 2+2 2 | 2+2 2 |
| | 5+6 2 | 5+6 2 | 5+6 2 | 5+6 2 | 3+3 2 | 3+3 2 |
| | 3+4 2 | 5+6 2 | 3+4 2 | 5+6 2 | 2+2 2 | 2+2 2 |
| | 3+4 2 | 3+4 2 | 3+4 2 | 3+4 2 | 3+3 2 | 2+2 2 |
| | 5+6 2 | 5+6 2 | 5+6 2 | 5+6 2 | 2+2 2 | 3+3 2 |
| | 3+4 2 | 3+4 2 | 3+4 2 | 3+4 2 | 2+2 2 | 3+3 2 |
| | 5+6 2 | 5+6 2 | 5+6 2 | 5+6 2 | 3+3 2 | 2+2 2 |
| | 5+6 2 | 5+6 2 | 5+6 2 | 5+6 2 | 2+2 2 | 2+2 2 |
| | 5+6 2 | 3+4 2 | 5+6 2 | 3+4 2 | 3+3 2 | 2+2 2 |
| | 3+4 2 | 3+4 2 | 5+6 2 | 3+4 2 | 2+2 2 | 3+3 2 |
| | 5+6 2 | 5+6 2 | 3+4 2 | 5+6 2 | 3+3 2 | 2+2 2 |
| | 3+4 2 | 3+4 2 | 3+4 2 | 3+4 2 | 2+2 2 | 3+3 2 |
| | 5+6 2 | 5+6 2 | 3+4 2 | 5+6 2 | 3+3 2 | 2+2 2 |
| | 3+4 2 | 5+6 2 | 5+6 2 | 3+4 2 | 3+3 2 | 3+3 2 |
| | 5+6 2 | 3+4 2 | 5+6 2 | 5+6 2 | 2+2 2 | 2+2 2 |
| | 3+4 2 | 3+4 2 | 3+4 2 | 3+4 2 | 2+2 2 | 3+3 2 |
| | 5+6 2 | 3+4 2 | 5+6 2 | 5+6 2 | 3+3 2 | 2+2 2 |
| | 3+4 2 | 5+6 2 | 3+4 2 | 5+6 2 | 2+2 2 | 3+3 2 |
| | 5+6 2 | 5+6 2 | 5+6 2 | 3+4 2 | 3+3 2 | 2+2 2 |
| | 3+4 2 | 3+4 2 | 3+4 2 | 5+6 2 | 2+2 2 | 3+3 2 |
| | 5+6 2 | 5+6 2 | 5+6 2 | 3+4 2 | 3+3 2 | 2+2 2 |
| | 3+4 2 | 3+4 2 | 3+4 2 | 5+6 2 | 2+2 2 | 3+3 2 |
| | 5+6 2 | 5+6 2 | 5+6 2 | 3+4 2 | 3+3 2 | 2+2 2 |
| | 5+6 2 | 3+4 2 | 3+4 2 | 5+6 2 | 2+2 2 | 3+3 2 |
| | 3+4 2 | 5+6 2 | 5+6 2 | 3+4 2 | 3+3 2 | 2+2 2 |
| | 3+4 2 | 3+4 2 | 3+4 2 | 3+4 2 | 3+3 2 | 3+3 2 |
| | 5+6 2 | 5+6 2 | 5+6 2 | 5+6 2 | 2+2 2 | 2+2 2 |
| | 3+4 2 | 3+4 2 | 3+4 2 | 5+6 2 | 2+2 2 | 3+3 2 |
| | 5+6 2 | 5+6 2 | 3+4 2 | 3+4 2 | 3+3 2 | 2+2 2 |
| | 3+4 2 | 3+4 2 | 5+6 2 | 5+6 2 | 2+2 2 | 2+2 2 |
| | 5+6 2 | 5+6 2 | 5+6 2 | 3+4 2 | 3+3 2 | 3+3 2 |
| | 3+4 2 | 5+6 2 | 3+4 2 | 5+6 2 | 2+2 2 | 3+3 2 |
| | 5+6 2 | 3+4 2 | 5+6 2 | 3+4 2 | 3+3 2 | 2+2 2 |
| | 3+4 2 | 5+6 2 | 3+4 2 | 5+6 2 | 2+2 2 | 3+3 2 |
| | 3+4 2 | 3+4 2 | 5+6 2 | 3+4 2 | 3+3 2 | 2+2 2 |
| | 5+6 2 | 5+6 2 | 3+4 2 | 5+6 2 | 2+2 2 | 3+3 2 |
| | 5+6 2 | 3+4 2 | 5+6 2 | 3+4 2 | 3+3 2 | 2+2 2 |
| | 3+4 2 | 5+6 2 | 3+4 2 | 5+6 2 | 2+2 2 | 3+3 2 |
| | 5+6 2 | 3+4 2 | 5+6 2 | 3+4 2 | 3+3 2 | 2+2 2 |
| | 3+4 2 | 3+4 2 | 3+4 2 | 5+6 2 | 3+3 2 | 3+3 2 |
| | 5+6 2 | 5+6 2 | 5+6 2 | 3+4 2 | 2+2 2 | 2+2 2 |
| | 3+4 2 | 3+4 2 | 3+4 2 | 3+4 2 | 3+3 2 | 3+3 2 |
| | 5+6 2 | 5+6 2 | 3+4 2 | 5+6 2 | 2+2 2 | 2+2 2 |
| | 3+4 2 | 5+6 2 | 5+6 2 | 5+6 2 | 2+2 2 | 3+3 2 |

Continued on next column

**TABLE XXIV– continued**

| TQC | ICSD | NoE | SF | | | |
|---|---|---|---|---|---|---|
| | 3+4 2 | 3+4 2 | 5+6 2 | 3+4 2 | 3+3 2 | 2+2 2 |
| | 5+6 2 | 5+6 2 | 5+6 2 | 5+6 2 | 2+2 2 | 3+3 2 |
| | 5+6 2 | 3+4 2 | 3+4 2 | 3+4 2 | 3+3 2 | 2+2 2 |
| | 3+4 2 | 3+4 2 | 3+4 2 | 3+4 2 | 2+2 2 | 3+3 2 |
| | 5+6 2 | 5+6 2 | 5+6 2 | 5+6 2 | 3+3 2 | 3+3 2 |
| | 3+4 2 | 5+6 2 | 3+4 2 | 5+6 2 | 2+2 2 | 2+2 2 |
| | 3+4 2 | 3+4 2 | 3+4 2 | 3+4 2 | 3+3 2 | 2+2 2 |
| | 5+6 2 | 5+6 2 | 5+6 2 | 5+6 2 | 2+2 2 | 3+3 2 |
| | 3+4 2 | 5+6 2 | 3+4 2 | 5+6 2 | 3+3 2 | 2+2 2 |
| | 5+6 2 | 3+4 2 | 5+6 2 | 3+4 2 | 2+2 2 | 3+3 2 |
| | 3+4 2 | 5+6 2 | 3+4 2 | 3+4 2 | 3+3 2 | 3+3 2 |
| | 5+6 2 | 3+4 2 | 5+6 2 | 5+6 2 | 2+2 2 | 2+2 2 |
| | 5+6 2 | 3+4 2 | 5+6 2 | 3+4 2 | 2+2 2 | 3+3 2 |
| | 5+6 2 | 5+6 2 | 3+4 2 | 5+6 2 | 3+3 2 | 2+2 2 |
| | 3+4 2 | 3+4 2 | 3+4 2 | 3+4 2 | 3+3 2 | 2+2 2 |
| | 3+4 2 | 3+4 2 | 5+6 2 | 3+4 2 | 2+2 2 | 3+3 2 |
| | 3+4 2 | 5+6 2 | 3+4 2 | 5+6 2 | 3+3 2 | 3+3 2 |
| | 5+6 2 | 3+4 2 | 5+6 2 | 5+6 2 | 3+3 2 | 2+2 2 |
| | 5+6 2 | 5+6 2 | 3+4 2 | 3+4 2 | 2+2 2 | 3+3 2 |
| | 3+4 2 | 3+4 2 | 5+6 2 | 5+6 2 | 2+2 2 | 2+2 2 |
| | 3+4 2 | 5+6 2 | 3+4 2 | 3+4 2 | 3+3 2 | 3+3 2 |
| | 5+6 2 | 5+6 2 | 5+6 2 | 5+6 2 | 2+2 2 | 3+3 2 |
| | 3+4 2 | 3+4 2 | 5+6 2 | 3+4 2 | 3+3 2 | 2+2 2 |
| | 5+6 2 | 5+6 2 | 3+4 2 | 5+6 2 | 2+2 2 | 3+3 2 |
| | 3+4 2 | 3+4 2 | 3+4 2 | 3+4 2 | 3+3 2 | 2+2 2 |
| | 3+4 2 | 5+6 2 | 3+4 2 | 5+6 2 | 3+3 2 | 3+3 2 |
| | 5+6 2 | 3+4 2 | 5+6 2 | 3+4 2 | 2+2 2 | 2+2 2 |
| | 3+4 2 | 5+6 2 | 3+4 2 | 5+6 2 | 2+2 2 | 2+2 2 |
| | 5+6 2 | 3+4 2 | 5+6 2 | 3+4 2 | 3+3 2 | 2+2 2 |
| | 5+6 2 | 5+6 2 | 3+4 2 | 5+6 2 | 2+2 2 | 3+3 2 |
| | 3+4 2 | 5+6 2 | 3+4 2 | 3+4 2 | 3+3 2 | 3+3 2 |
| | 5+6 2 | 3+4 2 | 5+6 2 | 5+6 2 | 2+2 2 | 2+2 2 |
| | 3+4 2 | 3+4 2 | 3+4 2 | 3+4 2 | 3+3 2 | 2+2 2 |
| | 5+6 2 | 3+4 2 | 5+6 2 | 5+6 2 | 3+3 2 | 3+3 2 |
| | 3+4 2 | 3+4 2 | 3+4 2 | 3+4 2 | 2+2 2 | 2+2 2 |
| | 5+6 2 | 5+6 2 | 5+6 2 | 5+6 2 | 3+3 2 | 3+3 2 |
| | 3+4 2 | 5+6 2 | 5+6 2 | 3+4 2 | 2+2 2 | 2+2 2 |
| | 5+6 2 | 3+4 2 | 5+6 2 | 5+6 2 | 3+3 2 | 3+3 2 |
| | 3+4 2 | 5+6 2 | 3+4 2 | 3+4 2 | 2+2 2 | 2+2 2 |
| | 5+6 2 | 3+4 2 | 5+6 2 | 5+6 2 | 3+3 2 | 3+3 2 |
| | 5+6 2 | 3+4 2 | 5+6 2 | 3+4 2 | 2+2 2 | 2+2 2 |
| 12.2.170 | 63501 | 84 | B4La2Ni5 | | | |
| | -A | -GM | -M | -Y | -L | -V |
| | 5+6 2 | 3+4 2 | 3+4 2 | 5+6 2 | 3+3 2 | 2+2 2 |
| | 3+4 2 | 5+6 2 | 5+6 2 | 3+4 2 | 2+2 2 | 3+3 2 |
| | 5+6 2 | 3+4 2 | 3+4 2 | 5+6 2 | 3+3 2 | 2+2 2 |
| | 3+4 2 | 5+6 2 | 5+6 2 | 3+4 2 | 2+2 2 | 3+3 2 |
| | 5+6 2 | 3+4 2 | 3+4 2 | 5+6 2 | 3+3 2 | 2+2 2 |
| | 5+6 2 | 3+4 2 | 5+6 2 | 3+4 2 | 2+2 2 | 3+3 2 |
| | 3+4 2 | 5+6 2 | 3+4 2 | 5+6 2 | 3+3 2 | 2+2 2 |
| | 3+4 2 | 5+6 2 | 5+6 2 | 3+4 2 | 2+2 2 | 3+3 2 |
| | 3+4 2 | 3+4 2 | 3+4 2 | 3+4 2 | 2+2 2 | 2+2 2 |
| | 5+6 2 | 5+6 2 | 5+6 2 | 5+6 2 | 3+3 2 | 3+3 2 |
| | 3+4 2 | 3+4 2 | 3+4 2 | 3+4 2 | 2+2 2 | 2+2 2 |
| | 5+6 2 | 5+6 2 | 5+6 2 | 5+6 2 | 3+3 2 | 3+3 2 |
| | 3+4 2 | 3+4 2 | 5+6 2 | 5+6 2 | 3+3 2 | 3+3 2 |
| | 5+6 2 | 5+6 2 | 5+6 2 | 3+4 2 | 2+2 2 | 3+3 2 |
| | 5+6 2 | 3+4 2 | 3+4 2 | 5+6 2 | 2+2 2 | 2+2 2 |
| | 5+6 2 | 3+4 2 | 3+4 2 | 3+4 2 | 3+3 2 | 2+2 2 |
| | 3+4 2 | 5+6 2 | 3+4 2 | 3+4 2 | 2+2 2 | 3+3 2 |
| | 3+4 2 | 3+4 2 | 5+6 2 | 5+6 2 | 2+2 2 | 2+2 2 |

Continued on next column

**TABLE XXIV– continued**

| TQC | ICSD | NoE | SF | | | |
|---|---|---|---|---|---|---|
| | 5+6 2 | 3+4 2 | 3+4 2 | 3+4 2 | 3+3 2 | 3+3 2 |
| | 3+4 2 | 5+6 2 | 5+6 2 | 3+4 2 | 3+3 2 | 2+2 2 |
| | 3+4 2 | 5+6 2 | 5+6 2 | 3+4 2 | 2+2 2 | 3+3 2 |
| | 5+6 2 | 3+4 2 | 5+6 2 | 5+6 2 | 3+3 2 | 3+3 2 |
| | 3+4 2 | 5+6 2 | 3+4 2 | 3+4 2 | 2+2 2 | 2+2 2 |
| | 3+4 2 | 3+4 2 | 3+4 2 | 5+6 2 | 2+2 2 | 3+3 2 |
| | 5+6 2 | 5+6 2 | 5+6 2 | 3+4 2 | 3+3 2 | 2+2 2 |
| | 5+6 2 | 3+4 2 | 5+6 2 | 5+6 2 | 2+2 2 | 3+3 2 |
| | 5+6 2 | 3+4 2 | 5+6 2 | 3+4 2 | 3+3 2 | 3+3 2 |
| | 3+4 2 | 5+6 2 | 5+6 2 | 3+4 2 | 2+2 2 | 2+2 2 |
| | 5+6 2 | 3+4 2 | 3+4 2 | 5+6 2 | 3+3 2 | 3+3 2 |
| | 3+4 2 | 5+6 2 | 3+4 2 | 5+6 2 | 2+2 2 | 2+2 2 |
| | 5+6 2 | 5+6 2 | 5+6 2 | 5+6 2 | 2+2 2 | 2+2 2 |
| | 3+4 2 | 3+4 2 | 3+4 2 | 3+4 2 | 3+3 2 | 3+3 2 |
| | 5+6 2 | 5+6 2 | 5+6 2 | 3+4 2 | 2+2 2 | 3+3 2 |
| | 3+4 2 | 3+4 2 | 3+4 2 | 5+6 2 | 3+3 2 | 2+2 2 |
| | 5+6 2 | 5+6 2 | 5+6 2 | 3+4 2 | 2+2 2 | 2+2 2 |
| | 5+6 2 | 3+4 2 | 5+6 2 | 5+6 2 | 3+3 2 | 3+3 2 |
| | 3+4 2 | 3+4 2 | 3+4 2 | 5+6 2 | 2+2 2 | 2+2 2 |
| | 5+6 2 | 3+4 2 | 5+6 2 | 3+4 2 | 2+2 2 | 3+3 2 |
| | 3+4 2 | 3+4 2 | 5+6 2 | 3+4 2 | 2+2 2 | 3+3 2 |
| | 5+6 2 | 5+6 2 | 3+4 2 | 3+4 2 | 3+3 2 | 3+3 2 |
| | 5+6 2 | 3+4 2 | 5+6 2 | 5+6 2 | 2+2 2 | 3+3 2 |
| | 3+4 2 | 3+4 2 | 5+6 2 | 3+4 2 | 3+3 2 | 2+2 2 |
| 12.2.171 | 170608 | 84 | B4La2Ni5 | | | |
| | -A | -GM | -M | -Y | -L | -V |
| | 5+6 2 | 3+4 2 | 3+4 2 | 5+6 2 | 3+3 2 | 2+2 2 |
| | 3+4 2 | 5+6 2 | 5+6 2 | 3+4 2 | 2+2 2 | 3+3 2 |
| | 5+6 2 | 3+4 2 | 3+4 2 | 5+6 2 | 3+3 2 | 2+2 2 |
| | 3+4 2 | 5+6 2 | 5+6 2 | 3+4 2 | 2+2 2 | 3+3 2 |
| | 5+6 2 | 3+4 2 | 3+4 2 | 5+6 2 | 3+3 2 | 2+2 2 |
| | 5+6 2 | 3+4 2 | 5+6 2 | 3+4 2 | 2+2 2 | 3+3 2 |
| | 3+4 2 | 5+6 2 | 3+4 2 | 5+6 2 | 3+3 2 | 2+2 2 |
| | 3+4 2 | 5+6 2 | 5+6 2 | 3+4 2 | 2+2 2 | 3+3 2 |
| | 3+4 2 | 3+4 2 | 3+4 2 | 3+4 2 | 2+2 2 | 2+2 2 |
| | 5+6 2 | 5+6 2 | 5+6 2 | 5+6 2 | 3+3 2 | 3+3 2 |
| | 3+4 2 | 3+4 2 | 3+4 2 | 3+4 2 | 2+2 2 | 2+2 2 |
| | 5+6 2 | 5+6 2 | 5+6 2 | 5+6 2 | 3+3 2 | 3+3 2 |
| | 3+4 2 | 3+4 2 | 5+6 2 | 5+6 2 | 3+3 2 | 3+3 2 |
| | 5+6 2 | 5+6 2 | 5+6 2 | 3+4 2 | 2+2 2 | 3+3 2 |
| | 5+6 2 | 3+4 2 | 3+4 2 | 5+6 2 | 2+2 2 | 2+2 2 |
| | 5+6 2 | 3+4 2 | 3+4 2 | 3+4 2 | 3+3 2 | 2+2 2 |
| | 3+4 2 | 5+6 2 | 3+4 2 | 3+4 2 | 2+2 2 | 3+3 2 |
| | 3+4 2 | 3+4 2 | 5+6 2 | 5+6 2 | 2+2 2 | 2+2 2 |
| | 5+6 2 | 3+4 2 | 3+4 2 | 3+4 2 | 3+3 2 | 3+3 2 |
| | 3+4 2 | 5+6 2 | 5+6 2 | 3+4 2 | 3+3 2 | 2+2 2 |
| | 3+4 2 | 5+6 2 | 5+6 2 | 3+4 2 | 2+2 2 | 3+3 2 |
| | 5+6 2 | 3+4 2 | 5+6 2 | 5+6 2 | 3+3 2 | 3+3 2 |
| | 3+4 2 | 5+6 2 | 3+4 2 | 3+4 2 | 2+2 2 | 2+2 2 |
| | 3+4 2 | 3+4 2 | 3+4 2 | 5+6 2 | 2+2 2 | 3+3 2 |
| | 5+6 2 | 5+6 2 | 5+6 2 | 3+4 2 | 3+3 2 | 2+2 2 |
| | 5+6 2 | 3+4 2 | 5+6 2 | 5+6 2 | 2+2 2 | 3+3 2 |
| | 5+6 2 | 3+4 2 | 5+6 2 | 3+4 2 | 3+3 2 | 3+3 2 |
| | 3+4 2 | 5+6 2 | 5+6 2 | 3+4 2 | 2+2 2 | 2+2 2 |
| | 5+6 2 | 3+4 2 | 3+4 2 | 5+6 2 | 3+3 2 | 3+3 2 |
| | 3+4 2 | 5+6 2 | 3+4 2 | 5+6 2 | 2+2 2 | 2+2 2 |
| | 5+6 2 | 5+6 2 | 5+6 2 | 3+4 2 | 2+2 2 | 2+2 2 |
| | 3+4 2 | 3+4 2 | 3+4 2 | 5+6 2 | 3+3 2 | 3+3 2 |
| | 5+6 2 | 5+6 2 | 5+6 2 | 3+4 2 | 2+2 2 | 3+3 2 |
| | 3+4 2 | 3+4 2 | 3+4 2 | 5+6 2 | 3+3 2 | 2+2 2 |
| | 5+6 2 | 5+6 2 | 5+6 2 | 3+4 2 | 2+2 2 | 2+2 2 |

Continued on next column

**TABLE XXIV– continued**

| TQC | ICSD | NoE | SF | | | |
|---|---|---|---|---|---|---|
| | 5+6 2 | 3+4 2 | 5+6 2 | 5+6 2 | 3+3 2 | 3+3 2 |
| | 3+4 2 | 3+4 2 | 3+4 2 | 5+6 2 | 2+2 2 | 2+2 2 |
| | 5+6 2 | 3+4 2 | 5+6 2 | 3+4 2 | 2+2 2 | 3+3 2 |
| | 3+4 2 | 3+4 2 | 5+6 2 | 3+4 2 | 2+2 2 | 3+3 2 |
| | 5+6 2 | 5+6 2 | 3+4 2 | 3+4 2 | 3+3 2 | 3+3 2 |
| | 5+6 2 | 3+4 2 | 5+6 2 | 5+6 2 | 2+2 2 | 3+3 2 |
| | 3+4 2 | 3+4 2 | 5+6 2 | 3+4 2 | 3+3 2 | 2+2 2 |
| 12.2.172 | 615319 | 52 | B2Rh3Tb1 | | | |
| | -A | -GM | -M | -Y | -L | -V |
| | 3+4 2 | 3+4 2 | 3+4 2 | 3+4 2 | 2+2 2 | 2+2 2 |
| | 5+6 2 | 5+6 2 | 5+6 2 | 5+6 2 | 3+3 2 | 3+3 2 |
| | 5+6 2 | 5+6 2 | 5+6 2 | 5+6 2 | 3+3 2 | 3+3 2 |
| | 5+6 2 | 5+6 2 | 5+6 2 | 5+6 2 | 3+3 2 | 3+3 2 |
| | 5+6 2 | 3+4 2 | 3+4 2 | 5+6 2 | 2+2 2 | 3+3 2 |
| | 5+6 2 | 3+4 2 | 5+6 2 | 3+4 2 | 3+3 2 | 2+2 2 |
| | 3+4 2 | 3+4 2 | 3+4 2 | 5+6 2 | 2+2 2 | 3+3 2 |
| | 5+6 2 | 5+6 2 | 3+4 2 | 5+6 2 | 2+2 2 | 3+3 2 |
| | 5+6 2 | 3+4 2 | 5+6 2 | 5+6 2 | 3+3 2 | 3+3 2 |
| | 5+6 2 | 3+4 2 | 3+4 2 | 3+4 2 | 2+2 2 | 3+3 2 |
| | 5+6 2 | 3+4 2 | 3+4 2 | 5+6 2 | 2+2 2 | 2+2 2 |
| | 5+6 2 | 3+4 2 | 5+6 2 | 3+4 2 | 3+3 2 | 2+2 2 |
| | 5+6 2 | 3+4 2 | 3+4 2 | 3+4 2 | 2+2 2 | 2+2 2 |
| | 3+4 2 | 3+4 2 | 3+4 2 | 3+4 2 | 2+2 2 | 2+2 2 |
| | 5+6 2 | 3+4 2 | 5+6 2 | 5+6 2 | 3+3 2 | 3+3 2 |
| | 3+4 2 | 3+4 2 | 5+6 2 | 5+6 2 | 3+3 2 | 3+3 2 |
| | 5+6 2 | 3+4 2 | 5+6 2 | 5+6 2 | 2+2 2 | 3+3 2 |
| | 3+4 2 | 3+4 2 | 3+4 2 | 3+4 2 | 3+3 2 | 2+2 2 |
| | 5+6 2 | 3+4 2 | 3+4 2 | 5+6 2 | 2+2 2 | 3+3 2 |
| | 5+6 2 | 3+4 2 | 3+4 2 | 3+4 2 | 2+2 2 | 2+2 2 |
| | 5+6 2 | 5+6 2 | 5+6 2 | 5+6 2 | 3+3 2 | 3+3 2 |
| | 5+6 2 | 5+6 2 | 5+6 2 | 5+6 2 | 3+3 2 | 3+3 2 |
| | 5+6 2 | 5+6 2 | 5+6 2 | 5+6 2 | 3+3 2 | 3+3 2 |
| | 5+6 2 | 5+6 2 | 3+4 2 | 5+6 2 | 2+2 2 | 3+3 2 |
| | 5+6 2 | 5+6 2 | 5+6 2 | 5+6 2 | 3+3 2 | 3+3 2 |
| | 3+4 2 | 5+6 2 | 3+4 2 | 5+6 2 | 2+2 2 | 3+3 2 |
| 12.2.173 | 614701 | 58 | B2Lu1Rh3 | | | |
| | -A | -GM | -M | -Y | -L | -V |
| | 5+6 2 | 3+4 2 | 5+6 2 | 3+4 2 | 3+3 2 | 2+2 2 |
| | 3+4 2 | 5+6 2 | 3+4 2 | 5+6 2 | 2+2 2 | 3+3 2 |
| | 3+4 2 | 5+6 2 | 3+4 2 | 5+6 2 | 2+2 2 | 3+3 2 |
| | 3+4 2 | 5+6 2 | 3+4 2 | 5+6 2 | 2+2 2 | 3+3 2 |
| | 3+4 2 | 3+4 2 | 5+6 2 | 5+6 2 | 3+3 2 | 3+3 2 |
| | 3+4 2 | 3+4 2 | 3+4 2 | 3+4 2 | 2+2 2 | 2+2 2 |
| | 5+6 2 | 3+4 2 | 5+6 2 | 5+6 2 | 3+3 2 | 3+3 2 |
| | 3+4 2 | 5+6 2 | 3+4 2 | 5+6 2 | 2+2 2 | 3+3 2 |
| | 3+4 2 | 3+4 2 | 3+4 2 | 5+6 2 | 2+2 2 | 3+3 2 |
| | 3+4 2 | 3+4 2 | 3+4 2 | 5+6 2 | 3+3 2 | 3+3 2 |
| | 3+4 2 | 5+6 2 | 5+6 2 | 5+6 2 | 2+2 2 | 3+3 2 |
| | 3+4 2 | 5+6 2 | 3+4 2 | 3+4 2 | 2+2 2 | 2+2 2 |
| | 3+4 2 | 5+6 2 | 5+6 2 | 5+6 2 | 3+3 2 | 3+3 2 |
| | 3+4 2 | 3+4 2 | 5+6 2 | 5+6 2 | 3+3 2 | 3+3 2 |
| | 3+4 2 | 3+4 2 | 3+4 2 | 5+6 2 | 2+2 2 | 3+3 2 |
| | 3+4 2 | 3+4 2 | 3+4 2 | 5+6 2 | 2+2 2 | 3+3 2 |
| | 3+4 2 | 5+6 2 | 3+4 2 | 3+4 2 | 2+2 2 | 2+2 2 |
| | 3+4 2 | 5+6 2 | 3+4 2 | 5+6 2 | 2+2 2 | 3+3 2 |
| | 3+4 2 | 5+6 2 | 3+4 2 | 3+4 2 | 2+2 2 | 2+2 2 |
| | 3+4 2 | 5+6 2 | 5+6 2 | 5+6 2 | 3+3 2 | 2+2 2 |
| | 5+6 2 | 3+4 2 | 5+6 2 | 3+4 2 | 3+3 2 | 2+2 2 |
| | 3+4 2 | 3+4 2 | 3+4 2 | 5+6 2 | 2+2 2 | 3+3 2 |
| | 5+6 2 | 3+4 2 | 3+4 2 | 5+6 2 | 2+2 2 | 3+3 2 |
| | 5+6 2 | 3+4 2 | 3+4 2 | 5+6 2 | 2+2 2 | 3+3 2 |

Continued on next column

**TABLE XXIV– continued**

| TQC | ICSD | NoE | SF | | | |
|---|---|---|---|---|---|---|
| | 3+4 2 | 3+4 2 | 5+6 2 | 3+4 2 | 3+3 2 | 2+2 2 |
| | 3+4 2 | 5+6 2 | 5+6 2 | 3+4 2 | 3+3 2 | 3+3 2 |
| | 3+4 2 | 3+4 2 | 5+6 2 | 5+6 2 | 3+3 2 | 2+2 2 |
| | 3+4 2 | 5+6 2 | 5+6 2 | 5+6 2 | 3+3 2 | 3+3 2 |
| | 5+6 2 | 3+4 2 | 5+6 2 | 5+6 2 | 3+3 2 | 3+3 2 |
| 12.2.174 | 614578 | 44 | B2Ir3Y1 | | | |
| | -A | -GM | -M | -Y | -L | -V |
| | 5+6 2 | 3+4 2 | 5+6 2 | 3+4 2 | 3+3 2 | 2+2 2 |
| | 3+4 2 | 5+6 2 | 3+4 2 | 5+6 2 | 2+2 2 | 3+3 2 |
| | 3+4 2 | 5+6 2 | 3+4 2 | 5+6 2 | 2+2 2 | 3+3 2 |
| | 3+4 2 | 5+6 2 | 3+4 2 | 5+6 2 | 2+2 2 | 3+3 2 |
| | 3+4 2 | 3+4 2 | 5+6 2 | 5+6 2 | 3+3 2 | 3+3 2 |
| | 3+4 2 | 3+4 2 | 3+4 2 | 3+4 2 | 2+2 2 | 2+2 2 |
| | 5+6 2 | 3+4 2 | 5+6 2 | 5+6 2 | 3+3 2 | 3+3 2 |
| | 3+4 2 | 5+6 2 | 3+4 2 | 5+6 2 | 2+2 2 | 3+3 2 |
| | 3+4 2 | 3+4 2 | 5+6 2 | 5+6 2 | 3+3 2 | 3+3 2 |
| | 3+4 2 | 3+4 2 | 5+6 2 | 3+4 2 | 3+3 2 | 2+2 2 |
| | 3+4 2 | 3+4 2 | 5+6 2 | 5+6 2 | 3+3 2 | 3+3 2 |
| | 3+4 2 | 3+4 2 | 3+4 2 | 3+4 2 | 2+2 2 | 2+2 2 |
| | 5+6 2 | 3+4 2 | 5+6 2 | 3+4 2 | 3+3 2 | 2+2 2 |
| | 3+4 2 | 3+4 2 | 5+6 2 | 3+4 2 | 3+3 2 | 2+2 2 |
| | 5+6 2 | 3+4 2 | 3+4 2 | 5+6 2 | 2+2 2 | 3+3 2 |
| | 3+4 2 | 3+4 2 | 3+4 2 | 5+6 2 | 2+2 2 | 3+3 2 |
| | 5+6 2 | 5+6 2 | 3+4 2 | 3+4 2 | 2+2 2 | 2+2 2 |
| | 3+4 2 | 3+4 2 | 5+6 2 | 5+6 2 | 3+3 2 | 3+3 2 |
| | 3+4 2 | 5+6 2 | 5+6 2 | 3+4 2 | 3+3 2 | 2+2 2 |
| | 3+4 2 | 3+4 2 | 5+6 2 | 5+6 2 | 3+3 2 | 3+3 2 |
| | 5+6 2 | 3+4 2 | 5+6 2 | 5+6 2 | 3+3 2 | 3+3 2 |
| | 5+6 2 | 3+4 2 | 3+4 2 | 5+6 2 | 2+2 2 | 3+3 2 |
| 12.2.175 | 614571 | 56 | B2Ir3Tm1 | | | |
| | -A | -GM | -M | -Y | -L | -V |
| | 5+6 2 | 3+4 2 | 5+6 2 | 3+4 2 | 3+3 2 | 2+2 2 |
| | 3+4 2 | 5+6 2 | 3+4 2 | 5+6 2 | 2+2 2 | 3+3 2 |
| | 3+4 2 | 5+6 2 | 3+4 2 | 5+6 2 | 2+2 2 | 3+3 2 |
| | 3+4 2 | 5+6 2 | 3+4 2 | 5+6 2 | 2+2 2 | 3+3 2 |
| | 3+4 2 | 3+4 2 | 5+6 2 | 5+6 2 | 3+3 2 | 3+3 2 |
| | 3+4 2 | 3+4 2 | 3+4 2 | 3+4 2 | 2+2 2 | 2+2 2 |
| | 5+6 2 | 3+4 2 | 5+6 2 | 5+6 2 | 3+3 2 | 3+3 2 |
| | 3+4 2 | 5+6 2 | 3+4 2 | 5+6 2 | 2+2 2 | 3+3 2 |
| | 3+4 2 | 3+4 2 | 5+6 2 | 5+6 2 | 3+3 2 | 3+3 2 |
| | 3+4 2 | 3+4 2 | 5+6 2 | 3+4 2 | 3+3 2 | 2+2 2 |
| | 3+4 2 | 3+4 2 | 5+6 2 | 5+6 2 | 3+3 2 | 3+3 2 |
| | 3+4 2 | 3+4 2 | 3+4 2 | 3+4 2 | 2+2 2 | 2+2 2 |
| | 5+6 2 | 3+4 2 | 5+6 2 | 3+4 2 | 3+3 2 | 2+2 2 |
| | 3+4 2 | 3+4 2 | 3+4 2 | 3+4 2 | 2+2 2 | 2+2 2 |
| | 5+6 2 | 3+4 2 | 5+6 2 | 5+6 2 | 3+3 2 | 3+3 2 |
| | 3+4 2 | 3+4 2 | 3+4 2 | 5+6 2 | 2+2 2 | 3+3 2 |
| | 5+6 2 | 5+6 2 | 3+4 2 | 5+6 2 | 2+2 2 | 3+3 2 |
| | 3+4 2 | 3+4 2 | 5+6 2 | 3+4 2 | 3+3 2 | 2+2 2 |
| | 3+4 2 | 5+6 2 | 3+4 2 | 5+6 2 | 2+2 2 | 3+3 2 |
| | 3+4 2 | 5+6 2 | 3+4 2 | 3+4 2 | 2+2 2 | 2+2 2 |
| | 3+4 2 | 5+6 2 | 5+6 2 | 5+6 2 | 3+3 2 | 3+3 2 |
| | 3+4 2 | 5+6 2 | 3+4 2 | 5+6 2 | 2+2 2 | 3+3 2 |
| | 3+4 2 | 3+4 2 | 5+6 2 | 5+6 2 | 3+3 2 | 3+3 2 |
| | 3+4 2 | 3+4 2 | 3+4 2 | 5+6 2 | 2+2 2 | 3+3 2 |
| | 5+6 2 | 3+4 2 | 5+6 2 | 5+6 2 | 3+3 2 | 3+3 2 |
| | 5+6 2 | 5+6 2 | 3+4 2 | 5+6 2 | 2+2 2 | 3+3 2 |
| | 3+4 2 | 5+6 2 | 3+4 2 | 5+6 2 | 2+2 2 | 3+3 2 |
| | 3+4 2 | 5+6 2 | 3+4 2 | 5+6 2 | 2+2 2 | 3+3 2 |
| 12.2.176 | 614566 | 52 | B2Ir3Tb1 | | | |
| | -A | -GM | -M | -Y | -L | -V |

Continued on next column

**TABLE XXIV– continued**

| TQC | ICSD | NoE | SF | | | |
|---|---|---|---|---|---|---|
| | 5+6 2 | 3+4 2 | 5+6 2 | 3+4 2 | 3+3 2 | 2+2 2 |
| | 3+4 2 | 5+6 2 | 3+4 2 | 5+6 2 | 2+2 2 | 3+3 2 |
| | 3+4 2 | 5+6 2 | 3+4 2 | 5+6 2 | 2+2 2 | 3+3 2 |
| | 3+4 2 | 5+6 2 | 3+4 2 | 5+6 2 | 2+2 2 | 3+3 2 |
| | 3+4 2 | 3+4 2 | 5+6 2 | 5+6 2 | 3+3 2 | 3+3 2 |
| | 3+4 2 | 3+4 2 | 3+4 2 | 3+4 2 | 2+2 2 | 2+2 2 |
| | 5+6 2 | 3+4 2 | 5+6 2 | 5+6 2 | 3+3 2 | 3+3 2 |
| | 3+4 2 | 5+6 2 | 3+4 2 | 5+6 2 | 2+2 2 | 3+3 2 |
| | 3+4 2 | 3+4 2 | 5+6 2 | 5+6 2 | 3+3 2 | 3+3 2 |
| | 3+4 2 | 3+4 2 | 5+6 2 | 3+4 2 | 3+3 2 | 2+2 2 |
| | 3+4 2 | 3+4 2 | 5+6 2 | 5+6 2 | 3+3 2 | 3+3 2 |
| | 3+4 2 | 3+4 2 | 3+4 2 | 3+4 2 | 2+2 2 | 2+2 2 |
| | 5+6 2 | 3+4 2 | 5+6 2 | 3+4 2 | 3+3 2 | 2+2 2 |
| | 3+4 2 | 3+4 2 | 3+4 2 | 3+4 2 | 2+2 2 | 2+2 2 |
| | 5+6 2 | 3+4 2 | 5+6 2 | 5+6 2 | 3+3 2 | 3+3 2 |
| | 3+4 2 | 3+4 2 | 3+4 2 | 5+6 2 | 2+2 2 | 3+3 2 |
| | 5+6 2 | 3+4 2 | 5+6 2 | 3+4 2 | 3+3 2 | 2+2 2 |
| | 3+4 2 | 5+6 2 | 3+4 2 | 5+6 2 | 2+2 2 | 3+3 2 |
| | 3+4 2 | 3+4 2 | 5+6 2 | 5+6 2 | 3+3 2 | 3+3 2 |
| | 3+4 2 | 5+6 2 | 5+6 2 | 3+4 2 | 3+3 2 | 2+2 2 |
| | 3+4 2 | 5+6 2 | 3+4 2 | 5+6 2 | 2+2 2 | 3+3 2 |
| | 5+6 2 | 5+6 2 | 3+4 2 | 5+6 2 | 2+2 2 | 3+3 2 |
| | 3+4 2 | 5+6 2 | 3+4 2 | 5+6 2 | 2+2 2 | 3+3 2 |
| | 3+4 2 | 3+4 2 | 3+4 2 | 5+6 2 | 2+2 2 | 3+3 2 |
| | 3+4 2 | 3+4 2 | 5+6 2 | 5+6 2 | 3+3 2 | 3+3 2 |
| | 5+6 2 | 5+6 2 | 3+4 2 | 5+6 2 | 2+2 2 | 3+3 2 |
| 12.2.177 | 614561 | 36 | B2Ir3Sc1 | | | |
| | -A | -GM | -M | -Y | -L | -V |
| | 5+6 2 | 3+4 2 | 3+4 2 | 5+6 2 | 2+2 2 | 3+3 2 |
| | 5+6 2 | 3+4 2 | 5+6 2 | 3+4 2 | 3+3 2 | 2+2 2 |
| | 3+4 2 | 3+4 2 | 3+4 2 | 5+6 2 | 2+2 2 | 3+3 2 |
| | 5+6 2 | 3+4 2 | 5+6 2 | 5+6 2 | 3+3 2 | 3+3 2 |
| | 5+6 2 | 3+4 2 | 3+4 2 | 5+6 2 | 2+2 2 | 3+3 2 |
| | 5+6 2 | 5+6 2 | 3+4 2 | 3+4 2 | 2+2 2 | 2+2 2 |
| | 5+6 2 | 3+4 2 | 3+4 2 | 5+6 2 | 2+2 2 | 3+3 2 |
| | 5+6 2 | 3+4 2 | 5+6 2 | 3+4 2 | 3+3 2 | 2+2 2 |
| | 3+4 2 | 3+4 2 | 3+4 2 | 3+4 2 | 2+2 2 | 2+2 2 |
| | 5+6 2 | 3+4 2 | 5+6 2 | 3+4 2 | 3+3 2 | 2+2 2 |
| | 3+4 2 | 3+4 2 | 3+4 2 | 5+6 2 | 2+2 2 | 3+3 2 |
| | 5+6 2 | 5+6 2 | 5+6 2 | 5+6 2 | 3+3 2 | 3+3 2 |
| | 3+4 2 | 3+4 2 | 5+6 2 | 3+4 2 | 3+3 2 | 2+2 2 |
| | 5+6 2 | 5+6 2 | 3+4 2 | 5+6 2 | 2+2 2 | 3+3 2 |
| | 5+6 2 | 3+4 2 | 3+4 2 | 5+6 2 | 2+2 2 | 2+2 2 |
| | 5+6 2 | 3+4 2 | 3+4 2 | 3+4 2 | 2+2 2 | 3+3 2 |
| | 3+4 2 | 3+4 2 | 3+4 2 | 5+6 2 | 2+2 2 | 3+3 2 |
| | 3+4 2 | 3+4 2 | 5+6 2 | 5+6 2 | 3+3 2 | 3+3 2 |
| 12.2.178 | 614531 | 58 | B2Ir3Lu1 | | | |
| | -A | -GM | -M | -Y | -L | -V |
| | 5+6 2 | 3+4 2 | 5+6 2 | 3+4 2 | 3+3 2 | 2+2 2 |
| | 3+4 2 | 5+6 2 | 3+4 2 | 5+6 2 | 2+2 2 | 3+3 2 |
| | 3+4 2 | 5+6 2 | 3+4 2 | 5+6 2 | 2+2 2 | 3+3 2 |
| | 3+4 2 | 5+6 2 | 3+4 2 | 5+6 2 | 2+2 2 | 3+3 2 |
| | 3+4 2 | 3+4 2 | 5+6 2 | 5+6 2 | 3+3 2 | 3+3 2 |
| | 3+4 2 | 3+4 2 | 3+4 2 | 3+4 2 | 2+2 2 | 2+2 2 |
| | 5+6 2 | 3+4 2 | 5+6 2 | 5+6 2 | 3+3 2 | 3+3 2 |
| | 3+4 2 | 5+6 2 | 3+4 2 | 5+6 2 | 2+2 2 | 3+3 2 |
| | 3+4 2 | 3+4 2 | 5+6 2 | 5+6 2 | 3+3 2 | 2+2 2 |
| | 3+4 2 | 3+4 2 | 3+4 2 | 3+4 2 | 2+2 2 | 3+3 2 |
| | 3+4 2 | 3+4 2 | 3+4 2 | 5+6 2 | 2+2 2 | 3+3 2 |
| | 3+4 2 | 5+6 2 | 3+4 2 | 5+6 2 | 2+2 2 | 3+3 2 |
| | 3+4 2 | 5+6 2 | 5+6 2 | 5+6 2 | 3+3 2 | 3+3 2 |

Continued on next column

**TABLE XXIV– continued**

| TQC | ICSD | NoE | SF | | | |
|---|---|---|---|---|---|---|
| | 3+4 2 | 5+6 2 | 5+6 2 | 5+6 2 | 3+3 2 | 3+3 2 |
| | 3+4 2 | 3+4 2 | 3+4 2 | 3+4 2 | 2+2 2 | 2+2 2 |
| | 3+4 2 | 3+4 2 | 3+4 2 | 3+4 2 | 2+2 2 | 2+2 2 |
| | 3+4 2 | 5+6 2 | 3+4 2 | 3+4 2 | 2+2 2 | 2+2 2 |
| | 3+4 2 | 5+6 2 | 3+4 2 | 5+6 2 | 2+2 2 | 3+3 2 |
| | 3+4 2 | 5+6 2 | 3+4 2 | 5+6 2 | 2+2 2 | 3+3 2 |
| | 5+6 2 | 5+6 2 | 5+6 2 | 5+6 2 | 3+3 2 | 3+3 2 |
| | 3+4 2 | 3+4 2 | 5+6 2 | 5+6 2 | 3+3 2 | 3+3 2 |
| | 5+6 2 | 3+4 2 | 3+4 2 | 5+6 2 | 2+2 2 | 3+3 2 |
| | 3+4 2 | 3+4 2 | 3+4 2 | 5+6 2 | 2+2 2 | 3+3 2 |
| | 5+6 2 | 5+6 2 | 3+4 2 | 3+4 2 | 2+2 2 | 2+2 2 |
| | 3+4 2 | 3+4 2 | 5+6 2 | 5+6 2 | 3+3 2 | 3+3 2 |
| | 3+4 2 | 5+6 2 | 5+6 2 | 3+4 2 | 3+3 2 | 2+2 2 |
| | 3+4 2 | 3+4 2 | 5+6 2 | 5+6 2 | 3+3 2 | 3+3 2 |
| | 5+6 2 | 3+4 2 | 5+6 2 | 5+6 2 | 3+3 2 | 3+3 2 |
| | 5+6 2 | 3+4 2 | 3+4 2 | 5+6 2 | 2+2 2 | 3+3 2 |
| 12.2.179 | 614472 | 54 | B2Ho1Ir3 | | | |
| | -A | -GM | -M | -Y | -L | -V |
| | 5+6 2 | 3+4 2 | 5+6 2 | 3+4 2 | 3+3 2 | 2+2 2 |
| | 3+4 2 | 5+6 2 | 3+4 2 | 5+6 2 | 2+2 2 | 3+3 2 |
| | 3+4 2 | 5+6 2 | 3+4 2 | 5+6 2 | 2+2 2 | 3+3 2 |
| | 3+4 2 | 5+6 2 | 3+4 2 | 5+6 2 | 2+2 2 | 3+3 2 |
| | 3+4 2 | 3+4 2 | 5+6 2 | 5+6 2 | 3+3 2 | 3+3 2 |
| | 3+4 2 | 3+4 2 | 3+4 2 | 3+4 2 | 2+2 2 | 2+2 2 |
| | 5+6 2 | 3+4 2 | 5+6 2 | 5+6 2 | 3+3 2 | 3+3 2 |
| | 3+4 2 | 5+6 2 | 3+4 2 | 5+6 2 | 2+2 2 | 3+3 2 |
| | 3+4 2 | 3+4 2 | 5+6 2 | 5+6 2 | 3+3 2 | 3+3 2 |
| | 3+4 2 | 3+4 2 | 5+6 2 | 3+4 2 | 3+3 2 | 2+2 2 |
| | 3+4 2 | 3+4 2 | 5+6 2 | 5+6 2 | 3+3 2 | 3+3 2 |
| | 3+4 2 | 3+4 2 | 3+4 2 | 3+4 2 | 2+2 2 | 2+2 2 |
| | 5+6 2 | 3+4 2 | 5+6 2 | 3+4 2 | 3+3 2 | 2+2 2 |
| | 3+4 2 | 3+4 2 | 3+4 2 | 3+4 2 | 2+2 2 | 2+2 2 |
| | 5+6 2 | 3+4 2 | 5+6 2 | 5+6 2 | 3+3 2 | 3+3 2 |
| | 3+4 2 | 3+4 2 | 3+4 2 | 5+6 2 | 2+2 2 | 3+3 2 |
| | 5+6 2 | 3+4 2 | 3+4 2 | 5+6 2 | 2+2 2 | 3+3 2 |
| | 3+4 2 | 5+6 2 | 5+6 2 | 3+4 2 | 3+3 2 | 2+2 2 |
| | 3+4 2 | 5+6 2 | 5+6 2 | 5+6 2 | 3+3 2 | 3+3 2 |
| | 3+4 2 | 3+4 2 | 3+4 2 | 3+4 2 | 2+2 2 | 2+2 2 |
| | 3+4 2 | 5+6 2 | 3+4 2 | 5+6 2 | 2+2 2 | 3+3 2 |
| | 3+4 2 | 5+6 2 | 3+4 2 | 5+6 2 | 2+2 2 | 3+3 2 |
| | 3+4 2 | 5+6 2 | 5+6 2 | 5+6 2 | 3+3 2 | 3+3 2 |
| | 3+4 2 | 3+4 2 | 3+4 2 | 5+6 2 | 2+2 2 | 3+3 2 |
| | 5+6 2 | 3+4 2 | 5+6 2 | 5+6 2 | 3+3 2 | 3+3 2 |
| | 5+6 2 | 5+6 2 | 3+4 2 | 5+6 2 | 2+2 2 | 3+3 2 |
| | 3+4 2 | 5+6 2 | 3+4 2 | 5+6 2 | 2+2 2 | 3+3 2 |
| 12.2.180 | 8058 | 74 | Au2P3 | | | |
| | -A | -GM | -M | -Y | -L | -V |
| | 3+4 2 | 3+4 2 | 3+4 2 | 3+4 2 | 2+2 2 | 2+2 2 |
| | 5+6 2 | 5+6 2 | 5+6 2 | 5+6 2 | 3+3 2 | 3+3 2 |
| | 5+6 2 | 5+6 2 | 5+6 2 | 5+6 2 | 3+3 2 | 3+3 2 |
| | 3+4 2 | 3+4 2 | 3+4 2 | 3+4 2 | 2+2 2 | 2+2 2 |
| | 5+6 2 | 3+4 2 | 3+4 2 | 3+4 2 | 3+3 2 | 2+2 2 |
| | 3+4 2 | 3+4 2 | 5+6 2 | 5+6 2 | 2+2 2 | 3+3 2 |
| | 5+6 2 | 3+4 2 | 3+4 2 | 3+4 2 | 2+2 2 | 2+2 2 |
| | 5+6 2 | 5+6 2 | 3+4 2 | 5+6 2 | 3+3 2 | 3+3 2 |
| | 3+4 2 | 5+6 2 | 5+6 2 | 5+6 2 | 2+2 2 | 2+2 2 |
| | 5+6 2 | 5+6 2 | 3+4 2 | 5+6 2 | 3+3 2 | 3+3 2 |
| | 5+6 2 | 3+4 2 | 3+4 2 | 5+6 2 | 2+2 2 | 3+3 2 |
| | 3+4 2 | 3+4 2 | 5+6 2 | 3+4 2 | 3+3 2 | 2+2 2 |
| | 5+6 2 | 3+4 2 | 3+4 2 | 5+6 2 | 2+2 2 | 2+2 2 |
| | 3+4 2 | 3+4 2 | 5+6 2 | 5+6 2 | 3+3 2 | 3+3 2 |

Continued on next column

**TABLE XXIV– continued**

| TQC | ICSD | NoE | SF | | | |
|---|---|---|---|---|---|---|
| | 5+6 2 | 3+4 2 | 3+4 2 | 5+6 2 | 3+3 2 | 2+2 2 |
| | 5+6 2 | 5+6 2 | 3+4 2 | 3+4 2 | 3+3 2 | 3+3 2 |
| | 5+6 2 | 3+4 2 | 5+6 2 | 3+4 2 | 2+2 2 | 2+2 2 |
| | 3+4 2 | 3+4 2 | 3+4 2 | 5+6 2 | 2+2 2 | 3+3 2 |
| | 5+6 2 | 5+6 2 | 3+4 2 | 5+6 2 | 3+3 2 | 3+3 2 |
| | 3+4 2 | 3+4 2 | 3+4 2 | 3+4 2 | 2+2 2 | 2+2 2 |
| | 5+6 2 | 5+6 2 | 5+6 2 | 5+6 2 | 3+3 2 | 2+2 2 |
| | 5+6 2 | 3+4 2 | 3+4 2 | 3+4 2 | 2+2 2 | 3+3 2 |
| | 3+4 2 | 3+4 2 | 5+6 2 | 5+6 2 | 3+3 2 | 2+2 2 |
| | 5+6 2 | 5+6 2 | 3+4 2 | 5+6 2 | 2+2 2 | 3+3 2 |
| | 5+6 2 | 3+4 2 | 5+6 2 | 5+6 2 | 3+3 2 | 2+2 2 |
| | 3+4 2 | 3+4 2 | 3+4 2 | 3+4 2 | 2+2 2 | 3+3 2 |
| | 5+6 2 | 5+6 2 | 3+4 2 | 5+6 2 | 3+3 2 | 2+2 2 |
| | 5+6 2 | 3+4 2 | 3+4 2 | 5+6 2 | 3+3 2 | 3+3 2 |
| | 3+4 2 | 3+4 2 | 5+6 2 | 3+4 2 | 2+2 2 | 2+2 2 |
| | 5+6 2 | 5+6 2 | 3+4 2 | 5+6 2 | 3+3 2 | 3+3 2 |
| | 5+6 2 | 3+4 2 | 5+6 2 | 5+6 2 | 2+2 2 | 2+2 2 |
| | 5+6 2 | 3+4 2 | 3+4 2 | 3+4 2 | 3+3 2 | 3+3 2 |
| | 3+4 2 | 3+4 2 | 5+6 2 | 5+6 2 | 2+2 2 | 3+3 2 |
| | 5+6 2 | 5+6 2 | 3+4 2 | 3+4 2 | 2+2 2 | 2+2 2 |
| | 3+4 2 | 5+6 2 | 5+6 2 | 3+4 2 | 3+3 2 | 2+2 2 |
| | 5+6 2 | 3+4 2 | 3+4 2 | 5+6 2 | 2+2 2 | 3+3 2 |
| | 3+4 2 | 3+4 2 | 3+4 2 | 5+6 2 | 3+3 2 | 2+2 2 |
| 12.2.181 | 429212 | 90 | Au2Ni4Si7 | | | |
| | -A | -GM | -M | -Y | -L | -V |
| | 5+6 2 | 3+4 2 | 5+6 2 | 3+4 2 | 3+3 2 | 2+2 2 |
| | 3+4 2 | 5+6 2 | 3+4 2 | 5+6 2 | 2+2 2 | 3+3 2 |
| | 3+4 2 | 3+4 2 | 3+4 2 | 5+6 2 | 2+2 2 | 2+2 2 |
| | 5+6 2 | 5+6 2 | 5+6 2 | 3+4 2 | 3+3 2 | 2+2 2 |
| | 5+6 2 | 3+4 2 | 3+4 2 | 3+4 2 | 3+3 2 | 3+3 2 |
| | 3+4 2 | 3+4 2 | 5+6 2 | 5+6 2 | 2+2 2 | 2+2 2 |
| | 3+4 2 | 5+6 2 | 5+6 2 | 3+4 2 | 2+2 2 | 3+3 2 |
| | 5+6 2 | 5+6 2 | 3+4 2 | 3+4 2 | 2+2 2 | 3+3 2 |
| | 5+6 2 | 3+4 2 | 3+4 2 | 5+6 2 | 3+3 2 | 3+3 2 |
| | 3+4 2 | 5+6 2 | 5+6 2 | 3+4 2 | 3+3 2 | 2+2 2 |
| | 5+6 2 | 3+4 2 | 5+6 2 | 5+6 2 | 2+2 2 | 3+3 2 |
| | 3+4 2 | 5+6 2 | 3+4 2 | 5+6 2 | 3+3 2 | 2+2 2 |
| | 3+4 2 | 3+4 2 | 5+6 2 | 3+4 2 | 2+2 2 | 3+3 2 |
| | 5+6 2 | 5+6 2 | 3+4 2 | 3+4 2 | 3+3 2 | 2+2 2 |
| | 5+6 2 | 5+6 2 | 5+6 2 | 5+6 2 | 2+2 2 | 3+3 2 |
| | 3+4 2 | 3+4 2 | 3+4 2 | 5+6 2 | 3+3 2 | 2+2 2 |
| | 3+4 2 | 3+4 2 | 5+6 2 | 5+6 2 | 3+3 2 | 3+3 2 |
| | 5+6 2 | 3+4 2 | 5+6 2 | 5+6 2 | 2+2 2 | 2+2 2 |
| | 5+6 2 | 5+6 2 | 3+4 2 | 3+4 2 | 2+2 2 | 2+2 2 |
| | 3+4 2 | 5+6 2 | 3+4 2 | 5+6 2 | 3+3 2 | 2+2 2 |
| | 5+6 2 | 3+4 2 | 5+6 2 | 3+4 2 | 3+3 2 | 3+3 2 |
| | 5+6 2 | 3+4 2 | 3+4 2 | 5+6 2 | 2+2 2 | 2+2 2 |
| | 5+6 2 | 5+6 2 | 3+4 2 | 5+6 2 | 3+3 2 | 3+3 2 |
| | 3+4 2 | 5+6 2 | 5+6 2 | 3+4 2 | 2+2 2 | 2+2 2 |
| | 3+4 2 | 3+4 2 | 5+6 2 | 3+4 2 | 3+3 2 | 3+3 2 |
| | 3+4 2 | 5+6 2 | 3+4 2 | 3+4 2 | 3+3 2 | 3+3 2 |
| | 5+6 2 | 3+4 2 | 3+4 2 | 5+6 2 | 2+2 2 | 3+3 2 |
| | 3+4 2 | 3+4 2 | 5+6 2 | 5+6 2 | 3+3 2 | 2+2 2 |
| | 5+6 2 | 5+6 2 | 3+4 2 | 3+4 2 | 2+2 2 | 2+2 2 |
| | 5+6 2 | 3+4 2 | 5+6 2 | 3+4 2 | 3+3 2 | 3+3 2 |
| | 3+4 2 | 5+6 2 | 3+4 2 | 3+4 2 | 2+2 2 | 2+2 2 |
| | 3+4 2 | 3+4 2 | 3+4 2 | 5+6 2 | 3+3 2 | 3+3 2 |
| | 3+4 2 | 3+4 2 | 5+6 2 | 5+6 2 | 2+2 2 | 3+3 2 |
| | 3+4 2 | 5+6 2 | 5+6 2 | 3+4 2 | 3+3 2 | 2+2 2 |
| | 5+6 2 | 5+6 2 | 3+4 2 | 5+6 2 | 2+2 2 | 3+3 2 |
| | 5+6 2 | 3+4 2 | 5+6 2 | 3+4 2 | 3+3 2 | 2+2 2 |
| | 3+4 2 | 5+6 2 | 3+4 2 | 5+6 2 | 3+3 2 | 2+2 2 |
| | 5+6 2 | 3+4 2 | 3+4 2 | 5+6 2 | 2+2 2 | 3+3 2 |
| | 3+4 2 | 5+6 2 | 5+6 2 | 3+4 2 | 3+3 2 | 2+2 2 |
| | 5+6 2 | 5+6 2 | 3+4 2 | 5+6 2 | 2+2 2 | 3+3 2 |
| | 5+6 2 | 3+4 2 | 5+6 2 | 3+4 2 | 2+2 2 | 2+2 2 |
| | 3+4 2 | 3+4 2 | 5+6 2 | 5+6 2 | 3+3 2 | 3+3 2 |
| | 5+6 2 | 5+6 2 | 5+6 2 | 3+4 2 | 2+2 2 | 3+3 2 |
| | 3+4 2 | 3+4 2 | 3+4 2 | 3+4 2 | 3+3 2 | 2+2 2 |
| | 5+6 2 | 3+4 2 | 5+6 2 | 3+4 2 | 2+2 2 | 2+2 2 |
| 12.2.182 | 425402 | 106 | Au1O1P2Pr2 | | | |
| | -A | -GM | -M | -Y | -L | -V |
| | 5+6 2 | 3+4 2 | 3+4 2 | 5+6 2 | 2+2 2 | 3+3 2 |
| | 3+4 2 | 5+6 2 | 5+6 2 | 3+4 2 | 3+3 2 | 2+2 2 |
| | 3+4 2 | 3+4 2 | 5+6 2 | 3+4 2 | 2+2 2 | 3+3 2 |
| | 5+6 2 | 5+6 2 | 3+4 2 | 5+6 2 | 3+3 2 | 2+2 2 |
| | 5+6 2 | 3+4 2 | 3+4 2 | 5+6 2 | 2+2 2 | 3+3 2 |
| | 3+4 2 | 5+6 2 | 5+6 2 | 3+4 2 | 3+3 2 | 2+2 2 |
| | 5+6 2 | 3+4 2 | 3+4 2 | 5+6 2 | 2+2 2 | 3+3 2 |
| | 3+4 2 | 5+6 2 | 5+6 2 | 3+4 2 | 3+3 2 | 2+2 2 |
| | 5+6 2 | 3+4 2 | 3+4 2 | 5+6 2 | 3+3 2 | 3+3 2 |
| | 3+4 2 | 5+6 2 | 5+6 2 | 3+4 2 | 2+2 2 | 2+2 2 |
| | 5+6 2 | 3+4 2 | 3+4 2 | 5+6 2 | 3+3 2 | 2+2 2 |
| | 5+6 2 | 3+4 2 | 5+6 2 | 5+6 2 | 2+2 2 | 3+3 2 |
| | 3+4 2 | 5+6 2 | 3+4 2 | 3+4 2 | 2+2 2 | 3+3 2 |
| | 3+4 2 | 3+4 2 | 5+6 2 | 3+4 2 | 3+3 2 | 2+2 2 |
| | 5+6 2 | 5+6 2 | 3+4 2 | 5+6 2 | 2+2 2 | 3+3 2 |
| | 3+4 2 | 5+6 2 | 5+6 2 | 3+4 2 | 3+3 2 | 2+2 2 |
| | 3+4 2 | 5+6 2 | 5+6 2 | 3+4 2 | 3+3 2 | 3+3 2 |
| | 5+6 2 | 3+4 2 | 3+4 2 | 5+6 2 | 2+2 2 | 2+2 2 |
| | 5+6 2 | 3+4 2 | 5+6 2 | 3+4 2 | 3+3 2 | 2+2 2 |
| | 3+4 2 | 3+4 2 | 3+4 2 | 5+6 2 | 3+3 2 | 3+3 2 |
| | 5+6 2 | 5+6 2 | 5+6 2 | 3+4 2 | 2+2 2 | 2+2 2 |
| | 3+4 2 | 5+6 2 | 3+4 2 | 5+6 2 | 2+2 2 | 3+3 2 |
| | 3+4 2 | 3+4 2 | 5+6 2 | 5+6 2 | 3+3 2 | 2+2 2 |
| | 5+6 2 | 5+6 2 | 3+4 2 | 3+4 2 | 2+2 2 | 3+3 2 |
| | 3+4 2 | 3+4 2 | 5+6 2 | 5+6 2 | 3+3 2 | 2+2 2 |
| | 5+6 2 | 5+6 2 | 3+4 2 | 3+4 2 | 2+2 2 | 3+3 2 |
| | 3+4 2 | 3+4 2 | 5+6 2 | 5+6 2 | 2+2 2 | 2+2 2 |
| | 5+6 2 | 5+6 2 | 5+6 2 | 3+4 2 | 3+3 2 | 3+3 2 |
| | 3+4 2 | 5+6 2 | 3+4 2 | 3+4 2 | 3+3 2 | 2+2 2 |
| | 5+6 2 | 3+4 2 | 3+4 2 | 5+6 2 | 2+2 2 | 3+3 2 |
| | 3+4 2 | 3+4 2 | 5+6 2 | 3+4 2 | 2+2 2 | 3+3 2 |
| | 5+6 2 | 3+4 2 | 3+4 2 | 5+6 2 | 3+3 2 | 2+2 2 |
| | 5+6 2 | 5+6 2 | 5+6 2 | 5+6 2 | 3+3 2 | 2+2 2 |
| | 3+4 2 | 3+4 2 | 3+4 2 | 3+4 2 | 2+2 2 | 3+3 2 |
| | 5+6 2 | 5+6 2 | 3+4 2 | 5+6 2 | 2+2 2 | 2+2 2 |
| | 5+6 2 | 3+4 2 | 3+4 2 | 3+4 2 | 3+3 2 | 3+3 2 |
| | 3+4 2 | 5+6 2 | 5+6 2 | 5+6 2 | 2+2 2 | 2+2 2 |
| | 3+4 2 | 5+6 2 | 5+6 2 | 3+4 2 | 3+3 2 | 3+3 2 |
| | 5+6 2 | 3+4 2 | 3+4 2 | 5+6 2 | 2+2 2 | 2+2 2 |
| | 3+4 2 | 5+6 2 | 5+6 2 | 3+4 2 | 3+3 2 | 3+3 2 |
| | 3+4 2 | 5+6 2 | 3+4 2 | 5+6 2 | 2+2 2 | 3+3 2 |
| | 5+6 2 | 3+4 2 | 5+6 2 | 3+4 2 | 3+3 2 | 2+2 2 |
| | 3+4 2 | 3+4 2 | 5+6 2 | 5+6 2 | 2+2 2 | 3+3 2 |
| | 3+4 2 | 3+4 2 | 5+6 2 | 5+6 2 | 3+3 2 | 2+2 2 |
| | 5+6 2 | 5+6 2 | 3+4 2 | 3+4 2 | 2+2 2 | 3+3 2 |
| | 5+6 2 | 5+6 2 | 3+4 2 | 3+4 2 | 3+3 2 | 2+2 2 |
| | 3+4 2 | 5+6 2 | 3+4 2 | 3+4 2 | 3+3 2 | 2+2 2 |
| | 5+6 2 | 3+4 2 | 5+6 2 | 5+6 2 | 3+3 2 | 3+3 2 |
| | 5+6 2 | 3+4 2 | 5+6 2 | 3+4 2 | 2+2 2 | 2+2 2 |
| | 3+4 2 | 3+4 2 | 5+6 2 | 5+6 2 | 2+2 2 | 3+3 2 |

Continued on next column

**TABLE XXIV– continued**

| TQC | ICSD | NoE | SF | | | |
|---|---|---|---|---|---|---|
| | 5+6 2 | 5+6 2 | 3+4 2 | 3+4 2 | 3+3 2 | 3+3 2 |
| | 3+4 2 | 5+6 2 | 3+4 2 | 5+6 2 | 2+2 2 | 2+2 2 |
| | 5+6 2 | 3+4 2 | 3+4 2 | 3+4 2 | 3+3 2 | 2+2 2 |
| 12.2.183 | 41831 | 50 | As3Sr1 | | | |
| | -A | -GM | -M | -Y | -L | -V |
| | 3+4 2 | 3+4 2 | 3+4 2 | 3+4 2 | 2+2 2 | 2+2 2 |
| | 5+6 2 | 5+6 2 | 5+6 2 | 5+6 2 | 3+3 2 | 3+3 2 |
| | 3+4 2 | 3+4 2 | 5+6 2 | 3+4 2 | 2+2 2 | 2+2 2 |
| | 5+6 2 | 5+6 2 | 3+4 2 | 5+6 2 | 3+3 2 | 3+3 2 |
| | 5+6 2 | 3+4 2 | 5+6 2 | 3+4 2 | 3+3 2 | 2+2 2 |
| | 3+4 2 | 3+4 2 | 3+4 2 | 5+6 2 | 2+2 2 | 3+3 2 |
| | 3+4 2 | 5+6 2 | 5+6 2 | 3+4 2 | 2+2 2 | 2+2 2 |
| | 5+6 2 | 5+6 2 | 3+4 2 | 5+6 2 | 3+3 2 | 3+3 2 |
| | 3+4 2 | 3+4 2 | 3+4 2 | 3+4 2 | 3+3 2 | 3+3 2 |
| | 5+6 2 | 3+4 2 | 5+6 2 | 5+6 2 | 2+2 2 | 2+2 2 |
| | 5+6 2 | 5+6 2 | 3+4 2 | 5+6 2 | 3+3 2 | 3+3 2 |
| | 3+4 2 | 3+4 2 | 3+4 2 | 5+6 2 | 2+2 2 | 2+2 2 |
| | 5+6 2 | 5+6 2 | 5+6 2 | 3+4 2 | 3+3 2 | 3+3 2 |
| | 3+4 2 | 5+6 2 | 5+6 2 | 3+4 2 | 2+2 2 | 2+2 2 |
| | 3+4 2 | 3+4 2 | 3+4 2 | 3+4 2 | 2+2 2 | 3+3 2 |
| | 5+6 2 | 5+6 2 | 5+6 2 | 5+6 2 | 3+3 2 | 2+2 2 |
| | 5+6 2 | 5+6 2 | 5+6 2 | 5+6 2 | 3+3 2 | 3+3 2 |
| | 3+4 2 | 3+4 2 | 3+4 2 | 3+4 2 | 2+2 2 | 2+2 2 |
| | 3+4 2 | 3+4 2 | 5+6 2 | 3+4 2 | 3+3 2 | 2+2 2 |
| | 5+6 2 | 5+6 2 | 3+4 2 | 5+6 2 | 2+2 2 | 3+3 2 |
| | 3+4 2 | 5+6 2 | 3+4 2 | 5+6 2 | 2+2 2 | 2+2 2 |
| | 3+4 2 | 3+4 2 | 5+6 2 | 3+4 2 | 3+3 2 | 3+3 2 |
| | 5+6 2 | 3+4 2 | 3+4 2 | 3+4 2 | 3+3 2 | 3+3 2 |
| | 5+6 2 | 3+4 2 | 3+4 2 | 5+6 2 | 2+2 2 | 3+3 2 |
| | 5+6 2 | 3+4 2 | 5+6 2 | 5+6 2 | 3+3 2 | 2+2 2 |
| 12.2.184 | 80375 | 86 | As3Mn4 | | | |
| | -A | -GM | -M | -Y | -L | -V |
| | 5+6 2 | 3+4 2 | 3+4 2 | 3+4 2 | 2+2 2 | 3+3 2 |
| | 3+4 2 | 5+6 2 | 5+6 2 | 5+6 2 | 3+3 2 | 2+2 2 |
| | 3+4 2 | 3+4 2 | 3+4 2 | 5+6 2 | 3+3 2 | 2+2 2 |
| | 3+4 2 | 5+6 2 | 5+6 2 | 3+4 2 | 2+2 2 | 3+3 2 |
| | 5+6 2 | 3+4 2 | 5+6 2 | 5+6 2 | 2+2 2 | 2+2 2 |
| | 5+6 2 | 5+6 2 | 3+4 2 | 3+4 2 | 3+3 2 | 3+3 2 |
| | 5+6 2 | 3+4 2 | 3+4 2 | 3+4 2 | 3+3 2 | 2+2 2 |
| | 3+4 2 | 3+4 2 | 5+6 2 | 3+4 2 | 2+2 2 | 3+3 2 |
| | 5+6 2 | 5+6 2 | 5+6 2 | 5+6 2 | 3+3 2 | 3+3 2 |
| | 3+4 2 | 3+4 2 | 5+6 2 | 5+6 2 | 2+2 2 | 3+3 2 |
| | 3+4 2 | 5+6 2 | 3+4 2 | 3+4 2 | 2+2 2 | 2+2 2 |
| | 5+6 2 | 5+6 2 | 3+4 2 | 5+6 2 | 3+3 2 | 3+3 2 |
| | 5+6 2 | 3+4 2 | 3+4 2 | 3+4 2 | 2+2 2 | 2+2 2 |
| | 3+4 2 | 3+4 2 | 5+6 2 | 3+4 2 | 3+3 2 | 3+3 2 |
| | 3+4 2 | 5+6 2 | 3+4 2 | 5+6 2 | 2+2 2 | 2+2 2 |
| | 5+6 2 | 3+4 2 | 5+6 2 | 3+4 2 | 3+3 2 | 3+3 2 |
| | 3+4 2 | 3+4 2 | 5+6 2 | 5+6 2 | 3+3 2 | 2+2 2 |
| | 5+6 2 | 3+4 2 | 3+4 2 | 5+6 2 | 2+2 2 | 2+2 2 |
| | 3+4 2 | 3+4 2 | 5+6 2 | 3+4 2 | 2+2 2 | 3+3 2 |
| | 5+6 2 | 5+6 2 | 5+6 2 | 3+4 2 | 3+3 2 | 2+2 2 |
| | 3+4 2 | 3+4 2 | 3+4 2 | 3+4 2 | 2+2 2 | 3+3 2 |
| | 5+6 2 | 5+6 2 | 3+4 2 | 5+6 2 | 3+3 2 | 2+2 2 |
| | 5+6 2 | 5+6 2 | 5+6 2 | 3+4 2 | 2+2 2 | 2+2 2 |
| | 3+4 2 | 3+4 2 | 5+6 2 | 5+6 2 | 3+3 2 | 3+3 2 |
| | 3+4 2 | 5+6 2 | 3+4 2 | 3+4 2 | 3+3 2 | 3+3 2 |
| | 5+6 2 | 5+6 2 | 3+4 2 | 5+6 2 | 2+2 2 | 3+3 2 |
| | 5+6 2 | 3+4 2 | 5+6 2 | 5+6 2 | 2+2 2 | 3+3 2 |
| | 3+4 2 | 5+6 2 | 5+6 2 | 3+4 2 | 3+3 2 | 2+2 2 |
| | 5+6 2 | 3+4 2 | 3+4 2 | 5+6 2 | 2+2 2 | 3+3 2 |

Continued on next column

**TABLE XXIV– continued**

| TQC | ICSD | NoE | SF | | | |
|---|---|---|---|---|---|---|
| | 3+4 2 | 3+4 2 | 3+4 2 | 3+4 2 | 3+3 2 | 2+2 2 |
| | 3+4 2 | 5+6 2 | 3+4 2 | 5+6 2 | 2+2 2 | 2+2 2 |
| | 5+6 2 | 3+4 2 | 5+6 2 | 3+4 2 | 3+3 2 | 3+3 2 |
| | 3+4 2 | 3+4 2 | 5+6 2 | 5+6 2 | 2+2 2 | 3+3 2 |
| | 3+4 2 | 3+4 2 | 3+4 2 | 3+4 2 | 3+3 2 | 2+2 2 |
| | 3+4 2 | 5+6 2 | 3+4 2 | 5+6 2 | 2+2 2 | 3+3 2 |
| | 5+6 2 | 3+4 2 | 3+4 2 | 3+4 2 | 3+3 2 | 2+2 2 |
| | 5+6 2 | 3+4 2 | 5+6 2 | 5+6 2 | 3+3 2 | 3+3 2 |
| | 3+4 2 | 5+6 2 | 3+4 2 | 5+6 2 | 2+2 2 | 2+2 2 |
| | 5+6 2 | 3+4 2 | 5+6 2 | 3+4 2 | 2+2 2 | 3+3 2 |
| | 5+6 2 | 5+6 2 | 5+6 2 | 3+4 2 | 2+2 2 | 3+3 2 |
| | 3+4 2 | 3+4 2 | 3+4 2 | 3+4 2 | 3+3 2 | 3+3 2 |
| | 3+4 2 | 3+4 2 | 3+4 2 | 5+6 2 | 3+3 2 | 3+3 2 |
| | 3+4 2 | 5+6 2 | 5+6 2 | 3+4 2 | 3+3 2 | 2+2 2 |
| 12.2.185 | 86402 | 50 | As3Ba1 | | | |
| | -A | -GM | -M | -Y | -L | -V |
| | 5+6 2 | 3+4 2 | 3+4 2 | 3+4 2 | 3+3 2 | 3+3 2 |
| | 3+4 2 | 5+6 2 | 5+6 2 | 5+6 2 | 2+2 2 | 2+2 2 |
| | 3+4 2 | 3+4 2 | 5+6 2 | 5+6 2 | 2+2 2 | 2+2 2 |
| | 5+6 2 | 5+6 2 | 3+4 2 | 3+4 2 | 3+3 2 | 3+3 2 |
| | 3+4 2 | 3+4 2 | 5+6 2 | 3+4 2 | 2+2 2 | 2+2 2 |
| | 5+6 2 | 3+4 2 | 3+4 2 | 5+6 2 | 3+3 2 | 3+3 2 |
| | 5+6 2 | 3+4 2 | 3+4 2 | 5+6 2 | 3+3 2 | 3+3 2 |
| | 3+4 2 | 5+6 2 | 5+6 2 | 3+4 2 | 2+2 2 | 2+2 2 |
| | 3+4 2 | 5+6 2 | 3+4 2 | 3+4 2 | 2+2 2 | 2+2 2 |
| | 3+4 2 | 3+4 2 | 5+6 2 | 5+6 2 | 3+3 2 | 3+3 2 |
| | 5+6 2 | 5+6 2 | 5+6 2 | 5+6 2 | 2+2 2 | 2+2 2 |
| | 3+4 2 | 3+4 2 | 3+4 2 | 3+4 2 | 3+3 2 | 3+3 2 |
| | 5+6 2 | 5+6 2 | 5+6 2 | 5+6 2 | 2+2 2 | 2+2 2 |
| | 5+6 2 | 5+6 2 | 3+4 2 | 3+4 2 | 3+3 2 | 3+3 2 |
| | 3+4 2 | 3+4 2 | 3+4 2 | 3+4 2 | 3+3 2 | 2+2 2 |
| | 5+6 2 | 5+6 2 | 5+6 2 | 5+6 2 | 2+2 2 | 3+3 2 |
| | 5+6 2 | 5+6 2 | 5+6 2 | 5+6 2 | 2+2 2 | 2+2 2 |
| | 3+4 2 | 3+4 2 | 3+4 2 | 3+4 2 | 3+3 2 | 3+3 2 |
| | 3+4 2 | 3+4 2 | 3+4 2 | 3+4 2 | 2+2 2 | 3+3 2 |
| | 5+6 2 | 5+6 2 | 5+6 2 | 5+6 2 | 3+3 2 | 2+2 2 |
| | 3+4 2 | 5+6 2 | 3+4 2 | 5+6 2 | 3+3 2 | 3+3 2 |
| | 5+6 2 | 3+4 2 | 3+4 2 | 3+4 2 | 2+2 2 | 2+2 2 |
| | 3+4 2 | 3+4 2 | 5+6 2 | 3+4 2 | 2+2 2 | 2+2 2 |
| | 3+4 2 | 3+4 2 | 5+6 2 | 5+6 2 | 2+2 2 | 2+2 2 |
| | 5+6 2 | 3+4 2 | 5+6 2 | 5+6 2 | 3+3 2 | 3+3 2 |
| 12.2.186 | 69001 | 70 | As2Co2O7 | | | |
| | -A | -GM | -M | -Y | -L | -V |
| | 3+4 2 | 3+4 2 | 3+4 2 | 3+4 2 | 3+3 2 | 3+3 2 |
| | 5+6 2 | 5+6 2 | 5+6 2 | 5+6 2 | 2+2 2 | 2+2 2 |
| | 3+4 2 | 3+4 2 | 3+4 2 | 3+4 2 | 3+3 2 | 3+3 2 |
| | 5+6 2 | 5+6 2 | 5+6 2 | 5+6 2 | 3+3 2 | 3+3 2 |
| | 3+4 2 | 3+4 2 | 3+4 2 | 5+6 2 | 2+2 2 | 2+2 2 |
| | 5+6 2 | 3+4 2 | 3+4 2 | 3+4 2 | 3+3 2 | 2+2 2 |
| | 3+4 2 | 5+6 2 | 5+6 2 | 3+4 2 | 2+2 2 | 3+3 2 |
| | 5+6 2 | 5+6 2 | 5+6 2 | 5+6 2 | 2+2 2 | 2+2 2 |
| | 3+4 2 | 3+4 2 | 3+4 2 | 3+4 2 | 3+3 2 | 3+3 2 |
| | 5+6 2 | 5+6 2 | 5+6 2 | 5+6 2 | 2+2 2 | 2+2 2 |
| | 5+6 2 | 3+4 2 | 3+4 2 | 5+6 2 | 3+3 2 | 2+2 2 |
| | 5+6 2 | 5+6 2 | 5+6 2 | 5+6 2 | 2+2 2 | 2+2 2 |
| | 3+4 2 | 3+4 2 | 5+6 2 | 3+4 2 | 2+2 2 | 3+3 2 |
| | 3+4 2 | 5+6 2 | 3+4 2 | 3+4 2 | 3+3 2 | 3+3 2 |
| | 3+4 2 | 5+6 2 | 3+4 2 | 3+4 2 | 3+3 2 | 2+2 2 |
| | 3+4 2 | 3+4 2 | 5+6 2 | 5+6 2 | 2+2 2 | 3+3 2 |
| | 5+6 2 | 5+6 2 | 5+6 2 | 5+6 2 | 2+2 2 | 2+2 2 |
| | 5+6 2 | 3+4 2 | 5+6 2 | 5+6 2 | 2+2 2 | 3+3 2 |

Continued on next column

**TABLE XXIV– continued**

| TQC | ICSD | | NoE | SF | | |
|---|---|---|---|---|---|---|
| | 5+6 2 | 3+4 2 | 3+4 2 | 3+4 2 | 3+3 2 | 2+2 2 |
| | 5+6 2 | 3+4 2 | 3+4 2 | 5+6 2 | 2+2 2 | 2+2 2 |
| | 3+4 2 | 5+6 2 | 5+6 2 | 3+4 2 | 3+3 2 | 3+3 2 |
| | 3+4 2 | 5+6 2 | 5+6 2 | 5+6 2 | 2+2 2 | 2+2 2 |
| | 5+6 2 | 3+4 2 | 3+4 2 | 3+4 2 | 3+3 2 | 3+3 2 |
| | 5+6 2 | 5+6 2 | 5+6 2 | 3+4 2 | 2+2 2 | 2+2 2 |
| | 3+4 2 | 5+6 2 | 3+4 2 | 5+6 2 | 3+3 2 | 3+3 2 |
| | 3+4 2 | 5+6 2 | 5+6 2 | 5+6 2 | 3+3 2 | 2+2 2 |
| | 5+6 2 | 3+4 2 | 5+6 2 | 5+6 2 | 2+2 2 | 2+2 2 |
| | 5+6 2 | 5+6 2 | 3+4 2 | 3+4 2 | 2+2 2 | 3+3 2 |
| | 5+6 2 | 3+4 2 | 5+6 2 | 3+4 2 | 2+2 2 | 3+3 2 |
| | 3+4 2 | 5+6 2 | 3+4 2 | 5+6 2 | 3+3 2 | 2+2 2 |
| | 5+6 2 | 3+4 2 | 3+4 2 | 3+4 2 | 3+3 2 | 3+3 2 |
| | 3+4 2 | 5+6 2 | 5+6 2 | 5+6 2 | 2+2 2 | 2+2 2 |
| | 3+4 2 | 3+4 2 | 5+6 2 | 3+4 2 | 3+3 2 | 3+3 2 |
| | 5+6 2 | 5+6 2 | 3+4 2 | 5+6 2 | 2+2 2 | 2+2 2 |
| | 3+4 2 | 5+6 2 | 3+4 2 | 3+4 2 | 3+3 2 | 3+3 2 |
| 12.2.187 | 239291 | | 110 | As1Pd5 | | |
| | -A | -GM | -M | -Y | -L | -V |
| | 3+4 2 | 3+4 2 | 3+4 2 | 3+4 2 | 3+3 2 | 3+3 2 |
| | 5+6 2 | 5+6 2 | 5+6 2 | 5+6 2 | 2+2 2 | 2+2 2 |
| | 3+4 2 | 3+4 2 | 5+6 2 | 5+6 2 | 2+2 2 | 3+3 2 |
| | 5+6 2 | 5+6 2 | 3+4 2 | 3+4 2 | 3+3 2 | 2+2 2 |
| | 5+6 2 | 5+6 2 | 3+4 2 | 5+6 2 | 2+2 2 | 2+2 2 |
| | 5+6 2 | 3+4 2 | 5+6 2 | 3+4 2 | 3+3 2 | 3+3 2 |
| | 3+4 2 | 3+4 2 | 5+6 2 | 3+4 2 | 2+2 2 | 2+2 2 |
| | 5+6 2 | 5+6 2 | 3+4 2 | 5+6 2 | 3+3 2 | 3+3 2 |
| | 3+4 2 | 3+4 2 | 3+4 2 | 5+6 2 | 2+2 2 | 2+2 2 |
| | 5+6 2 | 5+6 2 | 3+4 2 | 5+6 2 | 3+3 2 | 3+3 2 |
| | 3+4 2 | 5+6 2 | 5+6 2 | 3+4 2 | 3+3 2 | 3+3 2 |
| | 3+4 2 | 3+4 2 | 5+6 2 | 5+6 2 | 3+3 2 | 3+3 2 |
| | 3+4 2 | 3+4 2 | 3+4 2 | 3+4 2 | 2+2 2 | 2+2 2 |
| | 5+6 2 | 5+6 2 | 3+4 2 | 5+6 2 | 2+2 2 | 2+2 2 |
| | 5+6 2 | 5+6 2 | 5+6 2 | 3+4 2 | 2+2 2 | 3+3 2 |
| | 3+4 2 | 3+4 2 | 3+4 2 | 3+4 2 | 3+3 2 | 2+2 2 |
| | 5+6 2 | 5+6 2 | 5+6 2 | 3+4 2 | 2+2 2 | 3+3 2 |
| | 3+4 2 | 3+4 2 | 5+6 2 | 5+6 2 | 3+3 2 | 2+2 2 |
| | 5+6 2 | 5+6 2 | 3+4 2 | 3+4 2 | 2+2 2 | 3+3 2 |
| | 3+4 2 | 5+6 2 | 5+6 2 | 5+6 2 | 3+3 2 | 2+2 2 |
| | 5+6 2 | 3+4 2 | 3+4 2 | 5+6 2 | 2+2 2 | 2+2 2 |
| | 5+6 2 | 5+6 2 | 5+6 2 | 5+6 2 | 3+3 2 | 3+3 2 |
| | 3+4 2 | 3+4 2 | 3+4 2 | 3+4 2 | 3+3 2 | 2+2 2 |
| | 5+6 2 | 5+6 2 | 5+6 2 | 3+4 2 | 2+2 2 | 3+3 2 |
| | 3+4 2 | 3+4 2 | 3+4 2 | 5+6 2 | 2+2 2 | 2+2 2 |
| | 3+4 2 | 5+6 2 | 5+6 2 | 5+6 2 | 3+3 2 | 3+3 2 |
| | 3+4 2 | 3+4 2 | 5+6 2 | 3+4 2 | 3+3 2 | 3+3 2 |
| | 5+6 2 | 3+4 2 | 3+4 2 | 3+4 2 | 3+3 2 | 2+2 2 |
| | 3+4 2 | 3+4 2 | 5+6 2 | 3+4 2 | 2+2 2 | 3+3 2 |
| | 5+6 2 | 3+4 2 | 3+4 2 | 3+4 2 | 2+2 2 | 2+2 2 |
| | 3+4 2 | 5+6 2 | 5+6 2 | 5+6 2 | 2+2 2 | 3+3 2 |
| | 3+4 2 | 3+4 2 | 5+6 2 | 3+4 2 | 3+3 2 | 2+2 2 |
| | 5+6 2 | 5+6 2 | 3+4 2 | 5+6 2 | 2+2 2 | 3+3 2 |
| | 5+6 2 | 3+4 2 | 5+6 2 | 5+6 2 | 3+3 2 | 2+2 2 |
| | 3+4 2 | 5+6 2 | 3+4 2 | 3+4 2 | 3+3 2 | 2+2 2 |
| | 3+4 2 | 3+4 2 | 3+4 2 | 5+6 2 | 2+2 2 | 3+3 2 |
| | 5+6 2 | 5+6 2 | 5+6 2 | 5+6 2 | 3+3 2 | 2+2 2 |
| | 3+4 2 | 3+4 2 | 3+4 2 | 3+4 2 | 2+2 2 | 3+3 2 |
| | 5+6 2 | 3+4 2 | 5+6 2 | 3+4 2 | 3+3 2 | 2+2 2 |
| | 5+6 2 | 5+6 2 | 3+4 2 | 5+6 2 | 2+2 2 | 2+2 2 |
| | 3+4 2 | 3+4 2 | 5+6 2 | 3+4 2 | 3+3 2 | 3+3 2 |
| | 3+4 2 | 5+6 2 | 3+4 2 | 5+6 2 | 2+2 2 | 3+3 2 |
| | 3+4 2 | 3+4 2 | 3+4 2 | 5+6 2 | 3+3 2 | 2+2 2 |
| | 5+6 2 | 3+4 2 | 5+6 2 | 3+4 2 | 2+2 2 | 3+3 2 |
| | 5+6 2 | 5+6 2 | 3+4 2 | 3+4 2 | 2+2 2 | 3+3 2 |
| | 3+4 2 | 5+6 2 | 5+6 2 | 5+6 2 | 3+3 2 | 2+2 2 |
| | 5+6 2 | 3+4 2 | 5+6 2 | 3+4 2 | 3+3 2 | 3+3 2 |
| | 5+6 2 | 5+6 2 | 3+4 2 | 5+6 2 | 2+2 2 | 2+2 2 |
| | 3+4 2 | 3+4 2 | 5+6 2 | 5+6 2 | 2+2 2 | 3+3 2 |
| | 3+4 2 | 3+4 2 | 3+4 2 | 3+4 2 | 3+3 2 | 2+2 2 |
| | 5+6 2 | 5+6 2 | 5+6 2 | 3+4 2 | 3+3 2 | 3+3 2 |
| | 5+6 2 | 3+4 2 | 5+6 2 | 5+6 2 | 2+2 2 | 2+2 2 |
| | 5+6 2 | 5+6 2 | 3+4 2 | 3+4 2 | 2+2 2 | 3+3 2 |
| | 3+4 2 | 5+6 2 | 5+6 2 | 3+4 2 | 3+3 2 | 2+2 2 |
| | 5+6 2 | 3+4 2 | 3+4 2 | 5+6 2 | 3+3 2 | 2+2 2 |
| 12.2.188 | 417968 | | 52 | Al2Ge4Sr3 | | |
| | -A | -GM | -M | -Y | -L | -V |
| | 3+4 2 | 3+4 2 | 3+4 2 | 3+4 2 | 3+3 2 | 3+3 2 |
| | 5+6 2 | 3+4 2 | 5+6 2 | 3+4 2 | 3+3 2 | 2+2 2 |
| | 3+4 2 | 5+6 2 | 3+4 2 | 5+6 2 | 2+2 2 | 3+3 2 |
| | 5+6 2 | 5+6 2 | 5+6 2 | 5+6 2 | 2+2 2 | 2+2 2 |
| | 5+6 2 | 3+4 2 | 5+6 2 | 3+4 2 | 3+3 2 | 2+2 2 |
| | 3+4 2 | 5+6 2 | 3+4 2 | 5+6 2 | 2+2 2 | 3+3 2 |
| | 5+6 2 | 5+6 2 | 5+6 2 | 5+6 2 | 2+2 2 | 2+2 2 |
| | 5+6 2 | 3+4 2 | 5+6 2 | 3+4 2 | 2+2 2 | 2+2 2 |
| | 5+6 2 | 5+6 2 | 5+6 2 | 5+6 2 | 3+3 2 | 2+2 2 |
| | 3+4 2 | 5+6 2 | 3+4 2 | 5+6 2 | 2+2 2 | 3+3 2 |
| | 3+4 2 | 3+4 2 | 3+4 2 | 3+4 2 | 3+3 2 | 2+2 2 |
| | 5+6 2 | 5+6 2 | 5+6 2 | 5+6 2 | 2+2 2 | 3+3 2 |
| | 3+4 2 | 3+4 2 | 5+6 2 | 5+6 2 | 3+3 2 | 3+3 2 |
| | 5+6 2 | 3+4 2 | 3+4 2 | 5+6 2 | 2+2 2 | 2+2 2 |
| | 3+4 2 | 5+6 2 | 5+6 2 | 3+4 2 | 2+2 2 | 3+3 2 |
| | 5+6 2 | 5+6 2 | 3+4 2 | 3+4 2 | 3+3 2 | 2+2 2 |
| | 3+4 2 | 3+4 2 | 5+6 2 | 3+4 2 | 3+3 2 | 3+3 2 |
| | 5+6 2 | 5+6 2 | 3+4 2 | 5+6 2 | 2+2 2 | 2+2 2 |
| | 5+6 2 | 3+4 2 | 3+4 2 | 5+6 2 | 3+3 2 | 2+2 2 |
| | 5+6 2 | 3+4 2 | 3+4 2 | 3+4 2 | 2+2 2 | 3+3 2 |
| | 3+4 2 | 3+4 2 | 5+6 2 | 5+6 2 | 2+2 2 | 3+3 2 |
| | 5+6 2 | 5+6 2 | 3+4 2 | 5+6 2 | 3+3 2 | 2+2 2 |
| | 3+4 2 | 3+4 2 | 5+6 2 | 3+4 2 | 2+2 2 | 3+3 2 |
| | 3+4 2 | 3+4 2 | 5+6 2 | 3+4 2 | 3+3 2 | 3+3 2 |
| | 3+4 2 | 5+6 2 | 5+6 2 | 5+6 2 | 3+3 2 | 2+2 2 |
| | 3+4 2 | 5+6 2 | 5+6 2 | 3+4 2 | 2+2 2 | 2+2 2 |
| 12.2.189 | 417659 | | 168 | Al2Ba5Ge7 | | |
| | -A | -GM | -M | -Y | -L | -V |
| | 5+6 2 | 3+4 2 | 3+4 2 | 3+4 2 | 2+2 2 | 2+2 2 |
| | 3+4 2 | 5+6 2 | 5+6 2 | 5+6 2 | 3+3 2 | 3+3 2 |
| | 5+6 2 | 3+4 2 | 3+4 2 | 5+6 2 | 3+3 2 | 2+2 2 |
| | 3+4 2 | 3+4 2 | 5+6 2 | 3+4 2 | 2+2 2 | 3+3 2 |
| | 3+4 2 | 5+6 2 | 5+6 2 | 5+6 2 | 2+2 2 | 2+2 2 |
| | 5+6 2 | 5+6 2 | 3+4 2 | 3+4 2 | 3+3 2 | 3+3 2 |
| | 3+4 2 | 3+4 2 | 3+4 2 | 3+4 2 | 2+2 2 | 2+2 2 |
| | 5+6 2 | 5+6 2 | 5+6 2 | 5+6 2 | 3+3 2 | 3+3 2 |
| | 3+4 2 | 3+4 2 | 5+6 2 | 5+6 2 | 2+2 2 | 2+2 2 |
| | 5+6 2 | 5+6 2 | 3+4 2 | 3+4 2 | 3+3 2 | 3+3 2 |
| | 3+4 2 | 5+6 2 | 5+6 2 | 3+4 2 | 3+3 2 | 2+2 2 |
| | 5+6 2 | 3+4 2 | 3+4 2 | 5+6 2 | 2+2 2 | 3+3 2 |
| | 5+6 2 | 3+4 2 | 3+4 2 | 5+6 2 | 3+3 2 | 2+2 2 |
| | 3+4 2 | 3+4 2 | 5+6 2 | 3+4 2 | 2+2 2 | 3+3 2 |
| | 5+6 2 | 5+6 2 | 5+6 2 | 5+6 2 | 2+2 2 | 2+2 2 |
| | 3+4 2 | 5+6 2 | 3+4 2 | 3+4 2 | 3+3 2 | 3+3 2 |
| | 5+6 2 | 3+4 2 | 3+4 2 | 3+4 2 | 3+3 2 | 2+2 2 |
| | 3+4 2 | 5+6 2 | 5+6 2 | 5+6 2 | 2+2 2 | 3+3 2 |

Continued on next column

**TABLE XXIV– continued**

| TQC | ICSD | | NoE | SF | | |
|---|---|---|---|---|---|---|
| | 3+4 2 | 3+4 2 | 5+6 2 | 5+6 2 | 2+2 2 | 2+2 2 |
| | 5+6 2 | 5+6 2 | 3+4 2 | 3+4 2 | 3+3 2 | 3+3 2 |
| | 5+6 2 | 3+4 2 | 5+6 2 | 3+4 2 | 2+2 2 | 3+3 2 |
| | 3+4 2 | 5+6 2 | 3+4 2 | 5+6 2 | 3+3 2 | 2+2 2 |
| | 3+4 2 | 5+6 2 | 3+4 2 | 3+4 2 | 2+2 2 | 2+2 2 |
| | 5+6 2 | 3+4 2 | 5+6 2 | 5+6 2 | 3+3 2 | 3+3 2 |
| | 3+4 2 | 5+6 2 | 3+4 2 | 3+4 2 | 2+2 2 | 3+3 2 |
| | 5+6 2 | 3+4 2 | 5+6 2 | 5+6 2 | 3+3 2 | 2+2 2 |
| | 5+6 2 | 5+6 2 | 5+6 2 | 3+4 2 | 3+3 2 | 2+2 2 |
| | 3+4 2 | 3+4 2 | 3+4 2 | 5+6 2 | 2+2 2 | 3+3 2 |
| | 3+4 2 | 3+4 2 | 5+6 2 | 5+6 2 | 3+3 2 | 2+2 2 |
| | 5+6 2 | 3+4 2 | 3+4 2 | 3+4 2 | 2+2 2 | 3+3 2 |
| | 5+6 2 | 5+6 2 | 3+4 2 | 5+6 2 | 3+3 2 | 2+2 2 |
| | 5+6 2 | 3+4 2 | 5+6 2 | 3+4 2 | 2+2 2 | 3+3 2 |
| | 3+4 2 | 5+6 2 | 3+4 2 | 5+6 2 | 3+3 2 | 2+2 2 |
| | 3+4 2 | 3+4 2 | 3+4 2 | 3+4 2 | 2+2 2 | 3+3 2 |
| | 3+4 2 | 5+6 2 | 5+6 2 | 3+4 2 | 3+3 2 | 2+2 2 |
| | 5+6 2 | 3+4 2 | 5+6 2 | 5+6 2 | 2+2 2 | 3+3 2 |
| | 3+4 2 | 5+6 2 | 3+4 2 | 5+6 2 | 3+3 2 | 2+2 2 |
| | 3+4 2 | 3+4 2 | 5+6 2 | 3+4 2 | 2+2 2 | 3+3 2 |
| | 5+6 2 | 5+6 2 | 5+6 2 | 5+6 2 | 3+3 2 | 3+3 2 |
| | 5+6 2 | 5+6 2 | 3+4 2 | 3+4 2 | 2+2 2 | 2+2 2 |
| | 5+6 2 | 3+4 2 | 5+6 2 | 3+4 2 | 3+3 2 | 2+2 2 |
| | 3+4 2 | 5+6 2 | 3+4 2 | 5+6 2 | 2+2 2 | 3+3 2 |
| | 3+4 2 | 3+4 2 | 3+4 2 | 3+4 2 | 3+3 2 | 3+3 2 |
| | 5+6 2 | 5+6 2 | 5+6 2 | 5+6 2 | 2+2 2 | 2+2 2 |
| | 5+6 2 | 5+6 2 | 3+4 2 | 5+6 2 | 2+2 2 | 2+2 2 |
| | 3+4 2 | 3+4 2 | 3+4 2 | 3+4 2 | 3+3 2 | 3+3 2 |
| | 3+4 2 | 5+6 2 | 5+6 2 | 3+4 2 | 2+2 2 | 3+3 2 |
| | 5+6 2 | 3+4 2 | 5+6 2 | 5+6 2 | 3+3 2 | 2+2 2 |
| | 5+6 2 | 5+6 2 | 5+6 2 | 5+6 2 | 2+2 2 | 3+3 2 |
| | 5+6 2 | 3+4 2 | 5+6 2 | 3+4 2 | 3+3 2 | 2+2 2 |
| | 3+4 2 | 5+6 2 | 3+4 2 | 5+6 2 | 3+3 2 | 3+3 2 |
| | 3+4 2 | 3+4 2 | 3+4 2 | 3+4 2 | 2+2 2 | 2+2 2 |
| | 3+4 2 | 5+6 2 | 5+6 2 | 3+4 2 | 2+2 2 | 2+2 2 |
| | 5+6 2 | 3+4 2 | 3+4 2 | 5+6 2 | 3+3 2 | 3+3 2 |
| | 3+4 2 | 3+4 2 | 3+4 2 | 3+4 2 | 2+2 2 | 2+2 2 |
| | 5+6 2 | 5+6 2 | 5+6 2 | 3+4 2 | 3+3 2 | 3+3 2 |
| | 3+4 2 | 3+4 2 | 3+4 2 | 5+6 2 | 3+3 2 | 3+3 2 |
| | 5+6 2 | 3+4 2 | 5+6 2 | 5+6 2 | 2+2 2 | 2+2 2 |
| | 3+4 2 | 5+6 2 | 3+4 2 | 5+6 2 | 3+3 2 | 2+2 2 |
| | 3+4 2 | 5+6 2 | 5+6 2 | 3+4 2 | 2+2 2 | 3+3 2 |
| | 5+6 2 | 5+6 2 | 5+6 2 | 3+4 2 | 2+2 2 | 2+2 2 |
| | 5+6 2 | 3+4 2 | 3+4 2 | 5+6 2 | 3+3 2 | 3+3 2 |
| | 5+6 2 | 3+4 2 | 5+6 2 | 5+6 2 | 2+2 2 | 2+2 2 |
| | 3+4 2 | 5+6 2 | 5+6 2 | 3+4 2 | 3+3 2 | 3+3 2 |
| | 5+6 2 | 5+6 2 | 5+6 2 | 5+6 2 | 2+2 2 | 3+3 2 |
| | 5+6 2 | 3+4 2 | 3+4 2 | 5+6 2 | 3+3 2 | 2+2 2 |
| | 3+4 2 | 5+6 2 | 3+4 2 | 3+4 2 | 2+2 2 | 3+3 2 |
| | 3+4 2 | 3+4 2 | 5+6 2 | 5+6 2 | 3+3 2 | 2+2 2 |
| | 5+6 2 | 5+6 2 | 3+4 2 | 3+4 2 | 2+2 2 | 3+3 2 |
| | 3+4 2 | 3+4 2 | 5+6 2 | 3+4 2 | 3+3 2 | 2+2 2 |
| | 5+6 2 | 3+4 2 | 3+4 2 | 3+4 2 | 3+3 2 | 3+3 2 |
| | 5+6 2 | 5+6 2 | 3+4 2 | 3+4 2 | 2+2 2 | 3+3 2 |
| | 3+4 2 | 5+6 2 | 3+4 2 | 5+6 2 | 3+3 2 | 2+2 2 |
| | 5+6 2 | 5+6 2 | 5+6 2 | 5+6 2 | 2+2 2 | 3+3 2 |
| | 3+4 2 | 3+4 2 | 5+6 2 | 5+6 2 | 2+2 2 | 2+2 2 |
| | 5+6 2 | 5+6 2 | 3+4 2 | 3+4 2 | 3+3 2 | 3+3 2 |
| | 3+4 2 | 3+4 2 | 3+4 2 | 3+4 2 | 2+2 2 | 2+2 2 |
| | 3+4 2 | 5+6 2 | 5+6 2 | 3+4 2 | 2+2 2 | 2+2 2 |
| | 5+6 2 | 3+4 2 | 5+6 2 | 5+6 2 | 3+3 2 | 3+3 2 |

Continued on next column

**TABLE XXIV– continued**

| TQC | ICSD | | NoE | SF | | |
|---|---|---|---|---|---|---|
| | 3+4 2 | 5+6 2 | 3+4 2 | 5+6 2 | 3+3 2 | 2+2 2 |
| | 3+4 2 | 3+4 2 | 3+4 2 | 3+4 2 | 3+3 2 | 3+3 2 |
| | 5+6 2 | 3+4 2 | 5+6 2 | 3+4 2 | 2+2 2 | 2+2 2 |
| | 5+6 2 | 3+4 2 | 5+6 2 | 5+6 2 | 2+2 2 | 3+3 2 |
| | 3+4 2 | 3+4 2 | 3+4 2 | 5+6 2 | 2+2 2 | 2+2 2 |
| 12.2.190 | 20632 | | 48 | Al1Ce1Co1 | | |
| | -A | -GM | -M | -Y | -L | -V |
| | 5+6 2 | 3+4 2 | 3+4 2 | 5+6 2 | 2+2 2 | 3+3 2 |
| | 3+4 2 | 5+6 2 | 5+6 2 | 3+4 2 | 3+3 2 | 2+2 2 |
| | 5+6 2 | 3+4 2 | 3+4 2 | 5+6 2 | 3+3 2 | 3+3 2 |
| | 3+4 2 | 5+6 2 | 5+6 2 | 3+4 2 | 2+2 2 | 2+2 2 |
| | 5+6 2 | 3+4 2 | 3+4 2 | 3+4 2 | 2+2 2 | 3+3 2 |
| | 3+4 2 | 3+4 2 | 5+6 2 | 5+6 2 | 3+3 2 | 2+2 2 |
| | 5+6 2 | 5+6 2 | 3+4 2 | 5+6 2 | 3+3 2 | 3+3 2 |
| | 3+4 2 | 5+6 2 | 5+6 2 | 3+4 2 | 2+2 2 | 2+2 2 |
| | 3+4 2 | 3+4 2 | 5+6 2 | 3+4 2 | 2+2 2 | 2+2 2 |
| | 5+6 2 | 5+6 2 | 3+4 2 | 5+6 2 | 3+3 2 | 3+3 2 |
| | 3+4 2 | 3+4 2 | 5+6 2 | 3+4 2 | 2+2 2 | 2+2 2 |
| | 5+6 2 | 5+6 2 | 3+4 2 | 3+4 2 | 3+3 2 | 3+3 2 |
| | 5+6 2 | 5+6 2 | 5+6 2 | 5+6 2 | 2+2 2 | 2+2 2 |
| | 3+4 2 | 3+4 2 | 3+4 2 | 3+4 2 | 3+3 2 | 3+3 2 |
| | 5+6 2 | 5+6 2 | 3+4 2 | 5+6 2 | 3+3 2 | 3+3 2 |
| | 3+4 2 | 3+4 2 | 5+6 2 | 5+6 2 | 2+2 2 | 2+2 2 |
| | 5+6 2 | 5+6 2 | 3+4 2 | 3+4 2 | 3+3 2 | 3+3 2 |
| | 3+4 2 | 5+6 2 | 3+4 2 | 5+6 2 | 2+2 2 | 2+2 2 |
| | 5+6 2 | 3+4 2 | 5+6 2 | 5+6 2 | 3+3 2 | 2+2 2 |
| | 3+4 2 | 3+4 2 | 5+6 2 | 3+4 2 | 2+2 2 | 2+2 2 |
| | 3+4 2 | 3+4 2 | 5+6 2 | 3+4 2 | 2+2 2 | 3+3 2 |
| | 3+4 2 | 3+4 2 | 5+6 2 | 5+6 2 | 3+3 2 | 2+2 2 |
| | 5+6 2 | 3+4 2 | 3+4 2 | 3+4 2 | 2+2 2 | 3+3 2 |
| | 3+4 2 | 5+6 2 | 5+6 2 | 5+6 2 | 3+3 2 | 2+2 2 |
| 12.2.191 | 416360 | | 198 | Ag5Cl5Hg4P8 | | |
| | -A | -GM | -M | -Y | -L | -V |
| | 5+6 2 | 3+4 2 | 5+6 2 | 3+4 2 | 2+2 2 | 3+3 2 |
| | 3+4 2 | 5+6 2 | 3+4 2 | 5+6 2 | 3+3 2 | 2+2 2 |
| | 3+4 2 | 5+6 2 | 3+4 2 | 5+6 2 | 3+3 2 | 2+2 2 |
| | 5+6 2 | 3+4 2 | 5+6 2 | 3+4 2 | 2+2 2 | 3+3 2 |
| | 3+4 2 | 5+6 2 | 3+4 2 | 5+6 2 | 3+3 2 | 2+2 2 |
| | 5+6 2 | 3+4 2 | 5+6 2 | 3+4 2 | 2+2 2 | 3+3 2 |
| | 5+6 2 | 3+4 2 | 5+6 2 | 3+4 2 | 2+2 2 | 3+3 2 |
| | 3+4 2 | 5+6 2 | 3+4 2 | 5+6 2 | 3+3 2 | 2+2 2 |
| | 3+4 2 | 3+4 2 | 3+4 2 | 5+6 2 | 2+2 2 | 3+3 2 |
| | 5+6 2 | 5+6 2 | 5+6 2 | 3+4 2 | 3+3 2 | 2+2 2 |
| | 5+6 2 | 3+4 2 | 5+6 2 | 3+4 2 | 3+3 2 | 2+2 2 |
| | 3+4 2 | 3+4 2 | 3+4 2 | 5+6 2 | 2+2 2 | 3+3 2 |
| | 5+6 2 | 5+6 2 | 5+6 2 | 3+4 2 | 3+3 2 | 2+2 2 |
| | 3+4 2 | 5+6 2 | 3+4 2 | 5+6 2 | 2+2 2 | 3+3 2 |
| | 5+6 2 | 3+4 2 | 5+6 2 | 3+4 2 | 3+3 2 | 3+3 2 |
| | 3+4 2 | 5+6 2 | 3+4 2 | 5+6 2 | 2+2 2 | 2+2 2 |
| | 3+4 2 | 5+6 2 | 5+6 2 | 3+4 2 | 2+2 2 | 3+3 2 |
| | 5+6 2 | 3+4 2 | 3+4 2 | 5+6 2 | 3+3 2 | 2+2 2 |
| | 5+6 2 | 5+6 2 | 5+6 2 | 3+4 2 | 2+2 2 | 2+2 2 |
| | 3+4 2 | 3+4 2 | 3+4 2 | 5+6 2 | 3+3 2 | 3+3 2 |
| | 5+6 2 | 5+6 2 | 5+6 2 | 3+4 2 | 3+3 2 | 3+3 2 |
| | 3+4 2 | 3+4 2 | 3+4 2 | 5+6 2 | 2+2 2 | 2+2 2 |
| | 5+6 2 | 3+4 2 | 5+6 2 | 3+4 2 | 3+3 2 | 2+2 2 |
| | 3+4 2 | 5+6 2 | 3+4 2 | 5+6 2 | 2+2 2 | 3+3 2 |
| | 5+6 2 | 3+4 2 | 5+6 2 | 3+4 2 | 3+3 2 | 2+2 2 |
| | 5+6 2 | 3+4 2 | 3+4 2 | 5+6 2 | 2+2 2 | 2+2 2 |
| | 3+4 2 | 3+4 2 | 5+6 2 | 3+4 2 | 3+3 2 | 3+3 2 |
| | 3+4 2 | 5+6 2 | 3+4 2 | 5+6 2 | 2+2 2 | 3+3 2 |

Continued on next column

**TABLE XXIV– continued**

| TQC | ICSD NoE SF |
|---|---|
| | 5+6 2 5+6 2 5+6 2 3+4 2 2+2 2 3+3 2 |
| | 5+6 2 3+4 2 3+4 2 5+6 2 2+2 2 2+2 2 |
| | 3+4 2 5+6 2 5+6 2 5+6 2 3+3 2 2+2 2 |
| | 5+6 2 5+6 2 3+4 2 3+4 2 3+3 2 2+2 2 |
| | 3+4 2 3+4 2 3+4 2 3+4 2 3+3 2 3+3 2 |
| | 3+4 2 3+4 2 5+6 2 5+6 2 2+2 2 2+2 2 |
| | 5+6 2 5+6 2 3+4 2 5+6 2 3+3 2 3+3 2 |
| | 3+4 2 5+6 2 5+6 2 3+4 2 2+2 2 3+3 2 |
| | 3+4 2 5+6 2 3+4 2 5+6 2 3+3 2 2+2 2 |
| | 5+6 2 3+4 2 3+4 2 3+4 2 2+2 2 3+3 2 |
| | 5+6 2 3+4 2 5+6 2 3+4 2 3+3 2 2+2 2 |
| | 3+4 2 5+6 2 5+6 2 5+6 2 2+2 2 3+3 2 |
| | 3+4 2 3+4 2 3+4 2 3+4 2 3+3 2 2+2 2 |
| | 3+4 2 5+6 2 3+4 2 5+6 2 2+2 2 2+2 2 |
| | 5+6 2 3+4 2 5+6 2 3+4 2 3+3 2 3+3 2 |
| | 3+4 2 5+6 2 5+6 2 5+6 2 2+2 2 2+2 2 |
| | 5+6 2 3+4 2 3+4 2 3+4 2 2+2 2 3+3 2 |
| | 3+4 2 5+6 2 3+4 2 5+6 2 3+3 2 2+2 2 |
| | 5+6 2 3+4 2 5+6 2 3+4 2 2+2 2 3+3 2 |
| | 5+6 2 3+4 2 3+4 2 5+6 2 2+2 2 2+2 2 |
| | 3+4 2 3+4 2 5+6 2 3+4 2 3+3 2 3+3 2 |
| | 3+4 2 3+4 2 3+4 2 3+4 2 2+2 2 2+2 2 |
| | 3+4 2 5+6 2 5+6 2 3+4 2 3+3 2 2+2 2 |
| | 5+6 2 3+4 2 3+4 2 5+6 2 2+2 2 3+3 2 |
| | 3+4 2 5+6 2 5+6 2 3+4 2 2+2 2 2+2 2 |
| | 5+6 2 3+4 2 3+4 2 5+6 2 3+3 2 2+2 2 |
| | 5+6 2 5+6 2 5+6 2 3+4 2 3+3 2 3+3 2 |
| | 3+4 2 3+4 2 3+4 2 3+4 2 2+2 2 3+3 2 |
| | 3+4 2 3+4 2 3+4 2 5+6 2 2+2 2 2+2 2 |
| | 5+6 2 3+4 2 5+6 2 3+4 2 3+3 2 2+2 2 |
| | 3+4 2 3+4 2 3+4 2 5+6 2 2+2 2 2+2 2 |
| | 5+6 2 5+6 2 3+4 2 3+4 2 3+3 2 3+3 2 |
| | 3+4 2 3+4 2 5+6 2 3+4 2 3+3 2 2+2 2 |
| | 5+6 2 5+6 2 5+6 2 3+4 2 2+2 2 3+3 2 |
| | 3+4 2 3+4 2 3+4 2 5+6 2 3+3 2 2+2 2 |
| | 5+6 2 5+6 2 5+6 2 3+4 2 3+3 2 2+2 2 |
| | 5+6 2 3+4 2 3+4 2 5+6 2 2+2 2 3+3 2 |
| | 3+4 2 5+6 2 5+6 2 3+4 2 2+2 2 2+2 2 |
| | 5+6 2 3+4 2 5+6 2 5+6 2 3+3 2 3+3 2 |
| | 3+4 2 5+6 2 5+6 2 3+4 2 3+3 2 3+3 2 |
| | 5+6 2 5+6 2 5+6 2 5+6 2 2+2 2 2+2 2 |
| | 5+6 2 3+4 2 3+4 2 3+4 2 3+3 2 2+2 2 |
| | 5+6 2 5+6 2 5+6 2 5+6 2 3+3 2 3+3 2 |
| | 5+6 2 3+4 2 3+4 2 3+4 2 3+3 2 2+2 2 |
| | 3+4 2 3+4 2 5+6 2 3+4 2 2+2 2 2+2 2 |
| | 5+6 2 5+6 2 5+6 2 5+6 2 3+3 2 3+3 2 |
| | 5+6 2 3+4 2 3+4 2 5+6 2 2+2 2 3+3 2 |
| | 3+4 2 5+6 2 5+6 2 3+4 2 3+3 2 2+2 2 |
| | 3+4 2 3+4 2 3+4 2 3+4 2 3+3 2 2+2 2 |
| | 5+6 2 5+6 2 5+6 2 5+6 2 2+2 2 3+3 2 |
| | 3+4 2 3+4 2 5+6 2 5+6 2 3+3 2 3+3 2 |
| | 5+6 2 5+6 2 3+4 2 3+4 2 2+2 2 2+2 2 |
| | 5+6 2 5+6 2 3+4 2 5+6 2 3+3 2 3+3 2 |
| | 5+6 2 5+6 2 5+6 2 3+4 2 3+3 2 2+2 2 |
| | 3+4 2 3+4 2 3+4 2 5+6 2 2+2 2 2+2 2 |
| | 3+4 2 3+4 2 5+6 2 3+4 2 2+2 2 2+2 2 |
| | 3+4 2 3+4 2 3+4 2 3+4 2 2+2 2 3+3 2 |
| | 5+6 2 3+4 2 5+6 2 3+4 2 2+2 2 2+2 2 |
| | 3+4 2 3+4 2 3+4 2 3+4 2 3+3 2 2+2 2 |
| | 3+4 2 5+6 2 3+4 2 3+4 2 2+2 2 3+3 2 |
| | 3+4 2 3+4 2 5+6 2 5+6 2 3+3 2 2+2 2 |

Continued on next column

**TABLE XXIV– continued**

| TQC | ICSD NoE SF |
|---|---|
| | 5+6 2 3+4 2 3+4 2 3+4 2 2+2 2 2+2 2 |
| | 3+4 2 3+4 2 3+4 2 5+6 2 3+3 2 3+3 2 |
| | 5+6 2 5+6 2 3+4 2 3+4 2 2+2 2 2+2 2 |
| | 3+4 2 3+4 2 5+6 2 5+6 2 3+3 2 3+3 2 |
| | 5+6 2 5+6 2 5+6 2 3+4 2 2+2 2 2+2 2 |
| | 3+4 2 3+4 2 3+4 2 3+4 2 2+2 2 2+2 2 |
| | 5+6 2 3+4 2 5+6 2 5+6 2 3+3 2 3+3 2 |
| | 3+4 2 5+6 2 5+6 2 5+6 2 3+3 2 3+3 2 |
| | 5+6 2 5+6 2 5+6 2 5+6 2 2+2 2 2+2 2 |
| | 5+6 2 3+4 2 3+4 2 3+4 2 3+3 2 3+3 2 |
| 12.2.192 | 27743 186 Ag1Er1O8W2 |
| | -A -GM -M -Y -L -V |
| | 3+4 2 3+4 2 3+4 2 3+4 2 2+2 2 2+2 2 |
| | 5+6 2 5+6 2 5+6 2 5+6 2 3+3 2 3+3 2 |
| | 3+4 2 3+4 2 3+4 2 3+4 2 2+2 2 2+2 2 |
| | 5+6 2 5+6 2 5+6 2 5+6 2 3+3 2 3+3 2 |
| | 5+6 2 3+4 2 5+6 2 5+6 2 3+3 2 3+3 2 |
| | 3+4 2 5+6 2 3+4 2 3+4 2 2+2 2 2+2 2 |
| | 3+4 2 3+4 2 3+4 2 3+4 2 2+2 2 2+2 2 |
| | 5+6 2 5+6 2 5+6 2 5+6 2 3+3 2 3+3 2 |
| | 5+6 2 3+4 2 5+6 2 3+4 2 2+2 2 3+3 2 |
| | 3+4 2 5+6 2 3+4 2 5+6 2 3+3 2 2+2 2 |
| | 3+4 2 5+6 2 3+4 2 5+6 2 3+3 2 2+2 2 |
| | 5+6 2 3+4 2 5+6 2 3+4 2 2+2 2 3+3 2 |
| | 3+4 2 5+6 2 3+4 2 3+4 2 2+2 2 2+2 2 |
| | 5+6 2 3+4 2 5+6 2 5+6 2 3+3 2 2+2 2 |
| | 3+4 2 5+6 2 3+4 2 5+6 2 3+3 2 3+3 2 |
| | 5+6 2 3+4 2 5+6 2 3+4 2 2+2 2 3+3 2 |
| | 5+6 2 3+4 2 5+6 2 3+4 2 2+2 2 2+2 2 |
| | 3+4 2 5+6 2 3+4 2 5+6 2 3+3 2 3+3 2 |
| | 5+6 2 3+4 2 5+6 2 5+6 2 2+2 2 2+2 2 |
| | 3+4 2 5+6 2 3+4 2 3+4 2 3+3 2 3+3 2 |
| | 3+4 2 3+4 2 5+6 2 5+6 2 3+3 2 3+3 2 |
| | 5+6 2 5+6 2 3+4 2 5+6 2 2+2 2 2+2 2 |
| | 3+4 2 3+4 2 3+4 2 3+4 2 3+3 2 3+3 2 |
| | 5+6 2 5+6 2 5+6 2 3+4 2 2+2 2 2+2 2 |
| | 5+6 2 3+4 2 5+6 2 3+4 2 2+2 2 2+2 2 |
| | 3+4 2 3+4 2 3+4 2 5+6 2 3+3 2 3+3 2 |
| | 5+6 2 5+6 2 3+4 2 5+6 2 2+2 2 3+3 2 |
| | 5+6 2 3+4 2 5+6 2 5+6 2 3+3 2 2+2 2 |
| | 3+4 2 5+6 2 5+6 2 3+4 2 2+2 2 3+3 2 |
| | 3+4 2 3+4 2 5+6 2 3+4 2 3+3 2 2+2 2 |
| | 5+6 2 5+6 2 3+4 2 3+4 2 2+2 2 3+3 2 |
| | 3+4 2 5+6 2 5+6 2 5+6 2 3+3 2 2+2 2 |
| | 5+6 2 5+6 2 3+4 2 3+4 2 2+2 2 2+2 2 |
| | 3+4 2 3+4 2 3+4 2 5+6 2 3+3 2 3+3 2 |
| | 5+6 2 3+4 2 5+6 2 5+6 2 2+2 2 3+3 2 |
| | 3+4 2 5+6 2 3+4 2 3+4 2 3+3 2 2+2 2 |
| | 3+4 2 3+4 2 3+4 2 5+6 2 2+2 2 3+3 2 |
| | 5+6 2 3+4 2 5+6 2 3+4 2 3+3 2 3+3 2 |
| | 5+6 2 3+4 2 3+4 2 3+4 2 3+3 2 2+2 2 |
| | 3+4 2 5+6 2 3+4 2 5+6 2 2+2 2 2+2 2 |
| | 5+6 2 5+6 2 5+6 2 3+4 2 2+2 2 3+3 2 |
| | 3+4 2 5+6 2 5+6 2 5+6 2 3+3 2 2+2 2 |
| | 3+4 2 3+4 2 3+4 2 3+4 2 2+2 2 3+3 2 |
| | 5+6 2 5+6 2 5+6 2 5+6 2 3+3 2 2+2 2 |
| | 3+4 2 3+4 2 3+4 2 3+4 2 2+2 2 3+3 2 |
| | 5+6 2 3+4 2 5+6 2 3+4 2 3+3 2 2+2 2 |
| | 5+6 2 5+6 2 3+4 2 5+6 2 2+2 2 2+2 2 |
| | 3+4 2 5+6 2 3+4 2 3+4 2 3+3 2 3+3 2 |
| | 5+6 2 3+4 2 5+6 2 5+6 2 2+2 2 3+3 2 |

Continued on next column

**TABLE XXIV– continued**

| TQC | ICSD | NoE | SF | | | |
|---|---|---|---|---|---|---|
| | 5+6 2 | 5+6 2 | 5+6 2 | 3+4 2 | 3+3 2 | 3+3 2 |
| | 3+4 2 | 3+4 2 | 3+4 2 | 5+6 2 | 2+2 2 | 2+2 2 |
| | 3+4 2 | 3+4 2 | 5+6 2 | 3+4 2 | 3+3 2 | 2+2 2 |
| | 3+4 2 | 5+6 2 | 3+4 2 | 5+6 2 | 2+2 2 | 2+2 2 |
| | 3+4 2 | 5+6 2 | 5+6 2 | 5+6 2 | 3+3 2 | 3+3 2 |
| | 5+6 2 | 5+6 2 | 3+4 2 | 3+4 2 | 3+3 2 | 2+2 2 |
| | 5+6 2 | 3+4 2 | 5+6 2 | 5+6 2 | 2+2 2 | 3+3 2 |
| | 5+6 2 | 3+4 2 | 3+4 2 | 3+4 2 | 2+2 2 | 3+3 2 |
| | 3+4 2 | 5+6 2 | 5+6 2 | 5+6 2 | 2+2 2 | 2+2 2 |
| | 3+4 2 | 3+4 2 | 3+4 2 | 3+4 2 | 3+3 2 | 3+3 2 |
| | 5+6 2 | 5+6 2 | 5+6 2 | 5+6 2 | 3+3 2 | 2+2 2 |
| | 3+4 2 | 3+4 2 | 3+4 2 | 3+4 2 | 2+2 2 | 3+3 2 |
| | 3+4 2 | 5+6 2 | 5+6 2 | 5+6 2 | 3+3 2 | 2+2 2 |
| | 5+6 2 | 3+4 2 | 5+6 2 | 3+4 2 | 3+3 2 | 2+2 2 |
| | 5+6 2 | 5+6 2 | 5+6 2 | 5+6 2 | 2+2 2 | 2+2 2 |
| | 3+4 2 | 3+4 2 | 3+4 2 | 3+4 2 | 3+3 2 | 3+3 2 |
| | 5+6 2 | 3+4 2 | 3+4 2 | 5+6 2 | 2+2 2 | 3+3 2 |
| | 3+4 2 | 5+6 2 | 5+6 2 | 5+6 2 | 2+2 2 | 2+2 2 |
| | 3+4 2 | 5+6 2 | 5+6 2 | 3+4 2 | 3+3 2 | 3+3 2 |
| | 5+6 2 | 3+4 2 | 3+4 2 | 5+6 2 | 2+2 2 | 2+2 2 |
| | 5+6 2 | 5+6 2 | 3+4 2 | 3+4 2 | 2+2 2 | 3+3 2 |
| | 5+6 2 | 5+6 2 | 5+6 2 | 3+4 2 | 3+3 2 | 2+2 2 |
| | 3+4 2 | 3+4 2 | 3+4 2 | 5+6 2 | 3+3 2 | 3+3 2 |
| | 5+6 2 | 5+6 2 | 5+6 2 | 3+4 2 | 2+2 2 | 2+2 2 |
| | 3+4 2 | 3+4 2 | 3+4 2 | 5+6 2 | 3+3 2 | 3+3 2 |
| | 3+4 2 | 5+6 2 | 5+6 2 | 5+6 2 | 2+2 2 | 3+3 2 |
| | 5+6 2 | 3+4 2 | 3+4 2 | 3+4 2 | 3+3 2 | 2+2 2 |
| | 3+4 2 | 3+4 2 | 3+4 2 | 5+6 2 | 3+3 2 | 2+2 2 |
| | 5+6 2 | 5+6 2 | 5+6 2 | 3+4 2 | 2+2 2 | 3+3 2 |
| | 5+6 2 | 3+4 2 | 5+6 2 | 5+6 2 | 2+2 2 | 2+2 2 |
| | 5+6 2 | 5+6 2 | 3+4 2 | 3+4 2 | 3+3 2 | 2+2 2 |
| | 3+4 2 | 3+4 2 | 5+6 2 | 5+6 2 | 2+2 2 | 3+3 2 |
| | 3+4 2 | 5+6 2 | 3+4 2 | 3+4 2 | 3+3 2 | 3+3 2 |
| | 5+6 2 | 3+4 2 | 5+6 2 | 3+4 2 | 3+3 2 | 2+2 2 |
| | 3+4 2 | 5+6 2 | 3+4 2 | 5+6 2 | 2+2 2 | 3+3 2 |
| | 5+6 2 | 3+4 2 | 5+6 2 | 3+4 2 | 3+3 2 | 3+3 2 |
| | 3+4 2 | 3+4 2 | 3+4 2 | 5+6 2 | 2+2 2 | 2+2 2 |
| | 5+6 2 | 5+6 2 | 3+4 2 | 3+4 2 | 2+2 2 | 3+3 2 |
| | 3+4 2 | 5+6 2 | 5+6 2 | 5+6 2 | 3+3 2 | 2+2 2 |
| | 3+4 2 | 5+6 2 | 3+4 2 | 3+4 2 | 3+3 2 | 2+2 2 |
| | 5+6 2 | 3+4 2 | 3+4 2 | 3+4 2 | 2+2 2 | 3+3 2 |
| | 3+4 2 | 3+4 2 | 5+6 2 | 5+6 2 | 2+2 2 | 2+2 2 |
| | 5+6 2 | 5+6 2 | 5+6 2 | 5+6 2 | 3+3 2 | 3+3 2 |
| | 5+6 2 | 5+6 2 | 5+6 2 | 3+4 2 | 3+3 2 | 2+2 2 |

The End

TABLE XXV: The computed irreducible representations for the valence bands for the compounds of SEBR and NLC in space group 13. See the list of tables.

| TQC | ICSD | NoE | SF | | | | |
|---|---|---|---|---|---|---|---|
| 13.2.1 | 415459 | 104 | O4Ta1Tm1 | | | | |
| | -A | -B | -C | -D | -E | -GM | -Y | -Z |
| | 2+2 4 | 2+2 4 | 3+4 2 | 2+2 4 | 2+2 4 | 3+4 2 | 3+4 2 | 3+4 2 |
| | 2+2 4 | 2+2 4 | 5+6 2 | 2+2 4 | 2+2 4 | 5+6 2 | 5+6 2 | 5+6 2 |
| | 2+2 4 | 2+2 4 | 3+4 2 | 2+2 4 | 2+2 4 | 3+4 2 | 3+4 2 | 3+4 2 |
| | 2+2 4 | 2+2 4 | 5+6 2 | 2+2 4 | 2+2 4 | 5+6 2 | 5+6 2 | 5+6 2 |
| | 2+2 4 | 2+2 4 | 3+4 2 | 2+2 4 | 2+2 4 | 3+4 2 | 3+4 2 | 3+4 2 |
| | 2+2 4 | 2+2 4 | 5+6 2 | 2+2 4 | 2+2 4 | 5+6 2 | 5+6 2 | 5+6 2 |
| | 2+2 4 | 2+2 4 | 3+4 2 | 2+2 4 | 2+2 4 | 3+4 2 | 3+4 2 | 3+4 2 |

Continued on next column

**TABLE XXV– continued**

| TQC | ICSD | NoE | SF | | | | | |
|---|---|---|---|---|---|---|---|---|
| | 2+2 4 | 2+2 4 | 5+6 2 | 2+2 4 | 2+2 4 | 5+6 2 | 5+6 2 | 5+6 2 |
| | 2+2 4 | 2+2 4 | 3+4 2 | 2+2 4 | 2+2 4 | 3+4 2 | 5+6 2 | 5+6 2 |
| | 2+2 4 | 2+2 4 | 3+4 2 | 2+2 4 | 2+2 4 | 5+6 2 | 3+4 2 | 3+4 2 |
| | 2+2 4 | 2+2 4 | 5+6 2 | 2+2 4 | 2+2 4 | 3+4 2 | 5+6 2 | 5+6 2 |
| | 2+2 4 | 2+2 4 | 5+6 2 | 2+2 4 | 2+2 4 | 5+6 2 | 3+4 2 | 3+4 2 |
| | 2+2 4 | 2+2 4 | 3+4 2 | 2+2 4 | 2+2 4 | 5+6 2 | 5+6 2 | 3+4 2 |
| | 2+2 4 | 2+2 4 | 5+6 2 | 2+2 4 | 2+2 4 | 3+4 2 | 3+4 2 | 5+6 2 |
| | 2+2 4 | 2+2 4 | 5+6 2 | 2+2 4 | 2+2 4 | 3+4 2 | 5+6 2 | 5+6 2 |
| | 2+2 4 | 2+2 4 | 3+4 2 | 2+2 4 | 2+2 4 | 5+6 2 | 3+4 2 | 3+4 2 |
| | 2+2 4 | 2+2 4 | 3+4 2 | 2+2 4 | 2+2 4 | 3+4 2 | 3+4 2 | 3+4 2 |
| | 2+2 4 | 2+2 4 | 5+6 2 | 2+2 4 | 2+2 4 | 5+6 2 | 5+6 2 | 5+6 2 |
| | 2+2 4 | 2+2 4 | 3+4 2 | 2+2 4 | 2+2 4 | 5+6 2 | 5+6 2 | 5+6 2 |
| | 2+2 4 | 2+2 4 | 5+6 2 | 2+2 4 | 2+2 4 | 3+4 2 | 5+6 2 | 5+6 2 |
| | 2+2 4 | 2+2 4 | 3+4 2 | 2+2 4 | 2+2 4 | 3+4 2 | 3+4 2 | 3+4 2 |
| | 2+2 4 | 2+2 4 | 5+6 2 | 2+2 4 | 2+2 4 | 3+4 2 | 5+6 2 | 5+6 2 |
| | 2+2 4 | 2+2 4 | 3+4 2 | 2+2 4 | 2+2 4 | 3+4 2 | 3+4 2 | 3+4 2 |
| | 2+2 4 | 2+2 4 | 3+4 2 | 2+2 4 | 2+2 4 | 3+4 2 | 3+4 2 | 3+4 2 |
| | 2+2 4 | 2+2 4 | 5+6 2 | 2+2 4 | 2+2 4 | 5+6 2 | 5+6 2 | 3+4 2 |
| | 2+2 4 | 2+2 4 | 5+6 2 | 2+2 4 | 2+2 4 | 5+6 2 | 3+4 2 | 5+6 2 |
| | | | 5+6 2 | | | 5+6 2 | 5+6 2 | 3+4 2 |
| | | | 3+4 2 | | | 3+4 2 | 3+4 2 | 5+6 2 |
| | | | 3+4 2 | | | 3+4 2 | 3+4 2 | 5+6 2 |
| | | | 5+6 2 | | | 5+6 2 | 5+6 2 | 3+4 2 |
| | | | 3+4 2 | | | 3+4 2 | 3+4 2 | 5+6 2 |
| | | | 5+6 2 | | | 5+6 2 | 5+6 2 | 5+6 2 |
| | | | 5+6 2 | | | 5+6 2 | 5+6 2 | 3+4 2 |
| | | | 3+4 2 | | | 3+4 2 | 3+4 2 | 3+4 2 |
| | | | 3+4 2 | | | 5+6 2 | 5+6 2 | 5+6 2 |
| | | | 5+6 2 | | | 3+4 2 | 3+4 2 | 5+6 2 |
| | | | 3+4 2 | | | 5+6 2 | 5+6 2 | 3+4 2 |
| | | | 5+6 2 | | | 3+4 2 | 3+4 2 | 3+4 2 |
| | | | 5+6 2 | | | 5+6 2 | 3+4 2 | 3+4 2 |
| | | | 3+4 2 | | | 3+4 2 | 5+6 2 | 5+6 2 |
| | | | 3+4 2 | | | 3+4 2 | 5+6 2 | 5+6 2 |
| | | | 5+6 2 | | | 5+6 2 | 3+4 2 | 3+4 2 |
| | | | 3+4 2 | | | 3+4 2 | 3+4 2 | 3+4 2 |
| | | | 5+6 2 | | | 5+6 2 | 5+6 2 | 5+6 2 |
| | | | 3+4 2 | | | 5+6 2 | 3+4 2 | 3+4 2 |
| | | | 5+6 2 | | | 5+6 2 | 5+6 2 | 5+6 2 |
| | | | 3+4 2 | | | 3+4 2 | 3+4 2 | 3+4 2 |
| | | | 5+6 2 | | | 3+4 2 | 3+4 2 | 5+6 2 |
| | | | 3+4 2 | | | 5+6 2 | 5+6 2 | 3+4 2 |
| | | | 5+6 2 | | | 3+4 2 | 3+4 2 | 5+6 2 |
| | | | 3+4 2 | | | 3+4 2 | 5+6 2 | 3+4 2 |
| | | | 5+6 2 | | | 5+6 2 | 5+6 2 | 5+6 2 |
| 13.2.2 | 415434 | 96 | O4Ta1Tb1 | | | | | |
| | -A | -B | -C | -D | -E | -GM | -Y | -Z |
| | 2+2 4 | 2+2 4 | 3+4 2 | 2+2 4 | 2+2 4 | 3+4 2 | 3+4 2 | 3+4 2 |
| | 2+2 4 | 2+2 4 | 5+6 2 | 2+2 4 | 2+2 4 | 5+6 2 | 5+6 2 | 5+6 2 |
| | 2+2 4 | 2+2 4 | 3+4 2 | 2+2 4 | 2+2 4 | 3+4 2 | 3+4 2 | 3+4 2 |
| | 2+2 4 | 2+2 4 | 5+6 2 | 2+2 4 | 2+2 4 | 5+6 2 | 5+6 2 | 5+6 2 |
| | 2+2 4 | 2+2 4 | 3+4 2 | 2+2 4 | 2+2 4 | 3+4 2 | 3+4 2 | 5+6 2 |
| | 2+2 4 | 2+2 4 | 5+6 2 | 2+2 4 | 2+2 4 | 5+6 2 | 5+6 2 | 3+4 2 |
| | 2+2 4 | 2+2 4 | 3+4 2 | 2+2 4 | 2+2 4 | 3+4 2 | 3+4 2 | 5+6 2 |
| | 2+2 4 | 2+2 4 | 5+6 2 | 2+2 4 | 2+2 4 | 5+6 2 | 5+6 2 | 3+4 2 |
| | 2+2 4 | 2+2 4 | 3+4 2 | 2+2 4 | 2+2 4 | 3+4 2 | 5+6 2 | 5+6 2 |
| | 2+2 4 | 2+2 4 | 3+4 2 | 2+2 4 | 2+2 4 | 5+6 2 | 3+4 2 | 3+4 2 |
| | 2+2 4 | 2+2 4 | 5+6 2 | 2+2 4 | 2+2 4 | 3+4 2 | 5+6 2 | 3+4 2 |
| | 2+2 4 | 2+2 4 | 5+6 2 | 2+2 4 | 2+2 4 | 5+6 2 | 3+4 2 | 5+6 2 |
| | 2+2 4 | 2+2 4 | 3+4 2 | 2+2 4 | 2+2 4 | 5+6 2 | 5+6 2 | 5+6 2 |
| | 2+2 4 | 2+2 4 | 5+6 2 | 2+2 4 | 2+2 4 | 3+4 2 | 3+4 2 | 3+4 2 |

Continued on next column

**TABLE XXV– continued**

| TQC | ICSD | | NoE | SF | | | | |
|---|---|---|---|---|---|---|---|---|
| | 2+2 4 | 2+2 4 | 5+6 2 | 2+2 4 | 2+2 4 | 5+6 2 | 5+6 2 | 5+6 2 |
| | 2+2 4 | 2+2 4 | 3+4 2 | 2+2 4 | 2+2 4 | 3+4 2 | 3+4 2 | 3+4 2 |
| | 2+2 4 | 2+2 4 | 3+4 2 | 2+2 4 | 2+2 4 | 3+4 2 | 3+4 2 | 3+4 2 |
| | 2+2 4 | 2+2 4 | 5+6 2 | 2+2 4 | 2+2 4 | 5+6 2 | 5+6 2 | 5+6 2 |
| | 2+2 4 | 2+2 4 | 3+4 2 | 2+2 4 | 2+2 4 | 5+6 2 | 5+6 2 | 5+6 2 |
| | 2+2 4 | 2+2 4 | 3+4 2 | 2+2 4 | 2+2 4 | 3+4 2 | 5+6 2 | 5+6 2 |
| | 2+2 4 | 2+2 4 | 5+6 2 | 2+2 4 | 2+2 4 | 3+4 2 | 3+4 2 | 3+4 2 |
| | 2+2 4 | 2+2 4 | 5+6 2 | 2+2 4 | 2+2 4 | 3+4 2 | 5+6 2 | 5+6 2 |
| | 2+2 4 | 2+2 4 | 3+4 2 | 2+2 4 | 2+2 4 | 3+4 2 | 3+4 2 | 3+4 2 |
| | 2+2 4 | 2+2 4 | 3+4 2 | 2+2 4 | 2+2 4 | 3+4 2 | 3+4 2 | 3+4 2 |
| | | | 5+6 2 | | | 5+6 2 | 5+6 2 | 5+6 2 |
| | | | 5+6 2 | | | 5+6 2 | 3+4 2 | 3+4 2 |
| | | | 5+6 2 | | | 5+6 2 | 5+6 2 | 3+4 2 |
| | | | 3+4 2 | | | 3+4 2 | 3+4 2 | 5+6 2 |
| | | | 3+4 2 | | | 3+4 2 | 3+4 2 | 5+6 2 |
| | | | 5+6 2 | | | 5+6 2 | 5+6 2 | 3+4 2 |
| | | | 3+4 2 | | | 5+6 2 | 3+4 2 | 5+6 2 |
| | | | 5+6 2 | | | 3+4 2 | 5+6 2 | 5+6 2 |
| | | | 5+6 2 | | | 5+6 2 | 3+4 2 | 3+4 2 |
| | | | 3+4 2 | | | 3+4 2 | 5+6 2 | 3+4 2 |
| | | | 3+4 2 | | | 3+4 2 | 5+6 2 | 5+6 2 |
| | | | 5+6 2 | | | 5+6 2 | 3+4 2 | 5+6 2 |
| | | | 5+6 2 | | | 5+6 2 | 5+6 2 | 3+4 2 |
| | | | 3+4 2 | | | 3+4 2 | 3+4 2 | 3+4 2 |
| | | | 3+4 2 | | | 5+6 2 | 3+4 2 | 3+4 2 |
| | | | 5+6 2 | | | 5+6 2 | 5+6 2 | 5+6 2 |
| | | | 5+6 2 | | | 3+4 2 | 3+4 2 | 5+6 2 |
| | | | 3+4 2 | | | 3+4 2 | 3+4 2 | 3+4 2 |
| | | | 3+4 2 | | | 3+4 2 | 5+6 2 | 3+4 2 |
| | | | 5+6 2 | | | 5+6 2 | 3+4 2 | 3+4 2 |
| | | | 3+4 2 | | | 5+6 2 | 5+6 2 | 5+6 2 |
| | | | 5+6 2 | | | 5+6 2 | 5+6 2 | 5+6 2 |
| | | | 5+6 2 | | | 3+4 2 | 3+4 2 | 3+4 2 |
| | | | 3+4 2 | | | 3+4 2 | 3+4 2 | 5+6 2 |
| 13.2.3 | 174299 | | 72 | O4Sn3 | | | | |
| | -A | -B | -C | -D | -E | -GM | -Y | -Z |
| | 2+2 4 | 2+2 4 | 3+4 2 | 2+2 4 | 2+2 4 | 3+4 2 | 5+6 2 | 5+6 2 |
| | 2+2 4 | 2+2 4 | 5+6 2 | 2+2 4 | 2+2 4 | 5+6 2 | 3+4 2 | 3+4 2 |
| | 2+2 4 | 2+2 4 | 5+6 2 | 2+2 4 | 2+2 4 | 5+6 2 | 3+4 2 | 3+4 2 |
| | 2+2 4 | 2+2 4 | 3+4 2 | 2+2 4 | 2+2 4 | 3+4 2 | 5+6 2 | 5+6 2 |
| | 2+2 4 | 2+2 4 | 3+4 2 | 2+2 4 | 2+2 4 | 3+4 2 | 5+6 2 | 5+6 2 |
| | 2+2 4 | 2+2 4 | 5+6 2 | 2+2 4 | 2+2 4 | 5+6 2 | 3+4 2 | 3+4 2 |
| | 2+2 4 | 2+2 4 | 3+4 2 | 2+2 4 | 2+2 4 | 3+4 2 | 5+6 2 | 5+6 2 |
| | 2+2 4 | 2+2 4 | 5+6 2 | 2+2 4 | 2+2 4 | 5+6 2 | 3+4 2 | 3+4 2 |
| | 2+2 4 | 2+2 4 | 5+6 2 | 2+2 4 | 2+2 4 | 5+6 2 | 3+4 2 | 3+4 2 |
| | 2+2 4 | 2+2 4 | 5+6 2 | 2+2 4 | 2+2 4 | 3+4 2 | 5+6 2 | 5+6 2 |
| | 2+2 4 | 2+2 4 | 5+6 2 | 2+2 4 | 2+2 4 | 3+4 2 | 3+4 2 | 3+4 2 |
| | 2+2 4 | 2+2 4 | 3+4 2 | 2+2 4 | 2+2 4 | 5+6 2 | 5+6 2 | 3+4 2 |
| | 2+2 4 | 2+2 4 | 3+4 2 | 2+2 4 | 2+2 4 | 3+4 2 | 5+6 2 | 5+6 2 |
| | 2+2 4 | 2+2 4 | 3+4 2 | 2+2 4 | 2+2 4 | 3+4 2 | 3+4 2 | 5+6 2 |
| | 2+2 4 | 2+2 4 | 3+4 2 | 2+2 4 | 2+2 4 | 5+6 2 | 5+6 2 | 5+6 2 |
| | 2+2 4 | 2+2 4 | 3+4 2 | 2+2 4 | 2+2 4 | 3+4 2 | 3+4 2 | 5+6 2 |
| | 2+2 4 | 2+2 4 | 5+6 2 | 2+2 4 | 2+2 4 | 5+6 2 | 5+6 2 | 3+4 2 |
| | 2+2 4 | 2+2 4 | 5+6 2 | 2+2 4 | 2+2 4 | 5+6 2 | 3+4 2 | 5+6 2 |
| | | | 3+4 2 | | | 3+4 2 | 3+4 2 | 3+4 2 |
| | | | 5+6 2 | | | 5+6 2 | 5+6 2 | 3+4 2 |
| | | | 5+6 2 | | | 5+6 2 | 3+4 2 | 3+4 2 |
| | | | 5+6 2 | | | 3+4 2 | 5+6 2 | 5+6 2 |
| | | | 3+4 2 | | | 3+4 2 | 3+4 2 | 5+6 2 |
| | | | 3+4 2 | | | 3+4 2 | 5+6 2 | 3+4 2 |
| | | | 3+4 2 | | | 5+6 2 | 5+6 2 | 5+6 2 |
| | | | 5+6 2 | | | 3+4 2 | 3+4 2 | 5+6 2 |
| | | | 3+4 2 | | | 5+6 2 | 3+4 2 | 3+4 2 |
| | | | 5+6 2 | | | 3+4 2 | 5+6 2 | 5+6 2 |
| | | | 5+6 2 | | | 3+4 2 | 5+6 2 | 3+4 2 |
| | | | 3+4 2 | | | 5+6 2 | 3+4 2 | 3+4 2 |
| | | | 5+6 2 | | | 5+6 2 | 3+4 2 | 5+6 2 |
| | | | 3+4 2 | | | 5+6 2 | 3+4 2 | 3+4 2 |
| | | | 3+4 2 | | | 5+6 2 | 5+6 2 | 3+4 2 |
| | | | 5+6 2 | | | 5+6 2 | 5+6 2 | 3+4 2 |
| | | | 3+4 2 | | | 3+4 2 | 5+6 2 | 5+6 2 |
| | | | 5+6 2 | | | 5+6 2 | 3+4 2 | 3+4 2 |
| 13.2.4 | 16685 | | 80 | Ni1O4W1 | | | | |
| | -A | -B | -C | -D | -E | -GM | -Y | -Z |
| | 2+2 4 | 2+2 4 | 3+4 2 | 2+2 4 | 2+2 4 | 3+4 2 | 3+4 2 | 3+4 2 |
| | 2+2 4 | 2+2 4 | 3+4 2 | 2+2 4 | 2+2 4 | 5+6 2 | 5+6 2 | 5+6 2 |
| | 2+2 4 | 2+2 4 | 5+6 2 | 2+2 4 | 2+2 4 | 3+4 2 | 3+4 2 | 3+4 2 |
| | 2+2 4 | 2+2 4 | 5+6 2 | 2+2 4 | 2+2 4 | 5+6 2 | 5+6 2 | 5+6 2 |
| | 2+2 4 | 2+2 4 | 5+6 2 | 2+2 4 | 2+2 4 | 5+6 2 | 5+6 2 | 3+4 2 |
| | 2+2 4 | 2+2 4 | 3+4 2 | 2+2 4 | 2+2 4 | 3+4 2 | 3+4 2 | 5+6 2 |
| | 2+2 4 | 2+2 4 | 5+6 2 | 2+2 4 | 2+2 4 | 3+4 2 | 5+6 2 | 3+4 2 |
| | 2+2 4 | 2+2 4 | 3+4 2 | 2+2 4 | 2+2 4 | 5+6 2 | 3+4 2 | 5+6 2 |
| | 2+2 4 | 2+2 4 | 3+4 2 | 2+2 4 | 2+2 4 | 3+4 2 | 3+4 2 | 3+4 2 |
| | 2+2 4 | 2+2 4 | 3+4 2 | 2+2 4 | 2+2 4 | 3+4 2 | 5+6 2 | 5+6 2 |
| | 2+2 4 | 2+2 4 | 5+6 2 | 2+2 4 | 2+2 4 | 3+4 2 | 3+4 2 | 3+4 2 |
| | 2+2 4 | 2+2 4 | 5+6 2 | 2+2 4 | 2+2 4 | 5+6 2 | 3+4 2 | 5+6 2 |
| | 2+2 4 | 2+2 4 | 3+4 2 | 2+2 4 | 2+2 4 | 5+6 2 | 5+6 2 | 3+4 2 |
| | 2+2 4 | 2+2 4 | 5+6 2 | 2+2 4 | 2+2 4 | 5+6 2 | 3+4 2 | 5+6 2 |
| | 2+2 4 | 2+2 4 | 5+6 2 | 2+2 4 | 2+2 4 | 3+4 2 | 5+6 2 | 3+4 2 |
| | 2+2 4 | 2+2 4 | 3+4 2 | 2+2 4 | 2+2 4 | 5+6 2 | 3+4 2 | 5+6 2 |
| | 2+2 4 | 2+2 4 | 3+4 2 | 2+2 4 | 2+2 4 | 3+4 2 | 5+6 2 | 3+4 2 |
| | 2+2 4 | 2+2 4 | 5+6 2 | 2+2 4 | 2+2 4 | 3+4 2 | 5+6 2 | 5+6 2 |
| | 2+2 4 | 2+2 4 | 3+4 2 | 2+2 4 | 2+2 4 | 5+6 2 | 3+4 2 | 3+4 2 |
| | 2+2 4 | 2+2 4 | 3+4 2 | 2+2 4 | 2+2 4 | 5+6 2 | 5+6 2 | 5+6 2 |
| | | | 5+6 2 | | | 3+4 2 | 3+4 2 | 3+4 2 |
| | | | 5+6 2 | | | 5+6 2 | 5+6 2 | 3+4 2 |
| | | | 3+4 2 | | | 3+4 2 | 3+4 2 | 5+6 2 |
| | | | 5+6 2 | | | 3+4 2 | 5+6 2 | 3+4 2 |
| | | | 5+6 2 | | | 5+6 2 | 3+4 2 | 5+6 2 |
| | | | 3+4 2 | | | 3+4 2 | 5+6 2 | 3+4 2 |
| | | | 3+4 2 | | | 5+6 2 | 3+4 2 | 5+6 2 |
| | | | 5+6 2 | | | 3+4 2 | 5+6 2 | 3+4 2 |
| | | | 3+4 2 | | | 5+6 2 | 5+6 2 | 5+6 2 |
| | | | 5+6 2 | | | 5+6 2 | 5+6 2 | 5+6 2 |
| | | | 5+6 2 | | | 5+6 2 | 3+4 2 | 5+6 2 |
| | | | 3+4 2 | | | 5+6 2 | 3+4 2 | 3+4 2 |
| | | | 5+6 2 | | | 3+4 2 | 5+6 2 | 5+6 2 |
| | | | 3+4 2 | | | 3+4 2 | 3+4 2 | 3+4 2 |
| | | | 3+4 2 | | | 3+4 2 | 5+6 2 | 5+6 2 |
| | | | 3+4 2 | | | 3+4 2 | 3+4 2 | 5+6 2 |
| | | | 5+6 2 | | | 3+4 2 | 5+6 2 | 3+4 2 |
| | | | 3+4 2 | | | 5+6 2 | 3+4 2 | 5+6 2 |
| | | | 5+6 2 | | | 5+6 2 | 5+6 2 | 3+4 2 |
| | | | 5+6 2 | | | 3+4 2 | 3+4 2 | 3+4 2 |
| 13.2.5 | 15852 | | 80 | Ni1O4W1 | | | | |
| | -A | -B | -C | -D | -E | -GM | -Y | -Z |
| | 2+2 4 | 2+2 4 | 3+4 2 | 2+2 4 | 2+2 4 | 3+4 2 | 3+4 2 | 3+4 2 |
| | 2+2 4 | 2+2 4 | 3+4 2 | 2+2 4 | 2+2 4 | 5+6 2 | 5+6 2 | 5+6 2 |
| | 2+2 4 | 2+2 4 | 5+6 2 | 2+2 4 | 2+2 4 | 3+4 2 | 5+6 2 | 3+4 2 |
| | 2+2 4 | 2+2 4 | 5+6 2 | 2+2 4 | 2+2 4 | 5+6 2 | 3+4 2 | 5+6 2 |
| | 2+2 4 | 2+2 4 | 5+6 2 | 2+2 4 | 2+2 4 | 3+4 2 | 3+4 2 | 5+6 2 |
| | 2+2 4 | 2+2 4 | 3+4 2 | 2+2 4 | 2+2 4 | 5+6 2 | 5+6 2 | 3+4 2 |

Continued on next column

**TABLE XXV– continued**

| TQC | ICSD | NoE | SF | | | | | |
|---|---|---|---|---|---|---|---|---|
| | 2+2 4 | 2+2 4 | 5+6 2 | 2+2 4 | 2+2 4 | 3+4 2 | 5+6 2 | 3+4 2 |
| | 2+2 4 | 2+2 4 | 3+4 2 | 2+2 4 | 2+2 4 | 5+6 2 | 3+4 2 | 5+6 2 |
| | 2+2 4 | 2+2 4 | 3+4 2 | 2+2 4 | 2+2 4 | 3+4 2 | 3+4 2 | 5+6 2 |
| | 2+2 4 | 2+2 4 | 5+6 2 | 2+2 4 | 2+2 4 | 5+6 2 | 5+6 2 | 3+4 2 |
| | 2+2 4 | 2+2 4 | 3+4 2 | 2+2 4 | 2+2 4 | 3+4 2 | 3+4 2 | 5+6 2 |
| | 2+2 4 | 2+2 4 | 5+6 2 | 2+2 4 | 2+2 4 | 3+4 2 | 5+6 2 | 3+4 2 |
| | 2+2 4 | 2+2 4 | 3+4 2 | 2+2 4 | 2+2 4 | 3+4 2 | 5+6 2 | 3+4 2 |
| | 2+2 4 | 2+2 4 | 3+4 2 | 2+2 4 | 2+2 4 | 5+6 2 | 3+4 2 | 5+6 2 |
| | 2+2 4 | 2+2 4 | 5+6 2 | 2+2 4 | 2+2 4 | 5+6 2 | 3+4 2 | 3+4 2 |
| | 2+2 4 | 2+2 4 | 5+6 2 | 2+2 4 | 2+2 4 | 3+4 2 | 3+4 2 | 5+6 2 |
| | 2+2 4 | 2+2 4 | 3+4 2 | 2+2 4 | 2+2 4 | 5+6 2 | 5+6 2 | 3+4 2 |
| | 2+2 4 | 2+2 4 | 5+6 2 | 2+2 4 | 2+2 4 | 3+4 2 | 5+6 2 | 5+6 2 |
| | 2+2 4 | 2+2 4 | 3+4 2 | 2+2 4 | 2+2 4 | 5+6 2 | 3+4 2 | 5+6 2 |
| | 2+2 4 | 2+2 4 | 5+6 2 | 2+2 4 | 2+2 4 | 5+6 2 | 3+4 2 | 3+4 2 |
| | | | 5+6 2 | | | 3+4 2 | 5+6 2 | 3+4 2 |
| | | | 3+4 2 | | | 3+4 2 | 3+4 2 | 5+6 2 |
| | | | 3+4 2 | | | 3+4 2 | 5+6 2 | 3+4 2 |
| | | | 5+6 2 | | | 5+6 2 | 3+4 2 | 5+6 2 |
| | | | 3+4 2 | | | 5+6 2 | 5+6 2 | 3+4 2 |
| | | | 3+4 2 | | | 3+4 2 | 5+6 2 | 5+6 2 |
| | | | 5+6 2 | | | 5+6 2 | 5+6 2 | 5+6 2 |
| | | | 5+6 2 | | | 3+4 2 | 5+6 2 | 3+4 2 |
| | | | 5+6 2 | | | 5+6 2 | 3+4 2 | 5+6 2 |
| | | | 5+6 2 | | | 5+6 2 | 5+6 2 | 3+4 2 |
| | | | 3+4 2 | | | 3+4 2 | 3+4 2 | 3+4 2 |
| | | | 3+4 2 | | | 5+6 2 | 3+4 2 | 5+6 2 |
| | | | 3+4 2 | | | 3+4 2 | 3+4 2 | 5+6 2 |
| | | | 5+6 2 | | | 5+6 2 | 5+6 2 | 3+4 2 |
| | | | 3+4 2 | | | 3+4 2 | 5+6 2 | 5+6 2 |
| | | | 5+6 2 | | | 3+4 2 | 3+4 2 | 5+6 2 |
| | | | 5+6 2 | | | 5+6 2 | 5+6 2 | 3+4 2 |
| | | | 3+4 2 | | | 5+6 2 | 3+4 2 | 3+4 2 |
| | | | 3+4 2 | | | 3+4 2 | 3+4 2 | 3+4 2 |
| | | | 3+4 2 | | | 3+4 2 | 3+4 2 | 5+6 2 |
| 13.2.6 | 27701 | | 96 | Mo8O23 | | | | |
| | -A | -B | -C | -D | -E | -GM | -Y | -Z |
| | 2+2 4 | 2+2 4 | 3+4 2 | 2+2 4 | 2+2 4 | 3+4 2 | 5+6 2 | 5+6 2 |
| | 2+2 4 | 2+2 4 | 3+4 2 | 2+2 4 | 2+2 4 | 3+4 2 | 3+4 2 | 5+6 2 |
| | 2+2 4 | 2+2 4 | 5+6 2 | 2+2 4 | 2+2 4 | 5+6 2 | 5+6 2 | 3+4 2 |
| | 2+2 4 | 2+2 4 | 5+6 2 | 2+2 4 | 2+2 4 | 3+4 2 | 3+4 2 | 3+4 2 |
| | 2+2 4 | 2+2 4 | 5+6 2 | 2+2 4 | 2+2 4 | 3+4 2 | 5+6 2 | 3+4 2 |
| | 2+2 4 | 2+2 4 | 5+6 2 | 2+2 4 | 2+2 4 | 5+6 2 | 3+4 2 | 5+6 2 |
| | 2+2 4 | 2+2 4 | 3+4 2 | 2+2 4 | 2+2 4 | 5+6 2 | 3+4 2 | 5+6 2 |
| | 2+2 4 | 2+2 4 | 5+6 2 | 2+2 4 | 2+2 4 | 5+6 2 | 5+6 2 | 3+4 2 |
| | 2+2 4 | 2+2 4 | 3+4 2 | 2+2 4 | 2+2 4 | 3+4 2 | 5+6 2 | 5+6 2 |
| | 2+2 4 | 2+2 4 | 3+4 2 | 2+2 4 | 2+2 4 | 5+6 2 | 5+6 2 | 3+4 2 |
| | 2+2 4 | 2+2 4 | 3+4 2 | 2+2 4 | 2+2 4 | 3+4 2 | 3+4 2 | 5+6 2 |
| | 2+2 4 | 2+2 4 | 5+6 2 | 2+2 4 | 2+2 4 | 5+6 2 | 3+4 2 | 5+6 2 |
| | 2+2 4 | 2+2 4 | 3+4 2 | 2+2 4 | 2+2 4 | 5+6 2 | 3+4 2 | 3+4 2 |
| | 2+2 4 | 2+2 4 | 5+6 2 | 2+2 4 | 2+2 4 | 5+6 2 | 3+4 2 | 3+4 2 |
| | 2+2 4 | 2+2 4 | 5+6 2 | 2+2 4 | 2+2 4 | 3+4 2 | 3+4 2 | 5+6 2 |
| | 2+2 4 | 2+2 4 | 3+4 2 | 2+2 4 | 2+2 4 | 3+4 2 | 5+6 2 | 3+4 2 |
| | 2+2 4 | 2+2 4 | 5+6 2 | 2+2 4 | 2+2 4 | 3+4 2 | 5+6 2 | 3+4 2 |
| | 2+2 4 | 2+2 4 | 5+6 2 | 2+2 4 | 2+2 4 | 3+4 2 | 5+6 2 | 5+6 2 |
| | 2+2 4 | 2+2 4 | 5+6 2 | 2+2 4 | 2+2 4 | 5+6 2 | 5+6 2 | 3+4 2 |
| | 2+2 4 | 2+2 4 | 3+4 2 | 2+2 4 | 2+2 4 | 3+4 2 | 3+4 2 | 5+6 2 |
| | 2+2 4 | 2+2 4 | 3+4 2 | 2+2 4 | 2+2 4 | 5+6 2 | 3+4 2 | 5+6 2 |
| | 2+2 4 | 2+2 4 | 5+6 2 | 2+2 4 | 2+2 4 | 5+6 2 | 3+4 2 | 3+4 2 |
| | 2+2 4 | 2+2 4 | 3+4 2 | 2+2 4 | 2+2 4 | 5+6 2 | 5+6 2 | 3+4 2 |
| | 2+2 4 | 2+2 4 | 5+6 2 | 2+2 4 | 2+2 4 | 3+4 2 | 5+6 2 | 3+4 2 |
| | | | 3+4 2 | | | 3+4 2 | 3+4 2 | 5+6 2 |

Continued on next column

**TABLE XXV– continued**

| TQC | ICSD | NoE | SF | | | | | |
|---|---|---|---|---|---|---|---|---|
| | | | 5+6 2 | | | 5+6 2 | 5+6 2 | 5+6 2 |
| | | | 3+4 2 | | | 3+4 2 | 3+4 2 | 3+4 2 |
| | | | 3+4 2 | | | 5+6 2 | 3+4 2 | 3+4 2 |
| | | | 5+6 2 | | | 3+4 2 | 5+6 2 | 5+6 2 |
| | | | 3+4 2 | | | 5+6 2 | 5+6 2 | 3+4 2 |
| | | | 5+6 2 | | | 3+4 2 | 3+4 2 | 5+6 2 |
| | | | 5+6 2 | | | 5+6 2 | 5+6 2 | 3+4 2 |
| | | | 3+4 2 | | | 5+6 2 | 5+6 2 | 5+6 2 |
| | | | 3+4 2 | | | 3+4 2 | 3+4 2 | 3+4 2 |
| | | | 5+6 2 | | | 3+4 2 | 3+4 2 | 5+6 2 |
| | | | 3+4 2 | | | 5+6 2 | 5+6 2 | 5+6 2 |
| | | | 3+4 2 | | | 5+6 2 | 3+4 2 | 3+4 2 |
| | | | 3+4 2 | | | 3+4 2 | 3+4 2 | 3+4 2 |
| | | | 5+6 2 | | | 3+4 2 | 5+6 2 | 5+6 2 |
| | | | 3+4 2 | | | 5+6 2 | 5+6 2 | 5+6 2 |
| | | | 5+6 2 | | | 3+4 2 | 3+4 2 | 5+6 2 |
| | | | 5+6 2 | | | 3+4 2 | 5+6 2 | 3+4 2 |
| | | | 3+4 2 | | | 3+4 2 | 3+4 2 | 3+4 2 |
| | | | 5+6 2 | | | 5+6 2 | 5+6 2 | 5+6 2 |
| | | | 3+4 2 | | | 5+6 2 | 5+6 2 | 5+6 2 |
| | | | 5+6 2 | | | 3+4 2 | 3+4 2 | 3+4 2 |
| | | | 3+4 2 | | | 5+6 2 | 5+6 2 | 3+4 2 |
| | | | 5+6 2 | | | 5+6 2 | 3+4 2 | 3+4 2 |
| 13.2.7 | 81060 | | 80 | Mo1Ni1O4 | | | | |
| | -A | -B | -C | -D | -E | -GM | -Y | -Z |
| | 2+2 4 | 2+2 4 | 5+6 2 | 2+2 4 | 2+2 4 | 3+4 2 | 5+6 2 | 5+6 2 |
| | 2+2 4 | 2+2 4 | 5+6 2 | 2+2 4 | 2+2 4 | 5+6 2 | 3+4 2 | 3+4 2 |
| | 2+2 4 | 2+2 4 | 3+4 2 | 2+2 4 | 2+2 4 | 3+4 2 | 5+6 2 | 3+4 2 |
| | 2+2 4 | 2+2 4 | 3+4 2 | 2+2 4 | 2+2 4 | 5+6 2 | 3+4 2 | 5+6 2 |
| | 2+2 4 | 2+2 4 | 5+6 2 | 2+2 4 | 2+2 4 | 3+4 2 | 5+6 2 | 3+4 2 |
| | 2+2 4 | 2+2 4 | 3+4 2 | 2+2 4 | 2+2 4 | 5+6 2 | 3+4 2 | 5+6 2 |
| | 2+2 4 | 2+2 4 | 3+4 2 | 2+2 4 | 2+2 4 | 3+4 2 | 3+4 2 | 3+4 2 |
| | 2+2 4 | 2+2 4 | 5+6 2 | 2+2 4 | 2+2 4 | 5+6 2 | 5+6 2 | 5+6 2 |
| | 2+2 4 | 2+2 4 | 5+6 2 | 2+2 4 | 2+2 4 | 3+4 2 | 5+6 2 | 5+6 2 |
| | 2+2 4 | 2+2 4 | 3+4 2 | 2+2 4 | 2+2 4 | 5+6 2 | 3+4 2 | 5+6 2 |
| | 2+2 4 | 2+2 4 | 3+4 2 | 2+2 4 | 2+2 4 | 3+4 2 | 3+4 2 | 3+4 2 |
| | 2+2 4 | 2+2 4 | 5+6 2 | 2+2 4 | 2+2 4 | 3+4 2 | 5+6 2 | 3+4 2 |
| | 2+2 4 | 2+2 4 | 5+6 2 | 2+2 4 | 2+2 4 | 3+4 2 | 5+6 2 | 3+4 2 |
| | 2+2 4 | 2+2 4 | 5+6 2 | 2+2 4 | 2+2 4 | 5+6 2 | 3+4 2 | 3+4 2 |
| | 2+2 4 | 2+2 4 | 3+4 2 | 2+2 4 | 2+2 4 | 5+6 2 | 5+6 2 | 5+6 2 |
| | 2+2 4 | 2+2 4 | 3+4 2 | 2+2 4 | 2+2 4 | 3+4 2 | 5+6 2 | 5+6 2 |
| | 2+2 4 | 2+2 4 | 5+6 2 | 2+2 4 | 2+2 4 | 5+6 2 | 3+4 2 | 3+4 2 |
| | 2+2 4 | 2+2 4 | 3+4 2 | 2+2 4 | 2+2 4 | 5+6 2 | 5+6 2 | 5+6 2 |
| | 2+2 4 | 2+2 4 | 5+6 2 | 2+2 4 | 2+2 4 | 3+4 2 | 3+4 2 | 5+6 2 |
| | 2+2 4 | 2+2 4 | 5+6 2 | 2+2 4 | 2+2 4 | 3+4 2 | 5+6 2 | 3+4 2 |
| | | | 3+4 2 | | | 5+6 2 | 3+4 2 | 3+4 2 |
| | | | 3+4 2 | | | 3+4 2 | 5+6 2 | 3+4 2 |
| | | | 5+6 2 | | | 3+4 2 | 3+4 2 | 5+6 2 |
| | | | 5+6 2 | | | 5+6 2 | 5+6 2 | 5+6 2 |
| | | | 3+4 2 | | | 3+4 2 | 3+4 2 | 3+4 2 |
| | | | 5+6 2 | | | 5+6 2 | 3+4 2 | 5+6 2 |
| | | | 3+4 2 | | | 5+6 2 | 5+6 2 | 5+6 2 |
| | | | 3+4 2 | | | 5+6 2 | 3+4 2 | 3+4 2 |
| | | | 3+4 2 | | | 3+4 2 | 3+4 2 | 3+4 2 |
| | | | 5+6 2 | | | 5+6 2 | 3+4 2 | 5+6 2 |
| | | | 3+4 2 | | | 3+4 2 | 5+6 2 | 3+4 2 |
| | | | 5+6 2 | | | 5+6 2 | 5+6 2 | 5+6 2 |
| | | | 5+6 2 | | | 3+4 2 | 3+4 2 | 5+6 2 |
| | | | 3+4 2 | | | 5+6 2 | 5+6 2 | 3+4 2 |
| | | | 5+6 2 | | | 3+4 2 | 3+4 2 | 5+6 2 |
| | | | 3+4 2 | | | 3+4 2 | 5+6 2 | 5+6 2 |

Continued on next column

**TABLE XXV– continued**

| TQC | ICSD | | NoE | SF | | | | |
|---|---|---|---|---|---|---|---|---|
| | | | 3+4 2 | | | 5+6 2 | 3+4 2 | 3+4 2 |
| | | | 5+6 2 | | | 5+6 2 | 5+6 2 | 3+4 2 |
| | | | 5+6 2 | | | 3+4 2 | 5+6 2 | 5+6 2 |
| | | | 5+6 2 | | | 3+4 2 | 3+4 2 | 3+4 2 |
| 13.2.8 | 170685 | | 76 | Mn1O4Re1 | | | | |
| | -A | -B | -C | -D | -E | -GM | -Y | -Z |
| | 2+2 4 | 2+2 4 | 3+4 2 | 2+2 4 | 2+2 4 | 3+4 2 | 3+4 2 | 3+4 2 |
| | 2+2 4 | 2+2 4 | 3+4 2 | 2+2 4 | 2+2 4 | 5+6 2 | 5+6 2 | 5+6 2 |
| | 2+2 4 | 2+2 4 | 5+6 2 | 2+2 4 | 2+2 4 | 3+4 2 | 5+6 2 | 3+4 2 |
| | 2+2 4 | 2+2 4 | 5+6 2 | 2+2 4 | 2+2 4 | 5+6 2 | 3+4 2 | 5+6 2 |
| | 2+2 4 | 2+2 4 | 5+6 2 | 2+2 4 | 2+2 4 | 3+4 2 | 3+4 2 | 5+6 2 |
| | 2+2 4 | 2+2 4 | 3+4 2 | 2+2 4 | 2+2 4 | 5+6 2 | 5+6 2 | 3+4 2 |
| | 2+2 4 | 2+2 4 | 5+6 2 | 2+2 4 | 2+2 4 | 3+4 2 | 5+6 2 | 3+4 2 |
| | 2+2 4 | 2+2 4 | 3+4 2 | 2+2 4 | 2+2 4 | 5+6 2 | 3+4 2 | 5+6 2 |
| | 2+2 4 | 2+2 4 | 3+4 2 | 2+2 4 | 2+2 4 | 3+4 2 | 5+6 2 | 3+4 2 |
| | 2+2 4 | 2+2 4 | 5+6 2 | 2+2 4 | 2+2 4 | 5+6 2 | 3+4 2 | 5+6 2 |
| | 2+2 4 | 2+2 4 | 3+4 2 | 2+2 4 | 2+2 4 | 3+4 2 | 3+4 2 | 3+4 2 |
| | 2+2 4 | 2+2 4 | 5+6 2 | 2+2 4 | 2+2 4 | 3+4 2 | 5+6 2 | 5+6 2 |
| | 2+2 4 | 2+2 4 | 3+4 2 | 2+2 4 | 2+2 4 | 3+4 2 | 5+6 2 | 3+4 2 |
| | 2+2 4 | 2+2 4 | 3+4 2 | 2+2 4 | 2+2 4 | 5+6 2 | 3+4 2 | 5+6 2 |
| | 2+2 4 | 2+2 4 | 5+6 2 | 2+2 4 | 2+2 4 | 5+6 2 | 3+4 2 | 3+4 2 |
| | 2+2 4 | 2+2 4 | 3+4 2 | 2+2 4 | 2+2 4 | 3+4 2 | 5+6 2 | 5+6 2 |
| | 2+2 4 | 2+2 4 | 5+6 2 | 2+2 4 | 2+2 4 | 5+6 2 | 3+4 2 | 3+4 2 |
| | 2+2 4 | 2+2 4 | 5+6 2 | 2+2 4 | 2+2 4 | 3+4 2 | 3+4 2 | 5+6 2 |
| | 2+2 4 | 2+2 4 | 3+4 2 | 2+2 4 | 2+2 4 | 5+6 2 | 5+6 2 | 3+4 2 |
| | | | 5+6 2 | | | 3+4 2 | 3+4 2 | 5+6 2 |
| | | | 3+4 2 | | | 5+6 2 | 3+4 2 | 3+4 2 |
| | | | 5+6 2 | | | 3+4 2 | 5+6 2 | 5+6 2 |
| | | | 3+4 2 | | | 3+4 2 | 5+6 2 | 3+4 2 |
| | | | 3+4 2 | | | 5+6 2 | 3+4 2 | 3+4 2 |
| | | | 5+6 2 | | | 3+4 2 | 5+6 2 | 5+6 2 |
| | | | 3+4 2 | | | 5+6 2 | 5+6 2 | 5+6 2 |
| | | | 5+6 2 | | | 5+6 2 | 5+6 2 | 5+6 2 |
| | | | 5+6 2 | | | 5+6 2 | 5+6 2 | 3+4 2 |
| | | | 5+6 2 | | | 3+4 2 | 3+4 2 | 3+4 2 |
| | | | 5+6 2 | | | 5+6 2 | 5+6 2 | 5+6 2 |
| | | | 3+4 2 | | | 5+6 2 | 3+4 2 | 5+6 2 |
| | | | 3+4 2 | | | 3+4 2 | 3+4 2 | 3+4 2 |
| | | | 3+4 2 | | | 3+4 2 | 3+4 2 | 3+4 2 |
| | | | 5+6 2 | | | 3+4 2 | 5+6 2 | 5+6 2 |
| | | | 5+6 2 | | | 3+4 2 | 3+4 2 | 5+6 2 |
| | | | 3+4 2 | | | 5+6 2 | 5+6 2 | 5+6 2 |
| | | | 3+4 2 | | | 5+6 2 | 5+6 2 | 3+4 2 |
| | | | 3+4 2 | | | 5+6 2 | 3+4 2 | 3+4 2 |
| 13.2.9 | 415436 | | 100 | Ho1O4Ta1 | | | | |
| | -A | -B | -C | -D | -E | -GM | -Y | -Z |
| | 2+2 4 | 2+2 4 | 3+4 2 | 2+2 4 | 2+2 4 | 3+4 2 | 3+4 2 | 3+4 2 |
| | 2+2 4 | 2+2 4 | 5+6 2 | 2+2 4 | 2+2 4 | 5+6 2 | 5+6 2 | 5+6 2 |
| | 2+2 4 | 2+2 4 | 3+4 2 | 2+2 4 | 2+2 4 | 3+4 2 | 3+4 2 | 3+4 2 |
| | 2+2 4 | 2+2 4 | 5+6 2 | 2+2 4 | 2+2 4 | 5+6 2 | 5+6 2 | 5+6 2 |
| | 2+2 4 | 2+2 4 | 3+4 2 | 2+2 4 | 2+2 4 | 3+4 2 | 3+4 2 | 3+4 2 |
| | 2+2 4 | 2+2 4 | 5+6 2 | 2+2 4 | 2+2 4 | 5+6 2 | 5+6 2 | 5+6 2 |
| | 2+2 4 | 2+2 4 | 3+4 2 | 2+2 4 | 2+2 4 | 3+4 2 | 3+4 2 | 3+4 2 |
| | 2+2 4 | 2+2 4 | 5+6 2 | 2+2 4 | 2+2 4 | 5+6 2 | 5+6 2 | 5+6 2 |
| | 2+2 4 | 2+2 4 | 3+4 2 | 2+2 4 | 2+2 4 | 3+4 2 | 5+6 2 | 5+6 2 |
| | 2+2 4 | 2+2 4 | 3+4 2 | 2+2 4 | 2+2 4 | 5+6 2 | 3+4 2 | 3+4 2 |
| | 2+2 4 | 2+2 4 | 5+6 2 | 2+2 4 | 2+2 4 | 3+4 2 | 5+6 2 | 5+6 2 |
| | 2+2 4 | 2+2 4 | 5+6 2 | 2+2 4 | 2+2 4 | 5+6 2 | 3+4 2 | 3+4 2 |
| | 2+2 4 | 2+2 4 | 3+4 2 | 2+2 4 | 2+2 4 | 5+6 2 | 5+6 2 | 5+6 2 |
| | 2+2 4 | 2+2 4 | 5+6 2 | 2+2 4 | 2+2 4 | 3+4 2 | 3+4 2 | 3+4 2 |
| | 2+2 4 | 2+2 4 | 5+6 2 | 2+2 4 | 2+2 4 | 3+4 2 | 5+6 2 | 5+6 2 |
| | 2+2 4 | 2+2 4 | 3+4 2 | 2+2 4 | 2+2 4 | 5+6 2 | 3+4 2 | 3+4 2 |
| | 2+2 4 | 2+2 4 | 3+4 2 | 2+2 4 | 2+2 4 | 3+4 2 | 3+4 2 | 3+4 2 |
| | 2+2 4 | 2+2 4 | 5+6 2 | 2+2 4 | 2+2 4 | 5+6 2 | 5+6 2 | 5+6 2 |
| | 2+2 4 | 2+2 4 | 3+4 2 | 2+2 4 | 2+2 4 | 5+6 2 | 5+6 2 | 5+6 2 |
| | 2+2 4 | 2+2 4 | 5+6 2 | 2+2 4 | 2+2 4 | 3+4 2 | 5+6 2 | 5+6 2 |
| | 2+2 4 | 2+2 4 | 3+4 2 | 2+2 4 | 2+2 4 | 3+4 2 | 3+4 2 | 3+4 2 |
| | 2+2 4 | 2+2 4 | 5+6 2 | 2+2 4 | 2+2 4 | 3+4 2 | 5+6 2 | 5+6 2 |
| | 2+2 4 | 2+2 4 | 3+4 2 | 2+2 4 | 2+2 4 | 3+4 2 | 3+4 2 | 3+4 2 |
| | 2+2 4 | 2+2 4 | 3+4 2 | 2+2 4 | 2+2 4 | 5+6 2 | 3+4 2 | 3+4 2 |
| | 2+2 4 | 2+2 4 | 5+6 2 | 2+2 4 | 2+2 4 | 3+4 2 | 5+6 2 | 5+6 2 |
| | | | 5+6 2 | | | 5+6 2 | 3+4 2 | 3+4 2 |
| | | | 5+6 2 | | | 5+6 2 | 5+6 2 | 3+4 2 |
| | | | 3+4 2 | | | 3+4 2 | 3+4 2 | 5+6 2 |
| | | | 5+6 2 | | | 3+4 2 | 5+6 2 | 5+6 2 |
| | | | 3+4 2 | | | 5+6 2 | 3+4 2 | 3+4 2 |
| | | | 3+4 2 | | | 5+6 2 | 3+4 2 | 5+6 2 |
| | | | 5+6 2 | | | 3+4 2 | 5+6 2 | 5+6 2 |
| | | | 5+6 2 | | | 5+6 2 | 5+6 2 | 3+4 2 |
| | | | 3+4 2 | | | 3+4 2 | 3+4 2 | 3+4 2 |
| | | | 3+4 2 | | | 3+4 2 | 5+6 2 | 5+6 2 |
| | | | 5+6 2 | | | 5+6 2 | 3+4 2 | 5+6 2 |
| | | | 3+4 2 | | | 3+4 2 | 5+6 2 | 3+4 2 |
| | | | 5+6 2 | | | 5+6 2 | 3+4 2 | 3+4 2 |
| | | | 3+4 2 | | | 5+6 2 | 3+4 2 | 5+6 2 |
| | | | 5+6 2 | | | 5+6 2 | 5+6 2 | 3+4 2 |
| | | | 3+4 2 | | | 3+4 2 | 3+4 2 | 5+6 2 |
| | | | 5+6 2 | | | 3+4 2 | 3+4 2 | 3+4 2 |
| | | | 3+4 2 | | | 3+4 2 | 5+6 2 | 3+4 2 |
| | | | 5+6 2 | | | 5+6 2 | 5+6 2 | 5+6 2 |
| | | | 3+4 2 | | | 5+6 2 | 3+4 2 | 3+4 2 |
| | | | 5+6 2 | | | 5+6 2 | 5+6 2 | 5+6 2 |
| | | | 3+4 2 | | | 3+4 2 | 3+4 2 | 3+4 2 |
| | | | 5+6 2 | | | 3+4 2 | 3+4 2 | 5+6 2 |
| | | | 3+4 2 | | | 5+6 2 | 5+6 2 | 3+4 2 |
| | | | 5+6 2 | | | 3+4 2 | 3+4 2 | 5+6 2 |
| 13.2.10 | 155660 | | 100 | Ho1O4Ta1 | | | | |
| | -A | -B | -C | -D | -E | -GM | -Y | -Z |
| | 2+2 4 | 2+2 4 | 3+4 2 | 2+2 4 | 2+2 4 | 3+4 2 | 5+6 2 | 5+6 2 |
| | 2+2 4 | 2+2 4 | 5+6 2 | 2+2 4 | 2+2 4 | 5+6 2 | 3+4 2 | 3+4 2 |
| | 2+2 4 | 2+2 4 | 3+4 2 | 2+2 4 | 2+2 4 | 3+4 2 | 5+6 2 | 5+6 2 |
| | 2+2 4 | 2+2 4 | 5+6 2 | 2+2 4 | 2+2 4 | 5+6 2 | 3+4 2 | 3+4 2 |
| | 2+2 4 | 2+2 4 | 3+4 2 | 2+2 4 | 2+2 4 | 3+4 2 | 5+6 2 | 3+4 2 |
| | 2+2 4 | 2+2 4 | 5+6 2 | 2+2 4 | 2+2 4 | 5+6 2 | 3+4 2 | 5+6 2 |
| | 2+2 4 | 2+2 4 | 3+4 2 | 2+2 4 | 2+2 4 | 3+4 2 | 5+6 2 | 5+6 2 |
| | 2+2 4 | 2+2 4 | 5+6 2 | 2+2 4 | 2+2 4 | 5+6 2 | 3+4 2 | 3+4 2 |
| | 2+2 4 | 2+2 4 | 3+4 2 | 2+2 4 | 2+2 4 | 3+4 2 | 3+4 2 | 3+4 2 |
| | 2+2 4 | 2+2 4 | 3+4 2 | 2+2 4 | 2+2 4 | 5+6 2 | 5+6 2 | 5+6 2 |
| | 2+2 4 | 2+2 4 | 5+6 2 | 2+2 4 | 2+2 4 | 3+4 2 | 3+4 2 | 3+4 2 |
| | 2+2 4 | 2+2 4 | 5+6 2 | 2+2 4 | 2+2 4 | 5+6 2 | 5+6 2 | 5+6 2 |
| | 2+2 4 | 2+2 4 | 3+4 2 | 2+2 4 | 2+2 4 | 5+6 2 | 3+4 2 | 5+6 2 |
| | 2+2 4 | 2+2 4 | 5+6 2 | 2+2 4 | 2+2 4 | 3+4 2 | 3+4 2 | 3+4 2 |
| | 2+2 4 | 2+2 4 | 5+6 2 | 2+2 4 | 2+2 4 | 5+6 2 | 5+6 2 | 3+4 2 |
| | 2+2 4 | 2+2 4 | 3+4 2 | 2+2 4 | 2+2 4 | 3+4 2 | 5+6 2 | 5+6 2 |
| | 2+2 4 | 2+2 4 | 5+6 2 | 2+2 4 | 2+2 4 | 3+4 2 | 5+6 2 | 5+6 2 |
| | 2+2 4 | 2+2 4 | 3+4 2 | 2+2 4 | 2+2 4 | 5+6 2 | 3+4 2 | 3+4 2 |
| | 2+2 4 | 2+2 4 | 3+4 2 | 2+2 4 | 2+2 4 | 5+6 2 | 3+4 2 | 3+4 2 |
| | 2+2 4 | 2+2 4 | 3+4 2 | 2+2 4 | 2+2 4 | 3+4 2 | 3+4 2 | 3+4 2 |
| | 2+2 4 | 2+2 4 | 5+6 2 | 2+2 4 | 2+2 4 | 3+4 2 | 5+6 2 | 3+4 2 |
| | 2+2 4 | 2+2 4 | 5+6 2 | 2+2 4 | 2+2 4 | 3+4 2 | 3+4 2 | 5+6 2 |
| | 2+2 4 | 2+2 4 | 3+4 2 | 2+2 4 | 2+2 4 | 3+4 2 | 5+6 2 | 5+6 2 |
| | 2+2 4 | 2+2 4 | 3+4 2 | 2+2 4 | 2+2 4 | 5+6 2 | 5+6 2 | 5+6 2 |

Continued on next column

Continued on next column

**TABLE XXV– continued**

| TQC | ICSD | NoE | SF | | | | |
|---|---|---|---|---|---|---|---|
| | 2+2 4 | 2+2 4 | 5+6 2 | 2+2 4 | 2+2 4 | 3+4 2 | 5+6 2 | 5+6 2 |
| | | | 5+6 2 | | | 5+6 2 | 3+4 2 | 3+4 2 |
| | | | 3+4 2 | | | 3+4 2 | 3+4 2 | 5+6 2 |
| | | | 5+6 2 | | | 3+4 2 | 5+6 2 | 3+4 2 |
| | | | 3+4 2 | | | 5+6 2 | 5+6 2 | 5+6 2 |
| | | | 5+6 2 | | | 5+6 2 | 3+4 2 | 3+4 2 |
| | | | 3+4 2 | | | 5+6 2 | 5+6 2 | 3+4 2 |
| | | | 5+6 2 | | | 3+4 2 | 5+6 2 | 3+4 2 |
| | | | 3+4 2 | | | 3+4 2 | 3+4 2 | 5+6 2 |
| | | | 5+6 2 | | | 5+6 2 | 3+4 2 | 5+6 2 |
| | | | 3+4 2 | | | 5+6 2 | 3+4 2 | 5+6 2 |
| | | | 5+6 2 | | | 3+4 2 | 5+6 2 | 5+6 2 |
| | | | 5+6 2 | | | 3+4 2 | 3+4 2 | 3+4 2 |
| | | | 3+4 2 | | | 5+6 2 | 3+4 2 | 3+4 2 |
| | | | 5+6 2 | | | 5+6 2 | 5+6 2 | 5+6 2 |
| | | | 3+4 2 | | | 5+6 2 | 5+6 2 | 3+4 2 |
| | | | 3+4 2 | | | 3+4 2 | 5+6 2 | 5+6 2 |
| | | | 5+6 2 | | | 3+4 2 | 3+4 2 | 3+4 2 |
| | | | 5+6 2 | | | 3+4 2 | 5+6 2 | 5+6 2 |
| | | | 3+4 2 | | | 5+6 2 | 3+4 2 | 3+4 2 |
| | | | 5+6 2 | | | 5+6 2 | 3+4 2 | 5+6 2 |
| | | | 3+4 2 | | | 5+6 2 | 5+6 2 | 3+4 2 |
| | | | 3+4 2 | | | 3+4 2 | 5+6 2 | 5+6 2 |
| | | | 5+6 2 | | | 3+4 2 | 5+6 2 | 3+4 2 |
| | | | 3+4 2 | | | 3+4 2 | 3+4 2 | 3+4 2 |
| | | | 5+6 2 | | | 5+6 2 | 5+6 2 | 5+6 2 |
| 13.2.11 | 261836 | 164 | Ho1Li1O8W2 | | | | | |
| | -A | -B | -C | -D | -E | -GM | -Y | -Z |
| | 2+2 4 | 2+2 4 | 3+4 2 | 2+2 4 | 2+2 4 | 3+4 2 | 3+4 2 | 3+4 2 |
| | 2+2 4 | 2+2 4 | 5+6 2 | 2+2 4 | 2+2 4 | 5+6 2 | 5+6 2 | 5+6 2 |
| | 2+2 4 | 2+2 4 | 3+4 2 | 2+2 4 | 2+2 4 | 3+4 2 | 3+4 2 | 3+4 2 |
| | 2+2 4 | 2+2 4 | 5+6 2 | 2+2 4 | 2+2 4 | 5+6 2 | 5+6 2 | 5+6 2 |
| | 2+2 4 | 2+2 4 | 3+4 2 | 2+2 4 | 2+2 4 | 3+4 2 | 3+4 2 | 3+4 2 |
| | 2+2 4 | 2+2 4 | 5+6 2 | 2+2 4 | 2+2 4 | 5+6 2 | 5+6 2 | 5+6 2 |
| | 2+2 4 | 2+2 4 | 3+4 2 | 2+2 4 | 2+2 4 | 3+4 2 | 3+4 2 | 3+4 2 |
| | 2+2 4 | 2+2 4 | 5+6 2 | 2+2 4 | 2+2 4 | 5+6 2 | 5+6 2 | 5+6 2 |
| | 2+2 4 | 2+2 4 | 3+4 2 | 2+2 4 | 2+2 4 | 3+4 2 | 3+4 2 | 3+4 2 |
| | 2+2 4 | 2+2 4 | 5+6 2 | 2+2 4 | 2+2 4 | 3+4 2 | 5+6 2 | 3+4 2 |
| | 2+2 4 | 2+2 4 | 3+4 2 | 2+2 4 | 2+2 4 | 5+6 2 | 5+6 2 | 5+6 2 |
| | 2+2 4 | 2+2 4 | 5+6 2 | 2+2 4 | 2+2 4 | 3+4 2 | 3+4 2 | 3+4 2 |
| | 2+2 4 | 2+2 4 | 3+4 2 | 2+2 4 | 2+2 4 | 5+6 2 | 5+6 2 | 5+6 2 |
| | 2+2 4 | 2+2 4 | 5+6 2 | 2+2 4 | 2+2 4 | 5+6 2 | 3+4 2 | 5+6 2 |
| | 2+2 4 | 2+2 4 | 5+6 2 | 2+2 4 | 2+2 4 | 3+4 2 | 3+4 2 | 3+4 2 |
| | 2+2 4 | 2+2 4 | 3+4 2 | 2+2 4 | 2+2 4 | 5+6 2 | 5+6 2 | 5+6 2 |
| | 2+2 4 | 2+2 4 | 5+6 2 | 2+2 4 | 2+2 4 | 3+4 2 | 5+6 2 | 3+4 2 |
| | 2+2 4 | 2+2 4 | 3+4 2 | 2+2 4 | 2+2 4 | 5+6 2 | 3+4 2 | 5+6 2 |
| | 2+2 4 | 2+2 4 | 3+4 2 | 2+2 4 | 2+2 4 | 3+4 2 | 5+6 2 | 5+6 2 |
| | 2+2 4 | 2+2 4 | 5+6 2 | 2+2 4 | 2+2 4 | 5+6 2 | 3+4 2 | 3+4 2 |
| | 2+2 4 | 2+2 4 | 3+4 2 | 2+2 4 | 2+2 4 | 3+4 2 | 3+4 2 | 3+4 2 |
| | 2+2 4 | 2+2 4 | 5+6 2 | 2+2 4 | 2+2 4 | 5+6 2 | 5+6 2 | 5+6 2 |
| | 2+2 4 | 2+2 4 | 3+4 2 | 2+2 4 | 2+2 4 | 3+4 2 | 5+6 2 | 5+6 2 |
| | 2+2 4 | 2+2 4 | 5+6 2 | 2+2 4 | 2+2 4 | 5+6 2 | 3+4 2 | 3+4 2 |
| | 2+2 4 | 2+2 4 | 3+4 2 | 2+2 4 | 2+2 4 | 3+4 2 | 3+4 2 | 3+4 2 |
| | 2+2 4 | 2+2 4 | 5+6 2 | 2+2 4 | 2+2 4 | 5+6 2 | 5+6 2 | 5+6 2 |
| | 2+2 4 | 2+2 4 | 3+4 2 | 2+2 4 | 2+2 4 | 3+4 2 | 3+4 2 | 3+4 2 |
| | 2+2 4 | 2+2 4 | 5+6 2 | 2+2 4 | 2+2 4 | 3+4 2 | 5+6 2 | 5+6 2 |
| | 2+2 4 | 2+2 4 | 3+4 2 | 2+2 4 | 2+2 4 | 5+6 2 | 3+4 2 | 3+4 2 |
| | 2+2 4 | 2+2 4 | 5+6 2 | 2+2 4 | 2+2 4 | 3+4 2 | 5+6 2 | 3+4 2 |
| | 2+2 4 | 2+2 4 | 3+4 2 | 2+2 4 | 2+2 4 | 3+4 2 | 3+4 2 | 3+4 2 |
| | 2+2 4 | 2+2 4 | 5+6 2 | 2+2 4 | 2+2 4 | 3+4 2 | 5+6 2 | 5+6 2 |
| | 2+2 4 | 2+2 4 | 3+4 2 | 2+2 4 | 2+2 4 | 5+6 2 | 3+4 2 | 3+4 2 |

Continued on next column

**TABLE XXV– continued**

| TQC | ICSD | NoE | SF | | | | |
|---|---|---|---|---|---|---|---|
| | 2+2 4 | 2+2 4 | 5+6 2 | 2+2 4 | 2+2 4 | 5+6 2 | 3+4 2 | 5+6 2 |
| | 2+2 4 | 2+2 4 | 5+6 2 | 2+2 4 | 2+2 4 | 3+4 2 | 5+6 2 | 3+4 2 |
| | 2+2 4 | 2+2 4 | 3+4 2 | 2+2 4 | 2+2 4 | 3+4 2 | 5+6 2 | 5+6 2 |
| | 2+2 4 | 2+2 4 | 5+6 2 | 2+2 4 | 2+2 4 | 5+6 2 | 5+6 2 | 3+4 2 |
| | 2+2 4 | 2+2 4 | 3+4 2 | 2+2 4 | 2+2 4 | 5+6 2 | 5+6 2 | 5+6 2 |
| | 2+2 4 | 2+2 4 | 5+6 2 | 2+2 4 | 2+2 4 | 5+6 2 | 3+4 2 | 5+6 2 |
| | 2+2 4 | 2+2 4 | 3+4 2 | 2+2 4 | 2+2 4 | 3+4 2 | 3+4 2 | 3+4 2 |
| | 2+2 4 | 2+2 4 | 3+4 2 | 2+2 4 | 2+2 4 | 3+4 2 | 3+4 2 | 5+6 2 |
| | | | 5+6 2 | | | 5+6 2 | 5+6 2 | 5+6 2 |
| | | | 3+4 2 | | | 5+6 2 | 3+4 2 | 3+4 2 |
| | | | 3+4 2 | | | 3+4 2 | 5+6 2 | 5+6 2 |
| | | | 5+6 2 | | | 5+6 2 | 3+4 2 | 5+6 2 |
| | | | 3+4 2 | | | 3+4 2 | 3+4 2 | 3+4 2 |
| | | | 5+6 2 | | | 3+4 2 | 5+6 2 | 3+4 2 |
| | | | 5+6 2 | | | 5+6 2 | 3+4 2 | 5+6 2 |
| | | | 3+4 2 | | | 3+4 2 | 3+4 2 | 3+4 2 |
| | | | 5+6 2 | | | 5+6 2 | 5+6 2 | 5+6 2 |
| | | | 3+4 2 | | | 3+4 2 | 5+6 2 | 3+4 2 |
| | | | 5+6 2 | | | 3+4 2 | 5+6 2 | 3+4 2 |
| | | | 5+6 2 | | | 5+6 2 | 3+4 2 | 5+6 2 |
| | | | 3+4 2 | | | 5+6 2 | 3+4 2 | 3+4 2 |
| | | | 3+4 2 | | | 3+4 2 | 5+6 2 | 3+4 2 |
| | | | 5+6 2 | | | 5+6 2 | 5+6 2 | 5+6 2 |
| | | | 5+6 2 | | | 5+6 2 | 5+6 2 | 3+4 2 |
| | | | 3+4 2 | | | 5+6 2 | 5+6 2 | 5+6 2 |
| | | | 3+4 2 | | | 3+4 2 | 5+6 2 | 5+6 2 |
| | | | 5+6 2 | | | 3+4 2 | 3+4 2 | 3+4 2 |
| | | | 3+4 2 | | | 3+4 2 | 5+6 2 | 5+6 2 |
| | | | 5+6 2 | | | 5+6 2 | 3+4 2 | 5+6 2 |
| | | | 3+4 2 | | | 5+6 2 | 5+6 2 | 3+4 2 |
| | | | 5+6 2 | | | 3+4 2 | 3+4 2 | 5+6 2 |
| | | | 5+6 2 | | | 5+6 2 | 3+4 2 | 3+4 2 |
| | | | 5+6 2 | | | 3+4 2 | 5+6 2 | 5+6 2 |
| | | | 5+6 2 | | | 5+6 2 | 3+4 2 | 3+4 2 |
| | | | 5+6 2 | | | 5+6 2 | 5+6 2 | 5+6 2 |
| | | | 3+4 2 | | | 5+6 2 | 3+4 2 | 5+6 2 |
| | | | 3+4 2 | | | 5+6 2 | 5+6 2 | 3+4 2 |
| | | | 3+4 2 | | | 3+4 2 | 3+4 2 | 5+6 2 |
| | | | 3+4 2 | | | 3+4 2 | 3+4 2 | 3+4 2 |
| | | | 5+6 2 | | | 3+4 2 | 5+6 2 | 5+6 2 |
| | | | 3+4 2 | | | 5+6 2 | 3+4 2 | 3+4 2 |
| | | | 3+4 2 | | | 3+4 2 | 3+4 2 | 3+4 2 |
| | | | 5+6 2 | | | 5+6 2 | 5+6 2 | 5+6 2 |
| | | | 3+4 2 | | | 3+4 2 | 5+6 2 | 3+4 2 |
| | | | 5+6 2 | | | 5+6 2 | 3+4 2 | 5+6 2 |
| | | | 5+6 2 | | | 3+4 2 | 5+6 2 | 3+4 2 |
| | | | 3+4 2 | | | 5+6 2 | 3+4 2 | 5+6 2 |
| | | | 3+4 2 | | | 3+4 2 | 5+6 2 | 3+4 2 |
| | | | 5+6 2 | | | 3+4 2 | 3+4 2 | 5+6 2 |
| 13.2.12 | 36623 | 184 | Cu1Ho1O8W2 | | | | | |
| | -A | -B | -C | -D | -E | -GM | -Y | -Z |
| | 2+2 4 | 2+2 4 | 3+4 2 | 2+2 4 | 2+2 4 | 3+4 2 | 3+4 2 | 3+4 2 |
| | 2+2 4 | 2+2 4 | 5+6 2 | 2+2 4 | 2+2 4 | 5+6 2 | 5+6 2 | 5+6 2 |
| | 2+2 4 | 2+2 4 | 3+4 2 | 2+2 4 | 2+2 4 | 3+4 2 | 3+4 2 | 3+4 2 |
| | 2+2 4 | 2+2 4 | 5+6 2 | 2+2 4 | 2+2 4 | 5+6 2 | 5+6 2 | 5+6 2 |
| | 2+2 4 | 2+2 4 | 3+4 2 | 2+2 4 | 2+2 4 | 3+4 2 | 3+4 2 | 3+4 2 |
| | 2+2 4 | 2+2 4 | 5+6 2 | 2+2 4 | 2+2 4 | 5+6 2 | 5+6 2 | 5+6 2 |
| | 2+2 4 | 2+2 4 | 3+4 2 | 2+2 4 | 2+2 4 | 3+4 2 | 3+4 2 | 3+4 2 |
| | 2+2 4 | 2+2 4 | 5+6 2 | 2+2 4 | 2+2 4 | 5+6 2 | 5+6 2 | 5+6 2 |
| | 2+2 4 | 2+2 4 | 3+4 2 | 2+2 4 | 2+2 4 | 3+4 2 | 3+4 2 | 3+4 2 |
| | 2+2 4 | 2+2 4 | 5+6 2 | 2+2 4 | 2+2 4 | 3+4 2 | 5+6 2 | 3+4 2 |

Continued on next column

**TABLE XXV– continued**

| TQC | ICSD | | NoE | SF | | | | |
|---|---|---|---|---|---|---|---|---|
| | 2+2 4 | 2+2 4 | 3+4 2 | 2+2 4 | 2+2 4 | 5+6 2 | 5+6 2 | 5+6 2 |
| | 2+2 4 | 2+2 4 | 5+6 2 | 2+2 4 | 2+2 4 | 3+4 2 | 3+4 2 | 3+4 2 |
| | 2+2 4 | 2+2 4 | 3+4 2 | 2+2 4 | 2+2 4 | 5+6 2 | 3+4 2 | 5+6 2 |
| | 2+2 4 | 2+2 4 | 5+6 2 | 2+2 4 | 2+2 4 | 5+6 2 | 5+6 2 | 5+6 2 |
| | 2+2 4 | 2+2 4 | 5+6 2 | 2+2 4 | 2+2 4 | 3+4 2 | 5+6 2 | 5+6 2 |
| | 2+2 4 | 2+2 4 | 3+4 2 | 2+2 4 | 2+2 4 | 5+6 2 | 3+4 2 | 3+4 2 |
| | 2+2 4 | 2+2 4 | 5+6 2 | 2+2 4 | 2+2 4 | 3+4 2 | 5+6 2 | 3+4 2 |
| | 2+2 4 | 2+2 4 | 3+4 2 | 2+2 4 | 2+2 4 | 5+6 2 | 3+4 2 | 5+6 2 |
| | 2+2 4 | 2+2 4 | 3+4 2 | 2+2 4 | 2+2 4 | 3+4 2 | 3+4 2 | 3+4 2 |
| | 2+2 4 | 2+2 4 | 3+4 2 | 2+2 4 | 2+2 4 | 5+6 2 | 5+6 2 | 5+6 2 |
| | 2+2 4 | 2+2 4 | 5+6 2 | 2+2 4 | 2+2 4 | 3+4 2 | 3+4 2 | 3+4 2 |
| | 2+2 4 | 2+2 4 | 5+6 2 | 2+2 4 | 2+2 4 | 5+6 2 | 5+6 2 | 5+6 2 |
| | 2+2 4 | 2+2 4 | 3+4 2 | 2+2 4 | 2+2 4 | 3+4 2 | 3+4 2 | 3+4 2 |
| | 2+2 4 | 2+2 4 | 5+6 2 | 2+2 4 | 2+2 4 | 5+6 2 | 5+6 2 | 5+6 2 |
| | 2+2 4 | 2+2 4 | 5+6 2 | 2+2 4 | 2+2 4 | 3+4 2 | 5+6 2 | 3+4 2 |
| | 2+2 4 | 2+2 4 | 3+4 2 | 2+2 4 | 2+2 4 | 5+6 2 | 3+4 2 | 5+6 2 |
| | 2+2 4 | 2+2 4 | 3+4 2 | 2+2 4 | 2+2 4 | 3+4 2 | 5+6 2 | 3+4 2 |
| | 2+2 4 | 2+2 4 | 5+6 2 | 2+2 4 | 2+2 4 | 3+4 2 | 3+4 2 | 3+4 2 |
| | 2+2 4 | 2+2 4 | 3+4 2 | 2+2 4 | 2+2 4 | 5+6 2 | 3+4 2 | 5+6 2 |
| | 2+2 4 | 2+2 4 | 5+6 2 | 2+2 4 | 2+2 4 | 3+4 2 | 3+4 2 | 3+4 2 |
| | 2+2 4 | 2+2 4 | 5+6 2 | 2+2 4 | 2+2 4 | 3+4 2 | 5+6 2 | 3+4 2 |
| | 2+2 4 | 2+2 4 | 3+4 2 | 2+2 4 | 2+2 4 | 5+6 2 | 5+6 2 | 5+6 2 |
| | 2+2 4 | 2+2 4 | 3+4 2 | 2+2 4 | 2+2 4 | 3+4 2 | 3+4 2 | 3+4 2 |
| | 2+2 4 | 2+2 4 | 5+6 2 | 2+2 4 | 2+2 4 | 3+4 2 | 3+4 2 | 5+6 2 |
| | 2+2 4 | 2+2 4 | 5+6 2 | 2+2 4 | 2+2 4 | 5+6 2 | 5+6 2 | 3+4 2 |
| | 2+2 4 | 2+2 4 | 3+4 2 | 2+2 4 | 2+2 4 | 3+4 2 | 5+6 2 | 5+6 2 |
| | 2+2 4 | 2+2 4 | 3+4 2 | 2+2 4 | 2+2 4 | 5+6 2 | 3+4 2 | 3+4 2 |
| | 2+2 4 | 2+2 4 | 5+6 2 | 2+2 4 | 2+2 4 | 5+6 2 | 5+6 2 | 5+6 2 |
| | 2+2 4 | 2+2 4 | 5+6 2 | 2+2 4 | 2+2 4 | 3+4 2 | 3+4 2 | 3+4 2 |
| | 2+2 4 | 2+2 4 | 3+4 2 | 2+2 4 | 2+2 4 | 5+6 2 | 5+6 2 | 5+6 2 |
| | 2+2 4 | 2+2 4 | 3+4 2 | 2+2 4 | 2+2 4 | 3+4 2 | 5+6 2 | 3+4 2 |
| | 2+2 4 | 2+2 4 | 3+4 2 | 2+2 4 | 2+2 4 | 5+6 2 | 3+4 2 | 5+6 2 |
| | 2+2 4 | 2+2 4 | 5+6 2 | 2+2 4 | 2+2 4 | 5+6 2 | 3+4 2 | 5+6 2 |
| | 2+2 4 | 2+2 4 | 3+4 2 | 2+2 4 | 2+2 4 | 3+4 2 | 5+6 2 | 5+6 2 |
| | 2+2 4 | 2+2 4 | 5+6 2 | 2+2 4 | 2+2 4 | 3+4 2 | 3+4 2 | 5+6 2 |
| | 2+2 4 | 2+2 4 | 5+6 2 | 2+2 4 | 2+2 4 | 5+6 2 | 3+4 2 | 3+4 2 |
| | | | 3+4 2 | | | 5+6 2 | 5+6 2 | 5+6 2 |
| | | | 5+6 2 | | | 3+4 2 | 5+6 2 | 3+4 2 |
| | | | 5+6 2 | | | 3+4 2 | 3+4 2 | 3+4 2 |
| | | | 3+4 2 | | | 5+6 2 | 3+4 2 | 5+6 2 |
| | | | 3+4 2 | | | 3+4 2 | 5+6 2 | 3+4 2 |
| | | | 5+6 2 | | | 3+4 2 | 5+6 2 | 3+4 2 |
| | | | 5+6 2 | | | 5+6 2 | 3+4 2 | 5+6 2 |
| | | | 3+4 2 | | | 3+4 2 | 5+6 2 | 3+4 2 |
| | | | 5+6 2 | | | 5+6 2 | 5+6 2 | 5+6 2 |
| | | | 3+4 2 | | | 5+6 2 | 3+4 2 | 3+4 2 |
| | | | 5+6 2 | | | 5+6 2 | 5+6 2 | 5+6 2 |
| | | | 3+4 2 | | | 3+4 2 | 5+6 2 | 3+4 2 |
| | | | 5+6 2 | | | 5+6 2 | 3+4 2 | 3+4 2 |
| | | | 3+4 2 | | | 3+4 2 | 5+6 2 | 5+6 2 |
| | | | 3+4 2 | | | 3+4 2 | 5+6 2 | 5+6 2 |
| | | | 3+4 2 | | | 5+6 2 | 3+4 2 | 3+4 2 |
| | | | 5+6 2 | | | 5+6 2 | 3+4 2 | 5+6 2 |
| | | | 5+6 2 | | | 5+6 2 | 5+6 2 | 5+6 2 |
| | | | 3+4 2 | | | 3+4 2 | 3+4 2 | 5+6 2 |
| | | | 3+4 2 | | | 5+6 2 | 5+6 2 | 3+4 2 |
| | | | 5+6 2 | | | 5+6 2 | 3+4 2 | 3+4 2 |
| | | | 5+6 2 | | | 5+6 2 | 3+4 2 | 5+6 2 |
| | | | 5+6 2 | | | 3+4 2 | 5+6 2 | 3+4 2 |
| | | | 3+4 2 | | | 3+4 2 | 5+6 2 | 5+6 2 |
| | | | 3+4 2 | | | 5+6 2 | 3+4 2 | 5+6 2 |

Continued on next column

**TABLE XXV– continued**

| TQC | ICSD | | NoE | SF | | | | |
|---|---|---|---|---|---|---|---|---|
| | | | 5+6 2 | | | 3+4 2 | 3+4 2 | 3+4 2 |
| | | | 3+4 2 | | | 3+4 2 | 5+6 2 | 3+4 2 |
| | | | 5+6 2 | | | 5+6 2 | 3+4 2 | 5+6 2 |
| | | | 5+6 2 | | | 3+4 2 | 3+4 2 | 3+4 2 |
| | | | 3+4 2 | | | 5+6 2 | 5+6 2 | 5+6 2 |
| | | | 3+4 2 | | | 3+4 2 | 3+4 2 | 3+4 2 |
| | | | 5+6 2 | | | 5+6 2 | 5+6 2 | 5+6 2 |
| | | | 3+4 2 | | | 3+4 2 | 3+4 2 | 3+4 2 |
| | | | 5+6 2 | | | 5+6 2 | 5+6 2 | 5+6 2 |
| | | | 3+4 2 | | | 3+4 2 | 3+4 2 | 3+4 2 |
| | | | 5+6 2 | | | 5+6 2 | 5+6 2 | 5+6 2 |
| | | | 3+4 2 | | | 3+4 2 | 3+4 2 | 3+4 2 |
| | | | 5+6 2 | | | 5+6 2 | 5+6 2 | 5+6 2 |
| | | | 5+6 2 | | | 3+4 2 | 5+6 2 | 3+4 2 |
| | | | 5+6 2 | | | 3+4 2 | 3+4 2 | 5+6 2 |
| | | | 3+4 2 | | | 5+6 2 | 3+4 2 | 3+4 2 |
| | | | 5+6 2 | | | 5+6 2 | 5+6 2 | 5+6 2 |
| | | | 3+4 2 | | | 3+4 2 | 5+6 2 | 3+4 2 |
| | | | 3+4 2 | | | 5+6 2 | 3+4 2 | 5+6 2 |
| | | | 5+6 2 | | | 3+4 2 | 3+4 2 | 5+6 2 |
| | | | 3+4 2 | | | 3+4 2 | 5+6 2 | 3+4 2 |
| 13.2.13 | 67168 | | 132 | Cl7Cs1Ti2 | | | | |
| | -A | -B | -C | -D | -E | -GM | -Y | -Z |
| | 2+2 4 | 2+2 4 | 3+4 2 | 2+2 4 | 2+2 4 | 3+4 2 | 3+4 2 | 3+4 2 |
| | 2+2 4 | 2+2 4 | 5+6 2 | 2+2 4 | 2+2 4 | 5+6 2 | 5+6 2 | 5+6 2 |
| | 2+2 4 | 2+2 4 | 3+4 2 | 2+2 4 | 2+2 4 | 3+4 2 | 3+4 2 | 3+4 2 |
| | 2+2 4 | 2+2 4 | 5+6 2 | 2+2 4 | 2+2 4 | 3+4 2 | 3+4 2 | 5+6 2 |
| | 2+2 4 | 2+2 4 | 3+4 2 | 2+2 4 | 2+2 4 | 5+6 2 | 5+6 2 | 3+4 2 |
| | 2+2 4 | 2+2 4 | 5+6 2 | 2+2 4 | 2+2 4 | 3+4 2 | 5+6 2 | 5+6 2 |
| | 2+2 4 | 2+2 4 | 5+6 2 | 2+2 4 | 2+2 4 | 5+6 2 | 3+4 2 | 5+6 2 |
| | 2+2 4 | 2+2 4 | 5+6 2 | 2+2 4 | 2+2 4 | 3+4 2 | 5+6 2 | 5+6 2 |
| | 2+2 4 | 2+2 4 | 3+4 2 | 2+2 4 | 2+2 4 | 5+6 2 | 3+4 2 | 3+4 2 |
| | 2+2 4 | 2+2 4 | 3+4 2 | 2+2 4 | 2+2 4 | 5+6 2 | 5+6 2 | 3+4 2 |
| | 2+2 4 | 2+2 4 | 3+4 2 | 2+2 4 | 2+2 4 | 5+6 2 | 5+6 2 | 3+4 2 |
| | 2+2 4 | 2+2 4 | 5+6 2 | 2+2 4 | 2+2 4 | 3+4 2 | 3+4 2 | 5+6 2 |
| | 2+2 4 | 2+2 4 | 5+6 2 | 2+2 4 | 2+2 4 | 5+6 2 | 5+6 2 | 3+4 2 |
| | 2+2 4 | 2+2 4 | 3+4 2 | 2+2 4 | 2+2 4 | 3+4 2 | 3+4 2 | 3+4 2 |
| | 2+2 4 | 2+2 4 | 5+6 2 | 2+2 4 | 2+2 4 | 5+6 2 | 5+6 2 | 5+6 2 |
| | 2+2 4 | 2+2 4 | 3+4 2 | 2+2 4 | 2+2 4 | 3+4 2 | 3+4 2 | 5+6 2 |
| | 2+2 4 | 2+2 4 | 3+4 2 | 2+2 4 | 2+2 4 | 3+4 2 | 3+4 2 | 5+6 2 |
| | 2+2 4 | 2+2 4 | 5+6 2 | 2+2 4 | 2+2 4 | 5+6 2 | 5+6 2 | 3+4 2 |
| | 2+2 4 | 2+2 4 | 3+4 2 | 2+2 4 | 2+2 4 | 3+4 2 | 3+4 2 | 5+6 2 |
| | 2+2 4 | 2+2 4 | 5+6 2 | 2+2 4 | 2+2 4 | 5+6 2 | 5+6 2 | 3+4 2 |
| | 2+2 4 | 2+2 4 | 3+4 2 | 2+2 4 | 2+2 4 | 3+4 2 | 3+4 2 | 3+4 2 |
| | 2+2 4 | 2+2 4 | 5+6 2 | 2+2 4 | 2+2 4 | 5+6 2 | 5+6 2 | 5+6 2 |
| | 2+2 4 | 2+2 4 | 3+4 2 | 2+2 4 | 2+2 4 | 5+6 2 | 5+6 2 | 3+4 2 |
| | 2+2 4 | 2+2 4 | 5+6 2 | 2+2 4 | 2+2 4 | 5+6 2 | 5+6 2 | 5+6 2 |
| | 2+2 4 | 2+2 4 | 3+4 2 | 2+2 4 | 2+2 4 | 3+4 2 | 3+4 2 | 3+4 2 |
| | 2+2 4 | 2+2 4 | 5+6 2 | 2+2 4 | 2+2 4 | 3+4 2 | 5+6 2 | 5+6 2 |
| | 2+2 4 | 2+2 4 | 3+4 2 | 2+2 4 | 2+2 4 | 3+4 2 | 3+4 2 | 3+4 2 |
| | 2+2 4 | 2+2 4 | 3+4 2 | 2+2 4 | 2+2 4 | 5+6 2 | 3+4 2 | 5+6 2 |
| | 2+2 4 | 2+2 4 | 5+6 2 | 2+2 4 | 2+2 4 | 3+4 2 | 3+4 2 | 3+4 2 |
| | 2+2 4 | 2+2 4 | 3+4 2 | 2+2 4 | 2+2 4 | 3+4 2 | 3+4 2 | 5+6 2 |
| | 2+2 4 | 2+2 4 | 5+6 2 | 2+2 4 | 2+2 4 | 3+4 2 | 5+6 2 | 3+4 2 |
| | 2+2 4 | 2+2 4 | 5+6 2 | 2+2 4 | 2+2 4 | 5+6 2 | 3+4 2 | 5+6 2 |
| | 2+2 4 | 2+2 4 | 5+6 2 | 2+2 4 | 2+2 4 | 5+6 2 | 3+4 2 | 5+6 2 |
| | | | 3+4 2 | | | 3+4 2 | 3+4 2 | 3+4 2 |
| | | | 3+4 2 | | | 3+4 2 | 5+6 2 | 3+4 2 |
| | | | 5+6 2 | | | 5+6 2 | 5+6 2 | 5+6 2 |
| | | | 3+4 2 | | | 5+6 2 | 5+6 2 | 3+4 2 |
| | | | 5+6 2 | | | 3+4 2 | 3+4 2 | 5+6 2 |

Continued on next column

**TABLE XXV– continued**

| TQC | ICSD | NoE | SF | | |
|---|---|---|---|---|---|
| | | 5+6 2 | 3+4 2 | 3+4 2 | 3+4 2 |
| | | 5+6 2 | 3+4 2 | 3+4 2 | 5+6 2 |
| | | 3+4 2 | 5+6 2 | 5+6 2 | 5+6 2 |
| | | 5+6 2 | 3+4 2 | 5+6 2 | 5+6 2 |
| | | 5+6 2 | 5+6 2 | 3+4 2 | 3+4 2 |
| | | 5+6 2 | 5+6 2 | 5+6 2 | 5+6 2 |
| | | 3+4 2 | 3+4 2 | 3+4 2 | 3+4 2 |
| | | 3+4 2 | 5+6 2 | 5+6 2 | 5+6 2 |
| | | 3+4 2 | 3+4 2 | 3+4 2 | 3+4 2 |
| | | 3+4 2 | 5+6 2 | 5+6 2 | 3+4 2 |
| | | 3+4 2 | 3+4 2 | 5+6 2 | 5+6 2 |
| | | 5+6 2 | 5+6 2 | 5+6 2 | 3+4 2 |
| | | 3+4 2 | 5+6 2 | 3+4 2 | 5+6 2 |
| | | 5+6 2 | 3+4 2 | 3+4 2 | 5+6 2 |
| | | 5+6 2 | 3+4 2 | 3+4 2 | 3+4 2 |
| | | 3+4 2 | 3+4 2 | 5+6 2 | 3+4 2 |
| | | 5+6 2 | 5+6 2 | 5+6 2 | 3+4 2 |
| | | 5+6 2 | 3+4 2 | 3+4 2 | 5+6 2 |
| | | 3+4 2 | 5+6 2 | 3+4 2 | 3+4 2 |
| | | 3+4 2 | 5+6 2 | 5+6 2 | 3+4 2 |
| | | 3+4 2 | 3+4 2 | 5+6 2 | 5+6 2 |
| | | 5+6 2 | 5+6 2 | 5+6 2 | 3+4 2 |
| | | 3+4 2 | 5+6 2 | 5+6 2 | 5+6 2 |
| | | 5+6 2 | 5+6 2 | 5+6 2 | 5+6 2 |
| | | 3+4 2 | 5+6 2 | 3+4 2 | 3+4 2 |
| | | 5+6 2 | 3+4 2 | 3+4 2 | 5+6 2 |
| | | 3+4 2 | 3+4 2 | 3+4 2 | 3+4 2 |
| | | 3+4 2 | 5+6 2 | 5+6 2 | 3+4 2 |

The End

TABLE XXVI: The computed irreducible representations for the valence bands for the compounds of SEBR and NLC in space group 14. See the list of tables.

| TQC | ICSD | NoE | SF | | | | | |
|---|---|---|---|---|---|---|---|---|
| 14.2.1 | 96154 | 112 | Si4Ti3U2 | | | | | |
| | -A | -B | -C | -D | -E | -GM | -Y | -Z |
| | 2+2 4 | 2+2 4 | 2+2 4 | 3+3 2 | 3+3 2 | 3+4 2 | 3+4 2 | 2+2 4 |
| | 2+2 4 | 2+2 4 | 2+2 4 | 4+4 2 | 4+4 2 | 3+4 2 | 3+4 2 | 2+2 4 |
| | 2+2 4 | 2+2 4 | 2+2 4 | 5+5 2 | 5+5 2 | 5+6 2 | 5+6 2 | 2+2 4 |
| | 2+2 4 | 2+2 4 | 2+2 4 | 6+6 2 | 6+6 2 | 5+6 2 | 5+6 2 | 2+2 4 |
| | 2+2 4 | 2+2 4 | 2+2 4 | 3+3 2 | 3+3 2 | 3+4 2 | 3+4 2 | 2+2 4 |
| | 2+2 4 | 2+2 4 | 2+2 4 | 4+4 2 | 4+4 2 | 5+6 2 | 5+6 2 | 2+2 4 |
| | 2+2 4 | 2+2 4 | 2+2 4 | 6+6 2 | 6+6 2 | 3+4 2 | 3+4 2 | 2+2 4 |
| | 2+2 4 | 2+2 4 | 2+2 4 | 5+5 2 | 5+5 2 | 5+6 2 | 5+6 2 | 2+2 4 |
| | 2+2 4 | 2+2 4 | 2+2 4 | 3+3 2 | 3+3 2 | 3+4 2 | 5+6 2 | 2+2 4 |
| | 2+2 4 | 2+2 4 | 2+2 4 | 5+5 2 | 5+5 2 | 5+6 2 | 3+4 2 | 2+2 4 |
| | 2+2 4 | 2+2 4 | 2+2 4 | 4+4 2 | 4+4 2 | 3+4 2 | 3+4 2 | 2+2 4 |
| | 2+2 4 | 2+2 4 | 2+2 4 | 6+6 2 | 6+6 2 | 3+4 2 | 3+4 2 | 2+2 4 |
| | 2+2 4 | 2+2 4 | 2+2 4 | 4+4 2 | 4+4 2 | 5+6 2 | 5+6 2 | 2+2 4 |
| | 2+2 4 | 2+2 4 | 2+2 4 | 3+3 2 | 3+3 2 | 5+6 2 | 5+6 2 | 2+2 4 |
| | 2+2 4 | 2+2 4 | 2+2 4 | 6+6 2 | 6+6 2 | 3+4 2 | 3+4 2 | 2+2 4 |
| | 2+2 4 | 2+2 4 | 2+2 4 | 5+5 2 | 5+5 2 | 5+6 2 | 5+6 2 | 2+2 4 |
| | 2+2 4 | 2+2 4 | 2+2 4 | 4+4 2 | 5+5 2 | 3+4 2 | 5+6 2 | 2+2 4 |
| | 2+2 4 | 2+2 4 | 2+2 4 | 3+3 2 | 6+6 2 | 3+4 2 | 3+4 2 | 2+2 4 |
| | 2+2 4 | 2+2 4 | 2+2 4 | 6+6 2 | 4+4 2 | 3+4 2 | 5+6 2 | 2+2 4 |
| | 2+2 4 | 2+2 4 | 2+2 4 | 5+5 2 | 3+3 2 | 5+6 2 | 3+4 2 | 2+2 4 |
| | 2+2 4 | 2+2 4 | 2+2 4 | 3+3 2 | 6+6 2 | 3+4 2 | 5+6 2 | 2+2 4 |
| | 2+2 4 | 2+2 4 | 2+2 4 | 4+4 2 | 5+5 2 | 5+6 2 | 3+4 2 | 2+2 4 |
| | 2+2 4 | 2+2 4 | 2+2 4 | 6+6 2 | 3+3 2 | 5+6 2 | 5+6 2 | 2+2 4 |

Continued on next column

**TABLE XXVI– continued**

| TQC | ICSD | NoE | SF | | | | | |
|---|---|---|---|---|---|---|---|---|
| | 2+2 4 | 2+2 4 | 2+2 4 | 5+5 2 | 4+4 2 | 5+6 2 | 3+4 2 | 2+2 4 |
| | 2+2 4 | 2+2 4 | 2+2 4 | 5+5 2 | 3+3 2 | 3+4 2 | 5+6 2 | 2+2 4 |
| | 2+2 4 | 2+2 4 | 2+2 4 | 6+6 2 | 4+4 2 | 5+6 2 | 5+6 2 | 2+2 4 |
| | 2+2 4 | 2+2 4 | 2+2 4 | 4+4 2 | 5+5 2 | 5+6 2 | 5+6 2 | 2+2 4 |
| | 2+2 4 | 2+2 4 | 2+2 4 | 6+6 2 | 6+6 2 | 3+4 2 | 5+6 2 | 2+2 4 |
| | | | | 3+3 2 | 3+3 2 | 3+4 2 | 3+4 2 | |
| | | | | 5+5 2 | 4+4 2 | 3+4 2 | 3+4 2 | |
| | | | | 3+3 2 | 6+6 2 | 3+4 2 | 3+4 2 | |
| | | | | 4+4 2 | 5+5 2 | 5+6 2 | 5+6 2 | |
| | | | | 6+6 2 | 3+3 2 | 5+6 2 | 5+6 2 | |
| | | | | 5+5 2 | 4+4 2 | 3+4 2 | 3+4 2 | |
| | | | | 3+3 2 | 6+6 2 | 3+4 2 | 3+4 2 | |
| | | | | 6+6 2 | 5+5 2 | 5+6 2 | 3+4 2 | |
| | | | | 5+5 2 | 6+6 2 | 3+4 2 | 5+6 2 | |
| | | | | 4+4 2 | 5+5 2 | 5+6 2 | 5+6 2 | |
| | | | | 4+4 2 | 5+5 2 | 3+4 2 | 3+4 2 | |
| | | | | 3+3 2 | 6+6 2 | 5+6 2 | 5+6 2 | |
| | | | | 5+5 2 | 4+4 2 | 3+4 2 | 3+4 2 | |
| | | | | 6+6 2 | 3+3 2 | 3+4 2 | 3+4 2 | |
| | | | | 4+4 2 | 3+3 2 | 5+6 2 | 5+6 2 | |
| | | | | 3+3 2 | 4+4 2 | 5+6 2 | 5+6 2 | |
| | | | | 5+5 2 | 6+6 2 | 5+6 2 | 5+6 2 | |
| | | | | 6+6 2 | 5+5 2 | 3+4 2 | 3+4 2 | |
| | | | | 4+4 2 | 3+3 2 | 3+4 2 | 5+6 2 | |
| | | | | 3+3 2 | 4+4 2 | 3+4 2 | 3+4 2 | |
| | | | | 5+5 2 | 5+5 2 | 3+4 2 | 5+6 2 | |
| | | | | 6+6 2 | 3+3 2 | 3+4 2 | 5+6 2 | |
| | | | | 6+6 2 | 6+6 2 | 5+6 2 | 3+4 2 | |
| | | | | 5+5 2 | 4+4 2 | 3+4 2 | 5+6 2 | |
| | | | | 4+4 2 | 4+4 2 | 5+6 2 | 3+4 2 | |
| | | | | 3+3 2 | 3+3 2 | 5+6 2 | 5+6 2 | |
| | | | | 5+5 2 | 5+5 2 | 5+6 2 | 3+4 2 | |
| | | | | 6+6 2 | 6+6 2 | 3+4 2 | 3+4 2 | |
| 14.2.2 | 57480 | 144 | Re2Si1 | | | | | |
| | -A | -B | -C | -D | -E | -GM | -Y | -Z |
| | 2+2 4 | 2+2 4 | 2+2 4 | 6+6 2 | 6+6 2 | 3+4 2 | 3+4 2 | 2+2 4 |
| | 2+2 4 | 2+2 4 | 2+2 4 | 5+5 2 | 5+5 2 | 3+4 2 | 3+4 2 | 2+2 4 |
| | 2+2 4 | 2+2 4 | 2+2 4 | 3+3 2 | 3+3 2 | 5+6 2 | 5+6 2 | 2+2 4 |
| | 2+2 4 | 2+2 4 | 2+2 4 | 4+4 2 | 4+4 2 | 5+6 2 | 5+6 2 | 2+2 4 |
| | 2+2 4 | 2+2 4 | 2+2 4 | 6+6 2 | 4+4 2 | 5+6 2 | 5+6 2 | 2+2 4 |
| | 2+2 4 | 2+2 4 | 2+2 4 | 5+5 2 | 3+3 2 | 3+4 2 | 5+6 2 | 2+2 4 |
| | 2+2 4 | 2+2 4 | 2+2 4 | 4+4 2 | 5+5 2 | 3+4 2 | 3+4 2 | 2+2 4 |
| | 2+2 4 | 2+2 4 | 2+2 4 | 3+3 2 | 6+6 2 | 5+6 2 | 3+4 2 | 2+2 4 |
| | 2+2 4 | 2+2 4 | 2+2 4 | 3+3 2 | 6+6 2 | 5+6 2 | 3+4 2 | 2+2 4 |
| | 2+2 4 | 2+2 4 | 2+2 4 | 4+4 2 | 5+5 2 | 3+4 2 | 3+4 2 | 2+2 4 |
| | 2+2 4 | 2+2 4 | 2+2 4 | 5+5 2 | 3+3 2 | 3+4 2 | 5+6 2 | 2+2 4 |
| | 2+2 4 | 2+2 4 | 2+2 4 | 6+6 2 | 4+4 2 | 5+6 2 | 5+6 2 | 2+2 4 |
| | 2+2 4 | 2+2 4 | 2+2 4 | 4+4 2 | 3+3 2 | 3+4 2 | 5+6 2 | 2+2 4 |
| | 2+2 4 | 2+2 4 | 2+2 4 | 3+3 2 | 4+4 2 | 5+6 2 | 3+4 2 | 2+2 4 |
| | 2+2 4 | 2+2 4 | 2+2 4 | 6+6 2 | 5+5 2 | 3+4 2 | 3+4 2 | 2+2 4 |
| | 2+2 4 | 2+2 4 | 2+2 4 | 5+5 2 | 6+6 2 | 5+6 2 | 5+6 2 | 2+2 4 |
| | 2+2 4 | 2+2 4 | 2+2 4 | 5+5 2 | 4+4 2 | 3+4 2 | 3+4 2 | 2+2 4 |
| | 2+2 4 | 2+2 4 | 2+2 4 | 6+6 2 | 3+3 2 | 3+4 2 | 5+6 2 | 2+2 4 |
| | 2+2 4 | 2+2 4 | 2+2 4 | 4+4 2 | 5+5 2 | 3+4 2 | 5+6 2 | 2+2 4 |
| | 2+2 4 | 2+2 4 | 2+2 4 | 3+3 2 | 6+6 2 | 5+6 2 | 3+4 2 | 2+2 4 |
| | 2+2 4 | 2+2 4 | 2+2 4 | 6+6 2 | 4+4 2 | 5+6 2 | 3+4 2 | 2+2 4 |
| | 2+2 4 | 2+2 4 | 2+2 4 | 5+5 2 | 3+3 2 | 3+4 2 | 5+6 2 | 2+2 4 |
| | 2+2 4 | 2+2 4 | 2+2 4 | 5+5 2 | 5+5 2 | 3+4 2 | 3+4 2 | 2+2 4 |
| | 2+2 4 | 2+2 4 | 2+2 4 | 6+6 2 | 6+6 2 | 5+6 2 | 5+6 2 | 2+2 4 |
| | 2+2 4 | 2+2 4 | 2+2 4 | 3+3 2 | 5+5 2 | 5+6 2 | 5+6 2 | 2+2 4 |
| | 2+2 4 | 2+2 4 | 2+2 4 | 4+4 2 | 4+4 2 | 3+4 2 | 5+6 2 | 2+2 4 |

Continued on next column

**TABLE XXVI– continued**

| TQC | ICSD | NoE | SF | | | | | |
|---|---|---|---|---|---|---|---|---|
| | 2+2 4 | 2+2 4 | 2+2 4 | 4+4 2 | 3+3 2 | 5+6 2 | 3+4 2 | 2+2 4 |
| | 2+2 4 | 2+2 4 | 2+2 4 | 3+3 2 | 6+6 2 | 3+4 2 | 5+6 2 | 2+2 4 |
| | 2+2 4 | 2+2 4 | 2+2 4 | 5+5 2 | 3+3 2 | 3+4 2 | 5+6 2 | 2+2 4 |
| | 2+2 4 | 2+2 4 | 2+2 4 | 6+6 2 | 4+4 2 | 5+6 2 | 3+4 2 | 2+2 4 |
| | 2+2 4 | 2+2 4 | 2+2 4 | 5+5 2 | 6+6 2 | 5+6 2 | 3+4 2 | 2+2 4 |
| | 2+2 4 | 2+2 4 | 2+2 4 | 6+6 2 | 5+5 2 | 3+4 2 | 5+6 2 | 2+2 4 |
| | 2+2 4 | 2+2 4 | 2+2 4 | 3+3 2 | 4+4 2 | 5+6 2 | 5+6 2 | 2+2 4 |
| | 2+2 4 | 2+2 4 | 2+2 4 | 4+4 2 | 6+6 2 | 5+6 2 | 3+4 2 | 2+2 4 |
| | 2+2 4 | 2+2 4 | 2+2 4 | 3+3 2 | 3+3 2 | 5+6 2 | 3+4 2 | 2+2 4 |
| | 2+2 4 | 2+2 4 | 2+2 4 | 4+4 2 | 5+5 2 | 3+4 2 | 3+4 2 | 2+2 4 |
| | | | | 6+6 2 | 3+3 2 | 3+4 2 | 5+6 2 | |
| | | | | 5+5 2 | 4+4 2 | 3+4 2 | 3+4 2 | |
| | | | | 4+4 2 | 6+6 2 | 3+4 2 | 3+4 2 | |
| | | | | 3+3 2 | 5+5 2 | 5+6 2 | 5+6 2 | |
| | | | | 5+5 2 | 5+5 2 | 5+6 2 | 5+6 2 | |
| | | | | 6+6 2 | 6+6 2 | 3+4 2 | 3+4 2 | |
| | | | | 4+4 2 | 6+6 2 | 5+6 2 | 3+4 2 | |
| | | | | 3+3 2 | 4+4 2 | 5+6 2 | 5+6 2 | |
| | | | | 6+6 2 | 3+3 2 | 5+6 2 | 3+4 2 | |
| | | | | 5+5 2 | 5+5 2 | 5+6 2 | 5+6 2 | |
| | | | | 5+5 2 | 4+4 2 | 3+4 2 | 5+6 2 | |
| | | | | 6+6 2 | 3+3 2 | 3+4 2 | 3+4 2 | |
| | | | | 3+3 2 | 4+4 2 | 3+4 2 | 5+6 2 | |
| | | | | 4+4 2 | 3+3 2 | 3+4 2 | 3+4 2 | |
| | | | | 6+6 2 | 5+5 2 | 5+6 2 | 5+6 2 | |
| | | | | 5+5 2 | 6+6 2 | 5+6 2 | 3+4 2 | |
| | | | | 3+3 2 | 6+6 2 | 3+4 2 | 5+6 2 | |
| | | | | 4+4 2 | 5+5 2 | 5+6 2 | 3+4 2 | |
| | | | | 4+4 2 | 4+4 2 | 3+4 2 | 5+6 2 | |
| | | | | 3+3 2 | 3+3 2 | 5+6 2 | 3+4 2 | |
| | | | | 5+5 2 | 6+6 2 | 5+6 2 | 3+4 2 | |
| | | | | 6+6 2 | 5+5 2 | 3+4 2 | 5+6 2 | |
| | | | | 3+3 2 | 3+3 2 | 3+4 2 | 3+4 2 | |
| | | | | 4+4 2 | 4+4 2 | 3+4 2 | 5+6 2 | |
| | | | | 6+6 2 | 3+3 2 | 3+4 2 | 3+4 2 | |
| | | | | 5+5 2 | 4+4 2 | 5+6 2 | 5+6 2 | |
| | | | | 4+4 2 | 5+5 2 | 3+4 2 | 5+6 2 | |
| | | | | 3+3 2 | 6+6 2 | 5+6 2 | 3+4 2 | |
| | | | | 3+3 2 | 6+6 2 | 3+4 2 | 3+4 2 | |
| | | | | 4+4 2 | 5+5 2 | 5+6 2 | 3+4 2 | |
| | | | | 5+5 2 | 3+3 2 | 3+4 2 | 5+6 2 | |
| | | | | 6+6 2 | 4+4 2 | 5+6 2 | 3+4 2 | |
| | | | | 4+4 2 | 5+5 2 | 5+6 2 | 5+6 2 | |
| | | | | 3+3 2 | 6+6 2 | 3+4 2 | 5+6 2 | |
| | | | | 6+6 2 | 3+3 2 | 3+4 2 | 3+4 2 | |
| | | | | 5+5 2 | 4+4 2 | 3+4 2 | 5+6 2 | |
| 14.2.3 | 9402 | 140 | O4Te1V1 | | | | | |
| | -A | -B | -C | -D | -E | -GM | -Y | -Z |
| | 2+2 4 | 2+2 4 | 2+2 4 | 3+3 2 | 3+3 2 | 3+4 2 | 3+4 2 | 2+2 4 |
| | 2+2 4 | 2+2 4 | 2+2 4 | 4+4 2 | 4+4 2 | 5+6 2 | 5+6 2 | 2+2 4 |
| | 2+2 4 | 2+2 4 | 2+2 4 | 5+5 2 | 5+5 2 | 3+4 2 | 3+4 2 | 2+2 4 |
| | 2+2 4 | 2+2 4 | 2+2 4 | 6+6 2 | 6+6 2 | 5+6 2 | 5+6 2 | 2+2 4 |
| | 2+2 4 | 2+2 4 | 2+2 4 | 5+5 2 | 3+3 2 | 3+4 2 | 5+6 2 | 2+2 4 |
| | 2+2 4 | 2+2 4 | 2+2 4 | 6+6 2 | 4+4 2 | 3+4 2 | 5+6 2 | 2+2 4 |
| | 2+2 4 | 2+2 4 | 2+2 4 | 3+3 2 | 5+5 2 | 5+6 2 | 3+4 2 | 2+2 4 |
| | 2+2 4 | 2+2 4 | 2+2 4 | 4+4 2 | 6+6 2 | 5+6 2 | 5+6 2 | 2+2 4 |
| | 2+2 4 | 2+2 4 | 2+2 4 | 5+5 2 | 4+4 2 | 3+4 2 | 3+4 2 | 2+2 4 |
| | 2+2 4 | 2+2 4 | 2+2 4 | 6+6 2 | 3+3 2 | 3+4 2 | 3+4 2 | 2+2 4 |
| | 2+2 4 | 2+2 4 | 2+2 4 | 3+3 2 | 5+5 2 | 5+6 2 | 3+4 2 | 2+2 4 |
| | 2+2 4 | 2+2 4 | 2+2 4 | 4+4 2 | 6+6 2 | 5+6 2 | 5+6 2 | 2+2 4 |
| | 2+2 4 | 2+2 4 | 2+2 4 | 3+3 2 | 6+6 2 | 3+4 2 | 5+6 2 | 2+2 4 |

Continued on next column

**TABLE XXVI– continued**

| TQC | ICSD | NoE | SF | | | | | |
|---|---|---|---|---|---|---|---|---|
| | 2+2 4 | 2+2 4 | 2+2 4 | 4+4 2 | 5+5 2 | 3+4 2 | 5+6 2 | 2+2 4 |
| | 2+2 4 | 2+2 4 | 2+2 4 | 6+6 2 | 3+3 2 | 5+6 2 | 3+4 2 | 2+2 4 |
| | 2+2 4 | 2+2 4 | 2+2 4 | 5+5 2 | 4+4 2 | 5+6 2 | 3+4 2 | 2+2 4 |
| | 2+2 4 | 2+2 4 | 2+2 4 | 4+4 2 | 6+6 2 | 5+6 2 | 5+6 2 | 2+2 4 |
| | 2+2 4 | 2+2 4 | 2+2 4 | 3+3 2 | 5+5 2 | 3+4 2 | 3+4 2 | 2+2 4 |
| | 2+2 4 | 2+2 4 | 2+2 4 | 6+6 2 | 4+4 2 | 3+4 2 | 5+6 2 | 2+2 4 |
| | 2+2 4 | 2+2 4 | 2+2 4 | 5+5 2 | 3+3 2 | 5+6 2 | 3+4 2 | 2+2 4 |
| | 2+2 4 | 2+2 4 | 2+2 4 | 4+4 2 | 4+4 2 | 3+4 2 | 3+4 2 | 2+2 4 |
| | 2+2 4 | 2+2 4 | 2+2 4 | 6+6 2 | 3+3 2 | 5+6 2 | 5+6 2 | 2+2 4 |
| | 2+2 4 | 2+2 4 | 2+2 4 | 5+5 2 | 6+6 2 | 5+6 2 | 5+6 2 | 2+2 4 |
| | 2+2 4 | 2+2 4 | 2+2 4 | 3+3 2 | 5+5 2 | 3+4 2 | 3+4 2 | 2+2 4 |
| | 2+2 4 | 2+2 4 | 2+2 4 | 3+3 2 | 6+6 2 | 3+4 2 | 3+4 2 | 2+2 4 |
| | 2+2 4 | 2+2 4 | 2+2 4 | 4+4 2 | 5+5 2 | 5+6 2 | 5+6 2 | 2+2 4 |
| | 2+2 4 | 2+2 4 | 2+2 4 | 6+6 2 | 4+4 2 | 3+4 2 | 3+4 2 | 2+2 4 |
| | 2+2 4 | 2+2 4 | 2+2 4 | 5+5 2 | 3+3 2 | 5+6 2 | 5+6 2 | 2+2 4 |
| | 2+2 4 | 2+2 4 | 2+2 4 | 3+3 2 | 3+3 2 | 3+4 2 | 3+4 2 | 2+2 4 |
| | 2+2 4 | 2+2 4 | 2+2 4 | 4+4 2 | 5+5 2 | 3+4 2 | 5+6 2 | 2+2 4 |
| | 2+2 4 | 2+2 4 | 2+2 4 | 5+5 2 | 6+6 2 | 5+6 2 | 5+6 2 | 2+2 4 |
| | 2+2 4 | 2+2 4 | 2+2 4 | 6+6 2 | 4+4 2 | 3+4 2 | 5+6 2 | 2+2 4 |
| | 2+2 4 | 2+2 4 | 2+2 4 | 3+3 2 | 3+3 2 | 5+6 2 | 5+6 2 | 2+2 4 |
| | 2+2 4 | 2+2 4 | 2+2 4 | 4+4 2 | 4+4 2 | 3+4 2 | 3+4 2 | 2+2 4 |
| | 2+2 4 | 2+2 4 | 2+2 4 | 5+5 2 | 5+5 2 | 5+6 2 | 3+4 2 | 2+2 4 |
| | | | | 6+6 2 | 6+6 2 | 5+6 2 | 3+4 2 | |
| | | | | 5+5 2 | 5+5 2 | 3+4 2 | 5+6 2 | |
| | | | | 6+6 2 | 6+6 2 | 5+6 2 | 5+6 2 | |
| | | | | 5+5 2 | 3+3 2 | 3+4 2 | 3+4 2 | |
| | | | | 6+6 2 | 4+4 2 | 3+4 2 | 5+6 2 | |
| | | | | 3+3 2 | 3+3 2 | 3+4 2 | 5+6 2 | |
| | | | | 4+4 2 | 4+4 2 | 5+6 2 | 3+4 2 | |
| | | | | 3+3 2 | 5+5 2 | 5+6 2 | 3+4 2 | |
| | | | | 4+4 2 | 6+6 2 | 3+4 2 | 3+4 2 | |
| | | | | 5+5 2 | 5+5 2 | 5+6 2 | 5+6 2 | |
| | | | | 3+3 2 | 6+6 2 | 3+4 2 | 5+6 2 | |
| | | | | 4+4 2 | 3+3 2 | 3+4 2 | 3+4 2 | |
| | | | | 6+6 2 | 4+4 2 | 5+6 2 | 3+4 2 | |
| | | | | 5+5 2 | 3+3 2 | 3+4 2 | 5+6 2 | |
| | | | | 6+6 2 | 4+4 2 | 5+6 2 | 5+6 2 | |
| | | | | 3+3 2 | 6+6 2 | 3+4 2 | 3+4 2 | |
| | | | | 4+4 2 | 5+5 2 | 5+6 2 | 5+6 2 | |
| | | | | 3+3 2 | 5+5 2 | 3+4 2 | 3+4 2 | |
| | | | | 4+4 2 | 6+6 2 | 5+6 2 | 5+6 2 | |
| | | | | 6+6 2 | 3+3 2 | 3+4 2 | 3+4 2 | |
| | | | | 5+5 2 | 4+4 2 | 5+6 2 | 3+4 2 | |
| | | | | 5+5 2 | 5+5 2 | 5+6 2 | 3+4 2 | |
| | | | | 6+6 2 | 6+6 2 | 3+4 2 | 5+6 2 | |
| | | | | 4+4 2 | 3+3 2 | 5+6 2 | 3+4 2 | |
| | | | | 3+3 2 | 4+4 2 | 5+6 2 | 5+6 2 | |
| | | | | 5+5 2 | 3+3 2 | 3+4 2 | 5+6 2 | |
| | | | | 6+6 2 | 4+4 2 | 3+4 2 | 5+6 2 | |
| | | | | 3+3 2 | 5+5 2 | 5+6 2 | 3+4 2 | |
| | | | | 4+4 2 | 6+6 2 | 3+4 2 | 3+4 2 | |
| | | | | 6+6 2 | 3+3 2 | 5+6 2 | 5+6 2 | |
| | | | | 5+5 2 | 4+4 2 | 3+4 2 | 3+4 2 | |
| | | | | 3+3 2 | 5+5 2 | 5+6 2 | 5+6 2 | |
| | | | | 4+4 2 | 6+6 2 | 5+6 2 | 3+4 2 | |
| | | | | 3+3 2 | 6+6 2 | 3+4 2 | 5+6 2 | |
| | | | | 4+4 2 | 5+5 2 | 5+6 2 | 3+4 2 | |
| 14.2.4 | 8217 | 72 | O2W1 | | | | | |
| | -A | -B | -C | -D | -E | -GM | -Y | -Z |
| | 2+2 4 | 2+2 4 | 2+2 4 | 4+4 2 | 4+4 2 | 3+4 2 | 3+4 2 | 2+2 4 |
| | 2+2 4 | 2+2 4 | 2+2 4 | 3+3 2 | 3+3 2 | 5+6 2 | 5+6 2 | 2+2 4 |

Continued on next column

**TABLE XXVI– continued**

| TQC | ICSD | NoE | SF | | | | | |
|---|---|---|---|---|---|---|---|---|
| | 2+2 4 | 2+2 4 | 2+2 4 | 6+6 2 | 6+6 2 | 5+6 2 | 5+6 2 | 2+2 4 |
| | 2+2 4 | 2+2 4 | 2+2 4 | 5+5 2 | 5+5 2 | 3+4 2 | 3+4 2 | 2+2 4 |
| | 2+2 4 | 2+2 4 | 2+2 4 | 5+5 2 | 5+5 2 | 5+6 2 | 5+6 2 | 2+2 4 |
| | 2+2 4 | 2+2 4 | 2+2 4 | 6+6 2 | 6+6 2 | 3+4 2 | 3+4 2 | 2+2 4 |
| | 2+2 4 | 2+2 4 | 2+2 4 | 4+4 2 | 4+4 2 | 5+6 2 | 5+6 2 | 2+2 4 |
| | 2+2 4 | 2+2 4 | 2+2 4 | 3+3 2 | 3+3 2 | 3+4 2 | 3+4 2 | 2+2 4 |
| | 2+2 4 | 2+2 4 | 2+2 4 | 4+4 2 | 5+5 2 | 3+4 2 | 5+6 2 | 2+2 4 |
| | 2+2 4 | 2+2 4 | 2+2 4 | 3+3 2 | 6+6 2 | 3+4 2 | 5+6 2 | 2+2 4 |
| | 2+2 4 | 2+2 4 | 2+2 4 | 5+5 2 | 3+3 2 | 5+6 2 | 3+4 2 | 2+2 4 |
| | 2+2 4 | 2+2 4 | 2+2 4 | 6+6 2 | 4+4 2 | 5+6 2 | 3+4 2 | 2+2 4 |
| | 2+2 4 | 2+2 4 | 2+2 4 | 6+6 2 | 6+6 2 | 5+6 2 | 5+6 2 | 2+2 4 |
| | 2+2 4 | 2+2 4 | 2+2 4 | 5+5 2 | 5+5 2 | 5+6 2 | 3+4 2 | 2+2 4 |
| | 2+2 4 | 2+2 4 | 2+2 4 | 3+3 2 | 3+3 2 | 3+4 2 | 5+6 2 | 2+2 4 |
| | 2+2 4 | 2+2 4 | 2+2 4 | 4+4 2 | 4+4 2 | 3+4 2 | 3+4 2 | 2+2 4 |
| | 2+2 4 | 2+2 4 | 2+2 4 | 3+3 2 | 6+6 2 | 3+4 2 | 5+6 2 | 2+2 4 |
| | 2+2 4 | 2+2 4 | 2+2 4 | 4+4 2 | 5+5 2 | 3+4 2 | 5+6 2 | 2+2 4 |
| | | | | 3+3 2 | 4+4 2 | 3+4 2 | 3+4 2 | |
| | | | | 4+4 2 | 3+3 2 | 5+6 2 | 5+6 2 | |
| | | | | 6+6 2 | 3+3 2 | 5+6 2 | 3+4 2 | |
| | | | | 5+5 2 | 6+6 2 | 5+6 2 | 3+4 2 | |
| | | | | 5+5 2 | 4+4 2 | 3+4 2 | 5+6 2 | |
| | | | | 6+6 2 | 5+5 2 | 5+6 2 | 5+6 2 | |
| | | | | 3+3 2 | 6+6 2 | 3+4 2 | 3+4 2 | |
| | | | | 4+4 2 | 5+5 2 | 5+6 2 | 5+6 2 | |
| | | | | 5+5 2 | 3+3 2 | 3+4 2 | 3+4 2 | |
| | | | | 6+6 2 | 4+4 2 | 3+4 2 | 5+6 2 | |
| | | | | 4+4 2 | 5+5 2 | 5+6 2 | 3+4 2 | |
| | | | | 3+3 2 | 6+6 2 | 5+6 2 | 3+4 2 | |
| | | | | 6+6 2 | 4+4 2 | 5+6 2 | 3+4 2 | |
| | | | | 5+5 2 | 3+3 2 | 3+4 2 | 5+6 2 | |
| | | | | 4+4 2 | 4+4 2 | 3+4 2 | 3+4 2 | |
| | | | | 3+3 2 | 3+3 2 | 3+4 2 | 3+4 2 | |
| | | | | 6+6 2 | 6+6 2 | 5+6 2 | 5+6 2 | |
| | | | | 5+5 2 | 5+5 2 | 5+6 2 | 3+4 2 | |
| 14.2.5 | 80829 | 72 | O2W1 | | | | | |
| | -A | -B | -C | -D | -E | -GM | -Y | -Z |
| | 2+2 4 | 2+2 4 | 2+2 4 | 3+3 2 | 3+3 2 | 3+4 2 | 3+4 2 | 2+2 4 |
| | 2+2 4 | 2+2 4 | 2+2 4 | 4+4 2 | 4+4 2 | 5+6 2 | 5+6 2 | 2+2 4 |
| | 2+2 4 | 2+2 4 | 2+2 4 | 5+5 2 | 5+5 2 | 5+6 2 | 5+6 2 | 2+2 4 |
| | 2+2 4 | 2+2 4 | 2+2 4 | 6+6 2 | 6+6 2 | 3+4 2 | 3+4 2 | 2+2 4 |
| | 2+2 4 | 2+2 4 | 2+2 4 | 6+6 2 | 5+5 2 | 5+6 2 | 5+6 2 | 2+2 4 |
| | 2+2 4 | 2+2 4 | 2+2 4 | 5+5 2 | 6+6 2 | 3+4 2 | 3+4 2 | 2+2 4 |
| | 2+2 4 | 2+2 4 | 2+2 4 | 3+3 2 | 3+3 2 | 5+6 2 | 5+6 2 | 2+2 4 |
| | 2+2 4 | 2+2 4 | 2+2 4 | 4+4 2 | 4+4 2 | 3+4 2 | 3+4 2 | 2+2 4 |
| | 2+2 4 | 2+2 4 | 2+2 4 | 6+6 2 | 3+3 2 | 3+4 2 | 5+6 2 | 2+2 4 |
| | 2+2 4 | 2+2 4 | 2+2 4 | 5+5 2 | 4+4 2 | 3+4 2 | 5+6 2 | 2+2 4 |
| | 2+2 4 | 2+2 4 | 2+2 4 | 4+4 2 | 6+6 2 | 5+6 2 | 3+4 2 | 2+2 4 |
| | 2+2 4 | 2+2 4 | 2+2 4 | 3+3 2 | 5+5 2 | 5+6 2 | 3+4 2 | 2+2 4 |
| | 2+2 4 | 2+2 4 | 2+2 4 | 5+5 2 | 5+5 2 | 5+6 2 | 5+6 2 | 2+2 4 |
| | 2+2 4 | 2+2 4 | 2+2 4 | 6+6 2 | 6+6 2 | 5+6 2 | 3+4 2 | 2+2 4 |
| | 2+2 4 | 2+2 4 | 2+2 4 | 4+4 2 | 4+4 2 | 3+4 2 | 5+6 2 | 2+2 4 |
| | 2+2 4 | 2+2 4 | 2+2 4 | 3+3 2 | 3+3 2 | 3+4 2 | 3+4 2 | 2+2 4 |
| | 2+2 4 | 2+2 4 | 2+2 4 | 5+5 2 | 4+4 2 | 3+4 2 | 5+6 2 | 2+2 4 |
| | 2+2 4 | 2+2 4 | 2+2 4 | 6+6 2 | 3+3 2 | 3+4 2 | 5+6 2 | 2+2 4 |
| | | | | 3+3 2 | 4+4 2 | 3+4 2 | 3+4 2 | |
| | | | | 4+4 2 | 3+3 2 | 5+6 2 | 5+6 2 | |
| | | | | 4+4 2 | 5+5 2 | 5+6 2 | 3+4 2 | |
| | | | | 5+5 2 | 6+6 2 | 5+6 2 | 3+4 2 | |
| | | | | 6+6 2 | 6+6 2 | 3+4 2 | 5+6 2 | |
| | | | | 3+3 2 | 5+5 2 | 5+6 2 | 5+6 2 | |
| | | | | 5+5 2 | 4+4 2 | 3+4 2 | 3+4 2 | |
| | | | | 6+6 2 | 3+3 2 | 5+6 2 | 5+6 2 | |
| | | | | 4+4 2 | 5+5 2 | 3+4 2 | 3+4 2 | |
| | | | | 3+3 2 | 6+6 2 | 3+4 2 | 5+6 2 | |
| | | | | 6+6 2 | 3+3 2 | 5+6 2 | 3+4 2 | |
| | | | | 5+5 2 | 4+4 2 | 5+6 2 | 3+4 2 | |
| | | | | 3+3 2 | 5+5 2 | 5+6 2 | 3+4 2 | |
| | | | | 4+4 2 | 6+6 2 | 3+4 2 | 5+6 2 | |
| | | | | 3+3 2 | 3+3 2 | 3+4 2 | 3+4 2 | |
| | | | | 4+4 2 | 4+4 2 | 3+4 2 | 3+4 2 | |
| | | | | 5+5 2 | 5+5 2 | 5+6 2 | 5+6 2 | |
| | | | | 6+6 2 | 6+6 2 | 5+6 2 | 3+4 2 | |
| 14.2.6 | 67381 | 264 | O1S2Yb2 | | | | | |
| | -A | -B | -C | -D | -E | -GM | -Y | -Z |
| | 2+2 4 | 2+2 4 | 2+2 4 | 3+3 2 | 3+3 2 | 3+4 2 | 3+4 2 | 2+2 4 |
| | 2+2 4 | 2+2 4 | 2+2 4 | 4+4 2 | 4+4 2 | 5+6 2 | 5+6 2 | 2+2 4 |
| | 2+2 4 | 2+2 4 | 2+2 4 | 5+5 2 | 5+5 2 | 3+4 2 | 3+4 2 | 2+2 4 |
| | 2+2 4 | 2+2 4 | 2+2 4 | 6+6 2 | 6+6 2 | 5+6 2 | 5+6 2 | 2+2 4 |
| | 2+2 4 | 2+2 4 | 2+2 4 | 3+3 2 | 3+3 2 | 3+4 2 | 3+4 2 | 2+2 4 |
| | 2+2 4 | 2+2 4 | 2+2 4 | 4+4 2 | 4+4 2 | 5+6 2 | 5+6 2 | 2+2 4 |
| | 2+2 4 | 2+2 4 | 2+2 4 | 5+5 2 | 5+5 2 | 3+4 2 | 3+4 2 | 2+2 4 |
| | 2+2 4 | 2+2 4 | 2+2 4 | 6+6 2 | 6+6 2 | 5+6 2 | 5+6 2 | 2+2 4 |
| | 2+2 4 | 2+2 4 | 2+2 4 | 4+4 2 | 4+4 2 | 3+4 2 | 3+4 2 | 2+2 4 |
| | 2+2 4 | 2+2 4 | 2+2 4 | 3+3 2 | 3+3 2 | 5+6 2 | 5+6 2 | 2+2 4 |
| | 2+2 4 | 2+2 4 | 2+2 4 | 6+6 2 | 6+6 2 | 5+6 2 | 5+6 2 | 2+2 4 |
| | 2+2 4 | 2+2 4 | 2+2 4 | 5+5 2 | 5+5 2 | 3+4 2 | 3+4 2 | 2+2 4 |
| | 2+2 4 | 2+2 4 | 2+2 4 | 5+5 2 | 3+3 2 | 3+4 2 | 5+6 2 | 2+2 4 |
| | 2+2 4 | 2+2 4 | 2+2 4 | 6+6 2 | 4+4 2 | 5+6 2 | 3+4 2 | 2+2 4 |
| | 2+2 4 | 2+2 4 | 2+2 4 | 3+3 2 | 5+5 2 | 3+4 2 | 5+6 2 | 2+2 4 |
| | 2+2 4 | 2+2 4 | 2+2 4 | 4+4 2 | 6+6 2 | 5+6 2 | 3+4 2 | 2+2 4 |
| | 2+2 4 | 2+2 4 | 2+2 4 | 3+3 2 | 3+3 2 | 3+4 2 | 3+4 2 | 2+2 4 |
| | 2+2 4 | 2+2 4 | 2+2 4 | 4+4 2 | 4+4 2 | 5+6 2 | 5+6 2 | 2+2 4 |
| | 2+2 4 | 2+2 4 | 2+2 4 | 5+5 2 | 5+5 2 | 3+4 2 | 3+4 2 | 2+2 4 |
| | 2+2 4 | 2+2 4 | 2+2 4 | 6+6 2 | 6+6 2 | 5+6 2 | 5+6 2 | 2+2 4 |
| | 2+2 4 | 2+2 4 | 2+2 4 | 4+4 2 | 4+4 2 | 5+6 2 | 5+6 2 | 2+2 4 |
| | 2+2 4 | 2+2 4 | 2+2 4 | 6+6 2 | 6+6 2 | 3+4 2 | 3+4 2 | 2+2 4 |
| | 2+2 4 | 2+2 4 | 2+2 4 | 5+5 2 | 5+5 2 | 3+4 2 | 3+4 2 | 2+2 4 |
| | 2+2 4 | 2+2 4 | 2+2 4 | 3+3 2 | 3+3 2 | 5+6 2 | 5+6 2 | 2+2 4 |
| | 2+2 4 | 2+2 4 | 2+2 4 | 6+6 2 | 6+6 2 | 3+4 2 | 3+4 2 | 2+2 4 |
| | 2+2 4 | 2+2 4 | 2+2 4 | 5+5 2 | 5+5 2 | 5+6 2 | 5+6 2 | 2+2 4 |
| | 2+2 4 | 2+2 4 | 2+2 4 | 3+3 2 | 3+3 2 | 5+6 2 | 5+6 2 | 2+2 4 |
| | 2+2 4 | 2+2 4 | 2+2 4 | 4+4 2 | 4+4 2 | 3+4 2 | 3+4 2 | 2+2 4 |
| | 2+2 4 | 2+2 4 | 2+2 4 | 6+6 2 | 5+5 2 | 5+6 2 | 5+6 2 | 2+2 4 |
| | 2+2 4 | 2+2 4 | 2+2 4 | 5+5 2 | 6+6 2 | 3+4 2 | 3+4 2 | 2+2 4 |
| | 2+2 4 | 2+2 4 | 2+2 4 | 4+4 2 | 4+4 2 | 3+4 2 | 5+6 2 | 2+2 4 |
| | 2+2 4 | 2+2 4 | 2+2 4 | 3+3 2 | 3+3 2 | 5+6 2 | 3+4 2 | 2+2 4 |
| | 2+2 4 | 2+2 4 | 2+2 4 | 6+6 2 | 6+6 2 | 3+4 2 | 5+6 2 | 2+2 4 |
| | 2+2 4 | 2+2 4 | 2+2 4 | 5+5 2 | 5+5 2 | 5+6 2 | 3+4 2 | 2+2 4 |
| | 2+2 4 | 2+2 4 | 2+2 4 | 4+4 2 | 4+4 2 | 5+6 2 | 5+6 2 | 2+2 4 |
| | 2+2 4 | 2+2 4 | 2+2 4 | 3+3 2 | 3+3 2 | 3+4 2 | 3+4 2 | 2+2 4 |
| | 2+2 4 | 2+2 4 | 2+2 4 | 5+5 2 | 5+5 2 | 3+4 2 | 3+4 2 | 2+2 4 |
| | 2+2 4 | 2+2 4 | 2+2 4 | 6+6 2 | 6+6 2 | 5+6 2 | 5+6 2 | 2+2 4 |
| | 2+2 4 | 2+2 4 | 2+2 4 | 3+3 2 | 3+3 2 | 3+4 2 | 3+4 2 | 2+2 4 |
| | 2+2 4 | 2+2 4 | 2+2 4 | 4+4 2 | 4+4 2 | 5+6 2 | 5+6 2 | 2+2 4 |
| | 2+2 4 | 2+2 4 | 2+2 4 | 4+4 2 | 5+5 2 | 3+4 2 | 5+6 2 | 2+2 4 |
| | 2+2 4 | 2+2 4 | 2+2 4 | 3+3 2 | 6+6 2 | 5+6 2 | 3+4 2 | 2+2 4 |
| | 2+2 4 | 2+2 4 | 2+2 4 | 5+5 2 | 4+4 2 | 3+4 2 | 5+6 2 | 2+2 4 |
| | 2+2 4 | 2+2 4 | 2+2 4 | 6+6 2 | 3+3 2 | 5+6 2 | 3+4 2 | 2+2 4 |
| | 2+2 4 | 2+2 4 | 2+2 4 | 5+5 2 | 5+5 2 | 5+6 2 | 5+6 2 | 2+2 4 |
| | 2+2 4 | 2+2 4 | 2+2 4 | 6+6 2 | 6+6 2 | 3+4 2 | 3+4 2 | 2+2 4 |
| | 2+2 4 | 2+2 4 | 2+2 4 | 3+3 2 | 3+3 2 | 5+6 2 | 5+6 2 | 2+2 4 |
| | 2+2 4 | 2+2 4 | 2+2 4 | 4+4 2 | 4+4 2 | 3+4 2 | 3+4 2 | 2+2 4 |

Continued on next column

**TABLE XXVI– continued**

| TQC | ICSD | NoE | SF | | | | | |
|---|---|---|---|---|---|---|---|---|
| | 2+2 4 | 2+2 4 | 2+2 4 | 4+4 2 | 4+4 2 | 3+4 2 | 3+4 2 | 2+2 4 |
| | 2+2 4 | 2+2 4 | 2+2 4 | 3+3 2 | 3+3 2 | 5+6 2 | 5+6 2 | 2+2 4 |
| | 2+2 4 | 2+2 4 | 2+2 4 | 5+5 2 | 5+5 2 | 5+6 2 | 5+6 2 | 2+2 4 |
| | 2+2 4 | 2+2 4 | 2+2 4 | 6+6 2 | 6+6 2 | 3+4 2 | 3+4 2 | 2+2 4 |
| | 2+2 4 | 2+2 4 | 2+2 4 | 3+3 2 | 3+3 2 | 3+4 2 | 3+4 2 | 2+2 4 |
| | 2+2 4 | 2+2 4 | 2+2 4 | 5+5 2 | 5+5 2 | 5+6 2 | 5+6 2 | 2+2 4 |
| | 2+2 4 | 2+2 4 | 2+2 4 | 6+6 2 | 6+6 2 | 3+4 2 | 3+4 2 | 2+2 4 |
| | 2+2 4 | 2+2 4 | 2+2 4 | 4+4 2 | 4+4 2 | 5+6 2 | 5+6 2 | 2+2 4 |
| | 2+2 4 | 2+2 4 | 2+2 4 | 4+4 2 | 4+4 2 | 3+4 2 | 3+4 2 | 2+2 4 |
| | 2+2 4 | 2+2 4 | 2+2 4 | 6+6 2 | 6+6 2 | 5+6 2 | 5+6 2 | 2+2 4 |
| | 2+2 4 | 2+2 4 | 2+2 4 | 5+5 2 | 5+5 2 | 5+6 2 | 5+6 2 | 2+2 4 |
| | 2+2 4 | 2+2 4 | 2+2 4 | 3+3 2 | 3+3 2 | 3+4 2 | 3+4 2 | 2+2 4 |
| | 2+2 4 | 2+2 4 | 2+2 4 | 4+4 2 | 4+4 2 | 3+4 2 | 3+4 2 | 2+2 4 |
| | 2+2 4 | 2+2 4 | 2+2 4 | 3+3 2 | 3+3 2 | 5+6 2 | 5+6 2 | 2+2 4 |
| | 2+2 4 | 2+2 4 | 2+2 4 | 6+6 2 | 6+6 2 | 3+4 2 | 3+4 2 | 2+2 4 |
| | 2+2 4 | 2+2 4 | 2+2 4 | 5+5 2 | 5+5 2 | 5+6 2 | 5+6 2 | 2+2 4 |
| | 2+2 4 | 2+2 4 | 2+2 4 | 4+4 2 | 4+4 2 | 3+4 2 | 3+4 2 | 2+2 4 |
| | 2+2 4 | 2+2 4 | 2+2 4 | 3+3 2 | 3+3 2 | 3+4 2 | 3+4 2 | 2+2 4 |
| | | | | 5+5 2 | 5+5 2 | 5+6 2 | 5+6 2 | |
| | | | | 3+3 2 | 3+3 2 | 5+6 2 | 5+6 2 | |
| | | | | 6+6 2 | 6+6 2 | 3+4 2 | 3+4 2 | |
| | | | | 4+4 2 | 4+4 2 | 5+6 2 | 5+6 2 | |
| | | | | 6+6 2 | 6+6 2 | 3+4 2 | 3+4 2 | |
| | | | | 5+5 2 | 5+5 2 | 5+6 2 | 5+6 2 | |
| | | | | 4+4 2 | 4+4 2 | 3+4 2 | 3+4 2 | |
| | | | | 3+3 2 | 3+3 2 | 5+6 2 | 5+6 2 | |
| | | | | 5+5 2 | 5+5 2 | 5+6 2 | 3+4 2 | |
| | | | | 6+6 2 | 6+6 2 | 3+4 2 | 5+6 2 | |
| | | | | 6+6 2 | 6+6 2 | 3+4 2 | 3+4 2 | |
| | | | | 3+3 2 | 3+3 2 | 5+6 2 | 5+6 2 | |
| | | | | 4+4 2 | 4+4 2 | 5+6 2 | 5+6 2 | |
| | | | | 5+5 2 | 5+5 2 | 3+4 2 | 3+4 2 | |
| | | | | 4+4 2 | 4+4 2 | 3+4 2 | 3+4 2 | |
| | | | | 5+5 2 | 5+5 2 | 5+6 2 | 5+6 2 | |
| | | | | 3+3 2 | 3+3 2 | 5+6 2 | 3+4 2 | |
| | | | | 6+6 2 | 6+6 2 | 3+4 2 | 5+6 2 | |
| | | | | 3+3 2 | 3+3 2 | 3+4 2 | 3+4 2 | |
| | | | | 5+5 2 | 5+5 2 | 5+6 2 | 5+6 2 | |
| | | | | 6+6 2 | 6+6 2 | 3+4 2 | 3+4 2 | |
| | | | | 4+4 2 | 4+4 2 | 5+6 2 | 5+6 2 | |
| | | | | 6+6 2 | 6+6 2 | 3+4 2 | 3+4 2 | |
| | | | | 3+3 2 | 5+5 2 | 5+6 2 | 5+6 2 | |
| | | | | 5+5 2 | 3+3 2 | 5+6 2 | 5+6 2 | |
| | | | | 4+4 2 | 4+4 2 | 3+4 2 | 3+4 2 | |
| | | | | 5+5 2 | 5+5 2 | 3+4 2 | 3+4 2 | |
| | | | | 6+6 2 | 6+6 2 | 5+6 2 | 5+6 2 | |
| | | | | 3+3 2 | 3+3 2 | 3+4 2 | 3+4 2 | |
| | | | | 4+4 2 | 4+4 2 | 5+6 2 | 5+6 2 | |
| | | | | 6+6 2 | 6+6 2 | 3+4 2 | 3+4 2 | |
| | | | | 5+5 2 | 5+5 2 | 3+4 2 | 3+4 2 | |
| | | | | 3+3 2 | 3+3 2 | 5+6 2 | 5+6 2 | |
| | | | | 4+4 2 | 4+4 2 | 3+4 2 | 3+4 2 | |
| | | | | 5+5 2 | 5+5 2 | 5+6 2 | 5+6 2 | |
| | | | | 6+6 2 | 6+6 2 | 3+4 2 | 3+4 2 | |
| | | | | 3+3 2 | 3+3 2 | 5+6 2 | 5+6 2 | |
| | | | | 4+4 2 | 4+4 2 | 5+6 2 | 3+4 2 | |
| | | | | 5+5 2 | 5+5 2 | 3+4 2 | 5+6 2 | |
| | | | | 6+6 2 | 6+6 2 | 5+6 2 | 5+6 2 | |
| | | | | 4+4 2 | 4+4 2 | 3+4 2 | 3+4 2 | |
| | | | | 3+3 2 | 3+3 2 | 5+6 2 | 5+6 2 | |
| | | | | 6+6 2 | 6+6 2 | 3+4 2 | 3+4 2 | |
| | | | | 5+5 2 | 5+5 2 | 5+6 2 | 5+6 2 | |
| | | | | 4+4 2 | 4+4 2 | 5+6 2 | 3+4 2 | |
| | | | | 3+3 2 | 3+3 2 | 3+4 2 | 5+6 2 | |
| | | | | 3+3 2 | 6+6 2 | 5+6 2 | 3+4 2 | |
| | | | | 4+4 2 | 5+5 2 | 3+4 2 | 5+6 2 | |
| | | | | 6+6 2 | 3+3 2 | 3+4 2 | 3+4 2 | |
| | | | | 5+5 2 | 4+4 2 | 5+6 2 | 5+6 2 | |
| | | | | 4+4 2 | 5+5 2 | 5+6 2 | 3+4 2 | |
| | | | | 5+5 2 | 4+4 2 | 3+4 2 | 5+6 2 | |
| | | | | 3+3 2 | 6+6 2 | 3+4 2 | 5+6 2 | |
| | | | | 6+6 2 | 3+3 2 | 5+6 2 | 3+4 2 | |
| | | | | 4+4 2 | 5+5 2 | 5+6 2 | 3+4 2 | |
| | | | | 3+3 2 | 6+6 2 | 3+4 2 | 5+6 2 | |
| | | | | 5+5 2 | 4+4 2 | 3+4 2 | 5+6 2 | |
| | | | | 6+6 2 | 3+3 2 | 5+6 2 | 3+4 2 | |
| | | | | 4+4 2 | 5+5 2 | 3+4 2 | 5+6 2 | |
| | | | | 3+3 2 | 6+6 2 | 3+4 2 | 3+4 2 | |
| | | | | 6+6 2 | 3+3 2 | 5+6 2 | 5+6 2 | |
| | | | | 5+5 2 | 4+4 2 | 3+4 2 | 5+6 2 | |
| | | | | 3+3 2 | 6+6 2 | 5+6 2 | 3+4 2 | |
| | | | | 4+4 2 | 5+5 2 | 3+4 2 | 5+6 2 | |
| | | | | 5+5 2 | 4+4 2 | 3+4 2 | 3+4 2 | |
| | | | | 6+6 2 | 3+3 2 | 5+6 2 | 5+6 2 | |
| 14.2.7 | 281705 | 176 | Ni3O8P2 | | | | | |
| | -A | -B | -C | -D | -E | -GM | -Y | -Z |
| | 2+2 4 | 2+2 4 | 2+2 4 | 3+3 2 | 5+5 2 | 3+4 2 | 3+4 2 | 2+2 4 |
| | 2+2 4 | 2+2 4 | 2+2 4 | 4+4 2 | 6+6 2 | 3+4 2 | 5+6 2 | 2+2 4 |
| | 2+2 4 | 2+2 4 | 2+2 4 | 5+5 2 | 3+3 2 | 5+6 2 | 3+4 2 | 2+2 4 |
| | 2+2 4 | 2+2 4 | 2+2 4 | 6+6 2 | 4+4 2 | 5+6 2 | 5+6 2 | 2+2 4 |
| | 2+2 4 | 2+2 4 | 2+2 4 | 5+5 2 | 5+5 2 | 3+4 2 | 3+4 2 | 2+2 4 |
| | 2+2 4 | 2+2 4 | 2+2 4 | 6+6 2 | 6+6 2 | 5+6 2 | 5+6 2 | 2+2 4 |
| | 2+2 4 | 2+2 4 | 2+2 4 | 3+3 2 | 3+3 2 | 3+4 2 | 5+6 2 | 2+2 4 |
| | 2+2 4 | 2+2 4 | 2+2 4 | 4+4 2 | 4+4 2 | 5+6 2 | 3+4 2 | 2+2 4 |
| | 2+2 4 | 2+2 4 | 2+2 4 | 5+5 2 | 3+3 2 | 5+6 2 | 3+4 2 | 2+2 4 |
| | 2+2 4 | 2+2 4 | 2+2 4 | 6+6 2 | 4+4 2 | 3+4 2 | 3+4 2 | 2+2 4 |
| | 2+2 4 | 2+2 4 | 2+2 4 | 3+3 2 | 5+5 2 | 5+6 2 | 5+6 2 | 2+2 4 |
| | 2+2 4 | 2+2 4 | 2+2 4 | 4+4 2 | 6+6 2 | 3+4 2 | 5+6 2 | 2+2 4 |
| | 2+2 4 | 2+2 4 | 2+2 4 | 5+5 2 | 5+5 2 | 5+6 2 | 5+6 2 | 2+2 4 |
| | 2+2 4 | 2+2 4 | 2+2 4 | 6+6 2 | 6+6 2 | 5+6 2 | 3+4 2 | 2+2 4 |
| | 2+2 4 | 2+2 4 | 2+2 4 | 3+3 2 | 3+3 2 | 3+4 2 | 5+6 2 | 2+2 4 |
| | 2+2 4 | 2+2 4 | 2+2 4 | 4+4 2 | 4+4 2 | 3+4 2 | 3+4 2 | 2+2 4 |
| | 2+2 4 | 2+2 4 | 2+2 4 | 3+3 2 | 5+5 2 | 3+4 2 | 3+4 2 | 2+2 4 |
| | 2+2 4 | 2+2 4 | 2+2 4 | 4+4 2 | 6+6 2 | 3+4 2 | 5+6 2 | 2+2 4 |
| | 2+2 4 | 2+2 4 | 2+2 4 | 5+5 2 | 3+3 2 | 5+6 2 | 3+4 2 | 2+2 4 |
| | 2+2 4 | 2+2 4 | 2+2 4 | 6+6 2 | 4+4 2 | 5+6 2 | 5+6 2 | 2+2 4 |
| | 2+2 4 | 2+2 4 | 2+2 4 | 4+4 2 | 6+6 2 | 3+4 2 | 3+4 2 | 2+2 4 |
| | 2+2 4 | 2+2 4 | 2+2 4 | 3+3 2 | 5+5 2 | 3+4 2 | 5+6 2 | 2+2 4 |
| | 2+2 4 | 2+2 4 | 2+2 4 | 3+3 2 | 4+4 2 | 5+6 2 | 3+4 2 | 2+2 4 |
| | 2+2 4 | 2+2 4 | 2+2 4 | 4+4 2 | 3+3 2 | 5+6 2 | 5+6 2 | 2+2 4 |
| | 2+2 4 | 2+2 4 | 2+2 4 | 6+6 2 | 5+5 2 | 3+4 2 | 3+4 2 | 2+2 4 |
| | 2+2 4 | 2+2 4 | 2+2 4 | 5+5 2 | 6+6 2 | 5+6 2 | 5+6 2 | 2+2 4 |
| | 2+2 4 | 2+2 4 | 2+2 4 | 5+5 2 | 3+3 2 | 3+4 2 | 5+6 2 | 2+2 4 |
| | 2+2 4 | 2+2 4 | 2+2 4 | 6+6 2 | 4+4 2 | 3+4 2 | 3+4 2 | 2+2 4 |
| | 2+2 4 | 2+2 4 | 2+2 4 | 3+3 2 | 6+6 2 | 3+4 2 | 3+4 2 | 2+2 4 |
| | 2+2 4 | 2+2 4 | 2+2 4 | 4+4 2 | 5+5 2 | 5+6 2 | 3+4 2 | 2+2 4 |
| | 2+2 4 | 2+2 4 | 2+2 4 | 5+5 2 | 4+4 2 | 5+6 2 | 5+6 2 | 2+2 4 |
| | 2+2 4 | 2+2 4 | 2+2 4 | 6+6 2 | 3+3 2 | 5+6 2 | 5+6 2 | 2+2 4 |
| | 2+2 4 | 2+2 4 | 2+2 4 | 6+6 2 | 5+5 2 | 5+6 2 | 3+4 2 | 2+2 4 |
| | 2+2 4 | 2+2 4 | 2+2 4 | 5+5 2 | 6+6 2 | 3+4 2 | 3+4 2 | 2+2 4 |
| | 2+2 4 | 2+2 4 | 2+2 4 | 5+5 2 | 3+3 2 | 5+6 2 | 5+6 2 | 2+2 4 |
| | 2+2 4 | 2+2 4 | 2+2 4 | 6+6 2 | 4+4 2 | 3+4 2 | 3+4 2 | 2+2 4 |

Continued on next column

**TABLE XXVI– continued**

| TQC | ICSD | NoE | SF | | | | |
|---|---|---|---|---|---|---|---|
| | 2+2 4 | 2+2 4 | 2+2 4 | 3+3 2 | 5+5 2 | 3+4 2 | 5+6 2 | 2+2 4 |
| | 2+2 4 | 2+2 4 | 2+2 4 | 4+4 2 | 6+6 2 | 3+4 2 | 5+6 2 | 2+2 4 |
| | 2+2 4 | 2+2 4 | 2+2 4 | 3+3 2 | 3+3 2 | 5+6 2 | 3+4 2 | 2+2 4 |
| | 2+2 4 | 2+2 4 | 2+2 4 | 4+4 2 | 4+4 2 | 5+6 2 | 3+4 2 | 2+2 4 |
| | 2+2 4 | 2+2 4 | 2+2 4 | 6+6 2 | 5+5 2 | 3+4 2 | 5+6 2 | 2+2 4 |
| | 2+2 4 | 2+2 4 | 2+2 4 | 5+5 2 | 3+3 2 | 3+4 2 | 5+6 2 | 2+2 4 |
| | 2+2 4 | 2+2 4 | 2+2 4 | 4+4 2 | 4+4 2 | 5+6 2 | 3+4 2 | 2+2 4 |
| | 2+2 4 | 2+2 4 | 2+2 4 | 3+3 2 | 6+6 2 | 3+4 2 | 5+6 2 | 2+2 4 |
| | | | | 3+3 2 | 3+3 2 | 3+4 2 | 5+6 2 | |
| | | | | 4+4 2 | 4+4 2 | 5+6 2 | 3+4 2 | |
| | | | | 6+6 2 | 5+5 2 | 3+4 2 | 3+4 2 | |
| | | | | 5+5 2 | 6+6 2 | 3+4 2 | 3+4 2 | |
| | | | | 5+5 2 | 5+5 2 | 3+4 2 | 5+6 2 | |
| | | | | 6+6 2 | 6+6 2 | 5+6 2 | 5+6 2 | |
| | | | | 3+3 2 | 3+3 2 | 5+6 2 | 3+4 2 | |
| | | | | 4+4 2 | 4+4 2 | 5+6 2 | 3+4 2 | |
| | | | | 6+6 2 | 6+6 2 | 3+4 2 | 3+4 2 | |
| | | | | 5+5 2 | 5+5 2 | 5+6 2 | 5+6 2 | |
| | | | | 6+6 2 | 3+3 2 | 5+6 2 | 3+4 2 | |
| | | | | 3+3 2 | 4+4 2 | 5+6 2 | 5+6 2 | |
| | | | | 4+4 2 | 5+5 2 | 5+6 2 | 3+4 2 | |
| | | | | 5+5 2 | 6+6 2 | 5+6 2 | 5+6 2 | |
| | | | | 3+3 2 | 4+4 2 | 3+4 2 | 5+6 2 | |
| | | | | 4+4 2 | 3+3 2 | 3+4 2 | 3+4 2 | |
| | | | | 5+5 2 | 5+5 2 | 3+4 2 | 5+6 2 | |
| | | | | 6+6 2 | 6+6 2 | 5+6 2 | 5+6 2 | |
| | | | | 3+3 2 | 4+4 2 | 5+6 2 | 3+4 2 | |
| | | | | 4+4 2 | 3+3 2 | 3+4 2 | 5+6 2 | |
| | | | | 5+5 2 | 5+5 2 | 3+4 2 | 3+4 2 | |
| | | | | 6+6 2 | 6+6 2 | 3+4 2 | 3+4 2 | |
| | | | | 6+6 2 | 6+6 2 | 3+4 2 | 3+4 2 | |
| | | | | 5+5 2 | 5+5 2 | 3+4 2 | 3+4 2 | |
| | | | | 4+4 2 | 5+5 2 | 3+4 2 | 3+4 2 | |
| | | | | 3+3 2 | 6+6 2 | 3+4 2 | 5+6 2 | |
| | | | | 5+5 2 | 3+3 2 | 5+6 2 | 3+4 2 | |
| | | | | 6+6 2 | 4+4 2 | 3+4 2 | 5+6 2 | |
| | | | | 3+3 2 | 6+6 2 | 5+6 2 | 3+4 2 | |
| | | | | 6+6 2 | 5+5 2 | 3+4 2 | 5+6 2 | |
| | | | | 5+5 2 | 4+4 2 | 5+6 2 | 3+4 2 | |
| | | | | 4+4 2 | 3+3 2 | 3+4 2 | 5+6 2 | |
| | | | | 5+5 2 | 5+5 2 | 5+6 2 | 3+4 2 | |
| | | | | 6+6 2 | 6+6 2 | 3+4 2 | 3+4 2 | |
| | | | | 4+4 2 | 5+5 2 | 5+6 2 | 3+4 2 | |
| | | | | 3+3 2 | 6+6 2 | 3+4 2 | 5+6 2 | |
| | | | | 6+6 2 | 3+3 2 | 3+4 2 | 3+4 2 | |
| | | | | 5+5 2 | 4+4 2 | 5+6 2 | 5+6 2 | |
| | | | | 5+5 2 | 5+5 2 | 5+6 2 | 3+4 2 | |
| | | | | 6+6 2 | 6+6 2 | 3+4 2 | 5+6 2 | |
| | | | | 6+6 2 | 4+4 2 | 3+4 2 | 3+4 2 | |
| | | | | 5+5 2 | 3+3 2 | 5+6 2 | 5+6 2 | |
| | | | | 3+3 2 | 5+5 2 | 3+4 2 | 3+4 2 | |
| | | | | 4+4 2 | 6+6 2 | 5+6 2 | 3+4 2 | |
| 14.2.8 | 100194 | 144 | Ni2O7P2 | | | | | |
| | -A | -B | -C | -D | -E | -GM | -Y | -Z |
| | 2+2 4 | 2+2 4 | 2+2 4 | 5+5 2 | 5+5 2 | 3+4 2 | 3+4 2 | 2+2 4 |
| | 2+2 4 | 2+2 4 | 2+2 4 | 6+6 2 | 6+6 2 | 3+4 2 | 3+4 2 | 2+2 4 |
| | 2+2 4 | 2+2 4 | 2+2 4 | 3+3 2 | 3+3 2 | 5+6 2 | 5+6 2 | 2+2 4 |
| | 2+2 4 | 2+2 4 | 2+2 4 | 4+4 2 | 4+4 2 | 5+6 2 | 5+6 2 | 2+2 4 |
| | 2+2 4 | 2+2 4 | 2+2 4 | 5+5 2 | 5+5 2 | 3+4 2 | 3+4 2 | 2+2 4 |
| | 2+2 4 | 2+2 4 | 2+2 4 | 6+6 2 | 6+6 2 | 3+4 2 | 3+4 2 | 2+2 4 |
| | 2+2 4 | 2+2 4 | 2+2 4 | 4+4 2 | 3+3 2 | 3+4 2 | 5+6 2 | 2+2 4 |

Continued on next column

**TABLE XXVI– continued**

| TQC | ICSD | NoE | SF | | | | | |
|---|---|---|---|---|---|---|---|---|
| | 2+2 4 | 2+2 4 | 2+2 4 | 3+3 2 | 4+4 2 | 5+6 2 | 3+4 2 | 2+2 4 |
| | 2+2 4 | 2+2 4 | 2+2 4 | 5+5 2 | 6+6 2 | 3+4 2 | 3+4 2 | 2+2 4 |
| | 2+2 4 | 2+2 4 | 2+2 4 | 6+6 2 | 5+5 2 | 5+6 2 | 5+6 2 | 2+2 4 |
| | 2+2 4 | 2+2 4 | 2+2 4 | 3+3 2 | 5+5 2 | 3+4 2 | 5+6 2 | 2+2 4 |
| | 2+2 4 | 2+2 4 | 2+2 4 | 4+4 2 | 6+6 2 | 5+6 2 | 5+6 2 | 2+2 4 |
| | 2+2 4 | 2+2 4 | 2+2 4 | 6+6 2 | 3+3 2 | 3+4 2 | 3+4 2 | 2+2 4 |
| | 2+2 4 | 2+2 4 | 2+2 4 | 5+5 2 | 4+4 2 | 5+6 2 | 3+4 2 | 2+2 4 |
| | 2+2 4 | 2+2 4 | 2+2 4 | 3+3 2 | 3+3 2 | 5+6 2 | 5+6 2 | 2+2 4 |
| | 2+2 4 | 2+2 4 | 2+2 4 | 4+4 2 | 4+4 2 | 5+6 2 | 5+6 2 | 2+2 4 |
| | 2+2 4 | 2+2 4 | 2+2 4 | 5+5 2 | 5+5 2 | 3+4 2 | 3+4 2 | 2+2 4 |
| | 2+2 4 | 2+2 4 | 2+2 4 | 6+6 2 | 6+6 2 | 3+4 2 | 5+6 2 | 2+2 4 |
| | 2+2 4 | 2+2 4 | 2+2 4 | 3+3 2 | 3+3 2 | 5+6 2 | 5+6 2 | 2+2 4 |
| | 2+2 4 | 2+2 4 | 2+2 4 | 4+4 2 | 4+4 2 | 5+6 2 | 5+6 2 | 2+2 4 |
| | 2+2 4 | 2+2 4 | 2+2 4 | 4+4 2 | 6+6 2 | 5+6 2 | 3+4 2 | 2+2 4 |
| | 2+2 4 | 2+2 4 | 2+2 4 | 3+3 2 | 5+5 2 | 3+4 2 | 5+6 2 | 2+2 4 |
| | 2+2 4 | 2+2 4 | 2+2 4 | 3+3 2 | 3+3 2 | 5+6 2 | 3+4 2 | 2+2 4 |
| | 2+2 4 | 2+2 4 | 2+2 4 | 4+4 2 | 4+4 2 | 5+6 2 | 5+6 2 | 2+2 4 |
| | 2+2 4 | 2+2 4 | 2+2 4 | 5+5 2 | 4+4 2 | 3+4 2 | 5+6 2 | 2+2 4 |
| | 2+2 4 | 2+2 4 | 2+2 4 | 6+6 2 | 3+3 2 | 3+4 2 | 3+4 2 | 2+2 4 |
| | 2+2 4 | 2+2 4 | 2+2 4 | 5+5 2 | 5+5 2 | 5+6 2 | 3+4 2 | 2+2 4 |
| | 2+2 4 | 2+2 4 | 2+2 4 | 6+6 2 | 6+6 2 | 3+4 2 | 3+4 2 | 2+2 4 |
| | 2+2 4 | 2+2 4 | 2+2 4 | 3+3 2 | 5+5 2 | 5+6 2 | 5+6 2 | 2+2 4 |
| | 2+2 4 | 2+2 4 | 2+2 4 | 4+4 2 | 6+6 2 | 3+4 2 | 5+6 2 | 2+2 4 |
| | 2+2 4 | 2+2 4 | 2+2 4 | 5+5 2 | 3+3 2 | 5+6 2 | 3+4 2 | 2+2 4 |
| | 2+2 4 | 2+2 4 | 2+2 4 | 6+6 2 | 4+4 2 | 5+6 2 | 3+4 2 | 2+2 4 |
| | 2+2 4 | 2+2 4 | 2+2 4 | 3+3 2 | 3+3 2 | 3+4 2 | 5+6 2 | 2+2 4 |
| | 2+2 4 | 2+2 4 | 2+2 4 | 4+4 2 | 4+4 2 | 3+4 2 | 5+6 2 | 2+2 4 |
| | 2+2 4 | 2+2 4 | 2+2 4 | 4+4 2 | 3+3 2 | 5+6 2 | 3+4 2 | 2+2 4 |
| | 2+2 4 | 2+2 4 | 2+2 4 | 3+3 2 | 4+4 2 | 5+6 2 | 5+6 2 | 2+2 4 |
| | | | | 5+5 2 | 5+5 2 | 3+4 2 | 3+4 2 | |
| | | | | 6+6 2 | 6+6 2 | 3+4 2 | 5+6 2 | |
| | | | | 4+4 2 | 4+4 2 | 5+6 2 | 5+6 2 | |
| | | | | 3+3 2 | 3+3 2 | 3+4 2 | 3+4 2 | |
| | | | | 6+6 2 | 6+6 2 | 5+6 2 | 5+6 2 | |
| | | | | 5+5 2 | 5+5 2 | 3+4 2 | 3+4 2 | |
| | | | | 5+5 2 | 3+3 2 | 3+4 2 | 3+4 2 | |
| | | | | 6+6 2 | 4+4 2 | 3+4 2 | 5+6 2 | |
| | | | | 3+3 2 | 5+5 2 | 5+6 2 | 5+6 2 | |
| | | | | 4+4 2 | 6+6 2 | 5+6 2 | 3+4 2 | |
| | | | | 3+3 2 | 3+3 2 | 5+6 2 | 5+6 2 | |
| | | | | 4+4 2 | 4+4 2 | 5+6 2 | 5+6 2 | |
| | | | | 5+5 2 | 5+5 2 | 5+6 2 | 3+4 2 | |
| | | | | 6+6 2 | 6+6 2 | 3+4 2 | 3+4 2 | |
| | | | | 3+3 2 | 4+4 2 | 3+4 2 | 3+4 2 | |
| | | | | 4+4 2 | 3+3 2 | 5+6 2 | 5+6 2 | |
| | | | | 3+3 2 | 5+5 2 | 5+6 2 | 3+4 2 | |
| | | | | 4+4 2 | 6+6 2 | 5+6 2 | 5+6 2 | |
| | | | | 5+5 2 | 4+4 2 | 3+4 2 | 5+6 2 | |
| | | | | 6+6 2 | 3+3 2 | 5+6 2 | 5+6 2 | |
| | | | | 6+6 2 | 3+3 2 | 3+4 2 | 5+6 2 | |
| | | | | 5+5 2 | 4+4 2 | 5+6 2 | 3+4 2 | |
| | | | | 3+3 2 | 6+6 2 | 5+6 2 | 5+6 2 | |
| | | | | 4+4 2 | 5+5 2 | 3+4 2 | 3+4 2 | |
| | | | | 5+5 2 | 6+6 2 | 3+4 2 | 5+6 2 | |
| | | | | 6+6 2 | 5+5 2 | 3+4 2 | 3+4 2 | |
| | | | | 3+3 2 | 3+3 2 | 5+6 2 | 5+6 2 | |
| | | | | 4+4 2 | 4+4 2 | 3+4 2 | 3+4 2 | |
| | | | | 5+5 2 | 5+5 2 | 5+6 2 | 3+4 2 | |
| | | | | 6+6 2 | 6+6 2 | 3+4 2 | 5+6 2 | |
| | | | | 4+4 2 | 4+4 2 | 5+6 2 | 3+4 2 | |
| | | | | 3+3 2 | 3+3 2 | 5+6 2 | 5+6 2 | |

Continued on next column

**TABLE XXVI– continued**

| TQC | ICSD | NoE | SF | | | | | | | |
|---|---|---|---|---|---|---|---|---|---|---|
| | | | | | | 5+5 2 | 5+5 2 | 5+6 2 | 3+4 2 | |
| | | | | | | 6+6 2 | 6+6 2 | 3+4 2 | 3+4 2 | |
| | | | | | | 4+4 2 | 4+4 2 | 5+6 2 | 5+6 2 | |
| | | | | | | 3+3 2 | 3+3 2 | 3+4 2 | 3+4 2 | |
| 14.2.9 | 246365 | 180 | Ni1O10P2V2 | | | | | | | |
| | | | -A | -B | -C | -D | -E | -GM | -Y | -Z |
| | | | 2+2 4 | 2+2 4 | 2+2 4 | 5+5 2 | 3+3 2 | 3+4 2 | 3+4 2 | 2+2 4 |
| | | | 2+2 4 | 2+2 4 | 2+2 4 | 6+6 2 | 4+4 2 | 3+4 2 | 5+6 2 | 2+2 4 |
| | | | 2+2 4 | 2+2 4 | 2+2 4 | 3+3 2 | 5+5 2 | 5+6 2 | 5+6 2 | 2+2 4 |
| | | | 2+2 4 | 2+2 4 | 2+2 4 | 4+4 2 | 6+6 2 | 5+6 2 | 3+4 2 | 2+2 4 |
| | | | 2+2 4 | 2+2 4 | 2+2 4 | 3+3 2 | 3+3 2 | 5+6 2 | 3+4 2 | 2+2 4 |
| | | | 2+2 4 | 2+2 4 | 2+2 4 | 4+4 2 | 4+4 2 | 3+4 2 | 5+6 2 | 2+2 4 |
| | | | 2+2 4 | 2+2 4 | 2+2 4 | 5+5 2 | 6+6 2 | 5+6 2 | 5+6 2 | 2+2 4 |
| | | | 2+2 4 | 2+2 4 | 2+2 4 | 6+6 2 | 5+5 2 | 3+4 2 | 3+4 2 | 2+2 4 |
| | | | 2+2 4 | 2+2 4 | 2+2 4 | 5+5 2 | 5+5 2 | 3+4 2 | 3+4 2 | 2+2 4 |
| | | | 2+2 4 | 2+2 4 | 2+2 4 | 6+6 2 | 6+6 2 | 5+6 2 | 5+6 2 | 2+2 4 |
| | | | 2+2 4 | 2+2 4 | 2+2 4 | 4+4 2 | 3+3 2 | 3+4 2 | 5+6 2 | 2+2 4 |
| | | | 2+2 4 | 2+2 4 | 2+2 4 | 3+3 2 | 4+4 2 | 5+6 2 | 3+4 2 | 2+2 4 |
| | | | 2+2 4 | 2+2 4 | 2+2 4 | 3+3 2 | 5+5 2 | 3+4 2 | 3+4 2 | 2+2 4 |
| | | | 2+2 4 | 2+2 4 | 2+2 4 | 5+5 2 | 6+6 2 | 5+6 2 | 3+4 2 | 2+2 4 |
| | | | 2+2 4 | 2+2 4 | 2+2 4 | 6+6 2 | 3+3 2 | 3+4 2 | 5+6 2 | 2+2 4 |
| | | | 2+2 4 | 2+2 4 | 2+2 4 | 4+4 2 | 4+4 2 | 5+6 2 | 5+6 2 | 2+2 4 |
| | | | 2+2 4 | 2+2 4 | 2+2 4 | 3+3 2 | 3+3 2 | 3+4 2 | 3+4 2 | 2+2 4 |
| | | | 2+2 4 | 2+2 4 | 2+2 4 | 4+4 2 | 4+4 2 | 5+6 2 | 5+6 2 | 2+2 4 |
| | | | 2+2 4 | 2+2 4 | 2+2 4 | 5+5 2 | 5+5 2 | 3+4 2 | 3+4 2 | 2+2 4 |
| | | | 2+2 4 | 2+2 4 | 2+2 4 | 6+6 2 | 6+6 2 | 5+6 2 | 5+6 2 | 2+2 4 |
| | | | 2+2 4 | 2+2 4 | 2+2 4 | 3+3 2 | 3+3 2 | 3+4 2 | 3+4 2 | 2+2 4 |
| | | | 2+2 4 | 2+2 4 | 2+2 4 | 4+4 2 | 4+4 2 | 5+6 2 | 5+6 2 | 2+2 4 |
| | | | 2+2 4 | 2+2 4 | 2+2 4 | 5+5 2 | 5+5 2 | 5+6 2 | 5+6 2 | 2+2 4 |
| | | | 2+2 4 | 2+2 4 | 2+2 4 | 6+6 2 | 6+6 2 | 3+4 2 | 3+4 2 | 2+2 4 |
| | | | 2+2 4 | 2+2 4 | 2+2 4 | 4+4 2 | 4+4 2 | 3+4 2 | 3+4 2 | 2+2 4 |
| | | | 2+2 4 | 2+2 4 | 2+2 4 | 3+3 2 | 3+3 2 | 5+6 2 | 5+6 2 | 2+2 4 |
| | | | 2+2 4 | 2+2 4 | 2+2 4 | 6+6 2 | 6+6 2 | 5+6 2 | 5+6 2 | 2+2 4 |
| | | | 2+2 4 | 2+2 4 | 2+2 4 | 5+5 2 | 5+5 2 | 3+4 2 | 3+4 2 | 2+2 4 |
| | | | 2+2 4 | 2+2 4 | 2+2 4 | 3+3 2 | 5+5 2 | 3+4 2 | 3+4 2 | 2+2 4 |
| | | | 2+2 4 | 2+2 4 | 2+2 4 | 4+4 2 | 6+6 2 | 3+4 2 | 3+4 2 | 2+2 4 |
| | | | 2+2 4 | 2+2 4 | 2+2 4 | 5+5 2 | 3+3 2 | 3+4 2 | 5+6 2 | 2+2 4 |
| | | | 2+2 4 | 2+2 4 | 2+2 4 | 6+6 2 | 4+4 2 | 5+6 2 | 5+6 2 | 2+2 4 |
| | | | 2+2 4 | 2+2 4 | 2+2 4 | 3+3 2 | 3+3 2 | 5+6 2 | 3+4 2 | 2+2 4 |
| | | | 2+2 4 | 2+2 4 | 2+2 4 | 4+4 2 | 4+4 2 | 5+6 2 | 3+4 2 | 2+2 4 |
| | | | 2+2 4 | 2+2 4 | 2+2 4 | 5+5 2 | 5+5 2 | 3+4 2 | 5+6 2 | 2+2 4 |
| | | | 2+2 4 | 2+2 4 | 2+2 4 | 6+6 2 | 6+6 2 | 5+6 2 | 5+6 2 | 2+2 4 |
| | | | 2+2 4 | 2+2 4 | 2+2 4 | 3+3 2 | 3+3 2 | 3+4 2 | 3+4 2 | 2+2 4 |
| | | | 2+2 4 | 2+2 4 | 2+2 4 | 4+4 2 | 4+4 2 | 5+6 2 | 5+6 2 | 2+2 4 |
| | | | 2+2 4 | 2+2 4 | 2+2 4 | 3+3 2 | 5+5 2 | 3+4 2 | 3+4 2 | 2+2 4 |
| | | | 2+2 4 | 2+2 4 | 2+2 4 | 4+4 2 | 6+6 2 | 5+6 2 | 5+6 2 | 2+2 4 |
| | | | 2+2 4 | 2+2 4 | 2+2 4 | 5+5 2 | 4+4 2 | 3+4 2 | 3+4 2 | 2+2 4 |
| | | | 2+2 4 | 2+2 4 | 2+2 4 | 6+6 2 | 3+3 2 | 5+6 2 | 5+6 2 | 2+2 4 |
| | | | 2+2 4 | 2+2 4 | 2+2 4 | 5+5 2 | 3+3 2 | 3+4 2 | 3+4 2 | 2+2 4 |
| | | | 2+2 4 | 2+2 4 | 2+2 4 | 6+6 2 | 4+4 2 | 3+4 2 | 3+4 2 | 2+2 4 |
| | | | 2+2 4 | 2+2 4 | 2+2 4 | 5+5 2 | 5+5 2 | 3+4 2 | 5+6 2 | 2+2 4 |
| | | | | | | 4+4 2 | 6+6 2 | 3+4 2 | 5+6 2 | |
| | | | | | | 6+6 2 | 5+5 2 | 5+6 2 | 3+4 2 | |
| | | | | | | 3+3 2 | 6+6 2 | 3+4 2 | 3+4 2 | |
| | | | | | | 3+3 2 | 3+3 2 | 5+6 2 | 5+6 2 | |
| | | | | | | 4+4 2 | 4+4 2 | 5+6 2 | 5+6 2 | |
| | | | | | | 5+5 2 | 5+5 2 | 5+6 2 | 5+6 2 | |
| | | | | | | 6+6 2 | 6+6 2 | 3+4 2 | 3+4 2 | |
| | | | | | | 6+6 2 | 3+3 2 | 3+4 2 | 3+4 2 | |
| | | | | | | 5+5 2 | 4+4 2 | 3+4 2 | 5+6 2 | |
| | | | | | | 3+3 2 | 6+6 2 | 5+6 2 | 3+4 2 | |

Continued on next column

**TABLE XXVI– continued**

| TQC | ICSD | NoE | SF | | | | | | | |
|---|---|---|---|---|---|---|---|---|---|---|
| | | | | | | 4+4 2 | 5+5 2 | 5+6 2 | 5+6 2 | |
| | | | | | | 5+5 2 | 3+3 2 | 3+4 2 | 5+6 2 | |
| | | | | | | 6+6 2 | 4+4 2 | 5+6 2 | 3+4 2 | |
| | | | | | | 3+3 2 | 6+6 2 | 5+6 2 | 3+4 2 | |
| | | | | | | 4+4 2 | 5+5 2 | 3+4 2 | 5+6 2 | |
| | | | | | | 4+4 2 | 4+4 2 | 3+4 2 | 3+4 2 | |
| | | | | | | 3+3 2 | 3+3 2 | 5+6 2 | 3+4 2 | |
| | | | | | | 6+6 2 | 6+6 2 | 3+4 2 | 5+6 2 | |
| | | | | | | 5+5 2 | 5+5 2 | 5+6 2 | 3+4 2 | |
| | | | | | | 3+3 2 | 3+3 2 | 5+6 2 | 5+6 2 | |
| | | | | | | 4+4 2 | 4+4 2 | 5+6 2 | 3+4 2 | |
| | | | | | | 5+5 2 | 4+4 2 | 3+4 2 | 3+4 2 | |
| | | | | | | 6+6 2 | 3+3 2 | 5+6 2 | 5+6 2 | |
| | | | | | | 3+3 2 | 5+5 2 | 3+4 2 | 5+6 2 | |
| | | | | | | 4+4 2 | 6+6 2 | 5+6 2 | 3+4 2 | |
| | | | | | | 6+6 2 | 5+5 2 | 3+4 2 | 5+6 2 | |
| | | | | | | 5+5 2 | 6+6 2 | 5+6 2 | 5+6 2 | |
| | | | | | | 3+3 2 | 3+3 2 | 3+4 2 | 3+4 2 | |
| | | | | | | 4+4 2 | 4+4 2 | 5+6 2 | 5+6 2 | |
| | | | | | | 5+5 2 | 5+5 2 | 3+4 2 | 3+4 2 | |
| | | | | | | 6+6 2 | 6+6 2 | 5+6 2 | 5+6 2 | |
| | | | | | | 3+3 2 | 6+6 2 | 5+6 2 | 5+6 2 | |
| | | | | | | 4+4 2 | 5+5 2 | 3+4 2 | 3+4 2 | |
| | | | | | | 5+5 2 | 4+4 2 | 5+6 2 | 3+4 2 | |
| | | | | | | 6+6 2 | 3+3 2 | 3+4 2 | 5+6 2 | |
| | | | | | | 4+4 2 | 4+4 2 | 3+4 2 | 3+4 2 | |
| | | | | | | 3+3 2 | 3+3 2 | 3+4 2 | 3+4 2 | |
| | | | | | | 3+3 2 | 3+3 2 | 3+4 2 | 3+4 2 | |
| | | | | | | 4+4 2 | 4+4 2 | 3+4 2 | 3+4 2 | |
| | | | | | | 3+3 2 | 3+3 2 | 3+4 2 | 3+4 2 | |
| | | | | | | 4+4 2 | 4+4 2 | 3+4 2 | 3+4 2 | |
| | | | | | | 3+3 2 | 3+3 2 | 3+4 2 | 3+4 2 | |
| | | | | | | 4+4 2 | 4+4 2 | 3+4 2 | 3+4 2 | |
| | | | | | | 5+5 2 | 4+4 2 | 3+4 2 | 5+6 2 | |
| | | | | | | 6+6 2 | 3+3 2 | 3+4 2 | 3+4 2 | |
| 14.2.10 | 80000 | 124 | Mo3Si4U2 | | | | | | | |
| | | | -A | -B | -C | -D | -E | -GM | -Y | -Z |
| | | | 2+2 4 | 2+2 4 | 2+2 4 | 3+3 2 | 3+3 2 | 3+4 2 | 3+4 2 | 2+2 4 |
| | | | 2+2 4 | 2+2 4 | 2+2 4 | 4+4 2 | 4+4 2 | 3+4 2 | 5+6 2 | 2+2 4 |
| | | | 2+2 4 | 2+2 4 | 2+2 4 | 5+5 2 | 5+5 2 | 5+6 2 | 3+4 2 | 2+2 4 |
| | | | 2+2 4 | 2+2 4 | 2+2 4 | 6+6 2 | 6+6 2 | 5+6 2 | 5+6 2 | 2+2 4 |
| | | | 2+2 4 | 2+2 4 | 2+2 4 | 3+3 2 | 3+3 2 | 3+4 2 | 3+4 2 | 2+2 4 |
| | | | 2+2 4 | 2+2 4 | 2+2 4 | 4+4 2 | 4+4 2 | 5+6 2 | 5+6 2 | 2+2 4 |
| | | | 2+2 4 | 2+2 4 | 2+2 4 | 5+5 2 | 5+5 2 | 3+4 2 | 3+4 2 | 2+2 4 |
| | | | 2+2 4 | 2+2 4 | 2+2 4 | 6+6 2 | 6+6 2 | 5+6 2 | 5+6 2 | 2+2 4 |
| | | | 2+2 4 | 2+2 4 | 2+2 4 | 4+4 2 | 4+4 2 | 3+4 2 | 3+4 2 | 2+2 4 |
| | | | 2+2 4 | 2+2 4 | 2+2 4 | 6+6 2 | 6+6 2 | 5+6 2 | 5+6 2 | 2+2 4 |
| | | | 2+2 4 | 2+2 4 | 2+2 4 | 3+3 2 | 3+3 2 | 3+4 2 | 3+4 2 | 2+2 4 |
| | | | 2+2 4 | 2+2 4 | 2+2 4 | 5+5 2 | 5+5 2 | 3+4 2 | 3+4 2 | 2+2 4 |
| | | | 2+2 4 | 2+2 4 | 2+2 4 | 3+3 2 | 3+3 2 | 5+6 2 | 5+6 2 | 2+2 4 |
| | | | 2+2 4 | 2+2 4 | 2+2 4 | 4+4 2 | 5+5 2 | 5+6 2 | 5+6 2 | 2+2 4 |
| | | | 2+2 4 | 2+2 4 | 2+2 4 | 5+5 2 | 4+4 2 | 3+4 2 | 3+4 2 | 2+2 4 |
| | | | 2+2 4 | 2+2 4 | 2+2 4 | 6+6 2 | 6+6 2 | 5+6 2 | 5+6 2 | 2+2 4 |
| | | | 2+2 4 | 2+2 4 | 2+2 4 | 3+3 2 | 6+6 2 | 3+4 2 | 3+4 2 | 2+2 4 |
| | | | 2+2 4 | 2+2 4 | 2+2 4 | 5+5 2 | 5+5 2 | 3+4 2 | 5+6 2 | 2+2 4 |
| | | | 2+2 4 | 2+2 4 | 2+2 4 | 4+4 2 | 5+5 2 | 3+4 2 | 5+6 2 | 2+2 4 |
| | | | 2+2 4 | 2+2 4 | 2+2 4 | 6+6 2 | 6+6 2 | 3+4 2 | 3+4 2 | 2+2 4 |
| | | | 2+2 4 | 2+2 4 | 2+2 4 | 6+6 2 | 3+3 2 | 5+6 2 | 5+6 2 | 2+2 4 |
| | | | 2+2 4 | 2+2 4 | 2+2 4 | 5+5 2 | 4+4 2 | 5+6 2 | 3+4 2 | 2+2 4 |
| | | | 2+2 4 | 2+2 4 | 2+2 4 | 4+4 2 | 4+4 2 | 5+6 2 | 3+4 2 | 2+2 4 |
| | | | 2+2 4 | 2+2 4 | 2+2 4 | 3+3 2 | 3+3 2 | 5+6 2 | 5+6 2 | 2+2 4 |

Continued on next column

**TABLE XXVI– continued**

| TQC | ICSD | NoE | SF | | | | | |
|---|---|---|---|---|---|---|---|---|
| | 2+2 4 | 2+2 4 | 2+2 4 | 5+5 2 | 4+4 2 | 3+4 2 | 5+6 2 | 2+2 4 |
| | 2+2 4 | 2+2 4 | 2+2 4 | 6+6 2 | 3+3 2 | 3+4 2 | 5+6 2 | 2+2 4 |
| | 2+2 4 | 2+2 4 | 2+2 4 | 6+6 2 | 3+3 2 | 5+6 2 | 5+6 2 | 2+2 4 |
| | 2+2 4 | 2+2 4 | 2+2 4 | 5+5 2 | 4+4 2 | 5+6 2 | 3+4 2 | 2+2 4 |
| | 2+2 4 | 2+2 4 | 2+2 4 | 3+3 2 | 5+5 2 | 3+4 2 | 5+6 2 | 2+2 4 |
| | 2+2 4 | 2+2 4 | 2+2 4 | 4+4 2 | 6+6 2 | 3+4 2 | 3+4 2 | 2+2 4 |
| | 2+2 4 | 2+2 4 | 2+2 4 | 5+5 2 | 6+6 2 | 3+4 2 | 5+6 2 | 2+2 4 |
| | | | | 6+6 2 | 5+5 2 | 3+4 2 | 3+4 2 | |
| | | | | 3+3 2 | 6+6 2 | 5+6 2 | 5+6 2 | |
| | | | | 4+4 2 | 5+5 2 | 3+4 2 | 5+6 2 | |
| | | | | 5+5 2 | 4+4 2 | 5+6 2 | 3+4 2 | |
| | | | | 6+6 2 | 3+3 2 | 3+4 2 | 5+6 2 | |
| | | | | 4+4 2 | 5+5 2 | 3+4 2 | 3+4 2 | |
| | | | | 3+3 2 | 6+6 2 | 5+6 2 | 5+6 2 | |
| | | | | 4+4 2 | 3+3 2 | 3+4 2 | 3+4 2 | |
| | | | | 6+6 2 | 4+4 2 | 5+6 2 | 3+4 2 | |
| | | | | 3+3 2 | 4+4 2 | 3+4 2 | 3+4 2 | |
| | | | | 5+5 2 | 6+6 2 | 5+6 2 | 5+6 2 | |
| | | | | 6+6 2 | 5+5 2 | 3+4 2 | 5+6 2 | |
| | | | | 5+5 2 | 3+3 2 | 3+4 2 | 5+6 2 | |
| | | | | 6+6 2 | 5+5 2 | 5+6 2 | 5+6 2 | |
| | | | | 5+5 2 | 3+3 2 | 3+4 2 | 5+6 2 | |
| | | | | 4+4 2 | 6+6 2 | 5+6 2 | 3+4 2 | |
| | | | | 3+3 2 | 4+4 2 | 3+4 2 | 3+4 2 | |
| | | | | 5+5 2 | 3+3 2 | 3+4 2 | 5+6 2 | |
| | | | | 6+6 2 | 4+4 2 | 5+6 2 | 3+4 2 | |
| | | | | 4+4 2 | 6+6 2 | 3+4 2 | 3+4 2 | |
| | | | | 3+3 2 | 5+5 2 | 5+6 2 | 5+6 2 | |
| | | | | 3+3 2 | 3+3 2 | 5+6 2 | 5+6 2 | |
| | | | | 4+4 2 | 4+4 2 | 3+4 2 | 3+4 2 | |
| | | | | 6+6 2 | 4+4 2 | 3+4 2 | 3+4 2 | |
| | | | | 5+5 2 | 3+3 2 | 5+6 2 | 5+6 2 | |
| | | | | 5+5 2 | 6+6 2 | 5+6 2 | 3+4 2 | |
| | | | | 6+6 2 | 5+5 2 | 5+6 2 | 5+6 2 | |
| | | | | 3+3 2 | 4+4 2 | 5+6 2 | 5+6 2 | |
| | | | | 4+4 2 | 6+6 2 | 3+4 2 | 3+4 2 | |
| | | | | 5+5 2 | 5+5 2 | 5+6 2 | 5+6 2 | |
| | | | | 6+6 2 | 3+3 2 | 3+4 2 | 5+6 2 | |
| 14.2.11 | 291320 | 172 | Mo1O6Sr2Yb1 | | | | | |
| | -A | -B | -C | -D | -E | -GM | -Y | -Z |
| | 2+2 4 | 2+2 4 | 2+2 4 | 3+3 2 | 5+5 2 | 3+4 2 | 5+6 2 | 2+2 4 |
| | 2+2 4 | 2+2 4 | 2+2 4 | 4+4 2 | 6+6 2 | 3+4 2 | 5+6 2 | 2+2 4 |
| | 2+2 4 | 2+2 4 | 2+2 4 | 3+3 2 | 3+3 2 | 3+4 2 | 3+4 2 | 2+2 4 |
| | 2+2 4 | 2+2 4 | 2+2 4 | 4+4 2 | 4+4 2 | 5+6 2 | 5+6 2 | 2+2 4 |
| | 2+2 4 | 2+2 4 | 2+2 4 | 5+5 2 | 5+5 2 | 3+4 2 | 3+4 2 | 2+2 4 |
| | 2+2 4 | 2+2 4 | 2+2 4 | 6+6 2 | 6+6 2 | 5+6 2 | 5+6 2 | 2+2 4 |
| | 2+2 4 | 2+2 4 | 2+2 4 | 5+5 2 | 3+3 2 | 5+6 2 | 3+4 2 | 2+2 4 |
| | 2+2 4 | 2+2 4 | 2+2 4 | 6+6 2 | 4+4 2 | 5+6 2 | 3+4 2 | 2+2 4 |
| | 2+2 4 | 2+2 4 | 2+2 4 | 6+6 2 | 4+4 2 | 5+6 2 | 3+4 2 | 2+2 4 |
| | 2+2 4 | 2+2 4 | 2+2 4 | 5+5 2 | 3+3 2 | 5+6 2 | 3+4 2 | 2+2 4 |
| | 2+2 4 | 2+2 4 | 2+2 4 | 5+5 2 | 3+3 2 | 5+6 2 | 3+4 2 | 2+2 4 |
| | 2+2 4 | 2+2 4 | 2+2 4 | 6+6 2 | 4+4 2 | 5+6 2 | 3+4 2 | 2+2 4 |
| | 2+2 4 | 2+2 4 | 2+2 4 | 3+3 2 | 5+5 2 | 3+4 2 | 5+6 2 | 2+2 4 |
| | 2+2 4 | 2+2 4 | 2+2 4 | 4+4 2 | 6+6 2 | 3+4 2 | 3+4 2 | 2+2 4 |
| | 2+2 4 | 2+2 4 | 2+2 4 | 6+6 2 | 6+6 2 | 3+4 2 | 5+6 2 | 2+2 4 |
| | 2+2 4 | 2+2 4 | 2+2 4 | 5+5 2 | 5+5 2 | 3+4 2 | 3+4 2 | 2+2 4 |
| | 2+2 4 | 2+2 4 | 2+2 4 | 4+4 2 | 5+5 2 | 3+4 2 | 5+6 2 | 2+2 4 |
| | 2+2 4 | 2+2 4 | 2+2 4 | 3+3 2 | 6+6 2 | 3+4 2 | 5+6 2 | 2+2 4 |
| | 2+2 4 | 2+2 4 | 2+2 4 | 5+5 2 | 3+3 2 | 5+6 2 | 3+4 2 | 2+2 4 |
| | 2+2 4 | 2+2 4 | 2+2 4 | 3+3 2 | 4+4 2 | 5+6 2 | 5+6 2 | 2+2 4 |
| | 2+2 4 | 2+2 4 | 2+2 4 | 4+4 2 | 3+3 2 | 5+6 2 | 5+6 2 | 2+2 4 |

Continued on next column

**TABLE XXVI– continued**

| TQC | ICSD | NoE | SF | | | | | |
|---|---|---|---|---|---|---|---|---|
| | 2+2 4 | 2+2 4 | 2+2 4 | 6+6 2 | 4+4 2 | 5+6 2 | 3+4 2 | 2+2 4 |
| | 2+2 4 | 2+2 4 | 2+2 4 | 6+6 2 | 4+4 2 | 5+6 2 | 3+4 2 | 2+2 4 |
| | 2+2 4 | 2+2 4 | 2+2 4 | 5+5 2 | 3+3 2 | 3+4 2 | 3+4 2 | 2+2 4 |
| | 2+2 4 | 2+2 4 | 2+2 4 | 3+3 2 | 6+6 2 | 5+6 2 | 3+4 2 | 2+2 4 |
| | 2+2 4 | 2+2 4 | 2+2 4 | 6+6 2 | 5+5 2 | 3+4 2 | 5+6 2 | 2+2 4 |
| | 2+2 4 | 2+2 4 | 2+2 4 | 5+5 2 | 4+4 2 | 5+6 2 | 5+6 2 | 2+2 4 |
| | 2+2 4 | 2+2 4 | 2+2 4 | 4+4 2 | 3+3 2 | 5+6 2 | 3+4 2 | 2+2 4 |
| | 2+2 4 | 2+2 4 | 2+2 4 | 5+5 2 | 5+5 2 | 3+4 2 | 3+4 2 | 2+2 4 |
| | 2+2 4 | 2+2 4 | 2+2 4 | 4+4 2 | 6+6 2 | 3+4 2 | 5+6 2 | 2+2 4 |
| | 2+2 4 | 2+2 4 | 2+2 4 | 6+6 2 | 3+3 2 | 5+6 2 | 3+4 2 | 2+2 4 |
| | 2+2 4 | 2+2 4 | 2+2 4 | 3+3 2 | 4+4 2 | 3+4 2 | 5+6 2 | 2+2 4 |
| | 2+2 4 | 2+2 4 | 2+2 4 | 6+6 2 | 5+5 2 | 5+6 2 | 3+4 2 | 2+2 4 |
| | 2+2 4 | 2+2 4 | 2+2 4 | 3+3 2 | 6+6 2 | 5+6 2 | 5+6 2 | 2+2 4 |
| | 2+2 4 | 2+2 4 | 2+2 4 | 5+5 2 | 4+4 2 | 3+4 2 | 3+4 2 | 2+2 4 |
| | 2+2 4 | 2+2 4 | 2+2 4 | 4+4 2 | 3+3 2 | 5+6 2 | 5+6 2 | 2+2 4 |
| | 2+2 4 | 2+2 4 | 2+2 4 | 6+6 2 | 5+5 2 | 3+4 2 | 3+4 2 | 2+2 4 |
| | 2+2 4 | 2+2 4 | 2+2 4 | 5+5 2 | 6+6 2 | 3+4 2 | 3+4 2 | 2+2 4 |
| | 2+2 4 | 2+2 4 | 2+2 4 | 5+5 2 | 6+6 2 | 3+4 2 | 3+4 2 | 2+2 4 |
| | 2+2 4 | 2+2 4 | 2+2 4 | 6+6 2 | 5+5 2 | 3+4 2 | 3+4 2 | 2+2 4 |
| | 2+2 4 | 2+2 4 | 2+2 4 | 6+6 2 | 5+5 2 | 3+4 2 | 3+4 2 | 2+2 4 |
| | 2+2 4 | 2+2 4 | 2+2 4 | 5+5 2 | 6+6 2 | 3+4 2 | 3+4 2 | 2+2 4 |
| | 2+2 4 | 2+2 4 | 2+2 4 | 6+6 2 | 5+5 2 | 3+4 2 | 3+4 2 | 2+2 4 |
| | | | | 5+5 2 | 6+6 2 | 3+4 2 | 3+4 2 | |
| | | | | 6+6 2 | 5+5 2 | 3+4 2 | 3+4 2 | |
| | | | | 5+5 2 | 6+6 2 | 3+4 2 | 3+4 2 | |
| | | | | 4+4 2 | 5+5 2 | 3+4 2 | 5+6 2 | |
| | | | | 3+3 2 | 6+6 2 | 3+4 2 | 5+6 2 | |
| | | | | 5+5 2 | 3+3 2 | 5+6 2 | 3+4 2 | |
| | | | | 6+6 2 | 4+4 2 | 5+6 2 | 3+4 2 | |
| | | | | 3+3 2 | 4+4 2 | 5+6 2 | 5+6 2 | |
| | | | | 4+4 2 | 3+3 2 | 5+6 2 | 5+6 2 | |
| | | | | 3+3 2 | 4+4 2 | 5+6 2 | 5+6 2 | |
| | | | | 4+4 2 | 3+3 2 | 5+6 2 | 5+6 2 | |
| | | | | 3+3 2 | 3+3 2 | 5+6 2 | 5+6 2 | |
| | | | | 4+4 2 | 4+4 2 | 3+4 2 | 5+6 2 | |
| | | | | 4+4 2 | 5+5 2 | 3+4 2 | 5+6 2 | |
| | | | | 3+3 2 | 6+6 2 | 5+6 2 | 5+6 2 | |
| | | | | 4+4 2 | 4+4 2 | 5+6 2 | 5+6 2 | |
| | | | | 3+3 2 | 3+3 2 | 5+6 2 | 5+6 2 | |
| | | | | 3+3 2 | 5+5 2 | 5+6 2 | 5+6 2 | |
| | | | | 4+4 2 | 6+6 2 | 5+6 2 | 3+4 2 | |
| | | | | 4+4 2 | 3+3 2 | 5+6 2 | 5+6 2 | |
| | | | | 3+3 2 | 4+4 2 | 5+6 2 | 5+6 2 | |
| | | | | 6+6 2 | 3+3 2 | 5+6 2 | 5+6 2 | |
| | | | | 5+5 2 | 4+4 2 | 5+6 2 | 3+4 2 | |
| | | | | 4+4 2 | 4+4 2 | 5+6 2 | 5+6 2 | |
| | | | | 5+5 2 | 3+3 2 | 5+6 2 | 3+4 2 | |
| | | | | 3+3 2 | 3+3 2 | 3+4 2 | 5+6 2 | |
| | | | | 6+6 2 | 4+4 2 | 3+4 2 | 3+4 2 | |
| | | | | 5+5 2 | 5+5 2 | 3+4 2 | 3+4 2 | |
| | | | | 6+6 2 | 6+6 2 | 3+4 2 | 3+4 2 | |
| | | | | 6+6 2 | 4+4 2 | 5+6 2 | 3+4 2 | |
| | | | | 5+5 2 | 3+3 2 | 5+6 2 | 3+4 2 | |
| | | | | 5+5 2 | 3+3 2 | 5+6 2 | 3+4 2 | |
| | | | | 6+6 2 | 4+4 2 | 5+6 2 | 3+4 2 | |
| | | | | 5+5 2 | 3+3 2 | 5+6 2 | 3+4 2 | |
| | | | | 6+6 2 | 4+4 2 | 5+6 2 | 3+4 2 | |
| | | | | 5+5 2 | 3+3 2 | 5+6 2 | 3+4 2 | |
| | | | | 6+6 2 | 4+4 2 | 5+6 2 | 3+4 2 | |
| | | | | 6+6 2 | 4+4 2 | 3+4 2 | 3+4 2 | |
| | | | | 5+5 2 | 3+3 2 | 5+6 2 | 3+4 2 | |

Continued on next column

**TABLE XXVI– continued**

| TQC | ICSD | | NoE | SF | | | | |
|---|---|---|---|---|---|---|---|---|
| | | | | 6+6 2 | 3+3 2 | 5+6 2 | 3+4 2 | |
| | | | | 5+5 2 | 6+6 2 | 5+6 2 | 3+4 2 | |
| | | | | 5+5 2 | 4+4 2 | 5+6 2 | 3+4 2 | |
| | | | | 6+6 2 | 5+5 2 | 5+6 2 | 3+4 2 | |
| 14.2.12 | 80830 | | 72 | Mo1O2 | | | | |
| | -A | -B | -C | -D | -E | -GM | -Y | -Z |
| | 2+2 4 | 2+2 4 | 2+2 4 | 5+5 2 | 3+3 2 | 3+4 2 | 3+4 2 | 2+2 4 |
| | 2+2 4 | 2+2 4 | 2+2 4 | 6+6 2 | 4+4 2 | 5+6 2 | 5+6 2 | 2+2 4 |
| | 2+2 4 | 2+2 4 | 2+2 4 | 3+3 2 | 5+5 2 | 5+6 2 | 5+6 2 | 2+2 4 |
| | 2+2 4 | 2+2 4 | 2+2 4 | 4+4 2 | 6+6 2 | 3+4 2 | 3+4 2 | 2+2 4 |
| | 2+2 4 | 2+2 4 | 2+2 4 | 6+6 2 | 5+5 2 | 5+6 2 | 5+6 2 | 2+2 4 |
| | 2+2 4 | 2+2 4 | 2+2 4 | 5+5 2 | 6+6 2 | 5+6 2 | 5+6 2 | 2+2 4 |
| | 2+2 4 | 2+2 4 | 2+2 4 | 3+3 2 | 3+3 2 | 3+4 2 | 3+4 2 | 2+2 4 |
| | 2+2 4 | 2+2 4 | 2+2 4 | 4+4 2 | 4+4 2 | 3+4 2 | 3+4 2 | 2+2 4 |
| | 2+2 4 | 2+2 4 | 2+2 4 | 6+6 2 | 3+3 2 | 3+4 2 | 5+6 2 | 2+2 4 |
| | 2+2 4 | 2+2 4 | 2+2 4 | 5+5 2 | 4+4 2 | 3+4 2 | 5+6 2 | 2+2 4 |
| | 2+2 4 | 2+2 4 | 2+2 4 | 4+4 2 | 5+5 2 | 5+6 2 | 3+4 2 | 2+2 4 |
| | 2+2 4 | 2+2 4 | 2+2 4 | 3+3 2 | 6+6 2 | 5+6 2 | 5+6 2 | 2+2 4 |
| | 2+2 4 | 2+2 4 | 2+2 4 | 5+5 2 | 6+6 2 | 3+4 2 | 3+4 2 | 2+2 4 |
| | 2+2 4 | 2+2 4 | 2+2 4 | 6+6 2 | 5+5 2 | 5+6 2 | 3+4 2 | 2+2 4 |
| | 2+2 4 | 2+2 4 | 2+2 4 | 4+4 2 | 4+4 2 | 5+6 2 | 5+6 2 | 2+2 4 |
| | 2+2 4 | 2+2 4 | 2+2 4 | 3+3 2 | 3+3 2 | 3+4 2 | 3+4 2 | 2+2 4 |
| | 2+2 4 | 2+2 4 | 2+2 4 | 5+5 2 | 4+4 2 | 3+4 2 | 5+6 2 | 2+2 4 |
| | 2+2 4 | 2+2 4 | 2+2 4 | 6+6 2 | 3+3 2 | 3+4 2 | 5+6 2 | 2+2 4 |
| | | | | 3+3 2 | 4+4 2 | 3+4 2 | 3+4 2 | |
| | | | | 4+4 2 | 3+3 2 | 5+6 2 | 3+4 2 | |
| | | | | 4+4 2 | 5+5 2 | 5+6 2 | 5+6 2 | |
| | | | | 3+3 2 | 6+6 2 | 5+6 2 | 3+4 2 | |
| | | | | 5+5 2 | 6+6 2 | 5+6 2 | 5+6 2 | |
| | | | | 6+6 2 | 5+5 2 | 3+4 2 | 5+6 2 | |
| | | | | 5+5 2 | 4+4 2 | 3+4 2 | 5+6 2 | |
| | | | | 6+6 2 | 3+3 2 | 5+6 2 | 3+4 2 | |
| | | | | 4+4 2 | 6+6 2 | 3+4 2 | 3+4 2 | |
| | | | | 3+3 2 | 5+5 2 | 3+4 2 | 5+6 2 | |
| | | | | 5+5 2 | 3+3 2 | 5+6 2 | 3+4 2 | |
| | | | | 6+6 2 | 4+4 2 | 5+6 2 | 3+4 2 | |
| | | | | 3+3 2 | 5+5 2 | 5+6 2 | 3+4 2 | |
| | | | | 4+4 2 | 6+6 2 | 3+4 2 | 5+6 2 | |
| | | | | 3+3 2 | 3+3 2 | 3+4 2 | 3+4 2 | |
| | | | | 4+4 2 | 4+4 2 | 3+4 2 | 3+4 2 | |
| | | | | 5+5 2 | 5+5 2 | 5+6 2 | 5+6 2 | |
| | | | | 6+6 2 | 6+6 2 | 5+6 2 | 3+4 2 | |
| 14.2.13 | 23722 | | 72 | Mo1O2 | | | | |
| | -A | -B | -C | -D | -E | -GM | -Y | -Z |
| | 2+2 4 | 2+2 4 | 2+2 4 | 5+5 2 | 3+3 2 | 3+4 2 | 3+4 2 | 2+2 4 |
| | 2+2 4 | 2+2 4 | 2+2 4 | 6+6 2 | 4+4 2 | 5+6 2 | 5+6 2 | 2+2 4 |
| | 2+2 4 | 2+2 4 | 2+2 4 | 3+3 2 | 5+5 2 | 5+6 2 | 5+6 2 | 2+2 4 |
| | 2+2 4 | 2+2 4 | 2+2 4 | 4+4 2 | 6+6 2 | 3+4 2 | 3+4 2 | 2+2 4 |
| | 2+2 4 | 2+2 4 | 2+2 4 | 6+6 2 | 5+5 2 | 5+6 2 | 5+6 2 | 2+2 4 |
| | 2+2 4 | 2+2 4 | 2+2 4 | 5+5 2 | 6+6 2 | 5+6 2 | 5+6 2 | 2+2 4 |
| | 2+2 4 | 2+2 4 | 2+2 4 | 3+3 2 | 3+3 2 | 3+4 2 | 3+4 2 | 2+2 4 |
| | 2+2 4 | 2+2 4 | 2+2 4 | 4+4 2 | 4+4 2 | 3+4 2 | 3+4 2 | 2+2 4 |
| | 2+2 4 | 2+2 4 | 2+2 4 | 6+6 2 | 3+3 2 | 3+4 2 | 5+6 2 | 2+2 4 |
| | 2+2 4 | 2+2 4 | 2+2 4 | 5+5 2 | 4+4 2 | 3+4 2 | 5+6 2 | 2+2 4 |
| | 2+2 4 | 2+2 4 | 2+2 4 | 4+4 2 | 5+5 2 | 5+6 2 | 3+4 2 | 2+2 4 |
| | 2+2 4 | 2+2 4 | 2+2 4 | 3+3 2 | 6+6 2 | 5+6 2 | 5+6 2 | 2+2 4 |
| | 2+2 4 | 2+2 4 | 2+2 4 | 5+5 2 | 6+6 2 | 3+4 2 | 3+4 2 | 2+2 4 |
| | 2+2 4 | 2+2 4 | 2+2 4 | 6+6 2 | 5+5 2 | 5+6 2 | 3+4 2 | 2+2 4 |
| | 2+2 4 | 2+2 4 | 2+2 4 | 4+4 2 | 4+4 2 | 5+6 2 | 5+6 2 | 2+2 4 |
| | 2+2 4 | 2+2 4 | 2+2 4 | 3+3 2 | 3+3 2 | 3+4 2 | 3+4 2 | 2+2 4 |
| | 2+2 4 | 2+2 4 | 2+2 4 | 5+5 2 | 4+4 2 | 3+4 2 | 5+6 2 | 2+2 4 |

Continued on next column

**TABLE XXVI– continued**

| TQC | ICSD | | NoE | SF | | | | |
|---|---|---|---|---|---|---|---|---|
| | 2+2 4 | 2+2 4 | 2+2 4 | 6+6 2 | 3+3 2 | 3+4 2 | 5+6 2 | 2+2 4 |
| | | | | 3+3 2 | 4+4 2 | 3+4 2 | 3+4 2 | |
| | | | | 4+4 2 | 3+3 2 | 5+6 2 | 3+4 2 | |
| | | | | 4+4 2 | 5+5 2 | 5+6 2 | 5+6 2 | |
| | | | | 3+3 2 | 6+6 2 | 5+6 2 | 3+4 2 | |
| | | | | 5+5 2 | 6+6 2 | 3+4 2 | 5+6 2 | |
| | | | | 6+6 2 | 5+5 2 | 5+6 2 | 5+6 2 | |
| | | | | 5+5 2 | 4+4 2 | 3+4 2 | 5+6 2 | |
| | | | | 6+6 2 | 3+3 2 | 5+6 2 | 3+4 2 | |
| | | | | 4+4 2 | 6+6 2 | 3+4 2 | 3+4 2 | |
| | | | | 3+3 2 | 5+5 2 | 3+4 2 | 5+6 2 | |
| | | | | 5+5 2 | 3+3 2 | 5+6 2 | 3+4 2 | |
| | | | | 6+6 2 | 4+4 2 | 5+6 2 | 3+4 2 | |
| | | | | 3+3 2 | 5+5 2 | 5+6 2 | 3+4 2 | |
| | | | | 4+4 2 | 6+6 2 | 3+4 2 | 5+6 2 | |
| | | | | 3+3 2 | 3+3 2 | 3+4 2 | 3+4 2 | |
| | | | | 4+4 2 | 4+4 2 | 3+4 2 | 3+4 2 | |
| | | | | 5+5 2 | 5+5 2 | 5+6 2 | 5+6 2 | |
| | | | | 6+6 2 | 6+6 2 | 5+6 2 | 3+4 2 | |
| 14.2.14 | 174231 | | 76 | Ir1P2 | | | | |
| | -A | -B | -C | -D | -E | -GM | -Y | -Z |
| | 2+2 4 | 2+2 4 | 2+2 4 | 3+3 2 | 3+3 2 | 3+4 2 | 3+4 2 | 2+2 4 |
| | 2+2 4 | 2+2 4 | 2+2 4 | 4+4 2 | 4+4 2 | 3+4 2 | 5+6 2 | 2+2 4 |
| | 2+2 4 | 2+2 4 | 2+2 4 | 5+5 2 | 6+6 2 | 5+6 2 | 3+4 2 | 2+2 4 |
| | 2+2 4 | 2+2 4 | 2+2 4 | 6+6 2 | 5+5 2 | 5+6 2 | 5+6 2 | 2+2 4 |
| | 2+2 4 | 2+2 4 | 2+2 4 | 6+6 2 | 5+5 2 | 5+6 2 | 5+6 2 | 2+2 4 |
| | 2+2 4 | 2+2 4 | 2+2 4 | 5+5 2 | 6+6 2 | 3+4 2 | 5+6 2 | 2+2 4 |
| | 2+2 4 | 2+2 4 | 2+2 4 | 3+3 2 | 4+4 2 | 5+6 2 | 3+4 2 | 2+2 4 |
| | 2+2 4 | 2+2 4 | 2+2 4 | 4+4 2 | 3+3 2 | 3+4 2 | 3+4 2 | 2+2 4 |
| | 2+2 4 | 2+2 4 | 2+2 4 | 4+4 2 | 3+3 2 | 3+4 2 | 3+4 2 | 2+2 4 |
| | 2+2 4 | 2+2 4 | 2+2 4 | 3+3 2 | 4+4 2 | 5+6 2 | 5+6 2 | 2+2 4 |
| | 2+2 4 | 2+2 4 | 2+2 4 | 6+6 2 | 5+5 2 | 5+6 2 | 5+6 2 | 2+2 4 |
| | 2+2 4 | 2+2 4 | 2+2 4 | 5+5 2 | 6+6 2 | 3+4 2 | 3+4 2 | 2+2 4 |
| | 2+2 4 | 2+2 4 | 2+2 4 | 3+3 2 | 4+4 2 | 3+4 2 | 3+4 2 | 2+2 4 |
| | 2+2 4 | 2+2 4 | 2+2 4 | 5+5 2 | 3+3 2 | 3+4 2 | 5+6 2 | 2+2 4 |
| | 2+2 4 | 2+2 4 | 2+2 4 | 6+6 2 | 6+6 2 | 5+6 2 | 3+4 2 | 2+2 4 |
| | 2+2 4 | 2+2 4 | 2+2 4 | 4+4 2 | 5+5 2 | 5+6 2 | 5+6 2 | 2+2 4 |
| | 2+2 4 | 2+2 4 | 2+2 4 | 4+4 2 | 6+6 2 | 3+4 2 | 3+4 2 | 2+2 4 |
| | 2+2 4 | 2+2 4 | 2+2 4 | 3+3 2 | 5+5 2 | 3+4 2 | 5+6 2 | 2+2 4 |
| | 2+2 4 | 2+2 4 | 2+2 4 | 6+6 2 | 4+4 2 | 3+4 2 | 3+4 2 | 2+2 4 |
| | | | | 5+5 2 | 3+3 2 | 5+6 2 | 5+6 2 | |
| | | | | 3+3 2 | 5+5 2 | 5+6 2 | 5+6 2 | |
| | | | | 4+4 2 | 6+6 2 | 3+4 2 | 3+4 2 | |
| | | | | 6+6 2 | 3+3 2 | 3+4 2 | 3+4 2 | |
| | | | | 5+5 2 | 4+4 2 | 5+6 2 | 5+6 2 | |
| | | | | 5+5 2 | 6+6 2 | 5+6 2 | 5+6 2 | |
| | | | | 4+4 2 | 4+4 2 | 3+4 2 | 3+4 2 | |
| | | | | 3+3 2 | 3+3 2 | 5+6 2 | 5+6 2 | |
| | | | | 6+6 2 | 5+5 2 | 3+4 2 | 3+4 2 | |
| | | | | 4+4 2 | 5+5 2 | 3+4 2 | 5+6 2 | |
| | | | | 3+3 2 | 6+6 2 | 5+6 2 | 3+4 2 | |
| | | | | 6+6 2 | 5+5 2 | 3+4 2 | 5+6 2 | |
| | | | | 5+5 2 | 4+4 2 | 5+6 2 | 3+4 2 | |
| | | | | 6+6 2 | 6+6 2 | 3+4 2 | 5+6 2 | |
| | | | | 3+3 2 | 3+3 2 | 3+4 2 | 3+4 2 | |
| | | | | 4+4 2 | 6+6 2 | 5+6 2 | 3+4 2 | |
| | | | | 5+5 2 | 5+5 2 | 3+4 2 | 5+6 2 | |
| | | | | 5+5 2 | 3+3 2 | 5+6 2 | 5+6 2 | |
| | | | | 6+6 2 | 4+4 2 | 5+6 2 | 3+4 2 | |
| 14.2.15 | 171490 | | 168 | Ir1O6Sr2Tb1 | | | | |
| | -A | -B | -C | -D | -E | -GM | -Y | -Z |

Continued on next column

**TABLE XXVI– continued**

| TQC | ICSD | NoE | SF | | | | | |
|---|---|---|---|---|---|---|---|---|
| | 2+2 4 | 2+2 4 | 2+2 4 | 5+5 2 | 5+5 2 | 3+4 2 | 3+4 2 | 2+2 4 |
| | 2+2 4 | 2+2 4 | 2+2 4 | 6+6 2 | 6+6 2 | 3+4 2 | 3+4 2 | 2+2 4 |
| | 2+2 4 | 2+2 4 | 2+2 4 | 3+3 2 | 3+3 2 | 3+4 2 | 5+6 2 | 2+2 4 |
| | 2+2 4 | 2+2 4 | 2+2 4 | 4+4 2 | 4+4 2 | 5+6 2 | 5+6 2 | 2+2 4 |
| | 2+2 4 | 2+2 4 | 2+2 4 | 5+5 2 | 5+5 2 | 3+4 2 | 3+4 2 | 2+2 4 |
| | 2+2 4 | 2+2 4 | 2+2 4 | 6+6 2 | 6+6 2 | 5+6 2 | 3+4 2 | 2+2 4 |
| | 2+2 4 | 2+2 4 | 2+2 4 | 3+3 2 | 4+4 2 | 5+6 2 | 5+6 2 | 2+2 4 |
| | 2+2 4 | 2+2 4 | 2+2 4 | 4+4 2 | 3+3 2 | 5+6 2 | 5+6 2 | 2+2 4 |
| | 2+2 4 | 2+2 4 | 2+2 4 | 4+4 2 | 3+3 2 | 5+6 2 | 5+6 2 | 2+2 4 |
| | 2+2 4 | 2+2 4 | 2+2 4 | 3+3 2 | 4+4 2 | 5+6 2 | 5+6 2 | 2+2 4 |
| | 2+2 4 | 2+2 4 | 2+2 4 | 3+3 2 | 4+4 2 | 5+6 2 | 5+6 2 | 2+2 4 |
| | 2+2 4 | 2+2 4 | 2+2 4 | 4+4 2 | 3+3 2 | 5+6 2 | 5+6 2 | 2+2 4 |
| | 2+2 4 | 2+2 4 | 2+2 4 | 6+6 2 | 5+5 2 | 3+4 2 | 3+4 2 | 2+2 4 |
| | 2+2 4 | 2+2 4 | 2+2 4 | 5+5 2 | 6+6 2 | 3+4 2 | 3+4 2 | 2+2 4 |
| | 2+2 4 | 2+2 4 | 2+2 4 | 6+6 2 | 5+5 2 | 3+4 2 | 3+4 2 | 2+2 4 |
| | 2+2 4 | 2+2 4 | 2+2 4 | 5+5 2 | 6+6 2 | 3+4 2 | 5+6 2 | 2+2 4 |
| | 2+2 4 | 2+2 4 | 2+2 4 | 4+4 2 | 5+5 2 | 3+4 2 | 3+4 2 | 2+2 4 |
| | 2+2 4 | 2+2 4 | 2+2 4 | 3+3 2 | 6+6 2 | 3+4 2 | 5+6 2 | 2+2 4 |
| | 2+2 4 | 2+2 4 | 2+2 4 | 5+5 2 | 4+4 2 | 5+6 2 | 3+4 2 | 2+2 4 |
| | 2+2 4 | 2+2 4 | 2+2 4 | 6+6 2 | 3+3 2 | 5+6 2 | 3+4 2 | 2+2 4 |
| | 2+2 4 | 2+2 4 | 2+2 4 | 3+3 2 | 4+4 2 | 5+6 2 | 5+6 2 | 2+2 4 |
| | 2+2 4 | 2+2 4 | 2+2 4 | 4+4 2 | 3+3 2 | 3+4 2 | 5+6 2 | 2+2 4 |
| | 2+2 4 | 2+2 4 | 2+2 4 | 4+4 2 | 5+5 2 | 5+6 2 | 5+6 2 | 2+2 4 |
| | 2+2 4 | 2+2 4 | 2+2 4 | 3+3 2 | 6+6 2 | 3+4 2 | 3+4 2 | 2+2 4 |
| | 2+2 4 | 2+2 4 | 2+2 4 | 5+5 2 | 3+3 2 | 5+6 2 | 5+6 2 | 2+2 4 |
| | 2+2 4 | 2+2 4 | 2+2 4 | 6+6 2 | 4+4 2 | 5+6 2 | 3+4 2 | 2+2 4 |
| | 2+2 4 | 2+2 4 | 2+2 4 | 4+4 2 | 3+3 2 | 5+6 2 | 5+6 2 | 2+2 4 |
| | 2+2 4 | 2+2 4 | 2+2 4 | 3+3 2 | 4+4 2 | 5+6 2 | 5+6 2 | 2+2 4 |
| | 2+2 4 | 2+2 4 | 2+2 4 | 6+6 2 | 6+6 2 | 3+4 2 | 5+6 2 | 2+2 4 |
| | 2+2 4 | 2+2 4 | 2+2 4 | 3+3 2 | 5+5 2 | 3+4 2 | 3+4 2 | 2+2 4 |
| | 2+2 4 | 2+2 4 | 2+2 4 | 5+5 2 | 3+3 2 | 3+4 2 | 5+6 2 | 2+2 4 |
| | 2+2 4 | 2+2 4 | 2+2 4 | 4+4 2 | 4+4 2 | 5+6 2 | 3+4 2 | 2+2 4 |
| | 2+2 4 | 2+2 4 | 2+2 4 | 5+5 2 | 5+5 2 | 3+4 2 | 5+6 2 | 2+2 4 |
| | 2+2 4 | 2+2 4 | 2+2 4 | 4+4 2 | 6+6 2 | 5+6 2 | 3+4 2 | 2+2 4 |
| | 2+2 4 | 2+2 4 | 2+2 4 | 6+6 2 | 4+4 2 | 5+6 2 | 3+4 2 | 2+2 4 |
| | 2+2 4 | 2+2 4 | 2+2 4 | 3+3 2 | 3+3 2 | 5+6 2 | 5+6 2 | 2+2 4 |
| | 2+2 4 | 2+2 4 | 2+2 4 | 4+4 2 | 5+5 2 | 3+4 2 | 5+6 2 | 2+2 4 |
| | 2+2 4 | 2+2 4 | 2+2 4 | 3+3 2 | 6+6 2 | 3+4 2 | 5+6 2 | 2+2 4 |
| | 2+2 4 | 2+2 4 | 2+2 4 | 4+4 2 | 5+5 2 | 3+4 2 | 5+6 2 | 2+2 4 |
| | 2+2 4 | 2+2 4 | 2+2 4 | 3+3 2 | 6+6 2 | 3+4 2 | 5+6 2 | 2+2 4 |
| | 2+2 4 | 2+2 4 | 2+2 4 | 3+3 2 | 6+6 2 | 3+4 2 | 5+6 2 | 2+2 4 |
| | 2+2 4 | 2+2 4 | 2+2 4 | 4+4 2 | 5+5 2 | 3+4 2 | 5+6 2 | 2+2 4 |
| | | | | 3+3 2 | 6+6 2 | 3+4 2 | 5+6 2 | |
| | | | | 4+4 2 | 5+5 2 | 3+4 2 | 5+6 2 | |
| | | | | 4+4 2 | 5+5 2 | 3+4 2 | 5+6 2 | |
| | | | | 3+3 2 | 6+6 2 | 3+4 2 | 5+6 2 | |
| | | | | 3+3 2 | 6+6 2 | 3+4 2 | 5+6 2 | |
| | | | | 4+4 2 | 5+5 2 | 3+4 2 | 5+6 2 | |
| | | | | 6+6 2 | 4+4 2 | 5+6 2 | 3+4 2 | |
| | | | | 5+5 2 | 3+3 2 | 5+6 2 | 3+4 2 | |
| | | | | 5+5 2 | 3+3 2 | 5+6 2 | 3+4 2 | |
| | | | | 6+6 2 | 4+4 2 | 5+6 2 | 3+4 2 | |
| | | | | 5+5 2 | 4+4 2 | 5+6 2 | 3+4 2 | |
| | | | | 6+6 2 | 3+3 2 | 5+6 2 | 3+4 2 | |
| | | | | 5+5 2 | 3+3 2 | 5+6 2 | 3+4 2 | |
| | | | | 6+6 2 | 4+4 2 | 3+4 2 | 3+4 2 | |
| | | | | 5+5 2 | 6+6 2 | 3+4 2 | 3+4 2 | |
| | | | | 6+6 2 | 5+5 2 | 5+6 2 | 3+4 2 | |
| | | | | 6+6 2 | 5+5 2 | 5+6 2 | 3+4 2 | |
| | | | | 5+5 2 | 6+6 2 | 5+6 2 | 3+4 2 | |
| | | | | 5+5 2 | 3+3 2 | 5+6 2 | 3+4 2 | |

Continued on next column

**TABLE XXVI– continued**

| TQC | ICSD | NoE | SF | | | | | |
|---|---|---|---|---|---|---|---|---|
| | | | | 6+6 2 | 4+4 2 | 5+6 2 | 3+4 2 | |
| | | | | 5+5 2 | 3+3 2 | 5+6 2 | 5+6 2 | |
| | | | | 6+6 2 | 4+4 2 | 5+6 2 | 5+6 2 | |
| | | | | 4+4 2 | 4+4 2 | 5+6 2 | 3+4 2 | |
| | | | | 3+3 2 | 3+3 2 | 5+6 2 | 3+4 2 | |
| | | | | 6+6 2 | 3+3 2 | 5+6 2 | 5+6 2 | |
| | | | | 5+5 2 | 4+4 2 | 5+6 2 | 5+6 2 | |
| | | | | 3+3 2 | 3+3 2 | 3+4 2 | 5+6 2 | |
| | | | | 4+4 2 | 4+4 2 | 3+4 2 | 5+6 2 | |
| | | | | 4+4 2 | 6+6 2 | 3+4 2 | 3+4 2 | |
| | | | | 3+3 2 | 5+5 2 | 3+4 2 | 3+4 2 | |
| | | | | 4+4 2 | 5+5 2 | 3+4 2 | 5+6 2 | |
| | | | | 3+3 2 | 6+6 2 | 3+4 2 | 5+6 2 | |
| | | | | 3+3 2 | 3+3 2 | 3+4 2 | 5+6 2 | |
| | | | | 4+4 2 | 4+4 2 | 5+6 2 | 5+6 2 | |
| | | | | 3+3 2 | 3+3 2 | 5+6 2 | 5+6 2 | |
| | | | | 4+4 2 | 4+4 2 | 5+6 2 | 5+6 2 | |
| | | | | 3+3 2 | 4+4 2 | 5+6 2 | 5+6 2 | |
| | | | | 4+4 2 | 3+3 2 | 5+6 2 | 5+6 2 | |
| | | | | 4+4 2 | 6+6 2 | 5+6 2 | 5+6 2 | |
| | | | | 3+3 2 | 5+5 2 | 3+4 2 | 5+6 2 | |
| | | | | 4+4 2 | 4+4 2 | 3+4 2 | 5+6 2 | |
| | | | | 3+3 2 | 3+3 2 | 5+6 2 | 5+6 2 | |
| 14.2.16 | 153133 | 168 | Ir1O6Sr2Tb1 | | | | | |
| | -A | -B | -C | -D | -E | -GM | -Y | -Z |
| | 2+2 4 | 2+2 4 | 2+2 4 | 5+5 2 | 3+3 2 | 3+4 2 | 5+6 2 | 2+2 4 |
| | 2+2 4 | 2+2 4 | 2+2 4 | 6+6 2 | 4+4 2 | 3+4 2 | 5+6 2 | 2+2 4 |
| | 2+2 4 | 2+2 4 | 2+2 4 | 3+3 2 | 5+5 2 | 3+4 2 | 3+4 2 | 2+2 4 |
| | 2+2 4 | 2+2 4 | 2+2 4 | 4+4 2 | 6+6 2 | 5+6 2 | 5+6 2 | 2+2 4 |
| | 2+2 4 | 2+2 4 | 2+2 4 | 5+5 2 | 3+3 2 | 3+4 2 | 3+4 2 | 2+2 4 |
| | 2+2 4 | 2+2 4 | 2+2 4 | 6+6 2 | 4+4 2 | 5+6 2 | 5+6 2 | 2+2 4 |
| | 2+2 4 | 2+2 4 | 2+2 4 | 3+3 2 | 5+5 2 | 5+6 2 | 3+4 2 | 2+2 4 |
| | 2+2 4 | 2+2 4 | 2+2 4 | 4+4 2 | 6+6 2 | 5+6 2 | 3+4 2 | 2+2 4 |
| | 2+2 4 | 2+2 4 | 2+2 4 | 3+3 2 | 6+6 2 | 5+6 2 | 3+4 2 | 2+2 4 |
| | 2+2 4 | 2+2 4 | 2+2 4 | 4+4 2 | 5+5 2 | 5+6 2 | 3+4 2 | 2+2 4 |
| | 2+2 4 | 2+2 4 | 2+2 4 | 4+4 2 | 5+5 2 | 5+6 2 | 3+4 2 | 2+2 4 |
| | 2+2 4 | 2+2 4 | 2+2 4 | 3+3 2 | 6+6 2 | 5+6 2 | 3+4 2 | 2+2 4 |
| | 2+2 4 | 2+2 4 | 2+2 4 | 5+5 2 | 3+3 2 | 3+4 2 | 5+6 2 | 2+2 4 |
| | 2+2 4 | 2+2 4 | 2+2 4 | 6+6 2 | 4+4 2 | 3+4 2 | 5+6 2 | 2+2 4 |
| | 2+2 4 | 2+2 4 | 2+2 4 | 6+6 2 | 3+3 2 | 3+4 2 | 5+6 2 | 2+2 4 |
| | 2+2 4 | 2+2 4 | 2+2 4 | 5+5 2 | 4+4 2 | 3+4 2 | 5+6 2 | 2+2 4 |
| | 2+2 4 | 2+2 4 | 2+2 4 | 5+5 2 | 3+3 2 | 3+4 2 | 5+6 2 | 2+2 4 |
| | 2+2 4 | 2+2 4 | 2+2 4 | 6+6 2 | 4+4 2 | 3+4 2 | 5+6 2 | 2+2 4 |
| | 2+2 4 | 2+2 4 | 2+2 4 | 3+3 2 | 5+5 2 | 5+6 2 | 3+4 2 | 2+2 4 |
| | 2+2 4 | 2+2 4 | 2+2 4 | 4+4 2 | 6+6 2 | 5+6 2 | 3+4 2 | 2+2 4 |
| | 2+2 4 | 2+2 4 | 2+2 4 | 3+3 2 | 5+5 2 | 5+6 2 | 3+4 2 | 2+2 4 |
| | 2+2 4 | 2+2 4 | 2+2 4 | 4+4 2 | 6+6 2 | 3+4 2 | 3+4 2 | 2+2 4 |
| | 2+2 4 | 2+2 4 | 2+2 4 | 5+5 2 | 4+4 2 | 5+6 2 | 5+6 2 | 2+2 4 |
| | 2+2 4 | 2+2 4 | 2+2 4 | 6+6 2 | 3+3 2 | 3+4 2 | 5+6 2 | 2+2 4 |
| | 2+2 4 | 2+2 4 | 2+2 4 | 4+4 2 | 6+6 2 | 5+6 2 | 3+4 2 | 2+2 4 |
| | 2+2 4 | 2+2 4 | 2+2 4 | 3+3 2 | 5+5 2 | 5+6 2 | 3+4 2 | 2+2 4 |
| | 2+2 4 | 2+2 4 | 2+2 4 | 4+4 2 | 5+5 2 | 5+6 2 | 3+4 2 | 2+2 4 |
| | 2+2 4 | 2+2 4 | 2+2 4 | 3+3 2 | 6+6 2 | 5+6 2 | 3+4 2 | 2+2 4 |
| | 2+2 4 | 2+2 4 | 2+2 4 | 6+6 2 | 4+4 2 | 3+4 2 | 5+6 2 | 2+2 4 |
| | 2+2 4 | 2+2 4 | 2+2 4 | 3+3 2 | 3+3 2 | 3+4 2 | 3+4 2 | 2+2 4 |
| | 2+2 4 | 2+2 4 | 2+2 4 | 5+5 2 | 6+6 2 | 3+4 2 | 5+6 2 | 2+2 4 |
| | 2+2 4 | 2+2 4 | 2+2 4 | 4+4 2 | 5+5 2 | 5+6 2 | 3+4 2 | 2+2 4 |
| | 2+2 4 | 2+2 4 | 2+2 4 | 5+5 2 | 3+3 2 | 5+6 2 | 5+6 2 | 2+2 4 |
| | 2+2 4 | 2+2 4 | 2+2 4 | 4+4 2 | 4+4 2 | 3+4 2 | 3+4 2 | 2+2 4 |
| | 2+2 4 | 2+2 4 | 2+2 4 | 6+6 2 | 5+5 2 | 5+6 2 | 5+6 2 | 2+2 4 |
| | 2+2 4 | 2+2 4 | 2+2 4 | 3+3 2 | 6+6 2 | 5+6 2 | 3+4 2 | 2+2 4 |

Continued on next column

**TABLE XXVI– continued**

| TQC | ICSD | | NoE | SF | | | | |
|---|---|---|---|---|---|---|---|---|
| | 2+2 4 | 2+2 4 | 2+2 4 | 4+4 2 | 4+4 2 | 3+4 2 | 3+4 2 | 2+2 4 |
| | 2+2 4 | 2+2 4 | 2+2 4 | 3+3 2 | 3+3 2 | 3+4 2 | 3+4 2 | 2+2 4 |
| | 2+2 4 | 2+2 4 | 2+2 4 | 3+3 2 | 4+4 2 | 3+4 2 | 3+4 2 | 2+2 4 |
| | 2+2 4 | 2+2 4 | 2+2 4 | 4+4 2 | 3+3 2 | 3+4 2 | 3+4 2 | 2+2 4 |
| | 2+2 4 | 2+2 4 | 2+2 4 | 3+3 2 | 3+3 2 | 3+4 2 | 3+4 2 | 2+2 4 |
| | 2+2 4 | 2+2 4 | 2+2 4 | 4+4 2 | 4+4 2 | 3+4 2 | 3+4 2 | 2+2 4 |
| | | | | 3+3 2 | 3+3 2 | 3+4 2 | 3+4 2 | |
| | | | | 4+4 2 | 4+4 2 | 3+4 2 | 3+4 2 | |
| | | | | 4+4 2 | 4+4 2 | 3+4 2 | 3+4 2 | |
| | | | | 3+3 2 | 3+3 2 | 3+4 2 | 3+4 2 | |
| | | | | 3+3 2 | 3+3 2 | 3+4 2 | 5+6 2 | |
| | | | | 4+4 2 | 4+4 2 | 3+4 2 | 3+4 2 | |
| | | | | 6+6 2 | 5+5 2 | 5+6 2 | 5+6 2 | |
| | | | | 5+5 2 | 6+6 2 | 5+6 2 | 3+4 2 | |
| | | | | 5+5 2 | 6+6 2 | 5+6 2 | 5+6 2 | |
| | | | | 6+6 2 | 5+5 2 | 5+6 2 | 5+6 2 | |
| | | | | 6+6 2 | 6+6 2 | 5+6 2 | 5+6 2 | |
| | | | | 5+5 2 | 5+5 2 | 5+6 2 | 5+6 2 | |
| | | | | 5+5 2 | 4+4 2 | 3+4 2 | 5+6 2 | |
| | | | | 6+6 2 | 3+3 2 | 3+4 2 | 5+6 2 | |
| | | | | 5+5 2 | 5+5 2 | 5+6 2 | 5+6 2 | |
| | | | | 6+6 2 | 6+6 2 | 5+6 2 | 5+6 2 | |
| | | | | 6+6 2 | 3+3 2 | 5+6 2 | 5+6 2 | |
| | | | | 5+5 2 | 4+4 2 | 5+6 2 | 5+6 2 | |
| | | | | 4+4 2 | 6+6 2 | 5+6 2 | 5+6 2 | |
| | | | | 5+5 2 | 5+5 2 | 5+6 2 | 3+4 2 | |
| | | | | 6+6 2 | 5+5 2 | 5+6 2 | 5+6 2 | |
| | | | | 3+3 2 | 6+6 2 | 5+6 2 | 3+4 2 | |
| | | | | 3+3 2 | 5+5 2 | 5+6 2 | 3+4 2 | |
| | | | | 6+6 2 | 6+6 2 | 5+6 2 | 5+6 2 | |
| | | | | 5+5 2 | 6+6 2 | 5+6 2 | 3+4 2 | |
| | | | | 4+4 2 | 5+5 2 | 5+6 2 | 3+4 2 | |
| | | | | 3+3 2 | 5+5 2 | 3+4 2 | 3+4 2 | |
| | | | | 4+4 2 | 6+6 2 | 3+4 2 | 5+6 2 | |
| | | | | 5+5 2 | 3+3 2 | 3+4 2 | 5+6 2 | |
| | | | | 6+6 2 | 4+4 2 | 3+4 2 | 5+6 2 | |
| | | | | 4+4 2 | 4+4 2 | 3+4 2 | 3+4 2 | |
| | | | | 3+3 2 | 3+3 2 | 3+4 2 | 3+4 2 | |
| | | | | 4+4 2 | 6+6 2 | 3+4 2 | 3+4 2 | |
| | | | | 3+3 2 | 5+5 2 | 5+6 2 | 3+4 2 | |
| | | | | 3+3 2 | 5+5 2 | 5+6 2 | 3+4 2 | |
| | | | | 4+4 2 | 6+6 2 | 5+6 2 | 3+4 2 | |
| | | | | 3+3 2 | 5+5 2 | 3+4 2 | 3+4 2 | |
| | | | | 4+4 2 | 6+6 2 | 5+6 2 | 3+4 2 | |
| | | | | 4+4 2 | 3+3 2 | 5+6 2 | 3+4 2 | |
| | | | | 3+3 2 | 4+4 2 | 5+6 2 | 3+4 2 | |
| | | | | 4+4 2 | 5+5 2 | 3+4 2 | 3+4 2 | |
| | | | | 3+3 2 | 6+6 2 | 5+6 2 | 3+4 2 | |
| 14.2.17 | 47158 | | 100 | H2Li1N1O3 | | | | |
| | -A | -B | -C | -D | -E | -GM | -Y | -Z |
| | 2+2 4 | 2+2 4 | 2+2 4 | 5+5 2 | 3+3 2 | 3+4 2 | 3+4 2 | 2+2 4 |
| | 2+2 4 | 2+2 4 | 2+2 4 | 6+6 2 | 4+4 2 | 5+6 2 | 5+6 2 | 2+2 4 |
| | 2+2 4 | 2+2 4 | 2+2 4 | 3+3 2 | 5+5 2 | 3+4 2 | 3+4 2 | 2+2 4 |
| | 2+2 4 | 2+2 4 | 2+2 4 | 4+4 2 | 6+6 2 | 5+6 2 | 5+6 2 | 2+2 4 |
| | 2+2 4 | 2+2 4 | 2+2 4 | 5+5 2 | 3+3 2 | 3+4 2 | 5+6 2 | 2+2 4 |
| | 2+2 4 | 2+2 4 | 2+2 4 | 6+6 2 | 4+4 2 | 5+6 2 | 3+4 2 | 2+2 4 |
| | 2+2 4 | 2+2 4 | 2+2 4 | 3+3 2 | 5+5 2 | 3+4 2 | 3+4 2 | 2+2 4 |
| | 2+2 4 | 2+2 4 | 2+2 4 | 4+4 2 | 6+6 2 | 5+6 2 | 5+6 2 | 2+2 4 |
| | 2+2 4 | 2+2 4 | 2+2 4 | 3+3 2 | 3+3 2 | 3+4 2 | 3+4 2 | 2+2 4 |
| | 2+2 4 | 2+2 4 | 2+2 4 | 4+4 2 | 4+4 2 | 3+4 2 | 3+4 2 | 2+2 4 |
| | 2+2 4 | 2+2 4 | 2+2 4 | 5+5 2 | 5+5 2 | 5+6 2 | 5+6 2 | 2+2 4 |
| | 2+2 4 | 2+2 4 | 2+2 4 | 6+6 2 | 6+6 2 | 5+6 2 | 5+6 2 | 2+2 4 |
| | 2+2 4 | 2+2 4 | 2+2 4 | 5+5 2 | 3+3 2 | 3+4 2 | 3+4 2 | 2+2 4 |
| | 2+2 4 | 2+2 4 | 2+2 4 | 6+6 2 | 4+4 2 | 3+4 2 | 5+6 2 | 2+2 4 |
| | 2+2 4 | 2+2 4 | 2+2 4 | 3+3 2 | 5+5 2 | 5+6 2 | 3+4 2 | 2+2 4 |
| | 2+2 4 | 2+2 4 | 2+2 4 | 4+4 2 | 6+6 2 | 5+6 2 | 5+6 2 | 2+2 4 |
| | 2+2 4 | 2+2 4 | 2+2 4 | 3+3 2 | 3+3 2 | 3+4 2 | 3+4 2 | 2+2 4 |
| | 2+2 4 | 2+2 4 | 2+2 4 | 4+4 2 | 4+4 2 | 3+4 2 | 3+4 2 | 2+2 4 |
| | 2+2 4 | 2+2 4 | 2+2 4 | 5+5 2 | 5+5 2 | 5+6 2 | 5+6 2 | 2+2 4 |
| | 2+2 4 | 2+2 4 | 2+2 4 | 6+6 2 | 6+6 2 | 3+4 2 | 5+6 2 | 2+2 4 |
| | 2+2 4 | 2+2 4 | 2+2 4 | 3+3 2 | 3+3 2 | 3+4 2 | 3+4 2 | 2+2 4 |
| | 2+2 4 | 2+2 4 | 2+2 4 | 4+4 2 | 4+4 2 | 5+6 2 | 3+4 2 | 2+2 4 |
| | 2+2 4 | 2+2 4 | 2+2 4 | 5+5 2 | 5+5 2 | 5+6 2 | 5+6 2 | 2+2 4 |
| | 2+2 4 | 2+2 4 | 2+2 4 | 6+6 2 | 6+6 2 | 5+6 2 | 5+6 2 | 2+2 4 |
| | 2+2 4 | 2+2 4 | 2+2 4 | 3+3 2 | 3+3 2 | 3+4 2 | 3+4 2 | 2+2 4 |
| | | | | 4+4 2 | 4+4 2 | 3+4 2 | 3+4 2 | |
| | | | | 5+5 2 | 5+5 2 | 5+6 2 | 5+6 2 | |
| | | | | 6+6 2 | 6+6 2 | 5+6 2 | 5+6 2 | |
| | | | | 3+3 2 | 3+3 2 | 3+4 2 | 3+4 2 | |
| | | | | 4+4 2 | 4+4 2 | 3+4 2 | 3+4 2 | |
| | | | | 6+6 2 | 3+3 2 | 3+4 2 | 5+6 2 | |
| | | | | 5+5 2 | 4+4 2 | 5+6 2 | 3+4 2 | |
| | | | | 4+4 2 | 5+5 2 | 3+4 2 | 5+6 2 | |
| | | | | 3+3 2 | 6+6 2 | 5+6 2 | 3+4 2 | |
| | | | | 3+3 2 | 5+5 2 | 5+6 2 | 5+6 2 | |
| | | | | 4+4 2 | 6+6 2 | 3+4 2 | 3+4 2 | |
| | | | | 5+5 2 | 3+3 2 | 5+6 2 | 5+6 2 | |
| | | | | 6+6 2 | 4+4 2 | 3+4 2 | 3+4 2 | |
| | | | | 5+5 2 | 5+5 2 | 3+4 2 | 5+6 2 | |
| | | | | 6+6 2 | 6+6 2 | 5+6 2 | 5+6 2 | |
| | | | | 3+3 2 | 4+4 2 | 5+6 2 | 3+4 2 | |
| | | | | 4+4 2 | 3+3 2 | 3+4 2 | 3+4 2 | |
| | | | | 5+5 2 | 5+5 2 | 5+6 2 | 5+6 2 | |
| | | | | 6+6 2 | 6+6 2 | 5+6 2 | 5+6 2 | |
| | | | | 3+3 2 | 3+3 2 | 3+4 2 | 3+4 2 | |
| | | | | 4+4 2 | 4+4 2 | 3+4 2 | 3+4 2 | |
| | | | | 5+5 2 | 6+6 2 | 5+6 2 | 5+6 2 | |
| | | | | 6+6 2 | 5+5 2 | 3+4 2 | 5+6 2 | |
| | | | | 4+4 2 | 3+3 2 | 5+6 2 | 3+4 2 | |
| | | | | 3+3 2 | 4+4 2 | 5+6 2 | 3+4 2 | |
| 14.2.18 | 84949 | | 80 | H1Mn1O2 | | | | |
| | -A | -B | -C | -D | -E | -GM | -Y | -Z |
| | 2+2 4 | 2+2 4 | 2+2 4 | 5+5 2 | 3+3 2 | 3+4 2 | 5+6 2 | 2+2 4 |
| | 2+2 4 | 2+2 4 | 2+2 4 | 6+6 2 | 4+4 2 | 3+4 2 | 3+4 2 | 2+2 4 |
| | 2+2 4 | 2+2 4 | 2+2 4 | 3+3 2 | 5+5 2 | 5+6 2 | 5+6 2 | 2+2 4 |
| | 2+2 4 | 2+2 4 | 2+2 4 | 4+4 2 | 6+6 2 | 5+6 2 | 3+4 2 | 2+2 4 |
| | 2+2 4 | 2+2 4 | 2+2 4 | 3+3 2 | 3+3 2 | 3+4 2 | 3+4 2 | 2+2 4 |
| | 2+2 4 | 2+2 4 | 2+2 4 | 4+4 2 | 4+4 2 | 3+4 2 | 3+4 2 | 2+2 4 |
| | 2+2 4 | 2+2 4 | 2+2 4 | 5+5 2 | 5+5 2 | 5+6 2 | 5+6 2 | 2+2 4 |
| | 2+2 4 | 2+2 4 | 2+2 4 | 6+6 2 | 6+6 2 | 5+6 2 | 5+6 2 | 2+2 4 |
| | 2+2 4 | 2+2 4 | 2+2 4 | 5+5 2 | 3+3 2 | 3+4 2 | 5+6 2 | 2+2 4 |
| | 2+2 4 | 2+2 4 | 2+2 4 | 6+6 2 | 4+4 2 | 5+6 2 | 3+4 2 | 2+2 4 |
| | 2+2 4 | 2+2 4 | 2+2 4 | 4+4 2 | 5+5 2 | 3+4 2 | 3+4 2 | 2+2 4 |
| | 2+2 4 | 2+2 4 | 2+2 4 | 3+3 2 | 6+6 2 | 5+6 2 | 5+6 2 | 2+2 4 |
| | 2+2 4 | 2+2 4 | 2+2 4 | 3+3 2 | 5+5 2 | 3+4 2 | 3+4 2 | 2+2 4 |
| | 2+2 4 | 2+2 4 | 2+2 4 | 4+4 2 | 6+6 2 | 5+6 2 | 5+6 2 | 2+2 4 |
| | 2+2 4 | 2+2 4 | 2+2 4 | 5+5 2 | 3+3 2 | 5+6 2 | 3+4 2 | 2+2 4 |
| | 2+2 4 | 2+2 4 | 2+2 4 | 6+6 2 | 4+4 2 | 3+4 2 | 5+6 2 | 2+2 4 |
| | 2+2 4 | 2+2 4 | 2+2 4 | 6+6 2 | 3+3 2 | 3+4 2 | 5+6 2 | 2+2 4 |
| | 2+2 4 | 2+2 4 | 2+2 4 | 5+5 2 | 4+4 2 | 5+6 2 | 3+4 2 | 2+2 4 |
| | 2+2 4 | 2+2 4 | 2+2 4 | 3+3 2 | 5+5 2 | 3+4 2 | 3+4 2 | 2+2 4 |
| | 2+2 4 | 2+2 4 | 2+2 4 | 4+4 2 | 6+6 2 | 3+4 2 | 5+6 2 | 2+2 4 |

Continued on next column

**TABLE XXVI– continued**

| TQC | ICSD | | NoE | SF | | | | |
|---|---|---|---|---|---|---|---|---|
| | | | | 3+3 2 | 3+3 2 | 3+4 2 | 5+6 2 | |
| | | | | 4+4 2 | 4+4 2 | 5+6 2 | 3+4 2 | |
| | | | | 5+5 2 | 3+3 2 | 5+6 2 | 3+4 2 | |
| | | | | 6+6 2 | 4+4 2 | 3+4 2 | 5+6 2 | |
| | | | | 3+3 2 | 5+5 2 | 3+4 2 | 5+6 2 | |
| | | | | 4+4 2 | 6+6 2 | 5+6 2 | 3+4 2 | |
| | | | | 5+5 2 | 3+3 2 | 5+6 2 | 3+4 2 | |
| | | | | 6+6 2 | 4+4 2 | 3+4 2 | 5+6 2 | |
| | | | | 4+4 2 | 5+5 2 | 5+6 2 | 3+4 2 | |
| | | | | 3+3 2 | 6+6 2 | 3+4 2 | 5+6 2 | |
| | | | | 6+6 2 | 5+5 2 | 5+6 2 | 5+6 2 | |
| | | | | 5+5 2 | 6+6 2 | 5+6 2 | 3+4 2 | |
| | | | | 6+6 2 | 3+3 2 | 3+4 2 | 5+6 2 | |
| | | | | 5+5 2 | 4+4 2 | 3+4 2 | 3+4 2 | |
| | | | | 3+3 2 | 6+6 2 | 5+6 2 | 3+4 2 | |
| | | | | 4+4 2 | 5+5 2 | 5+6 2 | 5+6 2 | |
| | | | | 3+3 2 | 3+3 2 | 5+6 2 | 3+4 2 | |
| | | | | 4+4 2 | 4+4 2 | 3+4 2 | 3+4 2 | |
| | | | | 5+5 2 | 6+6 2 | 5+6 2 | 5+6 2 | |
| | | | | 6+6 2 | 5+5 2 | 3+4 2 | 3+4 2 | |
| 14.2.19 | 98698 | | 100 | H1Cu1F1O1 | | | | |
| | -A | -B | -C | -D | -E | -GM | -Y | -Z |
| | 2+2 4 | 2+2 4 | 2+2 4 | 3+3 2 | 5+5 2 | 3+4 2 | 3+4 2 | 2+2 4 |
| | 2+2 4 | 2+2 4 | 2+2 4 | 4+4 2 | 6+6 2 | 3+4 2 | 5+6 2 | 2+2 4 |
| | 2+2 4 | 2+2 4 | 2+2 4 | 5+5 2 | 3+3 2 | 5+6 2 | 3+4 2 | 2+2 4 |
| | 2+2 4 | 2+2 4 | 2+2 4 | 6+6 2 | 4+4 2 | 5+6 2 | 5+6 2 | 2+2 4 |
| | 2+2 4 | 2+2 4 | 2+2 4 | 5+5 2 | 5+5 2 | 3+4 2 | 3+4 2 | 2+2 4 |
| | 2+2 4 | 2+2 4 | 2+2 4 | 6+6 2 | 6+6 2 | 5+6 2 | 3+4 2 | 2+2 4 |
| | 2+2 4 | 2+2 4 | 2+2 4 | 3+3 2 | 3+3 2 | 3+4 2 | 5+6 2 | 2+2 4 |
| | 2+2 4 | 2+2 4 | 2+2 4 | 4+4 2 | 4+4 2 | 5+6 2 | 5+6 2 | 2+2 4 |
| | 2+2 4 | 2+2 4 | 2+2 4 | 5+5 2 | 5+5 2 | 3+4 2 | 3+4 2 | 2+2 4 |
| | 2+2 4 | 2+2 4 | 2+2 4 | 6+6 2 | 6+6 2 | 3+4 2 | 3+4 2 | 2+2 4 |
| | 2+2 4 | 2+2 4 | 2+2 4 | 3+3 2 | 3+3 2 | 5+6 2 | 5+6 2 | 2+2 4 |
| | 2+2 4 | 2+2 4 | 2+2 4 | 4+4 2 | 4+4 2 | 5+6 2 | 5+6 2 | 2+2 4 |
| | 2+2 4 | 2+2 4 | 2+2 4 | 5+5 2 | 3+3 2 | 3+4 2 | 5+6 2 | 2+2 4 |
| | 2+2 4 | 2+2 4 | 2+2 4 | 6+6 2 | 4+4 2 | 5+6 2 | 3+4 2 | 2+2 4 |
| | 2+2 4 | 2+2 4 | 2+2 4 | 3+3 2 | 5+5 2 | 5+6 2 | 3+4 2 | 2+2 4 |
| | 2+2 4 | 2+2 4 | 2+2 4 | 4+4 2 | 6+6 2 | 3+4 2 | 5+6 2 | 2+2 4 |
| | 2+2 4 | 2+2 4 | 2+2 4 | 6+6 2 | 5+5 2 | 3+4 2 | 3+4 2 | 2+2 4 |
| | 2+2 4 | 2+2 4 | 2+2 4 | 5+5 2 | 6+6 2 | 3+4 2 | 3+4 2 | 2+2 4 |
| | 2+2 4 | 2+2 4 | 2+2 4 | 4+4 2 | 4+4 2 | 5+6 2 | 5+6 2 | 2+2 4 |
| | 2+2 4 | 2+2 4 | 2+2 4 | 3+3 2 | 3+3 2 | 5+6 2 | 5+6 2 | 2+2 4 |
| | 2+2 4 | 2+2 4 | 2+2 4 | 5+5 2 | 5+5 2 | 3+4 2 | 3+4 2 | 2+2 4 |
| | 2+2 4 | 2+2 4 | 2+2 4 | 6+6 2 | 6+6 2 | 3+4 2 | 3+4 2 | 2+2 4 |
| | 2+2 4 | 2+2 4 | 2+2 4 | 3+3 2 | 3+3 2 | 3+4 2 | 3+4 2 | 2+2 4 |
| | 2+2 4 | 2+2 4 | 2+2 4 | 4+4 2 | 4+4 2 | 5+6 2 | 5+6 2 | 2+2 4 |
| | 2+2 4 | 2+2 4 | 2+2 4 | 3+3 2 | 3+3 2 | 3+4 2 | 3+4 2 | 2+2 4 |
| | | | | 4+4 2 | 4+4 2 | 5+6 2 | 5+6 2 | |
| | | | | 6+6 2 | 5+5 2 | 3+4 2 | 3+4 2 | |
| | | | | 5+5 2 | 6+6 2 | 5+6 2 | 5+6 2 | |
| | | | | 4+4 2 | 4+4 2 | 3+4 2 | 5+6 2 | |
| | | | | 3+3 2 | 3+3 2 | 5+6 2 | 3+4 2 | |
| | | | | 5+5 2 | 5+5 2 | 5+6 2 | 5+6 2 | |
| | | | | 6+6 2 | 6+6 2 | 5+6 2 | 5+6 2 | |
| | | | | 6+6 2 | 6+6 2 | 3+4 2 | 3+4 2 | |
| | | | | 5+5 2 | 5+5 2 | 5+6 2 | 5+6 2 | |
| | | | | 4+4 2 | 3+3 2 | 3+4 2 | 3+4 2 | |
| | | | | 3+3 2 | 4+4 2 | 5+6 2 | 5+6 2 | |
| | | | | 4+4 2 | 4+4 2 | 3+4 2 | 3+4 2 | |
| | | | | 3+3 2 | 3+3 2 | 5+6 2 | 5+6 2 | |
| | | | | 5+5 2 | 5+5 2 | 3+4 2 | 3+4 2 | |

Continued on next column

**TABLE XXVI– continued**

| TQC | ICSD | | NoE | SF | | | | |
|---|---|---|---|---|---|---|---|---|
| | | | | 6+6 2 | 6+6 2 | 5+6 2 | 5+6 2 | |
| | | | | 3+3 2 | 3+3 2 | 3+4 2 | 3+4 2 | |
| | | | | 4+4 2 | 4+4 2 | 5+6 2 | 5+6 2 | |
| | | | | 6+6 2 | 6+6 2 | 3+4 2 | 3+4 2 | |
| | | | | 5+5 2 | 5+5 2 | 5+6 2 | 5+6 2 | |
| | | | | 4+4 2 | 5+5 2 | 3+4 2 | 3+4 2 | |
| | | | | 3+3 2 | 4+4 2 | 5+6 2 | 5+6 2 | |
| | | | | 5+5 2 | 6+6 2 | 5+6 2 | 3+4 2 | |
| | | | | 6+6 2 | 3+3 2 | 5+6 2 | 5+6 2 | |
| | | | | 3+3 2 | 3+3 2 | 3+4 2 | 5+6 2 | |
| | | | | 4+4 2 | 4+4 2 | 3+4 2 | 3+4 2 | |
| 14.2.20 | 91088 | | 100 | H1Cl1Cu1O1 | | | | |
| | -A | -B | -C | -D | -E | -GM | -Y | -Z |
| | 2+2 4 | 2+2 4 | 2+2 4 | 5+5 2 | 3+3 2 | 3+4 2 | 5+6 2 | 2+2 4 |
| | 2+2 4 | 2+2 4 | 2+2 4 | 6+6 2 | 4+4 2 | 3+4 2 | 5+6 2 | 2+2 4 |
| | 2+2 4 | 2+2 4 | 2+2 4 | 3+3 2 | 5+5 2 | 5+6 2 | 3+4 2 | 2+2 4 |
| | 2+2 4 | 2+2 4 | 2+2 4 | 4+4 2 | 6+6 2 | 5+6 2 | 3+4 2 | 2+2 4 |
| | 2+2 4 | 2+2 4 | 2+2 4 | 3+3 2 | 5+5 2 | 3+4 2 | 3+4 2 | 2+2 4 |
| | 2+2 4 | 2+2 4 | 2+2 4 | 4+4 2 | 6+6 2 | 3+4 2 | 5+6 2 | 2+2 4 |
| | 2+2 4 | 2+2 4 | 2+2 4 | 5+5 2 | 3+3 2 | 5+6 2 | 5+6 2 | 2+2 4 |
| | 2+2 4 | 2+2 4 | 2+2 4 | 6+6 2 | 4+4 2 | 5+6 2 | 3+4 2 | 2+2 4 |
| | 2+2 4 | 2+2 4 | 2+2 4 | 3+3 2 | 3+3 2 | 3+4 2 | 3+4 2 | 2+2 4 |
| | 2+2 4 | 2+2 4 | 2+2 4 | 4+4 2 | 4+4 2 | 3+4 2 | 5+6 2 | 2+2 4 |
| | 2+2 4 | 2+2 4 | 2+2 4 | 6+6 2 | 5+5 2 | 5+6 2 | 5+6 2 | 2+2 4 |
| | 2+2 4 | 2+2 4 | 2+2 4 | 5+5 2 | 6+6 2 | 5+6 2 | 3+4 2 | 2+2 4 |
| | 2+2 4 | 2+2 4 | 2+2 4 | 3+3 2 | 4+4 2 | 3+4 2 | 5+6 2 | 2+2 4 |
| | 2+2 4 | 2+2 4 | 2+2 4 | 4+4 2 | 3+3 2 | 5+6 2 | 3+4 2 | 2+2 4 |
| | 2+2 4 | 2+2 4 | 2+2 4 | 5+5 2 | 5+5 2 | 5+6 2 | 3+4 2 | 2+2 4 |
| | 2+2 4 | 2+2 4 | 2+2 4 | 6+6 2 | 6+6 2 | 3+4 2 | 5+6 2 | 2+2 4 |
| | 2+2 4 | 2+2 4 | 2+2 4 | 5+5 2 | 3+3 2 | 3+4 2 | 5+6 2 | 2+2 4 |
| | 2+2 4 | 2+2 4 | 2+2 4 | 6+6 2 | 4+4 2 | 3+4 2 | 3+4 2 | 2+2 4 |
| | 2+2 4 | 2+2 4 | 2+2 4 | 5+5 2 | 3+3 2 | 3+4 2 | 3+4 2 | 2+2 4 |
| | 2+2 4 | 2+2 4 | 2+2 4 | 6+6 2 | 4+4 2 | 5+6 2 | 5+6 2 | 2+2 4 |
| | 2+2 4 | 2+2 4 | 2+2 4 | 3+3 2 | 5+5 2 | 5+6 2 | 5+6 2 | 2+2 4 |
| | 2+2 4 | 2+2 4 | 2+2 4 | 4+4 2 | 6+6 2 | 3+4 2 | 5+6 2 | 2+2 4 |
| | 2+2 4 | 2+2 4 | 2+2 4 | 3+3 2 | 6+6 2 | 5+6 2 | 3+4 2 | 2+2 4 |
| | 2+2 4 | 2+2 4 | 2+2 4 | 4+4 2 | 5+5 2 | 3+4 2 | 3+4 2 | 2+2 4 |
| | 2+2 4 | 2+2 4 | 2+2 4 | 4+4 2 | 4+4 2 | 5+6 2 | 5+6 2 | 2+2 4 |
| | | | | 3+3 2 | 3+3 2 | 5+6 2 | 3+4 2 | |
| | | | | 6+6 2 | 5+5 2 | 3+4 2 | 3+4 2 | |
| | | | | 5+5 2 | 6+6 2 | 3+4 2 | 5+6 2 | |
| | | | | 3+3 2 | 6+6 2 | 3+4 2 | 5+6 2 | |
| | | | | 4+4 2 | 5+5 2 | 5+6 2 | 5+6 2 | |
| | | | | 6+6 2 | 3+3 2 | 5+6 2 | 3+4 2 | |
| | | | | 5+5 2 | 4+4 2 | 3+4 2 | 3+4 2 | |
| | | | | 3+3 2 | 6+6 2 | 5+6 2 | 5+6 2 | |
| | | | | 4+4 2 | 3+3 2 | 3+4 2 | 3+4 2 | |
| | | | | 6+6 2 | 5+5 2 | 5+6 2 | 5+6 2 | |
| | | | | 5+5 2 | 4+4 2 | 3+4 2 | 3+4 2 | |
| | | | | 3+3 2 | 6+6 2 | 5+6 2 | 3+4 2 | |
| | | | | 4+4 2 | 5+5 2 | 3+4 2 | 5+6 2 | |
| | | | | 5+5 2 | 4+4 2 | 5+6 2 | 5+6 2 | |
| | | | | 6+6 2 | 3+3 2 | 3+4 2 | 3+4 2 | |
| | | | | 6+6 2 | 3+3 2 | 5+6 2 | 5+6 2 | |
| | | | | 5+5 2 | 4+4 2 | 3+4 2 | 5+6 2 | |
| | | | | 3+3 2 | 5+5 2 | 5+6 2 | 3+4 2 | |
| | | | | 4+4 2 | 6+6 2 | 5+6 2 | 3+4 2 | |
| | | | | 4+4 2 | 4+4 2 | 5+6 2 | 5+6 2 | |
| | | | | 3+3 2 | 3+3 2 | 3+4 2 | 3+4 2 | |
| | | | | 5+5 2 | 5+5 2 | 3+4 2 | 3+4 2 | |
| | | | | 6+6 2 | 6+6 2 | 5+6 2 | 5+6 2 | |

Continued on next column

**TABLE XXVI– continued**

| TQC | ICSD | NoE | SF | | | | | |
|---|---|---|---|---|---|---|---|---|
| | | | | 3+3 2 | 5+5 2 | 5+6 2 | 3+4 2 | |
| | | | | 4+4 2 | 6+6 2 | 3+4 2 | 5+6 2 | |
| 14.2.21 | 38384 | 100 | H1Cl1Cu1O1 | | | | | |
| | -A | -B | -C | -D | -E | -GM | -Y | -Z |
| | 2+2 4 | 2+2 4 | 2+2 4 | 5+5 2 | 5+5 2 | 3+4 2 | 3+4 2 | 2+2 4 |
| | 2+2 4 | 2+2 4 | 2+2 4 | 6+6 2 | 6+6 2 | 3+4 2 | 3+4 2 | 2+2 4 |
| | 2+2 4 | 2+2 4 | 2+2 4 | 3+3 2 | 3+3 2 | 5+6 2 | 5+6 2 | 2+2 4 |
| | 2+2 4 | 2+2 4 | 2+2 4 | 4+4 2 | 4+4 2 | 5+6 2 | 5+6 2 | 2+2 4 |
| | 2+2 4 | 2+2 4 | 2+2 4 | 3+3 2 | 3+3 2 | 3+4 2 | 5+6 2 | 2+2 4 |
| | 2+2 4 | 2+2 4 | 2+2 4 | 4+4 2 | 4+4 2 | 3+4 2 | 3+4 2 | 2+2 4 |
| | 2+2 4 | 2+2 4 | 2+2 4 | 5+5 2 | 5+5 2 | 5+6 2 | 3+4 2 | 2+2 4 |
| | 2+2 4 | 2+2 4 | 2+2 4 | 6+6 2 | 6+6 2 | 5+6 2 | 5+6 2 | 2+2 4 |
| | 2+2 4 | 2+2 4 | 2+2 4 | 5+5 2 | 5+5 2 | 3+4 2 | 3+4 2 | 2+2 4 |
| | 2+2 4 | 2+2 4 | 2+2 4 | 6+6 2 | 6+6 2 | 3+4 2 | 3+4 2 | 2+2 4 |
| | 2+2 4 | 2+2 4 | 2+2 4 | 3+3 2 | 3+3 2 | 5+6 2 | 5+6 2 | 2+2 4 |
| | 2+2 4 | 2+2 4 | 2+2 4 | 4+4 2 | 4+4 2 | 5+6 2 | 5+6 2 | 2+2 4 |
| | 2+2 4 | 2+2 4 | 2+2 4 | 6+6 2 | 3+3 2 | 3+4 2 | 3+4 2 | 2+2 4 |
| | 2+2 4 | 2+2 4 | 2+2 4 | 5+5 2 | 4+4 2 | 5+6 2 | 5+6 2 | 2+2 4 |
| | 2+2 4 | 2+2 4 | 2+2 4 | 3+3 2 | 6+6 2 | 5+6 2 | 5+6 2 | 2+2 4 |
| | 2+2 4 | 2+2 4 | 2+2 4 | 4+4 2 | 5+5 2 | 3+4 2 | 3+4 2 | 2+2 4 |
| | 2+2 4 | 2+2 4 | 2+2 4 | 5+5 2 | 5+5 2 | 3+4 2 | 3+4 2 | 2+2 4 |
| | 2+2 4 | 2+2 4 | 2+2 4 | 6+6 2 | 6+6 2 | 3+4 2 | 3+4 2 | 2+2 4 |
| | 2+2 4 | 2+2 4 | 2+2 4 | 4+4 2 | 4+4 2 | 5+6 2 | 5+6 2 | 2+2 4 |
| | 2+2 4 | 2+2 4 | 2+2 4 | 3+3 2 | 3+3 2 | 5+6 2 | 5+6 2 | 2+2 4 |
| | 2+2 4 | 2+2 4 | 2+2 4 | 6+6 2 | 6+6 2 | 3+4 2 | 3+4 2 | 2+2 4 |
| | 2+2 4 | 2+2 4 | 2+2 4 | 5+5 2 | 5+5 2 | 3+4 2 | 3+4 2 | 2+2 4 |
| | 2+2 4 | 2+2 4 | 2+2 4 | 4+4 2 | 4+4 2 | 5+6 2 | 5+6 2 | 2+2 4 |
| | 2+2 4 | 2+2 4 | 2+2 4 | 3+3 2 | 3+3 2 | 5+6 2 | 5+6 2 | 2+2 4 |
| | 2+2 4 | 2+2 4 | 2+2 4 | 3+3 2 | 6+6 2 | 3+4 2 | 3+4 2 | 2+2 4 |
| | | | | 4+4 2 | 5+5 2 | 3+4 2 | 5+6 2 | |
| | | | | 5+5 2 | 3+3 2 | 5+6 2 | 3+4 2 | |
| | | | | 6+6 2 | 4+4 2 | 3+4 2 | 5+6 2 | |
| | | | | 3+3 2 | 3+3 2 | 3+4 2 | 3+4 2 | |
| | | | | 4+4 2 | 4+4 2 | 5+6 2 | 3+4 2 | |
| | | | | 5+5 2 | 5+5 2 | 5+6 2 | 5+6 2 | |
| | | | | 6+6 2 | 6+6 2 | 3+4 2 | 5+6 2 | |
| | | | | 4+4 2 | 3+3 2 | 5+6 2 | 3+4 2 | |
| | | | | 3+3 2 | 4+4 2 | 3+4 2 | 5+6 2 | |
| | | | | 5+5 2 | 5+5 2 | 5+6 2 | 3+4 2 | |
| | | | | 6+6 2 | 6+6 2 | 5+6 2 | 5+6 2 | |
| | | | | 3+3 2 | 3+3 2 | 3+4 2 | 3+4 2 | |
| | | | | 4+4 2 | 4+4 2 | 3+4 2 | 5+6 2 | |
| | | | | 6+6 2 | 6+6 2 | 5+6 2 | 3+4 2 | |
| | | | | 5+5 2 | 5+5 2 | 5+6 2 | 5+6 2 | |
| | | | | 4+4 2 | 4+4 2 | 3+4 2 | 3+4 2 | |
| | | | | 3+3 2 | 3+3 2 | 3+4 2 | 3+4 2 | |
| | | | | 5+5 2 | 5+5 2 | 5+6 2 | 5+6 2 | |
| | | | | 6+6 2 | 6+6 2 | 5+6 2 | 5+6 2 | |
| | | | | 3+3 2 | 6+6 2 | 5+6 2 | 5+6 2 | |
| | | | | 4+4 2 | 5+5 2 | 3+4 2 | 3+4 2 | |
| | | | | 6+6 2 | 3+3 2 | 3+4 2 | 5+6 2 | |
| | | | | 5+5 2 | 4+4 2 | 5+6 2 | 3+4 2 | |
| | | | | 4+4 2 | 4+4 2 | 5+6 2 | 5+6 2 | |
| | | | | 3+3 2 | 3+3 2 | 3+4 2 | 5+6 2 | |
| 14.2.22 | 80002 | 124 | Ge4Mo3U2 | | | | | |
| | -A | -B | -C | -D | -E | -GM | -Y | -Z |
| | 2+2 4 | 2+2 4 | 2+2 4 | 3+3 2 | 3+3 2 | 3+4 2 | 3+4 2 | 2+2 4 |
| | 2+2 4 | 2+2 4 | 2+2 4 | 4+4 2 | 4+4 2 | 3+4 2 | 3+4 2 | 2+2 4 |
| | 2+2 4 | 2+2 4 | 2+2 4 | 5+5 2 | 5+5 2 | 5+6 2 | 5+6 2 | 2+2 4 |
| | 2+2 4 | 2+2 4 | 2+2 4 | 6+6 2 | 6+6 2 | 5+6 2 | 5+6 2 | 2+2 4 |
| | 2+2 4 | 2+2 4 | 2+2 4 | 4+4 2 | 4+4 2 | 3+4 2 | 3+4 2 | 2+2 4 |
| | 2+2 4 | 2+2 4 | 2+2 4 | 3+3 2 | 3+3 2 | 5+6 2 | 5+6 2 | 2+2 4 |
| | 2+2 4 | 2+2 4 | 2+2 4 | 5+5 2 | 6+6 2 | 3+4 2 | 3+4 2 | 2+2 4 |
| | 2+2 4 | 2+2 4 | 2+2 4 | 6+6 2 | 5+5 2 | 5+6 2 | 5+6 2 | 2+2 4 |
| | 2+2 4 | 2+2 4 | 2+2 4 | 3+3 2 | 3+3 2 | 3+4 2 | 3+4 2 | 2+2 4 |
| | 2+2 4 | 2+2 4 | 2+2 4 | 5+5 2 | 5+5 2 | 5+6 2 | 5+6 2 | 2+2 4 |
| | 2+2 4 | 2+2 4 | 2+2 4 | 4+4 2 | 4+4 2 | 3+4 2 | 3+4 2 | 2+2 4 |
| | 2+2 4 | 2+2 4 | 2+2 4 | 6+6 2 | 6+6 2 | 3+4 2 | 3+4 2 | 2+2 4 |
| | 2+2 4 | 2+2 4 | 2+2 4 | 4+4 2 | 4+4 2 | 5+6 2 | 5+6 2 | 2+2 4 |
| | 2+2 4 | 2+2 4 | 2+2 4 | 3+3 2 | 6+6 2 | 5+6 2 | 5+6 2 | 2+2 4 |
| | 2+2 4 | 2+2 4 | 2+2 4 | 6+6 2 | 3+3 2 | 3+4 2 | 3+4 2 | 2+2 4 |
| | 2+2 4 | 2+2 4 | 2+2 4 | 5+5 2 | 5+5 2 | 5+6 2 | 5+6 2 | 2+2 4 |
| | 2+2 4 | 2+2 4 | 2+2 4 | 4+4 2 | 5+5 2 | 3+4 2 | 5+6 2 | 2+2 4 |
| | 2+2 4 | 2+2 4 | 2+2 4 | 3+3 2 | 6+6 2 | 3+4 2 | 3+4 2 | 2+2 4 |
| | 2+2 4 | 2+2 4 | 2+2 4 | 6+6 2 | 6+6 2 | 3+4 2 | 5+6 2 | 2+2 4 |
| | 2+2 4 | 2+2 4 | 2+2 4 | 5+5 2 | 5+5 2 | 3+4 2 | 3+4 2 | 2+2 4 |
| | 2+2 4 | 2+2 4 | 2+2 4 | 5+5 2 | 4+4 2 | 5+6 2 | 5+6 2 | 2+2 4 |
| | 2+2 4 | 2+2 4 | 2+2 4 | 6+6 2 | 3+3 2 | 5+6 2 | 3+4 2 | 2+2 4 |
| | 2+2 4 | 2+2 4 | 2+2 4 | 3+3 2 | 3+3 2 | 5+6 2 | 3+4 2 | 2+2 4 |
| | 2+2 4 | 2+2 4 | 2+2 4 | 4+4 2 | 4+4 2 | 5+6 2 | 5+6 2 | 2+2 4 |
| | 2+2 4 | 2+2 4 | 2+2 4 | 6+6 2 | 3+3 2 | 3+4 2 | 5+6 2 | 2+2 4 |
| | 2+2 4 | 2+2 4 | 2+2 4 | 5+5 2 | 4+4 2 | 5+6 2 | 5+6 2 | 2+2 4 |
| | 2+2 4 | 2+2 4 | 2+2 4 | 5+5 2 | 4+4 2 | 3+4 2 | 3+4 2 | 2+2 4 |
| | 2+2 4 | 2+2 4 | 2+2 4 | 6+6 2 | 3+3 2 | 5+6 2 | 5+6 2 | 2+2 4 |
| | 2+2 4 | 2+2 4 | 2+2 4 | 4+4 2 | 6+6 2 | 3+4 2 | 5+6 2 | 2+2 4 |
| | 2+2 4 | 2+2 4 | 2+2 4 | 3+3 2 | 5+5 2 | 3+4 2 | 3+4 2 | 2+2 4 |
| | 2+2 4 | 2+2 4 | 2+2 4 | 6+6 2 | 5+5 2 | 3+4 2 | 5+6 2 | 2+2 4 |
| | | | | 5+5 2 | 6+6 2 | 3+4 2 | 3+4 2 | |
| | | | | 4+4 2 | 5+5 2 | 5+6 2 | 5+6 2 | |
| | | | | 3+3 2 | 6+6 2 | 3+4 2 | 5+6 2 | |
| | | | | 6+6 2 | 3+3 2 | 5+6 2 | 3+4 2 | |
| | | | | 5+5 2 | 4+4 2 | 3+4 2 | 5+6 2 | |
| | | | | 3+3 2 | 6+6 2 | 3+4 2 | 3+4 2 | |
| | | | | 4+4 2 | 5+5 2 | 5+6 2 | 3+4 2 | |
| | | | | 3+3 2 | 4+4 2 | 3+4 2 | 5+6 2 | |
| | | | | 4+4 2 | 3+3 2 | 3+4 2 | 3+4 2 | |
| | | | | 5+5 2 | 5+5 2 | 5+6 2 | 5+6 2 | |
| | | | | 6+6 2 | 6+6 2 | 5+6 2 | 3+4 2 | |
| | | | | 5+5 2 | 3+3 2 | 3+4 2 | 5+6 2 | |
| | | | | 6+6 2 | 4+4 2 | 5+6 2 | 5+6 2 | |
| | | | | 6+6 2 | 6+6 2 | 3+4 2 | 5+6 2 | |
| | | | | 5+5 2 | 5+5 2 | 5+6 2 | 3+4 2 | |
| | | | | 3+3 2 | 4+4 2 | 3+4 2 | 5+6 2 | |
| | | | | 4+4 2 | 3+3 2 | 3+4 2 | 3+4 2 | |
| | | | | 6+6 2 | 3+3 2 | 5+6 2 | 3+4 2 | |
| | | | | 5+5 2 | 4+4 2 | 3+4 2 | 5+6 2 | |
| | | | | 3+3 2 | 5+5 2 | 3+4 2 | 3+4 2 | |
| | | | | 4+4 2 | 6+6 2 | 5+6 2 | 5+6 2 | |
| | | | | 3+3 2 | 4+4 2 | 5+6 2 | 5+6 2 | |
| | | | | 4+4 2 | 3+3 2 | 3+4 2 | 3+4 2 | |
| | | | | 5+5 2 | 4+4 2 | 3+4 2 | 3+4 2 | |
| | | | | 6+6 2 | 3+3 2 | 5+6 2 | 5+6 2 | |
| | | | | 6+6 2 | 5+5 2 | 5+6 2 | 3+4 2 | |
| | | | | 5+5 2 | 6+6 2 | 5+6 2 | 5+6 2 | |
| | | | | 4+4 2 | 3+3 2 | 3+4 2 | 5+6 2 | |
| | | | | 3+3 2 | 4+4 2 | 5+6 2 | 3+4 2 | |
| | | | | 6+6 2 | 5+5 2 | 5+6 2 | 5+6 2 | |
| | | | | 5+5 2 | 6+6 2 | 3+4 2 | 5+6 2 | |
| 14.2.23 | 416421 | 156 | Ge2La1Pt2 | | | | | |
| | -A | -B | -C | -D | -E | -GM | -Y | -Z |
| | 2+2 4 | 2+2 4 | 2+2 4 | 3+3 2 | 3+3 2 | 3+4 2 | 3+4 2 | 2+2 4 |
| | 2+2 4 | 2+2 4 | 2+2 4 | 4+4 2 | 4+4 2 | 5+6 2 | 5+6 2 | 2+2 4 |

Continued on next column

Continued on next column

**TABLE XXVI– continued**

| TQC | ICSD | | NoE | SF | | | | |
|---|---|---|---|---|---|---|---|---|
| | 2+2 4 | 2+2 4 | 2+2 4 | 5+5 2 | 5+5 2 | 3+4 2 | 3+4 2 | 2+2 4 |
| | 2+2 4 | 2+2 4 | 2+2 4 | 6+6 2 | 6+6 2 | 5+6 2 | 5+6 2 | 2+2 4 |
| | 2+2 4 | 2+2 4 | 2+2 4 | 4+4 2 | 3+3 2 | 3+4 2 | 3+4 2 | 2+2 4 |
| | 2+2 4 | 2+2 4 | 2+2 4 | 5+5 2 | 4+4 2 | 5+6 2 | 5+6 2 | 2+2 4 |
| | 2+2 4 | 2+2 4 | 2+2 4 | 6+6 2 | 5+5 2 | 3+4 2 | 3+4 2 | 2+2 4 |
| | 2+2 4 | 2+2 4 | 2+2 4 | 3+3 2 | 6+6 2 | 5+6 2 | 5+6 2 | 2+2 4 |
| | 2+2 4 | 2+2 4 | 2+2 4 | 3+3 2 | 3+3 2 | 3+4 2 | 3+4 2 | 2+2 4 |
| | 2+2 4 | 2+2 4 | 2+2 4 | 5+5 2 | 5+5 2 | 5+6 2 | 5+6 2 | 2+2 4 |
| | 2+2 4 | 2+2 4 | 2+2 4 | 6+6 2 | 6+6 2 | 3+4 2 | 3+4 2 | 2+2 4 |
| | 2+2 4 | 2+2 4 | 2+2 4 | 4+4 2 | 4+4 2 | 5+6 2 | 5+6 2 | 2+2 4 |
| | 2+2 4 | 2+2 4 | 2+2 4 | 3+3 2 | 4+4 2 | 3+4 2 | 3+4 2 | 2+2 4 |
| | 2+2 4 | 2+2 4 | 2+2 4 | 4+4 2 | 6+6 2 | 5+6 2 | 5+6 2 | 2+2 4 |
| | 2+2 4 | 2+2 4 | 2+2 4 | 6+6 2 | 3+3 2 | 3+4 2 | 3+4 2 | 2+2 4 |
| | 2+2 4 | 2+2 4 | 2+2 4 | 5+5 2 | 5+5 2 | 5+6 2 | 5+6 2 | 2+2 4 |
| | 2+2 4 | 2+2 4 | 2+2 4 | 3+3 2 | 3+3 2 | 3+4 2 | 3+4 2 | 2+2 4 |
| | 2+2 4 | 2+2 4 | 2+2 4 | 4+4 2 | 4+4 2 | 3+4 2 | 5+6 2 | 2+2 4 |
| | 2+2 4 | 2+2 4 | 2+2 4 | 6+6 2 | 6+6 2 | 3+4 2 | 3+4 2 | 2+2 4 |
| | 2+2 4 | 2+2 4 | 2+2 4 | 5+5 2 | 5+5 2 | 5+6 2 | 5+6 2 | 2+2 4 |
| | 2+2 4 | 2+2 4 | 2+2 4 | 4+4 2 | 4+4 2 | 3+4 2 | 5+6 2 | 2+2 4 |
| | 2+2 4 | 2+2 4 | 2+2 4 | 5+5 2 | 3+3 2 | 5+6 2 | 3+4 2 | 2+2 4 |
| | 2+2 4 | 2+2 4 | 2+2 4 | 3+3 2 | 5+5 2 | 5+6 2 | 3+4 2 | 2+2 4 |
| | 2+2 4 | 2+2 4 | 2+2 4 | 6+6 2 | 6+6 2 | 5+6 2 | 5+6 2 | 2+2 4 |
| | 2+2 4 | 2+2 4 | 2+2 4 | 4+4 2 | 6+6 2 | 5+6 2 | 5+6 2 | 2+2 4 |
| | 2+2 4 | 2+2 4 | 2+2 4 | 3+3 2 | 5+5 2 | 3+4 2 | 3+4 2 | 2+2 4 |
| | 2+2 4 | 2+2 4 | 2+2 4 | 6+6 2 | 3+3 2 | 3+4 2 | 3+4 2 | 2+2 4 |
| | 2+2 4 | 2+2 4 | 2+2 4 | 5+5 2 | 4+4 2 | 3+4 2 | 3+4 2 | 2+2 4 |
| | 2+2 4 | 2+2 4 | 2+2 4 | 3+3 2 | 3+3 2 | 5+6 2 | 5+6 2 | 2+2 4 |
| | 2+2 4 | 2+2 4 | 2+2 4 | 6+6 2 | 4+4 2 | 3+4 2 | 5+6 2 | 2+2 4 |
| | 2+2 4 | 2+2 4 | 2+2 4 | 4+4 2 | 6+6 2 | 3+4 2 | 5+6 2 | 2+2 4 |
| | 2+2 4 | 2+2 4 | 2+2 4 | 5+5 2 | 5+5 2 | 5+6 2 | 3+4 2 | 2+2 4 |
| | 2+2 4 | 2+2 4 | 2+2 4 | 5+5 2 | 4+4 2 | 5+6 2 | 3+4 2 | 2+2 4 |
| | 2+2 4 | 2+2 4 | 2+2 4 | 4+4 2 | 5+5 2 | 5+6 2 | 3+4 2 | 2+2 4 |
| | 2+2 4 | 2+2 4 | 2+2 4 | 6+6 2 | 3+3 2 | 3+4 2 | 5+6 2 | 2+2 4 |
| | 2+2 4 | 2+2 4 | 2+2 4 | 5+5 2 | 4+4 2 | 3+4 2 | 5+6 2 | 2+2 4 |
| | 2+2 4 | 2+2 4 | 2+2 4 | 3+3 2 | 6+6 2 | 3+4 2 | 3+4 2 | 2+2 4 |
| | 2+2 4 | 2+2 4 | 2+2 4 | 4+4 2 | 5+5 2 | 5+6 2 | 3+4 2 | 2+2 4 |
| | 2+2 4 | 2+2 4 | 2+2 4 | 6+6 2 | 3+3 2 | 5+6 2 | 5+6 2 | 2+2 4 |
| | | | | 3+3 2 | 6+6 2 | 3+4 2 | 3+4 2 | |
| | | | | 6+6 2 | 4+4 2 | 5+6 2 | 3+4 2 | |
| | | | | 5+5 2 | 3+3 2 | 3+4 2 | 3+4 2 | |
| | | | | 4+4 2 | 6+6 2 | 5+6 2 | 5+6 2 | |
| | | | | 3+3 2 | 5+5 2 | 3+4 2 | 5+6 2 | |
| | | | | 6+6 2 | 3+3 2 | 5+6 2 | 5+6 2 | |
| | | | | 5+5 2 | 4+4 2 | 5+6 2 | 3+4 2 | |
| | | | | 3+3 2 | 5+5 2 | 3+4 2 | 5+6 2 | |
| | | | | 4+4 2 | 6+6 2 | 5+6 2 | 5+6 2 | |
| | | | | 5+5 2 | 4+4 2 | 5+6 2 | 3+4 2 | |
| | | | | 6+6 2 | 5+5 2 | 3+4 2 | 3+4 2 | |
| | | | | 3+3 2 | 3+3 2 | 3+4 2 | 5+6 2 | |
| | | | | 4+4 2 | 6+6 2 | 3+4 2 | 5+6 2 | |
| | | | | 5+5 2 | 4+4 2 | 5+6 2 | 3+4 2 | |
| | | | | 4+4 2 | 6+6 2 | 3+4 2 | 5+6 2 | |
| | | | | 6+6 2 | 3+3 2 | 3+4 2 | 3+4 2 | |
| | | | | 5+5 2 | 5+5 2 | 5+6 2 | 5+6 2 | |
| | | | | 6+6 2 | 3+3 2 | 3+4 2 | 3+4 2 | |
| | | | | 3+3 2 | 5+5 2 | 5+6 2 | 5+6 2 | |
| | | | | 4+4 2 | 4+4 2 | 3+4 2 | 3+4 2 | |
| | | | | 3+3 2 | 6+6 2 | 5+6 2 | 3+4 2 | |
| | | | | 6+6 2 | 3+3 2 | 5+6 2 | 5+6 2 | |
| | | | | 5+5 2 | 4+4 2 | 3+4 2 | 5+6 2 | |
| | | | | 3+3 2 | 6+6 2 | 5+6 2 | 5+6 2 | |

Continued on next column

**TABLE XXVI– continued**

| TQC | ICSD | | NoE | SF | | | | |
|---|---|---|---|---|---|---|---|---|
| | | | | 6+6 2 | 5+5 2 | 3+4 2 | 3+4 2 | |
| | | | | 4+4 2 | 3+3 2 | 5+6 2 | 5+6 2 | |
| | | | | 3+3 2 | 4+4 2 | 3+4 2 | 3+4 2 | |
| | | | | 5+5 2 | 6+6 2 | 3+4 2 | 5+6 2 | |
| | | | | 4+4 2 | 5+5 2 | 5+6 2 | 3+4 2 | |
| | | | | 6+6 2 | 6+6 2 | 3+4 2 | 5+6 2 | |
| | | | | 3+3 2 | 3+3 2 | 3+4 2 | 3+4 2 | |
| | | | | 5+5 2 | 5+5 2 | 5+6 2 | 5+6 2 | |
| | | | | 4+4 2 | 4+4 2 | 3+4 2 | 3+4 2 | |
| | | | | 4+4 2 | 6+6 2 | 3+4 2 | 3+4 2 | |
| | | | | 3+3 2 | 3+3 2 | 5+6 2 | 5+6 2 | |
| | | | | 5+5 2 | 5+5 2 | 3+4 2 | 5+6 2 | |
| | | | | 6+6 2 | 4+4 2 | 5+6 2 | 3+4 2 | |
| | | | | 3+3 2 | 6+6 2 | 5+6 2 | 5+6 2 | |
| | | | | 4+4 2 | 5+5 2 | 3+4 2 | 3+4 2 | |
| 14.2.24 | 421335 | | 120 | Gd1S2 | | | | |
| | -A | -B | -C | -D | -E | -GM | -Y | -Z |
| | 2+2 4 | 2+2 4 | 2+2 4 | 3+3 2 | 3+3 2 | 3+4 2 | 5+6 2 | 2+2 4 |
| | 2+2 4 | 2+2 4 | 2+2 4 | 4+4 2 | 4+4 2 | 5+6 2 | 3+4 2 | 2+2 4 |
| | 2+2 4 | 2+2 4 | 2+2 4 | 5+5 2 | 5+5 2 | 3+4 2 | 3+4 2 | 2+2 4 |
| | 2+2 4 | 2+2 4 | 2+2 4 | 6+6 2 | 6+6 2 | 5+6 2 | 5+6 2 | 2+2 4 |
| | 2+2 4 | 2+2 4 | 2+2 4 | 4+4 2 | 5+5 2 | 3+4 2 | 3+4 2 | 2+2 4 |
| | 2+2 4 | 2+2 4 | 2+2 4 | 5+5 2 | 6+6 2 | 5+6 2 | 5+6 2 | 2+2 4 |
| | 2+2 4 | 2+2 4 | 2+2 4 | 3+3 2 | 3+3 2 | 3+4 2 | 5+6 2 | 2+2 4 |
| | 2+2 4 | 2+2 4 | 2+2 4 | 6+6 2 | 4+4 2 | 5+6 2 | 3+4 2 | 2+2 4 |
| | 2+2 4 | 2+2 4 | 2+2 4 | 6+6 2 | 6+6 2 | 3+4 2 | 3+4 2 | 2+2 4 |
| | 2+2 4 | 2+2 4 | 2+2 4 | 5+5 2 | 5+5 2 | 5+6 2 | 5+6 2 | 2+2 4 |
| | 2+2 4 | 2+2 4 | 2+2 4 | 3+3 2 | 4+4 2 | 3+4 2 | 3+4 2 | 2+2 4 |
| | 2+2 4 | 2+2 4 | 2+2 4 | 4+4 2 | 3+3 2 | 5+6 2 | 5+6 2 | 2+2 4 |
| | 2+2 4 | 2+2 4 | 2+2 4 | 3+3 2 | 6+6 2 | 3+4 2 | 3+4 2 | 2+2 4 |
| | 2+2 4 | 2+2 4 | 2+2 4 | 6+6 2 | 3+3 2 | 5+6 2 | 5+6 2 | 2+2 4 |
| | 2+2 4 | 2+2 4 | 2+2 4 | 4+4 2 | 5+5 2 | 3+4 2 | 5+6 2 | 2+2 4 |
| | 2+2 4 | 2+2 4 | 2+2 4 | 5+5 2 | 4+4 2 | 5+6 2 | 3+4 2 | 2+2 4 |
| | 2+2 4 | 2+2 4 | 2+2 4 | 5+5 2 | 5+5 2 | 3+4 2 | 3+4 2 | 2+2 4 |
| | 2+2 4 | 2+2 4 | 2+2 4 | 6+6 2 | 6+6 2 | 3+4 2 | 3+4 2 | 2+2 4 |
| | 2+2 4 | 2+2 4 | 2+2 4 | 4+4 2 | 3+3 2 | 3+4 2 | 5+6 2 | 2+2 4 |
| | 2+2 4 | 2+2 4 | 2+2 4 | 3+3 2 | 5+5 2 | 5+6 2 | 5+6 2 | 2+2 4 |
| | 2+2 4 | 2+2 4 | 2+2 4 | 6+6 2 | 6+6 2 | 5+6 2 | 3+4 2 | 2+2 4 |
| | 2+2 4 | 2+2 4 | 2+2 4 | 5+5 2 | 4+4 2 | 3+4 2 | 3+4 2 | 2+2 4 |
| | 2+2 4 | 2+2 4 | 2+2 4 | 4+4 2 | 4+4 2 | 5+6 2 | 5+6 2 | 2+2 4 |
| | 2+2 4 | 2+2 4 | 2+2 4 | 3+3 2 | 3+3 2 | 5+6 2 | 5+6 2 | 2+2 4 |
| | 2+2 4 | 2+2 4 | 2+2 4 | 6+6 2 | 6+6 2 | 3+4 2 | 3+4 2 | 2+2 4 |
| | 2+2 4 | 2+2 4 | 2+2 4 | 5+5 2 | 5+5 2 | 5+6 2 | 5+6 2 | 2+2 4 |
| | 2+2 4 | 2+2 4 | 2+2 4 | 3+3 2 | 3+3 2 | 5+6 2 | 5+6 2 | 2+2 4 |
| | 2+2 4 | 2+2 4 | 2+2 4 | 4+4 2 | 4+4 2 | 3+4 2 | 3+4 2 | 2+2 4 |
| | 2+2 4 | 2+2 4 | 2+2 4 | 3+3 2 | 4+4 2 | 3+4 2 | 3+4 2 | 2+2 4 |
| | 2+2 4 | 2+2 4 | 2+2 4 | 4+4 2 | 3+3 2 | 3+4 2 | 5+6 2 | 2+2 4 |
| | | | | 6+6 2 | 6+6 2 | 5+6 2 | 5+6 2 | |
| | | | | 5+5 2 | 3+3 2 | 5+6 2 | 3+4 2 | |
| | | | | 3+3 2 | 5+5 2 | 5+6 2 | 5+6 2 | |
| | | | | 4+4 2 | 4+4 2 | 3+4 2 | 5+6 2 | |
| | | | | 5+5 2 | 5+5 2 | 5+6 2 | 3+4 2 | |
| | | | | 6+6 2 | 6+6 2 | 3+4 2 | 5+6 2 | |
| | | | | 4+4 2 | 6+6 2 | 3+4 2 | 5+6 2 | |
| | | | | 3+3 2 | 5+5 2 | 3+4 2 | 3+4 2 | |
| | | | | 5+5 2 | 3+3 2 | 3+4 2 | 5+6 2 | |
| | | | | 6+6 2 | 4+4 2 | 3+4 2 | 3+4 2 | |
| | | | | 4+4 2 | 5+5 2 | 5+6 2 | 3+4 2 | |
| | | | | 3+3 2 | 6+6 2 | 3+4 2 | 3+4 2 | |
| | | | | 5+5 2 | 4+4 2 | 3+4 2 | 3+4 2 | |
| | | | | 6+6 2 | 3+3 2 | 5+6 2 | 3+4 2 | |

Continued on next column

**TABLE XXVI– continued**

| TQC | ICSD | NoE | SF | | | | | |
|---|---|---|---|---|---|---|---|---|
| | | | | 5+5 2 | 5+5 2 | 5+6 2 | 5+6 2 | |
| | | | | 4+4 2 | 4+4 2 | 5+6 2 | 3+4 2 | |
| | | | | 6+6 2 | 6+6 2 | 3+4 2 | 5+6 2 | |
| | | | | 3+3 2 | 3+3 2 | 5+6 2 | 5+6 2 | |
| | | | | 4+4 2 | 4+4 2 | 3+4 2 | 3+4 2 | |
| | | | | 5+5 2 | 5+5 2 | 5+6 2 | 5+6 2 | |
| | | | | 3+3 2 | 3+3 2 | 3+4 2 | 3+4 2 | |
| | | | | 6+6 2 | 6+6 2 | 5+6 2 | 3+4 2 | |
| | | | | 3+3 2 | 5+5 2 | 3+4 2 | 5+6 2 | |
| | | | | 4+4 2 | 6+6 2 | 5+6 2 | 5+6 2 | |
| | | | | 6+6 2 | 4+4 2 | 3+4 2 | 3+4 2 | |
| | | | | 5+5 2 | 3+3 2 | 3+4 2 | 5+6 2 | |
| | | | | 6+6 2 | 6+6 2 | 5+6 2 | 3+4 2 | |
| | | | | 5+5 2 | 5+5 2 | 5+6 2 | 3+4 2 | |
| | | | | 3+3 2 | 3+3 2 | 3+4 2 | 5+6 2 | |
| | | | | 4+4 2 | 4+4 2 | 5+6 2 | 5+6 2 | |
| 14.2.25 | 62281 | 96 | Ga1La1Ni1 | | | | | |
| | -A | -B | -C | -D | -E | -GM | -Y | -Z |
| | 2+2 4 | 2+2 4 | 2+2 4 | 3+3 2 | 3+3 2 | 3+4 2 | 3+4 2 | 2+2 4 |
| | 2+2 4 | 2+2 4 | 2+2 4 | 4+4 2 | 4+4 2 | 3+4 2 | 3+4 2 | 2+2 4 |
| | 2+2 4 | 2+2 4 | 2+2 4 | 5+5 2 | 5+5 2 | 5+6 2 | 5+6 2 | 2+2 4 |
| | 2+2 4 | 2+2 4 | 2+2 4 | 6+6 2 | 6+6 2 | 5+6 2 | 5+6 2 | 2+2 4 |
| | 2+2 4 | 2+2 4 | 2+2 4 | 3+3 2 | 3+3 2 | 3+4 2 | 3+4 2 | 2+2 4 |
| | 2+2 4 | 2+2 4 | 2+2 4 | 4+4 2 | 4+4 2 | 3+4 2 | 3+4 2 | 2+2 4 |
| | 2+2 4 | 2+2 4 | 2+2 4 | 5+5 2 | 5+5 2 | 5+6 2 | 5+6 2 | 2+2 4 |
| | 2+2 4 | 2+2 4 | 2+2 4 | 6+6 2 | 6+6 2 | 5+6 2 | 5+6 2 | 2+2 4 |
| | 2+2 4 | 2+2 4 | 2+2 4 | 4+4 2 | 4+4 2 | 3+4 2 | 3+4 2 | 2+2 4 |
| | 2+2 4 | 2+2 4 | 2+2 4 | 3+3 2 | 3+3 2 | 3+4 2 | 3+4 2 | 2+2 4 |
| | 2+2 4 | 2+2 4 | 2+2 4 | 6+6 2 | 6+6 2 | 5+6 2 | 5+6 2 | 2+2 4 |
| | 2+2 4 | 2+2 4 | 2+2 4 | 5+5 2 | 5+5 2 | 3+4 2 | 5+6 2 | 2+2 4 |
| | 2+2 4 | 2+2 4 | 2+2 4 | 3+3 2 | 3+3 2 | 5+6 2 | 3+4 2 | 2+2 4 |
| | 2+2 4 | 2+2 4 | 2+2 4 | 4+4 2 | 6+6 2 | 3+4 2 | 5+6 2 | 2+2 4 |
| | 2+2 4 | 2+2 4 | 2+2 4 | 5+5 2 | 4+4 2 | 5+6 2 | 3+4 2 | 2+2 4 |
| | 2+2 4 | 2+2 4 | 2+2 4 | 6+6 2 | 5+5 2 | 5+6 2 | 5+6 2 | 2+2 4 |
| | 2+2 4 | 2+2 4 | 2+2 4 | 6+6 2 | 3+3 2 | 3+4 2 | 3+4 2 | 2+2 4 |
| | 2+2 4 | 2+2 4 | 2+2 4 | 5+5 2 | 4+4 2 | 3+4 2 | 5+6 2 | 2+2 4 |
| | 2+2 4 | 2+2 4 | 2+2 4 | 3+3 2 | 5+5 2 | 5+6 2 | 5+6 2 | 2+2 4 |
| | 2+2 4 | 2+2 4 | 2+2 4 | 4+4 2 | 6+6 2 | 5+6 2 | 3+4 2 | 2+2 4 |
| | 2+2 4 | 2+2 4 | 2+2 4 | 6+6 2 | 6+6 2 | 3+4 2 | 3+4 2 | 2+2 4 |
| | 2+2 4 | 2+2 4 | 2+2 4 | 5+5 2 | 5+5 2 | 5+6 2 | 5+6 2 | 2+2 4 |
| | 2+2 4 | 2+2 4 | 2+2 4 | 3+3 2 | 4+4 2 | 3+4 2 | 5+6 2 | 2+2 4 |
| | 2+2 4 | 2+2 4 | 2+2 4 | 4+4 2 | 3+3 2 | 5+6 2 | 5+6 2 | 2+2 4 |
| | | | | 4+4 2 | 4+4 2 | 5+6 2 | 3+4 2 | |
| | | | | 3+3 2 | 3+3 2 | 3+4 2 | 5+6 2 | |
| | | | | 5+5 2 | 5+5 2 | 5+6 2 | 3+4 2 | |
| | | | | 6+6 2 | 6+6 2 | 5+6 2 | 3+4 2 | |
| | | | | 3+3 2 | 4+4 2 | 3+4 2 | 5+6 2 | |
| | | | | 4+4 2 | 3+3 2 | 3+4 2 | 3+4 2 | |
| | | | | 5+5 2 | 6+6 2 | 5+6 2 | 5+6 2 | |
| | | | | 6+6 2 | 4+4 2 | 5+6 2 | 5+6 2 | |
| | | | | 4+4 2 | 5+5 2 | 3+4 2 | 5+6 2 | |
| | | | | 3+3 2 | 3+3 2 | 5+6 2 | 5+6 2 | |
| | | | | 5+5 2 | 6+6 2 | 3+4 2 | 3+4 2 | |
| | | | | 6+6 2 | 5+5 2 | 5+6 2 | 3+4 2 | |
| | | | | 4+4 2 | 3+3 2 | 3+4 2 | 3+4 2 | |
| | | | | 3+3 2 | 4+4 2 | 5+6 2 | 5+6 2 | |
| | | | | 6+6 2 | 5+5 2 | 3+4 2 | 5+6 2 | |
| | | | | 5+5 2 | 6+6 2 | 3+4 2 | 3+4 2 | |
| | | | | 5+5 2 | 3+3 2 | 3+4 2 | 3+4 2 | |
| | | | | 6+6 2 | 4+4 2 | 3+4 2 | 3+4 2 | |
| | | | | 3+3 2 | 6+6 2 | 5+6 2 | 5+6 2 | |

Continued on next column

**TABLE XXVI– continued**

| TQC | ICSD | NoE | SF | | | | | |
|---|---|---|---|---|---|---|---|---|
| | | | | 4+4 2 | 5+5 2 | 5+6 2 | 3+4 2 | |
| | | | | 5+5 2 | 3+3 2 | 3+4 2 | 5+6 2 | |
| | | | | 6+6 2 | 4+4 2 | 3+4 2 | 5+6 2 | |
| | | | | 5+5 2 | 6+6 2 | 5+6 2 | 3+4 2 | |
| | | | | 6+6 2 | 5+5 2 | 3+4 2 | 3+4 2 | |
| 14.2.26 | 189438 | 136 | Fe2O3 | | | | | |
| | -A | -B | -C | -D | -E | -GM | -Y | -Z |
| | 2+2 4 | 2+2 4 | 2+2 4 | 5+5 2 | 3+3 2 | 3+4 2 | 5+6 2 | 2+2 4 |
| | 2+2 4 | 2+2 4 | 2+2 4 | 6+6 2 | 4+4 2 | 3+4 2 | 5+6 2 | 2+2 4 |
| | 2+2 4 | 2+2 4 | 2+2 4 | 3+3 2 | 5+5 2 | 5+6 2 | 3+4 2 | 2+2 4 |
| | 2+2 4 | 2+2 4 | 2+2 4 | 4+4 2 | 6+6 2 | 3+4 2 | 3+4 2 | 2+2 4 |
| | 2+2 4 | 2+2 4 | 2+2 4 | 5+5 2 | 3+3 2 | 3+4 2 | 3+4 2 | 2+2 4 |
| | 2+2 4 | 2+2 4 | 2+2 4 | 6+6 2 | 4+4 2 | 5+6 2 | 5+6 2 | 2+2 4 |
| | 2+2 4 | 2+2 4 | 2+2 4 | 4+4 2 | 6+6 2 | 5+6 2 | 5+6 2 | 2+2 4 |
| | 2+2 4 | 2+2 4 | 2+2 4 | 3+3 2 | 5+5 2 | 3+4 2 | 3+4 2 | 2+2 4 |
| | 2+2 4 | 2+2 4 | 2+2 4 | 3+3 2 | 3+3 2 | 3+4 2 | 5+6 2 | 2+2 4 |
| | 2+2 4 | 2+2 4 | 2+2 4 | 4+4 2 | 4+4 2 | 5+6 2 | 3+4 2 | 2+2 4 |
| | 2+2 4 | 2+2 4 | 2+2 4 | 6+6 2 | 5+5 2 | 5+6 2 | 5+6 2 | 2+2 4 |
| | 2+2 4 | 2+2 4 | 2+2 4 | 5+5 2 | 6+6 2 | 5+6 2 | 3+4 2 | 2+2 4 |
| | 2+2 4 | 2+2 4 | 2+2 4 | 5+5 2 | 3+3 2 | 3+4 2 | 5+6 2 | 2+2 4 |
| | 2+2 4 | 2+2 4 | 2+2 4 | 6+6 2 | 4+4 2 | 3+4 2 | 5+6 2 | 2+2 4 |
| | 2+2 4 | 2+2 4 | 2+2 4 | 3+3 2 | 3+3 2 | 3+4 2 | 3+4 2 | 2+2 4 |
| | 2+2 4 | 2+2 4 | 2+2 4 | 4+4 2 | 4+4 2 | 3+4 2 | 3+4 2 | 2+2 4 |
| | 2+2 4 | 2+2 4 | 2+2 4 | 6+6 2 | 3+3 2 | 3+4 2 | 5+6 2 | 2+2 4 |
| | 2+2 4 | 2+2 4 | 2+2 4 | 5+5 2 | 4+4 2 | 3+4 2 | 3+4 2 | 2+2 4 |
| | 2+2 4 | 2+2 4 | 2+2 4 | 3+3 2 | 5+5 2 | 5+6 2 | 5+6 2 | 2+2 4 |
| | 2+2 4 | 2+2 4 | 2+2 4 | 4+4 2 | 6+6 2 | 3+4 2 | 3+4 2 | 2+2 4 |
| | 2+2 4 | 2+2 4 | 2+2 4 | 5+5 2 | 3+3 2 | 5+6 2 | 5+6 2 | 2+2 4 |
| | 2+2 4 | 2+2 4 | 2+2 4 | 6+6 2 | 4+4 2 | 3+4 2 | 3+4 2 | 2+2 4 |
| | 2+2 4 | 2+2 4 | 2+2 4 | 3+3 2 | 5+5 2 | 3+4 2 | 5+6 2 | 2+2 4 |
| | 2+2 4 | 2+2 4 | 2+2 4 | 4+4 2 | 6+6 2 | 5+6 2 | 5+6 2 | 2+2 4 |
| | 2+2 4 | 2+2 4 | 2+2 4 | 5+5 2 | 3+3 2 | 3+4 2 | 3+4 2 | 2+2 4 |
| | 2+2 4 | 2+2 4 | 2+2 4 | 6+6 2 | 4+4 2 | 5+6 2 | 5+6 2 | 2+2 4 |
| | 2+2 4 | 2+2 4 | 2+2 4 | 3+3 2 | 5+5 2 | 5+6 2 | 5+6 2 | 2+2 4 |
| | 2+2 4 | 2+2 4 | 2+2 4 | 4+4 2 | 6+6 2 | 3+4 2 | 3+4 2 | 2+2 4 |
| | 2+2 4 | 2+2 4 | 2+2 4 | 5+5 2 | 5+5 2 | 5+6 2 | 5+6 2 | 2+2 4 |
| | 2+2 4 | 2+2 4 | 2+2 4 | 6+6 2 | 6+6 2 | 5+6 2 | 3+4 2 | 2+2 4 |
| | 2+2 4 | 2+2 4 | 2+2 4 | 5+5 2 | 3+3 2 | 3+4 2 | 3+4 2 | 2+2 4 |
| | 2+2 4 | 2+2 4 | 2+2 4 | 6+6 2 | 4+4 2 | 3+4 2 | 3+4 2 | 2+2 4 |
| | 2+2 4 | 2+2 4 | 2+2 4 | 4+4 2 | 5+5 2 | 3+4 2 | 5+6 2 | 2+2 4 |
| | 2+2 4 | 2+2 4 | 2+2 4 | 3+3 2 | 6+6 2 | 5+6 2 | 5+6 2 | 2+2 4 |
| | | | | 3+3 2 | 3+3 2 | 5+6 2 | 3+4 2 | |
| | | | | 4+4 2 | 4+4 2 | 3+4 2 | 3+4 2 | |
| | | | | 4+4 2 | 6+6 2 | 5+6 2 | 5+6 2 | |
| | | | | 3+3 2 | 5+5 2 | 5+6 2 | 3+4 2 | |
| | | | | 5+5 2 | 3+3 2 | 3+4 2 | 3+4 2 | |
| | | | | 6+6 2 | 4+4 2 | 3+4 2 | 5+6 2 | |
| | | | | 6+6 2 | 5+5 2 | 3+4 2 | 5+6 2 | |
| | | | | 5+5 2 | 6+6 2 | 5+6 2 | 3+4 2 | |
| | | | | 3+3 2 | 4+4 2 | 5+6 2 | 3+4 2 | |
| | | | | 4+4 2 | 3+3 2 | 5+6 2 | 5+6 2 | |
| | | | | 4+4 2 | 6+6 2 | 5+6 2 | 3+4 2 | |
| | | | | 3+3 2 | 5+5 2 | 5+6 2 | 3+4 2 | |
| | | | | 5+5 2 | 5+5 2 | 5+6 2 | 5+6 2 | |
| | | | | 6+6 2 | 6+6 2 | 5+6 2 | 5+6 2 | |
| | | | | 6+6 2 | 3+3 2 | 3+4 2 | 3+4 2 | |
| | | | | 5+5 2 | 4+4 2 | 3+4 2 | 5+6 2 | |
| | | | | 3+3 2 | 4+4 2 | 3+4 2 | 5+6 2 | |
| | | | | 4+4 2 | 3+3 2 | 3+4 2 | 3+4 2 | |
| | | | | 6+6 2 | 3+3 2 | 3+4 2 | 5+6 2 | |
| | | | | 5+5 2 | 4+4 2 | 3+4 2 | 3+4 2 | |

Continued on next column

**TABLE XXVI– continued**

| TQC | ICSD | NoE | SF | | | | | |
|---|---|---|---|---|---|---|---|---|
| | | | | 3+3 2 | 3+3 2 | 3+4 2 | 3+4 2 | |
| | | | | 4+4 2 | 4+4 2 | 3+4 2 | 5+6 2 | |
| | | | | 6+6 2 | 4+4 2 | 5+6 2 | 5+6 2 | |
| | | | | 3+3 2 | 3+3 2 | 5+6 2 | 3+4 2 | |
| | | | | 4+4 2 | 5+5 2 | 3+4 2 | 5+6 2 | |
| | | | | 5+5 2 | 6+6 2 | 5+6 2 | 3+4 2 | |
| | | | | 3+3 2 | 4+4 2 | 3+4 2 | 3+4 2 | |
| | | | | 4+4 2 | 3+3 2 | 3+4 2 | 5+6 2 | |
| | | | | 5+5 2 | 5+5 2 | 3+4 2 | 3+4 2 | |
| | | | | 6+6 2 | 6+6 2 | 3+4 2 | 5+6 2 | |
| | | | | 4+4 2 | 3+3 2 | 5+6 2 | 3+4 2 | |
| | | | | 3+3 2 | 4+4 2 | 5+6 2 | 5+6 2 | |
| | | | | 5+5 2 | 6+6 2 | 3+4 2 | 3+4 2 | |
| | | | | 6+6 2 | 5+5 2 | 5+6 2 | 3+4 2 | |
| 14.2.27 | 154372 | 164 | Fe2K4O5 | | | | | |
| | -A | -B | -C | -D | -E | -GM | -Y | -Z |
| | 2+2 4 | 2+2 4 | 2+2 4 | 3+3 2 | 3+3 2 | 3+4 2 | 3+4 2 | 2+2 4 |
| | 2+2 4 | 2+2 4 | 2+2 4 | 4+4 2 | 4+4 2 | 5+6 2 | 5+6 2 | 2+2 4 |
| | 2+2 4 | 2+2 4 | 2+2 4 | 5+5 2 | 5+5 2 | 3+4 2 | 3+4 2 | 2+2 4 |
| | 2+2 4 | 2+2 4 | 2+2 4 | 6+6 2 | 6+6 2 | 5+6 2 | 5+6 2 | 2+2 4 |
| | 2+2 4 | 2+2 4 | 2+2 4 | 5+5 2 | 3+3 2 | 3+4 2 | 5+6 2 | 2+2 4 |
| | 2+2 4 | 2+2 4 | 2+2 4 | 6+6 2 | 4+4 2 | 3+4 2 | 5+6 2 | 2+2 4 |
| | 2+2 4 | 2+2 4 | 2+2 4 | 3+3 2 | 5+5 2 | 5+6 2 | 3+4 2 | 2+2 4 |
| | 2+2 4 | 2+2 4 | 2+2 4 | 4+4 2 | 6+6 2 | 5+6 2 | 3+4 2 | 2+2 4 |
| | 2+2 4 | 2+2 4 | 2+2 4 | 3+3 2 | 5+5 2 | 3+4 2 | 5+6 2 | 2+2 4 |
| | 2+2 4 | 2+2 4 | 2+2 4 | 4+4 2 | 6+6 2 | 5+6 2 | 3+4 2 | 2+2 4 |
| | 2+2 4 | 2+2 4 | 2+2 4 | 5+5 2 | 3+3 2 | 3+4 2 | 5+6 2 | 2+2 4 |
| | 2+2 4 | 2+2 4 | 2+2 4 | 6+6 2 | 4+4 2 | 3+4 2 | 5+6 2 | 2+2 4 |
| | 2+2 4 | 2+2 4 | 2+2 4 | 3+3 2 | 5+5 2 | 5+6 2 | 3+4 2 | 2+2 4 |
| | 2+2 4 | 2+2 4 | 2+2 4 | 4+4 2 | 6+6 2 | 3+4 2 | 5+6 2 | 2+2 4 |
| | 2+2 4 | 2+2 4 | 2+2 4 | 3+3 2 | 3+3 2 | 3+4 2 | 3+4 2 | 2+2 4 |
| | 2+2 4 | 2+2 4 | 2+2 4 | 4+4 2 | 4+4 2 | 5+6 2 | 5+6 2 | 2+2 4 |
| | 2+2 4 | 2+2 4 | 2+2 4 | 5+5 2 | 5+5 2 | 5+6 2 | 5+6 2 | 2+2 4 |
| | 2+2 4 | 2+2 4 | 2+2 4 | 6+6 2 | 6+6 2 | 3+4 2 | 3+4 2 | 2+2 4 |
| | 2+2 4 | 2+2 4 | 2+2 4 | 5+5 2 | 4+4 2 | 3+4 2 | 3+4 2 | 2+2 4 |
| | 2+2 4 | 2+2 4 | 2+2 4 | 6+6 2 | 3+3 2 | 5+6 2 | 5+6 2 | 2+2 4 |
| | 2+2 4 | 2+2 4 | 2+2 4 | 4+4 2 | 5+5 2 | 3+4 2 | 5+6 2 | 2+2 4 |
| | 2+2 4 | 2+2 4 | 2+2 4 | 3+3 2 | 6+6 2 | 3+4 2 | 3+4 2 | 2+2 4 |
| | 2+2 4 | 2+2 4 | 2+2 4 | 6+6 2 | 4+4 2 | 5+6 2 | 3+4 2 | 2+2 4 |
| | 2+2 4 | 2+2 4 | 2+2 4 | 5+5 2 | 3+3 2 | 3+4 2 | 5+6 2 | 2+2 4 |
| | 2+2 4 | 2+2 4 | 2+2 4 | 3+3 2 | 6+6 2 | 5+6 2 | 3+4 2 | 2+2 4 |
| | 2+2 4 | 2+2 4 | 2+2 4 | 4+4 2 | 5+5 2 | 3+4 2 | 3+4 2 | 2+2 4 |
| | 2+2 4 | 2+2 4 | 2+2 4 | 6+6 2 | 3+3 2 | 5+6 2 | 5+6 2 | 2+2 4 |
| | 2+2 4 | 2+2 4 | 2+2 4 | 5+5 2 | 4+4 2 | 3+4 2 | 5+6 2 | 2+2 4 |
| | 2+2 4 | 2+2 4 | 2+2 4 | 3+3 2 | 5+5 2 | 5+6 2 | 3+4 2 | 2+2 4 |
| | 2+2 4 | 2+2 4 | 2+2 4 | 5+5 2 | 6+6 2 | 3+4 2 | 3+4 2 | 2+2 4 |
| | 2+2 4 | 2+2 4 | 2+2 4 | 4+4 2 | 3+3 2 | 5+6 2 | 5+6 2 | 2+2 4 |
| | 2+2 4 | 2+2 4 | 2+2 4 | 6+6 2 | 4+4 2 | 3+4 2 | 3+4 2 | 2+2 4 |
| | 2+2 4 | 2+2 4 | 2+2 4 | 5+5 2 | 6+6 2 | 5+6 2 | 5+6 2 | 2+2 4 |
| | 2+2 4 | 2+2 4 | 2+2 4 | 4+4 2 | 5+5 2 | 3+4 2 | 3+4 2 | 2+2 4 |
| | 2+2 4 | 2+2 4 | 2+2 4 | 6+6 2 | 5+5 2 | 3+4 2 | 5+6 2 | 2+2 4 |
| | 2+2 4 | 2+2 4 | 2+2 4 | 3+3 2 | 4+4 2 | 5+6 2 | 3+4 2 | 2+2 4 |
| | 2+2 4 | 2+2 4 | 2+2 4 | 4+4 2 | 6+6 2 | 5+6 2 | 5+6 2 | 2+2 4 |
| | 2+2 4 | 2+2 4 | 2+2 4 | 3+3 2 | 3+3 2 | 5+6 2 | 5+6 2 | 2+2 4 |
| | 2+2 4 | 2+2 4 | 2+2 4 | 5+5 2 | 4+4 2 | 5+6 2 | 5+6 2 | 2+2 4 |
| | 2+2 4 | 2+2 4 | 2+2 4 | 6+6 2 | 3+3 2 | 5+6 2 | 3+4 2 | 2+2 4 |
| | 2+2 4 | 2+2 4 | 2+2 4 | 3+3 2 | 5+5 2 | 3+4 2 | 5+6 2 | 2+2 4 |
| | | | | 4+4 2 | 6+6 2 | 3+4 2 | 3+4 2 | |
| | | | | 5+5 2 | 4+4 2 | 5+6 2 | 3+4 2 | |
| | | | | 6+6 2 | 3+3 2 | 3+4 2 | 5+6 2 | |
| | | | | 3+3 2 | 5+5 2 | 5+6 2 | 3+4 2 | |

Continued on next column

**TABLE XXVI– continued**

| TQC | ICSD | NoE | SF | | | | | |
|---|---|---|---|---|---|---|---|---|
| | | | | 4+4 2 | 6+6 2 | 5+6 2 | 3+4 2 | |
| | | | | 5+5 2 | 3+3 2 | 5+6 2 | 3+4 2 | |
| | | | | 6+6 2 | 4+4 2 | 3+4 2 | 5+6 2 | |
| | | | | 5+5 2 | 4+4 2 | 5+6 2 | 3+4 2 | |
| | | | | 6+6 2 | 3+3 2 | 3+4 2 | 3+4 2 | |
| | | | | 3+3 2 | 5+5 2 | 5+6 2 | 3+4 2 | |
| | | | | 4+4 2 | 6+6 2 | 5+6 2 | 5+6 2 | |
| | | | | 5+5 2 | 3+3 2 | 3+4 2 | 5+6 2 | |
| | | | | 6+6 2 | 4+4 2 | 5+6 2 | 5+6 2 | |
| | | | | 5+5 2 | 4+4 2 | 3+4 2 | 3+4 2 | |
| | | | | 6+6 2 | 3+3 2 | 5+6 2 | 5+6 2 | |
| | | | | 4+4 2 | 6+6 2 | 3+4 2 | 3+4 2 | |
| | | | | 3+3 2 | 5+5 2 | 3+4 2 | 5+6 2 | |
| | | | | 4+4 2 | 3+3 2 | 5+6 2 | 3+4 2 | |
| | | | | 3+3 2 | 4+4 2 | 5+6 2 | 3+4 2 | |
| | | | | 6+6 2 | 5+5 2 | 5+6 2 | 3+4 2 | |
| | | | | 5+5 2 | 6+6 2 | 5+6 2 | 3+4 2 | |
| | | | | 6+6 2 | 3+3 2 | 3+4 2 | 5+6 2 | |
| | | | | 5+5 2 | 4+4 2 | 3+4 2 | 5+6 2 | |
| | | | | 3+3 2 | 6+6 2 | 3+4 2 | 5+6 2 | |
| | | | | 4+4 2 | 5+5 2 | 3+4 2 | 3+4 2 | |
| | | | | 4+4 2 | 4+4 2 | 5+6 2 | 3+4 2 | |
| | | | | 3+3 2 | 3+3 2 | 5+6 2 | 5+6 2 | |
| | | | | 5+5 2 | 5+5 2 | 5+6 2 | 3+4 2 | |
| | | | | 6+6 2 | 6+6 2 | 3+4 2 | 3+4 2 | |
| | | | | 6+6 2 | 3+3 2 | 5+6 2 | 5+6 2 | |
| | | | | 5+5 2 | 4+4 2 | 5+6 2 | 3+4 2 | |
| | | | | 4+4 2 | 5+5 2 | 3+4 2 | 5+6 2 | |
| | | | | 3+3 2 | 6+6 2 | 3+4 2 | 5+6 2 | |
| | | | | 3+3 2 | 6+6 2 | 5+6 2 | 3+4 2 | |
| | | | | 4+4 2 | 5+5 2 | 3+4 2 | 5+6 2 | |
| | | | | 6+6 2 | 3+3 2 | 3+4 2 | 5+6 2 | |
| | | | | 5+5 2 | 4+4 2 | 5+6 2 | 3+4 2 | |
| | | | | 4+4 2 | 5+5 2 | 3+4 2 | 5+6 2 | |
| | | | | 3+3 2 | 4+4 2 | 3+4 2 | 5+6 2 | |
| | | | | 5+5 2 | 6+6 2 | 5+6 2 | 3+4 2 | |
| | | | | 6+6 2 | 3+3 2 | 3+4 2 | 5+6 2 | |
| 14.2.28 | 419955 | 216 | F1O3Se1Tm1 | | | | | |
| | -A | -B | -C | -D | -E | -GM | -Y | -Z |
| | 2+2 4 | 2+2 4 | 2+2 4 | 3+3 2 | 3+3 2 | 3+4 2 | 3+4 2 | 2+2 4 |
| | 2+2 4 | 2+2 4 | 2+2 4 | 4+4 2 | 4+4 2 | 5+6 2 | 5+6 2 | 2+2 4 |
| | 2+2 4 | 2+2 4 | 2+2 4 | 5+5 2 | 5+5 2 | 3+4 2 | 3+4 2 | 2+2 4 |
| | 2+2 4 | 2+2 4 | 2+2 4 | 6+6 2 | 6+6 2 | 5+6 2 | 5+6 2 | 2+2 4 |
| | 2+2 4 | 2+2 4 | 2+2 4 | 3+3 2 | 3+3 2 | 3+4 2 | 3+4 2 | 2+2 4 |
| | 2+2 4 | 2+2 4 | 2+2 4 | 6+6 2 | 6+6 2 | 5+6 2 | 5+6 2 | 2+2 4 |
| | 2+2 4 | 2+2 4 | 2+2 4 | 4+4 2 | 4+4 2 | 3+4 2 | 3+4 2 | 2+2 4 |
| | 2+2 4 | 2+2 4 | 2+2 4 | 5+5 2 | 5+5 2 | 5+6 2 | 5+6 2 | 2+2 4 |
| | 2+2 4 | 2+2 4 | 2+2 4 | 5+5 2 | 5+5 2 | 3+4 2 | 3+4 2 | 2+2 4 |
| | 2+2 4 | 2+2 4 | 2+2 4 | 6+6 2 | 6+6 2 | 5+6 2 | 5+6 2 | 2+2 4 |
| | 2+2 4 | 2+2 4 | 2+2 4 | 4+4 2 | 4+4 2 | 3+4 2 | 3+4 2 | 2+2 4 |
| | 2+2 4 | 2+2 4 | 2+2 4 | 3+3 2 | 3+3 2 | 5+6 2 | 5+6 2 | 2+2 4 |
| | 2+2 4 | 2+2 4 | 2+2 4 | 3+3 2 | 3+3 2 | 3+4 2 | 5+6 2 | 2+2 4 |
| | 2+2 4 | 2+2 4 | 2+2 4 | 5+5 2 | 5+5 2 | 5+6 2 | 3+4 2 | 2+2 4 |
| | 2+2 4 | 2+2 4 | 2+2 4 | 4+4 2 | 6+6 2 | 3+4 2 | 3+4 2 | 2+2 4 |
| | 2+2 4 | 2+2 4 | 2+2 4 | 6+6 2 | 4+4 2 | 3+4 2 | 5+6 2 | 2+2 4 |
| | 2+2 4 | 2+2 4 | 2+2 4 | 4+4 2 | 4+4 2 | 5+6 2 | 3+4 2 | 2+2 4 |
| | 2+2 4 | 2+2 4 | 2+2 4 | 6+6 2 | 6+6 2 | 5+6 2 | 5+6 2 | 2+2 4 |
| | 2+2 4 | 2+2 4 | 2+2 4 | 3+3 2 | 3+3 2 | 3+4 2 | 3+4 2 | 2+2 4 |
| | 2+2 4 | 2+2 4 | 2+2 4 | 5+5 2 | 5+5 2 | 5+6 2 | 5+6 2 | 2+2 4 |
| | 2+2 4 | 2+2 4 | 2+2 4 | 4+4 2 | 3+3 2 | 3+4 2 | 3+4 2 | 2+2 4 |
| | 2+2 4 | 2+2 4 | 2+2 4 | 3+3 2 | 5+5 2 | 5+6 2 | 5+6 2 | 2+2 4 |

Continued on next column

**TABLE XXVI– continued**

| TQC | ICSD | NoE | SF |
|---|---|---|---|
| | 2+2 4 2+2 4 2+2 4 6+6 2 6+6 2 3+4 2 3+4 2 2+2 4 | | |
| | 2+2 4 2+2 4 2+2 4 5+5 2 4+4 2 5+6 2 5+6 2 2+2 4 | | |
| | 2+2 4 2+2 4 2+2 4 3+3 2 4+4 2 3+4 2 3+4 2 2+2 4 | | |
| | 2+2 4 2+2 4 2+2 4 4+4 2 3+3 2 5+6 2 5+6 2 2+2 4 | | |
| | 2+2 4 2+2 4 2+2 4 5+5 2 5+5 2 5+6 2 5+6 2 2+2 4 | | |
| | 2+2 4 2+2 4 2+2 4 6+6 2 6+6 2 3+4 2 3+4 2 2+2 4 | | |
| | 2+2 4 2+2 4 2+2 4 6+6 2 6+6 2 3+4 2 5+6 2 2+2 4 | | |
| | 2+2 4 2+2 4 2+2 4 5+5 2 5+5 2 5+6 2 3+4 2 2+2 4 | | |
| | 2+2 4 2+2 4 2+2 4 4+4 2 3+3 2 5+6 2 5+6 2 2+2 4 | | |
| | 2+2 4 2+2 4 2+2 4 3+3 2 4+4 2 3+4 2 3+4 2 2+2 4 | | |
| | 2+2 4 2+2 4 2+2 4 5+5 2 3+3 2 3+4 2 5+6 2 2+2 4 | | |
| | 2+2 4 2+2 4 2+2 4 6+6 2 4+4 2 5+6 2 3+4 2 2+2 4 | | |
| | 2+2 4 2+2 4 2+2 4 4+4 2 5+5 2 3+4 2 3+4 2 2+2 4 | | |
| | 2+2 4 2+2 4 2+2 4 3+3 2 6+6 2 5+6 2 5+6 2 2+2 4 | | |
| | 2+2 4 2+2 4 2+2 4 3+3 2 3+3 2 3+4 2 5+6 2 2+2 4 | | |
| | 2+2 4 2+2 4 2+2 4 6+6 2 4+4 2 5+6 2 3+4 2 2+2 4 | | |
| | 2+2 4 2+2 4 2+2 4 4+4 2 5+5 2 5+6 2 5+6 2 2+2 4 | | |
| | 2+2 4 2+2 4 2+2 4 5+5 2 6+6 2 3+4 2 3+4 2 2+2 4 | | |
| | 2+2 4 2+2 4 2+2 4 4+4 2 4+4 2 5+6 2 3+4 2 2+2 4 | | |
| | 2+2 4 2+2 4 2+2 4 3+3 2 3+3 2 3+4 2 5+6 2 2+2 4 | | |
| | 2+2 4 2+2 4 2+2 4 5+5 2 5+5 2 5+6 2 3+4 2 2+2 4 | | |
| | 2+2 4 2+2 4 2+2 4 6+6 2 6+6 2 3+4 2 3+4 2 2+2 4 | | |
| | 2+2 4 2+2 4 2+2 4 5+5 2 4+4 2 3+4 2 5+6 2 2+2 4 | | |
| | 2+2 4 2+2 4 2+2 4 6+6 2 3+3 2 5+6 2 5+6 2 2+2 4 | | |
| | 2+2 4 2+2 4 2+2 4 3+3 2 6+6 2 3+4 2 5+6 2 2+2 4 | | |
| | 2+2 4 2+2 4 2+2 4 4+4 2 5+5 2 3+4 2 3+4 2 2+2 4 | | |
| | 2+2 4 2+2 4 2+2 4 3+3 2 6+6 2 5+6 2 5+6 2 2+2 4 | | |
| | 2+2 4 2+2 4 2+2 4 4+4 2 4+4 2 5+6 2 3+4 2 2+2 4 | | |
| | 2+2 4 2+2 4 2+2 4 5+5 2 3+3 2 3+4 2 3+4 2 2+2 4 | | |
| | 2+2 4 2+2 4 2+2 4 6+6 2 5+5 2 3+4 2 5+6 2 2+2 4 | | |
| | 2+2 4 2+2 4 2+2 4 6+6 2 5+5 2 5+6 2 3+4 2 2+2 4 | | |
| | 2+2 4 2+2 4 2+2 4 5+5 2 6+6 2 3+4 2 3+4 2 2+2 4 | | |
| | | | 4+4 2 4+4 2 5+6 2 5+6 2 |
| | | | 3+3 2 3+3 2 5+6 2 5+6 2 |
| | | | 3+3 2 5+5 2 3+4 2 3+4 2 |
| | | | 5+5 2 6+6 2 5+6 2 3+4 2 |
| | | | 4+4 2 6+6 2 3+4 2 5+6 2 |
| | | | 6+6 2 5+5 2 3+4 2 5+6 2 |
| | | | 6+6 2 3+3 2 3+4 2 3+4 2 |
| | | | 3+3 2 4+4 2 5+6 2 5+6 2 |
| | | | 5+5 2 4+4 2 5+6 2 3+4 2 |
| | | | 4+4 2 3+3 2 5+6 2 5+6 2 |
| | | | 5+5 2 4+4 2 3+4 2 5+6 2 |
| | | | 4+4 2 3+3 2 3+4 2 3+4 2 |
| | | | 6+6 2 6+6 2 5+6 2 3+4 2 |
| | | | 3+3 2 5+5 2 5+6 2 5+6 2 |
| | | | 4+4 2 5+5 2 5+6 2 3+4 2 |
| | | | 3+3 2 6+6 2 3+4 2 5+6 2 |
| | | | 6+6 2 4+4 2 5+6 2 3+4 2 |
| | | | 5+5 2 3+3 2 3+4 2 5+6 2 |
| | | | 3+3 2 3+3 2 3+4 2 3+4 2 |
| | | | 4+4 2 4+4 2 3+4 2 5+6 2 |
| | | | 5+5 2 5+5 2 5+6 2 3+4 2 |
| | | | 6+6 2 6+6 2 3+4 2 5+6 2 |
| | | | 4+4 2 3+3 2 5+6 2 3+4 2 |
| | | | 3+3 2 4+4 2 3+4 2 5+6 2 |
| | | | 6+6 2 5+5 2 5+6 2 3+4 2 |
| | | | 5+5 2 6+6 2 5+6 2 5+6 2 |
| | | | 5+5 2 3+3 2 3+4 2 3+4 2 |
| | | | 3+3 2 4+4 2 3+4 2 5+6 2 |
| | | | 4+4 2 6+6 2 5+6 2 3+4 2 |

Continued on next column

**TABLE XXVI– continued**

| TQC | ICSD | NoE | SF |
|---|---|---|---|
| | | | 6+6 2 5+5 2 5+6 2 5+6 2 |
| | | | 3+3 2 4+4 2 3+4 2 3+4 2 |
| | | | 4+4 2 3+3 2 3+4 2 5+6 2 |
| | | | 6+6 2 5+5 2 5+6 2 3+4 2 |
| | | | 5+5 2 6+6 2 5+6 2 3+4 2 |
| | | | 6+6 2 3+3 2 3+4 2 5+6 2 |
| | | | 4+4 2 4+4 2 3+4 2 5+6 2 |
| | | | 3+3 2 6+6 2 5+6 2 3+4 2 |
| | | | 5+5 2 5+5 2 3+4 2 5+6 2 |
| | | | 6+6 2 4+4 2 5+6 2 5+6 2 |
| | | | 5+5 2 3+3 2 5+6 2 3+4 2 |
| | | | 3+3 2 6+6 2 3+4 2 3+4 2 |
| | | | 4+4 2 5+5 2 5+6 2 5+6 2 |
| | | | 4+4 2 3+3 2 3+4 2 3+4 2 |
| | | | 3+3 2 4+4 2 3+4 2 5+6 2 |
| | | | 5+5 2 5+5 2 5+6 2 3+4 2 |
| | | | 4+4 2 4+4 2 5+6 2 5+6 2 |
| | | | 5+5 2 6+6 2 3+4 2 5+6 2 |
| | | | 3+3 2 3+3 2 3+4 2 3+4 2 |
| | | | 6+6 2 5+5 2 5+6 2 3+4 2 |
| | | | 6+6 2 6+6 2 5+6 2 5+6 2 |
| | | | 3+3 2 4+4 2 3+4 2 3+4 2 |
| | | | 4+4 2 3+3 2 3+4 2 3+4 2 |
| | | | 5+5 2 5+5 2 5+6 2 5+6 2 |
| | | | 6+6 2 6+6 2 3+4 2 5+6 2 |
| 14.2.29 | 412767 | 160 | Eu2P2Se6 |
| | -A -B -C -D -E -GM -Y -Z | | |
| | 2+2 4 2+2 4 2+2 4 3+3 2 3+3 2 3+4 2 3+4 2 2+2 4 | | |
| | 2+2 4 2+2 4 2+2 4 4+4 2 4+4 2 3+4 2 3+4 2 2+2 4 | | |
| | 2+2 4 2+2 4 2+2 4 5+5 2 5+5 2 5+6 2 5+6 2 2+2 4 | | |
| | 2+2 4 2+2 4 2+2 4 6+6 2 6+6 2 5+6 2 5+6 2 2+2 4 | | |
| | 2+2 4 2+2 4 2+2 4 3+3 2 5+5 2 3+4 2 3+4 2 2+2 4 | | |
| | 2+2 4 2+2 4 2+2 4 4+4 2 6+6 2 5+6 2 5+6 2 2+2 4 | | |
| | 2+2 4 2+2 4 2+2 4 5+5 2 3+3 2 3+4 2 3+4 2 2+2 4 | | |
| | 2+2 4 2+2 4 2+2 4 6+6 2 4+4 2 5+6 2 5+6 2 2+2 4 | | |
| | 2+2 4 2+2 4 2+2 4 5+5 2 4+4 2 5+6 2 5+6 2 2+2 4 | | |
| | 2+2 4 2+2 4 2+2 4 4+4 2 3+3 2 3+4 2 3+4 2 2+2 4 | | |
| | 2+2 4 2+2 4 2+2 4 6+6 2 5+5 2 3+4 2 5+6 2 2+2 4 | | |
| | 2+2 4 2+2 4 2+2 4 3+3 2 6+6 2 3+4 2 3+4 2 2+2 4 | | |
| | 2+2 4 2+2 4 2+2 4 3+3 2 5+5 2 5+6 2 3+4 2 2+2 4 | | |
| | 2+2 4 2+2 4 2+2 4 4+4 2 6+6 2 5+6 2 5+6 2 2+2 4 | | |
| | 2+2 4 2+2 4 2+2 4 6+6 2 3+3 2 3+4 2 5+6 2 2+2 4 | | |
| | 2+2 4 2+2 4 2+2 4 5+5 2 4+4 2 5+6 2 3+4 2 2+2 4 | | |
| | 2+2 4 2+2 4 2+2 4 5+5 2 3+3 2 3+4 2 5+6 2 2+2 4 | | |
| | 2+2 4 2+2 4 2+2 4 6+6 2 4+4 2 3+4 2 5+6 2 2+2 4 | | |
| | 2+2 4 2+2 4 2+2 4 4+4 2 6+6 2 5+6 2 3+4 2 2+2 4 | | |
| | 2+2 4 2+2 4 2+2 4 3+3 2 5+5 2 5+6 2 3+4 2 2+2 4 | | |
| | 2+2 4 2+2 4 2+2 4 3+3 2 5+5 2 5+6 2 3+4 2 2+2 4 | | |
| | 2+2 4 2+2 4 2+2 4 4+4 2 6+6 2 3+4 2 3+4 2 2+2 4 | | |
| | 2+2 4 2+2 4 2+2 4 3+3 2 5+5 2 5+6 2 3+4 2 2+2 4 | | |
| | 2+2 4 2+2 4 2+2 4 4+4 2 6+6 2 5+6 2 5+6 2 2+2 4 | | |
| | 2+2 4 2+2 4 2+2 4 6+6 2 4+4 2 5+6 2 5+6 2 2+2 4 | | |
| | 2+2 4 2+2 4 2+2 4 5+5 2 3+3 2 3+4 2 3+4 2 2+2 4 | | |
| | 2+2 4 2+2 4 2+2 4 5+5 2 3+3 2 3+4 2 5+6 2 2+2 4 | | |
| | 2+2 4 2+2 4 2+2 4 6+6 2 4+4 2 3+4 2 5+6 2 2+2 4 | | |
| | 2+2 4 2+2 4 2+2 4 5+5 2 3+3 2 3+4 2 5+6 2 2+2 4 | | |
| | 2+2 4 2+2 4 2+2 4 6+6 2 4+4 2 3+4 2 5+6 2 2+2 4 | | |
| | 2+2 4 2+2 4 2+2 4 3+3 2 6+6 2 5+6 2 3+4 2 2+2 4 | | |
| | 2+2 4 2+2 4 2+2 4 4+4 2 5+5 2 5+6 2 3+4 2 2+2 4 | | |
| | 2+2 4 2+2 4 2+2 4 5+5 2 4+4 2 3+4 2 3+4 2 2+2 4 | | |
| | 2+2 4 2+2 4 2+2 4 6+6 2 3+3 2 3+4 2 5+6 2 2+2 4 | | |

Continued on next column

**TABLE XXVI– continued**

| TQC | ICSD | NoE | SF | | | | |
|---|---|---|---|---|---|---|---|
| | 2+2 4 | 2+2 4 | 2+2 4 | 3+3 2 | 5+5 2 | 5+6 2 | 3+4 2 | 2+2 4 |
| | 2+2 4 | 2+2 4 | 2+2 4 | 4+4 2 | 6+6 2 | 5+6 2 | 5+6 2 | 2+2 4 |
| | 2+2 4 | 2+2 4 | 2+2 4 | 4+4 2 | 6+6 2 | 5+6 2 | 3+4 2 | 2+2 4 |
| | 2+2 4 | 2+2 4 | 2+2 4 | 3+3 2 | 5+5 2 | 5+6 2 | 3+4 2 | 2+2 4 |
| | 2+2 4 | 2+2 4 | 2+2 4 | 6+6 2 | 4+4 2 | 3+4 2 | 5+6 2 | 2+2 4 |
| | 2+2 4 | 2+2 4 | 2+2 4 | 5+5 2 | 3+3 2 | 3+4 2 | 5+6 2 | 2+2 4 |
| | | | | 5+5 2 | 3+3 2 | 3+4 2 | 5+6 2 | |
| | | | | 6+6 2 | 4+4 2 | 3+4 2 | 5+6 2 | |
| | | | | 3+3 2 | 5+5 2 | 3+4 2 | 3+4 2 | |
| | | | | 4+4 2 | 6+6 2 | 5+6 2 | 5+6 2 | |
| | | | | 5+5 2 | 4+4 2 | 5+6 2 | 3+4 2 | |
| | | | | 6+6 2 | 5+5 2 | 3+4 2 | 5+6 2 | |
| | | | | 4+4 2 | 6+6 2 | 3+4 2 | 5+6 2 | |
| | | | | 3+3 2 | 3+3 2 | 5+6 2 | 3+4 2 | |
| | | | | 6+6 2 | 6+6 2 | 3+4 2 | 3+4 2 | |
| | | | | 5+5 2 | 5+5 2 | 3+4 2 | 5+6 2 | |
| | | | | 4+4 2 | 3+3 2 | 5+6 2 | 3+4 2 | |
| | | | | 3+3 2 | 6+6 2 | 3+4 2 | 3+4 2 | |
| | | | | 4+4 2 | 5+5 2 | 5+6 2 | 5+6 2 | |
| | | | | 3+3 2 | 4+4 2 | 5+6 2 | 3+4 2 | |
| | | | | 6+6 2 | 3+3 2 | 5+6 2 | 5+6 2 | |
| | | | | 5+5 2 | 4+4 2 | 3+4 2 | 3+4 2 | |
| | | | | 5+5 2 | 6+6 2 | 3+4 2 | 3+4 2 | |
| | | | | 6+6 2 | 5+5 2 | 5+6 2 | 5+6 2 | |
| | | | | 3+3 2 | 3+3 2 | 5+6 2 | 3+4 2 | |
| | | | | 4+4 2 | 4+4 2 | 5+6 2 | 5+6 2 | |
| | | | | 3+3 2 | 5+5 2 | 3+4 2 | 5+6 2 | |
| | | | | 5+5 2 | 6+6 2 | 3+4 2 | 3+4 2 | |
| | | | | 6+6 2 | 3+3 2 | 5+6 2 | 5+6 2 | |
| | | | | 4+4 2 | 4+4 2 | 3+4 2 | 3+4 2 | |
| | | | | 6+6 2 | 4+4 2 | 5+6 2 | 5+6 2 | |
| | | | | 5+5 2 | 3+3 2 | 3+4 2 | 5+6 2 | |
| | | | | 3+3 2 | 5+5 2 | 5+6 2 | 5+6 2 | |
| | | | | 4+4 2 | 6+6 2 | 3+4 2 | 3+4 2 | |
| | | | | 5+5 2 | 4+4 2 | 3+4 2 | 3+4 2 | |
| | | | | 6+6 2 | 3+3 2 | 5+6 2 | 5+6 2 | |
| | | | | 5+5 2 | 6+6 2 | 3+4 2 | 3+4 2 | |
| | | | | 3+3 2 | 5+5 2 | 5+6 2 | 5+6 2 | |
| | | | | 6+6 2 | 4+4 2 | 3+4 2 | 3+4 2 | |
| | | | | 4+4 2 | 3+3 2 | 5+6 2 | 3+4 2 | |
| | | | | 3+3 2 | 6+6 2 | 3+4 2 | 5+6 2 | |
| | | | | 6+6 2 | 5+5 2 | 5+6 2 | 5+6 2 | |
| | | | | 4+4 2 | 3+3 2 | 3+4 2 | 3+4 2 | |
| | | | | 5+5 2 | 4+4 2 | 5+6 2 | 5+6 2 | |
| | | | | 4+4 2 | 6+6 2 | 5+6 2 | 3+4 2 | |
| | | | | 3+3 2 | 5+5 2 | 5+6 2 | 5+6 2 | |
| 14.2.30 | 412763 | 160 | Eu2P2S6 | | | | | |
| | -A | -B | -C | -D | -E | -GM | -Y | -Z |
| | 2+2 4 | 2+2 4 | 2+2 4 | 3+3 2 | 3+3 2 | 3+4 2 | 3+4 2 | 2+2 4 |
| | 2+2 4 | 2+2 4 | 2+2 4 | 4+4 2 | 4+4 2 | 3+4 2 | 3+4 2 | 2+2 4 |
| | 2+2 4 | 2+2 4 | 2+2 4 | 5+5 2 | 5+5 2 | 5+6 2 | 5+6 2 | 2+2 4 |
| | 2+2 4 | 2+2 4 | 2+2 4 | 6+6 2 | 6+6 2 | 5+6 2 | 5+6 2 | 2+2 4 |
| | 2+2 4 | 2+2 4 | 2+2 4 | 3+3 2 | 6+6 2 | 3+4 2 | 3+4 2 | 2+2 4 |
| | 2+2 4 | 2+2 4 | 2+2 4 | 4+4 2 | 5+5 2 | 5+6 2 | 5+6 2 | 2+2 4 |
| | 2+2 4 | 2+2 4 | 2+2 4 | 5+5 2 | 3+3 2 | 3+4 2 | 3+4 2 | 2+2 4 |
| | 2+2 4 | 2+2 4 | 2+2 4 | 6+6 2 | 4+4 2 | 5+6 2 | 5+6 2 | 2+2 4 |
| | 2+2 4 | 2+2 4 | 2+2 4 | 5+5 2 | 4+4 2 | 3+4 2 | 5+6 2 | 2+2 4 |
| | 2+2 4 | 2+2 4 | 2+2 4 | 6+6 2 | 3+3 2 | 5+6 2 | 5+6 2 | 2+2 4 |
| | 2+2 4 | 2+2 4 | 2+2 4 | 4+4 2 | 5+5 2 | 3+4 2 | 3+4 2 | 2+2 4 |
| | 2+2 4 | 2+2 4 | 2+2 4 | 3+3 2 | 6+6 2 | 3+4 2 | 3+4 2 | 2+2 4 |
| | 2+2 4 | 2+2 4 | 2+2 4 | 3+3 2 | 5+5 2 | 5+6 2 | 3+4 2 | 2+2 4 |

Continued on next column

**TABLE XXVI– continued**

| TQC | ICSD | NoE | SF | | | | |
|---|---|---|---|---|---|---|---|
| | 2+2 4 | 2+2 4 | 2+2 4 | 6+6 2 | 6+6 2 | 5+6 2 | 5+6 2 | 2+2 4 |
| | 2+2 4 | 2+2 4 | 2+2 4 | 4+4 2 | 3+3 2 | 3+4 2 | 5+6 2 | 2+2 4 |
| | 2+2 4 | 2+2 4 | 2+2 4 | 5+5 2 | 4+4 2 | 5+6 2 | 3+4 2 | 2+2 4 |
| | 2+2 4 | 2+2 4 | 2+2 4 | 5+5 2 | 3+3 2 | 3+4 2 | 5+6 2 | 2+2 4 |
| | 2+2 4 | 2+2 4 | 2+2 4 | 6+6 2 | 4+4 2 | 3+4 2 | 5+6 2 | 2+2 4 |
| | 2+2 4 | 2+2 4 | 2+2 4 | 4+4 2 | 6+6 2 | 5+6 2 | 3+4 2 | 2+2 4 |
| | 2+2 4 | 2+2 4 | 2+2 4 | 3+3 2 | 5+5 2 | 5+6 2 | 3+4 2 | 2+2 4 |
| | 2+2 4 | 2+2 4 | 2+2 4 | 3+3 2 | 5+5 2 | 5+6 2 | 3+4 2 | 2+2 4 |
| | 2+2 4 | 2+2 4 | 2+2 4 | 4+4 2 | 6+6 2 | 5+6 2 | 3+4 2 | 2+2 4 |
| | 2+2 4 | 2+2 4 | 2+2 4 | 3+3 2 | 5+5 2 | 5+6 2 | 3+4 2 | 2+2 4 |
| | 2+2 4 | 2+2 4 | 2+2 4 | 4+4 2 | 6+6 2 | 3+4 2 | 3+4 2 | 2+2 4 |
| | 2+2 4 | 2+2 4 | 2+2 4 | 6+6 2 | 4+4 2 | 5+6 2 | 5+6 2 | 2+2 4 |
| | 2+2 4 | 2+2 4 | 2+2 4 | 5+5 2 | 3+3 2 | 3+4 2 | 5+6 2 | 2+2 4 |
| | 2+2 4 | 2+2 4 | 2+2 4 | 5+5 2 | 4+4 2 | 3+4 2 | 5+6 2 | 2+2 4 |
| | 2+2 4 | 2+2 4 | 2+2 4 | 6+6 2 | 3+3 2 | 3+4 2 | 5+6 2 | 2+2 4 |
| | 2+2 4 | 2+2 4 | 2+2 4 | 5+5 2 | 3+3 2 | 3+4 2 | 5+6 2 | 2+2 4 |
| | 2+2 4 | 2+2 4 | 2+2 4 | 6+6 2 | 4+4 2 | 3+4 2 | 5+6 2 | 2+2 4 |
| | 2+2 4 | 2+2 4 | 2+2 4 | 3+3 2 | 6+6 2 | 5+6 2 | 3+4 2 | 2+2 4 |
| | 2+2 4 | 2+2 4 | 2+2 4 | 4+4 2 | 5+5 2 | 5+6 2 | 3+4 2 | 2+2 4 |
| | 2+2 4 | 2+2 4 | 2+2 4 | 5+5 2 | 4+4 2 | 3+4 2 | 5+6 2 | 2+2 4 |
| | 2+2 4 | 2+2 4 | 2+2 4 | 6+6 2 | 3+3 2 | 3+4 2 | 3+4 2 | 2+2 4 |
| | 2+2 4 | 2+2 4 | 2+2 4 | 4+4 2 | 5+5 2 | 5+6 2 | 3+4 2 | 2+2 4 |
| | 2+2 4 | 2+2 4 | 2+2 4 | 3+3 2 | 6+6 2 | 5+6 2 | 5+6 2 | 2+2 4 |
| | 2+2 4 | 2+2 4 | 2+2 4 | 4+4 2 | 5+5 2 | 5+6 2 | 3+4 2 | 2+2 4 |
| | 2+2 4 | 2+2 4 | 2+2 4 | 3+3 2 | 6+6 2 | 5+6 2 | 3+4 2 | 2+2 4 |
| | 2+2 4 | 2+2 4 | 2+2 4 | 5+5 2 | 3+3 2 | 3+4 2 | 5+6 2 | 2+2 4 |
| | 2+2 4 | 2+2 4 | 2+2 4 | 6+6 2 | 4+4 2 | 3+4 2 | 5+6 2 | 2+2 4 |
| | | | | 6+6 2 | 3+3 2 | 3+4 2 | 5+6 2 | |
| | | | | 5+5 2 | 4+4 2 | 3+4 2 | 5+6 2 | |
| | | | | 3+3 2 | 5+5 2 | 3+4 2 | 3+4 2 | |
| | | | | 4+4 2 | 6+6 2 | 5+6 2 | 5+6 2 | |
| | | | | 5+5 2 | 5+5 2 | 5+6 2 | 3+4 2 | |
| | | | | 6+6 2 | 6+6 2 | 3+4 2 | 5+6 2 | |
| | | | | 3+3 2 | 4+4 2 | 3+4 2 | 5+6 2 | |
| | | | | 4+4 2 | 3+3 2 | 5+6 2 | 3+4 2 | |
| | | | | 5+5 2 | 5+5 2 | 5+6 2 | 3+4 2 | |
| | | | | 6+6 2 | 6+6 2 | 3+4 2 | 5+6 2 | |
| | | | | 4+4 2 | 4+4 2 | 5+6 2 | 3+4 2 | |
| | | | | 3+3 2 | 3+3 2 | 5+6 2 | 3+4 2 | |
| | | | | 4+4 2 | 6+6 2 | 5+6 2 | 3+4 2 | |
| | | | | 3+3 2 | 5+5 2 | 3+4 2 | 5+6 2 | |
| | | | | 3+3 2 | 6+6 2 | 3+4 2 | 3+4 2 | |
| | | | | 6+6 2 | 5+5 2 | 3+4 2 | 3+4 2 | |
| | | | | 5+5 2 | 3+3 2 | 5+6 2 | 5+6 2 | |
| | | | | 4+4 2 | 4+4 2 | 3+4 2 | 3+4 2 | |
| | | | | 5+5 2 | 3+3 2 | 5+6 2 | 5+6 2 | |
| | | | | 6+6 2 | 4+4 2 | 5+6 2 | 3+4 2 | |
| | | | | 3+3 2 | 5+5 2 | 5+6 2 | 5+6 2 | |
| | | | | 4+4 2 | 6+6 2 | 5+6 2 | 5+6 2 | |
| | | | | 5+5 2 | 3+3 2 | 3+4 2 | 5+6 2 | |
| | | | | 6+6 2 | 4+4 2 | 3+4 2 | 3+4 2 | |
| | | | | 6+6 2 | 4+4 2 | 3+4 2 | 5+6 2 | |
| | | | | 5+5 2 | 3+3 2 | 3+4 2 | 5+6 2 | |
| | | | | 3+3 2 | 5+5 2 | 5+6 2 | 3+4 2 | |
| | | | | 4+4 2 | 6+6 2 | 3+4 2 | 5+6 2 | |
| | | | | 5+5 2 | 4+4 2 | 3+4 2 | 3+4 2 | |
| | | | | 6+6 2 | 3+3 2 | 5+6 2 | 5+6 2 | |
| | | | | 3+3 2 | 5+5 2 | 3+4 2 | 3+4 2 | |
| | | | | 4+4 2 | 6+6 2 | 5+6 2 | 5+6 2 | |
| | | | | 5+5 2 | 3+3 2 | 3+4 2 | 3+4 2 | |
| | | | | 6+6 2 | 4+4 2 | 5+6 2 | 5+6 2 | |

Continued on next column

**TABLE XXVI– continued**

| TQC | ICSD | NoE | SF |
|---|---|---|---|
| | | | 3+3 2 5+5 2 3+4 2 3+4 2 |
| | | | 6+6 2 6+6 2 5+6 2 5+6 2 |
| | | | 4+4 2 3+3 2 3+4 2 5+6 2 |
| | | | 5+5 2 4+4 2 5+6 2 3+4 2 |
| | | | 4+4 2 6+6 2 5+6 2 3+4 2 |
| | | | 3+3 2 5+5 2 5+6 2 5+6 2 |
| 14.2.31 | 416422 | 180 | Eu1Ge2Pt2 |
| | | | -A -B -C -D -E -GM -Y -Z |
| | | | 2+2 4 2+2 4 2+2 4 3+3 2 3+3 2 3+4 2 3+4 2 2+2 4 |
| | | | 2+2 4 2+2 4 2+2 4 4+4 2 4+4 2 5+6 2 5+6 2 2+2 4 |
| | | | 2+2 4 2+2 4 2+2 4 5+5 2 5+5 2 3+4 2 3+4 2 2+2 4 |
| | | | 2+2 4 2+2 4 2+2 4 6+6 2 6+6 2 5+6 2 5+6 2 2+2 4 |
| | | | 2+2 4 2+2 4 2+2 4 3+3 2 3+3 2 3+4 2 3+4 2 2+2 4 |
| | | | 2+2 4 2+2 4 2+2 4 4+4 2 4+4 2 5+6 2 5+6 2 2+2 4 |
| | | | 2+2 4 2+2 4 2+2 4 5+5 2 5+5 2 3+4 2 3+4 2 2+2 4 |
| | | | 2+2 4 2+2 4 2+2 4 6+6 2 6+6 2 5+6 2 5+6 2 2+2 4 |
| | | | 2+2 4 2+2 4 2+2 4 3+3 2 3+3 2 3+4 2 3+4 2 2+2 4 |
| | | | 2+2 4 2+2 4 2+2 4 5+5 2 5+5 2 5+6 2 5+6 2 2+2 4 |
| | | | 2+2 4 2+2 4 2+2 4 6+6 2 6+6 2 3+4 2 3+4 2 2+2 4 |
| | | | 2+2 4 2+2 4 2+2 4 4+4 2 4+4 2 5+6 2 5+6 2 2+2 4 |
| | | | 2+2 4 2+2 4 2+2 4 3+3 2 4+4 2 3+4 2 3+4 2 2+2 4 |
| | | | 2+2 4 2+2 4 2+2 4 4+4 2 6+6 2 5+6 2 5+6 2 2+2 4 |
| | | | 2+2 4 2+2 4 2+2 4 5+5 2 3+3 2 3+4 2 3+4 2 2+2 4 |
| | | | 2+2 4 2+2 4 2+2 4 6+6 2 5+5 2 5+6 2 5+6 2 2+2 4 |
| | | | 2+2 4 2+2 4 2+2 4 4+4 2 4+4 2 3+4 2 3+4 2 2+2 4 |
| | | | 2+2 4 2+2 4 2+2 4 3+3 2 3+3 2 3+4 2 5+6 2 2+2 4 |
| | | | 2+2 4 2+2 4 2+2 4 5+5 2 6+6 2 3+4 2 5+6 2 2+2 4 |
| | | | 2+2 4 2+2 4 2+2 4 6+6 2 5+5 2 5+6 2 3+4 2 2+2 4 |
| | | | 2+2 4 2+2 4 2+2 4 4+4 2 4+4 2 3+4 2 5+6 2 2+2 4 |
| | | | 2+2 4 2+2 4 2+2 4 5+5 2 3+3 2 5+6 2 3+4 2 2+2 4 |
| | | | 2+2 4 2+2 4 2+2 4 3+3 2 5+5 2 5+6 2 3+4 2 2+2 4 |
| | | | 2+2 4 2+2 4 2+2 4 6+6 2 6+6 2 5+6 2 5+6 2 2+2 4 |
| | | | 2+2 4 2+2 4 2+2 4 4+4 2 6+6 2 5+6 2 5+6 2 2+2 4 |
| | | | 2+2 4 2+2 4 2+2 4 3+3 2 5+5 2 3+4 2 3+4 2 2+2 4 |
| | | | 2+2 4 2+2 4 2+2 4 6+6 2 3+3 2 5+6 2 3+4 2 2+2 4 |
| | | | 2+2 4 2+2 4 2+2 4 5+5 2 4+4 2 3+4 2 5+6 2 2+2 4 |
| | | | 2+2 4 2+2 4 2+2 4 3+3 2 3+3 2 3+4 2 3+4 2 2+2 4 |
| | | | 2+2 4 2+2 4 2+2 4 6+6 2 4+4 2 3+4 2 5+6 2 2+2 4 |
| | | | 2+2 4 2+2 4 2+2 4 4+4 2 6+6 2 5+6 2 5+6 2 2+2 4 |
| | | | 2+2 4 2+2 4 2+2 4 5+5 2 5+5 2 3+4 2 3+4 2 2+2 4 |
| | | | 2+2 4 2+2 4 2+2 4 5+5 2 4+4 2 5+6 2 3+4 2 2+2 4 |
| | | | 2+2 4 2+2 4 2+2 4 6+6 2 3+3 2 5+6 2 5+6 2 2+2 4 |
| | | | 2+2 4 2+2 4 2+2 4 4+4 2 5+5 2 3+4 2 3+4 2 2+2 4 |
| | | | 2+2 4 2+2 4 2+2 4 5+5 2 4+4 2 3+4 2 5+6 2 2+2 4 |
| | | | 2+2 4 2+2 4 2+2 4 3+3 2 6+6 2 5+6 2 3+4 2 2+2 4 |
| | | | 2+2 4 2+2 4 2+2 4 4+4 2 5+5 2 3+4 2 3+4 2 2+2 4 |
| | | | 2+2 4 2+2 4 2+2 4 6+6 2 3+3 2 3+4 2 5+6 2 2+2 4 |
| | | | 2+2 4 2+2 4 2+2 4 5+5 2 6+6 2 5+6 2 3+4 2 2+2 4 |
| | | | 2+2 4 2+2 4 2+2 4 3+3 2 4+4 2 5+6 2 3+4 2 2+2 4 |
| | | | 2+2 4 2+2 4 2+2 4 6+6 2 3+3 2 3+4 2 3+4 2 2+2 4 |
| | | | 2+2 4 2+2 4 2+2 4 4+4 2 5+5 2 5+6 2 5+6 2 2+2 4 |
| | | | 2+2 4 2+2 4 2+2 4 3+3 2 3+3 2 3+4 2 5+6 2 2+2 4 |
| | | | 2+2 4 2+2 4 2+2 4 6+6 2 6+6 2 5+6 2 5+6 2 2+2 4 |
| | | | 5+5 2 4+4 2 3+4 2 3+4 2 |
| | | | 3+3 2 5+5 2 5+6 2 5+6 2 |
| | | | 4+4 2 6+6 2 3+4 2 3+4 2 |
| | | | 6+6 2 4+4 2 5+6 2 5+6 2 |
| | | | 3+3 2 3+3 2 5+6 2 5+6 2 |
| | | | 5+5 2 5+5 2 3+4 2 3+4 2 |
| | | | 4+4 2 6+6 2 3+4 2 5+6 2 |
| | | | 5+5 2 3+3 2 5+6 2 3+4 2 |

Continued on next column

**TABLE XXVI– continued**

| TQC | ICSD | NoE | SF |
|---|---|---|---|
| | | | 6+6 2 4+4 2 3+4 2 5+6 2 |
| | | | 4+4 2 6+6 2 3+4 2 3+4 2 |
| | | | 6+6 2 5+5 2 3+4 2 5+6 2 |
| | | | 5+5 2 3+3 2 5+6 2 3+4 2 |
| | | | 3+3 2 4+4 2 5+6 2 5+6 2 |
| | | | 4+4 2 5+5 2 3+4 2 3+4 2 |
| | | | 3+3 2 6+6 2 5+6 2 3+4 2 |
| | | | 6+6 2 3+3 2 5+6 2 5+6 2 |
| | | | 5+5 2 4+4 2 3+4 2 5+6 2 |
| | | | 3+3 2 6+6 2 5+6 2 5+6 2 |
| | | | 4+4 2 5+5 2 3+4 2 3+4 2 |
| | | | 6+6 2 3+3 2 5+6 2 5+6 2 |
| | | | 3+3 2 4+4 2 3+4 2 3+4 2 |
| | | | 5+5 2 5+5 2 5+6 2 5+6 2 |
| | | | 4+4 2 6+6 2 3+4 2 3+4 2 |
| | | | 6+6 2 6+6 2 3+4 2 5+6 2 |
| | | | 3+3 2 3+3 2 5+6 2 3+4 2 |
| | | | 5+5 2 5+5 2 3+4 2 5+6 2 |
| | | | 4+4 2 4+4 2 3+4 2 3+4 2 |
| | | | 3+3 2 6+6 2 3+4 2 3+4 2 |
| | | | 4+4 2 5+5 2 5+6 2 5+6 2 |
| | | | 5+5 2 3+3 2 3+4 2 3+4 2 |
| | | | 6+6 2 4+4 2 5+6 2 5+6 2 |
| | | | 3+3 2 6+6 2 5+6 2 5+6 2 |
| | | | 4+4 2 5+5 2 3+4 2 3+4 2 |
| | | | 5+5 2 4+4 2 5+6 2 5+6 2 |
| | | | 4+4 2 6+6 2 3+4 2 3+4 2 |
| | | | 6+6 2 3+3 2 5+6 2 5+6 2 |
| | | | 3+3 2 5+5 2 5+6 2 3+4 2 |
| | | | 6+6 2 3+3 2 3+4 2 5+6 2 |
| | | | 4+4 2 6+6 2 3+4 2 3+4 2 |
| | | | 5+5 2 5+5 2 5+6 2 5+6 2 |
| | | | 3+3 2 4+4 2 3+4 2 3+4 2 |
| | | | 5+5 2 3+3 2 5+6 2 5+6 2 |
| | | | 6+6 2 4+4 2 3+4 2 3+4 2 |
| | | | 3+3 2 6+6 2 5+6 2 5+6 2 |
| | | | 4+4 2 5+5 2 5+6 2 3+4 2 |
| 14.2.32 | 401556 | 124 | Eu1Ge1Pd1 |
| | | | -A -B -C -D -E -GM -Y -Z |
| | | | 2+2 4 2+2 4 2+2 4 3+3 2 5+5 2 3+4 2 5+6 2 2+2 4 |
| | | | 2+2 4 2+2 4 2+2 4 4+4 2 6+6 2 3+4 2 3+4 2 2+2 4 |
| | | | 2+2 4 2+2 4 2+2 4 5+5 2 3+3 2 5+6 2 5+6 2 2+2 4 |
| | | | 2+2 4 2+2 4 2+2 4 6+6 2 4+4 2 5+6 2 3+4 2 2+2 4 |
| | | | 2+2 4 2+2 4 2+2 4 3+3 2 5+5 2 3+4 2 5+6 2 2+2 4 |
| | | | 2+2 4 2+2 4 2+2 4 4+4 2 6+6 2 3+4 2 3+4 2 2+2 4 |
| | | | 2+2 4 2+2 4 2+2 4 5+5 2 3+3 2 5+6 2 5+6 2 2+2 4 |
| | | | 2+2 4 2+2 4 2+2 4 6+6 2 4+4 2 5+6 2 3+4 2 2+2 4 |
| | | | 2+2 4 2+2 4 2+2 4 3+3 2 6+6 2 3+4 2 5+6 2 2+2 4 |
| | | | 2+2 4 2+2 4 2+2 4 4+4 2 5+5 2 3+4 2 3+4 2 2+2 4 |
| | | | 2+2 4 2+2 4 2+2 4 5+5 2 3+3 2 5+6 2 5+6 2 2+2 4 |
| | | | 2+2 4 2+2 4 2+2 4 6+6 2 4+4 2 3+4 2 3+4 2 2+2 4 |
| | | | 2+2 4 2+2 4 2+2 4 3+3 2 5+5 2 5+6 2 5+6 2 2+2 4 |
| | | | 2+2 4 2+2 4 2+2 4 4+4 2 4+4 2 3+4 2 3+4 2 2+2 4 |
| | | | 2+2 4 2+2 4 2+2 4 6+6 2 3+3 2 5+6 2 5+6 2 2+2 4 |
| | | | 2+2 4 2+2 4 2+2 4 5+5 2 6+6 2 5+6 2 3+4 2 2+2 4 |
| | | | 2+2 4 2+2 4 2+2 4 5+5 2 5+5 2 3+4 2 3+4 2 2+2 4 |
| | | | 2+2 4 2+2 4 2+2 4 6+6 2 6+6 2 3+4 2 5+6 2 2+2 4 |
| | | | 2+2 4 2+2 4 2+2 4 3+3 2 3+3 2 5+6 2 3+4 2 2+2 4 |
| | | | 2+2 4 2+2 4 2+2 4 4+4 2 4+4 2 5+6 2 5+6 2 2+2 4 |
| | | | 2+2 4 2+2 4 2+2 4 6+6 2 3+3 2 3+4 2 5+6 2 2+2 4 |
| | | | 2+2 4 2+2 4 2+2 4 5+5 2 4+4 2 5+6 2 3+4 2 2+2 4 |

Continued on next column

**TABLE XXVI– continued**

| TQC | ICSD | NoE | SF | | | | | |
|---|---|---|---|---|---|---|---|---|
| | 2+2 4 | 2+2 4 | 2+2 4 | 4+4 2 | 6+6 2 | 3+4 2 | 3+4 2 | 2+2 4 |
| | 2+2 4 | 2+2 4 | 2+2 4 | 3+3 2 | 5+5 2 | 5+6 2 | 3+4 2 | 2+2 4 |
| | 2+2 4 | 2+2 4 | 2+2 4 | 3+3 2 | 6+6 2 | 5+6 2 | 5+6 2 | 2+2 4 |
| | 2+2 4 | 2+2 4 | 2+2 4 | 4+4 2 | 5+5 2 | 3+4 2 | 3+4 2 | 2+2 4 |
| | 2+2 4 | 2+2 4 | 2+2 4 | 5+5 2 | 3+3 2 | 5+6 2 | 5+6 2 | 2+2 4 |
| | 2+2 4 | 2+2 4 | 2+2 4 | 6+6 2 | 4+4 2 | 3+4 2 | 5+6 2 | 2+2 4 |
| | 2+2 4 | 2+2 4 | 2+2 4 | 4+4 2 | 6+6 2 | 5+6 2 | 3+4 2 | 2+2 4 |
| | 2+2 4 | 2+2 4 | 2+2 4 | 3+3 2 | 5+5 2 | 3+4 2 | 5+6 2 | 2+2 4 |
| | 2+2 4 | 2+2 4 | 2+2 4 | 5+5 2 | 4+4 2 | 5+6 2 | 3+4 2 | 2+2 4 |
| | | | | 6+6 2 | 3+3 2 | 3+4 2 | 5+6 2 | |
| | | | | 3+3 2 | 4+4 2 | 3+4 2 | 3+4 2 | |
| | | | | 6+6 2 | 3+3 2 | 5+6 2 | 5+6 2 | |
| | | | | 4+4 2 | 6+6 2 | 3+4 2 | 3+4 2 | |
| | | | | 5+5 2 | 5+5 2 | 5+6 2 | 5+6 2 | |
| | | | | 6+6 2 | 3+3 2 | 3+4 2 | 3+4 2 | |
| | | | | 4+4 2 | 4+4 2 | 5+6 2 | 5+6 2 | |
| | | | | 5+5 2 | 5+5 2 | 5+6 2 | 3+4 2 | |
| | | | | 3+3 2 | 6+6 2 | 3+4 2 | 5+6 2 | |
| | | | | 6+6 2 | 5+5 2 | 3+4 2 | 3+4 2 | |
| | | | | 5+5 2 | 6+6 2 | 3+4 2 | 5+6 2 | |
| | | | | 4+4 2 | 4+4 2 | 5+6 2 | 5+6 2 | |
| | | | | 3+3 2 | 3+3 2 | 5+6 2 | 3+4 2 | |
| | | | | 5+5 2 | 3+3 2 | 3+4 2 | 3+4 2 | |
| | | | | 6+6 2 | 4+4 2 | 3+4 2 | 3+4 2 | |
| | | | | 6+6 2 | 5+5 2 | 3+4 2 | 5+6 2 | |
| | | | | 5+5 2 | 6+6 2 | 3+4 2 | 5+6 2 | |
| | | | | 3+3 2 | 5+5 2 | 5+6 2 | 5+6 2 | |
| | | | | 4+4 2 | 6+6 2 | 5+6 2 | 3+4 2 | |
| | | | | 3+3 2 | 3+3 2 | 5+6 2 | 3+4 2 | |
| | | | | 6+6 2 | 4+4 2 | 5+6 2 | 5+6 2 | |
| | | | | 4+4 2 | 4+4 2 | 3+4 2 | 3+4 2 | |
| | | | | 5+5 2 | 3+3 2 | 5+6 2 | 5+6 2 | |
| | | | | 4+4 2 | 5+5 2 | 3+4 2 | 3+4 2 | |
| | | | | 3+3 2 | 6+6 2 | 3+4 2 | 5+6 2 | |
| | | | | 5+5 2 | 4+4 2 | 3+4 2 | 5+6 2 | |
| | | | | 6+6 2 | 5+5 2 | 5+6 2 | 3+4 2 | |
| | | | | 5+5 2 | 3+3 2 | 3+4 2 | 5+6 2 | |
| | | | | 6+6 2 | 6+6 2 | 5+6 2 | 3+4 2 | |
| | | | | 4+4 2 | 6+6 2 | 3+4 2 | 3+4 2 | |
| | | | | 3+3 2 | 5+5 2 | 5+6 2 | 5+6 2 | |
| 14.2.33 | 67379 | 248 | Er2O1S2 | | | | | |
| | -A | -B | -C | -D | -E | -GM | -Y | -Z |
| | 2+2 4 | 2+2 4 | 2+2 4 | 3+3 2 | 3+3 2 | 3+4 2 | 5+6 2 | 2+2 4 |
| | 2+2 4 | 2+2 4 | 2+2 4 | 4+4 2 | 4+4 2 | 3+4 2 | 3+4 2 | 2+2 4 |
| | 2+2 4 | 2+2 4 | 2+2 4 | 5+5 2 | 5+5 2 | 5+6 2 | 3+4 2 | 2+2 4 |
| | 2+2 4 | 2+2 4 | 2+2 4 | 6+6 2 | 6+6 2 | 5+6 2 | 5+6 2 | 2+2 4 |
| | 2+2 4 | 2+2 4 | 2+2 4 | 3+3 2 | 3+3 2 | 3+4 2 | 3+4 2 | 2+2 4 |
| | 2+2 4 | 2+2 4 | 2+2 4 | 4+4 2 | 4+4 2 | 3+4 2 | 3+4 2 | 2+2 4 |
| | 2+2 4 | 2+2 4 | 2+2 4 | 5+5 2 | 5+5 2 | 5+6 2 | 5+6 2 | 2+2 4 |
| | 2+2 4 | 2+2 4 | 2+2 4 | 6+6 2 | 6+6 2 | 5+6 2 | 5+6 2 | 2+2 4 |
| | 2+2 4 | 2+2 4 | 2+2 4 | 6+6 2 | 3+3 2 | 3+4 2 | 3+4 2 | 2+2 4 |
| | 2+2 4 | 2+2 4 | 2+2 4 | 3+3 2 | 6+6 2 | 5+6 2 | 5+6 2 | 2+2 4 |
| | 2+2 4 | 2+2 4 | 2+2 4 | 5+5 2 | 4+4 2 | 3+4 2 | 5+6 2 | 2+2 4 |
| | 2+2 4 | 2+2 4 | 2+2 4 | 4+4 2 | 5+5 2 | 5+6 2 | 3+4 2 | 2+2 4 |
| | 2+2 4 | 2+2 4 | 2+2 4 | 4+4 2 | 4+4 2 | 3+4 2 | 3+4 2 | 2+2 4 |
| | 2+2 4 | 2+2 4 | 2+2 4 | 3+3 2 | 3+3 2 | 5+6 2 | 5+6 2 | 2+2 4 |
| | 2+2 4 | 2+2 4 | 2+2 4 | 6+6 2 | 6+6 2 | 3+4 2 | 3+4 2 | 2+2 4 |
| | 2+2 4 | 2+2 4 | 2+2 4 | 5+5 2 | 5+5 2 | 5+6 2 | 5+6 2 | 2+2 4 |
| | 2+2 4 | 2+2 4 | 2+2 4 | 3+3 2 | 3+3 2 | 3+4 2 | 3+4 2 | 2+2 4 |
| | 2+2 4 | 2+2 4 | 2+2 4 | 4+4 2 | 4+4 2 | 5+6 2 | 5+6 2 | 2+2 4 |
| | 2+2 4 | 2+2 4 | 2+2 4 | 5+5 2 | 5+5 2 | 5+6 2 | 5+6 2 | 2+2 4 |

Continued on next column

**TABLE XXVI– continued**

| TQC | ICSD | NoE | SF | | | | | |
|---|---|---|---|---|---|---|---|---|
| | 2+2 4 | 2+2 4 | 2+2 4 | 6+6 2 | 6+6 2 | 3+4 2 | 3+4 2 | 2+2 4 |
| | 2+2 4 | 2+2 4 | 2+2 4 | 4+4 2 | 4+4 2 | 3+4 2 | 5+6 2 | 2+2 4 |
| | 2+2 4 | 2+2 4 | 2+2 4 | 3+3 2 | 3+3 2 | 5+6 2 | 3+4 2 | 2+2 4 |
| | 2+2 4 | 2+2 4 | 2+2 4 | 6+6 2 | 5+5 2 | 3+4 2 | 5+6 2 | 2+2 4 |
| | 2+2 4 | 2+2 4 | 2+2 4 | 5+5 2 | 6+6 2 | 5+6 2 | 3+4 2 | 2+2 4 |
| | 2+2 4 | 2+2 4 | 2+2 4 | 6+6 2 | 3+3 2 | 3+4 2 | 5+6 2 | 2+2 4 |
| | 2+2 4 | 2+2 4 | 2+2 4 | 3+3 2 | 6+6 2 | 3+4 2 | 3+4 2 | 2+2 4 |
| | 2+2 4 | 2+2 4 | 2+2 4 | 5+5 2 | 4+4 2 | 5+6 2 | 3+4 2 | 2+2 4 |
| | 2+2 4 | 2+2 4 | 2+2 4 | 4+4 2 | 5+5 2 | 3+4 2 | 5+6 2 | 2+2 4 |
| | 2+2 4 | 2+2 4 | 2+2 4 | 4+4 2 | 4+4 2 | 5+6 2 | 3+4 2 | 2+2 4 |
| | 2+2 4 | 2+2 4 | 2+2 4 | 3+3 2 | 3+3 2 | 3+4 2 | 3+4 2 | 2+2 4 |
| | 2+2 4 | 2+2 4 | 2+2 4 | 5+5 2 | 5+5 2 | 5+6 2 | 5+6 2 | 2+2 4 |
| | 2+2 4 | 2+2 4 | 2+2 4 | 6+6 2 | 6+6 2 | 5+6 2 | 5+6 2 | 2+2 4 |
| | 2+2 4 | 2+2 4 | 2+2 4 | 6+6 2 | 6+6 2 | 3+4 2 | 3+4 2 | 2+2 4 |
| | 2+2 4 | 2+2 4 | 2+2 4 | 4+4 2 | 4+4 2 | 5+6 2 | 5+6 2 | 2+2 4 |
| | 2+2 4 | 2+2 4 | 2+2 4 | 3+3 2 | 3+3 2 | 5+6 2 | 5+6 2 | 2+2 4 |
| | 2+2 4 | 2+2 4 | 2+2 4 | 5+5 2 | 5+5 2 | 3+4 2 | 3+4 2 | 2+2 4 |
| | 2+2 4 | 2+2 4 | 2+2 4 | 4+4 2 | 5+5 2 | 3+4 2 | 5+6 2 | 2+2 4 |
| | 2+2 4 | 2+2 4 | 2+2 4 | 5+5 2 | 6+6 2 | 3+4 2 | 3+4 2 | 2+2 4 |
| | 2+2 4 | 2+2 4 | 2+2 4 | 6+6 2 | 4+4 2 | 5+6 2 | 3+4 2 | 2+2 4 |
| | 2+2 4 | 2+2 4 | 2+2 4 | 3+3 2 | 3+3 2 | 3+4 2 | 5+6 2 | 2+2 4 |
| | 2+2 4 | 2+2 4 | 2+2 4 | 5+5 2 | 6+6 2 | 5+6 2 | 3+4 2 | 2+2 4 |
| | 2+2 4 | 2+2 4 | 2+2 4 | 3+3 2 | 5+5 2 | 5+6 2 | 5+6 2 | 2+2 4 |
| | 2+2 4 | 2+2 4 | 2+2 4 | 6+6 2 | 3+3 2 | 3+4 2 | 3+4 2 | 2+2 4 |
| | 2+2 4 | 2+2 4 | 2+2 4 | 4+4 2 | 4+4 2 | 5+6 2 | 5+6 2 | 2+2 4 |
| | 2+2 4 | 2+2 4 | 2+2 4 | 3+3 2 | 3+3 2 | 3+4 2 | 3+4 2 | 2+2 4 |
| | 2+2 4 | 2+2 4 | 2+2 4 | 4+4 2 | 4+4 2 | 5+6 2 | 5+6 2 | 2+2 4 |
| | 2+2 4 | 2+2 4 | 2+2 4 | 6+6 2 | 6+6 2 | 5+6 2 | 5+6 2 | 2+2 4 |
| | 2+2 4 | 2+2 4 | 2+2 4 | 5+5 2 | 5+5 2 | 5+6 2 | 5+6 2 | 2+2 4 |
| | 2+2 4 | 2+2 4 | 2+2 4 | 5+5 2 | 5+5 2 | 3+4 2 | 3+4 2 | 2+2 4 |
| | 2+2 4 | 2+2 4 | 2+2 4 | 6+6 2 | 6+6 2 | 3+4 2 | 3+4 2 | 2+2 4 |
| | 2+2 4 | 2+2 4 | 2+2 4 | 3+3 2 | 3+3 2 | 3+4 2 | 3+4 2 | 2+2 4 |
| | 2+2 4 | 2+2 4 | 2+2 4 | 4+4 2 | 4+4 2 | 5+6 2 | 5+6 2 | 2+2 4 |
| | 2+2 4 | 2+2 4 | 2+2 4 | 4+4 2 | 3+3 2 | 5+6 2 | 3+4 2 | 2+2 4 |
| | 2+2 4 | 2+2 4 | 2+2 4 | 3+3 2 | 4+4 2 | 3+4 2 | 5+6 2 | 2+2 4 |
| | 2+2 4 | 2+2 4 | 2+2 4 | 5+5 2 | 5+5 2 | 3+4 2 | 3+4 2 | 2+2 4 |
| | 2+2 4 | 2+2 4 | 2+2 4 | 6+6 2 | 6+6 2 | 3+4 2 | 5+6 2 | 2+2 4 |
| | 2+2 4 | 2+2 4 | 2+2 4 | 4+4 2 | 5+5 2 | 5+6 2 | 3+4 2 | 2+2 4 |
| | 2+2 4 | 2+2 4 | 2+2 4 | 5+5 2 | 6+6 2 | 3+4 2 | 5+6 2 | 2+2 4 |
| | 2+2 4 | 2+2 4 | 2+2 4 | 3+3 2 | 4+4 2 | 3+4 2 | 5+6 2 | 2+2 4 |
| | 2+2 4 | 2+2 4 | 2+2 4 | 6+6 2 | 3+3 2 | 5+6 2 | 3+4 2 | 2+2 4 |
| | 2+2 4 | 2+2 4 | 2+2 4 | 4+4 2 | 5+5 2 | 5+6 2 | 5+6 2 | 2+2 4 |
| | 2+2 4 | 2+2 4 | 2+2 4 | 3+3 2 | 3+3 2 | 3+4 2 | 3+4 2 | 2+2 4 |
| | | | | 6+6 2 | 4+4 2 | 3+4 2 | 5+6 2 | |
| | | | | 5+5 2 | 6+6 2 | 5+6 2 | 3+4 2 | |
| | | | | 3+3 2 | 6+6 2 | 3+4 2 | 5+6 2 | |
| | | | | 4+4 2 | 5+5 2 | 5+6 2 | 3+4 2 | |
| | | | | 6+6 2 | 4+4 2 | 3+4 2 | 3+4 2 | |
| | | | | 5+5 2 | 3+3 2 | 5+6 2 | 3+4 2 | |
| | | | | 3+3 2 | 5+5 2 | 3+4 2 | 5+6 2 | |
| | | | | 4+4 2 | 6+6 2 | 5+6 2 | 5+6 2 | |
| | | | | 5+5 2 | 3+3 2 | 3+4 2 | 3+4 2 | |
| | | | | 6+6 2 | 4+4 2 | 5+6 2 | 3+4 2 | |
| | | | | 5+5 2 | 6+6 2 | 3+4 2 | 3+4 2 | |
| | | | | 6+6 2 | 5+5 2 | 5+6 2 | 5+6 2 | |
| | | | | 3+3 2 | 6+6 2 | 3+4 2 | 5+6 2 | |
| | | | | 4+4 2 | 3+3 2 | 5+6 2 | 5+6 2 | |
| | | | | 6+6 2 | 4+4 2 | 3+4 2 | 5+6 2 | |
| | | | | 3+3 2 | 5+5 2 | 5+6 2 | 3+4 2 | |
| | | | | 4+4 2 | 3+3 2 | 5+6 2 | 5+6 2 | |
| | | | | 5+5 2 | 4+4 2 | 3+4 2 | 3+4 2 | |

Continued on next column

**TABLE XXVI– continued**

| TQC | ICSD | | NoE | SF | | | | |
|---|---|---|---|---|---|---|---|---|
| | | | | 4+4 2 | 3+3 2 | 3+4 2 | 3+4 2 | |
| | | | | 3+3 2 | 4+4 2 | 5+6 2 | 5+6 2 | |
| | | | | 5+5 2 | 5+5 2 | 3+4 2 | 3+4 2 | |
| | | | | 6+6 2 | 6+6 2 | 5+6 2 | 5+6 2 | |
| | | | | 3+3 2 | 3+3 2 | 5+6 2 | 3+4 2 | |
| | | | | 5+5 2 | 4+4 2 | 3+4 2 | 5+6 2 | |
| | | | | 6+6 2 | 5+5 2 | 3+4 2 | 3+4 2 | |
| | | | | 4+4 2 | 6+6 2 | 5+6 2 | 5+6 2 | |
| | | | | 4+4 2 | 3+3 2 | 3+4 2 | 5+6 2 | |
| | | | | 3+3 2 | 5+5 2 | 5+6 2 | 3+4 2 | |
| | | | | 6+6 2 | 6+6 2 | 5+6 2 | 3+4 2 | |
| | | | | 5+5 2 | 4+4 2 | 3+4 2 | 3+4 2 | |
| | | | | 4+4 2 | 3+3 2 | 3+4 2 | 5+6 2 | |
| | | | | 6+6 2 | 5+5 2 | 5+6 2 | 5+6 2 | |
| | | | | 3+3 2 | 4+4 2 | 3+4 2 | 3+4 2 | |
| | | | | 5+5 2 | 6+6 2 | 5+6 2 | 5+6 2 | |
| | | | | 4+4 2 | 5+5 2 | 3+4 2 | 3+4 2 | |
| | | | | 6+6 2 | 6+6 2 | 5+6 2 | 5+6 2 | |
| | | | | 5+5 2 | 4+4 2 | 3+4 2 | 3+4 2 | |
| | | | | 3+3 2 | 5+5 2 | 5+6 2 | 5+6 2 | |
| | | | | 5+5 2 | 6+6 2 | 5+6 2 | 5+6 2 | |
| | | | | 3+3 2 | 3+3 2 | 3+4 2 | 3+4 2 | |
| | | | | 6+6 2 | 4+4 2 | 5+6 2 | 5+6 2 | |
| | | | | 4+4 2 | 3+3 2 | 5+6 2 | 3+4 2 | |
| | | | | 3+3 2 | 3+3 2 | 3+4 2 | 3+4 2 | |
| | | | | 4+4 2 | 4+4 2 | 3+4 2 | 3+4 2 | |
| | | | | 6+6 2 | 6+6 2 | 5+6 2 | 5+6 2 | |
| | | | | 5+5 2 | 5+5 2 | 5+6 2 | 5+6 2 | |
| | | | | 3+3 2 | 3+3 2 | 5+6 2 | 3+4 2 | |
| | | | | 4+4 2 | 4+4 2 | 3+4 2 | 5+6 2 | |
| | | | | 5+5 2 | 5+5 2 | 3+4 2 | 3+4 2 | |
| | | | | 6+6 2 | 6+6 2 | 3+4 2 | 3+4 2 | |
| | | | | 4+4 2 | 4+4 2 | 5+6 2 | 5+6 2 | |
| | | | | 5+5 2 | 3+3 2 | 3+4 2 | 5+6 2 | |
| | | | | 6+6 2 | 5+5 2 | 3+4 2 | 3+4 2 | |
| | | | | 3+3 2 | 6+6 2 | 5+6 2 | 5+6 2 | |
| | | | | 4+4 2 | 5+5 2 | 3+4 2 | 3+4 2 | |
| | | | | 3+3 2 | 4+4 2 | 5+6 2 | 5+6 2 | |
| | | | | 5+5 2 | 6+6 2 | 3+4 2 | 3+4 2 | |
| | | | | 6+6 2 | 3+3 2 | 5+6 2 | 5+6 2 | |
| | | | | 3+3 2 | 4+4 2 | 3+4 2 | 3+4 2 | |
| | | | | 4+4 2 | 6+6 2 | 5+6 2 | 3+4 2 | |
| | | | | 5+5 2 | 5+5 2 | 3+4 2 | 5+6 2 | |
| | | | | 6+6 2 | 3+3 2 | 5+6 2 | 5+6 2 | |
| 14.2.34 | 69772 | | 172 | Er1O6Ru1Sr2 | | | | |
| | -A | -B | -C | -D | -E | -GM | -Y | -Z |
| | 2+2 4 | 2+2 4 | 2+2 4 | 3+3 2 | 5+5 2 | 3+4 2 | 5+6 2 | 2+2 4 |
| | 2+2 4 | 2+2 4 | 2+2 4 | 4+4 2 | 6+6 2 | 3+4 2 | 5+6 2 | 2+2 4 |
| | 2+2 4 | 2+2 4 | 2+2 4 | 3+3 2 | 3+3 2 | 3+4 2 | 3+4 2 | 2+2 4 |
| | 2+2 4 | 2+2 4 | 2+2 4 | 4+4 2 | 4+4 2 | 5+6 2 | 5+6 2 | 2+2 4 |
| | 2+2 4 | 2+2 4 | 2+2 4 | 5+5 2 | 5+5 2 | 3+4 2 | 3+4 2 | 2+2 4 |
| | 2+2 4 | 2+2 4 | 2+2 4 | 6+6 2 | 6+6 2 | 5+6 2 | 5+6 2 | 2+2 4 |
| | 2+2 4 | 2+2 4 | 2+2 4 | 5+5 2 | 3+3 2 | 5+6 2 | 3+4 2 | 2+2 4 |
| | 2+2 4 | 2+2 4 | 2+2 4 | 6+6 2 | 4+4 2 | 5+6 2 | 3+4 2 | 2+2 4 |
| | 2+2 4 | 2+2 4 | 2+2 4 | 6+6 2 | 4+4 2 | 5+6 2 | 3+4 2 | 2+2 4 |
| | 2+2 4 | 2+2 4 | 2+2 4 | 5+5 2 | 3+3 2 | 5+6 2 | 3+4 2 | 2+2 4 |
| | 2+2 4 | 2+2 4 | 2+2 4 | 6+6 2 | 3+3 2 | 5+6 2 | 3+4 2 | 2+2 4 |
| | 2+2 4 | 2+2 4 | 2+2 4 | 5+5 2 | 4+4 2 | 5+6 2 | 3+4 2 | 2+2 4 |
| | 2+2 4 | 2+2 4 | 2+2 4 | 3+3 2 | 5+5 2 | 3+4 2 | 5+6 2 | 2+2 4 |
| | 2+2 4 | 2+2 4 | 2+2 4 | 5+5 2 | 6+6 2 | 3+4 2 | 3+4 2 | 2+2 4 |
| | 2+2 4 | 2+2 4 | 2+2 4 | 4+4 2 | 5+5 2 | 3+4 2 | 5+6 2 | 2+2 4 |

Continued on next column

**TABLE XXVI– continued**

| TQC | ICSD | | NoE | SF | | | | |
|---|---|---|---|---|---|---|---|---|
| | 2+2 4 | 2+2 4 | 2+2 4 | 6+6 2 | 6+6 2 | 3+4 2 | 3+4 2 | 2+2 4 |
| | 2+2 4 | 2+2 4 | 2+2 4 | 4+4 2 | 5+5 2 | 3+4 2 | 5+6 2 | 2+2 4 |
| | 2+2 4 | 2+2 4 | 2+2 4 | 3+3 2 | 6+6 2 | 3+4 2 | 5+6 2 | 2+2 4 |
| | 2+2 4 | 2+2 4 | 2+2 4 | 3+3 2 | 4+4 2 | 5+6 2 | 5+6 2 | 2+2 4 |
| | 2+2 4 | 2+2 4 | 2+2 4 | 4+4 2 | 3+3 2 | 5+6 2 | 5+6 2 | 2+2 4 |
| | 2+2 4 | 2+2 4 | 2+2 4 | 5+5 2 | 3+3 2 | 5+6 2 | 3+4 2 | 2+2 4 |
| | 2+2 4 | 2+2 4 | 2+2 4 | 6+6 2 | 4+4 2 | 5+6 2 | 3+4 2 | 2+2 4 |
| | 2+2 4 | 2+2 4 | 2+2 4 | 6+6 2 | 4+4 2 | 5+6 2 | 3+4 2 | 2+2 4 |
| | 2+2 4 | 2+2 4 | 2+2 4 | 5+5 2 | 3+3 2 | 5+6 2 | 3+4 2 | 2+2 4 |
| | 2+2 4 | 2+2 4 | 2+2 4 | 6+6 2 | 6+6 2 | 3+4 2 | 3+4 2 | 2+2 4 |
| | 2+2 4 | 2+2 4 | 2+2 4 | 3+3 2 | 5+5 2 | 5+6 2 | 5+6 2 | 2+2 4 |
| | 2+2 4 | 2+2 4 | 2+2 4 | 5+5 2 | 3+3 2 | 3+4 2 | 5+6 2 | 2+2 4 |
| | 2+2 4 | 2+2 4 | 2+2 4 | 4+4 2 | 4+4 2 | 5+6 2 | 3+4 2 | 2+2 4 |
| | 2+2 4 | 2+2 4 | 2+2 4 | 5+5 2 | 6+6 2 | 3+4 2 | 3+4 2 | 2+2 4 |
| | 2+2 4 | 2+2 4 | 2+2 4 | 4+4 2 | 5+5 2 | 3+4 2 | 5+6 2 | 2+2 4 |
| | 2+2 4 | 2+2 4 | 2+2 4 | 6+6 2 | 4+4 2 | 5+6 2 | 3+4 2 | 2+2 4 |
| | 2+2 4 | 2+2 4 | 2+2 4 | 3+3 2 | 3+3 2 | 3+4 2 | 5+6 2 | 2+2 4 |
| | 2+2 4 | 2+2 4 | 2+2 4 | 6+6 2 | 5+5 2 | 5+6 2 | 3+4 2 | 2+2 4 |
| | 2+2 4 | 2+2 4 | 2+2 4 | 3+3 2 | 3+3 2 | 5+6 2 | 5+6 2 | 2+2 4 |
| | 2+2 4 | 2+2 4 | 2+2 4 | 5+5 2 | 6+6 2 | 3+4 2 | 3+4 2 | 2+2 4 |
| | 2+2 4 | 2+2 4 | 2+2 4 | 4+4 2 | 4+4 2 | 5+6 2 | 5+6 2 | 2+2 4 |
| | 2+2 4 | 2+2 4 | 2+2 4 | 6+6 2 | 5+5 2 | 3+4 2 | 3+4 2 | 2+2 4 |
| | 2+2 4 | 2+2 4 | 2+2 4 | 5+5 2 | 6+6 2 | 3+4 2 | 3+4 2 | 2+2 4 |
| | 2+2 4 | 2+2 4 | 2+2 4 | 5+5 2 | 6+6 2 | 3+4 2 | 3+4 2 | 2+2 4 |
| | 2+2 4 | 2+2 4 | 2+2 4 | 6+6 2 | 5+5 2 | 3+4 2 | 3+4 2 | 2+2 4 |
| | 2+2 4 | 2+2 4 | 2+2 4 | 5+5 2 | 5+5 2 | 3+4 2 | 3+4 2 | 2+2 4 |
| | 2+2 4 | 2+2 4 | 2+2 4 | 6+6 2 | 6+6 2 | 3+4 2 | 3+4 2 | 2+2 4 |
| | 2+2 4 | 2+2 4 | 2+2 4 | 6+6 2 | 5+5 2 | 3+4 2 | 3+4 2 | 2+2 4 |
| | | | | 5+5 2 | 6+6 2 | 3+4 2 | 3+4 2 | |
| | | | | 6+6 2 | 5+5 2 | 3+4 2 | 3+4 2 | |
| | | | | 5+5 2 | 6+6 2 | 3+4 2 | 3+4 2 | |
| | | | | 4+4 2 | 5+5 2 | 3+4 2 | 5+6 2 | |
| | | | | 3+3 2 | 6+6 2 | 3+4 2 | 5+6 2 | |
| | | | | 5+5 2 | 3+3 2 | 5+6 2 | 3+4 2 | |
| | | | | 6+6 2 | 4+4 2 | 5+6 2 | 3+4 2 | |
| | | | | 3+3 2 | 3+3 2 | 5+6 2 | 5+6 2 | |
| | | | | 4+4 2 | 4+4 2 | 5+6 2 | 5+6 2 | |
| | | | | 4+4 2 | 4+4 2 | 5+6 2 | 5+6 2 | |
| | | | | 3+3 2 | 3+3 2 | 5+6 2 | 5+6 2 | |
| | | | | 3+3 2 | 3+3 2 | 5+6 2 | 5+6 2 | |
| | | | | 4+4 2 | 4+4 2 | 5+6 2 | 5+6 2 | |
| | | | | 3+3 2 | 4+4 2 | 3+4 2 | 5+6 2 | |
| | | | | 4+4 2 | 6+6 2 | 3+4 2 | 5+6 2 | |
| | | | | 4+4 2 | 5+5 2 | 5+6 2 | 5+6 2 | |
| | | | | 3+3 2 | 3+3 2 | 5+6 2 | 5+6 2 | |
| | | | | 3+3 2 | 5+5 2 | 5+6 2 | 5+6 2 | |
| | | | | 4+4 2 | 6+6 2 | 5+6 2 | 5+6 2 | |
| | | | | 4+4 2 | 3+3 2 | 5+6 2 | 3+4 2 | |
| | | | | 3+3 2 | 4+4 2 | 5+6 2 | 5+6 2 | |
| | | | | 6+6 2 | 3+3 2 | 5+6 2 | 5+6 2 | |
| | | | | 5+5 2 | 4+4 2 | 5+6 2 | 3+4 2 | |
| | | | | 5+5 2 | 4+4 2 | 5+6 2 | 3+4 2 | |
| | | | | 6+6 2 | 3+3 2 | 5+6 2 | 5+6 2 | |
| | | | | 3+3 2 | 4+4 2 | 3+4 2 | 3+4 2 | |
| | | | | 4+4 2 | 3+3 2 | 3+4 2 | 5+6 2 | |
| | | | | 5+5 2 | 5+5 2 | 3+4 2 | 3+4 2 | |
| | | | | 6+6 2 | 6+6 2 | 3+4 2 | 3+4 2 | |
| | | | | 6+6 2 | 4+4 2 | 5+6 2 | 3+4 2 | |
| | | | | 5+5 2 | 3+3 2 | 5+6 2 | 3+4 2 | |
| | | | | 5+5 2 | 3+3 2 | 5+6 2 | 3+4 2 | |
| | | | | 6+6 2 | 4+4 2 | 5+6 2 | 3+4 2 | |

Continued on next column

**TABLE XXVI– continued**

| TQC | ICSD | | NoE | SF | | | | |
|---|---|---|---|---|---|---|---|---|
| | | | | 6+6 2 | 3+3 2 | 5+6 2 | 3+4 2 | |
| | | | | 5+5 2 | 4+4 2 | 5+6 2 | 3+4 2 | |
| | | | | 5+5 2 | 6+6 2 | 3+4 2 | 3+4 2 | |
| | | | | 6+6 2 | 5+5 2 | 3+4 2 | 3+4 2 | |
| | | | | 6+6 2 | 3+3 2 | 3+4 2 | 3+4 2 | |
| | | | | 5+5 2 | 4+4 2 | 3+4 2 | 3+4 2 | |
| | | | | 6+6 2 | 4+4 2 | 5+6 2 | 3+4 2 | |
| | | | | 5+5 2 | 3+3 2 | 5+6 2 | 3+4 2 | |
| | | | | 5+5 2 | 3+3 2 | 5+6 2 | 3+4 2 | |
| | | | | 6+6 2 | 4+4 2 | 5+6 2 | 3+4 2 | |
| 14.2.35 | 69771 | | 172 | Er1O6Ru1Sr2 | | | | |
| | -A | -B | -C | -D | -E | -GM | -Y | -Z |
| | 2+2 4 | 2+2 4 | 2+2 4 | 3+3 2 | 5+5 2 | 3+4 2 | 5+6 2 | 2+2 4 |
| | 2+2 4 | 2+2 4 | 2+2 4 | 4+4 2 | 6+6 2 | 3+4 2 | 5+6 2 | 2+2 4 |
| | 2+2 4 | 2+2 4 | 2+2 4 | 3+3 2 | 3+3 2 | 3+4 2 | 3+4 2 | 2+2 4 |
| | 2+2 4 | 2+2 4 | 2+2 4 | 4+4 2 | 4+4 2 | 5+6 2 | 5+6 2 | 2+2 4 |
| | 2+2 4 | 2+2 4 | 2+2 4 | 5+5 2 | 5+5 2 | 3+4 2 | 3+4 2 | 2+2 4 |
| | 2+2 4 | 2+2 4 | 2+2 4 | 6+6 2 | 6+6 2 | 5+6 2 | 5+6 2 | 2+2 4 |
| | 2+2 4 | 2+2 4 | 2+2 4 | 5+5 2 | 3+3 2 | 5+6 2 | 3+4 2 | 2+2 4 |
| | 2+2 4 | 2+2 4 | 2+2 4 | 6+6 2 | 4+4 2 | 5+6 2 | 3+4 2 | 2+2 4 |
| | 2+2 4 | 2+2 4 | 2+2 4 | 6+6 2 | 4+4 2 | 5+6 2 | 3+4 2 | 2+2 4 |
| | 2+2 4 | 2+2 4 | 2+2 4 | 5+5 2 | 3+3 2 | 5+6 2 | 3+4 2 | 2+2 4 |
| | 2+2 4 | 2+2 4 | 2+2 4 | 6+6 2 | 3+3 2 | 5+6 2 | 3+4 2 | 2+2 4 |
| | 2+2 4 | 2+2 4 | 2+2 4 | 5+5 2 | 4+4 2 | 5+6 2 | 3+4 2 | 2+2 4 |
| | 2+2 4 | 2+2 4 | 2+2 4 | 3+3 2 | 5+5 2 | 3+4 2 | 5+6 2 | 2+2 4 |
| | 2+2 4 | 2+2 4 | 2+2 4 | 4+4 2 | 6+6 2 | 3+4 2 | 5+6 2 | 2+2 4 |
| | 2+2 4 | 2+2 4 | 2+2 4 | 5+5 2 | 5+5 2 | 3+4 2 | 3+4 2 | 2+2 4 |
| | 2+2 4 | 2+2 4 | 2+2 4 | 6+6 2 | 6+6 2 | 3+4 2 | 3+4 2 | 2+2 4 |
| | 2+2 4 | 2+2 4 | 2+2 4 | 4+4 2 | 5+5 2 | 3+4 2 | 5+6 2 | 2+2 4 |
| | 2+2 4 | 2+2 4 | 2+2 4 | 3+3 2 | 6+6 2 | 3+4 2 | 5+6 2 | 2+2 4 |
| | 2+2 4 | 2+2 4 | 2+2 4 | 3+3 2 | 4+4 2 | 5+6 2 | 5+6 2 | 2+2 4 |
| | 2+2 4 | 2+2 4 | 2+2 4 | 4+4 2 | 3+3 2 | 5+6 2 | 5+6 2 | 2+2 4 |
| | 2+2 4 | 2+2 4 | 2+2 4 | 5+5 2 | 3+3 2 | 5+6 2 | 3+4 2 | 2+2 4 |
| | 2+2 4 | 2+2 4 | 2+2 4 | 6+6 2 | 4+4 2 | 5+6 2 | 3+4 2 | 2+2 4 |
| | 2+2 4 | 2+2 4 | 2+2 4 | 6+6 2 | 4+4 2 | 5+6 2 | 3+4 2 | 2+2 4 |
| | 2+2 4 | 2+2 4 | 2+2 4 | 5+5 2 | 3+3 2 | 5+6 2 | 3+4 2 | 2+2 4 |
| | 2+2 4 | 2+2 4 | 2+2 4 | 6+6 2 | 6+6 2 | 3+4 2 | 5+6 2 | 2+2 4 |
| | 2+2 4 | 2+2 4 | 2+2 4 | 3+3 2 | 5+5 2 | 3+4 2 | 3+4 2 | 2+2 4 |
| | 2+2 4 | 2+2 4 | 2+2 4 | 4+4 2 | 4+4 2 | 5+6 2 | 5+6 2 | 2+2 4 |
| | 2+2 4 | 2+2 4 | 2+2 4 | 5+5 2 | 3+3 2 | 5+6 2 | 3+4 2 | 2+2 4 |
| | 2+2 4 | 2+2 4 | 2+2 4 | 5+5 2 | 6+6 2 | 3+4 2 | 3+4 2 | 2+2 4 |
| | 2+2 4 | 2+2 4 | 2+2 4 | 4+4 2 | 5+5 2 | 3+4 2 | 5+6 2 | 2+2 4 |
| | 2+2 4 | 2+2 4 | 2+2 4 | 6+6 2 | 4+4 2 | 5+6 2 | 3+4 2 | 2+2 4 |
| | 2+2 4 | 2+2 4 | 2+2 4 | 3+3 2 | 3+3 2 | 3+4 2 | 5+6 2 | 2+2 4 |
| | 2+2 4 | 2+2 4 | 2+2 4 | 6+6 2 | 5+5 2 | 5+6 2 | 3+4 2 | 2+2 4 |
| | 2+2 4 | 2+2 4 | 2+2 4 | 3+3 2 | 3+3 2 | 5+6 2 | 5+6 2 | 2+2 4 |
| | 2+2 4 | 2+2 4 | 2+2 4 | 5+5 2 | 6+6 2 | 3+4 2 | 3+4 2 | 2+2 4 |
| | 2+2 4 | 2+2 4 | 2+2 4 | 4+4 2 | 4+4 2 | 5+6 2 | 5+6 2 | 2+2 4 |
| | 2+2 4 | 2+2 4 | 2+2 4 | 6+6 2 | 5+5 2 | 3+4 2 | 3+4 2 | 2+2 4 |
| | 2+2 4 | 2+2 4 | 2+2 4 | 5+5 2 | 6+6 2 | 3+4 2 | 3+4 2 | 2+2 4 |
| | 2+2 4 | 2+2 4 | 2+2 4 | 5+5 2 | 6+6 2 | 3+4 2 | 3+4 2 | 2+2 4 |
| | 2+2 4 | 2+2 4 | 2+2 4 | 6+6 2 | 5+5 2 | 3+4 2 | 3+4 2 | 2+2 4 |
| | 2+2 4 | 2+2 4 | 2+2 4 | 5+5 2 | 5+5 2 | 3+4 2 | 3+4 2 | 2+2 4 |
| | 2+2 4 | 2+2 4 | 2+2 4 | 6+6 2 | 6+6 2 | 3+4 2 | 3+4 2 | 2+2 4 |
| | 2+2 4 | 2+2 4 | 2+2 4 | 6+6 2 | 5+5 2 | 3+4 2 | 3+4 2 | 2+2 4 |
| | | | | 5+5 2 | 6+6 2 | 3+4 2 | 3+4 2 | |
| | | | | 6+6 2 | 5+5 2 | 3+4 2 | 3+4 2 | |
| | | | | 5+5 2 | 6+6 2 | 3+4 2 | 3+4 2 | |
| | | | | 4+4 2 | 5+5 2 | 3+4 2 | 5+6 2 | |
| | | | | 3+3 2 | 6+6 2 | 3+4 2 | 5+6 2 | |
| | | | | 5+5 2 | 3+3 2 | 5+6 2 | 3+4 2 | |

Continued on next column

**TABLE XXVI– continued**

| TQC | ICSD | | NoE | SF | | | | |
|---|---|---|---|---|---|---|---|---|
| | | | | 6+6 2 | 4+4 2 | 5+6 2 | 3+4 2 | |
| | | | | 3+3 2 | 4+4 2 | 5+6 2 | 5+6 2 | |
| | | | | 4+4 2 | 3+3 2 | 5+6 2 | 5+6 2 | |
| | | | | 4+4 2 | 4+4 2 | 5+6 2 | 5+6 2 | |
| | | | | 3+3 2 | 3+3 2 | 5+6 2 | 5+6 2 | |
| | | | | 3+3 2 | 4+4 2 | 5+6 2 | 5+6 2 | |
| | | | | 4+4 2 | 3+3 2 | 3+4 2 | 5+6 2 | |
| | | | | 3+3 2 | 6+6 2 | 5+6 2 | 5+6 2 | |
| | | | | 4+4 2 | 5+5 2 | 3+4 2 | 5+6 2 | |
| | | | | 4+4 2 | 4+4 2 | 5+6 2 | 5+6 2 | |
| | | | | 3+3 2 | 3+3 2 | 5+6 2 | 5+6 2 | |
| | | | | 3+3 2 | 5+5 2 | 5+6 2 | 5+6 2 | |
| | | | | 4+4 2 | 6+6 2 | 5+6 2 | 3+4 2 | |
| | | | | 4+4 2 | 3+3 2 | 5+6 2 | 5+6 2 | |
| | | | | 3+3 2 | 4+4 2 | 5+6 2 | 5+6 2 | |
| | | | | 6+6 2 | 3+3 2 | 5+6 2 | 3+4 2 | |
| | | | | 5+5 2 | 4+4 2 | 5+6 2 | 5+6 2 | |
| | | | | 5+5 2 | 3+3 2 | 5+6 2 | 3+4 2 | |
| | | | | 6+6 2 | 4+4 2 | 5+6 2 | 3+4 2 | |
| | | | | 3+3 2 | 4+4 2 | 3+4 2 | 5+6 2 | |
| | | | | 4+4 2 | 3+3 2 | 3+4 2 | 3+4 2 | |
| | | | | 5+5 2 | 5+5 2 | 3+4 2 | 5+6 2 | |
| | | | | 6+6 2 | 6+6 2 | 3+4 2 | 3+4 2 | |
| | | | | 6+6 2 | 4+4 2 | 5+6 2 | 3+4 2 | |
| | | | | 5+5 2 | 3+3 2 | 5+6 2 | 3+4 2 | |
| | | | | 5+5 2 | 3+3 2 | 5+6 2 | 3+4 2 | |
| | | | | 6+6 2 | 4+4 2 | 5+6 2 | 3+4 2 | |
| | | | | 6+6 2 | 3+3 2 | 5+6 2 | 3+4 2 | |
| | | | | 5+5 2 | 4+4 2 | 5+6 2 | 3+4 2 | |
| | | | | 5+5 2 | 6+6 2 | 3+4 2 | 3+4 2 | |
| | | | | 6+6 2 | 5+5 2 | 3+4 2 | 3+4 2 | |
| | | | | 6+6 2 | 3+3 2 | 3+4 2 | 3+4 2 | |
| | | | | 5+5 2 | 4+4 2 | 3+4 2 | 3+4 2 | |
| | | | | 6+6 2 | 4+4 2 | 5+6 2 | 3+4 2 | |
| | | | | 5+5 2 | 3+3 2 | 5+6 2 | 3+4 2 | |
| | | | | 6+6 2 | 3+3 2 | 5+6 2 | 3+4 2 | |
| | | | | 5+5 2 | 4+4 2 | 5+6 2 | 3+4 2 | |
| 14.2.36 | 162113 | | 192 | Dy1Ni1O3 | | | | |
| | -A | -B | -C | -D | -E | -GM | -Y | -Z |
| | 2+2 4 | 2+2 4 | 2+2 4 | 3+3 2 | 3+3 2 | 3+4 2 | 3+4 2 | 2+2 4 |
| | 2+2 4 | 2+2 4 | 2+2 4 | 4+4 2 | 4+4 2 | 5+6 2 | 5+6 2 | 2+2 4 |
| | 2+2 4 | 2+2 4 | 2+2 4 | 5+5 2 | 5+5 2 | 3+4 2 | 3+4 2 | 2+2 4 |
| | 2+2 4 | 2+2 4 | 2+2 4 | 6+6 2 | 6+6 2 | 5+6 2 | 5+6 2 | 2+2 4 |
| | 2+2 4 | 2+2 4 | 2+2 4 | 3+3 2 | 4+4 2 | 3+4 2 | 3+4 2 | 2+2 4 |
| | 2+2 4 | 2+2 4 | 2+2 4 | 5+5 2 | 3+3 2 | 5+6 2 | 5+6 2 | 2+2 4 |
| | 2+2 4 | 2+2 4 | 2+2 4 | 6+6 2 | 6+6 2 | 3+4 2 | 5+6 2 | 2+2 4 |
| | 2+2 4 | 2+2 4 | 2+2 4 | 4+4 2 | 5+5 2 | 5+6 2 | 3+4 2 | 2+2 4 |
| | 2+2 4 | 2+2 4 | 2+2 4 | 6+6 2 | 5+5 2 | 5+6 2 | 3+4 2 | 2+2 4 |
| | 2+2 4 | 2+2 4 | 2+2 4 | 3+3 2 | 6+6 2 | 3+4 2 | 5+6 2 | 2+2 4 |
| | 2+2 4 | 2+2 4 | 2+2 4 | 5+5 2 | 3+3 2 | 3+4 2 | 3+4 2 | 2+2 4 |
| | 2+2 4 | 2+2 4 | 2+2 4 | 4+4 2 | 4+4 2 | 3+4 2 | 5+6 2 | 2+2 4 |
| | 2+2 4 | 2+2 4 | 2+2 4 | 4+4 2 | 4+4 2 | 5+6 2 | 5+6 2 | 2+2 4 |
| | 2+2 4 | 2+2 4 | 2+2 4 | 5+5 2 | 3+3 2 | 5+6 2 | 3+4 2 | 2+2 4 |
| | 2+2 4 | 2+2 4 | 2+2 4 | 6+6 2 | 6+6 2 | 5+6 2 | 5+6 2 | 2+2 4 |
| | 2+2 4 | 2+2 4 | 2+2 4 | 3+3 2 | 5+5 2 | 3+4 2 | 3+4 2 | 2+2 4 |
| | 2+2 4 | 2+2 4 | 2+2 4 | 5+5 2 | 3+3 2 | 3+4 2 | 3+4 2 | 2+2 4 |
| | 2+2 4 | 2+2 4 | 2+2 4 | 6+6 2 | 4+4 2 | 5+6 2 | 5+6 2 | 2+2 4 |
| | 2+2 4 | 2+2 4 | 2+2 4 | 3+3 2 | 5+5 2 | 3+4 2 | 3+4 2 | 2+2 4 |
| | 2+2 4 | 2+2 4 | 2+2 4 | 4+4 2 | 6+6 2 | 5+6 2 | 5+6 2 | 2+2 4 |
| | 2+2 4 | 2+2 4 | 2+2 4 | 3+3 2 | 5+5 2 | 5+6 2 | 3+4 2 | 2+2 4 |
| | 2+2 4 | 2+2 4 | 2+2 4 | 4+4 2 | 6+6 2 | 5+6 2 | 5+6 2 | 2+2 4 |

Continued on next column

**TABLE XXVI– continued**

| TQC | ICSD | NoE | SF | | | | |
|---|---|---|---|---|---|---|---|
| | 2+2 4 | 2+2 4 | 2+2 4 | 5+5 2 | 4+4 2 | 3+4 2 | 3+4 2 | 2+2 4 |
| | 2+2 4 | 2+2 4 | 2+2 4 | 6+6 2 | 3+3 2 | 3+4 2 | 5+6 2 | 2+2 4 |
| | 2+2 4 | 2+2 4 | 2+2 4 | 6+6 2 | 6+6 2 | 5+6 2 | 3+4 2 | 2+2 4 |
| | 2+2 4 | 2+2 4 | 2+2 4 | 3+3 2 | 5+5 2 | 3+4 2 | 5+6 2 | 2+2 4 |
| | 2+2 4 | 2+2 4 | 2+2 4 | 5+5 2 | 4+4 2 | 3+4 2 | 5+6 2 | 2+2 4 |
| | 2+2 4 | 2+2 4 | 2+2 4 | 4+4 2 | 3+3 2 | 5+6 2 | 3+4 2 | 2+2 4 |
| | 2+2 4 | 2+2 4 | 2+2 4 | 6+6 2 | 5+5 2 | 3+4 2 | 3+4 2 | 2+2 4 |
| | 2+2 4 | 2+2 4 | 2+2 4 | 5+5 2 | 6+6 2 | 3+4 2 | 5+6 2 | 2+2 4 |
| | 2+2 4 | 2+2 4 | 2+2 4 | 3+3 2 | 5+5 2 | 3+4 2 | 3+4 2 | 2+2 4 |
| | 2+2 4 | 2+2 4 | 2+2 4 | 4+4 2 | 6+6 2 | 3+4 2 | 5+6 2 | 2+2 4 |
| | 2+2 4 | 2+2 4 | 2+2 4 | 5+5 2 | 5+5 2 | 3+4 2 | 3+4 2 | 2+2 4 |
| | 2+2 4 | 2+2 4 | 2+2 4 | 6+6 2 | 6+6 2 | 5+6 2 | 3+4 2 | 2+2 4 |
| | 2+2 4 | 2+2 4 | 2+2 4 | 4+4 2 | 5+5 2 | 3+4 2 | 5+6 2 | 2+2 4 |
| | 2+2 4 | 2+2 4 | 2+2 4 | 3+3 2 | 6+6 2 | 3+4 2 | 3+4 2 | 2+2 4 |
| | 2+2 4 | 2+2 4 | 2+2 4 | 5+5 2 | 3+3 2 | 3+4 2 | 5+6 2 | 2+2 4 |
| | 2+2 4 | 2+2 4 | 2+2 4 | 6+6 2 | 4+4 2 | 5+6 2 | 5+6 2 | 2+2 4 |
| | 2+2 4 | 2+2 4 | 2+2 4 | 4+4 2 | 4+4 2 | 3+4 2 | 3+4 2 | 2+2 4 |
| | 2+2 4 | 2+2 4 | 2+2 4 | 3+3 2 | 3+3 2 | 5+6 2 | 3+4 2 | 2+2 4 |
| | 2+2 4 | 2+2 4 | 2+2 4 | 5+5 2 | 5+5 2 | 3+4 2 | 5+6 2 | 2+2 4 |
| | 2+2 4 | 2+2 4 | 2+2 4 | 6+6 2 | 6+6 2 | 5+6 2 | 5+6 2 | 2+2 4 |
| | 2+2 4 | 2+2 4 | 2+2 4 | 6+6 2 | 3+3 2 | 5+6 2 | 3+4 2 | 2+2 4 |
| | 2+2 4 | 2+2 4 | 2+2 4 | 5+5 2 | 4+4 2 | 3+4 2 | 5+6 2 | 2+2 4 |
| | 2+2 4 | 2+2 4 | 2+2 4 | 4+4 2 | 6+6 2 | 3+4 2 | 3+4 2 | 2+2 4 |
| | 2+2 4 | 2+2 4 | 2+2 4 | 3+3 2 | 5+5 2 | 5+6 2 | 3+4 2 | 2+2 4 |
| | 2+2 4 | 2+2 4 | 2+2 4 | 4+4 2 | 4+4 2 | 3+4 2 | 5+6 2 | 2+2 4 |
| | 2+2 4 | 2+2 4 | 2+2 4 | 3+3 2 | 3+3 2 | 5+6 2 | 3+4 2 | 2+2 4 |
| | | | | 6+6 2 | 3+3 2 | 3+4 2 | 5+6 2 | |
| | | | | 5+5 2 | 4+4 2 | 5+6 2 | 3+4 2 | |
| | | | | 3+3 2 | 6+6 2 | 5+6 2 | 5+6 2 | |
| | | | | 4+4 2 | 5+5 2 | 5+6 2 | 5+6 2 | |
| | | | | 4+4 2 | 5+5 2 | 3+4 2 | 3+4 2 | |
| | | | | 3+3 2 | 6+6 2 | 5+6 2 | 5+6 2 | |
| | | | | 5+5 2 | 3+3 2 | 3+4 2 | 5+6 2 | |
| | | | | 6+6 2 | 4+4 2 | 3+4 2 | 3+4 2 | |
| | | | | 5+5 2 | 4+4 2 | 5+6 2 | 5+6 2 | |
| | | | | 6+6 2 | 3+3 2 | 5+6 2 | 3+4 2 | |
| | | | | 4+4 2 | 5+5 2 | 5+6 2 | 5+6 2 | |
| | | | | 3+3 2 | 6+6 2 | 3+4 2 | 3+4 2 | |
| | | | | 4+4 2 | 4+4 2 | 3+4 2 | 5+6 2 | |
| | | | | 3+3 2 | 3+3 2 | 3+4 2 | 3+4 2 | |
| | | | | 5+5 2 | 3+3 2 | 3+4 2 | 5+6 2 | |
| | | | | 6+6 2 | 4+4 2 | 5+6 2 | 3+4 2 | |
| | | | | 6+6 2 | 6+6 2 | 3+4 2 | 3+4 2 | |
| | | | | 5+5 2 | 5+5 2 | 5+6 2 | 5+6 2 | |
| | | | | 3+3 2 | 6+6 2 | 5+6 2 | 3+4 2 | |
| | | | | 4+4 2 | 5+5 2 | 3+4 2 | 5+6 2 | |
| | | | | 5+5 2 | 5+5 2 | 5+6 2 | 3+4 2 | |
| | | | | 6+6 2 | 6+6 2 | 3+4 2 | 5+6 2 | |
| | | | | 4+4 2 | 5+5 2 | 3+4 2 | 3+4 2 | |
| | | | | 3+3 2 | 6+6 2 | 3+4 2 | 5+6 2 | |
| | | | | 6+6 2 | 5+5 2 | 3+4 2 | 3+4 2 | |
| | | | | 3+3 2 | 6+6 2 | 3+4 2 | 5+6 2 | |
| | | | | 5+5 2 | 3+3 2 | 3+4 2 | 3+4 2 | |
| | | | | 4+4 2 | 4+4 2 | 3+4 2 | 5+6 2 | |
| | | | | 5+5 2 | 6+6 2 | 5+6 2 | 3+4 2 | |
| | | | | 4+4 2 | 5+5 2 | 3+4 2 | 5+6 2 | |
| | | | | 6+6 2 | 3+3 2 | 3+4 2 | 5+6 2 | |
| | | | | 3+3 2 | 4+4 2 | 5+6 2 | 3+4 2 | |
| | | | | 5+5 2 | 3+3 2 | 5+6 2 | 5+6 2 | |
| | | | | 4+4 2 | 4+4 2 | 3+4 2 | 3+4 2 | |
| | | | | 6+6 2 | 6+6 2 | 5+6 2 | 3+4 2 | |

Continued on next column

**TABLE XXVI– continued**

| TQC | ICSD | NoE | SF | | | | |
|---|---|---|---|---|---|---|---|
| | | | | 3+3 2 | 5+5 2 | 5+6 2 | 5+6 2 | |
| | | | | 4+4 2 | 5+5 2 | 3+4 2 | 5+6 2 | |
| | | | | 5+5 2 | 6+6 2 | 5+6 2 | 3+4 2 | |
| | | | | 6+6 2 | 6+6 2 | 3+4 2 | 3+4 2 | |
| | | | | 3+3 2 | 5+5 2 | 3+4 2 | 5+6 2 | |
| | | | | 3+3 2 | 5+5 2 | 3+4 2 | 5+6 2 | |
| | | | | 4+4 2 | 6+6 2 | 3+4 2 | 3+4 2 | |
| | | | | 6+6 2 | 5+5 2 | 3+4 2 | 3+4 2 | |
| | | | | 5+5 2 | 6+6 2 | 3+4 2 | 5+6 2 | |
| | | | | 4+4 2 | 3+3 2 | 3+4 2 | 5+6 2 | |
| | | | | 5+5 2 | 4+4 2 | 5+6 2 | 3+4 2 | |
| | | | | 3+3 2 | 6+6 2 | 3+4 2 | 5+6 2 | |
| | | | | 6+6 2 | 5+5 2 | 3+4 2 | 3+4 2 | |
| 14.2.37 | 60653 | 208 | Cu2O4Se1 | | | | | |
| | -A | -B | -C | -D | -E | -GM | -Y | -Z |
| | 2+2 4 | 2+2 4 | 2+2 4 | 5+5 2 | 3+3 2 | 3+4 2 | 5+6 2 | 2+2 4 |
| | 2+2 4 | 2+2 4 | 2+2 4 | 6+6 2 | 4+4 2 | 3+4 2 | 5+6 2 | 2+2 4 |
| | 2+2 4 | 2+2 4 | 2+2 4 | 3+3 2 | 5+5 2 | 5+6 2 | 3+4 2 | 2+2 4 |
| | 2+2 4 | 2+2 4 | 2+2 4 | 4+4 2 | 6+6 2 | 5+6 2 | 3+4 2 | 2+2 4 |
| | 2+2 4 | 2+2 4 | 2+2 4 | 3+3 2 | 3+3 2 | 5+6 2 | 5+6 2 | 2+2 4 |
| | 2+2 4 | 2+2 4 | 2+2 4 | 4+4 2 | 4+4 2 | 3+4 2 | 3+4 2 | 2+2 4 |
| | 2+2 4 | 2+2 4 | 2+2 4 | 5+5 2 | 6+6 2 | 3+4 2 | 3+4 2 | 2+2 4 |
| | 2+2 4 | 2+2 4 | 2+2 4 | 6+6 2 | 5+5 2 | 5+6 2 | 5+6 2 | 2+2 4 |
| | 2+2 4 | 2+2 4 | 2+2 4 | 3+3 2 | 5+5 2 | 3+4 2 | 3+4 2 | 2+2 4 |
| | 2+2 4 | 2+2 4 | 2+2 4 | 4+4 2 | 6+6 2 | 3+4 2 | 3+4 2 | 2+2 4 |
| | 2+2 4 | 2+2 4 | 2+2 4 | 6+6 2 | 4+4 2 | 5+6 2 | 5+6 2 | 2+2 4 |
| | 2+2 4 | 2+2 4 | 2+2 4 | 5+5 2 | 3+3 2 | 5+6 2 | 5+6 2 | 2+2 4 |
| | 2+2 4 | 2+2 4 | 2+2 4 | 3+3 2 | 6+6 2 | 3+4 2 | 5+6 2 | 2+2 4 |
| | 2+2 4 | 2+2 4 | 2+2 4 | 4+4 2 | 3+3 2 | 3+4 2 | 3+4 2 | 2+2 4 |
| | 2+2 4 | 2+2 4 | 2+2 4 | 5+5 2 | 5+5 2 | 5+6 2 | 5+6 2 | 2+2 4 |
| | 2+2 4 | 2+2 4 | 2+2 4 | 6+6 2 | 4+4 2 | 5+6 2 | 3+4 2 | 2+2 4 |
| | 2+2 4 | 2+2 4 | 2+2 4 | 3+3 2 | 5+5 2 | 5+6 2 | 3+4 2 | 2+2 4 |
| | 2+2 4 | 2+2 4 | 2+2 4 | 4+4 2 | 6+6 2 | 3+4 2 | 5+6 2 | 2+2 4 |
| | 2+2 4 | 2+2 4 | 2+2 4 | 5+5 2 | 3+3 2 | 3+4 2 | 5+6 2 | 2+2 4 |
| | 2+2 4 | 2+2 4 | 2+2 4 | 6+6 2 | 4+4 2 | 5+6 2 | 3+4 2 | 2+2 4 |
| | 2+2 4 | 2+2 4 | 2+2 4 | 4+4 2 | 3+3 2 | 3+4 2 | 5+6 2 | 2+2 4 |
| | 2+2 4 | 2+2 4 | 2+2 4 | 3+3 2 | 4+4 2 | 3+4 2 | 3+4 2 | 2+2 4 |
| | 2+2 4 | 2+2 4 | 2+2 4 | 6+6 2 | 6+6 2 | 5+6 2 | 3+4 2 | 2+2 4 |
| | 2+2 4 | 2+2 4 | 2+2 4 | 5+5 2 | 5+5 2 | 3+4 2 | 5+6 2 | 2+2 4 |
| | 2+2 4 | 2+2 4 | 2+2 4 | 3+3 2 | 4+4 2 | 3+4 2 | 5+6 2 | 2+2 4 |
| | 2+2 4 | 2+2 4 | 2+2 4 | 4+4 2 | 3+3 2 | 3+4 2 | 3+4 2 | 2+2 4 |
| | 2+2 4 | 2+2 4 | 2+2 4 | 5+5 2 | 5+5 2 | 5+6 2 | 5+6 2 | 2+2 4 |
| | 2+2 4 | 2+2 4 | 2+2 4 | 6+6 2 | 6+6 2 | 3+4 2 | 5+6 2 | 2+2 4 |
| | 2+2 4 | 2+2 4 | 2+2 4 | 5+5 2 | 3+3 2 | 5+6 2 | 3+4 2 | 2+2 4 |
| | 2+2 4 | 2+2 4 | 2+2 4 | 6+6 2 | 4+4 2 | 5+6 2 | 3+4 2 | 2+2 4 |
| | 2+2 4 | 2+2 4 | 2+2 4 | 3+3 2 | 5+5 2 | 5+6 2 | 5+6 2 | 2+2 4 |
| | 2+2 4 | 2+2 4 | 2+2 4 | 4+4 2 | 6+6 2 | 5+6 2 | 3+4 2 | 2+2 4 |
| | 2+2 4 | 2+2 4 | 2+2 4 | 3+3 2 | 5+5 2 | 3+4 2 | 5+6 2 | 2+2 4 |
| | 2+2 4 | 2+2 4 | 2+2 4 | 4+4 2 | 6+6 2 | 3+4 2 | 3+4 2 | 2+2 4 |
| | 2+2 4 | 2+2 4 | 2+2 4 | 3+3 2 | 3+3 2 | 5+6 2 | 3+4 2 | 2+2 4 |
| | 2+2 4 | 2+2 4 | 2+2 4 | 4+4 2 | 4+4 2 | 3+4 2 | 5+6 2 | 2+2 4 |
| | 2+2 4 | 2+2 4 | 2+2 4 | 5+5 2 | 5+5 2 | 3+4 2 | 5+6 2 | 2+2 4 |
| | 2+2 4 | 2+2 4 | 2+2 4 | 6+6 2 | 6+6 2 | 5+6 2 | 5+6 2 | 2+2 4 |
| | 2+2 4 | 2+2 4 | 2+2 4 | 4+4 2 | 3+3 2 | 3+4 2 | 3+4 2 | 2+2 4 |
| | 2+2 4 | 2+2 4 | 2+2 4 | 3+3 2 | 4+4 2 | 3+4 2 | 3+4 2 | 2+2 4 |
| | 2+2 4 | 2+2 4 | 2+2 4 | 6+6 2 | 4+4 2 | 3+4 2 | 3+4 2 | 2+2 4 |
| | 2+2 4 | 2+2 4 | 2+2 4 | 5+5 2 | 3+3 2 | 5+6 2 | 5+6 2 | 2+2 4 |
| | 2+2 4 | 2+2 4 | 2+2 4 | 4+4 2 | 6+6 2 | 3+4 2 | 5+6 2 | 2+2 4 |
| | 2+2 4 | 2+2 4 | 2+2 4 | 3+3 2 | 5+5 2 | 5+6 2 | 3+4 2 | 2+2 4 |
| | 2+2 4 | 2+2 4 | 2+2 4 | 3+3 2 | 6+6 2 | 5+6 2 | 3+4 2 | 2+2 4 |
| | 2+2 4 | 2+2 4 | 2+2 4 | 4+4 2 | 5+5 2 | 5+6 2 | 5+6 2 | 2+2 4 |

Continued on next column

**TABLE XXVI– continued**

| TQC | ICSD | NoE | SF | | | | |
|---|---|---|---|---|---|---|---|
| | 2+2 4 | 2+2 4 | 2+2 4 | 6+6 2 | 3+3 2 | 3+4 2 | 3+4 2 | 2+2 4 |
| | 2+2 4 | 2+2 4 | 2+2 4 | 5+5 2 | 4+4 2 | 3+4 2 | 5+6 2 | 2+2 4 |
| | 2+2 4 | 2+2 4 | 2+2 4 | 5+5 2 | 4+4 2 | 5+6 2 | 5+6 2 | 2+2 4 |
| | 2+2 4 | 2+2 4 | 2+2 4 | 6+6 2 | 3+3 2 | 5+6 2 | 3+4 2 | 2+2 4 |
| | 2+2 4 | 2+2 4 | 2+2 4 | 6+6 2 | 5+5 2 | 3+4 2 | 3+4 2 | 2+2 4 |
| | 2+2 4 | 2+2 4 | 2+2 4 | 5+5 2 | 6+6 2 | 3+4 2 | 5+6 2 | 2+2 4 |
| | | | | 4+4 2 | 4+4 2 | 3+4 2 | 3+4 2 | |
| | | | | 3+3 2 | 3+3 2 | 5+6 2 | 5+6 2 | |
| | | | | 5+5 2 | 5+5 2 | 3+4 2 | 3+4 2 | |
| | | | | 6+6 2 | 6+6 2 | 5+6 2 | 5+6 2 | |
| | | | | 3+3 2 | 3+3 2 | 3+4 2 | 5+6 2 | |
| | | | | 4+4 2 | 4+4 2 | 3+4 2 | 3+4 2 | |
| | | | | 4+4 2 | 6+6 2 | 5+6 2 | 3+4 2 | |
| | | | | 3+3 2 | 5+5 2 | 3+4 2 | 3+4 2 | |
| | | | | 6+6 2 | 4+4 2 | 3+4 2 | 5+6 2 | |
| | | | | 5+5 2 | 3+3 2 | 5+6 2 | 5+6 2 | |
| | | | | 3+3 2 | 6+6 2 | 5+6 2 | 3+4 2 | |
| | | | | 4+4 2 | 5+5 2 | 5+6 2 | 5+6 2 | |
| | | | | 5+5 2 | 5+5 2 | 3+4 2 | 3+4 2 | |
| | | | | 6+6 2 | 6+6 2 | 3+4 2 | 5+6 2 | |
| | | | | 3+3 2 | 3+3 2 | 5+6 2 | 3+4 2 | |
| | | | | 4+4 2 | 4+4 2 | 5+6 2 | 5+6 2 | |
| | | | | 6+6 2 | 4+4 2 | 5+6 2 | 5+6 2 | |
| | | | | 5+5 2 | 3+3 2 | 3+4 2 | 3+4 2 | |
| | | | | 3+3 2 | 6+6 2 | 3+4 2 | 5+6 2 | |
| | | | | 4+4 2 | 5+5 2 | 3+4 2 | 3+4 2 | |
| | | | | 4+4 2 | 3+3 2 | 3+4 2 | 5+6 2 | |
| | | | | 3+3 2 | 6+6 2 | 3+4 2 | 3+4 2 | |
| | | | | 3+3 2 | 5+5 2 | 5+6 2 | 5+6 2 | |
| | | | | 4+4 2 | 4+4 2 | 3+4 2 | 3+4 2 | |
| | | | | 4+4 2 | 4+4 2 | 3+4 2 | 5+6 2 | |
| | | | | 3+3 2 | 5+5 2 | 3+4 2 | 3+4 2 | |
| | | | | 3+3 2 | 6+6 2 | 3+4 2 | 5+6 2 | |
| | | | | 4+4 2 | 3+3 2 | 3+4 2 | 3+4 2 | |
| | | | | 3+3 2 | 5+5 2 | 5+6 2 | 5+6 2 | |
| | | | | 4+4 2 | 6+6 2 | 3+4 2 | 3+4 2 | |
| | | | | 3+3 2 | 4+4 2 | 3+4 2 | 3+4 2 | |
| | | | | 4+4 2 | 3+3 2 | 5+6 2 | 5+6 2 | |
| | | | | 5+5 2 | 6+6 2 | 3+4 2 | 3+4 2 | |
| | | | | 6+6 2 | 5+5 2 | 5+6 2 | 5+6 2 | |
| | | | | 6+6 2 | 3+3 2 | 3+4 2 | 3+4 2 | |
| | | | | 5+5 2 | 4+4 2 | 3+4 2 | 5+6 2 | |
| | | | | 4+4 2 | 4+4 2 | 3+4 2 | 3+4 2 | |
| | | | | 3+3 2 | 5+5 2 | 3+4 2 | 5+6 2 | |
| | | | | 4+4 2 | 3+3 2 | 5+6 2 | 3+4 2 | |
| | | | | 3+3 2 | 6+6 2 | 3+4 2 | 5+6 2 | |
| | | | | 6+6 2 | 3+3 2 | 5+6 2 | 3+4 2 | |
| | | | | 3+3 2 | 4+4 2 | 3+4 2 | 5+6 2 | |
| | | | | 5+5 2 | 5+5 2 | 5+6 2 | 5+6 2 | |
| | | | | 4+4 2 | 6+6 2 | 3+4 2 | 3+4 2 | |
| | | | | 6+6 2 | 3+3 2 | 5+6 2 | 3+4 2 | |
| | | | | 5+5 2 | 4+4 2 | 3+4 2 | 5+6 2 | |
| | | | | 4+4 2 | 5+5 2 | 5+6 2 | 3+4 2 | |
| | | | | 3+3 2 | 6+6 2 | 3+4 2 | 5+6 2 | |
| | | | | 4+4 2 | 4+4 2 | 3+4 2 | 3+4 2 | |
| | | | | 3+3 2 | 3+3 2 | 3+4 2 | 5+6 2 | |
| | | | | 6+6 2 | 5+5 2 | 5+6 2 | 3+4 2 | |
| | | | | 5+5 2 | 6+6 2 | 3+4 2 | 5+6 2 | |
| 14.2.38 | 1672 | 164 | Cu1O4Te1 | | | | | |
| | -A | -B | -C | -D | -E | -GM | -Y | -Z |
| | 2+2 4 | 2+2 4 | 2+2 4 | 5+5 2 | 4+4 2 | 3+4 2 | 5+6 2 | 2+2 4 |

Continued on next column

**TABLE XXVI– continued**

| TQC | ICSD | NoE | SF | | | | |
|---|---|---|---|---|---|---|---|
| | 2+2 4 | 2+2 4 | 2+2 4 | 6+6 2 | 5+5 2 | 5+6 2 | 3+4 2 | 2+2 4 |
| | 2+2 4 | 2+2 4 | 2+2 4 | 3+3 2 | 6+6 2 | 3+4 2 | 3+4 2 | 2+2 4 |
| | 2+2 4 | 2+2 4 | 2+2 4 | 4+4 2 | 3+3 2 | 5+6 2 | 5+6 2 | 2+2 4 |
| | 2+2 4 | 2+2 4 | 2+2 4 | 5+5 2 | 6+6 2 | 3+4 2 | 3+4 2 | 2+2 4 |
| | 2+2 4 | 2+2 4 | 2+2 4 | 6+6 2 | 3+3 2 | 5+6 2 | 5+6 2 | 2+2 4 |
| | 2+2 4 | 2+2 4 | 2+2 4 | 3+3 2 | 5+5 2 | 5+6 2 | 3+4 2 | 2+2 4 |
| | 2+2 4 | 2+2 4 | 2+2 4 | 4+4 2 | 4+4 2 | 3+4 2 | 5+6 2 | 2+2 4 |
| | 2+2 4 | 2+2 4 | 2+2 4 | 3+3 2 | 4+4 2 | 3+4 2 | 3+4 2 | 2+2 4 |
| | 2+2 4 | 2+2 4 | 2+2 4 | 4+4 2 | 3+3 2 | 5+6 2 | 3+4 2 | 2+2 4 |
| | 2+2 4 | 2+2 4 | 2+2 4 | 6+6 2 | 5+5 2 | 5+6 2 | 5+6 2 | 2+2 4 |
| | 2+2 4 | 2+2 4 | 2+2 4 | 5+5 2 | 6+6 2 | 3+4 2 | 5+6 2 | 2+2 4 |
| | 2+2 4 | 2+2 4 | 2+2 4 | 5+5 2 | 5+5 2 | 3+4 2 | 3+4 2 | 2+2 4 |
| | 2+2 4 | 2+2 4 | 2+2 4 | 6+6 2 | 6+6 2 | 5+6 2 | 3+4 2 | 2+2 4 |
| | 2+2 4 | 2+2 4 | 2+2 4 | 4+4 2 | 4+4 2 | 3+4 2 | 5+6 2 | 2+2 4 |
| | 2+2 4 | 2+2 4 | 2+2 4 | 3+3 2 | 3+3 2 | 5+6 2 | 5+6 2 | 2+2 4 |
| | 2+2 4 | 2+2 4 | 2+2 4 | 3+3 2 | 5+5 2 | 3+4 2 | 3+4 2 | 2+2 4 |
| | 2+2 4 | 2+2 4 | 2+2 4 | 4+4 2 | 6+6 2 | 5+6 2 | 5+6 2 | 2+2 4 |
| | 2+2 4 | 2+2 4 | 2+2 4 | 5+5 2 | 4+4 2 | 3+4 2 | 5+6 2 | 2+2 4 |
| | 2+2 4 | 2+2 4 | 2+2 4 | 6+6 2 | 3+3 2 | 3+4 2 | 3+4 2 | 2+2 4 |
| | 2+2 4 | 2+2 4 | 2+2 4 | 6+6 2 | 5+5 2 | 5+6 2 | 3+4 2 | 2+2 4 |
| | 2+2 4 | 2+2 4 | 2+2 4 | 5+5 2 | 4+4 2 | 5+6 2 | 5+6 2 | 2+2 4 |
| | 2+2 4 | 2+2 4 | 2+2 4 | 4+4 2 | 6+6 2 | 3+4 2 | 3+4 2 | 2+2 4 |
| | 2+2 4 | 2+2 4 | 2+2 4 | 3+3 2 | 3+3 2 | 5+6 2 | 5+6 2 | 2+2 4 |
| | 2+2 4 | 2+2 4 | 2+2 4 | 4+4 2 | 3+3 2 | 3+4 2 | 3+4 2 | 2+2 4 |
| | 2+2 4 | 2+2 4 | 2+2 4 | 3+3 2 | 4+4 2 | 5+6 2 | 3+4 2 | 2+2 4 |
| | 2+2 4 | 2+2 4 | 2+2 4 | 6+6 2 | 6+6 2 | 3+4 2 | 5+6 2 | 2+2 4 |
| | 2+2 4 | 2+2 4 | 2+2 4 | 5+5 2 | 5+5 2 | 5+6 2 | 5+6 2 | 2+2 4 |
| | 2+2 4 | 2+2 4 | 2+2 4 | 5+5 2 | 3+3 2 | 3+4 2 | 3+4 2 | 2+2 4 |
| | 2+2 4 | 2+2 4 | 2+2 4 | 6+6 2 | 4+4 2 | 3+4 2 | 3+4 2 | 2+2 4 |
| | 2+2 4 | 2+2 4 | 2+2 4 | 3+3 2 | 6+6 2 | 3+4 2 | 3+4 2 | 2+2 4 |
| | 2+2 4 | 2+2 4 | 2+2 4 | 4+4 2 | 5+5 2 | 5+6 2 | 3+4 2 | 2+2 4 |
| | 2+2 4 | 2+2 4 | 2+2 4 | 6+6 2 | 5+5 2 | 3+4 2 | 5+6 2 | 2+2 4 |
| | 2+2 4 | 2+2 4 | 2+2 4 | 5+5 2 | 6+6 2 | 5+6 2 | 5+6 2 | 2+2 4 |
| | 2+2 4 | 2+2 4 | 2+2 4 | 4+4 2 | 4+4 2 | 3+4 2 | 5+6 2 | 2+2 4 |
| | 2+2 4 | 2+2 4 | 2+2 4 | 3+3 2 | 3+3 2 | 5+6 2 | 5+6 2 | 2+2 4 |
| | 2+2 4 | 2+2 4 | 2+2 4 | 4+4 2 | 3+3 2 | 5+6 2 | 3+4 2 | 2+2 4 |
| | 2+2 4 | 2+2 4 | 2+2 4 | 3+3 2 | 4+4 2 | 3+4 2 | 3+4 2 | 2+2 4 |
| | 2+2 4 | 2+2 4 | 2+2 4 | 6+6 2 | 5+5 2 | 3+4 2 | 3+4 2 | 2+2 4 |
| | 2+2 4 | 2+2 4 | 2+2 4 | 5+5 2 | 6+6 2 | 5+6 2 | 3+4 2 | 2+2 4 |
| | 2+2 4 | 2+2 4 | 2+2 4 | 3+3 2 | 3+3 2 | 3+4 2 | 5+6 2 | 2+2 4 |
| | | | | 5+5 2 | 4+4 2 | 5+6 2 | 5+6 2 | |
| | | | | 4+4 2 | 6+6 2 | 3+4 2 | 3+4 2 | |
| | | | | 6+6 2 | 5+5 2 | 3+4 2 | 3+4 2 | |
| | | | | 3+3 2 | 3+3 2 | 5+6 2 | 5+6 2 | |
| | | | | 4+4 2 | 4+4 2 | 3+4 2 | 5+6 2 | |
| | | | | 5+5 2 | 6+6 2 | 3+4 2 | 3+4 2 | |
| | | | | 6+6 2 | 5+5 2 | 3+4 2 | 3+4 2 | |
| | | | | 4+4 2 | 6+6 2 | 5+6 2 | 3+4 2 | |
| | | | | 3+3 2 | 5+5 2 | 5+6 2 | 3+4 2 | |
| | | | | 5+5 2 | 3+3 2 | 3+4 2 | 5+6 2 | |
| | | | | 6+6 2 | 4+4 2 | 5+6 2 | 5+6 2 | |
| | | | | 4+4 2 | 4+4 2 | 5+6 2 | 5+6 2 | |
| | | | | 3+3 2 | 3+3 2 | 5+6 2 | 5+6 2 | |
| | | | | 6+6 2 | 5+5 2 | 3+4 2 | 3+4 2 | |
| | | | | 5+5 2 | 6+6 2 | 3+4 2 | 3+4 2 | |
| | | | | 3+3 2 | 5+5 2 | 5+6 2 | 3+4 2 | |
| | | | | 4+4 2 | 6+6 2 | 3+4 2 | 5+6 2 | |
| | | | | 5+5 2 | 4+4 2 | 3+4 2 | 5+6 2 | |
| | | | | 6+6 2 | 3+3 2 | 5+6 2 | 3+4 2 | |
| | | | | 3+3 2 | 3+3 2 | 3+4 2 | 5+6 2 | |
| | | | | 5+5 2 | 6+6 2 | 3+4 2 | 5+6 2 | |

Continued on next column

**TABLE XXVI– continued**

| TQC | ICSD | | NoE | SF | | | | |
|---|---|---|---|---|---|---|---|---|
| | | | | 4+4 2 | 4+4 2 | 5+6 2 | 3+4 2 | |
| | | | | 6+6 2 | 5+5 2 | 5+6 2 | 3+4 2 | |
| | | | | 6+6 2 | 6+6 2 | 3+4 2 | 3+4 2 | |
| | | | | 5+5 2 | 5+5 2 | 3+4 2 | 5+6 2 | |
| | | | | 4+4 2 | 3+3 2 | 5+6 2 | 3+4 2 | |
| | | | | 3+3 2 | 4+4 2 | 3+4 2 | 5+6 2 | |
| | | | | 3+3 2 | 4+4 2 | 3+4 2 | 3+4 2 | |
| | | | | 4+4 2 | 5+5 2 | 5+6 2 | 3+4 2 | |
| | | | | 5+5 2 | 3+3 2 | 3+4 2 | 3+4 2 | |
| | | | | 6+6 2 | 6+6 2 | 3+4 2 | 3+4 2 | |
| | | | | 4+4 2 | 3+3 2 | 3+4 2 | 3+4 2 | |
| | | | | 3+3 2 | 4+4 2 | 3+4 2 | 3+4 2 | |
| | | | | 5+5 2 | 6+6 2 | 3+4 2 | 3+4 2 | |
| | | | | 6+6 2 | 5+5 2 | 3+4 2 | 3+4 2 | |
| | | | | 3+3 2 | 3+3 2 | 3+4 2 | 3+4 2 | |
| | | | | 4+4 2 | 4+4 2 | 3+4 2 | 3+4 2 | |
| | | | | 5+5 2 | 5+5 2 | 3+4 2 | 3+4 2 | |
| | | | | 6+6 2 | 6+6 2 | 3+4 2 | 3+4 2 | |
| | | | | 4+4 2 | 6+6 2 | 3+4 2 | 3+4 2 | |
| | | | | 3+3 2 | 5+5 2 | 3+4 2 | 3+4 2 | |
| 14.2.39 | 1671 | | 164 | Cu1O4Te1 | | | | |
| | -A | -B | -C | -D | -E | -GM | -Y | -Z |
| | 2+2 4 | 2+2 4 | 2+2 4 | 5+5 2 | 4+4 2 | 3+4 2 | 5+6 2 | 2+2 4 |
| | 2+2 4 | 2+2 4 | 2+2 4 | 6+6 2 | 5+5 2 | 5+6 2 | 3+4 2 | 2+2 4 |
| | 2+2 4 | 2+2 4 | 2+2 4 | 3+3 2 | 6+6 2 | 3+4 2 | 3+4 2 | 2+2 4 |
| | 2+2 4 | 2+2 4 | 2+2 4 | 4+4 2 | 3+3 2 | 5+6 2 | 5+6 2 | 2+2 4 |
| | 2+2 4 | 2+2 4 | 2+2 4 | 5+5 2 | 6+6 2 | 3+4 2 | 3+4 2 | 2+2 4 |
| | 2+2 4 | 2+2 4 | 2+2 4 | 6+6 2 | 3+3 2 | 5+6 2 | 5+6 2 | 2+2 4 |
| | 2+2 4 | 2+2 4 | 2+2 4 | 3+3 2 | 5+5 2 | 5+6 2 | 3+4 2 | 2+2 4 |
| | 2+2 4 | 2+2 4 | 2+2 4 | 4+4 2 | 4+4 2 | 3+4 2 | 5+6 2 | 2+2 4 |
| | 2+2 4 | 2+2 4 | 2+2 4 | 3+3 2 | 4+4 2 | 3+4 2 | 3+4 2 | 2+2 4 |
| | 2+2 4 | 2+2 4 | 2+2 4 | 4+4 2 | 3+3 2 | 5+6 2 | 3+4 2 | 2+2 4 |
| | 2+2 4 | 2+2 4 | 2+2 4 | 6+6 2 | 5+5 2 | 5+6 2 | 5+6 2 | 2+2 4 |
| | 2+2 4 | 2+2 4 | 2+2 4 | 5+5 2 | 6+6 2 | 3+4 2 | 5+6 2 | 2+2 4 |
| | 2+2 4 | 2+2 4 | 2+2 4 | 5+5 2 | 5+5 2 | 3+4 2 | 3+4 2 | 2+2 4 |
| | 2+2 4 | 2+2 4 | 2+2 4 | 6+6 2 | 6+6 2 | 5+6 2 | 3+4 2 | 2+2 4 |
| | 2+2 4 | 2+2 4 | 2+2 4 | 4+4 2 | 4+4 2 | 3+4 2 | 5+6 2 | 2+2 4 |
| | 2+2 4 | 2+2 4 | 2+2 4 | 3+3 2 | 3+3 2 | 5+6 2 | 5+6 2 | 2+2 4 |
| | 2+2 4 | 2+2 4 | 2+2 4 | 3+3 2 | 5+5 2 | 3+4 2 | 3+4 2 | 2+2 4 |
| | 2+2 4 | 2+2 4 | 2+2 4 | 4+4 2 | 6+6 2 | 5+6 2 | 5+6 2 | 2+2 4 |
| | 2+2 4 | 2+2 4 | 2+2 4 | 5+5 2 | 4+4 2 | 3+4 2 | 5+6 2 | 2+2 4 |
| | 2+2 4 | 2+2 4 | 2+2 4 | 6+6 2 | 3+3 2 | 3+4 2 | 3+4 2 | 2+2 4 |
| | 2+2 4 | 2+2 4 | 2+2 4 | 6+6 2 | 5+5 2 | 5+6 2 | 3+4 2 | 2+2 4 |
| | 2+2 4 | 2+2 4 | 2+2 4 | 5+5 2 | 4+4 2 | 5+6 2 | 5+6 2 | 2+2 4 |
| | 2+2 4 | 2+2 4 | 2+2 4 | 4+4 2 | 6+6 2 | 3+4 2 | 3+4 2 | 2+2 4 |
| | 2+2 4 | 2+2 4 | 2+2 4 | 3+3 2 | 3+3 2 | 5+6 2 | 5+6 2 | 2+2 4 |
| | 2+2 4 | 2+2 4 | 2+2 4 | 4+4 2 | 3+3 2 | 3+4 2 | 3+4 2 | 2+2 4 |
| | 2+2 4 | 2+2 4 | 2+2 4 | 3+3 2 | 4+4 2 | 5+6 2 | 3+4 2 | 2+2 4 |
| | 2+2 4 | 2+2 4 | 2+2 4 | 6+6 2 | 6+6 2 | 3+4 2 | 5+6 2 | 2+2 4 |
| | 2+2 4 | 2+2 4 | 2+2 4 | 5+5 2 | 5+5 2 | 5+6 2 | 5+6 2 | 2+2 4 |
| | 2+2 4 | 2+2 4 | 2+2 4 | 5+5 2 | 3+3 2 | 3+4 2 | 3+4 2 | 2+2 4 |
| | 2+2 4 | 2+2 4 | 2+2 4 | 6+6 2 | 4+4 2 | 3+4 2 | 3+4 2 | 2+2 4 |
| | 2+2 4 | 2+2 4 | 2+2 4 | 3+3 2 | 6+6 2 | 3+4 2 | 3+4 2 | 2+2 4 |
| | 2+2 4 | 2+2 4 | 2+2 4 | 4+4 2 | 5+5 2 | 5+6 2 | 3+4 2 | 2+2 4 |
| | 2+2 4 | 2+2 4 | 2+2 4 | 6+6 2 | 5+5 2 | 3+4 2 | 5+6 2 | 2+2 4 |
| | 2+2 4 | 2+2 4 | 2+2 4 | 5+5 2 | 6+6 2 | 5+6 2 | 5+6 2 | 2+2 4 |
| | 2+2 4 | 2+2 4 | 2+2 4 | 4+4 2 | 4+4 2 | 3+4 2 | 5+6 2 | 2+2 4 |
| | 2+2 4 | 2+2 4 | 2+2 4 | 3+3 2 | 3+3 2 | 5+6 2 | 5+6 2 | 2+2 4 |
| | 2+2 4 | 2+2 4 | 2+2 4 | 4+4 2 | 3+3 2 | 5+6 2 | 3+4 2 | 2+2 4 |
| | 2+2 4 | 2+2 4 | 2+2 4 | 3+3 2 | 4+4 2 | 3+4 2 | 3+4 2 | 2+2 4 |
| | 2+2 4 | 2+2 4 | 2+2 4 | 6+6 2 | 5+5 2 | 3+4 2 | 3+4 2 | 2+2 4 |

Continued on next column

| TQC | ICSD | | NoE | SF | | | | |
|---|---|---|---|---|---|---|---|---|
| | 2+2 4 | 2+2 4 | 2+2 4 | 5+5 2 | 6+6 2 | 5+6 2 | 3+4 2 | 2+2 4 |
| | 2+2 4 | 2+2 4 | 2+2 4 | 3+3 2 | 3+3 2 | 3+4 2 | 5+6 2 | 2+2 4 |
| | | | | 5+5 2 | 4+4 2 | 5+6 2 | 5+6 2 | |
| | | | | 4+4 2 | 6+6 2 | 3+4 2 | 3+4 2 | |
| | | | | 6+6 2 | 5+5 2 | 3+4 2 | 3+4 2 | |
| | | | | 3+3 2 | 3+3 2 | 5+6 2 | 5+6 2 | |
| | | | | 4+4 2 | 4+4 2 | 3+4 2 | 5+6 2 | |
| | | | | 5+5 2 | 6+6 2 | 3+4 2 | 3+4 2 | |
| | | | | 6+6 2 | 5+5 2 | 3+4 2 | 3+4 2 | |
| | | | | 4+4 2 | 6+6 2 | 5+6 2 | 3+4 2 | |
| | | | | 3+3 2 | 5+5 2 | 5+6 2 | 3+4 2 | |
| | | | | 5+5 2 | 3+3 2 | 3+4 2 | 5+6 2 | |
| | | | | 6+6 2 | 4+4 2 | 5+6 2 | 5+6 2 | |
| | | | | 4+4 2 | 4+4 2 | 5+6 2 | 5+6 2 | |
| | | | | 3+3 2 | 3+3 2 | 5+6 2 | 5+6 2 | |
| | | | | 6+6 2 | 5+5 2 | 3+4 2 | 3+4 2 | |
| | | | | 5+5 2 | 6+6 2 | 3+4 2 | 3+4 2 | |
| | | | | 3+3 2 | 5+5 2 | 5+6 2 | 3+4 2 | |
| | | | | 4+4 2 | 6+6 2 | 3+4 2 | 5+6 2 | |
| | | | | 5+5 2 | 4+4 2 | 3+4 2 | 5+6 2 | |
| | | | | 6+6 2 | 3+3 2 | 5+6 2 | 3+4 2 | |
| | | | | 3+3 2 | 3+3 2 | 3+4 2 | 5+6 2 | |
| | | | | 5+5 2 | 6+6 2 | 3+4 2 | 5+6 2 | |
| | | | | 4+4 2 | 4+4 2 | 5+6 2 | 3+4 2 | |
| | | | | 6+6 2 | 5+5 2 | 5+6 2 | 3+4 2 | |
| | | | | 6+6 2 | 6+6 2 | 3+4 2 | 3+4 2 | |
| | | | | 5+5 2 | 5+5 2 | 3+4 2 | 5+6 2 | |
| | | | | 4+4 2 | 3+3 2 | 5+6 2 | 3+4 2 | |
| | | | | 3+3 2 | 4+4 2 | 3+4 2 | 5+6 2 | |
| | | | | 3+3 2 | 4+4 2 | 3+4 2 | 3+4 2 | |
| | | | | 4+4 2 | 5+5 2 | 5+6 2 | 3+4 2 | |
| | | | | 5+5 2 | 3+3 2 | 3+4 2 | 3+4 2 | |
| | | | | 6+6 2 | 6+6 2 | 3+4 2 | 3+4 2 | |
| | | | | 4+4 2 | 3+3 2 | 3+4 2 | 3+4 2 | |
| | | | | 3+3 2 | 4+4 2 | 3+4 2 | 3+4 2 | |
| | | | | 5+5 2 | 6+6 2 | 3+4 2 | 3+4 2 | |
| | | | | 6+6 2 | 5+5 2 | 3+4 2 | 3+4 2 | |
| | | | | 3+3 2 | 3+3 2 | 3+4 2 | 3+4 2 | |
| | | | | 4+4 2 | 4+4 2 | 3+4 2 | 3+4 2 | |
| | | | | 5+5 2 | 5+5 2 | 3+4 2 | 3+4 2 | |
| | | | | 6+6 2 | 6+6 2 | 3+4 2 | 3+4 2 | |
| | | | | 4+4 2 | 6+6 2 | 3+4 2 | 3+4 2 | |
| | | | | 3+3 2 | 5+5 2 | 3+4 2 | 3+4 2 | |
| 14.2.40 | 202451 | | 140 | Cu1O3Te1 | | | | |
| | -A | -B | -C | -D | -E | -GM | -Y | -Z |
| | 2+2 4 | 2+2 4 | 2+2 4 | 3+3 2 | 3+3 2 | 3+4 2 | 5+6 2 | 2+2 4 |
| | 2+2 4 | 2+2 4 | 2+2 4 | 4+4 2 | 4+4 2 | 5+6 2 | 3+4 2 | 2+2 4 |
| | 2+2 4 | 2+2 4 | 2+2 4 | 6+6 2 | 6+6 2 | 3+4 2 | 5+6 2 | 2+2 4 |
| | 2+2 4 | 2+2 4 | 2+2 4 | 5+5 2 | 5+5 2 | 5+6 2 | 3+4 2 | 2+2 4 |
| | 2+2 4 | 2+2 4 | 2+2 4 | 4+4 2 | 5+5 2 | 3+4 2 | 3+4 2 | 2+2 4 |
| | 2+2 4 | 2+2 4 | 2+2 4 | 3+3 2 | 6+6 2 | 5+6 2 | 5+6 2 | 2+2 4 |
| | 2+2 4 | 2+2 4 | 2+2 4 | 5+5 2 | 4+4 2 | 3+4 2 | 5+6 2 | 2+2 4 |
| | 2+2 4 | 2+2 4 | 2+2 4 | 6+6 2 | 3+3 2 | 5+6 2 | 5+6 2 | 2+2 4 |
| | 2+2 4 | 2+2 4 | 2+2 4 | 4+4 2 | 5+5 2 | 3+4 2 | 3+4 2 | 2+2 4 |
| | 2+2 4 | 2+2 4 | 2+2 4 | 3+3 2 | 6+6 2 | 5+6 2 | 3+4 2 | 2+2 4 |
| | 2+2 4 | 2+2 4 | 2+2 4 | 5+5 2 | 4+4 2 | 5+6 2 | 5+6 2 | 2+2 4 |
| | 2+2 4 | 2+2 4 | 2+2 4 | 6+6 2 | 3+3 2 | 3+4 2 | 3+4 2 | 2+2 4 |
| | 2+2 4 | 2+2 4 | 2+2 4 | 3+3 2 | 6+6 2 | 3+4 2 | 3+4 2 | 2+2 4 |
| | 2+2 4 | 2+2 4 | 2+2 4 | 4+4 2 | 5+5 2 | 5+6 2 | 5+6 2 | 2+2 4 |
| | 2+2 4 | 2+2 4 | 2+2 4 | 6+6 2 | 3+3 2 | 3+4 2 | 5+6 2 | 2+2 4 |
| | 2+2 4 | 2+2 4 | 2+2 4 | 5+5 2 | 4+4 2 | 5+6 2 | 3+4 2 | 2+2 4 |

Continued on next column

**TABLE XXVI– continued**

| TQC | ICSD | NoE | SF | | | | | |
|---|---|---|---|---|---|---|---|---|
| | 2+2 4 | 2+2 4 | 2+2 4 | 3+3 2 | 6+6 2 | 3+4 2 | 3+4 2 | 2+2 4 |
| | 2+2 4 | 2+2 4 | 2+2 4 | 6+6 2 | 5+5 2 | 5+6 2 | 5+6 2 | 2+2 4 |
| | 2+2 4 | 2+2 4 | 2+2 4 | 4+4 2 | 3+3 2 | 3+4 2 | 5+6 2 | 2+2 4 |
| | 2+2 4 | 2+2 4 | 2+2 4 | 5+5 2 | 4+4 2 | 5+6 2 | 3+4 2 | 2+2 4 |
| | 2+2 4 | 2+2 4 | 2+2 4 | 3+3 2 | 5+5 2 | 5+6 2 | 3+4 2 | 2+2 4 |
| | 2+2 4 | 2+2 4 | 2+2 4 | 4+4 2 | 6+6 2 | 3+4 2 | 5+6 2 | 2+2 4 |
| | 2+2 4 | 2+2 4 | 2+2 4 | 5+5 2 | 4+4 2 | 5+6 2 | 3+4 2 | 2+2 4 |
| | 2+2 4 | 2+2 4 | 2+2 4 | 6+6 2 | 3+3 2 | 3+4 2 | 5+6 2 | 2+2 4 |
| | 2+2 4 | 2+2 4 | 2+2 4 | 5+5 2 | 5+5 2 | 3+4 2 | 3+4 2 | 2+2 4 |
| | 2+2 4 | 2+2 4 | 2+2 4 | 6+6 2 | 6+6 2 | 3+4 2 | 5+6 2 | 2+2 4 |
| | 2+2 4 | 2+2 4 | 2+2 4 | 4+4 2 | 3+3 2 | 5+6 2 | 5+6 2 | 2+2 4 |
| | 2+2 4 | 2+2 4 | 2+2 4 | 3+3 2 | 4+4 2 | 5+6 2 | 3+4 2 | 2+2 4 |
| | 2+2 4 | 2+2 4 | 2+2 4 | 3+3 2 | 3+3 2 | 3+4 2 | 3+4 2 | 2+2 4 |
| | 2+2 4 | 2+2 4 | 2+2 4 | 4+4 2 | 4+4 2 | 3+4 2 | 5+6 2 | 2+2 4 |
| | 2+2 4 | 2+2 4 | 2+2 4 | 5+5 2 | 6+6 2 | 5+6 2 | 5+6 2 | 2+2 4 |
| | 2+2 4 | 2+2 4 | 2+2 4 | 6+6 2 | 5+5 2 | 3+4 2 | 3+4 2 | 2+2 4 |
| | 2+2 4 | 2+2 4 | 2+2 4 | 3+3 2 | 3+3 2 | 3+4 2 | 5+6 2 | 2+2 4 |
| | 2+2 4 | 2+2 4 | 2+2 4 | 4+4 2 | 4+4 2 | 5+6 2 | 3+4 2 | 2+2 4 |
| | 2+2 4 | 2+2 4 | 2+2 4 | 5+5 2 | 6+6 2 | 3+4 2 | 5+6 2 | 2+2 4 |
| | | | | 6+6 2 | 5+5 2 | 5+6 2 | 3+4 2 | |
| | | | | 3+3 2 | 3+3 2 | 3+4 2 | 5+6 2 | |
| | | | | 4+4 2 | 4+4 2 | 5+6 2 | 3+4 2 | |
| | | | | 5+5 2 | 6+6 2 | 5+6 2 | 5+6 2 | |
| | | | | 6+6 2 | 5+5 2 | 5+6 2 | 3+4 2 | |
| | | | | 6+6 2 | 5+5 2 | 3+4 2 | 5+6 2 | |
| | | | | 5+5 2 | 6+6 2 | 5+6 2 | 3+4 2 | |
| | | | | 4+4 2 | 3+3 2 | 5+6 2 | 5+6 2 | |
| | | | | 3+3 2 | 4+4 2 | 3+4 2 | 3+4 2 | |
| | | | | 5+5 2 | 6+6 2 | 3+4 2 | 5+6 2 | |
| | | | | 6+6 2 | 5+5 2 | 5+6 2 | 5+6 2 | |
| | | | | 3+3 2 | 4+4 2 | 3+4 2 | 3+4 2 | |
| | | | | 4+4 2 | 3+3 2 | 5+6 2 | 3+4 2 | |
| | | | | 4+4 2 | 5+5 2 | 3+4 2 | 5+6 2 | |
| | | | | 3+3 2 | 6+6 2 | 5+6 2 | 3+4 2 | |
| | | | | 5+5 2 | 3+3 2 | 3+4 2 | 3+4 2 | |
| | | | | 6+6 2 | 4+4 2 | 5+6 2 | 5+6 2 | |
| | | | | 4+4 2 | 3+3 2 | 3+4 2 | 3+4 2 | |
| | | | | 3+3 2 | 4+4 2 | 5+6 2 | 5+6 2 | |
| | | | | 6+6 2 | 4+4 2 | 3+4 2 | 3+4 2 | |
| | | | | 5+5 2 | 3+3 2 | 3+4 2 | 5+6 2 | |
| | | | | 3+3 2 | 6+6 2 | 3+4 2 | 3+4 2 | |
| | | | | 4+4 2 | 5+5 2 | 3+4 2 | 5+6 2 | |
| | | | | 5+5 2 | 5+5 2 | 5+6 2 | 3+4 2 | |
| | | | | 6+6 2 | 6+6 2 | 5+6 2 | 3+4 2 | |
| | | | | 5+5 2 | 4+4 2 | 5+6 2 | 5+6 2 | |
| | | | | 4+4 2 | 3+3 2 | 3+4 2 | 5+6 2 | |
| | | | | 3+3 2 | 5+5 2 | 5+6 2 | 3+4 2 | |
| | | | | 6+6 2 | 6+6 2 | 3+4 2 | 5+6 2 | |
| | | | | 4+4 2 | 4+4 2 | 5+6 2 | 5+6 2 | |
| | | | | 3+3 2 | 3+3 2 | 5+6 2 | 3+4 2 | |
| | | | | 5+5 2 | 5+5 2 | 5+6 2 | 5+6 2 | |
| | | | | 6+6 2 | 6+6 2 | 3+4 2 | 3+4 2 | |
| | | | | 6+6 2 | 4+4 2 | 5+6 2 | 3+4 2 | |
| | | | | 5+5 2 | 3+3 2 | 3+4 2 | 3+4 2 | |
| 14.2.41 | 260194 | 152 | Cu1Np1Se2 | | | | | |
| | -A | -B | -C | -D | -E | -GM | -Y | -Z |
| | 2+2 4 | 2+2 4 | 2+2 4 | 3+3 2 | 5+5 2 | 3+4 2 | 5+6 2 | 2+2 4 |
| | 2+2 4 | 2+2 4 | 2+2 4 | 4+4 2 | 6+6 2 | 3+4 2 | 3+4 2 | 2+2 4 |
| | 2+2 4 | 2+2 4 | 2+2 4 | 5+5 2 | 3+3 2 | 5+6 2 | 5+6 2 | 2+2 4 |
| | 2+2 4 | 2+2 4 | 2+2 4 | 6+6 2 | 4+4 2 | 5+6 2 | 3+4 2 | 2+2 4 |
| | 2+2 4 | 2+2 4 | 2+2 4 | 3+3 2 | 5+5 2 | 3+4 2 | 5+6 2 | 2+2 4 |

Continued on next column

**TABLE XXVI– continued**

| TQC | ICSD | NoE | SF | | | | | |
|---|---|---|---|---|---|---|---|---|
| | 2+2 4 | 2+2 4 | 2+2 4 | 4+4 2 | 6+6 2 | 5+6 2 | 3+4 2 | 2+2 4 |
| | 2+2 4 | 2+2 4 | 2+2 4 | 5+5 2 | 3+3 2 | 3+4 2 | 5+6 2 | 2+2 4 |
| | 2+2 4 | 2+2 4 | 2+2 4 | 6+6 2 | 4+4 2 | 5+6 2 | 3+4 2 | 2+2 4 |
| | 2+2 4 | 2+2 4 | 2+2 4 | 6+6 2 | 3+3 2 | 5+6 2 | 5+6 2 | 2+2 4 |
| | 2+2 4 | 2+2 4 | 2+2 4 | 5+5 2 | 4+4 2 | 3+4 2 | 3+4 2 | 2+2 4 |
| | 2+2 4 | 2+2 4 | 2+2 4 | 3+3 2 | 6+6 2 | 3+4 2 | 5+6 2 | 2+2 4 |
| | 2+2 4 | 2+2 4 | 2+2 4 | 4+4 2 | 5+5 2 | 3+4 2 | 3+4 2 | 2+2 4 |
| | 2+2 4 | 2+2 4 | 2+2 4 | 4+4 2 | 4+4 2 | 3+4 2 | 5+6 2 | 2+2 4 |
| | 2+2 4 | 2+2 4 | 2+2 4 | 3+3 2 | 5+5 2 | 5+6 2 | 5+6 2 | 2+2 4 |
| | 2+2 4 | 2+2 4 | 2+2 4 | 5+5 2 | 6+6 2 | 5+6 2 | 3+4 2 | 2+2 4 |
| | 2+2 4 | 2+2 4 | 2+2 4 | 6+6 2 | 3+3 2 | 5+6 2 | 3+4 2 | 2+2 4 |
| | 2+2 4 | 2+2 4 | 2+2 4 | 3+3 2 | 3+3 2 | 3+4 2 | 3+4 2 | 2+2 4 |
| | 2+2 4 | 2+2 4 | 2+2 4 | 4+4 2 | 4+4 2 | 5+6 2 | 5+6 2 | 2+2 4 |
| | 2+2 4 | 2+2 4 | 2+2 4 | 6+6 2 | 6+6 2 | 5+6 2 | 5+6 2 | 2+2 4 |
| | 2+2 4 | 2+2 4 | 2+2 4 | 5+5 2 | 5+5 2 | 3+4 2 | 3+4 2 | 2+2 4 |
| | 2+2 4 | 2+2 4 | 2+2 4 | 5+5 2 | 5+5 2 | 5+6 2 | 5+6 2 | 2+2 4 |
| | 2+2 4 | 2+2 4 | 2+2 4 | 6+6 2 | 6+6 2 | 3+4 2 | 3+4 2 | 2+2 4 |
| | 2+2 4 | 2+2 4 | 2+2 4 | 4+4 2 | 4+4 2 | 5+6 2 | 3+4 2 | 2+2 4 |
| | 2+2 4 | 2+2 4 | 2+2 4 | 3+3 2 | 3+3 2 | 3+4 2 | 5+6 2 | 2+2 4 |
| | 2+2 4 | 2+2 4 | 2+2 4 | 6+6 2 | 6+6 2 | 3+4 2 | 3+4 2 | 2+2 4 |
| | 2+2 4 | 2+2 4 | 2+2 4 | 5+5 2 | 5+5 2 | 5+6 2 | 5+6 2 | 2+2 4 |
| | 2+2 4 | 2+2 4 | 2+2 4 | 4+4 2 | 3+3 2 | 3+4 2 | 3+4 2 | 2+2 4 |
| | 2+2 4 | 2+2 4 | 2+2 4 | 3+3 2 | 4+4 2 | 3+4 2 | 5+6 2 | 2+2 4 |
| | 2+2 4 | 2+2 4 | 2+2 4 | 4+4 2 | 5+5 2 | 5+6 2 | 5+6 2 | 2+2 4 |
| | 2+2 4 | 2+2 4 | 2+2 4 | 3+3 2 | 6+6 2 | 5+6 2 | 3+4 2 | 2+2 4 |
| | 2+2 4 | 2+2 4 | 2+2 4 | 5+5 2 | 3+3 2 | 5+6 2 | 3+4 2 | 2+2 4 |
| | 2+2 4 | 2+2 4 | 2+2 4 | 6+6 2 | 4+4 2 | 3+4 2 | 5+6 2 | 2+2 4 |
| | 2+2 4 | 2+2 4 | 2+2 4 | 4+4 2 | 4+4 2 | 3+4 2 | 3+4 2 | 2+2 4 |
| | 2+2 4 | 2+2 4 | 2+2 4 | 3+3 2 | 3+3 2 | 3+4 2 | 5+6 2 | 2+2 4 |
| | 2+2 4 | 2+2 4 | 2+2 4 | 6+6 2 | 4+4 2 | 5+6 2 | 5+6 2 | 2+2 4 |
| | 2+2 4 | 2+2 4 | 2+2 4 | 5+5 2 | 3+3 2 | 5+6 2 | 3+4 2 | 2+2 4 |
| | 2+2 4 | 2+2 4 | 2+2 4 | 6+6 2 | 5+5 2 | 3+4 2 | 5+6 2 | 2+2 4 |
| | 2+2 4 | 2+2 4 | 2+2 4 | 5+5 2 | 6+6 2 | 3+4 2 | 5+6 2 | 2+2 4 |
| | | | | 3+3 2 | 4+4 2 | 5+6 2 | 3+4 2 | |
| | | | | 4+4 2 | 3+3 2 | 3+4 2 | 3+4 2 | |
| | | | | 5+5 2 | 6+6 2 | 5+6 2 | 5+6 2 | |
| | | | | 6+6 2 | 5+5 2 | 3+4 2 | 5+6 2 | |
| | | | | 3+3 2 | 3+3 2 | 5+6 2 | 3+4 2 | |
| | | | | 4+4 2 | 5+5 2 | 3+4 2 | 3+4 2 | |
| | | | | 6+6 2 | 4+4 2 | 5+6 2 | 3+4 2 | |
| | | | | 5+5 2 | 6+6 2 | 3+4 2 | 5+6 2 | |
| | | | | 3+3 2 | 6+6 2 | 5+6 2 | 3+4 2 | |
| | | | | 4+4 2 | 5+5 2 | 3+4 2 | 5+6 2 | |
| | | | | 5+5 2 | 6+6 2 | 5+6 2 | 5+6 2 | |
| | | | | 6+6 2 | 5+5 2 | 5+6 2 | 3+4 2 | |
| | | | | 4+4 2 | 3+3 2 | 3+4 2 | 5+6 2 | |
| | | | | 3+3 2 | 4+4 2 | 5+6 2 | 3+4 2 | |
| | | | | 5+5 2 | 5+5 2 | 3+4 2 | 5+6 2 | |
| | | | | 6+6 2 | 4+4 2 | 3+4 2 | 3+4 2 | |
| | | | | 6+6 2 | 6+6 2 | 5+6 2 | 5+6 2 | |
| | | | | 4+4 2 | 3+3 2 | 3+4 2 | 3+4 2 | |
| | | | | 5+5 2 | 3+3 2 | 5+6 2 | 3+4 2 | |
| | | | | 3+3 2 | 4+4 2 | 3+4 2 | 3+4 2 | |
| | | | | 3+3 2 | 6+6 2 | 5+6 2 | 5+6 2 | |
| | | | | 4+4 2 | 5+5 2 | 5+6 2 | 5+6 2 | |
| | | | | 6+6 2 | 5+5 2 | 3+4 2 | 3+4 2 | |
| | | | | 5+5 2 | 3+3 2 | 3+4 2 | 5+6 2 | |
| | | | | 3+3 2 | 6+6 2 | 3+4 2 | 3+4 2 | |
| | | | | 4+4 2 | 4+4 2 | 5+6 2 | 5+6 2 | |
| | | | | 4+4 2 | 6+6 2 | 3+4 2 | 3+4 2 | |
| | | | | 3+3 2 | 5+5 2 | 5+6 2 | 3+4 2 | |

Continued on next column

**TABLE XXVI– continued**

| TQC | ICSD | NoE | SF | | | | | |
|---|---|---|---|---|---|---|---|---|
| | | | 6+6 2 | 4+4 2 | 5+6 2 | 5+6 2 | | |
| | | | 5+5 2 | 3+3 2 | 5+6 2 | 5+6 2 | | |
| | | | 5+5 2 | 6+6 2 | 3+4 2 | 3+4 2 | | |
| | | | 4+4 2 | 5+5 2 | 3+4 2 | 5+6 2 | | |
| | | | 3+3 2 | 3+3 2 | 5+6 2 | 3+4 2 | | |
| | | | 6+6 2 | 4+4 2 | 3+4 2 | 3+4 2 | | |
| | | | 5+5 2 | 4+4 2 | 3+4 2 | 5+6 2 | | |
| | | | 3+3 2 | 3+3 2 | 5+6 2 | 5+6 2 | | |
| | | | 4+4 2 | 5+5 2 | 3+4 2 | 3+4 2 | | |
| | | | 6+6 2 | 6+6 2 | 5+6 2 | 5+6 2 | | |
| 14.2.42 | 73398 | 164 | Cu1Na1O4P1 | | | | | |
| | -A | -B | -C | -D | -E | -GM | -Y | -Z |
| | 2+2 4 | 2+2 4 | 2+2 4 | 5+5 2 | 5+5 2 | 3+4 2 | 3+4 2 | 2+2 4 |
| | 2+2 4 | 2+2 4 | 2+2 4 | 6+6 2 | 6+6 2 | 5+6 2 | 5+6 2 | 2+2 4 |
| | 2+2 4 | 2+2 4 | 2+2 4 | 3+3 2 | 3+3 2 | 3+4 2 | 3+4 2 | 2+2 4 |
| | 2+2 4 | 2+2 4 | 2+2 4 | 4+4 2 | 4+4 2 | 5+6 2 | 5+6 2 | 2+2 4 |
| | 2+2 4 | 2+2 4 | 2+2 4 | 3+3 2 | 3+3 2 | 3+4 2 | 5+6 2 | 2+2 4 |
| | 2+2 4 | 2+2 4 | 2+2 4 | 4+4 2 | 4+4 2 | 3+4 2 | 3+4 2 | 2+2 4 |
| | 2+2 4 | 2+2 4 | 2+2 4 | 5+5 2 | 3+3 2 | 3+4 2 | 5+6 2 | 2+2 4 |
| | 2+2 4 | 2+2 4 | 2+2 4 | 6+6 2 | 4+4 2 | 5+6 2 | 3+4 2 | 2+2 4 |
| | 2+2 4 | 2+2 4 | 2+2 4 | 5+5 2 | 5+5 2 | 5+6 2 | 3+4 2 | 2+2 4 |
| | 2+2 4 | 2+2 4 | 2+2 4 | 6+6 2 | 6+6 2 | 5+6 2 | 5+6 2 | 2+2 4 |
| | 2+2 4 | 2+2 4 | 2+2 4 | 3+3 2 | 5+5 2 | 5+6 2 | 5+6 2 | 2+2 4 |
| | 2+2 4 | 2+2 4 | 2+2 4 | 4+4 2 | 6+6 2 | 3+4 2 | 3+4 2 | 2+2 4 |
| | 2+2 4 | 2+2 4 | 2+2 4 | 5+5 2 | 5+5 2 | 3+4 2 | 5+6 2 | 2+2 4 |
| | 2+2 4 | 2+2 4 | 2+2 4 | 6+6 2 | 6+6 2 | 5+6 2 | 3+4 2 | 2+2 4 |
| | 2+2 4 | 2+2 4 | 2+2 4 | 3+3 2 | 3+3 2 | 3+4 2 | 3+4 2 | 2+2 4 |
| | 2+2 4 | 2+2 4 | 2+2 4 | 4+4 2 | 4+4 2 | 5+6 2 | 5+6 2 | 2+2 4 |
| | 2+2 4 | 2+2 4 | 2+2 4 | 5+5 2 | 3+3 2 | 3+4 2 | 3+4 2 | 2+2 4 |
| | 2+2 4 | 2+2 4 | 2+2 4 | 6+6 2 | 4+4 2 | 3+4 2 | 3+4 2 | 2+2 4 |
| | 2+2 4 | 2+2 4 | 2+2 4 | 3+3 2 | 5+5 2 | 5+6 2 | 5+6 2 | 2+2 4 |
| | 2+2 4 | 2+2 4 | 2+2 4 | 4+4 2 | 6+6 2 | 5+6 2 | 5+6 2 | 2+2 4 |
| | 2+2 4 | 2+2 4 | 2+2 4 | 3+3 2 | 3+3 2 | 3+4 2 | 5+6 2 | 2+2 4 |
| | 2+2 4 | 2+2 4 | 2+2 4 | 4+4 2 | 4+4 2 | 3+4 2 | 3+4 2 | 2+2 4 |
| | 2+2 4 | 2+2 4 | 2+2 4 | 6+6 2 | 4+4 2 | 3+4 2 | 5+6 2 | 2+2 4 |
| | 2+2 4 | 2+2 4 | 2+2 4 | 5+5 2 | 3+3 2 | 3+4 2 | 3+4 2 | 2+2 4 |
| | 2+2 4 | 2+2 4 | 2+2 4 | 3+3 2 | 6+6 2 | 5+6 2 | 5+6 2 | 2+2 4 |
| | 2+2 4 | 2+2 4 | 2+2 4 | 4+4 2 | 5+5 2 | 5+6 2 | 3+4 2 | 2+2 4 |
| | 2+2 4 | 2+2 4 | 2+2 4 | 5+5 2 | 3+3 2 | 3+4 2 | 5+6 2 | 2+2 4 |
| | 2+2 4 | 2+2 4 | 2+2 4 | 6+6 2 | 4+4 2 | 5+6 2 | 3+4 2 | 2+2 4 |
| | 2+2 4 | 2+2 4 | 2+2 4 | 4+4 2 | 5+5 2 | 5+6 2 | 5+6 2 | 2+2 4 |
| | 2+2 4 | 2+2 4 | 2+2 4 | 3+3 2 | 6+6 2 | 3+4 2 | 5+6 2 | 2+2 4 |
| | 2+2 4 | 2+2 4 | 2+2 4 | 6+6 2 | 5+5 2 | 5+6 2 | 3+4 2 | 2+2 4 |
| | 2+2 4 | 2+2 4 | 2+2 4 | 5+5 2 | 6+6 2 | 5+6 2 | 3+4 2 | 2+2 4 |
| | 2+2 4 | 2+2 4 | 2+2 4 | 3+3 2 | 3+3 2 | 3+4 2 | 3+4 2 | 2+2 4 |
| | 2+2 4 | 2+2 4 | 2+2 4 | 4+4 2 | 4+4 2 | 3+4 2 | 3+4 2 | 2+2 4 |
| | 2+2 4 | 2+2 4 | 2+2 4 | 5+5 2 | 3+3 2 | 3+4 2 | 5+6 2 | 2+2 4 |
| | 2+2 4 | 2+2 4 | 2+2 4 | 6+6 2 | 4+4 2 | 3+4 2 | 3+4 2 | 2+2 4 |
| | 2+2 4 | 2+2 4 | 2+2 4 | 3+3 2 | 5+5 2 | 3+4 2 | 5+6 2 | 2+2 4 |
| | 2+2 4 | 2+2 4 | 2+2 4 | 4+4 2 | 6+6 2 | 5+6 2 | 3+4 2 | 2+2 4 |
| | 2+2 4 | 2+2 4 | 2+2 4 | 6+6 2 | 5+5 2 | 5+6 2 | 5+6 2 | 2+2 4 |
| | 2+2 4 | 2+2 4 | 2+2 4 | 5+5 2 | 6+6 2 | 3+4 2 | 5+6 2 | 2+2 4 |
| | 2+2 4 | 2+2 4 | 2+2 4 | 5+5 2 | 4+4 2 | 3+4 2 | 3+4 2 | 2+2 4 |
| | | | | 6+6 2 | 3+3 2 | 3+4 2 | 3+4 2 | |
| | | | | 3+3 2 | 5+5 2 | 3+4 2 | 5+6 2 | |
| | | | | 4+4 2 | 6+6 2 | 5+6 2 | 5+6 2 | |
| | | | | 4+4 2 | 4+4 2 | 5+6 2 | 3+4 2 | |
| | | | | 3+3 2 | 3+3 2 | 5+6 2 | 3+4 2 | |
| | | | | 6+6 2 | 6+6 2 | 5+6 2 | 5+6 2 | |
| | | | | 5+5 2 | 5+5 2 | 3+4 2 | 5+6 2 | |
| | | | | 5+5 2 | 4+4 2 | 3+4 2 | 3+4 2 | |

Continued on next column

**TABLE XXVI– continued**

| TQC | ICSD | NoE | SF | | | | | |
|---|---|---|---|---|---|---|---|---|
| | | | 6+6 2 | 3+3 2 | 5+6 2 | 5+6 2 | | |
| | | | 3+3 2 | 5+5 2 | 5+6 2 | 5+6 2 | | |
| | | | 4+4 2 | 6+6 2 | 3+4 2 | 3+4 2 | | |
| | | | 3+3 2 | 3+3 2 | 3+4 2 | 5+6 2 | | |
| | | | 4+4 2 | 4+4 2 | 3+4 2 | 5+6 2 | | |
| | | | 5+5 2 | 4+4 2 | 5+6 2 | 3+4 2 | | |
| | | | 6+6 2 | 3+3 2 | 5+6 2 | 3+4 2 | | |
| | | | 5+5 2 | 4+4 2 | 5+6 2 | 5+6 2 | | |
| | | | 6+6 2 | 3+3 2 | 3+4 2 | 3+4 2 | | |
| | | | 4+4 2 | 5+5 2 | 5+6 2 | 5+6 2 | | |
| | | | 3+3 2 | 6+6 2 | 5+6 2 | 3+4 2 | | |
| | | | 3+3 2 | 3+3 2 | 3+4 2 | 3+4 2 | | |
| | | | 4+4 2 | 4+4 2 | 5+6 2 | 5+6 2 | | |
| | | | 5+5 2 | 6+6 2 | 3+4 2 | 3+4 2 | | |
| | | | 6+6 2 | 5+5 2 | 5+6 2 | 5+6 2 | | |
| | | | 4+4 2 | 5+5 2 | 3+4 2 | 3+4 2 | | |
| | | | 3+3 2 | 6+6 2 | 5+6 2 | 3+4 2 | | |
| | | | 6+6 2 | 3+3 2 | 3+4 2 | 3+4 2 | | |
| | | | 4+4 2 | 4+4 2 | 3+4 2 | 3+4 2 | | |
| | | | 5+5 2 | 3+3 2 | 3+4 2 | 5+6 2 | | |
| | | | 3+3 2 | 4+4 2 | 3+4 2 | 5+6 2 | | |
| | | | 3+3 2 | 3+3 2 | 3+4 2 | 3+4 2 | | |
| | | | 4+4 2 | 4+4 2 | 3+4 2 | 3+4 2 | | |
| | | | 6+6 2 | 4+4 2 | 3+4 2 | 5+6 2 | | |
| | | | 5+5 2 | 3+3 2 | 3+4 2 | 3+4 2 | | |
| | | | 4+4 2 | 4+4 2 | 3+4 2 | 5+6 2 | | |
| | | | 3+3 2 | 3+3 2 | 3+4 2 | 5+6 2 | | |
| | | | 5+5 2 | 3+3 2 | 3+4 2 | 5+6 2 | | |
| | | | 6+6 2 | 4+4 2 | 3+4 2 | 3+4 2 | | |
| | | | 6+6 2 | 4+4 2 | 3+4 2 | 5+6 2 | | |
| | | | 5+5 2 | 3+3 2 | 3+4 2 | 5+6 2 | | |
| | | | 4+4 2 | 3+3 2 | 3+4 2 | 3+4 2 | | |
| | | | 3+3 2 | 4+4 2 | 3+4 2 | 5+6 2 | | |
| 14.2.43 | 411020 | 132 | Cu1Ge1O3 | | | | | |
| | -A | -B | -C | -D | -E | -GM | -Y | -Z |
| | 2+2 4 | 2+2 4 | 2+2 4 | 5+5 2 | 5+5 2 | 3+4 2 | 3+4 2 | 2+2 4 |
| | 2+2 4 | 2+2 4 | 2+2 4 | 6+6 2 | 6+6 2 | 3+4 2 | 3+4 2 | 2+2 4 |
| | 2+2 4 | 2+2 4 | 2+2 4 | 3+3 2 | 4+4 2 | 5+6 2 | 5+6 2 | 2+2 4 |
| | 2+2 4 | 2+2 4 | 2+2 4 | 4+4 2 | 3+3 2 | 5+6 2 | 5+6 2 | 2+2 4 |
| | 2+2 4 | 2+2 4 | 2+2 4 | 3+3 2 | 5+5 2 | 5+6 2 | 3+4 2 | 2+2 4 |
| | 2+2 4 | 2+2 4 | 2+2 4 | 4+4 2 | 6+6 2 | 3+4 2 | 5+6 2 | 2+2 4 |
| | 2+2 4 | 2+2 4 | 2+2 4 | 5+5 2 | 3+3 2 | 3+4 2 | 3+4 2 | 2+2 4 |
| | 2+2 4 | 2+2 4 | 2+2 4 | 6+6 2 | 4+4 2 | 5+6 2 | 5+6 2 | 2+2 4 |
| | 2+2 4 | 2+2 4 | 2+2 4 | 5+5 2 | 5+5 2 | 3+4 2 | 3+4 2 | 2+2 4 |
| | 2+2 4 | 2+2 4 | 2+2 4 | 6+6 2 | 6+6 2 | 3+4 2 | 3+4 2 | 2+2 4 |
| | 2+2 4 | 2+2 4 | 2+2 4 | 3+3 2 | 3+3 2 | 5+6 2 | 5+6 2 | 2+2 4 |
| | 2+2 4 | 2+2 4 | 2+2 4 | 4+4 2 | 4+4 2 | 5+6 2 | 5+6 2 | 2+2 4 |
| | 2+2 4 | 2+2 4 | 2+2 4 | 3+3 2 | 3+3 2 | 5+6 2 | 5+6 2 | 2+2 4 |
| | 2+2 4 | 2+2 4 | 2+2 4 | 4+4 2 | 4+4 2 | 5+6 2 | 5+6 2 | 2+2 4 |
| | 2+2 4 | 2+2 4 | 2+2 4 | 5+5 2 | 5+5 2 | 3+4 2 | 3+4 2 | 2+2 4 |
| | 2+2 4 | 2+2 4 | 2+2 4 | 6+6 2 | 6+6 2 | 3+4 2 | 3+4 2 | 2+2 4 |
| | 2+2 4 | 2+2 4 | 2+2 4 | 3+3 2 | 6+6 2 | 3+4 2 | 5+6 2 | 2+2 4 |
| | 2+2 4 | 2+2 4 | 2+2 4 | 4+4 2 | 5+5 2 | 5+6 2 | 3+4 2 | 2+2 4 |
| | 2+2 4 | 2+2 4 | 2+2 4 | 6+6 2 | 4+4 2 | 5+6 2 | 3+4 2 | 2+2 4 |
| | 2+2 4 | 2+2 4 | 2+2 4 | 5+5 2 | 3+3 2 | 3+4 2 | 5+6 2 | 2+2 4 |
| | 2+2 4 | 2+2 4 | 2+2 4 | 6+6 2 | 5+5 2 | 3+4 2 | 3+4 2 | 2+2 4 |
| | 2+2 4 | 2+2 4 | 2+2 4 | 5+5 2 | 6+6 2 | 3+4 2 | 3+4 2 | 2+2 4 |
| | 2+2 4 | 2+2 4 | 2+2 4 | 3+3 2 | 5+5 2 | 3+4 2 | 5+6 2 | 2+2 4 |
| | 2+2 4 | 2+2 4 | 2+2 4 | 4+4 2 | 6+6 2 | 5+6 2 | 3+4 2 | 2+2 4 |
| | 2+2 4 | 2+2 4 | 2+2 4 | 6+6 2 | 4+4 2 | 5+6 2 | 5+6 2 | 2+2 4 |
| | 2+2 4 | 2+2 4 | 2+2 4 | 5+5 2 | 3+3 2 | 3+4 2 | 3+4 2 | 2+2 4 |

Continued on next column

**TABLE XXVI– continued**

| TQC | ICSD | NoE | SF | | | | | |
|---|---|---|---|---|---|---|---|---|
| | 2+2 4 | 2+2 4 | 2+2 4 | 3+3 2 | 6+6 2 | 3+4 2 | 5+6 2 | 2+2 4 |
| | 2+2 4 | 2+2 4 | 2+2 4 | 4+4 2 | 5+5 2 | 5+6 2 | 3+4 2 | 2+2 4 |
| | 2+2 4 | 2+2 4 | 2+2 4 | 5+5 2 | 3+3 2 | 5+6 2 | 5+6 2 | 2+2 4 |
| | 2+2 4 | 2+2 4 | 2+2 4 | 6+6 2 | 4+4 2 | 3+4 2 | 5+6 2 | 2+2 4 |
| | 2+2 4 | 2+2 4 | 2+2 4 | 3+3 2 | 3+3 2 | 5+6 2 | 5+6 2 | 2+2 4 |
| | 2+2 4 | 2+2 4 | 2+2 4 | 4+4 2 | 4+4 2 | 3+4 2 | 3+4 2 | 2+2 4 |
| | 2+2 4 | 2+2 4 | 2+2 4 | 5+5 2 | 6+6 2 | 3+4 2 | 3+4 2 | 2+2 4 |
| | | | | 6+6 2 | 5+5 2 | 5+6 2 | 5+6 2 | |
| | | | | 4+4 2 | 3+3 2 | 5+6 2 | 5+6 2 | |
| | | | | 3+3 2 | 4+4 2 | 3+4 2 | 3+4 2 | |
| | | | | 4+4 2 | 4+4 2 | 3+4 2 | 5+6 2 | |
| | | | | 3+3 2 | 3+3 2 | 5+6 2 | 3+4 2 | |
| | | | | 5+5 2 | 6+6 2 | 3+4 2 | 5+6 2 | |
| | | | | 6+6 2 | 5+5 2 | 3+4 2 | 5+6 2 | |
| | | | | 5+5 2 | 3+3 2 | 5+6 2 | 5+6 2 | |
| | | | | 6+6 2 | 4+4 2 | 5+6 2 | 3+4 2 | |
| | | | | 3+3 2 | 5+5 2 | 5+6 2 | 3+4 2 | |
| | | | | 4+4 2 | 6+6 2 | 5+6 2 | 3+4 2 | |
| | | | | 3+3 2 | 3+3 2 | 3+4 2 | 3+4 2 | |
| | | | | 4+4 2 | 4+4 2 | 3+4 2 | 5+6 2 | |
| | | | | 6+6 2 | 5+5 2 | 5+6 2 | 5+6 2 | |
| | | | | 5+5 2 | 6+6 2 | 3+4 2 | 3+4 2 | |
| | | | | 3+3 2 | 3+3 2 | 5+6 2 | 3+4 2 | |
| | | | | 4+4 2 | 4+4 2 | 3+4 2 | 5+6 2 | |
| | | | | 5+5 2 | 5+5 2 | 3+4 2 | 3+4 2 | |
| | | | | 6+6 2 | 6+6 2 | 5+6 2 | 5+6 2 | |
| | | | | 6+6 2 | 4+4 2 | 5+6 2 | 3+4 2 | |
| | | | | 5+5 2 | 3+3 2 | 5+6 2 | 5+6 2 | |
| | | | | 4+4 2 | 5+5 2 | 5+6 2 | 3+4 2 | |
| | | | | 3+3 2 | 6+6 2 | 3+4 2 | 5+6 2 | |
| | | | | 5+5 2 | 4+4 2 | 3+4 2 | 3+4 2 | |
| | | | | 6+6 2 | 3+3 2 | 5+6 2 | 5+6 2 | |
| | | | | 4+4 2 | 6+6 2 | 3+4 2 | 3+4 2 | |
| | | | | 3+3 2 | 5+5 2 | 5+6 2 | 3+4 2 | |
| | | | | 5+5 2 | 3+3 2 | 5+6 2 | 5+6 2 | |
| | | | | 6+6 2 | 4+4 2 | 3+4 2 | 5+6 2 | |
| | | | | 3+3 2 | 5+5 2 | 5+6 2 | 3+4 2 | |
| | | | | 4+4 2 | 6+6 2 | 3+4 2 | 5+6 2 | |
| | | | | 5+5 2 | 6+6 2 | 5+6 2 | 3+4 2 | |
| | | | | 6+6 2 | 5+5 2 | 3+4 2 | 5+6 2 | |
| 14.2.44 | 59291 | 140 | Co2O7P2 | | | | | |
| | -A | -B | -C | -D | -E | -GM | -Y | -Z |
| | 2+2 4 | 2+2 4 | 2+2 4 | 5+5 2 | 5+5 2 | 3+4 2 | 3+4 2 | 2+2 4 |
| | 2+2 4 | 2+2 4 | 2+2 4 | 6+6 2 | 6+6 2 | 3+4 2 | 3+4 2 | 2+2 4 |
| | 2+2 4 | 2+2 4 | 2+2 4 | 3+3 2 | 3+3 2 | 5+6 2 | 5+6 2 | 2+2 4 |
| | 2+2 4 | 2+2 4 | 2+2 4 | 4+4 2 | 4+4 2 | 5+6 2 | 5+6 2 | 2+2 4 |
| | 2+2 4 | 2+2 4 | 2+2 4 | 5+5 2 | 5+5 2 | 3+4 2 | 3+4 2 | 2+2 4 |
| | 2+2 4 | 2+2 4 | 2+2 4 | 6+6 2 | 6+6 2 | 3+4 2 | 3+4 2 | 2+2 4 |
| | 2+2 4 | 2+2 4 | 2+2 4 | 4+4 2 | 3+3 2 | 3+4 2 | 5+6 2 | 2+2 4 |
| | 2+2 4 | 2+2 4 | 2+2 4 | 3+3 2 | 4+4 2 | 5+6 2 | 3+4 2 | 2+2 4 |
| | 2+2 4 | 2+2 4 | 2+2 4 | 5+5 2 | 5+5 2 | 3+4 2 | 5+6 2 | 2+2 4 |
| | 2+2 4 | 2+2 4 | 2+2 4 | 6+6 2 | 6+6 2 | 5+6 2 | 3+4 2 | 2+2 4 |
| | 2+2 4 | 2+2 4 | 2+2 4 | 3+3 2 | 5+5 2 | 3+4 2 | 5+6 2 | 2+2 4 |
| | 2+2 4 | 2+2 4 | 2+2 4 | 4+4 2 | 6+6 2 | 3+4 2 | 3+4 2 | 2+2 4 |
| | 2+2 4 | 2+2 4 | 2+2 4 | 5+5 2 | 3+3 2 | 5+6 2 | 5+6 2 | 2+2 4 |
| | 2+2 4 | 2+2 4 | 2+2 4 | 6+6 2 | 4+4 2 | 5+6 2 | 3+4 2 | 2+2 4 |
| | 2+2 4 | 2+2 4 | 2+2 4 | 3+3 2 | 3+3 2 | 5+6 2 | 5+6 2 | 2+2 4 |
| | 2+2 4 | 2+2 4 | 2+2 4 | 4+4 2 | 4+4 2 | 5+6 2 | 5+6 2 | 2+2 4 |
| | 2+2 4 | 2+2 4 | 2+2 4 | 5+5 2 | 5+5 2 | 3+4 2 | 3+4 2 | 2+2 4 |
| | 2+2 4 | 2+2 4 | 2+2 4 | 6+6 2 | 6+6 2 | 3+4 2 | 5+6 2 | 2+2 4 |
| | 2+2 4 | 2+2 4 | 2+2 4 | 3+3 2 | 3+3 2 | 5+6 2 | 5+6 2 | 2+2 4 |

Continued on next column

| TQC | ICSD | NoE | SF | | | | | |
|---|---|---|---|---|---|---|---|---|
| | 2+2 4 | 2+2 4 | 2+2 4 | 4+4 2 | 4+4 2 | 5+6 2 | 5+6 2 | 2+2 4 |
| | 2+2 4 | 2+2 4 | 2+2 4 | 4+4 2 | 6+6 2 | 5+6 2 | 3+4 2 | 2+2 4 |
| | 2+2 4 | 2+2 4 | 2+2 4 | 3+3 2 | 5+5 2 | 3+4 2 | 5+6 2 | 2+2 4 |
| | 2+2 4 | 2+2 4 | 2+2 4 | 3+3 2 | 3+3 2 | 5+6 2 | 3+4 2 | 2+2 4 |
| | 2+2 4 | 2+2 4 | 2+2 4 | 4+4 2 | 4+4 2 | 5+6 2 | 5+6 2 | 2+2 4 |
| | 2+2 4 | 2+2 4 | 2+2 4 | 5+5 2 | 4+4 2 | 3+4 2 | 5+6 2 | 2+2 4 |
| | 2+2 4 | 2+2 4 | 2+2 4 | 6+6 2 | 3+3 2 | 3+4 2 | 3+4 2 | 2+2 4 |
| | 2+2 4 | 2+2 4 | 2+2 4 | 5+5 2 | 5+5 2 | 5+6 2 | 3+4 2 | 2+2 4 |
| | 2+2 4 | 2+2 4 | 2+2 4 | 6+6 2 | 6+6 2 | 3+4 2 | 3+4 2 | 2+2 4 |
| | 2+2 4 | 2+2 4 | 2+2 4 | 3+3 2 | 5+5 2 | 5+6 2 | 5+6 2 | 2+2 4 |
| | 2+2 4 | 2+2 4 | 2+2 4 | 4+4 2 | 6+6 2 | 5+6 2 | 5+6 2 | 2+2 4 |
| | 2+2 4 | 2+2 4 | 2+2 4 | 5+5 2 | 3+3 2 | 3+4 2 | 3+4 2 | 2+2 4 |
| | 2+2 4 | 2+2 4 | 2+2 4 | 6+6 2 | 4+4 2 | 5+6 2 | 3+4 2 | 2+2 4 |
| | 2+2 4 | 2+2 4 | 2+2 4 | 3+3 2 | 3+3 2 | 3+4 2 | 5+6 2 | 2+2 4 |
| | 2+2 4 | 2+2 4 | 2+2 4 | 4+4 2 | 4+4 2 | 3+4 2 | 5+6 2 | 2+2 4 |
| | 2+2 4 | 2+2 4 | 2+2 4 | 3+3 2 | 3+3 2 | 5+6 2 | 3+4 2 | 2+2 4 |
| | | | | 4+4 2 | 4+4 2 | 5+6 2 | 5+6 2 | |
| | | | | 5+5 2 | 5+5 2 | 3+4 2 | 3+4 2 | |
| | | | | 6+6 2 | 6+6 2 | 3+4 2 | 5+6 2 | |
| | | | | 4+4 2 | 4+4 2 | 5+6 2 | 5+6 2 | |
| | | | | 3+3 2 | 3+3 2 | 3+4 2 | 3+4 2 | |
| | | | | 6+6 2 | 6+6 2 | 5+6 2 | 5+6 2 | |
| | | | | 5+5 2 | 5+5 2 | 3+4 2 | 3+4 2 | |
| | | | | 6+6 2 | 3+3 2 | 3+4 2 | 5+6 2 | |
| | | | | 5+5 2 | 4+4 2 | 3+4 2 | 3+4 2 | |
| | | | | 3+3 2 | 5+5 2 | 5+6 2 | 5+6 2 | |
| | | | | 4+4 2 | 6+6 2 | 5+6 2 | 3+4 2 | |
| | | | | 5+5 2 | 5+5 2 | 5+6 2 | 5+6 2 | |
| | | | | 6+6 2 | 6+6 2 | 5+6 2 | 5+6 2 | |
| | | | | 3+3 2 | 3+3 2 | 3+4 2 | 3+4 2 | |
| | | | | 4+4 2 | 4+4 2 | 5+6 2 | 3+4 2 | |
| | | | | 3+3 2 | 4+4 2 | 3+4 2 | 3+4 2 | |
| | | | | 4+4 2 | 3+3 2 | 5+6 2 | 5+6 2 | |
| | | | | 3+3 2 | 5+5 2 | 5+6 2 | 3+4 2 | |
| | | | | 4+4 2 | 6+6 2 | 5+6 2 | 5+6 2 | |
| | | | | 5+5 2 | 4+4 2 | 3+4 2 | 5+6 2 | |
| | | | | 6+6 2 | 3+3 2 | 5+6 2 | 5+6 2 | |
| | | | | 6+6 2 | 3+3 2 | 3+4 2 | 5+6 2 | |
| | | | | 5+5 2 | 4+4 2 | 5+6 2 | 5+6 2 | |
| | | | | 3+3 2 | 6+6 2 | 5+6 2 | 3+4 2 | |
| | | | | 4+4 2 | 5+5 2 | 3+4 2 | 3+4 2 | |
| | | | | 5+5 2 | 6+6 2 | 3+4 2 | 5+6 2 | |
| | | | | 6+6 2 | 5+5 2 | 5+6 2 | 3+4 2 | |
| | | | | 3+3 2 | 3+3 2 | 3+4 2 | 5+6 2 | |
| | | | | 4+4 2 | 4+4 2 | 3+4 2 | 3+4 2 | |
| | | | | 6+6 2 | 5+5 2 | 5+6 2 | 3+4 2 | |
| | | | | 5+5 2 | 6+6 2 | 5+6 2 | 5+6 2 | |
| | | | | 4+4 2 | 4+4 2 | 3+4 2 | 3+4 2 | |
| | | | | 3+3 2 | 3+3 2 | 5+6 2 | 5+6 2 | |
| | | | | 5+5 2 | 5+5 2 | 5+6 2 | 3+4 2 | |
| | | | | 6+6 2 | 6+6 2 | 3+4 2 | 3+4 2 | |
| 14.2.45 | 73738 | 104 | Co1Ta1Te2 | | | | | |
| | -A | -B | -C | -D | -E | -GM | -Y | -Z |
| | 2+2 4 | 2+2 4 | 2+2 4 | 4+4 2 | 5+5 2 | 3+4 2 | 5+6 2 | 2+2 4 |
| | 2+2 4 | 2+2 4 | 2+2 4 | 3+3 2 | 6+6 2 | 5+6 2 | 3+4 2 | 2+2 4 |
| | 2+2 4 | 2+2 4 | 2+2 4 | 5+5 2 | 4+4 2 | 5+6 2 | 3+4 2 | 2+2 4 |
| | 2+2 4 | 2+2 4 | 2+2 4 | 6+6 2 | 3+3 2 | 3+4 2 | 5+6 2 | 2+2 4 |
| | 2+2 4 | 2+2 4 | 2+2 4 | 5+5 2 | 4+4 2 | 5+6 2 | 3+4 2 | 2+2 4 |
| | 2+2 4 | 2+2 4 | 2+2 4 | 6+6 2 | 3+3 2 | 5+6 2 | 5+6 2 | 2+2 4 |
| | 2+2 4 | 2+2 4 | 2+2 4 | 3+3 2 | 5+5 2 | 3+4 2 | 3+4 2 | 2+2 4 |
| | 2+2 4 | 2+2 4 | 2+2 4 | 4+4 2 | 6+6 2 | 3+4 2 | 5+6 2 | 2+2 4 |

Continued on next column

**TABLE XXVI– continued**

| TQC | ICSD | NoE | SF | | | | |
|---|---|---|---|---|---|---|---|
| | 2+2 4 | 2+2 4 | 2+2 4 | 5+5 2 | 5+5 2 | 3+4 2 | 5+6 2 | 2+2 4 |
| | 2+2 4 | 2+2 4 | 2+2 4 | 6+6 2 | 6+6 2 | 3+4 2 | 3+4 2 | 2+2 4 |
| | 2+2 4 | 2+2 4 | 2+2 4 | 4+4 2 | 3+3 2 | 5+6 2 | 5+6 2 | 2+2 4 |
| | 2+2 4 | 2+2 4 | 2+2 4 | 3+3 2 | 4+4 2 | 5+6 2 | 3+4 2 | 2+2 4 |
| | 2+2 4 | 2+2 4 | 2+2 4 | 4+4 2 | 5+5 2 | 3+4 2 | 3+4 2 | 2+2 4 |
| | 2+2 4 | 2+2 4 | 2+2 4 | 3+3 2 | 6+6 2 | 5+6 2 | 5+6 2 | 2+2 4 |
| | 2+2 4 | 2+2 4 | 2+2 4 | 5+5 2 | 4+4 2 | 5+6 2 | 3+4 2 | 2+2 4 |
| | 2+2 4 | 2+2 4 | 2+2 4 | 6+6 2 | 3+3 2 | 3+4 2 | 3+4 2 | 2+2 4 |
| | 2+2 4 | 2+2 4 | 2+2 4 | 3+3 2 | 6+6 2 | 5+6 2 | 5+6 2 | 2+2 4 |
| | 2+2 4 | 2+2 4 | 2+2 4 | 4+4 2 | 5+5 2 | 3+4 2 | 3+4 2 | 2+2 4 |
| | 2+2 4 | 2+2 4 | 2+2 4 | 6+6 2 | 3+3 2 | 3+4 2 | 5+6 2 | 2+2 4 |
| | 2+2 4 | 2+2 4 | 2+2 4 | 5+5 2 | 4+4 2 | 5+6 2 | 5+6 2 | 2+2 4 |
| | 2+2 4 | 2+2 4 | 2+2 4 | 6+6 2 | 3+3 2 | 5+6 2 | 3+4 2 | 2+2 4 |
| | 2+2 4 | 2+2 4 | 2+2 4 | 5+5 2 | 4+4 2 | 3+4 2 | 5+6 2 | 2+2 4 |
| | 2+2 4 | 2+2 4 | 2+2 4 | 3+3 2 | 6+6 2 | 3+4 2 | 5+6 2 | 2+2 4 |
| | 2+2 4 | 2+2 4 | 2+2 4 | 4+4 2 | 3+3 2 | 5+6 2 | 3+4 2 | 2+2 4 |
| | 2+2 4 | 2+2 4 | 2+2 4 | 6+6 2 | 5+5 2 | 3+4 2 | 3+4 2 | 2+2 4 |
| | 2+2 4 | 2+2 4 | 2+2 4 | 5+5 2 | 4+4 2 | 5+6 2 | 3+4 2 | 2+2 4 |
| | | | | 3+3 2 | 6+6 2 | 3+4 2 | 5+6 2 | |
| | | | | 4+4 2 | 5+5 2 | 5+6 2 | 5+6 2 | |
| | | | | 5+5 2 | 4+4 2 | 3+4 2 | 3+4 2 | |
| | | | | 6+6 2 | 3+3 2 | 5+6 2 | 5+6 2 | |
| | | | | 4+4 2 | 5+5 2 | 3+4 2 | 5+6 2 | |
| | | | | 3+3 2 | 6+6 2 | 5+6 2 | 3+4 2 | |
| | | | | 6+6 2 | 4+4 2 | 5+6 2 | 3+4 2 | |
| | | | | 5+5 2 | 3+3 2 | 3+4 2 | 3+4 2 | |
| | | | | 4+4 2 | 5+5 2 | 5+6 2 | 3+4 2 | |
| | | | | 3+3 2 | 6+6 2 | 3+4 2 | 5+6 2 | |
| | | | | 5+5 2 | 4+4 2 | 5+6 2 | 5+6 2 | |
| | | | | 6+6 2 | 3+3 2 | 5+6 2 | 3+4 2 | |
| | | | | 3+3 2 | 5+5 2 | 3+4 2 | 5+6 2 | |
| | | | | 4+4 2 | 6+6 2 | 3+4 2 | 5+6 2 | |
| | | | | 4+4 2 | 5+5 2 | 3+4 2 | 3+4 2 | |
| | | | | 3+3 2 | 6+6 2 | 3+4 2 | 5+6 2 | |
| | | | | 4+4 2 | 3+3 2 | 5+6 2 | 5+6 2 | |
| | | | | 3+3 2 | 4+4 2 | 3+4 2 | 5+6 2 | |
| | | | | 5+5 2 | 5+5 2 | 3+4 2 | 3+4 2 | |
| | | | | 6+6 2 | 6+6 2 | 3+4 2 | 5+6 2 | |
| | | | | 4+4 2 | 5+5 2 | 3+4 2 | 5+6 2 | |
| | | | | 3+3 2 | 6+6 2 | 5+6 2 | 3+4 2 | |
| | | | | 5+5 2 | 3+3 2 | 5+6 2 | 5+6 2 | |
| | | | | 6+6 2 | 4+4 2 | 5+6 2 | 3+4 2 | |
| | | | | 5+5 2 | 4+4 2 | 3+4 2 | 3+4 2 | |
| | | | | 6+6 2 | 3+3 2 | 3+4 2 | 5+6 2 | |
| 14.2.46 | 63553 | 76 | Co1Sb2 | | | | | |
| | -A | -B | -C | -D | -E | -GM | -Y | -Z |
| | 2+2 4 | 2+2 4 | 2+2 4 | 3+3 2 | 5+5 2 | 3+4 2 | 5+6 2 | 2+2 4 |
| | 2+2 4 | 2+2 4 | 2+2 4 | 4+4 2 | 6+6 2 | 3+4 2 | 3+4 2 | 2+2 4 |
| | 2+2 4 | 2+2 4 | 2+2 4 | 5+5 2 | 3+3 2 | 5+6 2 | 5+6 2 | 2+2 4 |
| | 2+2 4 | 2+2 4 | 2+2 4 | 6+6 2 | 4+4 2 | 5+6 2 | 3+4 2 | 2+2 4 |
| | 2+2 4 | 2+2 4 | 2+2 4 | 5+5 2 | 3+3 2 | 5+6 2 | 3+4 2 | 2+2 4 |
| | 2+2 4 | 2+2 4 | 2+2 4 | 6+6 2 | 4+4 2 | 5+6 2 | 3+4 2 | 2+2 4 |
| | 2+2 4 | 2+2 4 | 2+2 4 | 3+3 2 | 5+5 2 | 3+4 2 | 5+6 2 | 2+2 4 |
| | 2+2 4 | 2+2 4 | 2+2 4 | 4+4 2 | 6+6 2 | 3+4 2 | 5+6 2 | 2+2 4 |
| | 2+2 4 | 2+2 4 | 2+2 4 | 6+6 2 | 6+6 2 | 3+4 2 | 5+6 2 | 2+2 4 |
| | 2+2 4 | 2+2 4 | 2+2 4 | 5+5 2 | 5+5 2 | 3+4 2 | 3+4 2 | 2+2 4 |
| | 2+2 4 | 2+2 4 | 2+2 4 | 3+3 2 | 4+4 2 | 5+6 2 | 3+4 2 | 2+2 4 |
| | 2+2 4 | 2+2 4 | 2+2 4 | 4+4 2 | 3+3 2 | 5+6 2 | 5+6 2 | 2+2 4 |
| | 2+2 4 | 2+2 4 | 2+2 4 | 4+4 2 | 4+4 2 | 3+4 2 | 3+4 2 | 2+2 4 |
| | 2+2 4 | 2+2 4 | 2+2 4 | 3+3 2 | 3+3 2 | 3+4 2 | 5+6 2 | 2+2 4 |
| | 2+2 4 | 2+2 4 | 2+2 4 | 5+5 2 | 5+5 2 | 5+6 2 | 3+4 2 | 2+2 4 |

Continued on next column

**TABLE XXVI– continued**

| TQC | ICSD | NoE | SF | | | | |
|---|---|---|---|---|---|---|---|
| | 2+2 4 | 2+2 4 | 2+2 4 | 6+6 2 | 6+6 2 | 5+6 2 | 5+6 2 | 2+2 4 |
| | 2+2 4 | 2+2 4 | 2+2 4 | 3+3 2 | 3+3 2 | 5+6 2 | 3+4 2 | 2+2 4 |
| | 2+2 4 | 2+2 4 | 2+2 4 | 4+4 2 | 4+4 2 | 3+4 2 | 3+4 2 | 2+2 4 |
| | 2+2 4 | 2+2 4 | 2+2 4 | 6+6 2 | 5+5 2 | 5+6 2 | 5+6 2 | 2+2 4 |
| | | | | 5+5 2 | 6+6 2 | 5+6 2 | 3+4 2 | |
| | | | | 4+4 2 | 3+3 2 | 3+4 2 | 3+4 2 | |
| | | | | 3+3 2 | 4+4 2 | 3+4 2 | 5+6 2 | |
| | | | | 6+6 2 | 3+3 2 | 5+6 2 | 3+4 2 | |
| | | | | 5+5 2 | 4+4 2 | 3+4 2 | 5+6 2 | |
| | | | | 4+4 2 | 5+5 2 | 3+4 2 | 5+6 2 | |
| | | | | 3+3 2 | 6+6 2 | 3+4 2 | 5+6 2 | |
| | | | | 5+5 2 | 4+4 2 | 3+4 2 | 3+4 2 | |
| | | | | 6+6 2 | 3+3 2 | 3+4 2 | 3+4 2 | |
| | | | | 3+3 2 | 6+6 2 | 5+6 2 | 5+6 2 | |
| | | | | 4+4 2 | 5+5 2 | 5+6 2 | 5+6 2 | |
| | | | | 5+5 2 | 3+3 2 | 3+4 2 | 3+4 2 | |
| | | | | 6+6 2 | 4+4 2 | 5+6 2 | 5+6 2 | |
| | | | | 3+3 2 | 6+6 2 | 5+6 2 | 5+6 2 | |
| | | | | 4+4 2 | 5+5 2 | 3+4 2 | 5+6 2 | |
| | | | | 5+5 2 | 3+3 2 | 3+4 2 | 3+4 2 | |
| | | | | 6+6 2 | 4+4 2 | 3+4 2 | 3+4 2 | |
| | | | | 6+6 2 | 6+6 2 | 5+6 2 | 5+6 2 | |
| | | | | 5+5 2 | 5+5 2 | 5+6 2 | 3+4 2 | |
| 14.2.47 | 73739 | 136 | Co1Nb1Te2 | | | | | |
| | -A | -B | -C | -D | -E | -GM | -Y | -Z |
| | 2+2 4 | 2+2 4 | 2+2 4 | 5+5 2 | 3+3 2 | 3+4 2 | 5+6 2 | 2+2 4 |
| | 2+2 4 | 2+2 4 | 2+2 4 | 6+6 2 | 4+4 2 | 5+6 2 | 3+4 2 | 2+2 4 |
| | 2+2 4 | 2+2 4 | 2+2 4 | 3+3 2 | 5+5 2 | 3+4 2 | 5+6 2 | 2+2 4 |
| | 2+2 4 | 2+2 4 | 2+2 4 | 4+4 2 | 6+6 2 | 5+6 2 | 3+4 2 | 2+2 4 |
| | 2+2 4 | 2+2 4 | 2+2 4 | 5+5 2 | 3+3 2 | 3+4 2 | 5+6 2 | 2+2 4 |
| | 2+2 4 | 2+2 4 | 2+2 4 | 6+6 2 | 4+4 2 | 5+6 2 | 3+4 2 | 2+2 4 |
| | 2+2 4 | 2+2 4 | 2+2 4 | 3+3 2 | 5+5 2 | 3+4 2 | 5+6 2 | 2+2 4 |
| | 2+2 4 | 2+2 4 | 2+2 4 | 4+4 2 | 6+6 2 | 5+6 2 | 3+4 2 | 2+2 4 |
| | 2+2 4 | 2+2 4 | 2+2 4 | 5+5 2 | 4+4 2 | 3+4 2 | 5+6 2 | 2+2 4 |
| | 2+2 4 | 2+2 4 | 2+2 4 | 3+3 2 | 6+6 2 | 5+6 2 | 3+4 2 | 2+2 4 |
| | 2+2 4 | 2+2 4 | 2+2 4 | 6+6 2 | 3+3 2 | 3+4 2 | 5+6 2 | 2+2 4 |
| | 2+2 4 | 2+2 4 | 2+2 4 | 4+4 2 | 5+5 2 | 3+4 2 | 5+6 2 | 2+2 4 |
| | 2+2 4 | 2+2 4 | 2+2 4 | 6+6 2 | 3+3 2 | 5+6 2 | 3+4 2 | 2+2 4 |
| | 2+2 4 | 2+2 4 | 2+2 4 | 5+5 2 | 4+4 2 | 5+6 2 | 3+4 2 | 2+2 4 |
| | 2+2 4 | 2+2 4 | 2+2 4 | 4+4 2 | 5+5 2 | 3+4 2 | 5+6 2 | 2+2 4 |
| | 2+2 4 | 2+2 4 | 2+2 4 | 3+3 2 | 6+6 2 | 5+6 2 | 3+4 2 | 2+2 4 |
| | 2+2 4 | 2+2 4 | 2+2 4 | 4+4 2 | 5+5 2 | 3+4 2 | 5+6 2 | 2+2 4 |
| | 2+2 4 | 2+2 4 | 2+2 4 | 3+3 2 | 6+6 2 | 5+6 2 | 3+4 2 | 2+2 4 |
| | 2+2 4 | 2+2 4 | 2+2 4 | 5+5 2 | 3+3 2 | 5+6 2 | 3+4 2 | 2+2 4 |
| | 2+2 4 | 2+2 4 | 2+2 4 | 6+6 2 | 4+4 2 | 3+4 2 | 5+6 2 | 2+2 4 |
| | 2+2 4 | 2+2 4 | 2+2 4 | 5+5 2 | 4+4 2 | 5+6 2 | 3+4 2 | 2+2 4 |
| | 2+2 4 | 2+2 4 | 2+2 4 | 6+6 2 | 3+3 2 | 5+6 2 | 5+6 2 | 2+2 4 |
| | 2+2 4 | 2+2 4 | 2+2 4 | 3+3 2 | 5+5 2 | 3+4 2 | 3+4 2 | 2+2 4 |
| | 2+2 4 | 2+2 4 | 2+2 4 | 4+4 2 | 6+6 2 | 3+4 2 | 5+6 2 | 2+2 4 |
| | 2+2 4 | 2+2 4 | 2+2 4 | 4+4 2 | 5+5 2 | 3+4 2 | 5+6 2 | 2+2 4 |
| | 2+2 4 | 2+2 4 | 2+2 4 | 3+3 2 | 6+6 2 | 3+4 2 | 3+4 2 | 2+2 4 |
| | 2+2 4 | 2+2 4 | 2+2 4 | 5+5 2 | 4+4 2 | 5+6 2 | 3+4 2 | 2+2 4 |
| | 2+2 4 | 2+2 4 | 2+2 4 | 6+6 2 | 3+3 2 | 5+6 2 | 5+6 2 | 2+2 4 |
| | 2+2 4 | 2+2 4 | 2+2 4 | 4+4 2 | 5+5 2 | 3+4 2 | 3+4 2 | 2+2 4 |
| | 2+2 4 | 2+2 4 | 2+2 4 | 3+3 2 | 6+6 2 | 5+6 2 | 5+6 2 | 2+2 4 |
| | 2+2 4 | 2+2 4 | 2+2 4 | 5+5 2 | 4+4 2 | 5+6 2 | 3+4 2 | 2+2 4 |
| | 2+2 4 | 2+2 4 | 2+2 4 | 6+6 2 | 3+3 2 | 3+4 2 | 3+4 2 | 2+2 4 |
| | 2+2 4 | 2+2 4 | 2+2 4 | 3+3 2 | 6+6 2 | 5+6 2 | 5+6 2 | 2+2 4 |
| | 2+2 4 | 2+2 4 | 2+2 4 | 4+4 2 | 5+5 2 | 3+4 2 | 3+4 2 | 2+2 4 |
| | | | | 6+6 2 | 3+3 2 | 3+4 2 | 5+6 2 | |
| | | | | 5+5 2 | 4+4 2 | 5+6 2 | 5+6 2 | |

Continued on next column

**TABLE XXVI– continued**

| TQC | ICSD | NoE | SF | | | | | |
|---|---|---|---|---|---|---|---|---|
| | | | | 6+6 2 | 3+3 2 | 5+6 2 | 3+4 2 | |
| | | | | 5+5 2 | 4+4 2 | 3+4 2 | 5+6 2 | |
| | | | | 3+3 2 | 6+6 2 | 3+4 2 | 5+6 2 | |
| | | | | 4+4 2 | 3+3 2 | 5+6 2 | 3+4 2 | |
| | | | | 6+6 2 | 5+5 2 | 3+4 2 | 3+4 2 | |
| | | | | 5+5 2 | 4+4 2 | 5+6 2 | 3+4 2 | |
| | | | | 3+3 2 | 6+6 2 | 5+6 2 | 5+6 2 | |
| | | | | 4+4 2 | 5+5 2 | 3+4 2 | 5+6 2 | |
| | | | | 5+5 2 | 4+4 2 | 3+4 2 | 5+6 2 | |
| | | | | 6+6 2 | 3+3 2 | 5+6 2 | 3+4 2 | |
| | | | | 4+4 2 | 5+5 2 | 3+4 2 | 5+6 2 | |
| | | | | 3+3 2 | 6+6 2 | 5+6 2 | 3+4 2 | |
| | | | | 6+6 2 | 3+3 2 | 3+4 2 | 3+4 2 | |
| | | | | 5+5 2 | 4+4 2 | 5+6 2 | 3+4 2 | |
| | | | | 4+4 2 | 5+5 2 | 5+6 2 | 5+6 2 | |
| | | | | 3+3 2 | 6+6 2 | 3+4 2 | 3+4 2 | |
| | | | | 5+5 2 | 4+4 2 | 5+6 2 | 5+6 2 | |
| | | | | 6+6 2 | 3+3 2 | 5+6 2 | 3+4 2 | |
| | | | | 4+4 2 | 5+5 2 | 3+4 2 | 5+6 2 | |
| | | | | 3+3 2 | 6+6 2 | 3+4 2 | 5+6 2 | |
| | | | | 4+4 2 | 5+5 2 | 3+4 2 | 5+6 2 | |
| | | | | 3+3 2 | 6+6 2 | 3+4 2 | 3+4 2 | |
| | | | | 4+4 2 | 3+3 2 | 5+6 2 | 5+6 2 | |
| | | | | 3+3 2 | 4+4 2 | 3+4 2 | 5+6 2 | |
| | | | | 6+6 2 | 5+5 2 | 3+4 2 | 5+6 2 | |
| | | | | 5+5 2 | 6+6 2 | 3+4 2 | 5+6 2 | |
| | | | | 4+4 2 | 5+5 2 | 3+4 2 | 3+4 2 | |
| | | | | 3+3 2 | 6+6 2 | 5+6 2 | 5+6 2 | |
| | | | | 6+6 2 | 3+3 2 | 5+6 2 | 3+4 2 | |
| | | | | 5+5 2 | 4+4 2 | 3+4 2 | 3+4 2 | |
| | | | | 5+5 2 | 4+4 2 | 5+6 2 | 3+4 2 | |
| | | | | 6+6 2 | 3+3 2 | 3+4 2 | 5+6 2 | |
| 14.2.48 | 250364 | 132 | Cl4V1 | | | | | |
| | -A | -B | -C | -D | -E | -GM | -Y | -Z |
| | 2+2 4 | 2+2 4 | 2+2 4 | 5+5 2 | 3+3 2 | 3+4 2 | 5+6 2 | 2+2 4 |
| | 2+2 4 | 2+2 4 | 2+2 4 | 6+6 2 | 4+4 2 | 5+6 2 | 3+4 2 | 2+2 4 |
| | 2+2 4 | 2+2 4 | 2+2 4 | 3+3 2 | 5+5 2 | 5+6 2 | 5+6 2 | 2+2 4 |
| | 2+2 4 | 2+2 4 | 2+2 4 | 4+4 2 | 6+6 2 | 3+4 2 | 3+4 2 | 2+2 4 |
| | 2+2 4 | 2+2 4 | 2+2 4 | 3+3 2 | 5+5 2 | 3+4 2 | 3+4 2 | 2+2 4 |
| | 2+2 4 | 2+2 4 | 2+2 4 | 4+4 2 | 6+6 2 | 3+4 2 | 5+6 2 | 2+2 4 |
| | 2+2 4 | 2+2 4 | 2+2 4 | 5+5 2 | 3+3 2 | 5+6 2 | 3+4 2 | 2+2 4 |
| | 2+2 4 | 2+2 4 | 2+2 4 | 6+6 2 | 4+4 2 | 3+4 2 | 5+6 2 | 2+2 4 |
| | 2+2 4 | 2+2 4 | 2+2 4 | 3+3 2 | 3+3 2 | 5+6 2 | 3+4 2 | 2+2 4 |
| | 2+2 4 | 2+2 4 | 2+2 4 | 4+4 2 | 4+4 2 | 5+6 2 | 5+6 2 | 2+2 4 |
| | 2+2 4 | 2+2 4 | 2+2 4 | 5+5 2 | 5+5 2 | 5+6 2 | 5+6 2 | 2+2 4 |
| | 2+2 4 | 2+2 4 | 2+2 4 | 6+6 2 | 6+6 2 | 3+4 2 | 3+4 2 | 2+2 4 |
| | 2+2 4 | 2+2 4 | 2+2 4 | 6+6 2 | 5+5 2 | 3+4 2 | 3+4 2 | 2+2 4 |
| | 2+2 4 | 2+2 4 | 2+2 4 | 5+5 2 | 6+6 2 | 5+6 2 | 3+4 2 | 2+2 4 |
| | 2+2 4 | 2+2 4 | 2+2 4 | 4+4 2 | 3+3 2 | 3+4 2 | 5+6 2 | 2+2 4 |
| | 2+2 4 | 2+2 4 | 2+2 4 | 3+3 2 | 4+4 2 | 5+6 2 | 5+6 2 | 2+2 4 |
| | 2+2 4 | 2+2 4 | 2+2 4 | 3+3 2 | 3+3 2 | 3+4 2 | 3+4 2 | 2+2 4 |
| | 2+2 4 | 2+2 4 | 2+2 4 | 4+4 2 | 4+4 2 | 3+4 2 | 3+4 2 | 2+2 4 |
| | 2+2 4 | 2+2 4 | 2+2 4 | 6+6 2 | 5+5 2 | 5+6 2 | 5+6 2 | 2+2 4 |
| | 2+2 4 | 2+2 4 | 2+2 4 | 5+5 2 | 6+6 2 | 5+6 2 | 5+6 2 | 2+2 4 |
| | 2+2 4 | 2+2 4 | 2+2 4 | 3+3 2 | 3+3 2 | 3+4 2 | 3+4 2 | 2+2 4 |
| | 2+2 4 | 2+2 4 | 2+2 4 | 4+4 2 | 4+4 2 | 3+4 2 | 3+4 2 | 2+2 4 |
| | 2+2 4 | 2+2 4 | 2+2 4 | 5+5 2 | 6+6 2 | 5+6 2 | 5+6 2 | 2+2 4 |
| | 2+2 4 | 2+2 4 | 2+2 4 | 6+6 2 | 5+5 2 | 3+4 2 | 5+6 2 | 2+2 4 |
| | 2+2 4 | 2+2 4 | 2+2 4 | 3+3 2 | 5+5 2 | 5+6 2 | 3+4 2 | 2+2 4 |
| | 2+2 4 | 2+2 4 | 2+2 4 | 4+4 2 | 3+3 2 | 5+6 2 | 3+4 2 | 2+2 4 |
| | 2+2 4 | 2+2 4 | 2+2 4 | 3+3 2 | 6+6 2 | 3+4 2 | 5+6 2 | 2+2 4 |

Continued on next column

**TABLE XXVI– continued**

| TQC | ICSD | NoE | SF | | | | | |
|---|---|---|---|---|---|---|---|---|
| | 2+2 4 | 2+2 4 | 2+2 4 | 4+4 2 | 4+4 2 | 3+4 2 | 5+6 2 | 2+2 4 |
| | 2+2 4 | 2+2 4 | 2+2 4 | 6+6 2 | 3+3 2 | 5+6 2 | 5+6 2 | 2+2 4 |
| | 2+2 4 | 2+2 4 | 2+2 4 | 5+5 2 | 4+4 2 | 3+4 2 | 5+6 2 | 2+2 4 |
| | 2+2 4 | 2+2 4 | 2+2 4 | 5+5 2 | 5+5 2 | 5+6 2 | 3+4 2 | 2+2 4 |
| | 2+2 4 | 2+2 4 | 2+2 4 | 6+6 2 | 6+6 2 | 3+4 2 | 3+4 2 | 2+2 4 |
| | 2+2 4 | 2+2 4 | 2+2 4 | 5+5 2 | 6+6 2 | 5+6 2 | 5+6 2 | 2+2 4 |
| | | | | 6+6 2 | 4+4 2 | 5+6 2 | 3+4 2 | |
| | | | | 4+4 2 | 5+5 2 | 3+4 2 | 3+4 2 | |
| | | | | 3+3 2 | 3+3 2 | 3+4 2 | 3+4 2 | |
| | | | | 5+5 2 | 5+5 2 | 5+6 2 | 5+6 2 | |
| | | | | 6+6 2 | 6+6 2 | 5+6 2 | 5+6 2 | |
| | | | | 3+3 2 | 4+4 2 | 3+4 2 | 5+6 2 | |
| | | | | 4+4 2 | 3+3 2 | 5+6 2 | 3+4 2 | |
| | | | | 6+6 2 | 3+3 2 | 3+4 2 | 5+6 2 | |
| | | | | 5+5 2 | 4+4 2 | 5+6 2 | 5+6 2 | |
| | | | | 5+5 2 | 5+5 2 | 5+6 2 | 3+4 2 | |
| | | | | 6+6 2 | 6+6 2 | 3+4 2 | 5+6 2 | |
| | | | | 4+4 2 | 6+6 2 | 3+4 2 | 3+4 2 | |
| | | | | 3+3 2 | 5+5 2 | 3+4 2 | 3+4 2 | |
| | | | | 3+3 2 | 3+3 2 | 3+4 2 | 5+6 2 | |
| | | | | 4+4 2 | 4+4 2 | 5+6 2 | 3+4 2 | |
| | | | | 5+5 2 | 4+4 2 | 3+4 2 | 5+6 2 | |
| | | | | 6+6 2 | 3+3 2 | 5+6 2 | 3+4 2 | |
| | | | | 4+4 2 | 5+5 2 | 5+6 2 | 5+6 2 | |
| | | | | 3+3 2 | 6+6 2 | 5+6 2 | 3+4 2 | |
| | | | | 4+4 2 | 4+4 2 | 3+4 2 | 3+4 2 | |
| | | | | 3+3 2 | 3+3 2 | 5+6 2 | 3+4 2 | |
| | | | | 5+5 2 | 5+5 2 | 3+4 2 | 5+6 2 | |
| | | | | 6+6 2 | 6+6 2 | 5+6 2 | 5+6 2 | |
| | | | | 4+4 2 | 5+5 2 | 3+4 2 | 5+6 2 | |
| | | | | 3+3 2 | 6+6 2 | 5+6 2 | 3+4 2 | |
| | | | | 6+6 2 | 3+3 2 | 3+4 2 | 5+6 2 | |
| | | | | 5+5 2 | 4+4 2 | 3+4 2 | 3+4 2 | |
| | | | | 5+5 2 | 6+6 2 | 5+6 2 | 5+6 2 | |
| | | | | 6+6 2 | 5+5 2 | 3+4 2 | 3+4 2 | |
| | | | | 3+3 2 | 3+3 2 | 5+6 2 | 3+4 2 | |
| | | | | 4+4 2 | 4+4 2 | 5+6 2 | 5+6 2 | |
| | | | | 3+3 2 | 6+6 2 | 5+6 2 | 3+4 2 | |
| | | | | 4+4 2 | 5+5 2 | 3+4 2 | 5+6 2 | |
| 14.2.49 | 422546 | 192 | Cl4Hg4I2S1 | | | | | |
| | -A | -B | -C | -D | -E | -GM | -Y | -Z |
| | 2+2 4 | 2+2 4 | 2+2 4 | 5+5 2 | 5+5 2 | 3+4 2 | 3+4 2 | 2+2 4 |
| | 2+2 4 | 2+2 4 | 2+2 4 | 6+6 2 | 6+6 2 | 3+4 2 | 3+4 2 | 2+2 4 |
| | 2+2 4 | 2+2 4 | 2+2 4 | 3+3 2 | 3+3 2 | 5+6 2 | 5+6 2 | 2+2 4 |
| | 2+2 4 | 2+2 4 | 2+2 4 | 4+4 2 | 4+4 2 | 5+6 2 | 5+6 2 | 2+2 4 |
| | 2+2 4 | 2+2 4 | 2+2 4 | 5+5 2 | 5+5 2 | 3+4 2 | 3+4 2 | 2+2 4 |
| | 2+2 4 | 2+2 4 | 2+2 4 | 6+6 2 | 6+6 2 | 3+4 2 | 3+4 2 | 2+2 4 |
| | 2+2 4 | 2+2 4 | 2+2 4 | 3+3 2 | 3+3 2 | 5+6 2 | 5+6 2 | 2+2 4 |
| | 2+2 4 | 2+2 4 | 2+2 4 | 4+4 2 | 4+4 2 | 5+6 2 | 5+6 2 | 2+2 4 |
| | 2+2 4 | 2+2 4 | 2+2 4 | 3+3 2 | 3+3 2 | 3+4 2 | 3+4 2 | 2+2 4 |
| | 2+2 4 | 2+2 4 | 2+2 4 | 4+4 2 | 4+4 2 | 3+4 2 | 3+4 2 | 2+2 4 |
| | 2+2 4 | 2+2 4 | 2+2 4 | 6+6 2 | 5+5 2 | 3+4 2 | 5+6 2 | 2+2 4 |
| | 2+2 4 | 2+2 4 | 2+2 4 | 5+5 2 | 6+6 2 | 3+4 2 | 3+4 2 | 2+2 4 |
| | 2+2 4 | 2+2 4 | 2+2 4 | 3+3 2 | 6+6 2 | 3+4 2 | 5+6 2 | 2+2 4 |
| | 2+2 4 | 2+2 4 | 2+2 4 | 4+4 2 | 4+4 2 | 3+4 2 | 3+4 2 | 2+2 4 |
| | 2+2 4 | 2+2 4 | 2+2 4 | 6+6 2 | 3+3 2 | 5+6 2 | 3+4 2 | 2+2 4 |
| | 2+2 4 | 2+2 4 | 2+2 4 | 5+5 2 | 5+5 2 | 5+6 2 | 3+4 2 | 2+2 4 |
| | 2+2 4 | 2+2 4 | 2+2 4 | 4+4 2 | 4+4 2 | 5+6 2 | 5+6 2 | 2+2 4 |
| | 2+2 4 | 2+2 4 | 2+2 4 | 3+3 2 | 3+3 2 | 5+6 2 | 5+6 2 | 2+2 4 |
| | 2+2 4 | 2+2 4 | 2+2 4 | 5+5 2 | 5+5 2 | 3+4 2 | 3+4 2 | 2+2 4 |
| | 2+2 4 | 2+2 4 | 2+2 4 | 6+6 2 | 6+6 2 | 3+4 2 | 3+4 2 | 2+2 4 |

Continued on next column

**TABLE XXVI– continued**

| TQC | ICSD | NoE | SF | | | | | |
|---|---|---|---|---|---|---|---|---|
| | 2+2 4 | 2+2 4 | 2+2 4 | 5+5 2 | 5+5 2 | 5+6 2 | 5+6 2 | 2+2 4 |
| | 2+2 4 | 2+2 4 | 2+2 4 | 6+6 2 | 6+6 2 | 5+6 2 | 5+6 2 | 2+2 4 |
| | 2+2 4 | 2+2 4 | 2+2 4 | 4+4 2 | 4+4 2 | 5+6 2 | 5+6 2 | 2+2 4 |
| | 2+2 4 | 2+2 4 | 2+2 4 | 3+3 2 | 3+3 2 | 5+6 2 | 5+6 2 | 2+2 4 |
| | 2+2 4 | 2+2 4 | 2+2 4 | 6+6 2 | 6+6 2 | 3+4 2 | 3+4 2 | 2+2 4 |
| | 2+2 4 | 2+2 4 | 2+2 4 | 5+5 2 | 5+5 2 | 3+4 2 | 3+4 2 | 2+2 4 |
| | 2+2 4 | 2+2 4 | 2+2 4 | 3+3 2 | 3+3 2 | 5+6 2 | 5+6 2 | 2+2 4 |
| | 2+2 4 | 2+2 4 | 2+2 4 | 4+4 2 | 4+4 2 | 5+6 2 | 3+4 2 | 2+2 4 |
| | 2+2 4 | 2+2 4 | 2+2 4 | 5+5 2 | 5+5 2 | 3+4 2 | 5+6 2 | 2+2 4 |
| | 2+2 4 | 2+2 4 | 2+2 4 | 6+6 2 | 6+6 2 | 3+4 2 | 3+4 2 | 2+2 4 |
| | 2+2 4 | 2+2 4 | 2+2 4 | 3+3 2 | 3+3 2 | 5+6 2 | 5+6 2 | 2+2 4 |
| | 2+2 4 | 2+2 4 | 2+2 4 | 4+4 2 | 4+4 2 | 5+6 2 | 3+4 2 | 2+2 4 |
| | 2+2 4 | 2+2 4 | 2+2 4 | 4+4 2 | 6+6 2 | 3+4 2 | 5+6 2 | 2+2 4 |
| | 2+2 4 | 2+2 4 | 2+2 4 | 6+6 2 | 4+4 2 | 5+6 2 | 5+6 2 | 2+2 4 |
| | 2+2 4 | 2+2 4 | 2+2 4 | 5+5 2 | 5+5 2 | 3+4 2 | 3+4 2 | 2+2 4 |
| | 2+2 4 | 2+2 4 | 2+2 4 | 3+3 2 | 3+3 2 | 5+6 2 | 5+6 2 | 2+2 4 |
| | 2+2 4 | 2+2 4 | 2+2 4 | 6+6 2 | 6+6 2 | 5+6 2 | 5+6 2 | 2+2 4 |
| | 2+2 4 | 2+2 4 | 2+2 4 | 5+5 2 | 5+5 2 | 3+4 2 | 3+4 2 | 2+2 4 |
| | 2+2 4 | 2+2 4 | 2+2 4 | 3+3 2 | 3+3 2 | 3+4 2 | 3+4 2 | 2+2 4 |
| | 2+2 4 | 2+2 4 | 2+2 4 | 4+4 2 | 4+4 2 | 5+6 2 | 5+6 2 | 2+2 4 |
| | 2+2 4 | 2+2 4 | 2+2 4 | 5+5 2 | 5+5 2 | 5+6 2 | 5+6 2 | 2+2 4 |
| | 2+2 4 | 2+2 4 | 2+2 4 | 6+6 2 | 6+6 2 | 3+4 2 | 3+4 2 | 2+2 4 |
| | 2+2 4 | 2+2 4 | 2+2 4 | 4+4 2 | 4+4 2 | 5+6 2 | 5+6 2 | 2+2 4 |
| | 2+2 4 | 2+2 4 | 2+2 4 | 3+3 2 | 3+3 2 | 3+4 2 | 3+4 2 | 2+2 4 |
| | 2+2 4 | 2+2 4 | 2+2 4 | 5+5 2 | 5+5 2 | 5+6 2 | 5+6 2 | 2+2 4 |
| | 2+2 4 | 2+2 4 | 2+2 4 | 6+6 2 | 6+6 2 | 3+4 2 | 3+4 2 | 2+2 4 |
| | 2+2 4 | 2+2 4 | 2+2 4 | 3+3 2 | 3+3 2 | 3+4 2 | 3+4 2 | 2+2 4 |
| | 2+2 4 | 2+2 4 | 2+2 4 | 4+4 2 | 4+4 2 | 3+4 2 | 5+6 2 | 2+2 4 |
| | | | | 5+5 2 | 5+5 2 | 5+6 2 | 3+4 2 | |
| | | | | 3+3 2 | 6+6 2 | 3+4 2 | 5+6 2 | |
| | | | | 6+6 2 | 3+3 2 | 5+6 2 | 3+4 2 | |
| | | | | 4+4 2 | 4+4 2 | 5+6 2 | 5+6 2 | |
| | | | | 4+4 2 | 6+6 2 | 5+6 2 | 5+6 2 | |
| | | | | 5+5 2 | 5+5 2 | 3+4 2 | 3+4 2 | |
| | | | | 6+6 2 | 4+4 2 | 3+4 2 | 3+4 2 | |
| | | | | 3+3 2 | 3+3 2 | 5+6 2 | 5+6 2 | |
| | | | | 3+3 2 | 3+3 2 | 5+6 2 | 5+6 2 | |
| | | | | 4+4 2 | 4+4 2 | 3+4 2 | 3+4 2 | |
| | | | | 4+4 2 | 3+3 2 | 5+6 2 | 5+6 2 | |
| | | | | 3+3 2 | 4+4 2 | 3+4 2 | 3+4 2 | |
| | | | | 6+6 2 | 6+6 2 | 5+6 2 | 3+4 2 | |
| | | | | 5+5 2 | 5+5 2 | 3+4 2 | 5+6 2 | |
| | | | | 6+6 2 | 6+6 2 | 5+6 2 | 5+6 2 | |
| | | | | 3+3 2 | 5+5 2 | 5+6 2 | 3+4 2 | |
| | | | | 5+5 2 | 3+3 2 | 3+4 2 | 5+6 2 | |
| | | | | 4+4 2 | 4+4 2 | 3+4 2 | 3+4 2 | |
| | | | | 6+6 2 | 6+6 2 | 3+4 2 | 3+4 2 | |
| | | | | 5+5 2 | 5+5 2 | 3+4 2 | 5+6 2 | |
| | | | | 4+4 2 | 6+6 2 | 3+4 2 | 3+4 2 | |
| | | | | 6+6 2 | 5+5 2 | 5+6 2 | 5+6 2 | |
| | | | | 3+3 2 | 3+3 2 | 5+6 2 | 3+4 2 | |
| | | | | 5+5 2 | 4+4 2 | 3+4 2 | 5+6 2 | |
| | | | | 3+3 2 | 4+4 2 | 5+6 2 | 3+4 2 | |
| | | | | 6+6 2 | 6+6 2 | 3+4 2 | 5+6 2 | |
| | | | | 4+4 2 | 5+5 2 | 5+6 2 | 3+4 2 | |
| | | | | 5+5 2 | 3+3 2 | 3+4 2 | 5+6 2 | |
| | | | | 5+5 2 | 5+5 2 | 5+6 2 | 3+4 2 | |
| | | | | 6+6 2 | 3+3 2 | 3+4 2 | 3+4 2 | |
| | | | | 4+4 2 | 4+4 2 | 3+4 2 | 5+6 2 | |
| | | | | 3+3 2 | 6+6 2 | 5+6 2 | 3+4 2 | |
| | | | | 4+4 2 | 6+6 2 | 5+6 2 | 5+6 2 | |
| | | | | 3+3 2 | 4+4 2 | 3+4 2 | 3+4 2 | |
| | | | | 5+5 2 | 5+5 2 | 5+6 2 | 5+6 2 | |
| | | | | 6+6 2 | 3+3 2 | 3+4 2 | 3+4 2 | |
| | | | | 3+3 2 | 4+4 2 | 5+6 2 | 5+6 2 | |
| | | | | 4+4 2 | 3+3 2 | 3+4 2 | 3+4 2 | |
| | | | | 5+5 2 | 5+5 2 | 5+6 2 | 5+6 2 | |
| | | | | 6+6 2 | 6+6 2 | 5+6 2 | 5+6 2 | |
| | | | | 4+4 2 | 3+3 2 | 3+4 2 | 3+4 2 | |
| | | | | 6+6 2 | 4+4 2 | 5+6 2 | 5+6 2 | |
| | | | | 3+3 2 | 5+5 2 | 3+4 2 | 5+6 2 | |
| | | | | 5+5 2 | 6+6 2 | 5+6 2 | 3+4 2 | |
| | | | | 5+5 2 | 5+5 2 | 3+4 2 | 5+6 2 | |
| | | | | 6+6 2 | 6+6 2 | 5+6 2 | 5+6 2 | |
| | | | | 6+6 2 | 5+5 2 | 5+6 2 | 3+4 2 | |
| | | | | 5+5 2 | 6+6 2 | 5+6 2 | 5+6 2 | |
| 14.2.50 | 4411 | 132 | Cl3Mo1O1 | | | | | |
| | -A | -B | -C | -D | -E | -GM | -Y | -Z |
| | 2+2 4 | 2+2 4 | 2+2 4 | 5+5 2 | 5+5 2 | 3+4 2 | 3+4 2 | 2+2 4 |
| | 2+2 4 | 2+2 4 | 2+2 4 | 6+6 2 | 6+6 2 | 3+4 2 | 3+4 2 | 2+2 4 |
| | 2+2 4 | 2+2 4 | 2+2 4 | 3+3 2 | 3+3 2 | 5+6 2 | 5+6 2 | 2+2 4 |
| | 2+2 4 | 2+2 4 | 2+2 4 | 4+4 2 | 4+4 2 | 5+6 2 | 5+6 2 | 2+2 4 |
| | 2+2 4 | 2+2 4 | 2+2 4 | 3+3 2 | 3+3 2 | 3+4 2 | 3+4 2 | 2+2 4 |
| | 2+2 4 | 2+2 4 | 2+2 4 | 4+4 2 | 4+4 2 | 5+6 2 | 5+6 2 | 2+2 4 |
| | 2+2 4 | 2+2 4 | 2+2 4 | 5+5 2 | 5+5 2 | 3+4 2 | 5+6 2 | 2+2 4 |
| | 2+2 4 | 2+2 4 | 2+2 4 | 6+6 2 | 6+6 2 | 5+6 2 | 3+4 2 | 2+2 4 |
| | 2+2 4 | 2+2 4 | 2+2 4 | 6+6 2 | 6+6 2 | 5+6 2 | 5+6 2 | 2+2 4 |
| | 2+2 4 | 2+2 4 | 2+2 4 | 3+3 2 | 5+5 2 | 3+4 2 | 3+4 2 | 2+2 4 |
| | 2+2 4 | 2+2 4 | 2+2 4 | 5+5 2 | 3+3 2 | 3+4 2 | 5+6 2 | 2+2 4 |
| | 2+2 4 | 2+2 4 | 2+2 4 | 4+4 2 | 4+4 2 | 5+6 2 | 3+4 2 | 2+2 4 |
| | 2+2 4 | 2+2 4 | 2+2 4 | 4+4 2 | 5+5 2 | 3+4 2 | 5+6 2 | 2+2 4 |
| | 2+2 4 | 2+2 4 | 2+2 4 | 3+3 2 | 6+6 2 | 3+4 2 | 3+4 2 | 2+2 4 |
| | 2+2 4 | 2+2 4 | 2+2 4 | 5+5 2 | 4+4 2 | 5+6 2 | 5+6 2 | 2+2 4 |
| | 2+2 4 | 2+2 4 | 2+2 4 | 6+6 2 | 3+3 2 | 5+6 2 | 3+4 2 | 2+2 4 |
| | 2+2 4 | 2+2 4 | 2+2 4 | 5+5 2 | 5+5 2 | 3+4 2 | 5+6 2 | 2+2 4 |
| | 2+2 4 | 2+2 4 | 2+2 4 | 6+6 2 | 6+6 2 | 5+6 2 | 3+4 2 | 2+2 4 |
| | 2+2 4 | 2+2 4 | 2+2 4 | 4+4 2 | 4+4 2 | 3+4 2 | 3+4 2 | 2+2 4 |
| | 2+2 4 | 2+2 4 | 2+2 4 | 3+3 2 | 3+3 2 | 5+6 2 | 5+6 2 | 2+2 4 |
| | 2+2 4 | 2+2 4 | 2+2 4 | 6+6 2 | 6+6 2 | 3+4 2 | 3+4 2 | 2+2 4 |
| | 2+2 4 | 2+2 4 | 2+2 4 | 4+4 2 | 5+5 2 | 3+4 2 | 3+4 2 | 2+2 4 |
| | 2+2 4 | 2+2 4 | 2+2 4 | 3+3 2 | 3+3 2 | 5+6 2 | 5+6 2 | 2+2 4 |
| | 2+2 4 | 2+2 4 | 2+2 4 | 5+5 2 | 4+4 2 | 3+4 2 | 3+4 2 | 2+2 4 |
| | 2+2 4 | 2+2 4 | 2+2 4 | 5+5 2 | 5+5 2 | 3+4 2 | 5+6 2 | 2+2 4 |
| | 2+2 4 | 2+2 4 | 2+2 4 | 6+6 2 | 6+6 2 | 5+6 2 | 3+4 2 | 2+2 4 |
| | 2+2 4 | 2+2 4 | 2+2 4 | 3+3 2 | 3+3 2 | 5+6 2 | 5+6 2 | 2+2 4 |
| | 2+2 4 | 2+2 4 | 2+2 4 | 4+4 2 | 4+4 2 | 5+6 2 | 5+6 2 | 2+2 4 |
| | 2+2 4 | 2+2 4 | 2+2 4 | 3+3 2 | 6+6 2 | 3+4 2 | 3+4 2 | 2+2 4 |
| | 2+2 4 | 2+2 4 | 2+2 4 | 4+4 2 | 3+3 2 | 5+6 2 | 5+6 2 | 2+2 4 |
| | 2+2 4 | 2+2 4 | 2+2 4 | 6+6 2 | 4+4 2 | 3+4 2 | 5+6 2 | 2+2 4 |
| | 2+2 4 | 2+2 4 | 2+2 4 | 5+5 2 | 5+5 2 | 5+6 2 | 3+4 2 | 2+2 4 |
| | 2+2 4 | 2+2 4 | 2+2 4 | 4+4 2 | 5+5 2 | 3+4 2 | 3+4 2 | 2+2 4 |
| | | | | 3+3 2 | 6+6 2 | 5+6 2 | 3+4 2 | |
| | | | | 5+5 2 | 3+3 2 | 3+4 2 | 5+6 2 | |
| | | | | 6+6 2 | 4+4 2 | 5+6 2 | 5+6 2 | |
| | | | | 4+4 2 | 6+6 2 | 3+4 2 | 5+6 2 | |
| | | | | 3+3 2 | 5+5 2 | 5+6 2 | 3+4 2 | |
| | | | | 5+5 2 | 3+3 2 | 3+4 2 | 5+6 2 | |
| | | | | 6+6 2 | 4+4 2 | 5+6 2 | 5+6 2 | |
| | | | | 4+4 2 | 5+5 2 | 3+4 2 | 3+4 2 | |
| | | | | 3+3 2 | 6+6 2 | 3+4 2 | 3+4 2 | |
| | | | | 6+6 2 | 4+4 2 | 5+6 2 | 5+6 2 | |
| | | | | 5+5 2 | 3+3 2 | 3+4 2 | 5+6 2 | |

Continued on next column

**TABLE XXVI– continued**

| TQC | ICSD | NoE | SF | | | | | |
|---|---|---|---|---|---|---|---|---|
| | | | | 3+3 2 | 6+6 2 | 5+6 2 | 3+4 2 | |
| | | | | 4+4 2 | 5+5 2 | 5+6 2 | 3+4 2 | |
| | | | | 6+6 2 | 3+3 2 | 3+4 2 | 5+6 2 | |
| | | | | 5+5 2 | 4+4 2 | 5+6 2 | 3+4 2 | |
| | | | | 3+3 2 | 4+4 2 | 3+4 2 | 5+6 2 | |
| | | | | 4+4 2 | 5+5 2 | 5+6 2 | 3+4 2 | |
| | | | | 6+6 2 | 3+3 2 | 3+4 2 | 5+6 2 | |
| | | | | 5+5 2 | 6+6 2 | 5+6 2 | 3+4 2 | |
| | | | | 3+3 2 | 3+3 2 | 5+6 2 | 5+6 2 | |
| | | | | 4+4 2 | 4+4 2 | 3+4 2 | 3+4 2 | |
| | | | | 6+6 2 | 5+5 2 | 5+6 2 | 3+4 2 | |
| | | | | 5+5 2 | 6+6 2 | 3+4 2 | 5+6 2 | |
| | | | | 6+6 2 | 6+6 2 | 5+6 2 | 3+4 2 | |
| | | | | 5+5 2 | 3+3 2 | 3+4 2 | 3+4 2 | |
| | | | | 4+4 2 | 5+5 2 | 3+4 2 | 5+6 2 | |
| | | | | 3+3 2 | 4+4 2 | 5+6 2 | 3+4 2 | |
| | | | | 3+3 2 | 3+3 2 | 5+6 2 | 5+6 2 | |
| | | | | 4+4 2 | 4+4 2 | 3+4 2 | 5+6 2 | |
| | | | | 5+5 2 | 5+5 2 | 3+4 2 | 3+4 2 | |
| | | | | 6+6 2 | 6+6 2 | 5+6 2 | 5+6 2 | |
| | | | | 5+5 2 | 4+4 2 | 5+6 2 | 5+6 2 | |
| | | | | 6+6 2 | 3+3 2 | 3+4 2 | 3+4 2 | |
| 14.2.51 | 16150 | 132 | Cl3Mo1O1 | | | | | |
| | -A | -B | -C | -D | -E | -GM | -Y | -Z |
| | 2+2 4 | 2+2 4 | 2+2 4 | 3+3 2 | 3+3 2 | 3+4 2 | 3+4 2 | 2+2 4 |
| | 2+2 4 | 2+2 4 | 2+2 4 | 4+4 2 | 4+4 2 | 3+4 2 | 3+4 2 | 2+2 4 |
| | 2+2 4 | 2+2 4 | 2+2 4 | 5+5 2 | 5+5 2 | 5+6 2 | 5+6 2 | 2+2 4 |
| | 2+2 4 | 2+2 4 | 2+2 4 | 6+6 2 | 6+6 2 | 5+6 2 | 5+6 2 | 2+2 4 |
| | 2+2 4 | 2+2 4 | 2+2 4 | 5+5 2 | 5+5 2 | 3+4 2 | 3+4 2 | 2+2 4 |
| | 2+2 4 | 2+2 4 | 2+2 4 | 6+6 2 | 6+6 2 | 5+6 2 | 5+6 2 | 2+2 4 |
| | 2+2 4 | 2+2 4 | 2+2 4 | 3+3 2 | 3+3 2 | 3+4 2 | 5+6 2 | 2+2 4 |
| | 2+2 4 | 2+2 4 | 2+2 4 | 4+4 2 | 4+4 2 | 5+6 2 | 3+4 2 | 2+2 4 |
| | 2+2 4 | 2+2 4 | 2+2 4 | 3+3 2 | 3+3 2 | 5+6 2 | 5+6 2 | 2+2 4 |
| | 2+2 4 | 2+2 4 | 2+2 4 | 4+4 2 | 4+4 2 | 3+4 2 | 3+4 2 | 2+2 4 |
| | 2+2 4 | 2+2 4 | 2+2 4 | 6+6 2 | 6+6 2 | 3+4 2 | 3+4 2 | 2+2 4 |
| | 2+2 4 | 2+2 4 | 2+2 4 | 5+5 2 | 5+5 2 | 5+6 2 | 5+6 2 | 2+2 4 |
| | 2+2 4 | 2+2 4 | 2+2 4 | 5+5 2 | 4+4 2 | 3+4 2 | 5+6 2 | 2+2 4 |
| | 2+2 4 | 2+2 4 | 2+2 4 | 6+6 2 | 3+3 2 | 3+4 2 | 5+6 2 | 2+2 4 |
| | 2+2 4 | 2+2 4 | 2+2 4 | 4+4 2 | 5+5 2 | 5+6 2 | 3+4 2 | 2+2 4 |
| | 2+2 4 | 2+2 4 | 2+2 4 | 3+3 2 | 6+6 2 | 5+6 2 | 3+4 2 | 2+2 4 |
| | 2+2 4 | 2+2 4 | 2+2 4 | 4+4 2 | 4+4 2 | 5+6 2 | 5+6 2 | 2+2 4 |
| | 2+2 4 | 2+2 4 | 2+2 4 | 3+3 2 | 3+3 2 | 3+4 2 | 3+4 2 | 2+2 4 |
| | 2+2 4 | 2+2 4 | 2+2 4 | 5+5 2 | 5+5 2 | 3+4 2 | 5+6 2 | 2+2 4 |
| | 2+2 4 | 2+2 4 | 2+2 4 | 6+6 2 | 6+6 2 | 3+4 2 | 3+4 2 | 2+2 4 |
| | 2+2 4 | 2+2 4 | 2+2 4 | 3+3 2 | 3+3 2 | 5+6 2 | 3+4 2 | 2+2 4 |
| | 2+2 4 | 2+2 4 | 2+2 4 | 4+4 2 | 4+4 2 | 5+6 2 | 3+4 2 | 2+2 4 |
| | 2+2 4 | 2+2 4 | 2+2 4 | 5+5 2 | 5+5 2 | 3+4 2 | 5+6 2 | 2+2 4 |
| | 2+2 4 | 2+2 4 | 2+2 4 | 6+6 2 | 6+6 2 | 3+4 2 | 3+4 2 | 2+2 4 |
| | 2+2 4 | 2+2 4 | 2+2 4 | 4+4 2 | 4+4 2 | 5+6 2 | 5+6 2 | 2+2 4 |
| | 2+2 4 | 2+2 4 | 2+2 4 | 3+3 2 | 3+3 2 | 3+4 2 | 3+4 2 | 2+2 4 |
| | 2+2 4 | 2+2 4 | 2+2 4 | 6+6 2 | 6+6 2 | 5+6 2 | 5+6 2 | 2+2 4 |
| | 2+2 4 | 2+2 4 | 2+2 4 | 5+5 2 | 5+5 2 | 5+6 2 | 3+4 2 | 2+2 4 |
| | 2+2 4 | 2+2 4 | 2+2 4 | 5+5 2 | 3+3 2 | 3+4 2 | 5+6 2 | 2+2 4 |
| | 2+2 4 | 2+2 4 | 2+2 4 | 3+3 2 | 4+4 2 | 5+6 2 | 5+6 2 | 2+2 4 |
| | 2+2 4 | 2+2 4 | 2+2 4 | 6+6 2 | 5+5 2 | 3+4 2 | 5+6 2 | 2+2 4 |
| | 2+2 4 | 2+2 4 | 2+2 4 | 4+4 2 | 6+6 2 | 5+6 2 | 3+4 2 | 2+2 4 |
| | 2+2 4 | 2+2 4 | 2+2 4 | 5+5 2 | 3+3 2 | 3+4 2 | 3+4 2 | 2+2 4 |
| | | | | 6+6 2 | 4+4 2 | 5+6 2 | 3+4 2 | |
| | | | | 4+4 2 | 6+6 2 | 3+4 2 | 5+6 2 | |
| | | | | 3+3 2 | 5+5 2 | 5+6 2 | 5+6 2 | |
| | | | | 5+5 2 | 3+3 2 | 3+4 2 | 5+6 2 | |

Continued on next column

**TABLE XXVI– continued**

| TQC | ICSD | NoE | SF | | | | | |
|---|---|---|---|---|---|---|---|---|
| | | | | 4+4 2 | 4+4 2 | 5+6 2 | 3+4 2 | |
| | | | | 6+6 2 | 5+5 2 | 3+4 2 | 5+6 2 | |
| | | | | 3+3 2 | 6+6 2 | 5+6 2 | 5+6 2 | |
| | | | | 5+5 2 | 4+4 2 | 3+4 2 | 3+4 2 | |
| | | | | 6+6 2 | 3+3 2 | 5+6 2 | 3+4 2 | |
| | | | | 3+3 2 | 5+5 2 | 3+4 2 | 5+6 2 | |
| | | | | 4+4 2 | 6+6 2 | 3+4 2 | 5+6 2 | |
| | | | | 6+6 2 | 3+3 2 | 5+6 2 | 3+4 2 | |
| | | | | 5+5 2 | 4+4 2 | 5+6 2 | 3+4 2 | |
| | | | | 3+3 2 | 6+6 2 | 3+4 2 | 5+6 2 | |
| | | | | 4+4 2 | 5+5 2 | 5+6 2 | 3+4 2 | |
| | | | | 5+5 2 | 5+5 2 | 3+4 2 | 5+6 2 | |
| | | | | 6+6 2 | 4+4 2 | 5+6 2 | 3+4 2 | |
| | | | | 3+3 2 | 6+6 2 | 3+4 2 | 5+6 2 | |
| | | | | 4+4 2 | 3+3 2 | 5+6 2 | 5+6 2 | |
| | | | | 6+6 2 | 5+5 2 | 5+6 2 | 3+4 2 | |
| | | | | 5+5 2 | 6+6 2 | 3+4 2 | 3+4 2 | |
| | | | | 3+3 2 | 3+3 2 | 3+4 2 | 3+4 2 | |
| | | | | 4+4 2 | 4+4 2 | 5+6 2 | 5+6 2 | |
| | | | | 3+3 2 | 3+3 2 | 5+6 2 | 3+4 2 | |
| | | | | 4+4 2 | 6+6 2 | 3+4 2 | 5+6 2 | |
| | | | | 5+5 2 | 5+5 2 | 3+4 2 | 3+4 2 | |
| | | | | 6+6 2 | 4+4 2 | 5+6 2 | 3+4 2 | |
| | | | | 5+5 2 | 6+6 2 | 3+4 2 | 5+6 2 | |
| | | | | 6+6 2 | 5+5 2 | 5+6 2 | 3+4 2 | |
| | | | | 4+4 2 | 4+4 2 | 3+4 2 | 5+6 2 | |
| | | | | 3+3 2 | 3+3 2 | 5+6 2 | 5+6 2 | |
| | | | | 4+4 2 | 5+5 2 | 5+6 2 | 5+6 2 | |
| | | | | 3+3 2 | 6+6 2 | 3+4 2 | 3+4 2 | |
| 14.2.52 | 27020 | 96 | Ce1Se2 | | | | | |
| | -A | -B | -C | -D | -E | -GM | -Y | -Z |
| | 2+2 4 | 2+2 4 | 2+2 4 | 3+3 2 | 3+3 2 | 3+4 2 | 3+4 2 | 2+2 4 |
| | 2+2 4 | 2+2 4 | 2+2 4 | 4+4 2 | 4+4 2 | 5+6 2 | 5+6 2 | 2+2 4 |
| | 2+2 4 | 2+2 4 | 2+2 4 | 5+5 2 | 5+5 2 | 3+4 2 | 3+4 2 | 2+2 4 |
| | 2+2 4 | 2+2 4 | 2+2 4 | 6+6 2 | 6+6 2 | 5+6 2 | 5+6 2 | 2+2 4 |
| | 2+2 4 | 2+2 4 | 2+2 4 | 6+6 2 | 3+3 2 | 3+4 2 | 5+6 2 | 2+2 4 |
| | 2+2 4 | 2+2 4 | 2+2 4 | 3+3 2 | 4+4 2 | 5+6 2 | 3+4 2 | 2+2 4 |
| | 2+2 4 | 2+2 4 | 2+2 4 | 5+5 2 | 5+5 2 | 3+4 2 | 3+4 2 | 2+2 4 |
| | 2+2 4 | 2+2 4 | 2+2 4 | 4+4 2 | 6+6 2 | 5+6 2 | 5+6 2 | 2+2 4 |
| | 2+2 4 | 2+2 4 | 2+2 4 | 5+5 2 | 4+4 2 | 3+4 2 | 5+6 2 | 2+2 4 |
| | 2+2 4 | 2+2 4 | 2+2 4 | 6+6 2 | 3+3 2 | 3+4 2 | 3+4 2 | 2+2 4 |
| | 2+2 4 | 2+2 4 | 2+2 4 | 3+3 2 | 6+6 2 | 5+6 2 | 5+6 2 | 2+2 4 |
| | 2+2 4 | 2+2 4 | 2+2 4 | 4+4 2 | 5+5 2 | 3+4 2 | 3+4 2 | 2+2 4 |
| | 2+2 4 | 2+2 4 | 2+2 4 | 5+5 2 | 4+4 2 | 5+6 2 | 5+6 2 | 2+2 4 |
| | 2+2 4 | 2+2 4 | 2+2 4 | 4+4 2 | 5+5 2 | 5+6 2 | 3+4 2 | 2+2 4 |
| | 2+2 4 | 2+2 4 | 2+2 4 | 6+6 2 | 3+3 2 | 3+4 2 | 3+4 2 | 2+2 4 |
| | 2+2 4 | 2+2 4 | 2+2 4 | 3+3 2 | 6+6 2 | 5+6 2 | 5+6 2 | 2+2 4 |
| | 2+2 4 | 2+2 4 | 2+2 4 | 6+6 2 | 3+3 2 | 3+4 2 | 5+6 2 | 2+2 4 |
| | 2+2 4 | 2+2 4 | 2+2 4 | 5+5 2 | 4+4 2 | 3+4 2 | 5+6 2 | 2+2 4 |
| | 2+2 4 | 2+2 4 | 2+2 4 | 3+3 2 | 5+5 2 | 5+6 2 | 3+4 2 | 2+2 4 |
| | 2+2 4 | 2+2 4 | 2+2 4 | 4+4 2 | 6+6 2 | 3+4 2 | 3+4 2 | 2+2 4 |
| | 2+2 4 | 2+2 4 | 2+2 4 | 5+5 2 | 3+3 2 | 5+6 2 | 5+6 2 | 2+2 4 |
| | 2+2 4 | 2+2 4 | 2+2 4 | 6+6 2 | 4+4 2 | 3+4 2 | 5+6 2 | 2+2 4 |
| | 2+2 4 | 2+2 4 | 2+2 4 | 3+3 2 | 6+6 2 | 5+6 2 | 3+4 2 | 2+2 4 |
| | 2+2 4 | 2+2 4 | 2+2 4 | 4+4 2 | 5+5 2 | 5+6 2 | 3+4 2 | 2+2 4 |
| | | | | 5+5 2 | 3+3 2 | 3+4 2 | 5+6 2 | |
| | | | | 6+6 2 | 4+4 2 | 5+6 2 | 3+4 2 | |
| | | | | 3+3 2 | 6+6 2 | 5+6 2 | 3+4 2 | |
| | | | | 4+4 2 | 5+5 2 | 3+4 2 | 5+6 2 | |
| | | | | 4+4 2 | 6+6 2 | 3+4 2 | 5+6 2 | |
| | | | | 3+3 2 | 5+5 2 | 3+4 2 | 3+4 2 | |

Continued on next column

**TABLE XXVI– continued**

| TQC | ICSD | | NoE | SF | | | | |
|---|---|---|---|---|---|---|---|---|
| | | | | 6+6 2 | 4+4 2 | 5+6 2 | 5+6 2 | |
| | | | | 5+5 2 | 3+3 2 | 3+4 2 | 3+4 2 | |
| | | | | 4+4 2 | 5+5 2 | 5+6 2 | 3+4 2 | |
| | | | | 3+3 2 | 6+6 2 | 5+6 2 | 5+6 2 | |
| | | | | 5+5 2 | 3+3 2 | 3+4 2 | 3+4 2 | |
| | | | | 6+6 2 | 4+4 2 | 5+6 2 | 3+4 2 | |
| | | | | 3+3 2 | 4+4 2 | 3+4 2 | 5+6 2 | |
| | | | | 4+4 2 | 5+5 2 | 3+4 2 | 3+4 2 | |
| | | | | 6+6 2 | 6+6 2 | 3+4 2 | 5+6 2 | |
| | | | | 5+5 2 | 3+3 2 | 3+4 2 | 3+4 2 | |
| | | | | 4+4 2 | 3+3 2 | 5+6 2 | 5+6 2 | |
| | | | | 5+5 2 | 4+4 2 | 3+4 2 | 5+6 2 | |
| | | | | 3+3 2 | 5+5 2 | 3+4 2 | 5+6 2 | |
| | | | | 6+6 2 | 6+6 2 | 5+6 2 | 5+6 2 | |
| | | | | 6+6 2 | 3+3 2 | 5+6 2 | 3+4 2 | |
| | | | | 5+5 2 | 4+4 2 | 5+6 2 | 5+6 2 | |
| | | | | 3+3 2 | 6+6 2 | 5+6 2 | 5+6 2 | |
| | | | | 4+4 2 | 5+5 2 | 5+6 2 | 3+4 2 | |
| 14.2.53 | 419346 | | 96 | Ce1S2 | | | | |
| | -A | -B | -C | -D | -E | -GM | -Y | -Z |
| | 2+2 4 | 2+2 4 | 2+2 4 | 3+3 2 | 5+5 2 | 3+4 2 | 5+6 2 | 2+2 4 |
| | 2+2 4 | 2+2 4 | 2+2 4 | 4+4 2 | 6+6 2 | 5+6 2 | 3+4 2 | 2+2 4 |
| | 2+2 4 | 2+2 4 | 2+2 4 | 5+5 2 | 3+3 2 | 3+4 2 | 3+4 2 | 2+2 4 |
| | 2+2 4 | 2+2 4 | 2+2 4 | 6+6 2 | 4+4 2 | 5+6 2 | 5+6 2 | 2+2 4 |
| | 2+2 4 | 2+2 4 | 2+2 4 | 6+6 2 | 6+6 2 | 3+4 2 | 3+4 2 | 2+2 4 |
| | 2+2 4 | 2+2 4 | 2+2 4 | 3+3 2 | 5+5 2 | 5+6 2 | 5+6 2 | 2+2 4 |
| | 2+2 4 | 2+2 4 | 2+2 4 | 5+5 2 | 3+3 2 | 3+4 2 | 3+4 2 | 2+2 4 |
| | 2+2 4 | 2+2 4 | 2+2 4 | 4+4 2 | 4+4 2 | 5+6 2 | 5+6 2 | 2+2 4 |
| | 2+2 4 | 2+2 4 | 2+2 4 | 5+5 2 | 5+5 2 | 3+4 2 | 3+4 2 | 2+2 4 |
| | 2+2 4 | 2+2 4 | 2+2 4 | 6+6 2 | 6+6 2 | 3+4 2 | 3+4 2 | 2+2 4 |
| | 2+2 4 | 2+2 4 | 2+2 4 | 3+3 2 | 3+3 2 | 5+6 2 | 5+6 2 | 2+2 4 |
| | 2+2 4 | 2+2 4 | 2+2 4 | 4+4 2 | 4+4 2 | 3+4 2 | 5+6 2 | 2+2 4 |
| | 2+2 4 | 2+2 4 | 2+2 4 | 5+5 2 | 5+5 2 | 5+6 2 | 3+4 2 | 2+2 4 |
| | 2+2 4 | 2+2 4 | 2+2 4 | 4+4 2 | 4+4 2 | 5+6 2 | 5+6 2 | 2+2 4 |
| | 2+2 4 | 2+2 4 | 2+2 4 | 6+6 2 | 6+6 2 | 3+4 2 | 5+6 2 | 2+2 4 |
| | 2+2 4 | 2+2 4 | 2+2 4 | 3+3 2 | 3+3 2 | 5+6 2 | 3+4 2 | 2+2 4 |
| | 2+2 4 | 2+2 4 | 2+2 4 | 6+6 2 | 6+6 2 | 3+4 2 | 3+4 2 | 2+2 4 |
| | 2+2 4 | 2+2 4 | 2+2 4 | 5+5 2 | 5+5 2 | 3+4 2 | 3+4 2 | 2+2 4 |
| | 2+2 4 | 2+2 4 | 2+2 4 | 3+3 2 | 4+4 2 | 3+4 2 | 5+6 2 | 2+2 4 |
| | 2+2 4 | 2+2 4 | 2+2 4 | 4+4 2 | 3+3 2 | 5+6 2 | 3+4 2 | 2+2 4 |
| | 2+2 4 | 2+2 4 | 2+2 4 | 5+5 2 | 6+6 2 | 5+6 2 | 5+6 2 | 2+2 4 |
| | 2+2 4 | 2+2 4 | 2+2 4 | 6+6 2 | 5+5 2 | 3+4 2 | 3+4 2 | 2+2 4 |
| | 2+2 4 | 2+2 4 | 2+2 4 | 3+3 2 | 3+3 2 | 5+6 2 | 5+6 2 | 2+2 4 |
| | 2+2 4 | 2+2 4 | 2+2 4 | 4+4 2 | 4+4 2 | 5+6 2 | 5+6 2 | 2+2 4 |
| | | | | 5+5 2 | 5+5 2 | 3+4 2 | 3+4 2 | |
| | | | | 6+6 2 | 6+6 2 | 5+6 2 | 5+6 2 | |
| | | | | 4+4 2 | 3+3 2 | 5+6 2 | 5+6 2 | |
| | | | | 3+3 2 | 4+4 2 | 3+4 2 | 3+4 2 | |
| | | | | 3+3 2 | 3+3 2 | 3+4 2 | 3+4 2 | |
| | | | | 4+4 2 | 4+4 2 | 3+4 2 | 5+6 2 | |
| | | | | 5+5 2 | 4+4 2 | 5+6 2 | 5+6 2 | |
| | | | | 6+6 2 | 5+5 2 | 5+6 2 | 5+6 2 | |
| | | | | 4+4 2 | 3+3 2 | 3+4 2 | 3+4 2 | |
| | | | | 3+3 2 | 6+6 2 | 5+6 2 | 5+6 2 | |
| | | | | 6+6 2 | 6+6 2 | 5+6 2 | 3+4 2 | |
| | | | | 5+5 2 | 5+5 2 | 3+4 2 | 5+6 2 | |
| | | | | 3+3 2 | 4+4 2 | 3+4 2 | 5+6 2 | |
| | | | | 4+4 2 | 3+3 2 | 3+4 2 | 3+4 2 | |
| | | | | 6+6 2 | 5+5 2 | 3+4 2 | 5+6 2 | |
| | | | | 5+5 2 | 6+6 2 | 3+4 2 | 3+4 2 | |
| | | | | 4+4 2 | 6+6 2 | 5+6 2 | 3+4 2 | |

Continued on next column

**TABLE XXVI– continued**

| TQC | ICSD | | NoE | SF | | | | |
|---|---|---|---|---|---|---|---|---|
| | | | | 3+3 2 | 5+5 2 | 3+4 2 | 3+4 2 | |
| | | | | 5+5 2 | 3+3 2 | 3+4 2 | 3+4 2 | |
| | | | | 6+6 2 | 4+4 2 | 5+6 2 | 3+4 2 | |
| | | | | 6+6 2 | 6+6 2 | 5+6 2 | 5+6 2 | |
| | | | | 5+5 2 | 5+5 2 | 5+6 2 | 3+4 2 | |
| | | | | 3+3 2 | 3+3 2 | 5+6 2 | 3+4 2 | |
| | | | | 4+4 2 | 4+4 2 | 5+6 2 | 5+6 2 | |
| 14.2.54 | 238150 | | 112 | Ca2Fe1O6Os1 | | | | |
| | -A | -B | -C | -D | -E | -GM | -Y | -Z |
| | 2+2 4 | 2+2 4 | 2+2 4 | 3+3 2 | 5+5 2 | 3+4 2 | 5+6 2 | 2+2 4 |
| | 2+2 4 | 2+2 4 | 2+2 4 | 4+4 2 | 6+6 2 | 3+4 2 | 5+6 2 | 2+2 4 |
| | 2+2 4 | 2+2 4 | 2+2 4 | 6+6 2 | 5+5 2 | 3+4 2 | 3+4 2 | 2+2 4 |
| | 2+2 4 | 2+2 4 | 2+2 4 | 5+5 2 | 6+6 2 | 3+4 2 | 3+4 2 | 2+2 4 |
| | 2+2 4 | 2+2 4 | 2+2 4 | 3+3 2 | 5+5 2 | 3+4 2 | 5+6 2 | 2+2 4 |
| | 2+2 4 | 2+2 4 | 2+2 4 | 4+4 2 | 6+6 2 | 5+6 2 | 5+6 2 | 2+2 4 |
| | 2+2 4 | 2+2 4 | 2+2 4 | 3+3 2 | 3+3 2 | 3+4 2 | 5+6 2 | 2+2 4 |
| | 2+2 4 | 2+2 4 | 2+2 4 | 4+4 2 | 4+4 2 | 5+6 2 | 5+6 2 | 2+2 4 |
| | 2+2 4 | 2+2 4 | 2+2 4 | 5+5 2 | 4+4 2 | 5+6 2 | 3+4 2 | 2+2 4 |
| | 2+2 4 | 2+2 4 | 2+2 4 | 6+6 2 | 3+3 2 | 5+6 2 | 3+4 2 | 2+2 4 |
| | 2+2 4 | 2+2 4 | 2+2 4 | 6+6 2 | 3+3 2 | 5+6 2 | 3+4 2 | 2+2 4 |
| | 2+2 4 | 2+2 4 | 2+2 4 | 5+5 2 | 4+4 2 | 5+6 2 | 3+4 2 | 2+2 4 |
| | 2+2 4 | 2+2 4 | 2+2 4 | 3+3 2 | 6+6 2 | 3+4 2 | 5+6 2 | 2+2 4 |
| | 2+2 4 | 2+2 4 | 2+2 4 | 4+4 2 | 5+5 2 | 3+4 2 | 5+6 2 | 2+2 4 |
| | 2+2 4 | 2+2 4 | 2+2 4 | 3+3 2 | 5+5 2 | 3+4 2 | 5+6 2 | 2+2 4 |
| | 2+2 4 | 2+2 4 | 2+2 4 | 4+4 2 | 6+6 2 | 3+4 2 | 5+6 2 | 2+2 4 |
| | 2+2 4 | 2+2 4 | 2+2 4 | 4+4 2 | 6+6 2 | 3+4 2 | 5+6 2 | 2+2 4 |
| | 2+2 4 | 2+2 4 | 2+2 4 | 3+3 2 | 5+5 2 | 3+4 2 | 5+6 2 | 2+2 4 |
| | 2+2 4 | 2+2 4 | 2+2 4 | 6+6 2 | 5+5 2 | 3+4 2 | 3+4 2 | 2+2 4 |
| | 2+2 4 | 2+2 4 | 2+2 4 | 5+5 2 | 6+6 2 | 3+4 2 | 3+4 2 | 2+2 4 |
| | 2+2 4 | 2+2 4 | 2+2 4 | 4+4 2 | 6+6 2 | 3+4 2 | 5+6 2 | 2+2 4 |
| | 2+2 4 | 2+2 4 | 2+2 4 | 3+3 2 | 5+5 2 | 3+4 2 | 5+6 2 | 2+2 4 |
| | 2+2 4 | 2+2 4 | 2+2 4 | 3+3 2 | 5+5 2 | 3+4 2 | 5+6 2 | 2+2 4 |
| | 2+2 4 | 2+2 4 | 2+2 4 | 4+4 2 | 6+6 2 | 3+4 2 | 5+6 2 | 2+2 4 |
| | 2+2 4 | 2+2 4 | 2+2 4 | 4+4 2 | 3+3 2 | 5+6 2 | 5+6 2 | 2+2 4 |
| | 2+2 4 | 2+2 4 | 2+2 4 | 3+3 2 | 4+4 2 | 5+6 2 | 5+6 2 | 2+2 4 |
| | 2+2 4 | 2+2 4 | 2+2 4 | 5+5 2 | 3+3 2 | 5+6 2 | 3+4 2 | 2+2 4 |
| | 2+2 4 | 2+2 4 | 2+2 4 | 6+6 2 | 4+4 2 | 5+6 2 | 3+4 2 | 2+2 4 |
| | | | | 6+6 2 | 3+3 2 | 5+6 2 | 3+4 2 | |
| | | | | 5+5 2 | 4+4 2 | 5+6 2 | 3+4 2 | |
| | | | | 5+5 2 | 4+4 2 | 5+6 2 | 3+4 2 | |
| | | | | 6+6 2 | 3+3 2 | 5+6 2 | 3+4 2 | |
| | | | | 5+5 2 | 4+4 2 | 3+4 2 | 3+4 2 | |
| | | | | 6+6 2 | 5+5 2 | 3+4 2 | 3+4 2 | |
| | | | | 5+5 2 | 6+6 2 | 5+6 2 | 3+4 2 | |
| | | | | 6+6 2 | 3+3 2 | 5+6 2 | 3+4 2 | |
| | | | | 5+5 2 | 6+6 2 | 5+6 2 | 5+6 2 | |
| | | | | 6+6 2 | 5+5 2 | 5+6 2 | 3+4 2 | |
| | | | | 6+6 2 | 3+3 2 | 5+6 2 | 3+4 2 | |
| | | | | 3+3 2 | 4+4 2 | 3+4 2 | 5+6 2 | |
| | | | | 4+4 2 | 3+3 2 | 3+4 2 | 3+4 2 | |
| | | | | 5+5 2 | 4+4 2 | 5+6 2 | 5+6 2 | |
| | | | | 3+3 2 | 4+4 2 | 5+6 2 | 3+4 2 | |
| | | | | 4+4 2 | 3+3 2 | 3+4 2 | 5+6 2 | |
| | | | | 5+5 2 | 3+3 2 | 3+4 2 | 3+4 2 | |
| | | | | 6+6 2 | 4+4 2 | 5+6 2 | 3+4 2 | |
| | | | | 4+4 2 | 6+6 2 | 5+6 2 | 5+6 2 | |
| | | | | 3+3 2 | 5+5 2 | 5+6 2 | 5+6 2 | |
| | | | | 6+6 2 | 6+6 2 | 3+4 2 | 3+4 2 | |
| | | | | 5+5 2 | 5+5 2 | 3+4 2 | 5+6 2 | |
| | | | | 4+4 2 | 6+6 2 | 3+4 2 | 3+4 2 | |
| | | | | 3+3 2 | 5+5 2 | 3+4 2 | 5+6 2 | |

Continued on next column

**TABLE XXVI– continued**

| TQC | ICSD | NoE | SF | | | | | |
|---|---|---|---|---|---|---|---|---|
| | | | | 5+5 2 | 5+5 2 | 3+4 2 | 3+4 2 | |
| | | | | 6+6 2 | 6+6 2 | 3+4 2 | 3+4 2 | |
| | | | | 5+5 2 | 5+5 2 | 3+4 2 | 3+4 2 | |
| | | | | 6+6 2 | 6+6 2 | 3+4 2 | 5+6 2 | |
| 14.2.55 | 238148 | 112 | Ca2Fe1O6Os1 | | | | | |
| | -A | -B | -C | -D | -E | -GM | -Y | -Z |
| | 2+2 4 | 2+2 4 | 2+2 4 | 3+3 2 | 5+5 2 | 3+4 2 | 5+6 2 | 2+2 4 |
| | 2+2 4 | 2+2 4 | 2+2 4 | 4+4 2 | 6+6 2 | 3+4 2 | 5+6 2 | 2+2 4 |
| | 2+2 4 | 2+2 4 | 2+2 4 | 6+6 2 | 5+5 2 | 3+4 2 | 3+4 2 | 2+2 4 |
| | 2+2 4 | 2+2 4 | 2+2 4 | 5+5 2 | 6+6 2 | 3+4 2 | 3+4 2 | 2+2 4 |
| | 2+2 4 | 2+2 4 | 2+2 4 | 3+3 2 | 5+5 2 | 3+4 2 | 5+6 2 | 2+2 4 |
| | 2+2 4 | 2+2 4 | 2+2 4 | 4+4 2 | 6+6 2 | 3+4 2 | 5+6 2 | 2+2 4 |
| | 2+2 4 | 2+2 4 | 2+2 4 | 3+3 2 | 3+3 2 | 5+6 2 | 5+6 2 | 2+2 4 |
| | 2+2 4 | 2+2 4 | 2+2 4 | 4+4 2 | 4+4 2 | 5+6 2 | 5+6 2 | 2+2 4 |
| | 2+2 4 | 2+2 4 | 2+2 4 | 6+6 2 | 4+4 2 | 5+6 2 | 3+4 2 | 2+2 4 |
| | 2+2 4 | 2+2 4 | 2+2 4 | 5+5 2 | 3+3 2 | 5+6 2 | 3+4 2 | 2+2 4 |
| | 2+2 4 | 2+2 4 | 2+2 4 | 5+5 2 | 3+3 2 | 5+6 2 | 3+4 2 | 2+2 4 |
| | 2+2 4 | 2+2 4 | 2+2 4 | 6+6 2 | 4+4 2 | 5+6 2 | 3+4 2 | 2+2 4 |
| | 2+2 4 | 2+2 4 | 2+2 4 | 4+4 2 | 5+5 2 | 3+4 2 | 5+6 2 | 2+2 4 |
| | 2+2 4 | 2+2 4 | 2+2 4 | 3+3 2 | 6+6 2 | 3+4 2 | 5+6 2 | 2+2 4 |
| | 2+2 4 | 2+2 4 | 2+2 4 | 3+3 2 | 6+6 2 | 3+4 2 | 5+6 2 | 2+2 4 |
| | 2+2 4 | 2+2 4 | 2+2 4 | 4+4 2 | 5+5 2 | 3+4 2 | 5+6 2 | 2+2 4 |
| | 2+2 4 | 2+2 4 | 2+2 4 | 4+4 2 | 5+5 2 | 3+4 2 | 5+6 2 | 2+2 4 |
| | 2+2 4 | 2+2 4 | 2+2 4 | 3+3 2 | 6+6 2 | 3+4 2 | 5+6 2 | 2+2 4 |
| | 2+2 4 | 2+2 4 | 2+2 4 | 6+6 2 | 5+5 2 | 3+4 2 | 3+4 2 | 2+2 4 |
| | 2+2 4 | 2+2 4 | 2+2 4 | 5+5 2 | 6+6 2 | 3+4 2 | 3+4 2 | 2+2 4 |
| | 2+2 4 | 2+2 4 | 2+2 4 | 4+4 2 | 6+6 2 | 3+4 2 | 5+6 2 | 2+2 4 |
| | 2+2 4 | 2+2 4 | 2+2 4 | 3+3 2 | 5+5 2 | 3+4 2 | 5+6 2 | 2+2 4 |
| | 2+2 4 | 2+2 4 | 2+2 4 | 3+3 2 | 5+5 2 | 3+4 2 | 5+6 2 | 2+2 4 |
| | 2+2 4 | 2+2 4 | 2+2 4 | 4+4 2 | 6+6 2 | 3+4 2 | 5+6 2 | 2+2 4 |
| | 2+2 4 | 2+2 4 | 2+2 4 | 4+4 2 | 3+3 2 | 5+6 2 | 5+6 2 | 2+2 4 |
| | 2+2 4 | 2+2 4 | 2+2 4 | 3+3 2 | 4+4 2 | 5+6 2 | 5+6 2 | 2+2 4 |
| | 2+2 4 | 2+2 4 | 2+2 4 | 5+5 2 | 4+4 2 | 5+6 2 | 3+4 2 | 2+2 4 |
| | 2+2 4 | 2+2 4 | 2+2 4 | 6+6 2 | 3+3 2 | 5+6 2 | 3+4 2 | 2+2 4 |
| | | | | 6+6 2 | 3+3 2 | 5+6 2 | 3+4 2 | |
| | | | | 5+5 2 | 4+4 2 | 5+6 2 | 3+4 2 | |
| | | | | 5+5 2 | 4+4 2 | 5+6 2 | 3+4 2 | |
| | | | | 6+6 2 | 3+3 2 | 5+6 2 | 3+4 2 | |
| | | | | 6+6 2 | 4+4 2 | 3+4 2 | 3+4 2 | |
| | | | | 5+5 2 | 3+3 2 | 3+4 2 | 3+4 2 | |
| | | | | 5+5 2 | 5+5 2 | 5+6 2 | 3+4 2 | |
| | | | | 6+6 2 | 6+6 2 | 5+6 2 | 3+4 2 | |
| | | | | 5+5 2 | 6+6 2 | 5+6 2 | 5+6 2 | |
| | | | | 6+6 2 | 5+5 2 | 5+6 2 | 3+4 2 | |
| | | | | 4+4 2 | 3+3 2 | 5+6 2 | 3+4 2 | |
| | | | | 3+3 2 | 4+4 2 | 3+4 2 | 3+4 2 | |
| | | | | 6+6 2 | 3+3 2 | 3+4 2 | 5+6 2 | |
| | | | | 3+3 2 | 4+4 2 | 5+6 2 | 5+6 2 | |
| | | | | 5+5 2 | 4+4 2 | 5+6 2 | 3+4 2 | |
| | | | | 4+4 2 | 3+3 2 | 3+4 2 | 5+6 2 | |
| | | | | 6+6 2 | 3+3 2 | 3+4 2 | 3+4 2 | |
| | | | | 5+5 2 | 4+4 2 | 5+6 2 | 3+4 2 | |
| | | | | 4+4 2 | 6+6 2 | 5+6 2 | 5+6 2 | |
| | | | | 3+3 2 | 5+5 2 | 5+6 2 | 5+6 2 | |
| | | | | 6+6 2 | 6+6 2 | 3+4 2 | 3+4 2 | |
| | | | | 5+5 2 | 5+5 2 | 3+4 2 | 5+6 2 | |
| | | | | 3+3 2 | 6+6 2 | 3+4 2 | 3+4 2 | |
| | | | | 4+4 2 | 5+5 2 | 3+4 2 | 3+4 2 | |
| | | | | 5+5 2 | 5+5 2 | 3+4 2 | 5+6 2 | |
| | | | | 6+6 2 | 6+6 2 | 3+4 2 | 3+4 2 | |
| | | | | 5+5 2 | 5+5 2 | 3+4 2 | 3+4 2 | |

Continued on next column

**TABLE XXVI– continued**

| TQC | ICSD | NoE | SF | | | | | |
|---|---|---|---|---|---|---|---|---|
| | | | | 6+6 2 | 6+6 2 | 3+4 2 | 5+6 2 | |
| 14.2.56 | 69790 | 64 | Ca1Pd1Si1 | | | | | |
| | -A | -B | -C | -D | -E | -GM | -Y | -Z |
| | 2+2 4 | 2+2 4 | 2+2 4 | 6+6 2 | 6+6 2 | 3+4 2 | 3+4 2 | 2+2 4 |
| | 2+2 4 | 2+2 4 | 2+2 4 | 5+5 2 | 5+5 2 | 3+4 2 | 3+4 2 | 2+2 4 |
| | 2+2 4 | 2+2 4 | 2+2 4 | 3+3 2 | 3+3 2 | 5+6 2 | 5+6 2 | 2+2 4 |
| | 2+2 4 | 2+2 4 | 2+2 4 | 4+4 2 | 4+4 2 | 5+6 2 | 5+6 2 | 2+2 4 |
| | 2+2 4 | 2+2 4 | 2+2 4 | 6+6 2 | 5+5 2 | 3+4 2 | 5+6 2 | 2+2 4 |
| | 2+2 4 | 2+2 4 | 2+2 4 | 5+5 2 | 6+6 2 | 5+6 2 | 3+4 2 | 2+2 4 |
| | 2+2 4 | 2+2 4 | 2+2 4 | 4+4 2 | 3+3 2 | 3+4 2 | 3+4 2 | 2+2 4 |
| | 2+2 4 | 2+2 4 | 2+2 4 | 3+3 2 | 4+4 2 | 5+6 2 | 3+4 2 | 2+2 4 |
| | 2+2 4 | 2+2 4 | 2+2 4 | 3+3 2 | 6+6 2 | 5+6 2 | 5+6 2 | 2+2 4 |
| | 2+2 4 | 2+2 4 | 2+2 4 | 4+4 2 | 5+5 2 | 3+4 2 | 3+4 2 | 2+2 4 |
| | 2+2 4 | 2+2 4 | 2+2 4 | 5+5 2 | 3+3 2 | 5+6 2 | 5+6 2 | 2+2 4 |
| | 2+2 4 | 2+2 4 | 2+2 4 | 6+6 2 | 4+4 2 | 3+4 2 | 3+4 2 | 2+2 4 |
| | 2+2 4 | 2+2 4 | 2+2 4 | 3+3 2 | 6+6 2 | 3+4 2 | 5+6 2 | 2+2 4 |
| | 2+2 4 | 2+2 4 | 2+2 4 | 4+4 2 | 5+5 2 | 5+6 2 | 5+6 2 | 2+2 4 |
| | 2+2 4 | 2+2 4 | 2+2 4 | 5+5 2 | 4+4 2 | 3+4 2 | 3+4 2 | 2+2 4 |
| | 2+2 4 | 2+2 4 | 2+2 4 | 6+6 2 | 3+3 2 | 5+6 2 | 5+6 2 | 2+2 4 |
| | | | | 3+3 2 | 3+3 2 | 3+4 2 | 5+6 2 | |
| | | | | 4+4 2 | 4+4 2 | 3+4 2 | 3+4 2 | |
| | | | | 6+6 2 | 5+5 2 | 5+6 2 | 5+6 2 | |
| | | | | 5+5 2 | 6+6 2 | 5+6 2 | 3+4 2 | |
| | | | | 6+6 2 | 6+6 2 | 3+4 2 | 3+4 2 | |
| | | | | 5+5 2 | 3+3 2 | 5+6 2 | 5+6 2 | |
| | | | | 4+4 2 | 5+5 2 | 3+4 2 | 3+4 2 | |
| | | | | 3+3 2 | 4+4 2 | 5+6 2 | 5+6 2 | |
| | | | | 6+6 2 | 5+5 2 | 3+4 2 | 3+4 2 | |
| | | | | 5+5 2 | 6+6 2 | 3+4 2 | 5+6 2 | |
| | | | | 4+4 2 | 4+4 2 | 5+6 2 | 5+6 2 | |
| | | | | 3+3 2 | 3+3 2 | 3+4 2 | 3+4 2 | |
| | | | | 5+5 2 | 4+4 2 | 3+4 2 | 3+4 2 | |
| | | | | 6+6 2 | 3+3 2 | 5+6 2 | 5+6 2 | |
| | | | | 6+6 2 | 5+5 2 | 3+4 2 | 5+6 2 | |
| | | | | 5+5 2 | 6+6 2 | 5+6 2 | 3+4 2 | |
| 14.2.57 | 86049 | 144 | C7Y4 | | | | | |
| | -A | -B | -C | -D | -E | -GM | -Y | -Z |
| | 2+2 4 | 2+2 4 | 2+2 4 | 5+5 2 | 3+3 2 | 3+4 2 | 3+4 2 | 2+2 4 |
| | 2+2 4 | 2+2 4 | 2+2 4 | 6+6 2 | 4+4 2 | 3+4 2 | 3+4 2 | 2+2 4 |
| | 2+2 4 | 2+2 4 | 2+2 4 | 3+3 2 | 5+5 2 | 5+6 2 | 5+6 2 | 2+2 4 |
| | 2+2 4 | 2+2 4 | 2+2 4 | 4+4 2 | 6+6 2 | 5+6 2 | 5+6 2 | 2+2 4 |
| | 2+2 4 | 2+2 4 | 2+2 4 | 3+3 2 | 3+3 2 | 3+4 2 | 3+4 2 | 2+2 4 |
| | 2+2 4 | 2+2 4 | 2+2 4 | 4+4 2 | 4+4 2 | 5+6 2 | 5+6 2 | 2+2 4 |
| | 2+2 4 | 2+2 4 | 2+2 4 | 5+5 2 | 5+5 2 | 3+4 2 | 3+4 2 | 2+2 4 |
| | 2+2 4 | 2+2 4 | 2+2 4 | 6+6 2 | 6+6 2 | 5+6 2 | 5+6 2 | 2+2 4 |
| | 2+2 4 | 2+2 4 | 2+2 4 | 5+5 2 | 3+3 2 | 3+4 2 | 5+6 2 | 2+2 4 |
| | 2+2 4 | 2+2 4 | 2+2 4 | 6+6 2 | 4+4 2 | 3+4 2 | 3+4 2 | 2+2 4 |
| | 2+2 4 | 2+2 4 | 2+2 4 | 3+3 2 | 5+5 2 | 5+6 2 | 3+4 2 | 2+2 4 |
| | 2+2 4 | 2+2 4 | 2+2 4 | 4+4 2 | 6+6 2 | 5+6 2 | 5+6 2 | 2+2 4 |
| | 2+2 4 | 2+2 4 | 2+2 4 | 3+3 2 | 5+5 2 | 3+4 2 | 5+6 2 | 2+2 4 |
| | 2+2 4 | 2+2 4 | 2+2 4 | 4+4 2 | 6+6 2 | 5+6 2 | 3+4 2 | 2+2 4 |
| | 2+2 4 | 2+2 4 | 2+2 4 | 6+6 2 | 3+3 2 | 3+4 2 | 5+6 2 | 2+2 4 |
| | 2+2 4 | 2+2 4 | 2+2 4 | 5+5 2 | 4+4 2 | 5+6 2 | 3+4 2 | 2+2 4 |
| | 2+2 4 | 2+2 4 | 2+2 4 | 6+6 2 | 4+4 2 | 3+4 2 | 5+6 2 | 2+2 4 |
| | 2+2 4 | 2+2 4 | 2+2 4 | 5+5 2 | 3+3 2 | 3+4 2 | 3+4 2 | 2+2 4 |
| | 2+2 4 | 2+2 4 | 2+2 4 | 3+3 2 | 6+6 2 | 5+6 2 | 5+6 2 | 2+2 4 |
| | 2+2 4 | 2+2 4 | 2+2 4 | 4+4 2 | 5+5 2 | 5+6 2 | 3+4 2 | 2+2 4 |
| | 2+2 4 | 2+2 4 | 2+2 4 | 5+5 2 | 6+6 2 | 3+4 2 | 5+6 2 | 2+2 4 |
| | 2+2 4 | 2+2 4 | 2+2 4 | 6+6 2 | 3+3 2 | 3+4 2 | 3+4 2 | 2+2 4 |
| | 2+2 4 | 2+2 4 | 2+2 4 | 3+3 2 | 5+5 2 | 5+6 2 | 5+6 2 | 2+2 4 |
| | 2+2 4 | 2+2 4 | 2+2 4 | 4+4 2 | 4+4 2 | 5+6 2 | 3+4 2 | 2+2 4 |

Continued on next column

**TABLE XXVI– continued**

| TQC | ICSD | NoE | SF | | | | |
|---|---|---|---|---|---|---|---|
| | 2+2 4 | 2+2 4 | 2+2 4 | 5+5 2 | 4+4 2 | 3+4 2 | 3+4 2 | 2+2 4 |
| | 2+2 4 | 2+2 4 | 2+2 4 | 6+6 2 | 5+5 2 | 5+6 2 | 5+6 2 | 2+2 4 |
| | 2+2 4 | 2+2 4 | 2+2 4 | 4+4 2 | 3+3 2 | 3+4 2 | 3+4 2 | 2+2 4 |
| | 2+2 4 | 2+2 4 | 2+2 4 | 3+3 2 | 6+6 2 | 5+6 2 | 5+6 2 | 2+2 4 |
| | 2+2 4 | 2+2 4 | 2+2 4 | 3+3 2 | 3+3 2 | 5+6 2 | 3+4 2 | 2+2 4 |
| | 2+2 4 | 2+2 4 | 2+2 4 | 4+4 2 | 6+6 2 | 3+4 2 | 5+6 2 | 2+2 4 |
| | 2+2 4 | 2+2 4 | 2+2 4 | 6+6 2 | 4+4 2 | 3+4 2 | 5+6 2 | 2+2 4 |
| | 2+2 4 | 2+2 4 | 2+2 4 | 5+5 2 | 5+5 2 | 5+6 2 | 3+4 2 | 2+2 4 |
| | 2+2 4 | 2+2 4 | 2+2 4 | 6+6 2 | 6+6 2 | 3+4 2 | 5+6 2 | 2+2 4 |
| | 2+2 4 | 2+2 4 | 2+2 4 | 5+5 2 | 3+3 2 | 5+6 2 | 3+4 2 | 2+2 4 |
| | 2+2 4 | 2+2 4 | 2+2 4 | 3+3 2 | 5+5 2 | 3+4 2 | 3+4 2 | 2+2 4 |
| | 2+2 4 | 2+2 4 | 2+2 4 | 4+4 2 | 4+4 2 | 5+6 2 | 5+6 2 | 2+2 4 |
| | | | | 5+5 2 | 4+4 2 | 3+4 2 | 5+6 2 | |
| | | | | 6+6 2 | 5+5 2 | 5+6 2 | 3+4 2 | |
| | | | | 4+4 2 | 3+3 2 | 5+6 2 | 3+4 2 | |
| | | | | 3+3 2 | 6+6 2 | 3+4 2 | 5+6 2 | |
| | | | | 3+3 2 | 5+5 2 | 3+4 2 | 5+6 2 | |
| | | | | 4+4 2 | 6+6 2 | 3+4 2 | 5+6 2 | |
| | | | | 5+5 2 | 3+3 2 | 5+6 2 | 3+4 2 | |
| | | | | 6+6 2 | 4+4 2 | 3+4 2 | 5+6 2 | |
| | | | | 4+4 2 | 6+6 2 | 3+4 2 | 3+4 2 | |
| | | | | 3+3 2 | 5+5 2 | 3+4 2 | 5+6 2 | |
| | | | | 5+5 2 | 4+4 2 | 5+6 2 | 5+6 2 | |
| | | | | 6+6 2 | 3+3 2 | 5+6 2 | 3+4 2 | |
| | | | | 3+3 2 | 5+5 2 | 5+6 2 | 5+6 2 | |
| | | | | 4+4 2 | 6+6 2 | 3+4 2 | 3+4 2 | |
| | | | | 6+6 2 | 4+4 2 | 3+4 2 | 5+6 2 | |
| | | | | 5+5 2 | 3+3 2 | 5+6 2 | 3+4 2 | |
| | | | | 3+3 2 | 6+6 2 | 3+4 2 | 3+4 2 | |
| | | | | 4+4 2 | 5+5 2 | 5+6 2 | 5+6 2 | |
| | | | | 5+5 2 | 3+3 2 | 3+4 2 | 3+4 2 | |
| | | | | 6+6 2 | 4+4 2 | 5+6 2 | 5+6 2 | |
| | | | | 4+4 2 | 6+6 2 | 5+6 2 | 3+4 2 | |
| | | | | 3+3 2 | 5+5 2 | 3+4 2 | 5+6 2 | |
| | | | | 6+6 2 | 5+5 2 | 5+6 2 | 3+4 2 | |
| | | | | 5+5 2 | 6+6 2 | 5+6 2 | 3+4 2 | |
| | | | | 3+3 2 | 4+4 2 | 3+4 2 | 3+4 2 | |
| | | | | 4+4 2 | 3+3 2 | 5+6 2 | 5+6 2 | |
| | | | | 5+5 2 | 3+3 2 | 3+4 2 | 5+6 2 | |
| | | | | 6+6 2 | 4+4 2 | 3+4 2 | 5+6 2 | |
| | | | | 4+4 2 | 6+6 2 | 3+4 2 | 3+4 2 | |
| | | | | 3+3 2 | 5+5 2 | 5+6 2 | 3+4 2 | |
| | | | | 5+5 2 | 3+3 2 | 5+6 2 | 3+4 2 | |
| | | | | 6+6 2 | 4+4 2 | 5+6 2 | 3+4 2 | |
| | | | | 5+5 2 | 4+4 2 | 5+6 2 | 5+6 2 | |
| | | | | 6+6 2 | 3+3 2 | 5+6 2 | 3+4 2 | |
| | | | | 5+5 2 | 3+3 2 | 5+6 2 | 3+4 2 | |
| | | | | 6+6 2 | 4+4 2 | 5+6 2 | 5+6 2 | |
| 14.2.58 | 54136 | 120 | Br3Cr1In1 | | | | | |
| | -A | -B | -C | -D | -E | -GM | -Y | -Z |
| | 2+2 4 | 2+2 4 | 2+2 4 | 5+5 2 | 5+5 2 | 3+4 2 | 3+4 2 | 2+2 4 |
| | 2+2 4 | 2+2 4 | 2+2 4 | 6+6 2 | 6+6 2 | 3+4 2 | 3+4 2 | 2+2 4 |
| | 2+2 4 | 2+2 4 | 2+2 4 | 3+3 2 | 3+3 2 | 5+6 2 | 5+6 2 | 2+2 4 |
| | 2+2 4 | 2+2 4 | 2+2 4 | 4+4 2 | 4+4 2 | 5+6 2 | 5+6 2 | 2+2 4 |
| | 2+2 4 | 2+2 4 | 2+2 4 | 3+3 2 | 5+5 2 | 3+4 2 | 5+6 2 | 2+2 4 |
| | 2+2 4 | 2+2 4 | 2+2 4 | 4+4 2 | 6+6 2 | 3+4 2 | 5+6 2 | 2+2 4 |
| | 2+2 4 | 2+2 4 | 2+2 4 | 5+5 2 | 3+3 2 | 5+6 2 | 3+4 2 | 2+2 4 |
| | 2+2 4 | 2+2 4 | 2+2 4 | 6+6 2 | 4+4 2 | 5+6 2 | 3+4 2 | 2+2 4 |
| | 2+2 4 | 2+2 4 | 2+2 4 | 5+5 2 | 3+3 2 | 3+4 2 | 3+4 2 | 2+2 4 |
| | 2+2 4 | 2+2 4 | 2+2 4 | 6+6 2 | 4+4 2 | 3+4 2 | 5+6 2 | 2+2 4 |
| | 2+2 4 | 2+2 4 | 2+2 4 | 3+3 2 | 5+5 2 | 5+6 2 | 5+6 2 | 2+2 4 |

Continued on next column

**TABLE XXVI– continued**

| TQC | ICSD | NoE | SF | | | | |
|---|---|---|---|---|---|---|---|
| | 2+2 4 | 2+2 4 | 2+2 4 | 4+4 2 | 6+6 2 | 5+6 2 | 3+4 2 | 2+2 4 |
| | 2+2 4 | 2+2 4 | 2+2 4 | 6+6 2 | 3+3 2 | 5+6 2 | 3+4 2 | 2+2 4 |
| | 2+2 4 | 2+2 4 | 2+2 4 | 5+5 2 | 4+4 2 | 3+4 2 | 3+4 2 | 2+2 4 |
| | 2+2 4 | 2+2 4 | 2+2 4 | 3+3 2 | 6+6 2 | 3+4 2 | 5+6 2 | 2+2 4 |
| | 2+2 4 | 2+2 4 | 2+2 4 | 4+4 2 | 5+5 2 | 5+6 2 | 5+6 2 | 2+2 4 |
| | 2+2 4 | 2+2 4 | 2+2 4 | 5+5 2 | 4+4 2 | 3+4 2 | 5+6 2 | 2+2 4 |
| | 2+2 4 | 2+2 4 | 2+2 4 | 6+6 2 | 5+5 2 | 5+6 2 | 3+4 2 | 2+2 4 |
| | 2+2 4 | 2+2 4 | 2+2 4 | 3+3 2 | 6+6 2 | 5+6 2 | 5+6 2 | 2+2 4 |
| | 2+2 4 | 2+2 4 | 2+2 4 | 4+4 2 | 3+3 2 | 3+4 2 | 3+4 2 | 2+2 4 |
| | 2+2 4 | 2+2 4 | 2+2 4 | 6+6 2 | 3+3 2 | 3+4 2 | 5+6 2 | 2+2 4 |
| | 2+2 4 | 2+2 4 | 2+2 4 | 5+5 2 | 4+4 2 | 5+6 2 | 3+4 2 | 2+2 4 |
| | 2+2 4 | 2+2 4 | 2+2 4 | 4+4 2 | 5+5 2 | 3+4 2 | 3+4 2 | 2+2 4 |
| | 2+2 4 | 2+2 4 | 2+2 4 | 3+3 2 | 6+6 2 | 3+4 2 | 5+6 2 | 2+2 4 |
| | 2+2 4 | 2+2 4 | 2+2 4 | 3+3 2 | 6+6 2 | 5+6 2 | 3+4 2 | 2+2 4 |
| | 2+2 4 | 2+2 4 | 2+2 4 | 4+4 2 | 4+4 2 | 3+4 2 | 3+4 2 | 2+2 4 |
| | 2+2 4 | 2+2 4 | 2+2 4 | 5+5 2 | 5+5 2 | 3+4 2 | 5+6 2 | 2+2 4 |
| | 2+2 4 | 2+2 4 | 2+2 4 | 6+6 2 | 3+3 2 | 3+4 2 | 3+4 2 | 2+2 4 |
| | 2+2 4 | 2+2 4 | 2+2 4 | 3+3 2 | 5+5 2 | 5+6 2 | 5+6 2 | 2+2 4 |
| | 2+2 4 | 2+2 4 | 2+2 4 | 4+4 2 | 6+6 2 | 5+6 2 | 5+6 2 | 2+2 4 |
| | | | | 5+5 2 | 3+3 2 | 3+4 2 | 3+4 2 | |
| | | | | 6+6 2 | 4+4 2 | 3+4 2 | 5+6 2 | |
| | | | | 6+6 2 | 5+5 2 | 5+6 2 | 3+4 2 | |
| | | | | 5+5 2 | 6+6 2 | 3+4 2 | 5+6 2 | |
| | | | | 4+4 2 | 3+3 2 | 3+4 2 | 3+4 2 | |
| | | | | 3+3 2 | 5+5 2 | 5+6 2 | 5+6 2 | |
| | | | | 6+6 2 | 6+6 2 | 5+6 2 | 5+6 2 | |
| | | | | 3+3 2 | 4+4 2 | 3+4 2 | 3+4 2 | |
| | | | | 5+5 2 | 3+3 2 | 5+6 2 | 3+4 2 | |
| | | | | 4+4 2 | 4+4 2 | 3+4 2 | 5+6 2 | |
| | | | | 5+5 2 | 5+5 2 | 5+6 2 | 3+4 2 | |
| | | | | 6+6 2 | 6+6 2 | 5+6 2 | 5+6 2 | |
| | | | | 3+3 2 | 3+3 2 | 3+4 2 | 3+4 2 | |
| | | | | 4+4 2 | 4+4 2 | 3+4 2 | 5+6 2 | |
| | | | | 5+5 2 | 6+6 2 | 5+6 2 | 5+6 2 | |
| | | | | 6+6 2 | 5+5 2 | 5+6 2 | 3+4 2 | |
| | | | | 4+4 2 | 3+3 2 | 3+4 2 | 5+6 2 | |
| | | | | 3+3 2 | 4+4 2 | 5+6 2 | 3+4 2 | |
| | | | | 6+6 2 | 6+6 2 | 5+6 2 | 3+4 2 | |
| | | | | 5+5 2 | 5+5 2 | 5+6 2 | 5+6 2 | |
| | | | | 3+3 2 | 4+4 2 | 5+6 2 | 5+6 2 | |
| | | | | 4+4 2 | 3+3 2 | 3+4 2 | 3+4 2 | |
| | | | | 3+3 2 | 4+4 2 | 5+6 2 | 3+4 2 | |
| | | | | 4+4 2 | 3+3 2 | 5+6 2 | 5+6 2 | |
| | | | | 5+5 2 | 6+6 2 | 3+4 2 | 5+6 2 | |
| | | | | 6+6 2 | 5+5 2 | 3+4 2 | 3+4 2 | |
| | | | | 6+6 2 | 4+4 2 | 3+4 2 | 3+4 2 | |
| | | | | 5+5 2 | 3+3 2 | 3+4 2 | 5+6 2 | |
| | | | | 3+3 2 | 5+5 2 | 3+4 2 | 5+6 2 | |
| | | | | 4+4 2 | 6+6 2 | 5+6 2 | 3+4 2 | |
| 14.2.59 | 172071 | 196 | Br2Cu2O6Sr2Te1 | | | | | |
| | -A | -B | -C | -D | -E | -GM | -Y | -Z |
| | 2+2 4 | 2+2 4 | 2+2 4 | 3+3 2 | 3+3 2 | 3+4 2 | 3+4 2 | 2+2 4 |
| | 2+2 4 | 2+2 4 | 2+2 4 | 4+4 2 | 4+4 2 | 5+6 2 | 3+4 2 | 2+2 4 |
| | 2+2 4 | 2+2 4 | 2+2 4 | 5+5 2 | 5+5 2 | 5+6 2 | 5+6 2 | 2+2 4 |
| | 2+2 4 | 2+2 4 | 2+2 4 | 6+6 2 | 6+6 2 | 3+4 2 | 5+6 2 | 2+2 4 |
| | 2+2 4 | 2+2 4 | 2+2 4 | 3+3 2 | 3+3 2 | 3+4 2 | 3+4 2 | 2+2 4 |
| | 2+2 4 | 2+2 4 | 2+2 4 | 4+4 2 | 4+4 2 | 3+4 2 | 3+4 2 | 2+2 4 |
| | 2+2 4 | 2+2 4 | 2+2 4 | 6+6 2 | 6+6 2 | 5+6 2 | 5+6 2 | 2+2 4 |
| | 2+2 4 | 2+2 4 | 2+2 4 | 5+5 2 | 5+5 2 | 5+6 2 | 5+6 2 | 2+2 4 |
| | 2+2 4 | 2+2 4 | 2+2 4 | 5+5 2 | 5+5 2 | 5+6 2 | 5+6 2 | 2+2 4 |
| | 2+2 4 | 2+2 4 | 2+2 4 | 6+6 2 | 6+6 2 | 5+6 2 | 5+6 2 | 2+2 4 |

Continued on next column

**TABLE XXVI– continued**

| TQC | ICSD | NoE | SF |
|---|---|---|---|
| | | | 2+2 4 2+2 4 2+2 4 6+6 2 6+6 2 5+6 2 5+6 2 2+2 4 |
| | | | 2+2 4 2+2 4 2+2 4 5+5 2 5+5 2 5+6 2 5+6 2 2+2 4 |
| | | | 2+2 4 2+2 4 2+2 4 3+3 2 3+3 2 3+4 2 3+4 2 2+2 4 |
| | | | 2+2 4 2+2 4 2+2 4 4+4 2 4+4 2 3+4 2 3+4 2 2+2 4 |
| | | | 2+2 4 2+2 4 2+2 4 3+3 2 3+3 2 3+4 2 3+4 2 2+2 4 |
| | | | 2+2 4 2+2 4 2+2 4 4+4 2 4+4 2 3+4 2 3+4 2 2+2 4 |
| | | | 2+2 4 2+2 4 2+2 4 5+5 2 5+5 2 3+4 2 5+6 2 2+2 4 |
| | | | 2+2 4 2+2 4 2+2 4 4+4 2 4+4 2 5+6 2 3+4 2 2+2 4 |
| | | | 2+2 4 2+2 4 2+2 4 6+6 2 6+6 2 5+6 2 5+6 2 2+2 4 |
| | | | 2+2 4 2+2 4 2+2 4 3+3 2 3+3 2 3+4 2 3+4 2 2+2 4 |
| | | | 2+2 4 2+2 4 2+2 4 4+4 2 4+4 2 3+4 2 3+4 2 2+2 4 |
| | | | 2+2 4 2+2 4 2+2 4 5+5 2 3+3 2 5+6 2 5+6 2 2+2 4 |
| | | | 2+2 4 2+2 4 2+2 4 3+3 2 5+5 2 5+6 2 5+6 2 2+2 4 |
| | | | 2+2 4 2+2 4 2+2 4 6+6 2 6+6 2 3+4 2 3+4 2 2+2 4 |
| | | | 2+2 4 2+2 4 2+2 4 6+6 2 6+6 2 3+4 2 5+6 2 2+2 4 |
| | | | 2+2 4 2+2 4 2+2 4 5+5 2 3+3 2 5+6 2 3+4 2 2+2 4 |
| | | | 2+2 4 2+2 4 2+2 4 3+3 2 5+5 2 5+6 2 5+6 2 2+2 4 |
| | | | 2+2 4 2+2 4 2+2 4 4+4 2 4+4 2 3+4 2 3+4 2 2+2 4 |
| | | | 2+2 4 2+2 4 2+2 4 6+6 2 3+3 2 3+4 2 5+6 2 2+2 4 |
| | | | 2+2 4 2+2 4 2+2 4 5+5 2 4+4 2 5+6 2 3+4 2 2+2 4 |
| | | | 2+2 4 2+2 4 2+2 4 3+3 2 5+5 2 3+4 2 5+6 2 2+2 4 |
| | | | 2+2 4 2+2 4 2+2 4 4+4 2 6+6 2 5+6 2 3+4 2 2+2 4 |
| | | | 2+2 4 2+2 4 2+2 4 4+4 2 4+4 2 3+4 2 3+4 2 2+2 4 |
| | | | 2+2 4 2+2 4 2+2 4 3+3 2 3+3 2 3+4 2 3+4 2 2+2 4 |
| | | | 2+2 4 2+2 4 2+2 4 5+5 2 5+5 2 5+6 2 5+6 2 2+2 4 |
| | | | 2+2 4 2+2 4 2+2 4 6+6 2 6+6 2 5+6 2 5+6 2 2+2 4 |
| | | | 2+2 4 2+2 4 2+2 4 6+6 2 6+6 2 5+6 2 5+6 2 2+2 4 |
| | | | 2+2 4 2+2 4 2+2 4 5+5 2 5+5 2 3+4 2 3+4 2 2+2 4 |
| | | | 2+2 4 2+2 4 2+2 4 5+5 2 5+5 2 5+6 2 5+6 2 2+2 4 |
| | | | 2+2 4 2+2 4 2+2 4 6+6 2 6+6 2 5+6 2 5+6 2 2+2 4 |
| | | | 2+2 4 2+2 4 2+2 4 3+3 2 3+3 2 5+6 2 5+6 2 2+2 4 |
| | | | 2+2 4 2+2 4 2+2 4 4+4 2 4+4 2 3+4 2 3+4 2 2+2 4 |
| | | | 2+2 4 2+2 4 2+2 4 6+6 2 6+6 2 5+6 2 5+6 2 2+2 4 |
| | | | 2+2 4 2+2 4 2+2 4 5+5 2 5+5 2 3+4 2 3+4 2 2+2 4 |
| | | | 2+2 4 2+2 4 2+2 4 4+4 2 4+4 2 5+6 2 5+6 2 2+2 4 |
| | | | 2+2 4 2+2 4 2+2 4 3+3 2 3+3 2 3+4 2 3+4 2 2+2 4 |
| | | | 2+2 4 2+2 4 2+2 4 3+3 2 3+3 2 3+4 2 3+4 2 2+2 4 |
| | | | 2+2 4 2+2 4 2+2 4 4+4 2 4+4 2 3+4 2 3+4 2 2+2 4 |
| | | | 2+2 4 2+2 4 2+2 4 3+3 2 3+3 2 5+6 2 5+6 2 2+2 4 |
| | | | 4+4 2 4+4 2 3+4 2 3+4 2 |
| | | | 5+5 2 5+5 2 5+6 2 5+6 2 |
| | | | 6+6 2 6+6 2 3+4 2 3+4 2 |
| | | | 3+3 2 6+6 2 3+4 2 3+4 2 |
| | | | 4+4 2 5+5 2 5+6 2 5+6 2 |
| | | | 6+6 2 3+3 2 3+4 2 5+6 2 |
| | | | 5+5 2 4+4 2 5+6 2 3+4 2 |
| | | | 5+5 2 5+5 2 5+6 2 5+6 2 |
| | | | 6+6 2 6+6 2 3+4 2 5+6 2 |
| | | | 3+3 2 3+3 2 5+6 2 3+4 2 |
| | | | 4+4 2 4+4 2 3+4 2 5+6 2 |
| | | | 5+5 2 5+5 2 5+6 2 3+4 2 |
| | | | 6+6 2 6+6 2 5+6 2 5+6 2 |
| | | | 4+4 2 6+6 2 3+4 2 3+4 2 |
| | | | 3+3 2 3+3 2 3+4 2 5+6 2 |
| | | | 6+6 2 4+4 2 3+4 2 3+4 2 |
| | | | 5+5 2 5+5 2 5+6 2 3+4 2 |
| | | | 6+6 2 3+3 2 3+4 2 5+6 2 |
| | | | 5+5 2 4+4 2 5+6 2 3+4 2 |
| | | | 5+5 2 6+6 2 3+4 2 5+6 2 |
| | | | 3+3 2 5+5 2 3+4 2 3+4 2 |
| | | | 4+4 2 4+4 2 5+6 2 5+6 2 |

Continued on next column

**TABLE XXVI– continued**

| TQC | ICSD | NoE | SF |
|---|---|---|---|
| | | | 6+6 2 3+3 2 3+4 2 3+4 2 |
| | | | 4+4 2 4+4 2 5+6 2 5+6 2 |
| | | | 3+3 2 5+5 2 3+4 2 3+4 2 |
| | | | 4+4 2 3+3 2 3+4 2 5+6 2 |
| | | | 3+3 2 6+6 2 5+6 2 3+4 2 |
| | | | 6+6 2 3+3 2 5+6 2 5+6 2 |
| | | | 5+5 2 4+4 2 3+4 2 3+4 2 |
| | | | 3+3 2 6+6 2 5+6 2 5+6 2 |
| | | | 4+4 2 5+5 2 5+6 2 3+4 2 |
| | | | 5+5 2 6+6 2 5+6 2 5+6 2 |
| | | | 6+6 2 4+4 2 3+4 2 3+4 2 |
| | | | 3+3 2 5+5 2 5+6 2 3+4 2 |
| | | | 4+4 2 3+3 2 3+4 2 5+6 2 |
| | | | 6+6 2 6+6 2 3+4 2 3+4 2 |
| | | | 5+5 2 5+5 2 3+4 2 5+6 2 |
| | | | 3+3 2 3+3 2 5+6 2 5+6 2 |
| | | | 4+4 2 4+4 2 5+6 2 3+4 2 |
| | | | 5+5 2 5+5 2 5+6 2 5+6 2 |
| | | | 6+6 2 6+6 2 5+6 2 3+4 2 |
| | | | 3+3 2 3+3 2 3+4 2 5+6 2 |
| | | | 4+4 2 4+4 2 3+4 2 3+4 2 |
| | | | 5+5 2 5+5 2 5+6 2 5+6 2 |
| | | | 6+6 2 6+6 2 5+6 2 5+6 2 |
| | | | 3+3 2 3+3 2 3+4 2 3+4 2 |
| | | | 4+4 2 4+4 2 3+4 2 3+4 2 |
| | | | 6+6 2 6+6 2 5+6 2 5+6 2 |
| | | | 5+5 2 5+5 2 3+4 2 3+4 2 |
| 14.2.60 | 75019 | 180 | Ba2P3Pd1 |
| | | | -A -B -C -D -E -GM -Y -Z |
| | | | 2+2 4 2+2 4 2+2 4 3+3 2 5+5 2 3+4 2 5+6 2 2+2 4 |
| | | | 2+2 4 2+2 4 2+2 4 4+4 2 6+6 2 3+4 2 5+6 2 2+2 4 |
| | | | 2+2 4 2+2 4 2+2 4 5+5 2 3+3 2 5+6 2 3+4 2 2+2 4 |
| | | | 2+2 4 2+2 4 2+2 4 6+6 2 4+4 2 5+6 2 3+4 2 2+2 4 |
| | | | 2+2 4 2+2 4 2+2 4 5+5 2 3+3 2 5+6 2 3+4 2 2+2 4 |
| | | | 2+2 4 2+2 4 2+2 4 6+6 2 4+4 2 5+6 2 3+4 2 2+2 4 |
| | | | 2+2 4 2+2 4 2+2 4 3+3 2 5+5 2 3+4 2 5+6 2 2+2 4 |
| | | | 2+2 4 2+2 4 2+2 4 4+4 2 6+6 2 3+4 2 5+6 2 2+2 4 |
| | | | 2+2 4 2+2 4 2+2 4 3+3 2 4+4 2 3+4 2 5+6 2 2+2 4 |
| | | | 2+2 4 2+2 4 2+2 4 6+6 2 6+6 2 5+6 2 3+4 2 2+2 4 |
| | | | 2+2 4 2+2 4 2+2 4 4+4 2 5+5 2 5+6 2 3+4 2 2+2 4 |
| | | | 2+2 4 2+2 4 2+2 4 5+5 2 3+3 2 3+4 2 5+6 2 2+2 4 |
| | | | 2+2 4 2+2 4 2+2 4 5+5 2 6+6 2 3+4 2 3+4 2 2+2 4 |
| | | | 2+2 4 2+2 4 2+2 4 4+4 2 3+3 2 3+4 2 5+6 2 2+2 4 |
| | | | 2+2 4 2+2 4 2+2 4 6+6 2 5+5 2 5+6 2 5+6 2 2+2 4 |
| | | | 2+2 4 2+2 4 2+2 4 3+3 2 4+4 2 5+6 2 3+4 2 2+2 4 |
| | | | 2+2 4 2+2 4 2+2 4 4+4 2 5+5 2 3+4 2 5+6 2 2+2 4 |
| | | | 2+2 4 2+2 4 2+2 4 5+5 2 3+3 2 5+6 2 3+4 2 2+2 4 |
| | | | 2+2 4 2+2 4 2+2 4 6+6 2 4+4 2 5+6 2 3+4 2 2+2 4 |
| | | | 2+2 4 2+2 4 2+2 4 3+3 2 6+6 2 3+4 2 5+6 2 2+2 4 |
| | | | 2+2 4 2+2 4 2+2 4 6+6 2 5+5 2 3+4 2 5+6 2 2+2 4 |
| | | | 2+2 4 2+2 4 2+2 4 3+3 2 4+4 2 3+4 2 3+4 2 2+2 4 |
| | | | 2+2 4 2+2 4 2+2 4 5+5 2 6+6 2 5+6 2 5+6 2 2+2 4 |
| | | | 2+2 4 2+2 4 2+2 4 4+4 2 3+3 2 5+6 2 3+4 2 2+2 4 |
| | | | 2+2 4 2+2 4 2+2 4 3+3 2 3+3 2 3+4 2 3+4 2 2+2 4 |
| | | | 2+2 4 2+2 4 2+2 4 4+4 2 6+6 2 5+6 2 5+6 2 2+2 4 |
| | | | 2+2 4 2+2 4 2+2 4 5+5 2 4+4 2 3+4 2 3+4 2 2+2 4 |
| | | | 2+2 4 2+2 4 2+2 4 6+6 2 5+5 2 3+4 2 5+6 2 2+2 4 |
| | | | 2+2 4 2+2 4 2+2 4 3+3 2 4+4 2 5+6 2 3+4 2 2+2 4 |
| | | | 2+2 4 2+2 4 2+2 4 6+6 2 5+5 2 5+6 2 3+4 2 2+2 4 |
| | | | 2+2 4 2+2 4 2+2 4 4+4 2 6+6 2 3+4 2 5+6 2 2+2 4 |
| | | | 2+2 4 2+2 4 2+2 4 5+5 2 3+3 2 3+4 2 5+6 2 2+2 4 |

Continued on next column

**TABLE XXVI– continued**

| TQC | ICSD | NoE | SF | | | | | |
|---|---|---|---|---|---|---|---|---|
| | 2+2 4 | 2+2 4 | 2+2 4 | 3+3 2 | 6+6 2 | 5+6 2 | 5+6 2 | 2+2 4 |
| | 2+2 4 | 2+2 4 | 2+2 4 | 4+4 2 | 4+4 2 | 5+6 2 | 3+4 2 | 2+2 4 |
| | 2+2 4 | 2+2 4 | 2+2 4 | 5+5 2 | 3+3 2 | 3+4 2 | 3+4 2 | 2+2 4 |
| | 2+2 4 | 2+2 4 | 2+2 4 | 6+6 2 | 5+5 2 | 5+6 2 | 5+6 2 | 2+2 4 |
| | 2+2 4 | 2+2 4 | 2+2 4 | 5+5 2 | 5+5 2 | 3+4 2 | 5+6 2 | 2+2 4 |
| | 2+2 4 | 2+2 4 | 2+2 4 | 6+6 2 | 6+6 2 | 5+6 2 | 3+4 2 | 2+2 4 |
| | 2+2 4 | 2+2 4 | 2+2 4 | 4+4 2 | 4+4 2 | 5+6 2 | 3+4 2 | 2+2 4 |
| | 2+2 4 | 2+2 4 | 2+2 4 | 3+3 2 | 3+3 2 | 3+4 2 | 5+6 2 | 2+2 4 |
| | 2+2 4 | 2+2 4 | 2+2 4 | 4+4 2 | 4+4 2 | 3+4 2 | 5+6 2 | 2+2 4 |
| | 2+2 4 | 2+2 4 | 2+2 4 | 3+3 2 | 3+3 2 | 5+6 2 | 3+4 2 | 2+2 4 |
| | 2+2 4 | 2+2 4 | 2+2 4 | 5+5 2 | 5+5 2 | 3+4 2 | 5+6 2 | 2+2 4 |
| | 2+2 4 | 2+2 4 | 2+2 4 | 6+6 2 | 6+6 2 | 5+6 2 | 3+4 2 | 2+2 4 |
| | 2+2 4 | 2+2 4 | 2+2 4 | 6+6 2 | 3+3 2 | 5+6 2 | 3+4 2 | 2+2 4 |
| | | | | 4+4 2 | 4+4 2 | 3+4 2 | 5+6 2 | |
| | | | | 5+5 2 | 5+5 2 | 3+4 2 | 5+6 2 | |
| | | | | 3+3 2 | 6+6 2 | 5+6 2 | 3+4 2 | |
| | | | | 5+5 2 | 5+5 2 | 3+4 2 | 5+6 2 | |
| | | | | 6+6 2 | 6+6 2 | 3+4 2 | 3+4 2 | |
| | | | | 4+4 2 | 3+3 2 | 5+6 2 | 5+6 2 | |
| | | | | 3+3 2 | 4+4 2 | 3+4 2 | 3+4 2 | |
| | | | | 5+5 2 | 6+6 2 | 5+6 2 | 5+6 2 | |
| | | | | 6+6 2 | 5+5 2 | 5+6 2 | 3+4 2 | |
| | | | | 3+3 2 | 4+4 2 | 3+4 2 | 5+6 2 | |
| | | | | 4+4 2 | 3+3 2 | 5+6 2 | 3+4 2 | |
| | | | | 6+6 2 | 3+3 2 | 3+4 2 | 5+6 2 | |
| | | | | 5+5 2 | 4+4 2 | 3+4 2 | 5+6 2 | |
| | | | | 3+3 2 | 6+6 2 | 5+6 2 | 3+4 2 | |
| | | | | 4+4 2 | 5+5 2 | 5+6 2 | 3+4 2 | |
| | | | | 6+6 2 | 3+3 2 | 3+4 2 | 5+6 2 | |
| | | | | 5+5 2 | 4+4 2 | 3+4 2 | 5+6 2 | |
| | | | | 4+4 2 | 5+5 2 | 5+6 2 | 3+4 2 | |
| | | | | 3+3 2 | 6+6 2 | 3+4 2 | 5+6 2 | |
| | | | | 4+4 2 | 5+5 2 | 3+4 2 | 3+4 2 | |
| | | | | 3+3 2 | 3+3 2 | 5+6 2 | 5+6 2 | |
| | | | | 6+6 2 | 6+6 2 | 5+6 2 | 3+4 2 | |
| | | | | 5+5 2 | 4+4 2 | 3+4 2 | 3+4 2 | |
| | | | | 3+3 2 | 6+6 2 | 5+6 2 | 5+6 2 | |
| | | | | 4+4 2 | 5+5 2 | 5+6 2 | 3+4 2 | |
| | | | | 5+5 2 | 4+4 2 | 3+4 2 | 5+6 2 | |
| | | | | 6+6 2 | 3+3 2 | 5+6 2 | 3+4 2 | |
| | | | | 3+3 2 | 6+6 2 | 3+4 2 | 5+6 2 | |
| | | | | 4+4 2 | 5+5 2 | 5+6 2 | 3+4 2 | |
| | | | | 5+5 2 | 4+4 2 | 3+4 2 | 5+6 2 | |
| | | | | 6+6 2 | 3+3 2 | 5+6 2 | 5+6 2 | |
| | | | | 5+5 2 | 4+4 2 | 3+4 2 | 3+4 2 | |
| | | | | 6+6 2 | 3+3 2 | 5+6 2 | 5+6 2 | |
| | | | | 4+4 2 | 6+6 2 | 3+4 2 | 3+4 2 | |
| | | | | 3+3 2 | 5+5 2 | 3+4 2 | 5+6 2 | |
| | | | | 5+5 2 | 3+3 2 | 5+6 2 | 3+4 2 | |
| | | | | 6+6 2 | 4+4 2 | 5+6 2 | 5+6 2 | |
| | | | | 3+3 2 | 5+5 2 | 5+6 2 | 3+4 2 | |
| | | | | 4+4 2 | 6+6 2 | 3+4 2 | 3+4 2 | |
| | | | | 3+3 2 | 4+4 2 | 3+4 2 | 3+4 2 | |
| | | | | 4+4 2 | 3+3 2 | 3+4 2 | 5+6 2 | |
| | | | | 5+5 2 | 6+6 2 | 5+6 2 | 5+6 2 | |
| | | | | 6+6 2 | 5+5 2 | 5+6 2 | 3+4 2 | |
| | | | | 3+3 2 | 3+3 2 | 3+4 2 | 5+6 2 | |
| | | | | 4+4 2 | 4+4 2 | 5+6 2 | 5+6 2 | |
| 14.2.61 | 50158 | 144 | Ba1Rh2Si2 | | | | | |
| | -A | -B | -C | -D | -E | -GM | -Y | -Z |
| | 2+2 4 | 2+2 4 | 2+2 4 | 5+5 2 | 5+5 2 | 3+4 2 | 3+4 2 | 2+2 4 |

Continued on next column

**TABLE XXVI– continued**

| TQC | ICSD | NoE | SF | | | | | |
|---|---|---|---|---|---|---|---|---|
| | 2+2 4 | 2+2 4 | 2+2 4 | 6+6 2 | 6+6 2 | 3+4 2 | 5+6 2 | 2+2 4 |
| | 2+2 4 | 2+2 4 | 2+2 4 | 3+3 2 | 3+3 2 | 5+6 2 | 3+4 2 | 2+2 4 |
| | 2+2 4 | 2+2 4 | 2+2 4 | 4+4 2 | 4+4 2 | 5+6 2 | 5+6 2 | 2+2 4 |
| | 2+2 4 | 2+2 4 | 2+2 4 | 5+5 2 | 5+5 2 | 3+4 2 | 3+4 2 | 2+2 4 |
| | 2+2 4 | 2+2 4 | 2+2 4 | 6+6 2 | 3+3 2 | 3+4 2 | 5+6 2 | 2+2 4 |
| | 2+2 4 | 2+2 4 | 2+2 4 | 3+3 2 | 6+6 2 | 5+6 2 | 3+4 2 | 2+2 4 |
| | 2+2 4 | 2+2 4 | 2+2 4 | 4+4 2 | 4+4 2 | 5+6 2 | 5+6 2 | 2+2 4 |
| | 2+2 4 | 2+2 4 | 2+2 4 | 6+6 2 | 4+4 2 | 3+4 2 | 3+4 2 | 2+2 4 |
| | 2+2 4 | 2+2 4 | 2+2 4 | 4+4 2 | 6+6 2 | 5+6 2 | 5+6 2 | 2+2 4 |
| | 2+2 4 | 2+2 4 | 2+2 4 | 5+5 2 | 5+5 2 | 3+4 2 | 3+4 2 | 2+2 4 |
| | 2+2 4 | 2+2 4 | 2+2 4 | 3+3 2 | 3+3 2 | 3+4 2 | 3+4 2 | 2+2 4 |
| | 2+2 4 | 2+2 4 | 2+2 4 | 5+5 2 | 5+5 2 | 5+6 2 | 5+6 2 | 2+2 4 |
| | 2+2 4 | 2+2 4 | 2+2 4 | 6+6 2 | 3+3 2 | 3+4 2 | 5+6 2 | 2+2 4 |
| | 2+2 4 | 2+2 4 | 2+2 4 | 3+3 2 | 6+6 2 | 5+6 2 | 3+4 2 | 2+2 4 |
| | 2+2 4 | 2+2 4 | 2+2 4 | 4+4 2 | 4+4 2 | 5+6 2 | 5+6 2 | 2+2 4 |
| | 2+2 4 | 2+2 4 | 2+2 4 | 3+3 2 | 3+3 2 | 3+4 2 | 3+4 2 | 2+2 4 |
| | 2+2 4 | 2+2 4 | 2+2 4 | 4+4 2 | 4+4 2 | 3+4 2 | 5+6 2 | 2+2 4 |
| | 2+2 4 | 2+2 4 | 2+2 4 | 4+4 2 | 5+5 2 | 3+4 2 | 3+4 2 | 2+2 4 |
| | 2+2 4 | 2+2 4 | 2+2 4 | 3+3 2 | 6+6 2 | 5+6 2 | 3+4 2 | 2+2 4 |
| | 2+2 4 | 2+2 4 | 2+2 4 | 5+5 2 | 4+4 2 | 5+6 2 | 3+4 2 | 2+2 4 |
| | 2+2 4 | 2+2 4 | 2+2 4 | 6+6 2 | 3+3 2 | 3+4 2 | 5+6 2 | 2+2 4 |
| | 2+2 4 | 2+2 4 | 2+2 4 | 5+5 2 | 6+6 2 | 5+6 2 | 5+6 2 | 2+2 4 |
| | 2+2 4 | 2+2 4 | 2+2 4 | 6+6 2 | 5+5 2 | 5+6 2 | 5+6 2 | 2+2 4 |
| | 2+2 4 | 2+2 4 | 2+2 4 | 4+4 2 | 4+4 2 | 3+4 2 | 5+6 2 | 2+2 4 |
| | 2+2 4 | 2+2 4 | 2+2 4 | 3+3 2 | 3+3 2 | 5+6 2 | 3+4 2 | 2+2 4 |
| | 2+2 4 | 2+2 4 | 2+2 4 | 6+6 2 | 6+6 2 | 5+6 2 | 3+4 2 | 2+2 4 |
| | 2+2 4 | 2+2 4 | 2+2 4 | 5+5 2 | 5+5 2 | 3+4 2 | 5+6 2 | 2+2 4 |
| | 2+2 4 | 2+2 4 | 2+2 4 | 5+5 2 | 3+3 2 | 5+6 2 | 3+4 2 | 2+2 4 |
| | 2+2 4 | 2+2 4 | 2+2 4 | 6+6 2 | 4+4 2 | 3+4 2 | 3+4 2 | 2+2 4 |
| | 2+2 4 | 2+2 4 | 2+2 4 | 4+4 2 | 6+6 2 | 5+6 2 | 3+4 2 | 2+2 4 |
| | 2+2 4 | 2+2 4 | 2+2 4 | 3+3 2 | 5+5 2 | 3+4 2 | 5+6 2 | 2+2 4 |
| | 2+2 4 | 2+2 4 | 2+2 4 | 6+6 2 | 4+4 2 | 3+4 2 | 3+4 2 | 2+2 4 |
| | 2+2 4 | 2+2 4 | 2+2 4 | 5+5 2 | 3+3 2 | 5+6 2 | 5+6 2 | 2+2 4 |
| | 2+2 4 | 2+2 4 | 2+2 4 | 5+5 2 | 3+3 2 | 3+4 2 | 5+6 2 | 2+2 4 |
| | 2+2 4 | 2+2 4 | 2+2 4 | 3+3 2 | 4+4 2 | 5+6 2 | 5+6 2 | 2+2 4 |
| | | | | 6+6 2 | 6+6 2 | 5+6 2 | 3+4 2 | |
| | | | | 4+4 2 | 5+5 2 | 3+4 2 | 3+4 2 | |
| | | | | 3+3 2 | 5+5 2 | 3+4 2 | 5+6 2 | |
| | | | | 4+4 2 | 6+6 2 | 5+6 2 | 5+6 2 | |
| | | | | 6+6 2 | 3+3 2 | 5+6 2 | 3+4 2 | |
| | | | | 3+3 2 | 4+4 2 | 3+4 2 | 5+6 2 | |
| | | | | 5+5 2 | 6+6 2 | 3+4 2 | 3+4 2 | |
| | | | | 4+4 2 | 5+5 2 | 3+4 2 | 5+6 2 | |
| | | | | 4+4 2 | 3+3 2 | 5+6 2 | 3+4 2 | |
| | | | | 3+3 2 | 5+5 2 | 5+6 2 | 5+6 2 | |
| | | | | 6+6 2 | 4+4 2 | 5+6 2 | 3+4 2 | |
| | | | | 4+4 2 | 6+6 2 | 3+4 2 | 5+6 2 | |
| | | | | 5+5 2 | 4+4 2 | 5+6 2 | 3+4 2 | |
| | | | | 3+3 2 | 3+3 2 | 3+4 2 | 5+6 2 | |
| | | | | 5+5 2 | 6+6 2 | 5+6 2 | 3+4 2 | |
| | | | | 3+3 2 | 5+5 2 | 3+4 2 | 3+4 2 | |
| | | | | 6+6 2 | 3+3 2 | 3+4 2 | 5+6 2 | |
| | | | | 4+4 2 | 5+5 2 | 5+6 2 | 3+4 2 | |
| | | | | 6+6 2 | 4+4 2 | 3+4 2 | 5+6 2 | |
| | | | | 5+5 2 | 6+6 2 | 5+6 2 | 3+4 2 | |
| | | | | 3+3 2 | 6+6 2 | 3+4 2 | 5+6 2 | |
| | | | | 4+4 2 | 5+5 2 | 5+6 2 | 5+6 2 | |
| | | | | 5+5 2 | 4+4 2 | 3+4 2 | 3+4 2 | |
| | | | | 4+4 2 | 3+3 2 | 5+6 2 | 5+6 2 | |
| | | | | 6+6 2 | 5+5 2 | 3+4 2 | 3+4 2 | |
| | | | | 3+3 2 | 6+6 2 | 3+4 2 | 5+6 2 | |

Continued on next column

**TABLE XXVI– continued**

| TQC | ICSD | | NoE | SF | | | | |
|---|---|---|---|---|---|---|---|---|
| | | | | 5+5 2 | 4+4 2 | 5+6 2 | 3+4 2 | |
| | | | | 6+6 2 | 3+3 2 | 3+4 2 | 5+6 2 | |
| | | | | 4+4 2 | 3+3 2 | 5+6 2 | 3+4 2 | |
| | | | | 3+3 2 | 4+4 2 | 3+4 2 | 5+6 2 | |
| | | | | 5+5 2 | 5+5 2 | 5+6 2 | 5+6 2 | |
| | | | | 6+6 2 | 6+6 2 | 5+6 2 | 5+6 2 | |
| | | | | 3+3 2 | 5+5 2 | 5+6 2 | 3+4 2 | |
| | | | | 4+4 2 | 6+6 2 | 3+4 2 | 5+6 2 | |
| | | | | 5+5 2 | 4+4 2 | 5+6 2 | 3+4 2 | |
| | | | | 6+6 2 | 3+3 2 | 3+4 2 | 5+6 2 | |
| 14.2.62 | 50160 | | 144 | Ba1Ga2Pt2 | | | | |
| | -A | -B | -C | -D | -E | -GM | -Y | -Z |
| | 2+2 4 | 2+2 4 | 2+2 4 | 5+5 2 | 5+5 2 | 3+4 2 | 3+4 2 | 2+2 4 |
| | 2+2 4 | 2+2 4 | 2+2 4 | 6+6 2 | 6+6 2 | 3+4 2 | 5+6 2 | 2+2 4 |
| | 2+2 4 | 2+2 4 | 2+2 4 | 3+3 2 | 3+3 2 | 5+6 2 | 3+4 2 | 2+2 4 |
| | 2+2 4 | 2+2 4 | 2+2 4 | 4+4 2 | 4+4 2 | 5+6 2 | 5+6 2 | 2+2 4 |
| | 2+2 4 | 2+2 4 | 2+2 4 | 5+5 2 | 5+5 2 | 3+4 2 | 3+4 2 | 2+2 4 |
| | 2+2 4 | 2+2 4 | 2+2 4 | 6+6 2 | 6+6 2 | 3+4 2 | 5+6 2 | 2+2 4 |
| | 2+2 4 | 2+2 4 | 2+2 4 | 3+3 2 | 3+3 2 | 5+6 2 | 3+4 2 | 2+2 4 |
| | 2+2 4 | 2+2 4 | 2+2 4 | 4+4 2 | 4+4 2 | 5+6 2 | 5+6 2 | 2+2 4 |
| | 2+2 4 | 2+2 4 | 2+2 4 | 6+6 2 | 6+6 2 | 3+4 2 | 3+4 2 | 2+2 4 |
| | 2+2 4 | 2+2 4 | 2+2 4 | 4+4 2 | 4+4 2 | 3+4 2 | 5+6 2 | 2+2 4 |
| | 2+2 4 | 2+2 4 | 2+2 4 | 5+5 2 | 5+5 2 | 5+6 2 | 3+4 2 | 2+2 4 |
| | 2+2 4 | 2+2 4 | 2+2 4 | 3+3 2 | 3+3 2 | 3+4 2 | 3+4 2 | 2+2 4 |
| | 2+2 4 | 2+2 4 | 2+2 4 | 5+5 2 | 5+5 2 | 5+6 2 | 5+6 2 | 2+2 4 |
| | 2+2 4 | 2+2 4 | 2+2 4 | 6+6 2 | 3+3 2 | 5+6 2 | 5+6 2 | 2+2 4 |
| | 2+2 4 | 2+2 4 | 2+2 4 | 3+3 2 | 6+6 2 | 3+4 2 | 3+4 2 | 2+2 4 |
| | 2+2 4 | 2+2 4 | 2+2 4 | 4+4 2 | 4+4 2 | 5+6 2 | 5+6 2 | 2+2 4 |
| | 2+2 4 | 2+2 4 | 2+2 4 | 3+3 2 | 3+3 2 | 3+4 2 | 3+4 2 | 2+2 4 |
| | 2+2 4 | 2+2 4 | 2+2 4 | 4+4 2 | 4+4 2 | 3+4 2 | 5+6 2 | 2+2 4 |
| | 2+2 4 | 2+2 4 | 2+2 4 | 5+5 2 | 5+5 2 | 3+4 2 | 3+4 2 | 2+2 4 |
| | 2+2 4 | 2+2 4 | 2+2 4 | 6+6 2 | 6+6 2 | 5+6 2 | 3+4 2 | 2+2 4 |
| | 2+2 4 | 2+2 4 | 2+2 4 | 3+3 2 | 4+4 2 | 5+6 2 | 5+6 2 | 2+2 4 |
| | 2+2 4 | 2+2 4 | 2+2 4 | 4+4 2 | 3+3 2 | 5+6 2 | 3+4 2 | 2+2 4 |
| | 2+2 4 | 2+2 4 | 2+2 4 | 5+5 2 | 6+6 2 | 3+4 2 | 5+6 2 | 2+2 4 |
| | 2+2 4 | 2+2 4 | 2+2 4 | 6+6 2 | 5+5 2 | 5+6 2 | 5+6 2 | 2+2 4 |
| | 2+2 4 | 2+2 4 | 2+2 4 | 3+3 2 | 6+6 2 | 3+4 2 | 5+6 2 | 2+2 4 |
| | 2+2 4 | 2+2 4 | 2+2 4 | 4+4 2 | 3+3 2 | 5+6 2 | 3+4 2 | 2+2 4 |
| | 2+2 4 | 2+2 4 | 2+2 4 | 6+6 2 | 5+5 2 | 5+6 2 | 3+4 2 | 2+2 4 |
| | 2+2 4 | 2+2 4 | 2+2 4 | 5+5 2 | 4+4 2 | 3+4 2 | 5+6 2 | 2+2 4 |
| | 2+2 4 | 2+2 4 | 2+2 4 | 5+5 2 | 3+3 2 | 3+4 2 | 5+6 2 | 2+2 4 |
| | 2+2 4 | 2+2 4 | 2+2 4 | 4+4 2 | 4+4 2 | 5+6 2 | 3+4 2 | 2+2 4 |
| | 2+2 4 | 2+2 4 | 2+2 4 | 6+6 2 | 5+5 2 | 3+4 2 | 3+4 2 | 2+2 4 |
| | 2+2 4 | 2+2 4 | 2+2 4 | 3+3 2 | 6+6 2 | 5+6 2 | 3+4 2 | 2+2 4 |
| | 2+2 4 | 2+2 4 | 2+2 4 | 6+6 2 | 4+4 2 | 5+6 2 | 5+6 2 | 2+2 4 |
| | 2+2 4 | 2+2 4 | 2+2 4 | 3+3 2 | 3+3 2 | 3+4 2 | 5+6 2 | 2+2 4 |
| | 2+2 4 | 2+2 4 | 2+2 4 | 5+5 2 | 6+6 2 | 5+6 2 | 3+4 2 | 2+2 4 |
| | 2+2 4 | 2+2 4 | 2+2 4 | 4+4 2 | 5+5 2 | 3+4 2 | 5+6 2 | 2+2 4 |
| | | | | 6+6 2 | 3+3 2 | 3+4 2 | 3+4 2 | |
| | | | | 5+5 2 | 4+4 2 | 5+6 2 | 5+6 2 | |
| | | | | 4+4 2 | 5+5 2 | 3+4 2 | 3+4 2 | |
| | | | | 3+3 2 | 6+6 2 | 5+6 2 | 5+6 2 | |
| | | | | 5+5 2 | 3+3 2 | 3+4 2 | 3+4 2 | |
| | | | | 3+3 2 | 4+4 2 | 5+6 2 | 5+6 2 | |
| | | | | 4+4 2 | 6+6 2 | 3+4 2 | 3+4 2 | |
| | | | | 6+6 2 | 5+5 2 | 5+6 2 | 3+4 2 | |
| | | | | 3+3 2 | 3+3 2 | 3+4 2 | 5+6 2 | |
| | | | | 4+4 2 | 4+4 2 | 5+6 2 | 5+6 2 | |
| | | | | 6+6 2 | 6+6 2 | 3+4 2 | 3+4 2 | |
| | | | | 5+5 2 | 5+5 2 | 5+6 2 | 5+6 2 | |
| | | | | 3+3 2 | 3+3 2 | 5+6 2 | 3+4 2 | |

Continued on next column

**TABLE XXVI– continued**

| TQC | ICSD | | NoE | SF | | | | |
|---|---|---|---|---|---|---|---|---|
| | | | | 4+4 2 | 4+4 2 | 3+4 2 | 3+4 2 | |
| | | | | 6+6 2 | 5+5 2 | 5+6 2 | 5+6 2 | |
| | | | | 5+5 2 | 6+6 2 | 3+4 2 | 5+6 2 | |
| | | | | 4+4 2 | 3+3 2 | 5+6 2 | 3+4 2 | |
| | | | | 3+3 2 | 4+4 2 | 3+4 2 | 5+6 2 | |
| | | | | 5+5 2 | 5+5 2 | 3+4 2 | 3+4 2 | |
| | | | | 6+6 2 | 6+6 2 | 5+6 2 | 3+4 2 | |
| | | | | 3+3 2 | 4+4 2 | 5+6 2 | 5+6 2 | |
| | | | | 5+5 2 | 3+3 2 | 3+4 2 | 5+6 2 | |
| | | | | 4+4 2 | 5+5 2 | 3+4 2 | 3+4 2 | |
| | | | | 6+6 2 | 6+6 2 | 5+6 2 | 3+4 2 | |
| | | | | 4+4 2 | 4+4 2 | 3+4 2 | 5+6 2 | |
| | | | | 3+3 2 | 6+6 2 | 5+6 2 | 3+4 2 | |
| | | | | 5+5 2 | 3+3 2 | 3+4 2 | 5+6 2 | |
| | | | | 6+6 2 | 5+5 2 | 3+4 2 | 5+6 2 | |
| | | | | 3+3 2 | 3+3 2 | 5+6 2 | 3+4 2 | |
| | | | | 4+4 2 | 4+4 2 | 5+6 2 | 5+6 2 | |
| | | | | 5+5 2 | 5+5 2 | 3+4 2 | 3+4 2 | |
| | | | | 6+6 2 | 6+6 2 | 5+6 2 | 5+6 2 | |
| | | | | 3+3 2 | 5+5 2 | 3+4 2 | 5+6 2 | |
| | | | | 4+4 2 | 6+6 2 | 5+6 2 | 5+6 2 | |
| | | | | 5+5 2 | 4+4 2 | 3+4 2 | 3+4 2 | |
| | | | | 6+6 2 | 3+3 2 | 5+6 2 | 5+6 2 | |
| 14.2.63 | 88450 | | 172 | B1Fe2O4 | | | | |
| | -A | -B | -C | -D | -E | -GM | -Y | -Z |
| | 2+2 4 | 2+2 4 | 2+2 4 | 5+5 2 | 3+3 2 | 3+4 2 | 3+4 2 | 2+2 4 |
| | 2+2 4 | 2+2 4 | 2+2 4 | 6+6 2 | 4+4 2 | 5+6 2 | 5+6 2 | 2+2 4 |
| | 2+2 4 | 2+2 4 | 2+2 4 | 3+3 2 | 5+5 2 | 5+6 2 | 5+6 2 | 2+2 4 |
| | 2+2 4 | 2+2 4 | 2+2 4 | 4+4 2 | 6+6 2 | 3+4 2 | 3+4 2 | 2+2 4 |
| | 2+2 4 | 2+2 4 | 2+2 4 | 3+3 2 | 3+3 2 | 3+4 2 | 5+6 2 | 2+2 4 |
| | 2+2 4 | 2+2 4 | 2+2 4 | 4+4 2 | 4+4 2 | 5+6 2 | 3+4 2 | 2+2 4 |
| | 2+2 4 | 2+2 4 | 2+2 4 | 5+5 2 | 5+5 2 | 5+6 2 | 5+6 2 | 2+2 4 |
| | 2+2 4 | 2+2 4 | 2+2 4 | 6+6 2 | 6+6 2 | 3+4 2 | 3+4 2 | 2+2 4 |
| | 2+2 4 | 2+2 4 | 2+2 4 | 3+3 2 | 3+3 2 | 3+4 2 | 5+6 2 | 2+2 4 |
| | 2+2 4 | 2+2 4 | 2+2 4 | 4+4 2 | 4+4 2 | 5+6 2 | 3+4 2 | 2+2 4 |
| | 2+2 4 | 2+2 4 | 2+2 4 | 5+5 2 | 5+5 2 | 3+4 2 | 3+4 2 | 2+2 4 |
| | 2+2 4 | 2+2 4 | 2+2 4 | 6+6 2 | 6+6 2 | 5+6 2 | 5+6 2 | 2+2 4 |
| | 2+2 4 | 2+2 4 | 2+2 4 | 3+3 2 | 3+3 2 | 3+4 2 | 3+4 2 | 2+2 4 |
| | 2+2 4 | 2+2 4 | 2+2 4 | 4+4 2 | 4+4 2 | 3+4 2 | 3+4 2 | 2+2 4 |
| | 2+2 4 | 2+2 4 | 2+2 4 | 5+5 2 | 5+5 2 | 5+6 2 | 5+6 2 | 2+2 4 |
| | 2+2 4 | 2+2 4 | 2+2 4 | 6+6 2 | 6+6 2 | 5+6 2 | 5+6 2 | 2+2 4 |
| | 2+2 4 | 2+2 4 | 2+2 4 | 5+5 2 | 5+5 2 | 3+4 2 | 3+4 2 | 2+2 4 |
| | 2+2 4 | 2+2 4 | 2+2 4 | 6+6 2 | 6+6 2 | 5+6 2 | 5+6 2 | 2+2 4 |
| | 2+2 4 | 2+2 4 | 2+2 4 | 3+3 2 | 3+3 2 | 5+6 2 | 5+6 2 | 2+2 4 |
| | 2+2 4 | 2+2 4 | 2+2 4 | 4+4 2 | 4+4 2 | 3+4 2 | 3+4 2 | 2+2 4 |
| | 2+2 4 | 2+2 4 | 2+2 4 | 3+3 2 | 3+3 2 | 5+6 2 | 3+4 2 | 2+2 4 |
| | 2+2 4 | 2+2 4 | 2+2 4 | 4+4 2 | 4+4 2 | 3+4 2 | 3+4 2 | 2+2 4 |
| | 2+2 4 | 2+2 4 | 2+2 4 | 5+5 2 | 5+5 2 | 3+4 2 | 5+6 2 | 2+2 4 |
| | 2+2 4 | 2+2 4 | 2+2 4 | 6+6 2 | 6+6 2 | 5+6 2 | 5+6 2 | 2+2 4 |
| | 2+2 4 | 2+2 4 | 2+2 4 | 5+5 2 | 3+3 2 | 3+4 2 | 3+4 2 | 2+2 4 |
| | 2+2 4 | 2+2 4 | 2+2 4 | 6+6 2 | 4+4 2 | 5+6 2 | 5+6 2 | 2+2 4 |
| | 2+2 4 | 2+2 4 | 2+2 4 | 3+3 2 | 5+5 2 | 3+4 2 | 5+6 2 | 2+2 4 |
| | 2+2 4 | 2+2 4 | 2+2 4 | 4+4 2 | 6+6 2 | 5+6 2 | 3+4 2 | 2+2 4 |
| | 2+2 4 | 2+2 4 | 2+2 4 | 3+3 2 | 5+5 2 | 3+4 2 | 3+4 2 | 2+2 4 |
| | 2+2 4 | 2+2 4 | 2+2 4 | 4+4 2 | 6+6 2 | 3+4 2 | 5+6 2 | 2+2 4 |
| | 2+2 4 | 2+2 4 | 2+2 4 | 5+5 2 | 3+3 2 | 5+6 2 | 3+4 2 | 2+2 4 |
| | 2+2 4 | 2+2 4 | 2+2 4 | 6+6 2 | 4+4 2 | 3+4 2 | 5+6 2 | 2+2 4 |
| | 2+2 4 | 2+2 4 | 2+2 4 | 5+5 2 | 4+4 2 | 5+6 2 | 3+4 2 | 2+2 4 |
| | 2+2 4 | 2+2 4 | 2+2 4 | 6+6 2 | 3+3 2 | 5+6 2 | 5+6 2 | 2+2 4 |
| | 2+2 4 | 2+2 4 | 2+2 4 | 3+3 2 | 5+5 2 | 3+4 2 | 3+4 2 | 2+2 4 |
| | 2+2 4 | 2+2 4 | 2+2 4 | 4+4 2 | 6+6 2 | 5+6 2 | 5+6 2 | 2+2 4 |

Continued on next column

**TABLE XXVI– continued**

| TQC | ICSD | NoE | SF | | | | |
|---|---|---|---|---|---|---|---|
| | 2+2 4 | 2+2 4 | 2+2 4 | 5+5 2 | 4+4 2 | 5+6 2 | 3+4 2 | 2+2 4 |
| | 2+2 4 | 2+2 4 | 2+2 4 | 6+6 2 | 3+3 2 | 3+4 2 | 5+6 2 | 2+2 4 |
| | 2+2 4 | 2+2 4 | 2+2 4 | 3+3 2 | 6+6 2 | 5+6 2 | 3+4 2 | 2+2 4 |
| | 2+2 4 | 2+2 4 | 2+2 4 | 4+4 2 | 5+5 2 | 3+4 2 | 3+4 2 | 2+2 4 |
| | 2+2 4 | 2+2 4 | 2+2 4 | 3+3 2 | 5+5 2 | 3+4 2 | 5+6 2 | 2+2 4 |
| | 2+2 4 | 2+2 4 | 2+2 4 | 4+4 2 | 6+6 2 | 3+4 2 | 5+6 2 | 2+2 4 |
| | 2+2 4 | 2+2 4 | 2+2 4 | 5+5 2 | 3+3 2 | 3+4 2 | 3+4 2 | 2+2 4 |
| | | | | 6+6 2 | 4+4 2 | 5+6 2 | 5+6 2 | |
| | | | | 3+3 2 | 5+5 2 | 3+4 2 | 5+6 2 | |
| | | | | 4+4 2 | 6+6 2 | 3+4 2 | 3+4 2 | |
| | | | | 5+5 2 | 3+3 2 | 5+6 2 | 3+4 2 | |
| | | | | 6+6 2 | 4+4 2 | 3+4 2 | 5+6 2 | |
| | | | | 5+5 2 | 3+3 2 | 5+6 2 | 5+6 2 | |
| | | | | 6+6 2 | 4+4 2 | 3+4 2 | 3+4 2 | |
| | | | | 3+3 2 | 5+5 2 | 5+6 2 | 3+4 2 | |
| | | | | 4+4 2 | 6+6 2 | 3+4 2 | 5+6 2 | |
| | | | | 3+3 2 | 3+3 2 | 5+6 2 | 3+4 2 | |
| | | | | 4+4 2 | 4+4 2 | 5+6 2 | 5+6 2 | |
| | | | | 3+3 2 | 5+5 2 | 5+6 2 | 5+6 2 | |
| | | | | 5+5 2 | 6+6 2 | 3+4 2 | 3+4 2 | |
| | | | | 6+6 2 | 5+5 2 | 3+4 2 | 3+4 2 | |
| | | | | 4+4 2 | 6+6 2 | 5+6 2 | 5+6 2 | |
| | | | | 5+5 2 | 3+3 2 | 5+6 2 | 5+6 2 | |
| | | | | 6+6 2 | 4+4 2 | 5+6 2 | 3+4 2 | |
| | | | | 5+5 2 | 3+3 2 | 3+4 2 | 5+6 2 | |
| | | | | 6+6 2 | 4+4 2 | 5+6 2 | 5+6 2 | |
| | | | | 3+3 2 | 5+5 2 | 3+4 2 | 3+4 2 | |
| | | | | 4+4 2 | 6+6 2 | 5+6 2 | 3+4 2 | |
| | | | | 3+3 2 | 5+5 2 | 5+6 2 | 3+4 2 | |
| | | | | 4+4 2 | 6+6 2 | 3+4 2 | 5+6 2 | |
| | | | | 5+5 2 | 3+3 2 | 5+6 2 | 5+6 2 | |
| | | | | 6+6 2 | 4+4 2 | 3+4 2 | 3+4 2 | |
| | | | | 3+3 2 | 5+5 2 | 5+6 2 | 3+4 2 | |
| | | | | 4+4 2 | 3+3 2 | 3+4 2 | 5+6 2 | |
| | | | | 5+5 2 | 6+6 2 | 3+4 2 | 3+4 2 | |
| | | | | 6+6 2 | 4+4 2 | 5+6 2 | 5+6 2 | |
| | | | | 4+4 2 | 4+4 2 | 3+4 2 | 5+6 2 | |
| | | | | 3+3 2 | 3+3 2 | 3+4 2 | 3+4 2 | |
| | | | | 5+5 2 | 6+6 2 | 5+6 2 | 3+4 2 | |
| | | | | 6+6 2 | 5+5 2 | 3+4 2 | 5+6 2 | |
| | | | | 3+3 2 | 6+6 2 | 3+4 2 | 3+4 2 | |
| | | | | 4+4 2 | 4+4 2 | 5+6 2 | 5+6 2 | |
| | | | | 6+6 2 | 3+3 2 | 3+4 2 | 5+6 2 | |
| | | | | 5+5 2 | 5+5 2 | 3+4 2 | 3+4 2 | |
| | | | | 4+4 2 | 5+5 2 | 5+6 2 | 3+4 2 | |
| | | | | 3+3 2 | 3+3 2 | 5+6 2 | 5+6 2 | |
| | | | | 5+5 2 | 6+6 2 | 3+4 2 | 5+6 2 | |
| | | | | 3+3 2 | 4+4 2 | 5+6 2 | 3+4 2 | |
| | | | | 6+6 2 | 5+5 2 | 5+6 2 | 5+6 2 | |
| | | | | 4+4 2 | 6+6 2 | 3+4 2 | 3+4 2 | |
| 14.2.64 | 86862 | 172 | B1Fe2O4 | | | | | |
| | -A | -B | -C | -D | -E | -GM | -Y | -Z |
| | 2+2 4 | 2+2 4 | 2+2 4 | 5+5 2 | 3+3 2 | 3+4 2 | 3+4 2 | 2+2 4 |
| | 2+2 4 | 2+2 4 | 2+2 4 | 6+6 2 | 4+4 2 | 5+6 2 | 5+6 2 | 2+2 4 |
| | 2+2 4 | 2+2 4 | 2+2 4 | 3+3 2 | 5+5 2 | 5+6 2 | 5+6 2 | 2+2 4 |
| | 2+2 4 | 2+2 4 | 2+2 4 | 4+4 2 | 6+6 2 | 3+4 2 | 3+4 2 | 2+2 4 |
| | 2+2 4 | 2+2 4 | 2+2 4 | 3+3 2 | 3+3 2 | 3+4 2 | 5+6 2 | 2+2 4 |
| | 2+2 4 | 2+2 4 | 2+2 4 | 4+4 2 | 4+4 2 | 5+6 2 | 3+4 2 | 2+2 4 |
| | 2+2 4 | 2+2 4 | 2+2 4 | 5+5 2 | 5+5 2 | 5+6 2 | 5+6 2 | 2+2 4 |
| | 2+2 4 | 2+2 4 | 2+2 4 | 6+6 2 | 6+6 2 | 3+4 2 | 3+4 2 | 2+2 4 |
| | 2+2 4 | 2+2 4 | 2+2 4 | 3+3 2 | 3+3 2 | 3+4 2 | 5+6 2 | 2+2 4 |

Continued on next column

**TABLE XXVI– continued**

| TQC | ICSD | NoE | SF | | | | |
|---|---|---|---|---|---|---|---|
| | 2+2 4 | 2+2 4 | 2+2 4 | 4+4 2 | 4+4 2 | 5+6 2 | 3+4 2 | 2+2 4 |
| | 2+2 4 | 2+2 4 | 2+2 4 | 5+5 2 | 5+5 2 | 3+4 2 | 3+4 2 | 2+2 4 |
| | 2+2 4 | 2+2 4 | 2+2 4 | 6+6 2 | 6+6 2 | 5+6 2 | 5+6 2 | 2+2 4 |
| | 2+2 4 | 2+2 4 | 2+2 4 | 3+3 2 | 3+3 2 | 3+4 2 | 3+4 2 | 2+2 4 |
| | 2+2 4 | 2+2 4 | 2+2 4 | 4+4 2 | 4+4 2 | 3+4 2 | 3+4 2 | 2+2 4 |
| | 2+2 4 | 2+2 4 | 2+2 4 | 5+5 2 | 5+5 2 | 5+6 2 | 5+6 2 | 2+2 4 |
| | 2+2 4 | 2+2 4 | 2+2 4 | 6+6 2 | 6+6 2 | 5+6 2 | 5+6 2 | 2+2 4 |
| | 2+2 4 | 2+2 4 | 2+2 4 | 5+5 2 | 5+5 2 | 3+4 2 | 3+4 2 | 2+2 4 |
| | 2+2 4 | 2+2 4 | 2+2 4 | 6+6 2 | 6+6 2 | 5+6 2 | 5+6 2 | 2+2 4 |
| | 2+2 4 | 2+2 4 | 2+2 4 | 3+3 2 | 3+3 2 | 5+6 2 | 5+6 2 | 2+2 4 |
| | 2+2 4 | 2+2 4 | 2+2 4 | 4+4 2 | 4+4 2 | 3+4 2 | 3+4 2 | 2+2 4 |
| | 2+2 4 | 2+2 4 | 2+2 4 | 3+3 2 | 3+3 2 | 5+6 2 | 3+4 2 | 2+2 4 |
| | 2+2 4 | 2+2 4 | 2+2 4 | 4+4 2 | 4+4 2 | 3+4 2 | 3+4 2 | 2+2 4 |
| | 2+2 4 | 2+2 4 | 2+2 4 | 5+5 2 | 5+5 2 | 3+4 2 | 5+6 2 | 2+2 4 |
| | 2+2 4 | 2+2 4 | 2+2 4 | 6+6 2 | 6+6 2 | 5+6 2 | 5+6 2 | 2+2 4 |
| | 2+2 4 | 2+2 4 | 2+2 4 | 5+5 2 | 3+3 2 | 3+4 2 | 3+4 2 | 2+2 4 |
| | 2+2 4 | 2+2 4 | 2+2 4 | 6+6 2 | 4+4 2 | 5+6 2 | 5+6 2 | 2+2 4 |
| | 2+2 4 | 2+2 4 | 2+2 4 | 3+3 2 | 5+5 2 | 3+4 2 | 5+6 2 | 2+2 4 |
| | 2+2 4 | 2+2 4 | 2+2 4 | 4+4 2 | 6+6 2 | 5+6 2 | 3+4 2 | 2+2 4 |
| | 2+2 4 | 2+2 4 | 2+2 4 | 3+3 2 | 5+5 2 | 3+4 2 | 3+4 2 | 2+2 4 |
| | 2+2 4 | 2+2 4 | 2+2 4 | 4+4 2 | 6+6 2 | 3+4 2 | 5+6 2 | 2+2 4 |
| | 2+2 4 | 2+2 4 | 2+2 4 | 5+5 2 | 3+3 2 | 5+6 2 | 3+4 2 | 2+2 4 |
| | 2+2 4 | 2+2 4 | 2+2 4 | 6+6 2 | 4+4 2 | 3+4 2 | 5+6 2 | 2+2 4 |
| | 2+2 4 | 2+2 4 | 2+2 4 | 5+5 2 | 4+4 2 | 5+6 2 | 3+4 2 | 2+2 4 |
| | 2+2 4 | 2+2 4 | 2+2 4 | 6+6 2 | 3+3 2 | 5+6 2 | 5+6 2 | 2+2 4 |
| | 2+2 4 | 2+2 4 | 2+2 4 | 3+3 2 | 5+5 2 | 3+4 2 | 3+4 2 | 2+2 4 |
| | 2+2 4 | 2+2 4 | 2+2 4 | 4+4 2 | 6+6 2 | 5+6 2 | 5+6 2 | 2+2 4 |
| | 2+2 4 | 2+2 4 | 2+2 4 | 5+5 2 | 4+4 2 | 5+6 2 | 3+4 2 | 2+2 4 |
| | 2+2 4 | 2+2 4 | 2+2 4 | 6+6 2 | 3+3 2 | 3+4 2 | 5+6 2 | 2+2 4 |
| | 2+2 4 | 2+2 4 | 2+2 4 | 3+3 2 | 6+6 2 | 5+6 2 | 3+4 2 | 2+2 4 |
| | 2+2 4 | 2+2 4 | 2+2 4 | 4+4 2 | 5+5 2 | 3+4 2 | 3+4 2 | 2+2 4 |
| | 2+2 4 | 2+2 4 | 2+2 4 | 3+3 2 | 5+5 2 | 3+4 2 | 5+6 2 | 2+2 4 |
| | 2+2 4 | 2+2 4 | 2+2 4 | 4+4 2 | 6+6 2 | 3+4 2 | 5+6 2 | 2+2 4 |
| | 2+2 4 | 2+2 4 | 2+2 4 | 5+5 2 | 3+3 2 | 3+4 2 | 3+4 2 | 2+2 4 |
| | | | | 6+6 2 | 4+4 2 | 5+6 2 | 5+6 2 | |
| | | | | 3+3 2 | 5+5 2 | 3+4 2 | 5+6 2 | |
| | | | | 4+4 2 | 6+6 2 | 3+4 2 | 3+4 2 | |
| | | | | 5+5 2 | 3+3 2 | 5+6 2 | 3+4 2 | |
| | | | | 6+6 2 | 4+4 2 | 3+4 2 | 5+6 2 | |
| | | | | 5+5 2 | 3+3 2 | 5+6 2 | 5+6 2 | |
| | | | | 6+6 2 | 4+4 2 | 3+4 2 | 3+4 2 | |
| | | | | 3+3 2 | 5+5 2 | 5+6 2 | 3+4 2 | |
| | | | | 4+4 2 | 6+6 2 | 3+4 2 | 5+6 2 | |
| | | | | 3+3 2 | 3+3 2 | 5+6 2 | 3+4 2 | |
| | | | | 4+4 2 | 4+4 2 | 5+6 2 | 5+6 2 | |
| | | | | 3+3 2 | 5+5 2 | 5+6 2 | 5+6 2 | |
| | | | | 5+5 2 | 6+6 2 | 3+4 2 | 3+4 2 | |
| | | | | 6+6 2 | 5+5 2 | 3+4 2 | 3+4 2 | |
| | | | | 4+4 2 | 6+6 2 | 5+6 2 | 5+6 2 | |
| | | | | 5+5 2 | 3+3 2 | 5+6 2 | 5+6 2 | |
| | | | | 6+6 2 | 4+4 2 | 5+6 2 | 3+4 2 | |
| | | | | 5+5 2 | 3+3 2 | 3+4 2 | 5+6 2 | |
| | | | | 6+6 2 | 4+4 2 | 5+6 2 | 5+6 2 | |
| | | | | 3+3 2 | 5+5 2 | 3+4 2 | 3+4 2 | |
| | | | | 4+4 2 | 6+6 2 | 5+6 2 | 3+4 2 | |
| | | | | 3+3 2 | 5+5 2 | 5+6 2 | 3+4 2 | |
| | | | | 4+4 2 | 6+6 2 | 3+4 2 | 5+6 2 | |
| | | | | 5+5 2 | 3+3 2 | 5+6 2 | 5+6 2 | |
| | | | | 6+6 2 | 4+4 2 | 3+4 2 | 3+4 2 | |
| | | | | 3+3 2 | 5+5 2 | 5+6 2 | 3+4 2 | |
| | | | | 4+4 2 | 3+3 2 | 3+4 2 | 5+6 2 | |

Continued on next column

**TABLE XXVI– continued**

| TQC | ICSD | NoE | SF | | | | |
|---|---|---|---|---|---|---|---|
| | | | | | | | |
| | | | 5+5 2 | 6+6 2 | 3+4 2 | 3+4 2 | |
| | | | 6+6 2 | 4+4 2 | 5+6 2 | 5+6 2 | |
| | | | 4+4 2 | 4+4 2 | 3+4 2 | 5+6 2 | |
| | | | 3+3 2 | 3+3 2 | 3+4 2 | 3+4 2 | |
| | | | 5+5 2 | 6+6 2 | 5+6 2 | 3+4 2 | |
| | | | 6+6 2 | 5+5 2 | 3+4 2 | 5+6 2 | |
| | | | 3+3 2 | 6+6 2 | 3+4 2 | 3+4 2 | |
| | | | 4+4 2 | 4+4 2 | 5+6 2 | 5+6 2 | |
| | | | 6+6 2 | 3+3 2 | 3+4 2 | 5+6 2 | |
| | | | 5+5 2 | 5+5 2 | 3+4 2 | 3+4 2 | |
| | | | 4+4 2 | 5+5 2 | 5+6 2 | 3+4 2 | |
| | | | 3+3 2 | 3+3 2 | 5+6 2 | 5+6 2 | |
| | | | 5+5 2 | 6+6 2 | 3+4 2 | 5+6 2 | |
| | | | 3+3 2 | 4+4 2 | 5+6 2 | 3+4 2 | |
| | | | 6+6 2 | 5+5 2 | 5+6 2 | 5+6 2 | |
| | | | 4+4 2 | 6+6 2 | 3+4 2 | 3+4 2 | |
| 14.2.65 | 237469 | 204 | Au4Ga2Na1 | | | | |
| -A | -B | -C | -D | -E | -GM | -Y | -Z |
| 2+2 4 | 2+2 4 | 2+2 4 | 5+5 2 | 5+5 2 | 3+4 2 | 5+6 2 | 2+2 4 |
| 2+2 4 | 2+2 4 | 2+2 4 | 6+6 2 | 6+6 2 | 3+4 2 | 3+4 2 | 2+2 4 |
| 2+2 4 | 2+2 4 | 2+2 4 | 4+4 2 | 6+6 2 | 3+4 2 | 5+6 2 | 2+2 4 |
| 2+2 4 | 2+2 4 | 2+2 4 | 3+3 2 | 5+5 2 | 5+6 2 | 3+4 2 | 2+2 4 |
| 2+2 4 | 2+2 4 | 2+2 4 | 5+5 2 | 3+3 2 | 3+4 2 | 3+4 2 | 2+2 4 |
| 2+2 4 | 2+2 4 | 2+2 4 | 6+6 2 | 4+4 2 | 5+6 2 | 3+4 2 | 2+2 4 |
| 2+2 4 | 2+2 4 | 2+2 4 | 4+4 2 | 4+4 2 | 5+6 2 | 5+6 2 | 2+2 4 |
| 2+2 4 | 2+2 4 | 2+2 4 | 3+3 2 | 3+3 2 | 5+6 2 | 5+6 2 | 2+2 4 |
| 2+2 4 | 2+2 4 | 2+2 4 | 6+6 2 | 5+5 2 | 5+6 2 | 3+4 2 | 2+2 4 |
| 2+2 4 | 2+2 4 | 2+2 4 | 4+4 2 | 6+6 2 | 5+6 2 | 5+6 2 | 2+2 4 |
| 2+2 4 | 2+2 4 | 2+2 4 | 5+5 2 | 3+3 2 | 3+4 2 | 5+6 2 | 2+2 4 |
| 2+2 4 | 2+2 4 | 2+2 4 | 3+3 2 | 4+4 2 | 3+4 2 | 3+4 2 | 2+2 4 |
| 2+2 4 | 2+2 4 | 2+2 4 | 6+6 2 | 4+4 2 | 3+4 2 | 3+4 2 | 2+2 4 |
| 2+2 4 | 2+2 4 | 2+2 4 | 5+5 2 | 3+3 2 | 5+6 2 | 5+6 2 | 2+2 4 |
| 2+2 4 | 2+2 4 | 2+2 4 | 3+3 2 | 5+5 2 | 3+4 2 | 5+6 2 | 2+2 4 |
| 2+2 4 | 2+2 4 | 2+2 4 | 4+4 2 | 6+6 2 | 5+6 2 | 3+4 2 | 2+2 4 |
| 2+2 4 | 2+2 4 | 2+2 4 | 4+4 2 | 5+5 2 | 5+6 2 | 5+6 2 | 2+2 4 |
| 2+2 4 | 2+2 4 | 2+2 4 | 5+5 2 | 6+6 2 | 3+4 2 | 3+4 2 | 2+2 4 |
| 2+2 4 | 2+2 4 | 2+2 4 | 3+3 2 | 3+3 2 | 3+4 2 | 3+4 2 | 2+2 4 |
| 2+2 4 | 2+2 4 | 2+2 4 | 6+6 2 | 4+4 2 | 5+6 2 | 5+6 2 | 2+2 4 |
| 2+2 4 | 2+2 4 | 2+2 4 | 3+3 2 | 6+6 2 | 3+4 2 | 3+4 2 | 2+2 4 |
| 2+2 4 | 2+2 4 | 2+2 4 | 6+6 2 | 5+5 2 | 5+6 2 | 5+6 2 | 2+2 4 |
| 2+2 4 | 2+2 4 | 2+2 4 | 5+5 2 | 3+3 2 | 3+4 2 | 5+6 2 | 2+2 4 |
| 2+2 4 | 2+2 4 | 2+2 4 | 4+4 2 | 4+4 2 | 5+6 2 | 3+4 2 | 2+2 4 |
| 2+2 4 | 2+2 4 | 2+2 4 | 3+3 2 | 6+6 2 | 3+4 2 | 3+4 2 | 2+2 4 |
| 2+2 4 | 2+2 4 | 2+2 4 | 5+5 2 | 5+5 2 | 3+4 2 | 5+6 2 | 2+2 4 |
| 2+2 4 | 2+2 4 | 2+2 4 | 6+6 2 | 4+4 2 | 5+6 2 | 5+6 2 | 2+2 4 |
| 2+2 4 | 2+2 4 | 2+2 4 | 4+4 2 | 3+3 2 | 3+4 2 | 3+4 2 | 2+2 4 |
| 2+2 4 | 2+2 4 | 2+2 4 | 3+3 2 | 6+6 2 | 5+6 2 | 3+4 2 | 2+2 4 |
| 2+2 4 | 2+2 4 | 2+2 4 | 6+6 2 | 5+5 2 | 3+4 2 | 5+6 2 | 2+2 4 |
| 2+2 4 | 2+2 4 | 2+2 4 | 5+5 2 | 3+3 2 | 5+6 2 | 5+6 2 | 2+2 4 |
| 2+2 4 | 2+2 4 | 2+2 4 | 4+4 2 | 4+4 2 | 5+6 2 | 3+4 2 | 2+2 4 |
| 2+2 4 | 2+2 4 | 2+2 4 | 4+4 2 | 6+6 2 | 5+6 2 | 3+4 2 | 2+2 4 |
| 2+2 4 | 2+2 4 | 2+2 4 | 3+3 2 | 3+3 2 | 3+4 2 | 5+6 2 | 2+2 4 |
| 2+2 4 | 2+2 4 | 2+2 4 | 5+5 2 | 4+4 2 | 3+4 2 | 5+6 2 | 2+2 4 |
| 2+2 4 | 2+2 4 | 2+2 4 | 6+6 2 | 5+5 2 | 5+6 2 | 3+4 2 | 2+2 4 |
| 2+2 4 | 2+2 4 | 2+2 4 | 4+4 2 | 4+4 2 | 3+4 2 | 5+6 2 | 2+2 4 |
| 2+2 4 | 2+2 4 | 2+2 4 | 3+3 2 | 5+5 2 | 5+6 2 | 3+4 2 | 2+2 4 |
| 2+2 4 | 2+2 4 | 2+2 4 | 5+5 2 | 3+3 2 | 3+4 2 | 5+6 2 | 2+2 4 |
| 2+2 4 | 2+2 4 | 2+2 4 | 6+6 2 | 6+6 2 | 5+6 2 | 3+4 2 | 2+2 4 |
| 2+2 4 | 2+2 4 | 2+2 4 | 3+3 2 | 3+3 2 | 3+4 2 | 5+6 2 | 2+2 4 |
| 2+2 4 | 2+2 4 | 2+2 4 | 5+5 2 | 5+5 2 | 5+6 2 | 3+4 2 | 2+2 4 |
| 2+2 4 | 2+2 4 | 2+2 4 | 6+6 2 | 4+4 2 | 3+4 2 | 5+6 2 | 2+2 4 |
| 2+2 4 | 2+2 4 | 2+2 4 | 4+4 2 | 6+6 2 | 5+6 2 | 3+4 2 | 2+2 4 |
| 2+2 4 | 2+2 4 | 2+2 4 | 3+3 2 | 5+5 2 | 3+4 2 | 5+6 2 | 2+2 4 |
| 2+2 4 | 2+2 4 | 2+2 4 | 4+4 2 | 6+6 2 | 5+6 2 | 3+4 2 | 2+2 4 |
| 2+2 4 | 2+2 4 | 2+2 4 | 5+5 2 | 4+4 2 | 3+4 2 | 5+6 2 | 2+2 4 |
| 2+2 4 | 2+2 4 | 2+2 4 | 6+6 2 | 3+3 2 | 5+6 2 | 3+4 2 | 2+2 4 |
| 2+2 4 | 2+2 4 | 2+2 4 | 4+4 2 | 5+5 2 | 3+4 2 | 5+6 2 | 2+2 4 |
| 2+2 4 | 2+2 4 | 2+2 4 | 3+3 2 | 6+6 2 | 5+6 2 | 3+4 2 | 2+2 4 |
| 2+2 4 | 2+2 4 | 2+2 4 | 6+6 2 | 3+3 2 | 3+4 2 | 5+6 2 | 2+2 4 |
| | | | 5+5 2 | 4+4 2 | 5+6 2 | 3+4 2 | |
| | | | 4+4 2 | 6+6 2 | 3+4 2 | 5+6 2 | |
| | | | 5+5 2 | 4+4 2 | 3+4 2 | 3+4 2 | |
| | | | 6+6 2 | 5+5 2 | 5+6 2 | 5+6 2 | |
| | | | 3+3 2 | 3+3 2 | 3+4 2 | 3+4 2 | |
| | | | 3+3 2 | 6+6 2 | 5+6 2 | 5+6 2 | |
| | | | 6+6 2 | 5+5 2 | 3+4 2 | 3+4 2 | |
| | | | 4+4 2 | 4+4 2 | 3+4 2 | 3+4 2 | |
| | | | 5+5 2 | 3+3 2 | 5+6 2 | 5+6 2 | |
| | | | 5+5 2 | 5+5 2 | 3+4 2 | 3+4 2 | |
| | | | 4+4 2 | 6+6 2 | 5+6 2 | 5+6 2 | |
| | | | 6+6 2 | 3+3 2 | 3+4 2 | 3+4 2 | |
| | | | 3+3 2 | 4+4 2 | 5+6 2 | 5+6 2 | |
| | | | 5+5 2 | 6+6 2 | 3+4 2 | 3+4 2 | |
| | | | 6+6 2 | 5+5 2 | 5+6 2 | 5+6 2 | |
| | | | 3+3 2 | 3+3 2 | 5+6 2 | 3+4 2 | |
| | | | 4+4 2 | 4+4 2 | 3+4 2 | 5+6 2 | |
| | | | 6+6 2 | 6+6 2 | 3+4 2 | 3+4 2 | |
| | | | 5+5 2 | 5+5 2 | 5+6 2 | 5+6 2 | |
| | | | 3+3 2 | 3+3 2 | 3+4 2 | 3+4 2 | |
| | | | 4+4 2 | 4+4 2 | 5+6 2 | 3+4 2 | |
| | | | 6+6 2 | 5+5 2 | 5+6 2 | 5+6 2 | |
| | | | 5+5 2 | 3+3 2 | 3+4 2 | 5+6 2 | |
| | | | 4+4 2 | 4+4 2 | 5+6 2 | 5+6 2 | |
| | | | 3+3 2 | 6+6 2 | 3+4 2 | 3+4 2 | |
| | | | 5+5 2 | 6+6 2 | 5+6 2 | 5+6 2 | |
| | | | 6+6 2 | 5+5 2 | 3+4 2 | 3+4 2 | |
| | | | 3+3 2 | 4+4 2 | 5+6 2 | 5+6 2 | |
| | | | 5+5 2 | 3+3 2 | 3+4 2 | 3+4 2 | |
| | | | 4+4 2 | 6+6 2 | 3+4 2 | 5+6 2 | |
| | | | 6+6 2 | 3+3 2 | 5+6 2 | 3+4 2 | |
| | | | 4+4 2 | 5+5 2 | 3+4 2 | 3+4 2 | |
| | | | 3+3 2 | 4+4 2 | 5+6 2 | 5+6 2 | |
| | | | 6+6 2 | 6+6 2 | 5+6 2 | 3+4 2 | |
| | | | 5+5 2 | 3+3 2 | 3+4 2 | 5+6 2 | |
| | | | 3+3 2 | 5+5 2 | 5+6 2 | 3+4 2 | |
| | | | 4+4 2 | 4+4 2 | 5+6 2 | 3+4 2 | |
| | | | 6+6 2 | 4+4 2 | 3+4 2 | 5+6 2 | |
| | | | 5+5 2 | 3+3 2 | 5+6 2 | 5+6 2 | |
| | | | 4+4 2 | 5+5 2 | 3+4 2 | 5+6 2 | |
| | | | 3+3 2 | 6+6 2 | 3+4 2 | 5+6 2 | |
| | | | 6+6 2 | 4+4 2 | 5+6 2 | 3+4 2 | |
| | | | 5+5 2 | 3+3 2 | 5+6 2 | 3+4 2 | |
| | | | 4+4 2 | 5+5 2 | 3+4 2 | 3+4 2 | |
| | | | 3+3 2 | 6+6 2 | 5+6 2 | 3+4 2 | |
| | | | 4+4 2 | 4+4 2 | 3+4 2 | 5+6 2 | |
| | | | 3+3 2 | 5+5 2 | 5+6 2 | 5+6 2 | |
| | | | 6+6 2 | 6+6 2 | 3+4 2 | 3+4 2 | |
| | | | 5+5 2 | 3+3 2 | 5+6 2 | 3+4 2 | |
| | | | 4+4 2 | 6+6 2 | 5+6 2 | 5+6 2 | |
| | | | 3+3 2 | 5+5 2 | 3+4 2 | 5+6 2 | |
| 14.2.66 | 58406 | 108 | Au4Ca5 | | | | |
| -A | -B | -C | -D | -E | -GM | -Y | -Z |

Continued on next column

**TABLE XXVI– continued**

| TQC | ICSD | NoE | SF | | | | | |
|---|---|---|---|---|---|---|---|---|
| | 2+2 4 | 2+2 4 | 2+2 4 | 6+6 2 | 6+6 2 | 3+4 2 | 3+4 2 | 2+2 4 |
| | 2+2 4 | 2+2 4 | 2+2 4 | 5+5 2 | 5+5 2 | 3+4 2 | 5+6 2 | 2+2 4 |
| | 2+2 4 | 2+2 4 | 2+2 4 | 3+3 2 | 5+5 2 | 3+4 2 | 3+4 2 | 2+2 4 |
| | 2+2 4 | 2+2 4 | 2+2 4 | 4+4 2 | 6+6 2 | 3+4 2 | 5+6 2 | 2+2 4 |
| | 2+2 4 | 2+2 4 | 2+2 4 | 4+4 2 | 3+3 2 | 5+6 2 | 5+6 2 | 2+2 4 |
| | 2+2 4 | 2+2 4 | 2+2 4 | 3+3 2 | 4+4 2 | 5+6 2 | 5+6 2 | 2+2 4 |
| | 2+2 4 | 2+2 4 | 2+2 4 | 6+6 2 | 3+3 2 | 3+4 2 | 3+4 2 | 2+2 4 |
| | 2+2 4 | 2+2 4 | 2+2 4 | 5+5 2 | 4+4 2 | 5+6 2 | 3+4 2 | 2+2 4 |
| | 2+2 4 | 2+2 4 | 2+2 4 | 4+4 2 | 3+3 2 | 5+6 2 | 3+4 2 | 2+2 4 |
| | 2+2 4 | 2+2 4 | 2+2 4 | 3+3 2 | 4+4 2 | 3+4 2 | 5+6 2 | 2+2 4 |
| | 2+2 4 | 2+2 4 | 2+2 4 | 5+5 2 | 5+5 2 | 5+6 2 | 5+6 2 | 2+2 4 |
| | 2+2 4 | 2+2 4 | 2+2 4 | 6+6 2 | 6+6 2 | 3+4 2 | 3+4 2 | 2+2 4 |
| | 2+2 4 | 2+2 4 | 2+2 4 | 4+4 2 | 4+4 2 | 5+6 2 | 5+6 2 | 2+2 4 |
| | 2+2 4 | 2+2 4 | 2+2 4 | 5+5 2 | 3+3 2 | 3+4 2 | 3+4 2 | 2+2 4 |
| | 2+2 4 | 2+2 4 | 2+2 4 | 3+3 2 | 5+5 2 | 5+6 2 | 3+4 2 | 2+2 4 |
| | 2+2 4 | 2+2 4 | 2+2 4 | 6+6 2 | 6+6 2 | 3+4 2 | 5+6 2 | 2+2 4 |
| | 2+2 4 | 2+2 4 | 2+2 4 | 4+4 2 | 5+5 2 | 5+6 2 | 3+4 2 | 2+2 4 |
| | 2+2 4 | 2+2 4 | 2+2 4 | 3+3 2 | 6+6 2 | 3+4 2 | 5+6 2 | 2+2 4 |
| | 2+2 4 | 2+2 4 | 2+2 4 | 6+6 2 | 6+6 2 | 3+4 2 | 3+4 2 | 2+2 4 |
| | 2+2 4 | 2+2 4 | 2+2 4 | 5+5 2 | 5+5 2 | 3+4 2 | 5+6 2 | 2+2 4 |
| | 2+2 4 | 2+2 4 | 2+2 4 | 4+4 2 | 4+4 2 | 5+6 2 | 5+6 2 | 2+2 4 |
| | 2+2 4 | 2+2 4 | 2+2 4 | 3+3 2 | 3+3 2 | 5+6 2 | 3+4 2 | 2+2 4 |
| | 2+2 4 | 2+2 4 | 2+2 4 | 5+5 2 | 4+4 2 | 5+6 2 | 5+6 2 | 2+2 4 |
| | 2+2 4 | 2+2 4 | 2+2 4 | 6+6 2 | 3+3 2 | 5+6 2 | 3+4 2 | 2+2 4 |
| | 2+2 4 | 2+2 4 | 2+2 4 | 6+6 2 | 3+3 2 | 3+4 2 | 3+4 2 | 2+2 4 |
| | 2+2 4 | 2+2 4 | 2+2 4 | 3+3 2 | 4+4 2 | 3+4 2 | 5+6 2 | 2+2 4 |
| | 2+2 4 | 2+2 4 | 2+2 4 | 4+4 2 | 5+5 2 | 3+4 2 | 3+4 2 | 2+2 4 |
| | | | | 5+5 2 | 6+6 2 | 5+6 2 | 3+4 2 | |
| | | | | 5+5 2 | 3+3 2 | 5+6 2 | 5+6 2 | |
| | | | | 4+4 2 | 4+4 2 | 3+4 2 | 5+6 2 | |
| | | | | 6+6 2 | 6+6 2 | 5+6 2 | 3+4 2 | |
| | | | | 3+3 2 | 5+5 2 | 3+4 2 | 5+6 2 | |
| | | | | 4+4 2 | 6+6 2 | 5+6 2 | 3+4 2 | |
| | | | | 3+3 2 | 4+4 2 | 3+4 2 | 3+4 2 | |
| | | | | 6+6 2 | 5+5 2 | 5+6 2 | 5+6 2 | |
| | | | | 5+5 2 | 3+3 2 | 5+6 2 | 5+6 2 | |
| | | | | 3+3 2 | 4+4 2 | 3+4 2 | 3+4 2 | |
| | | | | 5+5 2 | 3+3 2 | 5+6 2 | 5+6 2 | |
| | | | | 4+4 2 | 6+6 2 | 3+4 2 | 3+4 2 | |
| | | | | 6+6 2 | 5+5 2 | 3+4 2 | 3+4 2 | |
| | | | | 4+4 2 | 5+5 2 | 5+6 2 | 5+6 2 | |
| | | | | 5+5 2 | 6+6 2 | 3+4 2 | 5+6 2 | |
| | | | | 6+6 2 | 3+3 2 | 3+4 2 | 3+4 2 | |
| | | | | 3+3 2 | 6+6 2 | 3+4 2 | 5+6 2 | |
| | | | | 3+3 2 | 5+5 2 | 5+6 2 | 5+6 2 | |
| | | | | 4+4 2 | 4+4 2 | 5+6 2 | 3+4 2 | |
| | | | | 6+6 2 | 4+4 2 | 5+6 2 | 5+6 2 | |
| | | | | 5+5 2 | 3+3 2 | 5+6 2 | 3+4 2 | |
| | | | | 4+4 2 | 3+3 2 | 5+6 2 | 3+4 2 | |
| | | | | 3+3 2 | 4+4 2 | 3+4 2 | 3+4 2 | |
| | | | | 5+5 2 | 4+4 2 | 5+6 2 | 5+6 2 | |
| | | | | 6+6 2 | 3+3 2 | 3+4 2 | 5+6 2 | |
| | | | | 3+3 2 | 5+5 2 | 3+4 2 | 3+4 2 | |
| | | | | 4+4 2 | 6+6 2 | 3+4 2 | 5+6 2 | |
| 14.2.67 | 74948 | 180 | As2Ba1Cu2O8 | | | | | |
| | -A | -B | -C | -D | -E | -GM | -Y | -Z |
| | 2+2 4 | 2+2 4 | 2+2 4 | 3+3 2 | 3+3 2 | 3+4 2 | 3+4 2 | 2+2 4 |
| | 2+2 4 | 2+2 4 | 2+2 4 | 4+4 2 | 4+4 2 | 3+4 2 | 3+4 2 | 2+2 4 |
| | 2+2 4 | 2+2 4 | 2+2 4 | 5+5 2 | 3+3 2 | 3+4 2 | 5+6 2 | 2+2 4 |
| | 2+2 4 | 2+2 4 | 2+2 4 | 6+6 2 | 4+4 2 | 3+4 2 | 5+6 2 | 2+2 4 |
| | 2+2 4 | 2+2 4 | 2+2 4 | 3+3 2 | 5+5 2 | 5+6 2 | 3+4 2 | 2+2 4 |

Continued on next column

**TABLE XXVI– continued**

| TQC | ICSD | NoE | SF | | | | | |
|---|---|---|---|---|---|---|---|---|
| | 2+2 4 | 2+2 4 | 2+2 4 | 4+4 2 | 6+6 2 | 5+6 2 | 3+4 2 | 2+2 4 |
| | 2+2 4 | 2+2 4 | 2+2 4 | 5+5 2 | 5+5 2 | 3+4 2 | 3+4 2 | 2+2 4 |
| | 2+2 4 | 2+2 4 | 2+2 4 | 6+6 2 | 6+6 2 | 5+6 2 | 3+4 2 | 2+2 4 |
| | 2+2 4 | 2+2 4 | 2+2 4 | 3+3 2 | 3+3 2 | 5+6 2 | 5+6 2 | 2+2 4 |
| | 2+2 4 | 2+2 4 | 2+2 4 | 4+4 2 | 4+4 2 | 3+4 2 | 5+6 2 | 2+2 4 |
| | 2+2 4 | 2+2 4 | 2+2 4 | 6+6 2 | 6+6 2 | 3+4 2 | 5+6 2 | 2+2 4 |
| | 2+2 4 | 2+2 4 | 2+2 4 | 4+4 2 | 5+5 2 | 5+6 2 | 3+4 2 | 2+2 4 |
| | 2+2 4 | 2+2 4 | 2+2 4 | 5+5 2 | 4+4 2 | 5+6 2 | 5+6 2 | 2+2 4 |
| | 2+2 4 | 2+2 4 | 2+2 4 | 3+3 2 | 3+3 2 | 3+4 2 | 3+4 2 | 2+2 4 |
| | 2+2 4 | 2+2 4 | 2+2 4 | 3+3 2 | 3+3 2 | 5+6 2 | 5+6 2 | 2+2 4 |
| | 2+2 4 | 2+2 4 | 2+2 4 | 4+4 2 | 4+4 2 | 5+6 2 | 3+4 2 | 2+2 4 |
| | 2+2 4 | 2+2 4 | 2+2 4 | 5+5 2 | 5+5 2 | 3+4 2 | 5+6 2 | 2+2 4 |
| | 2+2 4 | 2+2 4 | 2+2 4 | 6+6 2 | 6+6 2 | 3+4 2 | 3+4 2 | 2+2 4 |
| | 2+2 4 | 2+2 4 | 2+2 4 | 6+6 2 | 5+5 2 | 5+6 2 | 5+6 2 | 2+2 4 |
| | 2+2 4 | 2+2 4 | 2+2 4 | 5+5 2 | 6+6 2 | 5+6 2 | 5+6 2 | 2+2 4 |
| | 2+2 4 | 2+2 4 | 2+2 4 | 5+5 2 | 6+6 2 | 5+6 2 | 5+6 2 | 2+2 4 |
| | 2+2 4 | 2+2 4 | 2+2 4 | 6+6 2 | 5+5 2 | 5+6 2 | 5+6 2 | 2+2 4 |
| | 2+2 4 | 2+2 4 | 2+2 4 | 5+5 2 | 6+6 2 | 5+6 2 | 5+6 2 | 2+2 4 |
| | 2+2 4 | 2+2 4 | 2+2 4 | 6+6 2 | 5+5 2 | 5+6 2 | 5+6 2 | 2+2 4 |
| | 2+2 4 | 2+2 4 | 2+2 4 | 6+6 2 | 5+5 2 | 3+4 2 | 3+4 2 | 2+2 4 |
| | 2+2 4 | 2+2 4 | 2+2 4 | 5+5 2 | 6+6 2 | 3+4 2 | 3+4 2 | 2+2 4 |
| | 2+2 4 | 2+2 4 | 2+2 4 | 4+4 2 | 4+4 2 | 5+6 2 | 5+6 2 | 2+2 4 |
| | 2+2 4 | 2+2 4 | 2+2 4 | 3+3 2 | 3+3 2 | 5+6 2 | 5+6 2 | 2+2 4 |
| | 2+2 4 | 2+2 4 | 2+2 4 | 4+4 2 | 4+4 2 | 3+4 2 | 5+6 2 | 2+2 4 |
| | 2+2 4 | 2+2 4 | 2+2 4 | 3+3 2 | 3+3 2 | 5+6 2 | 3+4 2 | 2+2 4 |
| | 2+2 4 | 2+2 4 | 2+2 4 | 6+6 2 | 5+5 2 | 3+4 2 | 3+4 2 | 2+2 4 |
| | 2+2 4 | 2+2 4 | 2+2 4 | 5+5 2 | 6+6 2 | 3+4 2 | 5+6 2 | 2+2 4 |
| | 2+2 4 | 2+2 4 | 2+2 4 | 4+4 2 | 3+3 2 | 3+4 2 | 3+4 2 | 2+2 4 |
| | 2+2 4 | 2+2 4 | 2+2 4 | 3+3 2 | 4+4 2 | 5+6 2 | 3+4 2 | 2+2 4 |
| | 2+2 4 | 2+2 4 | 2+2 4 | 5+5 2 | 6+6 2 | 5+6 2 | 5+6 2 | 2+2 4 |
| | 2+2 4 | 2+2 4 | 2+2 4 | 6+6 2 | 5+5 2 | 3+4 2 | 5+6 2 | 2+2 4 |
| | 2+2 4 | 2+2 4 | 2+2 4 | 6+6 2 | 3+3 2 | 5+6 2 | 5+6 2 | 2+2 4 |
| | 2+2 4 | 2+2 4 | 2+2 4 | 5+5 2 | 4+4 2 | 3+4 2 | 5+6 2 | 2+2 4 |
| | 2+2 4 | 2+2 4 | 2+2 4 | 3+3 2 | 5+5 2 | 5+6 2 | 3+4 2 | 2+2 4 |
| | 2+2 4 | 2+2 4 | 2+2 4 | 4+4 2 | 6+6 2 | 5+6 2 | 3+4 2 | 2+2 4 |
| | 2+2 4 | 2+2 4 | 2+2 4 | 5+5 2 | 5+5 2 | 3+4 2 | 3+4 2 | 2+2 4 |
| | 2+2 4 | 2+2 4 | 2+2 4 | 6+6 2 | 6+6 2 | 3+4 2 | 3+4 2 | 2+2 4 |
| | 2+2 4 | 2+2 4 | 2+2 4 | 4+4 2 | 4+4 2 | 5+6 2 | 3+4 2 | 2+2 4 |
| | 2+2 4 | 2+2 4 | 2+2 4 | 3+3 2 | 3+3 2 | 3+4 2 | 5+6 2 | 2+2 4 |
| | 2+2 4 | 2+2 4 | 2+2 4 | 4+4 2 | 5+5 2 | 5+6 2 | 5+6 2 | 2+2 4 |
| | | | | 3+3 2 | 6+6 2 | 5+6 2 | 3+4 2 | |
| | | | | 6+6 2 | 3+3 2 | 5+6 2 | 3+4 2 | |
| | | | | 5+5 2 | 4+4 2 | 3+4 2 | 3+4 2 | |
| | | | | 3+3 2 | 4+4 2 | 3+4 2 | 5+6 2 | |
| | | | | 4+4 2 | 3+3 2 | 3+4 2 | 5+6 2 | |
| | | | | 5+5 2 | 6+6 2 | 3+4 2 | 5+6 2 | |
| | | | | 6+6 2 | 5+5 2 | 5+6 2 | 3+4 2 | |
| | | | | 3+3 2 | 3+3 2 | 5+6 2 | 5+6 2 | |
| | | | | 4+4 2 | 4+4 2 | 3+4 2 | 3+4 2 | |
| | | | | 5+5 2 | 5+5 2 | 3+4 2 | 5+6 2 | |
| | | | | 6+6 2 | 6+6 2 | 3+4 2 | 5+6 2 | |
| | | | | 4+4 2 | 3+3 2 | 5+6 2 | 5+6 2 | |
| | | | | 3+3 2 | 4+4 2 | 5+6 2 | 3+4 2 | |
| | | | | 6+6 2 | 4+4 2 | 5+6 2 | 3+4 2 | |
| | | | | 5+5 2 | 3+3 2 | 3+4 2 | 5+6 2 | |
| | | | | 4+4 2 | 6+6 2 | 5+6 2 | 5+6 2 | |
| | | | | 3+3 2 | 5+5 2 | 3+4 2 | 3+4 2 | |
| | | | | 5+5 2 | 6+6 2 | 5+6 2 | 5+6 2 | |
| | | | | 6+6 2 | 5+5 2 | 5+6 2 | 3+4 2 | |
| | | | | 3+3 2 | 3+3 2 | 5+6 2 | 5+6 2 | |
| | | | | 4+4 2 | 4+4 2 | 3+4 2 | 3+4 2 | |

Continued on next column

**TABLE XXVI– continued**

| TQC | ICSD | | NoE | SF | | | |
|---|---|---|---|---|---|---|---|---|
| | | | | 6+6 2 | 6+6 2 | 5+6 2 | 3+4 2 | |
| | | | | 5+5 2 | 5+5 2 | 3+4 2 | 3+4 2 | |
| | | | | 4+4 2 | 4+4 2 | 3+4 2 | 5+6 2 | |
| | | | | 3+3 2 | 3+3 2 | 5+6 2 | 5+6 2 | |
| | | | | 6+6 2 | 6+6 2 | 3+4 2 | 3+4 2 | |
| | | | | 5+5 2 | 5+5 2 | 5+6 2 | 5+6 2 | |
| | | | | 5+5 2 | 3+3 2 | 3+4 2 | 5+6 2 | |
| | | | | 4+4 2 | 4+4 2 | 3+4 2 | 5+6 2 | |
| | | | | 6+6 2 | 5+5 2 | 5+6 2 | 3+4 2 | |
| | | | | 3+3 2 | 6+6 2 | 3+4 2 | 3+4 2 | |
| | | | | 5+5 2 | 6+6 2 | 3+4 2 | 3+4 2 | |
| | | | | 6+6 2 | 5+5 2 | 5+6 2 | 3+4 2 | |
| | | | | 4+4 2 | 3+3 2 | 3+4 2 | 5+6 2 | |
| | | | | 3+3 2 | 4+4 2 | 3+4 2 | 5+6 2 | |
| | | | | 6+6 2 | 6+6 2 | 5+6 2 | 3+4 2 | |
| | | | | 5+5 2 | 5+5 2 | 5+6 2 | 3+4 2 | |
| | | | | 4+4 2 | 3+3 2 | 3+4 2 | 5+6 2 | |
| | | | | 3+3 2 | 4+4 2 | 3+4 2 | 5+6 2 | |
| | | | | 6+6 2 | 5+5 2 | 5+6 2 | 3+4 2 | |
| | | | | 5+5 2 | 6+6 2 | 5+6 2 | 3+4 2 | |
| | | | | 3+3 2 | 4+4 2 | 5+6 2 | 5+6 2 | |
| | | | | 4+4 2 | 3+3 2 | 5+6 2 | 5+6 2 | |
| | | | | 3+3 2 | 5+5 2 | 5+6 2 | 3+4 2 | |
| | | | | 4+4 2 | 6+6 2 | 3+4 2 | 3+4 2 | |
| 14.2.68 | 73978 | | 148 | As1Fe1O4 | | | | |
| | -A | -B | -C | -D | -E | -GM | -Y | -Z |
| | 2+2 4 | 2+2 4 | 2+2 4 | 3+3 2 | 3+3 2 | 3+4 2 | 3+4 2 | 2+2 4 |
| | 2+2 4 | 2+2 4 | 2+2 4 | 4+4 2 | 4+4 2 | 5+6 2 | 5+6 2 | 2+2 4 |
| | 2+2 4 | 2+2 4 | 2+2 4 | 5+5 2 | 5+5 2 | 3+4 2 | 3+4 2 | 2+2 4 |
| | 2+2 4 | 2+2 4 | 2+2 4 | 6+6 2 | 6+6 2 | 5+6 2 | 5+6 2 | 2+2 4 |
| | 2+2 4 | 2+2 4 | 2+2 4 | 6+6 2 | 3+3 2 | 3+4 2 | 5+6 2 | 2+2 4 |
| | 2+2 4 | 2+2 4 | 2+2 4 | 5+5 2 | 4+4 2 | 3+4 2 | 5+6 2 | 2+2 4 |
| | 2+2 4 | 2+2 4 | 2+2 4 | 3+3 2 | 6+6 2 | 5+6 2 | 3+4 2 | 2+2 4 |
| | 2+2 4 | 2+2 4 | 2+2 4 | 4+4 2 | 5+5 2 | 3+4 2 | 3+4 2 | 2+2 4 |
| | 2+2 4 | 2+2 4 | 2+2 4 | 5+5 2 | 3+3 2 | 5+6 2 | 5+6 2 | 2+2 4 |
| | 2+2 4 | 2+2 4 | 2+2 4 | 6+6 2 | 4+4 2 | 5+6 2 | 3+4 2 | 2+2 4 |
| | 2+2 4 | 2+2 4 | 2+2 4 | 3+3 2 | 6+6 2 | 3+4 2 | 5+6 2 | 2+2 4 |
| | 2+2 4 | 2+2 4 | 2+2 4 | 4+4 2 | 5+5 2 | 3+4 2 | 3+4 2 | 2+2 4 |
| | 2+2 4 | 2+2 4 | 2+2 4 | 4+4 2 | 4+4 2 | 5+6 2 | 5+6 2 | 2+2 4 |
| | 2+2 4 | 2+2 4 | 2+2 4 | 3+3 2 | 3+3 2 | 5+6 2 | 3+4 2 | 2+2 4 |
| | 2+2 4 | 2+2 4 | 2+2 4 | 6+6 2 | 5+5 2 | 5+6 2 | 5+6 2 | 2+2 4 |
| | 2+2 4 | 2+2 4 | 2+2 4 | 5+5 2 | 6+6 2 | 3+4 2 | 3+4 2 | 2+2 4 |
| | 2+2 4 | 2+2 4 | 2+2 4 | 5+5 2 | 4+4 2 | 3+4 2 | 5+6 2 | 2+2 4 |
| | 2+2 4 | 2+2 4 | 2+2 4 | 6+6 2 | 3+3 2 | 3+4 2 | 5+6 2 | 2+2 4 |
| | 2+2 4 | 2+2 4 | 2+2 4 | 3+3 2 | 5+5 2 | 5+6 2 | 3+4 2 | 2+2 4 |
| | 2+2 4 | 2+2 4 | 2+2 4 | 4+4 2 | 6+6 2 | 5+6 2 | 3+4 2 | 2+2 4 |
| | 2+2 4 | 2+2 4 | 2+2 4 | 6+6 2 | 3+3 2 | 3+4 2 | 5+6 2 | 2+2 4 |
| | 2+2 4 | 2+2 4 | 2+2 4 | 5+5 2 | 4+4 2 | 3+4 2 | 3+4 2 | 2+2 4 |
| | 2+2 4 | 2+2 4 | 2+2 4 | 4+4 2 | 5+5 2 | 5+6 2 | 5+6 2 | 2+2 4 |
| | 2+2 4 | 2+2 4 | 2+2 4 | 3+3 2 | 6+6 2 | 3+4 2 | 5+6 2 | 2+2 4 |
| | 2+2 4 | 2+2 4 | 2+2 4 | 5+5 2 | 4+4 2 | 5+6 2 | 3+4 2 | 2+2 4 |
| | 2+2 4 | 2+2 4 | 2+2 4 | 6+6 2 | 3+3 2 | 3+4 2 | 5+6 2 | 2+2 4 |
| | 2+2 4 | 2+2 4 | 2+2 4 | 3+3 2 | 6+6 2 | 5+6 2 | 3+4 2 | 2+2 4 |
| | 2+2 4 | 2+2 4 | 2+2 4 | 4+4 2 | 5+5 2 | 3+4 2 | 5+6 2 | 2+2 4 |
| | 2+2 4 | 2+2 4 | 2+2 4 | 6+6 2 | 3+3 2 | 3+4 2 | 3+4 2 | 2+2 4 |
| | 2+2 4 | 2+2 4 | 2+2 4 | 5+5 2 | 4+4 2 | 5+6 2 | 3+4 2 | 2+2 4 |
| | 2+2 4 | 2+2 4 | 2+2 4 | 3+3 2 | 6+6 2 | 5+6 2 | 5+6 2 | 2+2 4 |
| | 2+2 4 | 2+2 4 | 2+2 4 | 4+4 2 | 5+5 2 | 5+6 2 | 3+4 2 | 2+2 4 |
| | 2+2 4 | 2+2 4 | 2+2 4 | 6+6 2 | 3+3 2 | 5+6 2 | 3+4 2 | 2+2 4 |
| | 2+2 4 | 2+2 4 | 2+2 4 | 5+5 2 | 4+4 2 | 3+4 2 | 5+6 2 | 2+2 4 |
| | 2+2 4 | 2+2 4 | 2+2 4 | 3+3 2 | 6+6 2 | 3+4 2 | 5+6 2 | 2+2 4 |

Continued on next column

**TABLE XXVI– continued**

| TQC | ICSD | | NoE | SF | | | | |
|---|---|---|---|---|---|---|---|---|
| | 2+2 4 | 2+2 4 | 2+2 4 | 4+4 2 | 5+5 2 | 5+6 2 | 3+4 2 | 2+2 4 |
| | 2+2 4 | 2+2 4 | 2+2 4 | 3+3 2 | 3+3 2 | 5+6 2 | 3+4 2 | 2+2 4 |
| | | | | 4+4 2 | 4+4 2 | 3+4 2 | 5+6 2 | |
| | | | | 5+5 2 | 6+6 2 | 3+4 2 | 5+6 2 | |
| | | | | 6+6 2 | 5+5 2 | 5+6 2 | 3+4 2 | |
| | | | | 3+3 2 | 5+5 2 | 3+4 2 | 3+4 2 | |
| | | | | 4+4 2 | 6+6 2 | 3+4 2 | 3+4 2 | |
| | | | | 5+5 2 | 3+3 2 | 3+4 2 | 5+6 2 | |
| | | | | 6+6 2 | 4+4 2 | 3+4 2 | 5+6 2 | |
| | | | | 3+3 2 | 6+6 2 | 5+6 2 | 5+6 2 | |
| | | | | 4+4 2 | 5+5 2 | 5+6 2 | 3+4 2 | |
| | | | | 5+5 2 | 4+4 2 | 3+4 2 | 3+4 2 | |
| | | | | 6+6 2 | 3+3 2 | 5+6 2 | 5+6 2 | |
| | | | | 3+3 2 | 5+5 2 | 5+6 2 | 3+4 2 | |
| | | | | 4+4 2 | 6+6 2 | 5+6 2 | 3+4 2 | |
| | | | | 6+6 2 | 3+3 2 | 3+4 2 | 5+6 2 | |
| | | | | 5+5 2 | 4+4 2 | 5+6 2 | 5+6 2 | |
| | | | | 3+3 2 | 5+5 2 | 5+6 2 | 3+4 2 | |
| | | | | 4+4 2 | 6+6 2 | 5+6 2 | 5+6 2 | |
| | | | | 5+5 2 | 3+3 2 | 5+6 2 | 5+6 2 | |
| | | | | 6+6 2 | 4+4 2 | 3+4 2 | 3+4 2 | |
| | | | | 5+5 2 | 4+4 2 | 3+4 2 | 3+4 2 | |
| | | | | 6+6 2 | 3+3 2 | 5+6 2 | 5+6 2 | |
| | | | | 3+3 2 | 5+5 2 | 3+4 2 | 5+6 2 | |
| | | | | 4+4 2 | 6+6 2 | 3+4 2 | 3+4 2 | |
| | | | | 5+5 2 | 3+3 2 | 5+6 2 | 3+4 2 | |
| | | | | 6+6 2 | 4+4 2 | 5+6 2 | 3+4 2 | |
| | | | | 4+4 2 | 5+5 2 | 3+4 2 | 5+6 2 | |
| | | | | 3+3 2 | 6+6 2 | 3+4 2 | 5+6 2 | |
| | | | | 4+4 2 | 5+5 2 | 3+4 2 | 5+6 2 | |
| | | | | 3+3 2 | 6+6 2 | 3+4 2 | 3+4 2 | |
| | | | | 6+6 2 | 4+4 2 | 5+6 2 | 5+6 2 | |
| | | | | 5+5 2 | 3+3 2 | 3+4 2 | 5+6 2 | |
| | | | | 4+4 2 | 6+6 2 | 5+6 2 | 3+4 2 | |
| | | | | 3+3 2 | 5+5 2 | 3+4 2 | 5+6 2 | |
| | | | | 5+5 2 | 3+3 2 | 3+4 2 | 3+4 2 | |
| | | | | 6+6 2 | 4+4 2 | 5+6 2 | 3+4 2 | |
| | | | | 4+4 2 | 6+6 2 | 5+6 2 | 5+6 2 | |
| | | | | 3+3 2 | 5+5 2 | 5+6 2 | 3+4 2 | |
| 14.2.69 | 73402 | | 112 | Ag2Te1 | | | | |
| | -A | -B | -C | -D | -E | -GM | -Y | -Z |
| | 2+2 4 | 2+2 4 | 2+2 4 | 3+3 2 | 5+5 2 | 3+4 2 | 5+6 2 | 2+2 4 |
| | 2+2 4 | 2+2 4 | 2+2 4 | 4+4 2 | 6+6 2 | 3+4 2 | 3+4 2 | 2+2 4 |
| | 2+2 4 | 2+2 4 | 2+2 4 | 5+5 2 | 4+4 2 | 5+6 2 | 3+4 2 | 2+2 4 |
| | 2+2 4 | 2+2 4 | 2+2 4 | 6+6 2 | 3+3 2 | 5+6 2 | 5+6 2 | 2+2 4 |
| | 2+2 4 | 2+2 4 | 2+2 4 | 6+6 2 | 6+6 2 | 5+6 2 | 3+4 2 | 2+2 4 |
| | 2+2 4 | 2+2 4 | 2+2 4 | 5+5 2 | 5+5 2 | 3+4 2 | 5+6 2 | 2+2 4 |
| | 2+2 4 | 2+2 4 | 2+2 4 | 3+3 2 | 3+3 2 | 3+4 2 | 5+6 2 | 2+2 4 |
| | 2+2 4 | 2+2 4 | 2+2 4 | 4+4 2 | 4+4 2 | 5+6 2 | 3+4 2 | 2+2 4 |
| | 2+2 4 | 2+2 4 | 2+2 4 | 3+3 2 | 4+4 2 | 5+6 2 | 5+6 2 | 2+2 4 |
| | 2+2 4 | 2+2 4 | 2+2 4 | 4+4 2 | 6+6 2 | 3+4 2 | 3+4 2 | 2+2 4 |
| | 2+2 4 | 2+2 4 | 2+2 4 | 6+6 2 | 3+3 2 | 5+6 2 | 3+4 2 | 2+2 4 |
| | 2+2 4 | 2+2 4 | 2+2 4 | 5+5 2 | 5+5 2 | 3+4 2 | 5+6 2 | 2+2 4 |
| | 2+2 4 | 2+2 4 | 2+2 4 | 4+4 2 | 5+5 2 | 5+6 2 | 3+4 2 | 2+2 4 |
| | 2+2 4 | 2+2 4 | 2+2 4 | 3+3 2 | 3+3 2 | 3+4 2 | 5+6 2 | 2+2 4 |
| | 2+2 4 | 2+2 4 | 2+2 4 | 6+6 2 | 4+4 2 | 5+6 2 | 3+4 2 | 2+2 4 |
| | 2+2 4 | 2+2 4 | 2+2 4 | 5+5 2 | 6+6 2 | 3+4 2 | 5+6 2 | 2+2 4 |
| | 2+2 4 | 2+2 4 | 2+2 4 | 3+3 2 | 6+6 2 | 3+4 2 | 3+4 2 | 2+2 4 |
| | 2+2 4 | 2+2 4 | 2+2 4 | 4+4 2 | 3+3 2 | 5+6 2 | 5+6 2 | 2+2 4 |
| | 2+2 4 | 2+2 4 | 2+2 4 | 5+5 2 | 5+5 2 | 3+4 2 | 5+6 2 | 2+2 4 |
| | 2+2 4 | 2+2 4 | 2+2 4 | 6+6 2 | 4+4 2 | 5+6 2 | 3+4 2 | 2+2 4 |

Continued on next column

**TABLE XXVI– continued**

| TQC | ICSD | NoE | SF | | | | | |
|---|---|---|---|---|---|---|---|---|
| | 2+2 4 | 2+2 4 | 2+2 4 | 4+4 2 | 3+3 2 | 3+4 2 | 5+6 2 | 2+2 4 |
| | 2+2 4 | 2+2 4 | 2+2 4 | 3+3 2 | 5+5 2 | 5+6 2 | 3+4 2 | 2+2 4 |
| | 2+2 4 | 2+2 4 | 2+2 4 | 5+5 2 | 4+4 2 | 3+4 2 | 3+4 2 | 2+2 4 |
| | 2+2 4 | 2+2 4 | 2+2 4 | 6+6 2 | 6+6 2 | 5+6 2 | 5+6 2 | 2+2 4 |
| | 2+2 4 | 2+2 4 | 2+2 4 | 4+4 2 | 5+5 2 | 3+4 2 | 5+6 2 | 2+2 4 |
| | 2+2 4 | 2+2 4 | 2+2 4 | 6+6 2 | 6+6 2 | 5+6 2 | 3+4 2 | 2+2 4 |
| | 2+2 4 | 2+2 4 | 2+2 4 | 5+5 2 | 4+4 2 | 3+4 2 | 3+4 2 | 2+2 4 |
| | 2+2 4 | 2+2 4 | 2+2 4 | 3+3 2 | 3+3 2 | 5+6 2 | 5+6 2 | 2+2 4 |
| | | | | 4+4 2 | 4+4 2 | 3+4 2 | 3+4 2 | |
| | | | | 3+3 2 | 3+3 2 | 5+6 2 | 3+4 2 | |
| | | | | 6+6 2 | 6+6 2 | 5+6 2 | 5+6 2 | |
| | | | | 5+5 2 | 5+5 2 | 3+4 2 | 5+6 2 | |
| | | | | 3+3 2 | 5+5 2 | 3+4 2 | 3+4 2 | |
| | | | | 4+4 2 | 4+4 2 | 5+6 2 | 5+6 2 | |
| | | | | 5+5 2 | 6+6 2 | 3+4 2 | 3+4 2 | |
| | | | | 6+6 2 | 3+3 2 | 5+6 2 | 3+4 2 | |
| | | | | 4+4 2 | 5+5 2 | 3+4 2 | 5+6 2 | |
| | | | | 3+3 2 | 6+6 2 | 3+4 2 | 5+6 2 | |
| | | | | 6+6 2 | 3+3 2 | 5+6 2 | 3+4 2 | |
| | | | | 5+5 2 | 4+4 2 | 3+4 2 | 5+6 2 | |
| | | | | 3+3 2 | 3+3 2 | 3+4 2 | 5+6 2 | |
| | | | | 4+4 2 | 6+6 2 | 5+6 2 | 3+4 2 | |
| | | | | 6+6 2 | 5+5 2 | 3+4 2 | 3+4 2 | |
| | | | | 5+5 2 | 4+4 2 | 5+6 2 | 5+6 2 | |
| | | | | 6+6 2 | 3+3 2 | 5+6 2 | 5+6 2 | |
| | | | | 5+5 2 | 4+4 2 | 3+4 2 | 3+4 2 | |
| | | | | 4+4 2 | 6+6 2 | 5+6 2 | 5+6 2 | |
| | | | | 3+3 2 | 5+5 2 | 3+4 2 | 3+4 2 | |
| | | | | 5+5 2 | 5+5 2 | 3+4 2 | 5+6 2 | |
| | | | | 3+3 2 | 6+6 2 | 5+6 2 | 3+4 2 | |
| | | | | 6+6 2 | 3+3 2 | 3+4 2 | 3+4 2 | |
| | | | | 4+4 2 | 4+4 2 | 5+6 2 | 5+6 2 | |
| | | | | 6+6 2 | 6+6 2 | 3+4 2 | 3+4 2 | |
| | | | | 5+5 2 | 5+5 2 | 5+6 2 | 5+6 2 | |
| | | | | 4+4 2 | 3+3 2 | 5+6 2 | 5+6 2 | |
| | | | | 3+3 2 | 4+4 2 | 3+4 2 | 3+4 2 | |
| 14.2.70 | 73230 | 112 | Ag2Te1 | | | | | |
| | -A | -B | -C | -D | -E | -GM | -Y | -Z |
| | 2+2 4 | 2+2 4 | 2+2 4 | 3+3 2 | 5+5 2 | 3+4 2 | 5+6 2 | 2+2 4 |
| | 2+2 4 | 2+2 4 | 2+2 4 | 4+4 2 | 6+6 2 | 3+4 2 | 3+4 2 | 2+2 4 |
| | 2+2 4 | 2+2 4 | 2+2 4 | 5+5 2 | 4+4 2 | 5+6 2 | 3+4 2 | 2+2 4 |
| | 2+2 4 | 2+2 4 | 2+2 4 | 6+6 2 | 3+3 2 | 5+6 2 | 5+6 2 | 2+2 4 |
| | 2+2 4 | 2+2 4 | 2+2 4 | 6+6 2 | 6+6 2 | 5+6 2 | 3+4 2 | 2+2 4 |
| | 2+2 4 | 2+2 4 | 2+2 4 | 5+5 2 | 5+5 2 | 3+4 2 | 5+6 2 | 2+2 4 |
| | 2+2 4 | 2+2 4 | 2+2 4 | 3+3 2 | 3+3 2 | 3+4 2 | 5+6 2 | 2+2 4 |
| | 2+2 4 | 2+2 4 | 2+2 4 | 4+4 2 | 4+4 2 | 5+6 2 | 3+4 2 | 2+2 4 |
| | 2+2 4 | 2+2 4 | 2+2 4 | 3+3 2 | 4+4 2 | 5+6 2 | 5+6 2 | 2+2 4 |
| | 2+2 4 | 2+2 4 | 2+2 4 | 4+4 2 | 6+6 2 | 3+4 2 | 3+4 2 | 2+2 4 |
| | 2+2 4 | 2+2 4 | 2+2 4 | 6+6 2 | 3+3 2 | 5+6 2 | 3+4 2 | 2+2 4 |
| | 2+2 4 | 2+2 4 | 2+2 4 | 5+5 2 | 5+5 2 | 3+4 2 | 5+6 2 | 2+2 4 |
| | 2+2 4 | 2+2 4 | 2+2 4 | 4+4 2 | 5+5 2 | 5+6 2 | 3+4 2 | 2+2 4 |
| | 2+2 4 | 2+2 4 | 2+2 4 | 3+3 2 | 4+4 2 | 3+4 2 | 5+6 2 | 2+2 4 |
| | 2+2 4 | 2+2 4 | 2+2 4 | 6+6 2 | 3+3 2 | 5+6 2 | 3+4 2 | 2+2 4 |
| | 2+2 4 | 2+2 4 | 2+2 4 | 5+5 2 | 6+6 2 | 3+4 2 | 5+6 2 | 2+2 4 |
| | 2+2 4 | 2+2 4 | 2+2 4 | 3+3 2 | 6+6 2 | 3+4 2 | 3+4 2 | 2+2 4 |
| | 2+2 4 | 2+2 4 | 2+2 4 | 4+4 2 | 3+3 2 | 5+6 2 | 5+6 2 | 2+2 4 |
| | 2+2 4 | 2+2 4 | 2+2 4 | 5+5 2 | 5+5 2 | 3+4 2 | 5+6 2 | 2+2 4 |
| | 2+2 4 | 2+2 4 | 2+2 4 | 6+6 2 | 4+4 2 | 5+6 2 | 3+4 2 | 2+2 4 |
| | 2+2 4 | 2+2 4 | 2+2 4 | 4+4 2 | 3+3 2 | 3+4 2 | 5+6 2 | 2+2 4 |
| | 2+2 4 | 2+2 4 | 2+2 4 | 3+3 2 | 5+5 2 | 5+6 2 | 3+4 2 | 2+2 4 |
| | 2+2 4 | 2+2 4 | 2+2 4 | 5+5 2 | 4+4 2 | 3+4 2 | 3+4 2 | 2+2 4 |

Continued on next column

**TABLE XXVI– continued**

| TQC | ICSD | NoE | SF | | | | | |
|---|---|---|---|---|---|---|---|---|
| | 2+2 4 | 2+2 4 | 2+2 4 | 6+6 2 | 6+6 2 | 5+6 2 | 5+6 2 | 2+2 4 |
| | 2+2 4 | 2+2 4 | 2+2 4 | 4+4 2 | 5+5 2 | 3+4 2 | 5+6 2 | 2+2 4 |
| | 2+2 4 | 2+2 4 | 2+2 4 | 6+6 2 | 4+4 2 | 5+6 2 | 3+4 2 | 2+2 4 |
| | 2+2 4 | 2+2 4 | 2+2 4 | 5+5 2 | 6+6 2 | 3+4 2 | 3+4 2 | 2+2 4 |
| | 2+2 4 | 2+2 4 | 2+2 4 | 3+3 2 | 3+3 2 | 5+6 2 | 5+6 2 | 2+2 4 |
| | | | | 4+4 2 | 4+4 2 | 3+4 2 | 3+4 2 | |
| | | | | 3+3 2 | 3+3 2 | 5+6 2 | 3+4 2 | |
| | | | | 5+5 2 | 5+5 2 | 5+6 2 | 5+6 2 | |
| | | | | 6+6 2 | 6+6 2 | 3+4 2 | 5+6 2 | |
| | | | | 3+3 2 | 4+4 2 | 3+4 2 | 3+4 2 | |
| | | | | 4+4 2 | 5+5 2 | 5+6 2 | 3+4 2 | |
| | | | | 5+5 2 | 3+3 2 | 3+4 2 | 5+6 2 | |
| | | | | 6+6 2 | 6+6 2 | 5+6 2 | 3+4 2 | |
| | | | | 4+4 2 | 5+5 2 | 3+4 2 | 5+6 2 | |
| | | | | 3+3 2 | 6+6 2 | 3+4 2 | 5+6 2 | |
| | | | | 6+6 2 | 3+3 2 | 5+6 2 | 3+4 2 | |
| | | | | 5+5 2 | 4+4 2 | 3+4 2 | 5+6 2 | |
| | | | | 3+3 2 | 3+3 2 | 3+4 2 | 5+6 2 | |
| | | | | 4+4 2 | 6+6 2 | 5+6 2 | 3+4 2 | |
| | | | | 6+6 2 | 5+5 2 | 3+4 2 | 3+4 2 | |
| | | | | 5+5 2 | 4+4 2 | 5+6 2 | 5+6 2 | |
| | | | | 6+6 2 | 3+3 2 | 5+6 2 | 5+6 2 | |
| | | | | 5+5 2 | 4+4 2 | 3+4 2 | 3+4 2 | |
| | | | | 4+4 2 | 6+6 2 | 5+6 2 | 5+6 2 | |
| | | | | 3+3 2 | 5+5 2 | 3+4 2 | 3+4 2 | |
| | | | | 5+5 2 | 5+5 2 | 3+4 2 | 5+6 2 | |
| | | | | 3+3 2 | 6+6 2 | 5+6 2 | 3+4 2 | |
| | | | | 6+6 2 | 3+3 2 | 3+4 2 | 3+4 2 | |
| | | | | 4+4 2 | 4+4 2 | 5+6 2 | 5+6 2 | |
| | | | | 6+6 2 | 6+6 2 | 3+4 2 | 3+4 2 | |
| | | | | 5+5 2 | 5+5 2 | 5+6 2 | 5+6 2 | |
| | | | | 4+4 2 | 3+3 2 | 5+6 2 | 5+6 2 | |
| | | | | 3+3 2 | 4+4 2 | 3+4 2 | 3+4 2 | |
| 14.2.71 | 165661 | 180 | Ag1Fe2S3 | | | | | |
| | -A | -B | -C | -D | -E | -GM | -Y | -Z |
| | 2+2 4 | 2+2 4 | 2+2 4 | 3+3 2 | 3+3 2 | 3+4 2 | 3+4 2 | 2+2 4 |
| | 2+2 4 | 2+2 4 | 2+2 4 | 4+4 2 | 4+4 2 | 3+4 2 | 5+6 2 | 2+2 4 |
| | 2+2 4 | 2+2 4 | 2+2 4 | 5+5 2 | 5+5 2 | 5+6 2 | 3+4 2 | 2+2 4 |
| | 2+2 4 | 2+2 4 | 2+2 4 | 6+6 2 | 6+6 2 | 5+6 2 | 5+6 2 | 2+2 4 |
| | 2+2 4 | 2+2 4 | 2+2 4 | 5+5 2 | 3+3 2 | 5+6 2 | 3+4 2 | 2+2 4 |
| | 2+2 4 | 2+2 4 | 2+2 4 | 6+6 2 | 4+4 2 | 3+4 2 | 3+4 2 | 2+2 4 |
| | 2+2 4 | 2+2 4 | 2+2 4 | 3+3 2 | 5+5 2 | 5+6 2 | 5+6 2 | 2+2 4 |
| | 2+2 4 | 2+2 4 | 2+2 4 | 4+4 2 | 6+6 2 | 5+6 2 | 3+4 2 | 2+2 4 |
| | 2+2 4 | 2+2 4 | 2+2 4 | 5+5 2 | 3+3 2 | 3+4 2 | 5+6 2 | 2+2 4 |
| | 2+2 4 | 2+2 4 | 2+2 4 | 6+6 2 | 4+4 2 | 5+6 2 | 3+4 2 | 2+2 4 |
| | 2+2 4 | 2+2 4 | 2+2 4 | 3+3 2 | 5+5 2 | 3+4 2 | 5+6 2 | 2+2 4 |
| | 2+2 4 | 2+2 4 | 2+2 4 | 4+4 2 | 6+6 2 | 3+4 2 | 5+6 2 | 2+2 4 |
| | 2+2 4 | 2+2 4 | 2+2 4 | 4+4 2 | 5+5 2 | 3+4 2 | 5+6 2 | 2+2 4 |
| | 2+2 4 | 2+2 4 | 2+2 4 | 3+3 2 | 6+6 2 | 5+6 2 | 3+4 2 | 2+2 4 |
| | 2+2 4 | 2+2 4 | 2+2 4 | 6+6 2 | 3+3 2 | 5+6 2 | 3+4 2 | 2+2 4 |
| | 2+2 4 | 2+2 4 | 2+2 4 | 5+5 2 | 4+4 2 | 3+4 2 | 5+6 2 | 2+2 4 |
| | 2+2 4 | 2+2 4 | 2+2 4 | 3+3 2 | 4+4 2 | 3+4 2 | 3+4 2 | 2+2 4 |
| | 2+2 4 | 2+2 4 | 2+2 4 | 4+4 2 | 3+3 2 | 5+6 2 | 5+6 2 | 2+2 4 |
| | 2+2 4 | 2+2 4 | 2+2 4 | 5+5 2 | 5+5 2 | 3+4 2 | 5+6 2 | 2+2 4 |
| | 2+2 4 | 2+2 4 | 2+2 4 | 6+6 2 | 6+6 2 | 5+6 2 | 3+4 2 | 2+2 4 |
| | 2+2 4 | 2+2 4 | 2+2 4 | 3+3 2 | 6+6 2 | 3+4 2 | 3+4 2 | 2+2 4 |
| | 2+2 4 | 2+2 4 | 2+2 4 | 4+4 2 | 5+5 2 | 5+6 2 | 5+6 2 | 2+2 4 |
| | 2+2 4 | 2+2 4 | 2+2 4 | 5+5 2 | 3+3 2 | 3+4 2 | 3+4 2 | 2+2 4 |
| | 2+2 4 | 2+2 4 | 2+2 4 | 6+6 2 | 4+4 2 | 3+4 2 | 5+6 2 | 2+2 4 |
| | 2+2 4 | 2+2 4 | 2+2 4 | 6+6 2 | 6+6 2 | 5+6 2 | 5+6 2 | 2+2 4 |
| | 2+2 4 | 2+2 4 | 2+2 4 | 4+4 2 | 3+3 2 | 5+6 2 | 3+4 2 | 2+2 4 |

Continued on next column

**TABLE XXVI– continued**

| TQC | ICSD | | NoE | | SF | | | |
|---|---|---|---|---|---|---|---|---|
| | 2+2 4 | 2+2 4 | 2+2 4 | 5+5 2 | 4+4 2 | 3+4 2 | 5+6 2 | 2+2 4 |
| | 2+2 4 | 2+2 4 | 2+2 4 | 3+3 2 | 5+5 2 | 5+6 2 | 3+4 2 | 2+2 4 |
| | 2+2 4 | 2+2 4 | 2+2 4 | 3+3 2 | 5+5 2 | 3+4 2 | 5+6 2 | 2+2 4 |
| | 2+2 4 | 2+2 4 | 2+2 4 | 4+4 2 | 4+4 2 | 3+4 2 | 3+4 2 | 2+2 4 |
| | 2+2 4 | 2+2 4 | 2+2 4 | 5+5 2 | 6+6 2 | 3+4 2 | 5+6 2 | 2+2 4 |
| | 2+2 4 | 2+2 4 | 2+2 4 | 6+6 2 | 3+3 2 | 5+6 2 | 3+4 2 | 2+2 4 |
| | 2+2 4 | 2+2 4 | 2+2 4 | 5+5 2 | 3+3 2 | 5+6 2 | 5+6 2 | 2+2 4 |
| | 2+2 4 | 2+2 4 | 2+2 4 | 6+6 2 | 4+4 2 | 5+6 2 | 3+4 2 | 2+2 4 |
| | 2+2 4 | 2+2 4 | 2+2 4 | 5+5 2 | 6+6 2 | 3+4 2 | 5+6 2 | 2+2 4 |
| | 2+2 4 | 2+2 4 | 2+2 4 | 6+6 2 | 5+5 2 | 5+6 2 | 3+4 2 | 2+2 4 |
| | 2+2 4 | 2+2 4 | 2+2 4 | 4+4 2 | 3+3 2 | 3+4 2 | 5+6 2 | 2+2 4 |
| | 2+2 4 | 2+2 4 | 2+2 4 | 3+3 2 | 4+4 2 | 5+6 2 | 3+4 2 | 2+2 4 |
| | 2+2 4 | 2+2 4 | 2+2 4 | 3+3 2 | 5+5 2 | 5+6 2 | 3+4 2 | 2+2 4 |
| | 2+2 4 | 2+2 4 | 2+2 4 | 4+4 2 | 6+6 2 | 3+4 2 | 5+6 2 | 2+2 4 |
| | 2+2 4 | 2+2 4 | 2+2 4 | 6+6 2 | 3+3 2 | 5+6 2 | 3+4 2 | 2+2 4 |
| | 2+2 4 | 2+2 4 | 2+2 4 | 5+5 2 | 4+4 2 | 3+4 2 | 5+6 2 | 2+2 4 |
| | 2+2 4 | 2+2 4 | 2+2 4 | 3+3 2 | 5+5 2 | 3+4 2 | 3+4 2 | 2+2 4 |
| | 2+2 4 | 2+2 4 | 2+2 4 | 4+4 2 | 6+6 2 | 5+6 2 | 5+6 2 | 2+2 4 |
| | 2+2 4 | 2+2 4 | 2+2 4 | 5+5 2 | 5+5 2 | 3+4 2 | 3+4 2 | 2+2 4 |
| | | | | 6+6 2 | 4+4 2 | 5+6 2 | 5+6 2 | |
| | | | | 3+3 2 | 6+6 2 | 3+4 2 | 3+4 2 | |
| | | | | 4+4 2 | 3+3 2 | 3+4 2 | 5+6 2 | |
| | | | | 6+6 2 | 6+6 2 | 5+6 2 | 3+4 2 | |
| | | | | 5+5 2 | 5+5 2 | 5+6 2 | 5+6 2 | |
| | | | | 3+3 2 | 3+3 2 | 5+6 2 | 5+6 2 | |
| | | | | 4+4 2 | 4+4 2 | 5+6 2 | 5+6 2 | |
| | | | | 3+3 2 | 5+5 2 | 3+4 2 | 3+4 2 | |
| | | | | 4+4 2 | 6+6 2 | 3+4 2 | 3+4 2 | |
| | | | | 5+5 2 | 4+4 2 | 3+4 2 | 5+6 2 | |
| | | | | 6+6 2 | 3+3 2 | 5+6 2 | 3+4 2 | |
| | | | | 3+3 2 | 6+6 2 | 3+4 2 | 5+6 2 | |
| | | | | 4+4 2 | 5+5 2 | 3+4 2 | 3+4 2 | |
| | | | | 6+6 2 | 4+4 2 | 5+6 2 | 5+6 2 | |
| | | | | 5+5 2 | 3+3 2 | 5+6 2 | 3+4 2 | |
| | | | | 4+4 2 | 5+5 2 | 3+4 2 | 5+6 2 | |
| | | | | 5+5 2 | 6+6 2 | 5+6 2 | 5+6 2 | |
| | | | | 3+3 2 | 4+4 2 | 3+4 2 | 3+4 2 | |
| | | | | 6+6 2 | 3+3 2 | 5+6 2 | 5+6 2 | |
| | | | | 3+3 2 | 6+6 2 | 3+4 2 | 3+4 2 | |
| | | | | 4+4 2 | 5+5 2 | 5+6 2 | 3+4 2 | |
| | | | | 5+5 2 | 3+3 2 | 5+6 2 | 3+4 2 | |
| | | | | 6+6 2 | 4+4 2 | 3+4 2 | 5+6 2 | |
| | | | | 3+3 2 | 3+3 2 | 3+4 2 | 3+4 2 | |
| | | | | 4+4 2 | 4+4 2 | 3+4 2 | 3+4 2 | |
| | | | | 5+5 2 | 5+5 2 | 5+6 2 | 5+6 2 | |
| | | | | 6+6 2 | 6+6 2 | 5+6 2 | 5+6 2 | |
| | | | | 4+4 2 | 3+3 2 | 3+4 2 | 3+4 2 | |
| | | | | 3+3 2 | 4+4 2 | 3+4 2 | 5+6 2 | |
| | | | | 3+3 2 | 5+5 2 | 5+6 2 | 3+4 2 | |
| | | | | 6+6 2 | 6+6 2 | 5+6 2 | 5+6 2 | |
| | | | | 5+5 2 | 3+3 2 | 3+4 2 | 3+4 2 | |
| | | | | 4+4 2 | 4+4 2 | 3+4 2 | 3+4 2 | |
| | | | | 5+5 2 | 6+6 2 | 5+6 2 | 5+6 2 | |
| | | | | 6+6 2 | 5+5 2 | 3+4 2 | 5+6 2 | |
| | | | | 4+4 2 | 4+4 2 | 5+6 2 | 3+4 2 | |
| | | | | 3+3 2 | 3+3 2 | 3+4 2 | 3+4 2 | |
| | | | | 6+6 2 | 6+6 2 | 3+4 2 | 5+6 2 | |
| | | | | 5+5 2 | 5+5 2 | 5+6 2 | 5+6 2 | |
| | | | | 3+3 2 | 3+3 2 | 5+6 2 | 3+4 2 | |
| | | | | 4+4 2 | 4+4 2 | 5+6 2 | 3+4 2 | |
| | | | | 5+5 2 | 4+4 2 | 3+4 2 | 5+6 2 | |

Continued on next column

**TABLE XXVI– continued**

| TQC | ICSD | | NoE | | SF | | | |
|---|---|---|---|---|---|---|---|---|
| | | | | 6+6 2 | 6+6 2 | 5+6 2 | 3+4 2 | |
| | | | | 6+6 2 | 5+5 2 | 5+6 2 | 5+6 2 | |
| | | | | 5+5 2 | 3+3 2 | 3+4 2 | 3+4 2 | |

The End

TABLE XXVII: The computed irreducible representations for the valence bands for the compounds of SEBR and NLC in space group 15. See the list of tables.

| TQC | ICSD | NoE | SF | | |
|---|---|---|---|---|---|
| 15.2.1 | 187497 | 56 | Sb2Te3 | | |
| | -A | -GM | -M | -Y | -L | -V |
| | 2+2 4 | 3+4 2 | 2+2 4 | 5+6 2 | 2+2 2 | 2+2 2 |
| | 2+2 4 | 3+4 2 | 2+2 4 | 3+4 2 | 2+2 2 | 3+3 2 |
| | 2+2 4 | 5+6 2 | 2+2 4 | 5+6 2 | 3+3 2 | 2+2 2 |
| | 2+2 4 | 3+4 2 | 2+2 4 | 3+4 2 | 3+3 2 | 3+3 2 |
| | 2+2 4 | 5+6 2 | 2+2 4 | 5+6 2 | 2+2 2 | 3+3 2 |
| | 2+2 4 | 5+6 2 | 2+2 4 | 3+4 2 | 3+3 2 | 2+2 2 |
| | 2+2 4 | 3+4 2 | 2+2 4 | 3+4 2 | 3+3 2 | 2+2 2 |
| | 2+2 4 | 5+6 2 | 2+2 4 | 5+6 2 | 2+2 2 | 3+3 2 |
| | 2+2 4 | 5+6 2 | 2+2 4 | 3+4 2 | 3+3 2 | 3+3 2 |
| | 2+2 4 | 3+4 2 | 2+2 4 | 5+6 2 | 2+2 2 | 2+2 2 |
| | 2+2 4 | 5+6 2 | 2+2 4 | 5+6 2 | 2+2 2 | 3+3 2 |
| | 2+2 4 | 5+6 2 | 2+2 4 | 3+4 2 | 3+3 2 | 2+2 2 |
| | 2+2 4 | 3+4 2 | 2+2 4 | 3+4 2 | 3+3 2 | 3+3 2 |
| | 2+2 4 | 3+4 2 | 2+2 4 | 5+6 2 | 2+2 2 | 2+2 2 |
| | | 3+4 2 | | 5+6 2 | 2+2 2 | 2+2 2 |
| | | 3+4 2 | | 3+4 2 | 3+3 2 | 3+3 2 |
| | | 5+6 2 | | 5+6 2 | 2+2 2 | 2+2 2 |
| | | 3+4 2 | | 3+4 2 | 3+3 2 | 3+3 2 |
| | | 5+6 2 | | 3+4 2 | 2+2 2 | 3+3 2 |
| | | 5+6 2 | | 5+6 2 | 3+3 2 | 2+2 2 |
| | | 3+4 2 | | 3+4 2 | 2+2 2 | 3+3 2 |
| | | 3+4 2 | | 5+6 2 | 3+3 2 | 2+2 2 |
| | | 5+6 2 | | 5+6 2 | 2+2 2 | 2+2 2 |
| | | 3+4 2 | | 3+4 2 | 3+3 2 | 3+3 2 |
| | | 3+4 2 | | 5+6 2 | 2+2 2 | 2+2 2 |
| | | 5+6 2 | | 3+4 2 | 2+2 2 | 3+3 2 |
| | | 3+4 2 | | 5+6 2 | 3+3 2 | 3+3 2 |
| | | 5+6 2 | | 5+6 2 | 3+3 2 | 3+3 2 |
| 15.2.2 | 185954 | 56 | Sb2Te3 | | |
| | -A | -GM | -M | -Y | -L | -V |
| | 2+2 4 | 3+4 2 | 2+2 4 | 3+4 2 | 2+2 2 | 3+3 2 |
| | 2+2 4 | 3+4 2 | 2+2 4 | 5+6 2 | 3+3 2 | 2+2 2 |
| | 2+2 4 | 5+6 2 | 2+2 4 | 5+6 2 | 2+2 2 | 2+2 2 |
| | 2+2 4 | 5+6 2 | 2+2 4 | 5+6 2 | 3+3 2 | 3+3 2 |
| | 2+2 4 | 3+4 2 | 2+2 4 | 3+4 2 | 2+2 2 | 3+3 2 |
| | 2+2 4 | 5+6 2 | 2+2 4 | 3+4 2 | 3+3 2 | 2+2 2 |
| | 2+2 4 | 3+4 2 | 2+2 4 | 3+4 2 | 3+3 2 | 2+2 2 |
| | 2+2 4 | 5+6 2 | 2+2 4 | 5+6 2 | 2+2 2 | 3+3 2 |
| | 2+2 4 | 3+4 2 | 2+2 4 | 3+4 2 | 3+3 2 | 3+3 2 |
| | 2+2 4 | 5+6 2 | 2+2 4 | 5+6 2 | 2+2 2 | 2+2 2 |
| | 2+2 4 | 3+4 2 | 2+2 4 | 5+6 2 | 2+2 2 | 3+3 2 |
| | 2+2 4 | 5+6 2 | 2+2 4 | 3+4 2 | 3+3 2 | 2+2 2 |
| | 2+2 4 | 5+6 2 | 2+2 4 | 5+6 2 | 2+2 2 | 2+2 2 |
| | 2+2 4 | 3+4 2 | 2+2 4 | 3+4 2 | 3+3 2 | 2+2 2 |
| | | 3+4 2 | | 3+4 2 | 2+2 2 | 3+3 2 |
| | | 5+6 2 | | 5+6 2 | 3+3 2 | 3+3 2 |
| | | 3+4 2 | | 3+4 2 | 2+2 2 | 3+3 2 |
| | | 5+6 2 | | 3+4 2 | 3+3 2 | 2+2 2 |

Continued on next column

**TABLE XXVII– continued**

| TQC | ICSD | | NoE | SF | | |
|---|---|---|---|---|---|---|
| | | 3+4 2 | | 5+6 2 | 3+3 2 | 3+3 2 |
| | | 3+4 2 | | 5+6 2 | 2+2 2 | 3+3 2 |
| | | 5+6 2 | | 3+4 2 | 2+2 2 | 2+2 2 |
| | | 3+4 2 | | 5+6 2 | 3+3 2 | 3+3 2 |
| | | 5+6 2 | | 3+4 2 | 2+2 2 | 2+2 2 |
| | | 3+4 2 | | 5+6 2 | 3+3 2 | 3+3 2 |
| | | 3+4 2 | | 3+4 2 | 2+2 2 | 2+2 2 |
| | | 5+6 2 | | 5+6 2 | 3+3 2 | 2+2 2 |
| | | 3+4 2 | | 5+6 2 | 2+2 2 | 3+3 2 |
| | | 5+6 2 | | 3+4 2 | 3+3 2 | 3+3 2 |
| 15.2.3 | 16041 | | 144 | S1Ta6 | | |
| | -A | -GM | -M | -Y | -L | -V |
| | 2+2 4 | 3+4 2 | 2+2 4 | 5+6 2 | 3+3 2 | 3+3 2 |
| | 2+2 4 | 3+4 2 | 2+2 4 | 3+4 2 | 2+2 2 | 3+3 2 |
| | 2+2 4 | 5+6 2 | 2+2 4 | 5+6 2 | 3+3 2 | 2+2 2 |
| | 2+2 4 | 5+6 2 | 2+2 4 | 3+4 2 | 2+2 2 | 2+2 2 |
| | 2+2 4 | 5+6 2 | 2+2 4 | 3+4 2 | 3+3 2 | 2+2 2 |
| | 2+2 4 | 3+4 2 | 2+2 4 | 5+6 2 | 2+2 2 | 3+3 2 |
| | 2+2 4 | 5+6 2 | 2+2 4 | 3+4 2 | 3+3 2 | 3+3 2 |
| | 2+2 4 | 3+4 2 | 2+2 4 | 5+6 2 | 2+2 2 | 2+2 2 |
| | 2+2 4 | 3+4 2 | 2+2 4 | 3+4 2 | 2+2 2 | 2+2 2 |
| | 2+2 4 | 5+6 2 | 2+2 4 | 5+6 2 | 2+2 2 | 3+3 2 |
| | 2+2 4 | 3+4 2 | 2+2 4 | 3+4 2 | 3+3 2 | 3+3 2 |
| | 2+2 4 | 5+6 2 | 2+2 4 | 5+6 2 | 3+3 2 | 2+2 2 |
| | 2+2 4 | 3+4 2 | 2+2 4 | 5+6 2 | 3+3 2 | 2+2 2 |
| | 2+2 4 | 5+6 2 | 2+2 4 | 5+6 2 | 2+2 2 | 3+3 2 |
| | 2+2 4 | 3+4 2 | 2+2 4 | 3+4 2 | 2+2 2 | 3+3 2 |
| | 2+2 4 | 5+6 2 | 2+2 4 | 3+4 2 | 3+3 2 | 2+2 2 |
| | 2+2 4 | 3+4 2 | 2+2 4 | 3+4 2 | 3+3 2 | 3+3 2 |
| | 2+2 4 | 5+6 2 | 2+2 4 | 5+6 2 | 2+2 2 | 3+3 2 |
| | 2+2 4 | 3+4 2 | 2+2 4 | 3+4 2 | 2+2 2 | 2+2 2 |
| | 2+2 4 | 5+6 2 | 2+2 4 | 5+6 2 | 3+3 2 | 2+2 2 |
| | 2+2 4 | 3+4 2 | 2+2 4 | 5+6 2 | 2+2 2 | 3+3 2 |
| | 2+2 4 | 5+6 2 | 2+2 4 | 3+4 2 | 2+2 2 | 2+2 2 |
| | 2+2 4 | 5+6 2 | 2+2 4 | 3+4 2 | 3+3 2 | 2+2 2 |
| | 2+2 4 | 3+4 2 | 2+2 4 | 3+4 2 | 3+3 2 | 3+3 2 |
| | 2+2 4 | 5+6 2 | 2+2 4 | 5+6 2 | 2+2 2 | 2+2 2 |
| | 2+2 4 | 3+4 2 | 2+2 4 | 5+6 2 | 2+2 2 | 2+2 2 |
| | 2+2 4 | 5+6 2 | 2+2 4 | 3+4 2 | 3+3 2 | 3+3 2 |
| | 2+2 4 | 3+4 2 | 2+2 4 | 5+6 2 | 3+3 2 | 3+3 2 |
| | 2+2 4 | 5+6 2 | 2+2 4 | 5+6 2 | 2+2 2 | 2+2 2 |
| | 2+2 4 | 5+6 2 | 2+2 4 | 3+4 2 | 3+3 2 | 2+2 2 |
| | 2+2 4 | 5+6 2 | 2+2 4 | 5+6 2 | 3+3 2 | 3+3 2 |
| | 2+2 4 | 3+4 2 | 2+2 4 | 3+4 2 | 2+2 2 | 3+3 2 |
| | 2+2 4 | 3+4 2 | 2+2 4 | 3+4 2 | 2+2 2 | 2+2 2 |
| | 2+2 4 | 5+6 2 | 2+2 4 | 5+6 2 | 2+2 2 | 2+2 2 |
| | 2+2 4 | 3+4 2 | 2+2 4 | 3+4 2 | 3+3 2 | 3+3 2 |
| | 2+2 4 | 3+4 2 | 2+2 4 | 5+6 2 | 3+3 2 | 3+3 2 |
| | | 3+4 2 | | 5+6 2 | 2+2 2 | 2+2 2 |
| | | 3+4 2 | | 3+4 2 | 3+3 2 | 3+3 2 |
| | | 5+6 2 | | 5+6 2 | 3+3 2 | 3+3 2 |
| | | 5+6 2 | | 3+4 2 | 3+3 2 | 2+2 2 |
| | | 3+4 2 | | 3+4 2 | 2+2 2 | 2+2 2 |
| | | 3+4 2 | | 5+6 2 | 2+2 2 | 2+2 2 |
| | | 3+4 2 | | 5+6 2 | 2+2 2 | 2+2 2 |
| | | 5+6 2 | | 3+4 2 | 3+3 2 | 3+3 2 |
| | | 5+6 2 | | 5+6 2 | 2+2 2 | 2+2 2 |
| | | 5+6 2 | | 3+4 2 | 3+3 2 | 3+3 2 |
| | | 5+6 2 | | 3+4 2 | 2+2 2 | 2+2 2 |
| | | 3+4 2 | | 5+6 2 | 3+3 2 | 3+3 2 |
| | | 5+6 2 | | 5+6 2 | 2+2 2 | 3+3 2 |

Continued on next column

**TABLE XXVII– continued**

| TQC | ICSD | | NoE | SF | | |
|---|---|---|---|---|---|---|
| | | 3+4 2 | | 3+4 2 | 3+3 2 | 2+2 2 |
| | | 3+4 2 | | 3+4 2 | 3+3 2 | 2+2 2 |
| | | 3+4 2 | | 5+6 2 | 2+2 2 | 3+3 2 |
| | | 3+4 2 | | 3+4 2 | 2+2 2 | 3+3 2 |
| | | 5+6 2 | | 5+6 2 | 3+3 2 | 2+2 2 |
| | | 3+4 2 | | 3+4 2 | 3+3 2 | 3+3 2 |
| | | 5+6 2 | | 5+6 2 | 2+2 2 | 2+2 2 |
| | | 5+6 2 | | 5+6 2 | 3+3 2 | 3+3 2 |
| | | 5+6 2 | | 3+4 2 | 2+2 2 | 3+3 2 |
| | | 3+4 2 | | 3+4 2 | 3+3 2 | 2+2 2 |
| | | 5+6 2 | | 5+6 2 | 2+2 2 | 2+2 2 |
| | | 3+4 2 | | 3+4 2 | 2+2 2 | 3+3 2 |
| | | 3+4 2 | | 5+6 2 | 3+3 2 | 3+3 2 |
| | | 3+4 2 | | 5+6 2 | 3+3 2 | 2+2 2 |
| | | 3+4 2 | | 3+4 2 | 2+2 2 | 2+2 2 |
| | | 5+6 2 | | 5+6 2 | 3+3 2 | 3+3 2 |
| | | 3+4 2 | | 3+4 2 | 2+2 2 | 3+3 2 |
| | | 5+6 2 | | 3+4 2 | 2+2 2 | 2+2 2 |
| | | 3+4 2 | | 3+4 2 | 3+3 2 | 3+3 2 |
| | | 5+6 2 | | 5+6 2 | 2+2 2 | 2+2 2 |
| | | 5+6 2 | | 5+6 2 | 3+3 2 | 3+3 2 |
| | | 5+6 2 | | 3+4 2 | 3+3 2 | 3+3 2 |
| | | 3+4 2 | | 5+6 2 | 2+2 2 | 2+2 2 |
| 15.2.4 | 24327 | | 120 | P2Pt5 | | |
| | -A | -GM | -M | -Y | -L | -V |
| | 2+2 4 | 3+4 2 | 2+2 4 | 5+6 2 | 3+3 2 | 3+3 2 |
| | 2+2 4 | 5+6 2 | 2+2 4 | 3+4 2 | 2+2 2 | 2+2 2 |
| | 2+2 4 | 5+6 2 | 2+2 4 | 3+4 2 | 3+3 2 | 2+2 2 |
| | 2+2 4 | 3+4 2 | 2+2 4 | 5+6 2 | 2+2 2 | 3+3 2 |
| | 2+2 4 | 3+4 2 | 2+2 4 | 5+6 2 | 3+3 2 | 2+2 2 |
| | 2+2 4 | 5+6 2 | 2+2 4 | 5+6 2 | 2+2 2 | 3+3 2 |
| | 2+2 4 | 3+4 2 | 2+2 4 | 3+4 2 | 3+3 2 | 3+3 2 |
| | 2+2 4 | 5+6 2 | 2+2 4 | 3+4 2 | 2+2 2 | 2+2 2 |
| | 2+2 4 | 5+6 2 | 2+2 4 | 3+4 2 | 2+2 2 | 2+2 2 |
| | 2+2 4 | 3+4 2 | 2+2 4 | 5+6 2 | 3+3 2 | 3+3 2 |
| | 2+2 4 | 5+6 2 | 2+2 4 | 3+4 2 | 2+2 2 | 2+2 2 |
| | 2+2 4 | 5+6 2 | 2+2 4 | 5+6 2 | 3+3 2 | 3+3 2 |
| | 2+2 4 | 3+4 2 | 2+2 4 | 3+4 2 | 2+2 2 | 3+3 2 |
| | 2+2 4 | 3+4 2 | 2+2 4 | 3+4 2 | 2+2 2 | 2+2 2 |
| | 2+2 4 | 5+6 2 | 2+2 4 | 5+6 2 | 3+3 2 | 2+2 2 |
| | 2+2 4 | 3+4 2 | 2+2 4 | 3+4 2 | 3+3 2 | 3+3 2 |
| | 2+2 4 | 3+4 2 | 2+2 4 | 3+4 2 | 2+2 2 | 3+3 2 |
| | 2+2 4 | 5+6 2 | 2+2 4 | 5+6 2 | 3+3 2 | 2+2 2 |
| | 2+2 4 | 5+6 2 | 2+2 4 | 5+6 2 | 2+2 2 | 2+2 2 |
| | 2+2 4 | 3+4 2 | 2+2 4 | 3+4 2 | 3+3 2 | 3+3 2 |
| | 2+2 4 | 3+4 2 | 2+2 4 | 5+6 2 | 3+3 2 | 2+2 2 |
| | 2+2 4 | 3+4 2 | 2+2 4 | 3+4 2 | 2+2 2 | 3+3 2 |
| | 2+2 4 | 5+6 2 | 2+2 4 | 5+6 2 | 2+2 2 | 2+2 2 |
| | 2+2 4 | 3+4 2 | 2+2 4 | 3+4 2 | 3+3 2 | 3+3 2 |
| | 2+2 4 | 5+6 2 | 2+2 4 | 3+4 2 | 3+3 2 | 2+2 2 |
| | 2+2 4 | 3+4 2 | 2+2 4 | 5+6 2 | 2+2 2 | 3+3 2 |
| | 2+2 4 | 3+4 2 | 2+2 4 | 5+6 2 | 2+2 2 | 2+2 2 |
| | 2+2 4 | 5+6 2 | 2+2 4 | 3+4 2 | 3+3 2 | 3+3 2 |
| | 2+2 4 | 5+6 2 | 2+2 4 | 5+6 2 | 3+3 2 | 2+2 2 |
| | 2+2 4 | 3+4 2 | 2+2 4 | 3+4 2 | 2+2 2 | 3+3 2 |
| | | 3+4 2 | | 5+6 2 | 3+3 2 | 2+2 2 |
| | | 5+6 2 | | 3+4 2 | 2+2 2 | 3+3 2 |
| | | 5+6 2 | | 5+6 2 | 3+3 2 | 3+3 2 |
| | | 5+6 2 | | 3+4 2 | 3+3 2 | 2+2 2 |
| | | 3+4 2 | | 5+6 2 | 2+2 2 | 3+3 2 |
| | | 3+4 2 | | 5+6 2 | 3+3 2 | 3+3 2 |

Continued on next column

**TABLE XXVII– continued**

| TQC | ICSD | | NoE | SF | | |
|---|---|---|---|---|---|---|
| | | 5+6 2 | | 5+6 2 | 2+2 2 | 2+2 2 |
| | | 3+4 2 | | 3+4 2 | 2+2 2 | 2+2 2 |
| | | 5+6 2 | | 5+6 2 | 3+3 2 | 3+3 2 |
| | | 5+6 2 | | 3+4 2 | 2+2 2 | 2+2 2 |
| | | 3+4 2 | | 3+4 2 | 3+3 2 | 3+3 2 |
| | | 3+4 2 | | 5+6 2 | 3+3 2 | 3+3 2 |
| | | 5+6 2 | | 5+6 2 | 2+2 2 | 2+2 2 |
| | | 5+6 2 | | 3+4 2 | 2+2 2 | 3+3 2 |
| | | 3+4 2 | | 5+6 2 | 3+3 2 | 2+2 2 |
| | | 3+4 2 | | 5+6 2 | 2+2 2 | 2+2 2 |
| | | 5+6 2 | | 3+4 2 | 3+3 2 | 3+3 2 |
| | | 3+4 2 | | 5+6 2 | 3+3 2 | 2+2 2 |
| | | 5+6 2 | | 3+4 2 | 2+2 2 | 3+3 2 |
| | | 5+6 2 | | 3+4 2 | 2+2 2 | 2+2 2 |
| | | 3+4 2 | | 5+6 2 | 3+3 2 | 3+3 2 |
| | | 5+6 2 | | 5+6 2 | 2+2 2 | 2+2 2 |
| | | 3+4 2 | | 3+4 2 | 3+3 2 | 3+3 2 |
| | | 5+6 2 | | 3+4 2 | 2+2 2 | 2+2 2 |
| | | 3+4 2 | | 5+6 2 | 2+2 2 | 3+3 2 |
| | | 3+4 2 | | 3+4 2 | 3+3 2 | 2+2 2 |
| | | 3+4 2 | | 3+4 2 | 3+3 2 | 2+2 2 |
| | | 5+6 2 | | 5+6 2 | 2+2 2 | 3+3 2 |
| | | 3+4 2 | | 3+4 2 | 3+3 2 | 2+2 2 |
| | | 3+4 2 | | 5+6 2 | 2+2 2 | 2+2 2 |
| 15.2.5 | 261031 | | 144 | O6Pb2Sb1Tm1 | | |
| | -A | -GM | -M | -Y | -L | -V |
| | 2+2 4 | 3+4 2 | 2+2 4 | 3+4 2 | 2+2 2 | 2+2 2 |
| | 2+2 4 | 5+6 2 | 2+2 4 | 5+6 2 | 3+3 2 | 3+3 2 |
| | 2+2 4 | 3+4 2 | 2+2 4 | 3+4 2 | 2+2 2 | 2+2 2 |
| | 2+2 4 | 5+6 2 | 2+2 4 | 5+6 2 | 3+3 2 | 3+3 2 |
| | 2+2 4 | 3+4 2 | 2+2 4 | 5+6 2 | 2+2 2 | 2+2 2 |
| | 2+2 4 | 5+6 2 | 2+2 4 | 3+4 2 | 3+3 2 | 3+3 2 |
| | 2+2 4 | 5+6 2 | 2+2 4 | 3+4 2 | 3+3 2 | 2+2 2 |
| | 2+2 4 | 3+4 2 | 2+2 4 | 5+6 2 | 2+2 2 | 3+3 2 |
| | 2+2 4 | 3+4 2 | 2+2 4 | 5+6 2 | 2+2 2 | 2+2 2 |
| | 2+2 4 | 3+4 2 | 2+2 4 | 5+6 2 | 2+2 2 | 3+3 2 |
| | 2+2 4 | 5+6 2 | 2+2 4 | 3+4 2 | 3+3 2 | 2+2 2 |
| | 2+2 4 | 5+6 2 | 2+2 4 | 3+4 2 | 3+3 2 | 3+3 2 |
| | 2+2 4 | 5+6 2 | 2+2 4 | 3+4 2 | 3+3 2 | 3+3 2 |
| | 2+2 4 | 5+6 2 | 2+2 4 | 3+4 2 | 3+3 2 | 2+2 2 |
| | 2+2 4 | 5+6 2 | 2+2 4 | 3+4 2 | 3+3 2 | 3+3 2 |
| | 2+2 4 | 5+6 2 | 2+2 4 | 5+6 2 | 2+2 2 | 2+2 2 |
| | 2+2 4 | 3+4 2 | 2+2 4 | 3+4 2 | 3+3 2 | 2+2 2 |
| | 2+2 4 | 3+4 2 | 2+2 4 | 5+6 2 | 2+2 2 | 3+3 2 |
| | 2+2 4 | 3+4 2 | 2+2 4 | 5+6 2 | 2+2 2 | 2+2 2 |
| | 2+2 4 | 3+4 2 | 2+2 4 | 5+6 2 | 2+2 2 | 3+3 2 |
| | 2+2 4 | 3+4 2 | 2+2 4 | 5+6 2 | 2+2 2 | 2+2 2 |
| | 2+2 4 | 3+4 2 | 2+2 4 | 5+6 2 | 2+2 2 | 3+3 2 |
| | 2+2 4 | 5+6 2 | 2+2 4 | 5+6 2 | 2+2 2 | 3+3 2 |
| | 2+2 4 | 3+4 2 | 2+2 4 | 3+4 2 | 3+3 2 | 2+2 2 |
| | 2+2 4 | 5+6 2 | 2+2 4 | 3+4 2 | 3+3 2 | 3+3 2 |
| | 2+2 4 | 3+4 2 | 2+2 4 | 5+6 2 | 2+2 2 | 2+2 2 |
| | 2+2 4 | 5+6 2 | 2+2 4 | 3+4 2 | 3+3 2 | 3+3 2 |
| | 2+2 4 | 5+6 2 | 2+2 4 | 3+4 2 | 3+3 2 | 2+2 2 |
| | 2+2 4 | 5+6 2 | 2+2 4 | 3+4 2 | 3+3 2 | 3+3 2 |
| | 2+2 4 | 5+6 2 | 2+2 4 | 3+4 2 | 3+3 2 | 2+2 2 |
| | 2+2 4 | 3+4 2 | 2+2 4 | 3+4 2 | 3+3 2 | 3+3 2 |
| | 2+2 4 | 5+6 2 | 2+2 4 | 3+4 2 | 3+3 2 | 2+2 2 |
| | 2+2 4 | 5+6 2 | 2+2 4 | 5+6 2 | 2+2 2 | 2+2 2 |
| | 2+2 4 | 3+4 2 | 2+2 4 | 5+6 2 | 2+2 2 | 2+2 2 |
| | 2+2 4 | 3+4 2 | 2+2 4 | 5+6 2 | 2+2 2 | 3+3 2 |
| | 2+2 4 | 3+4 2 | 2+2 4 | 3+4 2 | 3+3 2 | 3+3 2 |
| | | 3+4 2 | | 3+4 2 | 3+3 2 | 2+2 2 |
| | | 5+6 2 | | 5+6 2 | 2+2 2 | 3+3 2 |
| | | 5+6 2 | | 3+4 2 | 3+3 2 | 2+2 2 |
| | | 3+4 2 | | 3+4 2 | 2+2 2 | 3+3 2 |
| | | 3+4 2 | | 5+6 2 | 3+3 2 | 2+2 2 |
| | | 5+6 2 | | 5+6 2 | 2+2 2 | 3+3 2 |
| | | 5+6 2 | | 3+4 2 | 3+3 2 | 2+2 2 |
| | | 3+4 2 | | 5+6 2 | 2+2 2 | 3+3 2 |
| | | 5+6 2 | | 5+6 2 | 2+2 2 | 2+2 2 |
| | | 3+4 2 | | 5+6 2 | 3+3 2 | 3+3 2 |
| | | 5+6 2 | | 3+4 2 | 2+2 2 | 2+2 2 |
| | | 5+6 2 | | 5+6 2 | 2+2 2 | 3+3 2 |
| | | 5+6 2 | | 3+4 2 | 3+3 2 | 3+3 2 |
| | | 3+4 2 | | 3+4 2 | 3+3 2 | 2+2 2 |
| | | 5+6 2 | | 3+4 2 | 3+3 2 | 3+3 2 |
| | | 5+6 2 | | 5+6 2 | 3+3 2 | 2+2 2 |
| | | 3+4 2 | | 5+6 2 | 2+2 2 | 3+3 2 |
| | | 3+4 2 | | 3+4 2 | 2+2 2 | 2+2 2 |
| | | 3+4 2 | | 5+6 2 | 2+2 2 | 2+2 2 |
| | | 5+6 2 | | 5+6 2 | 2+2 2 | 3+3 2 |
| | | 5+6 2 | | 3+4 2 | 2+2 2 | 2+2 2 |
| | | 3+4 2 | | 3+4 2 | 3+3 2 | 3+3 2 |
| | | 5+6 2 | | 5+6 2 | 3+3 2 | 3+3 2 |
| | | 3+4 2 | | 5+6 2 | 2+2 2 | 2+2 2 |
| | | 5+6 2 | | 3+4 2 | 2+2 2 | 2+2 2 |
| | | 3+4 2 | | 5+6 2 | 3+3 2 | 3+3 2 |
| | | 3+4 2 | | 5+6 2 | 2+2 2 | 2+2 2 |
| | | 5+6 2 | | 3+4 2 | 3+3 2 | 3+3 2 |
| | | 3+4 2 | | 5+6 2 | 2+2 2 | 3+3 2 |
| | | 3+4 2 | | 3+4 2 | 3+3 2 | 2+2 2 |
| | | 3+4 2 | | 3+4 2 | 3+3 2 | 3+3 2 |
| | | 5+6 2 | | 3+4 2 | 2+2 2 | 2+2 2 |
| | | 3+4 2 | | 5+6 2 | 3+3 2 | 3+3 2 |
| | | 5+6 2 | | 5+6 2 | 3+3 2 | 2+2 2 |
| | | 5+6 2 | | 3+4 2 | 2+2 2 | 2+2 2 |
| | | 3+4 2 | | 5+6 2 | 3+3 2 | 3+3 2 |
| 15.2.6 | 35148 | | 84 | O5Ti3 | | |
| | -A | -GM | -M | -Y | -L | -V |
| | 2+2 4 | 3+4 2 | 2+2 4 | 3+4 2 | 3+3 2 | 2+2 2 |
| | 2+2 4 | 5+6 2 | 2+2 4 | 3+4 2 | 2+2 2 | 3+3 2 |
| | 2+2 4 | 3+4 2 | 2+2 4 | 3+4 2 | 3+3 2 | 3+3 2 |
| | 2+2 4 | 3+4 2 | 2+2 4 | 5+6 2 | 2+2 2 | 3+3 2 |
| | 2+2 4 | 3+4 2 | 2+2 4 | 5+6 2 | 2+2 2 | 3+3 2 |
| | 2+2 4 | 5+6 2 | 2+2 4 | 5+6 2 | 3+3 2 | 2+2 2 |
| | 2+2 4 | 3+4 2 | 2+2 4 | 5+6 2 | 2+2 2 | 2+2 2 |
| | 2+2 4 | 5+6 2 | 2+2 4 | 3+4 2 | 3+3 2 | 3+3 2 |
| | 2+2 4 | 5+6 2 | 2+2 4 | 3+4 2 | 2+2 2 | 2+2 2 |
| | 2+2 4 | 5+6 2 | 2+2 4 | 5+6 2 | 3+3 2 | 2+2 2 |
| | 2+2 4 | 3+4 2 | 2+2 4 | 3+4 2 | 2+2 2 | 3+3 2 |
| | 2+2 4 | 3+4 2 | 2+2 4 | 3+4 2 | 3+3 2 | 3+3 2 |
| | 2+2 4 | 5+6 2 | 2+2 4 | 3+4 2 | 3+3 2 | 3+3 2 |
| | 2+2 4 | 3+4 2 | 2+2 4 | 5+6 2 | 2+2 2 | 3+3 2 |
| | 2+2 4 | 5+6 2 | 2+2 4 | 5+6 2 | 3+3 2 | 2+2 2 |
| | 2+2 4 | 3+4 2 | 2+2 4 | 3+4 2 | 3+3 2 | 2+2 2 |
| | 2+2 4 | 3+4 2 | 2+2 4 | 5+6 2 | 2+2 2 | 2+2 2 |
| | 2+2 4 | 5+6 2 | 2+2 4 | 3+4 2 | 2+2 2 | 3+3 2 |
| | 2+2 4 | 3+4 2 | 2+2 4 | 5+6 2 | 3+3 2 | 2+2 2 |
| | 2+2 4 | 3+4 2 | 2+2 4 | 3+4 2 | 2+2 2 | 3+3 2 |
| | 2+2 4 | 5+6 2 | 2+2 4 | 3+4 2 | 3+3 2 | 2+2 2 |
| | | 3+4 2 | | 3+4 2 | 2+2 2 | 3+3 2 |

Continued on next column

**TABLE XXVII– continued**

| TQC | ICSD | | NoE | SF | | |
|---|---|---|---|---|---|---|
| | | 3+4 2 | | 5+6 2 | 2+2 2 | 2+2 2 |
| | | 3+4 2 | | 5+6 2 | 3+3 2 | 3+3 2 |
| | | 5+6 2 | | 3+4 2 | 2+2 2 | 3+3 2 |
| | | 5+6 2 | | 3+4 2 | 3+3 2 | 2+2 2 |
| | | 3+4 2 | | 5+6 2 | 2+2 2 | 2+2 2 |
| | | 3+4 2 | | 5+6 2 | 3+3 2 | 3+3 2 |
| | | 5+6 2 | | 3+4 2 | 3+3 2 | 2+2 2 |
| | | 5+6 2 | | 5+6 2 | 2+2 2 | 2+2 2 |
| | | 5+6 2 | | 3+4 2 | 3+3 2 | 3+3 2 |
| | | 3+4 2 | | 3+4 2 | 2+2 2 | 2+2 2 |
| | | 3+4 2 | | 5+6 2 | 2+2 2 | 3+3 2 |
| | | 5+6 2 | | 3+4 2 | 3+3 2 | 2+2 2 |
| | | 5+6 2 | | 5+6 2 | 2+2 2 | 3+3 2 |
| | | 5+6 2 | | 3+4 2 | 2+2 2 | 2+2 2 |
| | | 5+6 2 | | 5+6 2 | 3+3 2 | 3+3 2 |
| | | 3+4 2 | | 5+6 2 | 3+3 2 | 2+2 2 |
| | | 5+6 2 | | 5+6 2 | 2+2 2 | 3+3 2 |
| | | 5+6 2 | | 5+6 2 | 3+3 2 | 2+2 2 |
| | | 3+4 2 | | 3+4 2 | 3+3 2 | 3+3 2 |
| | | 5+6 2 | | 3+4 2 | 2+2 2 | 3+3 2 |
| 15.2.7 | 194464 | | 84 | O5Ti3 | | |
| | -A | -GM | -M | -Y | -L | -V |
| | 2+2 4 | 3+4 2 | 2+2 4 | 3+4 2 | 2+2 2 | 2+2 2 |
| | 2+2 4 | 5+6 2 | 2+2 4 | 3+4 2 | 3+3 2 | 3+3 2 |
| | 2+2 4 | 3+4 2 | 2+2 4 | 3+4 2 | 2+2 2 | 3+3 2 |
| | 2+2 4 | 3+4 2 | 2+2 4 | 5+6 2 | 3+3 2 | 3+3 2 |
| | 2+2 4 | 3+4 2 | 2+2 4 | 5+6 2 | 3+3 2 | 3+3 2 |
| | 2+2 4 | 5+6 2 | 2+2 4 | 5+6 2 | 2+2 2 | 2+2 2 |
| | 2+2 4 | 3+4 2 | 2+2 4 | 5+6 2 | 3+3 2 | 2+2 2 |
| | 2+2 4 | 5+6 2 | 2+2 4 | 3+4 2 | 2+2 2 | 3+3 2 |
| | 2+2 4 | 5+6 2 | 2+2 4 | 3+4 2 | 3+3 2 | 2+2 2 |
| | 2+2 4 | 5+6 2 | 2+2 4 | 5+6 2 | 2+2 2 | 2+2 2 |
| | 2+2 4 | 3+4 2 | 2+2 4 | 3+4 2 | 3+3 2 | 3+3 2 |
| | 2+2 4 | 3+4 2 | 2+2 4 | 3+4 2 | 2+2 2 | 3+3 2 |
| | 2+2 4 | 5+6 2 | 2+2 4 | 3+4 2 | 2+2 2 | 3+3 2 |
| | 2+2 4 | 3+4 2 | 2+2 4 | 5+6 2 | 3+3 2 | 3+3 2 |
| | 2+2 4 | 5+6 2 | 2+2 4 | 5+6 2 | 2+2 2 | 2+2 2 |
| | 2+2 4 | 3+4 2 | 2+2 4 | 3+4 2 | 2+2 2 | 2+2 2 |
| | 2+2 4 | 3+4 2 | 2+2 4 | 5+6 2 | 3+3 2 | 2+2 2 |
| | 2+2 4 | 5+6 2 | 2+2 4 | 3+4 2 | 3+3 2 | 2+2 2 |
| | 2+2 4 | 3+4 2 | 2+2 4 | 5+6 2 | 2+2 2 | 3+3 2 |
| | 2+2 4 | 3+4 2 | 2+2 4 | 3+4 2 | 3+3 2 | 3+3 2 |
| | 2+2 4 | 5+6 2 | 2+2 4 | 3+4 2 | 2+2 2 | 2+2 2 |
| | | 3+4 2 | | 3+4 2 | 3+3 2 | 3+3 2 |
| | | 3+4 2 | | 5+6 2 | 3+3 2 | 2+2 2 |
| | | 3+4 2 | | 5+6 2 | 2+2 2 | 3+3 2 |
| | | 5+6 2 | | 3+4 2 | 3+3 2 | 3+3 2 |
| | | 5+6 2 | | 3+4 2 | 2+2 2 | 2+2 2 |
| | | 3+4 2 | | 5+6 2 | 3+3 2 | 2+2 2 |
| | | 3+4 2 | | 5+6 2 | 2+2 2 | 3+3 2 |
| | | 5+6 2 | | 3+4 2 | 2+2 2 | 2+2 2 |
| | | 5+6 2 | | 5+6 2 | 3+3 2 | 2+2 2 |
| | | 5+6 2 | | 3+4 2 | 2+2 2 | 3+3 2 |
| | | 3+4 2 | | 3+4 2 | 3+3 2 | 2+2 2 |
| | | 3+4 2 | | 5+6 2 | 3+3 2 | 3+3 2 |
| | | 5+6 2 | | 5+6 2 | 2+2 2 | 2+2 2 |
| | | 5+6 2 | | 3+4 2 | 3+3 2 | 3+3 2 |
| | | 5+6 2 | | 5+6 2 | 3+3 2 | 2+2 2 |
| | | 5+6 2 | | 3+4 2 | 2+2 2 | 3+3 2 |
| | | 3+4 2 | | 5+6 2 | 2+2 2 | 2+2 2 |
| | | 5+6 2 | | 5+6 2 | 3+3 2 | 3+3 2 |

Continued on next column

**TABLE XXVII– continued**

| TQC | ICSD | | NoE | SF | | |
|---|---|---|---|---|---|---|
| | | 5+6 2 | | 5+6 2 | 2+2 2 | 2+2 2 |
| | | 3+4 2 | | 3+4 2 | 2+2 2 | 3+3 2 |
| | | 5+6 2 | | 3+4 2 | 3+3 2 | 3+3 2 |
| 15.2.8 | 109187 | | 96 | O4Ta1Tb1 | | |
| | -A | -GM | -M | -Y | -L | -V |
| | 2+2 4 | 3+4 2 | 2+2 4 | 3+4 2 | 2+2 2 | 2+2 2 |
| | 2+2 4 | 5+6 2 | 2+2 4 | 5+6 2 | 3+3 2 | 3+3 2 |
| | 2+2 4 | 3+4 2 | 2+2 4 | 5+6 2 | 2+2 2 | 2+2 2 |
| | 2+2 4 | 5+6 2 | 2+2 4 | 3+4 2 | 3+3 2 | 3+3 2 |
| | 2+2 4 | 3+4 2 | 2+2 4 | 5+6 2 | 2+2 2 | 3+3 2 |
| | 2+2 4 | 3+4 2 | 2+2 4 | 3+4 2 | 3+3 2 | 2+2 2 |
| | 2+2 4 | 5+6 2 | 2+2 4 | 3+4 2 | 2+2 2 | 2+2 2 |
| | 2+2 4 | 5+6 2 | 2+2 4 | 5+6 2 | 3+3 2 | 3+3 2 |
| | 2+2 4 | 3+4 2 | 2+2 4 | 3+4 2 | 2+2 2 | 2+2 2 |
| | 2+2 4 | 5+6 2 | 2+2 4 | 5+6 2 | 3+3 2 | 3+3 2 |
| | 2+2 4 | 5+6 2 | 2+2 4 | 5+6 2 | 2+2 2 | 2+2 2 |
| | 2+2 4 | 3+4 2 | 2+2 4 | 3+4 2 | 3+3 2 | 3+3 2 |
| | 2+2 4 | 5+6 2 | 2+2 4 | 3+4 2 | 3+3 2 | 3+3 2 |
| | 2+2 4 | 3+4 2 | 2+2 4 | 5+6 2 | 2+2 2 | 2+2 2 |
| | 2+2 4 | 5+6 2 | 2+2 4 | 3+4 2 | 3+3 2 | 3+3 2 |
| | 2+2 4 | 3+4 2 | 2+2 4 | 5+6 2 | 2+2 2 | 2+2 2 |
| | 2+2 4 | 3+4 2 | 2+2 4 | 5+6 2 | 2+2 2 | 3+3 2 |
| | 2+2 4 | 5+6 2 | 2+2 4 | 3+4 2 | 3+3 2 | 2+2 2 |
| | 2+2 4 | 5+6 2 | 2+2 4 | 3+4 2 | 2+2 2 | 2+2 2 |
| | 2+2 4 | 3+4 2 | 2+2 4 | 5+6 2 | 3+3 2 | 3+3 2 |
| | 2+2 4 | 3+4 2 | 2+2 4 | 3+4 2 | 3+3 2 | 3+3 2 |
| | 2+2 4 | 3+4 2 | 2+2 4 | 5+6 2 | 2+2 2 | 2+2 2 |
| | 2+2 4 | 3+4 2 | 2+2 4 | 3+4 2 | 2+2 2 | 3+3 2 |
| | 2+2 4 | 5+6 2 | 2+2 4 | 5+6 2 | 3+3 2 | 2+2 2 |
| | | 5+6 2 | | 5+6 2 | 2+2 2 | 3+3 2 |
| | | 5+6 2 | | 3+4 2 | 3+3 2 | 2+2 2 |
| | | 3+4 2 | | 3+4 2 | 2+2 2 | 2+2 2 |
| | | 3+4 2 | | 3+4 2 | 3+3 2 | 3+3 2 |
| | | 3+4 2 | | 5+6 2 | 2+2 2 | 3+3 2 |
| | | 5+6 2 | | 5+6 2 | 3+3 2 | 2+2 2 |
| | | 5+6 2 | | 5+6 2 | 3+3 2 | 2+2 2 |
| | | 3+4 2 | | 3+4 2 | 2+2 2 | 3+3 2 |
| | | 3+4 2 | | 3+4 2 | 3+3 2 | 3+3 2 |
| | | 5+6 2 | | 5+6 2 | 3+3 2 | 3+3 2 |
| | | 5+6 2 | | 5+6 2 | 2+2 2 | 2+2 2 |
| | | 3+4 2 | | 3+4 2 | 2+2 2 | 2+2 2 |
| | | 3+4 2 | | 3+4 2 | 2+2 2 | 3+3 2 |
| | | 5+6 2 | | 5+6 2 | 3+3 2 | 2+2 2 |
| | | 5+6 2 | | 3+4 2 | 3+3 2 | 3+3 2 |
| | | 5+6 2 | | 5+6 2 | 2+2 2 | 2+2 2 |
| | | 3+4 2 | | 3+4 2 | 2+2 2 | 2+2 2 |
| | | 3+4 2 | | 5+6 2 | 3+3 2 | 3+3 2 |
| | | 3+4 2 | | 5+6 2 | 2+2 2 | 2+2 2 |
| | | 5+6 2 | | 3+4 2 | 3+3 2 | 3+3 2 |
| | | 5+6 2 | | 5+6 2 | 2+2 2 | 2+2 2 |
| | | 5+6 2 | | 3+4 2 | 3+3 2 | 3+3 2 |
| | | 3+4 2 | | 3+4 2 | 2+2 2 | 2+2 2 |
| | | 3+4 2 | | 5+6 2 | 3+3 2 | 3+3 2 |
| 15.2.9 | 97442 | | 56 | O3V2 | | |
| | -A | -GM | -M | -Y | -L | -V |
| | 2+2 4 | 3+4 2 | 2+2 4 | 3+4 2 | 3+3 2 | 3+3 2 |
| | 2+2 4 | 5+6 2 | 2+2 4 | 5+6 2 | 2+2 2 | 2+2 2 |
| | 2+2 4 | 5+6 2 | 2+2 4 | 3+4 2 | 3+3 2 | 2+2 2 |
| | 2+2 4 | 5+6 2 | 2+2 4 | 5+6 2 | 2+2 2 | 3+3 2 |
| | 2+2 4 | 3+4 2 | 2+2 4 | 5+6 2 | 2+2 2 | 3+3 2 |
| | 2+2 4 | 3+4 2 | 2+2 4 | 3+4 2 | 3+3 2 | 2+2 2 |

Continued on next column

**TABLE XXVII– continued**

| TQC | ICSD | | NoE | | SF | |
|---|---|---|---|---|---|---|
| | 2+2 4 | 3+4 2 | 2+2 4 | 5+6 2 | 2+2 2 | 2+2 2 |
| | 2+2 4 | 5+6 2 | 2+2 4 | 3+4 2 | 3+3 2 | 3+3 2 |
| | 2+2 4 | 5+6 2 | 2+2 4 | 3+4 2 | 3+3 2 | 2+2 2 |
| | 2+2 4 | 5+6 2 | 2+2 4 | 5+6 2 | 2+2 2 | 3+3 2 |
| | 2+2 4 | 3+4 2 | 2+2 4 | 5+6 2 | 2+2 2 | 3+3 2 |
| | 2+2 4 | 3+4 2 | 2+2 4 | 3+4 2 | 3+3 2 | 2+2 2 |
| | 2+2 4 | 3+4 2 | 2+2 4 | 5+6 2 | 2+2 2 | 3+3 2 |
| | 2+2 4 | 3+4 2 | 2+2 4 | 3+4 2 | 3+3 2 | 2+2 2 |
| | | 5+6 2 | | 5+6 2 | 3+3 2 | 2+2 2 |
| | | 3+4 2 | | 3+4 2 | 2+2 2 | 3+3 2 |
| | | 5+6 2 | | 5+6 2 | 3+3 2 | 2+2 2 |
| | | 3+4 2 | | 3+4 2 | 2+2 2 | 3+3 2 |
| | | 5+6 2 | | 3+4 2 | 3+3 2 | 2+2 2 |
| | | 3+4 2 | | 3+4 2 | 3+3 2 | 3+3 2 |
| | | 3+4 2 | | 5+6 2 | 2+2 2 | 2+2 2 |
| | | 5+6 2 | | 3+4 2 | 3+3 2 | 3+3 2 |
| | | 5+6 2 | | 5+6 2 | 2+2 2 | 2+2 2 |
| | | 5+6 2 | | 5+6 2 | 2+2 2 | 3+3 2 |
| | | 3+4 2 | | 3+4 2 | 2+2 2 | 3+3 2 |
| | | 3+4 2 | | 5+6 2 | 3+3 2 | 2+2 2 |
| | | 3+4 2 | | 3+4 2 | 3+3 2 | 2+2 2 |
| | | 3+4 2 | | 3+4 2 | 3+3 2 | 3+3 2 |
| 15.2.10 | 95763 | | 56 | | O3V2 | |
| | -A | -GM | -M | -Y | -L | -V |
| | 2+2 4 | 3+4 2 | 2+2 4 | 3+4 2 | 3+3 2 | 3+3 2 |
| | 2+2 4 | 5+6 2 | 2+2 4 | 5+6 2 | 2+2 2 | 2+2 2 |
| | 2+2 4 | 5+6 2 | 2+2 4 | 5+6 2 | 3+3 2 | 2+2 2 |
| | 2+2 4 | 3+4 2 | 2+2 4 | 5+6 2 | 2+2 2 | 3+3 2 |
| | 2+2 4 | 5+6 2 | 2+2 4 | 3+4 2 | 2+2 2 | 3+3 2 |
| | 2+2 4 | 3+4 2 | 2+2 4 | 3+4 2 | 3+3 2 | 2+2 2 |
| | 2+2 4 | 3+4 2 | 2+2 4 | 5+6 2 | 2+2 2 | 3+3 2 |
| | 2+2 4 | 5+6 2 | 2+2 4 | 3+4 2 | 3+3 2 | 2+2 2 |
| | 2+2 4 | 5+6 2 | 2+2 4 | 3+4 2 | 3+3 2 | 2+2 2 |
| | 2+2 4 | 3+4 2 | 2+2 4 | 5+6 2 | 2+2 2 | 3+3 2 |
| | 2+2 4 | 5+6 2 | 2+2 4 | 5+6 2 | 2+2 2 | 2+2 2 |
| | 2+2 4 | 3+4 2 | 2+2 4 | 3+4 2 | 3+3 2 | 3+3 2 |
| | 2+2 4 | 3+4 2 | 2+2 4 | 5+6 2 | 2+2 2 | 3+3 2 |
| | 2+2 4 | 3+4 2 | 2+2 4 | 3+4 2 | 3+3 2 | 2+2 2 |
| | | 5+6 2 | | 5+6 2 | 3+3 2 | 2+2 2 |
| | | 3+4 2 | | 3+4 2 | 2+2 2 | 3+3 2 |
| | | 5+6 2 | | 5+6 2 | 3+3 2 | 2+2 2 |
| | | 3+4 2 | | 3+4 2 | 2+2 2 | 3+3 2 |
| | | 5+6 2 | | 3+4 2 | 3+3 2 | 2+2 2 |
| | | 3+4 2 | | 3+4 2 | 3+3 2 | 3+3 2 |
| | | 3+4 2 | | 5+6 2 | 2+2 2 | 2+2 2 |
| | | 5+6 2 | | 5+6 2 | 3+3 2 | 3+3 2 |
| | | 5+6 2 | | 3+4 2 | 2+2 2 | 2+2 2 |
| | | 5+6 2 | | 5+6 2 | 2+2 2 | 3+3 2 |
| | | 3+4 2 | | 3+4 2 | 2+2 2 | 3+3 2 |
| | | 3+4 2 | | 5+6 2 | 3+3 2 | 2+2 2 |
| | | 3+4 2 | | 3+4 2 | 3+3 2 | 2+2 2 |
| | | 3+4 2 | | 3+4 2 | 3+3 2 | 3+3 2 |
| 15.2.11 | 95762 | | 56 | | O3V2 | |
| | -A | -GM | -M | -Y | -L | -V |
| | 2+2 4 | 3+4 2 | 2+2 4 | 3+4 2 | 3+3 2 | 2+2 2 |
| | 2+2 4 | 5+6 2 | 2+2 4 | 5+6 2 | 2+2 2 | 3+3 2 |
| | 2+2 4 | 3+4 2 | 2+2 4 | 5+6 2 | 3+3 2 | 3+3 2 |
| | 2+2 4 | 5+6 2 | 2+2 4 | 5+6 2 | 2+2 2 | 2+2 2 |
| | 2+2 4 | 5+6 2 | 2+2 4 | 3+4 2 | 2+2 2 | 2+2 2 |
| | 2+2 4 | 3+4 2 | 2+2 4 | 3+4 2 | 3+3 2 | 3+3 2 |
| | 2+2 4 | 3+4 2 | 2+2 4 | 5+6 2 | 2+2 2 | 2+2 2 |

Continued on next column

**TABLE XXVII– continued**

| TQC | ICSD | | NoE | | SF | |
|---|---|---|---|---|---|---|
| | 2+2 4 | 5+6 2 | 2+2 4 | 3+4 2 | 3+3 2 | 3+3 2 |
| | 2+2 4 | 5+6 2 | 2+2 4 | 3+4 2 | 3+3 2 | 3+3 2 |
| | 2+2 4 | 3+4 2 | 2+2 4 | 5+6 2 | 2+2 2 | 2+2 2 |
| | 2+2 4 | 5+6 2 | 2+2 4 | 5+6 2 | 2+2 2 | 3+3 2 |
| | 2+2 4 | 3+4 2 | 2+2 4 | 3+4 2 | 3+3 2 | 2+2 2 |
| | 2+2 4 | 3+4 2 | 2+2 4 | 5+6 2 | 2+2 2 | 2+2 2 |
| | 2+2 4 | 5+6 2 | 2+2 4 | 3+4 2 | 3+3 2 | 3+3 2 |
| | | 3+4 2 | | 3+4 2 | 3+3 2 | 3+3 2 |
| | | 3+4 2 | | 5+6 2 | 2+2 2 | 2+2 2 |
| | | 5+6 2 | | 5+6 2 | 3+3 2 | 3+3 2 |
| | | 3+4 2 | | 3+4 2 | 2+2 2 | 2+2 2 |
| | | 5+6 2 | | 3+4 2 | 3+3 2 | 3+3 2 |
| | | 3+4 2 | | 5+6 2 | 3+3 2 | 2+2 2 |
| | | 3+4 2 | | 3+4 2 | 2+2 2 | 2+2 2 |
| | | 5+6 2 | | 5+6 2 | 3+3 2 | 3+3 2 |
| | | 5+6 2 | | 3+4 2 | 2+2 2 | 3+3 2 |
| | | 5+6 2 | | 5+6 2 | 2+2 2 | 2+2 2 |
| | | 3+4 2 | | 3+4 2 | 2+2 2 | 2+2 2 |
| | | 3+4 2 | | 5+6 2 | 3+3 2 | 3+3 2 |
| | | 3+4 2 | | 3+4 2 | 3+3 2 | 3+3 2 |
| | | 3+4 2 | | 3+4 2 | 3+3 2 | 2+2 2 |
| 15.2.12 | 6286 | | 56 | | O3V2 | |
| | -A | -GM | -M | -Y | -L | -V |
| | 2+2 4 | 3+4 2 | 2+2 4 | 3+4 2 | 3+3 2 | 3+3 2 |
| | 2+2 4 | 5+6 2 | 2+2 4 | 5+6 2 | 2+2 2 | 2+2 2 |
| | 2+2 4 | 5+6 2 | 2+2 4 | 3+4 2 | 3+3 2 | 2+2 2 |
| | 2+2 4 | 5+6 2 | 2+2 4 | 5+6 2 | 2+2 2 | 3+3 2 |
| | 2+2 4 | 3+4 2 | 2+2 4 | 5+6 2 | 2+2 2 | 3+3 2 |
| | 2+2 4 | 3+4 2 | 2+2 4 | 3+4 2 | 3+3 2 | 2+2 2 |
| | 2+2 4 | 3+4 2 | 2+2 4 | 5+6 2 | 2+2 2 | 3+3 2 |
| | 2+2 4 | 5+6 2 | 2+2 4 | 3+4 2 | 3+3 2 | 2+2 2 |
| | 2+2 4 | 5+6 2 | 2+2 4 | 3+4 2 | 3+3 2 | 2+2 2 |
| | 2+2 4 | 5+6 2 | 2+2 4 | 5+6 2 | 2+2 2 | 3+3 2 |
| | 2+2 4 | 3+4 2 | 2+2 4 | 5+6 2 | 2+2 2 | 3+3 2 |
| | 2+2 4 | 3+4 2 | 2+2 4 | 3+4 2 | 3+3 2 | 2+2 2 |
| | 2+2 4 | 3+4 2 | 2+2 4 | 5+6 2 | 2+2 2 | 3+3 2 |
| | 2+2 4 | 3+4 2 | 2+2 4 | 3+4 2 | 3+3 2 | 2+2 2 |
| | | 3+4 2 | | 3+4 2 | 3+3 2 | 2+2 2 |
| | | 5+6 2 | | 5+6 2 | 2+2 2 | 3+3 2 |
| | | 5+6 2 | | 3+4 2 | 3+3 2 | 2+2 2 |
| | | 3+4 2 | | 5+6 2 | 2+2 2 | 3+3 2 |
| | | 5+6 2 | | 3+4 2 | 3+3 2 | 2+2 2 |
| | | 3+4 2 | | 5+6 2 | 3+3 2 | 2+2 2 |
| | | 3+4 2 | | 3+4 2 | 3+3 2 | 3+3 2 |
| | | 5+6 2 | | 3+4 2 | 2+2 2 | 3+3 2 |
| | | 5+6 2 | | 5+6 2 | 2+2 2 | 3+3 2 |
| | | 5+6 2 | | 5+6 2 | 2+2 2 | 2+2 2 |
| | | 3+4 2 | | 3+4 2 | 2+2 2 | 3+3 2 |
| | | 3+4 2 | | 5+6 2 | 3+3 2 | 2+2 2 |
| | | 3+4 2 | | 3+4 2 | 3+3 2 | 2+2 2 |
| | | 5+6 2 | | 3+4 2 | 3+3 2 | 3+3 2 |
| 15.2.13 | 50703 | | 172 | | Ni2O8Te3 | |
| | -A | -GM | -M | -Y | -L | -V |
| | 2+2 4 | 3+4 2 | 2+2 4 | 5+6 2 | 2+2 2 | 2+2 2 |
| | 2+2 4 | 3+4 2 | 2+2 4 | 5+6 2 | 2+2 2 | 3+3 2 |
| | 2+2 4 | 5+6 2 | 2+2 4 | 3+4 2 | 3+3 2 | 2+2 2 |
| | 2+2 4 | 5+6 2 | 2+2 4 | 3+4 2 | 3+3 2 | 3+3 2 |
| | 2+2 4 | 3+4 2 | 2+2 4 | 5+6 2 | 2+2 2 | 3+3 2 |
| | 2+2 4 | 5+6 2 | 2+2 4 | 3+4 2 | 3+3 2 | 2+2 2 |
| | 2+2 4 | 3+4 2 | 2+2 4 | 5+6 2 | 2+2 2 | 3+3 2 |
| | 2+2 4 | 5+6 2 | 2+2 4 | 3+4 2 | 3+3 2 | 2+2 2 |

Continued on next column

**TABLE XXVII– continued**

| TQC | ICSD | | NoE | SF | | |
|---|---|---|---|---|---|---|
| | 2+2 4 | 5+6 2 | 2+2 4 | 3+4 2 | 3+3 2 | 3+3 2 |
| | 2+2 4 | 3+4 2 | 2+2 4 | 5+6 2 | 2+2 2 | 2+2 2 |
| | 2+2 4 | 3+4 2 | 2+2 4 | 5+6 2 | 2+2 2 | 2+2 2 |
| | 2+2 4 | 5+6 2 | 2+2 4 | 5+6 2 | 3+3 2 | 3+3 2 |
| | 2+2 4 | 3+4 2 | 2+2 4 | 5+6 2 | 2+2 2 | 3+3 2 |
| | 2+2 4 | 3+4 2 | 2+2 4 | 3+4 2 | 2+2 2 | 2+2 2 |
| | 2+2 4 | 5+6 2 | 2+2 4 | 3+4 2 | 3+3 2 | 2+2 2 |
| | 2+2 4 | 5+6 2 | 2+2 4 | 3+4 2 | 3+3 2 | 3+3 2 |
| | 2+2 4 | 5+6 2 | 2+2 4 | 3+4 2 | 3+3 2 | 2+2 2 |
| | 2+2 4 | 3+4 2 | 2+2 4 | 3+4 2 | 3+3 2 | 3+3 2 |
| | 2+2 4 | 5+6 2 | 2+2 4 | 5+6 2 | 2+2 2 | 2+2 2 |
| | 2+2 4 | 5+6 2 | 2+2 4 | 5+6 2 | 3+3 2 | 3+3 2 |
| | 2+2 4 | 3+4 2 | 2+2 4 | 3+4 2 | 2+2 2 | 3+3 2 |
| | 2+2 4 | 3+4 2 | 2+2 4 | 5+6 2 | 2+2 2 | 2+2 2 |
| | 2+2 4 | 3+4 2 | 2+2 4 | 5+6 2 | 2+2 2 | 2+2 2 |
| | 2+2 4 | 3+4 2 | 2+2 4 | 5+6 2 | 3+3 2 | 3+3 2 |
| | 2+2 4 | 5+6 2 | 2+2 4 | 3+4 2 | 2+2 2 | 2+2 2 |
| | 2+2 4 | 5+6 2 | 2+2 4 | 3+4 2 | 3+3 2 | 3+3 2 |
| | 2+2 4 | 3+4 2 | 2+2 4 | 3+4 2 | 3+3 2 | 2+2 2 |
| | 2+2 4 | 5+6 2 | 2+2 4 | 3+4 2 | 2+2 2 | 3+3 2 |
| | 2+2 4 | 3+4 2 | 2+2 4 | 5+6 2 | 2+2 2 | 2+2 2 |
| | 2+2 4 | 5+6 2 | 2+2 4 | 5+6 2 | 3+3 2 | 3+3 2 |
| | 2+2 4 | 3+4 2 | 2+2 4 | 5+6 2 | 2+2 2 | 3+3 2 |
| | 2+2 4 | 5+6 2 | 2+2 4 | 3+4 2 | 3+3 2 | 2+2 2 |
| | 2+2 4 | 3+4 2 | 2+2 4 | 5+6 2 | 2+2 2 | 2+2 2 |
| | 2+2 4 | 3+4 2 | 2+2 4 | 5+6 2 | 3+3 2 | 3+3 2 |
| | 2+2 4 | 5+6 2 | 2+2 4 | 3+4 2 | 2+2 2 | 2+2 2 |
| | 2+2 4 | 3+4 2 | 2+2 4 | 3+4 2 | 3+3 2 | 3+3 2 |
| | 2+2 4 | 5+6 2 | 2+2 4 | 3+4 2 | 2+2 2 | 2+2 2 |
| | 2+2 4 | 3+4 2 | 2+2 4 | 5+6 2 | 3+3 2 | 3+3 2 |
| | 2+2 4 | 5+6 2 | 2+2 4 | 3+4 2 | 3+3 2 | 2+2 2 |
| | 2+2 4 | 5+6 2 | 2+2 4 | 5+6 2 | 2+2 2 | 3+3 2 |
| | 2+2 4 | 3+4 2 | 2+2 4 | 5+6 2 | 3+3 2 | 3+3 2 |
| | 2+2 4 | 5+6 2 | 2+2 4 | 5+6 2 | 2+2 2 | 2+2 2 |
| | 2+2 4 | 3+4 2 | 2+2 4 | 3+4 2 | 2+2 2 | 3+3 2 |
| | | 5+6 2 | | 3+4 2 | 3+3 2 | 2+2 2 |
| | | 3+4 2 | | 5+6 2 | 3+3 2 | 2+2 2 |
| | | 5+6 2 | | 5+6 2 | 2+2 2 | 3+3 2 |
| | | 3+4 2 | | 5+6 2 | 3+3 2 | 2+2 2 |
| | | 3+4 2 | | 3+4 2 | 2+2 2 | 3+3 2 |
| | | 3+4 2 | | 3+4 2 | 3+3 2 | 2+2 2 |
| | | 3+4 2 | | 5+6 2 | 2+2 2 | 2+2 2 |
| | | 5+6 2 | | 3+4 2 | 2+2 2 | 3+3 2 |
| | | 5+6 2 | | 5+6 2 | 3+3 2 | 3+3 2 |
| | | 3+4 2 | | 5+6 2 | 3+3 2 | 3+3 2 |
| | | 5+6 2 | | 3+4 2 | 2+2 2 | 2+2 2 |
| | | 3+4 2 | | 5+6 2 | 2+2 2 | 3+3 2 |
| | | 5+6 2 | | 3+4 2 | 2+2 2 | 2+2 2 |
| | | 5+6 2 | | 3+4 2 | 3+3 2 | 3+3 2 |
| | | 3+4 2 | | 5+6 2 | 3+3 2 | 2+2 2 |
| | | 5+6 2 | | 5+6 2 | 2+2 2 | 3+3 2 |
| | | 5+6 2 | | 3+4 2 | 3+3 2 | 2+2 2 |
| | | 3+4 2 | | 5+6 2 | 2+2 2 | 3+3 2 |
| | | 3+4 2 | | 3+4 2 | 3+3 2 | 2+2 2 |
| | | 5+6 2 | | 3+4 2 | 2+2 2 | 3+3 2 |
| | | 5+6 2 | | 5+6 2 | 3+3 2 | 3+3 2 |
| | | 3+4 2 | | 3+4 2 | 2+2 2 | 2+2 2 |
| | | 5+6 2 | | 3+4 2 | 3+3 2 | 2+2 2 |
| | | 5+6 2 | | 3+4 2 | 3+3 2 | 3+3 2 |
| | | 3+4 2 | | 5+6 2 | 2+2 2 | 2+2 2 |
| | | 3+4 2 | | 5+6 2 | 3+3 2 | 2+2 2 |

Continued on next column

**TABLE XXVII– continued**

| TQC | ICSD | | NoE | SF | | |
|---|---|---|---|---|---|---|
| | | 5+6 2 | | 3+4 2 | 2+2 2 | 3+3 2 |
| | | 5+6 2 | | 3+4 2 | 2+2 2 | 2+2 2 |
| | | 3+4 2 | | 5+6 2 | 3+3 2 | 3+3 2 |
| | | 5+6 2 | | 5+6 2 | 2+2 2 | 2+2 2 |
| | | 3+4 2 | | 3+4 2 | 3+3 2 | 3+3 2 |
| | | 3+4 2 | | 5+6 2 | 2+2 2 | 3+3 2 |
| | | 5+6 2 | | 3+4 2 | 3+3 2 | 2+2 2 |
| | | 3+4 2 | | 5+6 2 | 2+2 2 | 3+3 2 |
| | | 5+6 2 | | 3+4 2 | 3+3 2 | 2+2 2 |
| | | 3+4 2 | | 5+6 2 | 2+2 2 | 2+2 2 |
| | | 5+6 2 | | 3+4 2 | 2+2 2 | 3+3 2 |
| | | 3+4 2 | | 5+6 2 | 3+3 2 | 2+2 2 |
| | | 5+6 2 | | 3+4 2 | 3+3 2 | 3+3 2 |
| | | 3+4 2 | | 3+4 2 | 2+2 2 | 3+3 2 |
| | | 5+6 2 | | 5+6 2 | 2+2 2 | 2+2 2 |
| | | 3+4 2 | | 5+6 2 | 3+3 2 | 3+3 2 |
| | | 5+6 2 | | 3+4 2 | 2+2 2 | 2+2 2 |
| 15.2.14 | 61148 | | 100 | Ni1Se5Ta2 | | |
| | -A | -GM | -M | -Y | -L | -V |
| | 2+2 4 | 3+4 2 | 2+2 4 | 5+6 2 | 2+2 2 | 2+2 2 |
| | 2+2 4 | 5+6 2 | 2+2 4 | 3+4 2 | 3+3 2 | 3+3 2 |
| | 2+2 4 | 3+4 2 | 2+2 4 | 5+6 2 | 2+2 2 | 3+3 2 |
| | 2+2 4 | 3+4 2 | 2+2 4 | 5+6 2 | 3+3 2 | 2+2 2 |
| | 2+2 4 | 5+6 2 | 2+2 4 | 3+4 2 | 2+2 2 | 2+2 2 |
| | 2+2 4 | 5+6 2 | 2+2 4 | 3+4 2 | 3+3 2 | 3+3 2 |
| | 2+2 4 | 3+4 2 | 2+2 4 | 5+6 2 | 2+2 2 | 2+2 2 |
| | 2+2 4 | 5+6 2 | 2+2 4 | 5+6 2 | 3+3 2 | 2+2 2 |
| | 2+2 4 | 5+6 2 | 2+2 4 | 3+4 2 | 2+2 2 | 3+3 2 |
| | 2+2 4 | 3+4 2 | 2+2 4 | 3+4 2 | 3+3 2 | 3+3 2 |
| | 2+2 4 | 3+4 2 | 2+2 4 | 3+4 2 | 2+2 2 | 3+3 2 |
| | 2+2 4 | 5+6 2 | 2+2 4 | 3+4 2 | 3+3 2 | 2+2 2 |
| | 2+2 4 | 5+6 2 | 2+2 4 | 5+6 2 | 3+3 2 | 3+3 2 |
| | 2+2 4 | 3+4 2 | 2+2 4 | 5+6 2 | 2+2 2 | 2+2 2 |
| | 2+2 4 | 5+6 2 | 2+2 4 | 3+4 2 | 3+3 2 | 2+2 2 |
| | 2+2 4 | 3+4 2 | 2+2 4 | 5+6 2 | 2+2 2 | 3+3 2 |
| | 2+2 4 | 3+4 2 | 2+2 4 | 5+6 2 | 3+3 2 | 2+2 2 |
| | 2+2 4 | 5+6 2 | 2+2 4 | 5+6 2 | 2+2 2 | 3+3 2 |
| | 2+2 4 | 3+4 2 | 2+2 4 | 5+6 2 | 3+3 2 | 2+2 2 |
| | 2+2 4 | 3+4 2 | 2+2 4 | 3+4 2 | 2+2 2 | 3+3 2 |
| | 2+2 4 | 5+6 2 | 2+2 4 | 3+4 2 | 2+2 2 | 2+2 2 |
| | 2+2 4 | 3+4 2 | 2+2 4 | 3+4 2 | 3+3 2 | 3+3 2 |
| | 2+2 4 | 5+6 2 | 2+2 4 | 3+4 2 | 3+3 2 | 2+2 2 |
| | 2+2 4 | 3+4 2 | 2+2 4 | 5+6 2 | 2+2 2 | 3+3 2 |
| | 2+2 4 | 5+6 2 | 2+2 4 | 5+6 2 | 2+2 2 | 2+2 2 |
| | | 3+4 2 | | 3+4 2 | 3+3 2 | 3+3 2 |
| | | 5+6 2 | | 5+6 2 | 2+2 2 | 2+2 2 |
| | | 3+4 2 | | 5+6 2 | 3+3 2 | 3+3 2 |
| | | 3+4 2 | | 5+6 2 | 2+2 2 | 2+2 2 |
| | | 5+6 2 | | 3+4 2 | 3+3 2 | 3+3 2 |
| | | 3+4 2 | | 5+6 2 | 2+2 2 | 2+2 2 |
| | | 5+6 2 | | 5+6 2 | 3+3 2 | 2+2 2 |
| | | 3+4 2 | | 3+4 2 | 2+2 2 | 3+3 2 |
| | | 5+6 2 | | 3+4 2 | 3+3 2 | 3+3 2 |
| | | 5+6 2 | | 3+4 2 | 2+2 2 | 2+2 2 |
| | | 3+4 2 | | 3+4 2 | 3+3 2 | 3+3 2 |
| | | 5+6 2 | | 5+6 2 | 2+2 2 | 3+3 2 |
| | | 3+4 2 | | 5+6 2 | 3+3 2 | 2+2 2 |
| | | 5+6 2 | | 3+4 2 | 2+2 2 | 3+3 2 |
| | | 3+4 2 | | 5+6 2 | 3+3 2 | 2+2 2 |
| | | 3+4 2 | | 5+6 2 | 3+3 2 | 3+3 2 |
| | | 5+6 2 | | 3+4 2 | 2+2 2 | 2+2 2 |

Continued on next column

**TABLE XXVII– continued**

| TQC | ICSD | | NoE | SF | | |
|---|---|---|---|---|---|---|
| | | 3+4 2 | | 5+6 2 | 2+2 2 | 3+3 2 |
| | | 5+6 2 | | 3+4 2 | 3+3 2 | 2+2 2 |
| | | 5+6 2 | | 3+4 2 | 2+2 2 | 3+3 2 |
| | | 3+4 2 | | 5+6 2 | 3+3 2 | 2+2 2 |
| | | 3+4 2 | | 5+6 2 | 3+3 2 | 2+2 2 |
| | | 5+6 2 | | 3+4 2 | 2+2 2 | 3+3 2 |
| | | 5+6 2 | | 3+4 2 | 2+2 2 | 3+3 2 |
| | | 3+4 2 | | 3+4 2 | 3+3 2 | 2+2 2 |
| 15.2.15 | 109181 | | 120 | Nb1O4Tm1 | | |
| | -A | -GM | -M | -Y | -L | -V |
| | 2+2 4 | 3+4 2 | 2+2 4 | 3+4 2 | 2+2 2 | 3+3 2 |
| | 2+2 4 | 5+6 2 | 2+2 4 | 5+6 2 | 3+3 2 | 2+2 2 |
| | 2+2 4 | 3+4 2 | 2+2 4 | 3+4 2 | 2+2 2 | 2+2 2 |
| | 2+2 4 | 5+6 2 | 2+2 4 | 5+6 2 | 3+3 2 | 3+3 2 |
| | 2+2 4 | 3+4 2 | 2+2 4 | 3+4 2 | 3+3 2 | 3+3 2 |
| | 2+2 4 | 5+6 2 | 2+2 4 | 5+6 2 | 2+2 2 | 2+2 2 |
| | 2+2 4 | 3+4 2 | 2+2 4 | 3+4 2 | 3+3 2 | 3+3 2 |
| | 2+2 4 | 5+6 2 | 2+2 4 | 5+6 2 | 2+2 2 | 2+2 2 |
| | 2+2 4 | 3+4 2 | 2+2 4 | 3+4 2 | 3+3 2 | 3+3 2 |
| | 2+2 4 | 5+6 2 | 2+2 4 | 5+6 2 | 2+2 2 | 2+2 2 |
| | 2+2 4 | 3+4 2 | 2+2 4 | 5+6 2 | 2+2 2 | 2+2 2 |
| | 2+2 4 | 5+6 2 | 2+2 4 | 3+4 2 | 3+3 2 | 3+3 2 |
| | 2+2 4 | 3+4 2 | 2+2 4 | 5+6 2 | 2+2 2 | 2+2 2 |
| | 2+2 4 | 3+4 2 | 2+2 4 | 3+4 2 | 3+3 2 | 3+3 2 |
| | 2+2 4 | 5+6 2 | 2+2 4 | 3+4 2 | 2+2 2 | 2+2 2 |
| | 2+2 4 | 5+6 2 | 2+2 4 | 5+6 2 | 3+3 2 | 3+3 2 |
| | 2+2 4 | 3+4 2 | 2+2 4 | 3+4 2 | 2+2 2 | 3+3 2 |
| | 2+2 4 | 5+6 2 | 2+2 4 | 5+6 2 | 3+3 2 | 2+2 2 |
| | 2+2 4 | 5+6 2 | 2+2 4 | 5+6 2 | 2+2 2 | 3+3 2 |
| | 2+2 4 | 3+4 2 | 2+2 4 | 3+4 2 | 3+3 2 | 2+2 2 |
| | 2+2 4 | 5+6 2 | 2+2 4 | 5+6 2 | 3+3 2 | 2+2 2 |
| | 2+2 4 | 5+6 2 | 2+2 4 | 3+4 2 | 2+2 2 | 3+3 2 |
| | 2+2 4 | 3+4 2 | 2+2 4 | 5+6 2 | 3+3 2 | 3+3 2 |
| | 2+2 4 | 3+4 2 | 2+2 4 | 3+4 2 | 2+2 2 | 2+2 2 |
| | 2+2 4 | 5+6 2 | 2+2 4 | 3+4 2 | 2+2 2 | 3+3 2 |
| | 2+2 4 | 3+4 2 | 2+2 4 | 5+6 2 | 3+3 2 | 2+2 2 |
| | 2+2 4 | 3+4 2 | 2+2 4 | 3+4 2 | 3+3 2 | 3+3 2 |
| | 2+2 4 | 3+4 2 | 2+2 4 | 3+4 2 | 2+2 2 | 2+2 2 |
| | 2+2 4 | 5+6 2 | 2+2 4 | 5+6 2 | 3+3 2 | 2+2 2 |
| | 2+2 4 | 3+4 2 | 2+2 4 | 5+6 2 | 2+2 2 | 3+3 2 |
| | | 3+4 2 | | 3+4 2 | 3+3 2 | 3+3 2 |
| | | 5+6 2 | | 5+6 2 | 2+2 2 | 2+2 2 |
| | | 5+6 2 | | 5+6 2 | 2+2 2 | 3+3 2 |
| | | 5+6 2 | | 3+4 2 | 3+3 2 | 2+2 2 |
| | | 3+4 2 | | 3+4 2 | 2+2 2 | 2+2 2 |
| | | 3+4 2 | | 3+4 2 | 3+3 2 | 3+3 2 |
| | | 3+4 2 | | 5+6 2 | 2+2 2 | 2+2 2 |
| | | 5+6 2 | | 5+6 2 | 3+3 2 | 3+3 2 |
| | | 5+6 2 | | 5+6 2 | 2+2 2 | 2+2 2 |
| | | 3+4 2 | | 3+4 2 | 3+3 2 | 3+3 2 |
| | | 3+4 2 | | 3+4 2 | 3+3 2 | 3+3 2 |
| | | 3+4 2 | | 5+6 2 | 2+2 2 | 2+2 2 |
| | | 5+6 2 | | 5+6 2 | 3+3 2 | 3+3 2 |
| | | 5+6 2 | | 3+4 2 | 2+2 2 | 2+2 2 |
| | | 5+6 2 | | 3+4 2 | 3+3 2 | 3+3 2 |
| | | 3+4 2 | | 5+6 2 | 2+2 2 | 3+3 2 |
| | | 5+6 2 | | 3+4 2 | 3+3 2 | 2+2 2 |
| | | 3+4 2 | | 5+6 2 | 2+2 2 | 2+2 2 |
| | | 3+4 2 | | 5+6 2 | 2+2 2 | 3+3 2 |
| | | 5+6 2 | | 3+4 2 | 3+3 2 | 2+2 2 |
| | | 5+6 2 | | 5+6 2 | 2+2 2 | 3+3 2 |

Continued on next column

**TABLE XXVII– continued**

| TQC | ICSD | | NoE | SF | | |
|---|---|---|---|---|---|---|
| | | 3+4 2 | | 3+4 2 | 3+3 2 | 2+2 2 |
| | | 5+6 2 | | 3+4 2 | 2+2 2 | 2+2 2 |
| | | 5+6 2 | | 5+6 2 | 3+3 2 | 3+3 2 |
| | | 3+4 2 | | 5+6 2 | 2+2 2 | 2+2 2 |
| | | 3+4 2 | | 3+4 2 | 3+3 2 | 3+3 2 |
| | | 3+4 2 | | 5+6 2 | 2+2 2 | 2+2 2 |
| | | 5+6 2 | | 3+4 2 | 3+3 2 | 3+3 2 |
| | | 5+6 2 | | 5+6 2 | 2+2 2 | 2+2 2 |
| | | 3+4 2 | | 3+4 2 | 3+3 2 | 3+3 2 |
| 15.2.16 | 109178 | | 112 | Nb1O4Tb1 | | |
| | -A | -GM | -M | -Y | -L | -V |
| | 2+2 4 | 3+4 2 | 2+2 4 | 3+4 2 | 2+2 2 | 3+3 2 |
| | 2+2 4 | 5+6 2 | 2+2 4 | 5+6 2 | 3+3 2 | 2+2 2 |
| | 2+2 4 | 3+4 2 | 2+2 4 | 5+6 2 | 2+2 2 | 2+2 2 |
| | 2+2 4 | 5+6 2 | 2+2 4 | 3+4 2 | 3+3 2 | 3+3 2 |
| | 2+2 4 | 3+4 2 | 2+2 4 | 3+4 2 | 3+3 2 | 3+3 2 |
| | 2+2 4 | 5+6 2 | 2+2 4 | 5+6 2 | 2+2 2 | 2+2 2 |
| | 2+2 4 | 3+4 2 | 2+2 4 | 3+4 2 | 3+3 2 | 3+3 2 |
| | 2+2 4 | 5+6 2 | 2+2 4 | 5+6 2 | 2+2 2 | 2+2 2 |
| | 2+2 4 | 3+4 2 | 2+2 4 | 3+4 2 | 3+3 2 | 3+3 2 |
| | 2+2 4 | 5+6 2 | 2+2 4 | 5+6 2 | 2+2 2 | 2+2 2 |
| | 2+2 4 | 3+4 2 | 2+2 4 | 5+6 2 | 2+2 2 | 2+2 2 |
| | 2+2 4 | 5+6 2 | 2+2 4 | 3+4 2 | 3+3 2 | 3+3 2 |
| | 2+2 4 | 3+4 2 | 2+2 4 | 5+6 2 | 2+2 2 | 3+3 2 |
| | 2+2 4 | 3+4 2 | 2+2 4 | 3+4 2 | 3+3 2 | 2+2 2 |
| | 2+2 4 | 5+6 2 | 2+2 4 | 3+4 2 | 2+2 2 | 2+2 2 |
| | 2+2 4 | 5+6 2 | 2+2 4 | 5+6 2 | 3+3 2 | 3+3 2 |
| | 2+2 4 | 3+4 2 | 2+2 4 | 3+4 2 | 2+2 2 | 3+3 2 |
| | 2+2 4 | 5+6 2 | 2+2 4 | 5+6 2 | 3+3 2 | 2+2 2 |
| | 2+2 4 | 5+6 2 | 2+2 4 | 5+6 2 | 2+2 2 | 2+2 2 |
| | 2+2 4 | 3+4 2 | 2+2 4 | 3+4 2 | 3+3 2 | 3+3 2 |
| | 2+2 4 | 5+6 2 | 2+2 4 | 5+6 2 | 3+3 2 | 3+3 2 |
| | 2+2 4 | 3+4 2 | 2+2 4 | 3+4 2 | 2+2 2 | 2+2 2 |
| | 2+2 4 | 5+6 2 | 2+2 4 | 5+6 2 | 3+3 2 | 3+3 2 |
| | 2+2 4 | 3+4 2 | 2+2 4 | 3+4 2 | 2+2 2 | 2+2 2 |
| | 2+2 4 | 3+4 2 | 2+2 4 | 5+6 2 | 3+3 2 | 3+3 2 |
| | 2+2 4 | 5+6 2 | 2+2 4 | 3+4 2 | 2+2 2 | 2+2 2 |
| | 2+2 4 | 3+4 2 | 2+2 4 | 3+4 2 | 3+3 2 | 3+3 2 |
| | 2+2 4 | 3+4 2 | 2+2 4 | 3+4 2 | 2+2 2 | 2+2 2 |
| | | 5+6 2 | | 5+6 2 | 3+3 2 | 3+3 2 |
| | | 3+4 2 | | 3+4 2 | 2+2 2 | 2+2 2 |
| | | 3+4 2 | | 5+6 2 | 3+3 2 | 3+3 2 |
| | | 5+6 2 | | 5+6 2 | 2+2 2 | 2+2 2 |
| | | 5+6 2 | | 5+6 2 | 2+2 2 | 3+3 2 |
| | | 5+6 2 | | 3+4 2 | 3+3 2 | 2+2 2 |
| | | 3+4 2 | | 3+4 2 | 2+2 2 | 2+2 2 |
| | | 3+4 2 | | 3+4 2 | 3+3 2 | 3+3 2 |
| | | 3+4 2 | | 5+6 2 | 2+2 2 | 3+3 2 |
| | | 5+6 2 | | 5+6 2 | 3+3 2 | 2+2 2 |
| | | 5+6 2 | | 5+6 2 | 3+3 2 | 2+2 2 |
| | | 3+4 2 | | 3+4 2 | 2+2 2 | 3+3 2 |
| | | 3+4 2 | | 3+4 2 | 3+3 2 | 3+3 2 |
| | | 5+6 2 | | 5+6 2 | 3+3 2 | 3+3 2 |
| | | 5+6 2 | | 5+6 2 | 2+2 2 | 2+2 2 |
| | | 3+4 2 | | 3+4 2 | 2+2 2 | 2+2 2 |
| | | 5+6 2 | | 3+4 2 | 2+2 2 | 3+3 2 |
| | | 3+4 2 | | 5+6 2 | 3+3 2 | 2+2 2 |
| | | 5+6 2 | | 5+6 2 | 3+3 2 | 3+3 2 |
| | | 5+6 2 | | 3+4 2 | 2+2 2 | 2+2 2 |
| | | 3+4 2 | | 3+4 2 | 2+2 2 | 2+2 2 |
| | | 3+4 2 | | 5+6 2 | 3+3 2 | 3+3 2 |

Continued on next column

**TABLE XXVII– continued**

| TQC | ICSD | | NoE | SF | | |
|---|---|---|---|---|---|---|
| | | 3+4 2 | | 5+6 2 | 2+2 2 | 2+2 2 |
| | | 5+6 2 | | 3+4 2 | 3+3 2 | 3+3 2 |
| | | 5+6 2 | | 5+6 2 | 2+2 2 | 2+2 2 |
| | | 5+6 2 | | 3+4 2 | 3+3 2 | 3+3 2 |
| | | 3+4 2 | | 3+4 2 | 2+2 2 | 2+2 2 |
| | | 3+4 2 | | 5+6 2 | 3+3 2 | 3+3 2 |
| 15.2.17 | 78342 | | 100 | Na1O6Si2V1 | | |
| | -A | -GM | -M | -Y | -L | -V |
| | 2+2 4 | 3+4 2 | 2+2 4 | 3+4 2 | 3+3 2 | 3+3 2 |
| | 2+2 4 | 5+6 2 | 2+2 4 | 5+6 2 | 2+2 2 | 2+2 2 |
| | 2+2 4 | 3+4 2 | 2+2 4 | 3+4 2 | 3+3 2 | 3+3 2 |
| | 2+2 4 | 5+6 2 | 2+2 4 | 5+6 2 | 2+2 2 | 2+2 2 |
| | 2+2 4 | 3+4 2 | 2+2 4 | 5+6 2 | 3+3 2 | 2+2 2 |
| | 2+2 4 | 3+4 2 | 2+2 4 | 3+4 2 | 2+2 2 | 3+3 2 |
| | 2+2 4 | 5+6 2 | 2+2 4 | 3+4 2 | 2+2 2 | 2+2 2 |
| | 2+2 4 | 5+6 2 | 2+2 4 | 5+6 2 | 3+3 2 | 3+3 2 |
| | 2+2 4 | 5+6 2 | 2+2 4 | 5+6 2 | 2+2 2 | 3+3 2 |
| | 2+2 4 | 3+4 2 | 2+2 4 | 3+4 2 | 3+3 2 | 2+2 2 |
| | 2+2 4 | 3+4 2 | 2+2 4 | 3+4 2 | 3+3 2 | 3+3 2 |
| | 2+2 4 | 5+6 2 | 2+2 4 | 5+6 2 | 2+2 2 | 2+2 2 |
| | 2+2 4 | 5+6 2 | 2+2 4 | 5+6 2 | 3+3 2 | 2+2 2 |
| | 2+2 4 | 3+4 2 | 2+2 4 | 3+4 2 | 2+2 2 | 3+3 2 |
| | 2+2 4 | 3+4 2 | 2+2 4 | 3+4 2 | 2+2 2 | 2+2 2 |
| | 2+2 4 | 3+4 2 | 2+2 4 | 5+6 2 | 3+3 2 | 3+3 2 |
| | 2+2 4 | 5+6 2 | 2+2 4 | 3+4 2 | 3+3 2 | 2+2 2 |
| | 2+2 4 | 5+6 2 | 2+2 4 | 3+4 2 | 2+2 2 | 3+3 2 |
| | 2+2 4 | 5+6 2 | 2+2 4 | 5+6 2 | 2+2 2 | 2+2 2 |
| | 2+2 4 | 3+4 2 | 2+2 4 | 5+6 2 | 3+3 2 | 3+3 2 |
| | 2+2 4 | 3+4 2 | 2+2 4 | 5+6 2 | 2+2 2 | 3+3 2 |
| | 2+2 4 | 5+6 2 | 2+2 4 | 5+6 2 | 3+3 2 | 2+2 2 |
| | 2+2 4 | 3+4 2 | 2+2 4 | 3+4 2 | 3+3 2 | 3+3 2 |
| | 2+2 4 | 5+6 2 | 2+2 4 | 3+4 2 | 2+2 2 | 2+2 2 |
| | 2+2 4 | 3+4 2 | 2+2 4 | 5+6 2 | 2+2 2 | 3+3 2 |
| | | 5+6 2 | | 3+4 2 | 3+3 2 | 2+2 2 |
| | | 5+6 2 | | 3+4 2 | 3+3 2 | 3+3 2 |
| | | 3+4 2 | | 5+6 2 | 3+3 2 | 2+2 2 |
| | | 3+4 2 | | 5+6 2 | 2+2 2 | 2+2 2 |
| | | 3+4 2 | | 3+4 2 | 2+2 2 | 3+3 2 |
| | | 5+6 2 | | 3+4 2 | 3+3 2 | 2+2 2 |
| | | 3+4 2 | | 3+4 2 | 2+2 2 | 3+3 2 |
| | | 3+4 2 | | 5+6 2 | 2+2 2 | 2+2 2 |
| | | 5+6 2 | | 5+6 2 | 3+3 2 | 3+3 2 |
| | | 3+4 2 | | 3+4 2 | 2+2 2 | 2+2 2 |
| | | 3+4 2 | | 5+6 2 | 3+3 2 | 3+3 2 |
| | | 5+6 2 | | 3+4 2 | 3+3 2 | 3+3 2 |
| | | 5+6 2 | | 5+6 2 | 2+2 2 | 2+2 2 |
| | | 3+4 2 | | 3+4 2 | 2+2 2 | 2+2 2 |
| | | 5+6 2 | | 5+6 2 | 3+3 2 | 3+3 2 |
| | | 3+4 2 | | 5+6 2 | 3+3 2 | 2+2 2 |
| | | 5+6 2 | | 3+4 2 | 2+2 2 | 2+2 2 |
| | | 5+6 2 | | 5+6 2 | 3+3 2 | 3+3 2 |
| | | 5+6 2 | | 3+4 2 | 2+2 2 | 3+3 2 |
| | | 5+6 2 | | 3+4 2 | 2+2 2 | 2+2 2 |
| | | 3+4 2 | | 5+6 2 | 3+3 2 | 3+3 2 |
| | | 3+4 2 | | 3+4 2 | 2+2 2 | 2+2 2 |
| | | 5+6 2 | | 5+6 2 | 3+3 2 | 3+3 2 |
| | | 3+4 2 | | 5+6 2 | 3+3 2 | 2+2 2 |
| | | 3+4 2 | | 3+4 2 | 2+2 2 | 3+3 2 |
| 15.2.18 | 163133 | | 100 | Na1O6Si2V1 | | |
| | -A | -GM | -M | -Y | -L | -V |
| | 2+2 4 | 3+4 2 | 2+2 4 | 3+4 2 | 2+2 2 | 3+3 2 |
| | 2+2 4 | 5+6 2 | 2+2 4 | 5+6 2 | 3+3 2 | 2+2 2 |
| | 2+2 4 | 3+4 2 | 2+2 4 | 3+4 2 | 2+2 2 | 3+3 2 |
| | 2+2 4 | 5+6 2 | 2+2 4 | 5+6 2 | 3+3 2 | 2+2 2 |
| | 2+2 4 | 3+4 2 | 2+2 4 | 3+4 2 | 2+2 2 | 2+2 2 |
| | 2+2 4 | 3+4 2 | 2+2 4 | 5+6 2 | 3+3 2 | 3+3 2 |
| | 2+2 4 | 5+6 2 | 2+2 4 | 3+4 2 | 3+3 2 | 2+2 2 |
| | 2+2 4 | 5+6 2 | 2+2 4 | 5+6 2 | 2+2 2 | 3+3 2 |
| | 2+2 4 | 5+6 2 | 2+2 4 | 5+6 2 | 3+3 2 | 3+3 2 |
| | 2+2 4 | 3+4 2 | 2+2 4 | 3+4 2 | 2+2 2 | 2+2 2 |
| | 2+2 4 | 5+6 2 | 2+2 4 | 3+4 2 | 3+3 2 | 3+3 2 |
| | 2+2 4 | 3+4 2 | 2+2 4 | 5+6 2 | 2+2 2 | 2+2 2 |
| | 2+2 4 | 5+6 2 | 2+2 4 | 5+6 2 | 2+2 2 | 2+2 2 |
| | 2+2 4 | 3+4 2 | 2+2 4 | 3+4 2 | 3+3 2 | 3+3 2 |
| | 2+2 4 | 3+4 2 | 2+2 4 | 3+4 2 | 2+2 2 | 2+2 2 |
| | 2+2 4 | 3+4 2 | 2+2 4 | 5+6 2 | 3+3 2 | 3+3 2 |
| | 2+2 4 | 5+6 2 | 2+2 4 | 3+4 2 | 2+2 2 | 2+2 2 |
| | 2+2 4 | 5+6 2 | 2+2 4 | 3+4 2 | 3+3 2 | 3+3 2 |
| | 2+2 4 | 3+4 2 | 2+2 4 | 5+6 2 | 3+3 2 | 2+2 2 |
| | 2+2 4 | 5+6 2 | 2+2 4 | 5+6 2 | 2+2 2 | 3+3 2 |
| | 2+2 4 | 3+4 2 | 2+2 4 | 5+6 2 | 3+3 2 | 3+3 2 |
| | 2+2 4 | 5+6 2 | 2+2 4 | 5+6 2 | 2+2 2 | 3+3 2 |
| | 2+2 4 | 3+4 2 | 2+2 4 | 3+4 2 | 2+2 2 | 2+2 2 |
| | 2+2 4 | 5+6 2 | 2+2 4 | 3+4 2 | 3+3 2 | 2+2 2 |
| | 2+2 4 | 3+4 2 | 2+2 4 | 5+6 2 | 3+3 2 | 3+3 2 |
| | | 5+6 2 | | 3+4 2 | 2+2 2 | 2+2 2 |
| | | 5+6 2 | | 3+4 2 | 2+2 2 | 3+3 2 |
| | | 3+4 2 | | 5+6 2 | 2+2 2 | 2+2 2 |
| | | 3+4 2 | | 5+6 2 | 3+3 2 | 2+2 2 |
| | | 5+6 2 | | 3+4 2 | 3+3 2 | 3+3 2 |
| | | 3+4 2 | | 3+4 2 | 2+2 2 | 2+2 2 |
| | | 3+4 2 | | 3+4 2 | 3+3 2 | 3+3 2 |
| | | 3+4 2 | | 5+6 2 | 3+3 2 | 3+3 2 |
| | | 5+6 2 | | 5+6 2 | 2+2 2 | 2+2 2 |
| | | 3+4 2 | | 3+4 2 | 3+3 2 | 2+2 2 |
| | | 3+4 2 | | 5+6 2 | 2+2 2 | 3+3 2 |
| | | 3+4 2 | | 3+4 2 | 3+3 2 | 3+3 2 |
| | | 5+6 2 | | 3+4 2 | 2+2 2 | 2+2 2 |
| | | 5+6 2 | | 5+6 2 | 3+3 2 | 3+3 2 |
| | | 3+4 2 | | 5+6 2 | 2+2 2 | 2+2 2 |
| | | 5+6 2 | | 5+6 2 | 2+2 2 | 2+2 2 |
| | | 5+6 2 | | 3+4 2 | 3+3 2 | 2+2 2 |
| | | 5+6 2 | | 5+6 2 | 2+2 2 | 3+3 2 |
| | | 5+6 2 | | 3+4 2 | 3+3 2 | 3+3 2 |
| | | 3+4 2 | | 3+4 2 | 3+3 2 | 3+3 2 |
| | | 5+6 2 | | 5+6 2 | 3+3 2 | 2+2 2 |
| | | 3+4 2 | | 3+4 2 | 2+2 2 | 2+2 2 |
| | | 5+6 2 | | 5+6 2 | 2+2 2 | 3+3 2 |
| | | 3+4 2 | | 5+6 2 | 2+2 2 | 2+2 2 |
| | | 3+4 2 | | 3+4 2 | 3+3 2 | 3+3 2 |
| 15.2.19 | 163132 | | 100 | Na1O6Si2V1 | | |
| | -A | -GM | -M | -Y | -L | -V |
| | 2+2 4 | 3+4 2 | 2+2 4 | 3+4 2 | 2+2 2 | 3+3 2 |
| | 2+2 4 | 5+6 2 | 2+2 4 | 5+6 2 | 3+3 2 | 2+2 2 |
| | 2+2 4 | 3+4 2 | 2+2 4 | 3+4 2 | 2+2 2 | 3+3 2 |
| | 2+2 4 | 5+6 2 | 2+2 4 | 5+6 2 | 3+3 2 | 2+2 2 |
| | 2+2 4 | 3+4 2 | 2+2 4 | 5+6 2 | 2+2 2 | 2+2 2 |
| | 2+2 4 | 3+4 2 | 2+2 4 | 3+4 2 | 3+3 2 | 3+3 2 |
| | 2+2 4 | 5+6 2 | 2+2 4 | 3+4 2 | 3+3 2 | 2+2 2 |
| | 2+2 4 | 5+6 2 | 2+2 4 | 5+6 2 | 2+2 2 | 3+3 2 |
| | 2+2 4 | 5+6 2 | 2+2 4 | 5+6 2 | 3+3 2 | 3+3 2 |
| | 2+2 4 | 3+4 2 | 2+2 4 | 3+4 2 | 2+2 2 | 2+2 2 |

Continued on next column

**TABLE XXVII– continued**

| TQC | ICSD | | NoE | | SF | |
|---|---|---|---|---|---|---|
| | 2+2 4 | 5+6 2 | 2+2 4 | 3+4 2 | 3+3 2 | 3+3 2 |
| | 2+2 4 | 3+4 2 | 2+2 4 | 5+6 2 | 2+2 2 | 2+2 2 |
| | 2+2 4 | 5+6 2 | 2+2 4 | 5+6 2 | 2+2 2 | 2+2 2 |
| | 2+2 4 | 3+4 2 | 2+2 4 | 3+4 2 | 3+3 2 | 3+3 2 |
| | 2+2 4 | 3+4 2 | 2+2 4 | 3+4 2 | 2+2 2 | 2+2 2 |
| | 2+2 4 | 3+4 2 | 2+2 4 | 5+6 2 | 3+3 2 | 3+3 2 |
| | 2+2 4 | 5+6 2 | 2+2 4 | 3+4 2 | 2+2 2 | 2+2 2 |
| | 2+2 4 | 5+6 2 | 2+2 4 | 3+4 2 | 3+3 2 | 3+3 2 |
| | 2+2 4 | 3+4 2 | 2+2 4 | 5+6 2 | 3+3 2 | 2+2 2 |
| | 2+2 4 | 5+6 2 | 2+2 4 | 5+6 2 | 2+2 2 | 3+3 2 |
| | 2+2 4 | 3+4 2 | 2+2 4 | 5+6 2 | 3+3 2 | 3+3 2 |
| | 2+2 4 | 5+6 2 | 2+2 4 | 5+6 2 | 2+2 2 | 3+3 2 |
| | 2+2 4 | 3+4 2 | 2+2 4 | 3+4 2 | 2+2 2 | 2+2 2 |
| | 2+2 4 | 5+6 2 | 2+2 4 | 3+4 2 | 3+3 2 | 2+2 2 |
| | 2+2 4 | 3+4 2 | 2+2 4 | 5+6 2 | 3+3 2 | 3+3 2 |
| | | 5+6 2 | | 3+4 2 | 2+2 2 | 2+2 2 |
| | | 5+6 2 | | 3+4 2 | 2+2 2 | 3+3 2 |
| | | 3+4 2 | | 5+6 2 | 2+2 2 | 2+2 2 |
| | | 3+4 2 | | 5+6 2 | 3+3 2 | 2+2 2 |
| | | 5+6 2 | | 3+4 2 | 3+3 2 | 3+3 2 |
| | | 3+4 2 | | 3+4 2 | 3+3 2 | 2+2 2 |
| | | 3+4 2 | | 3+4 2 | 2+2 2 | 3+3 2 |
| | | 3+4 2 | | 5+6 2 | 3+3 2 | 2+2 2 |
| | | 5+6 2 | | 5+6 2 | 2+2 2 | 3+3 2 |
| | | 3+4 2 | | 3+4 2 | 3+3 2 | 2+2 2 |
| | | 3+4 2 | | 5+6 2 | 2+2 2 | 3+3 2 |
| | | 3+4 2 | | 3+4 2 | 3+3 2 | 3+3 2 |
| | | 5+6 2 | | 5+6 2 | 2+2 2 | 2+2 2 |
| | | 5+6 2 | | 3+4 2 | 3+3 2 | 3+3 2 |
| | | 5+6 2 | | 5+6 2 | 2+2 2 | 2+2 2 |
| | | 3+4 2 | | 5+6 2 | 2+2 2 | 2+2 2 |
| | | 5+6 2 | | 3+4 2 | 3+3 2 | 3+3 2 |
| | | 5+6 2 | | 5+6 2 | 2+2 2 | 2+2 2 |
| | | 5+6 2 | | 3+4 2 | 3+3 2 | 3+3 2 |
| | | 5+6 2 | | 3+4 2 | 3+3 2 | 3+3 2 |
| | | 3+4 2 | | 5+6 2 | 3+3 2 | 2+2 2 |
| | | 3+4 2 | | 3+4 2 | 2+2 2 | 2+2 2 |
| | | 5+6 2 | | 5+6 2 | 2+2 2 | 3+3 2 |
| | | 3+4 2 | | 5+6 2 | 2+2 2 | 2+2 2 |
| | | 3+4 2 | | 3+4 2 | 3+3 2 | 3+3 2 |
| 15.2.20 | 163131 | | 100 | | Na1O6Si2V1 | |
| | -A | -GM | -M | -Y | -L | -V |
| | 2+2 4 | 3+4 2 | 2+2 4 | 3+4 2 | 3+3 2 | 3+3 2 |
| | 2+2 4 | 5+6 2 | 2+2 4 | 5+6 2 | 2+2 2 | 2+2 2 |
| | 2+2 4 | 3+4 2 | 2+2 4 | 3+4 2 | 3+3 2 | 3+3 2 |
| | 2+2 4 | 5+6 2 | 2+2 4 | 5+6 2 | 2+2 2 | 2+2 2 |
| | 2+2 4 | 3+4 2 | 2+2 4 | 3+4 2 | 3+3 2 | 2+2 2 |
| | 2+2 4 | 3+4 2 | 2+2 4 | 5+6 2 | 2+2 2 | 3+3 2 |
| | 2+2 4 | 5+6 2 | 2+2 4 | 3+4 2 | 2+2 2 | 2+2 2 |
| | 2+2 4 | 5+6 2 | 2+2 4 | 5+6 2 | 3+3 2 | 3+3 2 |
| | 2+2 4 | 5+6 2 | 2+2 4 | 5+6 2 | 2+2 2 | 3+3 2 |
| | 2+2 4 | 3+4 2 | 2+2 4 | 3+4 2 | 3+3 2 | 2+2 2 |
| | 2+2 4 | 5+6 2 | 2+2 4 | 3+4 2 | 2+2 2 | 3+3 2 |
| | 2+2 4 | 3+4 2 | 2+2 4 | 5+6 2 | 3+3 2 | 2+2 2 |
| | 2+2 4 | 5+6 2 | 2+2 4 | 5+6 2 | 3+3 2 | 2+2 2 |
| | 2+2 4 | 3+4 2 | 2+2 4 | 3+4 2 | 2+2 2 | 3+3 2 |
| | 2+2 4 | 3+4 2 | 2+2 4 | 3+4 2 | 3+3 2 | 2+2 2 |
| | 2+2 4 | 3+4 2 | 2+2 4 | 5+6 2 | 2+2 2 | 3+3 2 |
| | 2+2 4 | 5+6 2 | 2+2 4 | 3+4 2 | 3+3 2 | 2+2 2 |
| | 2+2 4 | 3+4 2 | 2+2 4 | 3+4 2 | 2+2 2 | 3+3 2 |
| | 2+2 4 | 5+6 2 | 2+2 4 | 5+6 2 | 2+2 2 | 2+2 2 |

Continued on next column

**TABLE XXVII– continued**

| TQC | ICSD | | NoE | | SF | |
|---|---|---|---|---|---|---|
| | 2+2 4 | 5+6 2 | 2+2 4 | 5+6 2 | 3+3 2 | 3+3 2 |
| | 2+2 4 | 3+4 2 | 2+2 4 | 5+6 2 | 2+2 2 | 3+3 2 |
| | 2+2 4 | 5+6 2 | 2+2 4 | 5+6 2 | 3+3 2 | 3+3 2 |
| | 2+2 4 | 3+4 2 | 2+2 4 | 3+4 2 | 3+3 2 | 2+2 2 |
| | 2+2 4 | 5+6 2 | 2+2 4 | 3+4 2 | 2+2 2 | 2+2 2 |
| | 2+2 4 | 3+4 2 | 2+2 4 | 5+6 2 | 2+2 2 | 3+3 2 |
| | | 5+6 2 | | 3+4 2 | 3+3 2 | 2+2 2 |
| | | 5+6 2 | | 3+4 2 | 3+3 2 | 3+3 2 |
| | | 3+4 2 | | 5+6 2 | 3+3 2 | 2+2 2 |
| | | 3+4 2 | | 5+6 2 | 2+2 2 | 2+2 2 |
| | | 5+6 2 | | 3+4 2 | 2+2 2 | 3+3 2 |
| | | 3+4 2 | | 3+4 2 | 2+2 2 | 2+2 2 |
| | | 3+4 2 | | 3+4 2 | 3+3 2 | 3+3 2 |
| | | 5+6 2 | | 5+6 2 | 2+2 2 | 2+2 2 |
| | | 3+4 2 | | 5+6 2 | 3+3 2 | 3+3 2 |
| | | 3+4 2 | | 3+4 2 | 2+2 2 | 2+2 2 |
| | | 3+4 2 | | 5+6 2 | 2+2 2 | 3+3 2 |
| | | 3+4 2 | | 3+4 2 | 3+3 2 | 3+3 2 |
| | | 5+6 2 | | 5+6 2 | 3+3 2 | 2+2 2 |
| | | 5+6 2 | | 3+4 2 | 2+2 2 | 3+3 2 |
| | | 3+4 2 | | 5+6 2 | 3+3 2 | 2+2 2 |
| | | 5+6 2 | | 3+4 2 | 3+3 2 | 2+2 2 |
| | | 5+6 2 | | 5+6 2 | 2+2 2 | 3+3 2 |
| | | 5+6 2 | | 5+6 2 | 3+3 2 | 2+2 2 |
| | | 5+6 2 | | 3+4 2 | 2+2 2 | 3+3 2 |
| | | 5+6 2 | | 3+4 2 | 2+2 2 | 3+3 2 |
| | | 3+4 2 | | 5+6 2 | 2+2 2 | 2+2 2 |
| | | 3+4 2 | | 3+4 2 | 3+3 2 | 2+2 2 |
| | | 5+6 2 | | 5+6 2 | 3+3 2 | 3+3 2 |
| | | 3+4 2 | | 5+6 2 | 3+3 2 | 2+2 2 |
| | | 3+4 2 | | 3+4 2 | 2+2 2 | 3+3 2 |
| 15.2.21 | 163130 | | 100 | | Na1O6Si2V1 | |
| | -A | -GM | -M | -Y | -L | -V |
| | 2+2 4 | 3+4 2 | 2+2 4 | 3+4 2 | 3+3 2 | 3+3 2 |
| | 2+2 4 | 5+6 2 | 2+2 4 | 5+6 2 | 2+2 2 | 2+2 2 |
| | 2+2 4 | 3+4 2 | 2+2 4 | 3+4 2 | 3+3 2 | 3+3 2 |
| | 2+2 4 | 5+6 2 | 2+2 4 | 5+6 2 | 2+2 2 | 2+2 2 |
| | 2+2 4 | 3+4 2 | 2+2 4 | 3+4 2 | 3+3 2 | 2+2 2 |
| | 2+2 4 | 3+4 2 | 2+2 4 | 5+6 2 | 2+2 2 | 3+3 2 |
| | 2+2 4 | 5+6 2 | 2+2 4 | 3+4 2 | 2+2 2 | 2+2 2 |
| | 2+2 4 | 5+6 2 | 2+2 4 | 5+6 2 | 3+3 2 | 3+3 2 |
| | 2+2 4 | 5+6 2 | 2+2 4 | 5+6 2 | 2+2 2 | 3+3 2 |
| | 2+2 4 | 3+4 2 | 2+2 4 | 3+4 2 | 3+3 2 | 2+2 2 |
| | 2+2 4 | 5+6 2 | 2+2 4 | 3+4 2 | 2+2 2 | 3+3 2 |
| | 2+2 4 | 3+4 2 | 2+2 4 | 5+6 2 | 3+3 2 | 2+2 2 |
| | 2+2 4 | 5+6 2 | 2+2 4 | 5+6 2 | 3+3 2 | 2+2 2 |
| | 2+2 4 | 3+4 2 | 2+2 4 | 3+4 2 | 2+2 2 | 3+3 2 |
| | 2+2 4 | 3+4 2 | 2+2 4 | 3+4 2 | 3+3 2 | 2+2 2 |
| | 2+2 4 | 3+4 2 | 2+2 4 | 5+6 2 | 2+2 2 | 3+3 2 |
| | 2+2 4 | 5+6 2 | 2+2 4 | 3+4 2 | 3+3 2 | 2+2 2 |
| | 2+2 4 | 3+4 2 | 2+2 4 | 3+4 2 | 2+2 2 | 3+3 2 |
| | 2+2 4 | 5+6 2 | 2+2 4 | 5+6 2 | 2+2 2 | 2+2 2 |
| | 2+2 4 | 5+6 2 | 2+2 4 | 5+6 2 | 3+3 2 | 3+3 2 |
| | 2+2 4 | 3+4 2 | 2+2 4 | 5+6 2 | 2+2 2 | 3+3 2 |
| | 2+2 4 | 5+6 2 | 2+2 4 | 5+6 2 | 3+3 2 | 3+3 2 |
| | 2+2 4 | 3+4 2 | 2+2 4 | 3+4 2 | 3+3 2 | 2+2 2 |
| | 2+2 4 | 5+6 2 | 2+2 4 | 3+4 2 | 2+2 2 | 2+2 2 |
| | 2+2 4 | 3+4 2 | 2+2 4 | 5+6 2 | 2+2 2 | 3+3 2 |
| | | 5+6 2 | | 3+4 2 | 3+3 2 | 2+2 2 |
| | | 5+6 2 | | 3+4 2 | 3+3 2 | 3+3 2 |
| | | 3+4 2 | | 5+6 2 | 3+3 2 | 2+2 2 |

Continued on next column

**TABLE XXVII– continued**

| TQC | ICSD | | NoE | SF | | |
|---|---|---|---|---|---|---|
| | | 3+4 2 | | 5+6 2 | 2+2 2 | 3+3 2 |
| | | 5+6 2 | | 3+4 2 | 2+2 2 | 2+2 2 |
| | | 3+4 2 | | 3+4 2 | 3+3 2 | 2+2 2 |
| | | 3+4 2 | | 3+4 2 | 2+2 2 | 3+3 2 |
| | | 5+6 2 | | 5+6 2 | 2+2 2 | 3+3 2 |
| | | 3+4 2 | | 5+6 2 | 3+3 2 | 2+2 2 |
| | | 3+4 2 | | 3+4 2 | 2+2 2 | 2+2 2 |
| | | 3+4 2 | | 5+6 2 | 2+2 2 | 3+3 2 |
| | | 3+4 2 | | 5+6 2 | 3+3 2 | 3+3 2 |
| | | 5+6 2 | | 3+4 2 | 3+3 2 | 2+2 2 |
| | | 5+6 2 | | 3+4 2 | 2+2 2 | 3+3 2 |
| | | 3+4 2 | | 5+6 2 | 3+3 2 | 2+2 2 |
| | | 5+6 2 | | 3+4 2 | 3+3 2 | 2+2 2 |
| | | 5+6 2 | | 5+6 2 | 2+2 2 | 3+3 2 |
| | | 5+6 2 | | 5+6 2 | 3+3 2 | 2+2 2 |
| | | 5+6 2 | | 3+4 2 | 2+2 2 | 3+3 2 |
| | | 3+4 2 | | 3+4 2 | 2+2 2 | 2+2 2 |
| | | 5+6 2 | | 5+6 2 | 2+2 2 | 3+3 2 |
| | | 3+4 2 | | 3+4 2 | 3+3 2 | 2+2 2 |
| | | 5+6 2 | | 5+6 2 | 3+3 2 | 3+3 2 |
| | | 3+4 2 | | 5+6 2 | 3+3 2 | 2+2 2 |
| | | 3+4 2 | | 3+4 2 | 2+2 2 | 3+3 2 |
| 15.2.22 | 163129 | | 100 | Na1O6Si2V1 | | |
| | -A | -GM | -M | -Y | -L | -V |
| | 2+2 4 | 3+4 2 | 2+2 4 | 3+4 2 | 2+2 2 | 3+3 2 |
| | 2+2 4 | 5+6 2 | 2+2 4 | 5+6 2 | 3+3 2 | 2+2 2 |
| | 2+2 4 | 3+4 2 | 2+2 4 | 3+4 2 | 2+2 2 | 3+3 2 |
| | 2+2 4 | 5+6 2 | 2+2 4 | 5+6 2 | 3+3 2 | 2+2 2 |
| | 2+2 4 | 3+4 2 | 2+2 4 | 5+6 2 | 2+2 2 | 2+2 2 |
| | 2+2 4 | 3+4 2 | 2+2 4 | 3+4 2 | 3+3 2 | 3+3 2 |
| | 2+2 4 | 5+6 2 | 2+2 4 | 3+4 2 | 3+3 2 | 2+2 2 |
| | 2+2 4 | 5+6 2 | 2+2 4 | 5+6 2 | 2+2 2 | 3+3 2 |
| | 2+2 4 | 5+6 2 | 2+2 4 | 5+6 2 | 3+3 2 | 3+3 2 |
| | 2+2 4 | 3+4 2 | 2+2 4 | 3+4 2 | 2+2 2 | 2+2 2 |
| | 2+2 4 | 5+6 2 | 2+2 4 | 3+4 2 | 3+3 2 | 3+3 2 |
| | 2+2 4 | 3+4 2 | 2+2 4 | 5+6 2 | 2+2 2 | 2+2 2 |
| | 2+2 4 | 5+6 2 | 2+2 4 | 5+6 2 | 2+2 2 | 2+2 2 |
| | 2+2 4 | 3+4 2 | 2+2 4 | 3+4 2 | 3+3 2 | 3+3 2 |
| | 2+2 4 | 3+4 2 | 2+2 4 | 3+4 2 | 2+2 2 | 2+2 2 |
| | 2+2 4 | 3+4 2 | 2+2 4 | 5+6 2 | 3+3 2 | 3+3 2 |
| | 2+2 4 | 5+6 2 | 2+2 4 | 3+4 2 | 2+2 2 | 2+2 2 |
| | 2+2 4 | 5+6 2 | 2+2 4 | 3+4 2 | 3+3 2 | 3+3 2 |
| | 2+2 4 | 3+4 2 | 2+2 4 | 5+6 2 | 3+3 2 | 2+2 2 |
| | 2+2 4 | 5+6 2 | 2+2 4 | 5+6 2 | 2+2 2 | 3+3 2 |
| | 2+2 4 | 3+4 2 | 2+2 4 | 5+6 2 | 3+3 2 | 3+3 2 |
| | 2+2 4 | 5+6 2 | 2+2 4 | 5+6 2 | 2+2 2 | 3+3 2 |
| | 2+2 4 | 3+4 2 | 2+2 4 | 3+4 2 | 2+2 2 | 2+2 2 |
| | 2+2 4 | 5+6 2 | 2+2 4 | 3+4 2 | 3+3 2 | 2+2 2 |
| | 2+2 4 | 3+4 2 | 2+2 4 | 5+6 2 | 3+3 2 | 3+3 2 |
| | | 5+6 2 | | 3+4 2 | 2+2 2 | 2+2 2 |
| | | 5+6 2 | | 3+4 2 | 2+2 2 | 3+3 2 |
| | | 3+4 2 | | 5+6 2 | 3+3 2 | 2+2 2 |
| | | 3+4 2 | | 5+6 2 | 2+2 2 | 3+3 2 |
| | | 3+4 2 | | 3+4 2 | 3+3 2 | 2+2 2 |
| | | 5+6 2 | | 3+4 2 | 2+2 2 | 2+2 2 |
| | | 3+4 2 | | 3+4 2 | 3+3 2 | 3+3 2 |
| | | 5+6 2 | | 5+6 2 | 3+3 2 | 3+3 2 |
| | | 3+4 2 | | 5+6 2 | 2+2 2 | 2+2 2 |
| | | 3+4 2 | | 3+4 2 | 3+3 2 | 2+2 2 |
| | | 3+4 2 | | 5+6 2 | 3+3 2 | 3+3 2 |
| | | 3+4 2 | | 3+4 2 | 2+2 2 | 3+3 2 |
| | | 5+6 2 | | 5+6 2 | 2+2 2 | 2+2 2 |
| | | 5+6 2 | | 3+4 2 | 3+3 2 | 3+3 2 |
| | | 3+4 2 | | 5+6 2 | 2+2 2 | 2+2 2 |
| | | 5+6 2 | | 3+4 2 | 2+2 2 | 2+2 2 |
| | | 5+6 2 | | 5+6 2 | 3+3 2 | 3+3 2 |
| | | 5+6 2 | | 5+6 2 | 2+2 2 | 2+2 2 |
| | | 5+6 2 | | 3+4 2 | 3+3 2 | 3+3 2 |
| | | 5+6 2 | | 3+4 2 | 3+3 2 | 2+2 2 |
| | | 3+4 2 | | 5+6 2 | 3+3 2 | 3+3 2 |
| | | 3+4 2 | | 3+4 2 | 2+2 2 | 2+2 2 |
| | | 5+6 2 | | 5+6 2 | 2+2 2 | 3+3 2 |
| | | 3+4 2 | | 5+6 2 | 2+2 2 | 2+2 2 |
| | | 3+4 2 | | 3+4 2 | 3+3 2 | 3+3 2 |
| 15.2.23 | 163128 | | 100 | Na1O6Si2V1 | | |
| | -A | -GM | -M | -Y | -L | -V |
| | 2+2 4 | 3+4 2 | 2+2 4 | 3+4 2 | 2+2 2 | 3+3 2 |
| | 2+2 4 | 5+6 2 | 2+2 4 | 5+6 2 | 3+3 2 | 2+2 2 |
| | 2+2 4 | 3+4 2 | 2+2 4 | 3+4 2 | 2+2 2 | 3+3 2 |
| | 2+2 4 | 5+6 2 | 2+2 4 | 5+6 2 | 3+3 2 | 2+2 2 |
| | 2+2 4 | 3+4 2 | 2+2 4 | 5+6 2 | 2+2 2 | 2+2 2 |
| | 2+2 4 | 3+4 2 | 2+2 4 | 3+4 2 | 3+3 2 | 3+3 2 |
| | 2+2 4 | 5+6 2 | 2+2 4 | 3+4 2 | 3+3 2 | 2+2 2 |
| | 2+2 4 | 5+6 2 | 2+2 4 | 5+6 2 | 2+2 2 | 3+3 2 |
| | 2+2 4 | 5+6 2 | 2+2 4 | 5+6 2 | 3+3 2 | 3+3 2 |
| | 2+2 4 | 3+4 2 | 2+2 4 | 3+4 2 | 2+2 2 | 2+2 2 |
| | 2+2 4 | 5+6 2 | 2+2 4 | 3+4 2 | 3+3 2 | 3+3 2 |
| | 2+2 4 | 3+4 2 | 2+2 4 | 5+6 2 | 2+2 2 | 2+2 2 |
| | 2+2 4 | 5+6 2 | 2+2 4 | 5+6 2 | 2+2 2 | 2+2 2 |
| | 2+2 4 | 3+4 2 | 2+2 4 | 3+4 2 | 3+3 2 | 3+3 2 |
| | 2+2 4 | 3+4 2 | 2+2 4 | 3+4 2 | 2+2 2 | 2+2 2 |
| | 2+2 4 | 3+4 2 | 2+2 4 | 5+6 2 | 3+3 2 | 3+3 2 |
| | 2+2 4 | 5+6 2 | 2+2 4 | 3+4 2 | 2+2 2 | 2+2 2 |
| | 2+2 4 | 5+6 2 | 2+2 4 | 3+4 2 | 3+3 2 | 3+3 2 |
| | 2+2 4 | 3+4 2 | 2+2 4 | 5+6 2 | 3+3 2 | 2+2 2 |
| | 2+2 4 | 5+6 2 | 2+2 4 | 5+6 2 | 2+2 2 | 3+3 2 |
| | 2+2 4 | 3+4 2 | 2+2 4 | 5+6 2 | 3+3 2 | 3+3 2 |
| | 2+2 4 | 5+6 2 | 2+2 4 | 5+6 2 | 2+2 2 | 3+3 2 |
| | 2+2 4 | 3+4 2 | 2+2 4 | 3+4 2 | 2+2 2 | 2+2 2 |
| | 2+2 4 | 5+6 2 | 2+2 4 | 3+4 2 | 3+3 2 | 2+2 2 |
| | 2+2 4 | 3+4 2 | 2+2 4 | 5+6 2 | 3+3 2 | 3+3 2 |
| | | 5+6 2 | | 3+4 2 | 2+2 2 | 2+2 2 |
| | | 5+6 2 | | 3+4 2 | 2+2 2 | 3+3 2 |
| | | 3+4 2 | | 5+6 2 | 2+2 2 | 2+2 2 |
| | | 3+4 2 | | 5+6 2 | 3+3 2 | 2+2 2 |
| | | 3+4 2 | | 3+4 2 | 3+3 2 | 3+3 2 |
| | | 5+6 2 | | 3+4 2 | 2+2 2 | 2+2 2 |
| | | 3+4 2 | | 3+4 2 | 3+3 2 | 3+3 2 |
| | | 5+6 2 | | 5+6 2 | 3+3 2 | 2+2 2 |
| | | 3+4 2 | | 5+6 2 | 2+2 2 | 3+3 2 |
| | | 3+4 2 | | 3+4 2 | 3+3 2 | 2+2 2 |
| | | 3+4 2 | | 5+6 2 | 3+3 2 | 3+3 2 |
| | | 3+4 2 | | 3+4 2 | 2+2 2 | 3+3 2 |
| | | 5+6 2 | | 5+6 2 | 2+2 2 | 2+2 2 |
| | | 5+6 2 | | 3+4 2 | 3+3 2 | 3+3 2 |
| | | 3+4 2 | | 5+6 2 | 2+2 2 | 2+2 2 |
| | | 5+6 2 | | 3+4 2 | 2+2 2 | 2+2 2 |
| | | 5+6 2 | | 5+6 2 | 3+3 2 | 3+3 2 |
| | | 5+6 2 | | 5+6 2 | 2+2 2 | 2+2 2 |
| | | 5+6 2 | | 3+4 2 | 3+3 2 | 3+3 2 |
| | | 5+6 2 | | 3+4 2 | 3+3 2 | 2+2 2 |
| | | 3+4 2 | | 5+6 2 | 3+3 2 | 3+3 2 |

Continued on next column

Continued on next column

**TABLE XXVII– continued**

| TQC | ICSD | | NoE | SF | | |
|---|---|---|---|---|---|---|
| | | 3+4 2 | | 3+4 2 | 2+2 2 | 2+2 2 |
| | | 5+6 2 | | 5+6 2 | 2+2 2 | 3+3 2 |
| | | 3+4 2 | | 5+6 2 | 2+2 2 | 2+2 2 |
| | | 3+4 2 | | 3+4 2 | 3+3 2 | 3+3 2 |
| 15.2.24 | 163127 | | 100 | Na1O6Si2V1 | | |
| | -A | -GM | -M | -Y | -L | -V |
| | 2+2 4 | 3+4 2 | 2+2 4 | 3+4 2 | 2+2 2 | 3+3 2 |
| | 2+2 4 | 5+6 2 | 2+2 4 | 5+6 2 | 3+3 2 | 2+2 2 |
| | 2+2 4 | 3+4 2 | 2+2 4 | 3+4 2 | 2+2 2 | 3+3 2 |
| | 2+2 4 | 5+6 2 | 2+2 4 | 5+6 2 | 3+3 2 | 2+2 2 |
| | 2+2 4 | 3+4 2 | 2+2 4 | 5+6 2 | 2+2 2 | 2+2 2 |
| | 2+2 4 | 3+4 2 | 2+2 4 | 3+4 2 | 3+3 2 | 3+3 2 |
| | 2+2 4 | 5+6 2 | 2+2 4 | 3+4 2 | 3+3 2 | 2+2 2 |
| | 2+2 4 | 5+6 2 | 2+2 4 | 5+6 2 | 2+2 2 | 3+3 2 |
| | 2+2 4 | 5+6 2 | 2+2 4 | 5+6 2 | 3+3 2 | 3+3 2 |
| | 2+2 4 | 3+4 2 | 2+2 4 | 3+4 2 | 2+2 2 | 2+2 2 |
| | 2+2 4 | 5+6 2 | 2+2 4 | 3+4 2 | 2+2 2 | 3+3 2 |
| | 2+2 4 | 3+4 2 | 2+2 4 | 5+6 2 | 3+3 2 | 2+2 2 |
| | 2+2 4 | 5+6 2 | 2+2 4 | 5+6 2 | 2+2 2 | 2+2 2 |
| | 2+2 4 | 3+4 2 | 2+2 4 | 3+4 2 | 3+3 2 | 3+3 2 |
| | 2+2 4 | 3+4 2 | 2+2 4 | 3+4 2 | 2+2 2 | 2+2 2 |
| | 2+2 4 | 3+4 2 | 2+2 4 | 5+6 2 | 3+3 2 | 3+3 2 |
| | 2+2 4 | 5+6 2 | 2+2 4 | 3+4 2 | 2+2 2 | 2+2 2 |
| | 2+2 4 | 5+6 2 | 2+2 4 | 3+4 2 | 3+3 2 | 3+3 2 |
| | 2+2 4 | 5+6 2 | 2+2 4 | 5+6 2 | 3+3 2 | 2+2 2 |
| | 2+2 4 | 3+4 2 | 2+2 4 | 5+6 2 | 2+2 2 | 3+3 2 |
| | 2+2 4 | 3+4 2 | 2+2 4 | 5+6 2 | 3+3 2 | 3+3 2 |
| | 2+2 4 | 5+6 2 | 2+2 4 | 5+6 2 | 2+2 2 | 3+3 2 |
| | 2+2 4 | 3+4 2 | 2+2 4 | 3+4 2 | 2+2 2 | 2+2 2 |
| | 2+2 4 | 5+6 2 | 2+2 4 | 3+4 2 | 3+3 2 | 2+2 2 |
| | 2+2 4 | 3+4 2 | 2+2 4 | 5+6 2 | 3+3 2 | 3+3 2 |
| | | 5+6 2 | | 3+4 2 | 2+2 2 | 2+2 2 |
| | | 5+6 2 | | 3+4 2 | 2+2 2 | 3+3 2 |
| | | 3+4 2 | | 5+6 2 | 2+2 2 | 2+2 2 |
| | | 3+4 2 | | 5+6 2 | 3+3 2 | 2+2 2 |
| | | 3+4 2 | | 3+4 2 | 3+3 2 | 3+3 2 |
| | | 5+6 2 | | 3+4 2 | 2+2 2 | 2+2 2 |
| | | 3+4 2 | | 3+4 2 | 3+3 2 | 3+3 2 |
| | | 3+4 2 | | 5+6 2 | 3+3 2 | 2+2 2 |
| | | 5+6 2 | | 5+6 2 | 2+2 2 | 3+3 2 |
| | | 3+4 2 | | 3+4 2 | 3+3 2 | 2+2 2 |
| | | 3+4 2 | | 5+6 2 | 2+2 2 | 3+3 2 |
| | | 5+6 2 | | 3+4 2 | 2+2 2 | 3+3 2 |
| | | 5+6 2 | | 5+6 2 | 3+3 2 | 2+2 2 |
| | | 3+4 2 | | 3+4 2 | 3+3 2 | 3+3 2 |
| | | 5+6 2 | | 5+6 2 | 2+2 2 | 2+2 2 |
| | | 3+4 2 | | 5+6 2 | 2+2 2 | 2+2 2 |
| | | 5+6 2 | | 3+4 2 | 3+3 2 | 2+2 2 |
| | | 5+6 2 | | 5+6 2 | 2+2 2 | 3+3 2 |
| | | 5+6 2 | | 3+4 2 | 3+3 2 | 3+3 2 |
| | | 5+6 2 | | 3+4 2 | 3+3 2 | 2+2 2 |
| | | 3+4 2 | | 5+6 2 | 2+2 2 | 3+3 2 |
| | | 3+4 2 | | 3+4 2 | 3+3 2 | 2+2 2 |
| | | 5+6 2 | | 5+6 2 | 2+2 2 | 3+3 2 |
| | | 3+4 2 | | 5+6 2 | 2+2 2 | 2+2 2 |
| | | 3+4 2 | | 3+4 2 | 3+3 2 | 3+3 2 |
| 15.2.25 | 163126 | | 100 | Na1O6Si2V1 | | |
| | -A | -GM | -M | -Y | -L | -V |
| | 2+2 4 | 3+4 2 | 2+2 4 | 3+4 2 | 3+3 2 | 3+3 2 |
| | 2+2 4 | 5+6 2 | 2+2 4 | 5+6 2 | 2+2 2 | 2+2 2 |
| | 2+2 4 | 3+4 2 | 2+2 4 | 3+4 2 | 3+3 2 | 3+3 2 |
| | 2+2 4 | 5+6 2 | 2+2 4 | 5+6 2 | 2+2 2 | 2+2 2 |
| | 2+2 4 | 3+4 2 | 2+2 4 | 5+6 2 | 3+3 2 | 2+2 2 |
| | 2+2 4 | 3+4 2 | 2+2 4 | 3+4 2 | 2+2 2 | 3+3 2 |
| | 2+2 4 | 5+6 2 | 2+2 4 | 3+4 2 | 2+2 2 | 2+2 2 |
| | 2+2 4 | 5+6 2 | 2+2 4 | 5+6 2 | 3+3 2 | 3+3 2 |
| | 2+2 4 | 5+6 2 | 2+2 4 | 5+6 2 | 2+2 2 | 3+3 2 |
| | 2+2 4 | 3+4 2 | 2+2 4 | 3+4 2 | 3+3 2 | 2+2 2 |
| | 2+2 4 | 5+6 2 | 2+2 4 | 3+4 2 | 2+2 2 | 3+3 2 |
| | 2+2 4 | 3+4 2 | 2+2 4 | 5+6 2 | 3+3 2 | 2+2 2 |
| | 2+2 4 | 5+6 2 | 2+2 4 | 5+6 2 | 3+3 2 | 2+2 2 |
| | 2+2 4 | 3+4 2 | 2+2 4 | 3+4 2 | 2+2 2 | 3+3 2 |
| | 2+2 4 | 3+4 2 | 2+2 4 | 3+4 2 | 2+2 2 | 2+2 2 |
| | 2+2 4 | 3+4 2 | 2+2 4 | 5+6 2 | 3+3 2 | 3+3 2 |
| | 2+2 4 | 5+6 2 | 2+2 4 | 3+4 2 | 3+3 2 | 2+2 2 |
| | 2+2 4 | 5+6 2 | 2+2 4 | 3+4 2 | 2+2 2 | 3+3 2 |
| | 2+2 4 | 3+4 2 | 2+2 4 | 5+6 2 | 2+2 2 | 2+2 2 |
| | 2+2 4 | 5+6 2 | 2+2 4 | 5+6 2 | 3+3 2 | 3+3 2 |
| | 2+2 4 | 3+4 2 | 2+2 4 | 5+6 2 | 2+2 2 | 3+3 2 |
| | 2+2 4 | 5+6 2 | 2+2 4 | 5+6 2 | 3+3 2 | 3+3 2 |
| | 2+2 4 | 3+4 2 | 2+2 4 | 3+4 2 | 3+3 2 | 2+2 2 |
| | 2+2 4 | 5+6 2 | 2+2 4 | 3+4 2 | 2+2 2 | 2+2 2 |
| | 2+2 4 | 3+4 2 | 2+2 4 | 5+6 2 | 2+2 2 | 3+3 2 |
| | | 5+6 2 | | 3+4 2 | 3+3 2 | 2+2 2 |
| | | 5+6 2 | | 3+4 2 | 3+3 2 | 3+3 2 |
| | | 3+4 2 | | 5+6 2 | 3+3 2 | 2+2 2 |
| | | 3+4 2 | | 5+6 2 | 2+2 2 | 2+2 2 |
| | | 3+4 2 | | 3+4 2 | 2+2 2 | 3+3 2 |
| | | 5+6 2 | | 3+4 2 | 3+3 2 | 2+2 2 |
| | | 3+4 2 | | 3+4 2 | 2+2 2 | 3+3 2 |
| | | 3+4 2 | | 5+6 2 | 2+2 2 | 2+2 2 |
| | | 5+6 2 | | 5+6 2 | 3+3 2 | 3+3 2 |
| | | 3+4 2 | | 3+4 2 | 2+2 2 | 2+2 2 |
| | | 3+4 2 | | 5+6 2 | 3+3 2 | 3+3 2 |
| | | 5+6 2 | | 3+4 2 | 2+2 2 | 3+3 2 |
| | | 3+4 2 | | 5+6 2 | 3+3 2 | 2+2 2 |
| | | 5+6 2 | | 3+4 2 | 2+2 2 | 3+3 2 |
| | | 5+6 2 | | 5+6 2 | 3+3 2 | 2+2 2 |
| | | 3+4 2 | | 5+6 2 | 3+3 2 | 2+2 2 |
| | | 5+6 2 | | 3+4 2 | 2+2 2 | 2+2 2 |
| | | 5+6 2 | | 5+6 2 | 3+3 2 | 3+3 2 |
| | | 5+6 2 | | 3+4 2 | 2+2 2 | 3+3 2 |
| | | 5+6 2 | | 3+4 2 | 2+2 2 | 3+3 2 |
| | | 3+4 2 | | 5+6 2 | 3+3 2 | 2+2 2 |
| | | 3+4 2 | | 3+4 2 | 2+2 2 | 2+2 2 |
| | | 5+6 2 | | 5+6 2 | 3+3 2 | 3+3 2 |
| | | 3+4 2 | | 5+6 2 | 3+3 2 | 2+2 2 |
| | | 3+4 2 | | 3+4 2 | 2+2 2 | 3+3 2 |
| 15.2.26 | 163125 | | 100 | Na1O6Si2V1 | | |
| | -A | -GM | -M | -Y | -L | -V |
| | 2+2 4 | 3+4 2 | 2+2 4 | 3+4 2 | 3+3 2 | 3+3 2 |
| | 2+2 4 | 5+6 2 | 2+2 4 | 5+6 2 | 2+2 2 | 2+2 2 |
| | 2+2 4 | 3+4 2 | 2+2 4 | 3+4 2 | 3+3 2 | 3+3 2 |
| | 2+2 4 | 5+6 2 | 2+2 4 | 5+6 2 | 2+2 2 | 2+2 2 |
| | 2+2 4 | 3+4 2 | 2+2 4 | 5+6 2 | 3+3 2 | 2+2 2 |
| | 2+2 4 | 3+4 2 | 2+2 4 | 3+4 2 | 2+2 2 | 3+3 2 |
| | 2+2 4 | 5+6 2 | 2+2 4 | 3+4 2 | 2+2 2 | 2+2 2 |
| | 2+2 4 | 5+6 2 | 2+2 4 | 5+6 2 | 3+3 2 | 3+3 2 |
| | 2+2 4 | 5+6 2 | 2+2 4 | 5+6 2 | 2+2 2 | 3+3 2 |
| | 2+2 4 | 3+4 2 | 2+2 4 | 3+4 2 | 3+3 2 | 2+2 2 |
| | 2+2 4 | 3+4 2 | 2+2 4 | 3+4 2 | 3+3 2 | 3+3 2 |
| | 2+2 4 | 5+6 2 | 2+2 4 | 5+6 2 | 2+2 2 | 2+2 2 |

Continued on next column

**TABLE XXVII– continued**

| TQC | ICSD | | NoE | | SF | |
|---|---|---|---|---|---|---|
| | 2+2 4 | 5+6 2 | 2+2 4 | 5+6 2 | 3+3 2 | 2+2 2 |
| | 2+2 4 | 3+4 2 | 2+2 4 | 3+4 2 | 2+2 2 | 3+3 2 |
| | 2+2 4 | 3+4 2 | 2+2 4 | 3+4 2 | 2+2 2 | 2+2 2 |
| | 2+2 4 | 3+4 2 | 2+2 4 | 5+6 2 | 3+3 2 | 3+3 2 |
| | 2+2 4 | 5+6 2 | 2+2 4 | 3+4 2 | 3+3 2 | 2+2 2 |
| | 2+2 4 | 5+6 2 | 2+2 4 | 3+4 2 | 2+2 2 | 3+3 2 |
| | 2+2 4 | 5+6 2 | 2+2 4 | 5+6 2 | 2+2 2 | 2+2 2 |
| | 2+2 4 | 3+4 2 | 2+2 4 | 5+6 2 | 3+3 2 | 3+3 2 |
| | 2+2 4 | 3+4 2 | 2+2 4 | 5+6 2 | 2+2 2 | 3+3 2 |
| | 2+2 4 | 5+6 2 | 2+2 4 | 5+6 2 | 3+3 2 | 2+2 2 |
| | 2+2 4 | 3+4 2 | 2+2 4 | 3+4 2 | 3+3 2 | 3+3 2 |
| | 2+2 4 | 5+6 2 | 2+2 4 | 3+4 2 | 2+2 2 | 2+2 2 |
| | 2+2 4 | 3+4 2 | 2+2 4 | 5+6 2 | 2+2 2 | 3+3 2 |
| | | 5+6 2 | | 3+4 2 | 3+3 2 | 2+2 2 |
| | | 5+6 2 | | 3+4 2 | 3+3 2 | 3+3 2 |
| | | 3+4 2 | | 5+6 2 | 3+3 2 | 2+2 2 |
| | | 3+4 2 | | 5+6 2 | 2+2 2 | 2+2 2 |
| | | 3+4 2 | | 3+4 2 | 2+2 2 | 3+3 2 |
| | | 5+6 2 | | 3+4 2 | 3+3 2 | 2+2 2 |
| | | 3+4 2 | | 3+4 2 | 2+2 2 | 3+3 2 |
| | | 3+4 2 | | 5+6 2 | 2+2 2 | 2+2 2 |
| | | 5+6 2 | | 5+6 2 | 3+3 2 | 3+3 2 |
| | | 3+4 2 | | 3+4 2 | 2+2 2 | 2+2 2 |
| | | 3+4 2 | | 5+6 2 | 3+3 2 | 3+3 2 |
| | | 5+6 2 | | 3+4 2 | 2+2 2 | 3+3 2 |
| | | 5+6 2 | | 5+6 2 | 3+3 2 | 2+2 2 |
| | | 3+4 2 | | 3+4 2 | 2+2 2 | 2+2 2 |
| | | 5+6 2 | | 5+6 2 | 3+3 2 | 3+3 2 |
| | | 3+4 2 | | 5+6 2 | 3+3 2 | 2+2 2 |
| | | 5+6 2 | | 3+4 2 | 2+2 2 | 2+2 2 |
| | | 5+6 2 | | 5+6 2 | 3+3 2 | 3+3 2 |
| | | 5+6 2 | | 3+4 2 | 2+2 2 | 3+3 2 |
| | | 5+6 2 | | 3+4 2 | 2+2 2 | 3+3 2 |
| | | 3+4 2 | | 5+6 2 | 3+3 2 | 2+2 2 |
| | | 3+4 2 | | 3+4 2 | 2+2 2 | 2+2 2 |
| | | 5+6 2 | | 5+6 2 | 3+3 2 | 3+3 2 |
| | | 3+4 2 | | 5+6 2 | 3+3 2 | 2+2 2 |
| | | 3+4 2 | | 3+4 2 | 2+2 2 | 3+3 2 |
| 15.2.27 | 105057 | | 52 | | Mo1P4 | |
| | -A | -GM | -M | -Y | -L | -V |
| | 2+2 4 | 3+4 2 | 2+2 4 | 5+6 2 | 2+2 2 | 2+2 2 |
| | 2+2 4 | 5+6 2 | 2+2 4 | 3+4 2 | 3+3 2 | 3+3 2 |
| | 2+2 4 | 3+4 2 | 2+2 4 | 5+6 2 | 2+2 2 | 3+3 2 |
| | 2+2 4 | 3+4 2 | 2+2 4 | 5+6 2 | 3+3 2 | 2+2 2 |
| | 2+2 4 | 5+6 2 | 2+2 4 | 3+4 2 | 2+2 2 | 2+2 2 |
| | 2+2 4 | 5+6 2 | 2+2 4 | 3+4 2 | 3+3 2 | 3+3 2 |
| | 2+2 4 | 3+4 2 | 2+2 4 | 5+6 2 | 2+2 2 | 3+3 2 |
| | 2+2 4 | 5+6 2 | 2+2 4 | 3+4 2 | 3+3 2 | 2+2 2 |
| | 2+2 4 | 5+6 2 | 2+2 4 | 3+4 2 | 3+3 2 | 3+3 2 |
| | 2+2 4 | 5+6 2 | 2+2 4 | 5+6 2 | 2+2 2 | 2+2 2 |
| | 2+2 4 | 3+4 2 | 2+2 4 | 5+6 2 | 2+2 2 | 3+3 2 |
| | 2+2 4 | 3+4 2 | 2+2 4 | 3+4 2 | 3+3 2 | 2+2 2 |
| | 2+2 4 | 3+4 2 | 2+2 4 | 3+4 2 | 3+3 2 | 3+3 2 |
| | | 3+4 2 | | 5+6 2 | 2+2 2 | 2+2 2 |
| | | 5+6 2 | | 5+6 2 | 3+3 2 | 3+3 2 |
| | | 5+6 2 | | 3+4 2 | 2+2 2 | 2+2 2 |
| | | 3+4 2 | | 5+6 2 | 2+2 2 | 2+2 2 |
| | | 5+6 2 | | 5+6 2 | 3+3 2 | 3+3 2 |
| | | 5+6 2 | | 3+4 2 | 2+2 2 | 2+2 2 |
| | | 3+4 2 | | 3+4 2 | 3+3 2 | 3+3 2 |
| | | 3+4 2 | | 5+6 2 | 2+2 2 | 2+2 2 |

Continued on next column

**TABLE XXVII– continued**

| TQC | ICSD | | NoE | | SF | |
|---|---|---|---|---|---|---|
| | | 3+4 2 | | 5+6 2 | 2+2 2 | 3+3 2 |
| | | 3+4 2 | | 3+4 2 | 3+3 2 | 2+2 2 |
| | | 3+4 2 | | 3+4 2 | 2+2 2 | 3+3 2 |
| | | 5+6 2 | | 5+6 2 | 3+3 2 | 2+2 2 |
| | | 5+6 2 | | 3+4 2 | 2+2 2 | 2+2 2 |
| 15.2.28 | 49802 | | 156 | | Mn2P12Ti1 | |
| | -A | -GM | -M | -Y | -L | -V |
| | 2+2 4 | 3+4 2 | 2+2 4 | 5+6 2 | 2+2 2 | 2+2 2 |
| | 2+2 4 | 3+4 2 | 2+2 4 | 3+4 2 | 2+2 2 | 3+3 2 |
| | 2+2 4 | 5+6 2 | 2+2 4 | 5+6 2 | 3+3 2 | 3+3 2 |
| | 2+2 4 | 3+4 2 | 2+2 4 | 3+4 2 | 3+3 2 | 2+2 2 |
| | 2+2 4 | 3+4 2 | 2+2 4 | 5+6 2 | 2+2 2 | 2+2 2 |
| | 2+2 4 | 5+6 2 | 2+2 4 | 3+4 2 | 2+2 2 | 3+3 2 |
| | 2+2 4 | 5+6 2 | 2+2 4 | 3+4 2 | 3+3 2 | 3+3 2 |
| | 2+2 4 | 5+6 2 | 2+2 4 | 5+6 2 | 3+3 2 | 2+2 2 |
| | 2+2 4 | 3+4 2 | 2+2 4 | 3+4 2 | 2+2 2 | 3+3 2 |
| | 2+2 4 | 3+4 2 | 2+2 4 | 3+4 2 | 3+3 2 | 2+2 2 |
| | 2+2 4 | 5+6 2 | 2+2 4 | 5+6 2 | 2+2 2 | 2+2 2 |
| | 2+2 4 | 3+4 2 | 2+2 4 | 5+6 2 | 3+3 2 | 2+2 2 |
| | 2+2 4 | 5+6 2 | 2+2 4 | 3+4 2 | 3+3 2 | 3+3 2 |
| | 2+2 4 | 3+4 2 | 2+2 4 | 3+4 2 | 2+2 2 | 2+2 2 |
| | 2+2 4 | 3+4 2 | 2+2 4 | 5+6 2 | 3+3 2 | 3+3 2 |
| | 2+2 4 | 5+6 2 | 2+2 4 | 5+6 2 | 2+2 2 | 2+2 2 |
| | 2+2 4 | 5+6 2 | 2+2 4 | 3+4 2 | 3+3 2 | 3+3 2 |
| | 2+2 4 | 3+4 2 | 2+2 4 | 5+6 2 | 2+2 2 | 3+3 2 |
| | 2+2 4 | 5+6 2 | 2+2 4 | 5+6 2 | 2+2 2 | 2+2 2 |
| | 2+2 4 | 5+6 2 | 2+2 4 | 3+4 2 | 3+3 2 | 3+3 2 |
| | 2+2 4 | 3+4 2 | 2+2 4 | 3+4 2 | 2+2 2 | 3+3 2 |
| | 2+2 4 | 5+6 2 | 2+2 4 | 5+6 2 | 2+2 2 | 2+2 2 |
| | 2+2 4 | 3+4 2 | 2+2 4 | 5+6 2 | 3+3 2 | 3+3 2 |
| | 2+2 4 | 5+6 2 | 2+2 4 | 3+4 2 | 3+3 2 | 2+2 2 |
| | 2+2 4 | 3+4 2 | 2+2 4 | 5+6 2 | 3+3 2 | 3+3 2 |
| | 2+2 4 | 5+6 2 | 2+2 4 | 3+4 2 | 2+2 2 | 2+2 2 |
| | 2+2 4 | 5+6 2 | 2+2 4 | 5+6 2 | 3+3 2 | 2+2 2 |
| | 2+2 4 | 3+4 2 | 2+2 4 | 3+4 2 | 2+2 2 | 3+3 2 |
| | 2+2 4 | 3+4 2 | 2+2 4 | 5+6 2 | 3+3 2 | 3+3 2 |
| | 2+2 4 | 5+6 2 | 2+2 4 | 3+4 2 | 2+2 2 | 2+2 2 |
| | 2+2 4 | 5+6 2 | 2+2 4 | 5+6 2 | 2+2 2 | 3+3 2 |
| | 2+2 4 | 3+4 2 | 2+2 4 | 3+4 2 | 3+3 2 | 2+2 2 |
| | 2+2 4 | 5+6 2 | 2+2 4 | 3+4 2 | 3+3 2 | 3+3 2 |
| | 2+2 4 | 5+6 2 | 2+2 4 | 5+6 2 | 2+2 2 | 2+2 2 |
| | 2+2 4 | 3+4 2 | 2+2 4 | 3+4 2 | 3+3 2 | 2+2 2 |
| | 2+2 4 | 3+4 2 | 2+2 4 | 5+6 2 | 2+2 2 | 3+3 2 |
| | 2+2 4 | 3+4 2 | 2+2 4 | 5+6 2 | 3+3 2 | 3+3 2 |
| | 2+2 4 | 3+4 2 | 2+2 4 | 3+4 2 | 3+3 2 | 3+3 2 |
| | 2+2 4 | 5+6 2 | 2+2 4 | 3+4 2 | 2+2 2 | 2+2 2 |
| | | 5+6 2 | | 5+6 2 | 3+3 2 | 2+2 2 |
| | | 3+4 2 | | 3+4 2 | 2+2 2 | 3+3 2 |
| | | 3+4 2 | | 5+6 2 | 2+2 2 | 2+2 2 |
| | | 5+6 2 | | 5+6 2 | 3+3 2 | 3+3 2 |
| | | 5+6 2 | | 3+4 2 | 2+2 2 | 2+2 2 |
| | | 3+4 2 | | 5+6 2 | 3+3 2 | 3+3 2 |
| | | 5+6 2 | | 3+4 2 | 2+2 2 | 2+2 2 |
| | | 3+4 2 | | 3+4 2 | 2+2 2 | 2+2 2 |
| | | 3+4 2 | | 3+4 2 | 3+3 2 | 3+3 2 |
| | | 5+6 2 | | 5+6 2 | 2+2 2 | 2+2 2 |
| | | 5+6 2 | | 5+6 2 | 3+3 2 | 2+2 2 |
| | | 3+4 2 | | 3+4 2 | 2+2 2 | 3+3 2 |
| | | 3+4 2 | | 5+6 2 | 3+3 2 | 3+3 2 |
| | | 5+6 2 | | 3+4 2 | 2+2 2 | 2+2 2 |
| | | 5+6 2 | | 5+6 2 | 2+2 2 | 3+3 2 |

Continued on next column

**TABLE XXVII– continued**

| TQC | ICSD | | NoE | SF | | |
|---|---|---|---|---|---|---|
| | | 5+6 2 | | 5+6 2 | 3+3 2 | 2+2 2 |
| | | 3+4 2 | | 3+4 2 | 2+2 2 | 3+3 2 |
| | | 3+4 2 | | 3+4 2 | 2+2 2 | 2+2 2 |
| | | 3+4 2 | | 5+6 2 | 3+3 2 | 3+3 2 |
| | | 5+6 2 | | 5+6 2 | 2+2 2 | 2+2 2 |
| | | 3+4 2 | | 3+4 2 | 3+3 2 | 2+2 2 |
| | | 3+4 2 | | 5+6 2 | 3+3 2 | 3+3 2 |
| | | 3+4 2 | | 5+6 2 | 2+2 2 | 2+2 2 |
| | | 5+6 2 | | 5+6 2 | 3+3 2 | 3+3 2 |
| | | 3+4 2 | | 3+4 2 | 2+2 2 | 3+3 2 |
| | | 5+6 2 | | 3+4 2 | 3+3 2 | 2+2 2 |
| | | 5+6 2 | | 5+6 2 | 2+2 2 | 2+2 2 |
| | | 3+4 2 | | 5+6 2 | 3+3 2 | 3+3 2 |
| | | 3+4 2 | | 3+4 2 | 2+2 2 | 3+3 2 |
| | | 5+6 2 | | 3+4 2 | 3+3 2 | 2+2 2 |
| | | 5+6 2 | | 5+6 2 | 2+2 2 | 2+2 2 |
| | | 3+4 2 | | 3+4 2 | 3+3 2 | 3+3 2 |
| | | 3+4 2 | | 3+4 2 | 2+2 2 | 2+2 2 |
| | | 3+4 2 | | 5+6 2 | 3+3 2 | 3+3 2 |
| | | 5+6 2 | | 3+4 2 | 3+3 2 | 2+2 2 |
| | | 3+4 2 | | 5+6 2 | 2+2 2 | 2+2 2 |
| | | 5+6 2 | | 5+6 2 | 3+3 2 | 3+3 2 |
| | | 5+6 2 | | 3+4 2 | 2+2 2 | 2+2 2 |
| | | 3+4 2 | | 5+6 2 | 3+3 2 | 3+3 2 |
| 15.2.29 | 78721 | | 112 | Li2O3Ru1 | | |
| | -A | -GM | -M | -Y | -L | -V |
| | 2+2 4 | 3+4 2 | 2+2 4 | 3+4 2 | 2+2 2 | 2+2 2 |
| | 2+2 4 | 3+4 2 | 2+2 4 | 3+4 2 | 3+3 2 | 3+3 2 |
| | 2+2 4 | 5+6 2 | 2+2 4 | 5+6 2 | 2+2 2 | 2+2 2 |
| | 2+2 4 | 5+6 2 | 2+2 4 | 5+6 2 | 3+3 2 | 3+3 2 |
| | 2+2 4 | 3+4 2 | 2+2 4 | 3+4 2 | 2+2 2 | 3+3 2 |
| | 2+2 4 | 3+4 2 | 2+2 4 | 5+6 2 | 2+2 2 | 2+2 2 |
| | 2+2 4 | 5+6 2 | 2+2 4 | 5+6 2 | 3+3 2 | 3+3 2 |
| | 2+2 4 | 5+6 2 | 2+2 4 | 3+4 2 | 3+3 2 | 2+2 2 |
| | 2+2 4 | 5+6 2 | 2+2 4 | 5+6 2 | 3+3 2 | 3+3 2 |
| | 2+2 4 | 5+6 2 | 2+2 4 | 3+4 2 | 2+2 2 | 2+2 2 |
| | 2+2 4 | 3+4 2 | 2+2 4 | 5+6 2 | 3+3 2 | 3+3 2 |
| | 2+2 4 | 3+4 2 | 2+2 4 | 3+4 2 | 2+2 2 | 2+2 2 |
| | 2+2 4 | 3+4 2 | 2+2 4 | 5+6 2 | 3+3 2 | 3+3 2 |
| | 2+2 4 | 5+6 2 | 2+2 4 | 5+6 2 | 2+2 2 | 2+2 2 |
| | 2+2 4 | 5+6 2 | 2+2 4 | 3+4 2 | 3+3 2 | 3+3 2 |
| | 2+2 4 | 5+6 2 | 2+2 4 | 3+4 2 | 2+2 2 | 2+2 2 |
| | 2+2 4 | 5+6 2 | 2+2 4 | 3+4 2 | 2+2 2 | 3+3 2 |
| | 2+2 4 | 3+4 2 | 2+2 4 | 5+6 2 | 3+3 2 | 2+2 2 |
| | 2+2 4 | 3+4 2 | 2+2 4 | 5+6 2 | 2+2 2 | 3+3 2 |
| | 2+2 4 | 5+6 2 | 2+2 4 | 3+4 2 | 3+3 2 | 2+2 2 |
| | 2+2 4 | 3+4 2 | 2+2 4 | 3+4 2 | 2+2 2 | 2+2 2 |
| | 2+2 4 | 5+6 2 | 2+2 4 | 5+6 2 | 3+3 2 | 3+3 2 |
| | 2+2 4 | 3+4 2 | 2+2 4 | 3+4 2 | 2+2 2 | 2+2 2 |
| | 2+2 4 | 3+4 2 | 2+2 4 | 3+4 2 | 2+2 2 | 3+3 2 |
| | 2+2 4 | 3+4 2 | 2+2 4 | 5+6 2 | 3+3 2 | 2+2 2 |
| | 2+2 4 | 3+4 2 | 2+2 4 | 5+6 2 | 3+3 2 | 3+3 2 |
| | 2+2 4 | 5+6 2 | 2+2 4 | 5+6 2 | 2+2 2 | 3+3 2 |
| | 2+2 4 | 5+6 2 | 2+2 4 | 5+6 2 | 2+2 2 | 3+3 2 |
| | | 3+4 2 | | 3+4 2 | 3+3 2 | 2+2 2 |
| | | 3+4 2 | | 3+4 2 | 2+2 2 | 2+2 2 |
| | | 3+4 2 | | 3+4 2 | 3+3 2 | 3+3 2 |
| | | 5+6 2 | | 5+6 2 | 2+2 2 | 2+2 2 |
| | | 5+6 2 | | 3+4 2 | 3+3 2 | 3+3 2 |
| | | 3+4 2 | | 5+6 2 | 2+2 2 | 2+2 2 |
| | | 3+4 2 | | 5+6 2 | 3+3 2 | 3+3 2 |

Continued on next column

**TABLE XXVII– continued**

| TQC | ICSD | | NoE | SF | | |
|---|---|---|---|---|---|---|
| | | 3+4 2 | | 3+4 2 | 3+3 2 | 2+2 2 |
| | | 5+6 2 | | 3+4 2 | 2+2 2 | 3+3 2 |
| | | 5+6 2 | | 3+4 2 | 2+2 2 | 2+2 2 |
| | | 5+6 2 | | 3+4 2 | 3+3 2 | 2+2 2 |
| | | 3+4 2 | | 3+4 2 | 2+2 2 | 2+2 2 |
| | | 3+4 2 | | 5+6 2 | 3+3 2 | 3+3 2 |
| | | 3+4 2 | | 5+6 2 | 2+2 2 | 3+3 2 |
| | | 3+4 2 | | 3+4 2 | 3+3 2 | 3+3 2 |
| | | 5+6 2 | | 5+6 2 | 2+2 2 | 3+3 2 |
| | | 5+6 2 | | 5+6 2 | 2+2 2 | 2+2 2 |
| | | 5+6 2 | | 3+4 2 | 3+3 2 | 2+2 2 |
| | | 5+6 2 | | 5+6 2 | 3+3 2 | 3+3 2 |
| | | 5+6 2 | | 5+6 2 | 3+3 2 | 2+2 2 |
| | | 3+4 2 | | 3+4 2 | 3+3 2 | 2+2 2 |
| | | 3+4 2 | | 3+4 2 | 2+2 2 | 2+2 2 |
| | | 5+6 2 | | 5+6 2 | 2+2 2 | 3+3 2 |
| | | 5+6 2 | | 5+6 2 | 3+3 2 | 3+3 2 |
| | | 3+4 2 | | 5+6 2 | 3+3 2 | 3+3 2 |
| | | 3+4 2 | | 3+4 2 | 3+3 2 | 2+2 2 |
| | | 3+4 2 | | 5+6 2 | 2+2 2 | 3+3 2 |
| | | 5+6 2 | | 3+4 2 | 2+2 2 | 2+2 2 |
| 15.2.30 | 180367 | | 112 | Li2O3Ru1 | | |
| | -A | -GM | -M | -Y | -L | -V |
| | 2+2 4 | 3+4 2 | 2+2 4 | 3+4 2 | 3+3 2 | 2+2 2 |
| | 2+2 4 | 3+4 2 | 2+2 4 | 3+4 2 | 2+2 2 | 3+3 2 |
| | 2+2 4 | 5+6 2 | 2+2 4 | 5+6 2 | 3+3 2 | 2+2 2 |
| | 2+2 4 | 5+6 2 | 2+2 4 | 5+6 2 | 3+3 2 | 3+3 2 |
| | 2+2 4 | 5+6 2 | 2+2 4 | 3+4 2 | 2+2 2 | 3+3 2 |
| | 2+2 4 | 5+6 2 | 2+2 4 | 5+6 2 | 2+2 2 | 2+2 2 |
| | 2+2 4 | 5+6 2 | 2+2 4 | 3+4 2 | 3+3 2 | 2+2 2 |
| | 2+2 4 | 5+6 2 | 2+2 4 | 5+6 2 | 2+2 2 | 3+3 2 |
| | 2+2 4 | 3+4 2 | 2+2 4 | 5+6 2 | 2+2 2 | 3+3 2 |
| | 2+2 4 | 3+4 2 | 2+2 4 | 5+6 2 | 3+3 2 | 2+2 2 |
| | 2+2 4 | 3+4 2 | 2+2 4 | 3+4 2 | 2+2 2 | 2+2 2 |
| | 2+2 4 | 3+4 2 | 2+2 4 | 3+4 2 | 3+3 2 | 3+3 2 |
| | 2+2 4 | 3+4 2 | 2+2 4 | 5+6 2 | 2+2 2 | 2+2 2 |
| | 2+2 4 | 5+6 2 | 2+2 4 | 3+4 2 | 3+3 2 | 3+3 2 |
| | 2+2 4 | 5+6 2 | 2+2 4 | 5+6 2 | 2+2 2 | 3+3 2 |
| | 2+2 4 | 5+6 2 | 2+2 4 | 3+4 2 | 3+3 2 | 2+2 2 |
| | 2+2 4 | 5+6 2 | 2+2 4 | 3+4 2 | 3+3 2 | 2+2 2 |
| | 2+2 4 | 3+4 2 | 2+2 4 | 3+4 2 | 3+3 2 | 3+3 2 |
| | 2+2 4 | 3+4 2 | 2+2 4 | 5+6 2 | 2+2 2 | 2+2 2 |
| | 2+2 4 | 3+4 2 | 2+2 4 | 3+4 2 | 2+2 2 | 3+3 2 |
| | 2+2 4 | 5+6 2 | 2+2 4 | 5+6 2 | 3+3 2 | 2+2 2 |
| | 2+2 4 | 5+6 2 | 2+2 4 | 5+6 2 | 2+2 2 | 3+3 2 |
| | 2+2 4 | 3+4 2 | 2+2 4 | 3+4 2 | 3+3 2 | 2+2 2 |
| | 2+2 4 | 3+4 2 | 2+2 4 | 3+4 2 | 3+3 2 | 3+3 2 |
| | 2+2 4 | 3+4 2 | 2+2 4 | 5+6 2 | 2+2 2 | 2+2 2 |
| | 2+2 4 | 3+4 2 | 2+2 4 | 5+6 2 | 3+3 2 | 3+3 2 |
| | 2+2 4 | 3+4 2 | 2+2 4 | 3+4 2 | 2+2 2 | 3+3 2 |
| | 2+2 4 | 3+4 2 | 2+2 4 | 5+6 2 | 2+2 2 | 2+2 2 |
| | | 5+6 2 | | 5+6 2 | 3+3 2 | 3+3 2 |
| | | 5+6 2 | | 3+4 2 | 3+3 2 | 3+3 2 |
| | | 5+6 2 | | 5+6 2 | 2+2 2 | 2+2 2 |
| | | 5+6 2 | | 3+4 2 | 3+3 2 | 2+2 2 |
| | | 3+4 2 | | 3+4 2 | 2+2 2 | 3+3 2 |
| | | 3+4 2 | | 5+6 2 | 2+2 2 | 2+2 2 |
| | | 3+4 2 | | 3+4 2 | 3+3 2 | 3+3 2 |
| | | 3+4 2 | | 5+6 2 | 2+2 2 | 2+2 2 |
| | | 5+6 2 | | 3+4 2 | 3+3 2 | 2+2 2 |
| | | 5+6 2 | | 3+4 2 | 3+3 2 | 3+3 2 |

Continued on next column

**TABLE XXVII– continued**

| TQC | ICSD | | NoE | SF | | |
|---|---|---|---|---|---|---|
| | | 5+6 2 | | 3+4 2 | 2+2 2 | 2+2 2 |
| | | 3+4 2 | | 5+6 2 | 3+3 2 | 3+3 2 |
| | | 3+4 2 | | 3+4 2 | 2+2 2 | 2+2 2 |
| | | 3+4 2 | | 5+6 2 | 3+3 2 | 3+3 2 |
| | | 3+4 2 | | 3+4 2 | 2+2 2 | 3+3 2 |
| | | 5+6 2 | | 5+6 2 | 3+3 2 | 2+2 2 |
| | | 5+6 2 | | 5+6 2 | 3+3 2 | 2+2 2 |
| | | 5+6 2 | | 3+4 2 | 2+2 2 | 3+3 2 |
| | | 5+6 2 | | 5+6 2 | 2+2 2 | 3+3 2 |
| | | 5+6 2 | | 5+6 2 | 2+2 2 | 2+2 2 |
| | | 3+4 2 | | 5+6 2 | 2+2 2 | 2+2 2 |
| | | 3+4 2 | | 3+4 2 | 3+3 2 | 3+3 2 |
| | | 5+6 2 | | 3+4 2 | 3+3 2 | 2+2 2 |
| | | 5+6 2 | | 5+6 2 | 2+2 2 | 3+3 2 |
| | | 3+4 2 | | 3+4 2 | 3+3 2 | 2+2 2 |
| | | 3+4 2 | | 5+6 2 | 2+2 2 | 3+3 2 |
| | | 3+4 2 | | 3+4 2 | 3+3 2 | 2+2 2 |
| | | 3+4 2 | | 5+6 2 | 2+2 2 | 3+3 2 |
| 15.2.31 | 55164 | | 100 | Li1O6Si2V1 | | |
| | -A | -GM | -M | -Y | -L | -V |
| | 2+2 4 | 3+4 2 | 2+2 4 | 3+4 2 | 3+3 2 | 3+3 2 |
| | 2+2 4 | 5+6 2 | 2+2 4 | 5+6 2 | 2+2 2 | 2+2 2 |
| | 2+2 4 | 5+6 2 | 2+2 4 | 5+6 2 | 2+2 2 | 2+2 2 |
| | 2+2 4 | 3+4 2 | 2+2 4 | 3+4 2 | 3+3 2 | 3+3 2 |
| | 2+2 4 | 3+4 2 | 2+2 4 | 5+6 2 | 3+3 2 | 2+2 2 |
| | 2+2 4 | 3+4 2 | 2+2 4 | 3+4 2 | 2+2 2 | 3+3 2 |
| | 2+2 4 | 5+6 2 | 2+2 4 | 3+4 2 | 2+2 2 | 2+2 2 |
| | 2+2 4 | 5+6 2 | 2+2 4 | 5+6 2 | 3+3 2 | 3+3 2 |
| | 2+2 4 | 5+6 2 | 2+2 4 | 5+6 2 | 2+2 2 | 3+3 2 |
| | 2+2 4 | 3+4 2 | 2+2 4 | 3+4 2 | 3+3 2 | 3+3 2 |
| | 2+2 4 | 5+6 2 | 2+2 4 | 3+4 2 | 3+3 2 | 2+2 2 |
| | 2+2 4 | 3+4 2 | 2+2 4 | 5+6 2 | 2+2 2 | 2+2 2 |
| | 2+2 4 | 3+4 2 | 2+2 4 | 5+6 2 | 2+2 2 | 3+3 2 |
| | 2+2 4 | 5+6 2 | 2+2 4 | 3+4 2 | 3+3 2 | 2+2 2 |
| | 2+2 4 | 3+4 2 | 2+2 4 | 3+4 2 | 3+3 2 | 2+2 2 |
| | 2+2 4 | 3+4 2 | 2+2 4 | 5+6 2 | 2+2 2 | 3+3 2 |
| | 2+2 4 | 5+6 2 | 2+2 4 | 3+4 2 | 3+3 2 | 2+2 2 |
| | 2+2 4 | 5+6 2 | 2+2 4 | 3+4 2 | 2+2 2 | 3+3 2 |
| | 2+2 4 | 3+4 2 | 2+2 4 | 5+6 2 | 2+2 2 | 2+2 2 |
| | 2+2 4 | 5+6 2 | 2+2 4 | 5+6 2 | 3+3 2 | 3+3 2 |
| | 2+2 4 | 3+4 2 | 2+2 4 | 5+6 2 | 2+2 2 | 3+3 2 |
| | 2+2 4 | 5+6 2 | 2+2 4 | 5+6 2 | 3+3 2 | 2+2 2 |
| | 2+2 4 | 3+4 2 | 2+2 4 | 3+4 2 | 3+3 2 | 3+3 2 |
| | 2+2 4 | 5+6 2 | 2+2 4 | 3+4 2 | 2+2 2 | 2+2 2 |
| | 2+2 4 | 5+6 2 | 2+2 4 | 5+6 2 | 2+2 2 | 3+3 2 |
| | | 3+4 2 | | 3+4 2 | 3+3 2 | 2+2 2 |
| | | 5+6 2 | | 3+4 2 | 3+3 2 | 3+3 2 |
| | | 3+4 2 | | 5+6 2 | 3+3 2 | 2+2 2 |
| | | 3+4 2 | | 5+6 2 | 2+2 2 | 2+2 2 |
| | | 5+6 2 | | 3+4 2 | 2+2 2 | 3+3 2 |
| | | 3+4 2 | | 3+4 2 | 3+3 2 | 2+2 2 |
| | | 3+4 2 | | 3+4 2 | 2+2 2 | 2+2 2 |
| | | 3+4 2 | | 5+6 2 | 2+2 2 | 3+3 2 |
| | | 5+6 2 | | 3+4 2 | 2+2 2 | 3+3 2 |
| | | 3+4 2 | | 5+6 2 | 3+3 2 | 2+2 2 |
| | | 3+4 2 | | 5+6 2 | 3+3 2 | 3+3 2 |
| | | 5+6 2 | | 3+4 2 | 2+2 2 | 2+2 2 |
| | | 5+6 2 | | 5+6 2 | 3+3 2 | 3+3 2 |
| | | 5+6 2 | | 3+4 2 | 2+2 2 | 2+2 2 |
| | | 3+4 2 | | 5+6 2 | 3+3 2 | 3+3 2 |
| | | 5+6 2 | | 5+6 2 | 3+3 2 | 2+2 2 |

Continued on next column

**TABLE XXVII– continued**

| TQC | ICSD | | NoE | SF | | |
|---|---|---|---|---|---|---|
| | | 5+6 2 | | 3+4 2 | 3+3 2 | 3+3 2 |
| | | 3+4 2 | | 5+6 2 | 2+2 2 | 2+2 2 |
| | | 5+6 2 | | 3+4 2 | 2+2 2 | 3+3 2 |
| | | 5+6 2 | | 5+6 2 | 2+2 2 | 3+3 2 |
| | | 3+4 2 | | 3+4 2 | 3+3 2 | 2+2 2 |
| | | 3+4 2 | | 3+4 2 | 2+2 2 | 2+2 2 |
| | | 5+6 2 | | 5+6 2 | 3+3 2 | 3+3 2 |
| | | 3+4 2 | | 5+6 2 | 3+3 2 | 2+2 2 |
| | | 3+4 2 | | 3+4 2 | 2+2 2 | 3+3 2 |
| 15.2.32 | 152074 | | 100 | Li1O6Si2V1 | | |
| | -A | -GM | -M | -Y | -L | -V |
| | 2+2 4 | 3+4 2 | 2+2 4 | 3+4 2 | 3+3 2 | 3+3 2 |
| | 2+2 4 | 5+6 2 | 2+2 4 | 5+6 2 | 2+2 2 | 2+2 2 |
| | 2+2 4 | 5+6 2 | 2+2 4 | 5+6 2 | 2+2 2 | 2+2 2 |
| | 2+2 4 | 3+4 2 | 2+2 4 | 3+4 2 | 3+3 2 | 3+3 2 |
| | 2+2 4 | 3+4 2 | 2+2 4 | 5+6 2 | 3+3 2 | 2+2 2 |
| | 2+2 4 | 3+4 2 | 2+2 4 | 3+4 2 | 2+2 2 | 3+3 2 |
| | 2+2 4 | 5+6 2 | 2+2 4 | 3+4 2 | 2+2 2 | 2+2 2 |
| | 2+2 4 | 5+6 2 | 2+2 4 | 5+6 2 | 3+3 2 | 3+3 2 |
| | 2+2 4 | 5+6 2 | 2+2 4 | 5+6 2 | 2+2 2 | 3+3 2 |
| | 2+2 4 | 3+4 2 | 2+2 4 | 3+4 2 | 3+3 2 | 3+3 2 |
| | 2+2 4 | 5+6 2 | 2+2 4 | 3+4 2 | 3+3 2 | 2+2 2 |
| | 2+2 4 | 3+4 2 | 2+2 4 | 5+6 2 | 2+2 2 | 2+2 2 |
| | 2+2 4 | 3+4 2 | 2+2 4 | 5+6 2 | 2+2 2 | 3+3 2 |
| | 2+2 4 | 5+6 2 | 2+2 4 | 3+4 2 | 3+3 2 | 2+2 2 |
| | 2+2 4 | 3+4 2 | 2+2 4 | 3+4 2 | 3+3 2 | 2+2 2 |
| | 2+2 4 | 3+4 2 | 2+2 4 | 5+6 2 | 2+2 2 | 3+3 2 |
| | 2+2 4 | 5+6 2 | 2+2 4 | 3+4 2 | 3+3 2 | 2+2 2 |
| | 2+2 4 | 5+6 2 | 2+2 4 | 3+4 2 | 2+2 2 | 3+3 2 |
| | 2+2 4 | 3+4 2 | 2+2 4 | 5+6 2 | 2+2 2 | 2+2 2 |
| | 2+2 4 | 5+6 2 | 2+2 4 | 5+6 2 | 3+3 2 | 3+3 2 |
| | 2+2 4 | 3+4 2 | 2+2 4 | 5+6 2 | 2+2 2 | 3+3 2 |
| | 2+2 4 | 5+6 2 | 2+2 4 | 5+6 2 | 3+3 2 | 2+2 2 |
| | 2+2 4 | 3+4 2 | 2+2 4 | 3+4 2 | 3+3 2 | 3+3 2 |
| | 2+2 4 | 5+6 2 | 2+2 4 | 3+4 2 | 2+2 2 | 2+2 2 |
| | 2+2 4 | 5+6 2 | 2+2 4 | 5+6 2 | 2+2 2 | 3+3 2 |
| | | 3+4 2 | | 3+4 2 | 3+3 2 | 2+2 2 |
| | | 5+6 2 | | 3+4 2 | 3+3 2 | 3+3 2 |
| | | 3+4 2 | | 5+6 2 | 3+3 2 | 2+2 2 |
| | | 3+4 2 | | 5+6 2 | 2+2 2 | 2+2 2 |
| | | 3+4 2 | | 3+4 2 | 2+2 2 | 3+3 2 |
| | | 5+6 2 | | 3+4 2 | 3+3 2 | 2+2 2 |
| | | 3+4 2 | | 3+4 2 | 2+2 2 | 2+2 2 |
| | | 3+4 2 | | 5+6 2 | 2+2 2 | 3+3 2 |
| | | 5+6 2 | | 3+4 2 | 2+2 2 | 3+3 2 |
| | | 3+4 2 | | 5+6 2 | 3+3 2 | 2+2 2 |
| | | 3+4 2 | | 5+6 2 | 3+3 2 | 3+3 2 |
| | | 5+6 2 | | 3+4 2 | 2+2 2 | 2+2 2 |
| | | 5+6 2 | | 5+6 2 | 3+3 2 | 3+3 2 |
| | | 5+6 2 | | 3+4 2 | 2+2 2 | 2+2 2 |
| | | 3+4 2 | | 5+6 2 | 3+3 2 | 3+3 2 |
| | | 5+6 2 | | 5+6 2 | 3+3 2 | 2+2 2 |
| | | 5+6 2 | | 3+4 2 | 3+3 2 | 3+3 2 |
| | | 3+4 2 | | 5+6 2 | 2+2 2 | 2+2 2 |
| | | 5+6 2 | | 3+4 2 | 2+2 2 | 3+3 2 |
| | | 5+6 2 | | 5+6 2 | 2+2 2 | 3+3 2 |
| | | 3+4 2 | | 3+4 2 | 3+3 2 | 2+2 2 |
| | | 3+4 2 | | 3+4 2 | 2+2 2 | 2+2 2 |
| | | 5+6 2 | | 5+6 2 | 3+3 2 | 3+3 2 |
| | | 3+4 2 | | 5+6 2 | 3+3 2 | 2+2 2 |
| | | 3+4 2 | | 3+4 2 | 2+2 2 | 3+3 2 |

Continued on next column

**TABLE XXVII– continued**

| TQC | ICSD | | NoE | | SF | |
|---|---|---|---|---|---|---|
| 15.2.33 | 416039 | | 184 | | K3Na1O8Ru2 | |
| | -A | -GM | -M | -Y | -L | -V |
| | 2+2 4 | 3+4 2 | 2+2 4 | 5+6 2 | 2+2 2 | 3+3 2 |
| | 2+2 4 | 3+4 2 | 2+2 4 | 5+6 2 | 2+2 2 | 2+2 2 |
| | 2+2 4 | 5+6 2 | 2+2 4 | 3+4 2 | 3+3 2 | 3+3 2 |
| | 2+2 4 | 5+6 2 | 2+2 4 | 3+4 2 | 3+3 2 | 2+2 2 |
| | 2+2 4 | 3+4 2 | 2+2 4 | 3+4 2 | 2+2 2 | 2+2 2 |
| | 2+2 4 | 5+6 2 | 2+2 4 | 5+6 2 | 3+3 2 | 3+3 2 |
| | 2+2 4 | 3+4 2 | 2+2 4 | 3+4 2 | 3+3 2 | 2+2 2 |
| | 2+2 4 | 5+6 2 | 2+2 4 | 5+6 2 | 2+2 2 | 3+3 2 |
| | 2+2 4 | 3+4 2 | 2+2 4 | 5+6 2 | 3+3 2 | 3+3 2 |
| | 2+2 4 | 5+6 2 | 2+2 4 | 3+4 2 | 2+2 2 | 2+2 2 |
| | 2+2 4 | 3+4 2 | 2+2 4 | 3+4 2 | 2+2 2 | 2+2 2 |
| | 2+2 4 | 3+4 2 | 2+2 4 | 3+4 2 | 3+3 2 | 2+2 2 |
| | 2+2 4 | 5+6 2 | 2+2 4 | 5+6 2 | 3+3 2 | 3+3 2 |
| | 2+2 4 | 5+6 2 | 2+2 4 | 5+6 2 | 2+2 2 | 3+3 2 |
| | 2+2 4 | 5+6 2 | 2+2 4 | 3+4 2 | 2+2 2 | 3+3 2 |
| | 2+2 4 | 3+4 2 | 2+2 4 | 5+6 2 | 3+3 2 | 2+2 2 |
| | 2+2 4 | 5+6 2 | 2+2 4 | 5+6 2 | 2+2 2 | 3+3 2 |
| | 2+2 4 | 3+4 2 | 2+2 4 | 3+4 2 | 3+3 2 | 2+2 2 |
| | 2+2 4 | 5+6 2 | 2+2 4 | 5+6 2 | 3+3 2 | 3+3 2 |
| | 2+2 4 | 3+4 2 | 2+2 4 | 5+6 2 | 2+2 2 | 2+2 2 |
| | 2+2 4 | 5+6 2 | 2+2 4 | 3+4 2 | 3+3 2 | 3+3 2 |
| | 2+2 4 | 3+4 2 | 2+2 4 | 3+4 2 | 2+2 2 | 2+2 2 |
| | 2+2 4 | 3+4 2 | 2+2 4 | 5+6 2 | 2+2 2 | 3+3 2 |
| | 2+2 4 | 3+4 2 | 2+2 4 | 5+6 2 | 3+3 2 | 2+2 2 |
| | 2+2 4 | 5+6 2 | 2+2 4 | 3+4 2 | 2+2 2 | 3+3 2 |
| | 2+2 4 | 5+6 2 | 2+2 4 | 3+4 2 | 3+3 2 | 2+2 2 |
| | 2+2 4 | 5+6 2 | 2+2 4 | 5+6 2 | 2+2 2 | 3+3 2 |
| | 2+2 4 | 3+4 2 | 2+2 4 | 3+4 2 | 3+3 2 | 2+2 2 |
| | 2+2 4 | 3+4 2 | 2+2 4 | 5+6 2 | 2+2 2 | 3+3 2 |
| | 2+2 4 | 3+4 2 | 2+2 4 | 5+6 2 | 3+3 2 | 3+3 2 |
| | 2+2 4 | 3+4 2 | 2+2 4 | 3+4 2 | 2+2 2 | 2+2 2 |
| | 2+2 4 | 3+4 2 | 2+2 4 | 3+4 2 | 3+3 2 | 2+2 2 |
| | 2+2 4 | 5+6 2 | 2+2 4 | 5+6 2 | 2+2 2 | 3+3 2 |
| | 2+2 4 | 5+6 2 | 2+2 4 | 5+6 2 | 2+2 2 | 2+2 2 |
| | 2+2 4 | 5+6 2 | 2+2 4 | 3+4 2 | 3+3 2 | 3+3 2 |
| | 2+2 4 | 5+6 2 | 2+2 4 | 3+4 2 | 3+3 2 | 2+2 2 |
| | 2+2 4 | 3+4 2 | 2+2 4 | 3+4 2 | 2+2 2 | 2+2 2 |
| | 2+2 4 | 5+6 2 | 2+2 4 | 5+6 2 | 3+3 2 | 3+3 2 |
| | 2+2 4 | 5+6 2 | 2+2 4 | 5+6 2 | 2+2 2 | 2+2 2 |
| | 2+2 4 | 3+4 2 | 2+2 4 | 3+4 2 | 3+3 2 | 3+3 2 |
| | 2+2 4 | 3+4 2 | 2+2 4 | 3+4 2 | 2+2 2 | 2+2 2 |
| | 2+2 4 | 5+6 2 | 2+2 4 | 3+4 2 | 3+3 2 | 3+3 2 |
| | 2+2 4 | 3+4 2 | 2+2 4 | 5+6 2 | 3+3 2 | 2+2 2 |
| | 2+2 4 | 5+6 2 | 2+2 4 | 5+6 2 | 2+2 2 | 3+3 2 |
| | 2+2 4 | 3+4 2 | 2+2 4 | 5+6 2 | 2+2 2 | 3+3 2 |
| | 2+2 4 | 5+6 2 | 2+2 4 | 3+4 2 | 3+3 2 | 2+2 2 |
| | | 3+4 2 | | 5+6 2 | 3+3 2 | 2+2 2 |
| | | 5+6 2 | | 3+4 2 | 2+2 2 | 3+3 2 |
| | | 5+6 2 | | 3+4 2 | 3+3 2 | 2+2 2 |
| | | 3+4 2 | | 5+6 2 | 2+2 2 | 3+3 2 |
| | | 5+6 2 | | 3+4 2 | 2+2 2 | 3+3 2 |
| | | 3+4 2 | | 5+6 2 | 3+3 2 | 2+2 2 |
| | | 3+4 2 | | 3+4 2 | 2+2 2 | 2+2 2 |
| | | 5+6 2 | | 5+6 2 | 3+3 2 | 3+3 2 |
| | | 3+4 2 | | 3+4 2 | 2+2 2 | 2+2 2 |
| | | 5+6 2 | | 5+6 2 | 3+3 2 | 3+3 2 |
| | | 3+4 2 | | 5+6 2 | 3+3 2 | 3+3 2 |
| | | 3+4 2 | | 5+6 2 | 2+2 2 | 3+3 2 |
| | | 5+6 2 | | 3+4 2 | 2+2 2 | 2+2 2 |

Continued on next column

**TABLE XXVII– continued**

| TQC | ICSD | | NoE | | SF | |
|---|---|---|---|---|---|---|
| | | 5+6 2 | | 3+4 2 | 3+3 2 | 2+2 2 |
| | | 3+4 2 | | 5+6 2 | 2+2 2 | 2+2 2 |
| | | 3+4 2 | | 3+4 2 | 2+2 2 | 3+3 2 |
| | | 5+6 2 | | 5+6 2 | 3+3 2 | 3+3 2 |
| | | 5+6 2 | | 3+4 2 | 3+3 2 | 2+2 2 |
| | | 5+6 2 | | 3+4 2 | 3+3 2 | 2+2 2 |
| | | 3+4 2 | | 5+6 2 | 2+2 2 | 3+3 2 |
| | | 5+6 2 | | 3+4 2 | 3+3 2 | 2+2 2 |
| | | 3+4 2 | | 5+6 2 | 2+2 2 | 3+3 2 |
| | | 3+4 2 | | 3+4 2 | 3+3 2 | 2+2 2 |
| | | 5+6 2 | | 5+6 2 | 2+2 2 | 3+3 2 |
| | | 3+4 2 | | 3+4 2 | 3+3 2 | 2+2 2 |
| | | 5+6 2 | | 5+6 2 | 3+3 2 | 3+3 2 |
| | | 5+6 2 | | 5+6 2 | 2+2 2 | 3+3 2 |
| | | 5+6 2 | | 3+4 2 | 2+2 2 | 2+2 2 |
| | | 3+4 2 | | 5+6 2 | 2+2 2 | 3+3 2 |
| | | 3+4 2 | | 3+4 2 | 3+3 2 | 2+2 2 |
| | | 5+6 2 | | 3+4 2 | 2+2 2 | 2+2 2 |
| | | 5+6 2 | | 3+4 2 | 3+3 2 | 3+3 2 |
| | | 3+4 2 | | 3+4 2 | 2+2 2 | 2+2 2 |
| | | 3+4 2 | | 5+6 2 | 3+3 2 | 2+2 2 |
| | | 3+4 2 | | 3+4 2 | 3+3 2 | 2+2 2 |
| | | 5+6 2 | | 5+6 2 | 2+2 2 | 3+3 2 |
| | | 5+6 2 | | 5+6 2 | 3+3 2 | 3+3 2 |
| | | 3+4 2 | | 5+6 2 | 2+2 2 | 3+3 2 |
| | | 5+6 2 | | 5+6 2 | 3+3 2 | 3+3 2 |
| | | 5+6 2 | | 5+6 2 | 2+2 2 | 2+2 2 |
| | | 3+4 2 | | 3+4 2 | 3+3 2 | 3+3 2 |
| | | 3+4 2 | | 3+4 2 | 2+2 2 | 2+2 2 |
| | | 3+4 2 | | 5+6 2 | 2+2 2 | 2+2 2 |
| | | 3+4 2 | | 3+4 2 | 3+3 2 | 3+3 2 |
| | | 3+4 2 | | 5+6 2 | 2+2 2 | 3+3 2 |
| | | 3+4 2 | | 3+4 2 | 2+2 2 | 2+2 2 |
| 15.2.34 | 192833 | | 100 | | Ho1O4V1 | |
| | -A | -GM | -M | -Y | -L | -V |
| | 2+2 4 | 3+4 2 | 2+2 4 | 3+4 2 | 2+2 2 | 3+3 2 |
| | 2+2 4 | 5+6 2 | 2+2 4 | 5+6 2 | 3+3 2 | 2+2 2 |
| | 2+2 4 | 3+4 2 | 2+2 4 | 5+6 2 | 2+2 2 | 3+3 2 |
| | 2+2 4 | 5+6 2 | 2+2 4 | 3+4 2 | 3+3 2 | 2+2 2 |
| | 2+2 4 | 3+4 2 | 2+2 4 | 5+6 2 | 2+2 2 | 3+3 2 |
| | 2+2 4 | 5+6 2 | 2+2 4 | 3+4 2 | 3+3 2 | 2+2 2 |
| | 2+2 4 | 3+4 2 | 2+2 4 | 5+6 2 | 2+2 2 | 3+3 2 |
| | 2+2 4 | 5+6 2 | 2+2 4 | 3+4 2 | 3+3 2 | 2+2 2 |
| | 2+2 4 | 3+4 2 | 2+2 4 | 5+6 2 | 2+2 2 | 3+3 2 |
| | 2+2 4 | 5+6 2 | 2+2 4 | 3+4 2 | 3+3 2 | 2+2 2 |
| | 2+2 4 | 5+6 2 | 2+2 4 | 3+4 2 | 2+2 2 | 3+3 2 |
| | 2+2 4 | 5+6 2 | 2+2 4 | 3+4 2 | 3+3 2 | 2+2 2 |
| | 2+2 4 | 3+4 2 | 2+2 4 | 5+6 2 | 3+3 2 | 2+2 2 |
| | 2+2 4 | 3+4 2 | 2+2 4 | 5+6 2 | 2+2 2 | 3+3 2 |
| | 2+2 4 | 5+6 2 | 2+2 4 | 5+6 2 | 3+3 2 | 2+2 2 |
| | 2+2 4 | 3+4 2 | 2+2 4 | 3+4 2 | 2+2 2 | 3+3 2 |
| | 2+2 4 | 3+4 2 | 2+2 4 | 3+4 2 | 2+2 2 | 2+2 2 |
| | 2+2 4 | 5+6 2 | 2+2 4 | 5+6 2 | 3+3 2 | 3+3 2 |
| | 2+2 4 | 3+4 2 | 2+2 4 | 5+6 2 | 3+3 2 | 3+3 2 |
| | 2+2 4 | 5+6 2 | 2+2 4 | 5+6 2 | 2+2 2 | 2+2 2 |
| | 2+2 4 | 3+4 2 | 2+2 4 | 3+4 2 | 3+3 2 | 2+2 2 |
| | 2+2 4 | 3+4 2 | 2+2 4 | 3+4 2 | 2+2 2 | 3+3 2 |
| | 2+2 4 | 3+4 2 | 2+2 4 | 3+4 2 | 3+3 2 | 3+3 2 |
| | 2+2 4 | 5+6 2 | 2+2 4 | 5+6 2 | 2+2 2 | 2+2 2 |
| | 2+2 4 | 5+6 2 | 2+2 4 | 5+6 2 | 2+2 2 | 2+2 2 |
| | | 5+6 2 | | 3+4 2 | 3+3 2 | 3+3 2 |

Continued on next column

**TABLE XXVII– continued**

| TQC | ICSD | | NoE | SF | | |
|---|---|---|---|---|---|---|
| | | 3+4 2 | | 5+6 2 | 2+2 2 | 3+3 2 |
| | | 3+4 2 | | 3+4 2 | 2+2 2 | 2+2 2 |
| | | 3+4 2 | | 3+4 2 | 3+3 2 | 2+2 2 |
| | | 5+6 2 | | 5+6 2 | 3+3 2 | 3+3 2 |
| | | 3+4 2 | | 3+4 2 | 3+3 2 | 3+3 2 |
| | | 5+6 2 | | 5+6 2 | 3+3 2 | 2+2 2 |
| | | 5+6 2 | | 5+6 2 | 2+2 2 | 2+2 2 |
| | | 5+6 2 | | 3+4 2 | 2+2 2 | 3+3 2 |
| | | 3+4 2 | | 3+4 2 | 2+2 2 | 2+2 2 |
| | | 3+4 2 | | 5+6 2 | 3+3 2 | 3+3 2 |
| | | 5+6 2 | | 3+4 2 | 2+2 2 | 2+2 2 |
| | | 5+6 2 | | 3+4 2 | 3+3 2 | 3+3 2 |
| | | 3+4 2 | | 5+6 2 | 3+3 2 | 2+2 2 |
| | | 5+6 2 | | 5+6 2 | 2+2 2 | 3+3 2 |
| | | 3+4 2 | | 3+4 2 | 2+2 2 | 3+3 2 |
| | | 3+4 2 | | 5+6 2 | 3+3 2 | 2+2 2 |
| | | 3+4 2 | | 3+4 2 | 2+2 2 | 3+3 2 |
| | | 5+6 2 | | 5+6 2 | 3+3 2 | 2+2 2 |
| | | 5+6 2 | | 3+4 2 | 2+2 2 | 3+3 2 |
| | | 5+6 2 | | 5+6 2 | 3+3 2 | 2+2 2 |
| | | 3+4 2 | | 3+4 2 | 2+2 2 | 3+3 2 |
| | | 3+4 2 | | 5+6 2 | 2+2 2 | 2+2 2 |
| | | 3+4 2 | | 5+6 2 | 3+3 2 | 3+3 2 |
| | | 5+6 2 | | 3+4 2 | 2+2 2 | 2+2 2 |
| 15.2.35 | 79499 | | 100 | Ho1O4Ta1 | | |
| | -A | -GM | -M | -Y | -L | -V |
| | 2+2 4 | 3+4 2 | 2+2 4 | 5+6 2 | 2+2 2 | 3+3 2 |
| | 2+2 4 | 5+6 2 | 2+2 4 | 3+4 2 | 3+3 2 | 2+2 2 |
| | 2+2 4 | 3+4 2 | 2+2 4 | 5+6 2 | 2+2 2 | 3+3 2 |
| | 2+2 4 | 5+6 2 | 2+2 4 | 3+4 2 | 3+3 2 | 2+2 2 |
| | 2+2 4 | 3+4 2 | 2+2 4 | 5+6 2 | 2+2 2 | 2+2 2 |
| | 2+2 4 | 3+4 2 | 2+2 4 | 3+4 2 | 3+3 2 | 3+3 2 |
| | 2+2 4 | 5+6 2 | 2+2 4 | 3+4 2 | 2+2 2 | 3+3 2 |
| | 2+2 4 | 5+6 2 | 2+2 4 | 5+6 2 | 3+3 2 | 2+2 2 |
| | 2+2 4 | 3+4 2 | 2+2 4 | 3+4 2 | 2+2 2 | 2+2 2 |
| | 2+2 4 | 5+6 2 | 2+2 4 | 5+6 2 | 3+3 2 | 3+3 2 |
| | 2+2 4 | 5+6 2 | 2+2 4 | 5+6 2 | 2+2 2 | 2+2 2 |
| | 2+2 4 | 3+4 2 | 2+2 4 | 3+4 2 | 3+3 2 | 3+3 2 |
| | 2+2 4 | 5+6 2 | 2+2 4 | 5+6 2 | 3+3 2 | 2+2 2 |
| | 2+2 4 | 3+4 2 | 2+2 4 | 3+4 2 | 2+2 2 | 3+3 2 |
| | 2+2 4 | 5+6 2 | 2+2 4 | 5+6 2 | 3+3 2 | 2+2 2 |
| | 2+2 4 | 3+4 2 | 2+2 4 | 3+4 2 | 2+2 2 | 3+3 2 |
| | 2+2 4 | 3+4 2 | 2+2 4 | 5+6 2 | 2+2 2 | 2+2 2 |
| | 2+2 4 | 5+6 2 | 2+2 4 | 3+4 2 | 3+3 2 | 3+3 2 |
| | 2+2 4 | 5+6 2 | 2+2 4 | 3+4 2 | 2+2 2 | 3+3 2 |
| | 2+2 4 | 3+4 2 | 2+2 4 | 5+6 2 | 3+3 2 | 2+2 2 |
| | 2+2 4 | 3+4 2 | 2+2 4 | 3+4 2 | 3+3 2 | 2+2 2 |
| | 2+2 4 | 3+4 2 | 2+2 4 | 3+4 2 | 2+2 2 | 2+2 2 |
| | 2+2 4 | 3+4 2 | 2+2 4 | 5+6 2 | 2+2 2 | 3+3 2 |
| | 2+2 4 | 5+6 2 | 2+2 4 | 5+6 2 | 3+3 2 | 3+3 2 |
| | 2+2 4 | 5+6 2 | 2+2 4 | 5+6 2 | 2+2 2 | 2+2 2 |
| | | 3+4 2 | | 3+4 2 | 3+3 2 | 3+3 2 |
| | | 5+6 2 | | 3+4 2 | 2+2 2 | 3+3 2 |
| | | 3+4 2 | | 3+4 2 | 3+3 2 | 2+2 2 |
| | | 3+4 2 | | 5+6 2 | 2+2 2 | 3+3 2 |
| | | 5+6 2 | | 5+6 2 | 3+3 2 | 2+2 2 |
| | | 5+6 2 | | 5+6 2 | 3+3 2 | 3+3 2 |
| | | 3+4 2 | | 3+4 2 | 2+2 2 | 2+2 2 |
| | | 5+6 2 | | 3+4 2 | 3+3 2 | 2+2 2 |
| | | 3+4 2 | | 5+6 2 | 3+3 2 | 2+2 2 |
| | | 3+4 2 | | 5+6 2 | 2+2 2 | 3+3 2 |

Continued on next column

**TABLE XXVII– continued**

| TQC | ICSD | | NoE | SF | | |
|---|---|---|---|---|---|---|
| | | 5+6 2 | | 3+4 2 | 2+2 2 | 3+3 2 |
| | | 3+4 2 | | 3+4 2 | 2+2 2 | 2+2 2 |
| | | 5+6 2 | | 5+6 2 | 3+3 2 | 3+3 2 |
| | | 5+6 2 | | 5+6 2 | 3+3 2 | 2+2 2 |
| | | 5+6 2 | | 3+4 2 | 2+2 2 | 3+3 2 |
| | | 3+4 2 | | 3+4 2 | 2+2 2 | 3+3 2 |
| | | 3+4 2 | | 5+6 2 | 3+3 2 | 2+2 2 |
| | | 5+6 2 | | 5+6 2 | 2+2 2 | 3+3 2 |
| | | 3+4 2 | | 3+4 2 | 3+3 2 | 2+2 2 |
| | | 5+6 2 | | 5+6 2 | 2+2 2 | 3+3 2 |
| | | 5+6 2 | | 3+4 2 | 3+3 2 | 2+2 2 |
| | | 3+4 2 | | 5+6 2 | 2+2 2 | 3+3 2 |
| | | 3+4 2 | | 3+4 2 | 3+3 2 | 2+2 2 |
| | | 3+4 2 | | 5+6 2 | 2+2 2 | 3+3 2 |
| | | 5+6 2 | | 3+4 2 | 3+3 2 | 2+2 2 |
| 15.2.36 | 153363 | | 100 | Ho1O4Ta1 | | |
| | -A | -GM | -M | -Y | -L | -V |
| | 2+2 4 | 3+4 2 | 2+2 4 | 5+6 2 | 2+2 2 | 3+3 2 |
| | 2+2 4 | 5+6 2 | 2+2 4 | 3+4 2 | 3+3 2 | 2+2 2 |
| | 2+2 4 | 3+4 2 | 2+2 4 | 5+6 2 | 2+2 2 | 3+3 2 |
| | 2+2 4 | 5+6 2 | 2+2 4 | 3+4 2 | 3+3 2 | 2+2 2 |
| | 2+2 4 | 3+4 2 | 2+2 4 | 5+6 2 | 2+2 2 | 2+2 2 |
| | 2+2 4 | 3+4 2 | 2+2 4 | 3+4 2 | 3+3 2 | 3+3 2 |
| | 2+2 4 | 5+6 2 | 2+2 4 | 3+4 2 | 2+2 2 | 3+3 2 |
| | 2+2 4 | 5+6 2 | 2+2 4 | 5+6 2 | 3+3 2 | 2+2 2 |
| | 2+2 4 | 3+4 2 | 2+2 4 | 3+4 2 | 2+2 2 | 2+2 2 |
| | 2+2 4 | 5+6 2 | 2+2 4 | 5+6 2 | 3+3 2 | 3+3 2 |
| | 2+2 4 | 5+6 2 | 2+2 4 | 5+6 2 | 2+2 2 | 2+2 2 |
| | 2+2 4 | 3+4 2 | 2+2 4 | 5+6 2 | 3+3 2 | 3+3 2 |
| | 2+2 4 | 5+6 2 | 2+2 4 | 3+4 2 | 3+3 2 | 2+2 2 |
| | 2+2 4 | 3+4 2 | 2+2 4 | 3+4 2 | 2+2 2 | 3+3 2 |
| | 2+2 4 | 5+6 2 | 2+2 4 | 5+6 2 | 3+3 2 | 2+2 2 |
| | 2+2 4 | 3+4 2 | 2+2 4 | 3+4 2 | 2+2 2 | 3+3 2 |
| | 2+2 4 | 3+4 2 | 2+2 4 | 5+6 2 | 2+2 2 | 2+2 2 |
| | 2+2 4 | 5+6 2 | 2+2 4 | 3+4 2 | 3+3 2 | 3+3 2 |
| | 2+2 4 | 5+6 2 | 2+2 4 | 3+4 2 | 3+3 2 | 3+3 2 |
| | 2+2 4 | 3+4 2 | 2+2 4 | 5+6 2 | 2+2 2 | 2+2 2 |
| | 2+2 4 | 3+4 2 | 2+2 4 | 3+4 2 | 3+3 2 | 2+2 2 |
| | 2+2 4 | 3+4 2 | 2+2 4 | 3+4 2 | 2+2 2 | 2+2 2 |
| | 2+2 4 | 3+4 2 | 2+2 4 | 5+6 2 | 2+2 2 | 3+3 2 |
| | 2+2 4 | 5+6 2 | 2+2 4 | 5+6 2 | 3+3 2 | 3+3 2 |
| | 2+2 4 | 5+6 2 | 2+2 4 | 5+6 2 | 2+2 2 | 2+2 2 |
| | | 5+6 2 | | 3+4 2 | 3+3 2 | 3+3 2 |
| | | 3+4 2 | | 3+4 2 | 2+2 2 | 3+3 2 |
| | | 3+4 2 | | 3+4 2 | 3+3 2 | 2+2 2 |
| | | 3+4 2 | | 5+6 2 | 2+2 2 | 3+3 2 |
| | | 5+6 2 | | 5+6 2 | 3+3 2 | 2+2 2 |
| | | 3+4 2 | | 5+6 2 | 3+3 2 | 3+3 2 |
| | | 5+6 2 | | 3+4 2 | 2+2 2 | 2+2 2 |
| | | 3+4 2 | | 3+4 2 | 3+3 2 | 2+2 2 |
| | | 5+6 2 | | 5+6 2 | 3+3 2 | 2+2 2 |
| | | 3+4 2 | | 5+6 2 | 2+2 2 | 3+3 2 |
| | | 5+6 2 | | 3+4 2 | 2+2 2 | 3+3 2 |
| | | 3+4 2 | | 3+4 2 | 3+3 2 | 2+2 2 |
| | | 5+6 2 | | 5+6 2 | 2+2 2 | 3+3 2 |
| | | 5+6 2 | | 5+6 2 | 3+3 2 | 2+2 2 |
| | | 5+6 2 | | 3+4 2 | 2+2 2 | 3+3 2 |
| | | 3+4 2 | | 3+4 2 | 2+2 2 | 3+3 2 |
| | | 3+4 2 | | 5+6 2 | 3+3 2 | 2+2 2 |
| | | 5+6 2 | | 5+6 2 | 2+2 2 | 3+3 2 |
| | | 3+4 2 | | 3+4 2 | 3+3 2 | 2+2 2 |

Continued on next column

**TABLE XXVII– continued**

| TQC | ICSD | | NoE | SF | | |
|---|---|---|---|---|---|---|
| | | 5+6 2 | | 5+6 2 | 2+2 2 | 3+3 2 |
| | | 5+6 2 | | 3+4 2 | 3+3 2 | 2+2 2 |
| | | 3+4 2 | | 5+6 2 | 2+2 2 | 3+3 2 |
| | | 3+4 2 | | 3+4 2 | 3+3 2 | 2+2 2 |
| | | 3+4 2 | | 5+6 2 | 2+2 2 | 3+3 2 |
| | | 5+6 2 | | 3+4 2 | 3+3 2 | 2+2 2 |
| 15.2.37 | 15271 | | 92 | H8O6Pt1 | | |
| | -A | -GM | -M | -Y | -L | -V |
| | 2+2 4 | 3+4 2 | 2+2 4 | 5+6 2 | 3+3 2 | 2+2 2 |
| | 2+2 4 | 3+4 2 | 2+2 4 | 5+6 2 | 3+3 2 | 3+3 2 |
| | 2+2 4 | 3+4 2 | 2+2 4 | 5+6 2 | 3+3 2 | 2+2 2 |
| | 2+2 4 | 3+4 2 | 2+2 4 | 3+4 2 | 3+3 2 | 3+3 2 |
| | 2+2 4 | 3+4 2 | 2+2 4 | 5+6 2 | 2+2 2 | 2+2 2 |
| | 2+2 4 | 3+4 2 | 2+2 4 | 5+6 2 | 3+3 2 | 3+3 2 |
| | 2+2 4 | 5+6 2 | 2+2 4 | 3+4 2 | 2+2 2 | 2+2 2 |
| | 2+2 4 | 5+6 2 | 2+2 4 | 5+6 2 | 3+3 2 | 3+3 2 |
| | 2+2 4 | 5+6 2 | 2+2 4 | 3+4 2 | 2+2 2 | 2+2 2 |
| | 2+2 4 | 5+6 2 | 2+2 4 | 3+4 2 | 2+2 2 | 3+3 2 |
| | 2+2 4 | 5+6 2 | 2+2 4 | 3+4 2 | 2+2 2 | 2+2 2 |
| | 2+2 4 | 5+6 2 | 2+2 4 | 3+4 2 | 2+2 2 | 3+3 2 |
| | 2+2 4 | 3+4 2 | 2+2 4 | 5+6 2 | 3+3 2 | 2+2 2 |
| | 2+2 4 | 3+4 2 | 2+2 4 | 5+6 2 | 3+3 2 | 3+3 2 |
| | 2+2 4 | 3+4 2 | 2+2 4 | 5+6 2 | 3+3 2 | 2+2 2 |
| | 2+2 4 | 3+4 2 | 2+2 4 | 5+6 2 | 3+3 2 | 3+3 2 |
| | 2+2 4 | 3+4 2 | 2+2 4 | 5+6 2 | 3+3 2 | 2+2 2 |
| | 2+2 4 | 3+4 2 | 2+2 4 | 5+6 2 | 3+3 2 | 3+3 2 |
| | 2+2 4 | 3+4 2 | 2+2 4 | 5+6 2 | 3+3 2 | 2+2 2 |
| | 2+2 4 | 3+4 2 | 2+2 4 | 5+6 2 | 3+3 2 | 3+3 2 |
| | 2+2 4 | 3+4 2 | 2+2 4 | 5+6 2 | 3+3 2 | 2+2 2 |
| | 2+2 4 | 3+4 2 | 2+2 4 | 5+6 2 | 3+3 2 | 3+3 2 |
| | 2+2 4 | 3+4 2 | 2+2 4 | 5+6 2 | 3+3 2 | 3+3 2 |
| | | 3+4 2 | | 5+6 2 | 3+3 2 | 2+2 2 |
| | | 5+6 2 | | 3+4 2 | 2+2 2 | 2+2 2 |
| | | 5+6 2 | | 3+4 2 | 2+2 2 | 3+3 2 |
| | | 5+6 2 | | 3+4 2 | 2+2 2 | 2+2 2 |
| | | 5+6 2 | | 3+4 2 | 2+2 2 | 3+3 2 |
| | | 5+6 2 | | 3+4 2 | 2+2 2 | 2+2 2 |
| | | 5+6 2 | | 3+4 2 | 2+2 2 | 3+3 2 |
| | | 5+6 2 | | 3+4 2 | 2+2 2 | 2+2 2 |
| | | 5+6 2 | | 3+4 2 | 2+2 2 | 3+3 2 |
| | | 5+6 2 | | 3+4 2 | 2+2 2 | 2+2 2 |
| | | 3+4 2 | | 3+4 2 | 2+2 2 | 3+3 2 |
| | | 3+4 2 | | 3+4 2 | 2+2 2 | 3+3 2 |
| | | 5+6 2 | | 3+4 2 | 3+3 2 | 2+2 2 |
| | | 3+4 2 | | 3+4 2 | 2+2 2 | 2+2 2 |
| | | 3+4 2 | | 5+6 2 | 2+2 2 | 3+3 2 |
| | | 5+6 2 | | 5+6 2 | 3+3 2 | 3+3 2 |
| | | 5+6 2 | | 3+4 2 | 3+3 2 | 2+2 2 |
| | | 3+4 2 | | 3+4 2 | 2+2 2 | 2+2 2 |
| | | 5+6 2 | | 5+6 2 | 3+3 2 | 2+2 2 |
| | | 5+6 2 | | 5+6 2 | 3+3 2 | 3+3 2 |
| | | 5+6 2 | | 5+6 2 | 2+2 2 | 3+3 2 |
| | | 3+4 2 | | 3+4 2 | 2+2 2 | 2+2 2 |
| | | 5+6 2 | | 3+4 2 | 3+3 2 | 3+3 2 |
| 15.2.38 | 14329 | | 140 | H4Ba1O8S2 | | |
| | -A | -GM | -M | -Y | -L | -V |
| | 2+2 4 | 3+4 2 | 2+2 4 | 3+4 2 | 2+2 2 | 2+2 2 |
| | 2+2 4 | 5+6 2 | 2+2 4 | 5+6 2 | 3+3 2 | 3+3 2 |
| | 2+2 4 | 3+4 2 | 2+2 4 | 5+6 2 | 3+3 2 | 3+3 2 |
| | 2+2 4 | 3+4 2 | 2+2 4 | 5+6 2 | 3+3 2 | 2+2 2 |
| | 2+2 4 | 5+6 2 | 2+2 4 | 3+4 2 | 2+2 2 | 2+2 2 |

Continued on next column

**TABLE XXVII– continued**

| TQC | ICSD | | NoE | SF | | |
|---|---|---|---|---|---|---|
| | 2+2 4 | 5+6 2 | 2+2 4 | 3+4 2 | 2+2 2 | 3+3 2 |
| | 2+2 4 | 5+6 2 | 2+2 4 | 3+4 2 | 2+2 2 | 3+3 2 |
| | 2+2 4 | 5+6 2 | 2+2 4 | 3+4 2 | 2+2 2 | 2+2 2 |
| | 2+2 4 | 5+6 2 | 2+2 4 | 3+4 2 | 2+2 2 | 2+2 2 |
| | 2+2 4 | 5+6 2 | 2+2 4 | 3+4 2 | 2+2 2 | 3+3 2 |
| | 2+2 4 | 3+4 2 | 2+2 4 | 5+6 2 | 3+3 2 | 2+2 2 |
| | 2+2 4 | 3+4 2 | 2+2 4 | 5+6 2 | 3+3 2 | 3+3 2 |
| | 2+2 4 | 3+4 2 | 2+2 4 | 5+6 2 | 3+3 2 | 2+2 2 |
| | 2+2 4 | 3+4 2 | 2+2 4 | 5+6 2 | 3+3 2 | 3+3 2 |
| | 2+2 4 | 3+4 2 | 2+2 4 | 3+4 2 | 3+3 2 | 2+2 2 |
| | 2+2 4 | 3+4 2 | 2+2 4 | 3+4 2 | 2+2 2 | 2+2 2 |
| | 2+2 4 | 5+6 2 | 2+2 4 | 5+6 2 | 2+2 2 | 3+3 2 |
| | 2+2 4 | 5+6 2 | 2+2 4 | 5+6 2 | 3+3 2 | 3+3 2 |
| | 2+2 4 | 3+4 2 | 2+2 4 | 5+6 2 | 3+3 2 | 3+3 2 |
| | 2+2 4 | 5+6 2 | 2+2 4 | 3+4 2 | 2+2 2 | 2+2 2 |
| | 2+2 4 | 3+4 2 | 2+2 4 | 5+6 2 | 3+3 2 | 2+2 2 |
| | 2+2 4 | 3+4 2 | 2+2 4 | 5+6 2 | 3+3 2 | 3+3 2 |
| | 2+2 4 | 5+6 2 | 2+2 4 | 3+4 2 | 2+2 2 | 2+2 2 |
| | 2+2 4 | 5+6 2 | 2+2 4 | 3+4 2 | 3+3 2 | 3+3 2 |
| | 2+2 4 | 3+4 2 | 2+2 4 | 5+6 2 | 2+2 2 | 3+3 2 |
| | 2+2 4 | 3+4 2 | 2+2 4 | 5+6 2 | 3+3 2 | 2+2 2 |
| | 2+2 4 | 5+6 2 | 2+2 4 | 3+4 2 | 2+2 2 | 3+3 2 |
| | 2+2 4 | 5+6 2 | 2+2 4 | 3+4 2 | 2+2 2 | 2+2 2 |
| | 2+2 4 | 3+4 2 | 2+2 4 | 5+6 2 | 3+3 2 | 2+2 2 |
| | 2+2 4 | 3+4 2 | 2+2 4 | 5+6 2 | 3+3 2 | 3+3 2 |
| | 2+2 4 | 5+6 2 | 2+2 4 | 3+4 2 | 2+2 2 | 2+2 2 |
| | 2+2 4 | 5+6 2 | 2+2 4 | 3+4 2 | 2+2 2 | 3+3 2 |
| | 2+2 4 | 3+4 2 | 2+2 4 | 5+6 2 | 3+3 2 | 2+2 2 |
| | 2+2 4 | 5+6 2 | 2+2 4 | 3+4 2 | 2+2 2 | 3+3 2 |
| | 2+2 4 | 5+6 2 | 2+2 4 | 5+6 2 | 2+2 2 | 3+3 2 |
| | | 3+4 2 | | 5+6 2 | 3+3 2 | 2+2 2 |
| | | 3+4 2 | | 3+4 2 | 3+3 2 | 2+2 2 |
| | | 3+4 2 | | 5+6 2 | 3+3 2 | 3+3 2 |
| | | 5+6 2 | | 3+4 2 | 2+2 2 | 3+3 2 |
| | | 5+6 2 | | 3+4 2 | 2+2 2 | 2+2 2 |
| | | 3+4 2 | | 5+6 2 | 3+3 2 | 3+3 2 |
| | | 3+4 2 | | 5+6 2 | 2+2 2 | 2+2 2 |
| | | 5+6 2 | | 5+6 2 | 3+3 2 | 3+3 2 |
| | | 3+4 2 | | 3+4 2 | 3+3 2 | 2+2 2 |
| | | 5+6 2 | | 5+6 2 | 2+2 2 | 3+3 2 |
| | | 3+4 2 | | 3+4 2 | 2+2 2 | 2+2 2 |
| | | 5+6 2 | | 3+4 2 | 3+3 2 | 3+3 2 |
| | | 5+6 2 | | 3+4 2 | 2+2 2 | 2+2 2 |
| | | 5+6 2 | | 3+4 2 | 2+2 2 | 3+3 2 |
| | | 5+6 2 | | 3+4 2 | 2+2 2 | 2+2 2 |
| | | 5+6 2 | | 3+4 2 | 2+2 2 | 2+2 2 |
| | | 5+6 2 | | 3+4 2 | 2+2 2 | 3+3 2 |
| | | 3+4 2 | | 5+6 2 | 3+3 2 | 2+2 2 |
| | | 3+4 2 | | 5+6 2 | 3+3 2 | 3+3 2 |
| | | 3+4 2 | | 5+6 2 | 3+3 2 | 3+3 2 |
| | | 3+4 2 | | 5+6 2 | 3+3 2 | 2+2 2 |
| | | 3+4 2 | | 3+4 2 | 3+3 2 | 2+2 2 |
| | | 3+4 2 | | 3+4 2 | 2+2 2 | 2+2 2 |
| | | 3+4 2 | | 3+4 2 | 2+2 2 | 3+3 2 |
| | | 5+6 2 | | 5+6 2 | 3+3 2 | 2+2 2 |
| | | 5+6 2 | | 5+6 2 | 3+3 2 | 3+3 2 |
| | | 3+4 2 | | 5+6 2 | 2+2 2 | 2+2 2 |
| | | 3+4 2 | | 3+4 2 | 3+3 2 | 3+3 2 |
| | | 5+6 2 | | 5+6 2 | 3+3 2 | 2+2 2 |
| | | 3+4 2 | | 5+6 2 | 3+3 2 | 2+2 2 |
| | | 5+6 2 | | 3+4 2 | 2+2 2 | 3+3 2 |

Continued on next column

**TABLE XXVII– continued**

| TQC | ICSD | | NoE | SF | | |
|---|---|---|---|---|---|---|
| | | 3+4 2 | | 3+4 2 | 2+2 2 | 3+3 2 |
| | | 3+4 2 | | 5+6 2 | 2+2 2 | 2+2 2 |
| | | 5+6 2 | | 5+6 2 | 3+3 2 | 3+3 2 |
| | | 5+6 2 | | 5+6 2 | 3+3 2 | 2+2 2 |
| 15.2.39 | 20138 | | 136 | H4Al2Ca1F4O4 | | |
| | -A | -GM | -M | -Y | -L | -V |
| | 2+2 4 | 3+4 2 | 2+2 4 | 3+4 2 | 2+2 2 | 3+3 2 |
| | 2+2 4 | 5+6 2 | 2+2 4 | 5+6 2 | 3+3 2 | 2+2 2 |
| | 2+2 4 | 3+4 2 | 2+2 4 | 3+4 2 | 3+3 2 | 3+3 2 |
| | 2+2 4 | 3+4 2 | 2+2 4 | 5+6 2 | 2+2 2 | 2+2 2 |
| | 2+2 4 | 5+6 2 | 2+2 4 | 3+4 2 | 3+3 2 | 3+3 2 |
| | 2+2 4 | 5+6 2 | 2+2 4 | 5+6 2 | 2+2 2 | 2+2 2 |
| | 2+2 4 | 5+6 2 | 2+2 4 | 3+4 2 | 2+2 2 | 3+3 2 |
| | 2+2 4 | 3+4 2 | 2+2 4 | 5+6 2 | 3+3 2 | 2+2 2 |
| | 2+2 4 | 3+4 2 | 2+2 4 | 3+4 2 | 2+2 2 | 2+2 2 |
| | 2+2 4 | 5+6 2 | 2+2 4 | 5+6 2 | 3+3 2 | 3+3 2 |
| | 2+2 4 | 3+4 2 | 2+2 4 | 3+4 2 | 3+3 2 | 3+3 2 |
| | 2+2 4 | 5+6 2 | 2+2 4 | 5+6 2 | 2+2 2 | 2+2 2 |
| | 2+2 4 | 3+4 2 | 2+2 4 | 3+4 2 | 3+3 2 | 3+3 2 |
| | 2+2 4 | 5+6 2 | 2+2 4 | 5+6 2 | 2+2 2 | 2+2 2 |
| | 2+2 4 | 5+6 2 | 2+2 4 | 5+6 2 | 2+2 2 | 2+2 2 |
| | 2+2 4 | 3+4 2 | 2+2 4 | 3+4 2 | 3+3 2 | 3+3 2 |
| | 2+2 4 | 3+4 2 | 2+2 4 | 5+6 2 | 3+3 2 | 2+2 2 |
| | 2+2 4 | 5+6 2 | 2+2 4 | 3+4 2 | 2+2 2 | 3+3 2 |
| | 2+2 4 | 3+4 2 | 2+2 4 | 5+6 2 | 3+3 2 | 2+2 2 |
| | 2+2 4 | 5+6 2 | 2+2 4 | 3+4 2 | 2+2 2 | 3+3 2 |
| | 2+2 4 | 5+6 2 | 2+2 4 | 3+4 2 | 2+2 2 | 2+2 2 |
| | 2+2 4 | 3+4 2 | 2+2 4 | 3+4 2 | 2+2 2 | 3+3 2 |
| | 2+2 4 | 5+6 2 | 2+2 4 | 5+6 2 | 3+3 2 | 2+2 2 |
| | 2+2 4 | 3+4 2 | 2+2 4 | 5+6 2 | 3+3 2 | 3+3 2 |
| | 2+2 4 | 3+4 2 | 2+2 4 | 5+6 2 | 3+3 2 | 3+3 2 |
| | 2+2 4 | 3+4 2 | 2+2 4 | 3+4 2 | 3+3 2 | 2+2 2 |
| | 2+2 4 | 3+4 2 | 2+2 4 | 5+6 2 | 2+2 2 | 2+2 2 |
| | 2+2 4 | 5+6 2 | 2+2 4 | 5+6 2 | 2+2 2 | 3+3 2 |
| | 2+2 4 | 5+6 2 | 2+2 4 | 3+4 2 | 3+3 2 | 3+3 2 |
| | 2+2 4 | 5+6 2 | 2+2 4 | 3+4 2 | 2+2 2 | 2+2 2 |
| | 2+2 4 | 5+6 2 | 2+2 4 | 3+4 2 | 2+2 2 | 2+2 2 |
| | 2+2 4 | 3+4 2 | 2+2 4 | 5+6 2 | 3+3 2 | 3+3 2 |
| | 2+2 4 | 3+4 2 | 2+2 4 | 5+6 2 | 2+2 2 | 2+2 2 |
| | 2+2 4 | 5+6 2 | 2+2 4 | 3+4 2 | 3+3 2 | 3+3 2 |
| | | 3+4 2 | | 3+4 2 | 2+2 2 | 2+2 2 |
| | | 5+6 2 | | 5+6 2 | 3+3 2 | 3+3 2 |
| | | 5+6 2 | | 3+4 2 | 3+3 2 | 3+3 2 |
| | | 3+4 2 | | 5+6 2 | 3+3 2 | 2+2 2 |
| | | 3+4 2 | | 5+6 2 | 2+2 2 | 2+2 2 |
| | | 5+6 2 | | 3+4 2 | 2+2 2 | 3+3 2 |
| | | 3+4 2 | | 5+6 2 | 3+3 2 | 3+3 2 |
| | | 5+6 2 | | 3+4 2 | 3+3 2 | 3+3 2 |
| | | 3+4 2 | | 3+4 2 | 2+2 2 | 2+2 2 |
| | | 5+6 2 | | 5+6 2 | 2+2 2 | 2+2 2 |
| | | 3+4 2 | | 3+4 2 | 2+2 2 | 3+3 2 |
| | | 5+6 2 | | 5+6 2 | 2+2 2 | 2+2 2 |
| | | 5+6 2 | | 3+4 2 | 3+3 2 | 2+2 2 |
| | | 3+4 2 | | 5+6 2 | 2+2 2 | 3+3 2 |
| | | 5+6 2 | | 5+6 2 | 3+3 2 | 2+2 2 |
| | | 3+4 2 | | 3+4 2 | 3+3 2 | 3+3 2 |
| | | 5+6 2 | | 3+4 2 | 2+2 2 | 2+2 2 |
| | | 3+4 2 | | 3+4 2 | 3+3 2 | 3+3 2 |
| | | 5+6 2 | | 5+6 2 | 2+2 2 | 3+3 2 |
| | | 5+6 2 | | 5+6 2 | 3+3 2 | 3+3 2 |
| | | 3+4 2 | | 3+4 2 | 2+2 2 | 2+2 2 |
| | | 3+4 2 | | 3+4 2 | 3+3 2 | 2+2 2 |
| | | 5+6 2 | | 5+6 2 | 3+3 2 | 3+3 2 |
| | | 3+4 2 | | 3+4 2 | 2+2 2 | 3+3 2 |
| | | 3+4 2 | | 5+6 2 | 3+3 2 | 2+2 2 |
| | | 3+4 2 | | 5+6 2 | 2+2 2 | 2+2 2 |
| | | 5+6 2 | | 5+6 2 | 2+2 2 | 3+3 2 |
| | | 3+4 2 | | 3+4 2 | 2+2 2 | 2+2 2 |
| | | 5+6 2 | | 3+4 2 | 3+3 2 | 2+2 2 |
| | | 5+6 2 | | 5+6 2 | 3+3 2 | 2+2 2 |
| | | 5+6 2 | | 3+4 2 | 2+2 2 | 3+3 2 |
| | | 3+4 2 | | 5+6 2 | 3+3 2 | 3+3 2 |
| | | 3+4 2 | | 3+4 2 | 2+2 2 | 3+3 2 |
| | | 5+6 2 | | 3+4 2 | 3+3 2 | 2+2 2 |
| 15.2.40 | 14176 | | 180 | H1Cr1Hg1O4.5 | | |
| | -A | -GM | -M | -Y | -L | -V |
| | 2+2 4 | 3+4 2 | 2+2 4 | 5+6 2 | 2+2 2 | 3+3 2 |
| | 2+2 4 | 3+4 2 | 2+2 4 | 5+6 2 | 3+3 2 | 2+2 2 |
| | 2+2 4 | 5+6 2 | 2+2 4 | 3+4 2 | 2+2 2 | 2+2 2 |
| | 2+2 4 | 5+6 2 | 2+2 4 | 3+4 2 | 3+3 2 | 3+3 2 |
| | 2+2 4 | 3+4 2 | 2+2 4 | 3+4 2 | 3+3 2 | 3+3 2 |
| | 2+2 4 | 5+6 2 | 2+2 4 | 5+6 2 | 2+2 2 | 2+2 2 |
| | 2+2 4 | 3+4 2 | 2+2 4 | 3+4 2 | 2+2 2 | 3+3 2 |
| | 2+2 4 | 5+6 2 | 2+2 4 | 5+6 2 | 3+3 2 | 2+2 2 |
| | 2+2 4 | 3+4 2 | 2+2 4 | 5+6 2 | 3+3 2 | 2+2 2 |
| | 2+2 4 | 3+4 2 | 2+2 4 | 3+4 2 | 2+2 2 | 2+2 2 |
| | 2+2 4 | 5+6 2 | 2+2 4 | 5+6 2 | 2+2 2 | 3+3 2 |
| | 2+2 4 | 5+6 2 | 2+2 4 | 3+4 2 | 3+3 2 | 3+3 2 |
| | 2+2 4 | 3+4 2 | 2+2 4 | 3+4 2 | 2+2 2 | 2+2 2 |
| | 2+2 4 | 3+4 2 | 2+2 4 | 3+4 2 | 3+3 2 | 2+2 2 |
| | 2+2 4 | 5+6 2 | 2+2 4 | 5+6 2 | 2+2 2 | 3+3 2 |
| | 2+2 4 | 5+6 2 | 2+2 4 | 5+6 2 | 3+3 2 | 3+3 2 |
| | 2+2 4 | 3+4 2 | 2+2 4 | 3+4 2 | 2+2 2 | 2+2 2 |
| | 2+2 4 | 5+6 2 | 2+2 4 | 5+6 2 | 3+3 2 | 3+3 2 |
| | 2+2 4 | 3+4 2 | 2+2 4 | 5+6 2 | 2+2 2 | 2+2 2 |
| | 2+2 4 | 3+4 2 | 2+2 4 | 5+6 2 | 2+2 2 | 3+3 2 |
| | 2+2 4 | 5+6 2 | 2+2 4 | 3+4 2 | 3+3 2 | 3+3 2 |
| | 2+2 4 | 5+6 2 | 2+2 4 | 3+4 2 | 3+3 2 | 2+2 2 |
| | 2+2 4 | 3+4 2 | 2+2 4 | 5+6 2 | 2+2 2 | 2+2 2 |
| | 2+2 4 | 3+4 2 | 2+2 4 | 3+4 2 | 3+3 2 | 3+3 2 |
| | 2+2 4 | 5+6 2 | 2+2 4 | 5+6 2 | 2+2 2 | 3+3 2 |
| | 2+2 4 | 5+6 2 | 2+2 4 | 3+4 2 | 3+3 2 | 2+2 2 |
| | 2+2 4 | 3+4 2 | 2+2 4 | 5+6 2 | 2+2 2 | 2+2 2 |
| | 2+2 4 | 3+4 2 | 2+2 4 | 5+6 2 | 3+3 2 | 3+3 2 |
| | 2+2 4 | 5+6 2 | 2+2 4 | 3+4 2 | 2+2 2 | 3+3 2 |
| | 2+2 4 | 5+6 2 | 2+2 4 | 3+4 2 | 3+3 2 | 2+2 2 |
| | 2+2 4 | 3+4 2 | 2+2 4 | 5+6 2 | 2+2 2 | 2+2 2 |
| | 2+2 4 | 3+4 2 | 2+2 4 | 3+4 2 | 3+3 2 | 3+3 2 |
| | 2+2 4 | 5+6 2 | 2+2 4 | 5+6 2 | 2+2 2 | 3+3 2 |
| | 2+2 4 | 5+6 2 | 2+2 4 | 3+4 2 | 3+3 2 | 2+2 2 |
| | 2+2 4 | 3+4 2 | 2+2 4 | 5+6 2 | 2+2 2 | 2+2 2 |
| | 2+2 4 | 3+4 2 | 2+2 4 | 5+6 2 | 2+2 2 | 3+3 2 |
| | 2+2 4 | 3+4 2 | 2+2 4 | 3+4 2 | 3+3 2 | 3+3 2 |
| | 2+2 4 | 5+6 2 | 2+2 4 | 3+4 2 | 3+3 2 | 2+2 2 |
| | 2+2 4 | 5+6 2 | 2+2 4 | 3+4 2 | 2+2 2 | 2+2 2 |
| | 2+2 4 | 5+6 2 | 2+2 4 | 5+6 2 | 2+2 2 | 3+3 2 |
| | 2+2 4 | 3+4 2 | 2+2 4 | 5+6 2 | 3+3 2 | 2+2 2 |
| | 2+2 4 | 5+6 2 | 2+2 4 | 3+4 2 | 3+3 2 | 3+3 2 |
| | 2+2 4 | 5+6 2 | 2+2 4 | 3+4 2 | 2+2 2 | 2+2 2 |
| | 2+2 4 | 3+4 2 | 2+2 4 | 5+6 2 | 2+2 2 | 3+3 2 |
| | 2+2 4 | 3+4 2 | 2+2 4 | 3+4 2 | 3+3 2 | 2+2 2 |
| | | 5+6 2 | | 5+6 2 | 3+3 2 | 3+3 2 |

Continued on next column

Continued on next column

**TABLE XXVII– continued**

| TQC | ICSD | | NoE | SF | | |
|---|---|---|---|---|---|---|
| | | 5+6 2 | | 5+6 2 | 2+2 2 | 2+2 2 |
| | | 5+6 2 | | 5+6 2 | 3+3 2 | 3+3 2 |
| | | 3+4 2 | | 3+4 2 | 2+2 2 | 2+2 2 |
| | | 3+4 2 | | 3+4 2 | 3+3 2 | 3+3 2 |
| | | 5+6 2 | | 5+6 2 | 2+2 2 | 2+2 2 |
| | | 3+4 2 | | 3+4 2 | 3+3 2 | 3+3 2 |
| | | 3+4 2 | | 5+6 2 | 3+3 2 | 2+2 2 |
| | | 5+6 2 | | 5+6 2 | 2+2 2 | 3+3 2 |
| | | 5+6 2 | | 5+6 2 | 2+2 2 | 2+2 2 |
| | | 3+4 2 | | 3+4 2 | 3+3 2 | 3+3 2 |
| | | 5+6 2 | | 3+4 2 | 2+2 2 | 2+2 2 |
| | | 3+4 2 | | 3+4 2 | 3+3 2 | 3+3 2 |
| | | 3+4 2 | | 3+4 2 | 3+3 2 | 3+3 2 |
| | | 3+4 2 | | 5+6 2 | 3+3 2 | 3+3 2 |
| | | 3+4 2 | | 3+4 2 | 2+2 2 | 2+2 2 |
| | | 5+6 2 | | 5+6 2 | 3+3 2 | 2+2 2 |
| | | 5+6 2 | | 5+6 2 | 2+2 2 | 3+3 2 |
| | | 3+4 2 | | 5+6 2 | 3+3 2 | 3+3 2 |
| | | 5+6 2 | | 3+4 2 | 2+2 2 | 2+2 2 |
| | | 3+4 2 | | 3+4 2 | 3+3 2 | 2+2 2 |
| | | 3+4 2 | | 3+4 2 | 3+3 2 | 3+3 2 |
| | | 5+6 2 | | 5+6 2 | 2+2 2 | 2+2 2 |
| | | 3+4 2 | | 5+6 2 | 3+3 2 | 3+3 2 |
| | | 5+6 2 | | 3+4 2 | 2+2 2 | 2+2 2 |
| | | 5+6 2 | | 3+4 2 | 2+2 2 | 3+3 2 |
| | | 3+4 2 | | 3+4 2 | 2+2 2 | 3+3 2 |
| | | 5+6 2 | | 5+6 2 | 3+3 2 | 2+2 2 |
| | | 5+6 2 | | 5+6 2 | 2+2 2 | 2+2 2 |
| | | 5+6 2 | | 3+4 2 | 2+2 2 | 2+2 2 |
| | | 3+4 2 | | 5+6 2 | 3+3 2 | 3+3 2 |
| | | 3+4 2 | | 3+4 2 | 2+2 2 | 2+2 2 |
| | | 5+6 2 | | 5+6 2 | 3+3 2 | 3+3 2 |
| | | 3+4 2 | | 3+4 2 | 2+2 2 | 2+2 2 |
| | | 3+4 2 | | 3+4 2 | 3+3 2 | 3+3 2 |
| | | 5+6 2 | | 5+6 2 | 2+2 2 | 2+2 2 |
| | | 5+6 2 | | 5+6 2 | 3+3 2 | 3+3 2 |
| | | 5+6 2 | | 3+4 2 | 2+2 2 | 2+2 2 |
| | | 3+4 2 | | 3+4 2 | 2+2 2 | 3+3 2 |
| | | 3+4 2 | | 5+6 2 | 3+3 2 | 2+2 2 |
| | | 5+6 2 | | 5+6 2 | 3+3 2 | 3+3 2 |
| | | 3+4 2 | | 5+6 2 | 2+2 2 | 3+3 2 |
| | | 5+6 2 | | 3+4 2 | 3+3 2 | 2+2 2 |
| | | 5+6 2 | | 3+4 2 | 3+3 2 | 2+2 2 |
| | | 5+6 2 | | 3+4 2 | 2+2 2 | 3+3 2 |
| 15.2.41 | 183508 | | 136 | H12Cl3Mg1O6Rb1 | | |
| | -A | -GM | -M | -Y | -L | -V |
| | 2+2 4 | 3+4 2 | 2+2 4 | 3+4 2 | 2+2 2 | 2+2 2 |
| | 2+2 4 | 5+6 2 | 2+2 4 | 5+6 2 | 3+3 2 | 3+3 2 |
| | 2+2 4 | 3+4 2 | 2+2 4 | 3+4 2 | 2+2 2 | 3+3 2 |
| | 2+2 4 | 3+4 2 | 2+2 4 | 3+4 2 | 3+3 2 | 3+3 2 |
| | 2+2 4 | 5+6 2 | 2+2 4 | 5+6 2 | 2+2 2 | 2+2 2 |
| | 2+2 4 | 5+6 2 | 2+2 4 | 5+6 2 | 3+3 2 | 2+2 2 |
| | 2+2 4 | 5+6 2 | 2+2 4 | 5+6 2 | 2+2 2 | 2+2 2 |
| | 2+2 4 | 5+6 2 | 2+2 4 | 5+6 2 | 3+3 2 | 2+2 2 |
| | 2+2 4 | 5+6 2 | 2+2 4 | 5+6 2 | 3+3 2 | 2+2 2 |
| | 2+2 4 | 5+6 2 | 2+2 4 | 5+6 2 | 2+2 2 | 2+2 2 |
| | 2+2 4 | 3+4 2 | 2+2 4 | 3+4 2 | 2+2 2 | 3+3 2 |
| | 2+2 4 | 3+4 2 | 2+2 4 | 3+4 2 | 3+3 2 | 3+3 2 |
| | 2+2 4 | 3+4 2 | 2+2 4 | 3+4 2 | 3+3 2 | 3+3 2 |
| | 2+2 4 | 3+4 2 | 2+2 4 | 3+4 2 | 2+2 2 | 3+3 2 |
| | 2+2 4 | 3+4 2 | 2+2 4 | 3+4 2 | 2+2 2 | 2+2 2 |
| | 2+2 4 | 3+4 2 | 2+2 4 | 3+4 2 | 3+3 2 | 2+2 2 |
| | 2+2 4 | 5+6 2 | 2+2 4 | 5+6 2 | 2+2 2 | 3+3 2 |
| | 2+2 4 | 3+4 2 | 2+2 4 | 3+4 2 | 3+3 2 | 2+2 2 |
| | 2+2 4 | 3+4 2 | 2+2 4 | 3+4 2 | 2+2 2 | 3+3 2 |
| | 2+2 4 | 5+6 2 | 2+2 4 | 5+6 2 | 3+3 2 | 2+2 2 |
| | 2+2 4 | 5+6 2 | 2+2 4 | 5+6 2 | 2+2 2 | 3+3 2 |
| | 2+2 4 | 3+4 2 | 2+2 4 | 3+4 2 | 3+3 2 | 2+2 2 |
| | 2+2 4 | 5+6 2 | 2+2 4 | 5+6 2 | 2+2 2 | 3+3 2 |
| | 2+2 4 | 3+4 2 | 2+2 4 | 3+4 2 | 3+3 2 | 2+2 2 |
| | 2+2 4 | 5+6 2 | 2+2 4 | 3+4 2 | 2+2 2 | 3+3 2 |
| | 2+2 4 | 3+4 2 | 2+2 4 | 5+6 2 | 3+3 2 | 2+2 2 |
| | 2+2 4 | 3+4 2 | 2+2 4 | 3+4 2 | 2+2 2 | 3+3 2 |
| | 2+2 4 | 3+4 2 | 2+2 4 | 3+4 2 | 3+3 2 | 3+3 2 |
| | 2+2 4 | 5+6 2 | 2+2 4 | 5+6 2 | 3+3 2 | 2+2 2 |
| | 2+2 4 | 5+6 2 | 2+2 4 | 3+4 2 | 2+2 2 | 3+3 2 |
| | 2+2 4 | 5+6 2 | 2+2 4 | 5+6 2 | 2+2 2 | 2+2 2 |
| | 2+2 4 | 5+6 2 | 2+2 4 | 5+6 2 | 3+3 2 | 3+3 2 |
| | 2+2 4 | 3+4 2 | 2+2 4 | 5+6 2 | 2+2 2 | 2+2 2 |
| | 2+2 4 | 5+6 2 | 2+2 4 | 3+4 2 | 3+3 2 | 3+3 2 |
| | | 5+6 2 | | 5+6 2 | 3+3 2 | 2+2 2 |
| | | 3+4 2 | | 5+6 2 | 2+2 2 | 3+3 2 |
| | | 3+4 2 | | 5+6 2 | 3+3 2 | 2+2 2 |
| | | 5+6 2 | | 3+4 2 | 2+2 2 | 3+3 2 |
| | | 3+4 2 | | 5+6 2 | 2+2 2 | 2+2 2 |
| | | 3+4 2 | | 5+6 2 | 3+3 2 | 3+3 2 |
| | | 5+6 2 | | 3+4 2 | 2+2 2 | 3+3 2 |
| | | 3+4 2 | | 3+4 2 | 3+3 2 | 3+3 2 |
| | | 5+6 2 | | 3+4 2 | 2+2 2 | 2+2 2 |
| | | 5+6 2 | | 5+6 2 | 3+3 2 | 2+2 2 |
| | | 5+6 2 | | 3+4 2 | 3+3 2 | 2+2 2 |
| | | 3+4 2 | | 5+6 2 | 2+2 2 | 3+3 2 |
| | | 3+4 2 | | 5+6 2 | 2+2 2 | 2+2 2 |
| | | 5+6 2 | | 3+4 2 | 3+3 2 | 3+3 2 |
| | | 3+4 2 | | 5+6 2 | 2+2 2 | 3+3 2 |
| | | 3+4 2 | | 3+4 2 | 3+3 2 | 2+2 2 |
| | | 5+6 2 | | 5+6 2 | 2+2 2 | 3+3 2 |
| | | 5+6 2 | | 3+4 2 | 3+3 2 | 3+3 2 |
| | | 3+4 2 | | 5+6 2 | 2+2 2 | 2+2 2 |
| | | 5+6 2 | | 5+6 2 | 3+3 2 | 3+3 2 |
| | | 5+6 2 | | 5+6 2 | 2+2 2 | 3+3 2 |
| | | 3+4 2 | | 5+6 2 | 3+3 2 | 3+3 2 |
| | | 5+6 2 | | 3+4 2 | 3+3 2 | 3+3 2 |
| | | 5+6 2 | | 3+4 2 | 2+2 2 | 3+3 2 |
| | | 3+4 2 | | 5+6 2 | 3+3 2 | 2+2 2 |
| | | 3+4 2 | | 3+4 2 | 2+2 2 | 2+2 2 |
| | | 5+6 2 | | 3+4 2 | 2+2 2 | 2+2 2 |
| | | 5+6 2 | | 5+6 2 | 3+3 2 | 2+2 2 |
| | | 3+4 2 | | 5+6 2 | 3+3 2 | 2+2 2 |
| | | 5+6 2 | | 5+6 2 | 2+2 2 | 3+3 2 |
| | | 5+6 2 | | 5+6 2 | 3+3 2 | 2+2 2 |
| | | 5+6 2 | | 3+4 2 | 2+2 2 | 2+2 2 |
| | | 3+4 2 | | 3+4 2 | 2+2 2 | 3+3 2 |
| | | 5+6 2 | | 5+6 2 | 3+3 2 | 3+3 2 |
| 15.2.42 | 192134 | | 136 | H12Br3Mg1O6Rb1 | | |
| | -A | -GM | -M | -Y | -L | -V |
| | 2+2 4 | 3+4 2 | 2+2 4 | 3+4 2 | 2+2 2 | 2+2 2 |
| | 2+2 4 | 5+6 2 | 2+2 4 | 5+6 2 | 3+3 2 | 3+3 2 |
| | 2+2 4 | 3+4 2 | 2+2 4 | 3+4 2 | 2+2 2 | 2+2 2 |
| | 2+2 4 | 3+4 2 | 2+2 4 | 3+4 2 | 3+3 2 | 2+2 2 |
| | 2+2 4 | 5+6 2 | 2+2 4 | 5+6 2 | 2+2 2 | 3+3 2 |
| | 2+2 4 | 5+6 2 | 2+2 4 | 5+6 2 | 3+3 2 | 3+3 2 |

Continued on next column

**TABLE XXVII– continued**

| TQC | ICSD | | NoE | | SF | |
|---|---|---|---|---|---|---|
| | 2+2 4 | 5+6 2 | 2+2 4 | 5+6 2 | 2+2 2 | 3+3 2 |
| | 2+2 4 | 5+6 2 | 2+2 4 | 5+6 2 | 3+3 2 | 3+3 2 |
| | 2+2 4 | 5+6 2 | 2+2 4 | 5+6 2 | 3+3 2 | 3+3 2 |
| | 2+2 4 | 5+6 2 | 2+2 4 | 5+6 2 | 2+2 2 | 3+3 2 |
| | 2+2 4 | 3+4 2 | 2+2 4 | 3+4 2 | 3+3 2 | 2+2 2 |
| | 2+2 4 | 3+4 2 | 2+2 4 | 3+4 2 | 2+2 2 | 2+2 2 |
| | 2+2 4 | 3+4 2 | 2+2 4 | 3+4 2 | 2+2 2 | 2+2 2 |
| | 2+2 4 | 3+4 2 | 2+2 4 | 3+4 2 | 3+3 2 | 2+2 2 |
| | 2+2 4 | 3+4 2 | 2+2 4 | 3+4 2 | 2+2 2 | 2+2 2 |
| | 2+2 4 | 5+6 2 | 2+2 4 | 5+6 2 | 3+3 2 | 3+3 2 |
| | 2+2 4 | 3+4 2 | 2+2 4 | 3+4 2 | 2+2 2 | 2+2 2 |
| | 2+2 4 | 5+6 2 | 2+2 4 | 5+6 2 | 3+3 2 | 3+3 2 |
| | 2+2 4 | 3+4 2 | 2+2 4 | 3+4 2 | 2+2 2 | 3+3 2 |
| | 2+2 4 | 3+4 2 | 2+2 4 | 3+4 2 | 3+3 2 | 3+3 2 |
| | 2+2 4 | 5+6 2 | 2+2 4 | 5+6 2 | 2+2 2 | 2+2 2 |
| | 2+2 4 | 3+4 2 | 2+2 4 | 3+4 2 | 3+3 2 | 3+3 2 |
| | 2+2 4 | 5+6 2 | 2+2 4 | 5+6 2 | 2+2 2 | 2+2 2 |
| | 2+2 4 | 3+4 2 | 2+2 4 | 3+4 2 | 3+3 2 | 3+3 2 |
| | 2+2 4 | 5+6 2 | 2+2 4 | 5+6 2 | 2+2 2 | 2+2 2 |
| | 2+2 4 | 3+4 2 | 2+2 4 | 3+4 2 | 3+3 2 | 3+3 2 |
| | 2+2 4 | 3+4 2 | 2+2 4 | 3+4 2 | 2+2 2 | 2+2 2 |
| | 2+2 4 | 3+4 2 | 2+2 4 | 3+4 2 | 3+3 2 | 2+2 2 |
| | 2+2 4 | 5+6 2 | 2+2 4 | 5+6 2 | 3+3 2 | 3+3 2 |
| | 2+2 4 | 5+6 2 | 2+2 4 | 3+4 2 | 2+2 2 | 2+2 2 |
| | 2+2 4 | 5+6 2 | 2+2 4 | 5+6 2 | 3+3 2 | 2+2 2 |
| | 2+2 4 | 3+4 2 | 2+2 4 | 5+6 2 | 2+2 2 | 3+3 2 |
| | 2+2 4 | 5+6 2 | 2+2 4 | 3+4 2 | 2+2 2 | 3+3 2 |
| | 2+2 4 | 5+6 2 | 2+2 4 | 5+6 2 | 3+3 2 | 2+2 2 |
| | | 3+4 2 | | 5+6 2 | 3+3 2 | 2+2 2 |
| | | 5+6 2 | | 5+6 2 | 2+2 2 | 3+3 2 |
| | | 5+6 2 | | 5+6 2 | 2+2 2 | 3+3 2 |
| | | 3+4 2 | | 5+6 2 | 3+3 2 | 2+2 2 |
| | | 5+6 2 | | 3+4 2 | 2+2 2 | 2+2 2 |
| | | 5+6 2 | | 5+6 2 | 3+3 2 | 3+3 2 |
| | | 3+4 2 | | 3+4 2 | 3+3 2 | 2+2 2 |
| | | 3+4 2 | | 5+6 2 | 2+2 2 | 3+3 2 |
| | | 3+4 2 | | 3+4 2 | 2+2 2 | 3+3 2 |
| | | 5+6 2 | | 5+6 2 | 3+3 2 | 2+2 2 |
| | | 5+6 2 | | 3+4 2 | 3+3 2 | 2+2 2 |
| | | 3+4 2 | | 3+4 2 | 2+2 2 | 3+3 2 |
| | | 3+4 2 | | 5+6 2 | 3+3 2 | 2+2 2 |
| | | 5+6 2 | | 3+4 2 | 2+2 2 | 2+2 2 |
| | | 5+6 2 | | 5+6 2 | 2+2 2 | 3+3 2 |
| | | 3+4 2 | | 3+4 2 | 3+3 2 | 3+3 2 |
| | | 5+6 2 | | 5+6 2 | 3+3 2 | 2+2 2 |
| | | 3+4 2 | | 3+4 2 | 2+2 2 | 2+2 2 |
| | | 5+6 2 | | 5+6 2 | 2+2 2 | 3+3 2 |
| | | 3+4 2 | | 5+6 2 | 3+3 2 | 2+2 2 |
| | | 5+6 2 | | 5+6 2 | 2+2 2 | 2+2 2 |
| | | 3+4 2 | | 5+6 2 | 3+3 2 | 2+2 2 |
| | | 3+4 2 | | 3+4 2 | 2+2 2 | 2+2 2 |
| | | 5+6 2 | | 3+4 2 | 3+3 2 | 3+3 2 |
| | | 5+6 2 | | 5+6 2 | 3+3 2 | 2+2 2 |
| | | 3+4 2 | | 3+4 2 | 2+2 2 | 3+3 2 |
| | | 5+6 2 | | 3+4 2 | 3+3 2 | 3+3 2 |
| | | 5+6 2 | | 5+6 2 | 2+2 2 | 3+3 2 |
| | | 5+6 2 | | 5+6 2 | 2+2 2 | 3+3 2 |
| | | 3+4 2 | | 5+6 2 | 3+3 2 | 3+3 2 |
| | | 5+6 2 | | 3+4 2 | 3+3 2 | 2+2 2 |
| | | 5+6 2 | | 5+6 2 | 2+2 2 | 3+3 2 |
| | | 3+4 2 | | 5+6 2 | 3+3 2 | 2+2 2 |

Continued on next column

**TABLE XXVII– continued**

| TQC | ICSD | | NoE | | SF | |
|---|---|---|---|---|---|---|
| | | 3+4 2 | | 3+4 2 | 2+2 2 | 3+3 2 |
| 15.2.43 | 237747 | | 104 | Ge2Mn1Na1O6 | | |
| | -A | -GM | -M | -Y | -L | -V |
| | 2+2 4 | 3+4 2 | 2+2 4 | 3+4 2 | 2+2 2 | 2+2 2 |
| | 2+2 4 | 5+6 2 | 2+2 4 | 5+6 2 | 3+3 2 | 3+3 2 |
| | 2+2 4 | 3+4 2 | 2+2 4 | 3+4 2 | 2+2 2 | 2+2 2 |
| | 2+2 4 | 5+6 2 | 2+2 4 | 5+6 2 | 3+3 2 | 3+3 2 |
| | 2+2 4 | 3+4 2 | 2+2 4 | 3+4 2 | 2+2 2 | 3+3 2 |
| | 2+2 4 | 5+6 2 | 2+2 4 | 5+6 2 | 3+3 2 | 2+2 2 |
| | 2+2 4 | 3+4 2 | 2+2 4 | 3+4 2 | 2+2 2 | 3+3 2 |
| | 2+2 4 | 5+6 2 | 2+2 4 | 5+6 2 | 3+3 2 | 2+2 2 |
| | 2+2 4 | 5+6 2 | 2+2 4 | 5+6 2 | 3+3 2 | 2+2 2 |
| | 2+2 4 | 3+4 2 | 2+2 4 | 3+4 2 | 2+2 2 | 3+3 2 |
| | 2+2 4 | 5+6 2 | 2+2 4 | 3+4 2 | 2+2 2 | 2+2 2 |
| | 2+2 4 | 3+4 2 | 2+2 4 | 5+6 2 | 3+3 2 | 3+3 2 |
| | 2+2 4 | 5+6 2 | 2+2 4 | 5+6 2 | 2+2 2 | 3+3 2 |
| | 2+2 4 | 3+4 2 | 2+2 4 | 3+4 2 | 3+3 2 | 2+2 2 |
| | 2+2 4 | 3+4 2 | 2+2 4 | 3+4 2 | 2+2 2 | 3+3 2 |
| | 2+2 4 | 5+6 2 | 2+2 4 | 5+6 2 | 3+3 2 | 2+2 2 |
| | 2+2 4 | 3+4 2 | 2+2 4 | 5+6 2 | 2+2 2 | 3+3 2 |
| | 2+2 4 | 5+6 2 | 2+2 4 | 3+4 2 | 3+3 2 | 2+2 2 |
| | 2+2 4 | 3+4 2 | 2+2 4 | 3+4 2 | 3+3 2 | 3+3 2 |
| | 2+2 4 | 5+6 2 | 2+2 4 | 5+6 2 | 2+2 2 | 2+2 2 |
| | 2+2 4 | 3+4 2 | 2+2 4 | 5+6 2 | 3+3 2 | 2+2 2 |
| | 2+2 4 | 5+6 2 | 2+2 4 | 5+6 2 | 2+2 2 | 3+3 2 |
| | 2+2 4 | 3+4 2 | 2+2 4 | 3+4 2 | 2+2 2 | 2+2 2 |
| | 2+2 4 | 3+4 2 | 2+2 4 | 3+4 2 | 3+3 2 | 3+3 2 |
| | 2+2 4 | 5+6 2 | 2+2 4 | 3+4 2 | 3+3 2 | 2+2 2 |
| | 2+2 4 | 5+6 2 | 2+2 4 | 5+6 2 | 2+2 2 | 3+3 2 |
| | | 5+6 2 | | 3+4 2 | 2+2 2 | 2+2 2 |
| | | 3+4 2 | | 5+6 2 | 3+3 2 | 3+3 2 |
| | | 5+6 2 | | 5+6 2 | 2+2 2 | 3+3 2 |
| | | 3+4 2 | | 3+4 2 | 3+3 2 | 2+2 2 |
| | | 3+4 2 | | 3+4 2 | 2+2 2 | 3+3 2 |
| | | 3+4 2 | | 3+4 2 | 3+3 2 | 2+2 2 |
| | | 5+6 2 | | 5+6 2 | 2+2 2 | 2+2 2 |
| | | 3+4 2 | | 3+4 2 | 3+3 2 | 3+3 2 |
| | | 5+6 2 | | 5+6 2 | 3+3 2 | 3+3 2 |
| | | 3+4 2 | | 3+4 2 | 2+2 2 | 2+2 2 |
| | | 3+4 2 | | 5+6 2 | 2+2 2 | 2+2 2 |
| | | 3+4 2 | | 3+4 2 | 3+3 2 | 3+3 2 |
| | | 5+6 2 | | 5+6 2 | 3+3 2 | 2+2 2 |
| | | 5+6 2 | | 3+4 2 | 2+2 2 | 3+3 2 |
| | | 3+4 2 | | 5+6 2 | 2+2 2 | 3+3 2 |
| | | 5+6 2 | | 3+4 2 | 3+3 2 | 2+2 2 |
| | | 5+6 2 | | 5+6 2 | 3+3 2 | 3+3 2 |
| | | 3+4 2 | | 5+6 2 | 2+2 2 | 3+3 2 |
| | | 5+6 2 | | 3+4 2 | 3+3 2 | 2+2 2 |
| | | 5+6 2 | | 5+6 2 | 2+2 2 | 2+2 2 |
| | | 3+4 2 | | 5+6 2 | 3+3 2 | 2+2 2 |
| | | 5+6 2 | | 3+4 2 | 2+2 2 | 3+3 2 |
| | | 3+4 2 | | 3+4 2 | 2+2 2 | 3+3 2 |
| | | 3+4 2 | | 5+6 2 | 3+3 2 | 2+2 2 |
| | | 5+6 2 | | 3+4 2 | 2+2 2 | 2+2 2 |
| | | 5+6 2 | | 5+6 2 | 3+3 2 | 2+2 2 |
| 15.2.44 | 237746 | | 104 | Ge2Mn1Na1O6 | | |
| | -A | -GM | -M | -Y | -L | -V |
| | 2+2 4 | 3+4 2 | 2+2 4 | 3+4 2 | 2+2 2 | 2+2 2 |
| | 2+2 4 | 5+6 2 | 2+2 4 | 5+6 2 | 3+3 2 | 3+3 2 |
| | 2+2 4 | 5+6 2 | 2+2 4 | 3+4 2 | 2+2 2 | 3+3 2 |
| | 2+2 4 | 3+4 2 | 2+2 4 | 5+6 2 | 3+3 2 | 2+2 2 |

Continued on next column

**TABLE XXVII– continued**

| TQC | ICSD | | NoE | SF | | |
|---|---|---|---|---|---|---|
| | 2+2 4 | 3+4 2 | 2+2 4 | 3+4 2 | 2+2 2 | 2+2 2 |
| | 2+2 4 | 3+4 2 | 2+2 4 | 3+4 2 | 3+3 2 | 3+3 2 |
| | 2+2 4 | 5+6 2 | 2+2 4 | 5+6 2 | 3+3 2 | 3+3 2 |
| | 2+2 4 | 5+6 2 | 2+2 4 | 5+6 2 | 2+2 2 | 2+2 2 |
| | 2+2 4 | 3+4 2 | 2+2 4 | 5+6 2 | 3+3 2 | 2+2 2 |
| | 2+2 4 | 5+6 2 | 2+2 4 | 3+4 2 | 2+2 2 | 3+3 2 |
| | 2+2 4 | 5+6 2 | 2+2 4 | 3+4 2 | 3+3 2 | 2+2 2 |
| | 2+2 4 | 3+4 2 | 2+2 4 | 5+6 2 | 2+2 2 | 3+3 2 |
| | 2+2 4 | 5+6 2 | 2+2 4 | 5+6 2 | 2+2 2 | 3+3 2 |
| | 2+2 4 | 3+4 2 | 2+2 4 | 3+4 2 | 3+3 2 | 2+2 2 |
| | 2+2 4 | 3+4 2 | 2+2 4 | 3+4 2 | 2+2 2 | 3+3 2 |
| | 2+2 4 | 5+6 2 | 2+2 4 | 5+6 2 | 3+3 2 | 2+2 2 |
| | 2+2 4 | 5+6 2 | 2+2 4 | 5+6 2 | 2+2 2 | 3+3 2 |
| | 2+2 4 | 3+4 2 | 2+2 4 | 3+4 2 | 3+3 2 | 2+2 2 |
| | 2+2 4 | 3+4 2 | 2+2 4 | 5+6 2 | 3+3 2 | 3+3 2 |
| | 2+2 4 | 3+4 2 | 2+2 4 | 3+4 2 | 2+2 2 | 2+2 2 |
| | 2+2 4 | 5+6 2 | 2+2 4 | 5+6 2 | 3+3 2 | 2+2 2 |
| | 2+2 4 | 5+6 2 | 2+2 4 | 5+6 2 | 2+2 2 | 3+3 2 |
| | 2+2 4 | 3+4 2 | 2+2 4 | 3+4 2 | 2+2 2 | 2+2 2 |
| | 2+2 4 | 3+4 2 | 2+2 4 | 3+4 2 | 3+3 2 | 3+3 2 |
| | 2+2 4 | 5+6 2 | 2+2 4 | 3+4 2 | 3+3 2 | 2+2 2 |
| | 2+2 4 | 5+6 2 | 2+2 4 | 5+6 2 | 2+2 2 | 3+3 2 |
| | | 5+6 2 | | 3+4 2 | 2+2 2 | 2+2 2 |
| | | 3+4 2 | | 5+6 2 | 3+3 2 | 3+3 2 |
| | | 5+6 2 | | 5+6 2 | 2+2 2 | 3+3 2 |
| | | 3+4 2 | | 3+4 2 | 3+3 2 | 3+3 2 |
| | | 3+4 2 | | 3+4 2 | 2+2 2 | 2+2 2 |
| | | 3+4 2 | | 3+4 2 | 3+3 2 | 2+2 2 |
| | | 5+6 2 | | 5+6 2 | 3+3 2 | 3+3 2 |
| | | 3+4 2 | | 3+4 2 | 2+2 2 | 2+2 2 |
| | | 5+6 2 | | 5+6 2 | 3+3 2 | 3+3 2 |
| | | 3+4 2 | | 3+4 2 | 2+2 2 | 2+2 2 |
| | | 3+4 2 | | 5+6 2 | 3+3 2 | 2+2 2 |
| | | 3+4 2 | | 3+4 2 | 2+2 2 | 2+2 2 |
| | | 5+6 2 | | 5+6 2 | 3+3 2 | 3+3 2 |
| | | 3+4 2 | | 5+6 2 | 2+2 2 | 3+3 2 |
| | | 5+6 2 | | 3+4 2 | 2+2 2 | 3+3 2 |
| | | 5+6 2 | | 3+4 2 | 3+3 2 | 2+2 2 |
| | | 5+6 2 | | 5+6 2 | 2+2 2 | 3+3 2 |
| | | 5+6 2 | | 5+6 2 | 3+3 2 | 3+3 2 |
| | | 3+4 2 | | 3+4 2 | 3+3 2 | 2+2 2 |
| | | 5+6 2 | | 5+6 2 | 2+2 2 | 2+2 2 |
| | | 3+4 2 | | 5+6 2 | 3+3 2 | 2+2 2 |
| | | 5+6 2 | | 3+4 2 | 2+2 2 | 3+3 2 |
| | | 3+4 2 | | 3+4 2 | 2+2 2 | 3+3 2 |
| | | 5+6 2 | | 5+6 2 | 3+3 2 | 2+2 2 |
| | | 3+4 2 | | 3+4 2 | 2+2 2 | 2+2 2 |
| | | 5+6 2 | | 5+6 2 | 3+3 2 | 2+2 2 |
| 15.2.45 | 430120 | | 220 | Gd1Ni8Si3 | | |
| | -A | -GM | -M | -Y | -L | -V |
| | 2+2 4 | 3+4 2 | 2+2 4 | 3+4 2 | 2+2 2 | 2+2 2 |
| | 2+2 4 | 5+6 2 | 2+2 4 | 5+6 2 | 3+3 2 | 3+3 2 |
| | 2+2 4 | 3+4 2 | 2+2 4 | 3+4 2 | 2+2 2 | 3+3 2 |
| | 2+2 4 | 5+6 2 | 2+2 4 | 5+6 2 | 3+3 2 | 2+2 2 |
| | 2+2 4 | 3+4 2 | 2+2 4 | 3+4 2 | 2+2 2 | 3+3 2 |
| | 2+2 4 | 5+6 2 | 2+2 4 | 5+6 2 | 3+3 2 | 2+2 2 |
| | 2+2 4 | 3+4 2 | 2+2 4 | 5+6 2 | 2+2 2 | 2+2 2 |
| | 2+2 4 | 5+6 2 | 2+2 4 | 3+4 2 | 3+3 2 | 3+3 2 |
| | 2+2 4 | 3+4 2 | 2+2 4 | 3+4 2 | 3+3 2 | 2+2 2 |
| | 2+2 4 | 5+6 2 | 2+2 4 | 5+6 2 | 2+2 2 | 3+3 2 |
| | 2+2 4 | 3+4 2 | 2+2 4 | 5+6 2 | 3+3 2 | 3+3 2 |
| | 2+2 4 | 3+4 2 | 2+2 4 | 5+6 2 | 3+3 2 | 2+2 2 |
| | 2+2 4 | 3+4 2 | 2+2 4 | 3+4 2 | 2+2 2 | 3+3 2 |
| | 2+2 4 | 5+6 2 | 2+2 4 | 5+6 2 | 3+3 2 | 2+2 2 |
| | 2+2 4 | 5+6 2 | 2+2 4 | 3+4 2 | 2+2 2 | 2+2 2 |
| | 2+2 4 | 5+6 2 | 2+2 4 | 3+4 2 | 2+2 2 | 3+3 2 |
| | 2+2 4 | 3+4 2 | 2+2 4 | 3+4 2 | 3+3 2 | 2+2 2 |
| | 2+2 4 | 3+4 2 | 2+2 4 | 5+6 2 | 2+2 2 | 2+2 2 |
| | 2+2 4 | 3+4 2 | 2+2 4 | 3+4 2 | 2+2 2 | 3+3 2 |
| | 2+2 4 | 5+6 2 | 2+2 4 | 5+6 2 | 3+3 2 | 2+2 2 |
| | 2+2 4 | 5+6 2 | 2+2 4 | 3+4 2 | 3+3 2 | 3+3 2 |
| | 2+2 4 | 5+6 2 | 2+2 4 | 3+4 2 | 2+2 2 | 2+2 2 |
| | 2+2 4 | 3+4 2 | 2+2 4 | 3+4 2 | 3+3 2 | 3+3 2 |
| | 2+2 4 | 3+4 2 | 2+2 4 | 5+6 2 | 2+2 2 | 2+2 2 |
| | 2+2 4 | 5+6 2 | 2+2 4 | 3+4 2 | 3+3 2 | 3+3 2 |
| | 2+2 4 | 3+4 2 | 2+2 4 | 3+4 2 | 2+2 2 | 2+2 2 |
| | 2+2 4 | 5+6 2 | 2+2 4 | 5+6 2 | 2+2 2 | 3+3 2 |
| | 2+2 4 | 5+6 2 | 2+2 4 | 5+6 2 | 2+2 2 | 2+2 2 |
| | 2+2 4 | 5+6 2 | 2+2 4 | 3+4 2 | 3+3 2 | 3+3 2 |
| | 2+2 4 | 3+4 2 | 2+2 4 | 5+6 2 | 2+2 2 | 2+2 2 |
| | 2+2 4 | 5+6 2 | 2+2 4 | 3+4 2 | 2+2 2 | 2+2 2 |
| | 2+2 4 | 3+4 2 | 2+2 4 | 3+4 2 | 3+3 2 | 2+2 2 |
| | 2+2 4 | 3+4 2 | 2+2 4 | 5+6 2 | 3+3 2 | 3+3 2 |
| | 2+2 4 | 3+4 2 | 2+2 4 | 3+4 2 | 2+2 2 | 3+3 2 |
| | 2+2 4 | 5+6 2 | 2+2 4 | 3+4 2 | 2+2 2 | 2+2 2 |
| | 2+2 4 | 3+4 2 | 2+2 4 | 5+6 2 | 3+3 2 | 3+3 2 |
| | 2+2 4 | 3+4 2 | 2+2 4 | 5+6 2 | 3+3 2 | 2+2 2 |
| | 2+2 4 | 5+6 2 | 2+2 4 | 3+4 2 | 3+3 2 | 2+2 2 |
| | 2+2 4 | 3+4 2 | 2+2 4 | 3+4 2 | 2+2 2 | 3+3 2 |
| | 2+2 4 | 5+6 2 | 2+2 4 | 5+6 2 | 2+2 2 | 2+2 2 |
| | 2+2 4 | 3+4 2 | 2+2 4 | 3+4 2 | 3+3 2 | 2+2 2 |
| | 2+2 4 | 3+4 2 | 2+2 4 | 3+4 2 | 2+2 2 | 2+2 2 |
| | 2+2 4 | 3+4 2 | 2+2 4 | 3+4 2 | 3+3 2 | 3+3 2 |
| | 2+2 4 | 5+6 2 | 2+2 4 | 5+6 2 | 2+2 2 | 3+3 2 |
| | 2+2 4 | 3+4 2 | 2+2 4 | 3+4 2 | 3+3 2 | 2+2 2 |
| | 2+2 4 | 5+6 2 | 2+2 4 | 5+6 2 | 2+2 2 | 3+3 2 |
| | 2+2 4 | 3+4 2 | 2+2 4 | 5+6 2 | 3+3 2 | 3+3 2 |
| | 2+2 4 | 5+6 2 | 2+2 4 | 3+4 2 | 2+2 2 | 2+2 2 |
| | 2+2 4 | 3+4 2 | 2+2 4 | 5+6 2 | 2+2 2 | 3+3 2 |
| | 2+2 4 | 3+4 2 | 2+2 4 | 3+4 2 | 3+3 2 | 2+2 2 |
| | 2+2 4 | 5+6 2 | 2+2 4 | 3+4 2 | 3+3 2 | 3+3 2 |
| | 2+2 4 | 5+6 2 | 2+2 4 | 5+6 2 | 2+2 2 | 2+2 2 |
| | 2+2 4 | 3+4 2 | 2+2 4 | 5+6 2 | 3+3 2 | 2+2 2 |
| | 2+2 4 | 5+6 2 | 2+2 4 | 5+6 2 | 2+2 2 | 3+3 2 |
| | 2+2 4 | 3+4 2 | 2+2 4 | 3+4 2 | 3+3 2 | 2+2 2 |
| | | 5+6 2 | | 5+6 2 | 2+2 2 | 3+3 2 |
| | | 3+4 2 | | 3+4 2 | 2+2 2 | 3+3 2 |
| | | 3+4 2 | | 5+6 2 | 2+2 2 | 2+2 2 |
| | | 5+6 2 | | 3+4 2 | 3+3 2 | 3+3 2 |
| | | 3+4 2 | | 3+4 2 | 3+3 2 | 3+3 2 |
| | | 3+4 2 | | 5+6 2 | 2+2 2 | 2+2 2 |
| | | 5+6 2 | | 3+4 2 | 3+3 2 | 2+2 2 |
| | | 3+4 2 | | 3+4 2 | 3+3 2 | 3+3 2 |
| | | 5+6 2 | | 5+6 2 | 3+3 2 | 2+2 2 |
| | | 5+6 2 | | 3+4 2 | 2+2 2 | 3+3 2 |
| | | 3+4 2 | | 5+6 2 | 2+2 2 | 2+2 2 |
| | | 5+6 2 | | 3+4 2 | 3+3 2 | 2+2 2 |
| | | 3+4 2 | | 5+6 2 | 2+2 2 | 3+3 2 |
| | | 5+6 2 | | 3+4 2 | 3+3 2 | 2+2 2 |
| | | 3+4 2 | | 5+6 2 | 2+2 2 | 3+3 2 |
| | | 3+4 2 | | 5+6 2 | 3+3 2 | 3+3 2 |
| | | 5+6 2 | | 3+4 2 | 2+2 2 | 2+2 2 |

Continued on next column

Continued on next column

**TABLE XXVII– continued**

| TQC | ICSD | | NoE | SF | | |
|---|---|---|---|---|---|---|
| | | 5+6 2 | | 3+4 2 | 3+3 2 | 3+3 2 |
| | | 5+6 2 | | 5+6 2 | 3+3 2 | 2+2 2 |
| | | 3+4 2 | | 5+6 2 | 2+2 2 | 2+2 2 |
| | | 5+6 2 | | 3+4 2 | 3+3 2 | 3+3 2 |
| | | 3+4 2 | | 5+6 2 | 2+2 2 | 2+2 2 |
| | | 3+4 2 | | 3+4 2 | 3+3 2 | 3+3 2 |
| | | 5+6 2 | | 5+6 2 | 2+2 2 | 2+2 2 |
| | | 5+6 2 | | 3+4 2 | 3+3 2 | 3+3 2 |
| | | 5+6 2 | | 5+6 2 | 2+2 2 | 2+2 2 |
| | | 3+4 2 | | 3+4 2 | 2+2 2 | 3+3 2 |
| | | 3+4 2 | | 5+6 2 | 3+3 2 | 2+2 2 |
| | | 3+4 2 | | 3+4 2 | 2+2 2 | 3+3 2 |
| | | 5+6 2 | | 5+6 2 | 3+3 2 | 2+2 2 |
| | | 5+6 2 | | 3+4 2 | 3+3 2 | 3+3 2 |
| | | 3+4 2 | | 3+4 2 | 2+2 2 | 2+2 2 |
| | | 3+4 2 | | 3+4 2 | 2+2 2 | 3+3 2 |
| | | 5+6 2 | | 5+6 2 | 3+3 2 | 2+2 2 |
| | | 3+4 2 | | 3+4 2 | 3+3 2 | 3+3 2 |
| | | 3+4 2 | | 5+6 2 | 2+2 2 | 2+2 2 |
| | | 5+6 2 | | 5+6 2 | 2+2 2 | 3+3 2 |
| | | 5+6 2 | | 3+4 2 | 3+3 2 | 2+2 2 |
| | | 3+4 2 | | 3+4 2 | 3+3 2 | 2+2 2 |
| | | 3+4 2 | | 5+6 2 | 2+2 2 | 2+2 2 |
| | | 3+4 2 | | 3+4 2 | 3+3 2 | 3+3 2 |
| | | 3+4 2 | | 5+6 2 | 3+3 2 | 3+3 2 |
| | | 5+6 2 | | 3+4 2 | 2+2 2 | 2+2 2 |
| | | 3+4 2 | | 3+4 2 | 3+3 2 | 2+2 2 |
| | | 5+6 2 | | 5+6 2 | 2+2 2 | 3+3 2 |
| | | 5+6 2 | | 3+4 2 | 3+3 2 | 2+2 2 |
| | | 3+4 2 | | 5+6 2 | 2+2 2 | 2+2 2 |
| | | 3+4 2 | | 3+4 2 | 3+3 2 | 3+3 2 |
| | | 3+4 2 | | 5+6 2 | 2+2 2 | 2+2 2 |
| | | 3+4 2 | | 5+6 2 | 3+3 2 | 3+3 2 |
| | | 5+6 2 | | 3+4 2 | 2+2 2 | 2+2 2 |
| | | 3+4 2 | | 5+6 2 | 3+3 2 | 3+3 2 |
| | | 3+4 2 | | 3+4 2 | 2+2 2 | 2+2 2 |
| | | 5+6 2 | | 5+6 2 | 2+2 2 | 3+3 2 |
| | | 3+4 2 | | 5+6 2 | 3+3 2 | 3+3 2 |
| 15.2.46 | 151766 | | 208 | Fe7S8 | | |
| | -A | -GM | -M | -Y | -L | -V |
| | 2+2 4 | 3+4 2 | 2+2 4 | 3+4 2 | 2+2 2 | 3+3 2 |
| | 2+2 4 | 3+4 2 | 2+2 4 | 3+4 2 | 3+3 2 | 2+2 2 |
| | 2+2 4 | 5+6 2 | 2+2 4 | 5+6 2 | 2+2 2 | 2+2 2 |
| | 2+2 4 | 5+6 2 | 2+2 4 | 5+6 2 | 3+3 2 | 3+3 2 |
| | 2+2 4 | 3+4 2 | 2+2 4 | 5+6 2 | 3+3 2 | 2+2 2 |
| | 2+2 4 | 5+6 2 | 2+2 4 | 3+4 2 | 2+2 2 | 3+3 2 |
| | 2+2 4 | 3+4 2 | 2+2 4 | 5+6 2 | 2+2 2 | 3+3 2 |
| | 2+2 4 | 5+6 2 | 2+2 4 | 5+6 2 | 3+3 2 | 2+2 2 |
| | 2+2 4 | 3+4 2 | 2+2 4 | 3+4 2 | 2+2 2 | 2+2 2 |
| | 2+2 4 | 3+4 2 | 2+2 4 | 3+4 2 | 3+3 2 | 3+3 2 |
| | 2+2 4 | 5+6 2 | 2+2 4 | 5+6 2 | 3+3 2 | 2+2 2 |
| | 2+2 4 | 3+4 2 | 2+2 4 | 3+4 2 | 2+2 2 | 3+3 2 |
| | 2+2 4 | 5+6 2 | 2+2 4 | 3+4 2 | 3+3 2 | 3+3 2 |
| | 2+2 4 | 5+6 2 | 2+2 4 | 5+6 2 | 2+2 2 | 2+2 2 |
| | 2+2 4 | 3+4 2 | 2+2 4 | 3+4 2 | 3+3 2 | 2+2 2 |
| | 2+2 4 | 5+6 2 | 2+2 4 | 5+6 2 | 2+2 2 | 3+3 2 |
| | 2+2 4 | 3+4 2 | 2+2 4 | 3+4 2 | 3+3 2 | 2+2 2 |
| | 2+2 4 | 5+6 2 | 2+2 4 | 5+6 2 | 2+2 2 | 3+3 2 |
| | 2+2 4 | 3+4 2 | 2+2 4 | 3+4 2 | 3+3 2 | 3+3 2 |
| | 2+2 4 | 5+6 2 | 2+2 4 | 3+4 2 | 2+2 2 | 2+2 2 |
| | 2+2 4 | 3+4 2 | 2+2 4 | 5+6 2 | 3+3 2 | 2+2 2 |

Continued on next column

**TABLE XXVII– continued**

| TQC | ICSD | | NoE | SF | | |
|---|---|---|---|---|---|---|
| | 2+2 4 | 5+6 2 | 2+2 4 | 5+6 2 | 2+2 2 | 3+3 2 |
| | 2+2 4 | 5+6 2 | 2+2 4 | 3+4 2 | 2+2 2 | 3+3 2 |
| | 2+2 4 | 3+4 2 | 2+2 4 | 5+6 2 | 3+3 2 | 2+2 2 |
| | 2+2 4 | 3+4 2 | 2+2 4 | 5+6 2 | 3+3 2 | 2+2 2 |
| | 2+2 4 | 5+6 2 | 2+2 4 | 3+4 2 | 2+2 2 | 3+3 2 |
| | 2+2 4 | 5+6 2 | 2+2 4 | 5+6 2 | 2+2 2 | 2+2 2 |
| | 2+2 4 | 3+4 2 | 2+2 4 | 3+4 2 | 3+3 2 | 3+3 2 |
| | 2+2 4 | 3+4 2 | 2+2 4 | 5+6 2 | 3+3 2 | 2+2 2 |
| | 2+2 4 | 5+6 2 | 2+2 4 | 5+6 2 | 2+2 2 | 3+3 2 |
| | 2+2 4 | 3+4 2 | 2+2 4 | 3+4 2 | 3+3 2 | 2+2 2 |
| | 2+2 4 | 3+4 2 | 2+2 4 | 5+6 2 | 2+2 2 | 3+3 2 |
| | 2+2 4 | 5+6 2 | 2+2 4 | 3+4 2 | 3+3 2 | 2+2 2 |
| | 2+2 4 | 3+4 2 | 2+2 4 | 3+4 2 | 2+2 2 | 3+3 2 |
| | 2+2 4 | 3+4 2 | 2+2 4 | 5+6 2 | 3+3 2 | 2+2 2 |
| | 2+2 4 | 5+6 2 | 2+2 4 | 3+4 2 | 2+2 2 | 3+3 2 |
| | 2+2 4 | 5+6 2 | 2+2 4 | 5+6 2 | 3+3 2 | 2+2 2 |
| | 2+2 4 | 5+6 2 | 2+2 4 | 3+4 2 | 2+2 2 | 3+3 2 |
| | 2+2 4 | 3+4 2 | 2+2 4 | 5+6 2 | 2+2 2 | 2+2 2 |
| | 2+2 4 | 3+4 2 | 2+2 4 | 5+6 2 | 3+3 2 | 3+3 2 |
| | 2+2 4 | 5+6 2 | 2+2 4 | 3+4 2 | 2+2 2 | 2+2 2 |
| | 2+2 4 | 3+4 2 | 2+2 4 | 3+4 2 | 3+3 2 | 3+3 2 |
| | 2+2 4 | 5+6 2 | 2+2 4 | 5+6 2 | 2+2 2 | 3+3 2 |
| | 2+2 4 | 5+6 2 | 2+2 4 | 3+4 2 | 3+3 2 | 2+2 2 |
| | 2+2 4 | 3+4 2 | 2+2 4 | 3+4 2 | 2+2 2 | 3+3 2 |
| | 2+2 4 | 3+4 2 | 2+2 4 | 5+6 2 | 2+2 2 | 2+2 2 |
| | 2+2 4 | 5+6 2 | 2+2 4 | 5+6 2 | 3+3 2 | 3+3 2 |
| | 2+2 4 | 3+4 2 | 2+2 4 | 3+4 2 | 3+3 2 | 2+2 2 |
| | 2+2 4 | 5+6 2 | 2+2 4 | 3+4 2 | 2+2 2 | 3+3 2 |
| | 2+2 4 | 5+6 2 | 2+2 4 | 5+6 2 | 2+2 2 | 3+3 2 |
| | 2+2 4 | 3+4 2 | 2+2 4 | 5+6 2 | 2+2 2 | 2+2 2 |
| | 2+2 4 | 5+6 2 | 2+2 4 | 3+4 2 | 3+3 2 | 3+3 2 |
| | | 3+4 2 | | 5+6 2 | 3+3 2 | 2+2 2 |
| | | 5+6 2 | | 3+4 2 | 3+3 2 | 2+2 2 |
| | | 3+4 2 | | 5+6 2 | 2+2 2 | 2+2 2 |
| | | 3+4 2 | | 5+6 2 | 3+3 2 | 2+2 2 |
| | | 3+4 2 | | 3+4 2 | 3+3 2 | 3+3 2 |
| | | 3+4 2 | | 3+4 2 | 2+2 2 | 3+3 2 |
| | | 3+4 2 | | 5+6 2 | 2+2 2 | 2+2 2 |
| | | 5+6 2 | | 3+4 2 | 3+3 2 | 3+3 2 |
| | | 5+6 2 | | 3+4 2 | 3+3 2 | 3+3 2 |
| | | 3+4 2 | | 5+6 2 | 2+2 2 | 2+2 2 |
| | | 5+6 2 | | 3+4 2 | 3+3 2 | 2+2 2 |
| | | 5+6 2 | | 5+6 2 | 2+2 2 | 3+3 2 |
| | | 3+4 2 | | 5+6 2 | 2+2 2 | 3+3 2 |
| | | 5+6 2 | | 3+4 2 | 3+3 2 | 2+2 2 |
| | | 3+4 2 | | 5+6 2 | 2+2 2 | 2+2 2 |
| | | 5+6 2 | | 3+4 2 | 3+3 2 | 3+3 2 |
| | | 5+6 2 | | 5+6 2 | 3+3 2 | 2+2 2 |
| | | 5+6 2 | | 3+4 2 | 2+2 2 | 3+3 2 |
| | | 5+6 2 | | 3+4 2 | 3+3 2 | 2+2 2 |
| | | 3+4 2 | | 5+6 2 | 2+2 2 | 2+2 2 |
| | | 3+4 2 | | 5+6 2 | 3+3 2 | 3+3 2 |
| | | 5+6 2 | | 3+4 2 | 2+2 2 | 3+3 2 |
| | | 3+4 2 | | 3+4 2 | 3+3 2 | 2+2 2 |
| | | 5+6 2 | | 5+6 2 | 3+3 2 | 3+3 2 |
| | | 3+4 2 | | 3+4 2 | 2+2 2 | 3+3 2 |
| | | 3+4 2 | | 3+4 2 | 3+3 2 | 2+2 2 |
| | | 3+4 2 | | 3+4 2 | 2+2 2 | 3+3 2 |
| | | 3+4 2 | | 5+6 2 | 3+3 2 | 2+2 2 |
| | | 5+6 2 | | 5+6 2 | 2+2 2 | 3+3 2 |
| | | 3+4 2 | | 5+6 2 | 2+2 2 | 2+2 2 |

Continued on next column

**TABLE XXVII– continued**

| TQC | ICSD | | NoE | SF | | |
|---|---|---|---|---|---|---|
| | | 3+4 2 | | 5+6 2 | 3+3 2 | 2+2 2 |
| | | 5+6 2 | | 3+4 2 | 2+2 2 | 3+3 2 |
| | | 3+4 2 | | 3+4 2 | 3+3 2 | 2+2 2 |
| | | 5+6 2 | | 5+6 2 | 2+2 2 | 3+3 2 |
| | | 5+6 2 | | 3+4 2 | 3+3 2 | 2+2 2 |
| | | 3+4 2 | | 3+4 2 | 3+3 2 | 3+3 2 |
| | | 3+4 2 | | 5+6 2 | 2+2 2 | 3+3 2 |
| | | 5+6 2 | | 3+4 2 | 2+2 2 | 2+2 2 |
| | | 5+6 2 | | 5+6 2 | 3+3 2 | 3+3 2 |
| | | 5+6 2 | | 5+6 2 | 2+2 2 | 3+3 2 |
| | | 3+4 2 | | 3+4 2 | 3+3 2 | 2+2 2 |
| | | 5+6 2 | | 3+4 2 | 2+2 2 | 3+3 2 |
| | | 3+4 2 | | 5+6 2 | 2+2 2 | 2+2 2 |
| | | 5+6 2 | | 3+4 2 | 3+3 2 | 3+3 2 |
| | | 5+6 2 | | 5+6 2 | 3+3 2 | 2+2 2 |
| | | 3+4 2 | | 5+6 2 | 2+2 2 | 3+3 2 |
| | | 5+6 2 | | 3+4 2 | 3+3 2 | 2+2 2 |
| | | 5+6 2 | | 3+4 2 | 2+2 2 | 3+3 2 |
| | | 3+4 2 | | 5+6 2 | 3+3 2 | 2+2 2 |
| | | 5+6 2 | | 3+4 2 | 2+2 2 | 3+3 2 |
| | | 3+4 2 | | 5+6 2 | 3+3 2 | 2+2 2 |
| | | 5+6 2 | | 5+6 2 | 2+2 2 | 3+3 2 |
| 15.2.47 | 24416 | | 100 | Fe2O5Ti1 | | |
| | -A | -GM | -M | -Y | -L | -V |
| | 2+2 4 | 3+4 2 | 2+2 4 | 3+4 2 | 2+2 2 | 3+3 2 |
| | 2+2 4 | 5+6 2 | 2+2 4 | 3+4 2 | 3+3 2 | 2+2 2 |
| | 2+2 4 | 3+4 2 | 2+2 4 | 5+6 2 | 3+3 2 | 2+2 2 |
| | 2+2 4 | 3+4 2 | 2+2 4 | 5+6 2 | 2+2 2 | 2+2 2 |
| | 2+2 4 | 3+4 2 | 2+2 4 | 3+4 2 | 2+2 2 | 3+3 2 |
| | 2+2 4 | 3+4 2 | 2+2 4 | 3+4 2 | 3+3 2 | 2+2 2 |
| | 2+2 4 | 5+6 2 | 2+2 4 | 5+6 2 | 3+3 2 | 3+3 2 |
| | 2+2 4 | 5+6 2 | 2+2 4 | 3+4 2 | 2+2 2 | 2+2 2 |
| | 2+2 4 | 5+6 2 | 2+2 4 | 5+6 2 | 3+3 2 | 3+3 2 |
| | 2+2 4 | 5+6 2 | 2+2 4 | 5+6 2 | 2+2 2 | 3+3 2 |
| | 2+2 4 | 3+4 2 | 2+2 4 | 3+4 2 | 2+2 2 | 2+2 2 |
| | 2+2 4 | 3+4 2 | 2+2 4 | 3+4 2 | 3+3 2 | 2+2 2 |
| | 2+2 4 | 5+6 2 | 2+2 4 | 5+6 2 | 2+2 2 | 2+2 2 |
| | 2+2 4 | 5+6 2 | 2+2 4 | 3+4 2 | 3+3 2 | 3+3 2 |
| | 2+2 4 | 3+4 2 | 2+2 4 | 3+4 2 | 2+2 2 | 2+2 2 |
| | 2+2 4 | 3+4 2 | 2+2 4 | 5+6 2 | 3+3 2 | 3+3 2 |
| | 2+2 4 | 3+4 2 | 2+2 4 | 5+6 2 | 2+2 2 | 3+3 2 |
| | 2+2 4 | 5+6 2 | 2+2 4 | 3+4 2 | 3+3 2 | 2+2 2 |
| | 2+2 4 | 3+4 2 | 2+2 4 | 3+4 2 | 3+3 2 | 3+3 2 |
| | 2+2 4 | 3+4 2 | 2+2 4 | 5+6 2 | 2+2 2 | 3+3 2 |
| | 2+2 4 | 5+6 2 | 2+2 4 | 3+4 2 | 2+2 2 | 2+2 2 |
| | 2+2 4 | 3+4 2 | 2+2 4 | 5+6 2 | 3+3 2 | 2+2 2 |
| | 2+2 4 | 5+6 2 | 2+2 4 | 3+4 2 | 2+2 2 | 3+3 2 |
| | 2+2 4 | 5+6 2 | 2+2 4 | 5+6 2 | 3+3 2 | 2+2 2 |
| | 2+2 4 | 3+4 2 | 2+2 4 | 3+4 2 | 3+3 2 | 2+2 2 |
| | | 3+4 2 | | 3+4 2 | 3+3 2 | 3+3 2 |
| | | 3+4 2 | | 5+6 2 | 2+2 2 | 2+2 2 |
| | | 3+4 2 | | 5+6 2 | 2+2 2 | 3+3 2 |
| | | 5+6 2 | | 3+4 2 | 2+2 2 | 3+3 2 |
| | | 5+6 2 | | 5+6 2 | 2+2 2 | 3+3 2 |
| | | 3+4 2 | | 3+4 2 | 3+3 2 | 2+2 2 |
| | | 5+6 2 | | 3+4 2 | 3+3 2 | 3+3 2 |
| | | 5+6 2 | | 5+6 2 | 3+3 2 | 2+2 2 |
| | | 3+4 2 | | 3+4 2 | 2+2 2 | 3+3 2 |
| | | 5+6 2 | | 5+6 2 | 3+3 2 | 2+2 2 |
| | | 5+6 2 | | 3+4 2 | 3+3 2 | 3+3 2 |
| | | 5+6 2 | | 5+6 2 | 2+2 2 | 2+2 2 |

Continued on next column

**TABLE XXVII– continued**

| TQC | ICSD | | NoE | SF | | |
|---|---|---|---|---|---|---|
| | | 3+4 2 | | 5+6 2 | 2+2 2 | 3+3 2 |
| | | 5+6 2 | | 5+6 2 | 3+3 2 | 2+2 2 |
| | | 5+6 2 | | 5+6 2 | 2+2 2 | 3+3 2 |
| | | 5+6 2 | | 3+4 2 | 2+2 2 | 2+2 2 |
| | | 3+4 2 | | 5+6 2 | 2+2 2 | 2+2 2 |
| | | 3+4 2 | | 5+6 2 | 3+3 2 | 3+3 2 |
| | | 3+4 2 | | 3+4 2 | 3+3 2 | 2+2 2 |
| | | 5+6 2 | | 3+4 2 | 3+3 2 | 3+3 2 |
| | | 3+4 2 | | 5+6 2 | 2+2 2 | 2+2 2 |
| | | 5+6 2 | | 3+4 2 | 3+3 2 | 3+3 2 |
| | | 3+4 2 | | 5+6 2 | 2+2 2 | 2+2 2 |
| | | 5+6 2 | | 3+4 2 | 3+3 2 | 2+2 2 |
| | | 3+4 2 | | 5+6 2 | 2+2 2 | 3+3 2 |
| 15.2.48 | 154291 | | 164 | Fe2O5Rb4 | | |
| | -A | -GM | -M | -Y | -L | -V |
| | 2+2 4 | 3+4 2 | 2+2 4 | 3+4 2 | 2+2 2 | 2+2 2 |
| | 2+2 4 | 3+4 2 | 2+2 4 | 5+6 2 | 2+2 2 | 3+3 2 |
| | 2+2 4 | 5+6 2 | 2+2 4 | 3+4 2 | 3+3 2 | 2+2 2 |
| | 2+2 4 | 5+6 2 | 2+2 4 | 5+6 2 | 3+3 2 | 3+3 2 |
| | 2+2 4 | 3+4 2 | 2+2 4 | 5+6 2 | 2+2 2 | 2+2 2 |
| | 2+2 4 | 5+6 2 | 2+2 4 | 3+4 2 | 3+3 2 | 3+3 2 |
| | 2+2 4 | 3+4 2 | 2+2 4 | 3+4 2 | 2+2 2 | 3+3 2 |
| | 2+2 4 | 5+6 2 | 2+2 4 | 5+6 2 | 3+3 2 | 2+2 2 |
| | 2+2 4 | 3+4 2 | 2+2 4 | 5+6 2 | 3+3 2 | 2+2 2 |
| | 2+2 4 | 3+4 2 | 2+2 4 | 5+6 2 | 3+3 2 | 3+3 2 |
| | 2+2 4 | 5+6 2 | 2+2 4 | 3+4 2 | 2+2 2 | 3+3 2 |
| | 2+2 4 | 3+4 2 | 2+2 4 | 5+6 2 | 3+3 2 | 2+2 2 |
| | 2+2 4 | 5+6 2 | 2+2 4 | 3+4 2 | 3+3 2 | 2+2 2 |
| | 2+2 4 | 3+4 2 | 2+2 4 | 5+6 2 | 2+2 2 | 3+3 2 |
| | 2+2 4 | 5+6 2 | 2+2 4 | 3+4 2 | 2+2 2 | 2+2 2 |
| | 2+2 4 | 3+4 2 | 2+2 4 | 5+6 2 | 2+2 2 | 3+3 2 |
| | 2+2 4 | 5+6 2 | 2+2 4 | 5+6 2 | 3+3 2 | 3+3 2 |
| | 2+2 4 | 3+4 2 | 2+2 4 | 3+4 2 | 3+3 2 | 2+2 2 |
| | 2+2 4 | 3+4 2 | 2+2 4 | 5+6 2 | 2+2 2 | 2+2 2 |
| | 2+2 4 | 3+4 2 | 2+2 4 | 3+4 2 | 3+3 2 | 3+3 2 |
| | 2+2 4 | 5+6 2 | 2+2 4 | 3+4 2 | 2+2 2 | 2+2 2 |
| | 2+2 4 | 5+6 2 | 2+2 4 | 5+6 2 | 3+3 2 | 2+2 2 |
| | 2+2 4 | 3+4 2 | 2+2 4 | 5+6 2 | 2+2 2 | 3+3 2 |
| | 2+2 4 | 5+6 2 | 2+2 4 | 3+4 2 | 3+3 2 | 3+3 2 |
| | 2+2 4 | 3+4 2 | 2+2 4 | 3+4 2 | 2+2 2 | 3+3 2 |
| | 2+2 4 | 5+6 2 | 2+2 4 | 5+6 2 | 3+3 2 | 2+2 2 |
| | 2+2 4 | 3+4 2 | 2+2 4 | 3+4 2 | 2+2 2 | 2+2 2 |
| | 2+2 4 | 5+6 2 | 2+2 4 | 5+6 2 | 2+2 2 | 3+3 2 |
| | 2+2 4 | 3+4 2 | 2+2 4 | 3+4 2 | 3+3 2 | 2+2 2 |
| | 2+2 4 | 3+4 2 | 2+2 4 | 5+6 2 | 3+3 2 | 2+2 2 |
| | 2+2 4 | 5+6 2 | 2+2 4 | 3+4 2 | 2+2 2 | 3+3 2 |
| | 2+2 4 | 3+4 2 | 2+2 4 | 5+6 2 | 3+3 2 | 3+3 2 |
| | 2+2 4 | 5+6 2 | 2+2 4 | 3+4 2 | 2+2 2 | 2+2 2 |
| | 2+2 4 | 3+4 2 | 2+2 4 | 5+6 2 | 2+2 2 | 3+3 2 |
| | 2+2 4 | 5+6 2 | 2+2 4 | 3+4 2 | 3+3 2 | 2+2 2 |
| | 2+2 4 | 3+4 2 | 2+2 4 | 3+4 2 | 3+3 2 | 3+3 2 |
| | 2+2 4 | 5+6 2 | 2+2 4 | 5+6 2 | 2+2 2 | 3+3 2 |
| | 2+2 4 | 3+4 2 | 2+2 4 | 5+6 2 | 3+3 2 | 3+3 2 |
| | 2+2 4 | 5+6 2 | 2+2 4 | 3+4 2 | 2+2 2 | 2+2 2 |
| | 2+2 4 | 3+4 2 | 2+2 4 | 5+6 2 | 3+3 2 | 2+2 2 |
| | 2+2 4 | 5+6 2 | 2+2 4 | 5+6 2 | 2+2 2 | 3+3 2 |
| | | 5+6 2 | | 3+4 2 | 3+3 2 | 2+2 2 |
| | | 5+6 2 | | 3+4 2 | 2+2 2 | 3+3 2 |
| | | 5+6 2 | | 3+4 2 | 2+2 2 | 2+2 2 |
| | | 3+4 2 | | 3+4 2 | 3+3 2 | 3+3 2 |
| | | 5+6 2 | | 5+6 2 | 2+2 2 | 2+2 2 |

Continued on next column

**TABLE XXVII– continued**

| TQC | ICSD | | NoE | SF | | |
|---|---|---|---|---|---|---|
| | | 3+4 2 | | 5+6 2 | 3+3 2 | 2+2 2 |
| | | 5+6 2 | | 3+4 2 | 2+2 2 | 3+3 2 |
| | | 5+6 2 | | 3+4 2 | 2+2 2 | 3+3 2 |
| | | 5+6 2 | | 3+4 2 | 2+2 2 | 2+2 2 |
| | | 3+4 2 | | 5+6 2 | 2+2 2 | 2+2 2 |
| | | 5+6 2 | | 3+4 2 | 3+3 2 | 3+3 2 |
| | | 3+4 2 | | 5+6 2 | 3+3 2 | 3+3 2 |
| | | 5+6 2 | | 3+4 2 | 2+2 2 | 2+2 2 |
| | | 5+6 2 | | 5+6 2 | 2+2 2 | 3+3 2 |
| | | 3+4 2 | | 3+4 2 | 3+3 2 | 2+2 2 |
| | | 5+6 2 | | 3+4 2 | 2+2 2 | 2+2 2 |
| | | 5+6 2 | | 3+4 2 | 3+3 2 | 3+3 2 |
| | | 3+4 2 | | 5+6 2 | 2+2 2 | 3+3 2 |
| | | 5+6 2 | | 5+6 2 | 3+3 2 | 2+2 2 |
| | | 3+4 2 | | 3+4 2 | 2+2 2 | 2+2 2 |
| | | 5+6 2 | | 3+4 2 | 3+3 2 | 3+3 2 |
| | | 3+4 2 | | 5+6 2 | 2+2 2 | 3+3 2 |
| | | 3+4 2 | | 3+4 2 | 2+2 2 | 2+2 2 |
| | | 5+6 2 | | 5+6 2 | 3+3 2 | 2+2 2 |
| | | 5+6 2 | | 5+6 2 | 3+3 2 | 3+3 2 |
| | | 3+4 2 | | 3+4 2 | 2+2 2 | 3+3 2 |
| | | 3+4 2 | | 5+6 2 | 3+3 2 | 2+2 2 |
| | | 5+6 2 | | 3+4 2 | 2+2 2 | 3+3 2 |
| | | 5+6 2 | | 5+6 2 | 3+3 2 | 3+3 2 |
| | | 5+6 2 | | 3+4 2 | 2+2 2 | 2+2 2 |
| | | 3+4 2 | | 3+4 2 | 2+2 2 | 2+2 2 |
| | | 3+4 2 | | 5+6 2 | 3+3 2 | 2+2 2 |
| | | 5+6 2 | | 3+4 2 | 2+2 2 | 3+3 2 |
| | | 3+4 2 | | 5+6 2 | 3+3 2 | 2+2 2 |
| | | 3+4 2 | | 5+6 2 | 3+3 2 | 3+3 2 |
| | | 3+4 2 | | 5+6 2 | 3+3 2 | 3+3 2 |
| | | 5+6 2 | | 3+4 2 | 3+3 2 | 2+2 2 |
| | | 3+4 2 | | 5+6 2 | 2+2 2 | 3+3 2 |
| | | 3+4 2 | | 5+6 2 | 3+3 2 | 2+2 2 |
| | | 3+4 2 | | 5+6 2 | 2+2 2 | 2+2 2 |
| | | 3+4 2 | | 3+4 2 | 3+3 2 | 3+3 2 |
| 15.2.49 | 71059 | | 104 | Fe2Li1N3Sr2 | | |
| | -A | -GM | -M | -Y | -L | -V |
| | 2+2 4 | 3+4 2 | 2+2 4 | 3+4 2 | 2+2 2 | 2+2 2 |
| | 2+2 4 | 5+6 2 | 2+2 4 | 5+6 2 | 3+3 2 | 3+3 2 |
| | 2+2 4 | 3+4 2 | 2+2 4 | 5+6 2 | 2+2 2 | 3+3 2 |
| | 2+2 4 | 5+6 2 | 2+2 4 | 3+4 2 | 3+3 2 | 2+2 2 |
| | 2+2 4 | 3+4 2 | 2+2 4 | 3+4 2 | 2+2 2 | 2+2 2 |
| | 2+2 4 | 5+6 2 | 2+2 4 | 5+6 2 | 3+3 2 | 3+3 2 |
| | 2+2 4 | 3+4 2 | 2+2 4 | 3+4 2 | 2+2 2 | 2+2 2 |
| | 2+2 4 | 5+6 2 | 2+2 4 | 5+6 2 | 3+3 2 | 3+3 2 |
| | 2+2 4 | 3+4 2 | 2+2 4 | 3+4 2 | 2+2 2 | 2+2 2 |
| | 2+2 4 | 5+6 2 | 2+2 4 | 5+6 2 | 3+3 2 | 3+3 2 |
| | 2+2 4 | 3+4 2 | 2+2 4 | 3+4 2 | 2+2 2 | 3+3 2 |
| | 2+2 4 | 5+6 2 | 2+2 4 | 5+6 2 | 3+3 2 | 2+2 2 |
| | 2+2 4 | 3+4 2 | 2+2 4 | 3+4 2 | 3+3 2 | 3+3 2 |
| | 2+2 4 | 5+6 2 | 2+2 4 | 5+6 2 | 2+2 2 | 2+2 2 |
| | 2+2 4 | 3+4 2 | 2+2 4 | 3+4 2 | 3+3 2 | 3+3 2 |
| | 2+2 4 | 5+6 2 | 2+2 4 | 5+6 2 | 2+2 2 | 2+2 2 |
| | 2+2 4 | 3+4 2 | 2+2 4 | 5+6 2 | 2+2 2 | 2+2 2 |
| | 2+2 4 | 3+4 2 | 2+2 4 | 5+6 2 | 2+2 2 | 3+3 2 |
| | 2+2 4 | 5+6 2 | 2+2 4 | 3+4 2 | 3+3 2 | 3+3 2 |
| | 2+2 4 | 5+6 2 | 2+2 4 | 3+4 2 | 3+3 2 | 2+2 2 |
| | 2+2 4 | 5+6 2 | 2+2 4 | 3+4 2 | 2+2 2 | 3+3 2 |
| | 2+2 4 | 3+4 2 | 2+2 4 | 5+6 2 | 3+3 2 | 2+2 2 |
| | 2+2 4 | 3+4 2 | 2+2 4 | 3+4 2 | 2+2 2 | 2+2 2 |
| | 2+2 4 | 5+6 2 | 2+2 4 | 5+6 2 | 3+3 2 | 3+3 2 |
| | 2+2 4 | 3+4 2 | 2+2 4 | 5+6 2 | 2+2 2 | 2+2 2 |
| | 2+2 4 | 5+6 2 | 2+2 4 | 3+4 2 | 3+3 2 | 3+3 2 |
| | | 5+6 2 | | 3+4 2 | 3+3 2 | 2+2 2 |
| | | 5+6 2 | | 3+4 2 | 2+2 2 | 3+3 2 |
| | | 5+6 2 | | 5+6 2 | 3+3 2 | 3+3 2 |
| | | 5+6 2 | | 3+4 2 | 3+3 2 | 2+2 2 |
| | | 5+6 2 | | 3+4 2 | 3+3 2 | 2+2 2 |
| | | 3+4 2 | | 3+4 2 | 2+2 2 | 3+3 2 |
| | | 3+4 2 | | 5+6 2 | 3+3 2 | 2+2 2 |
| | | 3+4 2 | | 5+6 2 | 3+3 2 | 3+3 2 |
| | | 3+4 2 | | 5+6 2 | 2+2 2 | 3+3 2 |
| | | 3+4 2 | | 5+6 2 | 2+2 2 | 2+2 2 |
| | | 5+6 2 | | 3+4 2 | 2+2 2 | 3+3 2 |
| | | 3+4 2 | | 5+6 2 | 2+2 2 | 2+2 2 |
| | | 3+4 2 | | 5+6 2 | 2+2 2 | 2+2 2 |
| | | 5+6 2 | | 3+4 2 | 3+3 2 | 3+3 2 |
| | | 3+4 2 | | 5+6 2 | 2+2 2 | 2+2 2 |
| | | 3+4 2 | | 5+6 2 | 3+3 2 | 3+3 2 |
| | | 3+4 2 | | 5+6 2 | 2+2 2 | 2+2 2 |
| | | 3+4 2 | | 3+4 2 | 2+2 2 | 3+3 2 |
| | | 5+6 2 | | 3+4 2 | 3+3 2 | 3+3 2 |
| | | 5+6 2 | | 5+6 2 | 2+2 2 | 2+2 2 |
| | | 5+6 2 | | 3+4 2 | 3+3 2 | 3+3 2 |
| | | 5+6 2 | | 3+4 2 | 2+2 2 | 2+2 2 |
| | | 3+4 2 | | 5+6 2 | 3+3 2 | 2+2 2 |
| | | 5+6 2 | | 3+4 2 | 3+3 2 | 3+3 2 |
| | | 3+4 2 | | 5+6 2 | 2+2 2 | 2+2 2 |
| | | 5+6 2 | | 5+6 2 | 2+2 2 | 3+3 2 |
| 15.2.50 | 41899 | | 132 | Fe1S5U2 | | |
| | -A | -GM | -M | -Y | -L | -V |
| | 2+2 4 | 3+4 2 | 2+2 4 | 3+4 2 | 2+2 2 | 2+2 2 |
| | 2+2 4 | 3+4 2 | 2+2 4 | 5+6 2 | 3+3 2 | 3+3 2 |
| | 2+2 4 | 5+6 2 | 2+2 4 | 3+4 2 | 2+2 2 | 2+2 2 |
| | 2+2 4 | 5+6 2 | 2+2 4 | 5+6 2 | 3+3 2 | 3+3 2 |
| | 2+2 4 | 3+4 2 | 2+2 4 | 5+6 2 | 3+3 2 | 2+2 2 |
| | 2+2 4 | 5+6 2 | 2+2 4 | 3+4 2 | 2+2 2 | 3+3 2 |
| | 2+2 4 | 3+4 2 | 2+2 4 | 3+4 2 | 3+3 2 | 2+2 2 |
| | 2+2 4 | 5+6 2 | 2+2 4 | 5+6 2 | 2+2 2 | 3+3 2 |
| | 2+2 4 | 5+6 2 | 2+2 4 | 5+6 2 | 3+3 2 | 3+3 2 |
| | 2+2 4 | 3+4 2 | 2+2 4 | 3+4 2 | 2+2 2 | 2+2 2 |
| | 2+2 4 | 3+4 2 | 2+2 4 | 3+4 2 | 3+3 2 | 2+2 2 |
| | 2+2 4 | 3+4 2 | 2+2 4 | 5+6 2 | 2+2 2 | 3+3 2 |
| | 2+2 4 | 5+6 2 | 2+2 4 | 3+4 2 | 3+3 2 | 2+2 2 |
| | 2+2 4 | 3+4 2 | 2+2 4 | 5+6 2 | 2+2 2 | 3+3 2 |
| | 2+2 4 | 5+6 2 | 2+2 4 | 3+4 2 | 3+3 2 | 2+2 2 |
| | 2+2 4 | 5+6 2 | 2+2 4 | 5+6 2 | 2+2 2 | 3+3 2 |
| | 2+2 4 | 3+4 2 | 2+2 4 | 5+6 2 | 3+3 2 | 2+2 2 |
| | 2+2 4 | 5+6 2 | 2+2 4 | 3+4 2 | 3+3 2 | 3+3 2 |
| | 2+2 4 | 3+4 2 | 2+2 4 | 3+4 2 | 2+2 2 | 3+3 2 |
| | 2+2 4 | 5+6 2 | 2+2 4 | 5+6 2 | 2+2 2 | 2+2 2 |
| | 2+2 4 | 5+6 2 | 2+2 4 | 5+6 2 | 2+2 2 | 3+3 2 |
| | 2+2 4 | 3+4 2 | 2+2 4 | 3+4 2 | 3+3 2 | 2+2 2 |
| | 2+2 4 | 5+6 2 | 2+2 4 | 3+4 2 | 2+2 2 | 3+3 2 |
| | 2+2 4 | 3+4 2 | 2+2 4 | 5+6 2 | 3+3 2 | 3+3 2 |
| | 2+2 4 | 3+4 2 | 2+2 4 | 5+6 2 | 2+2 2 | 2+2 2 |
| | 2+2 4 | 5+6 2 | 2+2 4 | 3+4 2 | 3+3 2 | 2+2 2 |
| | 2+2 4 | 3+4 2 | 2+2 4 | 5+6 2 | 2+2 2 | 3+3 2 |
| | 2+2 4 | 3+4 2 | 2+2 4 | 5+6 2 | 2+2 2 | 2+2 2 |
| | 2+2 4 | 5+6 2 | 2+2 4 | 3+4 2 | 3+3 2 | 3+3 2 |
| | 2+2 4 | 3+4 2 | 2+2 4 | 3+4 2 | 2+2 2 | 2+2 2 |

Continued on next column

Continued on next column

**TABLE XXVII– continued**

| TQC | ICSD | | NoE | | SF | |
|---|---|---|---|---|---|---|
| | 2+2 4 | 5+6 2 | 2+2 4 | 5+6 2 | 3+3 2 | 2+2 2 |
| | 2+2 4 | 3+4 2 | 2+2 4 | 3+4 2 | 2+2 2 | 3+3 2 |
| | 2+2 4 | 5+6 2 | 2+2 4 | 3+4 2 | 3+3 2 | 3+3 2 |
| | | 5+6 2 | | 5+6 2 | 3+3 2 | 2+2 2 |
| | | 3+4 2 | | 3+4 2 | 3+3 2 | 2+2 2 |
| | | 3+4 2 | | 5+6 2 | 2+2 2 | 3+3 2 |
| | | 3+4 2 | | 3+4 2 | 3+3 2 | 3+3 2 |
| | | 3+4 2 | | 5+6 2 | 2+2 2 | 2+2 2 |
| | | 5+6 2 | | 5+6 2 | 3+3 2 | 3+3 2 |
| | | 5+6 2 | | 3+4 2 | 2+2 2 | 2+2 2 |
| | | 3+4 2 | | 5+6 2 | 3+3 2 | 3+3 2 |
| | | 5+6 2 | | 5+6 2 | 2+2 2 | 3+3 2 |
| | | 5+6 2 | | 3+4 2 | 2+2 2 | 2+2 2 |
| | | 3+4 2 | | 3+4 2 | 2+2 2 | 2+2 2 |
| | | 3+4 2 | | 5+6 2 | 3+3 2 | 3+3 2 |
| | | 5+6 2 | | 5+6 2 | 3+3 2 | 2+2 2 |
| | | 5+6 2 | | 3+4 2 | 2+2 2 | 3+3 2 |
| | | 3+4 2 | | 5+6 2 | 3+3 2 | 2+2 2 |
| | | 3+4 2 | | 3+4 2 | 2+2 2 | 3+3 2 |
| | | 3+4 2 | | 5+6 2 | 3+3 2 | 2+2 2 |
| | | 3+4 2 | | 3+4 2 | 2+2 2 | 3+3 2 |
| | | 5+6 2 | | 3+4 2 | 3+3 2 | 2+2 2 |
| | | 5+6 2 | | 5+6 2 | 2+2 2 | 3+3 2 |
| | | 5+6 2 | | 3+4 2 | 2+2 2 | 2+2 2 |
| | | 5+6 2 | | 5+6 2 | 3+3 2 | 3+3 2 |
| | | 3+4 2 | | 3+4 2 | 3+3 2 | 2+2 2 |
| | | 3+4 2 | | 5+6 2 | 2+2 2 | 3+3 2 |
| | | 3+4 2 | | 5+6 2 | 2+2 2 | 2+2 2 |
| | | 5+6 2 | | 5+6 2 | 2+2 2 | 2+2 2 |
| | | 3+4 2 | | 5+6 2 | 2+2 2 | 3+3 2 |
| | | 3+4 2 | | 5+6 2 | 2+2 2 | 2+2 2 |
| | | 3+4 2 | | 5+6 2 | 2+2 2 | 3+3 2 |
| | | 3+4 2 | | 5+6 2 | 3+3 2 | 3+3 2 |
| | | 3+4 2 | | 3+4 2 | 2+2 2 | 2+2 2 |
| | | 5+6 2 | | 3+4 2 | 3+3 2 | 3+3 2 |
| | | 3+4 2 | | 5+6 2 | 2+2 2 | 2+2 2 |
| 15.2.51 | 15909 | | 32 | | F2Sn1 | |
| | -A | -GM | -M | -Y | -L | -V |
| | 2+2 4 | 3+4 2 | 2+2 4 | 3+4 2 | 3+3 2 | 2+2 2 |
| | 2+2 4 | 3+4 2 | 2+2 4 | 5+6 2 | 3+3 2 | 3+3 2 |
| | 2+2 4 | 5+6 2 | 2+2 4 | 3+4 2 | 2+2 2 | 3+3 2 |
| | 2+2 4 | 3+4 2 | 2+2 4 | 5+6 2 | 2+2 2 | 3+3 2 |
| | 2+2 4 | 5+6 2 | 2+2 4 | 3+4 2 | 2+2 2 | 2+2 2 |
| | 2+2 4 | 3+4 2 | 2+2 4 | 5+6 2 | 3+3 2 | 3+3 2 |
| | 2+2 4 | 5+6 2 | 2+2 4 | 3+4 2 | 3+3 2 | 3+3 2 |
| | 2+2 4 | 3+4 2 | 2+2 4 | 3+4 2 | 2+2 2 | 2+2 2 |
| | | 5+6 2 | | 5+6 2 | 2+2 2 | 2+2 2 |
| | | 5+6 2 | | 5+6 2 | 3+3 2 | 2+2 2 |
| | | 3+4 2 | | 3+4 2 | 2+2 2 | 2+2 2 |
| | | 3+4 2 | | 5+6 2 | 3+3 2 | 3+3 2 |
| | | 5+6 2 | | 5+6 2 | 3+3 2 | 2+2 2 |
| | | 5+6 2 | | 5+6 2 | 2+2 2 | 3+3 2 |
| | | 3+4 2 | | 3+4 2 | 2+2 2 | 2+2 2 |
| | | 5+6 2 | | 5+6 2 | 3+3 2 | 3+3 2 |
| 15.2.52 | 2739 | | 140 | | Er1Na1O2 | |
| | -A | -GM | -M | -Y | -L | -V |
| | 2+2 4 | 3+4 2 | 2+2 4 | 5+6 2 | 2+2 2 | 2+2 2 |
| | 2+2 4 | 5+6 2 | 2+2 4 | 3+4 2 | 3+3 2 | 3+3 2 |
| | 2+2 4 | 3+4 2 | 2+2 4 | 3+4 2 | 2+2 2 | 2+2 2 |
| | 2+2 4 | 5+6 2 | 2+2 4 | 5+6 2 | 3+3 2 | 3+3 2 |
| | 2+2 4 | 3+4 2 | 2+2 4 | 5+6 2 | 2+2 2 | 3+3 2 |

Continued on next column

**TABLE XXVII– continued**

| TQC | ICSD | | NoE | | SF | |
|---|---|---|---|---|---|---|
| | 2+2 4 | 5+6 2 | 2+2 4 | 3+4 2 | 3+3 2 | 2+2 2 |
| | 2+2 4 | 3+4 2 | 2+2 4 | 3+4 2 | 2+2 2 | 2+2 2 |
| | 2+2 4 | 5+6 2 | 2+2 4 | 5+6 2 | 3+3 2 | 3+3 2 |
| | 2+2 4 | 3+4 2 | 2+2 4 | 3+4 2 | 2+2 2 | 2+2 2 |
| | 2+2 4 | 5+6 2 | 2+2 4 | 5+6 2 | 3+3 2 | 3+3 2 |
| | 2+2 4 | 3+4 2 | 2+2 4 | 5+6 2 | 2+2 2 | 2+2 2 |
| | 2+2 4 | 5+6 2 | 2+2 4 | 3+4 2 | 3+3 2 | 3+3 2 |
| | 2+2 4 | 3+4 2 | 2+2 4 | 5+6 2 | 2+2 2 | 3+3 2 |
| | 2+2 4 | 5+6 2 | 2+2 4 | 3+4 2 | 3+3 2 | 2+2 2 |
| | 2+2 4 | 3+4 2 | 2+2 4 | 3+4 2 | 2+2 2 | 2+2 2 |
| | 2+2 4 | 5+6 2 | 2+2 4 | 5+6 2 | 3+3 2 | 3+3 2 |
| | 2+2 4 | 3+4 2 | 2+2 4 | 3+4 2 | 2+2 2 | 2+2 2 |
| | 2+2 4 | 5+6 2 | 2+2 4 | 5+6 2 | 3+3 2 | 3+3 2 |
| | 2+2 4 | 5+6 2 | 2+2 4 | 5+6 2 | 2+2 2 | 3+3 2 |
| | 2+2 4 | 3+4 2 | 2+2 4 | 3+4 2 | 3+3 2 | 2+2 2 |
| | 2+2 4 | 5+6 2 | 2+2 4 | 3+4 2 | 3+3 2 | 2+2 2 |
| | 2+2 4 | 3+4 2 | 2+2 4 | 5+6 2 | 2+2 2 | 3+3 2 |
| | 2+2 4 | 5+6 2 | 2+2 4 | 5+6 2 | 3+3 2 | 3+3 2 |
| | 2+2 4 | 3+4 2 | 2+2 4 | 3+4 2 | 2+2 2 | 2+2 2 |
| | 2+2 4 | 5+6 2 | 2+2 4 | 5+6 2 | 3+3 2 | 3+3 2 |
| | 2+2 4 | 3+4 2 | 2+2 4 | 3+4 2 | 2+2 2 | 3+3 2 |
| | 2+2 4 | 3+4 2 | 2+2 4 | 3+4 2 | 3+3 2 | 2+2 2 |
| | 2+2 4 | 5+6 2 | 2+2 4 | 3+4 2 | 2+2 2 | 2+2 2 |
| | 2+2 4 | 3+4 2 | 2+2 4 | 5+6 2 | 3+3 2 | 2+2 2 |
| | 2+2 4 | 5+6 2 | 2+2 4 | 5+6 2 | 2+2 2 | 3+3 2 |
| | 2+2 4 | 5+6 2 | 2+2 4 | 3+4 2 | 3+3 2 | 2+2 2 |
| | 2+2 4 | 3+4 2 | 2+2 4 | 5+6 2 | 2+2 2 | 3+3 2 |
| | 2+2 4 | 3+4 2 | 2+2 4 | 3+4 2 | 3+3 2 | 2+2 2 |
| | 2+2 4 | 3+4 2 | 2+2 4 | 5+6 2 | 2+2 2 | 3+3 2 |
| | 2+2 4 | 5+6 2 | 2+2 4 | 5+6 2 | 3+3 2 | 2+2 2 |
| | | 3+4 2 | | 3+4 2 | 3+3 2 | 3+3 2 |
| | | 3+4 2 | | 3+4 2 | 2+2 2 | 2+2 2 |
| | | 5+6 2 | | 5+6 2 | 2+2 2 | 3+3 2 |
| | | 5+6 2 | | 5+6 2 | 3+3 2 | 2+2 2 |
| | | 5+6 2 | | 3+4 2 | 2+2 2 | 3+3 2 |
| | | 3+4 2 | | 5+6 2 | 2+2 2 | 2+2 2 |
| | | 3+4 2 | | 3+4 2 | 3+3 2 | 3+3 2 |
| | | 5+6 2 | | 5+6 2 | 3+3 2 | 2+2 2 |
| | | 5+6 2 | | 3+4 2 | 2+2 2 | 3+3 2 |
| | | 3+4 2 | | 5+6 2 | 3+3 2 | 2+2 2 |
| | | 5+6 2 | | 3+4 2 | 2+2 2 | 3+3 2 |
| | | 3+4 2 | | 3+4 2 | 3+3 2 | 2+2 2 |
| | | 5+6 2 | | 5+6 2 | 2+2 2 | 3+3 2 |
| | | 3+4 2 | | 3+4 2 | 3+3 2 | 2+2 2 |
| | | 5+6 2 | | 3+4 2 | 2+2 2 | 3+3 2 |
| | | 3+4 2 | | 5+6 2 | 2+2 2 | 2+2 2 |
| | | 3+4 2 | | 5+6 2 | 3+3 2 | 3+3 2 |
| | | 5+6 2 | | 3+4 2 | 3+3 2 | 2+2 2 |
| | | 3+4 2 | | 5+6 2 | 2+2 2 | 3+3 2 |
| | | 5+6 2 | | 3+4 2 | 2+2 2 | 2+2 2 |
| | | 3+4 2 | | 5+6 2 | 3+3 2 | 3+3 2 |
| | | 5+6 2 | | 5+6 2 | 2+2 2 | 3+3 2 |
| | | 3+4 2 | | 3+4 2 | 2+2 2 | 2+2 2 |
| | | 5+6 2 | | 5+6 2 | 3+3 2 | 2+2 2 |
| | | 5+6 2 | | 3+4 2 | 3+3 2 | 3+3 2 |
| | | 3+4 2 | | 3+4 2 | 2+2 2 | 2+2 2 |
| | | 3+4 2 | | 5+6 2 | 3+3 2 | 2+2 2 |
| | | 5+6 2 | | 3+4 2 | 2+2 2 | 3+3 2 |
| | | 3+4 2 | | 5+6 2 | 3+3 2 | 3+3 2 |
| | | 5+6 2 | | 5+6 2 | 2+2 2 | 2+2 2 |
| | | 5+6 2 | | 3+4 2 | 3+3 2 | 3+3 2 |

Continued on next column

**TABLE XXVII– continued**

| TQC | ICSD | | NoE | SF | | |
|---|---|---|---|---|---|---|
| | | 3+4 2 | | 3+4 2 | 2+2 2 | 2+2 2 |
| | | 3+4 2 | | 5+6 2 | 2+2 2 | 3+3 2 |
| | | 5+6 2 | | 3+4 2 | 3+3 2 | 2+2 2 |
| | | 3+4 2 | | 5+6 2 | 3+3 2 | 2+2 2 |
| 15.2.53 | 94884 | | 92 | D7Al1Sr2 | | |
| | -A | -GM | -M | -Y | -L | -V |
| | 2+2 4 | 3+4 2 | 2+2 4 | 3+4 2 | 2+2 2 | 3+3 2 |
| | 2+2 4 | 3+4 2 | 2+2 4 | 5+6 2 | 3+3 2 | 2+2 2 |
| | 2+2 4 | 5+6 2 | 2+2 4 | 5+6 2 | 2+2 2 | 2+2 2 |
| | 2+2 4 | 5+6 2 | 2+2 4 | 3+4 2 | 3+3 2 | 3+3 2 |
| | 2+2 4 | 3+4 2 | 2+2 4 | 3+4 2 | 2+2 2 | 2+2 2 |
| | 2+2 4 | 3+4 2 | 2+2 4 | 5+6 2 | 3+3 2 | 2+2 2 |
| | 2+2 4 | 5+6 2 | 2+2 4 | 3+4 2 | 2+2 2 | 3+3 2 |
| | 2+2 4 | 5+6 2 | 2+2 4 | 5+6 2 | 3+3 2 | 3+3 2 |
| | 2+2 4 | 5+6 2 | 2+2 4 | 5+6 2 | 3+3 2 | 2+2 2 |
| | 2+2 4 | 3+4 2 | 2+2 4 | 3+4 2 | 2+2 2 | 3+3 2 |
| | 2+2 4 | 3+4 2 | 2+2 4 | 3+4 2 | 3+3 2 | 3+3 2 |
| | 2+2 4 | 5+6 2 | 2+2 4 | 5+6 2 | 2+2 2 | 2+2 2 |
| | 2+2 4 | 3+4 2 | 2+2 4 | 3+4 2 | 2+2 2 | 2+2 2 |
| | 2+2 4 | 3+4 2 | 2+2 4 | 5+6 2 | 3+3 2 | 3+3 2 |
| | 2+2 4 | 5+6 2 | 2+2 4 | 3+4 2 | 2+2 2 | 2+2 2 |
| | 2+2 4 | 5+6 2 | 2+2 4 | 5+6 2 | 3+3 2 | 3+3 2 |
| | 2+2 4 | 3+4 2 | 2+2 4 | 5+6 2 | 3+3 2 | 2+2 2 |
| | 2+2 4 | 5+6 2 | 2+2 4 | 3+4 2 | 2+2 2 | 3+3 2 |
| | 2+2 4 | 3+4 2 | 2+2 4 | 3+4 2 | 3+3 2 | 3+3 2 |
| | 2+2 4 | 3+4 2 | 2+2 4 | 5+6 2 | 2+2 2 | 2+2 2 |
| | 2+2 4 | 5+6 2 | 2+2 4 | 3+4 2 | 2+2 2 | 3+3 2 |
| | 2+2 4 | 5+6 2 | 2+2 4 | 5+6 2 | 3+3 2 | 2+2 2 |
| | 2+2 4 | 3+4 2 | 2+2 4 | 3+4 2 | 3+3 2 | 3+3 2 |
| | | 5+6 2 | | 5+6 2 | 2+2 2 | 2+2 2 |
| | | 3+4 2 | | 5+6 2 | 3+3 2 | 2+2 2 |
| | | 5+6 2 | | 3+4 2 | 2+2 2 | 3+3 2 |
| | | 3+4 2 | | 3+4 2 | 2+2 2 | 2+2 2 |
| | | 3+4 2 | | 3+4 2 | 3+3 2 | 3+3 2 |
| | | 5+6 2 | | 5+6 2 | 2+2 2 | 2+2 2 |
| | | 3+4 2 | | 5+6 2 | 3+3 2 | 3+3 2 |
| | | 5+6 2 | | 3+4 2 | 2+2 2 | 2+2 2 |
| | | 5+6 2 | | 5+6 2 | 3+3 2 | 3+3 2 |
| | | 3+4 2 | | 5+6 2 | 2+2 2 | 2+2 2 |
| | | 3+4 2 | | 5+6 2 | 3+3 2 | 3+3 2 |
| | | 5+6 2 | | 3+4 2 | 3+3 2 | 2+2 2 |
| | | 5+6 2 | | 3+4 2 | 2+2 2 | 3+3 2 |
| | | 3+4 2 | | 3+4 2 | 2+2 2 | 3+3 2 |
| | | 5+6 2 | | 5+6 2 | 3+3 2 | 2+2 2 |
| | | 3+4 2 | | 3+4 2 | 3+3 2 | 2+2 2 |
| | | 5+6 2 | | 5+6 2 | 2+2 2 | 3+3 2 |
| | | 3+4 2 | | 5+6 2 | 2+2 2 | 3+3 2 |
| | | 5+6 2 | | 5+6 2 | 3+3 2 | 2+2 2 |
| | | 3+4 2 | | 3+4 2 | 2+2 2 | 2+2 2 |
| | | 5+6 2 | | 5+6 2 | 3+3 2 | 3+3 2 |
| | | 3+4 2 | | 5+6 2 | 3+3 2 | 2+2 2 |
| | | 5+6 2 | | 3+4 2 | 3+3 2 | 3+3 2 |
| 15.2.54 | 185090 | | 172 | Cu6Sn5 | | |
| | -A | -GM | -M | -Y | -L | -V |
| | 2+2 4 | 3+4 2 | 2+2 4 | 3+4 2 | 3+3 2 | 3+3 2 |
| | 2+2 4 | 5+6 2 | 2+2 4 | 5+6 2 | 2+2 2 | 2+2 2 |
| | 2+2 4 | 3+4 2 | 2+2 4 | 5+6 2 | 2+2 2 | 3+3 2 |
| | 2+2 4 | 5+6 2 | 2+2 4 | 5+6 2 | 3+3 2 | 2+2 2 |
| | 2+2 4 | 3+4 2 | 2+2 4 | 3+4 2 | 3+3 2 | 3+3 2 |
| | 2+2 4 | 5+6 2 | 2+2 4 | 5+6 2 | 2+2 2 | 2+2 2 |
| | 2+2 4 | 3+4 2 | 2+2 4 | 3+4 2 | 3+3 2 | 2+2 2 |
| | 2+2 4 | 3+4 2 | 2+2 4 | 3+4 2 | 2+2 2 | 3+3 2 |
| | 2+2 4 | 5+6 2 | 2+2 4 | 5+6 2 | 2+2 2 | 3+3 2 |
| | 2+2 4 | 5+6 2 | 2+2 4 | 3+4 2 | 3+3 2 | 2+2 2 |
| | 2+2 4 | 3+4 2 | 2+2 4 | 3+4 2 | 3+3 2 | 2+2 2 |
| | 2+2 4 | 5+6 2 | 2+2 4 | 5+6 2 | 2+2 2 | 2+2 2 |
| | 2+2 4 | 5+6 2 | 2+2 4 | 3+4 2 | 2+2 2 | 3+3 2 |
| | 2+2 4 | 3+4 2 | 2+2 4 | 3+4 2 | 3+3 2 | 3+3 2 |
| | 2+2 4 | 3+4 2 | 2+2 4 | 5+6 2 | 3+3 2 | 2+2 2 |
| | 2+2 4 | 3+4 2 | 2+2 4 | 3+4 2 | 2+2 2 | 3+3 2 |
| | 2+2 4 | 3+4 2 | 2+2 4 | 5+6 2 | 3+3 2 | 2+2 2 |
| | 2+2 4 | 5+6 2 | 2+2 4 | 5+6 2 | 2+2 2 | 2+2 2 |
| | 2+2 4 | 5+6 2 | 2+2 4 | 3+4 2 | 3+3 2 | 3+3 2 |
| | 2+2 4 | 3+4 2 | 2+2 4 | 3+4 2 | 2+2 2 | 3+3 2 |
| | 2+2 4 | 3+4 2 | 2+2 4 | 5+6 2 | 2+2 2 | 2+2 2 |
| | 2+2 4 | 5+6 2 | 2+2 4 | 3+4 2 | 2+2 2 | 2+2 2 |
| | 2+2 4 | 3+4 2 | 2+2 4 | 3+4 2 | 3+3 2 | 2+2 2 |
| | 2+2 4 | 5+6 2 | 2+2 4 | 5+6 2 | 2+2 2 | 3+3 2 |
| | 2+2 4 | 3+4 2 | 2+2 4 | 5+6 2 | 3+3 2 | 2+2 2 |
| | 2+2 4 | 3+4 2 | 2+2 4 | 3+4 2 | 2+2 2 | 3+3 2 |
| | 2+2 4 | 5+6 2 | 2+2 4 | 3+4 2 | 3+3 2 | 2+2 2 |
| | 2+2 4 | 3+4 2 | 2+2 4 | 3+4 2 | 3+3 2 | 3+3 2 |
| | 2+2 4 | 3+4 2 | 2+2 4 | 5+6 2 | 2+2 2 | 2+2 2 |
| | 2+2 4 | 5+6 2 | 2+2 4 | 5+6 2 | 3+3 2 | 3+3 2 |
| | 2+2 4 | 5+6 2 | 2+2 4 | 3+4 2 | 3+3 2 | 2+2 2 |
| | 2+2 4 | 3+4 2 | 2+2 4 | 3+4 2 | 2+2 2 | 3+3 2 |
| | 2+2 4 | 5+6 2 | 2+2 4 | 5+6 2 | 3+3 2 | 2+2 2 |
| | 2+2 4 | 5+6 2 | 2+2 4 | 3+4 2 | 2+2 2 | 3+3 2 |
| | 2+2 4 | 3+4 2 | 2+2 4 | 5+6 2 | 2+2 2 | 2+2 2 |
| | 2+2 4 | 3+4 2 | 2+2 4 | 5+6 2 | 3+3 2 | 2+2 2 |
| | 2+2 4 | 3+4 2 | 2+2 4 | 5+6 2 | 2+2 2 | 3+3 2 |
| | 2+2 4 | 5+6 2 | 2+2 4 | 3+4 2 | 3+3 2 | 3+3 2 |
| | 2+2 4 | 3+4 2 | 2+2 4 | 5+6 2 | 3+3 2 | 2+2 2 |
| | 2+2 4 | 5+6 2 | 2+2 4 | 3+4 2 | 2+2 2 | 2+2 2 |
| | 2+2 4 | 3+4 2 | 2+2 4 | 3+4 2 | 3+3 2 | 3+3 2 |
| | 2+2 4 | 3+4 2 | 2+2 4 | 3+4 2 | 3+3 2 | 2+2 2 |
| | 2+2 4 | 5+6 2 | 2+2 4 | 5+6 2 | 2+2 2 | 3+3 2 |
| | | 3+4 2 | | 3+4 2 | 2+2 2 | 2+2 2 |
| | | 5+6 2 | | 5+6 2 | 2+2 2 | 3+3 2 |
| | | 5+6 2 | | 3+4 2 | 3+3 2 | 2+2 2 |
| | | 3+4 2 | | 5+6 2 | 3+3 2 | 3+3 2 |
| | | 3+4 2 | | 3+4 2 | 3+3 2 | 2+2 2 |
| | | 5+6 2 | | 5+6 2 | 2+2 2 | 2+2 2 |
| | | 3+4 2 | | 3+4 2 | 3+3 2 | 3+3 2 |
| | | 5+6 2 | | 3+4 2 | 2+2 2 | 3+3 2 |
| | | 3+4 2 | | 5+6 2 | 2+2 2 | 2+2 2 |
| | | 3+4 2 | | 3+4 2 | 3+3 2 | 3+3 2 |
| | | 5+6 2 | | 5+6 2 | 2+2 2 | 2+2 2 |
| | | 3+4 2 | | 3+4 2 | 2+2 2 | 2+2 2 |
| | | 5+6 2 | | 3+4 2 | 3+3 2 | 3+3 2 |
| | | 5+6 2 | | 3+4 2 | 3+3 2 | 2+2 2 |
| | | 3+4 2 | | 5+6 2 | 2+2 2 | 3+3 2 |
| | | 5+6 2 | | 5+6 2 | 3+3 2 | 2+2 2 |
| | | 3+4 2 | | 3+4 2 | 2+2 2 | 2+2 2 |
| | | 3+4 2 | | 5+6 2 | 3+3 2 | 2+2 2 |
| | | 5+6 2 | | 5+6 2 | 2+2 2 | 3+3 2 |
| | | 5+6 2 | | 3+4 2 | 2+2 2 | 2+2 2 |
| | | 3+4 2 | | 3+4 2 | 3+3 2 | 2+2 2 |
| | | 5+6 2 | | 5+6 2 | 3+3 2 | 3+3 2 |
| | | 3+4 2 | | 3+4 2 | 2+2 2 | 2+2 2 |
| | | 3+4 2 | | 3+4 2 | 3+3 2 | 3+3 2 |
| | | 3+4 2 | | 3+4 2 | 2+2 2 | 2+2 2 |

Continued on next column

Continued on next column

**TABLE XXVII– continued**

| TQC | ICSD | | NoE | SF | | |
|---|---|---|---|---|---|---|
| | | 5+6 2 | | 5+6 2 | 3+3 2 | 2+2 2 |
| | | 3+4 2 | | 3+4 2 | 2+2 2 | 2+2 2 |
| | | 3+4 2 | | 3+4 2 | 3+3 2 | 3+3 2 |
| | | 5+6 2 | | 3+4 2 | 2+2 2 | 2+2 2 |
| | | 5+6 2 | | 5+6 2 | 3+3 2 | 3+3 2 |
| | | 3+4 2 | | 5+6 2 | 2+2 2 | 2+2 2 |
| | | 3+4 2 | | 3+4 2 | 3+3 2 | 3+3 2 |
| | | 3+4 2 | | 5+6 2 | 2+2 2 | 3+3 2 |
| | | 5+6 2 | | 5+6 2 | 2+2 2 | 2+2 2 |
| | | 3+4 2 | | 3+4 2 | 3+3 2 | 3+3 2 |
| | | 5+6 2 | | 5+6 2 | 2+2 2 | 3+3 2 |
| | | 3+4 2 | | 3+4 2 | 2+2 2 | 2+2 2 |
| | | 3+4 2 | | 5+6 2 | 3+3 2 | 3+3 2 |
| | | 3+4 2 | | 3+4 2 | 3+3 2 | 2+2 2 |
| | | 5+6 2 | | 3+4 2 | 3+3 2 | 3+3 2 |
| | | 3+4 2 | | 3+4 2 | 2+2 2 | 2+2 2 |
| | | 5+6 2 | | 5+6 2 | 2+2 2 | 3+3 2 |
| | | 3+4 2 | | 5+6 2 | 3+3 2 | 2+2 2 |
| 15.2.55 | 185089 | | 172 | Cu6Sn5 | | |
| | -A | -GM | -M | -Y | -L | -V |
| | 2+2 4 | 3+4 2 | 2+2 4 | 3+4 2 | 2+2 2 | 3+3 2 |
| | 2+2 4 | 5+6 2 | 2+2 4 | 5+6 2 | 3+3 2 | 2+2 2 |
| | 2+2 4 | 3+4 2 | 2+2 4 | 5+6 2 | 3+3 2 | 3+3 2 |
| | 2+2 4 | 5+6 2 | 2+2 4 | 5+6 2 | 2+2 2 | 2+2 2 |
| | 2+2 4 | 3+4 2 | 2+2 4 | 3+4 2 | 2+2 2 | 3+3 2 |
| | 2+2 4 | 5+6 2 | 2+2 4 | 5+6 2 | 3+3 2 | 2+2 2 |
| | 2+2 4 | 3+4 2 | 2+2 4 | 3+4 2 | 2+2 2 | 2+2 2 |
| | 2+2 4 | 3+4 2 | 2+2 4 | 3+4 2 | 3+3 2 | 3+3 2 |
| | 2+2 4 | 5+6 2 | 2+2 4 | 5+6 2 | 3+3 2 | 3+3 2 |
| | 2+2 4 | 5+6 2 | 2+2 4 | 3+4 2 | 2+2 2 | 2+2 2 |
| | 2+2 4 | 3+4 2 | 2+2 4 | 3+4 2 | 2+2 2 | 2+2 2 |
| | 2+2 4 | 5+6 2 | 2+2 4 | 5+6 2 | 3+3 2 | 2+2 2 |
| | 2+2 4 | 5+6 2 | 2+2 4 | 3+4 2 | 3+3 2 | 3+3 2 |
| | 2+2 4 | 3+4 2 | 2+2 4 | 3+4 2 | 2+2 2 | 3+3 2 |
| | 2+2 4 | 3+4 2 | 2+2 4 | 5+6 2 | 2+2 2 | 2+2 2 |
| | 2+2 4 | 3+4 2 | 2+2 4 | 3+4 2 | 3+3 2 | 3+3 2 |
| | 2+2 4 | 3+4 2 | 2+2 4 | 5+6 2 | 2+2 2 | 2+2 2 |
| | 2+2 4 | 5+6 2 | 2+2 4 | 5+6 2 | 3+3 2 | 2+2 2 |
| | 2+2 4 | 5+6 2 | 2+2 4 | 3+4 2 | 2+2 2 | 3+3 2 |
| | 2+2 4 | 3+4 2 | 2+2 4 | 3+4 2 | 3+3 2 | 3+3 2 |
| | 2+2 4 | 3+4 2 | 2+2 4 | 5+6 2 | 3+3 2 | 2+2 2 |
| | 2+2 4 | 5+6 2 | 2+2 4 | 3+4 2 | 3+3 2 | 2+2 2 |
| | 2+2 4 | 3+4 2 | 2+2 4 | 3+4 2 | 2+2 2 | 2+2 2 |
| | 2+2 4 | 5+6 2 | 2+2 4 | 5+6 2 | 3+3 2 | 3+3 2 |
| | 2+2 4 | 3+4 2 | 2+2 4 | 5+6 2 | 2+2 2 | 2+2 2 |
| | 2+2 4 | 3+4 2 | 2+2 4 | 3+4 2 | 3+3 2 | 3+3 2 |
| | 2+2 4 | 5+6 2 | 2+2 4 | 3+4 2 | 2+2 2 | 2+2 2 |
| | 2+2 4 | 3+4 2 | 2+2 4 | 3+4 2 | 2+2 2 | 3+3 2 |
| | 2+2 4 | 3+4 2 | 2+2 4 | 5+6 2 | 3+3 2 | 2+2 2 |
| | 2+2 4 | 5+6 2 | 2+2 4 | 5+6 2 | 2+2 2 | 3+3 2 |
| | 2+2 4 | 5+6 2 | 2+2 4 | 3+4 2 | 2+2 2 | 2+2 2 |
| | 2+2 4 | 3+4 2 | 2+2 4 | 3+4 2 | 3+3 2 | 3+3 2 |
| | 2+2 4 | 5+6 2 | 2+2 4 | 5+6 2 | 2+2 2 | 2+2 2 |
| | 2+2 4 | 5+6 2 | 2+2 4 | 3+4 2 | 3+3 2 | 3+3 2 |
| | 2+2 4 | 3+4 2 | 2+2 4 | 5+6 2 | 3+3 2 | 2+2 2 |
| | 2+2 4 | 3+4 2 | 2+2 4 | 5+6 2 | 2+2 2 | 2+2 2 |
| | 2+2 4 | 3+4 2 | 2+2 4 | 5+6 2 | 3+3 2 | 3+3 2 |
| | 2+2 4 | 5+6 2 | 2+2 4 | 3+4 2 | 2+2 2 | 3+3 2 |
| | 2+2 4 | 3+4 2 | 2+2 4 | 5+6 2 | 2+2 2 | 2+2 2 |
| | 2+2 4 | 5+6 2 | 2+2 4 | 3+4 2 | 3+3 2 | 2+2 2 |
| | 2+2 4 | 3+4 2 | 2+2 4 | 3+4 2 | 2+2 2 | 3+3 2 |

Continued on next column

**TABLE XXVII– continued**

| TQC | ICSD | | NoE | SF | | |
|---|---|---|---|---|---|---|
| | 2+2 4 | 3+4 2 | 2+2 4 | 3+4 2 | 2+2 2 | 2+2 2 |
| | 2+2 4 | 5+6 2 | 2+2 4 | 5+6 2 | 3+3 2 | 3+3 2 |
| | | 3+4 2 | | 3+4 2 | 3+3 2 | 2+2 2 |
| | | 5+6 2 | | 5+6 2 | 3+3 2 | 3+3 2 |
| | | 5+6 2 | | 3+4 2 | 2+2 2 | 2+2 2 |
| | | 3+4 2 | | 5+6 2 | 2+2 2 | 3+3 2 |
| | | 3+4 2 | | 3+4 2 | 2+2 2 | 2+2 2 |
| | | 5+6 2 | | 5+6 2 | 3+3 2 | 2+2 2 |
| | | 3+4 2 | | 3+4 2 | 2+2 2 | 3+3 2 |
| | | 5+6 2 | | 3+4 2 | 3+3 2 | 3+3 2 |
| | | 3+4 2 | | 5+6 2 | 3+3 2 | 2+2 2 |
| | | 3+4 2 | | 3+4 2 | 2+2 2 | 3+3 2 |
| | | 5+6 2 | | 5+6 2 | 3+3 2 | 2+2 2 |
| | | 3+4 2 | | 3+4 2 | 3+3 2 | 2+2 2 |
| | | 5+6 2 | | 3+4 2 | 2+2 2 | 3+3 2 |
| | | 5+6 2 | | 3+4 2 | 2+2 2 | 2+2 2 |
| | | 3+4 2 | | 5+6 2 | 3+3 2 | 3+3 2 |
| | | 5+6 2 | | 5+6 2 | 2+2 2 | 2+2 2 |
| | | 3+4 2 | | 3+4 2 | 3+3 2 | 2+2 2 |
| | | 3+4 2 | | 5+6 2 | 2+2 2 | 2+2 2 |
| | | 5+6 2 | | 5+6 2 | 3+3 2 | 3+3 2 |
| | | 5+6 2 | | 3+4 2 | 3+3 2 | 2+2 2 |
| | | 3+4 2 | | 3+4 2 | 2+2 2 | 2+2 2 |
| | | 5+6 2 | | 5+6 2 | 2+2 2 | 3+3 2 |
| | | 3+4 2 | | 3+4 2 | 3+3 2 | 2+2 2 |
| | | 3+4 2 | | 3+4 2 | 2+2 2 | 3+3 2 |
| | | 3+4 2 | | 3+4 2 | 3+3 2 | 2+2 2 |
| | | 5+6 2 | | 5+6 2 | 2+2 2 | 2+2 2 |
| | | 3+4 2 | | 3+4 2 | 3+3 2 | 2+2 2 |
| | | 3+4 2 | | 3+4 2 | 2+2 2 | 3+3 2 |
| | | 5+6 2 | | 3+4 2 | 3+3 2 | 2+2 2 |
| | | 5+6 2 | | 5+6 2 | 2+2 2 | 3+3 2 |
| | | 3+4 2 | | 5+6 2 | 3+3 2 | 2+2 2 |
| | | 3+4 2 | | 3+4 2 | 2+2 2 | 3+3 2 |
| | | 3+4 2 | | 5+6 2 | 3+3 2 | 3+3 2 |
| | | 5+6 2 | | 5+6 2 | 3+3 2 | 2+2 2 |
| | | 3+4 2 | | 3+4 2 | 2+2 2 | 3+3 2 |
| | | 5+6 2 | | 5+6 2 | 3+3 2 | 3+3 2 |
| | | 3+4 2 | | 3+4 2 | 3+3 2 | 2+2 2 |
| | | 3+4 2 | | 5+6 2 | 2+2 2 | 3+3 2 |
| | | 3+4 2 | | 3+4 2 | 2+2 2 | 2+2 2 |
| | | 5+6 2 | | 3+4 2 | 2+2 2 | 3+3 2 |
| | | 3+4 2 | | 3+4 2 | 3+3 2 | 2+2 2 |
| | | 5+6 2 | | 5+6 2 | 3+3 2 | 3+3 2 |
| | | 3+4 2 | | 5+6 2 | 2+2 2 | 2+2 2 |
| 15.2.56 | 150124 | | 172 | Cu6Sn5 | | |
| | -A | -GM | -M | -Y | -L | -V |
| | 2+2 4 | 3+4 2 | 2+2 4 | 3+4 2 | 2+2 2 | 3+3 2 |
| | 2+2 4 | 5+6 2 | 2+2 4 | 5+6 2 | 3+3 2 | 2+2 2 |
| | 2+2 4 | 3+4 2 | 2+2 4 | 5+6 2 | 3+3 2 | 3+3 2 |
| | 2+2 4 | 5+6 2 | 2+2 4 | 5+6 2 | 2+2 2 | 2+2 2 |
| | 2+2 4 | 3+4 2 | 2+2 4 | 3+4 2 | 2+2 2 | 3+3 2 |
| | 2+2 4 | 5+6 2 | 2+2 4 | 5+6 2 | 3+3 2 | 2+2 2 |
| | 2+2 4 | 3+4 2 | 2+2 4 | 3+4 2 | 2+2 2 | 2+2 2 |
| | 2+2 4 | 3+4 2 | 2+2 4 | 3+4 2 | 3+3 2 | 3+3 2 |
| | 2+2 4 | 5+6 2 | 2+2 4 | 5+6 2 | 3+3 2 | 3+3 2 |
| | 2+2 4 | 5+6 2 | 2+2 4 | 3+4 2 | 2+2 2 | 2+2 2 |
| | 2+2 4 | 3+4 2 | 2+2 4 | 3+4 2 | 2+2 2 | 2+2 2 |
| | 2+2 4 | 5+6 2 | 2+2 4 | 5+6 2 | 3+3 2 | 2+2 2 |
| | 2+2 4 | 5+6 2 | 2+2 4 | 3+4 2 | 3+3 2 | 3+3 2 |
| | 2+2 4 | 3+4 2 | 2+2 4 | 3+4 2 | 2+2 2 | 3+3 2 |

Continued on next column

**TABLE XXVII– continued**

| TQC | ICSD | | NoE | | SF | |
|---|---|---|---|---|---|---|
| | 2+2 4 | 3+4 2 | 2+2 4 | 5+6 2 | 2+2 2 | 2+2 2 |
| | 2+2 4 | 3+4 2 | 2+2 4 | 3+4 2 | 3+3 2 | 3+3 2 |
| | 2+2 4 | 3+4 2 | 2+2 4 | 5+6 2 | 2+2 2 | 2+2 2 |
| | 2+2 4 | 5+6 2 | 2+2 4 | 5+6 2 | 3+3 2 | 2+2 2 |
| | 2+2 4 | 5+6 2 | 2+2 4 | 3+4 2 | 2+2 2 | 3+3 2 |
| | 2+2 4 | 3+4 2 | 2+2 4 | 3+4 2 | 3+3 2 | 3+3 2 |
| | 2+2 4 | 3+4 2 | 2+2 4 | 5+6 2 | 3+3 2 | 2+2 2 |
| | 2+2 4 | 5+6 2 | 2+2 4 | 3+4 2 | 3+3 2 | 2+2 2 |
| | 2+2 4 | 5+6 2 | 2+2 4 | 3+4 2 | 2+2 2 | 2+2 2 |
| | 2+2 4 | 3+4 2 | 2+2 4 | 5+6 2 | 3+3 2 | 3+3 2 |
| | 2+2 4 | 3+4 2 | 2+2 4 | 5+6 2 | 2+2 2 | 2+2 2 |
| | 2+2 4 | 3+4 2 | 2+2 4 | 3+4 2 | 3+3 2 | 3+3 2 |
| | 2+2 4 | 5+6 2 | 2+2 4 | 3+4 2 | 2+2 2 | 2+2 2 |
| | 2+2 4 | 3+4 2 | 2+2 4 | 3+4 2 | 2+2 2 | 3+3 2 |
| | 2+2 4 | 3+4 2 | 2+2 4 | 5+6 2 | 3+3 2 | 2+2 2 |
| | 2+2 4 | 5+6 2 | 2+2 4 | 5+6 2 | 2+2 2 | 3+3 2 |
| | 2+2 4 | 5+6 2 | 2+2 4 | 3+4 2 | 2+2 2 | 2+2 2 |
| | 2+2 4 | 3+4 2 | 2+2 4 | 3+4 2 | 3+3 2 | 3+3 2 |
| | 2+2 4 | 5+6 2 | 2+2 4 | 5+6 2 | 2+2 2 | 2+2 2 |
| | 2+2 4 | 5+6 2 | 2+2 4 | 3+4 2 | 3+3 2 | 3+3 2 |
| | 2+2 4 | 3+4 2 | 2+2 4 | 5+6 2 | 3+3 2 | 2+2 2 |
| | 2+2 4 | 3+4 2 | 2+2 4 | 5+6 2 | 2+2 2 | 2+2 2 |
| | 2+2 4 | 3+4 2 | 2+2 4 | 5+6 2 | 3+3 2 | 3+3 2 |
| | 2+2 4 | 5+6 2 | 2+2 4 | 3+4 2 | 2+2 2 | 3+3 2 |
| | 2+2 4 | 3+4 2 | 2+2 4 | 5+6 2 | 2+2 2 | 2+2 2 |
| | 2+2 4 | 5+6 2 | 2+2 4 | 3+4 2 | 3+3 2 | 2+2 2 |
| | 2+2 4 | 3+4 2 | 2+2 4 | 3+4 2 | 2+2 2 | 3+3 2 |
| | 2+2 4 | 3+4 2 | 2+2 4 | 3+4 2 | 2+2 2 | 2+2 2 |
| | 2+2 4 | 5+6 2 | 2+2 4 | 5+6 2 | 3+3 2 | 3+3 2 |
| | | 3+4 2 | | 3+4 2 | 3+3 2 | 2+2 2 |
| | | 5+6 2 | | 5+6 2 | 3+3 2 | 3+3 2 |
| | | 5+6 2 | | 3+4 2 | 2+2 2 | 2+2 2 |
| | | 3+4 2 | | 5+6 2 | 2+2 2 | 3+3 2 |
| | | 3+4 2 | | 3+4 2 | 2+2 2 | 2+2 2 |
| | | 5+6 2 | | 5+6 2 | 3+3 2 | 2+2 2 |
| | | 3+4 2 | | 3+4 2 | 2+2 2 | 3+3 2 |
| | | 5+6 2 | | 3+4 2 | 3+3 2 | 3+3 2 |
| | | 3+4 2 | | 5+6 2 | 3+3 2 | 2+2 2 |
| | | 3+4 2 | | 3+4 2 | 2+2 2 | 3+3 2 |
| | | 5+6 2 | | 5+6 2 | 3+3 2 | 2+2 2 |
| | | 3+4 2 | | 3+4 2 | 3+3 2 | 2+2 2 |
| | | 5+6 2 | | 3+4 2 | 2+2 2 | 3+3 2 |
| | | 5+6 2 | | 3+4 2 | 2+2 2 | 2+2 2 |
| | | 3+4 2 | | 5+6 2 | 3+3 2 | 3+3 2 |
| | | 5+6 2 | | 5+6 2 | 2+2 2 | 2+2 2 |
| | | 3+4 2 | | 3+4 2 | 3+3 2 | 2+2 2 |
| | | 3+4 2 | | 5+6 2 | 2+2 2 | 2+2 2 |
| | | 5+6 2 | | 5+6 2 | 3+3 2 | 3+3 2 |
| | | 5+6 2 | | 3+4 2 | 3+3 2 | 2+2 2 |
| | | 3+4 2 | | 3+4 2 | 2+2 2 | 3+3 2 |
| | | 5+6 2 | | 5+6 2 | 2+2 2 | 2+2 2 |
| | | 3+4 2 | | 3+4 2 | 3+3 2 | 2+2 2 |
| | | 3+4 2 | | 3+4 2 | 2+2 2 | 3+3 2 |
| | | 3+4 2 | | 3+4 2 | 3+3 2 | 2+2 2 |
| | | 5+6 2 | | 5+6 2 | 2+2 2 | 2+2 2 |
| | | 3+4 2 | | 3+4 2 | 3+3 2 | 2+2 2 |
| | | 3+4 2 | | 3+4 2 | 2+2 2 | 3+3 2 |
| | | 5+6 2 | | 3+4 2 | 3+3 2 | 2+2 2 |
| | | 5+6 2 | | 5+6 2 | 2+2 2 | 3+3 2 |
| | | 3+4 2 | | 5+6 2 | 3+3 2 | 2+2 2 |
| | | 3+4 2 | | 3+4 2 | 2+2 2 | 3+3 2 |

Continued on next column

**TABLE XXVII– continued**

| TQC | ICSD | | NoE | | SF | |
|---|---|---|---|---|---|---|
| | | 3+4 2 | | 5+6 2 | 3+3 2 | 3+3 2 |
| | | 5+6 2 | | 5+6 2 | 3+3 2 | 2+2 2 |
| | | 3+4 2 | | 3+4 2 | 2+2 2 | 3+3 2 |
| | | 5+6 2 | | 5+6 2 | 3+3 2 | 3+3 2 |
| | | 3+4 2 | | 3+4 2 | 3+3 2 | 2+2 2 |
| | | 3+4 2 | | 5+6 2 | 2+2 2 | 3+3 2 |
| | | 3+4 2 | | 3+4 2 | 2+2 2 | 3+3 2 |
| | | 5+6 2 | | 3+4 2 | 2+2 2 | 2+2 2 |
| | | 3+4 2 | | 3+4 2 | 3+3 2 | 2+2 2 |
| | | 5+6 2 | | 5+6 2 | 3+3 2 | 3+3 2 |
| | | 3+4 2 | | 5+6 2 | 2+2 2 | 2+2 2 |
| 15.2.57 | 106530 | | 172 | | Cu6Sn5 | |
| | -A | -GM | -M | -Y | -L | -V |
| | 2+2 4 | 3+4 2 | 2+2 4 | 3+4 2 | 3+3 2 | 3+3 2 |
| | 2+2 4 | 5+6 2 | 2+2 4 | 5+6 2 | 2+2 2 | 2+2 2 |
| | 2+2 4 | 3+4 2 | 2+2 4 | 5+6 2 | 2+2 2 | 3+3 2 |
| | 2+2 4 | 5+6 2 | 2+2 4 | 5+6 2 | 3+3 2 | 2+2 2 |
| | 2+2 4 | 3+4 2 | 2+2 4 | 3+4 2 | 3+3 2 | 3+3 2 |
| | 2+2 4 | 5+6 2 | 2+2 4 | 5+6 2 | 2+2 2 | 2+2 2 |
| | 2+2 4 | 3+4 2 | 2+2 4 | 3+4 2 | 3+3 2 | 2+2 2 |
| | 2+2 4 | 3+4 2 | 2+2 4 | 3+4 2 | 2+2 2 | 3+3 2 |
| | 2+2 4 | 5+6 2 | 2+2 4 | 5+6 2 | 2+2 2 | 3+3 2 |
| | 2+2 4 | 5+6 2 | 2+2 4 | 3+4 2 | 3+3 2 | 2+2 2 |
| | 2+2 4 | 3+4 2 | 2+2 4 | 3+4 2 | 3+3 2 | 2+2 2 |
| | 2+2 4 | 5+6 2 | 2+2 4 | 5+6 2 | 2+2 2 | 2+2 2 |
| | 2+2 4 | 5+6 2 | 2+2 4 | 3+4 2 | 2+2 2 | 3+3 2 |
| | 2+2 4 | 3+4 2 | 2+2 4 | 3+4 2 | 3+3 2 | 3+3 2 |
| | 2+2 4 | 3+4 2 | 2+2 4 | 5+6 2 | 3+3 2 | 2+2 2 |
| | 2+2 4 | 3+4 2 | 2+2 4 | 3+4 2 | 2+2 2 | 3+3 2 |
| | 2+2 4 | 3+4 2 | 2+2 4 | 5+6 2 | 3+3 2 | 2+2 2 |
| | 2+2 4 | 5+6 2 | 2+2 4 | 5+6 2 | 2+2 2 | 2+2 2 |
| | 2+2 4 | 5+6 2 | 2+2 4 | 3+4 2 | 3+3 2 | 3+3 2 |
| | 2+2 4 | 3+4 2 | 2+2 4 | 3+4 2 | 2+2 2 | 3+3 2 |
| | 2+2 4 | 3+4 2 | 2+2 4 | 5+6 2 | 2+2 2 | 2+2 2 |
| | 2+2 4 | 5+6 2 | 2+2 4 | 3+4 2 | 2+2 2 | 2+2 2 |
| | 2+2 4 | 5+6 2 | 2+2 4 | 3+4 2 | 3+3 2 | 2+2 2 |
| | 2+2 4 | 3+4 2 | 2+2 4 | 5+6 2 | 2+2 2 | 3+3 2 |
| | 2+2 4 | 3+4 2 | 2+2 4 | 5+6 2 | 3+3 2 | 2+2 2 |
| | 2+2 4 | 3+4 2 | 2+2 4 | 3+4 2 | 2+2 2 | 3+3 2 |
| | 2+2 4 | 5+6 2 | 2+2 4 | 3+4 2 | 3+3 2 | 2+2 2 |
| | 2+2 4 | 3+4 2 | 2+2 4 | 3+4 2 | 3+3 2 | 3+3 2 |
| | 2+2 4 | 3+4 2 | 2+2 4 | 5+6 2 | 2+2 2 | 2+2 2 |
| | 2+2 4 | 5+6 2 | 2+2 4 | 5+6 2 | 3+3 2 | 3+3 2 |
| | 2+2 4 | 5+6 2 | 2+2 4 | 3+4 2 | 3+3 2 | 2+2 2 |
| | 2+2 4 | 3+4 2 | 2+2 4 | 3+4 2 | 2+2 2 | 3+3 2 |
| | 2+2 4 | 5+6 2 | 2+2 4 | 5+6 2 | 2+2 2 | 2+2 2 |
| | 2+2 4 | 5+6 2 | 2+2 4 | 3+4 2 | 3+3 2 | 3+3 2 |
| | 2+2 4 | 3+4 2 | 2+2 4 | 5+6 2 | 2+2 2 | 2+2 2 |
| | 2+2 4 | 3+4 2 | 2+2 4 | 5+6 2 | 3+3 2 | 2+2 2 |
| | 2+2 4 | 3+4 2 | 2+2 4 | 5+6 2 | 2+2 2 | 3+3 2 |
| | 2+2 4 | 5+6 2 | 2+2 4 | 3+4 2 | 3+3 2 | 3+3 2 |
| | 2+2 4 | 3+4 2 | 2+2 4 | 5+6 2 | 3+3 2 | 2+2 2 |
| | 2+2 4 | 5+6 2 | 2+2 4 | 3+4 2 | 2+2 2 | 2+2 2 |
| | 2+2 4 | 3+4 2 | 2+2 4 | 3+4 2 | 3+3 2 | 3+3 2 |
| | 2+2 4 | 3+4 2 | 2+2 4 | 3+4 2 | 3+3 2 | 2+2 2 |
| | 2+2 4 | 3+4 2 | 2+2 4 | 5+6 2 | 2+2 2 | 3+3 2 |
| | | 5+6 2 | | 3+4 2 | 2+2 2 | 2+2 2 |
| | | 5+6 2 | | 5+6 2 | 2+2 2 | 3+3 2 |
| | | 5+6 2 | | 3+4 2 | 3+3 2 | 2+2 2 |
| | | 3+4 2 | | 5+6 2 | 3+3 2 | 3+3 2 |
| | | 3+4 2 | | 3+4 2 | 3+3 2 | 2+2 2 |

Continued on next column

**TABLE XXVII– continued**

| TQC | ICSD | | NoE | | SF | |
|---|---|---|---|---|---|---|
| | | 5+6 2 | | 5+6 2 | 2+2 2 | 2+2 2 |
| | | 3+4 2 | | 3+4 2 | 3+3 2 | 3+3 2 |
| | | 3+4 2 | | 3+4 2 | 2+2 2 | 3+3 2 |
| | | 5+6 2 | | 5+6 2 | 2+2 2 | 2+2 2 |
| | | 3+4 2 | | 3+4 2 | 3+3 2 | 3+3 2 |
| | | 5+6 2 | | 5+6 2 | 2+2 2 | 2+2 2 |
| | | 3+4 2 | | 3+4 2 | 2+2 2 | 2+2 2 |
| | | 5+6 2 | | 3+4 2 | 3+3 2 | 3+3 2 |
| | | 5+6 2 | | 3+4 2 | 3+3 2 | 2+2 2 |
| | | 3+4 2 | | 5+6 2 | 2+2 2 | 3+3 2 |
| | | 5+6 2 | | 5+6 2 | 3+3 2 | 2+2 2 |
| | | 3+4 2 | | 3+4 2 | 2+2 2 | 2+2 2 |
| | | 3+4 2 | | 5+6 2 | 3+3 2 | 2+2 2 |
| | | 5+6 2 | | 5+6 2 | 2+2 2 | 3+3 2 |
| | | 5+6 2 | | 3+4 2 | 2+2 2 | 2+2 2 |
| | | 3+4 2 | | 3+4 2 | 3+3 2 | 2+2 2 |
| | | 5+6 2 | | 3+4 2 | 3+3 2 | 3+3 2 |
| | | 3+4 2 | | 5+6 2 | 2+2 2 | 2+2 2 |
| | | 3+4 2 | | 3+4 2 | 3+3 2 | 3+3 2 |
| | | 3+4 2 | | 3+4 2 | 2+2 2 | 2+2 2 |
| | | 5+6 2 | | 5+6 2 | 3+3 2 | 2+2 2 |
| | | 3+4 2 | | 3+4 2 | 2+2 2 | 2+2 2 |
| | | 3+4 2 | | 3+4 2 | 3+3 2 | 3+3 2 |
| | | 5+6 2 | | 3+4 2 | 2+2 2 | 2+2 2 |
| | | 5+6 2 | | 5+6 2 | 3+3 2 | 3+3 2 |
| | | 3+4 2 | | 5+6 2 | 2+2 2 | 2+2 2 |
| | | 3+4 2 | | 3+4 2 | 3+3 2 | 3+3 2 |
| | | 3+4 2 | | 5+6 2 | 2+2 2 | 3+3 2 |
| | | 5+6 2 | | 5+6 2 | 2+2 2 | 2+2 2 |
| | | 3+4 2 | | 3+4 2 | 3+3 2 | 3+3 2 |
| | | 5+6 2 | | 5+6 2 | 2+2 2 | 3+3 2 |
| | | 3+4 2 | | 3+4 2 | 2+2 2 | 2+2 2 |
| | | 3+4 2 | | 5+6 2 | 3+3 2 | 3+3 2 |
| | | 3+4 2 | | 3+4 2 | 3+3 2 | 3+3 2 |
| | | 5+6 2 | | 3+4 2 | 3+3 2 | 2+2 2 |
| | | 3+4 2 | | 3+4 2 | 2+2 2 | 2+2 2 |
| | | 5+6 2 | | 5+6 2 | 2+2 2 | 3+3 2 |
| | | 3+4 2 | | 5+6 2 | 3+3 2 | 2+2 2 |
| 15.2.58 | 23479 | | 148 | | Cu2O7V2 | |
| | -A | -GM | -M | -Y | -L | -V |
| | 2+2 4 | 3+4 2 | 2+2 4 | 5+6 2 | 2+2 2 | 3+3 2 |
| | 2+2 4 | 5+6 2 | 2+2 4 | 3+4 2 | 3+3 2 | 2+2 2 |
| | 2+2 4 | 3+4 2 | 2+2 4 | 5+6 2 | 2+2 2 | 3+3 2 |
| | 2+2 4 | 3+4 2 | 2+2 4 | 5+6 2 | 2+2 2 | 2+2 2 |
| | 2+2 4 | 5+6 2 | 2+2 4 | 3+4 2 | 3+3 2 | 3+3 2 |
| | 2+2 4 | 5+6 2 | 2+2 4 | 3+4 2 | 3+3 2 | 2+2 2 |
| | 2+2 4 | 3+4 2 | 2+2 4 | 3+4 2 | 3+3 2 | 3+3 2 |
| | 2+2 4 | 3+4 2 | 2+2 4 | 3+4 2 | 2+2 2 | 3+3 2 |
| | 2+2 4 | 5+6 2 | 2+2 4 | 5+6 2 | 3+3 2 | 2+2 2 |
| | 2+2 4 | 3+4 2 | 2+2 4 | 5+6 2 | 2+2 2 | 2+2 2 |
| | 2+2 4 | 5+6 2 | 2+2 4 | 3+4 2 | 2+2 2 | 3+3 2 |
| | 2+2 4 | 5+6 2 | 2+2 4 | 3+4 2 | 2+2 2 | 2+2 2 |
| | 2+2 4 | 5+6 2 | 2+2 4 | 5+6 2 | 3+3 2 | 3+3 2 |
| | 2+2 4 | 3+4 2 | 2+2 4 | 5+6 2 | 3+3 2 | 2+2 2 |
| | 2+2 4 | 3+4 2 | 2+2 4 | 5+6 2 | 2+2 2 | 2+2 2 |
| | 2+2 4 | 5+6 2 | 2+2 4 | 3+4 2 | 2+2 2 | 3+3 2 |
| | 2+2 4 | 5+6 2 | 2+2 4 | 5+6 2 | 3+3 2 | 2+2 2 |
| | 2+2 4 | 3+4 2 | 2+2 4 | 3+4 2 | 3+3 2 | 3+3 2 |
| | 2+2 4 | 3+4 2 | 2+2 4 | 5+6 2 | 3+3 2 | 2+2 2 |
| | 2+2 4 | 5+6 2 | 2+2 4 | 3+4 2 | 2+2 2 | 3+3 2 |
| | 2+2 4 | 3+4 2 | 2+2 4 | 5+6 2 | 3+3 2 | 3+3 2 |
| | 2+2 4 | 3+4 2 | 2+2 4 | 3+4 2 | 2+2 2 | 2+2 2 |
| | 2+2 4 | 5+6 2 | 2+2 4 | 5+6 2 | 2+2 2 | 3+3 2 |
| | 2+2 4 | 5+6 2 | 2+2 4 | 5+6 2 | 3+3 2 | 2+2 2 |
| | 2+2 4 | 3+4 2 | 2+2 4 | 3+4 2 | 2+2 2 | 3+3 2 |
| | 2+2 4 | 3+4 2 | 2+2 4 | 3+4 2 | 3+3 2 | 2+2 2 |
| | 2+2 4 | 3+4 2 | 2+2 4 | 3+4 2 | 2+2 2 | 3+3 2 |
| | 2+2 4 | 5+6 2 | 2+2 4 | 3+4 2 | 2+2 2 | 2+2 2 |
| | 2+2 4 | 3+4 2 | 2+2 4 | 5+6 2 | 2+2 2 | 3+3 2 |
| | 2+2 4 | 5+6 2 | 2+2 4 | 5+6 2 | 3+3 2 | 3+3 2 |
| | 2+2 4 | 3+4 2 | 2+2 4 | 3+4 2 | 3+3 2 | 2+2 2 |
| | 2+2 4 | 5+6 2 | 2+2 4 | 5+6 2 | 3+3 2 | 2+2 2 |
| | 2+2 4 | 3+4 2 | 2+2 4 | 5+6 2 | 2+2 2 | 3+3 2 |
| | 2+2 4 | 5+6 2 | 2+2 4 | 3+4 2 | 3+3 2 | 3+3 2 |
| | 2+2 4 | 3+4 2 | 2+2 4 | 5+6 2 | 2+2 2 | 3+3 2 |
| | 2+2 4 | 5+6 2 | 2+2 4 | 3+4 2 | 3+3 2 | 2+2 2 |
| | 2+2 4 | 3+4 2 | 2+2 4 | 3+4 2 | 2+2 2 | 2+2 2 |
| | | 3+4 2 | | 5+6 2 | 3+3 2 | 3+3 2 |
| | | 5+6 2 | | 5+6 2 | 3+3 2 | 2+2 2 |
| | | 5+6 2 | | 3+4 2 | 2+2 2 | 2+2 2 |
| | | 5+6 2 | | 5+6 2 | 3+3 2 | 2+2 2 |
| | | 3+4 2 | | 3+4 2 | 2+2 2 | 3+3 2 |
| | | 3+4 2 | | 5+6 2 | 2+2 2 | 3+3 2 |
| | | 5+6 2 | | 3+4 2 | 2+2 2 | 2+2 2 |
| | | 5+6 2 | | 3+4 2 | 3+3 2 | 3+3 2 |
| | | 5+6 2 | | 5+6 2 | 2+2 2 | 2+2 2 |
| | | 3+4 2 | | 3+4 2 | 3+3 2 | 3+3 2 |
| | | 5+6 2 | | 5+6 2 | 3+3 2 | 2+2 2 |
| | | 3+4 2 | | 5+6 2 | 2+2 2 | 2+2 2 |
| | | 5+6 2 | | 3+4 2 | 3+3 2 | 3+3 2 |
| | | 3+4 2 | | 5+6 2 | 2+2 2 | 3+3 2 |
| | | 3+4 2 | | 3+4 2 | 3+3 2 | 2+2 2 |
| | | 5+6 2 | | 5+6 2 | 3+3 2 | 2+2 2 |
| | | 5+6 2 | | 3+4 2 | 2+2 2 | 3+3 2 |
| | | 5+6 2 | | 5+6 2 | 2+2 2 | 2+2 2 |
| | | 3+4 2 | | 3+4 2 | 3+3 2 | 3+3 2 |
| | | 5+6 2 | | 5+6 2 | 3+3 2 | 3+3 2 |
| | | 3+4 2 | | 3+4 2 | 2+2 2 | 2+2 2 |
| | | 3+4 2 | | 5+6 2 | 2+2 2 | 3+3 2 |
| | | 5+6 2 | | 3+4 2 | 3+3 2 | 2+2 2 |
| | | 5+6 2 | | 3+4 2 | 2+2 2 | 3+3 2 |
| | | 3+4 2 | | 3+4 2 | 3+3 2 | 2+2 2 |
| | | 5+6 2 | | 5+6 2 | 2+2 2 | 3+3 2 |
| | | 3+4 2 | | 5+6 2 | 3+3 2 | 2+2 2 |
| | | 3+4 2 | | 5+6 2 | 2+2 2 | 3+3 2 |
| | | 3+4 2 | | 3+4 2 | 3+3 2 | 3+3 2 |
| | | 5+6 2 | | 3+4 2 | 2+2 2 | 2+2 2 |
| | | 5+6 2 | | 5+6 2 | 3+3 2 | 2+2 2 |
| | | 3+4 2 | | 5+6 2 | 2+2 2 | 2+2 2 |
| | | 5+6 2 | | 5+6 2 | 3+3 2 | 3+3 2 |
| | | 3+4 2 | | 3+4 2 | 2+2 2 | 2+2 2 |
| | | 5+6 2 | | 3+4 2 | 3+3 2 | 3+3 2 |
| | | 5+6 2 | | 3+4 2 | 3+3 2 | 3+3 2 |
| | | 3+4 2 | | 5+6 2 | 3+3 2 | 2+2 2 |
| 15.2.59 | 158375 | | 148 | | Cu2O7V2 | |
| | -A | -GM | -M | -Y | -L | -V |
| | 2+2 4 | 3+4 2 | 2+2 4 | 5+6 2 | 2+2 2 | 3+3 2 |
| | 2+2 4 | 3+4 2 | 2+2 4 | 5+6 2 | 2+2 2 | 2+2 2 |
| | 2+2 4 | 5+6 2 | 2+2 4 | 3+4 2 | 3+3 2 | 3+3 2 |
| | 2+2 4 | 3+4 2 | 2+2 4 | 3+4 2 | 3+3 2 | 2+2 2 |
| | 2+2 4 | 5+6 2 | 2+2 4 | 5+6 2 | 2+2 2 | 3+3 2 |
| | 2+2 4 | 3+4 2 | 2+2 4 | 3+4 2 | 3+3 2 | 2+2 2 |

Continued on next column

**TABLE XXVII– continued**

| TQC | ICSD | NoE | SF | | | |
|---|---|---|---|---|---|---|
| | 2+2 4 | 5+6 2 | 2+2 4 | 3+4 2 | 3+3 2 | 3+3 2 |
| | 2+2 4 | 3+4 2 | 2+2 4 | 3+4 2 | 2+2 2 | 3+3 2 |
| | 2+2 4 | 5+6 2 | 2+2 4 | 5+6 2 | 3+3 2 | 2+2 2 |
| | 2+2 4 | 5+6 2 | 2+2 4 | 5+6 2 | 2+2 2 | 2+2 2 |
| | 2+2 4 | 5+6 2 | 2+2 4 | 3+4 2 | 2+2 2 | 2+2 2 |
| | 2+2 4 | 5+6 2 | 2+2 4 | 3+4 2 | 2+2 2 | 3+3 2 |
| | 2+2 4 | 3+4 2 | 2+2 4 | 5+6 2 | 3+3 2 | 3+3 2 |
| | 2+2 4 | 3+4 2 | 2+2 4 | 5+6 2 | 3+3 2 | 2+2 2 |
| | 2+2 4 | 3+4 2 | 2+2 4 | 5+6 2 | 2+2 2 | 2+2 2 |
| | 2+2 4 | 5+6 2 | 2+2 4 | 3+4 2 | 2+2 2 | 3+3 2 |
| | 2+2 4 | 3+4 2 | 2+2 4 | 5+6 2 | 3+3 2 | 2+2 2 |
| | 2+2 4 | 5+6 2 | 2+2 4 | 3+4 2 | 3+3 2 | 3+3 2 |
| | 2+2 4 | 3+4 2 | 2+2 4 | 5+6 2 | 2+2 2 | 3+3 2 |
| | 2+2 4 | 3+4 2 | 2+2 4 | 3+4 2 | 3+3 2 | 2+2 2 |
| | 2+2 4 | 3+4 2 | 2+2 4 | 5+6 2 | 2+2 2 | 3+3 2 |
| | 2+2 4 | 3+4 2 | 2+2 4 | 3+4 2 | 3+3 2 | 3+3 2 |
| | 2+2 4 | 5+6 2 | 2+2 4 | 5+6 2 | 2+2 2 | 2+2 2 |
| | 2+2 4 | 3+4 2 | 2+2 4 | 5+6 2 | 2+2 2 | 2+2 2 |
| | 2+2 4 | 5+6 2 | 2+2 4 | 3+4 2 | 3+3 2 | 3+3 2 |
| | 2+2 4 | 5+6 2 | 2+2 4 | 5+6 2 | 3+3 2 | 2+2 2 |
| | 2+2 4 | 3+4 2 | 2+2 4 | 3+4 2 | 2+2 2 | 3+3 2 |
| | 2+2 4 | 3+4 2 | 2+2 4 | 3+4 2 | 3+3 2 | 3+3 2 |
| | 2+2 4 | 5+6 2 | 2+2 4 | 5+6 2 | 2+2 2 | 2+2 2 |
| | 2+2 4 | 3+4 2 | 2+2 4 | 3+4 2 | 2+2 2 | 2+2 2 |
| | 2+2 4 | 5+6 2 | 2+2 4 | 5+6 2 | 3+3 2 | 3+3 2 |
| | 2+2 4 | 5+6 2 | 2+2 4 | 3+4 2 | 2+2 2 | 2+2 2 |
| | 2+2 4 | 3+4 2 | 2+2 4 | 5+6 2 | 3+3 2 | 3+3 2 |
| | 2+2 4 | 5+6 2 | 2+2 4 | 3+4 2 | 2+2 2 | 3+3 2 |
| | 2+2 4 | 3+4 2 | 2+2 4 | 5+6 2 | 3+3 2 | 3+3 2 |
| | 2+2 4 | 5+6 2 | 2+2 4 | 3+4 2 | 3+3 2 | 2+2 2 |
| | 2+2 4 | 5+6 2 | 2+2 4 | 5+6 2 | 2+2 2 | 2+2 2 |
| | | 5+6 2 | | 5+6 2 | 3+3 2 | 2+2 2 |
| | | 3+4 2 | | 3+4 2 | 3+3 2 | 3+3 2 |
| | | 3+4 2 | | 5+6 2 | 2+2 2 | 2+2 2 |
| | | 5+6 2 | | 3+4 2 | 3+3 2 | 2+2 2 |
| | | 5+6 2 | | 3+4 2 | 2+2 2 | 3+3 2 |
| | | 3+4 2 | | 5+6 2 | 2+2 2 | 2+2 2 |
| | | 3+4 2 | | 3+4 2 | 3+3 2 | 3+3 2 |
| | | 5+6 2 | | 3+4 2 | 2+2 2 | 3+3 2 |
| | | 5+6 2 | | 3+4 2 | 3+3 2 | 3+3 2 |
| | | 3+4 2 | | 5+6 2 | 3+3 2 | 2+2 2 |
| | | 5+6 2 | | 3+4 2 | 2+2 2 | 2+2 2 |
| | | 3+4 2 | | 5+6 2 | 3+3 2 | 3+3 2 |
| | | 3+4 2 | | 5+6 2 | 2+2 2 | 2+2 2 |
| | | 5+6 2 | | 5+6 2 | 3+3 2 | 3+3 2 |
| | | 5+6 2 | | 3+4 2 | 2+2 2 | 2+2 2 |
| | | 3+4 2 | | 5+6 2 | 2+2 2 | 2+2 2 |
| | | 5+6 2 | | 3+4 2 | 3+3 2 | 3+3 2 |
| | | 5+6 2 | | 3+4 2 | 3+3 2 | 2+2 2 |
| | | 3+4 2 | | 5+6 2 | 2+2 2 | 3+3 2 |
| | | 3+4 2 | | 5+6 2 | 3+3 2 | 2+2 2 |
| | | 5+6 2 | | 3+4 2 | 2+2 2 | 3+3 2 |
| | | 3+4 2 | | 5+6 2 | 3+3 2 | 2+2 2 |
| | | 5+6 2 | | 3+4 2 | 2+2 2 | 3+3 2 |
| | | 5+6 2 | | 3+4 2 | 3+3 2 | 3+3 2 |
| | | 3+4 2 | | 3+4 2 | 2+2 2 | 2+2 2 |
| | | 3+4 2 | | 5+6 2 | 2+2 2 | 3+3 2 |
| | | 5+6 2 | | 5+6 2 | 3+3 2 | 2+2 2 |
| | | 3+4 2 | | 3+4 2 | 2+2 2 | 2+2 2 |
| | | 5+6 2 | | 5+6 2 | 3+3 2 | 3+3 2 |
| | | 3+4 2 | | 3+4 2 | 2+2 2 | 3+3 2 |
| | | 5+6 2 | | 5+6 2 | 3+3 2 | 2+2 2 |
| | | 3+4 2 | | 5+6 2 | 2+2 2 | 2+2 2 |
| | | 5+6 2 | | 3+4 2 | 2+2 2 | 3+3 2 |
| | | 3+4 2 | | 5+6 2 | 3+3 2 | 3+3 2 |
| | | 5+6 2 | | 3+4 2 | 3+3 2 | 2+2 2 |
| | | 5+6 2 | | 3+4 2 | 3+3 2 | 3+3 2 |
| | | 3+4 2 | | 5+6 2 | 3+3 2 | 2+2 2 |
| 15.2.60 | 67316 | | 148 | Cu2O7P2 | | |
| | -A | -GM | -M | -Y | -L | -V |
| | 2+2 4 | 3+4 2 | 2+2 4 | 3+4 2 | 2+2 2 | 2+2 2 |
| | 2+2 4 | 5+6 2 | 2+2 4 | 5+6 2 | 3+3 2 | 3+3 2 |
| | 2+2 4 | 3+4 2 | 2+2 4 | 3+4 2 | 2+2 2 | 3+3 2 |
| | 2+2 4 | 5+6 2 | 2+2 4 | 5+6 2 | 3+3 2 | 2+2 2 |
| | 2+2 4 | 3+4 2 | 2+2 4 | 3+4 2 | 2+2 2 | 3+3 2 |
| | 2+2 4 | 5+6 2 | 2+2 4 | 5+6 2 | 3+3 2 | 2+2 2 |
| | 2+2 4 | 3+4 2 | 2+2 4 | 3+4 2 | 2+2 2 | 3+3 2 |
| | 2+2 4 | 3+4 2 | 2+2 4 | 3+4 2 | 3+3 2 | 2+2 2 |
| | 2+2 4 | 5+6 2 | 2+2 4 | 5+6 2 | 2+2 2 | 3+3 2 |
| | 2+2 4 | 5+6 2 | 2+2 4 | 5+6 2 | 3+3 2 | 2+2 2 |
| | 2+2 4 | 5+6 2 | 2+2 4 | 5+6 2 | 2+2 2 | 3+3 2 |
| | 2+2 4 | 3+4 2 | 2+2 4 | 3+4 2 | 3+3 2 | 3+3 2 |
| | 2+2 4 | 5+6 2 | 2+2 4 | 5+6 2 | 2+2 2 | 2+2 2 |
| | 2+2 4 | 3+4 2 | 2+2 4 | 3+4 2 | 3+3 2 | 2+2 2 |
| | 2+2 4 | 3+4 2 | 2+2 4 | 3+4 2 | 2+2 2 | 3+3 2 |
| | 2+2 4 | 5+6 2 | 2+2 4 | 5+6 2 | 3+3 2 | 2+2 2 |
| | 2+2 4 | 3+4 2 | 2+2 4 | 3+4 2 | 2+2 2 | 3+3 2 |
| | 2+2 4 | 3+4 2 | 2+2 4 | 3+4 2 | 3+3 2 | 2+2 2 |
| | 2+2 4 | 5+6 2 | 2+2 4 | 5+6 2 | 2+2 2 | 3+3 2 |
| | 2+2 4 | 5+6 2 | 2+2 4 | 5+6 2 | 3+3 2 | 2+2 2 |
| | 2+2 4 | 3+4 2 | 2+2 4 | 5+6 2 | 2+2 2 | 3+3 2 |
| | 2+2 4 | 3+4 2 | 2+2 4 | 3+4 2 | 3+3 2 | 3+3 2 |
| | 2+2 4 | 5+6 2 | 2+2 4 | 3+4 2 | 2+2 2 | 2+2 2 |
| | 2+2 4 | 3+4 2 | 2+2 4 | 5+6 2 | 2+2 2 | 3+3 2 |
| | 2+2 4 | 3+4 2 | 2+2 4 | 5+6 2 | 3+3 2 | 2+2 2 |
| | 2+2 4 | 5+6 2 | 2+2 4 | 3+4 2 | 3+3 2 | 2+2 2 |
| | 2+2 4 | 5+6 2 | 2+2 4 | 3+4 2 | 2+2 2 | 2+2 2 |
| | 2+2 4 | 5+6 2 | 2+2 4 | 5+6 2 | 3+3 2 | 3+3 2 |
| | 2+2 4 | 5+6 2 | 2+2 4 | 5+6 2 | 2+2 2 | 3+3 2 |
| | 2+2 4 | 5+6 2 | 2+2 4 | 5+6 2 | 2+2 2 | 2+2 2 |
| | 2+2 4 | 3+4 2 | 2+2 4 | 3+4 2 | 3+3 2 | 3+3 2 |
| | 2+2 4 | 3+4 2 | 2+2 4 | 5+6 2 | 3+3 2 | 3+3 2 |
| | 2+2 4 | 5+6 2 | 2+2 4 | 3+4 2 | 2+2 2 | 2+2 2 |
| | 2+2 4 | 3+4 2 | 2+2 4 | 3+4 2 | 3+3 2 | 3+3 2 |
| | 2+2 4 | 3+4 2 | 2+2 4 | 5+6 2 | 2+2 2 | 2+2 2 |
| | 2+2 4 | 3+4 2 | 2+2 4 | 3+4 2 | 3+3 2 | 2+2 2 |
| | 2+2 4 | 5+6 2 | 2+2 4 | 3+4 2 | 2+2 2 | 2+2 2 |
| | | 3+4 2 | | 5+6 2 | 3+3 2 | 3+3 2 |
| | | 5+6 2 | | 3+4 2 | 2+2 2 | 2+2 2 |
| | | 3+4 2 | | 5+6 2 | 3+3 2 | 2+2 2 |
| | | 5+6 2 | | 3+4 2 | 2+2 2 | 3+3 2 |
| | | 5+6 2 | | 5+6 2 | 3+3 2 | 2+2 2 |
| | | 3+4 2 | | 3+4 2 | 3+3 2 | 3+3 2 |
| | | 3+4 2 | | 5+6 2 | 2+2 2 | 3+3 2 |
| | | 3+4 2 | | 3+4 2 | 2+2 2 | 2+2 2 |
| | | 5+6 2 | | 5+6 2 | 3+3 2 | 2+2 2 |
| | | 5+6 2 | | 5+6 2 | 2+2 2 | 3+3 2 |
| | | 5+6 2 | | 3+4 2 | 3+3 2 | 3+3 2 |
| | | 5+6 2 | | 5+6 2 | 3+3 2 | 3+3 2 |
| | | 3+4 2 | | 5+6 2 | 2+2 2 | 3+3 2 |
| | | 3+4 2 | | 3+4 2 | 3+3 2 | 2+2 2 |
| | | 5+6 2 | | 5+6 2 | 2+2 2 | 3+3 2 |

Continued on next column

**TABLE XXVII– continued**

| TQC | ICSD | NoE | SF |
|---|---|---|---|
| | 5+6 2 | | 5+6 2 3+3 2 2+2 2 |
| | 3+4 2 | | 3+4 2 2+2 2 3+3 2 |
| | 5+6 2 | | 3+4 2 3+3 2 2+2 2 |
| | 3+4 2 | | 3+4 2 2+2 2 2+2 2 |
| | 3+4 2 | | 3+4 2 3+3 2 2+2 2 |
| | 3+4 2 | | 3+4 2 2+2 2 2+2 2 |
| | 5+6 2 | | 5+6 2 2+2 2 3+3 2 |
| | 5+6 2 | | 5+6 2 3+3 2 2+2 2 |
| | 3+4 2 | | 3+4 2 3+3 2 3+3 2 |
| | 3+4 2 | | 5+6 2 2+2 2 2+2 2 |
| | 5+6 2 | | 3+4 2 3+3 2 3+3 2 |
| | 5+6 2 | | 5+6 2 2+2 2 3+3 2 |
| | 3+4 2 | | 5+6 2 2+2 2 2+2 2 |
| | 5+6 2 | | 3+4 2 3+3 2 3+3 2 |
| | 3+4 2 | | 3+4 2 2+2 2 2+2 2 |
| | 5+6 2 | | 5+6 2 3+3 2 3+3 2 |
| | 5+6 2 | | 3+4 2 2+2 2 2+2 2 |
| | 5+6 2 | | 3+4 2 3+3 2 3+3 2 |
| | 3+4 2 | | 5+6 2 2+2 2 3+3 2 |
| | 5+6 2 | | 5+6 2 3+3 2 2+2 2 |
| | 3+4 2 | | 5+6 2 2+2 2 2+2 2 |
| | 3+4 2 | | 3+4 2 3+3 2 2+2 2 |
| 15.2.61 | 28279 | 148 | Cu2O7P2 |
| | -A -GM | -M | -Y -L -V |
| | 2+2 4 3+4 2 | 2+2 4 | 3+4 2 3+3 2 2+2 2 |
| | 2+2 4 5+6 2 | 2+2 4 | 5+6 2 2+2 2 3+3 2 |
| | 2+2 4 3+4 2 | 2+2 4 | 3+4 2 3+3 2 3+3 2 |
| | 2+2 4 5+6 2 | 2+2 4 | 5+6 2 2+2 2 2+2 2 |
| | 2+2 4 3+4 2 | 2+2 4 | 5+6 2 3+3 2 2+2 2 |
| | 2+2 4 5+6 2 | 2+2 4 | 3+4 2 2+2 2 3+3 2 |
| | 2+2 4 3+4 2 | 2+2 4 | 3+4 2 2+2 2 3+3 2 |
| | 2+2 4 3+4 2 | 2+2 4 | 3+4 2 3+3 2 3+3 2 |
| | 2+2 4 5+6 2 | 2+2 4 | 5+6 2 3+3 2 2+2 2 |
| | 2+2 4 5+6 2 | 2+2 4 | 5+6 2 2+2 2 2+2 2 |
| | 2+2 4 3+4 2 | 2+2 4 | 5+6 2 2+2 2 3+3 2 |
| | 2+2 4 5+6 2 | 2+2 4 | 3+4 2 3+3 2 3+3 2 |
| | 2+2 4 3+4 2 | 2+2 4 | 5+6 2 2+2 2 2+2 2 |
| | 2+2 4 5+6 2 | 2+2 4 | 3+4 2 3+3 2 2+2 2 |
| | 2+2 4 3+4 2 | 2+2 4 | 3+4 2 3+3 2 3+3 2 |
| | 2+2 4 5+6 2 | 2+2 4 | 5+6 2 2+2 2 2+2 2 |
| | 2+2 4 3+4 2 | 2+2 4 | 3+4 2 3+3 2 3+3 2 |
| | 2+2 4 3+4 2 | 2+2 4 | 3+4 2 2+2 2 2+2 2 |
| | 2+2 4 5+6 2 | 2+2 4 | 5+6 2 3+3 2 3+3 2 |
| | 2+2 4 5+6 2 | 2+2 4 | 5+6 2 2+2 2 2+2 2 |
| | 2+2 4 3+4 2 | 2+2 4 | 5+6 2 3+3 2 3+3 2 |
| | 2+2 4 3+4 2 | 2+2 4 | 3+4 2 2+2 2 3+3 2 |
| | 2+2 4 3+4 2 | 2+2 4 | 3+4 2 3+3 2 2+2 2 |
| | 2+2 4 5+6 2 | 2+2 4 | 5+6 2 2+2 2 3+3 2 |
| | 2+2 4 3+4 2 | 2+2 4 | 5+6 2 3+3 2 2+2 2 |
| | 2+2 4 5+6 2 | 2+2 4 | 3+4 2 2+2 2 2+2 2 |
| | 2+2 4 5+6 2 | 2+2 4 | 3+4 2 3+3 2 2+2 2 |
| | 2+2 4 5+6 2 | 2+2 4 | 5+6 2 2+2 2 3+3 2 |
| | 2+2 4 5+6 2 | 2+2 4 | 5+6 2 2+2 2 3+3 2 |
| | 2+2 4 5+6 2 | 2+2 4 | 5+6 2 3+3 2 3+3 2 |
| | 2+2 4 3+4 2 | 2+2 4 | 3+4 2 3+3 2 2+2 2 |
| | 2+2 4 3+4 2 | 2+2 4 | 5+6 2 2+2 2 3+3 2 |
| | 2+2 4 5+6 2 | 2+2 4 | 3+4 2 2+2 2 2+2 2 |
| | 2+2 4 3+4 2 | 2+2 4 | 3+4 2 3+3 2 2+2 2 |
| | 2+2 4 3+4 2 | 2+2 4 | 5+6 2 3+3 2 3+3 2 |
| | 2+2 4 5+6 2 | 2+2 4 | 3+4 2 2+2 2 2+2 2 |
| | 2+2 4 3+4 2 | 2+2 4 | 3+4 2 3+3 2 3+3 2 |

Continued on next column

**TABLE XXVII– continued**

| TQC | ICSD | NoE | SF |
|---|---|---|---|
| | 3+4 2 | | 5+6 2 2+2 2 2+2 2 |
| | 3+4 2 | | 3+4 2 3+3 2 3+3 2 |
| | 5+6 2 | | 5+6 2 2+2 2 2+2 2 |
| | 5+6 2 | | 3+4 2 3+3 2 2+2 2 |
| | 3+4 2 | | 3+4 2 2+2 2 2+2 2 |
| | 5+6 2 | | 5+6 2 2+2 2 3+3 2 |
| | 5+6 2 | | 5+6 2 3+3 2 3+3 2 |
| | 3+4 2 | | 3+4 2 2+2 2 2+2 2 |
| | 3+4 2 | | 5+6 2 3+3 2 3+3 2 |
| | 5+6 2 | | 5+6 2 3+3 2 2+2 2 |
| | 5+6 2 | | 5+6 2 2+2 2 3+3 2 |
| | 5+6 2 | | 3+4 2 2+2 2 3+3 2 |
| | 3+4 2 | | 5+6 2 3+3 2 3+3 2 |
| | 3+4 2 | | 3+4 2 2+2 2 2+2 2 |
| | 5+6 2 | | 5+6 2 3+3 2 2+2 2 |
| | 5+6 2 | | 5+6 2 2+2 2 3+3 2 |
| | 3+4 2 | | 3+4 2 3+3 2 3+3 2 |
| | 5+6 2 | | 3+4 2 2+2 2 2+2 2 |
| | 3+4 2 | | 3+4 2 3+3 2 2+2 2 |
| | 3+4 2 | | 3+4 2 2+2 2 2+2 2 |
| | 3+4 2 | | 3+4 2 3+3 2 2+2 2 |
| | 5+6 2 | | 5+6 2 3+3 2 3+3 2 |
| | 5+6 2 | | 5+6 2 2+2 2 2+2 2 |
| | 3+4 2 | | 3+4 2 3+3 2 3+3 2 |
| | 5+6 2 | | 5+6 2 2+2 2 2+2 2 |
| | 3+4 2 | | 5+6 2 3+3 2 3+3 2 |
| | 5+6 2 | | 3+4 2 2+2 2 3+3 2 |
| | 3+4 2 | | 5+6 2 3+3 2 2+2 2 |
| | 5+6 2 | | 3+4 2 2+2 2 3+3 2 |
| | 3+4 2 | | 3+4 2 2+2 2 2+2 2 |
| | 5+6 2 | | 5+6 2 3+3 2 3+3 2 |
| | 5+6 2 | | 3+4 2 3+3 2 2+2 2 |
| | 5+6 2 | | 3+4 2 3+3 2 3+3 2 |
| | 3+4 2 | | 5+6 2 2+2 2 3+3 2 |
| | 5+6 2 | | 5+6 2 2+2 2 2+2 2 |
| | 3+4 2 | | 5+6 2 2+2 2 2+2 2 |
| | 3+4 2 | | 3+4 2 3+3 2 2+2 2 |
| 15.2.62 | 157107 | 148 | Cu2O7P2 |
| | -A -GM | -M | -Y -L -V |
| | 2+2 4 3+4 2 | 2+2 4 | 3+4 2 2+2 2 2+2 2 |
| | 2+2 4 5+6 2 | 2+2 4 | 5+6 2 3+3 2 3+3 2 |
| | 2+2 4 3+4 2 | 2+2 4 | 3+4 2 2+2 2 3+3 2 |
| | 2+2 4 5+6 2 | 2+2 4 | 5+6 2 3+3 2 2+2 2 |
| | 2+2 4 3+4 2 | 2+2 4 | 3+4 2 2+2 2 3+3 2 |
| | 2+2 4 5+6 2 | 2+2 4 | 5+6 2 3+3 2 2+2 2 |
| | 2+2 4 3+4 2 | 2+2 4 | 3+4 2 2+2 2 3+3 2 |
| | 2+2 4 3+4 2 | 2+2 4 | 3+4 2 3+3 2 2+2 2 |
| | 2+2 4 5+6 2 | 2+2 4 | 5+6 2 2+2 2 3+3 2 |
| | 2+2 4 5+6 2 | 2+2 4 | 5+6 2 3+3 2 2+2 2 |
| | 2+2 4 5+6 2 | 2+2 4 | 5+6 2 2+2 2 3+3 2 |
| | 2+2 4 3+4 2 | 2+2 4 | 3+4 2 3+3 2 3+3 2 |
| | 2+2 4 5+6 2 | 2+2 4 | 5+6 2 2+2 2 2+2 2 |
| | 2+2 4 3+4 2 | 2+2 4 | 3+4 2 3+3 2 2+2 2 |
| | 2+2 4 3+4 2 | 2+2 4 | 3+4 2 2+2 2 3+3 2 |
| | 2+2 4 5+6 2 | 2+2 4 | 5+6 2 3+3 2 2+2 2 |
| | 2+2 4 3+4 2 | 2+2 4 | 3+4 2 2+2 2 3+3 2 |
| | 2+2 4 3+4 2 | 2+2 4 | 3+4 2 3+3 2 2+2 2 |
| | 2+2 4 5+6 2 | 2+2 4 | 5+6 2 2+2 2 3+3 2 |
| | 2+2 4 5+6 2 | 2+2 4 | 5+6 2 3+3 2 2+2 2 |
| | 2+2 4 3+4 2 | 2+2 4 | 5+6 2 2+2 2 3+3 2 |
| | 2+2 4 3+4 2 | 2+2 4 | 3+4 2 3+3 2 3+3 2 |

Continued on next column

**TABLE XXVII– continued**

| TQC | ICSD | | NoE | SF | | |
|---|---|---|---|---|---|---|
| | 2+2 4 | 5+6 2 | 2+2 4 | 3+4 2 | 2+2 2 | 2+2 2 |
| | 2+2 4 | 3+4 2 | 2+2 4 | 5+6 2 | 2+2 2 | 3+3 2 |
| | 2+2 4 | 3+4 2 | 2+2 4 | 5+6 2 | 3+3 2 | 2+2 2 |
| | 2+2 4 | 5+6 2 | 2+2 4 | 3+4 2 | 3+3 2 | 2+2 2 |
| | 2+2 4 | 5+6 2 | 2+2 4 | 3+4 2 | 2+2 2 | 2+2 2 |
| | 2+2 4 | 5+6 2 | 2+2 4 | 5+6 2 | 3+3 2 | 3+3 2 |
| | 2+2 4 | 5+6 2 | 2+2 4 | 5+6 2 | 2+2 2 | 3+3 2 |
| | 2+2 4 | 5+6 2 | 2+2 4 | 5+6 2 | 2+2 2 | 2+2 2 |
| | 2+2 4 | 3+4 2 | 2+2 4 | 3+4 2 | 3+3 2 | 3+3 2 |
| | 2+2 4 | 3+4 2 | 2+2 4 | 5+6 2 | 3+3 2 | 3+3 2 |
| | 2+2 4 | 5+6 2 | 2+2 4 | 3+4 2 | 2+2 2 | 2+2 2 |
| | 2+2 4 | 3+4 2 | 2+2 4 | 3+4 2 | 3+3 2 | 3+3 2 |
| | 2+2 4 | 3+4 2 | 2+2 4 | 5+6 2 | 2+2 2 | 2+2 2 |
| | 2+2 4 | 3+4 2 | 2+2 4 | 3+4 2 | 3+3 2 | 2+2 2 |
| | 2+2 4 | 5+6 2 | 2+2 4 | 3+4 2 | 2+2 2 | 2+2 2 |
| | | 3+4 2 | | 5+6 2 | 3+3 2 | 3+3 2 |
| | | 5+6 2 | | 3+4 2 | 2+2 2 | 2+2 2 |
| | | 3+4 2 | | 5+6 2 | 3+3 2 | 2+2 2 |
| | | 5+6 2 | | 3+4 2 | 2+2 2 | 3+3 2 |
| | | 5+6 2 | | 5+6 2 | 3+3 2 | 2+2 2 |
| | | 3+4 2 | | 3+4 2 | 3+3 2 | 3+3 2 |
| | | 3+4 2 | | 5+6 2 | 2+2 2 | 3+3 2 |
| | | 3+4 2 | | 3+4 2 | 2+2 2 | 2+2 2 |
| | | 5+6 2 | | 5+6 2 | 3+3 2 | 2+2 2 |
| | | 5+6 2 | | 5+6 2 | 2+2 2 | 3+3 2 |
| | | 5+6 2 | | 3+4 2 | 3+3 2 | 3+3 2 |
| | | 5+6 2 | | 5+6 2 | 3+3 2 | 3+3 2 |
| | | 3+4 2 | | 5+6 2 | 2+2 2 | 3+3 2 |
| | | 3+4 2 | | 3+4 2 | 3+3 2 | 2+2 2 |
| | | 5+6 2 | | 5+6 2 | 2+2 2 | 3+3 2 |
| | | 5+6 2 | | 5+6 2 | 3+3 2 | 2+2 2 |
| | | 3+4 2 | | 3+4 2 | 2+2 2 | 3+3 2 |
| | | 5+6 2 | | 3+4 2 | 3+3 2 | 2+2 2 |
| | | 3+4 2 | | 3+4 2 | 2+2 2 | 2+2 2 |
| | | 3+4 2 | | 3+4 2 | 3+3 2 | 2+2 2 |
| | | 3+4 2 | | 3+4 2 | 2+2 2 | 2+2 2 |
| | | 5+6 2 | | 5+6 2 | 3+3 2 | 3+3 2 |
| | | 5+6 2 | | 5+6 2 | 2+2 2 | 2+2 2 |
| | | 3+4 2 | | 3+4 2 | 3+3 2 | 3+3 2 |
| | | 3+4 2 | | 3+4 2 | 2+2 2 | 2+2 2 |
| | | 5+6 2 | | 5+6 2 | 3+3 2 | 3+3 2 |
| | | 5+6 2 | | 5+6 2 | 2+2 2 | 3+3 2 |
| | | 3+4 2 | | 5+6 2 | 2+2 2 | 2+2 2 |
| | | 5+6 2 | | 3+4 2 | 3+3 2 | 3+3 2 |
| | | 3+4 2 | | 3+4 2 | 2+2 2 | 2+2 2 |
| | | 5+6 2 | | 5+6 2 | 3+3 2 | 3+3 2 |
| | | 5+6 2 | | 3+4 2 | 2+2 2 | 2+2 2 |
| | | 5+6 2 | | 3+4 2 | 3+3 2 | 3+3 2 |
| | | 3+4 2 | | 5+6 2 | 2+2 2 | 3+3 2 |
| | | 5+6 2 | | 5+6 2 | 3+3 2 | 2+2 2 |
| | | 3+4 2 | | 5+6 2 | 2+2 2 | 2+2 2 |
| | | 3+4 2 | | 3+4 2 | 3+3 2 | 2+2 2 |
| 15.2.63 | 14369 | | 148 | Cu2O7P2 | | |
| | -A | -GM | -M | -Y | -L | -V |
| | 2+2 4 | 3+4 2 | 2+2 4 | 3+4 2 | 3+3 2 | 2+2 2 |
| | 2+2 4 | 5+6 2 | 2+2 4 | 5+6 2 | 2+2 2 | 3+3 2 |
| | 2+2 4 | 3+4 2 | 2+2 4 | 3+4 2 | 3+3 2 | 3+3 2 |
| | 2+2 4 | 5+6 2 | 2+2 4 | 5+6 2 | 2+2 2 | 2+2 2 |
| | 2+2 4 | 3+4 2 | 2+2 4 | 3+4 2 | 3+3 2 | 3+3 2 |
| | 2+2 4 | 5+6 2 | 2+2 4 | 5+6 2 | 2+2 2 | 2+2 2 |
| | 2+2 4 | 3+4 2 | 2+2 4 | 3+4 2 | 3+3 2 | 3+3 2 |

Continued on next column

**TABLE XXVII– continued**

| TQC | ICSD | | NoE | SF | | |
|---|---|---|---|---|---|---|
| | 2+2 4 | 3+4 2 | 2+2 4 | 5+6 2 | 2+2 2 | 2+2 2 |
| | 2+2 4 | 5+6 2 | 2+2 4 | 3+4 2 | 3+3 2 | 3+3 2 |
| | 2+2 4 | 5+6 2 | 2+2 4 | 5+6 2 | 2+2 2 | 2+2 2 |
| | 2+2 4 | 5+6 2 | 2+2 4 | 5+6 2 | 2+2 2 | 3+3 2 |
| | 2+2 4 | 3+4 2 | 2+2 4 | 3+4 2 | 3+3 2 | 3+3 2 |
| | 2+2 4 | 5+6 2 | 2+2 4 | 5+6 2 | 3+3 2 | 2+2 2 |
| | 2+2 4 | 3+4 2 | 2+2 4 | 3+4 2 | 2+2 2 | 2+2 2 |
| | 2+2 4 | 3+4 2 | 2+2 4 | 3+4 2 | 3+3 2 | 3+3 2 |
| | 2+2 4 | 5+6 2 | 2+2 4 | 5+6 2 | 2+2 2 | 2+2 2 |
| | 2+2 4 | 3+4 2 | 2+2 4 | 3+4 2 | 3+3 2 | 3+3 2 |
| | 2+2 4 | 3+4 2 | 2+2 4 | 3+4 2 | 2+2 2 | 2+2 2 |
| | 2+2 4 | 5+6 2 | 2+2 4 | 5+6 2 | 3+3 2 | 3+3 2 |
| | 2+2 4 | 5+6 2 | 2+2 4 | 5+6 2 | 2+2 2 | 2+2 2 |
| | 2+2 4 | 3+4 2 | 2+2 4 | 5+6 2 | 3+3 2 | 3+3 2 |
| | 2+2 4 | 3+4 2 | 2+2 4 | 3+4 2 | 2+2 2 | 3+3 2 |
| | 2+2 4 | 5+6 2 | 2+2 4 | 3+4 2 | 3+3 2 | 2+2 2 |
| | 2+2 4 | 3+4 2 | 2+2 4 | 5+6 2 | 3+3 2 | 3+3 2 |
| | 2+2 4 | 3+4 2 | 2+2 4 | 5+6 2 | 2+2 2 | 2+2 2 |
| | 2+2 4 | 5+6 2 | 2+2 4 | 3+4 2 | 2+2 2 | 2+2 2 |
| | 2+2 4 | 5+6 2 | 2+2 4 | 3+4 2 | 3+3 2 | 2+2 2 |
| | 2+2 4 | 5+6 2 | 2+2 4 | 5+6 2 | 2+2 2 | 3+3 2 |
| | 2+2 4 | 5+6 2 | 2+2 4 | 5+6 2 | 3+3 2 | 3+3 2 |
| | 2+2 4 | 5+6 2 | 2+2 4 | 5+6 2 | 3+3 2 | 2+2 2 |
| | 2+2 4 | 3+4 2 | 2+2 4 | 3+4 2 | 2+2 2 | 3+3 2 |
| | 2+2 4 | 3+4 2 | 2+2 4 | 5+6 2 | 2+2 2 | 3+3 2 |
| | 2+2 4 | 5+6 2 | 2+2 4 | 3+4 2 | 3+3 2 | 2+2 2 |
| | 2+2 4 | 3+4 2 | 2+2 4 | 3+4 2 | 2+2 2 | 3+3 2 |
| | 2+2 4 | 3+4 2 | 2+2 4 | 3+4 2 | 2+2 2 | 2+2 2 |
| | 2+2 4 | 3+4 2 | 2+2 4 | 5+6 2 | 3+3 2 | 2+2 2 |
| | 2+2 4 | 5+6 2 | 2+2 4 | 3+4 2 | 3+3 2 | 2+2 2 |
| | | 3+4 2 | | 5+6 2 | 2+2 2 | 3+3 2 |
| | | 5+6 2 | | 3+4 2 | 3+3 2 | 2+2 2 |
| | | 3+4 2 | | 5+6 2 | 2+2 2 | 2+2 2 |
| | | 5+6 2 | | 3+4 2 | 3+3 2 | 3+3 2 |
| | | 5+6 2 | | 5+6 2 | 2+2 2 | 2+2 2 |
| | | 3+4 2 | | 5+6 2 | 2+2 2 | 3+3 2 |
| | | 3+4 2 | | 3+4 2 | 3+3 2 | 3+3 2 |
| | | 3+4 2 | | 3+4 2 | 2+2 2 | 2+2 2 |
| | | 5+6 2 | | 5+6 2 | 3+3 2 | 2+2 2 |
| | | 5+6 2 | | 5+6 2 | 3+3 2 | 3+3 2 |
| | | 5+6 2 | | 3+4 2 | 2+2 2 | 3+3 2 |
| | | 5+6 2 | | 5+6 2 | 2+2 2 | 3+3 2 |
| | | 3+4 2 | | 5+6 2 | 3+3 2 | 3+3 2 |
| | | 3+4 2 | | 3+4 2 | 2+2 2 | 2+2 2 |
| | | 5+6 2 | | 3+4 2 | 3+3 2 | 3+3 2 |
| | | 3+4 2 | | 5+6 2 | 2+2 2 | 2+2 2 |
| | | 5+6 2 | | 5+6 2 | 3+3 2 | 3+3 2 |
| | | 5+6 2 | | 3+4 2 | 2+2 2 | 2+2 2 |
| | | 3+4 2 | | 3+4 2 | 3+3 2 | 2+2 2 |
| | | 3+4 2 | | 3+4 2 | 2+2 2 | 2+2 2 |
| | | 3+4 2 | | 3+4 2 | 3+3 2 | 2+2 2 |
| | | 5+6 2 | | 5+6 2 | 3+3 2 | 3+3 2 |
| | | 5+6 2 | | 5+6 2 | 2+2 2 | 2+2 2 |
| | | 3+4 2 | | 3+4 2 | 3+3 2 | 3+3 2 |
| | | 3+4 2 | | 5+6 2 | 2+2 2 | 2+2 2 |
| | | 5+6 2 | | 3+4 2 | 3+3 2 | 3+3 2 |
| | | 5+6 2 | | 5+6 2 | 2+2 2 | 3+3 2 |
| | | 3+4 2 | | 5+6 2 | 3+3 2 | 2+2 2 |
| | | 5+6 2 | | 3+4 2 | 2+2 2 | 3+3 2 |
| | | 3+4 2 | | 3+4 2 | 2+2 2 | 2+2 2 |
| | | 5+6 2 | | 5+6 2 | 3+3 2 | 3+3 2 |

Continued on next column

**TABLE XXVII– continued**

| TQC | -A | -GM | -M | -Y | -L | -V |
|---|---|---|---|---|---|---|
| | | 5+6 2 | | 3+4 2 | 3+3 2 | 2+2 2 |
| | | 5+6 2 | | 3+4 2 | 2+2 2 | 3+3 2 |
| | | 3+4 2 | | 5+6 2 | 3+3 2 | 3+3 2 |
| | | 5+6 2 | | 5+6 2 | 2+2 2 | 2+2 2 |
| | | 3+4 2 | | 5+6 2 | 3+3 2 | 2+2 2 |
| | | 3+4 2 | | 3+4 2 | 2+2 2 | 2+2 2 |
| 15.2.64 | 83490 | | 128 | Cu2Gd1O4 | | |
| | -A | -GM | -M | -Y | -L | -V |
| | 2+2 4 | 3+4 2 | 2+2 4 | 3+4 2 | 2+2 2 | 2+2 2 |
| | 2+2 4 | 5+6 2 | 2+2 4 | 5+6 2 | 3+3 2 | 3+3 2 |
| | 2+2 4 | 3+4 2 | 2+2 4 | 3+4 2 | 2+2 2 | 2+2 2 |
| | 2+2 4 | 5+6 2 | 2+2 4 | 5+6 2 | 3+3 2 | 3+3 2 |
| | 2+2 4 | 3+4 2 | 2+2 4 | 5+6 2 | 3+3 2 | 3+3 2 |
| | 2+2 4 | 3+4 2 | 2+2 4 | 3+4 2 | 2+2 2 | 2+2 2 |
| | 2+2 4 | 5+6 2 | 2+2 4 | 5+6 2 | 2+2 2 | 2+2 2 |
| | 2+2 4 | 5+6 2 | 2+2 4 | 3+4 2 | 3+3 2 | 3+3 2 |
| | 2+2 4 | 3+4 2 | 2+2 4 | 5+6 2 | 2+2 2 | 2+2 2 |
| | 2+2 4 | 5+6 2 | 2+2 4 | 3+4 2 | 3+3 2 | 3+3 2 |
| | 2+2 4 | 3+4 2 | 2+2 4 | 5+6 2 | 3+3 2 | 3+3 2 |
| | 2+2 4 | 5+6 2 | 2+2 4 | 3+4 2 | 2+2 2 | 2+2 2 |
| | 2+2 4 | 3+4 2 | 2+2 4 | 3+4 2 | 3+3 2 | 3+3 2 |
| | 2+2 4 | 5+6 2 | 2+2 4 | 5+6 2 | 2+2 2 | 2+2 2 |
| | 2+2 4 | 5+6 2 | 2+2 4 | 5+6 2 | 3+3 2 | 3+3 2 |
| | 2+2 4 | 3+4 2 | 2+2 4 | 3+4 2 | 2+2 2 | 2+2 2 |
| | 2+2 4 | 3+4 2 | 2+2 4 | 5+6 2 | 3+3 2 | 3+3 2 |
| | 2+2 4 | 3+4 2 | 2+2 4 | 5+6 2 | 3+3 2 | 3+3 2 |
| | 2+2 4 | 3+4 2 | 2+2 4 | 3+4 2 | 3+3 2 | 2+2 2 |
| | 2+2 4 | 3+4 2 | 2+2 4 | 5+6 2 | 2+2 2 | 3+3 2 |
| | 2+2 4 | 3+4 2 | 2+2 4 | 3+4 2 | 3+3 2 | 2+2 2 |
| | 2+2 4 | 3+4 2 | 2+2 4 | 3+4 2 | 3+3 2 | 3+3 2 |
| | 2+2 4 | 5+6 2 | 2+2 4 | 5+6 2 | 2+2 2 | 2+2 2 |
| | 2+2 4 | 5+6 2 | 2+2 4 | 5+6 2 | 3+3 2 | 3+3 2 |
| | 2+2 4 | 3+4 2 | 2+2 4 | 3+4 2 | 2+2 2 | 3+3 2 |
| | 2+2 4 | 3+4 2 | 2+2 4 | 5+6 2 | 3+3 2 | 2+2 2 |
| | 2+2 4 | 5+6 2 | 2+2 4 | 5+6 2 | 3+3 2 | 3+3 2 |
| | 2+2 4 | 3+4 2 | 2+2 4 | 3+4 2 | 2+2 2 | 3+3 2 |
| | 2+2 4 | 3+4 2 | 2+2 4 | 5+6 2 | 2+2 2 | 2+2 2 |
| | 2+2 4 | 3+4 2 | 2+2 4 | 3+4 2 | 3+3 2 | 2+2 2 |
| | 2+2 4 | 3+4 2 | 2+2 4 | 3+4 2 | 2+2 2 | 3+3 2 |
| | 2+2 4 | 5+6 2 | 2+2 4 | 5+6 2 | 3+3 2 | 2+2 2 |
| | | 5+6 2 | | 3+4 2 | 2+2 2 | 2+2 2 |
| | | 5+6 2 | | 3+4 2 | 2+2 2 | 3+3 2 |
| | | 3+4 2 | | 5+6 2 | 3+3 2 | 3+3 2 |
| | | 3+4 2 | | 5+6 2 | 3+3 2 | 3+3 2 |
| | | 5+6 2 | | 3+4 2 | 2+2 2 | 2+2 2 |
| | | 5+6 2 | | 3+4 2 | 3+3 2 | 3+3 2 |
| | | 3+4 2 | | 5+6 2 | 2+2 2 | 2+2 2 |
| | | 3+4 2 | | 3+4 2 | 3+3 2 | 3+3 2 |
| | | 3+4 2 | | 5+6 2 | 3+3 2 | 2+2 2 |
| | | 3+4 2 | | 3+4 2 | 2+2 2 | 3+3 2 |
| | | 3+4 2 | | 5+6 2 | 3+3 2 | 3+3 2 |
| | | 5+6 2 | | 5+6 2 | 3+3 2 | 3+3 2 |
| | | 3+4 2 | | 3+4 2 | 2+2 2 | 2+2 2 |
| | | 3+4 2 | | 5+6 2 | 3+3 2 | 2+2 2 |
| | | 3+4 2 | | 3+4 2 | 3+3 2 | 3+3 2 |
| | | 5+6 2 | | 5+6 2 | 2+2 2 | 2+2 2 |
| | | 5+6 2 | | 3+4 2 | 3+3 2 | 3+3 2 |
| | | 5+6 2 | | 5+6 2 | 2+2 2 | 2+2 2 |
| | | 3+4 2 | | 3+4 2 | 2+2 2 | 3+3 2 |
| | | 3+4 2 | | 3+4 2 | 3+3 2 | 3+3 2 |
| | | 3+4 2 | | 5+6 2 | 3+3 2 | 2+2 2 |

Continued on next column

**TABLE XXVII– continued**

| TQC | -A | -GM | -M | -Y | -L | -V |
|---|---|---|---|---|---|---|
| | | 3+4 2 | | 5+6 2 | 2+2 2 | 3+3 2 |
| | | 3+4 2 | | 3+4 2 | 3+3 2 | 3+3 2 |
| | | 3+4 2 | | 3+4 2 | 3+3 2 | 3+3 2 |
| | | 3+4 2 | | 3+4 2 | 3+3 2 | 2+2 2 |
| | | 3+4 2 | | 5+6 2 | 2+2 2 | 3+3 2 |
| | | 5+6 2 | | 3+4 2 | 2+2 2 | 2+2 2 |
| | | 3+4 2 | | 5+6 2 | 3+3 2 | 3+3 2 |
| | | 5+6 2 | | 3+4 2 | 2+2 2 | 2+2 2 |
| | | 5+6 2 | | 5+6 2 | 3+3 2 | 3+3 2 |
| | | 3+4 2 | | 3+4 2 | 3+3 2 | 3+3 2 |
| | | 3+4 2 | | 5+6 2 | 2+2 2 | 3+3 2 |
| 15.2.65 | 83494 | | 136 | Cu2Er1O4 | | |
| | -A | -GM | -M | -Y | -L | -V |
| | 2+2 4 | 3+4 2 | 2+2 4 | 3+4 2 | 2+2 2 | 2+2 2 |
| | 2+2 4 | 5+6 2 | 2+2 4 | 5+6 2 | 3+3 2 | 3+3 2 |
| | 2+2 4 | 3+4 2 | 2+2 4 | 3+4 2 | 2+2 2 | 2+2 2 |
| | 2+2 4 | 5+6 2 | 2+2 4 | 5+6 2 | 3+3 2 | 3+3 2 |
| | 2+2 4 | 3+4 2 | 2+2 4 | 3+4 2 | 2+2 2 | 2+2 2 |
| | 2+2 4 | 3+4 2 | 2+2 4 | 5+6 2 | 3+3 2 | 3+3 2 |
| | 2+2 4 | 5+6 2 | 2+2 4 | 3+4 2 | 2+2 2 | 2+2 2 |
| | 2+2 4 | 5+6 2 | 2+2 4 | 5+6 2 | 3+3 2 | 3+3 2 |
| | 2+2 4 | 3+4 2 | 2+2 4 | 5+6 2 | 2+2 2 | 2+2 2 |
| | 2+2 4 | 5+6 2 | 2+2 4 | 3+4 2 | 3+3 2 | 3+3 2 |
| | 2+2 4 | 3+4 2 | 2+2 4 | 5+6 2 | 3+3 2 | 3+3 2 |
| | 2+2 4 | 5+6 2 | 2+2 4 | 3+4 2 | 2+2 2 | 2+2 2 |
| | 2+2 4 | 5+6 2 | 2+2 4 | 3+4 2 | 3+3 2 | 3+3 2 |
| | 2+2 4 | 3+4 2 | 2+2 4 | 5+6 2 | 2+2 2 | 2+2 2 |
| | 2+2 4 | 5+6 2 | 2+2 4 | 5+6 2 | 3+3 2 | 3+3 2 |
| | 2+2 4 | 3+4 2 | 2+2 4 | 3+4 2 | 2+2 2 | 2+2 2 |
| | 2+2 4 | 3+4 2 | 2+2 4 | 3+4 2 | 3+3 2 | 3+3 2 |
| | 2+2 4 | 3+4 2 | 2+2 4 | 5+6 2 | 3+3 2 | 3+3 2 |
| | 2+2 4 | 3+4 2 | 2+2 4 | 5+6 2 | 2+2 2 | 2+2 2 |
| | 2+2 4 | 3+4 2 | 2+2 4 | 5+6 2 | 3+3 2 | 3+3 2 |
| | 2+2 4 | 3+4 2 | 2+2 4 | 3+4 2 | 3+3 2 | 2+2 2 |
| | 2+2 4 | 5+6 2 | 2+2 4 | 5+6 2 | 2+2 2 | 3+3 2 |
| | 2+2 4 | 3+4 2 | 2+2 4 | 3+4 2 | 3+3 2 | 2+2 2 |
| | 2+2 4 | 3+4 2 | 2+2 4 | 5+6 2 | 3+3 2 | 3+3 2 |
| | 2+2 4 | 5+6 2 | 2+2 4 | 5+6 2 | 2+2 2 | 2+2 2 |
| | 2+2 4 | 3+4 2 | 2+2 4 | 3+4 2 | 2+2 2 | 3+3 2 |
| | 2+2 4 | 3+4 2 | 2+2 4 | 3+4 2 | 3+3 2 | 3+3 2 |
| | 2+2 4 | 3+4 2 | 2+2 4 | 5+6 2 | 3+3 2 | 2+2 2 |
| | 2+2 4 | 5+6 2 | 2+2 4 | 3+4 2 | 2+2 2 | 3+3 2 |
| | 2+2 4 | 3+4 2 | 2+2 4 | 5+6 2 | 3+3 2 | 2+2 2 |
| | 2+2 4 | 3+4 2 | 2+2 4 | 3+4 2 | 2+2 2 | 2+2 2 |
| | 2+2 4 | 5+6 2 | 2+2 4 | 5+6 2 | 3+3 2 | 3+3 2 |
| | 2+2 4 | 5+6 2 | 2+2 4 | 3+4 2 | 2+2 2 | 2+2 2 |
| | 2+2 4 | 5+6 2 | 2+2 4 | 3+4 2 | 2+2 2 | 3+3 2 |
| | | 3+4 2 | | 5+6 2 | 3+3 2 | 3+3 2 |
| | | 5+6 2 | | 5+6 2 | 3+3 2 | 3+3 2 |
| | | 3+4 2 | | 3+4 2 | 2+2 2 | 2+2 2 |
| | | 3+4 2 | | 3+4 2 | 3+3 2 | 2+2 2 |
| | | 3+4 2 | | 5+6 2 | 2+2 2 | 3+3 2 |
| | | 3+4 2 | | 3+4 2 | 3+3 2 | 3+3 2 |
| | | 5+6 2 | | 3+4 2 | 3+3 2 | 2+2 2 |
| | | 3+4 2 | | 5+6 2 | 2+2 2 | 3+3 2 |
| | | 3+4 2 | | 5+6 2 | 3+3 2 | 2+2 2 |
| | | 3+4 2 | | 3+4 2 | 3+3 2 | 3+3 2 |
| | | 5+6 2 | | 5+6 2 | 2+2 2 | 3+3 2 |
| | | 3+4 2 | | 5+6 2 | 3+3 2 | 2+2 2 |
| | | 3+4 2 | | 3+4 2 | 2+2 2 | 2+2 2 |
| | | 5+6 2 | | 5+6 2 | 3+3 2 | 3+3 2 |

Continued on next column

**TABLE XXVII– continued**

| TQC | | ICSD | NoE | SF | | |
|---|---|---|---|---|---|---|
| | | 3+4 2 | | 3+4 2 | 2+2 2 | 3+3 2 |
| | | 5+6 2 | | 3+4 2 | 3+3 2 | 2+2 2 |
| | | 3+4 2 | | 5+6 2 | 3+3 2 | 3+3 2 |
| | | 3+4 2 | | 3+4 2 | 2+2 2 | 3+3 2 |
| | | 3+4 2 | | 3+4 2 | 2+2 2 | 3+3 2 |
| | | 3+4 2 | | 5+6 2 | 3+3 2 | 2+2 2 |
| | | 3+4 2 | | 5+6 2 | 3+3 2 | 3+3 2 |
| | | 5+6 2 | | 3+4 2 | 3+3 2 | 3+3 2 |
| | | 3+4 2 | | 5+6 2 | 2+2 2 | 2+2 2 |
| | | 5+6 2 | | 3+4 2 | 3+3 2 | 3+3 2 |
| | | 5+6 2 | | 5+6 2 | 2+2 2 | 2+2 2 |
| | | 3+4 2 | | 3+4 2 | 2+2 2 | 2+2 2 |
| | | 3+4 2 | | 3+4 2 | 3+3 2 | 3+3 2 |
| | | 5+6 2 | | 5+6 2 | 3+3 2 | 3+3 2 |
| | | 3+4 2 | | 3+4 2 | 3+3 2 | 3+3 2 |
| | | 3+4 2 | | 5+6 2 | 2+2 2 | 3+3 2 |
| | | 3+4 2 | | 3+4 2 | 3+3 2 | 2+2 2 |
| | | 3+4 2 | | 5+6 2 | 2+2 2 | 2+2 2 |
| | | 5+6 2 | | 3+4 2 | 2+2 2 | 2+2 2 |
| | | 5+6 2 | | 5+6 2 | 3+3 2 | 3+3 2 |
| 15.2.66 | 83491 | | 132 | Cu2Dy1O4 | | |
| | -A | -GM | -M | -Y | -L | -V |
| | 2+2 4 | 3+4 2 | 2+2 4 | 3+4 2 | 2+2 2 | 2+2 2 |
| | 2+2 4 | 5+6 2 | 2+2 4 | 5+6 2 | 3+3 2 | 3+3 2 |
| | 2+2 4 | 3+4 2 | 2+2 4 | 5+6 2 | 2+2 2 | 2+2 2 |
| | 2+2 4 | 5+6 2 | 2+2 4 | 3+4 2 | 3+3 2 | 3+3 2 |
| | 2+2 4 | 3+4 2 | 2+2 4 | 5+6 2 | 3+3 2 | 3+3 2 |
| | 2+2 4 | 3+4 2 | 2+2 4 | 3+4 2 | 2+2 2 | 2+2 2 |
| | 2+2 4 | 5+6 2 | 2+2 4 | 5+6 2 | 2+2 2 | 2+2 2 |
| | 2+2 4 | 5+6 2 | 2+2 4 | 3+4 2 | 3+3 2 | 3+3 2 |
| | 2+2 4 | 3+4 2 | 2+2 4 | 5+6 2 | 3+3 2 | 2+2 2 |
| | 2+2 4 | 3+4 2 | 2+2 4 | 3+4 2 | 2+2 2 | 3+3 2 |
| | 2+2 4 | 5+6 2 | 2+2 4 | 5+6 2 | 3+3 2 | 3+3 2 |
| | 2+2 4 | 5+6 2 | 2+2 4 | 3+4 2 | 2+2 2 | 2+2 2 |
| | 2+2 4 | 3+4 2 | 2+2 4 | 5+6 2 | 3+3 2 | 3+3 2 |
| | 2+2 4 | 5+6 2 | 2+2 4 | 3+4 2 | 2+2 2 | 2+2 2 |
| | 2+2 4 | 5+6 2 | 2+2 4 | 5+6 2 | 3+3 2 | 3+3 2 |
| | 2+2 4 | 3+4 2 | 2+2 4 | 3+4 2 | 2+2 2 | 2+2 2 |
| | 2+2 4 | 3+4 2 | 2+2 4 | 5+6 2 | 3+3 2 | 3+3 2 |
| | 2+2 4 | 3+4 2 | 2+2 4 | 5+6 2 | 3+3 2 | 3+3 2 |
| | 2+2 4 | 3+4 2 | 2+2 4 | 5+6 2 | 3+3 2 | 2+2 2 |
| | 2+2 4 | 3+4 2 | 2+2 4 | 3+4 2 | 2+2 2 | 3+3 2 |
| | 2+2 4 | 3+4 2 | 2+2 4 | 3+4 2 | 3+3 2 | 2+2 2 |
| | 2+2 4 | 3+4 2 | 2+2 4 | 5+6 2 | 3+3 2 | 2+2 2 |
| | 2+2 4 | 3+4 2 | 2+2 4 | 5+6 2 | 3+3 2 | 3+3 2 |
| | 2+2 4 | 5+6 2 | 2+2 4 | 3+4 2 | 2+2 2 | 3+3 2 |
| | 2+2 4 | 5+6 2 | 2+2 4 | 5+6 2 | 3+3 2 | 3+3 2 |
| | 2+2 4 | 3+4 2 | 2+2 4 | 5+6 2 | 2+2 2 | 2+2 2 |
| | 2+2 4 | 3+4 2 | 2+2 4 | 3+4 2 | 3+3 2 | 3+3 2 |
| | 2+2 4 | 5+6 2 | 2+2 4 | 5+6 2 | 2+2 2 | 2+2 2 |
| | 2+2 4 | 3+4 2 | 2+2 4 | 3+4 2 | 3+3 2 | 3+3 2 |
| | 2+2 4 | 3+4 2 | 2+2 4 | 3+4 2 | 2+2 2 | 2+2 2 |
| | 2+2 4 | 3+4 2 | 2+2 4 | 3+4 2 | 3+3 2 | 2+2 2 |
| | 2+2 4 | 5+6 2 | 2+2 4 | 5+6 2 | 2+2 2 | 3+3 2 |
| | 2+2 4 | 5+6 2 | 2+2 4 | 3+4 2 | 2+2 2 | 2+2 2 |
| | | 5+6 2 | | 5+6 2 | 3+3 2 | 3+3 2 |
| | | 3+4 2 | | 3+4 2 | 2+2 2 | 3+3 2 |
| | | 3+4 2 | | 5+6 2 | 3+3 2 | 3+3 2 |
| | | 5+6 2 | | 3+4 2 | 3+3 2 | 2+2 2 |
| | | 3+4 2 | | 3+4 2 | 2+2 2 | 3+3 2 |
| | | 3+4 2 | | 5+6 2 | 2+2 2 | 2+2 2 |
| | | 5+6 2 | | 5+6 2 | 3+3 2 | 3+3 2 |
| | | 3+4 2 | | 3+4 2 | 3+3 2 | 2+2 2 |
| | | 3+4 2 | | 3+4 2 | 2+2 2 | 3+3 2 |
| | | 3+4 2 | | 5+6 2 | 3+3 2 | 3+3 2 |
| | | 5+6 2 | | 5+6 2 | 3+3 2 | 2+2 2 |
| | | 3+4 2 | | 3+4 2 | 2+2 2 | 3+3 2 |
| | | 3+4 2 | | 5+6 2 | 3+3 2 | 2+2 2 |
| | | 5+6 2 | | 3+4 2 | 3+3 2 | 3+3 2 |
| | | 3+4 2 | | 5+6 2 | 2+2 2 | 2+2 2 |
| | | 5+6 2 | | 3+4 2 | 3+3 2 | 3+3 2 |
| | | 3+4 2 | | 5+6 2 | 2+2 2 | 2+2 2 |
| | | 5+6 2 | | 3+4 2 | 3+3 2 | 3+3 2 |
| | | 3+4 2 | | 3+4 2 | 2+2 2 | 3+3 2 |
| | | 3+4 2 | | 5+6 2 | 3+3 2 | 2+2 2 |
| | | 3+4 2 | | 3+4 2 | 2+2 2 | 3+3 2 |
| | | 3+4 2 | | 5+6 2 | 3+3 2 | 3+3 2 |
| | | 3+4 2 | | 3+4 2 | 3+3 2 | 3+3 2 |
| | | 3+4 2 | | 3+4 2 | 2+2 2 | 2+2 2 |
| | | 3+4 2 | | 5+6 2 | 3+3 2 | 2+2 2 |
| | | 5+6 2 | | 3+4 2 | 2+2 2 | 3+3 2 |
| | | 5+6 2 | | 5+6 2 | 2+2 2 | 3+3 2 |
| | | 3+4 2 | | 3+4 2 | 3+3 2 | 2+2 2 |
| | | 5+6 2 | | 5+6 2 | 3+3 2 | 3+3 2 |
| | | 3+4 2 | | 3+4 2 | 3+3 2 | 3+3 2 |
| | | 3+4 2 | | 5+6 2 | 2+2 2 | 3+3 2 |
| | | 3+4 2 | | 3+4 2 | 2+2 2 | 2+2 2 |
| | | 3+4 2 | | 5+6 2 | 3+3 2 | 2+2 2 |
| 15.2.67 | 2790 | | 52 | Cr1P4 | | |
| | -A | -GM | -M | -Y | -L | -V |
| | 2+2 4 | 3+4 2 | 2+2 4 | 5+6 2 | 3+3 2 | 3+3 2 |
| | 2+2 4 | 5+6 2 | 2+2 4 | 3+4 2 | 2+2 2 | 2+2 2 |
| | 2+2 4 | 3+4 2 | 2+2 4 | 5+6 2 | 3+3 2 | 3+3 2 |
| | 2+2 4 | 3+4 2 | 2+2 4 | 3+4 2 | 2+2 2 | 2+2 2 |
| | 2+2 4 | 5+6 2 | 2+2 4 | 5+6 2 | 3+3 2 | 2+2 2 |
| | 2+2 4 | 5+6 2 | 2+2 4 | 3+4 2 | 2+2 2 | 3+3 2 |
| | 2+2 4 | 3+4 2 | 2+2 4 | 5+6 2 | 3+3 2 | 3+3 2 |
| | 2+2 4 | 5+6 2 | 2+2 4 | 3+4 2 | 2+2 2 | 2+2 2 |
| | 2+2 4 | 5+6 2 | 2+2 4 | 3+4 2 | 3+3 2 | 3+3 2 |
| | 2+2 4 | 3+4 2 | 2+2 4 | 5+6 2 | 2+2 2 | 2+2 2 |
| | 2+2 4 | 5+6 2 | 2+2 4 | 5+6 2 | 2+2 2 | 2+2 2 |
| | 2+2 4 | 3+4 2 | 2+2 4 | 3+4 2 | 2+2 2 | 3+3 2 |
| | 2+2 4 | 3+4 2 | 2+2 4 | 3+4 2 | 3+3 2 | 3+3 2 |
| | | 3+4 2 | | 5+6 2 | 3+3 2 | 2+2 2 |
| | | 5+6 2 | | 5+6 2 | 2+2 2 | 3+3 2 |
| | | 5+6 2 | | 3+4 2 | 3+3 2 | 2+2 2 |
| | | 3+4 2 | | 5+6 2 | 3+3 2 | 2+2 2 |
| | | 3+4 2 | | 3+4 2 | 2+2 2 | 3+3 2 |
| | | 5+6 2 | | 5+6 2 | 3+3 2 | 2+2 2 |
| | | 5+6 2 | | 3+4 2 | 3+3 2 | 3+3 2 |
| | | 3+4 2 | | 5+6 2 | 2+2 2 | 2+2 2 |
| | | 3+4 2 | | 5+6 2 | 3+3 2 | 2+2 2 |
| | | 3+4 2 | | 3+4 2 | 2+2 2 | 3+3 2 |
| | | 3+4 2 | | 3+4 2 | 3+3 2 | 3+3 2 |
| | | 5+6 2 | | 5+6 2 | 2+2 2 | 2+2 2 |
| | | 5+6 2 | | 3+4 2 | 3+3 2 | 2+2 2 |
| 15.2.68 | 26814 | | 124 | Co1Ge1O3 | | |
| | -A | -GM | -M | -Y | -L | -V |
| | 2+2 4 | 3+4 2 | 2+2 4 | 3+4 2 | 2+2 2 | 3+3 2 |
| | 2+2 4 | 5+6 2 | 2+2 4 | 5+6 2 | 3+3 2 | 2+2 2 |
| | 2+2 4 | 3+4 2 | 2+2 4 | 5+6 2 | 2+2 2 | 2+2 2 |
| | 2+2 4 | 5+6 2 | 2+2 4 | 3+4 2 | 3+3 2 | 3+3 2 |

Continued on next column

**TABLE XXVII– continued**

| TQC | ICSD | | NoE | | SF | |
|---|---|---|---|---|---|---|
| | 2+2 4 | 3+4 2 | 2+2 4 | 3+4 2 | 3+3 2 | 3+3 2 |
| | 2+2 4 | 3+4 2 | 2+2 4 | 3+4 2 | 2+2 2 | 2+2 2 |
| | 2+2 4 | 5+6 2 | 2+2 4 | 5+6 2 | 3+3 2 | 2+2 2 |
| | 2+2 4 | 5+6 2 | 2+2 4 | 5+6 2 | 2+2 2 | 3+3 2 |
| | 2+2 4 | 5+6 2 | 2+2 4 | 5+6 2 | 3+3 2 | 2+2 2 |
| | 2+2 4 | 3+4 2 | 2+2 4 | 3+4 2 | 3+3 2 | 3+3 2 |
| | 2+2 4 | 3+4 2 | 2+2 4 | 3+4 2 | 2+2 2 | 3+3 2 |
| | 2+2 4 | 5+6 2 | 2+2 4 | 5+6 2 | 2+2 2 | 2+2 2 |
| | 2+2 4 | 5+6 2 | 2+2 4 | 5+6 2 | 2+2 2 | 2+2 2 |
| | 2+2 4 | 3+4 2 | 2+2 4 | 3+4 2 | 3+3 2 | 3+3 2 |
| | 2+2 4 | 3+4 2 | 2+2 4 | 3+4 2 | 2+2 2 | 2+2 2 |
| | 2+2 4 | 5+6 2 | 2+2 4 | 5+6 2 | 3+3 2 | 3+3 2 |
| | 2+2 4 | 3+4 2 | 2+2 4 | 5+6 2 | 2+2 2 | 2+2 2 |
| | 2+2 4 | 5+6 2 | 2+2 4 | 5+6 2 | 2+2 2 | 2+2 2 |
| | 2+2 4 | 3+4 2 | 2+2 4 | 3+4 2 | 3+3 2 | 3+3 2 |
| | 2+2 4 | 5+6 2 | 2+2 4 | 3+4 2 | 3+3 2 | 3+3 2 |
| | 2+2 4 | 5+6 2 | 2+2 4 | 3+4 2 | 2+2 2 | 3+3 2 |
| | 2+2 4 | 3+4 2 | 2+2 4 | 5+6 2 | 3+3 2 | 2+2 2 |
| | 2+2 4 | 3+4 2 | 2+2 4 | 5+6 2 | 2+2 2 | 3+3 2 |
| | 2+2 4 | 5+6 2 | 2+2 4 | 3+4 2 | 3+3 2 | 2+2 2 |
| | 2+2 4 | 3+4 2 | 2+2 4 | 5+6 2 | 3+3 2 | 2+2 2 |
| | 2+2 4 | 5+6 2 | 2+2 4 | 3+4 2 | 2+2 2 | 3+3 2 |
| | 2+2 4 | 5+6 2 | 2+2 4 | 3+4 2 | 3+3 2 | 2+2 2 |
| | 2+2 4 | 3+4 2 | 2+2 4 | 5+6 2 | 2+2 2 | 3+3 2 |
| | 2+2 4 | 3+4 2 | 2+2 4 | 3+4 2 | 3+3 2 | 3+3 2 |
| | 2+2 4 | 5+6 2 | 2+2 4 | 5+6 2 | 3+3 2 | 2+2 2 |
| | 2+2 4 | 3+4 2 | 2+2 4 | 5+6 2 | 2+2 2 | 2+2 2 |
| | | 5+6 2 | | 3+4 2 | 3+3 2 | 3+3 2 |
| | | 5+6 2 | | 3+4 2 | 2+2 2 | 3+3 2 |
| | | 3+4 2 | | 3+4 2 | 2+2 2 | 2+2 2 |
| | | 3+4 2 | | 5+6 2 | 3+3 2 | 3+3 2 |
| | | 3+4 2 | | 3+4 2 | 2+2 2 | 2+2 2 |
| | | 5+6 2 | | 5+6 2 | 2+2 2 | 3+3 2 |
| | | 3+4 2 | | 3+4 2 | 3+3 2 | 2+2 2 |
| | | 3+4 2 | | 5+6 2 | 3+3 2 | 2+2 2 |
| | | 5+6 2 | | 3+4 2 | 2+2 2 | 3+3 2 |
| | | 3+4 2 | | 5+6 2 | 3+3 2 | 3+3 2 |
| | | 5+6 2 | | 3+4 2 | 2+2 2 | 2+2 2 |
| | | 5+6 2 | | 3+4 2 | 3+3 2 | 3+3 2 |
| | | 3+4 2 | | 5+6 2 | 3+3 2 | 2+2 2 |
| | | 5+6 2 | | 5+6 2 | 2+2 2 | 3+3 2 |
| | | 3+4 2 | | 5+6 2 | 3+3 2 | 2+2 2 |
| | | 5+6 2 | | 5+6 2 | 2+2 2 | 3+3 2 |
| | | 5+6 2 | | 3+4 2 | 2+2 2 | 2+2 2 |
| | | 3+4 2 | | 5+6 2 | 2+2 2 | 2+2 2 |
| | | 3+4 2 | | 3+4 2 | 3+3 2 | 2+2 2 |
| | | 5+6 2 | | 3+4 2 | 2+2 2 | 3+3 2 |
| | | 5+6 2 | | 5+6 2 | 3+3 2 | 3+3 2 |
| | | 5+6 2 | | 3+4 2 | 2+2 2 | 2+2 2 |
| | | 3+4 2 | | 5+6 2 | 3+3 2 | 3+3 2 |
| | | 3+4 2 | | 3+4 2 | 2+2 2 | 2+2 2 |
| | | 3+4 2 | | 5+6 2 | 3+3 2 | 2+2 2 |
| | | 5+6 2 | | 3+4 2 | 2+2 2 | 3+3 2 |
| | | 5+6 2 | | 5+6 2 | 3+3 2 | 3+3 2 |
| | | 3+4 2 | | 3+4 2 | 3+3 2 | 2+2 2 |
| | | 5+6 2 | | 5+6 2 | 2+2 2 | 3+3 2 |
| | | 5+6 2 | | 3+4 2 | 3+3 2 | 2+2 2 |
| | | 5+6 2 | | 5+6 2 | 2+2 2 | 3+3 2 |
| 15.2.69 | 417870 | | 144 | | Cl8Ga2Ni1 | |
| | -A | -GM | -M | -Y | -L | -V |
| | 2+2 4 | 3+4 2 | 2+2 4 | 3+4 2 | 2+2 2 | 2+2 2 |

Continued on next column

**TABLE XXVII– continued**

| TQC | ICSD | | NoE | | SF | |
|---|---|---|---|---|---|---|
| | 2+2 4 | 5+6 2 | 2+2 4 | 3+4 2 | 3+3 2 | 3+3 2 |
| | 2+2 4 | 3+4 2 | 2+2 4 | 5+6 2 | 3+3 2 | 2+2 2 |
| | 2+2 4 | 5+6 2 | 2+2 4 | 5+6 2 | 2+2 2 | 3+3 2 |
| | 2+2 4 | 5+6 2 | 2+2 4 | 5+6 2 | 3+3 2 | 3+3 2 |
| | 2+2 4 | 3+4 2 | 2+2 4 | 3+4 2 | 2+2 2 | 2+2 2 |
| | 2+2 4 | 3+4 2 | 2+2 4 | 5+6 2 | 3+3 2 | 2+2 2 |
| | 2+2 4 | 5+6 2 | 2+2 4 | 3+4 2 | 2+2 2 | 3+3 2 |
| | 2+2 4 | 5+6 2 | 2+2 4 | 5+6 2 | 2+2 2 | 3+3 2 |
| | 2+2 4 | 5+6 2 | 2+2 4 | 5+6 2 | 3+3 2 | 3+3 2 |
| | 2+2 4 | 3+4 2 | 2+2 4 | 3+4 2 | 3+3 2 | 2+2 2 |
| | 2+2 4 | 3+4 2 | 2+2 4 | 3+4 2 | 2+2 2 | 2+2 2 |
| | 2+2 4 | 3+4 2 | 2+2 4 | 3+4 2 | 3+3 2 | 2+2 2 |
| | 2+2 4 | 5+6 2 | 2+2 4 | 5+6 2 | 2+2 2 | 2+2 2 |
| | 2+2 4 | 5+6 2 | 2+2 4 | 3+4 2 | 3+3 2 | 3+3 2 |
| | 2+2 4 | 3+4 2 | 2+2 4 | 5+6 2 | 2+2 2 | 3+3 2 |
| | 2+2 4 | 3+4 2 | 2+2 4 | 3+4 2 | 2+2 2 | 2+2 2 |
| | 2+2 4 | 5+6 2 | 2+2 4 | 5+6 2 | 3+3 2 | 3+3 2 |
| | 2+2 4 | 3+4 2 | 2+2 4 | 5+6 2 | 2+2 2 | 3+3 2 |
| | 2+2 4 | 5+6 2 | 2+2 4 | 3+4 2 | 3+3 2 | 2+2 2 |
| | 2+2 4 | 3+4 2 | 2+2 4 | 3+4 2 | 3+3 2 | 2+2 2 |
| | 2+2 4 | 5+6 2 | 2+2 4 | 5+6 2 | 2+2 2 | 3+3 2 |
| | 2+2 4 | 3+4 2 | 2+2 4 | 3+4 2 | 3+3 2 | 3+3 2 |
| | 2+2 4 | 5+6 2 | 2+2 4 | 5+6 2 | 2+2 2 | 2+2 2 |
| | 2+2 4 | 5+6 2 | 2+2 4 | 5+6 2 | 3+3 2 | 3+3 2 |
| | 2+2 4 | 3+4 2 | 2+2 4 | 3+4 2 | 3+3 2 | 2+2 2 |
| | 2+2 4 | 3+4 2 | 2+2 4 | 3+4 2 | 2+2 2 | 2+2 2 |
| | 2+2 4 | 3+4 2 | 2+2 4 | 3+4 2 | 2+2 2 | 3+3 2 |
| | 2+2 4 | 3+4 2 | 2+2 4 | 3+4 2 | 3+3 2 | 2+2 2 |
| | 2+2 4 | 5+6 2 | 2+2 4 | 5+6 2 | 2+2 2 | 3+3 2 |
| | 2+2 4 | 5+6 2 | 2+2 4 | 5+6 2 | 2+2 2 | 2+2 2 |
| | 2+2 4 | 5+6 2 | 2+2 4 | 3+4 2 | 3+3 2 | 3+3 2 |
| | 2+2 4 | 3+4 2 | 2+2 4 | 5+6 2 | 2+2 2 | 2+2 2 |
| | 2+2 4 | 5+6 2 | 2+2 4 | 5+6 2 | 3+3 2 | 3+3 2 |
| | 2+2 4 | 5+6 2 | 2+2 4 | 5+6 2 | 2+2 2 | 2+2 2 |
| | 2+2 4 | 3+4 2 | 2+2 4 | 3+4 2 | 3+3 2 | 3+3 2 |
| | | 5+6 2 | | 5+6 2 | 2+2 2 | 3+3 2 |
| | | 5+6 2 | | 3+4 2 | 3+3 2 | 2+2 2 |
| | | 3+4 2 | | 5+6 2 | 2+2 2 | 3+3 2 |
| | | 5+6 2 | | 5+6 2 | 3+3 2 | 2+2 2 |
| | | 3+4 2 | | 5+6 2 | 2+2 2 | 3+3 2 |
| | | 3+4 2 | | 3+4 2 | 3+3 2 | 2+2 2 |
| | | 5+6 2 | | 3+4 2 | 3+3 2 | 3+3 2 |
| | | 5+6 2 | | 5+6 2 | 2+2 2 | 2+2 2 |
| | | 3+4 2 | | 5+6 2 | 3+3 2 | 3+3 2 |
| | | 3+4 2 | | 3+4 2 | 2+2 2 | 2+2 2 |
| | | 5+6 2 | | 3+4 2 | 3+3 2 | 3+3 2 |
| | | 3+4 2 | | 5+6 2 | 2+2 2 | 3+3 2 |
| | | 3+4 2 | | 3+4 2 | 2+2 2 | 2+2 2 |
| | | 5+6 2 | | 3+4 2 | 3+3 2 | 3+3 2 |
| | | 3+4 2 | | 5+6 2 | 2+2 2 | 2+2 2 |
| | | 5+6 2 | | 3+4 2 | 3+3 2 | 2+2 2 |
| | | 5+6 2 | | 3+4 2 | 2+2 2 | 3+3 2 |
| | | 3+4 2 | | 5+6 2 | 3+3 2 | 2+2 2 |
| | | 5+6 2 | | 3+4 2 | 2+2 2 | 2+2 2 |
| | | 5+6 2 | | 5+6 2 | 3+3 2 | 3+3 2 |
| | | 3+4 2 | | 3+4 2 | 3+3 2 | 3+3 2 |
| | | 3+4 2 | | 5+6 2 | 2+2 2 | 2+2 2 |
| | | 3+4 2 | | 3+4 2 | 3+3 2 | 3+3 2 |
| | | 5+6 2 | | 5+6 2 | 2+2 2 | 2+2 2 |
| | | 3+4 2 | | 3+4 2 | 3+3 2 | 2+2 2 |
| | | 3+4 2 | | 5+6 2 | 2+2 2 | 3+3 2 |

Continued on next column

**TABLE XXVII– continued**

| TQC | ICSD | | NoE | SF | | |
|---|---|---|---|---|---|---|
| | | 5+6 2 | | 5+6 2 | 3+3 2 | 2+2 2 |
| | | 5+6 2 | | 3+4 2 | 2+2 2 | 3+3 2 |
| | | 3+4 2 | | 3+4 2 | 3+3 2 | 2+2 2 |
| | | 5+6 2 | | 5+6 2 | 2+2 2 | 3+3 2 |
| | | 5+6 2 | | 3+4 2 | 2+2 2 | 2+2 2 |
| | | 3+4 2 | | 5+6 2 | 3+3 2 | 3+3 2 |
| | | 5+6 2 | | 3+4 2 | 3+3 2 | 2+2 2 |
| | | 3+4 2 | | 5+6 2 | 2+2 2 | 3+3 2 |
| | | 5+6 2 | | 3+4 2 | 2+2 2 | 2+2 2 |
| | | 5+6 2 | | 5+6 2 | 3+3 2 | 3+3 2 |
| 15.2.70 | 56838 | | 164 | Cl3Cu1Rb1 | | |
| | -A | -GM | -M | -Y | -L | -V |
| | 2+2 4 | 3+4 2 | 2+2 4 | 5+6 2 | 2+2 2 | 2+2 2 |
| | 2+2 4 | 3+4 2 | 2+2 4 | 5+6 2 | 2+2 2 | 3+3 2 |
| | 2+2 4 | 5+6 2 | 2+2 4 | 3+4 2 | 3+3 2 | 3+3 2 |
| | 2+2 4 | 5+6 2 | 2+2 4 | 3+4 2 | 3+3 2 | 2+2 2 |
| | 2+2 4 | 3+4 2 | 2+2 4 | 3+4 2 | 3+3 2 | 2+2 2 |
| | 2+2 4 | 5+6 2 | 2+2 4 | 5+6 2 | 2+2 2 | 3+3 2 |
| | 2+2 4 | 3+4 2 | 2+2 4 | 3+4 2 | 3+3 2 | 2+2 2 |
| | 2+2 4 | 5+6 2 | 2+2 4 | 5+6 2 | 2+2 2 | 3+3 2 |
| | 2+2 4 | 5+6 2 | 2+2 4 | 5+6 2 | 2+2 2 | 2+2 2 |
| | 2+2 4 | 5+6 2 | 2+2 4 | 5+6 2 | 3+3 2 | 3+3 2 |
| | 2+2 4 | 3+4 2 | 2+2 4 | 3+4 2 | 2+2 2 | 3+3 2 |
| | 2+2 4 | 3+4 2 | 2+2 4 | 3+4 2 | 3+3 2 | 2+2 2 |
| | 2+2 4 | 5+6 2 | 2+2 4 | 5+6 2 | 3+3 2 | 3+3 2 |
| | 2+2 4 | 3+4 2 | 2+2 4 | 3+4 2 | 2+2 2 | 2+2 2 |
| | 2+2 4 | 5+6 2 | 2+2 4 | 3+4 2 | 3+3 2 | 3+3 2 |
| | 2+2 4 | 3+4 2 | 2+2 4 | 5+6 2 | 2+2 2 | 2+2 2 |
| | 2+2 4 | 3+4 2 | 2+2 4 | 5+6 2 | 2+2 2 | 3+3 2 |
| | 2+2 4 | 3+4 2 | 2+2 4 | 3+4 2 | 3+3 2 | 2+2 2 |
| | 2+2 4 | 5+6 2 | 2+2 4 | 5+6 2 | 2+2 2 | 2+2 2 |
| | 2+2 4 | 5+6 2 | 2+2 4 | 3+4 2 | 3+3 2 | 3+3 2 |
| | 2+2 4 | 3+4 2 | 2+2 4 | 5+6 2 | 2+2 2 | 3+3 2 |
| | 2+2 4 | 3+4 2 | 2+2 4 | 3+4 2 | 3+3 2 | 2+2 2 |
| | 2+2 4 | 5+6 2 | 2+2 4 | 5+6 2 | 2+2 2 | 3+3 2 |
| | 2+2 4 | 3+4 2 | 2+2 4 | 3+4 2 | 3+3 2 | 2+2 2 |
| | 2+2 4 | 5+6 2 | 2+2 4 | 5+6 2 | 2+2 2 | 3+3 2 |
| | 2+2 4 | 3+4 2 | 2+2 4 | 3+4 2 | 3+3 2 | 2+2 2 |
| | 2+2 4 | 5+6 2 | 2+2 4 | 5+6 2 | 2+2 2 | 3+3 2 |
| | 2+2 4 | 5+6 2 | 2+2 4 | 3+4 2 | 3+3 2 | 2+2 2 |
| | 2+2 4 | 3+4 2 | 2+2 4 | 3+4 2 | 3+3 2 | 2+2 2 |
| | 2+2 4 | 3+4 2 | 2+2 4 | 3+4 2 | 2+2 2 | 2+2 2 |
| | 2+2 4 | 5+6 2 | 2+2 4 | 5+6 2 | 3+3 2 | 2+2 2 |
| | 2+2 4 | 3+4 2 | 2+2 4 | 3+4 2 | 2+2 2 | 3+3 2 |
| | 2+2 4 | 5+6 2 | 2+2 4 | 3+4 2 | 2+2 2 | 2+2 2 |
| | 2+2 4 | 3+4 2 | 2+2 4 | 3+4 2 | 3+3 2 | 3+3 2 |
| | 2+2 4 | 3+4 2 | 2+2 4 | 5+6 2 | 2+2 2 | 2+2 2 |
| | 2+2 4 | 3+4 2 | 2+2 4 | 5+6 2 | 3+3 2 | 2+2 2 |
| | 2+2 4 | 5+6 2 | 2+2 4 | 3+4 2 | 3+3 2 | 3+3 2 |
| | 2+2 4 | 5+6 2 | 2+2 4 | 3+4 2 | 2+2 2 | 3+3 2 |
| | 2+2 4 | 3+4 2 | 2+2 4 | 5+6 2 | 2+2 2 | 2+2 2 |
| | 2+2 4 | 3+4 2 | 2+2 4 | 3+4 2 | 2+2 2 | 2+2 2 |
| | 2+2 4 | 5+6 2 | 2+2 4 | 5+6 2 | 3+3 2 | 3+3 2 |
| | | 3+4 2 | | 5+6 2 | 3+3 2 | 2+2 2 |
| | | 5+6 2 | | 3+4 2 | 2+2 2 | 3+3 2 |
| | | 3+4 2 | | 3+4 2 | 3+3 2 | 2+2 2 |
| | | 5+6 2 | | 5+6 2 | 3+3 2 | 3+3 2 |
| | | 5+6 2 | | 5+6 2 | 2+2 2 | 3+3 2 |
| | | 3+4 2 | | 3+4 2 | 3+3 2 | 2+2 2 |
| | | 5+6 2 | | 5+6 2 | 2+2 2 | 2+2 2 |
| | | 3+4 2 | | 5+6 2 | 3+3 2 | 3+3 2 |

Continued on next column

**TABLE XXVII– continued**

| TQC | ICSD | | NoE | SF | | |
|---|---|---|---|---|---|---|
| | | 5+6 2 | | 3+4 2 | 2+2 2 | 3+3 2 |
| | | 5+6 2 | | 3+4 2 | 3+3 2 | 2+2 2 |
| | | 3+4 2 | | 3+4 2 | 2+2 2 | 3+3 2 |
| | | 3+4 2 | | 5+6 2 | 3+3 2 | 2+2 2 |
| | | 3+4 2 | | 5+6 2 | 2+2 2 | 2+2 2 |
| | | 5+6 2 | | 3+4 2 | 3+3 2 | 3+3 2 |
| | | 3+4 2 | | 5+6 2 | 2+2 2 | 3+3 2 |
| | | 3+4 2 | | 5+6 2 | 3+3 2 | 3+3 2 |
| | | 3+4 2 | | 3+4 2 | 2+2 2 | 2+2 2 |
| | | 5+6 2 | | 5+6 2 | 2+2 2 | 2+2 2 |
| | | 3+4 2 | | 3+4 2 | 3+3 2 | 3+3 2 |
| | | 5+6 2 | | 3+4 2 | 2+2 2 | 2+2 2 |
| | | 5+6 2 | | 3+4 2 | 3+3 2 | 2+2 2 |
| | | 3+4 2 | | 3+4 2 | 2+2 2 | 2+2 2 |
| | | 5+6 2 | | 5+6 2 | 3+3 2 | 3+3 2 |
| | | 3+4 2 | | 3+4 2 | 3+3 2 | 2+2 2 |
| | | 3+4 2 | | 5+6 2 | 2+2 2 | 3+3 2 |
| | | 3+4 2 | | 3+4 2 | 2+2 2 | 2+2 2 |
| | | 5+6 2 | | 5+6 2 | 3+3 2 | 3+3 2 |
| | | 5+6 2 | | 3+4 2 | 3+3 2 | 2+2 2 |
| | | 5+6 2 | | 5+6 2 | 2+2 2 | 3+3 2 |
| | | 3+4 2 | | 3+4 2 | 3+3 2 | 2+2 2 |
| | | 5+6 2 | | 3+4 2 | 2+2 2 | 3+3 2 |
| | | 3+4 2 | | 5+6 2 | 3+3 2 | 2+2 2 |
| | | 3+4 2 | | 3+4 2 | 2+2 2 | 2+2 2 |
| | | 5+6 2 | | 5+6 2 | 2+2 2 | 3+3 2 |
| | | 3+4 2 | | 3+4 2 | 3+3 2 | 2+2 2 |
| | | 3+4 2 | | 3+4 2 | 3+3 2 | 2+2 2 |
| | | 3+4 2 | | 3+4 2 | 2+2 2 | 2+2 2 |
| | | 5+6 2 | | 5+6 2 | 3+3 2 | 3+3 2 |
| | | 3+4 2 | | 3+4 2 | 2+2 2 | 2+2 2 |
| | | 3+4 2 | | 3+4 2 | 2+2 2 | 2+2 2 |
| | | 5+6 2 | | 3+4 2 | 3+3 2 | 3+3 2 |
| 15.2.71 | 189154 | | 176 | Cl2Cu2O6Pb1Se2 | | |
| | -A | -GM | -M | -Y | -L | -V |
| | 2+2 4 | 3+4 2 | 2+2 4 | 3+4 2 | 2+2 2 | 2+2 2 |
| | 2+2 4 | 5+6 2 | 2+2 4 | 5+6 2 | 3+3 2 | 3+3 2 |
| | 2+2 4 | 5+6 2 | 2+2 4 | 5+6 2 | 3+3 2 | 3+3 2 |
| | 2+2 4 | 3+4 2 | 2+2 4 | 3+4 2 | 2+2 2 | 2+2 2 |
| | 2+2 4 | 5+6 2 | 2+2 4 | 5+6 2 | 2+2 2 | 3+3 2 |
| | 2+2 4 | 3+4 2 | 2+2 4 | 3+4 2 | 3+3 2 | 2+2 2 |
| | 2+2 4 | 3+4 2 | 2+2 4 | 3+4 2 | 2+2 2 | 2+2 2 |
| | 2+2 4 | 5+6 2 | 2+2 4 | 5+6 2 | 3+3 2 | 3+3 2 |
| | 2+2 4 | 3+4 2 | 2+2 4 | 3+4 2 | 2+2 2 | 2+2 2 |
| | 2+2 4 | 3+4 2 | 2+2 4 | 5+6 2 | 3+3 2 | 3+3 2 |
| | 2+2 4 | 5+6 2 | 2+2 4 | 5+6 2 | 2+2 2 | 2+2 2 |
| | 2+2 4 | 5+6 2 | 2+2 4 | 3+4 2 | 3+3 2 | 3+3 2 |
| | 2+2 4 | 3+4 2 | 2+2 4 | 3+4 2 | 2+2 2 | 3+3 2 |
| | 2+2 4 | 5+6 2 | 2+2 4 | 5+6 2 | 3+3 2 | 2+2 2 |
| | 2+2 4 | 3+4 2 | 2+2 4 | 5+6 2 | 2+2 2 | 2+2 2 |
| | 2+2 4 | 5+6 2 | 2+2 4 | 3+4 2 | 3+3 2 | 3+3 2 |
| | 2+2 4 | 3+4 2 | 2+2 4 | 5+6 2 | 2+2 2 | 2+2 2 |
| | 2+2 4 | 5+6 2 | 2+2 4 | 3+4 2 | 3+3 2 | 3+3 2 |
| | 2+2 4 | 3+4 2 | 2+2 4 | 3+4 2 | 2+2 2 | 2+2 2 |
| | 2+2 4 | 5+6 2 | 2+2 4 | 5+6 2 | 3+3 2 | 3+3 2 |
| | 2+2 4 | 3+4 2 | 2+2 4 | 5+6 2 | 2+2 2 | 3+3 2 |
| | 2+2 4 | 5+6 2 | 2+2 4 | 3+4 2 | 3+3 2 | 2+2 2 |
| | 2+2 4 | 3+4 2 | 2+2 4 | 5+6 2 | 2+2 2 | 2+2 2 |
| | 2+2 4 | 5+6 2 | 2+2 4 | 3+4 2 | 3+3 2 | 3+3 2 |
| | 2+2 4 | 3+4 2 | 2+2 4 | 5+6 2 | 2+2 2 | 3+3 2 |
| | 2+2 4 | 5+6 2 | 2+2 4 | 3+4 2 | 2+2 2 | 2+2 2 |

Continued on next column

**TABLE XXVII– continued**

| TQC | ICSD | | NoE | SF | | |
|---|---|---|---|---|---|---|
| | 2+2 4 | 5+6 2 | 2+2 4 | 5+6 2 | 3+3 2 | 3+3 2 |
| | 2+2 4 | 3+4 2 | 2+2 4 | 5+6 2 | 3+3 2 | 2+2 2 |
| | 2+2 4 | 3+4 2 | 2+2 4 | 3+4 2 | 2+2 2 | 3+3 2 |
| | 2+2 4 | 5+6 2 | 2+2 4 | 5+6 2 | 3+3 2 | 2+2 2 |
| | 2+2 4 | 3+4 2 | 2+2 4 | 3+4 2 | 3+3 2 | 2+2 2 |
| | 2+2 4 | 3+4 2 | 2+2 4 | 5+6 2 | 3+3 2 | 3+3 2 |
| | 2+2 4 | 5+6 2 | 2+2 4 | 3+4 2 | 2+2 2 | 3+3 2 |
| | 2+2 4 | 5+6 2 | 2+2 4 | 3+4 2 | 2+2 2 | 2+2 2 |
| | 2+2 4 | 3+4 2 | 2+2 4 | 5+6 2 | 2+2 2 | 3+3 2 |
| | 2+2 4 | 3+4 2 | 2+2 4 | 5+6 2 | 3+3 2 | 2+2 2 |
| | 2+2 4 | 3+4 2 | 2+2 4 | 3+4 2 | 2+2 2 | 2+2 2 |
| | 2+2 4 | 5+6 2 | 2+2 4 | 3+4 2 | 2+2 2 | 3+3 2 |
| | 2+2 4 | 3+4 2 | 2+2 4 | 5+6 2 | 2+2 2 | 3+3 2 |
| | 2+2 4 | 3+4 2 | 2+2 4 | 5+6 2 | 3+3 2 | 2+2 2 |
| | 2+2 4 | 5+6 2 | 2+2 4 | 3+4 2 | 2+2 2 | 2+2 2 |
| | 2+2 4 | 5+6 2 | 2+2 4 | 3+4 2 | 3+3 2 | 3+3 2 |
| | 2+2 4 | 3+4 2 | 2+2 4 | 5+6 2 | 2+2 2 | 2+2 2 |
| | 2+2 4 | 5+6 2 | 2+2 4 | 5+6 2 | 3+3 2 | 3+3 2 |
| | | 3+4 2 | | 3+4 2 | 2+2 2 | 2+2 2 |
| | | 3+4 2 | | 3+4 2 | 2+2 2 | 3+3 2 |
| | | 3+4 2 | | 5+6 2 | 3+3 2 | 2+2 2 |
| | | 5+6 2 | | 3+4 2 | 3+3 2 | 2+2 2 |
| | | 5+6 2 | | 5+6 2 | 2+2 2 | 3+3 2 |
| | | 5+6 2 | | 3+4 2 | 3+3 2 | 3+3 2 |
| | | 3+4 2 | | 5+6 2 | 3+3 2 | 2+2 2 |
| | | 3+4 2 | | 3+4 2 | 2+2 2 | 2+2 2 |
| | | 3+4 2 | | 5+6 2 | 3+3 2 | 3+3 2 |
| | | 5+6 2 | | 3+4 2 | 2+2 2 | 3+3 2 |
| | | 5+6 2 | | 5+6 2 | 3+3 2 | 3+3 2 |
| | | 5+6 2 | | 5+6 2 | 2+2 2 | 2+2 2 |
| | | 3+4 2 | | 5+6 2 | 2+2 2 | 2+2 2 |
| | | 3+4 2 | | 5+6 2 | 3+3 2 | 3+3 2 |
| | | 5+6 2 | | 3+4 2 | 2+2 2 | 3+3 2 |
| | | 3+4 2 | | 3+4 2 | 2+2 2 | 2+2 2 |
| | | 5+6 2 | | 5+6 2 | 3+3 2 | 3+3 2 |
| | | 5+6 2 | | 3+4 2 | 2+2 2 | 2+2 2 |
| | | 5+6 2 | | 5+6 2 | 3+3 2 | 3+3 2 |
| | | 3+4 2 | | 5+6 2 | 2+2 2 | 2+2 2 |
| | | 3+4 2 | | 5+6 2 | 3+3 2 | 3+3 2 |
| | | 3+4 2 | | 3+4 2 | 3+3 2 | 2+2 2 |
| | | 5+6 2 | | 3+4 2 | 2+2 2 | 2+2 2 |
| | | 5+6 2 | | 3+4 2 | 3+3 2 | 3+3 2 |
| | | 5+6 2 | | 5+6 2 | 3+3 2 | 3+3 2 |
| | | 3+4 2 | | 5+6 2 | 2+2 2 | 2+2 2 |
| | | 3+4 2 | | 3+4 2 | 2+2 2 | 3+3 2 |
| | | 5+6 2 | | 5+6 2 | 2+2 2 | 2+2 2 |
| | | 3+4 2 | | 3+4 2 | 3+3 2 | 2+2 2 |
| | | 3+4 2 | | 5+6 2 | 2+2 2 | 3+3 2 |
| | | 3+4 2 | | 3+4 2 | 2+2 2 | 2+2 2 |
| | | 5+6 2 | | 5+6 2 | 2+2 2 | 2+2 2 |
| | | 3+4 2 | | 5+6 2 | 3+3 2 | 3+3 2 |
| | | 3+4 2 | | 3+4 2 | 2+2 2 | 3+3 2 |
| | | 5+6 2 | | 5+6 2 | 3+3 2 | 2+2 2 |
| | | 3+4 2 | | 5+6 2 | 2+2 2 | 3+3 2 |
| | | 3+4 2 | | 3+4 2 | 2+2 2 | 2+2 2 |
| | | 5+6 2 | | 5+6 2 | 3+3 2 | 2+2 2 |
| | | 3+4 2 | | 5+6 2 | 2+2 2 | 3+3 2 |
| | | 5+6 2 | | 5+6 2 | 3+3 2 | 3+3 2 |
| | | 3+4 2 | | 3+4 2 | 2+2 2 | 3+3 2 |
| | | 3+4 2 | | 5+6 2 | 2+2 2 | 2+2 2 |
| | | 3+4 2 | | 3+4 2 | 2+2 2 | 3+3 2 |

Continued on next column

**TABLE XXVII– continued**

| TQC | ICSD | | NoE | SF | | |
|---|---|---|---|---|---|---|
| | | 3+4 2 | | 5+6 2 | 3+3 2 | 2+2 2 |
| 15.2.72 | 247292 | | 192 | Ce4Rh4Si3 | | |
| | -A | -GM | -M | -Y | -L | -V |
| | 2+2 4 | 3+4 2 | 2+2 4 | 3+4 2 | 2+2 2 | 2+2 2 |
| | 2+2 4 | 3+4 2 | 2+2 4 | 5+6 2 | 2+2 2 | 2+2 2 |
| | 2+2 4 | 3+4 2 | 2+2 4 | 5+6 2 | 3+3 2 | 3+3 2 |
| | 2+2 4 | 5+6 2 | 2+2 4 | 3+4 2 | 2+2 2 | 3+3 2 |
| | 2+2 4 | 5+6 2 | 2+2 4 | 5+6 2 | 3+3 2 | 2+2 2 |
| | 2+2 4 | 3+4 2 | 2+2 4 | 3+4 2 | 3+3 2 | 3+3 2 |
| | 2+2 4 | 5+6 2 | 2+2 4 | 3+4 2 | 2+2 2 | 2+2 2 |
| | 2+2 4 | 5+6 2 | 2+2 4 | 5+6 2 | 3+3 2 | 3+3 2 |
| | 2+2 4 | 3+4 2 | 2+2 4 | 3+4 2 | 2+2 2 | 2+2 2 |
| | 2+2 4 | 3+4 2 | 2+2 4 | 5+6 2 | 3+3 2 | 2+2 2 |
| | 2+2 4 | 5+6 2 | 2+2 4 | 5+6 2 | 2+2 2 | 3+3 2 |
| | 2+2 4 | 3+4 2 | 2+2 4 | 3+4 2 | 2+2 2 | 3+3 2 |
| | 2+2 4 | 5+6 2 | 2+2 4 | 5+6 2 | 3+3 2 | 2+2 2 |
| | 2+2 4 | 3+4 2 | 2+2 4 | 3+4 2 | 3+3 2 | 3+3 2 |
| | 2+2 4 | 5+6 2 | 2+2 4 | 3+4 2 | 2+2 2 | 2+2 2 |
| | 2+2 4 | 5+6 2 | 2+2 4 | 5+6 2 | 3+3 2 | 3+3 2 |
| | 2+2 4 | 3+4 2 | 2+2 4 | 3+4 2 | 2+2 2 | 2+2 2 |
| | 2+2 4 | 3+4 2 | 2+2 4 | 5+6 2 | 3+3 2 | 3+3 2 |
| | 2+2 4 | 5+6 2 | 2+2 4 | 5+6 2 | 2+2 2 | 2+2 2 |
| | 2+2 4 | 3+4 2 | 2+2 4 | 3+4 2 | 2+2 2 | 2+2 2 |
| | 2+2 4 | 5+6 2 | 2+2 4 | 5+6 2 | 3+3 2 | 3+3 2 |
| | 2+2 4 | 3+4 2 | 2+2 4 | 3+4 2 | 3+3 2 | 3+3 2 |
| | 2+2 4 | 5+6 2 | 2+2 4 | 3+4 2 | 2+2 2 | 2+2 2 |
| | 2+2 4 | 3+4 2 | 2+2 4 | 5+6 2 | 3+3 2 | 3+3 2 |
| | 2+2 4 | 3+4 2 | 2+2 4 | 5+6 2 | 2+2 2 | 2+2 2 |
| | 2+2 4 | 5+6 2 | 2+2 4 | 3+4 2 | 3+3 2 | 3+3 2 |
| | 2+2 4 | 5+6 2 | 2+2 4 | 3+4 2 | 2+2 2 | 2+2 2 |
| | 2+2 4 | 3+4 2 | 2+2 4 | 5+6 2 | 3+3 2 | 2+2 2 |
| | 2+2 4 | 5+6 2 | 2+2 4 | 3+4 2 | 2+2 2 | 3+3 2 |
| | 2+2 4 | 3+4 2 | 2+2 4 | 5+6 2 | 3+3 2 | 3+3 2 |
| | 2+2 4 | 5+6 2 | 2+2 4 | 3+4 2 | 2+2 2 | 2+2 2 |
| | 2+2 4 | 5+6 2 | 2+2 4 | 5+6 2 | 3+3 2 | 3+3 2 |
| | 2+2 4 | 3+4 2 | 2+2 4 | 5+6 2 | 3+3 2 | 3+3 2 |
| | 2+2 4 | 3+4 2 | 2+2 4 | 3+4 2 | 3+3 2 | 2+2 2 |
| | 2+2 4 | 5+6 2 | 2+2 4 | 3+4 2 | 2+2 2 | 3+3 2 |
| | 2+2 4 | 5+6 2 | 2+2 4 | 3+4 2 | 2+2 2 | 2+2 2 |
| | 2+2 4 | 3+4 2 | 2+2 4 | 5+6 2 | 2+2 2 | 3+3 2 |
| | 2+2 4 | 5+6 2 | 2+2 4 | 5+6 2 | 3+3 2 | 2+2 2 |
| | 2+2 4 | 3+4 2 | 2+2 4 | 5+6 2 | 3+3 2 | 3+3 2 |
| | 2+2 4 | 5+6 2 | 2+2 4 | 3+4 2 | 3+3 2 | 2+2 2 |
| | 2+2 4 | 3+4 2 | 2+2 4 | 3+4 2 | 2+2 2 | 3+3 2 |
| | 2+2 4 | 5+6 2 | 2+2 4 | 3+4 2 | 2+2 2 | 2+2 2 |
| | 2+2 4 | 3+4 2 | 2+2 4 | 5+6 2 | 2+2 2 | 2+2 2 |
| | 2+2 4 | 5+6 2 | 2+2 4 | 5+6 2 | 3+3 2 | 3+3 2 |
| | 2+2 4 | 5+6 2 | 2+2 4 | 5+6 2 | 2+2 2 | 2+2 2 |
| | 2+2 4 | 5+6 2 | 2+2 4 | 3+4 2 | 3+3 2 | 3+3 2 |
| | 2+2 4 | 3+4 2 | 2+2 4 | 5+6 2 | 2+2 2 | 2+2 2 |
| | 2+2 4 | 5+6 2 | 2+2 4 | 3+4 2 | 2+2 2 | 3+3 2 |
| | | 3+4 2 | | 3+4 2 | 3+3 2 | 3+3 2 |
| | | 3+4 2 | | 5+6 2 | 3+3 2 | 2+2 2 |
| | | 5+6 2 | | 3+4 2 | 2+2 2 | 2+2 2 |
| | | 5+6 2 | | 5+6 2 | 2+2 2 | 3+3 2 |
| | | 3+4 2 | | 5+6 2 | 3+3 2 | 2+2 2 |
| | | 5+6 2 | | 3+4 2 | 2+2 2 | 2+2 2 |
| | | 3+4 2 | | 5+6 2 | 3+3 2 | 3+3 2 |
| | | 5+6 2 | | 3+4 2 | 3+3 2 | 3+3 2 |
| | | 3+4 2 | | 3+4 2 | 3+3 2 | 2+2 2 |
| | | 5+6 2 | | 5+6 2 | 2+2 2 | 3+3 2 |

Continued on next column

**TABLE XXVII– continued**

| TQC | ICSD | | NoE | SF | | |
|---|---|---|---|---|---|---|
| | | 5+6 2 | | 3+4 2 | 3+3 2 | 3+3 2 |
| | | 3+4 2 | | 5+6 2 | 2+2 2 | 2+2 2 |
| | | 3+4 2 | | 5+6 2 | 3+3 2 | 2+2 2 |
| | | 3+4 2 | | 3+4 2 | 2+2 2 | 3+3 2 |
| | | 5+6 2 | | 5+6 2 | 2+2 2 | 2+2 2 |
| | | 3+4 2 | | 3+4 2 | 3+3 2 | 3+3 2 |
| | | 5+6 2 | | 5+6 2 | 3+3 2 | 3+3 2 |
| | | 3+4 2 | | 3+4 2 | 2+2 2 | 2+2 2 |
| | | 5+6 2 | | 5+6 2 | 3+3 2 | 2+2 2 |
| | | 5+6 2 | | 5+6 2 | 2+2 2 | 3+3 2 |
| | | 3+4 2 | | 3+4 2 | 2+2 2 | 2+2 2 |
| | | 5+6 2 | | 5+6 2 | 3+3 2 | 3+3 2 |
| | | 5+6 2 | | 3+4 2 | 2+2 2 | 2+2 2 |
| | | 5+6 2 | | 3+4 2 | 3+3 2 | 3+3 2 |
| | | 3+4 2 | | 5+6 2 | 2+2 2 | 2+2 2 |
| | | 5+6 2 | | 5+6 2 | 3+3 2 | 2+2 2 |
| | | 3+4 2 | | 3+4 2 | 3+3 2 | 3+3 2 |
| | | 3+4 2 | | 3+4 2 | 2+2 2 | 2+2 2 |
| | | 3+4 2 | | 5+6 2 | 2+2 2 | 3+3 2 |
| | | 3+4 2 | | 3+4 2 | 3+3 2 | 3+3 2 |
| | | 5+6 2 | | 5+6 2 | 3+3 2 | 3+3 2 |
| | | 5+6 2 | | 3+4 2 | 2+2 2 | 2+2 2 |
| | | 3+4 2 | | 5+6 2 | 2+2 2 | 3+3 2 |
| | | 3+4 2 | | 3+4 2 | 3+3 2 | 2+2 2 |
| | | 3+4 2 | | 3+4 2 | 3+3 2 | 3+3 2 |
| | | 5+6 2 | | 5+6 2 | 2+2 2 | 2+2 2 |
| | | 3+4 2 | | 3+4 2 | 3+3 2 | 3+3 2 |
| | | 3+4 2 | | 3+4 2 | 2+2 2 | 3+3 2 |
| | | 5+6 2 | | 5+6 2 | 3+3 2 | 2+2 2 |
| | | 3+4 2 | | 3+4 2 | 3+3 2 | 2+2 2 |
| | | 3+4 2 | | 5+6 2 | 3+3 2 | 3+3 2 |
| | | 3+4 2 | | 5+6 2 | 2+2 2 | 3+3 2 |
| | | 5+6 2 | | 5+6 2 | 3+3 2 | 2+2 2 |
| | | 5+6 2 | | 5+6 2 | 2+2 2 | 3+3 2 |
| | | 5+6 2 | | 3+4 2 | 2+2 2 | 2+2 2 |
| | | 3+4 2 | | 3+4 2 | 3+3 2 | 3+3 2 |
| | | 5+6 2 | | 3+4 2 | 2+2 2 | 2+2 2 |
| | | 3+4 2 | | 5+6 2 | 3+3 2 | 2+2 2 |
| 15.2.73 | 95789 | | 52 | Ca2Ir2Si1 | | |
| | -A | -GM | -M | -Y | -L | -V |
| | 2+2 4 | 3+4 2 | 2+2 4 | 3+4 2 | 3+3 2 | 3+3 2 |
| | 2+2 4 | 5+6 2 | 2+2 4 | 5+6 2 | 2+2 2 | 2+2 2 |
| | 2+2 4 | 3+4 2 | 2+2 4 | 5+6 2 | 3+3 2 | 3+3 2 |
| | 2+2 4 | 5+6 2 | 2+2 4 | 3+4 2 | 3+3 2 | 2+2 2 |
| | 2+2 4 | 5+6 2 | 2+2 4 | 5+6 2 | 2+2 2 | 2+2 2 |
| | 2+2 4 | 3+4 2 | 2+2 4 | 5+6 2 | 2+2 2 | 2+2 2 |
| | 2+2 4 | 5+6 2 | 2+2 4 | 3+4 2 | 2+2 2 | 3+3 2 |
| | 2+2 4 | 3+4 2 | 2+2 4 | 3+4 2 | 3+3 2 | 3+3 2 |
| | 2+2 4 | 5+6 2 | 2+2 4 | 3+4 2 | 2+2 2 | 2+2 2 |
| | 2+2 4 | 3+4 2 | 2+2 4 | 5+6 2 | 3+3 2 | 3+3 2 |
| | 2+2 4 | 3+4 2 | 2+2 4 | 3+4 2 | 2+2 2 | 3+3 2 |
| | 2+2 4 | 5+6 2 | 2+2 4 | 5+6 2 | 3+3 2 | 2+2 2 |
| | 2+2 4 | 3+4 2 | 2+2 4 | 5+6 2 | 2+2 2 | 2+2 2 |
| | | 5+6 2 | | 3+4 2 | 3+3 2 | 3+3 2 |
| | | 3+4 2 | | 5+6 2 | 3+3 2 | 3+3 2 |
| | | 5+6 2 | | 3+4 2 | 2+2 2 | 2+2 2 |
| | | 5+6 2 | | 3+4 2 | 2+2 2 | 3+3 2 |
| | | 3+4 2 | | 5+6 2 | 3+3 2 | 2+2 2 |
| | | 5+6 2 | | 3+4 2 | 2+2 2 | 2+2 2 |
| | | 3+4 2 | | 5+6 2 | 3+3 2 | 3+3 2 |
| | | 5+6 2 | | 3+4 2 | 2+2 2 | 2+2 2 |

Continued on next column

**TABLE XXVII– continued**

| TQC | ICSD | | NoE | SF | | |
|---|---|---|---|---|---|---|
| | | 3+4 2 | | 5+6 2 | 2+2 2 | 3+3 2 |
| | | 3+4 2 | | 3+4 2 | 3+3 2 | 2+2 2 |
| | | 3+4 2 | | 5+6 2 | 3+3 2 | 3+3 2 |
| | | 3+4 2 | | 3+4 2 | 2+2 2 | 2+2 2 |
| | | 5+6 2 | | 3+4 2 | 3+3 2 | 3+3 2 |
| 15.2.74 | 82216 | | 112 | Ca1Ni1O6Si2 | | |
| | -A | -GM | -M | -Y | -L | -V |
| | 2+2 4 | 3+4 2 | 2+2 4 | 3+4 2 | 2+2 2 | 3+3 2 |
| | 2+2 4 | 5+6 2 | 2+2 4 | 5+6 2 | 3+3 2 | 2+2 2 |
| | 2+2 4 | 3+4 2 | 2+2 4 | 3+4 2 | 2+2 2 | 3+3 2 |
| | 2+2 4 | 5+6 2 | 2+2 4 | 5+6 2 | 3+3 2 | 2+2 2 |
| | 2+2 4 | 3+4 2 | 2+2 4 | 3+4 2 | 2+2 2 | 2+2 2 |
| | 2+2 4 | 3+4 2 | 2+2 4 | 5+6 2 | 3+3 2 | 3+3 2 |
| | 2+2 4 | 5+6 2 | 2+2 4 | 3+4 2 | 2+2 2 | 2+2 2 |
| | 2+2 4 | 5+6 2 | 2+2 4 | 5+6 2 | 3+3 2 | 3+3 2 |
| | 2+2 4 | 5+6 2 | 2+2 4 | 5+6 2 | 3+3 2 | 2+2 2 |
| | 2+2 4 | 3+4 2 | 2+2 4 | 3+4 2 | 3+3 2 | 3+3 2 |
| | 2+2 4 | 3+4 2 | 2+2 4 | 3+4 2 | 2+2 2 | 3+3 2 |
| | 2+2 4 | 5+6 2 | 2+2 4 | 5+6 2 | 2+2 2 | 2+2 2 |
| | 2+2 4 | 5+6 2 | 2+2 4 | 5+6 2 | 2+2 2 | 2+2 2 |
| | 2+2 4 | 3+4 2 | 2+2 4 | 3+4 2 | 3+3 2 | 3+3 2 |
| | 2+2 4 | 3+4 2 | 2+2 4 | 5+6 2 | 2+2 2 | 3+3 2 |
| | 2+2 4 | 3+4 2 | 2+2 4 | 3+4 2 | 3+3 2 | 2+2 2 |
| | 2+2 4 | 5+6 2 | 2+2 4 | 3+4 2 | 2+2 2 | 2+2 2 |
| | 2+2 4 | 5+6 2 | 2+2 4 | 3+4 2 | 3+3 2 | 3+3 2 |
| | 2+2 4 | 5+6 2 | 2+2 4 | 5+6 2 | 3+3 2 | 2+2 2 |
| | 2+2 4 | 3+4 2 | 2+2 4 | 5+6 2 | 2+2 2 | 3+3 2 |
| | 2+2 4 | 3+4 2 | 2+2 4 | 5+6 2 | 2+2 2 | 3+3 2 |
| | 2+2 4 | 3+4 2 | 2+2 4 | 3+4 2 | 3+3 2 | 2+2 2 |
| | 2+2 4 | 5+6 2 | 2+2 4 | 5+6 2 | 2+2 2 | 3+3 2 |
| | 2+2 4 | 5+6 2 | 2+2 4 | 3+4 2 | 3+3 2 | 2+2 2 |
| | 2+2 4 | 3+4 2 | 2+2 4 | 3+4 2 | 3+3 2 | 3+3 2 |
| | 2+2 4 | 5+6 2 | 2+2 4 | 5+6 2 | 2+2 2 | 2+2 2 |
| | 2+2 4 | 5+6 2 | 2+2 4 | 3+4 2 | 3+3 2 | 2+2 2 |
| | 2+2 4 | 3+4 2 | 2+2 4 | 5+6 2 | 2+2 2 | 3+3 2 |
| | | 3+4 2 | | 5+6 2 | 2+2 2 | 2+2 2 |
| | | 3+4 2 | | 3+4 2 | 3+3 2 | 3+3 2 |
| | | 5+6 2 | | 3+4 2 | 2+2 2 | 2+2 2 |
| | | 3+4 2 | | 5+6 2 | 3+3 2 | 3+3 2 |
| | | 5+6 2 | | 3+4 2 | 3+3 2 | 3+3 2 |
| | | 3+4 2 | | 5+6 2 | 2+2 2 | 2+2 2 |
| | | 5+6 2 | | 3+4 2 | 2+2 2 | 3+3 2 |
| | | 3+4 2 | | 5+6 2 | 3+3 2 | 2+2 2 |
| | | 5+6 2 | | 3+4 2 | 3+3 2 | 3+3 2 |
| | | 3+4 2 | | 5+6 2 | 2+2 2 | 2+2 2 |
| | | 3+4 2 | | 3+4 2 | 3+3 2 | 2+2 2 |
| | | 5+6 2 | | 5+6 2 | 2+2 2 | 3+3 2 |
| | | 5+6 2 | | 5+6 2 | 2+2 2 | 2+2 2 |
| | | 5+6 2 | | 3+4 2 | 3+3 2 | 2+2 2 |
| | | 3+4 2 | | 3+4 2 | 3+3 2 | 3+3 2 |
| | | 5+6 2 | | 3+4 2 | 2+2 2 | 3+3 2 |
| | | 3+4 2 | | 5+6 2 | 2+2 2 | 2+2 2 |
| | | 5+6 2 | | 5+6 2 | 3+3 2 | 3+3 2 |
| | | 3+4 2 | | 5+6 2 | 3+3 2 | 2+2 2 |
| | | 5+6 2 | | 3+4 2 | 2+2 2 | 3+3 2 |
| | | 3+4 2 | | 3+4 2 | 2+2 2 | 2+2 2 |
| | | 3+4 2 | | 5+6 2 | 3+3 2 | 3+3 2 |
| | | 5+6 2 | | 5+6 2 | 2+2 2 | 3+3 2 |
| | | 5+6 2 | | 3+4 2 | 3+3 2 | 2+2 2 |
| | | 5+6 2 | | 5+6 2 | 2+2 2 | 3+3 2 |
| | | 3+4 2 | | 3+4 2 | 3+3 2 | 2+2 2 |

Continued on next column

**TABLE XXVII– continued**

| TQC | ICSD | | NoE | SF | | |
|---|---|---|---|---|---|---|
| | | 5+6 2 | | 5+6 2 | 3+3 2 | 3+3 2 |
| | | 5+6 2 | | 3+4 2 | 2+2 2 | 2+2 2 |
| 15.2.75 | 69381 | | 128 | Ca1Ni1O6Si2 | | |
| | -A | -GM | -M | -Y | -L | -V |
| | 2+2 4 | 3+4 2 | 2+2 4 | 3+4 2 | 2+2 2 | 2+2 2 |
| | 2+2 4 | 5+6 2 | 2+2 4 | 5+6 2 | 3+3 2 | 3+3 2 |
| | 2+2 4 | 3+4 2 | 2+2 4 | 3+4 2 | 2+2 2 | 3+3 2 |
| | 2+2 4 | 5+6 2 | 2+2 4 | 5+6 2 | 3+3 2 | 2+2 2 |
| | 2+2 4 | 3+4 2 | 2+2 4 | 3+4 2 | 2+2 2 | 3+3 2 |
| | 2+2 4 | 5+6 2 | 2+2 4 | 5+6 2 | 3+3 2 | 2+2 2 |
| | 2+2 4 | 3+4 2 | 2+2 4 | 3+4 2 | 3+3 2 | 3+3 2 |
| | 2+2 4 | 3+4 2 | 2+2 4 | 5+6 2 | 2+2 2 | 2+2 2 |
| | 2+2 4 | 5+6 2 | 2+2 4 | 3+4 2 | 3+3 2 | 3+3 2 |
| | 2+2 4 | 5+6 2 | 2+2 4 | 5+6 2 | 2+2 2 | 2+2 2 |
| | 2+2 4 | 5+6 2 | 2+2 4 | 5+6 2 | 3+3 2 | 2+2 2 |
| | 2+2 4 | 3+4 2 | 2+2 4 | 3+4 2 | 2+2 2 | 3+3 2 |
| | 2+2 4 | 3+4 2 | 2+2 4 | 3+4 2 | 2+2 2 | 2+2 2 |
| | 2+2 4 | 5+6 2 | 2+2 4 | 5+6 2 | 3+3 2 | 2+2 2 |
| | 2+2 4 | 3+4 2 | 2+2 4 | 3+4 2 | 3+3 2 | 3+3 2 |
| | 2+2 4 | 5+6 2 | 2+2 4 | 5+6 2 | 2+2 2 | 3+3 2 |
| | 2+2 4 | 5+6 2 | 2+2 4 | 5+6 2 | 3+3 2 | 2+2 2 |
| | 2+2 4 | 3+4 2 | 2+2 4 | 3+4 2 | 2+2 2 | 3+3 2 |
| | 2+2 4 | 3+4 2 | 2+2 4 | 3+4 2 | 3+3 2 | 3+3 2 |
| | 2+2 4 | 5+6 2 | 2+2 4 | 5+6 2 | 2+2 2 | 2+2 2 |
| | 2+2 4 | 5+6 2 | 2+2 4 | 5+6 2 | 2+2 2 | 2+2 2 |
| | 2+2 4 | 3+4 2 | 2+2 4 | 3+4 2 | 3+3 2 | 3+3 2 |
| | 2+2 4 | 3+4 2 | 2+2 4 | 3+4 2 | 2+2 2 | 2+2 2 |
| | 2+2 4 | 3+4 2 | 2+2 4 | 5+6 2 | 3+3 2 | 3+3 2 |
| | 2+2 4 | 5+6 2 | 2+2 4 | 3+4 2 | 2+2 2 | 2+2 2 |
| | 2+2 4 | 5+6 2 | 2+2 4 | 3+4 2 | 3+3 2 | 3+3 2 |
| | 2+2 4 | 5+6 2 | 2+2 4 | 5+6 2 | 3+3 2 | 2+2 2 |
| | 2+2 4 | 3+4 2 | 2+2 4 | 5+6 2 | 2+2 2 | 3+3 2 |
| | 2+2 4 | 3+4 2 | 2+2 4 | 5+6 2 | 2+2 2 | 3+3 2 |
| | 2+2 4 | 5+6 2 | 2+2 4 | 3+4 2 | 3+3 2 | 2+2 2 |
| | 2+2 4 | 3+4 2 | 2+2 4 | 5+6 2 | 2+2 2 | 3+3 2 |
| | 2+2 4 | 5+6 2 | 2+2 4 | 3+4 2 | 3+3 2 | 3+3 2 |
| | | 3+4 2 | | 5+6 2 | 3+3 2 | 2+2 2 |
| | | 5+6 2 | | 3+4 2 | 2+2 2 | 2+2 2 |
| | | 5+6 2 | | 3+4 2 | 3+3 2 | 2+2 2 |
| | | 3+4 2 | | 5+6 2 | 2+2 2 | 3+3 2 |
| | | 3+4 2 | | 3+4 2 | 2+2 2 | 2+2 2 |
| | | 3+4 2 | | 5+6 2 | 3+3 2 | 3+3 2 |
| | | 5+6 2 | | 3+4 2 | 2+2 2 | 2+2 2 |
| | | 3+4 2 | | 5+6 2 | 3+3 2 | 3+3 2 |
| | | 5+6 2 | | 3+4 2 | 3+3 2 | 3+3 2 |
| | | 3+4 2 | | 5+6 2 | 2+2 2 | 2+2 2 |
| | | 3+4 2 | | 3+4 2 | 2+2 2 | 3+3 2 |
| | | 3+4 2 | | 5+6 2 | 3+3 2 | 2+2 2 |
| | | 5+6 2 | | 3+4 2 | 3+3 2 | 3+3 2 |
| | | 5+6 2 | | 5+6 2 | 3+3 2 | 2+2 2 |
| | | 3+4 2 | | 3+4 2 | 2+2 2 | 2+2 2 |
| | | 5+6 2 | | 5+6 2 | 2+2 2 | 3+3 2 |
| | | 5+6 2 | | 5+6 2 | 3+3 2 | 2+2 2 |
| | | 3+4 2 | | 3+4 2 | 2+2 2 | 2+2 2 |
| | | 5+6 2 | | 3+4 2 | 3+3 2 | 3+3 2 |
| | | 5+6 2 | | 5+6 2 | 2+2 2 | 3+3 2 |
| | | 3+4 2 | | 3+4 2 | 2+2 2 | 2+2 2 |
| | | 3+4 2 | | 5+6 2 | 3+3 2 | 3+3 2 |
| | | 5+6 2 | | 3+4 2 | 3+3 2 | 2+2 2 |
| | | 5+6 2 | | 5+6 2 | 2+2 2 | 3+3 2 |
| | | 3+4 2 | | 3+4 2 | 2+2 2 | 2+2 2 |

Continued on next column

**TABLE XXVII– continued**

| TQC | ICSD | | NoE | SF | | |
|---|---|---|---|---|---|---|
| | | 3+4 2 | | 5+6 2 | 3+3 2 | 3+3 2 |
| | | 5+6 2 | | 5+6 2 | 2+2 2 | 3+3 2 |
| | | 5+6 2 | | 3+4 2 | 3+3 2 | 2+2 2 |
| | | 5+6 2 | | 5+6 2 | 2+2 2 | 3+3 2 |
| | | 3+4 2 | | 3+4 2 | 3+3 2 | 2+2 2 |
| | | 5+6 2 | | 3+4 2 | 3+3 2 | 3+3 2 |
| | | 5+6 2 | | 5+6 2 | 2+2 2 | 2+2 2 |
| 15.2.76 | 202244 | | 112 | Ca1Ni1O6Si2 | | |
| | -A | -GM | -M | -Y | -L | -V |
| | 2+2 4 | 3+4 2 | 2+2 4 | 3+4 2 | 2+2 2 | 3+3 2 |
| | 2+2 4 | 5+6 2 | 2+2 4 | 5+6 2 | 3+3 2 | 2+2 2 |
| | 2+2 4 | 3+4 2 | 2+2 4 | 3+4 2 | 2+2 2 | 3+3 2 |
| | 2+2 4 | 5+6 2 | 2+2 4 | 5+6 2 | 3+3 2 | 2+2 2 |
| | 2+2 4 | 3+4 2 | 2+2 4 | 5+6 2 | 2+2 2 | 2+2 2 |
| | 2+2 4 | 3+4 2 | 2+2 4 | 3+4 2 | 3+3 2 | 3+3 2 |
| | 2+2 4 | 5+6 2 | 2+2 4 | 3+4 2 | 2+2 2 | 2+2 2 |
| | 2+2 4 | 5+6 2 | 2+2 4 | 5+6 2 | 3+3 2 | 3+3 2 |
| | 2+2 4 | 5+6 2 | 2+2 4 | 5+6 2 | 3+3 2 | 2+2 2 |
| | 2+2 4 | 3+4 2 | 2+2 4 | 3+4 2 | 3+3 2 | 3+3 2 |
| | 2+2 4 | 3+4 2 | 2+2 4 | 3+4 2 | 2+2 2 | 3+3 2 |
| | 2+2 4 | 5+6 2 | 2+2 4 | 5+6 2 | 2+2 2 | 2+2 2 |
| | 2+2 4 | 5+6 2 | 2+2 4 | 5+6 2 | 2+2 2 | 2+2 2 |
| | 2+2 4 | 3+4 2 | 2+2 4 | 3+4 2 | 3+3 2 | 3+3 2 |
| | 2+2 4 | 3+4 2 | 2+2 4 | 3+4 2 | 2+2 2 | 2+2 2 |
| | 2+2 4 | 3+4 2 | 2+2 4 | 5+6 2 | 3+3 2 | 3+3 2 |
| | 2+2 4 | 5+6 2 | 2+2 4 | 3+4 2 | 2+2 2 | 2+2 2 |
| | 2+2 4 | 5+6 2 | 2+2 4 | 3+4 2 | 3+3 2 | 3+3 2 |
| | 2+2 4 | 5+6 2 | 2+2 4 | 5+6 2 | 3+3 2 | 2+2 2 |
| | 2+2 4 | 3+4 2 | 2+2 4 | 5+6 2 | 2+2 2 | 3+3 2 |
| | 2+2 4 | 3+4 2 | 2+2 4 | 5+6 2 | 2+2 2 | 3+3 2 |
| | 2+2 4 | 5+6 2 | 2+2 4 | 3+4 2 | 3+3 2 | 2+2 2 |
| | 2+2 4 | 3+4 2 | 2+2 4 | 5+6 2 | 2+2 2 | 3+3 2 |
| | 2+2 4 | 5+6 2 | 2+2 4 | 3+4 2 | 3+3 2 | 3+3 2 |
| | 2+2 4 | 3+4 2 | 2+2 4 | 3+4 2 | 3+3 2 | 2+2 2 |
| | 2+2 4 | 5+6 2 | 2+2 4 | 5+6 2 | 2+2 2 | 2+2 2 |
| | 2+2 4 | 5+6 2 | 2+2 4 | 3+4 2 | 3+3 2 | 2+2 2 |
| | 2+2 4 | 3+4 2 | 2+2 4 | 5+6 2 | 2+2 2 | 2+2 2 |
| | | 3+4 2 | | 3+4 2 | 2+2 2 | 3+3 2 |
| | | 3+4 2 | | 5+6 2 | 3+3 2 | 3+3 2 |
| | | 5+6 2 | | 3+4 2 | 2+2 2 | 2+2 2 |
| | | 3+4 2 | | 5+6 2 | 3+3 2 | 3+3 2 |
| | | 5+6 2 | | 3+4 2 | 3+3 2 | 3+3 2 |
| | | 3+4 2 | | 5+6 2 | 2+2 2 | 2+2 2 |
| | | 3+4 2 | | 3+4 2 | 2+2 2 | 3+3 2 |
| | | 3+4 2 | | 5+6 2 | 3+3 2 | 2+2 2 |
| | | 5+6 2 | | 3+4 2 | 3+3 2 | 3+3 2 |
| | | 5+6 2 | | 5+6 2 | 2+2 2 | 2+2 2 |
| | | 3+4 2 | | 3+4 2 | 3+3 2 | 2+2 2 |
| | | 5+6 2 | | 5+6 2 | 2+2 2 | 3+3 2 |
| | | 5+6 2 | | 5+6 2 | 3+3 2 | 2+2 2 |
| | | 3+4 2 | | 3+4 2 | 2+2 2 | 2+2 2 |
| | | 5+6 2 | | 3+4 2 | 3+3 2 | 3+3 2 |
| | | 5+6 2 | | 5+6 2 | 2+2 2 | 3+3 2 |
| | | 3+4 2 | | 3+4 2 | 2+2 2 | 2+2 2 |
| | | 3+4 2 | | 5+6 2 | 3+3 2 | 3+3 2 |
| | | 5+6 2 | | 3+4 2 | 3+3 2 | 2+2 2 |
| | | 5+6 2 | | 5+6 2 | 2+2 2 | 3+3 2 |
| | | 3+4 2 | | 3+4 2 | 2+2 2 | 2+2 2 |
| | | 3+4 2 | | 5+6 2 | 3+3 2 | 3+3 2 |
| | | 5+6 2 | | 5+6 2 | 2+2 2 | 3+3 2 |
| | | 5+6 2 | | 3+4 2 | 3+3 2 | 2+2 2 |

Continued on next column

**TABLE XXVII– continued**

| TQC | ICSD | | NoE | SF | | |
|---|---|---|---|---|---|---|
| | | 5+6 2 | | 5+6 2 | 2+2 2 | 3+3 2 |
| | | 3+4 2 | | 3+4 2 | 3+3 2 | 2+2 2 |
| | | 5+6 2 | | 3+4 2 | 3+3 2 | 3+3 2 |
| | | 5+6 2 | | 5+6 2 | 2+2 2 | 2+2 2 |
| 15.2.77 | 158299 | | 180 | C2H8Cl3Cu1N1 | | |
| | -A | -GM | -M | -Y | -L | -V |
| | 2+2 4 | 3+4 2 | 2+2 4 | 3+4 2 | 2+2 2 | 2+2 2 |
| | 2+2 4 | 5+6 2 | 2+2 4 | 3+4 2 | 3+3 2 | 3+3 2 |
| | 2+2 4 | 3+4 2 | 2+2 4 | 5+6 2 | 2+2 2 | 2+2 2 |
| | 2+2 4 | 5+6 2 | 2+2 4 | 5+6 2 | 3+3 2 | 3+3 2 |
| | 2+2 4 | 3+4 2 | 2+2 4 | 3+4 2 | 2+2 2 | 2+2 2 |
| | 2+2 4 | 3+4 2 | 2+2 4 | 3+4 2 | 3+3 2 | 2+2 2 |
| | 2+2 4 | 5+6 2 | 2+2 4 | 5+6 2 | 2+2 2 | 3+3 2 |
| | 2+2 4 | 5+6 2 | 2+2 4 | 5+6 2 | 3+3 2 | 3+3 2 |
| | 2+2 4 | 5+6 2 | 2+2 4 | 5+6 2 | 3+3 2 | 3+3 2 |
| | 2+2 4 | 3+4 2 | 2+2 4 | 3+4 2 | 2+2 2 | 2+2 2 |
| | 2+2 4 | 3+4 2 | 2+2 4 | 5+6 2 | 3+3 2 | 3+3 2 |
| | 2+2 4 | 5+6 2 | 2+2 4 | 3+4 2 | 2+2 2 | 2+2 2 |
| | 2+2 4 | 3+4 2 | 2+2 4 | 3+4 2 | 2+2 2 | 3+3 2 |
| | 2+2 4 | 3+4 2 | 2+2 4 | 3+4 2 | 3+3 2 | 3+3 2 |
| | 2+2 4 | 5+6 2 | 2+2 4 | 5+6 2 | 2+2 2 | 2+2 2 |
| | 2+2 4 | 5+6 2 | 2+2 4 | 5+6 2 | 3+3 2 | 2+2 2 |
| | 2+2 4 | 3+4 2 | 2+2 4 | 5+6 2 | 2+2 2 | 2+2 2 |
| | 2+2 4 | 3+4 2 | 2+2 4 | 3+4 2 | 2+2 2 | 3+3 2 |
| | 2+2 4 | 5+6 2 | 2+2 4 | 5+6 2 | 3+3 2 | 2+2 2 |
| | 2+2 4 | 5+6 2 | 2+2 4 | 3+4 2 | 3+3 2 | 3+3 2 |
| | 2+2 4 | 3+4 2 | 2+2 4 | 5+6 2 | 2+2 2 | 3+3 2 |
| | 2+2 4 | 5+6 2 | 2+2 4 | 3+4 2 | 2+2 2 | 2+2 2 |
| | 2+2 4 | 3+4 2 | 2+2 4 | 5+6 2 | 3+3 2 | 3+3 2 |
| | 2+2 4 | 5+6 2 | 2+2 4 | 3+4 2 | 3+3 2 | 2+2 2 |
| | 2+2 4 | 5+6 2 | 2+2 4 | 5+6 2 | 2+2 2 | 3+3 2 |
| | 2+2 4 | 5+6 2 | 2+2 4 | 5+6 2 | 3+3 2 | 3+3 2 |
| | 2+2 4 | 3+4 2 | 2+2 4 | 3+4 2 | 2+2 2 | 2+2 2 |
| | 2+2 4 | 3+4 2 | 2+2 4 | 3+4 2 | 3+3 2 | 2+2 2 |
| | 2+2 4 | 3+4 2 | 2+2 4 | 3+4 2 | 3+3 2 | 2+2 2 |
| | 2+2 4 | 5+6 2 | 2+2 4 | 5+6 2 | 2+2 2 | 2+2 2 |
| | 2+2 4 | 5+6 2 | 2+2 4 | 5+6 2 | 3+3 2 | 3+3 2 |
| | 2+2 4 | 3+4 2 | 2+2 4 | 3+4 2 | 2+2 2 | 3+3 2 |
| | 2+2 4 | 3+4 2 | 2+2 4 | 3+4 2 | 2+2 2 | 2+2 2 |
| | 2+2 4 | 3+4 2 | 2+2 4 | 3+4 2 | 3+3 2 | 2+2 2 |
| | 2+2 4 | 5+6 2 | 2+2 4 | 5+6 2 | 2+2 2 | 3+3 2 |
| | 2+2 4 | 3+4 2 | 2+2 4 | 5+6 2 | 3+3 2 | 2+2 2 |
| | 2+2 4 | 5+6 2 | 2+2 4 | 3+4 2 | 2+2 2 | 2+2 2 |
| | 2+2 4 | 3+4 2 | 2+2 4 | 3+4 2 | 3+3 2 | 3+3 2 |
| | 2+2 4 | 5+6 2 | 2+2 4 | 5+6 2 | 2+2 2 | 3+3 2 |
| | 2+2 4 | 5+6 2 | 2+2 4 | 5+6 2 | 3+3 2 | 3+3 2 |
| | 2+2 4 | 3+4 2 | 2+2 4 | 5+6 2 | 2+2 2 | 2+2 2 |
| | 2+2 4 | 3+4 2 | 2+2 4 | 3+4 2 | 2+2 2 | 3+3 2 |
| | 2+2 4 | 5+6 2 | 2+2 4 | 3+4 2 | 3+3 2 | 2+2 2 |
| | 2+2 4 | 5+6 2 | 2+2 4 | 5+6 2 | 3+3 2 | 3+3 2 |
| | 2+2 4 | 3+4 2 | 2+2 4 | 3+4 2 | 3+3 2 | 2+2 2 |
| | | 3+4 2 | | 5+6 2 | 2+2 2 | 3+3 2 |
| | | 5+6 2 | | 3+4 2 | 2+2 2 | 2+2 2 |
| | | 3+4 2 | | 3+4 2 | 3+3 2 | 2+2 2 |
| | | 5+6 2 | | 5+6 2 | 2+2 2 | 3+3 2 |
| | | 3+4 2 | | 3+4 2 | 3+3 2 | 2+2 2 |
| | | 5+6 2 | | 5+6 2 | 3+3 2 | 3+3 2 |
| | | 3+4 2 | | 3+4 2 | 2+2 2 | 2+2 2 |
| | | 5+6 2 | | 5+6 2 | 3+3 2 | 3+3 2 |
| | | 3+4 2 | | 3+4 2 | 2+2 2 | 3+3 2 |
| | | 5+6 2 | | 5+6 2 | 2+2 2 | 2+2 2 |

Continued on next column

**TABLE XXVII– continued**

| TQC | ICSD | | NoE | SF | | |
|---|---|---|---|---|---|---|
| | | 3+4 2 | | 3+4 2 | 3+3 2 | 2+2 2 |
| | | 3+4 2 | | 3+4 2 | 2+2 2 | 3+3 2 |
| | | 3+4 2 | | 5+6 2 | 3+3 2 | 3+3 2 |
| | | 5+6 2 | | 3+4 2 | 3+3 2 | 2+2 2 |
| | | 5+6 2 | | 5+6 2 | 2+2 2 | 2+2 2 |
| | | 5+6 2 | | 5+6 2 | 3+3 2 | 3+3 2 |
| | | 3+4 2 | | 3+4 2 | 2+2 2 | 2+2 2 |
| | | 5+6 2 | | 5+6 2 | 2+2 2 | 3+3 2 |
| | | 3+4 2 | | 3+4 2 | 3+3 2 | 2+2 2 |
| | | 5+6 2 | | 5+6 2 | 2+2 2 | 3+3 2 |
| | | 5+6 2 | | 5+6 2 | 3+3 2 | 3+3 2 |
| | | 5+6 2 | | 5+6 2 | 3+3 2 | 3+3 2 |
| | | 5+6 2 | | 5+6 2 | 2+2 2 | 3+3 2 |
| | | 3+4 2 | | 3+4 2 | 3+3 2 | 2+2 2 |
| | | 3+4 2 | | 3+4 2 | 2+2 2 | 2+2 2 |
| | | 3+4 2 | | 3+4 2 | 2+2 2 | 2+2 2 |
| | | 5+6 2 | | 5+6 2 | 3+3 2 | 3+3 2 |
| | | 3+4 2 | | 3+4 2 | 3+3 2 | 2+2 2 |
| | | 3+4 2 | | 5+6 2 | 2+2 2 | 3+3 2 |
| | | 5+6 2 | | 3+4 2 | 3+3 2 | 2+2 2 |
| | | 5+6 2 | | 5+6 2 | 2+2 2 | 3+3 2 |
| | | 3+4 2 | | 3+4 2 | 2+2 2 | 3+3 2 |
| | | 3+4 2 | | 3+4 2 | 3+3 2 | 2+2 2 |
| | | 5+6 2 | | 5+6 2 | 3+3 2 | 3+3 2 |
| | | 5+6 2 | | 3+4 2 | 2+2 2 | 2+2 2 |
| | | 3+4 2 | | 5+6 2 | 2+2 2 | 3+3 2 |
| | | 3+4 2 | | 3+4 2 | 3+3 2 | 2+2 2 |
| | | 5+6 2 | | 5+6 2 | 3+3 2 | 3+3 2 |
| | | 5+6 2 | | 5+6 2 | 2+2 2 | 3+3 2 |
| | | 3+4 2 | | 3+4 2 | 3+3 2 | 2+2 2 |
| | | 5+6 2 | | 5+6 2 | 2+2 2 | 2+2 2 |
| | | 3+4 2 | | 5+6 2 | 3+3 2 | 3+3 2 |
| | | 5+6 2 | | 3+4 2 | 2+2 2 | 2+2 2 |
| | | 5+6 2 | | 3+4 2 | 2+2 2 | 2+2 2 |
| | | 5+6 2 | | 5+6 2 | 2+2 2 | 3+3 2 |
| 15.2.78 | 423887 | | 96 | C2Fe5 | | |
| | -A | -GM | -M | -Y | -L | -V |
| | 2+2 4 | 3+4 2 | 2+2 4 | 3+4 2 | 2+2 2 | 3+3 2 |
| | 2+2 4 | 5+6 2 | 2+2 4 | 5+6 2 | 2+2 2 | 2+2 2 |
| | 2+2 4 | 5+6 2 | 2+2 4 | 3+4 2 | 3+3 2 | 2+2 2 |
| | 2+2 4 | 3+4 2 | 2+2 4 | 5+6 2 | 3+3 2 | 3+3 2 |
| | 2+2 4 | 3+4 2 | 2+2 4 | 5+6 2 | 3+3 2 | 3+3 2 |
| | 2+2 4 | 5+6 2 | 2+2 4 | 3+4 2 | 2+2 2 | 2+2 2 |
| | 2+2 4 | 3+4 2 | 2+2 4 | 5+6 2 | 2+2 2 | 2+2 2 |
| | 2+2 4 | 5+6 2 | 2+2 4 | 3+4 2 | 3+3 2 | 2+2 2 |
| | 2+2 4 | 3+4 2 | 2+2 4 | 3+4 2 | 2+2 2 | 3+3 2 |
| | 2+2 4 | 5+6 2 | 2+2 4 | 5+6 2 | 3+3 2 | 3+3 2 |
| | 2+2 4 | 3+4 2 | 2+2 4 | 5+6 2 | 3+3 2 | 2+2 2 |
| | 2+2 4 | 5+6 2 | 2+2 4 | 5+6 2 | 2+2 2 | 3+3 2 |
| | 2+2 4 | 3+4 2 | 2+2 4 | 5+6 2 | 3+3 2 | 2+2 2 |
| | 2+2 4 | 5+6 2 | 2+2 4 | 3+4 2 | 2+2 2 | 3+3 2 |
| | 2+2 4 | 5+6 2 | 2+2 4 | 3+4 2 | 3+3 2 | 3+3 2 |
| | 2+2 4 | 3+4 2 | 2+2 4 | 3+4 2 | 2+2 2 | 2+2 2 |
| | 2+2 4 | 3+4 2 | 2+2 4 | 5+6 2 | 3+3 2 | 3+3 2 |
| | 2+2 4 | 5+6 2 | 2+2 4 | 3+4 2 | 2+2 2 | 2+2 2 |
| | 2+2 4 | 5+6 2 | 2+2 4 | 3+4 2 | 3+3 2 | 3+3 2 |
| | 2+2 4 | 3+4 2 | 2+2 4 | 5+6 2 | 3+3 2 | 2+2 2 |
| | 2+2 4 | 5+6 2 | 2+2 4 | 5+6 2 | 3+3 2 | 3+3 2 |
| | 2+2 4 | 3+4 2 | 2+2 4 | 5+6 2 | 2+2 2 | 2+2 2 |
| | 2+2 4 | 5+6 2 | 2+2 4 | 3+4 2 | 3+3 2 | 3+3 2 |
| | 2+2 4 | 5+6 2 | 2+2 4 | 3+4 2 | 2+2 2 | 3+3 2 |

Continued on next column

**TABLE XXVII– continued**

| TQC | ICSD | | NoE | SF | | |
|---|---|---|---|---|---|---|
| | | 3+4 2 | | 3+4 2 | 2+2 2 | 2+2 2 |
| | | 3+4 2 | | 3+4 2 | 3+3 2 | 3+3 2 |
| | | 3+4 2 | | 5+6 2 | 2+2 2 | 2+2 2 |
| | | 3+4 2 | | 5+6 2 | 2+2 2 | 3+3 2 |
| | | 5+6 2 | | 3+4 2 | 3+3 2 | 2+2 2 |
| | | 3+4 2 | | 3+4 2 | 2+2 2 | 2+2 2 |
| | | 5+6 2 | | 5+6 2 | 3+3 2 | 3+3 2 |
| | | 3+4 2 | | 3+4 2 | 2+2 2 | 2+2 2 |
| | | 5+6 2 | | 5+6 2 | 3+3 2 | 3+3 2 |
| | | 5+6 2 | | 5+6 2 | 3+3 2 | 2+2 2 |
| | | 3+4 2 | | 3+4 2 | 2+2 2 | 3+3 2 |
| | | 3+4 2 | | 5+6 2 | 2+2 2 | 2+2 2 |
| | | 3+4 2 | | 3+4 2 | 2+2 2 | 2+2 2 |
| | | 5+6 2 | | 5+6 2 | 3+3 2 | 3+3 2 |
| | | 3+4 2 | | 3+4 2 | 2+2 2 | 3+3 2 |
| | | 3+4 2 | | 5+6 2 | 3+3 2 | 2+2 2 |
| | | 5+6 2 | | 5+6 2 | 2+2 2 | 2+2 2 |
| | | 3+4 2 | | 3+4 2 | 3+3 2 | 2+2 2 |
| | | 5+6 2 | | 5+6 2 | 2+2 2 | 3+3 2 |
| | | 5+6 2 | | 5+6 2 | 3+3 2 | 3+3 2 |
| | | 3+4 2 | | 3+4 2 | 2+2 2 | 2+2 2 |
| | | 3+4 2 | | 3+4 2 | 2+2 2 | 2+2 2 |
| | | 5+6 2 | | 3+4 2 | 3+3 2 | 3+3 2 |
| | | 3+4 2 | | 3+4 2 | 2+2 2 | 3+3 2 |
| 15.2.79 | 423886 | | 96 | C2Fe5 | | |
| | -A | -GM | -M | -Y | -L | -V |
| | 2+2 4 | 3+4 2 | 2+2 4 | 3+4 2 | 3+3 2 | 3+3 2 |
| | 2+2 4 | 5+6 2 | 2+2 4 | 5+6 2 | 3+3 2 | 2+2 2 |
| | 2+2 4 | 5+6 2 | 2+2 4 | 3+4 2 | 2+2 2 | 2+2 2 |
| | 2+2 4 | 3+4 2 | 2+2 4 | 5+6 2 | 2+2 2 | 3+3 2 |
| | 2+2 4 | 3+4 2 | 2+2 4 | 5+6 2 | 2+2 2 | 3+3 2 |
| | 2+2 4 | 5+6 2 | 2+2 4 | 3+4 2 | 3+3 2 | 2+2 2 |
| | 2+2 4 | 3+4 2 | 2+2 4 | 3+4 2 | 3+3 2 | 2+2 2 |
| | 2+2 4 | 5+6 2 | 2+2 4 | 5+6 2 | 2+2 2 | 2+2 2 |
| | 2+2 4 | 3+4 2 | 2+2 4 | 3+4 2 | 3+3 2 | 3+3 2 |
| | 2+2 4 | 5+6 2 | 2+2 4 | 5+6 2 | 2+2 2 | 3+3 2 |
| | 2+2 4 | 3+4 2 | 2+2 4 | 5+6 2 | 2+2 2 | 2+2 2 |
| | 2+2 4 | 5+6 2 | 2+2 4 | 5+6 2 | 3+3 2 | 3+3 2 |
| | 2+2 4 | 3+4 2 | 2+2 4 | 5+6 2 | 2+2 2 | 2+2 2 |
| | 2+2 4 | 5+6 2 | 2+2 4 | 3+4 2 | 2+2 2 | 3+3 2 |
| | 2+2 4 | 5+6 2 | 2+2 4 | 3+4 2 | 3+3 2 | 3+3 2 |
| | 2+2 4 | 3+4 2 | 2+2 4 | 3+4 2 | 3+3 2 | 2+2 2 |
| | 2+2 4 | 3+4 2 | 2+2 4 | 5+6 2 | 2+2 2 | 3+3 2 |
| | 2+2 4 | 5+6 2 | 2+2 4 | 3+4 2 | 2+2 2 | 2+2 2 |
| | 2+2 4 | 5+6 2 | 2+2 4 | 3+4 2 | 3+3 2 | 2+2 2 |
| | 2+2 4 | 3+4 2 | 2+2 4 | 5+6 2 | 2+2 2 | 3+3 2 |
| | 2+2 4 | 5+6 2 | 2+2 4 | 5+6 2 | 2+2 2 | 3+3 2 |
| | 2+2 4 | 3+4 2 | 2+2 4 | 5+6 2 | 3+3 2 | 3+3 2 |
| | 2+2 4 | 5+6 2 | 2+2 4 | 3+4 2 | 2+2 2 | 2+2 2 |
| | 2+2 4 | 5+6 2 | 2+2 4 | 3+4 2 | 3+3 2 | 3+3 2 |
| | | 3+4 2 | | 3+4 2 | 3+3 2 | 2+2 2 |
| | | 3+4 2 | | 3+4 2 | 2+2 2 | 3+3 2 |
| | | 3+4 2 | | 5+6 2 | 3+3 2 | 2+2 2 |
| | | 3+4 2 | | 5+6 2 | 3+3 2 | 3+3 2 |
| | | 5+6 2 | | 3+4 2 | 2+2 2 | 2+2 2 |
| | | 3+4 2 | | 5+6 2 | 3+3 2 | 2+2 2 |
| | | 5+6 2 | | 3+4 2 | 2+2 2 | 3+3 2 |
| | | 3+4 2 | | 3+4 2 | 3+3 2 | 2+2 2 |
| | | 5+6 2 | | 5+6 2 | 2+2 2 | 3+3 2 |
| | | 5+6 2 | | 5+6 2 | 2+2 2 | 2+2 2 |
| | | 3+4 2 | | 3+4 2 | 3+3 2 | 3+3 2 |

Continued on next column

**TABLE XXVII– continued**

| TQC | ICSD | | NoE | SF | | |
|---|---|---|---|---|---|---|
| | | 3+4 2 | | 5+6 2 | 3+3 2 | 2+2 2 |
| | | 3+4 2 | | 3+4 2 | 3+3 2 | 2+2 2 |
| | | 5+6 2 | | 5+6 2 | 2+2 2 | 3+3 2 |
| | | 3+4 2 | | 3+4 2 | 3+3 2 | 3+3 2 |
| | | 3+4 2 | | 5+6 2 | 2+2 2 | 2+2 2 |
| | | 5+6 2 | | 5+6 2 | 3+3 2 | 2+2 2 |
| | | 3+4 2 | | 3+4 2 | 2+2 2 | 3+3 2 |
| | | 5+6 2 | | 5+6 2 | 3+3 2 | 2+2 2 |
| | | 3+4 2 | | 5+6 2 | 2+2 2 | 3+3 2 |
| | | 5+6 2 | | 3+4 2 | 3+3 2 | 2+2 2 |
| | | 3+4 2 | | 3+4 2 | 3+3 2 | 2+2 2 |
| | | 5+6 2 | | 3+4 2 | 2+2 2 | 3+3 2 |
| | | 3+4 2 | | 3+4 2 | 3+3 2 | 3+3 2 |
| 15.2.80 | 423885 | | 96 | C2Fe5 | | |
| | -A | -GM | -M | -Y | -L | -V |
| | 2+2 4 | 3+4 2 | 2+2 4 | 3+4 2 | 3+3 2 | 3+3 2 |
| | 2+2 4 | 5+6 2 | 2+2 4 | 5+6 2 | 3+3 2 | 2+2 2 |
| | 2+2 4 | 5+6 2 | 2+2 4 | 3+4 2 | 2+2 2 | 2+2 2 |
| | 2+2 4 | 3+4 2 | 2+2 4 | 5+6 2 | 2+2 2 | 3+3 2 |
| | 2+2 4 | 3+4 2 | 2+2 4 | 5+6 2 | 2+2 2 | 3+3 2 |
| | 2+2 4 | 5+6 2 | 2+2 4 | 3+4 2 | 3+3 2 | 2+2 2 |
| | 2+2 4 | 3+4 2 | 2+2 4 | 3+4 2 | 3+3 2 | 2+2 2 |
| | 2+2 4 | 5+6 2 | 2+2 4 | 5+6 2 | 2+2 2 | 2+2 2 |
| | 2+2 4 | 3+4 2 | 2+2 4 | 3+4 2 | 3+3 2 | 3+3 2 |
| | 2+2 4 | 5+6 2 | 2+2 4 | 5+6 2 | 2+2 2 | 3+3 2 |
| | 2+2 4 | 3+4 2 | 2+2 4 | 5+6 2 | 2+2 2 | 2+2 2 |
| | 2+2 4 | 5+6 2 | 2+2 4 | 5+6 2 | 3+3 2 | 3+3 2 |
| | 2+2 4 | 3+4 2 | 2+2 4 | 5+6 2 | 2+2 2 | 2+2 2 |
| | 2+2 4 | 5+6 2 | 2+2 4 | 3+4 2 | 2+2 2 | 3+3 2 |
| | 2+2 4 | 5+6 2 | 2+2 4 | 3+4 2 | 3+3 2 | 2+2 2 |
| | 2+2 4 | 3+4 2 | 2+2 4 | 3+4 2 | 3+3 2 | 3+3 2 |
| | 2+2 4 | 3+4 2 | 2+2 4 | 5+6 2 | 2+2 2 | 3+3 2 |
| | 2+2 4 | 5+6 2 | 2+2 4 | 3+4 2 | 2+2 2 | 2+2 2 |
| | 2+2 4 | 5+6 2 | 2+2 4 | 3+4 2 | 3+3 2 | 2+2 2 |
| | 2+2 4 | 3+4 2 | 2+2 4 | 5+6 2 | 2+2 2 | 3+3 2 |
| | 2+2 4 | 5+6 2 | 2+2 4 | 5+6 2 | 2+2 2 | 3+3 2 |
| | 2+2 4 | 3+4 2 | 2+2 4 | 5+6 2 | 3+3 2 | 3+3 2 |
| | 2+2 4 | 5+6 2 | 2+2 4 | 3+4 2 | 2+2 2 | 2+2 2 |
| | 2+2 4 | 5+6 2 | 2+2 4 | 3+4 2 | 3+3 2 | 3+3 2 |
| | | 3+4 2 | | 3+4 2 | 3+3 2 | 2+2 2 |
| | | 3+4 2 | | 3+4 2 | 2+2 2 | 3+3 2 |
| | | 3+4 2 | | 5+6 2 | 3+3 2 | 2+2 2 |
| | | 3+4 2 | | 5+6 2 | 2+2 2 | 3+3 2 |
| | | 5+6 2 | | 3+4 2 | 3+3 2 | 2+2 2 |
| | | 3+4 2 | | 5+6 2 | 3+3 2 | 2+2 2 |
| | | 5+6 2 | | 3+4 2 | 3+3 2 | 3+3 2 |
| | | 3+4 2 | | 3+4 2 | 2+2 2 | 2+2 2 |
| | | 5+6 2 | | 5+6 2 | 2+2 2 | 3+3 2 |
| | | 5+6 2 | | 5+6 2 | 2+2 2 | 2+2 2 |
| | | 3+4 2 | | 3+4 2 | 3+3 2 | 3+3 2 |
| | | 3+4 2 | | 5+6 2 | 3+3 2 | 2+2 2 |
| | | 3+4 2 | | 3+4 2 | 3+3 2 | 2+2 2 |
| | | 5+6 2 | | 5+6 2 | 2+2 2 | 3+3 2 |
| | | 3+4 2 | | 3+4 2 | 3+3 2 | 3+3 2 |
| | | 3+4 2 | | 5+6 2 | 2+2 2 | 2+2 2 |
| | | 5+6 2 | | 5+6 2 | 3+3 2 | 2+2 2 |
| | | 3+4 2 | | 3+4 2 | 2+2 2 | 3+3 2 |
| | | 5+6 2 | | 5+6 2 | 3+3 2 | 2+2 2 |
| | | 3+4 2 | | 5+6 2 | 2+2 2 | 3+3 2 |
| | | 5+6 2 | | 3+4 2 | 3+3 2 | 2+2 2 |
| | | 3+4 2 | | 3+4 2 | 3+3 2 | 2+2 2 |

Continued on next column

**TABLE XXVII– continued**

| TQC | ICSD | | NoE | SF | | |
|---|---|---|---|---|---|---|
| | | 5+6 2 | | 3+4 2 | 2+2 2 | 3+3 2 |
| | | 3+4 2 | | 3+4 2 | 3+3 2 | 3+3 2 |
| 15.2.81 | 183902 | | 120 | Bi1Mn1O3 | | |
| | -A | -GM | -M | -Y | -L | -V |
| | 2+2 4 | 3+4 2 | 2+2 4 | 5+6 2 | 2+2 2 | 2+2 2 |
| | 2+2 4 | 5+6 2 | 2+2 4 | 3+4 2 | 3+3 2 | 2+2 2 |
| | 2+2 4 | 3+4 2 | 2+2 4 | 3+4 2 | 2+2 2 | 3+3 2 |
| | 2+2 4 | 5+6 2 | 2+2 4 | 5+6 2 | 3+3 2 | 3+3 2 |
| | 2+2 4 | 3+4 2 | 2+2 4 | 5+6 2 | 2+2 2 | 3+3 2 |
| | 2+2 4 | 3+4 2 | 2+2 4 | 3+4 2 | 3+3 2 | 2+2 2 |
| | 2+2 4 | 5+6 2 | 2+2 4 | 3+4 2 | 2+2 2 | 2+2 2 |
| | 2+2 4 | 5+6 2 | 2+2 4 | 5+6 2 | 3+3 2 | 3+3 2 |
| | 2+2 4 | 3+4 2 | 2+2 4 | 3+4 2 | 2+2 2 | 2+2 2 |
| | 2+2 4 | 3+4 2 | 2+2 4 | 5+6 2 | 3+3 2 | 2+2 2 |
| | 2+2 4 | 5+6 2 | 2+2 4 | 5+6 2 | 3+3 2 | 3+3 2 |
| | 2+2 4 | 5+6 2 | 2+2 4 | 3+4 2 | 2+2 2 | 3+3 2 |
| | 2+2 4 | 5+6 2 | 2+2 4 | 5+6 2 | 2+2 2 | 3+3 2 |
| | 2+2 4 | 5+6 2 | 2+2 4 | 5+6 2 | 3+3 2 | 3+3 2 |
| | 2+2 4 | 3+4 2 | 2+2 4 | 3+4 2 | 2+2 2 | 2+2 2 |
| | 2+2 4 | 3+4 2 | 2+2 4 | 3+4 2 | 3+3 2 | 2+2 2 |
| | 2+2 4 | 3+4 2 | 2+2 4 | 5+6 2 | 2+2 2 | 3+3 2 |
| | 2+2 4 | 3+4 2 | 2+2 4 | 5+6 2 | 3+3 2 | 2+2 2 |
| | 2+2 4 | 3+4 2 | 2+2 4 | 3+4 2 | 2+2 2 | 2+2 2 |
| | 2+2 4 | 3+4 2 | 2+2 4 | 3+4 2 | 2+2 2 | 3+3 2 |
| | 2+2 4 | 5+6 2 | 2+2 4 | 5+6 2 | 3+3 2 | 3+3 2 |
| | 2+2 4 | 5+6 2 | 2+2 4 | 5+6 2 | 2+2 2 | 2+2 2 |
| | 2+2 4 | 3+4 2 | 2+2 4 | 3+4 2 | 3+3 2 | 3+3 2 |
| | 2+2 4 | 5+6 2 | 2+2 4 | 5+6 2 | 2+2 2 | 2+2 2 |
| | 2+2 4 | 3+4 2 | 2+2 4 | 3+4 2 | 3+3 2 | 2+2 2 |
| | 2+2 4 | 5+6 2 | 2+2 4 | 5+6 2 | 2+2 2 | 3+3 2 |
| | 2+2 4 | 3+4 2 | 2+2 4 | 3+4 2 | 2+2 2 | 2+2 2 |
| | 2+2 4 | 5+6 2 | 2+2 4 | 3+4 2 | 3+3 2 | 3+3 2 |
| | 2+2 4 | 3+4 2 | 2+2 4 | 5+6 2 | 3+3 2 | 2+2 2 |
| | 2+2 4 | 5+6 2 | 2+2 4 | 5+6 2 | 2+2 2 | 3+3 2 |
| | | 3+4 2 | | 3+4 2 | 3+3 2 | 3+3 2 |
| | | 3+4 2 | | 5+6 2 | 2+2 2 | 2+2 2 |
| | | 5+6 2 | | 3+4 2 | 3+3 2 | 2+2 2 |
| | | 3+4 2 | | 5+6 2 | 3+3 2 | 3+3 2 |
| | | 5+6 2 | | 3+4 2 | 2+2 2 | 2+2 2 |
| | | 5+6 2 | | 5+6 2 | 2+2 2 | 2+2 2 |
| | | 3+4 2 | | 5+6 2 | 3+3 2 | 3+3 2 |
| | | 5+6 2 | | 3+4 2 | 3+3 2 | 3+3 2 |
| | | 5+6 2 | | 3+4 2 | 2+2 2 | 2+2 2 |
| | | 5+6 2 | | 3+4 2 | 2+2 2 | 3+3 2 |
| | | 3+4 2 | | 3+4 2 | 3+3 2 | 2+2 2 |
| | | 3+4 2 | | 5+6 2 | 2+2 2 | 3+3 2 |
| | | 5+6 2 | | 5+6 2 | 3+3 2 | 2+2 2 |
| | | 3+4 2 | | 5+6 2 | 2+2 2 | 2+2 2 |
| | | 3+4 2 | | 5+6 2 | 3+3 2 | 3+3 2 |
| | | 3+4 2 | | 3+4 2 | 2+2 2 | 2+2 2 |
| | | 5+6 2 | | 3+4 2 | 2+2 2 | 3+3 2 |
| | | 5+6 2 | | 5+6 2 | 3+3 2 | 3+3 2 |
| | | 3+4 2 | | 3+4 2 | 3+3 2 | 2+2 2 |
| | | 5+6 2 | | 3+4 2 | 3+3 2 | 3+3 2 |
| | | 5+6 2 | | 5+6 2 | 2+2 2 | 3+3 2 |
| | | 5+6 2 | | 3+4 2 | 3+3 2 | 2+2 2 |
| | | 3+4 2 | | 5+6 2 | 2+2 2 | 2+2 2 |
| | | 3+4 2 | | 5+6 2 | 2+2 2 | 3+3 2 |
| | | 3+4 2 | | 5+6 2 | 2+2 2 | 2+2 2 |
| | | 3+4 2 | | 5+6 2 | 3+3 2 | 3+3 2 |
| | | 5+6 2 | | 3+4 2 | 2+2 2 | 2+2 2 |

Continued on next column

**TABLE XXVII– continued**

| TQC | ICSD | | NoE | SF | | |
|---|---|---|---|---|---|---|
| | | 3+4 2 | | 5+6 2 | 3+3 2 | 3+3 2 |
| | | 5+6 2 | | 3+4 2 | 2+2 2 | 2+2 2 |
| | | 5+6 2 | | 5+6 2 | 2+2 2 | 3+3 2 |
| 15.2.82 | 174405 | | 116 | Bi1Cr1O3 | | |
| | -A | -GM | -M | -Y | -L | -V |
| | 2+2 4 | 3+4 2 | 2+2 4 | 5+6 2 | 2+2 2 | 2+2 2 |
| | 2+2 4 | 5+6 2 | 2+2 4 | 3+4 2 | 3+3 2 | 3+3 2 |
| | 2+2 4 | 3+4 2 | 2+2 4 | 5+6 2 | 2+2 2 | 3+3 2 |
| | 2+2 4 | 3+4 2 | 2+2 4 | 3+4 2 | 3+3 2 | 2+2 2 |
| | 2+2 4 | 5+6 2 | 2+2 4 | 5+6 2 | 2+2 2 | 2+2 2 |
| | 2+2 4 | 5+6 2 | 2+2 4 | 3+4 2 | 3+3 2 | 3+3 2 |
| | 2+2 4 | 3+4 2 | 2+2 4 | 3+4 2 | 2+2 2 | 2+2 2 |
| | 2+2 4 | 5+6 2 | 2+2 4 | 5+6 2 | 3+3 2 | 3+3 2 |
| | 2+2 4 | 3+4 2 | 2+2 4 | 3+4 2 | 2+2 2 | 2+2 2 |
| | 2+2 4 | 3+4 2 | 2+2 4 | 5+6 2 | 3+3 2 | 2+2 2 |
| | 2+2 4 | 5+6 2 | 2+2 4 | 5+6 2 | 3+3 2 | 3+3 2 |
| | 2+2 4 | 5+6 2 | 2+2 4 | 3+4 2 | 2+2 2 | 3+3 2 |
| | 2+2 4 | 3+4 2 | 2+2 4 | 5+6 2 | 2+2 2 | 3+3 2 |
| | 2+2 4 | 5+6 2 | 2+2 4 | 5+6 2 | 3+3 2 | 3+3 2 |
| | 2+2 4 | 5+6 2 | 2+2 4 | 3+4 2 | 2+2 2 | 2+2 2 |
| | 2+2 4 | 3+4 2 | 2+2 4 | 3+4 2 | 3+3 2 | 2+2 2 |
| | 2+2 4 | 3+4 2 | 2+2 4 | 5+6 2 | 2+2 2 | 2+2 2 |
| | 2+2 4 | 3+4 2 | 2+2 4 | 5+6 2 | 3+3 2 | 3+3 2 |
| | 2+2 4 | 3+4 2 | 2+2 4 | 3+4 2 | 2+2 2 | 2+2 2 |
| | 2+2 4 | 3+4 2 | 2+2 4 | 5+6 2 | 2+2 2 | 3+3 2 |
| | 2+2 4 | 5+6 2 | 2+2 4 | 3+4 2 | 3+3 2 | 3+3 2 |
| | 2+2 4 | 3+4 2 | 2+2 4 | 5+6 2 | 2+2 2 | 3+3 2 |
| | 2+2 4 | 5+6 2 | 2+2 4 | 3+4 2 | 2+2 2 | 2+2 2 |
| | 2+2 4 | 5+6 2 | 2+2 4 | 5+6 2 | 3+3 2 | 2+2 2 |
| | 2+2 4 | 3+4 2 | 2+2 4 | 5+6 2 | 2+2 2 | 2+2 2 |
| | 2+2 4 | 5+6 2 | 2+2 4 | 3+4 2 | 3+3 2 | 3+3 2 |
| | 2+2 4 | 5+6 2 | 2+2 4 | 3+4 2 | 2+2 2 | 2+2 2 |
| | 2+2 4 | 3+4 2 | 2+2 4 | 5+6 2 | 3+3 2 | 3+3 2 |
| | 2+2 4 | 3+4 2 | 2+2 4 | 3+4 2 | 2+2 2 | 2+2 2 |
| | | 3+4 2 | | 3+4 2 | 3+3 2 | 3+3 2 |
| | | 5+6 2 | | 5+6 2 | 3+3 2 | 2+2 2 |
| | | 3+4 2 | | 5+6 2 | 2+2 2 | 3+3 2 |
| | | 5+6 2 | | 3+4 2 | 3+3 2 | 2+2 2 |
| | | 5+6 2 | | 3+4 2 | 2+2 2 | 2+2 2 |
| | | 3+4 2 | | 5+6 2 | 3+3 2 | 3+3 2 |
| | | 5+6 2 | | 5+6 2 | 2+2 2 | 3+3 2 |
| | | 3+4 2 | | 5+6 2 | 3+3 2 | 2+2 2 |
| | | 5+6 2 | | 3+4 2 | 3+3 2 | 3+3 2 |
| | | 5+6 2 | | 3+4 2 | 2+2 2 | 2+2 2 |
| | | 5+6 2 | | 3+4 2 | 2+2 2 | 3+3 2 |
| | | 3+4 2 | | 3+4 2 | 3+3 2 | 2+2 2 |
| | | 3+4 2 | | 5+6 2 | 2+2 2 | 3+3 2 |
| | | 5+6 2 | | 5+6 2 | 2+2 2 | 2+2 2 |
| | | 3+4 2 | | 5+6 2 | 3+3 2 | 3+3 2 |
| | | 5+6 2 | | 5+6 2 | 2+2 2 | 2+2 2 |
| | | 5+6 2 | | 5+6 2 | 3+3 2 | 2+2 2 |
| | | 3+4 2 | | 3+4 2 | 3+3 2 | 3+3 2 |
| | | 5+6 2 | | 3+4 2 | 2+2 2 | 3+3 2 |
| | | 5+6 2 | | 3+4 2 | 3+3 2 | 2+2 2 |
| | | 3+4 2 | | 3+4 2 | 2+2 2 | 3+3 2 |
| | | 5+6 2 | | 5+6 2 | 3+3 2 | 3+3 2 |
| | | 3+4 2 | | 3+4 2 | 3+3 2 | 2+2 2 |
| | | 3+4 2 | | 5+6 2 | 2+2 2 | 2+2 2 |
| | | 3+4 2 | | 5+6 2 | 2+2 2 | 3+3 2 |
| | | 3+4 2 | | 5+6 2 | 3+3 2 | 2+2 2 |
| | | 5+6 2 | | 3+4 2 | 2+2 2 | 3+3 2 |

Continued on next column

**TABLE XXVII– continued**

| TQC | ICSD | | NoE | SF | | |
|---|---|---|---|---|---|---|
| | | 3+4 2 | | 5+6 2 | 2+2 2 | 2+2 2 |
| | | 3+4 2 | | 3+4 2 | 3+3 2 | 3+3 2 |
| 15.2.83 | 174404 | | 116 | Bi1Cr1O3 | | |
| | -A | -GM | -M | -Y | -L | -V |
| | 2+2 4 | 3+4 2 | 2+2 4 | 5+6 2 | 2+2 2 | 2+2 2 |
| | 2+2 4 | 5+6 2 | 2+2 4 | 3+4 2 | 3+3 2 | 3+3 2 |
| | 2+2 4 | 3+4 2 | 2+2 4 | 5+6 2 | 2+2 2 | 3+3 2 |
| | 2+2 4 | 3+4 2 | 2+2 4 | 3+4 2 | 3+3 2 | 2+2 2 |
| | 2+2 4 | 5+6 2 | 2+2 4 | 5+6 2 | 2+2 2 | 2+2 2 |
| | 2+2 4 | 5+6 2 | 2+2 4 | 3+4 2 | 3+3 2 | 3+3 2 |
| | 2+2 4 | 3+4 2 | 2+2 4 | 3+4 2 | 2+2 2 | 2+2 2 |
| | 2+2 4 | 3+4 2 | 2+2 4 | 5+6 2 | 3+3 2 | 3+3 2 |
| | 2+2 4 | 5+6 2 | 2+2 4 | 3+4 2 | 2+2 2 | 2+2 2 |
| | 2+2 4 | 3+4 2 | 2+2 4 | 5+6 2 | 3+3 2 | 2+2 2 |
| | 2+2 4 | 5+6 2 | 2+2 4 | 5+6 2 | 3+3 2 | 3+3 2 |
| | 2+2 4 | 5+6 2 | 2+2 4 | 3+4 2 | 2+2 2 | 3+3 2 |
| | 2+2 4 | 3+4 2 | 2+2 4 | 5+6 2 | 2+2 2 | 3+3 2 |
| | 2+2 4 | 5+6 2 | 2+2 4 | 3+4 2 | 3+3 2 | 3+3 2 |
| | 2+2 4 | 5+6 2 | 2+2 4 | 5+6 2 | 2+2 2 | 2+2 2 |
| | 2+2 4 | 3+4 2 | 2+2 4 | 3+4 2 | 3+3 2 | 2+2 2 |
| | 2+2 4 | 3+4 2 | 2+2 4 | 5+6 2 | 2+2 2 | 2+2 2 |
| | 2+2 4 | 3+4 2 | 2+2 4 | 5+6 2 | 3+3 2 | 3+3 2 |
| | 2+2 4 | 3+4 2 | 2+2 4 | 3+4 2 | 2+2 2 | 2+2 2 |
| | 2+2 4 | 3+4 2 | 2+2 4 | 5+6 2 | 2+2 2 | 3+3 2 |
| | 2+2 4 | 5+6 2 | 2+2 4 | 3+4 2 | 3+3 2 | 3+3 2 |
| | 2+2 4 | 3+4 2 | 2+2 4 | 5+6 2 | 2+2 2 | 3+3 2 |
| | 2+2 4 | 5+6 2 | 2+2 4 | 3+4 2 | 2+2 2 | 2+2 2 |
| | 2+2 4 | 5+6 2 | 2+2 4 | 5+6 2 | 3+3 2 | 2+2 2 |
| | 2+2 4 | 3+4 2 | 2+2 4 | 5+6 2 | 2+2 2 | 2+2 2 |
| | 2+2 4 | 5+6 2 | 2+2 4 | 3+4 2 | 3+3 2 | 3+3 2 |
| | 2+2 4 | 5+6 2 | 2+2 4 | 3+4 2 | 2+2 2 | 2+2 2 |
| | 2+2 4 | 3+4 2 | 2+2 4 | 3+4 2 | 3+3 2 | 3+3 2 |
| | 2+2 4 | 3+4 2 | 2+2 4 | 5+6 2 | 2+2 2 | 2+2 2 |
| | | 3+4 2 | | 3+4 2 | 3+3 2 | 3+3 2 |
| | | 5+6 2 | | 5+6 2 | 3+3 2 | 2+2 2 |
| | | 3+4 2 | | 5+6 2 | 2+2 2 | 3+3 2 |
| | | 5+6 2 | | 3+4 2 | 3+3 2 | 2+2 2 |
| | | 5+6 2 | | 3+4 2 | 2+2 2 | 2+2 2 |
| | | 3+4 2 | | 5+6 2 | 3+3 2 | 3+3 2 |
| | | 5+6 2 | | 5+6 2 | 2+2 2 | 3+3 2 |
| | | 3+4 2 | | 3+4 2 | 3+3 2 | 2+2 2 |
| | | 5+6 2 | | 5+6 2 | 2+2 2 | 3+3 2 |
| | | 5+6 2 | | 3+4 2 | 3+3 2 | 2+2 2 |
| | | 5+6 2 | | 3+4 2 | 2+2 2 | 3+3 2 |
| | | 3+4 2 | | 5+6 2 | 3+3 2 | 3+3 2 |
| | | 3+4 2 | | 3+4 2 | 2+2 2 | 2+2 2 |
| | | 5+6 2 | | 5+6 2 | 2+2 2 | 2+2 2 |
| | | 3+4 2 | | 5+6 2 | 3+3 2 | 3+3 2 |
| | | 5+6 2 | | 5+6 2 | 2+2 2 | 2+2 2 |
| | | 5+6 2 | | 3+4 2 | 3+3 2 | 2+2 2 |
| | | 3+4 2 | | 5+6 2 | 3+3 2 | 3+3 2 |
| | | 5+6 2 | | 3+4 2 | 2+2 2 | 3+3 2 |
| | | 5+6 2 | | 3+4 2 | 3+3 2 | 2+2 2 |
| | | 3+4 2 | | 3+4 2 | 2+2 2 | 3+3 2 |
| | | 5+6 2 | | 5+6 2 | 3+3 2 | 3+3 2 |
| | | 3+4 2 | | 3+4 2 | 3+3 2 | 2+2 2 |
| | | 3+4 2 | | 5+6 2 | 2+2 2 | 2+2 2 |
| | | 3+4 2 | | 5+6 2 | 2+2 2 | 3+3 2 |
| | | 3+4 2 | | 5+6 2 | 3+3 2 | 2+2 2 |
| | | 5+6 2 | | 3+4 2 | 2+2 2 | 3+3 2 |
| | | 3+4 2 | | 5+6 2 | 2+2 2 | 2+2 2 |

Continued on next column

**TABLE XXVII– continued**

| TQC | ICSD | | NoE | SF | | |
|---|---|---|---|---|---|---|
| | | 3+4 2 | | 5+6 2 | 3+3 2 | 3+3 2 |
| 15.2.84 | 160458 | | 116 | Bi1Cr1O3 | | |
| | -A | -GM | -M | -Y | -L | -V |
| | 2+2 4 | 3+4 2 | 2+2 4 | 5+6 2 | 2+2 2 | 2+2 2 |
| | 2+2 4 | 5+6 2 | 2+2 4 | 3+4 2 | 3+3 2 | 3+3 2 |
| | 2+2 4 | 3+4 2 | 2+2 4 | 5+6 2 | 2+2 2 | 2+2 2 |
| | 2+2 4 | 5+6 2 | 2+2 4 | 3+4 2 | 3+3 2 | 3+3 2 |
| | 2+2 4 | 3+4 2 | 2+2 4 | 5+6 2 | 2+2 2 | 3+3 2 |
| | 2+2 4 | 5+6 2 | 2+2 4 | 3+4 2 | 3+3 2 | 2+2 2 |
| | 2+2 4 | 3+4 2 | 2+2 4 | 3+4 2 | 2+2 2 | 2+2 2 |
| | 2+2 4 | 3+4 2 | 2+2 4 | 5+6 2 | 2+2 2 | 2+2 2 |
| | 2+2 4 | 5+6 2 | 2+2 4 | 3+4 2 | 3+3 2 | 3+3 2 |
| | 2+2 4 | 3+4 2 | 2+2 4 | 5+6 2 | 3+3 2 | 2+2 2 |
| | 2+2 4 | 5+6 2 | 2+2 4 | 3+4 2 | 3+3 2 | 3+3 2 |
| | 2+2 4 | 5+6 2 | 2+2 4 | 5+6 2 | 2+2 2 | 3+3 2 |
| | 2+2 4 | 3+4 2 | 2+2 4 | 5+6 2 | 2+2 2 | 3+3 2 |
| | 2+2 4 | 5+6 2 | 2+2 4 | 3+4 2 | 3+3 2 | 3+3 2 |
| | 2+2 4 | 5+6 2 | 2+2 4 | 5+6 2 | 2+2 2 | 2+2 2 |
| | 2+2 4 | 3+4 2 | 2+2 4 | 3+4 2 | 3+3 2 | 2+2 2 |
| | 2+2 4 | 3+4 2 | 2+2 4 | 5+6 2 | 2+2 2 | 2+2 2 |
| | 2+2 4 | 3+4 2 | 2+2 4 | 5+6 2 | 3+3 2 | 3+3 2 |
| | 2+2 4 | 3+4 2 | 2+2 4 | 3+4 2 | 2+2 2 | 2+2 2 |
| | 2+2 4 | 3+4 2 | 2+2 4 | 5+6 2 | 2+2 2 | 3+3 2 |
| | 2+2 4 | 5+6 2 | 2+2 4 | 3+4 2 | 3+3 2 | 3+3 2 |
| | 2+2 4 | 5+6 2 | 2+2 4 | 3+4 2 | 2+2 2 | 3+3 2 |
| | 2+2 4 | 3+4 2 | 2+2 4 | 5+6 2 | 2+2 2 | 2+2 2 |
| | 2+2 4 | 5+6 2 | 2+2 4 | 5+6 2 | 3+3 2 | 2+2 2 |
| | 2+2 4 | 3+4 2 | 2+2 4 | 5+6 2 | 2+2 2 | 2+2 2 |
| | 2+2 4 | 5+6 2 | 2+2 4 | 3+4 2 | 3+3 2 | 3+3 2 |
| | 2+2 4 | 5+6 2 | 2+2 4 | 3+4 2 | 2+2 2 | 2+2 2 |
| | 2+2 4 | 3+4 2 | 2+2 4 | 3+4 2 | 3+3 2 | 3+3 2 |
| | 2+2 4 | 3+4 2 | 2+2 4 | 5+6 2 | 2+2 2 | 2+2 2 |
| | | 3+4 2 | | 3+4 2 | 3+3 2 | 3+3 2 |
| | | 5+6 2 | | 5+6 2 | 3+3 2 | 2+2 2 |
| | | 3+4 2 | | 5+6 2 | 3+3 2 | 3+3 2 |
| | | 5+6 2 | | 3+4 2 | 2+2 2 | 2+2 2 |
| | | 5+6 2 | | 3+4 2 | 2+2 2 | 2+2 2 |
| | | 3+4 2 | | 5+6 2 | 3+3 2 | 3+3 2 |
| | | 5+6 2 | | 5+6 2 | 2+2 2 | 3+3 2 |
| | | 3+4 2 | | 5+6 2 | 3+3 2 | 2+2 2 |
| | | 5+6 2 | | 3+4 2 | 3+3 2 | 3+3 2 |
| | | 5+6 2 | | 3+4 2 | 2+2 2 | 2+2 2 |
| | | 5+6 2 | | 3+4 2 | 2+2 2 | 3+3 2 |
| | | 3+4 2 | | 5+6 2 | 3+3 2 | 3+3 2 |
| | | 3+4 2 | | 3+4 2 | 2+2 2 | 2+2 2 |
| | | 5+6 2 | | 5+6 2 | 2+2 2 | 2+2 2 |
| | | 3+4 2 | | 5+6 2 | 3+3 2 | 3+3 2 |
| | | 5+6 2 | | 5+6 2 | 2+2 2 | 2+2 2 |
| | | 5+6 2 | | 5+6 2 | 3+3 2 | 2+2 2 |
| | | 3+4 2 | | 3+4 2 | 3+3 2 | 3+3 2 |
| | | 5+6 2 | | 3+4 2 | 2+2 2 | 3+3 2 |
| | | 5+6 2 | | 3+4 2 | 3+3 2 | 2+2 2 |
| | | 3+4 2 | | 3+4 2 | 2+2 2 | 3+3 2 |
| | | 5+6 2 | | 5+6 2 | 3+3 2 | 3+3 2 |
| | | 3+4 2 | | 3+4 2 | 3+3 2 | 2+2 2 |
| | | 3+4 2 | | 5+6 2 | 2+2 2 | 2+2 2 |
| | | 3+4 2 | | 5+6 2 | 2+2 2 | 3+3 2 |
| | | 3+4 2 | | 5+6 2 | 3+3 2 | 2+2 2 |
| | | 5+6 2 | | 3+4 2 | 2+2 2 | 3+3 2 |
| | | 3+4 2 | | 5+6 2 | 2+2 2 | 2+2 2 |
| | | 3+4 2 | | 5+6 2 | 3+3 2 | 3+3 2 |

Continued on next column

**TABLE XXVII– continued**

| TQC | ICSD | NoE | SF | | | |
|---|---|---|---|---|---|---|
| 15.2.85 | 160457 | 116 | Bi1Cr1O3 | | | |
| | -A | -GM | -M | -Y | -L | -V |
| | 2+2 4 | 3+4 2 | 2+2 4 | 5+6 2 | 2+2 2 | 2+2 2 |
| | 2+2 4 | 5+6 2 | 2+2 4 | 3+4 2 | 3+3 2 | 3+3 2 |
| | 2+2 4 | 3+4 2 | 2+2 4 | 5+6 2 | 2+2 2 | 2+2 2 |
| | 2+2 4 | 5+6 2 | 2+2 4 | 3+4 2 | 3+3 2 | 3+3 2 |
| | 2+2 4 | 3+4 2 | 2+2 4 | 5+6 2 | 2+2 2 | 3+3 2 |
| | 2+2 4 | 5+6 2 | 2+2 4 | 3+4 2 | 3+3 2 | 2+2 2 |
| | 2+2 4 | 3+4 2 | 2+2 4 | 3+4 2 | 2+2 2 | 2+2 2 |
| | 2+2 4 | 3+4 2 | 2+2 4 | 5+6 2 | 2+2 2 | 2+2 2 |
| | 2+2 4 | 5+6 2 | 2+2 4 | 3+4 2 | 3+3 2 | 3+3 2 |
| | 2+2 4 | 3+4 2 | 2+2 4 | 5+6 2 | 3+3 2 | 2+2 2 |
| | 2+2 4 | 5+6 2 | 2+2 4 | 3+4 2 | 3+3 2 | 3+3 2 |
| | 2+2 4 | 5+6 2 | 2+2 4 | 5+6 2 | 2+2 2 | 3+3 2 |
| | 2+2 4 | 3+4 2 | 2+2 4 | 5+6 2 | 2+2 2 | 3+3 2 |
| | 2+2 4 | 5+6 2 | 2+2 4 | 5+6 2 | 3+3 2 | 3+3 2 |
| | 2+2 4 | 5+6 2 | 2+2 4 | 3+4 2 | 2+2 2 | 2+2 2 |
| | 2+2 4 | 3+4 2 | 2+2 4 | 3+4 2 | 3+3 2 | 2+2 2 |
| | 2+2 4 | 3+4 2 | 2+2 4 | 5+6 2 | 2+2 2 | 2+2 2 |
| | 2+2 4 | 3+4 2 | 2+2 4 | 5+6 2 | 3+3 2 | 3+3 2 |
| | 2+2 4 | 3+4 2 | 2+2 4 | 3+4 2 | 2+2 2 | 2+2 2 |
| | 2+2 4 | 3+4 2 | 2+2 4 | 5+6 2 | 2+2 2 | 3+3 2 |
| | 2+2 4 | 5+6 2 | 2+2 4 | 3+4 2 | 3+3 2 | 3+3 2 |
| | 2+2 4 | 5+6 2 | 2+2 4 | 3+4 2 | 2+2 2 | 3+3 2 |
| | 2+2 4 | 3+4 2 | 2+2 4 | 5+6 2 | 2+2 2 | 2+2 2 |
| | 2+2 4 | 5+6 2 | 2+2 4 | 5+6 2 | 3+3 2 | 2+2 2 |
| | 2+2 4 | 3+4 2 | 2+2 4 | 5+6 2 | 2+2 2 | 2+2 2 |
| | 2+2 4 | 5+6 2 | 2+2 4 | 3+4 2 | 3+3 2 | 3+3 2 |
| | 2+2 4 | 5+6 2 | 2+2 4 | 3+4 2 | 2+2 2 | 2+2 2 |
| | 2+2 4 | 3+4 2 | 2+2 4 | 3+4 2 | 3+3 2 | 3+3 2 |
| | 2+2 4 | 3+4 2 | 2+2 4 | 5+6 2 | 2+2 2 | 2+2 2 |
| | | 3+4 2 | | 3+4 2 | 3+3 2 | 3+3 2 |
| | | 5+6 2 | | 5+6 2 | 3+3 2 | 2+2 2 |
| | | 3+4 2 | | 5+6 2 | 3+3 2 | 3+3 2 |
| | | 5+6 2 | | 3+4 2 | 2+2 2 | 2+2 2 |
| | | 5+6 2 | | 3+4 2 | 2+2 2 | 2+2 2 |
| | | 3+4 2 | | 5+6 2 | 3+3 2 | 3+3 2 |
| | | 5+6 2 | | 5+6 2 | 2+2 2 | 3+3 2 |
| | | 3+4 2 | | 5+6 2 | 3+3 2 | 2+2 2 |
| | | 5+6 2 | | 3+4 2 | 3+3 2 | 3+3 2 |
| | | 5+6 2 | | 3+4 2 | 2+2 2 | 2+2 2 |
| | | 5+6 2 | | 3+4 2 | 2+2 2 | 3+3 2 |
| | | 3+4 2 | | 5+6 2 | 3+3 2 | 3+3 2 |
| | | 3+4 2 | | 3+4 2 | 2+2 2 | 2+2 2 |
| | | 5+6 2 | | 5+6 2 | 2+2 2 | 2+2 2 |
| | | 3+4 2 | | 5+6 2 | 3+3 2 | 3+3 2 |
| | | 5+6 2 | | 5+6 2 | 2+2 2 | 2+2 2 |
| | | 5+6 2 | | 5+6 2 | 3+3 2 | 2+2 2 |
| | | 3+4 2 | | 3+4 2 | 3+3 2 | 3+3 2 |
| | | 5+6 2 | | 3+4 2 | 2+2 2 | 3+3 2 |
| | | 5+6 2 | | 3+4 2 | 3+3 2 | 2+2 2 |
| | | 3+4 2 | | 3+4 2 | 2+2 2 | 3+3 2 |
| | | 5+6 2 | | 5+6 2 | 3+3 2 | 3+3 2 |
| | | 3+4 2 | | 3+4 2 | 3+3 2 | 2+2 2 |
| | | 3+4 2 | | 5+6 2 | 2+2 2 | 2+2 2 |
| | | 3+4 2 | | 5+6 2 | 2+2 2 | 3+3 2 |
| | | 3+4 2 | | 5+6 2 | 3+3 2 | 2+2 2 |
| | | 5+6 2 | | 3+4 2 | 2+2 2 | 3+3 2 |
| | | 3+4 2 | | 5+6 2 | 2+2 2 | 2+2 2 |
| | | 3+4 2 | | 5+6 2 | 3+3 2 | 3+3 2 |
| 15.2.86 | 160456 | 116 | Bi1Cr1O3 | | | |
| | -A | -GM | -M | -Y | -L | -V |
| | 2+2 4 | 3+4 2 | 2+2 4 | 5+6 2 | 2+2 2 | 2+2 2 |
| | 2+2 4 | 5+6 2 | 2+2 4 | 3+4 2 | 3+3 2 | 3+3 2 |
| | 2+2 4 | 3+4 2 | 2+2 4 | 5+6 2 | 2+2 2 | 2+2 2 |
| | 2+2 4 | 5+6 2 | 2+2 4 | 3+4 2 | 3+3 2 | 3+3 2 |
| | 2+2 4 | 3+4 2 | 2+2 4 | 5+6 2 | 2+2 2 | 3+3 2 |
| | 2+2 4 | 5+6 2 | 2+2 4 | 3+4 2 | 3+3 2 | 2+2 2 |
| | 2+2 4 | 3+4 2 | 2+2 4 | 3+4 2 | 2+2 2 | 2+2 2 |
| | 2+2 4 | 3+4 2 | 2+2 4 | 5+6 2 | 2+2 2 | 2+2 2 |
| | 2+2 4 | 5+6 2 | 2+2 4 | 3+4 2 | 3+3 2 | 3+3 2 |
| | 2+2 4 | 3+4 2 | 2+2 4 | 5+6 2 | 3+3 2 | 2+2 2 |
| | 2+2 4 | 5+6 2 | 2+2 4 | 3+4 2 | 3+3 2 | 3+3 2 |
| | 2+2 4 | 5+6 2 | 2+2 4 | 5+6 2 | 2+2 2 | 3+3 2 |
| | 2+2 4 | 3+4 2 | 2+2 4 | 5+6 2 | 2+2 2 | 3+3 2 |
| | 2+2 4 | 5+6 2 | 2+2 4 | 3+4 2 | 3+3 2 | 3+3 2 |
| | 2+2 4 | 5+6 2 | 2+2 4 | 5+6 2 | 2+2 2 | 2+2 2 |
| | 2+2 4 | 3+4 2 | 2+2 4 | 3+4 2 | 3+3 2 | 2+2 2 |
| | 2+2 4 | 3+4 2 | 2+2 4 | 5+6 2 | 2+2 2 | 2+2 2 |
| | 2+2 4 | 3+4 2 | 2+2 4 | 5+6 2 | 3+3 2 | 3+3 2 |
| | 2+2 4 | 3+4 2 | 2+2 4 | 3+4 2 | 2+2 2 | 2+2 2 |
| | 2+2 4 | 3+4 2 | 2+2 4 | 5+6 2 | 2+2 2 | 3+3 2 |
| | 2+2 4 | 5+6 2 | 2+2 4 | 3+4 2 | 3+3 2 | 3+3 2 |
| | 2+2 4 | 5+6 2 | 2+2 4 | 3+4 2 | 2+2 2 | 2+2 2 |
| | 2+2 4 | 3+4 2 | 2+2 4 | 5+6 2 | 2+2 2 | 3+3 2 |
| | 2+2 4 | 5+6 2 | 2+2 4 | 5+6 2 | 3+3 2 | 2+2 2 |
| | 2+2 4 | 3+4 2 | 2+2 4 | 5+6 2 | 2+2 2 | 2+2 2 |
| | 2+2 4 | 5+6 2 | 2+2 4 | 3+4 2 | 3+3 2 | 3+3 2 |
| | 2+2 4 | 5+6 2 | 2+2 4 | 3+4 2 | 2+2 2 | 2+2 2 |
| | 2+2 4 | 3+4 2 | 2+2 4 | 3+4 2 | 3+3 2 | 3+3 2 |
| | 2+2 4 | 3+4 2 | 2+2 4 | 5+6 2 | 2+2 2 | 2+2 2 |
| | | 3+4 2 | | 3+4 2 | 3+3 2 | 3+3 2 |
| | | 5+6 2 | | 5+6 2 | 3+3 2 | 2+2 2 |
| | | 3+4 2 | | 5+6 2 | 3+3 2 | 3+3 2 |
| | | 5+6 2 | | 3+4 2 | 2+2 2 | 2+2 2 |
| | | 5+6 2 | | 3+4 2 | 2+2 2 | 2+2 2 |
| | | 3+4 2 | | 5+6 2 | 3+3 2 | 3+3 2 |
| | | 5+6 2 | | 5+6 2 | 2+2 2 | 3+3 2 |
| | | 3+4 2 | | 5+6 2 | 3+3 2 | 2+2 2 |
| | | 5+6 2 | | 3+4 2 | 3+3 2 | 3+3 2 |
| | | 5+6 2 | | 3+4 2 | 2+2 2 | 2+2 2 |
| | | 5+6 2 | | 3+4 2 | 2+2 2 | 3+3 2 |
| | | 3+4 2 | | 5+6 2 | 3+3 2 | 3+3 2 |
| | | 3+4 2 | | 3+4 2 | 2+2 2 | 2+2 2 |
| | | 5+6 2 | | 5+6 2 | 2+2 2 | 2+2 2 |
| | | 3+4 2 | | 5+6 2 | 3+3 2 | 3+3 2 |
| | | 5+6 2 | | 5+6 2 | 2+2 2 | 2+2 2 |
| | | 5+6 2 | | 5+6 2 | 3+3 2 | 2+2 2 |
| | | 3+4 2 | | 3+4 2 | 3+3 2 | 3+3 2 |
| | | 5+6 2 | | 3+4 2 | 2+2 2 | 3+3 2 |
| | | 5+6 2 | | 3+4 2 | 3+3 2 | 2+2 2 |
| | | 3+4 2 | | 3+4 2 | 2+2 2 | 3+3 2 |
| | | 5+6 2 | | 5+6 2 | 3+3 2 | 2+2 2 |
| | | 3+4 2 | | 3+4 2 | 3+3 2 | 3+3 2 |
| | | 3+4 2 | | 5+6 2 | 2+2 2 | 2+2 2 |
| | | 3+4 2 | | 5+6 2 | 2+2 2 | 3+3 2 |
| | | 3+4 2 | | 5+6 2 | 3+3 2 | 2+2 2 |
| | | 5+6 2 | | 3+4 2 | 2+2 2 | 3+3 2 |
| | | 3+4 2 | | 5+6 2 | 2+2 2 | 2+2 2 |
| | | 3+4 2 | | 5+6 2 | 3+3 2 | 3+3 2 |
| 15.2.87 | 160455 | 116 | Bi1Cr1O3 | | | |
| | -A | -GM | -M | -Y | -L | -V |

Continued on next column

**TABLE XXVII– continued**

| TQC | ICSD | | NoE | SF | | |
|---|---|---|---|---|---|---|
| | 2+2 4 | 3+4 2 | 2+2 4 | 5+6 2 | 2+2 2 | 2+2 2 |
| | 2+2 4 | 5+6 2 | 2+2 4 | 3+4 2 | 3+3 2 | 3+3 2 |
| | 2+2 4 | 3+4 2 | 2+2 4 | 5+6 2 | 2+2 2 | 2+2 2 |
| | 2+2 4 | 5+6 2 | 2+2 4 | 3+4 2 | 3+3 2 | 3+3 2 |
| | 2+2 4 | 3+4 2 | 2+2 4 | 5+6 2 | 2+2 2 | 3+3 2 |
| | 2+2 4 | 5+6 2 | 2+2 4 | 3+4 2 | 3+3 2 | 2+2 2 |
| | 2+2 4 | 3+4 2 | 2+2 4 | 3+4 2 | 2+2 2 | 2+2 2 |
| | 2+2 4 | 3+4 2 | 2+2 4 | 5+6 2 | 2+2 2 | 2+2 2 |
| | 2+2 4 | 5+6 2 | 2+2 4 | 3+4 2 | 3+3 2 | 3+3 2 |
| | 2+2 4 | 3+4 2 | 2+2 4 | 5+6 2 | 3+3 2 | 2+2 2 |
| | 2+2 4 | 5+6 2 | 2+2 4 | 3+4 2 | 3+3 2 | 3+3 2 |
| | 2+2 4 | 5+6 2 | 2+2 4 | 5+6 2 | 2+2 2 | 3+3 2 |
| | 2+2 4 | 3+4 2 | 2+2 4 | 5+6 2 | 2+2 2 | 3+3 2 |
| | 2+2 4 | 5+6 2 | 2+2 4 | 3+4 2 | 3+3 2 | 3+3 2 |
| | 2+2 4 | 5+6 2 | 2+2 4 | 5+6 2 | 2+2 2 | 2+2 2 |
| | 2+2 4 | 3+4 2 | 2+2 4 | 3+4 2 | 3+3 2 | 2+2 2 |
| | 2+2 4 | 3+4 2 | 2+2 4 | 5+6 2 | 2+2 2 | 2+2 2 |
| | 2+2 4 | 3+4 2 | 2+2 4 | 5+6 2 | 3+3 2 | 3+3 2 |
| | 2+2 4 | 3+4 2 | 2+2 4 | 3+4 2 | 2+2 2 | 2+2 2 |
| | 2+2 4 | 3+4 2 | 2+2 4 | 5+6 2 | 2+2 2 | 3+3 2 |
| | 2+2 4 | 5+6 2 | 2+2 4 | 3+4 2 | 3+3 2 | 3+3 2 |
| | 2+2 4 | 5+6 2 | 2+2 4 | 3+4 2 | 2+2 2 | 2+2 2 |
| | 2+2 4 | 3+4 2 | 2+2 4 | 5+6 2 | 2+2 2 | 3+3 2 |
| | 2+2 4 | 5+6 2 | 2+2 4 | 5+6 2 | 3+3 2 | 2+2 2 |
| | 2+2 4 | 3+4 2 | 2+2 4 | 5+6 2 | 2+2 2 | 2+2 2 |
| | 2+2 4 | 5+6 2 | 2+2 4 | 3+4 2 | 3+3 2 | 3+3 2 |
| | 2+2 4 | 5+6 2 | 2+2 4 | 3+4 2 | 2+2 2 | 2+2 2 |
| | 2+2 4 | 3+4 2 | 2+2 4 | 3+4 2 | 3+3 2 | 3+3 2 |
| | 2+2 4 | 3+4 2 | 2+2 4 | 5+6 2 | 2+2 2 | 2+2 2 |
| | | 3+4 2 | | 3+4 2 | 3+3 2 | 3+3 2 |
| | | 5+6 2 | | 5+6 2 | 3+3 2 | 2+2 2 |
| | | 3+4 2 | | 5+6 2 | 3+3 2 | 3+3 2 |
| | | 5+6 2 | | 3+4 2 | 2+2 2 | 2+2 2 |
| | | 5+6 2 | | 3+4 2 | 2+2 2 | 2+2 2 |
| | | 3+4 2 | | 5+6 2 | 3+3 2 | 3+3 2 |
| | | 5+6 2 | | 5+6 2 | 2+2 2 | 3+3 2 |
| | | 3+4 2 | | 5+6 2 | 3+3 2 | 2+2 2 |
| | | 5+6 2 | | 3+4 2 | 3+3 2 | 3+3 2 |
| | | 5+6 2 | | 3+4 2 | 2+2 2 | 2+2 2 |
| | | 5+6 2 | | 3+4 2 | 2+2 2 | 3+3 2 |
| | | 3+4 2 | | 5+6 2 | 3+3 2 | 3+3 2 |
| | | 3+4 2 | | 3+4 2 | 2+2 2 | 2+2 2 |
| | | 5+6 2 | | 5+6 2 | 2+2 2 | 2+2 2 |
| | | 3+4 2 | | 5+6 2 | 3+3 2 | 3+3 2 |
| | | 5+6 2 | | 5+6 2 | 2+2 2 | 2+2 2 |
| | | 5+6 2 | | 5+6 2 | 3+3 2 | 2+2 2 |
| | | 3+4 2 | | 3+4 2 | 3+3 2 | 3+3 2 |
| | | 5+6 2 | | 3+4 2 | 2+2 2 | 3+3 2 |
| | | 5+6 2 | | 3+4 2 | 3+3 2 | 2+2 2 |
| | | 3+4 2 | | 3+4 2 | 2+2 2 | 3+3 2 |
| | | 5+6 2 | | 5+6 2 | 3+3 2 | 3+3 2 |
| | | 3+4 2 | | 3+4 2 | 3+3 2 | 2+2 2 |
| | | 3+4 2 | | 5+6 2 | 2+2 2 | 2+2 2 |
| | | 3+4 2 | | 5+6 2 | 2+2 2 | 3+3 2 |
| | | 3+4 2 | | 5+6 2 | 3+3 2 | 2+2 2 |
| | | 5+6 2 | | 3+4 2 | 2+2 2 | 3+3 2 |
| | | 3+4 2 | | 5+6 2 | 2+2 2 | 2+2 2 |
| | | 3+4 2 | | 5+6 2 | 3+3 2 | 3+3 2 |
| 15.2.88 | 245254 | | 224 | Ba3Ir2O9Sr1 | | |
| | -A | -GM | -M | -Y | -L | -V |
| | 2+2 4 | 3+4 2 | 2+2 4 | 3+4 2 | 2+2 2 | 2+2 2 |

Continued on next column

**TABLE XXVII– continued**

| TQC | ICSD | | NoE | SF | | |
|---|---|---|---|---|---|---|
| | 2+2 4 | 3+4 2 | 2+2 4 | 3+4 2 | 3+3 2 | 2+2 2 |
| | 2+2 4 | 3+4 2 | 2+2 4 | 5+6 2 | 2+2 2 | 3+3 2 |
| | 2+2 4 | 3+4 2 | 2+2 4 | 3+4 2 | 3+3 2 | 3+3 2 |
| | 2+2 4 | 5+6 2 | 2+2 4 | 5+6 2 | 2+2 2 | 2+2 2 |
| | 2+2 4 | 5+6 2 | 2+2 4 | 3+4 2 | 3+3 2 | 2+2 2 |
| | 2+2 4 | 3+4 2 | 2+2 4 | 3+4 2 | 2+2 2 | 2+2 2 |
| | 2+2 4 | 5+6 2 | 2+2 4 | 5+6 2 | 3+3 2 | 3+3 2 |
| | 2+2 4 | 5+6 2 | 2+2 4 | 5+6 2 | 3+3 2 | 3+3 2 |
| | 2+2 4 | 3+4 2 | 2+2 4 | 3+4 2 | 2+2 2 | 2+2 2 |
| | 2+2 4 | 5+6 2 | 2+2 4 | 5+6 2 | 2+2 2 | 3+3 2 |
| | 2+2 4 | 5+6 2 | 2+2 4 | 5+6 2 | 3+3 2 | 3+3 2 |
| | 2+2 4 | 3+4 2 | 2+2 4 | 3+4 2 | 2+2 2 | 2+2 2 |
| | 2+2 4 | 5+6 2 | 2+2 4 | 5+6 2 | 3+3 2 | 3+3 2 |
| | 2+2 4 | 5+6 2 | 2+2 4 | 5+6 2 | 3+3 2 | 3+3 2 |
| | 2+2 4 | 3+4 2 | 2+2 4 | 3+4 2 | 2+2 2 | 2+2 2 |
| | 2+2 4 | 3+4 2 | 2+2 4 | 3+4 2 | 3+3 2 | 3+3 2 |
| | 2+2 4 | 5+6 2 | 2+2 4 | 5+6 2 | 2+2 2 | 2+2 2 |
| | 2+2 4 | 5+6 2 | 2+2 4 | 5+6 2 | 2+2 2 | 3+3 2 |
| | 2+2 4 | 5+6 2 | 2+2 4 | 5+6 2 | 3+3 2 | 3+3 2 |
| | 2+2 4 | 3+4 2 | 2+2 4 | 3+4 2 | 2+2 2 | 3+3 2 |
| | 2+2 4 | 5+6 2 | 2+2 4 | 5+6 2 | 3+3 2 | 2+2 2 |
| | 2+2 4 | 3+4 2 | 2+2 4 | 3+4 2 | 3+3 2 | 2+2 2 |
| | 2+2 4 | 3+4 2 | 2+2 4 | 3+4 2 | 2+2 2 | 2+2 2 |
| | 2+2 4 | 3+4 2 | 2+2 4 | 3+4 2 | 3+3 2 | 2+2 2 |
| | 2+2 4 | 3+4 2 | 2+2 4 | 3+4 2 | 2+2 2 | 2+2 2 |
| | 2+2 4 | 5+6 2 | 2+2 4 | 5+6 2 | 3+3 2 | 3+3 2 |
| | 2+2 4 | 5+6 2 | 2+2 4 | 5+6 2 | 2+2 2 | 3+3 2 |
| | 2+2 4 | 5+6 2 | 2+2 4 | 5+6 2 | 2+2 2 | 3+3 2 |
| | 2+2 4 | 5+6 2 | 2+2 4 | 5+6 2 | 3+3 2 | 3+3 2 |
| | 2+2 4 | 5+6 2 | 2+2 4 | 5+6 2 | 2+2 2 | 3+3 2 |
| | 2+2 4 | 5+6 2 | 2+2 4 | 5+6 2 | 3+3 2 | 3+3 2 |
| | 2+2 4 | 3+4 2 | 2+2 4 | 5+6 2 | 2+2 2 | 2+2 2 |
| | 2+2 4 | 3+4 2 | 2+2 4 | 3+4 2 | 3+3 2 | 3+3 2 |
| | 2+2 4 | 5+6 2 | 2+2 4 | 5+6 2 | 2+2 2 | 3+3 2 |
| | 2+2 4 | 5+6 2 | 2+2 4 | 3+4 2 | 3+3 2 | 2+2 2 |
| | 2+2 4 | 3+4 2 | 2+2 4 | 3+4 2 | 3+3 2 | 2+2 2 |
| | 2+2 4 | 5+6 2 | 2+2 4 | 5+6 2 | 2+2 2 | 3+3 2 |
| | 2+2 4 | 3+4 2 | 2+2 4 | 5+6 2 | 2+2 2 | 3+3 2 |
| | 2+2 4 | 3+4 2 | 2+2 4 | 5+6 2 | 3+3 2 | 2+2 2 |
| | 2+2 4 | 3+4 2 | 2+2 4 | 3+4 2 | 2+2 2 | 3+3 2 |
| | 2+2 4 | 5+6 2 | 2+2 4 | 3+4 2 | 3+3 2 | 2+2 2 |
| | 2+2 4 | 3+4 2 | 2+2 4 | 5+6 2 | 3+3 2 | 3+3 2 |
| | 2+2 4 | 5+6 2 | 2+2 4 | 3+4 2 | 2+2 2 | 2+2 2 |
| | 2+2 4 | 5+6 2 | 2+2 4 | 5+6 2 | 2+2 2 | 3+3 2 |
| | 2+2 4 | 5+6 2 | 2+2 4 | 3+4 2 | 3+3 2 | 2+2 2 |
| | 2+2 4 | 3+4 2 | 2+2 4 | 3+4 2 | 3+3 2 | 2+2 2 |
| | 2+2 4 | 5+6 2 | 2+2 4 | 5+6 2 | 2+2 2 | 3+3 2 |
| | 2+2 4 | 5+6 2 | 2+2 4 | 3+4 2 | 2+2 2 | 2+2 2 |
| | 2+2 4 | 3+4 2 | 2+2 4 | 5+6 2 | 3+3 2 | 3+3 2 |
| | 2+2 4 | 5+6 2 | 2+2 4 | 3+4 2 | 3+3 2 | 2+2 2 |
| | 2+2 4 | 3+4 2 | 2+2 4 | 5+6 2 | 2+2 2 | 3+3 2 |
| | 2+2 4 | 3+4 2 | 2+2 4 | 3+4 2 | 2+2 2 | 2+2 2 |
| | 2+2 4 | 5+6 2 | 2+2 4 | 5+6 2 | 3+3 2 | 3+3 2 |
| | 2+2 4 | 3+4 2 | 2+2 4 | 3+4 2 | 3+3 2 | 2+2 2 |
| | 2+2 4 | 5+6 2 | 2+2 4 | 3+4 2 | 2+2 2 | 2+2 2 |
| | | 3+4 2 | | 5+6 2 | 2+2 2 | 3+3 2 |
| | | 5+6 2 | | 5+6 2 | 3+3 2 | 2+2 2 |
| | | 3+4 2 | | 3+4 2 | 2+2 2 | 3+3 2 |
| | | 3+4 2 | | 3+4 2 | 3+3 2 | 2+2 2 |
| | | 5+6 2 | | 5+6 2 | 2+2 2 | 3+3 2 |
| | | 5+6 2 | | 3+4 2 | 3+3 2 | 2+2 2 |

Continued on next column

**TABLE XXVII– continued**

| TQC | ICSD | | NoE | | SF | |
|---|---|---|---|---|---|---|
| | | 3+4 2 | | 5+6 2 | 2+2 2 | 3+3 2 |
| | | 5+6 2 | | 3+4 2 | 3+3 2 | 3+3 2 |
| | | 3+4 2 | | 5+6 2 | 3+3 2 | 2+2 2 |
| | | 5+6 2 | | 5+6 2 | 2+2 2 | 3+3 2 |
| | | 3+4 2 | | 3+4 2 | 3+3 2 | 2+2 2 |
| | | 3+4 2 | | 3+4 2 | 2+2 2 | 2+2 2 |
| | | 5+6 2 | | 5+6 2 | 2+2 2 | 3+3 2 |
| | | 3+4 2 | | 3+4 2 | 3+3 2 | 2+2 2 |
| | | 5+6 2 | | 5+6 2 | 2+2 2 | 3+3 2 |
| | | 3+4 2 | | 3+4 2 | 3+3 2 | 2+2 2 |
| | | 3+4 2 | | 5+6 2 | 3+3 2 | 3+3 2 |
| | | 5+6 2 | | 5+6 2 | 2+2 2 | 3+3 2 |
| | | 5+6 2 | | 3+4 2 | 3+3 2 | 2+2 2 |
| | | 3+4 2 | | 3+4 2 | 2+2 2 | 2+2 2 |
| | | 5+6 2 | | 5+6 2 | 3+3 2 | 3+3 2 |
| | | 3+4 2 | | 3+4 2 | 2+2 2 | 2+2 2 |
| | | 3+4 2 | | 5+6 2 | 3+3 2 | 3+3 2 |
| | | 5+6 2 | | 3+4 2 | 2+2 2 | 2+2 2 |
| | | 3+4 2 | | 5+6 2 | 2+2 2 | 3+3 2 |
| | | 5+6 2 | | 5+6 2 | 3+3 2 | 3+3 2 |
| | | 5+6 2 | | 5+6 2 | 2+2 2 | 2+2 2 |
| | | 5+6 2 | | 3+4 2 | 3+3 2 | 3+3 2 |
| | | 3+4 2 | | 3+4 2 | 3+3 2 | 2+2 2 |
| | | 5+6 2 | | 3+4 2 | 2+2 2 | 2+2 2 |
| | | 5+6 2 | | 5+6 2 | 2+2 2 | 3+3 2 |
| | | 3+4 2 | | 3+4 2 | 3+3 2 | 3+3 2 |
| | | 3+4 2 | | 3+4 2 | 2+2 2 | 2+2 2 |
| | | 3+4 2 | | 5+6 2 | 3+3 2 | 2+2 2 |
| | | 5+6 2 | | 3+4 2 | 3+3 2 | 3+3 2 |
| | | 3+4 2 | | 3+4 2 | 2+2 2 | 2+2 2 |
| | | 5+6 2 | | 5+6 2 | 2+2 2 | 3+3 2 |
| | | 3+4 2 | | 5+6 2 | 3+3 2 | 2+2 2 |
| | | 5+6 2 | | 3+4 2 | 3+3 2 | 2+2 2 |
| | | 5+6 2 | | 5+6 2 | 2+2 2 | 3+3 2 |
| | | 3+4 2 | | 5+6 2 | 2+2 2 | 2+2 2 |
| | | 5+6 2 | | 3+4 2 | 3+3 2 | 3+3 2 |
| | | 3+4 2 | | 5+6 2 | 3+3 2 | 3+3 2 |
| | | 3+4 2 | | 5+6 2 | 2+2 2 | 3+3 2 |
| | | 5+6 2 | | 5+6 2 | 3+3 2 | 3+3 2 |
| | | 5+6 2 | | 3+4 2 | 2+2 2 | 2+2 2 |
| | | 5+6 2 | | 3+4 2 | 2+2 2 | 3+3 2 |
| | | 3+4 2 | | 5+6 2 | 3+3 2 | 2+2 2 |
| | | 3+4 2 | | 3+4 2 | 2+2 2 | 2+2 2 |
| | | 3+4 2 | | 3+4 2 | 3+3 2 | 2+2 2 |
| | | 5+6 2 | | 5+6 2 | 2+2 2 | 3+3 2 |
| | | 5+6 2 | | 5+6 2 | 3+3 2 | 3+3 2 |
| | | 3+4 2 | | 3+4 2 | 2+2 2 | 2+2 2 |
| | | 3+4 2 | | 5+6 2 | 3+3 2 | 3+3 2 |
| | | 3+4 2 | | 3+4 2 | 2+2 2 | 3+3 2 |
| | | 5+6 2 | | 5+6 2 | 3+3 2 | 2+2 2 |
| 15.2.89 | 245253 | | 224 | | Ba3Ca1Ir2O9 | |
| | -A | -GM | -M | -Y | -L | -V |
| | 2+2 4 | 3+4 2 | 2+2 4 | 3+4 2 | 2+2 2 | 2+2 2 |
| | 2+2 4 | 3+4 2 | 2+2 4 | 3+4 2 | 3+3 2 | 2+2 2 |
| | 2+2 4 | 3+4 2 | 2+2 4 | 5+6 2 | 2+2 2 | 3+3 2 |
| | 2+2 4 | 3+4 2 | 2+2 4 | 5+6 2 | 3+3 2 | 3+3 2 |
| | 2+2 4 | 5+6 2 | 2+2 4 | 3+4 2 | 2+2 2 | 2+2 2 |
| | 2+2 4 | 5+6 2 | 2+2 4 | 3+4 2 | 3+3 2 | 2+2 2 |
| | 2+2 4 | 3+4 2 | 2+2 4 | 3+4 2 | 2+2 2 | 2+2 2 |
| | 2+2 4 | 5+6 2 | 2+2 4 | 5+6 2 | 3+3 2 | 3+3 2 |
| | 2+2 4 | 5+6 2 | 2+2 4 | 5+6 2 | 2+2 2 | 3+3 2 |
| | 2+2 4 | 5+6 2 | 2+2 4 | 5+6 2 | 3+3 2 | 3+3 2 |
| | 2+2 4 | 5+6 2 | 2+2 4 | 5+6 2 | 3+3 2 | 3+3 2 |
| | 2+2 4 | 5+6 2 | 2+2 4 | 5+6 2 | 2+2 2 | 3+3 2 |
| | 2+2 4 | 5+6 2 | 2+2 4 | 5+6 2 | 2+2 2 | 3+3 2 |
| | 2+2 4 | 5+6 2 | 2+2 4 | 5+6 2 | 3+3 2 | 3+3 2 |
| | 2+2 4 | 3+4 2 | 2+2 4 | 3+4 2 | 2+2 2 | 2+2 2 |
| | 2+2 4 | 5+6 2 | 2+2 4 | 5+6 2 | 3+3 2 | 3+3 2 |
| | 2+2 4 | 3+4 2 | 2+2 4 | 3+4 2 | 2+2 2 | 2+2 2 |
| | 2+2 4 | 3+4 2 | 2+2 4 | 5+6 2 | 3+3 2 | 2+2 2 |
| | 2+2 4 | 5+6 2 | 2+2 4 | 3+4 2 | 3+3 2 | 3+3 2 |
| | 2+2 4 | 3+4 2 | 2+2 4 | 3+4 2 | 2+2 2 | 2+2 2 |
| | 2+2 4 | 3+4 2 | 2+2 4 | 3+4 2 | 2+2 2 | 2+2 2 |
| | 2+2 4 | 5+6 2 | 2+2 4 | 5+6 2 | 3+3 2 | 3+3 2 |
| | 2+2 4 | 3+4 2 | 2+2 4 | 3+4 2 | 3+3 2 | 2+2 2 |
| | 2+2 4 | 3+4 2 | 2+2 4 | 3+4 2 | 2+2 2 | 3+3 2 |
| | 2+2 4 | 5+6 2 | 2+2 4 | 5+6 2 | 2+2 2 | 2+2 2 |
| | 2+2 4 | 5+6 2 | 2+2 4 | 3+4 2 | 3+3 2 | 2+2 2 |
| | 2+2 4 | 3+4 2 | 2+2 4 | 3+4 2 | 3+3 2 | 3+3 2 |
| | 2+2 4 | 3+4 2 | 2+2 4 | 5+6 2 | 2+2 2 | 2+2 2 |
| | 2+2 4 | 5+6 2 | 2+2 4 | 5+6 2 | 3+3 2 | 3+3 2 |
| | 2+2 4 | 5+6 2 | 2+2 4 | 5+6 2 | 2+2 2 | 3+3 2 |
| | 2+2 4 | 5+6 2 | 2+2 4 | 5+6 2 | 2+2 2 | 3+3 2 |
| | 2+2 4 | 5+6 2 | 2+2 4 | 5+6 2 | 3+3 2 | 3+3 2 |
| | 2+2 4 | 3+4 2 | 2+2 4 | 5+6 2 | 3+3 2 | 3+3 2 |
| | 2+2 4 | 3+4 2 | 2+2 4 | 5+6 2 | 2+2 2 | 3+3 2 |
| | 2+2 4 | 5+6 2 | 2+2 4 | 3+4 2 | 2+2 2 | 2+2 2 |
| | 2+2 4 | 5+6 2 | 2+2 4 | 3+4 2 | 3+3 2 | 2+2 2 |
| | 2+2 4 | 5+6 2 | 2+2 4 | 5+6 2 | 2+2 2 | 3+3 2 |
| | 2+2 4 | 3+4 2 | 2+2 4 | 3+4 2 | 3+3 2 | 2+2 2 |
| | 2+2 4 | 3+4 2 | 2+2 4 | 5+6 2 | 2+2 2 | 3+3 2 |
| | 2+2 4 | 3+4 2 | 2+2 4 | 5+6 2 | 3+3 2 | 3+3 2 |
| | 2+2 4 | 3+4 2 | 2+2 4 | 3+4 2 | 2+2 2 | 2+2 2 |
| | 2+2 4 | 5+6 2 | 2+2 4 | 3+4 2 | 3+3 2 | 2+2 2 |
| | 2+2 4 | 3+4 2 | 2+2 4 | 5+6 2 | 3+3 2 | 3+3 2 |
| | 2+2 4 | 5+6 2 | 2+2 4 | 5+6 2 | 2+2 2 | 2+2 2 |
| | 2+2 4 | 5+6 2 | 2+2 4 | 3+4 2 | 2+2 2 | 3+3 2 |
| | 2+2 4 | 5+6 2 | 2+2 4 | 3+4 2 | 3+3 2 | 2+2 2 |
| | 2+2 4 | 5+6 2 | 2+2 4 | 3+4 2 | 3+3 2 | 3+3 2 |
| | 2+2 4 | 3+4 2 | 2+2 4 | 5+6 2 | 2+2 2 | 2+2 2 |
| | 2+2 4 | 5+6 2 | 2+2 4 | 3+4 2 | 2+2 2 | 2+2 2 |
| | 2+2 4 | 3+4 2 | 2+2 4 | 5+6 2 | 3+3 2 | 3+3 2 |
| | 2+2 4 | 5+6 2 | 2+2 4 | 3+4 2 | 2+2 2 | 2+2 2 |
| | 2+2 4 | 3+4 2 | 2+2 4 | 5+6 2 | 3+3 2 | 3+3 2 |
| | 2+2 4 | 3+4 2 | 2+2 4 | 3+4 2 | 3+3 2 | 2+2 2 |
| | 2+2 4 | 5+6 2 | 2+2 4 | 5+6 2 | 2+2 2 | 3+3 2 |
| | 2+2 4 | 3+4 2 | 2+2 4 | 3+4 2 | 2+2 2 | 2+2 2 |
| | 2+2 4 | 5+6 2 | 2+2 4 | 5+6 2 | 3+3 2 | 3+3 2 |
| | | 3+4 2 | | 3+4 2 | 3+3 2 | 2+2 2 |
| | | 5+6 2 | | 5+6 2 | 2+2 2 | 3+3 2 |
| | | 3+4 2 | | 3+4 2 | 2+2 2 | 2+2 2 |
| | | 3+4 2 | | 3+4 2 | 3+3 2 | 2+2 2 |
| | | 5+6 2 | | 3+4 2 | 2+2 2 | 3+3 2 |
| | | 5+6 2 | | 5+6 2 | 3+3 2 | 2+2 2 |
| | | 3+4 2 | | 5+6 2 | 3+3 2 | 3+3 2 |
| | | 5+6 2 | | 5+6 2 | 2+2 2 | 3+3 2 |
| | | 3+4 2 | | 3+4 2 | 3+3 2 | 2+2 2 |
| | | 5+6 2 | | 5+6 2 | 2+2 2 | 3+3 2 |
| | | 3+4 2 | | 3+4 2 | 3+3 2 | 2+2 2 |
| | | 3+4 2 | | 3+4 2 | 2+2 2 | 3+3 2 |
| | | 5+6 2 | | 5+6 2 | 2+2 2 | 2+2 2 |
| | | 3+4 2 | | 3+4 2 | 3+3 2 | 3+3 2 |

Continued on next column

**TABLE XXVII– continued**

| TQC | ICSD | | NoE | SF | | |
|---|---|---|---|---|---|---|
| | | 5+6 2 | | 5+6 2 | 2+2 2 | 2+2 2 |
| | | 3+4 2 | | 3+4 2 | 3+3 2 | 3+3 2 |
| | | 3+4 2 | | 5+6 2 | 3+3 2 | 2+2 2 |
| | | 5+6 2 | | 5+6 2 | 2+2 2 | 3+3 2 |
| | | 5+6 2 | | 3+4 2 | 3+3 2 | 2+2 2 |
| | | 5+6 2 | | 3+4 2 | 2+2 2 | 3+3 2 |
| | | 3+4 2 | | 5+6 2 | 3+3 2 | 2+2 2 |
| | | 5+6 2 | | 5+6 2 | 2+2 2 | 3+3 2 |
| | | 3+4 2 | | 5+6 2 | 2+2 2 | 2+2 2 |
| | | 3+4 2 | | 3+4 2 | 3+3 2 | 3+3 2 |
| | | 5+6 2 | | 5+6 2 | 3+3 2 | 2+2 2 |
| | | 3+4 2 | | 3+4 2 | 2+2 2 | 2+2 2 |
| | | 5+6 2 | | 3+4 2 | 2+2 2 | 3+3 2 |
| | | 5+6 2 | | 3+4 2 | 3+3 2 | 3+3 2 |
| | | 3+4 2 | | 5+6 2 | 2+2 2 | 2+2 2 |
| | | 5+6 2 | | 3+4 2 | 3+3 2 | 2+2 2 |
| | | 3+4 2 | | 5+6 2 | 3+3 2 | 2+2 2 |
| | | 5+6 2 | | 3+4 2 | 2+2 2 | 3+3 2 |
| | | 5+6 2 | | 5+6 2 | 2+2 2 | 2+2 2 |
| | | 3+4 2 | | 3+4 2 | 3+3 2 | 3+3 2 |
| | | 5+6 2 | | 3+4 2 | 2+2 2 | 2+2 2 |
| | | 5+6 2 | | 5+6 2 | 3+3 2 | 3+3 2 |
| | | 3+4 2 | | 5+6 2 | 2+2 2 | 2+2 2 |
| | | 3+4 2 | | 3+4 2 | 3+3 2 | 3+3 2 |
| | | 5+6 2 | | 3+4 2 | 2+2 2 | 2+2 2 |
| | | 5+6 2 | | 5+6 2 | 3+3 2 | 3+3 2 |
| | | 3+4 2 | | 5+6 2 | 3+3 2 | 3+3 2 |
| | | 5+6 2 | | 5+6 2 | 2+2 2 | 3+3 2 |
| | | 3+4 2 | | 5+6 2 | 3+3 2 | 3+3 2 |
| | | 5+6 2 | | 3+4 2 | 2+2 2 | 2+2 2 |
| | | 3+4 2 | | 5+6 2 | 2+2 2 | 3+3 2 |
| | | 3+4 2 | | 3+4 2 | 3+3 2 | 2+2 2 |
| | | 5+6 2 | | 5+6 2 | 2+2 2 | 3+3 2 |
| | | 3+4 2 | | 3+4 2 | 3+3 2 | 2+2 2 |
| | | 3+4 2 | | 3+4 2 | 3+3 2 | 2+2 2 |
| | | 3+4 2 | | 3+4 2 | 2+2 2 | 2+2 2 |
| | | 5+6 2 | | 5+6 2 | 2+2 2 | 3+3 2 |
| | | 3+4 2 | | 5+6 2 | 3+3 2 | 3+3 2 |
| | | 5+6 2 | | 5+6 2 | 3+3 2 | 3+3 2 |
| | | 3+4 2 | | 3+4 2 | 2+2 2 | 3+3 2 |
| | | 3+4 2 | | 5+6 2 | 2+2 2 | 2+2 2 |
| | | 5+6 2 | | 5+6 2 | 3+3 2 | 2+2 2 |
| 15.2.90 | 71060 | | 104 | Ba2Fe2Li1N3 | | |
| | -A | -GM | -M | -Y | -L | -V |
| | 2+2 4 | 3+4 2 | 2+2 4 | 3+4 2 | 2+2 2 | 2+2 2 |
| | 2+2 4 | 5+6 2 | 2+2 4 | 5+6 2 | 3+3 2 | 3+3 2 |
| | 2+2 4 | 3+4 2 | 2+2 4 | 5+6 2 | 2+2 2 | 3+3 2 |
| | 2+2 4 | 5+6 2 | 2+2 4 | 3+4 2 | 3+3 2 | 2+2 2 |
| | 2+2 4 | 3+4 2 | 2+2 4 | 5+6 2 | 3+3 2 | 3+3 2 |
| | 2+2 4 | 5+6 2 | 2+2 4 | 3+4 2 | 2+2 2 | 2+2 2 |
| | 2+2 4 | 3+4 2 | 2+2 4 | 5+6 2 | 3+3 2 | 2+2 2 |
| | 2+2 4 | 5+6 2 | 2+2 4 | 3+4 2 | 2+2 2 | 3+3 2 |
| | 2+2 4 | 3+4 2 | 2+2 4 | 5+6 2 | 2+2 2 | 2+2 2 |
| | 2+2 4 | 5+6 2 | 2+2 4 | 3+4 2 | 3+3 2 | 3+3 2 |
| | 2+2 4 | 3+4 2 | 2+2 4 | 3+4 2 | 3+3 2 | 3+3 2 |
| | 2+2 4 | 5+6 2 | 2+2 4 | 5+6 2 | 2+2 2 | 2+2 2 |
| | 2+2 4 | 3+4 2 | 2+2 4 | 5+6 2 | 3+3 2 | 2+2 2 |
| | 2+2 4 | 5+6 2 | 2+2 4 | 3+4 2 | 2+2 2 | 3+3 2 |
| | 2+2 4 | 3+4 2 | 2+2 4 | 3+4 2 | 2+2 2 | 3+3 2 |
| | 2+2 4 | 3+4 2 | 2+2 4 | 5+6 2 | 3+3 2 | 2+2 2 |
| | 2+2 4 | 5+6 2 | 2+2 4 | 3+4 2 | 2+2 2 | 3+3 2 |
| | 2+2 4 | 5+6 2 | 2+2 4 | 5+6 2 | 3+3 2 | 2+2 2 |
| | 2+2 4 | 3+4 2 | 2+2 4 | 3+4 2 | 2+2 2 | 2+2 2 |
| | 2+2 4 | 3+4 2 | 2+2 4 | 3+4 2 | 3+3 2 | 2+2 2 |
| | 2+2 4 | 5+6 2 | 2+2 4 | 5+6 2 | 3+3 2 | 3+3 2 |
| | 2+2 4 | 5+6 2 | 2+2 4 | 5+6 2 | 2+2 2 | 3+3 2 |
| | 2+2 4 | 3+4 2 | 2+2 4 | 5+6 2 | 3+3 2 | 2+2 2 |
| | 2+2 4 | 5+6 2 | 2+2 4 | 5+6 2 | 2+2 2 | 3+3 2 |
| | 2+2 4 | 3+4 2 | 2+2 4 | 3+4 2 | 3+3 2 | 3+3 2 |
| | 2+2 4 | 5+6 2 | 2+2 4 | 3+4 2 | 2+2 2 | 2+2 2 |
| | | 5+6 2 | | 3+4 2 | 2+2 2 | 2+2 2 |
| | | 5+6 2 | | 3+4 2 | 2+2 2 | 3+3 2 |
| | | 5+6 2 | | 3+4 2 | 3+3 2 | 3+3 2 |
| | | 5+6 2 | | 3+4 2 | 2+2 2 | 2+2 2 |
| | | 3+4 2 | | 5+6 2 | 2+2 2 | 2+2 2 |
| | | 3+4 2 | | 5+6 2 | 3+3 2 | 3+3 2 |
| | | 5+6 2 | | 5+6 2 | 2+2 2 | 3+3 2 |
| | | 3+4 2 | | 5+6 2 | 3+3 2 | 2+2 2 |
| | | 5+6 2 | | 5+6 2 | 2+2 2 | 3+3 2 |
| | | 3+4 2 | | 3+4 2 | 3+3 2 | 2+2 2 |
| | | 3+4 2 | | 5+6 2 | 3+3 2 | 2+2 2 |
| | | 3+4 2 | | 3+4 2 | 3+3 2 | 3+3 2 |
| | | 3+4 2 | | 5+6 2 | 3+3 2 | 2+2 2 |
| | | 5+6 2 | | 3+4 2 | 2+2 2 | 3+3 2 |
| | | 3+4 2 | | 5+6 2 | 3+3 2 | 2+2 2 |
| | | 3+4 2 | | 5+6 2 | 3+3 2 | 3+3 2 |
| | | 3+4 2 | | 5+6 2 | 2+2 2 | 2+2 2 |
| | | 3+4 2 | | 5+6 2 | 3+3 2 | 3+3 2 |
| | | 5+6 2 | | 3+4 2 | 3+3 2 | 3+3 2 |
| | | 5+6 2 | | 3+4 2 | 2+2 2 | 2+2 2 |
| | | 5+6 2 | | 3+4 2 | 2+2 2 | 3+3 2 |
| | | 5+6 2 | | 3+4 2 | 2+2 2 | 2+2 2 |
| | | 3+4 2 | | 3+4 2 | 3+3 2 | 2+2 2 |
| | | 5+6 2 | | 5+6 2 | 2+2 2 | 3+3 2 |
| | | 3+4 2 | | 5+6 2 | 3+3 2 | 2+2 2 |
| | | 5+6 2 | | 5+6 2 | 3+3 2 | 3+3 2 |
| 15.2.91 | 86064 | | 156 | Ba1Cu1N1 | | |
| | -A | -GM | -M | -Y | -L | -V |
| | 2+2 4 | 3+4 2 | 2+2 4 | 5+6 2 | 3+3 2 | 2+2 2 |
| | 2+2 4 | 3+4 2 | 2+2 4 | 3+4 2 | 3+3 2 | 3+3 2 |
| | 2+2 4 | 5+6 2 | 2+2 4 | 5+6 2 | 2+2 2 | 2+2 2 |
| | 2+2 4 | 5+6 2 | 2+2 4 | 3+4 2 | 2+2 2 | 3+3 2 |
| | 2+2 4 | 5+6 2 | 2+2 4 | 5+6 2 | 2+2 2 | 2+2 2 |
| | 2+2 4 | 3+4 2 | 2+2 4 | 3+4 2 | 3+3 2 | 3+3 2 |
| | 2+2 4 | 3+4 2 | 2+2 4 | 3+4 2 | 2+2 2 | 2+2 2 |
| | 2+2 4 | 3+4 2 | 2+2 4 | 5+6 2 | 3+3 2 | 3+3 2 |
| | 2+2 4 | 5+6 2 | 2+2 4 | 3+4 2 | 2+2 2 | 2+2 2 |
| | 2+2 4 | 3+4 2 | 2+2 4 | 5+6 2 | 3+3 2 | 2+2 2 |
| | 2+2 4 | 5+6 2 | 2+2 4 | 3+4 2 | 2+2 2 | 3+3 2 |
| | 2+2 4 | 5+6 2 | 2+2 4 | 5+6 2 | 3+3 2 | 3+3 2 |
| | 2+2 4 | 3+4 2 | 2+2 4 | 3+4 2 | 3+3 2 | 2+2 2 |
| | 2+2 4 | 5+6 2 | 2+2 4 | 5+6 2 | 2+2 2 | 3+3 2 |
| | 2+2 4 | 3+4 2 | 2+2 4 | 3+4 2 | 2+2 2 | 2+2 2 |
| | 2+2 4 | 3+4 2 | 2+2 4 | 5+6 2 | 3+3 2 | 3+3 2 |
| | 2+2 4 | 5+6 2 | 2+2 4 | 3+4 2 | 3+3 2 | 2+2 2 |
| | 2+2 4 | 3+4 2 | 2+2 4 | 3+4 2 | 2+2 2 | 2+2 2 |
| | 2+2 4 | 3+4 2 | 2+2 4 | 5+6 2 | 3+3 2 | 3+3 2 |
| | 2+2 4 | 5+6 2 | 2+2 4 | 5+6 2 | 3+3 2 | 2+2 2 |
| | 2+2 4 | 5+6 2 | 2+2 4 | 3+4 2 | 2+2 2 | 3+3 2 |
| | 2+2 4 | 3+4 2 | 2+2 4 | 5+6 2 | 2+2 2 | 3+3 2 |
| | 2+2 4 | 5+6 2 | 2+2 4 | 5+6 2 | 3+3 2 | 2+2 2 |
| | 2+2 4 | 5+6 2 | 2+2 4 | 3+4 2 | 2+2 2 | 3+3 2 |

Continued on next column

Continued on next column

**TABLE XXVII– continued**

| TQC | ICSD | | NoE | | SF | |
|---|---|---|---|---|---|---|
| | 2+2 4 | 3+4 2 | 2+2 4 | 3+4 2 | 3+3 2 | 2+2 2 |
| | 2+2 4 | 3+4 2 | 2+2 4 | 5+6 2 | 2+2 2 | 3+3 2 |
| | 2+2 4 | 5+6 2 | 2+2 4 | 3+4 2 | 2+2 2 | 2+2 2 |
| | 2+2 4 | 3+4 2 | 2+2 4 | 5+6 2 | 3+3 2 | 3+3 2 |
| | 2+2 4 | 5+6 2 | 2+2 4 | 3+4 2 | 2+2 2 | 2+2 2 |
| | 2+2 4 | 3+4 2 | 2+2 4 | 3+4 2 | 3+3 2 | 2+2 2 |
| | 2+2 4 | 5+6 2 | 2+2 4 | 5+6 2 | 2+2 2 | 3+3 2 |
| | 2+2 4 | 5+6 2 | 2+2 4 | 5+6 2 | 3+3 2 | 3+3 2 |
| | 2+2 4 | 3+4 2 | 2+2 4 | 5+6 2 | 2+2 2 | 3+3 2 |
| | 2+2 4 | 3+4 2 | 2+2 4 | 5+6 2 | 2+2 2 | 2+2 2 |
| | 2+2 4 | 3+4 2 | 2+2 4 | 5+6 2 | 2+2 2 | 2+2 2 |
| | 2+2 4 | 3+4 2 | 2+2 4 | 5+6 2 | 2+2 2 | 3+3 2 |
| | 2+2 4 | 5+6 2 | 2+2 4 | 5+6 2 | 2+2 2 | 2+2 2 |
| | 2+2 4 | 3+4 2 | 2+2 4 | 3+4 2 | 3+3 2 | 3+3 2 |
| | 2+2 4 | 3+4 2 | 2+2 4 | 5+6 2 | 3+3 2 | 3+3 2 |
| | | 5+6 2 | | 3+4 2 | 2+2 2 | 3+3 2 |
| | | 3+4 2 | | 5+6 2 | 3+3 2 | 2+2 2 |
| | | 5+6 2 | | 5+6 2 | 2+2 2 | 2+2 2 |
| | | 5+6 2 | | 3+4 2 | 3+3 2 | 2+2 2 |
| | | 3+4 2 | | 3+4 2 | 2+2 2 | 3+3 2 |
| | | 5+6 2 | | 3+4 2 | 3+3 2 | 3+3 2 |
| | | 3+4 2 | | 5+6 2 | 2+2 2 | 2+2 2 |
| | | 3+4 2 | | 3+4 2 | 2+2 2 | 3+3 2 |
| | | 5+6 2 | | 5+6 2 | 2+2 2 | 2+2 2 |
| | | 3+4 2 | | 5+6 2 | 2+2 2 | 3+3 2 |
| | | 5+6 2 | | 3+4 2 | 3+3 2 | 3+3 2 |
| | | 3+4 2 | | 5+6 2 | 2+2 2 | 2+2 2 |
| | | 5+6 2 | | 5+6 2 | 3+3 2 | 3+3 2 |
| | | 3+4 2 | | 5+6 2 | 2+2 2 | 2+2 2 |
| | | 5+6 2 | | 3+4 2 | 3+3 2 | 2+2 2 |
| | | 3+4 2 | | 5+6 2 | 2+2 2 | 3+3 2 |
| | | 5+6 2 | | 3+4 2 | 3+3 2 | 2+2 2 |
| | | 3+4 2 | | 5+6 2 | 2+2 2 | 3+3 2 |
| | | 5+6 2 | | 3+4 2 | 3+3 2 | 2+2 2 |
| | | 3+4 2 | | 5+6 2 | 3+3 2 | 3+3 2 |
| | | 3+4 2 | | 3+4 2 | 2+2 2 | 2+2 2 |
| | | 3+4 2 | | 5+6 2 | 2+2 2 | 3+3 2 |
| | | 5+6 2 | | 3+4 2 | 3+3 2 | 2+2 2 |
| | | 3+4 2 | | 5+6 2 | 2+2 2 | 3+3 2 |
| | | 5+6 2 | | 5+6 2 | 2+2 2 | 2+2 2 |
| | | 3+4 2 | | 5+6 2 | 3+3 2 | 3+3 2 |
| | | 3+4 2 | | 3+4 2 | 2+2 2 | 2+2 2 |
| | | 5+6 2 | | 3+4 2 | 3+3 2 | 3+3 2 |
| | | 5+6 2 | | 5+6 2 | 2+2 2 | 3+3 2 |
| | | 5+6 2 | | 3+4 2 | 3+3 2 | 2+2 2 |
| | | 3+4 2 | | 5+6 2 | 2+2 2 | 3+3 2 |
| | | 3+4 2 | | 3+4 2 | 3+3 2 | 2+2 2 |
| | | 3+4 2 | | 5+6 2 | 2+2 2 | 3+3 2 |
| | | 3+4 2 | | 5+6 2 | 2+2 2 | 2+2 2 |
| | | 5+6 2 | | 5+6 2 | 2+2 2 | 3+3 2 |
| | | 3+4 2 | | 5+6 2 | 3+3 2 | 2+2 2 |
| | | 5+6 2 | | 5+6 2 | 2+2 2 | 3+3 2 |
| | | 5+6 2 | | 5+6 2 | 3+3 2 | 3+3 2 |
| | | 5+6 2 | | 3+4 2 | 2+2 2 | 3+3 2 |
| 15.2.92 | 418385 | | 80 | | B2Ni1O4 | |
| | -A | -GM | -M | -Y | -L | -V |
| | 2+2 4 | 3+4 2 | 2+2 4 | 5+6 2 | 3+3 2 | 2+2 2 |
| | 2+2 4 | 3+4 2 | 2+2 4 | 5+6 2 | 3+3 2 | 3+3 2 |
| | 2+2 4 | 5+6 2 | 2+2 4 | 3+4 2 | 2+2 2 | 2+2 2 |
| | 2+2 4 | 3+4 2 | 2+2 4 | 3+4 2 | 2+2 2 | 3+3 2 |
| | 2+2 4 | 3+4 2 | 2+2 4 | 3+4 2 | 3+3 2 | 2+2 2 |

Continued on next column

**TABLE XXVII– continued**

| TQC | ICSD | | NoE | | SF | |
|---|---|---|---|---|---|---|
| | 2+2 4 | 5+6 2 | 2+2 4 | 5+6 2 | 3+3 2 | 3+3 2 |
| | 2+2 4 | 5+6 2 | 2+2 4 | 5+6 2 | 2+2 2 | 2+2 2 |
| | 2+2 4 | 5+6 2 | 2+2 4 | 3+4 2 | 2+2 2 | 3+3 2 |
| | 2+2 4 | 5+6 2 | 2+2 4 | 3+4 2 | 2+2 2 | 3+3 2 |
| | 2+2 4 | 3+4 2 | 2+2 4 | 5+6 2 | 2+2 2 | 2+2 2 |
| | 2+2 4 | 3+4 2 | 2+2 4 | 5+6 2 | 3+3 2 | 3+3 2 |
| | 2+2 4 | 5+6 2 | 2+2 4 | 3+4 2 | 3+3 2 | 2+2 2 |
| | 2+2 4 | 3+4 2 | 2+2 4 | 5+6 2 | 3+3 2 | 3+3 2 |
| | 2+2 4 | 3+4 2 | 2+2 4 | 5+6 2 | 3+3 2 | 2+2 2 |
| | 2+2 4 | 5+6 2 | 2+2 4 | 3+4 2 | 3+3 2 | 2+2 2 |
| | 2+2 4 | 3+4 2 | 2+2 4 | 5+6 2 | 2+2 2 | 3+3 2 |
| | 2+2 4 | 5+6 2 | 2+2 4 | 5+6 2 | 2+2 2 | 2+2 2 |
| | 2+2 4 | 3+4 2 | 2+2 4 | 3+4 2 | 3+3 2 | 3+3 2 |
| | 2+2 4 | 5+6 2 | 2+2 4 | 3+4 2 | 2+2 2 | 2+2 2 |
| | 2+2 4 | 3+4 2 | 2+2 4 | 3+4 2 | 3+3 2 | 3+3 2 |
| | | 5+6 2 | | 5+6 2 | 2+2 2 | 2+2 2 |
| | | 5+6 2 | | 3+4 2 | 2+2 2 | 3+3 2 |
| | | 3+4 2 | | 5+6 2 | 3+3 2 | 2+2 2 |
| | | 3+4 2 | | 5+6 2 | 2+2 2 | 3+3 2 |
| | | 3+4 2 | | 3+4 2 | 2+2 2 | 3+3 2 |
| | | 5+6 2 | | 3+4 2 | 3+3 2 | 2+2 2 |
| | | 5+6 2 | | 3+4 2 | 3+3 2 | 3+3 2 |
| | | 3+4 2 | | 5+6 2 | 2+2 2 | 2+2 2 |
| | | 5+6 2 | | 3+4 2 | 2+2 2 | 2+2 2 |
| | | 3+4 2 | | 3+4 2 | 3+3 2 | 3+3 2 |
| | | 5+6 2 | | 5+6 2 | 2+2 2 | 2+2 2 |
| | | 5+6 2 | | 5+6 2 | 3+3 2 | 3+3 2 |
| | | 3+4 2 | | 3+4 2 | 3+3 2 | 3+3 2 |
| | | 3+4 2 | | 3+4 2 | 2+2 2 | 3+3 2 |
| | | 5+6 2 | | 5+6 2 | 2+2 2 | 2+2 2 |
| | | 5+6 2 | | 3+4 2 | 3+3 2 | 2+2 2 |
| | | 3+4 2 | | 5+6 2 | 2+2 2 | 3+3 2 |
| | | 5+6 2 | | 5+6 2 | 3+3 2 | 2+2 2 |
| | | 3+4 2 | | 5+6 2 | 2+2 2 | 2+2 2 |
| | | 3+4 2 | | 3+4 2 | 3+3 2 | 3+3 2 |
| 15.2.93 | 58403 | | 64 | | Au2Ca5 | |
| | -A | -GM | -M | -Y | -L | -V |
| | 2+2 4 | 3+4 2 | 2+2 4 | 3+4 2 | 3+3 2 | 3+3 2 |
| | 2+2 4 | 5+6 2 | 2+2 4 | 5+6 2 | 2+2 2 | 2+2 2 |
| | 2+2 4 | 3+4 2 | 2+2 4 | 3+4 2 | 2+2 2 | 3+3 2 |
| | 2+2 4 | 3+4 2 | 2+2 4 | 5+6 2 | 3+3 2 | 2+2 2 |
| | 2+2 4 | 5+6 2 | 2+2 4 | 5+6 2 | 3+3 2 | 3+3 2 |
| | 2+2 4 | 3+4 2 | 2+2 4 | 3+4 2 | 2+2 2 | 2+2 2 |
| | 2+2 4 | 5+6 2 | 2+2 4 | 3+4 2 | 2+2 2 | 2+2 2 |
| | 2+2 4 | 5+6 2 | 2+2 4 | 5+6 2 | 3+3 2 | 3+3 2 |
| | 2+2 4 | 3+4 2 | 2+2 4 | 3+4 2 | 3+3 2 | 3+3 2 |
| | 2+2 4 | 5+6 2 | 2+2 4 | 5+6 2 | 3+3 2 | 2+2 2 |
| | 2+2 4 | 3+4 2 | 2+2 4 | 3+4 2 | 2+2 2 | 2+2 2 |
| | 2+2 4 | 5+6 2 | 2+2 4 | 5+6 2 | 2+2 2 | 3+3 2 |
| | 2+2 4 | 3+4 2 | 2+2 4 | 3+4 2 | 3+3 2 | 2+2 2 |
| | 2+2 4 | 3+4 2 | 2+2 4 | 5+6 2 | 2+2 2 | 3+3 2 |
| | 2+2 4 | 5+6 2 | 2+2 4 | 3+4 2 | 2+2 2 | 3+3 2 |
| | 2+2 4 | 5+6 2 | 2+2 4 | 5+6 2 | 3+3 2 | 2+2 2 |
| | | 3+4 2 | | 3+4 2 | 2+2 2 | 3+3 2 |
| | | 3+4 2 | | 5+6 2 | 3+3 2 | 2+2 2 |
| | | 5+6 2 | | 3+4 2 | 3+3 2 | 3+3 2 |
| | | 3+4 2 | | 5+6 2 | 2+2 2 | 2+2 2 |
| | | 5+6 2 | | 3+4 2 | 2+2 2 | 3+3 2 |
| | | 3+4 2 | | 5+6 2 | 3+3 2 | 2+2 2 |
| | | 5+6 2 | | 3+4 2 | 3+3 2 | 3+3 2 |
| | | 5+6 2 | | 5+6 2 | 2+2 2 | 2+2 2 |

Continued on next column

**TABLE XXVII– continued**

| TQC | ICSD | | NoE | SF | | |
|---|---|---|---|---|---|---|
| | | 3+4 2 | | 5+6 2 | 2+2 2 | 3+3 2 |
| | | 5+6 2 | | 3+4 2 | 3+3 2 | 2+2 2 |
| | | 3+4 2 | | 5+6 2 | 2+2 2 | 2+2 2 |
| | | 3+4 2 | | 3+4 2 | 3+3 2 | 3+3 2 |
| | | 5+6 2 | | 3+4 2 | 3+3 2 | 2+2 2 |
| | | 5+6 2 | | 5+6 2 | 2+2 2 | 3+3 2 |
| | | 3+4 2 | | 5+6 2 | 2+2 2 | 3+3 2 |
| | | 3+4 2 | | 5+6 2 | 3+3 2 | 2+2 2 |
| 15.2.94 | 417872 | | 144 | Al2Cl8Ni1 | | |
| | -A | -GM | -M | -Y | -L | -V |
| | 2+2 4 | 3+4 2 | 2+2 4 | 3+4 2 | 2+2 2 | 2+2 2 |
| | 2+2 4 | 5+6 2 | 2+2 4 | 5+6 2 | 3+3 2 | 3+3 2 |
| | 2+2 4 | 3+4 2 | 2+2 4 | 3+4 2 | 3+3 2 | 2+2 2 |
| | 2+2 4 | 5+6 2 | 2+2 4 | 5+6 2 | 2+2 2 | 3+3 2 |
| | 2+2 4 | 5+6 2 | 2+2 4 | 5+6 2 | 3+3 2 | 3+3 2 |
| | 2+2 4 | 3+4 2 | 2+2 4 | 5+6 2 | 2+2 2 | 2+2 2 |
| | 2+2 4 | 3+4 2 | 2+2 4 | 3+4 2 | 3+3 2 | 3+3 2 |
| | 2+2 4 | 5+6 2 | 2+2 4 | 3+4 2 | 2+2 2 | 2+2 2 |
| | 2+2 4 | 5+6 2 | 2+2 4 | 5+6 2 | 2+2 2 | 3+3 2 |
| | 2+2 4 | 5+6 2 | 2+2 4 | 5+6 2 | 3+3 2 | 3+3 2 |
| | 2+2 4 | 3+4 2 | 2+2 4 | 3+4 2 | 3+3 2 | 2+2 2 |
| | 2+2 4 | 3+4 2 | 2+2 4 | 3+4 2 | 2+2 2 | 2+2 2 |
| | 2+2 4 | 3+4 2 | 2+2 4 | 3+4 2 | 3+3 2 | 2+2 2 |
| | 2+2 4 | 5+6 2 | 2+2 4 | 5+6 2 | 2+2 2 | 2+2 2 |
| | 2+2 4 | 5+6 2 | 2+2 4 | 3+4 2 | 3+3 2 | 3+3 2 |
| | 2+2 4 | 3+4 2 | 2+2 4 | 5+6 2 | 2+2 2 | 3+3 2 |
| | 2+2 4 | 3+4 2 | 2+2 4 | 3+4 2 | 2+2 2 | 2+2 2 |
| | 2+2 4 | 5+6 2 | 2+2 4 | 5+6 2 | 3+3 2 | 3+3 2 |
| | 2+2 4 | 3+4 2 | 2+2 4 | 5+6 2 | 2+2 2 | 3+3 2 |
| | 2+2 4 | 5+6 2 | 2+2 4 | 3+4 2 | 3+3 2 | 2+2 2 |
| | 2+2 4 | 3+4 2 | 2+2 4 | 3+4 2 | 3+3 2 | 2+2 2 |
| | 2+2 4 | 5+6 2 | 2+2 4 | 5+6 2 | 2+2 2 | 3+3 2 |
| | 2+2 4 | 5+6 2 | 2+2 4 | 3+4 2 | 3+3 2 | 3+3 2 |
| | 2+2 4 | 3+4 2 | 2+2 4 | 5+6 2 | 2+2 2 | 2+2 2 |
| | 2+2 4 | 5+6 2 | 2+2 4 | 5+6 2 | 3+3 2 | 3+3 2 |
| | 2+2 4 | 3+4 2 | 2+2 4 | 3+4 2 | 2+2 2 | 2+2 2 |
| | 2+2 4 | 3+4 2 | 2+2 4 | 3+4 2 | 3+3 2 | 2+2 2 |
| | 2+2 4 | 3+4 2 | 2+2 4 | 3+4 2 | 2+2 2 | 3+3 2 |
| | 2+2 4 | 3+4 2 | 2+2 4 | 5+6 2 | 3+3 2 | 2+2 2 |
| | 2+2 4 | 5+6 2 | 2+2 4 | 3+4 2 | 2+2 2 | 3+3 2 |
| | 2+2 4 | 5+6 2 | 2+2 4 | 5+6 2 | 3+3 2 | 2+2 2 |
| | 2+2 4 | 5+6 2 | 2+2 4 | 3+4 2 | 2+2 2 | 2+2 2 |
| | 2+2 4 | 3+4 2 | 2+2 4 | 5+6 2 | 2+2 2 | 3+3 2 |
| | 2+2 4 | 5+6 2 | 2+2 4 | 5+6 2 | 3+3 2 | 3+3 2 |
| | 2+2 4 | 3+4 2 | 2+2 4 | 3+4 2 | 2+2 2 | 2+2 2 |
| | 2+2 4 | 5+6 2 | 2+2 4 | 5+6 2 | 3+3 2 | 3+3 2 |
| | | 5+6 2 | | 3+4 2 | 2+2 2 | 3+3 2 |
| | | 5+6 2 | | 5+6 2 | 3+3 2 | 2+2 2 |
| | | 3+4 2 | | 5+6 2 | 2+2 2 | 3+3 2 |
| | | 5+6 2 | | 5+6 2 | 3+3 2 | 2+2 2 |
| | | 3+4 2 | | 5+6 2 | 2+2 2 | 3+3 2 |
| | | 3+4 2 | | 3+4 2 | 3+3 2 | 2+2 2 |
| | | 3+4 2 | | 3+4 2 | 2+2 2 | 3+3 2 |
| | | 5+6 2 | | 5+6 2 | 3+3 2 | 2+2 2 |
| | | 5+6 2 | | 5+6 2 | 3+3 2 | 3+3 2 |
| | | 3+4 2 | | 3+4 2 | 2+2 2 | 2+2 2 |
| | | 5+6 2 | | 3+4 2 | 3+3 2 | 3+3 2 |
| | | 3+4 2 | | 5+6 2 | 2+2 2 | 3+3 2 |
| | | 3+4 2 | | 3+4 2 | 2+2 2 | 2+2 2 |
| | | 5+6 2 | | 3+4 2 | 3+3 2 | 3+3 2 |
| | | 3+4 2 | | 3+4 2 | 2+2 2 | 2+2 2 |

Continued on next column

**TABLE XXVII– continued**

| TQC | ICSD | | NoE | SF | | |
|---|---|---|---|---|---|---|
| | | 3+4 2 | | 3+4 2 | 2+2 2 | 2+2 2 |
| | | 5+6 2 | | 5+6 2 | 3+3 2 | 3+3 2 |
| | | 5+6 2 | | 5+6 2 | 3+3 2 | 2+2 2 |
| | | 5+6 2 | | 3+4 2 | 2+2 2 | 2+2 2 |
| | | 3+4 2 | | 3+4 2 | 3+3 2 | 3+3 2 |
| | | 5+6 2 | | 5+6 2 | 3+3 2 | 2+2 2 |
| | | 3+4 2 | | 5+6 2 | 2+2 2 | 3+3 2 |
| | | 3+4 2 | | 3+4 2 | 3+3 2 | 3+3 2 |
| | | 5+6 2 | | 5+6 2 | 2+2 2 | 2+2 2 |
| | | 3+4 2 | | 3+4 2 | 3+3 2 | 2+2 2 |
| | | 5+6 2 | | 5+6 2 | 2+2 2 | 3+3 2 |
| | | 3+4 2 | | 5+6 2 | 3+3 2 | 2+2 2 |
| | | 5+6 2 | | 3+4 2 | 2+2 2 | 3+3 2 |
| | | 3+4 2 | | 3+4 2 | 3+3 2 | 2+2 2 |
| | | 5+6 2 | | 5+6 2 | 2+2 2 | 3+3 2 |
| | | 5+6 2 | | 5+6 2 | 2+2 2 | 2+2 2 |
| | | 3+4 2 | | 3+4 2 | 3+3 2 | 3+3 2 |
| | | 5+6 2 | | 5+6 2 | 3+3 2 | 3+3 2 |
| | | 3+4 2 | | 3+4 2 | 2+2 2 | 2+2 2 |
| | | 5+6 2 | | 3+4 2 | 2+2 2 | 2+2 2 |
| | | 5+6 2 | | 5+6 2 | 3+3 2 | 3+3 2 |
| 15.2.95 | 251835 | | 188 | Al1B2La2Os2 | | |
| | -A | -GM | -M | -Y | -L | -V |
| | 2+2 4 | 3+4 2 | 2+2 4 | 3+4 2 | 2+2 2 | 3+3 2 |
| | 2+2 4 | 5+6 2 | 2+2 4 | 5+6 2 | 3+3 2 | 2+2 2 |
| | 2+2 4 | 3+4 2 | 2+2 4 | 3+4 2 | 2+2 2 | 3+3 2 |
| | 2+2 4 | 5+6 2 | 2+2 4 | 5+6 2 | 3+3 2 | 2+2 2 |
| | 2+2 4 | 3+4 2 | 2+2 4 | 5+6 2 | 2+2 2 | 3+3 2 |
| | 2+2 4 | 3+4 2 | 2+2 4 | 5+6 2 | 2+2 2 | 2+2 2 |
| | 2+2 4 | 5+6 2 | 2+2 4 | 3+4 2 | 3+3 2 | 2+2 2 |
| | 2+2 4 | 5+6 2 | 2+2 4 | 3+4 2 | 3+3 2 | 3+3 2 |
| | 2+2 4 | 3+4 2 | 2+2 4 | 3+4 2 | 2+2 2 | 3+3 2 |
| | 2+2 4 | 5+6 2 | 2+2 4 | 5+6 2 | 3+3 2 | 2+2 2 |
| | 2+2 4 | 3+4 2 | 2+2 4 | 3+4 2 | 2+2 2 | 3+3 2 |
| | 2+2 4 | 3+4 2 | 2+2 4 | 5+6 2 | 3+3 2 | 2+2 2 |
| | 2+2 4 | 5+6 2 | 2+2 4 | 5+6 2 | 2+2 2 | 3+3 2 |
| | 2+2 4 | 3+4 2 | 2+2 4 | 5+6 2 | 2+2 2 | 2+2 2 |
| | 2+2 4 | 5+6 2 | 2+2 4 | 3+4 2 | 3+3 2 | 3+3 2 |
| | 2+2 4 | 5+6 2 | 2+2 4 | 3+4 2 | 3+3 2 | 2+2 2 |
| | 2+2 4 | 3+4 2 | 2+2 4 | 3+4 2 | 2+2 2 | 3+3 2 |
| | 2+2 4 | 5+6 2 | 2+2 4 | 5+6 2 | 3+3 2 | 2+2 2 |
| | 2+2 4 | 3+4 2 | 2+2 4 | 3+4 2 | 2+2 2 | 3+3 2 |
| | 2+2 4 | 3+4 2 | 2+2 4 | 5+6 2 | 3+3 2 | 2+2 2 |
| | 2+2 4 | 5+6 2 | 2+2 4 | 3+4 2 | 2+2 2 | 3+3 2 |
| | 2+2 4 | 3+4 2 | 2+2 4 | 5+6 2 | 3+3 2 | 2+2 2 |
| | 2+2 4 | 5+6 2 | 2+2 4 | 5+6 2 | 2+2 2 | 3+3 2 |
| | 2+2 4 | 3+4 2 | 2+2 4 | 3+4 2 | 3+3 2 | 2+2 2 |
| | 2+2 4 | 5+6 2 | 2+2 4 | 5+6 2 | 2+2 2 | 3+3 2 |
| | 2+2 4 | 5+6 2 | 2+2 4 | 3+4 2 | 3+3 2 | 2+2 2 |
| | 2+2 4 | 3+4 2 | 2+2 4 | 5+6 2 | 2+2 2 | 3+3 2 |
| | 2+2 4 | 3+4 2 | 2+2 4 | 3+4 2 | 2+2 2 | 2+2 2 |
| | 2+2 4 | 5+6 2 | 2+2 4 | 3+4 2 | 3+3 2 | 2+2 2 |
| | 2+2 4 | 5+6 2 | 2+2 4 | 5+6 2 | 2+2 2 | 3+3 2 |
| | 2+2 4 | 5+6 2 | 2+2 4 | 5+6 2 | 3+3 2 | 3+3 2 |
| | 2+2 4 | 3+4 2 | 2+2 4 | 3+4 2 | 3+3 2 | 2+2 2 |
| | 2+2 4 | 3+4 2 | 2+2 4 | 5+6 2 | 3+3 2 | 2+2 2 |
| | 2+2 4 | 3+4 2 | 2+2 4 | 5+6 2 | 3+3 2 | 3+3 2 |
| | 2+2 4 | 5+6 2 | 2+2 4 | 3+4 2 | 2+2 2 | 2+2 2 |
| | 2+2 4 | 5+6 2 | 2+2 4 | 3+4 2 | 2+2 2 | 3+3 2 |
| | 2+2 4 | 5+6 2 | 2+2 4 | 3+4 2 | 3+3 2 | 2+2 2 |
| | 2+2 4 | 3+4 2 | 2+2 4 | 3+4 2 | 2+2 2 | 3+3 2 |

Continued on next column

**TABLE XXVII– continued**

| TQC | ICSD | NoE | SF |
|---|---|---|---|
| | 2+2 4 3+4 2 | 2+2 4 5+6 2 | 2+2 2 2+2 2 |
| | 2+2 4 5+6 2 | 2+2 4 5+6 2 | 3+3 2 3+3 2 |
| | 2+2 4 3+4 2 | 2+2 4 3+4 2 | 2+2 2 2+2 2 |
| | 2+2 4 3+4 2 | 2+2 4 3+4 2 | 3+3 2 2+2 2 |
| | 2+2 4 5+6 2 | 2+2 4 5+6 2 | 2+2 2 3+3 2 |
| | 2+2 4 5+6 2 | 2+2 4 5+6 2 | 3+3 2 3+3 2 |
| | 2+2 4 3+4 2 | 2+2 4 5+6 2 | 2+2 2 3+3 2 |
| | 2+2 4 5+6 2 | 2+2 4 5+6 2 | 3+3 2 2+2 2 |
| | 2+2 4 3+4 2 | 2+2 4 3+4 2 | 2+2 2 2+2 2 |
| | 5+6 2 | 3+4 2 | 3+3 2 3+3 2 |
| | 5+6 2 | 3+4 2 | 2+2 2 3+3 2 |
| | 5+6 2 | 3+4 2 | 3+3 2 2+2 2 |
| | 3+4 2 | 3+4 2 | 2+2 2 3+3 2 |
| | 5+6 2 | 5+6 2 | 3+3 2 2+2 2 |
| | 3+4 2 | 3+4 2 | 2+2 2 3+3 2 |
| | 5+6 2 | 5+6 2 | 3+3 2 2+2 2 |
| | 3+4 2 | 5+6 2 | 3+3 2 2+2 2 |
| | 3+4 2 | 3+4 2 | 2+2 2 3+3 2 |
| | 3+4 2 | 5+6 2 | 3+3 2 2+2 2 |
| | 5+6 2 | 3+4 2 | 2+2 2 3+3 2 |
| | 3+4 2 | 5+6 2 | 2+2 2 2+2 2 |
| | 5+6 2 | 3+4 2 | 3+3 2 3+3 2 |
| | 3+4 2 | 5+6 2 | 3+3 2 2+2 2 |
| | 5+6 2 | 3+4 2 | 2+2 2 3+3 2 |
| | 3+4 2 | 5+6 2 | 3+3 2 2+2 2 |
| | 5+6 2 | 5+6 2 | 3+3 2 3+3 2 |
| | 3+4 2 | 5+6 2 | 2+2 2 2+2 2 |
| | 5+6 2 | 3+4 2 | 2+2 2 2+2 2 |
| | 5+6 2 | 5+6 2 | 3+3 2 3+3 2 |
| | 3+4 2 | 3+4 2 | 2+2 2 2+2 2 |
| | 5+6 2 | 3+4 2 | 3+3 2 3+3 2 |
| | 3+4 2 | 3+4 2 | 2+2 2 3+3 2 |
| | 3+4 2 | 5+6 2 | 3+3 2 2+2 2 |
| | 5+6 2 | 3+4 2 | 3+3 2 2+2 2 |
| | 5+6 2 | 5+6 2 | 2+2 2 2+2 2 |
| | 5+6 2 | 3+4 2 | 3+3 2 3+3 2 |
| | 3+4 2 | 5+6 2 | 2+2 2 3+3 2 |
| | 3+4 2 | 3+4 2 | 2+2 2 3+3 2 |
| | 5+6 2 | 5+6 2 | 3+3 2 2+2 2 |
| | 3+4 2 | 3+4 2 | 3+3 2 2+2 2 |
| | 5+6 2 | 5+6 2 | 2+2 2 2+2 2 |
| | 3+4 2 | 3+4 2 | 2+2 2 3+3 2 |
| | 5+6 2 | 5+6 2 | 3+3 2 3+3 2 |
| | 3+4 2 | 5+6 2 | 2+2 2 2+2 2 |
| | 5+6 2 | 3+4 2 | 2+2 2 3+3 2 |
| | 5+6 2 | 5+6 2 | 3+3 2 2+2 2 |
| | 3+4 2 | 5+6 2 | 3+3 2 3+3 2 |
| | 3+4 2 | 3+4 2 | 2+2 2 2+2 2 |
| | 3+4 2 | 5+6 2 | 3+3 2 3+3 2 |
| | 5+6 2 | 3+4 2 | 3+3 2 2+2 2 |
| | 3+4 2 | 5+6 2 | 2+2 2 3+3 2 |
| | 5+6 2 | 5+6 2 | 3+3 2 2+2 2 |
| | 5+6 2 | 3+4 2 | 2+2 2 3+3 2 |
| | 3+4 2 | 3+4 2 | 2+2 2 3+3 2 |
| | 3+4 2 | 5+6 2 | 3+3 2 2+2 2 |
| | 3+4 2 | 3+4 2 | 3+3 2 3+3 2 |

The End

TABLE XXVIII: The computed irreducible representations for the valence bands for the compounds of SEBR and NLC in space group 47. See the list of tables.

| TQC | ICSD | NoE | SF |
|---|---|---|---|
| 47.2.1 | 86427 | 64 | Cu1O5Sr2Tl1 |
| | -GM -R | -S -T -U | -X -Y -Z |
| | 5 2 6 2 | 5 2 5 2 5 2 | 6 2 6 2 6 2 |
| | 6 2 5 2 | 6 2 6 2 6 2 | 5 2 5 2 5 2 |
| | 5 2 5 2 | 6 2 6 2 6 2 | 5 2 5 2 6 2 |
| | 5 2 5 2 | 6 2 5 2 5 2 | 6 2 6 2 6 2 |
| | 6 2 6 2 | 6 2 6 2 6 2 | 6 2 6 2 6 2 |
| | 5 2 5 2 | 5 2 5 2 5 2 | 5 2 5 2 5 2 |
| | 5 2 5 2 | 5 2 6 2 6 2 | 6 2 6 2 5 2 |
| | 6 2 6 2 | 5 2 5 2 5 2 | 5 2 5 2 6 2 |
| | 5 2 5 2 | 6 2 6 2 6 2 | 6 2 6 2 5 2 |
| | 6 2 6 2 | 5 2 5 2 5 2 | 5 2 5 2 5 2 |
| | 6 2 6 2 | 5 2 6 2 5 2 | 6 2 6 2 6 2 |
| | 5 2 5 2 | 6 2 5 2 6 2 | 6 2 6 2 5 2 |
| | 5 2 5 2 | 6 2 6 2 6 2 | 5 2 5 2 6 2 |
| | 5 2 6 2 | 5 2 6 2 6 2 | 5 2 5 2 5 2 |
| | 6 2 6 2 | 5 2 5 2 5 2 | 5 2 5 2 5 2 |
| | 6 2 6 2 | 5 2 6 2 6 2 | 5 2 5 2 5 2 |
| | 6 2 5 2 | 5 2 6 2 6 2 | 5 2 5 2 6 2 |
| | 5 2 6 2 | 5 2 5 2 6 2 | 6 2 6 2 6 2 |
| | 5 2 6 2 | 5 2 6 2 5 2 | 6 2 6 2 5 2 |
| | 5 2 6 2 | 6 2 6 2 6 2 | 5 2 5 2 6 2 |
| | 6 2 6 2 | 6 2 5 2 5 2 | 6 2 6 2 6 2 |
| | 5 2 6 2 | 6 2 5 2 5 2 | 5 2 5 2 6 2 |
| | 5 2 5 2 | 5 2 6 2 6 2 | 6 2 6 2 6 2 |
| | 6 2 5 2 | 5 2 5 2 5 2 | 6 2 6 2 5 2 |
| | 6 2 6 2 | 5 2 6 2 6 2 | 5 2 5 2 6 2 |
| | 6 2 6 2 | 6 2 6 2 6 2 | 6 2 6 2 6 2 |
| | 6 2 6 2 | 6 2 5 2 5 2 | 5 2 5 2 5 2 |
| | 5 2 6 2 | 5 2 5 2 5 2 | 5 2 5 2 5 2 |
| | 6 2 6 2 | 6 2 6 2 5 2 | 5 2 5 2 6 2 |
| | 5 2 6 2 | 5 2 6 2 6 2 | 5 2 5 2 6 2 |
| | 6 2 6 2 | 5 2 5 2 5 2 | 5 2 5 2 5 2 |
| | 6 2 6 2 | 5 2 5 2 6 2 | 5 2 5 2 5 2 |
| 47.2.2 | 390001 | 206 | Ba4Cu6O13Y2 |
| | -GM -R | -S -T -U | -X -Y -Z |
| | 5 2 5 2 | 6 2 6 2 5 2 | 6 2 5 2 6 2 |
| | 6 2 6 2 | 5 2 5 2 6 2 | 5 2 6 2 5 2 |
| | 5 2 5 2 | 5 2 6 2 6 2 | 6 2 6 2 5 2 |
| | 6 2 6 2 | 6 2 5 2 5 2 | 5 2 5 2 6 2 |
| | 6 2 6 2 | 6 2 5 2 5 2 | 5 2 5 2 6 2 |
| | 5 2 5 2 | 5 2 6 2 6 2 | 6 2 6 2 5 2 |
| | 5 2 5 2 | 6 2 6 2 5 2 | 6 2 5 2 6 2 |
| | 6 2 6 2 | 5 2 5 2 6 2 | 5 2 6 2 5 2 |
| | 5 2 5 2 | 6 2 6 2 5 2 | 6 2 5 2 6 2 |
| | 6 2 6 2 | 5 2 5 2 6 2 | 5 2 6 2 5 2 |
| | 5 2 5 2 | 5 2 6 2 5 2 | 6 2 5 2 6 2 |
| | 6 2 6 2 | 6 2 5 2 6 2 | 5 2 6 2 5 2 |
| | 5 2 6 2 | 5 2 5 2 5 2 | 6 2 6 2 6 2 |
| | 6 2 5 2 | 6 2 6 2 5 2 | 6 2 5 2 5 2 |
| | 6 2 6 2 | 5 2 6 2 6 2 | 5 2 6 2 5 2 |
| | 5 2 5 2 | 6 2 6 2 6 2 | 5 2 5 2 5 2 |
| | 5 2 5 2 | 5 2 5 2 5 2 | 5 2 6 2 6 2 |
| | 6 2 6 2 | 5 2 5 2 6 2 | 6 2 5 2 5 2 |
| | 5 2 5 2 | 6 2 5 2 6 2 | 5 2 6 2 6 2 |
| | 6 2 6 2 | 6 2 6 2 5 2 | 6 2 5 2 5 2 |
| | 5 2 5 2 | 5 2 5 2 6 2 | 5 2 6 2 5 2 |
| | 6 2 5 2 | 6 2 6 2 5 2 | 6 2 5 2 6 2 |

Continued on next column

**TABLE XXVIII– continued**

| TQC | ICSD | NoE | SF |
|---|---|---|---|
| | 5 2 6 2 | 5 2 6 2 6 2 | 5 2 6 2 6 2 |
| | 6 2 6 2 | 5 2 5 2 6 2 | 6 2 5 2 6 2 |
| | 5 2 5 2 | 6 2 6 2 5 2 | 6 2 6 2 5 2 |
| | 5 2 6 2 | 6 2 5 2 5 2 | 5 2 5 2 6 2 |
| | 6 2 5 2 | 5 2 6 2 6 2 | 6 2 6 2 5 2 |
| | 6 2 6 2 | 6 2 5 2 5 2 | 5 2 5 2 6 2 |
| | 5 2 5 2 | 5 2 6 2 6 2 | 6 2 6 2 5 2 |
| | 6 2 6 2 | 6 2 5 2 5 2 | 5 2 5 2 6 2 |
| | 5 2 5 2 | 5 2 6 2 6 2 | 6 2 6 2 5 2 |
| | 6 2 6 2 | 6 2 5 2 5 2 | 5 2 5 2 6 2 |
| | 5 2 5 2 | 5 2 6 2 6 2 | 6 2 6 2 5 2 |
| | 6 2 6 2 | 6 2 5 2 5 2 | 5 2 5 2 6 2 |
| | 5 2 5 2 | 5 2 6 2 6 2 | 6 2 6 2 5 2 |
| | 6 2 6 2 | 6 2 5 2 5 2 | 5 2 5 2 6 2 |
| | 5 2 5 2 | 5 2 6 2 6 2 | 6 2 6 2 5 2 |
| | 5 2 5 2 | 6 2 5 2 6 2 | 5 2 6 2 6 2 |
| | 6 2 6 2 | 5 2 6 2 5 2 | 6 2 5 2 5 2 |
| | 5 2 6 2 | 5 2 6 2 5 2 | 5 2 5 2 6 2 |
| | 5 2 5 2 | 6 2 5 2 6 2 | 5 2 6 2 6 2 |
| | 5 2 6 2 | 6 2 6 2 5 2 | 6 2 5 2 6 2 |
| | 5 2 5 2 | 6 2 5 2 6 2 | 5 2 5 2 5 2 |
| | 6 2 6 2 | 5 2 5 2 6 2 | 5 2 6 2 5 2 |
| | 5 2 6 2 | 5 2 6 2 5 2 | 6 2 6 2 5 2 |
| | 6 2 5 2 | 6 2 6 2 5 2 | 5 2 5 2 5 2 |
| | 6 2 5 2 | 5 2 5 2 5 2 | 5 2 5 2 5 2 |
| | 6 2 6 2 | 5 2 6 2 5 2 | 6 2 6 2 5 2 |
| | 6 2 5 2 | 6 2 5 2 5 2 | 6 2 6 2 5 2 |
| | 5 2 6 2 | 6 2 5 2 5 2 | 6 2 5 2 6 2 |
| | 5 2 6 2 | 5 2 6 2 5 2 | 6 2 6 2 6 2 |
| | 6 2 5 2 | 6 2 5 2 6 2 | 5 2 5 2 6 2 |
| | 5 2 6 2 | 5 2 5 2 5 2 | 5 2 5 2 6 2 |
| | 5 2 5 2 | 6 2 6 2 6 2 | 5 2 6 2 6 2 |
| | 6 2 6 2 | 6 2 5 2 5 2 | 6 2 5 2 6 2 |
| | 5 2 6 2 | 5 2 6 2 6 2 | 6 2 5 2 5 2 |
| | 5 2 5 2 | 6 2 5 2 5 2 | 5 2 6 2 6 2 |
| | 6 2 6 2 | 5 2 6 2 6 2 | 6 2 6 2 5 2 |
| | 6 2 5 2 | 6 2 5 2 5 2 | 5 2 6 2 6 2 |
| | 5 2 6 2 | 6 2 5 2 6 2 | 5 2 6 2 5 2 |
| | 5 2 5 2 | 5 2 6 2 5 2 | 5 2 5 2 5 2 |
| | 6 2 5 2 | 6 2 5 2 5 2 | 6 2 5 2 5 2 |
| | 5 2 6 2 | 5 2 6 2 6 2 | 6 2 6 2 5 2 |
| | 6 2 5 2 | 5 2 6 2 6 2 | 5 2 5 2 5 2 |
| | 6 2 6 2 | 6 2 5 2 6 2 | 5 2 6 2 5 2 |
| | 6 2 5 2 | 6 2 6 2 5 2 | 5 2 6 2 6 2 |
| | 5 2 6 2 | 5 2 5 2 5 2 | 5 2 5 2 6 2 |
| | 5 2 6 2 | 6 2 6 2 5 2 | 6 2 5 2 5 2 |
| | 5 2 6 2 | 5 2 6 2 6 2 | 6 2 6 2 5 2 |
| | 6 2 5 2 | 6 2 6 2 5 2 | 6 2 5 2 6 2 |
| | 6 2 6 2 | 6 2 5 2 6 2 | 5 2 6 2 6 2 |
| | 5 2 5 2 | 5 2 5 2 6 2 | 5 2 6 2 6 2 |
| | 5 2 6 2 | 6 2 6 2 6 2 | 5 2 5 2 5 2 |
| | 5 2 6 2 | 5 2 6 2 5 2 | 5 2 6 2 6 2 |
| | 6 2 5 2 | 6 2 5 2 5 2 | 5 2 5 2 5 2 |
| | 6 2 5 2 | 6 2 6 2 6 2 | 5 2 6 2 5 2 |
| | 5 2 5 2 | 5 2 5 2 5 2 | 6 2 6 2 6 2 |
| | 6 2 6 2 | 5 2 6 2 6 2 | 5 2 5 2 5 2 |
| | 5 2 5 2 | 6 2 5 2 5 2 | 6 2 5 2 5 2 |
| | 5 2 6 2 | 5 2 6 2 5 2 | 5 2 6 2 6 2 |
| | 5 2 5 2 | 5 2 5 2 5 2 | 6 2 5 2 5 2 |
| | 6 2 5 2 | 5 2 6 2 5 2 | 6 2 6 2 6 2 |
| | 6 2 6 2 | 6 2 5 2 6 2 | 6 2 5 2 5 2 |

Continued on next column

| TQC | ICSD | NoE | SF |
|---|---|---|---|
| | 5 2 5 2 | 5 2 5 2 6 2 | 5 2 5 2 6 2 |
| | 6 2 6 2 | 6 2 6 2 6 2 | 6 2 5 2 5 2 |
| | 5 2 5 2 | 5 2 5 2 5 2 | 5 2 6 2 5 2 |
| | 6 2 5 2 | 5 2 5 2 5 2 | 6 2 5 2 6 2 |
| | 6 2 6 2 | 6 2 5 2 6 2 | 6 2 6 2 5 2 |
| | 5 2 5 2 | 6 2 5 2 5 2 | 5 2 6 2 6 2 |
| | 5 2 5 2 | 5 2 6 2 5 2 | 6 2 5 2 5 2 |
| | 6 2 6 2 | 5 2 6 2 5 2 | 5 2 5 2 5 2 |
| | 5 2 6 2 | 6 2 6 2 6 2 | 5 2 5 2 5 2 |
| | 5 2 6 2 | 6 2 6 2 6 2 | 6 2 5 2 6 2 |
| | 6 2 6 2 | 5 2 5 2 6 2 | 5 2 6 2 6 2 |
| | 5 2 6 2 | 6 2 5 2 5 2 | 5 2 6 2 5 2 |
| | 5 2 5 2 | 6 2 6 2 5 2 | 5 2 5 2 6 2 |
| | 6 2 5 2 | 5 2 5 2 6 2 | 5 2 6 2 5 2 |
| | 5 2 6 2 | 6 2 6 2 6 2 | 6 2 6 2 6 2 |
| | 6 2 6 2 | 6 2 5 2 5 2 | 5 2 5 2 6 2 |
| | 6 2 5 2 | 5 2 6 2 5 2 | 5 2 6 2 5 2 |
| | 5 2 5 2 | 5 2 5 2 5 2 | 6 2 5 2 6 2 |
| | 6 2 6 2 | 5 2 6 2 5 2 | 5 2 6 2 6 2 |
| | 5 2 5 2 | 6 2 5 2 5 2 | 5 2 5 2 5 2 |
| 47.2.3 | 77745 | 106 | Ba2Cu3O7Y1 |
| | -GM -R | -S -T -U | -X -Y -Z |
| | 5 2 6 2 | 5 2 5 2 5 2 | 6 2 6 2 6 2 |
| | 5 2 5 2 | 5 2 6 2 6 2 | 6 2 6 2 5 2 |
| | 6 2 6 2 | 6 2 5 2 5 2 | 5 2 5 2 6 2 |
| | 6 2 5 2 | 6 2 6 2 6 2 | 5 2 5 2 5 2 |
| | 6 2 5 2 | 6 2 6 2 6 2 | 5 2 5 2 5 2 |
| | 6 2 5 2 | 6 2 6 2 6 2 | 5 2 5 2 5 2 |
| | 5 2 6 2 | 5 2 5 2 5 2 | 6 2 6 2 6 2 |
| | 6 2 6 2 | 5 2 5 2 5 2 | 6 2 6 2 5 2 |
| | 5 2 5 2 | 6 2 6 2 5 2 | 5 2 5 2 5 2 |
| | 5 2 5 2 | 6 2 5 2 6 2 | 5 2 5 2 6 2 |
| | 6 2 6 2 | 5 2 6 2 6 2 | 5 2 5 2 5 2 |
| | 6 2 5 2 | 6 2 6 2 6 2 | 6 2 6 2 6 2 |
| | 5 2 6 2 | 6 2 6 2 5 2 | 5 2 6 2 5 2 |
| | 6 2 5 2 | 5 2 5 2 6 2 | 6 2 5 2 6 2 |
| | 5 2 6 2 | 6 2 6 2 5 2 | 5 2 6 2 5 2 |
| | 6 2 5 2 | 5 2 5 2 6 2 | 6 2 5 2 6 2 |
| | 5 2 6 2 | 6 2 6 2 5 2 | 5 2 6 2 5 2 |
| | 6 2 5 2 | 5 2 5 2 6 2 | 6 2 5 2 6 2 |
| | 5 2 6 2 | 6 2 6 2 5 2 | 5 2 6 2 5 2 |
| | 5 2 6 2 | 5 2 5 2 6 2 | 5 2 5 2 6 2 |
| | 5 2 5 2 | 6 2 6 2 5 2 | 6 2 5 2 6 2 |
| | 5 2 5 2 | 5 2 5 2 5 2 | 5 2 6 2 5 2 |
| | 5 2 6 2 | 5 2 5 2 5 2 | 6 2 5 2 6 2 |
| | 6 2 6 2 | 5 2 5 2 5 2 | 6 2 6 2 5 2 |
| | 6 2 5 2 | 6 2 5 2 5 2 | 6 2 5 2 5 2 |
| | 5 2 5 2 | 5 2 5 2 6 2 | 5 2 6 2 6 2 |
| | 6 2 5 2 | 6 2 5 2 6 2 | 5 2 6 2 6 2 |
| | 6 2 5 2 | 5 2 6 2 6 2 | 5 2 5 2 6 2 |
| | 5 2 5 2 | 6 2 6 2 6 2 | 6 2 5 2 5 2 |
| | 5 2 5 2 | 6 2 6 2 6 2 | 6 2 5 2 5 2 |
| | 6 2 6 2 | 5 2 5 2 5 2 | 5 2 5 2 5 2 |
| | 6 2 5 2 | 5 2 6 2 6 2 | 5 2 6 2 5 2 |
| | 5 2 6 2 | 6 2 5 2 5 2 | 6 2 5 2 6 2 |
| | 5 2 6 2 | 5 2 6 2 5 2 | 5 2 6 2 5 2 |
| | 6 2 5 2 | 5 2 5 2 5 2 | 6 2 5 2 5 2 |
| | 5 2 5 2 | 5 2 5 2 6 2 | 5 2 6 2 6 2 |
| | 5 2 6 2 | 5 2 6 2 5 2 | 6 2 6 2 6 2 |
| | 5 2 6 2 | 5 2 5 2 6 2 | 5 2 5 2 5 2 |
| | 6 2 5 2 | 6 2 6 2 5 2 | 5 2 5 2 5 2 |

Continued on next column

**TABLE XXVIII– continued**

| TQC | ICSD | NoE | SF |
|---|---|---|---|
| | 6 2 6 2 | 6 2 5 2 6 2 | 6 2 5 2 5 2 |
| | 6 2 5 2 | 5 2 6 2 5 2 | 5 2 6 2 6 2 |
| | 5 2 6 2 | 5 2 5 2 6 2 | 5 2 5 2 6 2 |
| | 5 2 5 2 | 6 2 6 2 5 2 | 6 2 6 2 6 2 |
| | 6 2 6 2 | 6 2 5 2 6 2 | 6 2 6 2 5 2 |
| | 5 2 5 2 | 6 2 5 2 5 2 | 6 2 6 2 6 2 |
| | 5 2 5 2 | 5 2 5 2 5 2 | 6 2 5 2 6 2 |
| | 6 2 6 2 | 6 2 6 2 6 2 | 6 2 5 2 5 2 |
| | 6 2 6 2 | 6 2 6 2 5 2 | 5 2 6 2 5 2 |
| | 6 2 5 2 | 5 2 6 2 6 2 | 5 2 5 2 5 2 |
| | 6 2 5 2 | 5 2 5 2 5 2 | 5 2 5 2 6 2 |
| | 5 2 5 2 | 5 2 6 2 6 2 | 5 2 5 2 6 2 |
| | 6 2 5 2 | 5 2 5 2 6 2 | 6 2 5 2 5 2 |
| | 6 2 5 2 | 5 2 5 2 6 2 | 5 2 6 2 6 2 |
| 47.2.4 | 63335 | 106 | Ba2Cu3O7Y1 |
| | -GM -R | -S -T -U | -X -Y -Z |
| | 5 2 6 2 | 5 2 5 2 5 2 | 6 2 6 2 6 2 |
| | 5 2 5 2 | 5 2 6 2 6 2 | 6 2 6 2 5 2 |
| | 6 2 6 2 | 6 2 5 2 5 2 | 5 2 5 2 6 2 |
| | 6 2 5 2 | 6 2 6 2 6 2 | 5 2 5 2 5 2 |
| | 6 2 5 2 | 6 2 6 2 6 2 | 5 2 5 2 5 2 |
| | 6 2 5 2 | 6 2 6 2 6 2 | 5 2 5 2 5 2 |
| | 5 2 6 2 | 5 2 5 2 5 2 | 6 2 6 2 6 2 |
| | 6 2 6 2 | 5 2 5 2 5 2 | 6 2 6 2 5 2 |
| | 5 2 5 2 | 6 2 6 2 5 2 | 5 2 5 2 5 2 |
| | 5 2 5 2 | 5 2 5 2 6 2 | 5 2 5 2 6 2 |
| | 6 2 5 2 | 6 2 6 2 6 2 | 5 2 5 2 5 2 |
| | 6 2 6 2 | 6 2 6 2 6 2 | 6 2 6 2 6 2 |
| | 5 2 6 2 | 6 2 6 2 5 2 | 5 2 6 2 5 2 |
| | 6 2 5 2 | 5 2 5 2 6 2 | 6 2 5 2 6 2 |
| | 5 2 6 2 | 6 2 6 2 5 2 | 5 2 6 2 5 2 |
| | 6 2 5 2 | 5 2 5 2 6 2 | 6 2 5 2 6 2 |
| | 5 2 6 2 | 6 2 6 2 5 2 | 5 2 6 2 5 2 |
| | 6 2 5 2 | 5 2 5 2 6 2 | 6 2 5 2 6 2 |
| | 5 2 6 2 | 6 2 6 2 5 2 | 5 2 6 2 5 2 |
| | 5 2 6 2 | 5 2 6 2 6 2 | 5 2 5 2 6 2 |
| | 5 2 5 2 | 6 2 5 2 5 2 | 6 2 6 2 5 2 |
| | 5 2 5 2 | 5 2 5 2 5 2 | 5 2 5 2 6 2 |
| | 5 2 6 2 | 5 2 5 2 5 2 | 6 2 5 2 6 2 |
| | 6 2 6 2 | 5 2 5 2 5 2 | 6 2 5 2 5 2 |
| | 6 2 5 2 | 6 2 5 2 5 2 | 6 2 6 2 5 2 |
| | 5 2 5 2 | 5 2 5 2 6 2 | 5 2 6 2 6 2 |
| | 6 2 5 2 | 6 2 5 2 6 2 | 5 2 6 2 6 2 |
| | 5 2 5 2 | 5 2 6 2 5 2 | 5 2 5 2 5 2 |
| | 6 2 5 2 | 6 2 6 2 6 2 | 6 2 5 2 6 2 |
| | 5 2 5 2 | 6 2 6 2 6 2 | 5 2 5 2 5 2 |
| | 6 2 6 2 | 5 2 5 2 6 2 | 6 2 5 2 5 2 |
| | 6 2 5 2 | 5 2 6 2 5 2 | 5 2 6 2 5 2 |
| | 5 2 6 2 | 5 2 6 2 6 2 | 6 2 5 2 6 2 |
| | 5 2 6 2 | 6 2 5 2 5 2 | 5 2 6 2 5 2 |
| | 5 2 5 2 | 5 2 5 2 5 2 | 6 2 5 2 6 2 |
| | 6 2 5 2 | 5 2 5 2 6 2 | 6 2 6 2 5 2 |
| | 5 2 6 2 | 5 2 6 2 5 2 | 5 2 6 2 5 2 |
| | 5 2 6 2 | 5 2 5 2 6 2 | 5 2 5 2 6 2 |
| | 6 2 5 2 | 6 2 6 2 5 2 | 5 2 5 2 5 2 |
| | 6 2 6 2 | 5 2 5 2 6 2 | 6 2 5 2 5 2 |
| | 6 2 6 2 | 6 2 6 2 5 2 | 5 2 6 2 6 2 |
| | 5 2 5 2 | 6 2 6 2 6 2 | 5 2 5 2 6 2 |
| | 5 2 5 2 | 5 2 5 2 5 2 | 6 2 6 2 5 2 |
| | 6 2 6 2 | 6 2 5 2 6 2 | 6 2 6 2 6 2 |
| | 5 2 5 2 | 6 2 5 2 5 2 | 6 2 6 2 6 2 |

Continued on next column

**TABLE XXVIII– continued**

| TQC | ICSD | NoE | SF |
|---|---|---|---|
| | 5 2 5 2 | 5 2 6 2 5 2 | 6 2 5 2 6 2 |
| | 6 2 6 2 | 6 2 5 2 6 2 | 6 2 5 2 5 2 |
| | 6 2 6 2 | 6 2 6 2 5 2 | 5 2 6 2 5 2 |
| | 6 2 5 2 | 5 2 6 2 6 2 | 5 2 5 2 5 2 |
| | 6 2 5 2 | 5 2 5 2 5 2 | 5 2 5 2 6 2 |
| | 5 2 5 2 | 5 2 6 2 6 2 | 5 2 5 2 6 2 |
| | 6 2 5 2 | 5 2 5 2 6 2 | 6 2 5 2 5 2 |
| | 6 2 6 2 | 5 2 5 2 6 2 | 5 2 6 2 6 2 |
| 47.2.5 | 262809 | 106 | Ba2Cu3O6.93Y1 |
| | -GM -R | -S -T -U | -X -Y -Z |
| | 5 2 6 2 | 5 2 5 2 5 2 | 6 2 6 2 6 2 |
| | 5 2 5 2 | 5 2 6 2 6 2 | 6 2 6 2 5 2 |
| | 6 2 6 2 | 6 2 5 2 5 2 | 5 2 5 2 6 2 |
| | 6 2 5 2 | 6 2 6 2 6 2 | 5 2 5 2 5 2 |
| | 6 2 5 2 | 6 2 6 2 6 2 | 5 2 5 2 5 2 |
| | 6 2 5 2 | 6 2 6 2 6 2 | 5 2 5 2 5 2 |
| | 5 2 6 2 | 5 2 5 2 5 2 | 6 2 6 2 6 2 |
| | 5 2 6 2 | 5 2 5 2 5 2 | 5 2 6 2 5 2 |
| | 6 2 5 2 | 6 2 6 2 5 2 | 6 2 5 2 5 2 |
| | 5 2 6 2 | 6 2 5 2 6 2 | 5 2 5 2 6 2 |
| | 6 2 5 2 | 5 2 6 2 6 2 | 5 2 5 2 5 2 |
| | 6 2 5 2 | 6 2 6 2 6 2 | 6 2 6 2 6 2 |
| | 5 2 6 2 | 6 2 6 2 5 2 | 5 2 6 2 5 2 |
| | 6 2 5 2 | 5 2 5 2 6 2 | 6 2 5 2 6 2 |
| | 5 2 6 2 | 6 2 6 2 5 2 | 5 2 6 2 5 2 |
| | 6 2 5 2 | 5 2 5 2 6 2 | 6 2 5 2 6 2 |
| | 5 2 6 2 | 6 2 6 2 5 2 | 5 2 6 2 5 2 |
| | 6 2 5 2 | 5 2 5 2 6 2 | 6 2 5 2 6 2 |
| | 5 2 6 2 | 6 2 6 2 5 2 | 5 2 6 2 5 2 |
| | 5 2 6 2 | 5 2 5 2 6 2 | 5 2 5 2 6 2 |
| | 5 2 5 2 | 6 2 6 2 5 2 | 6 2 5 2 5 2 |
| | 5 2 5 2 | 5 2 5 2 5 2 | 5 2 6 2 6 2 |
| | 5 2 6 2 | 5 2 5 2 5 2 | 6 2 5 2 6 2 |
| | 6 2 6 2 | 5 2 5 2 5 2 | 6 2 5 2 5 2 |
| | 6 2 5 2 | 6 2 5 2 5 2 | 6 2 6 2 5 2 |
| | 5 2 5 2 | 5 2 5 2 6 2 | 5 2 6 2 6 2 |
| | 6 2 5 2 | 6 2 5 2 6 2 | 5 2 6 2 6 2 |
| | 5 2 5 2 | 5 2 6 2 5 2 | 5 2 5 2 5 2 |
| | 6 2 5 2 | 6 2 6 2 6 2 | 6 2 5 2 6 2 |
| | 5 2 5 2 | 6 2 5 2 6 2 | 6 2 5 2 5 2 |
| | 5 2 5 2 | 5 2 6 2 6 2 | 5 2 5 2 5 2 |
| | 6 2 6 2 | 5 2 5 2 5 2 | 5 2 6 2 6 2 |
| | 5 2 6 2 | 5 2 6 2 6 2 | 6 2 5 2 5 2 |
| | 6 2 6 2 | 6 2 6 2 5 2 | 5 2 5 2 5 2 |
| | 5 2 5 2 | 5 2 5 2 5 2 | 5 2 6 2 5 2 |
| | 5 2 5 2 | 5 2 5 2 5 2 | 6 2 6 2 6 2 |
| | 6 2 6 2 | 5 2 6 2 6 2 | 6 2 5 2 5 2 |
| | 6 2 6 2 | 5 2 5 2 5 2 | 5 2 6 2 6 2 |
| | 5 2 5 2 | 6 2 6 2 6 2 | 5 2 5 2 5 2 |
| | 6 2 6 2 | 5 2 5 2 6 2 | 6 2 5 2 5 2 |
| | 6 2 6 2 | 6 2 6 2 5 2 | 5 2 6 2 6 2 |
| | 5 2 5 2 | 6 2 6 2 6 2 | 6 2 6 2 6 2 |
| | 6 2 5 2 | 5 2 5 2 5 2 | 5 2 5 2 6 2 |
| | 5 2 6 2 | 6 2 5 2 6 2 | 6 2 6 2 5 2 |
| | 5 2 5 2 | 6 2 5 2 5 2 | 6 2 6 2 6 2 |
| | 5 2 5 2 | 5 2 5 2 5 2 | 6 2 5 2 6 2 |
| | 6 2 6 2 | 6 2 6 2 5 2 | 5 2 5 2 5 2 |
| | 6 2 5 2 | 6 2 6 2 6 2 | 6 2 5 2 5 2 |
| | 6 2 6 2 | 5 2 5 2 5 2 | 5 2 6 2 5 2 |
| | 6 2 5 2 | 5 2 6 2 6 2 | 5 2 5 2 6 2 |
| | 5 2 5 2 | 5 2 6 2 6 2 | 5 2 5 2 6 2 |

Continued on next column

**TABLE XXVIII– continued**

| TQC | ICSD | NoE | SF |
|---|---|---|---|
| | 6 2 5 2 | 5 2 5 2 6 2 | 6 2 5 2 5 2 |
| | 6 2 5 2 | 5 2 5 2 6 2 | 5 2 5 2 6 2 |
| 47.2.6 | 73896 | 206 | Ba2Cu3O6.5Y1 |
| | -GM -R | -S -T -U | -X -Y -Z |
| | 5 2 5 2 | 6 2 6 2 5 2 | 6 2 5 2 6 2 |
| | 6 2 6 2 | 5 2 5 2 6 2 | 5 2 6 2 5 2 |
| | 5 2 5 2 | 5 2 6 2 6 2 | 6 2 6 2 5 2 |
| | 6 2 6 2 | 6 2 5 2 5 2 | 5 2 5 2 6 2 |
| | 6 2 6 2 | 6 2 5 2 5 2 | 5 2 5 2 6 2 |
| | 5 2 5 2 | 5 2 6 2 6 2 | 6 2 6 2 5 2 |
| | 5 2 5 2 | 6 2 6 2 5 2 | 6 2 5 2 6 2 |
| | 6 2 6 2 | 5 2 5 2 6 2 | 5 2 6 2 5 2 |
| | 5 2 5 2 | 6 2 6 2 5 2 | 6 2 5 2 6 2 |
| | 6 2 6 2 | 5 2 5 2 6 2 | 5 2 6 2 5 2 |
| | 5 2 5 2 | 6 2 6 2 5 2 | 6 2 5 2 6 2 |
| | 6 2 6 2 | 5 2 5 2 6 2 | 5 2 6 2 5 2 |
| | 5 2 6 2 | 5 2 5 2 5 2 | 6 2 6 2 6 2 |
| | 6 2 5 2 | 6 2 6 2 5 2 | 6 2 5 2 5 2 |
| | 6 2 6 2 | 5 2 6 2 6 2 | 5 2 6 2 5 2 |
| | 5 2 5 2 | 5 2 5 2 6 2 | 5 2 5 2 5 2 |
| | 5 2 5 2 | 6 2 6 2 5 2 | 5 2 6 2 6 2 |
| | 5 2 6 2 | 5 2 5 2 6 2 | 6 2 5 2 5 2 |
| | 6 2 5 2 | 6 2 5 2 5 2 | 5 2 6 2 5 2 |
| | 6 2 6 2 | 6 2 6 2 6 2 | 5 2 5 2 6 2 |
| | 5 2 5 2 | 6 2 5 2 6 2 | 6 2 6 2 6 2 |
| | 6 2 6 2 | 5 2 6 2 5 2 | 6 2 5 2 5 2 |
| | 5 2 5 2 | 5 2 6 2 6 2 | 5 2 6 2 6 2 |
| | 6 2 6 2 | 5 2 5 2 6 2 | 6 2 5 2 6 2 |
| | 5 2 5 2 | 6 2 6 2 5 2 | 6 2 6 2 5 2 |
| | 5 2 6 2 | 6 2 5 2 5 2 | 5 2 5 2 6 2 |
| | 6 2 5 2 | 5 2 6 2 6 2 | 6 2 6 2 5 2 |
| | 6 2 6 2 | 6 2 5 2 5 2 | 5 2 5 2 6 2 |
| | 5 2 5 2 | 5 2 6 2 6 2 | 6 2 6 2 5 2 |
| | 6 2 6 2 | 6 2 5 2 5 2 | 5 2 5 2 6 2 |
| | 5 2 5 2 | 5 2 6 2 6 2 | 6 2 6 2 5 2 |
| | 6 2 6 2 | 6 2 5 2 5 2 | 5 2 5 2 6 2 |
| | 5 2 5 2 | 5 2 6 2 6 2 | 6 2 6 2 5 2 |
| | 6 2 6 2 | 6 2 5 2 5 2 | 5 2 5 2 6 2 |
| | 5 2 5 2 | 5 2 6 2 6 2 | 6 2 6 2 5 2 |
| | 6 2 6 2 | 6 2 5 2 5 2 | 5 2 5 2 6 2 |
| | 5 2 5 2 | 5 2 6 2 6 2 | 6 2 6 2 5 2 |
| | 5 2 5 2 | 6 2 5 2 6 2 | 5 2 6 2 6 2 |
| | 6 2 6 2 | 5 2 6 2 5 2 | 6 2 5 2 5 2 |
| | 5 2 6 2 | 5 2 6 2 5 2 | 5 2 5 2 6 2 |
| | 5 2 5 2 | 6 2 5 2 6 2 | 5 2 6 2 6 2 |
| | 5 2 6 2 | 6 2 5 2 5 2 | 6 2 5 2 6 2 |
| | 5 2 5 2 | 6 2 6 2 6 2 | 5 2 5 2 5 2 |
| | 6 2 6 2 | 5 2 5 2 6 2 | 5 2 6 2 5 2 |
| | 5 2 6 2 | 5 2 6 2 5 2 | 6 2 6 2 5 2 |
| | 6 2 5 2 | 6 2 6 2 5 2 | 5 2 5 2 5 2 |
| | 6 2 5 2 | 5 2 5 2 5 2 | 5 2 6 2 5 2 |
| | 6 2 6 2 | 5 2 5 2 5 2 | 6 2 6 2 5 2 |
| | 6 2 5 2 | 6 2 6 2 5 2 | 6 2 5 2 5 2 |
| | 5 2 6 2 | 6 2 5 2 5 2 | 6 2 5 2 6 2 |
| | 5 2 6 2 | 5 2 6 2 5 2 | 6 2 6 2 6 2 |
| | 5 2 5 2 | 6 2 5 2 6 2 | 5 2 5 2 6 2 |
| | 6 2 6 2 | 5 2 5 2 5 2 | 5 2 5 2 6 2 |
| | 5 2 5 2 | 6 2 6 2 6 2 | 5 2 6 2 6 2 |
| | 6 2 6 2 | 6 2 5 2 5 2 | 6 2 5 2 6 2 |
| | 5 2 5 2 | 5 2 6 2 6 2 | 6 2 5 2 5 2 |
| | 5 2 6 2 | 6 2 5 2 5 2 | 5 2 6 2 6 2 |
| | 6 2 5 2 | 5 2 5 2 6 2 | 6 2 6 2 5 2 |
| | 6 2 6 2 | 6 2 6 2 5 2 | 5 2 6 2 6 2 |
| | 5 2 6 2 | 6 2 5 2 6 2 | 5 2 5 2 5 2 |
| | 5 2 5 2 | 5 2 6 2 5 2 | 5 2 6 2 5 2 |
| | 6 2 5 2 | 6 2 5 2 5 2 | 6 2 5 2 5 2 |
| | 5 2 5 2 | 5 2 6 2 6 2 | 6 2 5 2 5 2 |
| | 6 2 6 2 | 5 2 6 2 6 2 | 5 2 6 2 5 2 |
| | 6 2 6 2 | 6 2 5 2 5 2 | 5 2 6 2 5 2 |
| | 6 2 5 2 | 6 2 6 2 6 2 | 5 2 6 2 6 2 |
| | 5 2 6 2 | 5 2 5 2 5 2 | 5 2 5 2 6 2 |
| | 5 2 6 2 | 6 2 6 2 5 2 | 6 2 5 2 5 2 |
| | 5 2 6 2 | 5 2 6 2 6 2 | 6 2 6 2 6 2 |
| | 6 2 5 2 | 6 2 6 2 6 2 | 6 2 5 2 5 2 |
| | 6 2 6 2 | 6 2 5 2 5 2 | 5 2 6 2 6 2 |
| | 5 2 6 2 | 5 2 5 2 6 2 | 5 2 6 2 6 2 |
| | 5 2 5 2 | 6 2 6 2 6 2 | 5 2 5 2 5 2 |
| | 5 2 6 2 | 5 2 6 2 5 2 | 5 2 6 2 6 2 |
| | 6 2 5 2 | 6 2 5 2 5 2 | 5 2 5 2 5 2 |
| | 6 2 5 2 | 5 2 5 2 6 2 | 6 2 6 2 5 2 |
| | 5 2 5 2 | 6 2 6 2 5 2 | 5 2 6 2 6 2 |
| | 6 2 5 2 | 5 2 6 2 6 2 | 5 2 5 2 5 2 |
| | 5 2 6 2 | 6 2 5 2 5 2 | 5 2 5 2 5 2 |
| | 5 2 6 2 | 5 2 6 2 5 2 | 6 2 6 2 6 2 |
| | 5 2 5 2 | 5 2 5 2 6 2 | 6 2 5 2 5 2 |
| | 6 2 5 2 | 5 2 5 2 5 2 | 6 2 5 2 6 2 |
| | 5 2 5 2 | 6 2 6 2 5 2 | 6 2 6 2 6 2 |
| | 6 2 6 2 | 5 2 5 2 6 2 | 5 2 5 2 5 2 |
| | 6 2 5 2 | 6 2 6 2 6 2 | 6 2 5 2 5 2 |
| | 6 2 6 2 | 5 2 5 2 5 2 | 5 2 6 2 5 2 |
| | 5 2 6 2 | 5 2 5 2 5 2 | 6 2 5 2 6 2 |
| | 5 2 5 2 | 6 2 5 2 6 2 | 6 2 6 2 5 2 |
| | 6 2 5 2 | 6 2 5 2 5 2 | 6 2 6 2 5 2 |
| | 5 2 6 2 | 5 2 6 2 5 2 | 5 2 5 2 6 2 |
| | 5 2 6 2 | 5 2 6 2 5 2 | 5 2 5 2 5 2 |
| | 6 2 5 2 | 6 2 6 2 6 2 | 5 2 5 2 6 2 |
| | 5 2 6 2 | 6 2 6 2 6 2 | 6 2 5 2 5 2 |
| | 6 2 6 2 | 5 2 5 2 6 2 | 5 2 6 2 6 2 |
| | 5 2 6 2 | 6 2 5 2 5 2 | 5 2 6 2 5 2 |
| | 6 2 5 2 | 6 2 6 2 5 2 | 6 2 5 2 6 2 |
| | 5 2 5 2 | 5 2 5 2 6 2 | 5 2 6 2 5 2 |
| | 6 2 6 2 | 6 2 6 2 6 2 | 5 2 6 2 6 2 |
| | 5 2 6 2 | 6 2 5 2 5 2 | 5 2 6 2 6 2 |
| | 6 2 5 2 | 5 2 6 2 5 2 | 5 2 5 2 5 2 |
| | 5 2 5 2 | 5 2 5 2 5 2 | 6 2 5 2 6 2 |
| | 6 2 6 2 | 5 2 6 2 5 2 | 5 2 6 2 6 2 |
| | 5 2 5 2 | 6 2 5 2 5 2 | 5 2 5 2 5 2 |
| 47.2.7 | 68044 | 116 | Ba2Cu3Ho1O7 |
| | -GM -R | -S -T -U | -X -Y -Z |
| | 5 2 5 2 | 6 2 5 2 6 2 | 5 2 6 2 6 2 |
| | 5 2 6 2 | 6 2 6 2 5 2 | 5 2 6 2 5 2 |
| | 6 2 5 2 | 5 2 5 2 6 2 | 6 2 5 2 6 2 |
| | 6 2 6 2 | 5 2 6 2 5 2 | 6 2 5 2 5 2 |
| | 6 2 6 2 | 5 2 6 2 5 2 | 6 2 5 2 5 2 |
| | 6 2 6 2 | 5 2 6 2 5 2 | 6 2 5 2 5 2 |
| | 5 2 5 2 | 6 2 5 2 6 2 | 5 2 6 2 6 2 |
| | 6 2 5 2 | 6 2 5 2 6 2 | 5 2 6 2 5 2 |
| | 5 2 6 2 | 5 2 6 2 6 2 | 6 2 5 2 6 2 |
| | 5 2 6 2 | 5 2 6 2 5 2 | 6 2 5 2 5 2 |
| | 6 2 5 2 | 5 2 5 2 5 2 | 6 2 5 2 5 2 |
| | 6 2 6 2 | 6 2 6 2 5 2 | 5 2 6 2 6 2 |
| | 5 2 5 2 | 5 2 6 2 6 2 | 6 2 6 2 5 2 |

Continued on next column

**TABLE XXVIII– continued**

| TQC | ICSD | NoE | SF |
|---|---|---|---|
| | 6 2 6 2 | 6 2 5 2 5 2 | 5 2 5 2 6 2 |
| | 5 2 5 2 | 5 2 6 2 6 2 | 6 2 6 2 5 2 |
| | 6 2 6 2 | 6 2 5 2 5 2 | 5 2 5 2 6 2 |
| | 5 2 5 2 | 5 2 6 2 6 2 | 6 2 6 2 5 2 |
| | 6 2 6 2 | 6 2 5 2 5 2 | 5 2 5 2 6 2 |
| | 5 2 5 2 | 5 2 6 2 6 2 | 6 2 6 2 5 2 |
| | 5 2 5 2 | 6 2 6 2 5 2 | 6 2 5 2 6 2 |
| | 5 2 6 2 | 5 2 5 2 6 2 | 5 2 6 2 6 2 |
| | 5 2 6 2 | 6 2 5 2 6 2 | 5 2 5 2 6 2 |
| | 6 2 5 2 | 6 2 5 2 6 2 | 6 2 5 2 5 2 |
| | 6 2 5 2 | 6 2 5 2 6 2 | 5 2 6 2 5 2 |
| | 5 2 6 2 | 5 2 5 2 6 2 | 5 2 6 2 5 2 |
| | 5 2 6 2 | 5 2 5 2 5 2 | 6 2 6 2 6 2 |
| | 6 2 6 2 | 6 2 6 2 5 2 | 6 2 5 2 6 2 |
| | 6 2 6 2 | 5 2 5 2 5 2 | 6 2 5 2 6 2 |
| | 5 2 6 2 | 6 2 6 2 5 2 | 5 2 5 2 5 2 |
| | 6 2 5 2 | 5 2 6 2 5 2 | 5 2 5 2 5 2 |
| | 5 2 6 2 | 6 2 6 2 5 2 | 6 2 6 2 5 2 |
| | 6 2 5 2 | 6 2 6 2 6 2 | 5 2 6 2 5 2 |
| | 5 2 5 2 | 6 2 5 2 6 2 | 6 2 5 2 6 2 |
| | 6 2 6 2 | 5 2 5 2 6 2 | 6 2 5 2 5 2 |
| | 5 2 6 2 | 6 2 5 2 6 2 | 5 2 5 2 6 2 |
| | 6 2 6 2 | 6 2 5 2 5 2 | 6 2 6 2 5 2 |
| | 5 2 6 2 | 5 2 6 2 6 2 | 6 2 6 2 5 2 |
| | 5 2 5 2 | 6 2 6 2 5 2 | 5 2 5 2 6 2 |
| | 6 2 5 2 | 6 2 5 2 5 2 | 5 2 5 2 5 2 |
| | 5 2 6 2 | 5 2 5 2 6 2 | 6 2 5 2 5 2 |
| | 5 2 6 2 | 5 2 6 2 6 2 | 6 2 6 2 6 2 |
| | 6 2 5 2 | 6 2 5 2 5 2 | 6 2 5 2 5 2 |
| | 5 2 6 2 | 5 2 6 2 6 2 | 5 2 6 2 6 2 |
| | 6 2 5 2 | 6 2 5 2 5 2 | 5 2 6 2 6 2 |
| | 5 2 5 2 | 5 2 5 2 6 2 | 5 2 6 2 5 2 |
| | 6 2 6 2 | 5 2 6 2 6 2 | 6 2 5 2 6 2 |
| | 6 2 6 2 | 6 2 6 2 5 2 | 5 2 5 2 5 2 |
| | 5 2 6 2 | 5 2 6 2 5 2 | 6 2 5 2 6 2 |
| | 6 2 5 2 | 5 2 5 2 5 2 | 6 2 5 2 5 2 |
| | 6 2 6 2 | 5 2 6 2 6 2 | 6 2 5 2 5 2 |
| | 6 2 5 2 | 5 2 6 2 5 2 | 5 2 6 2 5 2 |
| | 6 2 6 2 | 6 2 5 2 5 2 | 6 2 5 2 5 2 |
| | 6 2 6 2 | 6 2 6 2 6 2 | 6 2 5 2 6 2 |
| | 5 2 6 2 | 6 2 6 2 5 2 | 6 2 5 2 6 2 |
| | 6 2 6 2 | 5 2 5 2 6 2 | 5 2 5 2 5 2 |
| | 6 2 6 2 | 5 2 5 2 5 2 | 6 2 6 2 6 2 |
| | 5 2 6 2 | 6 2 5 2 5 2 | 6 2 5 2 5 2 |
| | 5 2 6 2 | 5 2 6 2 5 2 | 6 2 5 2 5 2 |
| 47.2.8 | 94046 | 160 Ba2Co4O11Tb2 | |
| | -GM -R | -S -T -U | -X -Y -Z |
| | 5 2 6 2 | 5 2 5 2 5 2 | 6 2 6 2 6 2 |
| | 6 2 5 2 | 6 2 6 2 6 2 | 5 2 5 2 5 2 |
| | 5 2 6 2 | 6 2 5 2 6 2 | 6 2 5 2 5 2 |
| | 6 2 5 2 | 5 2 6 2 5 2 | 5 2 6 2 6 2 |
| | 5 2 5 2 | 5 2 5 2 5 2 | 6 2 6 2 6 2 |
| | 6 2 6 2 | 6 2 6 2 6 2 | 5 2 5 2 5 2 |
| | 5 2 5 2 | 6 2 5 2 6 2 | 5 2 6 2 6 2 |
| | 6 2 6 2 | 5 2 6 2 5 2 | 6 2 5 2 5 2 |
| | 5 2 5 2 | 5 2 5 2 6 2 | 6 2 6 2 6 2 |
| | 6 2 6 2 | 6 2 6 2 5 2 | 5 2 5 2 5 2 |
| | 5 2 5 2 | 5 2 5 2 5 2 | 5 2 6 2 6 2 |
| | 5 2 6 2 | 6 2 6 2 6 2 | 6 2 5 2 5 2 |
| | 6 2 5 2 | 5 2 6 2 6 2 | 5 2 5 2 5 2 |
| | 6 2 5 2 | 5 2 5 2 5 2 | 5 2 5 2 6 2 |

Continued on next column

**TABLE XXVIII– continued**

| TQC | ICSD | NoE | SF |
|---|---|---|---|
| | 5 2 6 2 | 6 2 5 2 6 2 | 6 2 5 2 5 2 |
| | 5 2 6 2 | 5 2 5 2 6 2 | 5 2 6 2 6 2 |
| | 5 2 5 2 | 6 2 6 2 5 2 | 5 2 6 2 6 2 |
| | 6 2 6 2 | 6 2 6 2 5 2 | 5 2 6 2 5 2 |
| | 5 2 5 2 | 5 2 5 2 6 2 | 6 2 5 2 6 2 |
| | 6 2 6 2 | 6 2 5 2 5 2 | 5 2 5 2 6 2 |
| | 5 2 5 2 | 5 2 6 2 6 2 | 6 2 6 2 5 2 |
| | 6 2 6 2 | 6 2 6 2 6 2 | 6 2 6 2 6 2 |
| | 5 2 5 2 | 5 2 5 2 5 2 | 5 2 5 2 5 2 |
| | 6 2 6 2 | 6 2 6 2 6 2 | 6 2 6 2 6 2 |
| | 5 2 5 2 | 6 2 5 2 5 2 | 6 2 5 2 5 2 |
| | 6 2 6 2 | 5 2 5 2 6 2 | 5 2 6 2 5 2 |
| | 5 2 5 2 | 5 2 6 2 5 2 | 5 2 5 2 6 2 |
| | 6 2 6 2 | 5 2 6 2 5 2 | 6 2 5 2 5 2 |
| | 5 2 6 2 | 6 2 6 2 5 2 | 5 2 6 2 5 2 |
| | 6 2 5 2 | 6 2 5 2 6 2 | 5 2 5 2 6 2 |
| | 6 2 5 2 | 5 2 5 2 6 2 | 6 2 6 2 5 2 |
| | 6 2 6 2 | 5 2 6 2 5 2 | 5 2 5 2 6 2 |
| | 6 2 6 2 | 5 2 6 2 5 2 | 6 2 5 2 6 2 |
| | 6 2 6 2 | 6 2 6 2 5 2 | 6 2 6 2 5 2 |
| | 5 2 5 2 | 6 2 5 2 6 2 | 6 2 6 2 5 2 |
| | 5 2 5 2 | 6 2 6 2 5 2 | 5 2 6 2 5 2 |
| | 5 2 5 2 | 5 2 5 2 6 2 | 6 2 5 2 6 2 |
| | 6 2 6 2 | 6 2 5 2 6 2 | 6 2 5 2 6 2 |
| | 6 2 5 2 | 5 2 6 2 6 2 | 6 2 6 2 5 2 |
| | 6 2 6 2 | 5 2 5 2 6 2 | 6 2 6 2 6 2 |
| | 6 2 5 2 | 6 2 6 2 5 2 | 5 2 6 2 5 2 |
| | 5 2 6 2 | 5 2 5 2 5 2 | 6 2 5 2 5 2 |
| | 5 2 6 2 | 6 2 6 2 6 2 | 6 2 6 2 5 2 |
| | 5 2 5 2 | 6 2 5 2 5 2 | 6 2 6 2 6 2 |
| | 6 2 6 2 | 5 2 6 2 6 2 | 5 2 6 2 6 2 |
| | 5 2 6 2 | 6 2 6 2 6 2 | 6 2 5 2 5 2 |
| | 6 2 5 2 | 5 2 5 2 5 2 | 5 2 5 2 6 2 |
| | 6 2 6 2 | 6 2 5 2 6 2 | 5 2 6 2 5 2 |
| | 6 2 5 2 | 6 2 5 2 5 2 | 6 2 6 2 6 2 |
| | 6 2 5 2 | 6 2 6 2 5 2 | 5 2 6 2 6 2 |
| | 5 2 6 2 | 5 2 6 2 5 2 | 6 2 6 2 5 2 |
| | 5 2 5 2 | 6 2 6 2 5 2 | 6 2 5 2 5 2 |
| | 6 2 6 2 | 5 2 5 2 6 2 | 5 2 5 2 6 2 |
| | 5 2 5 2 | 5 2 5 2 5 2 | 5 2 5 2 5 2 |
| | 6 2 6 2 | 6 2 6 2 5 2 | 6 2 5 2 6 2 |
| | 5 2 5 2 | 6 2 6 2 6 2 | 6 2 6 2 6 2 |
| | 6 2 6 2 | 5 2 6 2 6 2 | 6 2 6 2 5 2 |
| | 6 2 6 2 | 6 2 5 2 5 2 | 6 2 6 2 6 2 |
| | 6 2 5 2 | 5 2 5 2 6 2 | 5 2 5 2 5 2 |
| | 6 2 6 2 | 6 2 6 2 5 2 | 6 2 5 2 5 2 |
| | 5 2 5 2 | 5 2 5 2 6 2 | 5 2 5 2 6 2 |
| | 6 2 6 2 | 6 2 6 2 5 2 | 5 2 6 2 6 2 |
| | 5 2 5 2 | 6 2 5 2 6 2 | 5 2 5 2 5 2 |
| | 5 2 6 2 | 5 2 6 2 5 2 | 5 2 5 2 6 2 |
| | 5 2 5 2 | 6 2 5 2 6 2 | 6 2 6 2 5 2 |
| | 5 2 6 2 | 5 2 5 2 5 2 | 6 2 6 2 6 2 |
| | 5 2 5 2 | 5 2 6 2 5 2 | 5 2 5 2 5 2 |
| | 6 2 5 2 | 5 2 5 2 6 2 | 6 2 6 2 6 2 |
| | 5 2 6 2 | 5 2 6 2 5 2 | 5 2 5 2 5 2 |
| | 6 2 5 2 | 6 2 6 2 6 2 | 5 2 5 2 6 2 |
| | 6 2 6 2 | 5 2 5 2 5 2 | 5 2 5 2 6 2 |
| | 5 2 6 2 | 5 2 6 2 6 2 | 5 2 5 2 5 2 |
| | 5 2 6 2 | 6 2 5 2 5 2 | 6 2 6 2 5 2 |
| | 5 2 5 2 | 6 2 6 2 6 2 | 6 2 5 2 6 2 |
| | 6 2 5 2 | 5 2 5 2 6 2 | 5 2 6 2 6 2 |

Continued on next column

**TABLE XXVIII– continued**

| TQC | ICSD NoE | SF | |
|---|---|---|---|
| | 5 2 5 2 | 6 2 5 2 6 2 | 5 2 6 2 5 2 |
| | 6 2 6 2 | 6 2 6 2 5 2 | 6 2 5 2 5 2 |
| | 6 2 6 2 | 6 2 5 2 5 2 | 6 2 6 2 5 2 |
| | 5 2 5 2 | 6 2 5 2 5 2 | 5 2 5 2 6 2 |
| | 6 2 6 2 | 5 2 5 2 6 2 | 6 2 6 2 5 2 |
| 47.2.9 | 94045 160 | Ba2Co4O11Tb2 | |
| | -GM -R | -S -T -U | -X -Y -Z |
| | 5 2 6 2 | 5 2 5 2 5 2 | 6 2 6 2 6 2 |
| | 6 2 5 2 | 6 2 6 2 6 2 | 5 2 5 2 5 2 |
| | 5 2 5 2 | 6 2 6 2 5 2 | 6 2 5 2 6 2 |
| | 6 2 6 2 | 5 2 5 2 6 2 | 5 2 6 2 5 2 |
| | 5 2 6 2 | 5 2 5 2 5 2 | 6 2 6 2 6 2 |
| | 6 2 5 2 | 6 2 6 2 6 2 | 5 2 5 2 5 2 |
| | 5 2 5 2 | 6 2 5 2 6 2 | 5 2 6 2 6 2 |
| | 6 2 6 2 | 5 2 6 2 5 2 | 6 2 5 2 5 2 |
| | 5 2 6 2 | 5 2 5 2 5 2 | 6 2 6 2 6 2 |
| | 6 2 5 2 | 6 2 6 2 6 2 | 5 2 5 2 5 2 |
| | 5 2 5 2 | 5 2 5 2 5 2 | 5 2 6 2 6 2 |
| | 5 2 6 2 | 5 2 6 2 6 2 | 5 2 5 2 5 2 |
| | 6 2 5 2 | 5 2 5 2 6 2 | 6 2 5 2 6 2 |
| | 5 2 5 2 | 6 2 6 2 5 2 | 6 2 5 2 5 2 |
| | 6 2 6 2 | 6 2 6 2 6 2 | 5 2 5 2 5 2 |
| | 5 2 6 2 | 5 2 5 2 5 2 | 5 2 6 2 6 2 |
| | 6 2 5 2 | 6 2 5 2 6 2 | 5 2 6 2 6 2 |
| | 5 2 6 2 | 6 2 6 2 5 2 | 5 2 6 2 5 2 |
| | 5 2 5 2 | 5 2 5 2 6 2 | 6 2 5 2 6 2 |
| | 6 2 6 2 | 6 2 5 2 5 2 | 5 2 5 2 5 2 |
| | 5 2 5 2 | 5 2 6 2 6 2 | 6 2 6 2 6 2 |
| | 6 2 5 2 | 6 2 6 2 6 2 | 6 2 6 2 6 2 |
| | 5 2 6 2 | 5 2 5 2 5 2 | 5 2 5 2 5 2 |
| | 6 2 6 2 | 6 2 6 2 6 2 | 6 2 6 2 6 2 |
| | 5 2 5 2 | 6 2 5 2 5 2 | 6 2 5 2 5 2 |
| | 6 2 5 2 | 5 2 5 2 5 2 | 5 2 6 2 5 2 |
| | 5 2 6 2 | 5 2 6 2 6 2 | 5 2 5 2 6 2 |
| | 6 2 6 2 | 5 2 6 2 5 2 | 6 2 5 2 5 2 |
| | 6 2 6 2 | 6 2 6 2 5 2 | 5 2 6 2 5 2 |
| | 5 2 5 2 | 6 2 5 2 6 2 | 6 2 5 2 6 2 |
| | 6 2 5 2 | 5 2 6 2 6 2 | 5 2 6 2 5 2 |
| | 6 2 6 2 | 6 2 5 2 5 2 | 5 2 5 2 6 2 |
| | 6 2 5 2 | 5 2 6 2 5 2 | 6 2 6 2 6 2 |
| | 5 2 6 2 | 5 2 5 2 5 2 | 6 2 5 2 5 2 |
| | 5 2 6 2 | 6 2 6 2 6 2 | 6 2 6 2 5 2 |
| | 6 2 6 2 | 5 2 5 2 6 2 | 6 2 6 2 6 2 |
| | 6 2 5 2 | 6 2 6 2 5 2 | 5 2 6 2 5 2 |
| | 5 2 5 2 | 6 2 5 2 6 2 | 6 2 5 2 5 2 |
| | 6 2 5 2 | 5 2 6 2 6 2 | 6 2 5 2 6 2 |
| | 6 2 5 2 | 6 2 5 2 5 2 | 6 2 6 2 6 2 |
| | 6 2 6 2 | 5 2 6 2 5 2 | 6 2 6 2 5 2 |
| | 5 2 6 2 | 6 2 6 2 6 2 | 6 2 5 2 5 2 |
| | 5 2 6 2 | 5 2 5 2 6 2 | 6 2 6 2 6 2 |
| | 6 2 5 2 | 5 2 5 2 5 2 | 5 2 6 2 5 2 |
| | 5 2 6 2 | 6 2 6 2 6 2 | 5 2 6 2 6 2 |
| | 6 2 6 2 | 5 2 6 2 5 2 | 6 2 5 2 6 2 |
| | 6 2 5 2 | 6 2 5 2 6 2 | 5 2 5 2 5 2 |
| | 5 2 6 2 | 6 2 5 2 6 2 | 5 2 6 2 6 2 |
| | 6 2 5 2 | 6 2 5 2 5 2 | 6 2 6 2 5 2 |
| | 6 2 5 2 | 6 2 6 2 5 2 | 5 2 6 2 5 2 |
| | 5 2 6 2 | 5 2 6 2 5 2 | 6 2 6 2 6 2 |
| | 5 2 5 2 | 6 2 5 2 5 2 | 5 2 5 2 6 2 |
| | 6 2 6 2 | 5 2 6 2 6 2 | 6 2 5 2 5 2 |
| | 5 2 5 2 | 6 2 5 2 5 2 | 5 2 5 2 6 2 |
| | 6 2 6 2 | 5 2 6 2 6 2 | 6 2 5 2 5 2 |
| | 6 2 5 2 | 6 2 6 2 5 2 | 6 2 6 2 5 2 |
| | 5 2 6 2 | 6 2 6 2 6 2 | 6 2 6 2 6 2 |
| | 6 2 6 2 | 5 2 5 2 5 2 | 6 2 6 2 6 2 |
| | 6 2 5 2 | 5 2 5 2 6 2 | 5 2 5 2 5 2 |
| | 5 2 6 2 | 6 2 6 2 5 2 | 6 2 5 2 5 2 |
| | 5 2 5 2 | 5 2 5 2 6 2 | 5 2 6 2 6 2 |
| | 6 2 6 2 | 6 2 5 2 5 2 | 5 2 5 2 6 2 |
| | 6 2 6 2 | 6 2 6 2 6 2 | 5 2 5 2 5 2 |
| | 5 2 5 2 | 5 2 6 2 6 2 | 5 2 5 2 6 2 |
| | 5 2 5 2 | 6 2 5 2 5 2 | 6 2 6 2 5 2 |
| | 5 2 6 2 | 5 2 6 2 5 2 | 6 2 6 2 5 2 |
| | 5 2 5 2 | 5 2 5 2 5 2 | 5 2 5 2 6 2 |
| | 5 2 6 2 | 5 2 5 2 6 2 | 6 2 6 2 5 2 |
| | 6 2 5 2 | 5 2 6 2 5 2 | 5 2 5 2 6 2 |
| | 6 2 5 2 | 5 2 6 2 6 2 | 5 2 5 2 5 2 |
| | 6 2 6 2 | 6 2 5 2 6 2 | 5 2 5 2 6 2 |
| | 5 2 6 2 | 5 2 5 2 5 2 | 6 2 5 2 5 2 |
| | 5 2 5 2 | 5 2 6 2 6 2 | 5 2 6 2 6 2 |
| | 5 2 5 2 | 6 2 5 2 5 2 | 6 2 5 2 6 2 |
| | 6 2 6 2 | 6 2 6 2 6 2 | 5 2 6 2 6 2 |
| | 5 2 6 2 | 6 2 6 2 5 2 | 5 2 5 2 5 2 |
| | 6 2 5 2 | 6 2 5 2 6 2 | 6 2 6 2 5 2 |
| | 6 2 5 2 | 6 2 5 2 5 2 | 6 2 6 2 5 2 |
| | 5 2 6 2 | 5 2 5 2 5 2 | 5 2 5 2 6 2 |
| | 6 2 6 2 | 6 2 5 2 6 2 | 6 2 6 2 5 2 |
| 47.2.10 | 96380 150 | Ba1Co2Nd1O5.5 | |
| | -GM -R | -S -T -U | -X -Y -Z |
| | 5 2 6 2 | 5 2 5 2 5 2 | 6 2 6 2 6 2 |
| | 6 2 5 2 | 6 2 6 2 6 2 | 5 2 5 2 5 2 |
| | 5 2 5 2 | 6 2 6 2 5 2 | 6 2 5 2 6 2 |
| | 6 2 6 2 | 5 2 5 2 6 2 | 5 2 6 2 5 2 |
| | 5 2 5 2 | 6 2 5 2 6 2 | 5 2 6 2 6 2 |
| | 6 2 6 2 | 5 2 6 2 5 2 | 6 2 5 2 5 2 |
| | 5 2 5 2 | 6 2 5 2 6 2 | 5 2 6 2 6 2 |
| | 6 2 6 2 | 5 2 6 2 5 2 | 6 2 5 2 5 2 |
| | 5 2 5 2 | 5 2 5 2 6 2 | 5 2 6 2 6 2 |
| | 5 2 6 2 | 6 2 6 2 5 2 | 6 2 5 2 5 2 |
| | 6 2 5 2 | 5 2 6 2 6 2 | 5 2 5 2 6 2 |
| | 5 2 6 2 | 5 2 5 2 6 2 | 6 2 6 2 5 2 |
| | 6 2 6 2 | 6 2 5 2 5 2 | 5 2 5 2 6 2 |
| | 5 2 5 2 | 5 2 6 2 5 2 | 6 2 5 2 5 2 |
| | 5 2 5 2 | 6 2 6 2 6 2 | 5 2 6 2 5 2 |
| | 5 2 6 2 | 5 2 5 2 5 2 | 5 2 6 2 6 2 |
| | 6 2 5 2 | 6 2 5 2 6 2 | 5 2 5 2 6 2 |
| | 6 2 6 2 | 5 2 6 2 5 2 | 6 2 5 2 5 2 |
| | 5 2 5 2 | 6 2 5 2 6 2 | 5 2 6 2 6 2 |
| | 6 2 6 2 | 5 2 5 2 5 2 | 6 2 5 2 6 2 |
| | 5 2 5 2 | 6 2 6 2 6 2 | 5 2 6 2 5 2 |
| | 6 2 5 2 | 6 2 6 2 5 2 | 6 2 6 2 6 2 |
| | 5 2 6 2 | 5 2 5 2 6 2 | 5 2 5 2 5 2 |
| | 6 2 5 2 | 6 2 6 2 6 2 | 6 2 6 2 6 2 |
| | 5 2 6 2 | 6 2 5 2 5 2 | 6 2 5 2 5 2 |
| | 6 2 5 2 | 5 2 5 2 5 2 | 5 2 6 2 5 2 |
| | 5 2 6 2 | 5 2 6 2 6 2 | 5 2 5 2 6 2 |
| | 6 2 6 2 | 5 2 6 2 5 2 | 6 2 5 2 5 2 |
| | 6 2 6 2 | 6 2 6 2 5 2 | 5 2 6 2 5 2 |
| | 5 2 5 2 | 6 2 5 2 6 2 | 6 2 6 2 6 2 |
| | 6 2 5 2 | 5 2 5 2 6 2 | 5 2 5 2 5 2 |
| | 6 2 6 2 | 6 2 6 2 5 2 | 5 2 5 2 6 2 |
| | 6 2 5 2 | 5 2 6 2 5 2 | 6 2 5 2 5 2 |

Continued on next column

**TABLE XXVIII– continued**

| TQC | ICSD | NoE | SF |
|---|---|---|---|
| | 5 2 6 2 | 5 2 5 2 6 2 | 6 2 6 2 5 2 |
| | 5 2 6 2 | 6 2 6 2 5 2 | 6 2 6 2 6 2 |
| | 6 2 5 2 | 6 2 5 2 6 2 | 6 2 6 2 6 2 |
| | 6 2 6 2 | 5 2 6 2 5 2 | 5 2 6 2 5 2 |
| | 6 2 5 2 | 6 2 5 2 6 2 | 6 2 5 2 5 2 |
| | 5 2 5 2 | 6 2 6 2 6 2 | 6 2 5 2 6 2 |
| | 6 2 5 2 | 5 2 5 2 5 2 | 6 2 6 2 6 2 |
| | 6 2 6 2 | 5 2 6 2 5 2 | 6 2 6 2 5 2 |
| | 5 2 6 2 | 6 2 6 2 6 2 | 5 2 5 2 5 2 |
| | 5 2 6 2 | 5 2 5 2 6 2 | 6 2 6 2 6 2 |
| | 6 2 5 2 | 5 2 5 2 5 2 | 5 2 6 2 5 2 |
| | 5 2 5 2 | 5 2 6 2 6 2 | 5 2 6 2 6 2 |
| | 5 2 6 2 | 6 2 6 2 5 2 | 6 2 5 2 6 2 |
| | 6 2 6 2 | 6 2 5 2 6 2 | 6 2 5 2 5 2 |
| | 6 2 5 2 | 6 2 5 2 6 2 | 5 2 6 2 6 2 |
| | 6 2 6 2 | 6 2 5 2 5 2 | 6 2 6 2 5 2 |
| | 6 2 5 2 | 6 2 6 2 5 2 | 6 2 6 2 5 2 |
| | 5 2 6 2 | 5 2 6 2 5 2 | 5 2 5 2 6 2 |
| | 6 2 5 2 | 6 2 5 2 5 2 | 5 2 5 2 5 2 |
| | 5 2 6 2 | 5 2 6 2 6 2 | 6 2 6 2 6 2 |
| | 5 2 5 2 | 5 2 5 2 5 2 | 5 2 5 2 6 2 |
| | 6 2 6 2 | 6 2 6 2 5 2 | 6 2 5 2 5 2 |
| | 5 2 6 2 | 5 2 6 2 6 2 | 6 2 6 2 6 2 |
| | 6 2 5 2 | 6 2 5 2 5 2 | 5 2 6 2 5 2 |
| | 6 2 5 2 | 6 2 6 2 6 2 | 6 2 5 2 6 2 |
| | 6 2 6 2 | 5 2 5 2 6 2 | 6 2 6 2 5 2 |
| | 5 2 6 2 | 6 2 6 2 5 2 | 6 2 5 2 5 2 |
| | 5 2 5 2 | 5 2 5 2 6 2 | 5 2 5 2 6 2 |
| | 6 2 6 2 | 5 2 6 2 5 2 | 5 2 6 2 6 2 |
| | 6 2 5 2 | 5 2 6 2 5 2 | 5 2 5 2 5 2 |
| | 5 2 5 2 | 5 2 5 2 6 2 | 6 2 5 2 6 2 |
| | 5 2 6 2 | 6 2 6 2 6 2 | 5 2 5 2 5 2 |
| | 5 2 5 2 | 5 2 5 2 5 2 | 5 2 5 2 6 2 |
| | 6 2 6 2 | 6 2 5 2 6 2 | 5 2 5 2 5 2 |
| | 5 2 5 2 | 6 2 6 2 5 2 | 5 2 6 2 5 2 |
| | 5 2 6 2 | 5 2 5 2 6 2 | 6 2 6 2 6 2 |
| | 5 2 6 2 | 6 2 6 2 5 2 | 5 2 6 2 6 2 |
| | 6 2 5 2 | 6 2 5 2 6 2 | 6 2 5 2 5 2 |
| | 5 2 6 2 | 5 2 5 2 5 2 | 6 2 6 2 5 2 |
| | 6 2 6 2 | 6 2 5 2 5 2 | 6 2 5 2 5 2 |
| | 6 2 5 2 | 6 2 6 2 6 2 | 5 2 6 2 6 2 |
| | 6 2 5 2 | 5 2 5 2 5 2 | 6 2 6 2 6 2 |
| 47.2.11 | 96377 | 150 | Ba1Co2Nd1O5.5 |
| | -GM -R | -S -T -U | -X -Y -Z |
| | 5 2 6 2 | 5 2 5 2 5 2 | 6 2 6 2 6 2 |
| | 6 2 5 2 | 6 2 6 2 6 2 | 5 2 5 2 5 2 |
| | 5 2 5 2 | 6 2 6 2 5 2 | 6 2 5 2 6 2 |
| | 6 2 6 2 | 5 2 5 2 6 2 | 5 2 6 2 5 2 |
| | 5 2 5 2 | 6 2 5 2 6 2 | 5 2 6 2 6 2 |
| | 6 2 6 2 | 5 2 6 2 5 2 | 6 2 5 2 5 2 |
| | 5 2 5 2 | 6 2 5 2 6 2 | 5 2 6 2 6 2 |
| | 6 2 6 2 | 5 2 6 2 5 2 | 6 2 5 2 5 2 |
| | 5 2 5 2 | 5 2 5 2 6 2 | 5 2 6 2 6 2 |
| | 5 2 6 2 | 5 2 6 2 5 2 | 6 2 5 2 5 2 |
| | 6 2 5 2 | 5 2 6 2 6 2 | 5 2 5 2 6 2 |
| | 6 2 6 2 | 6 2 5 2 6 2 | 6 2 6 2 6 2 |
| | 5 2 6 2 | 6 2 5 2 5 2 | 5 2 5 2 5 2 |
| | 5 2 5 2 | 5 2 6 2 5 2 | 5 2 5 2 5 2 |
| | 5 2 5 2 | 6 2 6 2 6 2 | 6 2 6 2 5 2 |
| | 5 2 6 2 | 5 2 5 2 5 2 | 5 2 6 2 6 2 |
| | 6 2 5 2 | 6 2 5 2 6 2 | 5 2 5 2 6 2 |

Continued on next column

**TABLE XXVIII– continued**

| TQC | ICSD | NoE | SF |
|---|---|---|---|
| | 6 2 6 2 | 5 2 6 2 5 2 | 6 2 5 2 5 2 |
| | 5 2 5 2 | 6 2 5 2 6 2 | 5 2 6 2 6 2 |
| | 6 2 6 2 | 5 2 5 2 5 2 | 6 2 5 2 6 2 |
| | 5 2 5 2 | 6 2 6 2 6 2 | 5 2 6 2 5 2 |
| | 6 2 5 2 | 6 2 6 2 5 2 | 6 2 6 2 6 2 |
| | 5 2 6 2 | 5 2 5 2 6 2 | 5 2 5 2 5 2 |
| | 6 2 5 2 | 6 2 6 2 6 2 | 6 2 6 2 6 2 |
| | 5 2 6 2 | 6 2 5 2 5 2 | 6 2 5 2 5 2 |
| | 6 2 5 2 | 5 2 5 2 5 2 | 5 2 6 2 5 2 |
| | 5 2 6 2 | 5 2 6 2 6 2 | 5 2 5 2 6 2 |
| | 6 2 6 2 | 5 2 6 2 5 2 | 6 2 5 2 5 2 |
| | 6 2 6 2 | 6 2 6 2 5 2 | 5 2 6 2 5 2 |
| | 5 2 5 2 | 6 2 5 2 6 2 | 6 2 6 2 6 2 |
| | 6 2 5 2 | 5 2 5 2 6 2 | 5 2 5 2 5 2 |
| | 6 2 6 2 | 6 2 6 2 5 2 | 5 2 5 2 5 2 |
| | 6 2 5 2 | 5 2 6 2 5 2 | 6 2 6 2 6 2 |
| | 5 2 6 2 | 5 2 5 2 6 2 | 6 2 5 2 5 2 |
| | 5 2 6 2 | 6 2 6 2 5 2 | 6 2 6 2 6 2 |
| | 6 2 5 2 | 6 2 5 2 6 2 | 6 2 6 2 6 2 |
| | 6 2 6 2 | 5 2 6 2 5 2 | 6 2 6 2 5 2 |
| | 6 2 5 2 | 6 2 5 2 6 2 | 5 2 5 2 5 2 |
| | 5 2 5 2 | 6 2 6 2 6 2 | 6 2 5 2 6 2 |
| | 6 2 5 2 | 5 2 5 2 5 2 | 6 2 6 2 6 2 |
| | 6 2 6 2 | 5 2 6 2 5 2 | 6 2 5 2 5 2 |
| | 5 2 6 2 | 6 2 6 2 6 2 | 5 2 6 2 5 2 |
| | 5 2 6 2 | 5 2 5 2 6 2 | 6 2 6 2 5 2 |
| | 6 2 5 2 | 5 2 5 2 5 2 | 5 2 6 2 6 2 |
| | 5 2 5 2 | 5 2 6 2 6 2 | 5 2 6 2 6 2 |
| | 5 2 6 2 | 6 2 6 2 5 2 | 6 2 5 2 5 2 |
| | 6 2 5 2 | 6 2 5 2 6 2 | 6 2 5 2 6 2 |
| | 6 2 6 2 | 6 2 5 2 6 2 | 5 2 6 2 6 2 |
| | 6 2 6 2 | 6 2 5 2 5 2 | 6 2 6 2 5 2 |
| | 6 2 5 2 | 6 2 6 2 5 2 | 6 2 6 2 5 2 |
| | 5 2 6 2 | 5 2 6 2 5 2 | 5 2 5 2 6 2 |
| | 6 2 5 2 | 6 2 5 2 5 2 | 5 2 5 2 5 2 |
| | 5 2 6 2 | 5 2 6 2 6 2 | 6 2 6 2 6 2 |
| | 5 2 5 2 | 5 2 5 2 5 2 | 5 2 5 2 6 2 |
| | 6 2 6 2 | 6 2 6 2 5 2 | 6 2 5 2 5 2 |
| | 5 2 6 2 | 5 2 6 2 6 2 | 6 2 6 2 6 2 |
| | 6 2 5 2 | 6 2 5 2 6 2 | 5 2 6 2 5 2 |
| | 6 2 5 2 | 6 2 6 2 5 2 | 6 2 5 2 6 2 |
| | 6 2 6 2 | 5 2 5 2 6 2 | 6 2 6 2 5 2 |
| | 5 2 6 2 | 6 2 6 2 5 2 | 6 2 5 2 5 2 |
| | 5 2 5 2 | 5 2 5 2 6 2 | 5 2 6 2 6 2 |
| | 6 2 6 2 | 5 2 6 2 5 2 | 5 2 5 2 6 2 |
| | 6 2 5 2 | 5 2 6 2 5 2 | 5 2 5 2 5 2 |
| | 5 2 5 2 | 5 2 5 2 6 2 | 6 2 5 2 6 2 |
| | 5 2 6 2 | 5 2 6 2 6 2 | 5 2 5 2 5 2 |
| | 6 2 5 2 | 6 2 5 2 6 2 | 5 2 5 2 6 2 |
| | 5 2 6 2 | 6 2 5 2 5 2 | 5 2 5 2 5 2 |
| | 5 2 5 2 | 6 2 6 2 5 2 | 5 2 6 2 5 2 |
| | 5 2 6 2 | 5 2 5 2 6 2 | 6 2 6 2 6 2 |
| | 5 2 6 2 | 6 2 6 2 5 2 | 5 2 6 2 6 2 |
| | 6 2 5 2 | 6 2 5 2 6 2 | 6 2 5 2 5 2 |
| | 5 2 6 2 | 5 2 5 2 5 2 | 6 2 6 2 5 2 |
| | 6 2 5 2 | 6 2 5 2 5 2 | 6 2 5 2 5 2 |
| | 6 2 6 2 | 6 2 6 2 6 2 | 5 2 6 2 6 2 |
| | 6 2 5 2 | 6 2 6 2 5 2 | 6 2 6 2 6 2 |
| 47.2.12 | 187120 | 150 | Ba1Co2Nd1O5.5 |
| | -GM -R | -S -T -U | -X -Y -Z |
| | 5 2 6 2 | 5 2 5 2 5 2 | 6 2 6 2 6 2 |

Continued on next column

**TABLE XXVIII– continued**

| TQC | ICSD | NoE | SF |
|---|---|---|---|
| | 6 2 5 2 | 6 2 6 2 6 2 | 5 2 5 2 5 2 |
| | 5 2 5 2 | 6 2 6 2 5 2 | 6 2 5 2 6 2 |
| | 6 2 6 2 | 5 2 5 2 6 2 | 5 2 6 2 5 2 |
| | 5 2 5 2 | 6 2 5 2 6 2 | 5 2 6 2 6 2 |
| | 6 2 6 2 | 5 2 6 2 5 2 | 6 2 5 2 5 2 |
| | 5 2 5 2 | 6 2 5 2 6 2 | 5 2 6 2 6 2 |
| | 6 2 6 2 | 5 2 6 2 5 2 | 6 2 5 2 5 2 |
| | 5 2 5 2 | 5 2 5 2 6 2 | 5 2 6 2 6 2 |
| | 5 2 5 2 | 6 2 6 2 5 2 | 6 2 5 2 5 2 |
| | 6 2 6 2 | 5 2 6 2 6 2 | 5 2 5 2 6 2 |
| | 6 2 6 2 | 5 2 5 2 5 2 | 6 2 6 2 5 2 |
| | 5 2 6 2 | 6 2 5 2 6 2 | 5 2 5 2 6 2 |
| | 5 2 5 2 | 5 2 5 2 5 2 | 5 2 5 2 5 2 |
| | 5 2 5 2 | 6 2 6 2 6 2 | 6 2 6 2 5 2 |
| | 5 2 6 2 | 5 2 6 2 5 2 | 5 2 6 2 6 2 |
| | 6 2 5 2 | 6 2 5 2 6 2 | 5 2 5 2 6 2 |
| | 6 2 6 2 | 5 2 6 2 5 2 | 6 2 5 2 5 2 |
| | 5 2 5 2 | 6 2 5 2 6 2 | 5 2 6 2 6 2 |
| | 6 2 6 2 | 5 2 5 2 5 2 | 6 2 5 2 6 2 |
| | 5 2 5 2 | 6 2 6 2 6 2 | 5 2 6 2 5 2 |
| | 6 2 5 2 | 6 2 6 2 5 2 | 6 2 6 2 6 2 |
| | 5 2 6 2 | 5 2 5 2 6 2 | 5 2 5 2 5 2 |
| | 6 2 5 2 | 6 2 6 2 6 2 | 6 2 6 2 6 2 |
| | 5 2 6 2 | 6 2 5 2 5 2 | 6 2 5 2 5 2 |
| | 6 2 5 2 | 5 2 5 2 5 2 | 5 2 6 2 5 2 |
| | 5 2 6 2 | 5 2 6 2 6 2 | 5 2 5 2 6 2 |
| | 6 2 6 2 | 5 2 6 2 5 2 | 6 2 5 2 5 2 |
| | 6 2 6 2 | 6 2 6 2 5 2 | 5 2 6 2 5 2 |
| | 5 2 5 2 | 6 2 5 2 6 2 | 6 2 5 2 6 2 |
| | 6 2 5 2 | 5 2 5 2 6 2 | 5 2 6 2 5 2 |
| | 6 2 6 2 | 6 2 6 2 5 2 | 5 2 5 2 6 2 |
| | 6 2 5 2 | 5 2 6 2 5 2 | 6 2 6 2 5 2 |
| | 5 2 6 2 | 5 2 5 2 6 2 | 6 2 5 2 6 2 |
| | 5 2 6 2 | 6 2 6 2 5 2 | 6 2 6 2 5 2 |
| | 6 2 5 2 | 5 2 5 2 6 2 | 6 2 6 2 6 2 |
| | 6 2 6 2 | 6 2 6 2 5 2 | 6 2 6 2 5 2 |
| | 6 2 5 2 | 6 2 5 2 6 2 | 5 2 5 2 5 2 |
| | 5 2 5 2 | 6 2 6 2 6 2 | 6 2 5 2 6 2 |
| | 6 2 5 2 | 5 2 5 2 5 2 | 6 2 6 2 6 2 |
| | 6 2 6 2 | 5 2 6 2 5 2 | 6 2 6 2 5 2 |
| | 5 2 6 2 | 6 2 6 2 6 2 | 5 2 5 2 5 2 |
| | 5 2 6 2 | 5 2 5 2 6 2 | 6 2 6 2 6 2 |
| | 6 2 5 2 | 5 2 5 2 5 2 | 5 2 6 2 5 2 |
| | 5 2 5 2 | 5 2 6 2 6 2 | 6 2 6 2 6 2 |
| | 6 2 6 2 | 6 2 6 2 5 2 | 5 2 5 2 6 2 |
| | 5 2 6 2 | 6 2 5 2 6 2 | 6 2 5 2 5 2 |
| | 6 2 5 2 | 6 2 5 2 6 2 | 5 2 6 2 6 2 |
| | 6 2 6 2 | 6 2 5 2 5 2 | 6 2 6 2 5 2 |
| | 6 2 5 2 | 6 2 6 2 5 2 | 6 2 5 2 5 2 |
| | 5 2 6 2 | 5 2 6 2 5 2 | 5 2 6 2 6 2 |
| | 6 2 5 2 | 6 2 5 2 5 2 | 5 2 5 2 5 2 |
| | 5 2 6 2 | 5 2 6 2 6 2 | 6 2 6 2 6 2 |
| | 5 2 5 2 | 5 2 5 2 5 2 | 5 2 5 2 6 2 |
| | 6 2 6 2 | 5 2 6 2 5 2 | 6 2 5 2 5 2 |
| | 5 2 6 2 | 6 2 6 2 6 2 | 6 2 6 2 6 2 |
| | 6 2 5 2 | 6 2 5 2 6 2 | 6 2 6 2 5 2 |
| | 6 2 5 2 | 6 2 6 2 5 2 | 5 2 5 2 6 2 |
| | 6 2 6 2 | 5 2 5 2 6 2 | 6 2 6 2 5 2 |
| | 5 2 6 2 | 6 2 6 2 5 2 | 6 2 5 2 5 2 |
| | 5 2 5 2 | 5 2 5 2 6 2 | 5 2 5 2 6 2 |
| | 6 2 6 2 | 5 2 6 2 5 2 | 5 2 6 2 6 2 |

Continued on next column

**TABLE XXVIII– continued**

| TQC | ICSD | NoE | SF |
|---|---|---|---|
| | 6 2 5 2 | 5 2 6 2 5 2 | 5 2 5 2 5 2 |
| | 5 2 5 2 | 5 2 5 2 6 2 | 6 2 5 2 6 2 |
| | 5 2 6 2 | 6 2 6 2 6 2 | 5 2 5 2 5 2 |
| | 6 2 5 2 | 5 2 5 2 5 2 | 5 2 5 2 6 2 |
| | 5 2 6 2 | 6 2 5 2 6 2 | 5 2 5 2 5 2 |
| | 5 2 5 2 | 6 2 6 2 5 2 | 5 2 6 2 5 2 |
| | 5 2 6 2 | 5 2 5 2 6 2 | 6 2 6 2 6 2 |
| | 5 2 6 2 | 6 2 6 2 5 2 | 5 2 6 2 6 2 |
| | 6 2 5 2 | 6 2 5 2 6 2 | 6 2 5 2 5 2 |
| | 6 2 6 2 | 5 2 5 2 5 2 | 6 2 6 2 5 2 |
| | 5 2 5 2 | 6 2 5 2 5 2 | 6 2 5 2 5 2 |
| | 6 2 6 2 | 6 2 6 2 6 2 | 5 2 6 2 6 2 |
| | 5 2 5 2 | 5 2 5 2 6 2 | 6 2 6 2 6 2 |
| 47.2.13 | 71640 | 114 | B3Ir4Zn1 |
| | -GM -R | -S -T -U | -X -Y -Z |
| | 5 2 5 2 | 5 2 5 2 5 2 | 5 2 5 2 5 2 |
| | 6 2 6 2 | 6 2 6 2 6 2 | 6 2 6 2 6 2 |
| | 5 2 5 2 | 5 2 6 2 5 2 | 5 2 5 2 6 2 |
| | 6 2 5 2 | 6 2 5 2 5 2 | 6 2 6 2 5 2 |
| | 5 2 6 2 | 6 2 6 2 6 2 | 5 2 5 2 5 2 |
| | 6 2 6 2 | 5 2 5 2 6 2 | 6 2 6 2 6 2 |
| | 5 2 5 2 | 6 2 5 2 5 2 | 5 2 5 2 6 2 |
| | 6 2 5 2 | 5 2 6 2 6 2 | 6 2 5 2 5 2 |
| | 5 2 5 2 | 5 2 6 2 5 2 | 6 2 6 2 5 2 |
| | 5 2 6 2 | 6 2 5 2 6 2 | 5 2 6 2 6 2 |
| | 5 2 5 2 | 5 2 5 2 5 2 | 5 2 5 2 5 2 |
| | 5 2 5 2 | 6 2 6 2 5 2 | 5 2 5 2 6 2 |
| | 5 2 5 2 | 5 2 5 2 5 2 | 6 2 5 2 5 2 |
| | 5 2 5 2 | 6 2 6 2 6 2 | 6 2 6 2 6 2 |
| | 5 2 6 2 | 6 2 6 2 5 2 | 5 2 5 2 5 2 |
| | 5 2 6 2 | 5 2 5 2 5 2 | 6 2 5 2 5 2 |
| | 5 2 5 2 | 5 2 5 2 5 2 | 5 2 5 2 6 2 |
| | 5 2 5 2 | 6 2 6 2 5 2 | 6 2 5 2 6 2 |
| | 5 2 5 2 | 5 2 5 2 5 2 | 5 2 5 2 5 2 |
| | 6 2 5 2 | 6 2 6 2 5 2 | 6 2 5 2 6 2 |
| | 5 2 5 2 | 5 2 5 2 5 2 | 5 2 5 2 6 2 |
| | 5 2 5 2 | 6 2 6 2 5 2 | 6 2 5 2 5 2 |
| | 6 2 6 2 | 5 2 6 2 6 2 | 5 2 5 2 6 2 |
| | 6 2 5 2 | 6 2 5 2 5 2 | 6 2 6 2 5 2 |
| | 6 2 6 2 | 6 2 5 2 5 2 | 5 2 5 2 5 2 |
| | 5 2 6 2 | 5 2 6 2 5 2 | 6 2 6 2 6 2 |
| | 5 2 5 2 | 5 2 6 2 6 2 | 6 2 5 2 5 2 |
| | 5 2 5 2 | 6 2 5 2 5 2 | 5 2 5 2 6 2 |
| | 6 2 6 2 | 5 2 5 2 6 2 | 6 2 6 2 6 2 |
| | 5 2 5 2 | 6 2 6 2 5 2 | 5 2 5 2 5 2 |
| | 6 2 6 2 | 6 2 6 2 6 2 | 5 2 6 2 5 2 |
| | 5 2 5 2 | 5 2 5 2 6 2 | 6 2 5 2 6 2 |
| | 6 2 6 2 | 5 2 5 2 6 2 | 6 2 6 2 5 2 |
| | 5 2 6 2 | 6 2 6 2 6 2 | 5 2 6 2 6 2 |
| | 6 2 5 2 | 5 2 6 2 5 2 | 5 2 5 2 6 2 |
| | 5 2 5 2 | 6 2 5 2 6 2 | 6 2 6 2 5 2 |
| | 6 2 6 2 | 6 2 5 2 5 2 | 6 2 6 2 6 2 |
| | 5 2 5 2 | 5 2 6 2 5 2 | 5 2 5 2 5 2 |
| | 6 2 5 2 | 6 2 6 2 6 2 | 5 2 5 2 5 2 |
| | 6 2 6 2 | 5 2 5 2 6 2 | 6 2 6 2 6 2 |
| | 5 2 6 2 | 6 2 5 2 5 2 | 6 2 6 2 6 2 |
| | 5 2 6 2 | 5 2 6 2 5 2 | 5 2 6 2 5 2 |
| | 6 2 5 2 | 5 2 6 2 6 2 | 6 2 5 2 6 2 |
| | 6 2 6 2 | 6 2 6 2 6 2 | 5 2 6 2 5 2 |
| | 5 2 6 2 | 5 2 5 2 5 2 | 5 2 5 2 6 2 |
| | 5 2 6 2 | 6 2 5 2 6 2 | 6 2 6 2 5 2 |

Continued on next column

**TABLE XXVIII– continued**

| TQC | ICSD | NoE | SF |
|---|---|---|---|
| | 6 2 5 2 | 6 2 5 2 5 2 | 5 2 5 2 6 2 |
| | 6 2 5 2 | 5 2 6 2 6 2 | 6 2 6 2 5 2 |
| | 6 2 5 2 | 5 2 6 2 6 2 | 5 2 5 2 5 2 |
| | 5 2 6 2 | 6 2 5 2 5 2 | 6 2 6 2 6 2 |
| | 6 2 5 2 | 6 2 5 2 6 2 | 6 2 5 2 5 2 |
| | 5 2 5 2 | 5 2 6 2 5 2 | 5 2 6 2 6 2 |
| | 6 2 6 2 | 5 2 6 2 5 2 | 6 2 5 2 5 2 |
| | 5 2 5 2 | 6 2 5 2 6 2 | 5 2 6 2 6 2 |
| | 6 2 6 2 | 5 2 5 2 5 2 | 6 2 5 2 5 2 |
| | 5 2 5 2 | 6 2 6 2 6 2 | 5 2 6 2 6 2 |
| | 6 2 6 2 | 6 2 5 2 5 2 | 6 2 5 2 5 2 |

The End

TABLE XXIX: The computed irreducible representations for the valence bands for the compounds of SEBR and NLC in space group 51. See the list of tables.

| TQC | ICSD | NoE | SF | | | | | |
|---|---|---|---|---|---|---|---|---|
| 51.2.1 | 248052 | 208 | La2Se9U2 | | | | | |
| | -GM | -R | -S | -T | -U | -X | -Y | -Z |
| | 5 2 | 3+4 4 | 3+4 4 | 5 2 | 3+4 4 | 3+4 4 | 5 2 | 5 2 |
| | 6 2 | 3+4 4 | 3+4 4 | 6 2 | 3+4 4 | 3+4 4 | 6 2 | 6 2 |
| | 5 2 | 3+4 4 | 3+4 4 | 5 2 | 3+4 4 | 3+4 4 | 5 2 | 5 2 |
| | 6 2 | 3+4 4 | 3+4 4 | 6 2 | 3+4 4 | 3+4 4 | 6 2 | 6 2 |
| | 5 2 | 3+4 4 | 3+4 4 | 5 2 | 3+4 4 | 3+4 4 | 5 2 | 5 2 |
| | 6 2 | 3+4 4 | 3+4 4 | 6 2 | 3+4 4 | 3+4 4 | 6 2 | 6 2 |
| | 5 2 | 3+4 4 | 3+4 4 | 5 2 | 3+4 4 | 3+4 4 | 5 2 | 5 2 |
| | 6 2 | 3+4 4 | 3+4 4 | 6 2 | 3+4 4 | 3+4 4 | 6 2 | 6 2 |
| | 5 2 | 3+4 4 | 3+4 4 | 5 2 | 3+4 4 | 3+4 4 | 5 2 | 5 2 |
| | 6 2 | 3+4 4 | 3+4 4 | 6 2 | 3+4 4 | 3+4 4 | 6 2 | 6 2 |
| | 5 2 | 3+4 4 | 3+4 4 | 5 2 | 3+4 4 | 3+4 4 | 5 2 | 5 2 |
| | 6 2 | 3+4 4 | 3+4 4 | 6 2 | 3+4 4 | 3+4 4 | 6 2 | 6 2 |
| | 5 2 | 3+4 4 | 3+4 4 | 5 2 | 3+4 4 | 3+4 4 | 5 2 | 5 2 |
| | 6 2 | 3+4 4 | 3+4 4 | 6 2 | 3+4 4 | 3+4 4 | 6 2 | 6 2 |
| | 5 2 | 3+4 4 | 3+4 4 | 5 2 | 3+4 4 | 3+4 4 | 6 2 | 6 2 |
| | 6 2 | 3+4 4 | 3+4 4 | 6 2 | 3+4 4 | 3+4 4 | 5 2 | 5 2 |
| | 5 2 | 3+4 4 | 3+4 4 | 5 2 | 3+4 4 | 3+4 4 | 5 2 | 6 2 |
| | 5 2 | 3+4 4 | 3+4 4 | 6 2 | 3+4 4 | 3+4 4 | 5 2 | 5 2 |
| | 6 2 | 3+4 4 | 3+4 4 | 5 2 | 3+4 4 | 3+4 4 | 6 2 | 5 2 |
| | 5 2 | 3+4 4 | 3+4 4 | 6 2 | 3+4 4 | 3+4 4 | 5 2 | 6 2 |
| | 6 2 | 3+4 4 | 3+4 4 | 5 2 | 3+4 4 | 3+4 4 | 6 2 | 5 2 |
| | 6 2 | 3+4 4 | 3+4 4 | 6 2 | 3+4 4 | 3+4 4 | 6 2 | 6 2 |
| | 6 2 | 3+4 4 | 3+4 4 | 5 2 | 3+4 4 | 3+4 4 | 5 2 | 5 2 |
| | 5 2 | 3+4 4 | 3+4 4 | 6 2 | 3+4 4 | 3+4 4 | 6 2 | 6 2 |
| | 6 2 | 3+4 4 | 3+4 4 | 5 2 | 3+4 4 | 3+4 4 | 5 2 | 5 2 |
| | 5 2 | 3+4 4 | 3+4 4 | 6 2 | 3+4 4 | 3+4 4 | 6 2 | 6 2 |
| | 5 2 | 3+4 4 | 3+4 4 | 5 2 | 3+4 4 | 3+4 4 | 6 2 | 6 2 |
| | 6 2 | 3+4 4 | 3+4 4 | 6 2 | 3+4 4 | 3+4 4 | 6 2 | 5 2 |
| | 5 2 | 3+4 4 | 3+4 4 | 6 2 | 3+4 4 | 3+4 4 | 5 2 | 6 2 |
| | 6 2 | 3+4 4 | 3+4 4 | 5 2 | 3+4 4 | 3+4 4 | 5 2 | 5 2 |
| | 5 2 | 3+4 4 | 3+4 4 | 5 2 | 3+4 4 | 3+4 4 | 6 2 | 6 2 |
| | 6 2 | 3+4 4 | 3+4 4 | 6 2 | 3+4 4 | 3+4 4 | 5 2 | 5 2 |
| | 5 2 | 3+4 4 | 3+4 4 | 5 2 | 3+4 4 | 3+4 4 | 5 2 | 5 2 |
| | 5 2 | 3+4 4 | 3+4 4 | 5 2 | 3+4 4 | 3+4 4 | 5 2 | 6 2 |
| | 5 2 | 3+4 4 | 3+4 4 | 6 2 | 3+4 4 | 3+4 4 | 6 2 | 5 2 |
| | 6 2 | 3+4 4 | 3+4 4 | 5 2 | 3+4 4 | 3+4 4 | 6 2 | 6 2 |
| | 5 2 | 3+4 4 | 3+4 4 | 6 2 | 3+4 4 | 3+4 4 | 5 2 | 6 2 |
| | 6 2 | 3+4 4 | 3+4 4 | 5 2 | 3+4 4 | 3+4 4 | 6 2 | 6 2 |
| | 5 2 | 3+4 4 | 3+4 4 | 6 2 | 3+4 4 | 3+4 4 | 5 2 | 5 2 |
| | 6 2 | 3+4 4 | 3+4 4 | 6 2 | 3+4 4 | 3+4 4 | 5 2 | 5 2 |

Continued on next column

**TABLE XXIX– continued**

| TQC | -GM | -R | -S | -T | -U | -X | -Y | -Z |
|---|---|---|---|---|---|---|---|---|
| | 6 2 | 3+4 4 | 3+4 4 | 5 2 | 3+4 4 | 3+4 4 | 6 2 | 6 2 |
| | 5 2 | 3+4 4 | 3+4 4 | 6 2 | 3+4 4 | 3+4 4 | 5 2 | 5 2 |
| | 5 2 | 3+4 4 | 3+4 4 | 5 2 | 3+4 4 | 3+4 4 | 6 2 | 5 2 |
| | 5 2 | 3+4 4 | 3+4 4 | 6 2 | 3+4 4 | 3+4 4 | 6 2 | 6 2 |
| | 6 2 | 3+4 4 | 3+4 4 | 6 2 | 3+4 4 | 3+4 4 | 6 2 | 6 2 |
| | 6 2 | 3+4 4 | 3+4 4 | 5 2 | 3+4 4 | 3+4 4 | 5 2 | 5 2 |
| | 6 2 | 3+4 4 | 3+4 4 | 5 2 | 3+4 4 | 3+4 4 | 5 2 | 5 2 |
| | 6 2 | 3+4 4 | 3+4 4 | 6 2 | 3+4 4 | 3+4 4 | 5 2 | 6 2 |
| | 5 2 | 3+4 4 | 3+4 4 | 5 2 | 3+4 4 | 3+4 4 | 6 2 | 6 2 |
| | 6 2 | 3+4 4 | 3+4 4 | 6 2 | 3+4 4 | 3+4 4 | 6 2 | 5 2 |
| | 5 2 | 3+4 4 | 3+4 4 | 5 2 | 3+4 4 | 3+4 4 | 5 2 | 5 2 |
| | 5 2 | 3+4 4 | 3+4 4 | 6 2 | 3+4 4 | 3+4 4 | 6 2 | 6 2 |
| | 6 2 | | | 6 2 | | | 6 2 | 6 2 |
| | 5 2 | | | 5 2 | | | 6 2 | 5 2 |
| | 5 2 | | | 5 2 | | | 5 2 | 6 2 |
| | 6 2 | | | 6 2 | | | 5 2 | 5 2 |
| | 6 2 | | | 6 2 | | | 5 2 | 5 2 |
| | 6 2 | | | 5 2 | | | 5 2 | 6 2 |
| | 5 2 | | | 6 2 | | | 5 2 | 5 2 |
| | 6 2 | | | 5 2 | | | 6 2 | 6 2 |
| | 5 2 | | | 6 2 | | | 6 2 | 6 2 |
| | 6 2 | | | 5 2 | | | 6 2 | 6 2 |
| | 6 2 | | | 5 2 | | | 6 2 | 5 2 |
| | 6 2 | | | 6 2 | | | 6 2 | 6 2 |
| | 5 2 | | | 5 2 | | | 5 2 | 5 2 |
| | 6 2 | | | 6 2 | | | 5 2 | 5 2 |
| | 5 2 | | | 6 2 | | | 5 2 | 6 2 |
| | 5 2 | | | 5 2 | | | 6 2 | 5 2 |
| | 5 2 | | | 5 2 | | | 5 2 | 5 2 |
| | 6 2 | | | 6 2 | | | 5 2 | 6 2 |
| | 5 2 | | | 6 2 | | | 6 2 | 6 2 |
| | 6 2 | | | 5 2 | | | 6 2 | 5 2 |
| | 6 2 | | | 5 2 | | | 6 2 | 6 2 |
| | 5 2 | | | 6 2 | | | 6 2 | 5 2 |
| | 5 2 | | | 6 2 | | | 5 2 | 5 2 |
| | 5 2 | | | 5 2 | | | 6 2 | 6 2 |
| | 6 2 | | | 5 2 | | | 5 2 | 6 2 |
| | 6 2 | | | 6 2 | | | 5 2 | 6 2 |
| | 5 2 | | | 5 2 | | | 5 2 | 5 2 |
| | 6 2 | | | 6 2 | | | 6 2 | 6 2 |
| | 5 2 | | | 6 2 | | | 5 2 | 5 2 |
| | 5 2 | | | 5 2 | | | 6 2 | 5 2 |
| | 5 2 | | | 6 2 | | | 5 2 | 5 2 |
| | 6 2 | | | 5 2 | | | 5 2 | 6 2 |
| | 5 2 | | | 6 2 | | | 6 2 | 5 2 |
| | 6 2 | | | 5 2 | | | 5 2 | 6 2 |
| | 5 2 | | | 5 2 | | | 6 2 | 5 2 |
| | 5 2 | | | 6 2 | | | 6 2 | 5 2 |
| | 5 2 | | | 5 2 | | | 5 2 | 6 2 |
| | 6 2 | | | 6 2 | | | 6 2 | 6 2 |
| | 5 2 | | | 6 2 | | | 5 2 | 5 2 |
| | 6 2 | | | 5 2 | | | 5 2 | 6 2 |
| | 6 2 | | | 5 2 | | | 6 2 | 6 2 |
| | 6 2 | | | 6 2 | | | 5 2 | 5 2 |
| | 5 2 | | | 5 2 | | | 6 2 | 5 2 |
| | 6 2 | | | 6 2 | | | 6 2 | 6 2 |
| | 5 2 | | | 6 2 | | | 5 2 | 6 2 |
| | 6 2 | | | 5 2 | | | 6 2 | 5 2 |
| | 5 2 | | | 6 2 | | | 5 2 | 6 2 |
| | 6 2 | | | 6 2 | | | 5 2 | 5 2 |
| | 5 2 | | | 5 2 | | | 6 2 | 6 2 |

Continued on next column

**TABLE XXIX– continued**

| TQC | ICSD | NoE | SF | | | | | |
|---|---|---|---|---|---|---|---|---|
| | 6 2 | | | 5 2 | | | 5 2 | 5 2 |
| | 5 2 | | | 5 2 | | | 6 2 | 6 2 |
| | 6 2 | | | 6 2 | | | 6 2 | 5 2 |
| 51.2.2 | 106800 | 68 | In1La1Ni2 | | | | | |
| | -GM | -R | -S | -T | -U | -X | -Y | -Z |
| | 5 2 | 3+4 4 | 3+4 4 | 6 2 | 3+4 4 | 3+4 4 | 5 2 | 6 2 |
| | 6 2 | 3+4 4 | 3+4 4 | 5 2 | 3+4 4 | 3+4 4 | 6 2 | 5 2 |
| | 5 2 | 3+4 4 | 3+4 4 | 6 2 | 3+4 4 | 3+4 4 | 5 2 | 6 2 |
| | 6 2 | 3+4 4 | 3+4 4 | 5 2 | 3+4 4 | 3+4 4 | 6 2 | 5 2 |
| | 5 2 | 3+4 4 | 3+4 4 | 5 2 | 3+4 4 | 3+4 4 | 6 2 | 6 2 |
| | 6 2 | 3+4 4 | 3+4 4 | 6 2 | 3+4 4 | 3+4 4 | 5 2 | 5 2 |
| | 5 2 | 3+4 4 | 3+4 4 | 6 2 | 3+4 4 | 3+4 4 | 5 2 | 6 2 |
| | 6 2 | 3+4 4 | 3+4 4 | 5 2 | 3+4 4 | 3+4 4 | 6 2 | 5 2 |
| | 5 2 | 3+4 4 | 3+4 4 | 5 2 | 3+4 4 | 3+4 4 | 6 2 | 6 2 |
| | 6 2 | 3+4 4 | 3+4 4 | 6 2 | 3+4 4 | 3+4 4 | 5 2 | 5 2 |
| | 5 2 | 3+4 4 | 3+4 4 | 6 2 | 3+4 4 | 3+4 4 | 5 2 | 5 2 |
| | 5 2 | 3+4 4 | 3+4 4 | 5 2 | 3+4 4 | 3+4 4 | 6 2 | 5 2 |
| | 5 2 | 3+4 4 | 3+4 4 | 6 2 | 3+4 4 | 3+4 4 | 6 2 | 5 2 |
| | 5 2 | 3+4 4 | 3+4 4 | 6 2 | 3+4 4 | 3+4 4 | 6 2 | 6 2 |
| | 5 2 | 3+4 4 | 3+4 4 | 6 2 | 3+4 4 | 3+4 4 | 6 2 | 5 2 |
| | 5 2 | 3+4 4 | 3+4 4 | 5 2 | 3+4 4 | 3+4 4 | 6 2 | 5 2 |
| | 6 2 | 3+4 4 | 3+4 4 | 5 2 | 3+4 4 | 3+4 4 | 6 2 | 5 2 |
| | 6 2 | | | 6 2 | | | 5 2 | 6 2 |
| | 6 2 | | | 5 2 | | | 5 2 | 6 2 |
| | 5 2 | | | 6 2 | | | 6 2 | 5 2 |
| | 5 2 | | | 5 2 | | | 6 2 | 5 2 |
| | 5 2 | | | 6 2 | | | 6 2 | 6 2 |
| | 6 2 | | | 6 2 | | | 5 2 | 5 2 |
| | 6 2 | | | 5 2 | | | 6 2 | 6 2 |
| | 5 2 | | | 6 2 | | | 5 2 | 6 2 |
| | 5 2 | | | 6 2 | | | 5 2 | 5 2 |
| | 5 2 | | | 6 2 | | | 6 2 | 5 2 |
| | 5 2 | | | 5 2 | | | 6 2 | 5 2 |
| | 5 2 | | | 6 2 | | | 6 2 | 5 2 |
| | 6 2 | | | 5 2 | | | 6 2 | 5 2 |
| | 5 2 | | | 5 2 | | | 6 2 | 6 2 |
| | 5 2 | | | 6 2 | | | 6 2 | 5 2 |
| | 5 2 | | | 6 2 | | | 6 2 | 5 2 |
| | 5 2 | | | 6 2 | | | 5 2 | 6 2 |
| 51.2.3 | 370040 | 80 | Co2In1Tb1 | | | | | |
| | -GM | -R | -S | -T | -U | -X | -Y | -Z |
| | 5 2 | 3+4 4 | 3+4 4 | 5 2 | 3+4 4 | 3+4 4 | 6 2 | 6 2 |
| | 6 2 | 3+4 4 | 3+4 4 | 6 2 | 3+4 4 | 3+4 4 | 5 2 | 5 2 |
| | 5 2 | 3+4 4 | 3+4 4 | 5 2 | 3+4 4 | 3+4 4 | 6 2 | 6 2 |
| | 6 2 | 3+4 4 | 3+4 4 | 6 2 | 3+4 4 | 3+4 4 | 5 2 | 5 2 |
| | 5 2 | 3+4 4 | 3+4 4 | 5 2 | 3+4 4 | 3+4 4 | 6 2 | 6 2 |
| | 6 2 | 3+4 4 | 3+4 4 | 6 2 | 3+4 4 | 3+4 4 | 5 2 | 5 2 |
| | 5 2 | 3+4 4 | 3+4 4 | 5 2 | 3+4 4 | 3+4 4 | 6 2 | 6 2 |
| | 6 2 | 3+4 4 | 3+4 4 | 6 2 | 3+4 4 | 3+4 4 | 5 2 | 5 2 |
| | 5 2 | 3+4 4 | 3+4 4 | 6 2 | 3+4 4 | 3+4 4 | 5 2 | 6 2 |
| | 6 2 | 3+4 4 | 3+4 4 | 5 2 | 3+4 4 | 3+4 4 | 6 2 | 5 2 |
| | 5 2 | 3+4 4 | 3+4 4 | 5 2 | 3+4 4 | 3+4 4 | 6 2 | 5 2 |
| | 5 2 | 3+4 4 | 3+4 4 | 6 2 | 3+4 4 | 3+4 4 | 5 2 | 5 2 |
| | 5 2 | 3+4 4 | 3+4 4 | 5 2 | 3+4 4 | 3+4 4 | 5 2 | 6 2 |
| | 5 2 | 3+4 4 | 3+4 4 | 5 2 | 3+4 4 | 3+4 4 | 5 2 | 5 2 |
| | 5 2 | 3+4 4 | 3+4 4 | 5 2 | 3+4 4 | 3+4 4 | 5 2 | 5 2 |
| | 5 2 | 3+4 4 | 3+4 4 | 6 2 | 3+4 4 | 3+4 4 | 5 2 | 5 2 |
| | 6 2 | 3+4 4 | 3+4 4 | 5 2 | 3+4 4 | 3+4 4 | 5 2 | 5 2 |
| | 5 2 | 3+4 4 | 3+4 4 | 6 2 | 3+4 4 | 3+4 4 | 5 2 | 5 2 |
| | 6 2 | 3+4 4 | 3+4 4 | 5 2 | 3+4 4 | 3+4 4 | 6 2 | 6 2 |
| | 5 2 | 3+4 4 | 3+4 4 | 6 2 | 3+4 4 | 3+4 4 | 5 2 | 6 2 |

Continued on next column

| TQC | ICSD | NoE | SF | | | | | |
|---|---|---|---|---|---|---|---|---|
| | 6 2 | | | 5 2 | | | 5 2 | 5 2 |
| | 5 2 | | | 5 2 | | | 5 2 | 6 2 |
| | 6 2 | | | 6 2 | | | 6 2 | 5 2 |
| | 6 2 | | | 6 2 | | | 6 2 | 6 2 |
| | 5 2 | | | 5 2 | | | 6 2 | 6 2 |
| | 6 2 | | | 6 2 | | | 6 2 | 6 2 |
| | 5 2 | | | 5 2 | | | 5 2 | 5 2 |
| | 6 2 | | | 5 2 | | | 5 2 | 5 2 |
| | 5 2 | | | 5 2 | | | 5 2 | 6 2 |
| | 6 2 | | | 6 2 | | | 6 2 | 5 2 |
| | 5 2 | | | 6 2 | | | 6 2 | 6 2 |
| | 5 2 | | | 5 2 | | | 5 2 | 5 2 |
| | 5 2 | | | 6 2 | | | 6 2 | 6 2 |
| | 5 2 | | | 5 2 | | | 5 2 | 5 2 |
| | 6 2 | | | 5 2 | | | 5 2 | 5 2 |
| | 5 2 | | | 6 2 | | | 5 2 | 5 2 |
| | 5 2 | | | 5 2 | | | 5 2 | 5 2 |
| | 6 2 | | | 5 2 | | | 6 2 | 5 2 |
| | 6 2 | | | 6 2 | | | 5 2 | 6 2 |
| | 5 2 | | | 5 2 | | | 6 2 | 6 2 |
| 51.2.4 | 421074 | 84 | Ca2Ge2Ni3 | | | | | |
| | -GM | -R | -S | -T | -U | -X | -Y | -Z |
| | 5 2 | 3+4 4 | 3+4 4 | 5 2 | 3+4 4 | 3+4 4 | 6 2 | 6 2 |
| | 5 2 | 3+4 4 | 3+4 4 | 5 2 | 3+4 4 | 3+4 4 | 6 2 | 6 2 |
| | 6 2 | 3+4 4 | 3+4 4 | 6 2 | 3+4 4 | 3+4 4 | 5 2 | 5 2 |
| | 5 2 | 3+4 4 | 3+4 4 | 5 2 | 3+4 4 | 3+4 4 | 6 2 | 6 2 |
| | 6 2 | 3+4 4 | 3+4 4 | 6 2 | 3+4 4 | 3+4 4 | 5 2 | 5 2 |
| | 6 2 | 3+4 4 | 3+4 4 | 6 2 | 3+4 4 | 3+4 4 | 5 2 | 5 2 |
| | 6 2 | 3+4 4 | 3+4 4 | 6 2 | 3+4 4 | 3+4 4 | 5 2 | 5 2 |
| | 5 2 | 3+4 4 | 3+4 4 | 5 2 | 3+4 4 | 3+4 4 | 6 2 | 6 2 |
| | 5 2 | 3+4 4 | 3+4 4 | 6 2 | 3+4 4 | 3+4 4 | 6 2 | 5 2 |
| | 6 2 | 3+4 4 | 3+4 4 | 5 2 | 3+4 4 | 3+4 4 | 5 2 | 5 2 |
| | 6 2 | 3+4 4 | 3+4 4 | 6 2 | 3+4 4 | 3+4 4 | 6 2 | 5 2 |
| | 6 2 | 3+4 4 | 3+4 4 | 5 2 | 3+4 4 | 3+4 4 | 5 2 | 5 2 |
| | 6 2 | 3+4 4 | 3+4 4 | 6 2 | 3+4 4 | 3+4 4 | 5 2 | 6 2 |
| | 5 2 | 3+4 4 | 3+4 4 | 6 2 | 3+4 4 | 3+4 4 | 5 2 | 5 2 |
| | 5 2 | 3+4 4 | 3+4 4 | 6 2 | 3+4 4 | 3+4 4 | 5 2 | 6 2 |
| | 5 2 | 3+4 4 | 3+4 4 | 6 2 | 3+4 4 | 3+4 4 | 6 2 | 6 2 |
| | 5 2 | 3+4 4 | 3+4 4 | 5 2 | 3+4 4 | 3+4 4 | 6 2 | 6 2 |
| | 6 2 | 3+4 4 | 3+4 4 | 6 2 | 3+4 4 | 3+4 4 | 5 2 | 5 2 |
| | 5 2 | 3+4 4 | 3+4 4 | 5 2 | 3+4 4 | 3+4 4 | 6 2 | 6 2 |
| | 6 2 | 3+4 4 | 3+4 4 | 5 2 | 3+4 4 | 3+4 4 | 5 2 | 5 2 |
| | 5 2 | 3+4 4 | 3+4 4 | 6 2 | 3+4 4 | 3+4 4 | 6 2 | 6 2 |
| | 6 2 | | | 5 2 | | | 6 2 | 5 2 |
| | 5 2 | | | 5 2 | | | 5 2 | 6 2 |
| | 6 2 | | | 5 2 | | | 5 2 | 5 2 |
| | 6 2 | | | 6 2 | | | 6 2 | 5 2 |
| | 5 2 | | | 6 2 | | | 5 2 | 6 2 |
| | 6 2 | | | 5 2 | | | 6 2 | 5 2 |
| | 5 2 | | | 5 2 | | | 5 2 | 6 2 |
| | 6 2 | | | 6 2 | | | 6 2 | 6 2 |
| | 6 2 | | | 5 2 | | | 6 2 | 5 2 |
| | 5 2 | | | 6 2 | | | 5 2 | 5 2 |
| | 6 2 | | | 6 2 | | | 6 2 | 6 2 |
| | 5 2 | | | 5 2 | | | 5 2 | 6 2 |
| | 6 2 | | | 6 2 | | | 6 2 | 6 2 |
| | 5 2 | | | 6 2 | | | 5 2 | 5 2 |
| | 5 2 | | | 5 2 | | | 6 2 | 6 2 |
| | 6 2 | | | 5 2 | | | 6 2 | 6 2 |
| | 5 2 | | | 5 2 | | | 5 2 | 5 2 |
| | 5 2 | | | 6 2 | | | 5 2 | 5 2 |

Continued on next column

**TABLE XXIX– continued**

| TQC | ICSD | NoE | SF | | | | | |
|---|---|---|---|---|---|---|---|---|
| | 6 2 | | | 5 2 | | | 6 2 | 6 2 |
| | 6 2 | | | 5 2 | | | 5 2 | 5 2 |
| | 5 2 | | | 6 2 | | | 5 2 | 6 2 |
| 51.2.5 | 194877 | 96 | C3Al1Y3 | | | | | |
| | -GM | -R | -S | -T | -U | -X | -Y | -Z |
| | 5 2 | 3+4 4 | 3+4 4 | 6 2 | 3+4 4 | 3+4 4 | 5 2 | 6 2 |
| | 6 2 | 3+4 4 | 3+4 4 | 5 2 | 3+4 4 | 3+4 4 | 6 2 | 5 2 |
| | 5 2 | 3+4 4 | 3+4 4 | 6 2 | 3+4 4 | 3+4 4 | 6 2 | 5 2 |
| | 6 2 | 3+4 4 | 3+4 4 | 5 2 | 3+4 4 | 3+4 4 | 5 2 | 6 2 |
| | 5 2 | 3+4 4 | 3+4 4 | 6 2 | 3+4 4 | 3+4 4 | 6 2 | 5 2 |
| | 6 2 | 3+4 4 | 3+4 4 | 5 2 | 3+4 4 | 3+4 4 | 5 2 | 6 2 |
| | 5 2 | 3+4 4 | 3+4 4 | 6 2 | 3+4 4 | 3+4 4 | 5 2 | 6 2 |
| | 6 2 | 3+4 4 | 3+4 4 | 5 2 | 3+4 4 | 3+4 4 | 6 2 | 5 2 |
| | 5 2 | 3+4 4 | 3+4 4 | 6 2 | 3+4 4 | 3+4 4 | 6 2 | 5 2 |
| | 6 2 | 3+4 4 | 3+4 4 | 5 2 | 3+4 4 | 3+4 4 | 5 2 | 6 2 |
| | 5 2 | 3+4 4 | 3+4 4 | 6 2 | 3+4 4 | 3+4 4 | 6 2 | 5 2 |
| | 6 2 | 3+4 4 | 3+4 4 | 5 2 | 3+4 4 | 3+4 4 | 5 2 | 6 2 |
| | 5 2 | 3+4 4 | 3+4 4 | 6 2 | 3+4 4 | 3+4 4 | 5 2 | 6 2 |
| | 6 2 | 3+4 4 | 3+4 4 | 5 2 | 3+4 4 | 3+4 4 | 6 2 | 5 2 |
| | 6 2 | 3+4 4 | 3+4 4 | 6 2 | 3+4 4 | 3+4 4 | 5 2 | 5 2 |
| | 5 2 | 3+4 4 | 3+4 4 | 5 2 | 3+4 4 | 3+4 4 | 6 2 | 6 2 |
| | 6 2 | 3+4 4 | 3+4 4 | 5 2 | 3+4 4 | 3+4 4 | 5 2 | 5 2 |
| | 5 2 | 3+4 4 | 3+4 4 | 6 2 | 3+4 4 | 3+4 4 | 6 2 | 6 2 |
| | 5 2 | 3+4 4 | 3+4 4 | 6 2 | 3+4 4 | 3+4 4 | 5 2 | 5 2 |
| | 6 2 | 3+4 4 | 3+4 4 | 5 2 | 3+4 4 | 3+4 4 | 6 2 | 6 2 |
| | 5 2 | 3+4 4 | 3+4 4 | 6 2 | 3+4 4 | 3+4 4 | 6 2 | 5 2 |
| | 6 2 | 3+4 4 | 3+4 4 | 5 2 | 3+4 4 | 3+4 4 | 5 2 | 6 2 |
| | 5 2 | 3+4 4 | 3+4 4 | 6 2 | 3+4 4 | 3+4 4 | 6 2 | 6 2 |
| | 6 2 | 3+4 4 | 3+4 4 | 5 2 | 3+4 4 | 3+4 4 | 5 2 | 5 2 |
| | 5 2 | | | 5 2 | | | 6 2 | 6 2 |
| | 5 2 | | | 5 2 | | | 6 2 | 6 2 |
| | 5 2 | | | 5 2 | | | 5 2 | 5 2 |
| | 6 2 | | | 6 2 | | | 6 2 | 6 2 |
| | 6 2 | | | 6 2 | | | 5 2 | 5 2 |
| | 6 2 | | | 6 2 | | | 5 2 | 5 2 |
| | 6 2 | | | 6 2 | | | 5 2 | 5 2 |
| | 5 2 | | | 5 2 | | | 5 2 | 5 2 |
| | 6 2 | | | 6 2 | | | 5 2 | 6 2 |
| | 6 2 | | | 6 2 | | | 5 2 | 5 2 |
| | 5 2 | | | 6 2 | | | 5 2 | 6 2 |
| | 5 2 | | | 5 2 | | | 6 2 | 6 2 |
| | 6 2 | | | 6 2 | | | 6 2 | 5 2 |
| | 6 2 | | | 5 2 | | | 6 2 | 5 2 |
| | 5 2 | | | 6 2 | | | 6 2 | 5 2 |
| | 6 2 | | | 5 2 | | | 5 2 | 6 2 |
| | 5 2 | | | 5 2 | | | 5 2 | 5 2 |
| | 6 2 | | | 5 2 | | | 6 2 | 6 2 |
| | 5 2 | | | 5 2 | | | 6 2 | 5 2 |
| | 6 2 | | | 6 2 | | | 6 2 | 6 2 |
| | 5 2 | | | 5 2 | | | 5 2 | 5 2 |
| | 5 2 | | | 6 2 | | | 5 2 | 6 2 |
| | 5 2 | | | 6 2 | | | 5 2 | 5 2 |
| | 6 2 | | | 5 2 | | | 6 2 | 6 2 |
| 51.2.6 | 69767 | 48 | C1B2U1 | | | | | |
| | -GM | -R | -S | -T | -U | -X | -Y | -Z |
| | 5 2 | 3+4 4 | 3+4 4 | 6 2 | 3+4 4 | 3+4 4 | 5 2 | 6 2 |
| | 6 2 | 3+4 4 | 3+4 4 | 5 2 | 3+4 4 | 3+4 4 | 6 2 | 5 2 |
| | 5 2 | 3+4 4 | 3+4 4 | 6 2 | 3+4 4 | 3+4 4 | 5 2 | 6 2 |
| | 6 2 | 3+4 4 | 3+4 4 | 5 2 | 3+4 4 | 3+4 4 | 6 2 | 5 2 |
| | 5 2 | 3+4 4 | 3+4 4 | 5 2 | 3+4 4 | 3+4 4 | 6 2 | 6 2 |
| | 5 2 | 3+4 4 | 3+4 4 | 6 2 | 3+4 4 | 3+4 4 | 5 2 | 5 2 |

Continued on next column

**TABLE XXIX– continued**

| TQC | ICSD | NoE | SF | | | | | |
|---|---|---|---|---|---|---|---|---|
| | 6 2 | 3+4 4 | 3+4 4 | 6 2 | 3+4 4 | 3+4 4 | 5 2 | 6 2 |
| | 6 2 | 3+4 4 | 3+4 4 | 5 2 | 3+4 4 | 3+4 4 | 6 2 | 5 2 |
| | 5 2 | 3+4 4 | 3+4 4 | 5 2 | 3+4 4 | 3+4 4 | 6 2 | 6 2 |
| | 5 2 | 3+4 4 | 3+4 4 | 5 2 | 3+4 4 | 3+4 4 | 6 2 | 6 2 |
| | 6 2 | 3+4 4 | 3+4 4 | 6 2 | 3+4 4 | 3+4 4 | 5 2 | 5 2 |
| | 6 2 | 3+4 4 | 3+4 4 | 6 2 | 3+4 4 | 3+4 4 | 6 2 | 5 2 |
| | 6 2 | | | 5 2 | | | 5 2 | 5 2 |
| | 5 2 | | | 6 2 | | | 5 2 | 6 2 |
| | 5 2 | | | 6 2 | | | 6 2 | 5 2 |
| | 6 2 | | | 6 2 | | | 5 2 | 5 2 |
| | 5 2 | | | 6 2 | | | 6 2 | 5 2 |
| | 6 2 | | | 5 2 | | | 6 2 | 5 2 |
| | 6 2 | | | 6 2 | | | 5 2 | 6 2 |
| | 6 2 | | | 5 2 | | | 5 2 | 5 2 |
| | 5 2 | | | 6 2 | | | 5 2 | 6 2 |
| | 5 2 | | | 6 2 | | | 5 2 | 5 2 |
| | 6 2 | | | 5 2 | | | 5 2 | 6 2 |
| | 6 2 | | | 6 2 | | | 6 2 | 6 2 |
| 51.2.7 | 247569 | 148 | Ba1Fe2Gd1O5 | | | | | |
| | -GM | -R | -S | -T | -U | -X | -Y | -Z |
| | 5 2 | 3+4 4 | 3+4 4 | 6 2 | 3+4 4 | 3+4 4 | 5 2 | 6 2 |
| | 5 2 | 3+4 4 | 3+4 4 | 6 2 | 3+4 4 | 3+4 4 | 5 2 | 6 2 |
| | 5 2 | 3+4 4 | 3+4 4 | 5 2 | 3+4 4 | 3+4 4 | 5 2 | 5 2 |
| | 5 2 | 3+4 4 | 3+4 4 | 5 2 | 3+4 4 | 3+4 4 | 5 2 | 5 2 |
| | 6 2 | 3+4 4 | 3+4 4 | 5 2 | 3+4 4 | 3+4 4 | 6 2 | 5 2 |
| | 6 2 | 3+4 4 | 3+4 4 | 5 2 | 3+4 4 | 3+4 4 | 6 2 | 5 2 |
| | 6 2 | 3+4 4 | 3+4 4 | 5 2 | 3+4 4 | 3+4 4 | 6 2 | 5 2 |
| | 6 2 | 3+4 4 | 3+4 4 | 5 2 | 3+4 4 | 3+4 4 | 6 2 | 5 2 |
| | 6 2 | 3+4 4 | 3+4 4 | 5 2 | 3+4 4 | 3+4 4 | 6 2 | 5 2 |
| | 5 2 | 3+4 4 | 3+4 4 | 5 2 | 3+4 4 | 3+4 4 | 6 2 | 6 2 |
| | 6 2 | 3+4 4 | 3+4 4 | 6 2 | 3+4 4 | 3+4 4 | 6 2 | 6 2 |
| | 6 2 | 3+4 4 | 3+4 4 | 6 2 | 3+4 4 | 3+4 4 | 5 2 | 5 2 |
| | 5 2 | 3+4 4 | 3+4 4 | 5 2 | 3+4 4 | 3+4 4 | 5 2 | 5 2 |
| | 5 2 | 3+4 4 | 3+4 4 | 6 2 | 3+4 4 | 3+4 4 | 5 2 | 6 2 |
| | 5 2 | 3+4 4 | 3+4 4 | 6 2 | 3+4 4 | 3+4 4 | 5 2 | 6 2 |
| | 5 2 | 3+4 4 | 3+4 4 | 6 2 | 3+4 4 | 3+4 4 | 5 2 | 6 2 |
| | 6 2 | 3+4 4 | 3+4 4 | 5 2 | 3+4 4 | 3+4 4 | 6 2 | 5 2 |
| | 6 2 | 3+4 4 | 3+4 4 | 5 2 | 3+4 4 | 3+4 4 | 6 2 | 5 2 |
| | 6 2 | 3+4 4 | 3+4 4 | 5 2 | 3+4 4 | 3+4 4 | 6 2 | 5 2 |
| | 6 2 | 3+4 4 | 3+4 4 | 5 2 | 3+4 4 | 3+4 4 | 6 2 | 5 2 |
| | 6 2 | 3+4 4 | 3+4 4 | 6 2 | 3+4 4 | 3+4 4 | 6 2 | 6 2 |
| | 6 2 | 3+4 4 | 3+4 4 | 6 2 | 3+4 4 | 3+4 4 | 6 2 | 6 2 |
| | 6 2 | 3+4 4 | 3+4 4 | 6 2 | 3+4 4 | 3+4 4 | 6 2 | 6 2 |
| | 6 2 | 3+4 4 | 3+4 4 | 6 2 | 3+4 4 | 3+4 4 | 6 2 | 6 2 |
| | 6 2 | 3+4 4 | 3+4 4 | 6 2 | 3+4 4 | 3+4 4 | 6 2 | 6 2 |
| | 6 2 | 3+4 4 | 3+4 4 | 6 2 | 3+4 4 | 3+4 4 | 6 2 | 6 2 |
| | 5 2 | 3+4 4 | 3+4 4 | 6 2 | 3+4 4 | 3+4 4 | 5 2 | 5 2 |
| | 6 2 | 3+4 4 | 3+4 4 | 5 2 | 3+4 4 | 3+4 4 | 6 2 | 6 2 |
| | 5 2 | 3+4 4 | 3+4 4 | 5 2 | 3+4 4 | 3+4 4 | 5 2 | 6 2 |
| | 5 2 | 3+4 4 | 3+4 4 | 6 2 | 3+4 4 | 3+4 4 | 6 2 | 6 2 |
| | 5 2 | 3+4 4 | 3+4 4 | 6 2 | 3+4 4 | 3+4 4 | 5 2 | 6 2 |
| | 5 2 | 3+4 4 | 3+4 4 | 5 2 | 3+4 4 | 3+4 4 | 5 2 | 5 2 |
| | 6 2 | 3+4 4 | 3+4 4 | 6 2 | 3+4 4 | 3+4 4 | 5 2 | 6 2 |
| | 5 2 | 3+4 4 | 3+4 4 | 5 2 | 3+4 4 | 3+4 4 | 6 2 | 5 2 |
| | 6 2 | 3+4 4 | 3+4 4 | 5 2 | 3+4 4 | 3+4 4 | 5 2 | 5 2 |
| | 6 2 | 3+4 4 | 3+4 4 | 6 2 | 3+4 4 | 3+4 4 | 6 2 | 5 2 |
| | 5 2 | 3+4 4 | 3+4 4 | 5 2 | 3+4 4 | 3+4 4 | 5 2 | 6 2 |
| | 5 2 | | | 6 2 | | | 5 2 | 6 2 |
| | 6 2 | | | 6 2 | | | 6 2 | 6 2 |
| | 5 2 | | | 6 2 | | | 5 2 | 5 2 |
| | 5 2 | | | 5 2 | | | 6 2 | 5 2 |

Continued on next column

**TABLE XXIX– continued**

| TQC | ICSD | NoE | SF | | | | | |
|---|---|---|---|---|---|---|---|---|
| | 5 2 | | | 5 2 | | | 6 2 | 6 2 |
| | 6 2 | | | 5 2 | | | 5 2 | 5 2 |
| | 5 2 | | | 5 2 | | | 6 2 | 6 2 |
| | 6 2 | | | 5 2 | | | 5 2 | 5 2 |
| | 6 2 | | | 6 2 | | | 6 2 | 5 2 |
| | 5 2 | | | 5 2 | | | 5 2 | 5 2 |
| | 6 2 | | | 6 2 | | | 6 2 | 6 2 |
| | 6 2 | | | 6 2 | | | 5 2 | 6 2 |
| | 6 2 | | | 5 2 | | | 6 2 | 6 2 |
| | 6 2 | | | 6 2 | | | 5 2 | 5 2 |
| | 6 2 | | | 6 2 | | | 6 2 | 6 2 |
| | 5 2 | | | 5 2 | | | 6 2 | 5 2 |
| | 5 2 | | | 6 2 | | | 6 2 | 6 2 |
| | 6 2 | | | 6 2 | | | 5 2 | 5 2 |
| | 6 2 | | | 5 2 | | | 6 2 | 5 2 |
| | 5 2 | | | 5 2 | | | 5 2 | 5 2 |
| | 6 2 | | | 5 2 | | | 6 2 | 6 2 |
| | 6 2 | | | 6 2 | | | 6 2 | 5 2 |
| | 5 2 | | | 5 2 | | | 6 2 | 5 2 |
| | 6 2 | | | 5 2 | | | 6 2 | 5 2 |
| | 5 2 | | | 6 2 | | | 6 2 | 5 2 |
| | 6 2 | | | 5 2 | | | 6 2 | 6 2 |
| | 6 2 | | | 5 2 | | | 6 2 | 5 2 |
| | 6 2 | | | 5 2 | | | 6 2 | 5 2 |
| | 6 2 | | | 6 2 | | | 5 2 | 5 2 |
| | 6 2 | | | 5 2 | | | 5 2 | 5 2 |
| | 6 2 | | | 5 2 | | | 6 2 | 6 2 |
| | 5 2 | | | 6 2 | | | 5 2 | 6 2 |
| | 6 2 | | | 5 2 | | | 6 2 | 5 2 |
| | 5 2 | | | 5 2 | | | 6 2 | 5 2 |
| | 6 2 | | | 6 2 | | | 5 2 | 5 2 |
| | 5 2 | | | 5 2 | | | 6 2 | 5 2 |
| | 6 2 | | | 5 2 | | | 6 2 | 6 2 |
| 51.2.8 | 96391 | 148 | Ba1Fe2Gd1O4.999 | | | | | |
| | -GM | -R | -S | -T | -U | -X | -Y | -Z |
| | 5 2 | 3+4 4 | 3+4 4 | 6 2 | 3+4 4 | 3+4 4 | 5 2 | 6 2 |
| | 5 2 | 3+4 4 | 3+4 4 | 6 2 | 3+4 4 | 3+4 4 | 5 2 | 6 2 |
| | 6 2 | 3+4 4 | 3+4 4 | 5 2 | 3+4 4 | 3+4 4 | 6 2 | 5 2 |
| | 6 2 | 3+4 4 | 3+4 4 | 5 2 | 3+4 4 | 3+4 4 | 6 2 | 5 2 |
| | 5 2 | 3+4 4 | 3+4 4 | 5 2 | 3+4 4 | 3+4 4 | 5 2 | 5 2 |
| | 5 2 | 3+4 4 | 3+4 4 | 5 2 | 3+4 4 | 3+4 4 | 5 2 | 5 2 |
| | 6 2 | 3+4 4 | 3+4 4 | 5 2 | 3+4 4 | 3+4 4 | 6 2 | 5 2 |
| | 6 2 | 3+4 4 | 3+4 4 | 5 2 | 3+4 4 | 3+4 4 | 6 2 | 5 2 |
| | 6 2 | 3+4 4 | 3+4 4 | 5 2 | 3+4 4 | 3+4 4 | 6 2 | 5 2 |
| | 6 2 | 3+4 4 | 3+4 4 | 5 2 | 3+4 4 | 3+4 4 | 6 2 | 5 2 |
| | 5 2 | 3+4 4 | 3+4 4 | 6 2 | 3+4 4 | 3+4 4 | 5 2 | 6 2 |
| | 6 2 | 3+4 4 | 3+4 4 | 6 2 | 3+4 4 | 3+4 4 | 5 2 | 6 2 |
| | 5 2 | 3+4 4 | 3+4 4 | 6 2 | 3+4 4 | 3+4 4 | 6 2 | 5 2 |
| | 6 2 | 3+4 4 | 3+4 4 | 6 2 | 3+4 4 | 3+4 4 | 5 2 | 6 2 |
| | 5 2 | 3+4 4 | 3+4 4 | 5 2 | 3+4 4 | 3+4 4 | 5 2 | 5 2 |
| | 5 2 | 3+4 4 | 3+4 4 | 5 2 | 3+4 4 | 3+4 4 | 6 2 | 6 2 |
| | 6 2 | 3+4 4 | 3+4 4 | 5 2 | 3+4 4 | 3+4 4 | 6 2 | 5 2 |
| | 6 2 | 3+4 4 | 3+4 4 | 5 2 | 3+4 4 | 3+4 4 | 6 2 | 6 2 |
| | 6 2 | 3+4 4 | 3+4 4 | 6 2 | 3+4 4 | 3+4 4 | 6 2 | 5 2 |
| | 5 2 | 3+4 4 | 3+4 4 | 5 2 | 3+4 4 | 3+4 4 | 5 2 | 5 2 |
| | 6 2 | 3+4 4 | 3+4 4 | 5 2 | 3+4 4 | 3+4 4 | 6 2 | 5 2 |
| | 6 2 | 3+4 4 | 3+4 4 | 5 2 | 3+4 4 | 3+4 4 | 6 2 | 5 2 |
| | 6 2 | 3+4 4 | 3+4 4 | 5 2 | 3+4 4 | 3+4 4 | 6 2 | 5 2 |
| | 6 2 | 3+4 4 | 3+4 4 | 5 2 | 3+4 4 | 3+4 4 | 6 2 | 5 2 |
| | 6 2 | 3+4 4 | 3+4 4 | 5 2 | 3+4 4 | 3+4 4 | 6 2 | 5 2 |
| | 6 2 | 3+4 4 | 3+4 4 | 5 2 | 3+4 4 | 3+4 4 | 6 2 | 5 2 |

Continued on next column

**TABLE XXIX– continued**

| TQC | ICSD | NoE | SF | | | | | |
|---|---|---|---|---|---|---|---|---|
| | 6 2 | 3+4 4 | 3+4 4 | 5 2 | 3+4 4 | 3+4 4 | 6 2 | 5 2 |
| | 6 2 | 3+4 4 | 3+4 4 | 5 2 | 3+4 4 | 3+4 4 | 6 2 | 5 2 |
| | 6 2 | 3+4 4 | 3+4 4 | 5 2 | 3+4 4 | 3+4 4 | 6 2 | 5 2 |
| | 6 2 | 3+4 4 | 3+4 4 | 5 2 | 3+4 4 | 3+4 4 | 6 2 | 5 2 |
| | 6 2 | 3+4 4 | 3+4 4 | 5 2 | 3+4 4 | 3+4 4 | 6 2 | 5 2 |
| | 6 2 | 3+4 4 | 3+4 4 | 5 2 | 3+4 4 | 3+4 4 | 6 2 | 5 2 |
| | 6 2 | 3+4 4 | 3+4 4 | 5 2 | 3+4 4 | 3+4 4 | 6 2 | 5 2 |
| | 6 2 | 3+4 4 | 3+4 4 | 5 2 | 3+4 4 | 3+4 4 | 6 2 | 5 2 |
| | 6 2 | 3+4 4 | 3+4 4 | 6 2 | 3+4 4 | 3+4 4 | 6 2 | 6 2 |
| | 6 2 | 3+4 4 | 3+4 4 | 6 2 | 3+4 4 | 3+4 4 | 6 2 | 6 2 |
| | 6 2 | 3+4 4 | 3+4 4 | 6 2 | 3+4 4 | 3+4 4 | 6 2 | 6 2 |
| | 6 2 | | | 6 2 | | | 6 2 | 6 2 |
| | 6 2 | | | 6 2 | | | 6 2 | 6 2 |
| | 6 2 | | | 6 2 | | | 6 2 | 6 2 |
| | 5 2 | | | 6 2 | | | 5 2 | 6 2 |
| | 5 2 | | | 5 2 | | | 5 2 | 6 2 |
| | 5 2 | | | 6 2 | | | 6 2 | 5 2 |
| | 5 2 | | | 5 2 | | | 6 2 | 6 2 |
| | 5 2 | | | 6 2 | | | 5 2 | 6 2 |
| | 6 2 | | | 5 2 | | | 5 2 | 6 2 |
| | 5 2 | | | 6 2 | | | 5 2 | 5 2 |
| | 6 2 | | | 6 2 | | | 6 2 | 6 2 |
| | 6 2 | | | 5 2 | | | 5 2 | 5 2 |
| | 5 2 | | | 6 2 | | | 5 2 | 5 2 |
| | 5 2 | | | 5 2 | | | 5 2 | 6 2 |
| | 5 2 | | | 6 2 | | | 5 2 | 6 2 |
| | 5 2 | | | 5 2 | | | 6 2 | 6 2 |
| | 6 2 | | | 6 2 | | | 6 2 | 5 2 |
| | 6 2 | | | 5 2 | | | 5 2 | 6 2 |
| | 6 2 | | | 6 2 | | | 6 2 | 5 2 |
| | 6 2 | | | 5 2 | | | 6 2 | 5 2 |
| | 5 2 | | | 5 2 | | | 6 2 | 5 2 |
| | 5 2 | | | 5 2 | | | 6 2 | 5 2 |
| | 6 2 | | | 5 2 | | | 5 2 | 5 2 |
| | 5 2 | | | 6 2 | | | 6 2 | 5 2 |
| | 6 2 | | | 5 2 | | | 5 2 | 6 2 |
| | 6 2 | | | 6 2 | | | 5 2 | 6 2 |
| | 5 2 | | | 5 2 | | | 6 2 | 5 2 |
| | 6 2 | | | 6 2 | | | 5 2 | 6 2 |
| | 5 2 | | | 6 2 | | | 6 2 | 6 2 |
| | 6 2 | | | 6 2 | | | 5 2 | 6 2 |
| | 6 2 | | | 5 2 | | | 6 2 | 5 2 |
| | 6 2 | | | 6 2 | | | 6 2 | 5 2 |
| | 6 2 | | | 5 2 | | | 6 2 | 5 2 |
| | 5 2 | | | 6 2 | | | 5 2 | 6 2 |
| | 6 2 | | | 6 2 | | | 6 2 | 5 2 |
| | 5 2 | | | 6 2 | | | 6 2 | 5 2 |
| | 5 2 | | | 5 2 | | | 5 2 | 6 2 |
| 51.2.9 | 99611 | 152 | Ba1Dy1Fe2O4.996 | | | | | |
| | -GM | -R | -S | -T | -U | -X | -Y | -Z |
| | 5 2 | 3+4 4 | 3+4 4 | 6 2 | 3+4 4 | 3+4 4 | 5 2 | 6 2 |
| | 5 2 | 3+4 4 | 3+4 4 | 6 2 | 3+4 4 | 3+4 4 | 5 2 | 6 2 |
| | 5 2 | 3+4 4 | 3+4 4 | 5 2 | 3+4 4 | 3+4 4 | 5 2 | 5 2 |
| | 5 2 | 3+4 4 | 3+4 4 | 5 2 | 3+4 4 | 3+4 4 | 5 2 | 5 2 |
| | 6 2 | 3+4 4 | 3+4 4 | 5 2 | 3+4 4 | 3+4 4 | 6 2 | 5 2 |
| | 6 2 | 3+4 4 | 3+4 4 | 5 2 | 3+4 4 | 3+4 4 | 6 2 | 5 2 |
| | 6 2 | 3+4 4 | 3+4 4 | 5 2 | 3+4 4 | 3+4 4 | 6 2 | 5 2 |
| | 6 2 | 3+4 4 | 3+4 4 | 5 2 | 3+4 4 | 3+4 4 | 6 2 | 5 2 |
| | 6 2 | 3+4 4 | 3+4 4 | 5 2 | 3+4 4 | 3+4 4 | 6 2 | 5 2 |
| | 6 2 | 3+4 4 | 3+4 4 | 5 2 | 3+4 4 | 3+4 4 | 6 2 | 5 2 |
| | 5 2 | 3+4 4 | 3+4 4 | 6 2 | 3+4 4 | 3+4 4 | 6 2 | 6 2 |

Continued on next column

**TABLE XXIX– continued**

| TQC | ICSD | | NoE | | SF | | | |
|---|---|---|---|---|---|---|---|---|
| | 6 2 | 3+4 4 | 3+4 4 | 6 2 | 3+4 4 | 3+4 4 | 5 2 | 6 2 |
| | 5 2 | 3+4 4 | 3+4 4 | 5 2 | 3+4 4 | 3+4 4 | 5 2 | 6 2 |
| | 5 2 | 3+4 4 | 3+4 4 | 6 2 | 3+4 4 | 3+4 4 | 5 2 | 5 2 |
| | 5 2 | 3+4 4 | 3+4 4 | 6 2 | 3+4 4 | 3+4 4 | 5 2 | 6 2 |
| | 6 2 | 3+4 4 | 3+4 4 | 6 2 | 3+4 4 | 3+4 4 | 5 2 | 5 2 |
| | 5 2 | 3+4 4 | 3+4 4 | 5 2 | 3+4 4 | 3+4 4 | 6 2 | 6 2 |
| | 6 2 | 3+4 4 | 3+4 4 | 5 2 | 3+4 4 | 3+4 4 | 6 2 | 5 2 |
| | 6 2 | 3+4 4 | 3+4 4 | 5 2 | 3+4 4 | 3+4 4 | 6 2 | 5 2 |
| | 6 2 | 3+4 4 | 3+4 4 | 5 2 | 3+4 4 | 3+4 4 | 6 2 | 5 2 |
| | 6 2 | 3+4 4 | 3+4 4 | 6 2 | 3+4 4 | 3+4 4 | 6 2 | 6 2 |
| | 6 2 | 3+4 4 | 3+4 4 | 6 2 | 3+4 4 | 3+4 4 | 6 2 | 6 2 |
| | 6 2 | 3+4 4 | 3+4 4 | 6 2 | 3+4 4 | 3+4 4 | 6 2 | 6 2 |
| | 6 2 | 3+4 4 | 3+4 4 | 6 2 | 3+4 4 | 3+4 4 | 6 2 | 6 2 |
| | 6 2 | 3+4 4 | 3+4 4 | 6 2 | 3+4 4 | 3+4 4 | 6 2 | 6 2 |
| | 6 2 | 3+4 4 | 3+4 4 | 6 2 | 3+4 4 | 3+4 4 | 6 2 | 6 2 |
| | 5 2 | 3+4 4 | 3+4 4 | 6 2 | 3+4 4 | 3+4 4 | 5 2 | 5 2 |
| | 6 2 | 3+4 4 | 3+4 4 | 5 2 | 3+4 4 | 3+4 4 | 6 2 | 6 2 |
| | 5 2 | 3+4 4 | 3+4 4 | 5 2 | 3+4 4 | 3+4 4 | 5 2 | 6 2 |
| | 5 2 | 3+4 4 | 3+4 4 | 6 2 | 3+4 4 | 3+4 4 | 6 2 | 6 2 |
| | 5 2 | 3+4 4 | 3+4 4 | 6 2 | 3+4 4 | 3+4 4 | 5 2 | 6 2 |
| | 5 2 | 3+4 4 | 3+4 4 | 5 2 | 3+4 4 | 3+4 4 | 5 2 | 5 2 |
| | 6 2 | 3+4 4 | 3+4 4 | 6 2 | 3+4 4 | 3+4 4 | 5 2 | 5 2 |
| | 5 2 | 3+4 4 | 3+4 4 | 5 2 | 3+4 4 | 3+4 4 | 6 2 | 6 2 |
| | 6 2 | 3+4 4 | 3+4 4 | 6 2 | 3+4 4 | 3+4 4 | 5 2 | 5 2 |
| | 6 2 | 3+4 4 | 3+4 4 | 5 2 | 3+4 4 | 3+4 4 | 6 2 | 6 2 |
| | 5 2 | 3+4 4 | 3+4 4 | 5 2 | 3+4 4 | 3+4 4 | 5 2 | 5 2 |
| | 5 2 | 3+4 4 | 3+4 4 | 6 2 | 3+4 4 | 3+4 4 | 5 2 | 6 2 |
| | 6 2 | | | 6 2 | | | 6 2 | 6 2 |
| | 5 2 | | | 6 2 | | | 5 2 | 5 2 |
| | 5 2 | | | 5 2 | | | 6 2 | 5 2 |
| | 5 2 | | | 5 2 | | | 6 2 | 6 2 |
| | 6 2 | | | 5 2 | | | 5 2 | 5 2 |
| | 6 2 | | | 5 2 | | | 6 2 | 6 2 |
| | 6 2 | | | 5 2 | | | 6 2 | 5 2 |
| | 5 2 | | | 5 2 | | | 5 2 | 5 2 |
| | 5 2 | | | 6 2 | | | 6 2 | 5 2 |
| | 6 2 | | | 6 2 | | | 5 2 | 6 2 |
| | 6 2 | | | 6 2 | | | 5 2 | 6 2 |
| | 6 2 | | | 5 2 | | | 6 2 | 6 2 |
| | 6 2 | | | 6 2 | | | 5 2 | 5 2 |
| | 6 2 | | | 6 2 | | | 6 2 | 6 2 |
| | 5 2 | | | 5 2 | | | 6 2 | 5 2 |
| | 5 2 | | | 6 2 | | | 6 2 | 6 2 |
| | 6 2 | | | 6 2 | | | 5 2 | 5 2 |
| | 6 2 | | | 5 2 | | | 6 2 | 5 2 |
| | 5 2 | | | 5 2 | | | 5 2 | 5 2 |
| | 6 2 | | | 5 2 | | | 6 2 | 6 2 |
| | 6 2 | | | 5 2 | | | 6 2 | 5 2 |
| | 6 2 | | | 5 2 | | | 6 2 | 5 2 |
| | 6 2 | | | 5 2 | | | 6 2 | 5 2 |
| | 6 2 | | | 5 2 | | | 6 2 | 5 2 |
| | 6 2 | | | 6 2 | | | 6 2 | 5 2 |
| | 6 2 | | | 6 2 | | | 6 2 | 5 2 |
| | 5 2 | | | 5 2 | | | 6 2 | 5 2 |
| | 5 2 | | | 6 2 | | | 5 2 | 6 2 |
| | 6 2 | | | 5 2 | | | 5 2 | 5 2 |
| | 6 2 | | | 5 2 | | | 6 2 | 6 2 |
| | 5 2 | | | 6 2 | | | 5 2 | 6 2 |
| | 6 2 | | | 5 2 | | | 6 2 | 5 2 |
| | 6 2 | | | 5 2 | | | 6 2 | 5 2 |
| | 5 2 | | | 5 2 | | | 6 2 | 5 2 |
| | 5 2 | | | 5 2 | | | 6 2 | 5 2 |
| | 6 2 | | | 6 2 | | | 5 2 | 5 2 |
| | 6 2 | | | 5 2 | | | 6 2 | 5 2 |
| | 6 2 | | | 6 2 | | | 5 2 | 5 2 |
| 51.2.10 | 107495 | | 264 | | B3Rh7Zn5 | | | |
| | -GM | -R | -S | -T | -U | -X | -Y | -Z |
| | 5 2 | 3+4 4 | 3+4 4 | 6 2 | 3+4 4 | 3+4 4 | 5 2 | 6 2 |
| | 5 2 | 3+4 4 | 3+4 4 | 5 2 | 3+4 4 | 3+4 4 | 5 2 | 5 2 |
| | 5 2 | 3+4 4 | 3+4 4 | 6 2 | 3+4 4 | 3+4 4 | 6 2 | 6 2 |
| | 6 2 | 3+4 4 | 3+4 4 | 6 2 | 3+4 4 | 3+4 4 | 5 2 | 6 2 |
| | 6 2 | 3+4 4 | 3+4 4 | 5 2 | 3+4 4 | 3+4 4 | 6 2 | 5 2 |
| | 6 2 | 3+4 4 | 3+4 4 | 5 2 | 3+4 4 | 3+4 4 | 6 2 | 5 2 |
| | 6 2 | 3+4 4 | 3+4 4 | 6 2 | 3+4 4 | 3+4 4 | 5 2 | 5 2 |
| | 5 2 | 3+4 4 | 3+4 4 | 5 2 | 3+4 4 | 3+4 4 | 5 2 | 6 2 |
| | 5 2 | 3+4 4 | 3+4 4 | 6 2 | 3+4 4 | 3+4 4 | 6 2 | 5 2 |
| | 5 2 | 3+4 4 | 3+4 4 | 6 2 | 3+4 4 | 3+4 4 | 6 2 | 6 2 |
| | 5 2 | 3+4 4 | 3+4 4 | 5 2 | 3+4 4 | 3+4 4 | 6 2 | 6 2 |
| | 6 2 | 3+4 4 | 3+4 4 | 5 2 | 3+4 4 | 3+4 4 | 5 2 | 5 2 |
| | 5 2 | 3+4 4 | 3+4 4 | 6 2 | 3+4 4 | 3+4 4 | 5 2 | 6 2 |
| | 6 2 | 3+4 4 | 3+4 4 | 6 2 | 3+4 4 | 3+4 4 | 6 2 | 6 2 |
| | 6 2 | 3+4 4 | 3+4 4 | 5 2 | 3+4 4 | 3+4 4 | 5 2 | 5 2 |
| | 6 2 | 3+4 4 | 3+4 4 | 5 2 | 3+4 4 | 3+4 4 | 6 2 | 6 2 |
| | 5 2 | 3+4 4 | 3+4 4 | 6 2 | 3+4 4 | 3+4 4 | 6 2 | 5 2 |
| | 5 2 | 3+4 4 | 3+4 4 | 5 2 | 3+4 4 | 3+4 4 | 6 2 | 5 2 |
| | 6 2 | 3+4 4 | 3+4 4 | 6 2 | 3+4 4 | 3+4 4 | 5 2 | 6 2 |
| | 6 2 | 3+4 4 | 3+4 4 | 6 2 | 3+4 4 | 3+4 4 | 5 2 | 5 2 |
| | 5 2 | 3+4 4 | 3+4 4 | 6 2 | 3+4 4 | 3+4 4 | 5 2 | 6 2 |
| | 5 2 | 3+4 4 | 3+4 4 | 5 2 | 3+4 4 | 3+4 4 | 6 2 | 5 2 |
| | 6 2 | 3+4 4 | 3+4 4 | 5 2 | 3+4 4 | 3+4 4 | 5 2 | 6 2 |
| | 5 2 | 3+4 4 | 3+4 4 | 6 2 | 3+4 4 | 3+4 4 | 5 2 | 5 2 |
| | 6 2 | 3+4 4 | 3+4 4 | 5 2 | 3+4 4 | 3+4 4 | 5 2 | 5 2 |
| | 6 2 | 3+4 4 | 3+4 4 | 6 2 | 3+4 4 | 3+4 4 | 6 2 | 6 2 |
| | 5 2 | 3+4 4 | 3+4 4 | 5 2 | 3+4 4 | 3+4 4 | 5 2 | 5 2 |
| | 5 2 | 3+4 4 | 3+4 4 | 5 2 | 3+4 4 | 3+4 4 | 6 2 | 6 2 |
| | 5 2 | 3+4 4 | 3+4 4 | 6 2 | 3+4 4 | 3+4 4 | 5 2 | 6 2 |
| | 6 2 | 3+4 4 | 3+4 4 | 5 2 | 3+4 4 | 3+4 4 | 6 2 | 6 2 |
| | 5 2 | 3+4 4 | 3+4 4 | 6 2 | 3+4 4 | 3+4 4 | 5 2 | 5 2 |
| | 6 2 | 3+4 4 | 3+4 4 | 5 2 | 3+4 4 | 3+4 4 | 6 2 | 5 2 |
| | 6 2 | 3+4 4 | 3+4 4 | 6 2 | 3+4 4 | 3+4 4 | 6 2 | 6 2 |
| | 5 2 | 3+4 4 | 3+4 4 | 5 2 | 3+4 4 | 3+4 4 | 5 2 | 6 2 |
| | 6 2 | 3+4 4 | 3+4 4 | 6 2 | 3+4 4 | 3+4 4 | 6 2 | 5 2 |
| | 5 2 | 3+4 4 | 3+4 4 | 5 2 | 3+4 4 | 3+4 4 | 5 2 | 6 2 |
| | 6 2 | 3+4 4 | 3+4 4 | 6 2 | 3+4 4 | 3+4 4 | 6 2 | 5 2 |
| | 6 2 | 3+4 4 | 3+4 4 | 5 2 | 3+4 4 | 3+4 4 | 5 2 | 6 2 |
| | 5 2 | 3+4 4 | 3+4 4 | 6 2 | 3+4 4 | 3+4 4 | 6 2 | 5 2 |
| | 6 2 | 3+4 4 | 3+4 4 | 5 2 | 3+4 4 | 3+4 4 | 5 2 | 6 2 |
| | 5 2 | 3+4 4 | 3+4 4 | 6 2 | 3+4 4 | 3+4 4 | 5 2 | 5 2 |
| | 6 2 | 3+4 4 | 3+4 4 | 6 2 | 3+4 4 | 3+4 4 | 6 2 | 5 2 |
| | 5 2 | 3+4 4 | 3+4 4 | 6 2 | 3+4 4 | 3+4 4 | 5 2 | 6 2 |
| | 5 2 | 3+4 4 | 3+4 4 | 5 2 | 3+4 4 | 3+4 4 | 6 2 | 6 2 |
| | 6 2 | 3+4 4 | 3+4 4 | 6 2 | 3+4 4 | 3+4 4 | 6 2 | 5 2 |
| | 6 2 | 3+4 4 | 3+4 4 | 5 2 | 3+4 4 | 3+4 4 | 5 2 | 6 2 |
| | 5 2 | 3+4 4 | 3+4 4 | 5 2 | 3+4 4 | 3+4 4 | 6 2 | 5 2 |
| | 5 2 | 3+4 4 | 3+4 4 | 6 2 | 3+4 4 | 3+4 4 | 6 2 | 5 2 |
| | 6 2 | 3+4 4 | 3+4 4 | 5 2 | 3+4 4 | 3+4 4 | 6 2 | 6 2 |
| | 5 2 | 3+4 4 | 3+4 4 | 6 2 | 3+4 4 | 3+4 4 | 5 2 | 5 2 |
| | 6 2 | 3+4 4 | 3+4 4 | 6 2 | 3+4 4 | 3+4 4 | 5 2 | 6 2 |
| | 6 2 | 3+4 4 | 3+4 4 | 6 2 | 3+4 4 | 3+4 4 | 6 2 | 6 2 |
| | 6 2 | 3+4 4 | 3+4 4 | 5 2 | 3+4 4 | 3+4 4 | 6 2 | 5 2 |
| | 5 2 | 3+4 4 | 3+4 4 | 5 2 | 3+4 4 | 3+4 4 | 5 2 | 5 2 |
| | 5 2 | 3+4 4 | 3+4 4 | 5 2 | 3+4 4 | 3+4 4 | 5 2 | 5 2 |

Continued on next column

Continued on next column

**TABLE XXIX– continued**

| TQC | ICSD | NoE | SF | | | | | |
|---|---|---|---|---|---|---|---|---|
| | 6 2 | 3+4 4 | 3+4 4 | 6 2 | 3+4 4 | 3+4 4 | 5 2 | 6 2 |
| | 5 2 | 3+4 4 | 3+4 4 | 6 2 | 3+4 4 | 3+4 4 | 6 2 | 5 2 |
| | 5 2 | 3+4 4 | 3+4 4 | 5 2 | 3+4 4 | 3+4 4 | 5 2 | 5 2 |
| | 6 2 | 3+4 4 | 3+4 4 | 5 2 | 3+4 4 | 3+4 4 | 5 2 | 6 2 |
| | 5 2 | 3+4 4 | 3+4 4 | 6 2 | 3+4 4 | 3+4 4 | 5 2 | 5 2 |
| | 5 2 | 3+4 4 | 3+4 4 | 6 2 | 3+4 4 | 3+4 4 | 6 2 | 6 2 |
| | 6 2 | 3+4 4 | 3+4 4 | 6 2 | 3+4 4 | 3+4 4 | 5 2 | 6 2 |
| | 5 2 | 3+4 4 | 3+4 4 | 5 2 | 3+4 4 | 3+4 4 | 6 2 | 6 2 |
| | 5 2 | 3+4 4 | 3+4 4 | 6 2 | 3+4 4 | 3+4 4 | 5 2 | 6 2 |
| | 6 2 | 3+4 4 | 3+4 4 | 6 2 | 3+4 4 | 3+4 4 | 6 2 | 5 2 |
| | 6 2 | 3+4 4 | 3+4 4 | 5 2 | 3+4 4 | 3+4 4 | 6 2 | 6 2 |
| | 6 2 | | | 5 2 | | | 5 2 | 6 2 |
| | 5 2 | | | 5 2 | | | 6 2 | 5 2 |
| | 5 2 | | | 6 2 | | | 6 2 | 5 2 |
| | 6 2 | | | 5 2 | | | 5 2 | 6 2 |
| | 5 2 | | | 5 2 | | | 6 2 | 6 2 |
| | 6 2 | | | 6 2 | | | 5 2 | 5 2 |
| | 6 2 | | | 5 2 | | | 5 2 | 6 2 |
| | 5 2 | | | 6 2 | | | 5 2 | 6 2 |
| | 5 2 | | | 6 2 | | | 6 2 | 5 2 |
| | 5 2 | | | 5 2 | | | 6 2 | 6 2 |
| | 5 2 | | | 5 2 | | | 6 2 | 5 2 |
| | 6 2 | | | 6 2 | | | 5 2 | 6 2 |
| | 5 2 | | | 6 2 | | | 6 2 | 6 2 |
| | 6 2 | | | 6 2 | | | 5 2 | 6 2 |
| | 6 2 | | | 6 2 | | | 6 2 | 5 2 |
| | 5 2 | | | 5 2 | | | 5 2 | 6 2 |
| | 5 2 | | | 6 2 | | | 5 2 | 5 2 |
| | 5 2 | | | 5 2 | | | 6 2 | 6 2 |
| | 6 2 | | | 5 2 | | | 5 2 | 5 2 |
| | 5 2 | | | 6 2 | | | 6 2 | 6 2 |
| | 6 2 | | | 6 2 | | | 5 2 | 5 2 |
| | 5 2 | | | 5 2 | | | 6 2 | 5 2 |
| | 5 2 | | | 6 2 | | | 5 2 | 6 2 |
| | 6 2 | | | 5 2 | | | 5 2 | 5 2 |
| | 6 2 | | | 6 2 | | | 5 2 | 6 2 |
| | 5 2 | | | 5 2 | | | 6 2 | 6 2 |
| | 6 2 | | | 6 2 | | | 6 2 | 5 2 |
| | 6 2 | | | 6 2 | | | 5 2 | 5 2 |
| | 5 2 | | | 5 2 | | | 5 2 | 6 2 |
| | 5 2 | | | 5 2 | | | 6 2 | 5 2 |
| | 6 2 | | | 6 2 | | | 6 2 | 6 2 |
| | 6 2 | | | 6 2 | | | 5 2 | 5 2 |
| | 6 2 | | | 6 2 | | | 5 2 | 5 2 |
| | 6 2 | | | 5 2 | | | 6 2 | 5 2 |
| | 5 2 | | | 6 2 | | | 5 2 | 6 2 |
| | 5 2 | | | 5 2 | | | 6 2 | 5 2 |
| | 6 2 | | | 5 2 | | | 5 2 | 6 2 |
| | 5 2 | | | 5 2 | | | 6 2 | 6 2 |
| | 6 2 | | | 6 2 | | | 5 2 | 6 2 |
| | 5 2 | | | 6 2 | | | 6 2 | 5 2 |
| | 5 2 | | | 5 2 | | | 6 2 | 6 2 |
| | 5 2 | | | 5 2 | | | 5 2 | 5 2 |
| | 5 2 | | | 6 2 | | | 6 2 | 6 2 |
| | 5 2 | | | 5 2 | | | 6 2 | 5 2 |
| | 6 2 | | | 6 2 | | | 5 2 | 6 2 |
| | 5 2 | | | 5 2 | | | 5 2 | 5 2 |
| | 6 2 | | | 6 2 | | | 5 2 | 6 2 |
| | 5 2 | | | 5 2 | | | 5 2 | 5 2 |
| | 6 2 | | | 5 2 | | | 5 2 | 6 2 |
| | 6 2 | | | 6 2 | | | 5 2 | 6 2 |

Continued on next column

**TABLE XXIX– continued**

| TQC | ICSD | NoE | SF | | | | | |
|---|---|---|---|---|---|---|---|---|
| | 6 2 | | | 6 2 | | | 6 2 | 6 2 |
| | 5 2 | | | 6 2 | | | 5 2 | 5 2 |
| | 5 2 | | | 6 2 | | | 6 2 | 6 2 |
| | 5 2 | | | 6 2 | | | 5 2 | 5 2 |
| | 6 2 | | | 5 2 | | | 6 2 | 6 2 |
| | 6 2 | | | 6 2 | | | 5 2 | 5 2 |
| | 5 2 | | | 5 2 | | | 6 2 | 6 2 |
| | 5 2 | | | 5 2 | | | 5 2 | 6 2 |
| | 5 2 | | | 5 2 | | | 6 2 | 5 2 |
| | 6 2 | | | 6 2 | | | 6 2 | 5 2 |
| | 6 2 | | | 6 2 | | | 6 2 | 5 2 |
| | 5 2 | | | 6 2 | | | 5 2 | 6 2 |
| | 5 2 | | | 5 2 | | | 5 2 | 5 2 |
| | 6 2 | | | 5 2 | | | 5 2 | 6 2 |
| | 6 2 | | | 6 2 | | | 6 2 | 5 2 |
| | 5 2 | | | 6 2 | | | 5 2 | 6 2 |
| 51.2.11 | 107494 | 264 | B3Ir7Zn5 | | | | | |
| | -GM | -R | -S | -T | -U | -X | -Y | -Z |
| | 5 2 | 3+4 4 | 3+4 4 | 6 2 | 3+4 4 | 3+4 4 | 5 2 | 6 2 |
| | 5 2 | 3+4 4 | 3+4 4 | 6 2 | 3+4 4 | 3+4 4 | 5 2 | 5 2 |
| | 5 2 | 3+4 4 | 3+4 4 | 5 2 | 3+4 4 | 3+4 4 | 5 2 | 6 2 |
| | 6 2 | 3+4 4 | 3+4 4 | 6 2 | 3+4 4 | 3+4 4 | 6 2 | 6 2 |
| | 6 2 | 3+4 4 | 3+4 4 | 5 2 | 3+4 4 | 3+4 4 | 6 2 | 5 2 |
| | 6 2 | 3+4 4 | 3+4 4 | 6 2 | 3+4 4 | 3+4 4 | 5 2 | 5 2 |
| | 6 2 | 3+4 4 | 3+4 4 | 5 2 | 3+4 4 | 3+4 4 | 6 2 | 6 2 |
| | 5 2 | 3+4 4 | 3+4 4 | 6 2 | 3+4 4 | 3+4 4 | 6 2 | 5 2 |
| | 5 2 | 3+4 4 | 3+4 4 | 5 2 | 3+4 4 | 3+4 4 | 5 2 | 5 2 |
| | 5 2 | 3+4 4 | 3+4 4 | 6 2 | 3+4 4 | 3+4 4 | 5 2 | 5 2 |
| | 5 2 | 3+4 4 | 3+4 4 | 5 2 | 3+4 4 | 3+4 4 | 6 2 | 6 2 |
| | 6 2 | 3+4 4 | 3+4 4 | 6 2 | 3+4 4 | 3+4 4 | 6 2 | 6 2 |
| | 6 2 | 3+4 4 | 3+4 4 | 5 2 | 3+4 4 | 3+4 4 | 5 2 | 6 2 |
| | 5 2 | 3+4 4 | 3+4 4 | 6 2 | 3+4 4 | 3+4 4 | 5 2 | 6 2 |
| | 6 2 | 3+4 4 | 3+4 4 | 6 2 | 3+4 4 | 3+4 4 | 6 2 | 6 2 |
| | 5 2 | 3+4 4 | 3+4 4 | 5 2 | 3+4 4 | 3+4 4 | 5 2 | 5 2 |
| | 6 2 | 3+4 4 | 3+4 4 | 5 2 | 3+4 4 | 3+4 4 | 5 2 | 6 2 |
| | 5 2 | 3+4 4 | 3+4 4 | 6 2 | 3+4 4 | 3+4 4 | 5 2 | 6 2 |
| | 6 2 | 3+4 4 | 3+4 4 | 5 2 | 3+4 4 | 3+4 4 | 6 2 | 5 2 |
| | 5 2 | 3+4 4 | 3+4 4 | 6 2 | 3+4 4 | 3+4 4 | 6 2 | 5 2 |
| | 5 2 | 3+4 4 | 3+4 4 | 6 2 | 3+4 4 | 3+4 4 | 5 2 | 5 2 |
| | 6 2 | 3+4 4 | 3+4 4 | 5 2 | 3+4 4 | 3+4 4 | 5 2 | 6 2 |
| | 5 2 | 3+4 4 | 3+4 4 | 6 2 | 3+4 4 | 3+4 4 | 6 2 | 6 2 |
| | 6 2 | 3+4 4 | 3+4 4 | 5 2 | 3+4 4 | 3+4 4 | 6 2 | 5 2 |
| | 5 2 | 3+4 4 | 3+4 4 | 5 2 | 3+4 4 | 3+4 4 | 5 2 | 5 2 |
| | 6 2 | 3+4 4 | 3+4 4 | 6 2 | 3+4 4 | 3+4 4 | 5 2 | 5 2 |
| | 5 2 | 3+4 4 | 3+4 4 | 5 2 | 3+4 4 | 3+4 4 | 5 2 | 5 2 |
| | 6 2 | 3+4 4 | 3+4 4 | 6 2 | 3+4 4 | 3+4 4 | 6 2 | 6 2 |
| | 5 2 | 3+4 4 | 3+4 4 | 5 2 | 3+4 4 | 3+4 4 | 5 2 | 6 2 |
| | 6 2 | 3+4 4 | 3+4 4 | 6 2 | 3+4 4 | 3+4 4 | 6 2 | 6 2 |
| | 5 2 | 3+4 4 | 3+4 4 | 5 2 | 3+4 4 | 3+4 4 | 5 2 | 5 2 |
| | 6 2 | 3+4 4 | 3+4 4 | 5 2 | 3+4 4 | 3+4 4 | 6 2 | 5 2 |
| | 6 2 | 3+4 4 | 3+4 4 | 6 2 | 3+4 4 | 3+4 4 | 5 2 | 6 2 |
| | 5 2 | 3+4 4 | 3+4 4 | 6 2 | 3+4 4 | 3+4 4 | 6 2 | 6 2 |
| | 6 2 | 3+4 4 | 3+4 4 | 5 2 | 3+4 4 | 3+4 4 | 6 2 | 5 2 |
| | 5 2 | 3+4 4 | 3+4 4 | 6 2 | 3+4 4 | 3+4 4 | 5 2 | 6 2 |
| | 6 2 | 3+4 4 | 3+4 4 | 6 2 | 3+4 4 | 3+4 4 | 6 2 | 6 2 |
| | 6 2 | 3+4 4 | 3+4 4 | 5 2 | 3+4 4 | 3+4 4 | 5 2 | 5 2 |
| | 5 2 | 3+4 4 | 3+4 4 | 6 2 | 3+4 4 | 3+4 4 | 6 2 | 6 2 |
| | 5 2 | 3+4 4 | 3+4 4 | 6 2 | 3+4 4 | 3+4 4 | 5 2 | 5 2 |
| | 5 2 | 3+4 4 | 3+4 4 | 5 2 | 3+4 4 | 3+4 4 | 6 2 | 5 2 |
| | 6 2 | 3+4 4 | 3+4 4 | 6 2 | 3+4 4 | 3+4 4 | 6 2 | 6 2 |
| | 6 2 | 3+4 4 | 3+4 4 | 5 2 | 3+4 4 | 3+4 4 | 5 2 | 5 2 |

Continued on next column

**TABLE XXIX– continued**

| TQC | ICSD | | NoE | | SF | | | |
|---|---|---|---|---|---|---|---|---|
| | 5 2 | 3+4 4 | 3+4 4 | 6 2 | 3+4 4 | 3+4 4 | 6 2 | 5 2 |
| | 6 2 | 3+4 4 | 3+4 4 | 6 2 | 3+4 4 | 3+4 4 | 6 2 | 6 2 |
| | 5 2 | 3+4 4 | 3+4 4 | 5 2 | 3+4 4 | 3+4 4 | 5 2 | 6 2 |
| | 5 2 | 3+4 4 | 3+4 4 | 5 2 | 3+4 4 | 3+4 4 | 5 2 | 5 2 |
| | 6 2 | 3+4 4 | 3+4 4 | 5 2 | 3+4 4 | 3+4 4 | 6 2 | 6 2 |
| | 5 2 | 3+4 4 | 3+4 4 | 6 2 | 3+4 4 | 3+4 4 | 6 2 | 5 2 |
| | 5 2 | 3+4 4 | 3+4 4 | 5 2 | 3+4 4 | 3+4 4 | 5 2 | 5 2 |
| | 6 2 | 3+4 4 | 3+4 4 | 5 2 | 3+4 4 | 3+4 4 | 6 2 | 6 2 |
| | 6 2 | 3+4 4 | 3+4 4 | 6 2 | 3+4 4 | 3+4 4 | 5 2 | 6 2 |
| | 6 2 | 3+4 4 | 3+4 4 | 6 2 | 3+4 4 | 3+4 4 | 6 2 | 5 2 |
| | 5 2 | 3+4 4 | 3+4 4 | 5 2 | 3+4 4 | 3+4 4 | 5 2 | 6 2 |
| | 6 2 | 3+4 4 | 3+4 4 | 6 2 | 3+4 4 | 3+4 4 | 5 2 | 5 2 |
| | 5 2 | 3+4 4 | 3+4 4 | 5 2 | 3+4 4 | 3+4 4 | 6 2 | 6 2 |
| | 6 2 | 3+4 4 | 3+4 4 | 5 2 | 3+4 4 | 3+4 4 | 6 2 | 5 2 |
| | 5 2 | 3+4 4 | 3+4 4 | 5 2 | 3+4 4 | 3+4 4 | 5 2 | 6 2 |
| | 5 2 | 3+4 4 | 3+4 4 | 6 2 | 3+4 4 | 3+4 4 | 6 2 | 5 2 |
| | 5 2 | 3+4 4 | 3+4 4 | 6 2 | 3+4 4 | 3+4 4 | 5 2 | 5 2 |
| | 6 2 | 3+4 4 | 3+4 4 | 5 2 | 3+4 4 | 3+4 4 | 6 2 | 6 2 |
| | 5 2 | 3+4 4 | 3+4 4 | 6 2 | 3+4 4 | 3+4 4 | 5 2 | 5 2 |
| | 6 2 | 3+4 4 | 3+4 4 | 5 2 | 3+4 4 | 3+4 4 | 6 2 | 6 2 |
| | 5 2 | 3+4 4 | 3+4 4 | 6 2 | 3+4 4 | 3+4 4 | 5 2 | 6 2 |
| | 6 2 | 3+4 4 | 3+4 4 | 6 2 | 3+4 4 | 3+4 4 | 6 2 | 6 2 |
| | 6 2 | 3+4 4 | 3+4 4 | 6 2 | 3+4 4 | 3+4 4 | 5 2 | 5 2 |
| | 6 2 | | | 5 2 | | | 6 2 | 6 2 |
| | 5 2 | | | 5 2 | | | 5 2 | 5 2 |
| | 5 2 | | | 6 2 | | | 6 2 | 6 2 |
| | 6 2 | | | 5 2 | | | 5 2 | 5 2 |
| | 5 2 | | | 5 2 | | | 6 2 | 6 2 |
| | 6 2 | | | 6 2 | | | 5 2 | 5 2 |
| | 6 2 | | | 6 2 | | | 5 2 | 6 2 |
| | 5 2 | | | 5 2 | | | 5 2 | 5 2 |
| | 5 2 | | | 5 2 | | | 6 2 | 6 2 |
| | 5 2 | | | 6 2 | | | 6 2 | 6 2 |
| | 5 2 | | | 6 2 | | | 6 2 | 5 2 |
| | 6 2 | | | 5 2 | | | 6 2 | 6 2 |
| | 5 2 | | | 6 2 | | | 5 2 | 6 2 |
| | 6 2 | | | 6 2 | | | 5 2 | 6 2 |
| | 6 2 | | | 5 2 | | | 5 2 | 5 2 |
| | 5 2 | | | 6 2 | | | 5 2 | 6 2 |
| | 5 2 | | | 6 2 | | | 6 2 | 5 2 |
| | 5 2 | | | 5 2 | | | 6 2 | 5 2 |
| | 6 2 | | | 5 2 | | | 5 2 | 6 2 |
| | 6 2 | | | 6 2 | | | 5 2 | 5 2 |
| | 5 2 | | | 6 2 | | | 6 2 | 5 2 |
| | 5 2 | | | 5 2 | | | 5 2 | 6 2 |
| | 6 2 | | | 6 2 | | | 6 2 | 6 2 |
| | 5 2 | | | 5 2 | | | 5 2 | 6 2 |
| | 6 2 | | | 6 2 | | | 5 2 | 5 2 |
| | 5 2 | | | 6 2 | | | 6 2 | 6 2 |
| | 6 2 | | | 5 2 | | | 6 2 | 5 2 |
| | 6 2 | | | 6 2 | | | 5 2 | 5 2 |
| | 5 2 | | | 5 2 | | | 6 2 | 6 2 |
| | 6 2 | | | 5 2 | | | 5 2 | 5 2 |
| | 5 2 | | | 6 2 | | | 6 2 | 5 2 |
| | 6 2 | | | 6 2 | | | 5 2 | 6 2 |
| | 5 2 | | | 5 2 | | | 6 2 | 5 2 |
| | 6 2 | | | 6 2 | | | 5 2 | 6 2 |
| | 5 2 | | | 5 2 | | | 5 2 | 5 2 |
| | 6 2 | | | 5 2 | | | 6 2 | 5 2 |
| | 6 2 | | | 6 2 | | | 5 2 | 6 2 |
| | 5 2 | | | 5 2 | | | 6 2 | 6 2 |

Continued on next column

**TABLE XXIX– continued**

| TQC | ICSD | | NoE | | SF | | | |
|---|---|---|---|---|---|---|---|---|
| | 5 2 | | | 6 2 | | | 5 2 | 5 2 |
| | 6 2 | | | 6 2 | | | 6 2 | 5 2 |
| | 5 2 | | | 5 2 | | | 6 2 | 6 2 |
| | 5 2 | | | 6 2 | | | 5 2 | 6 2 |
| | 5 2 | | | 5 2 | | | 6 2 | 5 2 |
| | 6 2 | | | 6 2 | | | 5 2 | 6 2 |
| | 5 2 | | | 5 2 | | | 6 2 | 5 2 |
| | 5 2 | | | 5 2 | | | 5 2 | 6 2 |
| | 6 2 | | | 5 2 | | | 5 2 | 6 2 |
| | 5 2 | | | 6 2 | | | 5 2 | 5 2 |
| | 6 2 | | | 6 2 | | | 6 2 | 6 2 |
| | 6 2 | | | 6 2 | | | 5 2 | 6 2 |
| | 6 2 | | | 5 2 | | | 5 2 | 5 2 |
| | 5 2 | | | 6 2 | | | 6 2 | 6 2 |
| | 5 2 | | | 5 2 | | | 5 2 | 5 2 |
| | 6 2 | | | 6 2 | | | 6 2 | 5 2 |
| | 5 2 | | | 5 2 | | | 5 2 | 6 2 |
| | 5 2 | | | 6 2 | | | 6 2 | 6 2 |
| | 6 2 | | | 5 2 | | | 6 2 | 6 2 |
| | 5 2 | | | 6 2 | | | 5 2 | 5 2 |
| | 6 2 | | | 5 2 | | | 6 2 | 5 2 |
| | 5 2 | | | 6 2 | | | 5 2 | 5 2 |
| | 6 2 | | | 5 2 | | | 6 2 | 6 2 |
| | 5 2 | | | 6 2 | | | 6 2 | 5 2 |
| | 6 2 | | | 6 2 | | | 5 2 | 6 2 |
| | 5 2 | | | 5 2 | | | 6 2 | 6 2 |
| | 5 2 | | | 6 2 | | | 5 2 | 5 2 |
| | 6 2 | | | 6 2 | | | 5 2 | 6 2 |
| 51.2.12 | 429116 | | 96 | | B1Ru1Ta1 | | | |
| | -GM | -R | -S | -T | -U | -X | -Y | -Z |
| | 5 2 | 3+4 4 | 3+4 4 | 6 2 | 3+4 4 | 3+4 4 | 6 2 | 5 2 |
| | 5 2 | 3+4 4 | 3+4 4 | 6 2 | 3+4 4 | 3+4 4 | 6 2 | 5 2 |
| | 6 2 | 3+4 4 | 3+4 4 | 5 2 | 3+4 4 | 3+4 4 | 6 2 | 6 2 |
| | 5 2 | 3+4 4 | 3+4 4 | 6 2 | 3+4 4 | 3+4 4 | 5 2 | 5 2 |
| | 6 2 | 3+4 4 | 3+4 4 | 5 2 | 3+4 4 | 3+4 4 | 5 2 | 6 2 |
| | 6 2 | 3+4 4 | 3+4 4 | 5 2 | 3+4 4 | 3+4 4 | 5 2 | 6 2 |
| | 6 2 | 3+4 4 | 3+4 4 | 5 2 | 3+4 4 | 3+4 4 | 5 2 | 5 2 |
| | 5 2 | 3+4 4 | 3+4 4 | 5 2 | 3+4 4 | 3+4 4 | 5 2 | 6 2 |
| | 5 2 | 3+4 4 | 3+4 4 | 6 2 | 3+4 4 | 3+4 4 | 6 2 | 6 2 |
| | 6 2 | 3+4 4 | 3+4 4 | 5 2 | 3+4 4 | 3+4 4 | 6 2 | 6 2 |
| | 5 2 | 3+4 4 | 3+4 4 | 6 2 | 3+4 4 | 3+4 4 | 5 2 | 5 2 |
| | 6 2 | 3+4 4 | 3+4 4 | 6 2 | 3+4 4 | 3+4 4 | 6 2 | 5 2 |
| | 6 2 | 3+4 4 | 3+4 4 | 6 2 | 3+4 4 | 3+4 4 | 5 2 | 6 2 |
| | 5 2 | 3+4 4 | 3+4 4 | 6 2 | 3+4 4 | 3+4 4 | 6 2 | 6 2 |
| | 6 2 | 3+4 4 | 3+4 4 | 6 2 | 3+4 4 | 3+4 4 | 6 2 | 5 2 |
| | 5 2 | 3+4 4 | 3+4 4 | 5 2 | 3+4 4 | 3+4 4 | 5 2 | 5 2 |
| | 5 2 | 3+4 4 | 3+4 4 | 5 2 | 3+4 4 | 3+4 4 | 6 2 | 6 2 |
| | 6 2 | 3+4 4 | 3+4 4 | 5 2 | 3+4 4 | 3+4 4 | 5 2 | 5 2 |
| | 5 2 | 3+4 4 | 3+4 4 | 6 2 | 3+4 4 | 3+4 4 | 6 2 | 6 2 |
| | 6 2 | 3+4 4 | 3+4 4 | 6 2 | 3+4 4 | 3+4 4 | 5 2 | 5 2 |
| | 5 2 | 3+4 4 | 3+4 4 | 5 2 | 3+4 4 | 3+4 4 | 6 2 | 5 2 |
| | 6 2 | 3+4 4 | 3+4 4 | 6 2 | 3+4 4 | 3+4 4 | 6 2 | 5 2 |
| | 6 2 | 3+4 4 | 3+4 4 | 6 2 | 3+4 4 | 3+4 4 | 5 2 | 5 2 |
| | 6 2 | 3+4 4 | 3+4 4 | 5 2 | 3+4 4 | 3+4 4 | 6 2 | 5 2 |
| | 5 2 | | | 6 2 | | | 5 2 | 6 2 |
| | 5 2 | | | 6 2 | | | 5 2 | 6 2 |
| | 6 2 | | | 5 2 | | | 6 2 | 6 2 |
| | 6 2 | | | 6 2 | | | 5 2 | 6 2 |
| | 5 2 | | | 5 2 | | | 5 2 | 5 2 |
| | 5 2 | | | 5 2 | | | 5 2 | 6 2 |
| | 6 2 | | | 5 2 | | | 6 2 | 5 2 |

Continued on next column

**TABLE XXIX– continued**

| TQC | ICSD | NoE | SF | | | | | |
|---|---|---|---|---|---|---|---|---|
| | 5 2 | | | 5 2 | | | 5 2 | 6 2 |
| | 5 2 | | | 6 2 | | | 6 2 | 6 2 |
| | 5 2 | | | 6 2 | | | 6 2 | 5 2 |
| | 6 2 | | | 6 2 | | | 5 2 | 5 2 |
| | 5 2 | | | 6 2 | | | 6 2 | 5 2 |
| | 6 2 | | | 5 2 | | | 5 2 | 6 2 |
| | 5 2 | | | 5 2 | | | 6 2 | 6 2 |
| | 5 2 | | | 5 2 | | | 5 2 | 5 2 |
| | 6 2 | | | 6 2 | | | 6 2 | 5 2 |
| | 5 2 | | | 5 2 | | | 6 2 | 6 2 |
| | 6 2 | | | 5 2 | | | 5 2 | 6 2 |
| | 5 2 | | | 6 2 | | | 6 2 | 5 2 |
| | 6 2 | | | 5 2 | | | 6 2 | 6 2 |
| | 5 2 | | | 6 2 | | | 6 2 | 5 2 |
| | 5 2 | | | 5 2 | | | 5 2 | 6 2 |
| | 5 2 | | | 6 2 | | | 6 2 | 6 2 |
| | 6 2 | | | 5 2 | | | 6 2 | 6 2 |
| 51.2.13 | 426468 | 144 | B1Nb1Ru1 | | | | | |
| | -GM | -R | -S | -T | -U | -X | -Y | -Z |
| | 5 2 | 3+4 4 | 3+4 4 | 5 2 | 3+4 4 | 3+4 4 | 5 2 | 5 2 |
| | 6 2 | 3+4 4 | 3+4 4 | 6 2 | 3+4 4 | 3+4 4 | 6 2 | 6 2 |
| | 5 2 | 3+4 4 | 3+4 4 | 5 2 | 3+4 4 | 3+4 4 | 5 2 | 5 2 |
| | 5 2 | 3+4 4 | 3+4 4 | 5 2 | 3+4 4 | 3+4 4 | 5 2 | 5 2 |
| | 6 2 | 3+4 4 | 3+4 4 | 6 2 | 3+4 4 | 3+4 4 | 6 2 | 6 2 |
| | 6 2 | 3+4 4 | 3+4 4 | 6 2 | 3+4 4 | 3+4 4 | 6 2 | 6 2 |
| | 6 2 | 3+4 4 | 3+4 4 | 6 2 | 3+4 4 | 3+4 4 | 6 2 | 6 2 |
| | 5 2 | 3+4 4 | 3+4 4 | 5 2 | 3+4 4 | 3+4 4 | 5 2 | 5 2 |
| | 5 2 | 3+4 4 | 3+4 4 | 5 2 | 3+4 4 | 3+4 4 | 5 2 | 5 2 |
| | 5 2 | 3+4 4 | 3+4 4 | 5 2 | 3+4 4 | 3+4 4 | 5 2 | 5 2 |
| | 6 2 | 3+4 4 | 3+4 4 | 6 2 | 3+4 4 | 3+4 4 | 6 2 | 6 2 |
| | 6 2 | 3+4 4 | 3+4 4 | 6 2 | 3+4 4 | 3+4 4 | 6 2 | 6 2 |
| | 6 2 | 3+4 4 | 3+4 4 | 5 2 | 3+4 4 | 3+4 4 | 5 2 | 6 2 |
| | 5 2 | 3+4 4 | 3+4 4 | 6 2 | 3+4 4 | 3+4 4 | 6 2 | 5 2 |
| | 5 2 | 3+4 4 | 3+4 4 | 6 2 | 3+4 4 | 3+4 4 | 6 2 | 5 2 |
| | 5 2 | 3+4 4 | 3+4 4 | 5 2 | 3+4 4 | 3+4 4 | 5 2 | 6 2 |
| | 6 2 | 3+4 4 | 3+4 4 | 5 2 | 3+4 4 | 3+4 4 | 5 2 | 5 2 |
| | 5 2 | 3+4 4 | 3+4 4 | 6 2 | 3+4 4 | 3+4 4 | 6 2 | 5 2 |
| | 6 2 | 3+4 4 | 3+4 4 | 5 2 | 3+4 4 | 3+4 4 | 5 2 | 6 2 |
| | 6 2 | 3+4 4 | 3+4 4 | 6 2 | 3+4 4 | 3+4 4 | 6 2 | 6 2 |
| | 5 2 | 3+4 4 | 3+4 4 | 5 2 | 3+4 4 | 3+4 4 | 5 2 | 5 2 |
| | 5 2 | 3+4 4 | 3+4 4 | 5 2 | 3+4 4 | 3+4 4 | 5 2 | 5 2 |
| | 6 2 | 3+4 4 | 3+4 4 | 6 2 | 3+4 4 | 3+4 4 | 6 2 | 6 2 |
| | 6 2 | 3+4 4 | 3+4 4 | 6 2 | 3+4 4 | 3+4 4 | 6 2 | 6 2 |
| | 5 2 | 3+4 4 | 3+4 4 | 6 2 | 3+4 4 | 3+4 4 | 6 2 | 5 2 |
| | 5 2 | 3+4 4 | 3+4 4 | 6 2 | 3+4 4 | 3+4 4 | 6 2 | 5 2 |
| | 6 2 | 3+4 4 | 3+4 4 | 6 2 | 3+4 4 | 3+4 4 | 6 2 | 6 2 |
| | 5 2 | 3+4 4 | 3+4 4 | 5 2 | 3+4 4 | 3+4 4 | 5 2 | 5 2 |
| | 6 2 | 3+4 4 | 3+4 4 | 5 2 | 3+4 4 | 3+4 4 | 5 2 | 6 2 |
| | 6 2 | 3+4 4 | 3+4 4 | 5 2 | 3+4 4 | 3+4 4 | 5 2 | 6 2 |
| | 6 2 | 3+4 4 | 3+4 4 | 6 2 | 3+4 4 | 3+4 4 | 6 2 | 5 2 |
| | 5 2 | 3+4 4 | 3+4 4 | 5 2 | 3+4 4 | 3+4 4 | 5 2 | 6 2 |
| | 6 2 | 3+4 4 | 3+4 4 | 5 2 | 3+4 4 | 3+4 4 | 5 2 | 6 2 |
| | 5 2 | 3+4 4 | 3+4 4 | 5 2 | 3+4 4 | 3+4 4 | 6 2 | 6 2 |
| | 5 2 | 3+4 4 | 3+4 4 | 6 2 | 3+4 4 | 3+4 4 | 5 2 | 5 2 |
| | 6 2 | 3+4 4 | 3+4 4 | 6 2 | 3+4 4 | 3+4 4 | 5 2 | 5 2 |
| | 6 2 | | | 6 2 | | | 6 2 | 6 2 |
| | 5 2 | | | 6 2 | | | 6 2 | 6 2 |
| | 6 2 | | | 6 2 | | | 6 2 | 5 2 |
| | 5 2 | | | 5 2 | | | 5 2 | 5 2 |
| | 5 2 | | | 5 2 | | | 6 2 | 6 2 |
| | 6 2 | | | 6 2 | | | 5 2 | 5 2 |

Continued on next column

**TABLE XXIX– continued**

| TQC | ICSD | NoE | SF | | | | | |
|---|---|---|---|---|---|---|---|---|
| | 5 2 | | | 5 2 | | | 5 2 | 6 2 |
| | 6 2 | | | 6 2 | | | 6 2 | 5 2 |
| | 5 2 | | | 6 2 | | | 6 2 | 5 2 |
| | 6 2 | | | 5 2 | | | 6 2 | 5 2 |
| | 6 2 | | | 6 2 | | | 5 2 | 5 2 |
| | 5 2 | | | 5 2 | | | 6 2 | 5 2 |
| | 6 2 | | | 6 2 | | | 5 2 | 6 2 |
| | 6 2 | | | 5 2 | | | 5 2 | 6 2 |
| | 5 2 | | | 5 2 | | | 5 2 | 6 2 |
| | 6 2 | | | 6 2 | | | 5 2 | 6 2 |
| | 5 2 | | | 6 2 | | | 6 2 | 5 2 |
| | 5 2 | | | 5 2 | | | 5 2 | 6 2 |
| | 6 2 | | | 5 2 | | | 6 2 | 6 2 |
| | 5 2 | | | 6 2 | | | 6 2 | 5 2 |
| | 5 2 | | | 5 2 | | | 5 2 | 6 2 |
| | 5 2 | | | 6 2 | | | 6 2 | 5 2 |
| | 6 2 | | | 6 2 | | | 5 2 | 6 2 |
| | 5 2 | | | 6 2 | | | 6 2 | 5 2 |
| | 6 2 | | | 5 2 | | | 6 2 | 5 2 |
| | 5 2 | | | 5 2 | | | 5 2 | 6 2 |
| | 5 2 | | | 5 2 | | | 6 2 | 5 2 |
| | 5 2 | | | 6 2 | | | 5 2 | 5 2 |
| | 6 2 | | | 5 2 | | | 6 2 | 6 2 |
| | 6 2 | | | 5 2 | | | 6 2 | 6 2 |
| | 6 2 | | | 6 2 | | | 5 2 | 6 2 |
| | 5 2 | | | 5 2 | | | 6 2 | 5 2 |
| | 5 2 | | | 6 2 | | | 6 2 | 5 2 |
| | 5 2 | | | 5 2 | | | 5 2 | 6 2 |
| | 6 2 | | | 6 2 | | | 6 2 | 6 2 |
| | 5 2 | | | 5 2 | | | 6 2 | 6 2 |
| 51.2.14 | 194435 | 144 | B1Nb1Ru1 | | | | | |
| | -GM | -R | -S | -T | -U | -X | -Y | -Z |
| | 5 2 | 3+4 4 | 3+4 4 | 5 2 | 3+4 4 | 3+4 4 | 5 2 | 5 2 |
| | 6 2 | 3+4 4 | 3+4 4 | 6 2 | 3+4 4 | 3+4 4 | 6 2 | 6 2 |
| | 5 2 | 3+4 4 | 3+4 4 | 6 2 | 3+4 4 | 3+4 4 | 5 2 | 6 2 |
| | 5 2 | 3+4 4 | 3+4 4 | 6 2 | 3+4 4 | 3+4 4 | 5 2 | 6 2 |
| | 6 2 | 3+4 4 | 3+4 4 | 5 2 | 3+4 4 | 3+4 4 | 6 2 | 5 2 |
| | 6 2 | 3+4 4 | 3+4 4 | 5 2 | 3+4 4 | 3+4 4 | 6 2 | 5 2 |
| | 6 2 | 3+4 4 | 3+4 4 | 5 2 | 3+4 4 | 3+4 4 | 6 2 | 5 2 |
| | 5 2 | 3+4 4 | 3+4 4 | 6 2 | 3+4 4 | 3+4 4 | 5 2 | 6 2 |
| | 5 2 | 3+4 4 | 3+4 4 | 6 2 | 3+4 4 | 3+4 4 | 5 2 | 6 2 |
| | 5 2 | 3+4 4 | 3+4 4 | 6 2 | 3+4 4 | 3+4 4 | 5 2 | 6 2 |
| | 6 2 | 3+4 4 | 3+4 4 | 5 2 | 3+4 4 | 3+4 4 | 6 2 | 5 2 |
| | 6 2 | 3+4 4 | 3+4 4 | 5 2 | 3+4 4 | 3+4 4 | 6 2 | 5 2 |
| | 6 2 | 3+4 4 | 3+4 4 | 6 2 | 3+4 4 | 3+4 4 | 5 2 | 5 2 |
| | 5 2 | 3+4 4 | 3+4 4 | 5 2 | 3+4 4 | 3+4 4 | 6 2 | 6 2 |
| | 5 2 | 3+4 4 | 3+4 4 | 5 2 | 3+4 4 | 3+4 4 | 6 2 | 6 2 |
| | 5 2 | 3+4 4 | 3+4 4 | 6 2 | 3+4 4 | 3+4 4 | 5 2 | 5 2 |
| | 6 2 | 3+4 4 | 3+4 4 | 6 2 | 3+4 4 | 3+4 4 | 5 2 | 6 2 |
| | 5 2 | 3+4 4 | 3+4 4 | 5 2 | 3+4 4 | 3+4 4 | 6 2 | 6 2 |
| | 6 2 | 3+4 4 | 3+4 4 | 6 2 | 3+4 4 | 3+4 4 | 5 2 | 5 2 |
| | 6 2 | 3+4 4 | 3+4 4 | 5 2 | 3+4 4 | 3+4 4 | 6 2 | 5 2 |
| | 5 2 | 3+4 4 | 3+4 4 | 6 2 | 3+4 4 | 3+4 4 | 5 2 | 6 2 |
| | 5 2 | 3+4 4 | 3+4 4 | 6 2 | 3+4 4 | 3+4 4 | 5 2 | 6 2 |
| | 6 2 | 3+4 4 | 3+4 4 | 5 2 | 3+4 4 | 3+4 4 | 6 2 | 5 2 |
| | 6 2 | 3+4 4 | 3+4 4 | 5 2 | 3+4 4 | 3+4 4 | 6 2 | 5 2 |
| | 5 2 | 3+4 4 | 3+4 4 | 5 2 | 3+4 4 | 3+4 4 | 6 2 | 6 2 |
| | 5 2 | 3+4 4 | 3+4 4 | 5 2 | 3+4 4 | 3+4 4 | 6 2 | 6 2 |
| | 6 2 | 3+4 4 | 3+4 4 | 5 2 | 3+4 4 | 3+4 4 | 6 2 | 5 2 |
| | 5 2 | 3+4 4 | 3+4 4 | 6 2 | 3+4 4 | 3+4 4 | 5 2 | 6 2 |
| | 6 2 | 3+4 4 | 3+4 4 | 6 2 | 3+4 4 | 3+4 4 | 5 2 | 5 2 |

Continued on next column

**TABLE XXIX– continued**

| TQC | ICSD | NoE | SF | | | | | |
|---|---|---|---|---|---|---|---|---|
| | 6 2 | 3+4 4 | 3+4 4 | 6 2 | 3+4 4 | 3+4 4 | 5 2 | 5 2 |
| | 6 2 | 3+4 4 | 3+4 4 | 5 2 | 3+4 4 | 3+4 4 | 6 2 | 6 2 |
| | 5 2 | 3+4 4 | 3+4 4 | 6 2 | 3+4 4 | 3+4 4 | 5 2 | 5 2 |
| | 6 2 | 3+4 4 | 3+4 4 | 6 2 | 3+4 4 | 3+4 4 | 6 2 | 5 2 |
| | 5 2 | 3+4 4 | 3+4 4 | 6 2 | 3+4 4 | 3+4 4 | 5 2 | 5 2 |
| | 5 2 | 3+4 4 | 3+4 4 | 5 2 | 3+4 4 | 3+4 4 | 5 2 | 6 2 |
| | 6 2 | 3+4 4 | 3+4 4 | 5 2 | 3+4 4 | 3+4 4 | 5 2 | 6 2 |
| | 6 2 | | | 5 2 | | | 6 2 | 5 2 |
| | 5 2 | | | 5 2 | | | 6 2 | 5 2 |
| | 6 2 | | | 5 2 | | | 6 2 | 6 2 |
| | 5 2 | | | 6 2 | | | 5 2 | 5 2 |
| | 5 2 | | | 6 2 | | | 6 2 | 6 2 |
| | 6 2 | | | 5 2 | | | 5 2 | 6 2 |
| | 5 2 | | | 6 2 | | | 5 2 | 5 2 |
| | 6 2 | | | 5 2 | | | 6 2 | 6 2 |
| | 5 2 | | | 5 2 | | | 6 2 | 6 2 |
| | 6 2 | | | 5 2 | | | 6 2 | 6 2 |
| | 6 2 | | | 6 2 | | | 5 2 | 6 2 |
| | 5 2 | | | 6 2 | | | 6 2 | 6 2 |
| | 6 2 | | | 5 2 | | | 5 2 | 5 2 |
| | 6 2 | | | 6 2 | | | 5 2 | 5 2 |
| | 5 2 | | | 6 2 | | | 5 2 | 5 2 |
| | 6 2 | | | 5 2 | | | 5 2 | 5 2 |
| | 5 2 | | | 5 2 | | | 6 2 | 6 2 |
| | 5 2 | | | 6 2 | | | 5 2 | 5 2 |
| | 6 2 | | | 6 2 | | | 6 2 | 5 2 |
| | 5 2 | | | 5 2 | | | 6 2 | 6 2 |
| | 5 2 | | | 5 2 | | | 5 2 | 5 2 |
| | 5 2 | | | 6 2 | | | 6 2 | 6 2 |
| | 6 2 | | | 5 2 | | | 5 2 | 5 2 |
| | 5 2 | | | 5 2 | | | 6 2 | 6 2 |
| | 6 2 | | | 6 2 | | | 6 2 | 6 2 |
| | 5 2 | | | 6 2 | | | 5 2 | 6 2 |
| | 5 2 | | | 6 2 | | | 6 2 | 5 2 |
| | 5 2 | | | 5 2 | | | 6 2 | 6 2 |
| | 6 2 | | | 6 2 | | | 5 2 | 5 2 |
| | 6 2 | | | 6 2 | | | 6 2 | 5 2 |
| | 6 2 | | | 5 2 | | | 5 2 | 5 2 |
| | 5 2 | | | 6 2 | | | 6 2 | 6 2 |
| | 5 2 | | | 5 2 | | | 6 2 | 6 2 |
| | 5 2 | | | 6 2 | | | 5 2 | 5 2 |
| | 5 2 | | | 5 2 | | | 6 2 | 5 2 |
| | 6 2 | | | 6 2 | | | 5 2 | 5 2 |
| 51.2.15 | 249924 | 108 | Au3Rb2Tl1 | | | | | |
| | -GM | -R | -S | -T | -U | -X | -Y | -Z |
| | 5 2 | 3+4 4 | 3+4 4 | 6 2 | 3+4 4 | 3+4 4 | 5 2 | 6 2 |
| | 6 2 | 3+4 4 | 3+4 4 | 5 2 | 3+4 4 | 3+4 4 | 6 2 | 5 2 |
| | 5 2 | 3+4 4 | 3+4 4 | 6 2 | 3+4 4 | 3+4 4 | 6 2 | 5 2 |
| | 6 2 | 3+4 4 | 3+4 4 | 5 2 | 3+4 4 | 3+4 4 | 5 2 | 6 2 |
| | 5 2 | 3+4 4 | 3+4 4 | 6 2 | 3+4 4 | 3+4 4 | 5 2 | 6 2 |
| | 6 2 | 3+4 4 | 3+4 4 | 5 2 | 3+4 4 | 3+4 4 | 6 2 | 5 2 |
| | 5 2 | 3+4 4 | 3+4 4 | 6 2 | 3+4 4 | 3+4 4 | 6 2 | 5 2 |
| | 6 2 | 3+4 4 | 3+4 4 | 5 2 | 3+4 4 | 3+4 4 | 5 2 | 6 2 |
| | 5 2 | 3+4 4 | 3+4 4 | 6 2 | 3+4 4 | 3+4 4 | 5 2 | 6 2 |
| | 5 2 | 3+4 4 | 3+4 4 | 6 2 | 3+4 4 | 3+4 4 | 6 2 | 5 2 |
| | 6 2 | 3+4 4 | 3+4 4 | 5 2 | 3+4 4 | 3+4 4 | 5 2 | 6 2 |
| | 5 2 | 3+4 4 | 3+4 4 | 6 2 | 3+4 4 | 3+4 4 | 6 2 | 5 2 |
| | 6 2 | 3+4 4 | 3+4 4 | 5 2 | 3+4 4 | 3+4 4 | 6 2 | 5 2 |
| | 5 2 | 3+4 4 | 3+4 4 | 6 2 | 3+4 4 | 3+4 4 | 5 2 | 5 2 |
| | 6 2 | 3+4 4 | 3+4 4 | 5 2 | 3+4 4 | 3+4 4 | 6 2 | 6 2 |
| | 6 2 | 3+4 4 | 3+4 4 | 5 2 | 3+4 4 | 3+4 4 | 5 2 | 6 2 |

Continued on next column

**TABLE XXIX– continued**

| TQC | ICSD | NoE | SF | | | | | |
|---|---|---|---|---|---|---|---|---|
| | 5 2 | 3+4 4 | 3+4 4 | 5 2 | 3+4 4 | 3+4 4 | 6 2 | 6 2 |
| | 5 2 | 3+4 4 | 3+4 4 | 5 2 | 3+4 4 | 3+4 4 | 5 2 | 5 2 |
| | 6 2 | 3+4 4 | 3+4 4 | 6 2 | 3+4 4 | 3+4 4 | 5 2 | 5 2 |
| | 5 2 | 3+4 4 | 3+4 4 | 5 2 | 3+4 4 | 3+4 4 | 5 2 | 5 2 |
| | 5 2 | 3+4 4 | 3+4 4 | 5 2 | 3+4 4 | 3+4 4 | 5 2 | 5 2 |
| | 5 2 | 3+4 4 | 3+4 4 | 5 2 | 3+4 4 | 3+4 4 | 5 2 | 6 2 |
| | 5 2 | 3+4 4 | 3+4 4 | 6 2 | 3+4 4 | 3+4 4 | 6 2 | 5 2 |
| | 5 2 | 3+4 4 | 3+4 4 | 5 2 | 3+4 4 | 3+4 4 | 5 2 | 5 2 |
| | 5 2 | 3+4 4 | 3+4 4 | 5 2 | 3+4 4 | 3+4 4 | 5 2 | 5 2 |
| | 6 2 | 3+4 4 | 3+4 4 | 5 2 | 3+4 4 | 3+4 4 | 5 2 | 5 2 |
| | 5 2 | 3+4 4 | 3+4 4 | 5 2 | 3+4 4 | 3+4 4 | 6 2 | 6 2 |
| | 6 2 | | | 6 2 | | | 5 2 | 5 2 |
| | 5 2 | | | 6 2 | | | 6 2 | 6 2 |
| | 6 2 | | | 5 2 | | | 5 2 | 6 2 |
| | 6 2 | | | 6 2 | | | 6 2 | 5 2 |
| | 5 2 | | | 6 2 | | | 6 2 | 5 2 |
| | 6 2 | | | 5 2 | | | 5 2 | 6 2 |
| | 6 2 | | | 6 2 | | | 6 2 | 6 2 |
| | 5 2 | | | 5 2 | | | 5 2 | 5 2 |
| | 5 2 | | | 6 2 | | | 5 2 | 6 2 |
| | 6 2 | | | 5 2 | | | 6 2 | 5 2 |
| | 6 2 | | | 6 2 | | | 6 2 | 6 2 |
| | 5 2 | | | 5 2 | | | 6 2 | 5 2 |
| | 5 2 | | | 5 2 | | | 5 2 | 5 2 |
| | 6 2 | | | 6 2 | | | 5 2 | 6 2 |
| | 5 2 | | | 6 2 | | | 6 2 | 6 2 |
| | 5 2 | | | 5 2 | | | 5 2 | 5 2 |
| | 6 2 | | | 5 2 | | | 5 2 | 5 2 |
| | 5 2 | | | 6 2 | | | 6 2 | 6 2 |
| | 5 2 | | | 5 2 | | | 5 2 | 5 2 |
| | 6 2 | | | 5 2 | | | 5 2 | 5 2 |
| | 5 2 | | | 5 2 | | | 6 2 | 5 2 |
| | 5 2 | | | 5 2 | | | 5 2 | 5 2 |
| | 5 2 | | | 5 2 | | | 5 2 | 5 2 |
| | 6 2 | | | 6 2 | | | 5 2 | 5 2 |
| | 5 2 | | | 5 2 | | | 5 2 | 6 2 |
| | 5 2 | | | 5 2 | | | 6 2 | 6 2 |
| | 5 2 | | | 6 2 | | | 5 2 | 5 2 |
| 51.2.16 | 16325 | 60 | Au1I1Te2 | | | | | |
| | -GM | -R | -S | -T | -U | -X | -Y | -Z |
| | 5 2 | 3+4 4 | 3+4 4 | 6 2 | 3+4 4 | 3+4 4 | 6 2 | 5 2 |
| | 5 2 | 3+4 4 | 3+4 4 | 6 2 | 3+4 4 | 3+4 4 | 6 2 | 5 2 |
| | 6 2 | 3+4 4 | 3+4 4 | 6 2 | 3+4 4 | 3+4 4 | 5 2 | 6 2 |
| | 5 2 | 3+4 4 | 3+4 4 | 5 2 | 3+4 4 | 3+4 4 | 6 2 | 5 2 |
| | 6 2 | 3+4 4 | 3+4 4 | 5 2 | 3+4 4 | 3+4 4 | 5 2 | 6 2 |
| | 6 2 | 3+4 4 | 3+4 4 | 5 2 | 3+4 4 | 3+4 4 | 5 2 | 6 2 |
| | 5 2 | 3+4 4 | 3+4 4 | 6 2 | 3+4 4 | 3+4 4 | 5 2 | 6 2 |
| | 5 2 | 3+4 4 | 3+4 4 | 6 2 | 3+4 4 | 3+4 4 | 5 2 | 6 2 |
| | 5 2 | 3+4 4 | 3+4 4 | 6 2 | 3+4 4 | 3+4 4 | 5 2 | 6 2 |
| | 5 2 | 3+4 4 | 3+4 4 | 6 2 | 3+4 4 | 3+4 4 | 5 2 | 6 2 |
| | 5 2 | 3+4 4 | 3+4 4 | 6 2 | 3+4 4 | 3+4 4 | 5 2 | 6 2 |
| | 5 2 | 3+4 4 | 3+4 4 | 6 2 | 3+4 4 | 3+4 4 | 5 2 | 6 2 |
| | 5 2 | 3+4 4 | 3+4 4 | 5 2 | 3+4 4 | 3+4 4 | 5 2 | 5 2 |
| | 5 2 | 3+4 4 | 3+4 4 | 5 2 | 3+4 4 | 3+4 4 | 5 2 | 5 2 |
| | 5 2 | 3+4 4 | 3+4 4 | 6 2 | 3+4 4 | 3+4 4 | 6 2 | 6 2 |
| | 5 2 | | | 6 2 | | | 5 2 | 6 2 |
| | 6 2 | | | 5 2 | | | 5 2 | 6 2 |
| | 5 2 | | | 6 2 | | | 5 2 | 6 2 |
| | 5 2 | | | 6 2 | | | 6 2 | 6 2 |
| | 6 2 | | | 5 2 | | | 6 2 | 6 2 |
| | 5 2 | | | 6 2 | | | 5 2 | 5 2 |

Continued on next column

**TABLE XXIX– continued**

| TQC | -GM | -R | -S | -T | -U | -X | -Y | -Z |
|---|---|---|---|---|---|---|---|---|
| | 6 2 | | | 6 2 | | | 6 2 | 6 2 |
| | 6 2 | | | 6 2 | | | 6 2 | 5 2 |
| | 5 2 | | | 6 2 | | | 5 2 | 5 2 |
| | 6 2 | | | 5 2 | | | 6 2 | 5 2 |
| | 5 2 | | | 6 2 | | | 6 2 | 6 2 |
| | 6 2 | | | 5 2 | | | 5 2 | 5 2 |
| | 5 2 | | | 6 2 | | | 6 2 | 6 2 |
| | 5 2 | | | 5 2 | | | 6 2 | 5 2 |
| | 6 2 | | | 6 2 | | | 5 2 | 6 2 |
| 51.2.17 | ICSD: 9986 | NoE: 64 | SF: Al4Co1La1 | | | | | |
| | 5 2 | 3+4 4 | 3+4 4 | 6 2 | 3+4 4 | 3+4 4 | 5 2 | 6 2 |
| | 6 2 | 3+4 4 | 3+4 4 | 5 2 | 3+4 4 | 3+4 4 | 6 2 | 5 2 |
| | 5 2 | 3+4 4 | 3+4 4 | 6 2 | 3+4 4 | 3+4 4 | 5 2 | 6 2 |
| | 6 2 | 3+4 4 | 3+4 4 | 5 2 | 3+4 4 | 3+4 4 | 6 2 | 5 2 |
| | 5 2 | 3+4 4 | 3+4 4 | 6 2 | 3+4 4 | 3+4 4 | 5 2 | 6 2 |
| | 6 2 | 3+4 4 | 3+4 4 | 5 2 | 3+4 4 | 3+4 4 | 6 2 | 5 2 |
| | 5 2 | 3+4 4 | 3+4 4 | 6 2 | 3+4 4 | 3+4 4 | 5 2 | 6 2 |
| | 6 2 | 3+4 4 | 3+4 4 | 5 2 | 3+4 4 | 3+4 4 | 6 2 | 5 2 |
| | 5 2 | 3+4 4 | 3+4 4 | 6 2 | 3+4 4 | 3+4 4 | 6 2 | 5 2 |
| | 6 2 | 3+4 4 | 3+4 4 | 5 2 | 3+4 4 | 3+4 4 | 5 2 | 6 2 |
| | 5 2 | 3+4 4 | 3+4 4 | 5 2 | 3+4 4 | 3+4 4 | 6 2 | 5 2 |
| | 6 2 | 3+4 4 | 3+4 4 | 6 2 | 3+4 4 | 3+4 4 | 5 2 | 6 2 |
| | 5 2 | 3+4 4 | 3+4 4 | 5 2 | 3+4 4 | 3+4 4 | 5 2 | 5 2 |
| | 6 2 | 3+4 4 | 3+4 4 | 5 2 | 3+4 4 | 3+4 4 | 6 2 | 6 2 |
| | 5 2 | 3+4 4 | 3+4 4 | 6 2 | 3+4 4 | 3+4 4 | 6 2 | 6 2 |
| | 5 2 | 3+4 4 | 3+4 4 | 5 2 | 3+4 4 | 3+4 4 | 5 2 | 5 2 |
| | 6 2 | | | 6 2 | | | 5 2 | 6 2 |
| | 6 2 | | | 6 2 | | | 6 2 | 5 2 |
| | 5 2 | | | 5 2 | | | 6 2 | 6 2 |
| | 6 2 | | | 5 2 | | | 6 2 | 6 2 |
| | 6 2 | | | 5 2 | | | 5 2 | 6 2 |
| | 5 2 | | | 6 2 | | | 6 2 | 5 2 |
| | 6 2 | | | 5 2 | | | 5 2 | 5 2 |
| | 5 2 | | | 6 2 | | | 6 2 | 5 2 |
| | 5 2 | | | 6 2 | | | 5 2 | 6 2 |
| | 6 2 | | | 5 2 | | | 5 2 | 5 2 |
| | 6 2 | | | 6 2 | | | 5 2 | 6 2 |
| | 5 2 | | | 5 2 | | | 6 2 | 6 2 |
| | 6 2 | | | 6 2 | | | 5 2 | 6 2 |
| | 5 2 | | | 6 2 | | | 6 2 | 5 2 |
| | 5 2 | | | 5 2 | | | 6 2 | 6 2 |
| | 6 2 | | | 6 2 | | | 6 2 | 6 2 |

The End

TABLE XXX: The computed irreducible representations for the valence bands for the compounds of SEBR and NLC in space group 53. See the list of tables.

| TQC | -GM | -R | -S | -T | -U | -X | -Y | -Z |
|---|---|---|---|---|---|---|---|---|
| 53.1.1 | ICSD: 71063 | NoE: 108 | SF: Ni2Ta2Te4 | | | | | |
| | 5 2 | 6+8 2 | 3+4 4 | 3+4 4 | 6+8 2 | 3+4 4 | 5 2 | 3+4 4 |
| | 6 2 | 7+9 2 | 3+4 4 | 3+4 4 | 7+9 2 | 3+4 4 | 6 2 | 3+4 4 |
| | 5 2 | 3+5 2 | 3+4 4 | 3+4 4 | 2+4 2 | 3+4 4 | 5 2 | 3+4 4 |
| | 6 2 | 2+4 2 | 3+4 4 | 3+4 4 | 3+5 2 | 3+4 4 | 6 2 | 3+4 4 |
| | 6 2 | 3+5 2 | 3+4 4 | 3+4 4 | 3+5 2 | 3+4 4 | 6 2 | 3+4 4 |
| | 6 2 | 2+4 2 | 3+4 4 | 3+4 4 | 2+4 2 | 3+4 4 | 5 2 | 3+4 4 |
| | 5 2 | 7+9 2 | 3+4 4 | 3+4 4 | 6+8 2 | 3+4 4 | 5 2 | 3+4 4 |
| | 5 2 | 6+8 2 | 3+4 4 | 3+4 4 | 7+9 2 | 3+4 4 | 6 2 | 3+4 4 |

Continued on next column

**TABLE XXX– continued**

| TQC | -GM | -R | -S | -T | -U | -X | -Y | -Z |
|---|---|---|---|---|---|---|---|---|
| | 5 2 | 7+9 2 | 3+4 4 | 3+4 4 | 3+5 2 | 3+4 4 | 5 2 | 3+4 4 |
| | 5 2 | 6+8 2 | 3+4 4 | 3+4 4 | 6+8 2 | 3+4 4 | 6 2 | 3+4 4 |
| | 6 2 | 3+5 2 | 3+4 4 | 3+4 4 | 7+9 2 | 3+4 4 | 6 2 | 3+4 4 |
| | 6 2 | 2+4 2 | 3+4 4 | 3+4 4 | 2+4 2 | 3+4 4 | 5 2 | 3+4 4 |
| | 6 2 | 6+8 2 | 3+4 4 | 3+4 4 | 6+8 2 | 3+4 4 | 6 2 | 3+4 4 |
| | 5 2 | 7+9 2 | 3+4 4 | 3+4 4 | 7+9 2 | 3+4 4 | 5 2 | 3+4 4 |
| | 6 2 | 3+5 2 | 3+4 4 | 3+4 4 | 3+5 2 | 3+4 4 | 6 2 | 3+4 4 |
| | 5 2 | 2+4 2 | 3+4 4 | 3+4 4 | 2+4 2 | 3+4 4 | 5 2 | 3+4 4 |
| | 6 2 | 2+4 2 | 3+4 4 | 3+4 4 | 7+9 2 | 3+4 4 | 6 2 | 3+4 4 |
| | 5 2 | 3+5 2 | 3+4 4 | 3+4 4 | 6+8 2 | 3+4 4 | 5 2 | 3+4 4 |
| | 6 2 | 7+9 2 | 3+4 4 | 3+4 4 | 2+4 2 | 3+4 4 | 5 2 | 3+4 4 |
| | 6 2 | 2+4 2 | 3+4 4 | 3+4 4 | 3+5 2 | 3+4 4 | 6 2 | 3+4 4 |
| | 5 2 | 6+8 2 | 3+4 4 | 3+4 4 | 2+4 2 | 3+4 4 | 5 2 | 3+4 4 |
| | 5 2 | 3+5 2 | 3+4 4 | 3+4 4 | 3+5 2 | 3+4 4 | 6 2 | 3+4 4 |
| | 6 2 | 7+9 2 | 3+4 4 | 3+4 4 | 2+4 2 | 3+4 4 | 6 2 | 3+4 4 |
| | 6 2 | 2+4 2 | 3+4 4 | 3+4 4 | 3+5 2 | 3+4 4 | 5 2 | 3+4 4 |
| | 5 2 | 6+8 2 | 3+4 4 | 3+4 4 | 7+9 2 | 3+4 4 | 6 2 | 3+4 4 |
| | 6 2 | 3+5 2 | 3+4 4 | 3+4 4 | 6+8 2 | 3+4 4 | 6 2 | 3+4 4 |
| | 5 2 | 7+9 2 | 3+4 4 | 3+4 4 | 7+9 2 | 3+4 4 | 5 2 | 3+4 4 |
| | 5 2 | 6+8 2 | | | 6+8 2 | | 5 2 | |
| | 6 2 | 3+5 2 | | | 3+5 2 | | 5 2 | |
| | 5 2 | 2+4 2 | | | 6+8 2 | | 6 2 | |
| | 5 2 | 7+9 2 | | | 7+9 2 | | 5 2 | |
| | 6 2 | 6+8 2 | | | 2+4 2 | | 6 2 | |
| | 6 2 | 2+4 2 | | | 6+8 2 | | 5 2 | |
| | 5 2 | 3+5 2 | | | 7+9 2 | | 6 2 | |
| | 6 2 | 6+8 2 | | | 3+5 2 | | 6 2 | |
| | 6 2 | 7+9 2 | | | 2+4 2 | | 5 2 | |
| | 6 2 | 3+5 2 | | | 2+4 2 | | 6 2 | |
| | 5 2 | 2+4 2 | | | 3+5 2 | | 6 2 | |
| | 5 2 | 6+8 2 | | | 6+8 2 | | 5 2 | |
| | 5 2 | 7+9 2 | | | 7+9 2 | | 5 2 | |
| | 5 2 | 6+8 2 | | | 6+8 2 | | 5 2 | |
| | 5 2 | 7+9 2 | | | 7+9 2 | | 5 2 | |
| | 6 2 | 6+8 2 | | | 6+8 2 | | 6 2 | |
| | 5 2 | 7+9 2 | | | 7+9 2 | | 5 2 | |
| | 5 2 | 6+8 2 | | | 2+4 2 | | 6 2 | |
| | 5 2 | 7+9 2 | | | 3+5 2 | | 5 2 | |
| | 5 2 | 3+5 2 | | | 6+8 2 | | 5 2 | |
| | 6 2 | 2+4 2 | | | 7+9 2 | | 5 2 | |
| | 6 2 | 2+4 2 | | | 2+4 2 | | 6 2 | |
| | 5 2 | 3+5 2 | | | 3+5 2 | | 6 2 | |
| | 6 2 | 3+5 2 | | | 3+5 2 | | 5 2 | |
| | 5 2 | 2+4 2 | | | 2+4 2 | | 6 2 | |
| | 6 2 | 3+5 2 | | | 3+5 2 | | 6 2 | |
| | 6 2 | 2+4 2 | | | 2+4 2 | | 6 2 | |
| 53.1.2 | ICSD: 71520 | NoE: 132 | SF: Fe1Nb1Te2 | | | | | |
| | 5 2 | 2+4 2 | 3+4 4 | 3+4 4 | 2+4 2 | 3+4 4 | 5 2 | 3+4 4 |
| | 6 2 | 3+5 2 | 3+4 4 | 3+4 4 | 3+5 2 | 3+4 4 | 6 2 | 3+4 4 |
| | 5 2 | 6+8 2 | 3+4 4 | 3+4 4 | 6+8 2 | 3+4 4 | 5 2 | 3+4 4 |
| | 6 2 | 7+9 2 | 3+4 4 | 3+4 4 | 7+9 2 | 3+4 4 | 6 2 | 3+4 4 |
| | 5 2 | 2+4 2 | 3+4 4 | 3+4 4 | 2+4 2 | 3+4 4 | 5 2 | 3+4 4 |
| | 6 2 | 3+5 2 | 3+4 4 | 3+4 4 | 3+5 2 | 3+4 4 | 6 2 | 3+4 4 |
| | 5 2 | 6+8 2 | 3+4 4 | 3+4 4 | 6+8 2 | 3+4 4 | 5 2 | 3+4 4 |
| | 6 2 | 7+9 2 | 3+4 4 | 3+4 4 | 7+9 2 | 3+4 4 | 6 2 | 3+4 4 |
| | 5 2 | 2+4 2 | 3+4 4 | 3+4 4 | 2+4 2 | 3+4 4 | 5 2 | 3+4 4 |
| | 6 2 | 6+8 2 | 3+4 4 | 3+4 4 | 6+8 2 | 3+4 4 | 6 2 | 3+4 4 |
| | 5 2 | 3+5 2 | 3+4 4 | 3+4 4 | 3+5 2 | 3+4 4 | 5 2 | 3+4 4 |
| | 5 2 | 7+9 2 | 3+4 4 | 3+4 4 | 7+9 2 | 3+4 4 | 5 2 | 3+4 4 |
| | 6 2 | 3+5 2 | 3+4 4 | 3+4 4 | 3+5 2 | 3+4 4 | 6 2 | 3+4 4 |

Continued on next column

**TABLE XXX– continued**

| TQC | ICSD | NoE | SF | | | | | |
|---|---|---|---|---|---|---|---|---|
| | 6 2 | 2+4 2 | 3+4 4 | 3+4 4 | 2+4 2 | 3+4 4 | 6 2 | 3+4 4 |
| | 5 2 | 7+9 2 | 3+4 4 | 3+4 4 | 7+9 2 | 3+4 4 | 5 2 | 3+4 4 |
| | 6 2 | 6+8 2 | 3+4 4 | 3+4 4 | 6+8 2 | 3+4 4 | 6 2 | 3+4 4 |
| | 5 2 | 7+9 2 | 3+4 4 | 3+4 4 | 7+9 2 | 3+4 4 | 5 2 | 3+4 4 |
| | 6 2 | 6+8 2 | 3+4 4 | 3+4 4 | 6+8 2 | 3+4 4 | 6 2 | 3+4 4 |
| | 5 2 | 2+4 2 | 3+4 4 | 3+4 4 | 2+4 2 | 3+4 4 | 6 2 | 3+4 4 |
| | 6 2 | 3+5 2 | 3+4 4 | 3+4 4 | 3+5 2 | 3+4 4 | 5 2 | 3+4 4 |
| | 6 2 | 2+4 2 | 3+4 4 | 3+4 4 | 2+4 2 | 3+4 4 | 5 2 | 3+4 4 |
| | 6 2 | 3+5 2 | 3+4 4 | 3+4 4 | 3+5 2 | 3+4 4 | 6 2 | 3+4 4 |
| | 5 2 | 6+8 2 | 3+4 4 | 3+4 4 | 6+8 2 | 3+4 4 | 5 2 | 3+4 4 |
| | 5 2 | 7+9 2 | 3+4 4 | 3+4 4 | 7+9 2 | 3+4 4 | 6 2 | 3+4 4 |
| | 5 2 | 6+8 2 | 3+4 4 | 3+4 4 | 7+9 2 | 3+4 4 | 5 2 | 3+4 4 |
| | 5 2 | 7+9 2 | 3+4 4 | 3+4 4 | 6+8 2 | 3+4 4 | 6 2 | 3+4 4 |
| | 6 2 | 7+9 2 | 3+4 4 | 3+4 4 | 2+4 2 | 3+4 4 | 6 2 | 3+4 4 |
| | 6 2 | 6+8 2 | 3+4 4 | 3+4 4 | 3+5 2 | 3+4 4 | 5 2 | 3+4 4 |
| | 5 2 | 3+5 2 | 3+4 4 | 3+4 4 | 7+9 2 | 3+4 4 | 6 2 | 3+4 4 |
| | 6 2 | 2+4 2 | 3+4 4 | 3+4 4 | 6+8 2 | 3+4 4 | 5 2 | 3+4 4 |
| | 6 2 | 2+4 2 | 3+4 4 | 3+4 4 | 2+4 2 | 3+4 4 | 6 2 | 3+4 4 |
| | 5 2 | 3+5 2 | 3+4 4 | 3+4 4 | 3+5 2 | 3+4 4 | 6 2 | 3+4 4 |
| | 6 2 | 2+4 2 | 3+4 4 | 3+4 4 | 6+8 2 | 3+4 4 | 5 2 | 3+4 4 |
| | 6 2 | 3+5 2 | | | 7+9 2 | | 5 2 | |
| | 5 2 | 6+8 2 | | | 3+5 2 | | 5 2 | |
| | 6 2 | 3+5 2 | | | 2+4 2 | | 6 2 | |
| | 5 2 | 7+9 2 | | | 2+4 2 | | 5 2 | |
| | 5 2 | 2+4 2 | | | 3+5 2 | | 6 2 | |
| | 6 2 | 6+8 2 | | | 6+8 2 | | 5 2 | |
| | 5 2 | 7+9 2 | | | 3+5 2 | | 6 2 | |
| | 6 2 | 3+5 2 | | | 7+9 2 | | 6 2 | |
| | 5 2 | 2+4 2 | | | 2+4 2 | | 6 2 | |
| | 6 2 | 6+8 2 | | | 6+8 2 | | 5 2 | |
| | 6 2 | 7+9 2 | | | 7+9 2 | | 5 2 | |
| | 5 2 | 2+4 2 | | | 7+9 2 | | 5 2 | |
| | 5 2 | 3+5 2 | | | 6+8 2 | | 5 2 | |
| | 5 2 | 7+9 2 | | | 2+4 2 | | 5 2 | |
| | 6 2 | 6+8 2 | | | 3+5 2 | | 6 2 | |
| | 6 2 | 3+5 2 | | | 3+5 2 | | 6 2 | |
| | 5 2 | 2+4 2 | | | 2+4 2 | | 6 2 | |
| | 6 2 | 7+9 2 | | | 7+9 2 | | 6 2 | |
| | 5 2 | 6+8 2 | | | 6+8 2 | | 5 2 | |
| | 6 2 | 2+4 2 | | | 2+4 2 | | 6 2 | |
| | 6 2 | 3+5 2 | | | 3+5 2 | | 6 2 | |
| | 5 2 | 7+9 2 | | | 7+9 2 | | 5 2 | |
| | 5 2 | 6+8 2 | | | 6+8 2 | | 5 2 | |
| | 5 2 | 6+8 2 | | | 7+9 2 | | 5 2 | |
| | 5 2 | 7+9 2 | | | 6+8 2 | | 6 2 | |
| | 6 2 | 7+9 2 | | | 3+5 2 | | 5 2 | |
| | 5 2 | 6+8 2 | | | 2+4 2 | | 5 2 | |
| | 5 2 | 3+5 2 | | | 7+9 2 | | 5 2 | |
| | 5 2 | 2+4 2 | | | 6+8 2 | | 5 2 | |
| | 5 2 | 6+8 2 | | | 6+8 2 | | 6 2 | |
| | 6 2 | 7+9 2 | | | 7+9 2 | | 6 2 | |
| | 5 2 | 3+5 2 | | | 2+4 2 | | 5 2 | |
| | 6 2 | 2+4 2 | | | 3+5 2 | | 6 2 | |

The End

TABLE XXXI: The computed irreducible representations for the valence bands for the compounds of SEBR and NLC in space group 55. See the list of tables.

| TQC | ICSD | NoE | SF |
|---|---|---|---|

Continued on next column

**TABLE XXXI– continued**

| TQC | ICSD | NoE | SF | | | | | |
|---|---|---|---|---|---|---|---|---|
| 55.2.1 | 404193 | 96 | Se2Ti9 | | | | | |
| | -GM | -R | -S | -T | -U | -X | -Y | -Z |
| | 5 2 | 5+5 4 | 6+6 4 | 3+4 4 | 3+4 4 | 3+4 4 | 3+4 4 | 5 2 |
| | 6 2 | 6+6 4 | 5+5 4 | 3+4 4 | 3+4 4 | 3+4 4 | 3+4 4 | 6 2 |
| | 5 2 | 6+6 4 | 5+5 4 | 3+4 4 | 3+4 4 | 3+4 4 | 3+4 4 | 5 2 |
| | 6 2 | 5+5 4 | 6+6 4 | 3+4 4 | 3+4 4 | 3+4 4 | 3+4 4 | 6 2 |
| | 6 2 | 6+6 4 | 6+6 4 | 3+4 4 | 3+4 4 | 3+4 4 | 3+4 4 | 5 2 |
| | 5 2 | 5+5 4 | 5+5 4 | 3+4 4 | 3+4 4 | 3+4 4 | 3+4 4 | 6 2 |
| | 5 2 | 5+5 4 | 5+5 4 | 3+4 4 | 3+4 4 | 3+4 4 | 3+4 4 | 5 2 |
| | 5 2 | 6+6 4 | 5+5 4 | 3+4 4 | 3+4 4 | 3+4 4 | 3+4 4 | 6 2 |
| | 6 2 | 5+5 4 | 6+6 4 | 3+4 4 | 3+4 4 | 3+4 4 | 3+4 4 | 5 2 |
| | 6 2 | 5+5 4 | 6+6 4 | 3+4 4 | 3+4 4 | 3+4 4 | 3+4 4 | 6 2 |
| | 6 2 | 6+6 4 | 5+5 4 | 3+4 4 | 3+4 4 | 3+4 4 | 3+4 4 | 5 2 |
| | 5 2 | 6+6 4 | 6+6 4 | 3+4 4 | 3+4 4 | 3+4 4 | 3+4 4 | 6 2 |
| | 5 2 | 5+5 4 | 5+5 4 | 3+4 4 | 3+4 4 | 3+4 4 | 3+4 4 | 6 2 |
| | 6 2 | 5+5 4 | 6+6 4 | 3+4 4 | 3+4 4 | 3+4 4 | 3+4 4 | 5 2 |
| | 5 2 | 5+5 4 | 5+5 4 | 3+4 4 | 3+4 4 | 3+4 4 | 3+4 4 | 5 2 |
| | 5 2 | 6+6 4 | 5+5 4 | 3+4 4 | 3+4 4 | 3+4 4 | 3+4 4 | 6 2 |
| | 6 2 | 5+5 4 | 6+6 4 | 3+4 4 | 3+4 4 | 3+4 4 | 3+4 4 | 5 2 |
| | 5 2 | 6+6 4 | 5+5 4 | 3+4 4 | 3+4 4 | 3+4 4 | 3+4 4 | 5 2 |
| | 6 2 | 6+6 4 | 5+5 4 | 3+4 4 | 3+4 4 | 3+4 4 | 3+4 4 | 6 2 |
| | 6 2 | 5+5 4 | 6+6 4 | 3+4 4 | 3+4 4 | 3+4 4 | 3+4 4 | 5 2 |
| | 5 2 | 5+5 4 | 5+5 4 | 3+4 4 | 3+4 4 | 3+4 4 | 3+4 4 | 6 2 |
| | 6 2 | 6+6 4 | 6+6 4 | 3+4 4 | 3+4 4 | 3+4 4 | 3+4 4 | 5 2 |
| | 5 2 | 5+5 4 | 6+6 4 | 3+4 4 | 3+4 4 | 3+4 4 | 3+4 4 | 6 2 |
| | 5 2 | 6+6 4 | 5+5 4 | 3+4 4 | 3+4 4 | 3+4 4 | 3+4 4 | 5 2 |
| | 5 2 | | | | | | | 6 2 |
| | 6 2 | | | | | | | 5 2 |
| | 5 2 | | | | | | | 6 2 |
| | 5 2 | | | | | | | 5 2 |
| | 6 2 | | | | | | | 6 2 |
| | 5 2 | | | | | | | 5 2 |
| | 6 2 | | | | | | | 5 2 |
| | 5 2 | | | | | | | 6 2 |
| | 5 2 | | | | | | | 6 2 |
| | 6 2 | | | | | | | 5 2 |
| | 6 2 | | | | | | | 6 2 |
| | 5 2 | | | | | | | 5 2 |
| | 6 2 | | | | | | | 5 2 |
| | 5 2 | | | | | | | 6 2 |
| | 5 2 | | | | | | | 5 2 |
| | 6 2 | | | | | | | 6 2 |
| | 5 2 | | | | | | | 5 2 |
| | 6 2 | | | | | | | 6 2 |
| | 5 2 | | | | | | | 6 2 |
| | 6 2 | | | | | | | 5 2 |
| | 6 2 | | | | | | | 5 2 |
| | 5 2 | | | | | | | 6 2 |
| | 6 2 | | | | | | | 5 2 |
| | 5 2 | | | | | | | 6 2 |
| 55.2.2 | 249665 | 120 | Pr2S5Sn1 | | | | | |
| | -GM | -R | -S | -T | -U | -X | -Y | -Z |
| | 5 2 | 5+5 4 | 6+6 4 | 3+4 4 | 3+4 4 | 3+4 4 | 3+4 4 | 6 2 |
| | 5 2 | 6+6 4 | 5+5 4 | 3+4 4 | 3+4 4 | 3+4 4 | 3+4 4 | 6 2 |
| | 6 2 | 5+5 4 | 6+6 4 | 3+4 4 | 3+4 4 | 3+4 4 | 3+4 4 | 5 2 |
| | 6 2 | 6+6 4 | 5+5 4 | 3+4 4 | 3+4 4 | 3+4 4 | 3+4 4 | 5 2 |
| | 5 2 | 6+6 4 | 6+6 4 | 3+4 4 | 3+4 4 | 3+4 4 | 3+4 4 | 6 2 |
| | 5 2 | 5+5 4 | 5+5 4 | 3+4 4 | 3+4 4 | 3+4 4 | 3+4 4 | 5 2 |
| | 6 2 | 5+5 4 | 6+6 4 | 3+4 4 | 3+4 4 | 3+4 4 | 3+4 4 | 6 2 |
| | 6 2 | 6+6 4 | 5+5 4 | 3+4 4 | 3+4 4 | 3+4 4 | 3+4 4 | 5 2 |
| | 5 2 | 6+6 4 | 5+5 4 | 3+4 4 | 3+4 4 | 3+4 4 | 3+4 4 | 5 2 |

Continued on next column

**TABLE XXXI– continued**

| TQC | ICSD | NoE | SF | | | | | |
|---|---|---|---|---|---|---|---|---|
| | 5 2 | 5+5 4 | 6+6 4 | 3+4 4 | 3+4 4 | 3+4 4 | 3+4 4 | 6 2 |
| | 6 2 | 6+6 4 | 6+6 4 | 3+4 4 | 3+4 4 | 3+4 4 | 3+4 4 | 5 2 |
| | 6 2 | 5+5 4 | 5+5 4 | 3+4 4 | 3+4 4 | 3+4 4 | 3+4 4 | 6 2 |
| | 5 2 | 6+6 4 | 6+6 4 | 3+4 4 | 3+4 4 | 3+4 4 | 3+4 4 | 6 2 |
| | 6 2 | 5+5 4 | 5+5 4 | 3+4 4 | 3+4 4 | 3+4 4 | 3+4 4 | 5 2 |
| | 6 2 | 6+6 4 | 6+6 4 | 3+4 4 | 3+4 4 | 3+4 4 | 3+4 4 | 6 2 |
| | 5 2 | 6+6 4 | 5+5 4 | 3+4 4 | 3+4 4 | 3+4 4 | 3+4 4 | 5 2 |
| | 5 2 | 5+5 4 | 6+6 4 | 3+4 4 | 3+4 4 | 3+4 4 | 3+4 4 | 6 2 |
| | 5 2 | 6+6 4 | 5+5 4 | 3+4 4 | 3+4 4 | 3+4 4 | 3+4 4 | 5 2 |
| | 6 2 | 5+5 4 | 6+6 4 | 3+4 4 | 3+4 4 | 3+4 4 | 3+4 4 | 6 2 |
| | 6 2 | 5+5 4 | 5+5 4 | 3+4 4 | 3+4 4 | 3+4 4 | 3+4 4 | 6 2 |
| | 6 2 | 5+5 4 | 5+5 4 | 3+4 4 | 3+4 4 | 3+4 4 | 3+4 4 | 5 2 |
| | 6 2 | 6+6 4 | 6+6 4 | 3+4 4 | 3+4 4 | 3+4 4 | 3+4 4 | 6 2 |
| | 5 2 | 6+6 4 | 5+5 4 | 3+4 4 | 3+4 4 | 3+4 4 | 3+4 4 | 5 2 |
| | 5 2 | 5+5 4 | 6+6 4 | 3+4 4 | 3+4 4 | 3+4 4 | 3+4 4 | 5 2 |
| | 5 2 | 5+5 4 | 5+5 4 | 3+4 4 | 3+4 4 | 3+4 4 | 3+4 4 | 5 2 |
| | 5 2 | 6+6 4 | 6+6 4 | 3+4 4 | 3+4 4 | 3+4 4 | 3+4 4 | 5 2 |
| | 5 2 | 5+5 4 | 5+5 4 | 3+4 4 | 3+4 4 | 3+4 4 | 3+4 4 | 5 2 |
| | 5 2 | 5+5 4 | 5+5 4 | 3+4 4 | 3+4 4 | 3+4 4 | 3+4 4 | 5 2 |
| | 6 2 | 6+6 4 | 5+5 4 | 3+4 4 | 3+4 4 | 3+4 4 | 3+4 4 | 6 2 |
| | 6 2 | 5+5 4 | 6+6 4 | 3+4 4 | 3+4 4 | 3+4 4 | 3+4 4 | 6 2 |
| | 6 2 | | | | | | | 6 2 |
| | 5 2 | | | | | | | 6 2 |
| | 6 2 | | | | | | | 5 2 |
| | 6 2 | | | | | | | 6 2 |
| | 6 2 | | | | | | | 6 2 |
| | 6 2 | | | | | | | 6 2 |
| | 5 2 | | | | | | | 5 2 |
| | 6 2 | | | | | | | 5 2 |
| | 6 2 | | | | | | | 6 2 |
| | 5 2 | | | | | | | 6 2 |
| | 6 2 | | | | | | | 6 2 |
| | 5 2 | | | | | | | 5 2 |
| | 6 2 | | | | | | | 6 2 |
| | 5 2 | | | | | | | 5 2 |
| | 6 2 | | | | | | | 6 2 |
| | 5 2 | | | | | | | 5 2 |
| | 6 2 | | | | | | | 5 2 |
| | 6 2 | | | | | | | 6 2 |
| | 5 2 | | | | | | | 6 2 |
| | 5 2 | | | | | | | 6 2 |
| | 5 2 | | | | | | | 5 2 |
| | 5 2 | | | | | | | 6 2 |
| | 6 2 | | | | | | | 5 2 |
| | 6 2 | | | | | | | 5 2 |
| | 6 2 | | | | | | | 6 2 |
| | 6 2 | | | | | | | 6 2 |
| | 5 2 | | | | | | | 5 2 |
| | 5 2 | | | | | | | 5 2 |
| | 5 2 | | | | | | | 5 2 |
| | 6 2 | | | | | | | 5 2 |
| 55.2.3 | 97282 | 72 | O4Pb3 | | | | | |
| | -GM | -R | -S | -T | -U | -X | -Y | -Z |
| | 5 2 | 6+6 4 | 6+6 4 | 3+4 4 | 3+4 4 | 3+4 4 | 3+4 4 | 5 2 |
| | 6 2 | 5+5 4 | 5+5 4 | 3+4 4 | 3+4 4 | 3+4 4 | 3+4 4 | 6 2 |
| | 5 2 | 5+5 4 | 6+6 4 | 3+4 4 | 3+4 4 | 3+4 4 | 3+4 4 | 5 2 |
| | 6 2 | 6+6 4 | 5+5 4 | 3+4 4 | 3+4 4 | 3+4 4 | 3+4 4 | 6 2 |
| | 5 2 | 5+5 4 | 6+6 4 | 3+4 4 | 3+4 4 | 3+4 4 | 3+4 4 | 6 2 |
| | 6 2 | 6+6 4 | 5+5 4 | 3+4 4 | 3+4 4 | 3+4 4 | 3+4 4 | 6 2 |
| | 6 2 | 6+6 4 | 6+6 4 | 3+4 4 | 3+4 4 | 3+4 4 | 3+4 4 | 5 2 |
| | 5 2 | 5+5 4 | 5+5 4 | 3+4 4 | 3+4 4 | 3+4 4 | 3+4 4 | 5 2 |

Continued on next column

**TABLE XXXI– continued**

| TQC | ICSD | NoE | SF | | | | | |
|---|---|---|---|---|---|---|---|---|
| | 5 2 | 5+5 4 | 5+5 4 | 3+4 4 | 3+4 4 | 3+4 4 | 3+4 4 | 6 2 |
| | 5 2 | 5+5 4 | 6+6 4 | 3+4 4 | 3+4 4 | 3+4 4 | 3+4 4 | 6 2 |
| | 6 2 | 6+6 4 | 5+5 4 | 3+4 4 | 3+4 4 | 3+4 4 | 3+4 4 | 5 2 |
| | 6 2 | 6+6 4 | 6+6 4 | 3+4 4 | 3+4 4 | 3+4 4 | 3+4 4 | 5 2 |
| | 5 2 | 6+6 4 | 5+5 4 | 3+4 4 | 3+4 4 | 3+4 4 | 3+4 4 | 5 2 |
| | 5 2 | 5+5 4 | 6+6 4 | 3+4 4 | 3+4 4 | 3+4 4 | 3+4 4 | 6 2 |
| | 6 2 | 6+6 4 | 5+5 4 | 3+4 4 | 3+4 4 | 3+4 4 | 3+4 4 | 5 2 |
| | 6 2 | 5+5 4 | 5+5 4 | 3+4 4 | 3+4 4 | 3+4 4 | 3+4 4 | 6 2 |
| | 6 2 | 6+6 4 | 6+6 4 | 3+4 4 | 3+4 4 | 3+4 4 | 3+4 4 | 6 2 |
| | 6 2 | 5+5 4 | 6+6 4 | 3+4 4 | 3+4 4 | 3+4 4 | 3+4 4 | 5 2 |
| | 5 2 | | | | | | | 6 2 |
| | 6 2 | | | | | | | 6 2 |
| | 5 2 | | | | | | | 5 2 |
| | 5 2 | | | | | | | 6 2 |
| | 5 2 | | | | | | | 5 2 |
| | 6 2 | | | | | | | 5 2 |
| | 5 2 | | | | | | | 5 2 |
| | 5 2 | | | | | | | 6 2 |
| | 6 2 | | | | | | | 5 2 |
| | 5 2 | | | | | | | 6 2 |
| | 6 2 | | | | | | | 5 2 |
| | 6 2 | | | | | | | 6 2 |
| | 6 2 | | | | | | | 5 2 |
| | 5 2 | | | | | | | 5 2 |
| | 5 2 | | | | | | | 6 2 |
| | 6 2 | | | | | | | 6 2 |
| | 5 2 | | | | | | | 6 2 |
| | 6 2 | | | | | | | 6 2 |
| 55.2.4 | 249666 | 124 | Nd2S5Sn1 | | | | | |
| | -GM | -R | -S | -T | -U | -X | -Y | -Z |
| | 5 2 | 5+5 4 | 6+6 4 | 3+4 4 | 3+4 4 | 3+4 4 | 3+4 4 | 6 2 |
| | 5 2 | 6+6 4 | 5+5 4 | 3+4 4 | 3+4 4 | 3+4 4 | 3+4 4 | 6 2 |
| | 6 2 | 5+5 4 | 6+6 4 | 3+4 4 | 3+4 4 | 3+4 4 | 3+4 4 | 5 2 |
| | 6 2 | 6+6 4 | 5+5 4 | 3+4 4 | 3+4 4 | 3+4 4 | 3+4 4 | 5 2 |
| | 5 2 | 6+6 4 | 6+6 4 | 3+4 4 | 3+4 4 | 3+4 4 | 3+4 4 | 6 2 |
| | 5 2 | 5+5 4 | 5+5 4 | 3+4 4 | 3+4 4 | 3+4 4 | 3+4 4 | 5 2 |
| | 6 2 | 5+5 4 | 6+6 4 | 3+4 4 | 3+4 4 | 3+4 4 | 3+4 4 | 6 2 |
| | 6 2 | 6+6 4 | 5+5 4 | 3+4 4 | 3+4 4 | 3+4 4 | 3+4 4 | 5 2 |
| | 5 2 | 6+6 4 | 5+5 4 | 3+4 4 | 3+4 4 | 3+4 4 | 3+4 4 | 5 2 |
| | 5 2 | 5+5 4 | 6+6 4 | 3+4 4 | 3+4 4 | 3+4 4 | 3+4 4 | 6 2 |
| | 6 2 | 6+6 4 | 6+6 4 | 3+4 4 | 3+4 4 | 3+4 4 | 3+4 4 | 5 2 |
| | 6 2 | 5+5 4 | 5+5 4 | 3+4 4 | 3+4 4 | 3+4 4 | 3+4 4 | 6 2 |
| | 5 2 | 6+6 4 | 6+6 4 | 3+4 4 | 3+4 4 | 3+4 4 | 3+4 4 | 6 2 |
| | 6 2 | 5+5 4 | 5+5 4 | 3+4 4 | 3+4 4 | 3+4 4 | 3+4 4 | 5 2 |
| | 6 2 | 6+6 4 | 6+6 4 | 3+4 4 | 3+4 4 | 3+4 4 | 3+4 4 | 6 2 |
| | 5 2 | 6+6 4 | 5+5 4 | 3+4 4 | 3+4 4 | 3+4 4 | 3+4 4 | 5 2 |
| | 5 2 | 5+5 4 | 6+6 4 | 3+4 4 | 3+4 4 | 3+4 4 | 3+4 4 | 6 2 |
| | 5 2 | 6+6 4 | 5+5 4 | 3+4 4 | 3+4 4 | 3+4 4 | 3+4 4 | 6 2 |
| | 6 2 | 5+5 4 | 6+6 4 | 3+4 4 | 3+4 4 | 3+4 4 | 3+4 4 | 5 2 |
| | 6 2 | 5+5 4 | 5+5 4 | 3+4 4 | 3+4 4 | 3+4 4 | 3+4 4 | 6 2 |
| | 6 2 | 5+5 4 | 5+5 4 | 3+4 4 | 3+4 4 | 3+4 4 | 3+4 4 | 5 2 |
| | 6 2 | 6+6 4 | 6+6 4 | 3+4 4 | 3+4 4 | 3+4 4 | 3+4 4 | 6 2 |
| | 5 2 | 6+6 4 | 5+5 4 | 3+4 4 | 3+4 4 | 3+4 4 | 3+4 4 | 5 2 |
| | 5 2 | 5+5 4 | 6+6 4 | 3+4 4 | 3+4 4 | 3+4 4 | 3+4 4 | 5 2 |
| | 5 2 | 6+6 4 | 5+5 4 | 3+4 4 | 3+4 4 | 3+4 4 | 3+4 4 | 5 2 |
| | 5 2 | 5+5 4 | 6+6 4 | 3+4 4 | 3+4 4 | 3+4 4 | 3+4 4 | 5 2 |
| | 5 2 | 5+5 4 | 5+5 4 | 3+4 4 | 3+4 4 | 3+4 4 | 3+4 4 | 5 2 |
| | 5 2 | 5+5 4 | 5+5 4 | 3+4 4 | 3+4 4 | 3+4 4 | 3+4 4 | 5 2 |
| | 6 2 | 6+6 4 | 6+6 4 | 3+4 4 | 3+4 4 | 3+4 4 | 3+4 4 | 6 2 |
| | 6 2 | 5+5 4 | 5+5 4 | 3+4 4 | 3+4 4 | 3+4 4 | 3+4 4 | 6 2 |
| | 6 2 | 6+6 4 | 6+6 4 | 3+4 4 | 3+4 4 | 3+4 4 | 3+4 4 | 6 2 |

Continued on next column

**TABLE XXXI– continued**

| TQC | ICSD | NoE | SF | | | | | |
|---|---|---|---|---|---|---|---|---|
| | 5 2 | | | | | | | 6 2 |
| | 6 2 | | | | | | | 5 2 |
| | 6 2 | | | | | | | 6 2 |
| | 6 2 | | | | | | | 6 2 |
| | 6 2 | | | | | | | 6 2 |
| | 5 2 | | | | | | | 5 2 |
| | 6 2 | | | | | | | 5 2 |
| | 5 2 | | | | | | | 6 2 |
| | 6 2 | | | | | | | 6 2 |
| | 5 2 | | | | | | | 6 2 |
| | 6 2 | | | | | | | 5 2 |
| | 6 2 | | | | | | | 5 2 |
| | 5 2 | | | | | | | 6 2 |
| | 6 2 | | | | | | | 6 2 |
| | 5 2 | | | | | | | 5 2 |
| | 6 2 | | | | | | | 5 2 |
| | 6 2 | | | | | | | 6 2 |
| | 5 2 | | | | | | | 6 2 |
| | 5 2 | | | | | | | 6 2 |
| | 5 2 | | | | | | | 5 2 |
| | 5 2 | | | | | | | 6 2 |
| | 6 2 | | | | | | | 5 2 |
| | 6 2 | | | | | | | 5 2 |
| | 6 2 | | | | | | | 6 2 |
| | 6 2 | | | | | | | 6 2 |
| | 5 2 | | | | | | | 5 2 |
| | 5 2 | | | | | | | 5 2 |
| | 5 2 | | | | | | | 5 2 |
| | 6 2 | | | | | | | 5 2 |
| | 5 2 | | | | | | | 6 2 |
| | 5 2 | | | | | | | 6 2 |
| 55.2.5 | 90183 | 128 | Mn2O5Sr2 | | | | | |
| | -GM | -R | -S | -T | -U | -X | -Y | -Z |
| | 5 2 | 6+6 4 | 5+5 4 | 3+4 4 | 3+4 4 | 3+4 4 | 3+4 4 | 6 2 |
| | 6 2 | 5+5 4 | 6+6 4 | 3+4 4 | 3+4 4 | 3+4 4 | 3+4 4 | 5 2 |
| | 5 2 | 5+5 4 | 5+5 4 | 3+4 4 | 3+4 4 | 3+4 4 | 3+4 4 | 6 2 |
| | 6 2 | 6+6 4 | 6+6 4 | 3+4 4 | 3+4 4 | 3+4 4 | 3+4 4 | 5 2 |
| | 5 2 | 5+5 4 | 5+5 4 | 3+4 4 | 3+4 4 | 3+4 4 | 3+4 4 | 5 2 |
| | 6 2 | 6+6 4 | 6+6 4 | 3+4 4 | 3+4 4 | 3+4 4 | 3+4 4 | 6 2 |
| | 5 2 | 5+5 4 | 5+5 4 | 3+4 4 | 3+4 4 | 3+4 4 | 3+4 4 | 5 2 |
| | 5 2 | 6+6 4 | 6+6 4 | 3+4 4 | 3+4 4 | 3+4 4 | 3+4 4 | 6 2 |
| | 6 2 | 5+5 4 | 5+5 4 | 3+4 4 | 3+4 4 | 3+4 4 | 3+4 4 | 5 2 |
| | 5 2 | 6+6 4 | 5+5 4 | 3+4 4 | 3+4 4 | 3+4 4 | 3+4 4 | 6 2 |
| | 6 2 | 6+6 4 | 6+6 4 | 3+4 4 | 3+4 4 | 3+4 4 | 3+4 4 | 5 2 |
| | 5 2 | 5+5 4 | 6+6 4 | 3+4 4 | 3+4 4 | 3+4 4 | 3+4 4 | 5 2 |
| | 6 2 | 5+5 4 | 5+5 4 | 3+4 4 | 3+4 4 | 3+4 4 | 3+4 4 | 6 2 |
| | 5 2 | 6+6 4 | 6+6 4 | 3+4 4 | 3+4 4 | 3+4 4 | 3+4 4 | 5 2 |
| | 6 2 | 5+5 4 | 5+5 4 | 3+4 4 | 3+4 4 | 3+4 4 | 3+4 4 | 6 2 |
| | 5 2 | 6+6 4 | 6+6 4 | 3+4 4 | 3+4 4 | 3+4 4 | 3+4 4 | 5 2 |
| | 6 2 | 6+6 4 | 6+6 4 | 3+4 4 | 3+4 4 | 3+4 4 | 3+4 4 | 5 2 |
| | 5 2 | 5+5 4 | 6+6 4 | 3+4 4 | 3+4 4 | 3+4 4 | 3+4 4 | 6 2 |
| | 5 2 | 5+5 4 | 5+5 4 | 3+4 4 | 3+4 4 | 3+4 4 | 3+4 4 | 6 2 |
| | 6 2 | 5+5 4 | 5+5 4 | 3+4 4 | 3+4 4 | 3+4 4 | 3+4 4 | 6 2 |
| | 6 2 | 6+6 4 | 6+6 4 | 3+4 4 | 3+4 4 | 3+4 4 | 3+4 4 | 5 2 |
| | 5 2 | 6+6 4 | 6+6 4 | 3+4 4 | 3+4 4 | 3+4 4 | 3+4 4 | 5 2 |
| | 6 2 | 6+6 4 | 5+5 4 | 3+4 4 | 3+4 4 | 3+4 4 | 3+4 4 | 5 2 |
| | 6 2 | 5+5 4 | 5+5 4 | 3+4 4 | 3+4 4 | 3+4 4 | 3+4 4 | 6 2 |
| | 5 2 | 6+6 4 | 5+5 4 | 3+4 4 | 3+4 4 | 3+4 4 | 3+4 4 | 6 2 |
| | 5 2 | 6+6 4 | 6+6 4 | 3+4 4 | 3+4 4 | 3+4 4 | 3+4 4 | 5 2 |
| | 6 2 | 5+5 4 | 6+6 4 | 3+4 4 | 3+4 4 | 3+4 4 | 3+4 4 | 6 2 |
| | 6 2 | 6+6 4 | 6+6 4 | 3+4 4 | 3+4 4 | 3+4 4 | 3+4 4 | 5 2 |

Continued on next column

**TABLE XXXI– continued**

| TQC | ICSD | NoE | SF | | | | | |
|---|---|---|---|---|---|---|---|---|
| | 5 2 | 5+5 4 | 5+5 4 | 3+4 4 | 3+4 4 | 3+4 4 | 3+4 4 | 6 2 |
| | 6 2 | 6+6 4 | 6+6 4 | 3+4 4 | 3+4 4 | 3+4 4 | 3+4 4 | 5 2 |
| | 5 2 | 5+5 4 | 5+5 4 | 3+4 4 | 3+4 4 | 3+4 4 | 3+4 4 | 6 2 |
| | 5 2 | 6+6 4 | 5+5 4 | 3+4 4 | 3+4 4 | 3+4 4 | 3+4 4 | 6 2 |
| | 6 2 | | | | | | | 6 2 |
| | 6 2 | | | | | | | 5 2 |
| | 6 2 | | | | | | | 5 2 |
| | 6 2 | | | | | | | 6 2 |
| | 5 2 | | | | | | | 5 2 |
| | 6 2 | | | | | | | 5 2 |
| | 5 2 | | | | | | | 6 2 |
| | 6 2 | | | | | | | 6 2 |
| | 6 2 | | | | | | | 6 2 |
| | 5 2 | | | | | | | 6 2 |
| | 5 2 | | | | | | | 6 2 |
| | 6 2 | | | | | | | 5 2 |
| | 5 2 | | | | | | | 6 2 |
| | 5 2 | | | | | | | 5 2 |
| | 6 2 | | | | | | | 5 2 |
| | 5 2 | | | | | | | 6 2 |
| | 5 2 | | | | | | | 6 2 |
| | 6 2 | | | | | | | 6 2 |
| | 6 2 | | | | | | | 5 2 |
| | 6 2 | | | | | | | 6 2 |
| | 5 2 | | | | | | | 5 2 |
| | 6 2 | | | | | | | 5 2 |
| | 6 2 | | | | | | | 6 2 |
| | 6 2 | | | | | | | 6 2 |
| | 5 2 | | | | | | | 5 2 |
| | 5 2 | | | | | | | 5 2 |
| | 5 2 | | | | | | | 5 2 |
| | 6 2 | | | | | | | 6 2 |
| | 5 2 | | | | | | | 6 2 |
| | 6 2 | | | | | | | 6 2 |
| | 6 2 | | | | | | | 5 2 |
| | 5 2 | | | | | | | 5 2 |
| 55.2.6 | 417818 | 128 | Mn2O5Sr2 | | | | | |
| | -GM | -R | -S | -T | -U | -X | -Y | -Z |
| | 5 2 | 5+5 4 | 5+5 4 | 3+4 4 | 3+4 4 | 3+4 4 | 3+4 4 | 5 2 |
| | 6 2 | 6+6 4 | 6+6 4 | 3+4 4 | 3+4 4 | 3+4 4 | 3+4 4 | 6 2 |
| | 5 2 | 6+6 4 | 5+5 4 | 3+4 4 | 3+4 4 | 3+4 4 | 3+4 4 | 5 2 |
| | 6 2 | 5+5 4 | 6+6 4 | 3+4 4 | 3+4 4 | 3+4 4 | 3+4 4 | 6 2 |
| | 6 2 | 6+6 4 | 5+5 4 | 3+4 4 | 3+4 4 | 3+4 4 | 3+4 4 | 6 2 |
| | 5 2 | 5+5 4 | 6+6 4 | 3+4 4 | 3+4 4 | 3+4 4 | 3+4 4 | 5 2 |
| | 5 2 | 6+6 4 | 5+5 4 | 3+4 4 | 3+4 4 | 3+4 4 | 3+4 4 | 6 2 |
| | 5 2 | 5+5 4 | 6+6 4 | 3+4 4 | 3+4 4 | 3+4 4 | 3+4 4 | 5 2 |
| | 6 2 | 6+6 4 | 5+5 4 | 3+4 4 | 3+4 4 | 3+4 4 | 3+4 4 | 6 2 |
| | 5 2 | 5+5 4 | 5+5 4 | 3+4 4 | 3+4 4 | 3+4 4 | 3+4 4 | 5 2 |
| | 6 2 | 5+5 4 | 6+6 4 | 3+4 4 | 3+4 4 | 3+4 4 | 3+4 4 | 6 2 |
| | 6 2 | 6+6 4 | 6+6 4 | 3+4 4 | 3+4 4 | 3+4 4 | 3+4 4 | 6 2 |
| | 5 2 | 6+6 4 | 5+5 4 | 3+4 4 | 3+4 4 | 3+4 4 | 3+4 4 | 5 2 |
| | 5 2 | 5+5 4 | 6+6 4 | 3+4 4 | 3+4 4 | 3+4 4 | 3+4 4 | 6 2 |
| | 6 2 | 6+6 4 | 5+5 4 | 3+4 4 | 3+4 4 | 3+4 4 | 3+4 4 | 5 2 |
| | 5 2 | 5+5 4 | 6+6 4 | 3+4 4 | 3+4 4 | 3+4 4 | 3+4 4 | 6 2 |
| | 6 2 | 5+5 4 | 6+6 4 | 3+4 4 | 3+4 4 | 3+4 4 | 3+4 4 | 6 2 |
| | 5 2 | 6+6 4 | 6+6 4 | 3+4 4 | 3+4 4 | 3+4 4 | 3+4 4 | 5 2 |
| | 5 2 | 6+6 4 | 5+5 4 | 3+4 4 | 3+4 4 | 3+4 4 | 3+4 4 | 5 2 |
| | 6 2 | 6+6 4 | 5+5 4 | 3+4 4 | 3+4 4 | 3+4 4 | 3+4 4 | 5 2 |
| | 6 2 | 5+5 4 | 6+6 4 | 3+4 4 | 3+4 4 | 3+4 4 | 3+4 4 | 6 2 |
| | 5 2 | 5+5 4 | 6+6 4 | 3+4 4 | 3+4 4 | 3+4 4 | 3+4 4 | 6 2 |
| | 6 2 | 5+5 4 | 5+5 4 | 3+4 4 | 3+4 4 | 3+4 4 | 3+4 4 | 5 2 |

Continued on next column

**TABLE XXXI– continued**

| TQC | ICSD | NoE | SF | | | | | |
|---|---|---|---|---|---|---|---|---|
| | 6 2 | 6+6 4 | 5+5 4 | 3+4 4 | 3+4 4 | 3+4 4 | 3+4 4 | 6 2 |
| | 5 2 | 5+5 4 | 5+5 4 | 3+4 4 | 3+4 4 | 3+4 4 | 3+4 4 | 5 2 |
| | 5 2 | 5+5 4 | 6+6 4 | 3+4 4 | 3+4 4 | 3+4 4 | 3+4 4 | 6 2 |
| | 6 2 | 6+6 4 | 6+6 4 | 3+4 4 | 3+4 4 | 3+4 4 | 3+4 4 | 5 2 |
| | 6 2 | 5+5 4 | 6+6 4 | 3+4 4 | 3+4 4 | 3+4 4 | 3+4 4 | 6 2 |
| | 5 2 | 6+6 4 | 5+5 4 | 3+4 4 | 3+4 4 | 3+4 4 | 3+4 4 | 5 2 |
| | 6 2 | 5+5 4 | 6+6 4 | 3+4 4 | 3+4 4 | 3+4 4 | 3+4 4 | 6 2 |
| | 5 2 | 6+6 4 | 5+5 4 | 3+4 4 | 3+4 4 | 3+4 4 | 3+4 4 | 5 2 |
| | 5 2 | 5+5 4 | 5+5 4 | 3+4 4 | 3+4 4 | 3+4 4 | 3+4 4 | 5 2 |
| | 6 2 | | | | | | | 6 2 |
| | 6 2 | | | | | | | 5 2 |
| | 6 2 | | | | | | | 6 2 |
| | 6 2 | | | | | | | 5 2 |
| | 5 2 | | | | | | | 6 2 |
| | 5 2 | | | | | | | 6 2 |
| | 6 2 | | | | | | | 5 2 |
| | 6 2 | | | | | | | 5 2 |
| | 6 2 | | | | | | | 5 2 |
| | 5 2 | | | | | | | 5 2 |
| | 5 2 | | | | | | | 5 2 |
| | 5 2 | | | | | | | 6 2 |
| | 6 2 | | | | | | | 5 2 |
| | 5 2 | | | | | | | 6 2 |
| | 6 2 | | | | | | | 6 2 |
| | 5 2 | | | | | | | 5 2 |
| | 5 2 | | | | | | | 5 2 |
| | 6 2 | | | | | | | 5 2 |
| | 6 2 | | | | | | | 6 2 |
| | 6 2 | | | | | | | 5 2 |
| | 5 2 | | | | | | | 6 2 |
| | 6 2 | | | | | | | 6 2 |
| | 6 2 | | | | | | | 5 2 |
| | 6 2 | | | | | | | 5 2 |
| | 5 2 | | | | | | | 6 2 |
| | 5 2 | | | | | | | 6 2 |
| | 5 2 | | | | | | | 6 2 |
| | 6 2 | | | | | | | 5 2 |
| | 5 2 | | | | | | | 5 2 |
| | 6 2 | | | | | | | 5 2 |
| | 6 2 | | | | | | | 6 2 |
| | 5 2 | | | | | | | 6 2 |
| 55.2.7 | 417817 | 128 | Mn2O5Sr2 | | | | | |
| | -GM | -R | -S | -T | -U | -X | -Y | -Z |
| | 5 2 | 6+6 4 | 6+6 4 | 3+4 4 | 3+4 4 | 3+4 4 | 3+4 4 | 5 2 |
| | 6 2 | 5+5 4 | 5+5 4 | 3+4 4 | 3+4 4 | 3+4 4 | 3+4 4 | 6 2 |
| | 5 2 | 5+5 4 | 6+6 4 | 3+4 4 | 3+4 4 | 3+4 4 | 3+4 4 | 5 2 |
| | 6 2 | 6+6 4 | 5+5 4 | 3+4 4 | 3+4 4 | 3+4 4 | 3+4 4 | 6 2 |
| | 5 2 | 5+5 4 | 6+6 4 | 3+4 4 | 3+4 4 | 3+4 4 | 3+4 4 | 6 2 |
| | 6 2 | 6+6 4 | 5+5 4 | 3+4 4 | 3+4 4 | 3+4 4 | 3+4 4 | 5 2 |
| | 5 2 | 5+5 4 | 6+6 4 | 3+4 4 | 3+4 4 | 3+4 4 | 3+4 4 | 6 2 |
| | 5 2 | 6+6 4 | 5+5 4 | 3+4 4 | 3+4 4 | 3+4 4 | 3+4 4 | 5 2 |
| | 6 2 | 5+5 4 | 6+6 4 | 3+4 4 | 3+4 4 | 3+4 4 | 3+4 4 | 6 2 |
| | 5 2 | 6+6 4 | 6+6 4 | 3+4 4 | 3+4 4 | 3+4 4 | 3+4 4 | 5 2 |
| | 6 2 | 6+6 4 | 5+5 4 | 3+4 4 | 3+4 4 | 3+4 4 | 3+4 4 | 6 2 |
| | 5 2 | 5+5 4 | 5+5 4 | 3+4 4 | 3+4 4 | 3+4 4 | 3+4 4 | 6 2 |
| | 6 2 | 5+5 4 | 6+6 4 | 3+4 4 | 3+4 4 | 3+4 4 | 3+4 4 | 5 2 |
| | 5 2 | 6+6 4 | 5+5 4 | 3+4 4 | 3+4 4 | 3+4 4 | 3+4 4 | 6 2 |
| | 6 2 | 5+5 4 | 6+6 4 | 3+4 4 | 3+4 4 | 3+4 4 | 3+4 4 | 5 2 |
| | 5 2 | 6+6 4 | 5+5 4 | 3+4 4 | 3+4 4 | 3+4 4 | 3+4 4 | 6 2 |
| | 6 2 | 6+6 4 | 5+5 4 | 3+4 4 | 3+4 4 | 3+4 4 | 3+4 4 | 6 2 |
| | 5 2 | 5+5 4 | 5+5 4 | 3+4 4 | 3+4 4 | 3+4 4 | 3+4 4 | 5 2 |

Continued on next column

**TABLE XXXI– continued**

| TQC | ICSD | NoE | SF | | | | | |
|---|---|---|---|---|---|---|---|---|
| | 5 2 | 5+5 4 | 6+6 4 | 3+4 4 | 3+4 4 | 3+4 4 | 3+4 4 | 5 2 |
| | 6 2 | 5+5 4 | 5+5 4 | 3+4 4 | 3+4 4 | 3+4 4 | 3+4 4 | 5 2 |
| | 6 2 | 6+6 4 | 6+6 4 | 3+4 4 | 3+4 4 | 3+4 4 | 3+4 4 | 6 2 |
| | 5 2 | 6+6 4 | 5+5 4 | 3+4 4 | 3+4 4 | 3+4 4 | 3+4 4 | 6 2 |
| | 6 2 | 6+6 4 | 6+6 4 | 3+4 4 | 3+4 4 | 3+4 4 | 3+4 4 | 5 2 |
| | 6 2 | 5+5 4 | 6+6 4 | 3+4 4 | 3+4 4 | 3+4 4 | 3+4 4 | 6 2 |
| | 5 2 | 6+6 4 | 6+6 4 | 3+4 4 | 3+4 4 | 3+4 4 | 3+4 4 | 5 2 |
| | 5 2 | 6+6 4 | 5+5 4 | 3+4 4 | 3+4 4 | 3+4 4 | 3+4 4 | 6 2 |
| | 6 2 | 5+5 4 | 5+5 4 | 3+4 4 | 3+4 4 | 3+4 4 | 3+4 4 | 5 2 |
| | 6 2 | 6+6 4 | 5+5 4 | 3+4 4 | 3+4 4 | 3+4 4 | 3+4 4 | 6 2 |
| | 5 2 | 5+5 4 | 6+6 4 | 3+4 4 | 3+4 4 | 3+4 4 | 3+4 4 | 5 2 |
| | 6 2 | 6+6 4 | 5+5 4 | 3+4 4 | 3+4 4 | 3+4 4 | 3+4 4 | 6 2 |
| | 5 2 | 5+5 4 | 6+6 4 | 3+4 4 | 3+4 4 | 3+4 4 | 3+4 4 | 5 2 |
| | 5 2 | 6+6 4 | 6+6 4 | 3+4 4 | 3+4 4 | 3+4 4 | 3+4 4 | 5 2 |
| | 6 2 | | | | | | | 6 2 |
| | 6 2 | | | | | | | 5 2 |
| | 6 2 | | | | | | | 6 2 |
| | 6 2 | | | | | | | 5 2 |
| | 5 2 | | | | | | | 6 2 |
| | 5 2 | | | | | | | 6 2 |
| | 6 2 | | | | | | | 5 2 |
| | 6 2 | | | | | | | 5 2 |
| | 6 2 | | | | | | | 5 2 |
| | 5 2 | | | | | | | 5 2 |
| | 5 2 | | | | | | | 5 2 |
| | 5 2 | | | | | | | 6 2 |
| | 6 2 | | | | | | | 5 2 |
| | 5 2 | | | | | | | 6 2 |
| | 6 2 | | | | | | | 6 2 |
| | 5 2 | | | | | | | 5 2 |
| | 5 2 | | | | | | | 5 2 |
| | 6 2 | | | | | | | 5 2 |
| | 6 2 | | | | | | | 6 2 |
| | 6 2 | | | | | | | 5 2 |
| | 5 2 | | | | | | | 6 2 |
| | 6 2 | | | | | | | 6 2 |
| | 6 2 | | | | | | | 5 2 |
| | 6 2 | | | | | | | 5 2 |
| | 5 2 | | | | | | | 6 2 |
| | 5 2 | | | | | | | 6 2 |
| | 5 2 | | | | | | | 6 2 |
| | 6 2 | | | | | | | 5 2 |
| | 5 2 | | | | | | | 5 2 |
| | 6 2 | | | | | | | 5 2 |
| | 6 2 | | | | | | | 6 2 |
| | 5 2 | | | | | | | 6 2 |
| 55.2.8 | 417816 | 128 | Mn2O5Sr2 | | | | | |
| | -GM | -R | -S | -T | -U | -X | -Y | -Z |
| | 5 2 | 5+5 4 | 5+5 4 | 3+4 4 | 3+4 4 | 3+4 4 | 3+4 4 | 5 2 |
| | 6 2 | 6+6 4 | 6+6 4 | 3+4 4 | 3+4 4 | 3+4 4 | 3+4 4 | 6 2 |
| | 5 2 | 6+6 4 | 5+5 4 | 3+4 4 | 3+4 4 | 3+4 4 | 3+4 4 | 5 2 |
| | 6 2 | 5+5 4 | 6+6 4 | 3+4 4 | 3+4 4 | 3+4 4 | 3+4 4 | 6 2 |
| | 6 2 | 6+6 4 | 5+5 4 | 3+4 4 | 3+4 4 | 3+4 4 | 3+4 4 | 6 2 |
| | 5 2 | 5+5 4 | 6+6 4 | 3+4 4 | 3+4 4 | 3+4 4 | 3+4 4 | 5 2 |
| | 5 2 | 6+6 4 | 5+5 4 | 3+4 4 | 3+4 4 | 3+4 4 | 3+4 4 | 6 2 |
| | 5 2 | 5+5 4 | 6+6 4 | 3+4 4 | 3+4 4 | 3+4 4 | 3+4 4 | 5 2 |
| | 6 2 | 6+6 4 | 5+5 4 | 3+4 4 | 3+4 4 | 3+4 4 | 3+4 4 | 6 2 |
| | 5 2 | 5+5 4 | 5+5 4 | 3+4 4 | 3+4 4 | 3+4 4 | 3+4 4 | 5 2 |
| | 6 2 | 5+5 4 | 6+6 4 | 3+4 4 | 3+4 4 | 3+4 4 | 3+4 4 | 6 2 |
| | 5 2 | 6+6 4 | 6+6 4 | 3+4 4 | 3+4 4 | 3+4 4 | 3+4 4 | 6 2 |
| | 6 2 | 6+6 4 | 5+5 4 | 3+4 4 | 3+4 4 | 3+4 4 | 3+4 4 | 5 2 |

Continued on next column

**TABLE XXXI– continued**

| TQC | ICSD | NoE | SF | | | | | |
|---|---|---|---|---|---|---|---|---|
| | 5 2 | 5+5 4 | 6+6 4 | 3+4 4 | 3+4 4 | 3+4 4 | 3+4 4 | 6 2 |
| | 6 2 | 6+6 4 | 5+5 4 | 3+4 4 | 3+4 4 | 3+4 4 | 3+4 4 | 5 2 |
| | 5 2 | 5+5 4 | 6+6 4 | 3+4 4 | 3+4 4 | 3+4 4 | 3+4 4 | 6 2 |
| | 6 2 | 5+5 4 | 6+6 4 | 3+4 4 | 3+4 4 | 3+4 4 | 3+4 4 | 6 2 |
| | 5 2 | 6+6 4 | 6+6 4 | 3+4 4 | 3+4 4 | 3+4 4 | 3+4 4 | 5 2 |
| | 6 2 | 6+6 4 | 5+5 4 | 3+4 4 | 3+4 4 | 3+4 4 | 3+4 4 | 5 2 |
| | 5 2 | 6+6 4 | 5+5 4 | 3+4 4 | 3+4 4 | 3+4 4 | 3+4 4 | 5 2 |
| | 6 2 | 5+5 4 | 6+6 4 | 3+4 4 | 3+4 4 | 3+4 4 | 3+4 4 | 6 2 |
| | 5 2 | 5+5 4 | 6+6 4 | 3+4 4 | 3+4 4 | 3+4 4 | 3+4 4 | 6 2 |
| | 6 2 | 5+5 4 | 5+5 4 | 3+4 4 | 3+4 4 | 3+4 4 | 3+4 4 | 5 2 |
| | 6 2 | 6+6 4 | 5+5 4 | 3+4 4 | 3+4 4 | 3+4 4 | 3+4 4 | 6 2 |
| | 5 2 | 5+5 4 | 5+5 4 | 3+4 4 | 3+4 4 | 3+4 4 | 3+4 4 | 5 2 |
| | 5 2 | 5+5 4 | 6+6 4 | 3+4 4 | 3+4 4 | 3+4 4 | 3+4 4 | 6 2 |
| | 6 2 | 6+6 4 | 6+6 4 | 3+4 4 | 3+4 4 | 3+4 4 | 3+4 4 | 5 2 |
| | 6 2 | 5+5 4 | 6+6 4 | 3+4 4 | 3+4 4 | 3+4 4 | 3+4 4 | 6 2 |
| | 5 2 | 6+6 4 | 5+5 4 | 3+4 4 | 3+4 4 | 3+4 4 | 3+4 4 | 5 2 |
| | 6 2 | 5+5 4 | 6+6 4 | 3+4 4 | 3+4 4 | 3+4 4 | 3+4 4 | 6 2 |
| | 5 2 | 6+6 4 | 5+5 4 | 3+4 4 | 3+4 4 | 3+4 4 | 3+4 4 | 5 2 |
| | 5 2 | 5+5 4 | 5+5 4 | 3+4 4 | 3+4 4 | 3+4 4 | 3+4 4 | 5 2 |
| | 6 2 | | | | | | | 6 2 |
| | 6 2 | | | | | | | 5 2 |
| | 6 2 | | | | | | | 6 2 |
| | 6 2 | | | | | | | 5 2 |
| | 5 2 | | | | | | | 6 2 |
| | 5 2 | | | | | | | 6 2 |
| | 6 2 | | | | | | | 5 2 |
| | 6 2 | | | | | | | 5 2 |
| | 6 2 | | | | | | | 5 2 |
| | 5 2 | | | | | | | 5 2 |
| | 5 2 | | | | | | | 5 2 |
| | 5 2 | | | | | | | 6 2 |
| | 6 2 | | | | | | | 5 2 |
| | 5 2 | | | | | | | 6 2 |
| | 6 2 | | | | | | | 6 2 |
| | 5 2 | | | | | | | 5 2 |
| | 5 2 | | | | | | | 5 2 |
| | 6 2 | | | | | | | 5 2 |
| | 6 2 | | | | | | | 6 2 |
| | 6 2 | | | | | | | 5 2 |
| | 5 2 | | | | | | | 6 2 |
| | 6 2 | | | | | | | 6 2 |
| | 6 2 | | | | | | | 5 2 |
| | 6 2 | | | | | | | 5 2 |
| | 5 2 | | | | | | | 6 2 |
| | 5 2 | | | | | | | 6 2 |
| | 5 2 | | | | | | | 6 2 |
| | 6 2 | | | | | | | 5 2 |
| | 5 2 | | | | | | | 5 2 |
| | 6 2 | | | | | | | 5 2 |
| | 6 2 | | | | | | | 6 2 |
| | 5 2 | | | | | | | 6 2 |
| 55.2.9 | 417815 | 128 | Mn2O5Sr2 | | | | | |
| | -GM | -R | -S | -T | -U | -X | -Y | -Z |
| | 5 2 | 5+5 4 | 5+5 4 | 3+4 4 | 3+4 4 | 3+4 4 | 3+4 4 | 5 2 |
| | 6 2 | 6+6 4 | 6+6 4 | 3+4 4 | 3+4 4 | 3+4 4 | 3+4 4 | 6 2 |
| | 5 2 | 6+6 4 | 5+5 4 | 3+4 4 | 3+4 4 | 3+4 4 | 3+4 4 | 5 2 |
| | 6 2 | 5+5 4 | 6+6 4 | 3+4 4 | 3+4 4 | 3+4 4 | 3+4 4 | 6 2 |
| | 5 2 | 6+6 4 | 5+5 4 | 3+4 4 | 3+4 4 | 3+4 4 | 3+4 4 | 6 2 |
| | 6 2 | 5+5 4 | 6+6 4 | 3+4 4 | 3+4 4 | 3+4 4 | 3+4 4 | 5 2 |
| | 5 2 | 6+6 4 | 5+5 4 | 3+4 4 | 3+4 4 | 3+4 4 | 3+4 4 | 6 2 |
| | 5 2 | 5+5 4 | 6+6 4 | 3+4 4 | 3+4 4 | 3+4 4 | 3+4 4 | 5 2 |

Continued on next column

**TABLE XXXI– continued**

| TQC | ICSD | NoE | SF | | | | | |
|---|---|---|---|---|---|---|---|---|
| | 6 2 | 6+6 4 | 5+5 4 | 3+4 4 | 3+4 4 | 3+4 4 | 3+4 4 | 6 2 |
| | 5 2 | 5+5 4 | 5+5 4 | 3+4 4 | 3+4 4 | 3+4 4 | 3+4 4 | 5 2 |
| | 6 2 | 5+5 4 | 6+6 4 | 3+4 4 | 3+4 4 | 3+4 4 | 3+4 4 | 6 2 |
| | 5 2 | 6+6 4 | 6+6 4 | 3+4 4 | 3+4 4 | 3+4 4 | 3+4 4 | 6 2 |
| | 6 2 | 6+6 4 | 5+5 4 | 3+4 4 | 3+4 4 | 3+4 4 | 3+4 4 | 5 2 |
| | 5 2 | 5+5 4 | 6+6 4 | 3+4 4 | 3+4 4 | 3+4 4 | 3+4 4 | 6 2 |
| | 6 2 | 6+6 4 | 5+5 4 | 3+4 4 | 3+4 4 | 3+4 4 | 3+4 4 | 5 2 |
| | 5 2 | 5+5 4 | 6+6 4 | 3+4 4 | 3+4 4 | 3+4 4 | 3+4 4 | 6 2 |
| | 6 2 | 5+5 4 | 6+6 4 | 3+4 4 | 3+4 4 | 3+4 4 | 3+4 4 | 6 2 |
| | 5 2 | 6+6 4 | 6+6 4 | 3+4 4 | 3+4 4 | 3+4 4 | 3+4 4 | 5 2 |
| | 5 2 | 6+6 4 | 5+5 4 | 3+4 4 | 3+4 4 | 3+4 4 | 3+4 4 | 5 2 |
| | 6 2 | 6+6 4 | 6+6 4 | 3+4 4 | 3+4 4 | 3+4 4 | 3+4 4 | 5 2 |
| | 6 2 | 5+5 4 | 5+5 4 | 3+4 4 | 3+4 4 | 3+4 4 | 3+4 4 | 6 2 |
| | 5 2 | 5+5 4 | 6+6 4 | 3+4 4 | 3+4 4 | 3+4 4 | 3+4 4 | 6 2 |
| | 6 2 | 5+5 4 | 5+5 4 | 3+4 4 | 3+4 4 | 3+4 4 | 3+4 4 | 5 2 |
| | 6 2 | 6+6 4 | 5+5 4 | 3+4 4 | 3+4 4 | 3+4 4 | 3+4 4 | 6 2 |
| | 5 2 | 5+5 4 | 5+5 4 | 3+4 4 | 3+4 4 | 3+4 4 | 3+4 4 | 5 2 |
| | 5 2 | 5+5 4 | 6+6 4 | 3+4 4 | 3+4 4 | 3+4 4 | 3+4 4 | 6 2 |
| | 6 2 | 6+6 4 | 6+6 4 | 3+4 4 | 3+4 4 | 3+4 4 | 3+4 4 | 5 2 |
| | 6 2 | 5+5 4 | 6+6 4 | 3+4 4 | 3+4 4 | 3+4 4 | 3+4 4 | 6 2 |
| | 5 2 | 6+6 4 | 5+5 4 | 3+4 4 | 3+4 4 | 3+4 4 | 3+4 4 | 5 2 |
| | 6 2 | 5+5 4 | 6+6 4 | 3+4 4 | 3+4 4 | 3+4 4 | 3+4 4 | 6 2 |
| | 5 2 | 6+6 4 | 5+5 4 | 3+4 4 | 3+4 4 | 3+4 4 | 3+4 4 | 5 2 |
| | 5 2 | 5+5 4 | 5+5 4 | 3+4 4 | 3+4 4 | 3+4 4 | 3+4 4 | 5 2 |
| | 6 2 | | | | | | | 6 2 |
| | 6 2 | | | | | | | 5 2 |
| | 6 2 | | | | | | | 6 2 |
| | 6 2 | | | | | | | 5 2 |
| | 5 2 | | | | | | | 6 2 |
| | 5 2 | | | | | | | 6 2 |
| | 6 2 | | | | | | | 5 2 |
| | 6 2 | | | | | | | 5 2 |
| | 6 2 | | | | | | | 5 2 |
| | 5 2 | | | | | | | 5 2 |
| | 5 2 | | | | | | | 5 2 |
| | 6 2 | | | | | | | 6 2 |
| | 5 2 | | | | | | | 5 2 |
| | 5 2 | | | | | | | 6 2 |
| | 6 2 | | | | | | | 6 2 |
| | 5 2 | | | | | | | 5 2 |
| | 5 2 | | | | | | | 5 2 |
| | 6 2 | | | | | | | 5 2 |
| | 6 2 | | | | | | | 6 2 |
| | 6 2 | | | | | | | 5 2 |
| | 5 2 | | | | | | | 6 2 |
| | 6 2 | | | | | | | 6 2 |
| | 6 2 | | | | | | | 5 2 |
| | 6 2 | | | | | | | 5 2 |
| | 5 2 | | | | | | | 6 2 |
| | 5 2 | | | | | | | 6 2 |
| | 5 2 | | | | | | | 6 2 |
| | 6 2 | | | | | | | 5 2 |
| | 5 2 | | | | | | | 5 2 |
| | 6 2 | | | | | | | 5 2 |
| | 6 2 | | | | | | | 6 2 |
| | 5 2 | | | | | | | 6 2 |
| 55.2.10 | 417814 | 128 | Mn2O5Sr2 | | | | | |
| | -GM | -R | -S | -T | -U | -X | -Y | -Z |
| | 5 2 | 6+6 4 | 6+6 4 | 3+4 4 | 3+4 4 | 3+4 4 | 3+4 4 | 5 2 |
| | 6 2 | 5+5 4 | 5+5 4 | 3+4 4 | 3+4 4 | 3+4 4 | 3+4 4 | 6 2 |
| | 5 2 | 5+5 4 | 6+6 4 | 3+4 4 | 3+4 4 | 3+4 4 | 3+4 4 | 5 2 |

Continued on next column

**TABLE XXXI– continued**

| TQC | ICSD | NoE | SF | | | | | |
|---|---|---|---|---|---|---|---|---|
| | 6 2 | 6+6 4 | 5+5 4 | 3+4 4 | 3+4 4 | 3+4 4 | 3+4 4 | 6 2 |
| | 5 2 | 5+5 4 | 6+6 4 | 3+4 4 | 3+4 4 | 3+4 4 | 3+4 4 | 6 2 |
| | 6 2 | 6+6 4 | 5+5 4 | 3+4 4 | 3+4 4 | 3+4 4 | 3+4 4 | 5 2 |
| | 5 2 | 5+5 4 | 6+6 4 | 3+4 4 | 3+4 4 | 3+4 4 | 3+4 4 | 6 2 |
| | 5 2 | 6+6 4 | 5+5 4 | 3+4 4 | 3+4 4 | 3+4 4 | 3+4 4 | 5 2 |
| | 6 2 | 5+5 4 | 6+6 4 | 3+4 4 | 3+4 4 | 3+4 4 | 3+4 4 | 6 2 |
| | 5 2 | 6+6 4 | 6+6 4 | 3+4 4 | 3+4 4 | 3+4 4 | 3+4 4 | 5 2 |
| | 6 2 | 6+6 4 | 5+5 4 | 3+4 4 | 3+4 4 | 3+4 4 | 3+4 4 | 6 2 |
| | 6 2 | 5+5 4 | 5+5 4 | 3+4 4 | 3+4 4 | 3+4 4 | 3+4 4 | 6 2 |
| | 5 2 | 5+5 4 | 6+6 4 | 3+4 4 | 3+4 4 | 3+4 4 | 3+4 4 | 5 2 |
| | 5 2 | 6+6 4 | 5+5 4 | 3+4 4 | 3+4 4 | 3+4 4 | 3+4 4 | 6 2 |
| | 6 2 | 5+5 4 | 6+6 4 | 3+4 4 | 3+4 4 | 3+4 4 | 3+4 4 | 5 2 |
| | 5 2 | 6+6 4 | 5+5 4 | 3+4 4 | 3+4 4 | 3+4 4 | 3+4 4 | 6 2 |
| | 6 2 | 6+6 4 | 5+5 4 | 3+4 4 | 3+4 4 | 3+4 4 | 3+4 4 | 6 2 |
| | 5 2 | 5+5 4 | 5+5 4 | 3+4 4 | 3+4 4 | 3+4 4 | 3+4 4 | 5 2 |
| | 5 2 | 5+5 4 | 6+6 4 | 3+4 4 | 3+4 4 | 3+4 4 | 3+4 4 | 5 2 |
| | 6 2 | 5+5 4 | 6+6 4 | 3+4 4 | 3+4 4 | 3+4 4 | 3+4 4 | 5 2 |
| | 6 2 | 6+6 4 | 5+5 4 | 3+4 4 | 3+4 4 | 3+4 4 | 3+4 4 | 6 2 |
| | 5 2 | 6+6 4 | 5+5 4 | 3+4 4 | 3+4 4 | 3+4 4 | 3+4 4 | 6 2 |
| | 6 2 | 6+6 4 | 6+6 4 | 3+4 4 | 3+4 4 | 3+4 4 | 3+4 4 | 5 2 |
| | 6 2 | 5+5 4 | 6+6 4 | 3+4 4 | 3+4 4 | 3+4 4 | 3+4 4 | 6 2 |
| | 5 2 | 6+6 4 | 6+6 4 | 3+4 4 | 3+4 4 | 3+4 4 | 3+4 4 | 5 2 |
| | 5 2 | 6+6 4 | 5+5 4 | 3+4 4 | 3+4 4 | 3+4 4 | 3+4 4 | 6 2 |
| | 6 2 | 5+5 4 | 5+5 4 | 3+4 4 | 3+4 4 | 3+4 4 | 3+4 4 | 5 2 |
| | 6 2 | 6+6 4 | 5+5 4 | 3+4 4 | 3+4 4 | 3+4 4 | 3+4 4 | 6 2 |
| | 5 2 | 5+5 4 | 6+6 4 | 3+4 4 | 3+4 4 | 3+4 4 | 3+4 4 | 5 2 |
| | 6 2 | 6+6 4 | 5+5 4 | 3+4 4 | 3+4 4 | 3+4 4 | 3+4 4 | 6 2 |
| | 5 2 | 5+5 4 | 6+6 4 | 3+4 4 | 3+4 4 | 3+4 4 | 3+4 4 | 5 2 |
| | 5 2 | 6+6 4 | 6+6 4 | 3+4 4 | 3+4 4 | 3+4 4 | 3+4 4 | 5 2 |
| | 6 2 | | | | | | | 6 2 |
| | 6 2 | | | | | | | 5 2 |
| | 6 2 | | | | | | | 6 2 |
| | 6 2 | | | | | | | 5 2 |
| | 5 2 | | | | | | | 6 2 |
| | 5 2 | | | | | | | 6 2 |
| | 6 2 | | | | | | | 5 2 |
| | 6 2 | | | | | | | 5 2 |
| | 6 2 | | | | | | | 5 2 |
| | 5 2 | | | | | | | 5 2 |
| | 5 2 | | | | | | | 5 2 |
| | 5 2 | | | | | | | 6 2 |
| | 6 2 | | | | | | | 5 2 |
| | 5 2 | | | | | | | 6 2 |
| | 6 2 | | | | | | | 6 2 |
| | 5 2 | | | | | | | 5 2 |
| | 5 2 | | | | | | | 5 2 |
| | 6 2 | | | | | | | 5 2 |
| | 6 2 | | | | | | | 6 2 |
| | 6 2 | | | | | | | 5 2 |
| | 5 2 | | | | | | | 6 2 |
| | 6 2 | | | | | | | 6 2 |
| | 6 2 | | | | | | | 5 2 |
| | 6 2 | | | | | | | 5 2 |
| | 5 2 | | | | | | | 6 2 |
| | 5 2 | | | | | | | 6 2 |
| | 5 2 | | | | | | | 6 2 |
| | 6 2 | | | | | | | 5 2 |
| | 5 2 | | | | | | | 5 2 |
| | 6 2 | | | | | | | 5 2 |
| | 6 2 | | | | | | | 6 2 |
| | 5 2 | | | | | | | 6 2 |

Continued on next column

**TABLE XXXI– continued**

| TQC | ICSD | NoE | SF | | | | | |
|---|---|---|---|---|---|---|---|---|
| 55.2.11 | 417813 | 128 | Mn2O5Sr2 | | | | | |
| | -GM | -R | -S | -T | -U | -X | -Y | -Z |
| | 5 2 | 5+5 4 | 5+5 4 | 3+4 4 | 3+4 4 | 3+4 4 | 3+4 4 | 5 2 |
| | 6 2 | 6+6 4 | 6+6 4 | 3+4 4 | 3+4 4 | 3+4 4 | 3+4 4 | 6 2 |
| | 5 2 | 6+6 4 | 5+5 4 | 3+4 4 | 3+4 4 | 3+4 4 | 3+4 4 | 5 2 |
| | 6 2 | 5+5 4 | 6+6 4 | 3+4 4 | 3+4 4 | 3+4 4 | 3+4 4 | 6 2 |
| | 5 2 | 6+6 4 | 5+5 4 | 3+4 4 | 3+4 4 | 3+4 4 | 3+4 4 | 6 2 |
| | 6 2 | 5+5 4 | 6+6 4 | 3+4 4 | 3+4 4 | 3+4 4 | 3+4 4 | 5 2 |
| | 5 2 | 6+6 4 | 5+5 4 | 3+4 4 | 3+4 4 | 3+4 4 | 3+4 4 | 6 2 |
| | 5 2 | 5+5 4 | 6+6 4 | 3+4 4 | 3+4 4 | 3+4 4 | 3+4 4 | 5 2 |
| | 6 2 | 6+6 4 | 5+5 4 | 3+4 4 | 3+4 4 | 3+4 4 | 3+4 4 | 6 2 |
| | 5 2 | 5+5 4 | 5+5 4 | 3+4 4 | 3+4 4 | 3+4 4 | 3+4 4 | 5 2 |
| | 6 2 | 5+5 4 | 6+6 4 | 3+4 4 | 3+4 4 | 3+4 4 | 3+4 4 | 6 2 |
| | 5 2 | 6+6 4 | 6+6 4 | 3+4 4 | 3+4 4 | 3+4 4 | 3+4 4 | 6 2 |
| | 6 2 | 6+6 4 | 5+5 4 | 3+4 4 | 3+4 4 | 3+4 4 | 3+4 4 | 5 2 |
| | 5 2 | 5+5 4 | 6+6 4 | 3+4 4 | 3+4 4 | 3+4 4 | 3+4 4 | 6 2 |
| | 6 2 | 6+6 4 | 5+5 4 | 3+4 4 | 3+4 4 | 3+4 4 | 3+4 4 | 5 2 |
| | 5 2 | 5+5 4 | 6+6 4 | 3+4 4 | 3+4 4 | 3+4 4 | 3+4 4 | 6 2 |
| | 6 2 | 5+5 4 | 6+6 4 | 3+4 4 | 3+4 4 | 3+4 4 | 3+4 4 | 6 2 |
| | 5 2 | 6+6 4 | 6+6 4 | 3+4 4 | 3+4 4 | 3+4 4 | 3+4 4 | 5 2 |
| | 5 2 | 6+6 4 | 5+5 4 | 3+4 4 | 3+4 4 | 3+4 4 | 3+4 4 | 5 2 |
| | 6 2 | 6+6 4 | 6+6 4 | 3+4 4 | 3+4 4 | 3+4 4 | 3+4 4 | 5 2 |
| | 6 2 | 5+5 4 | 5+5 4 | 3+4 4 | 3+4 4 | 3+4 4 | 3+4 4 | 6 2 |
| | 5 2 | 5+5 4 | 6+6 4 | 3+4 4 | 3+4 4 | 3+4 4 | 3+4 4 | 6 2 |
| | 6 2 | 5+5 4 | 5+5 4 | 3+4 4 | 3+4 4 | 3+4 4 | 3+4 4 | 5 2 |
| | 6 2 | 6+6 4 | 5+5 4 | 3+4 4 | 3+4 4 | 3+4 4 | 3+4 4 | 6 2 |
| | 5 2 | 5+5 4 | 5+5 4 | 3+4 4 | 3+4 4 | 3+4 4 | 3+4 4 | 5 2 |
| | 5 2 | 5+5 4 | 6+6 4 | 3+4 4 | 3+4 4 | 3+4 4 | 3+4 4 | 6 2 |
| | 6 2 | 6+6 4 | 6+6 4 | 3+4 4 | 3+4 4 | 3+4 4 | 3+4 4 | 5 2 |
| | 6 2 | 5+5 4 | 6+6 4 | 3+4 4 | 3+4 4 | 3+4 4 | 3+4 4 | 6 2 |
| | 5 2 | 6+6 4 | 5+5 4 | 3+4 4 | 3+4 4 | 3+4 4 | 3+4 4 | 5 2 |
| | 6 2 | 5+5 4 | 6+6 4 | 3+4 4 | 3+4 4 | 3+4 4 | 3+4 4 | 6 2 |
| | 5 2 | 6+6 4 | 5+5 4 | 3+4 4 | 3+4 4 | 3+4 4 | 3+4 4 | 5 2 |
| | 5 2 | 5+5 4 | 5+5 4 | 3+4 4 | 3+4 4 | 3+4 4 | 3+4 4 | 5 2 |
| | 6 2 | | | | | | | 6 2 |
| | 6 2 | | | | | | | 5 2 |
| | 6 2 | | | | | | | 6 2 |
| | 6 2 | | | | | | | 5 2 |
| | 5 2 | | | | | | | 6 2 |
| | 5 2 | | | | | | | 6 2 |
| | 6 2 | | | | | | | 5 2 |
| | 6 2 | | | | | | | 5 2 |
| | 6 2 | | | | | | | 5 2 |
| | 5 2 | | | | | | | 5 2 |
| | 5 2 | | | | | | | 5 2 |
| | 5 2 | | | | | | | 6 2 |
| | 6 2 | | | | | | | 5 2 |
| | 5 2 | | | | | | | 6 2 |
| | 6 2 | | | | | | | 6 2 |
| | 5 2 | | | | | | | 5 2 |
| | 5 2 | | | | | | | 5 2 |
| | 6 2 | | | | | | | 5 2 |
| | 6 2 | | | | | | | 6 2 |
| | 6 2 | | | | | | | 5 2 |
| | 5 2 | | | | | | | 6 2 |
| | 6 2 | | | | | | | 6 2 |
| | 6 2 | | | | | | | 5 2 |
| | 6 2 | | | | | | | 5 2 |
| | 5 2 | | | | | | | 6 2 |
| | 5 2 | | | | | | | 6 2 |
| | 5 2 | | | | | | | 6 2 |

Continued on next column

**TABLE XXXI– continued**

| TQC | ICSD | NoE | SF | | | | | |
|---|---|---|---|---|---|---|---|---|
| | 6 2 | | | | | | | 5 2 |
| | 5 2 | | | | | | | 5 2 |
| | 6 2 | | | | | | | 6 2 |
| | 6 2 | | | | | | | 5 2 |
| | 5 2 | | | | | | | 6 2 |
| 55.2.12 | 417812 | 128 | Mn2O5Sr2 | | | | | |
| | -GM | -R | -S | -T | -U | -X | -Y | -Z |
| | 5 2 | 6+6 4 | 6+6 4 | 3+4 4 | 3+4 4 | 3+4 4 | 3+4 4 | 5 2 |
| | 6 2 | 5+5 4 | 5+5 4 | 3+4 4 | 3+4 4 | 3+4 4 | 3+4 4 | 6 2 |
| | 5 2 | 5+5 4 | 6+6 4 | 3+4 4 | 3+4 4 | 3+4 4 | 3+4 4 | 5 2 |
| | 6 2 | 6+6 4 | 5+5 4 | 3+4 4 | 3+4 4 | 3+4 4 | 3+4 4 | 6 2 |
| | 5 2 | 5+5 4 | 6+6 4 | 3+4 4 | 3+4 4 | 3+4 4 | 3+4 4 | 6 2 |
| | 6 2 | 6+6 4 | 5+5 4 | 3+4 4 | 3+4 4 | 3+4 4 | 3+4 4 | 5 2 |
| | 5 2 | 5+5 4 | 6+6 4 | 3+4 4 | 3+4 4 | 3+4 4 | 3+4 4 | 6 2 |
| | 5 2 | 6+6 4 | 5+5 4 | 3+4 4 | 3+4 4 | 3+4 4 | 3+4 4 | 5 2 |
| | 6 2 | 5+5 4 | 6+6 4 | 3+4 4 | 3+4 4 | 3+4 4 | 3+4 4 | 6 2 |
| | 5 2 | 6+6 4 | 6+6 4 | 3+4 4 | 3+4 4 | 3+4 4 | 3+4 4 | 5 2 |
| | 6 2 | 6+6 4 | 5+5 4 | 3+4 4 | 3+4 4 | 3+4 4 | 3+4 4 | 6 2 |
| | 5 2 | 5+5 4 | 5+5 4 | 3+4 4 | 3+4 4 | 3+4 4 | 3+4 4 | 6 2 |
| | 6 2 | 5+5 4 | 6+6 4 | 3+4 4 | 3+4 4 | 3+4 4 | 3+4 4 | 5 2 |
| | 5 2 | 6+6 4 | 5+5 4 | 3+4 4 | 3+4 4 | 3+4 4 | 3+4 4 | 6 2 |
| | 6 2 | 5+5 4 | 6+6 4 | 3+4 4 | 3+4 4 | 3+4 4 | 3+4 4 | 5 2 |
| | 5 2 | 6+6 4 | 5+5 4 | 3+4 4 | 3+4 4 | 3+4 4 | 3+4 4 | 6 2 |
| | 6 2 | 6+6 4 | 5+5 4 | 3+4 4 | 3+4 4 | 3+4 4 | 3+4 4 | 6 2 |
| | 5 2 | 5+5 4 | 5+5 4 | 3+4 4 | 3+4 4 | 3+4 4 | 3+4 4 | 5 2 |
| | 6 2 | 5+5 4 | 6+6 4 | 3+4 4 | 3+4 4 | 3+4 4 | 3+4 4 | 5 2 |
| | 5 2 | 5+5 4 | 5+5 4 | 3+4 4 | 3+4 4 | 3+4 4 | 3+4 4 | 5 2 |
| | 6 2 | 6+6 4 | 6+6 4 | 3+4 4 | 3+4 4 | 3+4 4 | 3+4 4 | 6 2 |
| | 5 2 | 6+6 4 | 5+5 4 | 3+4 4 | 3+4 4 | 3+4 4 | 3+4 4 | 6 2 |
| | 6 2 | 6+6 4 | 6+6 4 | 3+4 4 | 3+4 4 | 3+4 4 | 3+4 4 | 5 2 |
| | 6 2 | 5+5 4 | 6+6 4 | 3+4 4 | 3+4 4 | 3+4 4 | 3+4 4 | 6 2 |
| | 5 2 | 6+6 4 | 6+6 4 | 3+4 4 | 3+4 4 | 3+4 4 | 3+4 4 | 5 2 |
| | 5 2 | 6+6 4 | 5+5 4 | 3+4 4 | 3+4 4 | 3+4 4 | 3+4 4 | 6 2 |
| | 6 2 | 5+5 4 | 5+5 4 | 3+4 4 | 3+4 4 | 3+4 4 | 3+4 4 | 5 2 |
| | 6 2 | 6+6 4 | 5+5 4 | 3+4 4 | 3+4 4 | 3+4 4 | 3+4 4 | 6 2 |
| | 5 2 | 5+5 4 | 6+6 4 | 3+4 4 | 3+4 4 | 3+4 4 | 3+4 4 | 5 2 |
| | 6 2 | 6+6 4 | 5+5 4 | 3+4 4 | 3+4 4 | 3+4 4 | 3+4 4 | 6 2 |
| | 5 2 | 5+5 4 | 6+6 4 | 3+4 4 | 3+4 4 | 3+4 4 | 3+4 4 | 5 2 |
| | 5 2 | 6+6 4 | 6+6 4 | 3+4 4 | 3+4 4 | 3+4 4 | 3+4 4 | 5 2 |
| | 6 2 | | | | | | | 6 2 |
| | 6 2 | | | | | | | 5 2 |
| | 6 2 | | | | | | | 6 2 |
| | 6 2 | | | | | | | 5 2 |
| | 5 2 | | | | | | | 6 2 |
| | 5 2 | | | | | | | 6 2 |
| | 6 2 | | | | | | | 5 2 |
| | 6 2 | | | | | | | 5 2 |
| | 6 2 | | | | | | | 5 2 |
| | 5 2 | | | | | | | 5 2 |
| | 5 2 | | | | | | | 5 2 |
| | 6 2 | | | | | | | 6 2 |
| | 5 2 | | | | | | | 5 2 |
| | 5 2 | | | | | | | 6 2 |
| | 6 2 | | | | | | | 6 2 |
| | 5 2 | | | | | | | 5 2 |
| | 5 2 | | | | | | | 5 2 |
| | 6 2 | | | | | | | 5 2 |
| | 6 2 | | | | | | | 6 2 |
| | 6 2 | | | | | | | 5 2 |
| | 5 2 | | | | | | | 6 2 |
| | 6 2 | | | | | | | 6 2 |

Continued on next column

**TABLE XXXI– continued**

| TQC | ICSD | NoE | SF | | | | | |
|---|---|---|---|---|---|---|---|---|
| | 6 2 | | | | | | | 5 2 |
| | 6 2 | | | | | | | 5 2 |
| | 5 2 | | | | | | | 6 2 |
| | 5 2 | | | | | | | 6 2 |
| | 5 2 | | | | | | | 6 2 |
| | 6 2 | | | | | | | 5 2 |
| | 5 2 | | | | | | | 5 2 |
| | 6 2 | | | | | | | 6 2 |
| | 6 2 | | | | | | | 5 2 |
| | 5 2 | | | | | | | 6 2 |
| 55.2.13 | 417811 | 128 | Mn2O5Sr2 | | | | | |
| | -GM | -R | -S | -T | -U | -X | -Y | -Z |
| | 5 2 | 6+6 4 | 6+6 4 | 3+4 4 | 3+4 4 | 3+4 4 | 3+4 4 | 5 2 |
| | 6 2 | 5+5 4 | 5+5 4 | 3+4 4 | 3+4 4 | 3+4 4 | 3+4 4 | 6 2 |
| | 5 2 | 5+5 4 | 6+6 4 | 3+4 4 | 3+4 4 | 3+4 4 | 3+4 4 | 5 2 |
| | 6 2 | 6+6 4 | 5+5 4 | 3+4 4 | 3+4 4 | 3+4 4 | 3+4 4 | 6 2 |
| | 5 2 | 5+5 4 | 6+6 4 | 3+4 4 | 3+4 4 | 3+4 4 | 3+4 4 | 6 2 |
| | 6 2 | 6+6 4 | 5+5 4 | 3+4 4 | 3+4 4 | 3+4 4 | 3+4 4 | 5 2 |
| | 5 2 | 5+5 4 | 6+6 4 | 3+4 4 | 3+4 4 | 3+4 4 | 3+4 4 | 6 2 |
| | 5 2 | 6+6 4 | 5+5 4 | 3+4 4 | 3+4 4 | 3+4 4 | 3+4 4 | 5 2 |
| | 6 2 | 5+5 4 | 6+6 4 | 3+4 4 | 3+4 4 | 3+4 4 | 3+4 4 | 6 2 |
| | 5 2 | 6+6 4 | 6+6 4 | 3+4 4 | 3+4 4 | 3+4 4 | 3+4 4 | 5 2 |
| | 6 2 | 6+6 4 | 5+5 4 | 3+4 4 | 3+4 4 | 3+4 4 | 3+4 4 | 6 2 |
| | 5 2 | 5+5 4 | 5+5 4 | 3+4 4 | 3+4 4 | 3+4 4 | 3+4 4 | 6 2 |
| | 6 2 | 5+5 4 | 6+6 4 | 3+4 4 | 3+4 4 | 3+4 4 | 3+4 4 | 5 2 |
| | 5 2 | 6+6 4 | 5+5 4 | 3+4 4 | 3+4 4 | 3+4 4 | 3+4 4 | 6 2 |
| | 6 2 | 5+5 4 | 6+6 4 | 3+4 4 | 3+4 4 | 3+4 4 | 3+4 4 | 5 2 |
| | 5 2 | 6+6 4 | 5+5 4 | 3+4 4 | 3+4 4 | 3+4 4 | 3+4 4 | 6 2 |
| | 6 2 | 6+6 4 | 5+5 4 | 3+4 4 | 3+4 4 | 3+4 4 | 3+4 4 | 6 2 |
| | 5 2 | 5+5 4 | 5+5 4 | 3+4 4 | 3+4 4 | 3+4 4 | 3+4 4 | 5 2 |
| | 6 2 | 5+5 4 | 6+6 4 | 3+4 4 | 3+4 4 | 3+4 4 | 3+4 4 | 5 2 |
| | 5 2 | 5+5 4 | 5+5 4 | 3+4 4 | 3+4 4 | 3+4 4 | 3+4 4 | 5 2 |
| | 6 2 | 6+6 4 | 6+6 4 | 3+4 4 | 3+4 4 | 3+4 4 | 3+4 4 | 6 2 |
| | 5 2 | 6+6 4 | 5+5 4 | 3+4 4 | 3+4 4 | 3+4 4 | 3+4 4 | 6 2 |
| | 6 2 | 6+6 4 | 6+6 4 | 3+4 4 | 3+4 4 | 3+4 4 | 3+4 4 | 5 2 |
| | 6 2 | 5+5 4 | 6+6 4 | 3+4 4 | 3+4 4 | 3+4 4 | 3+4 4 | 6 2 |
| | 5 2 | 6+6 4 | 6+6 4 | 3+4 4 | 3+4 4 | 3+4 4 | 3+4 4 | 5 2 |
| | 5 2 | 6+6 4 | 5+5 4 | 3+4 4 | 3+4 4 | 3+4 4 | 3+4 4 | 6 2 |
| | 6 2 | 5+5 4 | 5+5 4 | 3+4 4 | 3+4 4 | 3+4 4 | 3+4 4 | 5 2 |
| | 6 2 | 6+6 4 | 5+5 4 | 3+4 4 | 3+4 4 | 3+4 4 | 3+4 4 | 6 2 |
| | 5 2 | 5+5 4 | 6+6 4 | 3+4 4 | 3+4 4 | 3+4 4 | 3+4 4 | 5 2 |
| | 6 2 | 6+6 4 | 5+5 4 | 3+4 4 | 3+4 4 | 3+4 4 | 3+4 4 | 6 2 |
| | 5 2 | 5+5 4 | 6+6 4 | 3+4 4 | 3+4 4 | 3+4 4 | 3+4 4 | 5 2 |
| | 5 2 | 6+6 4 | 6+6 4 | 3+4 4 | 3+4 4 | 3+4 4 | 3+4 4 | 5 2 |
| | 6 2 | | | | | | | 6 2 |
| | 6 2 | | | | | | | 5 2 |
| | 6 2 | | | | | | | 6 2 |
| | 6 2 | | | | | | | 5 2 |
| | 5 2 | | | | | | | 6 2 |
| | 5 2 | | | | | | | 6 2 |
| | 6 2 | | | | | | | 5 2 |
| | 6 2 | | | | | | | 5 2 |
| | 6 2 | | | | | | | 5 2 |
| | 5 2 | | | | | | | 5 2 |
| | 5 2 | | | | | | | 5 2 |
| | 6 2 | | | | | | | 6 2 |
| | 5 2 | | | | | | | 5 2 |
| | 5 2 | | | | | | | 6 2 |
| | 6 2 | | | | | | | 6 2 |
| | 5 2 | | | | | | | 5 2 |
| | 5 2 | | | | | | | 5 2 |

Continued on next column

**TABLE XXXI– continued**

| TQC | ICSD | NoE | SF | | | | | |
|---|---|---|---|---|---|---|---|---|
| | 6 2 | | | | | | | 5 2 |
| | 6 2 | | | | | | | 6 2 |
| | 6 2 | | | | | | | 5 2 |
| | 5 2 | | | | | | | 6 2 |
| | 6 2 | | | | | | | 6 2 |
| | 6 2 | | | | | | | 5 2 |
| | 6 2 | | | | | | | 5 2 |
| | 5 2 | | | | | | | 6 2 |
| | 5 2 | | | | | | | 6 2 |
| | 5 2 | | | | | | | 6 2 |
| | 6 2 | | | | | | | 5 2 |
| | 5 2 | | | | | | | 5 2 |
| | 6 2 | | | | | | | 6 2 |
| | 6 2 | | | | | | | 5 2 |
| | 5 2 | | | | | | | 6 2 |
| 55.2.14 | 417810 | 128 | Mn2O5Sr2 | | | | | |
| | -GM | -R | -S | -T | -U | -X | -Y | -Z |
| | 5 2 | 6+6 4 | 6+6 4 | 3+4 4 | 3+4 4 | 3+4 4 | 3+4 4 | 5 2 |
| | 6 2 | 5+5 4 | 5+5 4 | 3+4 4 | 3+4 4 | 3+4 4 | 3+4 4 | 6 2 |
| | 5 2 | 5+5 4 | 6+6 4 | 3+4 4 | 3+4 4 | 3+4 4 | 3+4 4 | 5 2 |
| | 6 2 | 6+6 4 | 5+5 4 | 3+4 4 | 3+4 4 | 3+4 4 | 3+4 4 | 6 2 |
| | 5 2 | 5+5 4 | 6+6 4 | 3+4 4 | 3+4 4 | 3+4 4 | 3+4 4 | 6 2 |
| | 6 2 | 6+6 4 | 5+5 4 | 3+4 4 | 3+4 4 | 3+4 4 | 3+4 4 | 5 2 |
| | 5 2 | 5+5 4 | 6+6 4 | 3+4 4 | 3+4 4 | 3+4 4 | 3+4 4 | 6 2 |
| | 5 2 | 6+6 4 | 5+5 4 | 3+4 4 | 3+4 4 | 3+4 4 | 3+4 4 | 5 2 |
| | 6 2 | 5+5 4 | 6+6 4 | 3+4 4 | 3+4 4 | 3+4 4 | 3+4 4 | 6 2 |
| | 5 2 | 6+6 4 | 6+6 4 | 3+4 4 | 3+4 4 | 3+4 4 | 3+4 4 | 5 2 |
| | 6 2 | 6+6 4 | 5+5 4 | 3+4 4 | 3+4 4 | 3+4 4 | 3+4 4 | 6 2 |
| | 5 2 | 5+5 4 | 5+5 4 | 3+4 4 | 3+4 4 | 3+4 4 | 3+4 4 | 6 2 |
| | 6 2 | 5+5 4 | 6+6 4 | 3+4 4 | 3+4 4 | 3+4 4 | 3+4 4 | 5 2 |
| | 5 2 | 6+6 4 | 5+5 4 | 3+4 4 | 3+4 4 | 3+4 4 | 3+4 4 | 6 2 |
| | 6 2 | 5+5 4 | 6+6 4 | 3+4 4 | 3+4 4 | 3+4 4 | 3+4 4 | 5 2 |
| | 5 2 | 6+6 4 | 5+5 4 | 3+4 4 | 3+4 4 | 3+4 4 | 3+4 4 | 6 2 |
| | 6 2 | 6+6 4 | 5+5 4 | 3+4 4 | 3+4 4 | 3+4 4 | 3+4 4 | 6 2 |
| | 5 2 | 5+5 4 | 5+5 4 | 3+4 4 | 3+4 4 | 3+4 4 | 3+4 4 | 5 2 |
| | 6 2 | 5+5 4 | 6+6 4 | 3+4 4 | 3+4 4 | 3+4 4 | 3+4 4 | 5 2 |
| | 5 2 | 5+5 4 | 5+5 4 | 3+4 4 | 3+4 4 | 3+4 4 | 3+4 4 | 5 2 |
| | 6 2 | 6+6 4 | 6+6 4 | 3+4 4 | 3+4 4 | 3+4 4 | 3+4 4 | 6 2 |
| | 5 2 | 6+6 4 | 5+5 4 | 3+4 4 | 3+4 4 | 3+4 4 | 3+4 4 | 6 2 |
| | 6 2 | 6+6 4 | 6+6 4 | 3+4 4 | 3+4 4 | 3+4 4 | 3+4 4 | 5 2 |
| | 6 2 | 5+5 4 | 6+6 4 | 3+4 4 | 3+4 4 | 3+4 4 | 3+4 4 | 6 2 |
| | 5 2 | 6+6 4 | 6+6 4 | 3+4 4 | 3+4 4 | 3+4 4 | 3+4 4 | 5 2 |
| | 5 2 | 6+6 4 | 5+5 4 | 3+4 4 | 3+4 4 | 3+4 4 | 3+4 4 | 6 2 |
| | 6 2 | 5+5 4 | 5+5 4 | 3+4 4 | 3+4 4 | 3+4 4 | 3+4 4 | 5 2 |
| | 6 2 | 6+6 4 | 5+5 4 | 3+4 4 | 3+4 4 | 3+4 4 | 3+4 4 | 6 2 |
| | 5 2 | 5+5 4 | 6+6 4 | 3+4 4 | 3+4 4 | 3+4 4 | 3+4 4 | 5 2 |
| | 6 2 | 6+6 4 | 5+5 4 | 3+4 4 | 3+4 4 | 3+4 4 | 3+4 4 | 6 2 |
| | 5 2 | 5+5 4 | 6+6 4 | 3+4 4 | 3+4 4 | 3+4 4 | 3+4 4 | 5 2 |
| | 5 2 | 6+6 4 | 6+6 4 | 3+4 4 | 3+4 4 | 3+4 4 | 3+4 4 | 5 2 |
| | 6 2 | | | | | | | 6 2 |
| | 6 2 | | | | | | | 5 2 |
| | 6 2 | | | | | | | 6 2 |
| | 6 2 | | | | | | | 5 2 |
| | 5 2 | | | | | | | 6 2 |
| | 6 2 | | | | | | | 5 2 |
| | 5 2 | | | | | | | 5 2 |
| | 6 2 | | | | | | | 6 2 |
| | 6 2 | | | | | | | 5 2 |
| | 5 2 | | | | | | | 5 2 |
| | 5 2 | | | | | | | 5 2 |
| | 6 2 | | | | | | | 6 2 |

Continued on next column

**TABLE XXXI– continued**

| TQC | ICSD | NoE | SF | | | | | |
|---|---|---|---|---|---|---|---|---|
| | 5 2 | | | | | | | 5 2 |
| | 5 2 | | | | | | | 6 2 |
| | 6 2 | | | | | | | 6 2 |
| | 5 2 | | | | | | | 5 2 |
| | 5 2 | | | | | | | 5 2 |
| | 6 2 | | | | | | | 5 2 |
| | 6 2 | | | | | | | 6 2 |
| | 6 2 | | | | | | | 5 2 |
| | 5 2 | | | | | | | 6 2 |
| | 6 2 | | | | | | | 6 2 |
| | 6 2 | | | | | | | 5 2 |
| | 6 2 | | | | | | | 5 2 |
| | 5 2 | | | | | | | 6 2 |
| | 5 2 | | | | | | | 6 2 |
| | 5 2 | | | | | | | 6 2 |
| | 6 2 | | | | | | | 5 2 |
| | 5 2 | | | | | | | 5 2 |
| | 6 2 | | | | | | | 6 2 |
| | 6 2 | | | | | | | 5 2 |
| | 5 2 | | | | | | | 6 2 |
| 55.2.15 | 417809 | 128 | Mn2O5Sr2 | | | | | |
| | -GM | -R | -S | -T | -U | -X | -Y | -Z |
| | 5 2 | 5+5 4 | 5+5 4 | 3+4 4 | 3+4 4 | 3+4 4 | 3+4 4 | 5 2 |
| | 6 2 | 6+6 4 | 6+6 4 | 3+4 4 | 3+4 4 | 3+4 4 | 3+4 4 | 6 2 |
| | 5 2 | 6+6 4 | 5+5 4 | 3+4 4 | 3+4 4 | 3+4 4 | 3+4 4 | 5 2 |
| | 6 2 | 5+5 4 | 6+6 4 | 3+4 4 | 3+4 4 | 3+4 4 | 3+4 4 | 6 2 |
| | 5 2 | 6+6 4 | 5+5 4 | 3+4 4 | 3+4 4 | 3+4 4 | 3+4 4 | 6 2 |
| | 6 2 | 5+5 4 | 6+6 4 | 3+4 4 | 3+4 4 | 3+4 4 | 3+4 4 | 5 2 |
| | 5 2 | 6+6 4 | 5+5 4 | 3+4 4 | 3+4 4 | 3+4 4 | 3+4 4 | 6 2 |
| | 5 2 | 5+5 4 | 6+6 4 | 3+4 4 | 3+4 4 | 3+4 4 | 3+4 4 | 5 2 |
| | 6 2 | 6+6 4 | 5+5 4 | 3+4 4 | 3+4 4 | 3+4 4 | 3+4 4 | 6 2 |
| | 5 2 | 5+5 4 | 5+5 4 | 3+4 4 | 3+4 4 | 3+4 4 | 3+4 4 | 5 2 |
| | 6 2 | 5+5 4 | 6+6 4 | 3+4 4 | 3+4 4 | 3+4 4 | 3+4 4 | 6 2 |
| | 5 2 | 6+6 4 | 6+6 4 | 3+4 4 | 3+4 4 | 3+4 4 | 3+4 4 | 6 2 |
| | 6 2 | 6+6 4 | 5+5 4 | 3+4 4 | 3+4 4 | 3+4 4 | 3+4 4 | 5 2 |
| | 5 2 | 5+5 4 | 6+6 4 | 3+4 4 | 3+4 4 | 3+4 4 | 3+4 4 | 6 2 |
| | 6 2 | 6+6 4 | 5+5 4 | 3+4 4 | 3+4 4 | 3+4 4 | 3+4 4 | 5 2 |
| | 5 2 | 5+5 4 | 6+6 4 | 3+4 4 | 3+4 4 | 3+4 4 | 3+4 4 | 6 2 |
| | 6 2 | 5+5 4 | 6+6 4 | 3+4 4 | 3+4 4 | 3+4 4 | 3+4 4 | 6 2 |
| | 5 2 | 6+6 4 | 6+6 4 | 3+4 4 | 3+4 4 | 3+4 4 | 3+4 4 | 5 2 |
| | 6 2 | 6+6 4 | 5+5 4 | 3+4 4 | 3+4 4 | 3+4 4 | 3+4 4 | 5 2 |
| | 5 2 | 6+6 4 | 6+6 4 | 3+4 4 | 3+4 4 | 3+4 4 | 3+4 4 | 5 2 |
| | 6 2 | 5+5 4 | 5+5 4 | 3+4 4 | 3+4 4 | 3+4 4 | 3+4 4 | 6 2 |
| | 5 2 | 5+5 4 | 6+6 4 | 3+4 4 | 3+4 4 | 3+4 4 | 3+4 4 | 6 2 |
| | 6 2 | 5+5 4 | 5+5 4 | 3+4 4 | 3+4 4 | 3+4 4 | 3+4 4 | 5 2 |
| | 6 2 | 6+6 4 | 5+5 4 | 3+4 4 | 3+4 4 | 3+4 4 | 3+4 4 | 6 2 |
| | 5 2 | 5+5 4 | 5+5 4 | 3+4 4 | 3+4 4 | 3+4 4 | 3+4 4 | 5 2 |
| | 5 2 | 5+5 4 | 6+6 4 | 3+4 4 | 3+4 4 | 3+4 4 | 3+4 4 | 6 2 |
| | 6 2 | 6+6 4 | 6+6 4 | 3+4 4 | 3+4 4 | 3+4 4 | 3+4 4 | 5 2 |
| | 6 2 | 5+5 4 | 6+6 4 | 3+4 4 | 3+4 4 | 3+4 4 | 3+4 4 | 6 2 |
| | 5 2 | 6+6 4 | 5+5 4 | 3+4 4 | 3+4 4 | 3+4 4 | 3+4 4 | 5 2 |
| | 6 2 | 5+5 4 | 6+6 4 | 3+4 4 | 3+4 4 | 3+4 4 | 3+4 4 | 6 2 |
| | 5 2 | 6+6 4 | 5+5 4 | 3+4 4 | 3+4 4 | 3+4 4 | 3+4 4 | 5 2 |
| | 5 2 | 5+5 4 | 5+5 4 | 3+4 4 | 3+4 4 | 3+4 4 | 3+4 4 | 5 2 |
| | 6 2 | | | | | | | 6 2 |
| | 6 2 | | | | | | | 5 2 |
| | 6 2 | | | | | | | 6 2 |
| | 6 2 | | | | | | | 5 2 |
| | 5 2 | | | | | | | 6 2 |
| | 6 2 | | | | | | | 5 2 |
| | 5 2 | | | | | | | 5 2 |

Continued on next column

**TABLE XXXI– continued**

| TQC | ICSD | NoE | SF | | | | | |
|---|---|---|---|---|---|---|---|---|
| | 6 2 | | | | | | | 6 2 |
| | 6 2 | | | | | | | 5 2 |
| | 5 2 | | | | | | | 5 2 |
| | 5 2 | | | | | | | 5 2 |
| | 6 2 | | | | | | | 6 2 |
| | 5 2 | | | | | | | 5 2 |
| | 5 2 | | | | | | | 6 2 |
| | 6 2 | | | | | | | 6 2 |
| | 5 2 | | | | | | | 5 2 |
| | 5 2 | | | | | | | 5 2 |
| | 6 2 | | | | | | | 5 2 |
| | 6 2 | | | | | | | 6 2 |
| | 6 2 | | | | | | | 5 2 |
| | 5 2 | | | | | | | 6 2 |
| | 6 2 | | | | | | | 6 2 |
| | 6 2 | | | | | | | 5 2 |
| | 6 2 | | | | | | | 5 2 |
| | 5 2 | | | | | | | 6 2 |
| | 5 2 | | | | | | | 6 2 |
| | 5 2 | | | | | | | 6 2 |
| | 6 2 | | | | | | | 5 2 |
| | 5 2 | | | | | | | 5 2 |
| | 6 2 | | | | | | | 6 2 |
| | 6 2 | | | | | | | 5 2 |
| | 5 2 | | | | | | | 6 2 |
| 55.2.16 | 417808 | 128 | Mn2O5Sr2 | | | | | |
| | -GM | -R | -S | -T | -U | -X | -Y | -Z |
| | 5 2 | 5+5 4 | 5+5 4 | 3+4 4 | 3+4 4 | 3+4 4 | 3+4 4 | 5 2 |
| | 6 2 | 6+6 4 | 6+6 4 | 3+4 4 | 3+4 4 | 3+4 4 | 3+4 4 | 6 2 |
| | 5 2 | 6+6 4 | 5+5 4 | 3+4 4 | 3+4 4 | 3+4 4 | 3+4 4 | 5 2 |
| | 6 2 | 5+5 4 | 6+6 4 | 3+4 4 | 3+4 4 | 3+4 4 | 3+4 4 | 6 2 |
| | 5 2 | 6+6 4 | 5+5 4 | 3+4 4 | 3+4 4 | 3+4 4 | 3+4 4 | 6 2 |
| | 6 2 | 5+5 4 | 6+6 4 | 3+4 4 | 3+4 4 | 3+4 4 | 3+4 4 | 5 2 |
| | 5 2 | 6+6 4 | 5+5 4 | 3+4 4 | 3+4 4 | 3+4 4 | 3+4 4 | 6 2 |
| | 5 2 | 5+5 4 | 6+6 4 | 3+4 4 | 3+4 4 | 3+4 4 | 3+4 4 | 5 2 |
| | 6 2 | 6+6 4 | 5+5 4 | 3+4 4 | 3+4 4 | 3+4 4 | 3+4 4 | 6 2 |
| | 5 2 | 5+5 4 | 5+5 4 | 3+4 4 | 3+4 4 | 3+4 4 | 3+4 4 | 5 2 |
| | 6 2 | 5+5 4 | 6+6 4 | 3+4 4 | 3+4 4 | 3+4 4 | 3+4 4 | 6 2 |
| | 5 2 | 6+6 4 | 6+6 4 | 3+4 4 | 3+4 4 | 3+4 4 | 3+4 4 | 6 2 |
| | 6 2 | 6+6 4 | 5+5 4 | 3+4 4 | 3+4 4 | 3+4 4 | 3+4 4 | 5 2 |
| | 5 2 | 5+5 4 | 6+6 4 | 3+4 4 | 3+4 4 | 3+4 4 | 3+4 4 | 6 2 |
| | 6 2 | 6+6 4 | 5+5 4 | 3+4 4 | 3+4 4 | 3+4 4 | 3+4 4 | 5 2 |
| | 5 2 | 5+5 4 | 6+6 4 | 3+4 4 | 3+4 4 | 3+4 4 | 3+4 4 | 6 2 |
| | 6 2 | 5+5 4 | 6+6 4 | 3+4 4 | 3+4 4 | 3+4 4 | 3+4 4 | 6 2 |
| | 5 2 | 6+6 4 | 6+6 4 | 3+4 4 | 3+4 4 | 3+4 4 | 3+4 4 | 5 2 |
| | 6 2 | 6+6 4 | 5+5 4 | 3+4 4 | 3+4 4 | 3+4 4 | 3+4 4 | 5 2 |
| | 5 2 | 6+6 4 | 6+6 4 | 3+4 4 | 3+4 4 | 3+4 4 | 3+4 4 | 5 2 |
| | 6 2 | 5+5 4 | 5+5 4 | 3+4 4 | 3+4 4 | 3+4 4 | 3+4 4 | 6 2 |
| | 5 2 | 5+5 4 | 6+6 4 | 3+4 4 | 3+4 4 | 3+4 4 | 3+4 4 | 6 2 |
| | 6 2 | 5+5 4 | 5+5 4 | 3+4 4 | 3+4 4 | 3+4 4 | 3+4 4 | 5 2 |
| | 6 2 | 6+6 4 | 5+5 4 | 3+4 4 | 3+4 4 | 3+4 4 | 3+4 4 | 6 2 |
| | 5 2 | 5+5 4 | 5+5 4 | 3+4 4 | 3+4 4 | 3+4 4 | 3+4 4 | 5 2 |
| | 5 2 | 5+5 4 | 6+6 4 | 3+4 4 | 3+4 4 | 3+4 4 | 3+4 4 | 6 2 |
| | 6 2 | 6+6 4 | 6+6 4 | 3+4 4 | 3+4 4 | 3+4 4 | 3+4 4 | 5 2 |
| | 6 2 | 5+5 4 | 6+6 4 | 3+4 4 | 3+4 4 | 3+4 4 | 3+4 4 | 6 2 |
| | 5 2 | 6+6 4 | 5+5 4 | 3+4 4 | 3+4 4 | 3+4 4 | 3+4 4 | 5 2 |
| | 6 2 | 5+5 4 | 6+6 4 | 3+4 4 | 3+4 4 | 3+4 4 | 3+4 4 | 6 2 |
| | 5 2 | 6+6 4 | 5+5 4 | 3+4 4 | 3+4 4 | 3+4 4 | 3+4 4 | 5 2 |
| | 5 2 | 5+5 4 | 5+5 4 | 3+4 4 | 3+4 4 | 3+4 4 | 3+4 4 | 5 2 |
| | 6 2 | | | | | | | 5 2 |
| | 6 2 | | | | | | | 6 2 |

Continued on next column

**TABLE XXXI– continued**

| TQC | ICSD | NoE | SF | | | | | |
|---|---|---|---|---|---|---|---|---|
| | 6 2 | | | | | | | 6 2 |
| | 6 2 | | | | | | | 5 2 |
| | 5 2 | | | | | | | 6 2 |
| | 6 2 | | | | | | | 5 2 |
| | 5 2 | | | | | | | 6 2 |
| | 6 2 | | | | | | | 5 2 |
| | 6 2 | | | | | | | 5 2 |
| | 5 2 | | | | | | | 5 2 |
| | 5 2 | | | | | | | 5 2 |
| | 6 2 | | | | | | | 6 2 |
| | 5 2 | | | | | | | 5 2 |
| | 5 2 | | | | | | | 6 2 |
| | 6 2 | | | | | | | 6 2 |
| | 5 2 | | | | | | | 5 2 |
| | 5 2 | | | | | | | 5 2 |
| | 6 2 | | | | | | | 5 2 |
| | 6 2 | | | | | | | 6 2 |
| | 6 2 | | | | | | | 5 2 |
| | 5 2 | | | | | | | 6 2 |
| | 6 2 | | | | | | | 6 2 |
| | 6 2 | | | | | | | 5 2 |
| | 6 2 | | | | | | | 5 2 |
| | 5 2 | | | | | | | 6 2 |
| | 5 2 | | | | | | | 6 2 |
| | 5 2 | | | | | | | 6 2 |
| | 6 2 | | | | | | | 5 2 |
| | 5 2 | | | | | | | 6 2 |
| | 6 2 | | | | | | | 5 2 |
| | 6 2 | | | | | | | 5 2 |
| | 5 2 | | | | | | | 6 2 |
| 55.2.17 | 417807 | 128 | Mn2O5Sr2 | | | | | |
| | -GM | -R | -S | -T | -U | -X | -Y | -Z |
| | 5 2 | 5+5 4 | 5+5 4 | 3+4 4 | 3+4 4 | 3+4 4 | 3+4 4 | 5 2 |
| | 6 2 | 6+6 4 | 6+6 4 | 3+4 4 | 3+4 4 | 3+4 4 | 3+4 4 | 6 2 |
| | 5 2 | 6+6 4 | 5+5 4 | 3+4 4 | 3+4 4 | 3+4 4 | 3+4 4 | 5 2 |
| | 6 2 | 5+5 4 | 6+6 4 | 3+4 4 | 3+4 4 | 3+4 4 | 3+4 4 | 6 2 |
| | 5 2 | 6+6 4 | 5+5 4 | 3+4 4 | 3+4 4 | 3+4 4 | 3+4 4 | 6 2 |
| | 6 2 | 5+5 4 | 6+6 4 | 3+4 4 | 3+4 4 | 3+4 4 | 3+4 4 | 5 2 |
| | 5 2 | 6+6 4 | 5+5 4 | 3+4 4 | 3+4 4 | 3+4 4 | 3+4 4 | 6 2 |
| | 5 2 | 5+5 4 | 6+6 4 | 3+4 4 | 3+4 4 | 3+4 4 | 3+4 4 | 5 2 |
| | 6 2 | 6+6 4 | 5+5 4 | 3+4 4 | 3+4 4 | 3+4 4 | 3+4 4 | 6 2 |
| | 5 2 | 5+5 4 | 5+5 4 | 3+4 4 | 3+4 4 | 3+4 4 | 3+4 4 | 5 2 |
| | 6 2 | 5+5 4 | 6+6 4 | 3+4 4 | 3+4 4 | 3+4 4 | 3+4 4 | 6 2 |
| | 5 2 | 6+6 4 | 6+6 4 | 3+4 4 | 3+4 4 | 3+4 4 | 3+4 4 | 6 2 |
| | 6 2 | 6+6 4 | 5+5 4 | 3+4 4 | 3+4 4 | 3+4 4 | 3+4 4 | 5 2 |
| | 5 2 | 5+5 4 | 6+6 4 | 3+4 4 | 3+4 4 | 3+4 4 | 3+4 4 | 6 2 |
| | 6 2 | 6+6 4 | 5+5 4 | 3+4 4 | 3+4 4 | 3+4 4 | 3+4 4 | 5 2 |
| | 5 2 | 5+5 4 | 6+6 4 | 3+4 4 | 3+4 4 | 3+4 4 | 3+4 4 | 6 2 |
| | 6 2 | 5+5 4 | 6+6 4 | 3+4 4 | 3+4 4 | 3+4 4 | 3+4 4 | 6 2 |
| | 5 2 | 6+6 4 | 6+6 4 | 3+4 4 | 3+4 4 | 3+4 4 | 3+4 4 | 5 2 |
| | 6 2 | 6+6 4 | 5+5 4 | 3+4 4 | 3+4 4 | 3+4 4 | 3+4 4 | 5 2 |
| | 5 2 | 6+6 4 | 6+6 4 | 3+4 4 | 3+4 4 | 3+4 4 | 3+4 4 | 5 2 |
| | 6 2 | 5+5 4 | 5+5 4 | 3+4 4 | 3+4 4 | 3+4 4 | 3+4 4 | 6 2 |
| | 5 2 | 5+5 4 | 6+6 4 | 3+4 4 | 3+4 4 | 3+4 4 | 3+4 4 | 6 2 |
| | 6 2 | 5+5 4 | 5+5 4 | 3+4 4 | 3+4 4 | 3+4 4 | 3+4 4 | 5 2 |
| | 6 2 | 6+6 4 | 5+5 4 | 3+4 4 | 3+4 4 | 3+4 4 | 3+4 4 | 6 2 |
| | 5 2 | 5+5 4 | 5+5 4 | 3+4 4 | 3+4 4 | 3+4 4 | 3+4 4 | 5 2 |
| | 5 2 | 5+5 4 | 6+6 4 | 3+4 4 | 3+4 4 | 3+4 4 | 3+4 4 | 6 2 |
| | 6 2 | 6+6 4 | 6+6 4 | 3+4 4 | 3+4 4 | 3+4 4 | 3+4 4 | 5 2 |
| | 6 2 | 5+5 4 | 6+6 4 | 3+4 4 | 3+4 4 | 3+4 4 | 3+4 4 | 6 2 |
| | 5 2 | 6+6 4 | 5+5 4 | 3+4 4 | 3+4 4 | 3+4 4 | 3+4 4 | 5 2 |

Continued on next column

**TABLE XXXI– continued**

| TQC | ICSD | NoE | SF | | | | | |
|---|---|---|---|---|---|---|---|---|
| | 6 2 | 5+5 4 | 6+6 4 | 3+4 4 | 3+4 4 | 3+4 4 | 3+4 4 | 6 2 |
| | 5 2 | 6+6 4 | 5+5 4 | 3+4 4 | 3+4 4 | 3+4 4 | 3+4 4 | 5 2 |
| | 5 2 | 5+5 4 | 5+5 4 | 3+4 4 | 3+4 4 | 3+4 4 | 3+4 4 | 5 2 |
| | 6 2 | | | | | | | 5 2 |
| | 6 2 | | | | | | | 6 2 |
| | 6 2 | | | | | | | 6 2 |
| | 6 2 | | | | | | | 5 2 |
| | 5 2 | | | | | | | 6 2 |
| | 6 2 | | | | | | | 5 2 |
| | 5 2 | | | | | | | 5 2 |
| | 6 2 | | | | | | | 6 2 |
| | 6 2 | | | | | | | 5 2 |
| | 5 2 | | | | | | | 5 2 |
| | 5 2 | | | | | | | 5 2 |
| | 6 2 | | | | | | | 6 2 |
| | 5 2 | | | | | | | 5 2 |
| | 5 2 | | | | | | | 6 2 |
| | 6 2 | | | | | | | 6 2 |
| | 5 2 | | | | | | | 5 2 |
| | 5 2 | | | | | | | 5 2 |
| | 6 2 | | | | | | | 5 2 |
| | 6 2 | | | | | | | 6 2 |
| | 6 2 | | | | | | | 5 2 |
| | 5 2 | | | | | | | 6 2 |
| | 6 2 | | | | | | | 6 2 |
| | 6 2 | | | | | | | 5 2 |
| | 6 2 | | | | | | | 5 2 |
| | 5 2 | | | | | | | 6 2 |
| | 5 2 | | | | | | | 6 2 |
| | 5 2 | | | | | | | 6 2 |
| | 6 2 | | | | | | | 5 2 |
| | 5 2 | | | | | | | 6 2 |
| | 6 2 | | | | | | | 5 2 |
| | 6 2 | | | | | | | 5 2 |
| | 5 2 | | | | | | | 6 2 |
| 55.2.18 | 417806 | 128 | Mn2O5Sr2 | | | | | |
| | -GM | -R | -S | -T | -U | -X | -Y | -Z |
| | 5 2 | 5+5 4 | 5+5 4 | 3+4 4 | 3+4 4 | 3+4 4 | 3+4 4 | 5 2 |
| | 6 2 | 6+6 4 | 6+6 4 | 3+4 4 | 3+4 4 | 3+4 4 | 3+4 4 | 6 2 |
| | 5 2 | 6+6 4 | 5+5 4 | 3+4 4 | 3+4 4 | 3+4 4 | 3+4 4 | 5 2 |
| | 6 2 | 5+5 4 | 6+6 4 | 3+4 4 | 3+4 4 | 3+4 4 | 3+4 4 | 6 2 |
| | 5 2 | 6+6 4 | 5+5 4 | 3+4 4 | 3+4 4 | 3+4 4 | 3+4 4 | 6 2 |
| | 6 2 | 5+5 4 | 6+6 4 | 3+4 4 | 3+4 4 | 3+4 4 | 3+4 4 | 5 2 |
| | 5 2 | 6+6 4 | 5+5 4 | 3+4 4 | 3+4 4 | 3+4 4 | 3+4 4 | 6 2 |
| | 5 2 | 5+5 4 | 6+6 4 | 3+4 4 | 3+4 4 | 3+4 4 | 3+4 4 | 5 2 |
| | 6 2 | 6+6 4 | 5+5 4 | 3+4 4 | 3+4 4 | 3+4 4 | 3+4 4 | 6 2 |
| | 5 2 | 5+5 4 | 5+5 4 | 3+4 4 | 3+4 4 | 3+4 4 | 3+4 4 | 5 2 |
| | 6 2 | 5+5 4 | 6+6 4 | 3+4 4 | 3+4 4 | 3+4 4 | 3+4 4 | 6 2 |
| | 5 2 | 6+6 4 | 6+6 4 | 3+4 4 | 3+4 4 | 3+4 4 | 3+4 4 | 6 2 |
| | 6 2 | 6+6 4 | 5+5 4 | 3+4 4 | 3+4 4 | 3+4 4 | 3+4 4 | 5 2 |
| | 5 2 | 5+5 4 | 6+6 4 | 3+4 4 | 3+4 4 | 3+4 4 | 3+4 4 | 6 2 |
| | 6 2 | 6+6 4 | 5+5 4 | 3+4 4 | 3+4 4 | 3+4 4 | 3+4 4 | 5 2 |
| | 5 2 | 5+5 4 | 6+6 4 | 3+4 4 | 3+4 4 | 3+4 4 | 3+4 4 | 6 2 |
| | 6 2 | 5+5 4 | 6+6 4 | 3+4 4 | 3+4 4 | 3+4 4 | 3+4 4 | 6 2 |
| | 5 2 | 6+6 4 | 6+6 4 | 3+4 4 | 3+4 4 | 3+4 4 | 3+4 4 | 5 2 |
| | 6 2 | 6+6 4 | 5+5 4 | 3+4 4 | 3+4 4 | 3+4 4 | 3+4 4 | 5 2 |
| | 5 2 | 6+6 4 | 6+6 4 | 3+4 4 | 3+4 4 | 3+4 4 | 3+4 4 | 5 2 |
| | 6 2 | 5+5 4 | 5+5 4 | 3+4 4 | 3+4 4 | 3+4 4 | 3+4 4 | 6 2 |
| | 5 2 | 5+5 4 | 6+6 4 | 3+4 4 | 3+4 4 | 3+4 4 | 3+4 4 | 6 2 |
| | 6 2 | 5+5 4 | 5+5 4 | 3+4 4 | 3+4 4 | 3+4 4 | 3+4 4 | 6 2 |
| | 6 2 | 6+6 4 | 5+5 4 | 3+4 4 | 3+4 4 | 3+4 4 | 3+4 4 | 5 2 |

Continued on next column

**TABLE XXXI– continued**

| TQC | ICSD | NoE | SF | | | | | |
|---|---|---|---|---|---|---|---|---|
| | 5 2 | 5+5 4 | 5+5 4 | 3+4 4 | 3+4 4 | 3+4 4 | 3+4 4 | 5 2 |
| | 5 2 | 5+5 4 | 6+6 4 | 3+4 4 | 3+4 4 | 3+4 4 | 3+4 4 | 6 2 |
| | 6 2 | 6+6 4 | 6+6 4 | 3+4 4 | 3+4 4 | 3+4 4 | 3+4 4 | 5 2 |
| | 6 2 | 5+5 4 | 6+6 4 | 3+4 4 | 3+4 4 | 3+4 4 | 3+4 4 | 6 2 |
| | 5 2 | 6+6 4 | 5+5 4 | 3+4 4 | 3+4 4 | 3+4 4 | 3+4 4 | 5 2 |
| | 6 2 | 5+5 4 | 6+6 4 | 3+4 4 | 3+4 4 | 3+4 4 | 3+4 4 | 6 2 |
| | 5 2 | 6+6 4 | 5+5 4 | 3+4 4 | 3+4 4 | 3+4 4 | 3+4 4 | 5 2 |
| | 5 2 | 5+5 4 | 5+5 4 | 3+4 4 | 3+4 4 | 3+4 4 | 3+4 4 | 5 2 |
| | 6 2 | | | | | | | 5 2 |
| | 6 2 | | | | | | | 6 2 |
| | 6 2 | | | | | | | 6 2 |
| | 6 2 | | | | | | | 5 2 |
| | 5 2 | | | | | | | 6 2 |
| | 6 2 | | | | | | | 5 2 |
| | 5 2 | | | | | | | 5 2 |
| | 6 2 | | | | | | | 6 2 |
| | 6 2 | | | | | | | 5 2 |
| | 5 2 | | | | | | | 5 2 |
| | 5 2 | | | | | | | 5 2 |
| | 6 2 | | | | | | | 6 2 |
| | 5 2 | | | | | | | 5 2 |
| | 5 2 | | | | | | | 6 2 |
| | 6 2 | | | | | | | 6 2 |
| | 5 2 | | | | | | | 5 2 |
| | 5 2 | | | | | | | 5 2 |
| | 6 2 | | | | | | | 5 2 |
| | 6 2 | | | | | | | 6 2 |
| | 6 2 | | | | | | | 5 2 |
| | 5 2 | | | | | | | 6 2 |
| | 6 2 | | | | | | | 6 2 |
| | 6 2 | | | | | | | 5 2 |
| | 6 2 | | | | | | | 5 2 |
| | 5 2 | | | | | | | 6 2 |
| | 5 2 | | | | | | | 6 2 |
| | 5 2 | | | | | | | 6 2 |
| | 6 2 | | | | | | | 5 2 |
| | 5 2 | | | | | | | 6 2 |
| | 6 2 | | | | | | | 5 2 |
| | 6 2 | | | | | | | 5 2 |
| | 5 2 | | | | | | | 6 2 |
| 55.2.19 | 391127 | 100 | In4Rh2Zr5 | | | | | |
| | -GM | -R | -S | -T | -U | -X | -Y | -Z |
| | 5 2 | 6+6 4 | 6+6 4 | 3+4 4 | 3+4 4 | 3+4 4 | 3+4 4 | 6 2 |
| | 5 2 | 5+5 4 | 5+5 4 | 3+4 4 | 3+4 4 | 3+4 4 | 3+4 4 | 5 2 |
| | 6 2 | 6+6 4 | 5+5 4 | 3+4 4 | 3+4 4 | 3+4 4 | 3+4 4 | 5 2 |
| | 6 2 | 5+5 4 | 6+6 4 | 3+4 4 | 3+4 4 | 3+4 4 | 3+4 4 | 6 2 |
| | 5 2 | 6+6 4 | 6+6 4 | 3+4 4 | 3+4 4 | 3+4 4 | 3+4 4 | 6 2 |
| | 5 2 | 5+5 4 | 5+5 4 | 3+4 4 | 3+4 4 | 3+4 4 | 3+4 4 | 6 2 |
| | 6 2 | 5+5 4 | 5+5 4 | 3+4 4 | 3+4 4 | 3+4 4 | 3+4 4 | 5 2 |
| | 6 2 | 6+6 4 | 6+6 4 | 3+4 4 | 3+4 4 | 3+4 4 | 3+4 4 | 5 2 |
| | 5 2 | 6+6 4 | 6+6 4 | 3+4 4 | 3+4 4 | 3+4 4 | 3+4 4 | 5 2 |
| | 6 2 | 6+6 4 | 5+5 4 | 3+4 4 | 3+4 4 | 3+4 4 | 3+4 4 | 5 2 |
| | 5 2 | 5+5 4 | 6+6 4 | 3+4 4 | 3+4 4 | 3+4 4 | 3+4 4 | 6 2 |
| | 5 2 | 5+5 4 | 5+5 4 | 3+4 4 | 3+4 4 | 3+4 4 | 3+4 4 | 6 2 |
| | 6 2 | 6+6 4 | 6+6 4 | 3+4 4 | 3+4 4 | 3+4 4 | 3+4 4 | 5 2 |
| | 5 2 | 6+6 4 | 5+5 4 | 3+4 4 | 3+4 4 | 3+4 4 | 3+4 4 | 5 2 |
| | 6 2 | 5+5 4 | 6+6 4 | 3+4 4 | 3+4 4 | 3+4 4 | 3+4 4 | 6 2 |
| | 5 2 | 6+6 4 | 5+5 4 | 3+4 4 | 3+4 4 | 3+4 4 | 3+4 4 | 5 2 |
| | 6 2 | 6+6 4 | 6+6 4 | 3+4 4 | 3+4 4 | 3+4 4 | 3+4 4 | 6 2 |
| | 5 2 | 5+5 4 | 5+5 4 | 3+4 4 | 3+4 4 | 3+4 4 | 3+4 4 | 6 2 |
| | 6 2 | 5+5 4 | 6+6 4 | 3+4 4 | 3+4 4 | 3+4 4 | 3+4 4 | 5 2 |

Continued on next column

**TABLE XXXI– continued**

| TQC | ICSD | NoE | SF | | | | | |
|---|---|---|---|---|---|---|---|---|
| | 5 2 | 6+6 4 | 5+5 4 | 3+4 4 | 3+4 4 | 3+4 4 | 3+4 4 | 6 2 |
| | 6 2 | 5+5 4 | 5+5 4 | 3+4 4 | 3+4 4 | 3+4 4 | 3+4 4 | 5 2 |
| | 5 2 | 6+6 4 | 6+6 4 | 3+4 4 | 3+4 4 | 3+4 4 | 3+4 4 | 5 2 |
| | 6 2 | 6+6 4 | 6+6 4 | 3+4 4 | 3+4 4 | 3+4 4 | 3+4 4 | 6 2 |
| | 5 2 | 5+5 4 | 5+5 4 | 3+4 4 | 3+4 4 | 3+4 4 | 3+4 4 | 6 2 |
| | 6 2 | 5+5 4 | 6+6 4 | 3+4 4 | 3+4 4 | 3+4 4 | 3+4 4 | 5 2 |
| | 5 2 | | | | | | | 6 2 |
| | 6 2 | | | | | | | 5 2 |
| | 6 2 | | | | | | | 5 2 |
| | 5 2 | | | | | | | 6 2 |
| | 6 2 | | | | | | | 6 2 |
| | 5 2 | | | | | | | 5 2 |
| | 6 2 | | | | | | | 6 2 |
| | 6 2 | | | | | | | 5 2 |
| | 5 2 | | | | | | | 6 2 |
| | 5 2 | | | | | | | 6 2 |
| | 6 2 | | | | | | | 5 2 |
| | 5 2 | | | | | | | 5 2 |
| | 5 2 | | | | | | | 5 2 |
| | 5 2 | | | | | | | 6 2 |
| | 5 2 | | | | | | | 6 2 |
| | 6 2 | | | | | | | 6 2 |
| | 6 2 | | | | | | | 5 2 |
| | 5 2 | | | | | | | 6 2 |
| | 5 2 | | | | | | | 5 2 |
| | 5 2 | | | | | | | 5 2 |
| | 6 2 | | | | | | | 6 2 |
| | 6 2 | | | | | | | 5 2 |
| | 5 2 | | | | | | | 6 2 |
| | 5 2 | | | | | | | 5 2 |
| | 6 2 | | | | | | | 5 2 |
| 55.2.20 | 193223 | 132 | In1La2Si2 | | | | | |
| | -GM | -R | -S | -T | -U | -X | -Y | -Z |
| | 5 2 | 5+5 4 | 5+5 4 | 3+4 4 | 3+4 4 | 3+4 4 | 3+4 4 | 6 2 |
| | 5 2 | 6+6 4 | 6+6 4 | 3+4 4 | 3+4 4 | 3+4 4 | 3+4 4 | 6 2 |
| | 6 2 | 5+5 4 | 6+6 4 | 3+4 4 | 3+4 4 | 3+4 4 | 3+4 4 | 5 2 |
| | 6 2 | 6+6 4 | 5+5 4 | 3+4 4 | 3+4 4 | 3+4 4 | 3+4 4 | 5 2 |
| | 6 2 | 5+5 4 | 6+6 4 | 3+4 4 | 3+4 4 | 3+4 4 | 3+4 4 | 5 2 |
| | 5 2 | 6+6 4 | 5+5 4 | 3+4 4 | 3+4 4 | 3+4 4 | 3+4 4 | 6 2 |
| | 5 2 | 6+6 4 | 5+5 4 | 3+4 4 | 3+4 4 | 3+4 4 | 3+4 4 | 5 2 |
| | 6 2 | 5+5 4 | 6+6 4 | 3+4 4 | 3+4 4 | 3+4 4 | 3+4 4 | 6 2 |
| | 5 2 | 5+5 4 | 6+6 4 | 3+4 4 | 3+4 4 | 3+4 4 | 3+4 4 | 6 2 |
| | 5 2 | 6+6 4 | 5+5 4 | 3+4 4 | 3+4 4 | 3+4 4 | 3+4 4 | 6 2 |
| | 6 2 | 6+6 4 | 5+5 4 | 3+4 4 | 3+4 4 | 3+4 4 | 3+4 4 | 5 2 |
| | 6 2 | 5+5 4 | 6+6 4 | 3+4 4 | 3+4 4 | 3+4 4 | 3+4 4 | 5 2 |
| | 5 2 | 5+5 4 | 6+6 4 | 3+4 4 | 3+4 4 | 3+4 4 | 3+4 4 | 6 2 |
| | 6 2 | 6+6 4 | 5+5 4 | 3+4 4 | 3+4 4 | 3+4 4 | 3+4 4 | 5 2 |
| | 5 2 | 6+6 4 | 5+5 4 | 3+4 4 | 3+4 4 | 3+4 4 | 3+4 4 | 6 2 |
| | 6 2 | 5+5 4 | 6+6 4 | 3+4 4 | 3+4 4 | 3+4 4 | 3+4 4 | 5 2 |
| | 5 2 | 5+5 4 | 5+5 4 | 3+4 4 | 3+4 4 | 3+4 4 | 3+4 4 | 6 2 |
| | 5 2 | 6+6 4 | 6+6 4 | 3+4 4 | 3+4 4 | 3+4 4 | 3+4 4 | 6 2 |
| | 6 2 | 5+5 4 | 5+5 4 | 3+4 4 | 3+4 4 | 3+4 4 | 3+4 4 | 5 2 |
| | 5 2 | 6+6 4 | 6+6 4 | 3+4 4 | 3+4 4 | 3+4 4 | 3+4 4 | 5 2 |
| | 6 2 | 5+5 4 | 5+5 4 | 3+4 4 | 3+4 4 | 3+4 4 | 3+4 4 | 6 2 |
| | 6 2 | 6+6 4 | 6+6 4 | 3+4 4 | 3+4 4 | 3+4 4 | 3+4 4 | 5 2 |
| | 5 2 | 6+6 4 | 6+6 4 | 3+4 4 | 3+4 4 | 3+4 4 | 3+4 4 | 6 2 |
| | 5 2 | 6+6 4 | 5+5 4 | 3+4 4 | 3+4 4 | 3+4 4 | 3+4 4 | 5 2 |
| | 6 2 | 5+5 4 | 6+6 4 | 3+4 4 | 3+4 4 | 3+4 4 | 3+4 4 | 6 2 |
| | 5 2 | 5+5 4 | 5+5 4 | 3+4 4 | 3+4 4 | 3+4 4 | 3+4 4 | 5 2 |
| | 6 2 | 5+5 4 | 5+5 4 | 3+4 4 | 3+4 4 | 3+4 4 | 3+4 4 | 6 2 |
| | 6 2 | 6+6 4 | 6+6 4 | 3+4 4 | 3+4 4 | 3+4 4 | 3+4 4 | 5 2 |

Continued on next column

**TABLE XXXI– continued**

| TQC | ICSD | NoE | SF | | | | | |
|---|---|---|---|---|---|---|---|---|
| | 6 2 | 6+6 4 | 5+5 4 | 3+4 4 | 3+4 4 | 3+4 4 | 3+4 4 | 5 2 |
| | 5 2 | 5+5 4 | 6+6 4 | 3+4 4 | 3+4 4 | 3+4 4 | 3+4 4 | 6 2 |
| | 5 2 | 5+5 4 | 6+6 4 | 3+4 4 | 3+4 4 | 3+4 4 | 3+4 4 | 6 2 |
| | 6 2 | 6+6 4 | 5+5 4 | 3+4 4 | 3+4 4 | 3+4 4 | 3+4 4 | 5 2 |
| | 5 2 | 5+5 4 | 6+6 4 | 3+4 4 | 3+4 4 | 3+4 4 | 3+4 4 | 5 2 |
| | 5 2 | | | | | | | 5 2 |
| | 6 2 | | | | | | | 6 2 |
| | 6 2 | | | | | | | 6 2 |
| | 5 2 | | | | | | | 5 2 |
| | 5 2 | | | | | | | 5 2 |
| | 6 2 | | | | | | | 6 2 |
| | 5 2 | | | | | | | 5 2 |
| | 6 2 | | | | | | | 6 2 |
| | 6 2 | | | | | | | 5 2 |
| | 5 2 | | | | | | | 6 2 |
| | 6 2 | | | | | | | 6 2 |
| | 5 2 | | | | | | | 6 2 |
| | 5 2 | | | | | | | 5 2 |
| | 6 2 | | | | | | | 6 2 |
| | 6 2 | | | | | | | 5 2 |
| | 6 2 | | | | | | | 5 2 |
| | 5 2 | | | | | | | 6 2 |
| | 6 2 | | | | | | | 5 2 |
| | 5 2 | | | | | | | 5 2 |
| | 5 2 | | | | | | | 5 2 |
| | 6 2 | | | | | | | 6 2 |
| | 5 2 | | | | | | | 6 2 |
| | 6 2 | | | | | | | 5 2 |
| | 6 2 | | | | | | | 5 2 |
| | 5 2 | | | | | | | 6 2 |
| | 5 2 | | | | | | | 6 2 |
| | 6 2 | | | | | | | 5 2 |
| | 5 2 | | | | | | | 5 2 |
| | 5 2 | | | | | | | 5 2 |
| | 6 2 | | | | | | | 6 2 |
| | 5 2 | | | | | | | 6 2 |
| | 5 2 | | | | | | | 5 2 |
| | 5 2 | | | | | | | 6 2 |
| 55.2.21 | 412706 | 100 | Hf5In4Rh2 | | | | | |
| | -GM | -R | -S | -T | -U | -X | -Y | -Z |
| | 5 2 | 5+5 4 | 5+5 4 | 3+4 4 | 3+4 4 | 3+4 4 | 3+4 4 | 6 2 |
| | 5 2 | 6+6 4 | 6+6 4 | 3+4 4 | 3+4 4 | 3+4 4 | 3+4 4 | 5 2 |
| | 6 2 | 5+5 4 | 6+6 4 | 3+4 4 | 3+4 4 | 3+4 4 | 3+4 4 | 6 2 |
| | 6 2 | 6+6 4 | 5+5 4 | 3+4 4 | 3+4 4 | 3+4 4 | 3+4 4 | 5 2 |
| | 5 2 | 5+5 4 | 5+5 4 | 3+4 4 | 3+4 4 | 3+4 4 | 3+4 4 | 5 2 |
| | 5 2 | 6+6 4 | 6+6 4 | 3+4 4 | 3+4 4 | 3+4 4 | 3+4 4 | 6 2 |
| | 6 2 | 6+6 4 | 6+6 4 | 3+4 4 | 3+4 4 | 3+4 4 | 3+4 4 | 6 2 |
| | 6 2 | 5+5 4 | 5+5 4 | 3+4 4 | 3+4 4 | 3+4 4 | 3+4 4 | 5 2 |
| | 5 2 | 5+5 4 | 5+5 4 | 3+4 4 | 3+4 4 | 3+4 4 | 3+4 4 | 5 2 |
| | 5 2 | 5+5 4 | 6+6 4 | 3+4 4 | 3+4 4 | 3+4 4 | 3+4 4 | 5 2 |
| | 6 2 | 6+6 4 | 5+5 4 | 3+4 4 | 3+4 4 | 3+4 4 | 3+4 4 | 6 2 |
| | 6 2 | 5+5 4 | 6+6 4 | 3+4 4 | 3+4 4 | 3+4 4 | 3+4 4 | 6 2 |
| | 5 2 | 6+6 4 | 5+5 4 | 3+4 4 | 3+4 4 | 3+4 4 | 3+4 4 | 5 2 |
| | 5 2 | 6+6 4 | 6+6 4 | 3+4 4 | 3+4 4 | 3+4 4 | 3+4 4 | 5 2 |
| | 6 2 | 5+5 4 | 5+5 4 | 3+4 4 | 3+4 4 | 3+4 4 | 3+4 4 | 6 2 |
| | 5 2 | 5+5 4 | 6+6 4 | 3+4 4 | 3+4 4 | 3+4 4 | 3+4 4 | 6 2 |
| | 5 2 | 6+6 4 | 5+5 4 | 3+4 4 | 3+4 4 | 3+4 4 | 3+4 4 | 5 2 |
| | 6 2 | 5+5 4 | 6+6 4 | 3+4 4 | 3+4 4 | 3+4 4 | 3+4 4 | 6 2 |
| | 6 2 | 6+6 4 | 5+5 4 | 3+4 4 | 3+4 4 | 3+4 4 | 3+4 4 | 6 2 |
| | 6 2 | 5+5 4 | 6+6 4 | 3+4 4 | 3+4 4 | 3+4 4 | 3+4 4 | 5 2 |
| | 5 2 | 5+5 4 | 6+6 4 | 3+4 4 | 3+4 4 | 3+4 4 | 3+4 4 | 5 2 |

Continued on next column

**TABLE XXXI– continued**

| TQC | ICSD | NoE | SF | | | | | |
|---|---|---|---|---|---|---|---|---|
| | 5 2 | 6+6 4 | 5+5 4 | 3+4 4 | 3+4 4 | 3+4 4 | 3+4 4 | 6 2 |
| | 6 2 | 5+5 4 | 5+5 4 | 3+4 4 | 3+4 4 | 3+4 4 | 3+4 4 | 6 2 |
| | 5 2 | 6+6 4 | 6+6 4 | 3+4 4 | 3+4 4 | 3+4 4 | 3+4 4 | 5 2 |
| | 6 2 | 6+6 4 | 5+5 4 | 3+4 4 | 3+4 4 | 3+4 4 | 3+4 4 | 5 2 |
| | 6 2 | | | | | | | 6 2 |
| | 5 2 | | | | | | | 5 2 |
| | 6 2 | | | | | | | 5 2 |
| | 6 2 | | | | | | | 6 2 |
| | 5 2 | | | | | | | 6 2 |
| | 6 2 | | | | | | | 5 2 |
| | 5 2 | | | | | | | 6 2 |
| | 6 2 | | | | | | | 5 2 |
| | 5 2 | | | | | | | 6 2 |
| | 5 2 | | | | | | | 6 2 |
| | 6 2 | | | | | | | 5 2 |
| | 5 2 | | | | | | | 6 2 |
| | 5 2 | | | | | | | 5 2 |
| | 5 2 | | | | | | | 5 2 |
| | 5 2 | | | | | | | 5 2 |
| | 6 2 | | | | | | | 6 2 |
| | 6 2 | | | | | | | 6 2 |
| | 5 2 | | | | | | | 6 2 |
| | 5 2 | | | | | | | 5 2 |
| | 6 2 | | | | | | | 5 2 |
| | 5 2 | | | | | | | 5 2 |
| | 6 2 | | | | | | | 6 2 |
| | 5 2 | | | | | | | 6 2 |
| | 5 2 | | | | | | | 5 2 |
| | 6 2 | | | | | | | 6 2 |
| 55.2.22 | 22046 | 84 | Ge1Mn2O4 | | | | | |
| | -GM | -R | -S | -T | -U | -X | -Y | -Z |
| | 5 2 | 6+6 4 | 5+5 4 | 3+4 4 | 3+4 4 | 3+4 4 | 3+4 4 | 6 2 |
| | 5 2 | 5+5 4 | 6+6 4 | 3+4 4 | 3+4 4 | 3+4 4 | 3+4 4 | 5 2 |
| | 6 2 | 6+6 4 | 6+6 4 | 3+4 4 | 3+4 4 | 3+4 4 | 3+4 4 | 6 2 |
| | 6 2 | 5+5 4 | 5+5 4 | 3+4 4 | 3+4 4 | 3+4 4 | 3+4 4 | 6 2 |
| | 6 2 | 5+5 4 | 5+5 4 | 3+4 4 | 3+4 4 | 3+4 4 | 3+4 4 | 5 2 |
| | 6 2 | 6+6 4 | 6+6 4 | 3+4 4 | 3+4 4 | 3+4 4 | 3+4 4 | 6 2 |
| | 5 2 | 6+6 4 | 5+5 4 | 3+4 4 | 3+4 4 | 3+4 4 | 3+4 4 | 5 2 |
| | 5 2 | 5+5 4 | 6+6 4 | 3+4 4 | 3+4 4 | 3+4 4 | 3+4 4 | 5 2 |
| | 5 2 | 6+6 4 | 6+6 4 | 3+4 4 | 3+4 4 | 3+4 4 | 3+4 4 | 5 2 |
| | 5 2 | 6+6 4 | 6+6 4 | 3+4 4 | 3+4 4 | 3+4 4 | 3+4 4 | 5 2 |
| | 6 2 | 5+5 4 | 5+5 4 | 3+4 4 | 3+4 4 | 3+4 4 | 3+4 4 | 6 2 |
| | 5 2 | 6+6 4 | 5+5 4 | 3+4 4 | 3+4 4 | 3+4 4 | 3+4 4 | 6 2 |
| | 6 2 | 5+5 4 | 6+6 4 | 3+4 4 | 3+4 4 | 3+4 4 | 3+4 4 | 6 2 |
| | 6 2 | 5+5 4 | 5+5 4 | 3+4 4 | 3+4 4 | 3+4 4 | 3+4 4 | 5 2 |
| | 5 2 | 5+5 4 | 6+6 4 | 3+4 4 | 3+4 4 | 3+4 4 | 3+4 4 | 6 2 |
| | 6 2 | 6+6 4 | 5+5 4 | 3+4 4 | 3+4 4 | 3+4 4 | 3+4 4 | 5 2 |
| | 6 2 | 6+6 4 | 6+6 4 | 3+4 4 | 3+4 4 | 3+4 4 | 3+4 4 | 6 2 |
| | 5 2 | 5+5 4 | 5+5 4 | 3+4 4 | 3+4 4 | 3+4 4 | 3+4 4 | 6 2 |
| | 6 2 | 5+5 4 | 6+6 4 | 3+4 4 | 3+4 4 | 3+4 4 | 3+4 4 | 6 2 |
| | 5 2 | 6+6 4 | 5+5 4 | 3+4 4 | 3+4 4 | 3+4 4 | 3+4 4 | 5 2 |
| | 6 2 | 6+6 4 | 6+6 4 | 3+4 4 | 3+4 4 | 3+4 4 | 3+4 4 | 6 2 |
| | 5 2 | | | | | | | 5 2 |
| | 6 2 | | | | | | | 6 2 |
| | 5 2 | | | | | | | 6 2 |
| | 5 2 | | | | | | | 5 2 |
| | 6 2 | | | | | | | 6 2 |
| | 6 2 | | | | | | | 6 2 |
| | 5 2 | | | | | | | 5 2 |
| | 6 2 | | | | | | | 5 2 |
| | 5 2 | | | | | | | 5 2 |

Continued on next column

**TABLE XXXI– continued**

| TQC | ICSD | NoE | SF | | | | | |
|---|---|---|---|---|---|---|---|---|
| | 5 2 | | | | | | | 5 2 |
| | 6 2 | | | | | | | 5 2 |
| | 5 2 | | | | | | | 5 2 |
| | 6 2 | | | | | | | 5 2 |
| | 5 2 | | | | | | | 6 2 |
| | 6 2 | | | | | | | 6 2 |
| | 6 2 | | | | | | | 6 2 |
| | 6 2 | | | | | | | 6 2 |
| | 5 2 | | | | | | | 5 2 |
| | 5 2 | | | | | | | 5 2 |
| | 6 2 | | | | | | | 5 2 |
| | 5 2 | | | | | | | 6 2 |
| 55.2.23 | 23226 | 64 | Ga2Mg1 | | | | | |
| | -GM | -R | -S | -T | -U | -X | -Y | -Z |
| | 5 2 | 5+5 4 | 5+5 4 | 3+4 4 | 3+4 4 | 3+4 4 | 3+4 4 | 6 2 |
| | 5 2 | 6+6 4 | 6+6 4 | 3+4 4 | 3+4 4 | 3+4 4 | 3+4 4 | 5 2 |
| | 6 2 | 5+5 4 | 6+6 4 | 3+4 4 | 3+4 4 | 3+4 4 | 3+4 4 | 6 2 |
| | 6 2 | 5+5 4 | 5+5 4 | 3+4 4 | 3+4 4 | 3+4 4 | 3+4 4 | 5 2 |
| | 5 2 | 6+6 4 | 6+6 4 | 3+4 4 | 3+4 4 | 3+4 4 | 3+4 4 | 5 2 |
| | 5 2 | 6+6 4 | 5+5 4 | 3+4 4 | 3+4 4 | 3+4 4 | 3+4 4 | 5 2 |
| | 6 2 | 6+6 4 | 5+5 4 | 3+4 4 | 3+4 4 | 3+4 4 | 3+4 4 | 6 2 |
| | 6 2 | 5+5 4 | 6+6 4 | 3+4 4 | 3+4 4 | 3+4 4 | 3+4 4 | 6 2 |
| | 5 2 | 5+5 4 | 6+6 4 | 3+4 4 | 3+4 4 | 3+4 4 | 3+4 4 | 6 2 |
| | 6 2 | 6+6 4 | 5+5 4 | 3+4 4 | 3+4 4 | 3+4 4 | 3+4 4 | 5 2 |
| | 5 2 | 5+5 4 | 5+5 4 | 3+4 4 | 3+4 4 | 3+4 4 | 3+4 4 | 6 2 |
| | 6 2 | 6+6 4 | 6+6 4 | 3+4 4 | 3+4 4 | 3+4 4 | 3+4 4 | 6 2 |
| | 5 2 | 5+5 4 | 6+6 4 | 3+4 4 | 3+4 4 | 3+4 4 | 3+4 4 | 5 2 |
| | 5 2 | 6+6 4 | 5+5 4 | 3+4 4 | 3+4 4 | 3+4 4 | 3+4 4 | 5 2 |
| | 5 2 | 5+5 4 | 5+5 4 | 3+4 4 | 3+4 4 | 3+4 4 | 3+4 4 | 5 2 |
| | 6 2 | 5+5 4 | 5+5 4 | 3+4 4 | 3+4 4 | 3+4 4 | 3+4 4 | 6 2 |
| | 6 2 | | | | | | | 5 2 |
| | 6 2 | | | | | | | 6 2 |
| | 5 2 | | | | | | | 5 2 |
| | 6 2 | | | | | | | 6 2 |
| | 5 2 | | | | | | | 5 2 |
| | 6 2 | | | | | | | 6 2 |
| | 6 2 | | | | | | | 6 2 |
| | 5 2 | | | | | | | 5 2 |
| | 5 2 | | | | | | | 6 2 |
| | 6 2 | | | | | | | 5 2 |
| | 5 2 | | | | | | | 5 2 |
| | 6 2 | | | | | | | 6 2 |
| | 5 2 | | | | | | | 6 2 |
| | 5 2 | | | | | | | 5 2 |
| | 6 2 | | | | | | | 5 2 |
| | 5 2 | | | | | | | 5 2 |
| 55.2.24 | 426620 | 164 | F3O1Rb1V1 | | | | | |
| | -GM | -R | -S | -T | -U | -X | -Y | -Z |
| | 5 2 | 5+5 4 | 5+5 4 | 3+4 4 | 3+4 4 | 3+4 4 | 3+4 4 | 5 2 |
| | 6 2 | 6+6 4 | 6+6 4 | 3+4 4 | 3+4 4 | 3+4 4 | 3+4 4 | 6 2 |
| | 5 2 | 5+5 4 | 5+5 4 | 3+4 4 | 3+4 4 | 3+4 4 | 3+4 4 | 5 2 |
| | 6 2 | 6+6 4 | 6+6 4 | 3+4 4 | 3+4 4 | 3+4 4 | 3+4 4 | 6 2 |
| | 5 2 | 6+6 4 | 5+5 4 | 3+4 4 | 3+4 4 | 3+4 4 | 3+4 4 | 5 2 |
| | 5 2 | 5+5 4 | 6+6 4 | 3+4 4 | 3+4 4 | 3+4 4 | 3+4 4 | 5 2 |
| | 6 2 | 5+5 4 | 6+6 4 | 3+4 4 | 3+4 4 | 3+4 4 | 3+4 4 | 6 2 |
| | 6 2 | 6+6 4 | 5+5 4 | 3+4 4 | 3+4 4 | 3+4 4 | 3+4 4 | 6 2 |
| | 5 2 | 5+5 4 | 6+6 4 | 3+4 4 | 3+4 4 | 3+4 4 | 3+4 4 | 6 2 |
| | 5 2 | 6+6 4 | 5+5 4 | 3+4 4 | 3+4 4 | 3+4 4 | 3+4 4 | 6 2 |
| | 6 2 | 6+6 4 | 5+5 4 | 3+4 4 | 3+4 4 | 3+4 4 | 3+4 4 | 5 2 |
| | 6 2 | 5+5 4 | 6+6 4 | 3+4 4 | 3+4 4 | 3+4 4 | 3+4 4 | 5 2 |
| | 6 2 | 6+6 4 | 5+5 4 | 3+4 4 | 3+4 4 | 3+4 4 | 3+4 4 | 5 2 |

Continued on next column

**TABLE XXXI– continued**

| TQC | ICSD | NoE | SF | | | | | |
|---|---|---|---|---|---|---|---|---|
| | 6 2 | 5+5 4 | 6+6 4 | 3+4 4 | 3+4 4 | 3+4 4 | 3+4 4 | 5 2 |
| | 5 2 | 6+6 4 | 5+5 4 | 3+4 4 | 3+4 4 | 3+4 4 | 3+4 4 | 6 2 |
| | 5 2 | 5+5 4 | 6+6 4 | 3+4 4 | 3+4 4 | 3+4 4 | 3+4 4 | 6 2 |
| | 5 2 | 6+6 4 | 6+6 4 | 3+4 4 | 3+4 4 | 3+4 4 | 3+4 4 | 6 2 |
| | 5 2 | 5+5 4 | 5+5 4 | 3+4 4 | 3+4 4 | 3+4 4 | 3+4 4 | 6 2 |
| | 6 2 | 5+5 4 | 5+5 4 | 3+4 4 | 3+4 4 | 3+4 4 | 3+4 4 | 5 2 |
| | 6 2 | 6+6 4 | 6+6 4 | 3+4 4 | 3+4 4 | 3+4 4 | 3+4 4 | 5 2 |
| | 5 2 | 6+6 4 | 5+5 4 | 3+4 4 | 3+4 4 | 3+4 4 | 3+4 4 | 6 2 |
| | 6 2 | 5+5 4 | 5+5 4 | 3+4 4 | 3+4 4 | 3+4 4 | 3+4 4 | 5 2 |
| | 5 2 | 6+6 4 | 6+6 4 | 3+4 4 | 3+4 4 | 3+4 4 | 3+4 4 | 6 2 |
| | 6 2 | 5+5 4 | 6+6 4 | 3+4 4 | 3+4 4 | 3+4 4 | 3+4 4 | 5 2 |
| | 5 2 | 6+6 4 | 5+5 4 | 3+4 4 | 3+4 4 | 3+4 4 | 3+4 4 | 6 2 |
| | 6 2 | 6+6 4 | 6+6 4 | 3+4 4 | 3+4 4 | 3+4 4 | 3+4 4 | 5 2 |
| | 5 2 | 5+5 4 | 5+5 4 | 3+4 4 | 3+4 4 | 3+4 4 | 3+4 4 | 6 2 |
| | 6 2 | 5+5 4 | 6+6 4 | 3+4 4 | 3+4 4 | 3+4 4 | 3+4 4 | 5 2 |
| | 5 2 | 5+5 4 | 5+5 4 | 3+4 4 | 3+4 4 | 3+4 4 | 3+4 4 | 5 2 |
| | 6 2 | 6+6 4 | 6+6 4 | 3+4 4 | 3+4 4 | 3+4 4 | 3+4 4 | 6 2 |
| | 5 2 | 5+5 4 | 5+5 4 | 3+4 4 | 3+4 4 | 3+4 4 | 3+4 4 | 6 2 |
| | 6 2 | 6+6 4 | 6+6 4 | 3+4 4 | 3+4 4 | 3+4 4 | 3+4 4 | 5 2 |
| | 6 2 | 6+6 4 | 5+5 4 | 3+4 4 | 3+4 4 | 3+4 4 | 3+4 4 | 6 2 |
| | 6 2 | 5+5 4 | 6+6 4 | 3+4 4 | 3+4 4 | 3+4 4 | 3+4 4 | 6 2 |
| | 5 2 | 6+6 4 | 6+6 4 | 3+4 4 | 3+4 4 | 3+4 4 | 3+4 4 | 5 2 |
| | 5 2 | 5+5 4 | 5+5 4 | 3+4 4 | 3+4 4 | 3+4 4 | 3+4 4 | 5 2 |
| | 5 2 | 5+5 4 | 6+6 4 | 3+4 4 | 3+4 4 | 3+4 4 | 3+4 4 | 5 2 |
| | 5 2 | 6+6 4 | 5+5 4 | 3+4 4 | 3+4 4 | 3+4 4 | 3+4 4 | 5 2 |
| | 6 2 | 5+5 4 | 6+6 4 | 3+4 4 | 3+4 4 | 3+4 4 | 3+4 4 | 6 2 |
| | 5 2 | 6+6 4 | 5+5 4 | 3+4 4 | 3+4 4 | 3+4 4 | 3+4 4 | 6 2 |
| | 5 2 | 5+5 4 | 6+6 4 | 3+4 4 | 3+4 4 | 3+4 4 | 3+4 4 | 6 2 |
| | 6 2 | | | | | | | 5 2 |
| | 5 2 | | | | | | | 6 2 |
| | 5 2 | | | | | | | 6 2 |
| | 6 2 | | | | | | | 6 2 |
| | 6 2 | | | | | | | 5 2 |
| | 5 2 | | | | | | | 5 2 |
| | 6 2 | | | | | | | 5 2 |
| | 6 2 | | | | | | | 6 2 |
| | 5 2 | | | | | | | 6 2 |
| | 6 2 | | | | | | | 5 2 |
| | 5 2 | | | | | | | 6 2 |
| | 6 2 | | | | | | | 6 2 |
| | 6 2 | | | | | | | 5 2 |
| | 6 2 | | | | | | | 5 2 |
| | 5 2 | | | | | | | 6 2 |
| | 6 2 | | | | | | | 5 2 |
| | 5 2 | | | | | | | 5 2 |
| | 5 2 | | | | | | | 5 2 |
| | 6 2 | | | | | | | 5 2 |
| | 5 2 | | | | | | | 5 2 |
| | 6 2 | | | | | | | 6 2 |
| | 6 2 | | | | | | | 6 2 |
| | 5 2 | | | | | | | 6 2 |
| | 6 2 | | | | | | | 6 2 |
| | 5 2 | | | | | | | 5 2 |
| | 5 2 | | | | | | | 6 2 |
| | 6 2 | | | | | | | 6 2 |
| | 5 2 | | | | | | | 6 2 |
| | 6 2 | | | | | | | 5 2 |
| | 6 2 | | | | | | | 5 2 |
| | 5 2 | | | | | | | 5 2 |
| | 5 2 | | | | | | | 5 2 |
| | 5 2 | | | | | | | 5 2 |

Continued on next column

**TABLE XXXI– continued**

| TQC | ICSD | NoE | SF | | | | | |
|---|---|---|---|---|---|---|---|---|
| | 6 2 | | | | | | | 6 2 |
| | 6 2 | | | | | | | 5 2 |
| | 6 2 | | | | | | | 6 2 |
| | 5 2 | | | | | | | 6 2 |
| | 5 2 | | | | | | | 6 2 |
| | 6 2 | | | | | | | 5 2 |
| | 6 2 | | | | | | | 5 2 |
| | 6 2 | | | | | | | 6 2 |
| 55.2.25 | 249403 | 164 | F3O1Rb1V1 | | | | | |
| | -GM | -R | -S | -T | -U | -X | -Y | -Z |
| | 5 2 | 5+5 4 | 5+5 4 | 3+4 4 | 3+4 4 | 3+4 4 | 3+4 4 | 5 2 |
| | 6 2 | 6+6 4 | 6+6 4 | 3+4 4 | 3+4 4 | 3+4 4 | 3+4 4 | 6 2 |
| | 5 2 | 5+5 4 | 6+6 4 | 3+4 4 | 3+4 4 | 3+4 4 | 3+4 4 | 5 2 |
| | 6 2 | 6+6 4 | 5+5 4 | 3+4 4 | 3+4 4 | 3+4 4 | 3+4 4 | 6 2 |
| | 5 2 | 6+6 4 | 6+6 4 | 3+4 4 | 3+4 4 | 3+4 4 | 3+4 4 | 6 2 |
| | 5 2 | 5+5 4 | 5+5 4 | 3+4 4 | 3+4 4 | 3+4 4 | 3+4 4 | 6 2 |
| | 6 2 | 5+5 4 | 5+5 4 | 3+4 4 | 3+4 4 | 3+4 4 | 3+4 4 | 5 2 |
| | 6 2 | 6+6 4 | 6+6 4 | 3+4 4 | 3+4 4 | 3+4 4 | 3+4 4 | 5 2 |
| | 5 2 | 5+5 4 | 5+5 4 | 3+4 4 | 3+4 4 | 3+4 4 | 3+4 4 | 5 2 |
| | 5 2 | 6+6 4 | 6+6 4 | 3+4 4 | 3+4 4 | 3+4 4 | 3+4 4 | 5 2 |
| | 6 2 | 6+6 4 | 5+5 4 | 3+4 4 | 3+4 4 | 3+4 4 | 3+4 4 | 6 2 |
| | 6 2 | 5+5 4 | 6+6 4 | 3+4 4 | 3+4 4 | 3+4 4 | 3+4 4 | 6 2 |
| | 6 2 | 6+6 4 | 5+5 4 | 3+4 4 | 3+4 4 | 3+4 4 | 3+4 4 | 6 2 |
| | 5 2 | 5+5 4 | 6+6 4 | 3+4 4 | 3+4 4 | 3+4 4 | 3+4 4 | 6 2 |
| | 6 2 | 6+6 4 | 5+5 4 | 3+4 4 | 3+4 4 | 3+4 4 | 3+4 4 | 5 2 |
| | 5 2 | 5+5 4 | 6+6 4 | 3+4 4 | 3+4 4 | 3+4 4 | 3+4 4 | 5 2 |
| | 5 2 | 6+6 4 | 5+5 4 | 3+4 4 | 3+4 4 | 3+4 4 | 3+4 4 | 5 2 |
| | 5 2 | 5+5 4 | 6+6 4 | 3+4 4 | 3+4 4 | 3+4 4 | 3+4 4 | 5 2 |
| | 6 2 | 5+5 4 | 6+6 4 | 3+4 4 | 3+4 4 | 3+4 4 | 3+4 4 | 6 2 |
| | 6 2 | 6+6 4 | 5+5 4 | 3+4 4 | 3+4 4 | 3+4 4 | 3+4 4 | 6 2 |
| | 5 2 | 6+6 4 | 6+6 4 | 3+4 4 | 3+4 4 | 3+4 4 | 3+4 4 | 6 2 |
| | 6 2 | 6+6 4 | 6+6 4 | 3+4 4 | 3+4 4 | 3+4 4 | 3+4 4 | 5 2 |
| | 5 2 | 5+5 4 | 5+5 4 | 3+4 4 | 3+4 4 | 3+4 4 | 3+4 4 | 5 2 |
| | 6 2 | 5+5 4 | 5+5 4 | 3+4 4 | 3+4 4 | 3+4 4 | 3+4 4 | 6 2 |
| | 5 2 | 6+6 4 | 6+6 4 | 3+4 4 | 3+4 4 | 3+4 4 | 3+4 4 | 6 2 |
| | 6 2 | 5+5 4 | 5+5 4 | 3+4 4 | 3+4 4 | 3+4 4 | 3+4 4 | 5 2 |
| | 5 2 | 6+6 4 | 5+5 4 | 3+4 4 | 3+4 4 | 3+4 4 | 3+4 4 | 5 2 |
| | 6 2 | 5+5 4 | 6+6 4 | 3+4 4 | 3+4 4 | 3+4 4 | 3+4 4 | 6 2 |
| | 6 2 | 5+5 4 | 6+6 4 | 3+4 4 | 3+4 4 | 3+4 4 | 3+4 4 | 6 2 |
| | 5 2 | 5+5 4 | 5+5 4 | 3+4 4 | 3+4 4 | 3+4 4 | 3+4 4 | 5 2 |
| | 5 2 | 6+6 4 | 5+5 4 | 3+4 4 | 3+4 4 | 3+4 4 | 3+4 4 | 5 2 |
| | 6 2 | 6+6 4 | 6+6 4 | 3+4 4 | 3+4 4 | 3+4 4 | 3+4 4 | 6 2 |
| | 6 2 | 6+6 4 | 6+6 4 | 3+4 4 | 3+4 4 | 3+4 4 | 3+4 4 | 5 2 |
| | 6 2 | 5+5 4 | 5+5 4 | 3+4 4 | 3+4 4 | 3+4 4 | 3+4 4 | 5 2 |
| | 5 2 | 6+6 4 | 6+6 4 | 3+4 4 | 3+4 4 | 3+4 4 | 3+4 4 | 6 2 |
| | 5 2 | 5+5 4 | 5+5 4 | 3+4 4 | 3+4 4 | 3+4 4 | 3+4 4 | 6 2 |
| | 5 2 | 5+5 4 | 6+6 4 | 3+4 4 | 3+4 4 | 3+4 4 | 3+4 4 | 6 2 |
| | 5 2 | 6+6 4 | 5+5 4 | 3+4 4 | 3+4 4 | 3+4 4 | 3+4 4 | 6 2 |
| | 6 2 | 5+5 4 | 5+5 4 | 3+4 4 | 3+4 4 | 3+4 4 | 3+4 4 | 5 2 |
| | 5 2 | 6+6 4 | 6+6 4 | 3+4 4 | 3+4 4 | 3+4 4 | 3+4 4 | 5 2 |
| | 5 2 | 6+6 4 | 5+5 4 | 3+4 4 | 3+4 4 | 3+4 4 | 3+4 4 | 5 2 |
| | 6 2 | | | | | | | 5 2 |
| | 5 2 | | | | | | | 6 2 |
| | 5 2 | | | | | | | 5 2 |
| | 6 2 | | | | | | | 5 2 |
| | 6 2 | | | | | | | 6 2 |
| | 6 2 | | | | | | | 6 2 |
| | 5 2 | | | | | | | 6 2 |
| | 6 2 | | | | | | | 5 2 |
| | 5 2 | | | | | | | 5 2 |
| | 5 2 | | | | | | | 6 2 |

Continued on next column

**TABLE XXXI– continued**

| TQC | ICSD | NoE | SF | | | | | |
|---|---|---|---|---|---|---|---|---|
| | 6 2 | | | | | | | 5 2 |
| | 5 2 | | | | | | | 5 2 |
| | 6 2 | | | | | | | 6 2 |
| | 6 2 | | | | | | | 6 2 |
| | 6 2 | | | | | | | 5 2 |
| | 5 2 | | | | | | | 6 2 |
| | 6 2 | | | | | | | 6 2 |
| | 6 2 | | | | | | | 5 2 |
| | 5 2 | | | | | | | 6 2 |
| | 5 2 | | | | | | | 6 2 |
| | 5 2 | | | | | | | 6 2 |
| | 6 2 | | | | | | | 5 2 |
| | 6 2 | | | | | | | 5 2 |
| | 5 2 | | | | | | | 5 2 |
| | 6 2 | | | | | | | 5 2 |
| | 6 2 | | | | | | | 6 2 |
| | 5 2 | | | | | | | 5 2 |
| | 5 2 | | | | | | | 6 2 |
| | 6 2 | | | | | | | 6 2 |
| | 5 2 | | | | | | | 5 2 |
| | 6 2 | | | | | | | 6 2 |
| | 5 2 | | | | | | | 6 2 |
| | 5 2 | | | | | | | 6 2 |
| | 6 2 | | | | | | | 5 2 |
| | 6 2 | | | | | | | 5 2 |
| | 5 2 | | | | | | | 5 2 |
| | 5 2 | | | | | | | 5 2 |
| | 6 2 | | | | | | | 6 2 |
| | 6 2 | | | | | | | 6 2 |
| | 6 2 | | | | | | | 6 2 |
| | 6 2 | | | | | | | 5 2 |
| 55.2.26 | 82389 | 148 | Cu2La2O5 | | | | | |
| | -GM | -R | -S | -T | -U | -X | -Y | -Z |
| | 5 2 | 6+6 4 | 5+5 4 | 3+4 4 | 3+4 4 | 3+4 4 | 3+4 4 | 6 2 |
| | 6 2 | 5+5 4 | 6+6 4 | 3+4 4 | 3+4 4 | 3+4 4 | 3+4 4 | 5 2 |
| | 5 2 | 5+5 4 | 5+5 4 | 3+4 4 | 3+4 4 | 3+4 4 | 3+4 4 | 6 2 |
| | 6 2 | 6+6 4 | 6+6 4 | 3+4 4 | 3+4 4 | 3+4 4 | 3+4 4 | 5 2 |
| | 5 2 | 5+5 4 | 5+5 4 | 3+4 4 | 3+4 4 | 3+4 4 | 3+4 4 | 5 2 |
| | 6 2 | 6+6 4 | 6+6 4 | 3+4 4 | 3+4 4 | 3+4 4 | 3+4 4 | 6 2 |
| | 5 2 | 5+5 4 | 5+5 4 | 3+4 4 | 3+4 4 | 3+4 4 | 3+4 4 | 5 2 |
| | 5 2 | 6+6 4 | 5+5 4 | 3+4 4 | 3+4 4 | 3+4 4 | 3+4 4 | 6 2 |
| | 6 2 | 5+5 4 | 6+6 4 | 3+4 4 | 3+4 4 | 3+4 4 | 3+4 4 | 5 2 |
| | 5 2 | 6+6 4 | 5+5 4 | 3+4 4 | 3+4 4 | 3+4 4 | 3+4 4 | 6 2 |
| | 6 2 | 5+5 4 | 6+6 4 | 3+4 4 | 3+4 4 | 3+4 4 | 3+4 4 | 5 2 |
| | 6 2 | 6+6 4 | 6+6 4 | 3+4 4 | 3+4 4 | 3+4 4 | 3+4 4 | 5 2 |
| | 5 2 | 5+5 4 | 5+5 4 | 3+4 4 | 3+4 4 | 3+4 4 | 3+4 4 | 6 2 |
| | 5 2 | 6+6 4 | 6+6 4 | 3+4 4 | 3+4 4 | 3+4 4 | 3+4 4 | 5 2 |
| | 6 2 | 5+5 4 | 6+6 4 | 3+4 4 | 3+4 4 | 3+4 4 | 3+4 4 | 6 2 |
| | 5 2 | 6+6 4 | 6+6 4 | 3+4 4 | 3+4 4 | 3+4 4 | 3+4 4 | 5 2 |
| | 6 2 | 6+6 4 | 5+5 4 | 3+4 4 | 3+4 4 | 3+4 4 | 3+4 4 | 5 2 |
| | 5 2 | 5+5 4 | 5+5 4 | 3+4 4 | 3+4 4 | 3+4 4 | 3+4 4 | 6 2 |
| | 6 2 | 5+5 4 | 6+6 4 | 3+4 4 | 3+4 4 | 3+4 4 | 3+4 4 | 6 2 |
| | 5 2 | 5+5 4 | 5+5 4 | 3+4 4 | 3+4 4 | 3+4 4 | 3+4 4 | 6 2 |
| | 6 2 | 6+6 4 | 6+6 4 | 3+4 4 | 3+4 4 | 3+4 4 | 3+4 4 | 5 2 |
| | 5 2 | 6+6 4 | 6+6 4 | 3+4 4 | 3+4 4 | 3+4 4 | 3+4 4 | 5 2 |
| | 6 2 | 6+6 4 | 5+5 4 | 3+4 4 | 3+4 4 | 3+4 4 | 3+4 4 | 6 2 |
| | 5 2 | 5+5 4 | 6+6 4 | 3+4 4 | 3+4 4 | 3+4 4 | 3+4 4 | 5 2 |
| | 6 2 | 6+6 4 | 5+5 4 | 3+4 4 | 3+4 4 | 3+4 4 | 3+4 4 | 6 2 |
| | 5 2 | 6+6 4 | 5+5 4 | 3+4 4 | 3+4 4 | 3+4 4 | 3+4 4 | 5 2 |
| | 6 2 | 5+5 4 | 6+6 4 | 3+4 4 | 3+4 4 | 3+4 4 | 3+4 4 | 6 2 |
| | 6 2 | 5+5 4 | 5+5 4 | 3+4 4 | 3+4 4 | 3+4 4 | 3+4 4 | 5 2 |
| | 5 2 | 6+6 4 | 6+6 4 | 3+4 4 | 3+4 4 | 3+4 4 | 3+4 4 | 6 2 |
| | 6 2 | 6+6 4 | 6+6 4 | 3+4 4 | 3+4 4 | 3+4 4 | 3+4 4 | 5 2 |
| | 6 2 | 5+5 4 | 5+5 4 | 3+4 4 | 3+4 4 | 3+4 4 | 3+4 4 | 6 2 |
| | 6 2 | 6+6 4 | 5+5 4 | 3+4 4 | 3+4 4 | 3+4 4 | 3+4 4 | 6 2 |
| | 5 2 | 5+5 4 | 6+6 4 | 3+4 4 | 3+4 4 | 3+4 4 | 3+4 4 | 6 2 |
| | 6 2 | 6+6 4 | 5+5 4 | 3+4 4 | 3+4 4 | 3+4 4 | 3+4 4 | 5 2 |
| | 5 2 | 5+5 4 | 6+6 4 | 3+4 4 | 3+4 4 | 3+4 4 | 3+4 4 | 5 2 |
| | 5 2 | 6+6 4 | 6+6 4 | 3+4 4 | 3+4 4 | 3+4 4 | 3+4 4 | 5 2 |
| | 6 2 | 5+5 4 | 5+5 4 | 3+4 4 | 3+4 4 | 3+4 4 | 3+4 4 | 6 2 |
| | 6 2 | | | | | | | 6 2 |
| | 6 2 | | | | | | | 6 2 |
| | 6 2 | | | | | | | 6 2 |
| | 5 2 | | | | | | | 5 2 |
| | 5 2 | | | | | | | 6 2 |
| | 5 2 | | | | | | | 5 2 |
| | 6 2 | | | | | | | 6 2 |
| | 5 2 | | | | | | | 6 2 |
| | 5 2 | | | | | | | 5 2 |
| | 5 2 | | | | | | | 5 2 |
| | 5 2 | | | | | | | 6 2 |
| | 6 2 | | | | | | | 6 2 |
| | 6 2 | | | | | | | 5 2 |
| | 6 2 | | | | | | | 6 2 |
| | 5 2 | | | | | | | 6 2 |
| | 5 2 | | | | | | | 5 2 |
| | 5 2 | | | | | | | 5 2 |
| | 6 2 | | | | | | | 5 2 |
| | 6 2 | | | | | | | 5 2 |
| | 5 2 | | | | | | | 5 2 |
| | 6 2 | | | | | | | 6 2 |
| | 5 2 | | | | | | | 6 2 |
| | 6 2 | | | | | | | 6 2 |
| | 5 2 | | | | | | | 6 2 |
| | 5 2 | | | | | | | 5 2 |
| | 6 2 | | | | | | | 6 2 |
| | 6 2 | | | | | | | 5 2 |
| | 6 2 | | | | | | | 6 2 |
| | 5 2 | | | | | | | 5 2 |
| | 6 2 | | | | | | | 6 2 |
| | 6 2 | | | | | | | 5 2 |
| | 5 2 | | | | | | | 6 2 |
| | 6 2 | | | | | | | 5 2 |
| | 6 2 | | | | | | | 6 2 |
| | 6 2 | | | | | | | 6 2 |
| | 5 2 | | | | | | | 5 2 |
| | 6 2 | | | | | | | 5 2 |
| 55.2.27 | 82388 | 148 | Cu2La2O5 | | | | | |
| | -GM | -R | -S | -T | -U | -X | -Y | -Z |
| | 5 2 | 6+6 4 | 5+5 4 | 3+4 4 | 3+4 4 | 3+4 4 | 3+4 4 | 6 2 |
| | 6 2 | 5+5 4 | 6+6 4 | 3+4 4 | 3+4 4 | 3+4 4 | 3+4 4 | 5 2 |
| | 5 2 | 5+5 4 | 5+5 4 | 3+4 4 | 3+4 4 | 3+4 4 | 3+4 4 | 6 2 |
| | 6 2 | 6+6 4 | 6+6 4 | 3+4 4 | 3+4 4 | 3+4 4 | 3+4 4 | 5 2 |
| | 5 2 | 5+5 4 | 5+5 4 | 3+4 4 | 3+4 4 | 3+4 4 | 3+4 4 | 5 2 |
| | 6 2 | 6+6 4 | 6+6 4 | 3+4 4 | 3+4 4 | 3+4 4 | 3+4 4 | 6 2 |
| | 5 2 | 5+5 4 | 5+5 4 | 3+4 4 | 3+4 4 | 3+4 4 | 3+4 4 | 5 2 |
| | 5 2 | 6+6 4 | 5+5 4 | 3+4 4 | 3+4 4 | 3+4 4 | 3+4 4 | 6 2 |
| | 6 2 | 5+5 4 | 6+6 4 | 3+4 4 | 3+4 4 | 3+4 4 | 3+4 4 | 5 2 |
| | 5 2 | 6+6 4 | 5+5 4 | 3+4 4 | 3+4 4 | 3+4 4 | 3+4 4 | 6 2 |
| | 6 2 | 6+6 4 | 6+6 4 | 3+4 4 | 3+4 4 | 3+4 4 | 3+4 4 | 5 2 |
| | 6 2 | 5+5 4 | 6+6 4 | 3+4 4 | 3+4 4 | 3+4 4 | 3+4 4 | 5 2 |
| | 5 2 | 5+5 4 | 5+5 4 | 3+4 4 | 3+4 4 | 3+4 4 | 3+4 4 | 6 2 |

Continued on next column

**TABLE XXXI– continued**

| TQC | ICSD | NoE | SF | | | | | |
|---|---|---|---|---|---|---|---|---|
| | 5 2 | 6+6 4 | 6+6 4 | 3+4 4 | 3+4 4 | 3+4 4 | 3+4 4 | 5 2 |
| | 6 2 | 5+5 4 | 6+6 4 | 3+4 4 | 3+4 4 | 3+4 4 | 3+4 4 | 6 2 |
| | 5 2 | 6+6 4 | 6+6 4 | 3+4 4 | 3+4 4 | 3+4 4 | 3+4 4 | 5 2 |
| | 5 2 | 6+6 4 | 5+5 4 | 3+4 4 | 3+4 4 | 3+4 4 | 3+4 4 | 5 2 |
| | 6 2 | 5+5 4 | 5+5 4 | 3+4 4 | 3+4 4 | 3+4 4 | 3+4 4 | 6 2 |
| | 5 2 | 5+5 4 | 6+6 4 | 3+4 4 | 3+4 4 | 3+4 4 | 3+4 4 | 6 2 |
| | 6 2 | 5+5 4 | 5+5 4 | 3+4 4 | 3+4 4 | 3+4 4 | 3+4 4 | 6 2 |
| | 6 2 | 6+6 4 | 6+6 4 | 3+4 4 | 3+4 4 | 3+4 4 | 3+4 4 | 5 2 |
| | 5 2 | 6+6 4 | 6+6 4 | 3+4 4 | 3+4 4 | 3+4 4 | 3+4 4 | 5 2 |
| | 6 2 | 6+6 4 | 5+5 4 | 3+4 4 | 3+4 4 | 3+4 4 | 3+4 4 | 6 2 |
| | 5 2 | 5+5 4 | 6+6 4 | 3+4 4 | 3+4 4 | 3+4 4 | 3+4 4 | 5 2 |
| | 6 2 | 6+6 4 | 5+5 4 | 3+4 4 | 3+4 4 | 3+4 4 | 3+4 4 | 6 2 |
| | 5 2 | 6+6 4 | 5+5 4 | 3+4 4 | 3+4 4 | 3+4 4 | 3+4 4 | 5 2 |
| | 6 2 | 5+5 4 | 6+6 4 | 3+4 4 | 3+4 4 | 3+4 4 | 3+4 4 | 6 2 |
| | 6 2 | 5+5 4 | 5+5 4 | 3+4 4 | 3+4 4 | 3+4 4 | 3+4 4 | 5 2 |
| | 5 2 | 6+6 4 | 6+6 4 | 3+4 4 | 3+4 4 | 3+4 4 | 3+4 4 | 6 2 |
| | 6 2 | 6+6 4 | 6+6 4 | 3+4 4 | 3+4 4 | 3+4 4 | 3+4 4 | 5 2 |
| | 6 2 | 5+5 4 | 5+5 4 | 3+4 4 | 3+4 4 | 3+4 4 | 3+4 4 | 6 2 |
| | 6 2 | 6+6 4 | 5+5 4 | 3+4 4 | 3+4 4 | 3+4 4 | 3+4 4 | 6 2 |
| | 5 2 | 5+5 4 | 6+6 4 | 3+4 4 | 3+4 4 | 3+4 4 | 3+4 4 | 6 2 |
| | 6 2 | 6+6 4 | 5+5 4 | 3+4 4 | 3+4 4 | 3+4 4 | 3+4 4 | 5 2 |
| | 5 2 | 5+5 4 | 6+6 4 | 3+4 4 | 3+4 4 | 3+4 4 | 3+4 4 | 5 2 |
| | 5 2 | 6+6 4 | 6+6 4 | 3+4 4 | 3+4 4 | 3+4 4 | 3+4 4 | 5 2 |
| | 6 2 | 5+5 4 | 5+5 4 | 3+4 4 | 3+4 4 | 3+4 4 | 3+4 4 | 6 2 |
| | 6 2 | | | | | | | 6 2 |
| | 6 2 | | | | | | | 6 2 |
| | 6 2 | | | | | | | 6 2 |
| | 5 2 | | | | | | | 5 2 |
| | 5 2 | | | | | | | 6 2 |
| | 5 2 | | | | | | | 5 2 |
| | 6 2 | | | | | | | 6 2 |
| | 5 2 | | | | | | | 6 2 |
| | 5 2 | | | | | | | 5 2 |
| | 5 2 | | | | | | | 5 2 |
| | 5 2 | | | | | | | 6 2 |
| | 6 2 | | | | | | | 6 2 |
| | 6 2 | | | | | | | 5 2 |
| | 6 2 | | | | | | | 6 2 |
| | 5 2 | | | | | | | 6 2 |
| | 5 2 | | | | | | | 5 2 |
| | 5 2 | | | | | | | 5 2 |
| | 6 2 | | | | | | | 5 2 |
| | 6 2 | | | | | | | 5 2 |
| | 5 2 | | | | | | | 5 2 |
| | 6 2 | | | | | | | 6 2 |
| | 5 2 | | | | | | | 6 2 |
| | 6 2 | | | | | | | 6 2 |
| | 5 2 | | | | | | | 6 2 |
| | 5 2 | | | | | | | 5 2 |
| | 6 2 | | | | | | | 6 2 |
| | 6 2 | | | | | | | 5 2 |
| | 6 2 | | | | | | | 6 2 |
| | 5 2 | | | | | | | 5 2 |
| | 6 2 | | | | | | | 6 2 |
| | 6 2 | | | | | | | 5 2 |
| | 5 2 | | | | | | | 6 2 |
| | 6 2 | | | | | | | 5 2 |
| | 6 2 | | | | | | | 6 2 |
| | 6 2 | | | | | | | 6 2 |
| | 5 2 | | | | | | | 5 2 |
| | 6 2 | | | | | | | 5 2 |

Continued on next column

**TABLE XXXI– continued**

| TQC | ICSD | NoE | SF | | | | | |
|---|---|---|---|---|---|---|---|---|
| 55.2.28 | 73399 | 148 | Cu2La2O5 | | | | | |
| | -GM | -R | -S | -T | -U | -X | -Y | -Z |
| | 5 2 | 6+6 4 | 5+5 4 | 3+4 4 | 3+4 4 | 3+4 4 | 3+4 4 | 6 2 |
| | 6 2 | 5+5 4 | 6+6 4 | 3+4 4 | 3+4 4 | 3+4 4 | 3+4 4 | 5 2 |
| | 5 2 | 5+5 4 | 5+5 4 | 3+4 4 | 3+4 4 | 3+4 4 | 3+4 4 | 6 2 |
| | 6 2 | 6+6 4 | 6+6 4 | 3+4 4 | 3+4 4 | 3+4 4 | 3+4 4 | 5 2 |
| | 5 2 | 5+5 4 | 5+5 4 | 3+4 4 | 3+4 4 | 3+4 4 | 3+4 4 | 5 2 |
| | 6 2 | 6+6 4 | 6+6 4 | 3+4 4 | 3+4 4 | 3+4 4 | 3+4 4 | 6 2 |
| | 5 2 | 5+5 4 | 5+5 4 | 3+4 4 | 3+4 4 | 3+4 4 | 3+4 4 | 5 2 |
| | 5 2 | 6+6 4 | 5+5 4 | 3+4 4 | 3+4 4 | 3+4 4 | 3+4 4 | 6 2 |
| | 5 2 | 5+5 4 | 6+6 4 | 3+4 4 | 3+4 4 | 3+4 4 | 3+4 4 | 5 2 |
| | 6 2 | 6+6 4 | 5+5 4 | 3+4 4 | 3+4 4 | 3+4 4 | 3+4 4 | 6 2 |
| | 6 2 | 5+5 4 | 6+6 4 | 3+4 4 | 3+4 4 | 3+4 4 | 3+4 4 | 5 2 |
| | 6 2 | 6+6 4 | 6+6 4 | 3+4 4 | 3+4 4 | 3+4 4 | 3+4 4 | 5 2 |
| | 5 2 | 5+5 4 | 5+5 4 | 3+4 4 | 3+4 4 | 3+4 4 | 3+4 4 | 6 2 |
| | 5 2 | 6+6 4 | 6+6 4 | 3+4 4 | 3+4 4 | 3+4 4 | 3+4 4 | 5 2 |
| | 6 2 | 5+5 4 | 6+6 4 | 3+4 4 | 3+4 4 | 3+4 4 | 3+4 4 | 6 2 |
| | 5 2 | 6+6 4 | 6+6 4 | 3+4 4 | 3+4 4 | 3+4 4 | 3+4 4 | 5 2 |
| | 6 2 | 6+6 4 | 5+5 4 | 3+4 4 | 3+4 4 | 3+4 4 | 3+4 4 | 5 2 |
| | 5 2 | 5+5 4 | 5+5 4 | 3+4 4 | 3+4 4 | 3+4 4 | 3+4 4 | 6 2 |
| | 6 2 | 5+5 4 | 6+6 4 | 3+4 4 | 3+4 4 | 3+4 4 | 3+4 4 | 6 2 |
| | 5 2 | 5+5 4 | 5+5 4 | 3+4 4 | 3+4 4 | 3+4 4 | 3+4 4 | 6 2 |
| | 6 2 | 6+6 4 | 6+6 4 | 3+4 4 | 3+4 4 | 3+4 4 | 3+4 4 | 5 2 |
| | 5 2 | 6+6 4 | 6+6 4 | 3+4 4 | 3+4 4 | 3+4 4 | 3+4 4 | 5 2 |
| | 6 2 | 6+6 4 | 5+5 4 | 3+4 4 | 3+4 4 | 3+4 4 | 3+4 4 | 6 2 |
| | 5 2 | 5+5 4 | 6+6 4 | 3+4 4 | 3+4 4 | 3+4 4 | 3+4 4 | 5 2 |
| | 6 2 | 6+6 4 | 5+5 4 | 3+4 4 | 3+4 4 | 3+4 4 | 3+4 4 | 6 2 |
| | 5 2 | 6+6 4 | 5+5 4 | 3+4 4 | 3+4 4 | 3+4 4 | 3+4 4 | 5 2 |
| | 6 2 | 5+5 4 | 6+6 4 | 3+4 4 | 3+4 4 | 3+4 4 | 3+4 4 | 6 2 |
| | 6 2 | 5+5 4 | 5+5 4 | 3+4 4 | 3+4 4 | 3+4 4 | 3+4 4 | 5 2 |
| | 5 2 | 6+6 4 | 6+6 4 | 3+4 4 | 3+4 4 | 3+4 4 | 3+4 4 | 6 2 |
| | 6 2 | 6+6 4 | 6+6 4 | 3+4 4 | 3+4 4 | 3+4 4 | 3+4 4 | 5 2 |
| | 6 2 | 5+5 4 | 5+5 4 | 3+4 4 | 3+4 4 | 3+4 4 | 3+4 4 | 6 2 |
| | 6 2 | 6+6 4 | 5+5 4 | 3+4 4 | 3+4 4 | 3+4 4 | 3+4 4 | 6 2 |
| | 5 2 | 5+5 4 | 6+6 4 | 3+4 4 | 3+4 4 | 3+4 4 | 3+4 4 | 6 2 |
| | 6 2 | 6+6 4 | 5+5 4 | 3+4 4 | 3+4 4 | 3+4 4 | 3+4 4 | 5 2 |
| | 5 2 | 5+5 4 | 6+6 4 | 3+4 4 | 3+4 4 | 3+4 4 | 3+4 4 | 5 2 |
| | 5 2 | 6+6 4 | 6+6 4 | 3+4 4 | 3+4 4 | 3+4 4 | 3+4 4 | 5 2 |
| | 6 2 | 5+5 4 | 5+5 4 | 3+4 4 | 3+4 4 | 3+4 4 | 3+4 4 | 6 2 |
| | 6 2 | | | | | | | 6 2 |
| | 6 2 | | | | | | | 6 2 |
| | 5 2 | | | | | | | 6 2 |
| | 6 2 | | | | | | | 6 2 |
| | 5 2 | | | | | | | 5 2 |
| | 5 2 | | | | | | | 6 2 |
| | 6 2 | | | | | | | 6 2 |
| | 5 2 | | | | | | | 5 2 |
| | 5 2 | | | | | | | 5 2 |
| | 5 2 | | | | | | | 5 2 |
| | 6 2 | | | | | | | 6 2 |
| | 6 2 | | | | | | | 6 2 |
| | 5 2 | | | | | | | 5 2 |
| | 6 2 | | | | | | | 6 2 |
| | 5 2 | | | | | | | 6 2 |
| | 5 2 | | | | | | | 5 2 |
| | 5 2 | | | | | | | 5 2 |
| | 6 2 | | | | | | | 5 2 |
| | 6 2 | | | | | | | 5 2 |
| | 6 2 | | | | | | | 5 2 |
| | 5 2 | | | | | | | 6 2 |
| | 6 2 | | | | | | | 6 2 |

Continued on next column

**TABLE XXXI– continued**

| TQC | ICSD | NoE | SF | | | | | |
|---|---|---|---|---|---|---|---|---|
| | 5 2 | | | | | | | 6 2 |
| | 5 2 | | | | | | | 6 2 |
| | 5 2 | | | | | | | 5 2 |
| | 6 2 | | | | | | | 6 2 |
| | 6 2 | | | | | | | 5 2 |
| | 5 2 | | | | | | | 6 2 |
| | 6 2 | | | | | | | 5 2 |
| | 6 2 | | | | | | | 6 2 |
| | 5 2 | | | | | | | 6 2 |
| | 6 2 | | | | | | | 5 2 |
| | 6 2 | | | | | | | 6 2 |
| | 6 2 | | | | | | | 5 2 |
| | 6 2 | | | | | | | 6 2 |
| | 5 2 | | | | | | | 5 2 |
| | 6 2 | | | | | | | 5 2 |
| 55.2.29 | 82019 | 144 | Cu1O2.5Sr1 | | | | | |
| | -GM | -R | -S | -T | -U | -X | -Y | -Z |
| | 5 2 | 5+5 4 | 5+5 4 | 3+4 4 | 3+4 4 | 3+4 4 | 3+4 4 | 5 2 |
| | 5 2 | 6+6 4 | 6+6 4 | 3+4 4 | 3+4 4 | 3+4 4 | 3+4 4 | 5 2 |
| | 6 2 | 6+6 4 | 5+5 4 | 3+4 4 | 3+4 4 | 3+4 4 | 3+4 4 | 6 2 |
| | 6 2 | 5+5 4 | 6+6 4 | 3+4 4 | 3+4 4 | 3+4 4 | 3+4 4 | 6 2 |
| | 6 2 | 6+6 4 | 5+5 4 | 3+4 4 | 3+4 4 | 3+4 4 | 3+4 4 | 6 2 |
| | 5 2 | 5+5 4 | 6+6 4 | 3+4 4 | 3+4 4 | 3+4 4 | 3+4 4 | 5 2 |
| | 5 2 | 6+6 4 | 5+5 4 | 3+4 4 | 3+4 4 | 3+4 4 | 3+4 4 | 6 2 |
| | 6 2 | 5+5 4 | 6+6 4 | 3+4 4 | 3+4 4 | 3+4 4 | 3+4 4 | 6 2 |
| | 5 2 | 6+6 4 | 5+5 4 | 3+4 4 | 3+4 4 | 3+4 4 | 3+4 4 | 5 2 |
| | 5 2 | 5+5 4 | 6+6 4 | 3+4 4 | 3+4 4 | 3+4 4 | 3+4 4 | 5 2 |
| | 6 2 | 5+5 4 | 5+5 4 | 3+4 4 | 3+4 4 | 3+4 4 | 3+4 4 | 6 2 |
| | 6 2 | 6+6 4 | 6+6 4 | 3+4 4 | 3+4 4 | 3+4 4 | 3+4 4 | 5 2 |
| | 5 2 | 6+6 4 | 5+5 4 | 3+4 4 | 3+4 4 | 3+4 4 | 3+4 4 | 6 2 |
| | 5 2 | 5+5 4 | 6+6 4 | 3+4 4 | 3+4 4 | 3+4 4 | 3+4 4 | 6 2 |
| | 6 2 | 6+6 4 | 5+5 4 | 3+4 4 | 3+4 4 | 3+4 4 | 3+4 4 | 5 2 |
| | 5 2 | 5+5 4 | 6+6 4 | 3+4 4 | 3+4 4 | 3+4 4 | 3+4 4 | 5 2 |
| | 6 2 | 6+6 4 | 6+6 4 | 3+4 4 | 3+4 4 | 3+4 4 | 3+4 4 | 6 2 |
| | 5 2 | 5+5 4 | 5+5 4 | 3+4 4 | 3+4 4 | 3+4 4 | 3+4 4 | 6 2 |
| | 6 2 | 6+6 4 | 6+6 4 | 3+4 4 | 3+4 4 | 3+4 4 | 3+4 4 | 5 2 |
| | 5 2 | 6+6 4 | 5+5 4 | 3+4 4 | 3+4 4 | 3+4 4 | 3+4 4 | 5 2 |
| | 6 2 | 5+5 4 | 6+6 4 | 3+4 4 | 3+4 4 | 3+4 4 | 3+4 4 | 5 2 |
| | 5 2 | 5+5 4 | 5+5 4 | 3+4 4 | 3+4 4 | 3+4 4 | 3+4 4 | 6 2 |
| | 6 2 | 5+5 4 | 6+6 4 | 3+4 4 | 3+4 4 | 3+4 4 | 3+4 4 | 5 2 |
| | 6 2 | 6+6 4 | 6+6 4 | 3+4 4 | 3+4 4 | 3+4 4 | 3+4 4 | 6 2 |
| | 5 2 | 5+5 4 | 5+5 4 | 3+4 4 | 3+4 4 | 3+4 4 | 3+4 4 | 6 2 |
| | 5 2 | 6+6 4 | 5+5 4 | 3+4 4 | 3+4 4 | 3+4 4 | 3+4 4 | 6 2 |
| | 6 2 | 5+5 4 | 6+6 4 | 3+4 4 | 3+4 4 | 3+4 4 | 3+4 4 | 5 2 |
| | 5 2 | 6+6 4 | 5+5 4 | 3+4 4 | 3+4 4 | 3+4 4 | 3+4 4 | 6 2 |
| | 6 2 | 5+5 4 | 5+5 4 | 3+4 4 | 3+4 4 | 3+4 4 | 3+4 4 | 5 2 |
| | 5 2 | 5+5 4 | 6+6 4 | 3+4 4 | 3+4 4 | 3+4 4 | 3+4 4 | 6 2 |
| | 6 2 | 6+6 4 | 6+6 4 | 3+4 4 | 3+4 4 | 3+4 4 | 3+4 4 | 5 2 |
| | 5 2 | 6+6 4 | 5+5 4 | 3+4 4 | 3+4 4 | 3+4 4 | 3+4 4 | 6 2 |
| | 5 2 | 5+5 4 | 6+6 4 | 3+4 4 | 3+4 4 | 3+4 4 | 3+4 4 | 5 2 |
| | 6 2 | 5+5 4 | 5+5 4 | 3+4 4 | 3+4 4 | 3+4 4 | 3+4 4 | 6 2 |
| | 6 2 | 6+6 4 | 6+6 4 | 3+4 4 | 3+4 4 | 3+4 4 | 3+4 4 | 6 2 |
| | 6 2 | 5+5 4 | 6+6 4 | 3+4 4 | 3+4 4 | 3+4 4 | 3+4 4 | 5 2 |
| | 6 2 | | | | | | | 6 2 |
| | 5 2 | | | | | | | 5 2 |
| | 6 2 | | | | | | | 5 2 |
| | 6 2 | | | | | | | 5 2 |
| | 5 2 | | | | | | | 5 2 |
| | 6 2 | | | | | | | 5 2 |
| | 5 2 | | | | | | | 6 2 |
| | 6 2 | | | | | | | 5 2 |

Continued on next column

**TABLE XXXI– continued**

| TQC | ICSD | NoE | SF | | | | | |
|---|---|---|---|---|---|---|---|---|
| | 5 2 | | | | | | | 6 2 |
| | 5 2 | | | | | | | 6 2 |
| | 6 2 | | | | | | | 5 2 |
| | 5 2 | | | | | | | 6 2 |
| | 5 2 | | | | | | | 6 2 |
| | 6 2 | | | | | | | 5 2 |
| | 5 2 | | | | | | | 5 2 |
| | 6 2 | | | | | | | 6 2 |
| | 5 2 | | | | | | | 5 2 |
| | 5 2 | | | | | | | 5 2 |
| | 6 2 | | | | | | | 6 2 |
| | 5 2 | | | | | | | 6 2 |
| | 6 2 | | | | | | | 5 2 |
| | 5 2 | | | | | | | 6 2 |
| | 6 2 | | | | | | | 5 2 |
| | 6 2 | | | | | | | 6 2 |
| | 6 2 | | | | | | | 5 2 |
| | 5 2 | | | | | | | 5 2 |
| | 6 2 | | | | | | | 6 2 |
| | 6 2 | | | | | | | 5 2 |
| | 5 2 | | | | | | | 5 2 |
| | 6 2 | | | | | | | 6 2 |
| | 6 2 | | | | | | | 6 2 |
| | 5 2 | | | | | | | 5 2 |
| | 5 2 | | | | | | | 6 2 |
| | 6 2 | | | | | | | 5 2 |
| | 6 2 | | | | | | | 5 2 |
| | 5 2 | | | | | | | 5 2 |
| 55.2.30 | 171414 | 184 | Ce5In4Pt2 | | | | | |
| | -GM | -R | -S | -T | -U | -X | -Y | -Z |
| | 5 2 | 5+5 4 | 5+5 4 | 3+4 4 | 3+4 4 | 3+4 4 | 3+4 4 | 5 2 |
| | 5 2 | 6+6 4 | 6+6 4 | 3+4 4 | 3+4 4 | 3+4 4 | 3+4 4 | 5 2 |
| | 5 2 | 5+5 4 | 5+5 4 | 3+4 4 | 3+4 4 | 3+4 4 | 3+4 4 | 5 2 |
| | 6 2 | 6+6 4 | 6+6 4 | 3+4 4 | 3+4 4 | 3+4 4 | 3+4 4 | 6 2 |
| | 5 2 | 5+5 4 | 5+5 4 | 3+4 4 | 3+4 4 | 3+4 4 | 3+4 4 | 5 2 |
| | 6 2 | 6+6 4 | 6+6 4 | 3+4 4 | 3+4 4 | 3+4 4 | 3+4 4 | 6 2 |
| | 5 2 | 6+6 4 | 5+5 4 | 3+4 4 | 3+4 4 | 3+4 4 | 3+4 4 | 5 2 |
| | 5 2 | 5+5 4 | 6+6 4 | 3+4 4 | 3+4 4 | 3+4 4 | 3+4 4 | 5 2 |
| | 6 2 | 6+6 4 | 6+6 4 | 3+4 4 | 3+4 4 | 3+4 4 | 3+4 4 | 6 2 |
| | 6 2 | 5+5 4 | 5+5 4 | 3+4 4 | 3+4 4 | 3+4 4 | 3+4 4 | 6 2 |
| | 6 2 | 6+6 4 | 6+6 4 | 3+4 4 | 3+4 4 | 3+4 4 | 3+4 4 | 6 2 |
| | 6 2 | 6+6 4 | 5+5 4 | 3+4 4 | 3+4 4 | 3+4 4 | 3+4 4 | 6 2 |
| | 5 2 | 5+5 4 | 6+6 4 | 3+4 4 | 3+4 4 | 3+4 4 | 3+4 4 | 5 2 |
| | 5 2 | 5+5 4 | 6+6 4 | 3+4 4 | 3+4 4 | 3+4 4 | 3+4 4 | 5 2 |
| | 6 2 | 6+6 4 | 5+5 4 | 3+4 4 | 3+4 4 | 3+4 4 | 3+4 4 | 6 2 |
| | 5 2 | 6+6 4 | 6+6 4 | 3+4 4 | 3+4 4 | 3+4 4 | 3+4 4 | 5 2 |
| | 6 2 | 5+5 4 | 5+5 4 | 3+4 4 | 3+4 4 | 3+4 4 | 3+4 4 | 6 2 |
| | 5 2 | 6+6 4 | 6+6 4 | 3+4 4 | 3+4 4 | 3+4 4 | 3+4 4 | 5 2 |
| | 6 2 | 5+5 4 | 5+5 4 | 3+4 4 | 3+4 4 | 3+4 4 | 3+4 4 | 6 2 |
| | 6 2 | 6+6 4 | 6+6 4 | 3+4 4 | 3+4 4 | 3+4 4 | 3+4 4 | 6 2 |
| | 6 2 | 5+5 4 | 5+5 4 | 3+4 4 | 3+4 4 | 3+4 4 | 3+4 4 | 6 2 |
| | 6 2 | 6+6 4 | 6+6 4 | 3+4 4 | 3+4 4 | 3+4 4 | 3+4 4 | 6 2 |
| | 5 2 | 5+5 4 | 6+6 4 | 3+4 4 | 3+4 4 | 3+4 4 | 3+4 4 | 5 2 |
| | 5 2 | 6+6 4 | 5+5 4 | 3+4 4 | 3+4 4 | 3+4 4 | 3+4 4 | 5 2 |
| | 6 2 | 5+5 4 | 5+5 4 | 3+4 4 | 3+4 4 | 3+4 4 | 3+4 4 | 6 2 |
| | 5 2 | 6+6 4 | 6+6 4 | 3+4 4 | 3+4 4 | 3+4 4 | 3+4 4 | 6 2 |
| | 6 2 | 6+6 4 | 6+6 4 | 3+4 4 | 3+4 4 | 3+4 4 | 3+4 4 | 5 2 |
| | 6 2 | 5+5 4 | 5+5 4 | 3+4 4 | 3+4 4 | 3+4 4 | 3+4 4 | 5 2 |
| | 5 2 | 5+5 4 | 5+5 4 | 3+4 4 | 3+4 4 | 3+4 4 | 3+4 4 | 6 2 |
| | 6 2 | 6+6 4 | 6+6 4 | 3+4 4 | 3+4 4 | 3+4 4 | 3+4 4 | 6 2 |
| | 6 2 | 5+5 4 | 6+6 4 | 3+4 4 | 3+4 4 | 3+4 4 | 3+4 4 | 6 2 |

Continued on next column

**TABLE XXXI– continued**

| TQC | ICSD | NoE | SF | | | | | |
|---|---|---|---|---|---|---|---|---|
| | 5 2 | 6+6 4 | 5+5 4 | 3+4 4 | 3+4 4 | 3+4 4 | 3+4 4 | 5 2 |
| | 5 2 | 5+5 4 | 5+5 4 | 3+4 4 | 3+4 4 | 3+4 4 | 3+4 4 | 5 2 |
| | 5 2 | 5+5 4 | 6+6 4 | 3+4 4 | 3+4 4 | 3+4 4 | 3+4 4 | 6 2 |
| | 6 2 | 6+6 4 | 5+5 4 | 3+4 4 | 3+4 4 | 3+4 4 | 3+4 4 | 5 2 |
| | 5 2 | 5+5 4 | 6+6 4 | 3+4 4 | 3+4 4 | 3+4 4 | 3+4 4 | 6 2 |
| | 6 2 | 6+6 4 | 5+5 4 | 3+4 4 | 3+4 4 | 3+4 4 | 3+4 4 | 6 2 |
| | 6 2 | 5+5 4 | 6+6 4 | 3+4 4 | 3+4 4 | 3+4 4 | 3+4 4 | 5 2 |
| | 6 2 | 6+6 4 | 5+5 4 | 3+4 4 | 3+4 4 | 3+4 4 | 3+4 4 | 6 2 |
| | 6 2 | 5+5 4 | 6+6 4 | 3+4 4 | 3+4 4 | 3+4 4 | 3+4 4 | 6 2 |
| | 5 2 | 6+6 4 | 5+5 4 | 3+4 4 | 3+4 4 | 3+4 4 | 3+4 4 | 5 2 |
| | 6 2 | 5+5 4 | 6+6 4 | 3+4 4 | 3+4 4 | 3+4 4 | 3+4 4 | 6 2 |
| | 5 2 | 6+6 4 | 5+5 4 | 3+4 4 | 3+4 4 | 3+4 4 | 3+4 4 | 5 2 |
| | 6 2 | 5+5 4 | 6+6 4 | 3+4 4 | 3+4 4 | 3+4 4 | 3+4 4 | 5 2 |
| | 5 2 | 6+6 4 | 5+5 4 | 3+4 4 | 3+4 4 | 3+4 4 | 3+4 4 | 6 2 |
| | 5 2 | 6+6 4 | 6+6 4 | 3+4 4 | 3+4 4 | 3+4 4 | 3+4 4 | 6 2 |
| | 6 2 | | | | | | | 6 2 |
| | 6 2 | | | | | | | 5 2 |
| | 6 2 | | | | | | | 5 2 |
| | 5 2 | | | | | | | 6 2 |
| | 5 2 | | | | | | | 5 2 |
| | 5 2 | | | | | | | 6 2 |
| | 6 2 | | | | | | | 5 2 |
| | 6 2 | | | | | | | 6 2 |
| | 5 2 | | | | | | | 5 2 |
| | 6 2 | | | | | | | 6 2 |
| | 5 2 | | | | | | | 6 2 |
| | 6 2 | | | | | | | 5 2 |
| | 5 2 | | | | | | | 5 2 |
| | 6 2 | | | | | | | 5 2 |
| | 5 2 | | | | | | | 6 2 |
| | 6 2 | | | | | | | 5 2 |
| | 5 2 | | | | | | | 5 2 |
| | 5 2 | | | | | | | 6 2 |
| | 6 2 | | | | | | | 5 2 |
| | 6 2 | | | | | | | 6 2 |
| | 5 2 | | | | | | | 6 2 |
| | 6 2 | | | | | | | 5 2 |
| | 5 2 | | | | | | | 6 2 |
| | 5 2 | | | | | | | 6 2 |
| | 6 2 | | | | | | | 5 2 |
| | 6 2 | | | | | | | 5 2 |
| | 5 2 | | | | | | | 6 2 |
| | 5 2 | | | | | | | 6 2 |
| | 6 2 | | | | | | | 5 2 |
| | 6 2 | | | | | | | 6 2 |
| | 5 2 | | | | | | | 5 2 |
| | 5 2 | | | | | | | 5 2 |
| | 6 2 | | | | | | | 6 2 |
| | 6 2 | | | | | | | 5 2 |
| | 5 2 | | | | | | | 6 2 |
| | 5 2 | | | | | | | 5 2 |
| | 5 2 | | | | | | | 6 2 |
| | 5 2 | | | | | | | 6 2 |
| | 6 2 | | | | | | | 6 2 |
| | 5 2 | | | | | | | 5 2 |
| | 5 2 | | | | | | | 6 2 |
| | 6 2 | | | | | | | 5 2 |
| | 6 2 | | | | | | | 5 2 |
| | 6 2 | | | | | | | 6 2 |
| | 5 2 | | | | | | | 5 2 |
| | 6 2 | | | | | | | 5 2 |

Continued on next column

**TABLE XXXI– continued**

| TQC | ICSD | NoE | SF | | | | | |
|---|---|---|---|---|---|---|---|---|
| 55.2.31 | 2578 | 180 | Ce4O4S3 | | | | | |
| | -GM | -R | -S | -T | -U | -X | -Y | -Z |
| | 5 2 | 5+5 4 | 5+5 4 | 3+4 4 | 3+4 4 | 3+4 4 | 3+4 4 | 5 2 |
| | 6 2 | 6+6 4 | 6+6 4 | 3+4 4 | 3+4 4 | 3+4 4 | 3+4 4 | 6 2 |
| | 5 2 | 5+5 4 | 5+5 4 | 3+4 4 | 3+4 4 | 3+4 4 | 3+4 4 | 5 2 |
| | 6 2 | 6+6 4 | 6+6 4 | 3+4 4 | 3+4 4 | 3+4 4 | 3+4 4 | 6 2 |
| | 5 2 | 5+5 4 | 6+6 4 | 3+4 4 | 3+4 4 | 3+4 4 | 3+4 4 | 5 2 |
| | 5 2 | 6+6 4 | 5+5 4 | 3+4 4 | 3+4 4 | 3+4 4 | 3+4 4 | 5 2 |
| | 6 2 | 5+5 4 | 5+5 4 | 3+4 4 | 3+4 4 | 3+4 4 | 3+4 4 | 6 2 |
| | 6 2 | 6+6 4 | 6+6 4 | 3+4 4 | 3+4 4 | 3+4 4 | 3+4 4 | 6 2 |
| | 6 2 | 5+5 4 | 5+5 4 | 3+4 4 | 3+4 4 | 3+4 4 | 3+4 4 | 5 2 |
| | 5 2 | 6+6 4 | 6+6 4 | 3+4 4 | 3+4 4 | 3+4 4 | 3+4 4 | 6 2 |
| | 6 2 | 6+6 4 | 6+6 4 | 3+4 4 | 3+4 4 | 3+4 4 | 3+4 4 | 5 2 |
| | 5 2 | 5+5 4 | 5+5 4 | 3+4 4 | 3+4 4 | 3+4 4 | 3+4 4 | 6 2 |
| | 5 2 | 5+5 4 | 5+5 4 | 3+4 4 | 3+4 4 | 3+4 4 | 3+4 4 | 5 2 |
| | 5 2 | 6+6 4 | 6+6 4 | 3+4 4 | 3+4 4 | 3+4 4 | 3+4 4 | 5 2 |
| | 6 2 | 5+5 4 | 6+6 4 | 3+4 4 | 3+4 4 | 3+4 4 | 3+4 4 | 6 2 |
| | 6 2 | 6+6 4 | 5+5 4 | 3+4 4 | 3+4 4 | 3+4 4 | 3+4 4 | 6 2 |
| | 5 2 | 6+6 4 | 6+6 4 | 3+4 4 | 3+4 4 | 3+4 4 | 3+4 4 | 5 2 |
| | 5 2 | 5+5 4 | 5+5 4 | 3+4 4 | 3+4 4 | 3+4 4 | 3+4 4 | 5 2 |
| | 6 2 | 5+5 4 | 6+6 4 | 3+4 4 | 3+4 4 | 3+4 4 | 3+4 4 | 6 2 |
| | 6 2 | 6+6 4 | 5+5 4 | 3+4 4 | 3+4 4 | 3+4 4 | 3+4 4 | 6 2 |
| | 6 2 | 6+6 4 | 6+6 4 | 3+4 4 | 3+4 4 | 3+4 4 | 3+4 4 | 5 2 |
| | 6 2 | 5+5 4 | 5+5 4 | 3+4 4 | 3+4 4 | 3+4 4 | 3+4 4 | 5 2 |
| | 5 2 | 5+5 4 | 6+6 4 | 3+4 4 | 3+4 4 | 3+4 4 | 3+4 4 | 6 2 |
| | 5 2 | 6+6 4 | 5+5 4 | 3+4 4 | 3+4 4 | 3+4 4 | 3+4 4 | 6 2 |
| | 6 2 | 6+6 4 | 6+6 4 | 3+4 4 | 3+4 4 | 3+4 4 | 3+4 4 | 5 2 |
| | 5 2 | 5+5 4 | 5+5 4 | 3+4 4 | 3+4 4 | 3+4 4 | 3+4 4 | 6 2 |
| | 6 2 | 5+5 4 | 5+5 4 | 3+4 4 | 3+4 4 | 3+4 4 | 3+4 4 | 5 2 |
| | 5 2 | 5+5 4 | 5+5 4 | 3+4 4 | 3+4 4 | 3+4 4 | 3+4 4 | 5 2 |
| | 6 2 | 6+6 4 | 6+6 4 | 3+4 4 | 3+4 4 | 3+4 4 | 3+4 4 | 6 2 |
| | 5 2 | 6+6 4 | 6+6 4 | 3+4 4 | 3+4 4 | 3+4 4 | 3+4 4 | 5 2 |
| | 5 2 | 5+5 4 | 6+6 4 | 3+4 4 | 3+4 4 | 3+4 4 | 3+4 4 | 6 2 |
| | 6 2 | 5+5 4 | 5+5 4 | 3+4 4 | 3+4 4 | 3+4 4 | 3+4 4 | 6 2 |
| | 6 2 | 6+6 4 | 6+6 4 | 3+4 4 | 3+4 4 | 3+4 4 | 3+4 4 | 6 2 |
| | 6 2 | 5+5 4 | 5+5 4 | 3+4 4 | 3+4 4 | 3+4 4 | 3+4 4 | 6 2 |
| | 5 2 | 6+6 4 | 6+6 4 | 3+4 4 | 3+4 4 | 3+4 4 | 3+4 4 | 5 2 |
| | 5 2 | 6+6 4 | 5+5 4 | 3+4 4 | 3+4 4 | 3+4 4 | 3+4 4 | 5 2 |
| | 5 2 | 5+5 4 | 5+5 4 | 3+4 4 | 3+4 4 | 3+4 4 | 3+4 4 | 6 2 |
| | 5 2 | 6+6 4 | 6+6 4 | 3+4 4 | 3+4 4 | 3+4 4 | 3+4 4 | 6 2 |
| | 6 2 | 5+5 4 | 6+6 4 | 3+4 4 | 3+4 4 | 3+4 4 | 3+4 4 | 5 2 |
| | 6 2 | 6+6 4 | 5+5 4 | 3+4 4 | 3+4 4 | 3+4 4 | 3+4 4 | 5 2 |
| | 6 2 | 6+6 4 | 6+6 4 | 3+4 4 | 3+4 4 | 3+4 4 | 3+4 4 | 5 2 |
| | 5 2 | 5+5 4 | 5+5 4 | 3+4 4 | 3+4 4 | 3+4 4 | 3+4 4 | 6 2 |
| | 5 2 | 6+6 4 | 5+5 4 | 3+4 4 | 3+4 4 | 3+4 4 | 3+4 4 | 5 2 |
| | 6 2 | 6+6 4 | 5+5 4 | 3+4 4 | 3+4 4 | 3+4 4 | 3+4 4 | 6 2 |
| | 5 2 | 5+5 4 | 6+6 4 | 3+4 4 | 3+4 4 | 3+4 4 | 3+4 4 | 6 2 |
| | 5 2 | | | | | | | 6 2 |
| | 5 2 | | | | | | | 6 2 |
| | 6 2 | | | | | | | 6 2 |
| | 5 2 | | | | | | | 5 2 |
| | 6 2 | | | | | | | 6 2 |
| | 5 2 | | | | | | | 5 2 |
| | 5 2 | | | | | | | 6 2 |
| | 6 2 | | | | | | | 5 2 |
| | 5 2 | | | | | | | 5 2 |
| | 5 2 | | | | | | | 6 2 |
| | 5 2 | | | | | | | 5 2 |
| | 6 2 | | | | | | | 5 2 |
| | 5 2 | | | | | | | 5 2 |
| | 6 2 | | | | | | | 5 2 |

Continued on next column

**TABLE XXXI– continued**

| TQC | ICSD | NoE | SF | | | | | |
|---|---|---|---|---|---|---|---|---|
| | 6 2 | | | | | | | 6 2 |
| | 6 2 | | | | | | | 5 2 |
| | 6 2 | | | | | | | 6 2 |
| | 6 2 | | | | | | | 5 2 |
| | 6 2 | | | | | | | 5 2 |
| | 6 2 | | | | | | | 5 2 |
| | 5 2 | | | | | | | 6 2 |
| | 6 2 | | | | | | | 5 2 |
| | 5 2 | | | | | | | 6 2 |
| | 6 2 | | | | | | | 5 2 |
| | 5 2 | | | | | | | 6 2 |
| | 5 2 | | | | | | | 6 2 |
| | 6 2 | | | | | | | 5 2 |
| | 5 2 | | | | | | | 6 2 |
| | 6 2 | | | | | | | 5 2 |
| | 6 2 | | | | | | | 6 2 |
| | 5 2 | | | | | | | 6 2 |
| | 5 2 | | | | | | | 5 2 |
| | 5 2 | | | | | | | 6 2 |
| | 6 2 | | | | | | | 5 2 |
| | 6 2 | | | | | | | 5 2 |
| | 6 2 | | | | | | | 5 2 |
| | 6 2 | | | | | | | 6 2 |
| | 5 2 | | | | | | | 5 2 |
| | 6 2 | | | | | | | 5 2 |
| | 5 2 | | | | | | | 6 2 |
| | 6 2 | | | | | | | 5 2 |
| | 6 2 | | | | | | | 6 2 |
| | 6 2 | | | | | | | 5 2 |
| | 5 2 | | | | | | | 6 2 |
| | 6 2 | | | | | | | 6 2 |
| 55.2.32 | 102075 | 312 | Cd7Th6 | | | | | |
| | -GM | -R | -S | -T | -U | -X | -Y | -Z |
| | 5 2 | 6+6 4 | 5+5 4 | 3+4 4 | 3+4 4 | 3+4 4 | 3+4 4 | 6 2 |
| | 5 2 | 5+5 4 | 6+6 4 | 3+4 4 | 3+4 4 | 3+4 4 | 3+4 4 | 6 2 |
| | 6 2 | 6+6 4 | 6+6 4 | 3+4 4 | 3+4 4 | 3+4 4 | 3+4 4 | 5 2 |
| | 6 2 | 5+5 4 | 5+5 4 | 3+4 4 | 3+4 4 | 3+4 4 | 3+4 4 | 5 2 |
| | 5 2 | 5+5 4 | 6+6 4 | 3+4 4 | 3+4 4 | 3+4 4 | 3+4 4 | 5 2 |
| | 5 2 | 6+6 4 | 5+5 4 | 3+4 4 | 3+4 4 | 3+4 4 | 3+4 4 | 5 2 |
| | 6 2 | 6+6 4 | 5+5 4 | 3+4 4 | 3+4 4 | 3+4 4 | 3+4 4 | 6 2 |
| | 6 2 | 5+5 4 | 6+6 4 | 3+4 4 | 3+4 4 | 3+4 4 | 3+4 4 | 6 2 |
| | 5 2 | 6+6 4 | 6+6 4 | 3+4 4 | 3+4 4 | 3+4 4 | 3+4 4 | 6 2 |
| | 5 2 | 5+5 4 | 6+6 4 | 3+4 4 | 3+4 4 | 3+4 4 | 3+4 4 | 6 2 |
| | 6 2 | 5+5 4 | 5+5 4 | 3+4 4 | 3+4 4 | 3+4 4 | 3+4 4 | 5 2 |
| | 6 2 | 6+6 4 | 5+5 4 | 3+4 4 | 3+4 4 | 3+4 4 | 3+4 4 | 5 2 |
| | 5 2 | 5+5 4 | 6+6 4 | 3+4 4 | 3+4 4 | 3+4 4 | 3+4 4 | 6 2 |
| | 5 2 | 6+6 4 | 5+5 4 | 3+4 4 | 3+4 4 | 3+4 4 | 3+4 4 | 6 2 |
| | 6 2 | 6+6 4 | 6+6 4 | 3+4 4 | 3+4 4 | 3+4 4 | 3+4 4 | 5 2 |
| | 6 2 | 5+5 4 | 5+5 4 | 3+4 4 | 3+4 4 | 3+4 4 | 3+4 4 | 5 2 |
| | 5 2 | 5+5 4 | 6+6 4 | 3+4 4 | 3+4 4 | 3+4 4 | 3+4 4 | 5 2 |
| | 5 2 | 6+6 4 | 5+5 4 | 3+4 4 | 3+4 4 | 3+4 4 | 3+4 4 | 5 2 |
| | 5 2 | 5+5 4 | 6+6 4 | 3+4 4 | 3+4 4 | 3+4 4 | 3+4 4 | 6 2 |
| | 5 2 | 6+6 4 | 5+5 4 | 3+4 4 | 3+4 4 | 3+4 4 | 3+4 4 | 6 2 |
| | 6 2 | 6+6 4 | 6+6 4 | 3+4 4 | 3+4 4 | 3+4 4 | 3+4 4 | 6 2 |
| | 6 2 | 5+5 4 | 6+6 4 | 3+4 4 | 3+4 4 | 3+4 4 | 3+4 4 | 6 2 |
| | 6 2 | 5+5 4 | 5+5 4 | 3+4 4 | 3+4 4 | 3+4 4 | 3+4 4 | 5 2 |
| | 6 2 | 6+6 4 | 5+5 4 | 3+4 4 | 3+4 4 | 3+4 4 | 3+4 4 | 5 2 |
| | 5 2 | 5+5 4 | 5+5 4 | 3+4 4 | 3+4 4 | 3+4 4 | 3+4 4 | 6 2 |
| | 5 2 | 5+5 4 | 6+6 4 | 3+4 4 | 3+4 4 | 3+4 4 | 3+4 4 | 5 2 |
| | 5 2 | 6+6 4 | 5+5 4 | 3+4 4 | 3+4 4 | 3+4 4 | 3+4 4 | 6 2 |
| | 6 2 | 6+6 4 | 6+6 4 | 3+4 4 | 3+4 4 | 3+4 4 | 3+4 4 | 5 2 |
| | 5 2 | 5+5 4 | 5+5 4 | 3+4 4 | 3+4 4 | 3+4 4 | 3+4 4 | 5 2 |
| | 6 2 | 5+5 4 | 5+5 4 | 3+4 4 | 3+4 4 | 3+4 4 | 3+4 4 | 6 2 |
| | 6 2 | 5+5 4 | 6+6 4 | 3+4 4 | 3+4 4 | 3+4 4 | 3+4 4 | 5 2 |
| | 5 2 | 6+6 4 | 5+5 4 | 3+4 4 | 3+4 4 | 3+4 4 | 3+4 4 | 6 2 |
| | 5 2 | 5+5 4 | 6+6 4 | 3+4 4 | 3+4 4 | 3+4 4 | 3+4 4 | 6 2 |
| | 6 2 | 6+6 4 | 5+5 4 | 3+4 4 | 3+4 4 | 3+4 4 | 3+4 4 | 6 2 |
| | 6 2 | 6+6 4 | 6+6 4 | 3+4 4 | 3+4 4 | 3+4 4 | 3+4 4 | 5 2 |
| | 5 2 | 5+5 4 | 5+5 4 | 3+4 4 | 3+4 4 | 3+4 4 | 3+4 4 | 5 2 |
| | 5 2 | 6+6 4 | 6+6 4 | 3+4 4 | 3+4 4 | 3+4 4 | 3+4 4 | 6 2 |
| | 6 2 | 5+5 4 | 5+5 4 | 3+4 4 | 3+4 4 | 3+4 4 | 3+4 4 | 6 2 |
| | 6 2 | 5+5 4 | 6+6 4 | 3+4 4 | 3+4 4 | 3+4 4 | 3+4 4 | 5 2 |
| | 6 2 | 6+6 4 | 5+5 4 | 3+4 4 | 3+4 4 | 3+4 4 | 3+4 4 | 5 2 |
| | 5 2 | 5+5 4 | 5+5 4 | 3+4 4 | 3+4 4 | 3+4 4 | 3+4 4 | 5 2 |
| | 5 2 | 6+6 4 | 6+6 4 | 3+4 4 | 3+4 4 | 3+4 4 | 3+4 4 | 5 2 |
| | 5 2 | 5+5 4 | 5+5 4 | 3+4 4 | 3+4 4 | 3+4 4 | 3+4 4 | 6 2 |
| | 5 2 | 6+6 4 | 6+6 4 | 3+4 4 | 3+4 4 | 3+4 4 | 3+4 4 | 6 2 |
| | 6 2 | 5+5 4 | 5+5 4 | 3+4 4 | 3+4 4 | 3+4 4 | 3+4 4 | 6 2 |
| | 6 2 | 6+6 4 | 5+5 4 | 3+4 4 | 3+4 4 | 3+4 4 | 3+4 4 | 6 2 |
| | 6 2 | 5+5 4 | 6+6 4 | 3+4 4 | 3+4 4 | 3+4 4 | 3+4 4 | 5 2 |
| | 6 2 | 6+6 4 | 5+5 4 | 3+4 4 | 3+4 4 | 3+4 4 | 3+4 4 | 5 2 |
| | 5 2 | 5+5 4 | 6+6 4 | 3+4 4 | 3+4 4 | 3+4 4 | 3+4 4 | 5 2 |
| | 5 2 | 5+5 4 | 5+5 4 | 3+4 4 | 3+4 4 | 3+4 4 | 3+4 4 | 5 2 |
| | 6 2 | 6+6 4 | 6+6 4 | 3+4 4 | 3+4 4 | 3+4 4 | 3+4 4 | 6 2 |
| | 6 2 | 5+5 4 | 6+6 4 | 3+4 4 | 3+4 4 | 3+4 4 | 3+4 4 | 5 2 |
| | 5 2 | 6+6 4 | 5+5 4 | 3+4 4 | 3+4 4 | 3+4 4 | 3+4 4 | 6 2 |
| | 6 2 | 5+5 4 | 5+5 4 | 3+4 4 | 3+4 4 | 3+4 4 | 3+4 4 | 5 2 |
| | 5 2 | 6+6 4 | 6+6 4 | 3+4 4 | 3+4 4 | 3+4 4 | 3+4 4 | 6 2 |
| | 5 2 | 5+5 4 | 5+5 4 | 3+4 4 | 3+4 4 | 3+4 4 | 3+4 4 | 5 2 |
| | 5 2 | 6+6 4 | 5+5 4 | 3+4 4 | 3+4 4 | 3+4 4 | 3+4 4 | 5 2 |
| | 5 2 | 5+5 4 | 6+6 4 | 3+4 4 | 3+4 4 | 3+4 4 | 3+4 4 | 5 2 |
| | 6 2 | 5+5 4 | 5+5 4 | 3+4 4 | 3+4 4 | 3+4 4 | 3+4 4 | 5 2 |
| | 6 2 | 5+5 4 | 5+5 4 | 3+4 4 | 3+4 4 | 3+4 4 | 3+4 4 | 6 2 |
| | 6 2 | 6+6 4 | 6+6 4 | 3+4 4 | 3+4 4 | 3+4 4 | 3+4 4 | 5 2 |
| | 5 2 | 6+6 4 | 5+5 4 | 3+4 4 | 3+4 4 | 3+4 4 | 3+4 4 | 6 2 |
| | 6 2 | 5+5 4 | 6+6 4 | 3+4 4 | 3+4 4 | 3+4 4 | 3+4 4 | 5 2 |
| | 5 2 | 5+5 4 | 5+5 4 | 3+4 4 | 3+4 4 | 3+4 4 | 3+4 4 | 6 2 |
| | 5 2 | 6+6 4 | 6+6 4 | 3+4 4 | 3+4 4 | 3+4 4 | 3+4 4 | 5 2 |
| | 5 2 | 5+5 4 | 5+5 4 | 3+4 4 | 3+4 4 | 3+4 4 | 3+4 4 | 6 2 |
| | 5 2 | 5+5 4 | 6+6 4 | 3+4 4 | 3+4 4 | 3+4 4 | 3+4 4 | 5 2 |
| | 6 2 | 6+6 4 | 5+5 4 | 3+4 4 | 3+4 4 | 3+4 4 | 3+4 4 | 6 2 |
| | 6 2 | 6+6 4 | 6+6 4 | 3+4 4 | 3+4 4 | 3+4 4 | 3+4 4 | 5 2 |
| | 5 2 | 6+6 4 | 5+5 4 | 3+4 4 | 3+4 4 | 3+4 4 | 3+4 4 | 6 2 |
| | 6 2 | 5+5 4 | 5+5 4 | 3+4 4 | 3+4 4 | 3+4 4 | 3+4 4 | 5 2 |
| | 5 2 | 6+6 4 | 6+6 4 | 3+4 4 | 3+4 4 | 3+4 4 | 3+4 4 | 6 2 |
| | 6 2 | 5+5 4 | 6+6 4 | 3+4 4 | 3+4 4 | 3+4 4 | 3+4 4 | 5 2 |
| | 5 2 | 6+6 4 | 5+5 4 | 3+4 4 | 3+4 4 | 3+4 4 | 3+4 4 | 6 2 |
| | 6 2 | 6+6 4 | 6+6 4 | 3+4 4 | 3+4 4 | 3+4 4 | 3+4 4 | 5 2 |
| | 5 2 | 5+5 4 | 6+6 4 | 3+4 4 | 3+4 4 | 3+4 4 | 3+4 4 | 6 2 |
| | 5 2 | 5+5 4 | 6+6 4 | 3+4 4 | 3+4 4 | 3+4 4 | 3+4 4 | 5 2 |
| | 6 2 | 5+5 4 | 5+5 4 | 3+4 4 | 3+4 4 | 3+4 4 | 3+4 4 | 6 2 |
| | 5 2 | | | | | | | 5 2 |
| | 6 2 | | | | | | | 6 2 |
| | 5 2 | | | | | | | 5 2 |
| | 6 2 | | | | | | | 6 2 |
| | 5 2 | | | | | | | 5 2 |
| | 6 2 | | | | | | | 5 2 |
| | 5 2 | | | | | | | 5 2 |
| | 5 2 | | | | | | | 6 2 |
| | 6 2 | | | | | | | 6 2 |
| | 6 2 | | | | | | | 6 2 |
| | 5 2 | | | | | | | 5 2 |

Continued on next column

Continued on next column

**TABLE XXXI– continued**

| TQC | ICSD | NoE | SF | | | | | |
|---|---|---|---|---|---|---|---|---|
| | 6 2 | | | | | | | 5 2 |
| | 5 2 | | | | | | | 5 2 |
| | 5 2 | | | | | | | 6 2 |
| | 6 2 | | | | | | | 6 2 |
| | 5 2 | | | | | | | 5 2 |
| | 6 2 | | | | | | | 6 2 |
| | 5 2 | | | | | | | 6 2 |
| | 5 2 | | | | | | | 5 2 |
| | 6 2 | | | | | | | 6 2 |
| | 5 2 | | | | | | | 5 2 |
| | 6 2 | | | | | | | 5 2 |
| | 6 2 | | | | | | | 6 2 |
| | 5 2 | | | | | | | 5 2 |
| | 5 2 | | | | | | | 6 2 |
| | 5 2 | | | | | | | 6 2 |
| | 5 2 | | | | | | | 5 2 |
| | 6 2 | | | | | | | 5 2 |
| | 6 2 | | | | | | | 5 2 |
| | 6 2 | | | | | | | 6 2 |
| | 5 2 | | | | | | | 5 2 |
| | 5 2 | | | | | | | 5 2 |
| | 6 2 | | | | | | | 6 2 |
| | 5 2 | | | | | | | 5 2 |
| | 5 2 | | | | | | | 6 2 |
| | 6 2 | | | | | | | 6 2 |
| | 5 2 | | | | | | | 5 2 |
| | 5 2 | | | | | | | 5 2 |
| | 6 2 | | | | | | | 5 2 |
| | 5 2 | | | | | | | 5 2 |
| | 5 2 | | | | | | | 5 2 |
| | 6 2 | | | | | | | 6 2 |
| | 5 2 | | | | | | | 6 2 |
| | 6 2 | | | | | | | 5 2 |
| | 6 2 | | | | | | | 6 2 |
| | 5 2 | | | | | | | 5 2 |
| | 5 2 | | | | | | | 5 2 |
| | 5 2 | | | | | | | 6 2 |
| | 5 2 | | | | | | | 5 2 |
| | 6 2 | | | | | | | 6 2 |
| | 6 2 | | | | | | | 5 2 |
| | 5 2 | | | | | | | 6 2 |
| | 6 2 | | | | | | | 5 2 |
| | 6 2 | | | | | | | 5 2 |
| | 6 2 | | | | | | | 6 2 |
| | 5 2 | | | | | | | 6 2 |
| | 6 2 | | | | | | | 5 2 |
| | 6 2 | | | | | | | 6 2 |
| | 6 2 | | | | | | | 5 2 |
| | 6 2 | | | | | | | 6 2 |
| | 5 2 | | | | | | | 5 2 |
| | 5 2 | | | | | | | 6 2 |
| | 6 2 | | | | | | | 5 2 |
| | 6 2 | | | | | | | 6 2 |
| | 5 2 | | | | | | | 5 2 |
| | 5 2 | | | | | | | 6 2 |
| | 5 2 | | | | | | | 6 2 |
| | 6 2 | | | | | | | 6 2 |
| | 6 2 | | | | | | | 6 2 |
| | 6 2 | | | | | | | 5 2 |
| | 5 2 | | | | | | | 5 2 |
| | 6 2 | | | | | | | 5 2 |

Continued on next column

**TABLE XXXI– continued**

| TQC | ICSD | NoE | SF | | | | | |
|---|---|---|---|---|---|---|---|---|
| | 5 2 | | | | | | | 6 2 |
| | 6 2 | | | | | | | 5 2 |
| | 5 2 | | | | | | | 6 2 |
| | 5 2 | | | | | | | 6 2 |
| | 5 2 | | | | | | | 6 2 |
| | 6 2 | | | | | | | 5 2 |
| 55.2.33 | 36466 | 172 | Ca5Ga2Sb6 | | | | | |
| | -GM | -R | -S | -T | -U | -X | -Y | -Z |
| | 5 2 | 5+5 4 | 5+5 4 | 3+4 4 | 3+4 4 | 3+4 4 | 3+4 4 | 5 2 |
| | 5 2 | 6+6 4 | 6+6 4 | 3+4 4 | 3+4 4 | 3+4 4 | 3+4 4 | 5 2 |
| | 6 2 | 5+5 4 | 6+6 4 | 3+4 4 | 3+4 4 | 3+4 4 | 3+4 4 | 6 2 |
| | 6 2 | 5+5 4 | 5+5 4 | 3+4 4 | 3+4 4 | 3+4 4 | 3+4 4 | 6 2 |
| | 5 2 | 6+6 4 | 6+6 4 | 3+4 4 | 3+4 4 | 3+4 4 | 3+4 4 | 6 2 |
| | 5 2 | 6+6 4 | 5+5 4 | 3+4 4 | 3+4 4 | 3+4 4 | 3+4 4 | 6 2 |
| | 5 2 | 5+5 4 | 6+6 4 | 3+4 4 | 3+4 4 | 3+4 4 | 3+4 4 | 5 2 |
| | 5 2 | 6+6 4 | 5+5 4 | 3+4 4 | 3+4 4 | 3+4 4 | 3+4 4 | 5 2 |
| | 6 2 | 5+5 4 | 5+5 4 | 3+4 4 | 3+4 4 | 3+4 4 | 3+4 4 | 6 2 |
| | 6 2 | 6+6 4 | 6+6 4 | 3+4 4 | 3+4 4 | 3+4 4 | 3+4 4 | 6 2 |
| | 5 2 | 5+5 4 | 5+5 4 | 3+4 4 | 3+4 4 | 3+4 4 | 3+4 4 | 6 2 |
| | 5 2 | 6+6 4 | 6+6 4 | 3+4 4 | 3+4 4 | 3+4 4 | 3+4 4 | 6 2 |
| | 6 2 | 6+6 4 | 5+5 4 | 3+4 4 | 3+4 4 | 3+4 4 | 3+4 4 | 5 2 |
| | 6 2 | 5+5 4 | 6+6 4 | 3+4 4 | 3+4 4 | 3+4 4 | 3+4 4 | 5 2 |
| | 6 2 | 6+6 4 | 5+5 4 | 3+4 4 | 3+4 4 | 3+4 4 | 3+4 4 | 5 2 |
| | 6 2 | 6+6 4 | 5+5 4 | 3+4 4 | 3+4 4 | 3+4 4 | 3+4 4 | 5 2 |
| | 5 2 | 5+5 4 | 5+5 4 | 3+4 4 | 3+4 4 | 3+4 4 | 3+4 4 | 5 2 |
| | 6 2 | 6+6 4 | 6+6 4 | 3+4 4 | 3+4 4 | 3+4 4 | 3+4 4 | 6 2 |
| | 5 2 | 5+5 4 | 6+6 4 | 3+4 4 | 3+4 4 | 3+4 4 | 3+4 4 | 5 2 |
| | 6 2 | 6+6 4 | 5+5 4 | 3+4 4 | 3+4 4 | 3+4 4 | 3+4 4 | 6 2 |
| | 5 2 | 5+5 4 | 5+5 4 | 3+4 4 | 3+4 4 | 3+4 4 | 3+4 4 | 5 2 |
| | 6 2 | 6+6 4 | 6+6 4 | 3+4 4 | 3+4 4 | 3+4 4 | 3+4 4 | 5 2 |
| | 5 2 | 5+5 4 | 5+5 4 | 3+4 4 | 3+4 4 | 3+4 4 | 3+4 4 | 6 2 |
| | 6 2 | 5+5 4 | 6+6 4 | 3+4 4 | 3+4 4 | 3+4 4 | 3+4 4 | 6 2 |
| | 5 2 | 6+6 4 | 5+5 4 | 3+4 4 | 3+4 4 | 3+4 4 | 3+4 4 | 6 2 |
| | 5 2 | 6+6 4 | 6+6 4 | 3+4 4 | 3+4 4 | 3+4 4 | 3+4 4 | 6 2 |
| | 6 2 | 5+5 4 | 6+6 4 | 3+4 4 | 3+4 4 | 3+4 4 | 3+4 4 | 5 2 |
| | 6 2 | 6+6 4 | 5+5 4 | 3+4 4 | 3+4 4 | 3+4 4 | 3+4 4 | 5 2 |
| | 6 2 | 6+6 4 | 5+5 4 | 3+4 4 | 3+4 4 | 3+4 4 | 3+4 4 | 5 2 |
| | 6 2 | 5+5 4 | 6+6 4 | 3+4 4 | 3+4 4 | 3+4 4 | 3+4 4 | 5 2 |
| | 6 2 | 5+5 4 | 5+5 4 | 3+4 4 | 3+4 4 | 3+4 4 | 3+4 4 | 5 2 |
| | 6 2 | 6+6 4 | 6+6 4 | 3+4 4 | 3+4 4 | 3+4 4 | 3+4 4 | 5 2 |
| | 5 2 | 6+6 4 | 5+5 4 | 3+4 4 | 3+4 4 | 3+4 4 | 3+4 4 | 5 2 |
| | 5 2 | 5+5 4 | 6+6 4 | 3+4 4 | 3+4 4 | 3+4 4 | 3+4 4 | 6 2 |
| | 6 2 | 5+5 4 | 6+6 4 | 3+4 4 | 3+4 4 | 3+4 4 | 3+4 4 | 6 2 |
| | 6 2 | 6+6 4 | 5+5 4 | 3+4 4 | 3+4 4 | 3+4 4 | 3+4 4 | 5 2 |
| | 5 2 | 6+6 4 | 6+6 4 | 3+4 4 | 3+4 4 | 3+4 4 | 3+4 4 | 6 2 |
| | 5 2 | 5+5 4 | 6+6 4 | 3+4 4 | 3+4 4 | 3+4 4 | 3+4 4 | 6 2 |
| | 6 2 | 5+5 4 | 5+5 4 | 3+4 4 | 3+4 4 | 3+4 4 | 3+4 4 | 5 2 |
| | 6 2 | 6+6 4 | 5+5 4 | 3+4 4 | 3+4 4 | 3+4 4 | 3+4 4 | 5 2 |
| | 5 2 | 5+5 4 | 6+6 4 | 3+4 4 | 3+4 4 | 3+4 4 | 3+4 4 | 5 2 |
| | 5 2 | 5+5 4 | 5+5 4 | 3+4 4 | 3+4 4 | 3+4 4 | 3+4 4 | 5 2 |
| | 6 2 | 6+6 4 | 6+6 4 | 3+4 4 | 3+4 4 | 3+4 4 | 3+4 4 | 6 2 |
| | 6 2 | | | | | | | 6 2 |
| | 5 2 | | | | | | | 6 2 |
| | 5 2 | | | | | | | 5 2 |
| | 5 2 | | | | | | | 5 2 |
| | 6 2 | | | | | | | 5 2 |
| | 6 2 | | | | | | | 6 2 |
| | 5 2 | | | | | | | 5 2 |
| | 6 2 | | | | | | | 6 2 |
| | 6 2 | | | | | | | 6 2 |
| | 5 2 | | | | | | | 5 2 |

Continued on next column

**TABLE XXXI– continued**

| TQC | ICSD | | NoE | | SF | | | |
|---|---|---|---|---|---|---|---|---|
| | 6 2 | | | | | | | 6 2 |
| | 5 2 | | | | | | | 5 2 |
| | 6 2 | | | | | | | 6 2 |
| | 5 2 | | | | | | | 6 2 |
| | 5 2 | | | | | | | 6 2 |
| | 6 2 | | | | | | | 6 2 |
| | 6 2 | | | | | | | 5 2 |
| | 5 2 | | | | | | | 6 2 |
| | 5 2 | | | | | | | 5 2 |
| | 6 2 | | | | | | | 5 2 |
| | 6 2 | | | | | | | 5 2 |
| | 5 2 | | | | | | | 6 2 |
| | 5 2 | | | | | | | 5 2 |
| | 6 2 | | | | | | | 6 2 |
| | 6 2 | | | | | | | 5 2 |
| | 5 2 | | | | | | | 6 2 |
| | 5 2 | | | | | | | 6 2 |
| | 6 2 | | | | | | | 6 2 |
| | 5 2 | | | | | | | 5 2 |
| | 6 2 | | | | | | | 5 2 |
| | 6 2 | | | | | | | 5 2 |
| | 6 2 | | | | | | | 5 2 |
| | 5 2 | | | | | | | 6 2 |
| | 5 2 | | | | | | | 5 2 |
| | 5 2 | | | | | | | 6 2 |
| | 6 2 | | | | | | | 5 2 |
| | 5 2 | | | | | | | 6 2 |
| | 6 2 | | | | | | | 5 2 |
| | 6 2 | | | | | | | 6 2 |
| | 5 2 | | | | | | | 5 2 |
| | 5 2 | | | | | | | 5 2 |
| | 6 2 | | | | | | | 5 2 |
| | 5 2 | | | | | | | 6 2 |
| 55.2.34 | 35130 | | 96 | | Ca2Mn2O5 | | | |
| | -GM | -R | -S | -T | -U | -X | -Y | -Z |
| | 5 2 | 5+5 4 | 5+5 4 | 3+4 4 | 3+4 4 | 3+4 4 | 3+4 4 | 5 2 |
| | 6 2 | 6+6 4 | 6+6 4 | 3+4 4 | 3+4 4 | 3+4 4 | 3+4 4 | 6 2 |
| | 5 2 | 5+5 4 | 5+5 4 | 3+4 4 | 3+4 4 | 3+4 4 | 3+4 4 | 6 2 |
| | 6 2 | 6+6 4 | 6+6 4 | 3+4 4 | 3+4 4 | 3+4 4 | 3+4 4 | 5 2 |
| | 5 2 | 5+5 4 | 5+5 4 | 3+4 4 | 3+4 4 | 3+4 4 | 3+4 4 | 6 2 |
| | 6 2 | 6+6 4 | 6+6 4 | 3+4 4 | 3+4 4 | 3+4 4 | 3+4 4 | 5 2 |
| | 5 2 | 5+5 4 | 5+5 4 | 3+4 4 | 3+4 4 | 3+4 4 | 3+4 4 | 5 2 |
| | 6 2 | 6+6 4 | 6+6 4 | 3+4 4 | 3+4 4 | 3+4 4 | 3+4 4 | 6 2 |
| | 5 2 | 6+6 4 | 6+6 4 | 3+4 4 | 3+4 4 | 3+4 4 | 3+4 4 | 5 2 |
| | 5 2 | 5+5 4 | 6+6 4 | 3+4 4 | 3+4 4 | 3+4 4 | 3+4 4 | 5 2 |
| | 6 2 | 5+5 4 | 5+5 4 | 3+4 4 | 3+4 4 | 3+4 4 | 3+4 4 | 6 2 |
| | 6 2 | 5+5 4 | 5+5 4 | 3+4 4 | 3+4 4 | 3+4 4 | 3+4 4 | 5 2 |
| | 5 2 | 6+6 4 | 6+6 4 | 3+4 4 | 3+4 4 | 3+4 4 | 3+4 4 | 5 2 |
| | 6 2 | 6+6 4 | 6+6 4 | 3+4 4 | 3+4 4 | 3+4 4 | 3+4 4 | 6 2 |
| | 5 2 | 6+6 4 | 5+5 4 | 3+4 4 | 3+4 4 | 3+4 4 | 3+4 4 | 6 2 |
| | 6 2 | 5+5 4 | 5+5 4 | 3+4 4 | 3+4 4 | 3+4 4 | 3+4 4 | 6 2 |
| | 5 2 | 6+6 4 | 5+5 4 | 3+4 4 | 3+4 4 | 3+4 4 | 3+4 4 | 5 2 |
| | 6 2 | 6+6 4 | 6+6 4 | 3+4 4 | 3+4 4 | 3+4 4 | 3+4 4 | 6 2 |
| | 6 2 | 5+5 4 | 6+6 4 | 3+4 4 | 3+4 4 | 3+4 4 | 3+4 4 | 6 2 |
| | 6 2 | 6+6 4 | 6+6 4 | 3+4 4 | 3+4 4 | 3+4 4 | 3+4 4 | 5 2 |
| | 5 2 | 5+5 4 | 5+5 4 | 3+4 4 | 3+4 4 | 3+4 4 | 3+4 4 | 5 2 |
| | 6 2 | 6+6 4 | 6+6 4 | 3+4 4 | 3+4 4 | 3+4 4 | 3+4 4 | 6 2 |
| | 6 2 | 5+5 4 | 5+5 4 | 3+4 4 | 3+4 4 | 3+4 4 | 3+4 4 | 6 2 |
| | 5 2 | 6+6 4 | 5+5 4 | 3+4 4 | 3+4 4 | 3+4 4 | 3+4 4 | 5 2 |
| | 6 2 | | | | | | | 6 2 |
| | 5 2 | | | | | | | 6 2 |

Continued on next column

**TABLE XXXI– continued**

| TQC | ICSD | | NoE | | SF | | | |
|---|---|---|---|---|---|---|---|---|
| | 5 2 | | | | | | | 5 2 |
| | 5 2 | | | | | | | 6 2 |
| | 6 2 | | | | | | | 5 2 |
| | 5 2 | | | | | | | 6 2 |
| | 5 2 | | | | | | | 5 2 |
| | 6 2 | | | | | | | 6 2 |
| | 5 2 | | | | | | | 6 2 |
| | 6 2 | | | | | | | 5 2 |
| | 5 2 | | | | | | | 6 2 |
| | 6 2 | | | | | | | 6 2 |
| | 6 2 | | | | | | | 5 2 |
| | 6 2 | | | | | | | 5 2 |
| | 6 2 | | | | | | | 6 2 |
| | 6 2 | | | | | | | 6 2 |
| | 5 2 | | | | | | | 5 2 |
| | 5 2 | | | | | | | 5 2 |
| | 5 2 | | | | | | | 5 2 |
| | 6 2 | | | | | | | 6 2 |
| | 5 2 | | | | | | | 6 2 |
| | 6 2 | | | | | | | 5 2 |
| | 5 2 | | | | | | | 6 2 |
| | 6 2 | | | | | | | 5 2 |
| 55.2.35 | 35219 | | 128 | | Ca1Mn1O2.5 | | | |
| | -GM | -R | -S | -T | -U | -X | -Y | -Z |
| | 5 2 | 6+6 4 | 5+5 4 | 3+4 4 | 3+4 4 | 3+4 4 | 3+4 4 | 6 2 |
| | 6 2 | 5+5 4 | 6+6 4 | 3+4 4 | 3+4 4 | 3+4 4 | 3+4 4 | 5 2 |
| | 5 2 | 5+5 4 | 5+5 4 | 3+4 4 | 3+4 4 | 3+4 4 | 3+4 4 | 6 2 |
| | 6 2 | 6+6 4 | 6+6 4 | 3+4 4 | 3+4 4 | 3+4 4 | 3+4 4 | 5 2 |
| | 5 2 | 5+5 4 | 5+5 4 | 3+4 4 | 3+4 4 | 3+4 4 | 3+4 4 | 5 2 |
| | 6 2 | 6+6 4 | 6+6 4 | 3+4 4 | 3+4 4 | 3+4 4 | 3+4 4 | 6 2 |
| | 5 2 | 6+6 4 | 6+6 4 | 3+4 4 | 3+4 4 | 3+4 4 | 3+4 4 | 5 2 |
| | 6 2 | 5+5 4 | 5+5 4 | 3+4 4 | 3+4 4 | 3+4 4 | 3+4 4 | 6 2 |
| | 6 2 | 5+5 4 | 5+5 4 | 3+4 4 | 3+4 4 | 3+4 4 | 3+4 4 | 6 2 |
| | 5 2 | 6+6 4 | 6+6 4 | 3+4 4 | 3+4 4 | 3+4 4 | 3+4 4 | 5 2 |
| | 5 2 | 5+5 4 | 5+5 4 | 3+4 4 | 3+4 4 | 3+4 4 | 3+4 4 | 5 2 |
| | 6 2 | 6+6 4 | 6+6 4 | 3+4 4 | 3+4 4 | 3+4 4 | 3+4 4 | 6 2 |
| | 6 2 | 5+5 4 | 5+5 4 | 3+4 4 | 3+4 4 | 3+4 4 | 3+4 4 | 6 2 |
| | 5 2 | 6+6 4 | 6+6 4 | 3+4 4 | 3+4 4 | 3+4 4 | 3+4 4 | 5 2 |
| | 6 2 | 5+5 4 | 5+5 4 | 3+4 4 | 3+4 4 | 3+4 4 | 3+4 4 | 6 2 |
| | 5 2 | 6+6 4 | 6+6 4 | 3+4 4 | 3+4 4 | 3+4 4 | 3+4 4 | 5 2 |
| | 5 2 | 5+5 4 | 6+6 4 | 3+4 4 | 3+4 4 | 3+4 4 | 3+4 4 | 6 2 |
| | 6 2 | 6+6 4 | 6+6 4 | 3+4 4 | 3+4 4 | 3+4 4 | 3+4 4 | 5 2 |
| | 6 2 | 5+5 4 | 5+5 4 | 3+4 4 | 3+4 4 | 3+4 4 | 3+4 4 | 6 2 |
| | 5 2 | 5+5 4 | 5+5 4 | 3+4 4 | 3+4 4 | 3+4 4 | 3+4 4 | 5 2 |
| | 6 2 | 6+6 4 | 6+6 4 | 3+4 4 | 3+4 4 | 3+4 4 | 3+4 4 | 6 2 |
| | 5 2 | 6+6 4 | 6+6 4 | 3+4 4 | 3+4 4 | 3+4 4 | 3+4 4 | 5 2 |
| | 5 2 | 6+6 4 | 5+5 4 | 3+4 4 | 3+4 4 | 3+4 4 | 3+4 4 | 5 2 |
| | 6 2 | 5+5 4 | 5+5 4 | 3+4 4 | 3+4 4 | 3+4 4 | 3+4 4 | 6 2 |
| | 5 2 | 6+6 4 | 5+5 4 | 3+4 4 | 3+4 4 | 3+4 4 | 3+4 4 | 5 2 |
| | 5 2 | 6+6 4 | 6+6 4 | 3+4 4 | 3+4 4 | 3+4 4 | 3+4 4 | 5 2 |
| | 6 2 | 5+5 4 | 6+6 4 | 3+4 4 | 3+4 4 | 3+4 4 | 3+4 4 | 6 2 |
| | 5 2 | 6+6 4 | 6+6 4 | 3+4 4 | 3+4 4 | 3+4 4 | 3+4 4 | 5 2 |
| | 6 2 | 5+5 4 | 5+5 4 | 3+4 4 | 3+4 4 | 3+4 4 | 3+4 4 | 6 2 |
| | 6 2 | 6+6 4 | 6+6 4 | 3+4 4 | 3+4 4 | 3+4 4 | 3+4 4 | 5 2 |
| | 5 2 | 5+5 4 | 5+5 4 | 3+4 4 | 3+4 4 | 3+4 4 | 3+4 4 | 5 2 |
| | 6 2 | 6+6 4 | 5+5 4 | 3+4 4 | 3+4 4 | 3+4 4 | 3+4 4 | 6 2 |
| | 5 2 | | | | | | | 6 2 |
| | 6 2 | | | | | | | 6 2 |
| | 6 2 | | | | | | | 5 2 |
| | 6 2 | | | | | | | 6 2 |
| | 5 2 | | | | | | | 6 2 |

Continued on next column

**TABLE XXXI– continued**

| TQC | ICSD | NoE | SF | | | | | |
|---|---|---|---|---|---|---|---|---|
| | 5 2 | | | | | | | 6 2 |
| | 5 2 | | | | | | | 5 2 |
| | 6 2 | | | | | | | 5 2 |
| | 5 2 | | | | | | | 6 2 |
| | 6 2 | | | | | | | 6 2 |
| | 6 2 | | | | | | | 5 2 |
| | 5 2 | | | | | | | 5 2 |
| | 5 2 | | | | | | | 6 2 |
| | 6 2 | | | | | | | 5 2 |
| | 5 2 | | | | | | | 6 2 |
| | 5 2 | | | | | | | 6 2 |
| | 6 2 | | | | | | | 6 2 |
| | 5 2 | | | | | | | 5 2 |
| | 6 2 | | | | | | | 6 2 |
| | 6 2 | | | | | | | 5 2 |
| | 6 2 | | | | | | | 5 2 |
| | 6 2 | | | | | | | 6 2 |
| | 6 2 | | | | | | | 6 2 |
| | 6 2 | | | | | | | 6 2 |
| | 5 2 | | | | | | | 5 2 |
| | 5 2 | | | | | | | 5 2 |
| | 5 2 | | | | | | | 5 2 |
| | 6 2 | | | | | | | 6 2 |
| | 6 2 | | | | | | | 6 2 |
| | 5 2 | | | | | | | 6 2 |
| | 6 2 | | | | | | | 5 2 |
| | 5 2 | | | | | | | 6 2 |
| 55.2.36 | 84869 | 128 | C5Y4 | | | | | |
| | -GM | -R | -S | -T | -U | -X | -Y | -Z |
| | 5 2 | 5+5 4 | 5+5 4 | 3+4 4 | 3+4 4 | 3+4 4 | 3+4 4 | 5 2 |
| | 6 2 | 6+6 4 | 6+6 4 | 3+4 4 | 3+4 4 | 3+4 4 | 3+4 4 | 6 2 |
| | 5 2 | 6+6 4 | 5+5 4 | 3+4 4 | 3+4 4 | 3+4 4 | 3+4 4 | 5 2 |
| | 6 2 | 5+5 4 | 6+6 4 | 3+4 4 | 3+4 4 | 3+4 4 | 3+4 4 | 6 2 |
| | 5 2 | 5+5 4 | 6+6 4 | 3+4 4 | 3+4 4 | 3+4 4 | 3+4 4 | 6 2 |
| | 5 2 | 6+6 4 | 5+5 4 | 3+4 4 | 3+4 4 | 3+4 4 | 3+4 4 | 6 2 |
| | 6 2 | 5+5 4 | 5+5 4 | 3+4 4 | 3+4 4 | 3+4 4 | 3+4 4 | 5 2 |
| | 6 2 | 6+6 4 | 6+6 4 | 3+4 4 | 3+4 4 | 3+4 4 | 3+4 4 | 5 2 |
| | 5 2 | 5+5 4 | 5+5 4 | 3+4 4 | 3+4 4 | 3+4 4 | 3+4 4 | 5 2 |
| | 6 2 | 6+6 4 | 6+6 4 | 3+4 4 | 3+4 4 | 3+4 4 | 3+4 4 | 6 2 |
| | 6 2 | 5+5 4 | 6+6 4 | 3+4 4 | 3+4 4 | 3+4 4 | 3+4 4 | 5 2 |
| | 5 2 | 6+6 4 | 5+5 4 | 3+4 4 | 3+4 4 | 3+4 4 | 3+4 4 | 6 2 |
| | 5 2 | 6+6 4 | 5+5 4 | 3+4 4 | 3+4 4 | 3+4 4 | 3+4 4 | 6 2 |
| | 5 2 | 5+5 4 | 6+6 4 | 3+4 4 | 3+4 4 | 3+4 4 | 3+4 4 | 6 2 |
| | 6 2 | 6+6 4 | 5+5 4 | 3+4 4 | 3+4 4 | 3+4 4 | 3+4 4 | 5 2 |
| | 6 2 | 5+5 4 | 6+6 4 | 3+4 4 | 3+4 4 | 3+4 4 | 3+4 4 | 5 2 |
| | 6 2 | 5+5 4 | 5+5 4 | 3+4 4 | 3+4 4 | 3+4 4 | 3+4 4 | 5 2 |
| | 5 2 | 6+6 4 | 6+6 4 | 3+4 4 | 3+4 4 | 3+4 4 | 3+4 4 | 6 2 |
| | 6 2 | 5+5 4 | 5+5 4 | 3+4 4 | 3+4 4 | 3+4 4 | 3+4 4 | 5 2 |
| | 5 2 | 5+5 4 | 6+6 4 | 3+4 4 | 3+4 4 | 3+4 4 | 3+4 4 | 6 2 |
| | 5 2 | 6+6 4 | 5+5 4 | 3+4 4 | 3+4 4 | 3+4 4 | 3+4 4 | 6 2 |
| | 6 2 | 6+6 4 | 6+6 4 | 3+4 4 | 3+4 4 | 3+4 4 | 3+4 4 | 5 2 |
| | 5 2 | 5+5 4 | 5+5 4 | 3+4 4 | 3+4 4 | 3+4 4 | 3+4 4 | 5 2 |
| | 6 2 | 6+6 4 | 6+6 4 | 3+4 4 | 3+4 4 | 3+4 4 | 3+4 4 | 6 2 |
| | 5 2 | 5+5 4 | 5+5 4 | 3+4 4 | 3+4 4 | 3+4 4 | 3+4 4 | 6 2 |
| | 5 2 | 5+5 4 | 6+6 4 | 3+4 4 | 3+4 4 | 3+4 4 | 3+4 4 | 6 2 |
| | 6 2 | 6+6 4 | 5+5 4 | 3+4 4 | 3+4 4 | 3+4 4 | 3+4 4 | 5 2 |
| | 5 2 | 6+6 4 | 6+6 4 | 3+4 4 | 3+4 4 | 3+4 4 | 3+4 4 | 6 2 |
| | 5 2 | 6+6 4 | 6+6 4 | 3+4 4 | 3+4 4 | 3+4 4 | 3+4 4 | 6 2 |
| | 6 2 | 6+6 4 | 6+6 4 | 3+4 4 | 3+4 4 | 3+4 4 | 3+4 4 | 5 2 |
| | 6 2 | 5+5 4 | 5+5 4 | 3+4 4 | 3+4 4 | 3+4 4 | 3+4 4 | 5 2 |
| | 6 2 | 6+6 4 | 6+6 4 | 3+4 4 | 3+4 4 | 3+4 4 | 3+4 4 | 5 2 |

Continued on next column

| TQC | ICSD | NoE | SF | | | | | |
|---|---|---|---|---|---|---|---|---|
| | 5 2 | | | | | | | 5 2 |
| | 6 2 | | | | | | | 6 2 |
| | 6 2 | | | | | | | 6 2 |
| | 5 2 | | | | | | | 5 2 |
| | 5 2 | | | | | | | 5 2 |
| | 5 2 | | | | | | | 5 2 |
| | 5 2 | | | | | | | 6 2 |
| | 6 2 | | | | | | | 5 2 |
| | 6 2 | | | | | | | 5 2 |
| | 5 2 | | | | | | | 6 2 |
| | 6 2 | | | | | | | 6 2 |
| | 5 2 | | | | | | | 5 2 |
| | 5 2 | | | | | | | 5 2 |
| | 6 2 | | | | | | | 6 2 |
| | 5 2 | | | | | | | 6 2 |
| | 6 2 | | | | | | | 5 2 |
| | 6 2 | | | | | | | 6 2 |
| | 5 2 | | | | | | | 5 2 |
| | 6 2 | | | | | | | 6 2 |
| | 5 2 | | | | | | | 5 2 |
| | 5 2 | | | | | | | 6 2 |
| | 6 2 | | | | | | | 5 2 |
| | 6 2 | | | | | | | 6 2 |
| | 6 2 | | | | | | | 6 2 |
| | 6 2 | | | | | | | 6 2 |
| | 6 2 | | | | | | | 6 2 |
| | 6 2 | | | | | | | 6 2 |
| | 6 2 | | | | | | | 5 2 |
| | 5 2 | | | | | | | 6 2 |
| | 6 2 | | | | | | | 5 2 |
| | 5 2 | | | | | | | 5 2 |
| | 5 2 | | | | | | | 6 2 |
| 55.2.37 | 23834 | 68 | C2B2Sc1 | | | | | |
| | -GM | -R | -S | -T | -U | -X | -Y | -Z |
| | 5 2 | 6+6 4 | 5+5 4 | 3+4 4 | 3+4 4 | 3+4 4 | 3+4 4 | 6 2 |
| | 5 2 | 5+5 4 | 6+6 4 | 3+4 4 | 3+4 4 | 3+4 4 | 3+4 4 | 6 2 |
| | 6 2 | 6+6 4 | 5+5 4 | 3+4 4 | 3+4 4 | 3+4 4 | 3+4 4 | 5 2 |
| | 5 2 | 5+5 4 | 6+6 4 | 3+4 4 | 3+4 4 | 3+4 4 | 3+4 4 | 6 2 |
| | 6 2 | 5+5 4 | 6+6 4 | 3+4 4 | 3+4 4 | 3+4 4 | 3+4 4 | 5 2 |
| | 5 2 | 6+6 4 | 5+5 4 | 3+4 4 | 3+4 4 | 3+4 4 | 3+4 4 | 6 2 |
| | 6 2 | 5+5 4 | 6+6 4 | 3+4 4 | 3+4 4 | 3+4 4 | 3+4 4 | 5 2 |
| | 6 2 | 5+5 4 | 5+5 4 | 3+4 4 | 3+4 4 | 3+4 4 | 3+4 4 | 5 2 |
| | 5 2 | 6+6 4 | 5+5 4 | 3+4 4 | 3+4 4 | 3+4 4 | 3+4 4 | 6 2 |
| | 6 2 | 6+6 4 | 6+6 4 | 3+4 4 | 3+4 4 | 3+4 4 | 3+4 4 | 5 2 |
| | 5 2 | 6+6 4 | 6+6 4 | 3+4 4 | 3+4 4 | 3+4 4 | 3+4 4 | 6 2 |
| | 6 2 | 5+5 4 | 5+5 4 | 3+4 4 | 3+4 4 | 3+4 4 | 3+4 4 | 5 2 |
| | 5 2 | 5+5 4 | 5+5 4 | 3+4 4 | 3+4 4 | 3+4 4 | 3+4 4 | 5 2 |
| | 6 2 | 6+6 4 | 6+6 4 | 3+4 4 | 3+4 4 | 3+4 4 | 3+4 4 | 6 2 |
| | 5 2 | 6+6 4 | 5+5 4 | 3+4 4 | 3+4 4 | 3+4 4 | 3+4 4 | 6 2 |
| | 6 2 | 5+5 4 | 6+6 4 | 3+4 4 | 3+4 4 | 3+4 4 | 3+4 4 | 5 2 |
| | 5 2 | 6+6 4 | 6+6 4 | 3+4 4 | 3+4 4 | 3+4 4 | 3+4 4 | 5 2 |
| | 6 2 | | | | | | | 5 2 |
| | 6 2 | | | | | | | 6 2 |
| | 5 2 | | | | | | | 6 2 |
| | 5 2 | | | | | | | 6 2 |
| | 6 2 | | | | | | | 6 2 |
| | 5 2 | | | | | | | 6 2 |
| | 6 2 | | | | | | | 5 2 |
| | 6 2 | | | | | | | 5 2 |
| | 5 2 | | | | | | | 5 2 |
| | 6 2 | | | | | | | 5 2 |

Continued on next column

**TABLE XXXI– continued**

| TQC | ICSD | NoE | SF | | | | | |
|---|---|---|---|---|---|---|---|---|
| | 5 2 | | | | | | | 6 2 |
| | 5 2 | | | | | | | 5 2 |
| | 5 2 | | | | | | | 6 2 |
| | 6 2 | | | | | | | 5 2 |
| | 5 2 | | | | | | | 6 2 |
| | 5 2 | | | | | | | 6 2 |
| | 5 2 | | | | | | | 6 2 |
| 55.2.38 | 423924 | 140 | C1B2Lu1 | | | | | |
| | -GM | -R | -S | -T | -U | -X | -Y | -Z |
| | 5 2 | 5+5 4 | 6+6 4 | 3+4 4 | 3+4 4 | 3+4 4 | 3+4 4 | 6 2 |
| | 6 2 | 6+6 4 | 5+5 4 | 3+4 4 | 3+4 4 | 3+4 4 | 3+4 4 | 5 2 |
| | 6 2 | 5+5 4 | 6+6 4 | 3+4 4 | 3+4 4 | 3+4 4 | 3+4 4 | 5 2 |
| | 5 2 | 6+6 4 | 5+5 4 | 3+4 4 | 3+4 4 | 3+4 4 | 3+4 4 | 6 2 |
| | 6 2 | 6+6 4 | 5+5 4 | 3+4 4 | 3+4 4 | 3+4 4 | 3+4 4 | 5 2 |
| | 5 2 | 5+5 4 | 6+6 4 | 3+4 4 | 3+4 4 | 3+4 4 | 3+4 4 | 6 2 |
| | 5 2 | 5+5 4 | 6+6 4 | 3+4 4 | 3+4 4 | 3+4 4 | 3+4 4 | 6 2 |
| | 6 2 | 6+6 4 | 5+5 4 | 3+4 4 | 3+4 4 | 3+4 4 | 3+4 4 | 5 2 |
| | 5 2 | 5+5 4 | 5+5 4 | 3+4 4 | 3+4 4 | 3+4 4 | 3+4 4 | 6 2 |
| | 6 2 | 6+6 4 | 6+6 4 | 3+4 4 | 3+4 4 | 3+4 4 | 3+4 4 | 6 2 |
| | 5 2 | 6+6 4 | 5+5 4 | 3+4 4 | 3+4 4 | 3+4 4 | 3+4 4 | 5 2 |
| | 6 2 | 5+5 4 | 6+6 4 | 3+4 4 | 3+4 4 | 3+4 4 | 3+4 4 | 5 2 |
| | 6 2 | 6+6 4 | 5+5 4 | 3+4 4 | 3+4 4 | 3+4 4 | 3+4 4 | 5 2 |
| | 5 2 | 5+5 4 | 6+6 4 | 3+4 4 | 3+4 4 | 3+4 4 | 3+4 4 | 6 2 |
| | 5 2 | 6+6 4 | 5+5 4 | 3+4 4 | 3+4 4 | 3+4 4 | 3+4 4 | 6 2 |
| | 6 2 | 5+5 4 | 6+6 4 | 3+4 4 | 3+4 4 | 3+4 4 | 3+4 4 | 5 2 |
| | 5 2 | 6+6 4 | 5+5 4 | 3+4 4 | 3+4 4 | 3+4 4 | 3+4 4 | 5 2 |
| | 5 2 | 5+5 4 | 6+6 4 | 3+4 4 | 3+4 4 | 3+4 4 | 3+4 4 | 5 2 |
| | 6 2 | 6+6 4 | 6+6 4 | 3+4 4 | 3+4 4 | 3+4 4 | 3+4 4 | 6 2 |
| | 6 2 | 6+6 4 | 5+5 4 | 3+4 4 | 3+4 4 | 3+4 4 | 3+4 4 | 6 2 |
| | 5 2 | 5+5 4 | 5+5 4 | 3+4 4 | 3+4 4 | 3+4 4 | 3+4 4 | 5 2 |
| | 5 2 | 5+5 4 | 6+6 4 | 3+4 4 | 3+4 4 | 3+4 4 | 3+4 4 | 5 2 |
| | 6 2 | 5+5 4 | 5+5 4 | 3+4 4 | 3+4 4 | 3+4 4 | 3+4 4 | 6 2 |
| | 6 2 | 6+6 4 | 6+6 4 | 3+4 4 | 3+4 4 | 3+4 4 | 3+4 4 | 6 2 |
| | 5 2 | 6+6 4 | 5+5 4 | 3+4 4 | 3+4 4 | 3+4 4 | 3+4 4 | 6 2 |
| | 5 2 | 5+5 4 | 6+6 4 | 3+4 4 | 3+4 4 | 3+4 4 | 3+4 4 | 5 2 |
| | 6 2 | 6+6 4 | 6+6 4 | 3+4 4 | 3+4 4 | 3+4 4 | 3+4 4 | 5 2 |
| | 5 2 | 5+5 4 | 5+5 4 | 3+4 4 | 3+4 4 | 3+4 4 | 3+4 4 | 6 2 |
| | 6 2 | 5+5 4 | 6+6 4 | 3+4 4 | 3+4 4 | 3+4 4 | 3+4 4 | 5 2 |
| | 5 2 | 6+6 4 | 5+5 4 | 3+4 4 | 3+4 4 | 3+4 4 | 3+4 4 | 6 2 |
| | 5 2 | 5+5 4 | 6+6 4 | 3+4 4 | 3+4 4 | 3+4 4 | 3+4 4 | 6 2 |
| | 6 2 | 6+6 4 | 5+5 4 | 3+4 4 | 3+4 4 | 3+4 4 | 3+4 4 | 5 2 |
| | 6 2 | 6+6 4 | 6+6 4 | 3+4 4 | 3+4 4 | 3+4 4 | 3+4 4 | 5 2 |
| | 5 2 | 6+6 4 | 5+5 4 | 3+4 4 | 3+4 4 | 3+4 4 | 3+4 4 | 6 2 |
| | 6 2 | 5+5 4 | 6+6 4 | 3+4 4 | 3+4 4 | 3+4 4 | 3+4 4 | 5 2 |
| | 6 2 | | | | | | | 6 2 |
| | 5 2 | | | | | | | 5 2 |
| | 5 2 | | | | | | | 6 2 |
| | 6 2 | | | | | | | 5 2 |
| | 5 2 | | | | | | | 5 2 |
| | 6 2 | | | | | | | 6 2 |
| | 6 2 | | | | | | | 5 2 |
| | 5 2 | | | | | | | 6 2 |
| | 6 2 | | | | | | | 6 2 |
| | 5 2 | | | | | | | 5 2 |
| | 6 2 | | | | | | | 5 2 |
| | 5 2 | | | | | | | 6 2 |
| | 6 2 | | | | | | | 5 2 |
| | 5 2 | | | | | | | 6 2 |
| | 5 2 | | | | | | | 5 2 |
| | 6 2 | | | | | | | 6 2 |
| | 6 2 | | | | | | | 6 2 |

Continued on next column

**TABLE XXXI– continued**

| TQC | ICSD | NoE | SF | | | | | |
|---|---|---|---|---|---|---|---|---|
| | 5 2 | | | | | | | 5 2 |
| | 5 2 | | | | | | | 5 2 |
| | 6 2 | | | | | | | 5 2 |
| | 6 2 | | | | | | | 6 2 |
| | 5 2 | | | | | | | 6 2 |
| | 6 2 | | | | | | | 5 2 |
| | 5 2 | | | | | | | 6 2 |
| | 6 2 | | | | | | | 6 2 |
| | 5 2 | | | | | | | 5 2 |
| | 6 2 | | | | | | | 5 2 |
| | 6 2 | | | | | | | 5 2 |
| | 5 2 | | | | | | | 6 2 |
| | 5 2 | | | | | | | 6 2 |
| | 6 2 | | | | | | | 6 2 |
| | 5 2 | | | | | | | 6 2 |
| | 5 2 | | | | | | | 5 2 |
| | 6 2 | | | | | | | 5 2 |
| | 6 2 | | | | | | | 5 2 |
| 55.2.39 | 412563 | 220 | Bi8Ca3Pd4 | | | | | |
| | -GM | -R | -S | -T | -U | -X | -Y | -Z |
| | 5 2 | 5+5 4 | 6+6 4 | 3+4 4 | 3+4 4 | 3+4 4 | 3+4 4 | 6 2 |
| | 5 2 | 5+5 4 | 5+5 4 | 3+4 4 | 3+4 4 | 3+4 4 | 3+4 4 | 6 2 |
| | 5 2 | 6+6 4 | 6+6 4 | 3+4 4 | 3+4 4 | 3+4 4 | 3+4 4 | 5 2 |
| | 5 2 | 6+6 4 | 5+5 4 | 3+4 4 | 3+4 4 | 3+4 4 | 3+4 4 | 5 2 |
| | 6 2 | 5+5 4 | 5+5 4 | 3+4 4 | 3+4 4 | 3+4 4 | 3+4 4 | 6 2 |
| | 6 2 | 6+6 4 | 6+6 4 | 3+4 4 | 3+4 4 | 3+4 4 | 3+4 4 | 6 2 |
| | 6 2 | 6+6 4 | 5+5 4 | 3+4 4 | 3+4 4 | 3+4 4 | 3+4 4 | 5 2 |
| | 6 2 | 6+6 4 | 5+5 4 | 3+4 4 | 3+4 4 | 3+4 4 | 3+4 4 | 5 2 |
| | 5 2 | 5+5 4 | 5+5 4 | 3+4 4 | 3+4 4 | 3+4 4 | 3+4 4 | 5 2 |
| | 5 2 | 6+6 4 | 6+6 4 | 3+4 4 | 3+4 4 | 3+4 4 | 3+4 4 | 5 2 |
| | 6 2 | 5+5 4 | 5+5 4 | 3+4 4 | 3+4 4 | 3+4 4 | 3+4 4 | 6 2 |
| | 6 2 | 6+6 4 | 6+6 4 | 3+4 4 | 3+4 4 | 3+4 4 | 3+4 4 | 6 2 |
| | 6 2 | 6+6 4 | 5+5 4 | 3+4 4 | 3+4 4 | 3+4 4 | 3+4 4 | 5 2 |
| | 6 2 | 6+6 4 | 6+6 4 | 3+4 4 | 3+4 4 | 3+4 4 | 3+4 4 | 5 2 |
| | 6 2 | 5+5 4 | 5+5 4 | 3+4 4 | 3+4 4 | 3+4 4 | 3+4 4 | 5 2 |
| | 6 2 | 5+5 4 | 6+6 4 | 3+4 4 | 3+4 4 | 3+4 4 | 3+4 4 | 5 2 |
| | 5 2 | 6+6 4 | 5+5 4 | 3+4 4 | 3+4 4 | 3+4 4 | 3+4 4 | 5 2 |
| | 5 2 | 5+5 4 | 6+6 4 | 3+4 4 | 3+4 4 | 3+4 4 | 3+4 4 | 5 2 |
| | 6 2 | 5+5 4 | 5+5 4 | 3+4 4 | 3+4 4 | 3+4 4 | 3+4 4 | 6 2 |
| | 6 2 | 6+6 4 | 6+6 4 | 3+4 4 | 3+4 4 | 3+4 4 | 3+4 4 | 6 2 |
| | 5 2 | 5+5 4 | 6+6 4 | 3+4 4 | 3+4 4 | 3+4 4 | 3+4 4 | 5 2 |
| | 5 2 | 5+5 4 | 5+5 4 | 3+4 4 | 3+4 4 | 3+4 4 | 3+4 4 | 5 2 |
| | 6 2 | 6+6 4 | 5+5 4 | 3+4 4 | 3+4 4 | 3+4 4 | 3+4 4 | 6 2 |
| | 6 2 | 6+6 4 | 6+6 4 | 3+4 4 | 3+4 4 | 3+4 4 | 3+4 4 | 6 2 |
| | 5 2 | 6+6 4 | 6+6 4 | 3+4 4 | 3+4 4 | 3+4 4 | 3+4 4 | 5 2 |
| | 5 2 | 5+5 4 | 5+5 4 | 3+4 4 | 3+4 4 | 3+4 4 | 3+4 4 | 6 2 |
| | 6 2 | 6+6 4 | 6+6 4 | 3+4 4 | 3+4 4 | 3+4 4 | 3+4 4 | 6 2 |
| | 6 2 | 5+5 4 | 5+5 4 | 3+4 4 | 3+4 4 | 3+4 4 | 3+4 4 | 6 2 |
| | 5 2 | 6+6 4 | 6+6 4 | 3+4 4 | 3+4 4 | 3+4 4 | 3+4 4 | 5 2 |
| | 5 2 | 6+6 4 | 6+6 4 | 3+4 4 | 3+4 4 | 3+4 4 | 3+4 4 | 6 2 |
| | 6 2 | 5+5 4 | 5+5 4 | 3+4 4 | 3+4 4 | 3+4 4 | 3+4 4 | 5 2 |
| | 6 2 | 5+5 4 | 5+5 4 | 3+4 4 | 3+4 4 | 3+4 4 | 3+4 4 | 6 2 |
| | 5 2 | 5+5 4 | 6+6 4 | 3+4 4 | 3+4 4 | 3+4 4 | 3+4 4 | 6 2 |
| | 6 2 | 6+6 4 | 5+5 4 | 3+4 4 | 3+4 4 | 3+4 4 | 3+4 4 | 5 2 |
| | 5 2 | 6+6 4 | 5+5 4 | 3+4 4 | 3+4 4 | 3+4 4 | 3+4 4 | 6 2 |
| | 5 2 | 5+5 4 | 6+6 4 | 3+4 4 | 3+4 4 | 3+4 4 | 3+4 4 | 5 2 |
| | 6 2 | 6+6 4 | 6+6 4 | 3+4 4 | 3+4 4 | 3+4 4 | 3+4 4 | 5 2 |
| | 5 2 | 5+5 4 | 5+5 4 | 3+4 4 | 3+4 4 | 3+4 4 | 3+4 4 | 5 2 |
| | 6 2 | 5+5 4 | 6+6 4 | 3+4 4 | 3+4 4 | 3+4 4 | 3+4 4 | 6 2 |
| | 6 2 | 6+6 4 | 5+5 4 | 3+4 4 | 3+4 4 | 3+4 4 | 3+4 4 | 5 2 |
| | 6 2 | 5+5 4 | 6+6 4 | 3+4 4 | 3+4 4 | 3+4 4 | 3+4 4 | 5 2 |

Continued on next column

**TABLE XXXI– continued**

| TQC | ICSD | NoE | SF | | | | | |
|---|---|---|---|---|---|---|---|---|
| | 5 2 | 5+5 4 | 5+5 4 | 3+4 4 | 3+4 4 | 3+4 4 | 3+4 4 | 6 2 |
| | 5 2 | 6+6 4 | 5+5 4 | 3+4 4 | 3+4 4 | 3+4 4 | 3+4 4 | 5 2 |
| | 6 2 | 5+5 4 | 6+6 4 | 3+4 4 | 3+4 4 | 3+4 4 | 3+4 4 | 5 2 |
| | 5 2 | 6+6 4 | 5+5 4 | 3+4 4 | 3+4 4 | 3+4 4 | 3+4 4 | 6 2 |
| | 6 2 | 6+6 4 | 6+6 4 | 3+4 4 | 3+4 4 | 3+4 4 | 3+4 4 | 6 2 |
| | 5 2 | 5+5 4 | 5+5 4 | 3+4 4 | 3+4 4 | 3+4 4 | 3+4 4 | 6 2 |
| | 5 2 | 6+6 4 | 6+6 4 | 3+4 4 | 3+4 4 | 3+4 4 | 3+4 4 | 6 2 |
| | 6 2 | 6+6 4 | 5+5 4 | 3+4 4 | 3+4 4 | 3+4 4 | 3+4 4 | 6 2 |
| | 5 2 | 5+5 4 | 6+6 4 | 3+4 4 | 3+4 4 | 3+4 4 | 3+4 4 | 5 2 |
| | 6 2 | 6+6 4 | 5+5 4 | 3+4 4 | 3+4 4 | 3+4 4 | 3+4 4 | 6 2 |
| | 5 2 | 5+5 4 | 6+6 4 | 3+4 4 | 3+4 4 | 3+4 4 | 3+4 4 | 5 2 |
| | 6 2 | 5+5 4 | 6+6 4 | 3+4 4 | 3+4 4 | 3+4 4 | 3+4 4 | 5 2 |
| | 6 2 | 6+6 4 | 5+5 4 | 3+4 4 | 3+4 4 | 3+4 4 | 3+4 4 | 6 2 |
| | 5 2 | 5+5 4 | 6+6 4 | 3+4 4 | 3+4 4 | 3+4 4 | 3+4 4 | 5 2 |
| | 6 2 | | | | | | | 5 2 |
| | 5 2 | | | | | | | 6 2 |
| | 6 2 | | | | | | | 6 2 |
| | 5 2 | | | | | | | 6 2 |
| | 6 2 | | | | | | | 5 2 |
| | 6 2 | | | | | | | 5 2 |
| | 5 2 | | | | | | | 6 2 |
| | 5 2 | | | | | | | 5 2 |
| | 6 2 | | | | | | | 5 2 |
| | 5 2 | | | | | | | 6 2 |
| | 6 2 | | | | | | | 5 2 |
| | 5 2 | | | | | | | 6 2 |
| | 5 2 | | | | | | | 6 2 |
| | 6 2 | | | | | | | 6 2 |
| | 5 2 | | | | | | | 5 2 |
| | 6 2 | | | | | | | 6 2 |
| | 6 2 | | | | | | | 5 2 |
| | 6 2 | | | | | | | 5 2 |
| | 5 2 | | | | | | | 6 2 |
| | 6 2 | | | | | | | 5 2 |
| | 6 2 | | | | | | | 5 2 |
| | 5 2 | | | | | | | 5 2 |
| | 6 2 | | | | | | | 6 2 |
| | 5 2 | | | | | | | 6 2 |
| | 6 2 | | | | | | | 5 2 |
| | 5 2 | | | | | | | 6 2 |
| | 6 2 | | | | | | | 5 2 |
| | 5 2 | | | | | | | 5 2 |
| | 5 2 | | | | | | | 5 2 |
| | 5 2 | | | | | | | 5 2 |
| | 5 2 | | | | | | | 5 2 |
| | 6 2 | | | | | | | 6 2 |
| | 5 2 | | | | | | | 5 2 |
| | 6 2 | | | | | | | 6 2 |
| | 5 2 | | | | | | | 6 2 |
| | 5 2 | | | | | | | 5 2 |
| | 6 2 | | | | | | | 6 2 |
| | 6 2 | | | | | | | 5 2 |
| | 5 2 | | | | | | | 6 2 |
| | 6 2 | | | | | | | 6 2 |
| | 6 2 | | | | | | | 6 2 |
| | 5 2 | | | | | | | 5 2 |
| | 6 2 | | | | | | | 6 2 |
| | 5 2 | | | | | | | 5 2 |
| | 6 2 | | | | | | | 6 2 |
| | 5 2 | | | | | | | 5 2 |
| | 5 2 | | | | | | | 5 2 |

Continued on next column

**TABLE XXXI– continued**

| TQC | ICSD | NoE | SF | | | | | |
|---|---|---|---|---|---|---|---|---|
| | 6 2 | | | | | | | 6 2 |
| | 5 2 | | | | | | | 6 2 |
| | 5 2 | | | | | | | 6 2 |
| | 6 2 | | | | | | | 5 2 |
| | 6 2 | | | | | | | 6 2 |
| | 6 2 | | | | | | | 5 2 |
| | 5 2 | | | | | | | 6 2 |
| | 6 2 | | | | | | | 5 2 |
| 55.2.40 | 415576 | 172 | Bi6In2Sr5 | | | | | |
| | -GM | -R | -S | -T | -U | -X | -Y | -Z |
| | 5 2 | 5+5 4 | 5+5 4 | 3+4 4 | 3+4 4 | 3+4 4 | 3+4 4 | 5 2 |
| | 5 2 | 6+6 4 | 6+6 4 | 3+4 4 | 3+4 4 | 3+4 4 | 3+4 4 | 5 2 |
| | 6 2 | 5+5 4 | 5+5 4 | 3+4 4 | 3+4 4 | 3+4 4 | 3+4 4 | 6 2 |
| | 6 2 | 6+6 4 | 6+6 4 | 3+4 4 | 3+4 4 | 3+4 4 | 3+4 4 | 6 2 |
| | 5 2 | 6+6 4 | 6+6 4 | 3+4 4 | 3+4 4 | 3+4 4 | 3+4 4 | 5 2 |
| | 6 2 | 5+5 4 | 5+5 4 | 3+4 4 | 3+4 4 | 3+4 4 | 3+4 4 | 6 2 |
| | 5 2 | 6+6 4 | 6+6 4 | 3+4 4 | 3+4 4 | 3+4 4 | 3+4 4 | 5 2 |
| | 6 2 | 5+5 4 | 5+5 4 | 3+4 4 | 3+4 4 | 3+4 4 | 3+4 4 | 6 2 |
| | 5 2 | 6+6 4 | 6+6 4 | 3+4 4 | 3+4 4 | 3+4 4 | 3+4 4 | 5 2 |
| | 5 2 | 5+5 4 | 5+5 4 | 3+4 4 | 3+4 4 | 3+4 4 | 3+4 4 | 5 2 |
| | 5 2 | 6+6 4 | 5+5 4 | 3+4 4 | 3+4 4 | 3+4 4 | 3+4 4 | 5 2 |
| | 5 2 | 5+5 4 | 6+6 4 | 3+4 4 | 3+4 4 | 3+4 4 | 3+4 4 | 5 2 |
| | 6 2 | 5+5 4 | 5+5 4 | 3+4 4 | 3+4 4 | 3+4 4 | 3+4 4 | 6 2 |
| | 6 2 | 6+6 4 | 6+6 4 | 3+4 4 | 3+4 4 | 3+4 4 | 3+4 4 | 6 2 |
| | 5 2 | 5+5 4 | 5+5 4 | 3+4 4 | 3+4 4 | 3+4 4 | 3+4 4 | 5 2 |
| | 6 2 | 6+6 4 | 6+6 4 | 3+4 4 | 3+4 4 | 3+4 4 | 3+4 4 | 6 2 |
| | 5 2 | 5+5 4 | 5+5 4 | 3+4 4 | 3+4 4 | 3+4 4 | 3+4 4 | 5 2 |
| | 6 2 | 5+5 4 | 6+6 4 | 3+4 4 | 3+4 4 | 3+4 4 | 3+4 4 | 6 2 |
| | 6 2 | 6+6 4 | 5+5 4 | 3+4 4 | 3+4 4 | 3+4 4 | 3+4 4 | 6 2 |
| | 6 2 | 5+5 4 | 5+5 4 | 3+4 4 | 3+4 4 | 3+4 4 | 3+4 4 | 6 2 |
| | 5 2 | 6+6 4 | 5+5 4 | 3+4 4 | 3+4 4 | 3+4 4 | 3+4 4 | 6 2 |
| | 5 2 | 6+6 4 | 6+6 4 | 3+4 4 | 3+4 4 | 3+4 4 | 3+4 4 | 6 2 |
| | 6 2 | 5+5 4 | 5+5 4 | 3+4 4 | 3+4 4 | 3+4 4 | 3+4 4 | 5 2 |
| | 6 2 | 5+5 4 | 6+6 4 | 3+4 4 | 3+4 4 | 3+4 4 | 3+4 4 | 5 2 |
| | 5 2 | 6+6 4 | 5+5 4 | 3+4 4 | 3+4 4 | 3+4 4 | 3+4 4 | 5 2 |
| | 5 2 | 5+5 4 | 6+6 4 | 3+4 4 | 3+4 4 | 3+4 4 | 3+4 4 | 5 2 |
| | 6 2 | 6+6 4 | 6+6 4 | 3+4 4 | 3+4 4 | 3+4 4 | 3+4 4 | 6 2 |
| | 6 2 | 5+5 4 | 5+5 4 | 3+4 4 | 3+4 4 | 3+4 4 | 3+4 4 | 6 2 |
| | 5 2 | 5+5 4 | 5+5 4 | 3+4 4 | 3+4 4 | 3+4 4 | 3+4 4 | 5 2 |
| | 6 2 | 6+6 4 | 6+6 4 | 3+4 4 | 3+4 4 | 3+4 4 | 3+4 4 | 6 2 |
| | 5 2 | 6+6 4 | 6+6 4 | 3+4 4 | 3+4 4 | 3+4 4 | 3+4 4 | 5 2 |
| | 6 2 | 5+5 4 | 5+5 4 | 3+4 4 | 3+4 4 | 3+4 4 | 3+4 4 | 6 2 |
| | 6 2 | 6+6 4 | 5+5 4 | 3+4 4 | 3+4 4 | 3+4 4 | 3+4 4 | 6 2 |
| | 6 2 | 5+5 4 | 6+6 4 | 3+4 4 | 3+4 4 | 3+4 4 | 3+4 4 | 6 2 |
| | 5 2 | 6+6 4 | 6+6 4 | 3+4 4 | 3+4 4 | 3+4 4 | 3+4 4 | 5 2 |
| | 6 2 | 5+5 4 | 5+5 4 | 3+4 4 | 3+4 4 | 3+4 4 | 3+4 4 | 6 2 |
| | 6 2 | 6+6 4 | 6+6 4 | 3+4 4 | 3+4 4 | 3+4 4 | 3+4 4 | 6 2 |
| | 6 2 | 5+5 4 | 5+5 4 | 3+4 4 | 3+4 4 | 3+4 4 | 3+4 4 | 6 2 |
| | 5 2 | 6+6 4 | 6+6 4 | 3+4 4 | 3+4 4 | 3+4 4 | 3+4 4 | 5 2 |
| | 6 2 | 5+5 4 | 5+5 4 | 3+4 4 | 3+4 4 | 3+4 4 | 3+4 4 | 6 2 |
| | 5 2 | 6+6 4 | 6+6 4 | 3+4 4 | 3+4 4 | 3+4 4 | 3+4 4 | 6 2 |
| | 5 2 | 5+5 4 | 5+5 4 | 3+4 4 | 3+4 4 | 3+4 4 | 3+4 4 | 6 2 |
| | 6 2 | 6+6 4 | 6+6 4 | 3+4 4 | 3+4 4 | 3+4 4 | 3+4 4 | 5 2 |
| | 5 2 | | | | | | | 6 2 |
| | 6 2 | | | | | | | 5 2 |
| | 5 2 | | | | | | | 6 2 |
| | 5 2 | | | | | | | 5 2 |
| | 6 2 | | | | | | | 5 2 |
| | 6 2 | | | | | | | 6 2 |
| | 5 2 | | | | | | | 6 2 |
| | 6 2 | | | | | | | 5 2 |

Continued on next column

**TABLE XXXI– continued**

| TQC | ICSD | NoE | SF | | | | | |
|---|---|---|---|---|---|---|---|---|
| | 6 2 | | | | | | | 5 2 |
| | 5 2 | | | | | | | 5 2 |
| | 6 2 | | | | | | | 6 2 |
| | 5 2 | | | | | | | 5 2 |
| | 6 2 | | | | | | | 6 2 |
| | 5 2 | | | | | | | 5 2 |
| | 5 2 | | | | | | | 5 2 |
| | 6 2 | | | | | | | 6 2 |
| | 6 2 | | | | | | | 6 2 |
| | 5 2 | | | | | | | 5 2 |
| | 5 2 | | | | | | | 6 2 |
| | 6 2 | | | | | | | 6 2 |
| | 6 2 | | | | | | | 5 2 |
| | 5 2 | | | | | | | 6 2 |
| | 6 2 | | | | | | | 5 2 |
| | 5 2 | | | | | | | 5 2 |
| | 5 2 | | | | | | | 6 2 |
| | 6 2 | | | | | | | 6 2 |
| | 5 2 | | | | | | | 5 2 |
| | 6 2 | | | | | | | 5 2 |
| | 6 2 | | | | | | | 6 2 |
| | 5 2 | | | | | | | 5 2 |
| | 5 2 | | | | | | | 5 2 |
| | 6 2 | | | | | | | 6 2 |
| | 6 2 | | | | | | | 6 2 |
| | 5 2 | | | | | | | 5 2 |
| | 6 2 | | | | | | | 6 2 |
| | 5 2 | | | | | | | 6 2 |
| | 5 2 | | | | | | | 5 2 |
| | 5 2 | | | | | | | 6 2 |
| | 6 2 | | | | | | | 5 2 |
| | 5 2 | | | | | | | 6 2 |
| | 6 2 | | | | | | | 5 2 |
| | 5 2 | | | | | | | 6 2 |
| | 5 2 | | | | | | | 6 2 |
| 55.2.41 | 262861 | 192 | Bi2Fe4O9 | | | | | |
| | -GM | -R | -S | -T | -U | -X | -Y | -Z |
| | 5 2 | 6+6 4 | 6+6 4 | 3+4 4 | 3+4 4 | 3+4 4 | 3+4 4 | 5 2 |
| | 5 2 | 5+5 4 | 5+5 4 | 3+4 4 | 3+4 4 | 3+4 4 | 3+4 4 | 6 2 |
| | 6 2 | 5+5 4 | 5+5 4 | 3+4 4 | 3+4 4 | 3+4 4 | 3+4 4 | 6 2 |
| | 6 2 | 6+6 4 | 5+5 4 | 3+4 4 | 3+4 4 | 3+4 4 | 3+4 4 | 6 2 |
| | 5 2 | 6+6 4 | 6+6 4 | 3+4 4 | 3+4 4 | 3+4 4 | 3+4 4 | 5 2 |
| | 5 2 | 5+5 4 | 6+6 4 | 3+4 4 | 3+4 4 | 3+4 4 | 3+4 4 | 6 2 |
| | 5 2 | 6+6 4 | 5+5 4 | 3+4 4 | 3+4 4 | 3+4 4 | 3+4 4 | 6 2 |
| | 6 2 | 6+6 4 | 6+6 4 | 3+4 4 | 3+4 4 | 3+4 4 | 3+4 4 | 5 2 |
| | 6 2 | 5+5 4 | 5+5 4 | 3+4 4 | 3+4 4 | 3+4 4 | 3+4 4 | 5 2 |
| | 5 2 | 5+5 4 | 5+5 4 | 3+4 4 | 3+4 4 | 3+4 4 | 3+4 4 | 6 2 |
| | 5 2 | 6+6 4 | 6+6 4 | 3+4 4 | 3+4 4 | 3+4 4 | 3+4 4 | 5 2 |
| | 5 2 | 5+5 4 | 6+6 4 | 3+4 4 | 3+4 4 | 3+4 4 | 3+4 4 | 6 2 |
| | 5 2 | 5+5 4 | 5+5 4 | 3+4 4 | 3+4 4 | 3+4 4 | 3+4 4 | 5 2 |
| | 6 2 | 5+5 4 | 6+6 4 | 3+4 4 | 3+4 4 | 3+4 4 | 3+4 4 | 6 2 |
| | 6 2 | 6+6 4 | 5+5 4 | 3+4 4 | 3+4 4 | 3+4 4 | 3+4 4 | 5 2 |
| | 6 2 | 6+6 4 | 6+6 4 | 3+4 4 | 3+4 4 | 3+4 4 | 3+4 4 | 6 2 |
| | 6 2 | 5+5 4 | 6+6 4 | 3+4 4 | 3+4 4 | 3+4 4 | 3+4 4 | 6 2 |
| | 5 2 | 5+5 4 | 5+5 4 | 3+4 4 | 3+4 4 | 3+4 4 | 3+4 4 | 5 2 |
| | 5 2 | 6+6 4 | 6+6 4 | 3+4 4 | 3+4 4 | 3+4 4 | 3+4 4 | 5 2 |
| | 6 2 | 6+6 4 | 5+5 4 | 3+4 4 | 3+4 4 | 3+4 4 | 3+4 4 | 6 2 |
| | 5 2 | 5+5 4 | 5+5 4 | 3+4 4 | 3+4 4 | 3+4 4 | 3+4 4 | 5 2 |
| | 6 2 | 6+6 4 | 6+6 4 | 3+4 4 | 3+4 4 | 3+4 4 | 3+4 4 | 6 2 |
| | 6 2 | 6+6 4 | 6+6 4 | 3+4 4 | 3+4 4 | 3+4 4 | 3+4 4 | 6 2 |
| | 5 2 | 5+5 4 | 5+5 4 | 3+4 4 | 3+4 4 | 3+4 4 | 3+4 4 | 5 2 |
| | 6 2 | 5+5 4 | 6+6 4 | 3+4 4 | 3+4 4 | 3+4 4 | 3+4 4 | 6 2 |
| | 5 2 | 6+6 4 | 6+6 4 | 3+4 4 | 3+4 4 | 3+4 4 | 3+4 4 | 5 2 |
| | 6 2 | 6+6 4 | 6+6 4 | 3+4 4 | 3+4 4 | 3+4 4 | 3+4 4 | 6 2 |
| | 5 2 | 5+5 4 | 5+5 4 | 3+4 4 | 3+4 4 | 3+4 4 | 3+4 4 | 5 2 |
| | 6 2 | 5+5 4 | 5+5 4 | 3+4 4 | 3+4 4 | 3+4 4 | 3+4 4 | 5 2 |
| | 5 2 | 6+6 4 | 5+5 4 | 3+4 4 | 3+4 4 | 3+4 4 | 3+4 4 | 6 2 |
| | 6 2 | 5+5 4 | 6+6 4 | 3+4 4 | 3+4 4 | 3+4 4 | 3+4 4 | 5 2 |
| | 6 2 | 6+6 4 | 5+5 4 | 3+4 4 | 3+4 4 | 3+4 4 | 3+4 4 | 6 2 |
| | 6 2 | 6+6 4 | 6+6 4 | 3+4 4 | 3+4 4 | 3+4 4 | 3+4 4 | 5 2 |
| | 5 2 | 5+5 4 | 6+6 4 | 3+4 4 | 3+4 4 | 3+4 4 | 3+4 4 | 5 2 |
| | 6 2 | 5+5 4 | 5+5 4 | 3+4 4 | 3+4 4 | 3+4 4 | 3+4 4 | 5 2 |
| | 5 2 | 6+6 4 | 5+5 4 | 3+4 4 | 3+4 4 | 3+4 4 | 3+4 4 | 6 2 |
| | 6 2 | 6+6 4 | 5+5 4 | 3+4 4 | 3+4 4 | 3+4 4 | 3+4 4 | 6 2 |
| | 6 2 | 5+5 4 | 6+6 4 | 3+4 4 | 3+4 4 | 3+4 4 | 3+4 4 | 5 2 |
| | 6 2 | 5+5 4 | 6+6 4 | 3+4 4 | 3+4 4 | 3+4 4 | 3+4 4 | 5 2 |
| | 6 2 | 5+5 4 | 6+6 4 | 3+4 4 | 3+4 4 | 3+4 4 | 3+4 4 | 6 2 |
| | 5 2 | 5+5 4 | 6+6 4 | 3+4 4 | 3+4 4 | 3+4 4 | 3+4 4 | 5 2 |
| | 5 2 | 6+6 4 | 5+5 4 | 3+4 4 | 3+4 4 | 3+4 4 | 3+4 4 | 5 2 |
| | 6 2 | 6+6 4 | 5+5 4 | 3+4 4 | 3+4 4 | 3+4 4 | 3+4 4 | 5 2 |
| | 5 2 | 5+5 4 | 5+5 4 | 3+4 4 | 3+4 4 | 3+4 4 | 3+4 4 | 6 2 |
| | 5 2 | 6+6 4 | 6+6 4 | 3+4 4 | 3+4 4 | 3+4 4 | 3+4 4 | 5 2 |
| | 5 2 | | | | | | | 6 2 |
| | 5 2 | | | | | | | 6 2 |
| | 6 2 | | | | | | | 5 2 |
| | 6 2 | | | | | | | 5 2 |
| | 5 2 | | | | | | | 5 2 |
| | 6 2 | | | | | | | 6 2 |
| | 6 2 | | | | | | | 6 2 |
| | 5 2 | | | | | | | 6 2 |
| | 6 2 | | | | | | | 5 2 |
| | 5 2 | | | | | | | 6 2 |
| | 6 2 | | | | | | | 5 2 |
| | 6 2 | | | | | | | 5 2 |
| | 5 2 | | | | | | | 6 2 |
| | 5 2 | | | | | | | 5 2 |
| | 5 2 | | | | | | | 6 2 |
| | 6 2 | | | | | | | 5 2 |
| | 6 2 | | | | | | | 6 2 |
| | 5 2 | | | | | | | 5 2 |
| | 6 2 | | | | | | | 6 2 |
| | 5 2 | | | | | | | 6 2 |
| | 6 2 | | | | | | | 5 2 |
| | 6 2 | | | | | | | 6 2 |
| | 6 2 | | | | | | | 5 2 |
| | 5 2 | | | | | | | 5 2 |
| | 6 2 | | | | | | | 5 2 |
| | 6 2 | | | | | | | 6 2 |
| | 6 2 | | | | | | | 5 2 |
| | 5 2 | | | | | | | 6 2 |
| | 5 2 | | | | | | | 6 2 |
| | 5 2 | | | | | | | 6 2 |
| | 5 2 | | | | | | | 6 2 |
| | 5 2 | | | | | | | 6 2 |
| | 5 2 | | | | | | | 6 2 |
| | 6 2 | | | | | | | 5 2 |
| | 5 2 | | | | | | | 6 2 |
| | 5 2 | | | | | | | 5 2 |
| | 6 2 | | | | | | | 5 2 |

Continued on next column

Continued on next column

**TABLE XXXI– continued**

| TQC | ICSD | NoE | SF | | | | | |
|---|---|---|---|---|---|---|---|---|
| | 5 2 | | | | | | | 6 2 |
| | 6 2 | | | | | | | 5 2 |
| | 5 2 | | | | | | | 6 2 |
| | 6 2 | | | | | | | 6 2 |
| | 5 2 | | | | | | | 5 2 |
| | 6 2 | | | | | | | 6 2 |
| | 6 2 | | | | | | | 5 2 |
| | 6 2 | | | | | | | 5 2 |
| | 5 2 | | | | | | | 6 2 |
| | 6 2 | | | | | | | 5 2 |
| | 6 2 | | | | | | | 5 2 |
| 55.2.42 | 186442 | 192 | Bi2Fe4O9 | | | | | |
| | -GM | -R | -S | -T | -U | -X | -Y | -Z |
| | 5 2 | 5+5 4 | 5+5 4 | 3+4 4 | 3+4 4 | 3+4 4 | 3+4 4 | 5 2 |
| | 6 2 | 6+6 4 | 6+6 4 | 3+4 4 | 3+4 4 | 3+4 4 | 3+4 4 | 6 2 |
| | 6 2 | 6+6 4 | 5+5 4 | 3+4 4 | 3+4 4 | 3+4 4 | 3+4 4 | 6 2 |
| | 5 2 | 5+5 4 | 6+6 4 | 3+4 4 | 3+4 4 | 3+4 4 | 3+4 4 | 6 2 |
| | 5 2 | 5+5 4 | 6+6 4 | 3+4 4 | 3+4 4 | 3+4 4 | 3+4 4 | 5 2 |
| | 5 2 | 6+6 4 | 5+5 4 | 3+4 4 | 3+4 4 | 3+4 4 | 3+4 4 | 6 2 |
| | 5 2 | 5+5 4 | 6+6 4 | 3+4 4 | 3+4 4 | 3+4 4 | 3+4 4 | 5 2 |
| | 6 2 | 5+5 4 | 5+5 4 | 3+4 4 | 3+4 4 | 3+4 4 | 3+4 4 | 6 2 |
| | 5 2 | 6+6 4 | 6+6 4 | 3+4 4 | 3+4 4 | 3+4 4 | 3+4 4 | 5 2 |
| | 5 2 | 6+6 4 | 6+6 4 | 3+4 4 | 3+4 4 | 3+4 4 | 3+4 4 | 6 2 |
| | 6 2 | 5+5 4 | 5+5 4 | 3+4 4 | 3+4 4 | 3+4 4 | 3+4 4 | 5 2 |
| | 5 2 | 6+6 4 | 5+5 4 | 3+4 4 | 3+4 4 | 3+4 4 | 3+4 4 | 6 2 |
| | 6 2 | 6+6 4 | 6+6 4 | 3+4 4 | 3+4 4 | 3+4 4 | 3+4 4 | 5 2 |
| | 6 2 | 6+6 4 | 6+6 4 | 3+4 4 | 3+4 4 | 3+4 4 | 3+4 4 | 6 2 |
| | 5 2 | 5+5 4 | 5+5 4 | 3+4 4 | 3+4 4 | 3+4 4 | 3+4 4 | 5 2 |
| | 6 2 | 5+5 4 | 5+5 4 | 3+4 4 | 3+4 4 | 3+4 4 | 3+4 4 | 6 2 |
| | 6 2 | 6+6 4 | 6+6 4 | 3+4 4 | 3+4 4 | 3+4 4 | 3+4 4 | 6 2 |
| | 5 2 | 5+5 4 | 5+5 4 | 3+4 4 | 3+4 4 | 3+4 4 | 3+4 4 | 5 2 |
| | 5 2 | 6+6 4 | 5+5 4 | 3+4 4 | 3+4 4 | 3+4 4 | 3+4 4 | 5 2 |
| | 6 2 | 5+5 4 | 6+6 4 | 3+4 4 | 3+4 4 | 3+4 4 | 3+4 4 | 6 2 |
| | 6 2 | 6+6 4 | 6+6 4 | 3+4 4 | 3+4 4 | 3+4 4 | 3+4 4 | 6 2 |
| | 5 2 | 5+5 4 | 5+5 4 | 3+4 4 | 3+4 4 | 3+4 4 | 3+4 4 | 5 2 |
| | 6 2 | 5+5 4 | 5+5 4 | 3+4 4 | 3+4 4 | 3+4 4 | 3+4 4 | 6 2 |
| | 5 2 | 6+6 4 | 6+6 4 | 3+4 4 | 3+4 4 | 3+4 4 | 3+4 4 | 5 2 |
| | 6 2 | 6+6 4 | 5+5 4 | 3+4 4 | 3+4 4 | 3+4 4 | 3+4 4 | 6 2 |
| | 5 2 | 5+5 4 | 5+5 4 | 3+4 4 | 3+4 4 | 3+4 4 | 3+4 4 | 6 2 |
| | 5 2 | 5+5 4 | 5+5 4 | 3+4 4 | 3+4 4 | 3+4 4 | 3+4 4 | 5 2 |
| | 6 2 | 6+6 4 | 6+6 4 | 3+4 4 | 3+4 4 | 3+4 4 | 3+4 4 | 5 2 |
| | 5 2 | 6+6 4 | 6+6 4 | 3+4 4 | 3+4 4 | 3+4 4 | 3+4 4 | 5 2 |
| | 6 2 | 6+6 4 | 6+6 4 | 3+4 4 | 3+4 4 | 3+4 4 | 3+4 4 | 5 2 |
| | 5 2 | 5+5 4 | 5+5 4 | 3+4 4 | 3+4 4 | 3+4 4 | 3+4 4 | 6 2 |
| | 6 2 | 6+6 4 | 5+5 4 | 3+4 4 | 3+4 4 | 3+4 4 | 3+4 4 | 6 2 |
| | 6 2 | 5+5 4 | 5+5 4 | 3+4 4 | 3+4 4 | 3+4 4 | 3+4 4 | 5 2 |
| | 5 2 | 6+6 4 | 6+6 4 | 3+4 4 | 3+4 4 | 3+4 4 | 3+4 4 | 5 2 |
| | 6 2 | 5+5 4 | 6+6 4 | 3+4 4 | 3+4 4 | 3+4 4 | 3+4 4 | 5 2 |
| | 6 2 | 6+6 4 | 5+5 4 | 3+4 4 | 3+4 4 | 3+4 4 | 3+4 4 | 5 2 |
| | 5 2 | 5+5 4 | 5+5 4 | 3+4 4 | 3+4 4 | 3+4 4 | 3+4 4 | 6 2 |
| | 6 2 | 6+6 4 | 6+6 4 | 3+4 4 | 3+4 4 | 3+4 4 | 3+4 4 | 6 2 |
| | 5 2 | 5+5 4 | 6+6 4 | 3+4 4 | 3+4 4 | 3+4 4 | 3+4 4 | 5 2 |
| | 6 2 | 5+5 4 | 6+6 4 | 3+4 4 | 3+4 4 | 3+4 4 | 3+4 4 | 6 2 |
| | 6 2 | 6+6 4 | 5+5 4 | 3+4 4 | 3+4 4 | 3+4 4 | 3+4 4 | 6 2 |
| | 6 2 | 6+6 4 | 5+5 4 | 3+4 4 | 3+4 4 | 3+4 4 | 3+4 4 | 5 2 |
| | 5 2 | 6+6 4 | 5+5 4 | 3+4 4 | 3+4 4 | 3+4 4 | 3+4 4 | 5 2 |
| | 6 2 | 5+5 4 | 6+6 4 | 3+4 4 | 3+4 4 | 3+4 4 | 3+4 4 | 5 2 |
| | 5 2 | 5+5 4 | 5+5 4 | 3+4 4 | 3+4 4 | 3+4 4 | 3+4 4 | 5 2 |
| | 6 2 | 6+6 4 | 6+6 4 | 3+4 4 | 3+4 4 | 3+4 4 | 3+4 4 | 6 2 |
| | 5 2 | 5+5 4 | 6+6 4 | 3+4 4 | 3+4 4 | 3+4 4 | 3+4 4 | 5 2 |
| | 5 2 | 6+6 4 | 5+5 4 | 3+4 4 | 3+4 4 | 3+4 4 | 3+4 4 | 5 2 |

Continued on next column

**TABLE XXXI– continued**

| TQC | ICSD | NoE | SF | | | | | |
|---|---|---|---|---|---|---|---|---|
| | 6 2 | | | | | | | 6 2 |
| | 5 2 | | | | | | | 6 2 |
| | 5 2 | | | | | | | 5 2 |
| | 6 2 | | | | | | | 5 2 |
| | 6 2 | | | | | | | 5 2 |
| | 6 2 | | | | | | | 6 2 |
| | 5 2 | | | | | | | 6 2 |
| | 5 2 | | | | | | | 6 2 |
| | 5 2 | | | | | | | 5 2 |
| | 6 2 | | | | | | | 6 2 |
| | 6 2 | | | | | | | 5 2 |
| | 6 2 | | | | | | | 6 2 |
| | 5 2 | | | | | | | 5 2 |
| | 5 2 | | | | | | | 5 2 |
| | 5 2 | | | | | | | 5 2 |
| | 5 2 | | | | | | | 6 2 |
| | 6 2 | | | | | | | 5 2 |
| | 6 2 | | | | | | | 6 2 |
| | 6 2 | | | | | | | 6 2 |
| | 5 2 | | | | | | | 5 2 |
| | 6 2 | | | | | | | 6 2 |
| | 6 2 | | | | | | | 6 2 |
| | 6 2 | | | | | | | 5 2 |
| | 5 2 | | | | | | | 5 2 |
| | 6 2 | | | | | | | 5 2 |
| | 5 2 | | | | | | | 5 2 |
| | 6 2 | | | | | | | 6 2 |
| | 6 2 | | | | | | | 6 2 |
| | 5 2 | | | | | | | 6 2 |
| | 5 2 | | | | | | | 6 2 |
| | 5 2 | | | | | | | 6 2 |
| | 5 2 | | | | | | | 6 2 |
| | 5 2 | | | | | | | 5 2 |
| | 6 2 | | | | | | | 6 2 |
| | 5 2 | | | | | | | 6 2 |
| | 5 2 | | | | | | | 5 2 |
| | 6 2 | | | | | | | 5 2 |
| | 5 2 | | | | | | | 6 2 |
| | 5 2 | | | | | | | 5 2 |
| | 6 2 | | | | | | | 6 2 |
| | 6 2 | | | | | | | 6 2 |
| | 6 2 | | | | | | | 5 2 |
| | 5 2 | | | | | | | 6 2 |
| | 6 2 | | | | | | | 5 2 |
| | 6 2 | | | | | | | 6 2 |
| | 6 2 | | | | | | | 5 2 |
| | 5 2 | | | | | | | 6 2 |
| | 6 2 | | | | | | | 5 2 |
| 55.2.43 | 186441 | 192 | Bi2Fe4O9 | | | | | |
| | -GM | -R | -S | -T | -U | -X | -Y | -Z |
| | 5 2 | 5+5 4 | 5+5 4 | 3+4 4 | 3+4 4 | 3+4 4 | 3+4 4 | 5 2 |
| | 6 2 | 6+6 4 | 6+6 4 | 3+4 4 | 3+4 4 | 3+4 4 | 3+4 4 | 6 2 |
| | 6 2 | 6+6 4 | 5+5 4 | 3+4 4 | 3+4 4 | 3+4 4 | 3+4 4 | 6 2 |
| | 5 2 | 5+5 4 | 6+6 4 | 3+4 4 | 3+4 4 | 3+4 4 | 3+4 4 | 6 2 |
| | 5 2 | 5+5 4 | 6+6 4 | 3+4 4 | 3+4 4 | 3+4 4 | 3+4 4 | 5 2 |
| | 5 2 | 6+6 4 | 5+5 4 | 3+4 4 | 3+4 4 | 3+4 4 | 3+4 4 | 6 2 |
| | 5 2 | 5+5 4 | 6+6 4 | 3+4 4 | 3+4 4 | 3+4 4 | 3+4 4 | 5 2 |
| | 6 2 | 5+5 4 | 5+5 4 | 3+4 4 | 3+4 4 | 3+4 4 | 3+4 4 | 6 2 |
| | 5 2 | 6+6 4 | 6+6 4 | 3+4 4 | 3+4 4 | 3+4 4 | 3+4 4 | 5 2 |
| | 6 2 | 6+6 4 | 6+6 4 | 3+4 4 | 3+4 4 | 3+4 4 | 3+4 4 | 6 2 |
| | 5 2 | 5+5 4 | 5+5 4 | 3+4 4 | 3+4 4 | 3+4 4 | 3+4 4 | 5 2 |

Continued on next column

**TABLE XXXI– continued**

| TQC | ICSD NoE SF |
|---|---|
| | 5 2 6+6 4 5+5 4 3+4 4 3+4 4 3+4 4 3+4 4 6 2 |
| | 6 2 6+6 4 6+6 4 3+4 4 3+4 4 3+4 4 3+4 4 5 2 |
| | 6 2 6+6 4 6+6 4 3+4 4 3+4 4 3+4 4 3+4 4 6 2 |
| | 5 2 5+5 4 5+5 4 3+4 4 3+4 4 3+4 4 3+4 4 5 2 |
| | 6 2 5+5 4 5+5 4 3+4 4 3+4 4 3+4 4 3+4 4 6 2 |
| | 6 2 6+6 4 5+5 4 3+4 4 3+4 4 3+4 4 3+4 4 6 2 |
| | 5 2 5+5 4 6+6 4 3+4 4 3+4 4 3+4 4 3+4 4 5 2 |
| | 5 2 6+6 4 5+5 4 3+4 4 3+4 4 3+4 4 3+4 4 5 2 |
| | 6 2 5+5 4 6+6 4 3+4 4 3+4 4 3+4 4 3+4 4 6 2 |
| | 6 2 6+6 4 6+6 4 3+4 4 3+4 4 3+4 4 3+4 4 6 2 |
| | 5 2 5+5 4 5+5 4 3+4 4 3+4 4 3+4 4 3+4 4 5 2 |
| | 6 2 5+5 4 5+5 4 3+4 4 3+4 4 3+4 4 3+4 4 6 2 |
| | 5 2 6+6 4 6+6 4 3+4 4 3+4 4 3+4 4 3+4 4 5 2 |
| | 6 2 6+6 4 5+5 4 3+4 4 3+4 4 3+4 4 3+4 4 6 2 |
| | 5 2 5+5 4 5+5 4 3+4 4 3+4 4 3+4 4 3+4 4 6 2 |
| | 5 2 5+5 4 5+5 4 3+4 4 3+4 4 3+4 4 3+4 4 5 2 |
| | 6 2 6+6 4 6+6 4 3+4 4 3+4 4 3+4 4 3+4 4 5 2 |
| | 5 2 6+6 4 6+6 4 3+4 4 3+4 4 3+4 4 3+4 4 5 2 |
| | 6 2 5+5 4 5+5 4 3+4 4 3+4 4 3+4 4 3+4 4 5 2 |
| | 5 2 6+6 4 6+6 4 3+4 4 3+4 4 3+4 4 3+4 4 6 2 |
| | 6 2 6+6 4 5+5 4 3+4 4 3+4 4 3+4 4 3+4 4 6 2 |
| | 5 2 5+5 4 5+5 4 3+4 4 3+4 4 3+4 4 3+4 4 5 2 |
| | 6 2 6+6 4 6+6 4 3+4 4 3+4 4 3+4 4 3+4 4 5 2 |
| | 6 2 5+5 4 6+6 4 3+4 4 3+4 4 3+4 4 3+4 4 5 2 |
| | 6 2 6+6 4 5+5 4 3+4 4 3+4 4 3+4 4 3+4 4 5 2 |
| | 5 2 5+5 4 5+5 4 3+4 4 3+4 4 3+4 4 3+4 4 6 2 |
| | 6 2 6+6 4 6+6 4 3+4 4 3+4 4 3+4 4 3+4 4 6 2 |
| | 5 2 5+5 4 6+6 4 3+4 4 3+4 4 3+4 4 3+4 4 5 2 |
| | 6 2 5+5 4 6+6 4 3+4 4 3+4 4 3+4 4 3+4 4 6 2 |
| | 6 2 6+6 4 5+5 4 3+4 4 3+4 4 3+4 4 3+4 4 6 2 |
| | 6 2 6+6 4 5+5 4 3+4 4 3+4 4 3+4 4 3+4 4 5 2 |
| | 5 2 6+6 4 5+5 4 3+4 4 3+4 4 3+4 4 3+4 4 5 2 |
| | 6 2 5+5 4 6+6 4 3+4 4 3+4 4 3+4 4 3+4 4 5 2 |
| | 5 2 5+5 4 5+5 4 3+4 4 3+4 4 3+4 4 3+4 4 5 2 |
| | 6 2 6+6 4 6+6 4 3+4 4 3+4 4 3+4 4 3+4 4 6 2 |
| | 5 2 5+5 4 6+6 4 3+4 4 3+4 4 3+4 4 3+4 4 5 2 |
| | 5 2 6+6 4 5+5 4 3+4 4 3+4 4 3+4 4 3+4 4 5 2 |
| | 6 2 6 2 |
| | 5 2 6 2 |
| | 5 2 5 2 |
| | 6 2 5 2 |
| | 5 2 5 2 |
| | 6 2 6 2 |
| | 6 2 6 2 |
| | 5 2 6 2 |
| | 5 2 5 2 |
| | 6 2 6 2 |
| | 6 2 5 2 |
| | 6 2 6 2 |
| | 5 2 5 2 |
| | 5 2 5 2 |
| | 5 2 5 2 |
| | 5 2 6 2 |
| | 6 2 5 2 |
| | 6 2 6 2 |
| | 6 2 6 2 |
| | 5 2 5 2 |
| | 6 2 6 2 |
| | 6 2 6 2 |
| | 6 2 5 2 |
| | 5 2 5 2 |

Continued on next column

**TABLE XXXI– continued**

| TQC | ICSD NoE SF |
|---|---|
| | 6 2 5 2 |
| | 5 2 5 2 |
| | 6 2 6 2 |
| | 6 2 6 2 |
| | 5 2 6 2 |
| | 5 2 6 2 |
| | 5 2 6 2 |
| | 5 2 6 2 |
| | 5 2 6 2 |
| | 6 2 5 2 |
| | 5 2 6 2 |
| | 6 2 5 2 |
| | 5 2 5 2 |
| | 5 2 6 2 |
| | 5 2 5 2 |
| | 6 2 6 2 |
| | 6 2 5 2 |
| | 5 2 6 2 |
| | 6 2 6 2 |
| | 6 2 5 2 |
| | 6 2 6 2 |
| | 6 2 5 2 |
| | 5 2 5 2 |
| | 6 2 6 2 |
| 55.2.44 | 186440 192 Bi2Fe4O9 |
| | -GM -R -S -T -U -X -Y -Z |
| | 5 2 5+5 4 5+5 4 3+4 4 3+4 4 3+4 4 3+4 4 5 2 |
| | 6 2 6+6 4 6+6 4 3+4 4 3+4 4 3+4 4 3+4 4 6 2 |
| | 5 2 6+6 4 5+5 4 3+4 4 3+4 4 3+4 4 3+4 4 6 2 |
| | 6 2 5+5 4 6+6 4 3+4 4 3+4 4 3+4 4 3+4 4 6 2 |
| | 5 2 5+5 4 6+6 4 3+4 4 3+4 4 3+4 4 3+4 4 5 2 |
| | 5 2 6+6 4 5+5 4 3+4 4 3+4 4 3+4 4 3+4 4 6 2 |
| | 6 2 5+5 4 6+6 4 3+4 4 3+4 4 3+4 4 3+4 4 5 2 |
| | 5 2 5+5 4 5+5 4 3+4 4 3+4 4 3+4 4 3+4 4 6 2 |
| | 5 2 6+6 4 6+6 4 3+4 4 3+4 4 3+4 4 3+4 4 5 2 |
| | 6 2 6+6 4 5+5 4 3+4 4 3+4 4 3+4 4 3+4 4 5 2 |
| | 5 2 5+5 4 6+6 4 3+4 4 3+4 4 3+4 4 3+4 4 6 2 |
| | 5 2 6+6 4 5+5 4 3+4 4 3+4 4 3+4 4 3+4 4 6 2 |
| | 6 2 6+6 4 6+6 4 3+4 4 3+4 4 3+4 4 3+4 4 5 2 |
| | 5 2 6+6 4 6+6 4 3+4 4 3+4 4 3+4 4 3+4 4 6 2 |
| | 6 2 5+5 4 5+5 4 3+4 4 3+4 4 3+4 4 3+4 4 5 2 |
| | 6 2 5+5 4 5+5 4 3+4 4 3+4 4 3+4 4 3+4 4 6 2 |
| | 6 2 6+6 4 5+5 4 3+4 4 3+4 4 3+4 4 3+4 4 6 2 |
| | 5 2 6+6 4 6+6 4 3+4 4 3+4 4 3+4 4 3+4 4 5 2 |
| | 5 2 5+5 4 5+5 4 3+4 4 3+4 4 3+4 4 3+4 4 5 2 |
| | 6 2 5+5 4 6+6 4 3+4 4 3+4 4 3+4 4 3+4 4 6 2 |
| | 6 2 6+6 4 6+6 4 3+4 4 3+4 4 3+4 4 3+4 4 6 2 |
| | 5 2 5+5 4 5+5 4 3+4 4 3+4 4 3+4 4 3+4 4 5 2 |
| | 6 2 5+5 4 6+6 4 3+4 4 3+4 4 3+4 4 3+4 4 6 2 |
| | 5 2 6+6 4 5+5 4 3+4 4 3+4 4 3+4 4 3+4 4 5 2 |
| | 6 2 6+6 4 5+5 4 3+4 4 3+4 4 3+4 4 3+4 4 6 2 |
| | 5 2 5+5 4 5+5 4 3+4 4 3+4 4 3+4 4 3+4 4 6 2 |
| | 5 2 5+5 4 5+5 4 3+4 4 3+4 4 3+4 4 3+4 4 5 2 |
| | 6 2 6+6 4 6+6 4 3+4 4 3+4 4 3+4 4 3+4 4 5 2 |
| | 6 2 6+6 4 6+6 4 3+4 4 3+4 4 3+4 4 3+4 4 5 2 |
| | 5 2 5+5 4 5+5 4 3+4 4 3+4 4 3+4 4 3+4 4 5 2 |
| | 5 2 6+6 4 5+5 4 3+4 4 3+4 4 3+4 4 3+4 4 6 2 |
| | 6 2 6+6 4 6+6 4 3+4 4 3+4 4 3+4 4 3+4 4 6 2 |
| | 5 2 5+5 4 5+5 4 3+4 4 3+4 4 3+4 4 3+4 4 5 2 |
| | 6 2 6+6 4 6+6 4 3+4 4 3+4 4 3+4 4 3+4 4 5 2 |
| | 6 2 5+5 4 6+6 4 3+4 4 3+4 4 3+4 4 3+4 4 6 2 |

Continued on next column

**TABLE XXXI– continued**

| TQC | ICSD NoE SF |
|---|---|
| | 6 2 5+5 4 5+5 4 3+4 4 3+4 4 3+4 4 3+4 4 5 2 |
| | 5 2 6+6 4 5+5 4 3+4 4 3+4 4 3+4 4 3+4 4 5 2 |
| | 6 2 6+6 4 6+6 4 3+4 4 3+4 4 3+4 4 3+4 4 6 2 |
| | 5 2 5+5 4 6+6 4 3+4 4 3+4 4 3+4 4 3+4 4 5 2 |
| | 6 2 5+5 4 6+6 4 3+4 4 3+4 4 3+4 4 3+4 4 6 2 |
| | 6 2 6+6 4 5+5 4 3+4 4 3+4 4 3+4 4 3+4 4 6 2 |
| | 6 2 6+6 4 5+5 4 3+4 4 3+4 4 3+4 4 3+4 4 5 2 |
| | 5 2 6+6 4 5+5 4 3+4 4 3+4 4 3+4 4 3+4 4 5 2 |
| | 6 2 5+5 4 5+5 4 3+4 4 3+4 4 3+4 4 3+4 4 5 2 |
| | 5 2 6+6 4 6+6 4 3+4 4 3+4 4 3+4 4 3+4 4 5 2 |
| | 6 2 5+5 4 6+6 4 3+4 4 3+4 4 3+4 4 3+4 4 5 2 |
| | 5 2 6+6 4 6+6 4 3+4 4 3+4 4 3+4 4 3+4 4 6 2 |
| | 5 2 5+5 4 5+5 4 3+4 4 3+4 4 3+4 4 3+4 4 5 2 |
| | 5 2 6 2 |
| | 5 2 6 2 |
| | 6 2 5 2 |
| | 6 2 5 2 |
| | 5 2 5 2 |
| | 6 2 6 2 |
| | 6 2 6 2 |
| | 5 2 6 2 |
| | 5 2 6 2 |
| | 6 2 5 2 |
| | 6 2 5 2 |
| | 6 2 6 2 |
| | 5 2 5 2 |
| | 5 2 5 2 |
| | 5 2 6 2 |
| | 6 2 5 2 |
| | 5 2 6 2 |
| | 6 2 5 2 |
| | 6 2 6 2 |
| | 5 2 6 2 |
| | 6 2 5 2 |
| | 6 2 6 2 |
| | 6 2 5 2 |
| | 5 2 5 2 |
| | 6 2 5 2 |
| | 6 2 5 2 |
| | 6 2 6 2 |
| | 5 2 6 2 |
| | 5 2 6 2 |
| | 5 2 6 2 |
| | 5 2 6 2 |
| | 5 2 6 2 |
| | 5 2 6 2 |
| | 6 2 5 2 |
| | 5 2 6 2 |
| | 6 2 5 2 |
| | 5 2 5 2 |
| | 5 2 6 2 |
| | 5 2 5 2 |
| | 6 2 6 2 |
| | 6 2 6 2 |
| | 5 2 5 2 |
| | 6 2 6 2 |
| | 6 2 5 2 |
| | 6 2 5 2 |
| | 6 2 5 2 |
| | 5 2 6 2 |
| | 6 2 5 2 |

Continued on next column

**TABLE XXXI– continued**

| TQC | ICSD NoE SF |
|---|---|
| 55.2.45 | 419361 180 B4Rh1Yb1 |
| | -GM -R -S -T -U -X -Y -Z |
| | 5 2 6+6 4 6+6 4 3+4 4 3+4 4 3+4 4 3+4 4 5 2 |
| | 6 2 5+5 4 5+5 4 3+4 4 3+4 4 3+4 4 3+4 4 6 2 |
| | 5 2 6+6 4 6+6 4 3+4 4 3+4 4 3+4 4 3+4 4 5 2 |
| | 6 2 5+5 4 5+5 4 3+4 4 3+4 4 3+4 4 3+4 4 6 2 |
| | 5 2 5+5 4 6+6 4 3+4 4 3+4 4 3+4 4 3+4 4 5 2 |
| | 6 2 6+6 4 5+5 4 3+4 4 3+4 4 3+4 4 3+4 4 6 2 |
| | 5 2 6+6 4 6+6 4 3+4 4 3+4 4 3+4 4 3+4 4 5 2 |
| | 6 2 5+5 4 5+5 4 3+4 4 3+4 4 3+4 4 3+4 4 6 2 |
| | 5 2 6+6 4 5+5 4 3+4 4 3+4 4 3+4 4 3+4 4 6 2 |
| | 6 2 5+5 4 6+6 4 3+4 4 3+4 4 3+4 4 3+4 4 5 2 |
| | 5 2 5+5 4 6+6 4 3+4 4 3+4 4 3+4 4 3+4 4 6 2 |
| | 6 2 6+6 4 5+5 4 3+4 4 3+4 4 3+4 4 3+4 4 5 2 |
| | 5 2 6+6 4 6+6 4 3+4 4 3+4 4 3+4 4 3+4 4 5 2 |
| | 6 2 5+5 4 5+5 4 3+4 4 3+4 4 3+4 4 3+4 4 6 2 |
| | 5 2 5+5 4 6+6 4 3+4 4 3+4 4 3+4 4 3+4 4 5 2 |
| | 6 2 6+6 4 5+5 4 3+4 4 3+4 4 3+4 4 3+4 4 6 2 |
| | 5 2 5+5 4 6+6 4 3+4 4 3+4 4 3+4 4 3+4 4 6 2 |
| | 5 2 6+6 4 5+5 4 3+4 4 3+4 4 3+4 4 3+4 4 6 2 |
| | 6 2 5+5 4 5+5 4 3+4 4 3+4 4 3+4 4 3+4 4 5 2 |
| | 6 2 5+5 4 6+6 4 3+4 4 3+4 4 3+4 4 3+4 4 6 2 |
| | 5 2 6+6 4 6+6 4 3+4 4 3+4 4 3+4 4 3+4 4 6 2 |
| | 6 2 6+6 4 5+5 4 3+4 4 3+4 4 3+4 4 3+4 4 5 2 |
| | 5 2 5+5 4 5+5 4 3+4 4 3+4 4 3+4 4 3+4 4 5 2 |
| | 6 2 6+6 4 6+6 4 3+4 4 3+4 4 3+4 4 3+4 4 6 2 |
| | 5 2 6+6 4 5+5 4 3+4 4 3+4 4 3+4 4 3+4 4 5 2 |
| | 5 2 5+5 4 6+6 4 3+4 4 3+4 4 3+4 4 3+4 4 6 2 |
| | 6 2 5+5 4 5+5 4 3+4 4 3+4 4 3+4 4 3+4 4 5 2 |
| | 5 2 6+6 4 6+6 4 3+4 4 3+4 4 3+4 4 3+4 4 5 2 |
| | 6 2 5+5 4 6+6 4 3+4 4 3+4 4 3+4 4 3+4 4 6 2 |
| | 5 2 6+6 4 5+5 4 3+4 4 3+4 4 3+4 4 3+4 4 5 2 |
| | 6 2 5+5 4 5+5 4 3+4 4 3+4 4 3+4 4 3+4 4 5 2 |
| | 6 2 6+6 4 6+6 4 3+4 4 3+4 4 3+4 4 3+4 4 6 2 |
| | 6 2 5+5 4 5+5 4 3+4 4 3+4 4 3+4 4 3+4 4 6 2 |
| | 6 2 6+6 4 6+6 4 3+4 4 3+4 4 3+4 4 3+4 4 5 2 |
| | 5 2 5+5 4 6+6 4 3+4 4 3+4 4 3+4 4 3+4 4 5 2 |
| | 5 2 6+6 4 5+5 4 3+4 4 3+4 4 3+4 4 3+4 4 5 2 |
| | 5 2 5+5 4 6+6 4 3+4 4 3+4 4 3+4 4 3+4 4 6 2 |
| | 5 2 5+5 4 5+5 4 3+4 4 3+4 4 3+4 4 3+4 4 5 2 |
| | 6 2 6+6 4 5+5 4 3+4 4 3+4 4 3+4 4 3+4 4 6 2 |
| | 6 2 5+5 4 6+6 4 3+4 4 3+4 4 3+4 4 3+4 4 6 2 |
| | 5 2 6+6 4 6+6 4 3+4 4 3+4 4 3+4 4 3+4 4 6 2 |
| | 6 2 5+5 4 5+5 4 3+4 4 3+4 4 3+4 4 3+4 4 6 2 |
| | 5 2 6+6 4 6+6 4 3+4 4 3+4 4 3+4 4 3+4 4 5 2 |
| | 5 2 6+6 4 5+5 4 3+4 4 3+4 4 3+4 4 3+4 4 5 2 |
| | 6 2 5+5 4 6+6 4 3+4 4 3+4 4 3+4 4 3+4 4 6 2 |
| | 5 2 6 2 |
| | 6 2 5 2 |
| | 5 2 6 2 |
| | 6 2 6 2 |
| | 5 2 5 2 |
| | 6 2 5 2 |
| | 5 2 5 2 |
| | 6 2 6 2 |
| | 5 2 5 2 |
| | 6 2 6 2 |
| | 5 2 6 2 |
| | 5 2 5 2 |
| | 6 2 6 2 |
| | 5 2 6 2 |

Continued on next column

**TABLE XXXI– continued**

| TQC | ICSD | NoE | SF | | | | | |
|---|---|---|---|---|---|---|---|---|
| | 6 2 | | | | | | | 5 2 |
| | 6 2 | | | | | | | 5 2 |
| | 5 2 | | | | | | | 5 2 |
| | 5 2 | | | | | | | 6 2 |
| | 6 2 | | | | | | | 5 2 |
| | 5 2 | | | | | | | 6 2 |
| | 6 2 | | | | | | | 6 2 |
| | 6 2 | | | | | | | 5 2 |
| | 5 2 | | | | | | | 6 2 |
| | 6 2 | | | | | | | 5 2 |
| | 6 2 | | | | | | | 6 2 |
| | 5 2 | | | | | | | 5 2 |
| | 6 2 | | | | | | | 5 2 |
| | 5 2 | | | | | | | 5 2 |
| | 5 2 | | | | | | | 6 2 |
| | 6 2 | | | | | | | 5 2 |
| | 5 2 | | | | | | | 6 2 |
| | 6 2 | | | | | | | 5 2 |
| | 5 2 | | | | | | | 6 2 |
| | 5 2 | | | | | | | 5 2 |
| | 6 2 | | | | | | | 5 2 |
| | 5 2 | | | | | | | 6 2 |
| | 6 2 | | | | | | | 5 2 |
| | 6 2 | | | | | | | 6 2 |
| | 5 2 | | | | | | | 5 2 |
| | 6 2 | | | | | | | 6 2 |
| | 6 2 | | | | | | | 5 2 |
| | 5 2 | | | | | | | 5 2 |
| | 6 2 | | | | | | | 6 2 |
| | 5 2 | | | | | | | 5 2 |
| | 5 2 | | | | | | | 6 2 |
| 55.2.46 | 419358 | 128 | B4Rh1Y1 | | | | | |
| | -GM | -R | -S | -T | -U | -X | -Y | -Z |
| | 5 2 | 5+5 4 | 5+5 4 | 3+4 4 | 3+4 4 | 3+4 4 | 3+4 4 | 5 2 |
| | 6 2 | 6+6 4 | 6+6 4 | 3+4 4 | 3+4 4 | 3+4 4 | 3+4 4 | 6 2 |
| | 5 2 | 5+5 4 | 5+5 4 | 3+4 4 | 3+4 4 | 3+4 4 | 3+4 4 | 5 2 |
| | 6 2 | 6+6 4 | 6+6 4 | 3+4 4 | 3+4 4 | 3+4 4 | 3+4 4 | 6 2 |
| | 5 2 | 6+6 4 | 5+5 4 | 3+4 4 | 3+4 4 | 3+4 4 | 3+4 4 | 5 2 |
| | 6 2 | 5+5 4 | 6+6 4 | 3+4 4 | 3+4 4 | 3+4 4 | 3+4 4 | 6 2 |
| | 5 2 | 5+5 4 | 5+5 4 | 3+4 4 | 3+4 4 | 3+4 4 | 3+4 4 | 5 2 |
| | 6 2 | 6+6 4 | 6+6 4 | 3+4 4 | 3+4 4 | 3+4 4 | 3+4 4 | 6 2 |
| | 5 2 | 5+5 4 | 6+6 4 | 3+4 4 | 3+4 4 | 3+4 4 | 3+4 4 | 6 2 |
| | 6 2 | 6+6 4 | 5+5 4 | 3+4 4 | 3+4 4 | 3+4 4 | 3+4 4 | 5 2 |
| | 5 2 | 6+6 4 | 5+5 4 | 3+4 4 | 3+4 4 | 3+4 4 | 3+4 4 | 5 2 |
| | 6 2 | 5+5 4 | 6+6 4 | 3+4 4 | 3+4 4 | 3+4 4 | 3+4 4 | 6 2 |
| | 5 2 | 5+5 4 | 5+5 4 | 3+4 4 | 3+4 4 | 3+4 4 | 3+4 4 | 5 2 |
| | 5 2 | 6+6 4 | 6+6 4 | 3+4 4 | 3+4 4 | 3+4 4 | 3+4 4 | 6 2 |
| | 6 2 | 6+6 4 | 5+5 4 | 3+4 4 | 3+4 4 | 3+4 4 | 3+4 4 | 5 2 |
| | 6 2 | 5+5 4 | 6+6 4 | 3+4 4 | 3+4 4 | 3+4 4 | 3+4 4 | 6 2 |
| | 5 2 | 6+6 4 | 5+5 4 | 3+4 4 | 3+4 4 | 3+4 4 | 3+4 4 | 6 2 |
| | 5 2 | 5+5 4 | 6+6 4 | 3+4 4 | 3+4 4 | 3+4 4 | 3+4 4 | 6 2 |
| | 6 2 | 6+6 4 | 5+5 4 | 3+4 4 | 3+4 4 | 3+4 4 | 3+4 4 | 5 2 |
| | 6 2 | 6+6 4 | 6+6 4 | 3+4 4 | 3+4 4 | 3+4 4 | 3+4 4 | 6 2 |
| | 5 2 | 5+5 4 | 5+5 4 | 3+4 4 | 3+4 4 | 3+4 4 | 3+4 4 | 5 2 |
| | 6 2 | 5+5 4 | 6+6 4 | 3+4 4 | 3+4 4 | 3+4 4 | 3+4 4 | 6 2 |
| | 5 2 | 6+6 4 | 6+6 4 | 3+4 4 | 3+4 4 | 3+4 4 | 3+4 4 | 5 2 |
| | 6 2 | 5+5 4 | 5+5 4 | 3+4 4 | 3+4 4 | 3+4 4 | 3+4 4 | 5 2 |
| | 5 2 | 5+5 4 | 5+5 4 | 3+4 4 | 3+4 4 | 3+4 4 | 3+4 4 | 6 2 |
| | 5 2 | 6+6 4 | 6+6 4 | 3+4 4 | 3+4 4 | 3+4 4 | 3+4 4 | 6 2 |
| | 6 2 | 5+5 4 | 6+6 4 | 3+4 4 | 3+4 4 | 3+4 4 | 3+4 4 | 5 2 |
| | 5 2 | 6+6 4 | 5+5 4 | 3+4 4 | 3+4 4 | 3+4 4 | 3+4 4 | 5 2 |

Continued on next column

**TABLE XXXI– continued**

| TQC | ICSD | NoE | SF | | | | | |
|---|---|---|---|---|---|---|---|---|
| | 6 2 | 6+6 4 | 5+5 4 | 3+4 4 | 3+4 4 | 3+4 4 | 3+4 4 | 6 2 |
| | 5 2 | 5+5 4 | 6+6 4 | 3+4 4 | 3+4 4 | 3+4 4 | 3+4 4 | 5 2 |
| | 6 2 | 6+6 4 | 5+5 4 | 3+4 4 | 3+4 4 | 3+4 4 | 3+4 4 | 5 2 |
| | 6 2 | 6+6 4 | 6+6 4 | 3+4 4 | 3+4 4 | 3+4 4 | 3+4 4 | 6 2 |
| | 6 2 | | | | | | | 6 2 |
| | 6 2 | | | | | | | 5 2 |
| | 5 2 | | | | | | | 5 2 |
| | 5 2 | | | | | | | 5 2 |
| | 5 2 | | | | | | | 6 2 |
| | 5 2 | | | | | | | 6 2 |
| | 6 2 | | | | | | | 6 2 |
| | 6 2 | | | | | | | 5 2 |
| | 5 2 | | | | | | | 6 2 |
| | 6 2 | | | | | | | 6 2 |
| | 5 2 | | | | | | | 5 2 |
| | 6 2 | | | | | | | 5 2 |
| | 5 2 | | | | | | | 6 2 |
| | 5 2 | | | | | | | 6 2 |
| | 6 2 | | | | | | | 5 2 |
| | 5 2 | | | | | | | 6 2 |
| | 5 2 | | | | | | | 6 2 |
| | 6 2 | | | | | | | 5 2 |
| | 6 2 | | | | | | | 5 2 |
| | 5 2 | | | | | | | 5 2 |
| | 6 2 | | | | | | | 5 2 |
| | 5 2 | | | | | | | 6 2 |
| | 6 2 | | | | | | | 6 2 |
| | 5 2 | | | | | | | 6 2 |
| | 6 2 | | | | | | | 5 2 |
| | 6 2 | | | | | | | 5 2 |
| | 5 2 | | | | | | | 6 2 |
| | 6 2 | | | | | | | 5 2 |
| | 5 2 | | | | | | | 6 2 |
| | 5 2 | | | | | | | 5 2 |
| | 5 2 | | | | | | | 5 2 |
| | 6 2 | | | | | | | 6 2 |
| 55.2.47 | 419360 | 176 | B4Rh1Tm1 | | | | | |
| | -GM | -R | -S | -T | -U | -X | -Y | -Z |
| | 5 2 | 6+6 4 | 6+6 4 | 3+4 4 | 3+4 4 | 3+4 4 | 3+4 4 | 5 2 |
| | 6 2 | 5+5 4 | 5+5 4 | 3+4 4 | 3+4 4 | 3+4 4 | 3+4 4 | 6 2 |
| | 5 2 | 6+6 4 | 6+6 4 | 3+4 4 | 3+4 4 | 3+4 4 | 3+4 4 | 5 2 |
| | 6 2 | 5+5 4 | 5+5 4 | 3+4 4 | 3+4 4 | 3+4 4 | 3+4 4 | 6 2 |
| | 5 2 | 5+5 4 | 6+6 4 | 3+4 4 | 3+4 4 | 3+4 4 | 3+4 4 | 5 2 |
| | 6 2 | 6+6 4 | 5+5 4 | 3+4 4 | 3+4 4 | 3+4 4 | 3+4 4 | 6 2 |
| | 5 2 | 6+6 4 | 6+6 4 | 3+4 4 | 3+4 4 | 3+4 4 | 3+4 4 | 5 2 |
| | 6 2 | 5+5 4 | 5+5 4 | 3+4 4 | 3+4 4 | 3+4 4 | 3+4 4 | 6 2 |
| | 5 2 | 6+6 4 | 5+5 4 | 3+4 4 | 3+4 4 | 3+4 4 | 3+4 4 | 6 2 |
| | 6 2 | 5+5 4 | 6+6 4 | 3+4 4 | 3+4 4 | 3+4 4 | 3+4 4 | 5 2 |
| | 5 2 | 5+5 4 | 6+6 4 | 3+4 4 | 3+4 4 | 3+4 4 | 3+4 4 | 6 2 |
| | 6 2 | 6+6 4 | 5+5 4 | 3+4 4 | 3+4 4 | 3+4 4 | 3+4 4 | 5 2 |
| | 5 2 | 6+6 4 | 6+6 4 | 3+4 4 | 3+4 4 | 3+4 4 | 3+4 4 | 5 2 |
| | 5 2 | 5+5 4 | 5+5 4 | 3+4 4 | 3+4 4 | 3+4 4 | 3+4 4 | 6 2 |
| | 6 2 | 5+5 4 | 6+6 4 | 3+4 4 | 3+4 4 | 3+4 4 | 3+4 4 | 5 2 |
| | 6 2 | 6+6 4 | 5+5 4 | 3+4 4 | 3+4 4 | 3+4 4 | 3+4 4 | 6 2 |
| | 5 2 | 5+5 4 | 6+6 4 | 3+4 4 | 3+4 4 | 3+4 4 | 3+4 4 | 6 2 |
| | 5 2 | 6+6 4 | 5+5 4 | 3+4 4 | 3+4 4 | 3+4 4 | 3+4 4 | 6 2 |
| | 6 2 | 5+5 4 | 5+5 4 | 3+4 4 | 3+4 4 | 3+4 4 | 3+4 4 | 5 2 |
| | 6 2 | 5+5 4 | 6+6 4 | 3+4 4 | 3+4 4 | 3+4 4 | 3+4 4 | 6 2 |
| | 5 2 | 6+6 4 | 6+6 4 | 3+4 4 | 3+4 4 | 3+4 4 | 3+4 4 | 6 2 |
| | 6 2 | 6+6 4 | 5+5 4 | 3+4 4 | 3+4 4 | 3+4 4 | 3+4 4 | 5 2 |
| | 5 2 | 5+5 4 | 5+5 4 | 3+4 4 | 3+4 4 | 3+4 4 | 3+4 4 | 5 2 |

Continued on next column

**TABLE XXXI– continued**

| TQC | ICSD NoE SF |
|---|---|
| | 6 2 6+6 4 6+6 4 3+4 4 3+4 4 3+4 4 3+4 4 6 2 |
| | 5 2 6+6 4 5+5 4 3+4 4 3+4 4 3+4 4 3+4 4 5 2 |
| | 5 2 5+5 4 6+6 4 3+4 4 3+4 4 3+4 4 3+4 4 6 2 |
| | 6 2 5+5 4 5+5 4 3+4 4 3+4 4 3+4 4 3+4 4 5 2 |
| | 5 2 6+6 4 6+6 4 3+4 4 3+4 4 3+4 4 3+4 4 5 2 |
| | 6 2 5+5 4 6+6 4 3+4 4 3+4 4 3+4 4 3+4 4 6 2 |
| | 5 2 6+6 4 5+5 4 3+4 4 3+4 4 3+4 4 3+4 4 5 2 |
| | 6 2 5+5 4 6+6 4 3+4 4 3+4 4 3+4 4 3+4 4 5 2 |
| | 6 2 5+5 4 5+5 4 3+4 4 3+4 4 3+4 4 3+4 4 6 2 |
| | 6 2 6+6 4 5+5 4 3+4 4 3+4 4 3+4 4 3+4 4 6 2 |
| | 6 2 6+6 4 6+6 4 3+4 4 3+4 4 3+4 4 3+4 4 5 2 |
| | 5 2 5+5 4 6+6 4 3+4 4 3+4 4 3+4 4 3+4 4 5 2 |
| | 5 2 6+6 4 5+5 4 3+4 4 3+4 4 3+4 4 3+4 4 5 2 |
| | 5 2 5+5 4 6+6 4 3+4 4 3+4 4 3+4 4 3+4 4 6 2 |
| | 5 2 5+5 4 5+5 4 3+4 4 3+4 4 3+4 4 3+4 4 5 2 |
| | 6 2 6+6 4 6+6 4 3+4 4 3+4 4 3+4 4 3+4 4 6 2 |
| | 6 2 5+5 4 5+5 4 3+4 4 3+4 4 3+4 4 3+4 4 6 2 |
| | 5 2 6+6 4 6+6 4 3+4 4 3+4 4 3+4 4 3+4 4 6 2 |
| | 6 2 5+5 4 5+5 4 3+4 4 3+4 4 3+4 4 3+4 4 6 2 |
| | 5 2 6+6 4 6+6 4 3+4 4 3+4 4 3+4 4 3+4 4 5 2 |
| | 5 2 6+6 4 5+5 4 3+4 4 3+4 4 3+4 4 3+4 4 5 2 |
| | 6 2 6 2 |
| | 5 2 6 2 |
| | 6 2 5 2 |
| | 5 2 6 2 |
| | 6 2 6 2 |
| | 5 2 5 2 |
| | 6 2 5 2 |
| | 5 2 5 2 |
| | 6 2 6 2 |
| | 5 2 5 2 |
| | 6 2 6 2 |
| | 5 2 6 2 |
| | 6 2 5 2 |
| | 6 2 6 2 |
| | 5 2 6 2 |
| | 5 2 5 2 |
| | 6 2 5 2 |
| | 5 2 6 2 |
| | 6 2 5 2 |
| | 5 2 5 2 |
| | 5 2 5 2 |
| | 6 2 6 2 |
| | 6 2 6 2 |
| | 5 2 5 2 |
| | 6 2 6 2 |
| | 6 2 5 2 |
| | 5 2 6 2 |
| | 6 2 5 2 |
| | 5 2 5 2 |
| | 5 2 5 2 |
| | 5 2 6 2 |
| | 6 2 6 2 |
| | 5 2 5 2 |
| | 6 2 5 2 |
| | 5 2 6 2 |
| | 6 2 5 2 |
| | 5 2 6 2 |
| | 6 2 6 2 |
| | 6 2 5 2 |
| | 5 2 5 2 |

Continued on next column

**TABLE XXXI– continued**

| TQC | ICSD NoE SF |
|---|---|
| | 6 2 6 2 |
| | 5 2 5 2 |
| | 6 2 6 2 |
| | 6 2 5 2 |
| 55.2.48 | 409820 184 B4Ni1Yb1 |
| | -GM -R -S -T -U -X -Y -Z |
| | 5 2 5+5 4 5+5 4 3+4 4 3+4 4 3+4 4 3+4 4 5 2 |
| | 6 2 6+6 4 6+6 4 3+4 4 3+4 4 3+4 4 3+4 4 6 2 |
| | 5 2 5+5 4 5+5 4 3+4 4 3+4 4 3+4 4 3+4 4 5 2 |
| | 6 2 6+6 4 6+6 4 3+4 4 3+4 4 3+4 4 3+4 4 6 2 |
| | 5 2 6+6 4 5+5 4 3+4 4 3+4 4 3+4 4 3+4 4 5 2 |
| | 6 2 5+5 4 6+6 4 3+4 4 3+4 4 3+4 4 3+4 4 6 2 |
| | 5 2 5+5 4 5+5 4 3+4 4 3+4 4 3+4 4 3+4 4 5 2 |
| | 6 2 6+6 4 6+6 4 3+4 4 3+4 4 3+4 4 3+4 4 6 2 |
| | 5 2 5+5 4 6+6 4 3+4 4 3+4 4 3+4 4 3+4 4 6 2 |
| | 6 2 6+6 4 5+5 4 3+4 4 3+4 4 3+4 4 3+4 4 5 2 |
| | 5 2 6+6 4 5+5 4 3+4 4 3+4 4 3+4 4 3+4 4 6 2 |
| | 6 2 5+5 4 6+6 4 3+4 4 3+4 4 3+4 4 3+4 4 5 2 |
| | 5 2 5+5 4 6+6 4 3+4 4 3+4 4 3+4 4 3+4 4 5 2 |
| | 6 2 6+6 4 5+5 4 3+4 4 3+4 4 3+4 4 3+4 4 6 2 |
| | 5 2 6+6 4 5+5 4 3+4 4 3+4 4 3+4 4 3+4 4 5 2 |
| | 6 2 5+5 4 6+6 4 3+4 4 3+4 4 3+4 4 3+4 4 6 2 |
| | 5 2 6+6 4 5+5 4 3+4 4 3+4 4 3+4 4 3+4 4 6 2 |
| | 5 2 5+5 4 5+5 4 3+4 4 3+4 4 3+4 4 3+4 4 6 2 |
| | 6 2 6+6 4 6+6 4 3+4 4 3+4 4 3+4 4 3+4 4 5 2 |
| | 6 2 5+5 4 6+6 4 3+4 4 3+4 4 3+4 4 3+4 4 5 2 |
| | 5 2 6+6 4 6+6 4 3+4 4 3+4 4 3+4 4 3+4 4 6 2 |
| | 6 2 5+5 4 6+6 4 3+4 4 3+4 4 3+4 4 3+4 4 6 2 |
| | 5 2 6+6 4 5+5 4 3+4 4 3+4 4 3+4 4 3+4 4 5 2 |
| | 6 2 5+5 4 5+5 4 3+4 4 3+4 4 3+4 4 3+4 4 5 2 |
| | 5 2 5+5 4 5+5 4 3+4 4 3+4 4 3+4 4 3+4 4 6 2 |
| | 5 2 6+6 4 6+6 4 3+4 4 3+4 4 3+4 4 3+4 4 5 2 |
| | 6 2 5+5 4 6+6 4 3+4 4 3+4 4 3+4 4 3+4 4 6 2 |
| | 5 2 6+6 4 5+5 4 3+4 4 3+4 4 3+4 4 3+4 4 5 2 |
| | 6 2 6+6 4 5+5 4 3+4 4 3+4 4 3+4 4 3+4 4 5 2 |
| | 6 2 5+5 4 6+6 4 3+4 4 3+4 4 3+4 4 3+4 4 6 2 |
| | 5 2 6+6 4 6+6 4 3+4 4 3+4 4 3+4 4 3+4 4 6 2 |
| | 6 2 5+5 4 5+5 4 3+4 4 3+4 4 3+4 4 3+4 4 5 2 |
| | 6 2 6+6 4 5+5 4 3+4 4 3+4 4 3+4 4 3+4 4 6 2 |
| | 6 2 5+5 4 6+6 4 3+4 4 3+4 4 3+4 4 3+4 4 5 2 |
| | 5 2 6+6 4 5+5 4 3+4 4 3+4 4 3+4 4 3+4 4 5 2 |
| | 5 2 5+5 4 5+5 4 3+4 4 3+4 4 3+4 4 3+4 4 6 2 |
| | 5 2 6+6 4 6+6 4 3+4 4 3+4 4 3+4 4 3+4 4 6 2 |
| | 5 2 6+6 4 6+6 4 3+4 4 3+4 4 3+4 4 3+4 4 6 2 |
| | 5 2 5+5 4 5+5 4 3+4 4 3+4 4 3+4 4 3+4 4 5 2 |
| | 6 2 6+6 4 6+6 4 3+4 4 3+4 4 3+4 4 3+4 4 6 2 |
| | 6 2 5+5 4 5+5 4 3+4 4 3+4 4 3+4 4 3+4 4 5 2 |
| | 6 2 5+5 4 6+6 4 3+4 4 3+4 4 3+4 4 3+4 4 6 2 |
| | 5 2 6+6 4 5+5 4 3+4 4 3+4 4 3+4 4 3+4 4 5 2 |
| | 6 2 5+5 4 6+6 4 3+4 4 3+4 4 3+4 4 3+4 4 5 2 |
| | 5 2 6+6 4 5+5 4 3+4 4 3+4 4 3+4 4 3+4 4 6 2 |
| | 5 2 5+5 4 6+6 4 3+4 4 3+4 4 3+4 4 3+4 4 6 2 |
| | 6 2 5 2 |
| | 5 2 6 2 |
| | 6 2 5 2 |
| | 5 2 6 2 |
| | 6 2 5 2 |
| | 5 2 5 2 |
| | 6 2 5 2 |
| | 5 2 6 2 |
| | 6 2 6 2 |

Continued on next column

**TABLE XXXI– continued**

| TQC | ICSD | NoE | SF | | | | | |
|---|---|---|---|---|---|---|---|---|
| | 6 2 | | | | | | | 6 2 |
| | 5 2 | | | | | | | 5 2 |
| | 5 2 | | | | | | | 6 2 |
| | 6 2 | | | | | | | 6 2 |
| | 5 2 | | | | | | | 5 2 |
| | 5 2 | | | | | | | 5 2 |
| | 6 2 | | | | | | | 6 2 |
| | 5 2 | | | | | | | 5 2 |
| | 6 2 | | | | | | | 6 2 |
| | 6 2 | | | | | | | 5 2 |
| | 5 2 | | | | | | | 5 2 |
| | 6 2 | | | | | | | 5 2 |
| | 6 2 | | | | | | | 6 2 |
| | 5 2 | | | | | | | 6 2 |
| | 6 2 | | | | | | | 5 2 |
| | 6 2 | | | | | | | 5 2 |
| | 5 2 | | | | | | | 6 2 |
| | 5 2 | | | | | | | 5 2 |
| | 5 2 | | | | | | | 5 2 |
| | 5 2 | | | | | | | 6 2 |
| | 5 2 | | | | | | | 6 2 |
| | 6 2 | | | | | | | 5 2 |
| | 6 2 | | | | | | | 6 2 |
| | 5 2 | | | | | | | 5 2 |
| | 6 2 | | | | | | | 5 2 |
| | 6 2 | | | | | | | 6 2 |
| | 5 2 | | | | | | | 5 2 |
| | 6 2 | | | | | | | 6 2 |
| | 6 2 | | | | | | | 5 2 |
| | 5 2 | | | | | | | 5 2 |
| | 5 2 | | | | | | | 6 2 |
| | 6 2 | | | | | | | 6 2 |
| | 6 2 | | | | | | | 5 2 |
| | 5 2 | | | | | | | 5 2 |
| | 5 2 | | | | | | | 6 2 |
| | 5 2 | | | | | | | 6 2 |
| | 6 2 | | | | | | | 6 2 |
| 55.2.49 | 425457 | 100 | B4Ni1Sc1 | | | | | |
| | -GM | -R | -S | -T | -U | -X | -Y | -Z |
| | 5 2 | 5+5 4 | 6+6 4 | 3+4 4 | 3+4 4 | 3+4 4 | 3+4 4 | 6 2 |
| | 5 2 | 6+6 4 | 5+5 4 | 3+4 4 | 3+4 4 | 3+4 4 | 3+4 4 | 6 2 |
| | 6 2 | 6+6 4 | 5+5 4 | 3+4 4 | 3+4 4 | 3+4 4 | 3+4 4 | 5 2 |
| | 6 2 | 5+5 4 | 6+6 4 | 3+4 4 | 3+4 4 | 3+4 4 | 3+4 4 | 5 2 |
| | 5 2 | 6+6 4 | 5+5 4 | 3+4 4 | 3+4 4 | 3+4 4 | 3+4 4 | 6 2 |
| | 5 2 | 5+5 4 | 6+6 4 | 3+4 4 | 3+4 4 | 3+4 4 | 3+4 4 | 6 2 |
| | 6 2 | 6+6 4 | 5+5 4 | 3+4 4 | 3+4 4 | 3+4 4 | 3+4 4 | 5 2 |
| | 6 2 | 5+5 4 | 6+6 4 | 3+4 4 | 3+4 4 | 3+4 4 | 3+4 4 | 5 2 |
| | 5 2 | 6+6 4 | 5+5 4 | 3+4 4 | 3+4 4 | 3+4 4 | 3+4 4 | 6 2 |
| | 5 2 | 5+5 4 | 5+5 4 | 3+4 4 | 3+4 4 | 3+4 4 | 3+4 4 | 5 2 |
| | 6 2 | 5+5 4 | 6+6 4 | 3+4 4 | 3+4 4 | 3+4 4 | 3+4 4 | 6 2 |
| | 5 2 | 6+6 4 | 6+6 4 | 3+4 4 | 3+4 4 | 3+4 4 | 3+4 4 | 5 2 |
| | 6 2 | 6+6 4 | 6+6 4 | 3+4 4 | 3+4 4 | 3+4 4 | 3+4 4 | 5 2 |
| | 6 2 | 5+5 4 | 6+6 4 | 3+4 4 | 3+4 4 | 3+4 4 | 3+4 4 | 6 2 |
| | 5 2 | 6+6 4 | 5+5 4 | 3+4 4 | 3+4 4 | 3+4 4 | 3+4 4 | 6 2 |
| | 6 2 | 5+5 4 | 5+5 4 | 3+4 4 | 3+4 4 | 3+4 4 | 3+4 4 | 5 2 |
| | 5 2 | 5+5 4 | 6+6 4 | 3+4 4 | 3+4 4 | 3+4 4 | 3+4 4 | 6 2 |
| | 6 2 | 6+6 4 | 5+5 4 | 3+4 4 | 3+4 4 | 3+4 4 | 3+4 4 | 5 2 |
| | 6 2 | 5+5 4 | 6+6 4 | 3+4 4 | 3+4 4 | 3+4 4 | 3+4 4 | 5 2 |
| | 5 2 | 6+6 4 | 5+5 4 | 3+4 4 | 3+4 4 | 3+4 4 | 3+4 4 | 6 2 |
| | 5 2 | 6+6 4 | 5+5 4 | 3+4 4 | 3+4 4 | 3+4 4 | 3+4 4 | 6 2 |
| | 5 2 | 5+5 4 | 5+5 4 | 3+4 4 | 3+4 4 | 3+4 4 | 3+4 4 | 6 2 |

Continued on next column

**TABLE XXXI– continued**

| TQC | ICSD | NoE | SF | | | | | |
|---|---|---|---|---|---|---|---|---|
| | 5 2 | 6+6 4 | 6+6 4 | 3+4 4 | 3+4 4 | 3+4 4 | 3+4 4 | 6 2 |
| | 6 2 | 6+6 4 | 6+6 4 | 3+4 4 | 3+4 4 | 3+4 4 | 3+4 4 | 5 2 |
| | 6 2 | 5+5 4 | 5+5 4 | 3+4 4 | 3+4 4 | 3+4 4 | 3+4 4 | 6 2 |
| | 6 2 | | | | | | | 5 2 |
| | 6 2 | | | | | | | 6 2 |
| | 5 2 | | | | | | | 5 2 |
| | 5 2 | | | | | | | 5 2 |
| | 5 2 | | | | | | | 6 2 |
| | 5 2 | | | | | | | 5 2 |
| | 6 2 | | | | | | | 5 2 |
| | 6 2 | | | | | | | 6 2 |
| | 5 2 | | | | | | | 6 2 |
| | 5 2 | | | | | | | 5 2 |
| | 6 2 | | | | | | | 5 2 |
| | 5 2 | | | | | | | 5 2 |
| | 6 2 | | | | | | | 6 2 |
| | 6 2 | | | | | | | 5 2 |
| | 5 2 | | | | | | | 6 2 |
| | 6 2 | | | | | | | 6 2 |
| | 6 2 | | | | | | | 6 2 |
| | 6 2 | | | | | | | 5 2 |
| | 5 2 | | | | | | | 5 2 |
| | 5 2 | | | | | | | 5 2 |
| | 5 2 | | | | | | | 5 2 |
| | 5 2 | | | | | | | 6 2 |
| | 6 2 | | | | | | | 6 2 |
| | 6 2 | | | | | | | 5 2 |
| | 5 2 | | | | | | | 5 2 |
| 55.2.50 | 61039 | 112 | B4Mg2Ru5 | | | | | |
| | -GM | -R | -S | -T | -U | -X | -Y | -Z |
| | 5 2 | 5+5 4 | 6+6 4 | 3+4 4 | 3+4 4 | 3+4 4 | 3+4 4 | 6 2 |
| | 5 2 | 5+5 4 | 5+5 4 | 3+4 4 | 3+4 4 | 3+4 4 | 3+4 4 | 6 2 |
| | 5 2 | 6+6 4 | 6+6 4 | 3+4 4 | 3+4 4 | 3+4 4 | 3+4 4 | 5 2 |
| | 6 2 | 6+6 4 | 5+5 4 | 3+4 4 | 3+4 4 | 3+4 4 | 3+4 4 | 5 2 |
| | 6 2 | 6+6 4 | 5+5 4 | 3+4 4 | 3+4 4 | 3+4 4 | 3+4 4 | 6 2 |
| | 5 2 | 5+5 4 | 6+6 4 | 3+4 4 | 3+4 4 | 3+4 4 | 3+4 4 | 5 2 |
| | 6 2 | 5+5 4 | 5+5 4 | 3+4 4 | 3+4 4 | 3+4 4 | 3+4 4 | 6 2 |
| | 6 2 | 5+5 4 | 5+5 4 | 3+4 4 | 3+4 4 | 3+4 4 | 3+4 4 | 5 2 |
| | 5 2 | 5+5 4 | 5+5 4 | 3+4 4 | 3+4 4 | 3+4 4 | 3+4 4 | 5 2 |
| | 5 2 | 6+6 4 | 6+6 4 | 3+4 4 | 3+4 4 | 3+4 4 | 3+4 4 | 5 2 |
| | 6 2 | 6+6 4 | 6+6 4 | 3+4 4 | 3+4 4 | 3+4 4 | 3+4 4 | 5 2 |
| | 6 2 | 5+5 4 | 5+5 4 | 3+4 4 | 3+4 4 | 3+4 4 | 3+4 4 | 6 2 |
| | 6 2 | 5+5 4 | 6+6 4 | 3+4 4 | 3+4 4 | 3+4 4 | 3+4 4 | 6 2 |
| | 5 2 | 6+6 4 | 5+5 4 | 3+4 4 | 3+4 4 | 3+4 4 | 3+4 4 | 6 2 |
| | 5 2 | 5+5 4 | 6+6 4 | 3+4 4 | 3+4 4 | 3+4 4 | 3+4 4 | 5 2 |
| | 5 2 | 5+5 4 | 5+5 4 | 3+4 4 | 3+4 4 | 3+4 4 | 3+4 4 | 6 2 |
| | 6 2 | 6+6 4 | 6+6 4 | 3+4 4 | 3+4 4 | 3+4 4 | 3+4 4 | 5 2 |
| | 6 2 | 6+6 4 | 5+5 4 | 3+4 4 | 3+4 4 | 3+4 4 | 3+4 4 | 5 2 |
| | 5 2 | 6+6 4 | 6+6 4 | 3+4 4 | 3+4 4 | 3+4 4 | 3+4 4 | 6 2 |
| | 5 2 | 5+5 4 | 5+5 4 | 3+4 4 | 3+4 4 | 3+4 4 | 3+4 4 | 5 2 |
| | 5 2 | 5+5 4 | 6+6 4 | 3+4 4 | 3+4 4 | 3+4 4 | 3+4 4 | 5 2 |
| | 5 2 | 6+6 4 | 6+6 4 | 3+4 4 | 3+4 4 | 3+4 4 | 3+4 4 | 5 2 |
| | 5 2 | 6+6 4 | 5+5 4 | 3+4 4 | 3+4 4 | 3+4 4 | 3+4 4 | 6 2 |
| | 6 2 | 5+5 4 | 5+5 4 | 3+4 4 | 3+4 4 | 3+4 4 | 3+4 4 | 6 2 |
| | 5 2 | 6+6 4 | 5+5 4 | 3+4 4 | 3+4 4 | 3+4 4 | 3+4 4 | 6 2 |
| | 5 2 | 5+5 4 | 6+6 4 | 3+4 4 | 3+4 4 | 3+4 4 | 3+4 4 | 6 2 |
| | 6 2 | 5+5 4 | 5+5 4 | 3+4 4 | 3+4 4 | 3+4 4 | 3+4 4 | 5 2 |
| | 6 2 | 6+6 4 | 6+6 4 | 3+4 4 | 3+4 4 | 3+4 4 | 3+4 4 | 6 2 |
| | 6 2 | | | | | | | 6 2 |
| | 5 2 | | | | | | | 5 2 |
| | 6 2 | | | | | | | 5 2 |

Continued on next column

**TABLE XXXI– continued**

| TQC | ICSD | NoE | SF | | | | | |
|---|---|---|---|---|---|---|---|---|
| | 5 2 | | | | | | | 5 2 |
| | 5 2 | | | | | | | 5 2 |
| | 6 2 | | | | | | | 6 2 |
| | 5 2 | | | | | | | 6 2 |
| | 6 2 | | | | | | | 5 2 |
| | 5 2 | | | | | | | 5 2 |
| | 5 2 | | | | | | | 5 2 |
| | 6 2 | | | | | | | 5 2 |
| | 5 2 | | | | | | | 5 2 |
| | 5 2 | | | | | | | 6 2 |
| | 6 2 | | | | | | | 5 2 |
| | 6 2 | | | | | | | 6 2 |
| | 5 2 | | | | | | | 5 2 |
| | 6 2 | | | | | | | 6 2 |
| | 5 2 | | | | | | | 6 2 |
| | 6 2 | | | | | | | 5 2 |
| | 5 2 | | | | | | | 6 2 |
| | 6 2 | | | | | | | 5 2 |
| | 5 2 | | | | | | | 5 2 |
| | 5 2 | | | | | | | 6 2 |
| | 5 2 | | | | | | | 5 2 |
| | 6 2 | | | | | | | 6 2 |
| | 5 2 | | | | | | | 5 2 |
| | 5 2 | | | | | | | 5 2 |
| | 6 2 | | | | | | | 6 2 |
| 55.2.51 | 190184 | 176 | B4Fe1Yb1 | | | | | |
| | -GM | -R | -S | -T | -U | -X | -Y | -Z |
| | 5 2 | 5+5 4 | 5+5 4 | 3+4 4 | 3+4 4 | 3+4 4 | 3+4 4 | 5 2 |
| | 6 2 | 6+6 4 | 6+6 4 | 3+4 4 | 3+4 4 | 3+4 4 | 3+4 4 | 6 2 |
| | 5 2 | 5+5 4 | 5+5 4 | 3+4 4 | 3+4 4 | 3+4 4 | 3+4 4 | 5 2 |
| | 6 2 | 6+6 4 | 6+6 4 | 3+4 4 | 3+4 4 | 3+4 4 | 3+4 4 | 6 2 |
| | 5 2 | 6+6 4 | 5+5 4 | 3+4 4 | 3+4 4 | 3+4 4 | 3+4 4 | 5 2 |
| | 6 2 | 5+5 4 | 6+6 4 | 3+4 4 | 3+4 4 | 3+4 4 | 3+4 4 | 6 2 |
| | 5 2 | 5+5 4 | 5+5 4 | 3+4 4 | 3+4 4 | 3+4 4 | 3+4 4 | 5 2 |
| | 6 2 | 6+6 4 | 6+6 4 | 3+4 4 | 3+4 4 | 3+4 4 | 3+4 4 | 6 2 |
| | 5 2 | 5+5 4 | 6+6 4 | 3+4 4 | 3+4 4 | 3+4 4 | 3+4 4 | 6 2 |
| | 6 2 | 6+6 4 | 5+5 4 | 3+4 4 | 3+4 4 | 3+4 4 | 3+4 4 | 5 2 |
| | 5 2 | 6+6 4 | 5+5 4 | 3+4 4 | 3+4 4 | 3+4 4 | 3+4 4 | 6 2 |
| | 6 2 | 5+5 4 | 6+6 4 | 3+4 4 | 3+4 4 | 3+4 4 | 3+4 4 | 5 2 |
| | 5 2 | 5+5 4 | 5+5 4 | 3+4 4 | 3+4 4 | 3+4 4 | 3+4 4 | 5 2 |
| | 5 2 | 6+6 4 | 6+6 4 | 3+4 4 | 3+4 4 | 3+4 4 | 3+4 4 | 6 2 |
| | 6 2 | 6+6 4 | 5+5 4 | 3+4 4 | 3+4 4 | 3+4 4 | 3+4 4 | 5 2 |
| | 6 2 | 5+5 4 | 6+6 4 | 3+4 4 | 3+4 4 | 3+4 4 | 3+4 4 | 6 2 |
| | 5 2 | 6+6 4 | 5+5 4 | 3+4 4 | 3+4 4 | 3+4 4 | 3+4 4 | 6 2 |
| | 5 2 | 5+5 4 | 5+5 4 | 3+4 4 | 3+4 4 | 3+4 4 | 3+4 4 | 6 2 |
| | 6 2 | 6+6 4 | 6+6 4 | 3+4 4 | 3+4 4 | 3+4 4 | 3+4 4 | 5 2 |
| | 6 2 | 6+6 4 | 6+6 4 | 3+4 4 | 3+4 4 | 3+4 4 | 3+4 4 | 5 2 |
| | 5 2 | 5+5 4 | 6+6 4 | 3+4 4 | 3+4 4 | 3+4 4 | 3+4 4 | 6 2 |
| | 5 2 | 5+5 4 | 6+6 4 | 3+4 4 | 3+4 4 | 3+4 4 | 3+4 4 | 6 2 |
| | 6 2 | 5+5 4 | 5+5 4 | 3+4 4 | 3+4 4 | 3+4 4 | 3+4 4 | 5 2 |
| | 6 2 | 6+6 4 | 5+5 4 | 3+4 4 | 3+4 4 | 3+4 4 | 3+4 4 | 5 2 |
| | 5 2 | 5+5 4 | 5+5 4 | 3+4 4 | 3+4 4 | 3+4 4 | 3+4 4 | 6 2 |
| | 5 2 | 6+6 4 | 6+6 4 | 3+4 4 | 3+4 4 | 3+4 4 | 3+4 4 | 5 2 |
| | 6 2 | 5+5 4 | 5+5 4 | 3+4 4 | 3+4 4 | 3+4 4 | 3+4 4 | 6 2 |
| | 5 2 | 6+6 4 | 6+6 4 | 3+4 4 | 3+4 4 | 3+4 4 | 3+4 4 | 5 2 |
| | 6 2 | 5+5 4 | 6+6 4 | 3+4 4 | 3+4 4 | 3+4 4 | 3+4 4 | 6 2 |
| | 6 2 | 6+6 4 | 5+5 4 | 3+4 4 | 3+4 4 | 3+4 4 | 3+4 4 | 5 2 |
| | 5 2 | 5+5 4 | 5+5 4 | 3+4 4 | 3+4 4 | 3+4 4 | 3+4 4 | 5 2 |
| | 6 2 | 6+6 4 | 6+6 4 | 3+4 4 | 3+4 4 | 3+4 4 | 3+4 4 | 6 2 |
| | 5 2 | 5+5 4 | 6+6 4 | 3+4 4 | 3+4 4 | 3+4 4 | 3+4 4 | 6 2 |
| | 6 2 | 6+6 4 | 5+5 4 | 3+4 4 | 3+4 4 | 3+4 4 | 3+4 4 | 5 2 |

Continued on next column

**TABLE XXXI– continued**

| TQC | ICSD | NoE | SF | | | | | |
|---|---|---|---|---|---|---|---|---|
| | 6 2 | 6+6 4 | 5+5 4 | 3+4 4 | 3+4 4 | 3+4 4 | 3+4 4 | 5 2 |
| | 5 2 | 5+5 4 | 6+6 4 | 3+4 4 | 3+4 4 | 3+4 4 | 3+4 4 | 6 2 |
| | 5 2 | 6+6 4 | 6+6 4 | 3+4 4 | 3+4 4 | 3+4 4 | 3+4 4 | 6 2 |
| | 5 2 | 6+6 4 | 5+5 4 | 3+4 4 | 3+4 4 | 3+4 4 | 3+4 4 | 6 2 |
| | 5 2 | 5+5 4 | 6+6 4 | 3+4 4 | 3+4 4 | 3+4 4 | 3+4 4 | 5 2 |
| | 6 2 | 5+5 4 | 5+5 4 | 3+4 4 | 3+4 4 | 3+4 4 | 3+4 4 | 6 2 |
| | 6 2 | 6+6 4 | 5+5 4 | 3+4 4 | 3+4 4 | 3+4 4 | 3+4 4 | 6 2 |
| | 6 2 | 5+5 4 | 6+6 4 | 3+4 4 | 3+4 4 | 3+4 4 | 3+4 4 | 5 2 |
| | 6 2 | 6+6 4 | 5+5 4 | 3+4 4 | 3+4 4 | 3+4 4 | 3+4 4 | 5 2 |
| | 5 2 | 5+5 4 | 6+6 4 | 3+4 4 | 3+4 4 | 3+4 4 | 3+4 4 | 6 2 |
| | 5 2 | | | | | | | 5 2 |
| | 6 2 | | | | | | | 6 2 |
| | 5 2 | | | | | | | 5 2 |
| | 5 2 | | | | | | | 6 2 |
| | 5 2 | | | | | | | 5 2 |
| | 6 2 | | | | | | | 6 2 |
| | 6 2 | | | | | | | 6 2 |
| | 5 2 | | | | | | | 6 2 |
| | 5 2 | | | | | | | 5 2 |
| | 6 2 | | | | | | | 5 2 |
| | 6 2 | | | | | | | 5 2 |
| | 5 2 | | | | | | | 6 2 |
| | 6 2 | | | | | | | 5 2 |
| | 5 2 | | | | | | | 6 2 |
| | 5 2 | | | | | | | 5 2 |
| | 5 2 | | | | | | | 6 2 |
| | 6 2 | | | | | | | 5 2 |
| | 5 2 | | | | | | | 5 2 |
| | 6 2 | | | | | | | 6 2 |
| | 6 2 | | | | | | | 6 2 |
| | 5 2 | | | | | | | 5 2 |
| | 6 2 | | | | | | | 6 2 |
| | 6 2 | | | | | | | 5 2 |
| | 5 2 | | | | | | | 5 2 |
| | 6 2 | | | | | | | 6 2 |
| | 5 2 | | | | | | | 6 2 |
| | 5 2 | | | | | | | 6 2 |
| | 6 2 | | | | | | | 6 2 |
| | 5 2 | | | | | | | 5 2 |
| | 6 2 | | | | | | | 5 2 |
| | 6 2 | | | | | | | 5 2 |
| | 5 2 | | | | | | | 5 2 |
| | 6 2 | | | | | | | 5 2 |
| | 5 2 | | | | | | | 5 2 |
| | 5 2 | | | | | | | 6 2 |
| | 6 2 | | | | | | | 6 2 |
| | 6 2 | | | | | | | 5 2 |
| | 5 2 | | | | | | | 6 2 |
| | 5 2 | | | | | | | 5 2 |
| | 6 2 | | | | | | | 5 2 |
| | 6 2 | | | | | | | 6 2 |
| | 5 2 | | | | | | | 5 2 |
| | 6 2 | | | | | | | 5 2 |
| | 6 2 | | | | | | | 6 2 |
| 55.2.52 | 409626 | 156 | B4Er1V1 | | | | | |
| | -GM | -R | -S | -T | -U | -X | -Y | -Z |
| | 5 2 | 6+6 4 | 5+5 4 | 3+4 4 | 3+4 4 | 3+4 4 | 3+4 4 | 6 2 |
| | 6 2 | 5+5 4 | 6+6 4 | 3+4 4 | 3+4 4 | 3+4 4 | 3+4 4 | 5 2 |
| | 5 2 | 6+6 4 | 5+5 4 | 3+4 4 | 3+4 4 | 3+4 4 | 3+4 4 | 6 2 |
| | 6 2 | 5+5 4 | 6+6 4 | 3+4 4 | 3+4 4 | 3+4 4 | 3+4 4 | 5 2 |
| | 5 2 | 5+5 4 | 5+5 4 | 3+4 4 | 3+4 4 | 3+4 4 | 3+4 4 | 6 2 |

Continued on next column

**TABLE XXXI– continued**

| TQC | ICSD | NoE | SF | | | | | |
|---|---|---|---|---|---|---|---|---|
| | 6 2 | 6+6 4 | 6+6 4 | 3+4 4 | 3+4 4 | 3+4 4 | 3+4 4 | 5 2 |
| | 5 2 | 6+6 4 | 5+5 4 | 3+4 4 | 3+4 4 | 3+4 4 | 3+4 4 | 6 2 |
| | 6 2 | 5+5 4 | 6+6 4 | 3+4 4 | 3+4 4 | 3+4 4 | 3+4 4 | 5 2 |
| | 5 2 | 6+6 4 | 6+6 4 | 3+4 4 | 3+4 4 | 3+4 4 | 3+4 4 | 5 2 |
| | 6 2 | 5+5 4 | 5+5 4 | 3+4 4 | 3+4 4 | 3+4 4 | 3+4 4 | 6 2 |
| | 5 2 | 5+5 4 | 5+5 4 | 3+4 4 | 3+4 4 | 3+4 4 | 3+4 4 | 5 2 |
| | 6 2 | 6+6 4 | 6+6 4 | 3+4 4 | 3+4 4 | 3+4 4 | 3+4 4 | 6 2 |
| | 5 2 | 6+6 4 | 5+5 4 | 3+4 4 | 3+4 4 | 3+4 4 | 3+4 4 | 6 2 |
| | 6 2 | 5+5 4 | 6+6 4 | 3+4 4 | 3+4 4 | 3+4 4 | 3+4 4 | 5 2 |
| | 5 2 | 5+5 4 | 5+5 4 | 3+4 4 | 3+4 4 | 3+4 4 | 3+4 4 | 6 2 |
| | 6 2 | 5+5 4 | 6+6 4 | 3+4 4 | 3+4 4 | 3+4 4 | 3+4 4 | 5 2 |
| | 5 2 | 6+6 4 | 5+5 4 | 3+4 4 | 3+4 4 | 3+4 4 | 3+4 4 | 5 2 |
| | 5 2 | 6+6 4 | 6+6 4 | 3+4 4 | 3+4 4 | 3+4 4 | 3+4 4 | 5 2 |
| | 6 2 | 6+6 4 | 5+5 4 | 3+4 4 | 3+4 4 | 3+4 4 | 3+4 4 | 6 2 |
| | 6 2 | 5+5 4 | 6+6 4 | 3+4 4 | 3+4 4 | 3+4 4 | 3+4 4 | 6 2 |
| | 5 2 | 5+5 4 | 6+6 4 | 3+4 4 | 3+4 4 | 3+4 4 | 3+4 4 | 5 2 |
| | 5 2 | 6+6 4 | 5+5 4 | 3+4 4 | 3+4 4 | 3+4 4 | 3+4 4 | 5 2 |
| | 6 2 | 6+6 4 | 6+6 4 | 3+4 4 | 3+4 4 | 3+4 4 | 3+4 4 | 6 2 |
| | 6 2 | 5+5 4 | 5+5 4 | 3+4 4 | 3+4 4 | 3+4 4 | 3+4 4 | 6 2 |
| | 5 2 | 6+6 4 | 6+6 4 | 3+4 4 | 3+4 4 | 3+4 4 | 3+4 4 | 5 2 |
| | 5 2 | 6+6 4 | 5+5 4 | 3+4 4 | 3+4 4 | 3+4 4 | 3+4 4 | 5 2 |
| | 6 2 | 5+5 4 | 6+6 4 | 3+4 4 | 3+4 4 | 3+4 4 | 3+4 4 | 6 2 |
| | 5 2 | 5+5 4 | 5+5 4 | 3+4 4 | 3+4 4 | 3+4 4 | 3+4 4 | 6 2 |
| | 6 2 | 6+6 4 | 6+6 4 | 3+4 4 | 3+4 4 | 3+4 4 | 3+4 4 | 5 2 |
| | 6 2 | 6+6 4 | 5+5 4 | 3+4 4 | 3+4 4 | 3+4 4 | 3+4 4 | 6 2 |
| | 5 2 | 5+5 4 | 6+6 4 | 3+4 4 | 3+4 4 | 3+4 4 | 3+4 4 | 6 2 |
| | 6 2 | 5+5 4 | 5+5 4 | 3+4 4 | 3+4 4 | 3+4 4 | 3+4 4 | 5 2 |
| | 5 2 | 6+6 4 | 5+5 4 | 3+4 4 | 3+4 4 | 3+4 4 | 3+4 4 | 5 2 |
| | 6 2 | 5+5 4 | 6+6 4 | 3+4 4 | 3+4 4 | 3+4 4 | 3+4 4 | 6 2 |
| | 6 2 | 6+6 4 | 5+5 4 | 3+4 4 | 3+4 4 | 3+4 4 | 3+4 4 | 6 2 |
| | 5 2 | 5+5 4 | 6+6 4 | 3+4 4 | 3+4 4 | 3+4 4 | 3+4 4 | 5 2 |
| | 5 2 | 6+6 4 | 6+6 4 | 3+4 4 | 3+4 4 | 3+4 4 | 3+4 4 | 5 2 |
| | 6 2 | 5+5 4 | 5+5 4 | 3+4 4 | 3+4 4 | 3+4 4 | 3+4 4 | 5 2 |
| | 5 2 | 6+6 4 | 5+5 4 | 3+4 4 | 3+4 4 | 3+4 4 | 3+4 4 | 5 2 |
| | 6 2 | | | | | | | 6 2 |
| | 5 2 | | | | | | | 6 2 |
| | 6 2 | | | | | | | 5 2 |
| | 6 2 | | | | | | | 5 2 |
| | 5 2 | | | | | | | 6 2 |
| | 6 2 | | | | | | | 5 2 |
| | 5 2 | | | | | | | 6 2 |
| | 5 2 | | | | | | | 6 2 |
| | 5 2 | | | | | | | 5 2 |
| | 5 2 | | | | | | | 6 2 |
| | 6 2 | | | | | | | 5 2 |
| | 6 2 | | | | | | | 5 2 |
| | 5 2 | | | | | | | 5 2 |
| | 5 2 | | | | | | | 6 2 |
| | 6 2 | | | | | | | 6 2 |
| | 6 2 | | | | | | | 6 2 |
| | 5 2 | | | | | | | 5 2 |
| | 6 2 | | | | | | | 6 2 |
| | 5 2 | | | | | | | 5 2 |
| | 5 2 | | | | | | | 6 2 |
| | 6 2 | | | | | | | 5 2 |
| | 5 2 | | | | | | | 5 2 |
| | 5 2 | | | | | | | 6 2 |
| | 6 2 | | | | | | | 6 2 |
| | 6 2 | | | | | | | 5 2 |
| | 5 2 | | | | | | | 6 2 |
| | 6 2 | | | | | | | 5 2 |

Continued on next column

**TABLE XXXI– continued**

| TQC | ICSD | NoE | SF | | | | | |
|---|---|---|---|---|---|---|---|---|
| | 6 2 | | | | | | | 5 2 |
| | 5 2 | | | | | | | 6 2 |
| | 6 2 | | | | | | | 5 2 |
| | 5 2 | | | | | | | 6 2 |
| | 5 2 | | | | | | | 6 2 |
| | 6 2 | | | | | | | 5 2 |
| | 6 2 | | | | | | | 5 2 |
| | 5 2 | | | | | | | 6 2 |
| | 5 2 | | | | | | | 5 2 |
| | 6 2 | | | | | | | 6 2 |
| | 5 2 | | | | | | | 6 2 |
| | 6 2 | | | | | | | 5 2 |
| 55.2.53 | 81542 | 120 | B4Cr1Th1 | | | | | |
| | -GM | -R | -S | -T | -U | -X | -Y | -Z |
| | 5 2 | 6+6 4 | 5+5 4 | 3+4 4 | 3+4 4 | 3+4 4 | 3+4 4 | 6 2 |
| | 6 2 | 5+5 4 | 6+6 4 | 3+4 4 | 3+4 4 | 3+4 4 | 3+4 4 | 5 2 |
| | 5 2 | 6+6 4 | 5+5 4 | 3+4 4 | 3+4 4 | 3+4 4 | 3+4 4 | 6 2 |
| | 6 2 | 5+5 4 | 6+6 4 | 3+4 4 | 3+4 4 | 3+4 4 | 3+4 4 | 5 2 |
| | 5 2 | 5+5 4 | 5+5 4 | 3+4 4 | 3+4 4 | 3+4 4 | 3+4 4 | 6 2 |
| | 6 2 | 6+6 4 | 6+6 4 | 3+4 4 | 3+4 4 | 3+4 4 | 3+4 4 | 5 2 |
| | 5 2 | 6+6 4 | 5+5 4 | 3+4 4 | 3+4 4 | 3+4 4 | 3+4 4 | 6 2 |
| | 6 2 | 5+5 4 | 6+6 4 | 3+4 4 | 3+4 4 | 3+4 4 | 3+4 4 | 5 2 |
| | 5 2 | 6+6 4 | 6+6 4 | 3+4 4 | 3+4 4 | 3+4 4 | 3+4 4 | 5 2 |
| | 6 2 | 5+5 4 | 5+5 4 | 3+4 4 | 3+4 4 | 3+4 4 | 3+4 4 | 6 2 |
| | 5 2 | 5+5 4 | 5+5 4 | 3+4 4 | 3+4 4 | 3+4 4 | 3+4 4 | 5 2 |
| | 6 2 | 6+6 4 | 6+6 4 | 3+4 4 | 3+4 4 | 3+4 4 | 3+4 4 | 6 2 |
| | 5 2 | 6+6 4 | 6+6 4 | 3+4 4 | 3+4 4 | 3+4 4 | 3+4 4 | 6 2 |
| | 6 2 | 5+5 4 | 5+5 4 | 3+4 4 | 3+4 4 | 3+4 4 | 3+4 4 | 5 2 |
| | 5 2 | 5+5 4 | 5+5 4 | 3+4 4 | 3+4 4 | 3+4 4 | 3+4 4 | 6 2 |
| | 6 2 | 5+5 4 | 6+6 4 | 3+4 4 | 3+4 4 | 3+4 4 | 3+4 4 | 5 2 |
| | 5 2 | 6+6 4 | 5+5 4 | 3+4 4 | 3+4 4 | 3+4 4 | 3+4 4 | 5 2 |
| | 5 2 | 6+6 4 | 5+5 4 | 3+4 4 | 3+4 4 | 3+4 4 | 3+4 4 | 5 2 |
| | 6 2 | 5+5 4 | 6+6 4 | 3+4 4 | 3+4 4 | 3+4 4 | 3+4 4 | 6 2 |
| | 6 2 | 6+6 4 | 6+6 4 | 3+4 4 | 3+4 4 | 3+4 4 | 3+4 4 | 6 2 |
| | 5 2 | 5+5 4 | 5+5 4 | 3+4 4 | 3+4 4 | 3+4 4 | 3+4 4 | 5 2 |
| | 5 2 | 6+6 4 | 6+6 4 | 3+4 4 | 3+4 4 | 3+4 4 | 3+4 4 | 5 2 |
| | 6 2 | 5+5 4 | 6+6 4 | 3+4 4 | 3+4 4 | 3+4 4 | 3+4 4 | 6 2 |
| | 6 2 | 6+6 4 | 5+5 4 | 3+4 4 | 3+4 4 | 3+4 4 | 3+4 4 | 6 2 |
| | 5 2 | 6+6 4 | 6+6 4 | 3+4 4 | 3+4 4 | 3+4 4 | 3+4 4 | 5 2 |
| | 5 2 | 5+5 4 | 5+5 4 | 3+4 4 | 3+4 4 | 3+4 4 | 3+4 4 | 5 2 |
| | 6 2 | 6+6 4 | 5+5 4 | 3+4 4 | 3+4 4 | 3+4 4 | 3+4 4 | 6 2 |
| | 5 2 | 5+5 4 | 6+6 4 | 3+4 4 | 3+4 4 | 3+4 4 | 3+4 4 | 5 2 |
| | 6 2 | 5+5 4 | 5+5 4 | 3+4 4 | 3+4 4 | 3+4 4 | 3+4 4 | 6 2 |
| | 6 2 | 6+6 4 | 5+5 4 | 3+4 4 | 3+4 4 | 3+4 4 | 3+4 4 | 6 2 |
| | 5 2 | | | | | | | 6 2 |
| | 6 2 | | | | | | | 5 2 |
| | 5 2 | | | | | | | 5 2 |
| | 6 2 | | | | | | | 6 2 |
| | 6 2 | | | | | | | 5 2 |
| | 5 2 | | | | | | | 6 2 |
| | 5 2 | | | | | | | 5 2 |
| | 6 2 | | | | | | | 6 2 |
| | 6 2 | | | | | | | 5 2 |
| | 6 2 | | | | | | | 5 2 |
| | 5 2 | | | | | | | 6 2 |
| | 5 2 | | | | | | | 5 2 |
| | 6 2 | | | | | | | 6 2 |
| | 5 2 | | | | | | | 5 2 |
| | 5 2 | | | | | | | 6 2 |
| | 5 2 | | | | | | | 5 2 |
| | 6 2 | | | | | | | 6 2 |

Continued on next column

**TABLE XXXI– continued**

| TQC | ICSD | NoE | SF | | | | | |
|---|---|---|---|---|---|---|---|---|
| | 5 2 | | | | | | | 5 2 |
| | 5 2 | | | | | | | 5 2 |
| | 6 2 | | | | | | | 5 2 |
| | 5 2 | | | | | | | 5 2 |
| | 6 2 | | | | | | | 6 2 |
| | 5 2 | | | | | | | 6 2 |
| | 6 2 | | | | | | | 5 2 |
| | 6 2 | | | | | | | 6 2 |
| | 5 2 | | | | | | | 6 2 |
| | 5 2 | | | | | | | 5 2 |
| | 6 2 | | | | | | | 6 2 |
| | 5 2 | | | | | | | 6 2 |
| | 6 2 | | | | | | | 5 2 |
| 55.2.54 | 250251 | 180 | B4Co1Yb1 | | | | | |
| | -GM | -R | -S | -T | -U | -X | -Y | -Z |
| | 5 2 | 5+5 4 | 5+5 4 | 3+4 4 | 3+4 4 | 3+4 4 | 3+4 4 | 5 2 |
| | 6 2 | 6+6 4 | 6+6 4 | 3+4 4 | 3+4 4 | 3+4 4 | 3+4 4 | 6 2 |
| | 5 2 | 5+5 4 | 5+5 4 | 3+4 4 | 3+4 4 | 3+4 4 | 3+4 4 | 5 2 |
| | 6 2 | 6+6 4 | 6+6 4 | 3+4 4 | 3+4 4 | 3+4 4 | 3+4 4 | 6 2 |
| | 5 2 | 6+6 4 | 5+5 4 | 3+4 4 | 3+4 4 | 3+4 4 | 3+4 4 | 5 2 |
| | 6 2 | 5+5 4 | 6+6 4 | 3+4 4 | 3+4 4 | 3+4 4 | 3+4 4 | 6 2 |
| | 5 2 | 5+5 4 | 5+5 4 | 3+4 4 | 3+4 4 | 3+4 4 | 3+4 4 | 5 2 |
| | 6 2 | 6+6 4 | 6+6 4 | 3+4 4 | 3+4 4 | 3+4 4 | 3+4 4 | 6 2 |
| | 5 2 | 5+5 4 | 6+6 4 | 3+4 4 | 3+4 4 | 3+4 4 | 3+4 4 | 6 2 |
| | 6 2 | 6+6 4 | 5+5 4 | 3+4 4 | 3+4 4 | 3+4 4 | 3+4 4 | 5 2 |
| | 5 2 | 6+6 4 | 5+5 4 | 3+4 4 | 3+4 4 | 3+4 4 | 3+4 4 | 6 2 |
| | 6 2 | 5+5 4 | 6+6 4 | 3+4 4 | 3+4 4 | 3+4 4 | 3+4 4 | 5 2 |
| | 5 2 | 5+5 4 | 5+5 4 | 3+4 4 | 3+4 4 | 3+4 4 | 3+4 4 | 5 2 |
| | 6 2 | 6+6 4 | 6+6 4 | 3+4 4 | 3+4 4 | 3+4 4 | 3+4 4 | 6 2 |
| | 5 2 | 6+6 4 | 5+5 4 | 3+4 4 | 3+4 4 | 3+4 4 | 3+4 4 | 5 2 |
| | 6 2 | 5+5 4 | 6+6 4 | 3+4 4 | 3+4 4 | 3+4 4 | 3+4 4 | 6 2 |
| | 5 2 | 6+6 4 | 5+5 4 | 3+4 4 | 3+4 4 | 3+4 4 | 3+4 4 | 6 2 |
| | 5 2 | 5+5 4 | 5+5 4 | 3+4 4 | 3+4 4 | 3+4 4 | 3+4 4 | 6 2 |
| | 6 2 | 6+6 4 | 6+6 4 | 3+4 4 | 3+4 4 | 3+4 4 | 3+4 4 | 5 2 |
| | 6 2 | 5+5 4 | 6+6 4 | 3+4 4 | 3+4 4 | 3+4 4 | 3+4 4 | 5 2 |
| | 5 2 | 6+6 4 | 6+6 4 | 3+4 4 | 3+4 4 | 3+4 4 | 3+4 4 | 6 2 |
| | 5 2 | 5+5 4 | 5+5 4 | 3+4 4 | 3+4 4 | 3+4 4 | 3+4 4 | 6 2 |
| | 6 2 | 5+5 4 | 6+6 4 | 3+4 4 | 3+4 4 | 3+4 4 | 3+4 4 | 5 2 |
| | 6 2 | 6+6 4 | 5+5 4 | 3+4 4 | 3+4 4 | 3+4 4 | 3+4 4 | 5 2 |
| | 5 2 | 5+5 4 | 5+5 4 | 3+4 4 | 3+4 4 | 3+4 4 | 3+4 4 | 6 2 |
| | 5 2 | 6+6 4 | 6+6 4 | 3+4 4 | 3+4 4 | 3+4 4 | 3+4 4 | 5 2 |
| | 6 2 | 5+5 4 | 5+5 4 | 3+4 4 | 3+4 4 | 3+4 4 | 3+4 4 | 6 2 |
| | 5 2 | 6+6 4 | 6+6 4 | 3+4 4 | 3+4 4 | 3+4 4 | 3+4 4 | 5 2 |
| | 6 2 | 6+6 4 | 5+5 4 | 3+4 4 | 3+4 4 | 3+4 4 | 3+4 4 | 5 2 |
| | 6 2 | 5+5 4 | 6+6 4 | 3+4 4 | 3+4 4 | 3+4 4 | 3+4 4 | 6 2 |
| | 5 2 | 6+6 4 | 5+5 4 | 3+4 4 | 3+4 4 | 3+4 4 | 3+4 4 | 5 2 |
| | 6 2 | 5+5 4 | 6+6 4 | 3+4 4 | 3+4 4 | 3+4 4 | 3+4 4 | 6 2 |
| | 5 2 | 5+5 4 | 6+6 4 | 3+4 4 | 3+4 4 | 3+4 4 | 3+4 4 | 6 2 |
| | 6 2 | 6+6 4 | 5+5 4 | 3+4 4 | 3+4 4 | 3+4 4 | 3+4 4 | 5 2 |
| | 6 2 | 6+6 4 | 5+5 4 | 3+4 4 | 3+4 4 | 3+4 4 | 3+4 4 | 5 2 |
| | 5 2 | 5+5 4 | 6+6 4 | 3+4 4 | 3+4 4 | 3+4 4 | 3+4 4 | 6 2 |
| | 5 2 | 6+6 4 | 5+5 4 | 3+4 4 | 3+4 4 | 3+4 4 | 3+4 4 | 6 2 |
| | 5 2 | 6+6 4 | 6+6 4 | 3+4 4 | 3+4 4 | 3+4 4 | 3+4 4 | 6 2 |
| | 6 2 | 5+5 4 | 5+5 4 | 3+4 4 | 3+4 4 | 3+4 4 | 3+4 4 | 5 2 |
| | 5 2 | 6+6 4 | 6+6 4 | 3+4 4 | 3+4 4 | 3+4 4 | 3+4 4 | 5 2 |
| | 6 2 | 5+5 4 | 5+5 4 | 3+4 4 | 3+4 4 | 3+4 4 | 3+4 4 | 6 2 |
| | 6 2 | 5+5 4 | 6+6 4 | 3+4 4 | 3+4 4 | 3+4 4 | 3+4 4 | 5 2 |
| | 6 2 | 6+6 4 | 5+5 4 | 3+4 4 | 3+4 4 | 3+4 4 | 3+4 4 | 6 2 |
| | 5 2 | 5+5 4 | 6+6 4 | 3+4 4 | 3+4 4 | 3+4 4 | 3+4 4 | 6 2 |
| | 5 2 | 6+6 4 | 5+5 4 | 3+4 4 | 3+4 4 | 3+4 4 | 3+4 4 | 5 2 |
| | 6 2 | | | | | | | 6 2 |

Continued on next column

**TABLE XXXI– continued**

| TQC | ICSD | NoE | SF | | | | | |
|---|---|---|---|---|---|---|---|---|
| | 5 2 | | | | | | | 5 2 |
| | 5 2 | | | | | | | 6 2 |
| | 5 2 | | | | | | | 5 2 |
| | 6 2 | | | | | | | 6 2 |
| | 6 2 | | | | | | | 6 2 |
| | 5 2 | | | | | | | 5 2 |
| | 5 2 | | | | | | | 5 2 |
| | 6 2 | | | | | | | 6 2 |
| | 6 2 | | | | | | | 5 2 |
| | 5 2 | | | | | | | 6 2 |
| | 5 2 | | | | | | | 5 2 |
| | 6 2 | | | | | | | 5 2 |
| | 5 2 | | | | | | | 6 2 |
| | 6 2 | | | | | | | 5 2 |
| | 5 2 | | | | | | | 6 2 |
| | 6 2 | | | | | | | 5 2 |
| | 5 2 | | | | | | | 5 2 |
| | 6 2 | | | | | | | 6 2 |
| | 6 2 | | | | | | | 6 2 |
| | 5 2 | | | | | | | 5 2 |
| | 5 2 | | | | | | | 5 2 |
| | 6 2 | | | | | | | 6 2 |
| | 6 2 | | | | | | | 6 2 |
| | 5 2 | | | | | | | 6 2 |
| | 6 2 | | | | | | | 5 2 |
| | 5 2 | | | | | | | 6 2 |
| | 6 2 | | | | | | | 5 2 |
| | 5 2 | | | | | | | 5 2 |
| | 6 2 | | | | | | | 5 2 |
| | 5 2 | | | | | | | 6 2 |
| | 6 2 | | | | | | | 5 2 |
| | 5 2 | | | | | | | 6 2 |
| | 5 2 | | | | | | | 5 2 |
| | 6 2 | | | | | | | 5 2 |
| | 6 2 | | | | | | | 6 2 |
| | 5 2 | | | | | | | 5 2 |
| | 6 2 | | | | | | | 6 2 |
| | 6 2 | | | | | | | 5 2 |
| | 5 2 | | | | | | | 6 2 |
| | 6 2 | | | | | | | 5 2 |
| | 6 2 | | | | | | | 5 2 |
| | 5 2 | | | | | | | 6 2 |
| | 5 2 | | | | | | | 5 2 |
| | 5 2 | | | | | | | 6 2 |
| 55.2.55 | 417873 | 156 | B4Co1Gd1 | | | | | |
| | -GM | -R | -S | -T | -U | -X | -Y | -Z |
| | 5 2 | 5+5 4 | 5+5 4 | 3+4 4 | 3+4 4 | 3+4 4 | 3+4 4 | 5 2 |
| | 6 2 | 6+6 4 | 6+6 4 | 3+4 4 | 3+4 4 | 3+4 4 | 3+4 4 | 6 2 |
| | 5 2 | 5+5 4 | 5+5 4 | 3+4 4 | 3+4 4 | 3+4 4 | 3+4 4 | 5 2 |
| | 6 2 | 6+6 4 | 6+6 4 | 3+4 4 | 3+4 4 | 3+4 4 | 3+4 4 | 6 2 |
| | 5 2 | 6+6 4 | 5+5 4 | 3+4 4 | 3+4 4 | 3+4 4 | 3+4 4 | 5 2 |
| | 6 2 | 5+5 4 | 6+6 4 | 3+4 4 | 3+4 4 | 3+4 4 | 3+4 4 | 6 2 |
| | 5 2 | 5+5 4 | 5+5 4 | 3+4 4 | 3+4 4 | 3+4 4 | 3+4 4 | 5 2 |
| | 6 2 | 6+6 4 | 6+6 4 | 3+4 4 | 3+4 4 | 3+4 4 | 3+4 4 | 6 2 |
| | 5 2 | 5+5 4 | 6+6 4 | 3+4 4 | 3+4 4 | 3+4 4 | 3+4 4 | 6 2 |
| | 6 2 | 6+6 4 | 5+5 4 | 3+4 4 | 3+4 4 | 3+4 4 | 3+4 4 | 5 2 |
| | 5 2 | 6+6 4 | 5+5 4 | 3+4 4 | 3+4 4 | 3+4 4 | 3+4 4 | 6 2 |
| | 6 2 | 5+5 4 | 6+6 4 | 3+4 4 | 3+4 4 | 3+4 4 | 3+4 4 | 5 2 |
| | 5 2 | 5+5 4 | 6+6 4 | 3+4 4 | 3+4 4 | 3+4 4 | 3+4 4 | 5 2 |
| | 6 2 | 6+6 4 | 5+5 4 | 3+4 4 | 3+4 4 | 3+4 4 | 3+4 4 | 6 2 |
| | 5 2 | 6+6 4 | 5+5 4 | 3+4 4 | 3+4 4 | 3+4 4 | 3+4 4 | 5 2 |

Continued on next column

**TABLE XXXI– continued**

| TQC | ICSD | NoE | SF | | | | | |
|---|---|---|---|---|---|---|---|---|
| | 6 2 | 6+6 4 | 6+6 4 | 3+4 4 | 3+4 4 | 3+4 4 | 3+4 4 | 6 2 |
| | 5 2 | 5+5 4 | 5+5 4 | 3+4 4 | 3+4 4 | 3+4 4 | 3+4 4 | 6 2 |
| | 5 2 | 5+5 4 | 5+5 4 | 3+4 4 | 3+4 4 | 3+4 4 | 3+4 4 | 6 2 |
| | 6 2 | 6+6 4 | 6+6 4 | 3+4 4 | 3+4 4 | 3+4 4 | 3+4 4 | 5 2 |
| | 6 2 | 5+5 4 | 6+6 4 | 3+4 4 | 3+4 4 | 3+4 4 | 3+4 4 | 5 2 |
| | 5 2 | 6+6 4 | 6+6 4 | 3+4 4 | 3+4 4 | 3+4 4 | 3+4 4 | 6 2 |
| | 6 2 | 5+5 4 | 6+6 4 | 3+4 4 | 3+4 4 | 3+4 4 | 3+4 4 | 5 2 |
| | 5 2 | 5+5 4 | 5+5 4 | 3+4 4 | 3+4 4 | 3+4 4 | 3+4 4 | 6 2 |
| | 6 2 | 6+6 4 | 5+5 4 | 3+4 4 | 3+4 4 | 3+4 4 | 3+4 4 | 5 2 |
| | 5 2 | 5+5 4 | 5+5 4 | 3+4 4 | 3+4 4 | 3+4 4 | 3+4 4 | 6 2 |
| | 5 2 | 6+6 4 | 6+6 4 | 3+4 4 | 3+4 4 | 3+4 4 | 3+4 4 | 5 2 |
| | 6 2 | 5+5 4 | 6+6 4 | 3+4 4 | 3+4 4 | 3+4 4 | 3+4 4 | 6 2 |
| | 5 2 | 6+6 4 | 5+5 4 | 3+4 4 | 3+4 4 | 3+4 4 | 3+4 4 | 5 2 |
| | 6 2 | 6+6 4 | 5+5 4 | 3+4 4 | 3+4 4 | 3+4 4 | 3+4 4 | 5 2 |
| | 5 2 | 5+5 4 | 6+6 4 | 3+4 4 | 3+4 4 | 3+4 4 | 3+4 4 | 6 2 |
| | 6 2 | 6+6 4 | 6+6 4 | 3+4 4 | 3+4 4 | 3+4 4 | 3+4 4 | 5 2 |
| | 6 2 | 6+6 4 | 5+5 4 | 3+4 4 | 3+4 4 | 3+4 4 | 3+4 4 | 6 2 |
| | 6 2 | 5+5 4 | 5+5 4 | 3+4 4 | 3+4 4 | 3+4 4 | 3+4 4 | 6 2 |
| | 6 2 | 5+5 4 | 6+6 4 | 3+4 4 | 3+4 4 | 3+4 4 | 3+4 4 | 5 2 |
| | 5 2 | 6+6 4 | 5+5 4 | 3+4 4 | 3+4 4 | 3+4 4 | 3+4 4 | 6 2 |
| | 5 2 | 5+5 4 | 6+6 4 | 3+4 4 | 3+4 4 | 3+4 4 | 3+4 4 | 5 2 |
| | 5 2 | 6+6 4 | 5+5 4 | 3+4 4 | 3+4 4 | 3+4 4 | 3+4 4 | 6 2 |
| | 5 2 | 6+6 4 | 6+6 4 | 3+4 4 | 3+4 4 | 3+4 4 | 3+4 4 | 6 2 |
| | 6 2 | 5+5 4 | 5+5 4 | 3+4 4 | 3+4 4 | 3+4 4 | 3+4 4 | 5 2 |
| | 6 2 | | | | | | | 5 2 |
| | 5 2 | | | | | | | 5 2 |
| | 6 2 | | | | | | | 6 2 |
| | 6 2 | | | | | | | 6 2 |
| | 5 2 | | | | | | | 5 2 |
| | 5 2 | | | | | | | 6 2 |
| | 6 2 | | | | | | | 6 2 |
| | 5 2 | | | | | | | 5 2 |
| | 5 2 | | | | | | | 6 2 |
| | 6 2 | | | | | | | 5 2 |
| | 5 2 | | | | | | | 6 2 |
| | 6 2 | | | | | | | 6 2 |
| | 5 2 | | | | | | | 5 2 |
| | 5 2 | | | | | | | 5 2 |
| | 6 2 | | | | | | | 5 2 |
| | 6 2 | | | | | | | 6 2 |
| | 5 2 | | | | | | | 5 2 |
| | 6 2 | | | | | | | 6 2 |
| | 5 2 | | | | | | | 6 2 |
| | 5 2 | | | | | | | 6 2 |
| | 6 2 | | | | | | | 6 2 |
| | 6 2 | | | | | | | 5 2 |
| | 5 2 | | | | | | | 5 2 |
| | 5 2 | | | | | | | 5 2 |
| | 6 2 | | | | | | | 5 2 |
| | 6 2 | | | | | | | 5 2 |
| | 5 2 | | | | | | | 6 2 |
| | 5 2 | | | | | | | 5 2 |
| | 6 2 | | | | | | | 6 2 |
| | 5 2 | | | | | | | 5 2 |
| | 6 2 | | | | | | | 5 2 |
| | 6 2 | | | | | | | 6 2 |
| | 5 2 | | | | | | | 5 2 |
| | 6 2 | | | | | | | 6 2 |
| | 5 2 | | | | | | | 5 2 |
| | 5 2 | | | | | | | 6 2 |
| | 5 2 | | | | | | | 5 2 |

Continued on next column

**TABLE XXXI– continued**

| TQC | ICSD | NoE | SF | | | | | |
|---|---|---|---|---|---|---|---|---|
| | 6 2 | | | | | | | 6 2 |
| | 6 2 | | | | | | | 6 2 |
| 55.2.56 | 78823 | 128 | B10Ni1Th2 | | | | | |
| | -GM | -R | -S | -T | -U | -X | -Y | -Z |
| | 5 2 | 5+5 4 | 6+6 4 | 3+4 4 | 3+4 4 | 3+4 4 | 3+4 4 | 6 2 |
| | 6 2 | 6+6 4 | 5+5 4 | 3+4 4 | 3+4 4 | 3+4 4 | 3+4 4 | 5 2 |
| | 5 2 | 5+5 4 | 6+6 4 | 3+4 4 | 3+4 4 | 3+4 4 | 3+4 4 | 6 2 |
| | 6 2 | 6+6 4 | 5+5 4 | 3+4 4 | 3+4 4 | 3+4 4 | 3+4 4 | 5 2 |
| | 5 2 | 6+6 4 | 6+6 4 | 3+4 4 | 3+4 4 | 3+4 4 | 3+4 4 | 6 2 |
| | 6 2 | 5+5 4 | 5+5 4 | 3+4 4 | 3+4 4 | 3+4 4 | 3+4 4 | 5 2 |
| | 5 2 | 5+5 4 | 6+6 4 | 3+4 4 | 3+4 4 | 3+4 4 | 3+4 4 | 6 2 |
| | 6 2 | 6+6 4 | 5+5 4 | 3+4 4 | 3+4 4 | 3+4 4 | 3+4 4 | 5 2 |
| | 5 2 | 5+5 4 | 5+5 4 | 3+4 4 | 3+4 4 | 3+4 4 | 3+4 4 | 5 2 |
| | 6 2 | 6+6 4 | 6+6 4 | 3+4 4 | 3+4 4 | 3+4 4 | 3+4 4 | 6 2 |
| | 5 2 | 5+5 4 | 6+6 4 | 3+4 4 | 3+4 4 | 3+4 4 | 3+4 4 | 5 2 |
| | 6 2 | 6+6 4 | 5+5 4 | 3+4 4 | 3+4 4 | 3+4 4 | 3+4 4 | 6 2 |
| | 5 2 | 6+6 4 | 5+5 4 | 3+4 4 | 3+4 4 | 3+4 4 | 3+4 4 | 6 2 |
| | 5 2 | 5+5 4 | 6+6 4 | 3+4 4 | 3+4 4 | 3+4 4 | 3+4 4 | 5 2 |
| | 6 2 | 6+6 4 | 5+5 4 | 3+4 4 | 3+4 4 | 3+4 4 | 3+4 4 | 6 2 |
| | 6 2 | 5+5 4 | 6+6 4 | 3+4 4 | 3+4 4 | 3+4 4 | 3+4 4 | 5 2 |
| | 5 2 | 5+5 4 | 5+5 4 | 3+4 4 | 3+4 4 | 3+4 4 | 3+4 4 | 5 2 |
| | 6 2 | 6+6 4 | 6+6 4 | 3+4 4 | 3+4 4 | 3+4 4 | 3+4 4 | 6 2 |
| | 6 2 | 5+5 4 | 5+5 4 | 3+4 4 | 3+4 4 | 3+4 4 | 3+4 4 | 6 2 |
| | 5 2 | 5+5 4 | 6+6 4 | 3+4 4 | 3+4 4 | 3+4 4 | 3+4 4 | 5 2 |
| | 5 2 | 6+6 4 | 5+5 4 | 3+4 4 | 3+4 4 | 3+4 4 | 3+4 4 | 6 2 |
| | 6 2 | 5+5 4 | 5+5 4 | 3+4 4 | 3+4 4 | 3+4 4 | 3+4 4 | 5 2 |
| | 5 2 | 6+6 4 | 5+5 4 | 3+4 4 | 3+4 4 | 3+4 4 | 3+4 4 | 5 2 |
| | 6 2 | 6+6 4 | 5+5 4 | 3+4 4 | 3+4 4 | 3+4 4 | 3+4 4 | 5 2 |
| | 5 2 | 5+5 4 | 6+6 4 | 3+4 4 | 3+4 4 | 3+4 4 | 3+4 4 | 5 2 |
| | 5 2 | 5+5 4 | 6+6 4 | 3+4 4 | 3+4 4 | 3+4 4 | 3+4 4 | 6 2 |
| | 5 2 | 5+5 4 | 5+5 4 | 3+4 4 | 3+4 4 | 3+4 4 | 3+4 4 | 6 2 |
| | 5 2 | 5+5 4 | 6+6 4 | 3+4 4 | 3+4 4 | 3+4 4 | 3+4 4 | 6 2 |
| | 6 2 | 6+6 4 | 5+5 4 | 3+4 4 | 3+4 4 | 3+4 4 | 3+4 4 | 5 2 |
| | 5 2 | 6+6 4 | 5+5 4 | 3+4 4 | 3+4 4 | 3+4 4 | 3+4 4 | 5 2 |
| | 6 2 | 5+5 4 | 6+6 4 | 3+4 4 | 3+4 4 | 3+4 4 | 3+4 4 | 6 2 |
| | 6 2 | 5+5 4 | 6+6 4 | 3+4 4 | 3+4 4 | 3+4 4 | 3+4 4 | 6 2 |
| | 6 2 | | | | | | | 5 2 |
| | 5 2 | | | | | | | 5 2 |
| | 6 2 | | | | | | | 5 2 |
| | 5 2 | | | | | | | 6 2 |
| | 6 2 | | | | | | | 6 2 |
| | 5 2 | | | | | | | 5 2 |
| | 6 2 | | | | | | | 6 2 |
| | 5 2 | | | | | | | 5 2 |
| | 6 2 | | | | | | | 5 2 |
| | 5 2 | | | | | | | 5 2 |
| | 6 2 | | | | | | | 5 2 |
| | 6 2 | | | | | | | 6 2 |
| | 5 2 | | | | | | | 6 2 |
| | 5 2 | | | | | | | 5 2 |
| | 5 2 | | | | | | | 6 2 |
| | 5 2 | | | | | | | 5 2 |
| | 5 2 | | | | | | | 5 2 |
| | 6 2 | | | | | | | 6 2 |
| | 5 2 | | | | | | | 5 2 |
| | 5 2 | | | | | | | 5 2 |
| | 6 2 | | | | | | | 5 2 |
| | 5 2 | | | | | | | 5 2 |
| | 5 2 | | | | | | | 6 2 |
| | 5 2 | | | | | | | 6 2 |
| | 6 2 | | | | | | | 5 2 |

Continued on next column

**TABLE XXXI– continued**

| TQC | ICSD | NoE | SF | | | | | |
|---|---|---|---|---|---|---|---|---|
| | 6 2 | | | | | | | 5 2 |
| | 5 2 | | | | | | | 5 2 |
| | 5 2 | | | | | | | 6 2 |
| | 5 2 | | | | | | | 6 2 |
| | 5 2 | | | | | | | 5 2 |
| | 6 2 | | | | | | | 5 2 |
| | 6 2 | | | | | | | 5 2 |
| 55.2.57 | 78821 | 124 | B10Fe1Th2 | | | | | |
| | -GM | -R | -S | -T | -U | -X | -Y | -Z |
| | 5 2 | 5+5 4 | 6+6 4 | 3+4 4 | 3+4 4 | 3+4 4 | 3+4 4 | 6 2 |
| | 6 2 | 6+6 4 | 5+5 4 | 3+4 4 | 3+4 4 | 3+4 4 | 3+4 4 | 5 2 |
| | 5 2 | 5+5 4 | 6+6 4 | 3+4 4 | 3+4 4 | 3+4 4 | 3+4 4 | 6 2 |
| | 6 2 | 6+6 4 | 5+5 4 | 3+4 4 | 3+4 4 | 3+4 4 | 3+4 4 | 5 2 |
| | 5 2 | 5+5 4 | 6+6 4 | 3+4 4 | 3+4 4 | 3+4 4 | 3+4 4 | 6 2 |
| | 6 2 | 6+6 4 | 5+5 4 | 3+4 4 | 3+4 4 | 3+4 4 | 3+4 4 | 5 2 |
| | 5 2 | 5+5 4 | 6+6 4 | 3+4 4 | 3+4 4 | 3+4 4 | 3+4 4 | 6 2 |
| | 6 2 | 6+6 4 | 5+5 4 | 3+4 4 | 3+4 4 | 3+4 4 | 3+4 4 | 5 2 |
| | 5 2 | 5+5 4 | 5+5 4 | 3+4 4 | 3+4 4 | 3+4 4 | 3+4 4 | 5 2 |
| | 6 2 | 6+6 4 | 6+6 4 | 3+4 4 | 3+4 4 | 3+4 4 | 3+4 4 | 6 2 |
| | 5 2 | 5+5 4 | 6+6 4 | 3+4 4 | 3+4 4 | 3+4 4 | 3+4 4 | 5 2 |
| | 6 2 | 6+6 4 | 5+5 4 | 3+4 4 | 3+4 4 | 3+4 4 | 3+4 4 | 6 2 |
| | 5 2 | 6+6 4 | 5+5 4 | 3+4 4 | 3+4 4 | 3+4 4 | 3+4 4 | 6 2 |
| | 5 2 | 5+5 4 | 6+6 4 | 3+4 4 | 3+4 4 | 3+4 4 | 3+4 4 | 5 2 |
| | 6 2 | 6+6 4 | 5+5 4 | 3+4 4 | 3+4 4 | 3+4 4 | 3+4 4 | 6 2 |
| | 6 2 | 5+5 4 | 6+6 4 | 3+4 4 | 3+4 4 | 3+4 4 | 3+4 4 | 5 2 |
| | 5 2 | 5+5 4 | 5+5 4 | 3+4 4 | 3+4 4 | 3+4 4 | 3+4 4 | 5 2 |
| | 6 2 | 6+6 4 | 6+6 4 | 3+4 4 | 3+4 4 | 3+4 4 | 3+4 4 | 6 2 |
| | 6 2 | 5+5 4 | 5+5 4 | 3+4 4 | 3+4 4 | 3+4 4 | 3+4 4 | 6 2 |
| | 5 2 | 6+6 4 | 6+6 4 | 3+4 4 | 3+4 4 | 3+4 4 | 3+4 4 | 5 2 |
| | 5 2 | 5+5 4 | 5+5 4 | 3+4 4 | 3+4 4 | 3+4 4 | 3+4 4 | 6 2 |
| | 6 2 | 5+5 4 | 5+5 4 | 3+4 4 | 3+4 4 | 3+4 4 | 3+4 4 | 5 2 |
| | 5 2 | 6+6 4 | 6+6 4 | 3+4 4 | 3+4 4 | 3+4 4 | 3+4 4 | 5 2 |
| | 6 2 | 6+6 4 | 5+5 4 | 3+4 4 | 3+4 4 | 3+4 4 | 3+4 4 | 5 2 |
| | 5 2 | 5+5 4 | 5+5 4 | 3+4 4 | 3+4 4 | 3+4 4 | 3+4 4 | 5 2 |
| | 5 2 | 6+6 4 | 6+6 4 | 3+4 4 | 3+4 4 | 3+4 4 | 3+4 4 | 6 2 |
| | 5 2 | 5+5 4 | 6+6 4 | 3+4 4 | 3+4 4 | 3+4 4 | 3+4 4 | 6 2 |
| | 5 2 | 6+6 4 | 5+5 4 | 3+4 4 | 3+4 4 | 3+4 4 | 3+4 4 | 6 2 |
| | 6 2 | 5+5 4 | 6+6 4 | 3+4 4 | 3+4 4 | 3+4 4 | 3+4 4 | 5 2 |
| | 5 2 | 5+5 4 | 5+5 4 | 3+4 4 | 3+4 4 | 3+4 4 | 3+4 4 | 5 2 |
| | 6 2 | 5+5 4 | 6+6 4 | 3+4 4 | 3+4 4 | 3+4 4 | 3+4 4 | 6 2 |
| | 6 2 | | | | | | | 6 2 |
| | 6 2 | | | | | | | 5 2 |
| | 5 2 | | | | | | | 5 2 |
| | 6 2 | | | | | | | 6 2 |
| | 6 2 | | | | | | | 5 2 |
| | 5 2 | | | | | | | 6 2 |
| | 5 2 | | | | | | | 5 2 |
| | 6 2 | | | | | | | 6 2 |
| | 6 2 | | | | | | | 5 2 |
| | 5 2 | | | | | | | 6 2 |
| | 5 2 | | | | | | | 5 2 |
| | 6 2 | | | | | | | 6 2 |
| | 6 2 | | | | | | | 5 2 |
| | 5 2 | | | | | | | 5 2 |
| | 5 2 | | | | | | | 6 2 |
| | 5 2 | | | | | | | 5 2 |
| | 6 2 | | | | | | | 5 2 |
| | 5 2 | | | | | | | 6 2 |
| | 5 2 | | | | | | | 5 2 |
| | 6 2 | | | | | | | 6 2 |
| | 5 2 | | | | | | | 6 2 |

Continued on next column

**TABLE XXXI– continued**

| TQC | ICSD | NoE | SF | | | | | |
|---|---|---|---|---|---|---|---|---|
| | 5 2 | | | | | | | 6 2 |
| | 6 2 | | | | | | | 5 2 |
| | 5 2 | | | | | | | 5 2 |
| | 6 2 | | | | | | | 5 2 |
| | 5 2 | | | | | | | 5 2 |
| | 5 2 | | | | | | | 6 2 |
| | 5 2 | | | | | | | 5 2 |
| | 5 2 | | | | | | | 5 2 |
| | 5 2 | | | | | | | 5 2 |
| | 6 2 | | | | | | | 5 2 |
| 55.2.58 | 56973 | 112 | As1Ga2Rh5 | | | | | |
| | -GM | -R | -S | -T | -U | -X | -Y | -Z |
| | 5 2 | 6+6 4 | 6+6 4 | 3+4 4 | 3+4 4 | 3+4 4 | 3+4 4 | 5 2 |
| | 5 2 | 5+5 4 | 5+5 4 | 3+4 4 | 3+4 4 | 3+4 4 | 3+4 4 | 5 2 |
| | 5 2 | 6+6 4 | 6+6 4 | 3+4 4 | 3+4 4 | 3+4 4 | 3+4 4 | 6 2 |
| | 6 2 | 5+5 4 | 5+5 4 | 3+4 4 | 3+4 4 | 3+4 4 | 3+4 4 | 6 2 |
| | 5 2 | 5+5 4 | 5+5 4 | 3+4 4 | 3+4 4 | 3+4 4 | 3+4 4 | 5 2 |
| | 6 2 | 5+5 4 | 5+5 4 | 3+4 4 | 3+4 4 | 3+4 4 | 3+4 4 | 5 2 |
| | 6 2 | 5+5 4 | 5+5 4 | 3+4 4 | 3+4 4 | 3+4 4 | 3+4 4 | 6 2 |
| | 6 2 | 6+6 4 | 6+6 4 | 3+4 4 | 3+4 4 | 3+4 4 | 3+4 4 | 6 2 |
| | 6 2 | 6+6 4 | 5+5 4 | 3+4 4 | 3+4 4 | 3+4 4 | 3+4 4 | 5 2 |
| | 6 2 | 6+6 4 | 6+6 4 | 3+4 4 | 3+4 4 | 3+4 4 | 3+4 4 | 6 2 |
| | 6 2 | 5+5 4 | 6+6 4 | 3+4 4 | 3+4 4 | 3+4 4 | 3+4 4 | 6 2 |
| | 5 2 | 6+6 4 | 5+5 4 | 3+4 4 | 3+4 4 | 3+4 4 | 3+4 4 | 6 2 |
| | 5 2 | 5+5 4 | 6+6 4 | 3+4 4 | 3+4 4 | 3+4 4 | 3+4 4 | 5 2 |
| | 6 2 | 5+5 4 | 5+5 4 | 3+4 4 | 3+4 4 | 3+4 4 | 3+4 4 | 5 2 |
| | 5 2 | 5+5 4 | 5+5 4 | 3+4 4 | 3+4 4 | 3+4 4 | 3+4 4 | 5 2 |
| | 5 2 | 6+6 4 | 6+6 4 | 3+4 4 | 3+4 4 | 3+4 4 | 3+4 4 | 5 2 |
| | 6 2 | 5+5 4 | 5+5 4 | 3+4 4 | 3+4 4 | 3+4 4 | 3+4 4 | 6 2 |
| | 5 2 | 6+6 4 | 5+5 4 | 3+4 4 | 3+4 4 | 3+4 4 | 3+4 4 | 6 2 |
| | 5 2 | 5+5 4 | 6+6 4 | 3+4 4 | 3+4 4 | 3+4 4 | 3+4 4 | 6 2 |
| | 5 2 | 5+5 4 | 6+6 4 | 3+4 4 | 3+4 4 | 3+4 4 | 3+4 4 | 5 2 |
| | 5 2 | 6+6 4 | 5+5 4 | 3+4 4 | 3+4 4 | 3+4 4 | 3+4 4 | 6 2 |
| | 6 2 | 6+6 4 | 5+5 4 | 3+4 4 | 3+4 4 | 3+4 4 | 3+4 4 | 5 2 |
| | 5 2 | 5+5 4 | 6+6 4 | 3+4 4 | 3+4 4 | 3+4 4 | 3+4 4 | 5 2 |
| | 6 2 | 6+6 4 | 6+6 4 | 3+4 4 | 3+4 4 | 3+4 4 | 3+4 4 | 5 2 |
| | 5 2 | 6+6 4 | 5+5 4 | 3+4 4 | 3+4 4 | 3+4 4 | 3+4 4 | 6 2 |
| | 5 2 | 5+5 4 | 6+6 4 | 3+4 4 | 3+4 4 | 3+4 4 | 3+4 4 | 5 2 |
| | 5 2 | 5+5 4 | 5+5 4 | 3+4 4 | 3+4 4 | 3+4 4 | 3+4 4 | 6 2 |
| | 6 2 | 5+5 4 | 5+5 4 | 3+4 4 | 3+4 4 | 3+4 4 | 3+4 4 | 5 2 |
| | 5 2 | | | | | | | 5 2 |
| | 6 2 | | | | | | | 5 2 |
| | 5 2 | | | | | | | 6 2 |
| | 5 2 | | | | | | | 6 2 |
| | 6 2 | | | | | | | 5 2 |
| | 5 2 | | | | | | | 6 2 |
| | 6 2 | | | | | | | 5 2 |
| | 6 2 | | | | | | | 6 2 |
| | 5 2 | | | | | | | 6 2 |
| | 5 2 | | | | | | | 6 2 |
| | 6 2 | | | | | | | 5 2 |
| | 5 2 | | | | | | | 5 2 |
| | 6 2 | | | | | | | 5 2 |
| | 6 2 | | | | | | | 5 2 |
| | 5 2 | | | | | | | 6 2 |
| | 5 2 | | | | | | | 6 2 |
| | 6 2 | | | | | | | 5 2 |
| | 5 2 | | | | | | | 5 2 |
| | 6 2 | | | | | | | 5 2 |
| | 5 2 | | | | | | | 5 2 |
| | 5 2 | | | | | | | 5 2 |

Continued on next column

**TABLE XXXI– continued**

| TQC | ICSD | NoE | SF | | | | | |
|---|---|---|---|---|---|---|---|---|
| | 6 2 | | | | | | | 6 2 |
| | 5 2 | | | | | | | 5 2 |
| | 5 2 | | | | | | | 6 2 |
| | 6 2 | | | | | | | 6 2 |
| | 5 2 | | | | | | | 6 2 |
| | 6 2 | | | | | | | 5 2 |
| | 6 2 | | | | | | | 6 2 |
| 55.2.59 | 36364 | 172 | Al2Bi6Ca5 | | | | | |
| | -GM | -R | -S | -T | -U | -X | -Y | -Z |
| | 5 2 | 5+5 4 | 5+5 4 | 3+4 4 | 3+4 4 | 3+4 4 | 3+4 4 | 5 2 |
| | 5 2 | 6+6 4 | 6+6 4 | 3+4 4 | 3+4 4 | 3+4 4 | 3+4 4 | 5 2 |
| | 6 2 | 5+5 4 | 5+5 4 | 3+4 4 | 3+4 4 | 3+4 4 | 3+4 4 | 6 2 |
| | 6 2 | 6+6 4 | 6+6 4 | 3+4 4 | 3+4 4 | 3+4 4 | 3+4 4 | 6 2 |
| | 5 2 | 6+6 4 | 6+6 4 | 3+4 4 | 3+4 4 | 3+4 4 | 3+4 4 | 5 2 |
| | 6 2 | 5+5 4 | 5+5 4 | 3+4 4 | 3+4 4 | 3+4 4 | 3+4 4 | 6 2 |
| | 5 2 | 6+6 4 | 6+6 4 | 3+4 4 | 3+4 4 | 3+4 4 | 3+4 4 | 5 2 |
| | 6 2 | 5+5 4 | 5+5 4 | 3+4 4 | 3+4 4 | 3+4 4 | 3+4 4 | 6 2 |
| | 5 2 | 6+6 4 | 6+6 4 | 3+4 4 | 3+4 4 | 3+4 4 | 3+4 4 | 5 2 |
| | 5 2 | 5+5 4 | 5+5 4 | 3+4 4 | 3+4 4 | 3+4 4 | 3+4 4 | 5 2 |
| | 5 2 | 6+6 4 | 6+6 4 | 3+4 4 | 3+4 4 | 3+4 4 | 3+4 4 | 5 2 |
| | 5 2 | 5+5 4 | 5+5 4 | 3+4 4 | 3+4 4 | 3+4 4 | 3+4 4 | 5 2 |
| | 6 2 | 5+5 4 | 5+5 4 | 3+4 4 | 3+4 4 | 3+4 4 | 3+4 4 | 6 2 |
| | 6 2 | 6+6 4 | 6+6 4 | 3+4 4 | 3+4 4 | 3+4 4 | 3+4 4 | 6 2 |
| | 5 2 | 5+5 4 | 5+5 4 | 3+4 4 | 3+4 4 | 3+4 4 | 3+4 4 | 5 2 |
| | 6 2 | 6+6 4 | 6+6 4 | 3+4 4 | 3+4 4 | 3+4 4 | 3+4 4 | 6 2 |
| | 5 2 | 5+5 4 | 5+5 4 | 3+4 4 | 3+4 4 | 3+4 4 | 3+4 4 | 5 2 |
| | 6 2 | 6+6 4 | 6+6 4 | 3+4 4 | 3+4 4 | 3+4 4 | 3+4 4 | 6 2 |
| | 6 2 | 5+5 4 | 5+5 4 | 3+4 4 | 3+4 4 | 3+4 4 | 3+4 4 | 6 2 |
| | 6 2 | 5+5 4 | 5+5 4 | 3+4 4 | 3+4 4 | 3+4 4 | 3+4 4 | 6 2 |
| | 5 2 | 6+6 4 | 5+5 4 | 3+4 4 | 3+4 4 | 3+4 4 | 3+4 4 | 5 2 |
| | 5 2 | 6+6 4 | 6+6 4 | 3+4 4 | 3+4 4 | 3+4 4 | 3+4 4 | 5 2 |
| | 5 2 | 5+5 4 | 5+5 4 | 3+4 4 | 3+4 4 | 3+4 4 | 3+4 4 | 5 2 |
| | 6 2 | 5+5 4 | 6+6 4 | 3+4 4 | 3+4 4 | 3+4 4 | 3+4 4 | 6 2 |
| | 6 2 | 6+6 4 | 5+5 4 | 3+4 4 | 3+4 4 | 3+4 4 | 3+4 4 | 6 2 |
| | 5 2 | 5+5 4 | 6+6 4 | 3+4 4 | 3+4 4 | 3+4 4 | 3+4 4 | 5 2 |
| | 6 2 | 6+6 4 | 6+6 4 | 3+4 4 | 3+4 4 | 3+4 4 | 3+4 4 | 6 2 |
| | 6 2 | 5+5 4 | 5+5 4 | 3+4 4 | 3+4 4 | 3+4 4 | 3+4 4 | 6 2 |
| | 5 2 | 5+5 4 | 5+5 4 | 3+4 4 | 3+4 4 | 3+4 4 | 3+4 4 | 5 2 |
| | 6 2 | 6+6 4 | 6+6 4 | 3+4 4 | 3+4 4 | 3+4 4 | 3+4 4 | 6 2 |
| | 5 2 | 6+6 4 | 5+5 4 | 3+4 4 | 3+4 4 | 3+4 4 | 3+4 4 | 5 2 |
| | 6 2 | 5+5 4 | 6+6 4 | 3+4 4 | 3+4 4 | 3+4 4 | 3+4 4 | 6 2 |
| | 5 2 | 5+5 4 | 5+5 4 | 3+4 4 | 3+4 4 | 3+4 4 | 3+4 4 | 5 2 |
| | 6 2 | 6+6 4 | 6+6 4 | 3+4 4 | 3+4 4 | 3+4 4 | 3+4 4 | 6 2 |
| | 5 2 | 6+6 4 | 6+6 4 | 3+4 4 | 3+4 4 | 3+4 4 | 3+4 4 | 5 2 |
| | 6 2 | 5+5 4 | 5+5 4 | 3+4 4 | 3+4 4 | 3+4 4 | 3+4 4 | 6 2 |
| | 6 2 | 6+6 4 | 6+6 4 | 3+4 4 | 3+4 4 | 3+4 4 | 3+4 4 | 6 2 |
| | 6 2 | 5+5 4 | 5+5 4 | 3+4 4 | 3+4 4 | 3+4 4 | 3+4 4 | 6 2 |
| | 6 2 | 6+6 4 | 6+6 4 | 3+4 4 | 3+4 4 | 3+4 4 | 3+4 4 | 6 2 |
| | 6 2 | 5+5 4 | 5+5 4 | 3+4 4 | 3+4 4 | 3+4 4 | 3+4 4 | 6 2 |
| | 5 2 | 6+6 4 | 6+6 4 | 3+4 4 | 3+4 4 | 3+4 4 | 3+4 4 | 6 2 |
| | 5 2 | 5+5 4 | 5+5 4 | 3+4 4 | 3+4 4 | 3+4 4 | 3+4 4 | 6 2 |
| | 6 2 | 6+6 4 | 6+6 4 | 3+4 4 | 3+4 4 | 3+4 4 | 3+4 4 | 5 2 |
| | 5 2 | | | | | | | 6 2 |
| | 6 2 | | | | | | | 5 2 |
| | 5 2 | | | | | | | 6 2 |
| | 5 2 | | | | | | | 5 2 |
| | 6 2 | | | | | | | 5 2 |
| | 5 2 | | | | | | | 6 2 |
| | 6 2 | | | | | | | 6 2 |
| | 6 2 | | | | | | | 5 2 |
| | 6 2 | | | | | | | 5 2 |

Continued on next column

**TABLE XXXI– continued**

| TQC | ICSD | NoE | SF | | | | | |
|---|---|---|---|---|---|---|---|---|
| | 5 2 | | | | | | | 5 2 |
| | 5 2 | | | | | | | 6 2 |
| | 6 2 | | | | | | | 5 2 |
| | 6 2 | | | | | | | 6 2 |
| | 5 2 | | | | | | | 5 2 |
| | 5 2 | | | | | | | 5 2 |
| | 6 2 | | | | | | | 6 2 |
| | 6 2 | | | | | | | 6 2 |
| | 5 2 | | | | | | | 5 2 |
| | 5 2 | | | | | | | 6 2 |
| | 6 2 | | | | | | | 6 2 |
| | 5 2 | | | | | | | 5 2 |
| | 6 2 | | | | | | | 6 2 |
| | 5 2 | | | | | | | 5 2 |
| | 6 2 | | | | | | | 6 2 |
| | 5 2 | | | | | | | 5 2 |
| | 5 2 | | | | | | | 6 2 |
| | 6 2 | | | | | | | 5 2 |
| | 6 2 | | | | | | | 6 2 |
| | 5 2 | | | | | | | 5 2 |
| | 6 2 | | | | | | | 5 2 |
| | 5 2 | | | | | | | 5 2 |
| | 6 2 | | | | | | | 6 2 |
| | 6 2 | | | | | | | 6 2 |
| | 5 2 | | | | | | | 6 2 |
| | 6 2 | | | | | | | 6 2 |
| | 5 2 | | | | | | | 5 2 |
| | 5 2 | | | | | | | 6 2 |
| | 5 2 | | | | | | | 5 2 |
| | 6 2 | | | | | | | 5 2 |
| | 5 2 | | | | | | | 6 2 |
| | 6 2 | | | | | | | 6 2 |
| | 5 2 | | | | | | | 5 2 |
| | 5 2 | | | | | | | 6 2 |
| 55.2.60 | 260796 | 156 | Al1B4Yb1 | | | | | |
| | -GM | -R | -S | -T | -U | -X | -Y | -Z |
| | 5 2 | 5+5 4 | 5+5 4 | 3+4 4 | 3+4 4 | 3+4 4 | 3+4 4 | 5 2 |
| | 6 2 | 6+6 4 | 6+6 4 | 3+4 4 | 3+4 4 | 3+4 4 | 3+4 4 | 6 2 |
| | 5 2 | 5+5 4 | 5+5 4 | 3+4 4 | 3+4 4 | 3+4 4 | 3+4 4 | 5 2 |
| | 6 2 | 6+6 4 | 6+6 4 | 3+4 4 | 3+4 4 | 3+4 4 | 3+4 4 | 6 2 |
| | 5 2 | 6+6 4 | 5+5 4 | 3+4 4 | 3+4 4 | 3+4 4 | 3+4 4 | 5 2 |
| | 6 2 | 5+5 4 | 6+6 4 | 3+4 4 | 3+4 4 | 3+4 4 | 3+4 4 | 6 2 |
| | 5 2 | 5+5 4 | 5+5 4 | 3+4 4 | 3+4 4 | 3+4 4 | 3+4 4 | 5 2 |
| | 6 2 | 6+6 4 | 6+6 4 | 3+4 4 | 3+4 4 | 3+4 4 | 3+4 4 | 6 2 |
| | 5 2 | 5+5 4 | 6+6 4 | 3+4 4 | 3+4 4 | 3+4 4 | 3+4 4 | 6 2 |
| | 6 2 | 6+6 4 | 5+5 4 | 3+4 4 | 3+4 4 | 3+4 4 | 3+4 4 | 5 2 |
| | 5 2 | 6+6 4 | 5+5 4 | 3+4 4 | 3+4 4 | 3+4 4 | 3+4 4 | 6 2 |
| | 6 2 | 6+6 4 | 6+6 4 | 3+4 4 | 3+4 4 | 3+4 4 | 3+4 4 | 5 2 |
| | 5 2 | 5+5 4 | 6+6 4 | 3+4 4 | 3+4 4 | 3+4 4 | 3+4 4 | 5 2 |
| | 5 2 | 5+5 4 | 5+5 4 | 3+4 4 | 3+4 4 | 3+4 4 | 3+4 4 | 6 2 |
| | 6 2 | 6+6 4 | 5+5 4 | 3+4 4 | 3+4 4 | 3+4 4 | 3+4 4 | 5 2 |
| | 6 2 | 5+5 4 | 6+6 4 | 3+4 4 | 3+4 4 | 3+4 4 | 3+4 4 | 6 2 |
| | 5 2 | 6+6 4 | 5+5 4 | 3+4 4 | 3+4 4 | 3+4 4 | 3+4 4 | 6 2 |
| | 5 2 | 5+5 4 | 5+5 4 | 3+4 4 | 3+4 4 | 3+4 4 | 3+4 4 | 6 2 |
| | 6 2 | 6+6 4 | 6+6 4 | 3+4 4 | 3+4 4 | 3+4 4 | 3+4 4 | 5 2 |
| | 6 2 | 5+5 4 | 6+6 4 | 3+4 4 | 3+4 4 | 3+4 4 | 3+4 4 | 5 2 |
| | 5 2 | 5+5 4 | 5+5 4 | 3+4 4 | 3+4 4 | 3+4 4 | 3+4 4 | 5 2 |
| | 6 2 | 6+6 4 | 6+6 4 | 3+4 4 | 3+4 4 | 3+4 4 | 3+4 4 | 6 2 |
| | 5 2 | 6+6 4 | 6+6 4 | 3+4 4 | 3+4 4 | 3+4 4 | 3+4 4 | 6 2 |
| | 6 2 | 5+5 4 | 5+5 4 | 3+4 4 | 3+4 4 | 3+4 4 | 3+4 4 | 5 2 |
| | 5 2 | 6+6 4 | 6+6 4 | 3+4 4 | 3+4 4 | 3+4 4 | 3+4 4 | 5 2 |

Continued on next column

**TABLE XXXI– continued**

| TQC | ICSD | NoE | SF | | | | | |
|---|---|---|---|---|---|---|---|---|
| | 5 2 | 5+5 4 | 5+5 4 | 3+4 4 | 3+4 4 | 3+4 4 | 3+4 4 | 5 2 |
| | 6 2 | 5+5 4 | 5+5 4 | 3+4 4 | 3+4 4 | 3+4 4 | 3+4 4 | 6 2 |
| | 5 2 | 6+6 4 | 6+6 4 | 3+4 4 | 3+4 4 | 3+4 4 | 3+4 4 | 6 2 |
| | 6 2 | 6+6 4 | 6+6 4 | 3+4 4 | 3+4 4 | 3+4 4 | 3+4 4 | 6 2 |
| | 6 2 | 5+5 4 | 5+5 4 | 3+4 4 | 3+4 4 | 3+4 4 | 3+4 4 | 5 2 |
| | 5 2 | 5+5 4 | 5+5 4 | 3+4 4 | 3+4 4 | 3+4 4 | 3+4 4 | 6 2 |
| | 6 2 | 6+6 4 | 6+6 4 | 3+4 4 | 3+4 4 | 3+4 4 | 3+4 4 | 6 2 |
| | 5 2 | 5+5 4 | 6+6 4 | 3+4 4 | 3+4 4 | 3+4 4 | 3+4 4 | 5 2 |
| | 6 2 | 6+6 4 | 5+5 4 | 3+4 4 | 3+4 4 | 3+4 4 | 3+4 4 | 6 2 |
| | 6 2 | 5+5 4 | 6+6 4 | 3+4 4 | 3+4 4 | 3+4 4 | 3+4 4 | 5 2 |
| | 5 2 | 6+6 4 | 5+5 4 | 3+4 4 | 3+4 4 | 3+4 4 | 3+4 4 | 5 2 |
| | 6 2 | 5+5 4 | 5+5 4 | 3+4 4 | 3+4 4 | 3+4 4 | 3+4 4 | 5 2 |
| | 5 2 | 5+5 4 | 6+6 4 | 3+4 4 | 3+4 4 | 3+4 4 | 3+4 4 | 6 2 |
| | 6 2 | 6+6 4 | 5+5 4 | 3+4 4 | 3+4 4 | 3+4 4 | 3+4 4 | 5 2 |
| | 5 2 | | | | | | | 6 2 |
| | 5 2 | | | | | | | 5 2 |
| | 5 2 | | | | | | | 5 2 |
| | 5 2 | | | | | | | 6 2 |
| | 6 2 | | | | | | | 6 2 |
| | 6 2 | | | | | | | 6 2 |
| | 5 2 | | | | | | | 5 2 |
| | 6 2 | | | | | | | 5 2 |
| | 5 2 | | | | | | | 6 2 |
| | 6 2 | | | | | | | 6 2 |
| | 6 2 | | | | | | | 6 2 |
| | 6 2 | | | | | | | 6 2 |
| | 5 2 | | | | | | | 5 2 |
| | 5 2 | | | | | | | 5 2 |
| | 5 2 | | | | | | | 5 2 |
| | 6 2 | | | | | | | 6 2 |
| | 6 2 | | | | | | | 5 2 |
| | 5 2 | | | | | | | 6 2 |
| | 5 2 | | | | | | | 5 2 |
| | 5 2 | | | | | | | 6 2 |
| | 5 2 | | | | | | | 5 2 |
| | 6 2 | | | | | | | 6 2 |
| | 5 2 | | | | | | | 5 2 |
| | 5 2 | | | | | | | 6 2 |
| | 6 2 | | | | | | | 6 2 |
| | 6 2 | | | | | | | 6 2 |
| | 6 2 | | | | | | | 5 2 |
| | 5 2 | | | | | | | 5 2 |
| | 6 2 | | | | | | | 6 2 |
| | 6 2 | | | | | | | 5 2 |
| | 5 2 | | | | | | | 6 2 |
| | 5 2 | | | | | | | 5 2 |
| | 6 2 | | | | | | | 6 2 |
| | 5 2 | | | | | | | 5 2 |
| | 6 2 | | | | | | | 5 2 |
| | 5 2 | | | | | | | 6 2 |
| | 5 2 | | | | | | | 6 2 |
| | 5 2 | | | | | | | 5 2 |
| | 6 2 | | | | | | | 5 2 |
| 55.2.61 | 167532 | 156 | Al1B4Yb1 | | | | | |
| | -GM | -R | -S | -T | -U | -X | -Y | -Z |
| | 5 2 | 5+5 4 | 5+5 4 | 3+4 4 | 3+4 4 | 3+4 4 | 3+4 4 | 5 2 |
| | 6 2 | 6+6 4 | 6+6 4 | 3+4 4 | 3+4 4 | 3+4 4 | 3+4 4 | 6 2 |
| | 5 2 | 5+5 4 | 5+5 4 | 3+4 4 | 3+4 4 | 3+4 4 | 3+4 4 | 5 2 |
| | 6 2 | 6+6 4 | 6+6 4 | 3+4 4 | 3+4 4 | 3+4 4 | 3+4 4 | 6 2 |
| | 5 2 | 6+6 4 | 5+5 4 | 3+4 4 | 3+4 4 | 3+4 4 | 3+4 4 | 5 2 |
| | 6 2 | 5+5 4 | 6+6 4 | 3+4 4 | 3+4 4 | 3+4 4 | 3+4 4 | 6 2 |

Continued on next column

**TABLE XXXI– continued**

| TQC | ICSD | NoE | SF | | | | | |
|---|---|---|---|---|---|---|---|---|
| | 5 2 | 5+5 4 | 5+5 4 | 3+4 4 | 3+4 4 | 3+4 4 | 3+4 4 | 5 2 |
| | 6 2 | 6+6 4 | 6+6 4 | 3+4 4 | 3+4 4 | 3+4 4 | 3+4 4 | 6 2 |
| | 5 2 | 5+5 4 | 6+6 4 | 3+4 4 | 3+4 4 | 3+4 4 | 3+4 4 | 6 2 |
| | 6 2 | 6+6 4 | 5+5 4 | 3+4 4 | 3+4 4 | 3+4 4 | 3+4 4 | 5 2 |
| | 5 2 | 6+6 4 | 5+5 4 | 3+4 4 | 3+4 4 | 3+4 4 | 3+4 4 | 6 2 |
| | 6 2 | 6+6 4 | 6+6 4 | 3+4 4 | 3+4 4 | 3+4 4 | 3+4 4 | 5 2 |
| | 5 2 | 5+5 4 | 6+6 4 | 3+4 4 | 3+4 4 | 3+4 4 | 3+4 4 | 5 2 |
| | 5 2 | 5+5 4 | 5+5 4 | 3+4 4 | 3+4 4 | 3+4 4 | 3+4 4 | 6 2 |
| | 6 2 | 6+6 4 | 5+5 4 | 3+4 4 | 3+4 4 | 3+4 4 | 3+4 4 | 5 2 |
| | 6 2 | 5+5 4 | 6+6 4 | 3+4 4 | 3+4 4 | 3+4 4 | 3+4 4 | 6 2 |
| | 5 2 | 6+6 4 | 5+5 4 | 3+4 4 | 3+4 4 | 3+4 4 | 3+4 4 | 6 2 |
| | 5 2 | 5+5 4 | 5+5 4 | 3+4 4 | 3+4 4 | 3+4 4 | 3+4 4 | 6 2 |
| | 6 2 | 6+6 4 | 6+6 4 | 3+4 4 | 3+4 4 | 3+4 4 | 3+4 4 | 5 2 |
| | 6 2 | 5+5 4 | 6+6 4 | 3+4 4 | 3+4 4 | 3+4 4 | 3+4 4 | 5 2 |
| | 5 2 | 5+5 4 | 5+5 4 | 3+4 4 | 3+4 4 | 3+4 4 | 3+4 4 | 5 2 |
| | 6 2 | 6+6 4 | 6+6 4 | 3+4 4 | 3+4 4 | 3+4 4 | 3+4 4 | 6 2 |
| | 5 2 | 6+6 4 | 6+6 4 | 3+4 4 | 3+4 4 | 3+4 4 | 3+4 4 | 6 2 |
| | 6 2 | 5+5 4 | 5+5 4 | 3+4 4 | 3+4 4 | 3+4 4 | 3+4 4 | 5 2 |
| | 5 2 | 6+6 4 | 6+6 4 | 3+4 4 | 3+4 4 | 3+4 4 | 3+4 4 | 5 2 |
| | 5 2 | 5+5 4 | 5+5 4 | 3+4 4 | 3+4 4 | 3+4 4 | 3+4 4 | 5 2 |
| | 6 2 | 5+5 4 | 5+5 4 | 3+4 4 | 3+4 4 | 3+4 4 | 3+4 4 | 6 2 |
| | 5 2 | 6+6 4 | 6+6 4 | 3+4 4 | 3+4 4 | 3+4 4 | 3+4 4 | 6 2 |
| | 6 2 | 6+6 4 | 6+6 4 | 3+4 4 | 3+4 4 | 3+4 4 | 3+4 4 | 6 2 |
| | 6 2 | 5+5 4 | 5+5 4 | 3+4 4 | 3+4 4 | 3+4 4 | 3+4 4 | 5 2 |
| | 5 2 | 5+5 4 | 5+5 4 | 3+4 4 | 3+4 4 | 3+4 4 | 3+4 4 | 6 2 |
| | 6 2 | 6+6 4 | 6+6 4 | 3+4 4 | 3+4 4 | 3+4 4 | 3+4 4 | 6 2 |
| | 5 2 | 5+5 4 | 6+6 4 | 3+4 4 | 3+4 4 | 3+4 4 | 3+4 4 | 5 2 |
| | 6 2 | 6+6 4 | 5+5 4 | 3+4 4 | 3+4 4 | 3+4 4 | 3+4 4 | 6 2 |
| | 6 2 | 5+5 4 | 6+6 4 | 3+4 4 | 3+4 4 | 3+4 4 | 3+4 4 | 5 2 |
| | 5 2 | 6+6 4 | 5+5 4 | 3+4 4 | 3+4 4 | 3+4 4 | 3+4 4 | 5 2 |
| | 6 2 | 5+5 4 | 5+5 4 | 3+4 4 | 3+4 4 | 3+4 4 | 3+4 4 | 5 2 |
| | 5 2 | 5+5 4 | 6+6 4 | 3+4 4 | 3+4 4 | 3+4 4 | 3+4 4 | 6 2 |
| | 6 2 | 6+6 4 | 5+5 4 | 3+4 4 | 3+4 4 | 3+4 4 | 3+4 4 | 5 2 |
| | 5 2 | | | | | | | 6 2 |
| | 5 2 | | | | | | | 5 2 |
| | 5 2 | | | | | | | 5 2 |
| | 5 2 | | | | | | | 6 2 |
| | 6 2 | | | | | | | 6 2 |
| | 6 2 | | | | | | | 6 2 |
| | 5 2 | | | | | | | 5 2 |
| | 6 2 | | | | | | | 5 2 |
| | 5 2 | | | | | | | 6 2 |
| | 6 2 | | | | | | | 6 2 |
| | 6 2 | | | | | | | 6 2 |
| | 6 2 | | | | | | | 6 2 |
| | 5 2 | | | | | | | 5 2 |
| | 5 2 | | | | | | | 5 2 |
| | 5 2 | | | | | | | 5 2 |
| | 6 2 | | | | | | | 6 2 |
| | 6 2 | | | | | | | 5 2 |
| | 5 2 | | | | | | | 6 2 |
| | 5 2 | | | | | | | 5 2 |
| | 5 2 | | | | | | | 6 2 |
| | 5 2 | | | | | | | 5 2 |
| | 6 2 | | | | | | | 6 2 |
| | 5 2 | | | | | | | 5 2 |
| | 5 2 | | | | | | | 6 2 |
| | 6 2 | | | | | | | 6 2 |
| | 6 2 | | | | | | | 6 2 |
| | 6 2 | | | | | | | 5 2 |
| | 5 2 | | | | | | | 5 2 |

Continued on next column

**TABLE XXXI– continued**

| TQC | ICSD | NoE | SF | | | | | |
|---|---|---|---|---|---|---|---|---|
| | 6 2 | | | | | | | 6 2 |
| | 6 2 | | | | | | | 5 2 |
| | 5 2 | | | | | | | 6 2 |
| | 5 2 | | | | | | | 5 2 |
| | 6 2 | | | | | | | 6 2 |
| | 5 2 | | | | | | | 5 2 |
| | 6 2 | | | | | | | 5 2 |
| | 5 2 | | | | | | | 6 2 |
| | 5 2 | | | | | | | 6 2 |
| | 5 2 | | | | | | | 5 2 |
| | 6 2 | | | | | | | 5 2 |
| 55.2.62 | 167534 | 160 | Al1B4Lu1 | | | | | |
| | -GM | -R | -S | -T | -U | -X | -Y | -Z |
| | 5 2 | 5+5 4 | 5+5 4 | 3+4 4 | 3+4 4 | 3+4 4 | 3+4 4 | 5 2 |
| | 6 2 | 6+6 4 | 6+6 4 | 3+4 4 | 3+4 4 | 3+4 4 | 3+4 4 | 6 2 |
| | 5 2 | 5+5 4 | 5+5 4 | 3+4 4 | 3+4 4 | 3+4 4 | 3+4 4 | 5 2 |
| | 6 2 | 6+6 4 | 6+6 4 | 3+4 4 | 3+4 4 | 3+4 4 | 3+4 4 | 6 2 |
| | 5 2 | 5+5 4 | 5+5 4 | 3+4 4 | 3+4 4 | 3+4 4 | 3+4 4 | 5 2 |
| | 6 2 | 6+6 4 | 6+6 4 | 3+4 4 | 3+4 4 | 3+4 4 | 3+4 4 | 6 2 |
| | 5 2 | 5+5 4 | 5+5 4 | 3+4 4 | 3+4 4 | 3+4 4 | 3+4 4 | 5 2 |
| | 6 2 | 6+6 4 | 6+6 4 | 3+4 4 | 3+4 4 | 3+4 4 | 3+4 4 | 6 2 |
| | 5 2 | 5+5 4 | 6+6 4 | 3+4 4 | 3+4 4 | 3+4 4 | 3+4 4 | 6 2 |
| | 6 2 | 6+6 4 | 5+5 4 | 3+4 4 | 3+4 4 | 3+4 4 | 3+4 4 | 5 2 |
| | 5 2 | 6+6 4 | 5+5 4 | 3+4 4 | 3+4 4 | 3+4 4 | 3+4 4 | 6 2 |
| | 6 2 | 5+5 4 | 6+6 4 | 3+4 4 | 3+4 4 | 3+4 4 | 3+4 4 | 5 2 |
| | 5 2 | 6+6 4 | 6+6 4 | 3+4 4 | 3+4 4 | 3+4 4 | 3+4 4 | 5 2 |
| | 6 2 | 5+5 4 | 5+5 4 | 3+4 4 | 3+4 4 | 3+4 4 | 3+4 4 | 6 2 |
| | 5 2 | 6+6 4 | 5+5 4 | 3+4 4 | 3+4 4 | 3+4 4 | 3+4 4 | 5 2 |
| | 6 2 | 5+5 4 | 6+6 4 | 3+4 4 | 3+4 4 | 3+4 4 | 3+4 4 | 6 2 |
| | 5 2 | 6+6 4 | 6+6 4 | 3+4 4 | 3+4 4 | 3+4 4 | 3+4 4 | 6 2 |
| | 5 2 | 5+5 4 | 5+5 4 | 3+4 4 | 3+4 4 | 3+4 4 | 3+4 4 | 6 2 |
| | 6 2 | 6+6 4 | 6+6 4 | 3+4 4 | 3+4 4 | 3+4 4 | 3+4 4 | 5 2 |
| | 6 2 | 5+5 4 | 5+5 4 | 3+4 4 | 3+4 4 | 3+4 4 | 3+4 4 | 5 2 |
| | 5 2 | 6+6 4 | 5+5 4 | 3+4 4 | 3+4 4 | 3+4 4 | 3+4 4 | 5 2 |
| | 5 2 | 5+5 4 | 6+6 4 | 3+4 4 | 3+4 4 | 3+4 4 | 3+4 4 | 6 2 |
| | 6 2 | 6+6 4 | 5+5 4 | 3+4 4 | 3+4 4 | 3+4 4 | 3+4 4 | 6 2 |
| | 6 2 | 5+5 4 | 6+6 4 | 3+4 4 | 3+4 4 | 3+4 4 | 3+4 4 | 5 2 |
| | 5 2 | 6+6 4 | 5+5 4 | 3+4 4 | 3+4 4 | 3+4 4 | 3+4 4 | 5 2 |
| | 5 2 | 5+5 4 | 6+6 4 | 3+4 4 | 3+4 4 | 3+4 4 | 3+4 4 | 5 2 |
| | 6 2 | 6+6 4 | 5+5 4 | 3+4 4 | 3+4 4 | 3+4 4 | 3+4 4 | 6 2 |
| | 5 2 | 5+5 4 | 6+6 4 | 3+4 4 | 3+4 4 | 3+4 4 | 3+4 4 | 6 2 |
| | 6 2 | 6+6 4 | 5+5 4 | 3+4 4 | 3+4 4 | 3+4 4 | 3+4 4 | 6 2 |
| | 6 2 | 5+5 4 | 6+6 4 | 3+4 4 | 3+4 4 | 3+4 4 | 3+4 4 | 5 2 |
| | 5 2 | 6+6 4 | 5+5 4 | 3+4 4 | 3+4 4 | 3+4 4 | 3+4 4 | 6 2 |
| | 6 2 | 5+5 4 | 5+5 4 | 3+4 4 | 3+4 4 | 3+4 4 | 3+4 4 | 6 2 |
| | 6 2 | 6+6 4 | 6+6 4 | 3+4 4 | 3+4 4 | 3+4 4 | 3+4 4 | 5 2 |
| | 5 2 | 5+5 4 | 6+6 4 | 3+4 4 | 3+4 4 | 3+4 4 | 3+4 4 | 6 2 |
| | 5 2 | 5+5 4 | 5+5 4 | 3+4 4 | 3+4 4 | 3+4 4 | 3+4 4 | 5 2 |
| | 5 2 | 6+6 4 | 6+6 4 | 3+4 4 | 3+4 4 | 3+4 4 | 3+4 4 | 6 2 |
| | 6 2 | 6+6 4 | 6+6 4 | 3+4 4 | 3+4 4 | 3+4 4 | 3+4 4 | 5 2 |
| | 6 2 | 5+5 4 | 5+5 4 | 3+4 4 | 3+4 4 | 3+4 4 | 3+4 4 | 5 2 |
| | 5 2 | 5+5 4 | 5+5 4 | 3+4 4 | 3+4 4 | 3+4 4 | 3+4 4 | 6 2 |
| | 5 2 | 6+6 4 | 6+6 4 | 3+4 4 | 3+4 4 | 3+4 4 | 3+4 4 | 5 2 |
| | 6 2 | | | | | | | 6 2 |
| | 6 2 | | | | | | | 6 2 |
| | 5 2 | | | | | | | 5 2 |
| | 6 2 | | | | | | | 5 2 |
| | 5 2 | | | | | | | 6 2 |
| | 6 2 | | | | | | | 6 2 |
| | 6 2 | | | | | | | 5 2 |
| | 5 2 | | | | | | | 5 2 |

Continued on next column

**TABLE XXXI– continued**

| TQC | ICSD | NoE | SF | | | | | |
|---|---|---|---|---|---|---|---|---|
| | 5 2 | | | | | | | 5 2 |
| | 5 2 | | | | | | | 6 2 |
| | 6 2 | | | | | | | 5 2 |
| | 5 2 | | | | | | | 6 2 |
| | 6 2 | | | | | | | 5 2 |
| | 5 2 | | | | | | | 6 2 |
| | 6 2 | | | | | | | 5 2 |
| | 6 2 | | | | | | | 6 2 |
| | 5 2 | | | | | | | 6 2 |
| | 5 2 | | | | | | | 5 2 |
| | 6 2 | | | | | | | 5 2 |
| | 5 2 | | | | | | | 6 2 |
| | 6 2 | | | | | | | 5 2 |
| | 6 2 | | | | | | | 6 2 |
| | 6 2 | | | | | | | 6 2 |
| | 5 2 | | | | | | | 6 2 |
| | 6 2 | | | | | | | 5 2 |
| | 6 2 | | | | | | | 5 2 |
| | 5 2 | | | | | | | 5 2 |
| | 5 2 | | | | | | | 6 2 |
| | 5 2 | | | | | | | 5 2 |
| | 6 2 | | | | | | | 5 2 |
| | 6 2 | | | | | | | 6 2 |
| | 5 2 | | | | | | | 6 2 |
| | 5 2 | | | | | | | 6 2 |
| | 5 2 | | | | | | | 5 2 |
| | 5 2 | | | | | | | 6 2 |
| | 5 2 | | | | | | | 5 2 |
| | 6 2 | | | | | | | 5 2 |
| | 5 2 | | | | | | | 6 2 |
| | 6 2 | | | | | | | 6 2 |
| | 5 2 | | | | | | | 6 2 |
| 55.2.63 | 260799 | 160 | Al1.33B5.33Lu1.33 | | | | | |
| | -GM | -R | -S | -T | -U | -X | -Y | -Z |
| | 5 2 | 5+5 4 | 5+5 4 | 3+4 4 | 3+4 4 | 3+4 4 | 3+4 4 | 5 2 |
| | 6 2 | 6+6 4 | 6+6 4 | 3+4 4 | 3+4 4 | 3+4 4 | 3+4 4 | 6 2 |
| | 5 2 | 5+5 4 | 5+5 4 | 3+4 4 | 3+4 4 | 3+4 4 | 3+4 4 | 5 2 |
| | 6 2 | 6+6 4 | 6+6 4 | 3+4 4 | 3+4 4 | 3+4 4 | 3+4 4 | 6 2 |
| | 5 2 | 5+5 4 | 5+5 4 | 3+4 4 | 3+4 4 | 3+4 4 | 3+4 4 | 5 2 |
| | 6 2 | 6+6 4 | 6+6 4 | 3+4 4 | 3+4 4 | 3+4 4 | 3+4 4 | 6 2 |
| | 5 2 | 5+5 4 | 5+5 4 | 3+4 4 | 3+4 4 | 3+4 4 | 3+4 4 | 5 2 |
| | 6 2 | 6+6 4 | 6+6 4 | 3+4 4 | 3+4 4 | 3+4 4 | 3+4 4 | 6 2 |
| | 5 2 | 5+5 4 | 6+6 4 | 3+4 4 | 3+4 4 | 3+4 4 | 3+4 4 | 6 2 |
| | 6 2 | 6+6 4 | 5+5 4 | 3+4 4 | 3+4 4 | 3+4 4 | 3+4 4 | 5 2 |
| | 5 2 | 6+6 4 | 5+5 4 | 3+4 4 | 3+4 4 | 3+4 4 | 3+4 4 | 6 2 |
| | 6 2 | 6+6 4 | 6+6 4 | 3+4 4 | 3+4 4 | 3+4 4 | 3+4 4 | 5 2 |
| | 5 2 | 5+5 4 | 6+6 4 | 3+4 4 | 3+4 4 | 3+4 4 | 3+4 4 | 5 2 |
| | 5 2 | 5+5 4 | 5+5 4 | 3+4 4 | 3+4 4 | 3+4 4 | 3+4 4 | 6 2 |
| | 6 2 | 6+6 4 | 5+5 4 | 3+4 4 | 3+4 4 | 3+4 4 | 3+4 4 | 5 2 |
| | 6 2 | 5+5 4 | 6+6 4 | 3+4 4 | 3+4 4 | 3+4 4 | 3+4 4 | 6 2 |
| | 5 2 | 6+6 4 | 6+6 4 | 3+4 4 | 3+4 4 | 3+4 4 | 3+4 4 | 6 2 |
| | 5 2 | 5+5 4 | 5+5 4 | 3+4 4 | 3+4 4 | 3+4 4 | 3+4 4 | 6 2 |
| | 6 2 | 6+6 4 | 6+6 4 | 3+4 4 | 3+4 4 | 3+4 4 | 3+4 4 | 5 2 |
| | 6 2 | 5+5 4 | 5+5 4 | 3+4 4 | 3+4 4 | 3+4 4 | 3+4 4 | 5 2 |
| | 5 2 | 6+6 4 | 5+5 4 | 3+4 4 | 3+4 4 | 3+4 4 | 3+4 4 | 5 2 |
| | 5 2 | 5+5 4 | 6+6 4 | 3+4 4 | 3+4 4 | 3+4 4 | 3+4 4 | 6 2 |
| | 6 2 | 6+6 4 | 5+5 4 | 3+4 4 | 3+4 4 | 3+4 4 | 3+4 4 | 6 2 |
| | 6 2 | 5+5 4 | 6+6 4 | 3+4 4 | 3+4 4 | 3+4 4 | 3+4 4 | 5 2 |
| | 5 2 | 6+6 4 | 5+5 4 | 3+4 4 | 3+4 4 | 3+4 4 | 3+4 4 | 5 2 |
| | 5 2 | 5+5 4 | 6+6 4 | 3+4 4 | 3+4 4 | 3+4 4 | 3+4 4 | 5 2 |
| | 6 2 | 6+6 4 | 5+5 4 | 3+4 4 | 3+4 4 | 3+4 4 | 3+4 4 | 6 2 |

Continued on next column

**TABLE XXXI– continued**

| TQC | ICSD | NoE | SF | | | | |
|---|---|---|---|---|---|---|---|
| | 5 2 | 5+5 4 | 6+6 4 | 3+4 4 | 3+4 4 | 3+4 4 | 3+4 4 | 6 2 |
| | 6 2 | 6+6 4 | 5+5 4 | 3+4 4 | 3+4 4 | 3+4 4 | 3+4 4 | 6 2 |
| | 6 2 | 5+5 4 | 6+6 4 | 3+4 4 | 3+4 4 | 3+4 4 | 3+4 4 | 5 2 |
| | 5 2 | 6+6 4 | 5+5 4 | 3+4 4 | 3+4 4 | 3+4 4 | 3+4 4 | 6 2 |
| | 6 2 | 5+5 4 | 5+5 4 | 3+4 4 | 3+4 4 | 3+4 4 | 3+4 4 | 6 2 |
| | 6 2 | 6+6 4 | 6+6 4 | 3+4 4 | 3+4 4 | 3+4 4 | 3+4 4 | 5 2 |
| | 5 2 | 5+5 4 | 6+6 4 | 3+4 4 | 3+4 4 | 3+4 4 | 3+4 4 | 6 2 |
| | 5 2 | 5+5 4 | 5+5 4 | 3+4 4 | 3+4 4 | 3+4 4 | 3+4 4 | 5 2 |
| | 5 2 | 6+6 4 | 6+6 4 | 3+4 4 | 3+4 4 | 3+4 4 | 3+4 4 | 6 2 |
| | 6 2 | 6+6 4 | 6+6 4 | 3+4 4 | 3+4 4 | 3+4 4 | 3+4 4 | 5 2 |
| | 6 2 | 5+5 4 | 5+5 4 | 3+4 4 | 3+4 4 | 3+4 4 | 3+4 4 | 6 2 |
| | 5 2 | 5+5 4 | 5+5 4 | 3+4 4 | 3+4 4 | 3+4 4 | 3+4 4 | 5 2 |
| | 6 2 | 6+6 4 | 6+6 4 | 3+4 4 | 3+4 4 | 3+4 4 | 3+4 4 | 5 2 |
| | 5 2 | | | | | | | 6 2 |
| | 6 2 | | | | | | | 6 2 |
| | 5 2 | | | | | | | 5 2 |
| | 6 2 | | | | | | | 6 2 |
| | 5 2 | | | | | | | 5 2 |
| | 6 2 | | | | | | | 6 2 |
| | 6 2 | | | | | | | 5 2 |
| | 5 2 | | | | | | | 5 2 |
| | 5 2 | | | | | | | 5 2 |
| | 5 2 | | | | | | | 6 2 |
| | 6 2 | | | | | | | 5 2 |
| | 5 2 | | | | | | | 6 2 |
| | 6 2 | | | | | | | 5 2 |
| | 5 2 | | | | | | | 5 2 |
| | 6 2 | | | | | | | 6 2 |
| | 6 2 | | | | | | | 6 2 |
| | 5 2 | | | | | | | 6 2 |
| | 6 2 | | | | | | | 5 2 |
| | 5 2 | | | | | | | 6 2 |
| | 5 2 | | | | | | | 5 2 |
| | 6 2 | | | | | | | 5 2 |
| | 6 2 | | | | | | | 6 2 |
| | 6 2 | | | | | | | 6 2 |
| | 5 2 | | | | | | | 5 2 |
| | 6 2 | | | | | | | 6 2 |
| | 6 2 | | | | | | | 5 2 |
| | 5 2 | | | | | | | 5 2 |
| | 5 2 | | | | | | | 6 2 |
| | 5 2 | | | | | | | 5 2 |
| | 6 2 | | | | | | | 5 2 |
| | 6 2 | | | | | | | 6 2 |
| | 5 2 | | | | | | | 6 2 |
| | 5 2 | | | | | | | 6 2 |
| | 5 2 | | | | | | | 5 2 |
| | 5 2 | | | | | | | 6 2 |
| | 5 2 | | | | | | | 5 2 |
| | 5 2 | | | | | | | 5 2 |
| | 6 2 | | | | | | | 6 2 |
| | 6 2 | | | | | | | 6 2 |
| | 5 2 | | | | | | | 6 2 |

The End

TABLE XXXII: The computed irreducible representations for the valence bands for the compounds of SEBR and NLC in space group 57. See the list of tables.

| TQC | ICSD | NoE | SF |
|---|---|---|---|

Continued on next column

**TABLE XXXII– continued**

| TQC | ICSD | NoE | SF | | | | | |
|---|---|---|---|---|---|---|---|---|
| 57.2.1 | 20933 | 112 | Fe2Hf1Si2 | | | | | |
| | -GM | -R | -S | -T | -U | -X | -Y | -Z |
| | 5 2 | 4+4 4 | 3+4 4 | 4+4 4 | 3+4 4 | 5 2 | 3+4 4 | 3+4 4 |
| | 6 2 | 3+3 4 | 3+4 4 | 3+3 4 | 3+4 4 | 6 2 | 3+4 4 | 3+4 4 |
| | 5 2 | 3+3 4 | 3+4 4 | 3+3 4 | 3+4 4 | 6 2 | 3+4 4 | 3+4 4 |
| | 5 2 | 4+4 4 | 3+4 4 | 4+4 4 | 3+4 4 | 6 2 | 3+4 4 | 3+4 4 |
| | 6 2 | 3+3 4 | 3+4 4 | 4+4 4 | 3+4 4 | 5 2 | 3+4 4 | 3+4 4 |
| | 6 2 | 4+4 4 | 3+4 4 | 3+3 4 | 3+4 4 | 5 2 | 3+4 4 | 3+4 4 |
| | 5 2 | 4+4 4 | 3+4 4 | 3+3 4 | 3+4 4 | 5 2 | 3+4 4 | 3+4 4 |
| | 6 2 | 3+3 4 | 3+4 4 | 4+4 4 | 3+4 4 | 6 2 | 3+4 4 | 3+4 4 |
| | 6 2 | 3+3 4 | 3+4 4 | 3+3 4 | 3+4 4 | 6 2 | 3+4 4 | 3+4 4 |
| | 5 2 | 4+4 4 | 3+4 4 | 4+4 4 | 3+4 4 | 5 2 | 3+4 4 | 3+4 4 |
| | 6 2 | 4+4 4 | 3+4 4 | 4+4 4 | 3+4 4 | 5 2 | 3+4 4 | 3+4 4 |
| | 5 2 | 3+3 4 | 3+4 4 | 3+3 4 | 3+4 4 | 6 2 | 3+4 4 | 3+4 4 |
| | 5 2 | 3+3 4 | 3+4 4 | 3+3 4 | 3+4 4 | 5 2 | 3+4 4 | 3+4 4 |
| | 6 2 | 4+4 4 | 3+4 4 | 4+4 4 | 3+4 4 | 6 2 | 3+4 4 | 3+4 4 |
| | 5 2 | 3+3 4 | 3+4 4 | 4+4 4 | 3+4 4 | 5 2 | 3+4 4 | 3+4 4 |
| | 5 2 | 4+4 4 | 3+4 4 | 3+3 4 | 3+4 4 | 6 2 | 3+4 4 | 3+4 4 |
| | 5 2 | 3+3 4 | 3+4 4 | 4+4 4 | 3+4 4 | 6 2 | 3+4 4 | 3+4 4 |
| | 6 2 | 4+4 4 | 3+4 4 | 3+3 4 | 3+4 4 | 6 2 | 3+4 4 | 3+4 4 |
| | 5 2 | 4+4 4 | 3+4 4 | 4+4 4 | 3+4 4 | 5 2 | 3+4 4 | 3+4 4 |
| | 6 2 | 3+3 4 | 3+4 4 | 3+3 4 | 3+4 4 | 6 2 | 3+4 4 | 3+4 4 |
| | 5 2 | 3+3 4 | 3+4 4 | 3+3 4 | 3+4 4 | 5 2 | 3+4 4 | 3+4 4 |
| | 5 2 | 4+4 4 | 3+4 4 | 4+4 4 | 3+4 4 | 6 2 | 3+4 4 | 3+4 4 |
| | 6 2 | 4+4 4 | 3+4 4 | 3+3 4 | 3+4 4 | 5 2 | 3+4 4 | 3+4 4 |
| | 5 2 | 3+3 4 | 3+4 4 | 4+4 4 | 3+4 4 | 5 2 | 3+4 4 | 3+4 4 |
| | 6 2 | 4+4 4 | 3+4 4 | 4+4 4 | 3+4 4 | 6 2 | 3+4 4 | 3+4 4 |
| | 5 2 | 3+3 4 | 3+4 4 | 3+3 4 | 3+4 4 | 6 2 | 3+4 4 | 3+4 4 |
| | 6 2 | 3+3 4 | 3+4 4 | 4+4 4 | 3+4 4 | 6 2 | 3+4 4 | 3+4 4 |
| | 5 2 | 4+4 4 | 3+4 4 | 3+3 4 | 3+4 4 | 5 2 | 3+4 4 | 3+4 4 |
| | 6 2 | | | | | 5 2 | | |
| | 5 2 | | | | | 6 2 | | |
| | 6 2 | | | | | 5 2 | | |
| | 5 2 | | | | | 6 2 | | |
| | 6 2 | | | | | 6 2 | | |
| | 6 2 | | | | | 5 2 | | |
| | 6 2 | | | | | 5 2 | | |
| | 5 2 | | | | | 6 2 | | |
| | 5 2 | | | | | 5 2 | | |
| | 5 2 | | | | | 6 2 | | |
| | 6 2 | | | | | 6 2 | | |
| | 5 2 | | | | | 5 2 | | |
| | 6 2 | | | | | 5 2 | | |
| | 6 2 | | | | | 6 2 | | |
| | 5 2 | | | | | 5 2 | | |
| | 5 2 | | | | | 6 2 | | |
| | 5 2 | | | | | 5 2 | | |
| | 6 2 | | | | | 5 2 | | |
| | 6 2 | | | | | 5 2 | | |
| | 5 2 | | | | | 5 2 | | |
| | 6 2 | | | | | 5 2 | | |
| | 5 2 | | | | | 5 2 | | |
| | 5 2 | | | | | 6 2 | | |
| | 6 2 | | | | | 5 2 | | |
| | 5 2 | | | | | 6 2 | | |
| | 6 2 | | | | | 5 2 | | |
| | 6 2 | | | | | 6 2 | | |
| | 5 2 | | | | | 6 2 | | |
| 57.2.2 | 85640 | 128 | Cr1La1Sb3 | | | | | |
| | -GM | -R | -S | -T | -U | -X | -Y | -Z |
| | 5 2 | 3+3 4 | 3+4 4 | 3+3 4 | 3+4 4 | 6 2 | 3+4 4 | 3+4 4 |

Continued on next column

**TABLE XXXII– continued**

| TQC | ICSD | NoE | SF | | | | | |
|---|---|---|---|---|---|---|---|---|
| | 6 2 | 4+4 4 | 3+4 4 | 4+4 4 | 3+4 4 | 5 2 | 3+4 4 | 3+4 4 |
| | 5 2 | 3+3 4 | 3+4 4 | 3+3 4 | 3+4 4 | 5 2 | 3+4 4 | 3+4 4 |
| | 6 2 | 4+4 4 | 3+4 4 | 4+4 4 | 3+4 4 | 6 2 | 3+4 4 | 3+4 4 |
| | 5 2 | 4+4 4 | 3+4 4 | 4+4 4 | 3+4 4 | 5 2 | 3+4 4 | 3+4 4 |
| | 6 2 | 3+3 4 | 3+4 4 | 3+3 4 | 3+4 4 | 6 2 | 3+4 4 | 3+4 4 |
| | 5 2 | 3+3 4 | 3+4 4 | 3+3 4 | 3+4 4 | 6 2 | 3+4 4 | 3+4 4 |
| | 6 2 | 4+4 4 | 3+4 4 | 4+4 4 | 3+4 4 | 5 2 | 3+4 4 | 3+4 4 |
| | 5 2 | 3+3 4 | 3+4 4 | 4+4 4 | 3+4 4 | 6 2 | 3+4 4 | 3+4 4 |
| | 6 2 | 4+4 4 | 3+4 4 | 3+3 4 | 3+4 4 | 5 2 | 3+4 4 | 3+4 4 |
| | 5 2 | 3+3 4 | 3+4 4 | 3+3 4 | 3+4 4 | 6 2 | 3+4 4 | 3+4 4 |
| | 6 2 | 4+4 4 | 3+4 4 | 4+4 4 | 3+4 4 | 5 2 | 3+4 4 | 3+4 4 |
| | 5 2 | 3+3 4 | 3+4 4 | 3+3 4 | 3+4 4 | 6 2 | 3+4 4 | 3+4 4 |
| | 6 2 | 4+4 4 | 3+4 4 | 4+4 4 | 3+4 4 | 5 2 | 3+4 4 | 3+4 4 |
| | 5 2 | 4+4 4 | 3+4 4 | 4+4 4 | 3+4 4 | 6 2 | 3+4 4 | 3+4 4 |
| | 6 2 | 3+3 4 | 3+4 4 | 3+3 4 | 3+4 4 | 5 2 | 3+4 4 | 3+4 4 |
| | 5 2 | 3+3 4 | 3+4 4 | 3+3 4 | 3+4 4 | 5 2 | 3+4 4 | 3+4 4 |
| | 5 2 | 4+4 4 | 3+4 4 | 4+4 4 | 3+4 4 | 6 2 | 3+4 4 | 3+4 4 |
| | 6 2 | 3+3 4 | 3+4 4 | 3+3 4 | 3+4 4 | 6 2 | 3+4 4 | 3+4 4 |
| | 6 2 | 4+4 4 | 3+4 4 | 4+4 4 | 3+4 4 | 6 2 | 3+4 4 | 3+4 4 |
| | 6 2 | 4+4 4 | 3+4 4 | 4+4 4 | 3+4 4 | 5 2 | 3+4 4 | 3+4 4 |
| | 5 2 | 3+3 4 | 3+4 4 | 3+3 4 | 3+4 4 | 5 2 | 3+4 4 | 3+4 4 |
| | 5 2 | 4+4 4 | 3+4 4 | 3+3 4 | 3+4 4 | 6 2 | 3+4 4 | 3+4 4 |
| | 5 2 | 3+3 4 | 3+4 4 | 4+4 4 | 3+4 4 | 5 2 | 3+4 4 | 3+4 4 |
| | 5 2 | 3+3 4 | 3+4 4 | 4+4 4 | 3+4 4 | 5 2 | 3+4 4 | 3+4 4 |
| | 6 2 | 4+4 4 | 3+4 4 | 3+3 4 | 3+4 4 | 6 2 | 3+4 4 | 3+4 4 |
| | 6 2 | 4+4 4 | 3+4 4 | 3+3 4 | 3+4 4 | 6 2 | 3+4 4 | 3+4 4 |
| | 6 2 | 3+3 4 | 3+4 4 | 4+4 4 | 3+4 4 | 5 2 | 3+4 4 | 3+4 4 |
| | 6 2 | 3+3 4 | 3+4 4 | 3+3 4 | 3+4 4 | 6 2 | 3+4 4 | 3+4 4 |
| | 5 2 | 4+4 4 | 3+4 4 | 4+4 4 | 3+4 4 | 5 2 | 3+4 4 | 3+4 4 |
| | 6 2 | 4+4 4 | 3+4 4 | 3+3 4 | 3+4 4 | 5 2 | 3+4 4 | 3+4 4 |
| | 5 2 | 3+3 4 | 3+4 4 | 4+4 4 | 3+4 4 | 6 2 | 3+4 4 | 3+4 4 |
| | 6 2 | | | | | 5 2 | | |
| | 5 2 | | | | | 6 2 | | |
| | 5 2 | | | | | 6 2 | | |
| | 6 2 | | | | | 6 2 | | |
| | 5 2 | | | | | 5 2 | | |
| | 5 2 | | | | | 5 2 | | |
| | 6 2 | | | | | 6 2 | | |
| | 5 2 | | | | | 6 2 | | |
| | 5 2 | | | | | 5 2 | | |
| | 6 2 | | | | | 5 2 | | |
| | 5 2 | | | | | 5 2 | | |
| | 6 2 | | | | | 6 2 | | |
| | 5 2 | | | | | 6 2 | | |
| | 6 2 | | | | | 6 2 | | |
| | 5 2 | | | | | 5 2 | | |
| | 5 2 | | | | | 5 2 | | |
| | 6 2 | | | | | 6 2 | | |
| | 5 2 | | | | | 6 2 | | |
| | 6 2 | | | | | 5 2 | | |
| | 6 2 | | | | | 5 2 | | |
| | 5 2 | | | | | 5 2 | | |
| | 5 2 | | | | | 6 2 | | |
| | 5 2 | | | | | 5 2 | | |
| | 6 2 | | | | | 5 2 | | |
| | 5 2 | | | | | 6 2 | | |
| | 5 2 | | | | | 5 2 | | |
| | 6 2 | | | | | 6 2 | | |
| | 6 2 | | | | | 6 2 | | |
| | 6 2 | | | | | 5 2 | | |
| | 5 2 | | | | | 6 2 | | |

Continued on next column

**TABLE XXXII– continued**

| TQC | ICSD | NoE | SF | | | | | |
|---|---|---|---|---|---|---|---|---|
| | 6 2 | | | | | 6 2 | | |
| | 5 2 | | | | | 5 2 | | |
| 57.2.3 | 85639 | 128 | Cr1La1Sb3 | | | | | |
| | -GM | -R | -S | -T | -U | -X | -Y | -Z |
| | 5 2 | 3+3 4 | 3+4 4 | 3+3 4 | 3+4 4 | 6 2 | 3+4 4 | 3+4 4 |
| | 6 2 | 4+4 4 | 3+4 4 | 4+4 4 | 3+4 4 | 5 2 | 3+4 4 | 3+4 4 |
| | 5 2 | 3+3 4 | 3+4 4 | 3+3 4 | 3+4 4 | 5 2 | 3+4 4 | 3+4 4 |
| | 6 2 | 4+4 4 | 3+4 4 | 4+4 4 | 3+4 4 | 6 2 | 3+4 4 | 3+4 4 |
| | 5 2 | 4+4 4 | 3+4 4 | 4+4 4 | 3+4 4 | 6 2 | 3+4 4 | 3+4 4 |
| | 6 2 | 3+3 4 | 3+4 4 | 3+3 4 | 3+4 4 | 5 2 | 3+4 4 | 3+4 4 |
| | 5 2 | 3+3 4 | 3+4 4 | 3+3 4 | 3+4 4 | 6 2 | 3+4 4 | 3+4 4 |
| | 6 2 | 4+4 4 | 3+4 4 | 4+4 4 | 3+4 4 | 5 2 | 3+4 4 | 3+4 4 |
| | 5 2 | 3+3 4 | 3+4 4 | 4+4 4 | 3+4 4 | 6 2 | 3+4 4 | 3+4 4 |
| | 6 2 | 4+4 4 | 3+4 4 | 3+3 4 | 3+4 4 | 5 2 | 3+4 4 | 3+4 4 |
| | 5 2 | 3+3 4 | 3+4 4 | 3+3 4 | 3+4 4 | 6 2 | 3+4 4 | 3+4 4 |
| | 6 2 | 4+4 4 | 3+4 4 | 4+4 4 | 3+4 4 | 5 2 | 3+4 4 | 3+4 4 |
| | 5 2 | 3+3 4 | 3+4 4 | 3+3 4 | 3+4 4 | 6 2 | 3+4 4 | 3+4 4 |
| | 6 2 | 4+4 4 | 3+4 4 | 4+4 4 | 3+4 4 | 5 2 | 3+4 4 | 3+4 4 |
| | 5 2 | 4+4 4 | 3+4 4 | 4+4 4 | 3+4 4 | 6 2 | 3+4 4 | 3+4 4 |
| | 6 2 | 3+3 4 | 3+4 4 | 3+3 4 | 3+4 4 | 5 2 | 3+4 4 | 3+4 4 |
| | 5 2 | 4+4 4 | 3+4 4 | 4+4 4 | 3+4 4 | 5 2 | 3+4 4 | 3+4 4 |
| | 5 2 | 3+3 4 | 3+4 4 | 3+3 4 | 3+4 4 | 6 2 | 3+4 4 | 3+4 4 |
| | 6 2 | 3+3 4 | 3+4 4 | 3+3 4 | 3+4 4 | 6 2 | 3+4 4 | 3+4 4 |
| | 6 2 | 4+4 4 | 3+4 4 | 4+4 4 | 3+4 4 | 5 2 | 3+4 4 | 3+4 4 |
| | 6 2 | 4+4 4 | 3+4 4 | 4+4 4 | 3+4 4 | 6 2 | 3+4 4 | 3+4 4 |
| | 5 2 | 3+3 4 | 3+4 4 | 3+3 4 | 3+4 4 | 5 2 | 3+4 4 | 3+4 4 |
| | 5 2 | 4+4 4 | 3+4 4 | 3+3 4 | 3+4 4 | 6 2 | 3+4 4 | 3+4 4 |
| | 5 2 | 3+3 4 | 3+4 4 | 4+4 4 | 3+4 4 | 5 2 | 3+4 4 | 3+4 4 |
| | 5 2 | 3+3 4 | 3+4 4 | 4+4 4 | 3+4 4 | 5 2 | 3+4 4 | 3+4 4 |
| | 6 2 | 4+4 4 | 3+4 4 | 3+3 4 | 3+4 4 | 6 2 | 3+4 4 | 3+4 4 |
| | 6 2 | 4+4 4 | 3+4 4 | 3+3 4 | 3+4 4 | 6 2 | 3+4 4 | 3+4 4 |
| | 6 2 | 3+3 4 | 3+4 4 | 4+4 4 | 3+4 4 | 5 2 | 3+4 4 | 3+4 4 |
| | 6 2 | 3+3 4 | 3+4 4 | 3+3 4 | 3+4 4 | 6 2 | 3+4 4 | 3+4 4 |
| | 5 2 | 4+4 4 | 3+4 4 | 4+4 4 | 3+4 4 | 5 2 | 3+4 4 | 3+4 4 |
| | 6 2 | 4+4 4 | 3+4 4 | 3+3 4 | 3+4 4 | 5 2 | 3+4 4 | 3+4 4 |
| | 5 2 | 3+3 4 | 3+4 4 | 4+4 4 | 3+4 4 | 6 2 | 3+4 4 | 3+4 4 |
| | 6 2 | | | | | 5 2 | | |
| | 5 2 | | | | | 6 2 | | |
| | 5 2 | | | | | 6 2 | | |
| | 6 2 | | | | | 6 2 | | |
| | 5 2 | | | | | 5 2 | | |
| | 5 2 | | | | | 5 2 | | |
| | 6 2 | | | | | 6 2 | | |
| | 5 2 | | | | | 6 2 | | |
| | 5 2 | | | | | 5 2 | | |
| | 6 2 | | | | | 5 2 | | |
| | 5 2 | | | | | 5 2 | | |
| | 5 2 | | | | | 6 2 | | |
| | 6 2 | | | | | 6 2 | | |
| | 5 2 | | | | | 6 2 | | |
| | 6 2 | | | | | 5 2 | | |
| | 5 2 | | | | | 5 2 | | |
| | 6 2 | | | | | 6 2 | | |
| | 5 2 | | | | | 6 2 | | |
| | 6 2 | | | | | 5 2 | | |
| | 6 2 | | | | | 5 2 | | |
| | 5 2 | | | | | 6 2 | | |
| | 5 2 | | | | | 5 2 | | |
| | 5 2 | | | | | 5 2 | | |
| | 6 2 | | | | | 5 2 | | |
| | 5 2 | | | | | 6 2 | | |

Continued on next column

**TABLE XXXII– continued**

| TQC | ICSD | NoE | SF | | | | |
|---|---|---|---|---|---|---|---|
| | 5 2 | | | | 5 2 | | |
| | 6 2 | | | | 6 2 | | |
| | 6 2 | | | | 6 2 | | |
| | 6 2 | | | | 6 2 | | |
| | 5 2 | | | | 5 2 | | |
| | 6 2 | | | | 6 2 | | |
| | 5 2 | | | | 5 2 | | |

| TQC | ICSD | NoE | SF | | | | | |
|---|---|---|---|---|---|---|---|---|
| 57.2.4 | 83907 | 128 | Cr1La1Sb3 | | | | | |
| | -GM | -R | -S | -T | -U | -X | -Y | -Z |
| | 5 2 | 3+3 4 | 3+4 4 | 3+3 4 | 3+4 4 | 5 2 | 3+4 4 | 3+4 4 |
| | 6 2 | 4+4 4 | 3+4 4 | 4+4 4 | 3+4 4 | 6 2 | 3+4 4 | 3+4 4 |
| | 5 2 | 4+4 4 | 3+4 4 | 4+4 4 | 3+4 4 | 5 2 | 3+4 4 | 3+4 4 |
| | 6 2 | 3+3 4 | 3+4 4 | 3+3 4 | 3+4 4 | 6 2 | 3+4 4 | 3+4 4 |
| | 5 2 | 3+3 4 | 3+4 4 | 3+3 4 | 3+4 4 | 5 2 | 3+4 4 | 3+4 4 |
| | 6 2 | 4+4 4 | 3+4 4 | 4+4 4 | 3+4 4 | 6 2 | 3+4 4 | 3+4 4 |
| | 5 2 | 4+4 4 | 3+4 4 | 4+4 4 | 3+4 4 | 6 2 | 3+4 4 | 3+4 4 |
| | 6 2 | 3+3 4 | 3+4 4 | 3+3 4 | 3+4 4 | 5 2 | 3+4 4 | 3+4 4 |
| | 5 2 | 4+4 4 | 3+4 4 | 4+4 4 | 3+4 4 | 6 2 | 3+4 4 | 3+4 4 |
| | 6 2 | 3+3 4 | 3+4 4 | 3+3 4 | 3+4 4 | 5 2 | 3+4 4 | 3+4 4 |
| | 5 2 | 4+4 4 | 3+4 4 | 4+4 4 | 3+4 4 | 6 2 | 3+4 4 | 3+4 4 |
| | 5 2 | 3+3 4 | 3+4 4 | 3+3 4 | 3+4 4 | 5 2 | 3+4 4 | 3+4 4 |
| | 6 2 | 3+3 4 | 3+4 4 | 3+3 4 | 3+4 4 | 6 2 | 3+4 4 | 3+4 4 |
| | 6 2 | 4+4 4 | 3+4 4 | 4+4 4 | 3+4 4 | 5 2 | 3+4 4 | 3+4 4 |
| | 5 2 | 4+4 4 | 3+4 4 | 4+4 4 | 3+4 4 | 6 2 | 3+4 4 | 3+4 4 |
| | 6 2 | 3+3 4 | 3+4 4 | 3+3 4 | 3+4 4 | 5 2 | 3+4 4 | 3+4 4 |
| | 5 2 | 4+4 4 | 3+4 4 | 3+3 4 | 3+4 4 | 5 2 | 3+4 4 | 3+4 4 |
| | 5 2 | 3+3 4 | 3+4 4 | 4+4 4 | 3+4 4 | 6 2 | 3+4 4 | 3+4 4 |
| | 6 2 | 4+4 4 | 3+4 4 | 4+4 4 | 3+4 4 | 6 2 | 3+4 4 | 3+4 4 |
| | 6 2 | 3+3 4 | 3+4 4 | 3+3 4 | 3+4 4 | 6 2 | 3+4 4 | 3+4 4 |
| | 5 2 | 4+4 4 | 3+4 4 | 4+4 4 | 3+4 4 | 5 2 | 3+4 4 | 3+4 4 |
| | 6 2 | 3+3 4 | 3+4 4 | 3+3 4 | 3+4 4 | 5 2 | 3+4 4 | 3+4 4 |
| | 5 2 | 3+3 4 | 3+4 4 | 3+3 4 | 3+4 4 | 5 2 | 3+4 4 | 3+4 4 |
| | 5 2 | 4+4 4 | 3+4 4 | 4+4 4 | 3+4 4 | 6 2 | 3+4 4 | 3+4 4 |
| | 6 2 | 3+3 4 | 3+4 4 | 4+4 4 | 3+4 4 | 6 2 | 3+4 4 | 3+4 4 |
| | 5 2 | 4+4 4 | 3+4 4 | 3+3 4 | 3+4 4 | 5 2 | 3+4 4 | 3+4 4 |
| | 6 2 | 4+4 4 | 3+4 4 | 3+3 4 | 3+4 4 | 6 2 | 3+4 4 | 3+4 4 |
| | 6 2 | 3+3 4 | 3+4 4 | 4+4 4 | 3+4 4 | 5 2 | 3+4 4 | 3+4 4 |
| | 5 2 | 3+3 4 | 3+4 4 | 3+3 4 | 3+4 4 | 5 2 | 3+4 4 | 3+4 4 |
| | 6 2 | 4+4 4 | 3+4 4 | 4+4 4 | 3+4 4 | 6 2 | 3+4 4 | 3+4 4 |
| | 6 2 | 4+4 4 | 3+4 4 | 3+3 4 | 3+4 4 | 5 2 | 3+4 4 | 3+4 4 |
| | 5 2 | 3+3 4 | 3+4 4 | 4+4 4 | 3+4 4 | 6 2 | 3+4 4 | 3+4 4 |
| | 6 2 | | | | | 5 2 | | |
| | 5 2 | | | | | 6 2 | | |
| | 6 2 | | | | | 6 2 | | |
| | 5 2 | | | | | 5 2 | | |
| | 5 2 | | | | | 5 2 | | |
| | 5 2 | | | | | 6 2 | | |
| | 6 2 | | | | | 6 2 | | |
| | 5 2 | | | | | 6 2 | | |
| | 6 2 | | | | | 5 2 | | |
| | 5 2 | | | | | 5 2 | | |
| | 5 2 | | | | | 6 2 | | |
| | 6 2 | | | | | 5 2 | | |
| | 6 2 | | | | | 6 2 | | |
| | 5 2 | | | | | 6 2 | | |
| | 6 2 | | | | | 5 2 | | |
| | 5 2 | | | | | 5 2 | | |
| | 6 2 | | | | | 6 2 | | |
| | 5 2 | | | | | 5 2 | | |
| | 5 2 | | | | | 6 2 | | |
| | 6 2 | | | | | 5 2 | | |

Continued on next column

**TABLE XXXII– continued**

| TQC | ICSD | NoE | SF | | | | |
|---|---|---|---|---|---|---|---|
| | 5 2 | | | | 6 2 | | |
| | 6 2 | | | | 5 2 | | |
| | 5 2 | | | | 5 2 | | |
| | 5 2 | | | | 5 2 | | |
| | 5 2 | | | | 6 2 | | |
| | 6 2 | | | | 6 2 | | |
| | 5 2 | | | | 5 2 | | |
| | 6 2 | | | | 5 2 | | |
| | 5 2 | | | | 6 2 | | |
| | 6 2 | | | | 6 2 | | |
| | 6 2 | | | | 6 2 | | |
| | 6 2 | | | | 6 2 | | |

| TQC | ICSD | NoE | SF | | | | | |
|---|---|---|---|---|---|---|---|---|
| 57.2.5 | 247291 | 248 | Ce3Rh2Si2 | | | | | |
| | -GM | -R | -S | -T | -U | -X | -Y | -Z |
| | 5 2 | 3+3 4 | 3+4 4 | 4+4 4 | 3+4 4 | 5 2 | 3+4 4 | 3+4 4 |
| | 5 2 | 4+4 4 | 3+4 4 | 3+3 4 | 3+4 4 | 5 2 | 3+4 4 | 3+4 4 |
| | 6 2 | 3+3 4 | 3+4 4 | 3+3 4 | 3+4 4 | 6 2 | 3+4 4 | 3+4 4 |
| | 6 2 | 4+4 4 | 3+4 4 | 4+4 4 | 3+4 4 | 5 2 | 3+4 4 | 3+4 4 |
| | 5 2 | 4+4 4 | 3+4 4 | 3+3 4 | 3+4 4 | 6 2 | 3+4 4 | 3+4 4 |
| | 6 2 | 3+3 4 | 3+4 4 | 4+4 4 | 3+4 4 | 6 2 | 3+4 4 | 3+4 4 |
| | 5 2 | 3+3 4 | 3+4 4 | 3+3 4 | 3+4 4 | 5 2 | 3+4 4 | 3+4 4 |
| | 5 2 | 4+4 4 | 3+4 4 | 4+4 4 | 3+4 4 | 5 2 | 3+4 4 | 3+4 4 |
| | 6 2 | 4+4 4 | 3+4 4 | 4+4 4 | 3+4 4 | 6 2 | 3+4 4 | 3+4 4 |
| | 6 2 | 3+3 4 | 3+4 4 | 3+3 4 | 3+4 4 | 5 2 | 3+4 4 | 3+4 4 |
| | 5 2 | 3+3 4 | 3+4 4 | 3+3 4 | 3+4 4 | 6 2 | 3+4 4 | 3+4 4 |
| | 6 2 | 4+4 4 | 3+4 4 | 4+4 4 | 3+4 4 | 6 2 | 3+4 4 | 3+4 4 |
| | 5 2 | 3+3 4 | 3+4 4 | 3+3 4 | 3+4 4 | 5 2 | 3+4 4 | 3+4 4 |
| | 5 2 | 4+4 4 | 3+4 4 | 4+4 4 | 3+4 4 | 5 2 | 3+4 4 | 3+4 4 |
| | 6 2 | 3+3 4 | 3+4 4 | 3+3 4 | 3+4 4 | 6 2 | 3+4 4 | 3+4 4 |
| | 5 2 | 4+4 4 | 3+4 4 | 4+4 4 | 3+4 4 | 5 2 | 3+4 4 | 3+4 4 |
| | 6 2 | 4+4 4 | 3+4 4 | 4+4 4 | 3+4 4 | 6 2 | 3+4 4 | 3+4 4 |
| | 6 2 | 3+3 4 | 3+4 4 | 3+3 4 | 3+4 4 | 5 2 | 3+4 4 | 3+4 4 |
| | 5 2 | 4+4 4 | 3+4 4 | 3+3 4 | 3+4 4 | 6 2 | 3+4 4 | 3+4 4 |
| | 6 2 | 3+3 4 | 3+4 4 | 4+4 4 | 3+4 4 | 5 2 | 3+4 4 | 3+4 4 |
| | 5 2 | 3+3 4 | 3+4 4 | 3+3 4 | 3+4 4 | 6 2 | 3+4 4 | 3+4 4 |
| | 6 2 | 4+4 4 | 3+4 4 | 4+4 4 | 3+4 4 | 5 2 | 3+4 4 | 3+4 4 |
| | 5 2 | 3+3 4 | 3+4 4 | 4+4 4 | 3+4 4 | 6 2 | 3+4 4 | 3+4 4 |
| | 6 2 | 4+4 4 | 3+4 4 | 3+3 4 | 3+4 4 | 6 2 | 3+4 4 | 3+4 4 |
| | 5 2 | 3+3 4 | 3+4 4 | 3+3 4 | 3+4 4 | 5 2 | 3+4 4 | 3+4 4 |
| | 5 2 | 4+4 4 | 3+4 4 | 4+4 4 | 3+4 4 | 6 2 | 3+4 4 | 3+4 4 |
| | 5 2 | 3+3 4 | 3+4 4 | 3+3 4 | 3+4 4 | 6 2 | 3+4 4 | 3+4 4 |
| | 6 2 | 4+4 4 | 3+4 4 | 4+4 4 | 3+4 4 | 5 2 | 3+4 4 | 3+4 4 |
| | 6 2 | 4+4 4 | 3+4 4 | 4+4 4 | 3+4 4 | 5 2 | 3+4 4 | 3+4 4 |
| | 5 2 | 3+3 4 | 3+4 4 | 3+3 4 | 3+4 4 | 5 2 | 3+4 4 | 3+4 4 |
| | 6 2 | 4+4 4 | 3+4 4 | 4+4 4 | 3+4 4 | 5 2 | 3+4 4 | 3+4 4 |
| | 6 2 | 3+3 4 | 3+4 4 | 3+3 4 | 3+4 4 | 6 2 | 3+4 4 | 3+4 4 |
| | 5 2 | 3+3 4 | 3+4 4 | 4+4 4 | 3+4 4 | 6 2 | 3+4 4 | 3+4 4 |
| | 5 2 | 4+4 4 | 3+4 4 | 3+3 4 | 3+4 4 | 5 2 | 3+4 4 | 3+4 4 |
| | 6 2 | 4+4 4 | 3+4 4 | 3+3 4 | 3+4 4 | 6 2 | 3+4 4 | 3+4 4 |
| | 6 2 | 3+3 4 | 3+4 4 | 4+4 4 | 3+4 4 | 6 2 | 3+4 4 | 3+4 4 |
| | 5 2 | 3+3 4 | 3+4 4 | 3+3 4 | 3+4 4 | 5 2 | 3+4 4 | 3+4 4 |
| | 6 2 | 4+4 4 | 3+4 4 | 4+4 4 | 3+4 4 | 5 2 | 3+4 4 | 3+4 4 |
| | 5 2 | 3+3 4 | 3+4 4 | 4+4 4 | 3+4 4 | 6 2 | 3+4 4 | 3+4 4 |
| | 6 2 | 4+4 4 | 3+4 4 | 3+3 4 | 3+4 4 | 5 2 | 3+4 4 | 3+4 4 |
| | 5 2 | 3+3 4 | 3+4 4 | 3+3 4 | 3+4 4 | 6 2 | 3+4 4 | 3+4 4 |
| | 6 2 | 4+4 4 | 3+4 4 | 4+4 4 | 3+4 4 | 5 2 | 3+4 4 | 3+4 4 |
| | 5 2 | 4+4 4 | 3+4 4 | 3+3 4 | 3+4 4 | 6 2 | 3+4 4 | 3+4 4 |
| | 5 2 | 3+3 4 | 3+4 4 | 4+4 4 | 3+4 4 | 5 2 | 3+4 4 | 3+4 4 |
| | 5 2 | 3+3 4 | 3+4 4 | 3+3 4 | 3+4 4 | 6 2 | 3+4 4 | 3+4 4 |
| | 6 2 | 4+4 4 | 3+4 4 | 4+4 4 | 3+4 4 | 5 2 | 3+4 4 | 3+4 4 |
| | 6 2 | 4+4 4 | 3+4 4 | 3+3 4 | 3+4 4 | 6 2 | 3+4 4 | 3+4 4 |

Continued on next column

**TABLE XXXII– continued**

| TQC | ICSD | NoE | SF | | | | | |
|---|---|---|---|---|---|---|---|---|
| | 6 2 | 3+3 4 | 3+4 4 | 4+4 4 | 3+4 4 | 6 2 | 3+4 4 | 3+4 4 |
| | 5 2 | 3+3 4 | 3+4 4 | 3+3 4 | 3+4 4 | 6 2 | 3+4 4 | 3+4 4 |
| | 5 2 | 4+4 4 | 3+4 4 | 4+4 4 | 3+4 4 | 6 2 | 3+4 4 | 3+4 4 |
| | 6 2 | 3+3 4 | 3+4 4 | 4+4 4 | 3+4 4 | 5 2 | 3+4 4 | 3+4 4 |
| | 6 2 | 4+4 4 | 3+4 4 | 3+3 4 | 3+4 4 | 5 2 | 3+4 4 | 3+4 4 |
| | 6 2 | 4+4 4 | 3+4 4 | 4+4 4 | 3+4 4 | 6 2 | 3+4 4 | 3+4 4 |
| | 5 2 | 3+3 4 | 3+4 4 | 3+3 4 | 3+4 4 | 5 2 | 3+4 4 | 3+4 4 |
| | 5 2 | 3+3 4 | 3+4 4 | 4+4 4 | 3+4 4 | 5 2 | 3+4 4 | 3+4 4 |
| | 6 2 | 4+4 4 | 3+4 4 | 3+3 4 | 3+4 4 | 6 2 | 3+4 4 | 3+4 4 |
| | 5 2 | 4+4 4 | 3+4 4 | 3+3 4 | 3+4 4 | 6 2 | 3+4 4 | 3+4 4 |
| | 6 2 | 3+3 4 | 3+4 4 | 4+4 4 | 3+4 4 | 6 2 | 3+4 4 | 3+4 4 |
| | 5 2 | 4+4 4 | 3+4 4 | 4+4 4 | 3+4 4 | 5 2 | 3+4 4 | 3+4 4 |
| | 6 2 | 3+3 4 | 3+4 4 | 3+3 4 | 3+4 4 | 5 2 | 3+4 4 | 3+4 4 |
| | 6 2 | 3+3 4 | 3+4 4 | 3+3 4 | 3+4 4 | 5 2 | 3+4 4 | 3+4 4 |
| | 6 2 | 4+4 4 | 3+4 4 | 4+4 4 | 3+4 4 | 6 2 | 3+4 4 | 3+4 4 |
| | 5 2 | | | | | 5 2 | | |
| | 5 2 | | | | | 6 2 | | |
| | 6 2 | | | | | 5 2 | | |
| | 6 2 | | | | | 5 2 | | |
| | 5 2 | | | | | 6 2 | | |
| | 6 2 | | | | | 6 2 | | |
| | 5 2 | | | | | 6 2 | | |
| | 5 2 | | | | | 5 2 | | |
| | 5 2 | | | | | 5 2 | | |
| | 6 2 | | | | | 5 2 | | |
| | 5 2 | | | | | 6 2 | | |
| | 6 2 | | | | | 5 2 | | |
| | 6 2 | | | | | 6 2 | | |
| | 5 2 | | | | | 5 2 | | |
| | 6 2 | | | | | 6 2 | | |
| | 5 2 | | | | | 5 2 | | |
| | 6 2 | | | | | 5 2 | | |
| | 5 2 | | | | | 6 2 | | |
| | 6 2 | | | | | 6 2 | | |
| | 6 2 | | | | | 6 2 | | |
| | 5 2 | | | | | 5 2 | | |
| | 6 2 | | | | | 5 2 | | |
| | 5 2 | | | | | 6 2 | | |
| | 6 2 | | | | | 5 2 | | |
| | 5 2 | | | | | 6 2 | | |
| | 6 2 | | | | | 5 2 | | |
| | 5 2 | | | | | 6 2 | | |
| | 6 2 | | | | | 6 2 | | |
| | 6 2 | | | | | 6 2 | | |
| | 5 2 | | | | | 5 2 | | |
| | 6 2 | | | | | 6 2 | | |
| | 5 2 | | | | | 5 2 | | |
| | 5 2 | | | | | 5 2 | | |
| | 5 2 | | | | | 6 2 | | |
| | 6 2 | | | | | 6 2 | | |
| | 5 2 | | | | | 5 2 | | |
| | 5 2 | | | | | 6 2 | | |
| | 6 2 | | | | | 5 2 | | |
| | 6 2 | | | | | 5 2 | | |
| | 6 2 | | | | | 6 2 | | |
| | 6 2 | | | | | 5 2 | | |
| | 5 2 | | | | | 6 2 | | |
| | 5 2 | | | | | 5 2 | | |
| | 5 2 | | | | | 6 2 | | |
| | 6 2 | | | | | 6 2 | | |
| | 5 2 | | | | | 6 2 | | |

Continued on next column

**TABLE XXXII– continued**

| TQC | ICSD | NoE | SF | | | | | |
|---|---|---|---|---|---|---|---|---|
| | 5 2 | | | | | 6 2 | | |
| | 5 2 | | | | | 5 2 | | |
| | 6 2 | | | | | 5 2 | | |
| | 5 2 | | | | | 6 2 | | |
| | 5 2 | | | | | 5 2 | | |
| | 5 2 | | | | | 6 2 | | |
| | 6 2 | | | | | 6 2 | | |
| | 5 2 | | | | | 5 2 | | |
| | 6 2 | | | | | 6 2 | | |
| | 6 2 | | | | | 6 2 | | |
| | 5 2 | | | | | 5 2 | | |
| | 6 2 | | | | | 5 2 | | |
| | 5 2 | | | | | 5 2 | | |
| | 6 2 | | | | | 6 2 | | |
| | 6 2 | | | | | 5 2 | | |
| | 5 2 | | | | | 6 2 | | |
| 57.2.6 | 76352 | 248 | Ce3Ge2Rh2 | | | | | |
| | -GM | -R | -S | -T | -U | -X | -Y | -Z |
| | 5 2 | 3+3 4 | 3+4 4 | 3+3 4 | 3+4 4 | 6 2 | 3+4 4 | 3+4 4 |
| | 6 2 | 4+4 4 | 3+4 4 | 4+4 4 | 3+4 4 | 6 2 | 3+4 4 | 3+4 4 |
| | 5 2 | 3+3 4 | 3+4 4 | 3+3 4 | 3+4 4 | 5 2 | 3+4 4 | 3+4 4 |
| | 6 2 | 4+4 4 | 3+4 4 | 4+4 4 | 3+4 4 | 6 2 | 3+4 4 | 3+4 4 |
| | 5 2 | 3+3 4 | 3+4 4 | 3+3 4 | 3+4 4 | 5 2 | 3+4 4 | 3+4 4 |
| | 6 2 | 4+4 4 | 3+4 4 | 4+4 4 | 3+4 4 | 6 2 | 3+4 4 | 3+4 4 |
| | 5 2 | 3+3 4 | 3+4 4 | 3+3 4 | 3+4 4 | 5 2 | 3+4 4 | 3+4 4 |
| | 5 2 | 4+4 4 | 3+4 4 | 4+4 4 | 3+4 4 | 5 2 | 3+4 4 | 3+4 4 |
| | 6 2 | 3+3 4 | 3+4 4 | 3+3 4 | 3+4 4 | 6 2 | 3+4 4 | 3+4 4 |
| | 5 2 | 4+4 4 | 3+4 4 | 4+4 4 | 3+4 4 | 6 2 | 3+4 4 | 3+4 4 |
| | 6 2 | 3+3 4 | 3+4 4 | 3+3 4 | 3+4 4 | 5 2 | 3+4 4 | 3+4 4 |
| | 6 2 | 4+4 4 | 3+4 4 | 4+4 4 | 3+4 4 | 5 2 | 3+4 4 | 3+4 4 |
| | 5 2 | 4+4 4 | 3+4 4 | 3+3 4 | 3+4 4 | 6 2 | 3+4 4 | 3+4 4 |
| | 5 2 | 3+3 4 | 3+4 4 | 4+4 4 | 3+4 4 | 6 2 | 3+4 4 | 3+4 4 |
| | 5 2 | 4+4 4 | 3+4 4 | 4+4 4 | 3+4 4 | 6 2 | 3+4 4 | 3+4 4 |
| | 6 2 | 3+3 4 | 3+4 4 | 3+3 4 | 3+4 4 | 5 2 | 3+4 4 | 3+4 4 |
| | 6 2 | 3+3 4 | 3+4 4 | 3+3 4 | 3+4 4 | 5 2 | 3+4 4 | 3+4 4 |
| | 6 2 | 4+4 4 | 3+4 4 | 4+4 4 | 3+4 4 | 6 2 | 3+4 4 | 3+4 4 |
| | 5 2 | 3+3 4 | 3+4 4 | 4+4 4 | 3+4 4 | 5 2 | 3+4 4 | 3+4 4 |
| | 6 2 | 4+4 4 | 3+4 4 | 3+3 4 | 3+4 4 | 6 2 | 3+4 4 | 3+4 4 |
| | 5 2 | 4+4 4 | 3+4 4 | 4+4 4 | 3+4 4 | 5 2 | 3+4 4 | 3+4 4 |
| | 6 2 | 3+3 4 | 3+4 4 | 3+3 4 | 3+4 4 | 6 2 | 3+4 4 | 3+4 4 |
| | 5 2 | 4+4 4 | 3+4 4 | 3+3 4 | 3+4 4 | 5 2 | 3+4 4 | 3+4 4 |
| | 6 2 | 3+3 4 | 3+4 4 | 4+4 4 | 3+4 4 | 5 2 | 3+4 4 | 3+4 4 |
| | 5 2 | 3+3 4 | 3+4 4 | 3+3 4 | 3+4 4 | 6 2 | 3+4 4 | 3+4 4 |
| | 5 2 | 4+4 4 | 3+4 4 | 4+4 4 | 3+4 4 | 5 2 | 3+4 4 | 3+4 4 |
| | 5 2 | 4+4 4 | 3+4 4 | 3+3 4 | 3+4 4 | 5 2 | 3+4 4 | 3+4 4 |
| | 6 2 | 3+3 4 | 3+4 4 | 4+4 4 | 3+4 4 | 6 2 | 3+4 4 | 3+4 4 |
| | 6 2 | 3+3 4 | 3+4 4 | 3+3 4 | 3+4 4 | 6 2 | 3+4 4 | 3+4 4 |
| | 5 2 | 4+4 4 | 3+4 4 | 4+4 4 | 3+4 4 | 6 2 | 3+4 4 | 3+4 4 |
| | 6 2 | 3+3 4 | 3+4 4 | 3+3 4 | 3+4 4 | 5 2 | 3+4 4 | 3+4 4 |
| | 6 2 | 4+4 4 | 3+4 4 | 4+4 4 | 3+4 4 | 6 2 | 3+4 4 | 3+4 4 |
| | 5 2 | 4+4 4 | 3+4 4 | 3+3 4 | 3+4 4 | 5 2 | 3+4 4 | 3+4 4 |
| | 5 2 | 3+3 4 | 3+4 4 | 4+4 4 | 3+4 4 | 5 2 | 3+4 4 | 3+4 4 |
| | 6 2 | 3+3 4 | 3+4 4 | 4+4 4 | 3+4 4 | 6 2 | 3+4 4 | 3+4 4 |
| | 6 2 | 4+4 4 | 3+4 4 | 3+3 4 | 3+4 4 | 5 2 | 3+4 4 | 3+4 4 |
| | 5 2 | 4+4 4 | 3+4 4 | 4+4 4 | 3+4 4 | 6 2 | 3+4 4 | 3+4 4 |
| | 6 2 | 3+3 4 | 3+4 4 | 3+3 4 | 3+4 4 | 6 2 | 3+4 4 | 3+4 4 |
| | 5 2 | 3+3 4 | 3+4 4 | 3+3 4 | 3+4 4 | 5 2 | 3+4 4 | 3+4 4 |
| | 6 2 | 4+4 4 | 3+4 4 | 4+4 4 | 3+4 4 | 6 2 | 3+4 4 | 3+4 4 |
| | 5 2 | 4+4 4 | 3+4 4 | 4+4 4 | 3+4 4 | 6 2 | 3+4 4 | 3+4 4 |
| | 5 2 | 3+3 4 | 3+4 4 | 3+3 4 | 3+4 4 | 5 2 | 3+4 4 | 3+4 4 |
| | 6 2 | 4+4 4 | 3+4 4 | 4+4 4 | 3+4 4 | 5 2 | 3+4 4 | 3+4 4 |

Continued on next column

**TABLE XXXII– continued**

| TQC | ICSD | NoE | SF |
|---|---|---|---|
| | 5 2 3+3 4 3+4 4 3+3 4 3+4 4 6 2 3+4 4 3+4 4 | | |
| | 5 2 4+4 4 3+4 4 4+4 4 3+4 4 5 2 3+4 4 3+4 4 | | |
| | 6 2 3+3 4 3+4 4 3+3 4 3+4 4 6 2 3+4 4 3+4 4 | | |
| | 6 2 3+3 4 3+4 4 4+4 4 3+4 4 5 2 3+4 4 3+4 4 | | |
| | 6 2 4+4 4 3+4 4 3+3 4 3+4 4 5 2 3+4 4 3+4 4 | | |
| | 5 2 4+4 4 3+4 4 4+4 4 3+4 4 5 2 3+4 4 3+4 4 | | |
| | 5 2 3+3 4 3+4 4 3+3 4 3+4 4 5 2 3+4 4 3+4 4 | | |
| | 6 2 4+4 4 3+4 4 3+3 4 3+4 4 6 2 3+4 4 3+4 4 | | |
| | 6 2 3+3 4 3+4 4 4+4 4 3+4 4 6 2 3+4 4 3+4 4 | | |
| | 6 2 3+3 4 3+4 4 3+3 4 3+4 4 6 2 3+4 4 3+4 4 | | |
| | 5 2 4+4 4 3+4 4 4+4 4 3+4 4 5 2 3+4 4 3+4 4 | | |
| | 5 2 3+3 4 3+4 4 3+3 4 3+4 4 6 2 3+4 4 3+4 4 | | |
| | 6 2 4+4 4 3+4 4 4+4 4 3+4 4 5 2 3+4 4 3+4 4 | | |
| | 5 2 3+3 4 3+4 4 3+3 4 3+4 4 5 2 3+4 4 3+4 4 | | |
| | 6 2 4+4 4 3+4 4 4+4 4 3+4 4 5 2 3+4 4 3+4 4 | | |
| | 5 2 3+3 4 3+4 4 3+3 4 3+4 4 6 2 3+4 4 3+4 4 | | |
| | 6 2 4+4 4 3+4 4 4+4 4 3+4 4 6 2 3+4 4 3+4 4 | | |
| | 6 2 4+4 4 3+4 4 4+4 4 3+4 4 6 2 3+4 4 3+4 4 | | |
| | 6 2 3+3 4 3+4 4 3+3 4 3+4 4 5 2 3+4 4 3+4 4 | | |
| | 5 2 5 2 | | |
| | 5 2 6 2 | | |
| | 6 2 6 2 | | |
| | 5 2 6 2 | | |
| | 6 2 5 2 | | |
| | 6 2 5 2 | | |
| | 5 2 5 2 | | |
| | 5 2 6 2 | | |
| | 6 2 6 2 | | |
| | 5 2 6 2 | | |
| | 5 2 5 2 | | |
| | 6 2 6 2 | | |
| | 6 2 5 2 | | |
| | 5 2 5 2 | | |
| | 6 2 6 2 | | |
| | 6 2 6 2 | | |
| | 5 2 6 2 | | |
| | 5 2 5 2 | | |
| | 6 2 5 2 | | |
| | 6 2 6 2 | | |
| | 5 2 6 2 | | |
| | 6 2 5 2 | | |
| | 5 2 6 2 | | |
| | 6 2 5 2 | | |
| | 6 2 6 2 | | |
| | 5 2 5 2 | | |
| | 5 2 5 2 | | |
| | 6 2 5 2 | | |
| | 6 2 6 2 | | |
| | 5 2 5 2 | | |
| | 5 2 5 2 | | |
| | 6 2 6 2 | | |
| | 5 2 6 2 | | |
| | 5 2 5 2 | | |
| | 6 2 6 2 | | |
| | 5 2 5 2 | | |
| | 6 2 5 2 | | |
| | 5 2 6 2 | | |
| | 6 2 6 2 | | |
| | 6 2 5 2 | | |
| | 6 2 6 2 | | |
| | 5 2 5 2 | | |

Continued on next column

**TABLE XXXII– continued**

| TQC | ICSD | NoE | SF |
|---|---|---|---|
| | 5 2 6 2 | | |
| | 5 2 5 2 | | |
| | 6 2 5 2 | | |
| | 5 2 5 2 | | |
| | 5 2 5 2 | | |
| | 6 2 5 2 | | |
| | 5 2 6 2 | | |
| | 5 2 6 2 | | |
| | 5 2 6 2 | | |
| | 6 2 5 2 | | |
| | 6 2 5 2 | | |
| | 5 2 6 2 | | |
| | 5 2 5 2 | | |
| | 6 2 6 2 | | |
| | 5 2 5 2 | | |
| | 6 2 6 2 | | |
| | 5 2 6 2 | | |
| | 6 2 5 2 | | |
| | 5 2 6 2 | | |
| | 6 2 5 2 | | |
| 57.2.7 | 190838 | 176 | Ce1Pd2Sn3 |
| | -GM -R -S -T -U -X -Y -Z | | |
| | 5 2 3+3 4 3+4 4 3+3 4 3+4 4 5 2 3+4 4 3+4 4 | | |
| | 6 2 4+4 4 3+4 4 4+4 4 3+4 4 6 2 3+4 4 3+4 4 | | |
| | 5 2 4+4 4 3+4 4 4+4 4 3+4 4 5 2 3+4 4 3+4 4 | | |
| | 6 2 3+3 4 3+4 4 3+3 4 3+4 4 6 2 3+4 4 3+4 4 | | |
| | 5 2 3+3 4 3+4 4 3+3 4 3+4 4 5 2 3+4 4 3+4 4 | | |
| | 6 2 4+4 4 3+4 4 4+4 4 3+4 4 6 2 3+4 4 3+4 4 | | |
| | 5 2 4+4 4 3+4 4 4+4 4 3+4 4 5 2 3+4 4 3+4 4 | | |
| | 6 2 3+3 4 3+4 4 3+3 4 3+4 4 6 2 3+4 4 3+4 4 | | |
| | 5 2 3+3 4 3+4 4 3+3 4 3+4 4 5 2 3+4 4 3+4 4 | | |
| | 6 2 4+4 4 3+4 4 4+4 4 3+4 4 6 2 3+4 4 3+4 4 | | |
| | 5 2 4+4 4 3+4 4 4+4 4 3+4 4 5 2 3+4 4 3+4 4 | | |
| | 6 2 3+3 4 3+4 4 3+3 4 3+4 4 6 2 3+4 4 3+4 4 | | |
| | 5 2 3+3 4 3+4 4 3+3 4 3+4 4 5 2 3+4 4 3+4 4 | | |
| | 6 2 4+4 4 3+4 4 4+4 4 3+4 4 6 2 3+4 4 3+4 4 | | |
| | 5 2 3+3 4 3+4 4 4+4 4 3+4 4 5 2 3+4 4 3+4 4 | | |
| | 6 2 4+4 4 3+4 4 3+3 4 3+4 4 6 2 3+4 4 3+4 4 | | |
| | 5 2 3+3 4 3+4 4 3+3 4 3+4 4 5 2 3+4 4 3+4 4 | | |
| | 5 2 4+4 4 3+4 4 4+4 4 3+4 4 6 2 3+4 4 3+4 4 | | |
| | 6 2 4+4 4 3+4 4 4+4 4 3+4 4 5 2 3+4 4 3+4 4 | | |
| | 6 2 3+3 4 3+4 4 3+3 4 3+4 4 6 2 3+4 4 3+4 4 | | |
| | 5 2 3+3 4 3+4 4 3+3 4 3+4 4 5 2 3+4 4 3+4 4 | | |
| | 6 2 4+4 4 3+4 4 4+4 4 3+4 4 5 2 3+4 4 3+4 4 | | |
| | 5 2 3+3 4 3+4 4 4+4 4 3+4 4 6 2 3+4 4 3+4 4 | | |
| | 6 2 4+4 4 3+4 4 3+3 4 3+4 4 6 2 3+4 4 3+4 4 | | |
| | 6 2 4+4 4 3+4 4 4+4 4 3+4 4 5 2 3+4 4 3+4 4 | | |
| | 5 2 3+3 4 3+4 4 3+3 4 3+4 4 5 2 3+4 4 3+4 4 | | |
| | 5 2 4+4 4 3+4 4 4+4 4 3+4 4 6 2 3+4 4 3+4 4 | | |
| | 6 2 3+3 4 3+4 4 3+3 4 3+4 4 6 2 3+4 4 3+4 4 | | |
| | 5 2 4+4 4 3+4 4 4+4 4 3+4 4 6 2 3+4 4 3+4 4 | | |
| | 6 2 3+3 4 3+4 4 3+3 4 3+4 4 5 2 3+4 4 3+4 4 | | |
| | 6 2 4+4 4 3+4 4 3+3 4 3+4 4 5 2 3+4 4 3+4 4 | | |
| | 5 2 3+3 4 3+4 4 4+4 4 3+4 4 6 2 3+4 4 3+4 4 | | |
| | 6 2 3+3 4 3+4 4 3+3 4 3+4 4 6 2 3+4 4 3+4 4 | | |
| | 5 2 4+4 4 3+4 4 4+4 4 3+4 4 5 2 3+4 4 3+4 4 | | |
| | 5 2 4+4 4 3+4 4 3+3 4 3+4 4 5 2 3+4 4 3+4 4 | | |
| | 5 2 3+3 4 3+4 4 4+4 4 3+4 4 6 2 3+4 4 3+4 4 | | |
| | 6 2 3+3 4 3+4 4 4+4 4 3+4 4 5 2 3+4 4 3+4 4 | | |
| | 5 2 4+4 4 3+4 4 3+3 4 3+4 4 5 2 3+4 4 3+4 4 | | |
| | 5 2 3+3 4 3+4 4 3+3 4 3+4 4 6 2 3+4 4 3+4 4 | | |

Continued on next column

**TABLE XXXII– continued**

| TQC | ICSD | NoE | SF | | | | | |
|---|---|---|---|---|---|---|---|---|
| | 6 2 | 4+4 4 | 3+4 4 | 4+4 4 | 3+4 4 | 6 2 | 3+4 4 | 3+4 4 |
| | 5 2 | 4+4 4 | 3+4 4 | 4+4 4 | 3+4 4 | 5 2 | 3+4 4 | 3+4 4 |
| | 6 2 | 3+3 4 | 3+4 4 | 3+3 4 | 3+4 4 | 5 2 | 3+4 4 | 3+4 4 |
| | 5 2 | 3+3 4 | 3+4 4 | 3+3 4 | 3+4 4 | 6 2 | 3+4 4 | 3+4 4 |
| | 5 2 | 4+4 4 | 3+4 4 | 4+4 4 | 3+4 4 | 6 2 | 3+4 4 | 3+4 4 |
| | 6 2 | | | | | 5 2 | | |
| | 5 2 | | | | | 6 2 | | |
| | 6 2 | | | | | 6 2 | | |
| | 5 2 | | | | | 5 2 | | |
| | 6 2 | | | | | 5 2 | | |
| | 6 2 | | | | | 6 2 | | |
| | 6 2 | | | | | 6 2 | | |
| | 5 2 | | | | | 5 2 | | |
| | 5 2 | | | | | 5 2 | | |
| | 5 2 | | | | | 6 2 | | |
| | 6 2 | | | | | 5 2 | | |
| | 6 2 | | | | | 6 2 | | |
| | 6 2 | | | | | 5 2 | | |
| | 5 2 | | | | | 6 2 | | |
| | 5 2 | | | | | 6 2 | | |
| | 6 2 | | | | | 5 2 | | |
| | 6 2 | | | | | 6 2 | | |
| | 5 2 | | | | | 5 2 | | |
| | 5 2 | | | | | 6 2 | | |
| | 6 2 | | | | | 5 2 | | |
| | 6 2 | | | | | 6 2 | | |
| | 5 2 | | | | | 5 2 | | |
| | 6 2 | | | | | 6 2 | | |
| | 5 2 | | | | | 5 2 | | |
| | 6 2 | | | | | 6 2 | | |
| | 5 2 | | | | | 5 2 | | |
| | 6 2 | | | | | 6 2 | | |
| | 5 2 | | | | | 5 2 | | |
| | 6 2 | | | | | 6 2 | | |
| | 5 2 | | | | | 5 2 | | |
| | 5 2 | | | | | 6 2 | | |
| | 5 2 | | | | | 5 2 | | |
| | 6 2 | | | | | 6 2 | | |
| | 6 2 | | | | | 5 2 | | |
| | 6 2 | | | | | 5 2 | | |
| | 5 2 | | | | | 6 2 | | |
| | 5 2 | | | | | 6 2 | | |
| | 6 2 | | | | | 5 2 | | |
| | 5 2 | | | | | 5 2 | | |
| | 6 2 | | | | | 5 2 | | |
| | 5 2 | | | | | 6 2 | | |
| | 5 2 | | | | | 6 2 | | |
| | 6 2 | | | | | 6 2 | | |
| | 6 2 | | | | | 5 2 | | |
| 57.2.8 | 93749 | 192 | Ca1Mn2O4 | | | | | |
| | -GM | -R | -S | -T | -U | -X | -Y | -Z |
| | 5 2 | 3+3 4 | 3+4 4 | 3+3 4 | 3+4 4 | 5 2 | 3+4 4 | 3+4 4 |
| | 5 2 | 4+4 4 | 3+4 4 | 4+4 4 | 3+4 4 | 5 2 | 3+4 4 | 3+4 4 |
| | 6 2 | 4+4 4 | 3+4 4 | 4+4 4 | 3+4 4 | 6 2 | 3+4 4 | 3+4 4 |
| | 6 2 | 3+3 4 | 3+4 4 | 3+3 4 | 3+4 4 | 6 2 | 3+4 4 | 3+4 4 |
| | 5 2 | 3+3 4 | 3+4 4 | 3+3 4 | 3+4 4 | 5 2 | 3+4 4 | 3+4 4 |
| | 6 2 | 4+4 4 | 3+4 4 | 4+4 4 | 3+4 4 | 6 2 | 3+4 4 | 3+4 4 |
| | 5 2 | 3+3 4 | 3+4 4 | 3+3 4 | 3+4 4 | 6 2 | 3+4 4 | 3+4 4 |
| | 6 2 | 4+4 4 | 3+4 4 | 4+4 4 | 3+4 4 | 5 2 | 3+4 4 | 3+4 4 |
| | 5 2 | 3+3 4 | 3+4 4 | 4+4 4 | 3+4 4 | 6 2 | 3+4 4 | 3+4 4 |
| | 5 2 | 4+4 4 | 3+4 4 | 3+3 4 | 3+4 4 | 5 2 | 3+4 4 | 3+4 4 |
| | 6 2 | 4+4 4 | 3+4 4 | 3+3 4 | 3+4 4 | 6 2 | 3+4 4 | 3+4 4 |
| | 6 2 | 3+3 4 | 3+4 4 | 4+4 4 | 3+4 4 | 5 2 | 3+4 4 | 3+4 4 |
| | 5 2 | 4+4 4 | 3+4 4 | 3+3 4 | 3+4 4 | 5 2 | 3+4 4 | 3+4 4 |
| | 5 2 | 3+3 4 | 3+4 4 | 4+4 4 | 3+4 4 | 6 2 | 3+4 4 | 3+4 4 |
| | 6 2 | 3+3 4 | 3+4 4 | 3+3 4 | 3+4 4 | 6 2 | 3+4 4 | 3+4 4 |
| | 6 2 | 4+4 4 | 3+4 4 | 4+4 4 | 3+4 4 | 5 2 | 3+4 4 | 3+4 4 |
| | 5 2 | 3+3 4 | 3+4 4 | 3+3 4 | 3+4 4 | 5 2 | 3+4 4 | 3+4 4 |
| | 5 2 | 4+4 4 | 3+4 4 | 4+4 4 | 3+4 4 | 5 2 | 3+4 4 | 3+4 4 |
| | 5 2 | 3+3 4 | 3+4 4 | 3+3 4 | 3+4 4 | 5 2 | 3+4 4 | 3+4 4 |
| | 5 2 | 4+4 4 | 3+4 4 | 4+4 4 | 3+4 4 | 5 2 | 3+4 4 | 3+4 4 |
| | 6 2 | 3+3 4 | 3+4 4 | 3+3 4 | 3+4 4 | 6 2 | 3+4 4 | 3+4 4 |
| | 6 2 | 4+4 4 | 3+4 4 | 4+4 4 | 3+4 4 | 6 2 | 3+4 4 | 3+4 4 |
| | 6 2 | 3+3 4 | 3+4 4 | 3+3 4 | 3+4 4 | 6 2 | 3+4 4 | 3+4 4 |
| | 6 2 | 4+4 4 | 3+4 4 | 4+4 4 | 3+4 4 | 6 2 | 3+4 4 | 3+4 4 |
| | 5 2 | 3+3 4 | 3+4 4 | 3+3 4 | 3+4 4 | 6 2 | 3+4 4 | 3+4 4 |
| | 5 2 | 4+4 4 | 3+4 4 | 4+4 4 | 3+4 4 | 5 2 | 3+4 4 | 3+4 4 |
| | 5 2 | 4+4 4 | 3+4 4 | 3+3 4 | 3+4 4 | 6 2 | 3+4 4 | 3+4 4 |
| | 6 2 | 3+3 4 | 3+4 4 | 4+4 4 | 3+4 4 | 5 2 | 3+4 4 | 3+4 4 |
| | 6 2 | 3+3 4 | 3+4 4 | 4+4 4 | 3+4 4 | 5 2 | 3+4 4 | 3+4 4 |
| | 6 2 | 4+4 4 | 3+4 4 | 3+3 4 | 3+4 4 | 6 2 | 3+4 4 | 3+4 4 |
| | 6 2 | 3+3 4 | 3+4 4 | 3+3 4 | 3+4 4 | 6 2 | 3+4 4 | 3+4 4 |
| | 5 2 | 4+4 4 | 3+4 4 | 4+4 4 | 3+4 4 | 5 2 | 3+4 4 | 3+4 4 |
| | 6 2 | 3+3 4 | 3+4 4 | 3+3 4 | 3+4 4 | 6 2 | 3+4 4 | 3+4 4 |
| | 5 2 | 4+4 4 | 3+4 4 | 4+4 4 | 3+4 4 | 5 2 | 3+4 4 | 3+4 4 |
| | 6 2 | 3+3 4 | 3+4 4 | 3+3 4 | 3+4 4 | 6 2 | 3+4 4 | 3+4 4 |
| | 5 2 | 4+4 4 | 3+4 4 | 4+4 4 | 3+4 4 | 5 2 | 3+4 4 | 3+4 4 |
| | 5 2 | 4+4 4 | 3+4 4 | 3+3 4 | 3+4 4 | 6 2 | 3+4 4 | 3+4 4 |
| | 6 2 | 3+3 4 | 3+4 4 | 4+4 4 | 3+4 4 | 5 2 | 3+4 4 | 3+4 4 |
| | 5 2 | 3+3 4 | 3+4 4 | 4+4 4 | 3+4 4 | 5 2 | 3+4 4 | 3+4 4 |
| | 5 2 | 4+4 4 | 3+4 4 | 3+3 4 | 3+4 4 | 6 2 | 3+4 4 | 3+4 4 |
| | 6 2 | 3+3 4 | 3+4 4 | 3+3 4 | 3+4 4 | 6 2 | 3+4 4 | 3+4 4 |
| | 6 2 | 4+4 4 | 3+4 4 | 4+4 4 | 3+4 4 | 5 2 | 3+4 4 | 3+4 4 |
| | 5 2 | 3+3 4 | 3+4 4 | 3+3 4 | 3+4 4 | 6 2 | 3+4 4 | 3+4 4 |
| | 5 2 | 4+4 4 | 3+4 4 | 4+4 4 | 3+4 4 | 6 2 | 3+4 4 | 3+4 4 |
| | 6 2 | 4+4 4 | 3+4 4 | 4+4 4 | 3+4 4 | 5 2 | 3+4 4 | 3+4 4 |
| | 5 2 | 3+3 4 | 3+4 4 | 3+3 4 | 3+4 4 | 6 2 | 3+4 4 | 3+4 4 |
| | 5 2 | 4+4 4 | 3+4 4 | 3+3 4 | 3+4 4 | 5 2 | 3+4 4 | 3+4 4 |
| | 6 2 | 3+3 4 | 3+4 4 | 4+4 4 | 3+4 4 | 5 2 | 3+4 4 | 3+4 4 |
| | 5 2 | | | | | 5 2 | | |
| | 6 2 | | | | | 5 2 | | |
| | 5 2 | | | | | 5 2 | | |
| | 6 2 | | | | | 6 2 | | |
| | 5 2 | | | | | 6 2 | | |
| | 5 2 | | | | | 5 2 | | |
| | 5 2 | | | | | 6 2 | | |
| | 6 2 | | | | | 6 2 | | |
| | 6 2 | | | | | 5 2 | | |
| | 6 2 | | | | | 5 2 | | |
| | 5 2 | | | | | 6 2 | | |
| | 5 2 | | | | | 6 2 | | |
| | 5 2 | | | | | 6 2 | | |
| | 5 2 | | | | | 6 2 | | |
| | 6 2 | | | | | 5 2 | | |
| | 6 2 | | | | | 5 2 | | |
| | 5 2 | | | | | 5 2 | | |
| | 6 2 | | | | | 5 2 | | |
| | 5 2 | | | | | 6 2 | | |
| | 5 2 | | | | | 6 2 | | |
| | 5 2 | | | | | 6 2 | | |
| | 6 2 | | | | | 5 2 | | |
| | 5 2 | | | | | 5 2 | | |
| | 6 2 | | | | | 5 2 | | |

Continued on next column

**TABLE XXXII– continued**

| TQC | ICSD | NoE | SF | | | | | |
|---|---|---|---|---|---|---|---|---|
| | 6 2 | | | | | 5 2 | | |
| | 6 2 | | | | | 5 2 | | |
| | 5 2 | | | | | 6 2 | | |
| | 5 2 | | | | | 6 2 | | |
| | 6 2 | | | | | 6 2 | | |
| | 6 2 | | | | | 6 2 | | |
| | 6 2 | | | | | 6 2 | | |
| | 6 2 | | | | | 5 2 | | |
| | 6 2 | | | | | 6 2 | | |
| | 6 2 | | | | | 6 2 | | |
| | 6 2 | | | | | 6 2 | | |
| | 5 2 | | | | | 5 2 | | |
| | 5 2 | | | | | 5 2 | | |
| | 5 2 | | | | | 6 2 | | |
| | 6 2 | | | | | 5 2 | | |
| | 5 2 | | | | | 6 2 | | |
| | 6 2 | | | | | 6 2 | | |
| | 5 2 | | | | | 5 2 | | |
| | 6 2 | | | | | 5 2 | | |
| | 5 2 | | | | | 6 2 | | |
| | 5 2 | | | | | 5 2 | | |
| | 6 2 | | | | | 6 2 | | |
| | 6 2 | | | | | 6 2 | | |
| | 5 2 | | | | | 5 2 | | |
| | 5 2 | | | | | 5 2 | | |
| 57.2.9 | 150592 | 184 | Al1Dy1 | | | | | |
| | -GM | -R | -S | -T | -U | -X | -Y | -Z |
| | 5 2 | 3+3 4 | 3+4 4 | 3+3 4 | 3+4 4 | 6 2 | 3+4 4 | 3+4 4 |
| | 5 2 | 4+4 4 | 3+4 4 | 4+4 4 | 3+4 4 | 5 2 | 3+4 4 | 3+4 4 |
| | 6 2 | 3+3 4 | 3+4 4 | 3+3 4 | 3+4 4 | 5 2 | 3+4 4 | 3+4 4 |
| | 5 2 | 4+4 4 | 3+4 4 | 4+4 4 | 3+4 4 | 5 2 | 3+4 4 | 3+4 4 |
| | 6 2 | 3+3 4 | 3+4 4 | 3+3 4 | 3+4 4 | 6 2 | 3+4 4 | 3+4 4 |
| | 6 2 | 4+4 4 | 3+4 4 | 4+4 4 | 3+4 4 | 5 2 | 3+4 4 | 3+4 4 |
| | 6 2 | 4+4 4 | 3+4 4 | 3+3 4 | 3+4 4 | 6 2 | 3+4 4 | 3+4 4 |
| | 5 2 | 3+3 4 | 3+4 4 | 4+4 4 | 3+4 4 | 6 2 | 3+4 4 | 3+4 4 |
| | 6 2 | 4+4 4 | 3+4 4 | 4+4 4 | 3+4 4 | 5 2 | 3+4 4 | 3+4 4 |
| | 5 2 | 3+3 4 | 3+4 4 | 3+3 4 | 3+4 4 | 5 2 | 3+4 4 | 3+4 4 |
| | 5 2 | 4+4 4 | 3+4 4 | 3+3 4 | 3+4 4 | 6 2 | 3+4 4 | 3+4 4 |
| | 5 2 | 3+3 4 | 3+4 4 | 4+4 4 | 3+4 4 | 5 2 | 3+4 4 | 3+4 4 |
| | 6 2 | 3+3 4 | 3+4 4 | 3+3 4 | 3+4 4 | 6 2 | 3+4 4 | 3+4 4 |
| | 5 2 | 4+4 4 | 3+4 4 | 4+4 4 | 3+4 4 | 6 2 | 3+4 4 | 3+4 4 |
| | 6 2 | 4+4 4 | 3+4 4 | 4+4 4 | 3+4 4 | 6 2 | 3+4 4 | 3+4 4 |
| | 6 2 | 3+3 4 | 3+4 4 | 3+3 4 | 3+4 4 | 5 2 | 3+4 4 | 3+4 4 |
| | 5 2 | 4+4 4 | 3+4 4 | 4+4 4 | 3+4 4 | 5 2 | 3+4 4 | 3+4 4 |
| | 5 2 | 3+3 4 | 3+4 4 | 3+3 4 | 3+4 4 | 5 2 | 3+4 4 | 3+4 4 |
| | 6 2 | 3+3 4 | 3+4 4 | 4+4 4 | 3+4 4 | 6 2 | 3+4 4 | 3+4 4 |
| | 5 2 | 4+4 4 | 3+4 4 | 3+3 4 | 3+4 4 | 6 2 | 3+4 4 | 3+4 4 |
| | 6 2 | 4+4 4 | 3+4 4 | 4+4 4 | 3+4 4 | 6 2 | 3+4 4 | 3+4 4 |
| | 5 2 | 3+3 4 | 3+4 4 | 3+3 4 | 3+4 4 | 5 2 | 3+4 4 | 3+4 4 |
| | 6 2 | 3+3 4 | 3+4 4 | 3+3 4 | 3+4 4 | 5 2 | 3+4 4 | 3+4 4 |
| | 5 2 | 4+4 4 | 3+4 4 | 4+4 4 | 3+4 4 | 5 2 | 3+4 4 | 3+4 4 |
| | 6 2 | 3+3 4 | 3+4 4 | 3+3 4 | 3+4 4 | 6 2 | 3+4 4 | 3+4 4 |
| | 5 2 | 4+4 4 | 3+4 4 | 4+4 4 | 3+4 4 | 5 2 | 3+4 4 | 3+4 4 |
| | 6 2 | 4+4 4 | 3+4 4 | 3+3 4 | 3+4 4 | 6 2 | 3+4 4 | 3+4 4 |
| | 5 2 | 3+3 4 | 3+4 4 | 4+4 4 | 3+4 4 | 5 2 | 3+4 4 | 3+4 4 |
| | 6 2 | 4+4 4 | 3+4 4 | 3+3 4 | 3+4 4 | 6 2 | 3+4 4 | 3+4 4 |
| | 6 2 | 3+3 4 | 3+4 4 | 4+4 4 | 3+4 4 | 5 2 | 3+4 4 | 3+4 4 |
| | 5 2 | 3+3 4 | 3+4 4 | 4+4 4 | 3+4 4 | 6 2 | 3+4 4 | 3+4 4 |
| | 6 2 | 4+4 4 | 3+4 4 | 3+3 4 | 3+4 4 | 6 2 | 3+4 4 | 3+4 4 |
| | 5 2 | 3+3 4 | 3+4 4 | 3+3 4 | 3+4 4 | 5 2 | 3+4 4 | 3+4 4 |
| | 6 2 | 4+4 4 | 3+4 4 | 4+4 4 | 3+4 4 | 6 2 | 3+4 4 | 3+4 4 |

Continued on next column

**TABLE XXXII– continued**

| TQC | ICSD | NoE | SF | | | | | |
|---|---|---|---|---|---|---|---|---|
| | 5 2 | 3+3 4 | 3+4 4 | 3+3 4 | 3+4 4 | 5 2 | 3+4 4 | 3+4 4 |
| | 6 2 | 4+4 4 | 3+4 4 | 4+4 4 | 3+4 4 | 6 2 | 3+4 4 | 3+4 4 |
| | 5 2 | 4+4 4 | 3+4 4 | 4+4 4 | 3+4 4 | 6 2 | 3+4 4 | 3+4 4 |
| | 6 2 | 3+3 4 | 3+4 4 | 3+3 4 | 3+4 4 | 5 2 | 3+4 4 | 3+4 4 |
| | 6 2 | 3+3 4 | 3+4 4 | 3+3 4 | 3+4 4 | 6 2 | 3+4 4 | 3+4 4 |
| | 5 2 | 4+4 4 | 3+4 4 | 4+4 4 | 3+4 4 | 5 2 | 3+4 4 | 3+4 4 |
| | 5 2 | 4+4 4 | 3+4 4 | 4+4 4 | 3+4 4 | 6 2 | 3+4 4 | 3+4 4 |
| | 6 2 | 3+3 4 | 3+4 4 | 3+3 4 | 3+4 4 | 5 2 | 3+4 4 | 3+4 4 |
| | 6 2 | 3+3 4 | 3+4 4 | 3+3 4 | 3+4 4 | 5 2 | 3+4 4 | 3+4 4 |
| | 5 2 | 4+4 4 | 3+4 4 | 4+4 4 | 3+4 4 | 5 2 | 3+4 4 | 3+4 4 |
| | 5 2 | 4+4 4 | 3+4 4 | 3+3 4 | 3+4 4 | 6 2 | 3+4 4 | 3+4 4 |
| | 6 2 | 3+3 4 | 3+4 4 | 4+4 4 | 3+4 4 | 6 2 | 3+4 4 | 3+4 4 |
| | 6 2 | | | | | 6 2 | | |
| | 5 2 | | | | | 6 2 | | |
| | 5 2 | | | | | 6 2 | | |
| | 6 2 | | | | | 5 2 | | |
| | 5 2 | | | | | 5 2 | | |
| | 6 2 | | | | | 6 2 | | |
| | 5 2 | | | | | 5 2 | | |
| | 6 2 | | | | | 5 2 | | |
| | 5 2 | | | | | 6 2 | | |
| | 5 2 | | | | | 6 2 | | |
| | 6 2 | | | | | 5 2 | | |
| | 5 2 | | | | | 6 2 | | |
| | 6 2 | | | | | 6 2 | | |
| | 6 2 | | | | | 5 2 | | |
| | 5 2 | | | | | 6 2 | | |
| | 6 2 | | | | | 5 2 | | |
| | 5 2 | | | | | 6 2 | | |
| | 6 2 | | | | | 5 2 | | |
| | 6 2 | | | | | 5 2 | | |
| | 5 2 | | | | | 5 2 | | |
| | 6 2 | | | | | 6 2 | | |
| | 5 2 | | | | | 5 2 | | |
| | 5 2 | | | | | 6 2 | | |
| | 6 2 | | | | | 5 2 | | |
| | 6 2 | | | | | 6 2 | | |
| | 5 2 | | | | | 6 2 | | |
| | 5 2 | | | | | 6 2 | | |
| | 6 2 | | | | | 5 2 | | |
| | 5 2 | | | | | 6 2 | | |
| | 5 2 | | | | | 5 2 | | |
| | 6 2 | | | | | 5 2 | | |
| | 6 2 | | | | | 6 2 | | |
| | 5 2 | | | | | 5 2 | | |
| | 6 2 | | | | | 6 2 | | |
| | 5 2 | | | | | 6 2 | | |
| | 5 2 | | | | | 5 2 | | |
| | 6 2 | | | | | 5 2 | | |
| | 5 2 | | | | | 6 2 | | |
| | 6 2 | | | | | 6 2 | | |
| | 6 2 | | | | | 5 2 | | |
| | 5 2 | | | | | 5 2 | | |
| | 5 2 | | | | | 6 2 | | |
| | 6 2 | | | | | 5 2 | | |
| | 5 2 | | | | | 6 2 | | |
| | 6 2 | | | | | 6 2 | | |
| | 6 2 | | | | | 5 2 | | |

The End

TABLE XXXIII: The computed irreducible representations for the valence bands for the compounds of SEBR and NLC in space group See the list of tables.

| TQC | ICSD | NoE | SF | | | | | |
|---|---|---|---|---|---|---|---|---|
| 58.2.1 | 42724 | 40 | Ni1Sb2 | | | | | |
| | -GM | -R | -S | -T | -U | -X | -Y | -Z |
| | 5 2 | 3+4 4 | 6+6 4 | 2+5 2 | 8+9 2 | 3+4 4 | 3+4 4 | 3+4 4 |
| | 5 2 | 3+4 4 | 5+5 4 | 3+4 2 | 6+7 2 | 3+4 4 | 3+4 4 | 3+4 4 |
| | 6 2 | 3+4 4 | 5+5 4 | 7+8 2 | 2+3 2 | 3+4 4 | 3+4 4 | 3+4 4 |
| | 6 2 | 3+4 4 | 5+5 4 | 6+9 2 | 4+5 2 | 3+4 4 | 3+4 4 | 3+4 4 |
| | 5 2 | 3+4 4 | 6+6 4 | 6+9 2 | 8+9 2 | 3+4 4 | 3+4 4 | 3+4 4 |
| | 5 2 | 3+4 4 | 5+5 4 | 7+8 2 | 6+7 2 | 3+4 4 | 3+4 4 | 3+4 4 |
| | 5 2 | 3+4 4 | 5+5 4 | 2+5 2 | 2+3 2 | 3+4 4 | 3+4 4 | 3+4 4 |
| | 5 2 | 3+4 4 | 5+5 4 | 3+4 2 | 4+5 2 | 3+4 4 | 3+4 4 | 3+4 4 |
| | 5 2 | 3+4 4 | 5+5 4 | 7+8 2 | 6+7 2 | 3+4 4 | 3+4 4 | 3+4 4 |
| | 6 2 | 3+4 4 | 6+6 4 | 6+9 2 | 8+9 2 | 3+4 4 | 3+4 4 | 3+4 4 |
| | 5 2 | | | 7+8 2 | 4+5 2 | | | |
| | 5 2 | | | 6+9 2 | 8+9 2 | | | |
| | 5 2 | | | 3+4 2 | 6+7 2 | | | |
| | 5 2 | | | 6+9 2 | 2+3 2 | | | |
| | 6 2 | | | 7+8 2 | 8+9 2 | | | |
| | 5 2 | | | 2+5 2 | 6+7 2 | | | |
| | 6 2 | | | 6+9 2 | 6+7 2 | | | |
| | 5 2 | | | 7+8 2 | 8+9 2 | | | |
| | 5 2 | | | 2+5 2 | 6+7 2 | | | |
| | 5 2 | | | 3+4 2 | 8+9 2 | | | |
| 58.2.2 | 414575 | 132 | Ge6Li2Sr4 | | | | | |
| | -GM | -R | -S | -T | -U | -X | -Y | -Z |
| | 5 2 | 3+4 4 | 5+5 4 | 6+9 2 | 6+7 2 | 3+4 4 | 3+4 4 | 3+4 4 |
| | 5 2 | 3+4 4 | 6+6 4 | 7+8 2 | 8+9 2 | 3+4 4 | 3+4 4 | 3+4 4 |
| | 6 2 | 3+4 4 | 6+6 4 | 2+5 2 | 2+3 2 | 3+4 4 | 3+4 4 | 3+4 4 |
| | 6 2 | 3+4 4 | 5+5 4 | 3+4 2 | 4+5 2 | 3+4 4 | 3+4 4 | 3+4 4 |
| | 5 2 | 3+4 4 | 5+5 4 | 6+9 2 | 2+3 2 | 3+4 4 | 3+4 4 | 3+4 4 |
| | 5 2 | 3+4 4 | 6+6 4 | 7+8 2 | 4+5 2 | 3+4 4 | 3+4 4 | 3+4 4 |
| | 6 2 | 3+4 4 | 6+6 4 | 2+5 2 | 6+7 2 | 3+4 4 | 3+4 4 | 3+4 4 |
| | 6 2 | 3+4 4 | 5+5 4 | 3+4 2 | 8+9 2 | 3+4 4 | 3+4 4 | 3+4 4 |
| | 5 2 | 3+4 4 | 5+5 4 | 6+9 2 | 6+7 2 | 3+4 4 | 3+4 4 | 3+4 4 |
| | 5 2 | 3+4 4 | 6+6 4 | 7+8 2 | 8+9 2 | 3+4 4 | 3+4 4 | 3+4 4 |
| | 6 2 | 3+4 4 | 5+5 4 | 2+5 2 | 2+3 2 | 3+4 4 | 3+4 4 | 3+4 4 |
| | 6 2 | 3+4 4 | 6+6 4 | 3+4 2 | 4+5 2 | 3+4 4 | 3+4 4 | 3+4 4 |
| | 5 2 | 3+4 4 | 6+6 4 | 6+9 2 | 2+3 2 | 3+4 4 | 3+4 4 | 3+4 4 |
| | 5 2 | 3+4 4 | 5+5 4 | 7+8 2 | 4+5 2 | 3+4 4 | 3+4 4 | 3+4 4 |
| | 6 2 | 3+4 4 | 6+6 4 | 2+5 2 | 6+7 2 | 3+4 4 | 3+4 4 | 3+4 4 |
| | 6 2 | 3+4 4 | 5+5 4 | 3+4 2 | 8+9 2 | 3+4 4 | 3+4 4 | 3+4 4 |
| | 5 2 | 3+4 4 | 5+5 4 | 2+5 2 | 2+3 2 | 3+4 4 | 3+4 4 | 3+4 4 |
| | 5 2 | 3+4 4 | 6+6 4 | 6+9 2 | 4+5 2 | 3+4 4 | 3+4 4 | 3+4 4 |
| | 6 2 | 3+4 4 | 6+6 4 | 3+4 2 | 6+7 2 | 3+4 4 | 3+4 4 | 3+4 4 |
| | 6 2 | 3+4 4 | 5+5 4 | 7+8 2 | 8+9 2 | 3+4 4 | 3+4 4 | 3+4 4 |
| | 5 2 | 3+4 4 | 5+5 4 | 6+9 2 | 6+7 2 | 3+4 4 | 3+4 4 | 3+4 4 |
| | 5 2 | 3+4 4 | 6+6 4 | 7+8 2 | 8+9 2 | 3+4 4 | 3+4 4 | 3+4 4 |
| | 6 2 | 3+4 4 | 6+6 4 | 2+5 2 | 4+5 2 | 3+4 4 | 3+4 4 | 3+4 4 |
| | 6 2 | 3+4 4 | 5+5 4 | 3+4 2 | 2+3 2 | 3+4 4 | 3+4 4 | 3+4 4 |
| | 5 2 | 3+4 4 | 6+6 4 | 7+8 2 | 2+3 2 | 3+4 4 | 3+4 4 | 3+4 4 |
| | 5 2 | 3+4 4 | 6+6 4 | 6+9 2 | 4+5 2 | 3+4 4 | 3+4 4 | 3+4 4 |
| | 6 2 | 3+4 4 | 5+5 4 | 3+4 2 | 6+7 2 | 3+4 4 | 3+4 4 | 3+4 4 |
| | 6 2 | 3+4 4 | 5+5 4 | 2+5 2 | 8+9 2 | 3+4 4 | 3+4 4 | 3+4 4 |
| | 5 2 | 3+4 4 | 6+6 4 | 7+8 2 | 2+3 2 | 3+4 4 | 3+4 4 | 3+4 4 |
| | 5 2 | 3+4 4 | 5+5 4 | 6+9 2 | 4+5 2 | 3+4 4 | 3+4 4 | 3+4 4 |
| | 6 2 | 3+4 4 | 5+5 4 | 2+5 2 | 8+9 2 | 3+4 4 | 3+4 4 | 3+4 4 |
| | 6 2 | 3+4 4 | 6+6 4 | 3+4 2 | 6+7 2 | 3+4 4 | 3+4 4 | 3+4 4 |
| | 5 2 | 3+4 4 | 5+5 4 | 2+5 2 | 2+3 2 | 3+4 4 | 3+4 4 | 3+4 4 |
| | 5 2 | | | 3+4 2 | 4+5 2 | | | |
| | 5 2 | | | 7+8 2 | 6+7 2 | | | |
| | 6 2 | | | 6+9 2 | 8+9 2 | | | |
| | 6 2 | | | 3+4 2 | 8+9 2 | | | |
| | 5 2 | | | 2+5 2 | 6+7 2 | | | |
| | 6 2 | | | 6+9 2 | 4+5 2 | | | |
| | 6 2 | | | 7+8 2 | 2+3 2 | | | |
| | 5 2 | | | 2+5 2 | 2+3 2 | | | |
| | 5 2 | | | 3+4 2 | 4+5 2 | | | |
| | 6 2 | | | 6+9 2 | 6+7 2 | | | |
| | 6 2 | | | 7+8 2 | 8+9 2 | | | |
| | 6 2 | | | 6+9 2 | 8+9 2 | | | |
| | 5 2 | | | 7+8 2 | 6+7 2 | | | |
| | 5 2 | | | 3+4 2 | 2+3 2 | | | |
| | 6 2 | | | 2+5 2 | 4+5 2 | | | |
| | 5 2 | | | 2+5 2 | 4+5 2 | | | |
| | 5 2 | | | 3+4 2 | 2+3 2 | | | |
| | 6 2 | | | 7+8 2 | 6+7 2 | | | |
| | 6 2 | | | 6+9 2 | 8+9 2 | | | |
| | 5 2 | | | 6+9 2 | 4+5 2 | | | |
| | 6 2 | | | 7+8 2 | 6+7 2 | | | |
| | 6 2 | | | 2+5 2 | 8+9 2 | | | |
| | 5 2 | | | 3+4 2 | 2+3 2 | | | |
| | 5 2 | | | 6+9 2 | 8+9 2 | | | |
| | 6 2 | | | 7+8 2 | 2+3 2 | | | |
| | 6 2 | | | 2+5 2 | 6+7 2 | | | |
| | 5 2 | | | 3+4 2 | 4+5 2 | | | |
| | 5 2 | | | 2+5 2 | 6+7 2 | | | |
| | 5 2 | | | 3+4 2 | 8+9 2 | | | |
| | 5 2 | | | 6+9 2 | 2+3 2 | | | |
| | 6 2 | | | 7+8 2 | 4+5 2 | | | |
| | 6 2 | | | 3+4 2 | 6+7 2 | | | |
| | 6 2 | | | 2+5 2 | 8+9 2 | | | |
| 58.2.3 | 409767 | 188 | Eu2Ge3Li1 | | | | | |
| | -GM | -R | -S | -T | -U | -X | -Y | -Z |
| | 5 2 | 3+4 4 | 5+5 4 | 2+5 2 | 2+3 2 | 3+4 4 | 3+4 4 | 3+4 4 |
| | 5 2 | 3+4 4 | 6+6 4 | 3+4 2 | 4+5 2 | 3+4 4 | 3+4 4 | 3+4 4 |
| | 6 2 | 3+4 4 | 6+6 4 | 6+9 2 | 6+7 2 | 3+4 4 | 3+4 4 | 3+4 4 |
| | 6 2 | 3+4 4 | 5+5 4 | 7+8 2 | 8+9 2 | 3+4 4 | 3+4 4 | 3+4 4 |
| | 5 2 | 3+4 4 | 5+5 4 | 6+9 2 | 2+3 2 | 3+4 4 | 3+4 4 | 3+4 4 |
| | 5 2 | 3+4 4 | 6+6 4 | 7+8 2 | 4+5 2 | 3+4 4 | 3+4 4 | 3+4 4 |
| | 6 2 | 3+4 4 | 6+6 4 | 2+5 2 | 6+7 2 | 3+4 4 | 3+4 4 | 3+4 4 |
| | 6 2 | 3+4 4 | 5+5 4 | 3+4 2 | 8+9 2 | 3+4 4 | 3+4 4 | 3+4 4 |
| | 5 2 | 3+4 4 | 5+5 4 | 6+9 2 | 6+7 2 | 3+4 4 | 3+4 4 | 3+4 4 |
| | 5 2 | 3+4 4 | 6+6 4 | 7+8 2 | 8+9 2 | 3+4 4 | 3+4 4 | 3+4 4 |
| | 6 2 | 3+4 4 | 5+5 4 | 2+5 2 | 2+3 2 | 3+4 4 | 3+4 4 | 3+4 4 |
| | 6 2 | 3+4 4 | 6+6 4 | 3+4 2 | 4+5 2 | 3+4 4 | 3+4 4 | 3+4 4 |
| | 5 2 | 3+4 4 | 6+6 4 | 6+9 2 | 2+3 2 | 3+4 4 | 3+4 4 | 3+4 4 |
| | 5 2 | 3+4 4 | 5+5 4 | 7+8 2 | 4+5 2 | 3+4 4 | 3+4 4 | 3+4 4 |
| | 6 2 | 3+4 4 | 6+6 4 | 2+5 2 | 6+7 2 | 3+4 4 | 3+4 4 | 3+4 4 |
| | 6 2 | 3+4 4 | 5+5 4 | 3+4 2 | 8+9 2 | 3+4 4 | 3+4 4 | 3+4 4 |
| | 5 2 | 3+4 4 | 5+5 4 | 6+9 2 | 2+3 2 | 3+4 4 | 3+4 4 | 3+4 4 |
| | 5 2 | 3+4 4 | 6+6 4 | 2+5 2 | 4+5 2 | 3+4 4 | 3+4 4 | 3+4 4 |
| | 5 2 | 3+4 4 | 6+6 4 | 3+4 2 | 6+7 2 | 3+4 4 | 3+4 4 | 3+4 4 |
| | 6 2 | 3+4 4 | 5+5 4 | 7+8 2 | 8+9 2 | 3+4 4 | 3+4 4 | 3+4 4 |
| | 6 2 | 3+4 4 | 5+5 4 | 6+9 2 | 6+7 2 | 3+4 4 | 3+4 4 | 3+4 4 |
| | 5 2 | 3+4 4 | 6+6 4 | 2+5 2 | 8+9 2 | 3+4 4 | 3+4 4 | 3+4 4 |
| | 6 2 | 3+4 4 | 6+6 4 | 7+8 2 | 4+5 2 | 3+4 4 | 3+4 4 | 3+4 4 |
| | 6 2 | 3+4 4 | 5+5 4 | 3+4 2 | 2+3 2 | 3+4 4 | 3+4 4 | 3+4 4 |
| | 5 2 | 3+4 4 | 6+6 4 | 7+8 2 | 2+3 2 | 3+4 4 | 3+4 4 | 3+4 4 |
| | 5 2 | 3+4 4 | 6+6 4 | 3+4 2 | 4+5 2 | 3+4 4 | 3+4 4 | 3+4 4 |
| | 6 2 | 3+4 4 | 5+5 4 | 6+9 2 | 6+7 2 | 3+4 4 | 3+4 4 | 3+4 4 |

Continued on next column

**TABLE XXXIII– continued**

| TQC | ICSD | NoE | SF | | | | | |
|---|---|---|---|---|---|---|---|---|
| | 6 2 | 3+4 4 | 5+5 4 | 2+5 2 | 8+9 2 | 3+4 4 | 3+4 4 | 3+4 4 |
| | 5 2 | 3+4 4 | 5+5 4 | 7+8 2 | 2+3 2 | 3+4 4 | 3+4 4 | 3+4 4 |
| | 5 2 | 3+4 4 | 6+6 4 | 6+9 2 | 4+5 2 | 3+4 4 | 3+4 4 | 3+4 4 |
| | 6 2 | 3+4 4 | 6+6 4 | 2+5 2 | 8+9 2 | 3+4 4 | 3+4 4 | 3+4 4 |
| | 6 2 | 3+4 4 | 5+5 4 | 3+4 2 | 6+7 2 | 3+4 4 | 3+4 4 | 3+4 4 |
| | 5 2 | 3+4 4 | 5+5 4 | 2+5 2 | 2+3 2 | 3+4 4 | 3+4 4 | 3+4 4 |
| | 5 2 | 3+4 4 | 6+6 4 | 3+4 2 | 4+5 2 | 3+4 4 | 3+4 4 | 3+4 4 |
| | 5 2 | 3+4 4 | 5+5 4 | 7+8 2 | 6+7 2 | 3+4 4 | 3+4 4 | 3+4 4 |
| | 6 2 | 3+4 4 | 6+6 4 | 6+9 2 | 8+9 2 | 3+4 4 | 3+4 4 | 3+4 4 |
| | 6 2 | 3+4 4 | 5+5 4 | 3+4 2 | 8+9 2 | 3+4 4 | 3+4 4 | 3+4 4 |
| | 5 2 | 3+4 4 | 6+6 4 | 2+5 2 | 6+7 2 | 3+4 4 | 3+4 4 | 3+4 4 |
| | 6 2 | 3+4 4 | 5+5 4 | 6+9 2 | 4+5 2 | 3+4 4 | 3+4 4 | 3+4 4 |
| | 6 2 | 3+4 4 | 6+6 4 | 7+8 2 | 2+3 2 | 3+4 4 | 3+4 4 | 3+4 4 |
| | 5 2 | 3+4 4 | 5+5 4 | 2+5 2 | 2+3 2 | 3+4 4 | 3+4 4 | 3+4 4 |
| | 5 2 | 3+4 4 | 6+6 4 | 3+4 2 | 4+5 2 | 3+4 4 | 3+4 4 | 3+4 4 |
| | 6 2 | 3+4 4 | 6+6 4 | 6+9 2 | 8+9 2 | 3+4 4 | 3+4 4 | 3+4 4 |
| | 6 2 | 3+4 4 | 5+5 4 | 7+8 2 | 6+7 2 | 3+4 4 | 3+4 4 | 3+4 4 |
| | 6 2 | 3+4 4 | 5+5 4 | 6+9 2 | 2+3 2 | 3+4 4 | 3+4 4 | 3+4 4 |
| | 5 2 | 3+4 4 | 6+6 4 | 7+8 2 | 4+5 2 | 3+4 4 | 3+4 4 | 3+4 4 |
| | 6 2 | 3+4 4 | 5+5 4 | 3+4 2 | 8+9 2 | 3+4 4 | 3+4 4 | 3+4 4 |
| | 5 2 | | | 2+5 2 | 6+7 2 | | | |
| | 5 2 | | | 2+5 2 | 4+5 2 | | | |
| | 5 2 | | | 3+4 2 | 2+3 2 | | | |
| | 6 2 | | | 6+9 2 | 8+9 2 | | | |
| | 6 2 | | | 7+8 2 | 6+7 2 | | | |
| | 5 2 | | | 6+9 2 | 6+7 2 | | | |
| | 6 2 | | | 7+8 2 | 8+9 2 | | | |
| | 5 2 | | | 3+4 2 | 4+5 2 | | | |
| | 6 2 | | | 2+5 2 | 2+3 2 | | | |
| | 6 2 | | | 7+8 2 | 2+3 2 | | | |
| | 5 2 | | | 6+9 2 | 6+7 2 | | | |
| | 6 2 | | | 2+5 2 | 8+9 2 | | | |
| | 5 2 | | | 3+4 2 | 4+5 2 | | | |
| | 5 2 | | | 2+5 2 | 6+7 2 | | | |
| | 5 2 | | | 3+4 2 | 8+9 2 | | | |
| | 5 2 | | | 7+8 2 | 6+7 2 | | | |
| | 6 2 | | | 6+9 2 | 8+9 2 | | | |
| | 6 2 | | | 3+4 2 | 2+3 2 | | | |
| | 5 2 | | | 2+5 2 | 4+5 2 | | | |
| | 6 2 | | | 2+5 2 | 2+3 2 | | | |
| | 6 2 | | | 3+4 2 | 4+5 2 | | | |
| | 5 2 | | | 7+8 2 | 6+7 2 | | | |
| | 5 2 | | | 6+9 2 | 8+9 2 | | | |
| | 6 2 | | | 2+5 2 | 2+3 2 | | | |
| | 6 2 | | | 3+4 2 | 4+5 2 | | | |
| | 6 2 | | | 6+9 2 | 8+9 2 | | | |
| | 5 2 | | | 7+8 2 | 4+5 2 | | | |
| | 5 2 | | | 2+5 2 | 2+3 2 | | | |
| | 6 2 | | | 3+4 2 | 6+7 2 | | | |
| | 5 2 | | | 6+9 2 | 2+3 2 | | | |
| | 6 2 | | | 7+8 2 | 8+9 2 | | | |
| | 5 2 | | | 2+5 2 | 4+5 2 | | | |
| | 5 2 | | | 3+4 2 | 6+7 2 | | | |
| | 6 2 | | | 6+9 2 | 8+9 2 | | | |
| | 6 2 | | | 7+8 2 | 2+3 2 | | | |
| | 5 2 | | | 6+9 2 | 6+7 2 | | | |
| | 6 2 | | | 7+8 2 | 4+5 2 | | | |
| | 5 2 | | | 3+4 2 | 4+5 2 | | | |
| | 5 2 | | | 2+5 2 | 6+7 2 | | | |
| | 6 2 | | | 6+9 2 | 8+9 2 | | | |
| | 5 2 | | | 7+8 2 | 2+3 2 | | | |

Continued on next column

**TABLE XXXIII– continued**

| TQC | ICSD | NoE | SF | | | | | |
|---|---|---|---|---|---|---|---|---|
| | 6 2 | | | 3+4 2 | 6+7 2 | | | |
| | 6 2 | | | 2+5 2 | 8+9 2 | | | |
| | 5 2 | | | 3+4 2 | 2+3 2 | | | |
| | 5 2 | | | 2+5 2 | 4+5 2 | | | |
| | 6 2 | | | 7+8 2 | 6+7 2 | | | |
| | 6 2 | | | 6+9 2 | 8+9 2 | | | |
| 58.2.4 | 42720 | 32 | Cr1Sb2 | | | | | |
| | -GM | -R | -S | -T | -U | -X | -Y | -Z |
| | 5 2 | 3+4 4 | 6+6 4 | 6+9 2 | 2+3 2 | 3+4 4 | 3+4 4 | 3+4 4 |
| | 5 2 | 3+4 4 | 5+5 4 | 7+8 2 | 4+5 2 | 3+4 4 | 3+4 4 | 3+4 4 |
| | 6 2 | 3+4 4 | 5+5 4 | 2+5 2 | 6+7 2 | 3+4 4 | 3+4 4 | 3+4 4 |
| | 6 2 | 3+4 4 | 6+6 4 | 3+4 2 | 8+9 2 | 3+4 4 | 3+4 4 | 3+4 4 |
| | 5 2 | 3+4 4 | 5+5 4 | 3+4 2 | 8+9 2 | 3+4 4 | 3+4 4 | 3+4 4 |
| | 5 2 | 3+4 4 | 6+6 4 | 2+5 2 | 6+7 2 | 3+4 4 | 3+4 4 | 3+4 4 |
| | 6 2 | 3+4 4 | 5+5 4 | 7+8 2 | 2+3 2 | 3+4 4 | 3+4 4 | 3+4 4 |
| | 5 2 | 3+4 4 | 5+5 4 | 6+9 2 | 4+5 2 | 3+4 4 | 3+4 4 | 3+4 4 |
| | 6 2 | | | 2+5 2 | 4+5 2 | | | |
| | 5 2 | | | 3+4 2 | 2+3 2 | | | |
| | 5 2 | | | 6+9 2 | 2+3 2 | | | |
| | 5 2 | | | 7+8 2 | 4+5 2 | | | |
| | 5 2 | | | 3+4 2 | 6+7 2 | | | |
| | 5 2 | | | 2+5 2 | 8+9 2 | | | |
| | 5 2 | | | 3+4 2 | 2+3 2 | | | |
| | 5 2 | | | 2+5 2 | 4+5 2 | | | |
| 58.2.5 | 42601 | 32 | Cr1Sb2 | | | | | |
| | -GM | -R | -S | -T | -U | -X | -Y | -Z |
| | 5 2 | 3+4 4 | 6+6 4 | 6+9 2 | 2+3 2 | 3+4 4 | 3+4 4 | 3+4 4 |
| | 5 2 | 3+4 4 | 5+5 4 | 7+8 2 | 4+5 2 | 3+4 4 | 3+4 4 | 3+4 4 |
| | 6 2 | 3+4 4 | 5+5 4 | 2+5 2 | 6+7 2 | 3+4 4 | 3+4 4 | 3+4 4 |
| | 6 2 | 3+4 4 | 6+6 4 | 3+4 2 | 8+9 2 | 3+4 4 | 3+4 4 | 3+4 4 |
| | 5 2 | 3+4 4 | 5+5 4 | 3+4 2 | 8+9 2 | 3+4 4 | 3+4 4 | 3+4 4 |
| | 5 2 | 3+4 4 | 6+6 4 | 2+5 2 | 6+7 2 | 3+4 4 | 3+4 4 | 3+4 4 |
| | 6 2 | 3+4 4 | 5+5 4 | 7+8 2 | 2+3 2 | 3+4 4 | 3+4 4 | 3+4 4 |
| | 5 2 | 3+4 4 | 5+5 4 | 6+9 2 | 4+5 2 | 3+4 4 | 3+4 4 | 3+4 4 |
| | 6 2 | | | 2+5 2 | 4+5 2 | | | |
| | 5 2 | | | 3+4 2 | 2+3 2 | | | |
| | 5 2 | | | 6+9 2 | 2+3 2 | | | |
| | 5 2 | | | 7+8 2 | 4+5 2 | | | |
| | 5 2 | | | 3+4 2 | 6+7 2 | | | |
| | 5 2 | | | 2+5 2 | 8+9 2 | | | |
| | 5 2 | | | 3+4 2 | 2+3 2 | | | |
| | 5 2 | | | 2+5 2 | 4+5 2 | | | |
| 58.2.6 | 81572 | 84 | C3Al2Th2 | | | | | |
| | -GM | -R | -S | -T | -U | -X | -Y | -Z |
| | 5 2 | 3+4 4 | 5+5 4 | 6+9 2 | 6+7 2 | 3+4 4 | 3+4 4 | 3+4 4 |
| | 6 2 | 3+4 4 | 6+6 4 | 7+8 2 | 8+9 2 | 3+4 4 | 3+4 4 | 3+4 4 |
| | 5 2 | 3+4 4 | 5+5 4 | 2+5 2 | 2+3 2 | 3+4 4 | 3+4 4 | 3+4 4 |
| | 6 2 | 3+4 4 | 6+6 4 | 3+4 2 | 4+5 2 | 3+4 4 | 3+4 4 | 3+4 4 |
| | 5 2 | 3+4 4 | 5+5 4 | 7+8 2 | 6+7 2 | 3+4 4 | 3+4 4 | 3+4 4 |
| | 6 2 | 3+4 4 | 6+6 4 | 6+9 2 | 8+9 2 | 3+4 4 | 3+4 4 | 3+4 4 |
| | 5 2 | 3+4 4 | 5+5 4 | 2+5 2 | 2+3 2 | 3+4 4 | 3+4 4 | 3+4 4 |
| | 6 2 | 3+4 4 | 6+6 4 | 3+4 2 | 4+5 2 | 3+4 4 | 3+4 4 | 3+4 4 |
| | 5 2 | 3+4 4 | 5+5 4 | 3+4 2 | 4+5 2 | 3+4 4 | 3+4 4 | 3+4 4 |
| | 6 2 | 3+4 4 | 6+6 4 | 2+5 2 | 2+3 2 | 3+4 4 | 3+4 4 | 3+4 4 |
| | 5 2 | 3+4 4 | 5+5 4 | 6+9 2 | 6+7 2 | 3+4 4 | 3+4 4 | 3+4 4 |
| | 6 2 | 3+4 4 | 6+6 4 | 7+8 2 | 8+9 2 | 3+4 4 | 3+4 4 | 3+4 4 |
| | 6 2 | 3+4 4 | 6+6 4 | 6+9 2 | 8+9 2 | 3+4 4 | 3+4 4 | 3+4 4 |
| | 5 2 | 3+4 4 | 5+5 4 | 7+8 2 | 6+7 2 | 3+4 4 | 3+4 4 | 3+4 4 |
| | 5 2 | 3+4 4 | 6+6 4 | 2+5 2 | 2+3 2 | 3+4 4 | 3+4 4 | 3+4 4 |
| | 6 2 | 3+4 4 | 6+6 4 | 3+4 2 | 4+5 2 | 3+4 4 | 3+4 4 | 3+4 4 |
| | 5 2 | 3+4 4 | 5+5 4 | 3+4 2 | 4+5 2 | 3+4 4 | 3+4 4 | 3+4 4 |

Continued on next column

**TABLE XXXIII– continued**

| TQC | ICSD | NoE | SF | | | | | |
|---|---|---|---|---|---|---|---|---|
| | 6 2 | 3+4 4 | 5+5 4 | 2+5 2 | 2+3 2 | 3+4 4 | 3+4 4 | 3+4 4 |
| | 6 2 | 3+4 4 | 6+6 4 | 7+8 2 | 8+9 2 | 3+4 4 | 3+4 4 | 3+4 4 |
| | 5 2 | 3+4 4 | 6+6 4 | 6+9 2 | 6+7 2 | 3+4 4 | 3+4 4 | 3+4 4 |
| | 5 2 | 3+4 4 | 5+5 4 | 2+5 2 | 2+3 2 | 3+4 4 | 3+4 4 | 3+4 4 |
| | 5 2 | | | 3+4 2 | 4+5 2 | | | |
| | 6 2 | | | 6+9 2 | 6+7 2 | | | |
| | 6 2 | | | 7+8 2 | 8+9 2 | | | |
| | 5 2 | | | 3+4 2 | 2+3 2 | | | |
| | 5 2 | | | 2+5 2 | 4+5 2 | | | |
| | 6 2 | | | 7+8 2 | 6+7 2 | | | |
| | 6 2 | | | 6+9 2 | 8+9 2 | | | |
| | 6 2 | | | 7+8 2 | 8+9 2 | | | |
| | 5 2 | | | 3+4 2 | 6+7 2 | | | |
| | 6 2 | | | 2+5 2 | 4+5 2 | | | |
| | 5 2 | | | 6+9 2 | 2+3 2 | | | |
| | 5 2 | | | 2+5 2 | 4+5 2 | | | |
| | 6 2 | | | 3+4 2 | 2+3 2 | | | |
| | 5 2 | | | 6+9 2 | 6+7 2 | | | |
| | 5 2 | | | 7+8 2 | 8+9 2 | | | |
| | 6 2 | | | 6+9 2 | 6+7 2 | | | |
| | 6 2 | | | 7+8 2 | 8+9 2 | | | |
| | 5 2 | | | 3+4 2 | 8+9 2 | | | |
| | 6 2 | | | 7+8 2 | 6+7 2 | | | |
| | 5 2 | | | 2+5 2 | 8+9 2 | | | |
| | 6 2 | | | 6+9 2 | 6+7 2 | | | |
| 58.2.7 | 425106 | 36 | B4Cr1 | | | | | |
| | -GM | -R | -S | -T | -U | -X | -Y | -Z |
| | 5 2 | 3+4 4 | 6+6 4 | 6+9 2 | 2+3 2 | 3+4 4 | 3+4 4 | 3+4 4 |
| | 5 2 | 3+4 4 | 5+5 4 | 7+8 2 | 4+5 2 | 3+4 4 | 3+4 4 | 3+4 4 |
| | 6 2 | 3+4 4 | 5+5 4 | 2+5 2 | 6+7 2 | 3+4 4 | 3+4 4 | 3+4 4 |
| | 6 2 | 3+4 4 | 6+6 4 | 3+4 2 | 8+9 2 | 3+4 4 | 3+4 4 | 3+4 4 |
| | 5 2 | 3+4 4 | 5+5 4 | 2+5 2 | 2+3 2 | 3+4 4 | 3+4 4 | 3+4 4 |
| | 5 2 | 3+4 4 | 5+5 4 | 3+4 2 | 4+5 2 | 3+4 4 | 3+4 4 | 3+4 4 |
| | 6 2 | 3+4 4 | 6+6 4 | 7+8 2 | 6+7 2 | 3+4 4 | 3+4 4 | 3+4 4 |
| | 6 2 | 3+4 4 | 5+5 4 | 6+9 2 | 8+9 2 | 3+4 4 | 3+4 4 | 3+4 4 |
| | 5 2 | 3+4 4 | 5+5 4 | 6+9 2 | 2+3 2 | 3+4 4 | 3+4 4 | 3+4 4 |
| | 5 2 | | | 7+8 2 | 4+5 2 | | | |
| | 5 2 | | | 2+5 2 | 6+7 2 | | | |
| | 5 2 | | | 3+4 2 | 8+9 2 | | | |
| | 6 2 | | | 3+4 2 | 2+3 2 | | | |
| | 5 2 | | | 2+5 2 | 4+5 2 | | | |
| | 5 2 | | | 6+9 2 | 4+5 2 | | | |
| | 5 2 | | | 7+8 2 | 2+3 2 | | | |
| | 5 2 | | | 2+5 2 | 2+3 2 | | | |
| | 5 2 | | | 3+4 2 | 4+5 2 | | | |
| 58.2.8 | 186851 | 36 | B4Cr1 | | | | | |
| | -GM | -R | -S | -T | -U | -X | -Y | -Z |
| | 5 2 | 3+4 4 | 6+6 4 | 6+9 2 | 2+3 2 | 3+4 4 | 3+4 4 | 3+4 4 |
| | 5 2 | 3+4 4 | 5+5 4 | 7+8 2 | 4+5 2 | 3+4 4 | 3+4 4 | 3+4 4 |
| | 6 2 | 3+4 4 | 5+5 4 | 2+5 2 | 6+7 2 | 3+4 4 | 3+4 4 | 3+4 4 |
| | 6 2 | 3+4 4 | 6+6 4 | 3+4 2 | 8+9 2 | 3+4 4 | 3+4 4 | 3+4 4 |
| | 5 2 | 3+4 4 | 5+5 4 | 2+5 2 | 2+3 2 | 3+4 4 | 3+4 4 | 3+4 4 |
| | 5 2 | 3+4 4 | 5+5 4 | 3+4 2 | 4+5 2 | 3+4 4 | 3+4 4 | 3+4 4 |
| | 6 2 | 3+4 4 | 6+6 4 | 7+8 2 | 6+7 2 | 3+4 4 | 3+4 4 | 3+4 4 |
| | 6 2 | 3+4 4 | 5+5 4 | 6+9 2 | 8+9 2 | 3+4 4 | 3+4 4 | 3+4 4 |
| | 5 2 | 3+4 4 | 5+5 4 | 6+9 2 | 2+3 2 | 3+4 4 | 3+4 4 | 3+4 4 |
| | 5 2 | | | 7+8 2 | 4+5 2 | | | |
| | 5 2 | | | 2+5 2 | 6+7 2 | | | |
| | 5 2 | | | 3+4 2 | 8+9 2 | | | |
| | 6 2 | | | 3+4 2 | 2+3 2 | | | |
| | 5 2 | | | 2+5 2 | 4+5 2 | | | |

Continued on next column

**TABLE XXXIII– continued**

| TQC | ICSD | NoE | SF | | | | | |
|---|---|---|---|---|---|---|---|---|
| | 5 2 | | | 6+9 2 | 4+5 2 | | | |
| | 5 2 | | | 7+8 2 | 2+3 2 | | | |
| | 5 2 | | | 2+5 2 | 2+3 2 | | | |
| | 5 2 | | | 3+4 2 | 4+5 2 | | | |
| 58.2.9 | 31387 | 144 | B2Ni3O6 | | | | | |
| | -GM | -R | -S | -T | -U | -X | -Y | -Z |
| | 5 2 | 3+4 4 | 5+5 4 | 6+9 2 | 6+7 2 | 3+4 4 | 3+4 4 | 3+4 4 |
| | 5 2 | 3+4 4 | 6+6 4 | 7+8 2 | 8+9 2 | 3+4 4 | 3+4 4 | 3+4 4 |
| | 6 2 | 3+4 4 | 6+6 4 | 2+5 2 | 2+3 2 | 3+4 4 | 3+4 4 | 3+4 4 |
| | 6 2 | 3+4 4 | 5+5 4 | 3+4 2 | 4+5 2 | 3+4 4 | 3+4 4 | 3+4 4 |
| | 6 2 | 3+4 4 | 5+5 4 | 2+5 2 | 2+3 2 | 3+4 4 | 3+4 4 | 3+4 4 |
| | 6 2 | 3+4 4 | 6+6 4 | 3+4 2 | 4+5 2 | 3+4 4 | 3+4 4 | 3+4 4 |
| | 6 2 | 3+4 4 | 6+6 4 | 6+9 2 | 6+7 2 | 3+4 4 | 3+4 4 | 3+4 4 |
| | 5 2 | 3+4 4 | 5+5 4 | 7+8 2 | 8+9 2 | 3+4 4 | 3+4 4 | 3+4 4 |
| | 5 2 | 3+4 4 | 6+6 4 | 6+9 2 | 2+3 2 | 3+4 4 | 3+4 4 | 3+4 4 |
| | 6 2 | 3+4 4 | 6+6 4 | 7+8 2 | 4+5 2 | 3+4 4 | 3+4 4 | 3+4 4 |
| | 5 2 | 3+4 4 | 5+5 4 | 2+5 2 | 6+7 2 | 3+4 4 | 3+4 4 | 3+4 4 |
| | 5 2 | 3+4 4 | 5+5 4 | 3+4 2 | 8+9 2 | 3+4 4 | 3+4 4 | 3+4 4 |
| | 5 2 | 3+4 4 | 5+5 4 | 6+9 2 | 4+5 2 | 3+4 4 | 3+4 4 | 3+4 4 |
| | 5 2 | 3+4 4 | 6+6 4 | 7+8 2 | 2+3 2 | 3+4 4 | 3+4 4 | 3+4 4 |
| | 5 2 | 3+4 4 | 5+5 4 | 2+5 2 | 6+7 2 | 3+4 4 | 3+4 4 | 3+4 4 |
| | 5 2 | 3+4 4 | 5+5 4 | 3+4 2 | 8+9 2 | 3+4 4 | 3+4 4 | 3+4 4 |
| | 5 2 | 3+4 4 | 6+6 4 | 2+5 2 | 6+7 2 | 3+4 4 | 3+4 4 | 3+4 4 |
| | 6 2 | 3+4 4 | 6+6 4 | 3+4 2 | 8+9 2 | 3+4 4 | 3+4 4 | 3+4 4 |
| | 6 2 | 3+4 4 | 5+5 4 | 7+8 2 | 2+3 2 | 3+4 4 | 3+4 4 | 3+4 4 |
| | 6 2 | 3+4 4 | 5+5 4 | 6+9 2 | 4+5 2 | 3+4 4 | 3+4 4 | 3+4 4 |
| | 5 2 | 3+4 4 | 6+6 4 | 3+4 2 | 6+7 2 | 3+4 4 | 3+4 4 | 3+4 4 |
| | 6 2 | 3+4 4 | 6+6 4 | 2+5 2 | 8+9 2 | 3+4 4 | 3+4 4 | 3+4 4 |
| | 6 2 | 3+4 4 | 5+5 4 | 7+8 2 | 2+3 2 | 3+4 4 | 3+4 4 | 3+4 4 |
| | 5 2 | 3+4 4 | 6+6 4 | 6+9 2 | 4+5 2 | 3+4 4 | 3+4 4 | 3+4 4 |
| | 6 2 | 3+4 4 | 5+5 4 | 2+5 2 | 6+7 2 | 3+4 4 | 3+4 4 | 3+4 4 |
| | 5 2 | 3+4 4 | 5+5 4 | 3+4 2 | 8+9 2 | 3+4 4 | 3+4 4 | 3+4 4 |
| | 6 2 | 3+4 4 | 6+6 4 | 6+9 2 | 4+5 2 | 3+4 4 | 3+4 4 | 3+4 4 |
| | 6 2 | 3+4 4 | 5+5 4 | 7+8 2 | 2+3 2 | 3+4 4 | 3+4 4 | 3+4 4 |
| | 6 2 | 3+4 4 | 5+5 4 | 2+5 2 | 2+3 2 | 3+4 4 | 3+4 4 | 3+4 4 |
| | 6 2 | 3+4 4 | 6+6 4 | 3+4 2 | 4+5 2 | 3+4 4 | 3+4 4 | 3+4 4 |
| | 6 2 | 3+4 4 | 5+5 4 | 7+8 2 | 4+5 2 | 3+4 4 | 3+4 4 | 3+4 4 |
| | 5 2 | 3+4 4 | 6+6 4 | 6+9 2 | 2+3 2 | 3+4 4 | 3+4 4 | 3+4 4 |
| | 5 2 | 3+4 4 | 5+5 4 | 3+4 2 | 4+5 2 | 3+4 4 | 3+4 4 | 3+4 4 |
| | 5 2 | 3+4 4 | 5+5 4 | 2+5 2 | 2+3 2 | 3+4 4 | 3+4 4 | 3+4 4 |
| | 6 2 | 3+4 4 | 6+6 4 | 7+8 2 | 8+9 2 | 3+4 4 | 3+4 4 | 3+4 4 |
| | 5 2 | 3+4 4 | 5+5 4 | 6+9 2 | 6+7 2 | 3+4 4 | 3+4 4 | 3+4 4 |
| | 5 2 | | | 3+4 2 | 8+9 2 | | | |
| | 5 2 | | | 2+5 2 | 6+7 2 | | | |
| | 5 2 | | | 7+8 2 | 2+3 2 | | | |
| | 5 2 | | | 6+9 2 | 4+5 2 | | | |
| | 6 2 | | | 2+5 2 | 6+7 2 | | | |
| | 6 2 | | | 3+4 2 | 8+9 2 | | | |
| | 6 2 | | | 6+9 2 | 2+3 2 | | | |
| | 5 2 | | | 7+8 2 | 4+5 2 | | | |
| | 6 2 | | | 2+5 2 | 6+7 2 | | | |
| | 5 2 | | | 3+4 2 | 8+9 2 | | | |
| | 6 2 | | | 6+9 2 | 8+9 2 | | | |
| | 6 2 | | | 7+8 2 | 6+7 2 | | | |
| | 5 2 | | | 3+4 2 | 2+3 2 | | | |
| | 5 2 | | | 2+5 2 | 4+5 2 | | | |
| | 5 2 | | | 3+4 2 | 4+5 2 | | | |
| | 5 2 | | | 2+5 2 | 2+3 2 | | | |
| | 5 2 | | | 2+5 2 | 4+5 2 | | | |
| | 5 2 | | | 7+8 2 | 6+7 2 | | | |
| | 6 2 | | | 3+4 2 | 2+3 2 | | | |

Continued on next column

**TABLE XXXIII– continued**

| TQC | ICSD | NoE | SF | | | | | |
|---|---|---|---|---|---|---|---|---|
| | 6 2 | | | 6+9 2 | 8+9 2 | | | |
| | 5 2 | | | 3+4 2 | 2+3 2 | | | |
| | 5 2 | | | 2+5 2 | 8+9 2 | | | |
| | 5 2 | | | 6+9 2 | 4+5 2 | | | |
| | 6 2 | | | 7+8 2 | 2+3 2 | | | |
| | 6 2 | | | 3+4 2 | 6+7 2 | | | |
| | 5 2 | | | 2+5 2 | 4+5 2 | | | |
| | 5 2 | | | 7+8 2 | 4+5 2 | | | |
| | 6 2 | | | 6+9 2 | 2+3 2 | | | |
| | 5 2 | | | 2+5 2 | 8+9 2 | | | |
| | 6 2 | | | 3+4 2 | 6+7 2 | | | |
| | 5 2 | | | 2+5 2 | 4+5 2 | | | |
| | 6 2 | | | 6+9 2 | 2+3 2 | | | |
| | 5 2 | | | 3+4 2 | 2+3 2 | | | |
| | 5 2 | | | 7+8 2 | 4+5 2 | | | |
| | 6 2 | | | 2+5 2 | 6+7 2 | | | |
| | 6 2 | | | 3+4 2 | 8+9 2 | | | |
| 58.2.10 | 2016 | 144 | B2Ni3O6 | | | | | |
| | -GM | -R | -S | -T | -U | -X | -Y | -Z |
| | 5 2 | 3+4 4 | 6+6 4 | 2+5 2 | 6+7 2 | 3+4 4 | 3+4 4 | 3+4 4 |
| | 5 2 | 3+4 4 | 5+5 4 | 3+4 2 | 8+9 2 | 3+4 4 | 3+4 4 | 3+4 4 |
| | 6 2 | 3+4 4 | 6+6 4 | 6+9 2 | 2+3 2 | 3+4 4 | 3+4 4 | 3+4 4 |
| | 6 2 | 3+4 4 | 5+5 4 | 7+8 2 | 4+5 2 | 3+4 4 | 3+4 4 | 3+4 4 |
| | 6 2 | 3+4 4 | 5+5 4 | 6+9 2 | 2+3 2 | 3+4 4 | 3+4 4 | 3+4 4 |
| | 6 2 | 3+4 4 | 6+6 4 | 7+8 2 | 4+5 2 | 3+4 4 | 3+4 4 | 3+4 4 |
| | 5 2 | 3+4 4 | 6+6 4 | 6+9 2 | 6+7 2 | 3+4 4 | 3+4 4 | 3+4 4 |
| | 6 2 | 3+4 4 | 5+5 4 | 7+8 2 | 8+9 2 | 3+4 4 | 3+4 4 | 3+4 4 |
| | 6 2 | 3+4 4 | 6+6 4 | 2+5 2 | 2+3 2 | 3+4 4 | 3+4 4 | 3+4 4 |
| | 5 2 | 3+4 4 | 6+6 4 | 3+4 2 | 4+5 2 | 3+4 4 | 3+4 4 | 3+4 4 |
| | 5 2 | 3+4 4 | 5+5 4 | 2+5 2 | 6+7 2 | 3+4 4 | 3+4 4 | 3+4 4 |
| | 5 2 | 3+4 4 | 5+5 4 | 3+4 2 | 8+9 2 | 3+4 4 | 3+4 4 | 3+4 4 |
| | 5 2 | 3+4 4 | 5+5 4 | 6+9 2 | 6+7 2 | 3+4 4 | 3+4 4 | 3+4 4 |
| | 5 2 | 3+4 4 | 6+6 4 | 7+8 2 | 8+9 2 | 3+4 4 | 3+4 4 | 3+4 4 |
| | 5 2 | 3+4 4 | 5+5 4 | 2+5 2 | 2+3 2 | 3+4 4 | 3+4 4 | 3+4 4 |
| | 5 2 | 3+4 4 | 5+5 4 | 3+4 2 | 4+5 2 | 3+4 4 | 3+4 4 | 3+4 4 |
| | 5 2 | 3+4 4 | 6+6 4 | 2+5 2 | 6+7 2 | 3+4 4 | 3+4 4 | 3+4 4 |
| | 6 2 | 3+4 4 | 6+6 4 | 3+4 2 | 8+9 2 | 3+4 4 | 3+4 4 | 3+4 4 |
| | 6 2 | 3+4 4 | 5+5 4 | 6+9 2 | 2+3 2 | 3+4 4 | 3+4 4 | 3+4 4 |
| | 6 2 | 3+4 4 | 5+5 4 | 7+8 2 | 4+5 2 | 3+4 4 | 3+4 4 | 3+4 4 |
| | 5 2 | 3+4 4 | 6+6 4 | 2+5 2 | 6+7 2 | 3+4 4 | 3+4 4 | 3+4 4 |
| | 6 2 | 3+4 4 | 6+6 4 | 3+4 2 | 8+9 2 | 3+4 4 | 3+4 4 | 3+4 4 |
| | 6 2 | 3+4 4 | 5+5 4 | 6+9 2 | 4+5 2 | 3+4 4 | 3+4 4 | 3+4 4 |
| | 5 2 | 3+4 4 | 6+6 4 | 7+8 2 | 2+3 2 | 3+4 4 | 3+4 4 | 3+4 4 |
| | 6 2 | 3+4 4 | 5+5 4 | 2+5 2 | 8+9 2 | 3+4 4 | 3+4 4 | 3+4 4 |
| | 5 2 | 3+4 4 | 5+5 4 | 3+4 2 | 6+7 2 | 3+4 4 | 3+4 4 | 3+4 4 |
| | 6 2 | 3+4 4 | 6+6 4 | 7+8 2 | 2+3 2 | 3+4 4 | 3+4 4 | 3+4 4 |
| | 6 2 | 3+4 4 | 5+5 4 | 6+9 2 | 4+5 2 | 3+4 4 | 3+4 4 | 3+4 4 |
| | 6 2 | 3+4 4 | 5+5 4 | 3+4 2 | 4+5 2 | 3+4 4 | 3+4 4 | 3+4 4 |
| | 6 2 | 3+4 4 | 6+6 4 | 2+5 2 | 2+3 2 | 3+4 4 | 3+4 4 | 3+4 4 |
| | 6 2 | 3+4 4 | 5+5 4 | 6+9 2 | 2+3 2 | 3+4 4 | 3+4 4 | 3+4 4 |
| | 5 2 | 3+4 4 | 6+6 4 | 7+8 2 | 4+5 2 | 3+4 4 | 3+4 4 | 3+4 4 |
| | 5 2 | 3+4 4 | 5+5 4 | 2+5 2 | 6+7 2 | 3+4 4 | 3+4 4 | 3+4 4 |
| | 5 2 | 3+4 4 | 5+5 4 | 3+4 2 | 8+9 2 | 3+4 4 | 3+4 4 | 3+4 4 |
| | 6 2 | 3+4 4 | 6+6 4 | 6+9 2 | 2+3 2 | 3+4 4 | 3+4 4 | 3+4 4 |
| | 5 2 | 3+4 4 | 5+5 4 | 7+8 2 | 4+5 2 | 3+4 4 | 3+4 4 | 3+4 4 |
| | 5 2 | | | 2+5 2 | 6+7 2 | | | |
| | 5 2 | | | 3+4 2 | 8+9 2 | | | |
| | 5 2 | | | 6+9 2 | 4+5 2 | | | |
| | 5 2 | | | 7+8 2 | 2+3 2 | | | |
| | 6 2 | | | 3+4 2 | 8+9 2 | | | |
| | 6 2 | | | 2+5 2 | 6+7 2 | | | |

Continued on next column

**TABLE XXXIII– continued**

| TQC | ICSD | NoE | SF | | | | | |
|---|---|---|---|---|---|---|---|---|
| | 5 2 | | | 6+9 2 | 4+5 2 | | | |
| | 6 2 | | | 7+8 2 | 2+3 2 | | | |
| | 5 2 | | | 3+4 2 | 8+9 2 | | | |
| | 6 2 | | | 2+5 2 | 6+7 2 | | | |
| | 6 2 | | | 7+8 2 | 6+7 2 | | | |
| | 6 2 | | | 6+9 2 | 8+9 2 | | | |
| | 5 2 | | | 2+5 2 | 4+5 2 | | | |
| | 5 2 | | | 3+4 2 | 2+3 2 | | | |
| | 5 2 | | | 2+5 2 | 2+3 2 | | | |
| | 5 2 | | | 3+4 2 | 4+5 2 | | | |
| | 5 2 | | | 3+4 2 | 2+3 2 | | | |
| | 5 2 | | | 6+9 2 | 8+9 2 | | | |
| | 6 2 | | | 2+5 2 | 4+5 2 | | | |
| | 6 2 | | | 7+8 2 | 6+7 2 | | | |
| | 5 2 | | | 2+5 2 | 4+5 2 | | | |
| | 5 2 | | | 3+4 2 | 6+7 2 | | | |
| | 5 2 | | | 7+8 2 | 2+3 2 | | | |
| | 6 2 | | | 6+9 2 | 8+9 2 | | | |
| | 6 2 | | | 2+5 2 | 4+5 2 | | | |
| | 5 2 | | | 3+4 2 | 2+3 2 | | | |
| | 5 2 | | | 6+9 2 | 2+3 2 | | | |
| | 6 2 | | | 7+8 2 | 4+5 2 | | | |
| | 5 2 | | | 3+4 2 | 6+7 2 | | | |
| | 6 2 | | | 2+5 2 | 8+9 2 | | | |
| | 5 2 | | | 3+4 2 | 2+3 2 | | | |
| | 6 2 | | | 2+5 2 | 4+5 2 | | | |
| | 5 2 | | | 7+8 2 | 4+5 2 | | | |
| | 5 2 | | | 6+9 2 | 2+3 2 | | | |
| | 6 2 | | | 3+4 2 | 8+9 2 | | | |
| | 6 2 | | | 2+5 2 | 6+7 2 | | | |
| 58.2.11 | 251066 | 260 | Au8Sn3Sr3 | | | | | |
| | -GM | -R | -S | -T | -U | -X | -Y | -Z |
| | 5 2 | 3+4 4 | 6+6 4 | 2+5 2 | 6+7 2 | 3+4 4 | 3+4 4 | 3+4 4 |
| | 5 2 | 3+4 4 | 5+5 4 | 3+4 2 | 8+9 2 | 3+4 4 | 3+4 4 | 3+4 4 |
| | 5 2 | 3+4 4 | 6+6 4 | 2+5 2 | 2+3 2 | 3+4 4 | 3+4 4 | 3+4 4 |
| | 5 2 | 3+4 4 | 5+5 4 | 3+4 2 | 4+5 2 | 3+4 4 | 3+4 4 | 3+4 4 |
| | 6 2 | 3+4 4 | 5+5 4 | 6+9 2 | 6+7 2 | 3+4 4 | 3+4 4 | 3+4 4 |
| | 6 2 | 3+4 4 | 6+6 4 | 7+8 2 | 8+9 2 | 3+4 4 | 3+4 4 | 3+4 4 |
| | 6 2 | 3+4 4 | 5+5 4 | 6+9 2 | 2+3 2 | 3+4 4 | 3+4 4 | 3+4 4 |
| | 6 2 | 3+4 4 | 5+5 4 | 7+8 2 | 4+5 2 | 3+4 4 | 3+4 4 | 3+4 4 |
| | 5 2 | 3+4 4 | 5+5 4 | 3+4 2 | 2+3 2 | 3+4 4 | 3+4 4 | 3+4 4 |
| | 6 2 | 3+4 4 | 6+6 4 | 6+9 2 | 4+5 2 | 3+4 4 | 3+4 4 | 3+4 4 |
| | 5 2 | 3+4 4 | 5+5 4 | 2+5 2 | 6+7 2 | 3+4 4 | 3+4 4 | 3+4 4 |
| | 6 2 | 3+4 4 | 6+6 4 | 7+8 2 | 8+9 2 | 3+4 4 | 3+4 4 | 3+4 4 |
| | 6 2 | 3+4 4 | 5+5 4 | 6+9 2 | 2+3 2 | 3+4 4 | 3+4 4 | 3+4 4 |
| | 6 2 | 3+4 4 | 6+6 4 | 7+8 2 | 4+5 2 | 3+4 4 | 3+4 4 | 3+4 4 |
| | 6 2 | 3+4 4 | 5+5 4 | 6+9 2 | 2+3 2 | 3+4 4 | 3+4 4 | 3+4 4 |
| | 6 2 | 3+4 4 | 6+6 4 | 7+8 2 | 4+5 2 | 3+4 4 | 3+4 4 | 3+4 4 |
| | 5 2 | 3+4 4 | 5+5 4 | 2+5 2 | 4+5 2 | 3+4 4 | 3+4 4 | 3+4 4 |
| | 6 2 | 3+4 4 | 6+6 4 | 3+4 2 | 8+9 2 | 3+4 4 | 3+4 4 | 3+4 4 |
| | 5 2 | 3+4 4 | 6+6 4 | 7+8 2 | 2+3 2 | 3+4 4 | 3+4 4 | 3+4 4 |
| | 6 2 | 3+4 4 | 5+5 4 | 6+9 2 | 6+7 2 | 3+4 4 | 3+4 4 | 3+4 4 |
| | 5 2 | 3+4 4 | 6+6 4 | 2+5 2 | 6+7 2 | 3+4 4 | 3+4 4 | 3+4 4 |
| | 6 2 | 3+4 4 | 6+6 4 | 6+9 2 | 2+3 2 | 3+4 4 | 3+4 4 | 3+4 4 |
| | 5 2 | 3+4 4 | 5+5 4 | 7+8 2 | 8+9 2 | 3+4 4 | 3+4 4 | 3+4 4 |
| | 6 2 | 3+4 4 | 5+5 4 | 3+4 2 | 4+5 2 | 3+4 4 | 3+4 4 | 3+4 4 |
| | 5 2 | 3+4 4 | 6+6 4 | 3+4 2 | 4+5 2 | 3+4 4 | 3+4 4 | 3+4 4 |
| | 5 2 | 3+4 4 | 5+5 4 | 2+5 2 | 2+3 2 | 3+4 4 | 3+4 4 | 3+4 4 |
| | 6 2 | 3+4 4 | 6+6 4 | 6+9 2 | 6+7 2 | 3+4 4 | 3+4 4 | 3+4 4 |
| | 6 2 | 3+4 4 | 5+5 4 | 7+8 2 | 8+9 2 | 3+4 4 | 3+4 4 | 3+4 4 |
| | 5 2 | 3+4 4 | 6+6 4 | 3+4 2 | 2+3 2 | 3+4 4 | 3+4 4 | 3+4 4 |

Continued on next column

**TABLE XXXIII– continued**

| TQC | ICSD | NoE | SF | | | | | |
|---|---|---|---|---|---|---|---|---|
| | 5 2 | 3+4 4 | 5+5 4 | 2+5 2 | 4+5 2 | 3+4 4 | 3+4 4 | 3+4 4 |
| | 6 2 | 3+4 4 | 6+6 4 | 7+8 2 | 6+7 2 | 3+4 4 | 3+4 4 | 3+4 4 |
| | 6 2 | 3+4 4 | 5+5 4 | 6+9 2 | 8+9 2 | 3+4 4 | 3+4 4 | 3+4 4 |
| | 5 2 | 3+4 4 | 6+6 4 | 2+5 2 | 8+9 2 | 3+4 4 | 3+4 4 | 3+4 4 |
| | 6 2 | 3+4 4 | 5+5 4 | 3+4 2 | 6+7 2 | 3+4 4 | 3+4 4 | 3+4 4 |
| | 6 2 | 3+4 4 | 6+6 4 | 7+8 2 | 4+5 2 | 3+4 4 | 3+4 4 | 3+4 4 |
| | 5 2 | 3+4 4 | 5+5 4 | 6+9 2 | 2+3 2 | 3+4 4 | 3+4 4 | 3+4 4 |
| | 5 2 | 3+4 4 | 6+6 4 | 2+5 2 | 8+9 2 | 3+4 4 | 3+4 4 | 3+4 4 |
| | 5 2 | 3+4 4 | 5+5 4 | 3+4 2 | 2+3 2 | 3+4 4 | 3+4 4 | 3+4 4 |
| | 6 2 | 3+4 4 | 6+6 4 | 6+9 2 | 6+7 2 | 3+4 4 | 3+4 4 | 3+4 4 |
| | 6 2 | 3+4 4 | 5+5 4 | 7+8 2 | 4+5 2 | 3+4 4 | 3+4 4 | 3+4 4 |
| | 5 2 | 3+4 4 | 6+6 4 | 3+4 2 | 4+5 2 | 3+4 4 | 3+4 4 | 3+4 4 |
| | 6 2 | 3+4 4 | 5+5 4 | 6+9 2 | 6+7 2 | 3+4 4 | 3+4 4 | 3+4 4 |
| | 5 2 | 3+4 4 | 6+6 4 | 2+5 2 | 8+9 2 | 3+4 4 | 3+4 4 | 3+4 4 |
| | 6 2 | 3+4 4 | 5+5 4 | 7+8 2 | 2+3 2 | 3+4 4 | 3+4 4 | 3+4 4 |
| | 5 2 | 3+4 4 | 6+6 4 | 6+9 2 | 4+5 2 | 3+4 4 | 3+4 4 | 3+4 4 |
| | 6 2 | 3+4 4 | 5+5 4 | 3+4 2 | 6+7 2 | 3+4 4 | 3+4 4 | 3+4 4 |
| | 6 2 | 3+4 4 | 6+6 4 | 7+8 2 | 8+9 2 | 3+4 4 | 3+4 4 | 3+4 4 |
| | 5 2 | 3+4 4 | 6+6 4 | 2+5 2 | 2+3 2 | 3+4 4 | 3+4 4 | 3+4 4 |
| | 5 2 | 3+4 4 | 5+5 4 | 6+9 2 | 8+9 2 | 3+4 4 | 3+4 4 | 3+4 4 |
| | 6 2 | 3+4 4 | 5+5 4 | 7+8 2 | 2+3 2 | 3+4 4 | 3+4 4 | 3+4 4 |
| | 5 2 | 3+4 4 | 6+6 4 | 2+5 2 | 6+7 2 | 3+4 4 | 3+4 4 | 3+4 4 |
| | 6 2 | 3+4 4 | 5+5 4 | 3+4 2 | 4+5 2 | 3+4 4 | 3+4 4 | 3+4 4 |
| | 5 2 | 3+4 4 | 6+6 4 | 2+5 2 | 2+3 2 | 3+4 4 | 3+4 4 | 3+4 4 |
| | 5 2 | 3+4 4 | 6+6 4 | 3+4 2 | 6+7 2 | 3+4 4 | 3+4 4 | 3+4 4 |
| | 6 2 | 3+4 4 | 5+5 4 | 7+8 2 | 8+9 2 | 3+4 4 | 3+4 4 | 3+4 4 |
| | 5 2 | 3+4 4 | 5+5 4 | 6+9 2 | 6+7 2 | 3+4 4 | 3+4 4 | 3+4 4 |
| | 5 2 | 3+4 4 | 5+5 4 | 3+4 2 | 8+9 2 | 3+4 4 | 3+4 4 | 3+4 4 |
| | 6 2 | 3+4 4 | 6+6 4 | 2+5 2 | 4+5 2 | 3+4 4 | 3+4 4 | 3+4 4 |
| | 5 2 | 3+4 4 | 5+5 4 | 6+9 2 | 2+3 2 | 3+4 4 | 3+4 4 | 3+4 4 |
| | 6 2 | 3+4 4 | 6+6 4 | 7+8 2 | 6+7 2 | 3+4 4 | 3+4 4 | 3+4 4 |
| | 6 2 | 3+4 4 | 5+5 4 | 2+5 2 | 8+9 2 | 3+4 4 | 3+4 4 | 3+4 4 |
| | 5 2 | 3+4 4 | 6+6 4 | 3+4 2 | 4+5 2 | 3+4 4 | 3+4 4 | 3+4 4 |
| | 5 2 | 3+4 4 | 6+6 4 | 7+8 2 | 4+5 2 | 3+4 4 | 3+4 4 | 3+4 4 |
| | 6 2 | 3+4 4 | 5+5 4 | 6+9 2 | 2+3 2 | 3+4 4 | 3+4 4 | 3+4 4 |
| | 6 2 | 3+4 4 | 6+6 4 | 2+5 2 | 6+7 2 | 3+4 4 | 3+4 4 | 3+4 4 |
| | 6 2 | | | 7+8 2 | 8+9 2 | | | |
| | 5 2 | | | 6+9 2 | 2+3 2 | | | |
| | 6 2 | | | 3+4 2 | 4+5 2 | | | |
| | 5 2 | | | 6+9 2 | 8+9 2 | | | |
| | 5 2 | | | 7+8 2 | 6+7 2 | | | |
| | 6 2 | | | 3+4 2 | 4+5 2 | | | |
| | 6 2 | | | 2+5 2 | 8+9 2 | | | |
| | 5 2 | | | 3+4 2 | 6+7 2 | | | |
| | 6 2 | | | 7+8 2 | 2+3 2 | | | |
| | 5 2 | | | 6+9 2 | 4+5 2 | | | |
| | 6 2 | | | 6+9 2 | 2+3 2 | | | |
| | 6 2 | | | 7+8 2 | 8+9 2 | | | |
| | 5 2 | | | 2+5 2 | 6+7 2 | | | |
| | 6 2 | | | 3+4 2 | 2+3 2 | | | |
| | 6 2 | | | 2+5 2 | 6+7 2 | | | |
| | 5 2 | | | 7+8 2 | 8+9 2 | | | |
| | 5 2 | | | 6+9 2 | 4+5 2 | | | |
| | 6 2 | | | 2+5 2 | 6+7 2 | | | |
| | 5 2 | | | 3+4 2 | 2+3 2 | | | |
| | 5 2 | | | 3+4 2 | 4+5 2 | | | |
| | 5 2 | | | 7+8 2 | 8+9 2 | | | |
| | 6 2 | | | 2+5 2 | 2+3 2 | | | |
| | 6 2 | | | 6+9 2 | 4+5 2 | | | |
| | 5 2 | | | 2+5 2 | 6+7 2 | | | |
| | 6 2 | | | 6+9 2 | 8+9 2 | | | |

Continued on next column

**TABLE XXXIII– continued**

| TQC | ICSD | NoE | SF | | | | | |
|---|---|---|---|---|---|---|---|---|
| | 5 2 | | | 3+4 2 | 2+3 2 | | | |
| | 5 2 | | | 7+8 2 | 4+5 2 | | | |
| | 6 2 | | | 6+9 2 | 6+7 2 | | | |
| | 5 2 | | | 7+8 2 | 8+9 2 | | | |
| | 5 2 | | | 3+4 2 | 4+5 2 | | | |
| | 6 2 | | | 2+5 2 | 8+9 2 | | | |
| | 6 2 | | | 6+9 2 | 6+7 2 | | | |
| | 5 2 | | | 2+5 2 | 2+3 2 | | | |
| | 6 2 | | | 7+8 2 | 4+5 2 | | | |
| | 5 2 | | | 3+4 2 | 2+3 2 | | | |
| | 6 2 | | | 2+5 2 | 6+7 2 | | | |
| | 5 2 | | | 6+9 2 | 8+9 2 | | | |
| | 6 2 | | | 7+8 2 | 2+3 2 | | | |
| | 6 2 | | | 3+4 2 | 4+5 2 | | | |
| | 5 2 | | | 3+4 2 | 8+9 2 | | | |
| | 5 2 | | | 2+5 2 | 2+3 2 | | | |
| | 6 2 | | | 6+9 2 | 4+5 2 | | | |
| | 5 2 | | | 7+8 2 | 6+7 2 | | | |
| | 5 2 | | | 2+5 2 | 4+5 2 | | | |
| | 6 2 | | | 7+8 2 | 8+9 2 | | | |
| | 5 2 | | | 6+9 2 | 6+7 2 | | | |
| | 6 2 | | | 3+4 2 | 2+3 2 | | | |
| | 6 2 | | | 6+9 2 | 8+9 2 | | | |
| | 6 2 | | | 7+8 2 | 4+5 2 | | | |
| | 5 2 | | | 3+4 2 | 6+7 2 | | | |
| | 5 2 | | | 2+5 2 | 2+3 2 | | | |
| | 6 2 | | | 3+4 2 | 2+3 2 | | | |
| | 6 2 | | | 7+8 2 | 4+5 2 | | | |
| | 5 2 | | | 6+9 2 | 6+7 2 | | | |
| | 5 2 | | | 2+5 2 | 8+9 2 | | | |
| | 6 2 | | | 6+9 2 | 2+3 2 | | | |
| | 6 2 | | | 7+8 2 | 4+5 2 | | | |
| | 6 2 | | | 3+4 2 | 4+5 2 | | | |
| | 6 2 | | | 2+5 2 | 2+3 2 | | | |
| | 5 2 | | | 3+4 2 | 8+9 2 | | | |
| | 6 2 | | | 2+5 2 | 6+7 2 | | | |
| | 5 2 | | | 6+9 2 | 6+7 2 | | | |
| | 5 2 | | | 7+8 2 | 8+9 2 | | | |
| | 6 2 | | | 7+8 2 | 6+7 2 | | | |
| | 5 2 | | | 6+9 2 | 8+9 2 | | | |
| 58.2.12 | 42605 | 40 | As2Ni1 | | | | | |
| | -GM | -R | -S | -T | -U | -X | -Y | -Z |
| | 5 2 | 3+4 4 | 6+6 4 | 6+9 2 | 2+3 2 | 3+4 4 | 3+4 4 | 3+4 4 |
| | 5 2 | 3+4 4 | 5+5 4 | 7+8 2 | 4+5 2 | 3+4 4 | 3+4 4 | 3+4 4 |
| | 6 2 | 3+4 4 | 5+5 4 | 2+5 2 | 8+9 2 | 3+4 4 | 3+4 4 | 3+4 4 |
| | 6 2 | 3+4 4 | 5+5 4 | 3+4 2 | 6+7 2 | 3+4 4 | 3+4 4 | 3+4 4 |
| | 5 2 | 3+4 4 | 6+6 4 | 3+4 2 | 2+3 2 | 3+4 4 | 3+4 4 | 3+4 4 |
| | 5 2 | 3+4 4 | 5+5 4 | 2+5 2 | 4+5 2 | 3+4 4 | 3+4 4 | 3+4 4 |
| | 5 2 | 3+4 4 | 5+5 4 | 2+5 2 | 8+9 2 | 3+4 4 | 3+4 4 | 3+4 4 |
| | 5 2 | 3+4 4 | 5+5 4 | 3+4 2 | 6+7 2 | 3+4 4 | 3+4 4 | 3+4 4 |
| | 5 2 | 3+4 4 | 5+5 4 | 7+8 2 | 4+5 2 | 3+4 4 | 3+4 4 | 3+4 4 |
| | 6 2 | 3+4 4 | 6+6 4 | 6+9 2 | 2+3 2 | 3+4 4 | 3+4 4 | 3+4 4 |
| | 5 2 | | | 2+5 2 | 2+3 2 | | | |
| | 5 2 | | | 3+4 2 | 4+5 2 | | | |
| | 5 2 | | | 3+4 2 | 6+7 2 | | | |
| | 5 2 | | | 2+5 2 | 8+9 2 | | | |
| | 6 2 | | | 6+9 2 | 2+3 2 | | | |
| | 5 2 | | | 7+8 2 | 4+5 2 | | | |
| | 6 2 | | | 3+4 2 | 4+5 2 | | | |
| | 5 2 | | | 2+5 2 | 2+3 2 | | | |
| | 5 2 | | | 6+9 2 | 2+3 2 | | | |

Continued on next column

**TABLE XXXIII– continued**

| TQC | ICSD | NoE | SF |
|---|---|---|---|
| | 5 2 | | 7+8 2 4+5 2 |

The End

TABLE XXXIV: The computed irreducible representations for the valence bands for the compounds of SEBR and NLC in space group 59. See the list of tables.

| TQC | -GM | -R | -S | -T | -U | -X | -Y | -Z |
|---|---|---|---|---|---|---|---|---|
| 59.1.1 | ICSD: 408080 | NoE: 200 | SF: Pd1Si1Y1 | | | | | |
| | 5 2 | 3+4 4 | 3+4 4 | 3+4 4 | 3+4 4 | 3+4 4 | 3+4 4 | 5 2 |
| | 6 2 | 3+4 4 | 3+4 4 | 3+4 4 | 3+4 4 | 3+4 4 | 3+4 4 | 6 2 |
| | 5 2 | 3+4 4 | 3+4 4 | 3+4 4 | 3+4 4 | 3+4 4 | 3+4 4 | 5 2 |
| | 5 2 | 3+4 4 | 3+4 4 | 3+4 4 | 3+4 4 | 3+4 4 | 3+4 4 | 5 2 |
| | 6 2 | 3+4 4 | 3+4 4 | 3+4 4 | 3+4 4 | 3+4 4 | 3+4 4 | 6 2 |
| | 6 2 | 3+4 4 | 3+4 4 | 3+4 4 | 3+4 4 | 3+4 4 | 3+4 4 | 6 2 |
| | 5 2 | 3+4 4 | 3+4 4 | 3+4 4 | 3+4 4 | 3+4 4 | 3+4 4 | 5 2 |
| | 6 2 | 3+4 4 | 3+4 4 | 3+4 4 | 3+4 4 | 3+4 4 | 3+4 4 | 6 2 |
| | 5 2 | 3+4 4 | 3+4 4 | 3+4 4 | 3+4 4 | 3+4 4 | 3+4 4 | 5 2 |
| | 6 2 | 3+4 4 | 3+4 4 | 3+4 4 | 3+4 4 | 3+4 4 | 3+4 4 | 6 2 |
| | 5 2 | 3+4 4 | 3+4 4 | 3+4 4 | 3+4 4 | 3+4 4 | 3+4 4 | 5 2 |
| | 5 2 | 3+4 4 | 3+4 4 | 3+4 4 | 3+4 4 | 3+4 4 | 3+4 4 | 5 2 |
| | 6 2 | 3+4 4 | 3+4 4 | 3+4 4 | 3+4 4 | 3+4 4 | 3+4 4 | 6 2 |
| | 6 2 | 3+4 4 | 3+4 4 | 3+4 4 | 3+4 4 | 3+4 4 | 3+4 4 | 6 2 |
| | 5 2 | 3+4 4 | 3+4 4 | 3+4 4 | 3+4 4 | 3+4 4 | 3+4 4 | 5 2 |
| | 6 2 | 3+4 4 | 3+4 4 | 3+4 4 | 3+4 4 | 3+4 4 | 3+4 4 | 6 2 |
| | 5 2 | 3+4 4 | 3+4 4 | 3+4 4 | 3+4 4 | 3+4 4 | 3+4 4 | 5 2 |
| | 6 2 | 3+4 4 | 3+4 4 | 3+4 4 | 3+4 4 | 3+4 4 | 3+4 4 | 6 2 |
| | 5 2 | 3+4 4 | 3+4 4 | 3+4 4 | 3+4 4 | 3+4 4 | 3+4 4 | 5 2 |
| | 5 2 | 3+4 4 | 3+4 4 | 3+4 4 | 3+4 4 | 3+4 4 | 3+4 4 | 5 2 |
| | 6 2 | 3+4 4 | 3+4 4 | 3+4 4 | 3+4 4 | 3+4 4 | 3+4 4 | 6 2 |
| | 5 2 | 3+4 4 | 3+4 4 | 3+4 4 | 3+4 4 | 3+4 4 | 3+4 4 | 5 2 |
| | 6 2 | 3+4 4 | 3+4 4 | 3+4 4 | 3+4 4 | 3+4 4 | 3+4 4 | 6 2 |
| | 5 2 | 3+4 4 | 3+4 4 | 3+4 4 | 3+4 4 | 3+4 4 | 3+4 4 | 5 2 |
| | 6 2 | 3+4 4 | 3+4 4 | 3+4 4 | 3+4 4 | 3+4 4 | 3+4 4 | 6 2 |
| | 5 2 | 3+4 4 | 3+4 4 | 3+4 4 | 3+4 4 | 3+4 4 | 3+4 4 | 5 2 |
| | 6 2 | 3+4 4 | 3+4 4 | 3+4 4 | 3+4 4 | 3+4 4 | 3+4 4 | 6 2 |
| | 6 2 | 3+4 4 | 3+4 4 | 3+4 4 | 3+4 4 | 3+4 4 | 3+4 4 | 6 2 |
| | 5 2 | 3+4 4 | 3+4 4 | 3+4 4 | 3+4 4 | 3+4 4 | 3+4 4 | 5 2 |
| | 6 2 | 3+4 4 | 3+4 4 | 3+4 4 | 3+4 4 | 3+4 4 | 3+4 4 | 6 2 |
| | 5 2 | 3+4 4 | 3+4 4 | 3+4 4 | 3+4 4 | 3+4 4 | 3+4 4 | 5 2 |
| | 6 2 | 3+4 4 | 3+4 4 | 3+4 4 | 3+4 4 | 3+4 4 | 3+4 4 | 6 2 |
| | 5 2 | 3+4 4 | 3+4 4 | 3+4 4 | 3+4 4 | 3+4 4 | 3+4 4 | 5 2 |
| | 6 2 | 3+4 4 | 3+4 4 | 3+4 4 | 3+4 4 | 3+4 4 | 3+4 4 | 6 2 |
| | 5 2 | 3+4 4 | 3+4 4 | 3+4 4 | 3+4 4 | 3+4 4 | 3+4 4 | 6 2 |
| | 5 2 | 3+4 4 | 3+4 4 | 3+4 4 | 3+4 4 | 3+4 4 | 3+4 4 | 6 2 |
| | 6 2 | 3+4 4 | 3+4 4 | 3+4 4 | 3+4 4 | 3+4 4 | 3+4 4 | 5 2 |
| | 6 2 | 3+4 4 | 3+4 4 | 3+4 4 | 3+4 4 | 3+4 4 | 3+4 4 | 5 2 |
| | 6 2 | 3+4 4 | 3+4 4 | 3+4 4 | 3+4 4 | 3+4 4 | 3+4 4 | 5 2 |
| | 5 2 | 3+4 4 | 3+4 4 | 3+4 4 | 3+4 4 | 3+4 4 | 3+4 4 | 6 2 |
| | 6 2 | 3+4 4 | 3+4 4 | 3+4 4 | 3+4 4 | 3+4 4 | 3+4 4 | 5 2 |
| | 5 2 | 3+4 4 | 3+4 4 | 3+4 4 | 3+4 4 | 3+4 4 | 3+4 4 | 6 2 |
| | 5 2 | 3+4 4 | 3+4 4 | 3+4 4 | 3+4 4 | 3+4 4 | 3+4 4 | 5 2 |
| | 5 2 | 3+4 4 | 3+4 4 | 3+4 4 | 3+4 4 | 3+4 4 | 3+4 4 | 6 2 |
| | 6 2 | 3+4 4 | 3+4 4 | 3+4 4 | 3+4 4 | 3+4 4 | 3+4 4 | 6 2 |
| | 6 2 | 3+4 4 | 3+4 4 | 3+4 4 | 3+4 4 | 3+4 4 | 3+4 4 | 5 2 |
| | 5 2 | 3+4 4 | 3+4 4 | 3+4 4 | 3+4 4 | 3+4 4 | 3+4 4 | 6 2 |
| | 6 2 | 3+4 4 | 3+4 4 | 3+4 4 | 3+4 4 | 3+4 4 | 3+4 4 | 6 2 |
| | 5 2 | 3+4 4 | 3+4 4 | 3+4 4 | 3+4 4 | 3+4 4 | 3+4 4 | 5 2 |
| | 6 2 | 3+4 4 | 3+4 4 | 3+4 4 | 3+4 4 | 3+4 4 | 3+4 4 | 5 2 |

Continued on next column

**TABLE XXXIV– continued**

| TQC | -GM | -R | -S | -T | -U | -X | -Y | -Z |
|---|---|---|---|---|---|---|---|---|
| | 5 2 | | | | | | | 5 2 |
| | 6 2 | | | | | | | 6 2 |
| | 6 2 | | | | | | | 6 2 |
| | 6 2 | | | | | | | 5 2 |
| | 5 2 | | | | | | | 6 2 |
| | 5 2 | | | | | | | 5 2 |
| | 5 2 | | | | | | | 5 2 |
| | 6 2 | | | | | | | 6 2 |
| | 5 2 | | | | | | | 6 2 |
| | 5 2 | | | | | | | 5 2 |
| | 6 2 | | | | | | | 5 2 |
| | 5 2 | | | | | | | 5 2 |
| | 6 2 | | | | | | | 6 2 |
| | 5 2 | | | | | | | 6 2 |
| | 6 2 | | | | | | | 5 2 |
| | 5 2 | | | | | | | 6 2 |
| | 6 2 | | | | | | | 5 2 |
| | 5 2 | | | | | | | 6 2 |
| | 5 2 | | | | | | | 6 2 |
| | 5 2 | | | | | | | 6 2 |
| | 6 2 | | | | | | | 6 2 |
| | 6 2 | | | | | | | 5 2 |
| | 6 2 | | | | | | | 5 2 |
| | 6 2 | | | | | | | 6 2 |
| | 5 2 | | | | | | | 5 2 |
| | 5 2 | | | | | | | 5 2 |
| | 6 2 | | | | | | | 6 2 |
| | 5 2 | | | | | | | 5 2 |
| | 6 2 | | | | | | | 5 2 |
| | 5 2 | | | | | | | 6 2 |
| | 6 2 | | | | | | | 5 2 |
| | 5 2 | | | | | | | 6 2 |
| | 6 2 | | | | | | | 6 2 |
| | 6 2 | | | | | | | 6 2 |
| | 5 2 | | | | | | | 6 2 |
| | 6 2 | | | | | | | 5 2 |
| | 5 2 | | | | | | | 5 2 |
| | 6 2 | | | | | | | 5 2 |
| | 6 2 | | | | | | | 6 2 |
| | 5 2 | | | | | | | 6 2 |
| | 5 2 | | | | | | | 6 2 |
| | 6 2 | | | | | | | 6 2 |
| | 5 2 | | | | | | | 5 2 |
| | 5 2 | | | | | | | 6 2 |
| | 5 2 | | | | | | | 5 2 |
| | 5 2 | | | | | | | 6 2 |
| | 6 2 | | | | | | | 6 2 |
| | 6 2 | | | | | | | 5 2 |
| | 6 2 | | | | | | | 5 2 |
| | 5 2 | | | | | | | 5 2 |
| 59.1.2 | ICSD: 154823 | NoE: 40 | SF: Li1Mn1O2 | | | | | |
| | 5 2 | 3+4 4 | 3+4 4 | 3+4 4 | 3+4 4 | 3+4 4 | 3+4 4 | 6 2 |
| | 6 2 | 3+4 4 | 3+4 4 | 3+4 4 | 3+4 4 | 3+4 4 | 3+4 4 | 5 2 |
| | 6 2 | 3+4 4 | 3+4 4 | 3+4 4 | 3+4 4 | 3+4 4 | 3+4 4 | 5 2 |
| | 5 2 | 3+4 4 | 3+4 4 | 3+4 4 | 3+4 4 | 3+4 4 | 3+4 4 | 6 2 |
| | 5 2 | 3+4 4 | 3+4 4 | 3+4 4 | 3+4 4 | 3+4 4 | 3+4 4 | 6 2 |
| | 5 2 | 3+4 4 | 3+4 4 | 3+4 4 | 3+4 4 | 3+4 4 | 3+4 4 | 5 2 |
| | 6 2 | 3+4 4 | 3+4 4 | 3+4 4 | 3+4 4 | 3+4 4 | 3+4 4 | 5 2 |
| | 5 2 | 3+4 4 | 3+4 4 | 3+4 4 | 3+4 4 | 3+4 4 | 3+4 4 | 6 2 |
| | 5 2 | 3+4 4 | 3+4 4 | 3+4 4 | 3+4 4 | 3+4 4 | 3+4 4 | 5 2 |

Continued on next column

**TABLE XXXIV– continued**

| TQC | ICSD | NoE | SF | | | | | |
|---|---|---|---|---|---|---|---|---|
| | 6 2 | 3+4 4 | 3+4 4 | 3+4 4 | 3+4 4 | 3+4 4 | 3+4 4 | 6 2 |
| | 6 2 | | | | | | | 6 2 |
| | 5 2 | | | | | | | 5 2 |
| | 5 2 | | | | | | | 6 2 |
| | 6 2 | | | | | | | 6 2 |
| | 6 2 | | | | | | | 5 2 |
| | 6 2 | | | | | | | 5 2 |
| | 6 2 | | | | | | | 6 2 |
| | 5 2 | | | | | | | 5 2 |
| | 5 2 | | | | | | | 6 2 |
| | 5 2 | | | | | | | 6 2 |
| 59.1.3 | 425285 | 68 | Ho1I1S1 | | | | | |
| | -GM | -R | -S | -T | -U | -X | -Y | -Z |
| | 5 2 | 3+4 4 | 3+4 4 | 3+4 4 | 3+4 4 | 3+4 4 | 3+4 4 | 5 2 |
| | 6 2 | 3+4 4 | 3+4 4 | 3+4 4 | 3+4 4 | 3+4 4 | 3+4 4 | 6 2 |
| | 5 2 | 3+4 4 | 3+4 4 | 3+4 4 | 3+4 4 | 3+4 4 | 3+4 4 | 5 2 |
| | 6 2 | 3+4 4 | 3+4 4 | 3+4 4 | 3+4 4 | 3+4 4 | 3+4 4 | 6 2 |
| | 5 2 | 3+4 4 | 3+4 4 | 3+4 4 | 3+4 4 | 3+4 4 | 3+4 4 | 5 2 |
| | 5 2 | 3+4 4 | 3+4 4 | 3+4 4 | 3+4 4 | 3+4 4 | 3+4 4 | 5 2 |
| | 6 2 | 3+4 4 | 3+4 4 | 3+4 4 | 3+4 4 | 3+4 4 | 3+4 4 | 6 2 |
| | 6 2 | 3+4 4 | 3+4 4 | 3+4 4 | 3+4 4 | 3+4 4 | 3+4 4 | 6 2 |
| | 5 2 | 3+4 4 | 3+4 4 | 3+4 4 | 3+4 4 | 3+4 4 | 3+4 4 | 5 2 |
| | 5 2 | 3+4 4 | 3+4 4 | 3+4 4 | 3+4 4 | 3+4 4 | 3+4 4 | 6 2 |
| | 6 2 | 3+4 4 | 3+4 4 | 3+4 4 | 3+4 4 | 3+4 4 | 3+4 4 | 6 2 |
| | 6 2 | 3+4 4 | 3+4 4 | 3+4 4 | 3+4 4 | 3+4 4 | 3+4 4 | 5 2 |
| | 6 2 | 3+4 4 | 3+4 4 | 3+4 4 | 3+4 4 | 3+4 4 | 3+4 4 | 6 2 |
| | 5 2 | 3+4 4 | 3+4 4 | 3+4 4 | 3+4 4 | 3+4 4 | 3+4 4 | 5 2 |
| | 5 2 | 3+4 4 | 3+4 4 | 3+4 4 | 3+4 4 | 3+4 4 | 3+4 4 | 5 2 |
| | 5 2 | 3+4 4 | 3+4 4 | 3+4 4 | 3+4 4 | 3+4 4 | 3+4 4 | 5 2 |
| | 5 2 | 3+4 4 | 3+4 4 | 3+4 4 | 3+4 4 | 3+4 4 | 3+4 4 | 6 2 |
| | 6 2 | | | | | | | 5 2 |
| | 5 2 | | | | | | | 6 2 |
| | 6 2 | | | | | | | 6 2 |
| | 5 2 | | | | | | | 6 2 |
| | 6 2 | | | | | | | 5 2 |
| | 6 2 | | | | | | | 6 2 |
| | 5 2 | | | | | | | 5 2 |
| | 5 2 | | | | | | | 5 2 |
| | 6 2 | | | | | | | 6 2 |
| | 6 2 | | | | | | | 5 2 |
| | 5 2 | | | | | | | 6 2 |
| | 6 2 | | | | | | | 5 2 |
| | 6 2 | | | | | | | 6 2 |
| | 5 2 | | | | | | | 5 2 |
| | 5 2 | | | | | | | 5 2 |
| | 6 2 | | | | | | | 6 2 |
| | 5 2 | | | | | | | 5 2 |
| 59.1.4 | 154397 | 180 | Ge5N2Sr6 | | | | | |
| | -GM | -R | -S | -T | -U | -X | -Y | -Z |
| | 5 2 | 3+4 4 | 3+4 4 | 3+4 4 | 3+4 4 | 3+4 4 | 3+4 4 | 5 2 |
| | 6 2 | 3+4 4 | 3+4 4 | 3+4 4 | 3+4 4 | 3+4 4 | 3+4 4 | 6 2 |
| | 5 2 | 3+4 4 | 3+4 4 | 3+4 4 | 3+4 4 | 3+4 4 | 3+4 4 | 5 2 |
| | 6 2 | 3+4 4 | 3+4 4 | 3+4 4 | 3+4 4 | 3+4 4 | 3+4 4 | 6 2 |
| | 5 2 | 3+4 4 | 3+4 4 | 3+4 4 | 3+4 4 | 3+4 4 | 3+4 4 | 6 2 |
| | 5 2 | 3+4 4 | 3+4 4 | 3+4 4 | 3+4 4 | 3+4 4 | 3+4 4 | 5 2 |
| | 6 2 | 3+4 4 | 3+4 4 | 3+4 4 | 3+4 4 | 3+4 4 | 3+4 4 | 6 2 |
| | 6 2 | 3+4 4 | 3+4 4 | 3+4 4 | 3+4 4 | 3+4 4 | 3+4 4 | 5 2 |
| | 5 2 | 3+4 4 | 3+4 4 | 3+4 4 | 3+4 4 | 3+4 4 | 3+4 4 | 5 2 |
| | 5 2 | 3+4 4 | 3+4 4 | 3+4 4 | 3+4 4 | 3+4 4 | 3+4 4 | 5 2 |
| | 6 2 | 3+4 4 | 3+4 4 | 3+4 4 | 3+4 4 | 3+4 4 | 3+4 4 | 6 2 |
| | 6 2 | 3+4 4 | 3+4 4 | 3+4 4 | 3+4 4 | 3+4 4 | 3+4 4 | 6 2 |

Continued on next column

**TABLE XXXIV– continued**

| TQC | ICSD | NoE | SF | | | | | |
|---|---|---|---|---|---|---|---|---|
| | 5 2 | 3+4 4 | 3+4 4 | 3+4 4 | 3+4 4 | 3+4 4 | 3+4 4 | 6 2 |
| | 6 2 | 3+4 4 | 3+4 4 | 3+4 4 | 3+4 4 | 3+4 4 | 3+4 4 | 5 2 |
| | 5 2 | 3+4 4 | 3+4 4 | 3+4 4 | 3+4 4 | 3+4 4 | 3+4 4 | 5 2 |
| | 6 2 | 3+4 4 | 3+4 4 | 3+4 4 | 3+4 4 | 3+4 4 | 3+4 4 | 6 2 |
| | 5 2 | 3+4 4 | 3+4 4 | 3+4 4 | 3+4 4 | 3+4 4 | 3+4 4 | 6 2 |
| | 6 2 | 3+4 4 | 3+4 4 | 3+4 4 | 3+4 4 | 3+4 4 | 3+4 4 | 5 2 |
| | 5 2 | 3+4 4 | 3+4 4 | 3+4 4 | 3+4 4 | 3+4 4 | 3+4 4 | 6 2 |
| | 6 2 | 3+4 4 | 3+4 4 | 3+4 4 | 3+4 4 | 3+4 4 | 3+4 4 | 5 2 |
| | 5 2 | 3+4 4 | 3+4 4 | 3+4 4 | 3+4 4 | 3+4 4 | 3+4 4 | 5 2 |
| | 5 2 | 3+4 4 | 3+4 4 | 3+4 4 | 3+4 4 | 3+4 4 | 3+4 4 | 5 2 |
| | 6 2 | 3+4 4 | 3+4 4 | 3+4 4 | 3+4 4 | 3+4 4 | 3+4 4 | 6 2 |
| | 6 2 | 3+4 4 | 3+4 4 | 3+4 4 | 3+4 4 | 3+4 4 | 3+4 4 | 6 2 |
| | 5 2 | 3+4 4 | 3+4 4 | 3+4 4 | 3+4 4 | 3+4 4 | 3+4 4 | 6 2 |
| | 6 2 | 3+4 4 | 3+4 4 | 3+4 4 | 3+4 4 | 3+4 4 | 3+4 4 | 5 2 |
| | 5 2 | 3+4 4 | 3+4 4 | 3+4 4 | 3+4 4 | 3+4 4 | 3+4 4 | 6 2 |
| | 6 2 | 3+4 4 | 3+4 4 | 3+4 4 | 3+4 4 | 3+4 4 | 3+4 4 | 5 2 |
| | 5 2 | 3+4 4 | 3+4 4 | 3+4 4 | 3+4 4 | 3+4 4 | 3+4 4 | 6 2 |
| | 6 2 | 3+4 4 | 3+4 4 | 3+4 4 | 3+4 4 | 3+4 4 | 3+4 4 | 5 2 |
| | 5 2 | 3+4 4 | 3+4 4 | 3+4 4 | 3+4 4 | 3+4 4 | 3+4 4 | 6 2 |
| | 6 2 | 3+4 4 | 3+4 4 | 3+4 4 | 3+4 4 | 3+4 4 | 3+4 4 | 5 2 |
| | 5 2 | 3+4 4 | 3+4 4 | 3+4 4 | 3+4 4 | 3+4 4 | 3+4 4 | 5 2 |
| | 6 2 | 3+4 4 | 3+4 4 | 3+4 4 | 3+4 4 | 3+4 4 | 3+4 4 | 6 2 |
| | 5 2 | 3+4 4 | 3+4 4 | 3+4 4 | 3+4 4 | 3+4 4 | 3+4 4 | 6 2 |
| | 6 2 | 3+4 4 | 3+4 4 | 3+4 4 | 3+4 4 | 3+4 4 | 3+4 4 | 5 2 |
| | 5 2 | 3+4 4 | 3+4 4 | 3+4 4 | 3+4 4 | 3+4 4 | 3+4 4 | 6 2 |
| | 6 2 | 3+4 4 | 3+4 4 | 3+4 4 | 3+4 4 | 3+4 4 | 3+4 4 | 5 2 |
| | 5 2 | 3+4 4 | 3+4 4 | 3+4 4 | 3+4 4 | 3+4 4 | 3+4 4 | 6 2 |
| | 6 2 | 3+4 4 | 3+4 4 | 3+4 4 | 3+4 4 | 3+4 4 | 3+4 4 | 5 2 |
| | 5 2 | 3+4 4 | 3+4 4 | 3+4 4 | 3+4 4 | 3+4 4 | 3+4 4 | 5 2 |
| | 5 2 | 3+4 4 | 3+4 4 | 3+4 4 | 3+4 4 | 3+4 4 | 3+4 4 | 5 2 |
| | 6 2 | 3+4 4 | 3+4 4 | 3+4 4 | 3+4 4 | 3+4 4 | 3+4 4 | 6 2 |
| | 6 2 | 3+4 4 | 3+4 4 | 3+4 4 | 3+4 4 | 3+4 4 | 3+4 4 | 6 2 |
| | 5 2 | 3+4 4 | 3+4 4 | 3+4 4 | 3+4 4 | 3+4 4 | 3+4 4 | 5 2 |
| | 5 2 | | | | | | | 5 2 |
| | 6 2 | | | | | | | 6 2 |
| | 6 2 | | | | | | | 6 2 |
| | 5 2 | | | | | | | 5 2 |
| | 6 2 | | | | | | | 6 2 |
| | 5 2 | | | | | | | 5 2 |
| | 6 2 | | | | | | | 6 2 |
| | 5 2 | | | | | | | 5 2 |
| | 5 2 | | | | | | | 5 2 |
| | 6 2 | | | | | | | 6 2 |
| | 6 2 | | | | | | | 6 2 |
| | 5 2 | | | | | | | 5 2 |
| | 6 2 | | | | | | | 6 2 |
| | 6 2 | | | | | | | 6 2 |
| | 5 2 | | | | | | | 6 2 |
| | 6 2 | | | | | | | 5 2 |
| | 5 2 | | | | | | | 5 2 |
| | 6 2 | | | | | | | 5 2 |
| | 5 2 | | | | | | | 6 2 |
| | 5 2 | | | | | | | 5 2 |
| | 6 2 | | | | | | | 6 2 |
| | 5 2 | | | | | | | 5 2 |
| | 5 2 | | | | | | | 5 2 |
| | 6 2 | | | | | | | 6 2 |
| | 6 2 | | | | | | | 5 2 |
| | 5 2 | | | | | | | 6 2 |
| | 5 2 | | | | | | | 6 2 |
| | 6 2 | | | | | | | 6 2 |

Continued on next column

**TABLE XXXIV– continued**

| TQC | ICSD | NoE | SF | | | | | |
|---|---|---|---|---|---|---|---|---|
| | 5 2 | | | | | | | 5 2 |
| | 6 2 | | | | | | | 6 2 |
| | 5 2 | | | | | | | 5 2 |
| | 6 2 | | | | | | | 5 2 |
| | 6 2 | | | | | | | 6 2 |
| | 5 2 | | | | | | | 5 2 |
| | 6 2 | | | | | | | 6 2 |
| | 6 2 | | | | | | | 6 2 |
| | 5 2 | | | | | | | 6 2 |
| | 6 2 | | | | | | | 5 2 |
| | 5 2 | | | | | | | 5 2 |
| | 5 2 | | | | | | | 6 2 |
| | 6 2 | | | | | | | 6 2 |
| | 6 2 | | | | | | | 5 2 |
| | 6 2 | | | | | | | 6 2 |
| | 5 2 | | | | | | | 6 2 |
| | 6 2 | | | | | | | 5 2 |
| 59.1.5 | 410357 | 216 | Ge1Pr1Pt1 | | | | | |
| | -GM | -R | -S | -T | -U | -X | -Y | -Z |
| | 5 2 | 3+4 4 | 3+4 4 | 3+4 4 | 3+4 4 | 3+4 4 | 3+4 4 | 6 2 |
| | 6 2 | 3+4 4 | 3+4 4 | 3+4 4 | 3+4 4 | 3+4 4 | 3+4 4 | 5 2 |
| | 5 2 | 3+4 4 | 3+4 4 | 3+4 4 | 3+4 4 | 3+4 4 | 3+4 4 | 5 2 |
| | 6 2 | 3+4 4 | 3+4 4 | 3+4 4 | 3+4 4 | 3+4 4 | 3+4 4 | 5 2 |
| | 5 2 | 3+4 4 | 3+4 4 | 3+4 4 | 3+4 4 | 3+4 4 | 3+4 4 | 6 2 |
| | 5 2 | 3+4 4 | 3+4 4 | 3+4 4 | 3+4 4 | 3+4 4 | 3+4 4 | 6 2 |
| | 6 2 | 3+4 4 | 3+4 4 | 3+4 4 | 3+4 4 | 3+4 4 | 3+4 4 | 5 2 |
| | 6 2 | 3+4 4 | 3+4 4 | 3+4 4 | 3+4 4 | 3+4 4 | 3+4 4 | 6 2 |
| | 5 2 | 3+4 4 | 3+4 4 | 3+4 4 | 3+4 4 | 3+4 4 | 3+4 4 | 5 2 |
| | 6 2 | 3+4 4 | 3+4 4 | 3+4 4 | 3+4 4 | 3+4 4 | 3+4 4 | 6 2 |
| | 5 2 | 3+4 4 | 3+4 4 | 3+4 4 | 3+4 4 | 3+4 4 | 3+4 4 | 5 2 |
| | 6 2 | 3+4 4 | 3+4 4 | 3+4 4 | 3+4 4 | 3+4 4 | 3+4 4 | 5 2 |
| | 5 2 | 3+4 4 | 3+4 4 | 3+4 4 | 3+4 4 | 3+4 4 | 3+4 4 | 6 2 |
| | 5 2 | 3+4 4 | 3+4 4 | 3+4 4 | 3+4 4 | 3+4 4 | 3+4 4 | 6 2 |
| | 6 2 | 3+4 4 | 3+4 4 | 3+4 4 | 3+4 4 | 3+4 4 | 3+4 4 | 5 2 |
| | 6 2 | 3+4 4 | 3+4 4 | 3+4 4 | 3+4 4 | 3+4 4 | 3+4 4 | 6 2 |
| | 5 2 | 3+4 4 | 3+4 4 | 3+4 4 | 3+4 4 | 3+4 4 | 3+4 4 | 5 2 |
| | 6 2 | 3+4 4 | 3+4 4 | 3+4 4 | 3+4 4 | 3+4 4 | 3+4 4 | 6 2 |
| | 5 2 | 3+4 4 | 3+4 4 | 3+4 4 | 3+4 4 | 3+4 4 | 3+4 4 | 5 2 |
| | 6 2 | 3+4 4 | 3+4 4 | 3+4 4 | 3+4 4 | 3+4 4 | 3+4 4 | 5 2 |
| | 5 2 | 3+4 4 | 3+4 4 | 3+4 4 | 3+4 4 | 3+4 4 | 3+4 4 | 6 2 |
| | 5 2 | 3+4 4 | 3+4 4 | 3+4 4 | 3+4 4 | 3+4 4 | 3+4 4 | 5 2 |
| | 6 2 | 3+4 4 | 3+4 4 | 3+4 4 | 3+4 4 | 3+4 4 | 3+4 4 | 6 2 |
| | 5 2 | 3+4 4 | 3+4 4 | 3+4 4 | 3+4 4 | 3+4 4 | 3+4 4 | 6 2 |
| | 5 2 | 3+4 4 | 3+4 4 | 3+4 4 | 3+4 4 | 3+4 4 | 3+4 4 | 5 2 |
| | 6 2 | 3+4 4 | 3+4 4 | 3+4 4 | 3+4 4 | 3+4 4 | 3+4 4 | 5 2 |
| | 6 2 | 3+4 4 | 3+4 4 | 3+4 4 | 3+4 4 | 3+4 4 | 3+4 4 | 5 2 |
| | 6 2 | 3+4 4 | 3+4 4 | 3+4 4 | 3+4 4 | 3+4 4 | 3+4 4 | 6 2 |
| | 5 2 | 3+4 4 | 3+4 4 | 3+4 4 | 3+4 4 | 3+4 4 | 3+4 4 | 6 2 |
| | 5 2 | 3+4 4 | 3+4 4 | 3+4 4 | 3+4 4 | 3+4 4 | 3+4 4 | 6 2 |
| | 6 2 | 3+4 4 | 3+4 4 | 3+4 4 | 3+4 4 | 3+4 4 | 3+4 4 | 5 2 |
| | 6 2 | 3+4 4 | 3+4 4 | 3+4 4 | 3+4 4 | 3+4 4 | 3+4 4 | 6 2 |
| | 5 2 | 3+4 4 | 3+4 4 | 3+4 4 | 3+4 4 | 3+4 4 | 3+4 4 | 5 2 |
| | 6 2 | 3+4 4 | 3+4 4 | 3+4 4 | 3+4 4 | 3+4 4 | 3+4 4 | 6 2 |
| | 5 2 | 3+4 4 | 3+4 4 | 3+4 4 | 3+4 4 | 3+4 4 | 3+4 4 | 6 2 |
| | 5 2 | 3+4 4 | 3+4 4 | 3+4 4 | 3+4 4 | 3+4 4 | 3+4 4 | 6 2 |
| | 6 2 | 3+4 4 | 3+4 4 | 3+4 4 | 3+4 4 | 3+4 4 | 3+4 4 | 5 2 |
| | 6 2 | 3+4 4 | 3+4 4 | 3+4 4 | 3+4 4 | 3+4 4 | 3+4 4 | 5 2 |
| | 5 2 | 3+4 4 | 3+4 4 | 3+4 4 | 3+4 4 | 3+4 4 | 3+4 4 | 5 2 |
| | 6 2 | 3+4 4 | 3+4 4 | 3+4 4 | 3+4 4 | 3+4 4 | 3+4 4 | 6 2 |
| | 6 2 | 3+4 4 | 3+4 4 | 3+4 4 | 3+4 4 | 3+4 4 | 3+4 4 | 6 2 |
| | 5 2 | 3+4 4 | 3+4 4 | 3+4 4 | 3+4 4 | 3+4 4 | 3+4 4 | 5 2 |
| | 5 2 | 3+4 4 | 3+4 4 | 3+4 4 | 3+4 4 | 3+4 4 | 3+4 4 | 6 2 |
| | 5 2 | 3+4 4 | 3+4 4 | 3+4 4 | 3+4 4 | 3+4 4 | 3+4 4 | 5 2 |
| | 6 2 | 3+4 4 | 3+4 4 | 3+4 4 | 3+4 4 | 3+4 4 | 3+4 4 | 6 2 |
| | 6 2 | 3+4 4 | 3+4 4 | 3+4 4 | 3+4 4 | 3+4 4 | 3+4 4 | 6 2 |
| | 6 2 | 3+4 4 | 3+4 4 | 3+4 4 | 3+4 4 | 3+4 4 | 3+4 4 | 5 2 |
| | 5 2 | 3+4 4 | 3+4 4 | 3+4 4 | 3+4 4 | 3+4 4 | 3+4 4 | 5 2 |
| | 6 2 | 3+4 4 | 3+4 4 | 3+4 4 | 3+4 4 | 3+4 4 | 3+4 4 | 6 2 |
| | 5 2 | 3+4 4 | 3+4 4 | 3+4 4 | 3+4 4 | 3+4 4 | 3+4 4 | 5 2 |
| | 5 2 | 3+4 4 | 3+4 4 | 3+4 4 | 3+4 4 | 3+4 4 | 3+4 4 | 6 2 |
| | 6 2 | 3+4 4 | 3+4 4 | 3+4 4 | 3+4 4 | 3+4 4 | 3+4 4 | 5 2 |
| | 5 2 | 3+4 4 | 3+4 4 | 3+4 4 | 3+4 4 | 3+4 4 | 3+4 4 | 6 2 |
| | 6 2 | 3+4 4 | 3+4 4 | 3+4 4 | 3+4 4 | 3+4 4 | 3+4 4 | 5 2 |
| | 5 2 | | | | | | | 6 2 |
| | 6 2 | | | | | | | 5 2 |
| | 5 2 | | | | | | | 6 2 |
| | 6 2 | | | | | | | 5 2 |
| | 5 2 | | | | | | | 6 2 |
| | 6 2 | | | | | | | 6 2 |
| | 5 2 | | | | | | | 6 2 |
| | 5 2 | | | | | | | 5 2 |
| | 6 2 | | | | | | | 5 2 |
| | 5 2 | | | | | | | 5 2 |
| | 6 2 | | | | | | | 5 2 |
| | 5 2 | | | | | | | 6 2 |
| | 6 2 | | | | | | | 6 2 |
| | 5 2 | | | | | | | 5 2 |
| | 6 2 | | | | | | | 6 2 |
| | 5 2 | | | | | | | 5 2 |
| | 6 2 | | | | | | | 6 2 |
| | 5 2 | | | | | | | 5 2 |
| | 6 2 | | | | | | | 6 2 |
| | 6 2 | | | | | | | 5 2 |
| | 5 2 | | | | | | | 6 2 |
| | 5 2 | | | | | | | 5 2 |
| | 6 2 | | | | | | | 6 2 |
| | 6 2 | | | | | | | 6 2 |
| | 5 2 | | | | | | | 5 2 |
| | 6 2 | | | | | | | 5 2 |
| | 5 2 | | | | | | | 5 2 |
| | 5 2 | | | | | | | 6 2 |
| | 6 2 | | | | | | | 6 2 |
| | 5 2 | | | | | | | 6 2 |
| | 6 2 | | | | | | | 6 2 |
| | 6 2 | | | | | | | 5 2 |
| | 6 2 | | | | | | | 5 2 |
| | 6 2 | | | | | | | 5 2 |
| | 5 2 | | | | | | | 6 2 |
| | 5 2 | | | | | | | 6 2 |
| | 5 2 | | | | | | | 5 2 |
| | 6 2 | | | | | | | 6 2 |
| | 5 2 | | | | | | | 6 2 |
| | 5 2 | | | | | | | 6 2 |
| | 5 2 | | | | | | | 5 2 |
| | 6 2 | | | | | | | 5 2 |
| | 5 2 | | | | | | | 6 2 |
| | 5 2 | | | | | | | 6 2 |
| | 6 2 | | | | | | | 5 2 |
| | 6 2 | | | | | | | 5 2 |
| | 5 2 | | | | | | | 5 2 |
| | 6 2 | | | | | | | 6 2 |
| | 6 2 | | | | | | | 6 2 |

Continued on next column

**TABLE XXXIV– continued**

| TQC | ICSD | NoE | SF | | | | | |
|---|---|---|---|---|---|---|---|---|
| | 6 2 | | | | | | | 5 2 |
| | 6 2 | | | | | | | 6 2 |
| | 6 2 | | | | | | | 5 2 |
| | 5 2 | | | | | | | 5 2 |
| | 6 2 | | | | | | | 6 2 |
| 59.1.6 | 195027 | 88 | Fe2O1Se2Sr1 | | | | | |
| | -GM | -R | -S | -T | -U | -X | -Y | -Z |
| | 5 2 | 3+4 4 | 3+4 4 | 3+4 4 | 3+4 4 | 3+4 4 | 3+4 4 | 5 2 |
| | 6 2 | 3+4 4 | 3+4 4 | 3+4 4 | 3+4 4 | 3+4 4 | 3+4 4 | 6 2 |
| | 5 2 | 3+4 4 | 3+4 4 | 3+4 4 | 3+4 4 | 3+4 4 | 3+4 4 | 6 2 |
| | 6 2 | 3+4 4 | 3+4 4 | 3+4 4 | 3+4 4 | 3+4 4 | 3+4 4 | 5 2 |
| | 5 2 | 3+4 4 | 3+4 4 | 3+4 4 | 3+4 4 | 3+4 4 | 3+4 4 | 5 2 |
| | 6 2 | 3+4 4 | 3+4 4 | 3+4 4 | 3+4 4 | 3+4 4 | 3+4 4 | 6 2 |
| | 5 2 | 3+4 4 | 3+4 4 | 3+4 4 | 3+4 4 | 3+4 4 | 3+4 4 | 5 2 |
| | 6 2 | 3+4 4 | 3+4 4 | 3+4 4 | 3+4 4 | 3+4 4 | 3+4 4 | 6 2 |
| | 6 2 | 3+4 4 | 3+4 4 | 3+4 4 | 3+4 4 | 3+4 4 | 3+4 4 | 6 2 |
| | 5 2 | 3+4 4 | 3+4 4 | 3+4 4 | 3+4 4 | 3+4 4 | 3+4 4 | 5 2 |
| | 5 2 | 3+4 4 | 3+4 4 | 3+4 4 | 3+4 4 | 3+4 4 | 3+4 4 | 6 2 |
| | 5 2 | 3+4 4 | 3+4 4 | 3+4 4 | 3+4 4 | 3+4 4 | 3+4 4 | 6 2 |
| | 6 2 | 3+4 4 | 3+4 4 | 3+4 4 | 3+4 4 | 3+4 4 | 3+4 4 | 5 2 |
| | 6 2 | 3+4 4 | 3+4 4 | 3+4 4 | 3+4 4 | 3+4 4 | 3+4 4 | 5 2 |
| | 5 2 | 3+4 4 | 3+4 4 | 3+4 4 | 3+4 4 | 3+4 4 | 3+4 4 | 6 2 |
| | 5 2 | 3+4 4 | 3+4 4 | 3+4 4 | 3+4 4 | 3+4 4 | 3+4 4 | 5 2 |
| | 6 2 | 3+4 4 | 3+4 4 | 3+4 4 | 3+4 4 | 3+4 4 | 3+4 4 | 5 2 |
| | 6 2 | 3+4 4 | 3+4 4 | 3+4 4 | 3+4 4 | 3+4 4 | 3+4 4 | 6 2 |
| | 5 2 | 3+4 4 | 3+4 4 | 3+4 4 | 3+4 4 | 3+4 4 | 3+4 4 | 6 2 |
| | 6 2 | 3+4 4 | 3+4 4 | 3+4 4 | 3+4 4 | 3+4 4 | 3+4 4 | 5 2 |
| | 5 2 | 3+4 4 | 3+4 4 | 3+4 4 | 3+4 4 | 3+4 4 | 3+4 4 | 6 2 |
| | 6 2 | 3+4 4 | 3+4 4 | 3+4 4 | 3+4 4 | 3+4 4 | 3+4 4 | 5 2 |
| | 5 2 | | | | | | | 6 2 |
| | 5 2 | | | | | | | 5 2 |
| | 6 2 | | | | | | | 5 2 |
| | 5 2 | | | | | | | 5 2 |
| | 5 2 | | | | | | | 6 2 |
| | 6 2 | | | | | | | 5 2 |
| | 6 2 | | | | | | | 5 2 |
| | 6 2 | | | | | | | 6 2 |
| | 5 2 | | | | | | | 6 2 |
| | 6 2 | | | | | | | 6 2 |
| | 5 2 | | | | | | | 6 2 |
| | 5 2 | | | | | | | 6 2 |
| | 5 2 | | | | | | | 6 2 |
| | 6 2 | | | | | | | 5 2 |
| | 5 2 | | | | | | | 6 2 |
| | 5 2 | | | | | | | 6 2 |
| | 6 2 | | | | | | | 5 2 |
| | 6 2 | | | | | | | 5 2 |
| | 6 2 | | | | | | | 6 2 |
| | 5 2 | | | | | | | 5 2 |
| | 6 2 | | | | | | | 6 2 |
| | 5 2 | | | | | | | 5 2 |
| 59.1.7 | 195026 | 88 | Fe2O1S2Sr1 | | | | | |
| | -GM | -R | -S | -T | -U | -X | -Y | -Z |
| | 5 2 | 3+4 4 | 3+4 4 | 3+4 4 | 3+4 4 | 3+4 4 | 3+4 4 | 5 2 |
| | 6 2 | 3+4 4 | 3+4 4 | 3+4 4 | 3+4 4 | 3+4 4 | 3+4 4 | 6 2 |
| | 5 2 | 3+4 4 | 3+4 4 | 3+4 4 | 3+4 4 | 3+4 4 | 3+4 4 | 6 2 |
| | 6 2 | 3+4 4 | 3+4 4 | 3+4 4 | 3+4 4 | 3+4 4 | 3+4 4 | 5 2 |
| | 5 2 | 3+4 4 | 3+4 4 | 3+4 4 | 3+4 4 | 3+4 4 | 3+4 4 | 5 2 |
| | 6 2 | 3+4 4 | 3+4 4 | 3+4 4 | 3+4 4 | 3+4 4 | 3+4 4 | 6 2 |
| | 5 2 | 3+4 4 | 3+4 4 | 3+4 4 | 3+4 4 | 3+4 4 | 3+4 4 | 5 2 |
| | 6 2 | 3+4 4 | 3+4 4 | 3+4 4 | 3+4 4 | 3+4 4 | 3+4 4 | 6 2 |

Continued on next column

**TABLE XXXIV– continued**

| TQC | ICSD | NoE | SF | | | | | |
|---|---|---|---|---|---|---|---|---|
| | 6 2 | 3+4 4 | 3+4 4 | 3+4 4 | 3+4 4 | 3+4 4 | 3+4 4 | 6 2 |
| | 5 2 | 3+4 4 | 3+4 4 | 3+4 4 | 3+4 4 | 3+4 4 | 3+4 4 | 5 2 |
| | 5 2 | 3+4 4 | 3+4 4 | 3+4 4 | 3+4 4 | 3+4 4 | 3+4 4 | 6 2 |
| | 5 2 | 3+4 4 | 3+4 4 | 3+4 4 | 3+4 4 | 3+4 4 | 3+4 4 | 6 2 |
| | 6 2 | 3+4 4 | 3+4 4 | 3+4 4 | 3+4 4 | 3+4 4 | 3+4 4 | 5 2 |
| | 6 2 | 3+4 4 | 3+4 4 | 3+4 4 | 3+4 4 | 3+4 4 | 3+4 4 | 5 2 |
| | 5 2 | 3+4 4 | 3+4 4 | 3+4 4 | 3+4 4 | 3+4 4 | 3+4 4 | 5 2 |
| | 6 2 | 3+4 4 | 3+4 4 | 3+4 4 | 3+4 4 | 3+4 4 | 3+4 4 | 6 2 |
| | 5 2 | 3+4 4 | 3+4 4 | 3+4 4 | 3+4 4 | 3+4 4 | 3+4 4 | 5 2 |
| | 6 2 | 3+4 4 | 3+4 4 | 3+4 4 | 3+4 4 | 3+4 4 | 3+4 4 | 6 2 |
| | 5 2 | 3+4 4 | 3+4 4 | 3+4 4 | 3+4 4 | 3+4 4 | 3+4 4 | 5 2 |
| | 6 2 | 3+4 4 | 3+4 4 | 3+4 4 | 3+4 4 | 3+4 4 | 3+4 4 | 6 2 |
| | 5 2 | 3+4 4 | 3+4 4 | 3+4 4 | 3+4 4 | 3+4 4 | 3+4 4 | 6 2 |
| | 6 2 | 3+4 4 | 3+4 4 | 3+4 4 | 3+4 4 | 3+4 4 | 3+4 4 | 5 2 |
| | 5 2 | | | | | | | 6 2 |
| | 5 2 | | | | | | | 5 2 |
| | 6 2 | | | | | | | 5 2 |
| | 5 2 | | | | | | | 5 2 |
| | 5 2 | | | | | | | 5 2 |
| | 6 2 | | | | | | | 5 2 |
| | 6 2 | | | | | | | 6 2 |
| | 6 2 | | | | | | | 6 2 |
| | 6 2 | | | | | | | 6 2 |
| | 5 2 | | | | | | | 6 2 |
| | 5 2 | | | | | | | 6 2 |
| | 5 2 | | | | | | | 6 2 |
| | 5 2 | | | | | | | 6 2 |
| | 6 2 | | | | | | | 5 2 |
| | 5 2 | | | | | | | 6 2 |
| | 5 2 | | | | | | | 6 2 |
| | 6 2 | | | | | | | 5 2 |
| | 6 2 | | | | | | | 5 2 |
| | 6 2 | | | | | | | 6 2 |
| | 5 2 | | | | | | | 5 2 |
| | 6 2 | | | | | | | 6 2 |
| | 5 2 | | | | | | | 5 2 |
| 59.1.8 | 195025 | 88 | Fe2O1S2Sr1 | | | | | |
| | -GM | -R | -S | -T | -U | -X | -Y | -Z |
| | 5 2 | 3+4 4 | 3+4 4 | 3+4 4 | 3+4 4 | 3+4 4 | 3+4 4 | 5 2 |
| | 6 2 | 3+4 4 | 3+4 4 | 3+4 4 | 3+4 4 | 3+4 4 | 3+4 4 | 6 2 |
| | 5 2 | 3+4 4 | 3+4 4 | 3+4 4 | 3+4 4 | 3+4 4 | 3+4 4 | 6 2 |
| | 6 2 | 3+4 4 | 3+4 4 | 3+4 4 | 3+4 4 | 3+4 4 | 3+4 4 | 5 2 |
| | 5 2 | 3+4 4 | 3+4 4 | 3+4 4 | 3+4 4 | 3+4 4 | 3+4 4 | 5 2 |
| | 6 2 | 3+4 4 | 3+4 4 | 3+4 4 | 3+4 4 | 3+4 4 | 3+4 4 | 6 2 |
| | 5 2 | 3+4 4 | 3+4 4 | 3+4 4 | 3+4 4 | 3+4 4 | 3+4 4 | 5 2 |
| | 6 2 | 3+4 4 | 3+4 4 | 3+4 4 | 3+4 4 | 3+4 4 | 3+4 4 | 6 2 |
| | 6 2 | 3+4 4 | 3+4 4 | 3+4 4 | 3+4 4 | 3+4 4 | 3+4 4 | 6 2 |
| | 5 2 | 3+4 4 | 3+4 4 | 3+4 4 | 3+4 4 | 3+4 4 | 3+4 4 | 5 2 |
| | 5 2 | 3+4 4 | 3+4 4 | 3+4 4 | 3+4 4 | 3+4 4 | 3+4 4 | 6 2 |
| | 5 2 | 3+4 4 | 3+4 4 | 3+4 4 | 3+4 4 | 3+4 4 | 3+4 4 | 6 2 |
| | 6 2 | 3+4 4 | 3+4 4 | 3+4 4 | 3+4 4 | 3+4 4 | 3+4 4 | 5 2 |
| | 6 2 | 3+4 4 | 3+4 4 | 3+4 4 | 3+4 4 | 3+4 4 | 3+4 4 | 5 2 |
| | 5 2 | 3+4 4 | 3+4 4 | 3+4 4 | 3+4 4 | 3+4 4 | 3+4 4 | 6 2 |
| | 5 2 | 3+4 4 | 3+4 4 | 3+4 4 | 3+4 4 | 3+4 4 | 3+4 4 | 5 2 |
| | 6 2 | 3+4 4 | 3+4 4 | 3+4 4 | 3+4 4 | 3+4 4 | 3+4 4 | 5 2 |
| | 6 2 | 3+4 4 | 3+4 4 | 3+4 4 | 3+4 4 | 3+4 4 | 3+4 4 | 6 2 |
| | 5 2 | 3+4 4 | 3+4 4 | 3+4 4 | 3+4 4 | 3+4 4 | 3+4 4 | 5 2 |
| | 6 2 | 3+4 4 | 3+4 4 | 3+4 4 | 3+4 4 | 3+4 4 | 3+4 4 | 6 2 |
| | 5 2 | 3+4 4 | 3+4 4 | 3+4 4 | 3+4 4 | 3+4 4 | 3+4 4 | 6 2 |
| | 6 2 | 3+4 4 | 3+4 4 | 3+4 4 | 3+4 4 | 3+4 4 | 3+4 4 | 5 2 |
| | 5 2 | | | | | | | 6 2 |

Continued on next column

**TABLE XXXIV– continued**

| TQC | ICSD | NoE | SF | | | | | |
|---|---|---|---|---|---|---|---|---|
| | 5 2 | | | | | | | 5 2 |
| | 6 2 | | | | | | | 5 2 |
| | 5 2 | | | | | | | 5 2 |
| | 5 2 | | | | | | | 5 2 |
| | 6 2 | | | | | | | 5 2 |
| | 6 2 | | | | | | | 6 2 |
| | 6 2 | | | | | | | 6 2 |
| | 6 2 | | | | | | | 6 2 |
| | 5 2 | | | | | | | 6 2 |
| | 5 2 | | | | | | | 6 2 |
| | 5 2 | | | | | | | 6 2 |
| | 5 2 | | | | | | | 6 2 |
| | 6 2 | | | | | | | 5 2 |
| | 5 2 | | | | | | | 6 2 |
| | 5 2 | | | | | | | 5 2 |
| | 6 2 | | | | | | | 6 2 |
| | 6 2 | | | | | | | 5 2 |
| | 6 2 | | | | | | | 6 2 |
| | 5 2 | | | | | | | 5 2 |
| | 6 2 | | | | | | | 6 2 |
| | 5 2 | | | | | | | 5 2 |
| 59.1.9 | 88203 | 128 | Er2Ge5 | | | | | |
| | -GM | -R | -S | -T | -U | -X | -Y | -Z |
| | 5 2 | 3+4 4 | 3+4 4 | 3+4 4 | 3+4 4 | 3+4 4 | 3+4 4 | 5 2 |
| | 6 2 | 3+4 4 | 3+4 4 | 3+4 4 | 3+4 4 | 3+4 4 | 3+4 4 | 6 2 |
| | 5 2 | 3+4 4 | 3+4 4 | 3+4 4 | 3+4 4 | 3+4 4 | 3+4 4 | 5 2 |
| | 6 2 | 3+4 4 | 3+4 4 | 3+4 4 | 3+4 4 | 3+4 4 | 3+4 4 | 6 2 |
| | 5 2 | 3+4 4 | 3+4 4 | 3+4 4 | 3+4 4 | 3+4 4 | 3+4 4 | 5 2 |
| | 6 2 | 3+4 4 | 3+4 4 | 3+4 4 | 3+4 4 | 3+4 4 | 3+4 4 | 6 2 |
| | 5 2 | 3+4 4 | 3+4 4 | 3+4 4 | 3+4 4 | 3+4 4 | 3+4 4 | 6 2 |
| | 6 2 | 3+4 4 | 3+4 4 | 3+4 4 | 3+4 4 | 3+4 4 | 3+4 4 | 5 2 |
| | 5 2 | 3+4 4 | 3+4 4 | 3+4 4 | 3+4 4 | 3+4 4 | 3+4 4 | 6 2 |
| | 6 2 | 3+4 4 | 3+4 4 | 3+4 4 | 3+4 4 | 3+4 4 | 3+4 4 | 5 2 |
| | 5 2 | 3+4 4 | 3+4 4 | 3+4 4 | 3+4 4 | 3+4 4 | 3+4 4 | 6 2 |
| | 6 2 | 3+4 4 | 3+4 4 | 3+4 4 | 3+4 4 | 3+4 4 | 3+4 4 | 5 2 |
| | 5 2 | 3+4 4 | 3+4 4 | 3+4 4 | 3+4 4 | 3+4 4 | 3+4 4 | 6 2 |
| | 6 2 | 3+4 4 | 3+4 4 | 3+4 4 | 3+4 4 | 3+4 4 | 3+4 4 | 5 2 |
| | 5 2 | 3+4 4 | 3+4 4 | 3+4 4 | 3+4 4 | 3+4 4 | 3+4 4 | 6 2 |
| | 6 2 | 3+4 4 | 3+4 4 | 3+4 4 | 3+4 4 | 3+4 4 | 3+4 4 | 5 2 |
| | 5 2 | 3+4 4 | 3+4 4 | 3+4 4 | 3+4 4 | 3+4 4 | 3+4 4 | 5 2 |
| | 5 2 | 3+4 4 | 3+4 4 | 3+4 4 | 3+4 4 | 3+4 4 | 3+4 4 | 6 2 |
| | 6 2 | 3+4 4 | 3+4 4 | 3+4 4 | 3+4 4 | 3+4 4 | 3+4 4 | 5 2 |
| | 5 2 | 3+4 4 | 3+4 4 | 3+4 4 | 3+4 4 | 3+4 4 | 3+4 4 | 6 2 |
| | 6 2 | 3+4 4 | 3+4 4 | 3+4 4 | 3+4 4 | 3+4 4 | 3+4 4 | 6 2 |
| | 6 2 | 3+4 4 | 3+4 4 | 3+4 4 | 3+4 4 | 3+4 4 | 3+4 4 | 5 2 |
| | 5 2 | 3+4 4 | 3+4 4 | 3+4 4 | 3+4 4 | 3+4 4 | 3+4 4 | 6 2 |
| | 6 2 | 3+4 4 | 3+4 4 | 3+4 4 | 3+4 4 | 3+4 4 | 3+4 4 | 6 2 |
| | 6 2 | 3+4 4 | 3+4 4 | 3+4 4 | 3+4 4 | 3+4 4 | 3+4 4 | 5 2 |
| | 5 2 | 3+4 4 | 3+4 4 | 3+4 4 | 3+4 4 | 3+4 4 | 3+4 4 | 5 2 |
| | 6 2 | 3+4 4 | 3+4 4 | 3+4 4 | 3+4 4 | 3+4 4 | 3+4 4 | 5 2 |
| | 5 2 | 3+4 4 | 3+4 4 | 3+4 4 | 3+4 4 | 3+4 4 | 3+4 4 | 5 2 |
| | 5 2 | 3+4 4 | 3+4 4 | 3+4 4 | 3+4 4 | 3+4 4 | 3+4 4 | 5 2 |
| | 5 2 | 3+4 4 | 3+4 4 | 3+4 4 | 3+4 4 | 3+4 4 | 3+4 4 | 5 2 |
| | 5 2 | 3+4 4 | 3+4 4 | 3+4 4 | 3+4 4 | 3+4 4 | 3+4 4 | 6 2 |
| | 6 2 | 3+4 4 | 3+4 4 | 3+4 4 | 3+4 4 | 3+4 4 | 3+4 4 | 6 2 |
| | 5 2 | | | | | | | 5 2 |
| | 5 2 | | | | | | | 6 2 |
| | 6 2 | | | | | | | 6 2 |
| | 5 2 | | | | | | | 6 2 |
| | 6 2 | | | | | | | 5 2 |
| | 6 2 | | | | | | | 6 2 |

Continued on next column

**TABLE XXXIV– continued**

| TQC | ICSD | NoE | SF | | | | | |
|---|---|---|---|---|---|---|---|---|
| | 5 2 | | | | | | | 6 2 |
| | 6 2 | | | | | | | 5 2 |
| | 5 2 | | | | | | | 6 2 |
| | 6 2 | | | | | | | 5 2 |
| | 5 2 | | | | | | | 5 2 |
| | 5 2 | | | | | | | 5 2 |
| | 6 2 | | | | | | | 6 2 |
| | 6 2 | | | | | | | 6 2 |
| | 5 2 | | | | | | | 5 2 |
| | 5 2 | | | | | | | 6 2 |
| | 6 2 | | | | | | | 5 2 |
| | 5 2 | | | | | | | 6 2 |
| | 6 2 | | | | | | | 6 2 |
| | 6 2 | | | | | | | 6 2 |
| | 5 2 | | | | | | | 6 2 |
| | 6 2 | | | | | | | 5 2 |
| | 5 2 | | | | | | | 6 2 |
| | 6 2 | | | | | | | 5 2 |
| | 6 2 | | | | | | | 5 2 |
| | 5 2 | | | | | | | 6 2 |
| | 5 2 | | | | | | | 6 2 |
| | 6 2 | | | | | | | 5 2 |
| | 5 2 | | | | | | | 5 2 |
| | 6 2 | | | | | | | 5 2 |
| | 5 2 | | | | | | | 6 2 |
| | 6 2 | | | | | | | 6 2 |
| 59.1.10 | 160543 | 120 | Dy2Sn5 | | | | | |
| | -GM | -R | -S | -T | -U | -X | -Y | -Z |
| | 5 2 | 3+4 4 | 3+4 4 | 3+4 4 | 3+4 4 | 3+4 4 | 3+4 4 | 5 2 |
| | 6 2 | 3+4 4 | 3+4 4 | 3+4 4 | 3+4 4 | 3+4 4 | 3+4 4 | 6 2 |
| | 5 2 | 3+4 4 | 3+4 4 | 3+4 4 | 3+4 4 | 3+4 4 | 3+4 4 | 5 2 |
| | 6 2 | 3+4 4 | 3+4 4 | 3+4 4 | 3+4 4 | 3+4 4 | 3+4 4 | 6 2 |
| | 5 2 | 3+4 4 | 3+4 4 | 3+4 4 | 3+4 4 | 3+4 4 | 3+4 4 | 5 2 |
| | 6 2 | 3+4 4 | 3+4 4 | 3+4 4 | 3+4 4 | 3+4 4 | 3+4 4 | 6 2 |
| | 5 2 | 3+4 4 | 3+4 4 | 3+4 4 | 3+4 4 | 3+4 4 | 3+4 4 | 5 2 |
| | 6 2 | 3+4 4 | 3+4 4 | 3+4 4 | 3+4 4 | 3+4 4 | 3+4 4 | 6 2 |
| | 5 2 | 3+4 4 | 3+4 4 | 3+4 4 | 3+4 4 | 3+4 4 | 3+4 4 | 5 2 |
| | 6 2 | 3+4 4 | 3+4 4 | 3+4 4 | 3+4 4 | 3+4 4 | 3+4 4 | 6 2 |
| | 5 2 | 3+4 4 | 3+4 4 | 3+4 4 | 3+4 4 | 3+4 4 | 3+4 4 | 5 2 |
| | 6 2 | 3+4 4 | 3+4 4 | 3+4 4 | 3+4 4 | 3+4 4 | 3+4 4 | 6 2 |
| | 5 2 | 3+4 4 | 3+4 4 | 3+4 4 | 3+4 4 | 3+4 4 | 3+4 4 | 5 2 |
| | 6 2 | 3+4 4 | 3+4 4 | 3+4 4 | 3+4 4 | 3+4 4 | 3+4 4 | 6 2 |
| | 5 2 | 3+4 4 | 3+4 4 | 3+4 4 | 3+4 4 | 3+4 4 | 3+4 4 | 5 2 |
| | 6 2 | 3+4 4 | 3+4 4 | 3+4 4 | 3+4 4 | 3+4 4 | 3+4 4 | 6 2 |
| | 5 2 | 3+4 4 | 3+4 4 | 3+4 4 | 3+4 4 | 3+4 4 | 3+4 4 | 6 2 |
| | 5 2 | 3+4 4 | 3+4 4 | 3+4 4 | 3+4 4 | 3+4 4 | 3+4 4 | 5 2 |
| | 6 2 | 3+4 4 | 3+4 4 | 3+4 4 | 3+4 4 | 3+4 4 | 3+4 4 | 5 2 |
| | 5 2 | 3+4 4 | 3+4 4 | 3+4 4 | 3+4 4 | 3+4 4 | 3+4 4 | 6 2 |
| | 6 2 | 3+4 4 | 3+4 4 | 3+4 4 | 3+4 4 | 3+4 4 | 3+4 4 | 5 2 |
| | 6 2 | 3+4 4 | 3+4 4 | 3+4 4 | 3+4 4 | 3+4 4 | 3+4 4 | 6 2 |
| | 5 2 | 3+4 4 | 3+4 4 | 3+4 4 | 3+4 4 | 3+4 4 | 3+4 4 | 5 2 |
| | 6 2 | 3+4 4 | 3+4 4 | 3+4 4 | 3+4 4 | 3+4 4 | 3+4 4 | 5 2 |
| | 6 2 | 3+4 4 | 3+4 4 | 3+4 4 | 3+4 4 | 3+4 4 | 3+4 4 | 6 2 |
| | 5 2 | 3+4 4 | 3+4 4 | 3+4 4 | 3+4 4 | 3+4 4 | 3+4 4 | 6 2 |
| | 5 2 | 3+4 4 | 3+4 4 | 3+4 4 | 3+4 4 | 3+4 4 | 3+4 4 | 6 2 |
| | 6 2 | 3+4 4 | 3+4 4 | 3+4 4 | 3+4 4 | 3+4 4 | 3+4 4 | 6 2 |
| | 5 2 | 3+4 4 | 3+4 4 | 3+4 4 | 3+4 4 | 3+4 4 | 3+4 4 | 5 2 |
| | 5 2 | 3+4 4 | 3+4 4 | 3+4 4 | 3+4 4 | 3+4 4 | 3+4 4 | 6 2 |
| | 5 2 | | | | | | | 6 2 |
| | 5 2 | | | | | | | 5 2 |
| | 6 2 | | | | | | | 5 2 |

Continued on next column

**TABLE XXXIV– continued**

| TQC | ICSD | NoE | SF | | | | | |
|---|---|---|---|---|---|---|---|---|
| | 6 2 | | | | | | | 5 2 |
| | 5 2 | | | | | | | 6 2 |
| | 6 2 | | | | | | | 5 2 |
| | 6 2 | | | | | | | 6 2 |
| | 5 2 | | | | | | | 5 2 |
| | 5 2 | | | | | | | 5 2 |
| | 6 2 | | | | | | | 5 2 |
| | 6 2 | | | | | | | 5 2 |
| | 5 2 | | | | | | | 6 2 |
| | 6 2 | | | | | | | 6 2 |
| | 5 2 | | | | | | | 6 2 |
| | 5 2 | | | | | | | 5 2 |
| | 6 2 | | | | | | | 6 2 |
| | 6 2 | | | | | | | 5 2 |
| | 5 2 | | | | | | | 6 2 |
| | 6 2 | | | | | | | 6 2 |
| | 5 2 | | | | | | | 5 2 |
| | 6 2 | | | | | | | 5 2 |
| | 5 2 | | | | | | | 5 2 |
| | 5 2 | | | | | | | 5 2 |
| | 6 2 | | | | | | | 6 2 |
| | 5 2 | | | | | | | 5 2 |
| | 6 2 | | | | | | | 6 2 |
| | 5 2 | | | | | | | 5 2 |
| | 6 2 | | | | | | | 5 2 |
| | 6 2 | | | | | | | 6 2 |
| | 5 2 | | | | | | | 6 2 |
| 59.1.11 | 72376 | 140 | Cr6P4U1 | | | | | |
| | -GM | -R | -S | -T | -U | -X | -Y | -Z |
| | 5 2 | 3+4 4 | 3+4 4 | 3+4 4 | 3+4 4 | 3+4 4 | 3+4 4 | 5 2 |
| | 6 2 | 3+4 4 | 3+4 4 | 3+4 4 | 3+4 4 | 3+4 4 | 3+4 4 | 6 2 |
| | 5 2 | 3+4 4 | 3+4 4 | 3+4 4 | 3+4 4 | 3+4 4 | 3+4 4 | 6 2 |
| | 6 2 | 3+4 4 | 3+4 4 | 3+4 4 | 3+4 4 | 3+4 4 | 3+4 4 | 5 2 |
| | 5 2 | 3+4 4 | 3+4 4 | 3+4 4 | 3+4 4 | 3+4 4 | 3+4 4 | 6 2 |
| | 6 2 | 3+4 4 | 3+4 4 | 3+4 4 | 3+4 4 | 3+4 4 | 3+4 4 | 5 2 |
| | 5 2 | 3+4 4 | 3+4 4 | 3+4 4 | 3+4 4 | 3+4 4 | 3+4 4 | 6 2 |
| | 6 2 | 3+4 4 | 3+4 4 | 3+4 4 | 3+4 4 | 3+4 4 | 3+4 4 | 5 2 |
| | 5 2 | 3+4 4 | 3+4 4 | 3+4 4 | 3+4 4 | 3+4 4 | 3+4 4 | 5 2 |
| | 6 2 | 3+4 4 | 3+4 4 | 3+4 4 | 3+4 4 | 3+4 4 | 3+4 4 | 6 2 |
| | 5 2 | 3+4 4 | 3+4 4 | 3+4 4 | 3+4 4 | 3+4 4 | 3+4 4 | 6 2 |
| | 6 2 | 3+4 4 | 3+4 4 | 3+4 4 | 3+4 4 | 3+4 4 | 3+4 4 | 5 2 |
| | 6 2 | 3+4 4 | 3+4 4 | 3+4 4 | 3+4 4 | 3+4 4 | 3+4 4 | 5 2 |
| | 5 2 | 3+4 4 | 3+4 4 | 3+4 4 | 3+4 4 | 3+4 4 | 3+4 4 | 6 2 |
| | 5 2 | 3+4 4 | 3+4 4 | 3+4 4 | 3+4 4 | 3+4 4 | 3+4 4 | 5 2 |
| | 6 2 | 3+4 4 | 3+4 4 | 3+4 4 | 3+4 4 | 3+4 4 | 3+4 4 | 6 2 |
| | 6 2 | 3+4 4 | 3+4 4 | 3+4 4 | 3+4 4 | 3+4 4 | 3+4 4 | 6 2 |
| | 5 2 | 3+4 4 | 3+4 4 | 3+4 4 | 3+4 4 | 3+4 4 | 3+4 4 | 5 2 |
| | 5 2 | 3+4 4 | 3+4 4 | 3+4 4 | 3+4 4 | 3+4 4 | 3+4 4 | 6 2 |
| | 5 2 | 3+4 4 | 3+4 4 | 3+4 4 | 3+4 4 | 3+4 4 | 3+4 4 | 5 2 |
| | 6 2 | 3+4 4 | 3+4 4 | 3+4 4 | 3+4 4 | 3+4 4 | 3+4 4 | 6 2 |
| | 5 2 | 3+4 4 | 3+4 4 | 3+4 4 | 3+4 4 | 3+4 4 | 3+4 4 | 5 2 |
| | 6 2 | 3+4 4 | 3+4 4 | 3+4 4 | 3+4 4 | 3+4 4 | 3+4 4 | 5 2 |
| | 6 2 | 3+4 4 | 3+4 4 | 3+4 4 | 3+4 4 | 3+4 4 | 3+4 4 | 5 2 |
| | 6 2 | 3+4 4 | 3+4 4 | 3+4 4 | 3+4 4 | 3+4 4 | 3+4 4 | 6 2 |
| | 5 2 | 3+4 4 | 3+4 4 | 3+4 4 | 3+4 4 | 3+4 4 | 3+4 4 | 5 2 |
| | 6 2 | 3+4 4 | 3+4 4 | 3+4 4 | 3+4 4 | 3+4 4 | 3+4 4 | 6 2 |
| | 6 2 | 3+4 4 | 3+4 4 | 3+4 4 | 3+4 4 | 3+4 4 | 3+4 4 | 6 2 |
| | 5 2 | 3+4 4 | 3+4 4 | 3+4 4 | 3+4 4 | 3+4 4 | 3+4 4 | 5 2 |
| | 5 2 | 3+4 4 | 3+4 4 | 3+4 4 | 3+4 4 | 3+4 4 | 3+4 4 | 5 2 |
| | 6 2 | 3+4 4 | 3+4 4 | 3+4 4 | 3+4 4 | 3+4 4 | 3+4 4 | 6 2 |
| | 5 2 | 3+4 4 | 3+4 4 | 3+4 4 | 3+4 4 | 3+4 4 | 3+4 4 | 5 2 |

Continued on next column

**TABLE XXXIV– continued**

| TQC | ICSD | NoE | SF | | | | | |
|---|---|---|---|---|---|---|---|---|
| | 6 2 | 3+4 4 | 3+4 4 | 3+4 4 | 3+4 4 | 3+4 4 | 3+4 4 | 6 2 |
| | 5 2 | 3+4 4 | 3+4 4 | 3+4 4 | 3+4 4 | 3+4 4 | 3+4 4 | 6 2 |
| | 5 2 | 3+4 4 | 3+4 4 | 3+4 4 | 3+4 4 | 3+4 4 | 3+4 4 | 5 2 |
| | 6 2 | | | | | | | 6 2 |
| | 6 2 | | | | | | | 6 2 |
| | 6 2 | | | | | | | 5 2 |
| | 6 2 | | | | | | | 6 2 |
| | 5 2 | | | | | | | 5 2 |
| | 6 2 | | | | | | | 6 2 |
| | 5 2 | | | | | | | 5 2 |
| | 5 2 | | | | | | | 6 2 |
| | 5 2 | | | | | | | 6 2 |
| | 6 2 | | | | | | | 5 2 |
| | 5 2 | | | | | | | 6 2 |
| | 6 2 | | | | | | | 5 2 |
| | 6 2 | | | | | | | 6 2 |
| | 5 2 | | | | | | | 5 2 |
| | 6 2 | | | | | | | 6 2 |
| | 5 2 | | | | | | | 5 2 |
| | 6 2 | | | | | | | 5 2 |
| | 5 2 | | | | | | | 5 2 |
| | 5 2 | | | | | | | 5 2 |
| | 5 2 | | | | | | | 6 2 |
| | 5 2 | | | | | | | 6 2 |
| | 6 2 | | | | | | | 5 2 |
| | 5 2 | | | | | | | 6 2 |
| | 5 2 | | | | | | | 5 2 |
| | 6 2 | | | | | | | 5 2 |
| | 6 2 | | | | | | | 5 2 |
| | 6 2 | | | | | | | 6 2 |
| | 6 2 | | | | | | | 5 2 |
| | 5 2 | | | | | | | 5 2 |
| | 5 2 | | | | | | | 6 2 |
| | 6 2 | | | | | | | 6 2 |
| | 5 2 | | | | | | | 5 2 |
| | 5 2 | | | | | | | 6 2 |
| | 5 2 | | | | | | | 6 2 |
| | 6 2 | | | | | | | 5 2 |
| 59.1.12 | 30701 | 216 | Co8La1P5 | | | | | |
| | -GM | -R | -S | -T | -U | -X | -Y | -Z |
| | 5 2 | 3+4 4 | 3+4 4 | 3+4 4 | 3+4 4 | 3+4 4 | 3+4 4 | 5 2 |
| | 6 2 | 3+4 4 | 3+4 4 | 3+4 4 | 3+4 4 | 3+4 4 | 3+4 4 | 6 2 |
| | 5 2 | 3+4 4 | 3+4 4 | 3+4 4 | 3+4 4 | 3+4 4 | 3+4 4 | 5 2 |
| | 6 2 | 3+4 4 | 3+4 4 | 3+4 4 | 3+4 4 | 3+4 4 | 3+4 4 | 6 2 |
| | 5 2 | 3+4 4 | 3+4 4 | 3+4 4 | 3+4 4 | 3+4 4 | 3+4 4 | 5 2 |
| | 6 2 | 3+4 4 | 3+4 4 | 3+4 4 | 3+4 4 | 3+4 4 | 3+4 4 | 6 2 |
| | 5 2 | 3+4 4 | 3+4 4 | 3+4 4 | 3+4 4 | 3+4 4 | 3+4 4 | 5 2 |
| | 6 2 | 3+4 4 | 3+4 4 | 3+4 4 | 3+4 4 | 3+4 4 | 3+4 4 | 6 2 |
| | 5 2 | 3+4 4 | 3+4 4 | 3+4 4 | 3+4 4 | 3+4 4 | 3+4 4 | 5 2 |
| | 6 2 | 3+4 4 | 3+4 4 | 3+4 4 | 3+4 4 | 3+4 4 | 3+4 4 | 6 2 |
| | 5 2 | 3+4 4 | 3+4 4 | 3+4 4 | 3+4 4 | 3+4 4 | 3+4 4 | 6 2 |
| | 5 2 | 3+4 4 | 3+4 4 | 3+4 4 | 3+4 4 | 3+4 4 | 3+4 4 | 5 2 |
| | 6 2 | 3+4 4 | 3+4 4 | 3+4 4 | 3+4 4 | 3+4 4 | 3+4 4 | 6 2 |
| | 6 2 | 3+4 4 | 3+4 4 | 3+4 4 | 3+4 4 | 3+4 4 | 3+4 4 | 5 2 |
| | 5 2 | 3+4 4 | 3+4 4 | 3+4 4 | 3+4 4 | 3+4 4 | 3+4 4 | 5 2 |
| | 6 2 | 3+4 4 | 3+4 4 | 3+4 4 | 3+4 4 | 3+4 4 | 3+4 4 | 6 2 |
| | 5 2 | 3+4 4 | 3+4 4 | 3+4 4 | 3+4 4 | 3+4 4 | 3+4 4 | 5 2 |
| | 6 2 | 3+4 4 | 3+4 4 | 3+4 4 | 3+4 4 | 3+4 4 | 3+4 4 | 6 2 |
| | 5 2 | 3+4 4 | 3+4 4 | 3+4 4 | 3+4 4 | 3+4 4 | 3+4 4 | 5 2 |
| | 5 2 | 3+4 4 | 3+4 4 | 3+4 4 | 3+4 4 | 3+4 4 | 3+4 4 | 6 2 |
| | 6 2 | 3+4 4 | 3+4 4 | 3+4 4 | 3+4 4 | 3+4 4 | 3+4 4 | 5 2 |

Continued on next column

**TABLE XXXIV– continued**

| TQC | ICSD | NoE | SF | | | | | |
|---|---|---|---|---|---|---|---|---|
| | 6 2 | 3+4 4 | 3+4 4 | 3+4 4 | 3+4 4 | 3+4 4 | 3+4 4 | 6 2 |
| | 5 2 | 3+4 4 | 3+4 4 | 3+4 4 | 3+4 4 | 3+4 4 | 3+4 4 | 5 2 |
| | 6 2 | 3+4 4 | 3+4 4 | 3+4 4 | 3+4 4 | 3+4 4 | 3+4 4 | 6 2 |
| | 5 2 | 3+4 4 | 3+4 4 | 3+4 4 | 3+4 4 | 3+4 4 | 3+4 4 | 6 2 |
| | 5 2 | 3+4 4 | 3+4 4 | 3+4 4 | 3+4 4 | 3+4 4 | 3+4 4 | 6 2 |
| | 6 2 | 3+4 4 | 3+4 4 | 3+4 4 | 3+4 4 | 3+4 4 | 3+4 4 | 5 2 |
| | 6 2 | 3+4 4 | 3+4 4 | 3+4 4 | 3+4 4 | 3+4 4 | 3+4 4 | 6 2 |
| | 6 2 | 3+4 4 | 3+4 4 | 3+4 4 | 3+4 4 | 3+4 4 | 3+4 4 | 5 2 |
| | 6 2 | 3+4 4 | 3+4 4 | 3+4 4 | 3+4 4 | 3+4 4 | 3+4 4 | 5 2 |
| | 5 2 | 3+4 4 | 3+4 4 | 3+4 4 | 3+4 4 | 3+4 4 | 3+4 4 | 5 2 |
| | 5 2 | 3+4 4 | 3+4 4 | 3+4 4 | 3+4 4 | 3+4 4 | 3+4 4 | 6 2 |
| | 6 2 | 3+4 4 | 3+4 4 | 3+4 4 | 3+4 4 | 3+4 4 | 3+4 4 | 5 2 |
| | 6 2 | 3+4 4 | 3+4 4 | 3+4 4 | 3+4 4 | 3+4 4 | 3+4 4 | 6 2 |
| | 6 2 | 3+4 4 | 3+4 4 | 3+4 4 | 3+4 4 | 3+4 4 | 3+4 4 | 6 2 |
| | 6 2 | 3+4 4 | 3+4 4 | 3+4 4 | 3+4 4 | 3+4 4 | 3+4 4 | 5 2 |
| | 5 2 | 3+4 4 | 3+4 4 | 3+4 4 | 3+4 4 | 3+4 4 | 3+4 4 | 6 2 |
| | 6 2 | 3+4 4 | 3+4 4 | 3+4 4 | 3+4 4 | 3+4 4 | 3+4 4 | 5 2 |
| | 5 2 | 3+4 4 | 3+4 4 | 3+4 4 | 3+4 4 | 3+4 4 | 3+4 4 | 6 2 |
| | 5 2 | 3+4 4 | 3+4 4 | 3+4 4 | 3+4 4 | 3+4 4 | 3+4 4 | 5 2 |
| | 6 2 | 3+4 4 | 3+4 4 | 3+4 4 | 3+4 4 | 3+4 4 | 3+4 4 | 5 2 |
| | 6 2 | 3+4 4 | 3+4 4 | 3+4 4 | 3+4 4 | 3+4 4 | 3+4 4 | 5 2 |
| | 5 2 | 3+4 4 | 3+4 4 | 3+4 4 | 3+4 4 | 3+4 4 | 3+4 4 | 5 2 |
| | 5 2 | 3+4 4 | 3+4 4 | 3+4 4 | 3+4 4 | 3+4 4 | 3+4 4 | 6 2 |
| | 5 2 | 3+4 4 | 3+4 4 | 3+4 4 | 3+4 4 | 3+4 4 | 3+4 4 | 5 2 |
| | 6 2 | 3+4 4 | 3+4 4 | 3+4 4 | 3+4 4 | 3+4 4 | 3+4 4 | 6 2 |
| | 5 2 | 3+4 4 | 3+4 4 | 3+4 4 | 3+4 4 | 3+4 4 | 3+4 4 | 6 2 |
| | 5 2 | 3+4 4 | 3+4 4 | 3+4 4 | 3+4 4 | 3+4 4 | 3+4 4 | 6 2 |
| | 6 2 | 3+4 4 | 3+4 4 | 3+4 4 | 3+4 4 | 3+4 4 | 3+4 4 | 5 2 |
| | 5 2 | 3+4 4 | 3+4 4 | 3+4 4 | 3+4 4 | 3+4 4 | 3+4 4 | 6 2 |
| | 5 2 | 3+4 4 | 3+4 4 | 3+4 4 | 3+4 4 | 3+4 4 | 3+4 4 | 6 2 |
| | 6 2 | 3+4 4 | 3+4 4 | 3+4 4 | 3+4 4 | 3+4 4 | 3+4 4 | 5 2 |
| | 5 2 | 3+4 4 | 3+4 4 | 3+4 4 | 3+4 4 | 3+4 4 | 3+4 4 | 6 2 |
| | 6 2 | 3+4 4 | 3+4 4 | 3+4 4 | 3+4 4 | 3+4 4 | 3+4 4 | 6 2 |
| | 6 2 | | | | | | | 5 2 |
| | 5 2 | | | | | | | 6 2 |
| | 5 2 | | | | | | | 5 2 |
| | 5 2 | | | | | | | 5 2 |
| | 6 2 | | | | | | | 5 2 |
| | 5 2 | | | | | | | 5 2 |
| | 6 2 | | | | | | | 6 2 |
| | 5 2 | | | | | | | 6 2 |
| | 6 2 | | | | | | | 5 2 |
| | 5 2 | | | | | | | 5 2 |
| | 5 2 | | | | | | | 6 2 |
| | 6 2 | | | | | | | 6 2 |
| | 5 2 | | | | | | | 5 2 |
| | 6 2 | | | | | | | 6 2 |
| | 6 2 | | | | | | | 5 2 |
| | 5 2 | | | | | | | 6 2 |
| | 5 2 | | | | | | | 6 2 |
| | 6 2 | | | | | | | 6 2 |
| | 5 2 | | | | | | | 5 2 |
| | 6 2 | | | | | | | 6 2 |
| | 6 2 | | | | | | | 6 2 |
| | 5 2 | | | | | | | 5 2 |
| | 6 2 | | | | | | | 5 2 |
| | 6 2 | | | | | | | 6 2 |
| | 5 2 | | | | | | | 6 2 |
| | 6 2 | | | | | | | 6 2 |
| | 6 2 | | | | | | | 5 2 |
| | 5 2 | | | | | | | 5 2 |
| | 5 2 | | | | | | | 5 2 |
| | 6 2 | | | | | | | 6 2 |
| | 5 2 | | | | | | | 5 2 |
| | 5 2 | | | | | | | 6 2 |
| | 5 2 | | | | | | | 6 2 |
| | 5 2 | | | | | | | 5 2 |
| | 6 2 | | | | | | | 5 2 |
| | 5 2 | | | | | | | 5 2 |
| | 6 2 | | | | | | | 5 2 |
| | 5 2 | | | | | | | 6 2 |
| | 5 2 | | | | | | | 6 2 |
| | 6 2 | | | | | | | 5 2 |
| | 5 2 | | | | | | | 6 2 |
| | 6 2 | | | | | | | 5 2 |
| | 5 2 | | | | | | | 6 2 |
| | 6 2 | | | | | | | 6 2 |
| | 5 2 | | | | | | | 5 2 |
| | 6 2 | | | | | | | 5 2 |
| | 5 2 | | | | | | | 5 2 |
| | 5 2 | | | | | | | 6 2 |
| | 6 2 | | | | | | | 6 2 |
| | 6 2 | | | | | | | 5 2 |
| | 6 2 | | | | | | | 5 2 |
| | 5 2 | | | | | | | 5 2 |
| | 6 2 | | | | | | | 6 2 |
| | 5 2 | | | | | | | 6 2 |
| 59.1.13 | 410358 | 208 | Ce1Ge1Pt1 | | | | | |
| | -GM | -R | -S | -T | -U | -X | -Y | -Z |
| | 5 2 | 3+4 4 | 3+4 4 | 3+4 4 | 3+4 4 | 3+4 4 | 3+4 4 | 6 2 |
| | 6 2 | 3+4 4 | 3+4 4 | 3+4 4 | 3+4 4 | 3+4 4 | 3+4 4 | 5 2 |
| | 5 2 | 3+4 4 | 3+4 4 | 3+4 4 | 3+4 4 | 3+4 4 | 3+4 4 | 5 2 |
| | 6 2 | 3+4 4 | 3+4 4 | 3+4 4 | 3+4 4 | 3+4 4 | 3+4 4 | 5 2 |
| | 5 2 | 3+4 4 | 3+4 4 | 3+4 4 | 3+4 4 | 3+4 4 | 3+4 4 | 6 2 |
| | 5 2 | 3+4 4 | 3+4 4 | 3+4 4 | 3+4 4 | 3+4 4 | 3+4 4 | 6 2 |
| | 6 2 | 3+4 4 | 3+4 4 | 3+4 4 | 3+4 4 | 3+4 4 | 3+4 4 | 5 2 |
| | 6 2 | 3+4 4 | 3+4 4 | 3+4 4 | 3+4 4 | 3+4 4 | 3+4 4 | 6 2 |
| | 5 2 | 3+4 4 | 3+4 4 | 3+4 4 | 3+4 4 | 3+4 4 | 3+4 4 | 5 2 |
| | 6 2 | 3+4 4 | 3+4 4 | 3+4 4 | 3+4 4 | 3+4 4 | 3+4 4 | 6 2 |
| | 6 2 | 3+4 4 | 3+4 4 | 3+4 4 | 3+4 4 | 3+4 4 | 3+4 4 | 5 2 |
| | 5 2 | 3+4 4 | 3+4 4 | 3+4 4 | 3+4 4 | 3+4 4 | 3+4 4 | 5 2 |
| | 5 2 | 3+4 4 | 3+4 4 | 3+4 4 | 3+4 4 | 3+4 4 | 3+4 4 | 6 2 |
| | 6 2 | 3+4 4 | 3+4 4 | 3+4 4 | 3+4 4 | 3+4 4 | 3+4 4 | 6 2 |
| | 5 2 | 3+4 4 | 3+4 4 | 3+4 4 | 3+4 4 | 3+4 4 | 3+4 4 | 5 2 |
| | 6 2 | 3+4 4 | 3+4 4 | 3+4 4 | 3+4 4 | 3+4 4 | 3+4 4 | 6 2 |
| | 5 2 | 3+4 4 | 3+4 4 | 3+4 4 | 3+4 4 | 3+4 4 | 3+4 4 | 5 2 |
| | 6 2 | 3+4 4 | 3+4 4 | 3+4 4 | 3+4 4 | 3+4 4 | 3+4 4 | 6 2 |
| | 5 2 | 3+4 4 | 3+4 4 | 3+4 4 | 3+4 4 | 3+4 4 | 3+4 4 | 5 2 |
| | 6 2 | 3+4 4 | 3+4 4 | 3+4 4 | 3+4 4 | 3+4 4 | 3+4 4 | 5 2 |
| | 5 2 | 3+4 4 | 3+4 4 | 3+4 4 | 3+4 4 | 3+4 4 | 3+4 4 | 6 2 |
| | 6 2 | 3+4 4 | 3+4 4 | 3+4 4 | 3+4 4 | 3+4 4 | 3+4 4 | 6 2 |
| | 5 2 | 3+4 4 | 3+4 4 | 3+4 4 | 3+4 4 | 3+4 4 | 3+4 4 | 5 2 |
| | 5 2 | 3+4 4 | 3+4 4 | 3+4 4 | 3+4 4 | 3+4 4 | 3+4 4 | 6 2 |
| | 5 2 | 3+4 4 | 3+4 4 | 3+4 4 | 3+4 4 | 3+4 4 | 3+4 4 | 5 2 |
| | 6 2 | 3+4 4 | 3+4 4 | 3+4 4 | 3+4 4 | 3+4 4 | 3+4 4 | 5 2 |
| | 6 2 | 3+4 4 | 3+4 4 | 3+4 4 | 3+4 4 | 3+4 4 | 3+4 4 | 6 2 |
| | 6 2 | 3+4 4 | 3+4 4 | 3+4 4 | 3+4 4 | 3+4 4 | 3+4 4 | 5 2 |
| | 5 2 | 3+4 4 | 3+4 4 | 3+4 4 | 3+4 4 | 3+4 4 | 3+4 4 | 6 2 |
| | 5 2 | 3+4 4 | 3+4 4 | 3+4 4 | 3+4 4 | 3+4 4 | 3+4 4 | 6 2 |
| | 6 2 | 3+4 4 | 3+4 4 | 3+4 4 | 3+4 4 | 3+4 4 | 3+4 4 | 5 2 |
| | 6 2 | 3+4 4 | 3+4 4 | 3+4 4 | 3+4 4 | 3+4 4 | 3+4 4 | 6 2 |
| | 5 2 | 3+4 4 | 3+4 4 | 3+4 4 | 3+4 4 | 3+4 4 | 3+4 4 | 5 2 |

Continued on next column

**TABLE XXXIV– continued**

| TQC | ICSD | NoE | SF | | | | | |
|---|---|---|---|---|---|---|---|---|
| | 6 2 | 3+4 4 | 3+4 4 | 3+4 4 | 3+4 4 | 3+4 4 | 3+4 4 | 6 2 |
| | 5 2 | 3+4 4 | 3+4 4 | 3+4 4 | 3+4 4 | 3+4 4 | 3+4 4 | 6 2 |
| | 5 2 | 3+4 4 | 3+4 4 | 3+4 4 | 3+4 4 | 3+4 4 | 3+4 4 | 6 2 |
| | 6 2 | 3+4 4 | 3+4 4 | 3+4 4 | 3+4 4 | 3+4 4 | 3+4 4 | 5 2 |
| | 6 2 | 3+4 4 | 3+4 4 | 3+4 4 | 3+4 4 | 3+4 4 | 3+4 4 | 5 2 |
| | 5 2 | 3+4 4 | 3+4 4 | 3+4 4 | 3+4 4 | 3+4 4 | 3+4 4 | 5 2 |
| | 6 2 | 3+4 4 | 3+4 4 | 3+4 4 | 3+4 4 | 3+4 4 | 3+4 4 | 6 2 |
| | 6 2 | 3+4 4 | 3+4 4 | 3+4 4 | 3+4 4 | 3+4 4 | 3+4 4 | 6 2 |
| | 5 2 | 3+4 4 | 3+4 4 | 3+4 4 | 3+4 4 | 3+4 4 | 3+4 4 | 5 2 |
| | 5 2 | 3+4 4 | 3+4 4 | 3+4 4 | 3+4 4 | 3+4 4 | 3+4 4 | 6 2 |
| | 5 2 | 3+4 4 | 3+4 4 | 3+4 4 | 3+4 4 | 3+4 4 | 3+4 4 | 5 2 |
| | 6 2 | 3+4 4 | 3+4 4 | 3+4 4 | 3+4 4 | 3+4 4 | 3+4 4 | 6 2 |
| | 6 2 | 3+4 4 | 3+4 4 | 3+4 4 | 3+4 4 | 3+4 4 | 3+4 4 | 6 2 |
| | 6 2 | 3+4 4 | 3+4 4 | 3+4 4 | 3+4 4 | 3+4 4 | 3+4 4 | 5 2 |
| | 5 2 | 3+4 4 | 3+4 4 | 3+4 4 | 3+4 4 | 3+4 4 | 3+4 4 | 5 2 |
| | 6 2 | 3+4 4 | 3+4 4 | 3+4 4 | 3+4 4 | 3+4 4 | 3+4 4 | 6 2 |
| | 5 2 | 3+4 4 | 3+4 4 | 3+4 4 | 3+4 4 | 3+4 4 | 3+4 4 | 5 2 |
| | 5 2 | 3+4 4 | 3+4 4 | 3+4 4 | 3+4 4 | 3+4 4 | 3+4 4 | 6 2 |
| | 6 2 | 3+4 4 | 3+4 4 | 3+4 4 | 3+4 4 | 3+4 4 | 3+4 4 | 5 2 |
| | 5 2 | | | | | | | 6 2 |
| | 5 2 | | | | | | | 5 2 |
| | 6 2 | | | | | | | 6 2 |
| | 6 2 | | | | | | | 5 2 |
| | 6 2 | | | | | | | 6 2 |
| | 5 2 | | | | | | | 5 2 |
| | 6 2 | | | | | | | 6 2 |
| | 5 2 | | | | | | | 6 2 |
| | 5 2 | | | | | | | 6 2 |
| | 5 2 | | | | | | | 5 2 |
| | 6 2 | | | | | | | 5 2 |
| | 5 2 | | | | | | | 5 2 |
| | 6 2 | | | | | | | 5 2 |
| | 5 2 | | | | | | | 6 2 |
| | 6 2 | | | | | | | 6 2 |
| | 5 2 | | | | | | | 5 2 |
| | 6 2 | | | | | | | 6 2 |
| | 5 2 | | | | | | | 5 2 |
| | 5 2 | | | | | | | 6 2 |
| | 6 2 | | | | | | | 6 2 |
| | 6 2 | | | | | | | 5 2 |
| | 6 2 | | | | | | | 5 2 |
| | 5 2 | | | | | | | 6 2 |
| | 6 2 | | | | | | | 5 2 |
| | 5 2 | | | | | | | 6 2 |
| | 6 2 | | | | | | | 6 2 |
| | 5 2 | | | | | | | 5 2 |
| | 6 2 | | | | | | | 5 2 |
| | 5 2 | | | | | | | 5 2 |
| | 5 2 | | | | | | | 6 2 |
| | 6 2 | | | | | | | 6 2 |
| | 6 2 | | | | | | | 6 2 |
| | 5 2 | | | | | | | 6 2 |
| | 6 2 | | | | | | | 5 2 |
| | 6 2 | | | | | | | 5 2 |
| | 6 2 | | | | | | | 5 2 |
| | 5 2 | | | | | | | 6 2 |
| | 5 2 | | | | | | | 6 2 |
| | 5 2 | | | | | | | 5 2 |
| | 6 2 | | | | | | | 6 2 |
| | 5 2 | | | | | | | 6 2 |
| | 5 2 | | | | | | | 6 2 |

Continued on next column

**TABLE XXXIV– continued**

| TQC | ICSD | NoE | SF | | | | | |
|---|---|---|---|---|---|---|---|---|
| | 5 2 | | | | | | | 5 2 |
| | 6 2 | | | | | | | 5 2 |
| | 5 2 | | | | | | | 6 2 |
| | 5 2 | | | | | | | 6 2 |
| | 6 2 | | | | | | | 5 2 |
| | 6 2 | | | | | | | 5 2 |
| | 5 2 | | | | | | | 5 2 |
| | 6 2 | | | | | | | 6 2 |
| | 6 2 | | | | | | | 6 2 |
| | 6 2 | | | | | | | 6 2 |
| 59.1.14 | 154398 | 180 | Ba6Ge5N2 | | | | | |
| | -GM | -R | -S | -T | -U | -X | -Y | -Z |
| | 5 2 | 3+4 4 | 3+4 4 | 3+4 4 | 3+4 4 | 3+4 4 | 3+4 4 | 5 2 |
| | 6 2 | 3+4 4 | 3+4 4 | 3+4 4 | 3+4 4 | 3+4 4 | 3+4 4 | 6 2 |
| | 5 2 | 3+4 4 | 3+4 4 | 3+4 4 | 3+4 4 | 3+4 4 | 3+4 4 | 6 2 |
| | 6 2 | 3+4 4 | 3+4 4 | 3+4 4 | 3+4 4 | 3+4 4 | 3+4 4 | 5 2 |
| | 5 2 | 3+4 4 | 3+4 4 | 3+4 4 | 3+4 4 | 3+4 4 | 3+4 4 | 6 2 |
| | 6 2 | 3+4 4 | 3+4 4 | 3+4 4 | 3+4 4 | 3+4 4 | 3+4 4 | 5 2 |
| | 5 2 | 3+4 4 | 3+4 4 | 3+4 4 | 3+4 4 | 3+4 4 | 3+4 4 | 6 2 |
| | 6 2 | 3+4 4 | 3+4 4 | 3+4 4 | 3+4 4 | 3+4 4 | 3+4 4 | 5 2 |
| | 5 2 | 3+4 4 | 3+4 4 | 3+4 4 | 3+4 4 | 3+4 4 | 3+4 4 | 5 2 |
| | 5 2 | 3+4 4 | 3+4 4 | 3+4 4 | 3+4 4 | 3+4 4 | 3+4 4 | 5 2 |
| | 6 2 | 3+4 4 | 3+4 4 | 3+4 4 | 3+4 4 | 3+4 4 | 3+4 4 | 6 2 |
| | 6 2 | 3+4 4 | 3+4 4 | 3+4 4 | 3+4 4 | 3+4 4 | 3+4 4 | 6 2 |
| | 5 2 | 3+4 4 | 3+4 4 | 3+4 4 | 3+4 4 | 3+4 4 | 3+4 4 | 5 2 |
| | 6 2 | 3+4 4 | 3+4 4 | 3+4 4 | 3+4 4 | 3+4 4 | 3+4 4 | 6 2 |
| | 5 2 | 3+4 4 | 3+4 4 | 3+4 4 | 3+4 4 | 3+4 4 | 3+4 4 | 5 2 |
| | 6 2 | 3+4 4 | 3+4 4 | 3+4 4 | 3+4 4 | 3+4 4 | 3+4 4 | 6 2 |
| | 5 2 | 3+4 4 | 3+4 4 | 3+4 4 | 3+4 4 | 3+4 4 | 3+4 4 | 6 2 |
| | 6 2 | 3+4 4 | 3+4 4 | 3+4 4 | 3+4 4 | 3+4 4 | 3+4 4 | 5 2 |
| | 5 2 | 3+4 4 | 3+4 4 | 3+4 4 | 3+4 4 | 3+4 4 | 3+4 4 | 5 2 |
| | 6 2 | 3+4 4 | 3+4 4 | 3+4 4 | 3+4 4 | 3+4 4 | 3+4 4 | 6 2 |
| | 5 2 | 3+4 4 | 3+4 4 | 3+4 4 | 3+4 4 | 3+4 4 | 3+4 4 | 5 2 |
| | 6 2 | 3+4 4 | 3+4 4 | 3+4 4 | 3+4 4 | 3+4 4 | 3+4 4 | 6 2 |
| | 5 2 | 3+4 4 | 3+4 4 | 3+4 4 | 3+4 4 | 3+4 4 | 3+4 4 | 5 2 |
| | 6 2 | 3+4 4 | 3+4 4 | 3+4 4 | 3+4 4 | 3+4 4 | 3+4 4 | 6 2 |
| | 5 2 | 3+4 4 | 3+4 4 | 3+4 4 | 3+4 4 | 3+4 4 | 3+4 4 | 5 2 |
| | 6 2 | 3+4 4 | 3+4 4 | 3+4 4 | 3+4 4 | 3+4 4 | 3+4 4 | 6 2 |
| | 5 2 | 3+4 4 | 3+4 4 | 3+4 4 | 3+4 4 | 3+4 4 | 3+4 4 | 6 2 |
| | 6 2 | 3+4 4 | 3+4 4 | 3+4 4 | 3+4 4 | 3+4 4 | 3+4 4 | 5 2 |
| | 5 2 | 3+4 4 | 3+4 4 | 3+4 4 | 3+4 4 | 3+4 4 | 3+4 4 | 6 2 |
| | 6 2 | 3+4 4 | 3+4 4 | 3+4 4 | 3+4 4 | 3+4 4 | 3+4 4 | 5 2 |
| | 5 2 | 3+4 4 | 3+4 4 | 3+4 4 | 3+4 4 | 3+4 4 | 3+4 4 | 5 2 |
| | 6 2 | 3+4 4 | 3+4 4 | 3+4 4 | 3+4 4 | 3+4 4 | 3+4 4 | 6 2 |
| | 5 2 | 3+4 4 | 3+4 4 | 3+4 4 | 3+4 4 | 3+4 4 | 3+4 4 | 5 2 |
| | 6 2 | 3+4 4 | 3+4 4 | 3+4 4 | 3+4 4 | 3+4 4 | 3+4 4 | 6 2 |
| | 5 2 | 3+4 4 | 3+4 4 | 3+4 4 | 3+4 4 | 3+4 4 | 3+4 4 | 5 2 |
| | 6 2 | 3+4 4 | 3+4 4 | 3+4 4 | 3+4 4 | 3+4 4 | 3+4 4 | 6 2 |
| | 5 2 | 3+4 4 | 3+4 4 | 3+4 4 | 3+4 4 | 3+4 4 | 3+4 4 | 5 2 |
| | 5 2 | 3+4 4 | 3+4 4 | 3+4 4 | 3+4 4 | 3+4 4 | 3+4 4 | 6 2 |
| | 6 2 | 3+4 4 | 3+4 4 | 3+4 4 | 3+4 4 | 3+4 4 | 3+4 4 | 6 2 |
| | 6 2 | 3+4 4 | 3+4 4 | 3+4 4 | 3+4 4 | 3+4 4 | 3+4 4 | 5 2 |
| | 5 2 | 3+4 4 | 3+4 4 | 3+4 4 | 3+4 4 | 3+4 4 | 3+4 4 | 5 2 |
| | 6 2 | 3+4 4 | 3+4 4 | 3+4 4 | 3+4 4 | 3+4 4 | 3+4 4 | 6 2 |
| | 5 2 | 3+4 4 | 3+4 4 | 3+4 4 | 3+4 4 | 3+4 4 | 3+4 4 | 5 2 |
| | 6 2 | 3+4 4 | 3+4 4 | 3+4 4 | 3+4 4 | 3+4 4 | 3+4 4 | 6 2 |
| | 5 2 | 3+4 4 | 3+4 4 | 3+4 4 | 3+4 4 | 3+4 4 | 3+4 4 | 5 2 |
| | 5 2 | | | | | | | 6 2 |
| | 6 2 | | | | | | | 5 2 |
| | 5 2 | | | | | | | 6 2 |
| | 6 2 | | | | | | | 5 2 |

Continued on next column

**TABLE XXXIV– continued**

| TQC | ICSD | NoE | SF | | | | | |
|---|---|---|---|---|---|---|---|---|
| | 6 2 | | | | | | | 6 2 |
| | 5 2 | | | | | | | 5 2 |
| | 6 2 | | | | | | | 6 2 |
| | 5 2 | | | | | | | 5 2 |
| | 5 2 | | | | | | | 5 2 |
| | 6 2 | | | | | | | 6 2 |
| | 6 2 | | | | | | | 6 2 |
| | 5 2 | | | | | | | 5 2 |
| | 6 2 | | | | | | | 6 2 |
| | 6 2 | | | | | | | 6 2 |
| | 6 2 | | | | | | | 6 2 |
| | 5 2 | | | | | | | 5 2 |
| | 5 2 | | | | | | | 5 2 |
| | 5 2 | | | | | | | 6 2 |
| | 6 2 | | | | | | | 5 2 |
| | 5 2 | | | | | | | 5 2 |
| | 6 2 | | | | | | | 6 2 |
| | 5 2 | | | | | | | 6 2 |
| | 6 2 | | | | | | | 5 2 |
| | 5 2 | | | | | | | 5 2 |
| | 6 2 | | | | | | | 5 2 |
| | 5 2 | | | | | | | 6 2 |
| | 6 2 | | | | | | | 6 2 |
| | 5 2 | | | | | | | 5 2 |
| | 5 2 | | | | | | | 6 2 |
| | 6 2 | | | | | | | 6 2 |
| | 5 2 | | | | | | | 5 2 |
| | 6 2 | | | | | | | 5 2 |
| | 6 2 | | | | | | | 6 2 |
| | 6 2 | | | | | | | 6 2 |
| | 6 2 | | | | | | | 5 2 |
| | 5 2 | | | | | | | 5 2 |
| | 5 2 | | | | | | | 6 2 |
| | 5 2 | | | | | | | 5 2 |
| | 6 2 | | | | | | | 6 2 |
| | 6 2 | | | | | | | 6 2 |
| | 5 2 | | | | | | | 6 2 |
| | 6 2 | | | | | | | 5 2 |
| | 6 2 | | | | | | | 6 2 |
| | 6 2 | | | | | | | 6 2 |
| | 5 2 | | | | | | | 5 2 |
| 59.1.15 | 421431 | 68 | Ba1Ge1Ni2 | | | | | |
| | -GM | -R | -S | -T | -U | -X | -Y | -Z |
| | 5 2 | 3+4 4 | 3+4 4 | 3+4 4 | 3+4 4 | 3+4 4 | 3+4 4 | 6 2 |
| | 6 2 | 3+4 4 | 3+4 4 | 3+4 4 | 3+4 4 | 3+4 4 | 3+4 4 | 5 2 |
| | 5 2 | 3+4 4 | 3+4 4 | 3+4 4 | 3+4 4 | 3+4 4 | 3+4 4 | 6 2 |
| | 6 2 | 3+4 4 | 3+4 4 | 3+4 4 | 3+4 4 | 3+4 4 | 3+4 4 | 5 2 |
| | 5 2 | 3+4 4 | 3+4 4 | 3+4 4 | 3+4 4 | 3+4 4 | 3+4 4 | 6 2 |
| | 5 2 | 3+4 4 | 3+4 4 | 3+4 4 | 3+4 4 | 3+4 4 | 3+4 4 | 5 2 |
| | 6 2 | 3+4 4 | 3+4 4 | 3+4 4 | 3+4 4 | 3+4 4 | 3+4 4 | 6 2 |
| | 6 2 | 3+4 4 | 3+4 4 | 3+4 4 | 3+4 4 | 3+4 4 | 3+4 4 | 5 2 |
| | 5 2 | 3+4 4 | 3+4 4 | 3+4 4 | 3+4 4 | 3+4 4 | 3+4 4 | 5 2 |
| | 6 2 | 3+4 4 | 3+4 4 | 3+4 4 | 3+4 4 | 3+4 4 | 3+4 4 | 6 2 |
| | 5 2 | 3+4 4 | 3+4 4 | 3+4 4 | 3+4 4 | 3+4 4 | 3+4 4 | 5 2 |
| | 6 2 | 3+4 4 | 3+4 4 | 3+4 4 | 3+4 4 | 3+4 4 | 3+4 4 | 6 2 |
| | 6 2 | 3+4 4 | 3+4 4 | 3+4 4 | 3+4 4 | 3+4 4 | 3+4 4 | 6 2 |
| | 6 2 | 3+4 4 | 3+4 4 | 3+4 4 | 3+4 4 | 3+4 4 | 3+4 4 | 6 2 |
| | 5 2 | 3+4 4 | 3+4 4 | 3+4 4 | 3+4 4 | 3+4 4 | 3+4 4 | 5 2 |
| | 5 2 | 3+4 4 | 3+4 4 | 3+4 4 | 3+4 4 | 3+4 4 | 3+4 4 | 5 2 |
| | 5 2 | 3+4 4 | 3+4 4 | 3+4 4 | 3+4 4 | 3+4 4 | 3+4 4 | 5 2 |
| | 6 2 | | | | | | | 6 2 |
| | 5 2 | | | | | | | 5 2 |
| | 5 2 | | | | | | | 5 2 |
| | 6 2 | | | | | | | 6 2 |
| | 6 2 | | | | | | | 6 2 |
| | 6 2 | | | | | | | 6 2 |
| | 5 2 | | | | | | | 6 2 |
| | 6 2 | | | | | | | 6 2 |
| | 5 2 | | | | | | | 5 2 |
| | 6 2 | | | | | | | 5 2 |
| | 5 2 | | | | | | | 6 2 |
| | 5 2 | | | | | | | 5 2 |
| | 6 2 | | | | | | | 5 2 |
| | 6 2 | | | | | | | 5 2 |
| | 5 2 | | | | | | | 6 2 |
| | 5 2 | | | | | | | 5 2 |
| | 5 2 | | | | | | | 5 2 |
| 59.1.16 | 428170 | 88 | Ba1Fe2O1S2 | | | | | |
| | -GM | -R | -S | -T | -U | -X | -Y | -Z |
| | 5 2 | 3+4 4 | 3+4 4 | 3+4 4 | 3+4 4 | 3+4 4 | 3+4 4 | 5 2 |
| | 6 2 | 3+4 4 | 3+4 4 | 3+4 4 | 3+4 4 | 3+4 4 | 3+4 4 | 6 2 |
| | 5 2 | 3+4 4 | 3+4 4 | 3+4 4 | 3+4 4 | 3+4 4 | 3+4 4 | 6 2 |
| | 6 2 | 3+4 4 | 3+4 4 | 3+4 4 | 3+4 4 | 3+4 4 | 3+4 4 | 5 2 |
| | 5 2 | 3+4 4 | 3+4 4 | 3+4 4 | 3+4 4 | 3+4 4 | 3+4 4 | 5 2 |
| | 6 2 | 3+4 4 | 3+4 4 | 3+4 4 | 3+4 4 | 3+4 4 | 3+4 4 | 6 2 |
| | 5 2 | 3+4 4 | 3+4 4 | 3+4 4 | 3+4 4 | 3+4 4 | 3+4 4 | 6 2 |
| | 5 2 | 3+4 4 | 3+4 4 | 3+4 4 | 3+4 4 | 3+4 4 | 3+4 4 | 5 2 |
| | 6 2 | 3+4 4 | 3+4 4 | 3+4 4 | 3+4 4 | 3+4 4 | 3+4 4 | 6 2 |
| | 6 2 | 3+4 4 | 3+4 4 | 3+4 4 | 3+4 4 | 3+4 4 | 3+4 4 | 5 2 |
| | 6 2 | 3+4 4 | 3+4 4 | 3+4 4 | 3+4 4 | 3+4 4 | 3+4 4 | 6 2 |
| | 5 2 | 3+4 4 | 3+4 4 | 3+4 4 | 3+4 4 | 3+4 4 | 3+4 4 | 5 2 |
| | 5 2 | 3+4 4 | 3+4 4 | 3+4 4 | 3+4 4 | 3+4 4 | 3+4 4 | 6 2 |
| | 6 2 | 3+4 4 | 3+4 4 | 3+4 4 | 3+4 4 | 3+4 4 | 3+4 4 | 5 2 |
| | 5 2 | 3+4 4 | 3+4 4 | 3+4 4 | 3+4 4 | 3+4 4 | 3+4 4 | 5 2 |
| | 6 2 | 3+4 4 | 3+4 4 | 3+4 4 | 3+4 4 | 3+4 4 | 3+4 4 | 6 2 |
| | 5 2 | 3+4 4 | 3+4 4 | 3+4 4 | 3+4 4 | 3+4 4 | 3+4 4 | 5 2 |
| | 6 2 | 3+4 4 | 3+4 4 | 3+4 4 | 3+4 4 | 3+4 4 | 3+4 4 | 6 2 |
| | 6 2 | 3+4 4 | 3+4 4 | 3+4 4 | 3+4 4 | 3+4 4 | 3+4 4 | 6 2 |
| | 5 2 | 3+4 4 | 3+4 4 | 3+4 4 | 3+4 4 | 3+4 4 | 3+4 4 | 5 2 |
| | 5 2 | 3+4 4 | 3+4 4 | 3+4 4 | 3+4 4 | 3+4 4 | 3+4 4 | 6 2 |
| | 5 2 | 3+4 4 | 3+4 4 | 3+4 4 | 3+4 4 | 3+4 4 | 3+4 4 | 5 2 |
| | 6 2 | | | | | | | 6 2 |
| | 5 2 | | | | | | | 5 2 |
| | 6 2 | | | | | | | 5 2 |
| | 5 2 | | | | | | | 5 2 |
| | 5 2 | | | | | | | 6 2 |
| | 6 2 | | | | | | | 6 2 |
| | 6 2 | | | | | | | 5 2 |
| | 6 2 | | | | | | | 5 2 |
| | 5 2 | | | | | | | 6 2 |
| | 6 2 | | | | | | | 6 2 |
| | 5 2 | | | | | | | 6 2 |
| | 5 2 | | | | | | | 6 2 |
| | 5 2 | | | | | | | 6 2 |
| | 6 2 | | | | | | | 5 2 |
| | 5 2 | | | | | | | 6 2 |
| | 5 2 | | | | | | | 5 2 |
| | 6 2 | | | | | | | 5 2 |
| | 6 2 | | | | | | | 6 2 |
| | 5 2 | | | | | | | 6 2 |
| | 6 2 | | | | | | | 6 2 |
| | 6 2 | | | | | | | 5 2 |

Continued on next column

Continued on next column

**TABLE XXXIV– continued**

| TQC | ICSD | NoE | SF | | | | | |
|---|---|---|---|---|---|---|---|---|
| | 5 2 | | | | | | | 5 2 |
| 59.1.17 | 421524 | 28 | B2Ru1 | | | | | |
| | -GM | -R | -S | -T | -U | -X | -Y | -Z |
| | 5 2 | 3+4 4 | 3+4 4 | 3+4 4 | 3+4 4 | 3+4 4 | 3+4 4 | 6 2 |
| | 6 2 | 3+4 4 | 3+4 4 | 3+4 4 | 3+4 4 | 3+4 4 | 3+4 4 | 5 2 |
| | 5 2 | 3+4 4 | 3+4 4 | 3+4 4 | 3+4 4 | 3+4 4 | 3+4 4 | 6 2 |
| | 6 2 | 3+4 4 | 3+4 4 | 3+4 4 | 3+4 4 | 3+4 4 | 3+4 4 | 5 2 |
| | 5 2 | 3+4 4 | 3+4 4 | 3+4 4 | 3+4 4 | 3+4 4 | 3+4 4 | 6 2 |
| | 6 2 | 3+4 4 | 3+4 4 | 3+4 4 | 3+4 4 | 3+4 4 | 3+4 4 | 6 2 |
| | 5 2 | 3+4 4 | 3+4 4 | 3+4 4 | 3+4 4 | 3+4 4 | 3+4 4 | 5 2 |
| | 5 2 | | | | | | | 6 2 |
| | 5 2 | | | | | | | 5 2 |
| | 5 2 | | | | | | | 5 2 |
| | 6 2 | | | | | | | 5 2 |
| | 5 2 | | | | | | | 6 2 |
| | 6 2 | | | | | | | 5 2 |
| | 6 2 | | | | | | | 5 2 |
| 59.1.18 | 31871 | 28 | B2Ru1 | | | | | |
| | -GM | -R | -S | -T | -U | -X | -Y | -Z |
| | 5 2 | 3+4 4 | 3+4 4 | 3+4 4 | 3+4 4 | 3+4 4 | 3+4 4 | 6 2 |
| | 6 2 | 3+4 4 | 3+4 4 | 3+4 4 | 3+4 4 | 3+4 4 | 3+4 4 | 5 2 |
| | 5 2 | 3+4 4 | 3+4 4 | 3+4 4 | 3+4 4 | 3+4 4 | 3+4 4 | 6 2 |
| | 6 2 | 3+4 4 | 3+4 4 | 3+4 4 | 3+4 4 | 3+4 4 | 3+4 4 | 5 2 |
| | 5 2 | 3+4 4 | 3+4 4 | 3+4 4 | 3+4 4 | 3+4 4 | 3+4 4 | 6 2 |
| | 6 2 | 3+4 4 | 3+4 4 | 3+4 4 | 3+4 4 | 3+4 4 | 3+4 4 | 6 2 |
| | 5 2 | 3+4 4 | 3+4 4 | 3+4 4 | 3+4 4 | 3+4 4 | 3+4 4 | 6 2 |
| | 5 2 | | | | | | | 5 2 |
| | 5 2 | | | | | | | 5 2 |
| | 5 2 | | | | | | | 5 2 |
| | 6 2 | | | | | | | 5 2 |
| | 5 2 | | | | | | | 6 2 |
| | 6 2 | | | | | | | 5 2 |
| | 6 2 | | | | | | | 5 2 |
| 59.1.19 | 421523 | 28 | B2Os1 | | | | | |
| | -GM | -R | -S | -T | -U | -X | -Y | -Z |
| | 5 2 | 3+4 4 | 3+4 4 | 3+4 4 | 3+4 4 | 3+4 4 | 3+4 4 | 6 2 |
| | 6 2 | 3+4 4 | 3+4 4 | 3+4 4 | 3+4 4 | 3+4 4 | 3+4 4 | 5 2 |
| | 5 2 | 3+4 4 | 3+4 4 | 3+4 4 | 3+4 4 | 3+4 4 | 3+4 4 | 6 2 |
| | 6 2 | 3+4 4 | 3+4 4 | 3+4 4 | 3+4 4 | 3+4 4 | 3+4 4 | 5 2 |
| | 5 2 | 3+4 4 | 3+4 4 | 3+4 4 | 3+4 4 | 3+4 4 | 3+4 4 | 6 2 |
| | 6 2 | 3+4 4 | 3+4 4 | 3+4 4 | 3+4 4 | 3+4 4 | 3+4 4 | 6 2 |
| | 5 2 | 3+4 4 | 3+4 4 | 3+4 4 | 3+4 4 | 3+4 4 | 3+4 4 | 5 2 |
| | 5 2 | | | | | | | 6 2 |
| | 5 2 | | | | | | | 5 2 |
| | 6 2 | | | | | | | 5 2 |
| | 5 2 | | | | | | | 5 2 |
| | 5 2 | | | | | | | 6 2 |
| | 6 2 | | | | | | | 5 2 |
| | 6 2 | | | | | | | 5 2 |
| 59.2.1 | 187622 | 28 | Ni1Si1 | | | | | |
| | -GM | -R | -S | -T | -U | -X | -Y | -Z |
| | 5 2 | 3+4 4 | 3+4 4 | 3+4 4 | 3+4 4 | 3+4 4 | 3+4 4 | 5 2 |
| | 6 2 | 3+4 4 | 3+4 4 | 3+4 4 | 3+4 4 | 3+4 4 | 3+4 4 | 6 2 |
| | 5 2 | 3+4 4 | 3+4 4 | 3+4 4 | 3+4 4 | 3+4 4 | 3+4 4 | 5 2 |
| | 6 2 | 3+4 4 | 3+4 4 | 3+4 4 | 3+4 4 | 3+4 4 | 3+4 4 | 5 2 |
| | 5 2 | 3+4 4 | 3+4 4 | 3+4 4 | 3+4 4 | 3+4 4 | 3+4 4 | 5 2 |
| | 5 2 | 3+4 4 | 3+4 4 | 3+4 4 | 3+4 4 | 3+4 4 | 3+4 4 | 6 2 |
| | 5 2 | 3+4 4 | 3+4 4 | 3+4 4 | 3+4 4 | 3+4 4 | 3+4 4 | 5 2 |
| | 6 2 | | | | | | | 6 2 |
| | 6 2 | | | | | | | 6 2 |
| | 5 2 | | | | | | | 6 2 |

Continued on next column

**TABLE XXXIV– continued**

| TQC | ICSD | NoE | SF | | | | | |
|---|---|---|---|---|---|---|---|---|
| | 6 2 | | | | | | | 6 2 |
| | 6 2 | | | | | | | 5 2 |
| | 5 2 | | | | | | | 5 2 |
| | 5 2 | | | | | | | 6 2 |
| 59.2.2 | 95640 | 40 | Li1Mn1O2 | | | | | |
| | -GM | -R | -S | -T | -U | -X | -Y | -Z |
| | 5 2 | 3+4 4 | 3+4 4 | 3+4 4 | 3+4 4 | 3+4 4 | 3+4 4 | 6 2 |
| | 6 2 | 3+4 4 | 3+4 4 | 3+4 4 | 3+4 4 | 3+4 4 | 3+4 4 | 5 2 |
| | 5 2 | 3+4 4 | 3+4 4 | 3+4 4 | 3+4 4 | 3+4 4 | 3+4 4 | 5 2 |
| | 6 2 | 3+4 4 | 3+4 4 | 3+4 4 | 3+4 4 | 3+4 4 | 3+4 4 | 6 2 |
| | 5 2 | 3+4 4 | 3+4 4 | 3+4 4 | 3+4 4 | 3+4 4 | 3+4 4 | 6 2 |
| | 6 2 | 3+4 4 | 3+4 4 | 3+4 4 | 3+4 4 | 3+4 4 | 3+4 4 | 5 2 |
| | 5 2 | 3+4 4 | 3+4 4 | 3+4 4 | 3+4 4 | 3+4 4 | 3+4 4 | 6 2 |
| | 5 2 | 3+4 4 | 3+4 4 | 3+4 4 | 3+4 4 | 3+4 4 | 3+4 4 | 5 2 |
| | 6 2 | 3+4 4 | 3+4 4 | 3+4 4 | 3+4 4 | 3+4 4 | 3+4 4 | 6 2 |
| | 5 2 | 3+4 4 | 3+4 4 | 3+4 4 | 3+4 4 | 3+4 4 | 3+4 4 | 5 2 |
| | 6 2 | | | | | | | 6 2 |
| | 5 2 | | | | | | | 5 2 |
| | 5 2 | | | | | | | 6 2 |
| | 6 2 | | | | | | | 6 2 |
| | 6 2 | | | | | | | 5 2 |
| | 6 2 | | | | | | | 5 2 |
| | 6 2 | | | | | | | 6 2 |
| | 5 2 | | | | | | | 5 2 |
| | 5 2 | | | | | | | 6 2 |
| | 5 2 | | | | | | | 5 2 |
| 59.2.3 | 408082 | 256 | Gd1Pd1Si1 | | | | | |
| | -GM | -R | -S | -T | -U | -X | -Y | -Z |
| | 5 2 | 3+4 4 | 3+4 4 | 3+4 4 | 3+4 4 | 3+4 4 | 3+4 4 | 5 2 |
| | 6 2 | 3+4 4 | 3+4 4 | 3+4 4 | 3+4 4 | 3+4 4 | 3+4 4 | 6 2 |
| | 5 2 | 3+4 4 | 3+4 4 | 3+4 4 | 3+4 4 | 3+4 4 | 3+4 4 | 6 2 |
| | 5 2 | 3+4 4 | 3+4 4 | 3+4 4 | 3+4 4 | 3+4 4 | 3+4 4 | 6 2 |
| | 6 2 | 3+4 4 | 3+4 4 | 3+4 4 | 3+4 4 | 3+4 4 | 3+4 4 | 5 2 |
| | 6 2 | 3+4 4 | 3+4 4 | 3+4 4 | 3+4 4 | 3+4 4 | 3+4 4 | 5 2 |
| | 5 2 | 3+4 4 | 3+4 4 | 3+4 4 | 3+4 4 | 3+4 4 | 3+4 4 | 5 2 |
| | 6 2 | 3+4 4 | 3+4 4 | 3+4 4 | 3+4 4 | 3+4 4 | 3+4 4 | 6 2 |
| | 5 2 | 3+4 4 | 3+4 4 | 3+4 4 | 3+4 4 | 3+4 4 | 3+4 4 | 6 2 |
| | 6 2 | 3+4 4 | 3+4 4 | 3+4 4 | 3+4 4 | 3+4 4 | 3+4 4 | 5 2 |
| | 5 2 | 3+4 4 | 3+4 4 | 3+4 4 | 3+4 4 | 3+4 4 | 3+4 4 | 6 2 |
| | 5 2 | 3+4 4 | 3+4 4 | 3+4 4 | 3+4 4 | 3+4 4 | 3+4 4 | 6 2 |
| | 6 2 | 3+4 4 | 3+4 4 | 3+4 4 | 3+4 4 | 3+4 4 | 3+4 4 | 5 2 |
| | 6 2 | 3+4 4 | 3+4 4 | 3+4 4 | 3+4 4 | 3+4 4 | 3+4 4 | 5 2 |
| | 5 2 | 3+4 4 | 3+4 4 | 3+4 4 | 3+4 4 | 3+4 4 | 3+4 4 | 5 2 |
| | 6 2 | 3+4 4 | 3+4 4 | 3+4 4 | 3+4 4 | 3+4 4 | 3+4 4 | 6 2 |
| | 5 2 | 3+4 4 | 3+4 4 | 3+4 4 | 3+4 4 | 3+4 4 | 3+4 4 | 6 2 |
| | 6 2 | 3+4 4 | 3+4 4 | 3+4 4 | 3+4 4 | 3+4 4 | 3+4 4 | 5 2 |
| | 5 2 | 3+4 4 | 3+4 4 | 3+4 4 | 3+4 4 | 3+4 4 | 3+4 4 | 6 2 |
| | 5 2 | 3+4 4 | 3+4 4 | 3+4 4 | 3+4 4 | 3+4 4 | 3+4 4 | 6 2 |
| | 6 2 | 3+4 4 | 3+4 4 | 3+4 4 | 3+4 4 | 3+4 4 | 3+4 4 | 6 2 |
| | 5 2 | 3+4 4 | 3+4 4 | 3+4 4 | 3+4 4 | 3+4 4 | 3+4 4 | 5 2 |
| | 6 2 | 3+4 4 | 3+4 4 | 3+4 4 | 3+4 4 | 3+4 4 | 3+4 4 | 5 2 |
| | 5 2 | 3+4 4 | 3+4 4 | 3+4 4 | 3+4 4 | 3+4 4 | 3+4 4 | 5 2 |
| | 6 2 | 3+4 4 | 3+4 4 | 3+4 4 | 3+4 4 | 3+4 4 | 3+4 4 | 6 2 |
| | 5 2 | 3+4 4 | 3+4 4 | 3+4 4 | 3+4 4 | 3+4 4 | 3+4 4 | 6 2 |
| | 6 2 | 3+4 4 | 3+4 4 | 3+4 4 | 3+4 4 | 3+4 4 | 3+4 4 | 5 2 |
| | 6 2 | 3+4 4 | 3+4 4 | 3+4 4 | 3+4 4 | 3+4 4 | 3+4 4 | 5 2 |
| | 5 2 | 3+4 4 | 3+4 4 | 3+4 4 | 3+4 4 | 3+4 4 | 3+4 4 | 6 2 |
| | 6 2 | 3+4 4 | 3+4 4 | 3+4 4 | 3+4 4 | 3+4 4 | 3+4 4 | 5 2 |
| | 5 2 | 3+4 4 | 3+4 4 | 3+4 4 | 3+4 4 | 3+4 4 | 3+4 4 | 6 2 |
| | 6 2 | 3+4 4 | 3+4 4 | 3+4 4 | 3+4 4 | 3+4 4 | 3+4 4 | 5 2 |
| | 5 2 | 3+4 4 | 3+4 4 | 3+4 4 | 3+4 4 | 3+4 4 | 3+4 4 | 6 2 |

Continued on next column

**TABLE XXXIV– continued**

| TQC | ICSD | NoE | SF | | | | | |
|---|---|---|---|---|---|---|---|---|
| | 6 2 | 3+4 4 | 3+4 4 | 3+4 4 | 3+4 4 | 3+4 4 | 3+4 4 | 5 2 |
| | 5 2 | 3+4 4 | 3+4 4 | 3+4 4 | 3+4 4 | 3+4 4 | 3+4 4 | 5 2 |
| | 5 2 | 3+4 4 | 3+4 4 | 3+4 4 | 3+4 4 | 3+4 4 | 3+4 4 | 5 2 |
| | 6 2 | 3+4 4 | 3+4 4 | 3+4 4 | 3+4 4 | 3+4 4 | 3+4 4 | 6 2 |
| | 6 2 | 3+4 4 | 3+4 4 | 3+4 4 | 3+4 4 | 3+4 4 | 3+4 4 | 6 2 |
| | 5 2 | 3+4 4 | 3+4 4 | 3+4 4 | 3+4 4 | 3+4 4 | 3+4 4 | 6 2 |
| | 6 2 | 3+4 4 | 3+4 4 | 3+4 4 | 3+4 4 | 3+4 4 | 3+4 4 | 5 2 |
| | 6 2 | 3+4 4 | 3+4 4 | 3+4 4 | 3+4 4 | 3+4 4 | 3+4 4 | 5 2 |
| | 5 2 | 3+4 4 | 3+4 4 | 3+4 4 | 3+4 4 | 3+4 4 | 3+4 4 | 6 2 |
| | 5 2 | 3+4 4 | 3+4 4 | 3+4 4 | 3+4 4 | 3+4 4 | 3+4 4 | 6 2 |
| | 5 2 | 3+4 4 | 3+4 4 | 3+4 4 | 3+4 4 | 3+4 4 | 3+4 4 | 5 2 |
| | 6 2 | 3+4 4 | 3+4 4 | 3+4 4 | 3+4 4 | 3+4 4 | 3+4 4 | 5 2 |
| | 6 2 | 3+4 4 | 3+4 4 | 3+4 4 | 3+4 4 | 3+4 4 | 3+4 4 | 6 2 |
| | 5 2 | 3+4 4 | 3+4 4 | 3+4 4 | 3+4 4 | 3+4 4 | 3+4 4 | 5 2 |
| | 5 2 | 3+4 4 | 3+4 4 | 3+4 4 | 3+4 4 | 3+4 4 | 3+4 4 | 5 2 |
| | 6 2 | 3+4 4 | 3+4 4 | 3+4 4 | 3+4 4 | 3+4 4 | 3+4 4 | 6 2 |
| | 6 2 | 3+4 4 | 3+4 4 | 3+4 4 | 3+4 4 | 3+4 4 | 3+4 4 | 6 2 |
| | 6 2 | 3+4 4 | 3+4 4 | 3+4 4 | 3+4 4 | 3+4 4 | 3+4 4 | 6 2 |
| | 5 2 | 3+4 4 | 3+4 4 | 3+4 4 | 3+4 4 | 3+4 4 | 3+4 4 | 5 2 |
| | 6 2 | 3+4 4 | 3+4 4 | 3+4 4 | 3+4 4 | 3+4 4 | 3+4 4 | 5 2 |
| | 6 2 | 3+4 4 | 3+4 4 | 3+4 4 | 3+4 4 | 3+4 4 | 3+4 4 | 5 2 |
| | 5 2 | 3+4 4 | 3+4 4 | 3+4 4 | 3+4 4 | 3+4 4 | 3+4 4 | 6 2 |
| | 5 2 | 3+4 4 | 3+4 4 | 3+4 4 | 3+4 4 | 3+4 4 | 3+4 4 | 6 2 |
| | 5 2 | 3+4 4 | 3+4 4 | 3+4 4 | 3+4 4 | 3+4 4 | 3+4 4 | 5 2 |
| | 6 2 | 3+4 4 | 3+4 4 | 3+4 4 | 3+4 4 | 3+4 4 | 3+4 4 | 6 2 |
| | 5 2 | 3+4 4 | 3+4 4 | 3+4 4 | 3+4 4 | 3+4 4 | 3+4 4 | 6 2 |
| | 5 2 | 3+4 4 | 3+4 4 | 3+4 4 | 3+4 4 | 3+4 4 | 3+4 4 | 5 2 |
| | 6 2 | 3+4 4 | 3+4 4 | 3+4 4 | 3+4 4 | 3+4 4 | 3+4 4 | 6 2 |
| | 5 2 | 3+4 4 | 3+4 4 | 3+4 4 | 3+4 4 | 3+4 4 | 3+4 4 | 5 2 |
| | 6 2 | 3+4 4 | 3+4 4 | 3+4 4 | 3+4 4 | 3+4 4 | 3+4 4 | 6 2 |
| | 5 2 | 3+4 4 | 3+4 4 | 3+4 4 | 3+4 4 | 3+4 4 | 3+4 4 | 5 2 |
| | 6 2 | | | | | | | 6 2 |
| | 5 2 | | | | | | | 5 2 |
| | 6 2 | | | | | | | 5 2 |
| | 5 2 | | | | | | | 5 2 |
| | 5 2 | | | | | | | 6 2 |
| | 5 2 | | | | | | | 5 2 |
| | 6 2 | | | | | | | 5 2 |
| | 6 2 | | | | | | | 6 2 |
| | 6 2 | | | | | | | 6 2 |
| | 6 2 | | | | | | | 5 2 |
| | 5 2 | | | | | | | 6 2 |
| | 6 2 | | | | | | | 6 2 |
| | 5 2 | | | | | | | 5 2 |
| | 6 2 | | | | | | | 6 2 |
| | 5 2 | | | | | | | 6 2 |
| | 6 2 | | | | | | | 5 2 |
| | 5 2 | | | | | | | 6 2 |
| | 5 2 | | | | | | | 5 2 |
| | 6 2 | | | | | | | 5 2 |
| | 6 2 | | | | | | | 5 2 |
| | 5 2 | | | | | | | 5 2 |
| | 6 2 | | | | | | | 6 2 |
| | 5 2 | | | | | | | 6 2 |
| | 6 2 | | | | | | | 6 2 |
| | 6 2 | | | | | | | 5 2 |
| | 5 2 | | | | | | | 5 2 |
| | 5 2 | | | | | | | 5 2 |
| | 5 2 | | | | | | | 5 2 |
| | 6 2 | | | | | | | 6 2 |
| | 5 2 | | | | | | | 6 2 |

Continued on next column

**TABLE XXXIV– continued**

| TQC | ICSD | NoE | SF | | | | | |
|---|---|---|---|---|---|---|---|---|
| | 6 2 | | | | | | | 5 2 |
| | 6 2 | | | | | | | 6 2 |
| | 5 2 | | | | | | | 5 2 |
| | 6 2 | | | | | | | 6 2 |
| | 5 2 | | | | | | | 5 2 |
| | 6 2 | | | | | | | 6 2 |
| | 6 2 | | | | | | | 6 2 |
| | 6 2 | | | | | | | 5 2 |
| | 5 2 | | | | | | | 6 2 |
| | 5 2 | | | | | | | 6 2 |
| | 6 2 | | | | | | | 5 2 |
| | 5 2 | | | | | | | 6 2 |
| | 5 2 | | | | | | | 5 2 |
| | 6 2 | | | | | | | 5 2 |
| | 5 2 | | | | | | | 6 2 |
| | 6 2 | | | | | | | 5 2 |
| | 5 2 | | | | | | | 6 2 |
| | 6 2 | | | | | | | 5 2 |
| | 6 2 | | | | | | | 6 2 |
| | 5 2 | | | | | | | 5 2 |
| | 5 2 | | | | | | | 5 2 |
| | 6 2 | | | | | | | 5 2 |
| | 5 2 | | | | | | | 6 2 |
| | 5 2 | | | | | | | 5 2 |
| | 5 2 | | | | | | | 6 2 |
| | 6 2 | | | | | | | 5 2 |
| | 5 2 | | | | | | | 6 2 |
| | 6 2 | | | | | | | 6 2 |
| | 6 2 | | | | | | | 5 2 |
| | 5 2 | | | | | | | 6 2 |
| | 6 2 | | | | | | | 6 2 |
| | 5 2 | | | | | | | 5 2 |
| | 6 2 | | | | | | | 6 2 |
| | 6 2 | | | | | | | 5 2 |
| 59.2.4 | 81303 | 152 | Fe1Hf8Te6 | | | | | |
| | -GM | -R | -S | -T | -U | -X | -Y | -Z |
| | 5 2 | 3+4 4 | 3+4 4 | 3+4 4 | 3+4 4 | 3+4 4 | 3+4 4 | 5 2 |
| | 6 2 | 3+4 4 | 3+4 4 | 3+4 4 | 3+4 4 | 3+4 4 | 3+4 4 | 6 2 |
| | 5 2 | 3+4 4 | 3+4 4 | 3+4 4 | 3+4 4 | 3+4 4 | 3+4 4 | 5 2 |
| | 6 2 | 3+4 4 | 3+4 4 | 3+4 4 | 3+4 4 | 3+4 4 | 3+4 4 | 6 2 |
| | 5 2 | 3+4 4 | 3+4 4 | 3+4 4 | 3+4 4 | 3+4 4 | 3+4 4 | 6 2 |
| | 5 2 | 3+4 4 | 3+4 4 | 3+4 4 | 3+4 4 | 3+4 4 | 3+4 4 | 6 2 |
| | 6 2 | 3+4 4 | 3+4 4 | 3+4 4 | 3+4 4 | 3+4 4 | 3+4 4 | 5 2 |
| | 6 2 | 3+4 4 | 3+4 4 | 3+4 4 | 3+4 4 | 3+4 4 | 3+4 4 | 5 2 |
| | 6 2 | 3+4 4 | 3+4 4 | 3+4 4 | 3+4 4 | 3+4 4 | 3+4 4 | 6 2 |
| | 5 2 | 3+4 4 | 3+4 4 | 3+4 4 | 3+4 4 | 3+4 4 | 3+4 4 | 5 2 |
| | 5 2 | 3+4 4 | 3+4 4 | 3+4 4 | 3+4 4 | 3+4 4 | 3+4 4 | 5 2 |
| | 6 2 | 3+4 4 | 3+4 4 | 3+4 4 | 3+4 4 | 3+4 4 | 3+4 4 | 6 2 |
| | 5 2 | 3+4 4 | 3+4 4 | 3+4 4 | 3+4 4 | 3+4 4 | 3+4 4 | 6 2 |
| | 6 2 | 3+4 4 | 3+4 4 | 3+4 4 | 3+4 4 | 3+4 4 | 3+4 4 | 5 2 |
| | 6 2 | 3+4 4 | 3+4 4 | 3+4 4 | 3+4 4 | 3+4 4 | 3+4 4 | 5 2 |
| | 5 2 | 3+4 4 | 3+4 4 | 3+4 4 | 3+4 4 | 3+4 4 | 3+4 4 | 6 2 |
| | 5 2 | 3+4 4 | 3+4 4 | 3+4 4 | 3+4 4 | 3+4 4 | 3+4 4 | 5 2 |
| | 6 2 | 3+4 4 | 3+4 4 | 3+4 4 | 3+4 4 | 3+4 4 | 3+4 4 | 6 2 |
| | 5 2 | 3+4 4 | 3+4 4 | 3+4 4 | 3+4 4 | 3+4 4 | 3+4 4 | 5 2 |
| | 6 2 | 3+4 4 | 3+4 4 | 3+4 4 | 3+4 4 | 3+4 4 | 3+4 4 | 5 2 |
| | 5 2 | 3+4 4 | 3+4 4 | 3+4 4 | 3+4 4 | 3+4 4 | 3+4 4 | 6 2 |
| | 5 2 | 3+4 4 | 3+4 4 | 3+4 4 | 3+4 4 | 3+4 4 | 3+4 4 | 6 2 |
| | 6 2 | 3+4 4 | 3+4 4 | 3+4 4 | 3+4 4 | 3+4 4 | 3+4 4 | 6 2 |
| | 5 2 | 3+4 4 | 3+4 4 | 3+4 4 | 3+4 4 | 3+4 4 | 3+4 4 | 5 2 |
| | 6 2 | 3+4 4 | 3+4 4 | 3+4 4 | 3+4 4 | 3+4 4 | 3+4 4 | 6 2 |

Continued on next column

**TABLE XXXIV– continued**

| TQC | ICSD | NoE | SF | | | | | |
|---|---|---|---|---|---|---|---|---|
| | 6 2 | 3+4 4 | 3+4 4 | 3+4 4 | 3+4 4 | 3+4 4 | 3+4 4 | 5 2 |
| | 6 2 | 3+4 4 | 3+4 4 | 3+4 4 | 3+4 4 | 3+4 4 | 3+4 4 | 5 2 |
| | 6 2 | 3+4 4 | 3+4 4 | 3+4 4 | 3+4 4 | 3+4 4 | 3+4 4 | 6 2 |
| | 5 2 | 3+4 4 | 3+4 4 | 3+4 4 | 3+4 4 | 3+4 4 | 3+4 4 | 6 2 |
| | 5 2 | 3+4 4 | 3+4 4 | 3+4 4 | 3+4 4 | 3+4 4 | 3+4 4 | 5 2 |
| | 5 2 | 3+4 4 | 3+4 4 | 3+4 4 | 3+4 4 | 3+4 4 | 3+4 4 | 5 2 |
| | 6 2 | 3+4 4 | 3+4 4 | 3+4 4 | 3+4 4 | 3+4 4 | 3+4 4 | 6 2 |
| | 5 2 | 3+4 4 | 3+4 4 | 3+4 4 | 3+4 4 | 3+4 4 | 3+4 4 | 6 2 |
| | 6 2 | 3+4 4 | 3+4 4 | 3+4 4 | 3+4 4 | 3+4 4 | 3+4 4 | 6 2 |
| | 5 2 | 3+4 4 | 3+4 4 | 3+4 4 | 3+4 4 | 3+4 4 | 3+4 4 | 5 2 |
| | 5 2 | 3+4 4 | 3+4 4 | 3+4 4 | 3+4 4 | 3+4 4 | 3+4 4 | 5 2 |
| | 6 2 | 3+4 4 | 3+4 4 | 3+4 4 | 3+4 4 | 3+4 4 | 3+4 4 | 6 2 |
| | 6 2 | 3+4 4 | 3+4 4 | 3+4 4 | 3+4 4 | 3+4 4 | 3+4 4 | 5 2 |
| | 5 2 | | | | | | | 5 2 |
| | 5 2 | | | | | | | 6 2 |
| | 6 2 | | | | | | | 6 2 |
| | 6 2 | | | | | | | 5 2 |
| | 5 2 | | | | | | | 6 2 |
| | 5 2 | | | | | | | 5 2 |
| | 6 2 | | | | | | | 5 2 |
| | 5 2 | | | | | | | 6 2 |
| | 6 2 | | | | | | | 5 2 |
| | 6 2 | | | | | | | 6 2 |
| | 5 2 | | | | | | | 6 2 |
| | 5 2 | | | | | | | 5 2 |
| | 6 2 | | | | | | | 6 2 |
| | 6 2 | | | | | | | 5 2 |
| | 5 2 | | | | | | | 5 2 |
| | 6 2 | | | | | | | 5 2 |
| | 6 2 | | | | | | | 6 2 |
| | 5 2 | | | | | | | 6 2 |
| | 6 2 | | | | | | | 5 2 |
| | 5 2 | | | | | | | 6 2 |
| | 6 2 | | | | | | | 5 2 |
| | 5 2 | | | | | | | 6 2 |
| | 5 2 | | | | | | | 6 2 |
| | 6 2 | | | | | | | 5 2 |
| | 6 2 | | | | | | | 5 2 |
| | 6 2 | | | | | | | 6 2 |
| | 5 2 | | | | | | | 5 2 |
| | 6 2 | | | | | | | 6 2 |
| | 5 2 | | | | | | | 5 2 |
| | 5 2 | | | | | | | 6 2 |
| | 6 2 | | | | | | | 5 2 |
| | 6 2 | | | | | | | 6 2 |
| | 5 2 | | | | | | | 5 2 |
| | 6 2 | | | | | | | 5 2 |
| | 5 2 | | | | | | | 6 2 |
| | 5 2 | | | | | | | 6 2 |
| | 5 2 | | | | | | | 6 2 |
| | 6 2 | | | | | | | 6 2 |
| 59.2.5 | 408081 | 288 | Er1Pd1Si1 | | | | | |
| | -GM | -R | -S | -T | -U | -X | -Y | -Z |
| | 5 2 | 3+4 4 | 3+4 4 | 3+4 4 | 3+4 4 | 3+4 4 | 3+4 4 | 5 2 |
| | 6 2 | 3+4 4 | 3+4 4 | 3+4 4 | 3+4 4 | 3+4 4 | 3+4 4 | 6 2 |
| | 5 2 | 3+4 4 | 3+4 4 | 3+4 4 | 3+4 4 | 3+4 4 | 3+4 4 | 5 2 |
| | 5 2 | 3+4 4 | 3+4 4 | 3+4 4 | 3+4 4 | 3+4 4 | 3+4 4 | 5 2 |
| | 6 2 | 3+4 4 | 3+4 4 | 3+4 4 | 3+4 4 | 3+4 4 | 3+4 4 | 6 2 |
| | 6 2 | 3+4 4 | 3+4 4 | 3+4 4 | 3+4 4 | 3+4 4 | 3+4 4 | 6 2 |
| | 5 2 | 3+4 4 | 3+4 4 | 3+4 4 | 3+4 4 | 3+4 4 | 3+4 4 | 5 2 |
| | 6 2 | 3+4 4 | 3+4 4 | 3+4 4 | 3+4 4 | 3+4 4 | 3+4 4 | 6 2 |

Continued on next column

**TABLE XXXIV– continued**

| TQC | ICSD | NoE | SF | | | | | |
|---|---|---|---|---|---|---|---|---|
| | 5 2 | 3+4 4 | 3+4 4 | 3+4 4 | 3+4 4 | 3+4 4 | 3+4 4 | 5 2 |
| | 6 2 | 3+4 4 | 3+4 4 | 3+4 4 | 3+4 4 | 3+4 4 | 3+4 4 | 6 2 |
| | 5 2 | 3+4 4 | 3+4 4 | 3+4 4 | 3+4 4 | 3+4 4 | 3+4 4 | 5 2 |
| | 5 2 | 3+4 4 | 3+4 4 | 3+4 4 | 3+4 4 | 3+4 4 | 3+4 4 | 5 2 |
| | 6 2 | 3+4 4 | 3+4 4 | 3+4 4 | 3+4 4 | 3+4 4 | 3+4 4 | 6 2 |
| | 6 2 | 3+4 4 | 3+4 4 | 3+4 4 | 3+4 4 | 3+4 4 | 3+4 4 | 6 2 |
| | 5 2 | 3+4 4 | 3+4 4 | 3+4 4 | 3+4 4 | 3+4 4 | 3+4 4 | 5 2 |
| | 6 2 | 3+4 4 | 3+4 4 | 3+4 4 | 3+4 4 | 3+4 4 | 3+4 4 | 6 2 |
| | 5 2 | 3+4 4 | 3+4 4 | 3+4 4 | 3+4 4 | 3+4 4 | 3+4 4 | 5 2 |
| | 6 2 | 3+4 4 | 3+4 4 | 3+4 4 | 3+4 4 | 3+4 4 | 3+4 4 | 6 2 |
| | 5 2 | 3+4 4 | 3+4 4 | 3+4 4 | 3+4 4 | 3+4 4 | 3+4 4 | 5 2 |
| | 5 2 | 3+4 4 | 3+4 4 | 3+4 4 | 3+4 4 | 3+4 4 | 3+4 4 | 5 2 |
| | 6 2 | 3+4 4 | 3+4 4 | 3+4 4 | 3+4 4 | 3+4 4 | 3+4 4 | 5 2 |
| | 5 2 | 3+4 4 | 3+4 4 | 3+4 4 | 3+4 4 | 3+4 4 | 3+4 4 | 6 2 |
| | 6 2 | 3+4 4 | 3+4 4 | 3+4 4 | 3+4 4 | 3+4 4 | 3+4 4 | 6 2 |
| | 5 2 | 3+4 4 | 3+4 4 | 3+4 4 | 3+4 4 | 3+4 4 | 3+4 4 | 6 2 |
| | 6 2 | 3+4 4 | 3+4 4 | 3+4 4 | 3+4 4 | 3+4 4 | 3+4 4 | 5 2 |
| | 5 2 | 3+4 4 | 3+4 4 | 3+4 4 | 3+4 4 | 3+4 4 | 3+4 4 | 5 2 |
| | 6 2 | 3+4 4 | 3+4 4 | 3+4 4 | 3+4 4 | 3+4 4 | 3+4 4 | 6 2 |
| | 6 2 | 3+4 4 | 3+4 4 | 3+4 4 | 3+4 4 | 3+4 4 | 3+4 4 | 6 2 |
| | 5 2 | 3+4 4 | 3+4 4 | 3+4 4 | 3+4 4 | 3+4 4 | 3+4 4 | 5 2 |
| | 6 2 | 3+4 4 | 3+4 4 | 3+4 4 | 3+4 4 | 3+4 4 | 3+4 4 | 6 2 |
| | 5 2 | 3+4 4 | 3+4 4 | 3+4 4 | 3+4 4 | 3+4 4 | 3+4 4 | 5 2 |
| | 6 2 | 3+4 4 | 3+4 4 | 3+4 4 | 3+4 4 | 3+4 4 | 3+4 4 | 6 2 |
| | 5 2 | 3+4 4 | 3+4 4 | 3+4 4 | 3+4 4 | 3+4 4 | 3+4 4 | 5 2 |
| | 6 2 | 3+4 4 | 3+4 4 | 3+4 4 | 3+4 4 | 3+4 4 | 3+4 4 | 6 2 |
| | 5 2 | 3+4 4 | 3+4 4 | 3+4 4 | 3+4 4 | 3+4 4 | 3+4 4 | 6 2 |
| | 5 2 | 3+4 4 | 3+4 4 | 3+4 4 | 3+4 4 | 3+4 4 | 3+4 4 | 6 2 |
| | 6 2 | 3+4 4 | 3+4 4 | 3+4 4 | 3+4 4 | 3+4 4 | 3+4 4 | 5 2 |
| | 6 2 | 3+4 4 | 3+4 4 | 3+4 4 | 3+4 4 | 3+4 4 | 3+4 4 | 5 2 |
| | 5 2 | 3+4 4 | 3+4 4 | 3+4 4 | 3+4 4 | 3+4 4 | 3+4 4 | 5 2 |
| | 6 2 | 3+4 4 | 3+4 4 | 3+4 4 | 3+4 4 | 3+4 4 | 3+4 4 | 6 2 |
| | 6 2 | 3+4 4 | 3+4 4 | 3+4 4 | 3+4 4 | 3+4 4 | 3+4 4 | 5 2 |
| | 5 2 | 3+4 4 | 3+4 4 | 3+4 4 | 3+4 4 | 3+4 4 | 3+4 4 | 6 2 |
| | 5 2 | 3+4 4 | 3+4 4 | 3+4 4 | 3+4 4 | 3+4 4 | 3+4 4 | 5 2 |
| | 5 2 | 3+4 4 | 3+4 4 | 3+4 4 | 3+4 4 | 3+4 4 | 3+4 4 | 6 2 |
| | 6 2 | 3+4 4 | 3+4 4 | 3+4 4 | 3+4 4 | 3+4 4 | 3+4 4 | 6 2 |
| | 6 2 | 3+4 4 | 3+4 4 | 3+4 4 | 3+4 4 | 3+4 4 | 3+4 4 | 5 2 |
| | 5 2 | 3+4 4 | 3+4 4 | 3+4 4 | 3+4 4 | 3+4 4 | 3+4 4 | 6 2 |
| | 6 2 | 3+4 4 | 3+4 4 | 3+4 4 | 3+4 4 | 3+4 4 | 3+4 4 | 6 2 |
| | 5 2 | 3+4 4 | 3+4 4 | 3+4 4 | 3+4 4 | 3+4 4 | 3+4 4 | 5 2 |
| | 5 2 | 3+4 4 | 3+4 4 | 3+4 4 | 3+4 4 | 3+4 4 | 3+4 4 | 5 2 |
| | 6 2 | 3+4 4 | 3+4 4 | 3+4 4 | 3+4 4 | 3+4 4 | 3+4 4 | 5 2 |
| | 6 2 | 3+4 4 | 3+4 4 | 3+4 4 | 3+4 4 | 3+4 4 | 3+4 4 | 6 2 |
| | 6 2 | 3+4 4 | 3+4 4 | 3+4 4 | 3+4 4 | 3+4 4 | 3+4 4 | 6 2 |
| | 6 2 | 3+4 4 | 3+4 4 | 3+4 4 | 3+4 4 | 3+4 4 | 3+4 4 | 6 2 |
| | 5 2 | 3+4 4 | 3+4 4 | 3+4 4 | 3+4 4 | 3+4 4 | 3+4 4 | 5 2 |
| | 5 2 | 3+4 4 | 3+4 4 | 3+4 4 | 3+4 4 | 3+4 4 | 3+4 4 | 5 2 |
| | 5 2 | 3+4 4 | 3+4 4 | 3+4 4 | 3+4 4 | 3+4 4 | 3+4 4 | 5 2 |
| | 6 2 | 3+4 4 | 3+4 4 | 3+4 4 | 3+4 4 | 3+4 4 | 3+4 4 | 6 2 |
| | 5 2 | 3+4 4 | 3+4 4 | 3+4 4 | 3+4 4 | 3+4 4 | 3+4 4 | 5 2 |
| | 5 2 | 3+4 4 | 3+4 4 | 3+4 4 | 3+4 4 | 3+4 4 | 3+4 4 | 6 2 |
| | 6 2 | 3+4 4 | 3+4 4 | 3+4 4 | 3+4 4 | 3+4 4 | 3+4 4 | 5 2 |
| | 5 2 | 3+4 4 | 3+4 4 | 3+4 4 | 3+4 4 | 3+4 4 | 3+4 4 | 5 2 |
| | 6 2 | 3+4 4 | 3+4 4 | 3+4 4 | 3+4 4 | 3+4 4 | 3+4 4 | 6 2 |
| | 5 2 | 3+4 4 | 3+4 4 | 3+4 4 | 3+4 4 | 3+4 4 | 3+4 4 | 6 2 |
| | 6 2 | 3+4 4 | 3+4 4 | 3+4 4 | 3+4 4 | 3+4 4 | 3+4 4 | 5 2 |
| | 5 2 | 3+4 4 | 3+4 4 | 3+4 4 | 3+4 4 | 3+4 4 | 3+4 4 | 6 2 |
| | 6 2 | 3+4 4 | 3+4 4 | 3+4 4 | 3+4 4 | 3+4 4 | 3+4 4 | 6 2 |
| | 5 2 | 3+4 4 | 3+4 4 | 3+4 4 | 3+4 4 | 3+4 4 | 3+4 4 | 6 2 |
| | 5 2 | 3+4 4 | 3+4 4 | 3+4 4 | 3+4 4 | 3+4 4 | 3+4 4 | 5 2 |

Continued on next column

**TABLE XXXIV– continued**

| TQC | ICSD | NoE | SF | | | | | |
|---|---|---|---|---|---|---|---|---|
| | 5 2 | 3+4 4 | 3+4 4 | 3+4 4 | 3+4 4 | 3+4 4 | 3+4 4 | 6 2 |
| | 6 2 | 3+4 4 | 3+4 4 | 3+4 4 | 3+4 4 | 3+4 4 | 3+4 4 | 6 2 |
| | 6 2 | 3+4 4 | 3+4 4 | 3+4 4 | 3+4 4 | 3+4 4 | 3+4 4 | 5 2 |
| | 6 2 | | | | | | | 5 2 |
| | 6 2 | | | | | | | 5 2 |
| | 5 2 | | | | | | | 6 2 |
| | 5 2 | | | | | | | 5 2 |
| | 6 2 | | | | | | | 5 2 |
| | 6 2 | | | | | | | 6 2 |
| | 5 2 | | | | | | | 5 2 |
| | 5 2 | | | | | | | 6 2 |
| | 5 2 | | | | | | | 5 2 |
| | 6 2 | | | | | | | 6 2 |
| | 6 2 | | | | | | | 6 2 |
| | 6 2 | | | | | | | 6 2 |
| | 5 2 | | | | | | | 6 2 |
| | 6 2 | | | | | | | 5 2 |
| | 5 2 | | | | | | | 5 2 |
| | 6 2 | | | | | | | 5 2 |
| | 5 2 | | | | | | | 6 2 |
| | 5 2 | | | | | | | 6 2 |
| | 5 2 | | | | | | | 5 2 |
| | 6 2 | | | | | | | 6 2 |
| | 6 2 | | | | | | | 5 2 |
| | 5 2 | | | | | | | 5 2 |
| | 6 2 | | | | | | | 6 2 |
| | 6 2 | | | | | | | 5 2 |
| | 5 2 | | | | | | | 6 2 |
| | 6 2 | | | | | | | 6 2 |
| | 5 2 | | | | | | | 5 2 |
| | 6 2 | | | | | | | 5 2 |
| | 6 2 | | | | | | | 6 2 |
| | 5 2 | | | | | | | 5 2 |
| | 6 2 | | | | | | | 6 2 |
| | 5 2 | | | | | | | 5 2 |
| | 6 2 | | | | | | | 6 2 |
| | 5 2 | | | | | | | 5 2 |
| | 5 2 | | | | | | | 6 2 |
| | 6 2 | | | | | | | 5 2 |
| | 5 2 | | | | | | | 6 2 |
| | 6 2 | | | | | | | 6 2 |
| | 5 2 | | | | | | | 5 2 |
| | 6 2 | | | | | | | 5 2 |
| | 5 2 | | | | | | | 6 2 |
| | 6 2 | | | | | | | 6 2 |
| | 5 2 | | | | | | | 6 2 |
| | 5 2 | | | | | | | 6 2 |
| | 6 2 | | | | | | | 5 2 |
| | 5 2 | | | | | | | 6 2 |
| | 5 2 | | | | | | | 5 2 |
| | 6 2 | | | | | | | 6 2 |
| | 6 2 | | | | | | | 5 2 |
| | 5 2 | | | | | | | 5 2 |
| | 6 2 | | | | | | | 6 2 |
| | 5 2 | | | | | | | 5 2 |
| | 6 2 | | | | | | | 5 2 |
| | 6 2 | | | | | | | 5 2 |
| | 5 2 | | | | | | | 6 2 |
| | 6 2 | | | | | | | 5 2 |
| | 5 2 | | | | | | | 6 2 |
| | 6 2 | | | | | | | 6 2 |
| | 6 2 | | | | | | | 5 2 |
| | 5 2 | | | | | | | 5 2 |
| | 6 2 | | | | | | | 5 2 |
| | 5 2 | | | | | | | 6 2 |
| | 6 2 | | | | | | | 5 2 |
| | 5 2 | | | | | | | 6 2 |
| | 6 2 | | | | | | | 5 2 |
| | 5 2 | | | | | | | 6 2 |
| | 6 2 | | | | | | | 5 2 |
| | 6 2 | | | | | | | 6 2 |
| | 5 2 | | | | | | | 6 2 |
| | 5 2 | | | | | | | 6 2 |
| | 5 2 | | | | | | | 6 2 |
| | 6 2 | | | | | | | 5 2 |
| 59.2.6 | 99042 | 100 | Cu2O3Sr1 | | | | | |
| | -GM | -R | -S | -T | -U | -X | -Y | -Z |
| | 5 2 | 3+4 4 | 3+4 4 | 3+4 4 | 3+4 4 | 3+4 4 | 3+4 4 | 5 2 |
| | 6 2 | 3+4 4 | 3+4 4 | 3+4 4 | 3+4 4 | 3+4 4 | 3+4 4 | 6 2 |
| | 5 2 | 3+4 4 | 3+4 4 | 3+4 4 | 3+4 4 | 3+4 4 | 3+4 4 | 5 2 |
| | 5 2 | 3+4 4 | 3+4 4 | 3+4 4 | 3+4 4 | 3+4 4 | 3+4 4 | 6 2 |
| | 6 2 | 3+4 4 | 3+4 4 | 3+4 4 | 3+4 4 | 3+4 4 | 3+4 4 | 5 2 |
| | 6 2 | 3+4 4 | 3+4 4 | 3+4 4 | 3+4 4 | 3+4 4 | 3+4 4 | 6 2 |
| | 6 2 | 3+4 4 | 3+4 4 | 3+4 4 | 3+4 4 | 3+4 4 | 3+4 4 | 5 2 |
| | 5 2 | 3+4 4 | 3+4 4 | 3+4 4 | 3+4 4 | 3+4 4 | 3+4 4 | 6 2 |
| | 5 2 | 3+4 4 | 3+4 4 | 3+4 4 | 3+4 4 | 3+4 4 | 3+4 4 | 6 2 |
| | 6 2 | 3+4 4 | 3+4 4 | 3+4 4 | 3+4 4 | 3+4 4 | 3+4 4 | 5 2 |
| | 5 2 | 3+4 4 | 3+4 4 | 3+4 4 | 3+4 4 | 3+4 4 | 3+4 4 | 6 2 |
| | 6 2 | 3+4 4 | 3+4 4 | 3+4 4 | 3+4 4 | 3+4 4 | 3+4 4 | 5 2 |
| | 6 2 | 3+4 4 | 3+4 4 | 3+4 4 | 3+4 4 | 3+4 4 | 3+4 4 | 6 2 |
| | 5 2 | 3+4 4 | 3+4 4 | 3+4 4 | 3+4 4 | 3+4 4 | 3+4 4 | 5 2 |
| | 5 2 | 3+4 4 | 3+4 4 | 3+4 4 | 3+4 4 | 3+4 4 | 3+4 4 | 5 2 |
| | 5 2 | 3+4 4 | 3+4 4 | 3+4 4 | 3+4 4 | 3+4 4 | 3+4 4 | 6 2 |
| | 6 2 | 3+4 4 | 3+4 4 | 3+4 4 | 3+4 4 | 3+4 4 | 3+4 4 | 5 2 |
| | 5 2 | 3+4 4 | 3+4 4 | 3+4 4 | 3+4 4 | 3+4 4 | 3+4 4 | 6 2 |
| | 5 2 | 3+4 4 | 3+4 4 | 3+4 4 | 3+4 4 | 3+4 4 | 3+4 4 | 6 2 |
| | 6 2 | 3+4 4 | 3+4 4 | 3+4 4 | 3+4 4 | 3+4 4 | 3+4 4 | 5 2 |
| | 5 2 | 3+4 4 | 3+4 4 | 3+4 4 | 3+4 4 | 3+4 4 | 3+4 4 | 6 2 |
| | 5 2 | 3+4 4 | 3+4 4 | 3+4 4 | 3+4 4 | 3+4 4 | 3+4 4 | 5 2 |
| | 5 2 | 3+4 4 | 3+4 4 | 3+4 4 | 3+4 4 | 3+4 4 | 3+4 4 | 6 2 |
| | 6 2 | 3+4 4 | 3+4 4 | 3+4 4 | 3+4 4 | 3+4 4 | 3+4 4 | 5 2 |
| | 6 2 | 3+4 4 | 3+4 4 | 3+4 4 | 3+4 4 | 3+4 4 | 3+4 4 | 5 2 |
| | 6 2 | | | | | | | 6 2 |
| | 6 2 | | | | | | | 5 2 |
| | 6 2 | | | | | | | 5 2 |
| | 5 2 | | | | | | | 6 2 |
| | 5 2 | | | | | | | 6 2 |
| | 6 2 | | | | | | | 5 2 |
| | 5 2 | | | | | | | 5 2 |
| | 5 2 | | | | | | | 6 2 |
| | 5 2 | | | | | | | 6 2 |
| | 6 2 | | | | | | | 6 2 |
| | 6 2 | | | | | | | 6 2 |
| | 6 2 | | | | | | | 5 2 |
| | 6 2 | | | | | | | 5 2 |
| | 5 2 | | | | | | | 6 2 |
| | 5 2 | | | | | | | 5 2 |
| | 5 2 | | | | | | | 6 2 |
| | 6 2 | | | | | | | 5 2 |
| | 6 2 | | | | | | | 6 2 |
| | 5 2 | | | | | | | 5 2 |
| | 6 2 | | | | | | | 6 2 |

Continued on next column

**TABLE XXXIV– continued**

| TQC | ICSD | NoE | SF | | | | |
|---|---|---|---|---|---|---|---|
| | 5 2 | | | | | | 5 2 |
| | 6 2 | | | | | | 6 2 |
| | 5 2 | | | | | | 5 2 |
| | 6 2 | | | | | | 6 2 |
| | 5 2 | | | | | | 5 2 |
| 59.2.7 | 155492 | 172 | Co1Na0.5O2 | | | | |
| | -GM -R | -S | -T | -U | -X | -Y | -Z |
| | 5 2 3+4 4 | 3+4 4 | 3+4 4 | 3+4 4 | 3+4 4 | 3+4 4 | 5 2 |
| | 5 2 3+4 4 | 3+4 4 | 3+4 4 | 3+4 4 | 3+4 4 | 3+4 4 | 5 2 |
| | 6 2 3+4 4 | 3+4 4 | 3+4 4 | 3+4 4 | 3+4 4 | 3+4 4 | 6 2 |
| | 5 2 3+4 4 | 3+4 4 | 3+4 4 | 3+4 4 | 3+4 4 | 3+4 4 | 6 2 |
| | 6 2 3+4 4 | 3+4 4 | 3+4 4 | 3+4 4 | 3+4 4 | 3+4 4 | 5 2 |
| | 5 2 3+4 4 | 3+4 4 | 3+4 4 | 3+4 4 | 3+4 4 | 3+4 4 | 5 2 |
| | 5 2 3+4 4 | 3+4 4 | 3+4 4 | 3+4 4 | 3+4 4 | 3+4 4 | 5 2 |
| | 5 2 3+4 4 | 3+4 4 | 3+4 4 | 3+4 4 | 3+4 4 | 3+4 4 | 6 2 |
| | 6 2 3+4 4 | 3+4 4 | 3+4 4 | 3+4 4 | 3+4 4 | 3+4 4 | 6 2 |
| | 6 2 3+4 4 | 3+4 4 | 3+4 4 | 3+4 4 | 3+4 4 | 3+4 4 | 5 2 |
| | 6 2 3+4 4 | 3+4 4 | 3+4 4 | 3+4 4 | 3+4 4 | 3+4 4 | 6 2 |
| | 6 2 3+4 4 | 3+4 4 | 3+4 4 | 3+4 4 | 3+4 4 | 3+4 4 | 6 2 |
| | 6 2 3+4 4 | 3+4 4 | 3+4 4 | 3+4 4 | 3+4 4 | 3+4 4 | 6 2 |
| | 5 2 3+4 4 | 3+4 4 | 3+4 4 | 3+4 4 | 3+4 4 | 3+4 4 | 6 2 |
| | 6 2 3+4 4 | 3+4 4 | 3+4 4 | 3+4 4 | 3+4 4 | 3+4 4 | 5 2 |
| | 5 2 3+4 4 | 3+4 4 | 3+4 4 | 3+4 4 | 3+4 4 | 3+4 4 | 5 2 |
| | 5 2 3+4 4 | 3+4 4 | 3+4 4 | 3+4 4 | 3+4 4 | 3+4 4 | 6 2 |
| | 5 2 3+4 4 | 3+4 4 | 3+4 4 | 3+4 4 | 3+4 4 | 3+4 4 | 6 2 |
| | 5 2 3+4 4 | 3+4 4 | 3+4 4 | 3+4 4 | 3+4 4 | 3+4 4 | 6 2 |
| | 6 2 3+4 4 | 3+4 4 | 3+4 4 | 3+4 4 | 3+4 4 | 3+4 4 | 6 2 |
| | 6 2 3+4 4 | 3+4 4 | 3+4 4 | 3+4 4 | 3+4 4 | 3+4 4 | 5 2 |
| | 5 2 3+4 4 | 3+4 4 | 3+4 4 | 3+4 4 | 3+4 4 | 3+4 4 | 6 2 |
| | 5 2 3+4 4 | 3+4 4 | 3+4 4 | 3+4 4 | 3+4 4 | 3+4 4 | 6 2 |
| | 5 2 3+4 4 | 3+4 4 | 3+4 4 | 3+4 4 | 3+4 4 | 3+4 4 | 5 2 |
| | 5 2 3+4 4 | 3+4 4 | 3+4 4 | 3+4 4 | 3+4 4 | 3+4 4 | 5 2 |
| | 6 2 3+4 4 | 3+4 4 | 3+4 4 | 3+4 4 | 3+4 4 | 3+4 4 | 6 2 |
| | 6 2 3+4 4 | 3+4 4 | 3+4 4 | 3+4 4 | 3+4 4 | 3+4 4 | 5 2 |
| | 5 2 3+4 4 | 3+4 4 | 3+4 4 | 3+4 4 | 3+4 4 | 3+4 4 | 5 2 |
| | 6 2 3+4 4 | 3+4 4 | 3+4 4 | 3+4 4 | 3+4 4 | 3+4 4 | 5 2 |
| | 5 2 3+4 4 | 3+4 4 | 3+4 4 | 3+4 4 | 3+4 4 | 3+4 4 | 5 2 |
| | 5 2 3+4 4 | 3+4 4 | 3+4 4 | 3+4 4 | 3+4 4 | 3+4 4 | 6 2 |
| | 6 2 3+4 4 | 3+4 4 | 3+4 4 | 3+4 4 | 3+4 4 | 3+4 4 | 5 2 |
| | 5 2 3+4 4 | 3+4 4 | 3+4 4 | 3+4 4 | 3+4 4 | 3+4 4 | 5 2 |
| | 5 2 3+4 4 | 3+4 4 | 3+4 4 | 3+4 4 | 3+4 4 | 3+4 4 | 5 2 |
| | 6 2 3+4 4 | 3+4 4 | 3+4 4 | 3+4 4 | 3+4 4 | 3+4 4 | 6 2 |
| | 6 2 3+4 4 | 3+4 4 | 3+4 4 | 3+4 4 | 3+4 4 | 3+4 4 | 6 2 |
| | 5 2 3+4 4 | 3+4 4 | 3+4 4 | 3+4 4 | 3+4 4 | 3+4 4 | 6 2 |
| | 5 2 3+4 4 | 3+4 4 | 3+4 4 | 3+4 4 | 3+4 4 | 3+4 4 | 6 2 |
| | 5 2 3+4 4 | 3+4 4 | 3+4 4 | 3+4 4 | 3+4 4 | 3+4 4 | 6 2 |
| | 5 2 3+4 4 | 3+4 4 | 3+4 4 | 3+4 4 | 3+4 4 | 3+4 4 | 6 2 |
| | 5 2 3+4 4 | 3+4 4 | 3+4 4 | 3+4 4 | 3+4 4 | 3+4 4 | 5 2 |
| | 5 2 3+4 4 | 3+4 4 | 3+4 4 | 3+4 4 | 3+4 4 | 3+4 4 | 6 2 |
| | 6 2 3+4 4 | 3+4 4 | 3+4 4 | 3+4 4 | 3+4 4 | 3+4 4 | 5 2 |
| | 5 2 | | | | | | 6 2 |
| | 6 2 | | | | | | 6 2 |
| | 5 2 | | | | | | 6 2 |
| | 5 2 | | | | | | 6 2 |
| | 6 2 | | | | | | 6 2 |
| | 5 2 | | | | | | 5 2 |
| | 6 2 | | | | | | 5 2 |
| | 6 2 | | | | | | 5 2 |
| | 6 2 | | | | | | 5 2 |
| | 6 2 | | | | | | 5 2 |
| | 5 2 | | | | | | 6 2 |

Continued on next column

**TABLE XXXIV– continued**

| TQC | ICSD | NoE | SF | | | | |
|---|---|---|---|---|---|---|---|
| | 6 2 | | | | | | 6 2 |
| | 6 2 | | | | | | 5 2 |
| | 5 2 | | | | | | 5 2 |
| | 6 2 | | | | | | 5 2 |
| | 6 2 | | | | | | 5 2 |
| | 6 2 | | | | | | 5 2 |
| | 6 2 | | | | | | 6 2 |
| | 6 2 | | | | | | 5 2 |
| | 6 2 | | | | | | 6 2 |
| | 6 2 | | | | | | 5 2 |
| | 5 2 | | | | | | 6 2 |
| | 5 2 | | | | | | 6 2 |
| | 5 2 | | | | | | 6 2 |
| | 5 2 | | | | | | 6 2 |
| | 5 2 | | | | | | 5 2 |
| | 5 2 | | | | | | 6 2 |
| | 5 2 | | | | | | 5 2 |
| | 5 2 | | | | | | 6 2 |
| | 6 2 | | | | | | 6 2 |
| | 6 2 | | | | | | 6 2 |
| | 5 2 | | | | | | 6 2 |
| | 5 2 | | | | | | 6 2 |
| | 5 2 | | | | | | 6 2 |
| | 5 2 | | | | | | 6 2 |
| | 6 2 | | | | | | 5 2 |
| | 6 2 | | | | | | 5 2 |
| | 5 2 | | | | | | 6 2 |
| | 5 2 | | | | | | 6 2 |
| | 5 2 | | | | | | 6 2 |
| | 6 2 | | | | | | 6 2 |
| | 5 2 | | | | | | 5 2 |
| | 6 2 | | | | | | 6 2 |
| 59.2.8 | 413690 | 108 | Br3Mo1 | | | | |
| | -GM -R | -S | -T | -U | -X | -Y | -Z |
| | 5 2 3+4 4 | 3+4 4 | 3+4 4 | 3+4 4 | 3+4 4 | 3+4 4 | 6 2 |
| | 6 2 3+4 4 | 3+4 4 | 3+4 4 | 3+4 4 | 3+4 4 | 3+4 4 | 5 2 |
| | 5 2 3+4 4 | 3+4 4 | 3+4 4 | 3+4 4 | 3+4 4 | 3+4 4 | 5 2 |
| | 5 2 3+4 4 | 3+4 4 | 3+4 4 | 3+4 4 | 3+4 4 | 3+4 4 | 5 2 |
| | 6 2 3+4 4 | 3+4 4 | 3+4 4 | 3+4 4 | 3+4 4 | 3+4 4 | 6 2 |
| | 5 2 3+4 4 | 3+4 4 | 3+4 4 | 3+4 4 | 3+4 4 | 3+4 4 | 6 2 |
| | 6 2 3+4 4 | 3+4 4 | 3+4 4 | 3+4 4 | 3+4 4 | 3+4 4 | 6 2 |
| | 5 2 3+4 4 | 3+4 4 | 3+4 4 | 3+4 4 | 3+4 4 | 3+4 4 | 5 2 |
| | 6 2 3+4 4 | 3+4 4 | 3+4 4 | 3+4 4 | 3+4 4 | 3+4 4 | 6 2 |
| | 6 2 3+4 4 | 3+4 4 | 3+4 4 | 3+4 4 | 3+4 4 | 3+4 4 | 5 2 |
| | 6 2 3+4 4 | 3+4 4 | 3+4 4 | 3+4 4 | 3+4 4 | 3+4 4 | 5 2 |
| | 5 2 3+4 4 | 3+4 4 | 3+4 4 | 3+4 4 | 3+4 4 | 3+4 4 | 6 2 |
| | 5 2 3+4 4 | 3+4 4 | 3+4 4 | 3+4 4 | 3+4 4 | 3+4 4 | 6 2 |
| | 6 2 3+4 4 | 3+4 4 | 3+4 4 | 3+4 4 | 3+4 4 | 3+4 4 | 5 2 |
| | 5 2 3+4 4 | 3+4 4 | 3+4 4 | 3+4 4 | 3+4 4 | 3+4 4 | 5 2 |
| | 5 2 3+4 4 | 3+4 4 | 3+4 4 | 3+4 4 | 3+4 4 | 3+4 4 | 6 2 |
| | 6 2 3+4 4 | 3+4 4 | 3+4 4 | 3+4 4 | 3+4 4 | 3+4 4 | 5 2 |
| | 5 2 3+4 4 | 3+4 4 | 3+4 4 | 3+4 4 | 3+4 4 | 3+4 4 | 6 2 |
| | 6 2 3+4 4 | 3+4 4 | 3+4 4 | 3+4 4 | 3+4 4 | 3+4 4 | 5 2 |
| | 6 2 3+4 4 | 3+4 4 | 3+4 4 | 3+4 4 | 3+4 4 | 3+4 4 | 6 2 |
| | 5 2 3+4 4 | 3+4 4 | 3+4 4 | 3+4 4 | 3+4 4 | 3+4 4 | 5 2 |
| | 5 2 3+4 4 | 3+4 4 | 3+4 4 | 3+4 4 | 3+4 4 | 3+4 4 | 6 2 |
| | 5 2 3+4 4 | 3+4 4 | 3+4 4 | 3+4 4 | 3+4 4 | 3+4 4 | 6 2 |
| | 5 2 3+4 4 | 3+4 4 | 3+4 4 | 3+4 4 | 3+4 4 | 3+4 4 | 5 2 |
| | 6 2 3+4 4 | 3+4 4 | 3+4 4 | 3+4 4 | 3+4 4 | 3+4 4 | 6 2 |
| | 6 2 3+4 4 | 3+4 4 | 3+4 4 | 3+4 4 | 3+4 4 | 3+4 4 | 6 2 |
| | 6 2 3+4 4 | 3+4 4 | 3+4 4 | 3+4 4 | 3+4 4 | 3+4 4 | 5 2 |

Continued on next column

**TABLE XXXIV– continued**

| TQC | ICSD | NoE | SF | | | | | |
|---|---|---|---|---|---|---|---|---|
| | 6 2 | | | | | | | 5 2 |
| | 6 2 | | | | | | | 6 2 |
| | 5 2 | | | | | | | 5 2 |
| | 6 2 | | | | | | | 5 2 |
| | 5 2 | | | | | | | 6 2 |
| | 6 2 | | | | | | | 5 2 |
| | 5 2 | | | | | | | 6 2 |
| | 6 2 | | | | | | | 5 2 |
| | 5 2 | | | | | | | 6 2 |
| | 6 2 | | | | | | | 5 2 |
| | 6 2 | | | | | | | 6 2 |
| | 5 2 | | | | | | | 6 2 |
| | 5 2 | | | | | | | 5 2 |
| | 5 2 | | | | | | | 5 2 |
| | 5 2 | | | | | | | 6 2 |
| | 5 2 | | | | | | | 6 2 |
| | 6 2 | | | | | | | 5 2 |
| | 5 2 | | | | | | | 5 2 |
| | 6 2 | | | | | | | 6 2 |
| | 6 2 | | | | | | | 5 2 |
| | 6 2 | | | | | | | 6 2 |
| | 6 2 | | | | | | | 5 2 |
| | 5 2 | | | | | | | 6 2 |
| | 6 2 | | | | | | | 6 2 |
| | 5 2 | | | | | | | 5 2 |
| | 6 2 | | | | | | | 6 2 |
| | 6 2 | | | | | | | 5 2 |
| 59.2.9 | 249645 | 108 | Au3K1Sn3 | | | | | |
| | -GM | -R | -S | -T | -U | -X | -Y | -Z |
| | 5 2 | 3+4 4 | 3+4 4 | 3+4 4 | 3+4 4 | 3+4 4 | 3+4 4 | 5 2 |
| | 6 2 | 3+4 4 | 3+4 4 | 3+4 4 | 3+4 4 | 3+4 4 | 3+4 4 | 6 2 |
| | 5 2 | 3+4 4 | 3+4 4 | 3+4 4 | 3+4 4 | 3+4 4 | 3+4 4 | 5 2 |
| | 6 2 | 3+4 4 | 3+4 4 | 3+4 4 | 3+4 4 | 3+4 4 | 3+4 4 | 6 2 |
| | 5 2 | 3+4 4 | 3+4 4 | 3+4 4 | 3+4 4 | 3+4 4 | 3+4 4 | 5 2 |
| | 6 2 | 3+4 4 | 3+4 4 | 3+4 4 | 3+4 4 | 3+4 4 | 3+4 4 | 6 2 |
| | 5 2 | 3+4 4 | 3+4 4 | 3+4 4 | 3+4 4 | 3+4 4 | 3+4 4 | 5 2 |
| | 6 2 | 3+4 4 | 3+4 4 | 3+4 4 | 3+4 4 | 3+4 4 | 3+4 4 | 6 2 |
| | 5 2 | 3+4 4 | 3+4 4 | 3+4 4 | 3+4 4 | 3+4 4 | 3+4 4 | 5 2 |
| | 6 2 | 3+4 4 | 3+4 4 | 3+4 4 | 3+4 4 | 3+4 4 | 3+4 4 | 6 2 |
| | 5 2 | 3+4 4 | 3+4 4 | 3+4 4 | 3+4 4 | 3+4 4 | 3+4 4 | 5 2 |
| | 5 2 | 3+4 4 | 3+4 4 | 3+4 4 | 3+4 4 | 3+4 4 | 3+4 4 | 6 2 |
| | 6 2 | 3+4 4 | 3+4 4 | 3+4 4 | 3+4 4 | 3+4 4 | 3+4 4 | 5 2 |
| | 6 2 | 3+4 4 | 3+4 4 | 3+4 4 | 3+4 4 | 3+4 4 | 3+4 4 | 6 2 |
| | 5 2 | 3+4 4 | 3+4 4 | 3+4 4 | 3+4 4 | 3+4 4 | 3+4 4 | 5 2 |
| | 5 2 | 3+4 4 | 3+4 4 | 3+4 4 | 3+4 4 | 3+4 4 | 3+4 4 | 6 2 |
| | 6 2 | 3+4 4 | 3+4 4 | 3+4 4 | 3+4 4 | 3+4 4 | 3+4 4 | 5 2 |
| | 6 2 | 3+4 4 | 3+4 4 | 3+4 4 | 3+4 4 | 3+4 4 | 3+4 4 | 5 2 |
| | 5 2 | 3+4 4 | 3+4 4 | 3+4 4 | 3+4 4 | 3+4 4 | 3+4 4 | 6 2 |
| | 6 2 | 3+4 4 | 3+4 4 | 3+4 4 | 3+4 4 | 3+4 4 | 3+4 4 | 5 2 |
| | 5 2 | 3+4 4 | 3+4 4 | 3+4 4 | 3+4 4 | 3+4 4 | 3+4 4 | 6 2 |
| | 5 2 | 3+4 4 | 3+4 4 | 3+4 4 | 3+4 4 | 3+4 4 | 3+4 4 | 6 2 |
| | 5 2 | 3+4 4 | 3+4 4 | 3+4 4 | 3+4 4 | 3+4 4 | 3+4 4 | 6 2 |
| | 6 2 | 3+4 4 | 3+4 4 | 3+4 4 | 3+4 4 | 3+4 4 | 3+4 4 | 5 2 |
| | 6 2 | 3+4 4 | 3+4 4 | 3+4 4 | 3+4 4 | 3+4 4 | 3+4 4 | 5 2 |
| | 5 2 | 3+4 4 | 3+4 4 | 3+4 4 | 3+4 4 | 3+4 4 | 3+4 4 | 6 2 |
| | 6 2 | 3+4 4 | 3+4 4 | 3+4 4 | 3+4 4 | 3+4 4 | 3+4 4 | 5 2 |
| | 6 2 | | | | | | | 6 2 |
| | 5 2 | | | | | | | 5 2 |
| | 5 2 | | | | | | | 6 2 |
| | 6 2 | | | | | | | 5 2 |
| | 5 2 | | | | | | | 6 2 |

Continued on next column

**TABLE XXXIV– continued**

| TQC | ICSD | NoE | SF | | | | | |
|---|---|---|---|---|---|---|---|---|
| | 6 2 | | | | | | | 6 2 |
| | 6 2 | | | | | | | 5 2 |
| | 5 2 | | | | | | | 5 2 |
| | 6 2 | | | | | | | 6 2 |
| | 5 2 | | | | | | | 5 2 |
| | 5 2 | | | | | | | 6 2 |
| | 6 2 | | | | | | | 5 2 |
| | 5 2 | | | | | | | 6 2 |
| | 6 2 | | | | | | | 5 2 |
| | 6 2 | | | | | | | 6 2 |
| | 6 2 | | | | | | | 5 2 |
| | 5 2 | | | | | | | 6 2 |
| | 6 2 | | | | | | | 6 2 |
| | 5 2 | | | | | | | 5 2 |
| | 5 2 | | | | | | | 6 2 |
| | 5 2 | | | | | | | 5 2 |
| | 5 2 | | | | | | | 5 2 |
| | 6 2 | | | | | | | 6 2 |
| | 5 2 | | | | | | | 5 2 |
| | 6 2 | | | | | | | 6 2 |
| | 6 2 | | | | | | | 5 2 |
| | 5 2 | | | | | | | 6 2 |
| 59.2.10 | 245679 | 104 | Au3In3Sr1 | | | | | |
| | -GM | -R | -S | -T | -U | -X | -Y | -Z |
| | 5 2 | 3+4 4 | 3+4 4 | 3+4 4 | 3+4 4 | 3+4 4 | 3+4 4 | 5 2 |
| | 6 2 | 3+4 4 | 3+4 4 | 3+4 4 | 3+4 4 | 3+4 4 | 3+4 4 | 6 2 |
| | 5 2 | 3+4 4 | 3+4 4 | 3+4 4 | 3+4 4 | 3+4 4 | 3+4 4 | 5 2 |
| | 6 2 | 3+4 4 | 3+4 4 | 3+4 4 | 3+4 4 | 3+4 4 | 3+4 4 | 6 2 |
| | 5 2 | 3+4 4 | 3+4 4 | 3+4 4 | 3+4 4 | 3+4 4 | 3+4 4 | 6 2 |
| | 6 2 | 3+4 4 | 3+4 4 | 3+4 4 | 3+4 4 | 3+4 4 | 3+4 4 | 5 2 |
| | 5 2 | 3+4 4 | 3+4 4 | 3+4 4 | 3+4 4 | 3+4 4 | 3+4 4 | 5 2 |
| | 6 2 | 3+4 4 | 3+4 4 | 3+4 4 | 3+4 4 | 3+4 4 | 3+4 4 | 6 2 |
| | 5 2 | 3+4 4 | 3+4 4 | 3+4 4 | 3+4 4 | 3+4 4 | 3+4 4 | 5 2 |
| | 5 2 | 3+4 4 | 3+4 4 | 3+4 4 | 3+4 4 | 3+4 4 | 3+4 4 | 6 2 |
| | 6 2 | 3+4 4 | 3+4 4 | 3+4 4 | 3+4 4 | 3+4 4 | 3+4 4 | 6 2 |
| | 5 2 | 3+4 4 | 3+4 4 | 3+4 4 | 3+4 4 | 3+4 4 | 3+4 4 | 5 2 |
| | 6 2 | 3+4 4 | 3+4 4 | 3+4 4 | 3+4 4 | 3+4 4 | 3+4 4 | 6 2 |
| | 5 2 | 3+4 4 | 3+4 4 | 3+4 4 | 3+4 4 | 3+4 4 | 3+4 4 | 5 2 |
| | 6 2 | 3+4 4 | 3+4 4 | 3+4 4 | 3+4 4 | 3+4 4 | 3+4 4 | 5 2 |
| | 5 2 | 3+4 4 | 3+4 4 | 3+4 4 | 3+4 4 | 3+4 4 | 3+4 4 | 6 2 |
| | 6 2 | 3+4 4 | 3+4 4 | 3+4 4 | 3+4 4 | 3+4 4 | 3+4 4 | 5 2 |
| | 6 2 | 3+4 4 | 3+4 4 | 3+4 4 | 3+4 4 | 3+4 4 | 3+4 4 | 6 2 |
| | 5 2 | 3+4 4 | 3+4 4 | 3+4 4 | 3+4 4 | 3+4 4 | 3+4 4 | 5 2 |
| | 5 2 | 3+4 4 | 3+4 4 | 3+4 4 | 3+4 4 | 3+4 4 | 3+4 4 | 6 2 |
| | 6 2 | 3+4 4 | 3+4 4 | 3+4 4 | 3+4 4 | 3+4 4 | 3+4 4 | 5 2 |
| | 5 2 | 3+4 4 | 3+4 4 | 3+4 4 | 3+4 4 | 3+4 4 | 3+4 4 | 6 2 |
| | 5 2 | 3+4 4 | 3+4 4 | 3+4 4 | 3+4 4 | 3+4 4 | 3+4 4 | 6 2 |
| | 6 2 | 3+4 4 | 3+4 4 | 3+4 4 | 3+4 4 | 3+4 4 | 3+4 4 | 5 2 |
| | 6 2 | 3+4 4 | 3+4 4 | 3+4 4 | 3+4 4 | 3+4 4 | 3+4 4 | 5 2 |
| | 5 2 | 3+4 4 | 3+4 4 | 3+4 4 | 3+4 4 | 3+4 4 | 3+4 4 | 6 2 |
| | 6 2 | | | | | | | 5 2 |
| | 6 2 | | | | | | | 6 2 |
| | 5 2 | | | | | | | 5 2 |
| | 5 2 | | | | | | | 6 2 |
| | 6 2 | | | | | | | 5 2 |
| | 5 2 | | | | | | | 6 2 |
| | 6 2 | | | | | | | 5 2 |
| | 6 2 | | | | | | | 6 2 |
| | 5 2 | | | | | | | 5 2 |
| | 5 2 | | | | | | | 6 2 |
| | 6 2 | | | | | | | 5 2 |

Continued on next column

**TABLE XXXIV– continued**

| TQC | ICSD | NoE | SF | | | | | |
|---|---|---|---|---|---|---|---|---|
| | 5 2 | | | | | | | 6 2 |
| | 6 2 | | | | | | | 5 2 |
| | 6 2 | | | | | | | 6 2 |
| | 5 2 | | | | | | | 5 2 |
| | 6 2 | | | | | | | 6 2 |
| | 5 2 | | | | | | | 5 2 |
| | 5 2 | | | | | | | 6 2 |
| | 6 2 | | | | | | | 6 2 |
| | 6 2 | | | | | | | 5 2 |
| | 5 2 | | | | | | | 5 2 |
| | 5 2 | | | | | | | 5 2 |
| | 5 2 | | | | | | | 6 2 |
| | 5 2 | | | | | | | 6 2 |
| | 6 2 | | | | | | | 5 2 |
| | 6 2 | | | | | | | 6 2 |
| 59.2.11 | 107445 | 108 | Au1Rb3Sn4 | | | | | |
| | -GM | -R | -S | -T | -U | -X | -Y | -Z |
| | 5 2 | 3+4 4 | 3+4 4 | 3+4 4 | 3+4 4 | 3+4 4 | 3+4 4 | 5 2 |
| | 5 2 | 3+4 4 | 3+4 4 | 3+4 4 | 3+4 4 | 3+4 4 | 3+4 4 | 5 2 |
| | 6 2 | 3+4 4 | 3+4 4 | 3+4 4 | 3+4 4 | 3+4 4 | 3+4 4 | 6 2 |
| | 6 2 | 3+4 4 | 3+4 4 | 3+4 4 | 3+4 4 | 3+4 4 | 3+4 4 | 6 2 |
| | 5 2 | 3+4 4 | 3+4 4 | 3+4 4 | 3+4 4 | 3+4 4 | 3+4 4 | 5 2 |
| | 6 2 | 3+4 4 | 3+4 4 | 3+4 4 | 3+4 4 | 3+4 4 | 3+4 4 | 6 2 |
| | 5 2 | 3+4 4 | 3+4 4 | 3+4 4 | 3+4 4 | 3+4 4 | 3+4 4 | 6 2 |
| | 6 2 | 3+4 4 | 3+4 4 | 3+4 4 | 3+4 4 | 3+4 4 | 3+4 4 | 5 2 |
| | 5 2 | 3+4 4 | 3+4 4 | 3+4 4 | 3+4 4 | 3+4 4 | 3+4 4 | 5 2 |
| | 6 2 | 3+4 4 | 3+4 4 | 3+4 4 | 3+4 4 | 3+4 4 | 3+4 4 | 6 2 |
| | 5 2 | 3+4 4 | 3+4 4 | 3+4 4 | 3+4 4 | 3+4 4 | 3+4 4 | 5 2 |
| | 5 2 | 3+4 4 | 3+4 4 | 3+4 4 | 3+4 4 | 3+4 4 | 3+4 4 | 5 2 |
| | 6 2 | 3+4 4 | 3+4 4 | 3+4 4 | 3+4 4 | 3+4 4 | 3+4 4 | 6 2 |
| | 6 2 | 3+4 4 | 3+4 4 | 3+4 4 | 3+4 4 | 3+4 4 | 3+4 4 | 6 2 |
| | 5 2 | 3+4 4 | 3+4 4 | 3+4 4 | 3+4 4 | 3+4 4 | 3+4 4 | 5 2 |
| | 5 2 | 3+4 4 | 3+4 4 | 3+4 4 | 3+4 4 | 3+4 4 | 3+4 4 | 5 2 |
| | 6 2 | 3+4 4 | 3+4 4 | 3+4 4 | 3+4 4 | 3+4 4 | 3+4 4 | 6 2 |
| | 6 2 | 3+4 4 | 3+4 4 | 3+4 4 | 3+4 4 | 3+4 4 | 3+4 4 | 6 2 |
| | 6 2 | 3+4 4 | 3+4 4 | 3+4 4 | 3+4 4 | 3+4 4 | 3+4 4 | 5 2 |
| | 5 2 | 3+4 4 | 3+4 4 | 3+4 4 | 3+4 4 | 3+4 4 | 3+4 4 | 6 2 |
| | 5 2 | 3+4 4 | 3+4 4 | 3+4 4 | 3+4 4 | 3+4 4 | 3+4 4 | 5 2 |
| | 6 2 | 3+4 4 | 3+4 4 | 3+4 4 | 3+4 4 | 3+4 4 | 3+4 4 | 6 2 |
| | 6 2 | 3+4 4 | 3+4 4 | 3+4 4 | 3+4 4 | 3+4 4 | 3+4 4 | 5 2 |
| | 5 2 | 3+4 4 | 3+4 4 | 3+4 4 | 3+4 4 | 3+4 4 | 3+4 4 | 6 2 |
| | 5 2 | 3+4 4 | 3+4 4 | 3+4 4 | 3+4 4 | 3+4 4 | 3+4 4 | 5 2 |
| | 6 2 | 3+4 4 | 3+4 4 | 3+4 4 | 3+4 4 | 3+4 4 | 3+4 4 | 6 2 |
| | 6 2 | 3+4 4 | 3+4 4 | 3+4 4 | 3+4 4 | 3+4 4 | 3+4 4 | 6 2 |
| | 5 2 | | | | | | | 5 2 |
| | 6 2 | | | | | | | 6 2 |
| | 5 2 | | | | | | | 5 2 |
| | 5 2 | | | | | | | 5 2 |
| | 6 2 | | | | | | | 6 2 |
| | 5 2 | | | | | | | 6 2 |
| | 6 2 | | | | | | | 5 2 |
| | 5 2 | | | | | | | 5 2 |
| | 6 2 | | | | | | | 6 2 |
| | 5 2 | | | | | | | 6 2 |
| | 6 2 | | | | | | | 5 2 |
| | 5 2 | | | | | | | 6 2 |
| | 6 2 | | | | | | | 5 2 |
| | 5 2 | | | | | | | 5 2 |
| | 6 2 | | | | | | | 6 2 |
| | 6 2 | | | | | | | 6 2 |
| | 5 2 | | | | | | | 5 2 |

Continued on next column

**TABLE XXXIV– continued**

| TQC | ICSD | NoE | SF | | | | | |
|---|---|---|---|---|---|---|---|---|
| | 6 2 | | | | | | | 5 2 |
| | 5 2 | | | | | | | 6 2 |
| | 6 2 | | | | | | | 5 2 |
| | 5 2 | | | | | | | 6 2 |
| | 6 2 | | | | | | | 6 2 |
| | 5 2 | | | | | | | 6 2 |
| | 5 2 | | | | | | | 5 2 |
| | 6 2 | | | | | | | 5 2 |
| | 5 2 | | | | | | | 6 2 |
| | 5 2 | | | | | | | 5 2 |
| 59.2.12 | 107447 | 108 | Au1Pb4Rb3 | | | | | |
| | -GM | -R | -S | -T | -U | -X | -Y | -Z |
| | 5 2 | 3+4 4 | 3+4 4 | 3+4 4 | 3+4 4 | 3+4 4 | 3+4 4 | 5 2 |
| | 5 2 | 3+4 4 | 3+4 4 | 3+4 4 | 3+4 4 | 3+4 4 | 3+4 4 | 5 2 |
| | 6 2 | 3+4 4 | 3+4 4 | 3+4 4 | 3+4 4 | 3+4 4 | 3+4 4 | 6 2 |
| | 6 2 | 3+4 4 | 3+4 4 | 3+4 4 | 3+4 4 | 3+4 4 | 3+4 4 | 6 2 |
| | 5 2 | 3+4 4 | 3+4 4 | 3+4 4 | 3+4 4 | 3+4 4 | 3+4 4 | 5 2 |
| | 6 2 | 3+4 4 | 3+4 4 | 3+4 4 | 3+4 4 | 3+4 4 | 3+4 4 | 6 2 |
| | 6 2 | 3+4 4 | 3+4 4 | 3+4 4 | 3+4 4 | 3+4 4 | 3+4 4 | 6 2 |
| | 5 2 | 3+4 4 | 3+4 4 | 3+4 4 | 3+4 4 | 3+4 4 | 3+4 4 | 5 2 |
| | 6 2 | 3+4 4 | 3+4 4 | 3+4 4 | 3+4 4 | 3+4 4 | 3+4 4 | 6 2 |
| | 5 2 | 3+4 4 | 3+4 4 | 3+4 4 | 3+4 4 | 3+4 4 | 3+4 4 | 5 2 |
| | 5 2 | 3+4 4 | 3+4 4 | 3+4 4 | 3+4 4 | 3+4 4 | 3+4 4 | 5 2 |
| | 5 2 | 3+4 4 | 3+4 4 | 3+4 4 | 3+4 4 | 3+4 4 | 3+4 4 | 5 2 |
| | 6 2 | 3+4 4 | 3+4 4 | 3+4 4 | 3+4 4 | 3+4 4 | 3+4 4 | 6 2 |
| | 6 2 | 3+4 4 | 3+4 4 | 3+4 4 | 3+4 4 | 3+4 4 | 3+4 4 | 6 2 |
| | 5 2 | 3+4 4 | 3+4 4 | 3+4 4 | 3+4 4 | 3+4 4 | 3+4 4 | 5 2 |
| | 5 2 | 3+4 4 | 3+4 4 | 3+4 4 | 3+4 4 | 3+4 4 | 3+4 4 | 5 2 |
| | 6 2 | 3+4 4 | 3+4 4 | 3+4 4 | 3+4 4 | 3+4 4 | 3+4 4 | 6 2 |
| | 6 2 | 3+4 4 | 3+4 4 | 3+4 4 | 3+4 4 | 3+4 4 | 3+4 4 | 6 2 |
| | 6 2 | 3+4 4 | 3+4 4 | 3+4 4 | 3+4 4 | 3+4 4 | 3+4 4 | 5 2 |
| | 5 2 | 3+4 4 | 3+4 4 | 3+4 4 | 3+4 4 | 3+4 4 | 3+4 4 | 6 2 |
| | 5 2 | 3+4 4 | 3+4 4 | 3+4 4 | 3+4 4 | 3+4 4 | 3+4 4 | 5 2 |
| | 6 2 | 3+4 4 | 3+4 4 | 3+4 4 | 3+4 4 | 3+4 4 | 3+4 4 | 6 2 |
| | 6 2 | 3+4 4 | 3+4 4 | 3+4 4 | 3+4 4 | 3+4 4 | 3+4 4 | 5 2 |
| | 5 2 | 3+4 4 | 3+4 4 | 3+4 4 | 3+4 4 | 3+4 4 | 3+4 4 | 6 2 |
| | 5 2 | 3+4 4 | 3+4 4 | 3+4 4 | 3+4 4 | 3+4 4 | 3+4 4 | 5 2 |
| | 6 2 | 3+4 4 | 3+4 4 | 3+4 4 | 3+4 4 | 3+4 4 | 3+4 4 | 6 2 |
| | 6 2 | 3+4 4 | 3+4 4 | 3+4 4 | 3+4 4 | 3+4 4 | 3+4 4 | 6 2 |
| | 5 2 | | | | | | | 6 2 |
| | 6 2 | | | | | | | 5 2 |
| | 5 2 | | | | | | | 5 2 |
| | 5 2 | | | | | | | 5 2 |
| | 6 2 | | | | | | | 6 2 |
| | 5 2 | | | | | | | 6 2 |
| | 6 2 | | | | | | | 5 2 |
| | 5 2 | | | | | | | 6 2 |
| | 6 2 | | | | | | | 5 2 |
| | 5 2 | | | | | | | 6 2 |
| | 6 2 | | | | | | | 5 2 |
| | 6 2 | | | | | | | 6 2 |
| | 5 2 | | | | | | | 5 2 |
| | 5 2 | | | | | | | 6 2 |
| | 6 2 | | | | | | | 5 2 |
| | 6 2 | | | | | | | 6 2 |
| | 5 2 | | | | | | | 5 2 |
| | 6 2 | | | | | | | 5 2 |
| | 5 2 | | | | | | | 6 2 |
| | 6 2 | | | | | | | 5 2 |
| | 5 2 | | | | | | | 6 2 |
| | 6 2 | | | | | | | 5 2 |

Continued on next column

**TABLE XXXIV– continued**

| TQC | ICSD | NoE | SF | | | | | |
|---|---|---|---|---|---|---|---|---|
| | 5 2 | | | | | | | 6 2 |
| | 5 2 | | | | | | | 6 2 |
| | 6 2 | | | | | | | 5 2 |
| | 5 2 | | | | | | | 5 2 |
| | 5 2 | | | | | | | 6 2 |
| 59.2.13 | 107448 | 108 | Au1Cs3Pb4 | | | | | |
| | -GM | -R | -S | -T | -U | -X | -Y | -Z |
| | 5 2 | 3+4 4 | 3+4 4 | 3+4 4 | 3+4 4 | 3+4 4 | 3+4 4 | 5 2 |
| | 5 2 | 3+4 4 | 3+4 4 | 3+4 4 | 3+4 4 | 3+4 4 | 3+4 4 | 5 2 |
| | 6 2 | 3+4 4 | 3+4 4 | 3+4 4 | 3+4 4 | 3+4 4 | 3+4 4 | 6 2 |
| | 6 2 | 3+4 4 | 3+4 4 | 3+4 4 | 3+4 4 | 3+4 4 | 3+4 4 | 6 2 |
| | 5 2 | 3+4 4 | 3+4 4 | 3+4 4 | 3+4 4 | 3+4 4 | 3+4 4 | 5 2 |
| | 6 2 | 3+4 4 | 3+4 4 | 3+4 4 | 3+4 4 | 3+4 4 | 3+4 4 | 6 2 |
| | 6 2 | 3+4 4 | 3+4 4 | 3+4 4 | 3+4 4 | 3+4 4 | 3+4 4 | 6 2 |
| | 5 2 | 3+4 4 | 3+4 4 | 3+4 4 | 3+4 4 | 3+4 4 | 3+4 4 | 5 2 |
| | 5 2 | 3+4 4 | 3+4 4 | 3+4 4 | 3+4 4 | 3+4 4 | 3+4 4 | 5 2 |
| | 6 2 | 3+4 4 | 3+4 4 | 3+4 4 | 3+4 4 | 3+4 4 | 3+4 4 | 6 2 |
| | 5 2 | 3+4 4 | 3+4 4 | 3+4 4 | 3+4 4 | 3+4 4 | 3+4 4 | 5 2 |
| | 5 2 | 3+4 4 | 3+4 4 | 3+4 4 | 3+4 4 | 3+4 4 | 3+4 4 | 5 2 |
| | 6 2 | 3+4 4 | 3+4 4 | 3+4 4 | 3+4 4 | 3+4 4 | 3+4 4 | 6 2 |
| | 6 2 | 3+4 4 | 3+4 4 | 3+4 4 | 3+4 4 | 3+4 4 | 3+4 4 | 6 2 |
| | 5 2 | 3+4 4 | 3+4 4 | 3+4 4 | 3+4 4 | 3+4 4 | 3+4 4 | 5 2 |
| | 6 2 | 3+4 4 | 3+4 4 | 3+4 4 | 3+4 4 | 3+4 4 | 3+4 4 | 6 2 |
| | 5 2 | 3+4 4 | 3+4 4 | 3+4 4 | 3+4 4 | 3+4 4 | 3+4 4 | 5 2 |
| | 5 2 | 3+4 4 | 3+4 4 | 3+4 4 | 3+4 4 | 3+4 4 | 3+4 4 | 5 2 |
| | 6 2 | 3+4 4 | 3+4 4 | 3+4 4 | 3+4 4 | 3+4 4 | 3+4 4 | 6 2 |
| | 6 2 | 3+4 4 | 3+4 4 | 3+4 4 | 3+4 4 | 3+4 4 | 3+4 4 | 6 2 |
| | 5 2 | 3+4 4 | 3+4 4 | 3+4 4 | 3+4 4 | 3+4 4 | 3+4 4 | 5 2 |
| | 6 2 | 3+4 4 | 3+4 4 | 3+4 4 | 3+4 4 | 3+4 4 | 3+4 4 | 6 2 |
| | 5 2 | 3+4 4 | 3+4 4 | 3+4 4 | 3+4 4 | 3+4 4 | 3+4 4 | 5 2 |
| | 6 2 | 3+4 4 | 3+4 4 | 3+4 4 | 3+4 4 | 3+4 4 | 3+4 4 | 6 2 |
| | 5 2 | 3+4 4 | 3+4 4 | 3+4 4 | 3+4 4 | 3+4 4 | 3+4 4 | 6 2 |
| | 6 2 | 3+4 4 | 3+4 4 | 3+4 4 | 3+4 4 | 3+4 4 | 3+4 4 | 5 2 |
| | 6 2 | 3+4 4 | 3+4 4 | 3+4 4 | 3+4 4 | 3+4 4 | 3+4 4 | 6 2 |
| | 5 2 | | | | | | | 5 2 |
| | 6 2 | | | | | | | 6 2 |
| | 5 2 | | | | | | | 5 2 |
| | 5 2 | | | | | | | 5 2 |
| | 6 2 | | | | | | | 6 2 |
| | 5 2 | | | | | | | 6 2 |
| | 6 2 | | | | | | | 5 2 |
| | 5 2 | | | | | | | 5 2 |
| | 6 2 | | | | | | | 6 2 |
| | 5 2 | | | | | | | 6 2 |
| | 6 2 | | | | | | | 5 2 |
| | 6 2 | | | | | | | 6 2 |
| | 5 2 | | | | | | | 5 2 |
| | 5 2 | | | | | | | 6 2 |
| | 6 2 | | | | | | | 5 2 |
| | 6 2 | | | | | | | 6 2 |
| | 5 2 | | | | | | | 5 2 |
| | 6 2 | | | | | | | 5 2 |
| | 5 2 | | | | | | | 6 2 |
| | 6 2 | | | | | | | 5 2 |
| | 5 2 | | | | | | | 6 2 |
| | 6 2 | | | | | | | 5 2 |
| | 5 2 | | | | | | | 6 2 |
| | 6 2 | | | | | | | 6 2 |
| | 5 2 | | | | | | | 5 2 |
| | 5 2 | | | | | | | 5 2 |
| | 5 2 | | | | | | | 6 2 |

Continued on next column

**TABLE XXXIV– continued**

| TQC | ICSD | NoE | SF | | | | | |
|---|---|---|---|---|---|---|---|---|
| 59.2.14 | 107525 | 200 | Al1Eu1 | | | | | |
| | -GM | -R | -S | -T | -U | -X | -Y | -Z |
| | 5 2 | 3+4 4 | 3+4 4 | 3+4 4 | 3+4 4 | 3+4 4 | 3+4 4 | 6 2 |
| | 6 2 | 3+4 4 | 3+4 4 | 3+4 4 | 3+4 4 | 3+4 4 | 3+4 4 | 5 2 |
| | 5 2 | 3+4 4 | 3+4 4 | 3+4 4 | 3+4 4 | 3+4 4 | 3+4 4 | 5 2 |
| | 6 2 | 3+4 4 | 3+4 4 | 3+4 4 | 3+4 4 | 3+4 4 | 3+4 4 | 5 2 |
| | 5 2 | 3+4 4 | 3+4 4 | 3+4 4 | 3+4 4 | 3+4 4 | 3+4 4 | 6 2 |
| | 6 2 | 3+4 4 | 3+4 4 | 3+4 4 | 3+4 4 | 3+4 4 | 3+4 4 | 6 2 |
| | 5 2 | 3+4 4 | 3+4 4 | 3+4 4 | 3+4 4 | 3+4 4 | 3+4 4 | 5 2 |
| | 5 2 | 3+4 4 | 3+4 4 | 3+4 4 | 3+4 4 | 3+4 4 | 3+4 4 | 5 2 |
| | 6 2 | 3+4 4 | 3+4 4 | 3+4 4 | 3+4 4 | 3+4 4 | 3+4 4 | 6 2 |
| | 6 2 | 3+4 4 | 3+4 4 | 3+4 4 | 3+4 4 | 3+4 4 | 3+4 4 | 6 2 |
| | 6 2 | 3+4 4 | 3+4 4 | 3+4 4 | 3+4 4 | 3+4 4 | 3+4 4 | 5 2 |
| | 5 2 | 3+4 4 | 3+4 4 | 3+4 4 | 3+4 4 | 3+4 4 | 3+4 4 | 6 2 |
| | 5 2 | 3+4 4 | 3+4 4 | 3+4 4 | 3+4 4 | 3+4 4 | 3+4 4 | 5 2 |
| | 6 2 | 3+4 4 | 3+4 4 | 3+4 4 | 3+4 4 | 3+4 4 | 3+4 4 | 6 2 |
| | 5 2 | 3+4 4 | 3+4 4 | 3+4 4 | 3+4 4 | 3+4 4 | 3+4 4 | 5 2 |
| | 5 2 | 3+4 4 | 3+4 4 | 3+4 4 | 3+4 4 | 3+4 4 | 3+4 4 | 6 2 |
| | 6 2 | 3+4 4 | 3+4 4 | 3+4 4 | 3+4 4 | 3+4 4 | 3+4 4 | 5 2 |
| | 5 2 | 3+4 4 | 3+4 4 | 3+4 4 | 3+4 4 | 3+4 4 | 3+4 4 | 6 2 |
| | 6 2 | 3+4 4 | 3+4 4 | 3+4 4 | 3+4 4 | 3+4 4 | 3+4 4 | 5 2 |
| | 6 2 | 3+4 4 | 3+4 4 | 3+4 4 | 3+4 4 | 3+4 4 | 3+4 4 | 6 2 |
| | 5 2 | 3+4 4 | 3+4 4 | 3+4 4 | 3+4 4 | 3+4 4 | 3+4 4 | 5 2 |
| | 6 2 | 3+4 4 | 3+4 4 | 3+4 4 | 3+4 4 | 3+4 4 | 3+4 4 | 6 2 |
| | 5 2 | 3+4 4 | 3+4 4 | 3+4 4 | 3+4 4 | 3+4 4 | 3+4 4 | 6 2 |
| | 5 2 | 3+4 4 | 3+4 4 | 3+4 4 | 3+4 4 | 3+4 4 | 3+4 4 | 5 2 |
| | 6 2 | 3+4 4 | 3+4 4 | 3+4 4 | 3+4 4 | 3+4 4 | 3+4 4 | 5 2 |
| | 5 2 | 3+4 4 | 3+4 4 | 3+4 4 | 3+4 4 | 3+4 4 | 3+4 4 | 5 2 |
| | 6 2 | 3+4 4 | 3+4 4 | 3+4 4 | 3+4 4 | 3+4 4 | 3+4 4 | 6 2 |
| | 6 2 | 3+4 4 | 3+4 4 | 3+4 4 | 3+4 4 | 3+4 4 | 3+4 4 | 5 2 |
| | 5 2 | 3+4 4 | 3+4 4 | 3+4 4 | 3+4 4 | 3+4 4 | 3+4 4 | 6 2 |
| | 6 2 | 3+4 4 | 3+4 4 | 3+4 4 | 3+4 4 | 3+4 4 | 3+4 4 | 5 2 |
| | 6 2 | 3+4 4 | 3+4 4 | 3+4 4 | 3+4 4 | 3+4 4 | 3+4 4 | 6 2 |
| | 5 2 | 3+4 4 | 3+4 4 | 3+4 4 | 3+4 4 | 3+4 4 | 3+4 4 | 6 2 |
| | 5 2 | 3+4 4 | 3+4 4 | 3+4 4 | 3+4 4 | 3+4 4 | 3+4 4 | 5 2 |
| | 5 2 | 3+4 4 | 3+4 4 | 3+4 4 | 3+4 4 | 3+4 4 | 3+4 4 | 6 2 |
| | 5 2 | 3+4 4 | 3+4 4 | 3+4 4 | 3+4 4 | 3+4 4 | 3+4 4 | 5 2 |
| | 6 2 | 3+4 4 | 3+4 4 | 3+4 4 | 3+4 4 | 3+4 4 | 3+4 4 | 5 2 |
| | 5 2 | 3+4 4 | 3+4 4 | 3+4 4 | 3+4 4 | 3+4 4 | 3+4 4 | 5 2 |
| | 6 2 | 3+4 4 | 3+4 4 | 3+4 4 | 3+4 4 | 3+4 4 | 3+4 4 | 6 2 |
| | 6 2 | 3+4 4 | 3+4 4 | 3+4 4 | 3+4 4 | 3+4 4 | 3+4 4 | 6 2 |
| | 6 2 | 3+4 4 | 3+4 4 | 3+4 4 | 3+4 4 | 3+4 4 | 3+4 4 | 6 2 |
| | 5 2 | 3+4 4 | 3+4 4 | 3+4 4 | 3+4 4 | 3+4 4 | 3+4 4 | 5 2 |
| | 6 2 | 3+4 4 | 3+4 4 | 3+4 4 | 3+4 4 | 3+4 4 | 3+4 4 | 6 2 |
| | 5 2 | 3+4 4 | 3+4 4 | 3+4 4 | 3+4 4 | 3+4 4 | 3+4 4 | 5 2 |
| | 6 2 | 3+4 4 | 3+4 4 | 3+4 4 | 3+4 4 | 3+4 4 | 3+4 4 | 6 2 |
| | 5 2 | 3+4 4 | 3+4 4 | 3+4 4 | 3+4 4 | 3+4 4 | 3+4 4 | 6 2 |
| | 5 2 | 3+4 4 | 3+4 4 | 3+4 4 | 3+4 4 | 3+4 4 | 3+4 4 | 5 2 |
| | 6 2 | 3+4 4 | 3+4 4 | 3+4 4 | 3+4 4 | 3+4 4 | 3+4 4 | 6 2 |
| | 6 2 | 3+4 4 | 3+4 4 | 3+4 4 | 3+4 4 | 3+4 4 | 3+4 4 | 5 2 |
| | 6 2 | 3+4 4 | 3+4 4 | 3+4 4 | 3+4 4 | 3+4 4 | 3+4 4 | 6 2 |
| | 5 2 | 3+4 4 | 3+4 4 | 3+4 4 | 3+4 4 | 3+4 4 | 3+4 4 | 5 2 |
| | 5 2 | | | | | | | 6 2 |
| | 6 2 | | | | | | | 5 2 |
| | 5 2 | | | | | | | 5 2 |
| | 6 2 | | | | | | | 5 2 |
| | 5 2 | | | | | | | 6 2 |
| | 6 2 | | | | | | | 6 2 |
| | 5 2 | | | | | | | 6 2 |
| | 5 2 | | | | | | | 5 2 |
| | 6 2 | | | | | | | 5 2 |

Continued on next column

**TABLE XXXIV– continued**

| TQC | ICSD | NoE | SF | |
|---|---|---|---|---|
| | 6 2 | | | 6 2 |
| | 5 2 | | | 6 2 |
| | 6 2 | | | 6 2 |
| | 5 2 | | | 6 2 |
| | 6 2 | | | 5 2 |
| | 5 2 | | | 5 2 |
| | 5 2 | | | 6 2 |
| | 6 2 | | | 5 2 |
| | 5 2 | | | 6 2 |
| | 6 2 | | | 5 2 |
| | 5 2 | | | 6 2 |
| | 6 2 | | | 5 2 |
| | 6 2 | | | 6 2 |
| | 6 2 | | | 5 2 |
| | 5 2 | | | 6 2 |
| | 6 2 | | | 5 2 |
| | 5 2 | | | 6 2 |
| | 6 2 | | | 5 2 |
| | 5 2 | | | 6 2 |
| | 5 2 | | | 5 2 |
| | 6 2 | | | 5 2 |
| | 5 2 | | | 6 2 |
| | 5 2 | | | 6 2 |
| | 6 2 | | | 5 2 |
| | 6 2 | | | 6 2 |
| | 6 2 | | | 5 2 |
| | 5 2 | | | 6 2 |
| | 5 2 | | | 5 2 |
| | 6 2 | | | 6 2 |
| | 6 2 | | | 5 2 |
| | 5 2 | | | 6 2 |
| | 5 2 | | | 6 2 |
| | 6 2 | | | 5 2 |
| | 5 2 | | | 6 2 |
| | 6 2 | | | 5 2 |
| | 6 2 | | | 5 2 |
| | 5 2 | | | 6 2 |
| | 5 2 | | | 6 2 |
| | 5 2 | | | 6 2 |
| | 6 2 | | | 5 2 |
| | 6 2 | | | 5 2 |

The End

TABLE XXXV: The computed irreducible representations for the valence bands for the compounds of SEBR and NLC in space group 60. See the list of tables.

| TQC | ICSD | NoE | SF | | | | | |
|---|---|---|---|---|---|---|---|---|
| 60.2.1 | 415270 | 64 | O2Pb1 | | | | | |
| | -GM | -R | -S | -T | -U | -X | -Y | -Z |
| | 5 2 | 4+4 4 | 6+7 2 | 4+4 4 | 4+4 4 | 3+4 4 | 3+4 4 | 3+4 4 |
| | 5 2 | 3+3 4 | 8+9 2 | 3+3 4 | 3+3 4 | 3+4 4 | 3+4 4 | 3+4 4 |
| | 5 2 | 3+3 4 | 2+3 2 | 4+4 4 | 3+3 4 | 3+4 4 | 3+4 4 | 3+4 4 |
| | 6 2 | 4+4 4 | 4+5 2 | 3+3 4 | 4+4 4 | 3+4 4 | 3+4 4 | 3+4 4 |
| | 6 2 | 4+4 4 | 6+7 2 | 3+3 4 | 3+3 4 | 3+4 4 | 3+4 4 | 3+4 4 |
| | 5 2 | 3+3 4 | 8+9 2 | 4+4 4 | 4+4 4 | 3+4 4 | 3+4 4 | 3+4 4 |
| | 6 2 | 3+3 4 | 4+5 2 | 3+3 4 | 3+3 4 | 3+4 4 | 3+4 4 | 3+4 4 |
| | 6 2 | 4+4 4 | 2+3 2 | 4+4 4 | 4+4 4 | 3+4 4 | 3+4 4 | 3+4 4 |
| | 5 2 | 4+4 4 | 2+3 2 | 4+4 4 | 4+4 4 | 3+4 4 | 3+4 4 | 3+4 4 |
| | 6 2 | 3+3 4 | 4+5 2 | 3+3 4 | 3+3 4 | 3+4 4 | 3+4 4 | 3+4 4 |

Continued on next column

**TABLE XXXV– continued**

| TQC | ICSD | NoE | SF | | | | | |
|---|---|---|---|---|---|---|---|---|
| | 6 2 | 4+4 4 | 8+9 2 | 4+4 4 | 4+4 4 | 3+4 4 | 3+4 4 | 3+4 4 |
| | 5 2 | 3+3 4 | 6+7 2 | 3+3 4 | 3+3 4 | 3+4 4 | 3+4 4 | 3+4 4 |
| | 5 2 | 3+3 4 | 8+9 2 | 4+4 4 | 4+4 4 | 3+4 4 | 3+4 4 | 3+4 4 |
| | 5 2 | 4+4 4 | 6+7 2 | 3+3 4 | 3+3 4 | 3+4 4 | 3+4 4 | 3+4 4 |
| | 6 2 | 4+4 4 | 6+7 2 | 3+3 4 | 3+3 4 | 3+4 4 | 3+4 4 | 3+4 4 |
| | 6 2 | 3+3 4 | 8+9 2 | 4+4 4 | 4+4 4 | 3+4 4 | 3+4 4 | 3+4 4 |
| | 5 2 | | 2+3 2 | | | | | |
| | 6 2 | | 4+5 2 | | | | | |
| | 5 2 | | 4+5 2 | | | | | |
| | 6 2 | | 2+3 2 | | | | | |
| | 5 2 | | 4+5 2 | | | | | |
| | 6 2 | | 2+3 2 | | | | | |
| | 5 2 | | 4+5 2 | | | | | |
| | 5 2 | | 2+3 2 | | | | | |
| | 5 2 | | 6+7 2 | | | | | |
| | 6 2 | | 8+9 2 | | | | | |
| | 6 2 | | 6+7 2 | | | | | |
| | 5 2 | | 8+9 2 | | | | | |
| | 6 2 | | 4+5 2 | | | | | |
| | 5 2 | | 2+3 2 | | | | | |
| | 5 2 | | 6+7 2 | | | | | |
| | 6 2 | | 8+9 2 | | | | | |
| 60.2.2 | 415269 | 64 | O2Pb1 | | | | | |
| | -GM | -R | -S | -T | -U | -X | -Y | -Z |
| | 5 2 | 4+4 4 | 6+7 2 | 4+4 4 | 3+3 4 | 3+4 4 | 3+4 4 | 3+4 4 |
| | 5 2 | 3+3 4 | 8+9 2 | 3+3 4 | 4+4 4 | 3+4 4 | 3+4 4 | 3+4 4 |
| | 5 2 | 3+3 4 | 2+3 2 | 4+4 4 | 4+4 4 | 3+4 4 | 3+4 4 | 3+4 4 |
| | 6 2 | 4+4 4 | 4+5 2 | 3+3 4 | 3+3 4 | 3+4 4 | 3+4 4 | 3+4 4 |
| | 6 2 | 4+4 4 | 6+7 2 | 3+3 4 | 4+4 4 | 3+4 4 | 3+4 4 | 3+4 4 |
| | 5 2 | 3+3 4 | 8+9 2 | 4+4 4 | 3+3 4 | 3+4 4 | 3+4 4 | 3+4 4 |
| | 6 2 | 3+3 4 | 4+5 2 | 3+3 4 | 4+4 4 | 3+4 4 | 3+4 4 | 3+4 4 |
| | 6 2 | 4+4 4 | 2+3 2 | 4+4 4 | 3+3 4 | 3+4 4 | 3+4 4 | 3+4 4 |
| | 5 2 | 4+4 4 | 2+3 2 | 4+4 4 | 3+3 4 | 3+4 4 | 3+4 4 | 3+4 4 |
| | 6 2 | 3+3 4 | 4+5 2 | 3+3 4 | 4+4 4 | 3+4 4 | 3+4 4 | 3+4 4 |
| | 6 2 | 4+4 4 | 8+9 2 | 4+4 4 | 3+3 4 | 3+4 4 | 3+4 4 | 3+4 4 |
| | 5 2 | 3+3 4 | 6+7 2 | 3+3 4 | 4+4 4 | 3+4 4 | 3+4 4 | 3+4 4 |
| | 5 2 | 3+3 4 | 8+9 2 | 4+4 4 | 3+3 4 | 3+4 4 | 3+4 4 | 3+4 4 |
| | 5 2 | 4+4 4 | 6+7 2 | 3+3 4 | 4+4 4 | 3+4 4 | 3+4 4 | 3+4 4 |
| | 6 2 | 4+4 4 | 6+7 2 | 3+3 4 | 4+4 4 | 3+4 4 | 3+4 4 | 3+4 4 |
| | 6 2 | 3+3 4 | 8+9 2 | 4+4 4 | 3+3 4 | 3+4 4 | 3+4 4 | 3+4 4 |
| | 5 2 | | 2+3 2 | | | | | |
| | 6 2 | | 4+5 2 | | | | | |
| | 5 2 | | 4+5 2 | | | | | |
| | 6 2 | | 2+3 2 | | | | | |
| | 5 2 | | 4+5 2 | | | | | |
| | 6 2 | | 2+3 2 | | | | | |
| | 5 2 | | 4+5 2 | | | | | |
| | 5 2 | | 2+3 2 | | | | | |
| | 5 2 | | 6+7 2 | | | | | |
| | 6 2 | | 8+9 2 | | | | | |
| | 6 2 | | 6+7 2 | | | | | |
| | 5 2 | | 8+9 2 | | | | | |
| | 6 2 | | 4+5 2 | | | | | |
| | 5 2 | | 2+3 2 | | | | | |
| | 5 2 | | 6+7 2 | | | | | |
| | 6 2 | | 8+9 2 | | | | | |
| 60.2.3 | 415268 | 64 | O2Pb1 | | | | | |
| | -GM | -R | -S | -T | -U | -X | -Y | -Z |
| | 5 2 | 4+4 4 | 6+7 2 | 4+4 4 | 4+4 4 | 3+4 4 | 3+4 4 | 3+4 4 |
| | 5 2 | 3+3 4 | 8+9 2 | 3+3 4 | 3+3 4 | 3+4 4 | 3+4 4 | 3+4 4 |
| | 5 2 | 3+3 4 | 2+3 2 | 4+4 4 | 3+3 4 | 3+4 4 | 3+4 4 | 3+4 4 |

Continued on next column

**TABLE XXXV– continued**

| TQC | ICSD | | NoE | SF | | | | |
|---|---|---|---|---|---|---|---|---|
| | 6 2 | 4+4 4 | 4+5 2 | 3+3 4 | 4+4 4 | 3+4 4 | 3+4 4 | 3+4 4 |
| | 6 2 | 4+4 4 | 6+7 2 | 3+3 4 | 3+3 4 | 3+4 4 | 3+4 4 | 3+4 4 |
| | 5 2 | 3+3 4 | 8+9 2 | 4+4 4 | 4+4 4 | 3+4 4 | 3+4 4 | 3+4 4 |
| | 6 2 | 3+3 4 | 4+5 2 | 3+3 4 | 3+3 4 | 3+4 4 | 3+4 4 | 3+4 4 |
| | 6 2 | 4+4 4 | 2+3 2 | 4+4 4 | 4+4 4 | 3+4 4 | 3+4 4 | 3+4 4 |
| | 5 2 | 4+4 4 | 2+3 2 | 4+4 4 | 4+4 4 | 3+4 4 | 3+4 4 | 3+4 4 |
| | 6 2 | 3+3 4 | 4+5 2 | 3+3 4 | 3+3 4 | 3+4 4 | 3+4 4 | 3+4 4 |
| | 6 2 | 4+4 4 | 8+9 2 | 4+4 4 | 4+4 4 | 3+4 4 | 3+4 4 | 3+4 4 |
| | 5 2 | 3+3 4 | 6+7 2 | 3+3 4 | 3+3 4 | 3+4 4 | 3+4 4 | 3+4 4 |
| | 5 2 | 3+3 4 | 8+9 2 | 4+4 4 | 4+4 4 | 3+4 4 | 3+4 4 | 3+4 4 |
| | 5 2 | 4+4 4 | 6+7 2 | 3+3 4 | 3+3 4 | 3+4 4 | 3+4 4 | 3+4 4 |
| | 6 2 | 4+4 4 | 6+7 2 | 3+3 4 | 3+3 4 | 3+4 4 | 3+4 4 | 3+4 4 |
| | 6 2 | 3+3 4 | 8+9 2 | 4+4 4 | 4+4 4 | 3+4 4 | 3+4 4 | 3+4 4 |
| | 5 2 | | 2+3 2 | | | | | |
| | 6 2 | | 4+5 2 | | | | | |
| | 5 2 | | 4+5 2 | | | | | |
| | 6 2 | | 2+3 2 | | | | | |
| | 5 2 | | 4+5 2 | | | | | |
| | 6 2 | | 2+3 2 | | | | | |
| | 5 2 | | 2+3 2 | | | | | |
| | 5 2 | | 4+5 2 | | | | | |
| | 5 2 | | 6+7 2 | | | | | |
| | 6 2 | | 8+9 2 | | | | | |
| | 6 2 | | 6+7 2 | | | | | |
| | 5 2 | | 8+9 2 | | | | | |
| | 6 2 | | 4+5 2 | | | | | |
| | 5 2 | | 2+3 2 | | | | | |
| | 5 2 | | 6+7 2 | | | | | |
| | 6 2 | | 8+9 2 | | | | | |
| 60.2.4 | 246148 | | 64 | C1Mo2 | | | | |
| | -GM | -R | -S | -T | -U | -X | -Y | -Z |
| | 5 2 | 4+4 4 | 8+9 2 | 3+3 4 | 3+3 4 | 3+4 4 | 3+4 4 | 3+4 4 |
| | 6 2 | 3+3 4 | 6+7 2 | 4+4 4 | 4+4 4 | 3+4 4 | 3+4 4 | 3+4 4 |
| | 6 2 | 4+4 4 | 4+5 2 | 4+4 4 | 4+4 4 | 3+4 4 | 3+4 4 | 3+4 4 |
| | 5 2 | 3+3 4 | 2+3 2 | 3+3 4 | 3+3 4 | 3+4 4 | 3+4 4 | 3+4 4 |
| | 5 2 | 4+4 4 | 6+7 2 | 4+4 4 | 3+3 4 | 3+4 4 | 3+4 4 | 3+4 4 |
| | 6 2 | 3+3 4 | 8+9 2 | 3+3 4 | 4+4 4 | 3+4 4 | 3+4 4 | 3+4 4 |
| | 5 2 | 3+3 4 | 4+5 2 | 3+3 4 | 4+4 4 | 3+4 4 | 3+4 4 | 3+4 4 |
| | 6 2 | 4+4 4 | 2+3 2 | 4+4 4 | 3+3 4 | 3+4 4 | 3+4 4 | 3+4 4 |
| | 6 2 | 4+4 4 | 4+5 2 | 3+3 4 | 3+3 4 | 3+4 4 | 3+4 4 | 3+4 4 |
| | 5 2 | 3+3 4 | 2+3 2 | 4+4 4 | 4+4 4 | 3+4 4 | 3+4 4 | 3+4 4 |
| | 6 2 | 4+4 4 | 6+7 2 | 4+4 4 | 4+4 4 | 3+4 4 | 3+4 4 | 3+4 4 |
| | 5 2 | 3+3 4 | 8+9 2 | 3+3 4 | 3+3 4 | 3+4 4 | 3+4 4 | 3+4 4 |
| | 6 2 | 4+4 4 | 2+3 2 | 3+3 4 | 4+4 4 | 3+4 4 | 3+4 4 | 3+4 4 |
| | 5 2 | 3+3 4 | 4+5 2 | 4+4 4 | 3+3 4 | 3+4 4 | 3+4 4 | 3+4 4 |
| | 6 2 | 4+4 4 | 8+9 2 | 3+3 4 | 4+4 4 | 3+4 4 | 3+4 4 | 3+4 4 |
| | 5 2 | 3+3 4 | 6+7 2 | 4+4 4 | 3+3 4 | 3+4 4 | 3+4 4 | 3+4 4 |
| | 6 2 | | 4+5 2 | | | | | |
| | 5 2 | | 2+3 2 | | | | | |
| | 6 2 | | 8+9 2 | | | | | |
| | 5 2 | | 6+7 2 | | | | | |
| | 5 2 | | 8+9 2 | | | | | |
| | 5 2 | | 6+7 2 | | | | | |
| | 6 2 | | 4+5 2 | | | | | |
| | 6 2 | | 2+3 2 | | | | | |
| | 6 2 | | 6+7 2 | | | | | |
| | 5 2 | | 8+9 2 | | | | | |
| | 5 2 | | 4+5 2 | | | | | |
| | 5 2 | | 2+3 2 | | | | | |
| | 6 2 | | 6+7 2 | | | | | |
| | 5 2 | | 8+9 2 | | | | | |

Continued on next column

**TABLE XXXV– continued**

| TQC | ICSD | | NoE | SF | | | | |
|---|---|---|---|---|---|---|---|---|
| | 6 2 | | 2+3 2 | | | | | |
| | 5 2 | | 4+5 2 | | | | | |
| 60.2.5 | 246147 | | 64 | C1Mo2 | | | | |
| | -GM | -R | -S | -T | -U | -X | -Y | -Z |
| | 5 2 | 4+4 4 | 8+9 2 | 3+3 4 | 3+3 4 | 3+4 4 | 3+4 4 | 3+4 4 |
| | 6 2 | 3+3 4 | 6+7 2 | 4+4 4 | 4+4 4 | 3+4 4 | 3+4 4 | 3+4 4 |
| | 6 2 | 4+4 4 | 4+5 2 | 4+4 4 | 4+4 4 | 3+4 4 | 3+4 4 | 3+4 4 |
| | 5 2 | 3+3 4 | 2+3 2 | 3+3 4 | 3+3 4 | 3+4 4 | 3+4 4 | 3+4 4 |
| | 5 2 | 4+4 4 | 6+7 2 | 4+4 4 | 3+3 4 | 3+4 4 | 3+4 4 | 3+4 4 |
| | 6 2 | 3+3 4 | 8+9 2 | 3+3 4 | 4+4 4 | 3+4 4 | 3+4 4 | 3+4 4 |
| | 5 2 | 3+3 4 | 4+5 2 | 3+3 4 | 4+4 4 | 3+4 4 | 3+4 4 | 3+4 4 |
| | 6 2 | 4+4 4 | 2+3 2 | 4+4 4 | 3+3 4 | 3+4 4 | 3+4 4 | 3+4 4 |
| | 6 2 | 4+4 4 | 4+5 2 | 3+3 4 | 3+3 4 | 3+4 4 | 3+4 4 | 3+4 4 |
| | 5 2 | 3+3 4 | 2+3 2 | 4+4 4 | 4+4 4 | 3+4 4 | 3+4 4 | 3+4 4 |
| | 6 2 | 4+4 4 | 6+7 2 | 4+4 4 | 4+4 4 | 3+4 4 | 3+4 4 | 3+4 4 |
| | 5 2 | 3+3 4 | 8+9 2 | 3+3 4 | 3+3 4 | 3+4 4 | 3+4 4 | 3+4 4 |
| | 6 2 | 4+4 4 | 2+3 2 | 3+3 4 | 4+4 4 | 3+4 4 | 3+4 4 | 3+4 4 |
| | 5 2 | 3+3 4 | 4+5 2 | 4+4 4 | 3+3 4 | 3+4 4 | 3+4 4 | 3+4 4 |
| | 6 2 | 4+4 4 | 8+9 2 | 3+3 4 | 4+4 4 | 3+4 4 | 3+4 4 | 3+4 4 |
| | 5 2 | 3+3 4 | 6+7 2 | 4+4 4 | 3+3 4 | 3+4 4 | 3+4 4 | 3+4 4 |
| | 6 2 | | 4+5 2 | | | | | |
| | 5 2 | | 2+3 2 | | | | | |
| | 6 2 | | 8+9 2 | | | | | |
| | 5 2 | | 6+7 2 | | | | | |
| | 5 2 | | 8+9 2 | | | | | |
| | 5 2 | | 6+7 2 | | | | | |
| | 6 2 | | 4+5 2 | | | | | |
| | 6 2 | | 2+3 2 | | | | | |
| | 6 2 | | 6+7 2 | | | | | |
| | 5 2 | | 8+9 2 | | | | | |
| | 5 2 | | 4+5 2 | | | | | |
| | 6 2 | | 2+3 2 | | | | | |
| | 5 2 | | 6+7 2 | | | | | |
| | 5 2 | | 8+9 2 | | | | | |
| | 6 2 | | 2+3 2 | | | | | |
| | 5 2 | | 4+5 2 | | | | | |
| 60.2.6 | 246146 | | 64 | C1Mo2 | | | | |
| | -GM | -R | -S | -T | -U | -X | -Y | -Z |
| | 5 2 | 4+4 4 | 8+9 2 | 3+3 4 | 3+3 4 | 3+4 4 | 3+4 4 | 3+4 4 |
| | 6 2 | 3+3 4 | 6+7 2 | 4+4 4 | 4+4 4 | 3+4 4 | 3+4 4 | 3+4 4 |
| | 6 2 | 4+4 4 | 4+5 2 | 4+4 4 | 4+4 4 | 3+4 4 | 3+4 4 | 3+4 4 |
| | 5 2 | 3+3 4 | 2+3 2 | 3+3 4 | 3+3 4 | 3+4 4 | 3+4 4 | 3+4 4 |
| | 5 2 | 4+4 4 | 6+7 2 | 4+4 4 | 3+3 4 | 3+4 4 | 3+4 4 | 3+4 4 |
| | 6 2 | 3+3 4 | 8+9 2 | 3+3 4 | 4+4 4 | 3+4 4 | 3+4 4 | 3+4 4 |
| | 5 2 | 3+3 4 | 4+5 2 | 3+3 4 | 4+4 4 | 3+4 4 | 3+4 4 | 3+4 4 |
| | 6 2 | 4+4 4 | 2+3 2 | 4+4 4 | 3+3 4 | 3+4 4 | 3+4 4 | 3+4 4 |
| | 6 2 | 4+4 4 | 4+5 2 | 3+3 4 | 4+4 4 | 3+4 4 | 3+4 4 | 3+4 4 |
| | 5 2 | 3+3 4 | 2+3 2 | 4+4 4 | 3+3 4 | 3+4 4 | 3+4 4 | 3+4 4 |
| | 6 2 | 4+4 4 | 6+7 2 | 4+4 4 | 3+3 4 | 3+4 4 | 3+4 4 | 3+4 4 |
| | 5 2 | 3+3 4 | 8+9 2 | 3+3 4 | 4+4 4 | 3+4 4 | 3+4 4 | 3+4 4 |
| | 6 2 | 4+4 4 | 2+3 2 | 3+3 4 | 4+4 4 | 3+4 4 | 3+4 4 | 3+4 4 |
| | 5 2 | 3+3 4 | 4+5 2 | 4+4 4 | 3+3 4 | 3+4 4 | 3+4 4 | 3+4 4 |
| | 6 2 | 3+3 4 | 8+9 2 | 3+3 4 | 4+4 4 | 3+4 4 | 3+4 4 | 3+4 4 |
| | 5 2 | 4+4 4 | 6+7 2 | 4+4 4 | 3+3 4 | 3+4 4 | 3+4 4 | 3+4 4 |
| | 6 2 | | 4+5 2 | | | | | |
| | 5 2 | | 2+3 2 | | | | | |
| | 6 2 | | 8+9 2 | | | | | |
| | 5 2 | | 6+7 2 | | | | | |
| | 5 2 | | 8+9 2 | | | | | |
| | 5 2 | | 6+7 2 | | | | | |
| | 6 2 | | 4+5 2 | | | | | |

Continued on next column

**TABLE XXXV– continued**

| TQC | ICSD | NoE | SF | | | | | |
|---|---|---|---|---|---|---|---|---|
| | 6 2 | | 2+3 2 | | | | | |
| | 6 2 | | 6+7 2 | | | | | |
| | 5 2 | | 8+9 2 | | | | | |
| | 5 2 | | 4+5 2 | | | | | |
| | 6 2 | | 2+3 2 | | | | | |
| | 5 2 | | 6+7 2 | | | | | |
| | 5 2 | | 8+9 2 | | | | | |
| | 6 2 | | 2+3 2 | | | | | |
| | 5 2 | | 4+5 2 | | | | | |
| 60.2.7 | 430298 | | 208 | Au2Pb1 | | | | |
| | -GM | -R | -S | -T | -U | -X | -Y | -Z |
| | 5 2 | 3+3 4 | 6+7 2 | 3+3 4 | 3+3 4 | 3+4 4 | 3+4 4 | 3+4 4 |
| | 6 2 | 4+4 4 | 8+9 2 | 4+4 4 | 4+4 4 | 3+4 4 | 3+4 4 | 3+4 4 |
| | 5 2 | 3+3 4 | 4+5 2 | 3+3 4 | 4+4 4 | 3+4 4 | 3+4 4 | 3+4 4 |
| | 6 2 | 4+4 4 | 2+3 2 | 4+4 4 | 3+3 4 | 3+4 4 | 3+4 4 | 3+4 4 |
| | 5 2 | 3+3 4 | 8+9 2 | 3+3 4 | 3+3 4 | 3+4 4 | 3+4 4 | 3+4 4 |
| | 6 2 | 4+4 4 | 6+7 2 | 4+4 4 | 4+4 4 | 3+4 4 | 3+4 4 | 3+4 4 |
| | 5 2 | 3+3 4 | 2+3 2 | 4+4 4 | 4+4 4 | 3+4 4 | 3+4 4 | 3+4 4 |
| | 5 2 | 4+4 4 | 4+5 2 | 3+3 4 | 3+3 4 | 3+4 4 | 3+4 4 | 3+4 4 |
| | 5 2 | 4+4 4 | 2+3 2 | 3+3 4 | 4+4 4 | 3+4 4 | 3+4 4 | 3+4 4 |
| | 6 2 | 3+3 4 | 4+5 2 | 4+4 4 | 3+3 4 | 3+4 4 | 3+4 4 | 3+4 4 |
| | 5 2 | 3+3 4 | 6+7 2 | 3+3 4 | 3+3 4 | 3+4 4 | 3+4 4 | 3+4 4 |
| | 6 2 | 4+4 4 | 2+3 2 | 4+4 4 | 4+4 4 | 3+4 4 | 3+4 4 | 3+4 4 |
| | 5 2 | 4+4 4 | 8+9 2 | 3+3 4 | 3+3 4 | 3+4 4 | 3+4 4 | 3+4 4 |
| | 5 2 | 3+3 4 | 4+5 2 | 4+4 4 | 4+4 4 | 3+4 4 | 3+4 4 | 3+4 4 |
| | 5 2 | 4+4 4 | 2+3 2 | 4+4 4 | 4+4 4 | 3+4 4 | 3+4 4 | 3+4 4 |
| | 6 2 | 3+3 4 | 4+5 2 | 3+3 4 | 3+3 4 | 3+4 4 | 3+4 4 | 3+4 4 |
| | 5 2 | 4+4 4 | 4+5 2 | 4+4 4 | 4+4 4 | 3+4 4 | 3+4 4 | 3+4 4 |
| | 6 2 | 3+3 4 | 2+3 2 | 3+3 4 | 3+3 4 | 3+4 4 | 3+4 4 | 3+4 4 |
| | 5 2 | 3+3 4 | 2+3 2 | 3+3 4 | 4+4 4 | 3+4 4 | 3+4 4 | 3+4 4 |
| | 5 2 | 4+4 4 | 4+5 2 | 4+4 4 | 3+3 4 | 3+4 4 | 3+4 4 | 3+4 4 |
| | 5 2 | 3+3 4 | 8+9 2 | 4+4 4 | 3+3 4 | 3+4 4 | 3+4 4 | 3+4 4 |
| | 6 2 | 4+4 4 | 4+5 2 | 3+3 4 | 4+4 4 | 3+4 4 | 3+4 4 | 3+4 4 |
| | 5 2 | 4+4 4 | 6+7 2 | 4+4 4 | 4+4 4 | 3+4 4 | 3+4 4 | 3+4 4 |
| | 6 2 | 3+3 4 | 2+3 2 | 3+3 4 | 3+3 4 | 3+4 4 | 3+4 4 | 3+4 4 |
| | 5 2 | 4+4 4 | 6+7 2 | 3+3 4 | 3+3 4 | 3+4 4 | 3+4 4 | 3+4 4 |
| | 6 2 | 3+3 4 | 8+9 2 | 4+4 4 | 4+4 4 | 3+4 4 | 3+4 4 | 3+4 4 |
| | 5 2 | 3+3 4 | 4+5 2 | 4+4 4 | 4+4 4 | 3+4 4 | 3+4 4 | 3+4 4 |
| | 5 2 | 4+4 4 | 2+3 2 | 3+3 4 | 3+3 4 | 3+4 4 | 3+4 4 | 3+4 4 |
| | 6 2 | 3+3 4 | 8+9 2 | 3+3 4 | 3+3 4 | 3+4 4 | 3+4 4 | 3+4 4 |
| | 5 2 | 4+4 4 | 6+7 2 | 4+4 4 | 4+4 4 | 3+4 4 | 3+4 4 | 3+4 4 |
| | 6 2 | 3+3 4 | 4+5 2 | 3+3 4 | 3+3 4 | 3+4 4 | 3+4 4 | 3+4 4 |
| | 6 2 | 4+4 4 | 2+3 2 | 4+4 4 | 4+4 4 | 3+4 4 | 3+4 4 | 3+4 4 |
| | 5 2 | 3+3 4 | 6+7 2 | 3+3 4 | 4+4 4 | 3+4 4 | 3+4 4 | 3+4 4 |
| | 5 2 | 4+4 4 | 8+9 2 | 4+4 4 | 3+3 4 | 3+4 4 | 3+4 4 | 3+4 4 |
| | 5 2 | 4+4 4 | 2+3 2 | 4+4 4 | 4+4 4 | 3+4 4 | 3+4 4 | 3+4 4 |
| | 6 2 | 3+3 4 | 4+5 2 | 3+3 4 | 3+3 4 | 3+4 4 | 3+4 4 | 3+4 4 |
| | 5 2 | 4+4 4 | 6+7 2 | 4+4 4 | 3+3 4 | 3+4 4 | 3+4 4 | 3+4 4 |
| | 5 2 | 3+3 4 | 8+9 2 | 3+3 4 | 4+4 4 | 3+4 4 | 3+4 4 | 3+4 4 |
| | 6 2 | 4+4 4 | 2+3 2 | 3+3 4 | 4+4 4 | 3+4 4 | 3+4 4 | 3+4 4 |
| | 6 2 | 3+3 4 | 4+5 2 | 4+4 4 | 3+3 4 | 3+4 4 | 3+4 4 | 3+4 4 |
| | 5 2 | 4+4 4 | 6+7 2 | 3+3 4 | 4+4 4 | 3+4 4 | 3+4 4 | 3+4 4 |
| | 5 2 | 3+3 4 | 2+3 2 | 4+4 4 | 3+3 4 | 3+4 4 | 3+4 4 | 3+4 4 |
| | 6 2 | 3+3 4 | 4+5 2 | 4+4 4 | 4+4 4 | 3+4 4 | 3+4 4 | 3+4 4 |
| | 5 2 | 4+4 4 | 8+9 2 | 3+3 4 | 3+3 4 | 3+4 4 | 3+4 4 | 3+4 4 |
| | 6 2 | 3+3 4 | 4+5 2 | 4+4 4 | 4+4 4 | 3+4 4 | 3+4 4 | 3+4 4 |
| | 5 2 | 4+4 4 | 2+3 2 | 3+3 4 | 3+3 4 | 3+4 4 | 3+4 4 | 3+4 4 |
| | 6 2 | 3+3 4 | 8+9 2 | 3+3 4 | 3+3 4 | 3+4 4 | 3+4 4 | 3+4 4 |
| | 6 2 | 4+4 4 | 6+7 2 | 4+4 4 | 4+4 4 | 3+4 4 | 3+4 4 | 3+4 4 |
| | 5 2 | 4+4 4 | 8+9 2 | 3+3 4 | 4+4 4 | 3+4 4 | 3+4 4 | 3+4 4 |
| | 5 2 | 3+3 4 | 2+3 2 | 4+4 4 | 3+3 4 | 3+4 4 | 3+4 4 | 3+4 4 |

Continued on next column

| TQC | ICSD | NoE | SF | | | | | |
|---|---|---|---|---|---|---|---|---|
| | 6 2 | 3+3 4 | 4+5 2 | 4+4 4 | 3+3 4 | 3+4 4 | 3+4 4 | 3+4 4 |
| | 6 2 | 4+4 4 | 4+5 2 | 3+3 4 | 4+4 4 | 3+4 4 | 3+4 4 | 3+4 4 |
| | 5 2 | | 6+7 2 | | | | | |
| | 5 2 | | 6+7 2 | | | | | |
| | 6 2 | | 2+3 2 | | | | | |
| | 5 2 | | 8+9 2 | | | | | |
| | 6 2 | | 2+3 2 | | | | | |
| | 5 2 | | 8+9 2 | | | | | |
| | 5 2 | | 6+7 2 | | | | | |
| | 6 2 | | 4+5 2 | | | | | |
| | 5 2 | | 8+9 2 | | | | | |
| | 6 2 | | 6+7 2 | | | | | |
| | 5 2 | | 2+3 2 | | | | | |
| | 5 2 | | 4+5 2 | | | | | |
| | 6 2 | | 2+3 2 | | | | | |
| | 6 2 | | 4+5 2 | | | | | |
| | 5 2 | | 4+5 2 | | | | | |
| | 5 2 | | 8+9 2 | | | | | |
| | 5 2 | | 2+3 2 | | | | | |
| | 6 2 | | 8+9 2 | | | | | |
| | 5 2 | | 4+5 2 | | | | | |
| | 5 2 | | 4+5 2 | | | | | |
| | 6 2 | | 2+3 2 | | | | | |
| | 5 2 | | 6+7 2 | | | | | |
| | 5 2 | | 4+5 2 | | | | | |
| | 6 2 | | 6+7 2 | | | | | |
| | 5 2 | | 2+3 2 | | | | | |
| | 6 2 | | 2+3 2 | | | | | |
| | 5 2 | | 2+3 2 | | | | | |
| | 6 2 | | 4+5 2 | | | | | |
| | 5 2 | | 2+3 2 | | | | | |
| | 5 2 | | 6+7 2 | | | | | |
| | 5 2 | | 4+5 2 | | | | | |
| | 6 2 | | 8+9 2 | | | | | |
| | 5 2 | | 2+3 2 | | | | | |
| | 6 2 | | 4+5 2 | | | | | |
| | 6 2 | | 6+7 2 | | | | | |
| | 5 2 | | 2+3 2 | | | | | |
| | 5 2 | | 4+5 2 | | | | | |
| | 6 2 | | 2+3 2 | | | | | |
| | 5 2 | | 8+9 2 | | | | | |
| | 5 2 | | 4+5 2 | | | | | |
| | 6 2 | | 6+7 2 | | | | | |
| | 5 2 | | 8+9 2 | | | | | |
| | 5 2 | | 4+5 2 | | | | | |
| | 5 2 | | 2+3 2 | | | | | |
| | 6 2 | | 8+9 2 | | | | | |
| | 6 2 | | 6+7 2 | | | | | |
| | 6 2 | | 6+7 2 | | | | | |
| | 5 2 | | 8+9 2 | | | | | |
| | 6 2 | | 4+5 2 | | | | | |
| | 5 2 | | 8+9 2 | | | | | |
| | 5 2 | | 6+7 2 | | | | | |
| | 6 2 | | 2+3 2 | | | | | |

The End

TABLE XXXVI: The computed irreducible representations for the valence bands for the compounds of SEBR and NLC in space group 61. See the list of tables.

| TQC | ICSD | NoE | SF | | | | | |
|---|---|---|---|---|---|---|---|---|
| 61.2.1 | 162404 | 160 | Pd1Sn1Te1 | | | | | |
| | -GM | -R | -S | -T | -U | -X | -Y | -Z |
| | 5 2 | 6+6 2 | 3+3 4 | 3+3 4 | 3+3 4 | 3+4 4 | 3+4 4 | 3+4 4 |
| | 6 2 | 8+8 2 | 4+4 4 | 4+4 4 | 4+4 4 | 3+4 4 | 3+4 4 | 3+4 4 |
| | 5 2 | 9+9 2 | 3+3 4 | 3+3 4 | 3+3 4 | 3+4 4 | 3+4 4 | 3+4 4 |
| | 6 2 | 7+7 2 | 4+4 4 | 4+4 4 | 4+4 4 | 3+4 4 | 3+4 4 | 3+4 4 |
| | 5 2 | 3+3 2 | 3+3 4 | 3+3 4 | 3+3 4 | 3+4 4 | 3+4 4 | 3+4 4 |
| | 6 2 | 4+4 2 | 4+4 4 | 4+4 4 | 4+4 4 | 3+4 4 | 3+4 4 | 3+4 4 |
| | 5 2 | 5+5 2 | 3+3 4 | 4+4 4 | 4+4 4 | 3+4 4 | 3+4 4 | 3+4 4 |
| | 6 2 | 2+2 2 | 4+4 4 | 3+3 4 | 3+3 4 | 3+4 4 | 3+4 4 | 3+4 4 |
| | 5 2 | 2+2 2 | 4+4 4 | 3+3 4 | 3+3 4 | 3+4 4 | 3+4 4 | 3+4 4 |
| | 5 2 | 5+5 2 | 3+3 4 | 4+4 4 | 4+4 4 | 3+4 4 | 3+4 4 | 3+4 4 |
| | 6 2 | 3+3 2 | 3+3 4 | 4+4 4 | 3+3 4 | 3+4 4 | 3+4 4 | 3+4 4 |
| | 6 2 | 4+4 2 | 4+4 4 | 3+3 4 | 4+4 4 | 3+4 4 | 3+4 4 | 3+4 4 |
| | 5 2 | 9+9 2 | 3+3 4 | 4+4 4 | 4+4 4 | 3+4 4 | 3+4 4 | 3+4 4 |
| | 6 2 | 8+8 2 | 4+4 4 | 3+3 4 | 3+3 4 | 3+4 4 | 3+4 4 | 3+4 4 |
| | 5 2 | 6+6 2 | 3+3 4 | 3+3 4 | 4+4 4 | 3+4 4 | 3+4 4 | 3+4 4 |
| | 6 2 | 7+7 2 | 4+4 4 | 4+4 4 | 3+3 4 | 3+4 4 | 3+4 4 | 3+4 4 |
| | 5 2 | 7+7 2 | 3+3 4 | 4+4 4 | 3+3 4 | 3+4 4 | 3+4 4 | 3+4 4 |
| | 6 2 | 6+6 2 | 4+4 4 | 3+3 4 | 4+4 4 | 3+4 4 | 3+4 4 | 3+4 4 |
| | 5 2 | 8+8 2 | 4+4 4 | 3+3 4 | 4+4 4 | 3+4 4 | 3+4 4 | 3+4 4 |
| | 6 2 | 9+9 2 | 3+3 4 | 4+4 4 | 3+3 4 | 3+4 4 | 3+4 4 | 3+4 4 |
| | 5 2 | 2+2 2 | 3+3 4 | 3+3 4 | 4+4 4 | 3+4 4 | 3+4 4 | 3+4 4 |
| | 6 2 | 3+3 2 | 4+4 4 | 4+4 4 | 3+3 4 | 3+4 4 | 3+4 4 | 3+4 4 |
| | 6 2 | 5+5 2 | 4+4 4 | 4+4 4 | 4+4 4 | 3+4 4 | 3+4 4 | 3+4 4 |
| | 6 2 | 4+4 2 | 3+3 4 | 3+3 4 | 3+3 4 | 3+4 4 | 3+4 4 | 3+4 4 |
| | 5 2 | 9+9 2 | 4+4 4 | 3+3 4 | 4+4 4 | 3+4 4 | 3+4 4 | 3+4 4 |
| | 6 2 | 6+6 2 | 3+3 4 | 4+4 4 | 3+3 4 | 3+4 4 | 3+4 4 | 3+4 4 |
| | 5 2 | 8+8 2 | 4+4 4 | 4+4 4 | 3+3 4 | 3+4 4 | 3+4 4 | 3+4 4 |
| | 5 2 | 7+7 2 | 3+3 4 | 3+3 4 | 4+4 4 | 3+4 4 | 3+4 4 | 3+4 4 |
| | 5 2 | 3+3 2 | 3+3 4 | 4+4 4 | 3+3 4 | 3+4 4 | 3+4 4 | 3+4 4 |
| | 6 2 | 2+2 2 | 4+4 4 | 3+3 4 | 4+4 4 | 3+4 4 | 3+4 4 | 3+4 4 |
| | 6 2 | 4+4 2 | 3+3 4 | 3+3 4 | 3+3 4 | 3+4 4 | 3+4 4 | 3+4 4 |
| | 5 2 | 5+5 2 | 4+4 4 | 4+4 4 | 4+4 4 | 3+4 4 | 3+4 4 | 3+4 4 |
| | 5 2 | 3+3 2 | 4+4 4 | 3+3 4 | 4+4 4 | 3+4 4 | 3+4 4 | 3+4 4 |
| | 5 2 | 2+2 2 | 3+3 4 | 4+4 4 | 3+3 4 | 3+4 4 | 3+4 4 | 3+4 4 |
| | 6 2 | 5+5 2 | 3+3 4 | 3+3 4 | 4+4 4 | 3+4 4 | 3+4 4 | 3+4 4 |
| | 6 2 | 4+4 2 | 4+4 4 | 4+4 4 | 3+3 4 | 3+4 4 | 3+4 4 | 3+4 4 |
| | 5 2 | 7+7 2 | 4+4 4 | 3+3 4 | 3+3 4 | 3+4 4 | 3+4 4 | 3+4 4 |
| | 5 2 | 8+8 2 | 3+3 4 | 4+4 4 | 4+4 4 | 3+4 4 | 3+4 4 | 3+4 4 |
| | 5 2 | 9+9 2 | 3+3 4 | 3+3 4 | 3+3 4 | 3+4 4 | 3+4 4 | 3+4 4 |
| | 6 2 | 6+6 2 | 4+4 4 | 4+4 4 | 4+4 4 | 3+4 4 | 3+4 4 | 3+4 4 |
| | 5 2 | 4+4 2 | | | | | | |
| | 6 2 | 5+5 2 | | | | | | |
| | 5 2 | 3+3 2 | | | | | | |
| | 5 2 | 2+2 2 | | | | | | |
| | 6 2 | 7+7 2 | | | | | | |
| | 6 2 | 9+9 2 | | | | | | |
| | 5 2 | 8+8 2 | | | | | | |
| | 6 2 | 6+6 2 | | | | | | |
| | 6 2 | 4+4 2 | | | | | | |
| | 5 2 | 5+5 2 | | | | | | |
| | 5 2 | 2+2 2 | | | | | | |
| | 6 2 | 3+3 2 | | | | | | |
| | 5 2 | 6+6 2 | | | | | | |
| | 6 2 | 9+9 2 | | | | | | |
| | 6 2 | 8+8 2 | | | | | | |
| | 5 2 | 7+7 2 | | | | | | |

Continued on next column

**TABLE XXXVI– continued**

| TQC | ICSD | NoE | SF | | | | | |
|---|---|---|---|---|---|---|---|---|
| | 6 2 | 3+3 2 | | | | | | |
| | 6 2 | 8+8 2 | | | | | | |
| | 5 2 | 4+4 2 | | | | | | |
| | 5 2 | 5+5 2 | | | | | | |
| | 5 2 | 2+2 2 | | | | | | |
| | 6 2 | 7+7 2 | | | | | | |
| | 5 2 | 6+6 2 | | | | | | |
| | 6 2 | 9+9 2 | | | | | | |
| | 6 2 | 2+2 2 | | | | | | |
| | 5 2 | 5+5 2 | | | | | | |
| | 6 2 | 4+4 2 | | | | | | |
| | 6 2 | 3+3 2 | | | | | | |
| | 5 2 | 9+9 2 | | | | | | |
| | 5 2 | 8+8 2 | | | | | | |
| | 6 2 | 6+6 2 | | | | | | |
| | 5 2 | 7+7 2 | | | | | | |
| | 6 2 | 2+2 2 | | | | | | |
| | 5 2 | 4+4 2 | | | | | | |
| | 6 2 | 5+5 2 | | | | | | |
| | 5 2 | 3+3 2 | | | | | | |
| | 6 2 | 9+9 2 | | | | | | |
| | 6 2 | 6+6 2 | | | | | | |
| | 5 2 | 7+7 2 | | | | | | |
| | 5 2 | 8+8 2 | | | | | | |
| 61.2.2 | 415968 | 152 | Au1Sn2 | | | | | |
| | -GM | -R | -S | -T | -U | -X | -Y | -Z |
| | 5 2 | 2+2 2 | 3+3 4 | 3+3 4 | 4+4 4 | 3+4 4 | 3+4 4 | 3+4 4 |
| | 6 2 | 3+3 2 | 4+4 4 | 4+4 4 | 3+3 4 | 3+4 4 | 3+4 4 | 3+4 4 |
| | 5 2 | 4+4 2 | 3+3 4 | 4+4 4 | 3+3 4 | 3+4 4 | 3+4 4 | 3+4 4 |
| | 5 2 | 5+5 2 | 4+4 4 | 3+3 4 | 4+4 4 | 3+4 4 | 3+4 4 | 3+4 4 |
| | 5 2 | 8+8 2 | 3+3 4 | 3+3 4 | 3+3 4 | 3+4 4 | 3+4 4 | 3+4 4 |
| | 6 2 | 7+7 2 | 4+4 4 | 4+4 4 | 4+4 4 | 3+4 4 | 3+4 4 | 3+4 4 |
| | 6 2 | 9+9 2 | 4+4 4 | 4+4 4 | 4+4 4 | 3+4 4 | 3+4 4 | 3+4 4 |
| | 6 2 | 6+6 2 | 3+3 4 | 3+3 4 | 3+3 4 | 3+4 4 | 3+4 4 | 3+4 4 |
| | 5 2 | 6+6 2 | 4+4 4 | 3+3 4 | 4+4 4 | 3+4 4 | 3+4 4 | 3+4 4 |
| | 6 2 | 9+9 2 | 3+3 4 | 4+4 4 | 3+3 4 | 3+4 4 | 3+4 4 | 3+4 4 |
| | 6 2 | 8+8 2 | 3+3 4 | 3+3 4 | 3+3 4 | 3+4 4 | 3+4 4 | 3+4 4 |
| | 5 2 | 7+7 2 | 4+4 4 | 4+4 4 | 4+4 4 | 3+4 4 | 3+4 4 | 3+4 4 |
| | 6 2 | 4+4 2 | 4+4 4 | 4+4 4 | 4+4 4 | 3+4 4 | 3+4 4 | 3+4 4 |
| | 6 2 | 5+5 2 | 3+3 4 | 3+3 4 | 3+3 4 | 3+4 4 | 3+4 4 | 3+4 4 |
| | 5 2 | 2+2 2 | 4+4 4 | 3+3 4 | 3+3 4 | 3+4 4 | 3+4 4 | 3+4 4 |
| | 6 2 | 3+3 2 | 3+3 4 | 4+4 4 | 4+4 4 | 3+4 4 | 3+4 4 | 3+4 4 |
| | 5 2 | 7+7 2 | 3+3 4 | 4+4 4 | 3+3 4 | 3+4 4 | 3+4 4 | 3+4 4 |
| | 6 2 | 8+8 2 | 4+4 4 | 3+3 4 | 4+4 4 | 3+4 4 | 3+4 4 | 3+4 4 |
| | 6 2 | 6+6 2 | 4+4 4 | 3+3 4 | 3+3 4 | 3+4 4 | 3+4 4 | 3+4 4 |
| | 5 2 | 3+3 2 | 3+3 4 | 4+4 4 | 4+4 4 | 3+4 4 | 3+4 4 | 3+4 4 |
| | 5 2 | 9+9 2 | 4+4 4 | 4+4 4 | 4+4 4 | 3+4 4 | 3+4 4 | 3+4 4 |
| | 5 2 | 2+2 2 | 3+3 4 | 3+3 4 | 3+3 4 | 3+4 4 | 3+4 4 | 3+4 4 |
| | 6 2 | 5+5 2 | 3+3 4 | 4+4 4 | 3+3 4 | 3+4 4 | 3+4 4 | 3+4 4 |
| | 5 2 | 4+4 2 | 4+4 4 | 3+3 4 | 4+4 4 | 3+4 4 | 3+4 4 | 3+4 4 |
| | 6 2 | 2+2 2 | 3+3 4 | 3+3 4 | 4+4 4 | 3+4 4 | 3+4 4 | 3+4 4 |
| | 5 2 | 4+4 2 | 4+4 4 | 4+4 4 | 3+3 4 | 3+4 4 | 3+4 4 | 3+4 4 |
| | 6 2 | 3+3 2 | 3+3 4 | 4+4 4 | 4+4 4 | 3+4 4 | 3+4 4 | 3+4 4 |
| | 5 2 | 5+5 2 | 4+4 4 | 3+3 4 | 3+3 4 | 3+4 4 | 3+4 4 | 3+4 4 |
| | 5 2 | 7+7 2 | 4+4 4 | 4+4 4 | 4+4 4 | 3+4 4 | 3+4 4 | 3+4 4 |
| | 6 2 | 9+9 2 | 3+3 4 | 3+3 4 | 3+3 4 | 3+4 4 | 3+4 4 | 3+4 4 |
| | 6 2 | 6+6 2 | 4+4 4 | 3+3 4 | 4+4 4 | 3+4 4 | 3+4 4 | 3+4 4 |
| | 5 2 | 8+8 2 | 3+3 4 | 4+4 4 | 3+3 4 | 3+4 4 | 3+4 4 | 3+4 4 |
| | 5 2 | 8+8 2 | 4+4 4 | 4+4 4 | 3+3 4 | 3+4 4 | 3+4 4 | 3+4 4 |
| | 6 2 | 4+4 2 | 3+3 4 | 3+3 4 | 4+4 4 | 3+4 4 | 3+4 4 | 3+4 4 |
| | 5 2 | 7+7 2 | 3+3 4 | 3+3 4 | 3+3 4 | 3+4 4 | 3+4 4 | 3+4 4 |

Continued on next column

**TABLE XXXVI– continued**

| TQC | ICSD NoE SF |
|---|---|
| | 6 2 6+6 2 4+4 4 4+4 4 4+4 4 3+4 4 3+4 4 3+4 4 |
| | 5 2 2+2 2 3+3 4 3+3 4 3+3 4 3+4 4 3+4 4 3+4 4 |
| | 6 2 5+5 2 4+4 4 4+4 4 4+4 4 3+4 4 3+4 4 3+4 4 |
| | 5 2 9+9 2 |
| | 6 2 3+3 2 |
| | 5 2 6+6 2 |
| | 5 2 9+9 2 |
| | 6 2 8+8 2 |
| | 6 2 3+3 2 |
| | 5 2 4+4 2 |
| | 6 2 5+5 2 |
| | 5 2 2+2 2 |
| | 5 2 7+7 2 |
| | 6 2 3+3 2 |
| | 6 2 2+2 2 |
| | 5 2 5+5 2 |
| | 6 2 4+4 2 |
| | 5 2 9+9 2 |
| | 6 2 7+7 2 |
| | 5 2 6+6 2 |
| | 5 2 8+8 2 |
| | 6 2 3+3 2 |
| | 5 2 4+4 2 |
| | 5 2 5+5 2 |
| | 5 2 2+2 2 |
| | 6 2 7+7 2 |
| | 6 2 6+6 2 |
| | 6 2 9+9 2 |
| | 6 2 8+8 2 |
| | 5 2 6+6 2 |
| | 6 2 7+7 2 |
| | 6 2 9+9 2 |
| | 5 2 8+8 2 |
| | 5 2 5+5 2 |
| | 5 2 2+2 2 |
| | 5 2 3+3 2 |
| | 6 2 4+4 2 |
| | 5 2 2+2 2 |
| | 5 2 4+4 2 |
| | 6 2 5+5 2 |
| | 6 2 3+3 2 |

The End

TABLE XXXVII: The computed irreducible representations for the valence bands for the compounds of SEBR and NLC in space group 62. See the list of tables.

| TQC | ICSD | NoE | SF | | | | |
|---|---|---|---|---|---|---|---|
| 62.2.1 | 418619 | 280 | Sr1Zn5 | | | | |
| | -GM | -R | -S | -T | -U | -X | -Y -Z |
| | 5 2 3+3 4 3+3 4 3+4 4 6+6 4 3+4 4 3+4 4 3+4 4 | | | | | | |
| | 5 2 4+4 4 4+4 4 3+4 4 5+5 4 3+4 4 3+4 4 3+4 4 | | | | | | |
| | 6 2 3+3 4 3+3 4 3+4 4 6+6 4 3+4 4 3+4 4 3+4 4 | | | | | | |
| | 6 2 4+4 4 4+4 4 3+4 4 5+5 4 3+4 4 3+4 4 3+4 4 | | | | | | |
| | 5 2 3+3 4 3+3 4 3+4 4 6+6 4 3+4 4 3+4 4 3+4 4 | | | | | | |
| | 5 2 4+4 4 4+4 4 3+4 4 5+5 4 3+4 4 3+4 4 3+4 4 | | | | | | |
| | 6 2 3+3 4 3+3 4 3+4 4 6+6 4 3+4 4 3+4 4 3+4 4 | | | | | | |
| | 6 2 4+4 4 4+4 4 3+4 4 5+5 4 3+4 4 3+4 4 3+4 4 | | | | | | |
| | 5 2 4+4 4 4+4 4 3+4 4 5+5 4 3+4 4 3+4 4 3+4 4 | | | | | | |
| | 5 2 3+3 4 3+3 4 3+4 4 6+6 4 3+4 4 3+4 4 3+4 4 | | | | | | |

Continued on next column

**TABLE XXXVII– continued**

| TQC | ICSD NoE SF |
|---|---|
| | 5 2 3+3 4 3+3 4 3+4 4 5+5 4 3+4 4 3+4 4 3+4 4 |
| | 5 2 4+4 4 4+4 4 3+4 4 6+6 4 3+4 4 3+4 4 3+4 4 |
| | 6 2 3+3 4 3+3 4 3+4 4 6+6 4 3+4 4 3+4 4 3+4 4 |
| | 6 2 4+4 4 4+4 4 3+4 4 5+5 4 3+4 4 3+4 4 3+4 4 |
| | 6 2 3+3 4 3+3 4 3+4 4 6+6 4 3+4 4 3+4 4 3+4 4 |
| | 6 2 4+4 4 4+4 4 3+4 4 5+5 4 3+4 4 3+4 4 3+4 4 |
| | 5 2 4+4 4 3+3 4 3+4 4 5+5 4 3+4 4 3+4 4 3+4 4 |
| | 6 2 3+3 4 4+4 4 3+4 4 6+6 4 3+4 4 3+4 4 3+4 4 |
| | 5 2 3+3 4 4+4 4 3+4 4 6+6 4 3+4 4 3+4 4 3+4 4 |
| | 6 2 4+4 4 3+3 4 3+4 4 5+5 4 3+4 4 3+4 4 3+4 4 |
| | 6 2 3+3 4 3+3 4 3+4 4 5+5 4 3+4 4 3+4 4 3+4 4 |
| | 5 2 4+4 4 4+4 4 3+4 4 6+6 4 3+4 4 3+4 4 3+4 4 |
| | 6 2 3+3 4 4+4 4 3+4 4 6+6 4 3+4 4 3+4 4 3+4 4 |
| | 5 2 4+4 4 3+3 4 3+4 4 5+5 4 3+4 4 3+4 4 3+4 4 |
| | 6 2 3+3 4 3+3 4 3+4 4 6+6 4 3+4 4 3+4 4 3+4 4 |
| | 5 2 4+4 4 4+4 4 3+4 4 5+5 4 3+4 4 3+4 4 3+4 4 |
| | 6 2 4+4 4 4+4 4 3+4 4 5+5 4 3+4 4 3+4 4 3+4 4 |
| | 5 2 3+3 4 3+3 4 3+4 4 6+6 4 3+4 4 3+4 4 3+4 4 |
| | 5 2 4+4 4 4+4 4 3+4 4 5+5 4 3+4 4 3+4 4 3+4 4 |
| | 5 2 3+3 4 3+3 4 3+4 4 6+6 4 3+4 4 3+4 4 3+4 4 |
| | 6 2 3+3 4 3+3 4 3+4 4 5+5 4 3+4 4 3+4 4 3+4 4 |
| | 5 2 4+4 4 4+4 4 3+4 4 5+5 4 3+4 4 3+4 4 3+4 4 |
| | 6 2 3+3 4 3+3 4 3+4 4 6+6 4 3+4 4 3+4 4 3+4 4 |
| | 6 2 4+4 4 4+4 4 3+4 4 6+6 4 3+4 4 3+4 4 3+4 4 |
| | 5 2 3+3 4 3+3 4 3+4 4 5+5 4 3+4 4 3+4 4 3+4 4 |
| | 6 2 4+4 4 4+4 4 3+4 4 6+6 4 3+4 4 3+4 4 3+4 4 |
| | 5 2 3+3 4 3+3 4 3+4 4 5+5 4 3+4 4 3+4 4 3+4 4 |
| | 6 2 4+4 4 4+4 4 3+4 4 6+6 4 3+4 4 3+4 4 3+4 4 |
| | 5 2 3+3 4 4+4 4 3+4 4 5+5 4 3+4 4 3+4 4 3+4 4 |
| | 5 2 4+4 4 3+3 4 3+4 4 6+6 4 3+4 4 3+4 4 3+4 4 |
| | 5 2 4+4 4 3+3 4 3+4 4 6+6 4 3+4 4 3+4 4 3+4 4 |
| | 6 2 3+3 4 4+4 4 3+4 4 5+5 4 3+4 4 3+4 4 3+4 4 |
| | 6 2 4+4 4 3+3 4 3+4 4 6+6 4 3+4 4 3+4 4 3+4 4 |
| | 6 2 3+3 4 4+4 4 3+4 4 5+5 4 3+4 4 3+4 4 3+4 4 |
| | 5 2 3+3 4 4+4 4 3+4 4 5+5 4 3+4 4 3+4 4 3+4 4 |
| | 5 2 4+4 4 3+3 4 3+4 4 6+6 4 3+4 4 3+4 4 3+4 4 |
| | 6 2 4+4 4 4+4 4 3+4 4 5+5 4 3+4 4 3+4 4 3+4 4 |
| | 6 2 3+3 4 3+3 4 3+4 4 6+6 4 3+4 4 3+4 4 3+4 4 |
| | 6 2 3+3 4 3+3 4 3+4 4 5+5 4 3+4 4 3+4 4 3+4 4 |
| | 5 2 4+4 4 4+4 4 3+4 4 6+6 4 3+4 4 3+4 4 3+4 4 |
| | 5 2 4+4 4 4+4 4 3+4 4 5+5 4 3+4 4 3+4 4 3+4 4 |
| | 6 2 3+3 4 3+3 4 3+4 4 6+6 4 3+4 4 3+4 4 3+4 4 |
| | 5 2 4+4 4 3+3 4 3+4 4 5+5 4 3+4 4 3+4 4 3+4 4 |
| | 6 2 3+3 4 4+4 4 3+4 4 6+6 4 3+4 4 3+4 4 3+4 4 |
| | 5 2 3+3 4 4+4 4 3+4 4 5+5 4 3+4 4 3+4 4 3+4 4 |
| | 6 2 4+4 4 3+3 4 3+4 4 6+6 4 3+4 4 3+4 4 3+4 4 |
| | 5 2 3+3 4 4+4 4 3+4 4 5+5 4 3+4 4 3+4 4 3+4 4 |
| | 6 2 4+4 4 3+3 4 3+4 4 6+6 4 3+4 4 3+4 4 3+4 4 |
| | 6 2 4+4 4 4+4 4 3+4 4 5+5 4 3+4 4 3+4 4 3+4 4 |
| | 5 2 3+3 4 3+3 4 3+4 4 6+6 4 3+4 4 3+4 4 3+4 4 |
| | 5 2 3+3 4 3+3 4 3+4 4 5+5 4 3+4 4 3+4 4 3+4 4 |
| | 6 2 4+4 4 4+4 4 3+4 4 6+6 4 3+4 4 3+4 4 3+4 4 |
| | 6 2 3+3 4 3+3 4 3+4 4 6+6 4 3+4 4 3+4 4 3+4 4 |
| | 5 2 4+4 4 4+4 4 3+4 4 5+5 4 3+4 4 3+4 4 3+4 4 |
| | 6 2 4+4 4 4+4 4 3+4 4 5+5 4 3+4 4 3+4 4 3+4 4 |
| | 5 2 3+3 4 3+3 4 3+4 4 5+5 4 3+4 4 3+4 4 3+4 4 |
| | 5 2 3+3 4 4+4 4 3+4 4 6+6 4 3+4 4 3+4 4 3+4 4 |
| | 6 2 4+4 4 3+3 4 3+4 4 6+6 4 3+4 4 3+4 4 3+4 4 |
| | 5 2 3+3 4 4+4 4 3+4 4 5+5 4 3+4 4 3+4 4 3+4 4 |
| | 6 2 4+4 4 3+3 4 3+4 4 5+5 4 3+4 4 3+4 4 3+4 4 |
| | 5 2 |

Continued on next column

**TABLE XXXVII– continued**

| TQC | ICSD | NoE | SF |
|---|---|---|---|
| | 5 2 | | |
| | 6 2 | | |
| | 6 2 | | |
| | 5 2 | | |
| | 6 2 | | |
| | 5 2 | | |
| | 6 2 | | |
| | 5 2 | | |
| | 5 2 | | |
| | 6 2 | | |
| | 5 2 | | |
| | 5 2 | | |
| | 6 2 | | |
| | 5 2 | | |
| | 6 2 | | |
| | 6 2 | | |
| | 6 2 | | |
| | 5 2 | | |
| | 6 2 | | |
| | 5 2 | | |
| | 5 2 | | |
| | 6 2 | | |
| | 5 2 | | |
| | 6 2 | | |
| | 5 2 | | |
| | 6 2 | | |
| | 6 2 | | |
| | 5 2 | | |
| | 5 2 | | |
| | 6 2 | | |
| | 5 2 | | |
| | 6 2 | | |
| | 6 2 | | |
| | 5 2 | | |
| | 5 2 | | |
| | 6 2 | | |
| | 5 2 | | |
| | 6 2 | | |
| | 5 2 | | |
| | 6 2 | | |
| | 5 2 | | |
| | 5 2 | | |
| | 6 2 | | |
| | 5 2 | | |
| | 6 2 | | |
| | 5 2 | | |
| | 6 2 | | |
| | 6 2 | | |
| | 6 2 | | |
| | 5 2 | | |
| | 6 2 | | |
| | 5 2 | | |
| | 5 2 | | |
| | 6 2 | | |
| | 6 2 | | |
| | 5 2 | | |
| | 6 2 | | |
| | 5 2 | | |
| | 6 2 | | |
| | 6 2 | | |
| | 5 2 | | |

Continued on next column

**TABLE XXXVII– continued**

| TQC | ICSD | | NoE | | SF | | | |
|---|---|---|---|---|---|---|---|---|
| | 5 2 | | | | | | | |
| | 6 2 | | | | | | | |
| | 5 2 | | | | | | | |
| | 5 2 | | | | | | | |
| | 6 2 | | | | | | | |
| | 5 2 | | | | | | | |
| | 5 2 | | | | | | | |
| | 6 2 | | | | | | | |
| 62.2.2 | 54980 | | 88 | | Sr1Zn1 | | | |
| | -GM | -R | -S | -T | -U | -X | -Y | -Z |
| | 5 2 | 3+3 4 | 3+3 4 | 3+4 4 | 5+5 4 | 3+4 4 | 3+4 4 | 3+4 4 |
| | 6 2 | 4+4 4 | 4+4 4 | 3+4 4 | 6+6 4 | 3+4 4 | 3+4 4 | 3+4 4 |
| | 5 2 | 3+3 4 | 3+3 4 | 3+4 4 | 5+5 4 | 3+4 4 | 3+4 4 | 3+4 4 |
| | 6 2 | 4+4 4 | 4+4 4 | 3+4 4 | 6+6 4 | 3+4 4 | 3+4 4 | 3+4 4 |
| | 5 2 | 4+4 4 | 4+4 4 | 3+4 4 | 5+5 4 | 3+4 4 | 3+4 4 | 3+4 4 |
| | 6 2 | 3+3 4 | 3+3 4 | 3+4 4 | 6+6 4 | 3+4 4 | 3+4 4 | 3+4 4 |
| | 5 2 | 3+3 4 | 3+3 4 | 3+4 4 | 5+5 4 | 3+4 4 | 3+4 4 | 3+4 4 |
| | 6 2 | 4+4 4 | 4+4 4 | 3+4 4 | 6+6 4 | 3+4 4 | 3+4 4 | 3+4 4 |
| | 5 2 | 3+3 4 | 3+3 4 | 3+4 4 | 6+6 4 | 3+4 4 | 3+4 4 | 3+4 4 |
| | 6 2 | 4+4 4 | 4+4 4 | 3+4 4 | 5+5 4 | 3+4 4 | 3+4 4 | 3+4 4 |
| | 5 2 | 4+4 4 | 3+3 4 | 3+4 4 | 6+6 4 | 3+4 4 | 3+4 4 | 3+4 4 |
| | 5 2 | 3+3 4 | 4+4 4 | 3+4 4 | 5+5 4 | 3+4 4 | 3+4 4 | 3+4 4 |
| | 6 2 | 3+3 4 | 3+3 4 | 3+4 4 | 5+5 4 | 3+4 4 | 3+4 4 | 3+4 4 |
| | 5 2 | 4+4 4 | 4+4 4 | 3+4 4 | 6+6 4 | 3+4 4 | 3+4 4 | 3+4 4 |
| | 6 2 | 3+3 4 | 4+4 4 | 3+4 4 | 5+5 4 | 3+4 4 | 3+4 4 | 3+4 4 |
| | 6 2 | 4+4 4 | 3+3 4 | 3+4 4 | 6+6 4 | 3+4 4 | 3+4 4 | 3+4 4 |
| | 5 2 | 3+3 4 | 3+3 4 | 3+4 4 | 6+6 4 | 3+4 4 | 3+4 4 | 3+4 4 |
| | 5 2 | 4+4 4 | 4+4 4 | 3+4 4 | 5+5 4 | 3+4 4 | 3+4 4 | 3+4 4 |
| | 6 2 | 4+4 4 | 3+3 4 | 3+4 4 | 6+6 4 | 3+4 4 | 3+4 4 | 3+4 4 |
| | 5 2 | 3+3 4 | 4+4 4 | 3+4 4 | 5+5 4 | 3+4 4 | 3+4 4 | 3+4 4 |
| | 5 2 | 3+3 4 | 4+4 4 | 3+4 4 | 5+5 4 | 3+4 4 | 3+4 4 | 3+4 4 |
| | 6 2 | 4+4 4 | 3+3 4 | 3+4 4 | 6+6 4 | 3+4 4 | 3+4 4 | 3+4 4 |
| | 6 2 | | | | | | | |
| | 6 2 | | | | | | | |
| | 5 2 | | | | | | | |
| | 6 2 | | | | | | | |
| | 6 2 | | | | | | | |
| | 5 2 | | | | | | | |
| | 6 2 | | | | | | | |
| | 5 2 | | | | | | | |
| | 5 2 | | | | | | | |
| | 6 2 | | | | | | | |
| | 5 2 | | | | | | | |
| | 5 2 | | | | | | | |
| | 6 2 | | | | | | | |
| | 6 2 | | | | | | | |
| | 5 2 | | | | | | | |
| | 5 2 | | | | | | | |
| | 6 2 | | | | | | | |
| | 6 2 | | | | | | | |
| | 6 2 | | | | | | | |
| | 5 2 | | | | | | | |
| | 5 2 | | | | | | | |
| | 5 2 | | | | | | | |
| 62.2.3 | 82295 | | 104 | | Se2U1 | | | |
| | -GM | -R | -S | -T | -U | -X | -Y | -Z |
| | 5 2 | 3+3 4 | 3+3 4 | 3+4 4 | 5+5 4 | 3+4 4 | 3+4 4 | 3+4 4 |
| | 6 2 | 4+4 4 | 4+4 4 | 3+4 4 | 6+6 4 | 3+4 4 | 3+4 4 | 3+4 4 |
| | 5 2 | 3+3 4 | 4+4 4 | 3+4 4 | 6+6 4 | 3+4 4 | 3+4 4 | 3+4 4 |
| | 6 2 | 4+4 4 | 3+3 4 | 3+4 4 | 5+5 4 | 3+4 4 | 3+4 4 | 3+4 4 |
| | 5 2 | 4+4 4 | 3+3 4 | 3+4 4 | 6+6 4 | 3+4 4 | 3+4 4 | 3+4 4 |

Continued on next column

**TABLE XXXVII– continued**

| TQC | ICSD NoE SF |
|---|---|
| | 6 2 3+3 4 4+4 4 3+4 4 5+5 4 3+4 4 3+4 4 3+4 4 |
| | 5 2 3+3 4 4+4 4 3+4 4 5+5 4 3+4 4 3+4 4 3+4 4 |
| | 6 2 4+4 4 3+3 4 3+4 4 6+6 4 3+4 4 3+4 4 3+4 4 |
| | 5 2 3+3 4 4+4 4 3+4 4 6+6 4 3+4 4 3+4 4 3+4 4 |
| | 6 2 4+4 4 3+3 4 3+4 4 5+5 4 3+4 4 3+4 4 3+4 4 |
| | 5 2 3+3 4 4+4 4 3+4 4 6+6 4 3+4 4 3+4 4 3+4 4 |
| | 5 2 4+4 4 3+3 4 3+4 4 5+5 4 3+4 4 3+4 4 3+4 4 |
| | 6 2 3+3 4 4+4 4 3+4 4 5+5 4 3+4 4 3+4 4 3+4 4 |
| | 5 2 4+4 4 3+3 4 3+4 4 6+6 4 3+4 4 3+4 4 3+4 4 |
| | 6 2 4+4 4 3+3 4 3+4 4 6+6 4 3+4 4 3+4 4 3+4 4 |
| | 6 2 3+3 4 4+4 4 3+4 4 6+6 4 3+4 4 3+4 4 3+4 4 |
| | 5 2 4+4 4 3+3 4 3+4 4 5+5 4 3+4 4 3+4 4 3+4 4 |
| | 6 2 3+3 4 4+4 4 3+4 4 5+5 4 3+4 4 3+4 4 3+4 4 |
| | 5 2 4+4 4 3+3 4 3+4 4 6+6 4 3+4 4 3+4 4 3+4 4 |
| | 6 2 3+3 4 4+4 4 3+4 4 5+5 4 3+4 4 3+4 4 3+4 4 |
| | 6 2 3+3 4 4+4 4 3+4 4 5+5 4 3+4 4 3+4 4 3+4 4 |
| | 5 2 4+4 4 3+3 4 3+4 4 6+6 4 3+4 4 3+4 4 3+4 4 |
| | 5 2 4+4 4 4+4 4 3+4 4 6+6 4 3+4 4 3+4 4 3+4 4 |
| | 6 2 3+3 4 3+3 4 3+4 4 5+5 4 3+4 4 3+4 4 3+4 4 |
| | 5 2 3+3 4 3+3 4 3+4 4 6+6 4 3+4 4 3+4 4 3+4 4 |
| | 6 2 4+4 4 4+4 4 3+4 4 5+5 4 3+4 4 3+4 4 3+4 4 |
| | 6 2 |
| | 5 2 |
| | 6 2 |
| | 5 2 |
| | 5 2 |
| | 6 2 |
| | 5 2 |
| | 5 2 |
| | 5 2 |
| | 5 2 |
| | 6 2 |
| | 5 2 |
| | 6 2 |
| | 5 2 |
| | 5 2 |
| | 6 2 |
| | 5 2 |
| | 6 2 |
| | 6 2 |
| | 6 2 |
| | 6 2 |
| | 6 2 |
| | 5 2 |
| | 6 2 |
| | 5 2 |
| | 5 2 |
| 62.2.4 | 10001 128 Sb2Sr1Zn1 |
| | -GM -R -S -T -U -X -Y -Z |
| | 5 2 3+3 4 3+3 4 3+4 4 5+5 4 3+4 4 3+4 4 3+4 4 |
| | 5 2 4+4 4 4+4 4 3+4 4 6+6 4 3+4 4 3+4 4 3+4 4 |
| | 6 2 3+3 4 3+3 4 3+4 4 5+5 4 3+4 4 3+4 4 3+4 4 |
| | 6 2 4+4 4 4+4 4 3+4 4 6+6 4 3+4 4 3+4 4 3+4 4 |
| | 5 2 3+3 4 3+3 4 3+4 4 5+5 4 3+4 4 3+4 4 3+4 4 |
| | 5 2 4+4 4 4+4 4 3+4 4 6+6 4 3+4 4 3+4 4 3+4 4 |
| | 6 2 3+3 4 3+3 4 3+4 4 5+5 4 3+4 4 3+4 4 3+4 4 |
| | 6 2 4+4 4 4+4 4 3+4 4 6+6 4 3+4 4 3+4 4 3+4 4 |
| | 5 2 3+3 4 3+3 4 3+4 4 5+5 4 3+4 4 3+4 4 3+4 4 |
| | 5 2 4+4 4 4+4 4 3+4 4 6+6 4 3+4 4 3+4 4 3+4 4 |
| | 6 2 3+3 4 3+3 4 3+4 4 5+5 4 3+4 4 3+4 4 3+4 4 |
| | 6 2 4+4 4 4+4 4 3+4 4 6+6 4 3+4 4 3+4 4 3+4 4 |

Continued on next column

**TABLE XXXVII– continued**

| TQC | ICSD NoE SF |
|---|---|
| | 5 2 3+3 4 3+3 4 3+4 4 5+5 4 3+4 4 3+4 4 3+4 4 |
| | 5 2 4+4 4 4+4 4 3+4 4 6+6 4 3+4 4 3+4 4 3+4 4 |
| | 6 2 3+3 4 3+3 4 3+4 4 5+5 4 3+4 4 3+4 4 3+4 4 |
| | 6 2 4+4 4 4+4 4 3+4 4 6+6 4 3+4 4 3+4 4 3+4 4 |
| | 5 2 3+3 4 3+3 4 3+4 4 5+5 4 3+4 4 3+4 4 3+4 4 |
| | 5 2 4+4 4 4+4 4 3+4 4 6+6 4 3+4 4 3+4 4 3+4 4 |
| | 6 2 3+3 4 3+3 4 3+4 4 5+5 4 3+4 4 3+4 4 3+4 4 |
| | 5 2 4+4 4 4+4 4 3+4 4 6+6 4 3+4 4 3+4 4 3+4 4 |
| | 6 2 3+3 4 3+3 4 3+4 4 5+5 4 3+4 4 3+4 4 3+4 4 |
| | 5 2 4+4 4 4+4 4 3+4 4 6+6 4 3+4 4 3+4 4 3+4 4 |
| | 5 2 3+3 4 3+3 4 3+4 4 6+6 4 3+4 4 3+4 4 3+4 4 |
| | 6 2 4+4 4 4+4 4 3+4 4 5+5 4 3+4 4 3+4 4 3+4 4 |
| | 5 2 3+3 4 4+4 4 3+4 4 6+6 4 3+4 4 3+4 4 3+4 4 |
| | 6 2 4+4 4 3+3 4 3+4 4 5+5 4 3+4 4 3+4 4 3+4 4 |
| | 6 2 3+3 4 3+3 4 3+4 4 6+6 4 3+4 4 3+4 4 3+4 4 |
| | 5 2 4+4 4 4+4 4 3+4 4 5+5 4 3+4 4 3+4 4 3+4 4 |
| | 6 2 3+3 4 4+4 4 3+4 4 5+5 4 3+4 4 3+4 4 3+4 4 |
| | 5 2 4+4 4 3+3 4 3+4 4 6+6 4 3+4 4 3+4 4 3+4 4 |
| | 6 2 4+4 4 3+3 4 3+4 4 5+5 4 3+4 4 3+4 4 3+4 4 |
| | 5 2 3+3 4 4+4 4 3+4 4 6+6 4 3+4 4 3+4 4 3+4 4 |
| | 5 2 |
| | 6 2 |
| | 5 2 |
| | 6 2 |
| | 5 2 |
| | 6 2 |
| | 5 2 |
| | 6 2 |
| | 5 2 |
| | 6 2 |
| | 6 2 |
| | 6 2 |
| | 6 2 |
| | 5 2 |
| | 6 2 |
| | 5 2 |
| | 5 2 |
| | 6 2 |
| | 5 2 |
| | 6 2 |
| | 5 2 |
| | 5 2 |
| | 6 2 |
| | 5 2 |
| | 5 2 |
| | 6 2 |
| | 5 2 |
| | 6 2 |
| | 5 2 |
| | 6 2 |
| | 6 2 |
| | 5 2 |
| 62.2.5 | 402457 256 S3Tm2 |
| | -GM -R -S -T -U -X -Y -Z |
| | 5 2 3+3 4 3+3 4 3+4 4 5+5 4 3+4 4 3+4 4 3+4 4 |
| | 5 2 4+4 4 4+4 4 3+4 4 6+6 4 3+4 4 3+4 4 3+4 4 |
| | 6 2 3+3 4 3+3 4 3+4 4 5+5 4 3+4 4 3+4 4 3+4 4 |
| | 6 2 4+4 4 4+4 4 3+4 4 6+6 4 3+4 4 3+4 4 3+4 4 |
| | 5 2 3+3 4 3+3 4 3+4 4 5+5 4 3+4 4 3+4 4 3+4 4 |
| | 5 2 4+4 4 4+4 4 3+4 4 6+6 4 3+4 4 3+4 4 3+4 4 |
| | 6 2 3+3 4 3+3 4 3+4 4 6+6 4 3+4 4 3+4 4 3+4 4 |

Continued on next column

**TABLE XXXVII– continued**

| TQC | ICSD | NoE | SF |
|---|---|---|---|
| | 6 2 4+4 4 4+4 4 3+4 4 5+5 4 3+4 4 3+4 4 3+4 4 | | |
| | 5 2 4+4 4 3+3 4 3+4 4 5+5 4 3+4 4 3+4 4 3+4 4 | | |
| | 6 2 3+3 4 4+4 4 3+4 4 6+6 4 3+4 4 3+4 4 3+4 4 | | |
| | 5 2 3+3 4 3+3 4 3+4 4 6+6 4 3+4 4 3+4 4 3+4 4 | | |
| | 6 2 4+4 4 4+4 4 3+4 4 5+5 4 3+4 4 3+4 4 3+4 4 | | |
| | 5 2 4+4 4 4+4 4 3+4 4 5+5 4 3+4 4 3+4 4 3+4 4 | | |
| | 5 2 3+3 4 3+3 4 3+4 4 6+6 4 3+4 4 3+4 4 3+4 4 | | |
| | 6 2 3+3 4 4+4 4 3+4 4 6+6 4 3+4 4 3+4 4 3+4 4 | | |
| | 6 2 4+4 4 3+3 4 3+4 4 5+5 4 3+4 4 3+4 4 3+4 4 | | |
| | 5 2 4+4 4 4+4 4 3+4 4 6+6 4 3+4 4 3+4 4 3+4 4 | | |
| | 6 2 3+3 4 3+3 4 3+4 4 5+5 4 3+4 4 3+4 4 3+4 4 | | |
| | 5 2 3+3 4 3+3 4 3+4 4 5+5 4 3+4 4 3+4 4 3+4 4 | | |
| | 5 2 4+4 4 4+4 4 3+4 4 6+6 4 3+4 4 3+4 4 3+4 4 | | |
| | 6 2 3+3 4 3+3 4 3+4 4 6+6 4 3+4 4 3+4 4 3+4 4 | | |
| | 5 2 4+4 4 4+4 4 3+4 4 5+5 4 3+4 4 3+4 4 3+4 4 | | |
| | 6 2 3+3 4 3+3 4 3+4 4 5+5 4 3+4 4 3+4 4 3+4 4 | | |
| | 6 2 4+4 4 4+4 4 3+4 4 6+6 4 3+4 4 3+4 4 3+4 4 | | |
| | 5 2 4+4 4 3+3 4 3+4 4 6+6 4 3+4 4 3+4 4 3+4 4 | | |
| | 5 2 3+3 4 4+4 4 3+4 4 5+5 4 3+4 4 3+4 4 3+4 4 | | |
| | 6 2 3+3 4 4+4 4 3+4 4 6+6 4 3+4 4 3+4 4 3+4 4 | | |
| | 5 2 4+4 4 3+3 4 3+4 4 5+5 4 3+4 4 3+4 4 3+4 4 | | |
| | 6 2 3+3 4 3+3 4 3+4 4 6+6 4 3+4 4 3+4 4 3+4 4 | | |
| | 6 2 4+4 4 4+4 4 3+4 4 5+5 4 3+4 4 3+4 4 3+4 4 | | |
| | 5 2 4+4 4 4+4 4 3+4 4 5+5 4 3+4 4 3+4 4 3+4 4 | | |
| | 6 2 3+3 4 3+3 4 3+4 4 6+6 4 3+4 4 3+4 4 3+4 4 | | |
| | 5 2 4+4 4 3+3 4 3+4 4 6+6 4 3+4 4 3+4 4 3+4 4 | | |
| | 5 2 3+3 4 4+4 4 3+4 4 5+5 4 3+4 4 3+4 4 3+4 4 | | |
| | 5 2 3+3 4 4+4 4 3+4 4 5+5 4 3+4 4 3+4 4 3+4 4 | | |
| | 6 2 4+4 4 3+3 4 3+4 4 6+6 4 3+4 4 3+4 4 3+4 4 | | |
| | 6 2 3+3 4 3+3 4 3+4 4 6+6 4 3+4 4 3+4 4 3+4 4 | | |
| | 6 2 4+4 4 4+4 4 3+4 4 5+5 4 3+4 4 3+4 4 3+4 4 | | |
| | 5 2 3+3 4 3+3 4 3+4 4 5+5 4 3+4 4 3+4 4 3+4 4 | | |
| | 6 2 4+4 4 4+4 4 3+4 4 6+6 4 3+4 4 3+4 4 3+4 4 | | |
| | 5 2 3+3 4 4+4 4 3+4 4 5+5 4 3+4 4 3+4 4 3+4 4 | | |
| | 6 2 4+4 4 3+3 4 3+4 4 6+6 4 3+4 4 3+4 4 3+4 4 | | |
| | 5 2 4+4 4 4+4 4 3+4 4 6+6 4 3+4 4 3+4 4 3+4 4 | | |
| | 6 2 3+3 4 3+3 4 3+4 4 5+5 4 3+4 4 3+4 4 3+4 4 | | |
| | 6 2 3+3 4 3+3 4 3+4 4 6+6 4 3+4 4 3+4 4 3+4 4 | | |
| | 5 2 4+4 4 4+4 4 3+4 4 5+5 4 3+4 4 3+4 4 3+4 4 | | |
| | 6 2 3+3 4 3+3 4 3+4 4 6+6 4 3+4 4 3+4 4 3+4 4 | | |
| | 5 2 4+4 4 4+4 4 3+4 4 5+5 4 3+4 4 3+4 4 3+4 4 | | |
| | 5 2 4+4 4 4+4 4 3+4 4 5+5 4 3+4 4 3+4 4 3+4 4 | | |
| | 5 2 3+3 4 3+3 4 3+4 4 6+6 4 3+4 4 3+4 4 3+4 4 | | |
| | 5 2 3+3 4 4+4 4 3+4 4 6+6 4 3+4 4 3+4 4 3+4 4 | | |
| | 6 2 4+4 4 3+3 4 3+4 4 5+5 4 3+4 4 3+4 4 3+4 4 | | |
| | 6 2 3+3 4 4+4 4 3+4 4 5+5 4 3+4 4 3+4 4 3+4 4 | | |
| | 5 2 4+4 4 3+3 4 3+4 4 6+6 4 3+4 4 3+4 4 3+4 4 | | |
| | 5 2 3+3 4 4+4 4 3+4 4 5+5 4 3+4 4 3+4 4 3+4 4 | | |
| | 6 2 4+4 4 3+3 4 3+4 4 6+6 4 3+4 4 3+4 4 3+4 4 | | |
| | 5 2 4+4 4 3+3 4 3+4 4 6+6 4 3+4 4 3+4 4 3+4 4 | | |
| | 5 2 3+3 4 4+4 4 3+4 4 5+5 4 3+4 4 3+4 4 3+4 4 | | |
| | 5 2 3+3 4 4+4 4 3+4 4 5+5 4 3+4 4 3+4 4 3+4 4 | | |
| | 6 2 4+4 4 3+3 4 3+4 4 6+6 4 3+4 4 3+4 4 3+4 4 | | |
| | 5 2 4+4 4 3+3 4 3+4 4 6+6 4 3+4 4 3+4 4 3+4 4 | | |
| | 5 2 3+3 4 4+4 4 3+4 4 5+5 4 3+4 4 3+4 4 3+4 4 | | |
| | 6 2 3+3 4 3+3 4 3+4 4 6+6 4 3+4 4 3+4 4 3+4 4 | | |
| | 5 2 4+4 4 4+4 4 3+4 4 5+5 4 3+4 4 3+4 4 3+4 4 | | |
| | 6 2 | | |
| | 6 2 | | |
| | 6 2 | | |
| | 6 2 | | |

Continued on next column

**TABLE XXXVII– continued**

| TQC | ICSD | NoE | SF |
|---|---|---|---|
| | 6 2 | | |
| | 5 2 | | |
| | 5 2 | | |
| | 6 2 | | |
| | 5 2 | | |
| | 6 2 | | |
| | 5 2 | | |
| | 6 2 | | |
| | 6 2 | | |
| | 5 2 | | |
| | 5 2 | | |
| | 6 2 | | |
| | 5 2 | | |
| | 6 2 | | |
| | 5 2 | | |
| | 6 2 | | |
| | 5 2 | | |
| | 6 2 | | |
| | 5 2 | | |
| | 6 2 | | |
| | 5 2 | | |
| | 5 2 | | |
| | 6 2 | | |
| | 6 2 | | |
| | 5 2 | | |
| | 5 2 | | |
| | 6 2 | | |
| | 5 2 | | |
| | 6 2 | | |
| | 5 2 | | |
| | 6 2 | | |
| | 5 2 | | |
| | 6 2 | | |
| | 6 2 | | |
| | 6 2 | | |
| | 6 2 | | |
| | 5 2 | | |
| | 6 2 | | |
| | 5 2 | | |
| | 6 2 | | |
| | 5 2 | | |
| | 5 2 | | |
| | 5 2 | | |
| | 6 2 | | |
| | 5 2 | | |
| | 5 2 | | |
| | 6 2 | | |
| | 6 2 | | |
| | 5 2 | | |
| | 5 2 | | |
| | 6 2 | | |
| | 5 2 | | |
| | 6 2 | | |
| | 5 2 | | |
| | 6 2 | | |
| | 5 2 | | |
| | 6 2 | | |
| | 6 2 | | |
| | 5 2 | | |
| | 5 2 | | |
| 62.2.6 | 42430 | 104 | S1Te1U1 |

Continued on next column

**TABLE XXXVII– continued**

| TQC | ICSD | NoE | SF | | | | | |
|---|---|---|---|---|---|---|---|---|
| | -GM | -R | -S | -T | -U | -X | -Y | -Z |
| | 5 2 | 3+3 4 | 3+3 4 | 3+4 4 | 5+5 4 | 3+4 4 | 3+4 4 | 3+4 4 |
| | 5 2 | 4+4 4 | 4+4 4 | 3+4 4 | 6+6 4 | 3+4 4 | 3+4 4 | 3+4 4 |
| | 6 2 | 3+3 4 | 4+4 4 | 3+4 4 | 5+5 4 | 3+4 4 | 3+4 4 | 3+4 4 |
| | 6 2 | 4+4 4 | 3+3 4 | 3+4 4 | 6+6 4 | 3+4 4 | 3+4 4 | 3+4 4 |
| | 5 2 | 3+3 4 | 3+3 4 | 3+4 4 | 5+5 4 | 3+4 4 | 3+4 4 | 3+4 4 |
| | 5 2 | 4+4 4 | 4+4 4 | 3+4 4 | 6+6 4 | 3+4 4 | 3+4 4 | 3+4 4 |
| | 6 2 | 4+4 4 | 4+4 4 | 3+4 4 | 5+5 4 | 3+4 4 | 3+4 4 | 3+4 4 |
| | 6 2 | 3+3 4 | 3+3 4 | 3+4 4 | 6+6 4 | 3+4 4 | 3+4 4 | 3+4 4 |
| | 5 2 | 4+4 4 | 4+4 4 | 3+4 4 | 5+5 4 | 3+4 4 | 3+4 4 | 3+4 4 |
| | 6 2 | 3+3 4 | 3+3 4 | 3+4 4 | 6+6 4 | 3+4 4 | 3+4 4 | 3+4 4 |
| | 5 2 | 3+3 4 | 3+3 4 | 3+4 4 | 6+6 4 | 3+4 4 | 3+4 4 | 3+4 4 |
| | 5 2 | 4+4 4 | 4+4 4 | 3+4 4 | 5+5 4 | 3+4 4 | 3+4 4 | 3+4 4 |
| | 6 2 | 3+3 4 | 4+4 4 | 3+4 4 | 6+6 4 | 3+4 4 | 3+4 4 | 3+4 4 |
| | 5 2 | 4+4 4 | 3+3 4 | 3+4 4 | 5+5 4 | 3+4 4 | 3+4 4 | 3+4 4 |
| | 6 2 | 3+3 4 | 3+3 4 | 3+4 4 | 5+5 4 | 3+4 4 | 3+4 4 | 3+4 4 |
| | 6 2 | 4+4 4 | 4+4 4 | 3+4 4 | 5+5 4 | 3+4 4 | 3+4 4 | 3+4 4 |
| | 5 2 | 3+3 4 | 3+3 4 | 3+4 4 | 6+6 4 | 3+4 4 | 3+4 4 | 3+4 4 |
| | 6 2 | 4+4 4 | 4+4 4 | 3+4 4 | 6+6 4 | 3+4 4 | 3+4 4 | 3+4 4 |
| | 5 2 | 4+4 4 | 3+3 4 | 3+4 4 | 6+6 4 | 3+4 4 | 3+4 4 | 3+4 4 |
| | 6 2 | 3+3 4 | 4+4 4 | 3+4 4 | 5+5 4 | 3+4 4 | 3+4 4 | 3+4 4 |
| | 5 2 | 4+4 4 | 4+4 4 | 3+4 4 | 6+6 4 | 3+4 4 | 3+4 4 | 3+4 4 |
| | 5 2 | 3+3 4 | 3+3 4 | 3+4 4 | 5+5 4 | 3+4 4 | 3+4 4 | 3+4 4 |
| | 6 2 | 3+3 4 | 3+3 4 | 3+4 4 | 6+6 4 | 3+4 4 | 3+4 4 | 3+4 4 |
| | 6 2 | 4+4 4 | 4+4 4 | 3+4 4 | 5+5 4 | 3+4 4 | 3+4 4 | 3+4 4 |
| | 5 2 | 4+4 4 | 3+3 4 | 3+4 4 | 5+5 4 | 3+4 4 | 3+4 4 | 3+4 4 |
| | 6 2 | 3+3 4 | 4+4 4 | 3+4 4 | 6+6 4 | 3+4 4 | 3+4 4 | 3+4 4 |
| | 5 2 | | | | | | | |
| | 6 2 | | | | | | | |
| | 6 2 | | | | | | | |
| | 5 2 | | | | | | | |
| | 5 2 | | | | | | | |
| | 6 2 | | | | | | | |
| | 5 2 | | | | | | | |
| | 5 2 | | | | | | | |
| | 5 2 | | | | | | | |
| | 5 2 | | | | | | | |
| | 6 2 | | | | | | | |
| | 5 2 | | | | | | | |
| | 6 2 | | | | | | | |
| | 5 2 | | | | | | | |
| | 5 2 | | | | | | | |
| | 6 2 | | | | | | | |
| | 6 2 | | | | | | | |
| | 5 2 | | | | | | | |
| | 6 2 | | | | | | | |
| | 6 2 | | | | | | | |
| | 6 2 | | | | | | | |
| | 5 2 | | | | | | | |
| | 6 2 | | | | | | | |
| | 5 2 | | | | | | | |
| | 6 2 | | | | | | | |
| | 5 2 | | | | | | | |
| 62.2.7 | 30398 | 176 | Ru4Si3 | | | | | |
| | -GM | -R | -S | -T | -U | -X | -Y | -Z |
| | 5 2 | 3+3 4 | 4+4 4 | 3+4 4 | 6+6 4 | 3+4 4 | 3+4 4 | 3+4 4 |
| | 5 2 | 4+4 4 | 3+3 4 | 3+4 4 | 5+5 4 | 3+4 4 | 3+4 4 | 3+4 4 |
| | 6 2 | 3+3 4 | 3+3 4 | 3+4 4 | 6+6 4 | 3+4 4 | 3+4 4 | 3+4 4 |
| | 6 2 | 4+4 4 | 4+4 4 | 3+4 4 | 5+5 4 | 3+4 4 | 3+4 4 | 3+4 4 |
| | 5 2 | 3+3 4 | 3+3 4 | 3+4 4 | 6+6 4 | 3+4 4 | 3+4 4 | 3+4 4 |
| | 6 2 | 4+4 4 | 4+4 4 | 3+4 4 | 5+5 4 | 3+4 4 | 3+4 4 | 3+4 4 |

Continued on next column

**TABLE XXXVII– continued**

| TQC | ICSD | NoE | SF | | | | | |
|---|---|---|---|---|---|---|---|---|
| | 5 2 | 3+3 4 | 3+3 4 | 3+4 4 | 5+5 4 | 3+4 4 | 3+4 4 | 3+4 4 |
| | 5 2 | 4+4 4 | 4+4 4 | 3+4 4 | 6+6 4 | 3+4 4 | 3+4 4 | 3+4 4 |
| | 6 2 | 3+3 4 | 4+4 4 | 3+4 4 | 6+6 4 | 3+4 4 | 3+4 4 | 3+4 4 |
| | 6 2 | 4+4 4 | 3+3 4 | 3+4 4 | 6+6 4 | 3+4 4 | 3+4 4 | 3+4 4 |
| | 5 2 | 4+4 4 | 3+3 4 | 3+4 4 | 5+5 4 | 3+4 4 | 3+4 4 | 3+4 4 |
| | 6 2 | 3+3 4 | 4+4 4 | 3+4 4 | 6+6 4 | 3+4 4 | 3+4 4 | 3+4 4 |
| | 6 2 | 3+3 4 | 4+4 4 | 3+4 4 | 5+5 4 | 3+4 4 | 3+4 4 | 3+4 4 |
| | 5 2 | 4+4 4 | 3+3 4 | 3+4 4 | 6+6 4 | 3+4 4 | 3+4 4 | 3+4 4 |
| | 5 2 | 3+3 4 | 3+3 4 | 3+4 4 | 5+5 4 | 3+4 4 | 3+4 4 | 3+4 4 |
| | 6 2 | 4+4 4 | 4+4 4 | 3+4 4 | 6+6 4 | 3+4 4 | 3+4 4 | 3+4 4 |
| | 5 2 | 4+4 4 | 3+3 4 | 3+4 4 | 5+5 4 | 3+4 4 | 3+4 4 | 3+4 4 |
| | 5 2 | 3+3 4 | 4+4 4 | 3+4 4 | 5+5 4 | 3+4 4 | 3+4 4 | 3+4 4 |
| | 6 2 | 3+3 4 | 3+3 4 | 3+4 4 | 6+6 4 | 3+4 4 | 3+4 4 | 3+4 4 |
| | 5 2 | 4+4 4 | 4+4 4 | 3+4 4 | 6+6 4 | 3+4 4 | 3+4 4 | 3+4 4 |
| | 5 2 | 3+3 4 | 4+4 4 | 3+4 4 | 5+5 4 | 3+4 4 | 3+4 4 | 3+4 4 |
| | 6 2 | 4+4 4 | 3+3 4 | 3+4 4 | 6+6 4 | 3+4 4 | 3+4 4 | 3+4 4 |
| | 5 2 | 4+4 4 | 3+3 4 | 3+4 4 | 5+5 4 | 3+4 4 | 3+4 4 | 3+4 4 |
| | 6 2 | 3+3 4 | 4+4 4 | 3+4 4 | 5+5 4 | 3+4 4 | 3+4 4 | 3+4 4 |
| | 6 2 | 4+4 4 | 3+3 4 | 3+4 4 | 6+6 4 | 3+4 4 | 3+4 4 | 3+4 4 |
| | 5 2 | 3+3 4 | 4+4 4 | 3+4 4 | 5+5 4 | 3+4 4 | 3+4 4 | 3+4 4 |
| | 6 2 | 4+4 4 | 4+4 4 | 3+4 4 | 5+5 4 | 3+4 4 | 3+4 4 | 3+4 4 |
| | 5 2 | 3+3 4 | 3+3 4 | 3+4 4 | 6+6 4 | 3+4 4 | 3+4 4 | 3+4 4 |
| | 6 2 | 3+3 4 | 4+4 4 | 3+4 4 | 6+6 4 | 3+4 4 | 3+4 4 | 3+4 4 |
| | 5 2 | 4+4 4 | 3+3 4 | 3+4 4 | 5+5 4 | 3+4 4 | 3+4 4 | 3+4 4 |
| | 6 2 | 3+3 4 | 4+4 4 | 3+4 4 | 5+5 4 | 3+4 4 | 3+4 4 | 3+4 4 |
| | 6 2 | 4+4 4 | 3+3 4 | 3+4 4 | 6+6 4 | 3+4 4 | 3+4 4 | 3+4 4 |
| | 5 2 | 4+4 4 | 4+4 4 | 3+4 4 | 5+5 4 | 3+4 4 | 3+4 4 | 3+4 4 |
| | 5 2 | 3+3 4 | 3+3 4 | 3+4 4 | 5+5 4 | 3+4 4 | 3+4 4 | 3+4 4 |
| | 5 2 | 3+3 4 | 3+3 4 | 3+4 4 | 6+6 4 | 3+4 4 | 3+4 4 | 3+4 4 |
| | 6 2 | 4+4 4 | 4+4 4 | 3+4 4 | 6+6 4 | 3+4 4 | 3+4 4 | 3+4 4 |
| | 5 2 | 4+4 4 | 4+4 4 | 3+4 4 | 5+5 4 | 3+4 4 | 3+4 4 | 3+4 4 |
| | 6 2 | 3+3 4 | 3+3 4 | 3+4 4 | 5+5 4 | 3+4 4 | 3+4 4 | 3+4 4 |
| | 5 2 | 4+4 4 | 4+4 4 | 3+4 4 | 6+6 4 | 3+4 4 | 3+4 4 | 3+4 4 |
| | 6 2 | 3+3 4 | 3+3 4 | 3+4 4 | 6+6 4 | 3+4 4 | 3+4 4 | 3+4 4 |
| | 5 2 | 3+3 4 | 3+3 4 | 3+4 4 | 5+5 4 | 3+4 4 | 3+4 4 | 3+4 4 |
| | 5 2 | 4+4 4 | 4+4 4 | 3+4 4 | 6+6 4 | 3+4 4 | 3+4 4 | 3+4 4 |
| | 6 2 | 4+4 4 | 3+3 4 | 3+4 4 | 5+5 4 | 3+4 4 | 3+4 4 | 3+4 4 |
| | 6 2 | 3+3 4 | 4+4 4 | 3+4 4 | 5+5 4 | 3+4 4 | 3+4 4 | 3+4 4 |
| | 6 2 | | | | | | | |
| | 5 2 | | | | | | | |
| | 6 2 | | | | | | | |
| | 6 2 | | | | | | | |
| | 5 2 | | | | | | | |
| | 5 2 | | | | | | | |
| | 6 2 | | | | | | | |
| | 5 2 | | | | | | | |
| | 6 2 | | | | | | | |
| | 5 2 | | | | | | | |
| | 5 2 | | | | | | | |
| | 6 2 | | | | | | | |
| | 6 2 | | | | | | | |
| | 5 2 | | | | | | | |
| | 6 2 | | | | | | | |
| | 5 2 | | | | | | | |
| | 6 2 | | | | | | | |
| | 5 2 | | | | | | | |
| | 5 2 | | | | | | | |
| | 6 2 | | | | | | | |
| | 6 2 | | | | | | | |
| | 6 2 | | | | | | | |
| | 5 2 | | | | | | | |

Continued on next column

**TABLE XXXVII– continued**

| TQC | ICSD | NoE | SF | | | | | |
|---|---|---|---|---|---|---|---|---|
| | 5 2 | | | | | | | |
| | 6 2 | | | | | | | |
| | 5 2 | | | | | | | |
| | 6 2 | | | | | | | |
| | 6 2 | | | | | | | |
| | 5 2 | | | | | | | |
| | 5 2 | | | | | | | |
| | 6 2 | | | | | | | |
| | 6 2 | | | | | | | |
| | 5 2 | | | | | | | |
| | 6 2 | | | | | | | |
| | 5 2 | | | | | | | |
| | 5 2 | | | | | | | |
| | 5 2 | | | | | | | |
| | 6 2 | | | | | | | |
| | 5 2 | | | | | | | |
| | 6 2 | | | | | | | |
| | 5 2 | | | | | | | |
| | 5 2 | | | | | | | |
| | 6 2 | | | | | | | |
| | 6 2 | | | | | | | |
| 62.2.8 | 421663 | 240 | Ru2Sn2Zn3 | | | | | |
| | -GM | -R | -S | -T | -U | -X | -Y | -Z |
| | 5 2 | 3+3 4 | 3+3 4 | 3+4 4 | 5+5 4 | 3+4 4 | 3+4 4 | 3+4 4 |
| | 6 2 | 4+4 4 | 4+4 4 | 3+4 4 | 6+6 4 | 3+4 4 | 3+4 4 | 3+4 4 |
| | 5 2 | 4+4 4 | 4+4 4 | 3+4 4 | 5+5 4 | 3+4 4 | 3+4 4 | 3+4 4 |
| | 6 2 | 3+3 4 | 3+3 4 | 3+4 4 | 6+6 4 | 3+4 4 | 3+4 4 | 3+4 4 |
| | 5 2 | 4+4 4 | 3+3 4 | 3+4 4 | 6+6 4 | 3+4 4 | 3+4 4 | 3+4 4 |
| | 5 2 | 3+3 4 | 4+4 4 | 3+4 4 | 5+5 4 | 3+4 4 | 3+4 4 | 3+4 4 |
| | 6 2 | 4+4 4 | 3+3 4 | 3+4 4 | 5+5 4 | 3+4 4 | 3+4 4 | 3+4 4 |
| | 5 2 | 3+3 4 | 4+4 4 | 3+4 4 | 6+6 4 | 3+4 4 | 3+4 4 | 3+4 4 |
| | 5 2 | 4+4 4 | 4+4 4 | 3+4 4 | 5+5 4 | 3+4 4 | 3+4 4 | 3+4 4 |
| | 6 2 | 3+3 4 | 3+3 4 | 3+4 4 | 6+6 4 | 3+4 4 | 3+4 4 | 3+4 4 |
| | 6 2 | 3+3 4 | 3+3 4 | 3+4 4 | 6+6 4 | 3+4 4 | 3+4 4 | 3+4 4 |
| | 6 2 | 4+4 4 | 4+4 4 | 3+4 4 | 5+5 4 | 3+4 4 | 3+4 4 | 3+4 4 |
| | 5 2 | 3+3 4 | 4+4 4 | 3+4 4 | 6+6 4 | 3+4 4 | 3+4 4 | 3+4 4 |
| | 5 2 | 4+4 4 | 3+3 4 | 3+4 4 | 5+5 4 | 3+4 4 | 3+4 4 | 3+4 4 |
| | 6 2 | 3+3 4 | 3+3 4 | 3+4 4 | 6+6 4 | 3+4 4 | 3+4 4 | 3+4 4 |
| | 5 2 | 4+4 4 | 4+4 4 | 3+4 4 | 5+5 4 | 3+4 4 | 3+4 4 | 3+4 4 |
| | 6 2 | 4+4 4 | 3+3 4 | 3+4 4 | 6+6 4 | 3+4 4 | 3+4 4 | 3+4 4 |
| | 6 2 | 3+3 4 | 4+4 4 | 3+4 4 | 5+5 4 | 3+4 4 | 3+4 4 | 3+4 4 |
| | 5 2 | 3+3 4 | 4+4 4 | 3+4 4 | 6+6 4 | 3+4 4 | 3+4 4 | 3+4 4 |
| | 6 2 | 4+4 4 | 3+3 4 | 3+4 4 | 5+5 4 | 3+4 4 | 3+4 4 | 3+4 4 |
| | 5 2 | 4+4 4 | 4+4 4 | 3+4 4 | 6+6 4 | 3+4 4 | 3+4 4 | 3+4 4 |
| | 6 2 | 3+3 4 | 3+3 4 | 3+4 4 | 5+5 4 | 3+4 4 | 3+4 4 | 3+4 4 |
| | 5 2 | 3+3 4 | 3+3 4 | 3+4 4 | 6+6 4 | 3+4 4 | 3+4 4 | 3+4 4 |
| | 5 2 | 4+4 4 | 4+4 4 | 3+4 4 | 5+5 4 | 3+4 4 | 3+4 4 | 3+4 4 |
| | 5 2 | 4+4 4 | 4+4 4 | 3+4 4 | 5+5 4 | 3+4 4 | 3+4 4 | 3+4 4 |
| | 6 2 | 3+3 4 | 3+3 4 | 3+4 4 | 6+6 4 | 3+4 4 | 3+4 4 | 3+4 4 |
| | 5 2 | 3+3 4 | 4+4 4 | 3+4 4 | 5+5 4 | 3+4 4 | 3+4 4 | 3+4 4 |
| | 5 2 | 4+4 4 | 3+3 4 | 3+4 4 | 6+6 4 | 3+4 4 | 3+4 4 | 3+4 4 |
| | 6 2 | 3+3 4 | 4+4 4 | 3+4 4 | 5+5 4 | 3+4 4 | 3+4 4 | 3+4 4 |
| | 6 2 | 4+4 4 | 3+3 4 | 3+4 4 | 6+6 4 | 3+4 4 | 3+4 4 | 3+4 4 |
| | 6 2 | 4+4 4 | 3+3 4 | 3+4 4 | 5+5 4 | 3+4 4 | 3+4 4 | 3+4 4 |
| | 6 2 | 3+3 4 | 4+4 4 | 3+4 4 | 5+5 4 | 3+4 4 | 3+4 4 | 3+4 4 |
| | 5 2 | 3+3 4 | 3+3 4 | 3+4 4 | 6+6 4 | 3+4 4 | 3+4 4 | 3+4 4 |
| | 6 2 | 4+4 4 | 4+4 4 | 3+4 4 | 6+6 4 | 3+4 4 | 3+4 4 | 3+4 4 |
| | 5 2 | 4+4 4 | 4+4 4 | 3+4 4 | 6+6 4 | 3+4 4 | 3+4 4 | 3+4 4 |
| | 6 2 | 3+3 4 | 3+3 4 | 3+4 4 | 5+5 4 | 3+4 4 | 3+4 4 | 3+4 4 |
| | 5 2 | 3+3 4 | 4+4 4 | 3+4 4 | 6+6 4 | 3+4 4 | 3+4 4 | 3+4 4 |
| | 5 2 | 4+4 4 | 3+3 4 | 3+4 4 | 5+5 4 | 3+4 4 | 3+4 4 | 3+4 4 |

Continued on next column

**TABLE XXXVII– continued**

| TQC | ICSD | NoE | SF | | | | | |
|---|---|---|---|---|---|---|---|---|
| | 6 2 | 4+4 4 | 3+3 4 | 3+4 4 | 6+6 4 | 3+4 4 | 3+4 4 | 3+4 4 |
| | 5 2 | 3+3 4 | 4+4 4 | 3+4 4 | 5+5 4 | 3+4 4 | 3+4 4 | 3+4 4 |
| | 5 2 | 3+3 4 | 4+4 4 | 3+4 4 | 5+5 4 | 3+4 4 | 3+4 4 | 3+4 4 |
| | 6 2 | 4+4 4 | 3+3 4 | 3+4 4 | 6+6 4 | 3+4 4 | 3+4 4 | 3+4 4 |
| | 6 2 | 4+4 4 | 4+4 4 | 3+4 4 | 5+5 4 | 3+4 4 | 3+4 4 | 3+4 4 |
| | 5 2 | 3+3 4 | 3+3 4 | 3+4 4 | 6+6 4 | 3+4 4 | 3+4 4 | 3+4 4 |
| | 5 2 | 3+3 4 | 3+3 4 | 3+4 4 | 5+5 4 | 3+4 4 | 3+4 4 | 3+4 4 |
| | 5 2 | 4+4 4 | 4+4 4 | 3+4 4 | 6+6 4 | 3+4 4 | 3+4 4 | 3+4 4 |
| | 6 2 | 3+3 4 | 3+3 4 | 3+4 4 | 6+6 4 | 3+4 4 | 3+4 4 | 3+4 4 |
| | 6 2 | 4+4 4 | 4+4 4 | 3+4 4 | 5+5 4 | 3+4 4 | 3+4 4 | 3+4 4 |
| | 5 2 | 3+3 4 | 4+4 4 | 3+4 4 | 6+6 4 | 3+4 4 | 3+4 4 | 3+4 4 |
| | 6 2 | 4+4 4 | 3+3 4 | 3+4 4 | 6+6 4 | 3+4 4 | 3+4 4 | 3+4 4 |
| | 6 2 | 3+3 4 | 3+3 4 | 3+4 4 | 5+5 4 | 3+4 4 | 3+4 4 | 3+4 4 |
| | 5 2 | 4+4 4 | 4+4 4 | 3+4 4 | 5+5 4 | 3+4 4 | 3+4 4 | 3+4 4 |
| | 6 2 | 4+4 4 | 4+4 4 | 3+4 4 | 5+5 4 | 3+4 4 | 3+4 4 | 3+4 4 |
| | 6 2 | 3+3 4 | 3+3 4 | 3+4 4 | 6+6 4 | 3+4 4 | 3+4 4 | 3+4 4 |
| | 5 2 | 4+4 4 | 3+3 4 | 3+4 4 | 5+5 4 | 3+4 4 | 3+4 4 | 3+4 4 |
| | 5 2 | 3+3 4 | 4+4 4 | 3+4 4 | 6+6 4 | 3+4 4 | 3+4 4 | 3+4 4 |
| | 6 2 | 4+4 4 | 3+3 4 | 3+4 4 | 5+5 4 | 3+4 4 | 3+4 4 | 3+4 4 |
| | 6 2 | 3+3 4 | 4+4 4 | 3+4 4 | 6+6 4 | 3+4 4 | 3+4 4 | 3+4 4 |
| | 6 2 | 3+3 4 | 4+4 4 | 3+4 4 | 5+5 4 | 3+4 4 | 3+4 4 | 3+4 4 |
| | 5 2 | 4+4 4 | 3+3 4 | 3+4 4 | 5+5 4 | 3+4 4 | 3+4 4 | 3+4 4 |
| | 6 2 | | | | | | | |
| | 5 2 | | | | | | | |
| | 6 2 | | | | | | | |
| | 6 2 | | | | | | | |
| | 6 2 | | | | | | | |
| | 5 2 | | | | | | | |
| | 5 2 | | | | | | | |
| | 5 2 | | | | | | | |
| | 5 2 | | | | | | | |
| | 6 2 | | | | | | | |
| | 5 2 | | | | | | | |
| | 6 2 | | | | | | | |
| | 5 2 | | | | | | | |
| | 6 2 | | | | | | | |
| | 5 2 | | | | | | | |
| | 6 2 | | | | | | | |
| | 5 2 | | | | | | | |
| | 5 2 | | | | | | | |
| | 6 2 | | | | | | | |
| | 6 2 | | | | | | | |
| | 5 2 | | | | | | | |
| | 6 2 | | | | | | | |
| | 5 2 | | | | | | | |
| | 6 2 | | | | | | | |
| | 6 2 | | | | | | | |
| | 5 2 | | | | | | | |
| | 5 2 | | | | | | | |
| | 5 2 | | | | | | | |
| | 6 2 | | | | | | | |
| | 5 2 | | | | | | | |
| | 5 2 | | | | | | | |
| | 6 2 | | | | | | | |
| | 5 2 | | | | | | | |
| | 6 2 | | | | | | | |
| | 5 2 | | | | | | | |
| | 6 2 | | | | | | | |
| | 5 2 | | | | | | | |
| | 5 2 | | | | | | | |
| | 5 2 | | | | | | | |

Continued on next column

**TABLE XXXVII– continued**

| TQC | ICSD | NoE | SF | | | | | |
|---|---|---|---|---|---|---|---|---|
| | 5 2 | | | | | | | |
| | 5 2 | | | | | | | |
| | 5 2 | | | | | | | |
| | 5 2 | | | | | | | |
| | 6 2 | | | | | | | |
| | 5 2 | | | | | | | |
| | 6 2 | | | | | | | |
| | 6 2 | | | | | | | |
| | 6 2 | | | | | | | |
| | 6 2 | | | | | | | |
| | 6 2 | | | | | | | |
| | 6 2 | | | | | | | |
| | 6 2 | | | | | | | |
| | 5 2 | | | | | | | |
| | 6 2 | | | | | | | |
| | 5 2 | | | | | | | |
| | 6 2 | | | | | | | |
| | 6 2 | | | | | | | |
| | 6 2 | | | | | | | |
| | 6 2 | | | | | | | |
| | 6 2 | | | | | | | |
| | 5 2 | | | | | | | |
| 62.2.9 | 421255 | 120 | Rh1Sc3Si3 | | | | | |
| | -GM | -R | -S | -T | -U | -X | -Y | -Z |
| | 5 2 | 3+3 4 | 3+3 4 | 3+4 4 | 5+5 4 | 3+4 4 | 3+4 4 | 3+4 4 |
| | 6 2 | 4+4 4 | 4+4 4 | 3+4 4 | 6+6 4 | 3+4 4 | 3+4 4 | 3+4 4 |
| | 6 2 | 3+3 4 | 3+3 4 | 3+4 4 | 6+6 4 | 3+4 4 | 3+4 4 | 3+4 4 |
| | 5 2 | 4+4 4 | 4+4 4 | 3+4 4 | 5+5 4 | 3+4 4 | 3+4 4 | 3+4 4 |
| | 5 2 | 4+4 4 | 3+3 4 | 3+4 4 | 5+5 4 | 3+4 4 | 3+4 4 | 3+4 4 |
| | 6 2 | 3+3 4 | 4+4 4 | 3+4 4 | 6+6 4 | 3+4 4 | 3+4 4 | 3+4 4 |
| | 6 2 | 3+3 4 | 4+4 4 | 3+4 4 | 6+6 4 | 3+4 4 | 3+4 4 | 3+4 4 |
| | 5 2 | 4+4 4 | 3+3 4 | 3+4 4 | 5+5 4 | 3+4 4 | 3+4 4 | 3+4 4 |
| | 5 2 | 3+3 4 | 3+3 4 | 3+4 4 | 5+5 4 | 3+4 4 | 3+4 4 | 3+4 4 |
| | 6 2 | 4+4 4 | 4+4 4 | 3+4 4 | 6+6 4 | 3+4 4 | 3+4 4 | 3+4 4 |
| | 6 2 | 3+3 4 | 4+4 4 | 3+4 4 | 6+6 4 | 3+4 4 | 3+4 4 | 3+4 4 |
| | 5 2 | 4+4 4 | 3+3 4 | 3+4 4 | 5+5 4 | 3+4 4 | 3+4 4 | 3+4 4 |
| | 6 2 | 3+3 4 | 3+3 4 | 3+4 4 | 6+6 4 | 3+4 4 | 3+4 4 | 3+4 4 |
| | 6 2 | 4+4 4 | 4+4 4 | 3+4 4 | 5+5 4 | 3+4 4 | 3+4 4 | 3+4 4 |
| | 5 2 | 4+4 4 | 3+3 4 | 3+4 4 | 6+6 4 | 3+4 4 | 3+4 4 | 3+4 4 |
| | 5 2 | 3+3 4 | 4+4 4 | 3+4 4 | 5+5 4 | 3+4 4 | 3+4 4 | 3+4 4 |
| | 5 2 | 4+4 4 | 4+4 4 | 3+4 4 | 6+6 4 | 3+4 4 | 3+4 4 | 3+4 4 |
| | 6 2 | 3+3 4 | 3+3 4 | 3+4 4 | 5+5 4 | 3+4 4 | 3+4 4 | 3+4 4 |
| | 5 2 | 3+3 4 | 3+3 4 | 3+4 4 | 5+5 4 | 3+4 4 | 3+4 4 | 3+4 4 |
| | 5 2 | 4+4 4 | 4+4 4 | 3+4 4 | 6+6 4 | 3+4 4 | 3+4 4 | 3+4 4 |
| | 6 2 | 3+3 4 | 4+4 4 | 3+4 4 | 5+5 4 | 3+4 4 | 3+4 4 | 3+4 4 |
| | 6 2 | 4+4 4 | 3+3 4 | 3+4 4 | 6+6 4 | 3+4 4 | 3+4 4 | 3+4 4 |
| | 5 2 | 4+4 4 | 4+4 4 | 3+4 4 | 5+5 4 | 3+4 4 | 3+4 4 | 3+4 4 |
| | 6 2 | 3+3 4 | 3+3 4 | 3+4 4 | 6+6 4 | 3+4 4 | 3+4 4 | 3+4 4 |
| | 5 2 | 3+3 4 | 3+3 4 | 3+4 4 | 6+6 4 | 3+4 4 | 3+4 4 | 3+4 4 |
| | 5 2 | 4+4 4 | 4+4 4 | 3+4 4 | 5+5 4 | 3+4 4 | 3+4 4 | 3+4 4 |
| | 6 2 | 3+3 4 | 3+3 4 | 3+4 4 | 6+6 4 | 3+4 4 | 3+4 4 | 3+4 4 |
| | 6 2 | 4+4 4 | 4+4 4 | 3+4 4 | 5+5 4 | 3+4 4 | 3+4 4 | 3+4 4 |
| | 5 2 | 3+3 4 | 3+3 4 | 3+4 4 | 6+6 4 | 3+4 4 | 3+4 4 | 3+4 4 |
| | 6 2 | 4+4 4 | 4+4 4 | 3+4 4 | 5+5 4 | 3+4 4 | 3+4 4 | 3+4 4 |
| | 5 2 | | | | | | | |
| | 6 2 | | | | | | | |
| | 5 2 | | | | | | | |
| | 5 2 | | | | | | | |
| | 5 2 | | | | | | | |
| | 6 2 | | | | | | | |
| | 6 2 | | | | | | | |
| | 6 2 | | | | | | | |

Continued on next column

**TABLE XXXVII– continued**

| TQC | ICSD | NoE | SF | | | | | |
|---|---|---|---|---|---|---|---|---|
| | 6 2 | | | | | | | |
| | 5 2 | | | | | | | |
| | 6 2 | | | | | | | |
| | 5 2 | | | | | | | |
| | 5 2 | | | | | | | |
| | 5 2 | | | | | | | |
| | 6 2 | | | | | | | |
| | 6 2 | | | | | | | |
| | 6 2 | | | | | | | |
| | 6 2 | | | | | | | |
| | 5 2 | | | | | | | |
| | 6 2 | | | | | | | |
| | 6 2 | | | | | | | |
| | 6 2 | | | | | | | |
| | 5 2 | | | | | | | |
| | 5 2 | | | | | | | |
| | 6 2 | | | | | | | |
| | 5 2 | | | | | | | |
| | 5 2 | | | | | | | |
| | 6 2 | | | | | | | |
| | 5 2 | | | | | | | |
| | 6 2 | | | | | | | |
| 62.2.10 | 15248 | 80 | Rh1Sc1Si2 | | | | | |
| | -GM | -R | -S | -T | -U | -X | -Y | -Z |
| | 5 2 | 3+3 4 | 3+3 4 | 3+4 4 | 6+6 4 | 3+4 4 | 3+4 4 | 3+4 4 |
| | 5 2 | 4+4 4 | 4+4 4 | 3+4 4 | 5+5 4 | 3+4 4 | 3+4 4 | 3+4 4 |
| | 6 2 | 4+4 4 | 4+4 4 | 3+4 4 | 5+5 4 | 3+4 4 | 3+4 4 | 3+4 4 |
| | 6 2 | 3+3 4 | 3+3 4 | 3+4 4 | 6+6 4 | 3+4 4 | 3+4 4 | 3+4 4 |
| | 6 2 | 3+3 4 | 4+4 4 | 3+4 4 | 6+6 4 | 3+4 4 | 3+4 4 | 3+4 4 |
| | 6 2 | 4+4 4 | 3+3 4 | 3+4 4 | 6+6 4 | 3+4 4 | 3+4 4 | 3+4 4 |
| | 5 2 | 4+4 4 | 3+3 4 | 3+4 4 | 5+5 4 | 3+4 4 | 3+4 4 | 3+4 4 |
| | 5 2 | 3+3 4 | 4+4 4 | 3+4 4 | 6+6 4 | 3+4 4 | 3+4 4 | 3+4 4 |
| | 5 2 | 4+4 4 | 3+3 4 | 3+4 4 | 5+5 4 | 3+4 4 | 3+4 4 | 3+4 4 |
| | 5 2 | 3+3 4 | 4+4 4 | 3+4 4 | 5+5 4 | 3+4 4 | 3+4 4 | 3+4 4 |
| | 6 2 | 3+3 4 | 3+3 4 | 3+4 4 | 5+5 4 | 3+4 4 | 3+4 4 | 3+4 4 |
| | 5 2 | 4+4 4 | 4+4 4 | 3+4 4 | 5+5 4 | 3+4 4 | 3+4 4 | 3+4 4 |
| | 6 2 | 4+4 4 | 4+4 4 | 3+4 4 | 6+6 4 | 3+4 4 | 3+4 4 | 3+4 4 |
| | 6 2 | 3+3 4 | 3+3 4 | 3+4 4 | 6+6 4 | 3+4 4 | 3+4 4 | 3+4 4 |
| | 6 2 | 4+4 4 | 3+3 4 | 3+4 4 | 5+5 4 | 3+4 4 | 3+4 4 | 3+4 4 |
| | 6 2 | 3+3 4 | 4+4 4 | 3+4 4 | 6+6 4 | 3+4 4 | 3+4 4 | 3+4 4 |
| | 5 2 | 4+4 4 | 4+4 4 | 3+4 4 | 5+5 4 | 3+4 4 | 3+4 4 | 3+4 4 |
| | 5 2 | 3+3 4 | 3+3 4 | 3+4 4 | 6+6 4 | 3+4 4 | 3+4 4 | 3+4 4 |
| | 6 2 | 3+3 4 | 4+4 4 | 3+4 4 | 5+5 4 | 3+4 4 | 3+4 4 | 3+4 4 |
| | 5 2 | 4+4 4 | 3+3 4 | 3+4 4 | 6+6 4 | 3+4 4 | 3+4 4 | 3+4 4 |
| | 5 2 | | | | | | | |
| | 5 2 | | | | | | | |
| | 5 2 | | | | | | | |
| | 6 2 | | | | | | | |
| | 5 2 | | | | | | | |
| | 6 2 | | | | | | | |
| | 6 2 | | | | | | | |
| | 5 2 | | | | | | | |
| | 5 2 | | | | | | | |
| | 6 2 | | | | | | | |
| | 6 2 | | | | | | | |
| | 5 2 | | | | | | | |
| | 6 2 | | | | | | | |
| | 5 2 | | | | | | | |
| | 5 2 | | | | | | | |
| | 6 2 | | | | | | | |
| | 6 2 | | | | | | | |

Continued on next column

**TABLE XXXVII– continued**

| TQC | ICSD | NoE | SF | | | | |
|---|---|---|---|---|---|---|---|
| | 5 2 | | | | | | |
| | 5 2 | | | | | | |
| | 6 2 | | | | | | |
| 62.2.11 | 159305 | 184 | Pt1Yb1Zn1 | | | | |
| | -GM | -R | -S | -T | -U | -X | -Y | -Z |
| | 5 2 | 3+3 4 | 3+3 4 | 3+4 4 | 6+6 4 | 3+4 4 | 3+4 4 | 3+4 4 |
| | 5 2 | 4+4 4 | 4+4 4 | 3+4 4 | 5+5 4 | 3+4 4 | 3+4 4 | 3+4 4 |
| | 6 2 | 3+3 4 | 3+3 4 | 3+4 4 | 6+6 4 | 3+4 4 | 3+4 4 | 3+4 4 |
| | 6 2 | 4+4 4 | 4+4 4 | 3+4 4 | 5+5 4 | 3+4 4 | 3+4 4 | 3+4 4 |
| | 5 2 | 4+4 4 | 3+3 4 | 3+4 4 | 6+6 4 | 3+4 4 | 3+4 4 | 3+4 4 |
| | 6 2 | 3+3 4 | 4+4 4 | 3+4 4 | 5+5 4 | 3+4 4 | 3+4 4 | 3+4 4 |
| | 5 2 | 3+3 4 | 4+4 4 | 3+4 4 | 6+6 4 | 3+4 4 | 3+4 4 | 3+4 4 |
| | 6 2 | 4+4 4 | 3+3 4 | 3+4 4 | 5+5 4 | 3+4 4 | 3+4 4 | 3+4 4 |
| | 5 2 | 4+4 4 | 3+3 4 | 3+4 4 | 5+5 4 | 3+4 4 | 3+4 4 | 3+4 4 |
| | 6 2 | 3+3 4 | 4+4 4 | 3+4 4 | 6+6 4 | 3+4 4 | 3+4 4 | 3+4 4 |
| | 5 2 | 3+3 4 | 4+4 4 | 3+4 4 | 5+5 4 | 3+4 4 | 3+4 4 | 3+4 4 |
| | 5 2 | 4+4 4 | 3+3 4 | 3+4 4 | 6+6 4 | 3+4 4 | 3+4 4 | 3+4 4 |
| | 6 2 | 4+4 4 | 4+4 4 | 3+4 4 | 5+5 4 | 3+4 4 | 3+4 4 | 3+4 4 |
| | 6 2 | 3+3 4 | 3+3 4 | 3+4 4 | 6+6 4 | 3+4 4 | 3+4 4 | 3+4 4 |
| | 5 2 | 3+3 4 | 3+3 4 | 3+4 4 | 5+5 4 | 3+4 4 | 3+4 4 | 3+4 4 |
| | 6 2 | 4+4 4 | 4+4 4 | 3+4 4 | 6+6 4 | 3+4 4 | 3+4 4 | 3+4 4 |
| | 5 2 | 3+3 4 | 3+3 4 | 3+4 4 | 5+5 4 | 3+4 4 | 3+4 4 | 3+4 4 |
| | 6 2 | 4+4 4 | 4+4 4 | 3+4 4 | 5+5 4 | 3+4 4 | 3+4 4 | 3+4 4 |
| | 6 2 | 3+3 4 | 4+4 4 | 3+4 4 | 6+6 4 | 3+4 4 | 3+4 4 | 3+4 4 |
| | 5 2 | 4+4 4 | 3+3 4 | 3+4 4 | 6+6 4 | 3+4 4 | 3+4 4 | 3+4 4 |
| | 5 2 | 4+4 4 | 4+4 4 | 3+4 4 | 6+6 4 | 3+4 4 | 3+4 4 | 3+4 4 |
| | 5 2 | 3+3 4 | 3+3 4 | 3+4 4 | 5+5 4 | 3+4 4 | 3+4 4 | 3+4 4 |
| | 6 2 | 4+4 4 | 3+3 4 | 3+4 4 | 5+5 4 | 3+4 4 | 3+4 4 | 3+4 4 |
| | 6 2 | 3+3 4 | 4+4 4 | 3+4 4 | 6+6 4 | 3+4 4 | 3+4 4 | 3+4 4 |
| | 5 2 | 3+3 4 | 3+3 4 | 3+4 4 | 5+5 4 | 3+4 4 | 3+4 4 | 3+4 4 |
| | 6 2 | 4+4 4 | 4+4 4 | 3+4 4 | 6+6 4 | 3+4 4 | 3+4 4 | 3+4 4 |
| | 5 2 | 4+4 4 | 4+4 4 | 3+4 4 | 5+5 4 | 3+4 4 | 3+4 4 | 3+4 4 |
| | 6 2 | 3+3 4 | 3+3 4 | 3+4 4 | 6+6 4 | 3+4 4 | 3+4 4 | 3+4 4 |
| | 5 2 | 3+3 4 | 4+4 4 | 3+4 4 | 5+5 4 | 3+4 4 | 3+4 4 | 3+4 4 |
| | 6 2 | 4+4 4 | 3+3 4 | 3+4 4 | 6+6 4 | 3+4 4 | 3+4 4 | 3+4 4 |
| | 6 2 | 4+4 4 | 4+4 4 | 3+4 4 | 5+5 4 | 3+4 4 | 3+4 4 | 3+4 4 |
| | 5 2 | 3+3 4 | 3+3 4 | 3+4 4 | 6+6 4 | 3+4 4 | 3+4 4 | 3+4 4 |
| | 5 2 | 3+3 4 | 4+4 4 | 3+4 4 | 5+5 4 | 3+4 4 | 3+4 4 | 3+4 4 |
| | 5 2 | 4+4 4 | 3+3 4 | 3+4 4 | 6+6 4 | 3+4 4 | 3+4 4 | 3+4 4 |
| | 5 2 | 3+3 4 | 3+3 4 | 3+4 4 | 6+6 4 | 3+4 4 | 3+4 4 | 3+4 4 |
| | 6 2 | 4+4 4 | 4+4 4 | 3+4 4 | 5+5 4 | 3+4 4 | 3+4 4 | 3+4 4 |
| | 6 2 | 4+4 4 | 4+4 4 | 3+4 4 | 5+5 4 | 3+4 4 | 3+4 4 | 3+4 4 |
| | 5 2 | 3+3 4 | 3+3 4 | 3+4 4 | 6+6 4 | 3+4 4 | 3+4 4 | 3+4 4 |
| | 6 2 | 3+3 4 | 4+4 4 | 3+4 4 | 5+5 4 | 3+4 4 | 3+4 4 | 3+4 4 |
| | 6 2 | 4+4 4 | 3+3 4 | 3+4 4 | 6+6 4 | 3+4 4 | 3+4 4 | 3+4 4 |
| | 6 2 | 3+3 4 | 3+3 4 | 3+4 4 | 5+5 4 | 3+4 4 | 3+4 4 | 3+4 4 |
| | 5 2 | 4+4 4 | 4+4 4 | 3+4 4 | 6+6 4 | 3+4 4 | 3+4 4 | 3+4 4 |
| | 6 2 | 4+4 4 | 3+3 4 | 3+4 4 | 6+6 4 | 3+4 4 | 3+4 4 | 3+4 4 |
| | 5 2 | 3+3 4 | 4+4 4 | 3+4 4 | 5+5 4 | 3+4 4 | 3+4 4 | 3+4 4 |
| | 5 2 | 4+4 4 | 4+4 4 | 3+4 4 | 5+5 4 | 3+4 4 | 3+4 4 | 3+4 4 |
| | 6 2 | 3+3 4 | 3+3 4 | 3+4 4 | 6+6 4 | 3+4 4 | 3+4 4 | 3+4 4 |
| | 5 2 | | | | | | | |
| | 6 2 | | | | | | | |
| | 5 2 | | | | | | | |
| | 6 2 | | | | | | | |
| | 5 2 | | | | | | | |
| | 5 2 | | | | | | | |
| | 5 2 | | | | | | | |
| | 6 2 | | | | | | | |
| | 5 2 | | | | | | | |
| | 6 2 | | | | | | | |

Continued on next column

**TABLE XXXVII– continued**

| TQC | ICSD | NoE | SF | | | | |
|---|---|---|---|---|---|---|---|
| | 6 2 | | | | | | |
| | 5 2 | | | | | | |
| | 5 2 | | | | | | |
| | 6 2 | | | | | | |
| | 6 2 | | | | | | |
| | 6 2 | | | | | | |
| | 5 2 | | | | | | |
| | 5 2 | | | | | | |
| | 6 2 | | | | | | |
| | 5 2 | | | | | | |
| | 5 2 | | | | | | |
| | 6 2 | | | | | | |
| | 6 2 | | | | | | |
| | 5 2 | | | | | | |
| | 6 2 | | | | | | |
| | 5 2 | | | | | | |
| | 6 2 | | | | | | |
| | 6 2 | | | | | | |
| | 5 2 | | | | | | |
| | 5 2 | | | | | | |
| | 6 2 | | | | | | |
| | 5 2 | | | | | | |
| | 6 2 | | | | | | |
| | 5 2 | | | | | | |
| | 6 2 | | | | | | |
| | 5 2 | | | | | | |
| | 6 2 | | | | | | |
| | 5 2 | | | | | | |
| | 6 2 | | | | | | |
| | 5 2 | | | | | | |
| | 5 2 | | | | | | |
| | 6 2 | | | | | | |
| | 6 2 | | | | | | |
| | 6 2 | | | | | | |
| | 5 2 | | | | | | |
| | 5 2 | | | | | | |
| 62.2.12 | 412012 | 96 | Pt1Sn1Sr1 | | | | |
| | -GM | -R | -S | -T | -U | -X | -Y | -Z |
| | 5 2 | 3+3 4 | 3+3 4 | 3+4 4 | 6+6 4 | 3+4 4 | 3+4 4 | 3+4 4 |
| | 5 2 | 4+4 4 | 4+4 4 | 3+4 4 | 5+5 4 | 3+4 4 | 3+4 4 | 3+4 4 |
| | 6 2 | 3+3 4 | 3+3 4 | 3+4 4 | 6+6 4 | 3+4 4 | 3+4 4 | 3+4 4 |
| | 6 2 | 4+4 4 | 4+4 4 | 3+4 4 | 5+5 4 | 3+4 4 | 3+4 4 | 3+4 4 |
| | 5 2 | 4+4 4 | 3+3 4 | 3+4 4 | 6+6 4 | 3+4 4 | 3+4 4 | 3+4 4 |
| | 5 2 | 3+3 4 | 4+4 4 | 3+4 4 | 5+5 4 | 3+4 4 | 3+4 4 | 3+4 4 |
| | 6 2 | 3+3 4 | 4+4 4 | 3+4 4 | 6+6 4 | 3+4 4 | 3+4 4 | 3+4 4 |
| | 6 2 | 4+4 4 | 3+3 4 | 3+4 4 | 5+5 4 | 3+4 4 | 3+4 4 | 3+4 4 |
| | 5 2 | 3+3 4 | 4+4 4 | 3+4 4 | 5+5 4 | 3+4 4 | 3+4 4 | 3+4 4 |
| | 5 2 | 4+4 4 | 3+3 4 | 3+4 4 | 6+6 4 | 3+4 4 | 3+4 4 | 3+4 4 |
| | 6 2 | 4+4 4 | 4+4 4 | 3+4 4 | 6+6 4 | 3+4 4 | 3+4 4 | 3+4 4 |
| | 5 2 | 3+3 4 | 3+3 4 | 3+4 4 | 5+5 4 | 3+4 4 | 3+4 4 | 3+4 4 |
| | 6 2 | 4+4 4 | 3+3 4 | 3+4 4 | 5+5 4 | 3+4 4 | 3+4 4 | 3+4 4 |
| | 6 2 | 3+3 4 | 4+4 4 | 3+4 4 | 6+6 4 | 3+4 4 | 3+4 4 | 3+4 4 |
| | 5 2 | 4+4 4 | 3+3 4 | 3+4 4 | 5+5 4 | 3+4 4 | 3+4 4 | 3+4 4 |
| | 6 2 | 3+3 4 | 4+4 4 | 3+4 4 | 6+6 4 | 3+4 4 | 3+4 4 | 3+4 4 |
| | 5 2 | 3+3 4 | 3+3 4 | 3+4 4 | 5+5 4 | 3+4 4 | 3+4 4 | 3+4 4 |
| | 5 2 | 4+4 4 | 4+4 4 | 3+4 4 | 6+6 4 | 3+4 4 | 3+4 4 | 3+4 4 |
| | 6 2 | 3+3 4 | 4+4 4 | 3+4 4 | 5+5 4 | 3+4 4 | 3+4 4 | 3+4 4 |
| | 6 2 | 4+4 4 | 3+3 4 | 3+4 4 | 6+6 4 | 3+4 4 | 3+4 4 | 3+4 4 |
| | 6 2 | 3+3 4 | 4+4 4 | 3+4 4 | 5+5 4 | 3+4 4 | 3+4 4 | 3+4 4 |
| | 5 2 | 4+4 4 | 3+3 4 | 3+4 4 | 6+6 4 | 3+4 4 | 3+4 4 | 3+4 4 |
| | 5 2 | 3+3 4 | 4+4 4 | 3+4 4 | 5+5 4 | 3+4 4 | 3+4 4 | 3+4 4 |

Continued on next column

**TABLE XXXVII– continued**

| TQC | ICSD | NoE | SF | | | | | |
|---|---|---|---|---|---|---|---|---|
| | 6 2 | 4+4 4 | 3+3 4 | 3+4 4 | 5+5 4 | 3+4 4 | 3+4 4 | 3+4 4 |
| | 5 2 | | | | | | | |
| | 6 2 | | | | | | | |
| | 5 2 | | | | | | | |
| | 5 2 | | | | | | | |
| | 6 2 | | | | | | | |
| | 5 2 | | | | | | | |
| | 6 2 | | | | | | | |
| | 5 2 | | | | | | | |
| | 6 2 | | | | | | | |
| | 6 2 | | | | | | | |
| | 5 2 | | | | | | | |
| | 6 2 | | | | | | | |
| | 5 2 | | | | | | | |
| | 5 2 | | | | | | | |
| | 6 2 | | | | | | | |
| | 6 2 | | | | | | | |
| | 5 2 | | | | | | | |
| | 6 2 | | | | | | | |
| | 5 2 | | | | | | | |
| | 6 2 | | | | | | | |
| | 5 2 | | | | | | | |
| | 5 2 | | | | | | | |
| | 6 2 | | | | | | | |
| | 5 2 | | | | | | | |
| 62.2.13 | 260792 | 72 | Pt1Si1Zr1 | | | | | |
| | -GM | -R | -S | -T | -U | -X | -Y | -Z |
| | 5 2 | 4+4 4 | 3+3 4 | 3+4 4 | 5+5 4 | 3+4 4 | 3+4 4 | 3+4 4 |
| | 5 2 | 3+3 4 | 4+4 4 | 3+4 4 | 6+6 4 | 3+4 4 | 3+4 4 | 3+4 4 |
| | 6 2 | 3+3 4 | 4+4 4 | 3+4 4 | 5+5 4 | 3+4 4 | 3+4 4 | 3+4 4 |
| | 6 2 | 4+4 4 | 3+3 4 | 3+4 4 | 6+6 4 | 3+4 4 | 3+4 4 | 3+4 4 |
| | 6 2 | 4+4 4 | 4+4 4 | 3+4 4 | 6+6 4 | 3+4 4 | 3+4 4 | 3+4 4 |
| | 5 2 | 3+3 4 | 3+3 4 | 3+4 4 | 5+5 4 | 3+4 4 | 3+4 4 | 3+4 4 |
| | 6 2 | 4+4 4 | 4+4 4 | 3+4 4 | 6+6 4 | 3+4 4 | 3+4 4 | 3+4 4 |
| | 5 2 | 3+3 4 | 3+3 4 | 3+4 4 | 5+5 4 | 3+4 4 | 3+4 4 | 3+4 4 |
| | 6 2 | 3+3 4 | 3+3 4 | 3+4 4 | 5+5 4 | 3+4 4 | 3+4 4 | 3+4 4 |
| | 5 2 | 4+4 4 | 4+4 4 | 3+4 4 | 6+6 4 | 3+4 4 | 3+4 4 | 3+4 4 |
| | 5 2 | 3+3 4 | 4+4 4 | 3+4 4 | 5+5 4 | 3+4 4 | 3+4 4 | 3+4 4 |
| | 6 2 | 4+4 4 | 3+3 4 | 3+4 4 | 6+6 4 | 3+4 4 | 3+4 4 | 3+4 4 |
| | 6 2 | 4+4 4 | 3+3 4 | 3+4 4 | 6+6 4 | 3+4 4 | 3+4 4 | 3+4 4 |
| | 5 2 | 3+3 4 | 4+4 4 | 3+4 4 | 5+5 4 | 3+4 4 | 3+4 4 | 3+4 4 |
| | 6 2 | 4+4 4 | 3+3 4 | 3+4 4 | 5+5 4 | 3+4 4 | 3+4 4 | 3+4 4 |
| | 6 2 | 3+3 4 | 4+4 4 | 3+4 4 | 5+5 4 | 3+4 4 | 3+4 4 | 3+4 4 |
| | 5 2 | 4+4 4 | 4+4 4 | 3+4 4 | 6+6 4 | 3+4 4 | 3+4 4 | 3+4 4 |
| | 5 2 | 3+3 4 | 3+3 4 | 3+4 4 | 6+6 4 | 3+4 4 | 3+4 4 | 3+4 4 |
| | 6 2 | | | | | | | |
| | 5 2 | | | | | | | |
| | 5 2 | | | | | | | |
| | 6 2 | | | | | | | |
| | 5 2 | | | | | | | |
| | 6 2 | | | | | | | |
| | 5 2 | | | | | | | |
| | 6 2 | | | | | | | |
| | 5 2 | | | | | | | |
| | 5 2 | | | | | | | |
| | 6 2 | | | | | | | |
| | 6 2 | | | | | | | |
| | 5 2 | | | | | | | |
| | 5 2 | | | | | | | |
| | 5 2 | | | | | | | |
| | 5 2 | | | | | | | |

Continued on next column

| TQC | ICSD | NoE | SF | | | | | |
|---|---|---|---|---|---|---|---|---|
| | 6 2 | | | | | | | |
| | 6 2 | | | | | | | |
| 62.2.14 | 260794 | 72 | Pt1Si1Ti1 | | | | | |
| | -GM | -R | -S | -T | -U | -X | -Y | -Z |
| | 5 2 | 4+4 4 | 3+3 4 | 3+4 4 | 6+6 4 | 3+4 4 | 3+4 4 | 3+4 4 |
| | 5 2 | 3+3 4 | 4+4 4 | 3+4 4 | 5+5 4 | 3+4 4 | 3+4 4 | 3+4 4 |
| | 6 2 | 3+3 4 | 4+4 4 | 3+4 4 | 6+6 4 | 3+4 4 | 3+4 4 | 3+4 4 |
| | 6 2 | 4+4 4 | 3+3 4 | 3+4 4 | 5+5 4 | 3+4 4 | 3+4 4 | 3+4 4 |
| | 6 2 | 4+4 4 | 4+4 4 | 3+4 4 | 5+5 4 | 3+4 4 | 3+4 4 | 3+4 4 |
| | 5 2 | 3+3 4 | 3+3 4 | 3+4 4 | 6+6 4 | 3+4 4 | 3+4 4 | 3+4 4 |
| | 6 2 | 4+4 4 | 3+3 4 | 3+4 4 | 5+5 4 | 3+4 4 | 3+4 4 | 3+4 4 |
| | 5 2 | 3+3 4 | 4+4 4 | 3+4 4 | 6+6 4 | 3+4 4 | 3+4 4 | 3+4 4 |
| | 6 2 | 3+3 4 | 3+3 4 | 3+4 4 | 6+6 4 | 3+4 4 | 3+4 4 | 3+4 4 |
| | 5 2 | 4+4 4 | 4+4 4 | 3+4 4 | 5+5 4 | 3+4 4 | 3+4 4 | 3+4 4 |
| | 5 2 | 3+3 4 | 4+4 4 | 3+4 4 | 6+6 4 | 3+4 4 | 3+4 4 | 3+4 4 |
| | 6 2 | 4+4 4 | 3+3 4 | 3+4 4 | 5+5 4 | 3+4 4 | 3+4 4 | 3+4 4 |
| | 6 2 | 4+4 4 | 3+3 4 | 3+4 4 | 5+5 4 | 3+4 4 | 3+4 4 | 3+4 4 |
| | 5 2 | 3+3 4 | 4+4 4 | 3+4 4 | 6+6 4 | 3+4 4 | 3+4 4 | 3+4 4 |
| | 6 2 | 4+4 4 | 3+3 4 | 3+4 4 | 6+6 4 | 3+4 4 | 3+4 4 | 3+4 4 |
| | 6 2 | 3+3 4 | 4+4 4 | 3+4 4 | 6+6 4 | 3+4 4 | 3+4 4 | 3+4 4 |
| | 5 2 | 3+3 4 | 3+3 4 | 3+4 4 | 5+5 4 | 3+4 4 | 3+4 4 | 3+4 4 |
| | 5 2 | 4+4 4 | 4+4 4 | 3+4 4 | 5+5 4 | 3+4 4 | 3+4 4 | 3+4 4 |
| | 6 2 | | | | | | | |
| | 5 2 | | | | | | | |
| | 5 2 | | | | | | | |
| | 5 2 | | | | | | | |
| | 6 2 | | | | | | | |
| | 6 2 | | | | | | | |
| | 5 2 | | | | | | | |
| | 6 2 | | | | | | | |
| | 5 2 | | | | | | | |
| | 5 2 | | | | | | | |
| | 6 2 | | | | | | | |
| | 6 2 | | | | | | | |
| | 5 2 | | | | | | | |
| | 5 2 | | | | | | | |
| | 5 2 | | | | | | | |
| | 5 2 | | | | | | | |
| | 6 2 | | | | | | | |
| | 6 2 | | | | | | | |
| 62.2.15 | 79238 | 56 | Pt1Si1 | | | | | |
| | -GM | -R | -S | -T | -U | -X | -Y | -Z |
| | 5 2 | 3+3 4 | 4+4 4 | 3+4 4 | 6+6 4 | 3+4 4 | 3+4 4 | 3+4 4 |
| | 5 2 | 4+4 4 | 3+3 4 | 3+4 4 | 5+5 4 | 3+4 4 | 3+4 4 | 3+4 4 |
| | 6 2 | 3+3 4 | 3+3 4 | 3+4 4 | 6+6 4 | 3+4 4 | 3+4 4 | 3+4 4 |
| | 6 2 | 4+4 4 | 4+4 4 | 3+4 4 | 5+5 4 | 3+4 4 | 3+4 4 | 3+4 4 |
| | 6 2 | 4+4 4 | 3+3 4 | 3+4 4 | 6+6 4 | 3+4 4 | 3+4 4 | 3+4 4 |
| | 5 2 | 3+3 4 | 4+4 4 | 3+4 4 | 5+5 4 | 3+4 4 | 3+4 4 | 3+4 4 |
| | 6 2 | 3+3 4 | 4+4 4 | 3+4 4 | 6+6 4 | 3+4 4 | 3+4 4 | 3+4 4 |
| | 5 2 | 4+4 4 | 3+3 4 | 3+4 4 | 6+6 4 | 3+4 4 | 3+4 4 | 3+4 4 |
| | 5 2 | 3+3 4 | 3+3 4 | 3+4 4 | 5+5 4 | 3+4 4 | 3+4 4 | 3+4 4 |
| | 6 2 | 4+4 4 | 4+4 4 | 3+4 4 | 5+5 4 | 3+4 4 | 3+4 4 | 3+4 4 |
| | 6 2 | 4+4 4 | 3+3 4 | 3+4 4 | 6+6 4 | 3+4 4 | 3+4 4 | 3+4 4 |
| | 5 2 | 3+3 4 | 4+4 4 | 3+4 4 | 5+5 4 | 3+4 4 | 3+4 4 | 3+4 4 |
| | 5 2 | 3+3 4 | 4+4 4 | 3+4 4 | 5+5 4 | 3+4 4 | 3+4 4 | 3+4 4 |
| | 5 2 | 4+4 4 | 3+3 4 | 3+4 4 | 6+6 4 | 3+4 4 | 3+4 4 | 3+4 4 |
| | 5 2 | | | | | | | |
| | 6 2 | | | | | | | |
| | 5 2 | | | | | | | |
| | 6 2 | | | | | | | |
| | 6 2 | | | | | | | |

Continued on next column

**TABLE XXXVII– continued**

| TQC | ICSD | NoE | SF | | | | | |
|---|---|---|---|---|---|---|---|---|
| | 6 2 | | | | | | | |
| | 5 2 | | | | | | | |
| | 6 2 | | | | | | | |
| | 5 2 | | | | | | | |
| | 6 2 | | | | | | | |
| | 5 2 | | | | | | | |
| | 5 2 | | | | | | | |
| | 5 2 | | | | | | | |
| | 5 2 | | | | | | | |
| 62.2.16 | 2623 | 56 | Pt1Si1 | | | | | |
| | -GM | -R | -S | -T | -U | -X | -Y | -Z |
| | 5 2 | 3+3 4 | 4+4 4 | 3+4 4 | 6+6 4 | 3+4 4 | 3+4 4 | 3+4 4 |
| | 5 2 | 4+4 4 | 3+3 4 | 3+4 4 | 5+5 4 | 3+4 4 | 3+4 4 | 3+4 4 |
| | 6 2 | 3+3 4 | 4+4 4 | 3+4 4 | 6+6 4 | 3+4 4 | 3+4 4 | 3+4 4 |
| | 6 2 | 4+4 4 | 3+3 4 | 3+4 4 | 5+5 4 | 3+4 4 | 3+4 4 | 3+4 4 |
| | 6 2 | 4+4 4 | 3+3 4 | 3+4 4 | 6+6 4 | 3+4 4 | 3+4 4 | 3+4 4 |
| | 5 2 | 3+3 4 | 4+4 4 | 3+4 4 | 5+5 4 | 3+4 4 | 3+4 4 | 3+4 4 |
| | 6 2 | 3+3 4 | 4+4 4 | 3+4 4 | 6+6 4 | 3+4 4 | 3+4 4 | 3+4 4 |
| | 5 2 | 4+4 4 | 3+3 4 | 3+4 4 | 6+6 4 | 3+4 4 | 3+4 4 | 3+4 4 |
| | 5 2 | 3+3 4 | 3+3 4 | 3+4 4 | 5+5 4 | 3+4 4 | 3+4 4 | 3+4 4 |
| | 6 2 | 4+4 4 | 4+4 4 | 3+4 4 | 5+5 4 | 3+4 4 | 3+4 4 | 3+4 4 |
| | 6 2 | 4+4 4 | 3+3 4 | 3+4 4 | 6+6 4 | 3+4 4 | 3+4 4 | 3+4 4 |
| | 5 2 | 3+3 4 | 4+4 4 | 3+4 4 | 5+5 4 | 3+4 4 | 3+4 4 | 3+4 4 |
| | 5 2 | 3+3 4 | 4+4 4 | 3+4 4 | 5+5 4 | 3+4 4 | 3+4 4 | 3+4 4 |
| | 5 2 | 4+4 4 | 3+3 4 | 3+4 4 | 6+6 4 | 3+4 4 | 3+4 4 | 3+4 4 |
| | 5 2 | | | | | | | |
| | 6 2 | | | | | | | |
| | 5 2 | | | | | | | |
| | 6 2 | | | | | | | |
| | 6 2 | | | | | | | |
| | 5 2 | | | | | | | |
| | 6 2 | | | | | | | |
| | 6 2 | | | | | | | |
| | 6 2 | | | | | | | |
| | 5 2 | | | | | | | |
| | 5 2 | | | | | | | |
| | 5 2 | | | | | | | |
| | 5 2 | | | | | | | |
| | 5 2 | | | | | | | |
| 62.2.17 | 105785 | 64 | Pt1Sc2 | | | | | |
| | -GM | -R | -S | -T | -U | -X | -Y | -Z |
| | 5 2 | 4+4 4 | 3+3 4 | 3+4 4 | 6+6 4 | 3+4 4 | 3+4 4 | 3+4 4 |
| | 5 2 | 3+3 4 | 4+4 4 | 3+4 4 | 5+5 4 | 3+4 4 | 3+4 4 | 3+4 4 |
| | 6 2 | 3+3 4 | 3+3 4 | 3+4 4 | 5+5 4 | 3+4 4 | 3+4 4 | 3+4 4 |
| | 6 2 | 4+4 4 | 4+4 4 | 3+4 4 | 6+6 4 | 3+4 4 | 3+4 4 | 3+4 4 |
| | 5 2 | 3+3 4 | 3+3 4 | 3+4 4 | 5+5 4 | 3+4 4 | 3+4 4 | 3+4 4 |
| | 6 2 | 4+4 4 | 4+4 4 | 3+4 4 | 6+6 4 | 3+4 4 | 3+4 4 | 3+4 4 |
| | 5 2 | 4+4 4 | 4+4 4 | 3+4 4 | 5+5 4 | 3+4 4 | 3+4 4 | 3+4 4 |
| | 6 2 | 3+3 4 | 3+3 4 | 3+4 4 | 6+6 4 | 3+4 4 | 3+4 4 | 3+4 4 |
| | 5 2 | 3+3 4 | 4+4 4 | 3+4 4 | 5+5 4 | 3+4 4 | 3+4 4 | 3+4 4 |
| | 6 2 | 4+4 4 | 3+3 4 | 3+4 4 | 6+6 4 | 3+4 4 | 3+4 4 | 3+4 4 |
| | 5 2 | 3+3 4 | 4+4 4 | 3+4 4 | 5+5 4 | 3+4 4 | 3+4 4 | 3+4 4 |
| | 6 2 | 4+4 4 | 3+3 4 | 3+4 4 | 6+6 4 | 3+4 4 | 3+4 4 | 3+4 4 |
| | 6 2 | 3+3 4 | 3+3 4 | 3+4 4 | 6+6 4 | 3+4 4 | 3+4 4 | 3+4 4 |
| | 5 2 | 4+4 4 | 4+4 4 | 3+4 4 | 5+5 4 | 3+4 4 | 3+4 4 | 3+4 4 |
| | 5 2 | 4+4 4 | 4+4 4 | 3+4 4 | 5+5 4 | 3+4 4 | 3+4 4 | 3+4 4 |
| | 6 2 | 3+3 4 | 3+3 4 | 3+4 4 | 5+5 4 | 3+4 4 | 3+4 4 | 3+4 4 |
| | 5 2 | | | | | | | |
| | 5 2 | | | | | | | |
| | 5 2 | | | | | | | |
| | 6 2 | | | | | | | |

Continued on next column

**TABLE XXXVII– continued**

| TQC | ICSD | NoE | SF | | | | | |
|---|---|---|---|---|---|---|---|---|
| | 6 2 | | | | | | | |
| | 5 2 | | | | | | | |
| | 6 2 | | | | | | | |
| | 6 2 | | | | | | | |
| | 5 2 | | | | | | | |
| | 6 2 | | | | | | | |
| | 5 2 | | | | | | | |
| | 5 2 | | | | | | | |
| | 6 2 | | | | | | | |
| | 6 2 | | | | | | | |
| | 5 2 | | | | | | | |
| | 6 2 | | | | | | | |
| 62.2.18 | 158364 | 96 | Pd2Sn1 | | | | | |
| | -GM | -R | -S | -T | -U | -X | -Y | -Z |
| | 5 2 | 3+3 4 | 3+3 4 | 3+4 4 | 5+5 4 | 3+4 4 | 3+4 4 | 3+4 4 |
| | 6 2 | 4+4 4 | 4+4 4 | 3+4 4 | 6+6 4 | 3+4 4 | 3+4 4 | 3+4 4 |
| | 5 2 | 3+3 4 | 3+3 4 | 3+4 4 | 6+6 4 | 3+4 4 | 3+4 4 | 3+4 4 |
| | 6 2 | 4+4 4 | 4+4 4 | 3+4 4 | 5+5 4 | 3+4 4 | 3+4 4 | 3+4 4 |
| | 6 2 | 4+4 4 | 4+4 4 | 3+4 4 | 5+5 4 | 3+4 4 | 3+4 4 | 3+4 4 |
| | 5 2 | 3+3 4 | 3+3 4 | 3+4 4 | 5+5 4 | 3+4 4 | 3+4 4 | 3+4 4 |
| | 5 2 | 3+3 4 | 4+4 4 | 3+4 4 | 6+6 4 | 3+4 4 | 3+4 4 | 3+4 4 |
| | 6 2 | 4+4 4 | 3+3 4 | 3+4 4 | 6+6 4 | 3+4 4 | 3+4 4 | 3+4 4 |
| | 5 2 | 4+4 4 | 3+3 4 | 3+4 4 | 5+5 4 | 3+4 4 | 3+4 4 | 3+4 4 |
| | 6 2 | 3+3 4 | 4+4 4 | 3+4 4 | 6+6 4 | 3+4 4 | 3+4 4 | 3+4 4 |
| | 6 2 | 3+3 4 | 3+3 4 | 3+4 4 | 5+5 4 | 3+4 4 | 3+4 4 | 3+4 4 |
| | 6 2 | 4+4 4 | 4+4 4 | 3+4 4 | 6+6 4 | 3+4 4 | 3+4 4 | 3+4 4 |
| | 5 2 | 3+3 4 | 4+4 4 | 3+4 4 | 5+5 4 | 3+4 4 | 3+4 4 | 3+4 4 |
| | 5 2 | 4+4 4 | 3+3 4 | 3+4 4 | 6+6 4 | 3+4 4 | 3+4 4 | 3+4 4 |
| | 6 2 | 3+3 4 | 3+3 4 | 3+4 4 | 6+6 4 | 3+4 4 | 3+4 4 | 3+4 4 |
| | 5 2 | 4+4 4 | 4+4 4 | 3+4 4 | 6+6 4 | 3+4 4 | 3+4 4 | 3+4 4 |
| | 5 2 | 3+3 4 | 4+4 4 | 3+4 4 | 5+5 4 | 3+4 4 | 3+4 4 | 3+4 4 |
| | 5 2 | 4+4 4 | 3+3 4 | 3+4 4 | 5+5 4 | 3+4 4 | 3+4 4 | 3+4 4 |
| | 6 2 | 4+4 4 | 4+4 4 | 3+4 4 | 6+6 4 | 3+4 4 | 3+4 4 | 3+4 4 |
| | 6 2 | 3+3 4 | 3+3 4 | 3+4 4 | 6+6 4 | 3+4 4 | 3+4 4 | 3+4 4 |
| | 5 2 | 4+4 4 | 4+4 4 | 3+4 4 | 5+5 4 | 3+4 4 | 3+4 4 | 3+4 4 |
| | 5 2 | 3+3 4 | 3+3 4 | 3+4 4 | 5+5 4 | 3+4 4 | 3+4 4 | 3+4 4 |
| | 6 2 | 4+4 4 | 3+3 4 | 3+4 4 | 5+5 4 | 3+4 4 | 3+4 4 | 3+4 4 |
| | 5 2 | 3+3 4 | 4+4 4 | 3+4 4 | 6+6 4 | 3+4 4 | 3+4 4 | 3+4 4 |
| | 5 2 | | | | | | | |
| | 6 2 | | | | | | | |
| | 6 2 | | | | | | | |
| | 6 2 | | | | | | | |
| | 5 2 | | | | | | | |
| | 6 2 | | | | | | | |
| | 5 2 | | | | | | | |
| | 6 2 | | | | | | | |
| | 6 2 | | | | | | | |
| | 5 2 | | | | | | | |
| | 5 2 | | | | | | | |
| | 5 2 | | | | | | | |
| | 5 2 | | | | | | | |
| | 6 2 | | | | | | | |
| | 5 2 | | | | | | | |
| | 6 2 | | | | | | | |
| | 5 2 | | | | | | | |
| | 6 2 | | | | | | | |
| | 5 2 | | | | | | | |
| | 6 2 | | | | | | | |
| | 5 2 | | | | | | | |
| | 6 2 | | | | | | | |
| | 5 2 | | | | | | | |

Continued on next column

**TABLE XXXVII– continued**

| TQC | ICSD | NoE | SF | | | | | |
|---|---|---|---|---|---|---|---|---|
| | 5 2 | | | | | | | |
| 62.2.19 | 159304 | 184 | Pd1Yb1Zn1 | | | | | |
| | -GM | -R | -S | -T | -U | -X | -Y | -Z |
| | 5 2 | 3+3 4 | 3+3 4 | 3+4 4 | 6+6 4 | 3+4 4 | 3+4 4 | 3+4 4 |
| | 5 2 | 4+4 4 | 4+4 4 | 3+4 4 | 5+5 4 | 3+4 4 | 3+4 4 | 3+4 4 |
| | 6 2 | 3+3 4 | 3+3 4 | 3+4 4 | 6+6 4 | 3+4 4 | 3+4 4 | 3+4 4 |
| | 6 2 | 4+4 4 | 4+4 4 | 3+4 4 | 5+5 4 | 3+4 4 | 3+4 4 | 3+4 4 |
| | 5 2 | 3+3 4 | 4+4 4 | 3+4 4 | 6+6 4 | 3+4 4 | 3+4 4 | 3+4 4 |
| | 6 2 | 4+4 4 | 3+3 4 | 3+4 4 | 5+5 4 | 3+4 4 | 3+4 4 | 3+4 4 |
| | 5 2 | 4+4 4 | 3+3 4 | 3+4 4 | 6+6 4 | 3+4 4 | 3+4 4 | 3+4 4 |
| | 6 2 | 3+3 4 | 4+4 4 | 3+4 4 | 5+5 4 | 3+4 4 | 3+4 4 | 3+4 4 |
| | 5 2 | 4+4 4 | 3+3 4 | 3+4 4 | 6+6 4 | 3+4 4 | 3+4 4 | 3+4 4 |
| | 6 2 | 3+3 4 | 4+4 4 | 3+4 4 | 5+5 4 | 3+4 4 | 3+4 4 | 3+4 4 |
| | 5 2 | 3+3 4 | 3+3 4 | 3+4 4 | 5+5 4 | 3+4 4 | 3+4 4 | 3+4 4 |
| | 5 2 | 4+4 4 | 4+4 4 | 3+4 4 | 6+6 4 | 3+4 4 | 3+4 4 | 3+4 4 |
| | 6 2 | 3+3 4 | 3+3 4 | 3+4 4 | 6+6 4 | 3+4 4 | 3+4 4 | 3+4 4 |
| | 6 2 | 4+4 4 | 4+4 4 | 3+4 4 | 5+5 4 | 3+4 4 | 3+4 4 | 3+4 4 |
| | 5 2 | 4+4 4 | 4+4 4 | 3+4 4 | 5+5 4 | 3+4 4 | 3+4 4 | 3+4 4 |
| | 6 2 | 3+3 4 | 3+3 4 | 3+4 4 | 6+6 4 | 3+4 4 | 3+4 4 | 3+4 4 |
| | 5 2 | 3+3 4 | 4+4 4 | 3+4 4 | 5+5 4 | 3+4 4 | 3+4 4 | 3+4 4 |
| | 6 2 | 4+4 4 | 3+3 4 | 3+4 4 | 6+6 4 | 3+4 4 | 3+4 4 | 3+4 4 |
| | 5 2 | 4+4 4 | 3+3 4 | 3+4 4 | 5+5 4 | 3+4 4 | 3+4 4 | 3+4 4 |
| | 6 2 | 3+3 4 | 4+4 4 | 3+4 4 | 6+6 4 | 3+4 4 | 3+4 4 | 3+4 4 |
| | 5 2 | 3+3 4 | 4+4 4 | 3+4 4 | 6+6 4 | 3+4 4 | 3+4 4 | 3+4 4 |
| | 5 2 | 4+4 4 | 3+3 4 | 3+4 4 | 5+5 4 | 3+4 4 | 3+4 4 | 3+4 4 |
| | 6 2 | 3+3 4 | 4+4 4 | 3+4 4 | 5+5 4 | 3+4 4 | 3+4 4 | 3+4 4 |
| | 6 2 | 4+4 4 | 3+3 4 | 3+4 4 | 6+6 4 | 3+4 4 | 3+4 4 | 3+4 4 |
| | 5 2 | 4+4 4 | 4+4 4 | 3+4 4 | 5+5 4 | 3+4 4 | 3+4 4 | 3+4 4 |
| | 6 2 | 3+3 4 | 3+3 4 | 3+4 4 | 6+6 4 | 3+4 4 | 3+4 4 | 3+4 4 |
| | 5 2 | 3+3 4 | 4+4 4 | 3+4 4 | 5+5 4 | 3+4 4 | 3+4 4 | 3+4 4 |
| | 6 2 | 4+4 4 | 3+3 4 | 3+4 4 | 6+6 4 | 3+4 4 | 3+4 4 | 3+4 4 |
| | 5 2 | 4+4 4 | 3+3 4 | 3+4 4 | 5+5 4 | 3+4 4 | 3+4 4 | 3+4 4 |
| | 5 2 | 3+3 4 | 4+4 4 | 3+4 4 | 6+6 4 | 3+4 4 | 3+4 4 | 3+4 4 |
| | 6 2 | 3+3 4 | 3+3 4 | 3+4 4 | 5+5 4 | 3+4 4 | 3+4 4 | 3+4 4 |
| | 5 2 | 4+4 4 | 4+4 4 | 3+4 4 | 6+6 4 | 3+4 4 | 3+4 4 | 3+4 4 |
| | 6 2 | 3+3 4 | 3+3 4 | 3+4 4 | 5+5 4 | 3+4 4 | 3+4 4 | 3+4 4 |
| | 5 2 | 4+4 4 | 4+4 4 | 3+4 4 | 6+6 4 | 3+4 4 | 3+4 4 | 3+4 4 |
| | 5 2 | 4+4 4 | 4+4 4 | 3+4 4 | 6+6 4 | 3+4 4 | 3+4 4 | 3+4 4 |
| | 6 2 | 3+3 4 | 3+3 4 | 3+4 4 | 5+5 4 | 3+4 4 | 3+4 4 | 3+4 4 |
| | 6 2 | 3+3 4 | 3+3 4 | 3+4 4 | 5+5 4 | 3+4 4 | 3+4 4 | 3+4 4 |
| | 5 2 | 4+4 4 | 4+4 4 | 3+4 4 | 6+6 4 | 3+4 4 | 3+4 4 | 3+4 4 |
| | 6 2 | 3+3 4 | 4+4 4 | 3+4 4 | 5+5 4 | 3+4 4 | 3+4 4 | 3+4 4 |
| | 6 2 | 4+4 4 | 3+3 4 | 3+4 4 | 5+5 4 | 3+4 4 | 3+4 4 | 3+4 4 |
| | 6 2 | 4+4 4 | 4+4 4 | 3+4 4 | 6+6 4 | 3+4 4 | 3+4 4 | 3+4 4 |
| | 5 2 | 3+3 4 | 3+3 4 | 3+4 4 | 6+6 4 | 3+4 4 | 3+4 4 | 3+4 4 |
| | 6 2 | 3+3 4 | 3+3 4 | 3+4 4 | 6+6 4 | 3+4 4 | 3+4 4 | 3+4 4 |
| | 5 2 | 4+4 4 | 4+4 4 | 3+4 4 | 5+5 4 | 3+4 4 | 3+4 4 | 3+4 4 |
| | 5 2 | 3+3 4 | 3+3 4 | 3+4 4 | 5+5 4 | 3+4 4 | 3+4 4 | 3+4 4 |
| | 6 2 | 4+4 4 | 4+4 4 | 3+4 4 | 6+6 4 | 3+4 4 | 3+4 4 | 3+4 4 |
| | 5 2 | | | | | | | |
| | 5 2 | | | | | | | |
| | 6 2 | | | | | | | |
| | 5 2 | | | | | | | |
| | 5 2 | | | | | | | |
| | 6 2 | | | | | | | |
| | 5 2 | | | | | | | |
| | 6 2 | | | | | | | |
| | 5 2 | | | | | | | |
| | 6 2 | | | | | | | |
| | 5 2 | | | | | | | |
| | 6 2 | | | | | | | |

Continued on next column

**TABLE XXXVII– continued**

| TQC | ICSD | NoE | SF | | | | | |
|---|---|---|---|---|---|---|---|---|
| | 6 2 | | | | | | | |
| | 5 2 | | | | | | | |
| | 6 2 | | | | | | | |
| | 6 2 | | | | | | | |
| | 5 2 | | | | | | | |
| | 5 2 | | | | | | | |
| | 6 2 | | | | | | | |
| | 5 2 | | | | | | | |
| | 5 2 | | | | | | | |
| | 6 2 | | | | | | | |
| | 6 2 | | | | | | | |
| | 5 2 | | | | | | | |
| | 6 2 | | | | | | | |
| | 5 2 | | | | | | | |
| | 6 2 | | | | | | | |
| | 6 2 | | | | | | | |
| | 5 2 | | | | | | | |
| | 5 2 | | | | | | | |
| | 6 2 | | | | | | | |
| | 5 2 | | | | | | | |
| | 5 2 | | | | | | | |
| | 6 2 | | | | | | | |
| | 6 2 | | | | | | | |
| | 6 2 | | | | | | | |
| | 5 2 | | | | | | | |
| | 6 2 | | | | | | | |
| | 5 2 | | | | | | | |
| | 5 2 | | | | | | | |
| | 5 2 | | | | | | | |
| | 6 2 | | | | | | | |
| | 6 2 | | | | | | | |
| | 6 2 | | | | | | | |
| | 5 2 | | | | | | | |
| | 5 2 | | | | | | | |
| 62.2.20 | 408408 | 152 | Pd1Sn1Yb1 | | | | | |
| | -GM | -R | -S | -T | -U | -X | -Y | -Z |
| | 5 2 | 3+3 4 | 3+3 4 | 3+4 4 | 6+6 4 | 3+4 4 | 3+4 4 | 3+4 4 |
| | 5 2 | 4+4 4 | 4+4 4 | 3+4 4 | 5+5 4 | 3+4 4 | 3+4 4 | 3+4 4 |
| | 6 2 | 3+3 4 | 3+3 4 | 3+4 4 | 6+6 4 | 3+4 4 | 3+4 4 | 3+4 4 |
| | 6 2 | 4+4 4 | 4+4 4 | 3+4 4 | 5+5 4 | 3+4 4 | 3+4 4 | 3+4 4 |
| | 5 2 | 4+4 4 | 3+3 4 | 3+4 4 | 6+6 4 | 3+4 4 | 3+4 4 | 3+4 4 |
| | 6 2 | 3+3 4 | 4+4 4 | 3+4 4 | 5+5 4 | 3+4 4 | 3+4 4 | 3+4 4 |
| | 5 2 | 3+3 4 | 4+4 4 | 3+4 4 | 6+6 4 | 3+4 4 | 3+4 4 | 3+4 4 |
| | 6 2 | 4+4 4 | 3+3 4 | 3+4 4 | 5+5 4 | 3+4 4 | 3+4 4 | 3+4 4 |
| | 5 2 | 3+3 4 | 4+4 4 | 3+4 4 | 5+5 4 | 3+4 4 | 3+4 4 | 3+4 4 |
| | 6 2 | 4+4 4 | 3+3 4 | 3+4 4 | 6+6 4 | 3+4 4 | 3+4 4 | 3+4 4 |
| | 5 2 | 4+4 4 | 3+3 4 | 3+4 4 | 6+6 4 | 3+4 4 | 3+4 4 | 3+4 4 |
| | 5 2 | 3+3 4 | 4+4 4 | 3+4 4 | 5+5 4 | 3+4 4 | 3+4 4 | 3+4 4 |
| | 6 2 | 4+4 4 | 3+3 4 | 3+4 4 | 5+5 4 | 3+4 4 | 3+4 4 | 3+4 4 |
| | 6 2 | 3+3 4 | 4+4 4 | 3+4 4 | 5+5 4 | 3+4 4 | 3+4 4 | 3+4 4 |
| | 5 2 | 4+4 4 | 3+3 4 | 3+4 4 | 6+6 4 | 3+4 4 | 3+4 4 | 3+4 4 |
| | 6 2 | 3+3 4 | 4+4 4 | 3+4 4 | 6+6 4 | 3+4 4 | 3+4 4 | 3+4 4 |
| | 5 2 | 3+3 4 | 3+3 4 | 3+4 4 | 5+5 4 | 3+4 4 | 3+4 4 | 3+4 4 |
| | 5 2 | 4+4 4 | 4+4 4 | 3+4 4 | 6+6 4 | 3+4 4 | 3+4 4 | 3+4 4 |
| | 6 2 | 3+3 4 | 4+4 4 | 3+4 4 | 5+5 4 | 3+4 4 | 3+4 4 | 3+4 4 |
| | 6 2 | 4+4 4 | 3+3 4 | 3+4 4 | 6+6 4 | 3+4 4 | 3+4 4 | 3+4 4 |
| | 6 2 | 4+4 4 | 4+4 4 | 3+4 4 | 5+5 4 | 3+4 4 | 3+4 4 | 3+4 4 |
| | 5 2 | 3+3 4 | 3+3 4 | 3+4 4 | 6+6 4 | 3+4 4 | 3+4 4 | 3+4 4 |
| | 6 2 | 3+3 4 | 4+4 4 | 3+4 4 | 5+5 4 | 3+4 4 | 3+4 4 | 3+4 4 |
| | 5 2 | 4+4 4 | 3+3 4 | 3+4 4 | 6+6 4 | 3+4 4 | 3+4 4 | 3+4 4 |
| | 5 2 | 3+3 4 | 3+3 4 | 3+4 4 | 5+5 4 | 3+4 4 | 3+4 4 | 3+4 4 |

Continued on next column

**TABLE XXXVII– continued**

| TQC | ICSD | NoE | SF | | | | | |
|---|---|---|---|---|---|---|---|---|
| | 5 2 | 4+4 4 | 4+4 4 | 3+4 4 | 6+6 4 | 3+4 4 | 3+4 4 | 3+4 4 |
| | 6 2 | 3+3 4 | 3+3 4 | 3+4 4 | 5+5 4 | 3+4 4 | 3+4 4 | 3+4 4 |
| | 5 2 | 4+4 4 | 4+4 4 | 3+4 4 | 6+6 4 | 3+4 4 | 3+4 4 | 3+4 4 |
| | 6 2 | 4+4 4 | 3+3 4 | 3+4 4 | 5+5 4 | 3+4 4 | 3+4 4 | 3+4 4 |
| | 5 2 | 3+3 4 | 4+4 4 | 3+4 4 | 5+5 4 | 3+4 4 | 3+4 4 | 3+4 4 |
| | 5 2 | 4+4 4 | 3+3 4 | 3+4 4 | 6+6 4 | 3+4 4 | 3+4 4 | 3+4 4 |
| | 6 2 | 3+3 4 | 4+4 4 | 3+4 4 | 6+6 4 | 3+4 4 | 3+4 4 | 3+4 4 |
| | 5 2 | 4+4 4 | 3+3 4 | 3+4 4 | 5+5 4 | 3+4 4 | 3+4 4 | 3+4 4 |
| | 6 2 | 3+3 4 | 4+4 4 | 3+4 4 | 6+6 4 | 3+4 4 | 3+4 4 | 3+4 4 |
| | 5 2 | 4+4 4 | 4+4 4 | 3+4 4 | 5+5 4 | 3+4 4 | 3+4 4 | 3+4 4 |
| | 6 2 | 3+3 4 | 3+3 4 | 3+4 4 | 6+6 4 | 3+4 4 | 3+4 4 | 3+4 4 |
| | 5 2 | 3+3 4 | 4+4 4 | 3+4 4 | 5+5 4 | 3+4 4 | 3+4 4 | 3+4 4 |
| | 6 2 | 4+4 4 | 3+3 4 | 3+4 4 | 5+5 4 | 3+4 4 | 3+4 4 | 3+4 4 |
| | 6 2 | | | | | | | |
| | 5 2 | | | | | | | |
| | 6 2 | | | | | | | |
| | 6 2 | | | | | | | |
| | 5 2 | | | | | | | |
| | 5 2 | | | | | | | |
| | 5 2 | | | | | | | |
| | 6 2 | | | | | | | |
| | 6 2 | | | | | | | |
| | 5 2 | | | | | | | |
| | 6 2 | | | | | | | |
| | 6 2 | | | | | | | |
| | 5 2 | | | | | | | |
| | 5 2 | | | | | | | |
| | 6 2 | | | | | | | |
| | 6 2 | | | | | | | |
| | 5 2 | | | | | | | |
| | 5 2 | | | | | | | |
| | 5 2 | | | | | | | |
| | 6 2 | | | | | | | |
| | 6 2 | | | | | | | |
| | 6 2 | | | | | | | |
| | 5 2 | | | | | | | |
| | 6 2 | | | | | | | |
| | 5 2 | | | | | | | |
| | 6 2 | | | | | | | |
| | 5 2 | | | | | | | |
| | 6 2 | | | | | | | |
| | 5 2 | | | | | | | |
| | 6 2 | | | | | | | |
| | 5 2 | | | | | | | |
| | 6 2 | | | | | | | |
| | 6 2 | | | | | | | |
| | 5 2 | | | | | | | |
| | 6 2 | | | | | | | |
| | 5 2 | | | | | | | |
| | 5 2 | | | | | | | |
| | 5 2 | | | | | | | |
| 62.2.21 | 151167 | 152 | Pd1Sn1Yb1 | | | | | |
| | -GM | -R | -S | -T | -U | -X | -Y | -Z |
| | 5 2 | 3+3 4 | 3+3 4 | 3+4 4 | 6+6 4 | 3+4 4 | 3+4 4 | 3+4 4 |
| | 5 2 | 4+4 4 | 4+4 4 | 3+4 4 | 5+5 4 | 3+4 4 | 3+4 4 | 3+4 4 |
| | 6 2 | 3+3 4 | 3+3 4 | 3+4 4 | 6+6 4 | 3+4 4 | 3+4 4 | 3+4 4 |
| | 6 2 | 4+4 4 | 4+4 4 | 3+4 4 | 5+5 4 | 3+4 4 | 3+4 4 | 3+4 4 |
| | 5 2 | 4+4 4 | 3+3 4 | 3+4 4 | 6+6 4 | 3+4 4 | 3+4 4 | 3+4 4 |
| | 6 2 | 3+3 4 | 4+4 4 | 3+4 4 | 5+5 4 | 3+4 4 | 3+4 4 | 3+4 4 |
| | 5 2 | 3+3 4 | 4+4 4 | 3+4 4 | 6+6 4 | 3+4 4 | 3+4 4 | 3+4 4 |
| | 6 2 | 4+4 4 | 3+3 4 | 3+4 4 | 5+5 4 | 3+4 4 | 3+4 4 | 3+4 4 |

Continued on next column

**TABLE XXXVII– continued**

| TQC | ICSD | NoE | SF | | | | | |
|---|---|---|---|---|---|---|---|---|
| | 5 2 | 3+3 4 | 4+4 4 | 3+4 4 | 5+5 4 | 3+4 4 | 3+4 4 | 3+4 4 |
| | 6 2 | 4+4 4 | 3+3 4 | 3+4 4 | 6+6 4 | 3+4 4 | 3+4 4 | 3+4 4 |
| | 5 2 | 4+4 4 | 3+3 4 | 3+4 4 | 6+6 4 | 3+4 4 | 3+4 4 | 3+4 4 |
| | 5 2 | 3+3 4 | 4+4 4 | 3+4 4 | 5+5 4 | 3+4 4 | 3+4 4 | 3+4 4 |
| | 6 2 | 4+4 4 | 3+3 4 | 3+4 4 | 5+5 4 | 3+4 4 | 3+4 4 | 3+4 4 |
| | 6 2 | 3+3 4 | 4+4 4 | 3+4 4 | 5+5 4 | 3+4 4 | 3+4 4 | 3+4 4 |
| | 5 2 | 4+4 4 | 3+3 4 | 3+4 4 | 6+6 4 | 3+4 4 | 3+4 4 | 3+4 4 |
| | 6 2 | 3+3 4 | 4+4 4 | 3+4 4 | 6+6 4 | 3+4 4 | 3+4 4 | 3+4 4 |
| | 5 2 | 3+3 4 | 3+3 4 | 3+4 4 | 5+5 4 | 3+4 4 | 3+4 4 | 3+4 4 |
| | 5 2 | 4+4 4 | 4+4 4 | 3+4 4 | 6+6 4 | 3+4 4 | 3+4 4 | 3+4 4 |
| | 6 2 | 3+3 4 | 4+4 4 | 3+4 4 | 5+5 4 | 3+4 4 | 3+4 4 | 3+4 4 |
| | 6 2 | 4+4 4 | 3+3 4 | 3+4 4 | 6+6 4 | 3+4 4 | 3+4 4 | 3+4 4 |
| | 6 2 | 4+4 4 | 4+4 4 | 3+4 4 | 5+5 4 | 3+4 4 | 3+4 4 | 3+4 4 |
| | 5 2 | 3+3 4 | 3+3 4 | 3+4 4 | 6+6 4 | 3+4 4 | 3+4 4 | 3+4 4 |
| | 6 2 | 3+3 4 | 4+4 4 | 3+4 4 | 5+5 4 | 3+4 4 | 3+4 4 | 3+4 4 |
| | 5 2 | 4+4 4 | 3+3 4 | 3+4 4 | 6+6 4 | 3+4 4 | 3+4 4 | 3+4 4 |
| | 5 2 | 3+3 4 | 3+3 4 | 3+4 4 | 5+5 4 | 3+4 4 | 3+4 4 | 3+4 4 |
| | 5 2 | 4+4 4 | 4+4 4 | 3+4 4 | 6+6 4 | 3+4 4 | 3+4 4 | 3+4 4 |
| | 6 2 | 3+3 4 | 3+3 4 | 3+4 4 | 5+5 4 | 3+4 4 | 3+4 4 | 3+4 4 |
| | 5 2 | 4+4 4 | 4+4 4 | 3+4 4 | 6+6 4 | 3+4 4 | 3+4 4 | 3+4 4 |
| | 6 2 | 4+4 4 | 3+3 4 | 3+4 4 | 5+5 4 | 3+4 4 | 3+4 4 | 3+4 4 |
| | 5 2 | 3+3 4 | 4+4 4 | 3+4 4 | 5+5 4 | 3+4 4 | 3+4 4 | 3+4 4 |
| | 5 2 | 4+4 4 | 3+3 4 | 3+4 4 | 6+6 4 | 3+4 4 | 3+4 4 | 3+4 4 |
| | 6 2 | 3+3 4 | 4+4 4 | 3+4 4 | 6+6 4 | 3+4 4 | 3+4 4 | 3+4 4 |
| | 5 2 | 4+4 4 | 3+3 4 | 3+4 4 | 5+5 4 | 3+4 4 | 3+4 4 | 3+4 4 |
| | 6 2 | 3+3 4 | 4+4 4 | 3+4 4 | 6+6 4 | 3+4 4 | 3+4 4 | 3+4 4 |
| | 5 2 | 4+4 4 | 4+4 4 | 3+4 4 | 5+5 4 | 3+4 4 | 3+4 4 | 3+4 4 |
| | 6 2 | 3+3 4 | 3+3 4 | 3+4 4 | 6+6 4 | 3+4 4 | 3+4 4 | 3+4 4 |
| | 5 2 | 3+3 4 | 4+4 4 | 3+4 4 | 5+5 4 | 3+4 4 | 3+4 4 | 3+4 4 |
| | 6 2 | 4+4 4 | 3+3 4 | 3+4 4 | 5+5 4 | 3+4 4 | 3+4 4 | 3+4 4 |
| | 6 2 | | | | | | | |
| | 5 2 | | | | | | | |
| | 6 2 | | | | | | | |
| | 6 2 | | | | | | | |
| | 5 2 | | | | | | | |
| | 5 2 | | | | | | | |
| | 5 2 | | | | | | | |
| | 6 2 | | | | | | | |
| | 6 2 | | | | | | | |
| | 5 2 | | | | | | | |
| | 6 2 | | | | | | | |
| | 6 2 | | | | | | | |
| | 5 2 | | | | | | | |
| | 5 2 | | | | | | | |
| | 6 2 | | | | | | | |
| | 6 2 | | | | | | | |
| | 5 2 | | | | | | | |
| | 5 2 | | | | | | | |
| | 5 2 | | | | | | | |
| | 6 2 | | | | | | | |
| | 6 2 | | | | | | | |
| | 6 2 | | | | | | | |
| | 5 2 | | | | | | | |
| | 6 2 | | | | | | | |
| | 5 2 | | | | | | | |
| | 6 2 | | | | | | | |
| | 5 2 | | | | | | | |
| | 6 2 | | | | | | | |
| | 5 2 | | | | | | | |
| | 6 2 | | | | | | | |
| | 5 2 | | | | | | | |

Continued on next column

**TABLE XXXVII– continued**

| TQC | ICSD | NoE | SF | | | | | |
|---|---|---|---|---|---|---|---|---|
| | 6 2 | | | | | | | |
| | 6 2 | | | | | | | |
| | 5 2 | | | | | | | |
| | 6 2 | | | | | | | |
| | 5 2 | | | | | | | |
| | 5 2 | | | | | | | |
| | 5 2 | | | | | | | |
| 62.2.22 | 109415 | 56 | Pd1Sn1 | | | | | |
| | -GM | -R | -S | -T | -U | -X | -Y | -Z |
| | 5 2 | 4+4 4 | 3+3 4 | 3+4 4 | 5+5 4 | 3+4 4 | 3+4 4 | 3+4 4 |
| | 5 2 | 3+3 4 | 4+4 4 | 3+4 4 | 6+6 4 | 3+4 4 | 3+4 4 | 3+4 4 |
| | 6 2 | 4+4 4 | 3+3 4 | 3+4 4 | 5+5 4 | 3+4 4 | 3+4 4 | 3+4 4 |
| | 6 2 | 3+3 4 | 4+4 4 | 3+4 4 | 6+6 4 | 3+4 4 | 3+4 4 | 3+4 4 |
| | 6 2 | 3+3 4 | 3+3 4 | 3+4 4 | 5+5 4 | 3+4 4 | 3+4 4 | 3+4 4 |
| | 5 2 | 4+4 4 | 4+4 4 | 3+4 4 | 5+5 4 | 3+4 4 | 3+4 4 | 3+4 4 |
| | 6 2 | 3+3 4 | 4+4 4 | 3+4 4 | 6+6 4 | 3+4 4 | 3+4 4 | 3+4 4 |
| | 5 2 | 4+4 4 | 3+3 4 | 3+4 4 | 6+6 4 | 3+4 4 | 3+4 4 | 3+4 4 |
| | 6 2 | 4+4 4 | 4+4 4 | 3+4 4 | 5+5 4 | 3+4 4 | 3+4 4 | 3+4 4 |
| | 5 2 | 3+3 4 | 3+3 4 | 3+4 4 | 6+6 4 | 3+4 4 | 3+4 4 | 3+4 4 |
| | 6 2 | 3+3 4 | 4+4 4 | 3+4 4 | 5+5 4 | 3+4 4 | 3+4 4 | 3+4 4 |
| | 5 2 | 4+4 4 | 3+3 4 | 3+4 4 | 6+6 4 | 3+4 4 | 3+4 4 | 3+4 4 |
| | 5 2 | 3+3 4 | 3+3 4 | 3+4 4 | 6+6 4 | 3+4 4 | 3+4 4 | 3+4 4 |
| | 5 2 | 4+4 4 | 4+4 4 | 3+4 4 | 5+5 4 | 3+4 4 | 3+4 4 | 3+4 4 |
| | 5 2 | | | | | | | |
| | 6 2 | | | | | | | |
| | 5 2 | | | | | | | |
| | 6 2 | | | | | | | |
| | 6 2 | | | | | | | |
| | 5 2 | | | | | | | |
| | 5 2 | | | | | | | |
| | 6 2 | | | | | | | |
| | 6 2 | | | | | | | |
| | 6 2 | | | | | | | |
| | 5 2 | | | | | | | |
| | 5 2 | | | | | | | |
| | 5 2 | | | | | | | |
| | 5 2 | | | | | | | |
| 62.2.23 | 79237 | 56 | Pd1Si1 | | | | | |
| | -GM | -R | -S | -T | -U | -X | -Y | -Z |
| | 5 2 | 4+4 4 | 3+3 4 | 3+4 4 | 5+5 4 | 3+4 4 | 3+4 4 | 3+4 4 |
| | 5 2 | 3+3 4 | 4+4 4 | 3+4 4 | 6+6 4 | 3+4 4 | 3+4 4 | 3+4 4 |
| | 6 2 | 4+4 4 | 3+3 4 | 3+4 4 | 5+5 4 | 3+4 4 | 3+4 4 | 3+4 4 |
| | 6 2 | 3+3 4 | 4+4 4 | 3+4 4 | 6+6 4 | 3+4 4 | 3+4 4 | 3+4 4 |
| | 6 2 | 3+3 4 | 4+4 4 | 3+4 4 | 5+5 4 | 3+4 4 | 3+4 4 | 3+4 4 |
| | 5 2 | 4+4 4 | 3+3 4 | 3+4 4 | 6+6 4 | 3+4 4 | 3+4 4 | 3+4 4 |
| | 6 2 | 4+4 4 | 4+4 4 | 3+4 4 | 5+5 4 | 3+4 4 | 3+4 4 | 3+4 4 |
| | 5 2 | 3+3 4 | 3+3 4 | 3+4 4 | 5+5 4 | 3+4 4 | 3+4 4 | 3+4 4 |
| | 5 2 | 3+3 4 | 4+4 4 | 3+4 4 | 6+6 4 | 3+4 4 | 3+4 4 | 3+4 4 |
| | 6 2 | 4+4 4 | 3+3 4 | 3+4 4 | 6+6 4 | 3+4 4 | 3+4 4 | 3+4 4 |
| | 6 2 | 3+3 4 | 4+4 4 | 3+4 4 | 5+5 4 | 3+4 4 | 3+4 4 | 3+4 4 |
| | 5 2 | 4+4 4 | 3+3 4 | 3+4 4 | 6+6 4 | 3+4 4 | 3+4 4 | 3+4 4 |
| | 5 2 | 4+4 4 | 3+3 4 | 3+4 4 | 6+6 4 | 3+4 4 | 3+4 4 | 3+4 4 |
| | 5 2 | 3+3 4 | 4+4 4 | 3+4 4 | 5+5 4 | 3+4 4 | 3+4 4 | 3+4 4 |
| | 5 2 | | | | | | | |
| | 6 2 | | | | | | | |
| | 5 2 | | | | | | | |
| | 6 2 | | | | | | | |
| | 6 2 | | | | | | | |
| | 5 2 | | | | | | | |
| | 6 2 | | | | | | | |
| | 6 2 | | | | | | | |

Continued on next column

**TABLE XXXVII– continued**

| TQC | ICSD | NoE | SF | | | | | |
|---|---|---|---|---|---|---|---|---|
| | 6 2 | | | | | | | |
| | 5 2 | | | | | | | |
| | 5 2 | | | | | | | |
| | 5 2 | | | | | | | |
| | 5 2 | | | | | | | |
| | 5 2 | | | | | | | |
| 62.2.24 | 15016 | 56 | Pd1Si1 | | | | | |
| | -GM | -R | -S | -T | -U | -X | -Y | -Z |
| | 5 2 | 3+3 4 | 4+4 4 | 3+4 4 | 6+6 4 | 3+4 4 | 3+4 4 | 3+4 4 |
| | 5 2 | 4+4 4 | 3+3 4 | 3+4 4 | 5+5 4 | 3+4 4 | 3+4 4 | 3+4 4 |
| | 6 2 | 3+3 4 | 4+4 4 | 3+4 4 | 6+6 4 | 3+4 4 | 3+4 4 | 3+4 4 |
| | 6 2 | 4+4 4 | 3+3 4 | 3+4 4 | 5+5 4 | 3+4 4 | 3+4 4 | 3+4 4 |
| | 6 2 | 4+4 4 | 3+3 4 | 3+4 4 | 6+6 4 | 3+4 4 | 3+4 4 | 3+4 4 |
| | 5 2 | 3+3 4 | 4+4 4 | 3+4 4 | 5+5 4 | 3+4 4 | 3+4 4 | 3+4 4 |
| | 6 2 | 3+3 4 | 3+3 4 | 3+4 4 | 6+6 4 | 3+4 4 | 3+4 4 | 3+4 4 |
| | 5 2 | 4+4 4 | 4+4 4 | 3+4 4 | 6+6 4 | 3+4 4 | 3+4 4 | 3+4 4 |
| | 5 2 | 4+4 4 | 3+3 4 | 3+4 4 | 5+5 4 | 3+4 4 | 3+4 4 | 3+4 4 |
| | 6 2 | 3+3 4 | 4+4 4 | 3+4 4 | 5+5 4 | 3+4 4 | 3+4 4 | 3+4 4 |
| | 6 2 | 4+4 4 | 3+3 4 | 3+4 4 | 6+6 4 | 3+4 4 | 3+4 4 | 3+4 4 |
| | 5 2 | 3+3 4 | 4+4 4 | 3+4 4 | 5+5 4 | 3+4 4 | 3+4 4 | 3+4 4 |
| | 5 2 | 3+3 4 | 4+4 4 | 3+4 4 | 5+5 4 | 3+4 4 | 3+4 4 | 3+4 4 |
| | 5 2 | 4+4 4 | 3+3 4 | 3+4 4 | 6+6 4 | 3+4 4 | 3+4 4 | 3+4 4 |
| | 5 2 | | | | | | | |
| | 6 2 | | | | | | | |
| | 5 2 | | | | | | | |
| | 6 2 | | | | | | | |
| | 6 2 | | | | | | | |
| | 5 2 | | | | | | | |
| | 6 2 | | | | | | | |
| | 6 2 | | | | | | | |
| | 6 2 | | | | | | | |
| | 5 2 | | | | | | | |
| | 5 2 | | | | | | | |
| | 5 2 | | | | | | | |
| | 5 2 | | | | | | | |
| | 5 2 | | | | | | | |
| 62.2.25 | 1918 | 104 | Pd1Pu1 | | | | | |
| | -GM | -R | -S | -T | -U | -X | -Y | -Z |
| | 5 2 | 3+3 4 | 3+3 4 | 3+4 4 | 5+5 4 | 3+4 4 | 3+4 4 | 3+4 4 |
| | 6 2 | 4+4 4 | 4+4 4 | 3+4 4 | 6+6 4 | 3+4 4 | 3+4 4 | 3+4 4 |
| | 6 2 | 3+3 4 | 3+3 4 | 3+4 4 | 5+5 4 | 3+4 4 | 3+4 4 | 3+4 4 |
| | 5 2 | 4+4 4 | 4+4 4 | 3+4 4 | 6+6 4 | 3+4 4 | 3+4 4 | 3+4 4 |
| | 6 2 | 4+4 4 | 4+4 4 | 3+4 4 | 5+5 4 | 3+4 4 | 3+4 4 | 3+4 4 |
| | 5 2 | 3+3 4 | 3+3 4 | 3+4 4 | 6+6 4 | 3+4 4 | 3+4 4 | 3+4 4 |
| | 5 2 | 3+3 4 | 3+3 4 | 3+4 4 | 5+5 4 | 3+4 4 | 3+4 4 | 3+4 4 |
| | 6 2 | 4+4 4 | 4+4 4 | 3+4 4 | 6+6 4 | 3+4 4 | 3+4 4 | 3+4 4 |
| | 5 2 | 3+3 4 | 4+4 4 | 3+4 4 | 5+5 4 | 3+4 4 | 3+4 4 | 3+4 4 |
| | 6 2 | 4+4 4 | 3+3 4 | 3+4 4 | 6+6 4 | 3+4 4 | 3+4 4 | 3+4 4 |
| | 5 2 | 4+4 4 | 3+3 4 | 3+4 4 | 6+6 4 | 3+4 4 | 3+4 4 | 3+4 4 |
| | 6 2 | 3+3 4 | 4+4 4 | 3+4 4 | 5+5 4 | 3+4 4 | 3+4 4 | 3+4 4 |
| | 5 2 | 4+4 4 | 3+3 4 | 3+4 4 | 6+6 4 | 3+4 4 | 3+4 4 | 3+4 4 |
| | 5 2 | 3+3 4 | 4+4 4 | 3+4 4 | 5+5 4 | 3+4 4 | 3+4 4 | 3+4 4 |
| | 6 2 | 4+4 4 | 4+4 4 | 3+4 4 | 5+5 4 | 3+4 4 | 3+4 4 | 3+4 4 |
| | 6 2 | 3+3 4 | 3+3 4 | 3+4 4 | 6+6 4 | 3+4 4 | 3+4 4 | 3+4 4 |
| | 5 2 | 3+3 4 | 3+3 4 | 3+4 4 | 5+5 4 | 3+4 4 | 3+4 4 | 3+4 4 |
| | 6 2 | 4+4 4 | 4+4 4 | 3+4 4 | 6+6 4 | 3+4 4 | 3+4 4 | 3+4 4 |
| | 5 2 | 3+3 4 | 4+4 4 | 3+4 4 | 5+5 4 | 3+4 4 | 3+4 4 | 3+4 4 |
| | 6 2 | 4+4 4 | 3+3 4 | 3+4 4 | 6+6 4 | 3+4 4 | 3+4 4 | 3+4 4 |
| | 5 2 | 4+4 4 | 4+4 4 | 3+4 4 | 6+6 4 | 3+4 4 | 3+4 4 | 3+4 4 |
| | 6 2 | 3+3 4 | 3+3 4 | 3+4 4 | 6+6 4 | 3+4 4 | 3+4 4 | 3+4 4 |
| | 5 2 | 4+4 4 | 4+4 4 | 3+4 4 | 5+5 4 | 3+4 4 | 3+4 4 | 3+4 4 |

Continued on next column

**TABLE XXXVII– continued**

| | | | | | | | | |
|---|---|---|---|---|---|---|---|---|
| TQC | ICSD | NoE | SF | | | | | |
| | 6 2 | 3+3 4 | 3+3 4 | 3+4 4 | 5+5 4 | 3+4 4 | 3+4 4 | 3+4 4 |
| | 5 2 | 3+3 4 | 3+3 4 | 3+4 4 | 6+6 4 | 3+4 4 | 3+4 4 | 3+4 4 |
| | 6 2 | 4+4 4 | 4+4 4 | 3+4 4 | 5+5 4 | 3+4 4 | 3+4 4 | 3+4 4 |
| | 6 2 | | | | | | | |
| | 6 2 | | | | | | | |
| | 5 2 | | | | | | | |
| | 5 2 | | | | | | | |
| | 5 2 | | | | | | | |
| | 6 2 | | | | | | | |
| | 5 2 | | | | | | | |
| | 5 2 | | | | | | | |
| | 6 2 | | | | | | | |
| | 6 2 | | | | | | | |
| | 5 2 | | | | | | | |
| | 6 2 | | | | | | | |
| | 5 2 | | | | | | | |
| | 6 2 | | | | | | | |
| | 6 2 | | | | | | | |
| | 5 2 | | | | | | | |
| | 5 2 | | | | | | | |
| | 5 2 | | | | | | | |
| | 6 2 | | | | | | | |
| | 5 2 | | | | | | | |
| | 6 2 | | | | | | | |
| | 5 2 | | | | | | | |
| | 6 2 | | | | | | | |
| | 5 2 | | | | | | | |
| | 5 2 | | | | | | | |
| | 6 2 | | | | | | | |
| 62.2.26 | 400334 | 56 | P2Zr1 | | | | | |
| | -GM | -R | -S | -T | -U | -X | -Y | -Z |
| | 5 2 | 3+3 4 | 3+3 4 | 3+4 4 | 6+6 4 | 3+4 4 | 3+4 4 | 3+4 4 |
| | 5 2 | 4+4 4 | 4+4 4 | 3+4 4 | 5+5 4 | 3+4 4 | 3+4 4 | 3+4 4 |
| | 6 2 | 4+4 4 | 3+3 4 | 3+4 4 | 6+6 4 | 3+4 4 | 3+4 4 | 3+4 4 |
| | 6 2 | 3+3 4 | 4+4 4 | 3+4 4 | 5+5 4 | 3+4 4 | 3+4 4 | 3+4 4 |
| | 5 2 | 3+3 4 | 3+3 4 | 3+4 4 | 5+5 4 | 3+4 4 | 3+4 4 | 3+4 4 |
| | 5 2 | 4+4 4 | 4+4 4 | 3+4 4 | 6+6 4 | 3+4 4 | 3+4 4 | 3+4 4 |
| | 6 2 | 4+4 4 | 3+3 4 | 3+4 4 | 6+6 4 | 3+4 4 | 3+4 4 | 3+4 4 |
| | 6 2 | 3+3 4 | 4+4 4 | 3+4 4 | 5+5 4 | 3+4 4 | 3+4 4 | 3+4 4 |
| | 5 2 | 3+3 4 | 4+4 4 | 3+4 4 | 6+6 4 | 3+4 4 | 3+4 4 | 3+4 4 |
| | 6 2 | 4+4 4 | 3+3 4 | 3+4 4 | 5+5 4 | 3+4 4 | 3+4 4 | 3+4 4 |
| | 6 2 | 3+3 4 | 3+3 4 | 3+4 4 | 5+5 4 | 3+4 4 | 3+4 4 | 3+4 4 |
| | 5 2 | 4+4 4 | 4+4 4 | 3+4 4 | 6+6 4 | 3+4 4 | 3+4 4 | 3+4 4 |
| | 5 2 | 3+3 4 | 4+4 4 | 3+4 4 | 6+6 4 | 3+4 4 | 3+4 4 | 3+4 4 |
| | 5 2 | 4+4 4 | 3+3 4 | 3+4 4 | 5+5 4 | 3+4 4 | 3+4 4 | 3+4 4 |
| | 5 2 | | | | | | | |
| | 5 2 | | | | | | | |
| | 6 2 | | | | | | | |
| | 5 2 | | | | | | | |
| | 5 2 | | | | | | | |
| | 6 2 | | | | | | | |
| | 6 2 | | | | | | | |
| | 5 2 | | | | | | | |
| | 6 2 | | | | | | | |
| | 5 2 | | | | | | | |
| | 6 2 | | | | | | | |
| | 5 2 | | | | | | | |
| | 6 2 | | | | | | | |
| | 6 2 | | | | | | | |
| 62.2.27 | 417911 | 72 | P1Pt1Sc1 | | | | | |
| | -GM | -R | -S | -T | -U | -X | -Y | -Z |

Continued on next column

**TABLE XXXVII– continued**

| | | | | | | | | |
|---|---|---|---|---|---|---|---|---|
| TQC | ICSD | NoE | SF | | | | | |
| | 5 2 | 4+4 4 | 3+3 4 | 3+4 4 | 5+5 4 | 3+4 4 | 3+4 4 | 3+4 4 |
| | 5 2 | 3+3 4 | 4+4 4 | 3+4 4 | 6+6 4 | 3+4 4 | 3+4 4 | 3+4 4 |
| | 6 2 | 4+4 4 | 3+3 4 | 3+4 4 | 6+6 4 | 3+4 4 | 3+4 4 | 3+4 4 |
| | 6 2 | 3+3 4 | 4+4 4 | 3+4 4 | 5+5 4 | 3+4 4 | 3+4 4 | 3+4 4 |
| | 6 2 | 4+4 4 | 3+3 4 | 3+4 4 | 5+5 4 | 3+4 4 | 3+4 4 | 3+4 4 |
| | 5 2 | 3+3 4 | 4+4 4 | 3+4 4 | 5+5 4 | 3+4 4 | 3+4 4 | 3+4 4 |
| | 6 2 | 4+4 4 | 4+4 4 | 3+4 4 | 6+6 4 | 3+4 4 | 3+4 4 | 3+4 4 |
| | 5 2 | 3+3 4 | 3+3 4 | 3+4 4 | 6+6 4 | 3+4 4 | 3+4 4 | 3+4 4 |
| | 5 2 | 3+3 4 | 3+3 4 | 3+4 4 | 5+5 4 | 3+4 4 | 3+4 4 | 3+4 4 |
| | 5 2 | 4+4 4 | 4+4 4 | 3+4 4 | 6+6 4 | 3+4 4 | 3+4 4 | 3+4 4 |
| | 5 2 | 3+3 4 | 4+4 4 | 3+4 4 | 5+5 4 | 3+4 4 | 3+4 4 | 3+4 4 |
| | 6 2 | 4+4 4 | 3+3 4 | 3+4 4 | 6+6 4 | 3+4 4 | 3+4 4 | 3+4 4 |
| | 5 2 | 3+3 4 | 4+4 4 | 3+4 4 | 5+5 4 | 3+4 4 | 3+4 4 | 3+4 4 |
| | 6 2 | 4+4 4 | 3+3 4 | 3+4 4 | 6+6 4 | 3+4 4 | 3+4 4 | 3+4 4 |
| | 5 2 | 4+4 4 | 4+4 4 | 3+4 4 | 5+5 4 | 3+4 4 | 3+4 4 | 3+4 4 |
| | 6 2 | 3+3 4 | 3+3 4 | 3+4 4 | 5+5 4 | 3+4 4 | 3+4 4 | 3+4 4 |
| | 6 2 | 4+4 4 | 3+3 4 | 3+4 4 | 6+6 4 | 3+4 4 | 3+4 4 | 3+4 4 |
| | 5 2 | 3+3 4 | 4+4 4 | 3+4 4 | 6+6 4 | 3+4 4 | 3+4 4 | 3+4 4 |
| | 6 2 | | | | | | | |
| | 5 2 | | | | | | | |
| | 5 2 | | | | | | | |
| | 6 2 | | | | | | | |
| | 6 2 | | | | | | | |
| | 6 2 | | | | | | | |
| | 5 2 | | | | | | | |
| | 6 2 | | | | | | | |
| | 5 2 | | | | | | | |
| | 5 2 | | | | | | | |
| | 6 2 | | | | | | | |
| | 5 2 | | | | | | | |
| | 6 2 | | | | | | | |
| | 5 2 | | | | | | | |
| | 6 2 | | | | | | | |
| | 5 2 | | | | | | | |
| | 6 2 | | | | | | | |
| | 5 2 | | | | | | | |
| 62.2.28 | 100624 | 160 | O4S1Sr1 | | | | | |
| | -GM | -R | -S | -T | -U | -X | -Y | -Z |
| | 5 2 | 3+3 4 | 3+3 4 | 3+4 4 | 5+5 4 | 3+4 4 | 3+4 4 | 3+4 4 |
| | 6 2 | 4+4 4 | 4+4 4 | 3+4 4 | 6+6 4 | 3+4 4 | 3+4 4 | 3+4 4 |
| | 6 2 | 3+3 4 | 3+3 4 | 3+4 4 | 6+6 4 | 3+4 4 | 3+4 4 | 3+4 4 |
| | 5 2 | 4+4 4 | 4+4 4 | 3+4 4 | 5+5 4 | 3+4 4 | 3+4 4 | 3+4 4 |
| | 5 2 | 3+3 4 | 4+4 4 | 3+4 4 | 5+5 4 | 3+4 4 | 3+4 4 | 3+4 4 |
| | 6 2 | 4+4 4 | 3+3 4 | 3+4 4 | 6+6 4 | 3+4 4 | 3+4 4 | 3+4 4 |
| | 5 2 | 4+4 4 | 3+3 4 | 3+4 4 | 5+5 4 | 3+4 4 | 3+4 4 | 3+4 4 |
| | 6 2 | 3+3 4 | 4+4 4 | 3+4 4 | 6+6 4 | 3+4 4 | 3+4 4 | 3+4 4 |
| | 5 2 | 4+4 4 | 3+3 4 | 3+4 4 | 6+6 4 | 3+4 4 | 3+4 4 | 3+4 4 |
| | 5 2 | 3+3 4 | 4+4 4 | 3+4 4 | 5+5 4 | 3+4 4 | 3+4 4 | 3+4 4 |
| | 6 2 | 3+3 4 | 3+3 4 | 3+4 4 | 5+5 4 | 3+4 4 | 3+4 4 | 3+4 4 |
| | 6 2 | 4+4 4 | 4+4 4 | 3+4 4 | 6+6 4 | 3+4 4 | 3+4 4 | 3+4 4 |
| | 6 2 | 3+3 4 | 4+4 4 | 3+4 4 | 6+6 4 | 3+4 4 | 3+4 4 | 3+4 4 |
| | 5 2 | 4+4 4 | 3+3 4 | 3+4 4 | 5+5 4 | 3+4 4 | 3+4 4 | 3+4 4 |
| | 5 2 | 3+3 4 | 3+3 4 | 3+4 4 | 5+5 4 | 3+4 4 | 3+4 4 | 3+4 4 |
| | 6 2 | 4+4 4 | 4+4 4 | 3+4 4 | 6+6 4 | 3+4 4 | 3+4 4 | 3+4 4 |
| | 5 2 | 4+4 4 | 4+4 4 | 3+4 4 | 6+6 4 | 3+4 4 | 3+4 4 | 3+4 4 |
| | 6 2 | 3+3 4 | 3+3 4 | 3+4 4 | 5+5 4 | 3+4 4 | 3+4 4 | 3+4 4 |
| | 5 2 | 4+4 4 | 3+3 4 | 3+4 4 | 5+5 4 | 3+4 4 | 3+4 4 | 3+4 4 |
| | 6 2 | 3+3 4 | 4+4 4 | 3+4 4 | 6+6 4 | 3+4 4 | 3+4 4 | 3+4 4 |
| | 6 2 | 3+3 4 | 3+3 4 | 3+4 4 | 5+5 4 | 3+4 4 | 3+4 4 | 3+4 4 |
| | 6 2 | 4+4 4 | 4+4 4 | 3+4 4 | 6+6 4 | 3+4 4 | 3+4 4 | 3+4 4 |
| | 5 2 | 3+3 4 | 3+3 4 | 3+4 4 | 6+6 4 | 3+4 4 | 3+4 4 | 3+4 4 |

Continued on next column

**TABLE XXXVII– continued**

| TQC | ICSD NoE SF |
|---|---|
| | 5 2 4+4 4 4+4 4 3+4 4 5+5 4 3+4 4 3+4 4 3+4 4 |
| | 5 2 4+4 4 4+4 4 3+4 4 5+5 4 3+4 4 3+4 4 3+4 4 |
| | 6 2 3+3 4 3+3 4 3+4 4 5+5 4 3+4 4 3+4 4 3+4 4 |
| | 6 2 3+3 4 3+3 4 3+4 4 6+6 4 3+4 4 3+4 4 3+4 4 |
| | 5 2 4+4 4 4+4 4 3+4 4 6+6 4 3+4 4 3+4 4 3+4 4 |
| | 5 2 3+3 4 3+3 4 3+4 4 5+5 4 3+4 4 3+4 4 3+4 4 |
| | 6 2 4+4 4 4+4 4 3+4 4 6+6 4 3+4 4 3+4 4 3+4 4 |
| | 5 2 3+3 4 4+4 4 3+4 4 5+5 4 3+4 4 3+4 4 3+4 4 |
| | 6 2 4+4 4 3+3 4 3+4 4 6+6 4 3+4 4 3+4 4 3+4 4 |
| | 5 2 3+3 4 3+3 4 3+4 4 5+5 4 3+4 4 3+4 4 3+4 4 |
| | 6 2 4+4 4 4+4 4 3+4 4 5+5 4 3+4 4 3+4 4 3+4 4 |
| | 5 2 4+4 4 3+3 4 3+4 4 6+6 4 3+4 4 3+4 4 3+4 4 |
| | 6 2 3+3 4 4+4 4 3+4 4 6+6 4 3+4 4 3+4 4 3+4 4 |
| | 6 2 4+4 4 4+4 4 3+4 4 6+6 4 3+4 4 3+4 4 3+4 4 |
| | 6 2 3+3 4 3+3 4 3+4 4 6+6 4 3+4 4 3+4 4 3+4 4 |
| | 5 2 3+3 4 3+3 4 3+4 4 5+5 4 3+4 4 3+4 4 3+4 4 |
| | 5 2 4+4 4 4+4 4 3+4 4 5+5 4 3+4 4 3+4 4 3+4 4 |
| | 5 2 |
| | 5 2 |
| | 6 2 |
| | 6 2 |
| | 5 2 |
| | 6 2 |
| | 5 2 |
| | 5 2 |
| | 5 2 |
| | 6 2 |
| | 5 2 |
| | 6 2 |
| | 5 2 |
| | 6 2 |
| | 6 2 |
| | 6 2 |
| | 5 2 |
| | 6 2 |
| | 5 2 |
| | 5 2 |
| | 6 2 |
| | 6 2 |
| | 5 2 |
| | 6 2 |
| | 6 2 |
| | 5 2 |
| | 5 2 |
| | 5 2 |
| | 6 2 |
| | 6 2 |
| | 5 2 |
| | 6 2 |
| | 5 2 |
| | 6 2 |
| | 5 2 |
| | 6 2 |
| | 5 2 |
| | 6 2 |
| | 5 2 |
| | 5 2 |
| 62.2.29 | 415748 200 O4Rb2Ru1 |
| | -GM -R -S -T -U -X -Y -Z |
| | 5 2 3+3 4 3+3 4 3+4 4 6+6 4 3+4 4 3+4 4 3+4 4 |
| | 5 2 4+4 4 4+4 4 3+4 4 5+5 4 3+4 4 3+4 4 3+4 4 |

Continued on next column

**TABLE XXXVII– continued**

| TQC | ICSD NoE SF |
|---|---|
| | 6 2 3+3 4 3+3 4 3+4 4 5+5 4 3+4 4 3+4 4 3+4 4 |
| | 6 2 4+4 4 4+4 4 3+4 4 6+6 4 3+4 4 3+4 4 3+4 4 |
| | 5 2 3+3 4 3+3 4 3+4 4 5+5 4 3+4 4 3+4 4 3+4 4 |
| | 6 2 4+4 4 4+4 4 3+4 4 6+6 4 3+4 4 3+4 4 3+4 4 |
| | 5 2 4+4 4 4+4 4 3+4 4 5+5 4 3+4 4 3+4 4 3+4 4 |
| | 6 2 3+3 4 3+3 4 3+4 4 6+6 4 3+4 4 3+4 4 3+4 4 |
| | 5 2 3+3 4 4+4 4 3+4 4 5+5 4 3+4 4 3+4 4 3+4 4 |
| | 5 2 4+4 4 3+3 4 3+4 4 6+6 4 3+4 4 3+4 4 3+4 4 |
| | 6 2 3+3 4 3+3 4 3+4 4 5+5 4 3+4 4 3+4 4 3+4 4 |
| | 6 2 4+4 4 4+4 4 3+4 4 6+6 4 3+4 4 3+4 4 3+4 4 |
| | 6 2 3+3 4 3+3 4 3+4 4 6+6 4 3+4 4 3+4 4 3+4 4 |
| | 5 2 4+4 4 4+4 4 3+4 4 5+5 4 3+4 4 3+4 4 3+4 4 |
| | 5 2 4+4 4 3+3 4 3+4 4 5+5 4 3+4 4 3+4 4 3+4 4 |
| | 6 2 3+3 4 4+4 4 3+4 4 6+6 4 3+4 4 3+4 4 3+4 4 |
| | 5 2 3+3 4 4+4 4 3+4 4 6+6 4 3+4 4 3+4 4 3+4 4 |
| | 6 2 4+4 4 3+3 4 3+4 4 6+6 4 3+4 4 3+4 4 3+4 4 |
| | 6 2 4+4 4 4+4 4 3+4 4 5+5 4 3+4 4 3+4 4 3+4 4 |
| | 5 2 3+3 4 3+3 4 3+4 4 5+5 4 3+4 4 3+4 4 3+4 4 |
| | 5 2 3+3 4 3+3 4 3+4 4 5+5 4 3+4 4 3+4 4 3+4 4 |
| | 6 2 4+4 4 4+4 4 3+4 4 6+6 4 3+4 4 3+4 4 3+4 4 |
| | 6 2 4+4 4 4+4 4 3+4 4 5+5 4 3+4 4 3+4 4 3+4 4 |
| | 5 2 3+3 4 3+3 4 3+4 4 6+6 4 3+4 4 3+4 4 3+4 4 |
| | 5 2 3+3 4 4+4 4 3+4 4 5+5 4 3+4 4 3+4 4 3+4 4 |
| | 5 2 4+4 4 3+3 4 3+4 4 6+6 4 3+4 4 3+4 4 3+4 4 |
| | 6 2 4+4 4 3+3 4 3+4 4 5+5 4 3+4 4 3+4 4 3+4 4 |
| | 6 2 3+3 4 4+4 4 3+4 4 6+6 4 3+4 4 3+4 4 3+4 4 |
| | 5 2 3+3 4 4+4 4 3+4 4 5+5 4 3+4 4 3+4 4 3+4 4 |
| | 6 2 4+4 4 3+3 4 3+4 4 6+6 4 3+4 4 3+4 4 3+4 4 |
| | 5 2 4+4 4 3+3 4 3+4 4 5+5 4 3+4 4 3+4 4 3+4 4 |
| | 6 2 3+3 4 4+4 4 3+4 4 6+6 4 3+4 4 3+4 4 3+4 4 |
| | 5 2 3+3 4 3+3 4 3+4 4 5+5 4 3+4 4 3+4 4 3+4 4 |
| | 5 2 4+4 4 4+4 4 3+4 4 6+6 4 3+4 4 3+4 4 3+4 4 |
| | 5 2 3+3 4 3+3 4 3+4 4 6+6 4 3+4 4 3+4 4 3+4 4 |
| | 6 2 4+4 4 4+4 4 3+4 4 5+5 4 3+4 4 3+4 4 3+4 4 |
| | 6 2 4+4 4 3+3 4 3+4 4 5+5 4 3+4 4 3+4 4 3+4 4 |
| | 5 2 3+3 4 4+4 4 3+4 4 5+5 4 3+4 4 3+4 4 3+4 4 |
| | 6 2 3+3 4 4+4 4 3+4 4 6+6 4 3+4 4 3+4 4 3+4 4 |
| | 6 2 4+4 4 3+3 4 3+4 4 6+6 4 3+4 4 3+4 4 3+4 4 |
| | 5 2 3+3 4 3+3 4 3+4 4 5+5 4 3+4 4 3+4 4 3+4 4 |
| | 6 2 4+4 4 4+4 4 3+4 4 6+6 4 3+4 4 3+4 4 3+4 4 |
| | 5 2 3+3 4 4+4 4 3+4 4 5+5 4 3+4 4 3+4 4 3+4 4 |
| | 6 2 4+4 4 3+3 4 3+4 4 5+5 4 3+4 4 3+4 4 3+4 4 |
| | 5 2 3+3 4 3+3 4 3+4 4 6+6 4 3+4 4 3+4 4 3+4 4 |
| | 6 2 4+4 4 4+4 4 3+4 4 6+6 4 3+4 4 3+4 4 3+4 4 |
| | 5 2 4+4 4 3+3 4 3+4 4 5+5 4 3+4 4 3+4 4 3+4 4 |
| | 6 2 3+3 4 4+4 4 3+4 4 6+6 4 3+4 4 3+4 4 3+4 4 |
| | 6 2 4+4 4 4+4 4 3+4 4 6+6 4 3+4 4 3+4 4 3+4 4 |
| | 5 2 3+3 4 3+3 4 3+4 4 5+5 4 3+4 4 3+4 4 3+4 4 |
| | 5 2 |
| | 6 2 |
| | 5 2 |
| | 5 2 |
| | 6 2 |
| | 5 2 |
| | 6 2 |
| | 5 2 |
| | 6 2 |
| | 6 2 |
| | 5 2 |
| | 6 2 |
| | 5 2 |

Continued on next column

**TABLE XXXVII– continued**

| TQC | ICSD | NoE | SF | | | | | |
|---|---|---|---|---|---|---|---|---|
| | 6 2 | | | | | | | |
| | 6 2 | | | | | | | |
| | 5 2 | | | | | | | |
| | 6 2 | | | | | | | |
| | 5 2 | | | | | | | |
| | 5 2 | | | | | | | |
| | 6 2 | | | | | | | |
| | 6 2 | | | | | | | |
| | 5 2 | | | | | | | |
| | 6 2 | | | | | | | |
| | 5 2 | | | | | | | |
| | 5 2 | | | | | | | |
| | 5 2 | | | | | | | |
| | 5 2 | | | | | | | |
| | 6 2 | | | | | | | |
| | 5 2 | | | | | | | |
| | 6 2 | | | | | | | |
| | 5 2 | | | | | | | |
| | 6 2 | | | | | | | |
| | 6 2 | | | | | | | |
| | 6 2 | | | | | | | |
| | 6 2 | | | | | | | |
| | 6 2 | | | | | | | |
| | 5 2 | | | | | | | |
| | 5 2 | | | | | | | |
| | 6 2 | | | | | | | |
| | 6 2 | | | | | | | |
| | 5 2 | | | | | | | |
| | 5 2 | | | | | | | |
| | 5 2 | | | | | | | |
| | 6 2 | | | | | | | |
| | 6 2 | | | | | | | |
| | 5 2 | | | | | | | |
| | 6 2 | | | | | | | |
| | 5 2 | | | | | | | |
| | 5 2 | | | | | | | |
| | 5 2 | | | | | | | |
| 62.2.30 | 100625 | 136 | O4Pb1S1 | | | | | |
| | -GM | -R | -S | -T | -U | -X | -Y | -Z |
| | 5 2 | 3+3 4 | 3+3 4 | 3+4 4 | 6+6 4 | 3+4 4 | 3+4 4 | 3+4 4 |
| | 5 2 | 4+4 4 | 4+4 4 | 3+4 4 | 5+5 4 | 3+4 4 | 3+4 4 | 3+4 4 |
| | 6 2 | 3+3 4 | 3+3 4 | 3+4 4 | 5+5 4 | 3+4 4 | 3+4 4 | 3+4 4 |
| | 6 2 | 4+4 4 | 4+4 4 | 3+4 4 | 6+6 4 | 3+4 4 | 3+4 4 | 3+4 4 |
| | 5 2 | 3+3 4 | 3+3 4 | 3+4 4 | 5+5 4 | 3+4 4 | 3+4 4 | 3+4 4 |
| | 5 2 | 4+4 4 | 4+4 4 | 3+4 4 | 6+6 4 | 3+4 4 | 3+4 4 | 3+4 4 |
| | 6 2 | 3+3 4 | 3+3 4 | 3+4 4 | 5+5 4 | 3+4 4 | 3+4 4 | 3+4 4 |
| | 6 2 | 4+4 4 | 4+4 4 | 3+4 4 | 6+6 4 | 3+4 4 | 3+4 4 | 3+4 4 |
| | 6 2 | 4+4 4 | 3+3 4 | 3+4 4 | 6+6 4 | 3+4 4 | 3+4 4 | 3+4 4 |
| | 6 2 | 3+3 4 | 4+4 4 | 3+4 4 | 5+5 4 | 3+4 4 | 3+4 4 | 3+4 4 |
| | 5 2 | 3+3 4 | 3+3 4 | 3+4 4 | 6+6 4 | 3+4 4 | 3+4 4 | 3+4 4 |
| | 5 2 | 4+4 4 | 4+4 4 | 3+4 4 | 5+5 4 | 3+4 4 | 3+4 4 | 3+4 4 |
| | 6 2 | 4+4 4 | 3+3 4 | 3+4 4 | 5+5 4 | 3+4 4 | 3+4 4 | 3+4 4 |
| | 6 2 | 3+3 4 | 4+4 4 | 3+4 4 | 6+6 4 | 3+4 4 | 3+4 4 | 3+4 4 |
| | 5 2 | 4+4 4 | 3+3 4 | 3+4 4 | 6+6 4 | 3+4 4 | 3+4 4 | 3+4 4 |
| | 5 2 | 3+3 4 | 4+4 4 | 3+4 4 | 5+5 4 | 3+4 4 | 3+4 4 | 3+4 4 |
| | 5 2 | 4+4 4 | 4+4 4 | 3+4 4 | 5+5 4 | 3+4 4 | 3+4 4 | 3+4 4 |
| | 5 2 | 3+3 4 | 3+3 4 | 3+4 4 | 6+6 4 | 3+4 4 | 3+4 4 | 3+4 4 |
| | 6 2 | 3+3 4 | 4+4 4 | 3+4 4 | 5+5 4 | 3+4 4 | 3+4 4 | 3+4 4 |
| | 6 2 | 4+4 4 | 3+3 4 | 3+4 4 | 6+6 4 | 3+4 4 | 3+4 4 | 3+4 4 |
| | 5 2 | 4+4 4 | 3+3 4 | 3+4 4 | 6+6 4 | 3+4 4 | 3+4 4 | 3+4 4 |
| | 6 2 | 3+3 4 | 4+4 4 | 3+4 4 | 5+5 4 | 3+4 4 | 3+4 4 | 3+4 4 |
| | 5 2 | 3+3 4 | 3+3 4 | 3+4 4 | 6+6 4 | 3+4 4 | 3+4 4 | 3+4 4 |
| | 6 2 | 4+4 4 | 4+4 4 | 3+4 4 | 5+5 4 | 3+4 4 | 3+4 4 | 3+4 4 |
| | 6 2 | 3+3 4 | 3+3 4 | 3+4 4 | 5+5 4 | 3+4 4 | 3+4 4 | 3+4 4 |
| | 5 2 | 4+4 4 | 4+4 4 | 3+4 4 | 6+6 4 | 3+4 4 | 3+4 4 | 3+4 4 |
| | 6 2 | 3+3 4 | 4+4 4 | 3+4 4 | 5+5 4 | 3+4 4 | 3+4 4 | 3+4 4 |
| | 5 2 | 4+4 4 | 3+3 4 | 3+4 4 | 6+6 4 | 3+4 4 | 3+4 4 | 3+4 4 |
| | 5 2 | 4+4 4 | 3+3 4 | 3+4 4 | 6+6 4 | 3+4 4 | 3+4 4 | 3+4 4 |
| | 5 2 | 3+3 4 | 4+4 4 | 3+4 4 | 5+5 4 | 3+4 4 | 3+4 4 | 3+4 4 |
| | 6 2 | 3+3 4 | 3+3 4 | 3+4 4 | 6+6 4 | 3+4 4 | 3+4 4 | 3+4 4 |
| | 5 2 | 4+4 4 | 4+4 4 | 3+4 4 | 5+5 4 | 3+4 4 | 3+4 4 | 3+4 4 |
| | 6 2 | 4+4 4 | 3+3 4 | 3+4 4 | 5+5 4 | 3+4 4 | 3+4 4 | 3+4 4 |
| | 6 2 | 3+3 4 | 4+4 4 | 3+4 4 | 6+6 4 | 3+4 4 | 3+4 4 | 3+4 4 |
| | 6 2 | | | | | | | |
| | 5 2 | | | | | | | |
| | 6 2 | | | | | | | |
| | 5 2 | | | | | | | |
| | 5 2 | | | | | | | |
| | 6 2 | | | | | | | |
| | 5 2 | | | | | | | |
| | 5 2 | | | | | | | |
| | 5 2 | | | | | | | |
| | 5 2 | | | | | | | |
| | 6 2 | | | | | | | |
| | 6 2 | | | | | | | |
| | 6 2 | | | | | | | |
| | 6 2 | | | | | | | |
| | 6 2 | | | | | | | |
| | 5 2 | | | | | | | |
| | 6 2 | | | | | | | |
| | 5 2 | | | | | | | |
| | 5 2 | | | | | | | |
| | 6 2 | | | | | | | |
| | 6 2 | | | | | | | |
| | 5 2 | | | | | | | |
| | 5 2 | | | | | | | |
| | 6 2 | | | | | | | |
| | 5 2 | | | | | | | |
| | 5 2 | | | | | | | |
| | 5 2 | | | | | | | |
| | 6 2 | | | | | | | |
| | 6 2 | | | | | | | |
| | 6 2 | | | | | | | |
| | 5 2 | | | | | | | |
| | 5 2 | | | | | | | |
| | 5 2 | | | | | | | |
| | 5 2 | | | | | | | |
| 62.2.31 | 185833 | 144 | O3Pr1V1 | | | | | |
| | -GM | -R | -S | -T | -U | -X | -Y | -Z |
| | 5 2 | 3+3 4 | 3+3 4 | 3+4 4 | 5+5 4 | 3+4 4 | 3+4 4 | 3+4 4 |
| | 6 2 | 4+4 4 | 4+4 4 | 3+4 4 | 6+6 4 | 3+4 4 | 3+4 4 | 3+4 4 |
| | 5 2 | 4+4 4 | 3+3 4 | 3+4 4 | 6+6 4 | 3+4 4 | 3+4 4 | 3+4 4 |
| | 6 2 | 3+3 4 | 4+4 4 | 3+4 4 | 5+5 4 | 3+4 4 | 3+4 4 | 3+4 4 |
| | 5 2 | 3+3 4 | 4+4 4 | 3+4 4 | 5+5 4 | 3+4 4 | 3+4 4 | 3+4 4 |
| | 6 2 | 4+4 4 | 3+3 4 | 3+4 4 | 6+6 4 | 3+4 4 | 3+4 4 | 3+4 4 |
| | 5 2 | 4+4 4 | 3+3 4 | 3+4 4 | 6+6 4 | 3+4 4 | 3+4 4 | 3+4 4 |
| | 6 2 | 3+3 4 | 4+4 4 | 3+4 4 | 5+5 4 | 3+4 4 | 3+4 4 | 3+4 4 |
| | 5 2 | 4+4 4 | 3+3 4 | 3+4 4 | 6+6 4 | 3+4 4 | 3+4 4 | 3+4 4 |
| | 6 2 | 3+3 4 | 4+4 4 | 3+4 4 | 5+5 4 | 3+4 4 | 3+4 4 | 3+4 4 |
| | 5 2 | 3+3 4 | 4+4 4 | 3+4 4 | 6+6 4 | 3+4 4 | 3+4 4 | 3+4 4 |
| | 5 2 | 4+4 4 | 3+3 4 | 3+4 4 | 5+5 4 | 3+4 4 | 3+4 4 | 3+4 4 |
| | 5 2 | 3+3 4 | 3+3 4 | 3+4 4 | 5+5 4 | 3+4 4 | 3+4 4 | 3+4 4 |

Continued on next column

**TABLE XXXVII– continued**

| TQC | ICSD NoE SF |
|---|---|
| | 6 2 4+4 4 4+4 4 3+4 4 6+6 4 3+4 4 3+4 4 3+4 4 |
| | 6 2 3+3 4 4+4 4 3+4 4 6+6 4 3+4 4 3+4 4 3+4 4 |
| | 6 2 4+4 4 3+3 4 3+4 4 6+6 4 3+4 4 3+4 4 3+4 4 |
| | 5 2 4+4 4 3+3 4 3+4 4 6+6 4 3+4 4 3+4 4 3+4 4 |
| | 6 2 3+3 4 4+4 4 3+4 4 6+6 4 3+4 4 3+4 4 3+4 4 |
| | 6 2 4+4 4 3+3 4 3+4 4 5+5 4 3+4 4 3+4 4 3+4 4 |
| | 5 2 3+3 4 4+4 4 3+4 4 5+5 4 3+4 4 3+4 4 3+4 4 |
| | 5 2 3+3 4 3+3 4 3+4 4 5+5 4 3+4 4 3+4 4 3+4 4 |
| | 6 2 4+4 4 4+4 4 3+4 4 6+6 4 3+4 4 3+4 4 3+4 4 |
| | 5 2 3+3 4 4+4 4 3+4 4 6+6 4 3+4 4 3+4 4 3+4 4 |
| | 6 2 4+4 4 3+3 4 3+4 4 5+5 4 3+4 4 3+4 4 3+4 4 |
| | 5 2 4+4 4 4+4 4 3+4 4 6+6 4 3+4 4 3+4 4 3+4 4 |
| | 6 2 3+3 4 3+3 4 3+4 4 5+5 4 3+4 4 3+4 4 3+4 4 |
| | 5 2 4+4 4 4+4 4 3+4 4 6+6 4 3+4 4 3+4 4 3+4 4 |
| | 6 2 3+3 4 3+3 4 3+4 4 5+5 4 3+4 4 3+4 4 3+4 4 |
| | 5 2 3+3 4 3+3 4 3+4 4 5+5 4 3+4 4 3+4 4 3+4 4 |
| | 5 2 4+4 4 4+4 4 3+4 4 5+5 4 3+4 4 3+4 4 3+4 4 |
| | 5 2 3+3 4 4+4 4 3+4 4 5+5 4 3+4 4 3+4 4 3+4 4 |
| | 5 2 4+4 4 3+3 4 3+4 4 6+6 4 3+4 4 3+4 4 3+4 4 |
| | 5 2 4+4 4 4+4 4 3+4 4 6+6 4 3+4 4 3+4 4 3+4 4 |
| | 5 2 3+3 4 3+3 4 3+4 4 6+6 4 3+4 4 3+4 4 3+4 4 |
| | 5 2 4+4 4 4+4 4 3+4 4 6+6 4 3+4 4 3+4 4 3+4 4 |
| | 6 2 3+3 4 3+3 4 3+4 4 5+5 4 3+4 4 3+4 4 3+4 4 |
| | 5 2 |
| | 6 2 |
| | 6 2 |
| | 5 2 |
| | 5 2 |
| | 6 2 |
| | 6 2 |
| | 5 2 |
| | 5 2 |
| | 6 2 |
| | 6 2 |
| | 5 2 |
| | 5 2 |
| | 6 2 |
| | 6 2 |
| | 6 2 |
| | 6 2 |
| | 6 2 |
| | 5 2 |
| | 5 2 |
| | 6 2 |
| | 5 2 |
| | 6 2 |
| | 6 2 |
| | 5 2 |
| | 6 2 |
| | 6 2 |
| | 6 2 |
| | 5 2 |
| | 5 2 |
| | 5 2 |
| | 5 2 |
| | 5 2 |
| | 5 2 |
| | 6 2 |
| | 5 2 |
| 62.2.32 | 172354 160 O3Pr1Rh1 |
| | -GM -R -S -T -U -X -Y -Z |

Continued on next column

**TABLE XXXVII– continued**

| TQC | ICSD NoE SF |
|---|---|
| | 5 2 3+3 4 3+3 4 3+4 4 5+5 4 3+4 4 3+4 4 3+4 4 |
| | 6 2 4+4 4 4+4 4 3+4 4 6+6 4 3+4 4 3+4 4 3+4 4 |
| | 5 2 4+4 4 3+3 4 3+4 4 6+6 4 3+4 4 3+4 4 3+4 4 |
| | 6 2 3+3 4 4+4 4 3+4 4 5+5 4 3+4 4 3+4 4 3+4 4 |
| | 5 2 3+3 4 3+3 4 3+4 4 5+5 4 3+4 4 3+4 4 3+4 4 |
| | 6 2 4+4 4 4+4 4 3+4 4 6+6 4 3+4 4 3+4 4 3+4 4 |
| | 5 2 4+4 4 4+4 4 3+4 4 6+6 4 3+4 4 3+4 4 3+4 4 |
| | 6 2 3+3 4 3+3 4 3+4 4 5+5 4 3+4 4 3+4 4 3+4 4 |
| | 5 2 4+4 4 4+4 4 3+4 4 6+6 4 3+4 4 3+4 4 3+4 4 |
| | 6 2 3+3 4 3+3 4 3+4 4 5+5 4 3+4 4 3+4 4 3+4 4 |
| | 5 2 3+3 4 3+3 4 3+4 4 6+6 4 3+4 4 3+4 4 3+4 4 |
| | 5 2 4+4 4 4+4 4 3+4 4 5+5 4 3+4 4 3+4 4 3+4 4 |
| | 5 2 3+3 4 4+4 4 3+4 4 5+5 4 3+4 4 3+4 4 3+4 4 |
| | 6 2 4+4 4 3+3 4 3+4 4 6+6 4 3+4 4 3+4 4 3+4 4 |
| | 5 2 3+3 4 4+4 4 3+4 4 6+6 4 3+4 4 3+4 4 3+4 4 |
| | 6 2 4+4 4 3+3 4 3+4 4 6+6 4 3+4 4 3+4 4 3+4 4 |
| | 6 2 4+4 4 3+3 4 3+4 4 6+6 4 3+4 4 3+4 4 3+4 4 |
| | 6 2 3+3 4 4+4 4 3+4 4 6+6 4 3+4 4 3+4 4 3+4 4 |
| | 6 2 4+4 4 4+4 4 3+4 4 6+6 4 3+4 4 3+4 4 3+4 4 |
| | 5 2 3+3 4 3+3 4 3+4 4 5+5 4 3+4 4 3+4 4 3+4 4 |
| | 5 2 3+3 4 4+4 4 3+4 4 5+5 4 3+4 4 3+4 4 3+4 4 |
| | 6 2 4+4 4 3+3 4 3+4 4 6+6 4 3+4 4 3+4 4 3+4 4 |
| | 5 2 3+3 4 3+3 4 3+4 4 5+5 4 3+4 4 3+4 4 3+4 4 |
| | 6 2 4+4 4 4+4 4 3+4 4 6+6 4 3+4 4 3+4 4 3+4 4 |
| | 6 2 4+4 4 3+3 4 3+4 4 6+6 4 3+4 4 3+4 4 3+4 4 |
| | 5 2 3+3 4 4+4 4 3+4 4 5+5 4 3+4 4 3+4 4 3+4 4 |
| | 5 2 4+4 4 3+3 4 3+4 4 5+5 4 3+4 4 3+4 4 3+4 4 |
| | 6 2 3+3 4 4+4 4 3+4 4 5+5 4 3+4 4 3+4 4 3+4 4 |
| | 5 2 4+4 4 4+4 4 3+4 4 5+5 4 3+4 4 3+4 4 3+4 4 |
| | 5 2 3+3 4 3+3 4 3+4 4 6+6 4 3+4 4 3+4 4 3+4 4 |
| | 5 2 3+3 4 3+3 4 3+4 4 5+5 4 3+4 4 3+4 4 3+4 4 |
| | 5 2 4+4 4 4+4 4 3+4 4 5+5 4 3+4 4 3+4 4 3+4 4 |
| | 5 2 4+4 4 3+3 4 3+4 4 6+6 4 3+4 4 3+4 4 3+4 4 |
| | 5 2 3+3 4 4+4 4 3+4 4 6+6 4 3+4 4 3+4 4 3+4 4 |
| | 5 2 4+4 4 4+4 4 3+4 4 6+6 4 3+4 4 3+4 4 3+4 4 |
| | 5 2 3+3 4 3+3 4 3+4 4 6+6 4 3+4 4 3+4 4 3+4 4 |
| | 6 2 3+3 4 4+4 4 3+4 4 6+6 4 3+4 4 3+4 4 3+4 4 |
| | 5 2 4+4 4 3+3 4 3+4 4 6+6 4 3+4 4 3+4 4 3+4 4 |
| | 5 2 4+4 4 4+4 4 3+4 4 6+6 4 3+4 4 3+4 4 3+4 4 |
| | 5 2 3+3 4 3+3 4 3+4 4 5+5 4 3+4 4 3+4 4 3+4 4 |
| | 6 2 |
| | 5 2 |
| | 6 2 |
| | 5 2 |
| | 5 2 |
| | 6 2 |
| | 6 2 |
| | 6 2 |
| | 6 2 |
| | 6 2 |
| | 6 2 |
| | 5 2 |
| | 6 2 |
| | 5 2 |
| | 5 2 |
| | 6 2 |
| | 6 2 |
| | 6 2 |
| | 6 2 |
| | 5 2 |
| | 6 2 |

Continued on next column

**TABLE XXXVII– continued**

| TQC | ICSD | NoE | SF | | | | | |
|---|---|---|---|---|---|---|---|---|
| | 6 2 | | | | | | | |
| | 6 2 | | | | | | | |
| | 5 2 | | | | | | | |
| | 6 2 | | | | | | | |
| | 5 2 | | | | | | | |
| | 5 2 | | | | | | | |
| | 5 2 | | | | | | | |
| | 5 2 | | | | | | | |
| | 5 2 | | | | | | | |
| | 5 2 | | | | | | | |
| | 5 2 | | | | | | | |
| | 5 2 | | | | | | | |
| | 5 2 | | | | | | | |
| | 5 2 | | | | | | | |
| | 5 2 | | | | | | | |
| | 5 2 | | | | | | | |
| | 5 2 | | | | | | | |
| | 5 2 | | | | | | | |
| | 6 2 | | | | | | | |
| 62.2.33 | 172348 | 160 | O3Pr1Rh1 | | | | | |
| | -GM | -R | -S | -T | -U | -X | -Y | -Z |
| | 5 2 | 3+3 4 | 3+3 4 | 3+4 4 | 5+5 4 | 3+4 4 | 3+4 4 | 3+4 4 |
| | 6 2 | 4+4 4 | 4+4 4 | 3+4 4 | 6+6 4 | 3+4 4 | 3+4 4 | 3+4 4 |
| | 5 2 | 4+4 4 | 3+3 4 | 3+4 4 | 6+6 4 | 3+4 4 | 3+4 4 | 3+4 4 |
| | 6 2 | 3+3 4 | 4+4 4 | 3+4 4 | 5+5 4 | 3+4 4 | 3+4 4 | 3+4 4 |
| | 5 2 | 3+3 4 | 3+3 4 | 3+4 4 | 5+5 4 | 3+4 4 | 3+4 4 | 3+4 4 |
| | 6 2 | 4+4 4 | 4+4 4 | 3+4 4 | 6+6 4 | 3+4 4 | 3+4 4 | 3+4 4 |
| | 5 2 | 4+4 4 | 4+4 4 | 3+4 4 | 6+6 4 | 3+4 4 | 3+4 4 | 3+4 4 |
| | 6 2 | 3+3 4 | 3+3 4 | 3+4 4 | 5+5 4 | 3+4 4 | 3+4 4 | 3+4 4 |
| | 5 2 | 4+4 4 | 4+4 4 | 3+4 4 | 6+6 4 | 3+4 4 | 3+4 4 | 3+4 4 |
| | 6 2 | 3+3 4 | 3+3 4 | 3+4 4 | 5+5 4 | 3+4 4 | 3+4 4 | 3+4 4 |
| | 5 2 | 3+3 4 | 3+3 4 | 3+4 4 | 6+6 4 | 3+4 4 | 3+4 4 | 3+4 4 |
| | 5 2 | 4+4 4 | 4+4 4 | 3+4 4 | 5+5 4 | 3+4 4 | 3+4 4 | 3+4 4 |
| | 5 2 | 3+3 4 | 4+4 4 | 3+4 4 | 5+5 4 | 3+4 4 | 3+4 4 | 3+4 4 |
| | 6 2 | 4+4 4 | 3+3 4 | 3+4 4 | 6+6 4 | 3+4 4 | 3+4 4 | 3+4 4 |
| | 5 2 | 3+3 4 | 4+4 4 | 3+4 4 | 6+6 4 | 3+4 4 | 3+4 4 | 3+4 4 |
| | 6 2 | 4+4 4 | 3+3 4 | 3+4 4 | 6+6 4 | 3+4 4 | 3+4 4 | 3+4 4 |
| | 6 2 | 4+4 4 | 3+3 4 | 3+4 4 | 6+6 4 | 3+4 4 | 3+4 4 | 3+4 4 |
| | 6 2 | 3+3 4 | 4+4 4 | 3+4 4 | 6+6 4 | 3+4 4 | 3+4 4 | 3+4 4 |
| | 6 2 | 4+4 4 | 4+4 4 | 3+4 4 | 6+6 4 | 3+4 4 | 3+4 4 | 3+4 4 |
| | 5 2 | 3+3 4 | 3+3 4 | 3+4 4 | 5+5 4 | 3+4 4 | 3+4 4 | 3+4 4 |
| | 5 2 | 3+3 4 | 4+4 4 | 3+4 4 | 5+5 4 | 3+4 4 | 3+4 4 | 3+4 4 |
| | 6 2 | 4+4 4 | 3+3 4 | 3+4 4 | 6+6 4 | 3+4 4 | 3+4 4 | 3+4 4 |
| | 5 2 | 3+3 4 | 3+3 4 | 3+4 4 | 5+5 4 | 3+4 4 | 3+4 4 | 3+4 4 |
| | 6 2 | 4+4 4 | 4+4 4 | 3+4 4 | 6+6 4 | 3+4 4 | 3+4 4 | 3+4 4 |
| | 6 2 | 4+4 4 | 3+3 4 | 3+4 4 | 6+6 4 | 3+4 4 | 3+4 4 | 3+4 4 |
| | 5 2 | 3+3 4 | 4+4 4 | 3+4 4 | 5+5 4 | 3+4 4 | 3+4 4 | 3+4 4 |
| | 5 2 | 4+4 4 | 3+3 4 | 3+4 4 | 5+5 4 | 3+4 4 | 3+4 4 | 3+4 4 |
| | 6 2 | 3+3 4 | 4+4 4 | 3+4 4 | 5+5 4 | 3+4 4 | 3+4 4 | 3+4 4 |
| | 5 2 | 4+4 4 | 4+4 4 | 3+4 4 | 5+5 4 | 3+4 4 | 3+4 4 | 3+4 4 |
| | 5 2 | 3+3 4 | 3+3 4 | 3+4 4 | 6+6 4 | 3+4 4 | 3+4 4 | 3+4 4 |
| | 5 2 | 3+3 4 | 3+3 4 | 3+4 4 | 5+5 4 | 3+4 4 | 3+4 4 | 3+4 4 |
| | 5 2 | 4+4 4 | 4+4 4 | 3+4 4 | 5+5 4 | 3+4 4 | 3+4 4 | 3+4 4 |
| | 5 2 | 4+4 4 | 3+3 4 | 3+4 4 | 6+6 4 | 3+4 4 | 3+4 4 | 3+4 4 |
| | 5 2 | 3+3 4 | 4+4 4 | 3+4 4 | 6+6 4 | 3+4 4 | 3+4 4 | 3+4 4 |
| | 5 2 | 4+4 4 | 4+4 4 | 3+4 4 | 6+6 4 | 3+4 4 | 3+4 4 | 3+4 4 |
| | 5 2 | 3+3 4 | 3+3 4 | 3+4 4 | 6+6 4 | 3+4 4 | 3+4 4 | 3+4 4 |
| | 6 2 | 3+3 4 | 4+4 4 | 3+4 4 | 6+6 4 | 3+4 4 | 3+4 4 | 3+4 4 |
| | 5 2 | 4+4 4 | 3+3 4 | 3+4 4 | 6+6 4 | 3+4 4 | 3+4 4 | 3+4 4 |
| | 5 2 | 4+4 4 | 4+4 4 | 3+4 4 | 6+6 4 | 3+4 4 | 3+4 4 | 3+4 4 |
| | 5 2 | 3+3 4 | 3+3 4 | 3+4 4 | 5+5 4 | 3+4 4 | 3+4 4 | 3+4 4 |

Continued on next column

| TQC | ICSD | NoE | SF | | | | | |
|---|---|---|---|---|---|---|---|---|
| | 6 2 | | | | | | | |
| | 5 2 | | | | | | | |
| | 6 2 | | | | | | | |
| | 5 2 | | | | | | | |
| | 5 2 | | | | | | | |
| | 6 2 | | | | | | | |
| | 6 2 | | | | | | | |
| | 6 2 | | | | | | | |
| | 6 2 | | | | | | | |
| | 6 2 | | | | | | | |
| | 6 2 | | | | | | | |
| | 5 2 | | | | | | | |
| | 6 2 | | | | | | | |
| | 5 2 | | | | | | | |
| | 5 2 | | | | | | | |
| | 6 2 | | | | | | | |
| | 6 2 | | | | | | | |
| | 6 2 | | | | | | | |
| | 6 2 | | | | | | | |
| | 5 2 | | | | | | | |
| | 6 2 | | | | | | | |
| | 6 2 | | | | | | | |
| | 6 2 | | | | | | | |
| | 5 2 | | | | | | | |
| | 6 2 | | | | | | | |
| | 5 2 | | | | | | | |
| | 5 2 | | | | | | | |
| | 5 2 | | | | | | | |
| | 5 2 | | | | | | | |
| | 5 2 | | | | | | | |
| | 5 2 | | | | | | | |
| | 5 2 | | | | | | | |
| | 5 2 | | | | | | | |
| | 5 2 | | | | | | | |
| | 5 2 | | | | | | | |
| | 5 2 | | | | | | | |
| | 5 2 | | | | | | | |
| | 5 2 | | | | | | | |
| | 5 2 | | | | | | | |
| | 6 2 | | | | | | | |
| 62.2.34 | 413386 | 328 | O1Si1Yb3 | | | | | |
| | -GM | -R | -S | -T | -U | -X | -Y | -Z |
| | 5 2 | 3+3 4 | 3+3 4 | 3+4 4 | 5+5 4 | 3+4 4 | 3+4 4 | 3+4 4 |
| | 6 2 | 4+4 4 | 4+4 4 | 3+4 4 | 6+6 4 | 3+4 4 | 3+4 4 | 3+4 4 |
| | 5 2 | 3+3 4 | 3+3 4 | 3+4 4 | 5+5 4 | 3+4 4 | 3+4 4 | 3+4 4 |
| | 6 2 | 3+3 4 | 4+4 4 | 3+4 4 | 5+5 4 | 3+4 4 | 3+4 4 | 3+4 4 |
| | 5 2 | 4+4 4 | 3+3 4 | 3+4 4 | 6+6 4 | 3+4 4 | 3+4 4 | 3+4 4 |
| | 5 2 | 4+4 4 | 4+4 4 | 3+4 4 | 6+6 4 | 3+4 4 | 3+4 4 | 3+4 4 |
| | 6 2 | 3+3 4 | 3+3 4 | 3+4 4 | 6+6 4 | 3+4 4 | 3+4 4 | 3+4 4 |
| | 6 2 | 4+4 4 | 4+4 4 | 3+4 4 | 5+5 4 | 3+4 4 | 3+4 4 | 3+4 4 |
| | 6 2 | 4+4 4 | 3+3 4 | 3+4 4 | 6+6 4 | 3+4 4 | 3+4 4 | 3+4 4 |
| | 6 2 | 3+3 4 | 4+4 4 | 3+4 4 | 6+6 4 | 3+4 4 | 3+4 4 | 3+4 4 |
| | 5 2 | 3+3 4 | 3+3 4 | 3+4 4 | 5+5 4 | 3+4 4 | 3+4 4 | 3+4 4 |
| | 5 2 | 4+4 4 | 4+4 4 | 3+4 4 | 5+5 4 | 3+4 4 | 3+4 4 | 3+4 4 |
| | 5 2 | 3+3 4 | 4+4 4 | 3+4 4 | 6+6 4 | 3+4 4 | 3+4 4 | 3+4 4 |
| | 6 2 | 4+4 4 | 3+3 4 | 3+4 4 | 5+5 4 | 3+4 4 | 3+4 4 | 3+4 4 |
| | 5 2 | 4+4 4 | 3+3 4 | 3+4 4 | 6+6 4 | 3+4 4 | 3+4 4 | 3+4 4 |
| | 6 2 | 3+3 4 | 4+4 4 | 3+4 4 | 6+6 4 | 3+4 4 | 3+4 4 | 3+4 4 |
| | 6 2 | 3+3 4 | 3+3 4 | 3+4 4 | 5+5 4 | 3+4 4 | 3+4 4 | 3+4 4 |
| | 6 2 | 4+4 4 | 4+4 4 | 3+4 4 | 5+5 4 | 3+4 4 | 3+4 4 | 3+4 4 |
| | 5 2 | 4+4 4 | 4+4 4 | 3+4 4 | 5+5 4 | 3+4 4 | 3+4 4 | 3+4 4 |

Continued on next column

**TABLE XXXVII– continued**

| TQC | ICSD | NoE | SF |
|---|---|---|---|
| | 5 2 3+3 4 3+3 4 3+4 4 5+5 4 3+4 4 3+4 4 3+4 4 | | |
| | 5 2 4+4 4 3+3 4 3+4 4 6+6 4 3+4 4 3+4 4 3+4 4 | | |
| | 5 2 3+3 4 4+4 4 3+4 4 6+6 4 3+4 4 3+4 4 3+4 4 | | |
| | 6 2 3+3 4 3+3 4 3+4 4 6+6 4 3+4 4 3+4 4 3+4 4 | | |
| | 6 2 4+4 4 4+4 4 3+4 4 5+5 4 3+4 4 3+4 4 3+4 4 | | |
| | 5 2 4+4 4 4+4 4 3+4 4 6+6 4 3+4 4 3+4 4 3+4 4 | | |
| | 5 2 3+3 4 3+3 4 3+4 4 6+6 4 3+4 4 3+4 4 3+4 4 | | |
| | 5 2 4+4 4 3+3 4 3+4 4 5+5 4 3+4 4 3+4 4 3+4 4 | | |
| | 6 2 3+3 4 4+4 4 3+4 4 6+6 4 3+4 4 3+4 4 3+4 4 | | |
| | 5 2 3+3 4 3+3 4 3+4 4 5+5 4 3+4 4 3+4 4 3+4 4 | | |
| | 6 2 4+4 4 4+4 4 3+4 4 5+5 4 3+4 4 3+4 4 3+4 4 | | |
| | 6 2 3+3 4 4+4 4 3+4 4 5+5 4 3+4 4 3+4 4 3+4 4 | | |
| | 6 2 4+4 4 3+3 4 3+4 4 5+5 4 3+4 4 3+4 4 3+4 4 | | |
| | 6 2 4+4 4 4+4 4 3+4 4 5+5 4 3+4 4 3+4 4 3+4 4 | | |
| | 5 2 3+3 4 3+3 4 3+4 4 5+5 4 3+4 4 3+4 4 3+4 4 | | |
| | 6 2 3+3 4 3+3 4 3+4 4 6+6 4 3+4 4 3+4 4 3+4 4 | | |
| | 5 2 4+4 4 4+4 4 3+4 4 5+5 4 3+4 4 3+4 4 3+4 4 | | |
| | 5 2 3+3 4 3+3 4 3+4 4 5+5 4 3+4 4 3+4 4 3+4 4 | | |
| | 5 2 4+4 4 4+4 4 3+4 4 6+6 4 3+4 4 3+4 4 3+4 4 | | |
| | 6 2 3+3 4 4+4 4 3+4 4 6+6 4 3+4 4 3+4 4 3+4 4 | | |
| | 6 2 4+4 4 3+3 4 3+4 4 5+5 4 3+4 4 3+4 4 3+4 4 | | |
| | 6 2 4+4 4 4+4 4 3+4 4 5+5 4 3+4 4 3+4 4 3+4 4 | | |
| | 5 2 3+3 4 3+3 4 3+4 4 6+6 4 3+4 4 3+4 4 3+4 4 | | |
| | 5 2 3+3 4 4+4 4 3+4 4 5+5 4 3+4 4 3+4 4 3+4 4 | | |
| | 6 2 4+4 4 3+3 4 3+4 4 5+5 4 3+4 4 3+4 4 3+4 4 | | |
| | 5 2 3+3 4 3+3 4 3+4 4 6+6 4 3+4 4 3+4 4 3+4 4 | | |
| | 5 2 4+4 4 4+4 4 3+4 4 6+6 4 3+4 4 3+4 4 3+4 4 | | |
| | 6 2 4+4 4 3+3 4 3+4 4 6+6 4 3+4 4 3+4 4 3+4 4 | | |
| | 6 2 3+3 4 4+4 4 3+4 4 6+6 4 3+4 4 3+4 4 3+4 4 | | |
| | 5 2 3+3 4 3+3 4 3+4 4 6+6 4 3+4 4 3+4 4 3+4 4 | | |
| | 5 2 4+4 4 4+4 4 3+4 4 5+5 4 3+4 4 3+4 4 3+4 4 | | |
| | 5 2 3+3 4 4+4 4 3+4 4 6+6 4 3+4 4 3+4 4 3+4 4 | | |
| | 5 2 4+4 4 3+3 4 3+4 4 5+5 4 3+4 4 3+4 4 3+4 4 | | |
| | 5 2 3+3 4 3+3 4 3+4 4 5+5 4 3+4 4 3+4 4 3+4 4 | | |
| | 6 2 4+4 4 4+4 4 3+4 4 6+6 4 3+4 4 3+4 4 3+4 4 | | |
| | 6 2 4+4 4 3+3 4 3+4 4 5+5 4 3+4 4 3+4 4 3+4 4 | | |
| | 5 2 3+3 4 4+4 4 3+4 4 5+5 4 3+4 4 3+4 4 3+4 4 | | |
| | 6 2 4+4 4 3+3 4 3+4 4 5+5 4 3+4 4 3+4 4 3+4 4 | | |
| | 6 2 3+3 4 4+4 4 3+4 4 6+6 4 3+4 4 3+4 4 3+4 4 | | |
| | 6 2 4+4 4 3+3 4 3+4 4 6+6 4 3+4 4 3+4 4 3+4 4 | | |
| | 6 2 3+3 4 4+4 4 3+4 4 6+6 4 3+4 4 3+4 4 3+4 4 | | |
| | 6 2 3+3 4 3+3 4 3+4 4 5+5 4 3+4 4 3+4 4 3+4 4 | | |
| | 6 2 4+4 4 4+4 4 3+4 4 6+6 4 3+4 4 3+4 4 3+4 4 | | |
| | 6 2 3+3 4 4+4 4 3+4 4 5+5 4 3+4 4 3+4 4 3+4 4 | | |
| | 6 2 4+4 4 3+3 4 3+4 4 5+5 4 3+4 4 3+4 4 3+4 4 | | |
| | 6 2 3+3 4 3+3 4 3+4 4 6+6 4 3+4 4 3+4 4 3+4 4 | | |
| | 6 2 4+4 4 4+4 4 3+4 4 5+5 4 3+4 4 3+4 4 3+4 4 | | |
| | 6 2 4+4 4 4+4 4 3+4 4 6+6 4 3+4 4 3+4 4 3+4 4 | | |
| | 6 2 3+3 4 3+3 4 3+4 4 6+6 4 3+4 4 3+4 4 3+4 4 | | |
| | 5 2 3+3 4 4+4 4 3+4 4 5+5 4 3+4 4 3+4 4 3+4 4 | | |
| | 6 2 4+4 4 3+3 4 3+4 4 6+6 4 3+4 4 3+4 4 3+4 4 | | |
| | 6 2 4+4 4 3+3 4 3+4 4 6+6 4 3+4 4 3+4 4 3+4 4 | | |
| | 5 2 3+3 4 4+4 4 3+4 4 5+5 4 3+4 4 3+4 4 3+4 4 | | |
| | 5 2 4+4 4 3+3 4 3+4 4 6+6 4 3+4 4 3+4 4 3+4 4 | | |
| | 6 2 3+3 4 4+4 4 3+4 4 6+6 4 3+4 4 3+4 4 3+4 4 | | |
| | 5 2 4+4 4 3+3 4 3+4 4 5+5 4 3+4 4 3+4 4 3+4 4 | | |
| | 5 2 3+3 4 4+4 4 3+4 4 5+5 4 3+4 4 3+4 4 3+4 4 | | |
| | 6 2 3+3 4 4+4 4 3+4 4 6+6 4 3+4 4 3+4 4 3+4 4 | | |
| | 5 2 4+4 4 3+3 4 3+4 4 5+5 4 3+4 4 3+4 4 3+4 4 | | |
| | 6 2 4+4 4 4+4 4 3+4 4 5+5 4 3+4 4 3+4 4 3+4 4 | | |
| | 5 2 3+3 4 3+3 4 3+4 4 5+5 4 3+4 4 3+4 4 3+4 4 | | |

Continued on next column

**TABLE XXXVII– continued**

| TQC | ICSD | NoE | SF |
|---|---|---|---|
| | 6 2 3+3 4 3+3 4 3+4 4 6+6 4 3+4 4 3+4 4 3+4 4 | | |
| | 6 2 4+4 4 4+4 4 3+4 4 5+5 4 3+4 4 3+4 4 3+4 4 | | |
| | 5 2 | | |
| | 6 2 | | |
| | 5 2 | | |
| | 5 2 | | |
| | 6 2 | | |
| | 5 2 | | |
| | 6 2 | | |
| | 6 2 | | |
| | 5 2 | | |
| | 5 2 | | |
| | 5 2 | | |
| | 5 2 | | |
| | 5 2 | | |
| | 6 2 | | |
| | 6 2 | | |
| | 5 2 | | |
| | 6 2 | | |
| | 6 2 | | |
| | 6 2 | | |
| | 5 2 | | |
| | 5 2 | | |
| | 6 2 | | |
| | 5 2 | | |
| | 5 2 | | |
| | 6 2 | | |
| | 6 2 | | |
| | 6 2 | | |
| | 5 2 | | |
| | 6 2 | | |
| | 5 2 | | |
| | 5 2 | | |
| | 6 2 | | |
| | 6 2 | | |
| | 6 2 | | |
| | 5 2 | | |
| | 5 2 | | |
| | 5 2 | | |
| | 5 2 | | |
| | 5 2 | | |
| | 5 2 | | |
| | 6 2 | | |
| | 5 2 | | |
| | 6 2 | | |
| | 6 2 | | |
| | 5 2 | | |
| | 6 2 | | |
| | 5 2 | | |
| | 6 2 | | |
| | 6 2 | | |
| | 6 2 | | |
| | 5 2 | | |
| | 6 2 | | |
| | 5 2 | | |
| | 5 2 | | |
| | 5 2 | | |
| | 6 2 | | |
| | 6 2 | | |
| | 5 2 | | |
| | 6 2 | | |

Continued on next column

**TABLE XXXVII– continued**

| TQC | ICSD | NoE | SF | | | | | |
|---|---|---|---|---|---|---|---|---|
| | 5 2 | | | | | | | |
| | 5 2 | | | | | | | |
| | 6 2 | | | | | | | |
| | 6 2 | | | | | | | |
| | 5 2 | | | | | | | |
| | 6 2 | | | | | | | |
| | 5 2 | | | | | | | |
| | 6 2 | | | | | | | |
| | 5 2 | | | | | | | |
| | 5 2 | | | | | | | |
| | 6 2 | | | | | | | |
| | 5 2 | | | | | | | |
| | 5 2 | | | | | | | |
| | 6 2 | | | | | | | |
| | 5 2 | | | | | | | |
| | 6 2 | | | | | | | |
| | 5 2 | | | | | | | |
| | 6 2 | | | | | | | |
| | 6 2 | | | | | | | |
| | 6 2 | | | | | | | |
| | 5 2 | | | | | | | |
| | 5 2 | | | | | | | |
| | 6 2 | | | | | | | |
| 62.2.35 | 74785 | 168 | Ni2P2Th1 | | | | | |
| | -GM | -R | -S | -T | -U | -X | -Y | -Z |
| | 5 2 | 3+3 4 | 3+3 4 | 3+4 4 | 5+5 4 | 3+4 4 | 3+4 4 | 3+4 4 |
| | 6 2 | 4+4 4 | 4+4 4 | 3+4 4 | 6+6 4 | 3+4 4 | 3+4 4 | 3+4 4 |
| | 5 2 | 3+3 4 | 4+4 4 | 3+4 4 | 6+6 4 | 3+4 4 | 3+4 4 | 3+4 4 |
| | 6 2 | 4+4 4 | 3+3 4 | 3+4 4 | 5+5 4 | 3+4 4 | 3+4 4 | 3+4 4 |
| | 6 2 | 4+4 4 | 3+3 4 | 3+4 4 | 6+6 4 | 3+4 4 | 3+4 4 | 3+4 4 |
| | 5 2 | 3+3 4 | 4+4 4 | 3+4 4 | 5+5 4 | 3+4 4 | 3+4 4 | 3+4 4 |
| | 5 2 | 3+3 4 | 4+4 4 | 3+4 4 | 6+6 4 | 3+4 4 | 3+4 4 | 3+4 4 |
| | 6 2 | 4+4 4 | 3+3 4 | 3+4 4 | 5+5 4 | 3+4 4 | 3+4 4 | 3+4 4 |
| | 5 2 | 4+4 4 | 3+3 4 | 3+4 4 | 6+6 4 | 3+4 4 | 3+4 4 | 3+4 4 |
| | 6 2 | 3+3 4 | 4+4 4 | 3+4 4 | 5+5 4 | 3+4 4 | 3+4 4 | 3+4 4 |
| | 5 2 | 3+3 4 | 4+4 4 | 3+4 4 | 6+6 4 | 3+4 4 | 3+4 4 | 3+4 4 |
| | 6 2 | 4+4 4 | 3+3 4 | 3+4 4 | 5+5 4 | 3+4 4 | 3+4 4 | 3+4 4 |
| | 6 2 | 3+3 4 | 4+4 4 | 3+4 4 | 5+5 4 | 3+4 4 | 3+4 4 | 3+4 4 |
| | 5 2 | 4+4 4 | 3+3 4 | 3+4 4 | 6+6 4 | 3+4 4 | 3+4 4 | 3+4 4 |
| | 5 2 | 4+4 4 | 3+3 4 | 3+4 4 | 5+5 4 | 3+4 4 | 3+4 4 | 3+4 4 |
| | 6 2 | 3+3 4 | 4+4 4 | 3+4 4 | 6+6 4 | 3+4 4 | 3+4 4 | 3+4 4 |
| | 5 2 | 3+3 4 | 3+3 4 | 3+4 4 | 6+6 4 | 3+4 4 | 3+4 4 | 3+4 4 |
| | 5 2 | 4+4 4 | 4+4 4 | 3+4 4 | 6+6 4 | 3+4 4 | 3+4 4 | 3+4 4 |
| | 6 2 | 4+4 4 | 4+4 4 | 3+4 4 | 6+6 4 | 3+4 4 | 3+4 4 | 3+4 4 |
| | 5 2 | 3+3 4 | 3+3 4 | 3+4 4 | 5+5 4 | 3+4 4 | 3+4 4 | 3+4 4 |
| | 6 2 | 3+3 4 | 3+3 4 | 3+4 4 | 5+5 4 | 3+4 4 | 3+4 4 | 3+4 4 |
| | 5 2 | 4+4 4 | 4+4 4 | 3+4 4 | 6+6 4 | 3+4 4 | 3+4 4 | 3+4 4 |
| | 6 2 | 4+4 4 | 3+3 4 | 3+4 4 | 5+5 4 | 3+4 4 | 3+4 4 | 3+4 4 |
| | 6 2 | 3+3 4 | 4+4 4 | 3+4 4 | 5+5 4 | 3+4 4 | 3+4 4 | 3+4 4 |
| | 5 2 | 3+3 4 | 4+4 4 | 3+4 4 | 5+5 4 | 3+4 4 | 3+4 4 | 3+4 4 |
| | 6 2 | 4+4 4 | 3+3 4 | 3+4 4 | 6+6 4 | 3+4 4 | 3+4 4 | 3+4 4 |
| | 5 2 | 3+3 4 | 3+3 4 | 3+4 4 | 6+6 4 | 3+4 4 | 3+4 4 | 3+4 4 |
| | 6 2 | 4+4 4 | 4+4 4 | 3+4 4 | 5+5 4 | 3+4 4 | 3+4 4 | 3+4 4 |
| | 5 2 | 4+4 4 | 3+3 4 | 3+4 4 | 6+6 4 | 3+4 4 | 3+4 4 | 3+4 4 |
| | 6 2 | 3+3 4 | 4+4 4 | 3+4 4 | 5+5 4 | 3+4 4 | 3+4 4 | 3+4 4 |
| | 5 2 | 3+3 4 | 4+4 4 | 3+4 4 | 6+6 4 | 3+4 4 | 3+4 4 | 3+4 4 |
| | 5 2 | 4+4 4 | 3+3 4 | 3+4 4 | 5+5 4 | 3+4 4 | 3+4 4 | 3+4 4 |
| | 5 2 | 4+4 4 | 4+4 4 | 3+4 4 | 6+6 4 | 3+4 4 | 3+4 4 | 3+4 4 |
| | 6 2 | 3+3 4 | 3+3 4 | 3+4 4 | 5+5 4 | 3+4 4 | 3+4 4 | 3+4 4 |
| | 6 2 | 4+4 4 | 3+3 4 | 3+4 4 | 5+5 4 | 3+4 4 | 3+4 4 | 3+4 4 |
| | 5 2 | 3+3 4 | 4+4 4 | 3+4 4 | 6+6 4 | 3+4 4 | 3+4 4 | 3+4 4 |

Continued on next column

**TABLE XXXVII– continued**

| TQC | ICSD | NoE | SF | | | | | |
|---|---|---|---|---|---|---|---|---|
| | 6 2 | 4+4 4 | 3+3 4 | 3+4 4 | 6+6 4 | 3+4 4 | 3+4 4 | 3+4 4 |
| | 6 2 | 3+3 4 | 4+4 4 | 3+4 4 | 5+5 4 | 3+4 4 | 3+4 4 | 3+4 4 |
| | 5 2 | 4+4 4 | 3+3 4 | 3+4 4 | 6+6 4 | 3+4 4 | 3+4 4 | 3+4 4 |
| | 5 2 | 3+3 4 | 4+4 4 | 3+4 4 | 5+5 4 | 3+4 4 | 3+4 4 | 3+4 4 |
| | 6 2 | 3+3 4 | 4+4 4 | 3+4 4 | 6+6 4 | 3+4 4 | 3+4 4 | 3+4 4 |
| | 6 2 | 4+4 4 | 3+3 4 | 3+4 4 | 5+5 4 | 3+4 4 | 3+4 4 | 3+4 4 |
| | 5 2 | | | | | | | |
| | 6 2 | | | | | | | |
| | 6 2 | | | | | | | |
| | 5 2 | | | | | | | |
| | 6 2 | | | | | | | |
| | 5 2 | | | | | | | |
| | 5 2 | | | | | | | |
| | 5 2 | | | | | | | |
| | 6 2 | | | | | | | |
| | 6 2 | | | | | | | |
| | 5 2 | | | | | | | |
| | 5 2 | | | | | | | |
| | 5 2 | | | | | | | |
| | 5 2 | | | | | | | |
| | 6 2 | | | | | | | |
| | 6 2 | | | | | | | |
| | 6 2 | | | | | | | |
| | 5 2 | | | | | | | |
| | 6 2 | | | | | | | |
| | 5 2 | | | | | | | |
| | 5 2 | | | | | | | |
| | 6 2 | | | | | | | |
| | 6 2 | | | | | | | |
| | 6 2 | | | | | | | |
| | 5 2 | | | | | | | |
| | 5 2 | | | | | | | |
| | 6 2 | | | | | | | |
| | 5 2 | | | | | | | |
| | 5 2 | | | | | | | |
| | 5 2 | | | | | | | |
| | 5 2 | | | | | | | |
| | 5 2 | | | | | | | |
| | 5 2 | | | | | | | |
| | 6 2 | | | | | | | |
| | 6 2 | | | | | | | |
| | 5 2 | | | | | | | |
| | 6 2 | | | | | | | |
| | 6 2 | | | | | | | |
| | 6 2 | | | | | | | |
| | 6 2 | | | | | | | |
| | 5 2 | | | | | | | |
| | 5 2 | | | | | | | |
| 62.2.36 | 64952 | 192 | Ni2O4Si1 | | | | | |
| | -GM | -R | -S | -T | -U | -X | -Y | -Z |
| | 5 2 | 3+3 4 | 3+3 4 | 3+4 4 | 5+5 4 | 3+4 4 | 3+4 4 | 3+4 4 |
| | 5 2 | 4+4 4 | 4+4 4 | 3+4 4 | 6+6 4 | 3+4 4 | 3+4 4 | 3+4 4 |
| | 6 2 | 3+3 4 | 3+3 4 | 3+4 4 | 6+6 4 | 3+4 4 | 3+4 4 | 3+4 4 |
| | 6 2 | 4+4 4 | 4+4 4 | 3+4 4 | 5+5 4 | 3+4 4 | 3+4 4 | 3+4 4 |
| | 6 2 | 3+3 4 | 3+3 4 | 3+4 4 | 6+6 4 | 3+4 4 | 3+4 4 | 3+4 4 |
| | 6 2 | 4+4 4 | 4+4 4 | 3+4 4 | 5+5 4 | 3+4 4 | 3+4 4 | 3+4 4 |
| | 5 2 | 3+3 4 | 3+3 4 | 3+4 4 | 6+6 4 | 3+4 4 | 3+4 4 | 3+4 4 |
| | 6 2 | 4+4 4 | 4+4 4 | 3+4 4 | 5+5 4 | 3+4 4 | 3+4 4 | 3+4 4 |
| | 5 2 | 3+3 4 | 3+3 4 | 3+4 4 | 6+6 4 | 3+4 4 | 3+4 4 | 3+4 4 |
| | 5 2 | 4+4 4 | 4+4 4 | 3+4 4 | 5+5 4 | 3+4 4 | 3+4 4 | 3+4 4 |
| | 6 2 | 3+3 4 | 3+3 4 | 3+4 4 | 5+5 4 | 3+4 4 | 3+4 4 | 3+4 4 |

Continued on next column

**TABLE XXXVII– continued**

| TQC | ICSD NoE SF |
|---|---|
| | 6 2 4+4 4 4+4 4 3+4 4 5+5 4 3+4 4 3+4 4 3+4 4 |
| | 5 2 3+3 4 3+3 4 3+4 4 6+6 4 3+4 4 3+4 4 3+4 4 |
| | 5 2 4+4 4 4+4 4 3+4 4 6+6 4 3+4 4 3+4 4 3+4 4 |
| | 5 2 3+3 4 4+4 4 3+4 4 6+6 4 3+4 4 3+4 4 3+4 4 |
| | 6 2 4+4 4 3+3 4 3+4 4 5+5 4 3+4 4 3+4 4 3+4 4 |
| | 5 2 3+3 4 3+3 4 3+4 4 6+6 4 3+4 4 3+4 4 3+4 4 |
| | 5 2 4+4 4 4+4 4 3+4 4 6+6 4 3+4 4 3+4 4 3+4 4 |
| | 5 2 3+3 4 3+3 4 3+4 4 6+6 4 3+4 4 3+4 4 3+4 4 |
| | 5 2 4+4 4 4+4 4 3+4 4 5+5 4 3+4 4 3+4 4 3+4 4 |
| | 6 2 3+3 4 3+3 4 3+4 4 5+5 4 3+4 4 3+4 4 3+4 4 |
| | 6 2 4+4 4 4+4 4 3+4 4 5+5 4 3+4 4 3+4 4 3+4 4 |
| | 5 2 3+3 4 4+4 4 3+4 4 6+6 4 3+4 4 3+4 4 3+4 4 |
| | 6 2 4+4 4 3+3 4 3+4 4 6+6 4 3+4 4 3+4 4 3+4 4 |
| | 6 2 3+3 4 3+3 4 3+4 4 5+5 4 3+4 4 3+4 4 3+4 4 |
| | 5 2 4+4 4 4+4 4 3+4 4 6+6 4 3+4 4 3+4 4 3+4 4 |
| | 6 2 4+4 4 3+3 4 3+4 4 5+5 4 3+4 4 3+4 4 3+4 4 |
| | 5 2 3+3 4 4+4 4 3+4 4 6+6 4 3+4 4 3+4 4 3+4 4 |
| | 5 2 3+3 4 3+3 4 3+4 4 6+6 4 3+4 4 3+4 4 3+4 4 |
| | 6 2 4+4 4 4+4 4 3+4 4 5+5 4 3+4 4 3+4 4 3+4 4 |
| | 6 2 3+3 4 4+4 4 3+4 4 5+5 4 3+4 4 3+4 4 3+4 4 |
| | 6 2 4+4 4 3+3 4 3+4 4 5+5 4 3+4 4 3+4 4 3+4 4 |
| | 6 2 4+4 4 4+4 4 3+4 4 6+6 4 3+4 4 3+4 4 3+4 4 |
| | 5 2 3+3 4 3+3 4 3+4 4 6+6 4 3+4 4 3+4 4 3+4 4 |
| | 5 2 3+3 4 3+3 4 3+4 4 6+6 4 3+4 4 3+4 4 3+4 4 |
| | 5 2 4+4 4 4+4 4 3+4 4 6+6 4 3+4 4 3+4 4 3+4 4 |
| | 6 2 3+3 4 3+3 4 3+4 4 6+6 4 3+4 4 3+4 4 3+4 4 |
| | 5 2 4+4 4 4+4 4 3+4 4 5+5 4 3+4 4 3+4 4 3+4 4 |
| | 6 2 4+4 4 3+3 4 3+4 4 6+6 4 3+4 4 3+4 4 3+4 4 |
| | 5 2 3+3 4 4+4 4 3+4 4 5+5 4 3+4 4 3+4 4 3+4 4 |
| | 6 2 4+4 4 3+3 4 3+4 4 6+6 4 3+4 4 3+4 4 3+4 4 |
| | 6 2 3+3 4 4+4 4 3+4 4 6+6 4 3+4 4 3+4 4 3+4 4 |
| | 5 2 3+3 4 4+4 4 3+4 4 5+5 4 3+4 4 3+4 4 3+4 4 |
| | 6 2 4+4 4 3+3 4 3+4 4 6+6 4 3+4 4 3+4 4 3+4 4 |
| | 5 2 3+3 4 3+3 4 3+4 4 6+6 4 3+4 4 3+4 4 3+4 4 |
| | 5 2 4+4 4 4+4 4 3+4 4 6+6 4 3+4 4 3+4 4 3+4 4 |
| | 5 2 3+3 4 3+3 4 3+4 4 6+6 4 3+4 4 3+4 4 3+4 4 |
| | 5 2 4+4 4 4+4 4 3+4 4 5+5 4 3+4 4 3+4 4 3+4 4 |
| | 6 2 |
| | 5 2 |
| | 6 2 |
| | 5 2 |
| | 5 2 |
| | 6 2 |
| | 6 2 |
| | 5 2 |
| | 5 2 |
| | 6 2 |
| | 6 2 |
| | 6 2 |
| | 5 2 |
| | 6 2 |
| | 6 2 |
| | 6 2 |
| | 5 2 |
| | 5 2 |
| | 5 2 |
| | 5 2 |
| | 5 2 |
| | 5 2 |
| | 5 2 |
| | 5 2 |

Continued on next column

**TABLE XXXVII– continued**

| TQC | ICSD NoE SF |
|---|---|
| | 5 2 |
| | 5 2 |
| | 6 2 |
| | 5 2 |
| | 5 2 |
| | 6 2 |
| | 5 2 |
| | 6 2 |
| | 6 2 |
| | 5 2 |
| | 6 2 |
| | 5 2 |
| | 5 2 |
| | 5 2 |
| | 5 2 |
| | 6 2 |
| | 5 2 |
| | 5 2 |
| | 6 2 |
| | 6 2 |
| | 5 2 |
| | 5 2 |
| | 6 2 |
| | 5 2 |
| 62.2.37 | 35675 192 Ni2O4Si1 |
| | -GM -R -S -T -U -X -Y -Z |
| | 5 2 3+3 4 3+3 4 3+4 4 5+5 4 3+4 4 3+4 4 3+4 4 |
| | 5 2 4+4 4 4+4 4 3+4 4 6+6 4 3+4 4 3+4 4 3+4 4 |
| | 6 2 3+3 4 3+3 4 3+4 4 6+6 4 3+4 4 3+4 4 3+4 4 |
| | 6 2 4+4 4 4+4 4 3+4 4 5+5 4 3+4 4 3+4 4 3+4 4 |
| | 6 2 3+3 4 3+3 4 3+4 4 6+6 4 3+4 4 3+4 4 3+4 4 |
| | 6 2 4+4 4 4+4 4 3+4 4 5+5 4 3+4 4 3+4 4 3+4 4 |
| | 5 2 3+3 4 3+3 4 3+4 4 6+6 4 3+4 4 3+4 4 3+4 4 |
| | 6 2 4+4 4 4+4 4 3+4 4 5+5 4 3+4 4 3+4 4 3+4 4 |
| | 5 2 3+3 4 3+3 4 3+4 4 6+6 4 3+4 4 3+4 4 3+4 4 |
| | 5 2 4+4 4 4+4 4 3+4 4 5+5 4 3+4 4 3+4 4 3+4 4 |
| | 6 2 3+3 4 3+3 4 3+4 4 5+5 4 3+4 4 3+4 4 3+4 4 |
| | 6 2 4+4 4 4+4 4 3+4 4 5+5 4 3+4 4 3+4 4 3+4 4 |
| | 5 2 3+3 4 3+3 4 3+4 4 6+6 4 3+4 4 3+4 4 3+4 4 |
| | 5 2 4+4 4 4+4 4 3+4 4 6+6 4 3+4 4 3+4 4 3+4 4 |
| | 5 2 3+3 4 4+4 4 3+4 4 6+6 4 3+4 4 3+4 4 3+4 4 |
| | 6 2 4+4 4 3+3 4 3+4 4 5+5 4 3+4 4 3+4 4 3+4 4 |
| | 5 2 3+3 4 3+3 4 3+4 4 6+6 4 3+4 4 3+4 4 3+4 4 |
| | 5 2 4+4 4 4+4 4 3+4 4 6+6 4 3+4 4 3+4 4 3+4 4 |
| | 5 2 3+3 4 3+3 4 3+4 4 6+6 4 3+4 4 3+4 4 3+4 4 |
| | 5 2 4+4 4 4+4 4 3+4 4 5+5 4 3+4 4 3+4 4 3+4 4 |
| | 6 2 3+3 4 3+3 4 3+4 4 5+5 4 3+4 4 3+4 4 3+4 4 |
| | 6 2 4+4 4 4+4 4 3+4 4 5+5 4 3+4 4 3+4 4 3+4 4 |
| | 5 2 4+4 4 3+3 4 3+4 4 6+6 4 3+4 4 3+4 4 3+4 4 |
| | 6 2 3+3 4 4+4 4 3+4 4 6+6 4 3+4 4 3+4 4 3+4 4 |
| | 6 2 3+3 4 3+3 4 3+4 4 5+5 4 3+4 4 3+4 4 3+4 4 |
| | 5 2 4+4 4 4+4 4 3+4 4 6+6 4 3+4 4 3+4 4 3+4 4 |
| | 6 2 4+4 4 3+3 4 3+4 4 5+5 4 3+4 4 3+4 4 3+4 4 |
| | 5 2 3+3 4 4+4 4 3+4 4 6+6 4 3+4 4 3+4 4 3+4 4 |
| | 5 2 3+3 4 3+3 4 3+4 4 6+6 4 3+4 4 3+4 4 3+4 4 |
| | 6 2 4+4 4 4+4 4 3+4 4 5+5 4 3+4 4 3+4 4 3+4 4 |
| | 6 2 3+3 4 4+4 4 3+4 4 5+5 4 3+4 4 3+4 4 3+4 4 |
| | 6 2 4+4 4 3+3 4 3+4 4 5+5 4 3+4 4 3+4 4 3+4 4 |
| | 6 2 4+4 4 4+4 4 3+4 4 6+6 4 3+4 4 3+4 4 3+4 4 |
| | 5 2 3+3 4 3+3 4 3+4 4 6+6 4 3+4 4 3+4 4 3+4 4 |
| | 5 2 3+3 4 3+3 4 3+4 4 6+6 4 3+4 4 3+4 4 3+4 4 |

Continued on next column

**TABLE XXXVII– continued**

| TQC | ICSD | NoE | SF | | | | | |
|---|---|---|---|---|---|---|---|---|
| | 6 2 | 4+4 4 | 4+4 4 | 3+4 4 | 6+6 4 | 3+4 4 | 3+4 4 | 3+4 4 |
| | 5 2 | 3+3 4 | 3+3 4 | 3+4 4 | 5+5 4 | 3+4 4 | 3+4 4 | 3+4 4 |
| | 5 2 | 4+4 4 | 4+4 4 | 3+4 4 | 6+6 4 | 3+4 4 | 3+4 4 | 3+4 4 |
| | 6 2 | 4+4 4 | 3+3 4 | 3+4 4 | 6+6 4 | 3+4 4 | 3+4 4 | 3+4 4 |
| | 5 2 | 3+3 4 | 4+4 4 | 3+4 4 | 5+5 4 | 3+4 4 | 3+4 4 | 3+4 4 |
| | 6 2 | 4+4 4 | 3+3 4 | 3+4 4 | 6+6 4 | 3+4 4 | 3+4 4 | 3+4 4 |
| | 6 2 | 3+3 4 | 4+4 4 | 3+4 4 | 6+6 4 | 3+4 4 | 3+4 4 | 3+4 4 |
| | 5 2 | 3+3 4 | 4+4 4 | 3+4 4 | 5+5 4 | 3+4 4 | 3+4 4 | 3+4 4 |
| | 6 2 | 4+4 4 | 3+3 4 | 3+4 4 | 6+6 4 | 3+4 4 | 3+4 4 | 3+4 4 |
| | 5 2 | 3+3 4 | 3+3 4 | 3+4 4 | 6+6 4 | 3+4 4 | 3+4 4 | 3+4 4 |
| | 5 2 | 4+4 4 | 4+4 4 | 3+4 4 | 6+6 4 | 3+4 4 | 3+4 4 | 3+4 4 |
| | 5 2 | 3+3 4 | 3+3 4 | 3+4 4 | 6+6 4 | 3+4 4 | 3+4 4 | 3+4 4 |
| | 5 2 | 4+4 4 | 4+4 4 | 3+4 4 | 5+5 4 | 3+4 4 | 3+4 4 | 3+4 4 |
| | 6 2 | | | | | | | |
| | 5 2 | | | | | | | |
| | 6 2 | | | | | | | |
| | 5 2 | | | | | | | |
| | 5 2 | | | | | | | |
| | 6 2 | | | | | | | |
| | 6 2 | | | | | | | |
| | 5 2 | | | | | | | |
| | 5 2 | | | | | | | |
| | 6 2 | | | | | | | |
| | 6 2 | | | | | | | |
| | 6 2 | | | | | | | |
| | 5 2 | | | | | | | |
| | 6 2 | | | | | | | |
| | 6 2 | | | | | | | |
| | 6 2 | | | | | | | |
| | 5 2 | | | | | | | |
| | 5 2 | | | | | | | |
| | 5 2 | | | | | | | |
| | 5 2 | | | | | | | |
| | 5 2 | | | | | | | |
| | 5 2 | | | | | | | |
| | 5 2 | | | | | | | |
| | 5 2 | | | | | | | |
| | 5 2 | | | | | | | |
| | 5 2 | | | | | | | |
| | 6 2 | | | | | | | |
| | 5 2 | | | | | | | |
| | 5 2 | | | | | | | |
| | 6 2 | | | | | | | |
| | 5 2 | | | | | | | |
| | 6 2 | | | | | | | |
| | 6 2 | | | | | | | |
| | 5 2 | | | | | | | |
| | 6 2 | | | | | | | |
| | 5 2 | | | | | | | |
| | 5 2 | | | | | | | |
| | 5 2 | | | | | | | |
| | 5 2 | | | | | | | |
| | 6 2 | | | | | | | |
| | 5 2 | | | | | | | |
| | 5 2 | | | | | | | |
| | 6 2 | | | | | | | |
| | 6 2 | | | | | | | |
| | 5 2 | | | | | | | |
| | 5 2 | | | | | | | |
| | 6 2 | | | | | | | |
| | 5 2 | | | | | | | |

Continued on next column

**TABLE XXXVII– continued**

| TQC | ICSD | NoE | SF | | | | | |
|---|---|---|---|---|---|---|---|---|
| 62.2.38 | 202376 | 192 | Ni2O4Si1 | | | | | |
| | -GM | -R | -S | -T | -U | -X | -Y | -Z |
| | 5 2 | 3+3 4 | 3+3 4 | 3+4 4 | 6+6 4 | 3+4 4 | 3+4 4 | 3+4 4 |
| | 5 2 | 4+4 4 | 4+4 4 | 3+4 4 | 5+5 4 | 3+4 4 | 3+4 4 | 3+4 4 |
| | 6 2 | 3+3 4 | 3+3 4 | 3+4 4 | 5+5 4 | 3+4 4 | 3+4 4 | 3+4 4 |
| | 6 2 | 4+4 4 | 4+4 4 | 3+4 4 | 6+6 4 | 3+4 4 | 3+4 4 | 3+4 4 |
| | 6 2 | 3+3 4 | 3+3 4 | 3+4 4 | 5+5 4 | 3+4 4 | 3+4 4 | 3+4 4 |
| | 6 2 | 4+4 4 | 4+4 4 | 3+4 4 | 6+6 4 | 3+4 4 | 3+4 4 | 3+4 4 |
| | 5 2 | 3+3 4 | 3+3 4 | 3+4 4 | 5+5 4 | 3+4 4 | 3+4 4 | 3+4 4 |
| | 6 2 | 4+4 4 | 4+4 4 | 3+4 4 | 6+6 4 | 3+4 4 | 3+4 4 | 3+4 4 |
| | 5 2 | 3+3 4 | 3+3 4 | 3+4 4 | 5+5 4 | 3+4 4 | 3+4 4 | 3+4 4 |
| | 5 2 | 4+4 4 | 4+4 4 | 3+4 4 | 6+6 4 | 3+4 4 | 3+4 4 | 3+4 4 |
| | 6 2 | 3+3 4 | 3+3 4 | 3+4 4 | 6+6 4 | 3+4 4 | 3+4 4 | 3+4 4 |
| | 6 2 | 4+4 4 | 4+4 4 | 3+4 4 | 6+6 4 | 3+4 4 | 3+4 4 | 3+4 4 |
| | 5 2 | 3+3 4 | 3+3 4 | 3+4 4 | 5+5 4 | 3+4 4 | 3+4 4 | 3+4 4 |
| | 5 2 | 4+4 4 | 4+4 4 | 3+4 4 | 5+5 4 | 3+4 4 | 3+4 4 | 3+4 4 |
| | 5 2 | 3+3 4 | 3+3 4 | 3+4 4 | 5+5 4 | 3+4 4 | 3+4 4 | 3+4 4 |
| | 6 2 | 4+4 4 | 4+4 4 | 3+4 4 | 6+6 4 | 3+4 4 | 3+4 4 | 3+4 4 |
| | 5 2 | 3+3 4 | 3+3 4 | 3+4 4 | 5+5 4 | 3+4 4 | 3+4 4 | 3+4 4 |
| | 5 2 | 4+4 4 | 4+4 4 | 3+4 4 | 5+5 4 | 3+4 4 | 3+4 4 | 3+4 4 |
| | 5 2 | 4+4 4 | 4+4 4 | 3+4 4 | 5+5 4 | 3+4 4 | 3+4 4 | 3+4 4 |
| | 5 2 | 3+3 4 | 3+3 4 | 3+4 4 | 6+6 4 | 3+4 4 | 3+4 4 | 3+4 4 |
| | 6 2 | 4+4 4 | 3+3 4 | 3+4 4 | 6+6 4 | 3+4 4 | 3+4 4 | 3+4 4 |
| | 6 2 | 3+3 4 | 4+4 4 | 3+4 4 | 6+6 4 | 3+4 4 | 3+4 4 | 3+4 4 |
| | 5 2 | 3+3 4 | 3+3 4 | 3+4 4 | 5+5 4 | 3+4 4 | 3+4 4 | 3+4 4 |
| | 6 2 | 4+4 4 | 4+4 4 | 3+4 4 | 5+5 4 | 3+4 4 | 3+4 4 | 3+4 4 |
| | 6 2 | 4+4 4 | 4+4 4 | 3+4 4 | 6+6 4 | 3+4 4 | 3+4 4 | 3+4 4 |
| | 5 2 | 3+3 4 | 3+3 4 | 3+4 4 | 5+5 4 | 3+4 4 | 3+4 4 | 3+4 4 |
| | 6 2 | 3+3 4 | 3+3 4 | 3+4 4 | 6+6 4 | 3+4 4 | 3+4 4 | 3+4 4 |
| | 5 2 | 4+4 4 | 4+4 4 | 3+4 4 | 5+5 4 | 3+4 4 | 3+4 4 | 3+4 4 |
| | 5 2 | 3+3 4 | 4+4 4 | 3+4 4 | 5+5 4 | 3+4 4 | 3+4 4 | 3+4 4 |
| | 6 2 | 4+4 4 | 3+3 4 | 3+4 4 | 6+6 4 | 3+4 4 | 3+4 4 | 3+4 4 |
| | 6 2 | 3+3 4 | 3+3 4 | 3+4 4 | 6+6 4 | 3+4 4 | 3+4 4 | 3+4 4 |
| | 6 2 | 4+4 4 | 4+4 4 | 3+4 4 | 6+6 4 | 3+4 4 | 3+4 4 | 3+4 4 |
| | 6 2 | 3+3 4 | 3+3 4 | 3+4 4 | 5+5 4 | 3+4 4 | 3+4 4 | 3+4 4 |
| | 5 2 | 4+4 4 | 4+4 4 | 3+4 4 | 5+5 4 | 3+4 4 | 3+4 4 | 3+4 4 |
| | 5 2 | 4+4 4 | 4+4 4 | 3+4 4 | 5+5 4 | 3+4 4 | 3+4 4 | 3+4 4 |
| | 5 2 | 3+3 4 | 3+3 4 | 3+4 4 | 5+5 4 | 3+4 4 | 3+4 4 | 3+4 4 |
| | 6 2 | 4+4 4 | 4+4 4 | 3+4 4 | 5+5 4 | 3+4 4 | 3+4 4 | 3+4 4 |
| | 5 2 | 3+3 4 | 3+3 4 | 3+4 4 | 6+6 4 | 3+4 4 | 3+4 4 | 3+4 4 |
| | 6 2 | 3+3 4 | 4+4 4 | 3+4 4 | 5+5 4 | 3+4 4 | 3+4 4 | 3+4 4 |
| | 5 2 | 4+4 4 | 3+3 4 | 3+4 4 | 6+6 4 | 3+4 4 | 3+4 4 | 3+4 4 |
| | 6 2 | 3+3 4 | 4+4 4 | 3+4 4 | 5+5 4 | 3+4 4 | 3+4 4 | 3+4 4 |
| | 6 2 | 4+4 4 | 3+3 4 | 3+4 4 | 5+5 4 | 3+4 4 | 3+4 4 | 3+4 4 |
| | 5 2 | 4+4 4 | 3+3 4 | 3+4 4 | 6+6 4 | 3+4 4 | 3+4 4 | 3+4 4 |
| | 6 2 | 3+3 4 | 4+4 4 | 3+4 4 | 5+5 4 | 3+4 4 | 3+4 4 | 3+4 4 |
| | 5 2 | 4+4 4 | 4+4 4 | 3+4 4 | 5+5 4 | 3+4 4 | 3+4 4 | 3+4 4 |
| | 5 2 | 3+3 4 | 3+3 4 | 3+4 4 | 5+5 4 | 3+4 4 | 3+4 4 | 3+4 4 |
| | 5 2 | 4+4 4 | 4+4 4 | 3+4 4 | 5+5 4 | 3+4 4 | 3+4 4 | 3+4 4 |
| | 5 2 | 3+3 4 | 3+3 4 | 3+4 4 | 6+6 4 | 3+4 4 | 3+4 4 | 3+4 4 |
| | 6 2 | | | | | | | |
| | 6 2 | | | | | | | |
| | 5 2 | | | | | | | |
| | 5 2 | | | | | | | |
| | 5 2 | | | | | | | |
| | 6 2 | | | | | | | |
| | 6 2 | | | | | | | |
| | 5 2 | | | | | | | |
| | 5 2 | | | | | | | |
| | 6 2 | | | | | | | |
| | 6 2 | | | | | | | |

Continued on next column

**TABLE XXXVII– continued**

| TQC | ICSD | NoE | SF | | | | | |
|---|---|---|---|---|---|---|---|---|
| | 6 2 | | | | | | | |
| | 5 2 | | | | | | | |
| | 6 2 | | | | | | | |
| | 6 2 | | | | | | | |
| | 6 2 | | | | | | | |
| | 5 2 | | | | | | | |
| | 5 2 | | | | | | | |
| | 5 2 | | | | | | | |
| | 5 2 | | | | | | | |
| | 5 2 | | | | | | | |
| | 5 2 | | | | | | | |
| | 5 2 | | | | | | | |
| | 5 2 | | | | | | | |
| | 5 2 | | | | | | | |
| | 5 2 | | | | | | | |
| | 6 2 | | | | | | | |
| | 5 2 | | | | | | | |
| | 5 2 | | | | | | | |
| | 6 2 | | | | | | | |
| | 5 2 | | | | | | | |
| | 6 2 | | | | | | | |
| | 6 2 | | | | | | | |
| | 5 2 | | | | | | | |
| | 6 2 | | | | | | | |
| | 5 2 | | | | | | | |
| | 5 2 | | | | | | | |
| | 5 2 | | | | | | | |
| | 5 2 | | | | | | | |
| | 6 2 | | | | | | | |
| | 5 2 | | | | | | | |
| | 5 2 | | | | | | | |
| | 6 2 | | | | | | | |
| | 6 2 | | | | | | | |
| | 5 2 | | | | | | | |
| | 5 2 | | | | | | | |
| | 6 2 | | | | | | | |
| | 5 2 | | | | | | | |
| 62.2.39 | 100645 | 192 | Ni2O4Si1 | | | | | |
| | -GM | -R | -S | -T | -U | -X | -Y | -Z |
| | 5 2 | 3+3 4 | 3+3 4 | 3+4 4 | 6+6 4 | 3+4 4 | 3+4 4 | 3+4 4 |
| | 5 2 | 4+4 4 | 4+4 4 | 3+4 4 | 5+5 4 | 3+4 4 | 3+4 4 | 3+4 4 |
| | 6 2 | 3+3 4 | 3+3 4 | 3+4 4 | 5+5 4 | 3+4 4 | 3+4 4 | 3+4 4 |
| | 6 2 | 4+4 4 | 4+4 4 | 3+4 4 | 6+6 4 | 3+4 4 | 3+4 4 | 3+4 4 |
| | 6 2 | 3+3 4 | 3+3 4 | 3+4 4 | 5+5 4 | 3+4 4 | 3+4 4 | 3+4 4 |
| | 6 2 | 4+4 4 | 4+4 4 | 3+4 4 | 6+6 4 | 3+4 4 | 3+4 4 | 3+4 4 |
| | 5 2 | 3+3 4 | 3+3 4 | 3+4 4 | 5+5 4 | 3+4 4 | 3+4 4 | 3+4 4 |
| | 6 2 | 4+4 4 | 4+4 4 | 3+4 4 | 6+6 4 | 3+4 4 | 3+4 4 | 3+4 4 |
| | 5 2 | 3+3 4 | 3+3 4 | 3+4 4 | 5+5 4 | 3+4 4 | 3+4 4 | 3+4 4 |
| | 5 2 | 4+4 4 | 4+4 4 | 3+4 4 | 6+6 4 | 3+4 4 | 3+4 4 | 3+4 4 |
| | 6 2 | 3+3 4 | 3+3 4 | 3+4 4 | 6+6 4 | 3+4 4 | 3+4 4 | 3+4 4 |
| | 5 2 | 4+4 4 | 4+4 4 | 3+4 4 | 6+6 4 | 3+4 4 | 3+4 4 | 3+4 4 |
| | 6 2 | 3+3 4 | 3+3 4 | 3+4 4 | 5+5 4 | 3+4 4 | 3+4 4 | 3+4 4 |
| | 5 2 | 4+4 4 | 4+4 4 | 3+4 4 | 5+5 4 | 3+4 4 | 3+4 4 | 3+4 4 |
| | 5 2 | 3+3 4 | 3+3 4 | 3+4 4 | 5+5 4 | 3+4 4 | 3+4 4 | 3+4 4 |
| | 6 2 | 4+4 4 | 4+4 4 | 3+4 4 | 6+6 4 | 3+4 4 | 3+4 4 | 3+4 4 |
| | 5 2 | 3+3 4 | 3+3 4 | 3+4 4 | 5+5 4 | 3+4 4 | 3+4 4 | 3+4 4 |
| | 5 2 | 4+4 4 | 4+4 4 | 3+4 4 | 5+5 4 | 3+4 4 | 3+4 4 | 3+4 4 |
| | 5 2 | 4+4 4 | 4+4 4 | 3+4 4 | 5+5 4 | 3+4 4 | 3+4 4 | 3+4 4 |
| | 6 2 | 3+3 4 | 3+3 4 | 3+4 4 | 6+6 4 | 3+4 4 | 3+4 4 | 3+4 4 |
| | 5 2 | 4+4 4 | 3+3 4 | 3+4 4 | 6+6 4 | 3+4 4 | 3+4 4 | 3+4 4 |
| | 6 2 | 3+3 4 | 4+4 4 | 3+4 4 | 6+6 4 | 3+4 4 | 3+4 4 | 3+4 4 |
| | 5 2 | 3+3 4 | 3+3 4 | 3+4 4 | 5+5 4 | 3+4 4 | 3+4 4 | 3+4 4 |
| | 6 2 | 4+4 4 | 4+4 4 | 3+4 4 | 5+5 4 | 3+4 4 | 3+4 4 | 3+4 4 |
| | 6 2 | 4+4 4 | 4+4 4 | 3+4 4 | 6+6 4 | 3+4 4 | 3+4 4 | 3+4 4 |
| | 5 2 | 3+3 4 | 3+3 4 | 3+4 4 | 5+5 4 | 3+4 4 | 3+4 4 | 3+4 4 |
| | 6 2 | 3+3 4 | 3+3 4 | 3+4 4 | 6+6 4 | 3+4 4 | 3+4 4 | 3+4 4 |
| | 5 2 | 4+4 4 | 4+4 4 | 3+4 4 | 5+5 4 | 3+4 4 | 3+4 4 | 3+4 4 |
| | 5 2 | 4+4 4 | 4+4 4 | 3+4 4 | 5+5 4 | 3+4 4 | 3+4 4 | 3+4 4 |
| | 6 2 | 3+3 4 | 3+3 4 | 3+4 4 | 6+6 4 | 3+4 4 | 3+4 4 | 3+4 4 |
| | 6 2 | 3+3 4 | 3+3 4 | 3+4 4 | 6+6 4 | 3+4 4 | 3+4 4 | 3+4 4 |
| | 6 2 | 4+4 4 | 4+4 4 | 3+4 4 | 6+6 4 | 3+4 4 | 3+4 4 | 3+4 4 |
| | 6 2 | 3+3 4 | 3+3 4 | 3+4 4 | 5+5 4 | 3+4 4 | 3+4 4 | 3+4 4 |
| | 5 2 | 4+4 4 | 4+4 4 | 3+4 4 | 5+5 4 | 3+4 4 | 3+4 4 | 3+4 4 |
| | 5 2 | 4+4 4 | 4+4 4 | 3+4 4 | 5+5 4 | 3+4 4 | 3+4 4 | 3+4 4 |
| | 6 2 | 3+3 4 | 3+3 4 | 3+4 4 | 5+5 4 | 3+4 4 | 3+4 4 | 3+4 4 |
| | 5 2 | 4+4 4 | 4+4 4 | 3+4 4 | 5+5 4 | 3+4 4 | 3+4 4 | 3+4 4 |
| | 6 2 | 3+3 4 | 3+3 4 | 3+4 4 | 6+6 4 | 3+4 4 | 3+4 4 | 3+4 4 |
| | 5 2 | 3+3 4 | 4+4 4 | 3+4 4 | 5+5 4 | 3+4 4 | 3+4 4 | 3+4 4 |
| | 6 2 | 4+4 4 | 3+3 4 | 3+4 4 | 6+6 4 | 3+4 4 | 3+4 4 | 3+4 4 |
| | 5 2 | 3+3 4 | 3+3 4 | 3+4 4 | 5+5 4 | 3+4 4 | 3+4 4 | 3+4 4 |
| | 6 2 | 4+4 4 | 4+4 4 | 3+4 4 | 5+5 4 | 3+4 4 | 3+4 4 | 3+4 4 |
| | 5 2 | 4+4 4 | 3+3 4 | 3+4 4 | 5+5 4 | 3+4 4 | 3+4 4 | 3+4 4 |
| | 6 2 | 3+3 4 | 4+4 4 | 3+4 4 | 6+6 4 | 3+4 4 | 3+4 4 | 3+4 4 |
| | 5 2 | 4+4 4 | 4+4 4 | 3+4 4 | 5+5 4 | 3+4 4 | 3+4 4 | 3+4 4 |
| | 5 2 | 3+3 4 | 3+3 4 | 3+4 4 | 5+5 4 | 3+4 4 | 3+4 4 | 3+4 4 |
| | 5 2 | 4+4 4 | 4+4 4 | 3+4 4 | 5+5 4 | 3+4 4 | 3+4 4 | 3+4 4 |
| | 5 2 | 3+3 4 | 3+3 4 | 3+4 4 | 6+6 4 | 3+4 4 | 3+4 4 | 3+4 4 |
| | 5 2 | | | | | | | |
| | 5 2 | | | | | | | |
| | 6 2 | | | | | | | |
| | 6 2 | | | | | | | |
| | 5 2 | | | | | | | |
| | 6 2 | | | | | | | |
| | 6 2 | | | | | | | |
| | 5 2 | | | | | | | |
| | 5 2 | | | | | | | |
| | 6 2 | | | | | | | |
| | 6 2 | | | | | | | |
| | 6 2 | | | | | | | |
| | 5 2 | | | | | | | |
| | 6 2 | | | | | | | |
| | 6 2 | | | | | | | |
| | 6 2 | | | | | | | |
| | 5 2 | | | | | | | |
| | 5 2 | | | | | | | |
| | 5 2 | | | | | | | |
| | 5 2 | | | | | | | |
| | 5 2 | | | | | | | |
| | 5 2 | | | | | | | |
| | 5 2 | | | | | | | |
| | 5 2 | | | | | | | |
| | 5 2 | | | | | | | |
| | 5 2 | | | | | | | |
| | 6 2 | | | | | | | |
| | 5 2 | | | | | | | |
| | 5 2 | | | | | | | |
| | 6 2 | | | | | | | |
| | 5 2 | | | | | | | |
| | 6 2 | | | | | | | |
| | 6 2 | | | | | | | |
| | 5 2 | | | | | | | |
| | 6 2 | | | | | | | |

Continued on next column

Continued on next column

**TABLE XXXVII– continued**

| TQC | ICSD | NoE | SF | | | | | |
|---|---|---|---|---|---|---|---|---|
| | 5 2 | | | | | | | |
| | 5 2 | | | | | | | |
| | 5 2 | | | | | | | |
| | 5 2 | | | | | | | |
| | 6 2 | | | | | | | |
| | 5 2 | | | | | | | |
| | 6 2 | | | | | | | |
| | 5 2 | | | | | | | |
| | 6 2 | | | | | | | |
| | 5 2 | | | | | | | |
| | 5 2 | | | | | | | |
| | 6 2 | | | | | | | |
| | 5 2 | | | | | | | |
| 62.2.40 | 100644 | 192 | Ni2O4Si1 | | | | | |
| | -GM | -R | -S | -T | -U | -X | -Y | -Z |
| | 5 2 | 3+3 4 | 3+3 4 | 3+4 4 | 5+5 4 | 3+4 4 | 3+4 4 | 3+4 4 |
| | 5 2 | 4+4 4 | 4+4 4 | 3+4 4 | 6+6 4 | 3+4 4 | 3+4 4 | 3+4 4 |
| | 6 2 | 3+3 4 | 3+3 4 | 3+4 4 | 6+6 4 | 3+4 4 | 3+4 4 | 3+4 4 |
| | 6 2 | 4+4 4 | 4+4 4 | 3+4 4 | 5+5 4 | 3+4 4 | 3+4 4 | 3+4 4 |
| | 6 2 | 3+3 4 | 3+3 4 | 3+4 4 | 6+6 4 | 3+4 4 | 3+4 4 | 3+4 4 |
| | 6 2 | 4+4 4 | 4+4 4 | 3+4 4 | 5+5 4 | 3+4 4 | 3+4 4 | 3+4 4 |
| | 5 2 | 3+3 4 | 3+3 4 | 3+4 4 | 6+6 4 | 3+4 4 | 3+4 4 | 3+4 4 |
| | 6 2 | 4+4 4 | 4+4 4 | 3+4 4 | 5+5 4 | 3+4 4 | 3+4 4 | 3+4 4 |
| | 5 2 | 3+3 4 | 3+3 4 | 3+4 4 | 6+6 4 | 3+4 4 | 3+4 4 | 3+4 4 |
| | 5 2 | 4+4 4 | 4+4 4 | 3+4 4 | 5+5 4 | 3+4 4 | 3+4 4 | 3+4 4 |
| | 6 2 | 3+3 4 | 3+3 4 | 3+4 4 | 5+5 4 | 3+4 4 | 3+4 4 | 3+4 4 |
| | 5 2 | 4+4 4 | 4+4 4 | 3+4 4 | 5+5 4 | 3+4 4 | 3+4 4 | 3+4 4 |
| | 6 2 | 3+3 4 | 3+3 4 | 3+4 4 | 6+6 4 | 3+4 4 | 3+4 4 | 3+4 4 |
| | 5 2 | 4+4 4 | 4+4 4 | 3+4 4 | 6+6 4 | 3+4 4 | 3+4 4 | 3+4 4 |
| | 5 2 | 3+3 4 | 4+4 4 | 3+4 4 | 6+6 4 | 3+4 4 | 3+4 4 | 3+4 4 |
| | 6 2 | 4+4 4 | 3+3 4 | 3+4 4 | 5+5 4 | 3+4 4 | 3+4 4 | 3+4 4 |
| | 5 2 | 3+3 4 | 3+3 4 | 3+4 4 | 6+6 4 | 3+4 4 | 3+4 4 | 3+4 4 |
| | 5 2 | 4+4 4 | 4+4 4 | 3+4 4 | 6+6 4 | 3+4 4 | 3+4 4 | 3+4 4 |
| | 5 2 | 3+3 4 | 3+3 4 | 3+4 4 | 6+6 4 | 3+4 4 | 3+4 4 | 3+4 4 |
| | 6 2 | 4+4 4 | 4+4 4 | 3+4 4 | 5+5 4 | 3+4 4 | 3+4 4 | 3+4 4 |
| | 5 2 | 3+3 4 | 3+3 4 | 3+4 4 | 5+5 4 | 3+4 4 | 3+4 4 | 3+4 4 |
| | 6 2 | 4+4 4 | 4+4 4 | 3+4 4 | 5+5 4 | 3+4 4 | 3+4 4 | 3+4 4 |
| | 5 2 | 4+4 4 | 4+4 4 | 3+4 4 | 6+6 4 | 3+4 4 | 3+4 4 | 3+4 4 |
| | 6 2 | 3+3 4 | 3+3 4 | 3+4 4 | 6+6 4 | 3+4 4 | 3+4 4 | 3+4 4 |
| | 6 2 | 3+3 4 | 3+3 4 | 3+4 4 | 5+5 4 | 3+4 4 | 3+4 4 | 3+4 4 |
| | 5 2 | 4+4 4 | 4+4 4 | 3+4 4 | 6+6 4 | 3+4 4 | 3+4 4 | 3+4 4 |
| | 6 2 | 4+4 4 | 3+3 4 | 3+4 4 | 5+5 4 | 3+4 4 | 3+4 4 | 3+4 4 |
| | 5 2 | 3+3 4 | 4+4 4 | 3+4 4 | 6+6 4 | 3+4 4 | 3+4 4 | 3+4 4 |
| | 5 2 | 3+3 4 | 3+3 4 | 3+4 4 | 6+6 4 | 3+4 4 | 3+4 4 | 3+4 4 |
| | 6 2 | 4+4 4 | 4+4 4 | 3+4 4 | 5+5 4 | 3+4 4 | 3+4 4 | 3+4 4 |
| | 6 2 | 3+3 4 | 4+4 4 | 3+4 4 | 5+5 4 | 3+4 4 | 3+4 4 | 3+4 4 |
| | 6 2 | 4+4 4 | 3+3 4 | 3+4 4 | 5+5 4 | 3+4 4 | 3+4 4 | 3+4 4 |
| | 6 2 | 4+4 4 | 4+4 4 | 3+4 4 | 6+6 4 | 3+4 4 | 3+4 4 | 3+4 4 |
| | 5 2 | 3+3 4 | 3+3 4 | 3+4 4 | 6+6 4 | 3+4 4 | 3+4 4 | 3+4 4 |
| | 5 2 | 3+3 4 | 3+3 4 | 3+4 4 | 6+6 4 | 3+4 4 | 3+4 4 | 3+4 4 |
| | 6 2 | 4+4 4 | 4+4 4 | 3+4 4 | 6+6 4 | 3+4 4 | 3+4 4 | 3+4 4 |
| | 5 2 | 3+3 4 | 3+3 4 | 3+4 4 | 6+6 4 | 3+4 4 | 3+4 4 | 3+4 4 |
| | 6 2 | 4+4 4 | 4+4 4 | 3+4 4 | 5+5 4 | 3+4 4 | 3+4 4 | 3+4 4 |
| | 5 2 | 4+4 4 | 3+3 4 | 3+4 4 | 6+6 4 | 3+4 4 | 3+4 4 | 3+4 4 |
| | 5 2 | 3+3 4 | 4+4 4 | 3+4 4 | 5+5 4 | 3+4 4 | 3+4 4 | 3+4 4 |
| | 6 2 | 4+4 4 | 3+3 4 | 3+4 4 | 6+6 4 | 3+4 4 | 3+4 4 | 3+4 4 |
| | 6 2 | 3+3 4 | 4+4 4 | 3+4 4 | 6+6 4 | 3+4 4 | 3+4 4 | 3+4 4 |
| | 5 2 | 3+3 4 | 4+4 4 | 3+4 4 | 5+5 4 | 3+4 4 | 3+4 4 | 3+4 4 |
| | 6 2 | 4+4 4 | 3+3 4 | 3+4 4 | 6+6 4 | 3+4 4 | 3+4 4 | 3+4 4 |
| | 5 2 | 3+3 4 | 3+3 4 | 3+4 4 | 6+6 4 | 3+4 4 | 3+4 4 | 3+4 4 |
| | 5 2 | 4+4 4 | 4+4 4 | 3+4 4 | 6+6 4 | 3+4 4 | 3+4 4 | 3+4 4 |

Continued on next column

| TQC | ICSD | NoE | SF | | | | | |
|---|---|---|---|---|---|---|---|---|
| | 5 2 | 3+3 4 | 3+3 4 | 3+4 4 | 6+6 4 | 3+4 4 | 3+4 4 | 3+4 4 |
| | 5 2 | 4+4 4 | 4+4 4 | 3+4 4 | 5+5 4 | 3+4 4 | 3+4 4 | 3+4 4 |
| | 5 2 | | | | | | | |
| | 6 2 | | | | | | | |
| | 5 2 | | | | | | | |
| | 6 2 | | | | | | | |
| | 5 2 | | | | | | | |
| | 6 2 | | | | | | | |
| | 6 2 | | | | | | | |
| | 5 2 | | | | | | | |
| | 5 2 | | | | | | | |
| | 6 2 | | | | | | | |
| | 6 2 | | | | | | | |
| | 6 2 | | | | | | | |
| | 5 2 | | | | | | | |
| | 6 2 | | | | | | | |
| | 6 2 | | | | | | | |
| | 6 2 | | | | | | | |
| | 5 2 | | | | | | | |
| | 5 2 | | | | | | | |
| | 5 2 | | | | | | | |
| | 5 2 | | | | | | | |
| | 5 2 | | | | | | | |
| | 5 2 | | | | | | | |
| | 5 2 | | | | | | | |
| | 5 2 | | | | | | | |
| | 5 2 | | | | | | | |
| | 5 2 | | | | | | | |
| | 6 2 | | | | | | | |
| | 5 2 | | | | | | | |
| | 5 2 | | | | | | | |
| | 6 2 | | | | | | | |
| | 5 2 | | | | | | | |
| | 6 2 | | | | | | | |
| | 6 2 | | | | | | | |
| | 5 2 | | | | | | | |
| | 6 2 | | | | | | | |
| | 5 2 | | | | | | | |
| | 5 2 | | | | | | | |
| | 5 2 | | | | | | | |
| | 5 2 | | | | | | | |
| | 6 2 | | | | | | | |
| | 5 2 | | | | | | | |
| | 6 2 | | | | | | | |
| | 5 2 | | | | | | | |
| | 6 2 | | | | | | | |
| | 5 2 | | | | | | | |
| | 5 2 | | | | | | | |
| | 6 2 | | | | | | | |
| | 5 2 | | | | | | | |
| 62.2.41 | 100643 | 192 | Ni2O4Si1 | | | | | |
| | -GM | -R | -S | -T | -U | -X | -Y | -Z |
| | 5 2 | 3+3 4 | 3+3 4 | 3+4 4 | 5+5 4 | 3+4 4 | 3+4 4 | 3+4 4 |
| | 5 2 | 4+4 4 | 4+4 4 | 3+4 4 | 6+6 4 | 3+4 4 | 3+4 4 | 3+4 4 |
| | 6 2 | 3+3 4 | 3+3 4 | 3+4 4 | 6+6 4 | 3+4 4 | 3+4 4 | 3+4 4 |
| | 6 2 | 4+4 4 | 4+4 4 | 3+4 4 | 5+5 4 | 3+4 4 | 3+4 4 | 3+4 4 |
| | 6 2 | 3+3 4 | 3+3 4 | 3+4 4 | 6+6 4 | 3+4 4 | 3+4 4 | 3+4 4 |
| | 6 2 | 4+4 4 | 4+4 4 | 3+4 4 | 5+5 4 | 3+4 4 | 3+4 4 | 3+4 4 |
| | 5 2 | 3+3 4 | 3+3 4 | 3+4 4 | 6+6 4 | 3+4 4 | 3+4 4 | 3+4 4 |
| | 6 2 | 4+4 4 | 4+4 4 | 3+4 4 | 5+5 4 | 3+4 4 | 3+4 4 | 3+4 4 |
| | 5 2 | 3+3 4 | 3+3 4 | 3+4 4 | 6+6 4 | 3+4 4 | 3+4 4 | 3+4 4 |

Continued on next column

**TABLE XXXVII– continued**

| TQC | ICSD NoE SF |
|---|---|
| | 5 2 4+4 4 4+4 4 3+4 4 5+5 4 3+4 4 3+4 4 3+4 4 |
| | 6 2 3+3 4 3+3 4 3+4 4 5+5 4 3+4 4 3+4 4 3+4 4 |
| | 6 2 4+4 4 4+4 4 3+4 4 5+5 4 3+4 4 3+4 4 3+4 4 |
| | 5 2 3+3 4 3+3 4 3+4 4 6+6 4 3+4 4 3+4 4 3+4 4 |
| | 5 2 4+4 4 4+4 4 3+4 4 6+6 4 3+4 4 3+4 4 3+4 4 |
| | 5 2 3+3 4 4+4 4 3+4 4 6+6 4 3+4 4 3+4 4 3+4 4 |
| | 6 2 4+4 4 3+3 4 3+4 4 5+5 4 3+4 4 3+4 4 3+4 4 |
| | 5 2 3+3 4 3+3 4 3+4 4 6+6 4 3+4 4 3+4 4 3+4 4 |
| | 5 2 4+4 4 4+4 4 3+4 4 6+6 4 3+4 4 3+4 4 3+4 4 |
| | 5 2 4+4 4 3+3 4 3+4 4 6+6 4 3+4 4 3+4 4 3+4 4 |
| | 5 2 3+3 4 4+4 4 3+4 4 5+5 4 3+4 4 3+4 4 3+4 4 |
| | 6 2 4+4 4 3+3 4 3+4 4 5+5 4 3+4 4 3+4 4 3+4 4 |
| | 6 2 3+3 4 4+4 4 3+4 4 5+5 4 3+4 4 3+4 4 3+4 4 |
| | 5 2 3+3 4 4+4 4 3+4 4 6+6 4 3+4 4 3+4 4 3+4 4 |
| | 6 2 4+4 4 3+3 4 3+4 4 6+6 4 3+4 4 3+4 4 3+4 4 |
| | 6 2 4+4 4 3+3 4 3+4 4 5+5 4 3+4 4 3+4 4 3+4 4 |
| | 5 2 3+3 4 4+4 4 3+4 4 6+6 4 3+4 4 3+4 4 3+4 4 |
| | 6 2 3+3 4 3+3 4 3+4 4 5+5 4 3+4 4 3+4 4 3+4 4 |
| | 5 2 4+4 4 4+4 4 3+4 4 6+6 4 3+4 4 3+4 4 3+4 4 |
| | 5 2 4+4 4 3+3 4 3+4 4 6+6 4 3+4 4 3+4 4 3+4 4 |
| | 6 2 3+3 4 4+4 4 3+4 4 5+5 4 3+4 4 3+4 4 3+4 4 |
| | 6 2 3+3 4 4+4 4 3+4 4 5+5 4 3+4 4 3+4 4 3+4 4 |
| | 6 2 4+4 4 3+3 4 3+4 4 5+5 4 3+4 4 3+4 4 3+4 4 |
| | 6 2 3+3 4 4+4 4 3+4 4 6+6 4 3+4 4 3+4 4 3+4 4 |
| | 5 2 4+4 4 3+3 4 3+4 4 6+6 4 3+4 4 3+4 4 3+4 4 |
| | 5 2 4+4 4 3+3 4 3+4 4 6+6 4 3+4 4 3+4 4 3+4 4 |
| | 5 2 3+3 4 4+4 4 3+4 4 6+6 4 3+4 4 3+4 4 3+4 4 |
| | 6 2 4+4 4 3+3 4 3+4 4 6+6 4 3+4 4 3+4 4 3+4 4 |
| | 5 2 3+3 4 4+4 4 3+4 4 5+5 4 3+4 4 3+4 4 3+4 4 |
| | 6 2 3+3 4 3+3 4 3+4 4 6+6 4 3+4 4 3+4 4 3+4 4 |
| | 5 2 4+4 4 4+4 4 3+4 4 5+5 4 3+4 4 3+4 4 3+4 4 |
| | 6 2 3+3 4 3+3 4 3+4 4 6+6 4 3+4 4 3+4 4 3+4 4 |
| | 6 2 4+4 4 4+4 4 3+4 4 6+6 4 3+4 4 3+4 4 3+4 4 |
| | 5 2 4+4 4 4+4 4 3+4 4 6+6 4 3+4 4 3+4 4 3+4 4 |
| | 6 2 3+3 4 3+3 4 3+4 4 5+5 4 3+4 4 3+4 4 3+4 4 |
| | 5 2 4+4 4 3+3 4 3+4 4 6+6 4 3+4 4 3+4 4 3+4 4 |
| | 5 2 3+3 4 4+4 4 3+4 4 6+6 4 3+4 4 3+4 4 3+4 4 |
| | 5 2 4+4 4 3+3 4 3+4 4 6+6 4 3+4 4 3+4 4 3+4 4 |
| | 5 2 3+3 4 4+4 4 3+4 4 5+5 4 3+4 4 3+4 4 3+4 4 |
| | 5 2 |
| | 6 2 |
| | 5 2 |
| | 6 2 |
| | 5 2 |
| | 6 2 |
| | 6 2 |
| | 5 2 |
| | 5 2 |
| | 6 2 |
| | 6 2 |
| | 6 2 |
| | 5 2 |
| | 6 2 |
| | 6 2 |
| | 6 2 |
| | 5 2 |
| | 5 2 |
| | 5 2 |
| | 5 2 |
| | 5 2 |
| | 5 2 |
| | 5 2 |
| | 5 2 |
| | 5 2 |
| | 5 2 |
| | 6 2 |
| | 5 2 |
| | 5 2 |
| | 6 2 |
| | 5 2 |
| | 6 2 |
| | 6 2 |
| | 5 2 |
| | 6 2 |
| | 5 2 |
| | 5 2 |
| | 5 2 |
| | 5 2 |
| | 6 2 |
| | 5 2 |
| | 5 2 |
| | 6 2 |
| | 6 2 |
| | 5 2 |
| | 5 2 |
| | 6 2 |
| | 5 2 |
| 62.2.42 | 100642 192 Ni2O4Si1 |
| | -GM -R -S -T -U -X -Y -Z |
| | 5 2 3+3 4 3+3 4 3+4 4 6+6 4 3+4 4 3+4 4 3+4 4 |
| | 5 2 4+4 4 4+4 4 3+4 4 5+5 4 3+4 4 3+4 4 3+4 4 |
| | 6 2 3+3 4 3+3 4 3+4 4 5+5 4 3+4 4 3+4 4 3+4 4 |
| | 6 2 4+4 4 4+4 4 3+4 4 6+6 4 3+4 4 3+4 4 3+4 4 |
| | 6 2 3+3 4 3+3 4 3+4 4 5+5 4 3+4 4 3+4 4 3+4 4 |
| | 6 2 4+4 4 4+4 4 3+4 4 6+6 4 3+4 4 3+4 4 3+4 4 |
| | 5 2 3+3 4 3+3 4 3+4 4 5+5 4 3+4 4 3+4 4 3+4 4 |
| | 6 2 4+4 4 4+4 4 3+4 4 6+6 4 3+4 4 3+4 4 3+4 4 |
| | 5 2 3+3 4 3+3 4 3+4 4 5+5 4 3+4 4 3+4 4 3+4 4 |
| | 5 2 4+4 4 4+4 4 3+4 4 6+6 4 3+4 4 3+4 4 3+4 4 |
| | 6 2 3+3 4 3+3 4 3+4 4 6+6 4 3+4 4 3+4 4 3+4 4 |
| | 5 2 4+4 4 4+4 4 3+4 4 6+6 4 3+4 4 3+4 4 3+4 4 |
| | 6 2 3+3 4 3+3 4 3+4 4 5+5 4 3+4 4 3+4 4 3+4 4 |
| | 5 2 4+4 4 4+4 4 3+4 4 5+5 4 3+4 4 3+4 4 3+4 4 |
| | 5 2 3+3 4 3+3 4 3+4 4 5+5 4 3+4 4 3+4 4 3+4 4 |
| | 6 2 4+4 4 4+4 4 3+4 4 6+6 4 3+4 4 3+4 4 3+4 4 |
| | 5 2 3+3 4 3+3 4 3+4 4 5+5 4 3+4 4 3+4 4 3+4 4 |
| | 5 2 4+4 4 4+4 4 3+4 4 5+5 4 3+4 4 3+4 4 3+4 4 |
| | 5 2 3+3 4 4+4 4 3+4 4 5+5 4 3+4 4 3+4 4 3+4 4 |
| | 5 2 4+4 4 3+3 4 3+4 4 6+6 4 3+4 4 3+4 4 3+4 4 |
| | 6 2 4+4 4 3+3 4 3+4 4 6+6 4 3+4 4 3+4 4 3+4 4 |
| | 6 2 3+3 4 4+4 4 3+4 4 6+6 4 3+4 4 3+4 4 3+4 4 |
| | 5 2 3+3 4 3+3 4 3+4 4 5+5 4 3+4 4 3+4 4 3+4 4 |
| | 6 2 4+4 4 4+4 4 3+4 4 5+5 4 3+4 4 3+4 4 3+4 4 |
| | 6 2 4+4 4 4+4 4 3+4 4 6+6 4 3+4 4 3+4 4 3+4 4 |
| | 5 2 3+3 4 3+3 4 3+4 4 5+5 4 3+4 4 3+4 4 3+4 4 |
| | 6 2 3+3 4 3+3 4 3+4 4 6+6 4 3+4 4 3+4 4 3+4 4 |
| | 5 2 4+4 4 4+4 4 3+4 4 5+5 4 3+4 4 3+4 4 3+4 4 |
| | 5 2 4+4 4 4+4 4 3+4 4 5+5 4 3+4 4 3+4 4 3+4 4 |
| | 6 2 3+3 4 3+3 4 3+4 4 6+6 4 3+4 4 3+4 4 3+4 4 |
| | 6 2 4+4 4 3+3 4 3+4 4 6+6 4 3+4 4 3+4 4 3+4 4 |
| | 6 2 3+3 4 4+4 4 3+4 4 6+6 4 3+4 4 3+4 4 3+4 4 |
| | 6 2 3+3 4 3+3 4 3+4 4 5+5 4 3+4 4 3+4 4 3+4 4 |

Continued on next column

**TABLE XXXVII– continued**

| TQC | ICSD | NoE | SF | | | | | |
|---|---|---|---|---|---|---|---|---|
| | 5 2 4+4 4 | 4+4 4 | 3+4 4 | 5+5 4 | 3+4 4 | 3+4 4 | 3+4 4 | |
| | 5 2 4+4 4 | 4+4 4 | 3+4 4 | 5+5 4 | 3+4 4 | 3+4 4 | 3+4 4 | |
| | 6 2 3+3 4 | 3+3 4 | 3+4 4 | 5+5 4 | 3+4 4 | 3+4 4 | 3+4 4 | |
| | 5 2 4+4 4 | 4+4 4 | 3+4 4 | 5+5 4 | 3+4 4 | 3+4 4 | 3+4 4 | |
| | 5 2 3+3 4 | 3+3 4 | 3+4 4 | 6+6 4 | 3+4 4 | 3+4 4 | 3+4 4 | |
| | 6 2 3+3 4 | 4+4 4 | 3+4 4 | 5+5 4 | 3+4 4 | 3+4 4 | 3+4 4 | |
| | 5 2 4+4 4 | 3+3 4 | 3+4 4 | 6+6 4 | 3+4 4 | 3+4 4 | 3+4 4 | |
| | 6 2 3+3 4 | 4+4 4 | 3+4 4 | 5+5 4 | 3+4 4 | 3+4 4 | 3+4 4 | |
| | 6 2 4+4 4 | 3+3 4 | 3+4 4 | 5+5 4 | 3+4 4 | 3+4 4 | 3+4 4 | |
| | 5 2 4+4 4 | 3+3 4 | 3+4 4 | 5+5 4 | 3+4 4 | 3+4 4 | 3+4 4 | |
| | 6 2 3+3 4 | 4+4 4 | 3+4 4 | 6+6 4 | 3+4 4 | 3+4 4 | 3+4 4 | |
| | 5 2 4+4 4 | 4+4 4 | 3+4 4 | 5+5 4 | 3+4 4 | 3+4 4 | 3+4 4 | |
| | 5 2 3+3 4 | 3+3 4 | 3+4 4 | 5+5 4 | 3+4 4 | 3+4 4 | 3+4 4 | |
| | 5 2 4+4 4 | 4+4 4 | 3+4 4 | 5+5 4 | 3+4 4 | 3+4 4 | 3+4 4 | |
| | 5 2 3+3 4 | 3+3 4 | 3+4 4 | 6+6 4 | 3+4 4 | 3+4 4 | 3+4 4 | |
| | 6 2 | | | | | | | |
| | 5 2 | | | | | | | |
| | 6 2 | | | | | | | |
| | 5 2 | | | | | | | |
| | 5 2 | | | | | | | |
| | 6 2 | | | | | | | |
| | 6 2 | | | | | | | |
| | 5 2 | | | | | | | |
| | 5 2 | | | | | | | |
| | 6 2 | | | | | | | |
| | 6 2 | | | | | | | |
| | 6 2 | | | | | | | |
| | 5 2 | | | | | | | |
| | 6 2 | | | | | | | |
| | 6 2 | | | | | | | |
| | 6 2 | | | | | | | |
| | 5 2 | | | | | | | |
| | 5 2 | | | | | | | |
| | 5 2 | | | | | | | |
| | 5 2 | | | | | | | |
| | 5 2 | | | | | | | |
| | 5 2 | | | | | | | |
| | 5 2 | | | | | | | |
| | 5 2 | | | | | | | |
| | 5 2 | | | | | | | |
| | 5 2 | | | | | | | |
| | 6 2 | | | | | | | |
| | 5 2 | | | | | | | |
| | 5 2 | | | | | | | |
| | 6 2 | | | | | | | |
| | 5 2 | | | | | | | |
| | 6 2 | | | | | | | |
| | 6 2 | | | | | | | |
| | 5 2 | | | | | | | |
| | 6 2 | | | | | | | |
| | 5 2 | | | | | | | |
| | 5 2 | | | | | | | |
| | 5 2 | | | | | | | |
| | 5 2 | | | | | | | |
| | 6 2 | | | | | | | |
| | 5 2 | | | | | | | |
| | 5 2 | | | | | | | |
| | 6 2 | | | | | | | |
| | 6 2 | | | | | | | |
| | 5 2 | | | | | | | |
| | 5 2 | | | | | | | |

Continued on next column

**TABLE XXXVII– continued**

| TQC | ICSD | NoE | SF | | | | | |
|---|---|---|---|---|---|---|---|---|
| | 6 2 | | | | | | | |
| | 5 2 | | | | | | | |
| 62.2.43 | 105403 | 176 | Ni1Th1 | | | | | |
| | -GM | -R | -S | -T | -U | -X | -Y | -Z |
| | 5 2 3+3 4 | 3+3 4 | 3+4 4 | 5+5 4 | 3+4 4 | 3+4 4 | 3+4 4 | |
| | 5 2 4+4 4 | 4+4 4 | 3+4 4 | 6+6 4 | 3+4 4 | 3+4 4 | 3+4 4 | |
| | 6 2 3+3 4 | 3+3 4 | 3+4 4 | 6+6 4 | 3+4 4 | 3+4 4 | 3+4 4 | |
| | 5 2 4+4 4 | 4+4 4 | 3+4 4 | 5+5 4 | 3+4 4 | 3+4 4 | 3+4 4 | |
| | 6 2 3+3 4 | 3+3 4 | 3+4 4 | 6+6 4 | 3+4 4 | 3+4 4 | 3+4 4 | |
| | 6 2 4+4 4 | 4+4 4 | 3+4 4 | 5+5 4 | 3+4 4 | 3+4 4 | 3+4 4 | |
| | 6 2 3+3 4 | 3+3 4 | 3+4 4 | 5+5 4 | 3+4 4 | 3+4 4 | 3+4 4 | |
| | 5 2 4+4 4 | 4+4 4 | 3+4 4 | 6+6 4 | 3+4 4 | 3+4 4 | 3+4 4 | |
| | 5 2 4+4 4 | 4+4 4 | 3+4 4 | 6+6 4 | 3+4 4 | 3+4 4 | 3+4 4 | |
| | 6 2 3+3 4 | 3+3 4 | 3+4 4 | 5+5 4 | 3+4 4 | 3+4 4 | 3+4 4 | |
| | 5 2 4+4 4 | 4+4 4 | 3+4 4 | 6+6 4 | 3+4 4 | 3+4 4 | 3+4 4 | |
| | 5 2 3+3 4 | 3+3 4 | 3+4 4 | 5+5 4 | 3+4 4 | 3+4 4 | 3+4 4 | |
| | 6 2 3+3 4 | 3+3 4 | 3+4 4 | 6+6 4 | 3+4 4 | 3+4 4 | 3+4 4 | |
| | 6 2 4+4 4 | 4+4 4 | 3+4 4 | 5+5 4 | 3+4 4 | 3+4 4 | 3+4 4 | |
| | 5 2 3+3 4 | 3+3 4 | 3+4 4 | 5+5 4 | 3+4 4 | 3+4 4 | 3+4 4 | |
| | 6 2 4+4 4 | 4+4 4 | 3+4 4 | 6+6 4 | 3+4 4 | 3+4 4 | 3+4 4 | |
| | 5 2 3+3 4 | 4+4 4 | 3+4 4 | 6+6 4 | 3+4 4 | 3+4 4 | 3+4 4 | |
| | 5 2 4+4 4 | 3+3 4 | 3+4 4 | 5+5 4 | 3+4 4 | 3+4 4 | 3+4 4 | |
| | 6 2 3+3 4 | 4+4 4 | 3+4 4 | 5+5 4 | 3+4 4 | 3+4 4 | 3+4 4 | |
| | 5 2 4+4 4 | 3+3 4 | 3+4 4 | 6+6 4 | 3+4 4 | 3+4 4 | 3+4 4 | |
| | 6 2 4+4 4 | 3+3 4 | 3+4 4 | 6+6 4 | 3+4 4 | 3+4 4 | 3+4 4 | |
| | 6 2 3+3 4 | 4+4 4 | 3+4 4 | 5+5 4 | 3+4 4 | 3+4 4 | 3+4 4 | |
| | 5 2 4+4 4 | 3+3 4 | 3+4 4 | 5+5 4 | 3+4 4 | 3+4 4 | 3+4 4 | |
| | 5 2 3+3 4 | 4+4 4 | 3+4 4 | 6+6 4 | 3+4 4 | 3+4 4 | 3+4 4 | |
| | 5 2 4+4 4 | 4+4 4 | 3+4 4 | 5+5 4 | 3+4 4 | 3+4 4 | 3+4 4 | |
| | 6 2 3+3 4 | 3+3 4 | 3+4 4 | 5+5 4 | 3+4 4 | 3+4 4 | 3+4 4 | |
| | 5 2 3+3 4 | 3+3 4 | 3+4 4 | 6+6 4 | 3+4 4 | 3+4 4 | 3+4 4 | |
| | 6 2 4+4 4 | 4+4 4 | 3+4 4 | 6+6 4 | 3+4 4 | 3+4 4 | 3+4 4 | |
| | 6 2 4+4 4 | 3+3 4 | 3+4 4 | 5+5 4 | 3+4 4 | 3+4 4 | 3+4 4 | |
| | 5 2 3+3 4 | 4+4 4 | 3+4 4 | 6+6 4 | 3+4 4 | 3+4 4 | 3+4 4 | |
| | 6 2 4+4 4 | 4+4 4 | 3+4 4 | 6+6 4 | 3+4 4 | 3+4 4 | 3+4 4 | |
| | 6 2 3+3 4 | 3+3 4 | 3+4 4 | 5+5 4 | 3+4 4 | 3+4 4 | 3+4 4 | |
| | 5 2 3+3 4 | 3+3 4 | 3+4 4 | 6+6 4 | 3+4 4 | 3+4 4 | 3+4 4 | |
| | 6 2 4+4 4 | 4+4 4 | 3+4 4 | 5+5 4 | 3+4 4 | 3+4 4 | 3+4 4 | |
| | 5 2 3+3 4 | 3+3 4 | 3+4 4 | 6+6 4 | 3+4 4 | 3+4 4 | 3+4 4 | |
| | 6 2 4+4 4 | 4+4 4 | 3+4 4 | 5+5 4 | 3+4 4 | 3+4 4 | 3+4 4 | |
| | 6 2 3+3 4 | 4+4 4 | 3+4 4 | 6+6 4 | 3+4 4 | 3+4 4 | 3+4 4 | |
| | 5 2 4+4 4 | 3+3 4 | 3+4 4 | 5+5 4 | 3+4 4 | 3+4 4 | 3+4 4 | |
| | 6 2 3+3 4 | 4+4 4 | 3+4 4 | 5+5 4 | 3+4 4 | 3+4 4 | 3+4 4 | |
| | 5 2 4+4 4 | 3+3 4 | 3+4 4 | 6+6 4 | 3+4 4 | 3+4 4 | 3+4 4 | |
| | 6 2 4+4 4 | 4+4 4 | 3+4 4 | 5+5 4 | 3+4 4 | 3+4 4 | 3+4 4 | |
| | 5 2 3+3 4 | 3+3 4 | 3+4 4 | 5+5 4 | 3+4 4 | 3+4 4 | 3+4 4 | |
| | 5 2 4+4 4 | 4+4 4 | 3+4 4 | 6+6 4 | 3+4 4 | 3+4 4 | 3+4 4 | |
| | 5 2 3+3 4 | 3+3 4 | 3+4 4 | 6+6 4 | 3+4 4 | 3+4 4 | 3+4 4 | |
| | 6 2 | | | | | | | |
| | 6 2 | | | | | | | |
| | 5 2 | | | | | | | |
| | 5 2 | | | | | | | |
| | 6 2 | | | | | | | |
| | 5 2 | | | | | | | |
| | 6 2 | | | | | | | |
| | 6 2 | | | | | | | |
| | 5 2 | | | | | | | |
| | 6 2 | | | | | | | |
| | 6 2 | | | | | | | |
| | 6 2 | | | | | | | |
| | 5 2 | | | | | | | |

Continued on next column

**TABLE XXXVII– continued**

| TQC | ICSD | NoE | SF | | | | | |
|---|---|---|---|---|---|---|---|---|
| | 6 2 | | | | | | | |
| | 5 2 | | | | | | | |
| | 5 2 | | | | | | | |
| | 6 2 | | | | | | | |
| | 5 2 | | | | | | | |
| | 6 2 | | | | | | | |
| | 5 2 | | | | | | | |
| | 5 2 | | | | | | | |
| | 6 2 | | | | | | | |
| | 5 2 | | | | | | | |
| | 5 2 | | | | | | | |
| | 6 2 | | | | | | | |
| | 6 2 | | | | | | | |
| | 6 2 | | | | | | | |
| | 5 2 | | | | | | | |
| | 5 2 | | | | | | | |
| | 6 2 | | | | | | | |
| | 6 2 | | | | | | | |
| | 5 2 | | | | | | | |
| | 6 2 | | | | | | | |
| | 5 2 | | | | | | | |
| | 6 2 | | | | | | | |
| | 5 2 | | | | | | | |
| | 6 2 | | | | | | | |
| | 5 2 | | | | | | | |
| | 6 2 | | | | | | | |
| | 6 2 | | | | | | | |
| | 5 2 | | | | | | | |
| | 6 2 | | | | | | | |
| | 6 2 | | | | | | | |
| | 5 2 | | | | | | | |
| 62.2.44 | 54360 | 152 | Ni1Sn1Yb1 | | | | | |
| | -GM | -R | -S | -T | -U | -X | -Y | -Z |
| | 5 2 | 3+3 4 | 3+3 4 | 3+4 4 | 5+5 4 | 3+4 4 | 3+4 4 | 3+4 4 |
| | 6 2 | 4+4 4 | 4+4 4 | 3+4 4 | 6+6 4 | 3+4 4 | 3+4 4 | 3+4 4 |
| | 5 2 | 3+3 4 | 3+3 4 | 3+4 4 | 5+5 4 | 3+4 4 | 3+4 4 | 3+4 4 |
| | 6 2 | 4+4 4 | 4+4 4 | 3+4 4 | 6+6 4 | 3+4 4 | 3+4 4 | 3+4 4 |
| | 5 2 | 4+4 4 | 4+4 4 | 3+4 4 | 5+5 4 | 3+4 4 | 3+4 4 | 3+4 4 |
| | 6 2 | 3+3 4 | 3+3 4 | 3+4 4 | 6+6 4 | 3+4 4 | 3+4 4 | 3+4 4 |
| | 5 2 | 3+3 4 | 3+3 4 | 3+4 4 | 5+5 4 | 3+4 4 | 3+4 4 | 3+4 4 |
| | 6 2 | 4+4 4 | 4+4 4 | 3+4 4 | 6+6 4 | 3+4 4 | 3+4 4 | 3+4 4 |
| | 5 2 | 3+3 4 | 3+3 4 | 3+4 4 | 6+6 4 | 3+4 4 | 3+4 4 | 3+4 4 |
| | 6 2 | 4+4 4 | 4+4 4 | 3+4 4 | 5+5 4 | 3+4 4 | 3+4 4 | 3+4 4 |
| | 5 2 | 4+4 4 | 4+4 4 | 3+4 4 | 5+5 4 | 3+4 4 | 3+4 4 | 3+4 4 |
| | 5 2 | 3+3 4 | 3+3 4 | 3+4 4 | 6+6 4 | 3+4 4 | 3+4 4 | 3+4 4 |
| | 6 2 | 4+4 4 | 4+4 4 | 3+4 4 | 6+6 4 | 3+4 4 | 3+4 4 | 3+4 4 |
| | 6 2 | 3+3 4 | 3+3 4 | 3+4 4 | 6+6 4 | 3+4 4 | 3+4 4 | 3+4 4 |
| | 5 2 | 4+4 4 | 3+3 4 | 3+4 4 | 5+5 4 | 3+4 4 | 3+4 4 | 3+4 4 |
| | 6 2 | 3+3 4 | 4+4 4 | 3+4 4 | 5+5 4 | 3+4 4 | 3+4 4 | 3+4 4 |
| | 5 2 | 3+3 4 | 4+4 4 | 3+4 4 | 6+6 4 | 3+4 4 | 3+4 4 | 3+4 4 |
| | 5 2 | 4+4 4 | 3+3 4 | 3+4 4 | 6+6 4 | 3+4 4 | 3+4 4 | 3+4 4 |
| | 6 2 | 4+4 4 | 3+3 4 | 3+4 4 | 5+5 4 | 3+4 4 | 3+4 4 | 3+4 4 |
| | 6 2 | 3+3 4 | 4+4 4 | 3+4 4 | 5+5 4 | 3+4 4 | 3+4 4 | 3+4 4 |
| | 6 2 | 3+3 4 | 3+3 4 | 3+4 4 | 6+6 4 | 3+4 4 | 3+4 4 | 3+4 4 |
| | 5 2 | 4+4 4 | 4+4 4 | 3+4 4 | 5+5 4 | 3+4 4 | 3+4 4 | 3+4 4 |
| | 5 2 | 3+3 4 | 4+4 4 | 3+4 4 | 6+6 4 | 3+4 4 | 3+4 4 | 3+4 4 |
| | 6 2 | 4+4 4 | 3+3 4 | 3+4 4 | 5+5 4 | 3+4 4 | 3+4 4 | 3+4 4 |
| | 5 2 | 3+3 4 | 4+4 4 | 3+4 4 | 6+6 4 | 3+4 4 | 3+4 4 | 3+4 4 |
| | 5 2 | 4+4 4 | 3+3 4 | 3+4 4 | 5+5 4 | 3+4 4 | 3+4 4 | 3+4 4 |
| | 5 2 | 4+4 4 | 3+3 4 | 3+4 4 | 6+6 4 | 3+4 4 | 3+4 4 | 3+4 4 |
| | 6 2 | 3+3 4 | 4+4 4 | 3+4 4 | 5+5 4 | 3+4 4 | 3+4 4 | 3+4 4 |

Continued on next column

**TABLE XXXVII– continued**

| TQC | ICSD | NoE | SF | | | | | |
|---|---|---|---|---|---|---|---|---|
| | 6 2 | 4+4 4 | 4+4 4 | 3+4 4 | 6+6 4 | 3+4 4 | 3+4 4 | 3+4 4 |
| | 5 2 | 3+3 4 | 3+3 4 | 3+4 4 | 6+6 4 | 3+4 4 | 3+4 4 | 3+4 4 |
| | 6 2 | 4+4 4 | 3+3 4 | 3+4 4 | 5+5 4 | 3+4 4 | 3+4 4 | 3+4 4 |
| | 6 2 | 3+3 4 | 4+4 4 | 3+4 4 | 5+5 4 | 3+4 4 | 3+4 4 | 3+4 4 |
| | 5 2 | 4+4 4 | 4+4 4 | 3+4 4 | 5+5 4 | 3+4 4 | 3+4 4 | 3+4 4 |
| | 6 2 | 3+3 4 | 3+3 4 | 3+4 4 | 6+6 4 | 3+4 4 | 3+4 4 | 3+4 4 |
| | 5 2 | 3+3 4 | 3+3 4 | 3+4 4 | 6+6 4 | 3+4 4 | 3+4 4 | 3+4 4 |
| | 5 2 | 4+4 4 | 4+4 4 | 3+4 4 | 5+5 4 | 3+4 4 | 3+4 4 | 3+4 4 |
| | 5 2 | 3+3 4 | 3+3 4 | 3+4 4 | 6+6 4 | 3+4 4 | 3+4 4 | 3+4 4 |
| | 5 2 | 4+4 4 | 4+4 4 | 3+4 4 | 6+6 4 | 3+4 4 | 3+4 4 | 3+4 4 |
| | 6 2 | | | | | | | |
| | 6 2 | | | | | | | |
| | 5 2 | | | | | | | |
| | 6 2 | | | | | | | |
| | 6 2 | | | | | | | |
| | 5 2 | | | | | | | |
| | 6 2 | | | | | | | |
| | 5 2 | | | | | | | |
| | 6 2 | | | | | | | |
| | 6 2 | | | | | | | |
| | 5 2 | | | | | | | |
| | 5 2 | | | | | | | |
| | 6 2 | | | | | | | |
| | 5 2 | | | | | | | |
| | 6 2 | | | | | | | |
| | 5 2 | | | | | | | |
| | 6 2 | | | | | | | |
| | 5 2 | | | | | | | |
| | 5 2 | | | | | | | |
| | 6 2 | | | | | | | |
| | 6 2 | | | | | | | |
| | 6 2 | | | | | | | |
| | 5 2 | | | | | | | |
| | 5 2 | | | | | | | |
| | 6 2 | | | | | | | |
| | 6 2 | | | | | | | |
| | 5 2 | | | | | | | |
| | 6 2 | | | | | | | |
| | 6 2 | | | | | | | |
| | 5 2 | | | | | | | |
| | 5 2 | | | | | | | |
| | 6 2 | | | | | | | |
| | 6 2 | | | | | | | |
| | 5 2 | | | | | | | |
| | 6 2 | | | | | | | |
| | 5 2 | | | | | | | |
| | 5 2 | | | | | | | |
| | 5 2 | | | | | | | |
| 62.2.45 | 18188 | 72 | Ni1Si1Ti1 | | | | | |
| | -GM | -R | -S | -T | -U | -X | -Y | -Z |
| | 5 2 | 3+3 4 | 3+3 4 | 3+4 4 | 6+6 4 | 3+4 4 | 3+4 4 | 3+4 4 |
| | 5 2 | 4+4 4 | 4+4 4 | 3+4 4 | 5+5 4 | 3+4 4 | 3+4 4 | 3+4 4 |
| | 6 2 | 3+3 4 | 4+4 4 | 3+4 4 | 6+6 4 | 3+4 4 | 3+4 4 | 3+4 4 |
| | 6 2 | 4+4 4 | 3+3 4 | 3+4 4 | 5+5 4 | 3+4 4 | 3+4 4 | 3+4 4 |
| | 5 2 | 4+4 4 | 3+3 4 | 3+4 4 | 5+5 4 | 3+4 4 | 3+4 4 | 3+4 4 |
| | 6 2 | 3+3 4 | 4+4 4 | 3+4 4 | 5+5 4 | 3+4 4 | 3+4 4 | 3+4 4 |
| | 5 2 | 4+4 4 | 4+4 4 | 3+4 4 | 6+6 4 | 3+4 4 | 3+4 4 | 3+4 4 |
| | 6 2 | 3+3 4 | 3+3 4 | 3+4 4 | 6+6 4 | 3+4 4 | 3+4 4 | 3+4 4 |
| | 5 2 | 4+4 4 | 4+4 4 | 3+4 4 | 6+6 4 | 3+4 4 | 3+4 4 | 3+4 4 |
| | 6 2 | 3+3 4 | 3+3 4 | 3+4 4 | 5+5 4 | 3+4 4 | 3+4 4 | 3+4 4 |
| | 5 2 | 3+3 4 | 4+4 4 | 3+4 4 | 6+6 4 | 3+4 4 | 3+4 4 | 3+4 4 |

Continued on next column

**TABLE XXXVII– continued**

| TQC | ICSD | NoE | SF | | | | | |
|---|---|---|---|---|---|---|---|---|
| | 6 2 | 4+4 4 | 3+3 4 | 3+4 4 | 5+5 4 | 3+4 4 | 3+4 4 | 3+4 4 |
| | 6 2 | 4+4 4 | 4+4 4 | 3+4 4 | 5+5 4 | 3+4 4 | 3+4 4 | 3+4 4 |
| | 6 2 | 3+3 4 | 3+3 4 | 3+4 4 | 6+6 4 | 3+4 4 | 3+4 4 | 3+4 4 |
| | 6 2 | 4+4 4 | 3+3 4 | 3+4 4 | 6+6 4 | 3+4 4 | 3+4 4 | 3+4 4 |
| | 5 2 | 3+3 4 | 4+4 4 | 3+4 4 | 6+6 4 | 3+4 4 | 3+4 4 | 3+4 4 |
| | 6 2 | 3+3 4 | 4+4 4 | 3+4 4 | 5+5 4 | 3+4 4 | 3+4 4 | 3+4 4 |
| | 5 2 | 4+4 4 | 3+3 4 | 3+4 4 | 6+6 4 | 3+4 4 | 3+4 4 | 3+4 4 |
| | 5 2 | | | | | | | |
| | 5 2 | | | | | | | |
| | 6 2 | | | | | | | |
| | 5 2 | | | | | | | |
| | 5 2 | | | | | | | |
| | 6 2 | | | | | | | |
| | 5 2 | | | | | | | |
| | 5 2 | | | | | | | |
| | 6 2 | | | | | | | |
| | 5 2 | | | | | | | |
| | 6 2 | | | | | | | |
| | 6 2 | | | | | | | |
| | 5 2 | | | | | | | |
| | 5 2 | | | | | | | |
| | 5 2 | | | | | | | |
| | 6 2 | | | | | | | |
| | 5 2 | | | | | | | |
| | 6 2 | | | | | | | |
| 62.2.46 | 95449 | 56 | Ni1Si1 | | | | | |
| | -GM | -R | -S | -T | -U | -X | -Y | -Z |
| | 5 2 | 3+3 4 | 4+4 4 | 3+4 4 | 6+6 4 | 3+4 4 | 3+4 4 | 3+4 4 |
| | 5 2 | 4+4 4 | 3+3 4 | 3+4 4 | 5+5 4 | 3+4 4 | 3+4 4 | 3+4 4 |
| | 6 2 | 4+4 4 | 4+4 4 | 3+4 4 | 6+6 4 | 3+4 4 | 3+4 4 | 3+4 4 |
| | 6 2 | 3+3 4 | 3+3 4 | 3+4 4 | 5+5 4 | 3+4 4 | 3+4 4 | 3+4 4 |
| | 5 2 | 4+4 4 | 3+3 4 | 3+4 4 | 6+6 4 | 3+4 4 | 3+4 4 | 3+4 4 |
| | 6 2 | 3+3 4 | 4+4 4 | 3+4 4 | 6+6 4 | 3+4 4 | 3+4 4 | 3+4 4 |
| | 6 2 | 3+3 4 | 4+4 4 | 3+4 4 | 5+5 4 | 3+4 4 | 3+4 4 | 3+4 4 |
| | 5 2 | 4+4 4 | 3+3 4 | 3+4 4 | 6+6 4 | 3+4 4 | 3+4 4 | 3+4 4 |
| | 5 2 | 3+3 4 | 3+3 4 | 3+4 4 | 5+5 4 | 3+4 4 | 3+4 4 | 3+4 4 |
| | 5 2 | 4+4 4 | 4+4 4 | 3+4 4 | 5+5 4 | 3+4 4 | 3+4 4 | 3+4 4 |
| | 6 2 | 4+4 4 | 3+3 4 | 3+4 4 | 6+6 4 | 3+4 4 | 3+4 4 | 3+4 4 |
| | 6 2 | 3+3 4 | 4+4 4 | 3+4 4 | 5+5 4 | 3+4 4 | 3+4 4 | 3+4 4 |
| | 5 2 | 3+3 4 | 4+4 4 | 3+4 4 | 5+5 4 | 3+4 4 | 3+4 4 | 3+4 4 |
| | 5 2 | 4+4 4 | 3+3 4 | 3+4 4 | 6+6 4 | 3+4 4 | 3+4 4 | 3+4 4 |
| | 5 2 | | | | | | | |
| | 5 2 | | | | | | | |
| | 6 2 | | | | | | | |
| | 6 2 | | | | | | | |
| | 6 2 | | | | | | | |
| | 5 2 | | | | | | | |
| | 6 2 | | | | | | | |
| | 6 2 | | | | | | | |
| | 6 2 | | | | | | | |
| | 5 2 | | | | | | | |
| | 5 2 | | | | | | | |
| | 5 2 | | | | | | | |
| | 5 2 | | | | | | | |
| | 5 2 | | | | | | | |
| 62.2.47 | 95448 | 56 | Ni1Si1 | | | | | |
| | -GM | -R | -S | -T | -U | -X | -Y | -Z |
| | 5 2 | 3+3 4 | 4+4 4 | 3+4 4 | 6+6 4 | 3+4 4 | 3+4 4 | 3+4 4 |
| | 5 2 | 4+4 4 | 3+3 4 | 3+4 4 | 5+5 4 | 3+4 4 | 3+4 4 | 3+4 4 |
| | 6 2 | 4+4 4 | 4+4 4 | 3+4 4 | 6+6 4 | 3+4 4 | 3+4 4 | 3+4 4 |
| | 6 2 | 3+3 4 | 3+3 4 | 3+4 4 | 5+5 4 | 3+4 4 | 3+4 4 | 3+4 4 |

Continued on next column

| TQC | ICSD | NoE | SF | | | | | |
|---|---|---|---|---|---|---|---|---|
| | 5 2 | 4+4 4 | 3+3 4 | 3+4 4 | 6+6 4 | 3+4 4 | 3+4 4 | 3+4 4 |
| | 6 2 | 3+3 4 | 4+4 4 | 3+4 4 | 6+6 4 | 3+4 4 | 3+4 4 | 3+4 4 |
| | 6 2 | 3+3 4 | 4+4 4 | 3+4 4 | 5+5 4 | 3+4 4 | 3+4 4 | 3+4 4 |
| | 5 2 | 4+4 4 | 3+3 4 | 3+4 4 | 6+6 4 | 3+4 4 | 3+4 4 | 3+4 4 |
| | 5 2 | 3+3 4 | 3+3 4 | 3+4 4 | 5+5 4 | 3+4 4 | 3+4 4 | 3+4 4 |
| | 6 2 | 4+4 4 | 4+4 4 | 3+4 4 | 5+5 4 | 3+4 4 | 3+4 4 | 3+4 4 |
| | 5 2 | 4+4 4 | 3+3 4 | 3+4 4 | 6+6 4 | 3+4 4 | 3+4 4 | 3+4 4 |
| | 6 2 | 3+3 4 | 4+4 4 | 3+4 4 | 5+5 4 | 3+4 4 | 3+4 4 | 3+4 4 |
| | 5 2 | 3+3 4 | 4+4 4 | 3+4 4 | 5+5 4 | 3+4 4 | 3+4 4 | 3+4 4 |
| | 5 2 | 4+4 4 | 3+3 4 | 3+4 4 | 6+6 4 | 3+4 4 | 3+4 4 | 3+4 4 |
| | 5 2 | | | | | | | |
| | 5 2 | | | | | | | |
| | 6 2 | | | | | | | |
| | 6 2 | | | | | | | |
| | 6 2 | | | | | | | |
| | 5 2 | | | | | | | |
| | 6 2 | | | | | | | |
| | 6 2 | | | | | | | |
| | 6 2 | | | | | | | |
| | 5 2 | | | | | | | |
| | 5 2 | | | | | | | |
| | 5 2 | | | | | | | |
| | 5 2 | | | | | | | |
| | 5 2 | | | | | | | |
| 62.2.48 | 190105 | 56 | Ni1Si1 | | | | | |
| | -GM | -R | -S | -T | -U | -X | -Y | -Z |
| | 5 2 | 3+3 4 | 4+4 4 | 3+4 4 | 6+6 4 | 3+4 4 | 3+4 4 | 3+4 4 |
| | 5 2 | 4+4 4 | 3+3 4 | 3+4 4 | 5+5 4 | 3+4 4 | 3+4 4 | 3+4 4 |
| | 6 2 | 4+4 4 | 4+4 4 | 3+4 4 | 6+6 4 | 3+4 4 | 3+4 4 | 3+4 4 |
| | 6 2 | 3+3 4 | 3+3 4 | 3+4 4 | 5+5 4 | 3+4 4 | 3+4 4 | 3+4 4 |
| | 5 2 | 4+4 4 | 3+3 4 | 3+4 4 | 6+6 4 | 3+4 4 | 3+4 4 | 3+4 4 |
| | 6 2 | 3+3 4 | 4+4 4 | 3+4 4 | 6+6 4 | 3+4 4 | 3+4 4 | 3+4 4 |
| | 6 2 | 3+3 4 | 4+4 4 | 3+4 4 | 5+5 4 | 3+4 4 | 3+4 4 | 3+4 4 |
| | 5 2 | 4+4 4 | 3+3 4 | 3+4 4 | 6+6 4 | 3+4 4 | 3+4 4 | 3+4 4 |
| | 5 2 | 3+3 4 | 3+3 4 | 3+4 4 | 5+5 4 | 3+4 4 | 3+4 4 | 3+4 4 |
| | 6 2 | 4+4 4 | 4+4 4 | 3+4 4 | 5+5 4 | 3+4 4 | 3+4 4 | 3+4 4 |
| | 5 2 | 4+4 4 | 3+3 4 | 3+4 4 | 6+6 4 | 3+4 4 | 3+4 4 | 3+4 4 |
| | 6 2 | 3+3 4 | 4+4 4 | 3+4 4 | 5+5 4 | 3+4 4 | 3+4 4 | 3+4 4 |
| | 5 2 | 3+3 4 | 4+4 4 | 3+4 4 | 5+5 4 | 3+4 4 | 3+4 4 | 3+4 4 |
| | 5 2 | 4+4 4 | 3+3 4 | 3+4 4 | 6+6 4 | 3+4 4 | 3+4 4 | 3+4 4 |
| | 5 2 | | | | | | | |
| | 5 2 | | | | | | | |
| | 6 2 | | | | | | | |
| | 6 2 | | | | | | | |
| | 6 2 | | | | | | | |
| | 5 2 | | | | | | | |
| | 6 2 | | | | | | | |
| | 6 2 | | | | | | | |
| | 6 2 | | | | | | | |
| | 5 2 | | | | | | | |
| | 5 2 | | | | | | | |
| | 5 2 | | | | | | | |
| | 5 2 | | | | | | | |
| | 5 2 | | | | | | | |
| 62.2.49 | 248408 | 168 | Ni1S3U1 | | | | | |
| | -GM | -R | -S | -T | -U | -X | -Y | -Z |
| | 5 2 | 3+3 4 | 3+3 4 | 3+4 4 | 5+5 4 | 3+4 4 | 3+4 4 | 3+4 4 |
| | 6 2 | 4+4 4 | 4+4 4 | 3+4 4 | 6+6 4 | 3+4 4 | 3+4 4 | 3+4 4 |
| | 5 2 | 3+3 4 | 3+3 4 | 3+4 4 | 6+6 4 | 3+4 4 | 3+4 4 | 3+4 4 |
| | 6 2 | 4+4 4 | 4+4 4 | 3+4 4 | 5+5 4 | 3+4 4 | 3+4 4 | 3+4 4 |
| | 5 2 | 3+3 4 | 4+4 4 | 3+4 4 | 5+5 4 | 3+4 4 | 3+4 4 | 3+4 4 |

Continued on next column

**TABLE XXXVII– continued**

| TQC | ICSD | NoE | SF |
|---|---|---|---|
| | 6 2 4+4 4 3+3 4 3+4 4 6+6 4 3+4 4 3+4 4 3+4 4 | | |
| | 5 2 4+4 4 3+3 4 3+4 4 6+6 4 3+4 4 3+4 4 3+4 4 | | |
| | 6 2 3+3 4 4+4 4 3+4 4 5+5 4 3+4 4 3+4 4 3+4 4 | | |
| | 6 2 3+3 4 4+4 4 3+4 4 6+6 4 3+4 4 3+4 4 3+4 4 | | |
| | 5 2 4+4 4 3+3 4 3+4 4 5+5 4 3+4 4 3+4 4 3+4 4 | | |
| | 5 2 4+4 4 3+3 4 3+4 4 5+5 4 3+4 4 3+4 4 3+4 4 | | |
| | 5 2 3+3 4 4+4 4 3+4 4 6+6 4 3+4 4 3+4 4 3+4 4 | | |
| | 6 2 3+3 4 4+4 4 3+4 4 5+5 4 3+4 4 3+4 4 3+4 4 | | |
| | 6 2 4+4 4 3+3 4 3+4 4 6+6 4 3+4 4 3+4 4 3+4 4 | | |
| | 5 2 4+4 4 3+3 4 3+4 4 5+5 4 3+4 4 3+4 4 3+4 4 | | |
| | 6 2 3+3 4 4+4 4 3+4 4 5+5 4 3+4 4 3+4 4 3+4 4 | | |
| | 5 2 4+4 4 4+4 4 3+4 4 5+5 4 3+4 4 3+4 4 3+4 4 | | |
| | 6 2 3+3 4 3+3 4 3+4 4 6+6 4 3+4 4 3+4 4 3+4 4 | | |
| | 6 2 4+4 4 4+4 4 3+4 4 6+6 4 3+4 4 3+4 4 3+4 4 | | |
| | 5 2 3+3 4 3+3 4 3+4 4 6+6 4 3+4 4 3+4 4 3+4 4 | | |
| | 6 2 3+3 4 3+3 4 3+4 4 6+6 4 3+4 4 3+4 4 3+4 4 | | |
| | 5 2 4+4 4 4+4 4 3+4 4 5+5 4 3+4 4 3+4 4 3+4 4 | | |
| | 5 2 4+4 4 4+4 4 3+4 4 6+6 4 3+4 4 3+4 4 3+4 4 | | |
| | 6 2 3+3 4 3+3 4 3+4 4 5+5 4 3+4 4 3+4 4 3+4 4 | | |
| | 6 2 3+3 4 3+3 4 3+4 4 5+5 4 3+4 4 3+4 4 3+4 4 | | |
| | 5 2 4+4 4 4+4 4 3+4 4 6+6 4 3+4 4 3+4 4 3+4 4 | | |
| | 5 2 3+3 4 3+3 4 3+4 4 5+5 4 3+4 4 3+4 4 3+4 4 | | |
| | 6 2 4+4 4 4+4 4 3+4 4 5+5 4 3+4 4 3+4 4 3+4 4 | | |
| | 5 2 4+4 4 4+4 4 3+4 4 6+6 4 3+4 4 3+4 4 3+4 4 | | |
| | 5 2 3+3 4 3+3 4 3+4 4 5+5 4 3+4 4 3+4 4 3+4 4 | | |
| | 5 2 3+3 4 4+4 4 3+4 4 6+6 4 3+4 4 3+4 4 3+4 4 | | |
| | 6 2 4+4 4 3+3 4 3+4 4 6+6 4 3+4 4 3+4 4 3+4 4 | | |
| | 5 2 4+4 4 3+3 4 3+4 4 5+5 4 3+4 4 3+4 4 3+4 4 | | |
| | 6 2 3+3 4 4+4 4 3+4 4 5+5 4 3+4 4 3+4 4 3+4 4 | | |
| | 5 2 3+3 4 3+3 4 3+4 4 5+5 4 3+4 4 3+4 4 3+4 4 | | |
| | 6 2 4+4 4 4+4 4 3+4 4 5+5 4 3+4 4 3+4 4 3+4 4 | | |
| | 6 2 3+3 4 4+4 4 3+4 4 5+5 4 3+4 4 3+4 4 3+4 4 | | |
| | 5 2 4+4 4 3+3 4 3+4 4 5+5 4 3+4 4 3+4 4 3+4 4 | | |
| | 5 2 4+4 4 3+3 4 3+4 4 5+5 4 3+4 4 3+4 4 3+4 4 | | |
| | 6 2 3+3 4 4+4 4 3+4 4 5+5 4 3+4 4 3+4 4 3+4 4 | | |
| | 5 2 4+4 4 3+3 4 3+4 4 5+5 4 3+4 4 3+4 4 3+4 4 | | |
| | 6 2 3+3 4 4+4 4 3+4 4 6+6 4 3+4 4 3+4 4 3+4 4 | | |
| | 5 2 | | |
| | 6 2 | | |
| | 5 2 | | |
| | 5 2 | | |
| | 5 2 | | |
| | 5 2 | | |
| | 6 2 | | |
| | 6 2 | | |
| | 5 2 | | |
| | 6 2 | | |
| | 5 2 | | |
| | 5 2 | | |
| | 6 2 | | |
| | 5 2 | | |
| | 5 2 | | |
| | 6 2 | | |
| | 6 2 | | |
| | 6 2 | | |
| | 5 2 | | |
| | 5 2 | | |
| | 5 2 | | |
| | 5 2 | | |
| | 5 2 | | |
| | 6 2 | | |

Continued on next column

**TABLE XXXVII– continued**

| TQC | ICSD | NoE | SF |
|---|---|---|---|
| | 5 2 | | |
| | 5 2 | | |
| | 5 2 | | |
| | 6 2 | | |
| | 5 2 | | |
| | 6 2 | | |
| | 5 2 | | |
| | 5 2 | | |
| | 5 2 | | |
| | 5 2 | | |
| | 5 2 | | |
| | 6 2 | | |
| | 5 2 | | |
| | 5 2 | | |
| | 5 2 | | |
| | 5 2 | | |
| | 6 2 | | |
| | 5 2 | | |
| 62.2.50 | 419759 | 104 | Ni1Pd1Te1 |
| | -GM -R -S -T -U -X -Y -Z | | |
| | 5 2 3+3 4 4+4 4 3+4 4 6+6 4 3+4 4 3+4 4 3+4 4 | | |
| | 6 2 4+4 4 3+3 4 3+4 4 5+5 4 3+4 4 3+4 4 3+4 4 | | |
| | 5 2 4+4 4 3+3 4 3+4 4 6+6 4 3+4 4 3+4 4 3+4 4 | | |
| | 6 2 3+3 4 4+4 4 3+4 4 5+5 4 3+4 4 3+4 4 3+4 4 | | |
| | 5 2 3+3 4 4+4 4 3+4 4 6+6 4 3+4 4 3+4 4 3+4 4 | | |
| | 5 2 4+4 4 3+3 4 3+4 4 5+5 4 3+4 4 3+4 4 3+4 4 | | |
| | 6 2 4+4 4 3+3 4 3+4 4 6+6 4 3+4 4 3+4 4 3+4 4 | | |
| | 6 2 3+3 4 4+4 4 3+4 4 5+5 4 3+4 4 3+4 4 3+4 4 | | |
| | 5 2 3+3 4 4+4 4 3+4 4 5+5 4 3+4 4 3+4 4 3+4 4 | | |
| | 6 2 4+4 4 3+3 4 3+4 4 5+5 4 3+4 4 3+4 4 3+4 4 | | |
| | 6 2 3+3 4 3+3 4 3+4 4 6+6 4 3+4 4 3+4 4 3+4 4 | | |
| | 5 2 4+4 4 4+4 4 3+4 4 6+6 4 3+4 4 3+4 4 3+4 4 | | |
| | 6 2 4+4 4 4+4 4 3+4 4 6+6 4 3+4 4 3+4 4 3+4 4 | | |
| | 6 2 3+3 4 3+3 4 3+4 4 5+5 4 3+4 4 3+4 4 3+4 4 | | |
| | 5 2 3+3 4 3+3 4 3+4 4 5+5 4 3+4 4 3+4 4 3+4 4 | | |
| | 6 2 4+4 4 4+4 4 3+4 4 6+6 4 3+4 4 3+4 4 3+4 4 | | |
| | 5 2 3+3 4 4+4 4 3+4 4 5+5 4 3+4 4 3+4 4 3+4 4 | | |
| | 6 2 4+4 4 3+3 4 3+4 4 6+6 4 3+4 4 3+4 4 3+4 4 | | |
| | 6 2 3+3 4 3+3 4 3+4 4 6+6 4 3+4 4 3+4 4 3+4 4 | | |
| | 5 2 4+4 4 4+4 4 3+4 4 5+5 4 3+4 4 3+4 4 3+4 4 | | |
| | 5 2 4+4 4 4+4 4 3+4 4 6+6 4 3+4 4 3+4 4 3+4 4 | | |
| | 6 2 3+3 4 3+3 4 3+4 4 5+5 4 3+4 4 3+4 4 3+4 4 | | |
| | 5 2 4+4 4 3+3 4 3+4 4 6+6 4 3+4 4 3+4 4 3+4 4 | | |
| | 5 2 3+3 4 4+4 4 3+4 4 5+5 4 3+4 4 3+4 4 3+4 4 | | |
| | 5 2 3+3 4 4+4 4 3+4 4 5+5 4 3+4 4 3+4 4 3+4 4 | | |
| | 5 2 4+4 4 3+3 4 3+4 4 6+6 4 3+4 4 3+4 4 3+4 4 | | |
| | 6 2 | | |
| | 6 2 | | |
| | 5 2 | | |
| | 6 2 | | |
| | 5 2 | | |
| | 5 2 | | |
| | 5 2 | | |
| | 6 2 | | |
| | 5 2 | | |
| | 5 2 | | |
| | 6 2 | | |
| | 5 2 | | |
| | 6 2 | | |
| | 5 2 | | |
| | 6 2 | | |

Continued on next column

**TABLE XXXVII– continued**

| TQC | ICSD | NoE | SF | | | | | |
|---|---|---|---|---|---|---|---|---|
| | 6 2 | | | | | | | |
| | 5 2 | | | | | | | |
| | 5 2 | | | | | | | |
| | 5 2 | | | | | | | |
| | 6 2 | | | | | | | |
| | 6 2 | | | | | | | |
| | 6 2 | | | | | | | |
| | 5 2 | | | | | | | |
| | 6 2 | | | | | | | |
| | 5 2 | | | | | | | |
| | 6 2 | | | | | | | |
| 62.2.51 | 50990 | 72 | Ni1P1Sc1 | | | | | |
| | -GM | -R | -S | -T | -U | -X | -Y | -Z |
| | 5 2 | 3+3 4 | 4+4 4 | 3+4 4 | 5+5 4 | 3+4 4 | 3+4 4 | 3+4 4 |
| | 6 2 | 4+4 4 | 3+3 4 | 3+4 4 | 6+6 4 | 3+4 4 | 3+4 4 | 3+4 4 |
| | 5 2 | 3+3 4 | 4+4 4 | 3+4 4 | 5+5 4 | 3+4 4 | 3+4 4 | 3+4 4 |
| | 6 2 | 4+4 4 | 3+3 4 | 3+4 4 | 6+6 4 | 3+4 4 | 3+4 4 | 3+4 4 |
| | 5 2 | 4+4 4 | 3+3 4 | 3+4 4 | 6+6 4 | 3+4 4 | 3+4 4 | 3+4 4 |
| | 6 2 | 3+3 4 | 4+4 4 | 3+4 4 | 6+6 4 | 3+4 4 | 3+4 4 | 3+4 4 |
| | 5 2 | 3+3 4 | 3+3 4 | 3+4 4 | 5+5 4 | 3+4 4 | 3+4 4 | 3+4 4 |
| | 6 2 | 4+4 4 | 4+4 4 | 3+4 4 | 5+5 4 | 3+4 4 | 3+4 4 | 3+4 4 |
| | 5 2 | 4+4 4 | 4+4 4 | 3+4 4 | 5+5 4 | 3+4 4 | 3+4 4 | 3+4 4 |
| | 5 2 | 3+3 4 | 3+3 4 | 3+4 4 | 6+6 4 | 3+4 4 | 3+4 4 | 3+4 4 |
| | 6 2 | 3+3 4 | 4+4 4 | 3+4 4 | 5+5 4 | 3+4 4 | 3+4 4 | 3+4 4 |
| | 6 2 | 4+4 4 | 3+3 4 | 3+4 4 | 6+6 4 | 3+4 4 | 3+4 4 | 3+4 4 |
| | 6 2 | 4+4 4 | 4+4 4 | 3+4 4 | 5+5 4 | 3+4 4 | 3+4 4 | 3+4 4 |
| | 6 2 | 3+3 4 | 3+3 4 | 3+4 4 | 6+6 4 | 3+4 4 | 3+4 4 | 3+4 4 |
| | 5 2 | 4+4 4 | 3+3 4 | 3+4 4 | 5+5 4 | 3+4 4 | 3+4 4 | 3+4 4 |
| | 5 2 | 3+3 4 | 4+4 4 | 3+4 4 | 5+5 4 | 3+4 4 | 3+4 4 | 3+4 4 |
| | 6 2 | 3+3 4 | 4+4 4 | 3+4 4 | 6+6 4 | 3+4 4 | 3+4 4 | 3+4 4 |
| | 5 2 | 4+4 4 | 3+3 4 | 3+4 4 | 6+6 4 | 3+4 4 | 3+4 4 | 3+4 4 |
| | 6 2 | | | | | | | |
| | 5 2 | | | | | | | |
| | 5 2 | | | | | | | |
| | 6 2 | | | | | | | |
| | 5 2 | | | | | | | |
| | 6 2 | | | | | | | |
| | 5 2 | | | | | | | |
| | 5 2 | | | | | | | |
| | 5 2 | | | | | | | |
| | 6 2 | | | | | | | |
| | 5 2 | | | | | | | |
| | 6 2 | | | | | | | |
| | 6 2 | | | | | | | |
| | 5 2 | | | | | | | |
| | 5 2 | | | | | | | |
| | 6 2 | | | | | | | |
| | 6 2 | | | | | | | |
| | 5 2 | | | | | | | |
| 62.2.52 | 189299 | 208 | Ni1O3Yb1 | | | | | |
| | -GM | -R | -S | -T | -U | -X | -Y | -Z |
| | 5 2 | 3+3 4 | 3+3 4 | 3+4 4 | 5+5 4 | 3+4 4 | 3+4 4 | 3+4 4 |
| | 6 2 | 4+4 4 | 4+4 4 | 3+4 4 | 6+6 4 | 3+4 4 | 3+4 4 | 3+4 4 |
| | 5 2 | 3+3 4 | 4+4 4 | 3+4 4 | 5+5 4 | 3+4 4 | 3+4 4 | 3+4 4 |
| | 6 2 | 4+4 4 | 3+3 4 | 3+4 4 | 6+6 4 | 3+4 4 | 3+4 4 | 3+4 4 |
| | 5 2 | 3+3 4 | 4+4 4 | 3+4 4 | 6+6 4 | 3+4 4 | 3+4 4 | 3+4 4 |
| | 6 2 | 4+4 4 | 3+3 4 | 3+4 4 | 5+5 4 | 3+4 4 | 3+4 4 | 3+4 4 |
| | 5 2 | 4+4 4 | 3+3 4 | 3+4 4 | 5+5 4 | 3+4 4 | 3+4 4 | 3+4 4 |
| | 6 2 | 3+3 4 | 4+4 4 | 3+4 4 | 6+6 4 | 3+4 4 | 3+4 4 | 3+4 4 |
| | 6 2 | 4+4 4 | 3+3 4 | 3+4 4 | 5+5 4 | 3+4 4 | 3+4 4 | 3+4 4 |
| | 5 2 | 3+3 4 | 4+4 4 | 3+4 4 | 6+6 4 | 3+4 4 | 3+4 4 | 3+4 4 |

Continued on next column

**TABLE XXXVII– continued**

| TQC | ICSD | NoE | SF | | | | | |
|---|---|---|---|---|---|---|---|---|
| | 5 2 | 4+4 4 | 3+3 4 | 3+4 4 | 6+6 4 | 3+4 4 | 3+4 4 | 3+4 4 |
| | 5 2 | 3+3 4 | 4+4 4 | 3+4 4 | 5+5 4 | 3+4 4 | 3+4 4 | 3+4 4 |
| | 6 2 | 3+3 4 | 4+4 4 | 3+4 4 | 6+6 4 | 3+4 4 | 3+4 4 | 3+4 4 |
| | 6 2 | 4+4 4 | 3+3 4 | 3+4 4 | 5+5 4 | 3+4 4 | 3+4 4 | 3+4 4 |
| | 6 2 | 3+3 4 | 4+4 4 | 3+4 4 | 6+6 4 | 3+4 4 | 3+4 4 | 3+4 4 |
| | 5 2 | 4+4 4 | 3+3 4 | 3+4 4 | 6+6 4 | 3+4 4 | 3+4 4 | 3+4 4 |
| | 5 2 | 4+4 4 | 3+3 4 | 3+4 4 | 6+6 4 | 3+4 4 | 3+4 4 | 3+4 4 |
| | 6 2 | 3+3 4 | 4+4 4 | 3+4 4 | 6+6 4 | 3+4 4 | 3+4 4 | 3+4 4 |
| | 5 2 | 3+3 4 | 3+3 4 | 3+4 4 | 5+5 4 | 3+4 4 | 3+4 4 | 3+4 4 |
| | 6 2 | 4+4 4 | 4+4 4 | 3+4 4 | 5+5 4 | 3+4 4 | 3+4 4 | 3+4 4 |
| | 6 2 | 3+3 4 | 3+3 4 | 3+4 4 | 6+6 4 | 3+4 4 | 3+4 4 | 3+4 4 |
| | 5 2 | 4+4 4 | 4+4 4 | 3+4 4 | 5+5 4 | 3+4 4 | 3+4 4 | 3+4 4 |
| | 6 2 | 3+3 4 | 3+3 4 | 3+4 4 | 6+6 4 | 3+4 4 | 3+4 4 | 3+4 4 |
| | 5 2 | 4+4 4 | 4+4 4 | 3+4 4 | 5+5 4 | 3+4 4 | 3+4 4 | 3+4 4 |
| | 6 2 | 4+4 4 | 4+4 4 | 3+4 4 | 5+5 4 | 3+4 4 | 3+4 4 | 3+4 4 |
| | 5 2 | 3+3 4 | 3+3 4 | 3+4 4 | 6+6 4 | 3+4 4 | 3+4 4 | 3+4 4 |
| | 5 2 | 4+4 4 | 3+3 4 | 3+4 4 | 6+6 4 | 3+4 4 | 3+4 4 | 3+4 4 |
| | 6 2 | 3+3 4 | 4+4 4 | 3+4 4 | 5+5 4 | 3+4 4 | 3+4 4 | 3+4 4 |
| | 5 2 | 4+4 4 | 3+3 4 | 3+4 4 | 5+5 4 | 3+4 4 | 3+4 4 | 3+4 4 |
| | 5 2 | 3+3 4 | 4+4 4 | 3+4 4 | 6+6 4 | 3+4 4 | 3+4 4 | 3+4 4 |
| | 5 2 | 3+3 4 | 3+3 4 | 3+4 4 | 6+6 4 | 3+4 4 | 3+4 4 | 3+4 4 |
| | 5 2 | 4+4 4 | 4+4 4 | 3+4 4 | 5+5 4 | 3+4 4 | 3+4 4 | 3+4 4 |
| | 5 2 | 4+4 4 | 3+3 4 | 3+4 4 | 5+5 4 | 3+4 4 | 3+4 4 | 3+4 4 |
| | 5 2 | 3+3 4 | 4+4 4 | 3+4 4 | 6+6 4 | 3+4 4 | 3+4 4 | 3+4 4 |
| | 6 2 | 4+4 4 | 4+4 4 | 3+4 4 | 6+6 4 | 3+4 4 | 3+4 4 | 3+4 4 |
| | 5 2 | 3+3 4 | 3+3 4 | 3+4 4 | 6+6 4 | 3+4 4 | 3+4 4 | 3+4 4 |
| | 5 2 | 4+4 4 | 4+4 4 | 3+4 4 | 5+5 4 | 3+4 4 | 3+4 4 | 3+4 4 |
| | 5 2 | 3+3 4 | 3+3 4 | 3+4 4 | 6+6 4 | 3+4 4 | 3+4 4 | 3+4 4 |
| | 5 2 | 3+3 4 | 3+3 4 | 3+4 4 | 5+5 4 | 3+4 4 | 3+4 4 | 3+4 4 |
| | 6 2 | 4+4 4 | 4+4 4 | 3+4 4 | 5+5 4 | 3+4 4 | 3+4 4 | 3+4 4 |
| | 6 2 | 4+4 4 | 3+3 4 | 3+4 4 | 6+6 4 | 3+4 4 | 3+4 4 | 3+4 4 |
| | 6 2 | 3+3 4 | 4+4 4 | 3+4 4 | 6+6 4 | 3+4 4 | 3+4 4 | 3+4 4 |
| | 6 2 | 3+3 4 | 3+3 4 | 3+4 4 | 6+6 4 | 3+4 4 | 3+4 4 | 3+4 4 |
| | 5 2 | 4+4 4 | 4+4 4 | 3+4 4 | 6+6 4 | 3+4 4 | 3+4 4 | 3+4 4 |
| | 5 2 | 3+3 4 | 4+4 4 | 3+4 4 | 6+6 4 | 3+4 4 | 3+4 4 | 3+4 4 |
| | 5 2 | 4+4 4 | 3+3 4 | 3+4 4 | 5+5 4 | 3+4 4 | 3+4 4 | 3+4 4 |
| | 5 2 | 4+4 4 | 3+3 4 | 3+4 4 | 6+6 4 | 3+4 4 | 3+4 4 | 3+4 4 |
| | 6 2 | 3+3 4 | 4+4 4 | 3+4 4 | 5+5 4 | 3+4 4 | 3+4 4 | 3+4 4 |
| | 6 2 | 4+4 4 | 4+4 4 | 3+4 4 | 5+5 4 | 3+4 4 | 3+4 4 | 3+4 4 |
| | 6 2 | 3+3 4 | 3+3 4 | 3+4 4 | 6+6 4 | 3+4 4 | 3+4 4 | 3+4 4 |
| | 6 2 | 3+3 4 | 3+3 4 | 3+4 4 | 5+5 4 | 3+4 4 | 3+4 4 | 3+4 4 |
| | 5 2 | 4+4 4 | 4+4 4 | 3+4 4 | 6+6 4 | 3+4 4 | 3+4 4 | 3+4 4 |
| | 6 2 | | | | | | | |
| | 5 2 | | | | | | | |
| | 6 2 | | | | | | | |
| | 5 2 | | | | | | | |
| | 5 2 | | | | | | | |
| | 6 2 | | | | | | | |
| | 6 2 | | | | | | | |
| | 6 2 | | | | | | | |
| | 5 2 | | | | | | | |
| | 5 2 | | | | | | | |
| | 5 2 | | | | | | | |
| | 5 2 | | | | | | | |
| | 6 2 | | | | | | | |
| | 5 2 | | | | | | | |
| | 6 2 | | | | | | | |
| | 5 2 | | | | | | | |
| | 5 2 | | | | | | | |
| | 5 2 | | | | | | | |
| | 5 2 | | | | | | | |

Continued on next column

**TABLE XXXVII– continued**

| TQC | ICSD | NoE | SF | | | | | |
|---|---|---|---|---|---|---|---|---|
| | 6 2 | | | | | | | |
| | 6 2 | | | | | | | |
| | 5 2 | | | | | | | |
| | 5 2 | | | | | | | |
| | 6 2 | | | | | | | |
| | 6 2 | | | | | | | |
| | 5 2 | | | | | | | |
| | 6 2 | | | | | | | |
| | 5 2 | | | | | | | |
| | 6 2 | | | | | | | |
| | 5 2 | | | | | | | |
| | 6 2 | | | | | | | |
| | 6 2 | | | | | | | |
| | 5 2 | | | | | | | |
| | 5 2 | | | | | | | |
| | 5 2 | | | | | | | |
| | 5 2 | | | | | | | |
| | 5 2 | | | | | | | |
| | 5 2 | | | | | | | |
| | 5 2 | | | | | | | |
| | 5 2 | | | | | | | |
| | 5 2 | | | | | | | |
| | 6 2 | | | | | | | |
| | 6 2 | | | | | | | |
| | 5 2 | | | | | | | |
| | 6 2 | | | | | | | |
| | 5 2 | | | | | | | |
| | 6 2 | | | | | | | |
| | 6 2 | | | | | | | |
| | 5 2 | | | | | | | |
| | 6 2 | | | | | | | |
| | 6 2 | | | | | | | |
| | 5 2 | | | | | | | |
| 62.2.53 | 72290 | 184 | Nd2S3 | | | | | |
| | -GM | -R | -S | -T | -U | -X | -Y | -Z |
| | 5 2 | 3+3 4 | 3+3 4 | 3+4 4 | 5+5 4 | 3+4 4 | 3+4 4 | 3+4 4 |
| | 6 2 | 4+4 4 | 4+4 4 | 3+4 4 | 6+6 4 | 3+4 4 | 3+4 4 | 3+4 4 |
| | 5 2 | 3+3 4 | 3+3 4 | 3+4 4 | 5+5 4 | 3+4 4 | 3+4 4 | 3+4 4 |
| | 6 2 | 4+4 4 | 4+4 4 | 3+4 4 | 6+6 4 | 3+4 4 | 3+4 4 | 3+4 4 |
| | 5 2 | 3+3 4 | 3+3 4 | 3+4 4 | 5+5 4 | 3+4 4 | 3+4 4 | 3+4 4 |
| | 5 2 | 4+4 4 | 4+4 4 | 3+4 4 | 6+6 4 | 3+4 4 | 3+4 4 | 3+4 4 |
| | 6 2 | 3+3 4 | 4+4 4 | 3+4 4 | 5+5 4 | 3+4 4 | 3+4 4 | 3+4 4 |
| | 6 2 | 4+4 4 | 3+3 4 | 3+4 4 | 6+6 4 | 3+4 4 | 3+4 4 | 3+4 4 |
| | 5 2 | 3+3 4 | 4+4 4 | 3+4 4 | 6+6 4 | 3+4 4 | 3+4 4 | 3+4 4 |
| | 6 2 | 4+4 4 | 3+3 4 | 3+4 4 | 5+5 4 | 3+4 4 | 3+4 4 | 3+4 4 |
| | 6 2 | 3+3 4 | 3+3 4 | 3+4 4 | 6+6 4 | 3+4 4 | 3+4 4 | 3+4 4 |
| | 5 2 | 4+4 4 | 4+4 4 | 3+4 4 | 5+5 4 | 3+4 4 | 3+4 4 | 3+4 4 |
| | 5 2 | 4+4 4 | 4+4 4 | 3+4 4 | 5+5 4 | 3+4 4 | 3+4 4 | 3+4 4 |
| | 5 2 | 3+3 4 | 3+3 4 | 3+4 4 | 6+6 4 | 3+4 4 | 3+4 4 | 3+4 4 |
| | 6 2 | 4+4 4 | 3+3 4 | 3+4 4 | 5+5 4 | 3+4 4 | 3+4 4 | 3+4 4 |
| | 6 2 | 3+3 4 | 4+4 4 | 3+4 4 | 6+6 4 | 3+4 4 | 3+4 4 | 3+4 4 |
| | 5 2 | 3+3 4 | 4+4 4 | 3+4 4 | 6+6 4 | 3+4 4 | 3+4 4 | 3+4 4 |
| | 6 2 | 4+4 4 | 3+3 4 | 3+4 4 | 5+5 4 | 3+4 4 | 3+4 4 | 3+4 4 |
| | 6 2 | 4+4 4 | 3+3 4 | 3+4 4 | 6+6 4 | 3+4 4 | 3+4 4 | 3+4 4 |
| | 5 2 | 3+3 4 | 4+4 4 | 3+4 4 | 5+5 4 | 3+4 4 | 3+4 4 | 3+4 4 |
| | 5 2 | 3+3 4 | 3+3 4 | 3+4 4 | 6+6 4 | 3+4 4 | 3+4 4 | 3+4 4 |
| | 5 2 | 4+4 4 | 4+4 4 | 3+4 4 | 5+5 4 | 3+4 4 | 3+4 4 | 3+4 4 |
| | 6 2 | 3+3 4 | 4+4 4 | 3+4 4 | 5+5 4 | 3+4 4 | 3+4 4 | 3+4 4 |
| | 6 2 | 4+4 4 | 3+3 4 | 3+4 4 | 6+6 4 | 3+4 4 | 3+4 4 | 3+4 4 |
| | 5 2 | 4+4 4 | 3+3 4 | 3+4 4 | 6+6 4 | 3+4 4 | 3+4 4 | 3+4 4 |
| | 5 2 | 3+3 4 | 4+4 4 | 3+4 4 | 6+6 4 | 3+4 4 | 3+4 4 | 3+4 4 |

Continued on next column

**TABLE XXXVII– continued**

| TQC | ICSD | NoE | SF | | | | | |
|---|---|---|---|---|---|---|---|---|
| | 6 2 | 3+3 4 | 4+4 4 | 3+4 4 | 5+5 4 | 3+4 4 | 3+4 4 | 3+4 4 |
| | 5 2 | 4+4 4 | 3+3 4 | 3+4 4 | 6+6 4 | 3+4 4 | 3+4 4 | 3+4 4 |
| | 6 2 | 4+4 4 | 3+3 4 | 3+4 4 | 5+5 4 | 3+4 4 | 3+4 4 | 3+4 4 |
| | 5 2 | 3+3 4 | 4+4 4 | 3+4 4 | 5+5 4 | 3+4 4 | 3+4 4 | 3+4 4 |
| | 6 2 | 3+3 4 | 4+4 4 | 3+4 4 | 6+6 4 | 3+4 4 | 3+4 4 | 3+4 4 |
| | 6 2 | 4+4 4 | 3+3 4 | 3+4 4 | 5+5 4 | 3+4 4 | 3+4 4 | 3+4 4 |
| | 5 2 | 4+4 4 | 3+3 4 | 3+4 4 | 6+6 4 | 3+4 4 | 3+4 4 | 3+4 4 |
| | 6 2 | 3+3 4 | 4+4 4 | 3+4 4 | 5+5 4 | 3+4 4 | 3+4 4 | 3+4 4 |
| | 5 2 | 3+3 4 | 4+4 4 | 3+4 4 | 5+5 4 | 3+4 4 | 3+4 4 | 3+4 4 |
| | 5 2 | 4+4 4 | 3+3 4 | 3+4 4 | 6+6 4 | 3+4 4 | 3+4 4 | 3+4 4 |
| | 6 2 | 4+4 4 | 4+4 4 | 3+4 4 | 5+5 4 | 3+4 4 | 3+4 4 | 3+4 4 |
| | 6 2 | 3+3 4 | 3+3 4 | 3+4 4 | 6+6 4 | 3+4 4 | 3+4 4 | 3+4 4 |
| | 6 2 | 3+3 4 | 3+3 4 | 3+4 4 | 6+6 4 | 3+4 4 | 3+4 4 | 3+4 4 |
| | 5 2 | 4+4 4 | 4+4 4 | 3+4 4 | 5+5 4 | 3+4 4 | 3+4 4 | 3+4 4 |
| | 6 2 | 3+3 4 | 3+3 4 | 3+4 4 | 5+5 4 | 3+4 4 | 3+4 4 | 3+4 4 |
| | 6 2 | 4+4 4 | 4+4 4 | 3+4 4 | 6+6 4 | 3+4 4 | 3+4 4 | 3+4 4 |
| | 5 2 | 4+4 4 | 3+3 4 | 3+4 4 | 5+5 4 | 3+4 4 | 3+4 4 | 3+4 4 |
| | 5 2 | 3+3 4 | 4+4 4 | 3+4 4 | 6+6 4 | 3+4 4 | 3+4 4 | 3+4 4 |
| | 5 2 | 3+3 4 | 4+4 4 | 3+4 4 | 5+5 4 | 3+4 4 | 3+4 4 | 3+4 4 |
| | 6 2 | 4+4 4 | 3+3 4 | 3+4 4 | 6+6 4 | 3+4 4 | 3+4 4 | 3+4 4 |
| | 6 2 | | | | | | | |
| | 5 2 | | | | | | | |
| | 6 2 | | | | | | | |
| | 5 2 | | | | | | | |
| | 5 2 | | | | | | | |
| | 5 2 | | | | | | | |
| | 5 2 | | | | | | | |
| | 6 2 | | | | | | | |
| | 6 2 | | | | | | | |
| | 5 2 | | | | | | | |
| | 5 2 | | | | | | | |
| | 6 2 | | | | | | | |
| | 5 2 | | | | | | | |
| | 6 2 | | | | | | | |
| | 6 2 | | | | | | | |
| | 5 2 | | | | | | | |
| | 5 2 | | | | | | | |
| | 6 2 | | | | | | | |
| | 6 2 | | | | | | | |
| | 5 2 | | | | | | | |
| | 6 2 | | | | | | | |
| | 6 2 | | | | | | | |
| | 5 2 | | | | | | | |
| | 5 2 | | | | | | | |
| | 5 2 | | | | | | | |
| | 5 2 | | | | | | | |
| | 5 2 | | | | | | | |
| | 6 2 | | | | | | | |
| | 5 2 | | | | | | | |
| | 6 2 | | | | | | | |
| | 6 2 | | | | | | | |
| | 6 2 | | | | | | | |
| | 6 2 | | | | | | | |
| | 6 2 | | | | | | | |
| | 6 2 | | | | | | | |
| | 5 2 | | | | | | | |
| | 6 2 | | | | | | | |
| | 5 2 | | | | | | | |
| | 5 2 | | | | | | | |
| | 6 2 | | | | | | | |
| | 5 2 | | | | | | | |

Continued on next column

**TABLE XXXVII– continued**

| TQC | ICSD | NoE | SF | | | | | |
|---|---|---|---|---|---|---|---|---|
| | 5 2 | | | | | | | |
| | 5 2 | | | | | | | |
| | 6 2 | | | | | | | |
| | 5 2 | | | | | | | |
| | 6 2 | | | | | | | |
| 62.2.54 | 415953 | 112 | Nd1Pt1Sn1 | | | | | |
| | -GM | -R | -S | -T | -U | -X | -Y | -Z |
| | 5 2 | 3+3 4 | 3+3 4 | 3+4 4 | 6+6 4 | 3+4 4 | 3+4 4 | 3+4 4 |
| | 6 2 | 4+4 4 | 4+4 4 | 3+4 4 | 5+5 4 | 3+4 4 | 3+4 4 | 3+4 4 |
| | 5 2 | 3+3 4 | 3+3 4 | 3+4 4 | 6+6 4 | 3+4 4 | 3+4 4 | 3+4 4 |
| | 6 2 | 4+4 4 | 4+4 4 | 3+4 4 | 5+5 4 | 3+4 4 | 3+4 4 | 3+4 4 |
| | 5 2 | 4+4 4 | 3+3 4 | 3+4 4 | 6+6 4 | 3+4 4 | 3+4 4 | 3+4 4 |
| | 6 2 | 3+3 4 | 4+4 4 | 3+4 4 | 5+5 4 | 3+4 4 | 3+4 4 | 3+4 4 |
| | 5 2 | 3+3 4 | 4+4 4 | 3+4 4 | 6+6 4 | 3+4 4 | 3+4 4 | 3+4 4 |
| | 6 2 | 4+4 4 | 3+3 4 | 3+4 4 | 5+5 4 | 3+4 4 | 3+4 4 | 3+4 4 |
| | 5 2 | 3+3 4 | 4+4 4 | 3+4 4 | 5+5 4 | 3+4 4 | 3+4 4 | 3+4 4 |
| | 6 2 | 4+4 4 | 3+3 4 | 3+4 4 | 6+6 4 | 3+4 4 | 3+4 4 | 3+4 4 |
| | 5 2 | 4+4 4 | 3+3 4 | 3+4 4 | 6+6 4 | 3+4 4 | 3+4 4 | 3+4 4 |
| | 5 2 | 3+3 4 | 4+4 4 | 3+4 4 | 5+5 4 | 3+4 4 | 3+4 4 | 3+4 4 |
| | 6 2 | 4+4 4 | 4+4 4 | 3+4 4 | 5+5 4 | 3+4 4 | 3+4 4 | 3+4 4 |
| | 5 2 | 3+3 4 | 3+3 4 | 3+4 4 | 6+6 4 | 3+4 4 | 3+4 4 | 3+4 4 |
| | 6 2 | 4+4 4 | 3+3 4 | 3+4 4 | 5+5 4 | 3+4 4 | 3+4 4 | 3+4 4 |
| | 6 2 | 3+3 4 | 4+4 4 | 3+4 4 | 6+6 4 | 3+4 4 | 3+4 4 | 3+4 4 |
| | 5 2 | 3+3 4 | 3+3 4 | 3+4 4 | 5+5 4 | 3+4 4 | 3+4 4 | 3+4 4 |
| | 5 2 | 4+4 4 | 4+4 4 | 3+4 4 | 6+6 4 | 3+4 4 | 3+4 4 | 3+4 4 |
| | 6 2 | 3+3 4 | 4+4 4 | 3+4 4 | 5+5 4 | 3+4 4 | 3+4 4 | 3+4 4 |
| | 6 2 | 4+4 4 | 3+3 4 | 3+4 4 | 6+6 4 | 3+4 4 | 3+4 4 | 3+4 4 |
| | 6 2 | 3+3 4 | 4+4 4 | 3+4 4 | 5+5 4 | 3+4 4 | 3+4 4 | 3+4 4 |
| | 5 2 | 4+4 4 | 3+3 4 | 3+4 4 | 6+6 4 | 3+4 4 | 3+4 4 | 3+4 4 |
| | 6 2 | 3+3 4 | 4+4 4 | 3+4 4 | 5+5 4 | 3+4 4 | 3+4 4 | 3+4 4 |
| | 5 2 | 4+4 4 | 3+3 4 | 3+4 4 | 5+5 4 | 3+4 4 | 3+4 4 | 3+4 4 |
| | 5 2 | 3+3 4 | 4+4 4 | 3+4 4 | 6+6 4 | 3+4 4 | 3+4 4 | 3+4 4 |
| | 6 2 | 4+4 4 | 3+3 4 | 3+4 4 | 6+6 4 | 3+4 4 | 3+4 4 | 3+4 4 |
| | 5 2 | 3+3 4 | 3+3 4 | 3+4 4 | 5+5 4 | 3+4 4 | 3+4 4 | 3+4 4 |
| | 5 2 | 4+4 4 | 4+4 4 | 3+4 4 | 6+6 4 | 3+4 4 | 3+4 4 | 3+4 4 |
| | 6 2 | | | | | | | |
| | 5 2 | | | | | | | |
| | 6 2 | | | | | | | |
| | 5 2 | | | | | | | |
| | 6 2 | | | | | | | |
| | 5 2 | | | | | | | |
| | 6 2 | | | | | | | |
| | 6 2 | | | | | | | |
| | 5 2 | | | | | | | |
| | 5 2 | | | | | | | |
| | 6 2 | | | | | | | |
| | 6 2 | | | | | | | |
| | 5 2 | | | | | | | |
| | 6 2 | | | | | | | |
| | 5 2 | | | | | | | |
| | 6 2 | | | | | | | |
| | 5 2 | | | | | | | |
| | 6 2 | | | | | | | |
| | 5 2 | | | | | | | |
| | 5 2 | | | | | | | |
| | 6 2 | | | | | | | |
| | 5 2 | | | | | | | |
| | 6 2 | | | | | | | |
| | 6 2 | | | | | | | |
| | 6 2 | | | | | | | |
| | 5 2 | | | | | | | |

Continued on next column

| TQC | ICSD | NoE | SF | | | | | |
|---|---|---|---|---|---|---|---|---|
| | 5 2 | | | | | | | |
| | 5 2 | | | | | | | |
| 62.2.55 | 82008 | 144 | Nd1O3Ti1 | | | | | |
| | -GM | -R | -S | -T | -U | -X | -Y | -Z |
| | 5 2 | 3+3 4 | 3+3 4 | 3+4 4 | 5+5 4 | 3+4 4 | 3+4 4 | 3+4 4 |
| | 6 2 | 4+4 4 | 4+4 4 | 3+4 4 | 6+6 4 | 3+4 4 | 3+4 4 | 3+4 4 |
| | 5 2 | 4+4 4 | 4+4 4 | 3+4 4 | 6+6 4 | 3+4 4 | 3+4 4 | 3+4 4 |
| | 6 2 | 3+3 4 | 3+3 4 | 3+4 4 | 5+5 4 | 3+4 4 | 3+4 4 | 3+4 4 |
| | 5 2 | 4+4 4 | 4+4 4 | 3+4 4 | 5+5 4 | 3+4 4 | 3+4 4 | 3+4 4 |
| | 6 2 | 3+3 4 | 3+3 4 | 3+4 4 | 6+6 4 | 3+4 4 | 3+4 4 | 3+4 4 |
| | 5 2 | 3+3 4 | 3+3 4 | 3+4 4 | 6+6 4 | 3+4 4 | 3+4 4 | 3+4 4 |
| | 6 2 | 4+4 4 | 4+4 4 | 3+4 4 | 5+5 4 | 3+4 4 | 3+4 4 | 3+4 4 |
| | 5 2 | 3+3 4 | 3+3 4 | 3+4 4 | 6+6 4 | 3+4 4 | 3+4 4 | 3+4 4 |
| | 6 2 | 4+4 4 | 4+4 4 | 3+4 4 | 5+5 4 | 3+4 4 | 3+4 4 | 3+4 4 |
| | 5 2 | 4+4 4 | 4+4 4 | 3+4 4 | 6+6 4 | 3+4 4 | 3+4 4 | 3+4 4 |
| | 5 2 | 3+3 4 | 3+3 4 | 3+4 4 | 5+5 4 | 3+4 4 | 3+4 4 | 3+4 4 |
| | 5 2 | 4+4 4 | 3+3 4 | 3+4 4 | 5+5 4 | 3+4 4 | 3+4 4 | 3+4 4 |
| | 6 2 | 3+3 4 | 4+4 4 | 3+4 4 | 6+6 4 | 3+4 4 | 3+4 4 | 3+4 4 |
| | 5 2 | 3+3 4 | 4+4 4 | 3+4 4 | 6+6 4 | 3+4 4 | 3+4 4 | 3+4 4 |
| | 6 2 | 4+4 4 | 3+3 4 | 3+4 4 | 6+6 4 | 3+4 4 | 3+4 4 | 3+4 4 |
| | 6 2 | 3+3 4 | 3+3 4 | 3+4 4 | 6+6 4 | 3+4 4 | 3+4 4 | 3+4 4 |
| | 6 2 | 4+4 4 | 4+4 4 | 3+4 4 | 6+6 4 | 3+4 4 | 3+4 4 | 3+4 4 |
| | 6 2 | 3+3 4 | 3+3 4 | 3+4 4 | 5+5 4 | 3+4 4 | 3+4 4 | 3+4 4 |
| | 5 2 | 4+4 4 | 4+4 4 | 3+4 4 | 5+5 4 | 3+4 4 | 3+4 4 | 3+4 4 |
| | 5 2 | 4+4 4 | 3+3 4 | 3+4 4 | 5+5 4 | 3+4 4 | 3+4 4 | 3+4 4 |
| | 6 2 | 3+3 4 | 4+4 4 | 3+4 4 | 6+6 4 | 3+4 4 | 3+4 4 | 3+4 4 |
| | 6 2 | 3+3 4 | 4+4 4 | 3+4 4 | 6+6 4 | 3+4 4 | 3+4 4 | 3+4 4 |
| | 5 2 | 4+4 4 | 3+3 4 | 3+4 4 | 5+5 4 | 3+4 4 | 3+4 4 | 3+4 4 |
| | 6 2 | 3+3 4 | 4+4 4 | 3+4 4 | 6+6 4 | 3+4 4 | 3+4 4 | 3+4 4 |
| | 5 2 | 4+4 4 | 3+3 4 | 3+4 4 | 6+6 4 | 3+4 4 | 3+4 4 | 3+4 4 |
| | 5 2 | 3+3 4 | 3+3 4 | 3+4 4 | 5+5 4 | 3+4 4 | 3+4 4 | 3+4 4 |
| | 6 2 | 4+4 4 | 4+4 4 | 3+4 4 | 5+5 4 | 3+4 4 | 3+4 4 | 3+4 4 |
| | 5 2 | 3+3 4 | 3+3 4 | 3+4 4 | 5+5 4 | 3+4 4 | 3+4 4 | 3+4 4 |
| | 5 2 | 4+4 4 | 4+4 4 | 3+4 4 | 5+5 4 | 3+4 4 | 3+4 4 | 3+4 4 |
| | 5 2 | 4+4 4 | 4+4 4 | 3+4 4 | 6+6 4 | 3+4 4 | 3+4 4 | 3+4 4 |
| | 5 2 | 3+3 4 | 3+3 4 | 3+4 4 | 5+5 4 | 3+4 4 | 3+4 4 | 3+4 4 |
| | 5 2 | 4+4 4 | 4+4 4 | 3+4 4 | 6+6 4 | 3+4 4 | 3+4 4 | 3+4 4 |
| | 6 2 | 3+3 4 | 3+3 4 | 3+4 4 | 6+6 4 | 3+4 4 | 3+4 4 | 3+4 4 |
| | 5 2 | 3+3 4 | 4+4 4 | 3+4 4 | 5+5 4 | 3+4 4 | 3+4 4 | 3+4 4 |
| | 5 2 | 4+4 4 | 3+3 4 | 3+4 4 | 6+6 4 | 3+4 4 | 3+4 4 | 3+4 4 |
| | 5 2 | | | | | | | |
| | 5 2 | | | | | | | |
| | 6 2 | | | | | | | |
| | 6 2 | | | | | | | |
| | 5 2 | | | | | | | |
| | 6 2 | | | | | | | |
| | 6 2 | | | | | | | |
| | 5 2 | | | | | | | |
| | 5 2 | | | | | | | |
| | 6 2 | | | | | | | |
| | 5 2 | | | | | | | |
| | 5 2 | | | | | | | |
| | 6 2 | | | | | | | |
| | 6 2 | | | | | | | |
| | 6 2 | | | | | | | |
| | 6 2 | | | | | | | |
| | 6 2 | | | | | | | |
| | 5 2 | | | | | | | |
| | 5 2 | | | | | | | |
| | 5 2 | | | | | | | |
| | 6 2 | | | | | | | |

Continued on next column

**TABLE XXXVII– continued**

| TQC | ICSD | NoE | SF | | | | | |
|---|---|---|---|---|---|---|---|---|
| | 6 2 | | | | | | | |
| | 6 2 | | | | | | | |
| | 5 2 | | | | | | | |
| | 6 2 | | | | | | | |
| | 6 2 | | | | | | | |
| | 6 2 | | | | | | | |
| | 6 2 | | | | | | | |
| | 5 2 | | | | | | | |
| | 5 2 | | | | | | | |
| | 5 2 | | | | | | | |
| | 5 2 | | | | | | | |
| | 5 2 | | | | | | | |
| | 5 2 | | | | | | | |
| | 6 2 | | | | | | | |
| | 5 2 | | | | | | | |
| 62.2.56 | 67723 | 168 | Nd1Ni1O3 | | | | | |
| | -GM | -R | -S | -T | -U | -X | -Y | -Z |
| | 5 2 | 3+3 4 | 3+3 4 | 3+4 4 | 5+5 4 | 3+4 4 | 3+4 4 | 3+4 4 |
| | 6 2 | 4+4 4 | 4+4 4 | 3+4 4 | 6+6 4 | 3+4 4 | 3+4 4 | 3+4 4 |
| | 5 2 | 3+3 4 | 3+3 4 | 3+4 4 | 6+6 4 | 3+4 4 | 3+4 4 | 3+4 4 |
| | 6 2 | 4+4 4 | 4+4 4 | 3+4 4 | 5+5 4 | 3+4 4 | 3+4 4 | 3+4 4 |
| | 5 2 | 3+3 4 | 4+4 4 | 3+4 4 | 5+5 4 | 3+4 4 | 3+4 4 | 3+4 4 |
| | 6 2 | 4+4 4 | 3+3 4 | 3+4 4 | 6+6 4 | 3+4 4 | 3+4 4 | 3+4 4 |
| | 5 2 | 4+4 4 | 3+3 4 | 3+4 4 | 6+6 4 | 3+4 4 | 3+4 4 | 3+4 4 |
| | 6 2 | 3+3 4 | 4+4 4 | 3+4 4 | 5+5 4 | 3+4 4 | 3+4 4 | 3+4 4 |
| | 6 2 | 4+4 4 | 3+3 4 | 3+4 4 | 5+5 4 | 3+4 4 | 3+4 4 | 3+4 4 |
| | 5 2 | 3+3 4 | 4+4 4 | 3+4 4 | 6+6 4 | 3+4 4 | 3+4 4 | 3+4 4 |
| | 5 2 | 4+4 4 | 4+4 4 | 3+4 4 | 6+6 4 | 3+4 4 | 3+4 4 | 3+4 4 |
| | 5 2 | 3+3 4 | 3+3 4 | 3+4 4 | 5+5 4 | 3+4 4 | 3+4 4 | 3+4 4 |
| | 6 2 | 3+3 4 | 3+3 4 | 3+4 4 | 5+5 4 | 3+4 4 | 3+4 4 | 3+4 4 |
| | 5 2 | 4+4 4 | 4+4 4 | 3+4 4 | 6+6 4 | 3+4 4 | 3+4 4 | 3+4 4 |
| | 6 2 | 3+3 4 | 4+4 4 | 3+4 4 | 6+6 4 | 3+4 4 | 3+4 4 | 3+4 4 |
| | 5 2 | 4+4 4 | 3+3 4 | 3+4 4 | 6+6 4 | 3+4 4 | 3+4 4 | 3+4 4 |
| | 6 2 | 4+4 4 | 3+3 4 | 3+4 4 | 6+6 4 | 3+4 4 | 3+4 4 | 3+4 4 |
| | 6 2 | 3+3 4 | 4+4 4 | 3+4 4 | 6+6 4 | 3+4 4 | 3+4 4 | 3+4 4 |
| | 6 2 | 3+3 4 | 3+3 4 | 3+4 4 | 5+5 4 | 3+4 4 | 3+4 4 | 3+4 4 |
| | 5 2 | 4+4 4 | 4+4 4 | 3+4 4 | 5+5 4 | 3+4 4 | 3+4 4 | 3+4 4 |
| | 5 2 | 3+3 4 | 3+3 4 | 3+4 4 | 6+6 4 | 3+4 4 | 3+4 4 | 3+4 4 |
| | 6 2 | 4+4 4 | 4+4 4 | 3+4 4 | 6+6 4 | 3+4 4 | 3+4 4 | 3+4 4 |
| | 5 2 | 3+3 4 | 3+3 4 | 3+4 4 | 5+5 4 | 3+4 4 | 3+4 4 | 3+4 4 |
| | 6 2 | 4+4 4 | 4+4 4 | 3+4 4 | 5+5 4 | 3+4 4 | 3+4 4 | 3+4 4 |
| | 5 2 | 4+4 4 | 4+4 4 | 3+4 4 | 6+6 4 | 3+4 4 | 3+4 4 | 3+4 4 |
| | 6 2 | 3+3 4 | 3+3 4 | 3+4 4 | 6+6 4 | 3+4 4 | 3+4 4 | 3+4 4 |
| | 5 2 | 4+4 4 | 3+3 4 | 3+4 4 | 5+5 4 | 3+4 4 | 3+4 4 | 3+4 4 |
| | 6 2 | 3+3 4 | 4+4 4 | 3+4 4 | 5+5 4 | 3+4 4 | 3+4 4 | 3+4 4 |
| | 5 2 | 3+3 4 | 3+3 4 | 3+4 4 | 5+5 4 | 3+4 4 | 3+4 4 | 3+4 4 |
| | 5 2 | 4+4 4 | 4+4 4 | 3+4 4 | 5+5 4 | 3+4 4 | 3+4 4 | 3+4 4 |
| | 5 2 | 3+3 4 | 3+3 4 | 3+4 4 | 5+5 4 | 3+4 4 | 3+4 4 | 3+4 4 |
| | 5 2 | 4+4 4 | 4+4 4 | 3+4 4 | 6+6 4 | 3+4 4 | 3+4 4 | 3+4 4 |
| | 5 2 | 4+4 4 | 4+4 4 | 3+4 4 | 6+6 4 | 3+4 4 | 3+4 4 | 3+4 4 |
| | 5 2 | 3+3 4 | 3+3 4 | 3+4 4 | 6+6 4 | 3+4 4 | 3+4 4 | 3+4 4 |
| | 5 2 | 3+3 4 | 4+4 4 | 3+4 4 | 6+6 4 | 3+4 4 | 3+4 4 | 3+4 4 |
| | 6 2 | 4+4 4 | 3+3 4 | 3+4 4 | 6+6 4 | 3+4 4 | 3+4 4 | 3+4 4 |
| | 6 2 | 3+3 4 | 3+3 4 | 3+4 4 | 6+6 4 | 3+4 4 | 3+4 4 | 3+4 4 |
| | 6 2 | 4+4 4 | 4+4 4 | 3+4 4 | 6+6 4 | 3+4 4 | 3+4 4 | 3+4 4 |
| | 5 2 | 3+3 4 | 3+3 4 | 3+4 4 | 6+6 4 | 3+4 4 | 3+4 4 | 3+4 4 |
| | 6 2 | 4+4 4 | 4+4 4 | 3+4 4 | 6+6 4 | 3+4 4 | 3+4 4 | 3+4 4 |
| | 6 2 | 3+3 4 | 4+4 4 | 3+4 4 | 5+5 4 | 3+4 4 | 3+4 4 | 3+4 4 |
| | 5 2 | 4+4 4 | 3+3 4 | 3+4 4 | 6+6 4 | 3+4 4 | 3+4 4 | 3+4 4 |
| | 5 2 | | | | | | | |
| | 5 2 | | | | | | | |
| | 5 2 | | | | | | | |
| | 6 2 | | | | | | | |
| | 6 2 | | | | | | | |
| | 5 2 | | | | | | | |
| | 6 2 | | | | | | | |
| | 5 2 | | | | | | | |
| | 6 2 | | | | | | | |
| | 6 2 | | | | | | | |
| | 6 2 | | | | | | | |
| | 6 2 | | | | | | | |
| | 5 2 | | | | | | | |
| | 5 2 | | | | | | | |
| | 5 2 | | | | | | | |
| | 6 2 | | | | | | | |
| | 6 2 | | | | | | | |
| | 6 2 | | | | | | | |
| | 5 2 | | | | | | | |
| | 5 2 | | | | | | | |
| | 5 2 | | | | | | | |
| | 6 2 | | | | | | | |
| | 5 2 | | | | | | | |
| | 5 2 | | | | | | | |
| | 6 2 | | | | | | | |
| | 6 2 | | | | | | | |
| | 5 2 | | | | | | | |
| | 5 2 | | | | | | | |
| | 5 2 | | | | | | | |
| | 5 2 | | | | | | | |
| | 5 2 | | | | | | | |
| | 5 2 | | | | | | | |
| | 5 2 | | | | | | | |
| | 5 2 | | | | | | | |
| | 5 2 | | | | | | | |
| | 5 2 | | | | | | | |
| | 5 2 | | | | | | | |
| | 6 2 | | | | | | | |
| | 5 2 | | | | | | | |
| | 5 2 | | | | | | | |
| | 5 2 | | | | | | | |
| 62.2.57 | 67722 | 168 | Nd1Ni1O3 | | | | | |
| | -GM | -R | -S | -T | -U | -X | -Y | -Z |
| | 5 2 | 3+3 4 | 3+3 4 | 3+4 4 | 5+5 4 | 3+4 4 | 3+4 4 | 3+4 4 |
| | 6 2 | 4+4 4 | 4+4 4 | 3+4 4 | 6+6 4 | 3+4 4 | 3+4 4 | 3+4 4 |
| | 5 2 | 3+3 4 | 3+3 4 | 3+4 4 | 6+6 4 | 3+4 4 | 3+4 4 | 3+4 4 |
| | 6 2 | 4+4 4 | 4+4 4 | 3+4 4 | 5+5 4 | 3+4 4 | 3+4 4 | 3+4 4 |
| | 5 2 | 3+3 4 | 4+4 4 | 3+4 4 | 5+5 4 | 3+4 4 | 3+4 4 | 3+4 4 |
| | 6 2 | 4+4 4 | 3+3 4 | 3+4 4 | 6+6 4 | 3+4 4 | 3+4 4 | 3+4 4 |
| | 5 2 | 4+4 4 | 3+3 4 | 3+4 4 | 6+6 4 | 3+4 4 | 3+4 4 | 3+4 4 |
| | 6 2 | 3+3 4 | 4+4 4 | 3+4 4 | 5+5 4 | 3+4 4 | 3+4 4 | 3+4 4 |
| | 6 2 | 4+4 4 | 3+3 4 | 3+4 4 | 5+5 4 | 3+4 4 | 3+4 4 | 3+4 4 |
| | 5 2 | 3+3 4 | 4+4 4 | 3+4 4 | 6+6 4 | 3+4 4 | 3+4 4 | 3+4 4 |
| | 5 2 | 4+4 4 | 4+4 4 | 3+4 4 | 6+6 4 | 3+4 4 | 3+4 4 | 3+4 4 |
| | 5 2 | 3+3 4 | 3+3 4 | 3+4 4 | 5+5 4 | 3+4 4 | 3+4 4 | 3+4 4 |
| | 6 2 | 3+3 4 | 3+3 4 | 3+4 4 | 5+5 4 | 3+4 4 | 3+4 4 | 3+4 4 |
| | 5 2 | 4+4 4 | 4+4 4 | 3+4 4 | 6+6 4 | 3+4 4 | 3+4 4 | 3+4 4 |
| | 6 2 | 3+3 4 | 4+4 4 | 3+4 4 | 6+6 4 | 3+4 4 | 3+4 4 | 3+4 4 |
| | 5 2 | 4+4 4 | 3+3 4 | 3+4 4 | 6+6 4 | 3+4 4 | 3+4 4 | 3+4 4 |
| | 6 2 | 4+4 4 | 3+3 4 | 3+4 4 | 6+6 4 | 3+4 4 | 3+4 4 | 3+4 4 |
| | 6 2 | 3+3 4 | 4+4 4 | 3+4 4 | 6+6 4 | 3+4 4 | 3+4 4 | 3+4 4 |
| | 6 2 | 3+3 4 | 3+3 4 | 3+4 4 | 5+5 4 | 3+4 4 | 3+4 4 | 3+4 4 |

Continued on next column

**TABLE XXXVII– continued**

| TQC | ICSD NoE SF |
|---|---|
| | 5 2 4+4 4 4+4 4 3+4 4 5+5 4 3+4 4 3+4 4 3+4 4 |
| | 5 2 3+3 4 3+3 4 3+4 4 6+6 4 3+4 4 3+4 4 3+4 4 |
| | 6 2 4+4 4 4+4 4 3+4 4 6+6 4 3+4 4 3+4 4 3+4 4 |
| | 5 2 3+3 4 3+3 4 3+4 4 5+5 4 3+4 4 3+4 4 3+4 4 |
| | 6 2 4+4 4 4+4 4 3+4 4 5+5 4 3+4 4 3+4 4 3+4 4 |
| | 5 2 4+4 4 4+4 4 3+4 4 6+6 4 3+4 4 3+4 4 3+4 4 |
| | 6 2 3+3 4 3+3 4 3+4 4 6+6 4 3+4 4 3+4 4 3+4 4 |
| | 5 2 4+4 4 3+3 4 3+4 4 5+5 4 3+4 4 3+4 4 3+4 4 |
| | 6 2 3+3 4 4+4 4 3+4 4 5+5 4 3+4 4 3+4 4 3+4 4 |
| | 5 2 3+3 4 3+3 4 3+4 4 5+5 4 3+4 4 3+4 4 3+4 4 |
| | 5 2 4+4 4 4+4 4 3+4 4 5+5 4 3+4 4 3+4 4 3+4 4 |
| | 5 2 3+3 4 3+3 4 3+4 4 5+5 4 3+4 4 3+4 4 3+4 4 |
| | 5 2 4+4 4 4+4 4 3+4 4 6+6 4 3+4 4 3+4 4 3+4 4 |
| | 5 2 4+4 4 4+4 4 3+4 4 6+6 4 3+4 4 3+4 4 3+4 4 |
| | 5 2 3+3 4 3+3 4 3+4 4 6+6 4 3+4 4 3+4 4 3+4 4 |
| | 5 2 3+3 4 4+4 4 3+4 4 6+6 4 3+4 4 3+4 4 3+4 4 |
| | 6 2 4+4 4 3+3 4 3+4 4 6+6 4 3+4 4 3+4 4 3+4 4 |
| | 6 2 3+3 4 3+3 4 3+4 4 6+6 4 3+4 4 3+4 4 3+4 4 |
| | 6 2 4+4 4 4+4 4 3+4 4 6+6 4 3+4 4 3+4 4 3+4 4 |
| | 5 2 3+3 4 4+4 4 3+4 4 6+6 4 3+4 4 3+4 4 3+4 4 |
| | 6 2 4+4 4 3+3 4 3+4 4 6+6 4 3+4 4 3+4 4 3+4 4 |
| | 6 2 3+3 4 3+3 4 3+4 4 5+5 4 3+4 4 3+4 4 3+4 4 |
| | 5 2 4+4 4 4+4 4 3+4 4 6+6 4 3+4 4 3+4 4 3+4 4 |
| | 5 2 |
| | 5 2 |
| | 5 2 |
| | 6 2 |
| | 6 2 |
| | 5 2 |
| | 6 2 |
| | 5 2 |
| | 6 2 |
| | 6 2 |
| | 6 2 |
| | 6 2 |
| | 5 2 |
| | 5 2 |
| | 5 2 |
| | 6 2 |
| | 6 2 |
| | 6 2 |
| | 5 2 |
| | 5 2 |
| | 5 2 |
| | 6 2 |
| | 5 2 |
| | 5 2 |
| | 6 2 |
| | 6 2 |
| | 5 2 |
| | 5 2 |
| | 5 2 |
| | 5 2 |
| | 5 2 |
| | 5 2 |
| | 5 2 |
| | 5 2 |
| | 5 2 |
| | 5 2 |
| | 5 2 |
| | 5 2 |
| | 5 2 |

Continued on next column

**TABLE XXXVII– continued**

| TQC | ICSD NoE SF |
|---|---|
| | 5 2 |
| | 6 2 |
| | 5 2 |
| | 5 2 |
| 62.2.58 | 75008 88 Nb1P1Zr1 |
| | -GM -R -S -T -U -X -Y -Z |
| | 5 2 3+3 4 3+3 4 3+4 4 5+5 4 3+4 4 3+4 4 3+4 4 |
| | 5 2 4+4 4 4+4 4 3+4 4 6+6 4 3+4 4 3+4 4 3+4 4 |
| | 6 2 3+3 4 3+3 4 3+4 4 5+5 4 3+4 4 3+4 4 3+4 4 |
| | 6 2 4+4 4 4+4 4 3+4 4 6+6 4 3+4 4 3+4 4 3+4 4 |
| | 5 2 3+3 4 3+3 4 3+4 4 5+5 4 3+4 4 3+4 4 3+4 4 |
| | 5 2 4+4 4 4+4 4 3+4 4 5+5 4 3+4 4 3+4 4 3+4 4 |
| | 6 2 3+3 4 3+3 4 3+4 4 6+6 4 3+4 4 3+4 4 3+4 4 |
| | 6 2 4+4 4 4+4 4 3+4 4 6+6 4 3+4 4 3+4 4 3+4 4 |
| | 5 2 3+3 4 3+3 4 3+4 4 5+5 4 3+4 4 3+4 4 3+4 4 |
| | 5 2 4+4 4 4+4 4 3+4 4 6+6 4 3+4 4 3+4 4 3+4 4 |
| | 5 2 3+3 4 3+3 4 3+4 4 5+5 4 3+4 4 3+4 4 3+4 4 |
| | 5 2 4+4 4 4+4 4 3+4 4 6+6 4 3+4 4 3+4 4 3+4 4 |
| | 6 2 4+4 4 4+4 4 3+4 4 6+6 4 3+4 4 3+4 4 3+4 4 |
| | 6 2 3+3 4 3+3 4 3+4 4 5+5 4 3+4 4 3+4 4 3+4 4 |
| | 6 2 3+3 4 4+4 4 3+4 4 6+6 4 3+4 4 3+4 4 3+4 4 |
| | 6 2 4+4 4 3+3 4 3+4 4 5+5 4 3+4 4 3+4 4 3+4 4 |
| | 5 2 4+4 4 3+3 4 3+4 4 6+6 4 3+4 4 3+4 4 3+4 4 |
| | 6 2 3+3 4 4+4 4 3+4 4 5+5 4 3+4 4 3+4 4 3+4 4 |
| | 6 2 4+4 4 3+3 4 3+4 4 6+6 4 3+4 4 3+4 4 3+4 4 |
| | 5 2 3+3 4 4+4 4 3+4 4 5+5 4 3+4 4 3+4 4 3+4 4 |
| | 5 2 3+3 4 4+4 4 3+4 4 5+5 4 3+4 4 3+4 4 3+4 4 |
| | 6 2 4+4 4 3+3 4 3+4 4 6+6 4 3+4 4 3+4 4 3+4 4 |
| | 6 2 |
| | 5 2 |
| | 5 2 |
| | 6 2 |
| | 6 2 |
| | 5 2 |
| | 6 2 |
| | 6 2 |
| | 5 2 |
| | 6 2 |
| | 6 2 |
| | 5 2 |
| | 6 2 |
| | 5 2 |
| | 5 2 |
| | 5 2 |
| | 5 2 |
| | 6 2 |
| | 5 2 |
| | 6 2 |
| | 5 2 |
| | 5 2 |
| 62.2.59 | 49728 112 Nb1Ni1P1 |
| | -GM -R -S -T -U -X -Y -Z |
| | 5 2 3+3 4 3+3 4 3+4 4 6+6 4 3+4 4 3+4 4 3+4 4 |
| | 5 2 4+4 4 4+4 4 3+4 4 5+5 4 3+4 4 3+4 4 3+4 4 |
| | 6 2 4+4 4 3+3 4 3+4 4 6+6 4 3+4 4 3+4 4 3+4 4 |
| | 6 2 3+3 4 4+4 4 3+4 4 5+5 4 3+4 4 3+4 4 3+4 4 |
| | 5 2 3+3 4 4+4 4 3+4 4 6+6 4 3+4 4 3+4 4 3+4 4 |
| | 5 2 4+4 4 3+3 4 3+4 4 5+5 4 3+4 4 3+4 4 3+4 4 |
| | 6 2 4+4 4 3+3 4 3+4 4 6+6 4 3+4 4 3+4 4 3+4 4 |
| | 6 2 3+3 4 4+4 4 3+4 4 5+5 4 3+4 4 3+4 4 3+4 4 |
| | 5 2 4+4 4 4+4 4 3+4 4 5+5 4 3+4 4 3+4 4 3+4 4 |

Continued on next column

**TABLE XXXVII– continued**

| TQC | ICSD | NoE | SF | | | | | |
|---|---|---|---|---|---|---|---|---|
| | 5 2 | 3+3 4 | 3+3 4 | 3+4 4 | 6+6 4 | 3+4 4 | 3+4 4 | 3+4 4 |
| | 5 2 | 3+3 4 | 3+3 4 | 3+4 4 | 5+5 4 | 3+4 4 | 3+4 4 | 3+4 4 |
| | 6 2 | 4+4 4 | 4+4 4 | 3+4 4 | 6+6 4 | 3+4 4 | 3+4 4 | 3+4 4 |
| | 6 2 | 3+3 4 | 3+3 4 | 3+4 4 | 6+6 4 | 3+4 4 | 3+4 4 | 3+4 4 |
| | 5 2 | 4+4 4 | 4+4 4 | 3+4 4 | 6+6 4 | 3+4 4 | 3+4 4 | 3+4 4 |
| | 6 2 | 3+3 4 | 3+3 4 | 3+4 4 | 5+5 4 | 3+4 4 | 3+4 4 | 3+4 4 |
| | 6 2 | 4+4 4 | 4+4 4 | 3+4 4 | 5+5 4 | 3+4 4 | 3+4 4 | 3+4 4 |
| | 5 2 | 3+3 4 | 4+4 4 | 3+4 4 | 5+5 4 | 3+4 4 | 3+4 4 | 3+4 4 |
| | 6 2 | 4+4 4 | 3+3 4 | 3+4 4 | 5+5 4 | 3+4 4 | 3+4 4 | 3+4 4 |
| | 5 2 | 4+4 4 | 3+3 4 | 3+4 4 | 6+6 4 | 3+4 4 | 3+4 4 | 3+4 4 |
| | 6 2 | 3+3 4 | 4+4 4 | 3+4 4 | 5+5 4 | 3+4 4 | 3+4 4 | 3+4 4 |
| | 5 2 | 4+4 4 | 4+4 4 | 3+4 4 | 6+6 4 | 3+4 4 | 3+4 4 | 3+4 4 |
| | 6 2 | 3+3 4 | 3+3 4 | 3+4 4 | 6+6 4 | 3+4 4 | 3+4 4 | 3+4 4 |
| | 5 2 | 3+3 4 | 4+4 4 | 3+4 4 | 5+5 4 | 3+4 4 | 3+4 4 | 3+4 4 |
| | 6 2 | 4+4 4 | 3+3 4 | 3+4 4 | 5+5 4 | 3+4 4 | 3+4 4 | 3+4 4 |
| | 5 2 | 3+3 4 | 4+4 4 | 3+4 4 | 6+6 4 | 3+4 4 | 3+4 4 | 3+4 4 |
| | 5 2 | 4+4 4 | 3+3 4 | 3+4 4 | 6+6 4 | 3+4 4 | 3+4 4 | 3+4 4 |
| | 6 2 | 3+3 4 | 4+4 4 | 3+4 4 | 5+5 4 | 3+4 4 | 3+4 4 | 3+4 4 |
| | 6 2 | 4+4 4 | 3+3 4 | 3+4 4 | 6+6 4 | 3+4 4 | 3+4 4 | 3+4 4 |
| | 6 2 | | | | | | | |
| | 6 2 | | | | | | | |
| | 6 2 | | | | | | | |
| | 5 2 | | | | | | | |
| | 5 2 | | | | | | | |
| | 6 2 | | | | | | | |
| | 5 2 | | | | | | | |
| | 5 2 | | | | | | | |
| | 6 2 | | | | | | | |
| | 5 2 | | | | | | | |
| | 6 2 | | | | | | | |
| | 5 2 | | | | | | | |
| | 5 2 | | | | | | | |
| | 6 2 | | | | | | | |
| | 5 2 | | | | | | | |
| | 5 2 | | | | | | | |
| | 6 2 | | | | | | | |
| | 5 2 | | | | | | | |
| | 5 2 | | | | | | | |
| | 6 2 | | | | | | | |
| | 6 2 | | | | | | | |
| | 5 2 | | | | | | | |
| | 5 2 | | | | | | | |
| | 6 2 | | | | | | | |
| | 5 2 | | | | | | | |
| | 5 2 | | | | | | | |
| | 6 2 | | | | | | | |
| | 6 2 | | | | | | | |
| 62.2.60 | 41149 | 80 | Mn1Rh1Si1 | | | | | |
| | -GM | -R | -S | -T | -U | -X | -Y | -Z |
| | 5 2 | 4+4 4 | 4+4 4 | 3+4 4 | 6+6 4 | 3+4 4 | 3+4 4 | 3+4 4 |
| | 6 2 | 3+3 4 | 3+3 4 | 3+4 4 | 5+5 4 | 3+4 4 | 3+4 4 | 3+4 4 |
| | 6 2 | 3+3 4 | 4+4 4 | 3+4 4 | 5+5 4 | 3+4 4 | 3+4 4 | 3+4 4 |
| | 5 2 | 4+4 4 | 3+3 4 | 3+4 4 | 6+6 4 | 3+4 4 | 3+4 4 | 3+4 4 |
| | 5 2 | 4+4 4 | 3+3 4 | 3+4 4 | 5+5 4 | 3+4 4 | 3+4 4 | 3+4 4 |
| | 6 2 | 3+3 4 | 4+4 4 | 3+4 4 | 5+5 4 | 3+4 4 | 3+4 4 | 3+4 4 |
| | 6 2 | 4+4 4 | 4+4 4 | 3+4 4 | 6+6 4 | 3+4 4 | 3+4 4 | 3+4 4 |
| | 5 2 | 3+3 4 | 3+3 4 | 3+4 4 | 6+6 4 | 3+4 4 | 3+4 4 | 3+4 4 |
| | 6 2 | 3+3 4 | 4+4 4 | 3+4 4 | 6+6 4 | 3+4 4 | 3+4 4 | 3+4 4 |
| | 5 2 | 4+4 4 | 3+3 4 | 3+4 4 | 5+5 4 | 3+4 4 | 3+4 4 | 3+4 4 |
| | 6 2 | 3+3 4 | 4+4 4 | 3+4 4 | 6+6 4 | 3+4 4 | 3+4 4 | 3+4 4 |
| | 5 2 | 4+4 4 | 3+3 4 | 3+4 4 | 5+5 4 | 3+4 4 | 3+4 4 | 3+4 4 |

Continued on next column

**TABLE XXXVII– continued**

| TQC | ICSD | NoE | SF | | | | | |
|---|---|---|---|---|---|---|---|---|
| | 6 2 | 4+4 4 | 3+3 4 | 3+4 4 | 5+5 4 | 3+4 4 | 3+4 4 | 3+4 4 |
| | 6 2 | 3+3 4 | 4+4 4 | 3+4 4 | 6+6 4 | 3+4 4 | 3+4 4 | 3+4 4 |
| | 5 2 | 4+4 4 | 3+3 4 | 3+4 4 | 6+6 4 | 3+4 4 | 3+4 4 | 3+4 4 |
| | 6 2 | 3+3 4 | 4+4 4 | 3+4 4 | 6+6 4 | 3+4 4 | 3+4 4 | 3+4 4 |
| | 5 2 | 4+4 4 | 3+3 4 | 3+4 4 | 5+5 4 | 3+4 4 | 3+4 4 | 3+4 4 |
| | 5 2 | 3+3 4 | 4+4 4 | 3+4 4 | 5+5 4 | 3+4 4 | 3+4 4 | 3+4 4 |
| | 5 2 | 4+4 4 | 3+3 4 | 3+4 4 | 6+6 4 | 3+4 4 | 3+4 4 | 3+4 4 |
| | 6 2 | 3+3 4 | 4+4 4 | 3+4 4 | 5+5 4 | 3+4 4 | 3+4 4 | 3+4 4 |
| | 6 2 | | | | | | | |
| | 5 2 | | | | | | | |
| | 5 2 | | | | | | | |
| | 6 2 | | | | | | | |
| | 6 2 | | | | | | | |
| | 5 2 | | | | | | | |
| | 5 2 | | | | | | | |
| | 5 2 | | | | | | | |
| | 6 2 | | | | | | | |
| | 5 2 | | | | | | | |
| | 5 2 | | | | | | | |
| | 6 2 | | | | | | | |
| | 5 2 | | | | | | | |
| | 5 2 | | | | | | | |
| | 5 2 | | | | | | | |
| | 6 2 | | | | | | | |
| | 5 2 | | | | | | | |
| | 5 2 | | | | | | | |
| | 6 2 | | | | | | | |
| | 6 2 | | | | | | | |
| 62.2.61 | 43399 | 48 | Mn1P1 | | | | | |
| | -GM | -R | -S | -T | -U | -X | -Y | -Z |
| | 5 2 | 3+3 4 | 4+4 4 | 3+4 4 | 6+6 4 | 3+4 4 | 3+4 4 | 3+4 4 |
| | 5 2 | 4+4 4 | 3+3 4 | 3+4 4 | 5+5 4 | 3+4 4 | 3+4 4 | 3+4 4 |
| | 6 2 | 4+4 4 | 3+3 4 | 3+4 4 | 6+6 4 | 3+4 4 | 3+4 4 | 3+4 4 |
| | 6 2 | 3+3 4 | 4+4 4 | 3+4 4 | 5+5 4 | 3+4 4 | 3+4 4 | 3+4 4 |
| | 6 2 | 4+4 4 | 3+3 4 | 3+4 4 | 6+6 4 | 3+4 4 | 3+4 4 | 3+4 4 |
| | 5 2 | 3+3 4 | 4+4 4 | 3+4 4 | 6+6 4 | 3+4 4 | 3+4 4 | 3+4 4 |
| | 5 2 | 3+3 4 | 3+3 4 | 3+4 4 | 5+5 4 | 3+4 4 | 3+4 4 | 3+4 4 |
| | 5 2 | 4+4 4 | 4+4 4 | 3+4 4 | 5+5 4 | 3+4 4 | 3+4 4 | 3+4 4 |
| | 6 2 | 3+3 4 | 3+3 4 | 3+4 4 | 6+6 4 | 3+4 4 | 3+4 4 | 3+4 4 |
| | 5 2 | 4+4 4 | 4+4 4 | 3+4 4 | 5+5 4 | 3+4 4 | 3+4 4 | 3+4 4 |
| | 5 2 | 4+4 4 | 4+4 4 | 3+4 4 | 5+5 4 | 3+4 4 | 3+4 4 | 3+4 4 |
| | 6 2 | 3+3 4 | 3+3 4 | 3+4 4 | 6+6 4 | 3+4 4 | 3+4 4 | 3+4 4 |
| | 6 2 | | | | | | | |
| | 5 2 | | | | | | | |
| | 5 2 | | | | | | | |
| | 6 2 | | | | | | | |
| | 5 2 | | | | | | | |
| | 6 2 | | | | | | | |
| | 5 2 | | | | | | | |
| | 5 2 | | | | | | | |
| | 5 2 | | | | | | | |
| | 6 2 | | | | | | | |
| | 6 2 | | | | | | | |
| | 6 2 | | | | | | | |
| 62.2.62 | 43246 | 48 | Mn1P1 | | | | | |
| | -GM | -R | -S | -T | -U | -X | -Y | -Z |
| | 5 2 | 3+3 4 | 4+4 4 | 3+4 4 | 6+6 4 | 3+4 4 | 3+4 4 | 3+4 4 |
| | 5 2 | 4+4 4 | 3+3 4 | 3+4 4 | 5+5 4 | 3+4 4 | 3+4 4 | 3+4 4 |
| | 6 2 | 4+4 4 | 3+3 4 | 3+4 4 | 6+6 4 | 3+4 4 | 3+4 4 | 3+4 4 |
| | 6 2 | 3+3 4 | 4+4 4 | 3+4 4 | 5+5 4 | 3+4 4 | 3+4 4 | 3+4 4 |
| | 6 2 | 4+4 4 | 3+3 4 | 3+4 4 | 6+6 4 | 3+4 4 | 3+4 4 | 3+4 4 |

Continued on next column

**TABLE XXXVII– continued**

| TQC | ICSD | NoE | SF | | | | | |
|---|---|---|---|---|---|---|---|---|
| | 5 2 | 3+3 4 | 4+4 4 | 3+4 4 | 6+6 4 | 3+4 4 | 3+4 4 | 3+4 4 |
| | 5 2 | 3+3 4 | 3+3 4 | 3+4 4 | 5+5 4 | 3+4 4 | 3+4 4 | 3+4 4 |
| | 5 2 | 4+4 4 | 4+4 4 | 3+4 4 | 5+5 4 | 3+4 4 | 3+4 4 | 3+4 4 |
| | 6 2 | 3+3 4 | 3+3 4 | 3+4 4 | 6+6 4 | 3+4 4 | 3+4 4 | 3+4 4 |
| | 5 2 | 4+4 4 | 4+4 4 | 3+4 4 | 5+5 4 | 3+4 4 | 3+4 4 | 3+4 4 |
| | 5 2 | 4+4 4 | 4+4 4 | 3+4 4 | 5+5 4 | 3+4 4 | 3+4 4 | 3+4 4 |
| | 6 2 | 3+3 4 | 3+3 4 | 3+4 4 | 6+6 4 | 3+4 4 | 3+4 4 | 3+4 4 |
| | 6 2 | | | | | | | |
| | 5 2 | | | | | | | |
| | 5 2 | | | | | | | |
| | 6 2 | | | | | | | |
| | 5 2 | | | | | | | |
| | 6 2 | | | | | | | |
| | 5 2 | | | | | | | |
| | 5 2 | | | | | | | |
| | 5 2 | | | | | | | |
| | 6 2 | | | | | | | |
| | 6 2 | | | | | | | |
| | 6 2 | | | | | | | |
| 62.2.63 | 185418 | 120 | Mn1O3V1 | | | | | |
| | -GM | -R | -S | -T | -U | -X | -Y | -Z |
| | 5 2 | 3+3 4 | 3+3 4 | 3+4 4 | 6+6 4 | 3+4 4 | 3+4 4 | 3+4 4 |
| | 5 2 | 4+4 4 | 4+4 4 | 3+4 4 | 6+6 4 | 3+4 4 | 3+4 4 | 3+4 4 |
| | 5 2 | 3+3 4 | 3+3 4 | 3+4 4 | 5+5 4 | 3+4 4 | 3+4 4 | 3+4 4 |
| | 6 2 | 4+4 4 | 4+4 4 | 3+4 4 | 6+6 4 | 3+4 4 | 3+4 4 | 3+4 4 |
| | 5 2 | 4+4 4 | 3+3 4 | 3+4 4 | 5+5 4 | 3+4 4 | 3+4 4 | 3+4 4 |
| | 5 2 | 3+3 4 | 4+4 4 | 3+4 4 | 5+5 4 | 3+4 4 | 3+4 4 | 3+4 4 |
| | 6 2 | 3+3 4 | 4+4 4 | 3+4 4 | 6+6 4 | 3+4 4 | 3+4 4 | 3+4 4 |
| | 5 2 | 4+4 4 | 3+3 4 | 3+4 4 | 6+6 4 | 3+4 4 | 3+4 4 | 3+4 4 |
| | 6 2 | 3+3 4 | 3+3 4 | 3+4 4 | 6+6 4 | 3+4 4 | 3+4 4 | 3+4 4 |
| | 6 2 | 4+4 4 | 4+4 4 | 3+4 4 | 6+6 4 | 3+4 4 | 3+4 4 | 3+4 4 |
| | 6 2 | 3+3 4 | 3+3 4 | 3+4 4 | 5+5 4 | 3+4 4 | 3+4 4 | 3+4 4 |
| | 6 2 | 4+4 4 | 4+4 4 | 3+4 4 | 5+5 4 | 3+4 4 | 3+4 4 | 3+4 4 |
| | 5 2 | 3+3 4 | 3+3 4 | 3+4 4 | 6+6 4 | 3+4 4 | 3+4 4 | 3+4 4 |
| | 5 2 | 4+4 4 | 4+4 4 | 3+4 4 | 5+5 4 | 3+4 4 | 3+4 4 | 3+4 4 |
| | 5 2 | 3+3 4 | 4+4 4 | 3+4 4 | 6+6 4 | 3+4 4 | 3+4 4 | 3+4 4 |
| | 5 2 | 4+4 4 | 3+3 4 | 3+4 4 | 5+5 4 | 3+4 4 | 3+4 4 | 3+4 4 |
| | 5 2 | 3+3 4 | 3+3 4 | 3+4 4 | 5+5 4 | 3+4 4 | 3+4 4 | 3+4 4 |
| | 5 2 | 4+4 4 | 4+4 4 | 3+4 4 | 6+6 4 | 3+4 4 | 3+4 4 | 3+4 4 |
| | 5 2 | 4+4 4 | 3+3 4 | 3+4 4 | 6+6 4 | 3+4 4 | 3+4 4 | 3+4 4 |
| | 5 2 | 3+3 4 | 4+4 4 | 3+4 4 | 5+5 4 | 3+4 4 | 3+4 4 | 3+4 4 |
| | 5 2 | 3+3 4 | 3+3 4 | 3+4 4 | 5+5 4 | 3+4 4 | 3+4 4 | 3+4 4 |
| | 5 2 | 4+4 4 | 4+4 4 | 3+4 4 | 6+6 4 | 3+4 4 | 3+4 4 | 3+4 4 |
| | 6 2 | 4+4 4 | 4+4 4 | 3+4 4 | 5+5 4 | 3+4 4 | 3+4 4 | 3+4 4 |
| | 6 2 | 3+3 4 | 3+3 4 | 3+4 4 | 5+5 4 | 3+4 4 | 3+4 4 | 3+4 4 |
| | 6 2 | 4+4 4 | 3+3 4 | 3+4 4 | 6+6 4 | 3+4 4 | 3+4 4 | 3+4 4 |
| | 6 2 | 3+3 4 | 4+4 4 | 3+4 4 | 6+6 4 | 3+4 4 | 3+4 4 | 3+4 4 |
| | 6 2 | 3+3 4 | 3+3 4 | 3+4 4 | 6+6 4 | 3+4 4 | 3+4 4 | 3+4 4 |
| | 6 2 | 4+4 4 | 4+4 4 | 3+4 4 | 5+5 4 | 3+4 4 | 3+4 4 | 3+4 4 |
| | 6 2 | 4+4 4 | 3+3 4 | 3+4 4 | 6+6 4 | 3+4 4 | 3+4 4 | 3+4 4 |
| | 5 2 | 3+3 4 | 4+4 4 | 3+4 4 | 5+5 4 | 3+4 4 | 3+4 4 | 3+4 4 |
| | 5 2 | | | | | | | |
| | 5 2 | | | | | | | |
| | 5 2 | | | | | | | |
| | 6 2 | | | | | | | |
| | 6 2 | | | | | | | |
| | 6 2 | | | | | | | |
| | 6 2 | | | | | | | |
| | 5 2 | | | | | | | |
| | 5 2 | | | | | | | |
| | 6 2 | | | | | | | |

Continued on next column

**TABLE XXXVII– continued**

| TQC | ICSD | NoE | SF | | | | | |
|---|---|---|---|---|---|---|---|---|
| | 5 2 | | | | | | | |
| | 5 2 | | | | | | | |
| | 6 2 | | | | | | | |
| | 6 2 | | | | | | | |
| | 6 2 | | | | | | | |
| | 6 2 | | | | | | | |
| | 6 2 | | | | | | | |
| | 6 2 | | | | | | | |
| | 5 2 | | | | | | | |
| | 5 2 | | | | | | | |
| | 5 2 | | | | | | | |
| | 5 2 | | | | | | | |
| | 5 2 | | | | | | | |
| | 6 2 | | | | | | | |
| | 6 2 | | | | | | | |
| | 5 2 | | | | | | | |
| | 5 2 | | | | | | | |
| | 6 2 | | | | | | | |
| | 6 2 | | | | | | | |
| | 6 2 | | | | | | | |
| 62.2.64 | 97084 | 152 | Mn1O3Pr1 | | | | | |
| | -GM | -R | -S | -T | -U | -X | -Y | -Z |
| | 5 2 | 3+3 4 | 3+3 4 | 3+4 4 | 5+5 4 | 3+4 4 | 3+4 4 | 3+4 4 |
| | 6 2 | 4+4 4 | 4+4 4 | 3+4 4 | 6+6 4 | 3+4 4 | 3+4 4 | 3+4 4 |
| | 5 2 | 3+3 4 | 3+3 4 | 3+4 4 | 6+6 4 | 3+4 4 | 3+4 4 | 3+4 4 |
| | 6 2 | 4+4 4 | 4+4 4 | 3+4 4 | 5+5 4 | 3+4 4 | 3+4 4 | 3+4 4 |
| | 5 2 | 3+3 4 | 4+4 4 | 3+4 4 | 5+5 4 | 3+4 4 | 3+4 4 | 3+4 4 |
| | 6 2 | 4+4 4 | 3+3 4 | 3+4 4 | 6+6 4 | 3+4 4 | 3+4 4 | 3+4 4 |
| | 5 2 | 4+4 4 | 3+3 4 | 3+4 4 | 6+6 4 | 3+4 4 | 3+4 4 | 3+4 4 |
| | 6 2 | 3+3 4 | 4+4 4 | 3+4 4 | 5+5 4 | 3+4 4 | 3+4 4 | 3+4 4 |
| | 6 2 | 4+4 4 | 3+3 4 | 3+4 4 | 6+6 4 | 3+4 4 | 3+4 4 | 3+4 4 |
| | 5 2 | 3+3 4 | 4+4 4 | 3+4 4 | 5+5 4 | 3+4 4 | 3+4 4 | 3+4 4 |
| | 5 2 | 3+3 4 | 4+4 4 | 3+4 4 | 6+6 4 | 3+4 4 | 3+4 4 | 3+4 4 |
| | 5 2 | 4+4 4 | 3+3 4 | 3+4 4 | 5+5 4 | 3+4 4 | 3+4 4 | 3+4 4 |
| | 5 2 | 3+3 4 | 3+3 4 | 3+4 4 | 5+5 4 | 3+4 4 | 3+4 4 | 3+4 4 |
| | 6 2 | 4+4 4 | 4+4 4 | 3+4 4 | 6+6 4 | 3+4 4 | 3+4 4 | 3+4 4 |
| | 6 2 | 3+3 4 | 4+4 4 | 3+4 4 | 6+6 4 | 3+4 4 | 3+4 4 | 3+4 4 |
| | 6 2 | 4+4 4 | 3+3 4 | 3+4 4 | 6+6 4 | 3+4 4 | 3+4 4 | 3+4 4 |
| | 5 2 | 4+4 4 | 3+3 4 | 3+4 4 | 6+6 4 | 3+4 4 | 3+4 4 | 3+4 4 |
| | 6 2 | 3+3 4 | 4+4 4 | 3+4 4 | 6+6 4 | 3+4 4 | 3+4 4 | 3+4 4 |
| | 6 2 | 3+3 4 | 3+3 4 | 3+4 4 | 5+5 4 | 3+4 4 | 3+4 4 | 3+4 4 |
| | 5 2 | 4+4 4 | 4+4 4 | 3+4 4 | 5+5 4 | 3+4 4 | 3+4 4 | 3+4 4 |
| | 5 2 | 3+3 4 | 3+3 4 | 3+4 4 | 5+5 4 | 3+4 4 | 3+4 4 | 3+4 4 |
| | 6 2 | 4+4 4 | 4+4 4 | 3+4 4 | 6+6 4 | 3+4 4 | 3+4 4 | 3+4 4 |
| | 5 2 | 3+3 4 | 3+3 4 | 3+4 4 | 6+6 4 | 3+4 4 | 3+4 4 | 3+4 4 |
| | 6 2 | 4+4 4 | 4+4 4 | 3+4 4 | 5+5 4 | 3+4 4 | 3+4 4 | 3+4 4 |
| | 5 2 | 4+4 4 | 4+4 4 | 3+4 4 | 5+5 4 | 3+4 4 | 3+4 4 | 3+4 4 |
| | 6 2 | 3+3 4 | 3+3 4 | 3+4 4 | 6+6 4 | 3+4 4 | 3+4 4 | 3+4 4 |
| | 5 2 | 4+4 4 | 3+3 4 | 3+4 4 | 6+6 4 | 3+4 4 | 3+4 4 | 3+4 4 |
| | 6 2 | 3+3 4 | 4+4 4 | 3+4 4 | 5+5 4 | 3+4 4 | 3+4 4 | 3+4 4 |
| | 5 2 | 3+3 4 | 3+3 4 | 3+4 4 | 5+5 4 | 3+4 4 | 3+4 4 | 3+4 4 |
| | 5 2 | 4+4 4 | 4+4 4 | 3+4 4 | 5+5 4 | 3+4 4 | 3+4 4 | 3+4 4 |
| | 5 2 | 3+3 4 | 3+3 4 | 3+4 4 | 5+5 4 | 3+4 4 | 3+4 4 | 3+4 4 |
| | 5 2 | 4+4 4 | 4+4 4 | 3+4 4 | 6+6 4 | 3+4 4 | 3+4 4 | 3+4 4 |
| | 5 2 | 4+4 4 | 4+4 4 | 3+4 4 | 6+6 4 | 3+4 4 | 3+4 4 | 3+4 4 |
| | 5 2 | 3+3 4 | 3+3 4 | 3+4 4 | 6+6 4 | 3+4 4 | 3+4 4 | 3+4 4 |
| | 6 2 | 4+4 4 | 4+4 4 | 3+4 4 | 6+6 4 | 3+4 4 | 3+4 4 | 3+4 4 |
| | 5 2 | 3+3 4 | 3+3 4 | 3+4 4 | 6+6 4 | 3+4 4 | 3+4 4 | 3+4 4 |
| | 6 2 | 3+3 4 | 3+3 4 | 3+4 4 | 5+5 4 | 3+4 4 | 3+4 4 | 3+4 4 |
| | 5 2 | 4+4 4 | 4+4 4 | 3+4 4 | 6+6 4 | 3+4 4 | 3+4 4 | 3+4 4 |
| | 6 2 | | | | | | | |

Continued on next column

**TABLE XXXVII– continued**

| TQC | ICSD NoE SF |
|---|---|
| | 6 2 |
| | 5 2 |
| | 6 2 |
| | 5 2 |
| | 5 2 |
| | 5 2 |
| | 6 2 |
| | 6 2 |
| | 6 2 |
| | 5 2 |
| | 5 2 |
| | 6 2 |
| | 6 2 |
| | 6 2 |
| | 6 2 |
| | 5 2 |
| | 5 2 |
| | 6 2 |
| | 5 2 |
| | 6 2 |
| | 6 2 |
| | 5 2 |
| | 6 2 |
| | 6 2 |
| | 6 2 |
| | 5 2 |
| | 5 2 |
| | 5 2 |
| | 5 2 |
| | 5 2 |
| | 5 2 |
| | 5 2 |
| | 5 2 |
| | 5 2 |
| | 5 2 |
| | 5 2 |
| | 6 2 |
| 62.2.65 | 97083 152 Mn1O3Pr1 |
| | -GM -R -S -T -U -X -Y -Z |
| | 5 2 3+3 4 3+3 4 3+4 4 5+5 4 3+4 4 3+4 4 3+4 4 |
| | 6 2 4+4 4 4+4 4 3+4 4 6+6 4 3+4 4 3+4 4 3+4 4 |
| | 5 2 3+3 4 3+3 4 3+4 4 6+6 4 3+4 4 3+4 4 3+4 4 |
| | 6 2 4+4 4 4+4 4 3+4 4 5+5 4 3+4 4 3+4 4 3+4 4 |
| | 5 2 3+3 4 4+4 4 3+4 4 5+5 4 3+4 4 3+4 4 3+4 4 |
| | 6 2 4+4 4 3+3 4 3+4 4 6+6 4 3+4 4 3+4 4 3+4 4 |
| | 5 2 4+4 4 3+3 4 3+4 4 6+6 4 3+4 4 3+4 4 3+4 4 |
| | 6 2 3+3 4 4+4 4 3+4 4 5+5 4 3+4 4 3+4 4 3+4 4 |
| | 6 2 4+4 4 3+3 4 3+4 4 6+6 4 3+4 4 3+4 4 3+4 4 |
| | 5 2 3+3 4 4+4 4 3+4 4 5+5 4 3+4 4 3+4 4 3+4 4 |
| | 5 2 3+3 4 4+4 4 3+4 4 6+6 4 3+4 4 3+4 4 3+4 4 |
| | 5 2 4+4 4 3+3 4 3+4 4 5+5 4 3+4 4 3+4 4 3+4 4 |
| | 5 2 3+3 4 3+3 4 3+4 4 5+5 4 3+4 4 3+4 4 3+4 4 |
| | 6 2 4+4 4 4+4 4 3+4 4 6+6 4 3+4 4 3+4 4 3+4 4 |
| | 6 2 3+3 4 4+4 4 3+4 4 6+6 4 3+4 4 3+4 4 3+4 4 |
| | 6 2 4+4 4 3+3 4 3+4 4 6+6 4 3+4 4 3+4 4 3+4 4 |
| | 5 2 4+4 4 3+3 4 3+4 4 6+6 4 3+4 4 3+4 4 3+4 4 |
| | 6 2 3+3 4 4+4 4 3+4 4 6+6 4 3+4 4 3+4 4 3+4 4 |
| | 6 2 3+3 4 3+3 4 3+4 4 5+5 4 3+4 4 3+4 4 3+4 4 |
| | 5 2 4+4 4 4+4 4 3+4 4 5+5 4 3+4 4 3+4 4 3+4 4 |
| | 5 2 3+3 4 3+3 4 3+4 4 5+5 4 3+4 4 3+4 4 3+4 4 |
| | 6 2 4+4 4 4+4 4 3+4 4 6+6 4 3+4 4 3+4 4 3+4 4 |
| | 5 2 3+3 4 4+4 4 3+4 4 6+6 4 3+4 4 3+4 4 3+4 4 |
| | 6 2 4+4 4 3+3 4 3+4 4 5+5 4 3+4 4 3+4 4 3+4 4 |
| | 5 2 4+4 4 4+4 4 3+4 4 5+5 4 3+4 4 3+4 4 3+4 4 |
| | 6 2 3+3 4 3+3 4 3+4 4 6+6 4 3+4 4 3+4 4 3+4 4 |
| | 5 2 4+4 4 3+3 4 3+4 4 6+6 4 3+4 4 3+4 4 3+4 4 |
| | 6 2 3+3 4 4+4 4 3+4 4 5+5 4 3+4 4 3+4 4 3+4 4 |
| | 5 2 3+3 4 3+3 4 3+4 4 5+5 4 3+4 4 3+4 4 3+4 4 |
| | 5 2 4+4 4 4+4 4 3+4 4 5+5 4 3+4 4 3+4 4 3+4 4 |
| | 5 2 3+3 4 3+3 4 3+4 4 5+5 4 3+4 4 3+4 4 3+4 4 |
| | 5 2 4+4 4 4+4 4 3+4 4 6+6 4 3+4 4 3+4 4 3+4 4 |
| | 5 2 4+4 4 4+4 4 3+4 4 6+6 4 3+4 4 3+4 4 3+4 4 |
| | 5 2 3+3 4 3+3 4 3+4 4 6+6 4 3+4 4 3+4 4 3+4 4 |
| | 6 2 4+4 4 4+4 4 3+4 4 6+6 4 3+4 4 3+4 4 3+4 4 |
| | 5 2 3+3 4 3+3 4 3+4 4 6+6 4 3+4 4 3+4 4 3+4 4 |
| | 6 2 3+3 4 3+3 4 3+4 4 5+5 4 3+4 4 3+4 4 3+4 4 |
| | 5 2 4+4 4 4+4 4 3+4 4 6+6 4 3+4 4 3+4 4 3+4 4 |
| | 6 2 |
| | 6 2 |
| | 5 2 |
| | 6 2 |
| | 5 2 |
| | 5 2 |
| | 5 2 |
| | 6 2 |
| | 6 2 |
| | 6 2 |
| | 5 2 |
| | 5 2 |
| | 6 2 |
| | 6 2 |
| | 6 2 |
| | 6 2 |
| | 5 2 |
| | 5 2 |
| | 6 2 |
| | 5 2 |
| | 6 2 |
| | 6 2 |
| | 5 2 |
| | 6 2 |
| | 6 2 |
| | 6 2 |
| | 5 2 |
| | 5 2 |
| | 5 2 |
| | 5 2 |
| | 5 2 |
| | 5 2 |
| | 5 2 |
| | 5 2 |
| | 5 2 |
| | 5 2 |
| | 5 2 |
| | 6 2 |
| 62.2.66 | 97082 152 Mn1O3Pr1 |
| | -GM -R -S -T -U -X -Y -Z |
| | 5 2 3+3 4 3+3 4 3+4 4 5+5 4 3+4 4 3+4 4 3+4 4 |
| | 6 2 4+4 4 4+4 4 3+4 4 6+6 4 3+4 4 3+4 4 3+4 4 |
| | 5 2 4+4 4 3+3 4 3+4 4 6+6 4 3+4 4 3+4 4 3+4 4 |
| | 6 2 3+3 4 4+4 4 3+4 4 5+5 4 3+4 4 3+4 4 3+4 4 |
| | 5 2 4+4 4 4+4 4 3+4 4 5+5 4 3+4 4 3+4 4 3+4 4 |

Continued on next column

**TABLE XXXVII– continued**

| TQC | ICSD | NoE | SF | | | | | |
|---|---|---|---|---|---|---|---|---|
| | 6 2 | 3+3 4 | 3+3 4 | 3+4 4 | 6+6 4 | 3+4 4 | 3+4 4 | 3+4 4 |
| | 5 2 | 3+3 4 | 3+3 4 | 3+4 4 | 6+6 4 | 3+4 4 | 3+4 4 | 3+4 4 |
| | 6 2 | 4+4 4 | 4+4 4 | 3+4 4 | 5+5 4 | 3+4 4 | 3+4 4 | 3+4 4 |
| | 6 2 | 3+3 4 | 3+3 4 | 3+4 4 | 6+6 4 | 3+4 4 | 3+4 4 | 3+4 4 |
| | 5 2 | 4+4 4 | 4+4 4 | 3+4 4 | 5+5 4 | 3+4 4 | 3+4 4 | 3+4 4 |
| | 5 2 | 4+4 4 | 4+4 4 | 3+4 4 | 6+6 4 | 3+4 4 | 3+4 4 | 3+4 4 |
| | 5 2 | 3+3 4 | 3+3 4 | 3+4 4 | 5+5 4 | 3+4 4 | 3+4 4 | 3+4 4 |
| | 5 2 | 4+4 4 | 3+3 4 | 3+4 4 | 5+5 4 | 3+4 4 | 3+4 4 | 3+4 4 |
| | 6 2 | 3+3 4 | 4+4 4 | 3+4 4 | 6+6 4 | 3+4 4 | 3+4 4 | 3+4 4 |
| | 6 2 | 4+4 4 | 4+4 4 | 3+4 4 | 6+6 4 | 3+4 4 | 3+4 4 | 3+4 4 |
| | 6 2 | 3+3 4 | 3+3 4 | 3+4 4 | 6+6 4 | 3+4 4 | 3+4 4 | 3+4 4 |
| | 5 2 | 3+3 4 | 4+4 4 | 3+4 4 | 6+6 4 | 3+4 4 | 3+4 4 | 3+4 4 |
| | 6 2 | 4+4 4 | 3+3 4 | 3+4 4 | 6+6 4 | 3+4 4 | 3+4 4 | 3+4 4 |
| | 6 2 | 3+3 4 | 3+3 4 | 3+4 4 | 5+5 4 | 3+4 4 | 3+4 4 | 3+4 4 |
| | 5 2 | 4+4 4 | 4+4 4 | 3+4 4 | 5+5 4 | 3+4 4 | 3+4 4 | 3+4 4 |
| | 5 2 | 4+4 4 | 4+4 4 | 3+4 4 | 5+5 4 | 3+4 4 | 3+4 4 | 3+4 4 |
| | 6 2 | 3+3 4 | 3+3 4 | 3+4 4 | 6+6 4 | 3+4 4 | 3+4 4 | 3+4 4 |
| | 6 2 | 4+4 4 | 4+4 4 | 3+4 4 | 6+6 4 | 3+4 4 | 3+4 4 | 3+4 4 |
| | 5 2 | 3+3 4 | 3+3 4 | 3+4 4 | 5+5 4 | 3+4 4 | 3+4 4 | 3+4 4 |
| | 5 2 | 3+3 4 | 4+4 4 | 3+4 4 | 5+5 4 | 3+4 4 | 3+4 4 | 3+4 4 |
| | 6 2 | 4+4 4 | 3+3 4 | 3+4 4 | 6+6 4 | 3+4 4 | 3+4 4 | 3+4 4 |
| | 5 2 | 3+3 4 | 3+3 4 | 3+4 4 | 6+6 4 | 3+4 4 | 3+4 4 | 3+4 4 |
| | 6 2 | 4+4 4 | 4+4 4 | 3+4 4 | 5+5 4 | 3+4 4 | 3+4 4 | 3+4 4 |
| | 5 2 | 3+3 4 | 4+4 4 | 3+4 4 | 5+5 4 | 3+4 4 | 3+4 4 | 3+4 4 |
| | 5 2 | 4+4 4 | 3+3 4 | 3+4 4 | 5+5 4 | 3+4 4 | 3+4 4 | 3+4 4 |
| | 5 2 | 4+4 4 | 3+3 4 | 3+4 4 | 6+6 4 | 3+4 4 | 3+4 4 | 3+4 4 |
| | 5 2 | 3+3 4 | 4+4 4 | 3+4 4 | 5+5 4 | 3+4 4 | 3+4 4 | 3+4 4 |
| | 5 2 | 3+3 4 | 4+4 4 | 3+4 4 | 6+6 4 | 3+4 4 | 3+4 4 | 3+4 4 |
| | 6 2 | 4+4 4 | 3+3 4 | 3+4 4 | 6+6 4 | 3+4 4 | 3+4 4 | 3+4 4 |
| | 5 2 | 3+3 4 | 4+4 4 | 3+4 4 | 6+6 4 | 3+4 4 | 3+4 4 | 3+4 4 |
| | 6 2 | 4+4 4 | 3+3 4 | 3+4 4 | 6+6 4 | 3+4 4 | 3+4 4 | 3+4 4 |
| | 5 2 | 4+4 4 | 3+3 4 | 3+4 4 | 6+6 4 | 3+4 4 | 3+4 4 | 3+4 4 |
| | 5 2 | 3+3 4 | 4+4 4 | 3+4 4 | 5+5 4 | 3+4 4 | 3+4 4 | 3+4 4 |
| | 6 2 | | | | | | | |
| | 5 2 | | | | | | | |
| | 6 2 | | | | | | | |
| | 5 2 | | | | | | | |
| | 6 2 | | | | | | | |
| | 5 2 | | | | | | | |
| | 5 2 | | | | | | | |
| | 6 2 | | | | | | | |
| | 5 2 | | | | | | | |
| | 6 2 | | | | | | | |
| | 6 2 | | | | | | | |
| | 5 2 | | | | | | | |
| | 6 2 | | | | | | | |
| | 6 2 | | | | | | | |
| | 5 2 | | | | | | | |
| | 5 2 | | | | | | | |
| | 6 2 | | | | | | | |
| | 6 2 | | | | | | | |
| | 6 2 | | | | | | | |
| | 5 2 | | | | | | | |
| | 6 2 | | | | | | | |
| | 5 2 | | | | | | | |
| | 6 2 | | | | | | | |
| | 6 2 | | | | | | | |
| | 6 2 | | | | | | | |
| | 6 2 | | | | | | | |
| | 5 2 | | | | | | | |
| | 5 2 | | | | | | | |

Continued on next column

**TABLE XXXVII– continued**

| TQC | ICSD | NoE | SF | | | | | |
|---|---|---|---|---|---|---|---|---|
| | 5 2 | | | | | | | |
| | 5 2 | | | | | | | |
| | 5 2 | | | | | | | |
| | 5 2 | | | | | | | |
| | 5 2 | | | | | | | |
| | 5 2 | | | | | | | |
| | 5 2 | | | | | | | |
| | 5 2 | | | | | | | |
| | 6 2 | | | | | | | |
| | 5 2 | | | | | | | |
| 62.2.67 | 84922 | 152 | Mn1O3Pr1 | | | | | |
| | -GM | -R | -S | -T | -U | -X | -Y | -Z |
| | 5 2 | 3+3 4 | 3+3 4 | 3+4 4 | 5+5 4 | 3+4 4 | 3+4 4 | 3+4 4 |
| | 6 2 | 4+4 4 | 4+4 4 | 3+4 4 | 6+6 4 | 3+4 4 | 3+4 4 | 3+4 4 |
| | 5 2 | 3+3 4 | 4+4 4 | 3+4 4 | 6+6 4 | 3+4 4 | 3+4 4 | 3+4 4 |
| | 6 2 | 4+4 4 | 3+3 4 | 3+4 4 | 5+5 4 | 3+4 4 | 3+4 4 | 3+4 4 |
| | 5 2 | 3+3 4 | 4+4 4 | 3+4 4 | 5+5 4 | 3+4 4 | 3+4 4 | 3+4 4 |
| | 6 2 | 4+4 4 | 3+3 4 | 3+4 4 | 6+6 4 | 3+4 4 | 3+4 4 | 3+4 4 |
| | 5 2 | 4+4 4 | 3+3 4 | 3+4 4 | 6+6 4 | 3+4 4 | 3+4 4 | 3+4 4 |
| | 6 2 | 3+3 4 | 4+4 4 | 3+4 4 | 5+5 4 | 3+4 4 | 3+4 4 | 3+4 4 |
| | 6 2 | 4+4 4 | 3+3 4 | 3+4 4 | 6+6 4 | 3+4 4 | 3+4 4 | 3+4 4 |
| | 5 2 | 3+3 4 | 4+4 4 | 3+4 4 | 5+5 4 | 3+4 4 | 3+4 4 | 3+4 4 |
| | 5 2 | 3+3 4 | 4+4 4 | 3+4 4 | 6+6 4 | 3+4 4 | 3+4 4 | 3+4 4 |
| | 5 2 | 4+4 4 | 3+3 4 | 3+4 4 | 5+5 4 | 3+4 4 | 3+4 4 | 3+4 4 |
| | 5 2 | 3+3 4 | 3+3 4 | 3+4 4 | 5+5 4 | 3+4 4 | 3+4 4 | 3+4 4 |
| | 6 2 | 4+4 4 | 4+4 4 | 3+4 4 | 6+6 4 | 3+4 4 | 3+4 4 | 3+4 4 |
| | 6 2 | 3+3 4 | 4+4 4 | 3+4 4 | 6+6 4 | 3+4 4 | 3+4 4 | 3+4 4 |
| | 6 2 | 4+4 4 | 3+3 4 | 3+4 4 | 6+6 4 | 3+4 4 | 3+4 4 | 3+4 4 |
| | 5 2 | 4+4 4 | 4+4 4 | 3+4 4 | 6+6 4 | 3+4 4 | 3+4 4 | 3+4 4 |
| | 6 2 | 3+3 4 | 3+3 4 | 3+4 4 | 6+6 4 | 3+4 4 | 3+4 4 | 3+4 4 |
| | 6 2 | 3+3 4 | 3+3 4 | 3+4 4 | 5+5 4 | 3+4 4 | 3+4 4 | 3+4 4 |
| | 5 2 | 4+4 4 | 4+4 4 | 3+4 4 | 5+5 4 | 3+4 4 | 3+4 4 | 3+4 4 |
| | 5 2 | 3+3 4 | 4+4 4 | 3+4 4 | 5+5 4 | 3+4 4 | 3+4 4 | 3+4 4 |
| | 6 2 | 4+4 4 | 3+3 4 | 3+4 4 | 6+6 4 | 3+4 4 | 3+4 4 | 3+4 4 |
| | 5 2 | 3+3 4 | 4+4 4 | 3+4 4 | 6+6 4 | 3+4 4 | 3+4 4 | 3+4 4 |
| | 6 2 | 4+4 4 | 3+3 4 | 3+4 4 | 5+5 4 | 3+4 4 | 3+4 4 | 3+4 4 |
| | 5 2 | 4+4 4 | 4+4 4 | 3+4 4 | 5+5 4 | 3+4 4 | 3+4 4 | 3+4 4 |
| | 6 2 | 3+3 4 | 3+3 4 | 3+4 4 | 6+6 4 | 3+4 4 | 3+4 4 | 3+4 4 |
| | 5 2 | 4+4 4 | 3+3 4 | 3+4 4 | 6+6 4 | 3+4 4 | 3+4 4 | 3+4 4 |
| | 6 2 | 3+3 4 | 4+4 4 | 3+4 4 | 5+5 4 | 3+4 4 | 3+4 4 | 3+4 4 |
| | 5 2 | 4+4 4 | 3+3 4 | 3+4 4 | 5+5 4 | 3+4 4 | 3+4 4 | 3+4 4 |
| | 5 2 | 3+3 4 | 4+4 4 | 3+4 4 | 5+5 4 | 3+4 4 | 3+4 4 | 3+4 4 |
| | 5 2 | 3+3 4 | 3+3 4 | 3+4 4 | 6+6 4 | 3+4 4 | 3+4 4 | 3+4 4 |
| | 5 2 | 4+4 4 | 4+4 4 | 3+4 4 | 5+5 4 | 3+4 4 | 3+4 4 | 3+4 4 |
| | 6 2 | 4+4 4 | 4+4 4 | 3+4 4 | 6+6 4 | 3+4 4 | 3+4 4 | 3+4 4 |
| | 5 2 | 3+3 4 | 3+3 4 | 3+4 4 | 6+6 4 | 3+4 4 | 3+4 4 | 3+4 4 |
| | 5 2 | 4+4 4 | 4+4 4 | 3+4 4 | 6+6 4 | 3+4 4 | 3+4 4 | 3+4 4 |
| | 6 2 | 3+3 4 | 3+3 4 | 3+4 4 | 6+6 4 | 3+4 4 | 3+4 4 | 3+4 4 |
| | 5 2 | 3+3 4 | 3+3 4 | 3+4 4 | 6+6 4 | 3+4 4 | 3+4 4 | 3+4 4 |
| | 5 2 | 4+4 4 | 4+4 4 | 3+4 4 | 5+5 4 | 3+4 4 | 3+4 4 | 3+4 4 |
| | 6 2 | | | | | | | |
| | 6 2 | | | | | | | |
| | 5 2 | | | | | | | |
| | 5 2 | | | | | | | |
| | 6 2 | | | | | | | |
| | 5 2 | | | | | | | |
| | 5 2 | | | | | | | |
| | 5 2 | | | | | | | |
| | 6 2 | | | | | | | |
| | 6 2 | | | | | | | |
| | 6 2 | | | | | | | |

Continued on next column

**TABLE XXXVII– continued**

| TQC | ICSD | NoE | SF | | | | | |
|---|---|---|---|---|---|---|---|---|
| | 5 2 | | | | | | | |
| | 6 2 | | | | | | | |
| | 6 2 | | | | | | | |
| | 5 2 | | | | | | | |
| | 6 2 | | | | | | | |
| | 5 2 | | | | | | | |
| | 6 2 | | | | | | | |
| | 5 2 | | | | | | | |
| | 6 2 | | | | | | | |
| | 6 2 | | | | | | | |
| | 5 2 | | | | | | | |
| | 6 2 | | | | | | | |
| | 6 2 | | | | | | | |
| | 6 2 | | | | | | | |
| | 6 2 | | | | | | | |
| | 5 2 | | | | | | | |
| | 5 2 | | | | | | | |
| | 5 2 | | | | | | | |
| | 5 2 | | | | | | | |
| | 5 2 | | | | | | | |
| | 5 2 | | | | | | | |
| | 5 2 | | | | | | | |
| | 5 2 | | | | | | | |
| | 5 2 | | | | | | | |
| | 5 2 | | | | | | | |
| | 6 2 | | | | | | | |
| | 5 2 | | | | | | | |
| 62.2.68 | 426614 | 88 | Mg1Pt1Sr1 | | | | | |
| | -GM | -R | -S | -T | -U | -X | -Y | -Z |
| | 5 2 | 3+3 4 | 3+3 4 | 3+4 4 | 6+6 4 | 3+4 4 | 3+4 4 | 3+4 4 |
| | 5 2 | 4+4 4 | 4+4 4 | 3+4 4 | 5+5 4 | 3+4 4 | 3+4 4 | 3+4 4 |
| | 6 2 | 3+3 4 | 3+3 4 | 3+4 4 | 6+6 4 | 3+4 4 | 3+4 4 | 3+4 4 |
| | 6 2 | 4+4 4 | 4+4 4 | 3+4 4 | 5+5 4 | 3+4 4 | 3+4 4 | 3+4 4 |
| | 5 2 | 4+4 4 | 3+3 4 | 3+4 4 | 6+6 4 | 3+4 4 | 3+4 4 | 3+4 4 |
| | 5 2 | 3+3 4 | 4+4 4 | 3+4 4 | 5+5 4 | 3+4 4 | 3+4 4 | 3+4 4 |
| | 6 2 | 3+3 4 | 4+4 4 | 3+4 4 | 6+6 4 | 3+4 4 | 3+4 4 | 3+4 4 |
| | 6 2 | 4+4 4 | 3+3 4 | 3+4 4 | 5+5 4 | 3+4 4 | 3+4 4 | 3+4 4 |
| | 5 2 | 3+3 4 | 4+4 4 | 3+4 4 | 5+5 4 | 3+4 4 | 3+4 4 | 3+4 4 |
| | 5 2 | 4+4 4 | 3+3 4 | 3+4 4 | 6+6 4 | 3+4 4 | 3+4 4 | 3+4 4 |
| | 6 2 | 3+3 4 | 4+4 4 | 3+4 4 | 6+6 4 | 3+4 4 | 3+4 4 | 3+4 4 |
| | 5 2 | 4+4 4 | 3+3 4 | 3+4 4 | 5+5 4 | 3+4 4 | 3+4 4 | 3+4 4 |
| | 6 2 | 4+4 4 | 3+3 4 | 3+4 4 | 5+5 4 | 3+4 4 | 3+4 4 | 3+4 4 |
| | 6 2 | 3+3 4 | 4+4 4 | 3+4 4 | 6+6 4 | 3+4 4 | 3+4 4 | 3+4 4 |
| | 5 2 | 4+4 4 | 3+3 4 | 3+4 4 | 5+5 4 | 3+4 4 | 3+4 4 | 3+4 4 |
| | 6 2 | 3+3 4 | 4+4 4 | 3+4 4 | 6+6 4 | 3+4 4 | 3+4 4 | 3+4 4 |
| | 5 2 | 4+4 4 | 4+4 4 | 3+4 4 | 5+5 4 | 3+4 4 | 3+4 4 | 3+4 4 |
| | 5 2 | 3+3 4 | 3+3 4 | 3+4 4 | 6+6 4 | 3+4 4 | 3+4 4 | 3+4 4 |
| | 6 2 | 3+3 4 | 3+3 4 | 3+4 4 | 5+5 4 | 3+4 4 | 3+4 4 | 3+4 4 |
| | 6 2 | 4+4 4 | 4+4 4 | 3+4 4 | 6+6 4 | 3+4 4 | 3+4 4 | 3+4 4 |
| | 6 2 | 3+3 4 | 4+4 4 | 3+4 4 | 5+5 4 | 3+4 4 | 3+4 4 | 3+4 4 |
| | 5 2 | 4+4 4 | 3+3 4 | 3+4 4 | 6+6 4 | 3+4 4 | 3+4 4 | 3+4 4 |
| | 5 2 | | | | | | | |
| | 6 2 | | | | | | | |
| | 5 2 | | | | | | | |
| | 6 2 | | | | | | | |
| | 5 2 | | | | | | | |
| | 6 2 | | | | | | | |
| | 6 2 | | | | | | | |
| | 5 2 | | | | | | | |
| | 5 2 | | | | | | | |
| | 5 2 | | | | | | | |

Continued on next column

**TABLE XXXVII– continued**

| TQC | ICSD | NoE | SF | | | | | |
|---|---|---|---|---|---|---|---|---|
| | 6 2 | | | | | | | |
| | 6 2 | | | | | | | |
| | 5 2 | | | | | | | |
| | 5 2 | | | | | | | |
| | 5 2 | | | | | | | |
| | 6 2 | | | | | | | |
| | 6 2 | | | | | | | |
| | 6 2 | | | | | | | |
| | 5 2 | | | | | | | |
| | 6 2 | | | | | | | |
| | 5 2 | | | | | | | |
| | 5 2 | | | | | | | |
| 62.2.69 | 97765 | 160 | Li1Ni1O4P1 | | | | | |
| | -GM | -R | -S | -T | -U | -X | -Y | -Z |
| | 5 2 | 3+3 4 | 3+3 4 | 3+4 4 | 6+6 4 | 3+4 4 | 3+4 4 | 3+4 4 |
| | 5 2 | 4+4 4 | 4+4 4 | 3+4 4 | 5+5 4 | 3+4 4 | 3+4 4 | 3+4 4 |
| | 6 2 | 3+3 4 | 3+3 4 | 3+4 4 | 5+5 4 | 3+4 4 | 3+4 4 | 3+4 4 |
| | 6 2 | 4+4 4 | 4+4 4 | 3+4 4 | 6+6 4 | 3+4 4 | 3+4 4 | 3+4 4 |
| | 6 2 | 3+3 4 | 3+3 4 | 3+4 4 | 6+6 4 | 3+4 4 | 3+4 4 | 3+4 4 |
| | 6 2 | 4+4 4 | 4+4 4 | 3+4 4 | 5+5 4 | 3+4 4 | 3+4 4 | 3+4 4 |
| | 5 2 | 3+3 4 | 3+3 4 | 3+4 4 | 5+5 4 | 3+4 4 | 3+4 4 | 3+4 4 |
| | 6 2 | 4+4 4 | 4+4 4 | 3+4 4 | 6+6 4 | 3+4 4 | 3+4 4 | 3+4 4 |
| | 5 2 | 3+3 4 | 3+3 4 | 3+4 4 | 6+6 4 | 3+4 4 | 3+4 4 | 3+4 4 |
| | 5 2 | 4+4 4 | 4+4 4 | 3+4 4 | 5+5 4 | 3+4 4 | 3+4 4 | 3+4 4 |
| | 5 2 | 3+3 4 | 3+3 4 | 3+4 4 | 6+6 4 | 3+4 4 | 3+4 4 | 3+4 4 |
| | 6 2 | 4+4 4 | 4+4 4 | 3+4 4 | 6+6 4 | 3+4 4 | 3+4 4 | 3+4 4 |
| | 6 2 | 4+4 4 | 3+3 4 | 3+4 4 | 5+5 4 | 3+4 4 | 3+4 4 | 3+4 4 |
| | 6 2 | 3+3 4 | 4+4 4 | 3+4 4 | 6+6 4 | 3+4 4 | 3+4 4 | 3+4 4 |
| | 5 2 | 3+3 4 | 3+3 4 | 3+4 4 | 5+5 4 | 3+4 4 | 3+4 4 | 3+4 4 |
| | 5 2 | 4+4 4 | 4+4 4 | 3+4 4 | 5+5 4 | 3+4 4 | 3+4 4 | 3+4 4 |
| | 5 2 | 4+4 4 | 3+3 4 | 3+4 4 | 5+5 4 | 3+4 4 | 3+4 4 | 3+4 4 |
| | 5 2 | 3+3 4 | 4+4 4 | 3+4 4 | 6+6 4 | 3+4 4 | 3+4 4 | 3+4 4 |
| | 6 2 | 4+4 4 | 4+4 4 | 3+4 4 | 5+5 4 | 3+4 4 | 3+4 4 | 3+4 4 |
| | 6 2 | 3+3 4 | 3+3 4 | 3+4 4 | 5+5 4 | 3+4 4 | 3+4 4 | 3+4 4 |
| | 5 2 | 4+4 4 | 4+4 4 | 3+4 4 | 6+6 4 | 3+4 4 | 3+4 4 | 3+4 4 |
| | 5 2 | 3+3 4 | 3+3 4 | 3+4 4 | 6+6 4 | 3+4 4 | 3+4 4 | 3+4 4 |
| | 6 2 | 4+4 4 | 3+3 4 | 3+4 4 | 6+6 4 | 3+4 4 | 3+4 4 | 3+4 4 |
| | 6 2 | 3+3 4 | 4+4 4 | 3+4 4 | 6+6 4 | 3+4 4 | 3+4 4 | 3+4 4 |
| | 5 2 | 3+3 4 | 4+4 4 | 3+4 4 | 5+5 4 | 3+4 4 | 3+4 4 | 3+4 4 |
| | 6 2 | 4+4 4 | 3+3 4 | 3+4 4 | 5+5 4 | 3+4 4 | 3+4 4 | 3+4 4 |
| | 5 2 | 4+4 4 | 4+4 4 | 3+4 4 | 6+6 4 | 3+4 4 | 3+4 4 | 3+4 4 |
| | 5 2 | 3+3 4 | 3+3 4 | 3+4 4 | 6+6 4 | 3+4 4 | 3+4 4 | 3+4 4 |
| | 6 2 | 3+3 4 | 3+3 4 | 3+4 4 | 5+5 4 | 3+4 4 | 3+4 4 | 3+4 4 |
| | 5 2 | 4+4 4 | 4+4 4 | 3+4 4 | 5+5 4 | 3+4 4 | 3+4 4 | 3+4 4 |
| | 6 2 | 3+3 4 | 3+3 4 | 3+4 4 | 5+5 4 | 3+4 4 | 3+4 4 | 3+4 4 |
| | 6 2 | 4+4 4 | 4+4 4 | 3+4 4 | 6+6 4 | 3+4 4 | 3+4 4 | 3+4 4 |
| | 6 2 | 4+4 4 | 4+4 4 | 3+4 4 | 6+6 4 | 3+4 4 | 3+4 4 | 3+4 4 |
| | 6 2 | 3+3 4 | 3+3 4 | 3+4 4 | 5+5 4 | 3+4 4 | 3+4 4 | 3+4 4 |
| | 5 2 | 4+4 4 | 3+3 4 | 3+4 4 | 6+6 4 | 3+4 4 | 3+4 4 | 3+4 4 |
| | 5 2 | 3+3 4 | 4+4 4 | 3+4 4 | 5+5 4 | 3+4 4 | 3+4 4 | 3+4 4 |
| | 6 2 | 4+4 4 | 3+3 4 | 3+4 4 | 6+6 4 | 3+4 4 | 3+4 4 | 3+4 4 |
| | 6 2 | 3+3 4 | 4+4 4 | 3+4 4 | 5+5 4 | 3+4 4 | 3+4 4 | 3+4 4 |
| | 6 2 | 3+3 4 | 4+4 4 | 3+4 4 | 5+5 4 | 3+4 4 | 3+4 4 | 3+4 4 |
| | 5 2 | 4+4 4 | 3+3 4 | 3+4 4 | 6+6 4 | 3+4 4 | 3+4 4 | 3+4 4 |
| | 5 2 | | | | | | | |
| | 5 2 | | | | | | | |
| | 5 2 | | | | | | | |
| | 6 2 | | | | | | | |
| | 6 2 | | | | | | | |
| | 5 2 | | | | | | | |
| | 5 2 | | | | | | | |

Continued on next column

**TABLE XXXVII– continued**

| TQC | ICSD | NoE | SF | | | | | |
|---|---|---|---|---|---|---|---|---|
| | 5 2 | | | | | | | |
| | 6 2 | | | | | | | |
| | 6 2 | | | | | | | |
| | 5 2 | | | | | | | |
| | 6 2 | | | | | | | |
| | 5 2 | | | | | | | |
| | 6 2 | | | | | | | |
| | 6 2 | | | | | | | |
| | 6 2 | | | | | | | |
| | 5 2 | | | | | | | |
| | 6 2 | | | | | | | |
| | 5 2 | | | | | | | |
| | 5 2 | | | | | | | |
| | 6 2 | | | | | | | |
| | 6 2 | | | | | | | |
| | 5 2 | | | | | | | |
| | 5 2 | | | | | | | |
| | 5 2 | | | | | | | |
| | 5 2 | | | | | | | |
| | 6 2 | | | | | | | |
| | 5 2 | | | | | | | |
| | 6 2 | | | | | | | |
| | 6 2 | | | | | | | |
| | 5 2 | | | | | | | |
| | 6 2 | | | | | | | |
| | 5 2 | | | | | | | |
| | 6 2 | | | | | | | |
| | 5 2 | | | | | | | |
| | 6 2 | | | | | | | |
| | 6 2 | | | | | | | |
| | 5 2 | | | | | | | |
| | 6 2 | | | | | | | |
| | 6 2 | | | | | | | |
| 62.2.70 | 72929 | 160 | Li1Ni1O4P1 | | | | | |
| | -GM | -R | -S | -T | -U | -X | -Y | -Z |
| | 5 2 | 3+3 4 | 3+3 4 | 3+4 4 | 6+6 4 | 3+4 4 | 3+4 4 | 3+4 4 |
| | 5 2 | 4+4 4 | 4+4 4 | 3+4 4 | 5+5 4 | 3+4 4 | 3+4 4 | 3+4 4 |
| | 6 2 | 3+3 4 | 3+3 4 | 3+4 4 | 5+5 4 | 3+4 4 | 3+4 4 | 3+4 4 |
| | 6 2 | 4+4 4 | 4+4 4 | 3+4 4 | 6+6 4 | 3+4 4 | 3+4 4 | 3+4 4 |
| | 6 2 | 3+3 4 | 4+4 4 | 3+4 4 | 5+5 4 | 3+4 4 | 3+4 4 | 3+4 4 |
| | 6 2 | 4+4 4 | 3+3 4 | 3+4 4 | 6+6 4 | 3+4 4 | 3+4 4 | 3+4 4 |
| | 5 2 | 3+3 4 | 3+3 4 | 3+4 4 | 5+5 4 | 3+4 4 | 3+4 4 | 3+4 4 |
| | 6 2 | 4+4 4 | 4+4 4 | 3+4 4 | 6+6 4 | 3+4 4 | 3+4 4 | 3+4 4 |
| | 5 2 | 3+3 4 | 3+3 4 | 3+4 4 | 6+6 4 | 3+4 4 | 3+4 4 | 3+4 4 |
| | 6 2 | 4+4 4 | 4+4 4 | 3+4 4 | 5+5 4 | 3+4 4 | 3+4 4 | 3+4 4 |
| | 5 2 | 3+3 4 | 3+3 4 | 3+4 4 | 6+6 4 | 3+4 4 | 3+4 4 | 3+4 4 |
| | 6 2 | 4+4 4 | 4+4 4 | 3+4 4 | 6+6 4 | 3+4 4 | 3+4 4 | 3+4 4 |
| | 5 2 | 4+4 4 | 3+3 4 | 3+4 4 | 5+5 4 | 3+4 4 | 3+4 4 | 3+4 4 |
| | 6 2 | 3+3 4 | 4+4 4 | 3+4 4 | 6+6 4 | 3+4 4 | 3+4 4 | 3+4 4 |
| | 5 2 | 3+3 4 | 3+3 4 | 3+4 4 | 5+5 4 | 3+4 4 | 3+4 4 | 3+4 4 |
| | 5 2 | 4+4 4 | 4+4 4 | 3+4 4 | 5+5 4 | 3+4 4 | 3+4 4 | 3+4 4 |
| | 5 2 | 4+4 4 | 3+3 4 | 3+4 4 | 5+5 4 | 3+4 4 | 3+4 4 | 3+4 4 |
| | 6 2 | 3+3 4 | 4+4 4 | 3+4 4 | 6+6 4 | 3+4 4 | 3+4 4 | 3+4 4 |
| | 5 2 | 4+4 4 | 3+3 4 | 3+4 4 | 5+5 4 | 3+4 4 | 3+4 4 | 3+4 4 |
| | 6 2 | 3+3 4 | 4+4 4 | 3+4 4 | 5+5 4 | 3+4 4 | 3+4 4 | 3+4 4 |
| | 5 2 | 4+4 4 | 4+4 4 | 3+4 4 | 6+6 4 | 3+4 4 | 3+4 4 | 3+4 4 |
| | 5 2 | 3+3 4 | 3+3 4 | 3+4 4 | 6+6 4 | 3+4 4 | 3+4 4 | 3+4 4 |
| | 6 2 | 4+4 4 | 3+3 4 | 3+4 4 | 6+6 4 | 3+4 4 | 3+4 4 | 3+4 4 |
| | 6 2 | 3+3 4 | 4+4 4 | 3+4 4 | 6+6 4 | 3+4 4 | 3+4 4 | 3+4 4 |
| | 6 2 | 3+3 4 | 4+4 4 | 3+4 4 | 5+5 4 | 3+4 4 | 3+4 4 | 3+4 4 |
| | 5 2 | 4+4 4 | 3+3 4 | 3+4 4 | 5+5 4 | 3+4 4 | 3+4 4 | 3+4 4 |

Continued on next column

**TABLE XXXVII– continued**

| TQC | ICSD | NoE | SF | | | | | |
|---|---|---|---|---|---|---|---|---|
| | 5 2 | 4+4 4 | 4+4 4 | 3+4 4 | 6+6 4 | 3+4 4 | 3+4 4 | 3+4 4 |
| | 5 2 | 3+3 4 | 3+3 4 | 3+4 4 | 6+6 4 | 3+4 4 | 3+4 4 | 3+4 4 |
| | 5 2 | 3+3 4 | 3+3 4 | 3+4 4 | 5+5 4 | 3+4 4 | 3+4 4 | 3+4 4 |
| | 6 2 | 4+4 4 | 4+4 4 | 3+4 4 | 5+5 4 | 3+4 4 | 3+4 4 | 3+4 4 |
| | 6 2 | 3+3 4 | 3+3 4 | 3+4 4 | 5+5 4 | 3+4 4 | 3+4 4 | 3+4 4 |
| | 6 2 | 4+4 4 | 4+4 4 | 3+4 4 | 6+6 4 | 3+4 4 | 3+4 4 | 3+4 4 |
| | 6 2 | 3+3 4 | 4+4 4 | 3+4 4 | 6+6 4 | 3+4 4 | 3+4 4 | 3+4 4 |
| | 6 2 | 4+4 4 | 3+3 4 | 3+4 4 | 5+5 4 | 3+4 4 | 3+4 4 | 3+4 4 |
| | 5 2 | 4+4 4 | 3+3 4 | 3+4 4 | 6+6 4 | 3+4 4 | 3+4 4 | 3+4 4 |
| | 5 2 | 3+3 4 | 4+4 4 | 3+4 4 | 5+5 4 | 3+4 4 | 3+4 4 | 3+4 4 |
| | 6 2 | 4+4 4 | 3+3 4 | 3+4 4 | 6+6 4 | 3+4 4 | 3+4 4 | 3+4 4 |
| | 6 2 | 3+3 4 | 4+4 4 | 3+4 4 | 5+5 4 | 3+4 4 | 3+4 4 | 3+4 4 |
| | 6 2 | 3+3 4 | 4+4 4 | 3+4 4 | 5+5 4 | 3+4 4 | 3+4 4 | 3+4 4 |
| | 5 2 | 4+4 4 | 3+3 4 | 3+4 4 | 6+6 4 | 3+4 4 | 3+4 4 | 3+4 4 |
| | 5 2 | | | | | | | |
| | 5 2 | | | | | | | |
| | 5 2 | | | | | | | |
| | 6 2 | | | | | | | |
| | 5 2 | | | | | | | |
| | 6 2 | | | | | | | |
| | 5 2 | | | | | | | |
| | 6 2 | | | | | | | |
| | 5 2 | | | | | | | |
| | 5 2 | | | | | | | |
| | 6 2 | | | | | | | |
| | 5 2 | | | | | | | |
| | 6 2 | | | | | | | |
| | 6 2 | | | | | | | |
| | 6 2 | | | | | | | |
| | 6 2 | | | | | | | |
| | 6 2 | | | | | | | |
| | 5 2 | | | | | | | |
| | 5 2 | | | | | | | |
| | 5 2 | | | | | | | |
| | 6 2 | | | | | | | |
| | 6 2 | | | | | | | |
| | 5 2 | | | | | | | |
| | 5 2 | | | | | | | |
| | 5 2 | | | | | | | |
| | 6 2 | | | | | | | |
| | 5 2 | | | | | | | |
| | 5 2 | | | | | | | |
| | 6 2 | | | | | | | |
| | 6 2 | | | | | | | |
| | 5 2 | | | | | | | |
| | 6 2 | | | | | | | |
| | 5 2 | | | | | | | |
| | 6 2 | | | | | | | |
| | 5 2 | | | | | | | |
| | 6 2 | | | | | | | |
| | 6 2 | | | | | | | |
| | 5 2 | | | | | | | |
| | 6 2 | | | | | | | |
| | 6 2 | | | | | | | |
| 62.2.71 | 402760 | 160 | Li1Ni1O4P1 | | | | | |
| | -GM | -R | -S | -T | -U | -X | -Y | -Z |
| | 5 2 | 3+3 4 | 3+3 4 | 3+4 4 | 6+6 4 | 3+4 4 | 3+4 4 | 3+4 4 |
| | 5 2 | 4+4 4 | 4+4 4 | 3+4 4 | 5+5 4 | 3+4 4 | 3+4 4 | 3+4 4 |
| | 6 2 | 3+3 4 | 3+3 4 | 3+4 4 | 5+5 4 | 3+4 4 | 3+4 4 | 3+4 4 |
| | 6 2 | 4+4 4 | 4+4 4 | 3+4 4 | 6+6 4 | 3+4 4 | 3+4 4 | 3+4 4 |
| | 6 2 | 3+3 4 | 3+3 4 | 3+4 4 | 5+5 4 | 3+4 4 | 3+4 4 | 3+4 4 |

Continued on next column

**TABLE XXXVII– continued**

| TQC | ICSD NoE SF |
|---|---|
| | 6 2 4+4 4 4+4 4 3+4 4 6+6 4 3+4 4 3+4 4 3+4 4 |
| | 5 2 3+3 4 3+3 4 3+4 4 5+5 4 3+4 4 3+4 4 3+4 4 |
| | 6 2 4+4 4 4+4 4 3+4 4 6+6 4 3+4 4 3+4 4 3+4 4 |
| | 5 2 3+3 4 3+3 4 3+4 4 6+6 4 3+4 4 3+4 4 3+4 4 |
| | 6 2 4+4 4 4+4 4 3+4 4 5+5 4 3+4 4 3+4 4 3+4 4 |
| | 5 2 3+3 4 3+3 4 3+4 4 6+6 4 3+4 4 3+4 4 3+4 4 |
| | 6 2 4+4 4 4+4 4 3+4 4 6+6 4 3+4 4 3+4 4 3+4 4 |
| | 5 2 4+4 4 3+3 4 3+4 4 5+5 4 3+4 4 3+4 4 3+4 4 |
| | 6 2 3+3 4 4+4 4 3+4 4 6+6 4 3+4 4 3+4 4 3+4 4 |
| | 5 2 3+3 4 3+3 4 3+4 4 5+5 4 3+4 4 3+4 4 3+4 4 |
| | 5 2 4+4 4 4+4 4 3+4 4 5+5 4 3+4 4 3+4 4 3+4 4 |
| | 5 2 4+4 4 3+3 4 3+4 4 5+5 4 3+4 4 3+4 4 3+4 4 |
| | 6 2 3+3 4 4+4 4 3+4 4 6+6 4 3+4 4 3+4 4 3+4 4 |
| | 5 2 4+4 4 3+3 4 3+4 4 5+5 4 3+4 4 3+4 4 3+4 4 |
| | 6 2 3+3 4 4+4 4 3+4 4 5+5 4 3+4 4 3+4 4 3+4 4 |
| | 5 2 4+4 4 4+4 4 3+4 4 6+6 4 3+4 4 3+4 4 3+4 4 |
| | 5 2 3+3 4 3+3 4 3+4 4 6+6 4 3+4 4 3+4 4 3+4 4 |
| | 6 2 4+4 4 3+3 4 3+4 4 6+6 4 3+4 4 3+4 4 3+4 4 |
| | 6 2 3+3 4 4+4 4 3+4 4 6+6 4 3+4 4 3+4 4 3+4 4 |
| | 5 2 3+3 4 4+4 4 3+4 4 5+5 4 3+4 4 3+4 4 3+4 4 |
| | 6 2 4+4 4 3+3 4 3+4 4 5+5 4 3+4 4 3+4 4 3+4 4 |
| | 5 2 4+4 4 4+4 4 3+4 4 6+6 4 3+4 4 3+4 4 3+4 4 |
| | 5 2 3+3 4 3+3 4 3+4 4 6+6 4 3+4 4 3+4 4 3+4 4 |
| | 5 2 3+3 4 3+3 4 3+4 4 5+5 4 3+4 4 3+4 4 3+4 4 |
| | 6 2 4+4 4 4+4 4 3+4 4 5+5 4 3+4 4 3+4 4 3+4 4 |
| | 6 2 3+3 4 3+3 4 3+4 4 5+5 4 3+4 4 3+4 4 3+4 4 |
| | 6 2 4+4 4 4+4 4 3+4 4 6+6 4 3+4 4 3+4 4 3+4 4 |
| | 6 2 3+3 4 4+4 4 3+4 4 6+6 4 3+4 4 3+4 4 3+4 4 |
| | 6 2 4+4 4 3+3 4 3+4 4 5+5 4 3+4 4 3+4 4 3+4 4 |
| | 5 2 4+4 4 3+3 4 3+4 4 6+6 4 3+4 4 3+4 4 3+4 4 |
| | 5 2 3+3 4 4+4 4 3+4 4 5+5 4 3+4 4 3+4 4 3+4 4 |
| | 6 2 4+4 4 3+3 4 3+4 4 6+6 4 3+4 4 3+4 4 3+4 4 |
| | 6 2 3+3 4 4+4 4 3+4 4 5+5 4 3+4 4 3+4 4 3+4 4 |
| | 6 2 3+3 4 4+4 4 3+4 4 5+5 4 3+4 4 3+4 4 3+4 4 |
| | 5 2 4+4 4 3+3 4 3+4 4 6+6 4 3+4 4 3+4 4 3+4 4 |
| | 5 2 |
| | 5 2 |
| | 5 2 |
| | 6 2 |
| | 5 2 |
| | 6 2 |
| | 5 2 |
| | 6 2 |
| | 5 2 |
| | 5 2 |
| | 6 2 |
| | 5 2 |
| | 6 2 |
| | 6 2 |
| | 6 2 |
| | 6 2 |
| | 6 2 |
| | 5 2 |
| | 5 2 |
| | 5 2 |
| | 6 2 |
| | 6 2 |
| | 5 2 |
| | 5 2 |
| | 5 2 |
| | 6 2 |
| | 5 2 |
| | 5 2 |
| | 6 2 |
| | 6 2 |
| | 5 2 |
| | 6 2 |
| | 5 2 |
| | 6 2 |
| | 5 2 |
| | 6 2 |
| | 6 2 |
| | 5 2 |
| | 6 2 |
| | 6 2 |
| 62.2.72 | 417912 104 La1P1Pt1 |
| | -GM -R -S -T -U -X -Y -Z |
| | 5 2 3+3 4 3+3 4 3+4 4 5+5 4 3+4 4 3+4 4 3+4 4 |
| | 5 2 4+4 4 4+4 4 3+4 4 6+6 4 3+4 4 3+4 4 3+4 4 |
| | 6 2 4+4 4 4+4 4 3+4 4 5+5 4 3+4 4 3+4 4 3+4 4 |
| | 6 2 3+3 4 3+3 4 3+4 4 6+6 4 3+4 4 3+4 4 3+4 4 |
| | 5 2 3+3 4 4+4 4 3+4 4 5+5 4 3+4 4 3+4 4 3+4 4 |
| | 5 2 4+4 4 3+3 4 3+4 4 6+6 4 3+4 4 3+4 4 3+4 4 |
| | 6 2 4+4 4 3+3 4 3+4 4 5+5 4 3+4 4 3+4 4 3+4 4 |
| | 6 2 3+3 4 4+4 4 3+4 4 6+6 4 3+4 4 3+4 4 3+4 4 |
| | 5 2 4+4 4 3+3 4 3+4 4 6+6 4 3+4 4 3+4 4 3+4 4 |
| | 5 2 3+3 4 4+4 4 3+4 4 5+5 4 3+4 4 3+4 4 3+4 4 |
| | 6 2 3+3 4 4+4 4 3+4 4 5+5 4 3+4 4 3+4 4 3+4 4 |
| | 5 2 4+4 4 3+3 4 3+4 4 6+6 4 3+4 4 3+4 4 3+4 4 |
| | 6 2 3+3 4 4+4 4 3+4 4 6+6 4 3+4 4 3+4 4 3+4 4 |
| | 5 2 4+4 4 3+3 4 3+4 4 6+6 4 3+4 4 3+4 4 3+4 4 |
| | 6 2 3+3 4 3+3 4 3+4 4 5+5 4 3+4 4 3+4 4 3+4 4 |
| | 6 2 4+4 4 4+4 4 3+4 4 5+5 4 3+4 4 3+4 4 3+4 4 |
| | 5 2 3+3 4 4+4 4 3+4 4 6+6 4 3+4 4 3+4 4 3+4 4 |
| | 5 2 4+4 4 3+3 4 3+4 4 5+5 4 3+4 4 3+4 4 3+4 4 |
| | 6 2 4+4 4 3+3 4 3+4 4 6+6 4 3+4 4 3+4 4 3+4 4 |
| | 6 2 3+3 4 4+4 4 3+4 4 5+5 4 3+4 4 3+4 4 3+4 4 |
| | 6 2 4+4 4 3+3 4 3+4 4 6+6 4 3+4 4 3+4 4 3+4 4 |
| | 5 2 3+3 4 4+4 4 3+4 4 5+5 4 3+4 4 3+4 4 3+4 4 |
| | 5 2 4+4 4 3+3 4 3+4 4 6+6 4 3+4 4 3+4 4 3+4 4 |
| | 6 2 3+3 4 4+4 4 3+4 4 6+6 4 3+4 4 3+4 4 3+4 4 |
| | 5 2 3+3 4 3+3 4 3+4 4 5+5 4 3+4 4 3+4 4 3+4 4 |
| | 5 2 4+4 4 4+4 4 3+4 4 5+5 4 3+4 4 3+4 4 3+4 4 |
| | 6 2 |
| | 5 2 |
| | 5 2 |
| | 6 2 |
| | 5 2 |
| | 6 2 |
| | 6 2 |
| | 6 2 |
| | 5 2 |
| | 6 2 |
| | 5 2 |
| | 6 2 |
| | 5 2 |
| | 6 2 |
| | 6 2 |
| | 5 2 |
| | 6 2 |
| | 5 2 |
| | 5 2 |

Continued on next column

Continued on next column

**TABLE XXXVII– continued**

| TQC | ICSD | NoE | SF | | | | | |
|---|---|---|---|---|---|---|---|---|
| | 6 2 | | | | | | | |
| | 5 2 | | | | | | | |
| | 5 2 | | | | | | | |
| | 6 2 | | | | | | | |
| | 5 2 | | | | | | | |
| | 6 2 | | | | | | | |
| | 5 2 | | | | | | | |
| 62.2.73 | 415749 | 200 | K2O4Ru1 | | | | | |
| | -GM | -R | -S | -T | -U | -X | -Y | -Z |
| | 5 2 | 3+3 4 | 3+3 4 | 3+4 4 | 6+6 4 | 3+4 4 | 3+4 4 | 3+4 4 |
| | 5 2 | 4+4 4 | 4+4 4 | 3+4 4 | 5+5 4 | 3+4 4 | 3+4 4 | 3+4 4 |
| | 6 2 | 3+3 4 | 3+3 4 | 3+4 4 | 5+5 4 | 3+4 4 | 3+4 4 | 3+4 4 |
| | 6 2 | 4+4 4 | 4+4 4 | 3+4 4 | 6+6 4 | 3+4 4 | 3+4 4 | 3+4 4 |
| | 5 2 | 3+3 4 | 3+3 4 | 3+4 4 | 5+5 4 | 3+4 4 | 3+4 4 | 3+4 4 |
| | 6 2 | 4+4 4 | 4+4 4 | 3+4 4 | 6+6 4 | 3+4 4 | 3+4 4 | 3+4 4 |
| | 5 2 | 4+4 4 | 3+3 4 | 3+4 4 | 6+6 4 | 3+4 4 | 3+4 4 | 3+4 4 |
| | 6 2 | 3+3 4 | 4+4 4 | 3+4 4 | 5+5 4 | 3+4 4 | 3+4 4 | 3+4 4 |
| | 5 2 | 3+3 4 | 3+3 4 | 3+4 4 | 5+5 4 | 3+4 4 | 3+4 4 | 3+4 4 |
| | 5 2 | 4+4 4 | 4+4 4 | 3+4 4 | 6+6 4 | 3+4 4 | 3+4 4 | 3+4 4 |
| | 6 2 | 3+3 4 | 4+4 4 | 3+4 4 | 5+5 4 | 3+4 4 | 3+4 4 | 3+4 4 |
| | 6 2 | 4+4 4 | 3+3 4 | 3+4 4 | 6+6 4 | 3+4 4 | 3+4 4 | 3+4 4 |
| | 6 2 | 3+3 4 | 4+4 4 | 3+4 4 | 6+6 4 | 3+4 4 | 3+4 4 | 3+4 4 |
| | 5 2 | 4+4 4 | 3+3 4 | 3+4 4 | 5+5 4 | 3+4 4 | 3+4 4 | 3+4 4 |
| | 5 2 | 3+3 4 | 3+3 4 | 3+4 4 | 6+6 4 | 3+4 4 | 3+4 4 | 3+4 4 |
| | 6 2 | 4+4 4 | 4+4 4 | 3+4 4 | 5+5 4 | 3+4 4 | 3+4 4 | 3+4 4 |
| | 5 2 | 4+4 4 | 3+3 4 | 3+4 4 | 6+6 4 | 3+4 4 | 3+4 4 | 3+4 4 |
| | 6 2 | 3+3 4 | 4+4 4 | 3+4 4 | 5+5 4 | 3+4 4 | 3+4 4 | 3+4 4 |
| | 6 2 | 3+3 4 | 3+3 4 | 3+4 4 | 5+5 4 | 3+4 4 | 3+4 4 | 3+4 4 |
| | 5 2 | 4+4 4 | 4+4 4 | 3+4 4 | 6+6 4 | 3+4 4 | 3+4 4 | 3+4 4 |
| | 5 2 | 3+3 4 | 4+4 4 | 3+4 4 | 5+5 4 | 3+4 4 | 3+4 4 | 3+4 4 |
| | 6 2 | 4+4 4 | 3+3 4 | 3+4 4 | 6+6 4 | 3+4 4 | 3+4 4 | 3+4 4 |
| | 6 2 | 4+4 4 | 3+3 4 | 3+4 4 | 5+5 4 | 3+4 4 | 3+4 4 | 3+4 4 |
| | 5 2 | 3+3 4 | 4+4 4 | 3+4 4 | 6+6 4 | 3+4 4 | 3+4 4 | 3+4 4 |
| | 5 2 | 3+3 4 | 3+3 4 | 3+4 4 | 5+5 4 | 3+4 4 | 3+4 4 | 3+4 4 |
| | 5 2 | 4+4 4 | 4+4 4 | 3+4 4 | 6+6 4 | 3+4 4 | 3+4 4 | 3+4 4 |
| | 6 2 | 4+4 4 | 4+4 4 | 3+4 4 | 5+5 4 | 3+4 4 | 3+4 4 | 3+4 4 |
| | 6 2 | 3+3 4 | 3+3 4 | 3+4 4 | 6+6 4 | 3+4 4 | 3+4 4 | 3+4 4 |
| | 5 2 | 3+3 4 | 3+3 4 | 3+4 4 | 5+5 4 | 3+4 4 | 3+4 4 | 3+4 4 |
| | 5 2 | 4+4 4 | 4+4 4 | 3+4 4 | 6+6 4 | 3+4 4 | 3+4 4 | 3+4 4 |
| | 6 2 | 3+3 4 | 3+3 4 | 3+4 4 | 6+6 4 | 3+4 4 | 3+4 4 | 3+4 4 |
| | 5 2 | 4+4 4 | 4+4 4 | 3+4 4 | 5+5 4 | 3+4 4 | 3+4 4 | 3+4 4 |
| | 6 2 | 4+4 4 | 4+4 4 | 3+4 4 | 6+6 4 | 3+4 4 | 3+4 4 | 3+4 4 |
| | 5 2 | 3+3 4 | 3+3 4 | 3+4 4 | 5+5 4 | 3+4 4 | 3+4 4 | 3+4 4 |
| | 6 2 | 3+3 4 | 3+3 4 | 3+4 4 | 6+6 4 | 3+4 4 | 3+4 4 | 3+4 4 |
| | 6 2 | 4+4 4 | 4+4 4 | 3+4 4 | 5+5 4 | 3+4 4 | 3+4 4 | 3+4 4 |
| | 5 2 | 4+4 4 | 4+4 4 | 3+4 4 | 5+5 4 | 3+4 4 | 3+4 4 | 3+4 4 |
| | 6 2 | 3+3 4 | 3+3 4 | 3+4 4 | 5+5 4 | 3+4 4 | 3+4 4 | 3+4 4 |
| | 5 2 | 3+3 4 | 3+3 4 | 3+4 4 | 6+6 4 | 3+4 4 | 3+4 4 | 3+4 4 |
| | 6 2 | 4+4 4 | 4+4 4 | 3+4 4 | 6+6 4 | 3+4 4 | 3+4 4 | 3+4 4 |
| | 5 2 | 3+3 4 | 3+3 4 | 3+4 4 | 5+5 4 | 3+4 4 | 3+4 4 | 3+4 4 |
| | 6 2 | 4+4 4 | 4+4 4 | 3+4 4 | 6+6 4 | 3+4 4 | 3+4 4 | 3+4 4 |
| | 5 2 | 3+3 4 | 3+3 4 | 3+4 4 | 5+5 4 | 3+4 4 | 3+4 4 | 3+4 4 |
| | 6 2 | 4+4 4 | 4+4 4 | 3+4 4 | 5+5 4 | 3+4 4 | 3+4 4 | 3+4 4 |
| | 5 2 | 3+3 4 | 3+3 4 | 3+4 4 | 6+6 4 | 3+4 4 | 3+4 4 | 3+4 4 |
| | 6 2 | 4+4 4 | 4+4 4 | 3+4 4 | 6+6 4 | 3+4 4 | 3+4 4 | 3+4 4 |
| | 5 2 | 4+4 4 | 4+4 4 | 3+4 4 | 5+5 4 | 3+4 4 | 3+4 4 | 3+4 4 |
| | 6 2 | 3+3 4 | 3+3 4 | 3+4 4 | 6+6 4 | 3+4 4 | 3+4 4 | 3+4 4 |
| | 5 2 | 3+3 4 | 3+3 4 | 3+4 4 | 6+6 4 | 3+4 4 | 3+4 4 | 3+4 4 |
| | 6 2 | 4+4 4 | 4+4 4 | 3+4 4 | 6+6 4 | 3+4 4 | 3+4 4 | 3+4 4 |
| | 5 2 | | | | | | | |
| | 6 2 | | | | | | | |

Continued on next column

**TABLE XXXVII– continued**

| TQC | ICSD | NoE | SF | | | | | |
|---|---|---|---|---|---|---|---|---|
| | 5 2 | | | | | | | |
| | 5 2 | | | | | | | |
| | 6 2 | | | | | | | |
| | 5 2 | | | | | | | |
| | 5 2 | | | | | | | |
| | 6 2 | | | | | | | |
| | 6 2 | | | | | | | |
| | 5 2 | | | | | | | |
| | 6 2 | | | | | | | |
| | 6 2 | | | | | | | |
| | 5 2 | | | | | | | |
| | 6 2 | | | | | | | |
| | 5 2 | | | | | | | |
| | 5 2 | | | | | | | |
| | 6 2 | | | | | | | |
| | 6 2 | | | | | | | |
| | 5 2 | | | | | | | |
| | 6 2 | | | | | | | |
| | 6 2 | | | | | | | |
| | 5 2 | | | | | | | |
| | 5 2 | | | | | | | |
| | 6 2 | | | | | | | |
| | 6 2 | | | | | | | |
| | 5 2 | | | | | | | |
| | 5 2 | | | | | | | |
| | 6 2 | | | | | | | |
| | 5 2 | | | | | | | |
| | 5 2 | | | | | | | |
| | 5 2 | | | | | | | |
| | 6 2 | | | | | | | |
| | 6 2 | | | | | | | |
| | 6 2 | | | | | | | |
| | 6 2 | | | | | | | |
| | 6 2 | | | | | | | |
| | 5 2 | | | | | | | |
| | 5 2 | | | | | | | |
| | 6 2 | | | | | | | |
| | 5 2 | | | | | | | |
| | 6 2 | | | | | | | |
| | 5 2 | | | | | | | |
| | 5 2 | | | | | | | |
| | 6 2 | | | | | | | |
| | 6 2 | | | | | | | |
| | 5 2 | | | | | | | |
| | 6 2 | | | | | | | |
| | 5 2 | | | | | | | |
| | 5 2 | | | | | | | |
| | 5 2 | | | | | | | |
| 62.2.74 | 427252 | 96 | Ir1Sn1Y1 | | | | | |
| | -GM | -R | -S | -T | -U | -X | -Y | -Z |
| | 5 2 | 3+3 4 | 3+3 4 | 3+4 4 | 5+5 4 | 3+4 4 | 3+4 4 | 3+4 4 |
| | 6 2 | 4+4 4 | 4+4 4 | 3+4 4 | 6+6 4 | 3+4 4 | 3+4 4 | 3+4 4 |
| | 5 2 | 4+4 4 | 4+4 4 | 3+4 4 | 5+5 4 | 3+4 4 | 3+4 4 | 3+4 4 |
| | 6 2 | 3+3 4 | 3+3 4 | 3+4 4 | 6+6 4 | 3+4 4 | 3+4 4 | 3+4 4 |
| | 5 2 | 3+3 4 | 3+3 4 | 3+4 4 | 5+5 4 | 3+4 4 | 3+4 4 | 3+4 4 |
| | 6 2 | 4+4 4 | 4+4 4 | 3+4 4 | 6+6 4 | 3+4 4 | 3+4 4 | 3+4 4 |
| | 5 2 | 4+4 4 | 3+3 4 | 3+4 4 | 5+5 4 | 3+4 4 | 3+4 4 | 3+4 4 |
| | 6 2 | 3+3 4 | 4+4 4 | 3+4 4 | 6+6 4 | 3+4 4 | 3+4 4 | 3+4 4 |
| | 5 2 | 3+3 4 | 4+4 4 | 3+4 4 | 6+6 4 | 3+4 4 | 3+4 4 | 3+4 4 |
| | 6 2 | 4+4 4 | 3+3 4 | 3+4 4 | 5+5 4 | 3+4 4 | 3+4 4 | 3+4 4 |
| | 5 2 | 3+3 4 | 3+3 4 | 3+4 4 | 5+5 4 | 3+4 4 | 3+4 4 | 3+4 4 |

Continued on next column

**TABLE XXXVII– continued**

| TQC | ICSD | NoE | SF | | | | | |
|---|---|---|---|---|---|---|---|---|
| | 5 2 | 4+4 4 | 4+4 4 | 3+4 4 | 6+6 4 | 3+4 4 | 3+4 4 | 3+4 4 |
| | 6 2 | 3+3 4 | 4+4 4 | 3+4 4 | 6+6 4 | 3+4 4 | 3+4 4 | 3+4 4 |
| | 6 2 | 4+4 4 | 3+3 4 | 3+4 4 | 6+6 4 | 3+4 4 | 3+4 4 | 3+4 4 |
| | 5 2 | 3+3 4 | 3+3 4 | 3+4 4 | 5+5 4 | 3+4 4 | 3+4 4 | 3+4 4 |
| | 6 2 | 4+4 4 | 4+4 4 | 3+4 4 | 5+5 4 | 3+4 4 | 3+4 4 | 3+4 4 |
| | 5 2 | 4+4 4 | 4+4 4 | 3+4 4 | 6+6 4 | 3+4 4 | 3+4 4 | 3+4 4 |
| | 5 2 | 3+3 4 | 3+3 4 | 3+4 4 | 5+5 4 | 3+4 4 | 3+4 4 | 3+4 4 |
| | 6 2 | 4+4 4 | 3+3 4 | 3+4 4 | 6+6 4 | 3+4 4 | 3+4 4 | 3+4 4 |
| | 6 2 | 3+3 4 | 4+4 4 | 3+4 4 | 5+5 4 | 3+4 4 | 3+4 4 | 3+4 4 |
| | 6 2 | 4+4 4 | 3+3 4 | 3+4 4 | 6+6 4 | 3+4 4 | 3+4 4 | 3+4 4 |
| | 5 2 | 3+3 4 | 4+4 4 | 3+4 4 | 5+5 4 | 3+4 4 | 3+4 4 | 3+4 4 |
| | 6 2 | 3+3 4 | 3+3 4 | 3+4 4 | 6+6 4 | 3+4 4 | 3+4 4 | 3+4 4 |
| | 5 2 | 4+4 4 | 4+4 4 | 3+4 4 | 6+6 4 | 3+4 4 | 3+4 4 | 3+4 4 |
| | 5 2 | | | | | | | |
| | 5 2 | | | | | | | |
| | 5 2 | | | | | | | |
| | 6 2 | | | | | | | |
| | 5 2 | | | | | | | |
| | 6 2 | | | | | | | |
| | 5 2 | | | | | | | |
| | 6 2 | | | | | | | |
| | 5 2 | | | | | | | |
| | 6 2 | | | | | | | |
| | 5 2 | | | | | | | |
| | 6 2 | | | | | | | |
| | 6 2 | | | | | | | |
| | 5 2 | | | | | | | |
| | 6 2 | | | | | | | |
| | 5 2 | | | | | | | |
| | 6 2 | | | | | | | |
| | 6 2 | | | | | | | |
| | 5 2 | | | | | | | |
| | 6 2 | | | | | | | |
| | 5 2 | | | | | | | |
| | 5 2 | | | | | | | |
| | 5 2 | | | | | | | |
| | 6 2 | | | | | | | |
| 62.2.75 | 427253 | 96 | Ir1Si1Y1 | | | | | |
| | -GM | -R | -S | -T | -U | -X | -Y | -Z |
| | 5 2 | 3+3 4 | 3+3 4 | 3+4 4 | 6+6 4 | 3+4 4 | 3+4 4 | 3+4 4 |
| | 5 2 | 4+4 4 | 4+4 4 | 3+4 4 | 5+5 4 | 3+4 4 | 3+4 4 | 3+4 4 |
| | 6 2 | 3+3 4 | 4+4 4 | 3+4 4 | 6+6 4 | 3+4 4 | 3+4 4 | 3+4 4 |
| | 6 2 | 4+4 4 | 3+3 4 | 3+4 4 | 5+5 4 | 3+4 4 | 3+4 4 | 3+4 4 |
| | 5 2 | 4+4 4 | 3+3 4 | 3+4 4 | 6+6 4 | 3+4 4 | 3+4 4 | 3+4 4 |
| | 6 2 | 3+3 4 | 4+4 4 | 3+4 4 | 5+5 4 | 3+4 4 | 3+4 4 | 3+4 4 |
| | 5 2 | 3+3 4 | 3+3 4 | 3+4 4 | 6+6 4 | 3+4 4 | 3+4 4 | 3+4 4 |
| | 6 2 | 4+4 4 | 4+4 4 | 3+4 4 | 5+5 4 | 3+4 4 | 3+4 4 | 3+4 4 |
| | 5 2 | 4+4 4 | 4+4 4 | 3+4 4 | 5+5 4 | 3+4 4 | 3+4 4 | 3+4 4 |
| | 5 2 | 3+3 4 | 3+3 4 | 3+4 4 | 6+6 4 | 3+4 4 | 3+4 4 | 3+4 4 |
| | 6 2 | 3+3 4 | 4+4 4 | 3+4 4 | 6+6 4 | 3+4 4 | 3+4 4 | 3+4 4 |
| | 5 2 | 4+4 4 | 3+3 4 | 3+4 4 | 5+5 4 | 3+4 4 | 3+4 4 | 3+4 4 |
| | 6 2 | 4+4 4 | 3+3 4 | 3+4 4 | 6+6 4 | 3+4 4 | 3+4 4 | 3+4 4 |
| | 5 2 | 3+3 4 | 4+4 4 | 3+4 4 | 5+5 4 | 3+4 4 | 3+4 4 | 3+4 4 |
| | 6 2 | 4+4 4 | 3+3 4 | 3+4 4 | 6+6 4 | 3+4 4 | 3+4 4 | 3+4 4 |
| | 6 2 | 3+3 4 | 4+4 4 | 3+4 4 | 5+5 4 | 3+4 4 | 3+4 4 | 3+4 4 |
| | 5 2 | 3+3 4 | 4+4 4 | 3+4 4 | 5+5 4 | 3+4 4 | 3+4 4 | 3+4 4 |
| | 5 2 | 4+4 4 | 3+3 4 | 3+4 4 | 6+6 4 | 3+4 4 | 3+4 4 | 3+4 4 |
| | 6 2 | 3+3 4 | 4+4 4 | 3+4 4 | 5+5 4 | 3+4 4 | 3+4 4 | 3+4 4 |
| | 6 2 | 4+4 4 | 3+3 4 | 3+4 4 | 6+6 4 | 3+4 4 | 3+4 4 | 3+4 4 |
| | 6 2 | 4+4 4 | 3+3 4 | 3+4 4 | 5+5 4 | 3+4 4 | 3+4 4 | 3+4 4 |
| | 5 2 | 3+3 4 | 4+4 4 | 3+4 4 | 6+6 4 | 3+4 4 | 3+4 4 | 3+4 4 |
| | 5 2 | 4+4 4 | 3+3 4 | 3+4 4 | 5+5 4 | 3+4 4 | 3+4 4 | 3+4 4 |
| | 6 2 | 3+3 4 | 4+4 4 | 3+4 4 | 5+5 4 | 3+4 4 | 3+4 4 | 3+4 4 |
| | 6 2 | | | | | | | |
| | 5 2 | | | | | | | |
| | 5 2 | | | | | | | |
| | 6 2 | | | | | | | |
| | 6 2 | | | | | | | |
| | 5 2 | | | | | | | |
| | 6 2 | | | | | | | |
| | 6 2 | | | | | | | |
| | 5 2 | | | | | | | |
| | 5 2 | | | | | | | |
| | 5 2 | | | | | | | |
| | 6 2 | | | | | | | |
| | 5 2 | | | | | | | |
| | 6 2 | | | | | | | |
| | 5 2 | | | | | | | |
| | 6 2 | | | | | | | |
| | 5 2 | | | | | | | |
| | 6 2 | | | | | | | |
| | 5 2 | | | | | | | |
| | 5 2 | | | | | | | |
| | 5 2 | | | | | | | |
| | 5 2 | | | | | | | |
| | 6 2 | | | | | | | |
| | 6 2 | | | | | | | |
| 62.2.76 | 411884 | 72 | Ir1Si1Ta1 | | | | | |
| | -GM | -R | -S | -T | -U | -X | -Y | -Z |
| | 5 2 | 3+3 4 | 3+3 4 | 3+4 4 | 5+5 4 | 3+4 4 | 3+4 4 | 3+4 4 |
| | 5 2 | 4+4 4 | 4+4 4 | 3+4 4 | 6+6 4 | 3+4 4 | 3+4 4 | 3+4 4 |
| | 6 2 | 4+4 4 | 3+3 4 | 3+4 4 | 5+5 4 | 3+4 4 | 3+4 4 | 3+4 4 |
| | 6 2 | 3+3 4 | 4+4 4 | 3+4 4 | 6+6 4 | 3+4 4 | 3+4 4 | 3+4 4 |
| | 6 2 | 4+4 4 | 4+4 4 | 3+4 4 | 6+6 4 | 3+4 4 | 3+4 4 | 3+4 4 |
| | 5 2 | 3+3 4 | 3+3 4 | 3+4 4 | 6+6 4 | 3+4 4 | 3+4 4 | 3+4 4 |
| | 5 2 | 4+4 4 | 3+3 4 | 3+4 4 | 5+5 4 | 3+4 4 | 3+4 4 | 3+4 4 |
| | 6 2 | 3+3 4 | 4+4 4 | 3+4 4 | 5+5 4 | 3+4 4 | 3+4 4 | 3+4 4 |
| | 5 2 | 3+3 4 | 3+3 4 | 3+4 4 | 5+5 4 | 3+4 4 | 3+4 4 | 3+4 4 |
| | 6 2 | 4+4 4 | 4+4 4 | 3+4 4 | 6+6 4 | 3+4 4 | 3+4 4 | 3+4 4 |
| | 5 2 | 3+3 4 | 3+3 4 | 3+4 4 | 5+5 4 | 3+4 4 | 3+4 4 | 3+4 4 |
| | 6 2 | 4+4 4 | 4+4 4 | 3+4 4 | 6+6 4 | 3+4 4 | 3+4 4 | 3+4 4 |
| | 6 2 | 4+4 4 | 4+4 4 | 3+4 4 | 5+5 4 | 3+4 4 | 3+4 4 | 3+4 4 |
| | 6 2 | 3+3 4 | 3+3 4 | 3+4 4 | 6+6 4 | 3+4 4 | 3+4 4 | 3+4 4 |
| | 6 2 | 4+4 4 | 4+4 4 | 3+4 4 | 5+5 4 | 3+4 4 | 3+4 4 | 3+4 4 |
| | 5 2 | 3+3 4 | 3+3 4 | 3+4 4 | 5+5 4 | 3+4 4 | 3+4 4 | 3+4 4 |
| | 6 2 | 4+4 4 | 4+4 4 | 3+4 4 | 6+6 4 | 3+4 4 | 3+4 4 | 3+4 4 |
| | 5 2 | 3+3 4 | 3+3 4 | 3+4 4 | 6+6 4 | 3+4 4 | 3+4 4 | 3+4 4 |
| | 5 2 | | | | | | | |
| | 6 2 | | | | | | | |
| | 5 2 | | | | | | | |
| | 5 2 | | | | | | | |
| | 5 2 | | | | | | | |
| | 6 2 | | | | | | | |
| | 5 2 | | | | | | | |
| | 6 2 | | | | | | | |
| | 5 2 | | | | | | | |
| | 6 2 | | | | | | | |
| | 5 2 | | | | | | | |
| | 5 2 | | | | | | | |
| | 5 2 | | | | | | | |
| | 6 2 | | | | | | | |
| | 5 2 | | | | | | | |

Continued on next column

**TABLE XXXVII– continued**

| TQC | ICSD | NoE | SF | | | | | |
|---|---|---|---|---|---|---|---|---|
| | 6 2 | | | | | | | |
| | 5 2 | | | | | | | |
| | 5 2 | | | | | | | |
| 62.2.77 | 421256 | 120 | Ir1Sc3Si3 | | | | | |
| | -GM | -R | -S | -T | -U | -X | -Y | -Z |
| | 5 2 | 3+3 4 | 3+3 4 | 3+4 4 | 6+6 4 | 3+4 4 | 3+4 4 | 3+4 4 |
| | 6 2 | 4+4 4 | 4+4 4 | 3+4 4 | 5+5 4 | 3+4 4 | 3+4 4 | 3+4 4 |
| | 6 2 | 4+4 4 | 4+4 4 | 3+4 4 | 5+5 4 | 3+4 4 | 3+4 4 | 3+4 4 |
| | 5 2 | 3+3 4 | 3+3 4 | 3+4 4 | 6+6 4 | 3+4 4 | 3+4 4 | 3+4 4 |
| | 6 2 | 4+4 4 | 3+3 4 | 3+4 4 | 6+6 4 | 3+4 4 | 3+4 4 | 3+4 4 |
| | 6 2 | 3+3 4 | 4+4 4 | 3+4 4 | 5+5 4 | 3+4 4 | 3+4 4 | 3+4 4 |
| | 5 2 | 3+3 4 | 4+4 4 | 3+4 4 | 5+5 4 | 3+4 4 | 3+4 4 | 3+4 4 |
| | 5 2 | 4+4 4 | 3+3 4 | 3+4 4 | 6+6 4 | 3+4 4 | 3+4 4 | 3+4 4 |
| | 5 2 | 3+3 4 | 3+3 4 | 3+4 4 | 6+6 4 | 3+4 4 | 3+4 4 | 3+4 4 |
| | 6 2 | 4+4 4 | 4+4 4 | 3+4 4 | 5+5 4 | 3+4 4 | 3+4 4 | 3+4 4 |
| | 6 2 | 3+3 4 | 4+4 4 | 3+4 4 | 6+6 4 | 3+4 4 | 3+4 4 | 3+4 4 |
| | 5 2 | 4+4 4 | 3+3 4 | 3+4 4 | 5+5 4 | 3+4 4 | 3+4 4 | 3+4 4 |
| | 6 2 | 4+4 4 | 3+3 4 | 3+4 4 | 5+5 4 | 3+4 4 | 3+4 4 | 3+4 4 |
| | 6 2 | 3+3 4 | 4+4 4 | 3+4 4 | 6+6 4 | 3+4 4 | 3+4 4 | 3+4 4 |
| | 5 2 | 4+4 4 | 3+3 4 | 3+4 4 | 5+5 4 | 3+4 4 | 3+4 4 | 3+4 4 |
| | 5 2 | 3+3 4 | 4+4 4 | 3+4 4 | 6+6 4 | 3+4 4 | 3+4 4 | 3+4 4 |
| | 5 2 | 4+4 4 | 4+4 4 | 3+4 4 | 5+5 4 | 3+4 4 | 3+4 4 | 3+4 4 |
| | 6 2 | 3+3 4 | 3+3 4 | 3+4 4 | 6+6 4 | 3+4 4 | 3+4 4 | 3+4 4 |
| | 5 2 | 3+3 4 | 4+4 4 | 3+4 4 | 6+6 4 | 3+4 4 | 3+4 4 | 3+4 4 |
| | 5 2 | 4+4 4 | 3+3 4 | 3+4 4 | 5+5 4 | 3+4 4 | 3+4 4 | 3+4 4 |
| | 6 2 | 3+3 4 | 4+4 4 | 3+4 4 | 6+6 4 | 3+4 4 | 3+4 4 | 3+4 4 |
| | 6 2 | 4+4 4 | 3+3 4 | 3+4 4 | 5+5 4 | 3+4 4 | 3+4 4 | 3+4 4 |
| | 5 2 | 4+4 4 | 3+3 4 | 3+4 4 | 6+6 4 | 3+4 4 | 3+4 4 | 3+4 4 |
| | 6 2 | 3+3 4 | 4+4 4 | 3+4 4 | 5+5 4 | 3+4 4 | 3+4 4 | 3+4 4 |
| | 5 2 | 3+3 4 | 3+3 4 | 3+4 4 | 5+5 4 | 3+4 4 | 3+4 4 | 3+4 4 |
| | 5 2 | 4+4 4 | 4+4 4 | 3+4 4 | 6+6 4 | 3+4 4 | 3+4 4 | 3+4 4 |
| | 6 2 | 3+3 4 | 3+3 4 | 3+4 4 | 5+5 4 | 3+4 4 | 3+4 4 | 3+4 4 |
| | 6 2 | 4+4 4 | 4+4 4 | 3+4 4 | 6+6 4 | 3+4 4 | 3+4 4 | 3+4 4 |
| | 5 2 | 3+3 4 | 3+3 4 | 3+4 4 | 5+5 4 | 3+4 4 | 3+4 4 | 3+4 4 |
| | 5 2 | 4+4 4 | 4+4 4 | 3+4 4 | 6+6 4 | 3+4 4 | 3+4 4 | 3+4 4 |
| | 6 2 | | | | | | | |
| | 6 2 | | | | | | | |
| | 5 2 | | | | | | | |
| | 5 2 | | | | | | | |
| | 6 2 | | | | | | | |
| | 5 2 | | | | | | | |
| | 6 2 | | | | | | | |
| | 6 2 | | | | | | | |
| | 6 2 | | | | | | | |
| | 5 2 | | | | | | | |
| | 6 2 | | | | | | | |
| | 5 2 | | | | | | | |
| | 5 2 | | | | | | | |
| | 5 2 | | | | | | | |
| | 6 2 | | | | | | | |
| | 6 2 | | | | | | | |
| | 6 2 | | | | | | | |
| | 6 2 | | | | | | | |
| | 5 2 | | | | | | | |
| | 6 2 | | | | | | | |
| | 5 2 | | | | | | | |
| | 6 2 | | | | | | | |
| | 6 2 | | | | | | | |
| | 5 2 | | | | | | | |
| | 5 2 | | | | | | | |
| | 6 2 | | | | | | | |

Continued on next column

**TABLE XXXVII– continued**

| TQC | ICSD | NoE | SF | | | | | |
|---|---|---|---|---|---|---|---|---|
| | 5 2 | | | | | | | |
| | 6 2 | | | | | | | |
| | 6 2 | | | | | | | |
| | 5 2 | | | | | | | |
| 62.2.78 | 420416 | 64 | Ir1Sc1Si1 | | | | | |
| | -GM | -R | -S | -T | -U | -X | -Y | -Z |
| | 5 2 | 4+4 4 | 4+4 4 | 3+4 4 | 5+5 4 | 3+4 4 | 3+4 4 | 3+4 4 |
| | 5 2 | 3+3 4 | 3+3 4 | 3+4 4 | 6+6 4 | 3+4 4 | 3+4 4 | 3+4 4 |
| | 6 2 | 3+3 4 | 4+4 4 | 3+4 4 | 6+6 4 | 3+4 4 | 3+4 4 | 3+4 4 |
| | 6 2 | 4+4 4 | 3+3 4 | 3+4 4 | 5+5 4 | 3+4 4 | 3+4 4 | 3+4 4 |
| | 6 2 | 4+4 4 | 3+3 4 | 3+4 4 | 6+6 4 | 3+4 4 | 3+4 4 | 3+4 4 |
| | 5 2 | 3+3 4 | 4+4 4 | 3+4 4 | 6+6 4 | 3+4 4 | 3+4 4 | 3+4 4 |
| | 6 2 | 4+4 4 | 3+3 4 | 3+4 4 | 5+5 4 | 3+4 4 | 3+4 4 | 3+4 4 |
| | 5 2 | 3+3 4 | 4+4 4 | 3+4 4 | 5+5 4 | 3+4 4 | 3+4 4 | 3+4 4 |
| | 6 2 | 3+3 4 | 4+4 4 | 3+4 4 | 5+5 4 | 3+4 4 | 3+4 4 | 3+4 4 |
| | 5 2 | 4+4 4 | 3+3 4 | 3+4 4 | 6+6 4 | 3+4 4 | 3+4 4 | 3+4 4 |
| | 5 2 | 3+3 4 | 4+4 4 | 3+4 4 | 5+5 4 | 3+4 4 | 3+4 4 | 3+4 4 |
| | 6 2 | 4+4 4 | 3+3 4 | 3+4 4 | 6+6 4 | 3+4 4 | 3+4 4 | 3+4 4 |
| | 6 2 | 4+4 4 | 3+3 4 | 3+4 4 | 6+6 4 | 3+4 4 | 3+4 4 | 3+4 4 |
| | 6 2 | 3+3 4 | 4+4 4 | 3+4 4 | 5+5 4 | 3+4 4 | 3+4 4 | 3+4 4 |
| | 5 2 | 4+4 4 | 3+3 4 | 3+4 4 | 5+5 4 | 3+4 4 | 3+4 4 | 3+4 4 |
| | 6 2 | 3+3 4 | 4+4 4 | 3+4 4 | 5+5 4 | 3+4 4 | 3+4 4 | 3+4 4 |
| | 5 2 | | | | | | | |
| | 5 2 | | | | | | | |
| | 5 2 | | | | | | | |
| | 6 2 | | | | | | | |
| | 5 2 | | | | | | | |
| | 5 2 | | | | | | | |
| | 6 2 | | | | | | | |
| | 5 2 | | | | | | | |
| | 6 2 | | | | | | | |
| | 5 2 | | | | | | | |
| | 6 2 | | | | | | | |
| | 5 2 | | | | | | | |
| | 5 2 | | | | | | | |
| | 6 2 | | | | | | | |
| | 5 2 | | | | | | | |
| | 6 2 | | | | | | | |
| 62.2.79 | 422623 | 152 | Ir1Sb1Yb1 | | | | | |
| | -GM | -R | -S | -T | -U | -X | -Y | -Z |
| | 5 2 | 3+3 4 | 3+3 4 | 3+4 4 | 6+6 4 | 3+4 4 | 3+4 4 | 3+4 4 |
| | 6 2 | 4+4 4 | 4+4 4 | 3+4 4 | 5+5 4 | 3+4 4 | 3+4 4 | 3+4 4 |
| | 5 2 | 3+3 4 | 3+3 4 | 3+4 4 | 6+6 4 | 3+4 4 | 3+4 4 | 3+4 4 |
| | 6 2 | 4+4 4 | 4+4 4 | 3+4 4 | 5+5 4 | 3+4 4 | 3+4 4 | 3+4 4 |
| | 5 2 | 4+4 4 | 3+3 4 | 3+4 4 | 6+6 4 | 3+4 4 | 3+4 4 | 3+4 4 |
| | 6 2 | 3+3 4 | 4+4 4 | 3+4 4 | 5+5 4 | 3+4 4 | 3+4 4 | 3+4 4 |
| | 5 2 | 3+3 4 | 4+4 4 | 3+4 4 | 6+6 4 | 3+4 4 | 3+4 4 | 3+4 4 |
| | 6 2 | 4+4 4 | 3+3 4 | 3+4 4 | 5+5 4 | 3+4 4 | 3+4 4 | 3+4 4 |
| | 5 2 | 3+3 4 | 3+3 4 | 3+4 4 | 5+5 4 | 3+4 4 | 3+4 4 | 3+4 4 |
| | 6 2 | 4+4 4 | 4+4 4 | 3+4 4 | 6+6 4 | 3+4 4 | 3+4 4 | 3+4 4 |
| | 5 2 | 4+4 4 | 3+3 4 | 3+4 4 | 6+6 4 | 3+4 4 | 3+4 4 | 3+4 4 |
| | 5 2 | 3+3 4 | 4+4 4 | 3+4 4 | 5+5 4 | 3+4 4 | 3+4 4 | 3+4 4 |
| | 6 2 | 4+4 4 | 4+4 4 | 3+4 4 | 5+5 4 | 3+4 4 | 3+4 4 | 3+4 4 |
| | 6 2 | 3+3 4 | 3+3 4 | 3+4 4 | 5+5 4 | 3+4 4 | 3+4 4 | 3+4 4 |
| | 5 2 | 4+4 4 | 4+4 4 | 3+4 4 | 6+6 4 | 3+4 4 | 3+4 4 | 3+4 4 |
| | 6 2 | 3+3 4 | 3+3 4 | 3+4 4 | 6+6 4 | 3+4 4 | 3+4 4 | 3+4 4 |
| | 5 2 | 3+3 4 | 3+3 4 | 3+4 4 | 5+5 4 | 3+4 4 | 3+4 4 | 3+4 4 |
| | 5 2 | 4+4 4 | 4+4 4 | 3+4 4 | 6+6 4 | 3+4 4 | 3+4 4 | 3+4 4 |
| | 6 2 | 3+3 4 | 3+3 4 | 3+4 4 | 5+5 4 | 3+4 4 | 3+4 4 | 3+4 4 |
| | 6 2 | 4+4 4 | 4+4 4 | 3+4 4 | 6+6 4 | 3+4 4 | 3+4 4 | 3+4 4 |
| | 6 2 | 3+3 4 | 4+4 4 | 3+4 4 | 5+5 4 | 3+4 4 | 3+4 4 | 3+4 4 |

Continued on next column

**TABLE XXXVII– continued**

| TQC | ICSD NoE SF |
|---|---|
| | 5 2 4+4 4 3+3 4 3+4 4 6+6 4 3+4 4 3+4 4 3+4 4 |
| | 5 2 3+3 4 4+4 4 3+4 4 5+5 4 3+4 4 3+4 4 3+4 4 |
| | 6 2 4+4 4 3+3 4 3+4 4 6+6 4 3+4 4 3+4 4 3+4 4 |
| | 5 2 3+3 4 4+4 4 3+4 4 5+5 4 3+4 4 3+4 4 3+4 4 |
| | 5 2 4+4 4 3+3 4 3+4 4 6+6 4 3+4 4 3+4 4 3+4 4 |
| | 5 2 3+3 4 4+4 4 3+4 4 5+5 4 3+4 4 3+4 4 3+4 4 |
| | 6 2 4+4 4 3+3 4 3+4 4 6+6 4 3+4 4 3+4 4 3+4 4 |
| | 6 2 4+4 4 4+4 4 3+4 4 5+5 4 3+4 4 3+4 4 3+4 4 |
| | 5 2 3+3 4 3+3 4 3+4 4 5+5 4 3+4 4 3+4 4 3+4 4 |
| | 6 2 4+4 4 3+3 4 3+4 4 6+6 4 3+4 4 3+4 4 3+4 4 |
| | 6 2 3+3 4 4+4 4 3+4 4 6+6 4 3+4 4 3+4 4 3+4 4 |
| | 5 2 4+4 4 3+3 4 3+4 4 6+6 4 3+4 4 3+4 4 3+4 4 |
| | 6 2 3+3 4 4+4 4 3+4 4 5+5 4 3+4 4 3+4 4 3+4 4 |
| | 5 2 3+3 4 4+4 4 3+4 4 5+5 4 3+4 4 3+4 4 3+4 4 |
| | 5 2 4+4 4 3+3 4 3+4 4 6+6 4 3+4 4 3+4 4 3+4 4 |
| | 6 2 3+3 4 4+4 4 3+4 4 5+5 4 3+4 4 3+4 4 3+4 4 |
| | 5 2 4+4 4 3+3 4 3+4 4 5+5 4 3+4 4 3+4 4 3+4 4 |
| | 5 2 |
| | 6 2 |
| | 5 2 |
| | 6 2 |
| | 6 2 |
| | 5 2 |
| | 6 2 |
| | 5 2 |
| | 6 2 |
| | 6 2 |
| | 5 2 |
| | 6 2 |
| | 5 2 |
| | 6 2 |
| | 5 2 |
| | 5 2 |
| | 6 2 |
| | 5 2 |
| | 6 2 |
| | 5 2 |
| | 6 2 |
| | 6 2 |
| | 5 2 |
| | 6 2 |
| | 5 2 |
| | 6 2 |
| | 5 2 |
| | 6 2 |
| | 6 2 |
| | 5 2 |
| | 5 2 |
| | 6 2 |
| | 5 2 |
| | 5 2 |
| | 6 2 |
| | 6 2 |
| | 5 2 |
| | 5 2 |
| 62.2.80 | 411882 104 Ir1Nb1Si1 |
| | -GM -R -S -T -U -X -Y -Z |
| | 5 2 3+3 4 3+3 4 3+4 4 5+5 4 3+4 4 3+4 4 3+4 4 |
| | 5 2 4+4 4 4+4 4 3+4 4 6+6 4 3+4 4 3+4 4 3+4 4 |
| | 6 2 3+3 4 3+3 4 3+4 4 5+5 4 3+4 4 3+4 4 3+4 4 |
| | 6 2 4+4 4 4+4 4 3+4 4 6+6 4 3+4 4 3+4 4 3+4 4 |

Continued on next column

**TABLE XXXVII– continued**

| TQC | ICSD NoE SF |
|---|---|
| | 5 2 4+4 4 4+4 4 3+4 4 5+5 4 3+4 4 3+4 4 3+4 4 |
| | 5 2 3+3 4 3+3 4 3+4 4 6+6 4 3+4 4 3+4 4 3+4 4 |
| | 6 2 3+3 4 3+3 4 3+4 4 5+5 4 3+4 4 3+4 4 3+4 4 |
| | 6 2 4+4 4 4+4 4 3+4 4 6+6 4 3+4 4 3+4 4 3+4 4 |
| | 5 2 4+4 4 3+3 4 3+4 4 6+6 4 3+4 4 3+4 4 3+4 4 |
| | 5 2 3+3 4 4+4 4 3+4 4 5+5 4 3+4 4 3+4 4 3+4 4 |
| | 6 2 3+3 4 3+3 4 3+4 4 6+6 4 3+4 4 3+4 4 3+4 4 |
| | 5 2 4+4 4 4+4 4 3+4 4 5+5 4 3+4 4 3+4 4 3+4 4 |
| | 6 2 4+4 4 4+4 4 3+4 4 5+5 4 3+4 4 3+4 4 3+4 4 |
| | 5 2 3+3 4 3+3 4 3+4 4 5+5 4 3+4 4 3+4 4 3+4 4 |
| | 6 2 4+4 4 4+4 4 3+4 4 6+6 4 3+4 4 3+4 4 3+4 4 |
| | 6 2 3+3 4 3+3 4 3+4 4 6+6 4 3+4 4 3+4 4 3+4 4 |
| | 5 2 3+3 4 3+3 4 3+4 4 6+6 4 3+4 4 3+4 4 3+4 4 |
| | 5 2 4+4 4 4+4 4 3+4 4 5+5 4 3+4 4 3+4 4 3+4 4 |
| | 6 2 3+3 4 3+3 4 3+4 4 6+6 4 3+4 4 3+4 4 3+4 4 |
| | 6 2 4+4 4 4+4 4 3+4 4 5+5 4 3+4 4 3+4 4 3+4 4 |
| | 6 2 4+4 4 4+4 4 3+4 4 6+6 4 3+4 4 3+4 4 3+4 4 |
| | 5 2 3+3 4 3+3 4 3+4 4 5+5 4 3+4 4 3+4 4 3+4 4 |
| | 5 2 4+4 4 4+4 4 3+4 4 6+6 4 3+4 4 3+4 4 3+4 4 |
| | 6 2 3+3 4 3+3 4 3+4 4 6+6 4 3+4 4 3+4 4 3+4 4 |
| | 5 2 4+4 4 4+4 4 3+4 4 5+5 4 3+4 4 3+4 4 3+4 4 |
| | 6 2 3+3 4 3+3 4 3+4 4 5+5 4 3+4 4 3+4 4 3+4 4 |
| | 5 2 |
| | 6 2 |
| | 6 2 |
| | 6 2 |
| | 5 2 |
| | 6 2 |
| | 6 2 |
| | 5 2 |
| | 5 2 |
| | 6 2 |
| | 5 2 |
| | 5 2 |
| | 5 2 |
| | 6 2 |
| | 5 2 |
| | 6 2 |
| | 5 2 |
| | 6 2 |
| | 5 2 |
| | 5 2 |
| | 5 2 |
| | 6 2 |
| | 5 2 |
| | 6 2 |
| | 5 2 |
| | 5 2 |
| 62.2.81 | 246548 80 Ir1Mn1Si1 |
| | -GM -R -S -T -U -X -Y -Z |
| | 5 2 4+4 4 4+4 4 3+4 4 6+6 4 3+4 4 3+4 4 3+4 4 |
| | 5 2 3+3 4 3+3 4 3+4 4 5+5 4 3+4 4 3+4 4 3+4 4 |
| | 6 2 3+3 4 4+4 4 3+4 4 5+5 4 3+4 4 3+4 4 3+4 4 |
| | 6 2 4+4 4 3+3 4 3+4 4 6+6 4 3+4 4 3+4 4 3+4 4 |
| | 6 2 4+4 4 3+3 4 3+4 4 5+5 4 3+4 4 3+4 4 3+4 4 |
| | 5 2 3+3 4 4+4 4 3+4 4 5+5 4 3+4 4 3+4 4 3+4 4 |
| | 6 2 4+4 4 3+3 4 3+4 4 6+6 4 3+4 4 3+4 4 3+4 4 |
| | 5 2 3+3 4 4+4 4 3+4 4 6+6 4 3+4 4 3+4 4 3+4 4 |
| | 5 2 3+3 4 4+4 4 3+4 4 6+6 4 3+4 4 3+4 4 3+4 4 |
| | 6 2 4+4 4 3+3 4 3+4 4 5+5 4 3+4 4 3+4 4 3+4 4 |
| | 5 2 3+3 4 4+4 4 3+4 4 6+6 4 3+4 4 3+4 4 3+4 4 |

Continued on next column

**TABLE XXXVII– continued**

| TQC | ICSD | NoE | SF | | | | | |
|---|---|---|---|---|---|---|---|---|
| | 6 2 | 4+4 4 | 3+3 4 | 3+4 4 | 5+5 4 | 3+4 4 | 3+4 4 | 3+4 4 |
| | 6 2 | 4+4 4 | 3+3 4 | 3+4 4 | 5+5 4 | 3+4 4 | 3+4 4 | 3+4 4 |
| | 6 2 | 3+3 4 | 4+4 4 | 3+4 4 | 6+6 4 | 3+4 4 | 3+4 4 | 3+4 4 |
| | 5 2 | 4+4 4 | 3+3 4 | 3+4 4 | 6+6 4 | 3+4 4 | 3+4 4 | 3+4 4 |
| | 6 2 | 3+3 4 | 4+4 4 | 3+4 4 | 6+6 4 | 3+4 4 | 3+4 4 | 3+4 4 |
| | 5 2 | 4+4 4 | 3+3 4 | 3+4 4 | 5+5 4 | 3+4 4 | 3+4 4 | 3+4 4 |
| | 5 2 | 3+3 4 | 4+4 4 | 3+4 4 | 5+5 4 | 3+4 4 | 3+4 4 | 3+4 4 |
| | 6 2 | 3+3 4 | 3+3 4 | 3+4 4 | 6+6 4 | 3+4 4 | 3+4 4 | 3+4 4 |
| | 5 2 | 4+4 4 | 4+4 4 | 3+4 4 | 5+5 4 | 3+4 4 | 3+4 4 | 3+4 4 |
| | 5 2 | | | | | | | |
| | 6 2 | | | | | | | |
| | 5 2 | | | | | | | |
| | 6 2 | | | | | | | |
| | 5 2 | | | | | | | |
| | 5 2 | | | | | | | |
| | 6 2 | | | | | | | |
| | 5 2 | | | | | | | |
| | 6 2 | | | | | | | |
| | 5 2 | | | | | | | |
| | 5 2 | | | | | | | |
| | 6 2 | | | | | | | |
| | 5 2 | | | | | | | |
| | 5 2 | | | | | | | |
| | 6 2 | | | | | | | |
| | 5 2 | | | | | | | |
| | 5 2 | | | | | | | |
| | 5 2 | | | | | | | |
| | 6 2 | | | | | | | |
| | 6 2 | | | | | | | |
| 62.2.82 | 246547 | 80 | Ir1Mn1Si1 | | | | | |
| | -GM | -R | -S | -T | -U | -X | -Y | -Z |
| | 5 2 | 4+4 4 | 4+4 4 | 3+4 4 | 6+6 4 | 3+4 4 | 3+4 4 | 3+4 4 |
| | 5 2 | 3+3 4 | 3+3 4 | 3+4 4 | 5+5 4 | 3+4 4 | 3+4 4 | 3+4 4 |
| | 6 2 | 3+3 4 | 4+4 4 | 3+4 4 | 5+5 4 | 3+4 4 | 3+4 4 | 3+4 4 |
| | 6 2 | 4+4 4 | 3+3 4 | 3+4 4 | 6+6 4 | 3+4 4 | 3+4 4 | 3+4 4 |
| | 6 2 | 4+4 4 | 3+3 4 | 3+4 4 | 5+5 4 | 3+4 4 | 3+4 4 | 3+4 4 |
| | 5 2 | 3+3 4 | 4+4 4 | 3+4 4 | 5+5 4 | 3+4 4 | 3+4 4 | 3+4 4 |
| | 6 2 | 4+4 4 | 3+3 4 | 3+4 4 | 6+6 4 | 3+4 4 | 3+4 4 | 3+4 4 |
| | 5 2 | 3+3 4 | 4+4 4 | 3+4 4 | 6+6 4 | 3+4 4 | 3+4 4 | 3+4 4 |
| | 5 2 | 3+3 4 | 4+4 4 | 3+4 4 | 6+6 4 | 3+4 4 | 3+4 4 | 3+4 4 |
| | 6 2 | 4+4 4 | 3+3 4 | 3+4 4 | 5+5 4 | 3+4 4 | 3+4 4 | 3+4 4 |
| | 5 2 | 3+3 4 | 4+4 4 | 3+4 4 | 6+6 4 | 3+4 4 | 3+4 4 | 3+4 4 |
| | 6 2 | 4+4 4 | 3+3 4 | 3+4 4 | 5+5 4 | 3+4 4 | 3+4 4 | 3+4 4 |
| | 6 2 | 4+4 4 | 3+3 4 | 3+4 4 | 5+5 4 | 3+4 4 | 3+4 4 | 3+4 4 |
| | 6 2 | 3+3 4 | 4+4 4 | 3+4 4 | 6+6 4 | 3+4 4 | 3+4 4 | 3+4 4 |
| | 5 2 | 4+4 4 | 3+3 4 | 3+4 4 | 6+6 4 | 3+4 4 | 3+4 4 | 3+4 4 |
| | 6 2 | 3+3 4 | 4+4 4 | 3+4 4 | 6+6 4 | 3+4 4 | 3+4 4 | 3+4 4 |
| | 5 2 | 4+4 4 | 3+3 4 | 3+4 4 | 5+5 4 | 3+4 4 | 3+4 4 | 3+4 4 |
| | 5 2 | 3+3 4 | 4+4 4 | 3+4 4 | 5+5 4 | 3+4 4 | 3+4 4 | 3+4 4 |
| | 6 2 | 3+3 4 | 3+3 4 | 3+4 4 | 6+6 4 | 3+4 4 | 3+4 4 | 3+4 4 |
| | 5 2 | 4+4 4 | 4+4 4 | 3+4 4 | 5+5 4 | 3+4 4 | 3+4 4 | 3+4 4 |
| | 5 2 | | | | | | | |
| | 6 2 | | | | | | | |
| | 5 2 | | | | | | | |
| | 6 2 | | | | | | | |
| | 5 2 | | | | | | | |
| | 5 2 | | | | | | | |
| | 6 2 | | | | | | | |
| | 5 2 | | | | | | | |
| | 6 2 | | | | | | | |
| | 5 2 | | | | | | | |

Continued on next column

**TABLE XXXVII– continued**

| TQC | ICSD | NoE | SF | | | | | |
|---|---|---|---|---|---|---|---|---|
| | 5 2 | | | | | | | |
| | 6 2 | | | | | | | |
| | 5 2 | | | | | | | |
| | 5 2 | | | | | | | |
| | 5 2 | | | | | | | |
| | 6 2 | | | | | | | |
| | 5 2 | | | | | | | |
| | 5 2 | | | | | | | |
| | 6 2 | | | | | | | |
| | 6 2 | | | | | | | |
| 62.2.83 | 413857 | 152 | Ir1Lu1Si1 | | | | | |
| | -GM | -R | -S | -T | -U | -X | -Y | -Z |
| | 5 2 | 3+3 4 | 3+3 4 | 3+4 4 | 6+6 4 | 3+4 4 | 3+4 4 | 3+4 4 |
| | 5 2 | 4+4 4 | 4+4 4 | 3+4 4 | 5+5 4 | 3+4 4 | 3+4 4 | 3+4 4 |
| | 6 2 | 3+3 4 | 3+3 4 | 3+4 4 | 6+6 4 | 3+4 4 | 3+4 4 | 3+4 4 |
| | 6 2 | 4+4 4 | 4+4 4 | 3+4 4 | 5+5 4 | 3+4 4 | 3+4 4 | 3+4 4 |
| | 5 2 | 4+4 4 | 4+4 4 | 3+4 4 | 6+6 4 | 3+4 4 | 3+4 4 | 3+4 4 |
| | 6 2 | 3+3 4 | 3+3 4 | 3+4 4 | 5+5 4 | 3+4 4 | 3+4 4 | 3+4 4 |
| | 5 2 | 3+3 4 | 3+3 4 | 3+4 4 | 6+6 4 | 3+4 4 | 3+4 4 | 3+4 4 |
| | 6 2 | 4+4 4 | 4+4 4 | 3+4 4 | 5+5 4 | 3+4 4 | 3+4 4 | 3+4 4 |
| | 5 2 | 3+3 4 | 3+3 4 | 3+4 4 | 5+5 4 | 3+4 4 | 3+4 4 | 3+4 4 |
| | 6 2 | 4+4 4 | 4+4 4 | 3+4 4 | 6+6 4 | 3+4 4 | 3+4 4 | 3+4 4 |
| | 5 2 | 4+4 4 | 3+3 4 | 3+4 4 | 6+6 4 | 3+4 4 | 3+4 4 | 3+4 4 |
| | 5 2 | 3+3 4 | 4+4 4 | 3+4 4 | 5+5 4 | 3+4 4 | 3+4 4 | 3+4 4 |
| | 6 2 | 3+3 4 | 4+4 4 | 3+4 4 | 5+5 4 | 3+4 4 | 3+4 4 | 3+4 4 |
| | 5 2 | 4+4 4 | 3+3 4 | 3+4 4 | 6+6 4 | 3+4 4 | 3+4 4 | 3+4 4 |
| | 6 2 | 3+3 4 | 4+4 4 | 3+4 4 | 5+5 4 | 3+4 4 | 3+4 4 | 3+4 4 |
| | 6 2 | 4+4 4 | 3+3 4 | 3+4 4 | 6+6 4 | 3+4 4 | 3+4 4 | 3+4 4 |
| | 5 2 | 3+3 4 | 4+4 4 | 3+4 4 | 5+5 4 | 3+4 4 | 3+4 4 | 3+4 4 |
| | 5 2 | 4+4 4 | 3+3 4 | 3+4 4 | 6+6 4 | 3+4 4 | 3+4 4 | 3+4 4 |
| | 6 2 | 4+4 4 | 3+3 4 | 3+4 4 | 6+6 4 | 3+4 4 | 3+4 4 | 3+4 4 |
| | 6 2 | 3+3 4 | 4+4 4 | 3+4 4 | 5+5 4 | 3+4 4 | 3+4 4 | 3+4 4 |
| | 6 2 | 4+4 4 | 4+4 4 | 3+4 4 | 6+6 4 | 3+4 4 | 3+4 4 | 3+4 4 |
| | 5 2 | 3+3 4 | 3+3 4 | 3+4 4 | 5+5 4 | 3+4 4 | 3+4 4 | 3+4 4 |
| | 6 2 | 3+3 4 | 4+4 4 | 3+4 4 | 5+5 4 | 3+4 4 | 3+4 4 | 3+4 4 |
| | 5 2 | 4+4 4 | 3+3 4 | 3+4 4 | 6+6 4 | 3+4 4 | 3+4 4 | 3+4 4 |
| | 6 2 | 4+4 4 | 3+3 4 | 3+4 4 | 5+5 4 | 3+4 4 | 3+4 4 | 3+4 4 |
| | 5 2 | 3+3 4 | 4+4 4 | 3+4 4 | 6+6 4 | 3+4 4 | 3+4 4 | 3+4 4 |
| | 5 2 | 3+3 4 | 4+4 4 | 3+4 4 | 6+6 4 | 3+4 4 | 3+4 4 | 3+4 4 |
| | 6 2 | 4+4 4 | 3+3 4 | 3+4 4 | 6+6 4 | 3+4 4 | 3+4 4 | 3+4 4 |
| | 6 2 | 4+4 4 | 3+3 4 | 3+4 4 | 5+5 4 | 3+4 4 | 3+4 4 | 3+4 4 |
| | 5 2 | 3+3 4 | 4+4 4 | 3+4 4 | 5+5 4 | 3+4 4 | 3+4 4 | 3+4 4 |
| | 5 2 | 3+3 4 | 3+3 4 | 3+4 4 | 5+5 4 | 3+4 4 | 3+4 4 | 3+4 4 |
| | 6 2 | 4+4 4 | 4+4 4 | 3+4 4 | 6+6 4 | 3+4 4 | 3+4 4 | 3+4 4 |
| | 6 2 | 3+3 4 | 3+3 4 | 3+4 4 | 5+5 4 | 3+4 4 | 3+4 4 | 3+4 4 |
| | 5 2 | 4+4 4 | 4+4 4 | 3+4 4 | 6+6 4 | 3+4 4 | 3+4 4 | 3+4 4 |
| | 5 2 | 4+4 4 | 4+4 4 | 3+4 4 | 5+5 4 | 3+4 4 | 3+4 4 | 3+4 4 |
| | 6 2 | 3+3 4 | 3+3 4 | 3+4 4 | 6+6 4 | 3+4 4 | 3+4 4 | 3+4 4 |
| | 5 2 | 4+4 4 | 4+4 4 | 3+4 4 | 5+5 4 | 3+4 4 | 3+4 4 | 3+4 4 |
| | 6 2 | 3+3 4 | 3+3 4 | 3+4 4 | 5+5 4 | 3+4 4 | 3+4 4 | 3+4 4 |
| | 5 2 | | | | | | | |
| | 6 2 | | | | | | | |
| | 6 2 | | | | | | | |
| | 5 2 | | | | | | | |
| | 6 2 | | | | | | | |
| | 5 2 | | | | | | | |
| | 5 2 | | | | | | | |
| | 6 2 | | | | | | | |
| | 5 2 | | | | | | | |
| | 6 2 | | | | | | | |
| | 5 2 | | | | | | | |

Continued on next column

**TABLE XXXVII– continued**

| TQC | ICSD | NoE | SF | | | | | |
|---|---|---|---|---|---|---|---|---|
| | 6 2 | | | | | | | |
| | 6 2 | | | | | | | |
| | 5 2 | | | | | | | |
| | 5 2 | | | | | | | |
| | 6 2 | | | | | | | |
| | 5 2 | | | | | | | |
| | 6 2 | | | | | | | |
| | 6 2 | | | | | | | |
| | 5 2 | | | | | | | |
| | 6 2 | | | | | | | |
| | 6 2 | | | | | | | |
| | 5 2 | | | | | | | |
| | 5 2 | | | | | | | |
| | 5 2 | | | | | | | |
| | 6 2 | | | | | | | |
| | 5 2 | | | | | | | |
| | 6 2 | | | | | | | |
| | 5 2 | | | | | | | |
| | 6 2 | | | | | | | |
| | 5 2 | | | | | | | |
| | 5 2 | | | | | | | |
| | 6 2 | | | | | | | |
| | 5 2 | | | | | | | |
| | 5 2 | | | | | | | |
| | 6 2 | | | | | | | |
| | 5 2 | | | | | | | |
| | 6 2 | | | | | | | |
| 62.2.84 | 240201 | 72 | In3Ir1 | | | | | |
| | -GM | -R | -S | -T | -U | -X | -Y | -Z |
| | 5 2 | 4+4 4 | 3+3 4 | 3+4 4 | 5+5 4 | 3+4 4 | 3+4 4 | 3+4 4 |
| | 6 2 | 3+3 4 | 4+4 4 | 3+4 4 | 6+6 4 | 3+4 4 | 3+4 4 | 3+4 4 |
| | 5 2 | 3+3 4 | 3+3 4 | 3+4 4 | 5+5 4 | 3+4 4 | 3+4 4 | 3+4 4 |
| | 6 2 | 4+4 4 | 4+4 4 | 3+4 4 | 6+6 4 | 3+4 4 | 3+4 4 | 3+4 4 |
| | 5 2 | 4+4 4 | 4+4 4 | 3+4 4 | 6+6 4 | 3+4 4 | 3+4 4 | 3+4 4 |
| | 6 2 | 3+3 4 | 3+3 4 | 3+4 4 | 5+5 4 | 3+4 4 | 3+4 4 | 3+4 4 |
| | 5 2 | 3+3 4 | 3+3 4 | 3+4 4 | 6+6 4 | 3+4 4 | 3+4 4 | 3+4 4 |
| | 6 2 | 4+4 4 | 4+4 4 | 3+4 4 | 5+5 4 | 3+4 4 | 3+4 4 | 3+4 4 |
| | 5 2 | 3+3 4 | 4+4 4 | 3+4 4 | 5+5 4 | 3+4 4 | 3+4 4 | 3+4 4 |
| | 6 2 | 4+4 4 | 3+3 4 | 3+4 4 | 5+5 4 | 3+4 4 | 3+4 4 | 3+4 4 |
| | 5 2 | 4+4 4 | 4+4 4 | 3+4 4 | 6+6 4 | 3+4 4 | 3+4 4 | 3+4 4 |
| | 5 2 | 3+3 4 | 3+3 4 | 3+4 4 | 5+5 4 | 3+4 4 | 3+4 4 | 3+4 4 |
| | 6 2 | 3+3 4 | 3+3 4 | 3+4 4 | 6+6 4 | 3+4 4 | 3+4 4 | 3+4 4 |
| | 5 2 | 4+4 4 | 4+4 4 | 3+4 4 | 5+5 4 | 3+4 4 | 3+4 4 | 3+4 4 |
| | 6 2 | 4+4 4 | 3+3 4 | 3+4 4 | 6+6 4 | 3+4 4 | 3+4 4 | 3+4 4 |
| | 5 2 | 3+3 4 | 4+4 4 | 3+4 4 | 5+5 4 | 3+4 4 | 3+4 4 | 3+4 4 |
| | 6 2 | 3+3 4 | 4+4 4 | 3+4 4 | 6+6 4 | 3+4 4 | 3+4 4 | 3+4 4 |
| | 6 2 | 4+4 4 | 3+3 4 | 3+4 4 | 6+6 4 | 3+4 4 | 3+4 4 | 3+4 4 |
| | 5 2 | | | | | | | |
| | 5 2 | | | | | | | |
| | 6 2 | | | | | | | |
| | 5 2 | | | | | | | |
| | 6 2 | | | | | | | |
| | 5 2 | | | | | | | |
| | 6 2 | | | | | | | |
| | 6 2 | | | | | | | |
| | 5 2 | | | | | | | |
| | 6 2 | | | | | | | |
| | 5 2 | | | | | | | |
| | 6 2 | | | | | | | |
| | 5 2 | | | | | | | |
| | 5 2 | | | | | | | |

Continued on next column

**TABLE XXXVII– continued**

| TQC | ICSD | NoE | SF | | | | | |
|---|---|---|---|---|---|---|---|---|
| | 6 2 | | | | | | | |
| | 5 2 | | | | | | | |
| | 6 2 | | | | | | | |
| | 5 2 | | | | | | | |
| 62.2.85 | 412913 | 144 | In1Rh1Yb1 | | | | | |
| | -GM | -R | -S | -T | -U | -X | -Y | -Z |
| | 5 2 | 3+3 4 | 3+3 4 | 3+4 4 | 5+5 4 | 3+4 4 | 3+4 4 | 3+4 4 |
| | 5 2 | 4+4 4 | 4+4 4 | 3+4 4 | 6+6 4 | 3+4 4 | 3+4 4 | 3+4 4 |
| | 6 2 | 3+3 4 | 3+3 4 | 3+4 4 | 5+5 4 | 3+4 4 | 3+4 4 | 3+4 4 |
| | 6 2 | 4+4 4 | 4+4 4 | 3+4 4 | 6+6 4 | 3+4 4 | 3+4 4 | 3+4 4 |
| | 5 2 | 3+3 4 | 4+4 4 | 3+4 4 | 5+5 4 | 3+4 4 | 3+4 4 | 3+4 4 |
| | 5 2 | 4+4 4 | 3+3 4 | 3+4 4 | 6+6 4 | 3+4 4 | 3+4 4 | 3+4 4 |
| | 6 2 | 4+4 4 | 3+3 4 | 3+4 4 | 5+5 4 | 3+4 4 | 3+4 4 | 3+4 4 |
| | 6 2 | 3+3 4 | 4+4 4 | 3+4 4 | 6+6 4 | 3+4 4 | 3+4 4 | 3+4 4 |
| | 5 2 | 4+4 4 | 3+3 4 | 3+4 4 | 6+6 4 | 3+4 4 | 3+4 4 | 3+4 4 |
| | 6 2 | 3+3 4 | 4+4 4 | 3+4 4 | 5+5 4 | 3+4 4 | 3+4 4 | 3+4 4 |
| | 5 2 | 3+3 4 | 4+4 4 | 3+4 4 | 5+5 4 | 3+4 4 | 3+4 4 | 3+4 4 |
| | 5 2 | 4+4 4 | 3+3 4 | 3+4 4 | 6+6 4 | 3+4 4 | 3+4 4 | 3+4 4 |
| | 6 2 | 3+3 4 | 4+4 4 | 3+4 4 | 6+6 4 | 3+4 4 | 3+4 4 | 3+4 4 |
| | 5 2 | 4+4 4 | 3+3 4 | 3+4 4 | 5+5 4 | 3+4 4 | 3+4 4 | 3+4 4 |
| | 6 2 | 3+3 4 | 4+4 4 | 3+4 4 | 6+6 4 | 3+4 4 | 3+4 4 | 3+4 4 |
| | 6 2 | 4+4 4 | 3+3 4 | 3+4 4 | 5+5 4 | 3+4 4 | 3+4 4 | 3+4 4 |
| | 5 2 | 3+3 4 | 3+3 4 | 3+4 4 | 5+5 4 | 3+4 4 | 3+4 4 | 3+4 4 |
| | 5 2 | 4+4 4 | 4+4 4 | 3+4 4 | 6+6 4 | 3+4 4 | 3+4 4 | 3+4 4 |
| | 6 2 | 4+4 4 | 3+3 4 | 3+4 4 | 6+6 4 | 3+4 4 | 3+4 4 | 3+4 4 |
| | 6 2 | 3+3 4 | 4+4 4 | 3+4 4 | 5+5 4 | 3+4 4 | 3+4 4 | 3+4 4 |
| | 6 2 | 4+4 4 | 3+3 4 | 3+4 4 | 6+6 4 | 3+4 4 | 3+4 4 | 3+4 4 |
| | 5 2 | 3+3 4 | 4+4 4 | 3+4 4 | 5+5 4 | 3+4 4 | 3+4 4 | 3+4 4 |
| | 5 2 | 4+4 4 | 4+4 4 | 3+4 4 | 6+6 4 | 3+4 4 | 3+4 4 | 3+4 4 |
| | 6 2 | 3+3 4 | 3+3 4 | 3+4 4 | 5+5 4 | 3+4 4 | 3+4 4 | 3+4 4 |
| | 5 2 | 4+4 4 | 3+3 4 | 3+4 4 | 5+5 4 | 3+4 4 | 3+4 4 | 3+4 4 |
| | 5 2 | 3+3 4 | 4+4 4 | 3+4 4 | 6+6 4 | 3+4 4 | 3+4 4 | 3+4 4 |
| | 5 2 | 3+3 4 | 3+3 4 | 3+4 4 | 6+6 4 | 3+4 4 | 3+4 4 | 3+4 4 |
| | 6 2 | 4+4 4 | 4+4 4 | 3+4 4 | 5+5 4 | 3+4 4 | 3+4 4 | 3+4 4 |
| | 5 2 | 3+3 4 | 4+4 4 | 3+4 4 | 6+6 4 | 3+4 4 | 3+4 4 | 3+4 4 |
| | 5 2 | 4+4 4 | 3+3 4 | 3+4 4 | 5+5 4 | 3+4 4 | 3+4 4 | 3+4 4 |
| | 6 2 | 3+3 4 | 4+4 4 | 3+4 4 | 6+6 4 | 3+4 4 | 3+4 4 | 3+4 4 |
| | 5 2 | 4+4 4 | 3+3 4 | 3+4 4 | 5+5 4 | 3+4 4 | 3+4 4 | 3+4 4 |
| | 6 2 | 4+4 4 | 3+3 4 | 3+4 4 | 5+5 4 | 3+4 4 | 3+4 4 | 3+4 4 |
| | 5 2 | 3+3 4 | 4+4 4 | 3+4 4 | 6+6 4 | 3+4 4 | 3+4 4 | 3+4 4 |
| | 6 2 | 3+3 4 | 3+3 4 | 3+4 4 | 6+6 4 | 3+4 4 | 3+4 4 | 3+4 4 |
| | 6 2 | 4+4 4 | 4+4 4 | 3+4 4 | 5+5 4 | 3+4 4 | 3+4 4 | 3+4 4 |
| | 6 2 | | | | | | | |
| | 6 2 | | | | | | | |
| | 5 2 | | | | | | | |
| | 5 2 | | | | | | | |
| | 6 2 | | | | | | | |
| | 6 2 | | | | | | | |
| | 5 2 | | | | | | | |
| | 5 2 | | | | | | | |
| | 6 2 | | | | | | | |
| | 5 2 | | | | | | | |
| | 5 2 | | | | | | | |
| | 6 2 | | | | | | | |
| | 6 2 | | | | | | | |
| | 5 2 | | | | | | | |
| | 6 2 | | | | | | | |
| | 5 2 | | | | | | | |
| | 6 2 | | | | | | | |
| | 6 2 | | | | | | | |
| | 5 2 | | | | | | | |

Continued on next column

**TABLE XXXVII– continued**

| TQC | ICSD | NoE | SF | | | | | |
|---|---|---|---|---|---|---|---|---|
| | 5 2 | | | | | | | |
| | 6 2 | | | | | | | |
| | 5 2 | | | | | | | |
| | 6 2 | | | | | | | |
| | 5 2 | | | | | | | |
| | 6 2 | | | | | | | |
| | 5 2 | | | | | | | |
| | 6 2 | | | | | | | |
| | 6 2 | | | | | | | |
| | 5 2 | | | | | | | |
| | 5 2 | | | | | | | |
| | 5 2 | | | | | | | |
| | 6 2 | | | | | | | |
| | 6 2 | | | | | | | |
| | 6 2 | | | | | | | |
| | 5 2 | | | | | | | |
| | 5 2 | | | | | | | |
| 62.2.86 | 72291 | 240 | Ho2S3 | | | | | |
| | -GM | -R | -S | -T | -U | -X | -Y | -Z |
| | 5 2 | 3+3 4 | 3+3 4 | 3+4 4 | 5+5 4 | 3+4 4 | 3+4 4 | 3+4 4 |
| | 5 2 | 4+4 4 | 4+4 4 | 3+4 4 | 6+6 4 | 3+4 4 | 3+4 4 | 3+4 4 |
| | 6 2 | 3+3 4 | 3+3 4 | 3+4 4 | 5+5 4 | 3+4 4 | 3+4 4 | 3+4 4 |
| | 6 2 | 4+4 4 | 4+4 4 | 3+4 4 | 6+6 4 | 3+4 4 | 3+4 4 | 3+4 4 |
| | 5 2 | 3+3 4 | 3+3 4 | 3+4 4 | 5+5 4 | 3+4 4 | 3+4 4 | 3+4 4 |
| | 5 2 | 4+4 4 | 4+4 4 | 3+4 4 | 6+6 4 | 3+4 4 | 3+4 4 | 3+4 4 |
| | 6 2 | 3+3 4 | 3+3 4 | 3+4 4 | 6+6 4 | 3+4 4 | 3+4 4 | 3+4 4 |
| | 6 2 | 4+4 4 | 4+4 4 | 3+4 4 | 5+5 4 | 3+4 4 | 3+4 4 | 3+4 4 |
| | 5 2 | 4+4 4 | 3+3 4 | 3+4 4 | 5+5 4 | 3+4 4 | 3+4 4 | 3+4 4 |
| | 6 2 | 3+3 4 | 4+4 4 | 3+4 4 | 6+6 4 | 3+4 4 | 3+4 4 | 3+4 4 |
| | 5 2 | 3+3 4 | 3+3 4 | 3+4 4 | 6+6 4 | 3+4 4 | 3+4 4 | 3+4 4 |
| | 6 2 | 4+4 4 | 4+4 4 | 3+4 4 | 5+5 4 | 3+4 4 | 3+4 4 | 3+4 4 |
| | 5 2 | 4+4 4 | 4+4 4 | 3+4 4 | 5+5 4 | 3+4 4 | 3+4 4 | 3+4 4 |
| | 5 2 | 3+3 4 | 3+3 4 | 3+4 4 | 6+6 4 | 3+4 4 | 3+4 4 | 3+4 4 |
| | 6 2 | 3+3 4 | 4+4 4 | 3+4 4 | 6+6 4 | 3+4 4 | 3+4 4 | 3+4 4 |
| | 6 2 | 4+4 4 | 3+3 4 | 3+4 4 | 5+5 4 | 3+4 4 | 3+4 4 | 3+4 4 |
| | 5 2 | 4+4 4 | 4+4 4 | 3+4 4 | 6+6 4 | 3+4 4 | 3+4 4 | 3+4 4 |
| | 6 2 | 3+3 4 | 3+3 4 | 3+4 4 | 5+5 4 | 3+4 4 | 3+4 4 | 3+4 4 |
| | 5 2 | 3+3 4 | 4+4 4 | 3+4 4 | 5+5 4 | 3+4 4 | 3+4 4 | 3+4 4 |
| | 5 2 | 4+4 4 | 3+3 4 | 3+4 4 | 6+6 4 | 3+4 4 | 3+4 4 | 3+4 4 |
| | 6 2 | 3+3 4 | 3+3 4 | 3+4 4 | 6+6 4 | 3+4 4 | 3+4 4 | 3+4 4 |
| | 5 2 | 4+4 4 | 4+4 4 | 3+4 4 | 5+5 4 | 3+4 4 | 3+4 4 | 3+4 4 |
| | 6 2 | 3+3 4 | 3+3 4 | 3+4 4 | 5+5 4 | 3+4 4 | 3+4 4 | 3+4 4 |
| | 6 2 | 4+4 4 | 4+4 4 | 3+4 4 | 6+6 4 | 3+4 4 | 3+4 4 | 3+4 4 |
| | 5 2 | 4+4 4 | 3+3 4 | 3+4 4 | 5+5 4 | 3+4 4 | 3+4 4 | 3+4 4 |
| | 5 2 | 3+3 4 | 4+4 4 | 3+4 4 | 6+6 4 | 3+4 4 | 3+4 4 | 3+4 4 |
| | 6 2 | 3+3 4 | 4+4 4 | 3+4 4 | 6+6 4 | 3+4 4 | 3+4 4 | 3+4 4 |
| | 5 2 | 4+4 4 | 3+3 4 | 3+4 4 | 5+5 4 | 3+4 4 | 3+4 4 | 3+4 4 |
| | 6 2 | 3+3 4 | 3+3 4 | 3+4 4 | 6+6 4 | 3+4 4 | 3+4 4 | 3+4 4 |
| | 6 2 | 4+4 4 | 4+4 4 | 3+4 4 | 5+5 4 | 3+4 4 | 3+4 4 | 3+4 4 |
| | 5 2 | 4+4 4 | 4+4 4 | 3+4 4 | 5+5 4 | 3+4 4 | 3+4 4 | 3+4 4 |
| | 6 2 | 3+3 4 | 3+3 4 | 3+4 4 | 6+6 4 | 3+4 4 | 3+4 4 | 3+4 4 |
| | 5 2 | 4+4 4 | 3+3 4 | 3+4 4 | 6+6 4 | 3+4 4 | 3+4 4 | 3+4 4 |
| | 5 2 | 3+3 4 | 4+4 4 | 3+4 4 | 5+5 4 | 3+4 4 | 3+4 4 | 3+4 4 |
| | 5 2 | 3+3 4 | 4+4 4 | 3+4 4 | 5+5 4 | 3+4 4 | 3+4 4 | 3+4 4 |
| | 6 2 | 4+4 4 | 3+3 4 | 3+4 4 | 6+6 4 | 3+4 4 | 3+4 4 | 3+4 4 |
| | 6 2 | 3+3 4 | 3+3 4 | 3+4 4 | 5+5 4 | 3+4 4 | 3+4 4 | 3+4 4 |
| | 6 2 | 4+4 4 | 4+4 4 | 3+4 4 | 6+6 4 | 3+4 4 | 3+4 4 | 3+4 4 |
| | 5 2 | 3+3 4 | 3+3 4 | 3+4 4 | 6+6 4 | 3+4 4 | 3+4 4 | 3+4 4 |
| | 6 2 | 4+4 4 | 4+4 4 | 3+4 4 | 5+5 4 | 3+4 4 | 3+4 4 | 3+4 4 |
| | 5 2 | 3+3 4 | 4+4 4 | 3+4 4 | 5+5 4 | 3+4 4 | 3+4 4 | 3+4 4 |
| | 6 2 | 4+4 4 | 3+3 4 | 3+4 4 | 6+6 4 | 3+4 4 | 3+4 4 | 3+4 4 |

Continued on next column

**TABLE XXXVII– continued**

| TQC | ICSD | NoE | SF | | | | | |
|---|---|---|---|---|---|---|---|---|
| | 5 2 | 3+3 4 | 3+3 4 | 3+4 4 | 5+5 4 | 3+4 4 | 3+4 4 | 3+4 4 |
| | 6 2 | 4+4 4 | 4+4 4 | 3+4 4 | 6+6 4 | 3+4 4 | 3+4 4 | 3+4 4 |
| | 6 2 | 3+3 4 | 3+3 4 | 3+4 4 | 6+6 4 | 3+4 4 | 3+4 4 | 3+4 4 |
| | 5 2 | 4+4 4 | 4+4 4 | 3+4 4 | 5+5 4 | 3+4 4 | 3+4 4 | 3+4 4 |
| | 6 2 | 3+3 4 | 3+3 4 | 3+4 4 | 6+6 4 | 3+4 4 | 3+4 4 | 3+4 4 |
| | 5 2 | 4+4 4 | 4+4 4 | 3+4 4 | 5+5 4 | 3+4 4 | 3+4 4 | 3+4 4 |
| | 5 2 | 4+4 4 | 4+4 4 | 3+4 4 | 5+5 4 | 3+4 4 | 3+4 4 | 3+4 4 |
| | 5 2 | 3+3 4 | 3+3 4 | 3+4 4 | 6+6 4 | 3+4 4 | 3+4 4 | 3+4 4 |
| | 5 2 | 3+3 4 | 4+4 4 | 3+4 4 | 5+5 4 | 3+4 4 | 3+4 4 | 3+4 4 |
| | 6 2 | 4+4 4 | 3+3 4 | 3+4 4 | 6+6 4 | 3+4 4 | 3+4 4 | 3+4 4 |
| | 6 2 | 3+3 4 | 4+4 4 | 3+4 4 | 5+5 4 | 3+4 4 | 3+4 4 | 3+4 4 |
| | 5 2 | 4+4 4 | 3+3 4 | 3+4 4 | 6+6 4 | 3+4 4 | 3+4 4 | 3+4 4 |
| | 5 2 | 3+3 4 | 4+4 4 | 3+4 4 | 5+5 4 | 3+4 4 | 3+4 4 | 3+4 4 |
| | 5 2 | 4+4 4 | 3+3 4 | 3+4 4 | 6+6 4 | 3+4 4 | 3+4 4 | 3+4 4 |
| | 6 2 | 4+4 4 | 3+3 4 | 3+4 4 | 6+6 4 | 3+4 4 | 3+4 4 | 3+4 4 |
| | 5 2 | 3+3 4 | 4+4 4 | 3+4 4 | 5+5 4 | 3+4 4 | 3+4 4 | 3+4 4 |
| | 5 2 | 4+4 4 | 4+4 4 | 3+4 4 | 5+5 4 | 3+4 4 | 3+4 4 | 3+4 4 |
| | 5 2 | 3+3 4 | 3+3 4 | 3+4 4 | 6+6 4 | 3+4 4 | 3+4 4 | 3+4 4 |
| | 6 2 | | | | | | | |
| | 5 2 | | | | | | | |
| | 6 2 | | | | | | | |
| | 6 2 | | | | | | | |
| | 5 2 | | | | | | | |
| | 6 2 | | | | | | | |
| | 6 2 | | | | | | | |
| | 6 2 | | | | | | | |
| | 6 2 | | | | | | | |
| | 5 2 | | | | | | | |
| | 5 2 | | | | | | | |
| | 5 2 | | | | | | | |
| | 6 2 | | | | | | | |
| | 5 2 | | | | | | | |
| | 6 2 | | | | | | | |
| | 6 2 | | | | | | | |
| | 5 2 | | | | | | | |
| | 6 2 | | | | | | | |
| | 6 2 | | | | | | | |
| | 5 2 | | | | | | | |
| | 5 2 | | | | | | | |
| | 6 2 | | | | | | | |
| | 5 2 | | | | | | | |
| | 6 2 | | | | | | | |
| | 5 2 | | | | | | | |
| | 6 2 | | | | | | | |
| | 5 2 | | | | | | | |
| | 6 2 | | | | | | | |
| | 5 2 | | | | | | | |
| | 6 2 | | | | | | | |
| | 5 2 | | | | | | | |
| | 6 2 | | | | | | | |
| | 5 2 | | | | | | | |
| | 5 2 | | | | | | | |
| | 6 2 | | | | | | | |
| | 6 2 | | | | | | | |
| | 5 2 | | | | | | | |
| | 5 2 | | | | | | | |
| | 6 2 | | | | | | | |
| | 5 2 | | | | | | | |
| | 6 2 | | | | | | | |
| | 6 2 | | | | | | | |
| | 6 2 | | | | | | | |

Continued on next column

**TABLE XXXVII– continued**

| TQC | ICSD | NoE | SF | | | | | |
|---|---|---|---|---|---|---|---|---|
| | 6 2 | | | | | | | |
| | 5 2 | | | | | | | |
| | 6 2 | | | | | | | |
| | 5 2 | | | | | | | |
| | 6 2 | | | | | | | |
| | 5 2 | | | | | | | |
| | 5 2 | | | | | | | |
| | 6 2 | | | | | | | |
| | 5 2 | | | | | | | |
| | 5 2 | | | | | | | |
| | 5 2 | | | | | | | |
| | 6 2 | | | | | | | |
| | 6 2 | | | | | | | |
| | 5 2 | | | | | | | |
| | 5 2 | | | | | | | |
| | 6 2 | | | | | | | |
| | 5 2 | | | | | | | |
| 62.2.87 | 99546 | 168 | Ho1O3Sc1 | | | | | |
| | -GM | -R | -S | -T | -U | -X | -Y | -Z |
| | 5 2 | 3+3 4 | 3+3 4 | 3+4 4 | 6+6 4 | 3+4 4 | 3+4 4 | 3+4 4 |
| | 6 2 | 4+4 4 | 4+4 4 | 3+4 4 | 5+5 4 | 3+4 4 | 3+4 4 | 3+4 4 |
| | 5 2 | 3+3 4 | 3+3 4 | 3+4 4 | 6+6 4 | 3+4 4 | 3+4 4 | 3+4 4 |
| | 6 2 | 4+4 4 | 4+4 4 | 3+4 4 | 5+5 4 | 3+4 4 | 3+4 4 | 3+4 4 |
| | 5 2 | 3+3 4 | 3+3 4 | 3+4 4 | 5+5 4 | 3+4 4 | 3+4 4 | 3+4 4 |
| | 6 2 | 4+4 4 | 4+4 4 | 3+4 4 | 6+6 4 | 3+4 4 | 3+4 4 | 3+4 4 |
| | 5 2 | 4+4 4 | 4+4 4 | 3+4 4 | 6+6 4 | 3+4 4 | 3+4 4 | 3+4 4 |
| | 6 2 | 3+3 4 | 3+3 4 | 3+4 4 | 5+5 4 | 3+4 4 | 3+4 4 | 3+4 4 |
| | 6 2 | 4+4 4 | 3+3 4 | 3+4 4 | 5+5 4 | 3+4 4 | 3+4 4 | 3+4 4 |
| | 5 2 | 3+3 4 | 4+4 4 | 3+4 4 | 5+5 4 | 3+4 4 | 3+4 4 | 3+4 4 |
| | 5 2 | 4+4 4 | 4+4 4 | 3+4 4 | 6+6 4 | 3+4 4 | 3+4 4 | 3+4 4 |
| | 5 2 | 3+3 4 | 3+3 4 | 3+4 4 | 5+5 4 | 3+4 4 | 3+4 4 | 3+4 4 |
| | 6 2 | 3+3 4 | 3+3 4 | 3+4 4 | 6+6 4 | 3+4 4 | 3+4 4 | 3+4 4 |
| | 6 2 | 4+4 4 | 4+4 4 | 3+4 4 | 6+6 4 | 3+4 4 | 3+4 4 | 3+4 4 |
| | 6 2 | 4+4 4 | 3+3 4 | 3+4 4 | 5+5 4 | 3+4 4 | 3+4 4 | 3+4 4 |
| | 5 2 | 3+3 4 | 4+4 4 | 3+4 4 | 5+5 4 | 3+4 4 | 3+4 4 | 3+4 4 |
| | 5 2 | 3+3 4 | 4+4 4 | 3+4 4 | 5+5 4 | 3+4 4 | 3+4 4 | 3+4 4 |
| | 5 2 | 4+4 4 | 3+3 4 | 3+4 4 | 6+6 4 | 3+4 4 | 3+4 4 | 3+4 4 |
| | 6 2 | 4+4 4 | 4+4 4 | 3+4 4 | 5+5 4 | 3+4 4 | 3+4 4 | 3+4 4 |
| | 5 2 | 3+3 4 | 3+3 4 | 3+4 4 | 6+6 4 | 3+4 4 | 3+4 4 | 3+4 4 |
| | 6 2 | 3+3 4 | 3+3 4 | 3+4 4 | 6+6 4 | 3+4 4 | 3+4 4 | 3+4 4 |
| | 6 2 | 4+4 4 | 4+4 4 | 3+4 4 | 5+5 4 | 3+4 4 | 3+4 4 | 3+4 4 |
| | 5 2 | 3+3 4 | 4+4 4 | 3+4 4 | 5+5 4 | 3+4 4 | 3+4 4 | 3+4 4 |
| | 5 2 | 4+4 4 | 3+3 4 | 3+4 4 | 6+6 4 | 3+4 4 | 3+4 4 | 3+4 4 |
| | 6 2 | 4+4 4 | 3+3 4 | 3+4 4 | 6+6 4 | 3+4 4 | 3+4 4 | 3+4 4 |
| | 5 2 | 3+3 4 | 4+4 4 | 3+4 4 | 5+5 4 | 3+4 4 | 3+4 4 | 3+4 4 |
| | 6 2 | 3+3 4 | 4+4 4 | 3+4 4 | 5+5 4 | 3+4 4 | 3+4 4 | 3+4 4 |
| | 6 2 | 4+4 4 | 3+3 4 | 3+4 4 | 6+6 4 | 3+4 4 | 3+4 4 | 3+4 4 |
| | 5 2 | 3+3 4 | 3+3 4 | 3+4 4 | 6+6 4 | 3+4 4 | 3+4 4 | 3+4 4 |
| | 5 2 | 4+4 4 | 4+4 4 | 3+4 4 | 5+5 4 | 3+4 4 | 3+4 4 | 3+4 4 |
| | 5 2 | 3+3 4 | 4+4 4 | 3+4 4 | 6+6 4 | 3+4 4 | 3+4 4 | 3+4 4 |
| | 5 2 | 4+4 4 | 3+3 4 | 3+4 4 | 6+6 4 | 3+4 4 | 3+4 4 | 3+4 4 |
| | 5 2 | 3+3 4 | 4+4 4 | 3+4 4 | 5+5 4 | 3+4 4 | 3+4 4 | 3+4 4 |
| | 6 2 | 4+4 4 | 3+3 4 | 3+4 4 | 6+6 4 | 3+4 4 | 3+4 4 | 3+4 4 |
| | 5 2 | 4+4 4 | 3+3 4 | 3+4 4 | 5+5 4 | 3+4 4 | 3+4 4 | 3+4 4 |
| | 5 2 | 3+3 4 | 4+4 4 | 3+4 4 | 6+6 4 | 3+4 4 | 3+4 4 | 3+4 4 |
| | 5 2 | 4+4 4 | 3+3 4 | 3+4 4 | 5+5 4 | 3+4 4 | 3+4 4 | 3+4 4 |
| | 5 2 | 3+3 4 | 4+4 4 | 3+4 4 | 6+6 4 | 3+4 4 | 3+4 4 | 3+4 4 |
| | 5 2 | 4+4 4 | 3+3 4 | 3+4 4 | 5+5 4 | 3+4 4 | 3+4 4 | 3+4 4 |
| | 6 2 | 3+3 4 | 4+4 4 | 3+4 4 | 6+6 4 | 3+4 4 | 3+4 4 | 3+4 4 |
| | 6 2 | 3+3 4 | 4+4 4 | 3+4 4 | 5+5 4 | 3+4 4 | 3+4 4 | 3+4 4 |
| | 6 2 | 4+4 4 | 3+3 4 | 3+4 4 | 6+6 4 | 3+4 4 | 3+4 4 | 3+4 4 |

Continued on next column

**TABLE XXXVII– continued**

| TQC | ICSD | NoE | SF | | | | | |
|---|---|---|---|---|---|---|---|---|
| | 6 2 | | | | | | | |
| | 6 2 | | | | | | | |
| | 6 2 | | | | | | | |
| | 5 2 | | | | | | | |
| | 5 2 | | | | | | | |
| | 6 2 | | | | | | | |
| | 5 2 | | | | | | | |
| | 5 2 | | | | | | | |
| | 6 2 | | | | | | | |
| | 5 2 | | | | | | | |
| | 5 2 | | | | | | | |
| | 6 2 | | | | | | | |
| | 6 2 | | | | | | | |
| | 5 2 | | | | | | | |
| | 6 2 | | | | | | | |
| | 6 2 | | | | | | | |
| | 5 2 | | | | | | | |
| | 6 2 | | | | | | | |
| | 6 2 | | | | | | | |
| | 6 2 | | | | | | | |
| | 6 2 | | | | | | | |
| | 5 2 | | | | | | | |
| | 5 2 | | | | | | | |
| | 5 2 | | | | | | | |
| | 6 2 | | | | | | | |
| | 6 2 | | | | | | | |
| | 5 2 | | | | | | | |
| | 6 2 | | | | | | | |
| | 5 2 | | | | | | | |
| | 6 2 | | | | | | | |
| | 5 2 | | | | | | | |
| | 6 2 | | | | | | | |
| | 6 2 | | | | | | | |
| | 6 2 | | | | | | | |
| | 5 2 | | | | | | | |
| | 5 2 | | | | | | | |
| | 5 2 | | | | | | | |
| | 6 2 | | | | | | | |
| | 5 2 | | | | | | | |
| | 5 2 | | | | | | | |
| | 6 2 | | | | | | | |
| | 6 2 | | | | | | | |
| 62.2.88 | 165196 | 184 | Ho1Mn1O3 | | | | | |
| | -GM | -R | -S | -T | -U | -X | -Y | -Z |
| | 5 2 | 3+3 4 | 3+3 4 | 3+4 4 | 5+5 4 | 3+4 4 | 3+4 4 | 3+4 4 |
| | 6 2 | 4+4 4 | 4+4 4 | 3+4 4 | 6+6 4 | 3+4 4 | 3+4 4 | 3+4 4 |
| | 5 2 | 3+3 4 | 4+4 4 | 3+4 4 | 5+5 4 | 3+4 4 | 3+4 4 | 3+4 4 |
| | 6 2 | 4+4 4 | 3+3 4 | 3+4 4 | 6+6 4 | 3+4 4 | 3+4 4 | 3+4 4 |
| | 5 2 | 3+3 4 | 4+4 4 | 3+4 4 | 6+6 4 | 3+4 4 | 3+4 4 | 3+4 4 |
| | 6 2 | 4+4 4 | 3+3 4 | 3+4 4 | 5+5 4 | 3+4 4 | 3+4 4 | 3+4 4 |
| | 5 2 | 4+4 4 | 3+3 4 | 3+4 4 | 5+5 4 | 3+4 4 | 3+4 4 | 3+4 4 |
| | 6 2 | 3+3 4 | 4+4 4 | 3+4 4 | 6+6 4 | 3+4 4 | 3+4 4 | 3+4 4 |
| | 6 2 | 3+3 4 | 4+4 4 | 3+4 4 | 5+5 4 | 3+4 4 | 3+4 4 | 3+4 4 |
| | 5 2 | 4+4 4 | 3+3 4 | 3+4 4 | 6+6 4 | 3+4 4 | 3+4 4 | 3+4 4 |
| | 5 2 | 4+4 4 | 3+3 4 | 3+4 4 | 6+6 4 | 3+4 4 | 3+4 4 | 3+4 4 |
| | 5 2 | 3+3 4 | 4+4 4 | 3+4 4 | 5+5 4 | 3+4 4 | 3+4 4 | 3+4 4 |
| | 6 2 | 3+3 4 | 4+4 4 | 3+4 4 | 6+6 4 | 3+4 4 | 3+4 4 | 3+4 4 |
| | 6 2 | 4+4 4 | 3+3 4 | 3+4 4 | 5+5 4 | 3+4 4 | 3+4 4 | 3+4 4 |
| | 5 2 | 3+3 4 | 4+4 4 | 3+4 4 | 6+6 4 | 3+4 4 | 3+4 4 | 3+4 4 |
| | 6 2 | 4+4 4 | 3+3 4 | 3+4 4 | 6+6 4 | 3+4 4 | 3+4 4 | 3+4 4 |
| | 5 2 | 4+4 4 | 3+3 4 | 3+4 4 | 6+6 4 | 3+4 4 | 3+4 4 | 3+4 4 |

Continued on next column

**TABLE XXXVII– continued**

| TQC | ICSD | NoE | SF |
|---|---|---|---|
| | 5 2 3+3 4 4+4 4 3+4 4 6+6 4 3+4 4 3+4 4 3+4 4 | | |
| | 6 2 4+4 4 3+3 4 3+4 4 5+5 4 3+4 4 3+4 4 3+4 4 | | |
| | 6 2 3+3 4 4+4 4 3+4 4 5+5 4 3+4 4 3+4 4 3+4 4 | | |
| | 6 2 3+3 4 3+3 4 3+4 4 5+5 4 3+4 4 3+4 4 3+4 4 | | |
| | 5 2 4+4 4 4+4 4 3+4 4 6+6 4 3+4 4 3+4 4 3+4 4 | | |
| | 6 2 4+4 4 4+4 4 3+4 4 5+5 4 3+4 4 3+4 4 3+4 4 | | |
| | 5 2 3+3 4 3+3 4 3+4 4 6+6 4 3+4 4 3+4 4 3+4 4 | | |
| | 5 2 3+3 4 4+4 4 3+4 4 5+5 4 3+4 4 3+4 4 3+4 4 | | |
| | 6 2 4+4 4 3+3 4 3+4 4 6+6 4 3+4 4 3+4 4 3+4 4 | | |
| | 5 2 4+4 4 3+3 4 3+4 4 6+6 4 3+4 4 3+4 4 3+4 4 | | |
| | 6 2 3+3 4 4+4 4 3+4 4 5+5 4 3+4 4 3+4 4 3+4 4 | | |
| | 5 2 4+4 4 3+3 4 3+4 4 5+5 4 3+4 4 3+4 4 3+4 4 | | |
| | 5 2 3+3 4 4+4 4 3+4 4 6+6 4 3+4 4 3+4 4 3+4 4 | | |
| | 5 2 3+3 4 3+3 4 3+4 4 5+5 4 3+4 4 3+4 4 3+4 4 | | |
| | 5 2 4+4 4 4+4 4 3+4 4 5+5 4 3+4 4 3+4 4 3+4 4 | | |
| | 6 2 3+3 4 3+3 4 3+4 4 6+6 4 3+4 4 3+4 4 3+4 4 | | |
| | 5 2 4+4 4 4+4 4 3+4 4 5+5 4 3+4 4 3+4 4 3+4 4 | | |
| | 5 2 4+4 4 4+4 4 3+4 4 6+6 4 3+4 4 3+4 4 3+4 4 | | |
| | 5 2 3+3 4 3+3 4 3+4 4 5+5 4 3+4 4 3+4 4 3+4 4 | | |
| | 5 2 4+4 4 4+4 4 3+4 4 6+6 4 3+4 4 3+4 4 3+4 4 | | |
| | 5 2 3+3 4 3+3 4 3+4 4 5+5 4 3+4 4 3+4 4 3+4 4 | | |
| | 6 2 4+4 4 3+3 4 3+4 4 6+6 4 3+4 4 3+4 4 3+4 4 | | |
| | 5 2 3+3 4 4+4 4 3+4 4 6+6 4 3+4 4 3+4 4 3+4 4 | | |
| | 6 2 3+3 4 4+4 4 3+4 4 6+6 4 3+4 4 3+4 4 3+4 4 | | |
| | 6 2 4+4 4 3+3 4 3+4 4 6+6 4 3+4 4 3+4 4 3+4 4 | | |
| | 5 2 3+3 4 4+4 4 3+4 4 6+6 4 3+4 4 3+4 4 3+4 4 | | |
| | 6 2 4+4 4 3+3 4 3+4 4 5+5 4 3+4 4 3+4 4 3+4 4 | | |
| | 6 2 3+3 4 3+3 4 3+4 4 5+5 4 3+4 4 3+4 4 3+4 4 | | |
| | 5 2 4+4 4 4+4 4 3+4 4 5+5 4 3+4 4 3+4 4 3+4 4 | | |
| | 6 2 | | |
| | 5 2 | | |
| | 5 2 | | |
| | 6 2 | | |
| | 6 2 | | |
| | 5 2 | | |
| | 6 2 | | |
| | 5 2 | | |
| | 6 2 | | |
| | 5 2 | | |
| | 5 2 | | |
| | 6 2 | | |
| | 6 2 | | |
| | 6 2 | | |
| | 6 2 | | |
| | 6 2 | | |
| | 6 2 | | |
| | 6 2 | | |
| | 5 2 | | |
| | 5 2 | | |
| | 5 2 | | |
| | 5 2 | | |
| | 6 2 | | |
| | 5 2 | | |
| | 6 2 | | |
| | 5 2 | | |
| | 6 2 | | |
| | 6 2 | | |
| | 6 2 | | |
| | 6 2 | | |
| | 5 2 | | |
| | 5 2 | | |

Continued on next column

**TABLE XXXVII– continued**

| TQC | ICSD | NoE | SF |
|---|---|---|---|
| | 5 2 | | |
| | 5 2 | | |
| | 5 2 | | |
| | 5 2 | | |
| | 5 2 | | |
| | 5 2 | | |
| | 5 2 | | |
| | 5 2 | | |
| | 6 2 | | |
| | 5 2 | | |
| | 5 2 | | |
| | 6 2 | | |
| | 5 2 | | |
| | 6 2 | | |
| 62.2.89 | 82727 | 88 | Hf3P2 |
| | -GM -R -S -T -U -X -Y -Z | | |
| | 5 2 3+3 4 4+4 4 3+4 4 6+6 4 3+4 4 3+4 4 3+4 4 | | |
| | 5 2 4+4 4 3+3 4 3+4 4 5+5 4 3+4 4 3+4 4 3+4 4 | | |
| | 6 2 3+3 4 4+4 4 3+4 4 5+5 4 3+4 4 3+4 4 3+4 4 | | |
| | 5 2 4+4 4 3+3 4 3+4 4 6+6 4 3+4 4 3+4 4 3+4 4 | | |
| | 6 2 3+3 4 3+3 4 3+4 4 5+5 4 3+4 4 3+4 4 3+4 4 | | |
| | 6 2 4+4 4 4+4 4 3+4 4 6+6 4 3+4 4 3+4 4 3+4 4 | | |
| | 6 2 4+4 4 4+4 4 3+4 4 6+6 4 3+4 4 3+4 4 3+4 4 | | |
| | 5 2 3+3 4 3+3 4 3+4 4 5+5 4 3+4 4 3+4 4 3+4 4 | | |
| | 5 2 3+3 4 3+3 4 3+4 4 5+5 4 3+4 4 3+4 4 3+4 4 | | |
| | 5 2 4+4 4 4+4 4 3+4 4 6+6 4 3+4 4 3+4 4 3+4 4 | | |
| | 6 2 4+4 4 3+3 4 3+4 4 5+5 4 3+4 4 3+4 4 3+4 4 | | |
| | 6 2 3+3 4 4+4 4 3+4 4 6+6 4 3+4 4 3+4 4 3+4 4 | | |
| | 6 2 3+3 4 4+4 4 3+4 4 6+6 4 3+4 4 3+4 4 3+4 4 | | |
| | 6 2 4+4 4 3+3 4 3+4 4 5+5 4 3+4 4 3+4 4 3+4 4 | | |
| | 5 2 3+3 4 3+3 4 3+4 4 6+6 4 3+4 4 3+4 4 3+4 4 | | |
| | 5 2 4+4 4 4+4 4 3+4 4 6+6 4 3+4 4 3+4 4 3+4 4 | | |
| | 5 2 4+4 4 3+3 4 3+4 4 5+5 4 3+4 4 3+4 4 3+4 4 | | |
| | 6 2 3+3 4 4+4 4 3+4 4 5+5 4 3+4 4 3+4 4 3+4 4 | | |
| | 6 2 4+4 4 4+4 4 3+4 4 6+6 4 3+4 4 3+4 4 3+4 4 | | |
| | 5 2 3+3 4 3+3 4 3+4 4 5+5 4 3+4 4 3+4 4 3+4 4 | | |
| | 6 2 3+3 4 4+4 4 3+4 4 6+6 4 3+4 4 3+4 4 3+4 4 | | |
| | 5 2 4+4 4 3+3 4 3+4 4 6+6 4 3+4 4 3+4 4 3+4 4 | | |
| | 5 2 | | |
| | 5 2 | | |
| | 6 2 | | |
| | 5 2 | | |
| | 6 2 | | |
| | 6 2 | | |
| | 5 2 | | |
| | 6 2 | | |
| | 6 2 | | |
| | 6 2 | | |
| | 6 2 | | |
| | 5 2 | | |
| | 5 2 | | |
| | 5 2 | | |
| | 6 2 | | |
| | 5 2 | | |
| | 6 2 | | |
| | 5 2 | | |
| | 6 2 | | |
| | 5 2 | | |
| | 5 2 | | |
| | 5 2 | | |
| 62.2.90 | 260839 | 176 | H2Cl1O2Tm1 |

Continued on next column

**TABLE XXXVII– continued**

| TQC | ICSD | NoE | SF | | | | | |
|---|---|---|---|---|---|---|---|---|
| | -GM | -R | -S | -T | -U | -X | -Y | -Z |
| | 5 2 | 3+3 4 | 3+3 4 | 3+4 4 | 5+5 4 | 3+4 4 | 3+4 4 | 3+4 4 |
| | 5 2 | 4+4 4 | 4+4 4 | 3+4 4 | 6+6 4 | 3+4 4 | 3+4 4 | 3+4 4 |
| | 6 2 | 3+3 4 | 3+3 4 | 3+4 4 | 6+6 4 | 3+4 4 | 3+4 4 | 3+4 4 |
| | 6 2 | 4+4 4 | 4+4 4 | 3+4 4 | 5+5 4 | 3+4 4 | 3+4 4 | 3+4 4 |
| | 5 2 | 3+3 4 | 3+3 4 | 3+4 4 | 5+5 4 | 3+4 4 | 3+4 4 | 3+4 4 |
| | 5 2 | 4+4 4 | 4+4 4 | 3+4 4 | 6+6 4 | 3+4 4 | 3+4 4 | 3+4 4 |
| | 6 2 | 3+3 4 | 3+3 4 | 3+4 4 | 5+5 4 | 3+4 4 | 3+4 4 | 3+4 4 |
| | 6 2 | 4+4 4 | 4+4 4 | 3+4 4 | 6+6 4 | 3+4 4 | 3+4 4 | 3+4 4 |
| | 5 2 | 3+3 4 | 3+3 4 | 3+4 4 | 6+6 4 | 3+4 4 | 3+4 4 | 3+4 4 |
| | 5 2 | 4+4 4 | 4+4 4 | 3+4 4 | 5+5 4 | 3+4 4 | 3+4 4 | 3+4 4 |
| | 6 2 | 3+3 4 | 3+3 4 | 3+4 4 | 5+5 4 | 3+4 4 | 3+4 4 | 3+4 4 |
| | 6 2 | 4+4 4 | 4+4 4 | 3+4 4 | 6+6 4 | 3+4 4 | 3+4 4 | 3+4 4 |
| | 5 2 | 3+3 4 | 4+4 4 | 3+4 4 | 5+5 4 | 3+4 4 | 3+4 4 | 3+4 4 |
| | 5 2 | 4+4 4 | 3+3 4 | 3+4 4 | 6+6 4 | 3+4 4 | 3+4 4 | 3+4 4 |
| | 6 2 | 3+3 4 | 3+3 4 | 3+4 4 | 5+5 4 | 3+4 4 | 3+4 4 | 3+4 4 |
| | 6 2 | 4+4 4 | 4+4 4 | 3+4 4 | 6+6 4 | 3+4 4 | 3+4 4 | 3+4 4 |
| | 5 2 | 4+4 4 | 4+4 4 | 3+4 4 | 5+5 4 | 3+4 4 | 3+4 4 | 3+4 4 |
| | 5 2 | 3+3 4 | 3+3 4 | 3+4 4 | 6+6 4 | 3+4 4 | 3+4 4 | 3+4 4 |
| | 6 2 | 3+3 4 | 4+4 4 | 3+4 4 | 6+6 4 | 3+4 4 | 3+4 4 | 3+4 4 |
| | 6 2 | 4+4 4 | 3+3 4 | 3+4 4 | 6+6 4 | 3+4 4 | 3+4 4 | 3+4 4 |
| | 6 2 | 4+4 4 | 3+3 4 | 3+4 4 | 5+5 4 | 3+4 4 | 3+4 4 | 3+4 4 |
| | 6 2 | 3+3 4 | 4+4 4 | 3+4 4 | 5+5 4 | 3+4 4 | 3+4 4 | 3+4 4 |
| | 5 2 | 4+4 4 | 4+4 4 | 3+4 4 | 6+6 4 | 3+4 4 | 3+4 4 | 3+4 4 |
| | 5 2 | 3+3 4 | 3+3 4 | 3+4 4 | 5+5 4 | 3+4 4 | 3+4 4 | 3+4 4 |
| | 6 2 | 3+3 4 | 3+3 4 | 3+4 4 | 5+5 4 | 3+4 4 | 3+4 4 | 3+4 4 |
| | 5 2 | 4+4 4 | 4+4 4 | 3+4 4 | 6+6 4 | 3+4 4 | 3+4 4 | 3+4 4 |
| | 6 2 | 3+3 4 | 4+4 4 | 3+4 4 | 5+5 4 | 3+4 4 | 3+4 4 | 3+4 4 |
| | 5 2 | 4+4 4 | 3+3 4 | 3+4 4 | 6+6 4 | 3+4 4 | 3+4 4 | 3+4 4 |
| | 5 2 | 3+3 4 | 3+3 4 | 3+4 4 | 5+5 4 | 3+4 4 | 3+4 4 | 3+4 4 |
| | 5 2 | 4+4 4 | 4+4 4 | 3+4 4 | 6+6 4 | 3+4 4 | 3+4 4 | 3+4 4 |
| | 5 2 | 3+3 4 | 3+3 4 | 3+4 4 | 6+6 4 | 3+4 4 | 3+4 4 | 3+4 4 |
| | 5 2 | 4+4 4 | 4+4 4 | 3+4 4 | 5+5 4 | 3+4 4 | 3+4 4 | 3+4 4 |
| | 6 2 | 4+4 4 | 3+3 4 | 3+4 4 | 5+5 4 | 3+4 4 | 3+4 4 | 3+4 4 |
| | 6 2 | 3+3 4 | 4+4 4 | 3+4 4 | 6+6 4 | 3+4 4 | 3+4 4 | 3+4 4 |
| | 6 2 | 4+4 4 | 4+4 4 | 3+4 4 | 5+5 4 | 3+4 4 | 3+4 4 | 3+4 4 |
| | 6 2 | 3+3 4 | 3+3 4 | 3+4 4 | 6+6 4 | 3+4 4 | 3+4 4 | 3+4 4 |
| | 6 2 | 3+3 4 | 3+3 4 | 3+4 4 | 6+6 4 | 3+4 4 | 3+4 4 | 3+4 4 |
| | 6 2 | 4+4 4 | 4+4 4 | 3+4 4 | 5+5 4 | 3+4 4 | 3+4 4 | 3+4 4 |
| | 5 2 | 3+3 4 | 3+3 4 | 3+4 4 | 5+5 4 | 3+4 4 | 3+4 4 | 3+4 4 |
| | 5 2 | 4+4 4 | 4+4 4 | 3+4 4 | 6+6 4 | 3+4 4 | 3+4 4 | 3+4 4 |
| | 5 2 | 3+3 4 | 4+4 4 | 3+4 4 | 6+6 4 | 3+4 4 | 3+4 4 | 3+4 4 |
| | 5 2 | 4+4 4 | 3+3 4 | 3+4 4 | 5+5 4 | 3+4 4 | 3+4 4 | 3+4 4 |
| | 5 2 | 4+4 4 | 4+4 4 | 3+4 4 | 5+5 4 | 3+4 4 | 3+4 4 | 3+4 4 |
| | 5 2 | 3+3 4 | 3+3 4 | 3+4 4 | 6+6 4 | 3+4 4 | 3+4 4 | 3+4 4 |
| | 5 2 | | | | | | | |
| | 5 2 | | | | | | | |
| | 6 2 | | | | | | | |
| | 6 2 | | | | | | | |
| | 6 2 | | | | | | | |
| | 5 2 | | | | | | | |
| | 5 2 | | | | | | | |
| | 6 2 | | | | | | | |
| | 5 2 | | | | | | | |
| | 6 2 | | | | | | | |
| | 6 2 | | | | | | | |
| | 6 2 | | | | | | | |
| | 6 2 | | | | | | | |
| | 5 2 | | | | | | | |
| | 6 2 | | | | | | | |
| | 5 2 | | | | | | | |

Continued on next column

**TABLE XXXVII– continued**

| TQC | ICSD | NoE | SF | | | | | |
|---|---|---|---|---|---|---|---|---|
| | 5 2 | | | | | | | |
| | 6 2 | | | | | | | |
| | 5 2 | | | | | | | |
| | 6 2 | | | | | | | |
| | 5 2 | | | | | | | |
| | 5 2 | | | | | | | |
| | 6 2 | | | | | | | |
| | 6 2 | | | | | | | |
| | 6 2 | | | | | | | |
| | 5 2 | | | | | | | |
| | 6 2 | | | | | | | |
| | 5 2 | | | | | | | |
| | 6 2 | | | | | | | |
| | 6 2 | | | | | | | |
| | 5 2 | | | | | | | |
| | 6 2 | | | | | | | |
| | 5 2 | | | | | | | |
| | 5 2 | | | | | | | |
| | 6 2 | | | | | | | |
| | 5 2 | | | | | | | |
| | 6 2 | | | | | | | |
| | 5 2 | | | | | | | |
| | 5 2 | | | | | | | |
| | 6 2 | | | | | | | |
| | 5 2 | | | | | | | |
| | 6 2 | | | | | | | |
| | 5 2 | | | | | | | |
| | 5 2 | | | | | | | |
| 62.2.91 | 42912 | 128 | Ge3Pt2 | | | | | |
| | -GM | -R | -S | -T | -U | -X | -Y | -Z |
| | 5 2 | 3+3 4 | 4+4 4 | 3+4 4 | 5+5 4 | 3+4 4 | 3+4 4 | 3+4 4 |
| | 5 2 | 4+4 4 | 3+3 4 | 3+4 4 | 6+6 4 | 3+4 4 | 3+4 4 | 3+4 4 |
| | 6 2 | 3+3 4 | 4+4 4 | 3+4 4 | 5+5 4 | 3+4 4 | 3+4 4 | 3+4 4 |
| | 5 2 | 4+4 4 | 3+3 4 | 3+4 4 | 6+6 4 | 3+4 4 | 3+4 4 | 3+4 4 |
| | 6 2 | 3+3 4 | 3+3 4 | 3+4 4 | 5+5 4 | 3+4 4 | 3+4 4 | 3+4 4 |
| | 5 2 | 4+4 4 | 4+4 4 | 3+4 4 | 6+6 4 | 3+4 4 | 3+4 4 | 3+4 4 |
| | 6 2 | 4+4 4 | 3+3 4 | 3+4 4 | 5+5 4 | 3+4 4 | 3+4 4 | 3+4 4 |
| | 6 2 | 3+3 4 | 4+4 4 | 3+4 4 | 6+6 4 | 3+4 4 | 3+4 4 | 3+4 4 |
| | 5 2 | 3+3 4 | 3+3 4 | 3+4 4 | 5+5 4 | 3+4 4 | 3+4 4 | 3+4 4 |
| | 5 2 | 4+4 4 | 4+4 4 | 3+4 4 | 6+6 4 | 3+4 4 | 3+4 4 | 3+4 4 |
| | 6 2 | 4+4 4 | 3+3 4 | 3+4 4 | 5+5 4 | 3+4 4 | 3+4 4 | 3+4 4 |
| | 6 2 | 3+3 4 | 4+4 4 | 3+4 4 | 6+6 4 | 3+4 4 | 3+4 4 | 3+4 4 |
| | 5 2 | 4+4 4 | 4+4 4 | 3+4 4 | 5+5 4 | 3+4 4 | 3+4 4 | 3+4 4 |
| | 6 2 | 3+3 4 | 3+3 4 | 3+4 4 | 6+6 4 | 3+4 4 | 3+4 4 | 3+4 4 |
| | 6 2 | 4+4 4 | 3+3 4 | 3+4 4 | 5+5 4 | 3+4 4 | 3+4 4 | 3+4 4 |
| | 5 2 | 3+3 4 | 4+4 4 | 3+4 4 | 5+5 4 | 3+4 4 | 3+4 4 | 3+4 4 |
| | 5 2 | 3+3 4 | 4+4 4 | 3+4 4 | 6+6 4 | 3+4 4 | 3+4 4 | 3+4 4 |
| | 6 2 | 4+4 4 | 3+3 4 | 3+4 4 | 5+5 4 | 3+4 4 | 3+4 4 | 3+4 4 |
| | 6 2 | 3+3 4 | 4+4 4 | 3+4 4 | 6+6 4 | 3+4 4 | 3+4 4 | 3+4 4 |
| | 5 2 | 4+4 4 | 3+3 4 | 3+4 4 | 6+6 4 | 3+4 4 | 3+4 4 | 3+4 4 |
| | 6 2 | 4+4 4 | 4+4 4 | 3+4 4 | 5+5 4 | 3+4 4 | 3+4 4 | 3+4 4 |
| | 5 2 | 3+3 4 | 3+3 4 | 3+4 4 | 6+6 4 | 3+4 4 | 3+4 4 | 3+4 4 |
| | 5 2 | 3+3 4 | 3+3 4 | 3+4 4 | 5+5 4 | 3+4 4 | 3+4 4 | 3+4 4 |
| | 6 2 | 4+4 4 | 4+4 4 | 3+4 4 | 6+6 4 | 3+4 4 | 3+4 4 | 3+4 4 |
| | 5 2 | 3+3 4 | 4+4 4 | 3+4 4 | 6+6 4 | 3+4 4 | 3+4 4 | 3+4 4 |
| | 5 2 | 4+4 4 | 3+3 4 | 3+4 4 | 5+5 4 | 3+4 4 | 3+4 4 | 3+4 4 |
| | 6 2 | 3+3 4 | 4+4 4 | 3+4 4 | 6+6 4 | 3+4 4 | 3+4 4 | 3+4 4 |
| | 6 2 | 4+4 4 | 3+3 4 | 3+4 4 | 5+5 4 | 3+4 4 | 3+4 4 | 3+4 4 |
| | 5 2 | 4+4 4 | 3+3 4 | 3+4 4 | 6+6 4 | 3+4 4 | 3+4 4 | 3+4 4 |
| | 5 2 | 3+3 4 | 4+4 4 | 3+4 4 | 5+5 4 | 3+4 4 | 3+4 4 | 3+4 4 |
| | 5 2 | 4+4 4 | 4+4 4 | 3+4 4 | 6+6 4 | 3+4 4 | 3+4 4 | 3+4 4 |

Continued on next column

**TABLE XXXVII– continued**

| TQC | ICSD | NoE | SF | | | | | |
|---|---|---|---|---|---|---|---|---|
| | 6 2 | 3+3 4 | 3+3 4 | 3+4 4 | 5+5 4 | 3+4 4 | 3+4 4 | 3+4 4 |
| | 5 2 | | | | | | | |
| | 6 2 | | | | | | | |
| | 6 2 | | | | | | | |
| | 5 2 | | | | | | | |
| | 6 2 | | | | | | | |
| | 6 2 | | | | | | | |
| | 5 2 | | | | | | | |
| | 5 2 | | | | | | | |
| | 6 2 | | | | | | | |
| | 5 2 | | | | | | | |
| | 6 2 | | | | | | | |
| | 5 2 | | | | | | | |
| | 5 2 | | | | | | | |
| | 6 2 | | | | | | | |
| | 5 2 | | | | | | | |
| | 6 2 | | | | | | | |
| | 5 2 | | | | | | | |
| | 6 2 | | | | | | | |
| | 6 2 | | | | | | | |
| | 5 2 | | | | | | | |
| | 5 2 | | | | | | | |
| | 6 2 | | | | | | | |
| | 6 2 | | | | | | | |
| | 6 2 | | | | | | | |
| | 5 2 | | | | | | | |
| | 5 2 | | | | | | | |
| | 6 2 | | | | | | | |
| | 5 2 | | | | | | | |
| | 6 2 | | | | | | | |
| | 5 2 | | | | | | | |
| | 5 2 | | | | | | | |
| | 6 2 | | | | | | | |
| 62.2.92 | 16822 | 56 | Ge2Mo1 | | | | | |
| | -GM | -R | -S | -T | -U | -X | -Y | -Z |
| | 5 2 | 3+3 4 | 3+3 4 | 3+4 4 | 6+6 4 | 3+4 4 | 3+4 4 | 3+4 4 |
| | 5 2 | 4+4 4 | 4+4 4 | 3+4 4 | 5+5 4 | 3+4 4 | 3+4 4 | 3+4 4 |
| | 6 2 | 3+3 4 | 3+3 4 | 3+4 4 | 6+6 4 | 3+4 4 | 3+4 4 | 3+4 4 |
| | 6 2 | 4+4 4 | 4+4 4 | 3+4 4 | 5+5 4 | 3+4 4 | 3+4 4 | 3+4 4 |
| | 5 2 | 3+3 4 | 4+4 4 | 3+4 4 | 5+5 4 | 3+4 4 | 3+4 4 | 3+4 4 |
| | 5 2 | 4+4 4 | 3+3 4 | 3+4 4 | 6+6 4 | 3+4 4 | 3+4 4 | 3+4 4 |
| | 6 2 | 3+3 4 | 4+4 4 | 3+4 4 | 6+6 4 | 3+4 4 | 3+4 4 | 3+4 4 |
| | 6 2 | 4+4 4 | 3+3 4 | 3+4 4 | 5+5 4 | 3+4 4 | 3+4 4 | 3+4 4 |
| | 5 2 | 4+4 4 | 3+3 4 | 3+4 4 | 6+6 4 | 3+4 4 | 3+4 4 | 3+4 4 |
| | 5 2 | 3+3 4 | 4+4 4 | 3+4 4 | 5+5 4 | 3+4 4 | 3+4 4 | 3+4 4 |
| | 6 2 | 4+4 4 | 4+4 4 | 3+4 4 | 5+5 4 | 3+4 4 | 3+4 4 | 3+4 4 |
| | 6 2 | 3+3 4 | 3+3 4 | 3+4 4 | 6+6 4 | 3+4 4 | 3+4 4 | 3+4 4 |
| | 5 2 | 3+3 4 | 3+3 4 | 3+4 4 | 6+6 4 | 3+4 4 | 3+4 4 | 3+4 4 |
| | 5 2 | 4+4 4 | 4+4 4 | 3+4 4 | 5+5 4 | 3+4 4 | 3+4 4 | 3+4 4 |
| | 5 2 | | | | | | | |
| | 5 2 | | | | | | | |
| | 6 2 | | | | | | | |
| | 5 2 | | | | | | | |
| | 5 2 | | | | | | | |
| | 6 2 | | | | | | | |
| | 6 2 | | | | | | | |
| | 5 2 | | | | | | | |
| | 6 2 | | | | | | | |
| | 6 2 | | | | | | | |
| | 5 2 | | | | | | | |
| | 6 2 | | | | | | | |

Continued on next column

**TABLE XXXVII– continued**

| TQC | ICSD | NoE | SF | | | | | |
|---|---|---|---|---|---|---|---|---|
| | 5 2 | | | | | | | |
| | 6 2 | | | | | | | |
| 62.2.93 | 429881 | 48 | Ge2Li3Na1 | | | | | |
| | -GM | -R | -S | -T | -U | -X | -Y | -Z |
| | 5 2 | 3+3 4 | 3+3 4 | 3+4 4 | 5+5 4 | 3+4 4 | 3+4 4 | 3+4 4 |
| | 6 2 | 4+4 4 | 4+4 4 | 3+4 4 | 6+6 4 | 3+4 4 | 3+4 4 | 3+4 4 |
| | 5 2 | 3+3 4 | 3+3 4 | 3+4 4 | 5+5 4 | 3+4 4 | 3+4 4 | 3+4 4 |
| | 6 2 | 4+4 4 | 4+4 4 | 3+4 4 | 6+6 4 | 3+4 4 | 3+4 4 | 3+4 4 |
| | 5 2 | 3+3 4 | 4+4 4 | 3+4 4 | 6+6 4 | 3+4 4 | 3+4 4 | 3+4 4 |
| | 6 2 | 4+4 4 | 3+3 4 | 3+4 4 | 5+5 4 | 3+4 4 | 3+4 4 | 3+4 4 |
| | 5 2 | 3+3 4 | 3+3 4 | 3+4 4 | 5+5 4 | 3+4 4 | 3+4 4 | 3+4 4 |
| | 6 2 | 4+4 4 | 4+4 4 | 3+4 4 | 6+6 4 | 3+4 4 | 3+4 4 | 3+4 4 |
| | 6 2 | 4+4 4 | 4+4 4 | 3+4 4 | 5+5 4 | 3+4 4 | 3+4 4 | 3+4 4 |
| | 5 2 | 3+3 4 | 3+3 4 | 3+4 4 | 6+6 4 | 3+4 4 | 3+4 4 | 3+4 4 |
| | 5 2 | 4+4 4 | 3+3 4 | 3+4 4 | 5+5 4 | 3+4 4 | 3+4 4 | 3+4 4 |
| | 5 2 | 3+3 4 | 4+4 4 | 3+4 4 | 6+6 4 | 3+4 4 | 3+4 4 | 3+4 4 |
| | 6 2 | | | | | | | |
| | 5 2 | | | | | | | |
| | 6 2 | | | | | | | |
| | 6 2 | | | | | | | |
| | 5 2 | | | | | | | |
| | 6 2 | | | | | | | |
| | 5 2 | | | | | | | |
| | 5 2 | | | | | | | |
| | 6 2 | | | | | | | |
| | 6 2 | | | | | | | |
| | 5 2 | | | | | | | |
| | 5 2 | | | | | | | |
| 62.2.94 | 280691 | 128 | Ge2In1Li1Sr2 | | | | | |
| | -GM | -R | -S | -T | -U | -X | -Y | -Z |
| | 5 2 | 3+3 4 | 3+3 4 | 3+4 4 | 5+5 4 | 3+4 4 | 3+4 4 | 3+4 4 |
| | 5 2 | 4+4 4 | 4+4 4 | 3+4 4 | 6+6 4 | 3+4 4 | 3+4 4 | 3+4 4 |
| | 6 2 | 3+3 4 | 3+3 4 | 3+4 4 | 5+5 4 | 3+4 4 | 3+4 4 | 3+4 4 |
| | 6 2 | 4+4 4 | 4+4 4 | 3+4 4 | 6+6 4 | 3+4 4 | 3+4 4 | 3+4 4 |
| | 5 2 | 3+3 4 | 3+3 4 | 3+4 4 | 5+5 4 | 3+4 4 | 3+4 4 | 3+4 4 |
| | 6 2 | 4+4 4 | 4+4 4 | 3+4 4 | 6+6 4 | 3+4 4 | 3+4 4 | 3+4 4 |
| | 5 2 | 3+3 4 | 3+3 4 | 3+4 4 | 5+5 4 | 3+4 4 | 3+4 4 | 3+4 4 |
| | 6 2 | 4+4 4 | 4+4 4 | 3+4 4 | 6+6 4 | 3+4 4 | 3+4 4 | 3+4 4 |
| | 5 2 | 3+3 4 | 4+4 4 | 3+4 4 | 5+5 4 | 3+4 4 | 3+4 4 | 3+4 4 |
| | 5 2 | 4+4 4 | 3+3 4 | 3+4 4 | 6+6 4 | 3+4 4 | 3+4 4 | 3+4 4 |
| | 6 2 | 3+3 4 | 3+3 4 | 3+4 4 | 5+5 4 | 3+4 4 | 3+4 4 | 3+4 4 |
| | 6 2 | 4+4 4 | 4+4 4 | 3+4 4 | 6+6 4 | 3+4 4 | 3+4 4 | 3+4 4 |
| | 5 2 | 3+3 4 | 3+3 4 | 3+4 4 | 5+5 4 | 3+4 4 | 3+4 4 | 3+4 4 |
| | 6 2 | 4+4 4 | 4+4 4 | 3+4 4 | 6+6 4 | 3+4 4 | 3+4 4 | 3+4 4 |
| | 5 2 | 3+3 4 | 3+3 4 | 3+4 4 | 5+5 4 | 3+4 4 | 3+4 4 | 3+4 4 |
| | 6 2 | 4+4 4 | 4+4 4 | 3+4 4 | 6+6 4 | 3+4 4 | 3+4 4 | 3+4 4 |
| | 5 2 | 3+3 4 | 3+3 4 | 3+4 4 | 5+5 4 | 3+4 4 | 3+4 4 | 3+4 4 |
| | 5 2 | 4+4 4 | 4+4 4 | 3+4 4 | 6+6 4 | 3+4 4 | 3+4 4 | 3+4 4 |
| | 6 2 | 3+3 4 | 3+3 4 | 3+4 4 | 6+6 4 | 3+4 4 | 3+4 4 | 3+4 4 |
| | 6 2 | 4+4 4 | 4+4 4 | 3+4 4 | 5+5 4 | 3+4 4 | 3+4 4 | 3+4 4 |
| | 5 2 | 3+3 4 | 4+4 4 | 3+4 4 | 6+6 4 | 3+4 4 | 3+4 4 | 3+4 4 |
| | 5 2 | 4+4 4 | 3+3 4 | 3+4 4 | 5+5 4 | 3+4 4 | 3+4 4 | 3+4 4 |
| | 6 2 | 3+3 4 | 3+3 4 | 3+4 4 | 5+5 4 | 3+4 4 | 3+4 4 | 3+4 4 |
| | 6 2 | 4+4 4 | 4+4 4 | 3+4 4 | 6+6 4 | 3+4 4 | 3+4 4 | 3+4 4 |
| | 5 2 | 3+3 4 | 3+3 4 | 3+4 4 | 6+6 4 | 3+4 4 | 3+4 4 | 3+4 4 |
| | 6 2 | 4+4 4 | 4+4 4 | 3+4 4 | 6+6 4 | 3+4 4 | 3+4 4 | 3+4 4 |
| | 5 2 | 3+3 4 | 4+4 4 | 3+4 4 | 5+5 4 | 3+4 4 | 3+4 4 | 3+4 4 |
| | 6 2 | 4+4 4 | 3+3 4 | 3+4 4 | 5+5 4 | 3+4 4 | 3+4 4 | 3+4 4 |
| | 5 2 | 4+4 4 | 3+3 4 | 3+4 4 | 6+6 4 | 3+4 4 | 3+4 4 | 3+4 4 |
| | 6 2 | 3+3 4 | 4+4 4 | 3+4 4 | 5+5 4 | 3+4 4 | 3+4 4 | 3+4 4 |
| | 5 2 | 3+3 4 | 4+4 4 | 3+4 4 | 5+5 4 | 3+4 4 | 3+4 4 | 3+4 4 |

Continued on next column

**TABLE XXXVII– continued**

| TQC | ICSD | NoE | SF | | | | | |
|---|---|---|---|---|---|---|---|---|
| | 6 2 | 4+4 4 | 3+3 4 | 3+4 4 | 6+6 4 | 3+4 4 | 3+4 4 | 3+4 4 |
| | 5 2 | | | | | | | |
| | 6 2 | | | | | | | |
| | 5 2 | | | | | | | |
| | 6 2 | | | | | | | |
| | 6 2 | | | | | | | |
| | 5 2 | | | | | | | |
| | 6 2 | | | | | | | |
| | 5 2 | | | | | | | |
| | 5 2 | | | | | | | |
| | 6 2 | | | | | | | |
| | 5 2 | | | | | | | |
| | 6 2 | | | | | | | |
| | 6 2 | | | | | | | |
| | 5 2 | | | | | | | |
| | 5 2 | | | | | | | |
| | 6 2 | | | | | | | |
| | 5 2 | | | | | | | |
| | 6 2 | | | | | | | |
| | 5 2 | | | | | | | |
| | 5 2 | | | | | | | |
| | 6 2 | | | | | | | |
| | 6 2 | | | | | | | |
| | 5 2 | | | | | | | |
| | 6 2 | | | | | | | |
| | 5 2 | | | | | | | |
| | 5 2 | | | | | | | |
| | 6 2 | | | | | | | |
| | 5 2 | | | | | | | |
| | 5 2 | | | | | | | |
| | 6 2 | | | | | | | |
| | 5 2 | | | | | | | |
| | 6 2 | | | | | | | |
| 62.2.95 | 44860 | 88 | Ge1Rh2 | | | | | |
| | -GM | -R | -S | -T | -U | -X | -Y | -Z |
| | 5 2 | 3+3 4 | 4+4 4 | 3+4 4 | 5+5 4 | 3+4 4 | 3+4 4 | 3+4 4 |
| | 6 2 | 4+4 4 | 3+3 4 | 3+4 4 | 6+6 4 | 3+4 4 | 3+4 4 | 3+4 4 |
| | 5 2 | 3+3 4 | 4+4 4 | 3+4 4 | 6+6 4 | 3+4 4 | 3+4 4 | 3+4 4 |
| | 6 2 | 4+4 4 | 3+3 4 | 3+4 4 | 5+5 4 | 3+4 4 | 3+4 4 | 3+4 4 |
| | 6 2 | 4+4 4 | 4+4 4 | 3+4 4 | 5+5 4 | 3+4 4 | 3+4 4 | 3+4 4 |
| | 5 2 | 3+3 4 | 3+3 4 | 3+4 4 | 5+5 4 | 3+4 4 | 3+4 4 | 3+4 4 |
| | 5 2 | 3+3 4 | 4+4 4 | 3+4 4 | 6+6 4 | 3+4 4 | 3+4 4 | 3+4 4 |
| | 5 2 | 4+4 4 | 3+3 4 | 3+4 4 | 6+6 4 | 3+4 4 | 3+4 4 | 3+4 4 |
| | 6 2 | 3+3 4 | 3+3 4 | 3+4 4 | 5+5 4 | 3+4 4 | 3+4 4 | 3+4 4 |
| | 6 2 | 4+4 4 | 4+4 4 | 3+4 4 | 6+6 4 | 3+4 4 | 3+4 4 | 3+4 4 |
| | 6 2 | 4+4 4 | 3+3 4 | 3+4 4 | 5+5 4 | 3+4 4 | 3+4 4 | 3+4 4 |
| | 6 2 | 3+3 4 | 4+4 4 | 3+4 4 | 6+6 4 | 3+4 4 | 3+4 4 | 3+4 4 |
| | 6 2 | 3+3 4 | 4+4 4 | 3+4 4 | 5+5 4 | 3+4 4 | 3+4 4 | 3+4 4 |
| | 5 2 | 4+4 4 | 3+3 4 | 3+4 4 | 6+6 4 | 3+4 4 | 3+4 4 | 3+4 4 |
| | 5 2 | 3+3 4 | 3+3 4 | 3+4 4 | 6+6 4 | 3+4 4 | 3+4 4 | 3+4 4 |
| | 5 2 | 4+4 4 | 4+4 4 | 3+4 4 | 5+5 4 | 3+4 4 | 3+4 4 | 3+4 4 |
| | 6 2 | 4+4 4 | 4+4 4 | 3+4 4 | 5+5 4 | 3+4 4 | 3+4 4 | 3+4 4 |
| | 5 2 | 3+3 4 | 3+3 4 | 3+4 4 | 6+6 4 | 3+4 4 | 3+4 4 | 3+4 4 |
| | 5 2 | 4+4 4 | 4+4 4 | 3+4 4 | 6+6 4 | 3+4 4 | 3+4 4 | 3+4 4 |
| | 6 2 | 3+3 4 | 3+3 4 | 3+4 4 | 6+6 4 | 3+4 4 | 3+4 4 | 3+4 4 |
| | 5 2 | 4+4 4 | 3+3 4 | 3+4 4 | 5+5 4 | 3+4 4 | 3+4 4 | 3+4 4 |
| | 5 2 | 3+3 4 | 4+4 4 | 3+4 4 | 5+5 4 | 3+4 4 | 3+4 4 | 3+4 4 |
| | 5 2 | | | | | | | |
| | 6 2 | | | | | | | |
| | 5 2 | | | | | | | |
| | 6 2 | | | | | | | |

Continued on next column

**TABLE XXXVII– continued**

| TQC | ICSD | NoE | SF | | | | | |
|---|---|---|---|---|---|---|---|---|
| | 5 2 | | | | | | | |
| | 6 2 | | | | | | | |
| | 6 2 | | | | | | | |
| | 5 2 | | | | | | | |
| | 6 2 | | | | | | | |
| | 5 2 | | | | | | | |
| | 5 2 | | | | | | | |
| | 6 2 | | | | | | | |
| | 6 2 | | | | | | | |
| | 5 2 | | | | | | | |
| | 5 2 | | | | | | | |
| | 5 2 | | | | | | | |
| | 5 2 | | | | | | | |
| | 6 2 | | | | | | | |
| | 6 2 | | | | | | | |
| | 5 2 | | | | | | | |
| | 6 2 | | | | | | | |
| | 5 2 | | | | | | | |
| 62.2.96 | 421092 | 72 | Ge1Rh1Ta1 | | | | | |
| | -GM | -R | -S | -T | -U | -X | -Y | -Z |
| | 5 2 | 3+3 4 | 4+4 4 | 3+4 4 | 6+6 4 | 3+4 4 | 3+4 4 | 3+4 4 |
| | 6 2 | 4+4 4 | 3+3 4 | 3+4 4 | 5+5 4 | 3+4 4 | 3+4 4 | 3+4 4 |
| | 5 2 | 4+4 4 | 3+3 4 | 3+4 4 | 6+6 4 | 3+4 4 | 3+4 4 | 3+4 4 |
| | 6 2 | 3+3 4 | 4+4 4 | 3+4 4 | 5+5 4 | 3+4 4 | 3+4 4 | 3+4 4 |
| | 5 2 | 3+3 4 | 3+3 4 | 3+4 4 | 5+5 4 | 3+4 4 | 3+4 4 | 3+4 4 |
| | 6 2 | 4+4 4 | 4+4 4 | 3+4 4 | 5+5 4 | 3+4 4 | 3+4 4 | 3+4 4 |
| | 6 2 | 4+4 4 | 3+3 4 | 3+4 4 | 6+6 4 | 3+4 4 | 3+4 4 | 3+4 4 |
| | 5 2 | 3+3 4 | 4+4 4 | 3+4 4 | 6+6 4 | 3+4 4 | 3+4 4 | 3+4 4 |
| | 5 2 | 4+4 4 | 3+3 4 | 3+4 4 | 6+6 4 | 3+4 4 | 3+4 4 | 3+4 4 |
| | 6 2 | 3+3 4 | 4+4 4 | 3+4 4 | 5+5 4 | 3+4 4 | 3+4 4 | 3+4 4 |
| | 5 2 | 3+3 4 | 3+3 4 | 3+4 4 | 6+6 4 | 3+4 4 | 3+4 4 | 3+4 4 |
| | 6 2 | 4+4 4 | 4+4 4 | 3+4 4 | 5+5 4 | 3+4 4 | 3+4 4 | 3+4 4 |
| | 6 2 | 4+4 4 | 4+4 4 | 3+4 4 | 5+5 4 | 3+4 4 | 3+4 4 | 3+4 4 |
| | 6 2 | 3+3 4 | 3+3 4 | 3+4 4 | 6+6 4 | 3+4 4 | 3+4 4 | 3+4 4 |
| | 6 2 | 4+4 4 | 4+4 4 | 3+4 4 | 6+6 4 | 3+4 4 | 3+4 4 | 3+4 4 |
| | 5 2 | 3+3 4 | 3+3 4 | 3+4 4 | 6+6 4 | 3+4 4 | 3+4 4 | 3+4 4 |
| | 6 2 | 3+3 4 | 4+4 4 | 3+4 4 | 5+5 4 | 3+4 4 | 3+4 4 | 3+4 4 |
| | 5 2 | 4+4 4 | 3+3 4 | 3+4 4 | 5+5 4 | 3+4 4 | 3+4 4 | 3+4 4 |
| | 5 2 | | | | | | | |
| | 6 2 | | | | | | | |
| | 5 2 | | | | | | | |
| | 5 2 | | | | | | | |
| | 5 2 | | | | | | | |
| | 6 2 | | | | | | | |
| | 5 2 | | | | | | | |
| | 5 2 | | | | | | | |
| | 6 2 | | | | | | | |
| | 5 2 | | | | | | | |
| | 6 2 | | | | | | | |
| | 5 2 | | | | | | | |
| | 6 2 | | | | | | | |
| | 5 2 | | | | | | | |
| | 5 2 | | | | | | | |
| | 5 2 | | | | | | | |
| | 6 2 | | | | | | | |
| | 6 2 | | | | | | | |
| 62.2.97 | 260793 | 72 | Ge1Pt1Zr1 | | | | | |
| | -GM | -R | -S | -T | -U | -X | -Y | -Z |
| | 5 2 | 3+3 4 | 3+3 4 | 3+4 4 | 6+6 4 | 3+4 4 | 3+4 4 | 3+4 4 |
| | 6 2 | 4+4 4 | 4+4 4 | 3+4 4 | 5+5 4 | 3+4 4 | 3+4 4 | 3+4 4 |
| | 5 2 | 4+4 4 | 3+3 4 | 3+4 4 | 6+6 4 | 3+4 4 | 3+4 4 | 3+4 4 |

Continued on next column

**TABLE XXXVII– continued**

| TQC | ICSD | NoE | SF | | | | | |
|---|---|---|---|---|---|---|---|---|
| | 6 2 | 3+3 4 | 4+4 4 | 3+4 4 | 5+5 4 | 3+4 4 | 3+4 4 | 3+4 4 |
| | 5 2 | 3+3 4 | 3+3 4 | 3+4 4 | 5+5 4 | 3+4 4 | 3+4 4 | 3+4 4 |
| | 6 2 | 4+4 4 | 4+4 4 | 3+4 4 | 6+6 4 | 3+4 4 | 3+4 4 | 3+4 4 |
| | 5 2 | 3+3 4 | 3+3 4 | 3+4 4 | 5+5 4 | 3+4 4 | 3+4 4 | 3+4 4 |
| | 6 2 | 4+4 4 | 4+4 4 | 3+4 4 | 6+6 4 | 3+4 4 | 3+4 4 | 3+4 4 |
| | 6 2 | 4+4 4 | 4+4 4 | 3+4 4 | 6+6 4 | 3+4 4 | 3+4 4 | 3+4 4 |
| | 5 2 | 3+3 4 | 3+3 4 | 3+4 4 | 5+5 4 | 3+4 4 | 3+4 4 | 3+4 4 |
| | 5 2 | 4+4 4 | 3+3 4 | 3+4 4 | 6+6 4 | 3+4 4 | 3+4 4 | 3+4 4 |
| | 6 2 | 3+3 4 | 4+4 4 | 3+4 4 | 5+5 4 | 3+4 4 | 3+4 4 | 3+4 4 |
| | 6 2 | 3+3 4 | 4+4 4 | 3+4 4 | 5+5 4 | 3+4 4 | 3+4 4 | 3+4 4 |
| | 6 2 | 4+4 4 | 3+3 4 | 3+4 4 | 6+6 4 | 3+4 4 | 3+4 4 | 3+4 4 |
| | 5 2 | 3+3 4 | 4+4 4 | 3+4 4 | 6+6 4 | 3+4 4 | 3+4 4 | 3+4 4 |
| | 6 2 | 4+4 4 | 3+3 4 | 3+4 4 | 6+6 4 | 3+4 4 | 3+4 4 | 3+4 4 |
| | 5 2 | 3+3 4 | 3+3 4 | 3+4 4 | 5+5 4 | 3+4 4 | 3+4 4 | 3+4 4 |
| | 5 2 | 4+4 4 | 4+4 4 | 3+4 4 | 5+5 4 | 3+4 4 | 3+4 4 | 3+4 4 |
| | 6 2 | | | | | | | |
| | 5 2 | | | | | | | |
| | 6 2 | | | | | | | |
| | 5 2 | | | | | | | |
| | 5 2 | | | | | | | |
| | 6 2 | | | | | | | |
| | 5 2 | | | | | | | |
| | 6 2 | | | | | | | |
| | 5 2 | | | | | | | |
| | 5 2 | | | | | | | |
| | 6 2 | | | | | | | |
| | 6 2 | | | | | | | |
| | 5 2 | | | | | | | |
| | 5 2 | | | | | | | |
| | 5 2 | | | | | | | |
| | 5 2 | | | | | | | |
| | 6 2 | | | | | | | |
| | 6 2 | | | | | | | |
| 62.2.98 | 419135 | 152 | Ge1Pt1Yb1 | | | | | |
| | -GM | -R | -S | -T | -U | -X | -Y | -Z |
| | 5 2 | 3+3 4 | 3+3 4 | 3+4 4 | 6+6 4 | 3+4 4 | 3+4 4 | 3+4 4 |
| | 6 2 | 4+4 4 | 4+4 4 | 3+4 4 | 5+5 4 | 3+4 4 | 3+4 4 | 3+4 4 |
| | 5 2 | 3+3 4 | 3+3 4 | 3+4 4 | 6+6 4 | 3+4 4 | 3+4 4 | 3+4 4 |
| | 6 2 | 4+4 4 | 4+4 4 | 3+4 4 | 5+5 4 | 3+4 4 | 3+4 4 | 3+4 4 |
| | 5 2 | 4+4 4 | 4+4 4 | 3+4 4 | 6+6 4 | 3+4 4 | 3+4 4 | 3+4 4 |
| | 6 2 | 3+3 4 | 3+3 4 | 3+4 4 | 5+5 4 | 3+4 4 | 3+4 4 | 3+4 4 |
| | 5 2 | 3+3 4 | 4+4 4 | 3+4 4 | 6+6 4 | 3+4 4 | 3+4 4 | 3+4 4 |
| | 6 2 | 4+4 4 | 3+3 4 | 3+4 4 | 5+5 4 | 3+4 4 | 3+4 4 | 3+4 4 |
| | 5 2 | 3+3 4 | 4+4 4 | 3+4 4 | 5+5 4 | 3+4 4 | 3+4 4 | 3+4 4 |
| | 6 2 | 4+4 4 | 3+3 4 | 3+4 4 | 6+6 4 | 3+4 4 | 3+4 4 | 3+4 4 |
| | 5 2 | 4+4 4 | 3+3 4 | 3+4 4 | 6+6 4 | 3+4 4 | 3+4 4 | 3+4 4 |
| | 5 2 | 3+3 4 | 4+4 4 | 3+4 4 | 5+5 4 | 3+4 4 | 3+4 4 | 3+4 4 |
| | 6 2 | 4+4 4 | 3+3 4 | 3+4 4 | 5+5 4 | 3+4 4 | 3+4 4 | 3+4 4 |
| | 6 2 | 3+3 4 | 4+4 4 | 3+4 4 | 6+6 4 | 3+4 4 | 3+4 4 | 3+4 4 |
| | 5 2 | 4+4 4 | 4+4 4 | 3+4 4 | 5+5 4 | 3+4 4 | 3+4 4 | 3+4 4 |
| | 6 2 | 3+3 4 | 3+3 4 | 3+4 4 | 6+6 4 | 3+4 4 | 3+4 4 | 3+4 4 |
| | 5 2 | 3+3 4 | 3+3 4 | 3+4 4 | 5+5 4 | 3+4 4 | 3+4 4 | 3+4 4 |
| | 5 2 | 4+4 4 | 4+4 4 | 3+4 4 | 6+6 4 | 3+4 4 | 3+4 4 | 3+4 4 |
| | 6 2 | 3+3 4 | 4+4 4 | 3+4 4 | 5+5 4 | 3+4 4 | 3+4 4 | 3+4 4 |
| | 6 2 | 4+4 4 | 3+3 4 | 3+4 4 | 6+6 4 | 3+4 4 | 3+4 4 | 3+4 4 |
| | 6 2 | 4+4 4 | 4+4 4 | 3+4 4 | 5+5 4 | 3+4 4 | 3+4 4 | 3+4 4 |
| | 5 2 | 3+3 4 | 3+3 4 | 3+4 4 | 6+6 4 | 3+4 4 | 3+4 4 | 3+4 4 |
| | 6 2 | 3+3 4 | 4+4 4 | 3+4 4 | 5+5 4 | 3+4 4 | 3+4 4 | 3+4 4 |
| | 5 2 | 4+4 4 | 3+3 4 | 3+4 4 | 6+6 4 | 3+4 4 | 3+4 4 | 3+4 4 |
| | 5 2 | 3+3 4 | 4+4 4 | 3+4 4 | 5+5 4 | 3+4 4 | 3+4 4 | 3+4 4 |
| | 6 2 | 4+4 4 | 3+3 4 | 3+4 4 | 6+6 4 | 3+4 4 | 3+4 4 | 3+4 4 |

Continued on next column

**TABLE XXXVII– continued**

| TQC | ICSD | NoE | SF | | | | | |
|---|---|---|---|---|---|---|---|---|
| | 5 2 | 4+4 4 | 3+3 4 | 3+4 4 | 5+5 4 | 3+4 4 | 3+4 4 | 3+4 4 |
| | 5 2 | 3+3 4 | 4+4 4 | 3+4 4 | 6+6 4 | 3+4 4 | 3+4 4 | 3+4 4 |
| | 6 2 | 4+4 4 | 4+4 4 | 3+4 4 | 5+5 4 | 3+4 4 | 3+4 4 | 3+4 4 |
| | 5 2 | 3+3 4 | 3+3 4 | 3+4 4 | 5+5 4 | 3+4 4 | 3+4 4 | 3+4 4 |
| | 6 2 | 4+4 4 | 3+3 4 | 3+4 4 | 6+6 4 | 3+4 4 | 3+4 4 | 3+4 4 |
| | 5 2 | 3+3 4 | 4+4 4 | 3+4 4 | 6+6 4 | 3+4 4 | 3+4 4 | 3+4 4 |
| | 6 2 | 4+4 4 | 3+3 4 | 3+4 4 | 6+6 4 | 3+4 4 | 3+4 4 | 3+4 4 |
| | 5 2 | 3+3 4 | 4+4 4 | 3+4 4 | 5+5 4 | 3+4 4 | 3+4 4 | 3+4 4 |
| | 6 2 | 3+3 4 | 3+3 4 | 3+4 4 | 5+5 4 | 3+4 4 | 3+4 4 | 3+4 4 |
| | 6 2 | 4+4 4 | 4+4 4 | 3+4 4 | 6+6 4 | 3+4 4 | 3+4 4 | 3+4 4 |
| | 5 2 | 3+3 4 | 4+4 4 | 3+4 4 | 5+5 4 | 3+4 4 | 3+4 4 | 3+4 4 |
| | 5 2 | 4+4 4 | 3+3 4 | 3+4 4 | 5+5 4 | 3+4 4 | 3+4 4 | 3+4 4 |
| | 6 2 | | | | | | | |
| | 5 2 | | | | | | | |
| | 6 2 | | | | | | | |
| | 5 2 | | | | | | | |
| | 6 2 | | | | | | | |
| | 5 2 | | | | | | | |
| | 5 2 | | | | | | | |
| | 6 2 | | | | | | | |
| | 6 2 | | | | | | | |
| | 6 2 | | | | | | | |
| | 5 2 | | | | | | | |
| | 6 2 | | | | | | | |
| | 5 2 | | | | | | | |
| | 6 2 | | | | | | | |
| | 5 2 | | | | | | | |
| | 5 2 | | | | | | | |
| | 6 2 | | | | | | | |
| | 5 2 | | | | | | | |
| | 6 2 | | | | | | | |
| | 5 2 | | | | | | | |
| | 6 2 | | | | | | | |
| | 6 2 | | | | | | | |
| | 5 2 | | | | | | | |
| | 6 2 | | | | | | | |
| | 6 2 | | | | | | | |
| | 5 2 | | | | | | | |
| | 5 2 | | | | | | | |
| | 6 2 | | | | | | | |
| | 5 2 | | | | | | | |
| | 6 2 | | | | | | | |
| | 5 2 | | | | | | | |
| | 6 2 | | | | | | | |
| | 5 2 | | | | | | | |
| | 6 2 | | | | | | | |
| | 6 2 | | | | | | | |
| | 5 2 | | | | | | | |
| | 5 2 | | | | | | | |
| | 5 2 | | | | | | | |
| 62.2.99 | 2624 | 56 | Ge1Pt1 | | | | | |
| | -GM | -R | -S | -T | -U | -X | -Y | -Z |
| | 5 2 | 3+3 4 | 4+4 4 | 3+4 4 | 6+6 4 | 3+4 4 | 3+4 4 | 3+4 4 |
| | 5 2 | 4+4 4 | 3+3 4 | 3+4 4 | 5+5 4 | 3+4 4 | 3+4 4 | 3+4 4 |
| | 6 2 | 3+3 4 | 3+3 4 | 3+4 4 | 6+6 4 | 3+4 4 | 3+4 4 | 3+4 4 |
| | 6 2 | 4+4 4 | 4+4 4 | 3+4 4 | 5+5 4 | 3+4 4 | 3+4 4 | 3+4 4 |
| | 6 2 | 4+4 4 | 3+3 4 | 3+4 4 | 6+6 4 | 3+4 4 | 3+4 4 | 3+4 4 |
| | 5 2 | 3+3 4 | 4+4 4 | 3+4 4 | 5+5 4 | 3+4 4 | 3+4 4 | 3+4 4 |
| | 6 2 | 4+4 4 | 3+3 4 | 3+4 4 | 6+6 4 | 3+4 4 | 3+4 4 | 3+4 4 |
| | 5 2 | 3+3 4 | 4+4 4 | 3+4 4 | 5+5 4 | 3+4 4 | 3+4 4 | 3+4 4 |
| | 6 2 | 3+3 4 | 3+3 4 | 3+4 4 | 6+6 4 | 3+4 4 | 3+4 4 | 3+4 4 |

Continued on next column

**TABLE XXXVII– continued**

| TQC | ICSD | NoE | SF | | | | | |
|---|---|---|---|---|---|---|---|---|
| | 5 2 | 4+4 4 | 4+4 4 | 3+4 4 | 5+5 4 | 3+4 4 | 3+4 4 | 3+4 4 |
| | 6 2 | 4+4 4 | 3+3 4 | 3+4 4 | 6+6 4 | 3+4 4 | 3+4 4 | 3+4 4 |
| | 5 2 | 3+3 4 | 4+4 4 | 3+4 4 | 5+5 4 | 3+4 4 | 3+4 4 | 3+4 4 |
| | 5 2 | 4+4 4 | 4+4 4 | 3+4 4 | 5+5 4 | 3+4 4 | 3+4 4 | 3+4 4 |
| | 5 2 | 3+3 4 | 3+3 4 | 3+4 4 | 6+6 4 | 3+4 4 | 3+4 4 | 3+4 4 |
| | 5 2 | | | | | | | |
| | 6 2 | | | | | | | |
| | 5 2 | | | | | | | |
| | 6 2 | | | | | | | |
| | 6 2 | | | | | | | |
| | 5 2 | | | | | | | |
| | 6 2 | | | | | | | |
| | 6 2 | | | | | | | |
| | 5 2 | | | | | | | |
| | 5 2 | | | | | | | |
| | 6 2 | | | | | | | |
| | 5 2 | | | | | | | |
| | 5 2 | | | | | | | |
| | 5 2 | | | | | | | |
| 62.2.100 | 409926 | 72 | Ge1Pd1Ti1 | | | | | |
| | -GM | -R | -S | -T | -U | -X | -Y | -Z |
| | 5 2 | 3+3 4 | 3+3 4 | 3+4 4 | 6+6 4 | 3+4 4 | 3+4 4 | 3+4 4 |
| | 6 2 | 4+4 4 | 4+4 4 | 3+4 4 | 5+5 4 | 3+4 4 | 3+4 4 | 3+4 4 |
| | 5 2 | 4+4 4 | 3+3 4 | 3+4 4 | 6+6 4 | 3+4 4 | 3+4 4 | 3+4 4 |
| | 6 2 | 3+3 4 | 4+4 4 | 3+4 4 | 5+5 4 | 3+4 4 | 3+4 4 | 3+4 4 |
| | 6 2 | 3+3 4 | 3+3 4 | 3+4 4 | 5+5 4 | 3+4 4 | 3+4 4 | 3+4 4 |
| | 5 2 | 4+4 4 | 4+4 4 | 3+4 4 | 5+5 4 | 3+4 4 | 3+4 4 | 3+4 4 |
| | 6 2 | 3+3 4 | 3+3 4 | 3+4 4 | 6+6 4 | 3+4 4 | 3+4 4 | 3+4 4 |
| | 5 2 | 4+4 4 | 4+4 4 | 3+4 4 | 6+6 4 | 3+4 4 | 3+4 4 | 3+4 4 |
| | 5 2 | 4+4 4 | 4+4 4 | 3+4 4 | 6+6 4 | 3+4 4 | 3+4 4 | 3+4 4 |
| | 6 2 | 3+3 4 | 3+3 4 | 3+4 4 | 5+5 4 | 3+4 4 | 3+4 4 | 3+4 4 |
| | 5 2 | 4+4 4 | 3+3 4 | 3+4 4 | 6+6 4 | 3+4 4 | 3+4 4 | 3+4 4 |
| | 6 2 | 3+3 4 | 4+4 4 | 3+4 4 | 5+5 4 | 3+4 4 | 3+4 4 | 3+4 4 |
| | 6 2 | 3+3 4 | 4+4 4 | 3+4 4 | 6+6 4 | 3+4 4 | 3+4 4 | 3+4 4 |
| | 6 2 | 4+4 4 | 3+3 4 | 3+4 4 | 5+5 4 | 3+4 4 | 3+4 4 | 3+4 4 |
| | 5 2 | 3+3 4 | 4+4 4 | 3+4 4 | 6+6 4 | 3+4 4 | 3+4 4 | 3+4 4 |
| | 5 2 | 4+4 4 | 3+3 4 | 3+4 4 | 6+6 4 | 3+4 4 | 3+4 4 | 3+4 4 |
| | 6 2 | 3+3 4 | 3+3 4 | 3+4 4 | 5+5 4 | 3+4 4 | 3+4 4 | 3+4 4 |
| | 5 2 | 4+4 4 | 4+4 4 | 3+4 4 | 5+5 4 | 3+4 4 | 3+4 4 | 3+4 4 |
| | 5 2 | | | | | | | |
| | 6 2 | | | | | | | |
| | 5 2 | | | | | | | |
| | 5 2 | | | | | | | |
| | 6 2 | | | | | | | |
| | 6 2 | | | | | | | |
| | 5 2 | | | | | | | |
| | 6 2 | | | | | | | |
| | 5 2 | | | | | | | |
| | 5 2 | | | | | | | |
| | 6 2 | | | | | | | |
| | 6 2 | | | | | | | |
| | 5 2 | | | | | | | |
| | 5 2 | | | | | | | |
| | 5 2 | | | | | | | |
| | 6 2 | | | | | | | |
| | 5 2 | | | | | | | |
| | 6 2 | | | | | | | |
| 62.2.101 | 413387 | 328 | Ge1O1Yb3 | | | | | |
| | -GM | -R | -S | -T | -U | -X | -Y | -Z |
| | 5 2 | 3+3 4 | 3+3 4 | 3+4 4 | 5+5 4 | 3+4 4 | 3+4 4 | 3+4 4 |
| | 6 2 | 4+4 4 | 4+4 4 | 3+4 4 | 6+6 4 | 3+4 4 | 3+4 4 | 3+4 4 |

Continued on next column

**TABLE XXXVII– continued**

| TQC | ICSD | NoE | SF | | | | | |
|---|---|---|---|---|---|---|---|---|
| | 6 2 | 3+3 4 | 3+3 4 | 3+4 4 | 5+5 4 | 3+4 4 | 3+4 4 | 3+4 4 |
| | 5 2 | 4+4 4 | 4+4 4 | 3+4 4 | 6+6 4 | 3+4 4 | 3+4 4 | 3+4 4 |
| | 5 2 | 3+3 4 | 3+3 4 | 3+4 4 | 6+6 4 | 3+4 4 | 3+4 4 | 3+4 4 |
| | 6 2 | 4+4 4 | 4+4 4 | 3+4 4 | 5+5 4 | 3+4 4 | 3+4 4 | 3+4 4 |
| | 6 2 | 3+3 4 | 3+3 4 | 3+4 4 | 6+6 4 | 3+4 4 | 3+4 4 | 3+4 4 |
| | 5 2 | 4+4 4 | 4+4 4 | 3+4 4 | 5+5 4 | 3+4 4 | 3+4 4 | 3+4 4 |
| | 6 2 | 4+4 4 | 3+3 4 | 3+4 4 | 5+5 4 | 3+4 4 | 3+4 4 | 3+4 4 |
| | 5 2 | 3+3 4 | 4+4 4 | 3+4 4 | 6+6 4 | 3+4 4 | 3+4 4 | 3+4 4 |
| | 6 2 | 3+3 4 | 4+4 4 | 3+4 4 | 5+5 4 | 3+4 4 | 3+4 4 | 3+4 4 |
| | 5 2 | 4+4 4 | 3+3 4 | 3+4 4 | 6+6 4 | 3+4 4 | 3+4 4 | 3+4 4 |
| | 5 2 | 4+4 4 | 3+3 4 | 3+4 4 | 5+5 4 | 3+4 4 | 3+4 4 | 3+4 4 |
| | 6 2 | 3+3 4 | 4+4 4 | 3+4 4 | 5+5 4 | 3+4 4 | 3+4 4 | 3+4 4 |
| | 6 2 | 4+4 4 | 4+4 4 | 3+4 4 | 6+6 4 | 3+4 4 | 3+4 4 | 3+4 4 |
| | 5 2 | 3+3 4 | 3+3 4 | 3+4 4 | 5+5 4 | 3+4 4 | 3+4 4 | 3+4 4 |
| | 6 2 | 3+3 4 | 3+3 4 | 3+4 4 | 6+6 4 | 3+4 4 | 3+4 4 | 3+4 4 |
| | 5 2 | 4+4 4 | 4+4 4 | 3+4 4 | 6+6 4 | 3+4 4 | 3+4 4 | 3+4 4 |
| | 5 2 | 3+3 4 | 3+3 4 | 3+4 4 | 6+6 4 | 3+4 4 | 3+4 4 | 3+4 4 |
| | 5 2 | 4+4 4 | 4+4 4 | 3+4 4 | 5+5 4 | 3+4 4 | 3+4 4 | 3+4 4 |
| | 6 2 | 3+3 4 | 3+3 4 | 3+4 4 | 6+6 4 | 3+4 4 | 3+4 4 | 3+4 4 |
| | 6 2 | 4+4 4 | 4+4 4 | 3+4 4 | 5+5 4 | 3+4 4 | 3+4 4 | 3+4 4 |
| | 5 2 | 4+4 4 | 4+4 4 | 3+4 4 | 5+5 4 | 3+4 4 | 3+4 4 | 3+4 4 |
| | 6 2 | 3+3 4 | 3+3 4 | 3+4 4 | 6+6 4 | 3+4 4 | 3+4 4 | 3+4 4 |
| | 5 2 | 3+3 4 | 3+3 4 | 3+4 4 | 5+5 4 | 3+4 4 | 3+4 4 | 3+4 4 |
| | 5 2 | 4+4 4 | 4+4 4 | 3+4 4 | 5+5 4 | 3+4 4 | 3+4 4 | 3+4 4 |
| | 6 2 | 3+3 4 | 4+4 4 | 3+4 4 | 6+6 4 | 3+4 4 | 3+4 4 | 3+4 4 |
| | 5 2 | 4+4 4 | 3+3 4 | 3+4 4 | 5+5 4 | 3+4 4 | 3+4 4 | 3+4 4 |
| | 6 2 | 4+4 4 | 4+4 4 | 3+4 4 | 6+6 4 | 3+4 4 | 3+4 4 | 3+4 4 |
| | 5 2 | 3+3 4 | 3+3 4 | 3+4 4 | 6+6 4 | 3+4 4 | 3+4 4 | 3+4 4 |
| | 6 2 | 4+4 4 | 4+4 4 | 3+4 4 | 6+6 4 | 3+4 4 | 3+4 4 | 3+4 4 |
| | 5 2 | 3+3 4 | 3+3 4 | 3+4 4 | 6+6 4 | 3+4 4 | 3+4 4 | 3+4 4 |
| | 6 2 | 3+3 4 | 3+3 4 | 3+4 4 | 6+6 4 | 3+4 4 | 3+4 4 | 3+4 4 |
| | 6 2 | 4+4 4 | 4+4 4 | 3+4 4 | 6+6 4 | 3+4 4 | 3+4 4 | 3+4 4 |
| | 6 2 | 4+4 4 | 4+4 4 | 3+4 4 | 5+5 4 | 3+4 4 | 3+4 4 | 3+4 4 |
| | 5 2 | 3+3 4 | 3+3 4 | 3+4 4 | 6+6 4 | 3+4 4 | 3+4 4 | 3+4 4 |
| | 6 2 | 4+4 4 | 4+4 4 | 3+4 4 | 6+6 4 | 3+4 4 | 3+4 4 | 3+4 4 |
| | 5 2 | 3+3 4 | 3+3 4 | 3+4 4 | 5+5 4 | 3+4 4 | 3+4 4 | 3+4 4 |
| | 5 2 | 3+3 4 | 3+3 4 | 3+4 4 | 6+6 4 | 3+4 4 | 3+4 4 | 3+4 4 |
| | 6 2 | 4+4 4 | 4+4 4 | 3+4 4 | 5+5 4 | 3+4 4 | 3+4 4 | 3+4 4 |
| | 5 2 | 4+4 4 | 3+3 4 | 3+4 4 | 6+6 4 | 3+4 4 | 3+4 4 | 3+4 4 |
| | 6 2 | 3+3 4 | 4+4 4 | 3+4 4 | 6+6 4 | 3+4 4 | 3+4 4 | 3+4 4 |
| | 6 2 | 4+4 4 | 3+3 4 | 3+4 4 | 5+5 4 | 3+4 4 | 3+4 4 | 3+4 4 |
| | 5 2 | 3+3 4 | 4+4 4 | 3+4 4 | 5+5 4 | 3+4 4 | 3+4 4 | 3+4 4 |
| | 5 2 | 4+4 4 | 3+3 4 | 3+4 4 | 6+6 4 | 3+4 4 | 3+4 4 | 3+4 4 |
| | 6 2 | 3+3 4 | 4+4 4 | 3+4 4 | 5+5 4 | 3+4 4 | 3+4 4 | 3+4 4 |
| | 6 2 | 4+4 4 | 3+3 4 | 3+4 4 | 5+5 4 | 3+4 4 | 3+4 4 | 3+4 4 |
| | 5 2 | 3+3 4 | 4+4 4 | 3+4 4 | 5+5 4 | 3+4 4 | 3+4 4 | 3+4 4 |
| | 5 2 | 4+4 4 | 3+3 4 | 3+4 4 | 6+6 4 | 3+4 4 | 3+4 4 | 3+4 4 |
| | 5 2 | 3+3 4 | 4+4 4 | 3+4 4 | 5+5 4 | 3+4 4 | 3+4 4 | 3+4 4 |
| | 5 2 | 4+4 4 | 3+3 4 | 3+4 4 | 6+6 4 | 3+4 4 | 3+4 4 | 3+4 4 |
| | 5 2 | 3+3 4 | 4+4 4 | 3+4 4 | 6+6 4 | 3+4 4 | 3+4 4 | 3+4 4 |
| | 5 2 | 3+3 4 | 3+3 4 | 3+4 4 | 6+6 4 | 3+4 4 | 3+4 4 | 3+4 4 |
| | 6 2 | 4+4 4 | 4+4 4 | 3+4 4 | 5+5 4 | 3+4 4 | 3+4 4 | 3+4 4 |
| | 6 2 | 3+3 4 | 4+4 4 | 3+4 4 | 6+6 4 | 3+4 4 | 3+4 4 | 3+4 4 |
| | 5 2 | 4+4 4 | 3+3 4 | 3+4 4 | 5+5 4 | 3+4 4 | 3+4 4 | 3+4 4 |
| | 6 2 | 3+3 4 | 4+4 4 | 3+4 4 | 6+6 4 | 3+4 4 | 3+4 4 | 3+4 4 |
| | 6 2 | 4+4 4 | 3+3 4 | 3+4 4 | 5+5 4 | 3+4 4 | 3+4 4 | 3+4 4 |
| | 6 2 | 4+4 4 | 3+3 4 | 3+4 4 | 5+5 4 | 3+4 4 | 3+4 4 | 3+4 4 |
| | 6 2 | 3+3 4 | 4+4 4 | 3+4 4 | 6+6 4 | 3+4 4 | 3+4 4 | 3+4 4 |
| | 6 2 | 3+3 4 | 3+3 4 | 3+4 4 | 6+6 4 | 3+4 4 | 3+4 4 | 3+4 4 |
| | 6 2 | 4+4 4 | 4+4 4 | 3+4 4 | 5+5 4 | 3+4 4 | 3+4 4 | 3+4 4 |
| | 6 2 | 3+3 4 | 4+4 4 | 3+4 4 | 5+5 4 | 3+4 4 | 3+4 4 | 3+4 4 |

Continued on next column

**TABLE XXXVII– continued**

| TQC | ICSD | NoE | SF | | | | | |
|---|---|---|---|---|---|---|---|---|
| | 6 2 | 4+4 4 | 3+3 4 | 3+4 4 | 5+5 4 | 3+4 4 | 3+4 4 | 3+4 4 |
| | 6 2 | 4+4 4 | 4+4 4 | 3+4 4 | 6+6 4 | 3+4 4 | 3+4 4 | 3+4 4 |
| | 6 2 | 3+3 4 | 3+3 4 | 3+4 4 | 5+5 4 | 3+4 4 | 3+4 4 | 3+4 4 |
| | 6 2 | 3+3 4 | 3+3 4 | 3+4 4 | 6+6 4 | 3+4 4 | 3+4 4 | 3+4 4 |
| | 6 2 | 4+4 4 | 4+4 4 | 3+4 4 | 5+5 4 | 3+4 4 | 3+4 4 | 3+4 4 |
| | 5 2 | 3+3 4 | 3+3 4 | 3+4 4 | 6+6 4 | 3+4 4 | 3+4 4 | 3+4 4 |
| | 6 2 | 4+4 4 | 4+4 4 | 3+4 4 | 5+5 4 | 3+4 4 | 3+4 4 | 3+4 4 |
| | 6 2 | 4+4 4 | 3+3 4 | 3+4 4 | 5+5 4 | 3+4 4 | 3+4 4 | 3+4 4 |
| | 5 2 | 3+3 4 | 4+4 4 | 3+4 4 | 5+5 4 | 3+4 4 | 3+4 4 | 3+4 4 |
| | 5 2 | 4+4 4 | 3+3 4 | 3+4 4 | 6+6 4 | 3+4 4 | 3+4 4 | 3+4 4 |
| | 6 2 | 3+3 4 | 4+4 4 | 3+4 4 | 6+6 4 | 3+4 4 | 3+4 4 | 3+4 4 |
| | 5 2 | 3+3 4 | 3+3 4 | 3+4 4 | 6+6 4 | 3+4 4 | 3+4 4 | 3+4 4 |
| | 5 2 | 4+4 4 | 4+4 4 | 3+4 4 | 6+6 4 | 3+4 4 | 3+4 4 | 3+4 4 |
| | 6 2 | 3+3 4 | 3+3 4 | 3+4 4 | 5+5 4 | 3+4 4 | 3+4 4 | 3+4 4 |
| | 6 2 | 4+4 4 | 4+4 4 | 3+4 4 | 6+6 4 | 3+4 4 | 3+4 4 | 3+4 4 |
| | 6 2 | 3+3 4 | 4+4 4 | 3+4 4 | 5+5 4 | 3+4 4 | 3+4 4 | 3+4 4 |
| | 5 2 | 4+4 4 | 3+3 4 | 3+4 4 | 6+6 4 | 3+4 4 | 3+4 4 | 3+4 4 |
| | 5 2 | 3+3 4 | 3+3 4 | 3+4 4 | 6+6 4 | 3+4 4 | 3+4 4 | 3+4 4 |
| | 5 2 | 4+4 4 | 4+4 4 | 3+4 4 | 5+5 4 | 3+4 4 | 3+4 4 | 3+4 4 |
| | 6 2 | | | | | | | |
| | 6 2 | | | | | | | |
| | 5 2 | | | | | | | |
| | 6 2 | | | | | | | |
| | 5 2 | | | | | | | |
| | 5 2 | | | | | | | |
| | 5 2 | | | | | | | |
| | 5 2 | | | | | | | |
| | 6 2 | | | | | | | |
| | 6 2 | | | | | | | |
| | 6 2 | | | | | | | |
| | 5 2 | | | | | | | |
| | 6 2 | | | | | | | |
| | 5 2 | | | | | | | |
| | 6 2 | | | | | | | |
| | 5 2 | | | | | | | |
| | 5 2 | | | | | | | |
| | 6 2 | | | | | | | |
| | 6 2 | | | | | | | |
| | 6 2 | | | | | | | |
| | 6 2 | | | | | | | |
| | 5 2 | | | | | | | |
| | 5 2 | | | | | | | |
| | 5 2 | | | | | | | |
| | 6 2 | | | | | | | |
| | 5 2 | | | | | | | |
| | 6 2 | | | | | | | |
| | 5 2 | | | | | | | |
| | 5 2 | | | | | | | |
| | 6 2 | | | | | | | |
| | 5 2 | | | | | | | |
| | 6 2 | | | | | | | |
| | 6 2 | | | | | | | |
| | 6 2 | | | | | | | |
| | 5 2 | | | | | | | |
| | 5 2 | | | | | | | |
| | 5 2 | | | | | | | |
| | 6 2 | | | | | | | |
| | 5 2 | | | | | | | |
| | 6 2 | | | | | | | |
| | 5 2 | | | | | | | |
| | 6 2 | | | | | | | |

Continued on next column

**TABLE XXXVII– continued**

| TQC | ICSD | NoE | SF | | | | | |
|---|---|---|---|---|---|---|---|---|
| | 5 2 | | | | | | | |
| | 5 2 | | | | | | | |
| | 5 2 | | | | | | | |
| | 6 2 | | | | | | | |
| | 6 2 | | | | | | | |
| | 5 2 | | | | | | | |
| | 6 2 | | | | | | | |
| | 5 2 | | | | | | | |
| | 5 2 | | | | | | | |
| | 6 2 | | | | | | | |
| | 5 2 | | | | | | | |
| | 6 2 | | | | | | | |
| | 5 2 | | | | | | | |
| | 6 2 | | | | | | | |
| | 5 2 | | | | | | | |
| | 6 2 | | | | | | | |
| | 5 2 | | | | | | | |
| | 6 2 | | | | | | | |
| | 5 2 | | | | | | | |
| | 6 2 | | | | | | | |
| | 5 2 | | | | | | | |
| | 6 2 | | | | | | | |
| | 5 2 | | | | | | | |
| | 6 2 | | | | | | | |
| | 5 2 | | | | | | | |
| | 5 2 | | | | | | | |
| | 6 2 | | | | | | | |
| | 6 2 | | | | | | | |
| | 5 2 | | | | | | | |
| | 6 2 | | | | | | | |
| | 5 2 | | | | | | | |
| | 6 2 | | | | | | | |
| | 5 2 | | | | | | | |
| | 5 2 | | | | | | | |
| | 6 2 | | | | | | | |
| | 6 2 | | | | | | | |
| | 5 2 | | | | | | | |
| | 6 2 | | | | | | | |
| | 6 2 | | | | | | | |
| | 6 2 | | | | | | | |
| 62.2.102 | 237302 | 64 | Ge1Mg1Ni1 | | | | | |
| | -GM | -R | -S | -T | -U | -X | -Y | -Z |
| | 5 2 | 3+3 4 | 4+4 4 | 3+4 4 | 6+6 4 | 3+4 4 | 3+4 4 | 3+4 4 |
| | 5 2 | 4+4 4 | 3+3 4 | 3+4 4 | 5+5 4 | 3+4 4 | 3+4 4 | 3+4 4 |
| | 6 2 | 4+4 4 | 3+3 4 | 3+4 4 | 5+5 4 | 3+4 4 | 3+4 4 | 3+4 4 |
| | 6 2 | 3+3 4 | 4+4 4 | 3+4 4 | 6+6 4 | 3+4 4 | 3+4 4 | 3+4 4 |
| | 5 2 | 4+4 4 | 4+4 4 | 3+4 4 | 5+5 4 | 3+4 4 | 3+4 4 | 3+4 4 |
| | 6 2 | 3+3 4 | 3+3 4 | 3+4 4 | 5+5 4 | 3+4 4 | 3+4 4 | 3+4 4 |
| | 5 2 | 3+3 4 | 3+3 4 | 3+4 4 | 6+6 4 | 3+4 4 | 3+4 4 | 3+4 4 |
| | 6 2 | 4+4 4 | 4+4 4 | 3+4 4 | 6+6 4 | 3+4 4 | 3+4 4 | 3+4 4 |
| | 5 2 | 3+3 4 | 4+4 4 | 3+4 4 | 6+6 4 | 3+4 4 | 3+4 4 | 3+4 4 |
| | 6 2 | 4+4 4 | 3+3 4 | 3+4 4 | 5+5 4 | 3+4 4 | 3+4 4 | 3+4 4 |
| | 6 2 | 4+4 4 | 4+4 4 | 3+4 4 | 6+6 4 | 3+4 4 | 3+4 4 | 3+4 4 |
| | 5 2 | 3+3 4 | 3+3 4 | 3+4 4 | 5+5 4 | 3+4 4 | 3+4 4 | 3+4 4 |
| | 6 2 | 3+3 4 | 3+3 4 | 3+4 4 | 6+6 4 | 3+4 4 | 3+4 4 | 3+4 4 |
| | 6 2 | 4+4 4 | 4+4 4 | 3+4 4 | 5+5 4 | 3+4 4 | 3+4 4 | 3+4 4 |
| | 5 2 | 4+4 4 | 3+3 4 | 3+4 4 | 6+6 4 | 3+4 4 | 3+4 4 | 3+4 4 |
| | 5 2 | 3+3 4 | 4+4 4 | 3+4 4 | 6+6 4 | 3+4 4 | 3+4 4 | 3+4 4 |
| | 5 2 | | | | | | | |
| | 5 2 | | | | | | | |
| | 6 2 | | | | | | | |

Continued on next column

**TABLE XXXVII– continued**

| TQC | ICSD | NoE | SF | | | | | |
|---|---|---|---|---|---|---|---|---|
| | 5 2 | | | | | | | |
| | 5 2 | | | | | | | |
| | 6 2 | | | | | | | |
| | 5 2 | | | | | | | |
| | 6 2 | | | | | | | |
| | 5 2 | | | | | | | |
| | 6 2 | | | | | | | |
| | 6 2 | | | | | | | |
| | 5 2 | | | | | | | |
| | 5 2 | | | | | | | |
| | 6 2 | | | | | | | |
| | 6 2 | | | | | | | |
| | 5 2 | | | | | | | |
| 62.2.103 | 428468 | 64 | Ge1Ir1Sc1 | | | | | |
| | -GM | -R | -S | -T | -U | -X | -Y | -Z |
| | 5 2 | 4+4 4 | 4+4 4 | 3+4 4 | 6+6 4 | 3+4 4 | 3+4 4 | 3+4 4 |
| | 5 2 | 3+3 4 | 3+3 4 | 3+4 4 | 5+5 4 | 3+4 4 | 3+4 4 | 3+4 4 |
| | 6 2 | 3+3 4 | 4+4 4 | 3+4 4 | 6+6 4 | 3+4 4 | 3+4 4 | 3+4 4 |
| | 6 2 | 4+4 4 | 3+3 4 | 3+4 4 | 5+5 4 | 3+4 4 | 3+4 4 | 3+4 4 |
| | 6 2 | 4+4 4 | 3+3 4 | 3+4 4 | 5+5 4 | 3+4 4 | 3+4 4 | 3+4 4 |
| | 5 2 | 3+3 4 | 4+4 4 | 3+4 4 | 5+5 4 | 3+4 4 | 3+4 4 | 3+4 4 |
| | 6 2 | 4+4 4 | 3+3 4 | 3+4 4 | 6+6 4 | 3+4 4 | 3+4 4 | 3+4 4 |
| | 5 2 | 3+3 4 | 4+4 4 | 3+4 4 | 6+6 4 | 3+4 4 | 3+4 4 | 3+4 4 |
| | 5 2 | 3+3 4 | 4+4 4 | 3+4 4 | 6+6 4 | 3+4 4 | 3+4 4 | 3+4 4 |
| | 6 2 | 4+4 4 | 3+3 4 | 3+4 4 | 5+5 4 | 3+4 4 | 3+4 4 | 3+4 4 |
| | 5 2 | 3+3 4 | 4+4 4 | 3+4 4 | 6+6 4 | 3+4 4 | 3+4 4 | 3+4 4 |
| | 6 2 | 4+4 4 | 3+3 4 | 3+4 4 | 5+5 4 | 3+4 4 | 3+4 4 | 3+4 4 |
| | 6 2 | 4+4 4 | 3+3 4 | 3+4 4 | 5+5 4 | 3+4 4 | 3+4 4 | 3+4 4 |
| | 6 2 | 3+3 4 | 4+4 4 | 3+4 4 | 6+6 4 | 3+4 4 | 3+4 4 | 3+4 4 |
| | 5 2 | 4+4 4 | 3+3 4 | 3+4 4 | 6+6 4 | 3+4 4 | 3+4 4 | 3+4 4 |
| | 5 2 | 3+3 4 | 4+4 4 | 3+4 4 | 6+6 4 | 3+4 4 | 3+4 4 | 3+4 4 |
| | 6 2 | | | | | | | |
| | 5 2 | | | | | | | |
| | 5 2 | | | | | | | |
| | 6 2 | | | | | | | |
| | 5 2 | | | | | | | |
| | 5 2 | | | | | | | |
| | 6 2 | | | | | | | |
| | 6 2 | | | | | | | |
| | 5 2 | | | | | | | |
| | 5 2 | | | | | | | |
| | 6 2 | | | | | | | |
| | 5 2 | | | | | | | |
| | 5 2 | | | | | | | |
| | 6 2 | | | | | | | |
| | 5 2 | | | | | | | |
| | 6 2 | | | | | | | |
| 62.2.104 | 411883 | 104 | Ge1Ir1Nb1 | | | | | |
| | -GM | -R | -S | -T | -U | -X | -Y | -Z |
| | 5 2 | 3+3 4 | 3+3 4 | 3+4 4 | 5+5 4 | 3+4 4 | 3+4 4 | 3+4 4 |
| | 5 2 | 4+4 4 | 4+4 4 | 3+4 4 | 6+6 4 | 3+4 4 | 3+4 4 | 3+4 4 |
| | 6 2 | 4+4 4 | 3+3 4 | 3+4 4 | 5+5 4 | 3+4 4 | 3+4 4 | 3+4 4 |
| | 6 2 | 3+3 4 | 4+4 4 | 3+4 4 | 6+6 4 | 3+4 4 | 3+4 4 | 3+4 4 |
| | 5 2 | 3+3 4 | 4+4 4 | 3+4 4 | 5+5 4 | 3+4 4 | 3+4 4 | 3+4 4 |
| | 5 2 | 4+4 4 | 3+3 4 | 3+4 4 | 6+6 4 | 3+4 4 | 3+4 4 | 3+4 4 |
| | 6 2 | 4+4 4 | 3+3 4 | 3+4 4 | 5+5 4 | 3+4 4 | 3+4 4 | 3+4 4 |
| | 6 2 | 3+3 4 | 4+4 4 | 3+4 4 | 6+6 4 | 3+4 4 | 3+4 4 | 3+4 4 |
| | 5 2 | 3+3 4 | 3+3 4 | 3+4 4 | 6+6 4 | 3+4 4 | 3+4 4 | 3+4 4 |
| | 6 2 | 4+4 4 | 4+4 4 | 3+4 4 | 5+5 4 | 3+4 4 | 3+4 4 | 3+4 4 |
| | 5 2 | 4+4 4 | 3+3 4 | 3+4 4 | 6+6 4 | 3+4 4 | 3+4 4 | 3+4 4 |
| | 5 2 | 3+3 4 | 4+4 4 | 3+4 4 | 5+5 4 | 3+4 4 | 3+4 4 | 3+4 4 |
| | 6 2 | 3+3 4 | 4+4 4 | 3+4 4 | 5+5 4 | 3+4 4 | 3+4 4 | 3+4 4 |
| | 5 2 | 4+4 4 | 3+3 4 | 3+4 4 | 5+5 4 | 3+4 4 | 3+4 4 | 3+4 4 |
| | 6 2 | 3+3 4 | 4+4 4 | 3+4 4 | 6+6 4 | 3+4 4 | 3+4 4 | 3+4 4 |
| | 6 2 | 4+4 4 | 3+3 4 | 3+4 4 | 6+6 4 | 3+4 4 | 3+4 4 | 3+4 4 |
| | 5 2 | 4+4 4 | 4+4 4 | 3+4 4 | 6+6 4 | 3+4 4 | 3+4 4 | 3+4 4 |
| | 6 2 | 3+3 4 | 3+3 4 | 3+4 4 | 5+5 4 | 3+4 4 | 3+4 4 | 3+4 4 |
| | 5 2 | 4+4 4 | 3+3 4 | 3+4 4 | 6+6 4 | 3+4 4 | 3+4 4 | 3+4 4 |
| | 6 2 | 3+3 4 | 4+4 4 | 3+4 4 | 5+5 4 | 3+4 4 | 3+4 4 | 3+4 4 |
| | 5 2 | 4+4 4 | 3+3 4 | 3+4 4 | 5+5 4 | 3+4 4 | 3+4 4 | 3+4 4 |
| | 6 2 | 3+3 4 | 4+4 4 | 3+4 4 | 6+6 4 | 3+4 4 | 3+4 4 | 3+4 4 |
| | 5 2 | 3+3 4 | 4+4 4 | 3+4 4 | 6+6 4 | 3+4 4 | 3+4 4 | 3+4 4 |
| | 6 2 | 4+4 4 | 3+3 4 | 3+4 4 | 6+6 4 | 3+4 4 | 3+4 4 | 3+4 4 |
| | 5 2 | 3+3 4 | 4+4 4 | 3+4 4 | 5+5 4 | 3+4 4 | 3+4 4 | 3+4 4 |
| | 6 2 | 4+4 4 | 3+3 4 | 3+4 4 | 5+5 4 | 3+4 4 | 3+4 4 | 3+4 4 |
| | 5 2 | | | | | | | |
| | 6 2 | | | | | | | |
| | 6 2 | | | | | | | |
| | 6 2 | | | | | | | |
| | 5 2 | | | | | | | |
| | 6 2 | | | | | | | |
| | 5 2 | | | | | | | |
| | 6 2 | | | | | | | |
| | 5 2 | | | | | | | |
| | 6 2 | | | | | | | |
| | 5 2 | | | | | | | |
| | 5 2 | | | | | | | |
| | 5 2 | | | | | | | |
| | 6 2 | | | | | | | |
| | 5 2 | | | | | | | |
| | 6 2 | | | | | | | |
| | 5 2 | | | | | | | |
| | 6 2 | | | | | | | |
| | 5 2 | | | | | | | |
| | 5 2 | | | | | | | |
| | 5 2 | | | | | | | |
| | 6 2 | | | | | | | |
| | 5 2 | | | | | | | |
| | 5 2 | | | | | | | |
| | 6 2 | | | | | | | |
| | 5 2 | | | | | | | |
| 62.2.105 | 15534 | 288 | Gd3Ni1Si2 | | | | | |
| | -GM | -R | -S | -T | -U | -X | -Y | -Z |
| | 5 2 | 3+3 4 | 3+3 4 | 3+4 4 | 5+5 4 | 3+4 4 | 3+4 4 | 3+4 4 |
| | 5 2 | 4+4 4 | 4+4 4 | 3+4 4 | 6+6 4 | 3+4 4 | 3+4 4 | 3+4 4 |
| | 6 2 | 3+3 4 | 3+3 4 | 3+4 4 | 5+5 4 | 3+4 4 | 3+4 4 | 3+4 4 |
| | 6 2 | 4+4 4 | 4+4 4 | 3+4 4 | 6+6 4 | 3+4 4 | 3+4 4 | 3+4 4 |
| | 5 2 | 3+3 4 | 3+3 4 | 3+4 4 | 6+6 4 | 3+4 4 | 3+4 4 | 3+4 4 |
| | 5 2 | 4+4 4 | 4+4 4 | 3+4 4 | 5+5 4 | 3+4 4 | 3+4 4 | 3+4 4 |
| | 6 2 | 3+3 4 | 3+3 4 | 3+4 4 | 5+5 4 | 3+4 4 | 3+4 4 | 3+4 4 |
| | 6 2 | 4+4 4 | 4+4 4 | 3+4 4 | 6+6 4 | 3+4 4 | 3+4 4 | 3+4 4 |
| | 5 2 | 3+3 4 | 3+3 4 | 3+4 4 | 5+5 4 | 3+4 4 | 3+4 4 | 3+4 4 |
| | 5 2 | 4+4 4 | 4+4 4 | 3+4 4 | 6+6 4 | 3+4 4 | 3+4 4 | 3+4 4 |
| | 6 2 | 3+3 4 | 4+4 4 | 3+4 4 | 6+6 4 | 3+4 4 | 3+4 4 | 3+4 4 |
| | 6 2 | 4+4 4 | 3+3 4 | 3+4 4 | 5+5 4 | 3+4 4 | 3+4 4 | 3+4 4 |
| | 5 2 | 4+4 4 | 3+3 4 | 3+4 4 | 6+6 4 | 3+4 4 | 3+4 4 | 3+4 4 |
| | 6 2 | 3+3 4 | 4+4 4 | 3+4 4 | 5+5 4 | 3+4 4 | 3+4 4 | 3+4 4 |
| | 5 2 | 4+4 4 | 3+3 4 | 3+4 4 | 5+5 4 | 3+4 4 | 3+4 4 | 3+4 4 |
| | 5 2 | 3+3 4 | 4+4 4 | 3+4 4 | 6+6 4 | 3+4 4 | 3+4 4 | 3+4 4 |
| | 6 2 | 4+4 4 | 3+3 4 | 3+4 4 | 5+5 4 | 3+4 4 | 3+4 4 | 3+4 4 |
| | 5 2 | 3+3 4 | 4+4 4 | 3+4 4 | 6+6 4 | 3+4 4 | 3+4 4 | 3+4 4 |
| | 6 2 | 3+3 4 | 3+3 4 | 3+4 4 | 6+6 4 | 3+4 4 | 3+4 4 | 3+4 4 |

Continued on next column

**TABLE XXXVII– continued**

| TQC | ICSD | NoE | SF |
|---|---|---|---|
| | 6 | 2 | 4+4 4 4+4 4 3+4 4 5+5 4 3+4 4 3+4 4 3+4 4 |
| | 5 | 2 | 3+3 4 4+4 4 3+4 4 5+5 4 3+4 4 3+4 4 3+4 4 |
| | 5 | 2 | 4+4 4 3+3 4 3+4 4 6+6 4 3+4 4 3+4 4 3+4 4 |
| | 6 | 2 | 3+3 4 4+4 4 3+4 4 6+6 4 3+4 4 3+4 4 3+4 4 |
| | 6 | 2 | 4+4 4 3+3 4 3+4 4 5+5 4 3+4 4 3+4 4 3+4 4 |
| | 5 | 2 | 4+4 4 3+3 4 3+4 4 6+6 4 3+4 4 3+4 4 3+4 4 |
| | 6 | 2 | 3+3 4 4+4 4 3+4 4 5+5 4 3+4 4 3+4 4 3+4 4 |
| | 5 | 2 | 3+3 4 3+3 4 3+4 4 5+5 4 3+4 4 3+4 4 3+4 4 |
| | 5 | 2 | 4+4 4 4+4 4 3+4 4 6+6 4 3+4 4 3+4 4 3+4 4 |
| | 6 | 2 | 3+3 4 3+3 4 3+4 4 5+5 4 3+4 4 3+4 4 3+4 4 |
| | 5 | 2 | 4+4 4 4+4 4 3+4 4 6+6 4 3+4 4 3+4 4 3+4 4 |
| | 6 | 2 | 4+4 4 3+3 4 3+4 4 5+5 4 3+4 4 3+4 4 3+4 4 |
| | 5 | 2 | 3+3 4 4+4 4 3+4 4 6+6 4 3+4 4 3+4 4 3+4 4 |
| | 6 | 2 | 3+3 4 4+4 4 3+4 4 6+6 4 3+4 4 3+4 4 3+4 4 |
| | 5 | 2 | 4+4 4 3+3 4 3+4 4 5+5 4 3+4 4 3+4 4 3+4 4 |
| | 6 | 2 | 3+3 4 4+4 4 3+4 4 5+5 4 3+4 4 3+4 4 3+4 4 |
| | 5 | 2 | 4+4 4 3+3 4 3+4 4 6+6 4 3+4 4 3+4 4 3+4 4 |
| | 6 | 2 | 4+4 4 4+4 4 3+4 4 6+6 4 3+4 4 3+4 4 3+4 4 |
| | 5 | 2 | 3+3 4 3+3 4 3+4 4 5+5 4 3+4 4 3+4 4 3+4 4 |
| | 5 | 2 | 4+4 4 3+3 4 3+4 4 6+6 4 3+4 4 3+4 4 3+4 4 |
| | 5 | 2 | 3+3 4 4+4 4 3+4 4 5+5 4 3+4 4 3+4 4 3+4 4 |
| | 6 | 2 | 4+4 4 3+3 4 3+4 4 6+6 4 3+4 4 3+4 4 3+4 4 |
| | 6 | 2 | 3+3 4 4+4 4 3+4 4 5+5 4 3+4 4 3+4 4 3+4 4 |
| | 5 | 2 | 3+3 4 4+4 4 3+4 4 5+5 4 3+4 4 3+4 4 3+4 4 |
| | 6 | 2 | 4+4 4 3+3 4 3+4 4 5+5 4 3+4 4 3+4 4 3+4 4 |
| | 6 | 2 | 3+3 4 4+4 4 3+4 4 6+6 4 3+4 4 3+4 4 3+4 4 |
| | 5 | 2 | 4+4 4 3+3 4 3+4 4 6+6 4 3+4 4 3+4 4 3+4 4 |
| | 6 | 2 | 3+3 4 4+4 4 3+4 4 5+5 4 3+4 4 3+4 4 3+4 4 |
| | 6 | 2 | 4+4 4 3+3 4 3+4 4 6+6 4 3+4 4 3+4 4 3+4 4 |
| | 5 | 2 | 4+4 4 3+3 4 3+4 4 6+6 4 3+4 4 3+4 4 3+4 4 |
| | 5 | 2 | 3+3 4 4+4 4 3+4 4 5+5 4 3+4 4 3+4 4 3+4 4 |
| | 6 | 2 | 4+4 4 3+3 4 3+4 4 5+5 4 3+4 4 3+4 4 3+4 4 |
| | 6 | 2 | 3+3 4 4+4 4 3+4 4 6+6 4 3+4 4 3+4 4 3+4 4 |
| | 5 | 2 | 4+4 4 3+3 4 3+4 4 5+5 4 3+4 4 3+4 4 3+4 4 |
| | 6 | 2 | 3+3 4 4+4 4 3+4 4 6+6 4 3+4 4 3+4 4 3+4 4 |
| | 6 | 2 | 4+4 4 3+3 4 3+4 4 5+5 4 3+4 4 3+4 4 3+4 4 |
| | 5 | 2 | 3+3 4 4+4 4 3+4 4 6+6 4 3+4 4 3+4 4 3+4 4 |
| | 6 | 2 | 3+3 4 4+4 4 3+4 4 6+6 4 3+4 4 3+4 4 3+4 4 |
| | 5 | 2 | 4+4 4 3+3 4 3+4 4 5+5 4 3+4 4 3+4 4 3+4 4 |
| | 5 | 2 | 4+4 4 4+4 4 3+4 4 6+6 4 3+4 4 3+4 4 3+4 4 |
| | 6 | 2 | 3+3 4 3+3 4 3+4 4 5+5 4 3+4 4 3+4 4 3+4 4 |
| | 5 | 2 | 4+4 4 3+3 4 3+4 4 5+5 4 3+4 4 3+4 4 3+4 4 |
| | 6 | 2 | 3+3 4 4+4 4 3+4 4 6+6 4 3+4 4 3+4 4 3+4 4 |
| | 6 | 2 | 4+4 4 4+4 4 3+4 4 5+5 4 3+4 4 3+4 4 3+4 4 |
| | 5 | 2 | 3+3 4 3+3 4 3+4 4 6+6 4 3+4 4 3+4 4 3+4 4 |
| | 5 | 2 | 3+3 4 3+3 4 3+4 4 5+5 4 3+4 4 3+4 4 3+4 4 |
| | 6 | 2 | 4+4 4 4+4 4 3+4 4 6+6 4 3+4 4 3+4 4 3+4 4 |
| | 6 | 2 | 4+4 4 3+3 4 3+4 4 6+6 4 3+4 4 3+4 4 3+4 4 |
| | 5 | 2 | 3+3 4 4+4 4 3+4 4 5+5 4 3+4 4 3+4 4 3+4 4 |
| | 6 | 2 | 3+3 4 3+3 4 3+4 4 5+5 4 3+4 4 3+4 4 3+4 4 |
| | 5 | 2 | 4+4 4 4+4 4 3+4 4 6+6 4 3+4 4 3+4 4 3+4 4 |
| | 6 | 2 | 3+3 4 3+3 4 3+4 4 5+5 4 3+4 4 3+4 4 3+4 4 |
| | 5 | 2 | 4+4 4 4+4 4 3+4 4 6+6 4 3+4 4 3+4 4 3+4 4 |
| | 5 | 2 | |
| | 5 | 2 | |
| | 6 | 2 | |
| | 6 | 2 | |
| | 6 | 2 | |
| | 5 | 2 | |
| | 6 | 2 | |
| | 5 | 2 | |

Continued on next column

**TABLE XXXVII– continued**

| TQC | ICSD | NoE | SF |
|---|---|---|---|
| | 6 | 2 | |
| | 5 | 2 | |
| | 5 | 2 | |
| | 6 | 2 | |
| | 6 | 2 | |
| | 6 | 2 | |
| | 6 | 2 | |
| | 6 | 2 | |
| | 5 | 2 | |
| | 5 | 2 | |
| | 5 | 2 | |
| | 6 | 2 | |
| | 5 | 2 | |
| | 5 | 2 | |
| | 6 | 2 | |
| | 5 | 2 | |
| | 6 | 2 | |
| | 5 | 2 | |
| | 6 | 2 | |
| | 5 | 2 | |
| | 6 | 2 | |
| | 5 | 2 | |
| | 6 | 2 | |
| | 5 | 2 | |
| | 6 | 2 | |
| | 5 | 2 | |
| | 6 | 2 | |
| | 5 | 2 | |
| | 6 | 2 | |
| | 5 | 2 | |
| | 6 | 2 | |
| | 6 | 2 | |
| | 5 | 2 | |
| | 5 | 2 | |
| | 6 | 2 | |
| | 5 | 2 | |
| | 6 | 2 | |
| | 6 | 2 | |
| | 5 | 2 | |
| | 6 | 2 | |
| | 5 | 2 | |
| | 6 | 2 | |
| | 5 | 2 | |
| | 6 | 2 | |
| | 5 | 2 | |
| | 6 | 2 | |
| | 5 | 2 | |
| | 5 | 2 | |
| | 6 | 2 | |
| | 6 | 2 | |
| | 5 | 2 | |
| | 5 | 2 | |
| | 6 | 2 | |
| | 6 | 2 | |
| | 5 | 2 | |
| | 6 | 2 | |
| | 5 | 2 | |
| | 6 | 2 | |
| | 5 | 2 | |
| | 6 | 2 | |
| | 5 | 2 | |

Continued on next column

**TABLE XXXVII– continued**

| TQC | ICSD | NoE | SF | | | | | |
|---|---|---|---|---|---|---|---|---|
| | 5 2 | | | | | | | |
| | 6 2 | | | | | | | |
| | 6 2 | | | | | | | |
| 62.2.106 | 280201 | 216 | Gd2O1Se2 | | | | | |
| | -GM | -R | -S | -T | -U | -X | -Y | -Z |
| | 5 2 | 3+3 4 | 3+3 4 | 3+4 4 | 6+6 4 | 3+4 4 | 3+4 4 | 3+4 4 |
| | 5 2 | 4+4 4 | 4+4 4 | 3+4 4 | 5+5 4 | 3+4 4 | 3+4 4 | 3+4 4 |
| | 6 2 | 3+3 4 | 3+3 4 | 3+4 4 | 6+6 4 | 3+4 4 | 3+4 4 | 3+4 4 |
| | 6 2 | 4+4 4 | 4+4 4 | 3+4 4 | 5+5 4 | 3+4 4 | 3+4 4 | 3+4 4 |
| | 5 2 | 4+4 4 | 3+3 4 | 3+4 4 | 6+6 4 | 3+4 4 | 3+4 4 | 3+4 4 |
| | 5 2 | 3+3 4 | 4+4 4 | 3+4 4 | 5+5 4 | 3+4 4 | 3+4 4 | 3+4 4 |
| | 6 2 | 4+4 4 | 4+4 4 | 3+4 4 | 6+6 4 | 3+4 4 | 3+4 4 | 3+4 4 |
| | 6 2 | 3+3 4 | 3+3 4 | 3+4 4 | 5+5 4 | 3+4 4 | 3+4 4 | 3+4 4 |
| | 5 2 | 3+3 4 | 3+3 4 | 3+4 4 | 6+6 4 | 3+4 4 | 3+4 4 | 3+4 4 |
| | 5 2 | 4+4 4 | 4+4 4 | 3+4 4 | 5+5 4 | 3+4 4 | 3+4 4 | 3+4 4 |
| | 6 2 | 3+3 4 | 3+3 4 | 3+4 4 | 6+6 4 | 3+4 4 | 3+4 4 | 3+4 4 |
| | 6 2 | 4+4 4 | 4+4 4 | 3+4 4 | 5+5 4 | 3+4 4 | 3+4 4 | 3+4 4 |
| | 5 2 | 4+4 4 | 3+3 4 | 3+4 4 | 6+6 4 | 3+4 4 | 3+4 4 | 3+4 4 |
| | 5 2 | 3+3 4 | 4+4 4 | 3+4 4 | 6+6 4 | 3+4 4 | 3+4 4 | 3+4 4 |
| | 6 2 | 3+3 4 | 3+3 4 | 3+4 4 | 5+5 4 | 3+4 4 | 3+4 4 | 3+4 4 |
| | 6 2 | 4+4 4 | 4+4 4 | 3+4 4 | 5+5 4 | 3+4 4 | 3+4 4 | 3+4 4 |
| | 5 2 | 3+3 4 | 3+3 4 | 3+4 4 | 5+5 4 | 3+4 4 | 3+4 4 | 3+4 4 |
| | 5 2 | 4+4 4 | 4+4 4 | 3+4 4 | 6+6 4 | 3+4 4 | 3+4 4 | 3+4 4 |
| | 6 2 | 3+3 4 | 3+3 4 | 3+4 4 | 5+5 4 | 3+4 4 | 3+4 4 | 3+4 4 |
| | 6 2 | 4+4 4 | 4+4 4 | 3+4 4 | 6+6 4 | 3+4 4 | 3+4 4 | 3+4 4 |
| | 5 2 | 4+4 4 | 4+4 4 | 3+4 4 | 5+5 4 | 3+4 4 | 3+4 4 | 3+4 4 |
| | 5 2 | 3+3 4 | 3+3 4 | 3+4 4 | 6+6 4 | 3+4 4 | 3+4 4 | 3+4 4 |
| | 6 2 | 3+3 4 | 3+3 4 | 3+4 4 | 5+5 4 | 3+4 4 | 3+4 4 | 3+4 4 |
| | 6 2 | 4+4 4 | 4+4 4 | 3+4 4 | 6+6 4 | 3+4 4 | 3+4 4 | 3+4 4 |
| | 5 2 | 4+4 4 | 3+3 4 | 3+4 4 | 6+6 4 | 3+4 4 | 3+4 4 | 3+4 4 |
| | 5 2 | 3+3 4 | 4+4 4 | 3+4 4 | 6+6 4 | 3+4 4 | 3+4 4 | 3+4 4 |
| | 5 2 | 3+3 4 | 3+3 4 | 3+4 4 | 5+5 4 | 3+4 4 | 3+4 4 | 3+4 4 |
| | 5 2 | 4+4 4 | 4+4 4 | 3+4 4 | 5+5 4 | 3+4 4 | 3+4 4 | 3+4 4 |
| | 6 2 | 3+3 4 | 4+4 4 | 3+4 4 | 6+6 4 | 3+4 4 | 3+4 4 | 3+4 4 |
| | 6 2 | 4+4 4 | 3+3 4 | 3+4 4 | 5+5 4 | 3+4 4 | 3+4 4 | 3+4 4 |
| | 6 2 | 4+4 4 | 3+3 4 | 3+4 4 | 6+6 4 | 3+4 4 | 3+4 4 | 3+4 4 |
| | 6 2 | 3+3 4 | 4+4 4 | 3+4 4 | 5+5 4 | 3+4 4 | 3+4 4 | 3+4 4 |
| | 6 2 | 3+3 4 | 3+3 4 | 3+4 4 | 5+5 4 | 3+4 4 | 3+4 4 | 3+4 4 |
| | 6 2 | 4+4 4 | 4+4 4 | 3+4 4 | 6+6 4 | 3+4 4 | 3+4 4 | 3+4 4 |
| | 5 2 | 3+3 4 | 4+4 4 | 3+4 4 | 5+5 4 | 3+4 4 | 3+4 4 | 3+4 4 |
| | 5 2 | 4+4 4 | 3+3 4 | 3+4 4 | 6+6 4 | 3+4 4 | 3+4 4 | 3+4 4 |
| | 5 2 | 3+3 4 | 3+3 4 | 3+4 4 | 5+5 4 | 3+4 4 | 3+4 4 | 3+4 4 |
| | 6 2 | 4+4 4 | 4+4 4 | 3+4 4 | 5+5 4 | 3+4 4 | 3+4 4 | 3+4 4 |
| | 5 2 | 4+4 4 | 3+3 4 | 3+4 4 | 6+6 4 | 3+4 4 | 3+4 4 | 3+4 4 |
| | 6 2 | 3+3 4 | 4+4 4 | 3+4 4 | 6+6 4 | 3+4 4 | 3+4 4 | 3+4 4 |
| | 5 2 | 4+4 4 | 3+3 4 | 3+4 4 | 6+6 4 | 3+4 4 | 3+4 4 | 3+4 4 |
| | 6 2 | 3+3 4 | 4+4 4 | 3+4 4 | 6+6 4 | 3+4 4 | 3+4 4 | 3+4 4 |
| | 5 2 | 4+4 4 | 3+3 4 | 3+4 4 | 5+5 4 | 3+4 4 | 3+4 4 | 3+4 4 |
| | 6 2 | 3+3 4 | 4+4 4 | 3+4 4 | 5+5 4 | 3+4 4 | 3+4 4 | 3+4 4 |
| | 6 2 | 3+3 4 | 4+4 4 | 3+4 4 | 5+5 4 | 3+4 4 | 3+4 4 | 3+4 4 |
| | 6 2 | 4+4 4 | 3+3 4 | 3+4 4 | 6+6 4 | 3+4 4 | 3+4 4 | 3+4 4 |
| | 5 2 | 3+3 4 | 3+3 4 | 3+4 4 | 6+6 4 | 3+4 4 | 3+4 4 | 3+4 4 |
| | 5 2 | 4+4 4 | 4+4 4 | 3+4 4 | 5+5 4 | 3+4 4 | 3+4 4 | 3+4 4 |
| | 6 2 | 3+3 4 | 3+3 4 | 3+4 4 | 5+5 4 | 3+4 4 | 3+4 4 | 3+4 4 |
| | 5 2 | 4+4 4 | 4+4 4 | 3+4 4 | 6+6 4 | 3+4 4 | 3+4 4 | 3+4 4 |
| | 5 2 | 3+3 4 | 3+3 4 | 3+4 4 | 5+5 4 | 3+4 4 | 3+4 4 | 3+4 4 |
| | 5 2 | 4+4 4 | 4+4 4 | 3+4 4 | 6+6 4 | 3+4 4 | 3+4 4 | 3+4 4 |
| | 5 2 | 3+3 4 | 4+4 4 | 3+4 4 | 6+6 4 | 3+4 4 | 3+4 4 | 3+4 4 |
| | 6 2 | 4+4 4 | 3+3 4 | 3+4 4 | 5+5 4 | 3+4 4 | 3+4 4 | 3+4 4 |
| | 6 2 | | | | | | | |
| | 6 2 | | | | | | | |
| | 5 2 | | | | | | | |
| | 5 2 | | | | | | | |
| | 6 2 | | | | | | | |
| | 5 2 | | | | | | | |
| | 5 2 | | | | | | | |
| | 6 2 | | | | | | | |
| | 5 2 | | | | | | | |
| | 6 2 | | | | | | | |
| | 6 2 | | | | | | | |
| | 5 2 | | | | | | | |
| | 5 2 | | | | | | | |
| | 5 2 | | | | | | | |
| | 6 2 | | | | | | | |
| | 6 2 | | | | | | | |
| | 5 2 | | | | | | | |
| | 5 2 | | | | | | | |
| | 6 2 | | | | | | | |
| | 6 2 | | | | | | | |
| | 5 2 | | | | | | | |
| | 5 2 | | | | | | | |
| | 6 2 | | | | | | | |
| | 6 2 | | | | | | | |
| | 6 2 | | | | | | | |
| | 6 2 | | | | | | | |
| | 5 2 | | | | | | | |
| | 5 2 | | | | | | | |
| | 5 2 | | | | | | | |
| | 5 2 | | | | | | | |
| | 6 2 | | | | | | | |
| | 6 2 | | | | | | | |
| | 6 2 | | | | | | | |
| | 5 2 | | | | | | | |
| | 5 2 | | | | | | | |
| | 6 2 | | | | | | | |
| | 5 2 | | | | | | | |
| | 6 2 | | | | | | | |
| | 5 2 | | | | | | | |
| | 6 2 | | | | | | | |
| | 5 2 | | | | | | | |
| | 6 2 | | | | | | | |
| | 6 2 | | | | | | | |
| | 5 2 | | | | | | | |
| | 6 2 | | | | | | | |
| | 5 2 | | | | | | | |
| | 5 2 | | | | | | | |
| | 6 2 | | | | | | | |
| | 6 2 | | | | | | | |
| | 6 2 | | | | | | | |
| | 5 2 | | | | | | | |
| | 5 2 | | | | | | | |
| | 5 2 | | | | | | | |
| | 6 2 | | | | | | | |
| 62.2.107 | 1375 | 232 | Gd1P1S1 | | | | | |
| | -GM | -R | -S | -T | -U | -X | -Y | -Z |
| | 5 2 | 3+3 4 | 3+3 4 | 3+4 4 | 5+5 4 | 3+4 4 | 3+4 4 | 3+4 4 |
| | 5 2 | 4+4 4 | 4+4 4 | 3+4 4 | 6+6 4 | 3+4 4 | 3+4 4 | 3+4 4 |
| | 6 2 | 3+3 4 | 3+3 4 | 3+4 4 | 5+5 4 | 3+4 4 | 3+4 4 | 3+4 4 |
| | 6 2 | 4+4 4 | 4+4 4 | 3+4 4 | 6+6 4 | 3+4 4 | 3+4 4 | 3+4 4 |
| | 5 2 | 3+3 4 | 3+3 4 | 3+4 4 | 5+5 4 | 3+4 4 | 3+4 4 | 3+4 4 |
| | 5 2 | 4+4 4 | 4+4 4 | 3+4 4 | 6+6 4 | 3+4 4 | 3+4 4 | 3+4 4 |
| | 6 2 | 3+3 4 | 3+3 4 | 3+4 4 | 6+6 4 | 3+4 4 | 3+4 4 | 3+4 4 |

Continued on next column

**TABLE XXXVII– continued**

| TQC | ICSD | NoE | SF |
|---|---|---|---|
| | 6 2 4+4 4 4+4 4 3+4 4 5+5 4 3+4 4 3+4 4 3+4 4 | | |
| | 5 2 3+3 4 3+3 4 3+4 4 5+5 4 3+4 4 3+4 4 3+4 4 | | |
| | 5 2 4+4 4 4+4 4 3+4 4 6+6 4 3+4 4 3+4 4 3+4 4 | | |
| | 6 2 3+3 4 3+3 4 3+4 4 6+6 4 3+4 4 3+4 4 3+4 4 | | |
| | 6 2 4+4 4 4+4 4 3+4 4 5+5 4 3+4 4 3+4 4 3+4 4 | | |
| | 5 2 3+3 4 3+3 4 3+4 4 5+5 4 3+4 4 3+4 4 3+4 4 | | |
| | 6 2 4+4 4 4+4 4 3+4 4 6+6 4 3+4 4 3+4 4 3+4 4 | | |
| | 5 2 4+4 4 3+3 4 3+4 4 6+6 4 3+4 4 3+4 4 3+4 4 | | |
| | 6 2 3+3 4 4+4 4 3+4 4 5+5 4 3+4 4 3+4 4 3+4 4 | | |
| | 5 2 3+3 4 3+3 4 3+4 4 6+6 4 3+4 4 3+4 4 3+4 4 | | |
| | 6 2 4+4 4 4+4 4 3+4 4 5+5 4 3+4 4 3+4 4 3+4 4 | | |
| | 5 2 4+4 4 3+3 4 3+4 4 6+6 4 3+4 4 3+4 4 3+4 4 | | |
| | 6 2 3+3 4 4+4 4 3+4 4 5+5 4 3+4 4 3+4 4 3+4 4 | | |
| | 5 2 3+3 4 4+4 4 3+4 4 6+6 4 3+4 4 3+4 4 3+4 4 | | |
| | 6 2 4+4 4 3+3 4 3+4 4 5+5 4 3+4 4 3+4 4 3+4 4 | | |
| | 5 2 3+3 4 3+3 4 3+4 4 6+6 4 3+4 4 3+4 4 3+4 4 | | |
| | 6 2 4+4 4 4+4 4 3+4 4 5+5 4 3+4 4 3+4 4 3+4 4 | | |
| | 5 2 3+3 4 3+3 4 3+4 4 5+5 4 3+4 4 3+4 4 3+4 4 | | |
| | 6 2 4+4 4 4+4 4 3+4 4 6+6 4 3+4 4 3+4 4 3+4 4 | | |
| | 5 2 3+3 4 3+3 4 3+4 4 5+5 4 3+4 4 3+4 4 3+4 4 | | |
| | 6 2 4+4 4 4+4 4 3+4 4 6+6 4 3+4 4 3+4 4 3+4 4 | | |
| | 5 2 3+3 4 3+3 4 3+4 4 5+5 4 3+4 4 3+4 4 3+4 4 | | |
| | 6 2 4+4 4 4+4 4 3+4 4 6+6 4 3+4 4 3+4 4 3+4 4 | | |
| | 5 2 3+3 4 3+3 4 3+4 4 6+6 4 3+4 4 3+4 4 3+4 4 | | |
| | 6 2 4+4 4 4+4 4 3+4 4 5+5 4 3+4 4 3+4 4 3+4 4 | | |
| | 5 2 3+3 4 4+4 4 3+4 4 6+6 4 3+4 4 3+4 4 3+4 4 | | |
| | 5 2 4+4 4 3+3 4 3+4 4 5+5 4 3+4 4 3+4 4 3+4 4 | | |
| | 6 2 3+3 4 3+3 4 3+4 4 5+5 4 3+4 4 3+4 4 3+4 4 | | |
| | 5 2 4+4 4 4+4 4 3+4 4 6+6 4 3+4 4 3+4 4 3+4 4 | | |
| | 6 2 4+4 4 3+3 4 3+4 4 6+6 4 3+4 4 3+4 4 3+4 4 | | |
| | 5 2 3+3 4 4+4 4 3+4 4 5+5 4 3+4 4 3+4 4 3+4 4 | | |
| | 6 2 3+3 4 3+3 4 3+4 4 6+6 4 3+4 4 3+4 4 3+4 4 | | |
| | 5 2 4+4 4 4+4 4 3+4 4 5+5 4 3+4 4 3+4 4 3+4 4 | | |
| | 6 2 3+3 4 3+3 4 3+4 4 5+5 4 3+4 4 3+4 4 3+4 4 | | |
| | 5 2 4+4 4 4+4 4 3+4 4 6+6 4 3+4 4 3+4 4 3+4 4 | | |
| | 6 2 3+3 4 3+3 4 3+4 4 6+6 4 3+4 4 3+4 4 3+4 4 | | |
| | 6 2 4+4 4 4+4 4 3+4 4 6+6 4 3+4 4 3+4 4 3+4 4 | | |
| | 5 2 3+3 4 3+3 4 3+4 4 5+5 4 3+4 4 3+4 4 3+4 4 | | |
| | 5 2 4+4 4 4+4 4 3+4 4 6+6 4 3+4 4 3+4 4 3+4 4 | | |
| | 5 2 3+3 4 3+3 4 3+4 4 5+5 4 3+4 4 3+4 4 3+4 4 | | |
| | 6 2 4+4 4 4+4 4 3+4 4 5+5 4 3+4 4 3+4 4 3+4 4 | | |
| | 6 2 4+4 4 3+3 4 3+4 4 6+6 4 3+4 4 3+4 4 3+4 4 | | |
| | 5 2 3+3 4 4+4 4 3+4 4 5+5 4 3+4 4 3+4 4 3+4 4 | | |
| | 6 2 3+3 4 3+3 4 3+4 4 5+5 4 3+4 4 3+4 4 3+4 4 | | |
| | 6 2 4+4 4 4+4 4 3+4 4 6+6 4 3+4 4 3+4 4 3+4 4 | | |
| | 6 2 4+4 4 3+3 4 3+4 4 5+5 4 3+4 4 3+4 4 3+4 4 | | |
| | 6 2 3+3 4 4+4 4 3+4 4 6+6 4 3+4 4 3+4 4 3+4 4 | | |
| | 5 2 4+4 4 4+4 4 3+4 4 6+6 4 3+4 4 3+4 4 3+4 4 | | |
| | 6 2 3+3 4 3+3 4 3+4 4 6+6 4 3+4 4 3+4 4 3+4 4 | | |
| | 6 2 3+3 4 4+4 4 3+4 4 5+5 4 3+4 4 3+4 4 3+4 4 | | |
| | 5 2 4+4 4 3+3 4 3+4 4 5+5 4 3+4 4 3+4 4 3+4 4 | | |
| | 5 2 | | |
| | 5 2 | | |
| | 5 2 | | |
| | 6 2 | | |
| | 5 2 | | |
| | 5 2 | | |
| | 5 2 | | |
| | 5 2 | | |
| | 6 2 | | |
| | 6 2 | | |
| | 6 2 | | |
| | 5 2 | | |
| | 6 2 | | |
| | 5 2 | | |
| | 6 2 | | |
| | 6 2 | | |
| | 5 2 | | |
| | 5 2 | | |
| | 6 2 | | |
| | 5 2 | | |
| | 6 2 | | |
| | 5 2 | | |
| | 5 2 | | |
| | 6 2 | | |
| | 5 2 | | |
| | 6 2 | | |
| | 5 2 | | |
| | 6 2 | | |
| | 5 2 | | |
| | 5 2 | | |
| | 6 2 | | |
| | 5 2 | | |
| | 5 2 | | |
| | 6 2 | | |
| | 6 2 | | |
| | 6 2 | | |
| | 5 2 | | |
| | 6 2 | | |
| | 5 2 | | |
| | 6 2 | | |
| | 5 2 | | |
| | 6 2 | | |
| | 5 2 | | |
| | 6 2 | | |
| | 5 2 | | |
| | 5 2 | | |
| | 6 2 | | |
| | 5 2 | | |
| | 5 2 | | |
| | 6 2 | | |
| | 6 2 | | |
| | 5 2 | | |
| | 6 2 | | |
| | 5 2 | | |
| | 6 2 | | |
| | 5 2 | | |
| | 5 2 | | |
| | 6 2 | | |
| 62.2.108 | 419793 | 120 | Gd1Ge1Ru1 |
| | -GM -R -S -T -U -X -Y -Z | | |
| | 5 2 3+3 4 3+3 4 3+4 4 5+5 4 3+4 4 3+4 4 3+4 4 | | |
| | 5 2 4+4 4 4+4 4 3+4 4 6+6 4 3+4 4 3+4 4 3+4 4 | | |
| | 6 2 3+3 4 3+3 4 3+4 4 5+5 4 3+4 4 3+4 4 3+4 4 | | |
| | 6 2 4+4 4 4+4 4 3+4 4 6+6 4 3+4 4 3+4 4 3+4 4 | | |
| | 5 2 4+4 4 3+3 4 3+4 4 5+5 4 3+4 4 3+4 4 3+4 4 | | |
| | 5 2 3+3 4 4+4 4 3+4 4 6+6 4 3+4 4 3+4 4 3+4 4 | | |
| | 6 2 3+3 4 4+4 4 3+4 4 5+5 4 3+4 4 3+4 4 3+4 4 | | |
| | 6 2 4+4 4 3+3 4 3+4 4 6+6 4 3+4 4 3+4 4 3+4 4 | | |
| | 5 2 3+3 4 4+4 4 3+4 4 6+6 4 3+4 4 3+4 4 3+4 4 | | |
| | 5 2 4+4 4 3+3 4 3+4 4 5+5 4 3+4 4 3+4 4 3+4 4 | | |
| | 6 2 4+4 4 3+3 4 3+4 4 5+5 4 3+4 4 3+4 4 3+4 4 | | |

Continued on next column

Continued on next column

**TABLE XXXVII– continued**

| TQC | ICSD | NoE | SF | | | | | |
|---|---|---|---|---|---|---|---|---|
| | 5 2 | 3+3 4 | 4+4 4 | 3+4 4 | 6+6 4 | 3+4 4 | 3+4 4 | 3+4 4 |
| | 6 2 | 3+3 4 | 3+3 4 | 3+4 4 | 6+6 4 | 3+4 4 | 3+4 4 | 3+4 4 |
| | 5 2 | 4+4 4 | 4+4 4 | 3+4 4 | 5+5 4 | 3+4 4 | 3+4 4 | 3+4 4 |
| | 6 2 | 3+3 4 | 4+4 4 | 3+4 4 | 5+5 4 | 3+4 4 | 3+4 4 | 3+4 4 |
| | 6 2 | 4+4 4 | 3+3 4 | 3+4 4 | 6+6 4 | 3+4 4 | 3+4 4 | 3+4 4 |
| | 5 2 | 3+3 4 | 3+3 4 | 3+4 4 | 6+6 4 | 3+4 4 | 3+4 4 | 3+4 4 |
| | 6 2 | 4+4 4 | 4+4 4 | 3+4 4 | 5+5 4 | 3+4 4 | 3+4 4 | 3+4 4 |
| | 5 2 | 4+4 4 | 3+3 4 | 3+4 4 | 6+6 4 | 3+4 4 | 3+4 4 | 3+4 4 |
| | 6 2 | 3+3 4 | 4+4 4 | 3+4 4 | 5+5 4 | 3+4 4 | 3+4 4 | 3+4 4 |
| | 5 2 | 4+4 4 | 4+4 4 | 3+4 4 | 6+6 4 | 3+4 4 | 3+4 4 | 3+4 4 |
| | 6 2 | 3+3 4 | 3+3 4 | 3+4 4 | 5+5 4 | 3+4 4 | 3+4 4 | 3+4 4 |
| | 5 2 | 4+4 4 | 4+4 4 | 3+4 4 | 6+6 4 | 3+4 4 | 3+4 4 | 3+4 4 |
| | 5 2 | 3+3 4 | 3+3 4 | 3+4 4 | 5+5 4 | 3+4 4 | 3+4 4 | 3+4 4 |
| | 6 2 | 3+3 4 | 3+3 4 | 3+4 4 | 6+6 4 | 3+4 4 | 3+4 4 | 3+4 4 |
| | 5 2 | 4+4 4 | 4+4 4 | 3+4 4 | 5+5 4 | 3+4 4 | 3+4 4 | 3+4 4 |
| | 6 2 | 4+4 4 | 3+3 4 | 3+4 4 | 6+6 4 | 3+4 4 | 3+4 4 | 3+4 4 |
| | 6 2 | 3+3 4 | 4+4 4 | 3+4 4 | 6+6 4 | 3+4 4 | 3+4 4 | 3+4 4 |
| | 6 2 | 3+3 4 | 4+4 4 | 3+4 4 | 5+5 4 | 3+4 4 | 3+4 4 | 3+4 4 |
| | 6 2 | 4+4 4 | 3+3 4 | 3+4 4 | 6+6 4 | 3+4 4 | 3+4 4 | 3+4 4 |
| | 5 2 | | | | | | | |
| | 5 2 | | | | | | | |
| | 6 2 | | | | | | | |
| | 5 2 | | | | | | | |
| | 5 2 | | | | | | | |
| | 6 2 | | | | | | | |
| | 5 2 | | | | | | | |
| | 6 2 | | | | | | | |
| | 6 2 | | | | | | | |
| | 5 2 | | | | | | | |
| | 5 2 | | | | | | | |
| | 6 2 | | | | | | | |
| | 5 2 | | | | | | | |
| | 5 2 | | | | | | | |
| | 6 2 | | | | | | | |
| | 5 2 | | | | | | | |
| | 6 2 | | | | | | | |
| | 6 2 | | | | | | | |
| | 6 2 | | | | | | | |
| | 6 2 | | | | | | | |
| | 5 2 | | | | | | | |
| | 5 2 | | | | | | | |
| | 5 2 | | | | | | | |
| | 6 2 | | | | | | | |
| | 5 2 | | | | | | | |
| | 5 2 | | | | | | | |
| | 5 2 | | | | | | | |
| | 6 2 | | | | | | | |
| | 5 2 | | | | | | | |
| | 6 2 | | | | | | | |
| 62.2.109 | 103904 | 224 | Ga2Pd5 | | | | | |
| | -GM | -R | -S | -T | -U | -X | -Y | -Z |
| | 5 2 | 3+3 4 | 3+3 4 | 3+4 4 | 5+5 4 | 3+4 4 | 3+4 4 | 3+4 4 |
| | 5 2 | 4+4 4 | 4+4 4 | 3+4 4 | 6+6 4 | 3+4 4 | 3+4 4 | 3+4 4 |
| | 6 2 | 3+3 4 | 3+3 4 | 3+4 4 | 5+5 4 | 3+4 4 | 3+4 4 | 3+4 4 |
| | 5 2 | 4+4 4 | 4+4 4 | 3+4 4 | 6+6 4 | 3+4 4 | 3+4 4 | 3+4 4 |
| | 6 2 | 3+3 4 | 3+3 4 | 3+4 4 | 6+6 4 | 3+4 4 | 3+4 4 | 3+4 4 |
| | 5 2 | 4+4 4 | 4+4 4 | 3+4 4 | 5+5 4 | 3+4 4 | 3+4 4 | 3+4 4 |
| | 6 2 | 3+3 4 | 4+4 4 | 3+4 4 | 6+6 4 | 3+4 4 | 3+4 4 | 3+4 4 |
| | 6 2 | 4+4 4 | 3+3 4 | 3+4 4 | 5+5 4 | 3+4 4 | 3+4 4 | 3+4 4 |
| | 6 2 | 4+4 4 | 3+3 4 | 3+4 4 | 5+5 4 | 3+4 4 | 3+4 4 | 3+4 4 |
| | 6 2 | 3+3 4 | 4+4 4 | 3+4 4 | 5+5 4 | 3+4 4 | 3+4 4 | 3+4 4 |
| | 5 2 | 3+3 4 | 3+3 4 | 3+4 4 | 6+6 4 | 3+4 4 | 3+4 4 | 3+4 4 |
| | 5 2 | 4+4 4 | 4+4 4 | 3+4 4 | 5+5 4 | 3+4 4 | 3+4 4 | 3+4 4 |
| | 5 2 | 4+4 4 | 3+3 4 | 3+4 4 | 6+6 4 | 3+4 4 | 3+4 4 | 3+4 4 |
| | 5 2 | 3+3 4 | 4+4 4 | 3+4 4 | 5+5 4 | 3+4 4 | 3+4 4 | 3+4 4 |
| | 6 2 | 3+3 4 | 3+3 4 | 3+4 4 | 6+6 4 | 3+4 4 | 3+4 4 | 3+4 4 |
| | 5 2 | 4+4 4 | 4+4 4 | 3+4 4 | 5+5 4 | 3+4 4 | 3+4 4 | 3+4 4 |
| | 5 2 | 4+4 4 | 4+4 4 | 3+4 4 | 6+6 4 | 3+4 4 | 3+4 4 | 3+4 4 |
| | 6 2 | 3+3 4 | 3+3 4 | 3+4 4 | 6+6 4 | 3+4 4 | 3+4 4 | 3+4 4 |
| | 6 2 | 3+3 4 | 4+4 4 | 3+4 4 | 5+5 4 | 3+4 4 | 3+4 4 | 3+4 4 |
| | 6 2 | 4+4 4 | 3+3 4 | 3+4 4 | 6+6 4 | 3+4 4 | 3+4 4 | 3+4 4 |
| | 6 2 | 4+4 4 | 3+3 4 | 3+4 4 | 5+5 4 | 3+4 4 | 3+4 4 | 3+4 4 |
| | 5 2 | 3+3 4 | 4+4 4 | 3+4 4 | 6+6 4 | 3+4 4 | 3+4 4 | 3+4 4 |
| | 6 2 | 4+4 4 | 3+3 4 | 3+4 4 | 5+5 4 | 3+4 4 | 3+4 4 | 3+4 4 |
| | 5 2 | 3+3 4 | 4+4 4 | 3+4 4 | 5+5 4 | 3+4 4 | 3+4 4 | 3+4 4 |
| | 6 2 | 4+4 4 | 4+4 4 | 3+4 4 | 6+6 4 | 3+4 4 | 3+4 4 | 3+4 4 |
| | 6 2 | 3+3 4 | 3+3 4 | 3+4 4 | 6+6 4 | 3+4 4 | 3+4 4 | 3+4 4 |
| | 5 2 | 3+3 4 | 4+4 4 | 3+4 4 | 5+5 4 | 3+4 4 | 3+4 4 | 3+4 4 |
| | 5 2 | 4+4 4 | 3+3 4 | 3+4 4 | 6+6 4 | 3+4 4 | 3+4 4 | 3+4 4 |
| | 6 2 | 3+3 4 | 3+3 4 | 3+4 4 | 6+6 4 | 3+4 4 | 3+4 4 | 3+4 4 |
| | 5 2 | 4+4 4 | 4+4 4 | 3+4 4 | 5+5 4 | 3+4 4 | 3+4 4 | 3+4 4 |
| | 5 2 | 3+3 4 | 3+3 4 | 3+4 4 | 6+6 4 | 3+4 4 | 3+4 4 | 3+4 4 |
| | 6 2 | 4+4 4 | 4+4 4 | 3+4 4 | 5+5 4 | 3+4 4 | 3+4 4 | 3+4 4 |
| | 5 2 | 3+3 4 | 4+4 4 | 3+4 4 | 6+6 4 | 3+4 4 | 3+4 4 | 3+4 4 |
| | 6 2 | 4+4 4 | 3+3 4 | 3+4 4 | 5+5 4 | 3+4 4 | 3+4 4 | 3+4 4 |
| | 5 2 | 3+3 4 | 3+3 4 | 3+4 4 | 6+6 4 | 3+4 4 | 3+4 4 | 3+4 4 |
| | 5 2 | 4+4 4 | 4+4 4 | 3+4 4 | 5+5 4 | 3+4 4 | 3+4 4 | 3+4 4 |
| | 6 2 | 4+4 4 | 4+4 4 | 3+4 4 | 6+6 4 | 3+4 4 | 3+4 4 | 3+4 4 |
| | 5 2 | 3+3 4 | 3+3 4 | 3+4 4 | 5+5 4 | 3+4 4 | 3+4 4 | 3+4 4 |
| | 6 2 | 4+4 4 | 4+4 4 | 3+4 4 | 6+6 4 | 3+4 4 | 3+4 4 | 3+4 4 |
| | 6 2 | 3+3 4 | 3+3 4 | 3+4 4 | 5+5 4 | 3+4 4 | 3+4 4 | 3+4 4 |
| | 5 2 | 4+4 4 | 3+3 4 | 3+4 4 | 6+6 4 | 3+4 4 | 3+4 4 | 3+4 4 |
| | 6 2 | 3+3 4 | 4+4 4 | 3+4 4 | 5+5 4 | 3+4 4 | 3+4 4 | 3+4 4 |
| | 5 2 | 3+3 4 | 3+3 4 | 3+4 4 | 6+6 4 | 3+4 4 | 3+4 4 | 3+4 4 |
| | 5 2 | 4+4 4 | 4+4 4 | 3+4 4 | 5+5 4 | 3+4 4 | 3+4 4 | 3+4 4 |
| | 6 2 | 4+4 4 | 3+3 4 | 3+4 4 | 5+5 4 | 3+4 4 | 3+4 4 | 3+4 4 |
| | 6 2 | 3+3 4 | 4+4 4 | 3+4 4 | 6+6 4 | 3+4 4 | 3+4 4 | 3+4 4 |
| | 6 2 | 4+4 4 | 3+3 4 | 3+4 4 | 6+6 4 | 3+4 4 | 3+4 4 | 3+4 4 |
| | 5 2 | 3+3 4 | 4+4 4 | 3+4 4 | 6+6 4 | 3+4 4 | 3+4 4 | 3+4 4 |
| | 5 2 | 4+4 4 | 4+4 4 | 3+4 4 | 5+5 4 | 3+4 4 | 3+4 4 | 3+4 4 |
| | 5 2 | 3+3 4 | 3+3 4 | 3+4 4 | 6+6 4 | 3+4 4 | 3+4 4 | 3+4 4 |
| | 6 2 | 3+3 4 | 3+3 4 | 3+4 4 | 5+5 4 | 3+4 4 | 3+4 4 | 3+4 4 |
| | 6 2 | 4+4 4 | 4+4 4 | 3+4 4 | 5+5 4 | 3+4 4 | 3+4 4 | 3+4 4 |
| | 5 2 | 3+3 4 | 4+4 4 | 3+4 4 | 6+6 4 | 3+4 4 | 3+4 4 | 3+4 4 |
| | 6 2 | 4+4 4 | 3+3 4 | 3+4 4 | 5+5 4 | 3+4 4 | 3+4 4 | 3+4 4 |
| | 5 2 | 3+3 4 | 3+3 4 | 3+4 4 | 6+6 4 | 3+4 4 | 3+4 4 | 3+4 4 |
| | 5 2 | 4+4 4 | 4+4 4 | 3+4 4 | 5+5 4 | 3+4 4 | 3+4 4 | 3+4 4 |
| | 5 2 | | | | | | | |
| | 6 2 | | | | | | | |
| | 5 2 | | | | | | | |
| | 6 2 | | | | | | | |
| | 6 2 | | | | | | | |
| | 5 2 | | | | | | | |
| | 6 2 | | | | | | | |
| | 5 2 | | | | | | | |
| | 6 2 | | | | | | | |
| | 6 2 | | | | | | | |
| | 5 2 | | | | | | | |
| | 5 2 | | | | | | | |
| | 6 2 | | | | | | | |
| | 5 2 | | | | | | | |
| | 6 2 | | | | | | | |

Continued on next column

**TABLE XXXVII– continued**

| TQC | ICSD | NoE | SF | | | | | |
|---|---|---|---|---|---|---|---|---|
| | 5 2 | | | | | | | |
| | 6 2 | | | | | | | |
| | 6 2 | | | | | | | |
| | 5 2 | | | | | | | |
| | 6 2 | | | | | | | |
| | 5 2 | | | | | | | |
| | 5 2 | | | | | | | |
| | 5 2 | | | | | | | |
| | 6 2 | | | | | | | |
| | 6 2 | | | | | | | |
| | 5 2 | | | | | | | |
| | 6 2 | | | | | | | |
| | 5 2 | | | | | | | |
| | 5 2 | | | | | | | |
| | 6 2 | | | | | | | |
| | 6 2 | | | | | | | |
| | 6 2 | | | | | | | |
| | 5 2 | | | | | | | |
| | 6 2 | | | | | | | |
| | 5 2 | | | | | | | |
| | 5 2 | | | | | | | |
| | 5 2 | | | | | | | |
| | 5 2 | | | | | | | |
| | 6 2 | | | | | | | |
| | 5 2 | | | | | | | |
| | 5 2 | | | | | | | |
| | 6 2 | | | | | | | |
| | 6 2 | | | | | | | |
| | 5 2 | | | | | | | |
| | 6 2 | | | | | | | |
| | 5 2 | | | | | | | |
| | 6 2 | | | | | | | |
| | 6 2 | | | | | | | |
| | 6 2 | | | | | | | |
| | 5 2 | | | | | | | |
| | 6 2 | | | | | | | |
| | 5 2 | | | | | | | |
| | 6 2 | | | | | | | |
| | 6 2 | | | | | | | |
| | 5 2 | | | | | | | |
| | 5 2 | | | | | | | |
| 62.2.110 | 420711 | 168 | Ga1Ho1O3 | | | | | |
| | -GM | -R | -S | -T | -U | -X | -Y | -Z |
| | 5 2 | 3+3 4 | 3+3 4 | 3+4 4 | 6+6 4 | 3+4 4 | 3+4 4 | 3+4 4 |
| | 6 2 | 4+4 4 | 4+4 4 | 3+4 4 | 5+5 4 | 3+4 4 | 3+4 4 | 3+4 4 |
| | 5 2 | 3+3 4 | 4+4 4 | 3+4 4 | 6+6 4 | 3+4 4 | 3+4 4 | 3+4 4 |
| | 6 2 | 4+4 4 | 3+3 4 | 3+4 4 | 5+5 4 | 3+4 4 | 3+4 4 | 3+4 4 |
| | 5 2 | 3+3 4 | 4+4 4 | 3+4 4 | 5+5 4 | 3+4 4 | 3+4 4 | 3+4 4 |
| | 6 2 | 4+4 4 | 3+3 4 | 3+4 4 | 6+6 4 | 3+4 4 | 3+4 4 | 3+4 4 |
| | 5 2 | 4+4 4 | 3+3 4 | 3+4 4 | 6+6 4 | 3+4 4 | 3+4 4 | 3+4 4 |
| | 6 2 | 3+3 4 | 4+4 4 | 3+4 4 | 5+5 4 | 3+4 4 | 3+4 4 | 3+4 4 |
| | 6 2 | 4+4 4 | 3+3 4 | 3+4 4 | 6+6 4 | 3+4 4 | 3+4 4 | 3+4 4 |
| | 5 2 | 3+3 4 | 4+4 4 | 3+4 4 | 5+5 4 | 3+4 4 | 3+4 4 | 3+4 4 |
| | 5 2 | 4+4 4 | 3+3 4 | 3+4 4 | 5+5 4 | 3+4 4 | 3+4 4 | 3+4 4 |
| | 5 2 | 3+3 4 | 4+4 4 | 3+4 4 | 6+6 4 | 3+4 4 | 3+4 4 | 3+4 4 |
| | 6 2 | 3+3 4 | 4+4 4 | 3+4 4 | 6+6 4 | 3+4 4 | 3+4 4 | 3+4 4 |
| | 6 2 | 4+4 4 | 3+3 4 | 3+4 4 | 5+5 4 | 3+4 4 | 3+4 4 | 3+4 4 |
| | 6 2 | 4+4 4 | 4+4 4 | 3+4 4 | 5+5 4 | 3+4 4 | 3+4 4 | 3+4 4 |
| | 5 2 | 3+3 4 | 3+3 4 | 3+4 4 | 5+5 4 | 3+4 4 | 3+4 4 | 3+4 4 |
| | 5 2 | 3+3 4 | 3+3 4 | 3+4 4 | 6+6 4 | 3+4 4 | 3+4 4 | 3+4 4 |
| | 6 2 | 4+4 4 | 4+4 4 | 3+4 4 | 6+6 4 | 3+4 4 | 3+4 4 | 3+4 4 |

Continued on next column

**TABLE XXXVII– continued**

| TQC | ICSD | NoE | SF | | | | | |
|---|---|---|---|---|---|---|---|---|
| | 6 2 | 4+4 4 | 3+3 4 | 3+4 4 | 5+5 4 | 3+4 4 | 3+4 4 | 3+4 4 |
| | 6 2 | 3+3 4 | 4+4 4 | 3+4 4 | 6+6 4 | 3+4 4 | 3+4 4 | 3+4 4 |
| | 5 2 | 4+4 4 | 4+4 4 | 3+4 4 | 6+6 4 | 3+4 4 | 3+4 4 | 3+4 4 |
| | 6 2 | 3+3 4 | 3+3 4 | 3+4 4 | 6+6 4 | 3+4 4 | 3+4 4 | 3+4 4 |
| | 5 2 | 4+4 4 | 4+4 4 | 3+4 4 | 6+6 4 | 3+4 4 | 3+4 4 | 3+4 4 |
| | 6 2 | 3+3 4 | 3+3 4 | 3+4 4 | 5+5 4 | 3+4 4 | 3+4 4 | 3+4 4 |
| | 5 2 | 3+3 4 | 4+4 4 | 3+4 4 | 5+5 4 | 3+4 4 | 3+4 4 | 3+4 4 |
| | 6 2 | 4+4 4 | 3+3 4 | 3+4 4 | 6+6 4 | 3+4 4 | 3+4 4 | 3+4 4 |
| | 5 2 | 3+3 4 | 3+3 4 | 3+4 4 | 5+5 4 | 3+4 4 | 3+4 4 | 3+4 4 |
| | 5 2 | 4+4 4 | 4+4 4 | 3+4 4 | 5+5 4 | 3+4 4 | 3+4 4 | 3+4 4 |
| | 5 2 | 3+3 4 | 3+3 4 | 3+4 4 | 6+6 4 | 3+4 4 | 3+4 4 | 3+4 4 |
| | 5 2 | 4+4 4 | 4+4 4 | 3+4 4 | 6+6 4 | 3+4 4 | 3+4 4 | 3+4 4 |
| | 6 2 | 4+4 4 | 3+3 4 | 3+4 4 | 5+5 4 | 3+4 4 | 3+4 4 | 3+4 4 |
| | 5 2 | 3+3 4 | 4+4 4 | 3+4 4 | 5+5 4 | 3+4 4 | 3+4 4 | 3+4 4 |
| | 6 2 | 3+3 4 | 4+4 4 | 3+4 4 | 6+6 4 | 3+4 4 | 3+4 4 | 3+4 4 |
| | 6 2 | 4+4 4 | 3+3 4 | 3+4 4 | 5+5 4 | 3+4 4 | 3+4 4 | 3+4 4 |
| | 6 2 | 4+4 4 | 3+3 4 | 3+4 4 | 6+6 4 | 3+4 4 | 3+4 4 | 3+4 4 |
| | 6 2 | 3+3 4 | 4+4 4 | 3+4 4 | 6+6 4 | 3+4 4 | 3+4 4 | 3+4 4 |
| | 5 2 | 3+3 4 | 4+4 4 | 3+4 4 | 5+5 4 | 3+4 4 | 3+4 4 | 3+4 4 |
| | 6 2 | 4+4 4 | 3+3 4 | 3+4 4 | 5+5 4 | 3+4 4 | 3+4 4 | 3+4 4 |
| | 5 2 | 4+4 4 | 3+3 4 | 3+4 4 | 6+6 4 | 3+4 4 | 3+4 4 | 3+4 4 |
| | 6 2 | 3+3 4 | 4+4 4 | 3+4 4 | 5+5 4 | 3+4 4 | 3+4 4 | 3+4 4 |
| | 6 2 | 4+4 4 | 3+3 4 | 3+4 4 | 6+6 4 | 3+4 4 | 3+4 4 | 3+4 4 |
| | 5 2 | 3+3 4 | 4+4 4 | 3+4 4 | 5+5 4 | 3+4 4 | 3+4 4 | 3+4 4 |
| | 5 2 | | | | | | | |
| | 6 2 | | | | | | | |
| | 6 2 | | | | | | | |
| | 5 2 | | | | | | | |
| | 6 2 | | | | | | | |
| | 5 2 | | | | | | | |
| | 5 2 | | | | | | | |
| | 5 2 | | | | | | | |
| | 5 2 | | | | | | | |
| | 6 2 | | | | | | | |
| | 6 2 | | | | | | | |
| | 6 2 | | | | | | | |
| | 5 2 | | | | | | | |
| | 5 2 | | | | | | | |
| | 5 2 | | | | | | | |
| | 5 2 | | | | | | | |
| | 6 2 | | | | | | | |
| | 5 2 | | | | | | | |
| | 6 2 | | | | | | | |
| | 5 2 | | | | | | | |
| | 6 2 | | | | | | | |
| | 6 2 | | | | | | | |
| | 5 2 | | | | | | | |
| | 6 2 | | | | | | | |
| | 5 2 | | | | | | | |
| | 5 2 | | | | | | | |
| | 6 2 | | | | | | | |
| | 5 2 | | | | | | | |
| | 6 2 | | | | | | | |
| | 5 2 | | | | | | | |
| | 6 2 | | | | | | | |
| | 6 2 | | | | | | | |
| | 5 2 | | | | | | | |
| | 6 2 | | | | | | | |
| | 5 2 | | | | | | | |
| | 6 2 | | | | | | | |
| | 5 2 | | | | | | | |

Continued on next column

**TABLE XXXVII– continued**

| TQC | ICSD | NoE | SF | | | | | |
|---|---|---|---|---|---|---|---|---|
| | 5 2 | | | | | | | |
| | 5 2 | | | | | | | |
| | 6 2 | | | | | | | |
| | 5 2 | | | | | | | |
| | 6 2 | | | | | | | |
| 62.2.111 | 55537 | 104 | Fe1Si1U1 | | | | | |
| | -GM | -R | -S | -T | -U | -X | -Y | -Z |
| | 5 2 | 3+3 4 | 3+3 4 | 3+4 4 | 6+6 4 | 3+4 4 | 3+4 4 | 3+4 4 |
| | 5 2 | 4+4 4 | 4+4 4 | 3+4 4 | 5+5 4 | 3+4 4 | 3+4 4 | 3+4 4 |
| | 6 2 | 4+4 4 | 3+3 4 | 3+4 4 | 6+6 4 | 3+4 4 | 3+4 4 | 3+4 4 |
| | 6 2 | 3+3 4 | 4+4 4 | 3+4 4 | 5+5 4 | 3+4 4 | 3+4 4 | 3+4 4 |
| | 5 2 | 3+3 4 | 4+4 4 | 3+4 4 | 6+6 4 | 3+4 4 | 3+4 4 | 3+4 4 |
| | 5 2 | 4+4 4 | 3+3 4 | 3+4 4 | 5+5 4 | 3+4 4 | 3+4 4 | 3+4 4 |
| | 6 2 | 4+4 4 | 3+3 4 | 3+4 4 | 6+6 4 | 3+4 4 | 3+4 4 | 3+4 4 |
| | 6 2 | 3+3 4 | 4+4 4 | 3+4 4 | 5+5 4 | 3+4 4 | 3+4 4 | 3+4 4 |
| | 5 2 | 4+4 4 | 3+3 4 | 3+4 4 | 5+5 4 | 3+4 4 | 3+4 4 | 3+4 4 |
| | 5 2 | 3+3 4 | 4+4 4 | 3+4 4 | 6+6 4 | 3+4 4 | 3+4 4 | 3+4 4 |
| | 6 2 | 3+3 4 | 4+4 4 | 3+4 4 | 5+5 4 | 3+4 4 | 3+4 4 | 3+4 4 |
| | 5 2 | 4+4 4 | 3+3 4 | 3+4 4 | 6+6 4 | 3+4 4 | 3+4 4 | 3+4 4 |
| | 5 2 | 3+3 4 | 4+4 4 | 3+4 4 | 5+5 4 | 3+4 4 | 3+4 4 | 3+4 4 |
| | 6 2 | 4+4 4 | 3+3 4 | 3+4 4 | 6+6 4 | 3+4 4 | 3+4 4 | 3+4 4 |
| | 6 2 | 3+3 4 | 4+4 4 | 3+4 4 | 6+6 4 | 3+4 4 | 3+4 4 | 3+4 4 |
| | 6 2 | 4+4 4 | 3+3 4 | 3+4 4 | 5+5 4 | 3+4 4 | 3+4 4 | 3+4 4 |
| | 5 2 | 3+3 4 | 3+3 4 | 3+4 4 | 5+5 4 | 3+4 4 | 3+4 4 | 3+4 4 |
| | 6 2 | 4+4 4 | 4+4 4 | 3+4 4 | 6+6 4 | 3+4 4 | 3+4 4 | 3+4 4 |
| | 5 2 | 4+4 4 | 3+3 4 | 3+4 4 | 5+5 4 | 3+4 4 | 3+4 4 | 3+4 4 |
| | 6 2 | 3+3 4 | 4+4 4 | 3+4 4 | 6+6 4 | 3+4 4 | 3+4 4 | 3+4 4 |
| | 5 2 | 3+3 4 | 4+4 4 | 3+4 4 | 5+5 4 | 3+4 4 | 3+4 4 | 3+4 4 |
| | 6 2 | 4+4 4 | 3+3 4 | 3+4 4 | 6+6 4 | 3+4 4 | 3+4 4 | 3+4 4 |
| | 6 2 | 3+3 4 | 3+3 4 | 3+4 4 | 5+5 4 | 3+4 4 | 3+4 4 | 3+4 4 |
| | 5 2 | 4+4 4 | 4+4 4 | 3+4 4 | 5+5 4 | 3+4 4 | 3+4 4 | 3+4 4 |
| | 5 2 | 4+4 4 | 4+4 4 | 3+4 4 | 6+6 4 | 3+4 4 | 3+4 4 | 3+4 4 |
| | 6 2 | 3+3 4 | 3+3 4 | 3+4 4 | 5+5 4 | 3+4 4 | 3+4 4 | 3+4 4 |
| | 5 2 | | | | | | | |
| | 6 2 | | | | | | | |
| | 6 2 | | | | | | | |
| | 6 2 | | | | | | | |
| | 5 2 | | | | | | | |
| | 5 2 | | | | | | | |
| | 6 2 | | | | | | | |
| | 6 2 | | | | | | | |
| | 5 2 | | | | | | | |
| | 5 2 | | | | | | | |
| | 5 2 | | | | | | | |
| | 6 2 | | | | | | | |
| | 6 2 | | | | | | | |
| | 5 2 | | | | | | | |
| | 5 2 | | | | | | | |
| | 6 2 | | | | | | | |
| | 5 2 | | | | | | | |
| | 5 2 | | | | | | | |
| | 6 2 | | | | | | | |
| | 5 2 | | | | | | | |
| | 5 2 | | | | | | | |
| | 5 2 | | | | | | | |
| | 6 2 | | | | | | | |
| | 6 2 | | | | | | | |
| | 6 2 | | | | | | | |
| | 5 2 | | | | | | | |
| 62.2.112 | 169317 | 56 | Fe1Se1 | | | | | |
| | -GM | -R | -S | -T | -U | -X | -Y | -Z |
| | 5 2 | 4+4 4 | 4+4 4 | 3+4 4 | 6+6 4 | 3+4 4 | 3+4 4 | 3+4 4 |
| | 5 2 | 3+3 4 | 3+3 4 | 3+4 4 | 5+5 4 | 3+4 4 | 3+4 4 | 3+4 4 |
| | 6 2 | 3+3 4 | 4+4 4 | 3+4 4 | 5+5 4 | 3+4 4 | 3+4 4 | 3+4 4 |
| | 6 2 | 4+4 4 | 3+3 4 | 3+4 4 | 6+6 4 | 3+4 4 | 3+4 4 | 3+4 4 |
| | 5 2 | 3+3 4 | 3+3 4 | 3+4 4 | 6+6 4 | 3+4 4 | 3+4 4 | 3+4 4 |
| | 6 2 | 4+4 4 | 4+4 4 | 3+4 4 | 6+6 4 | 3+4 4 | 3+4 4 | 3+4 4 |
| | 5 2 | 4+4 4 | 4+4 4 | 3+4 4 | 5+5 4 | 3+4 4 | 3+4 4 | 3+4 4 |
| | 5 2 | 3+3 4 | 3+3 4 | 3+4 4 | 5+5 4 | 3+4 4 | 3+4 4 | 3+4 4 |
| | 6 2 | 3+3 4 | 4+4 4 | 3+4 4 | 5+5 4 | 3+4 4 | 3+4 4 | 3+4 4 |
| | 5 2 | 4+4 4 | 3+3 4 | 3+4 4 | 6+6 4 | 3+4 4 | 3+4 4 | 3+4 4 |
| | 6 2 | 4+4 4 | 3+3 4 | 3+4 4 | 5+5 4 | 3+4 4 | 3+4 4 | 3+4 4 |
| | 5 2 | 3+3 4 | 4+4 4 | 3+4 4 | 6+6 4 | 3+4 4 | 3+4 4 | 3+4 4 |
| | 5 2 | 3+3 4 | 3+3 4 | 3+4 4 | 5+5 4 | 3+4 4 | 3+4 4 | 3+4 4 |
| | 6 2 | 4+4 4 | 4+4 4 | 3+4 4 | 6+6 4 | 3+4 4 | 3+4 4 | 3+4 4 |
| | 5 2 | | | | | | | |
| | 5 2 | | | | | | | |
| | 6 2 | | | | | | | |
| | 6 2 | | | | | | | |
| | 5 2 | | | | | | | |
| | 6 2 | | | | | | | |
| | 5 2 | | | | | | | |
| | 5 2 | | | | | | | |
| | 6 2 | | | | | | | |
| | 6 2 | | | | | | | |
| | 5 2 | | | | | | | |
| | 6 2 | | | | | | | |
| | 5 2 | | | | | | | |
| | 6 2 | | | | | | | |
| 62.2.113 | 169316 | 56 | Fe1Se1 | | | | | |
| | -GM | -R | -S | -T | -U | -X | -Y | -Z |
| | 5 2 | 3+3 4 | 4+4 4 | 3+4 4 | 6+6 4 | 3+4 4 | 3+4 4 | 3+4 4 |
| | 5 2 | 4+4 4 | 3+3 4 | 3+4 4 | 5+5 4 | 3+4 4 | 3+4 4 | 3+4 4 |
| | 6 2 | 4+4 4 | 4+4 4 | 3+4 4 | 6+6 4 | 3+4 4 | 3+4 4 | 3+4 4 |
| | 6 2 | 3+3 4 | 3+3 4 | 3+4 4 | 5+5 4 | 3+4 4 | 3+4 4 | 3+4 4 |
| | 5 2 | 4+4 4 | 3+3 4 | 3+4 4 | 6+6 4 | 3+4 4 | 3+4 4 | 3+4 4 |
| | 6 2 | 3+3 4 | 4+4 4 | 3+4 4 | 6+6 4 | 3+4 4 | 3+4 4 | 3+4 4 |
| | 5 2 | 3+3 4 | 4+4 4 | 3+4 4 | 5+5 4 | 3+4 4 | 3+4 4 | 3+4 4 |
| | 5 2 | 4+4 4 | 3+3 4 | 3+4 4 | 5+5 4 | 3+4 4 | 3+4 4 | 3+4 4 |
| | 6 2 | 4+4 4 | 4+4 4 | 3+4 4 | 5+5 4 | 3+4 4 | 3+4 4 | 3+4 4 |
| | 6 2 | 3+3 4 | 3+3 4 | 3+4 4 | 6+6 4 | 3+4 4 | 3+4 4 | 3+4 4 |
| | 5 2 | 3+3 4 | 3+3 4 | 3+4 4 | 5+5 4 | 3+4 4 | 3+4 4 | 3+4 4 |
| | 5 2 | 4+4 4 | 4+4 4 | 3+4 4 | 6+6 4 | 3+4 4 | 3+4 4 | 3+4 4 |
| | 5 2 | 4+4 4 | 4+4 4 | 3+4 4 | 5+5 4 | 3+4 4 | 3+4 4 | 3+4 4 |
| | 6 2 | 3+3 4 | 3+3 4 | 3+4 4 | 5+5 4 | 3+4 4 | 3+4 4 | 3+4 4 |
| | 5 2 | | | | | | | |
| | 5 2 | | | | | | | |
| | 6 2 | | | | | | | |
| | 6 2 | | | | | | | |
| | 5 2 | | | | | | | |
| | 6 2 | | | | | | | |
| | 5 2 | | | | | | | |
| | 6 2 | | | | | | | |
| | 5 2 | | | | | | | |
| | 6 2 | | | | | | | |
| | 6 2 | | | | | | | |
| | 5 2 | | | | | | | |
| | 5 2 | | | | | | | |
| | 6 2 | | | | | | | |
| 62.2.114 | 169314 | 56 | Fe1Se1 | | | | | |
| | -GM | -R | -S | -T | -U | -X | -Y | -Z |
| | 5 2 | 3+3 4 | 3+3 4 | 3+4 4 | 5+5 4 | 3+4 4 | 3+4 4 | 3+4 4 |

Continued on next column

**TABLE XXXVII– continued**

| TQC | ICSD | NoE | SF | | | | |
|---|---|---|---|---|---|---|---|
| | 6 2 | 4+4 4 | 4+4 4 | 3+4 4 | 6+6 4 | 3+4 4 | 3+4 4 | 3+4 4 |
| | 5 2 | 3+3 4 | 3+3 4 | 3+4 4 | 5+5 4 | 3+4 4 | 3+4 4 | 3+4 4 |
| | 6 2 | 4+4 4 | 4+4 4 | 3+4 4 | 6+6 4 | 3+4 4 | 3+4 4 | 3+4 4 |
| | 5 2 | 3+3 4 | 4+4 4 | 3+4 4 | 5+5 4 | 3+4 4 | 3+4 4 | 3+4 4 |
| | 6 2 | 4+4 4 | 3+3 4 | 3+4 4 | 5+5 4 | 3+4 4 | 3+4 4 | 3+4 4 |
| | 5 2 | 4+4 4 | 3+3 4 | 3+4 4 | 6+6 4 | 3+4 4 | 3+4 4 | 3+4 4 |
| | 6 2 | 3+3 4 | 4+4 4 | 3+4 4 | 6+6 4 | 3+4 4 | 3+4 4 | 3+4 4 |
| | 5 2 | 3+3 4 | 3+3 4 | 3+4 4 | 6+6 4 | 3+4 4 | 3+4 4 | 3+4 4 |
| | 6 2 | 4+4 4 | 4+4 4 | 3+4 4 | 5+5 4 | 3+4 4 | 3+4 4 | 3+4 4 |
| | 5 2 | 4+4 4 | 4+4 4 | 3+4 4 | 6+6 4 | 3+4 4 | 3+4 4 | 3+4 4 |
| | 5 2 | 3+3 4 | 3+3 4 | 3+4 4 | 5+5 4 | 3+4 4 | 3+4 4 | 3+4 4 |
| | 5 2 | 3+3 4 | 3+3 4 | 3+4 4 | 6+6 4 | 3+4 4 | 3+4 4 | 3+4 4 |
| | 6 2 | 4+4 4 | 4+4 4 | 3+4 4 | 5+5 4 | 3+4 4 | 3+4 4 | 3+4 4 |
| | 5 2 | | | | | | | |
| | 5 2 | | | | | | | |
| | 6 2 | | | | | | | |
| | 5 2 | | | | | | | |
| | 6 2 | | | | | | | |
| | 5 2 | | | | | | | |
| | 6 2 | | | | | | | |
| | 6 2 | | | | | | | |
| | 5 2 | | | | | | | |
| | 6 2 | | | | | | | |
| | 5 2 | | | | | | | |
| | 6 2 | | | | | | | |
| | 5 2 | | | | | | | |
| | 6 2 | | | | | | | |
| 62.2.115 | 169312 | 56 | Fe1Se1 | | | | | |
| | -GM | -R | -S | -T | -U | -X | -Y | -Z |
| | 5 2 | 3+3 4 | 4+4 4 | 3+4 4 | 6+6 4 | 3+4 4 | 3+4 4 | 3+4 4 |
| | 5 2 | 4+4 4 | 3+3 4 | 3+4 4 | 5+5 4 | 3+4 4 | 3+4 4 | 3+4 4 |
| | 6 2 | 3+3 4 | 4+4 4 | 3+4 4 | 6+6 4 | 3+4 4 | 3+4 4 | 3+4 4 |
| | 6 2 | 4+4 4 | 3+3 4 | 3+4 4 | 5+5 4 | 3+4 4 | 3+4 4 | 3+4 4 |
| | 5 2 | 3+3 4 | 3+3 4 | 3+4 4 | 6+6 4 | 3+4 4 | 3+4 4 | 3+4 4 |
| | 6 2 | 4+4 4 | 4+4 4 | 3+4 4 | 6+6 4 | 3+4 4 | 3+4 4 | 3+4 4 |
| | 5 2 | 4+4 4 | 4+4 4 | 3+4 4 | 5+5 4 | 3+4 4 | 3+4 4 | 3+4 4 |
| | 5 2 | 3+3 4 | 3+3 4 | 3+4 4 | 5+5 4 | 3+4 4 | 3+4 4 | 3+4 4 |
| | 6 2 | 3+3 4 | 4+4 4 | 3+4 4 | 5+5 4 | 3+4 4 | 3+4 4 | 3+4 4 |
| | 6 2 | 4+4 4 | 3+3 4 | 3+4 4 | 6+6 4 | 3+4 4 | 3+4 4 | 3+4 4 |
| | 5 2 | 4+4 4 | 3+3 4 | 3+4 4 | 5+5 4 | 3+4 4 | 3+4 4 | 3+4 4 |
| | 5 2 | 3+3 4 | 4+4 4 | 3+4 4 | 6+6 4 | 3+4 4 | 3+4 4 | 3+4 4 |
| | 5 2 | 3+3 4 | 4+4 4 | 3+4 4 | 5+5 4 | 3+4 4 | 3+4 4 | 3+4 4 |
| | 6 2 | 4+4 4 | 3+3 4 | 3+4 4 | 5+5 4 | 3+4 4 | 3+4 4 | 3+4 4 |
| | 5 2 | | | | | | | |
| | 5 2 | | | | | | | |
| | 6 2 | | | | | | | |
| | 6 2 | | | | | | | |
| | 5 2 | | | | | | | |
| | 6 2 | | | | | | | |
| | 5 2 | | | | | | | |
| | 6 2 | | | | | | | |
| | 5 2 | | | | | | | |
| | 6 2 | | | | | | | |
| | 6 2 | | | | | | | |
| | 5 2 | | | | | | | |
| | 5 2 | | | | | | | |
| | 6 2 | | | | | | | |
| 62.2.116 | 169310 | 56 | Fe1Se1 | | | | | |
| | -GM | -R | -S | -T | -U | -X | -Y | -Z |
| | 5 2 | 3+3 4 | 4+4 4 | 3+4 4 | 6+6 4 | 3+4 4 | 3+4 4 | 3+4 4 |
| | 5 2 | 4+4 4 | 3+3 4 | 3+4 4 | 5+5 4 | 3+4 4 | 3+4 4 | 3+4 4 |

Continued on next column

**TABLE XXXVII– continued**

| TQC | ICSD | NoE | SF | | | | |
|---|---|---|---|---|---|---|---|
| | 6 2 | 3+3 4 | 4+4 4 | 3+4 4 | 6+6 4 | 3+4 4 | 3+4 4 | 3+4 4 |
| | 6 2 | 4+4 4 | 3+3 4 | 3+4 4 | 5+5 4 | 3+4 4 | 3+4 4 | 3+4 4 |
| | 5 2 | 3+3 4 | 3+3 4 | 3+4 4 | 6+6 4 | 3+4 4 | 3+4 4 | 3+4 4 |
| | 6 2 | 4+4 4 | 4+4 4 | 3+4 4 | 6+6 4 | 3+4 4 | 3+4 4 | 3+4 4 |
| | 5 2 | 4+4 4 | 4+4 4 | 3+4 4 | 5+5 4 | 3+4 4 | 3+4 4 | 3+4 4 |
| | 5 2 | 3+3 4 | 3+3 4 | 3+4 4 | 5+5 4 | 3+4 4 | 3+4 4 | 3+4 4 |
| | 6 2 | 3+3 4 | 4+4 4 | 3+4 4 | 5+5 4 | 3+4 4 | 3+4 4 | 3+4 4 |
| | 6 2 | 4+4 4 | 3+3 4 | 3+4 4 | 6+6 4 | 3+4 4 | 3+4 4 | 3+4 4 |
| | 5 2 | 4+4 4 | 3+3 4 | 3+4 4 | 5+5 4 | 3+4 4 | 3+4 4 | 3+4 4 |
| | 5 2 | 3+3 4 | 4+4 4 | 3+4 4 | 6+6 4 | 3+4 4 | 3+4 4 | 3+4 4 |
| | 5 2 | 3+3 4 | 4+4 4 | 3+4 4 | 5+5 4 | 3+4 4 | 3+4 4 | 3+4 4 |
| | 6 2 | 4+4 4 | 3+3 4 | 3+4 4 | 5+5 4 | 3+4 4 | 3+4 4 | 3+4 4 |
| | 5 2 | | | | | | | |
| | 6 2 | | | | | | | |
| | 5 2 | | | | | | | |
| | 6 2 | | | | | | | |
| | 5 2 | | | | | | | |
| | 6 2 | | | | | | | |
| | 5 2 | | | | | | | |
| | 6 2 | | | | | | | |
| | 5 2 | | | | | | | |
| | 6 2 | | | | | | | |
| | 6 2 | | | | | | | |
| | 5 2 | | | | | | | |
| | 5 2 | | | | | | | |
| | 6 2 | | | | | | | |
| 62.2.117 | 169308 | 56 | Fe1Se1 | | | | | |
| | -GM | -R | -S | -T | -U | -X | -Y | -Z |
| | 5 2 | 3+3 4 | 3+3 4 | 3+4 4 | 6+6 4 | 3+4 4 | 3+4 4 | 3+4 4 |
| | 6 2 | 4+4 4 | 4+4 4 | 3+4 4 | 5+5 4 | 3+4 4 | 3+4 4 | 3+4 4 |
| | 5 2 | 4+4 4 | 3+3 4 | 3+4 4 | 5+5 4 | 3+4 4 | 3+4 4 | 3+4 4 |
| | 6 2 | 3+3 4 | 4+4 4 | 3+4 4 | 6+6 4 | 3+4 4 | 3+4 4 | 3+4 4 |
| | 6 2 | 4+4 4 | 4+4 4 | 3+4 4 | 5+5 4 | 3+4 4 | 3+4 4 | 3+4 4 |
| | 5 2 | 3+3 4 | 3+3 4 | 3+4 4 | 5+5 4 | 3+4 4 | 3+4 4 | 3+4 4 |
| | 5 2 | 3+3 4 | 3+3 4 | 3+4 4 | 6+6 4 | 3+4 4 | 3+4 4 | 3+4 4 |
| | 6 2 | 4+4 4 | 4+4 4 | 3+4 4 | 6+6 4 | 3+4 4 | 3+4 4 | 3+4 4 |
| | 5 2 | 4+4 4 | 3+3 4 | 3+4 4 | 6+6 4 | 3+4 4 | 3+4 4 | 3+4 4 |
| | 6 2 | 3+3 4 | 4+4 4 | 3+4 4 | 5+5 4 | 3+4 4 | 3+4 4 | 3+4 4 |
| | 5 2 | 3+3 4 | 4+4 4 | 3+4 4 | 6+6 4 | 3+4 4 | 3+4 4 | 3+4 4 |
| | 5 2 | 4+4 4 | 3+3 4 | 3+4 4 | 5+5 4 | 3+4 4 | 3+4 4 | 3+4 4 |
| | 5 2 | 4+4 4 | 3+3 4 | 3+4 4 | 6+6 4 | 3+4 4 | 3+4 4 | 3+4 4 |
| | 6 2 | 3+3 4 | 4+4 4 | 3+4 4 | 6+6 4 | 3+4 4 | 3+4 4 | 3+4 4 |
| | 5 2 | | | | | | | |
| | 6 2 | | | | | | | |
| | 5 2 | | | | | | | |
| | 5 2 | | | | | | | |
| | 6 2 | | | | | | | |
| | 5 2 | | | | | | | |
| | 6 2 | | | | | | | |
| | 6 2 | | | | | | | |
| | 5 2 | | | | | | | |
| | 6 2 | | | | | | | |
| | 5 2 | | | | | | | |
| | 6 2 | | | | | | | |
| | 6 2 | | | | | | | |
| | 5 2 | | | | | | | |
| 62.2.118 | 169305 | 56 | Fe1Se1 | | | | | |
| | -GM | -R | -S | -T | -U | -X | -Y | -Z |
| | 5 2 | 3+3 4 | 3+3 4 | 3+4 4 | 6+6 4 | 3+4 4 | 3+4 4 | 3+4 4 |
| | 6 2 | 4+4 4 | 4+4 4 | 3+4 4 | 5+5 4 | 3+4 4 | 3+4 4 | 3+4 4 |
| | 5 2 | 3+3 4 | 3+3 4 | 3+4 4 | 6+6 4 | 3+4 4 | 3+4 4 | 3+4 4 |

Continued on next column

**TABLE XXXVII– continued**

| TQC | ICSD NoE SF |
|---|---|
| | 6 2 4+4 4 4+4 4 3+4 4 5+5 4 3+4 4 3+4 4 3+4 4 |
| | 5 2 4+4 4 4+4 4 3+4 4 6+6 4 3+4 4 3+4 4 3+4 4 |
| | 6 2 3+3 4 3+3 4 3+4 4 6+6 4 3+4 4 3+4 4 3+4 4 |
| | 5 2 3+3 4 3+3 4 3+4 4 5+5 4 3+4 4 3+4 4 3+4 4 |
| | 6 2 4+4 4 4+4 4 3+4 4 5+5 4 3+4 4 3+4 4 3+4 4 |
| | 6 2 4+4 4 3+3 4 3+4 4 6+6 4 3+4 4 3+4 4 3+4 4 |
| | 5 2 3+3 4 4+4 4 3+4 4 5+5 4 3+4 4 3+4 4 3+4 4 |
| | 5 2 3+3 4 4+4 4 3+4 4 5+5 4 3+4 4 3+4 4 3+4 4 |
| | 5 2 4+4 4 3+3 4 3+4 4 6+6 4 3+4 4 3+4 4 3+4 4 |
| | 5 2 4+4 4 3+3 4 3+4 4 5+5 4 3+4 4 3+4 4 3+4 4 |
| | 6 2 3+3 4 4+4 4 3+4 4 6+6 4 3+4 4 3+4 4 3+4 4 |
| | 5 2 |
| | 5 2 |
| | 6 2 |
| | 5 2 |
| | 6 2 |
| | 6 2 |
| | 6 2 |
| | 5 2 |
| | 5 2 |
| | 6 2 |
| | 5 2 |
| | 6 2 |
| | 5 2 |
| | 6 2 |
| 62.2.119 | 169282 56 Fe1Se1 |
| | -GM -R -S -T -U -X -Y -Z |
| | 5 2 4+4 4 4+4 4 3+4 4 5+5 4 3+4 4 3+4 4 3+4 4 |
| | 5 2 3+3 4 3+3 4 3+4 4 6+6 4 3+4 4 3+4 4 3+4 4 |
| | 6 2 3+3 4 4+4 4 3+4 4 5+5 4 3+4 4 3+4 4 3+4 4 |
| | 6 2 4+4 4 3+3 4 3+4 4 6+6 4 3+4 4 3+4 4 3+4 4 |
| | 5 2 3+3 4 3+3 4 3+4 4 5+5 4 3+4 4 3+4 4 3+4 4 |
| | 6 2 4+4 4 4+4 4 3+4 4 5+5 4 3+4 4 3+4 4 3+4 4 |
| | 5 2 4+4 4 4+4 4 3+4 4 6+6 4 3+4 4 3+4 4 3+4 4 |
| | 6 2 3+3 4 3+3 4 3+4 4 6+6 4 3+4 4 3+4 4 3+4 4 |
| | 5 2 4+4 4 4+4 4 3+4 4 5+5 4 3+4 4 3+4 4 3+4 4 |
| | 5 2 3+3 4 3+3 4 3+4 4 6+6 4 3+4 4 3+4 4 3+4 4 |
| | 6 2 3+3 4 3+3 4 3+4 4 5+5 4 3+4 4 3+4 4 3+4 4 |
| | 5 2 4+4 4 4+4 4 3+4 4 6+6 4 3+4 4 3+4 4 3+4 4 |
| | 5 2 3+3 4 3+3 4 3+4 4 6+6 4 3+4 4 3+4 4 3+4 4 |
| | 6 2 4+4 4 4+4 4 3+4 4 5+5 4 3+4 4 3+4 4 3+4 4 |
| | 5 2 |
| | 6 2 |
| | 5 2 |
| | 5 2 |
| | 5 2 |
| | 5 2 |
| | 6 2 |
| | 6 2 |
| | 6 2 |
| | 6 2 |
| | 5 2 |
| | 6 2 |
| | 5 2 |
| | 6 2 |
| 62.2.120 | 169281 56 Fe1Se1 |
| | -GM -R -S -T -U -X -Y -Z |
| | 5 2 4+4 4 4+4 4 3+4 4 5+5 4 3+4 4 3+4 4 3+4 4 |
| | 5 2 3+3 4 3+3 4 3+4 4 6+6 4 3+4 4 3+4 4 3+4 4 |
| | 6 2 3+3 4 4+4 4 3+4 4 5+5 4 3+4 4 3+4 4 3+4 4 |
| | 6 2 4+4 4 3+3 4 3+4 4 6+6 4 3+4 4 3+4 4 3+4 4 |

Continued on next column

**TABLE XXXVII– continued**

| TQC | ICSD NoE SF |
|---|---|
| | 5 2 3+3 4 3+3 4 3+4 4 5+5 4 3+4 4 3+4 4 3+4 4 |
| | 6 2 4+4 4 4+4 4 3+4 4 5+5 4 3+4 4 3+4 4 3+4 4 |
| | 5 2 4+4 4 4+4 4 3+4 4 6+6 4 3+4 4 3+4 4 3+4 4 |
| | 6 2 3+3 4 3+3 4 3+4 4 6+6 4 3+4 4 3+4 4 3+4 4 |
| | 5 2 4+4 4 4+4 4 3+4 4 5+5 4 3+4 4 3+4 4 3+4 4 |
| | 5 2 3+3 4 3+3 4 3+4 4 6+6 4 3+4 4 3+4 4 3+4 4 |
| | 5 2 3+3 4 3+3 4 3+4 4 6+6 4 3+4 4 3+4 4 3+4 4 |
| | 6 2 4+4 4 4+4 4 3+4 4 5+5 4 3+4 4 3+4 4 3+4 4 |
| | 5 2 3+3 4 3+3 4 3+4 4 6+6 4 3+4 4 3+4 4 3+4 4 |
| | 6 2 4+4 4 4+4 4 3+4 4 5+5 4 3+4 4 3+4 4 3+4 4 |
| | 5 2 |
| | 6 2 |
| | 5 2 |
| | 5 2 |
| | 5 2 |
| | 5 2 |
| | 6 2 |
| | 6 2 |
| | 6 2 |
| | 6 2 |
| | 6 2 |
| | 5 2 |
| | 5 2 |
| | 6 2 |
| 62.2.121 | 169280 56 Fe1Se1 |
| | -GM -R -S -T -U -X -Y -Z |
| | 5 2 3+3 4 4+4 4 3+4 4 5+5 4 3+4 4 3+4 4 3+4 4 |
| | 5 2 4+4 4 3+3 4 3+4 4 6+6 4 3+4 4 3+4 4 3+4 4 |
| | 6 2 4+4 4 4+4 4 3+4 4 5+5 4 3+4 4 3+4 4 3+4 4 |
| | 6 2 3+3 4 3+3 4 3+4 4 6+6 4 3+4 4 3+4 4 3+4 4 |
| | 5 2 4+4 4 3+3 4 3+4 4 5+5 4 3+4 4 3+4 4 3+4 4 |
| | 6 2 3+3 4 4+4 4 3+4 4 5+5 4 3+4 4 3+4 4 3+4 4 |
| | 5 2 3+3 4 4+4 4 3+4 4 6+6 4 3+4 4 3+4 4 3+4 4 |
| | 6 2 4+4 4 3+3 4 3+4 4 6+6 4 3+4 4 3+4 4 3+4 4 |
| | 5 2 3+3 4 4+4 4 3+4 4 5+5 4 3+4 4 3+4 4 3+4 4 |
| | 5 2 4+4 4 3+3 4 3+4 4 6+6 4 3+4 4 3+4 4 3+4 4 |
| | 5 2 4+4 4 4+4 4 3+4 4 6+6 4 3+4 4 3+4 4 3+4 4 |
| | 6 2 3+3 4 3+3 4 3+4 4 5+5 4 3+4 4 3+4 4 3+4 4 |
| | 5 2 3+3 4 3+3 4 3+4 4 6+6 4 3+4 4 3+4 4 3+4 4 |
| | 6 2 4+4 4 4+4 4 3+4 4 5+5 4 3+4 4 3+4 4 3+4 4 |
| | 5 2 |
| | 6 2 |
| | 5 2 |
| | 5 2 |
| | 5 2 |
| | 6 2 |
| | 5 2 |
| | 6 2 |
| | 6 2 |
| | 6 2 |
| | 5 2 |
| | 6 2 |
| | 5 2 |
| | 6 2 |
| 62.2.122 | 169279 56 Fe1Se1 |
| | -GM -R -S -T -U -X -Y -Z |
| | 5 2 3+3 4 3+3 4 3+4 4 6+6 4 3+4 4 3+4 4 3+4 4 |
| | 5 2 4+4 4 4+4 4 3+4 4 5+5 4 3+4 4 3+4 4 3+4 4 |
| | 6 2 4+4 4 3+3 4 3+4 4 6+6 4 3+4 4 3+4 4 3+4 4 |
| | 6 2 3+3 4 4+4 4 3+4 4 5+5 4 3+4 4 3+4 4 3+4 4 |
| | 5 2 4+4 4 4+4 4 3+4 4 6+6 4 3+4 4 3+4 4 3+4 4 |

Continued on next column

**TABLE XXXVII– continued**

| TQC | ICSD | NoE | SF | | | | | |
|---|---|---|---|---|---|---|---|---|
| | 6 2 | 3+3 4 | 3+3 4 | 3+4 4 | 6+6 4 | 3+4 4 | 3+4 4 | 3+4 4 |
| | 5 2 | 3+3 4 | 3+3 4 | 3+4 4 | 5+5 4 | 3+4 4 | 3+4 4 | 3+4 4 |
| | 6 2 | 4+4 4 | 4+4 4 | 3+4 4 | 5+5 4 | 3+4 4 | 3+4 4 | 3+4 4 |
| | 5 2 | 3+3 4 | 3+3 4 | 3+4 4 | 6+6 4 | 3+4 4 | 3+4 4 | 3+4 4 |
| | 5 2 | 4+4 4 | 4+4 4 | 3+4 4 | 5+5 4 | 3+4 4 | 3+4 4 | 3+4 4 |
| | 6 2 | 4+4 4 | 4+4 4 | 3+4 4 | 5+5 4 | 3+4 4 | 3+4 4 | 3+4 4 |
| | 5 2 | 3+3 4 | 3+3 4 | 3+4 4 | 6+6 4 | 3+4 4 | 3+4 4 | 3+4 4 |
| | 5 2 | 4+4 4 | 4+4 4 | 3+4 4 | 5+5 4 | 3+4 4 | 3+4 4 | 3+4 4 |
| | 6 2 | 3+3 4 | 3+3 4 | 3+4 4 | 6+6 4 | 3+4 4 | 3+4 4 | 3+4 4 |
| | 5 2 | | | | | | | |
| | 6 2 | | | | | | | |
| | 5 2 | | | | | | | |
| | 5 2 | | | | | | | |
| | 6 2 | | | | | | | |
| | 5 2 | | | | | | | |
| | 6 2 | | | | | | | |
| | 5 2 | | | | | | | |
| | 6 2 | | | | | | | |
| | 6 2 | | | | | | | |
| | 6 2 | | | | | | | |
| | 5 2 | | | | | | | |
| | 5 2 | | | | | | | |
| | 6 2 | | | | | | | |
| 62.2.123 | 169278 | 56 | Fe1Se1 | | | | | |
| | -GM | -R | -S | -T | -U | -X | -Y | -Z |
| | 5 2 | 4+4 4 | 4+4 4 | 3+4 4 | 5+5 4 | 3+4 4 | 3+4 4 | 3+4 4 |
| | 5 2 | 3+3 4 | 3+3 4 | 3+4 4 | 6+6 4 | 3+4 4 | 3+4 4 | 3+4 4 |
| | 6 2 | 3+3 4 | 4+4 4 | 3+4 4 | 5+5 4 | 3+4 4 | 3+4 4 | 3+4 4 |
| | 6 2 | 4+4 4 | 3+3 4 | 3+4 4 | 6+6 4 | 3+4 4 | 3+4 4 | 3+4 4 |
| | 5 2 | 3+3 4 | 3+3 4 | 3+4 4 | 5+5 4 | 3+4 4 | 3+4 4 | 3+4 4 |
| | 6 2 | 4+4 4 | 4+4 4 | 3+4 4 | 5+5 4 | 3+4 4 | 3+4 4 | 3+4 4 |
| | 5 2 | 4+4 4 | 4+4 4 | 3+4 4 | 6+6 4 | 3+4 4 | 3+4 4 | 3+4 4 |
| | 6 2 | 3+3 4 | 3+3 4 | 3+4 4 | 6+6 4 | 3+4 4 | 3+4 4 | 3+4 4 |
| | 5 2 | 4+4 4 | 4+4 4 | 3+4 4 | 5+5 4 | 3+4 4 | 3+4 4 | 3+4 4 |
| | 5 2 | 3+3 4 | 3+3 4 | 3+4 4 | 6+6 4 | 3+4 4 | 3+4 4 | 3+4 4 |
| | 6 2 | 3+3 4 | 3+3 4 | 3+4 4 | 6+6 4 | 3+4 4 | 3+4 4 | 3+4 4 |
| | 5 2 | 4+4 4 | 4+4 4 | 3+4 4 | 5+5 4 | 3+4 4 | 3+4 4 | 3+4 4 |
| | 5 2 | 3+3 4 | 3+3 4 | 3+4 4 | 6+6 4 | 3+4 4 | 3+4 4 | 3+4 4 |
| | 6 2 | 4+4 4 | 4+4 4 | 3+4 4 | 5+5 4 | 3+4 4 | 3+4 4 | 3+4 4 |
| | 5 2 | | | | | | | |
| | 6 2 | | | | | | | |
| | 5 2 | | | | | | | |
| | 5 2 | | | | | | | |
| | 5 2 | | | | | | | |
| | 5 2 | | | | | | | |
| | 6 2 | | | | | | | |
| | 6 2 | | | | | | | |
| | 6 2 | | | | | | | |
| | 6 2 | | | | | | | |
| | 6 2 | | | | | | | |
| | 5 2 | | | | | | | |
| | 5 2 | | | | | | | |
| | 6 2 | | | | | | | |
| 62.2.124 | 169277 | 56 | Fe1Se1 | | | | | |
| | -GM | -R | -S | -T | -U | -X | -Y | -Z |
| | 5 2 | 4+4 4 | 3+3 4 | 3+4 4 | 6+6 4 | 3+4 4 | 3+4 4 | 3+4 4 |
| | 6 2 | 3+3 4 | 4+4 4 | 3+4 4 | 5+5 4 | 3+4 4 | 3+4 4 | 3+4 4 |
| | 5 2 | 3+3 4 | 3+3 4 | 3+4 4 | 6+6 4 | 3+4 4 | 3+4 4 | 3+4 4 |
| | 6 2 | 4+4 4 | 4+4 4 | 3+4 4 | 5+5 4 | 3+4 4 | 3+4 4 | 3+4 4 |
| | 5 2 | 3+3 4 | 4+4 4 | 3+4 4 | 6+6 4 | 3+4 4 | 3+4 4 | 3+4 4 |
| | 6 2 | 4+4 4 | 3+3 4 | 3+4 4 | 6+6 4 | 3+4 4 | 3+4 4 | 3+4 4 |

Continued on next column

**TABLE XXXVII– continued**

| TQC | ICSD | NoE | SF | | | | | |
|---|---|---|---|---|---|---|---|---|
| | 5 2 | 4+4 4 | 3+3 4 | 3+4 4 | 5+5 4 | 3+4 4 | 3+4 4 | 3+4 4 |
| | 6 2 | 3+3 4 | 4+4 4 | 3+4 4 | 5+5 4 | 3+4 4 | 3+4 4 | 3+4 4 |
| | 5 2 | 4+4 4 | 3+3 4 | 3+4 4 | 6+6 4 | 3+4 4 | 3+4 4 | 3+4 4 |
| | 5 2 | 3+3 4 | 4+4 4 | 3+4 4 | 5+5 4 | 3+4 4 | 3+4 4 | 3+4 4 |
| | 6 2 | 3+3 4 | 3+3 4 | 3+4 4 | 5+5 4 | 3+4 4 | 3+4 4 | 3+4 4 |
| | 5 2 | 4+4 4 | 4+4 4 | 3+4 4 | 6+6 4 | 3+4 4 | 3+4 4 | 3+4 4 |
| | 5 2 | 3+3 4 | 4+4 4 | 3+4 4 | 5+5 4 | 3+4 4 | 3+4 4 | 3+4 4 |
| | 6 2 | 4+4 4 | 3+3 4 | 3+4 4 | 6+6 4 | 3+4 4 | 3+4 4 | 3+4 4 |
| | 5 2 | | | | | | | |
| | 6 2 | | | | | | | |
| | 5 2 | | | | | | | |
| | 5 2 | | | | | | | |
| | 5 2 | | | | | | | |
| | 6 2 | | | | | | | |
| | 5 2 | | | | | | | |
| | 6 2 | | | | | | | |
| | 6 2 | | | | | | | |
| | 6 2 | | | | | | | |
| | 5 2 | | | | | | | |
| | 6 2 | | | | | | | |
| | 5 2 | | | | | | | |
| | 6 2 | | | | | | | |
| 62.2.125 | 169276 | 56 | Fe1Se1 | | | | | |
| | -GM | -R | -S | -T | -U | -X | -Y | -Z |
| | 5 2 | 3+3 4 | 3+3 4 | 3+4 4 | 6+6 4 | 3+4 4 | 3+4 4 | 3+4 4 |
| | 5 2 | 4+4 4 | 4+4 4 | 3+4 4 | 5+5 4 | 3+4 4 | 3+4 4 | 3+4 4 |
| | 6 2 | 4+4 4 | 3+3 4 | 3+4 4 | 6+6 4 | 3+4 4 | 3+4 4 | 3+4 4 |
| | 6 2 | 3+3 4 | 4+4 4 | 3+4 4 | 5+5 4 | 3+4 4 | 3+4 4 | 3+4 4 |
| | 5 2 | 3+3 4 | 4+4 4 | 3+4 4 | 6+6 4 | 3+4 4 | 3+4 4 | 3+4 4 |
| | 6 2 | 4+4 4 | 3+3 4 | 3+4 4 | 6+6 4 | 3+4 4 | 3+4 4 | 3+4 4 |
| | 5 2 | 4+4 4 | 3+3 4 | 3+4 4 | 5+5 4 | 3+4 4 | 3+4 4 | 3+4 4 |
| | 6 2 | 3+3 4 | 4+4 4 | 3+4 4 | 5+5 4 | 3+4 4 | 3+4 4 | 3+4 4 |
| | 5 2 | 4+4 4 | 3+3 4 | 3+4 4 | 6+6 4 | 3+4 4 | 3+4 4 | 3+4 4 |
| | 6 2 | 3+3 4 | 4+4 4 | 3+4 4 | 5+5 4 | 3+4 4 | 3+4 4 | 3+4 4 |
| | 5 2 | 3+3 4 | 4+4 4 | 3+4 4 | 5+5 4 | 3+4 4 | 3+4 4 | 3+4 4 |
| | 5 2 | 4+4 4 | 3+3 4 | 3+4 4 | 6+6 4 | 3+4 4 | 3+4 4 | 3+4 4 |
| | 5 2 | 3+3 4 | 3+3 4 | 3+4 4 | 5+5 4 | 3+4 4 | 3+4 4 | 3+4 4 |
| | 6 2 | 4+4 4 | 4+4 4 | 3+4 4 | 5+5 4 | 3+4 4 | 3+4 4 | 3+4 4 |
| | 5 2 | | | | | | | |
| | 6 2 | | | | | | | |
| | 5 2 | | | | | | | |
| | 6 2 | | | | | | | |
| | 5 2 | | | | | | | |
| | 5 2 | | | | | | | |
| | 5 2 | | | | | | | |
| | 6 2 | | | | | | | |
| | 6 2 | | | | | | | |
| | 6 2 | | | | | | | |
| | 6 2 | | | | | | | |
| | 5 2 | | | | | | | |
| | 6 2 | | | | | | | |
| | 5 2 | | | | | | | |
| 62.2.126 | 169275 | 56 | Fe1Se1 | | | | | |
| | -GM | -R | -S | -T | -U | -X | -Y | -Z |
| | 5 2 | 3+3 4 | 4+4 4 | 3+4 4 | 6+6 4 | 3+4 4 | 3+4 4 | 3+4 4 |
| | 5 2 | 4+4 4 | 3+3 4 | 3+4 4 | 5+5 4 | 3+4 4 | 3+4 4 | 3+4 4 |
| | 6 2 | 3+3 4 | 4+4 4 | 3+4 4 | 6+6 4 | 3+4 4 | 3+4 4 | 3+4 4 |
| | 6 2 | 4+4 4 | 3+3 4 | 3+4 4 | 5+5 4 | 3+4 4 | 3+4 4 | 3+4 4 |
| | 5 2 | 4+4 4 | 3+3 4 | 3+4 4 | 6+6 4 | 3+4 4 | 3+4 4 | 3+4 4 |
| | 6 2 | 3+3 4 | 4+4 4 | 3+4 4 | 6+6 4 | 3+4 4 | 3+4 4 | 3+4 4 |
| | 5 2 | 3+3 4 | 4+4 4 | 3+4 4 | 5+5 4 | 3+4 4 | 3+4 4 | 3+4 4 |

Continued on next column

**TABLE XXXVII– continued**

| TQC | ICSD | NoE | SF | | | | |
|---|---|---|---|---|---|---|---|
| | 5 2 | 4+4 4 | 3+3 4 | 3+4 4 | 5+5 4 | 3+4 4 | 3+4 4 | 3+4 4 |
| | 6 2 | 3+3 4 | 4+4 4 | 3+4 4 | 6+6 4 | 3+4 4 | 3+4 4 | 3+4 4 |
| | 6 2 | 4+4 4 | 3+3 4 | 3+4 4 | 5+5 4 | 3+4 4 | 3+4 4 | 3+4 4 |
| | 5 2 | 4+4 4 | 3+3 4 | 3+4 4 | 5+5 4 | 3+4 4 | 3+4 4 | 3+4 4 |
| | 5 2 | 3+3 4 | 4+4 4 | 3+4 4 | 6+6 4 | 3+4 4 | 3+4 4 | 3+4 4 |
| | 5 2 | 3+3 4 | 4+4 4 | 3+4 4 | 5+5 4 | 3+4 4 | 3+4 4 | 3+4 4 |
| | 6 2 | 4+4 4 | 3+3 4 | 3+4 4 | 5+5 4 | 3+4 4 | 3+4 4 | 3+4 4 |
| | 5 2 | | | | | | | |
| | 6 2 | | | | | | | |
| | 5 2 | | | | | | | |
| | 6 2 | | | | | | | |
| | 5 2 | | | | | | | |
| | 5 2 | | | | | | | |
| | 5 2 | | | | | | | |
| | 6 2 | | | | | | | |
| | 6 2 | | | | | | | |
| | 6 2 | | | | | | | |
| | 5 2 | | | | | | | |
| | 6 2 | | | | | | | |
| | 6 2 | | | | | | | |
| | 5 2 | | | | | | | |
| 62.2.127 | 169274 | 56 | Fe1Se1 | | | | | |
| | -GM | -R | -S | -T | -U | -X | -Y | -Z |
| | 5 2 | 3+3 4 | 4+4 4 | 3+4 4 | 6+6 4 | 3+4 4 | 3+4 4 | 3+4 4 |
| | 6 2 | 4+4 4 | 3+3 4 | 3+4 4 | 5+5 4 | 3+4 4 | 3+4 4 | 3+4 4 |
| | 6 2 | 3+3 4 | 4+4 4 | 3+4 4 | 6+6 4 | 3+4 4 | 3+4 4 | 3+4 4 |
| | 5 2 | 4+4 4 | 3+3 4 | 3+4 4 | 5+5 4 | 3+4 4 | 3+4 4 | 3+4 4 |
| | 5 2 | 4+4 4 | 3+3 4 | 3+4 4 | 6+6 4 | 3+4 4 | 3+4 4 | 3+4 4 |
| | 6 2 | 3+3 4 | 4+4 4 | 3+4 4 | 6+6 4 | 3+4 4 | 3+4 4 | 3+4 4 |
| | 5 2 | 3+3 4 | 4+4 4 | 3+4 4 | 5+5 4 | 3+4 4 | 3+4 4 | 3+4 4 |
| | 6 2 | 4+4 4 | 3+3 4 | 3+4 4 | 5+5 4 | 3+4 4 | 3+4 4 | 3+4 4 |
| | 6 2 | 3+3 4 | 4+4 4 | 3+4 4 | 6+6 4 | 3+4 4 | 3+4 4 | 3+4 4 |
| | 5 2 | 4+4 4 | 3+3 4 | 3+4 4 | 5+5 4 | 3+4 4 | 3+4 4 | 3+4 4 |
| | 5 2 | 4+4 4 | 3+3 4 | 3+4 4 | 5+5 4 | 3+4 4 | 3+4 4 | 3+4 4 |
| | 5 2 | 3+3 4 | 4+4 4 | 3+4 4 | 6+6 4 | 3+4 4 | 3+4 4 | 3+4 4 |
| | 5 2 | 4+4 4 | 4+4 4 | 3+4 4 | 5+5 4 | 3+4 4 | 3+4 4 | 3+4 4 |
| | 6 2 | 3+3 4 | 3+3 4 | 3+4 4 | 6+6 4 | 3+4 4 | 3+4 4 | 3+4 4 |
| | 5 2 | | | | | | | |
| | 6 2 | | | | | | | |
| | 5 2 | | | | | | | |
| | 6 2 | | | | | | | |
| | 5 2 | | | | | | | |
| | 5 2 | | | | | | | |
| | 5 2 | | | | | | | |
| | 6 2 | | | | | | | |
| | 6 2 | | | | | | | |
| | 5 2 | | | | | | | |
| | 6 2 | | | | | | | |
| | 6 2 | | | | | | | |
| | 6 2 | | | | | | | |
| | 5 2 | | | | | | | |
| 62.2.128 | 169273 | 56 | Fe1Se1 | | | | | |
| | -GM | -R | -S | -T | -U | -X | -Y | -Z |
| | 5 2 | 3+3 4 | 4+4 4 | 3+4 4 | 5+5 4 | 3+4 4 | 3+4 4 | 3+4 4 |
| | 6 2 | 4+4 4 | 3+3 4 | 3+4 4 | 6+6 4 | 3+4 4 | 3+4 4 | 3+4 4 |
| | 6 2 | 3+3 4 | 4+4 4 | 3+4 4 | 6+6 4 | 3+4 4 | 3+4 4 | 3+4 4 |
| | 5 2 | 4+4 4 | 3+3 4 | 3+4 4 | 5+5 4 | 3+4 4 | 3+4 4 | 3+4 4 |
| | 5 2 | 4+4 4 | 3+3 4 | 3+4 4 | 6+6 4 | 3+4 4 | 3+4 4 | 3+4 4 |
| | 6 2 | 3+3 4 | 4+4 4 | 3+4 4 | 5+5 4 | 3+4 4 | 3+4 4 | 3+4 4 |
| | 5 2 | 3+3 4 | 4+4 4 | 3+4 4 | 5+5 4 | 3+4 4 | 3+4 4 | 3+4 4 |
| | 6 2 | 4+4 4 | 3+3 4 | 3+4 4 | 6+6 4 | 3+4 4 | 3+4 4 | 3+4 4 |

Continued on next column

**TABLE XXXVII– continued**

| TQC | ICSD | NoE | SF | | | | | |
|---|---|---|---|---|---|---|---|---|
| | 6 2 | 3+3 4 | 4+4 4 | 3+4 4 | 5+5 4 | 3+4 4 | 3+4 4 | 3+4 4 |
| | 5 2 | 4+4 4 | 3+3 4 | 3+4 4 | 6+6 4 | 3+4 4 | 3+4 4 | 3+4 4 |
| | 5 2 | 4+4 4 | 3+3 4 | 3+4 4 | 5+5 4 | 3+4 4 | 3+4 4 | 3+4 4 |
| | 5 2 | 3+3 4 | 4+4 4 | 3+4 4 | 6+6 4 | 3+4 4 | 3+4 4 | 3+4 4 |
| | 5 2 | 4+4 4 | 4+4 4 | 3+4 4 | 5+5 4 | 3+4 4 | 3+4 4 | 3+4 4 |
| | 6 2 | 3+3 4 | 3+3 4 | 3+4 4 | 6+6 4 | 3+4 4 | 3+4 4 | 3+4 4 |
| | 6 2 | | | | | | | |
| | 5 2 | | | | | | | |
| | 5 2 | | | | | | | |
| | 6 2 | | | | | | | |
| | 5 2 | | | | | | | |
| | 5 2 | | | | | | | |
| | 5 2 | | | | | | | |
| | 6 2 | | | | | | | |
| | 5 2 | | | | | | | |
| | 6 2 | | | | | | | |
| | 6 2 | | | | | | | |
| | 6 2 | | | | | | | |
| | 5 2 | | | | | | | |
| | 6 2 | | | | | | | |
| 62.2.129 | 35010 | 56 | Fe1S1 | | | | | |
| | -GM | -R | -S | -T | -U | -X | -Y | -Z |
| | 5 2 | 3+3 4 | 4+4 4 | 3+4 4 | 6+6 4 | 3+4 4 | 3+4 4 | 3+4 4 |
| | 6 2 | 4+4 4 | 3+3 4 | 3+4 4 | 5+5 4 | 3+4 4 | 3+4 4 | 3+4 4 |
| | 5 2 | 3+3 4 | 3+3 4 | 3+4 4 | 6+6 4 | 3+4 4 | 3+4 4 | 3+4 4 |
| | 6 2 | 4+4 4 | 4+4 4 | 3+4 4 | 5+5 4 | 3+4 4 | 3+4 4 | 3+4 4 |
| | 5 2 | 4+4 4 | 3+3 4 | 3+4 4 | 6+6 4 | 3+4 4 | 3+4 4 | 3+4 4 |
| | 6 2 | 3+3 4 | 4+4 4 | 3+4 4 | 6+6 4 | 3+4 4 | 3+4 4 | 3+4 4 |
| | 5 2 | 3+3 4 | 3+3 4 | 3+4 4 | 5+5 4 | 3+4 4 | 3+4 4 | 3+4 4 |
| | 5 2 | 4+4 4 | 4+4 4 | 3+4 4 | 5+5 4 | 3+4 4 | 3+4 4 | 3+4 4 |
| | 6 2 | 3+3 4 | 4+4 4 | 3+4 4 | 6+6 4 | 3+4 4 | 3+4 4 | 3+4 4 |
| | 6 2 | 4+4 4 | 3+3 4 | 3+4 4 | 5+5 4 | 3+4 4 | 3+4 4 | 3+4 4 |
| | 5 2 | 4+4 4 | 4+4 4 | 3+4 4 | 5+5 4 | 3+4 4 | 3+4 4 | 3+4 4 |
| | 5 2 | 3+3 4 | 3+3 4 | 3+4 4 | 6+6 4 | 3+4 4 | 3+4 4 | 3+4 4 |
| | 5 2 | 3+3 4 | 3+3 4 | 3+4 4 | 5+5 4 | 3+4 4 | 3+4 4 | 3+4 4 |
| | 6 2 | 4+4 4 | 4+4 4 | 3+4 4 | 5+5 4 | 3+4 4 | 3+4 4 | 3+4 4 |
| | 5 2 | | | | | | | |
| | 6 2 | | | | | | | |
| | 6 2 | | | | | | | |
| | 6 2 | | | | | | | |
| | 5 2 | | | | | | | |
| | 5 2 | | | | | | | |
| | 6 2 | | | | | | | |
| | 5 2 | | | | | | | |
| | 5 2 | | | | | | | |
| | 6 2 | | | | | | | |
| | 6 2 | | | | | | | |
| | 5 2 | | | | | | | |
| | 5 2 | | | | | | | |
| | 6 2 | | | | | | | |
| 62.2.130 | 35009 | 56 | Fe1S1 | | | | | |
| | -GM | -R | -S | -T | -U | -X | -Y | -Z |
| | 5 2 | 3+3 4 | 4+4 4 | 3+4 4 | 6+6 4 | 3+4 4 | 3+4 4 | 3+4 4 |
| | 6 2 | 4+4 4 | 3+3 4 | 3+4 4 | 5+5 4 | 3+4 4 | 3+4 4 | 3+4 4 |
| | 5 2 | 3+3 4 | 3+3 4 | 3+4 4 | 6+6 4 | 3+4 4 | 3+4 4 | 3+4 4 |
| | 6 2 | 4+4 4 | 4+4 4 | 3+4 4 | 5+5 4 | 3+4 4 | 3+4 4 | 3+4 4 |
| | 5 2 | 4+4 4 | 3+3 4 | 3+4 4 | 6+6 4 | 3+4 4 | 3+4 4 | 3+4 4 |
| | 6 2 | 3+3 4 | 4+4 4 | 3+4 4 | 6+6 4 | 3+4 4 | 3+4 4 | 3+4 4 |
| | 5 2 | 3+3 4 | 3+3 4 | 3+4 4 | 5+5 4 | 3+4 4 | 3+4 4 | 3+4 4 |
| | 5 2 | 4+4 4 | 4+4 4 | 3+4 4 | 5+5 4 | 3+4 4 | 3+4 4 | 3+4 4 |
| | 6 2 | 3+3 4 | 4+4 4 | 3+4 4 | 6+6 4 | 3+4 4 | 3+4 4 | 3+4 4 |

Continued on next column

**TABLE XXXVII– continued**

| TQC | ICSD | NoE | SF | | | | | |
|---|---|---|---|---|---|---|---|---|
| | 6 2 | 4+4 4 | 3+3 4 | 3+4 4 | 5+5 4 | 3+4 4 | 3+4 4 | 3+4 4 |
| | 5 2 | 4+4 4 | 4+4 4 | 3+4 4 | 5+5 4 | 3+4 4 | 3+4 4 | 3+4 4 |
| | 5 2 | 3+3 4 | 3+3 4 | 3+4 4 | 6+6 4 | 3+4 4 | 3+4 4 | 3+4 4 |
| | 5 2 | 3+3 4 | 3+3 4 | 3+4 4 | 5+5 4 | 3+4 4 | 3+4 4 | 3+4 4 |
| | 6 2 | 4+4 4 | 4+4 4 | 3+4 4 | 5+5 4 | 3+4 4 | 3+4 4 | 3+4 4 |
| | 5 2 | | | | | | | |
| | 6 2 | | | | | | | |
| | 6 2 | | | | | | | |
| | 6 2 | | | | | | | |
| | 5 2 | | | | | | | |
| | 5 2 | | | | | | | |
| | 6 2 | | | | | | | |
| | 5 2 | | | | | | | |
| | 5 2 | | | | | | | |
| | 6 2 | | | | | | | |
| | 5 2 | | | | | | | |
| | 6 2 | | | | | | | |
| | 5 2 | | | | | | | |
| | 6 2 | | | | | | | |
| 62.2.131 | 35008 | 56 | Fe1S1 | | | | | |
| | -GM | -R | -S | -T | -U | -X | -Y | -Z |
| | 5 2 | 3+3 4 | 4+4 4 | 3+4 4 | 6+6 4 | 3+4 4 | 3+4 4 | 3+4 4 |
| | 6 2 | 4+4 4 | 3+3 4 | 3+4 4 | 5+5 4 | 3+4 4 | 3+4 4 | 3+4 4 |
| | 5 2 | 3+3 4 | 3+3 4 | 3+4 4 | 6+6 4 | 3+4 4 | 3+4 4 | 3+4 4 |
| | 6 2 | 4+4 4 | 4+4 4 | 3+4 4 | 5+5 4 | 3+4 4 | 3+4 4 | 3+4 4 |
| | 5 2 | 4+4 4 | 3+3 4 | 3+4 4 | 6+6 4 | 3+4 4 | 3+4 4 | 3+4 4 |
| | 6 2 | 3+3 4 | 4+4 4 | 3+4 4 | 5+5 4 | 3+4 4 | 3+4 4 | 3+4 4 |
| | 5 2 | 3+3 4 | 4+4 4 | 3+4 4 | 6+6 4 | 3+4 4 | 3+4 4 | 3+4 4 |
| | 5 2 | 4+4 4 | 3+3 4 | 3+4 4 | 5+5 4 | 3+4 4 | 3+4 4 | 3+4 4 |
| | 6 2 | 3+3 4 | 4+4 4 | 3+4 4 | 6+6 4 | 3+4 4 | 3+4 4 | 3+4 4 |
| | 6 2 | 4+4 4 | 3+3 4 | 3+4 4 | 5+5 4 | 3+4 4 | 3+4 4 | 3+4 4 |
| | 5 2 | 4+4 4 | 4+4 4 | 3+4 4 | 5+5 4 | 3+4 4 | 3+4 4 | 3+4 4 |
| | 5 2 | 3+3 4 | 3+3 4 | 3+4 4 | 6+6 4 | 3+4 4 | 3+4 4 | 3+4 4 |
| | 5 2 | 3+3 4 | 3+3 4 | 3+4 4 | 5+5 4 | 3+4 4 | 3+4 4 | 3+4 4 |
| | 5 2 | 4+4 4 | 4+4 4 | 3+4 4 | 6+6 4 | 3+4 4 | 3+4 4 | 3+4 4 |
| | 6 2 | | | | | | | |
| | 6 2 | | | | | | | |
| | 6 2 | | | | | | | |
| | 6 2 | | | | | | | |
| | 5 2 | | | | | | | |
| | 5 2 | | | | | | | |
| | 5 2 | | | | | | | |
| | 5 2 | | | | | | | |
| | 6 2 | | | | | | | |
| | 6 2 | | | | | | | |
| | 5 2 | | | | | | | |
| | 6 2 | | | | | | | |
| | 5 2 | | | | | | | |
| | 6 2 | | | | | | | |
| 62.2.132 | 49729 | 72 | Fe1P1Ta1 | | | | | |
| | -GM | -R | -S | -T | -U | -X | -Y | -Z |
| | 5 2 | 3+3 4 | 3+3 4 | 3+4 4 | 5+5 4 | 3+4 4 | 3+4 4 | 3+4 4 |
| | 6 2 | 4+4 4 | 4+4 4 | 3+4 4 | 6+6 4 | 3+4 4 | 3+4 4 | 3+4 4 |
| | 5 2 | 4+4 4 | 4+4 4 | 3+4 4 | 5+5 4 | 3+4 4 | 3+4 4 | 3+4 4 |
| | 6 2 | 3+3 4 | 3+3 4 | 3+4 4 | 6+6 4 | 3+4 4 | 3+4 4 | 3+4 4 |
| | 5 2 | 4+4 4 | 4+4 4 | 3+4 4 | 6+6 4 | 3+4 4 | 3+4 4 | 3+4 4 |
| | 6 2 | 3+3 4 | 3+3 4 | 3+4 4 | 6+6 4 | 3+4 4 | 3+4 4 | 3+4 4 |
| | 6 2 | 4+4 4 | 4+4 4 | 3+4 4 | 5+5 4 | 3+4 4 | 3+4 4 | 3+4 4 |
| | 5 2 | 3+3 4 | 3+3 4 | 3+4 4 | 5+5 4 | 3+4 4 | 3+4 4 | 3+4 4 |
| | 5 2 | 3+3 4 | 3+3 4 | 3+4 4 | 5+5 4 | 3+4 4 | 3+4 4 | 3+4 4 |
| | 5 2 | 4+4 4 | 4+4 4 | 3+4 4 | 6+6 4 | 3+4 4 | 3+4 4 | 3+4 4 |

Continued on next column

| TQC | ICSD | NoE | SF | | | | | |
|---|---|---|---|---|---|---|---|---|
| | 6 2 | 3+3 4 | 3+3 4 | 3+4 4 | 5+5 4 | 3+4 4 | 3+4 4 | 3+4 4 |
| | 6 2 | 4+4 4 | 4+4 4 | 3+4 4 | 6+6 4 | 3+4 4 | 3+4 4 | 3+4 4 |
| | 6 2 | 3+3 4 | 4+4 4 | 3+4 4 | 5+5 4 | 3+4 4 | 3+4 4 | 3+4 4 |
| | 6 2 | 4+4 4 | 3+3 4 | 3+4 4 | 6+6 4 | 3+4 4 | 3+4 4 | 3+4 4 |
| | 6 2 | 3+3 4 | 3+3 4 | 3+4 4 | 5+5 4 | 3+4 4 | 3+4 4 | 3+4 4 |
| | 5 2 | 4+4 4 | 4+4 4 | 3+4 4 | 5+5 4 | 3+4 4 | 3+4 4 | 3+4 4 |
| | 5 2 | 4+4 4 | 3+3 4 | 3+4 4 | 6+6 4 | 3+4 4 | 3+4 4 | 3+4 4 |
| | 6 2 | 3+3 4 | 4+4 4 | 3+4 4 | 6+6 4 | 3+4 4 | 3+4 4 | 3+4 4 |
| | 5 2 | | | | | | | |
| | 5 2 | | | | | | | |
| | 6 2 | | | | | | | |
| | 6 2 | | | | | | | |
| | 5 2 | | | | | | | |
| | 5 2 | | | | | | | |
| | 5 2 | | | | | | | |
| | 6 2 | | | | | | | |
| | 5 2 | | | | | | | |
| | 5 2 | | | | | | | |
| | 5 2 | | | | | | | |
| | 6 2 | | | | | | | |
| | 5 2 | | | | | | | |
| | 6 2 | | | | | | | |
| | 5 2 | | | | | | | |
| | 6 2 | | | | | | | |
| | 5 2 | | | | | | | |
| | 5 2 | | | | | | | |
| 62.2.133 | 153236 | 160 | Fe1Nd1O3 | | | | | |
| | -GM | -R | -S | -T | -U | -X | -Y | -Z |
| | 5 2 | 3+3 4 | 3+3 4 | 3+4 4 | 5+5 4 | 3+4 4 | 3+4 4 | 3+4 4 |
| | 5 2 | 4+4 4 | 4+4 4 | 3+4 4 | 6+6 4 | 3+4 4 | 3+4 4 | 3+4 4 |
| | 6 2 | 3+3 4 | 4+4 4 | 3+4 4 | 5+5 4 | 3+4 4 | 3+4 4 | 3+4 4 |
| | 6 2 | 4+4 4 | 3+3 4 | 3+4 4 | 6+6 4 | 3+4 4 | 3+4 4 | 3+4 4 |
| | 5 2 | 4+4 4 | 3+3 4 | 3+4 4 | 5+5 4 | 3+4 4 | 3+4 4 | 3+4 4 |
| | 5 2 | 3+3 4 | 4+4 4 | 3+4 4 | 6+6 4 | 3+4 4 | 3+4 4 | 3+4 4 |
| | 6 2 | 3+3 4 | 4+4 4 | 3+4 4 | 5+5 4 | 3+4 4 | 3+4 4 | 3+4 4 |
| | 6 2 | 4+4 4 | 3+3 4 | 3+4 4 | 6+6 4 | 3+4 4 | 3+4 4 | 3+4 4 |
| | 5 2 | 4+4 4 | 4+4 4 | 3+4 4 | 5+5 4 | 3+4 4 | 3+4 4 | 3+4 4 |
| | 5 2 | 3+3 4 | 3+3 4 | 3+4 4 | 5+5 4 | 3+4 4 | 3+4 4 | 3+4 4 |
| | 6 2 | 4+4 4 | 3+3 4 | 3+4 4 | 6+6 4 | 3+4 4 | 3+4 4 | 3+4 4 |
| | 5 2 | 3+3 4 | 4+4 4 | 3+4 4 | 6+6 4 | 3+4 4 | 3+4 4 | 3+4 4 |
| | 5 2 | 3+3 4 | 4+4 4 | 3+4 4 | 5+5 4 | 3+4 4 | 3+4 4 | 3+4 4 |
| | 6 2 | 4+4 4 | 3+3 4 | 3+4 4 | 6+6 4 | 3+4 4 | 3+4 4 | 3+4 4 |
| | 6 2 | 4+4 4 | 3+3 4 | 3+4 4 | 5+5 4 | 3+4 4 | 3+4 4 | 3+4 4 |
| | 6 2 | 3+3 4 | 4+4 4 | 3+4 4 | 5+5 4 | 3+4 4 | 3+4 4 | 3+4 4 |
| | 5 2 | 3+3 4 | 4+4 4 | 3+4 4 | 5+5 4 | 3+4 4 | 3+4 4 | 3+4 4 |
| | 5 2 | 4+4 4 | 3+3 4 | 3+4 4 | 5+5 4 | 3+4 4 | 3+4 4 | 3+4 4 |
| | 5 2 | 3+3 4 | 3+3 4 | 3+4 4 | 5+5 4 | 3+4 4 | 3+4 4 | 3+4 4 |
| | 6 2 | 4+4 4 | 4+4 4 | 3+4 4 | 5+5 4 | 3+4 4 | 3+4 4 | 3+4 4 |
| | 5 2 | 3+3 4 | 4+4 4 | 3+4 4 | 6+6 4 | 3+4 4 | 3+4 4 | 3+4 4 |
| | 6 2 | 4+4 4 | 3+3 4 | 3+4 4 | 6+6 4 | 3+4 4 | 3+4 4 | 3+4 4 |
| | 6 2 | 3+3 4 | 3+3 4 | 3+4 4 | 5+5 4 | 3+4 4 | 3+4 4 | 3+4 4 |
| | 6 2 | 4+4 4 | 4+4 4 | 3+4 4 | 6+6 4 | 3+4 4 | 3+4 4 | 3+4 4 |
| | 5 2 | 4+4 4 | 3+3 4 | 3+4 4 | 5+5 4 | 3+4 4 | 3+4 4 | 3+4 4 |
| | 6 2 | 3+3 4 | 4+4 4 | 3+4 4 | 6+6 4 | 3+4 4 | 3+4 4 | 3+4 4 |
| | 5 2 | 3+3 4 | 3+3 4 | 3+4 4 | 5+5 4 | 3+4 4 | 3+4 4 | 3+4 4 |
| | 6 2 | 4+4 4 | 4+4 4 | 3+4 4 | 6+6 4 | 3+4 4 | 3+4 4 | 3+4 4 |
| | 5 2 | 3+3 4 | 4+4 4 | 3+4 4 | 6+6 4 | 3+4 4 | 3+4 4 | 3+4 4 |
| | 5 2 | 4+4 4 | 3+3 4 | 3+4 4 | 5+5 4 | 3+4 4 | 3+4 4 | 3+4 4 |
| | 5 2 | 4+4 4 | 4+4 4 | 3+4 4 | 5+5 4 | 3+4 4 | 3+4 4 | 3+4 4 |
| | 5 2 | 3+3 4 | 3+3 4 | 3+4 4 | 6+6 4 | 3+4 4 | 3+4 4 | 3+4 4 |
| | 5 2 | 3+3 4 | 4+4 4 | 3+4 4 | 5+5 4 | 3+4 4 | 3+4 4 | 3+4 4 |

Continued on next column

**TABLE XXXVII– continued**

| TQC | ICSD | NoE | SF | | | | | |
|---|---|---|---|---|---|---|---|---|
| | 5 2 | 4+4 4 | 3+3 4 | 3+4 4 | 5+5 4 | 3+4 4 | 3+4 4 | 3+4 4 |
| | 5 2 | 4+4 4 | 4+4 4 | 3+4 4 | 6+6 4 | 3+4 4 | 3+4 4 | 3+4 4 |
| | 5 2 | 3+3 4 | 3+3 4 | 3+4 4 | 5+5 4 | 3+4 4 | 3+4 4 | 3+4 4 |
| | 6 2 | 3+3 4 | 3+3 4 | 3+4 4 | 6+6 4 | 3+4 4 | 3+4 4 | 3+4 4 |
| | 5 2 | 4+4 4 | 4+4 4 | 3+4 4 | 5+5 4 | 3+4 4 | 3+4 4 | 3+4 4 |
| | 6 2 | 4+4 4 | 3+3 4 | 3+4 4 | 6+6 4 | 3+4 4 | 3+4 4 | 3+4 4 |
| | 5 2 | 3+3 4 | 4+4 4 | 3+4 4 | 5+5 4 | 3+4 4 | 3+4 4 | 3+4 4 |
| | 6 2 | | | | | | | |
| | 5 2 | | | | | | | |
| | 6 2 | | | | | | | |
| | 5 2 | | | | | | | |
| | 5 2 | | | | | | | |
| | 6 2 | | | | | | | |
| | 6 2 | | | | | | | |
| | 5 2 | | | | | | | |
| | 5 2 | | | | | | | |
| | 5 2 | | | | | | | |
| | 6 2 | | | | | | | |
| | 6 2 | | | | | | | |
| | 5 2 | | | | | | | |
| | 5 2 | | | | | | | |
| | 6 2 | | | | | | | |
| | 6 2 | | | | | | | |
| | 6 2 | | | | | | | |
| | 6 2 | | | | | | | |
| | 5 2 | | | | | | | |
| | 5 2 | | | | | | | |
| | 5 2 | | | | | | | |
| | 5 2 | | | | | | | |
| | 5 2 | | | | | | | |
| | 6 2 | | | | | | | |
| | 5 2 | | | | | | | |
| | 6 2 | | | | | | | |
| | 5 2 | | | | | | | |
| | 6 2 | | | | | | | |
| | 6 2 | | | | | | | |
| | 5 2 | | | | | | | |
| | 6 2 | | | | | | | |
| | 5 2 | | | | | | | |
| | 5 2 | | | | | | | |
| | 5 2 | | | | | | | |
| | 6 2 | | | | | | | |
| | 5 2 | | | | | | | |
| | 5 2 | | | | | | | |
| | 5 2 | | | | | | | |
| | 6 2 | | | | | | | |
| | 5 2 | | | | | | | |
| 62.2.134 | 32756 | 200 | Fe1K2O4 | | | | | |
| | -GM | -R | -S | -T | -U | -X | -Y | -Z |
| | 5 2 | 3+3 4 | 3+3 4 | 3+4 4 | 6+6 4 | 3+4 4 | 3+4 4 | 3+4 4 |
| | 5 2 | 4+4 4 | 4+4 4 | 3+4 4 | 5+5 4 | 3+4 4 | 3+4 4 | 3+4 4 |
| | 6 2 | 3+3 4 | 3+3 4 | 3+4 4 | 5+5 4 | 3+4 4 | 3+4 4 | 3+4 4 |
| | 6 2 | 4+4 4 | 4+4 4 | 3+4 4 | 6+6 4 | 3+4 4 | 3+4 4 | 3+4 4 |
| | 5 2 | 3+3 4 | 3+3 4 | 3+4 4 | 5+5 4 | 3+4 4 | 3+4 4 | 3+4 4 |
| | 6 2 | 4+4 4 | 4+4 4 | 3+4 4 | 6+6 4 | 3+4 4 | 3+4 4 | 3+4 4 |
| | 5 2 | 3+3 4 | 3+3 4 | 3+4 4 | 5+5 4 | 3+4 4 | 3+4 4 | 3+4 4 |
| | 6 2 | 4+4 4 | 4+4 4 | 3+4 4 | 6+6 4 | 3+4 4 | 3+4 4 | 3+4 4 |
| | 5 2 | 3+3 4 | 3+3 4 | 3+4 4 | 5+5 4 | 3+4 4 | 3+4 4 | 3+4 4 |
| | 5 2 | 4+4 4 | 4+4 4 | 3+4 4 | 6+6 4 | 3+4 4 | 3+4 4 | 3+4 4 |
| | 6 2 | 4+4 4 | 4+4 4 | 3+4 4 | 5+5 4 | 3+4 4 | 3+4 4 | 3+4 4 |
| | 6 2 | 3+3 4 | 3+3 4 | 3+4 4 | 6+6 4 | 3+4 4 | 3+4 4 | 3+4 4 |

Continued on next column

**TABLE XXXVII– continued**

| TQC | ICSD | NoE | SF | | | | | |
|---|---|---|---|---|---|---|---|---|
| | 5 2 | 3+3 4 | 4+4 4 | 3+4 4 | 6+6 4 | 3+4 4 | 3+4 4 | 3+4 4 |
| | 6 2 | 4+4 4 | 3+3 4 | 3+4 4 | 5+5 4 | 3+4 4 | 3+4 4 | 3+4 4 |
| | 5 2 | 4+4 4 | 3+3 4 | 3+4 4 | 6+6 4 | 3+4 4 | 3+4 4 | 3+4 4 |
| | 6 2 | 3+3 4 | 4+4 4 | 3+4 4 | 6+6 4 | 3+4 4 | 3+4 4 | 3+4 4 |
| | 5 2 | 3+3 4 | 3+3 4 | 3+4 4 | 5+5 4 | 3+4 4 | 3+4 4 | 3+4 4 |
| | 6 2 | 4+4 4 | 4+4 4 | 3+4 4 | 5+5 4 | 3+4 4 | 3+4 4 | 3+4 4 |
| | 5 2 | 4+4 4 | 3+3 4 | 3+4 4 | 5+5 4 | 3+4 4 | 3+4 4 | 3+4 4 |
| | 6 2 | 3+3 4 | 4+4 4 | 3+4 4 | 6+6 4 | 3+4 4 | 3+4 4 | 3+4 4 |
| | 5 2 | 4+4 4 | 4+4 4 | 3+4 4 | 5+5 4 | 3+4 4 | 3+4 4 | 3+4 4 |
| | 6 2 | 3+3 4 | 3+3 4 | 3+4 4 | 6+6 4 | 3+4 4 | 3+4 4 | 3+4 4 |
| | 5 2 | 3+3 4 | 3+3 4 | 3+4 4 | 5+5 4 | 3+4 4 | 3+4 4 | 3+4 4 |
| | 6 2 | 4+4 4 | 4+4 4 | 3+4 4 | 6+6 4 | 3+4 4 | 3+4 4 | 3+4 4 |
| | 5 2 | 4+4 4 | 3+3 4 | 3+4 4 | 5+5 4 | 3+4 4 | 3+4 4 | 3+4 4 |
| | 5 2 | 3+3 4 | 4+4 4 | 3+4 4 | 6+6 4 | 3+4 4 | 3+4 4 | 3+4 4 |
| | 6 2 | 3+3 4 | 4+4 4 | 3+4 4 | 5+5 4 | 3+4 4 | 3+4 4 | 3+4 4 |
| | 6 2 | 4+4 4 | 3+3 4 | 3+4 4 | 6+6 4 | 3+4 4 | 3+4 4 | 3+4 4 |
| | 5 2 | 3+3 4 | 3+3 4 | 3+4 4 | 5+5 4 | 3+4 4 | 3+4 4 | 3+4 4 |
| | 5 2 | 4+4 4 | 4+4 4 | 3+4 4 | 6+6 4 | 3+4 4 | 3+4 4 | 3+4 4 |
| | 5 2 | 4+4 4 | 4+4 4 | 3+4 4 | 6+6 4 | 3+4 4 | 3+4 4 | 3+4 4 |
| | 6 2 | 3+3 4 | 3+3 4 | 3+4 4 | 5+5 4 | 3+4 4 | 3+4 4 | 3+4 4 |
| | 6 2 | 3+3 4 | 4+4 4 | 3+4 4 | 6+6 4 | 3+4 4 | 3+4 4 | 3+4 4 |
| | 5 2 | 4+4 4 | 3+3 4 | 3+4 4 | 5+5 4 | 3+4 4 | 3+4 4 | 3+4 4 |
| | 6 2 | 3+3 4 | 3+3 4 | 3+4 4 | 6+6 4 | 3+4 4 | 3+4 4 | 3+4 4 |
| | 6 2 | 4+4 4 | 4+4 4 | 3+4 4 | 5+5 4 | 3+4 4 | 3+4 4 | 3+4 4 |
| | 6 2 | 3+3 4 | 4+4 4 | 3+4 4 | 5+5 4 | 3+4 4 | 3+4 4 | 3+4 4 |
| | 5 2 | 4+4 4 | 3+3 4 | 3+4 4 | 5+5 4 | 3+4 4 | 3+4 4 | 3+4 4 |
| | 5 2 | 4+4 4 | 3+3 4 | 3+4 4 | 6+6 4 | 3+4 4 | 3+4 4 | 3+4 4 |
| | 6 2 | 3+3 4 | 4+4 4 | 3+4 4 | 6+6 4 | 3+4 4 | 3+4 4 | 3+4 4 |
| | 5 2 | 4+4 4 | 3+3 4 | 3+4 4 | 5+5 4 | 3+4 4 | 3+4 4 | 3+4 4 |
| | 6 2 | 3+3 4 | 4+4 4 | 3+4 4 | 6+6 4 | 3+4 4 | 3+4 4 | 3+4 4 |
| | 6 2 | 4+4 4 | 4+4 4 | 3+4 4 | 5+5 4 | 3+4 4 | 3+4 4 | 3+4 4 |
| | 5 2 | 3+3 4 | 3+3 4 | 3+4 4 | 5+5 4 | 3+4 4 | 3+4 4 | 3+4 4 |
| | 5 2 | 3+3 4 | 3+3 4 | 3+4 4 | 6+6 4 | 3+4 4 | 3+4 4 | 3+4 4 |
| | 6 2 | 4+4 4 | 4+4 4 | 3+4 4 | 6+6 4 | 3+4 4 | 3+4 4 | 3+4 4 |
| | 5 2 | 3+3 4 | 4+4 4 | 3+4 4 | 5+5 4 | 3+4 4 | 3+4 4 | 3+4 4 |
| | 6 2 | 4+4 4 | 3+3 4 | 3+4 4 | 6+6 4 | 3+4 4 | 3+4 4 | 3+4 4 |
| | 6 2 | 3+3 4 | 3+3 4 | 3+4 4 | 6+6 4 | 3+4 4 | 3+4 4 | 3+4 4 |
| | 5 2 | 4+4 4 | 4+4 4 | 3+4 4 | 5+5 4 | 3+4 4 | 3+4 4 | 3+4 4 |
| | 5 2 | | | | | | | |
| | 6 2 | | | | | | | |
| | 5 2 | | | | | | | |
| | 5 2 | | | | | | | |
| | 5 2 | | | | | | | |
| | 6 2 | | | | | | | |
| | 5 2 | | | | | | | |
| | 6 2 | | | | | | | |
| | 6 2 | | | | | | | |
| | 6 2 | | | | | | | |
| | 5 2 | | | | | | | |
| | 6 2 | | | | | | | |
| | 5 2 | | | | | | | |
| | 5 2 | | | | | | | |
| | 6 2 | | | | | | | |
| | 6 2 | | | | | | | |
| | 6 2 | | | | | | | |
| | 5 2 | | | | | | | |
| | 5 2 | | | | | | | |
| | 6 2 | | | | | | | |
| | 6 2 | | | | | | | |
| | 5 2 | | | | | | | |
| | 6 2 | | | | | | | |

Continued on next column

**TABLE XXXVII– continued**

| TQC | ICSD | NoE | SF | | | | | |
|---|---|---|---|---|---|---|---|---|
| | 5 2 | | | | | | | |
| | 5 2 | | | | | | | |
| | 6 2 | | | | | | | |
| | 5 2 | | | | | | | |
| | 5 2 | | | | | | | |
| | 5 2 | | | | | | | |
| | 6 2 | | | | | | | |
| | 5 2 | | | | | | | |
| | 6 2 | | | | | | | |
| | 6 2 | | | | | | | |
| | 6 2 | | | | | | | |
| | 6 2 | | | | | | | |
| | 6 2 | | | | | | | |
| | 5 2 | | | | | | | |
| | 5 2 | | | | | | | |
| | 5 2 | | | | | | | |
| | 6 2 | | | | | | | |
| | 5 2 | | | | | | | |
| | 6 2 | | | | | | | |
| | 5 2 | | | | | | | |
| | 6 2 | | | | | | | |
| | 5 2 | | | | | | | |
| | 6 2 | | | | | | | |
| | 6 2 | | | | | | | |
| | 5 2 | | | | | | | |
| | 5 2 | | | | | | | |
| | 5 2 | | | | | | | |
| 62.2.135 | 2876 | 200 | Fe1K2O4 | | | | | |
| | -GM | -R | -S | -T | -U | -X | -Y | -Z |
| | 5 2 | 3+3 4 | 3+3 4 | 3+4 4 | 5+5 4 | 3+4 4 | 3+4 4 | 3+4 4 |
| | 5 2 | 4+4 4 | 4+4 4 | 3+4 4 | 6+6 4 | 3+4 4 | 3+4 4 | 3+4 4 |
| | 6 2 | 3+3 4 | 3+3 4 | 3+4 4 | 5+5 4 | 3+4 4 | 3+4 4 | 3+4 4 |
| | 6 2 | 4+4 4 | 4+4 4 | 3+4 4 | 6+6 4 | 3+4 4 | 3+4 4 | 3+4 4 |
| | 5 2 | 3+3 4 | 3+3 4 | 3+4 4 | 6+6 4 | 3+4 4 | 3+4 4 | 3+4 4 |
| | 6 2 | 4+4 4 | 4+4 4 | 3+4 4 | 5+5 4 | 3+4 4 | 3+4 4 | 3+4 4 |
| | 5 2 | 4+4 4 | 4+4 4 | 3+4 4 | 6+6 4 | 3+4 4 | 3+4 4 | 3+4 4 |
| | 6 2 | 3+3 4 | 3+3 4 | 3+4 4 | 5+5 4 | 3+4 4 | 3+4 4 | 3+4 4 |
| | 5 2 | 3+3 4 | 3+3 4 | 3+4 4 | 6+6 4 | 3+4 4 | 3+4 4 | 3+4 4 |
| | 5 2 | 4+4 4 | 4+4 4 | 3+4 4 | 5+5 4 | 3+4 4 | 3+4 4 | 3+4 4 |
| | 6 2 | 3+3 4 | 3+3 4 | 3+4 4 | 5+5 4 | 3+4 4 | 3+4 4 | 3+4 4 |
| | 6 2 | 4+4 4 | 4+4 4 | 3+4 4 | 6+6 4 | 3+4 4 | 3+4 4 | 3+4 4 |
| | 5 2 | 3+3 4 | 3+3 4 | 3+4 4 | 5+5 4 | 3+4 4 | 3+4 4 | 3+4 4 |
| | 6 2 | 4+4 4 | 4+4 4 | 3+4 4 | 6+6 4 | 3+4 4 | 3+4 4 | 3+4 4 |
| | 5 2 | 3+3 4 | 3+3 4 | 3+4 4 | 5+5 4 | 3+4 4 | 3+4 4 | 3+4 4 |
| | 5 2 | 4+4 4 | 4+4 4 | 3+4 4 | 5+5 4 | 3+4 4 | 3+4 4 | 3+4 4 |
| | 6 2 | 4+4 4 | 3+3 4 | 3+4 4 | 6+6 4 | 3+4 4 | 3+4 4 | 3+4 4 |
| | 6 2 | 3+3 4 | 4+4 4 | 3+4 4 | 6+6 4 | 3+4 4 | 3+4 4 | 3+4 4 |
| | 6 2 | 3+3 4 | 4+4 4 | 3+4 4 | 5+5 4 | 3+4 4 | 3+4 4 | 3+4 4 |
| | 5 2 | 4+4 4 | 3+3 4 | 3+4 4 | 6+6 4 | 3+4 4 | 3+4 4 | 3+4 4 |
| | 5 2 | 3+3 4 | 3+3 4 | 3+4 4 | 5+5 4 | 3+4 4 | 3+4 4 | 3+4 4 |
| | 6 2 | 4+4 4 | 4+4 4 | 3+4 4 | 6+6 4 | 3+4 4 | 3+4 4 | 3+4 4 |
| | 5 2 | 4+4 4 | 3+3 4 | 3+4 4 | 6+6 4 | 3+4 4 | 3+4 4 | 3+4 4 |
| | 6 2 | 3+3 4 | 4+4 4 | 3+4 4 | 5+5 4 | 3+4 4 | 3+4 4 | 3+4 4 |
| | 5 2 | 3+3 4 | 4+4 4 | 3+4 4 | 5+5 4 | 3+4 4 | 3+4 4 | 3+4 4 |
| | 5 2 | 4+4 4 | 3+3 4 | 3+4 4 | 6+6 4 | 3+4 4 | 3+4 4 | 3+4 4 |
| | 6 2 | 4+4 4 | 3+3 4 | 3+4 4 | 6+6 4 | 3+4 4 | 3+4 4 | 3+4 4 |
| | 6 2 | 3+3 4 | 4+4 4 | 3+4 4 | 5+5 4 | 3+4 4 | 3+4 4 | 3+4 4 |
| | 5 2 | 3+3 4 | 4+4 4 | 3+4 4 | 6+6 4 | 3+4 4 | 3+4 4 | 3+4 4 |
| | 5 2 | 4+4 4 | 3+3 4 | 3+4 4 | 5+5 4 | 3+4 4 | 3+4 4 | 3+4 4 |
| | 5 2 | 3+3 4 | 3+3 4 | 3+4 4 | 5+5 4 | 3+4 4 | 3+4 4 | 3+4 4 |
| | 6 2 | 4+4 4 | 4+4 4 | 3+4 4 | 6+6 4 | 3+4 4 | 3+4 4 | 3+4 4 |

Continued on next column

**TABLE XXXVII– continued**

| TQC | ICSD | NoE | SF | | | | | |
|---|---|---|---|---|---|---|---|---|
| | 6 2 | 4+4 4 | 3+3 4 | 3+4 4 | 5+5 4 | 3+4 4 | 3+4 4 | 3+4 4 |
| | 5 2 | 3+3 4 | 4+4 4 | 3+4 4 | 6+6 4 | 3+4 4 | 3+4 4 | 3+4 4 |
| | 6 2 | 3+3 4 | 3+3 4 | 3+4 4 | 5+5 4 | 3+4 4 | 3+4 4 | 3+4 4 |
| | 6 2 | 4+4 4 | 4+4 4 | 3+4 4 | 6+6 4 | 3+4 4 | 3+4 4 | 3+4 4 |
| | 6 2 | 4+4 4 | 3+3 4 | 3+4 4 | 6+6 4 | 3+4 4 | 3+4 4 | 3+4 4 |
| | 5 2 | 3+3 4 | 4+4 4 | 3+4 4 | 6+6 4 | 3+4 4 | 3+4 4 | 3+4 4 |
| | 5 2 | 3+3 4 | 4+4 4 | 3+4 4 | 5+5 4 | 3+4 4 | 3+4 4 | 3+4 4 |
| | 6 2 | 4+4 4 | 3+3 4 | 3+4 4 | 5+5 4 | 3+4 4 | 3+4 4 | 3+4 4 |
| | 5 2 | 3+3 4 | 3+3 4 | 3+4 4 | 6+6 4 | 3+4 4 | 3+4 4 | 3+4 4 |
| | 6 2 | 4+4 4 | 4+4 4 | 3+4 4 | 5+5 4 | 3+4 4 | 3+4 4 | 3+4 4 |
| | 6 2 | 3+3 4 | 3+3 4 | 3+4 4 | 6+6 4 | 3+4 4 | 3+4 4 | 3+4 4 |
| | 5 2 | 4+4 4 | 4+4 4 | 3+4 4 | 6+6 4 | 3+4 4 | 3+4 4 | 3+4 4 |
| | 5 2 | 4+4 4 | 4+4 4 | 3+4 4 | 5+5 4 | 3+4 4 | 3+4 4 | 3+4 4 |
| | 6 2 | 3+3 4 | 3+3 4 | 3+4 4 | 5+5 4 | 3+4 4 | 3+4 4 | 3+4 4 |
| | 5 2 | 4+4 4 | 3+3 4 | 3+4 4 | 6+6 4 | 3+4 4 | 3+4 4 | 3+4 4 |
| | 6 2 | 3+3 4 | 4+4 4 | 3+4 4 | 5+5 4 | 3+4 4 | 3+4 4 | 3+4 4 |
| | 6 2 | 3+3 4 | 4+4 4 | 3+4 4 | 5+5 4 | 3+4 4 | 3+4 4 | 3+4 4 |
| | 5 2 | 4+4 4 | 3+3 4 | 3+4 4 | 6+6 4 | 3+4 4 | 3+4 4 | 3+4 4 |
| | 5 2 | | | | | | | |
| | 5 2 | | | | | | | |
| | 6 2 | | | | | | | |
| | 5 2 | | | | | | | |
| | 6 2 | | | | | | | |
| | 5 2 | | | | | | | |
| | 6 2 | | | | | | | |
| | 5 2 | | | | | | | |
| | 6 2 | | | | | | | |
| | 6 2 | | | | | | | |
| | 5 2 | | | | | | | |
| | 6 2 | | | | | | | |
| | 5 2 | | | | | | | |
| | 5 2 | | | | | | | |
| | 6 2 | | | | | | | |
| | 6 2 | | | | | | | |
| | 6 2 | | | | | | | |
| | 5 2 | | | | | | | |
| | 5 2 | | | | | | | |
| | 6 2 | | | | | | | |
| | 6 2 | | | | | | | |
| | 5 2 | | | | | | | |
| | 6 2 | | | | | | | |
| | 5 2 | | | | | | | |
| | 5 2 | | | | | | | |
| | 6 2 | | | | | | | |
| | 5 2 | | | | | | | |
| | 5 2 | | | | | | | |
| | 5 2 | | | | | | | |
| | 6 2 | | | | | | | |
| | 5 2 | | | | | | | |
| | 6 2 | | | | | | | |
| | 6 2 | | | | | | | |
| | 6 2 | | | | | | | |
| | 6 2 | | | | | | | |
| | 6 2 | | | | | | | |
| | 5 2 | | | | | | | |
| | 5 2 | | | | | | | |
| | 5 2 | | | | | | | |
| | 6 2 | | | | | | | |
| | 5 2 | | | | | | | |
| | 6 2 | | | | | | | |
| | 5 2 | | | | | | | |

Continued on next column

**TABLE XXXVII– continued**

| TQC | ICSD | NoE | SF | | | | | |
|---|---|---|---|---|---|---|---|---|
| | 6 2 | | | | | | | |
| | 5 2 | | | | | | | |
| | 6 2 | | | | | | | |
| | 6 2 | | | | | | | |
| | 5 2 | | | | | | | |
| | 5 2 | | | | | | | |
| | 5 2 | | | | | | | |
| 62.2.136 | 240891 | 136 | F2K1N1O1 | | | | | |
| | -GM | -R | -S | -T | -U | -X | -Y | -Z |
| | 5 2 | 3+3 4 | 3+3 4 | 3+4 4 | 6+6 4 | 3+4 4 | 3+4 4 | 3+4 4 |
| | 6 2 | 4+4 4 | 4+4 4 | 3+4 4 | 5+5 4 | 3+4 4 | 3+4 4 | 3+4 4 |
| | 5 2 | 3+3 4 | 3+3 4 | 3+4 4 | 5+5 4 | 3+4 4 | 3+4 4 | 3+4 4 |
| | 6 2 | 4+4 4 | 4+4 4 | 3+4 4 | 6+6 4 | 3+4 4 | 3+4 4 | 3+4 4 |
| | 5 2 | 3+3 4 | 3+3 4 | 3+4 4 | 6+6 4 | 3+4 4 | 3+4 4 | 3+4 4 |
| | 6 2 | 4+4 4 | 4+4 4 | 3+4 4 | 5+5 4 | 3+4 4 | 3+4 4 | 3+4 4 |
| | 6 2 | 3+3 4 | 3+3 4 | 3+4 4 | 5+5 4 | 3+4 4 | 3+4 4 | 3+4 4 |
| | 5 2 | 4+4 4 | 4+4 4 | 3+4 4 | 6+6 4 | 3+4 4 | 3+4 4 | 3+4 4 |
| | 5 2 | 3+3 4 | 4+4 4 | 3+4 4 | 6+6 4 | 3+4 4 | 3+4 4 | 3+4 4 |
| | 6 2 | 4+4 4 | 3+3 4 | 3+4 4 | 5+5 4 | 3+4 4 | 3+4 4 | 3+4 4 |
| | 5 2 | 4+4 4 | 3+3 4 | 3+4 4 | 6+6 4 | 3+4 4 | 3+4 4 | 3+4 4 |
| | 6 2 | 3+3 4 | 4+4 4 | 3+4 4 | 5+5 4 | 3+4 4 | 3+4 4 | 3+4 4 |
| | 5 2 | 3+3 4 | 4+4 4 | 3+4 4 | 6+6 4 | 3+4 4 | 3+4 4 | 3+4 4 |
| | 5 2 | 4+4 4 | 3+3 4 | 3+4 4 | 5+5 4 | 3+4 4 | 3+4 4 | 3+4 4 |
| | 6 2 | 3+3 4 | 4+4 4 | 3+4 4 | 5+5 4 | 3+4 4 | 3+4 4 | 3+4 4 |
| | 6 2 | 4+4 4 | 3+3 4 | 3+4 4 | 6+6 4 | 3+4 4 | 3+4 4 | 3+4 4 |
| | 5 2 | 3+3 4 | 3+3 4 | 3+4 4 | 5+5 4 | 3+4 4 | 3+4 4 | 3+4 4 |
| | 6 2 | 4+4 4 | 4+4 4 | 3+4 4 | 6+6 4 | 3+4 4 | 3+4 4 | 3+4 4 |
| | 5 2 | 3+3 4 | 3+3 4 | 3+4 4 | 5+5 4 | 3+4 4 | 3+4 4 | 3+4 4 |
| | 6 2 | 4+4 4 | 4+4 4 | 3+4 4 | 6+6 4 | 3+4 4 | 3+4 4 | 3+4 4 |
| | 5 2 | 3+3 4 | 3+3 4 | 3+4 4 | 5+5 4 | 3+4 4 | 3+4 4 | 3+4 4 |
| | 6 2 | 4+4 4 | 4+4 4 | 3+4 4 | 6+6 4 | 3+4 4 | 3+4 4 | 3+4 4 |
| | 5 2 | 3+3 4 | 3+3 4 | 3+4 4 | 6+6 4 | 3+4 4 | 3+4 4 | 3+4 4 |
| | 6 2 | 4+4 4 | 4+4 4 | 3+4 4 | 5+5 4 | 3+4 4 | 3+4 4 | 3+4 4 |
| | 5 2 | 3+3 4 | 3+3 4 | 3+4 4 | 6+6 4 | 3+4 4 | 3+4 4 | 3+4 4 |
| | 5 2 | 4+4 4 | 4+4 4 | 3+4 4 | 5+5 4 | 3+4 4 | 3+4 4 | 3+4 4 |
| | 6 2 | 3+3 4 | 3+3 4 | 3+4 4 | 5+5 4 | 3+4 4 | 3+4 4 | 3+4 4 |
| | 5 2 | 4+4 4 | 4+4 4 | 3+4 4 | 6+6 4 | 3+4 4 | 3+4 4 | 3+4 4 |
| | 6 2 | 3+3 4 | 3+3 4 | 3+4 4 | 5+5 4 | 3+4 4 | 3+4 4 | 3+4 4 |
| | 6 2 | 4+4 4 | 4+4 4 | 3+4 4 | 6+6 4 | 3+4 4 | 3+4 4 | 3+4 4 |
| | 5 2 | 3+3 4 | 3+3 4 | 3+4 4 | 5+5 4 | 3+4 4 | 3+4 4 | 3+4 4 |
| | 6 2 | 4+4 4 | 4+4 4 | 3+4 4 | 6+6 4 | 3+4 4 | 3+4 4 | 3+4 4 |
| | 5 2 | 3+3 4 | 3+3 4 | 3+4 4 | 6+6 4 | 3+4 4 | 3+4 4 | 3+4 4 |
| | 6 2 | 4+4 4 | 4+4 4 | 3+4 4 | 5+5 4 | 3+4 4 | 3+4 4 | 3+4 4 |
| | 6 2 | | | | | | | |
| | 5 2 | | | | | | | |
| | 6 2 | | | | | | | |
| | 5 2 | | | | | | | |
| | 5 2 | | | | | | | |
| | 6 2 | | | | | | | |
| | 6 2 | | | | | | | |
| | 5 2 | | | | | | | |
| | 5 2 | | | | | | | |
| | 6 2 | | | | | | | |
| | 5 2 | | | | | | | |
| | 6 2 | | | | | | | |
| | 5 2 | | | | | | | |
| | 6 2 | | | | | | | |
| | 5 2 | | | | | | | |
| | 6 2 | | | | | | | |
| | 5 2 | | | | | | | |
| | 6 2 | | | | | | | |

Continued on next column

**TABLE XXXVII– continued**

| TQC | ICSD | NoE | SF | | | | | |
|---|---|---|---|---|---|---|---|---|
| | 5 2 | | | | | | | |
| | 5 2 | | | | | | | |
| | 6 2 | | | | | | | |
| | 6 2 | | | | | | | |
| | 6 2 | | | | | | | |
| | 5 2 | | | | | | | |
| | 5 2 | | | | | | | |
| | 6 2 | | | | | | | |
| | 5 2 | | | | | | | |
| | 5 2 | | | | | | | |
| | 6 2 | | | | | | | |
| | 6 2 | | | | | | | |
| | 5 2 | | | | | | | |
| | 6 2 | | | | | | | |
| | 5 2 | | | | | | | |
| | 5 2 | | | | | | | |
| 62.2.137 | 2635 | 96 | F1Se1Y1 | | | | | |
| | -GM | -R | -S | -T | -U | -X | -Y | -Z |
| | 5 2 | 3+3 4 | 3+3 4 | 3+4 4 | 6+6 4 | 3+4 4 | 3+4 4 | 3+4 4 |
| | 5 2 | 4+4 4 | 4+4 4 | 3+4 4 | 5+5 4 | 3+4 4 | 3+4 4 | 3+4 4 |
| | 6 2 | 4+4 4 | 4+4 4 | 3+4 4 | 6+6 4 | 3+4 4 | 3+4 4 | 3+4 4 |
| | 6 2 | 3+3 4 | 3+3 4 | 3+4 4 | 5+5 4 | 3+4 4 | 3+4 4 | 3+4 4 |
| | 5 2 | 4+4 4 | 4+4 4 | 3+4 4 | 5+5 4 | 3+4 4 | 3+4 4 | 3+4 4 |
| | 5 2 | 3+3 4 | 3+3 4 | 3+4 4 | 6+6 4 | 3+4 4 | 3+4 4 | 3+4 4 |
| | 6 2 | 3+3 4 | 3+3 4 | 3+4 4 | 6+6 4 | 3+4 4 | 3+4 4 | 3+4 4 |
| | 6 2 | 4+4 4 | 4+4 4 | 3+4 4 | 5+5 4 | 3+4 4 | 3+4 4 | 3+4 4 |
| | 5 2 | 3+3 4 | 3+3 4 | 3+4 4 | 5+5 4 | 3+4 4 | 3+4 4 | 3+4 4 |
| | 5 2 | 4+4 4 | 4+4 4 | 3+4 4 | 6+6 4 | 3+4 4 | 3+4 4 | 3+4 4 |
| | 6 2 | 4+4 4 | 4+4 4 | 3+4 4 | 6+6 4 | 3+4 4 | 3+4 4 | 3+4 4 |
| | 5 2 | 3+3 4 | 3+3 4 | 3+4 4 | 5+5 4 | 3+4 4 | 3+4 4 | 3+4 4 |
| | 6 2 | 4+4 4 | 3+3 4 | 3+4 4 | 5+5 4 | 3+4 4 | 3+4 4 | 3+4 4 |
| | 5 2 | 3+3 4 | 4+4 4 | 3+4 4 | 6+6 4 | 3+4 4 | 3+4 4 | 3+4 4 |
| | 6 2 | 4+4 4 | 3+3 4 | 3+4 4 | 5+5 4 | 3+4 4 | 3+4 4 | 3+4 4 |
| | 5 2 | 3+3 4 | 4+4 4 | 3+4 4 | 6+6 4 | 3+4 4 | 3+4 4 | 3+4 4 |
| | 6 2 | 3+3 4 | 3+3 4 | 3+4 4 | 6+6 4 | 3+4 4 | 3+4 4 | 3+4 4 |
| | 5 2 | 4+4 4 | 4+4 4 | 3+4 4 | 5+5 4 | 3+4 4 | 3+4 4 | 3+4 4 |
| | 6 2 | 3+3 4 | 4+4 4 | 3+4 4 | 5+5 4 | 3+4 4 | 3+4 4 | 3+4 4 |
| | 6 2 | 4+4 4 | 3+3 4 | 3+4 4 | 6+6 4 | 3+4 4 | 3+4 4 | 3+4 4 |
| | 6 2 | 3+3 4 | 4+4 4 | 3+4 4 | 5+5 4 | 3+4 4 | 3+4 4 | 3+4 4 |
| | 6 2 | 4+4 4 | 3+3 4 | 3+4 4 | 6+6 4 | 3+4 4 | 3+4 4 | 3+4 4 |
| | 5 2 | 3+3 4 | 4+4 4 | 3+4 4 | 5+5 4 | 3+4 4 | 3+4 4 | 3+4 4 |
| | 5 2 | 4+4 4 | 3+3 4 | 3+4 4 | 6+6 4 | 3+4 4 | 3+4 4 | 3+4 4 |
| | 5 2 | | | | | | | |
| | 5 2 | | | | | | | |
| | 6 2 | | | | | | | |
| | 5 2 | | | | | | | |
| | 6 2 | | | | | | | |
| | 5 2 | | | | | | | |
| | 6 2 | | | | | | | |
| | 6 2 | | | | | | | |
| | 5 2 | | | | | | | |
| | 6 2 | | | | | | | |
| | 6 2 | | | | | | | |
| | 5 2 | | | | | | | |
| | 5 2 | | | | | | | |
| | 5 2 | | | | | | | |
| | 6 2 | | | | | | | |
| | 6 2 | | | | | | | |
| | 5 2 | | | | | | | |
| | 5 2 | | | | | | | |
| | 6 2 | | | | | | | |

Continued on next column

**TABLE XXXVII– continued**

| TQC | ICSD | NoE | SF | | | | | |
|---|---|---|---|---|---|---|---|---|
| | 6 2 | | | | | | | |
| | 5 2 | | | | | | | |
| | 6 2 | | | | | | | |
| | 5 2 | | | | | | | |
| | 5 2 | | | | | | | |
| 62.2.138 | 195382 | 136 | F1Ho1Se1 | | | | | |
| | -GM | -R | -S | -T | -U | -X | -Y | -Z |
| | 5 2 | 3+3 4 | 3+3 4 | 3+4 4 | 5+5 4 | 3+4 4 | 3+4 4 | 3+4 4 |
| | 5 2 | 4+4 4 | 4+4 4 | 3+4 4 | 6+6 4 | 3+4 4 | 3+4 4 | 3+4 4 |
| | 6 2 | 3+3 4 | 3+3 4 | 3+4 4 | 5+5 4 | 3+4 4 | 3+4 4 | 3+4 4 |
| | 6 2 | 4+4 4 | 4+4 4 | 3+4 4 | 6+6 4 | 3+4 4 | 3+4 4 | 3+4 4 |
| | 5 2 | 3+3 4 | 3+3 4 | 3+4 4 | 5+5 4 | 3+4 4 | 3+4 4 | 3+4 4 |
| | 5 2 | 4+4 4 | 4+4 4 | 3+4 4 | 6+6 4 | 3+4 4 | 3+4 4 | 3+4 4 |
| | 6 2 | 4+4 4 | 3+3 4 | 3+4 4 | 6+6 4 | 3+4 4 | 3+4 4 | 3+4 4 |
| | 6 2 | 3+3 4 | 4+4 4 | 3+4 4 | 5+5 4 | 3+4 4 | 3+4 4 | 3+4 4 |
| | 5 2 | 4+4 4 | 3+3 4 | 3+4 4 | 6+6 4 | 3+4 4 | 3+4 4 | 3+4 4 |
| | 5 2 | 3+3 4 | 4+4 4 | 3+4 4 | 5+5 4 | 3+4 4 | 3+4 4 | 3+4 4 |
| | 6 2 | 3+3 4 | 4+4 4 | 3+4 4 | 6+6 4 | 3+4 4 | 3+4 4 | 3+4 4 |
| | 6 2 | 4+4 4 | 3+3 4 | 3+4 4 | 5+5 4 | 3+4 4 | 3+4 4 | 3+4 4 |
| | 5 2 | 4+4 4 | 3+3 4 | 3+4 4 | 5+5 4 | 3+4 4 | 3+4 4 | 3+4 4 |
| | 5 2 | 3+3 4 | 4+4 4 | 3+4 4 | 5+5 4 | 3+4 4 | 3+4 4 | 3+4 4 |
| | 6 2 | 3+3 4 | 3+3 4 | 3+4 4 | 6+6 4 | 3+4 4 | 3+4 4 | 3+4 4 |
| | 6 2 | 4+4 4 | 4+4 4 | 3+4 4 | 6+6 4 | 3+4 4 | 3+4 4 | 3+4 4 |
| | 6 2 | 4+4 4 | 3+3 4 | 3+4 4 | 5+5 4 | 3+4 4 | 3+4 4 | 3+4 4 |
| | 6 2 | 3+3 4 | 4+4 4 | 3+4 4 | 6+6 4 | 3+4 4 | 3+4 4 | 3+4 4 |
| | 5 2 | 4+4 4 | 3+3 4 | 3+4 4 | 5+5 4 | 3+4 4 | 3+4 4 | 3+4 4 |
| | 5 2 | 3+3 4 | 4+4 4 | 3+4 4 | 6+6 4 | 3+4 4 | 3+4 4 | 3+4 4 |
| | 5 2 | 4+4 4 | 4+4 4 | 3+4 4 | 6+6 4 | 3+4 4 | 3+4 4 | 3+4 4 |
| | 6 2 | 3+3 4 | 3+3 4 | 3+4 4 | 5+5 4 | 3+4 4 | 3+4 4 | 3+4 4 |
| | 6 2 | 4+4 4 | 3+3 4 | 3+4 4 | 6+6 4 | 3+4 4 | 3+4 4 | 3+4 4 |
| | 5 2 | 3+3 4 | 4+4 4 | 3+4 4 | 5+5 4 | 3+4 4 | 3+4 4 | 3+4 4 |
| | 5 2 | 4+4 4 | 3+3 4 | 3+4 4 | 6+6 4 | 3+4 4 | 3+4 4 | 3+4 4 |
| | 5 2 | 3+3 4 | 4+4 4 | 3+4 4 | 5+5 4 | 3+4 4 | 3+4 4 | 3+4 4 |
| | 5 2 | 3+3 4 | 4+4 4 | 3+4 4 | 6+6 4 | 3+4 4 | 3+4 4 | 3+4 4 |
| | 5 2 | 4+4 4 | 3+3 4 | 3+4 4 | 5+5 4 | 3+4 4 | 3+4 4 | 3+4 4 |
| | 6 2 | 3+3 4 | 3+3 4 | 3+4 4 | 6+6 4 | 3+4 4 | 3+4 4 | 3+4 4 |
| | 6 2 | 4+4 4 | 4+4 4 | 3+4 4 | 5+5 4 | 3+4 4 | 3+4 4 | 3+4 4 |
| | 6 2 | 4+4 4 | 3+3 4 | 3+4 4 | 6+6 4 | 3+4 4 | 3+4 4 | 3+4 4 |
| | 6 2 | 3+3 4 | 4+4 4 | 3+4 4 | 6+6 4 | 3+4 4 | 3+4 4 | 3+4 4 |
| | 5 2 | 4+4 4 | 4+4 4 | 3+4 4 | 5+5 4 | 3+4 4 | 3+4 4 | 3+4 4 |
| | 5 2 | 3+3 4 | 3+3 4 | 3+4 4 | 6+6 4 | 3+4 4 | 3+4 4 | 3+4 4 |
| | 6 2 | | | | | | | |
| | 6 2 | | | | | | | |
| | 5 2 | | | | | | | |
| | 6 2 | | | | | | | |
| | 5 2 | | | | | | | |
| | 5 2 | | | | | | | |
| | 6 2 | | | | | | | |
| | 6 2 | | | | | | | |
| | 5 2 | | | | | | | |
| | 5 2 | | | | | | | |
| | 6 2 | | | | | | | |
| | 6 2 | | | | | | | |
| | 5 2 | | | | | | | |
| | 5 2 | | | | | | | |
| | 6 2 | | | | | | | |
| | 6 2 | | | | | | | |
| | 5 2 | | | | | | | |
| | 6 2 | | | | | | | |
| | 5 2 | | | | | | | |
| | 6 2 | | | | | | | |

Continued on next column

**TABLE XXXVII– continued**

| TQC | ICSD | NoE | SF | | | | | |
|---|---|---|---|---|---|---|---|---|
| | 5 2 | | | | | | | |
| | 5 2 | | | | | | | |
| | 6 2 | | | | | | | |
| | 5 2 | | | | | | | |
| | 6 2 | | | | | | | |
| | 6 2 | | | | | | | |
| | 5 2 | | | | | | | |
| | 5 2 | | | | | | | |
| | 5 2 | | | | | | | |
| | 6 2 | | | | | | | |
| | 6 2 | | | | | | | |
| | 6 2 | | | | | | | |
| | 6 2 | | | | | | | |
| | 5 2 | | | | | | | |
| 62.2.139 | 405049 | 120 | Eu1In1Pt1 | | | | | |
| | -GM | -R | -S | -T | -U | -X | -Y | -Z |
| | 5 2 | 3+3 4 | 3+3 4 | 3+4 4 | 5+5 4 | 3+4 4 | 3+4 4 | 3+4 4 |
| | 5 2 | 4+4 4 | 4+4 4 | 3+4 4 | 6+6 4 | 3+4 4 | 3+4 4 | 3+4 4 |
| | 6 2 | 3+3 4 | 3+3 4 | 3+4 4 | 5+5 4 | 3+4 4 | 3+4 4 | 3+4 4 |
| | 6 2 | 4+4 4 | 4+4 4 | 3+4 4 | 6+6 4 | 3+4 4 | 3+4 4 | 3+4 4 |
| | 5 2 | 4+4 4 | 3+3 4 | 3+4 4 | 5+5 4 | 3+4 4 | 3+4 4 | 3+4 4 |
| | 5 2 | 3+3 4 | 4+4 4 | 3+4 4 | 6+6 4 | 3+4 4 | 3+4 4 | 3+4 4 |
| | 6 2 | 3+3 4 | 4+4 4 | 3+4 4 | 5+5 4 | 3+4 4 | 3+4 4 | 3+4 4 |
| | 6 2 | 4+4 4 | 3+3 4 | 3+4 4 | 6+6 4 | 3+4 4 | 3+4 4 | 3+4 4 |
| | 5 2 | 3+3 4 | 4+4 4 | 3+4 4 | 6+6 4 | 3+4 4 | 3+4 4 | 3+4 4 |
| | 5 2 | 4+4 4 | 3+3 4 | 3+4 4 | 5+5 4 | 3+4 4 | 3+4 4 | 3+4 4 |
| | 6 2 | 4+4 4 | 3+3 4 | 3+4 4 | 5+5 4 | 3+4 4 | 3+4 4 | 3+4 4 |
| | 5 2 | 3+3 4 | 4+4 4 | 3+4 4 | 6+6 4 | 3+4 4 | 3+4 4 | 3+4 4 |
| | 6 2 | 4+4 4 | 3+3 4 | 3+4 4 | 6+6 4 | 3+4 4 | 3+4 4 | 3+4 4 |
| | 5 2 | 3+3 4 | 4+4 4 | 3+4 4 | 5+5 4 | 3+4 4 | 3+4 4 | 3+4 4 |
| | 6 2 | 4+4 4 | 3+3 4 | 3+4 4 | 6+6 4 | 3+4 4 | 3+4 4 | 3+4 4 |
| | 6 2 | 3+3 4 | 4+4 4 | 3+4 4 | 5+5 4 | 3+4 4 | 3+4 4 | 3+4 4 |
| | 5 2 | 4+4 4 | 3+3 4 | 3+4 4 | 6+6 4 | 3+4 4 | 3+4 4 | 3+4 4 |
| | 5 2 | 3+3 4 | 4+4 4 | 3+4 4 | 5+5 4 | 3+4 4 | 3+4 4 | 3+4 4 |
| | 6 2 | 3+3 4 | 4+4 4 | 3+4 4 | 6+6 4 | 3+4 4 | 3+4 4 | 3+4 4 |
| | 6 2 | 4+4 4 | 3+3 4 | 3+4 4 | 5+5 4 | 3+4 4 | 3+4 4 | 3+4 4 |
| | 6 2 | 3+3 4 | 4+4 4 | 3+4 4 | 6+6 4 | 3+4 4 | 3+4 4 | 3+4 4 |
| | 5 2 | 4+4 4 | 3+3 4 | 3+4 4 | 5+5 4 | 3+4 4 | 3+4 4 | 3+4 4 |
| | 6 2 | 3+3 4 | 4+4 4 | 3+4 4 | 6+6 4 | 3+4 4 | 3+4 4 | 3+4 4 |
| | 5 2 | 4+4 4 | 3+3 4 | 3+4 4 | 5+5 4 | 3+4 4 | 3+4 4 | 3+4 4 |
| | 5 2 | 4+4 4 | 4+4 4 | 3+4 4 | 6+6 4 | 3+4 4 | 3+4 4 | 3+4 4 |
| | 6 2 | 3+3 4 | 3+3 4 | 3+4 4 | 5+5 4 | 3+4 4 | 3+4 4 | 3+4 4 |
| | 5 2 | 3+3 4 | 3+3 4 | 3+4 4 | 6+6 4 | 3+4 4 | 3+4 4 | 3+4 4 |
| | 6 2 | 4+4 4 | 4+4 4 | 3+4 4 | 5+5 4 | 3+4 4 | 3+4 4 | 3+4 4 |
| | 6 2 | 3+3 4 | 3+3 4 | 3+4 4 | 6+6 4 | 3+4 4 | 3+4 4 | 3+4 4 |
| | 5 2 | 4+4 4 | 4+4 4 | 3+4 4 | 6+6 4 | 3+4 4 | 3+4 4 | 3+4 4 |
| | 5 2 | | | | | | | |
| | 6 2 | | | | | | | |
| | 5 2 | | | | | | | |
| | 5 2 | | | | | | | |
| | 6 2 | | | | | | | |
| | 5 2 | | | | | | | |
| | 6 2 | | | | | | | |
| | 5 2 | | | | | | | |
| | 6 2 | | | | | | | |
| | 6 2 | | | | | | | |
| | 5 2 | | | | | | | |
| | 6 2 | | | | | | | |
| | 5 2 | | | | | | | |
| | 5 2 | | | | | | | |
| | 6 2 | | | | | | | |

Continued on next column

**TABLE XXXVII– continued**

| TQC | ICSD | NoE | SF | | | | | |
|---|---|---|---|---|---|---|---|---|
| | 5 2 | | | | | | | |
| | 6 2 | | | | | | | |
| | 5 2 | | | | | | | |
| | 6 2 | | | | | | | |
| | 6 2 | | | | | | | |
| | 5 2 | | | | | | | |
| | 6 2 | | | | | | | |
| | 5 2 | | | | | | | |
| | 5 2 | | | | | | | |
| | 6 2 | | | | | | | |
| | 5 2 | | | | | | | |
| | 6 2 | | | | | | | |
| | 5 2 | | | | | | | |
| | 6 2 | | | | | | | |
| | 5 2 | | | | | | | |
| 62.2.140 | 404044 | 120 | Eu1In1Pd1 | | | | | |
| | -GM | -R | -S | -T | -U | -X | -Y | -Z |
| | 5 2 | 3+3 4 | 3+3 4 | 3+4 4 | 6+6 4 | 3+4 4 | 3+4 4 | 3+4 4 |
| | 5 2 | 4+4 4 | 4+4 4 | 3+4 4 | 5+5 4 | 3+4 4 | 3+4 4 | 3+4 4 |
| | 6 2 | 3+3 4 | 3+3 4 | 3+4 4 | 6+6 4 | 3+4 4 | 3+4 4 | 3+4 4 |
| | 6 2 | 4+4 4 | 4+4 4 | 3+4 4 | 5+5 4 | 3+4 4 | 3+4 4 | 3+4 4 |
| | 5 2 | 3+3 4 | 4+4 4 | 3+4 4 | 6+6 4 | 3+4 4 | 3+4 4 | 3+4 4 |
| | 5 2 | 4+4 4 | 3+3 4 | 3+4 4 | 5+5 4 | 3+4 4 | 3+4 4 | 3+4 4 |
| | 6 2 | 4+4 4 | 3+3 4 | 3+4 4 | 6+6 4 | 3+4 4 | 3+4 4 | 3+4 4 |
| | 6 2 | 3+3 4 | 4+4 4 | 3+4 4 | 5+5 4 | 3+4 4 | 3+4 4 | 3+4 4 |
| | 5 2 | 4+4 4 | 3+3 4 | 3+4 4 | 5+5 4 | 3+4 4 | 3+4 4 | 3+4 4 |
| | 5 2 | 3+3 4 | 4+4 4 | 3+4 4 | 6+6 4 | 3+4 4 | 3+4 4 | 3+4 4 |
| | 6 2 | 3+3 4 | 4+4 4 | 3+4 4 | 6+6 4 | 3+4 4 | 3+4 4 | 3+4 4 |
| | 5 2 | 4+4 4 | 3+3 4 | 3+4 4 | 5+5 4 | 3+4 4 | 3+4 4 | 3+4 4 |
| | 6 2 | 3+3 4 | 4+4 4 | 3+4 4 | 5+5 4 | 3+4 4 | 3+4 4 | 3+4 4 |
| | 5 2 | 4+4 4 | 3+3 4 | 3+4 4 | 6+6 4 | 3+4 4 | 3+4 4 | 3+4 4 |
| | 6 2 | 3+3 4 | 4+4 4 | 3+4 4 | 5+5 4 | 3+4 4 | 3+4 4 | 3+4 4 |
| | 6 2 | 4+4 4 | 3+3 4 | 3+4 4 | 6+6 4 | 3+4 4 | 3+4 4 | 3+4 4 |
| | 5 2 | 3+3 4 | 3+3 4 | 3+4 4 | 5+5 4 | 3+4 4 | 3+4 4 | 3+4 4 |
| | 5 2 | 4+4 4 | 4+4 4 | 3+4 4 | 6+6 4 | 3+4 4 | 3+4 4 | 3+4 4 |
| | 6 2 | 4+4 4 | 3+3 4 | 3+4 4 | 5+5 4 | 3+4 4 | 3+4 4 | 3+4 4 |
| | 6 2 | 3+3 4 | 4+4 4 | 3+4 4 | 6+6 4 | 3+4 4 | 3+4 4 | 3+4 4 |
| | 6 2 | 3+3 4 | 3+3 4 | 3+4 4 | 5+5 4 | 3+4 4 | 3+4 4 | 3+4 4 |
| | 5 2 | 4+4 4 | 4+4 4 | 3+4 4 | 6+6 4 | 3+4 4 | 3+4 4 | 3+4 4 |
| | 5 2 | 4+4 4 | 4+4 4 | 3+4 4 | 5+5 4 | 3+4 4 | 3+4 4 | 3+4 4 |
| | 6 2 | 3+3 4 | 3+3 4 | 3+4 4 | 6+6 4 | 3+4 4 | 3+4 4 | 3+4 4 |
| | 5 2 | 3+3 4 | 3+3 4 | 3+4 4 | 5+5 4 | 3+4 4 | 3+4 4 | 3+4 4 |
| | 6 2 | 4+4 4 | 4+4 4 | 3+4 4 | 6+6 4 | 3+4 4 | 3+4 4 | 3+4 4 |
| | 5 2 | 4+4 4 | 4+4 4 | 3+4 4 | 5+5 4 | 3+4 4 | 3+4 4 | 3+4 4 |
| | 5 2 | 3+3 4 | 3+3 4 | 3+4 4 | 6+6 4 | 3+4 4 | 3+4 4 | 3+4 4 |
| | 5 2 | 3+3 4 | 4+4 4 | 3+4 4 | 5+5 4 | 3+4 4 | 3+4 4 | 3+4 4 |
| | 6 2 | 4+4 4 | 3+3 4 | 3+4 4 | 5+5 4 | 3+4 4 | 3+4 4 | 3+4 4 |
| | 5 2 | | | | | | | |
| | 6 2 | | | | | | | |
| | 6 2 | | | | | | | |
| | 5 2 | | | | | | | |
| | 5 2 | | | | | | | |
| | 6 2 | | | | | | | |
| | 6 2 | | | | | | | |
| | 5 2 | | | | | | | |
| | 6 2 | | | | | | | |
| | 6 2 | | | | | | | |
| | 5 2 | | | | | | | |
| | 6 2 | | | | | | | |
| | 5 2 | | | | | | | |
| | 5 2 | | | | | | | |

Continued on next column

**TABLE XXXVII– continued**

| TQC | ICSD | NoE | SF | | | | | |
|---|---|---|---|---|---|---|---|---|
| | 6 2 | | | | | | | |
| | 5 2 | | | | | | | |
| | 6 2 | | | | | | | |
| | 5 2 | | | | | | | |
| | 6 2 | | | | | | | |
| | 6 2 | | | | | | | |
| | 5 2 | | | | | | | |
| | 5 2 | | | | | | | |
| | 6 2 | | | | | | | |
| | 5 2 | | | | | | | |
| | 6 2 | | | | | | | |
| | 6 2 | | | | | | | |
| | 5 2 | | | | | | | |
| | 5 2 | | | | | | | |
| | 6 2 | | | | | | | |
| | 5 2 | | | | | | | |
| 62.2.141 | 423115 | 176 | Er1Pt1Zn1 | | | | | |
| | -GM | -R | -S | -T | -U | -X | -Y | -Z |
| | 5 2 | 3+3 4 | 3+3 4 | 3+4 4 | 6+6 4 | 3+4 4 | 3+4 4 | 3+4 4 |
| | 6 2 | 4+4 4 | 4+4 4 | 3+4 4 | 5+5 4 | 3+4 4 | 3+4 4 | 3+4 4 |
| | 5 2 | 3+3 4 | 3+3 4 | 3+4 4 | 6+6 4 | 3+4 4 | 3+4 4 | 3+4 4 |
| | 6 2 | 4+4 4 | 4+4 4 | 3+4 4 | 5+5 4 | 3+4 4 | 3+4 4 | 3+4 4 |
| | 5 2 | 4+4 4 | 3+3 4 | 3+4 4 | 6+6 4 | 3+4 4 | 3+4 4 | 3+4 4 |
| | 6 2 | 3+3 4 | 4+4 4 | 3+4 4 | 5+5 4 | 3+4 4 | 3+4 4 | 3+4 4 |
| | 5 2 | 3+3 4 | 4+4 4 | 3+4 4 | 6+6 4 | 3+4 4 | 3+4 4 | 3+4 4 |
| | 6 2 | 4+4 4 | 3+3 4 | 3+4 4 | 5+5 4 | 3+4 4 | 3+4 4 | 3+4 4 |
| | 5 2 | 4+4 4 | 3+3 4 | 3+4 4 | 6+6 4 | 3+4 4 | 3+4 4 | 3+4 4 |
| | 6 2 | 3+3 4 | 4+4 4 | 3+4 4 | 5+5 4 | 3+4 4 | 3+4 4 | 3+4 4 |
| | 5 2 | 3+3 4 | 4+4 4 | 3+4 4 | 5+5 4 | 3+4 4 | 3+4 4 | 3+4 4 |
| | 5 2 | 4+4 4 | 3+3 4 | 3+4 4 | 6+6 4 | 3+4 4 | 3+4 4 | 3+4 4 |
| | 6 2 | 4+4 4 | 4+4 4 | 3+4 4 | 5+5 4 | 3+4 4 | 3+4 4 | 3+4 4 |
| | 6 2 | 3+3 4 | 3+3 4 | 3+4 4 | 6+6 4 | 3+4 4 | 3+4 4 | 3+4 4 |
| | 5 2 | 3+3 4 | 3+3 4 | 3+4 4 | 5+5 4 | 3+4 4 | 3+4 4 | 3+4 4 |
| | 6 2 | 4+4 4 | 4+4 4 | 3+4 4 | 6+6 4 | 3+4 4 | 3+4 4 | 3+4 4 |
| | 5 2 | 3+3 4 | 3+3 4 | 3+4 4 | 5+5 4 | 3+4 4 | 3+4 4 | 3+4 4 |
| | 6 2 | 4+4 4 | 4+4 4 | 3+4 4 | 5+5 4 | 3+4 4 | 3+4 4 | 3+4 4 |
| | 6 2 | 3+3 4 | 4+4 4 | 3+4 4 | 6+6 4 | 3+4 4 | 3+4 4 | 3+4 4 |
| | 5 2 | 4+4 4 | 3+3 4 | 3+4 4 | 6+6 4 | 3+4 4 | 3+4 4 | 3+4 4 |
| | 5 2 | 4+4 4 | 4+4 4 | 3+4 4 | 6+6 4 | 3+4 4 | 3+4 4 | 3+4 4 |
| | 5 2 | 3+3 4 | 3+3 4 | 3+4 4 | 5+5 4 | 3+4 4 | 3+4 4 | 3+4 4 |
| | 6 2 | 4+4 4 | 3+3 4 | 3+4 4 | 5+5 4 | 3+4 4 | 3+4 4 | 3+4 4 |
| | 6 2 | 3+3 4 | 4+4 4 | 3+4 4 | 6+6 4 | 3+4 4 | 3+4 4 | 3+4 4 |
| | 5 2 | 3+3 4 | 3+3 4 | 3+4 4 | 5+5 4 | 3+4 4 | 3+4 4 | 3+4 4 |
| | 6 2 | 4+4 4 | 4+4 4 | 3+4 4 | 6+6 4 | 3+4 4 | 3+4 4 | 3+4 4 |
| | 5 2 | 4+4 4 | 4+4 4 | 3+4 4 | 5+5 4 | 3+4 4 | 3+4 4 | 3+4 4 |
| | 6 2 | 3+3 4 | 3+3 4 | 3+4 4 | 6+6 4 | 3+4 4 | 3+4 4 | 3+4 4 |
| | 5 2 | 3+3 4 | 4+4 4 | 3+4 4 | 5+5 4 | 3+4 4 | 3+4 4 | 3+4 4 |
| | 6 2 | 4+4 4 | 3+3 4 | 3+4 4 | 6+6 4 | 3+4 4 | 3+4 4 | 3+4 4 |
| | 6 2 | 4+4 4 | 4+4 4 | 3+4 4 | 5+5 4 | 3+4 4 | 3+4 4 | 3+4 4 |
| | 5 2 | 3+3 4 | 3+3 4 | 3+4 4 | 6+6 4 | 3+4 4 | 3+4 4 | 3+4 4 |
| | 5 2 | 3+3 4 | 4+4 4 | 3+4 4 | 5+5 4 | 3+4 4 | 3+4 4 | 3+4 4 |
| | 5 2 | 4+4 4 | 3+3 4 | 3+4 4 | 6+6 4 | 3+4 4 | 3+4 4 | 3+4 4 |
| | 5 2 | 3+3 4 | 3+3 4 | 3+4 4 | 6+6 4 | 3+4 4 | 3+4 4 | 3+4 4 |
| | 6 2 | 4+4 4 | 4+4 4 | 3+4 4 | 5+5 4 | 3+4 4 | 3+4 4 | 3+4 4 |
| | 6 2 | 4+4 4 | 4+4 4 | 3+4 4 | 5+5 4 | 3+4 4 | 3+4 4 | 3+4 4 |
| | 5 2 | 3+3 4 | 3+3 4 | 3+4 4 | 6+6 4 | 3+4 4 | 3+4 4 | 3+4 4 |
| | 6 2 | 3+3 4 | 4+4 4 | 3+4 4 | 5+5 4 | 3+4 4 | 3+4 4 | 3+4 4 |
| | 6 2 | 4+4 4 | 3+3 4 | 3+4 4 | 6+6 4 | 3+4 4 | 3+4 4 | 3+4 4 |
| | 6 2 | 3+3 4 | 3+3 4 | 3+4 4 | 5+5 4 | 3+4 4 | 3+4 4 | 3+4 4 |
| | 5 2 | 4+4 4 | 4+4 4 | 3+4 4 | 6+6 4 | 3+4 4 | 3+4 4 | 3+4 4 |
| | 6 2 | 4+4 4 | 3+3 4 | 3+4 4 | 6+6 4 | 3+4 4 | 3+4 4 | 3+4 4 |

Continued on next column

**TABLE XXXVII– continued**

| TQC | ICSD NoE SF |
|---|---|
| | 5 2 3+3 4 4+4 4 3+4 4 5+5 4 3+4 4 3+4 4 3+4 4 |
| | 5 2 |
| | 6 2 |
| | 5 2 |
| | 6 2 |
| | 5 2 |
| | 6 2 |
| | 5 2 |
| | 5 2 |
| | 5 2 |
| | 6 2 |
| | 5 2 |
| | 6 2 |
| | 6 2 |
| | 5 2 |
| | 5 2 |
| | 6 2 |
| | 6 2 |
| | 6 2 |
| | 5 2 |
| | 5 2 |
| | 6 2 |
| | 5 2 |
| | 5 2 |
| | 6 2 |
| | 6 2 |
| | 5 2 |
| | 6 2 |
| | 5 2 |
| | 6 2 |
| | 6 2 |
| | 5 2 |
| | 5 2 |
| | 6 2 |
| | 6 2 |
| | 5 2 |
| | 5 2 |
| | 6 2 |
| | 5 2 |
| | 6 2 |
| | 5 2 |
| | 6 2 |
| | 5 2 |
| | 5 2 |
| | 6 2 |
| 62.2.142 | 27282 192 Er1Fe1O3 |
| | -GM -R -S -T -U -X -Y -Z |
| | 5 2 3+3 4 3+3 4 3+4 4 5+5 4 3+4 4 3+4 4 3+4 4 |
| | 6 2 4+4 4 4+4 4 3+4 4 6+6 4 3+4 4 3+4 4 3+4 4 |
| | 5 2 4+4 4 4+4 4 3+4 4 5+5 4 3+4 4 3+4 4 3+4 4 |
| | 6 2 3+3 4 3+3 4 3+4 4 6+6 4 3+4 4 3+4 4 3+4 4 |
| | 5 2 4+4 4 4+4 4 3+4 4 6+6 4 3+4 4 3+4 4 3+4 4 |
| | 6 2 3+3 4 3+3 4 3+4 4 5+5 4 3+4 4 3+4 4 3+4 4 |
| | 5 2 3+3 4 3+3 4 3+4 4 5+5 4 3+4 4 3+4 4 3+4 4 |
| | 6 2 4+4 4 4+4 4 3+4 4 6+6 4 3+4 4 3+4 4 3+4 4 |
| | 6 2 4+4 4 4+4 4 3+4 4 5+5 4 3+4 4 3+4 4 3+4 4 |
| | 5 2 3+3 4 3+3 4 3+4 4 6+6 4 3+4 4 3+4 4 3+4 4 |
| | 5 2 3+3 4 3+3 4 3+4 4 6+6 4 3+4 4 3+4 4 3+4 4 |
| | 5 2 4+4 4 4+4 4 3+4 4 5+5 4 3+4 4 3+4 4 3+4 4 |
| | 6 2 4+4 4 4+4 4 3+4 4 6+6 4 3+4 4 3+4 4 3+4 4 |
| | 6 2 3+3 4 3+3 4 3+4 4 5+5 4 3+4 4 3+4 4 3+4 4 |

Continued on next column

**TABLE XXXVII– continued**

| TQC | ICSD NoE SF |
|---|---|
| | 6 2 3+3 4 4+4 4 3+4 4 6+6 4 3+4 4 3+4 4 3+4 4 |
| | 5 2 4+4 4 3+3 4 3+4 4 6+6 4 3+4 4 3+4 4 3+4 4 |
| | 5 2 3+3 4 3+3 4 3+4 4 6+6 4 3+4 4 3+4 4 3+4 4 |
| | 6 2 4+4 4 4+4 4 3+4 4 6+6 4 3+4 4 3+4 4 3+4 4 |
| | 5 2 3+3 4 3+3 4 3+4 4 5+5 4 3+4 4 3+4 4 3+4 4 |
| | 6 2 4+4 4 4+4 4 3+4 4 5+5 4 3+4 4 3+4 4 3+4 4 |
| | 6 2 4+4 4 3+3 4 3+4 4 5+5 4 3+4 4 3+4 4 3+4 4 |
| | 5 2 3+3 4 4+4 4 3+4 4 6+6 4 3+4 4 3+4 4 3+4 4 |
| | 6 2 3+3 4 4+4 4 3+4 4 5+5 4 3+4 4 3+4 4 3+4 4 |
| | 5 2 4+4 4 3+3 4 3+4 4 6+6 4 3+4 4 3+4 4 3+4 4 |
| | 5 2 3+3 4 4+4 4 3+4 4 5+5 4 3+4 4 3+4 4 3+4 4 |
| | 6 2 4+4 4 3+3 4 3+4 4 6+6 4 3+4 4 3+4 4 3+4 4 |
| | 5 2 3+3 4 3+3 4 3+4 4 6+6 4 3+4 4 3+4 4 3+4 4 |
| | 6 2 4+4 4 4+4 4 3+4 4 5+5 4 3+4 4 3+4 4 3+4 4 |
| | 5 2 3+3 4 3+3 4 3+4 4 5+5 4 3+4 4 3+4 4 3+4 4 |
| | 5 2 4+4 4 4+4 4 3+4 4 6+6 4 3+4 4 3+4 4 3+4 4 |
| | 5 2 4+4 4 3+3 4 3+4 4 5+5 4 3+4 4 3+4 4 3+4 4 |
| | 5 2 3+3 4 4+4 4 3+4 4 5+5 4 3+4 4 3+4 4 3+4 4 |
| | 6 2 3+3 4 3+3 4 3+4 4 6+6 4 3+4 4 3+4 4 3+4 4 |
| | 5 2 4+4 4 4+4 4 3+4 4 5+5 4 3+4 4 3+4 4 3+4 4 |
| | 5 2 3+3 4 4+4 4 3+4 4 6+6 4 3+4 4 3+4 4 3+4 4 |
| | 5 2 4+4 4 3+3 4 3+4 4 5+5 4 3+4 4 3+4 4 3+4 4 |
| | 5 2 3+3 4 4+4 4 3+4 4 6+6 4 3+4 4 3+4 4 3+4 4 |
| | 5 2 4+4 4 3+3 4 3+4 4 5+5 4 3+4 4 3+4 4 3+4 4 |
| | 6 2 3+3 4 3+3 4 3+4 4 6+6 4 3+4 4 3+4 4 3+4 4 |
| | 6 2 4+4 4 4+4 4 3+4 4 6+6 4 3+4 4 3+4 4 3+4 4 |
| | 5 2 4+4 4 3+3 4 3+4 4 6+6 4 3+4 4 3+4 4 3+4 4 |
| | 6 2 3+3 4 4+4 4 3+4 4 6+6 4 3+4 4 3+4 4 3+4 4 |
| | 6 2 4+4 4 3+3 4 3+4 4 6+6 4 3+4 4 3+4 4 3+4 4 |
| | 6 2 3+3 4 4+4 4 3+4 4 5+5 4 3+4 4 3+4 4 3+4 4 |
| | 5 2 4+4 4 4+4 4 3+4 4 6+6 4 3+4 4 3+4 4 3+4 4 |
| | 6 2 3+3 4 3+3 4 3+4 4 5+5 4 3+4 4 3+4 4 3+4 4 |
| | 5 2 3+3 4 3+3 4 3+4 4 5+5 4 3+4 4 3+4 4 3+4 4 |
| | 5 2 4+4 4 4+4 4 3+4 4 6+6 4 3+4 4 3+4 4 3+4 4 |
| | 6 2 |
| | 5 2 |
| | 6 2 |
| | 6 2 |
| | 5 2 |
| | 5 2 |
| | 6 2 |
| | 5 2 |
| | 6 2 |
| | 6 2 |
| | 6 2 |
| | 5 2 |
| | 6 2 |
| | 6 2 |
| | 6 2 |
| | 6 2 |
| | 6 2 |
| | 5 2 |
| | 5 2 |
| | 5 2 |
| | 5 2 |
| | 6 2 |
| | 5 2 |
| | 6 2 |
| | 5 2 |
| | 6 2 |
| | 6 2 |
| | 6 2 |

Continued on next column

**TABLE XXXVII– continued**

| TQC | ICSD | NoE | SF | | | | | |
|---|---|---|---|---|---|---|---|---|
| | 5 2 | | | | | | | |
| | 6 2 | | | | | | | |
| | 5 2 | | | | | | | |
| | 5 2 | | | | | | | |
| | 5 2 | | | | | | | |
| | 5 2 | | | | | | | |
| | 5 2 | | | | | | | |
| | 5 2 | | | | | | | |
| | 5 2 | | | | | | | |
| | 5 2 | | | | | | | |
| | 5 2 | | | | | | | |
| | 5 2 | | | | | | | |
| | 5 2 | | | | | | | |
| | 6 2 | | | | | | | |
| | 5 2 | | | | | | | |
| | 6 2 | | | | | | | |
| | 5 2 | | | | | | | |
| | 6 2 | | | | | | | |
| | 6 2 | | | | | | | |
| | 5 2 | | | | | | | |
| | 6 2 | | | | | | | |
| 62.2.143 | 419 | 232 | Dy2Se3 | | | | | |
| | -GM | -R | -S | -T | -U | -X | -Y | -Z |
| | 5 2 | 3+3 4 | 3+3 4 | 3+4 4 | 5+5 4 | 3+4 4 | 3+4 4 | 3+4 4 |
| | 5 2 | 4+4 4 | 4+4 4 | 3+4 4 | 6+6 4 | 3+4 4 | 3+4 4 | 3+4 4 |
| | 6 2 | 3+3 4 | 3+3 4 | 3+4 4 | 5+5 4 | 3+4 4 | 3+4 4 | 3+4 4 |
| | 6 2 | 4+4 4 | 4+4 4 | 3+4 4 | 6+6 4 | 3+4 4 | 3+4 4 | 3+4 4 |
| | 5 2 | 3+3 4 | 3+3 4 | 3+4 4 | 5+5 4 | 3+4 4 | 3+4 4 | 3+4 4 |
| | 5 2 | 4+4 4 | 4+4 4 | 3+4 4 | 6+6 4 | 3+4 4 | 3+4 4 | 3+4 4 |
| | 6 2 | 3+3 4 | 3+3 4 | 3+4 4 | 6+6 4 | 3+4 4 | 3+4 4 | 3+4 4 |
| | 6 2 | 4+4 4 | 4+4 4 | 3+4 4 | 5+5 4 | 3+4 4 | 3+4 4 | 3+4 4 |
| | 5 2 | 3+3 4 | 4+4 4 | 3+4 4 | 5+5 4 | 3+4 4 | 3+4 4 | 3+4 4 |
| | 6 2 | 4+4 4 | 3+3 4 | 3+4 4 | 6+6 4 | 3+4 4 | 3+4 4 | 3+4 4 |
| | 5 2 | 4+4 4 | 4+4 4 | 3+4 4 | 6+6 4 | 3+4 4 | 3+4 4 | 3+4 4 |
| | 6 2 | 3+3 4 | 3+3 4 | 3+4 4 | 5+5 4 | 3+4 4 | 3+4 4 | 3+4 4 |
| | 5 2 | 3+3 4 | 3+3 4 | 3+4 4 | 5+5 4 | 3+4 4 | 3+4 4 | 3+4 4 |
| | 5 2 | 4+4 4 | 4+4 4 | 3+4 4 | 6+6 4 | 3+4 4 | 3+4 4 | 3+4 4 |
| | 6 2 | 4+4 4 | 3+3 4 | 3+4 4 | 6+6 4 | 3+4 4 | 3+4 4 | 3+4 4 |
| | 6 2 | 3+3 4 | 4+4 4 | 3+4 4 | 5+5 4 | 3+4 4 | 3+4 4 | 3+4 4 |
| | 5 2 | 3+3 4 | 3+3 4 | 3+4 4 | 6+6 4 | 3+4 4 | 3+4 4 | 3+4 4 |
| | 6 2 | 4+4 4 | 4+4 4 | 3+4 4 | 5+5 4 | 3+4 4 | 3+4 4 | 3+4 4 |
| | 5 2 | 3+3 4 | 3+3 4 | 3+4 4 | 5+5 4 | 3+4 4 | 3+4 4 | 3+4 4 |
| | 5 2 | 4+4 4 | 4+4 4 | 3+4 4 | 6+6 4 | 3+4 4 | 3+4 4 | 3+4 4 |
| | 6 2 | 4+4 4 | 4+4 4 | 3+4 4 | 6+6 4 | 3+4 4 | 3+4 4 | 3+4 4 |
| | 5 2 | 3+3 4 | 3+3 4 | 3+4 4 | 5+5 4 | 3+4 4 | 3+4 4 | 3+4 4 |
| | 6 2 | 4+4 4 | 4+4 4 | 3+4 4 | 5+5 4 | 3+4 4 | 3+4 4 | 3+4 4 |
| | 6 2 | 3+3 4 | 3+3 4 | 3+4 4 | 6+6 4 | 3+4 4 | 3+4 4 | 3+4 4 |
| | 5 2 | 3+3 4 | 4+4 4 | 3+4 4 | 5+5 4 | 3+4 4 | 3+4 4 | 3+4 4 |
| | 5 2 | 4+4 4 | 3+3 4 | 3+4 4 | 6+6 4 | 3+4 4 | 3+4 4 | 3+4 4 |
| | 6 2 | 4+4 4 | 3+3 4 | 3+4 4 | 6+6 4 | 3+4 4 | 3+4 4 | 3+4 4 |
| | 6 2 | 3+3 4 | 4+4 4 | 3+4 4 | 5+5 4 | 3+4 4 | 3+4 4 | 3+4 4 |
| | 5 2 | 3+3 4 | 4+4 4 | 3+4 4 | 6+6 4 | 3+4 4 | 3+4 4 | 3+4 4 |
| | 6 2 | 4+4 4 | 3+3 4 | 3+4 4 | 5+5 4 | 3+4 4 | 3+4 4 | 3+4 4 |
| | 5 2 | 4+4 4 | 4+4 4 | 3+4 4 | 5+5 4 | 3+4 4 | 3+4 4 | 3+4 4 |
| | 6 2 | 3+3 4 | 3+3 4 | 3+4 4 | 6+6 4 | 3+4 4 | 3+4 4 | 3+4 4 |
| | 5 2 | 3+3 4 | 3+3 4 | 3+4 4 | 6+6 4 | 3+4 4 | 3+4 4 | 3+4 4 |
| | 5 2 | 4+4 4 | 4+4 4 | 3+4 4 | 5+5 4 | 3+4 4 | 3+4 4 | 3+4 4 |
| | 5 2 | 4+4 4 | 4+4 4 | 3+4 4 | 5+5 4 | 3+4 4 | 3+4 4 | 3+4 4 |
| | 6 2 | 3+3 4 | 3+3 4 | 3+4 4 | 6+6 4 | 3+4 4 | 3+4 4 | 3+4 4 |
| | 6 2 | 4+4 4 | 3+3 4 | 3+4 4 | 5+5 4 | 3+4 4 | 3+4 4 | 3+4 4 |
| | 6 2 | 3+3 4 | 4+4 4 | 3+4 4 | 6+6 4 | 3+4 4 | 3+4 4 | 3+4 4 |
| | 5 2 | 4+4 4 | 4+4 4 | 3+4 4 | 6+6 4 | 3+4 4 | 3+4 4 | 3+4 4 |
| | 6 2 | 3+3 4 | 3+3 4 | 3+4 4 | 5+5 4 | 3+4 4 | 3+4 4 | 3+4 4 |
| | 5 2 | 4+4 4 | 3+3 4 | 3+4 4 | 5+5 4 | 3+4 4 | 3+4 4 | 3+4 4 |
| | 6 2 | 3+3 4 | 4+4 4 | 3+4 4 | 6+6 4 | 3+4 4 | 3+4 4 | 3+4 4 |
| | 5 2 | 3+3 4 | 3+3 4 | 3+4 4 | 5+5 4 | 3+4 4 | 3+4 4 | 3+4 4 |
| | 6 2 | 4+4 4 | 4+4 4 | 3+4 4 | 6+6 4 | 3+4 4 | 3+4 4 | 3+4 4 |
| | 6 2 | 3+3 4 | 4+4 4 | 3+4 4 | 6+6 4 | 3+4 4 | 3+4 4 | 3+4 4 |
| | 6 2 | 4+4 4 | 3+3 4 | 3+4 4 | 5+5 4 | 3+4 4 | 3+4 4 | 3+4 4 |
| | 5 2 | 4+4 4 | 3+3 4 | 3+4 4 | 6+6 4 | 3+4 4 | 3+4 4 | 3+4 4 |
| | 5 2 | 3+3 4 | 4+4 4 | 3+4 4 | 5+5 4 | 3+4 4 | 3+4 4 | 3+4 4 |
| | 5 2 | 3+3 4 | 3+3 4 | 3+4 4 | 5+5 4 | 3+4 4 | 3+4 4 | 3+4 4 |
| | 5 2 | 4+4 4 | 4+4 4 | 3+4 4 | 6+6 4 | 3+4 4 | 3+4 4 | 3+4 4 |
| | 5 2 | 4+4 4 | 3+3 4 | 3+4 4 | 5+5 4 | 3+4 4 | 3+4 4 | 3+4 4 |
| | 6 2 | 3+3 4 | 4+4 4 | 3+4 4 | 6+6 4 | 3+4 4 | 3+4 4 | 3+4 4 |
| | 6 2 | 4+4 4 | 3+3 4 | 3+4 4 | 5+5 4 | 3+4 4 | 3+4 4 | 3+4 4 |
| | 5 2 | 3+3 4 | 4+4 4 | 3+4 4 | 6+6 4 | 3+4 4 | 3+4 4 | 3+4 4 |
| | 5 2 | 3+3 4 | 3+3 4 | 3+4 4 | 5+5 4 | 3+4 4 | 3+4 4 | 3+4 4 |
| | 6 2 | 4+4 4 | 4+4 4 | 3+4 4 | 6+6 4 | 3+4 4 | 3+4 4 | 3+4 4 |
| | 5 2 | 3+3 4 | 3+3 4 | 3+4 4 | 6+6 4 | 3+4 4 | 3+4 4 | 3+4 4 |
| | 5 2 | 4+4 4 | 4+4 4 | 3+4 4 | 5+5 4 | 3+4 4 | 3+4 4 | 3+4 4 |
| | 6 2 | | | | | | | |
| | 5 2 | | | | | | | |
| | 5 2 | | | | | | | |
| | 5 2 | | | | | | | |
| | 6 2 | | | | | | | |
| | 6 2 | | | | | | | |
| | 5 2 | | | | | | | |
| | 6 2 | | | | | | | |
| | 6 2 | | | | | | | |
| | 6 2 | | | | | | | |
| | 6 2 | | | | | | | |
| | 5 2 | | | | | | | |
| | 5 2 | | | | | | | |
| | 5 2 | | | | | | | |
| | 6 2 | | | | | | | |
| | 5 2 | | | | | | | |
| | 5 2 | | | | | | | |
| | 6 2 | | | | | | | |
| | 6 2 | | | | | | | |
| | 6 2 | | | | | | | |
| | 6 2 | | | | | | | |
| | 5 2 | | | | | | | |
| | 5 2 | | | | | | | |
| | 6 2 | | | | | | | |
| | 6 2 | | | | | | | |
| | 5 2 | | | | | | | |
| | 6 2 | | | | | | | |
| | 5 2 | | | | | | | |
| | 5 2 | | | | | | | |
| | 6 2 | | | | | | | |
| | 5 2 | | | | | | | |
| | 6 2 | | | | | | | |
| | 5 2 | | | | | | | |
| | 6 2 | | | | | | | |
| | 5 2 | | | | | | | |
| | 6 2 | | | | | | | |
| | 5 2 | | | | | | | |
| | 6 2 | | | | | | | |
| | 5 2 | | | | | | | |
| | 5 2 | | | | | | | |
| | 6 2 | | | | | | | |

Continued on next column

**TABLE XXXVII– continued**

| TQC | ICSD | NoE | SF | | | | | |
|---|---|---|---|---|---|---|---|---|
| | 5 2 | | | | | | | |
| | 6 2 | | | | | | | |
| | 6 2 | | | | | | | |
| | 6 2 | | | | | | | |
| | 6 2 | | | | | | | |
| | 5 2 | | | | | | | |
| | 6 2 | | | | | | | |
| | 5 2 | | | | | | | |
| | 6 2 | | | | | | | |
| | 5 2 | | | | | | | |
| | 5 2 | | | | | | | |
| | 6 2 | | | | | | | |
| | 5 2 | | | | | | | |
| | 5 2 | | | | | | | |
| | 5 2 | | | | | | | |
| | 6 2 | | | | | | | |
| | 6 2 | | | | | | | |
| 62.2.144 | 80768 | 232 | Dy2S3 | | | | | |
| | -GM | -R | -S | -T | -U | -X | -Y | -Z |
| | 5 2 | 3+3 4 | 3+3 4 | 3+4 4 | 5+5 4 | 3+4 4 | 3+4 4 | 3+4 4 |
| | 5 2 | 4+4 4 | 4+4 4 | 3+4 4 | 6+6 4 | 3+4 4 | 3+4 4 | 3+4 4 |
| | 6 2 | 3+3 4 | 3+3 4 | 3+4 4 | 5+5 4 | 3+4 4 | 3+4 4 | 3+4 4 |
| | 6 2 | 4+4 4 | 4+4 4 | 3+4 4 | 6+6 4 | 3+4 4 | 3+4 4 | 3+4 4 |
| | 5 2 | 3+3 4 | 3+3 4 | 3+4 4 | 6+6 4 | 3+4 4 | 3+4 4 | 3+4 4 |
| | 5 2 | 4+4 4 | 4+4 4 | 3+4 4 | 5+5 4 | 3+4 4 | 3+4 4 | 3+4 4 |
| | 6 2 | 3+3 4 | 3+3 4 | 3+4 4 | 5+5 4 | 3+4 4 | 3+4 4 | 3+4 4 |
| | 6 2 | 4+4 4 | 4+4 4 | 3+4 4 | 6+6 4 | 3+4 4 | 3+4 4 | 3+4 4 |
| | 6 2 | 4+4 4 | 3+3 4 | 3+4 4 | 6+6 4 | 3+4 4 | 3+4 4 | 3+4 4 |
| | 5 2 | 3+3 4 | 4+4 4 | 3+4 4 | 5+5 4 | 3+4 4 | 3+4 4 | 3+4 4 |
| | 5 2 | 3+3 4 | 3+3 4 | 3+4 4 | 5+5 4 | 3+4 4 | 3+4 4 | 3+4 4 |
| | 6 2 | 4+4 4 | 4+4 4 | 3+4 4 | 6+6 4 | 3+4 4 | 3+4 4 | 3+4 4 |
| | 5 2 | 4+4 4 | 4+4 4 | 3+4 4 | 5+5 4 | 3+4 4 | 3+4 4 | 3+4 4 |
| | 5 2 | 3+3 4 | 3+3 4 | 3+4 4 | 6+6 4 | 3+4 4 | 3+4 4 | 3+4 4 |
| | 6 2 | 3+3 4 | 4+4 4 | 3+4 4 | 5+5 4 | 3+4 4 | 3+4 4 | 3+4 4 |
| | 6 2 | 4+4 4 | 3+3 4 | 3+4 4 | 6+6 4 | 3+4 4 | 3+4 4 | 3+4 4 |
| | 5 2 | 4+4 4 | 4+4 4 | 3+4 4 | 5+5 4 | 3+4 4 | 3+4 4 | 3+4 4 |
| | 6 2 | 3+3 4 | 3+3 4 | 3+4 4 | 6+6 4 | 3+4 4 | 3+4 4 | 3+4 4 |
| | 5 2 | 3+3 4 | 4+4 4 | 3+4 4 | 6+6 4 | 3+4 4 | 3+4 4 | 3+4 4 |
| | 5 2 | 4+4 4 | 3+3 4 | 3+4 4 | 5+5 4 | 3+4 4 | 3+4 4 | 3+4 4 |
| | 6 2 | 3+3 4 | 3+3 4 | 3+4 4 | 5+5 4 | 3+4 4 | 3+4 4 | 3+4 4 |
| | 5 2 | 4+4 4 | 4+4 4 | 3+4 4 | 6+6 4 | 3+4 4 | 3+4 4 | 3+4 4 |
| | 6 2 | 3+3 4 | 3+3 4 | 3+4 4 | 6+6 4 | 3+4 4 | 3+4 4 | 3+4 4 |
| | 6 2 | 4+4 4 | 4+4 4 | 3+4 4 | 5+5 4 | 3+4 4 | 3+4 4 | 3+4 4 |
| | 5 2 | 4+4 4 | 3+3 4 | 3+4 4 | 6+6 4 | 3+4 4 | 3+4 4 | 3+4 4 |
| | 5 2 | 3+3 4 | 4+4 4 | 3+4 4 | 5+5 4 | 3+4 4 | 3+4 4 | 3+4 4 |
| | 6 2 | 3+3 4 | 4+4 4 | 3+4 4 | 5+5 4 | 3+4 4 | 3+4 4 | 3+4 4 |
| | 5 2 | 4+4 4 | 3+3 4 | 3+4 4 | 6+6 4 | 3+4 4 | 3+4 4 | 3+4 4 |
| | 6 2 | 3+3 4 | 3+3 4 | 3+4 4 | 5+5 4 | 3+4 4 | 3+4 4 | 3+4 4 |
| | 6 2 | 4+4 4 | 4+4 4 | 3+4 4 | 6+6 4 | 3+4 4 | 3+4 4 | 3+4 4 |
| | 5 2 | 4+4 4 | 4+4 4 | 3+4 4 | 6+6 4 | 3+4 4 | 3+4 4 | 3+4 4 |
| | 6 2 | 3+3 4 | 3+3 4 | 3+4 4 | 5+5 4 | 3+4 4 | 3+4 4 | 3+4 4 |
| | 5 2 | 4+4 4 | 3+3 4 | 3+4 4 | 5+5 4 | 3+4 4 | 3+4 4 | 3+4 4 |
| | 5 2 | 3+3 4 | 4+4 4 | 3+4 4 | 6+6 4 | 3+4 4 | 3+4 4 | 3+4 4 |
| | 5 2 | 3+3 4 | 4+4 4 | 3+4 4 | 6+6 4 | 3+4 4 | 3+4 4 | 3+4 4 |
| | 6 2 | 4+4 4 | 3+3 4 | 3+4 4 | 5+5 4 | 3+4 4 | 3+4 4 | 3+4 4 |
| | 6 2 | 3+3 4 | 3+3 4 | 3+4 4 | 6+6 4 | 3+4 4 | 3+4 4 | 3+4 4 |
| | 6 2 | 4+4 4 | 4+4 4 | 3+4 4 | 5+5 4 | 3+4 4 | 3+4 4 | 3+4 4 |
| | 6 2 | 3+3 4 | 3+3 4 | 3+4 4 | 5+5 4 | 3+4 4 | 3+4 4 | 3+4 4 |
| | 5 2 | 4+4 4 | 4+4 4 | 3+4 4 | 6+6 4 | 3+4 4 | 3+4 4 | 3+4 4 |
| | 5 2 | 3+3 4 | 4+4 4 | 3+4 4 | 6+6 4 | 3+4 4 | 3+4 4 | 3+4 4 |
| | 6 2 | 4+4 4 | 3+3 4 | 3+4 4 | 5+5 4 | 3+4 4 | 3+4 4 | 3+4 4 |
| | 5 2 | 3+3 4 | 3+3 4 | 3+4 4 | 6+6 4 | 3+4 4 | 3+4 4 | 3+4 4 |
| | 6 2 | 4+4 4 | 4+4 4 | 3+4 4 | 5+5 4 | 3+4 4 | 3+4 4 | 3+4 4 |
| | 6 2 | 3+3 4 | 3+3 4 | 3+4 4 | 5+5 4 | 3+4 4 | 3+4 4 | 3+4 4 |
| | 5 2 | 4+4 4 | 4+4 4 | 3+4 4 | 6+6 4 | 3+4 4 | 3+4 4 | 3+4 4 |
| | 6 2 | 3+3 4 | 4+4 4 | 3+4 4 | 5+5 4 | 3+4 4 | 3+4 4 | 3+4 4 |
| | 5 2 | 4+4 4 | 3+3 4 | 3+4 4 | 6+6 4 | 3+4 4 | 3+4 4 | 3+4 4 |
| | 5 2 | 4+4 4 | 4+4 4 | 3+4 4 | 6+6 4 | 3+4 4 | 3+4 4 | 3+4 4 |
| | 5 2 | 3+3 4 | 3+3 4 | 3+4 4 | 5+5 4 | 3+4 4 | 3+4 4 | 3+4 4 |
| | 5 2 | 3+3 4 | 4+4 4 | 3+4 4 | 6+6 4 | 3+4 4 | 3+4 4 | 3+4 4 |
| | 6 2 | 4+4 4 | 3+3 4 | 3+4 4 | 5+5 4 | 3+4 4 | 3+4 4 | 3+4 4 |
| | 6 2 | 3+3 4 | 4+4 4 | 3+4 4 | 6+6 4 | 3+4 4 | 3+4 4 | 3+4 4 |
| | 5 2 | 4+4 4 | 3+3 4 | 3+4 4 | 5+5 4 | 3+4 4 | 3+4 4 | 3+4 4 |
| | 5 2 | 3+3 4 | 4+4 4 | 3+4 4 | 6+6 4 | 3+4 4 | 3+4 4 | 3+4 4 |
| | 5 2 | 4+4 4 | 3+3 4 | 3+4 4 | 5+5 4 | 3+4 4 | 3+4 4 | 3+4 4 |
| | 6 2 | 4+4 4 | 3+3 4 | 3+4 4 | 5+5 4 | 3+4 4 | 3+4 4 | 3+4 4 |
| | 5 2 | 3+3 4 | 4+4 4 | 3+4 4 | 6+6 4 | 3+4 4 | 3+4 4 | 3+4 4 |
| | 5 2 | | | | | | | |
| | 5 2 | | | | | | | |
| | 6 2 | | | | | | | |
| | 6 2 | | | | | | | |
| | 5 2 | | | | | | | |
| | 6 2 | | | | | | | |
| | 5 2 | | | | | | | |
| | 6 2 | | | | | | | |
| | 6 2 | | | | | | | |
| | 6 2 | | | | | | | |
| | 6 2 | | | | | | | |
| | 5 2 | | | | | | | |
| | 5 2 | | | | | | | |
| | 5 2 | | | | | | | |
| | 6 2 | | | | | | | |
| | 5 2 | | | | | | | |
| | 6 2 | | | | | | | |
| | 5 2 | | | | | | | |
| | 6 2 | | | | | | | |
| | 6 2 | | | | | | | |
| | 6 2 | | | | | | | |
| | 5 2 | | | | | | | |
| | 5 2 | | | | | | | |
| | 6 2 | | | | | | | |
| | 6 2 | | | | | | | |
| | 5 2 | | | | | | | |
| | 6 2 | | | | | | | |
| | 5 2 | | | | | | | |
| | 6 2 | | | | | | | |
| | 5 2 | | | | | | | |
| | 5 2 | | | | | | | |
| | 6 2 | | | | | | | |
| | 5 2 | | | | | | | |
| | 6 2 | | | | | | | |
| | 5 2 | | | | | | | |
| | 6 2 | | | | | | | |
| | 5 2 | | | | | | | |
| | 6 2 | | | | | | | |
| | 5 2 | | | | | | | |
| | 5 2 | | | | | | | |
| | 6 2 | | | | | | | |
| | 5 2 | | | | | | | |
| | 6 2 | | | | | | | |
| | 6 2 | | | | | | | |
| | 6 2 | | | | | | | |

Continued on next column

Continued on next column

**TABLE XXXVII– continued**

| TQC | ICSD | NoE | SF | | | | | |
|---|---|---|---|---|---|---|---|---|
| | 6 2 | | | | | | | |
| | 5 2 | | | | | | | |
| | 6 2 | | | | | | | |
| | 5 2 | | | | | | | |
| | 6 2 | | | | | | | |
| | 5 2 | | | | | | | |
| | 5 2 | | | | | | | |
| | 6 2 | | | | | | | |
| | 5 2 | | | | | | | |
| | 5 2 | | | | | | | |
| | 6 2 | | | | | | | |
| | 5 2 | | | | | | | |
| | 6 2 | | | | | | | |
| 62.2.145 | 66532 | 232 | Dy2S3 | | | | | |
| | -GM | -R | -S | -T | -U | -X | -Y | -Z |
| | 5 2 | 3+3 4 | 3+3 4 | 3+4 4 | 5+5 4 | 3+4 4 | 3+4 4 | 3+4 4 |
| | 6 2 | 4+4 4 | 4+4 4 | 3+4 4 | 6+6 4 | 3+4 4 | 3+4 4 | 3+4 4 |
| | 5 2 | 3+3 4 | 3+3 4 | 3+4 4 | 6+6 4 | 3+4 4 | 3+4 4 | 3+4 4 |
| | 6 2 | 4+4 4 | 4+4 4 | 3+4 4 | 5+5 4 | 3+4 4 | 3+4 4 | 3+4 4 |
| | 5 2 | 3+3 4 | 3+3 4 | 3+4 4 | 5+5 4 | 3+4 4 | 3+4 4 | 3+4 4 |
| | 5 2 | 4+4 4 | 4+4 4 | 3+4 4 | 6+6 4 | 3+4 4 | 3+4 4 | 3+4 4 |
| | 6 2 | 3+3 4 | 3+3 4 | 3+4 4 | 6+6 4 | 3+4 4 | 3+4 4 | 3+4 4 |
| | 6 2 | 4+4 4 | 4+4 4 | 3+4 4 | 5+5 4 | 3+4 4 | 3+4 4 | 3+4 4 |
| | 5 2 | 3+3 4 | 3+3 4 | 3+4 4 | 5+5 4 | 3+4 4 | 3+4 4 | 3+4 4 |
| | 6 2 | 4+4 4 | 4+4 4 | 3+4 4 | 6+6 4 | 3+4 4 | 3+4 4 | 3+4 4 |
| | 6 2 | 3+3 4 | 4+4 4 | 3+4 4 | 5+5 4 | 3+4 4 | 3+4 4 | 3+4 4 |
| | 5 2 | 4+4 4 | 3+3 4 | 3+4 4 | 6+6 4 | 3+4 4 | 3+4 4 | 3+4 4 |
| | 5 2 | 4+4 4 | 3+3 4 | 3+4 4 | 6+6 4 | 3+4 4 | 3+4 4 | 3+4 4 |
| | 5 2 | 3+3 4 | 4+4 4 | 3+4 4 | 5+5 4 | 3+4 4 | 3+4 4 | 3+4 4 |
| | 6 2 | 4+4 4 | 4+4 4 | 3+4 4 | 6+6 4 | 3+4 4 | 3+4 4 | 3+4 4 |
| | 6 2 | 3+3 4 | 3+3 4 | 3+4 4 | 5+5 4 | 3+4 4 | 3+4 4 | 3+4 4 |
| | 5 2 | 3+3 4 | 3+3 4 | 3+4 4 | 5+5 4 | 3+4 4 | 3+4 4 | 3+4 4 |
| | 6 2 | 4+4 4 | 4+4 4 | 3+4 4 | 6+6 4 | 3+4 4 | 3+4 4 | 3+4 4 |
| | 6 2 | 4+4 4 | 4+4 4 | 3+4 4 | 6+6 4 | 3+4 4 | 3+4 4 | 3+4 4 |
| | 5 2 | 3+3 4 | 3+3 4 | 3+4 4 | 5+5 4 | 3+4 4 | 3+4 4 | 3+4 4 |
| | 5 2 | 3+3 4 | 4+4 4 | 3+4 4 | 5+5 4 | 3+4 4 | 3+4 4 | 3+4 4 |
| | 5 2 | 4+4 4 | 3+3 4 | 3+4 4 | 6+6 4 | 3+4 4 | 3+4 4 | 3+4 4 |
| | 6 2 | 3+3 4 | 3+3 4 | 3+4 4 | 6+6 4 | 3+4 4 | 3+4 4 | 3+4 4 |
| | 6 2 | 4+4 4 | 4+4 4 | 3+4 4 | 5+5 4 | 3+4 4 | 3+4 4 | 3+4 4 |
| | 5 2 | 4+4 4 | 4+4 4 | 3+4 4 | 5+5 4 | 3+4 4 | 3+4 4 | 3+4 4 |
| | 5 2 | 3+3 4 | 3+3 4 | 3+4 4 | 6+6 4 | 3+4 4 | 3+4 4 | 3+4 4 |
| | 6 2 | 3+3 4 | 3+3 4 | 3+4 4 | 5+5 4 | 3+4 4 | 3+4 4 | 3+4 4 |
| | 5 2 | 4+4 4 | 4+4 4 | 3+4 4 | 6+6 4 | 3+4 4 | 3+4 4 | 3+4 4 |
| | 6 2 | 4+4 4 | 4+4 4 | 3+4 4 | 5+5 4 | 3+4 4 | 3+4 4 | 3+4 4 |
| | 5 2 | 3+3 4 | 3+3 4 | 3+4 4 | 6+6 4 | 3+4 4 | 3+4 4 | 3+4 4 |
| | 6 2 | 3+3 4 | 3+3 4 | 3+4 4 | 6+6 4 | 3+4 4 | 3+4 4 | 3+4 4 |
| | 6 2 | 4+4 4 | 4+4 4 | 3+4 4 | 5+5 4 | 3+4 4 | 3+4 4 | 3+4 4 |
| | 5 2 | 4+4 4 | 4+4 4 | 3+4 4 | 5+5 4 | 3+4 4 | 3+4 4 | 3+4 4 |
| | 6 2 | 3+3 4 | 3+3 4 | 3+4 4 | 6+6 4 | 3+4 4 | 3+4 4 | 3+4 4 |
| | 5 2 | 4+4 4 | 3+3 4 | 3+4 4 | 6+6 4 | 3+4 4 | 3+4 4 | 3+4 4 |
| | 6 2 | 3+3 4 | 4+4 4 | 3+4 4 | 5+5 4 | 3+4 4 | 3+4 4 | 3+4 4 |
| | 5 2 | 3+3 4 | 3+3 4 | 3+4 4 | 6+6 4 | 3+4 4 | 3+4 4 | 3+4 4 |
| | 6 2 | 4+4 4 | 4+4 4 | 3+4 4 | 5+5 4 | 3+4 4 | 3+4 4 | 3+4 4 |
| | 5 2 | 3+3 4 | 3+3 4 | 3+4 4 | 5+5 4 | 3+4 4 | 3+4 4 | 3+4 4 |
| | 6 2 | 4+4 4 | 4+4 4 | 3+4 4 | 6+6 4 | 3+4 4 | 3+4 4 | 3+4 4 |
| | 6 2 | 4+4 4 | 3+3 4 | 3+4 4 | 6+6 4 | 3+4 4 | 3+4 4 | 3+4 4 |
| | 5 2 | 3+3 4 | 4+4 4 | 3+4 4 | 5+5 4 | 3+4 4 | 3+4 4 | 3+4 4 |
| | 5 2 | 3+3 4 | 3+3 4 | 3+4 4 | 6+6 4 | 3+4 4 | 3+4 4 | 3+4 4 |
| | 6 2 | 4+4 4 | 4+4 4 | 3+4 4 | 5+5 4 | 3+4 4 | 3+4 4 | 3+4 4 |
| | 5 2 | 3+3 4 | 3+3 4 | 3+4 4 | 5+5 4 | 3+4 4 | 3+4 4 | 3+4 4 |
| | 6 2 | 4+4 4 | 4+4 4 | 3+4 4 | 6+6 4 | 3+4 4 | 3+4 4 | 3+4 4 |

Continued on next column

**TABLE XXXVII– continued**

| TQC | ICSD | NoE | SF | | | | | |
|---|---|---|---|---|---|---|---|---|
| | 6 2 | 3+3 4 | 3+3 4 | 3+4 4 | 6+6 4 | 3+4 4 | 3+4 4 | 3+4 4 |
| | 5 2 | 4+4 4 | 4+4 4 | 3+4 4 | 5+5 4 | 3+4 4 | 3+4 4 | 3+4 4 |
| | 6 2 | 3+3 4 | 4+4 4 | 3+4 4 | 6+6 4 | 3+4 4 | 3+4 4 | 3+4 4 |
| | 5 2 | 4+4 4 | 3+3 4 | 3+4 4 | 5+5 4 | 3+4 4 | 3+4 4 | 3+4 4 |
| | 5 2 | 4+4 4 | 4+4 4 | 3+4 4 | 6+6 4 | 3+4 4 | 3+4 4 | 3+4 4 |
| | 5 2 | 3+3 4 | 3+3 4 | 3+4 4 | 5+5 4 | 3+4 4 | 3+4 4 | 3+4 4 |
| | 5 2 | 3+3 4 | 3+3 4 | 3+4 4 | 6+6 4 | 3+4 4 | 3+4 4 | 3+4 4 |
| | 6 2 | 4+4 4 | 4+4 4 | 3+4 4 | 5+5 4 | 3+4 4 | 3+4 4 | 3+4 4 |
| | 5 2 | 3+3 4 | 3+3 4 | 3+4 4 | 6+6 4 | 3+4 4 | 3+4 4 | 3+4 4 |
| | 6 2 | 4+4 4 | 4+4 4 | 3+4 4 | 5+5 4 | 3+4 4 | 3+4 4 | 3+4 4 |
| | 6 2 | 3+3 4 | 3+3 4 | 3+4 4 | 5+5 4 | 3+4 4 | 3+4 4 | 3+4 4 |
| | 5 2 | 4+4 4 | 4+4 4 | 3+4 4 | 6+6 4 | 3+4 4 | 3+4 4 | 3+4 4 |
| | 5 2 | | | | | | | |
| | 5 2 | | | | | | | |
| | 6 2 | | | | | | | |
| | 5 2 | | | | | | | |
| | 6 2 | | | | | | | |
| | 5 2 | | | | | | | |
| | 6 2 | | | | | | | |
| | 6 2 | | | | | | | |
| | 6 2 | | | | | | | |
| | 5 2 | | | | | | | |
| | 5 2 | | | | | | | |
| | 6 2 | | | | | | | |
| | 5 2 | | | | | | | |
| | 5 2 | | | | | | | |
| | 6 2 | | | | | | | |
| | 5 2 | | | | | | | |
| | 5 2 | | | | | | | |
| | 6 2 | | | | | | | |
| | 6 2 | | | | | | | |
| | 6 2 | | | | | | | |
| | 6 2 | | | | | | | |
| | 5 2 | | | | | | | |
| | 6 2 | | | | | | | |
| | 5 2 | | | | | | | |
| | 5 2 | | | | | | | |
| | 6 2 | | | | | | | |
| | 5 2 | | | | | | | |
| | 6 2 | | | | | | | |
| | 5 2 | | | | | | | |
| | 6 2 | | | | | | | |
| | 5 2 | | | | | | | |
| | 5 2 | | | | | | | |
| | 6 2 | | | | | | | |
| | 6 2 | | | | | | | |
| | 5 2 | | | | | | | |
| | 5 2 | | | | | | | |
| | 6 2 | | | | | | | |
| | 5 2 | | | | | | | |
| | 6 2 | | | | | | | |
| | 6 2 | | | | | | | |
| | 5 2 | | | | | | | |
| | 6 2 | | | | | | | |
| | 5 2 | | | | | | | |
| | 6 2 | | | | | | | |
| | 6 2 | | | | | | | |
| | 6 2 | | | | | | | |
| | 5 2 | | | | | | | |
| | 6 2 | | | | | | | |
| | 5 2 | | | | | | | |

Continued on next column

**TABLE XXXVII– continued**

| TQC | ICSD | NoE | SF | | | | | |
|---|---|---|---|---|---|---|---|---|
| | 6 2 | | | | | | | |
| | 5 2 | | | | | | | |
| | 5 2 | | | | | | | |
| | 5 2 | | | | | | | |
| | 6 2 | | | | | | | |
| | 5 2 | | | | | | | |
| | 5 2 | | | | | | | |
| | 6 2 | | | | | | | |
| | 6 2 | | | | | | | |
| 62.2.146 | 66531 | 232 | Dy2S3 | | | | | |
| | -GM | -R | -S | -T | -U | -X | -Y | -Z |
| | 5 2 | 3+3 4 | 3+3 4 | 3+4 4 | 5+5 4 | 3+4 4 | 3+4 4 | 3+4 4 |
| | 5 2 | 4+4 4 | 4+4 4 | 3+4 4 | 6+6 4 | 3+4 4 | 3+4 4 | 3+4 4 |
| | 6 2 | 3+3 4 | 3+3 4 | 3+4 4 | 5+5 4 | 3+4 4 | 3+4 4 | 3+4 4 |
| | 6 2 | 4+4 4 | 4+4 4 | 3+4 4 | 6+6 4 | 3+4 4 | 3+4 4 | 3+4 4 |
| | 5 2 | 3+3 4 | 3+3 4 | 3+4 4 | 6+6 4 | 3+4 4 | 3+4 4 | 3+4 4 |
| | 5 2 | 4+4 4 | 4+4 4 | 3+4 4 | 5+5 4 | 3+4 4 | 3+4 4 | 3+4 4 |
| | 6 2 | 3+3 4 | 3+3 4 | 3+4 4 | 5+5 4 | 3+4 4 | 3+4 4 | 3+4 4 |
| | 6 2 | 4+4 4 | 4+4 4 | 3+4 4 | 6+6 4 | 3+4 4 | 3+4 4 | 3+4 4 |
| | 6 2 | 4+4 4 | 3+3 4 | 3+4 4 | 6+6 4 | 3+4 4 | 3+4 4 | 3+4 4 |
| | 5 2 | 3+3 4 | 4+4 4 | 3+4 4 | 5+5 4 | 3+4 4 | 3+4 4 | 3+4 4 |
| | 5 2 | 3+3 4 | 3+3 4 | 3+4 4 | 5+5 4 | 3+4 4 | 3+4 4 | 3+4 4 |
| | 6 2 | 4+4 4 | 4+4 4 | 3+4 4 | 6+6 4 | 3+4 4 | 3+4 4 | 3+4 4 |
| | 5 2 | 4+4 4 | 4+4 4 | 3+4 4 | 5+5 4 | 3+4 4 | 3+4 4 | 3+4 4 |
| | 5 2 | 3+3 4 | 3+3 4 | 3+4 4 | 6+6 4 | 3+4 4 | 3+4 4 | 3+4 4 |
| | 6 2 | 3+3 4 | 4+4 4 | 3+4 4 | 5+5 4 | 3+4 4 | 3+4 4 | 3+4 4 |
| | 6 2 | 4+4 4 | 3+3 4 | 3+4 4 | 6+6 4 | 3+4 4 | 3+4 4 | 3+4 4 |
| | 5 2 | 4+4 4 | 4+4 4 | 3+4 4 | 5+5 4 | 3+4 4 | 3+4 4 | 3+4 4 |
| | 6 2 | 3+3 4 | 3+3 4 | 3+4 4 | 6+6 4 | 3+4 4 | 3+4 4 | 3+4 4 |
| | 5 2 | 3+3 4 | 4+4 4 | 3+4 4 | 6+6 4 | 3+4 4 | 3+4 4 | 3+4 4 |
| | 5 2 | 4+4 4 | 3+3 4 | 3+4 4 | 5+5 4 | 3+4 4 | 3+4 4 | 3+4 4 |
| | 6 2 | 3+3 4 | 3+3 4 | 3+4 4 | 5+5 4 | 3+4 4 | 3+4 4 | 3+4 4 |
| | 5 2 | 4+4 4 | 4+4 4 | 3+4 4 | 6+6 4 | 3+4 4 | 3+4 4 | 3+4 4 |
| | 6 2 | 3+3 4 | 3+3 4 | 3+4 4 | 6+6 4 | 3+4 4 | 3+4 4 | 3+4 4 |
| | 6 2 | 4+4 4 | 4+4 4 | 3+4 4 | 5+5 4 | 3+4 4 | 3+4 4 | 3+4 4 |
| | 5 2 | 4+4 4 | 3+3 4 | 3+4 4 | 6+6 4 | 3+4 4 | 3+4 4 | 3+4 4 |
| | 5 2 | 3+3 4 | 4+4 4 | 3+4 4 | 5+5 4 | 3+4 4 | 3+4 4 | 3+4 4 |
| | 6 2 | 3+3 4 | 4+4 4 | 3+4 4 | 5+5 4 | 3+4 4 | 3+4 4 | 3+4 4 |
| | 5 2 | 4+4 4 | 3+3 4 | 3+4 4 | 6+6 4 | 3+4 4 | 3+4 4 | 3+4 4 |
| | 6 2 | 3+3 4 | 3+3 4 | 3+4 4 | 5+5 4 | 3+4 4 | 3+4 4 | 3+4 4 |
| | 6 2 | 4+4 4 | 4+4 4 | 3+4 4 | 6+6 4 | 3+4 4 | 3+4 4 | 3+4 4 |
| | 5 2 | 4+4 4 | 4+4 4 | 3+4 4 | 6+6 4 | 3+4 4 | 3+4 4 | 3+4 4 |
| | 6 2 | 3+3 4 | 3+3 4 | 3+4 4 | 5+5 4 | 3+4 4 | 3+4 4 | 3+4 4 |
| | 5 2 | 4+4 4 | 3+3 4 | 3+4 4 | 5+5 4 | 3+4 4 | 3+4 4 | 3+4 4 |
| | 5 2 | 3+3 4 | 4+4 4 | 3+4 4 | 6+6 4 | 3+4 4 | 3+4 4 | 3+4 4 |
| | 5 2 | 3+3 4 | 4+4 4 | 3+4 4 | 6+6 4 | 3+4 4 | 3+4 4 | 3+4 4 |
| | 6 2 | 4+4 4 | 3+3 4 | 3+4 4 | 5+5 4 | 3+4 4 | 3+4 4 | 3+4 4 |
| | 6 2 | 3+3 4 | 3+3 4 | 3+4 4 | 6+6 4 | 3+4 4 | 3+4 4 | 3+4 4 |
| | 6 2 | 4+4 4 | 4+4 4 | 3+4 4 | 5+5 4 | 3+4 4 | 3+4 4 | 3+4 4 |
| | 6 2 | 3+3 4 | 3+3 4 | 3+4 4 | 5+5 4 | 3+4 4 | 3+4 4 | 3+4 4 |
| | 5 2 | 4+4 4 | 4+4 4 | 3+4 4 | 6+6 4 | 3+4 4 | 3+4 4 | 3+4 4 |
| | 5 2 | 3+3 4 | 4+4 4 | 3+4 4 | 6+6 4 | 3+4 4 | 3+4 4 | 3+4 4 |
| | 6 2 | 4+4 4 | 3+3 4 | 3+4 4 | 5+5 4 | 3+4 4 | 3+4 4 | 3+4 4 |
| | 5 2 | 3+3 4 | 3+3 4 | 3+4 4 | 6+6 4 | 3+4 4 | 3+4 4 | 3+4 4 |
| | 6 2 | 4+4 4 | 4+4 4 | 3+4 4 | 5+5 4 | 3+4 4 | 3+4 4 | 3+4 4 |
| | 6 2 | 3+3 4 | 3+3 4 | 3+4 4 | 5+5 4 | 3+4 4 | 3+4 4 | 3+4 4 |
| | 5 2 | 4+4 4 | 4+4 4 | 3+4 4 | 6+6 4 | 3+4 4 | 3+4 4 | 3+4 4 |
| | 6 2 | 3+3 4 | 4+4 4 | 3+4 4 | 5+5 4 | 3+4 4 | 3+4 4 | 3+4 4 |
| | 5 2 | 4+4 4 | 3+3 4 | 3+4 4 | 6+6 4 | 3+4 4 | 3+4 4 | 3+4 4 |
| | 5 2 | 4+4 4 | 4+4 4 | 3+4 4 | 6+6 4 | 3+4 4 | 3+4 4 | 3+4 4 |
| | 5 2 | 3+3 4 | 3+3 4 | 3+4 4 | 5+5 4 | 3+4 4 | 3+4 4 | 3+4 4 |

Continued on next column

| TQC | ICSD | NoE | SF | | | | | |
|---|---|---|---|---|---|---|---|---|
| | 5 2 | 3+3 4 | 4+4 4 | 3+4 4 | 6+6 4 | 3+4 4 | 3+4 4 | 3+4 4 |
| | 6 2 | 4+4 4 | 3+3 4 | 3+4 4 | 5+5 4 | 3+4 4 | 3+4 4 | 3+4 4 |
| | 6 2 | 3+3 4 | 4+4 4 | 3+4 4 | 6+6 4 | 3+4 4 | 3+4 4 | 3+4 4 |
| | 5 2 | 4+4 4 | 3+3 4 | 3+4 4 | 5+5 4 | 3+4 4 | 3+4 4 | 3+4 4 |
| | 5 2 | 3+3 4 | 4+4 4 | 3+4 4 | 6+6 4 | 3+4 4 | 3+4 4 | 3+4 4 |
| | 5 2 | 4+4 4 | 3+3 4 | 3+4 4 | 5+5 4 | 3+4 4 | 3+4 4 | 3+4 4 |
| | 6 2 | 4+4 4 | 3+3 4 | 3+4 4 | 5+5 4 | 3+4 4 | 3+4 4 | 3+4 4 |
| | 5 2 | 3+3 4 | 4+4 4 | 3+4 4 | 6+6 4 | 3+4 4 | 3+4 4 | 3+4 4 |
| | 5 2 | | | | | | | |
| | 5 2 | | | | | | | |
| | 6 2 | | | | | | | |
| | 6 2 | | | | | | | |
| | 5 2 | | | | | | | |
| | 6 2 | | | | | | | |
| | 5 2 | | | | | | | |
| | 6 2 | | | | | | | |
| | 6 2 | | | | | | | |
| | 6 2 | | | | | | | |
| | 6 2 | | | | | | | |
| | 5 2 | | | | | | | |
| | 5 2 | | | | | | | |
| | 5 2 | | | | | | | |
| | 6 2 | | | | | | | |
| | 5 2 | | | | | | | |
| | 6 2 | | | | | | | |
| | 5 2 | | | | | | | |
| | 6 2 | | | | | | | |
| | 6 2 | | | | | | | |
| | 6 2 | | | | | | | |
| | 5 2 | | | | | | | |
| | 5 2 | | | | | | | |
| | 6 2 | | | | | | | |
| | 6 2 | | | | | | | |
| | 5 2 | | | | | | | |
| | 6 2 | | | | | | | |
| | 5 2 | | | | | | | |
| | 6 2 | | | | | | | |
| | 5 2 | | | | | | | |
| | 5 2 | | | | | | | |
| | 6 2 | | | | | | | |
| | 5 2 | | | | | | | |
| | 6 2 | | | | | | | |
| | 5 2 | | | | | | | |
| | 6 2 | | | | | | | |
| | 5 2 | | | | | | | |
| | 6 2 | | | | | | | |
| | 5 2 | | | | | | | |
| | 5 2 | | | | | | | |
| | 6 2 | | | | | | | |
| | 5 2 | | | | | | | |
| | 6 2 | | | | | | | |
| | 6 2 | | | | | | | |
| | 6 2 | | | | | | | |
| | 6 2 | | | | | | | |
| | 5 2 | | | | | | | |
| | 6 2 | | | | | | | |
| | 5 2 | | | | | | | |
| | 6 2 | | | | | | | |
| | 5 2 | | | | | | | |
| | 5 2 | | | | | | | |
| | 6 2 | | | | | | | |

Continued on next column

**TABLE XXXVII– continued**

| TQC | ICSD NoE SF |
|---|---|
| | 5 2 |
| | 5 2 |
| | 5 2 |
| | 6 2 |
| | 6 2 |
| 62.2.147 | 36603 232 Dy2O1S2 |
| | -GM -R -S -T -U -X -Y -Z |
| | 5 2 3+3 4 3+3 4 3+4 4 6+6 4 3+4 4 3+4 4 3+4 4 |
| | 5 2 4+4 4 4+4 4 3+4 4 5+5 4 3+4 4 3+4 4 3+4 4 |
| | 6 2 3+3 4 3+3 4 3+4 4 6+6 4 3+4 4 3+4 4 3+4 4 |
| | 6 2 4+4 4 4+4 4 3+4 4 5+5 4 3+4 4 3+4 4 3+4 4 |
| | 5 2 3+3 4 3+3 4 3+4 4 6+6 4 3+4 4 3+4 4 3+4 4 |
| | 5 2 4+4 4 4+4 4 3+4 4 5+5 4 3+4 4 3+4 4 3+4 4 |
| | 6 2 3+3 4 4+4 4 3+4 4 6+6 4 3+4 4 3+4 4 3+4 4 |
| | 6 2 4+4 4 3+3 4 3+4 4 5+5 4 3+4 4 3+4 4 3+4 4 |
| | 5 2 4+4 4 3+3 4 3+4 4 6+6 4 3+4 4 3+4 4 3+4 4 |
| | 5 2 3+3 4 4+4 4 3+4 4 5+5 4 3+4 4 3+4 4 3+4 4 |
| | 6 2 4+4 4 3+3 4 3+4 4 6+6 4 3+4 4 3+4 4 3+4 4 |
| | 6 2 3+3 4 4+4 4 3+4 4 5+5 4 3+4 4 3+4 4 3+4 4 |
| | 5 2 3+3 4 4+4 4 3+4 4 6+6 4 3+4 4 3+4 4 3+4 4 |
| | 5 2 4+4 4 3+3 4 3+4 4 6+6 4 3+4 4 3+4 4 3+4 4 |
| | 6 2 4+4 4 3+3 4 3+4 4 5+5 4 3+4 4 3+4 4 3+4 4 |
| | 6 2 3+3 4 4+4 4 3+4 4 5+5 4 3+4 4 3+4 4 3+4 4 |
| | 5 2 4+4 4 3+3 4 3+4 4 5+5 4 3+4 4 3+4 4 3+4 4 |
| | 5 2 3+3 4 4+4 4 3+4 4 6+6 4 3+4 4 3+4 4 3+4 4 |
| | 6 2 4+4 4 3+3 4 3+4 4 5+5 4 3+4 4 3+4 4 3+4 4 |
| | 6 2 3+3 4 4+4 4 3+4 4 6+6 4 3+4 4 3+4 4 3+4 4 |
| | 5 2 3+3 4 4+4 4 3+4 4 5+5 4 3+4 4 3+4 4 3+4 4 |
| | 5 2 4+4 4 3+3 4 3+4 4 6+6 4 3+4 4 3+4 4 3+4 4 |
| | 6 2 3+3 4 3+3 4 3+4 4 5+5 4 3+4 4 3+4 4 3+4 4 |
| | 6 2 4+4 4 4+4 4 3+4 4 6+6 4 3+4 4 3+4 4 3+4 4 |
| | 5 2 3+3 4 4+4 4 3+4 4 6+6 4 3+4 4 3+4 4 3+4 4 |
| | 5 2 4+4 4 3+3 4 3+4 4 6+6 4 3+4 4 3+4 4 3+4 4 |
| | 5 2 4+4 4 3+3 4 3+4 4 5+5 4 3+4 4 3+4 4 3+4 4 |
| | 5 2 3+3 4 4+4 4 3+4 4 5+5 4 3+4 4 3+4 4 3+4 4 |
| | 6 2 4+4 4 4+4 4 3+4 4 6+6 4 3+4 4 3+4 4 3+4 4 |
| | 6 2 3+3 4 3+3 4 3+4 4 6+6 4 3+4 4 3+4 4 3+4 4 |
| | 6 2 4+4 4 3+3 4 3+4 4 5+5 4 3+4 4 3+4 4 3+4 4 |
| | 6 2 3+3 4 4+4 4 3+4 4 5+5 4 3+4 4 3+4 4 3+4 4 |
| | 6 2 4+4 4 3+3 4 3+4 4 5+5 4 3+4 4 3+4 4 3+4 4 |
| | 6 2 3+3 4 4+4 4 3+4 4 6+6 4 3+4 4 3+4 4 3+4 4 |
| | 5 2 4+4 4 4+4 4 3+4 4 5+5 4 3+4 4 3+4 4 3+4 4 |
| | 5 2 3+3 4 3+3 4 3+4 4 6+6 4 3+4 4 3+4 4 3+4 4 |
| | 5 2 3+3 4 3+3 4 3+4 4 5+5 4 3+4 4 3+4 4 3+4 4 |
| | 6 2 4+4 4 4+4 4 3+4 4 5+5 4 3+4 4 3+4 4 3+4 4 |
| | 5 2 3+3 4 4+4 4 3+4 4 6+6 4 3+4 4 3+4 4 3+4 4 |
| | 6 2 4+4 4 3+3 4 3+4 4 6+6 4 3+4 4 3+4 4 3+4 4 |
| | 5 2 3+3 4 3+3 4 3+4 4 6+6 4 3+4 4 3+4 4 3+4 4 |
| | 6 2 4+4 4 4+4 4 3+4 4 5+5 4 3+4 4 3+4 4 3+4 4 |
| | 5 2 3+3 4 3+3 4 3+4 4 6+6 4 3+4 4 3+4 4 3+4 4 |
| | 6 2 4+4 4 4+4 4 3+4 4 5+5 4 3+4 4 3+4 4 3+4 4 |
| | 6 2 3+3 4 4+4 4 3+4 4 5+5 4 3+4 4 3+4 4 3+4 4 |
| | 6 2 4+4 4 3+3 4 3+4 4 6+6 4 3+4 4 3+4 4 3+4 4 |
| | 5 2 4+4 4 3+3 4 3+4 4 6+6 4 3+4 4 3+4 4 3+4 4 |
| | 6 2 3+3 4 4+4 4 3+4 4 5+5 4 3+4 4 3+4 4 3+4 4 |
| | 5 2 4+4 4 3+3 4 3+4 4 5+5 4 3+4 4 3+4 4 3+4 4 |
| | 5 2 3+3 4 4+4 4 3+4 4 6+6 4 3+4 4 3+4 4 3+4 4 |
| | 5 2 3+3 4 4+4 4 3+4 4 5+5 4 3+4 4 3+4 4 3+4 4 |
| | 5 2 4+4 4 3+3 4 3+4 4 6+6 4 3+4 4 3+4 4 3+4 4 |
| | 5 2 4+4 4 4+4 4 3+4 4 6+6 4 3+4 4 3+4 4 3+4 4 |
| | 6 2 3+3 4 3+3 4 3+4 4 5+5 4 3+4 4 3+4 4 3+4 4 |

Continued on next column

**TABLE XXXVII– continued**

| TQC | ICSD NoE SF |
|---|---|
| | 6 2 4+4 4 3+3 4 3+4 4 6+6 4 3+4 4 3+4 4 3+4 4 |
| | 6 2 3+3 4 4+4 4 3+4 4 5+5 4 3+4 4 3+4 4 3+4 4 |
| | 5 2 3+3 4 3+3 4 3+4 4 5+5 4 3+4 4 3+4 4 3+4 4 |
| | 5 2 4+4 4 4+4 4 3+4 4 6+6 4 3+4 4 3+4 4 3+4 4 |
| | 5 2 |
| | 5 2 |
| | 6 2 |
| | 6 2 |
| | 5 2 |
| | 6 2 |
| | 5 2 |
| | 6 2 |
| | 5 2 |
| | 6 2 |
| | 5 2 |
| | 6 2 |
| | 5 2 |
| | 5 2 |
| | 6 2 |
| | 6 2 |
| | 5 2 |
| | 5 2 |
| | 6 2 |
| | 6 2 |
| | 6 2 |
| | 6 2 |
| | 5 2 |
| | 5 2 |
| | 5 2 |
| | 6 2 |
| | 5 2 |
| | 6 2 |
| | 6 2 |
| | 5 2 |
| | 5 2 |
| | 6 2 |
| | 5 2 |
| | 6 2 |
| | 5 2 |
| | 6 2 |
| | 5 2 |
| | 6 2 |
| | 6 2 |
| | 5 2 |
| | 6 2 |
| | 5 2 |
| | 5 2 |
| | 6 2 |
| | 6 2 |
| | 6 2 |
| | 5 2 |
| | 5 2 |
| | 5 2 |
| | 5 2 |
| | 6 2 |
| | 6 2 |
| | 6 2 |
| | 5 2 |
| | 6 2 |
| | 5 2 |
| | 6 2 |

Continued on next column

**TABLE XXXVII– continued**

| TQC | ICSD | NoE | SF | | | | |
|---|---|---|---|---|---|---|---|
| | 5 2 | | | | | | |
| 62.2.148 | 180220 | 96 | Dy1Si1 | | | | |
| | -GM | -R | -S | -T | -U | -X | -Y | -Z |
| | 5 2 | 3+3 4 | 3+3 4 | 3+4 4 | 6+6 4 | 3+4 4 | 3+4 4 | 3+4 4 |
| | 6 2 | 4+4 4 | 4+4 4 | 3+4 4 | 5+5 4 | 3+4 4 | 3+4 4 | 3+4 4 |
| | 5 2 | 3+3 4 | 3+3 4 | 3+4 4 | 6+6 4 | 3+4 4 | 3+4 4 | 3+4 4 |
| | 6 2 | 4+4 4 | 4+4 4 | 3+4 4 | 5+5 4 | 3+4 4 | 3+4 4 | 3+4 4 |
| | 5 2 | 4+4 4 | 3+3 4 | 3+4 4 | 6+6 4 | 3+4 4 | 3+4 4 | 3+4 4 |
| | 6 2 | 3+3 4 | 4+4 4 | 3+4 4 | 5+5 4 | 3+4 4 | 3+4 4 | 3+4 4 |
| | 5 2 | 3+3 4 | 4+4 4 | 3+4 4 | 6+6 4 | 3+4 4 | 3+4 4 | 3+4 4 |
| | 6 2 | 4+4 4 | 3+3 4 | 3+4 4 | 5+5 4 | 3+4 4 | 3+4 4 | 3+4 4 |
| | 5 2 | 3+3 4 | 3+3 4 | 3+4 4 | 5+5 4 | 3+4 4 | 3+4 4 | 3+4 4 |
| | 6 2 | 4+4 4 | 4+4 4 | 3+4 4 | 6+6 4 | 3+4 4 | 3+4 4 | 3+4 4 |
| | 5 2 | 4+4 4 | 3+3 4 | 3+4 4 | 6+6 4 | 3+4 4 | 3+4 4 | 3+4 4 |
| | 5 2 | 3+3 4 | 4+4 4 | 3+4 4 | 5+5 4 | 3+4 4 | 3+4 4 | 3+4 4 |
| | 6 2 | 3+3 4 | 3+3 4 | 3+4 4 | 5+5 4 | 3+4 4 | 3+4 4 | 3+4 4 |
| | 6 2 | 4+4 4 | 4+4 4 | 3+4 4 | 5+5 4 | 3+4 4 | 3+4 4 | 3+4 4 |
| | 5 2 | 4+4 4 | 3+3 4 | 3+4 4 | 6+6 4 | 3+4 4 | 3+4 4 | 3+4 4 |
| | 6 2 | 3+3 4 | 4+4 4 | 3+4 4 | 6+6 4 | 3+4 4 | 3+4 4 | 3+4 4 |
| | 5 2 | 3+3 4 | 4+4 4 | 3+4 4 | 5+5 4 | 3+4 4 | 3+4 4 | 3+4 4 |
| | 5 2 | 4+4 4 | 3+3 4 | 3+4 4 | 6+6 4 | 3+4 4 | 3+4 4 | 3+4 4 |
| | 6 2 | 3+3 4 | 3+3 4 | 3+4 4 | 5+5 4 | 3+4 4 | 3+4 4 | 3+4 4 |
| | 6 2 | 4+4 4 | 4+4 4 | 3+4 4 | 5+5 4 | 3+4 4 | 3+4 4 | 3+4 4 |
| | 6 2 | 3+3 4 | 4+4 4 | 3+4 4 | 6+6 4 | 3+4 4 | 3+4 4 | 3+4 4 |
| | 5 2 | 4+4 4 | 3+3 4 | 3+4 4 | 5+5 4 | 3+4 4 | 3+4 4 | 3+4 4 |
| | 5 2 | 3+3 4 | 3+3 4 | 3+4 4 | 6+6 4 | 3+4 4 | 3+4 4 | 3+4 4 |
| | 5 2 | 4+4 4 | 4+4 4 | 3+4 4 | 5+5 4 | 3+4 4 | 3+4 4 | 3+4 4 |
| | 6 2 | | | | | | | |
| | 5 2 | | | | | | | |
| | 5 2 | | | | | | | |
| | 6 2 | | | | | | | |
| | 6 2 | | | | | | | |
| | 5 2 | | | | | | | |
| | 6 2 | | | | | | | |
| | 6 2 | | | | | | | |
| | 5 2 | | | | | | | |
| | 5 2 | | | | | | | |
| | 6 2 | | | | | | | |
| | 6 2 | | | | | | | |
| | 5 2 | | | | | | | |
| | 5 2 | | | | | | | |
| | 5 2 | | | | | | | |
| | 6 2 | | | | | | | |
| | 6 2 | | | | | | | |
| | 6 2 | | | | | | | |
| | 5 2 | | | | | | | |
| | 6 2 | | | | | | | |
| | 5 2 | | | | | | | |
| | 6 2 | | | | | | | |
| | 5 2 | | | | | | | |
| | 6 2 | | | | | | | |
| 62.2.149 | 280091 | 184 | Dy1Fe1O3 | | | | |
| | -GM | -R | -S | -T | -U | -X | -Y | -Z |
| | 5 2 | 3+3 4 | 3+3 4 | 3+4 4 | 5+5 4 | 3+4 4 | 3+4 4 | 3+4 4 |
| | 6 2 | 4+4 4 | 4+4 4 | 3+4 4 | 6+6 4 | 3+4 4 | 3+4 4 | 3+4 4 |
| | 5 2 | 3+3 4 | 4+4 4 | 3+4 4 | 5+5 4 | 3+4 4 | 3+4 4 | 3+4 4 |
| | 6 2 | 4+4 4 | 3+3 4 | 3+4 4 | 6+6 4 | 3+4 4 | 3+4 4 | 3+4 4 |
| | 5 2 | 3+3 4 | 4+4 4 | 3+4 4 | 6+6 4 | 3+4 4 | 3+4 4 | 3+4 4 |
| | 6 2 | 4+4 4 | 3+3 4 | 3+4 4 | 5+5 4 | 3+4 4 | 3+4 4 | 3+4 4 |
| | 5 2 | 4+4 4 | 3+3 4 | 3+4 4 | 5+5 4 | 3+4 4 | 3+4 4 | 3+4 4 |
| | 6 2 | 3+3 4 | 4+4 4 | 3+4 4 | 6+6 4 | 3+4 4 | 3+4 4 | 3+4 4 |

Continued on next column

**TABLE XXXVII– continued**

| TQC | ICSD | NoE | SF | | | | |
|---|---|---|---|---|---|---|---|
| | 6 2 | 3+3 4 | 4+4 4 | 3+4 4 | 5+5 4 | 3+4 4 | 3+4 4 | 3+4 4 |
| | 5 2 | 4+4 4 | 3+3 4 | 3+4 4 | 6+6 4 | 3+4 4 | 3+4 4 | 3+4 4 |
| | 5 2 | 4+4 4 | 3+3 4 | 3+4 4 | 6+6 4 | 3+4 4 | 3+4 4 | 3+4 4 |
| | 5 2 | 3+3 4 | 4+4 4 | 3+4 4 | 5+5 4 | 3+4 4 | 3+4 4 | 3+4 4 |
| | 6 2 | 3+3 4 | 4+4 4 | 3+4 4 | 6+6 4 | 3+4 4 | 3+4 4 | 3+4 4 |
| | 6 2 | 4+4 4 | 3+3 4 | 3+4 4 | 5+5 4 | 3+4 4 | 3+4 4 | 3+4 4 |
| | 6 2 | 3+3 4 | 4+4 4 | 3+4 4 | 6+6 4 | 3+4 4 | 3+4 4 | 3+4 4 |
| | 5 2 | 4+4 4 | 3+3 4 | 3+4 4 | 6+6 4 | 3+4 4 | 3+4 4 | 3+4 4 |
| | 5 2 | 4+4 4 | 3+3 4 | 3+4 4 | 6+6 4 | 3+4 4 | 3+4 4 | 3+4 4 |
| | 6 2 | 3+3 4 | 4+4 4 | 3+4 4 | 6+6 4 | 3+4 4 | 3+4 4 | 3+4 4 |
| | 5 2 | 3+3 4 | 3+3 4 | 3+4 4 | 5+5 4 | 3+4 4 | 3+4 4 | 3+4 4 |
| | 6 2 | 4+4 4 | 4+4 4 | 3+4 4 | 5+5 4 | 3+4 4 | 3+4 4 | 3+4 4 |
| | 6 2 | 3+3 4 | 3+3 4 | 3+4 4 | 5+5 4 | 3+4 4 | 3+4 4 | 3+4 4 |
| | 5 2 | 4+4 4 | 4+4 4 | 3+4 4 | 6+6 4 | 3+4 4 | 3+4 4 | 3+4 4 |
| | 6 2 | 4+4 4 | 4+4 4 | 3+4 4 | 6+6 4 | 3+4 4 | 3+4 4 | 3+4 4 |
| | 5 2 | 3+3 4 | 3+3 4 | 3+4 4 | 5+5 4 | 3+4 4 | 3+4 4 | 3+4 4 |
| | 5 2 | 4+4 4 | 4+4 4 | 3+4 4 | 5+5 4 | 3+4 4 | 3+4 4 | 3+4 4 |
| | 6 2 | 3+3 4 | 3+3 4 | 3+4 4 | 6+6 4 | 3+4 4 | 3+4 4 | 3+4 4 |
| | 5 2 | 4+4 4 | 3+3 4 | 3+4 4 | 6+6 4 | 3+4 4 | 3+4 4 | 3+4 4 |
| | 6 2 | 3+3 4 | 4+4 4 | 3+4 4 | 5+5 4 | 3+4 4 | 3+4 4 | 3+4 4 |
| | 5 2 | 4+4 4 | 3+3 4 | 3+4 4 | 5+5 4 | 3+4 4 | 3+4 4 | 3+4 4 |
| | 5 2 | 3+3 4 | 4+4 4 | 3+4 4 | 6+6 4 | 3+4 4 | 3+4 4 | 3+4 4 |
| | 5 2 | 3+3 4 | 3+3 4 | 3+4 4 | 5+5 4 | 3+4 4 | 3+4 4 | 3+4 4 |
| | 5 2 | 4+4 4 | 4+4 4 | 3+4 4 | 5+5 4 | 3+4 4 | 3+4 4 | 3+4 4 |
| | 6 2 | 4+4 4 | 3+3 4 | 3+4 4 | 6+6 4 | 3+4 4 | 3+4 4 | 3+4 4 |
| | 5 2 | 3+3 4 | 4+4 4 | 3+4 4 | 6+6 4 | 3+4 4 | 3+4 4 | 3+4 4 |
| | 5 2 | 4+4 4 | 4+4 4 | 3+4 4 | 5+5 4 | 3+4 4 | 3+4 4 | 3+4 4 |
| | 5 2 | 3+3 4 | 3+3 4 | 3+4 4 | 6+6 4 | 3+4 4 | 3+4 4 | 3+4 4 |
| | 5 2 | 3+3 4 | 4+4 4 | 3+4 4 | 5+5 4 | 3+4 4 | 3+4 4 | 3+4 4 |
| | 5 2 | 4+4 4 | 3+3 4 | 3+4 4 | 5+5 4 | 3+4 4 | 3+4 4 | 3+4 4 |
| | 6 2 | 4+4 4 | 3+3 4 | 3+4 4 | 6+6 4 | 3+4 4 | 3+4 4 | 3+4 4 |
| | 6 2 | 3+3 4 | 4+4 4 | 3+4 4 | 6+6 4 | 3+4 4 | 3+4 4 | 3+4 4 |
| | 5 2 | 3+3 4 | 3+3 4 | 3+4 4 | 6+6 4 | 3+4 4 | 3+4 4 | 3+4 4 |
| | 6 2 | 4+4 4 | 4+4 4 | 3+4 4 | 6+6 4 | 3+4 4 | 3+4 4 | 3+4 4 |
| | 6 2 | 3+3 4 | 3+3 4 | 3+4 4 | 6+6 4 | 3+4 4 | 3+4 4 | 3+4 4 |
| | 6 2 | 4+4 4 | 4+4 4 | 3+4 4 | 5+5 4 | 3+4 4 | 3+4 4 | 3+4 4 |
| | 5 2 | 3+3 4 | 4+4 4 | 3+4 4 | 6+6 4 | 3+4 4 | 3+4 4 | 3+4 4 |
| | 6 2 | 4+4 4 | 3+3 4 | 3+4 4 | 5+5 4 | 3+4 4 | 3+4 4 | 3+4 4 |
| | 5 2 | | | | | | | |
| | 6 2 | | | | | | | |
| | 5 2 | | | | | | | |
| | 5 2 | | | | | | | |
| | 6 2 | | | | | | | |
| | 6 2 | | | | | | | |
| | 5 2 | | | | | | | |
| | 6 2 | | | | | | | |
| | 5 2 | | | | | | | |
| | 5 2 | | | | | | | |
| | 6 2 | | | | | | | |
| | 6 2 | | | | | | | |
| | 6 2 | | | | | | | |
| | 5 2 | | | | | | | |
| | 6 2 | | | | | | | |
| | 6 2 | | | | | | | |
| | 6 2 | | | | | | | |
| | 6 2 | | | | | | | |
| | 5 2 | | | | | | | |
| | 5 2 | | | | | | | |
| | 5 2 | | | | | | | |
| | 5 2 | | | | | | | |
| | 5 2 | | | | | | | |

Continued on next column

**TABLE XXXVII– continued**

| TQC | ICSD | NoE | SF | | | | | |
|---|---|---|---|---|---|---|---|---|
| | 5 2 | | | | | | | |
| | 6 2 | | | | | | | |
| | 5 2 | | | | | | | |
| | 6 2 | | | | | | | |
| | 5 2 | | | | | | | |
| | 6 2 | | | | | | | |
| | 5 2 | | | | | | | |
| | 6 2 | | | | | | | |
| | 6 2 | | | | | | | |
| | 5 2 | | | | | | | |
| | 6 2 | | | | | | | |
| | 5 2 | | | | | | | |
| | 5 2 | | | | | | | |
| | 5 2 | | | | | | | |
| | 5 2 | | | | | | | |
| | 5 2 | | | | | | | |
| | 5 2 | | | | | | | |
| | 5 2 | | | | | | | |
| | 5 2 | | | | | | | |
| | 6 2 | | | | | | | |
| | 5 2 | | | | | | | |
| | 6 2 | | | | | | | |
| | 5 2 | | | | | | | |
| 62.2.150 | 27280 | 184 | Dy1Fe1O3 | | | | | |
| | -GM | -R | -S | -T | -U | -X | -Y | -Z |
| | 5 2 | 3+3 4 | 3+3 4 | 3+4 4 | 5+5 4 | 3+4 4 | 3+4 4 | 3+4 4 |
| | 6 2 | 4+4 4 | 4+4 4 | 3+4 4 | 6+6 4 | 3+4 4 | 3+4 4 | 3+4 4 |
| | 5 2 | 4+4 4 | 4+4 4 | 3+4 4 | 5+5 4 | 3+4 4 | 3+4 4 | 3+4 4 |
| | 6 2 | 3+3 4 | 3+3 4 | 3+4 4 | 6+6 4 | 3+4 4 | 3+4 4 | 3+4 4 |
| | 5 2 | 4+4 4 | 4+4 4 | 3+4 4 | 6+6 4 | 3+4 4 | 3+4 4 | 3+4 4 |
| | 6 2 | 3+3 4 | 3+3 4 | 3+4 4 | 5+5 4 | 3+4 4 | 3+4 4 | 3+4 4 |
| | 5 2 | 3+3 4 | 3+3 4 | 3+4 4 | 5+5 4 | 3+4 4 | 3+4 4 | 3+4 4 |
| | 6 2 | 4+4 4 | 4+4 4 | 3+4 4 | 6+6 4 | 3+4 4 | 3+4 4 | 3+4 4 |
| | 6 2 | 4+4 4 | 4+4 4 | 3+4 4 | 5+5 4 | 3+4 4 | 3+4 4 | 3+4 4 |
| | 5 2 | 3+3 4 | 3+3 4 | 3+4 4 | 6+6 4 | 3+4 4 | 3+4 4 | 3+4 4 |
| | 5 2 | 3+3 4 | 3+3 4 | 3+4 4 | 6+6 4 | 3+4 4 | 3+4 4 | 3+4 4 |
| | 5 2 | 4+4 4 | 4+4 4 | 3+4 4 | 5+5 4 | 3+4 4 | 3+4 4 | 3+4 4 |
| | 6 2 | 4+4 4 | 4+4 4 | 3+4 4 | 6+6 4 | 3+4 4 | 3+4 4 | 3+4 4 |
| | 6 2 | 3+3 4 | 3+3 4 | 3+4 4 | 5+5 4 | 3+4 4 | 3+4 4 | 3+4 4 |
| | 6 2 | 3+3 4 | 4+4 4 | 3+4 4 | 6+6 4 | 3+4 4 | 3+4 4 | 3+4 4 |
| | 5 2 | 4+4 4 | 3+3 4 | 3+4 4 | 6+6 4 | 3+4 4 | 3+4 4 | 3+4 4 |
| | 5 2 | 3+3 4 | 3+3 4 | 3+4 4 | 6+6 4 | 3+4 4 | 3+4 4 | 3+4 4 |
| | 6 2 | 4+4 4 | 4+4 4 | 3+4 4 | 6+6 4 | 3+4 4 | 3+4 4 | 3+4 4 |
| | 5 2 | 3+3 4 | 3+3 4 | 3+4 4 | 5+5 4 | 3+4 4 | 3+4 4 | 3+4 4 |
| | 6 2 | 4+4 4 | 4+4 4 | 3+4 4 | 5+5 4 | 3+4 4 | 3+4 4 | 3+4 4 |
| | 6 2 | 4+4 4 | 3+3 4 | 3+4 4 | 5+5 4 | 3+4 4 | 3+4 4 | 3+4 4 |
| | 5 2 | 3+3 4 | 4+4 4 | 3+4 4 | 6+6 4 | 3+4 4 | 3+4 4 | 3+4 4 |
| | 6 2 | 3+3 4 | 4+4 4 | 3+4 4 | 6+6 4 | 3+4 4 | 3+4 4 | 3+4 4 |
| | 5 2 | 4+4 4 | 3+3 4 | 3+4 4 | 5+5 4 | 3+4 4 | 3+4 4 | 3+4 4 |
| | 5 2 | 3+3 4 | 4+4 4 | 3+4 4 | 5+5 4 | 3+4 4 | 3+4 4 | 3+4 4 |
| | 6 2 | 4+4 4 | 3+3 4 | 3+4 4 | 6+6 4 | 3+4 4 | 3+4 4 | 3+4 4 |
| | 5 2 | 3+3 4 | 3+3 4 | 3+4 4 | 6+6 4 | 3+4 4 | 3+4 4 | 3+4 4 |
| | 6 2 | 4+4 4 | 4+4 4 | 3+4 4 | 5+5 4 | 3+4 4 | 3+4 4 | 3+4 4 |
| | 5 2 | 3+3 4 | 3+3 4 | 3+4 4 | 5+5 4 | 3+4 4 | 3+4 4 | 3+4 4 |
| | 5 2 | 4+4 4 | 4+4 4 | 3+4 4 | 6+6 4 | 3+4 4 | 3+4 4 | 3+4 4 |
| | 5 2 | 4+4 4 | 3+3 4 | 3+4 4 | 5+5 4 | 3+4 4 | 3+4 4 | 3+4 4 |
| | 5 2 | 3+3 4 | 4+4 4 | 3+4 4 | 5+5 4 | 3+4 4 | 3+4 4 | 3+4 4 |
| | 6 2 | 3+3 4 | 3+3 4 | 3+4 4 | 6+6 4 | 3+4 4 | 3+4 4 | 3+4 4 |
| | 5 2 | 4+4 4 | 4+4 4 | 3+4 4 | 6+6 4 | 3+4 4 | 3+4 4 | 3+4 4 |
| | 5 2 | 3+3 4 | 4+4 4 | 3+4 4 | 5+5 4 | 3+4 4 | 3+4 4 | 3+4 4 |
| | 5 2 | 4+4 4 | 3+3 4 | 3+4 4 | 6+6 4 | 3+4 4 | 3+4 4 | 3+4 4 |

Continued on next column

**TABLE XXXVII– continued**

| TQC | ICSD | NoE | SF | | | | | |
|---|---|---|---|---|---|---|---|---|
| | 5 2 | 4+4 4 | 4+4 4 | 3+4 4 | 5+5 4 | 3+4 4 | 3+4 4 | 3+4 4 |
| | 5 2 | 3+3 4 | 3+3 4 | 3+4 4 | 5+5 4 | 3+4 4 | 3+4 4 | 3+4 4 |
| | 6 2 | 3+3 4 | 3+3 4 | 3+4 4 | 6+6 4 | 3+4 4 | 3+4 4 | 3+4 4 |
| | 6 2 | 4+4 4 | 4+4 4 | 3+4 4 | 6+6 4 | 3+4 4 | 3+4 4 | 3+4 4 |
| | 5 2 | 4+4 4 | 3+3 4 | 3+4 4 | 6+6 4 | 3+4 4 | 3+4 4 | 3+4 4 |
| | 6 2 | 3+3 4 | 4+4 4 | 3+4 4 | 6+6 4 | 3+4 4 | 3+4 4 | 3+4 4 |
| | 6 2 | 4+4 4 | 3+3 4 | 3+4 4 | 6+6 4 | 3+4 4 | 3+4 4 | 3+4 4 |
| | 6 2 | 3+3 4 | 4+4 4 | 3+4 4 | 5+5 4 | 3+4 4 | 3+4 4 | 3+4 4 |
| | 5 2 | 4+4 4 | 4+4 4 | 3+4 4 | 6+6 4 | 3+4 4 | 3+4 4 | 3+4 4 |
| | 5 2 | 3+3 4 | 3+3 4 | 3+4 4 | 5+5 4 | 3+4 4 | 3+4 4 | 3+4 4 |
| | 6 2 | | | | | | | |
| | 6 2 | | | | | | | |
| | 5 2 | | | | | | | |
| | 5 2 | | | | | | | |
| | 6 2 | | | | | | | |
| | 6 2 | | | | | | | |
| | 5 2 | | | | | | | |
| | 6 2 | | | | | | | |
| | 5 2 | | | | | | | |
| | 5 2 | | | | | | | |
| | 6 2 | | | | | | | |
| | 6 2 | | | | | | | |
| | 6 2 | | | | | | | |
| | 5 2 | | | | | | | |
| | 6 2 | | | | | | | |
| | 6 2 | | | | | | | |
| | 6 2 | | | | | | | |
| | 6 2 | | | | | | | |
| | 5 2 | | | | | | | |
| | 5 2 | | | | | | | |
| | 5 2 | | | | | | | |
| | 5 2 | | | | | | | |
| | 5 2 | | | | | | | |
| | 5 2 | | | | | | | |
| | 6 2 | | | | | | | |
| | 5 2 | | | | | | | |
| | 6 2 | | | | | | | |
| | 5 2 | | | | | | | |
| | 6 2 | | | | | | | |
| | 5 2 | | | | | | | |
| | 6 2 | | | | | | | |
| | 6 2 | | | | | | | |
| | 5 2 | | | | | | | |
| | 6 2 | | | | | | | |
| | 5 2 | | | | | | | |
| | 5 2 | | | | | | | |
| | 5 2 | | | | | | | |
| | 5 2 | | | | | | | |
| | 5 2 | | | | | | | |
| | 5 2 | | | | | | | |
| | 5 2 | | | | | | | |
| | 5 2 | | | | | | | |
| | 6 2 | | | | | | | |
| | 5 2 | | | | | | | |
| | 6 2 | | | | | | | |
| | 5 2 | | | | | | | |
| 62.2.151 | 86391 | 72 | Cu1Sc1Si1 | | | | | |
| | -GM | -R | -S | -T | -U | -X | -Y | -Z |
| | 5 2 | 3+3 4 | 3+3 4 | 3+4 4 | 6+6 4 | 3+4 4 | 3+4 4 | 3+4 4 |
| | 5 2 | 4+4 4 | 4+4 4 | 3+4 4 | 5+5 4 | 3+4 4 | 3+4 4 | 3+4 4 |
| | 6 2 | 3+3 4 | 4+4 4 | 3+4 4 | 6+6 4 | 3+4 4 | 3+4 4 | 3+4 4 |

Continued on next column

**TABLE XXXVII– continued**

| TQC | ICSD | NoE | SF | | | | | |
|---|---|---|---|---|---|---|---|---|
| | 6 2 | 4+4 4 | 3+3 4 | 3+4 4 | 5+5 4 | 3+4 4 | 3+4 4 | 3+4 4 |
| | 6 2 | 4+4 4 | 3+3 4 | 3+4 4 | 5+5 4 | 3+4 4 | 3+4 4 | 3+4 4 |
| | 5 2 | 3+3 4 | 4+4 4 | 3+4 4 | 5+5 4 | 3+4 4 | 3+4 4 | 3+4 4 |
| | 5 2 | 4+4 4 | 4+4 4 | 3+4 4 | 6+6 4 | 3+4 4 | 3+4 4 | 3+4 4 |
| | 5 2 | 3+3 4 | 3+3 4 | 3+4 4 | 6+6 4 | 3+4 4 | 3+4 4 | 3+4 4 |
| | 6 2 | 3+3 4 | 4+4 4 | 3+4 4 | 6+6 4 | 3+4 4 | 3+4 4 | 3+4 4 |
| | 6 2 | 4+4 4 | 3+3 4 | 3+4 4 | 5+5 4 | 3+4 4 | 3+4 4 | 3+4 4 |
| | 5 2 | 3+3 4 | 4+4 4 | 3+4 4 | 6+6 4 | 3+4 4 | 3+4 4 | 3+4 4 |
| | 6 2 | 4+4 4 | 3+3 4 | 3+4 4 | 5+5 4 | 3+4 4 | 3+4 4 | 3+4 4 |
| | 6 2 | 4+4 4 | 3+3 4 | 3+4 4 | 6+6 4 | 3+4 4 | 3+4 4 | 3+4 4 |
| | 6 2 | 3+3 4 | 4+4 4 | 3+4 4 | 5+5 4 | 3+4 4 | 3+4 4 | 3+4 4 |
| | 5 2 | 4+4 4 | 3+3 4 | 3+4 4 | 6+6 4 | 3+4 4 | 3+4 4 | 3+4 4 |
| | 6 2 | 3+3 4 | 4+4 4 | 3+4 4 | 6+6 4 | 3+4 4 | 3+4 4 | 3+4 4 |
| | 5 2 | 3+3 4 | 3+3 4 | 3+4 4 | 5+5 4 | 3+4 4 | 3+4 4 | 3+4 4 |
| | 5 2 | 4+4 4 | 4+4 4 | 3+4 4 | 5+5 4 | 3+4 4 | 3+4 4 | 3+4 4 |
| | 5 2 | | | | | | | |
| | 5 2 | | | | | | | |
| | 6 2 | | | | | | | |
| | 5 2 | | | | | | | |
| | 6 2 | | | | | | | |
| | 5 2 | | | | | | | |
| | 6 2 | | | | | | | |
| | 6 2 | | | | | | | |
| | 5 2 | | | | | | | |
| | 5 2 | | | | | | | |
| | 6 2 | | | | | | | |
| | 6 2 | | | | | | | |
| | 5 2 | | | | | | | |
| | 5 2 | | | | | | | |
| | 5 2 | | | | | | | |
| | 6 2 | | | | | | | |
| | 5 2 | | | | | | | |
| | 6 2 | | | | | | | |
| 62.2.152 | 172850 | 184 | Cu1S2Tm1 | | | | | |
| | -GM | -R | -S | -T | -U | -X | -Y | -Z |
| | 5 2 | 3+3 4 | 3+3 4 | 3+4 4 | 5+5 4 | 3+4 4 | 3+4 4 | 3+4 4 |
| | 5 2 | 4+4 4 | 4+4 4 | 3+4 4 | 6+6 4 | 3+4 4 | 3+4 4 | 3+4 4 |
| | 6 2 | 4+4 4 | 4+4 4 | 3+4 4 | 6+6 4 | 3+4 4 | 3+4 4 | 3+4 4 |
| | 6 2 | 3+3 4 | 3+3 4 | 3+4 4 | 5+5 4 | 3+4 4 | 3+4 4 | 3+4 4 |
| | 5 2 | 3+3 4 | 3+3 4 | 3+4 4 | 6+6 4 | 3+4 4 | 3+4 4 | 3+4 4 |
| | 5 2 | 4+4 4 | 4+4 4 | 3+4 4 | 5+5 4 | 3+4 4 | 3+4 4 | 3+4 4 |
| | 6 2 | 3+3 4 | 3+3 4 | 3+4 4 | 6+6 4 | 3+4 4 | 3+4 4 | 3+4 4 |
| | 6 2 | 4+4 4 | 4+4 4 | 3+4 4 | 5+5 4 | 3+4 4 | 3+4 4 | 3+4 4 |
| | 5 2 | 3+3 4 | 3+3 4 | 3+4 4 | 6+6 4 | 3+4 4 | 3+4 4 | 3+4 4 |
| | 5 2 | 4+4 4 | 4+4 4 | 3+4 4 | 5+5 4 | 3+4 4 | 3+4 4 | 3+4 4 |
| | 6 2 | 4+4 4 | 4+4 4 | 3+4 4 | 5+5 4 | 3+4 4 | 3+4 4 | 3+4 4 |
| | 6 2 | 3+3 4 | 3+3 4 | 3+4 4 | 6+6 4 | 3+4 4 | 3+4 4 | 3+4 4 |
| | 5 2 | 4+4 4 | 3+3 4 | 3+4 4 | 5+5 4 | 3+4 4 | 3+4 4 | 3+4 4 |
| | 5 2 | 3+3 4 | 4+4 4 | 3+4 4 | 6+6 4 | 3+4 4 | 3+4 4 | 3+4 4 |
| | 6 2 | 3+3 4 | 3+3 4 | 3+4 4 | 6+6 4 | 3+4 4 | 3+4 4 | 3+4 4 |
| | 6 2 | 4+4 4 | 4+4 4 | 3+4 4 | 5+5 4 | 3+4 4 | 3+4 4 | 3+4 4 |
| | 5 2 | 3+3 4 | 4+4 4 | 3+4 4 | 6+6 4 | 3+4 4 | 3+4 4 | 3+4 4 |
| | 5 2 | 4+4 4 | 3+3 4 | 3+4 4 | 5+5 4 | 3+4 4 | 3+4 4 | 3+4 4 |
| | 6 2 | 3+3 4 | 3+3 4 | 3+4 4 | 6+6 4 | 3+4 4 | 3+4 4 | 3+4 4 |
| | 6 2 | 4+4 4 | 4+4 4 | 3+4 4 | 5+5 4 | 3+4 4 | 3+4 4 | 3+4 4 |
| | 6 2 | 3+3 4 | 3+3 4 | 3+4 4 | 6+6 4 | 3+4 4 | 3+4 4 | 3+4 4 |
| | 5 2 | 4+4 4 | 4+4 4 | 3+4 4 | 5+5 4 | 3+4 4 | 3+4 4 | 3+4 4 |
| | 6 2 | 3+3 4 | 3+3 4 | 3+4 4 | 5+5 4 | 3+4 4 | 3+4 4 | 3+4 4 |
| | 5 2 | 4+4 4 | 4+4 4 | 3+4 4 | 6+6 4 | 3+4 4 | 3+4 4 | 3+4 4 |
| | 5 2 | 3+3 4 | 4+4 4 | 3+4 4 | 5+5 4 | 3+4 4 | 3+4 4 | 3+4 4 |
| | 5 2 | 4+4 4 | 3+3 4 | 3+4 4 | 5+5 4 | 3+4 4 | 3+4 4 | 3+4 4 |

Continued on next column

| TQC | ICSD | NoE | SF | | | | | |
|---|---|---|---|---|---|---|---|---|
| | 6 2 | 4+4 4 | 3+3 4 | 3+4 4 | 6+6 4 | 3+4 4 | 3+4 4 | 3+4 4 |
| | 6 2 | 3+3 4 | 4+4 4 | 3+4 4 | 5+5 4 | 3+4 4 | 3+4 4 | 3+4 4 |
| | 6 2 | 4+4 4 | 4+4 4 | 3+4 4 | 6+6 4 | 3+4 4 | 3+4 4 | 3+4 4 |
| | 5 2 | 3+3 4 | 3+3 4 | 3+4 4 | 5+5 4 | 3+4 4 | 3+4 4 | 3+4 4 |
| | 6 2 | 3+3 4 | 3+3 4 | 3+4 4 | 6+6 4 | 3+4 4 | 3+4 4 | 3+4 4 |
| | 5 2 | 4+4 4 | 4+4 4 | 3+4 4 | 5+5 4 | 3+4 4 | 3+4 4 | 3+4 4 |
| | 5 2 | 3+3 4 | 4+4 4 | 3+4 4 | 6+6 4 | 3+4 4 | 3+4 4 | 3+4 4 |
| | 6 2 | 4+4 4 | 3+3 4 | 3+4 4 | 5+5 4 | 3+4 4 | 3+4 4 | 3+4 4 |
| | 5 2 | 3+3 4 | 3+3 4 | 3+4 4 | 6+6 4 | 3+4 4 | 3+4 4 | 3+4 4 |
| | 6 2 | 4+4 4 | 4+4 4 | 3+4 4 | 6+6 4 | 3+4 4 | 3+4 4 | 3+4 4 |
| | 6 2 | 3+3 4 | 3+3 4 | 3+4 4 | 5+5 4 | 3+4 4 | 3+4 4 | 3+4 4 |
| | 5 2 | 4+4 4 | 4+4 4 | 3+4 4 | 6+6 4 | 3+4 4 | 3+4 4 | 3+4 4 |
| | 6 2 | 3+3 4 | 4+4 4 | 3+4 4 | 5+5 4 | 3+4 4 | 3+4 4 | 3+4 4 |
| | 5 2 | 4+4 4 | 3+3 4 | 3+4 4 | 6+6 4 | 3+4 4 | 3+4 4 | 3+4 4 |
| | 5 2 | 3+3 4 | 3+3 4 | 3+4 4 | 6+6 4 | 3+4 4 | 3+4 4 | 3+4 4 |
| | 6 2 | 4+4 4 | 4+4 4 | 3+4 4 | 5+5 4 | 3+4 4 | 3+4 4 | 3+4 4 |
| | 5 2 | 3+3 4 | 4+4 4 | 3+4 4 | 6+6 4 | 3+4 4 | 3+4 4 | 3+4 4 |
| | 6 2 | 4+4 4 | 3+3 4 | 3+4 4 | 5+5 4 | 3+4 4 | 3+4 4 | 3+4 4 |
| | 5 2 | 4+4 4 | 3+3 4 | 3+4 4 | 6+6 4 | 3+4 4 | 3+4 4 | 3+4 4 |
| | 5 2 | 3+3 4 | 4+4 4 | 3+4 4 | 5+5 4 | 3+4 4 | 3+4 4 | 3+4 4 |
| | 6 2 | | | | | | | |
| | 6 2 | | | | | | | |
| | 6 2 | | | | | | | |
| | 5 2 | | | | | | | |
| | 6 2 | | | | | | | |
| | 5 2 | | | | | | | |
| | 5 2 | | | | | | | |
| | 5 2 | | | | | | | |
| | 5 2 | | | | | | | |
| | 5 2 | | | | | | | |
| | 6 2 | | | | | | | |
| | 6 2 | | | | | | | |
| | 6 2 | | | | | | | |
| | 6 2 | | | | | | | |
| | 5 2 | | | | | | | |
| | 5 2 | | | | | | | |
| | 5 2 | | | | | | | |
| | 5 2 | | | | | | | |
| | 6 2 | | | | | | | |
| | 6 2 | | | | | | | |
| | 5 2 | | | | | | | |
| | 5 2 | | | | | | | |
| | 5 2 | | | | | | | |
| | 6 2 | | | | | | | |
| | 6 2 | | | | | | | |
| | 5 2 | | | | | | | |
| | 5 2 | | | | | | | |
| | 6 2 | | | | | | | |
| | 6 2 | | | | | | | |
| | 6 2 | | | | | | | |
| | 6 2 | | | | | | | |
| | 5 2 | | | | | | | |
| | 6 2 | | | | | | | |
| | 6 2 | | | | | | | |
| | 5 2 | | | | | | | |
| | 5 2 | | | | | | | |
| | 5 2 | | | | | | | |
| | 6 2 | | | | | | | |
| | 6 2 | | | | | | | |
| | 5 2 | | | | | | | |
| | 5 2 | | | | | | | |

Continued on next column

**TABLE XXXVII– continued**

| TQC | ICSD | NoE | SF | | | | |
|---|---|---|---|---|---|---|---|
| | 5 2 | | | | | | |
| | 6 2 | | | | | | |
| | 5 2 | | | | | | |
| | 6 2 | | | | | | |
| | 6 2 | | | | | | |
| 62.2.153 | 172848 | 176 | Cu1Ho1S2 | | | | |
| | -GM | -R | -S | -T | -U | -X | -Y | -Z |
| | 5 2 | 3+3 4 | 3+3 4 | 3+4 4 | 6+6 4 | 3+4 4 | 3+4 4 | 3+4 4 |
| | 5 2 | 4+4 4 | 4+4 4 | 3+4 4 | 5+5 4 | 3+4 4 | 3+4 4 | 3+4 4 |
| | 6 2 | 4+4 4 | 4+4 4 | 3+4 4 | 6+6 4 | 3+4 4 | 3+4 4 | 3+4 4 |
| | 6 2 | 3+3 4 | 3+3 4 | 3+4 4 | 5+5 4 | 3+4 4 | 3+4 4 | 3+4 4 |
| | 5 2 | 3+3 4 | 3+3 4 | 3+4 4 | 6+6 4 | 3+4 4 | 3+4 4 | 3+4 4 |
| | 5 2 | 4+4 4 | 4+4 4 | 3+4 4 | 5+5 4 | 3+4 4 | 3+4 4 | 3+4 4 |
| | 6 2 | 4+4 4 | 4+4 4 | 3+4 4 | 6+6 4 | 3+4 4 | 3+4 4 | 3+4 4 |
| | 6 2 | 3+3 4 | 3+3 4 | 3+4 4 | 5+5 4 | 3+4 4 | 3+4 4 | 3+4 4 |
| | 5 2 | 3+3 4 | 3+3 4 | 3+4 4 | 6+6 4 | 3+4 4 | 3+4 4 | 3+4 4 |
| | 6 2 | 4+4 4 | 4+4 4 | 3+4 4 | 5+5 4 | 3+4 4 | 3+4 4 | 3+4 4 |
| | 5 2 | 4+4 4 | 4+4 4 | 3+4 4 | 5+5 4 | 3+4 4 | 3+4 4 | 3+4 4 |
| | 6 2 | 3+3 4 | 3+3 4 | 3+4 4 | 6+6 4 | 3+4 4 | 3+4 4 | 3+4 4 |
| | 5 2 | 4+4 4 | 3+3 4 | 3+4 4 | 5+5 4 | 3+4 4 | 3+4 4 | 3+4 4 |
| | 5 2 | 3+3 4 | 4+4 4 | 3+4 4 | 6+6 4 | 3+4 4 | 3+4 4 | 3+4 4 |
| | 6 2 | 3+3 4 | 3+3 4 | 3+4 4 | 6+6 4 | 3+4 4 | 3+4 4 | 3+4 4 |
| | 6 2 | 4+4 4 | 4+4 4 | 3+4 4 | 6+6 4 | 3+4 4 | 3+4 4 | 3+4 4 |
| | 5 2 | 3+3 4 | 4+4 4 | 3+4 4 | 5+5 4 | 3+4 4 | 3+4 4 | 3+4 4 |
| | 5 2 | 4+4 4 | 3+3 4 | 3+4 4 | 5+5 4 | 3+4 4 | 3+4 4 | 3+4 4 |
| | 6 2 | 3+3 4 | 3+3 4 | 3+4 4 | 6+6 4 | 3+4 4 | 3+4 4 | 3+4 4 |
| | 6 2 | 4+4 4 | 4+4 4 | 3+4 4 | 5+5 4 | 3+4 4 | 3+4 4 | 3+4 4 |
| | 6 2 | 3+3 4 | 3+3 4 | 3+4 4 | 5+5 4 | 3+4 4 | 3+4 4 | 3+4 4 |
| | 5 2 | 4+4 4 | 4+4 4 | 3+4 4 | 6+6 4 | 3+4 4 | 3+4 4 | 3+4 4 |
| | 6 2 | 3+3 4 | 3+3 4 | 3+4 4 | 5+5 4 | 3+4 4 | 3+4 4 | 3+4 4 |
| | 5 2 | 4+4 4 | 4+4 4 | 3+4 4 | 6+6 4 | 3+4 4 | 3+4 4 | 3+4 4 |
| | 5 2 | 3+3 4 | 4+4 4 | 3+4 4 | 5+5 4 | 3+4 4 | 3+4 4 | 3+4 4 |
| | 5 2 | 4+4 4 | 3+3 4 | 3+4 4 | 5+5 4 | 3+4 4 | 3+4 4 | 3+4 4 |
| | 6 2 | 4+4 4 | 3+3 4 | 3+4 4 | 6+6 4 | 3+4 4 | 3+4 4 | 3+4 4 |
| | 6 2 | 3+3 4 | 4+4 4 | 3+4 4 | 5+5 4 | 3+4 4 | 3+4 4 | 3+4 4 |
| | 6 2 | 3+3 4 | 4+4 4 | 3+4 4 | 6+6 4 | 3+4 4 | 3+4 4 | 3+4 4 |
| | 5 2 | 4+4 4 | 3+3 4 | 3+4 4 | 5+5 4 | 3+4 4 | 3+4 4 | 3+4 4 |
| | 6 2 | 4+4 4 | 3+3 4 | 3+4 4 | 6+6 4 | 3+4 4 | 3+4 4 | 3+4 4 |
| | 5 2 | 3+3 4 | 4+4 4 | 3+4 4 | 6+6 4 | 3+4 4 | 3+4 4 | 3+4 4 |
| | 5 2 | 3+3 4 | 3+3 4 | 3+4 4 | 5+5 4 | 3+4 4 | 3+4 4 | 3+4 4 |
| | 6 2 | 4+4 4 | 4+4 4 | 3+4 4 | 5+5 4 | 3+4 4 | 3+4 4 | 3+4 4 |
| | 5 2 | 3+3 4 | 3+3 4 | 3+4 4 | 6+6 4 | 3+4 4 | 3+4 4 | 3+4 4 |
| | 6 2 | 4+4 4 | 4+4 4 | 3+4 4 | 6+6 4 | 3+4 4 | 3+4 4 | 3+4 4 |
| | 6 2 | 4+4 4 | 3+3 4 | 3+4 4 | 5+5 4 | 3+4 4 | 3+4 4 | 3+4 4 |
| | 5 2 | 3+3 4 | 4+4 4 | 3+4 4 | 6+6 4 | 3+4 4 | 3+4 4 | 3+4 4 |
| | 6 2 | 3+3 4 | 3+3 4 | 3+4 4 | 5+5 4 | 3+4 4 | 3+4 4 | 3+4 4 |
| | 5 2 | 4+4 4 | 4+4 4 | 3+4 4 | 6+6 4 | 3+4 4 | 3+4 4 | 3+4 4 |
| | 6 2 | 3+3 4 | 3+3 4 | 3+4 4 | 6+6 4 | 3+4 4 | 3+4 4 | 3+4 4 |
| | 5 2 | 4+4 4 | 4+4 4 | 3+4 4 | 5+5 4 | 3+4 4 | 3+4 4 | 3+4 4 |
| | 5 2 | 3+3 4 | 4+4 4 | 3+4 4 | 6+6 4 | 3+4 4 | 3+4 4 | 3+4 4 |
| | 6 2 | 4+4 4 | 3+3 4 | 3+4 4 | 5+5 4 | 3+4 4 | 3+4 4 | 3+4 4 |
| | 5 2 | | | | | | | |
| | 5 2 | | | | | | | |
| | 6 2 | | | | | | | |
| | 6 2 | | | | | | | |
| | 6 2 | | | | | | | |
| | 5 2 | | | | | | | |
| | 6 2 | | | | | | | |
| | 5 2 | | | | | | | |
| | 5 2 | | | | | | | |
| | 5 2 | | | | | | | |

Continued on next column

| TQC | ICSD | NoE | SF | | | | |
|---|---|---|---|---|---|---|---|
| | 5 2 | | | | | | |
| | 5 2 | | | | | | |
| | 5 2 | | | | | | |
| | 6 2 | | | | | | |
| | 6 2 | | | | | | |
| | 5 2 | | | | | | |
| | 6 2 | | | | | | |
| | 5 2 | | | | | | |
| | 6 2 | | | | | | |
| | 6 2 | | | | | | |
| | 6 2 | | | | | | |
| | 5 2 | | | | | | |
| | 5 2 | | | | | | |
| | 5 2 | | | | | | |
| | 6 2 | | | | | | |
| | 6 2 | | | | | | |
| | 5 2 | | | | | | |
| | 5 2 | | | | | | |
| | 5 2 | | | | | | |
| | 6 2 | | | | | | |
| | 6 2 | | | | | | |
| | 6 2 | | | | | | |
| | 5 2 | | | | | | |
| | 6 2 | | | | | | |
| | 6 2 | | | | | | |
| | 6 2 | | | | | | |
| | 5 2 | | | | | | |
| | 5 2 | | | | | | |
| | 5 2 | | | | | | |
| | 6 2 | | | | | | |
| | 6 2 | | | | | | |
| | 5 2 | | | | | | |
| | 5 2 | | | | | | |
| | 5 2 | | | | | | |
| 62.2.154 | 33799 | 200 | Cs2O4Ru1 | | | | |
| | -GM | -R | -S | -T | -U | -X | -Y | -Z |
| | 5 2 | 3+3 4 | 3+3 4 | 3+4 4 | 6+6 4 | 3+4 4 | 3+4 4 | 3+4 4 |
| | 5 2 | 4+4 4 | 4+4 4 | 3+4 4 | 5+5 4 | 3+4 4 | 3+4 4 | 3+4 4 |
| | 6 2 | 3+3 4 | 3+3 4 | 3+4 4 | 5+5 4 | 3+4 4 | 3+4 4 | 3+4 4 |
| | 6 2 | 4+4 4 | 4+4 4 | 3+4 4 | 6+6 4 | 3+4 4 | 3+4 4 | 3+4 4 |
| | 6 2 | 3+3 4 | 3+3 4 | 3+4 4 | 5+5 4 | 3+4 4 | 3+4 4 | 3+4 4 |
| | 5 2 | 4+4 4 | 4+4 4 | 3+4 4 | 6+6 4 | 3+4 4 | 3+4 4 | 3+4 4 |
| | 5 2 | 3+3 4 | 3+3 4 | 3+4 4 | 5+5 4 | 3+4 4 | 3+4 4 | 3+4 4 |
| | 6 2 | 4+4 4 | 4+4 4 | 3+4 4 | 6+6 4 | 3+4 4 | 3+4 4 | 3+4 4 |
| | 5 2 | 4+4 4 | 3+3 4 | 3+4 4 | 5+5 4 | 3+4 4 | 3+4 4 | 3+4 4 |
| | 6 2 | 3+3 4 | 4+4 4 | 3+4 4 | 6+6 4 | 3+4 4 | 3+4 4 | 3+4 4 |
| | 6 2 | 3+3 4 | 3+3 4 | 3+4 4 | 5+5 4 | 3+4 4 | 3+4 4 | 3+4 4 |
| | 5 2 | 4+4 4 | 4+4 4 | 3+4 4 | 6+6 4 | 3+4 4 | 3+4 4 | 3+4 4 |
| | 5 2 | 4+4 4 | 4+4 4 | 3+4 4 | 6+6 4 | 3+4 4 | 3+4 4 | 3+4 4 |
| | 6 2 | 3+3 4 | 3+3 4 | 3+4 4 | 5+5 4 | 3+4 4 | 3+4 4 | 3+4 4 |
| | 5 2 | 3+3 4 | 3+3 4 | 3+4 4 | 5+5 4 | 3+4 4 | 3+4 4 | 3+4 4 |
| | 6 2 | 4+4 4 | 4+4 4 | 3+4 4 | 6+6 4 | 3+4 4 | 3+4 4 | 3+4 4 |
| | 5 2 | 4+4 4 | 3+3 4 | 3+4 4 | 6+6 4 | 3+4 4 | 3+4 4 | 3+4 4 |
| | 5 2 | 3+3 4 | 4+4 4 | 3+4 4 | 5+5 4 | 3+4 4 | 3+4 4 | 3+4 4 |
| | 6 2 | 3+3 4 | 3+3 4 | 3+4 4 | 6+6 4 | 3+4 4 | 3+4 4 | 3+4 4 |
| | 6 2 | 4+4 4 | 4+4 4 | 3+4 4 | 5+5 4 | 3+4 4 | 3+4 4 | 3+4 4 |
| | 5 2 | 3+3 4 | 4+4 4 | 3+4 4 | 5+5 4 | 3+4 4 | 3+4 4 | 3+4 4 |
| | 6 2 | 4+4 4 | 3+3 4 | 3+4 4 | 6+6 4 | 3+4 4 | 3+4 4 | 3+4 4 |
| | 5 2 | 4+4 4 | 4+4 4 | 3+4 4 | 5+5 4 | 3+4 4 | 3+4 4 | 3+4 4 |
| | 6 2 | 3+3 4 | 3+3 4 | 3+4 4 | 6+6 4 | 3+4 4 | 3+4 4 | 3+4 4 |
| | 5 2 | 3+3 4 | 3+3 4 | 3+4 4 | 5+5 4 | 3+4 4 | 3+4 4 | 3+4 4 |

Continued on next column

**TABLE XXXVII– continued**

| TQC | ICSD | NoE | SF |
|---|---|---|---|
| | 5 2 4+4 4 4+4 4 3+4 4 6+6 4 3+4 4 3+4 4 3+4 4 | | |
| | 6 2 4+4 4 3+3 4 3+4 4 5+5 4 3+4 4 3+4 4 3+4 4 | | |
| | 6 2 3+3 4 4+4 4 3+4 4 6+6 4 3+4 4 3+4 4 3+4 4 | | |
| | 5 2 3+3 4 3+3 4 3+4 4 6+6 4 3+4 4 3+4 4 3+4 4 | | |
| | 6 2 4+4 4 4+4 4 3+4 4 5+5 4 3+4 4 3+4 4 3+4 4 | | |
| | 5 2 3+3 4 4+4 4 3+4 4 5+5 4 3+4 4 3+4 4 3+4 4 | | |
| | 6 2 4+4 4 3+3 4 3+4 4 6+6 4 3+4 4 3+4 4 3+4 4 | | |
| | 5 2 3+3 4 4+4 4 3+4 4 5+5 4 3+4 4 3+4 4 3+4 4 | | |
| | 5 2 4+4 4 3+3 4 3+4 4 6+6 4 3+4 4 3+4 4 3+4 4 | | |
| | 5 2 3+3 4 4+4 4 3+4 4 5+5 4 3+4 4 3+4 4 3+4 4 | | |
| | 6 2 4+4 4 3+3 4 3+4 4 6+6 4 3+4 4 3+4 4 3+4 4 | | |
| | 6 2 3+3 4 4+4 4 3+4 4 5+5 4 3+4 4 3+4 4 3+4 4 | | |
| | 5 2 4+4 4 3+3 4 3+4 4 6+6 4 3+4 4 3+4 4 3+4 4 | | |
| | 6 2 4+4 4 3+3 4 3+4 4 5+5 4 3+4 4 3+4 4 3+4 4 | | |
| | 6 2 3+3 4 4+4 4 3+4 4 5+5 4 3+4 4 3+4 4 3+4 4 | | |
| | 5 2 4+4 4 4+4 4 3+4 4 6+6 4 3+4 4 3+4 4 3+4 4 | | |
| | 6 2 3+3 4 3+3 4 3+4 4 6+6 4 3+4 4 3+4 4 3+4 4 | | |
| | 6 2 4+4 4 3+3 4 3+4 4 5+5 4 3+4 4 3+4 4 3+4 4 | | |
| | 5 2 3+3 4 4+4 4 3+4 4 6+6 4 3+4 4 3+4 4 3+4 4 | | |
| | 5 2 3+3 4 3+3 4 3+4 4 5+5 4 3+4 4 3+4 4 3+4 4 | | |
| | 6 2 4+4 4 4+4 4 3+4 4 6+6 4 3+4 4 3+4 4 3+4 4 | | |
| | 6 2 3+3 4 4+4 4 3+4 4 5+5 4 3+4 4 3+4 4 3+4 4 | | |
| | 5 2 4+4 4 3+3 4 3+4 4 6+6 4 3+4 4 3+4 4 3+4 4 | | |
| | 5 2 3+3 4 3+3 4 3+4 4 6+6 4 3+4 4 3+4 4 3+4 4 | | |
| | 6 2 4+4 4 4+4 4 3+4 4 5+5 4 3+4 4 3+4 4 3+4 4 | | |
| | 5 2 | | |
| | 6 2 | | |
| | 5 2 | | |
| | 6 2 | | |
| | 6 2 | | |
| | 5 2 | | |
| | 6 2 | | |
| | 6 2 | | |
| | 5 2 | | |
| | 5 2 | | |
| | 6 2 | | |
| | 5 2 | | |
| | 6 2 | | |
| | 5 2 | | |
| | 6 2 | | |
| | 6 2 | | |
| | 5 2 | | |
| | 5 2 | | |
| | 5 2 | | |
| | 6 2 | | |
| | 6 2 | | |
| | 5 2 | | |
| | 5 2 | | |
| | 5 2 | | |
| | 6 2 | | |
| | 5 2 | | |
| | 6 2 | | |
| | 5 2 | | |
| | 6 2 | | |
| | 5 2 | | |
| | 5 2 | | |
| | 6 2 | | |
| | 6 2 | | |
| | 6 2 | | |
| | 6 2 | | |
| | 5 2 | | |

Continued on next column

**TABLE XXXVII– continued**

| TQC | ICSD | NoE | SF |
|---|---|---|---|
| | 6 2 | | |
| | 6 2 | | |
| | 5 2 | | |
| | 6 2 | | |
| | 5 2 | | |
| | 5 2 | | |
| | 6 2 | | |
| | 5 2 | | |
| | 5 2 | | |
| | 6 2 | | |
| | 6 2 | | |
| | 5 2 | | |
| | 5 2 | | |
| | 5 2 | | |
| 62.2.155 | 33861 | 200 | Cs2Fe1O4 |
| | -GM -R -S -T -U -X -Y -Z | | |
| | 5 2 3+3 4 3+3 4 3+4 4 5+5 4 3+4 4 3+4 4 3+4 4 | | |
| | 5 2 4+4 4 4+4 4 3+4 4 6+6 4 3+4 4 3+4 4 3+4 4 | | |
| | 6 2 3+3 4 3+3 4 3+4 4 6+6 4 3+4 4 3+4 4 3+4 4 | | |
| | 6 2 4+4 4 4+4 4 3+4 4 5+5 4 3+4 4 3+4 4 3+4 4 | | |
| | 6 2 3+3 4 3+3 4 3+4 4 5+5 4 3+4 4 3+4 4 3+4 4 | | |
| | 5 2 4+4 4 4+4 4 3+4 4 6+6 4 3+4 4 3+4 4 3+4 4 | | |
| | 5 2 3+3 4 3+3 4 3+4 4 6+6 4 3+4 4 3+4 4 3+4 4 | | |
| | 6 2 4+4 4 4+4 4 3+4 4 5+5 4 3+4 4 3+4 4 3+4 4 | | |
| | 6 2 4+4 4 3+3 4 3+4 4 5+5 4 3+4 4 3+4 4 3+4 4 | | |
| | 6 2 3+3 4 4+4 4 3+4 4 6+6 4 3+4 4 3+4 4 3+4 4 | | |
| | 5 2 3+3 4 3+3 4 3+4 4 6+6 4 3+4 4 3+4 4 3+4 4 | | |
| | 5 2 4+4 4 4+4 4 3+4 4 5+5 4 3+4 4 3+4 4 3+4 4 | | |
| | 5 2 3+3 4 3+3 4 3+4 4 5+5 4 3+4 4 3+4 4 3+4 4 | | |
| | 6 2 4+4 4 4+4 4 3+4 4 6+6 4 3+4 4 3+4 4 3+4 4 | | |
| | 5 2 4+4 4 3+3 4 3+4 4 6+6 4 3+4 4 3+4 4 3+4 4 | | |
| | 6 2 3+3 4 4+4 4 3+4 4 5+5 4 3+4 4 3+4 4 3+4 4 | | |
| | 5 2 3+3 4 4+4 4 3+4 4 5+5 4 3+4 4 3+4 4 3+4 4 | | |
| | 6 2 4+4 4 3+3 4 3+4 4 6+6 4 3+4 4 3+4 4 3+4 4 | | |
| | 5 2 4+4 4 4+4 4 3+4 4 5+5 4 3+4 4 3+4 4 3+4 4 | | |
| | 6 2 3+3 4 3+3 4 3+4 4 6+6 4 3+4 4 3+4 4 3+4 4 | | |
| | 5 2 4+4 4 3+3 4 3+4 4 5+5 4 3+4 4 3+4 4 3+4 4 | | |
| | 5 2 3+3 4 4+4 4 3+4 4 6+6 4 3+4 4 3+4 4 3+4 4 | | |
| | 6 2 4+4 4 3+3 4 3+4 4 6+6 4 3+4 4 3+4 4 3+4 4 | | |
| | 6 2 3+3 4 4+4 4 3+4 4 5+5 4 3+4 4 3+4 4 3+4 4 | | |
| | 5 2 4+4 4 3+3 4 3+4 4 5+5 4 3+4 4 3+4 4 3+4 4 | | |
| | 6 2 3+3 4 4+4 4 3+4 4 6+6 4 3+4 4 3+4 4 3+4 4 | | |
| | 5 2 3+3 4 4+4 4 3+4 4 5+5 4 3+4 4 3+4 4 3+4 4 | | |
| | 6 2 4+4 4 3+3 4 3+4 4 6+6 4 3+4 4 3+4 4 3+4 4 | | |
| | 5 2 3+3 4 3+3 4 3+4 4 5+5 4 3+4 4 3+4 4 3+4 4 | | |
| | 6 2 4+4 4 4+4 4 3+4 4 6+6 4 3+4 4 3+4 4 3+4 4 | | |
| | 5 2 4+4 4 3+3 4 3+4 4 6+6 4 3+4 4 3+4 4 3+4 4 | | |
| | 6 2 3+3 4 4+4 4 3+4 4 5+5 4 3+4 4 3+4 4 3+4 4 | | |
| | 5 2 3+3 4 3+3 4 3+4 4 6+6 4 3+4 4 3+4 4 3+4 4 | | |
| | 5 2 4+4 4 4+4 4 3+4 4 5+5 4 3+4 4 3+4 4 3+4 4 | | |
| | 6 2 4+4 4 3+3 4 3+4 4 6+6 4 3+4 4 3+4 4 3+4 4 | | |
| | 5 2 3+3 4 4+4 4 3+4 4 5+5 4 3+4 4 3+4 4 3+4 4 | | |
| | 5 2 3+3 4 3+3 4 3+4 4 6+6 4 3+4 4 3+4 4 3+4 4 | | |
| | 6 2 4+4 4 4+4 4 3+4 4 5+5 4 3+4 4 3+4 4 3+4 4 | | |
| | 6 2 4+4 4 3+3 4 3+4 4 6+6 4 3+4 4 3+4 4 3+4 4 | | |
| | 6 2 3+3 4 4+4 4 3+4 4 6+6 4 3+4 4 3+4 4 3+4 4 | | |
| | 6 2 3+3 4 3+3 4 3+4 4 5+5 4 3+4 4 3+4 4 3+4 4 | | |
| | 5 2 4+4 4 4+4 4 3+4 4 5+5 4 3+4 4 3+4 4 3+4 4 | | |
| | 6 2 3+3 4 4+4 4 3+4 4 6+6 4 3+4 4 3+4 4 3+4 4 | | |
| | 5 2 4+4 4 3+3 4 3+4 4 5+5 4 3+4 4 3+4 4 3+4 4 | | |
| | 5 2 4+4 4 4+4 4 3+4 4 6+6 4 3+4 4 3+4 4 3+4 4 | | |

Continued on next column

**TABLE XXXVII– continued**

| TQC | ICSD | NoE | SF | | | | | |
|---|---|---|---|---|---|---|---|---|
| | 6 2 | 3+3 4 | 3+3 4 | 3+4 4 | 5+5 4 | 3+4 4 | 3+4 4 | 3+4 4 |
| | 5 2 | 3+3 4 | 3+3 4 | 3+4 4 | 6+6 4 | 3+4 4 | 3+4 4 | 3+4 4 |
| | 6 2 | 4+4 4 | 4+4 4 | 3+4 4 | 5+5 4 | 3+4 4 | 3+4 4 | 3+4 4 |
| | 6 2 | 3+3 4 | 4+4 4 | 3+4 4 | 5+5 4 | 3+4 4 | 3+4 4 | 3+4 4 |
| | 5 2 | 4+4 4 | 3+3 4 | 3+4 4 | 6+6 4 | 3+4 4 | 3+4 4 | 3+4 4 |
| | 5 2 | | | | | | | |
| | 5 2 | | | | | | | |
| | 6 2 | | | | | | | |
| | 6 2 | | | | | | | |
| | 6 2 | | | | | | | |
| | 6 2 | | | | | | | |
| | 5 2 | | | | | | | |
| | 6 2 | | | | | | | |
| | 5 2 | | | | | | | |
| | 5 2 | | | | | | | |
| | 6 2 | | | | | | | |
| | 6 2 | | | | | | | |
| | 5 2 | | | | | | | |
| | 6 2 | | | | | | | |
| | 5 2 | | | | | | | |
| | 6 2 | | | | | | | |
| | 5 2 | | | | | | | |
| | 5 2 | | | | | | | |
| | 5 2 | | | | | | | |
| | 6 2 | | | | | | | |
| | 6 2 | | | | | | | |
| | 5 2 | | | | | | | |
| | 5 2 | | | | | | | |
| | 5 2 | | | | | | | |
| | 6 2 | | | | | | | |
| | 5 2 | | | | | | | |
| | 6 2 | | | | | | | |
| | 5 2 | | | | | | | |
| | 6 2 | | | | | | | |
| | 6 2 | | | | | | | |
| | 5 2 | | | | | | | |
| | 6 2 | | | | | | | |
| | 5 2 | | | | | | | |
| | 6 2 | | | | | | | |
| | 6 2 | | | | | | | |
| | 6 2 | | | | | | | |
| | 5 2 | | | | | | | |
| | 5 2 | | | | | | | |
| | 5 2 | | | | | | | |
| | 6 2 | | | | | | | |
| | 6 2 | | | | | | | |
| | 5 2 | | | | | | | |
| | 6 2 | | | | | | | |
| | 5 2 | | | | | | | |
| | 5 2 | | | | | | | |
| | 6 2 | | | | | | | |
| | 6 2 | | | | | | | |
| | 5 2 | | | | | | | |
| | 5 2 | | | | | | | |
| | 5 2 | | | | | | | |
| 62.2.156 | 192388 | 176 | Cr2Fe1O4 | | | | | |
| | -GM | -R | -S | -T | -U | -X | -Y | -Z |
| | 5 2 | 3+3 4 | 3+3 4 | 3+4 4 | 6+6 4 | 3+4 4 | 3+4 4 | 3+4 4 |
| | 5 2 | 4+4 4 | 4+4 4 | 3+4 4 | 5+5 4 | 3+4 4 | 3+4 4 | 3+4 4 |
| | 5 2 | 3+3 4 | 3+3 4 | 3+4 4 | 5+5 4 | 3+4 4 | 3+4 4 | 3+4 4 |
| | 5 2 | 4+4 4 | 4+4 4 | 3+4 4 | 6+6 4 | 3+4 4 | 3+4 4 | 3+4 4 |

Continued on next column

**TABLE XXXVII– continued**

| TQC | ICSD | NoE | SF | | | | | |
|---|---|---|---|---|---|---|---|---|
| | 6 2 | 3+3 4 | 3+3 4 | 3+4 4 | 6+6 4 | 3+4 4 | 3+4 4 | 3+4 4 |
| | 6 2 | 4+4 4 | 4+4 4 | 3+4 4 | 5+5 4 | 3+4 4 | 3+4 4 | 3+4 4 |
| | 6 2 | 3+3 4 | 3+3 4 | 3+4 4 | 6+6 4 | 3+4 4 | 3+4 4 | 3+4 4 |
| | 5 2 | 4+4 4 | 4+4 4 | 3+4 4 | 5+5 4 | 3+4 4 | 3+4 4 | 3+4 4 |
| | 6 2 | 3+3 4 | 3+3 4 | 3+4 4 | 5+5 4 | 3+4 4 | 3+4 4 | 3+4 4 |
| | 5 2 | 4+4 4 | 4+4 4 | 3+4 4 | 6+6 4 | 3+4 4 | 3+4 4 | 3+4 4 |
| | 6 2 | 3+3 4 | 3+3 4 | 3+4 4 | 6+6 4 | 3+4 4 | 3+4 4 | 3+4 4 |
| | 6 2 | 4+4 4 | 4+4 4 | 3+4 4 | 5+5 4 | 3+4 4 | 3+4 4 | 3+4 4 |
| | 6 2 | 3+3 4 | 3+3 4 | 3+4 4 | 6+6 4 | 3+4 4 | 3+4 4 | 3+4 4 |
| | 6 2 | 4+4 4 | 4+4 4 | 3+4 4 | 5+5 4 | 3+4 4 | 3+4 4 | 3+4 4 |
| | 5 2 | 3+3 4 | 3+3 4 | 3+4 4 | 6+6 4 | 3+4 4 | 3+4 4 | 3+4 4 |
| | 5 2 | 4+4 4 | 4+4 4 | 3+4 4 | 5+5 4 | 3+4 4 | 3+4 4 | 3+4 4 |
| | 6 2 | 3+3 4 | 3+3 4 | 3+4 4 | 6+6 4 | 3+4 4 | 3+4 4 | 3+4 4 |
| | 6 2 | 4+4 4 | 4+4 4 | 3+4 4 | 5+5 4 | 3+4 4 | 3+4 4 | 3+4 4 |
| | 5 2 | 3+3 4 | 3+3 4 | 3+4 4 | 5+5 4 | 3+4 4 | 3+4 4 | 3+4 4 |
| | 5 2 | 4+4 4 | 4+4 4 | 3+4 4 | 5+5 4 | 3+4 4 | 3+4 4 | 3+4 4 |
| | 5 2 | 3+3 4 | 4+4 4 | 3+4 4 | 6+6 4 | 3+4 4 | 3+4 4 | 3+4 4 |
| | 6 2 | 4+4 4 | 3+3 4 | 3+4 4 | 6+6 4 | 3+4 4 | 3+4 4 | 3+4 4 |
| | 6 2 | 3+3 4 | 3+3 4 | 3+4 4 | 5+5 4 | 3+4 4 | 3+4 4 | 3+4 4 |
| | 5 2 | 4+4 4 | 4+4 4 | 3+4 4 | 6+6 4 | 3+4 4 | 3+4 4 | 3+4 4 |
| | 5 2 | 3+3 4 | 4+4 4 | 3+4 4 | 6+6 4 | 3+4 4 | 3+4 4 | 3+4 4 |
| | 5 2 | 4+4 4 | 3+3 4 | 3+4 4 | 6+6 4 | 3+4 4 | 3+4 4 | 3+4 4 |
| | 5 2 | 3+3 4 | 3+3 4 | 3+4 4 | 5+5 4 | 3+4 4 | 3+4 4 | 3+4 4 |
| | 5 2 | 4+4 4 | 4+4 4 | 3+4 4 | 5+5 4 | 3+4 4 | 3+4 4 | 3+4 4 |
| | 6 2 | 3+3 4 | 3+3 4 | 3+4 4 | 5+5 4 | 3+4 4 | 3+4 4 | 3+4 4 |
| | 6 2 | 4+4 4 | 4+4 4 | 3+4 4 | 5+5 4 | 3+4 4 | 3+4 4 | 3+4 4 |
| | 5 2 | 4+4 4 | 3+3 4 | 3+4 4 | 6+6 4 | 3+4 4 | 3+4 4 | 3+4 4 |
| | 5 2 | 3+3 4 | 4+4 4 | 3+4 4 | 6+6 4 | 3+4 4 | 3+4 4 | 3+4 4 |
| | 5 2 | 4+4 4 | 3+3 4 | 3+4 4 | 5+5 4 | 3+4 4 | 3+4 4 | 3+4 4 |
| | 5 2 | 3+3 4 | 4+4 4 | 3+4 4 | 6+6 4 | 3+4 4 | 3+4 4 | 3+4 4 |
| | 5 2 | 3+3 4 | 3+3 4 | 3+4 4 | 6+6 4 | 3+4 4 | 3+4 4 | 3+4 4 |
| | 6 2 | 4+4 4 | 4+4 4 | 3+4 4 | 5+5 4 | 3+4 4 | 3+4 4 | 3+4 4 |
| | 6 2 | 3+3 4 | 4+4 4 | 3+4 4 | 5+5 4 | 3+4 4 | 3+4 4 | 3+4 4 |
| | 5 2 | 4+4 4 | 3+3 4 | 3+4 4 | 6+6 4 | 3+4 4 | 3+4 4 | 3+4 4 |
| | 6 2 | 3+3 4 | 4+4 4 | 3+4 4 | 6+6 4 | 3+4 4 | 3+4 4 | 3+4 4 |
| | 6 2 | 4+4 4 | 3+3 4 | 3+4 4 | 5+5 4 | 3+4 4 | 3+4 4 | 3+4 4 |
| | 6 2 | 3+3 4 | 4+4 4 | 3+4 4 | 6+6 4 | 3+4 4 | 3+4 4 | 3+4 4 |
| | 6 2 | 4+4 4 | 3+3 4 | 3+4 4 | 5+5 4 | 3+4 4 | 3+4 4 | 3+4 4 |
| | 6 2 | 4+4 4 | 3+3 4 | 3+4 4 | 6+6 4 | 3+4 4 | 3+4 4 | 3+4 4 |
| | 5 2 | 3+3 4 | 4+4 4 | 3+4 4 | 5+5 4 | 3+4 4 | 3+4 4 | 3+4 4 |
| | 6 2 | | | | | | | |
| | 5 2 | | | | | | | |
| | 5 2 | | | | | | | |
| | 5 2 | | | | | | | |
| | 5 2 | | | | | | | |
| | 6 2 | | | | | | | |
| | 6 2 | | | | | | | |
| | 5 2 | | | | | | | |
| | 5 2 | | | | | | | |
| | 5 2 | | | | | | | |
| | 6 2 | | | | | | | |
| | 5 2 | | | | | | | |
| | 6 2 | | | | | | | |
| | 6 2 | | | | | | | |
| | 6 2 | | | | | | | |
| | 5 2 | | | | | | | |
| | 6 2 | | | | | | | |
| | 6 2 | | | | | | | |
| | 6 2 | | | | | | | |
| | 6 2 | | | | | | | |
| | 5 2 | | | | | | | |

Continued on next column

**TABLE XXXVII– continued**

| TQC | ICSD | NoE | SF | | | | | |
|---|---|---|---|---|---|---|---|---|
| | 6 2 | | | | | | | |
| | 5 2 | | | | | | | |
| | 6 2 | | | | | | | |
| | 6 2 | | | | | | | |
| | 5 2 | | | | | | | |
| | 5 2 | | | | | | | |
| | 6 2 | | | | | | | |
| | 6 2 | | | | | | | |
| | 5 2 | | | | | | | |
| | 6 2 | | | | | | | |
| | 5 2 | | | | | | | |
| | 5 2 | | | | | | | |
| | 5 2 | | | | | | | |
| | 5 2 | | | | | | | |
| | 5 2 | | | | | | | |
| | 6 2 | | | | | | | |
| | 6 2 | | | | | | | |
| | 5 2 | | | | | | | |
| | 6 2 | | | | | | | |
| | 6 2 | | | | | | | |
| | 5 2 | | | | | | | |
| | 5 2 | | | | | | | |
| | 6 2 | | | | | | | |
| 62.2.157 | 80332 | 152 | Cr1Nd1O3 | | | | | |
| | -GM | -R | -S | -T | -U | -X | -Y | -Z |
| | 5 2 | 3+3 4 | 3+3 4 | 3+4 4 | 5+5 4 | 3+4 4 | 3+4 4 | 3+4 4 |
| | 6 2 | 4+4 4 | 4+4 4 | 3+4 4 | 6+6 4 | 3+4 4 | 3+4 4 | 3+4 4 |
| | 5 2 | 4+4 4 | 3+3 4 | 3+4 4 | 6+6 4 | 3+4 4 | 3+4 4 | 3+4 4 |
| | 6 2 | 3+3 4 | 4+4 4 | 3+4 4 | 5+5 4 | 3+4 4 | 3+4 4 | 3+4 4 |
| | 5 2 | 3+3 4 | 4+4 4 | 3+4 4 | 5+5 4 | 3+4 4 | 3+4 4 | 3+4 4 |
| | 6 2 | 4+4 4 | 3+3 4 | 3+4 4 | 6+6 4 | 3+4 4 | 3+4 4 | 3+4 4 |
| | 5 2 | 4+4 4 | 3+3 4 | 3+4 4 | 6+6 4 | 3+4 4 | 3+4 4 | 3+4 4 |
| | 6 2 | 3+3 4 | 4+4 4 | 3+4 4 | 5+5 4 | 3+4 4 | 3+4 4 | 3+4 4 |
| | 6 2 | 4+4 4 | 3+3 4 | 3+4 4 | 6+6 4 | 3+4 4 | 3+4 4 | 3+4 4 |
| | 5 2 | 3+3 4 | 4+4 4 | 3+4 4 | 5+5 4 | 3+4 4 | 3+4 4 | 3+4 4 |
| | 5 2 | 3+3 4 | 4+4 4 | 3+4 4 | 6+6 4 | 3+4 4 | 3+4 4 | 3+4 4 |
| | 5 2 | 4+4 4 | 3+3 4 | 3+4 4 | 5+5 4 | 3+4 4 | 3+4 4 | 3+4 4 |
| | 5 2 | 3+3 4 | 3+3 4 | 3+4 4 | 5+5 4 | 3+4 4 | 3+4 4 | 3+4 4 |
| | 6 2 | 4+4 4 | 4+4 4 | 3+4 4 | 6+6 4 | 3+4 4 | 3+4 4 | 3+4 4 |
| | 5 2 | 3+3 4 | 4+4 4 | 3+4 4 | 6+6 4 | 3+4 4 | 3+4 4 | 3+4 4 |
| | 6 2 | 4+4 4 | 3+3 4 | 3+4 4 | 6+6 4 | 3+4 4 | 3+4 4 | 3+4 4 |
| | 6 2 | 4+4 4 | 3+3 4 | 3+4 4 | 6+6 4 | 3+4 4 | 3+4 4 | 3+4 4 |
| | 6 2 | 3+3 4 | 4+4 4 | 3+4 4 | 6+6 4 | 3+4 4 | 3+4 4 | 3+4 4 |
| | 6 2 | 4+4 4 | 3+3 4 | 3+4 4 | 5+5 4 | 3+4 4 | 3+4 4 | 3+4 4 |
| | 5 2 | 3+3 4 | 4+4 4 | 3+4 4 | 5+5 4 | 3+4 4 | 3+4 4 | 3+4 4 |
| | 5 2 | 3+3 4 | 3+3 4 | 3+4 4 | 5+5 4 | 3+4 4 | 3+4 4 | 3+4 4 |
| | 6 2 | 4+4 4 | 4+4 4 | 3+4 4 | 6+6 4 | 3+4 4 | 3+4 4 | 3+4 4 |
| | 5 2 | 3+3 4 | 4+4 4 | 3+4 4 | 6+6 4 | 3+4 4 | 3+4 4 | 3+4 4 |
| | 6 2 | 4+4 4 | 3+3 4 | 3+4 4 | 5+5 4 | 3+4 4 | 3+4 4 | 3+4 4 |
| | 5 2 | 4+4 4 | 4+4 4 | 3+4 4 | 5+5 4 | 3+4 4 | 3+4 4 | 3+4 4 |
| | 6 2 | 3+3 4 | 3+3 4 | 3+4 4 | 6+6 4 | 3+4 4 | 3+4 4 | 3+4 4 |
| | 5 2 | 4+4 4 | 4+4 4 | 3+4 4 | 6+6 4 | 3+4 4 | 3+4 4 | 3+4 4 |
| | 6 2 | 3+3 4 | 3+3 4 | 3+4 4 | 5+5 4 | 3+4 4 | 3+4 4 | 3+4 4 |
| | 5 2 | 3+3 4 | 3+3 4 | 3+4 4 | 5+5 4 | 3+4 4 | 3+4 4 | 3+4 4 |
| | 5 2 | 4+4 4 | 4+4 4 | 3+4 4 | 5+5 4 | 3+4 4 | 3+4 4 | 3+4 4 |
| | 5 2 | 3+3 4 | 4+4 4 | 3+4 4 | 5+5 4 | 3+4 4 | 3+4 4 | 3+4 4 |
| | 5 2 | 4+4 4 | 3+3 4 | 3+4 4 | 6+6 4 | 3+4 4 | 3+4 4 | 3+4 4 |
| | 5 2 | 4+4 4 | 4+4 4 | 3+4 4 | 6+6 4 | 3+4 4 | 3+4 4 | 3+4 4 |
| | 5 2 | 3+3 4 | 3+3 4 | 3+4 4 | 6+6 4 | 3+4 4 | 3+4 4 | 3+4 4 |
| | 5 2 | 4+4 4 | 4+4 4 | 3+4 4 | 6+6 4 | 3+4 4 | 3+4 4 | 3+4 4 |
| | 6 2 | 3+3 4 | 3+3 4 | 3+4 4 | 6+6 4 | 3+4 4 | 3+4 4 | 3+4 4 |

Continued on next column

**TABLE XXXVII– continued**

| TQC | ICSD | NoE | SF | | | | | |
|---|---|---|---|---|---|---|---|---|
| | 5 2 | 3+3 4 | 3+3 4 | 3+4 4 | 5+5 4 | 3+4 4 | 3+4 4 | 3+4 4 |
| | 6 2 | 4+4 4 | 4+4 4 | 3+4 4 | 5+5 4 | 3+4 4 | 3+4 4 | 3+4 4 |
| | 6 2 | | | | | | | |
| | 5 2 | | | | | | | |
| | 6 2 | | | | | | | |
| | 5 2 | | | | | | | |
| | 6 2 | | | | | | | |
| | 5 2 | | | | | | | |
| | 5 2 | | | | | | | |
| | 6 2 | | | | | | | |
| | 6 2 | | | | | | | |
| | 5 2 | | | | | | | |
| | 5 2 | | | | | | | |
| | 6 2 | | | | | | | |
| | 6 2 | | | | | | | |
| | 6 2 | | | | | | | |
| | 6 2 | | | | | | | |
| | 6 2 | | | | | | | |
| | 5 2 | | | | | | | |
| | 5 2 | | | | | | | |
| | 6 2 | | | | | | | |
| | 5 2 | | | | | | | |
| | 6 2 | | | | | | | |
| | 6 2 | | | | | | | |
| | 5 2 | | | | | | | |
| | 6 2 | | | | | | | |
| | 6 2 | | | | | | | |
| | 6 2 | | | | | | | |
| | 5 2 | | | | | | | |
| | 5 2 | | | | | | | |
| | 5 2 | | | | | | | |
| | 5 2 | | | | | | | |
| | 5 2 | | | | | | | |
| | 5 2 | | | | | | | |
| | 5 2 | | | | | | | |
| | 5 2 | | | | | | | |
| | 6 2 | | | | | | | |
| | 5 2 | | | | | | | |
| | 6 2 | | | | | | | |
| | 5 2 | | | | | | | |
| 62.2.158 | 79349 | 152 | Cr1Nd1O3 | | | | | |
| | -GM | -R | -S | -T | -U | -X | -Y | -Z |
| | 5 2 | 3+3 4 | 3+3 4 | 3+4 4 | 5+5 4 | 3+4 4 | 3+4 4 | 3+4 4 |
| | 6 2 | 4+4 4 | 4+4 4 | 3+4 4 | 6+6 4 | 3+4 4 | 3+4 4 | 3+4 4 |
| | 5 2 | 4+4 4 | 3+3 4 | 3+4 4 | 6+6 4 | 3+4 4 | 3+4 4 | 3+4 4 |
| | 6 2 | 3+3 4 | 4+4 4 | 3+4 4 | 5+5 4 | 3+4 4 | 3+4 4 | 3+4 4 |
| | 5 2 | 3+3 4 | 4+4 4 | 3+4 4 | 5+5 4 | 3+4 4 | 3+4 4 | 3+4 4 |
| | 6 2 | 4+4 4 | 3+3 4 | 3+4 4 | 6+6 4 | 3+4 4 | 3+4 4 | 3+4 4 |
| | 5 2 | 4+4 4 | 3+3 4 | 3+4 4 | 6+6 4 | 3+4 4 | 3+4 4 | 3+4 4 |
| | 6 2 | 3+3 4 | 4+4 4 | 3+4 4 | 5+5 4 | 3+4 4 | 3+4 4 | 3+4 4 |
| | 6 2 | 4+4 4 | 3+3 4 | 3+4 4 | 6+6 4 | 3+4 4 | 3+4 4 | 3+4 4 |
| | 5 2 | 3+3 4 | 4+4 4 | 3+4 4 | 5+5 4 | 3+4 4 | 3+4 4 | 3+4 4 |
| | 5 2 | 3+3 4 | 4+4 4 | 3+4 4 | 6+6 4 | 3+4 4 | 3+4 4 | 3+4 4 |
| | 5 2 | 4+4 4 | 3+3 4 | 3+4 4 | 5+5 4 | 3+4 4 | 3+4 4 | 3+4 4 |
| | 5 2 | 3+3 4 | 3+3 4 | 3+4 4 | 5+5 4 | 3+4 4 | 3+4 4 | 3+4 4 |
| | 6 2 | 4+4 4 | 4+4 4 | 3+4 4 | 6+6 4 | 3+4 4 | 3+4 4 | 3+4 4 |
| | 5 2 | 3+3 4 | 4+4 4 | 3+4 4 | 6+6 4 | 3+4 4 | 3+4 4 | 3+4 4 |
| | 6 2 | 4+4 4 | 3+3 4 | 3+4 4 | 6+6 4 | 3+4 4 | 3+4 4 | 3+4 4 |
| | 6 2 | 4+4 4 | 3+3 4 | 3+4 4 | 6+6 4 | 3+4 4 | 3+4 4 | 3+4 4 |
| | 6 2 | 3+3 4 | 4+4 4 | 3+4 4 | 6+6 4 | 3+4 4 | 3+4 4 | 3+4 4 |
| | 6 2 | 4+4 4 | 3+3 4 | 3+4 4 | 5+5 4 | 3+4 4 | 3+4 4 | 3+4 4 |

Continued on next column

**TABLE XXXVII– continued**

| TQC | ICSD NoE SF |
|---|---|
| | 5 2 3+3 4 4+4 4 3+4 4 5+5 4 3+4 4 3+4 4 3+4 4 |
| | 5 2 3+3 4 3+3 4 3+4 4 5+5 4 3+4 4 3+4 4 3+4 4 |
| | 6 2 4+4 4 4+4 4 3+4 4 6+6 4 3+4 4 3+4 4 3+4 4 |
| | 5 2 3+3 4 4+4 4 3+4 4 6+6 4 3+4 4 3+4 4 3+4 4 |
| | 6 2 4+4 4 3+3 4 3+4 4 5+5 4 3+4 4 3+4 4 3+4 4 |
| | 5 2 4+4 4 4+4 4 3+4 4 5+5 4 3+4 4 3+4 4 3+4 4 |
| | 6 2 3+3 4 3+3 4 3+4 4 6+6 4 3+4 4 3+4 4 3+4 4 |
| | 5 2 4+4 4 4+4 4 3+4 4 6+6 4 3+4 4 3+4 4 3+4 4 |
| | 6 2 3+3 4 3+3 4 3+4 4 5+5 4 3+4 4 3+4 4 3+4 4 |
| | 5 2 3+3 4 3+3 4 3+4 4 5+5 4 3+4 4 3+4 4 3+4 4 |
| | 5 2 4+4 4 4+4 4 3+4 4 5+5 4 3+4 4 3+4 4 3+4 4 |
| | 5 2 3+3 4 4+4 4 3+4 4 5+5 4 3+4 4 3+4 4 3+4 4 |
| | 5 2 4+4 4 3+3 4 3+4 4 6+6 4 3+4 4 3+4 4 3+4 4 |
| | 5 2 4+4 4 4+4 4 3+4 4 6+6 4 3+4 4 3+4 4 3+4 4 |
| | 5 2 3+3 4 3+3 4 3+4 4 6+6 4 3+4 4 3+4 4 3+4 4 |
| | 5 2 4+4 4 4+4 4 3+4 4 6+6 4 3+4 4 3+4 4 3+4 4 |
| | 6 2 3+3 4 3+3 4 3+4 4 6+6 4 3+4 4 3+4 4 3+4 4 |
| | 5 2 3+3 4 3+3 4 3+4 4 5+5 4 3+4 4 3+4 4 3+4 4 |
| | 6 2 4+4 4 4+4 4 3+4 4 5+5 4 3+4 4 3+4 4 3+4 4 |
| | 6 2 |
| | 5 2 |
| | 6 2 |
| | 5 2 |
| | 6 2 |
| | 5 2 |
| | 5 2 |
| | 6 2 |
| | 6 2 |
| | 5 2 |
| | 5 2 |
| | 6 2 |
| | 6 2 |
| | 6 2 |
| | 6 2 |
| | 6 2 |
| | 5 2 |
| | 5 2 |
| | 6 2 |
| | 5 2 |
| | 6 2 |
| | 6 2 |
| | 5 2 |
| | 6 2 |
| | 6 2 |
| | 6 2 |
| | 5 2 |
| | 5 2 |
| | 5 2 |
| | 5 2 |
| | 5 2 |
| | 5 2 |
| | 5 2 |
| | 5 2 |
| | 6 2 |
| | 5 2 |
| | 6 2 |
| | 5 2 |
| 62.2.159 | 156326 152 Cr1Nd1O3 |
| | -GM -R -S -T -U -X -Y -Z |
| | 5 2 3+3 4 3+3 4 3+4 4 5+5 4 3+4 4 3+4 4 3+4 4 |
| | 6 2 4+4 4 4+4 4 3+4 4 6+6 4 3+4 4 3+4 4 3+4 4 |

Continued on next column

**TABLE XXXVII– continued**

| TQC | ICSD NoE SF |
|---|---|
| | 5 2 3+3 4 3+3 4 3+4 4 6+6 4 3+4 4 3+4 4 3+4 4 |
| | 6 2 4+4 4 4+4 4 3+4 4 5+5 4 3+4 4 3+4 4 3+4 4 |
| | 5 2 3+3 4 4+4 4 3+4 4 5+5 4 3+4 4 3+4 4 3+4 4 |
| | 6 2 4+4 4 3+3 4 3+4 4 6+6 4 3+4 4 3+4 4 3+4 4 |
| | 5 2 4+4 4 3+3 4 3+4 4 6+6 4 3+4 4 3+4 4 3+4 4 |
| | 6 2 3+3 4 4+4 4 3+4 4 5+5 4 3+4 4 3+4 4 3+4 4 |
| | 5 2 4+4 4 3+3 4 3+4 4 6+6 4 3+4 4 3+4 4 3+4 4 |
| | 6 2 3+3 4 4+4 4 3+4 4 5+5 4 3+4 4 3+4 4 3+4 4 |
| | 5 2 3+3 4 4+4 4 3+4 4 6+6 4 3+4 4 3+4 4 3+4 4 |
| | 5 2 4+4 4 3+3 4 3+4 4 5+5 4 3+4 4 3+4 4 3+4 4 |
| | 5 2 3+3 4 3+3 4 3+4 4 5+5 4 3+4 4 3+4 4 3+4 4 |
| | 6 2 4+4 4 4+4 4 3+4 4 6+6 4 3+4 4 3+4 4 3+4 4 |
| | 5 2 3+3 4 4+4 4 3+4 4 6+6 4 3+4 4 3+4 4 3+4 4 |
| | 6 2 4+4 4 3+3 4 3+4 4 6+6 4 3+4 4 3+4 4 3+4 4 |
| | 6 2 4+4 4 4+4 4 3+4 4 6+6 4 3+4 4 3+4 4 3+4 4 |
| | 6 2 3+3 4 3+3 4 3+4 4 6+6 4 3+4 4 3+4 4 3+4 4 |
| | 6 2 4+4 4 3+3 4 3+4 4 5+5 4 3+4 4 3+4 4 3+4 4 |
| | 5 2 3+3 4 4+4 4 3+4 4 5+5 4 3+4 4 3+4 4 3+4 4 |
| | 5 2 3+3 4 3+3 4 3+4 4 5+5 4 3+4 4 3+4 4 3+4 4 |
| | 6 2 4+4 4 4+4 4 3+4 4 6+6 4 3+4 4 3+4 4 3+4 4 |
| | 5 2 3+3 4 4+4 4 3+4 4 6+6 4 3+4 4 3+4 4 3+4 4 |
| | 6 2 4+4 4 3+3 4 3+4 4 5+5 4 3+4 4 3+4 4 3+4 4 |
| | 5 2 4+4 4 4+4 4 3+4 4 5+5 4 3+4 4 3+4 4 3+4 4 |
| | 6 2 3+3 4 3+3 4 3+4 4 6+6 4 3+4 4 3+4 4 3+4 4 |
| | 5 2 4+4 4 4+4 4 3+4 4 6+6 4 3+4 4 3+4 4 3+4 4 |
| | 6 2 3+3 4 3+3 4 3+4 4 5+5 4 3+4 4 3+4 4 3+4 4 |
| | 5 2 3+3 4 3+3 4 3+4 4 5+5 4 3+4 4 3+4 4 3+4 4 |
| | 5 2 4+4 4 4+4 4 3+4 4 5+5 4 3+4 4 3+4 4 3+4 4 |
| | 5 2 3+3 4 3+3 4 3+4 4 5+5 4 3+4 4 3+4 4 3+4 4 |
| | 5 2 4+4 4 4+4 4 3+4 4 6+6 4 3+4 4 3+4 4 3+4 4 |
| | 5 2 4+4 4 4+4 4 3+4 4 6+6 4 3+4 4 3+4 4 3+4 4 |
| | 5 2 3+3 4 3+3 4 3+4 4 6+6 4 3+4 4 3+4 4 3+4 4 |
| | 5 2 4+4 4 4+4 4 3+4 4 6+6 4 3+4 4 3+4 4 3+4 4 |
| | 5 2 3+3 4 3+3 4 3+4 4 6+6 4 3+4 4 3+4 4 3+4 4 |
| | 6 2 3+3 4 3+3 4 3+4 4 5+5 4 3+4 4 3+4 4 3+4 4 |
| | 6 2 4+4 4 4+4 4 3+4 4 5+5 4 3+4 4 3+4 4 3+4 4 |
| | 6 2 |
| | 5 2 |
| | 6 2 |
| | 5 2 |
| | 6 2 |
| | 5 2 |
| | 5 2 |
| | 6 2 |
| | 6 2 |
| | 5 2 |
| | 5 2 |
| | 6 2 |
| | 6 2 |
| | 6 2 |
| | 6 2 |
| | 6 2 |
| | 5 2 |
| | 6 2 |
| | 5 2 |
| | 5 2 |
| | 6 2 |
| | 6 2 |
| | 5 2 |
| | 6 2 |
| | 6 2 |

Continued on next column

**TABLE XXXVII– continued**

| TQC | ICSD | NoE | SF | | | | | |
|---|---|---|---|---|---|---|---|---|
| | 6 2 | | | | | | | |
| | 5 2 | | | | | | | |
| | 5 2 | | | | | | | |
| | 5 2 | | | | | | | |
| | 5 2 | | | | | | | |
| | 5 2 | | | | | | | |
| | 5 2 | | | | | | | |
| | 5 2 | | | | | | | |
| | 5 2 | | | | | | | |
| | 6 2 | | | | | | | |
| | 5 2 | | | | | | | |
| | 6 2 | | | | | | | |
| | 5 2 | | | | | | | |
| 62.2.160 | 156322 | 152 | Cr1Nd1O3 | | | | | |
| | -GM | -R | -S | -T | -U | -X | -Y | -Z |
| | 5 2 | 3+3 4 | 3+3 4 | 3+4 4 | 6+6 4 | 3+4 4 | 3+4 4 | 3+4 4 |
| | 6 2 | 4+4 4 | 4+4 4 | 3+4 4 | 5+5 4 | 3+4 4 | 3+4 4 | 3+4 4 |
| | 5 2 | 3+3 4 | 4+4 4 | 3+4 4 | 5+5 4 | 3+4 4 | 3+4 4 | 3+4 4 |
| | 6 2 | 4+4 4 | 3+3 4 | 3+4 4 | 6+6 4 | 3+4 4 | 3+4 4 | 3+4 4 |
| | 5 2 | 4+4 4 | 3+3 4 | 3+4 4 | 6+6 4 | 3+4 4 | 3+4 4 | 3+4 4 |
| | 6 2 | 3+3 4 | 4+4 4 | 3+4 4 | 5+5 4 | 3+4 4 | 3+4 4 | 3+4 4 |
| | 5 2 | 3+3 4 | 4+4 4 | 3+4 4 | 5+5 4 | 3+4 4 | 3+4 4 | 3+4 4 |
| | 6 2 | 4+4 4 | 3+3 4 | 3+4 4 | 6+6 4 | 3+4 4 | 3+4 4 | 3+4 4 |
| | 5 2 | 3+3 4 | 4+4 4 | 3+4 4 | 5+5 4 | 3+4 4 | 3+4 4 | 3+4 4 |
| | 6 2 | 4+4 4 | 3+3 4 | 3+4 4 | 6+6 4 | 3+4 4 | 3+4 4 | 3+4 4 |
| | 5 2 | 3+3 4 | 3+3 4 | 3+4 4 | 5+5 4 | 3+4 4 | 3+4 4 | 3+4 4 |
| | 5 2 | 4+4 4 | 4+4 4 | 3+4 4 | 6+6 4 | 3+4 4 | 3+4 4 | 3+4 4 |
| | 5 2 | 4+4 4 | 4+4 4 | 3+4 4 | 5+5 4 | 3+4 4 | 3+4 4 | 3+4 4 |
| | 6 2 | 3+3 4 | 3+3 4 | 3+4 4 | 6+6 4 | 3+4 4 | 3+4 4 | 3+4 4 |
| | 5 2 | 4+4 4 | 3+3 4 | 3+4 4 | 5+5 4 | 3+4 4 | 3+4 4 | 3+4 4 |
| | 6 2 | 3+3 4 | 4+4 4 | 3+4 4 | 5+5 4 | 3+4 4 | 3+4 4 | 3+4 4 |
| | 6 2 | 3+3 4 | 4+4 4 | 3+4 4 | 5+5 4 | 3+4 4 | 3+4 4 | 3+4 4 |
| | 6 2 | 4+4 4 | 3+3 4 | 3+4 4 | 5+5 4 | 3+4 4 | 3+4 4 | 3+4 4 |
| | 6 2 | 3+3 4 | 4+4 4 | 3+4 4 | 6+6 4 | 3+4 4 | 3+4 4 | 3+4 4 |
| | 5 2 | 4+4 4 | 3+3 4 | 3+4 4 | 6+6 4 | 3+4 4 | 3+4 4 | 3+4 4 |
| | 5 2 | 4+4 4 | 4+4 4 | 3+4 4 | 6+6 4 | 3+4 4 | 3+4 4 | 3+4 4 |
| | 6 2 | 3+3 4 | 3+3 4 | 3+4 4 | 5+5 4 | 3+4 4 | 3+4 4 | 3+4 4 |
| | 5 2 | 4+4 4 | 3+3 4 | 3+4 4 | 5+5 4 | 3+4 4 | 3+4 4 | 3+4 4 |
| | 6 2 | 3+3 4 | 4+4 4 | 3+4 4 | 6+6 4 | 3+4 4 | 3+4 4 | 3+4 4 |
| | 5 2 | 3+3 4 | 3+3 4 | 3+4 4 | 6+6 4 | 3+4 4 | 3+4 4 | 3+4 4 |
| | 6 2 | 4+4 4 | 4+4 4 | 3+4 4 | 5+5 4 | 3+4 4 | 3+4 4 | 3+4 4 |
| | 5 2 | 3+3 4 | 3+3 4 | 3+4 4 | 5+5 4 | 3+4 4 | 3+4 4 | 3+4 4 |
| | 6 2 | 4+4 4 | 4+4 4 | 3+4 4 | 6+6 4 | 3+4 4 | 3+4 4 | 3+4 4 |
| | 5 2 | 3+3 4 | 4+4 4 | 3+4 4 | 6+6 4 | 3+4 4 | 3+4 4 | 3+4 4 |
| | 5 2 | 4+4 4 | 3+3 4 | 3+4 4 | 6+6 4 | 3+4 4 | 3+4 4 | 3+4 4 |
| | 5 2 | 4+4 4 | 3+3 4 | 3+4 4 | 6+6 4 | 3+4 4 | 3+4 4 | 3+4 4 |
| | 5 2 | 3+3 4 | 4+4 4 | 3+4 4 | 5+5 4 | 3+4 4 | 3+4 4 | 3+4 4 |
| | 5 2 | 3+3 4 | 3+3 4 | 3+4 4 | 5+5 4 | 3+4 4 | 3+4 4 | 3+4 4 |
| | 5 2 | 4+4 4 | 4+4 4 | 3+4 4 | 5+5 4 | 3+4 4 | 3+4 4 | 3+4 4 |
| | 5 2 | 3+3 4 | 3+3 4 | 3+4 4 | 5+5 4 | 3+4 4 | 3+4 4 | 3+4 4 |
| | 6 2 | 4+4 4 | 4+4 4 | 3+4 4 | 6+6 4 | 3+4 4 | 3+4 4 | 3+4 4 |
| | 5 2 | 3+3 4 | 3+3 4 | 3+4 4 | 5+5 4 | 3+4 4 | 3+4 4 | 3+4 4 |
| | 6 2 | 4+4 4 | 4+4 4 | 3+4 4 | 6+6 4 | 3+4 4 | 3+4 4 | 3+4 4 |
| | 6 2 | | | | | | | |
| | 5 2 | | | | | | | |
| | 6 2 | | | | | | | |
| | 5 2 | | | | | | | |
| | 6 2 | | | | | | | |
| | 5 2 | | | | | | | |
| | 5 2 | | | | | | | |
| | 6 2 | | | | | | | |

Continued on next column

| TQC | ICSD | NoE | SF | | | | | |
|---|---|---|---|---|---|---|---|---|
| | 6 2 | | | | | | | |
| | 5 2 | | | | | | | |
| | 5 2 | | | | | | | |
| | 6 2 | | | | | | | |
| | 6 2 | | | | | | | |
| | 6 2 | | | | | | | |
| | 6 2 | | | | | | | |
| | 6 2 | | | | | | | |
| | 5 2 | | | | | | | |
| | 5 2 | | | | | | | |
| | 6 2 | | | | | | | |
| | 5 2 | | | | | | | |
| | 6 2 | | | | | | | |
| | 6 2 | | | | | | | |
| | 5 2 | | | | | | | |
| | 6 2 | | | | | | | |
| | 6 2 | | | | | | | |
| | 6 2 | | | | | | | |
| | 5 2 | | | | | | | |
| | 5 2 | | | | | | | |
| | 5 2 | | | | | | | |
| | 5 2 | | | | | | | |
| | 5 2 | | | | | | | |
| | 5 2 | | | | | | | |
| | 5 2 | | | | | | | |
| | 6 2 | | | | | | | |
| | 5 2 | | | | | | | |
| | 5 2 | | | | | | | |
| | 6 2 | | | | | | | |
| | 5 2 | | | | | | | |
| 62.2.161 | 156320 | 152 | Cr1Nd1O3 | | | | | |
| | -GM | -R | -S | -T | -U | -X | -Y | -Z |
| | 5 2 | 3+3 4 | 3+3 4 | 3+4 4 | 5+5 4 | 3+4 4 | 3+4 4 | 3+4 4 |
| | 6 2 | 4+4 4 | 4+4 4 | 3+4 4 | 6+6 4 | 3+4 4 | 3+4 4 | 3+4 4 |
| | 5 2 | 4+4 4 | 3+3 4 | 3+4 4 | 6+6 4 | 3+4 4 | 3+4 4 | 3+4 4 |
| | 6 2 | 3+3 4 | 4+4 4 | 3+4 4 | 5+5 4 | 3+4 4 | 3+4 4 | 3+4 4 |
| | 5 2 | 3+3 4 | 4+4 4 | 3+4 4 | 5+5 4 | 3+4 4 | 3+4 4 | 3+4 4 |
| | 6 2 | 4+4 4 | 3+3 4 | 3+4 4 | 6+6 4 | 3+4 4 | 3+4 4 | 3+4 4 |
| | 5 2 | 4+4 4 | 3+3 4 | 3+4 4 | 6+6 4 | 3+4 4 | 3+4 4 | 3+4 4 |
| | 6 2 | 3+3 4 | 4+4 4 | 3+4 4 | 5+5 4 | 3+4 4 | 3+4 4 | 3+4 4 |
| | 5 2 | 4+4 4 | 3+3 4 | 3+4 4 | 6+6 4 | 3+4 4 | 3+4 4 | 3+4 4 |
| | 6 2 | 3+3 4 | 4+4 4 | 3+4 4 | 5+5 4 | 3+4 4 | 3+4 4 | 3+4 4 |
| | 5 2 | 3+3 4 | 4+4 4 | 3+4 4 | 6+6 4 | 3+4 4 | 3+4 4 | 3+4 4 |
| | 5 2 | 4+4 4 | 3+3 4 | 3+4 4 | 5+5 4 | 3+4 4 | 3+4 4 | 3+4 4 |
| | 5 2 | 3+3 4 | 3+3 4 | 3+4 4 | 5+5 4 | 3+4 4 | 3+4 4 | 3+4 4 |
| | 6 2 | 4+4 4 | 4+4 4 | 3+4 4 | 6+6 4 | 3+4 4 | 3+4 4 | 3+4 4 |
| | 5 2 | 3+3 4 | 4+4 4 | 3+4 4 | 6+6 4 | 3+4 4 | 3+4 4 | 3+4 4 |
| | 6 2 | 4+4 4 | 3+3 4 | 3+4 4 | 6+6 4 | 3+4 4 | 3+4 4 | 3+4 4 |
| | 6 2 | 4+4 4 | 3+3 4 | 3+4 4 | 6+6 4 | 3+4 4 | 3+4 4 | 3+4 4 |
| | 6 2 | 3+3 4 | 4+4 4 | 3+4 4 | 6+6 4 | 3+4 4 | 3+4 4 | 3+4 4 |
| | 6 2 | 4+4 4 | 3+3 4 | 3+4 4 | 5+5 4 | 3+4 4 | 3+4 4 | 3+4 4 |
| | 5 2 | 3+3 4 | 4+4 4 | 3+4 4 | 5+5 4 | 3+4 4 | 3+4 4 | 3+4 4 |
| | 5 2 | 3+3 4 | 3+3 4 | 3+4 4 | 5+5 4 | 3+4 4 | 3+4 4 | 3+4 4 |
| | 6 2 | 4+4 4 | 4+4 4 | 3+4 4 | 6+6 4 | 3+4 4 | 3+4 4 | 3+4 4 |
| | 5 2 | 3+3 4 | 4+4 4 | 3+4 4 | 6+6 4 | 3+4 4 | 3+4 4 | 3+4 4 |
| | 6 2 | 4+4 4 | 3+3 4 | 3+4 4 | 5+5 4 | 3+4 4 | 3+4 4 | 3+4 4 |
| | 5 2 | 4+4 4 | 4+4 4 | 3+4 4 | 6+6 4 | 3+4 4 | 3+4 4 | 3+4 4 |
| | 6 2 | 3+3 4 | 3+3 4 | 3+4 4 | 5+5 4 | 3+4 4 | 3+4 4 | 3+4 4 |
| | 5 2 | 4+4 4 | 4+4 4 | 3+4 4 | 6+6 4 | 3+4 4 | 3+4 4 | 3+4 4 |
| | 6 2 | 3+3 4 | 3+3 4 | 3+4 4 | 5+5 4 | 3+4 4 | 3+4 4 | 3+4 4 |
| | 5 2 | 3+3 4 | 3+3 4 | 3+4 4 | 5+5 4 | 3+4 4 | 3+4 4 | 3+4 4 |

Continued on next column

**TABLE XXXVII– continued**

| TQC | ICSD | NoE | SF | | | | |
|---|---|---|---|---|---|---|---|
| | 5 2 | 4+4 4 | 4+4 4 | 3+4 4 | 5+5 4 | 3+4 4 | 3+4 4 | 3+4 4 |
| | 5 2 | 3+3 4 | 3+3 4 | 3+4 4 | 5+5 4 | 3+4 4 | 3+4 4 | 3+4 4 |
| | 5 2 | 4+4 4 | 4+4 4 | 3+4 4 | 6+6 4 | 3+4 4 | 3+4 4 | 3+4 4 |
| | 5 2 | 4+4 4 | 4+4 4 | 3+4 4 | 6+6 4 | 3+4 4 | 3+4 4 | 3+4 4 |
| | 5 2 | 3+3 4 | 3+3 4 | 3+4 4 | 6+6 4 | 3+4 4 | 3+4 4 | 3+4 4 |
| | 5 2 | 4+4 4 | 4+4 4 | 3+4 4 | 6+6 4 | 3+4 4 | 3+4 4 | 3+4 4 |
| | 6 2 | 3+3 4 | 3+3 4 | 3+4 4 | 6+6 4 | 3+4 4 | 3+4 4 | 3+4 4 |
| | 5 2 | 3+3 4 | 3+3 4 | 3+4 4 | 5+5 4 | 3+4 4 | 3+4 4 | 3+4 4 |
| | 6 2 | 4+4 4 | 4+4 4 | 3+4 4 | 5+5 4 | 3+4 4 | 3+4 4 | 3+4 4 |
| | 6 2 | | | | | | | |
| | 5 2 | | | | | | | |
| | 6 2 | | | | | | | |
| | 5 2 | | | | | | | |
| | 6 2 | | | | | | | |
| | 5 2 | | | | | | | |
| | 5 2 | | | | | | | |
| | 6 2 | | | | | | | |
| | 6 2 | | | | | | | |
| | 5 2 | | | | | | | |
| | 5 2 | | | | | | | |
| | 6 2 | | | | | | | |
| | 6 2 | | | | | | | |
| | 6 2 | | | | | | | |
| | 6 2 | | | | | | | |
| | 6 2 | | | | | | | |
| | 5 2 | | | | | | | |
| | 6 2 | | | | | | | |
| | 5 2 | | | | | | | |
| | 5 2 | | | | | | | |
| | 6 2 | | | | | | | |
| | 6 2 | | | | | | | |
| | 5 2 | | | | | | | |
| | 6 2 | | | | | | | |
| | 6 2 | | | | | | | |
| | 6 2 | | | | | | | |
| | 5 2 | | | | | | | |
| | 5 2 | | | | | | | |
| | 5 2 | | | | | | | |
| | 5 2 | | | | | | | |
| | 5 2 | | | | | | | |
| | 5 2 | | | | | | | |
| | 5 2 | | | | | | | |
| | 5 2 | | | | | | | |
| | 6 2 | | | | | | | |
| | 5 2 | | | | | | | |
| | 6 2 | | | | | | | |
| | 5 2 | | | | | | | |
| 62.2.162 | 251083 | 184 | Cr1Er1O3 | | | | |
| | -GM | -R | -S | -T | -U | -X | -Y | -Z |
| | 5 2 | 3+3 4 | 3+3 4 | 3+4 4 | 5+5 4 | 3+4 4 | 3+4 4 | 3+4 4 |
| | 6 2 | 4+4 4 | 4+4 4 | 3+4 4 | 6+6 4 | 3+4 4 | 3+4 4 | 3+4 4 |
| | 5 2 | 3+3 4 | 4+4 4 | 3+4 4 | 5+5 4 | 3+4 4 | 3+4 4 | 3+4 4 |
| | 6 2 | 4+4 4 | 3+3 4 | 3+4 4 | 6+6 4 | 3+4 4 | 3+4 4 | 3+4 4 |
| | 5 2 | 3+3 4 | 4+4 4 | 3+4 4 | 6+6 4 | 3+4 4 | 3+4 4 | 3+4 4 |
| | 6 2 | 4+4 4 | 3+3 4 | 3+4 4 | 5+5 4 | 3+4 4 | 3+4 4 | 3+4 4 |
| | 5 2 | 4+4 4 | 3+3 4 | 3+4 4 | 5+5 4 | 3+4 4 | 3+4 4 | 3+4 4 |
| | 6 2 | 3+3 4 | 4+4 4 | 3+4 4 | 6+6 4 | 3+4 4 | 3+4 4 | 3+4 4 |
| | 6 2 | 3+3 4 | 4+4 4 | 3+4 4 | 5+5 4 | 3+4 4 | 3+4 4 | 3+4 4 |
| | 5 2 | 4+4 4 | 3+3 4 | 3+4 4 | 6+6 4 | 3+4 4 | 3+4 4 | 3+4 4 |
| | 5 2 | 4+4 4 | 3+3 4 | 3+4 4 | 6+6 4 | 3+4 4 | 3+4 4 | 3+4 4 |
| | 5 2 | 3+3 4 | 4+4 4 | 3+4 4 | 5+5 4 | 3+4 4 | 3+4 4 | 3+4 4 |

Continued on next column

**TABLE XXXVII– continued**

| TQC | ICSD | NoE | SF | | | | |
|---|---|---|---|---|---|---|---|
| | 6 2 | 3+3 4 | 4+4 4 | 3+4 4 | 6+6 4 | 3+4 4 | 3+4 4 | 3+4 4 |
| | 6 2 | 4+4 4 | 3+3 4 | 3+4 4 | 5+5 4 | 3+4 4 | 3+4 4 | 3+4 4 |
| | 6 2 | 3+3 4 | 4+4 4 | 3+4 4 | 6+6 4 | 3+4 4 | 3+4 4 | 3+4 4 |
| | 5 2 | 4+4 4 | 3+3 4 | 3+4 4 | 6+6 4 | 3+4 4 | 3+4 4 | 3+4 4 |
| | 5 2 | 4+4 4 | 3+3 4 | 3+4 4 | 6+6 4 | 3+4 4 | 3+4 4 | 3+4 4 |
| | 5 2 | 3+3 4 | 4+4 4 | 3+4 4 | 6+6 4 | 3+4 4 | 3+4 4 | 3+4 4 |
| | 6 2 | 4+4 4 | 3+3 4 | 3+4 4 | 5+5 4 | 3+4 4 | 3+4 4 | 3+4 4 |
| | 6 2 | 3+3 4 | 4+4 4 | 3+4 4 | 5+5 4 | 3+4 4 | 3+4 4 | 3+4 4 |
| | 6 2 | 3+3 4 | 3+3 4 | 3+4 4 | 5+5 4 | 3+4 4 | 3+4 4 | 3+4 4 |
| | 5 2 | 4+4 4 | 4+4 4 | 3+4 4 | 6+6 4 | 3+4 4 | 3+4 4 | 3+4 4 |
| | 6 2 | 4+4 4 | 4+4 4 | 3+4 4 | 6+6 4 | 3+4 4 | 3+4 4 | 3+4 4 |
| | 5 2 | 3+3 4 | 3+3 4 | 3+4 4 | 5+5 4 | 3+4 4 | 3+4 4 | 3+4 4 |
| | 5 2 | 4+4 4 | 4+4 4 | 3+4 4 | 5+5 4 | 3+4 4 | 3+4 4 | 3+4 4 |
| | 6 2 | 3+3 4 | 3+3 4 | 3+4 4 | 6+6 4 | 3+4 4 | 3+4 4 | 3+4 4 |
| | 5 2 | 4+4 4 | 3+3 4 | 3+4 4 | 6+6 4 | 3+4 4 | 3+4 4 | 3+4 4 |
| | 6 2 | 3+3 4 | 4+4 4 | 3+4 4 | 5+5 4 | 3+4 4 | 3+4 4 | 3+4 4 |
| | 5 2 | 4+4 4 | 3+3 4 | 3+4 4 | 5+5 4 | 3+4 4 | 3+4 4 | 3+4 4 |
| | 5 2 | 3+3 4 | 4+4 4 | 3+4 4 | 6+6 4 | 3+4 4 | 3+4 4 | 3+4 4 |
| | 5 2 | 3+3 4 | 4+4 4 | 3+4 4 | 5+5 4 | 3+4 4 | 3+4 4 | 3+4 4 |
| | 5 2 | 4+4 4 | 3+3 4 | 3+4 4 | 5+5 4 | 3+4 4 | 3+4 4 | 3+4 4 |
| | 5 2 | 3+3 4 | 3+3 4 | 3+4 4 | 6+6 4 | 3+4 4 | 3+4 4 | 3+4 4 |
| | 5 2 | 4+4 4 | 4+4 4 | 3+4 4 | 5+5 4 | 3+4 4 | 3+4 4 | 3+4 4 |
| | 6 2 | 4+4 4 | 4+4 4 | 3+4 4 | 6+6 4 | 3+4 4 | 3+4 4 | 3+4 4 |
| | 5 2 | 3+3 4 | 3+3 4 | 3+4 4 | 5+5 4 | 3+4 4 | 3+4 4 | 3+4 4 |
| | 5 2 | 3+3 4 | 4+4 4 | 3+4 4 | 5+5 4 | 3+4 4 | 3+4 4 | 3+4 4 |
| | 5 2 | 4+4 4 | 3+3 4 | 3+4 4 | 6+6 4 | 3+4 4 | 3+4 4 | 3+4 4 |
| | 5 2 | 4+4 4 | 3+3 4 | 3+4 4 | 6+6 4 | 3+4 4 | 3+4 4 | 3+4 4 |
| | 6 2 | 3+3 4 | 4+4 4 | 3+4 4 | 6+6 4 | 3+4 4 | 3+4 4 | 3+4 4 |
| | 6 2 | 3+3 4 | 3+3 4 | 3+4 4 | 6+6 4 | 3+4 4 | 3+4 4 | 3+4 4 |
| | 6 2 | 4+4 4 | 4+4 4 | 3+4 4 | 6+6 4 | 3+4 4 | 3+4 4 | 3+4 4 |
| | 6 2 | 4+4 4 | 3+3 4 | 3+4 4 | 5+5 4 | 3+4 4 | 3+4 4 | 3+4 4 |
| | 6 2 | 3+3 4 | 4+4 4 | 3+4 4 | 5+5 4 | 3+4 4 | 3+4 4 | 3+4 4 |
| | 5 2 | 3+3 4 | 3+3 4 | 3+4 4 | 5+5 4 | 3+4 4 | 3+4 4 | 3+4 4 |
| | 5 2 | 4+4 4 | 4+4 4 | 3+4 4 | 6+6 4 | 3+4 4 | 3+4 4 | 3+4 4 |
| | 6 2 | | | | | | | |
| | 5 2 | | | | | | | |
| | 5 2 | | | | | | | |
| | 6 2 | | | | | | | |
| | 6 2 | | | | | | | |
| | 6 2 | | | | | | | |
| | 6 2 | | | | | | | |
| | 5 2 | | | | | | | |
| | 5 2 | | | | | | | |
| | 5 2 | | | | | | | |
| | 6 2 | | | | | | | |
| | 6 2 | | | | | | | |
| | 6 2 | | | | | | | |
| | 5 2 | | | | | | | |
| | 6 2 | | | | | | | |
| | 6 2 | | | | | | | |
| | 6 2 | | | | | | | |
| | 6 2 | | | | | | | |
| | 5 2 | | | | | | | |
| | 5 2 | | | | | | | |
| | 5 2 | | | | | | | |
| | 5 2 | | | | | | | |
| | 6 2 | | | | | | | |
| | 5 2 | | | | | | | |
| | 6 2 | | | | | | | |
| | 6 2 | | | | | | | |
| | 5 2 | | | | | | | |

Continued on next column

**TABLE XXXVII– continued**

| TQC | ICSD | NoE | SF | | | | | |
|---|---|---|---|---|---|---|---|---|
| | 6 2 | | | | | | | |
| | 6 2 | | | | | | | |
| | 6 2 | | | | | | | |
| | 5 2 | | | | | | | |
| | 5 2 | | | | | | | |
| | 5 2 | | | | | | | |
| | 5 2 | | | | | | | |
| | 5 2 | | | | | | | |
| | 5 2 | | | | | | | |
| | 5 2 | | | | | | | |
| | 5 2 | | | | | | | |
| | 6 2 | | | | | | | |
| | 5 2 | | | | | | | |
| | 5 2 | | | | | | | |
| | 6 2 | | | | | | | |
| | 5 2 | | | | | | | |
| | 6 2 | | | | | | | |
| | 6 2 | | | | | | | |
| | 6 2 | | | | | | | |
| 62.2.163 | 183759 | 184 | Cr1Er1O3 | | | | | |
| | -GM | -R | -S | -T | -U | -X | -Y | -Z |
| | 5 2 | 3+3 4 | 3+3 4 | 3+4 4 | 5+5 4 | 3+4 4 | 3+4 4 | 3+4 4 |
| | 6 2 | 4+4 4 | 4+4 4 | 3+4 4 | 6+6 4 | 3+4 4 | 3+4 4 | 3+4 4 |
| | 5 2 | 3+3 4 | 3+3 4 | 3+4 4 | 5+5 4 | 3+4 4 | 3+4 4 | 3+4 4 |
| | 6 2 | 4+4 4 | 4+4 4 | 3+4 4 | 6+6 4 | 3+4 4 | 3+4 4 | 3+4 4 |
| | 5 2 | 3+3 4 | 3+3 4 | 3+4 4 | 6+6 4 | 3+4 4 | 3+4 4 | 3+4 4 |
| | 6 2 | 4+4 4 | 4+4 4 | 3+4 4 | 5+5 4 | 3+4 4 | 3+4 4 | 3+4 4 |
| | 5 2 | 4+4 4 | 4+4 4 | 3+4 4 | 5+5 4 | 3+4 4 | 3+4 4 | 3+4 4 |
| | 6 2 | 3+3 4 | 3+3 4 | 3+4 4 | 6+6 4 | 3+4 4 | 3+4 4 | 3+4 4 |
| | 6 2 | 3+3 4 | 3+3 4 | 3+4 4 | 5+5 4 | 3+4 4 | 3+4 4 | 3+4 4 |
| | 5 2 | 4+4 4 | 4+4 4 | 3+4 4 | 6+6 4 | 3+4 4 | 3+4 4 | 3+4 4 |
| | 5 2 | 4+4 4 | 3+3 4 | 3+4 4 | 6+6 4 | 3+4 4 | 3+4 4 | 3+4 4 |
| | 5 2 | 3+3 4 | 4+4 4 | 3+4 4 | 5+5 4 | 3+4 4 | 3+4 4 | 3+4 4 |
| | 6 2 | 3+3 4 | 3+3 4 | 3+4 4 | 6+6 4 | 3+4 4 | 3+4 4 | 3+4 4 |
| | 6 2 | 4+4 4 | 4+4 4 | 3+4 4 | 5+5 4 | 3+4 4 | 3+4 4 | 3+4 4 |
| | 5 2 | 3+3 4 | 3+3 4 | 3+4 4 | 6+6 4 | 3+4 4 | 3+4 4 | 3+4 4 |
| | 6 2 | 4+4 4 | 4+4 4 | 3+4 4 | 6+6 4 | 3+4 4 | 3+4 4 | 3+4 4 |
| | 5 2 | 4+4 4 | 4+4 4 | 3+4 4 | 6+6 4 | 3+4 4 | 3+4 4 | 3+4 4 |
| | 5 2 | 3+3 4 | 3+3 4 | 3+4 4 | 6+6 4 | 3+4 4 | 3+4 4 | 3+4 4 |
| | 6 2 | 4+4 4 | 3+3 4 | 3+4 4 | 5+5 4 | 3+4 4 | 3+4 4 | 3+4 4 |
| | 6 2 | 3+3 4 | 4+4 4 | 3+4 4 | 5+5 4 | 3+4 4 | 3+4 4 | 3+4 4 |
| | 6 2 | 3+3 4 | 4+4 4 | 3+4 4 | 6+6 4 | 3+4 4 | 3+4 4 | 3+4 4 |
| | 5 2 | 4+4 4 | 3+3 4 | 3+4 4 | 5+5 4 | 3+4 4 | 3+4 4 | 3+4 4 |
| | 6 2 | 3+3 4 | 3+3 4 | 3+4 4 | 6+6 4 | 3+4 4 | 3+4 4 | 3+4 4 |
| | 5 2 | 4+4 4 | 4+4 4 | 3+4 4 | 5+5 4 | 3+4 4 | 3+4 4 | 3+4 4 |
| | 5 2 | 4+4 4 | 3+3 4 | 3+4 4 | 5+5 4 | 3+4 4 | 3+4 4 | 3+4 4 |
| | 6 2 | 3+3 4 | 4+4 4 | 3+4 4 | 6+6 4 | 3+4 4 | 3+4 4 | 3+4 4 |
| | 5 2 | 4+4 4 | 4+4 4 | 3+4 4 | 6+6 4 | 3+4 4 | 3+4 4 | 3+4 4 |
| | 6 2 | 3+3 4 | 3+3 4 | 3+4 4 | 5+5 4 | 3+4 4 | 3+4 4 | 3+4 4 |
| | 5 2 | 3+3 4 | 4+4 4 | 3+4 4 | 5+5 4 | 3+4 4 | 3+4 4 | 3+4 4 |
| | 5 2 | 4+4 4 | 3+3 4 | 3+4 4 | 6+6 4 | 3+4 4 | 3+4 4 | 3+4 4 |
| | 5 2 | 3+3 4 | 3+3 4 | 3+4 4 | 5+5 4 | 3+4 4 | 3+4 4 | 3+4 4 |
| | 5 2 | 4+4 4 | 4+4 4 | 3+4 4 | 5+5 4 | 3+4 4 | 3+4 4 | 3+4 4 |
| | 5 2 | 3+3 4 | 3+3 4 | 3+4 4 | 6+6 4 | 3+4 4 | 3+4 4 | 3+4 4 |
| | 6 2 | 4+4 4 | 4+4 4 | 3+4 4 | 5+5 4 | 3+4 4 | 3+4 4 | 3+4 4 |
| | 5 2 | 4+4 4 | 3+3 4 | 3+4 4 | 6+6 4 | 3+4 4 | 3+4 4 | 3+4 4 |
| | 5 2 | 3+3 4 | 4+4 4 | 3+4 4 | 5+5 4 | 3+4 4 | 3+4 4 | 3+4 4 |
| | 5 2 | 3+3 4 | 3+3 4 | 3+4 4 | 5+5 4 | 3+4 4 | 3+4 4 | 3+4 4 |
| | 5 2 | 4+4 4 | 4+4 4 | 3+4 4 | 6+6 4 | 3+4 4 | 3+4 4 | 3+4 4 |
| | 6 2 | 4+4 4 | 4+4 4 | 3+4 4 | 6+6 4 | 3+4 4 | 3+4 4 | 3+4 4 |
| | 5 2 | 3+3 4 | 3+3 4 | 3+4 4 | 6+6 4 | 3+4 4 | 3+4 4 | 3+4 4 |

Continued on next column

**TABLE XXXVII– continued**

| TQC | ICSD | NoE | SF | | | | | |
|---|---|---|---|---|---|---|---|---|
| | 6 2 | 3+3 4 | 3+3 4 | 3+4 4 | 6+6 4 | 3+4 4 | 3+4 4 | 3+4 4 |
| | 6 2 | 4+4 4 | 4+4 4 | 3+4 4 | 6+6 4 | 3+4 4 | 3+4 4 | 3+4 4 |
| | 6 2 | 4+4 4 | 4+4 4 | 3+4 4 | 5+5 4 | 3+4 4 | 3+4 4 | 3+4 4 |
| | 6 2 | 3+3 4 | 3+3 4 | 3+4 4 | 5+5 4 | 3+4 4 | 3+4 4 | 3+4 4 |
| | 6 2 | 4+4 4 | 4+4 4 | 3+4 4 | 5+5 4 | 3+4 4 | 3+4 4 | 3+4 4 |
| | 5 2 | 3+3 4 | 3+3 4 | 3+4 4 | 6+6 4 | 3+4 4 | 3+4 4 | 3+4 4 |
| | 5 2 | | | | | | | |
| | 5 2 | | | | | | | |
| | 5 2 | | | | | | | |
| | 6 2 | | | | | | | |
| | 6 2 | | | | | | | |
| | 6 2 | | | | | | | |
| | 5 2 | | | | | | | |
| | 5 2 | | | | | | | |
| | 6 2 | | | | | | | |
| | 5 2 | | | | | | | |
| | 6 2 | | | | | | | |
| | 6 2 | | | | | | | |
| | 6 2 | | | | | | | |
| | 5 2 | | | | | | | |
| | 6 2 | | | | | | | |
| | 6 2 | | | | | | | |
| | 6 2 | | | | | | | |
| | 6 2 | | | | | | | |
| | 5 2 | | | | | | | |
| | 5 2 | | | | | | | |
| | 5 2 | | | | | | | |
| | 5 2 | | | | | | | |
| | 6 2 | | | | | | | |
| | 5 2 | | | | | | | |
| | 6 2 | | | | | | | |
| | 6 2 | | | | | | | |
| | 5 2 | | | | | | | |
| | 6 2 | | | | | | | |
| | 6 2 | | | | | | | |
| | 6 2 | | | | | | | |
| | 5 2 | | | | | | | |
| | 5 2 | | | | | | | |
| | 5 2 | | | | | | | |
| | 5 2 | | | | | | | |
| | 5 2 | | | | | | | |
| | 5 2 | | | | | | | |
| | 5 2 | | | | | | | |
| | 5 2 | | | | | | | |
| | 6 2 | | | | | | | |
| | 5 2 | | | | | | | |
| | 5 2 | | | | | | | |
| | 6 2 | | | | | | | |
| | 5 2 | | | | | | | |
| | 6 2 | | | | | | | |
| | 6 2 | | | | | | | |
| | 6 2 | | | | | | | |
| 62.2.164 | 65751 | 184 | Co2O4Si1 | | | | | |
| | -GM | -R | -S | -T | -U | -X | -Y | -Z |
| | 5 2 | 3+3 4 | 3+3 4 | 3+4 4 | 6+6 4 | 3+4 4 | 3+4 4 | 3+4 4 |
| | 5 2 | 4+4 4 | 4+4 4 | 3+4 4 | 5+5 4 | 3+4 4 | 3+4 4 | 3+4 4 |
| | 6 2 | 3+3 4 | 3+3 4 | 3+4 4 | 5+5 4 | 3+4 4 | 3+4 4 | 3+4 4 |
| | 6 2 | 4+4 4 | 4+4 4 | 3+4 4 | 6+6 4 | 3+4 4 | 3+4 4 | 3+4 4 |
| | 6 2 | 3+3 4 | 3+3 4 | 3+4 4 | 5+5 4 | 3+4 4 | 3+4 4 | 3+4 4 |
| | 6 2 | 4+4 4 | 4+4 4 | 3+4 4 | 6+6 4 | 3+4 4 | 3+4 4 | 3+4 4 |
| | 5 2 | 3+3 4 | 3+3 4 | 3+4 4 | 5+5 4 | 3+4 4 | 3+4 4 | 3+4 4 |

Continued on next column

**TABLE XXXVII– continued**

| TQC | ICSD NoE SF |
|---|---|
| | 5 2 4+4 4 4+4 4 3+4 4 6+6 4 3+4 4 3+4 4 3+4 4 |
| | 6 2 3+3 4 3+3 4 3+4 4 5+5 4 3+4 4 3+4 4 3+4 4 |
| | 6 2 4+4 4 4+4 4 3+4 4 6+6 4 3+4 4 3+4 4 3+4 4 |
| | 5 2 3+3 4 3+3 4 3+4 4 6+6 4 3+4 4 3+4 4 3+4 4 |
| | 5 2 4+4 4 4+4 4 3+4 4 6+6 4 3+4 4 3+4 4 3+4 4 |
| | 5 2 3+3 4 3+3 4 3+4 4 5+5 4 3+4 4 3+4 4 3+4 4 |
| | 6 2 4+4 4 4+4 4 3+4 4 5+5 4 3+4 4 3+4 4 3+4 4 |
| | 5 2 3+3 4 3+3 4 3+4 4 5+5 4 3+4 4 3+4 4 3+4 4 |
| | 6 2 4+4 4 4+4 4 3+4 4 6+6 4 3+4 4 3+4 4 3+4 4 |
| | 5 2 3+3 4 3+3 4 3+4 4 5+5 4 3+4 4 3+4 4 3+4 4 |
| | 5 2 4+4 4 4+4 4 3+4 4 5+5 4 3+4 4 3+4 4 3+4 4 |
| | 6 2 4+4 4 4+4 4 3+4 4 5+5 4 3+4 4 3+4 4 3+4 4 |
| | 5 2 3+3 4 3+3 4 3+4 4 6+6 4 3+4 4 3+4 4 3+4 4 |
| | 6 2 4+4 4 3+3 4 3+4 4 6+6 4 3+4 4 3+4 4 3+4 4 |
| | 5 2 3+3 4 4+4 4 3+4 4 6+6 4 3+4 4 3+4 4 3+4 4 |
| | 5 2 3+3 4 3+3 4 3+4 4 5+5 4 3+4 4 3+4 4 3+4 4 |
| | 6 2 4+4 4 4+4 4 3+4 4 6+6 4 3+4 4 3+4 4 3+4 4 |
| | 6 2 4+4 4 4+4 4 3+4 4 5+5 4 3+4 4 3+4 4 3+4 4 |
| | 5 2 3+3 4 3+3 4 3+4 4 6+6 4 3+4 4 3+4 4 3+4 4 |
| | 6 2 3+3 4 3+3 4 3+4 4 5+5 4 3+4 4 3+4 4 3+4 4 |
| | 5 2 4+4 4 4+4 4 3+4 4 5+5 4 3+4 4 3+4 4 3+4 4 |
| | 5 2 3+3 4 4+4 4 3+4 4 5+5 4 3+4 4 3+4 4 3+4 4 |
| | 6 2 4+4 4 3+3 4 3+4 4 6+6 4 3+4 4 3+4 4 3+4 4 |
| | 6 2 3+3 4 3+3 4 3+4 4 6+6 4 3+4 4 3+4 4 3+4 4 |
| | 6 2 4+4 4 4+4 4 3+4 4 6+6 4 3+4 4 3+4 4 3+4 4 |
| | 6 2 3+3 4 3+3 4 3+4 4 5+5 4 3+4 4 3+4 4 3+4 4 |
| | 5 2 4+4 4 4+4 4 3+4 4 5+5 4 3+4 4 3+4 4 3+4 4 |
| | 6 2 4+4 4 4+4 4 3+4 4 5+5 4 3+4 4 3+4 4 3+4 4 |
| | 6 2 3+3 4 3+3 4 3+4 4 5+5 4 3+4 4 3+4 4 3+4 4 |
| | 5 2 4+4 4 4+4 4 3+4 4 6+6 4 3+4 4 3+4 4 3+4 4 |
| | 5 2 3+3 4 3+3 4 3+4 4 5+5 4 3+4 4 3+4 4 3+4 4 |
| | 5 2 3+3 4 4+4 4 3+4 4 6+6 4 3+4 4 3+4 4 3+4 4 |
| | 6 2 4+4 4 3+3 4 3+4 4 5+5 4 3+4 4 3+4 4 3+4 4 |
| | 6 2 3+3 4 3+3 4 3+4 4 5+5 4 3+4 4 3+4 4 3+4 4 |
| | 5 2 4+4 4 4+4 4 3+4 4 5+5 4 3+4 4 3+4 4 3+4 4 |
| | 5 2 4+4 4 3+3 4 3+4 4 6+6 4 3+4 4 3+4 4 3+4 4 |
| | 6 2 3+3 4 4+4 4 3+4 4 5+5 4 3+4 4 3+4 4 3+4 4 |
| | 5 2 4+4 4 4+4 4 3+4 4 5+5 4 3+4 4 3+4 4 3+4 4 |
| | 5 2 3+3 4 3+3 4 3+4 4 5+5 4 3+4 4 3+4 4 3+4 4 |
| | 5 2 |
| | 5 2 |
| | 5 2 |
| | 5 2 |
| | 6 2 |
| | 6 2 |
| | 5 2 |
| | 6 2 |
| | 6 2 |
| | 5 2 |
| | 5 2 |
| | 6 2 |
| | 6 2 |
| | 6 2 |
| | 6 2 |
| | 5 2 |
| | 6 2 |
| | 6 2 |
| | 5 2 |
| | 5 2 |
| | 5 2 |
| | 5 2 |

Continued on next column

**TABLE XXXVII– continued**

| TQC | ICSD NoE SF |
|---|---|
| | 5 2 |
| | 5 2 |
| | 5 2 |
| | 5 2 |
| | 5 2 |
| | 5 2 |
| | 6 2 |
| | 5 2 |
| | 6 2 |
| | 5 2 |
| | 6 2 |
| | 5 2 |
| | 6 2 |
| | 6 2 |
| | 5 2 |
| | 5 2 |
| | 5 2 |
| | 6 2 |
| | 5 2 |
| | 5 2 |
| | 5 2 |
| | 6 2 |
| | 5 2 |
| | 5 2 |
| 62.2.165 | 6248 184 Co2O4Si1 |
| | -GM -R -S -T -U -X -Y -Z |
| | 5 2 3+3 4 3+3 4 3+4 4 6+6 4 3+4 4 3+4 4 3+4 4 |
| | 5 2 4+4 4 4+4 4 3+4 4 5+5 4 3+4 4 3+4 4 3+4 4 |
| | 6 2 3+3 4 3+3 4 3+4 4 5+5 4 3+4 4 3+4 4 3+4 4 |
| | 6 2 4+4 4 4+4 4 3+4 4 6+6 4 3+4 4 3+4 4 3+4 4 |
| | 6 2 3+3 4 3+3 4 3+4 4 5+5 4 3+4 4 3+4 4 3+4 4 |
| | 6 2 4+4 4 4+4 4 3+4 4 6+6 4 3+4 4 3+4 4 3+4 4 |
| | 5 2 3+3 4 3+3 4 3+4 4 5+5 4 3+4 4 3+4 4 3+4 4 |
| | 5 2 4+4 4 4+4 4 3+4 4 6+6 4 3+4 4 3+4 4 3+4 4 |
| | 6 2 3+3 4 3+3 4 3+4 4 5+5 4 3+4 4 3+4 4 3+4 4 |
| | 6 2 4+4 4 4+4 4 3+4 4 6+6 4 3+4 4 3+4 4 3+4 4 |
| | 5 2 3+3 4 3+3 4 3+4 4 6+6 4 3+4 4 3+4 4 3+4 4 |
| | 5 2 4+4 4 4+4 4 3+4 4 6+6 4 3+4 4 3+4 4 3+4 4 |
| | 5 2 3+3 4 3+3 4 3+4 4 5+5 4 3+4 4 3+4 4 3+4 4 |
| | 6 2 4+4 4 4+4 4 3+4 4 5+5 4 3+4 4 3+4 4 3+4 4 |
| | 5 2 3+3 4 3+3 4 3+4 4 5+5 4 3+4 4 3+4 4 3+4 4 |
| | 6 2 4+4 4 4+4 4 3+4 4 6+6 4 3+4 4 3+4 4 3+4 4 |
| | 5 2 3+3 4 3+3 4 3+4 4 5+5 4 3+4 4 3+4 4 3+4 4 |
| | 5 2 4+4 4 4+4 4 3+4 4 5+5 4 3+4 4 3+4 4 3+4 4 |
| | 6 2 4+4 4 4+4 4 3+4 4 5+5 4 3+4 4 3+4 4 3+4 4 |
| | 5 2 3+3 4 3+3 4 3+4 4 6+6 4 3+4 4 3+4 4 3+4 4 |
| | 6 2 4+4 4 3+3 4 3+4 4 6+6 4 3+4 4 3+4 4 3+4 4 |
| | 5 2 3+3 4 4+4 4 3+4 4 6+6 4 3+4 4 3+4 4 3+4 4 |
| | 5 2 3+3 4 3+3 4 3+4 4 6+6 4 3+4 4 3+4 4 3+4 4 |
| | 6 2 4+4 4 4+4 4 3+4 4 5+5 4 3+4 4 3+4 4 3+4 4 |
| | 6 2 4+4 4 4+4 4 3+4 4 5+5 4 3+4 4 3+4 4 3+4 4 |
| | 5 2 3+3 4 3+3 4 3+4 4 6+6 4 3+4 4 3+4 4 3+4 4 |
| | 6 2 3+3 4 3+3 4 3+4 4 5+5 4 3+4 4 3+4 4 3+4 4 |
| | 5 2 4+4 4 4+4 4 3+4 4 5+5 4 3+4 4 3+4 4 3+4 4 |
| | 5 2 3+3 4 4+4 4 3+4 4 5+5 4 3+4 4 3+4 4 3+4 4 |
| | 6 2 4+4 4 3+3 4 3+4 4 6+6 4 3+4 4 3+4 4 3+4 4 |
| | 6 2 3+3 4 3+3 4 3+4 4 6+6 4 3+4 4 3+4 4 3+4 4 |
| | 6 2 4+4 4 4+4 4 3+4 4 6+6 4 3+4 4 3+4 4 3+4 4 |
| | 6 2 3+3 4 3+3 4 3+4 4 5+5 4 3+4 4 3+4 4 3+4 4 |
| | 5 2 4+4 4 4+4 4 3+4 4 5+5 4 3+4 4 3+4 4 3+4 4 |
| | 6 2 4+4 4 4+4 4 3+4 4 5+5 4 3+4 4 3+4 4 3+4 4 |

Continued on next column

**TABLE XXXVII– continued**

| TQC | ICSD NoE SF |
|---|---|
| | 6 2 3+3 4 3+3 4 3+4 4 5+5 4 3+4 4 3+4 4 3+4 4 |
| | 5 2 4+4 4 4+4 4 3+4 4 6+6 4 3+4 4 3+4 4 3+4 4 |
| | 5 2 3+3 4 3+3 4 3+4 4 5+5 4 3+4 4 3+4 4 3+4 4 |
| | 5 2 3+3 4 4+4 4 3+4 4 6+6 4 3+4 4 3+4 4 3+4 4 |
| | 6 2 4+4 4 3+3 4 3+4 4 5+5 4 3+4 4 3+4 4 3+4 4 |
| | 6 2 3+3 4 3+3 4 3+4 4 5+5 4 3+4 4 3+4 4 3+4 4 |
| | 5 2 4+4 4 4+4 4 3+4 4 5+5 4 3+4 4 3+4 4 3+4 4 |
| | 5 2 4+4 4 3+3 4 3+4 4 6+6 4 3+4 4 3+4 4 3+4 4 |
| | 6 2 3+3 4 4+4 4 3+4 4 5+5 4 3+4 4 3+4 4 3+4 4 |
| | 5 2 4+4 4 4+4 4 3+4 4 5+5 4 3+4 4 3+4 4 3+4 4 |
| | 5 2 3+3 4 3+3 4 3+4 4 5+5 4 3+4 4 3+4 4 3+4 4 |
| | 5 2 |
| | 5 2 |
| | 5 2 |
| | 5 2 |
| | 6 2 |
| | 6 2 |
| | 5 2 |
| | 6 2 |
| | 6 2 |
| | 5 2 |
| | 5 2 |
| | 6 2 |
| | 6 2 |
| | 6 2 |
| | 6 2 |
| | 6 2 |
| | 5 2 |
| | 6 2 |
| | 5 2 |
| | 5 2 |
| | 5 2 |
| | 5 2 |
| | 5 2 |
| | 5 2 |
| | 5 2 |
| | 5 2 |
| | 5 2 |
| | 6 2 |
| | 5 2 |
| | 5 2 |
| | 6 2 |
| | 5 2 |
| | 6 2 |
| | 5 2 |
| | 6 2 |
| | 6 2 |
| | 5 2 |
| | 5 2 |
| | 5 2 |
| | 6 2 |
| | 5 2 |
| | 5 2 |
| | 5 2 |
| | 6 2 |
| | 5 2 |
| | 5 2 |
| 62.2.166 | 50072 184 Co2O4Si1 |
| | -GM -R -S -T -U -X -Y -Z |
| | 5 2 3+3 4 3+3 4 3+4 4 6+6 4 3+4 4 3+4 4 3+4 4 |
| | 5 2 4+4 4 4+4 4 3+4 4 5+5 4 3+4 4 3+4 4 3+4 4 |
| | 6 2 3+3 4 3+3 4 3+4 4 5+5 4 3+4 4 3+4 4 3+4 4 |
| | 6 2 4+4 4 4+4 4 3+4 4 6+6 4 3+4 4 3+4 4 3+4 4 |
| | 6 2 3+3 4 3+3 4 3+4 4 5+5 4 3+4 4 3+4 4 3+4 4 |
| | 6 2 4+4 4 4+4 4 3+4 4 5+5 4 3+4 4 3+4 4 3+4 4 |
| | 5 2 3+3 4 3+3 4 3+4 4 6+6 4 3+4 4 3+4 4 3+4 4 |
| | 5 2 4+4 4 4+4 4 3+4 4 6+6 4 3+4 4 3+4 4 3+4 4 |
| | 6 2 3+3 4 3+3 4 3+4 4 5+5 4 3+4 4 3+4 4 3+4 4 |
| | 5 2 4+4 4 4+4 4 3+4 4 6+6 4 3+4 4 3+4 4 3+4 4 |
| | 6 2 3+3 4 3+3 4 3+4 4 6+6 4 3+4 4 3+4 4 3+4 4 |
| | 6 2 4+4 4 4+4 4 3+4 4 6+6 4 3+4 4 3+4 4 3+4 4 |
| | 5 2 3+3 4 3+3 4 3+4 4 5+5 4 3+4 4 3+4 4 3+4 4 |
| | 5 2 4+4 4 4+4 4 3+4 4 5+5 4 3+4 4 3+4 4 3+4 4 |
| | 5 2 3+3 4 3+3 4 3+4 4 5+5 4 3+4 4 3+4 4 3+4 4 |
| | 6 2 4+4 4 4+4 4 3+4 4 6+6 4 3+4 4 3+4 4 3+4 4 |
| | 5 2 3+3 4 3+3 4 3+4 4 5+5 4 3+4 4 3+4 4 3+4 4 |
| | 5 2 4+4 4 4+4 4 3+4 4 5+5 4 3+4 4 3+4 4 3+4 4 |
| | 5 2 4+4 4 4+4 4 3+4 4 5+5 4 3+4 4 3+4 4 3+4 4 |
| | 6 2 3+3 4 3+3 4 3+4 4 6+6 4 3+4 4 3+4 4 3+4 4 |
| | 6 2 4+4 4 3+3 4 3+4 4 6+6 4 3+4 4 3+4 4 3+4 4 |
| | 5 2 3+3 4 4+4 4 3+4 4 6+6 4 3+4 4 3+4 4 3+4 4 |
| | 5 2 3+3 4 3+3 4 3+4 4 6+6 4 3+4 4 3+4 4 3+4 4 |
| | 6 2 4+4 4 4+4 4 3+4 4 5+5 4 3+4 4 3+4 4 3+4 4 |
| | 6 2 4+4 4 4+4 4 3+4 4 5+5 4 3+4 4 3+4 4 3+4 4 |
| | 5 2 3+3 4 3+3 4 3+4 4 6+6 4 3+4 4 3+4 4 3+4 4 |
| | 6 2 3+3 4 4+4 4 3+4 4 5+5 4 3+4 4 3+4 4 3+4 4 |
| | 5 2 4+4 4 3+3 4 3+4 4 5+5 4 3+4 4 3+4 4 3+4 4 |
| | 5 2 3+3 4 4+4 4 3+4 4 5+5 4 3+4 4 3+4 4 3+4 4 |
| | 6 2 4+4 4 3+3 4 3+4 4 6+6 4 3+4 4 3+4 4 3+4 4 |
| | 6 2 3+3 4 3+3 4 3+4 4 6+6 4 3+4 4 3+4 4 3+4 4 |
| | 6 2 4+4 4 4+4 4 3+4 4 6+6 4 3+4 4 3+4 4 3+4 4 |
| | 6 2 3+3 4 3+3 4 3+4 4 5+5 4 3+4 4 3+4 4 3+4 4 |
| | 5 2 4+4 4 4+4 4 3+4 4 5+5 4 3+4 4 3+4 4 3+4 4 |
| | 6 2 4+4 4 4+4 4 3+4 4 5+5 4 3+4 4 3+4 4 3+4 4 |
| | 6 2 3+3 4 3+3 4 3+4 4 5+5 4 3+4 4 3+4 4 3+4 4 |
| | 5 2 4+4 4 4+4 4 3+4 4 6+6 4 3+4 4 3+4 4 3+4 4 |
| | 5 2 3+3 4 3+3 4 3+4 4 5+5 4 3+4 4 3+4 4 3+4 4 |
| | 5 2 3+3 4 4+4 4 3+4 4 6+6 4 3+4 4 3+4 4 3+4 4 |
| | 6 2 4+4 4 3+3 4 3+4 4 5+5 4 3+4 4 3+4 4 3+4 4 |
| | 6 2 3+3 4 3+3 4 3+4 4 5+5 4 3+4 4 3+4 4 3+4 4 |
| | 5 2 4+4 4 4+4 4 3+4 4 5+5 4 3+4 4 3+4 4 3+4 4 |
| | 5 2 3+3 4 3+3 4 3+4 4 6+6 4 3+4 4 3+4 4 3+4 4 |
| | 6 2 4+4 4 4+4 4 3+4 4 5+5 4 3+4 4 3+4 4 3+4 4 |
| | 5 2 3+3 4 4+4 4 3+4 4 5+5 4 3+4 4 3+4 4 3+4 4 |
| | 5 2 4+4 4 3+3 4 3+4 4 5+5 4 3+4 4 3+4 4 3+4 4 |
| | 5 2 |
| | 5 2 |
| | 5 2 |
| | 5 2 |
| | 6 2 |
| | 5 2 |
| | 6 2 |
| | 6 2 |
| | 6 2 |
| | 5 2 |
| | 5 2 |
| | 6 2 |
| | 6 2 |
| | 6 2 |
| | 6 2 |
| | 6 2 |
| | 5 2 |

Continued on next column

**TABLE XXXVII– continued**

| TQC | ICSD | NoE | SF | | | | | |
|---|---|---|---|---|---|---|---|---|
| | 6 2 | | | | | | | |
| | 5 2 | | | | | | | |
| | 5 2 | | | | | | | |
| | 5 2 | | | | | | | |
| | 5 2 | | | | | | | |
| | 5 2 | | | | | | | |
| | 5 2 | | | | | | | |
| | 5 2 | | | | | | | |
| | 5 2 | | | | | | | |
| | 5 2 | | | | | | | |
| | 6 2 | | | | | | | |
| | 5 2 | | | | | | | |
| | 5 2 | | | | | | | |
| | 5 2 | | | | | | | |
| | 6 2 | | | | | | | |
| | 6 2 | | | | | | | |
| | 5 2 | | | | | | | |
| | 6 2 | | | | | | | |
| | 6 2 | | | | | | | |
| | 5 2 | | | | | | | |
| | 5 2 | | | | | | | |
| | 5 2 | | | | | | | |
| | 6 2 | | | | | | | |
| | 5 2 | | | | | | | |
| | 5 2 | | | | | | | |
| | 5 2 | | | | | | | |
| | 6 2 | | | | | | | |
| | 5 2 | | | | | | | |
| | 5 2 | | | | | | | |
| 62.2.167 | 260092 | 184 | Co2O4Si1 | | | | | |
| | -GM | -R | -S | -T | -U | -X | -Y | -Z |
| | 5 2 | 3+3 4 | 3+3 4 | 3+4 4 | 6+6 4 | 3+4 4 | 3+4 4 | 3+4 4 |
| | 5 2 | 4+4 4 | 4+4 4 | 3+4 4 | 5+5 4 | 3+4 4 | 3+4 4 | 3+4 4 |
| | 6 2 | 3+3 4 | 3+3 4 | 3+4 4 | 5+5 4 | 3+4 4 | 3+4 4 | 3+4 4 |
| | 6 2 | 4+4 4 | 4+4 4 | 3+4 4 | 6+6 4 | 3+4 4 | 3+4 4 | 3+4 4 |
| | 6 2 | 3+3 4 | 3+3 4 | 3+4 4 | 5+5 4 | 3+4 4 | 3+4 4 | 3+4 4 |
| | 6 2 | 4+4 4 | 4+4 4 | 3+4 4 | 6+6 4 | 3+4 4 | 3+4 4 | 3+4 4 |
| | 5 2 | 3+3 4 | 3+3 4 | 3+4 4 | 5+5 4 | 3+4 4 | 3+4 4 | 3+4 4 |
| | 5 2 | 4+4 4 | 4+4 4 | 3+4 4 | 6+6 4 | 3+4 4 | 3+4 4 | 3+4 4 |
| | 6 2 | 3+3 4 | 3+3 4 | 3+4 4 | 5+5 4 | 3+4 4 | 3+4 4 | 3+4 4 |
| | 6 2 | 4+4 4 | 4+4 4 | 3+4 4 | 6+6 4 | 3+4 4 | 3+4 4 | 3+4 4 |
| | 5 2 | 3+3 4 | 3+3 4 | 3+4 4 | 6+6 4 | 3+4 4 | 3+4 4 | 3+4 4 |
| | 5 2 | 4+4 4 | 4+4 4 | 3+4 4 | 6+6 4 | 3+4 4 | 3+4 4 | 3+4 4 |
| | 5 2 | 3+3 4 | 3+3 4 | 3+4 4 | 5+5 4 | 3+4 4 | 3+4 4 | 3+4 4 |
| | 6 2 | 4+4 4 | 4+4 4 | 3+4 4 | 5+5 4 | 3+4 4 | 3+4 4 | 3+4 4 |
| | 5 2 | 3+3 4 | 3+3 4 | 3+4 4 | 5+5 4 | 3+4 4 | 3+4 4 | 3+4 4 |
| | 6 2 | 4+4 4 | 4+4 4 | 3+4 4 | 6+6 4 | 3+4 4 | 3+4 4 | 3+4 4 |
| | 5 2 | 3+3 4 | 3+3 4 | 3+4 4 | 5+5 4 | 3+4 4 | 3+4 4 | 3+4 4 |
| | 5 2 | 4+4 4 | 4+4 4 | 3+4 4 | 5+5 4 | 3+4 4 | 3+4 4 | 3+4 4 |
| | 6 2 | 4+4 4 | 4+4 4 | 3+4 4 | 5+5 4 | 3+4 4 | 3+4 4 | 3+4 4 |
| | 5 2 | 3+3 4 | 3+3 4 | 3+4 4 | 6+6 4 | 3+4 4 | 3+4 4 | 3+4 4 |
| | 6 2 | 4+4 4 | 3+3 4 | 3+4 4 | 6+6 4 | 3+4 4 | 3+4 4 | 3+4 4 |
| | 5 2 | 3+3 4 | 4+4 4 | 3+4 4 | 6+6 4 | 3+4 4 | 3+4 4 | 3+4 4 |
| | 5 2 | 3+3 4 | 3+3 4 | 3+4 4 | 5+5 4 | 3+4 4 | 3+4 4 | 3+4 4 |
| | 6 2 | 4+4 4 | 4+4 4 | 3+4 4 | 6+6 4 | 3+4 4 | 3+4 4 | 3+4 4 |
| | 6 2 | 4+4 4 | 4+4 4 | 3+4 4 | 5+5 4 | 3+4 4 | 3+4 4 | 3+4 4 |
| | 5 2 | 3+3 4 | 3+3 4 | 3+4 4 | 6+6 4 | 3+4 4 | 3+4 4 | 3+4 4 |
| | 6 2 | 3+3 4 | 4+4 4 | 3+4 4 | 5+5 4 | 3+4 4 | 3+4 4 | 3+4 4 |
| | 5 2 | 4+4 4 | 3+3 4 | 3+4 4 | 5+5 4 | 3+4 4 | 3+4 4 | 3+4 4 |
| | 5 2 | 3+3 4 | 4+4 4 | 3+4 4 | 5+5 4 | 3+4 4 | 3+4 4 | 3+4 4 |
| | 6 2 | 4+4 4 | 3+3 4 | 3+4 4 | 6+6 4 | 3+4 4 | 3+4 4 | 3+4 4 |
| | 6 2 | 3+3 4 | 3+3 4 | 3+4 4 | 6+6 4 | 3+4 4 | 3+4 4 | 3+4 4 |
| | 6 2 | 4+4 4 | 4+4 4 | 3+4 4 | 6+6 4 | 3+4 4 | 3+4 4 | 3+4 4 |
| | 6 2 | 3+3 4 | 3+3 4 | 3+4 4 | 5+5 4 | 3+4 4 | 3+4 4 | 3+4 4 |
| | 5 2 | 4+4 4 | 4+4 4 | 3+4 4 | 5+5 4 | 3+4 4 | 3+4 4 | 3+4 4 |
| | 6 2 | 4+4 4 | 4+4 4 | 3+4 4 | 5+5 4 | 3+4 4 | 3+4 4 | 3+4 4 |
| | 6 2 | 3+3 4 | 3+3 4 | 3+4 4 | 5+5 4 | 3+4 4 | 3+4 4 | 3+4 4 |
| | 5 2 | 4+4 4 | 4+4 4 | 3+4 4 | 6+6 4 | 3+4 4 | 3+4 4 | 3+4 4 |
| | 5 2 | 3+3 4 | 3+3 4 | 3+4 4 | 5+5 4 | 3+4 4 | 3+4 4 | 3+4 4 |
| | 5 2 | 3+3 4 | 4+4 4 | 3+4 4 | 6+6 4 | 3+4 4 | 3+4 4 | 3+4 4 |
| | 6 2 | 4+4 4 | 3+3 4 | 3+4 4 | 5+5 4 | 3+4 4 | 3+4 4 | 3+4 4 |
| | 6 2 | 3+3 4 | 3+3 4 | 3+4 4 | 5+5 4 | 3+4 4 | 3+4 4 | 3+4 4 |
| | 5 2 | 4+4 4 | 4+4 4 | 3+4 4 | 5+5 4 | 3+4 4 | 3+4 4 | 3+4 4 |
| | 5 2 | 4+4 4 | 3+3 4 | 3+4 4 | 6+6 4 | 3+4 4 | 3+4 4 | 3+4 4 |
| | 6 2 | 3+3 4 | 4+4 4 | 3+4 4 | 5+5 4 | 3+4 4 | 3+4 4 | 3+4 4 |
| | 5 2 | 4+4 4 | 4+4 4 | 3+4 4 | 5+5 4 | 3+4 4 | 3+4 4 | 3+4 4 |
| | 5 2 | 3+3 4 | 3+3 4 | 3+4 4 | 5+5 4 | 3+4 4 | 3+4 4 | 3+4 4 |
| | 5 2 | | | | | | | |
| | 5 2 | | | | | | | |
| | 5 2 | | | | | | | |
| | 5 2 | | | | | | | |
| | 6 2 | | | | | | | |
| | 6 2 | | | | | | | |
| | 5 2 | | | | | | | |
| | 6 2 | | | | | | | |
| | 6 2 | | | | | | | |
| | 5 2 | | | | | | | |
| | 5 2 | | | | | | | |
| | 6 2 | | | | | | | |
| | 6 2 | | | | | | | |
| | 6 2 | | | | | | | |
| | 6 2 | | | | | | | |
| | 5 2 | | | | | | | |
| | 6 2 | | | | | | | |
| | 6 2 | | | | | | | |
| | 5 2 | | | | | | | |
| | 5 2 | | | | | | | |
| | 5 2 | | | | | | | |
| | 5 2 | | | | | | | |
| | 5 2 | | | | | | | |
| | 5 2 | | | | | | | |
| | 5 2 | | | | | | | |
| | 5 2 | | | | | | | |
| | 5 2 | | | | | | | |
| | 5 2 | | | | | | | |
| | 6 2 | | | | | | | |
| | 5 2 | | | | | | | |
| | 6 2 | | | | | | | |
| | 5 2 | | | | | | | |
| | 6 2 | | | | | | | |
| | 5 2 | | | | | | | |
| | 6 2 | | | | | | | |
| | 6 2 | | | | | | | |
| | 5 2 | | | | | | | |
| | 5 2 | | | | | | | |
| | 5 2 | | | | | | | |
| | 6 2 | | | | | | | |
| | 5 2 | | | | | | | |
| | 5 2 | | | | | | | |
| | 5 2 | | | | | | | |
| | 6 2 | | | | | | | |
| | 5 2 | | | | | | | |

Continued on next column

**TABLE XXXVII– continued**

| TQC | ICSD | NoE | SF | | | | | |
|---|---|---|---|---|---|---|---|---|
| | 5 2 | | | | | | | |
| 62.2.168 | 260091 | 184 | Co2O4Si1 | | | | | |
| | -GM | -R | -S | -T | -U | -X | -Y | -Z |
| | 5 2 | 3+3 4 | 3+3 4 | 3+4 4 | 6+6 4 | 3+4 4 | 3+4 4 | 3+4 4 |
| | 5 2 | 4+4 4 | 4+4 4 | 3+4 4 | 5+5 4 | 3+4 4 | 3+4 4 | 3+4 4 |
| | 6 2 | 3+3 4 | 3+3 4 | 3+4 4 | 5+5 4 | 3+4 4 | 3+4 4 | 3+4 4 |
| | 6 2 | 4+4 4 | 4+4 4 | 3+4 4 | 6+6 4 | 3+4 4 | 3+4 4 | 3+4 4 |
| | 6 2 | 3+3 4 | 3+3 4 | 3+4 4 | 5+5 4 | 3+4 4 | 3+4 4 | 3+4 4 |
| | 6 2 | 4+4 4 | 4+4 4 | 3+4 4 | 6+6 4 | 3+4 4 | 3+4 4 | 3+4 4 |
| | 5 2 | 3+3 4 | 3+3 4 | 3+4 4 | 5+5 4 | 3+4 4 | 3+4 4 | 3+4 4 |
| | 5 2 | 4+4 4 | 4+4 4 | 3+4 4 | 6+6 4 | 3+4 4 | 3+4 4 | 3+4 4 |
| | 6 2 | 3+3 4 | 3+3 4 | 3+4 4 | 5+5 4 | 3+4 4 | 3+4 4 | 3+4 4 |
| | 6 2 | 4+4 4 | 4+4 4 | 3+4 4 | 6+6 4 | 3+4 4 | 3+4 4 | 3+4 4 |
| | 5 2 | 3+3 4 | 3+3 4 | 3+4 4 | 6+6 4 | 3+4 4 | 3+4 4 | 3+4 4 |
| | 5 2 | 4+4 4 | 4+4 4 | 3+4 4 | 6+6 4 | 3+4 4 | 3+4 4 | 3+4 4 |
| | 5 2 | 3+3 4 | 3+3 4 | 3+4 4 | 5+5 4 | 3+4 4 | 3+4 4 | 3+4 4 |
| | 6 2 | 4+4 4 | 4+4 4 | 3+4 4 | 5+5 4 | 3+4 4 | 3+4 4 | 3+4 4 |
| | 5 2 | 3+3 4 | 3+3 4 | 3+4 4 | 5+5 4 | 3+4 4 | 3+4 4 | 3+4 4 |
| | 6 2 | 4+4 4 | 4+4 4 | 3+4 4 | 6+6 4 | 3+4 4 | 3+4 4 | 3+4 4 |
| | 5 2 | 3+3 4 | 3+3 4 | 3+4 4 | 5+5 4 | 3+4 4 | 3+4 4 | 3+4 4 |
| | 5 2 | 4+4 4 | 4+4 4 | 3+4 4 | 5+5 4 | 3+4 4 | 3+4 4 | 3+4 4 |
| | 6 2 | 4+4 4 | 4+4 4 | 3+4 4 | 5+5 4 | 3+4 4 | 3+4 4 | 3+4 4 |
| | 5 2 | 3+3 4 | 3+3 4 | 3+4 4 | 6+6 4 | 3+4 4 | 3+4 4 | 3+4 4 |
| | 6 2 | 4+4 4 | 3+3 4 | 3+4 4 | 6+6 4 | 3+4 4 | 3+4 4 | 3+4 4 |
| | 5 2 | 3+3 4 | 4+4 4 | 3+4 4 | 6+6 4 | 3+4 4 | 3+4 4 | 3+4 4 |
| | 5 2 | 3+3 4 | 3+3 4 | 3+4 4 | 5+5 4 | 3+4 4 | 3+4 4 | 3+4 4 |
| | 6 2 | 4+4 4 | 4+4 4 | 3+4 4 | 6+6 4 | 3+4 4 | 3+4 4 | 3+4 4 |
| | 6 2 | 4+4 4 | 4+4 4 | 3+4 4 | 5+5 4 | 3+4 4 | 3+4 4 | 3+4 4 |
| | 5 2 | 3+3 4 | 3+3 4 | 3+4 4 | 6+6 4 | 3+4 4 | 3+4 4 | 3+4 4 |
| | 6 2 | 3+3 4 | 3+3 4 | 3+4 4 | 5+5 4 | 3+4 4 | 3+4 4 | 3+4 4 |
| | 5 2 | 4+4 4 | 4+4 4 | 3+4 4 | 5+5 4 | 3+4 4 | 3+4 4 | 3+4 4 |
| | 5 2 | 3+3 4 | 4+4 4 | 3+4 4 | 5+5 4 | 3+4 4 | 3+4 4 | 3+4 4 |
| | 6 2 | 4+4 4 | 3+3 4 | 3+4 4 | 6+6 4 | 3+4 4 | 3+4 4 | 3+4 4 |
| | 6 2 | 3+3 4 | 3+3 4 | 3+4 4 | 6+6 4 | 3+4 4 | 3+4 4 | 3+4 4 |
| | 6 2 | 4+4 4 | 4+4 4 | 3+4 4 | 6+6 4 | 3+4 4 | 3+4 4 | 3+4 4 |
| | 6 2 | 3+3 4 | 3+3 4 | 3+4 4 | 5+5 4 | 3+4 4 | 3+4 4 | 3+4 4 |
| | 5 2 | 4+4 4 | 4+4 4 | 3+4 4 | 5+5 4 | 3+4 4 | 3+4 4 | 3+4 4 |
| | 6 2 | 4+4 4 | 4+4 4 | 3+4 4 | 5+5 4 | 3+4 4 | 3+4 4 | 3+4 4 |
| | 6 2 | 3+3 4 | 3+3 4 | 3+4 4 | 5+5 4 | 3+4 4 | 3+4 4 | 3+4 4 |
| | 5 2 | 4+4 4 | 4+4 4 | 3+4 4 | 6+6 4 | 3+4 4 | 3+4 4 | 3+4 4 |
| | 5 2 | 3+3 4 | 3+3 4 | 3+4 4 | 5+5 4 | 3+4 4 | 3+4 4 | 3+4 4 |
| | 5 2 | 3+3 4 | 4+4 4 | 3+4 4 | 6+6 4 | 3+4 4 | 3+4 4 | 3+4 4 |
| | 6 2 | 4+4 4 | 3+3 4 | 3+4 4 | 5+5 4 | 3+4 4 | 3+4 4 | 3+4 4 |
| | 6 2 | 3+3 4 | 3+3 4 | 3+4 4 | 5+5 4 | 3+4 4 | 3+4 4 | 3+4 4 |
| | 5 2 | 4+4 4 | 4+4 4 | 3+4 4 | 5+5 4 | 3+4 4 | 3+4 4 | 3+4 4 |
| | 5 2 | 4+4 4 | 3+3 4 | 3+4 4 | 6+6 4 | 3+4 4 | 3+4 4 | 3+4 4 |
| | 6 2 | 3+3 4 | 4+4 4 | 3+4 4 | 5+5 4 | 3+4 4 | 3+4 4 | 3+4 4 |
| | 5 2 | 4+4 4 | 4+4 4 | 3+4 4 | 5+5 4 | 3+4 4 | 3+4 4 | 3+4 4 |
| | 5 2 | 3+3 4 | 3+3 4 | 3+4 4 | 5+5 4 | 3+4 4 | 3+4 4 | 3+4 4 |
| | 5 2 | | | | | | | |
| | 5 2 | | | | | | | |
| | 5 2 | | | | | | | |
| | 5 2 | | | | | | | |
| | 6 2 | | | | | | | |
| | 6 2 | | | | | | | |
| | 5 2 | | | | | | | |
| | 6 2 | | | | | | | |
| | 6 2 | | | | | | | |
| | 5 2 | | | | | | | |
| | 5 2 | | | | | | | |
| | 6 2 | | | | | | | |

Continued on next column

**TABLE XXXVII– continued**

| TQC | ICSD | NoE | SF | | | | | |
|---|---|---|---|---|---|---|---|---|
| | 6 2 | | | | | | | |
| | 6 2 | | | | | | | |
| | 6 2 | | | | | | | |
| | 5 2 | | | | | | | |
| | 6 2 | | | | | | | |
| | 6 2 | | | | | | | |
| | 5 2 | | | | | | | |
| | 5 2 | | | | | | | |
| | 5 2 | | | | | | | |
| | 5 2 | | | | | | | |
| | 5 2 | | | | | | | |
| | 5 2 | | | | | | | |
| | 5 2 | | | | | | | |
| | 5 2 | | | | | | | |
| | 5 2 | | | | | | | |
| | 5 2 | | | | | | | |
| | 6 2 | | | | | | | |
| | 5 2 | | | | | | | |
| | 6 2 | | | | | | | |
| | 5 2 | | | | | | | |
| | 6 2 | | | | | | | |
| | 5 2 | | | | | | | |
| | 6 2 | | | | | | | |
| | 6 2 | | | | | | | |
| | 5 2 | | | | | | | |
| | 5 2 | | | | | | | |
| | 5 2 | | | | | | | |
| | 6 2 | | | | | | | |
| | 5 2 | | | | | | | |
| | 5 2 | | | | | | | |
| | 5 2 | | | | | | | |
| | 6 2 | | | | | | | |
| | 5 2 | | | | | | | |
| | 5 2 | | | | | | | |
| 62.2.169 | 202255 | 184 | Co2O4Si1 | | | | | |
| | -GM | -R | -S | -T | -U | -X | -Y | -Z |
| | 5 2 | 3+3 4 | 3+3 4 | 3+4 4 | 6+6 4 | 3+4 4 | 3+4 4 | 3+4 4 |
| | 5 2 | 4+4 4 | 4+4 4 | 3+4 4 | 5+5 4 | 3+4 4 | 3+4 4 | 3+4 4 |
| | 6 2 | 3+3 4 | 3+3 4 | 3+4 4 | 5+5 4 | 3+4 4 | 3+4 4 | 3+4 4 |
| | 6 2 | 4+4 4 | 4+4 4 | 3+4 4 | 6+6 4 | 3+4 4 | 3+4 4 | 3+4 4 |
| | 6 2 | 3+3 4 | 3+3 4 | 3+4 4 | 5+5 4 | 3+4 4 | 3+4 4 | 3+4 4 |
| | 6 2 | 4+4 4 | 4+4 4 | 3+4 4 | 6+6 4 | 3+4 4 | 3+4 4 | 3+4 4 |
| | 5 2 | 3+3 4 | 3+3 4 | 3+4 4 | 5+5 4 | 3+4 4 | 3+4 4 | 3+4 4 |
| | 5 2 | 4+4 4 | 4+4 4 | 3+4 4 | 6+6 4 | 3+4 4 | 3+4 4 | 3+4 4 |
| | 6 2 | 3+3 4 | 3+3 4 | 3+4 4 | 5+5 4 | 3+4 4 | 3+4 4 | 3+4 4 |
| | 6 2 | 4+4 4 | 4+4 4 | 3+4 4 | 6+6 4 | 3+4 4 | 3+4 4 | 3+4 4 |
| | 5 2 | 3+3 4 | 3+3 4 | 3+4 4 | 6+6 4 | 3+4 4 | 3+4 4 | 3+4 4 |
| | 5 2 | 4+4 4 | 4+4 4 | 3+4 4 | 6+6 4 | 3+4 4 | 3+4 4 | 3+4 4 |
| | 5 2 | 3+3 4 | 3+3 4 | 3+4 4 | 5+5 4 | 3+4 4 | 3+4 4 | 3+4 4 |
| | 6 2 | 4+4 4 | 4+4 4 | 3+4 4 | 5+5 4 | 3+4 4 | 3+4 4 | 3+4 4 |
| | 5 2 | 3+3 4 | 3+3 4 | 3+4 4 | 5+5 4 | 3+4 4 | 3+4 4 | 3+4 4 |
| | 6 2 | 4+4 4 | 4+4 4 | 3+4 4 | 6+6 4 | 3+4 4 | 3+4 4 | 3+4 4 |
| | 5 2 | 3+3 4 | 3+3 4 | 3+4 4 | 5+5 4 | 3+4 4 | 3+4 4 | 3+4 4 |
| | 5 2 | 4+4 4 | 4+4 4 | 3+4 4 | 5+5 4 | 3+4 4 | 3+4 4 | 3+4 4 |
| | 6 2 | 4+4 4 | 4+4 4 | 3+4 4 | 5+5 4 | 3+4 4 | 3+4 4 | 3+4 4 |
| | 5 2 | 3+3 4 | 3+3 4 | 3+4 4 | 6+6 4 | 3+4 4 | 3+4 4 | 3+4 4 |
| | 6 2 | 4+4 4 | 3+3 4 | 3+4 4 | 6+6 4 | 3+4 4 | 3+4 4 | 3+4 4 |
| | 5 2 | 3+3 4 | 4+4 4 | 3+4 4 | 6+6 4 | 3+4 4 | 3+4 4 | 3+4 4 |
| | 5 2 | 3+3 4 | 3+3 4 | 3+4 4 | 5+5 4 | 3+4 4 | 3+4 4 | 3+4 4 |
| | 6 2 | 4+4 4 | 4+4 4 | 3+4 4 | 6+6 4 | 3+4 4 | 3+4 4 | 3+4 4 |
| | 6 2 | 4+4 4 | 4+4 4 | 3+4 4 | 5+5 4 | 3+4 4 | 3+4 4 | 3+4 4 |

Continued on next column

**TABLE XXXVII– continued**

| TQC | ICSD | NoE | SF |
|---|---|---|---|
| | 5 2 3+3 4 3+3 4 3+4 4 6+6 4 3+4 4 3+4 4 3+4 4 | | |
| | 6 2 3+3 4 4+4 4 3+4 4 5+5 4 3+4 4 3+4 4 3+4 4 | | |
| | 5 2 4+4 4 3+3 4 3+4 4 5+5 4 3+4 4 3+4 4 3+4 4 | | |
| | 5 2 3+3 4 4+4 4 3+4 4 5+5 4 3+4 4 3+4 4 3+4 4 | | |
| | 6 2 4+4 4 3+3 4 3+4 4 6+6 4 3+4 4 3+4 4 3+4 4 | | |
| | 6 2 3+3 4 3+3 4 3+4 4 6+6 4 3+4 4 3+4 4 3+4 4 | | |
| | 6 2 4+4 4 4+4 4 3+4 4 6+6 4 3+4 4 3+4 4 3+4 4 | | |
| | 6 2 3+3 4 3+3 4 3+4 4 5+5 4 3+4 4 3+4 4 3+4 4 | | |
| | 5 2 4+4 4 4+4 4 3+4 4 5+5 4 3+4 4 3+4 4 3+4 4 | | |
| | 6 2 4+4 4 4+4 4 3+4 4 5+5 4 3+4 4 3+4 4 3+4 4 | | |
| | 6 2 3+3 4 3+3 4 3+4 4 5+5 4 3+4 4 3+4 4 3+4 4 | | |
| | 5 2 4+4 4 4+4 4 3+4 4 6+6 4 3+4 4 3+4 4 3+4 4 | | |
| | 5 2 3+3 4 3+3 4 3+4 4 5+5 4 3+4 4 3+4 4 3+4 4 | | |
| | 5 2 3+3 4 4+4 4 3+4 4 6+6 4 3+4 4 3+4 4 3+4 4 | | |
| | 6 2 4+4 4 3+3 4 3+4 4 5+5 4 3+4 4 3+4 4 3+4 4 | | |
| | 6 2 3+3 4 3+3 4 3+4 4 5+5 4 3+4 4 3+4 4 3+4 4 | | |
| | 5 2 4+4 4 4+4 4 3+4 4 5+5 4 3+4 4 3+4 4 3+4 4 | | |
| | 5 2 4+4 4 3+3 4 3+4 4 6+6 4 3+4 4 3+4 4 3+4 4 | | |
| | 6 2 3+3 4 4+4 4 3+4 4 5+5 4 3+4 4 3+4 4 3+4 4 | | |
| | 5 2 4+4 4 4+4 4 3+4 4 5+5 4 3+4 4 3+4 4 3+4 4 | | |
| | 5 2 3+3 4 3+3 4 3+4 4 5+5 4 3+4 4 3+4 4 3+4 4 | | |
| | 5 2 | | |
| | 5 2 | | |
| | 5 2 | | |
| | 5 2 | | |
| | 6 2 | | |
| | 6 2 | | |
| | 5 2 | | |
| | 6 2 | | |
| | 6 2 | | |
| | 5 2 | | |
| | 5 2 | | |
| | 6 2 | | |
| | 6 2 | | |
| | 6 2 | | |
| | 6 2 | | |
| | 5 2 | | |
| | 6 2 | | |
| | 6 2 | | |
| | 5 2 | | |
| | 5 2 | | |
| | 5 2 | | |
| | 5 2 | | |
| | 5 2 | | |
| | 5 2 | | |
| | 5 2 | | |
| | 5 2 | | |
| | 5 2 | | |
| | 5 2 | | |
| | 6 2 | | |
| | 5 2 | | |
| | 5 2 | | |
| | 6 2 | | |
| | 6 2 | | |
| | 5 2 | | |
| | 6 2 | | |
| | 6 2 | | |
| | 5 2 | | |
| | 5 2 | | |
| | 5 2 | | |
| | 6 2 | | |
| | 5 2 | | |
| | 5 2 | | |
| | 5 2 | | |
| | 6 2 | | |
| | 5 2 | | |
| | 5 2 | | |
| 62.2.170 | 200705 | 184 | Co2O4Si1 |
| | -GM -R -S -T -U -X -Y -Z | | |
| | 5 2 3+3 4 3+3 4 3+4 4 6+6 4 3+4 4 3+4 4 3+4 4 | | |
| | 5 2 4+4 4 4+4 4 3+4 4 5+5 4 3+4 4 3+4 4 3+4 4 | | |
| | 6 2 3+3 4 3+3 4 3+4 4 5+5 4 3+4 4 3+4 4 3+4 4 | | |
| | 6 2 4+4 4 4+4 4 3+4 4 6+6 4 3+4 4 3+4 4 3+4 4 | | |
| | 6 2 3+3 4 3+3 4 3+4 4 5+5 4 3+4 4 3+4 4 3+4 4 | | |
| | 6 2 4+4 4 4+4 4 3+4 4 6+6 4 3+4 4 3+4 4 3+4 4 | | |
| | 5 2 3+3 4 3+3 4 3+4 4 5+5 4 3+4 4 3+4 4 3+4 4 | | |
| | 5 2 4+4 4 4+4 4 3+4 4 6+6 4 3+4 4 3+4 4 3+4 4 | | |
| | 6 2 3+3 4 3+3 4 3+4 4 5+5 4 3+4 4 3+4 4 3+4 4 | | |
| | 6 2 4+4 4 4+4 4 3+4 4 6+6 4 3+4 4 3+4 4 3+4 4 | | |
| | 5 2 3+3 4 3+3 4 3+4 4 6+6 4 3+4 4 3+4 4 3+4 4 | | |
| | 5 2 4+4 4 4+4 4 3+4 4 6+6 4 3+4 4 3+4 4 3+4 4 | | |
| | 6 2 3+3 4 3+3 4 3+4 4 5+5 4 3+4 4 3+4 4 3+4 4 | | |
| | 5 2 4+4 4 4+4 4 3+4 4 5+5 4 3+4 4 3+4 4 3+4 4 | | |
| | 5 2 4+4 4 3+3 4 3+4 4 5+5 4 3+4 4 3+4 4 3+4 4 | | |
| | 6 2 3+3 4 4+4 4 3+4 4 6+6 4 3+4 4 3+4 4 3+4 4 | | |
| | 5 2 4+4 4 3+3 4 3+4 4 5+5 4 3+4 4 3+4 4 3+4 4 | | |
| | 5 2 3+3 4 4+4 4 3+4 4 5+5 4 3+4 4 3+4 4 3+4 4 | | |
| | 6 2 3+3 4 4+4 4 3+4 4 5+5 4 3+4 4 3+4 4 3+4 4 | | |
| | 5 2 4+4 4 3+3 4 3+4 4 6+6 4 3+4 4 3+4 4 3+4 4 | | |
| | 6 2 3+3 4 3+3 4 3+4 4 6+6 4 3+4 4 3+4 4 3+4 4 | | |
| | 5 2 4+4 4 4+4 4 3+4 4 6+6 4 3+4 4 3+4 4 3+4 4 | | |
| | 5 2 4+4 4 3+3 4 3+4 4 6+6 4 3+4 4 3+4 4 3+4 4 | | |
| | 6 2 3+3 4 4+4 4 3+4 4 5+5 4 3+4 4 3+4 4 3+4 4 | | |
| | 6 2 3+3 4 4+4 4 3+4 4 5+5 4 3+4 4 3+4 4 3+4 4 | | |
| | 5 2 4+4 4 3+3 4 3+4 4 6+6 4 3+4 4 3+4 4 3+4 4 | | |
| | 6 2 4+4 4 3+3 4 3+4 4 5+5 4 3+4 4 3+4 4 3+4 4 | | |
| | 5 2 3+3 4 4+4 4 3+4 4 5+5 4 3+4 4 3+4 4 3+4 4 | | |
| | 5 2 3+3 4 4+4 4 3+4 4 5+5 4 3+4 4 3+4 4 3+4 4 | | |
| | 6 2 4+4 4 3+3 4 3+4 4 6+6 4 3+4 4 3+4 4 3+4 4 | | |
| | 6 2 3+3 4 3+3 4 3+4 4 6+6 4 3+4 4 3+4 4 3+4 4 | | |
| | 6 2 4+4 4 4+4 4 3+4 4 6+6 4 3+4 4 3+4 4 3+4 4 | | |
| | 6 2 4+4 4 3+3 4 3+4 4 5+5 4 3+4 4 3+4 4 3+4 4 | | |
| | 5 2 3+3 4 4+4 4 3+4 4 5+5 4 3+4 4 3+4 4 3+4 4 | | |
| | 6 2 3+3 4 4+4 4 3+4 4 5+5 4 3+4 4 3+4 4 3+4 4 | | |
| | 6 2 4+4 4 3+3 4 3+4 4 5+5 4 3+4 4 3+4 4 3+4 4 | | |
| | 5 2 3+3 4 4+4 4 3+4 4 6+6 4 3+4 4 3+4 4 3+4 4 | | |
| | 5 2 4+4 4 3+3 4 3+4 4 5+5 4 3+4 4 3+4 4 3+4 4 | | |
| | 5 2 4+4 4 4+4 4 3+4 4 6+6 4 3+4 4 3+4 4 3+4 4 | | |
| | 6 2 3+3 4 3+3 4 3+4 4 5+5 4 3+4 4 3+4 4 3+4 4 | | |
| | 6 2 4+4 4 3+3 4 3+4 4 5+5 4 3+4 4 3+4 4 3+4 4 | | |
| | 5 2 3+3 4 4+4 4 3+4 4 5+5 4 3+4 4 3+4 4 3+4 4 | | |
| | 5 2 3+3 4 3+3 4 3+4 4 6+6 4 3+4 4 3+4 4 3+4 4 | | |
| | 6 2 4+4 4 4+4 4 3+4 4 5+5 4 3+4 4 3+4 4 3+4 4 | | |
| | 5 2 3+3 4 4+4 4 3+4 4 5+5 4 3+4 4 3+4 4 3+4 4 | | |
| | 5 2 4+4 4 3+3 4 3+4 4 5+5 4 3+4 4 3+4 4 3+4 4 | | |
| | 5 2 | | |
| | 5 2 | | |
| | 5 2 | | |
| | 5 2 | | |
| | 6 2 | | |
| | 6 2 | | |
| | 5 2 | | |

Continued on next column

Continued on next column

**TABLE XXXVII– continued**

| TQC | ICSD | NoE | SF | | | | | |
|---|---|---|---|---|---|---|---|---|
| | 6 2 | | | | | | | |
| | 6 2 | | | | | | | |
| | 5 2 | | | | | | | |
| | 5 2 | | | | | | | |
| | 6 2 | | | | | | | |
| | 6 2 | | | | | | | |
| | 6 2 | | | | | | | |
| | 6 2 | | | | | | | |
| | 5 2 | | | | | | | |
| | 6 2 | | | | | | | |
| | 6 2 | | | | | | | |
| | 5 2 | | | | | | | |
| | 5 2 | | | | | | | |
| | 5 2 | | | | | | | |
| | 5 2 | | | | | | | |
| | 5 2 | | | | | | | |
| | 5 2 | | | | | | | |
| | 5 2 | | | | | | | |
| | 5 2 | | | | | | | |
| | 5 2 | | | | | | | |
| | 5 2 | | | | | | | |
| | 6 2 | | | | | | | |
| | 5 2 | | | | | | | |
| | 5 2 | | | | | | | |
| | 6 2 | | | | | | | |
| | 6 2 | | | | | | | |
| | 5 2 | | | | | | | |
| | 6 2 | | | | | | | |
| | 6 2 | | | | | | | |
| | 5 2 | | | | | | | |
| | 5 2 | | | | | | | |
| | 5 2 | | | | | | | |
| | 6 2 | | | | | | | |
| | 5 2 | | | | | | | |
| | 5 2 | | | | | | | |
| | 5 2 | | | | | | | |
| | 6 2 | | | | | | | |
| | 5 2 | | | | | | | |
| | 5 2 | | | | | | | |
| 62.2.171 | 192245 | 168 | Co1Y3 | | | | | |
| | -GM | -R | -S | -T | -U | -X | -Y | -Z |
| | 5 2 | 3+3 4 | 3+3 4 | 3+4 4 | 6+6 4 | 3+4 4 | 3+4 4 | 3+4 4 |
| | 5 2 | 4+4 4 | 4+4 4 | 3+4 4 | 5+5 4 | 3+4 4 | 3+4 4 | 3+4 4 |
| | 6 2 | 3+3 4 | 3+3 4 | 3+4 4 | 6+6 4 | 3+4 4 | 3+4 4 | 3+4 4 |
| | 6 2 | 4+4 4 | 4+4 4 | 3+4 4 | 5+5 4 | 3+4 4 | 3+4 4 | 3+4 4 |
| | 5 2 | 3+3 4 | 3+3 4 | 3+4 4 | 6+6 4 | 3+4 4 | 3+4 4 | 3+4 4 |
| | 6 2 | 4+4 4 | 4+4 4 | 3+4 4 | 5+5 4 | 3+4 4 | 3+4 4 | 3+4 4 |
| | 6 2 | 3+3 4 | 3+3 4 | 3+4 4 | 6+6 4 | 3+4 4 | 3+4 4 | 3+4 4 |
| | 5 2 | 4+4 4 | 4+4 4 | 3+4 4 | 6+6 4 | 3+4 4 | 3+4 4 | 3+4 4 |
| | 6 2 | 3+3 4 | 3+3 4 | 3+4 4 | 5+5 4 | 3+4 4 | 3+4 4 | 3+4 4 |
| | 5 2 | 4+4 4 | 4+4 4 | 3+4 4 | 5+5 4 | 3+4 4 | 3+4 4 | 3+4 4 |
| | 5 2 | 3+3 4 | 3+3 4 | 3+4 4 | 5+5 4 | 3+4 4 | 3+4 4 | 3+4 4 |
| | 6 2 | 4+4 4 | 4+4 4 | 3+4 4 | 6+6 4 | 3+4 4 | 3+4 4 | 3+4 4 |
| | 6 2 | 4+4 4 | 4+4 4 | 3+4 4 | 5+5 4 | 3+4 4 | 3+4 4 | 3+4 4 |
| | 5 2 | 3+3 4 | 3+3 4 | 3+4 4 | 6+6 4 | 3+4 4 | 3+4 4 | 3+4 4 |
| | 5 2 | 4+4 4 | 3+3 4 | 3+4 4 | 6+6 4 | 3+4 4 | 3+4 4 | 3+4 4 |
| | 6 2 | 3+3 4 | 4+4 4 | 3+4 4 | 5+5 4 | 3+4 4 | 3+4 4 | 3+4 4 |
| | 6 2 | 3+3 4 | 4+4 4 | 3+4 4 | 6+6 4 | 3+4 4 | 3+4 4 | 3+4 4 |
| | 5 2 | 4+4 4 | 3+3 4 | 3+4 4 | 5+5 4 | 3+4 4 | 3+4 4 | 3+4 4 |
| | 5 2 | 3+3 4 | 3+3 4 | 3+4 4 | 6+6 4 | 3+4 4 | 3+4 4 | 3+4 4 |
| | 5 2 | 4+4 4 | 4+4 4 | 3+4 4 | 6+6 4 | 3+4 4 | 3+4 4 | 3+4 4 |

Continued on next column

**TABLE XXXVII– continued**

| TQC | ICSD | NoE | SF | | | | | |
|---|---|---|---|---|---|---|---|---|
| | 6 2 | 3+3 4 | 4+4 4 | 3+4 4 | 5+5 4 | 3+4 4 | 3+4 4 | 3+4 4 |
| | 6 2 | 4+4 4 | 3+3 4 | 3+4 4 | 5+5 4 | 3+4 4 | 3+4 4 | 3+4 4 |
| | 5 2 | 3+3 4 | 3+3 4 | 3+4 4 | 5+5 4 | 3+4 4 | 3+4 4 | 3+4 4 |
| | 6 2 | 4+4 4 | 4+4 4 | 3+4 4 | 6+6 4 | 3+4 4 | 3+4 4 | 3+4 4 |
| | 5 2 | 3+3 4 | 3+3 4 | 3+4 4 | 6+6 4 | 3+4 4 | 3+4 4 | 3+4 4 |
| | 6 2 | 4+4 4 | 4+4 4 | 3+4 4 | 5+5 4 | 3+4 4 | 3+4 4 | 3+4 4 |
| | 5 2 | 4+4 4 | 4+4 4 | 3+4 4 | 5+5 4 | 3+4 4 | 3+4 4 | 3+4 4 |
| | 6 2 | 3+3 4 | 3+3 4 | 3+4 4 | 6+6 4 | 3+4 4 | 3+4 4 | 3+4 4 |
| | 5 2 | 4+4 4 | 3+3 4 | 3+4 4 | 6+6 4 | 3+4 4 | 3+4 4 | 3+4 4 |
| | 6 2 | 3+3 4 | 4+4 4 | 3+4 4 | 5+5 4 | 3+4 4 | 3+4 4 | 3+4 4 |
| | 5 2 | 3+3 4 | 4+4 4 | 3+4 4 | 5+5 4 | 3+4 4 | 3+4 4 | 3+4 4 |
| | 6 2 | 4+4 4 | 3+3 4 | 3+4 4 | 6+6 4 | 3+4 4 | 3+4 4 | 3+4 4 |
| | 5 2 | 3+3 4 | 4+4 4 | 3+4 4 | 5+5 4 | 3+4 4 | 3+4 4 | 3+4 4 |
| | 6 2 | 4+4 4 | 3+3 4 | 3+4 4 | 5+5 4 | 3+4 4 | 3+4 4 | 3+4 4 |
| | 5 2 | 3+3 4 | 4+4 4 | 3+4 4 | 6+6 4 | 3+4 4 | 3+4 4 | 3+4 4 |
| | 5 2 | 4+4 4 | 3+3 4 | 3+4 4 | 6+6 4 | 3+4 4 | 3+4 4 | 3+4 4 |
| | 6 2 | 3+3 4 | 4+4 4 | 3+4 4 | 6+6 4 | 3+4 4 | 3+4 4 | 3+4 4 |
| | 6 2 | 4+4 4 | 3+3 4 | 3+4 4 | 5+5 4 | 3+4 4 | 3+4 4 | 3+4 4 |
| | 6 2 | 4+4 4 | 3+3 4 | 3+4 4 | 6+6 4 | 3+4 4 | 3+4 4 | 3+4 4 |
| | 5 2 | 3+3 4 | 4+4 4 | 3+4 4 | 5+5 4 | 3+4 4 | 3+4 4 | 3+4 4 |
| | 5 2 | 3+3 4 | 4+4 4 | 3+4 4 | 5+5 4 | 3+4 4 | 3+4 4 | 3+4 4 |
| | 6 2 | 4+4 4 | 3+3 4 | 3+4 4 | 5+5 4 | 3+4 4 | 3+4 4 | 3+4 4 |
| | 5 2 | | | | | | | |
| | 5 2 | | | | | | | |
| | 5 2 | | | | | | | |
| | 6 2 | | | | | | | |
| | 6 2 | | | | | | | |
| | 6 2 | | | | | | | |
| | 5 2 | | | | | | | |
| | 6 2 | | | | | | | |
| | 5 2 | | | | | | | |
| | 6 2 | | | | | | | |
| | 5 2 | | | | | | | |
| | 6 2 | | | | | | | |
| | 6 2 | | | | | | | |
| | 6 2 | | | | | | | |
| | 6 2 | | | | | | | |
| | 5 2 | | | | | | | |
| | 5 2 | | | | | | | |
| | 6 2 | | | | | | | |
| | 6 2 | | | | | | | |
| | 5 2 | | | | | | | |
| | 5 2 | | | | | | | |
| | 5 2 | | | | | | | |
| | 5 2 | | | | | | | |
| | 5 2 | | | | | | | |
| | 6 2 | | | | | | | |
| | 6 2 | | | | | | | |
| | 5 2 | | | | | | | |
| | 6 2 | | | | | | | |
| | 6 2 | | | | | | | |
| | 5 2 | | | | | | | |
| | 5 2 | | | | | | | |
| | 6 2 | | | | | | | |
| | 5 2 | | | | | | | |
| | 5 2 | | | | | | | |
| | 6 2 | | | | | | | |
| | 5 2 | | | | | | | |
| | 5 2 | | | | | | | |
| | 5 2 | | | | | | | |
| | 6 2 | | | | | | | |

Continued on next column

**TABLE XXXVII– continued**

| TQC | ICSD | NoE | SF | | | | | |
|---|---|---|---|---|---|---|---|---|
| | 6 2 | | | | | | | |
| | 6 2 | | | | | | | |
| | 5 2 | | | | | | | |
| 62.2.172 | 192244 | 168 | Co1Y3 | | | | | |
| | -GM | -R | -S | -T | -U | -X | -Y | -Z |
| | 5 2 | 3+3 4 | 3+3 4 | 3+4 4 | 6+6 4 | 3+4 4 | 3+4 4 | 3+4 4 |
| | 5 2 | 4+4 4 | 4+4 4 | 3+4 4 | 5+5 4 | 3+4 4 | 3+4 4 | 3+4 4 |
| | 6 2 | 3+3 4 | 3+3 4 | 3+4 4 | 6+6 4 | 3+4 4 | 3+4 4 | 3+4 4 |
| | 6 2 | 4+4 4 | 4+4 4 | 3+4 4 | 5+5 4 | 3+4 4 | 3+4 4 | 3+4 4 |
| | 5 2 | 3+3 4 | 3+3 4 | 3+4 4 | 6+6 4 | 3+4 4 | 3+4 4 | 3+4 4 |
| | 6 2 | 4+4 4 | 4+4 4 | 3+4 4 | 5+5 4 | 3+4 4 | 3+4 4 | 3+4 4 |
| | 6 2 | 3+3 4 | 3+3 4 | 3+4 4 | 6+6 4 | 3+4 4 | 3+4 4 | 3+4 4 |
| | 5 2 | 4+4 4 | 4+4 4 | 3+4 4 | 6+6 4 | 3+4 4 | 3+4 4 | 3+4 4 |
| | 6 2 | 4+4 4 | 3+3 4 | 3+4 4 | 5+5 4 | 3+4 4 | 3+4 4 | 3+4 4 |
| | 5 2 | 3+3 4 | 4+4 4 | 3+4 4 | 5+5 4 | 3+4 4 | 3+4 4 | 3+4 4 |
| | 5 2 | 3+3 4 | 3+3 4 | 3+4 4 | 5+5 4 | 3+4 4 | 3+4 4 | 3+4 4 |
| | 6 2 | 4+4 4 | 4+4 4 | 3+4 4 | 6+6 4 | 3+4 4 | 3+4 4 | 3+4 4 |
| | 6 2 | 3+3 4 | 4+4 4 | 3+4 4 | 5+5 4 | 3+4 4 | 3+4 4 | 3+4 4 |
| | 5 2 | 4+4 4 | 3+3 4 | 3+4 4 | 6+6 4 | 3+4 4 | 3+4 4 | 3+4 4 |
| | 5 2 | 3+3 4 | 3+3 4 | 3+4 4 | 6+6 4 | 3+4 4 | 3+4 4 | 3+4 4 |
| | 6 2 | 4+4 4 | 4+4 4 | 3+4 4 | 5+5 4 | 3+4 4 | 3+4 4 | 3+4 4 |
| | 6 2 | 4+4 4 | 4+4 4 | 3+4 4 | 6+6 4 | 3+4 4 | 3+4 4 | 3+4 4 |
| | 5 2 | 3+3 4 | 3+3 4 | 3+4 4 | 5+5 4 | 3+4 4 | 3+4 4 | 3+4 4 |
| | 5 2 | 4+4 4 | 3+3 4 | 3+4 4 | 6+6 4 | 3+4 4 | 3+4 4 | 3+4 4 |
| | 5 2 | 3+3 4 | 4+4 4 | 3+4 4 | 6+6 4 | 3+4 4 | 3+4 4 | 3+4 4 |
| | 6 2 | 4+4 4 | 4+4 4 | 3+4 4 | 5+5 4 | 3+4 4 | 3+4 4 | 3+4 4 |
| | 6 2 | 3+3 4 | 3+3 4 | 3+4 4 | 5+5 4 | 3+4 4 | 3+4 4 | 3+4 4 |
| | 5 2 | 4+4 4 | 3+3 4 | 3+4 4 | 5+5 4 | 3+4 4 | 3+4 4 | 3+4 4 |
| | 6 2 | 3+3 4 | 4+4 4 | 3+4 4 | 6+6 4 | 3+4 4 | 3+4 4 | 3+4 4 |
| | 5 2 | 3+3 4 | 3+3 4 | 3+4 4 | 6+6 4 | 3+4 4 | 3+4 4 | 3+4 4 |
| | 6 2 | 4+4 4 | 4+4 4 | 3+4 4 | 5+5 4 | 3+4 4 | 3+4 4 | 3+4 4 |
| | 5 2 | 3+3 4 | 4+4 4 | 3+4 4 | 5+5 4 | 3+4 4 | 3+4 4 | 3+4 4 |
| | 6 2 | 4+4 4 | 3+3 4 | 3+4 4 | 6+6 4 | 3+4 4 | 3+4 4 | 3+4 4 |
| | 5 2 | 3+3 4 | 3+3 4 | 3+4 4 | 6+6 4 | 3+4 4 | 3+4 4 | 3+4 4 |
| | 6 2 | 4+4 4 | 4+4 4 | 3+4 4 | 5+5 4 | 3+4 4 | 3+4 4 | 3+4 4 |
| | 5 2 | 4+4 4 | 4+4 4 | 3+4 4 | 5+5 4 | 3+4 4 | 3+4 4 | 3+4 4 |
| | 6 2 | 3+3 4 | 3+3 4 | 3+4 4 | 6+6 4 | 3+4 4 | 3+4 4 | 3+4 4 |
| | 5 2 | 4+4 4 | 4+4 4 | 3+4 4 | 5+5 4 | 3+4 4 | 3+4 4 | 3+4 4 |
| | 6 2 | 3+3 4 | 3+3 4 | 3+4 4 | 5+5 4 | 3+4 4 | 3+4 4 | 3+4 4 |
| | 5 2 | 4+4 4 | 4+4 4 | 3+4 4 | 6+6 4 | 3+4 4 | 3+4 4 | 3+4 4 |
| | 5 2 | 3+3 4 | 3+3 4 | 3+4 4 | 6+6 4 | 3+4 4 | 3+4 4 | 3+4 4 |
| | 6 2 | 4+4 4 | 4+4 4 | 3+4 4 | 6+6 4 | 3+4 4 | 3+4 4 | 3+4 4 |
| | 6 2 | 3+3 4 | 3+3 4 | 3+4 4 | 5+5 4 | 3+4 4 | 3+4 4 | 3+4 4 |
| | 6 2 | 3+3 4 | 3+3 4 | 3+4 4 | 6+6 4 | 3+4 4 | 3+4 4 | 3+4 4 |
| | 5 2 | 4+4 4 | 4+4 4 | 3+4 4 | 5+5 4 | 3+4 4 | 3+4 4 | 3+4 4 |
| | 5 2 | 4+4 4 | 4+4 4 | 3+4 4 | 5+5 4 | 3+4 4 | 3+4 4 | 3+4 4 |
| | 6 2 | 3+3 4 | 3+3 4 | 3+4 4 | 5+5 4 | 3+4 4 | 3+4 4 | 3+4 4 |
| | 5 2 | | | | | | | |
| | 5 2 | | | | | | | |
| | 5 2 | | | | | | | |
| | 6 2 | | | | | | | |
| | 6 2 | | | | | | | |
| | 6 2 | | | | | | | |
| | 5 2 | | | | | | | |
| | 6 2 | | | | | | | |
| | 5 2 | | | | | | | |
| | 6 2 | | | | | | | |
| | 5 2 | | | | | | | |
| | 6 2 | | | | | | | |
| | 6 2 | | | | | | | |
| | 6 2 | | | | | | | |

Continued on next column

**TABLE XXXVII– continued**

| TQC | ICSD | NoE | SF | | | | | |
|---|---|---|---|---|---|---|---|---|
| | 6 2 | | | | | | | |
| | 5 2 | | | | | | | |
| | 5 2 | | | | | | | |
| | 6 2 | | | | | | | |
| | 6 2 | | | | | | | |
| | 5 2 | | | | | | | |
| | 5 2 | | | | | | | |
| | 5 2 | | | | | | | |
| | 5 2 | | | | | | | |
| | 5 2 | | | | | | | |
| | 6 2 | | | | | | | |
| | 6 2 | | | | | | | |
| | 5 2 | | | | | | | |
| | 6 2 | | | | | | | |
| | 6 2 | | | | | | | |
| | 5 2 | | | | | | | |
| | 5 2 | | | | | | | |
| | 6 2 | | | | | | | |
| | 5 2 | | | | | | | |
| | 5 2 | | | | | | | |
| | 6 2 | | | | | | | |
| | 5 2 | | | | | | | |
| | 5 2 | | | | | | | |
| | 5 2 | | | | | | | |
| | 6 2 | | | | | | | |
| | 6 2 | | | | | | | |
| | 6 2 | | | | | | | |
| | 5 2 | | | | | | | |
| 62.2.173 | 409847 | 72 | Co1Si1V1 | | | | | |
| | -GM | -R | -S | -T | -U | -X | -Y | -Z |
| | 5 2 | 3+3 4 | 3+3 4 | 3+4 4 | 5+5 4 | 3+4 4 | 3+4 4 | 3+4 4 |
| | 6 2 | 4+4 4 | 4+4 4 | 3+4 4 | 6+6 4 | 3+4 4 | 3+4 4 | 3+4 4 |
| | 5 2 | 3+3 4 | 4+4 4 | 3+4 4 | 5+5 4 | 3+4 4 | 3+4 4 | 3+4 4 |
| | 6 2 | 4+4 4 | 3+3 4 | 3+4 4 | 6+6 4 | 3+4 4 | 3+4 4 | 3+4 4 |
| | 5 2 | 3+3 4 | 3+3 4 | 3+4 4 | 6+6 4 | 3+4 4 | 3+4 4 | 3+4 4 |
| | 6 2 | 4+4 4 | 4+4 4 | 3+4 4 | 6+6 4 | 3+4 4 | 3+4 4 | 3+4 4 |
| | 5 2 | 4+4 4 | 4+4 4 | 3+4 4 | 5+5 4 | 3+4 4 | 3+4 4 | 3+4 4 |
| | 6 2 | 3+3 4 | 3+3 4 | 3+4 4 | 5+5 4 | 3+4 4 | 3+4 4 | 3+4 4 |
| | 5 2 | 4+4 4 | 3+3 4 | 3+4 4 | 5+5 4 | 3+4 4 | 3+4 4 | 3+4 4 |
| | 5 2 | 3+3 4 | 4+4 4 | 3+4 4 | 6+6 4 | 3+4 4 | 3+4 4 | 3+4 4 |
| | 6 2 | 3+3 4 | 4+4 4 | 3+4 4 | 5+5 4 | 3+4 4 | 3+4 4 | 3+4 4 |
| | 6 2 | 4+4 4 | 3+3 4 | 3+4 4 | 6+6 4 | 3+4 4 | 3+4 4 | 3+4 4 |
| | 6 2 | 4+4 4 | 4+4 4 | 3+4 4 | 5+5 4 | 3+4 4 | 3+4 4 | 3+4 4 |
| | 6 2 | 3+3 4 | 3+3 4 | 3+4 4 | 6+6 4 | 3+4 4 | 3+4 4 | 3+4 4 |
| | 6 2 | 4+4 4 | 3+3 4 | 3+4 4 | 5+5 4 | 3+4 4 | 3+4 4 | 3+4 4 |
| | 5 2 | 3+3 4 | 4+4 4 | 3+4 4 | 5+5 4 | 3+4 4 | 3+4 4 | 3+4 4 |
| | 5 2 | 4+4 4 | 3+3 4 | 3+4 4 | 6+6 4 | 3+4 4 | 3+4 4 | 3+4 4 |
| | 6 2 | 3+3 4 | 4+4 4 | 3+4 4 | 6+6 4 | 3+4 4 | 3+4 4 | 3+4 4 |
| | 6 2 | | | | | | | |
| | 5 2 | | | | | | | |
| | 5 2 | | | | | | | |
| | 5 2 | | | | | | | |
| | 5 2 | | | | | | | |
| | 5 2 | | | | | | | |
| | 6 2 | | | | | | | |
| | 5 2 | | | | | | | |
| | 6 2 | | | | | | | |
| | 6 2 | | | | | | | |
| | 5 2 | | | | | | | |
| | 5 2 | | | | | | | |
| | 6 2 | | | | | | | |

Continued on next column

**TABLE XXXVII– continued**

| TQC | ICSD | NoE | SF | | | | |
|---|---|---|---|---|---|---|---|
| | 5 2 | | | | | | |
| | 5 2 | | | | | | |
| | 5 2 | | | | | | |
| | 6 2 | | | | | | |
| | 5 2 | | | | | | |
| 62.2.174 | 49726 | 72 | Co1P1Zr1 | | | | |
| | -GM | -R | -S | -T | -U | -X | -Y | -Z |
| | 5 2 | 4+4 4 | 4+4 4 | 3+4 4 | 5+5 4 | 3+4 4 | 3+4 4 | 3+4 4 |
| | 6 2 | 3+3 4 | 3+3 4 | 3+4 4 | 6+6 4 | 3+4 4 | 3+4 4 | 3+4 4 |
| | 5 2 | 3+3 4 | 3+3 4 | 3+4 4 | 5+5 4 | 3+4 4 | 3+4 4 | 3+4 4 |
| | 6 2 | 4+4 4 | 4+4 4 | 3+4 4 | 6+6 4 | 3+4 4 | 3+4 4 | 3+4 4 |
| | 5 2 | 4+4 4 | 3+3 4 | 3+4 4 | 6+6 4 | 3+4 4 | 3+4 4 | 3+4 4 |
| | 6 2 | 3+3 4 | 4+4 4 | 3+4 4 | 6+6 4 | 3+4 4 | 3+4 4 | 3+4 4 |
| | 5 2 | 3+3 4 | 3+3 4 | 3+4 4 | 5+5 4 | 3+4 4 | 3+4 4 | 3+4 4 |
| | 6 2 | 4+4 4 | 4+4 4 | 3+4 4 | 5+5 4 | 3+4 4 | 3+4 4 | 3+4 4 |
| | 5 2 | 3+3 4 | 4+4 4 | 3+4 4 | 5+5 4 | 3+4 4 | 3+4 4 | 3+4 4 |
| | 5 2 | 4+4 4 | 3+3 4 | 3+4 4 | 6+6 4 | 3+4 4 | 3+4 4 | 3+4 4 |
| | 6 2 | 4+4 4 | 4+4 4 | 3+4 4 | 5+5 4 | 3+4 4 | 3+4 4 | 3+4 4 |
| | 6 2 | 3+3 4 | 3+3 4 | 3+4 4 | 6+6 4 | 3+4 4 | 3+4 4 | 3+4 4 |
| | 6 2 | 3+3 4 | 3+3 4 | 3+4 4 | 5+5 4 | 3+4 4 | 3+4 4 | 3+4 4 |
| | 6 2 | 4+4 4 | 4+4 4 | 3+4 4 | 6+6 4 | 3+4 4 | 3+4 4 | 3+4 4 |
| | 6 2 | 3+3 4 | 4+4 4 | 3+4 4 | 5+5 4 | 3+4 4 | 3+4 4 | 3+4 4 |
| | 5 2 | 4+4 4 | 3+3 4 | 3+4 4 | 5+5 4 | 3+4 4 | 3+4 4 | 3+4 4 |
| | 5 2 | 3+3 4 | 4+4 4 | 3+4 4 | 6+6 4 | 3+4 4 | 3+4 4 | 3+4 4 |
| | 6 2 | 4+4 4 | 3+3 4 | 3+4 4 | 6+6 4 | 3+4 4 | 3+4 4 | 3+4 4 |
| | 5 2 | | | | | | | |
| | 5 2 | | | | | | | |
| | 6 2 | | | | | | | |
| | 5 2 | | | | | | | |
| | 5 2 | | | | | | | |
| | 5 2 | | | | | | | |
| | 6 2 | | | | | | | |
| | 6 2 | | | | | | | |
| | 5 2 | | | | | | | |
| | 5 2 | | | | | | | |
| | 5 2 | | | | | | | |
| | 6 2 | | | | | | | |
| | 5 2 | | | | | | | |
| | 6 2 | | | | | | | |
| | 5 2 | | | | | | | |
| | 5 2 | | | | | | | |
| | 6 2 | | | | | | | |
| | 6 2 | | | | | | | |
| 62.2.175 | 43401 | 56 | Co1P1 | | | | |
| | -GM | -R | -S | -T | -U | -X | -Y | -Z |
| | 5 2 | 3+3 4 | 4+4 4 | 3+4 4 | 6+6 4 | 3+4 4 | 3+4 4 | 3+4 4 |
| | 5 2 | 4+4 4 | 3+3 4 | 3+4 4 | 5+5 4 | 3+4 4 | 3+4 4 | 3+4 4 |
| | 6 2 | 4+4 4 | 3+3 4 | 3+4 4 | 6+6 4 | 3+4 4 | 3+4 4 | 3+4 4 |
| | 6 2 | 3+3 4 | 4+4 4 | 3+4 4 | 5+5 4 | 3+4 4 | 3+4 4 | 3+4 4 |
| | 6 2 | 4+4 4 | 3+3 4 | 3+4 4 | 6+6 4 | 3+4 4 | 3+4 4 | 3+4 4 |
| | 5 2 | 3+3 4 | 4+4 4 | 3+4 4 | 6+6 4 | 3+4 4 | 3+4 4 | 3+4 4 |
| | 5 2 | 3+3 4 | 4+4 4 | 3+4 4 | 5+5 4 | 3+4 4 | 3+4 4 | 3+4 4 |
| | 5 2 | 4+4 4 | 3+3 4 | 3+4 4 | 5+5 4 | 3+4 4 | 3+4 4 | 3+4 4 |
| | 6 2 | 3+3 4 | 3+3 4 | 3+4 4 | 6+6 4 | 3+4 4 | 3+4 4 | 3+4 4 |
| | 5 2 | 4+4 4 | 4+4 4 | 3+4 4 | 5+5 4 | 3+4 4 | 3+4 4 | 3+4 4 |
| | 5 2 | 4+4 4 | 3+3 4 | 3+4 4 | 5+5 4 | 3+4 4 | 3+4 4 | 3+4 4 |
| | 6 2 | 3+3 4 | 4+4 4 | 3+4 4 | 6+6 4 | 3+4 4 | 3+4 4 | 3+4 4 |
| | 6 2 | 3+3 4 | 4+4 4 | 3+4 4 | 5+5 4 | 3+4 4 | 3+4 4 | 3+4 4 |
| | 5 2 | 4+4 4 | 3+3 4 | 3+4 4 | 6+6 4 | 3+4 4 | 3+4 4 | 3+4 4 |
| | 5 2 | | | | | | | |
| | 6 2 | | | | | | | |

Continued on next column

| TQC | ICSD | NoE | SF | | | | |
|---|---|---|---|---|---|---|---|
| | 5 2 | | | | | | |
| | 6 2 | | | | | | |
| | 5 2 | | | | | | |
| | 5 2 | | | | | | |
| | 5 2 | | | | | | |
| | 6 2 | | | | | | |
| | 6 2 | | | | | | |
| | 6 2 | | | | | | |
| | 6 2 | | | | | | |
| | 5 2 | | | | | | |
| | 5 2 | | | | | | |
| | 5 2 | | | | | | |
| 62.2.176 | 43249 | 56 | Co1P1 | | | | |
| | -GM | -R | -S | -T | -U | -X | -Y | -Z |
| | 5 2 | 3+3 4 | 4+4 4 | 3+4 4 | 6+6 4 | 3+4 4 | 3+4 4 | 3+4 4 |
| | 5 2 | 4+4 4 | 3+3 4 | 3+4 4 | 5+5 4 | 3+4 4 | 3+4 4 | 3+4 4 |
| | 6 2 | 4+4 4 | 3+3 4 | 3+4 4 | 6+6 4 | 3+4 4 | 3+4 4 | 3+4 4 |
| | 6 2 | 3+3 4 | 4+4 4 | 3+4 4 | 5+5 4 | 3+4 4 | 3+4 4 | 3+4 4 |
| | 5 2 | 4+4 4 | 3+3 4 | 3+4 4 | 6+6 4 | 3+4 4 | 3+4 4 | 3+4 4 |
| | 6 2 | 3+3 4 | 4+4 4 | 3+4 4 | 6+6 4 | 3+4 4 | 3+4 4 | 3+4 4 |
| | 5 2 | 4+4 4 | 4+4 4 | 3+4 4 | 5+5 4 | 3+4 4 | 3+4 4 | 3+4 4 |
| | 6 2 | 3+3 4 | 3+3 4 | 3+4 4 | 5+5 4 | 3+4 4 | 3+4 4 | 3+4 4 |
| | 5 2 | 3+3 4 | 3+3 4 | 3+4 4 | 6+6 4 | 3+4 4 | 3+4 4 | 3+4 4 |
| | 5 2 | 4+4 4 | 4+4 4 | 3+4 4 | 5+5 4 | 3+4 4 | 3+4 4 | 3+4 4 |
| | 6 2 | 4+4 4 | 3+3 4 | 3+4 4 | 5+5 4 | 3+4 4 | 3+4 4 | 3+4 4 |
| | 5 2 | 3+3 4 | 4+4 4 | 3+4 4 | 6+6 4 | 3+4 4 | 3+4 4 | 3+4 4 |
| | 6 2 | 3+3 4 | 4+4 4 | 3+4 4 | 5+5 4 | 3+4 4 | 3+4 4 | 3+4 4 |
| | 5 2 | 4+4 4 | 3+3 4 | 3+4 4 | 6+6 4 | 3+4 4 | 3+4 4 | 3+4 4 |
| | 5 2 | | | | | | | |
| | 5 2 | | | | | | | |
| | 6 2 | | | | | | | |
| | 6 2 | | | | | | | |
| | 5 2 | | | | | | | |
| | 5 2 | | | | | | | |
| | 5 2 | | | | | | | |
| | 6 2 | | | | | | | |
| | 6 2 | | | | | | | |
| | 6 2 | | | | | | | |
| | 6 2 | | | | | | | |
| | 5 2 | | | | | | | |
| | 5 2 | | | | | | | |
| | 5 2 | | | | | | | |
| 62.2.177 | 172046 | 184 | Co1O3Tb1 | | | | |
| | -GM | -R | -S | -T | -U | -X | -Y | -Z |
| | 5 2 | 3+3 4 | 3+3 4 | 3+4 4 | 5+5 4 | 3+4 4 | 3+4 4 | 3+4 4 |
| | 6 2 | 4+4 4 | 4+4 4 | 3+4 4 | 6+6 4 | 3+4 4 | 3+4 4 | 3+4 4 |
| | 5 2 | 4+4 4 | 4+4 4 | 3+4 4 | 5+5 4 | 3+4 4 | 3+4 4 | 3+4 4 |
| | 6 2 | 3+3 4 | 3+3 4 | 3+4 4 | 6+6 4 | 3+4 4 | 3+4 4 | 3+4 4 |
| | 5 2 | 4+4 4 | 4+4 4 | 3+4 4 | 6+6 4 | 3+4 4 | 3+4 4 | 3+4 4 |
| | 6 2 | 3+3 4 | 3+3 4 | 3+4 4 | 5+5 4 | 3+4 4 | 3+4 4 | 3+4 4 |
| | 5 2 | 3+3 4 | 3+3 4 | 3+4 4 | 5+5 4 | 3+4 4 | 3+4 4 | 3+4 4 |
| | 6 2 | 4+4 4 | 4+4 4 | 3+4 4 | 6+6 4 | 3+4 4 | 3+4 4 | 3+4 4 |
| | 6 2 | 4+4 4 | 3+3 4 | 3+4 4 | 5+5 4 | 3+4 4 | 3+4 4 | 3+4 4 |
| | 5 2 | 3+3 4 | 4+4 4 | 3+4 4 | 6+6 4 | 3+4 4 | 3+4 4 | 3+4 4 |
| | 5 2 | 3+3 4 | 4+4 4 | 3+4 4 | 6+6 4 | 3+4 4 | 3+4 4 | 3+4 4 |
| | 5 2 | 4+4 4 | 3+3 4 | 3+4 4 | 5+5 4 | 3+4 4 | 3+4 4 | 3+4 4 |
| | 6 2 | 4+4 4 | 4+4 4 | 3+4 4 | 6+6 4 | 3+4 4 | 3+4 4 | 3+4 4 |
| | 5 2 | 3+3 4 | 3+3 4 | 3+4 4 | 5+5 4 | 3+4 4 | 3+4 4 | 3+4 4 |
| | 6 2 | 3+3 4 | 4+4 4 | 3+4 4 | 6+6 4 | 3+4 4 | 3+4 4 | 3+4 4 |
| | 6 2 | 4+4 4 | 3+3 4 | 3+4 4 | 6+6 4 | 3+4 4 | 3+4 4 | 3+4 4 |
| | 5 2 | 3+3 4 | 3+3 4 | 3+4 4 | 6+6 4 | 3+4 4 | 3+4 4 | 3+4 4 |

Continued on next column

**TABLE XXXVII– continued**

| TQC | ICSD | NoE | SF |
|---|---|---|---|
| | 6 2 4+4 4 4+4 4 3+4 4 6+6 4 3+4 4 3+4 4 3+4 4 | | |
| | 5 2 4+4 4 3+3 4 3+4 4 5+5 4 3+4 4 3+4 4 3+4 4 | | |
| | 6 2 3+3 4 4+4 4 3+4 4 5+5 4 3+4 4 3+4 4 3+4 4 | | |
| | 6 2 4+4 4 3+3 4 3+4 4 5+5 4 3+4 4 3+4 4 3+4 4 | | |
| | 5 2 3+3 4 4+4 4 3+4 4 6+6 4 3+4 4 3+4 4 3+4 4 | | |
| | 6 2 4+4 4 4+4 4 3+4 4 6+6 4 3+4 4 3+4 4 3+4 4 | | |
| | 5 2 3+3 4 3+3 4 3+4 4 5+5 4 3+4 4 3+4 4 3+4 4 | | |
| | 6 2 3+3 4 4+4 4 3+4 4 5+5 4 3+4 4 3+4 4 3+4 4 | | |
| | 5 2 4+4 4 3+3 4 3+4 4 6+6 4 3+4 4 3+4 4 3+4 4 | | |
| | 5 2 3+3 4 3+3 4 3+4 4 6+6 4 3+4 4 3+4 4 3+4 4 | | |
| | 6 2 4+4 4 4+4 4 3+4 4 5+5 4 3+4 4 3+4 4 3+4 4 | | |
| | 5 2 3+3 4 3+3 4 3+4 4 5+5 4 3+4 4 3+4 4 3+4 4 | | |
| | 5 2 4+4 4 4+4 4 3+4 4 5+5 4 3+4 4 3+4 4 3+4 4 | | |
| | 5 2 4+4 4 3+3 4 3+4 4 5+5 4 3+4 4 3+4 4 3+4 4 | | |
| | 5 2 3+3 4 4+4 4 3+4 4 6+6 4 3+4 4 3+4 4 3+4 4 | | |
| | 5 2 3+3 4 4+4 4 3+4 4 6+6 4 3+4 4 3+4 4 3+4 4 | | |
| | 6 2 4+4 4 3+3 4 3+4 4 6+6 4 3+4 4 3+4 4 3+4 4 | | |
| | 5 2 4+4 4 4+4 4 3+4 4 6+6 4 3+4 4 3+4 4 3+4 4 | | |
| | 5 2 3+3 4 3+3 4 3+4 4 6+6 4 3+4 4 3+4 4 3+4 4 | | |
| | 5 2 4+4 4 4+4 4 3+4 4 5+5 4 3+4 4 3+4 4 3+4 4 | | |
| | 6 2 3+3 4 3+3 4 3+4 4 6+6 4 3+4 4 3+4 4 3+4 4 | | |
| | 6 2 3+3 4 3+3 4 3+4 4 5+5 4 3+4 4 3+4 4 3+4 4 | | |
| | 5 2 4+4 4 4+4 4 3+4 4 5+5 4 3+4 4 3+4 4 3+4 4 | | |
| | 6 2 4+4 4 3+3 4 3+4 4 6+6 4 3+4 4 3+4 4 3+4 4 | | |
| | 5 2 3+3 4 4+4 4 3+4 4 6+6 4 3+4 4 3+4 4 3+4 4 | | |
| | 6 2 3+3 4 3+3 4 3+4 4 6+6 4 3+4 4 3+4 4 3+4 4 | | |
| | 5 2 4+4 4 4+4 4 3+4 4 6+6 4 3+4 4 3+4 4 3+4 4 | | |
| | 5 2 4+4 4 4+4 4 3+4 4 6+6 4 3+4 4 3+4 4 3+4 4 | | |
| | 5 2 3+3 4 3+3 4 3+4 4 5+5 4 3+4 4 3+4 4 3+4 4 | | |
| | 6 2 | | |
| | 6 2 | | |
| | 5 2 | | |
| | 6 2 | | |
| | 6 2 | | |
| | 6 2 | | |
| | 6 2 | | |
| | 5 2 | | |
| | 5 2 | | |
| | 6 2 | | |
| | 5 2 | | |
| | 6 2 | | |
| | 6 2 | | |
| | 5 2 | | |
| | 6 2 | | |
| | 6 2 | | |
| | 6 2 | | |
| | 6 2 | | |
| | 5 2 | | |
| | 5 2 | | |
| | 5 2 | | |
| | 5 2 | | |
| | 5 2 | | |
| | 5 2 | | |
| | 5 2 | | |
| | 5 2 | | |
| | 5 2 | | |
| | 6 2 | | |
| | 5 2 | | |
| | 5 2 | | |
| | 6 2 | | |
| | 5 2 | | |

Continued on next column

**TABLE XXXVII– continued**

| TQC | ICSD | NoE | SF |
|---|---|---|---|
| | 6 2 | | |
| | 5 2 | | |
| | 6 2 | | |
| | 5 2 | | |
| | 6 2 | | |
| | 6 2 | | |
| | 5 2 | | |
| | 5 2 | | |
| | 5 2 | | |
| | 5 2 | | |
| | 5 2 | | |
| | 5 2 | | |
| | 6 2 | | |
| | 5 2 | | |
| 62.2.178 | 88738 | 160 | Co1O3Pr1 |
| | -GM -R -S -T -U -X -Y -Z | | |
| | 5 2 3+3 4 3+3 4 3+4 4 6+6 4 3+4 4 3+4 4 3+4 4 | | |
| | 6 2 4+4 4 4+4 4 3+4 4 5+5 4 3+4 4 3+4 4 3+4 4 | | |
| | 5 2 3+3 4 4+4 4 3+4 4 5+5 4 3+4 4 3+4 4 3+4 4 | | |
| | 6 2 4+4 4 3+3 4 3+4 4 6+6 4 3+4 4 3+4 4 3+4 4 | | |
| | 5 2 4+4 4 3+3 4 3+4 4 6+6 4 3+4 4 3+4 4 3+4 4 | | |
| | 6 2 3+3 4 4+4 4 3+4 4 5+5 4 3+4 4 3+4 4 3+4 4 | | |
| | 5 2 3+3 4 4+4 4 3+4 4 5+5 4 3+4 4 3+4 4 3+4 4 | | |
| | 6 2 4+4 4 3+3 4 3+4 4 6+6 4 3+4 4 3+4 4 3+4 4 | | |
| | 5 2 3+3 4 4+4 4 3+4 4 5+5 4 3+4 4 3+4 4 3+4 4 | | |
| | 6 2 4+4 4 3+3 4 3+4 4 6+6 4 3+4 4 3+4 4 3+4 4 | | |
| | 5 2 3+3 4 3+3 4 3+4 4 5+5 4 3+4 4 3+4 4 3+4 4 | | |
| | 5 2 4+4 4 4+4 4 3+4 4 6+6 4 3+4 4 3+4 4 3+4 4 | | |
| | 6 2 4+4 4 4+4 4 3+4 4 6+6 4 3+4 4 3+4 4 3+4 4 | | |
| | 5 2 3+3 4 3+3 4 3+4 4 5+5 4 3+4 4 3+4 4 3+4 4 | | |
| | 6 2 4+4 4 3+3 4 3+4 4 5+5 4 3+4 4 3+4 4 3+4 4 | | |
| | 6 2 3+3 4 4+4 4 3+4 4 5+5 4 3+4 4 3+4 4 3+4 4 | | |
| | 5 2 3+3 4 3+3 4 3+4 4 5+5 4 3+4 4 3+4 4 3+4 4 | | |
| | 6 2 4+4 4 4+4 4 3+4 4 5+5 4 3+4 4 3+4 4 3+4 4 | | |
| | 6 2 4+4 4 3+3 4 3+4 4 6+6 4 3+4 4 3+4 4 3+4 4 | | |
| | 5 2 3+3 4 4+4 4 3+4 4 6+6 4 3+4 4 3+4 4 3+4 4 | | |
| | 5 2 4+4 4 4+4 4 3+4 4 5+5 4 3+4 4 3+4 4 3+4 4 | | |
| | 6 2 3+3 4 3+3 4 3+4 4 6+6 4 3+4 4 3+4 4 3+4 4 | | |
| | 5 2 4+4 4 4+4 4 3+4 4 5+5 4 3+4 4 3+4 4 3+4 4 | | |
| | 6 2 3+3 4 3+3 4 3+4 4 6+6 4 3+4 4 3+4 4 3+4 4 | | |
| | 5 2 3+3 4 3+3 4 3+4 4 5+5 4 3+4 4 3+4 4 3+4 4 | | |
| | 6 2 4+4 4 4+4 4 3+4 4 6+6 4 3+4 4 3+4 4 3+4 4 | | |
| | 5 2 3+3 4 4+4 4 3+4 4 5+5 4 3+4 4 3+4 4 3+4 4 | | |
| | 6 2 4+4 4 3+3 4 3+4 4 6+6 4 3+4 4 3+4 4 3+4 4 | | |
| | 5 2 4+4 4 4+4 4 3+4 4 6+6 4 3+4 4 3+4 4 3+4 4 | | |
| | 5 2 3+3 4 3+3 4 3+4 4 6+6 4 3+4 4 3+4 4 3+4 4 | | |
| | 5 2 4+4 4 3+3 4 3+4 4 6+6 4 3+4 4 3+4 4 3+4 4 | | |
| | 5 2 3+3 4 4+4 4 3+4 4 5+5 4 3+4 4 3+4 4 3+4 4 | | |
| | 5 2 3+3 4 3+3 4 3+4 4 5+5 4 3+4 4 3+4 4 3+4 4 | | |
| | 5 2 4+4 4 4+4 4 3+4 4 5+5 4 3+4 4 3+4 4 3+4 4 | | |
| | 6 2 3+3 4 3+3 4 3+4 4 5+5 4 3+4 4 3+4 4 3+4 4 | | |
| | 5 2 4+4 4 4+4 4 3+4 4 5+5 4 3+4 4 3+4 4 3+4 4 | | |
| | 6 2 4+4 4 4+4 4 3+4 4 5+5 4 3+4 4 3+4 4 3+4 4 | | |
| | 6 2 3+3 4 3+3 4 3+4 4 5+5 4 3+4 4 3+4 4 3+4 4 | | |
| | 5 2 4+4 4 3+3 4 3+4 4 5+5 4 3+4 4 3+4 4 3+4 4 | | |
| | 6 2 3+3 4 4+4 4 3+4 4 6+6 4 3+4 4 3+4 4 3+4 4 | | |
| | 6 2 | | |
| | 5 2 | | |
| | 5 2 | | |
| | 5 2 | | |
| | 5 2 | | |

Continued on next column

**TABLE XXXVII– continued**

| TQC | ICSD | NoE | SF | | | | | |
|---|---|---|---|---|---|---|---|---|
| | 6 2 | | | | | | | |
| | 6 2 | | | | | | | |
| | 6 2 | | | | | | | |
| | 5 2 | | | | | | | |
| | 5 2 | | | | | | | |
| | 6 2 | | | | | | | |
| | 6 2 | | | | | | | |
| | 6 2 | | | | | | | |
| | 6 2 | | | | | | | |
| | 6 2 | | | | | | | |
| | 5 2 | | | | | | | |
| | 5 2 | | | | | | | |
| | 6 2 | | | | | | | |
| | 5 2 | | | | | | | |
| | 6 2 | | | | | | | |
| | 5 2 | | | | | | | |
| | 6 2 | | | | | | | |
| | 6 2 | | | | | | | |
| | 6 2 | | | | | | | |
| | 5 2 | | | | | | | |
| | 5 2 | | | | | | | |
| | 5 2 | | | | | | | |
| | 5 2 | | | | | | | |
| | 5 2 | | | | | | | |
| | 5 2 | | | | | | | |
| | 5 2 | | | | | | | |
| | 5 2 | | | | | | | |
| | 5 2 | | | | | | | |
| | 5 2 | | | | | | | |
| | 5 2 | | | | | | | |
| | 5 2 | | | | | | | |
| | 5 2 | | | | | | | |
| | 6 2 | | | | | | | |
| | 5 2 | | | | | | | |
| | 5 2 | | | | | | | |
| 62.2.179 | 94437 | 192 | Co1O2Rb3 | | | | | |
| | -GM | -R | -S | -T | -U | -X | -Y | -Z |
| | 5 2 | 3+3 4 | 3+3 4 | 3+4 4 | 6+6 4 | 3+4 4 | 3+4 4 | 3+4 4 |
| | 5 2 | 4+4 4 | 4+4 4 | 3+4 4 | 5+5 4 | 3+4 4 | 3+4 4 | 3+4 4 |
| | 6 2 | 3+3 4 | 3+3 4 | 3+4 4 | 6+6 4 | 3+4 4 | 3+4 4 | 3+4 4 |
| | 6 2 | 4+4 4 | 4+4 4 | 3+4 4 | 5+5 4 | 3+4 4 | 3+4 4 | 3+4 4 |
| | 5 2 | 3+3 4 | 3+3 4 | 3+4 4 | 6+6 4 | 3+4 4 | 3+4 4 | 3+4 4 |
| | 6 2 | 4+4 4 | 4+4 4 | 3+4 4 | 5+5 4 | 3+4 4 | 3+4 4 | 3+4 4 |
| | 5 2 | 3+3 4 | 3+3 4 | 3+4 4 | 6+6 4 | 3+4 4 | 3+4 4 | 3+4 4 |
| | 6 2 | 4+4 4 | 4+4 4 | 3+4 4 | 5+5 4 | 3+4 4 | 3+4 4 | 3+4 4 |
| | 5 2 | 3+3 4 | 3+3 4 | 3+4 4 | 5+5 4 | 3+4 4 | 3+4 4 | 3+4 4 |
| | 6 2 | 4+4 4 | 4+4 4 | 3+4 4 | 6+6 4 | 3+4 4 | 3+4 4 | 3+4 4 |
| | 5 2 | 3+3 4 | 3+3 4 | 3+4 4 | 5+5 4 | 3+4 4 | 3+4 4 | 3+4 4 |
| | 6 2 | 4+4 4 | 4+4 4 | 3+4 4 | 6+6 4 | 3+4 4 | 3+4 4 | 3+4 4 |
| | 6 2 | 4+4 4 | 4+4 4 | 3+4 4 | 5+5 4 | 3+4 4 | 3+4 4 | 3+4 4 |
| | 5 2 | 3+3 4 | 3+3 4 | 3+4 4 | 6+6 4 | 3+4 4 | 3+4 4 | 3+4 4 |
| | 6 2 | 4+4 4 | 3+3 4 | 3+4 4 | 6+6 4 | 3+4 4 | 3+4 4 | 3+4 4 |
| | 5 2 | 3+3 4 | 4+4 4 | 3+4 4 | 5+5 4 | 3+4 4 | 3+4 4 | 3+4 4 |
| | 5 2 | 4+4 4 | 3+3 4 | 3+4 4 | 6+6 4 | 3+4 4 | 3+4 4 | 3+4 4 |
| | 5 2 | 3+3 4 | 4+4 4 | 3+4 4 | 5+5 4 | 3+4 4 | 3+4 4 | 3+4 4 |
| | 6 2 | 3+3 4 | 4+4 4 | 3+4 4 | 5+5 4 | 3+4 4 | 3+4 4 | 3+4 4 |
| | 6 2 | 4+4 4 | 3+3 4 | 3+4 4 | 6+6 4 | 3+4 4 | 3+4 4 | 3+4 4 |
| | 6 2 | 3+3 4 | 4+4 4 | 3+4 4 | 6+6 4 | 3+4 4 | 3+4 4 | 3+4 4 |
| | 5 2 | 4+4 4 | 3+3 4 | 3+4 4 | 5+5 4 | 3+4 4 | 3+4 4 | 3+4 4 |
| | 6 2 | 3+3 4 | 4+4 4 | 3+4 4 | 6+6 4 | 3+4 4 | 3+4 4 | 3+4 4 |
| | 5 2 | 4+4 4 | 3+3 4 | 3+4 4 | 5+5 4 | 3+4 4 | 3+4 4 | 3+4 4 |
| | 5 2 | 3+3 4 | 4+4 4 | 3+4 4 | 6+6 4 | 3+4 4 | 3+4 4 | 3+4 4 |
| | 5 2 | 4+4 4 | 3+3 4 | 3+4 4 | 5+5 4 | 3+4 4 | 3+4 4 | 3+4 4 |
| | 6 2 | 4+4 4 | 3+3 4 | 3+4 4 | 6+6 4 | 3+4 4 | 3+4 4 | 3+4 4 |
| | 5 2 | 3+3 4 | 4+4 4 | 3+4 4 | 5+5 4 | 3+4 4 | 3+4 4 | 3+4 4 |
| | 6 2 | 3+3 4 | 3+3 4 | 3+4 4 | 5+5 4 | 3+4 4 | 3+4 4 | 3+4 4 |
| | 5 2 | 4+4 4 | 4+4 4 | 3+4 4 | 6+6 4 | 3+4 4 | 3+4 4 | 3+4 4 |
| | 6 2 | 3+3 4 | 3+3 4 | 3+4 4 | 6+6 4 | 3+4 4 | 3+4 4 | 3+4 4 |
| | 6 2 | 4+4 4 | 4+4 4 | 3+4 4 | 6+6 4 | 3+4 4 | 3+4 4 | 3+4 4 |
| | 5 2 | 4+4 4 | 3+3 4 | 3+4 4 | 5+5 4 | 3+4 4 | 3+4 4 | 3+4 4 |
| | 6 2 | 3+3 4 | 4+4 4 | 3+4 4 | 5+5 4 | 3+4 4 | 3+4 4 | 3+4 4 |
| | 5 2 | 3+3 4 | 4+4 4 | 3+4 4 | 5+5 4 | 3+4 4 | 3+4 4 | 3+4 4 |
| | 6 2 | 4+4 4 | 3+3 4 | 3+4 4 | 6+6 4 | 3+4 4 | 3+4 4 | 3+4 4 |
| | 6 2 | 3+3 4 | 4+4 4 | 3+4 4 | 5+5 4 | 3+4 4 | 3+4 4 | 3+4 4 |
| | 6 2 | 4+4 4 | 3+3 4 | 3+4 4 | 5+5 4 | 3+4 4 | 3+4 4 | 3+4 4 |
| | 5 2 | 4+4 4 | 3+3 4 | 3+4 4 | 6+6 4 | 3+4 4 | 3+4 4 | 3+4 4 |
| | 5 2 | 3+3 4 | 4+4 4 | 3+4 4 | 6+6 4 | 3+4 4 | 3+4 4 | 3+4 4 |
| | 5 2 | 3+3 4 | 3+3 4 | 3+4 4 | 5+5 4 | 3+4 4 | 3+4 4 | 3+4 4 |
| | 5 2 | 4+4 4 | 4+4 4 | 3+4 4 | 6+6 4 | 3+4 4 | 3+4 4 | 3+4 4 |
| | 6 2 | 3+3 4 | 3+3 4 | 3+4 4 | 5+5 4 | 3+4 4 | 3+4 4 | 3+4 4 |
| | 5 2 | 4+4 4 | 4+4 4 | 3+4 4 | 6+6 4 | 3+4 4 | 3+4 4 | 3+4 4 |
| | 6 2 | 3+3 4 | 3+3 4 | 3+4 4 | 5+5 4 | 3+4 4 | 3+4 4 | 3+4 4 |
| | 6 2 | 4+4 4 | 4+4 4 | 3+4 4 | 6+6 4 | 3+4 4 | 3+4 4 | 3+4 4 |
| | 5 2 | 3+3 4 | 4+4 4 | 3+4 4 | 6+6 4 | 3+4 4 | 3+4 4 | 3+4 4 |
| | 6 2 | 4+4 4 | 3+3 4 | 3+4 4 | 6+6 4 | 3+4 4 | 3+4 4 | 3+4 4 |
| | 5 2 | | | | | | | |
| | 5 2 | | | | | | | |
| | 6 2 | | | | | | | |
| | 6 2 | | | | | | | |
| | 5 2 | | | | | | | |
| | 6 2 | | | | | | | |
| | 6 2 | | | | | | | |
| | 5 2 | | | | | | | |
| | 5 2 | | | | | | | |
| | 5 2 | | | | | | | |
| | 6 2 | | | | | | | |
| | 6 2 | | | | | | | |
| | 6 2 | | | | | | | |
| | 6 2 | | | | | | | |
| | 5 2 | | | | | | | |
| | 6 2 | | | | | | | |
| | 6 2 | | | | | | | |
| | 5 2 | | | | | | | |
| | 5 2 | | | | | | | |
| | 5 2 | | | | | | | |
| | 5 2 | | | | | | | |
| | 5 2 | | | | | | | |
| | 6 2 | | | | | | | |
| | 5 2 | | | | | | | |
| | 6 2 | | | | | | | |
| | 5 2 | | | | | | | |
| | 5 2 | | | | | | | |
| | 6 2 | | | | | | | |
| | 6 2 | | | | | | | |
| | 5 2 | | | | | | | |
| | 6 2 | | | | | | | |
| | 6 2 | | | | | | | |
| | 5 2 | | | | | | | |
| | 5 2 | | | | | | | |
| | 6 2 | | | | | | | |
| | 6 2 | | | | | | | |
| | 5 2 | | | | | | | |

Continued on next column

**TABLE XXXVII– continued**

| TQC | ICSD | NoE | SF | | | | | |
|---|---|---|---|---|---|---|---|---|
| | 5 2 | | | | | | | |
| | 6 2 | | | | | | | |
| | 6 2 | | | | | | | |
| | 5 2 | | | | | | | |
| | 5 2 | | | | | | | |
| | 6 2 | | | | | | | |
| | 6 2 | | | | | | | |
| | 6 2 | | | | | | | |
| | 6 2 | | | | | | | |
| | 5 2 | | | | | | | |
| | 6 2 | | | | | | | |
| 62.2.180 | 2421 | 80 | Co1Mo1P1 | | | | | |
| | -GM | -R | -S | -T | -U | -X | -Y | -Z |
| | 5 2 | 3+3 4 | 4+4 4 | 3+4 4 | 5+5 4 | 3+4 4 | 3+4 4 | 3+4 4 |
| | 6 2 | 4+4 4 | 3+3 4 | 3+4 4 | 6+6 4 | 3+4 4 | 3+4 4 | 3+4 4 |
| | 5 2 | 4+4 4 | 3+3 4 | 3+4 4 | 5+5 4 | 3+4 4 | 3+4 4 | 3+4 4 |
| | 6 2 | 3+3 4 | 4+4 4 | 3+4 4 | 6+6 4 | 3+4 4 | 3+4 4 | 3+4 4 |
| | 5 2 | 4+4 4 | 3+3 4 | 3+4 4 | 6+6 4 | 3+4 4 | 3+4 4 | 3+4 4 |
| | 6 2 | 3+3 4 | 4+4 4 | 3+4 4 | 5+5 4 | 3+4 4 | 3+4 4 | 3+4 4 |
| | 6 2 | 3+3 4 | 3+3 4 | 3+4 4 | 6+6 4 | 3+4 4 | 3+4 4 | 3+4 4 |
| | 5 2 | 4+4 4 | 4+4 4 | 3+4 4 | 5+5 4 | 3+4 4 | 3+4 4 | 3+4 4 |
| | 5 2 | 3+3 4 | 4+4 4 | 3+4 4 | 5+5 4 | 3+4 4 | 3+4 4 | 3+4 4 |
| | 5 2 | 4+4 4 | 3+3 4 | 3+4 4 | 5+5 4 | 3+4 4 | 3+4 4 | 3+4 4 |
| | 6 2 | 3+3 4 | 3+3 4 | 3+4 4 | 6+6 4 | 3+4 4 | 3+4 4 | 3+4 4 |
| | 6 2 | 4+4 4 | 4+4 4 | 3+4 4 | 6+6 4 | 3+4 4 | 3+4 4 | 3+4 4 |
| | 6 2 | 3+3 4 | 3+3 4 | 3+4 4 | 6+6 4 | 3+4 4 | 3+4 4 | 3+4 4 |
| | 6 2 | 4+4 4 | 4+4 4 | 3+4 4 | 5+5 4 | 3+4 4 | 3+4 4 | 3+4 4 |
| | 6 2 | 3+3 4 | 4+4 4 | 3+4 4 | 5+5 4 | 3+4 4 | 3+4 4 | 3+4 4 |
| | 5 2 | 4+4 4 | 3+3 4 | 3+4 4 | 6+6 4 | 3+4 4 | 3+4 4 | 3+4 4 |
| | 5 2 | 4+4 4 | 4+4 4 | 3+4 4 | 5+5 4 | 3+4 4 | 3+4 4 | 3+4 4 |
| | 5 2 | 3+3 4 | 3+3 4 | 3+4 4 | 6+6 4 | 3+4 4 | 3+4 4 | 3+4 4 |
| | 6 2 | 4+4 4 | 3+3 4 | 3+4 4 | 5+5 4 | 3+4 4 | 3+4 4 | 3+4 4 |
| | 5 2 | 3+3 4 | 4+4 4 | 3+4 4 | 6+6 4 | 3+4 4 | 3+4 4 | 3+4 4 |
| | 6 2 | | | | | | | |
| | 5 2 | | | | | | | |
| | 6 2 | | | | | | | |
| | 5 2 | | | | | | | |
| | 5 2 | | | | | | | |
| | 5 2 | | | | | | | |
| | 5 2 | | | | | | | |
| | 6 2 | | | | | | | |
| | 6 2 | | | | | | | |
| | 5 2 | | | | | | | |
| | 5 2 | | | | | | | |
| | 6 2 | | | | | | | |
| | 5 2 | | | | | | | |
| | 6 2 | | | | | | | |
| | 5 2 | | | | | | | |
| | 5 2 | | | | | | | |
| | 5 2 | | | | | | | |
| | 5 2 | | | | | | | |
| | 6 2 | | | | | | | |
| | 6 2 | | | | | | | |
| 62.2.181 | 150531 | 168 | Co1La3 | | | | | |
| | -GM | -R | -S | -T | -U | -X | -Y | -Z |
| | 5 2 | 3+3 4 | 3+3 4 | 3+4 4 | 5+5 4 | 3+4 4 | 3+4 4 | 3+4 4 |
| | 5 2 | 4+4 4 | 4+4 4 | 3+4 4 | 6+6 4 | 3+4 4 | 3+4 4 | 3+4 4 |
| | 6 2 | 3+3 4 | 3+3 4 | 3+4 4 | 6+6 4 | 3+4 4 | 3+4 4 | 3+4 4 |
| | 5 2 | 4+4 4 | 4+4 4 | 3+4 4 | 6+6 4 | 3+4 4 | 3+4 4 | 3+4 4 |
| | 6 2 | 3+3 4 | 3+3 4 | 3+4 4 | 5+5 4 | 3+4 4 | 3+4 4 | 3+4 4 |
| | 6 2 | 4+4 4 | 4+4 4 | 3+4 4 | 5+5 4 | 3+4 4 | 3+4 4 | 3+4 4 |

Continued on next column

**TABLE XXXVII– continued**

| TQC | ICSD | NoE | SF | | | | | |
|---|---|---|---|---|---|---|---|---|
| | 6 2 | 4+4 4 | 4+4 4 | 3+4 4 | 6+6 4 | 3+4 4 | 3+4 4 | 3+4 4 |
| | 5 2 | 3+3 4 | 3+3 4 | 3+4 4 | 6+6 4 | 3+4 4 | 3+4 4 | 3+4 4 |
| | 5 2 | 3+3 4 | 3+3 4 | 3+4 4 | 5+5 4 | 3+4 4 | 3+4 4 | 3+4 4 |
| | 6 2 | 4+4 4 | 4+4 4 | 3+4 4 | 5+5 4 | 3+4 4 | 3+4 4 | 3+4 4 |
| | 6 2 | 3+3 4 | 3+3 4 | 3+4 4 | 5+5 4 | 3+4 4 | 3+4 4 | 3+4 4 |
| | 5 2 | 4+4 4 | 4+4 4 | 3+4 4 | 6+6 4 | 3+4 4 | 3+4 4 | 3+4 4 |
| | 5 2 | 3+3 4 | 4+4 4 | 3+4 4 | 6+6 4 | 3+4 4 | 3+4 4 | 3+4 4 |
| | 6 2 | 4+4 4 | 3+3 4 | 3+4 4 | 5+5 4 | 3+4 4 | 3+4 4 | 3+4 4 |
| | 5 2 | 4+4 4 | 3+3 4 | 3+4 4 | 6+6 4 | 3+4 4 | 3+4 4 | 3+4 4 |
| | 6 2 | 3+3 4 | 4+4 4 | 3+4 4 | 5+5 4 | 3+4 4 | 3+4 4 | 3+4 4 |
| | 6 2 | 3+3 4 | 4+4 4 | 3+4 4 | 6+6 4 | 3+4 4 | 3+4 4 | 3+4 4 |
| | 5 2 | 4+4 4 | 3+3 4 | 3+4 4 | 5+5 4 | 3+4 4 | 3+4 4 | 3+4 4 |
| | 5 2 | 4+4 4 | 4+4 4 | 3+4 4 | 6+6 4 | 3+4 4 | 3+4 4 | 3+4 4 |
| | 5 2 | 3+3 4 | 3+3 4 | 3+4 4 | 6+6 4 | 3+4 4 | 3+4 4 | 3+4 4 |
| | 6 2 | 4+4 4 | 3+3 4 | 3+4 4 | 5+5 4 | 3+4 4 | 3+4 4 | 3+4 4 |
| | 5 2 | 3+3 4 | 4+4 4 | 3+4 4 | 5+5 4 | 3+4 4 | 3+4 4 | 3+4 4 |
| | 6 2 | 3+3 4 | 4+4 4 | 3+4 4 | 6+6 4 | 3+4 4 | 3+4 4 | 3+4 4 |
| | 6 2 | 4+4 4 | 3+3 4 | 3+4 4 | 5+5 4 | 3+4 4 | 3+4 4 | 3+4 4 |
| | 5 2 | 4+4 4 | 3+3 4 | 3+4 4 | 5+5 4 | 3+4 4 | 3+4 4 | 3+4 4 |
| | 6 2 | 3+3 4 | 4+4 4 | 3+4 4 | 6+6 4 | 3+4 4 | 3+4 4 | 3+4 4 |
| | 5 2 | 4+4 4 | 4+4 4 | 3+4 4 | 6+6 4 | 3+4 4 | 3+4 4 | 3+4 4 |
| | 5 2 | 3+3 4 | 3+3 4 | 3+4 4 | 5+5 4 | 3+4 4 | 3+4 4 | 3+4 4 |
| | 6 2 | 3+3 4 | 3+3 4 | 3+4 4 | 6+6 4 | 3+4 4 | 3+4 4 | 3+4 4 |
| | 6 2 | 4+4 4 | 4+4 4 | 3+4 4 | 5+5 4 | 3+4 4 | 3+4 4 | 3+4 4 |
| | 5 2 | 4+4 4 | 3+3 4 | 3+4 4 | 5+5 4 | 3+4 4 | 3+4 4 | 3+4 4 |
| | 6 2 | 3+3 4 | 4+4 4 | 3+4 4 | 6+6 4 | 3+4 4 | 3+4 4 | 3+4 4 |
| | 5 2 | 3+3 4 | 4+4 4 | 3+4 4 | 5+5 4 | 3+4 4 | 3+4 4 | 3+4 4 |
| | 6 2 | 4+4 4 | 3+3 4 | 3+4 4 | 5+5 4 | 3+4 4 | 3+4 4 | 3+4 4 |
| | 5 2 | 4+4 4 | 4+4 4 | 3+4 4 | 6+6 4 | 3+4 4 | 3+4 4 | 3+4 4 |
| | 5 2 | 3+3 4 | 3+3 4 | 3+4 4 | 6+6 4 | 3+4 4 | 3+4 4 | 3+4 4 |
| | 6 2 | 4+4 4 | 4+4 4 | 3+4 4 | 5+5 4 | 3+4 4 | 3+4 4 | 3+4 4 |
| | 6 2 | 3+3 4 | 3+3 4 | 3+4 4 | 6+6 4 | 3+4 4 | 3+4 4 | 3+4 4 |
| | 6 2 | 3+3 4 | 3+3 4 | 3+4 4 | 5+5 4 | 3+4 4 | 3+4 4 | 3+4 4 |
| | 5 2 | 4+4 4 | 4+4 4 | 3+4 4 | 5+5 4 | 3+4 4 | 3+4 4 | 3+4 4 |
| | 5 2 | 3+3 4 | 3+3 4 | 3+4 4 | 6+6 4 | 3+4 4 | 3+4 4 | 3+4 4 |
| | 5 2 | 4+4 4 | 4+4 4 | 3+4 4 | 5+5 4 | 3+4 4 | 3+4 4 | 3+4 4 |
| | 6 2 | | | | | | | |
| | 5 2 | | | | | | | |
| | 5 2 | | | | | | | |
| | 6 2 | | | | | | | |
| | 6 2 | | | | | | | |
| | 6 2 | | | | | | | |
| | 5 2 | | | | | | | |
| | 6 2 | | | | | | | |
| | 6 2 | | | | | | | |
| | 5 2 | | | | | | | |
| | 6 2 | | | | | | | |
| | 5 2 | | | | | | | |
| | 6 2 | | | | | | | |
| | 6 2 | | | | | | | |
| | 5 2 | | | | | | | |
| | 5 2 | | | | | | | |
| | 5 2 | | | | | | | |
| | 6 2 | | | | | | | |
| | 6 2 | | | | | | | |
| | 5 2 | | | | | | | |
| | 6 2 | | | | | | | |
| | 5 2 | | | | | | | |
| | 5 2 | | | | | | | |
| | 6 2 | | | | | | | |
| | 5 2 | | | | | | | |

Continued on next column

**TABLE XXXVII– continued**

| TQC | ICSD | NoE | SF | | | | | |
|---|---|---|---|---|---|---|---|---|
| | 6 2 | | | | | | | |
| | 6 2 | | | | | | | |
| | 5 2 | | | | | | | |
| | 6 2 | | | | | | | |
| | 5 2 | | | | | | | |
| | 6 2 | | | | | | | |
| | 5 2 | | | | | | | |
| | 5 2 | | | | | | | |
| | 5 2 | | | | | | | |
| | 5 2 | | | | | | | |
| | 5 2 | | | | | | | |
| | 6 2 | | | | | | | |
| | 5 2 | | | | | | | |
| | 6 2 | | | | | | | |
| | 6 2 | | | | | | | |
| | 6 2 | | | | | | | |
| | 6 2 | | | | | | | |
| 62.2.182 | 73212 | 192 | Co1K3O2 | | | | | |
| | -GM | -R | -S | -T | -U | -X | -Y | -Z |
| | 5 2 | 3+3 4 | 3+3 4 | 3+4 4 | 6+6 4 | 3+4 4 | 3+4 4 | 3+4 4 |
| | 5 2 | 4+4 4 | 4+4 4 | 3+4 4 | 5+5 4 | 3+4 4 | 3+4 4 | 3+4 4 |
| | 6 2 | 3+3 4 | 3+3 4 | 3+4 4 | 6+6 4 | 3+4 4 | 3+4 4 | 3+4 4 |
| | 6 2 | 4+4 4 | 4+4 4 | 3+4 4 | 5+5 4 | 3+4 4 | 3+4 4 | 3+4 4 |
| | 5 2 | 3+3 4 | 3+3 4 | 3+4 4 | 6+6 4 | 3+4 4 | 3+4 4 | 3+4 4 |
| | 6 2 | 4+4 4 | 4+4 4 | 3+4 4 | 5+5 4 | 3+4 4 | 3+4 4 | 3+4 4 |
| | 5 2 | 3+3 4 | 3+3 4 | 3+4 4 | 6+6 4 | 3+4 4 | 3+4 4 | 3+4 4 |
| | 6 2 | 4+4 4 | 4+4 4 | 3+4 4 | 5+5 4 | 3+4 4 | 3+4 4 | 3+4 4 |
| | 5 2 | 3+3 4 | 3+3 4 | 3+4 4 | 5+5 4 | 3+4 4 | 3+4 4 | 3+4 4 |
| | 6 2 | 4+4 4 | 4+4 4 | 3+4 4 | 6+6 4 | 3+4 4 | 3+4 4 | 3+4 4 |
| | 5 2 | 3+3 4 | 3+3 4 | 3+4 4 | 5+5 4 | 3+4 4 | 3+4 4 | 3+4 4 |
| | 6 2 | 4+4 4 | 4+4 4 | 3+4 4 | 6+6 4 | 3+4 4 | 3+4 4 | 3+4 4 |
| | 6 2 | 4+4 4 | 3+3 4 | 3+4 4 | 5+5 4 | 3+4 4 | 3+4 4 | 3+4 4 |
| | 5 2 | 3+3 4 | 4+4 4 | 3+4 4 | 6+6 4 | 3+4 4 | 3+4 4 | 3+4 4 |
| | 6 2 | 4+4 4 | 3+3 4 | 3+4 4 | 6+6 4 | 3+4 4 | 3+4 4 | 3+4 4 |
| | 5 2 | 3+3 4 | 4+4 4 | 3+4 4 | 5+5 4 | 3+4 4 | 3+4 4 | 3+4 4 |
| | 5 2 | 4+4 4 | 3+3 4 | 3+4 4 | 6+6 4 | 3+4 4 | 3+4 4 | 3+4 4 |
| | 5 2 | 3+3 4 | 4+4 4 | 3+4 4 | 5+5 4 | 3+4 4 | 3+4 4 | 3+4 4 |
| | 6 2 | 4+4 4 | 4+4 4 | 3+4 4 | 5+5 4 | 3+4 4 | 3+4 4 | 3+4 4 |
| | 6 2 | 3+3 4 | 3+3 4 | 3+4 4 | 6+6 4 | 3+4 4 | 3+4 4 | 3+4 4 |
| | 6 2 | 3+3 4 | 4+4 4 | 3+4 4 | 6+6 4 | 3+4 4 | 3+4 4 | 3+4 4 |
| | 5 2 | 4+4 4 | 3+3 4 | 3+4 4 | 5+5 4 | 3+4 4 | 3+4 4 | 3+4 4 |
| | 6 2 | 3+3 4 | 3+3 4 | 3+4 4 | 6+6 4 | 3+4 4 | 3+4 4 | 3+4 4 |
| | 5 2 | 4+4 4 | 4+4 4 | 3+4 4 | 5+5 4 | 3+4 4 | 3+4 4 | 3+4 4 |
| | 6 2 | 3+3 4 | 4+4 4 | 3+4 4 | 6+6 4 | 3+4 4 | 3+4 4 | 3+4 4 |
| | 5 2 | 4+4 4 | 3+3 4 | 3+4 4 | 5+5 4 | 3+4 4 | 3+4 4 | 3+4 4 |
| | 6 2 | 3+3 4 | 4+4 4 | 3+4 4 | 5+5 4 | 3+4 4 | 3+4 4 | 3+4 4 |
| | 5 2 | 4+4 4 | 3+3 4 | 3+4 4 | 6+6 4 | 3+4 4 | 3+4 4 | 3+4 4 |
| | 5 2 | 3+3 4 | 3+3 4 | 3+4 4 | 5+5 4 | 3+4 4 | 3+4 4 | 3+4 4 |
| | 6 2 | 4+4 4 | 4+4 4 | 3+4 4 | 6+6 4 | 3+4 4 | 3+4 4 | 3+4 4 |
| | 5 2 | 3+3 4 | 3+3 4 | 3+4 4 | 6+6 4 | 3+4 4 | 3+4 4 | 3+4 4 |
| | 6 2 | 4+4 4 | 4+4 4 | 3+4 4 | 6+6 4 | 3+4 4 | 3+4 4 | 3+4 4 |
| | 5 2 | 4+4 4 | 4+4 4 | 3+4 4 | 5+5 4 | 3+4 4 | 3+4 4 | 3+4 4 |
| | 5 2 | 3+3 4 | 3+3 4 | 3+4 4 | 5+5 4 | 3+4 4 | 3+4 4 | 3+4 4 |
| | 6 2 | 3+3 4 | 4+4 4 | 3+4 4 | 5+5 4 | 3+4 4 | 3+4 4 | 3+4 4 |
| | 6 2 | 4+4 4 | 3+3 4 | 3+4 4 | 6+6 4 | 3+4 4 | 3+4 4 | 3+4 4 |
| | 6 2 | 3+3 4 | 4+4 4 | 3+4 4 | 5+5 4 | 3+4 4 | 3+4 4 | 3+4 4 |
| | 5 2 | 4+4 4 | 3+3 4 | 3+4 4 | 5+5 4 | 3+4 4 | 3+4 4 | 3+4 4 |
| | 5 2 | 4+4 4 | 3+3 4 | 3+4 4 | 6+6 4 | 3+4 4 | 3+4 4 | 3+4 4 |
| | 6 2 | 3+3 4 | 4+4 4 | 3+4 4 | 6+6 4 | 3+4 4 | 3+4 4 | 3+4 4 |
| | 5 2 | 3+3 4 | 3+3 4 | 3+4 4 | 5+5 4 | 3+4 4 | 3+4 4 | 3+4 4 |
| | 5 2 | 4+4 4 | 4+4 4 | 3+4 4 | 6+6 4 | 3+4 4 | 3+4 4 | 3+4 4 |

Continued on next column

**TABLE XXXVII– continued**

| TQC | ICSD | NoE | SF | | | | | |
|---|---|---|---|---|---|---|---|---|
| | 6 2 | 3+3 4 | 3+3 4 | 3+4 4 | 5+5 4 | 3+4 4 | 3+4 4 | 3+4 4 |
| | 5 2 | 4+4 4 | 4+4 4 | 3+4 4 | 6+6 4 | 3+4 4 | 3+4 4 | 3+4 4 |
| | 6 2 | 4+4 4 | 4+4 4 | 3+4 4 | 5+5 4 | 3+4 4 | 3+4 4 | 3+4 4 |
| | 5 2 | 3+3 4 | 3+3 4 | 3+4 4 | 6+6 4 | 3+4 4 | 3+4 4 | 3+4 4 |
| | 6 2 | 3+3 4 | 4+4 4 | 3+4 4 | 6+6 4 | 3+4 4 | 3+4 4 | 3+4 4 |
| | 5 2 | 4+4 4 | 3+3 4 | 3+4 4 | 6+6 4 | 3+4 4 | 3+4 4 | 3+4 4 |
| | 6 2 | | | | | | | |
| | 6 2 | | | | | | | |
| | 6 2 | | | | | | | |
| | 5 2 | | | | | | | |
| | 6 2 | | | | | | | |
| | 5 2 | | | | | | | |
| | 6 2 | | | | | | | |
| | 5 2 | | | | | | | |
| | 5 2 | | | | | | | |
| | 5 2 | | | | | | | |
| | 6 2 | | | | | | | |
| | 6 2 | | | | | | | |
| | 6 2 | | | | | | | |
| | 5 2 | | | | | | | |
| | 6 2 | | | | | | | |
| | 5 2 | | | | | | | |
| | 6 2 | | | | | | | |
| | 5 2 | | | | | | | |
| | 6 2 | | | | | | | |
| | 5 2 | | | | | | | |
| | 5 2 | | | | | | | |
| | 5 2 | | | | | | | |
| | 6 2 | | | | | | | |
| | 6 2 | | | | | | | |
| | 5 2 | | | | | | | |
| | 5 2 | | | | | | | |
| | 5 2 | | | | | | | |
| | 6 2 | | | | | | | |
| | 6 2 | | | | | | | |
| | 5 2 | | | | | | | |
| | 6 2 | | | | | | | |
| | 6 2 | | | | | | | |
| | 5 2 | | | | | | | |
| | 5 2 | | | | | | | |
| | 6 2 | | | | | | | |
| | 6 2 | | | | | | | |
| | 6 2 | | | | | | | |
| | 5 2 | | | | | | | |
| | 5 2 | | | | | | | |
| | 6 2 | | | | | | | |
| | 5 2 | | | | | | | |
| | 5 2 | | | | | | | |
| | 6 2 | | | | | | | |
| | 6 2 | | | | | | | |
| | 6 2 | | | | | | | |
| | 6 2 | | | | | | | |
| | 5 2 | | | | | | | |
| | 5 2 | | | | | | | |
| 62.2.183 | 172048 | 192 | Co1Ho1O3 | | | | | |
| | -GM | -R | -S | -T | -U | -X | -Y | -Z |
| | 5 2 | 3+3 4 | 3+3 4 | 3+4 4 | 5+5 4 | 3+4 4 | 3+4 4 | 3+4 4 |
| | 6 2 | 4+4 4 | 4+4 4 | 3+4 4 | 6+6 4 | 3+4 4 | 3+4 4 | 3+4 4 |
| | 5 2 | 3+3 4 | 4+4 4 | 3+4 4 | 5+5 4 | 3+4 4 | 3+4 4 | 3+4 4 |
| | 6 2 | 4+4 4 | 3+3 4 | 3+4 4 | 6+6 4 | 3+4 4 | 3+4 4 | 3+4 4 |
| | 5 2 | 3+3 4 | 4+4 4 | 3+4 4 | 6+6 4 | 3+4 4 | 3+4 4 | 3+4 4 |

Continued on next column

**TABLE XXXVII– continued**

| TQC | ICSD | NoE | SF |
|---|---|---|---|
| | 6 2 4+4 4 3+3 4 3+4 4 5+5 4 3+4 4 3+4 4 3+4 4 | | |
| | 5 2 4+4 4 3+3 4 3+4 4 5+5 4 3+4 4 3+4 4 3+4 4 | | |
| | 6 2 3+3 4 4+4 4 3+4 4 6+6 4 3+4 4 3+4 4 3+4 4 | | |
| | 6 2 3+3 4 3+3 4 3+4 4 5+5 4 3+4 4 3+4 4 3+4 4 | | |
| | 5 2 4+4 4 4+4 4 3+4 4 6+6 4 3+4 4 3+4 4 3+4 4 | | |
| | 5 2 4+4 4 3+3 4 3+4 4 6+6 4 3+4 4 3+4 4 3+4 4 | | |
| | 5 2 3+3 4 4+4 4 3+4 4 5+5 4 3+4 4 3+4 4 3+4 4 | | |
| | 6 2 3+3 4 4+4 4 3+4 4 6+6 4 3+4 4 3+4 4 3+4 4 | | |
| | 6 2 4+4 4 3+3 4 3+4 4 5+5 4 3+4 4 3+4 4 3+4 4 | | |
| | 6 2 3+3 4 4+4 4 3+4 4 6+6 4 3+4 4 3+4 4 3+4 4 | | |
| | 5 2 4+4 4 3+3 4 3+4 4 6+6 4 3+4 4 3+4 4 3+4 4 | | |
| | 5 2 4+4 4 3+3 4 3+4 4 6+6 4 3+4 4 3+4 4 3+4 4 | | |
| | 5 2 3+3 4 4+4 4 3+4 4 6+6 4 3+4 4 3+4 4 3+4 4 | | |
| | 6 2 3+3 4 3+3 4 3+4 4 5+5 4 3+4 4 3+4 4 3+4 4 | | |
| | 6 2 4+4 4 4+4 4 3+4 4 5+5 4 3+4 4 3+4 4 3+4 4 | | |
| | 6 2 3+3 4 3+3 4 3+4 4 5+5 4 3+4 4 3+4 4 3+4 4 | | |
| | 5 2 4+4 4 4+4 4 3+4 4 6+6 4 3+4 4 3+4 4 3+4 4 | | |
| | 6 2 3+3 4 4+4 4 3+4 4 6+6 4 3+4 4 3+4 4 3+4 4 | | |
| | 5 2 4+4 4 3+3 4 3+4 4 5+5 4 3+4 4 3+4 4 3+4 4 | | |
| | 5 2 4+4 4 4+4 4 3+4 4 5+5 4 3+4 4 3+4 4 3+4 4 | | |
| | 6 2 3+3 4 3+3 4 3+4 4 6+6 4 3+4 4 3+4 4 3+4 4 | | |
| | 5 2 4+4 4 3+3 4 3+4 4 6+6 4 3+4 4 3+4 4 3+4 4 | | |
| | 6 2 3+3 4 4+4 4 3+4 4 5+5 4 3+4 4 3+4 4 3+4 4 | | |
| | 5 2 4+4 4 3+3 4 3+4 4 5+5 4 3+4 4 3+4 4 3+4 4 | | |
| | 5 2 3+3 4 4+4 4 3+4 4 6+6 4 3+4 4 3+4 4 3+4 4 | | |
| | 5 2 3+3 4 3+3 4 3+4 4 5+5 4 3+4 4 3+4 4 3+4 4 | | |
| | 5 2 4+4 4 4+4 4 3+4 4 5+5 4 3+4 4 3+4 4 3+4 4 | | |
| | 5 2 4+4 4 4+4 4 3+4 4 6+6 4 3+4 4 3+4 4 3+4 4 | | |
| | 6 2 3+3 4 3+3 4 3+4 4 6+6 4 3+4 4 3+4 4 3+4 4 | | |
| | 5 2 3+3 4 4+4 4 3+4 4 6+6 4 3+4 4 3+4 4 3+4 4 | | |
| | 5 2 4+4 4 4+4 4 3+4 4 5+5 4 3+4 4 3+4 4 3+4 4 | | |
| | 5 2 4+4 4 3+3 4 3+4 4 6+6 4 3+4 4 3+4 4 3+4 4 | | |
| | 5 2 3+3 4 3+3 4 3+4 4 6+6 4 3+4 4 3+4 4 3+4 4 | | |
| | 6 2 4+4 4 3+3 4 3+4 4 5+5 4 3+4 4 3+4 4 3+4 4 | | |
| | 6 2 3+3 4 4+4 4 3+4 4 5+5 4 3+4 4 3+4 4 3+4 4 | | |
| | 5 2 3+3 4 3+3 4 3+4 4 6+6 4 3+4 4 3+4 4 3+4 4 | | |
| | 6 2 4+4 4 4+4 4 3+4 4 6+6 4 3+4 4 3+4 4 3+4 4 | | |
| | 6 2 3+3 4 3+3 4 3+4 4 6+6 4 3+4 4 3+4 4 3+4 4 | | |
| | 5 2 4+4 4 4+4 4 3+4 4 6+6 4 3+4 4 3+4 4 3+4 4 | | |
| | 5 2 3+3 4 4+4 4 3+4 4 5+5 4 3+4 4 3+4 4 3+4 4 | | |
| | 5 2 4+4 4 3+3 4 3+4 4 6+6 4 3+4 4 3+4 4 3+4 4 | | |
| | 6 2 4+4 4 3+3 4 3+4 4 5+5 4 3+4 4 3+4 4 3+4 4 | | |
| | 6 2 3+3 4 4+4 4 3+4 4 6+6 4 3+4 4 3+4 4 3+4 4 | | |
| | 5 2 | | |
| | 6 2 | | |
| | 6 2 | | |
| | 6 2 | | |
| | 6 2 | | |
| | 5 2 | | |
| | 5 2 | | |
| | 6 2 | | |
| | 5 2 | | |
| | 6 2 | | |
| | 6 2 | | |
| | 5 2 | | |
| | 6 2 | | |
| | 6 2 | | |
| | 6 2 | | |
| | 5 2 | | |
| | 6 2 | | |
| | 5 2 | | |

Continued on next column

**TABLE XXXVII– continued**

| TQC | ICSD | NoE | SF |
|---|---|---|---|
| | 5 2 | | |
| | 5 2 | | |
| | 5 2 | | |
| | 5 2 | | |
| | 5 2 | | |
| | 5 2 | | |
| | 6 2 | | |
| | 5 2 | | |
| | 6 2 | | |
| | 5 2 | | |
| | 6 2 | | |
| | 5 2 | | |
| | 6 2 | | |
| | 6 2 | | |
| | 5 2 | | |
| | 6 2 | | |
| | 5 2 | | |
| | 5 2 | | |
| | 5 2 | | |
| | 5 2 | | |
| | 5 2 | | |
| | 5 2 | | |
| | 5 2 | | |
| | 6 2 | | |
| | 5 2 | | |
| | 5 2 | | |
| | 6 2 | | |
| | 5 2 | | |
| | 6 2 | | |
| | 5 2 | | |
| 62.2.184 | 102428 | 120 | Co1Ga1Gd1 |
| | -GM -R -S -T -U -X -Y -Z | | |
| | 5 2 3+3 4 3+3 4 3+4 4 5+5 4 3+4 4 3+4 4 3+4 4 | | |
| | 5 2 4+4 4 4+4 4 3+4 4 6+6 4 3+4 4 3+4 4 3+4 4 | | |
| | 6 2 4+4 4 3+3 4 3+4 4 5+5 4 3+4 4 3+4 4 3+4 4 | | |
| | 6 2 3+3 4 4+4 4 3+4 4 6+6 4 3+4 4 3+4 4 3+4 4 | | |
| | 5 2 3+3 4 4+4 4 3+4 4 5+5 4 3+4 4 3+4 4 3+4 4 | | |
| | 5 2 4+4 4 3+3 4 3+4 4 6+6 4 3+4 4 3+4 4 3+4 4 | | |
| | 6 2 4+4 4 4+4 4 3+4 4 5+5 4 3+4 4 3+4 4 3+4 4 | | |
| | 6 2 3+3 4 3+3 4 3+4 4 6+6 4 3+4 4 3+4 4 3+4 4 | | |
| | 5 2 4+4 4 3+3 4 3+4 4 6+6 4 3+4 4 3+4 4 3+4 4 | | |
| | 6 2 3+3 4 4+4 4 3+4 4 5+5 4 3+4 4 3+4 4 3+4 4 | | |
| | 5 2 4+4 4 4+4 4 3+4 4 5+5 4 3+4 4 3+4 4 3+4 4 | | |
| | 6 2 3+3 4 3+3 4 3+4 4 6+6 4 3+4 4 3+4 4 3+4 4 | | |
| | 5 2 3+3 4 4+4 4 3+4 4 6+6 4 3+4 4 3+4 4 3+4 4 | | |
| | 5 2 4+4 4 3+3 4 3+4 4 6+6 4 3+4 4 3+4 4 3+4 4 | | |
| | 6 2 4+4 4 4+4 4 3+4 4 5+5 4 3+4 4 3+4 4 3+4 4 | | |
| | 6 2 3+3 4 3+3 4 3+4 4 5+5 4 3+4 4 3+4 4 3+4 4 | | |
| | 5 2 3+3 4 3+3 4 3+4 4 6+6 4 3+4 4 3+4 4 3+4 4 | | |
| | 5 2 4+4 4 4+4 4 3+4 4 5+5 4 3+4 4 3+4 4 3+4 4 | | |
| | 6 2 4+4 4 3+3 4 3+4 4 6+6 4 3+4 4 3+4 4 3+4 4 | | |
| | 6 2 3+3 4 4+4 4 3+4 4 5+5 4 3+4 4 3+4 4 3+4 4 | | |
| | 6 2 3+3 4 3+3 4 3+4 4 5+5 4 3+4 4 3+4 4 3+4 4 | | |
| | 5 2 4+4 4 4+4 4 3+4 4 6+6 4 3+4 4 3+4 4 3+4 4 | | |
| | 5 2 3+3 4 4+4 4 3+4 4 6+6 4 3+4 4 3+4 4 3+4 4 | | |
| | 6 2 4+4 4 3+3 4 3+4 4 5+5 4 3+4 4 3+4 4 3+4 4 | | |
| | 5 2 4+4 4 3+3 4 3+4 4 6+6 4 3+4 4 3+4 4 3+4 4 | | |
| | 5 2 3+3 4 4+4 4 3+4 4 5+5 4 3+4 4 3+4 4 3+4 4 | | |
| | 5 2 3+3 4 4+4 4 3+4 4 6+6 4 3+4 4 3+4 4 3+4 4 | | |
| | 6 2 4+4 4 3+3 4 3+4 4 5+5 4 3+4 4 3+4 4 3+4 4 | | |
| | 6 2 4+4 4 3+3 4 3+4 4 6+6 4 3+4 4 3+4 4 3+4 4 | | |

Continued on next column

**TABLE XXXVII– continued**

| TQC | ICSD | NoE | SF |
|---|---|---|---|
| | 6 2 3+3 4 4+4 4 3+4 4 5+5 4 3+4 4 3+4 4 3+4 4 | | |
| | 5 2 | | |
| | 6 2 | | |
| | 5 2 | | |
| | 5 2 | | |
| | 6 2 | | |
| | 5 2 | | |
| | 5 2 | | |
| | 6 2 | | |
| | 6 2 | | |
| | 6 2 | | |
| | 5 2 | | |
| | 5 2 | | |
| | 6 2 | | |
| | 6 2 | | |
| | 6 2 | | |
| | 5 2 | | |
| | 5 2 | | |
| | 6 2 | | |
| | 5 2 | | |
| | 6 2 | | |
| | 5 2 | | |
| | 6 2 | | |
| | 5 2 | | |
| | 6 2 | | |
| | 5 2 | | |
| | 6 2 | | |
| | 5 2 | | |
| | 6 2 | | |
| | 5 2 | | |
| | 5 2 | | |
| 62.2.185 | 48206 | 112 | Cl2Nd1 |
| | -GM -R -S -T -U -X -Y -Z | | |
| | 5 2 3+3 4 3+3 4 3+4 4 5+5 4 3+4 4 3+4 4 3+4 4 | | |
| | 5 2 4+4 4 4+4 4 3+4 4 6+6 4 3+4 4 3+4 4 3+4 4 | | |
| | 6 2 3+3 4 4+4 4 3+4 4 6+6 4 3+4 4 3+4 4 3+4 4 | | |
| | 6 2 4+4 4 3+3 4 3+4 4 5+5 4 3+4 4 3+4 4 3+4 4 | | |
| | 5 2 4+4 4 3+3 4 3+4 4 6+6 4 3+4 4 3+4 4 3+4 4 | | |
| | 6 2 3+3 4 4+4 4 3+4 4 5+5 4 3+4 4 3+4 4 3+4 4 | | |
| | 5 2 3+3 4 4+4 4 3+4 4 5+5 4 3+4 4 3+4 4 3+4 4 | | |
| | 6 2 4+4 4 3+3 4 3+4 4 6+6 4 3+4 4 3+4 4 3+4 4 | | |
| | 5 2 3+3 4 4+4 4 3+4 4 6+6 4 3+4 4 3+4 4 3+4 4 | | |
| | 6 2 4+4 4 3+3 4 3+4 4 5+5 4 3+4 4 3+4 4 3+4 4 | | |
| | 5 2 3+3 4 4+4 4 3+4 4 5+5 4 3+4 4 3+4 4 3+4 4 | | |
| | 5 2 4+4 4 3+3 4 3+4 4 6+6 4 3+4 4 3+4 4 3+4 4 | | |
| | 6 2 3+3 4 3+3 4 3+4 4 5+5 4 3+4 4 3+4 4 3+4 4 | | |
| | 5 2 4+4 4 4+4 4 3+4 4 6+6 4 3+4 4 3+4 4 3+4 4 | | |
| | 6 2 4+4 4 3+3 4 3+4 4 6+6 4 3+4 4 3+4 4 3+4 4 | | |
| | 6 2 3+3 4 4+4 4 3+4 4 6+6 4 3+4 4 3+4 4 3+4 4 | | |
| | 5 2 4+4 4 3+3 4 3+4 4 5+5 4 3+4 4 3+4 4 3+4 4 | | |
| | 6 2 3+3 4 4+4 4 3+4 4 5+5 4 3+4 4 3+4 4 3+4 4 | | |
| | 6 2 3+3 4 3+3 4 3+4 4 6+6 4 3+4 4 3+4 4 3+4 4 | | |
| | 5 2 4+4 4 4+4 4 3+4 4 5+5 4 3+4 4 3+4 4 3+4 4 | | |
| | 6 2 3+3 4 4+4 4 3+4 4 5+5 4 3+4 4 3+4 4 3+4 4 | | |
| | 6 2 4+4 4 3+3 4 3+4 4 6+6 4 3+4 4 3+4 4 3+4 4 | | |
| | 5 2 4+4 4 4+4 4 3+4 4 5+5 4 3+4 4 3+4 4 3+4 4 | | |
| | 5 2 3+3 4 3+3 4 3+4 4 6+6 4 3+4 4 3+4 4 3+4 4 | | |
| | 5 2 3+3 4 3+3 4 3+4 4 6+6 4 3+4 4 3+4 4 3+4 4 | | |
| | 6 2 4+4 4 4+4 4 3+4 4 5+5 4 3+4 4 3+4 4 3+4 4 | | |
| | 5 2 4+4 4 3+3 4 3+4 4 5+5 4 3+4 4 3+4 4 3+4 4 | | |
| | 6 2 3+3 4 4+4 4 3+4 4 6+6 4 3+4 4 3+4 4 3+4 4 | | |

Continued on next column

**TABLE XXXVII– continued**

| TQC | ICSD | NoE | SF |
|---|---|---|---|
| | 5 2 | | |
| | 6 2 | | |
| | 5 2 | | |
| | 5 2 | | |
| | 6 2 | | |
| | 5 2 | | |
| | 5 2 | | |
| | 5 2 | | |
| | 6 2 | | |
| | 6 2 | | |
| | 5 2 | | |
| | 5 2 | | |
| | 5 2 | | |
| | 6 2 | | |
| | 5 2 | | |
| | 6 2 | | |
| | 6 2 | | |
| | 6 2 | | |
| | 6 2 | | |
| | 6 2 | | |
| | 6 2 | | |
| | 5 2 | | |
| | 6 2 | | |
| | 5 2 | | |
| | 5 2 | | |
| | 5 2 | | |
| | 5 2 | | |
| | 6 2 | | |
| 62.2.186 | 157332 | 232 | Ce3In2Ru2 |
| | -GM -R -S -T -U -X -Y -Z | | |
| | 5 2 3+3 4 3+3 4 3+4 4 5+5 4 3+4 4 3+4 4 3+4 4 | | |
| | 6 2 4+4 4 4+4 4 3+4 4 6+6 4 3+4 4 3+4 4 3+4 4 | | |
| | 5 2 3+3 4 3+3 4 3+4 4 6+6 4 3+4 4 3+4 4 3+4 4 | | |
| | 6 2 4+4 4 4+4 4 3+4 4 5+5 4 3+4 4 3+4 4 3+4 4 | | |
| | 5 2 3+3 4 3+3 4 3+4 4 6+6 4 3+4 4 3+4 4 3+4 4 | | |
| | 5 2 4+4 4 4+4 4 3+4 4 5+5 4 3+4 4 3+4 4 3+4 4 | | |
| | 6 2 4+4 4 3+3 4 3+4 4 6+6 4 3+4 4 3+4 4 3+4 4 | | |
| | 6 2 3+3 4 4+4 4 3+4 4 5+5 4 3+4 4 3+4 4 3+4 4 | | |
| | 5 2 3+3 4 3+3 4 3+4 4 6+6 4 3+4 4 3+4 4 3+4 4 | | |
| | 6 2 4+4 4 4+4 4 3+4 4 5+5 4 3+4 4 3+4 4 3+4 4 | | |
| | 5 2 4+4 4 4+4 4 3+4 4 6+6 4 3+4 4 3+4 4 3+4 4 | | |
| | 6 2 3+3 4 3+3 4 3+4 4 5+5 4 3+4 4 3+4 4 3+4 4 | | |
| | 6 2 3+3 4 4+4 4 3+4 4 6+6 4 3+4 4 3+4 4 3+4 4 | | |
| | 5 2 4+4 4 3+3 4 3+4 4 5+5 4 3+4 4 3+4 4 3+4 4 | | |
| | 5 2 3+3 4 3+3 4 3+4 4 6+6 4 3+4 4 3+4 4 3+4 4 | | |
| | 6 2 4+4 4 4+4 4 3+4 4 5+5 4 3+4 4 3+4 4 3+4 4 | | |
| | 5 2 3+3 4 4+4 4 3+4 4 6+6 4 3+4 4 3+4 4 3+4 4 | | |
| | 5 2 4+4 4 3+3 4 3+4 4 5+5 4 3+4 4 3+4 4 3+4 4 | | |
| | 6 2 4+4 4 3+3 4 3+4 4 6+6 4 3+4 4 3+4 4 3+4 4 | | |
| | 6 2 3+3 4 4+4 4 3+4 4 5+5 4 3+4 4 3+4 4 3+4 4 | | |
| | 5 2 4+4 4 3+3 4 3+4 4 6+6 4 3+4 4 3+4 4 3+4 4 | | |
| | 6 2 3+3 4 4+4 4 3+4 4 5+5 4 3+4 4 3+4 4 3+4 4 | | |
| | 5 2 4+4 4 4+4 4 3+4 4 6+6 4 3+4 4 3+4 4 3+4 4 | | |
| | 6 2 3+3 4 3+3 4 3+4 4 5+5 4 3+4 4 3+4 4 3+4 4 | | |
| | 5 2 4+4 4 3+3 4 3+4 4 5+5 4 3+4 4 3+4 4 3+4 4 | | |
| | 6 2 3+3 4 4+4 4 3+4 4 6+6 4 3+4 4 3+4 4 3+4 4 | | |
| | 5 2 3+3 4 3+3 4 3+4 4 5+5 4 3+4 4 3+4 4 3+4 4 | | |
| | 6 2 4+4 4 4+4 4 3+4 4 6+6 4 3+4 4 3+4 4 3+4 4 | | |
| | 5 2 3+3 4 4+4 4 3+4 4 6+6 4 3+4 4 3+4 4 3+4 4 | | |
| | 6 2 4+4 4 3+3 4 3+4 4 5+5 4 3+4 4 3+4 4 3+4 4 | | |
| | 5 2 4+4 4 3+3 4 3+4 4 5+5 4 3+4 4 3+4 4 3+4 4 | | |

Continued on next column

**TABLE XXXVII– continued**

| TQC | ICSD | NoE | SF |
|---|---|---|---|
| | 6 2 3+3 4 4+4 4 3+4 4 6+6 4 3+4 4 3+4 4 3+4 4 | | |
| | 5 2 3+3 4 4+4 4 3+4 4 5+5 4 3+4 4 3+4 4 3+4 4 | | |
| | 5 2 4+4 4 3+3 4 3+4 4 6+6 4 3+4 4 3+4 4 3+4 4 | | |
| | 6 2 4+4 4 3+3 4 3+4 4 5+5 4 3+4 4 3+4 4 3+4 4 | | |
| | 6 2 3+3 4 4+4 4 3+4 4 6+6 4 3+4 4 3+4 4 3+4 4 | | |
| | 5 2 3+3 4 3+3 4 3+4 4 5+5 4 3+4 4 3+4 4 3+4 4 | | |
| | 5 2 4+4 4 4+4 4 3+4 4 5+5 4 3+4 4 3+4 4 3+4 4 | | |
| | 6 2 3+3 4 3+3 4 3+4 4 6+6 4 3+4 4 3+4 4 3+4 4 | | |
| | 5 2 4+4 4 4+4 4 3+4 4 6+6 4 3+4 4 3+4 4 3+4 4 | | |
| | 6 2 3+3 4 4+4 4 3+4 4 6+6 4 3+4 4 3+4 4 3+4 4 | | |
| | 5 2 4+4 4 3+3 4 3+4 4 5+5 4 3+4 4 3+4 4 3+4 4 | | |
| | 6 2 4+4 4 3+3 4 3+4 4 6+6 4 3+4 4 3+4 4 3+4 4 | | |
| | 6 2 3+3 4 4+4 4 3+4 4 5+5 4 3+4 4 3+4 4 3+4 4 | | |
| | 5 2 3+3 4 4+4 4 3+4 4 5+5 4 3+4 4 3+4 4 3+4 4 | | |
| | 5 2 4+4 4 3+3 4 3+4 4 5+5 4 3+4 4 3+4 4 3+4 4 | | |
| | 6 2 4+4 4 4+4 4 3+4 4 6+6 4 3+4 4 3+4 4 3+4 4 | | |
| | 6 2 3+3 4 3+3 4 3+4 4 6+6 4 3+4 4 3+4 4 3+4 4 | | |
| | 5 2 4+4 4 3+3 4 3+4 4 6+6 4 3+4 4 3+4 4 3+4 4 | | |
| | 5 2 3+3 4 4+4 4 3+4 4 5+5 4 3+4 4 3+4 4 3+4 4 | | |
| | 6 2 3+3 4 4+4 4 3+4 4 5+5 4 3+4 4 3+4 4 3+4 4 | | |
| | 6 2 4+4 4 3+3 4 3+4 4 5+5 4 3+4 4 3+4 4 3+4 4 | | |
| | 5 2 3+3 4 4+4 4 3+4 4 6+6 4 3+4 4 3+4 4 3+4 4 | | |
| | 5 2 4+4 4 3+3 4 3+4 4 6+6 4 3+4 4 3+4 4 3+4 4 | | |
| | 6 2 3+3 4 3+3 4 3+4 4 5+5 4 3+4 4 3+4 4 3+4 4 | | |
| | 6 2 4+4 4 4+4 4 3+4 4 5+5 4 3+4 4 3+4 4 3+4 4 | | |
| | 6 2 3+3 4 3+3 4 3+4 4 6+6 4 3+4 4 3+4 4 3+4 4 | | |
| | 6 2 4+4 4 4+4 4 3+4 4 5+5 4 3+4 4 3+4 4 3+4 4 | | |
| | 5 2 | | |
| | 5 2 | | |
| | 5 2 | | |
| | 5 2 | | |
| | 5 2 | | |
| | 6 2 | | |
| | 6 2 | | |
| | 6 2 | | |
| | 6 2 | | |
| | 5 2 | | |
| | 5 2 | | |
| | 5 2 | | |
| | 5 2 | | |
| | 6 2 | | |
| | 5 2 | | |
| | 6 2 | | |
| | 6 2 | | |
| | 6 2 | | |
| | 6 2 | | |
| | 5 2 | | |
| | 5 2 | | |
| | 6 2 | | |
| | 5 2 | | |
| | 6 2 | | |
| | 5 2 | | |
| | 5 2 | | |
| | 6 2 | | |
| | 5 2 | | |
| | 5 2 | | |
| | 5 2 | | |
| | 6 2 | | |
| | 6 2 | | |
| | 6 2 | | |
| | 5 2 | | |
| | 5 2 | | |
| | 5 2 | | |
| | 6 2 | | |
| | 6 2 | | |
| | 5 2 | | |
| | 6 2 | | |
| | 5 2 | | |
| | 6 2 | | |
| | 6 2 | | |
| | 6 2 | | |
| | 5 2 | | |
| | 6 2 | | |
| | 5 2 | | |
| | 6 2 | | |
| | 5 2 | | |
| | 6 2 | | |
| | 6 2 | | |
| | 5 2 | | |
| | 5 2 | | |
| | 5 2 | | |
| | 6 2 | | |
| | 5 2 | | |
| | 6 2 | | |
| | 6 2 | | |
| 62.2.187 | 428424 | 168 | Ce2Te3 |
| | -GM -R -S -T -U -X -Y -Z | | |
| | 5 2 3+3 4 3+3 4 3+4 4 5+5 4 3+4 4 3+4 4 3+4 4 | | |
| | 5 2 4+4 4 4+4 4 3+4 4 6+6 4 3+4 4 3+4 4 3+4 4 | | |
| | 6 2 3+3 4 3+3 4 3+4 4 5+5 4 3+4 4 3+4 4 3+4 4 | | |
| | 6 2 4+4 4 4+4 4 3+4 4 6+6 4 3+4 4 3+4 4 3+4 4 | | |
| | 5 2 3+3 4 3+3 4 3+4 4 6+6 4 3+4 4 3+4 4 3+4 4 | | |
| | 5 2 4+4 4 4+4 4 3+4 4 5+5 4 3+4 4 3+4 4 3+4 4 | | |
| | 6 2 3+3 4 4+4 4 3+4 4 5+5 4 3+4 4 3+4 4 3+4 4 | | |
| | 6 2 4+4 4 3+3 4 3+4 4 6+6 4 3+4 4 3+4 4 3+4 4 | | |
| | 6 2 4+4 4 3+3 4 3+4 4 5+5 4 3+4 4 3+4 4 3+4 4 | | |
| | 5 2 3+3 4 4+4 4 3+4 4 6+6 4 3+4 4 3+4 4 3+4 4 | | |
| | 5 2 3+3 4 3+3 4 3+4 4 5+5 4 3+4 4 3+4 4 3+4 4 | | |
| | 6 2 4+4 4 4+4 4 3+4 4 6+6 4 3+4 4 3+4 4 3+4 4 | | |
| | 5 2 3+3 4 4+4 4 3+4 4 5+5 4 3+4 4 3+4 4 3+4 4 | | |
| | 5 2 4+4 4 3+3 4 3+4 4 6+6 4 3+4 4 3+4 4 3+4 4 | | |
| | 6 2 3+3 4 4+4 4 3+4 4 5+5 4 3+4 4 3+4 4 3+4 4 | | |
| | 6 2 4+4 4 3+3 4 3+4 4 6+6 4 3+4 4 3+4 4 3+4 4 | | |
| | 5 2 4+4 4 4+4 4 3+4 4 5+5 4 3+4 4 3+4 4 3+4 4 | | |
| | 6 2 3+3 4 3+3 4 3+4 4 6+6 4 3+4 4 3+4 4 3+4 4 | | |
| | 5 2 3+3 4 4+4 4 3+4 4 6+6 4 3+4 4 3+4 4 3+4 4 | | |
| | 5 2 4+4 4 3+3 4 3+4 4 5+5 4 3+4 4 3+4 4 3+4 4 | | |
| | 6 2 3+3 4 3+3 4 3+4 4 6+6 4 3+4 4 3+4 4 3+4 4 | | |
| | 5 2 4+4 4 4+4 4 3+4 4 5+5 4 3+4 4 3+4 4 3+4 4 | | |
| | 6 2 3+3 4 4+4 4 3+4 4 5+5 4 3+4 4 3+4 4 3+4 4 | | |
| | 6 2 4+4 4 3+3 4 3+4 4 6+6 4 3+4 4 3+4 4 3+4 4 | | |
| | 5 2 4+4 4 3+3 4 3+4 4 6+6 4 3+4 4 3+4 4 3+4 4 | | |
| | 5 2 3+3 4 4+4 4 3+4 4 5+5 4 3+4 4 3+4 4 3+4 4 | | |
| | 6 2 3+3 4 3+3 4 3+4 4 5+5 4 3+4 4 3+4 4 3+4 4 | | |
| | 6 2 4+4 4 4+4 4 3+4 4 6+6 4 3+4 4 3+4 4 3+4 4 | | |
| | 5 2 4+4 4 3+3 4 3+4 4 5+5 4 3+4 4 3+4 4 3+4 4 | | |
| | 6 2 3+3 4 4+4 4 3+4 4 6+6 4 3+4 4 3+4 4 3+4 4 | | |
| | 5 2 3+3 4 3+3 4 3+4 4 6+6 4 3+4 4 3+4 4 3+4 4 | | |
| | 6 2 4+4 4 4+4 4 3+4 4 5+5 4 3+4 4 3+4 4 3+4 4 | | |
| | 5 2 3+3 4 4+4 4 3+4 4 5+5 4 3+4 4 3+4 4 3+4 4 | | |
| | 5 2 4+4 4 3+3 4 3+4 4 6+6 4 3+4 4 3+4 4 3+4 4 | | |
| | 5 2 4+4 4 3+3 4 3+4 4 6+6 4 3+4 4 3+4 4 3+4 4 | | |

Continued on next column

**TABLE XXXVII– continued**

| TQC | ICSD | NoE | SF | | | | |
|---|---|---|---|---|---|---|---|
| | 6 2 | 3+3 4 | 4+4 4 | 3+4 4 | 5+5 4 | 3+4 4 | 3+4 4 | 3+4 4 |
| | 6 2 | 4+4 4 | 4+4 4 | 3+4 4 | 6+6 4 | 3+4 4 | 3+4 4 | 3+4 4 |
| | 6 2 | 3+3 4 | 3+3 4 | 3+4 4 | 5+5 4 | 3+4 4 | 3+4 4 | 3+4 4 |
| | 5 2 | 3+3 4 | 4+4 4 | 3+4 4 | 5+5 4 | 3+4 4 | 3+4 4 | 3+4 4 |
| | 6 2 | 4+4 4 | 3+3 4 | 3+4 4 | 6+6 4 | 3+4 4 | 3+4 4 | 3+4 4 |
| | 5 2 | 3+3 4 | 3+3 4 | 3+4 4 | 5+5 4 | 3+4 4 | 3+4 4 | 3+4 4 |
| | 6 2 | 4+4 4 | 4+4 4 | 3+4 4 | 6+6 4 | 3+4 4 | 3+4 4 | 3+4 4 |
| | 5 2 | | | | | | | |
| | 6 2 | | | | | | | |
| | 6 2 | | | | | | | |
| | 5 2 | | | | | | | |
| | 6 2 | | | | | | | |
| | 5 2 | | | | | | | |
| | 5 2 | | | | | | | |
| | 5 2 | | | | | | | |
| | 5 2 | | | | | | | |
| | 6 2 | | | | | | | |
| | 6 2 | | | | | | | |
| | 5 2 | | | | | | | |
| | 5 2 | | | | | | | |
| | 5 2 | | | | | | | |
| | 6 2 | | | | | | | |
| | 5 2 | | | | | | | |
| | 6 2 | | | | | | | |
| | 5 2 | | | | | | | |
| | 5 2 | | | | | | | |
| | 6 2 | | | | | | | |
| | 5 2 | | | | | | | |
| | 6 2 | | | | | | | |
| | 6 2 | | | | | | | |
| | 6 2 | | | | | | | |
| | 5 2 | | | | | | | |
| | 6 2 | | | | | | | |
| | 5 2 | | | | | | | |
| | 6 2 | | | | | | | |
| | 5 2 | | | | | | | |
| | 5 2 | | | | | | | |
| | 6 2 | | | | | | | |
| | 5 2 | | | | | | | |
| | 6 2 | | | | | | | |
| | 5 2 | | | | | | | |
| | 6 2 | | | | | | | |
| | 6 2 | | | | | | | |
| | 6 2 | | | | | | | |
| | 6 2 | | | | | | | |
| | 6 2 | | | | | | | |
| | 6 2 | | | | | | | |
| | 5 2 | | | | | | | |
| | 6 2 | | | | | | | |
| 62.2.188 | 106433 | 64 | Ce1Si1 | | | | |
| | -GM | -R | -S | -T | -U | -X | -Y | -Z |
| | 5 2 | 3+3 4 | 3+3 4 | 3+4 4 | 6+6 4 | 3+4 4 | 3+4 4 | 3+4 4 |
| | 6 2 | 4+4 4 | 4+4 4 | 3+4 4 | 5+5 4 | 3+4 4 | 3+4 4 | 3+4 4 |
| | 5 2 | 4+4 4 | 4+4 4 | 3+4 4 | 6+6 4 | 3+4 4 | 3+4 4 | 3+4 4 |
| | 6 2 | 3+3 4 | 3+3 4 | 3+4 4 | 5+5 4 | 3+4 4 | 3+4 4 | 3+4 4 |
| | 5 2 | 3+3 4 | 3+3 4 | 3+4 4 | 6+6 4 | 3+4 4 | 3+4 4 | 3+4 4 |
| | 6 2 | 4+4 4 | 4+4 4 | 3+4 4 | 5+5 4 | 3+4 4 | 3+4 4 | 3+4 4 |
| | 5 2 | 4+4 4 | 3+3 4 | 3+4 4 | 6+6 4 | 3+4 4 | 3+4 4 | 3+4 4 |
| | 6 2 | 3+3 4 | 4+4 4 | 3+4 4 | 5+5 4 | 3+4 4 | 3+4 4 | 3+4 4 |
| | 5 2 | 4+4 4 | 4+4 4 | 3+4 4 | 5+5 4 | 3+4 4 | 3+4 4 | 3+4 4 |
| | 6 2 | 3+3 4 | 3+3 4 | 3+4 4 | 6+6 4 | 3+4 4 | 3+4 4 | 3+4 4 |

Continued on next column

| TQC | ICSD | NoE | SF | | | | |
|---|---|---|---|---|---|---|---|
| | 5 2 | 3+3 4 | 4+4 4 | 3+4 4 | 6+6 4 | 3+4 4 | 3+4 4 | 3+4 4 |
| | 5 2 | 4+4 4 | 3+3 4 | 3+4 4 | 5+5 4 | 3+4 4 | 3+4 4 | 3+4 4 |
| | 6 2 | 4+4 4 | 4+4 4 | 3+4 4 | 5+5 4 | 3+4 4 | 3+4 4 | 3+4 4 |
| | 6 2 | 3+3 4 | 3+3 4 | 3+4 4 | 5+5 4 | 3+4 4 | 3+4 4 | 3+4 4 |
| | 5 2 | 3+3 4 | 3+3 4 | 3+4 4 | 6+6 4 | 3+4 4 | 3+4 4 | 3+4 4 |
| | 6 2 | 4+4 4 | 4+4 4 | 3+4 4 | 5+5 4 | 3+4 4 | 3+4 4 | 3+4 4 |
| | 5 2 | | | | | | | |
| | 5 2 | | | | | | | |
| | 6 2 | | | | | | | |
| | 6 2 | | | | | | | |
| | 6 2 | | | | | | | |
| | 5 2 | | | | | | | |
| | 5 2 | | | | | | | |
| | 5 2 | | | | | | | |
| | 6 2 | | | | | | | |
| | 5 2 | | | | | | | |
| | 6 2 | | | | | | | |
| | 5 2 | | | | | | | |
| | 6 2 | | | | | | | |
| | 5 2 | | | | | | | |
| | 5 2 | | | | | | | |
| | 6 2 | | | | | | | |
| 62.2.189 | 418546 | 136 | Ce1Pt1Zn1 | | | | |
| | -GM | -R | -S | -T | -U | -X | -Y | -Z |
| | 5 2 | 3+3 4 | 3+3 4 | 3+4 4 | 6+6 4 | 3+4 4 | 3+4 4 | 3+4 4 |
| | 6 2 | 4+4 4 | 4+4 4 | 3+4 4 | 5+5 4 | 3+4 4 | 3+4 4 | 3+4 4 |
| | 5 2 | 3+3 4 | 3+3 4 | 3+4 4 | 6+6 4 | 3+4 4 | 3+4 4 | 3+4 4 |
| | 6 2 | 4+4 4 | 4+4 4 | 3+4 4 | 5+5 4 | 3+4 4 | 3+4 4 | 3+4 4 |
| | 5 2 | 4+4 4 | 3+3 4 | 3+4 4 | 6+6 4 | 3+4 4 | 3+4 4 | 3+4 4 |
| | 6 2 | 3+3 4 | 4+4 4 | 3+4 4 | 5+5 4 | 3+4 4 | 3+4 4 | 3+4 4 |
| | 5 2 | 3+3 4 | 4+4 4 | 3+4 4 | 6+6 4 | 3+4 4 | 3+4 4 | 3+4 4 |
| | 6 2 | 4+4 4 | 3+3 4 | 3+4 4 | 5+5 4 | 3+4 4 | 3+4 4 | 3+4 4 |
| | 5 2 | 3+3 4 | 3+3 4 | 3+4 4 | 6+6 4 | 3+4 4 | 3+4 4 | 3+4 4 |
| | 6 2 | 4+4 4 | 4+4 4 | 3+4 4 | 5+5 4 | 3+4 4 | 3+4 4 | 3+4 4 |
| | 5 2 | 4+4 4 | 4+4 4 | 3+4 4 | 5+5 4 | 3+4 4 | 3+4 4 | 3+4 4 |
| | 5 2 | 3+3 4 | 3+3 4 | 3+4 4 | 6+6 4 | 3+4 4 | 3+4 4 | 3+4 4 |
| | 6 2 | 4+4 4 | 4+4 4 | 3+4 4 | 6+6 4 | 3+4 4 | 3+4 4 | 3+4 4 |
| | 6 2 | 3+3 4 | 3+3 4 | 3+4 4 | 5+5 4 | 3+4 4 | 3+4 4 | 3+4 4 |
| | 5 2 | 3+3 4 | 3+3 4 | 3+4 4 | 5+5 4 | 3+4 4 | 3+4 4 | 3+4 4 |
| | 6 2 | 4+4 4 | 4+4 4 | 3+4 4 | 6+6 4 | 3+4 4 | 3+4 4 | 3+4 4 |
| | 5 2 | 3+3 4 | 3+3 4 | 3+4 4 | 5+5 4 | 3+4 4 | 3+4 4 | 3+4 4 |
| | 6 2 | 4+4 4 | 4+4 4 | 3+4 4 | 6+6 4 | 3+4 4 | 3+4 4 | 3+4 4 |
| | 6 2 | 3+3 4 | 4+4 4 | 3+4 4 | 5+5 4 | 3+4 4 | 3+4 4 | 3+4 4 |
| | 5 2 | 4+4 4 | 3+3 4 | 3+4 4 | 6+6 4 | 3+4 4 | 3+4 4 | 3+4 4 |
| | 5 2 | 4+4 4 | 3+3 4 | 3+4 4 | 5+5 4 | 3+4 4 | 3+4 4 | 3+4 4 |
| | 6 2 | 3+3 4 | 4+4 4 | 3+4 4 | 6+6 4 | 3+4 4 | 3+4 4 | 3+4 4 |
| | 5 2 | 4+4 4 | 3+3 4 | 3+4 4 | 5+5 4 | 3+4 4 | 3+4 4 | 3+4 4 |
| | 5 2 | 3+3 4 | 4+4 4 | 3+4 4 | 6+6 4 | 3+4 4 | 3+4 4 | 3+4 4 |
| | 6 2 | 3+3 4 | 3+3 4 | 3+4 4 | 5+5 4 | 3+4 4 | 3+4 4 | 3+4 4 |
| | 6 2 | 4+4 4 | 4+4 4 | 3+4 4 | 6+6 4 | 3+4 4 | 3+4 4 | 3+4 4 |
| | 5 2 | 4+4 4 | 3+3 4 | 3+4 4 | 5+5 4 | 3+4 4 | 3+4 4 | 3+4 4 |
| | 6 2 | 3+3 4 | 4+4 4 | 3+4 4 | 6+6 4 | 3+4 4 | 3+4 4 | 3+4 4 |
| | 5 2 | 4+4 4 | 4+4 4 | 3+4 4 | 5+5 4 | 3+4 4 | 3+4 4 | 3+4 4 |
| | 6 2 | 3+3 4 | 3+3 4 | 3+4 4 | 6+6 4 | 3+4 4 | 3+4 4 | 3+4 4 |
| | 5 2 | 3+3 4 | 4+4 4 | 3+4 4 | 5+5 4 | 3+4 4 | 3+4 4 | 3+4 4 |
| | 5 2 | 4+4 4 | 3+3 4 | 3+4 4 | 6+6 4 | 3+4 4 | 3+4 4 | 3+4 4 |
| | 6 2 | 3+3 4 | 4+4 4 | 3+4 4 | 5+5 4 | 3+4 4 | 3+4 4 | 3+4 4 |
| | 5 2 | 4+4 4 | 3+3 4 | 3+4 4 | 5+5 4 | 3+4 4 | 3+4 4 | 3+4 4 |
| | 5 2 | | | | | | | |
| | 6 2 | | | | | | | |
| | 6 2 | | | | | | | |

Continued on next column

**TABLE XXXVII– continued**

| TQC | ICSD | NoE | SF | | | | | |
|---|---|---|---|---|---|---|---|---|
| | 5 2 | | | | | | | |
| | 6 2 | | | | | | | |
| | 6 2 | | | | | | | |
| | 5 2 | | | | | | | |
| | 6 2 | | | | | | | |
| | 5 2 | | | | | | | |
| | 6 2 | | | | | | | |
| | 6 2 | | | | | | | |
| | 5 2 | | | | | | | |
| | 5 2 | | | | | | | |
| | 6 2 | | | | | | | |
| | 5 2 | | | | | | | |
| | 5 2 | | | | | | | |
| | 6 2 | | | | | | | |
| | 5 2 | | | | | | | |
| | 6 2 | | | | | | | |
| | 6 2 | | | | | | | |
| | 5 2 | | | | | | | |
| | 6 2 | | | | | | | |
| | 5 2 | | | | | | | |
| | 5 2 | | | | | | | |
| | 6 2 | | | | | | | |
| | 6 2 | | | | | | | |
| | 6 2 | | | | | | | |
| | 5 2 | | | | | | | |
| | 5 2 | | | | | | | |
| | 5 2 | | | | | | | |
| | 6 2 | | | | | | | |
| | 5 2 | | | | | | | |
| | 5 2 | | | | | | | |
| | 6 2 | | | | | | | |
| 62.2.190 | 420207 | 136 | Ce1Pd1Zn1 | | | | | |
| | -GM | -R | -S | -T | -U | -X | -Y | -Z |
| | 5 2 | 3+3 4 | 3+3 4 | 3+4 4 | 6+6 4 | 3+4 4 | 3+4 4 | 3+4 4 |
| | 6 2 | 4+4 4 | 4+4 4 | 3+4 4 | 5+5 4 | 3+4 4 | 3+4 4 | 3+4 4 |
| | 5 2 | 3+3 4 | 3+3 4 | 3+4 4 | 6+6 4 | 3+4 4 | 3+4 4 | 3+4 4 |
| | 6 2 | 4+4 4 | 4+4 4 | 3+4 4 | 5+5 4 | 3+4 4 | 3+4 4 | 3+4 4 |
| | 5 2 | 4+4 4 | 3+3 4 | 3+4 4 | 6+6 4 | 3+4 4 | 3+4 4 | 3+4 4 |
| | 6 2 | 3+3 4 | 4+4 4 | 3+4 4 | 5+5 4 | 3+4 4 | 3+4 4 | 3+4 4 |
| | 5 2 | 3+3 4 | 4+4 4 | 3+4 4 | 6+6 4 | 3+4 4 | 3+4 4 | 3+4 4 |
| | 6 2 | 4+4 4 | 3+3 4 | 3+4 4 | 5+5 4 | 3+4 4 | 3+4 4 | 3+4 4 |
| | 5 2 | 3+3 4 | 3+3 4 | 3+4 4 | 6+6 4 | 3+4 4 | 3+4 4 | 3+4 4 |
| | 6 2 | 4+4 4 | 4+4 4 | 3+4 4 | 5+5 4 | 3+4 4 | 3+4 4 | 3+4 4 |
| | 5 2 | 4+4 4 | 4+4 4 | 3+4 4 | 5+5 4 | 3+4 4 | 3+4 4 | 3+4 4 |
| | 5 2 | 3+3 4 | 3+3 4 | 3+4 4 | 6+6 4 | 3+4 4 | 3+4 4 | 3+4 4 |
| | 6 2 | 4+4 4 | 3+3 4 | 3+4 4 | 6+6 4 | 3+4 4 | 3+4 4 | 3+4 4 |
| | 6 2 | 3+3 4 | 4+4 4 | 3+4 4 | 5+5 4 | 3+4 4 | 3+4 4 | 3+4 4 |
| | 5 2 | 3+3 4 | 3+3 4 | 3+4 4 | 5+5 4 | 3+4 4 | 3+4 4 | 3+4 4 |
| | 6 2 | 4+4 4 | 4+4 4 | 3+4 4 | 6+6 4 | 3+4 4 | 3+4 4 | 3+4 4 |
| | 6 2 | 3+3 4 | 3+3 4 | 3+4 4 | 5+5 4 | 3+4 4 | 3+4 4 | 3+4 4 |
| | 5 2 | 4+4 4 | 4+4 4 | 3+4 4 | 6+6 4 | 3+4 4 | 3+4 4 | 3+4 4 |
| | 6 2 | 3+3 4 | 4+4 4 | 3+4 4 | 5+5 4 | 3+4 4 | 3+4 4 | 3+4 4 |
| | 5 2 | 4+4 4 | 3+3 4 | 3+4 4 | 6+6 4 | 3+4 4 | 3+4 4 | 3+4 4 |
| | 5 2 | 4+4 4 | 3+3 4 | 3+4 4 | 5+5 4 | 3+4 4 | 3+4 4 | 3+4 4 |
| | 6 2 | 3+3 4 | 4+4 4 | 3+4 4 | 6+6 4 | 3+4 4 | 3+4 4 | 3+4 4 |
| | 5 2 | 4+4 4 | 3+3 4 | 3+4 4 | 6+6 4 | 3+4 4 | 3+4 4 | 3+4 4 |
| | 6 2 | 3+3 4 | 4+4 4 | 3+4 4 | 5+5 4 | 3+4 4 | 3+4 4 | 3+4 4 |
| | 6 2 | 3+3 4 | 3+3 4 | 3+4 4 | 5+5 4 | 3+4 4 | 3+4 4 | 3+4 4 |
| | 5 2 | 4+4 4 | 4+4 4 | 3+4 4 | 6+6 4 | 3+4 4 | 3+4 4 | 3+4 4 |
| | 6 2 | 4+4 4 | 3+3 4 | 3+4 4 | 5+5 4 | 3+4 4 | 3+4 4 | 3+4 4 |
| | 5 2 | 3+3 4 | 4+4 4 | 3+4 4 | 6+6 4 | 3+4 4 | 3+4 4 | 3+4 4 |

Continued on next column

**TABLE XXXVII– continued**

| TQC | ICSD | NoE | SF | | | | | |
|---|---|---|---|---|---|---|---|---|
| | 5 2 | 3+3 4 | 4+4 4 | 3+4 4 | 5+5 4 | 3+4 4 | 3+4 4 | 3+4 4 |
| | 5 2 | 4+4 4 | 3+3 4 | 3+4 4 | 6+6 4 | 3+4 4 | 3+4 4 | 3+4 4 |
| | 6 2 | 4+4 4 | 4+4 4 | 3+4 4 | 5+5 4 | 3+4 4 | 3+4 4 | 3+4 4 |
| | 6 2 | 3+3 4 | 3+3 4 | 3+4 4 | 6+6 4 | 3+4 4 | 3+4 4 | 3+4 4 |
| | 5 2 | 3+3 4 | 4+4 4 | 3+4 4 | 5+5 4 | 3+4 4 | 3+4 4 | 3+4 4 |
| | 5 2 | 4+4 4 | 3+3 4 | 3+4 4 | 5+5 4 | 3+4 4 | 3+4 4 | 3+4 4 |
| | 5 2 | | | | | | | |
| | 6 2 | | | | | | | |
| | 6 2 | | | | | | | |
| | 5 2 | | | | | | | |
| | 6 2 | | | | | | | |
| | 6 2 | | | | | | | |
| | 5 2 | | | | | | | |
| | 5 2 | | | | | | | |
| | 6 2 | | | | | | | |
| | 5 2 | | | | | | | |
| | 6 2 | | | | | | | |
| | 6 2 | | | | | | | |
| | 5 2 | | | | | | | |
| | 5 2 | | | | | | | |
| | 6 2 | | | | | | | |
| | 5 2 | | | | | | | |
| | 6 2 | | | | | | | |
| | 5 2 | | | | | | | |
| | 6 2 | | | | | | | |
| | 6 2 | | | | | | | |
| | 5 2 | | | | | | | |
| | 6 2 | | | | | | | |
| | 5 2 | | | | | | | |
| | 5 2 | | | | | | | |
| | 6 2 | | | | | | | |
| | 6 2 | | | | | | | |
| | 6 2 | | | | | | | |
| | 5 2 | | | | | | | |
| | 5 2 | | | | | | | |
| | 5 2 | | | | | | | |
| | 6 2 | | | | | | | |
| | 5 2 | | | | | | | |
| | 6 2 | | | | | | | |
| | 5 2 | | | | | | | |
| 62.2.191 | 409972 | 248 | Ce1P2Pt4 | | | | | |
| | -GM | -R | -S | -T | -U | -X | -Y | -Z |
| | 5 2 | 3+3 4 | 3+3 4 | 3+4 4 | 5+5 4 | 3+4 4 | 3+4 4 | 3+4 4 |
| | 6 2 | 4+4 4 | 4+4 4 | 3+4 4 | 6+6 4 | 3+4 4 | 3+4 4 | 3+4 4 |
| | 5 2 | 3+3 4 | 3+3 4 | 3+4 4 | 5+5 4 | 3+4 4 | 3+4 4 | 3+4 4 |
| | 6 2 | 4+4 4 | 4+4 4 | 3+4 4 | 6+6 4 | 3+4 4 | 3+4 4 | 3+4 4 |
| | 5 2 | 3+3 4 | 3+3 4 | 3+4 4 | 6+6 4 | 3+4 4 | 3+4 4 | 3+4 4 |
| | 6 2 | 4+4 4 | 4+4 4 | 3+4 4 | 5+5 4 | 3+4 4 | 3+4 4 | 3+4 4 |
| | 5 2 | 3+3 4 | 3+3 4 | 3+4 4 | 6+6 4 | 3+4 4 | 3+4 4 | 3+4 4 |
| | 6 2 | 4+4 4 | 4+4 4 | 3+4 4 | 5+5 4 | 3+4 4 | 3+4 4 | 3+4 4 |
| | 5 2 | 3+3 4 | 3+3 4 | 3+4 4 | 5+5 4 | 3+4 4 | 3+4 4 | 3+4 4 |
| | 6 2 | 4+4 4 | 4+4 4 | 3+4 4 | 6+6 4 | 3+4 4 | 3+4 4 | 3+4 4 |
| | 5 2 | 3+3 4 | 3+3 4 | 3+4 4 | 5+5 4 | 3+4 4 | 3+4 4 | 3+4 4 |
| | 6 2 | 4+4 4 | 4+4 4 | 3+4 4 | 6+6 4 | 3+4 4 | 3+4 4 | 3+4 4 |
| | 5 2 | 3+3 4 | 3+3 4 | 3+4 4 | 6+6 4 | 3+4 4 | 3+4 4 | 3+4 4 |
| | 6 2 | 4+4 4 | 4+4 4 | 3+4 4 | 5+5 4 | 3+4 4 | 3+4 4 | 3+4 4 |
| | 5 2 | 3+3 4 | 4+4 4 | 3+4 4 | 6+6 4 | 3+4 4 | 3+4 4 | 3+4 4 |
| | 6 2 | 4+4 4 | 3+3 4 | 3+4 4 | 5+5 4 | 3+4 4 | 3+4 4 | 3+4 4 |
| | 5 2 | 4+4 4 | 3+3 4 | 3+4 4 | 6+6 4 | 3+4 4 | 3+4 4 | 3+4 4 |
| | 5 2 | 3+3 4 | 4+4 4 | 3+4 4 | 5+5 4 | 3+4 4 | 3+4 4 | 3+4 4 |
| | 6 2 | 4+4 4 | 3+3 4 | 3+4 4 | 6+6 4 | 3+4 4 | 3+4 4 | 3+4 4 |

Continued on next column

**TABLE XXXVII– continued**

| TQC | ICSD NoE SF |
|---|---|
| | 6 2 3+3 4 4+4 4 3+4 4 5+5 4 3+4 4 3+4 4 3+4 4 |
| | 5 2 3+3 4 4+4 4 3+4 4 6+6 4 3+4 4 3+4 4 3+4 4 |
| | 6 2 4+4 4 3+3 4 3+4 4 6+6 4 3+4 4 3+4 4 3+4 4 |
| | 6 2 3+3 4 3+3 4 3+4 4 5+5 4 3+4 4 3+4 4 3+4 4 |
| | 5 2 4+4 4 4+4 4 3+4 4 5+5 4 3+4 4 3+4 4 3+4 4 |
| | 5 2 4+4 4 4+4 4 3+4 4 6+6 4 3+4 4 3+4 4 3+4 4 |
| | 6 2 3+3 4 3+3 4 3+4 4 6+6 4 3+4 4 3+4 4 3+4 4 |
| | 5 2 4+4 4 4+4 4 3+4 4 5+5 4 3+4 4 3+4 4 3+4 4 |
| | 5 2 3+3 4 3+3 4 3+4 4 5+5 4 3+4 4 3+4 4 3+4 4 |
| | 6 2 3+3 4 3+3 4 3+4 4 6+6 4 3+4 4 3+4 4 3+4 4 |
| | 5 2 4+4 4 4+4 4 3+4 4 5+5 4 3+4 4 3+4 4 3+4 4 |
| | 6 2 3+3 4 4+4 4 3+4 4 5+5 4 3+4 4 3+4 4 3+4 4 |
| | 6 2 4+4 4 3+3 4 3+4 4 6+6 4 3+4 4 3+4 4 3+4 4 |
| | 5 2 4+4 4 3+3 4 3+4 4 5+5 4 3+4 4 3+4 4 3+4 4 |
| | 6 2 3+3 4 4+4 4 3+4 4 6+6 4 3+4 4 3+4 4 3+4 4 |
| | 6 2 3+3 4 3+3 4 3+4 4 5+5 4 3+4 4 3+4 4 3+4 4 |
| | 6 2 4+4 4 4+4 4 3+4 4 6+6 4 3+4 4 3+4 4 3+4 4 |
| | 5 2 4+4 4 4+4 4 3+4 4 5+5 4 3+4 4 3+4 4 3+4 4 |
| | 5 2 3+3 4 3+3 4 3+4 4 5+5 4 3+4 4 3+4 4 3+4 4 |
| | 6 2 4+4 4 3+3 4 3+4 4 6+6 4 3+4 4 3+4 4 3+4 4 |
| | 5 2 3+3 4 4+4 4 3+4 4 6+6 4 3+4 4 3+4 4 3+4 4 |
| | 6 2 3+3 4 4+4 4 3+4 4 5+5 4 3+4 4 3+4 4 3+4 4 |
| | 6 2 4+4 4 3+3 4 3+4 4 6+6 4 3+4 4 3+4 4 3+4 4 |
| | 5 2 4+4 4 4+4 4 3+4 4 5+5 4 3+4 4 3+4 4 3+4 4 |
| | 5 2 3+3 4 3+3 4 3+4 4 6+6 4 3+4 4 3+4 4 3+4 4 |
| | 6 2 3+3 4 3+3 4 3+4 4 5+5 4 3+4 4 3+4 4 3+4 4 |
| | 5 2 4+4 4 4+4 4 3+4 4 6+6 4 3+4 4 3+4 4 3+4 4 |
| | 5 2 4+4 4 3+3 4 3+4 4 6+6 4 3+4 4 3+4 4 3+4 4 |
| | 6 2 3+3 4 4+4 4 3+4 4 5+5 4 3+4 4 3+4 4 3+4 4 |
| | 6 2 4+4 4 4+4 4 3+4 4 5+5 4 3+4 4 3+4 4 3+4 4 |
| | 5 2 3+3 4 3+3 4 3+4 4 6+6 4 3+4 4 3+4 4 3+4 4 |
| | 6 2 3+3 4 4+4 4 3+4 4 5+5 4 3+4 4 3+4 4 3+4 4 |
| | 5 2 4+4 4 3+3 4 3+4 4 6+6 4 3+4 4 3+4 4 3+4 4 |
| | 6 2 3+3 4 4+4 4 3+4 4 5+5 4 3+4 4 3+4 4 3+4 4 |
| | 5 2 4+4 4 3+3 4 3+4 4 6+6 4 3+4 4 3+4 4 3+4 4 |
| | 6 2 4+4 4 4+4 4 3+4 4 5+5 4 3+4 4 3+4 4 3+4 4 |
| | 5 2 3+3 4 3+3 4 3+4 4 6+6 4 3+4 4 3+4 4 3+4 4 |
| | 6 2 3+3 4 3+3 4 3+4 4 6+6 4 3+4 4 3+4 4 3+4 4 |
| | 5 2 4+4 4 4+4 4 3+4 4 5+5 4 3+4 4 3+4 4 3+4 4 |
| | 6 2 4+4 4 3+3 4 3+4 4 5+5 4 3+4 4 3+4 4 3+4 4 |
| | 5 2 3+3 4 4+4 4 3+4 4 6+6 4 3+4 4 3+4 4 3+4 4 |
| | 6 2 3+3 4 4+4 4 3+4 4 5+5 4 3+4 4 3+4 4 3+4 4 |
| | 5 2 4+4 4 3+3 4 3+4 4 6+6 4 3+4 4 3+4 4 3+4 4 |
| | 6 2 |
| | 6 2 |
| | 5 2 |
| | 5 2 |
| | 6 2 |
| | 5 2 |
| | 6 2 |
| | 5 2 |
| | 6 2 |
| | 5 2 |
| | 5 2 |
| | 6 2 |
| | 5 2 |
| | 5 2 |
| | 6 2 |
| | 5 2 |
| | 5 2 |
| | 6 2 |

Continued on next column

**TABLE XXXVII– continued**

| TQC | ICSD NoE SF |
|---|---|
| | 5 2 |
| | 6 2 |
| | 5 2 |
| | 6 2 |
| | 5 2 |
| | 6 2 |
| | 5 2 |
| | 6 2 |
| | 6 2 |
| | 6 2 |
| | 5 2 |
| | 6 2 |
| | 5 2 |
| | 5 2 |
| | 5 2 |
| | 6 2 |
| | 5 2 |
| | 6 2 |
| | 6 2 |
| | 5 2 |
| | 5 2 |
| | 6 2 |
| | 5 2 |
| | 6 2 |
| | 5 2 |
| | 6 2 |
| | 6 2 |
| | 5 2 |
| | 6 2 |
| | 5 2 |
| | 5 2 |
| | 6 2 |
| | 5 2 |
| | 5 2 |
| | 6 2 |
| | 6 2 |
| | 6 2 |
| | 5 2 |
| | 5 2 |
| | 6 2 |
| | 5 2 |
| | 6 2 |
| | 6 2 |
| | 5 2 |
| 62.2.192 | 54733 176 Ce1In4Pd2 |
| | -GM -R -S -T -U -X -Y -Z |
| | 5 2 3+3 4 3+3 4 3+4 4 5+5 4 3+4 4 3+4 4 3+4 4 |
| | 5 2 4+4 4 4+4 4 3+4 4 6+6 4 3+4 4 3+4 4 3+4 4 |
| | 6 2 3+3 4 3+3 4 3+4 4 5+5 4 3+4 4 3+4 4 3+4 4 |
| | 6 2 4+4 4 4+4 4 3+4 4 6+6 4 3+4 4 3+4 4 3+4 4 |
| | 5 2 3+3 4 4+4 4 3+4 4 5+5 4 3+4 4 3+4 4 3+4 4 |
| | 6 2 4+4 4 3+3 4 3+4 4 6+6 4 3+4 4 3+4 4 3+4 4 |
| | 5 2 4+4 4 3+3 4 3+4 4 5+5 4 3+4 4 3+4 4 3+4 4 |
| | 6 2 3+3 4 4+4 4 3+4 4 6+6 4 3+4 4 3+4 4 3+4 4 |
| | 5 2 4+4 4 3+3 4 3+4 4 6+6 4 3+4 4 3+4 4 3+4 4 |
| | 6 2 3+3 4 4+4 4 3+4 4 5+5 4 3+4 4 3+4 4 3+4 4 |
| | 5 2 3+3 4 4+4 4 3+4 4 5+5 4 3+4 4 3+4 4 3+4 4 |
| | 6 2 4+4 4 3+3 4 3+4 4 6+6 4 3+4 4 3+4 4 3+4 4 |
| | 5 2 3+3 4 3+3 4 3+4 4 6+6 4 3+4 4 3+4 4 3+4 4 |
| | 6 2 4+4 4 4+4 4 3+4 4 5+5 4 3+4 4 3+4 4 3+4 4 |
| | 5 2 3+3 4 3+3 4 3+4 4 5+5 4 3+4 4 3+4 4 3+4 4 |

Continued on next column

**TABLE XXXVII– continued**

| TQC | ICSD | NoE | SF |
|---|---|---|---|
| | 6 2 4+4 4 4+4 4 3+4 4 6+6 4 3+4 4 3+4 4 3+4 4 | | |
| | 5 2 3+3 4 4+4 4 3+4 4 6+6 4 3+4 4 3+4 4 3+4 4 | | |
| | 6 2 4+4 4 3+3 4 3+4 4 5+5 4 3+4 4 3+4 4 3+4 4 | | |
| | 5 2 4+4 4 3+3 4 3+4 4 6+6 4 3+4 4 3+4 4 3+4 4 | | |
| | 5 2 3+3 4 4+4 4 3+4 4 5+5 4 3+4 4 3+4 4 3+4 4 | | |
| | 6 2 3+3 4 4+4 4 3+4 4 5+5 4 3+4 4 3+4 4 3+4 4 | | |
| | 5 2 4+4 4 3+3 4 3+4 4 6+6 4 3+4 4 3+4 4 3+4 4 | | |
| | 6 2 3+3 4 4+4 4 3+4 4 5+5 4 3+4 4 3+4 4 3+4 4 | | |
| | 5 2 4+4 4 3+3 4 3+4 4 6+6 4 3+4 4 3+4 4 3+4 4 | | |
| | 6 2 3+3 4 3+3 4 3+4 4 6+6 4 3+4 4 3+4 4 3+4 4 | | |
| | 5 2 4+4 4 4+4 4 3+4 4 5+5 4 3+4 4 3+4 4 3+4 4 | | |
| | 6 2 3+3 4 3+3 4 3+4 4 5+5 4 3+4 4 3+4 4 3+4 4 | | |
| | 5 2 4+4 4 4+4 4 3+4 4 6+6 4 3+4 4 3+4 4 3+4 4 | | |
| | 6 2 4+4 4 4+4 4 3+4 4 5+5 4 3+4 4 3+4 4 3+4 4 | | |
| | 6 2 3+3 4 3+3 4 3+4 4 6+6 4 3+4 4 3+4 4 3+4 4 | | |
| | 5 2 4+4 4 3+3 4 3+4 4 6+6 4 3+4 4 3+4 4 3+4 4 | | |
| | 6 2 3+3 4 4+4 4 3+4 4 5+5 4 3+4 4 3+4 4 3+4 4 | | |
| | 6 2 3+3 4 4+4 4 3+4 4 5+5 4 3+4 4 3+4 4 3+4 4 | | |
| | 5 2 4+4 4 3+3 4 3+4 4 6+6 4 3+4 4 3+4 4 3+4 4 | | |
| | 6 2 4+4 4 4+4 4 3+4 4 5+5 4 3+4 4 3+4 4 3+4 4 | | |
| | 5 2 3+3 4 3+3 4 3+4 4 6+6 4 3+4 4 3+4 4 3+4 4 | | |
| | 5 2 3+3 4 4+4 4 3+4 4 6+6 4 3+4 4 3+4 4 3+4 4 | | |
| | 6 2 4+4 4 3+3 4 3+4 4 5+5 4 3+4 4 3+4 4 3+4 4 | | |
| | 5 2 3+3 4 3+3 4 3+4 4 6+6 4 3+4 4 3+4 4 3+4 4 | | |
| | 5 2 4+4 4 4+4 4 3+4 4 6+6 4 3+4 4 3+4 4 3+4 4 | | |
| | 5 2 3+3 4 4+4 4 3+4 4 5+5 4 3+4 4 3+4 4 3+4 4 | | |
| | 6 2 4+4 4 3+3 4 3+4 4 5+5 4 3+4 4 3+4 4 3+4 4 | | |
| | 5 2 3+3 4 3+3 4 3+4 4 5+5 4 3+4 4 3+4 4 3+4 4 | | |
| | 5 2 4+4 4 4+4 4 3+4 4 6+6 4 3+4 4 3+4 4 3+4 4 | | |
| | 6 2 | | |
| | 5 2 | | |
| | 6 2 | | |
| | 6 2 | | |
| | 5 2 | | |
| | 6 2 | | |
| | 5 2 | | |
| | 6 2 | | |
| | 5 2 | | |
| | 6 2 | | |
| | 5 2 | | |
| | 5 2 | | |
| | 6 2 | | |
| | 6 2 | | |
| | 5 2 | | |
| | 5 2 | | |
| | 6 2 | | |
| | 5 2 | | |
| | 6 2 | | |
| | 6 2 | | |
| | 5 2 | | |
| | 6 2 | | |
| | 6 2 | | |
| | 6 2 | | |
| | 5 2 | | |
| | 5 2 | | |
| | 6 2 | | |
| | 5 2 | | |
| | 6 2 | | |
| | 5 2 | | |
| | 5 2 | | |
| | 5 2 | | |

Continued on next column

**TABLE XXXVII– continued**

| TQC | ICSD | NoE | SF |
|---|---|---|---|
| | 6 2 | | |
| | 5 2 | | |
| | 6 2 | | |
| | 5 2 | | |
| | 6 2 | | |
| | 5 2 | | |
| | 6 2 | | |
| | 6 2 | | |
| | 5 2 | | |
| | 6 2 | | |
| | 5 2 | | |
| | 6 2 | | |
| 62.2.193 | 99240 | 192 | Ce1In2Pd3 |
| | -GM -R -S -T -U -X -Y -Z | | |
| | 5 2 3+3 4 3+3 4 3+4 4 5+5 4 3+4 4 3+4 4 3+4 4 | | |
| | 6 2 4+4 4 4+4 4 3+4 4 6+6 4 3+4 4 3+4 4 3+4 4 | | |
| | 5 2 3+3 4 4+4 4 3+4 4 5+5 4 3+4 4 3+4 4 3+4 4 | | |
| | 6 2 4+4 4 3+3 4 3+4 4 6+6 4 3+4 4 3+4 4 3+4 4 | | |
| | 5 2 4+4 4 3+3 4 3+4 4 5+5 4 3+4 4 3+4 4 3+4 4 | | |
| | 6 2 3+3 4 4+4 4 3+4 4 6+6 4 3+4 4 3+4 4 3+4 4 | | |
| | 5 2 3+3 4 3+3 4 3+4 4 5+5 4 3+4 4 3+4 4 3+4 4 | | |
| | 6 2 4+4 4 4+4 4 3+4 4 6+6 4 3+4 4 3+4 4 3+4 4 | | |
| | 5 2 3+3 4 3+3 4 3+4 4 6+6 4 3+4 4 3+4 4 3+4 4 | | |
| | 6 2 4+4 4 4+4 4 3+4 4 5+5 4 3+4 4 3+4 4 3+4 4 | | |
| | 5 2 3+3 4 3+3 4 3+4 4 6+6 4 3+4 4 3+4 4 3+4 4 | | |
| | 6 2 4+4 4 4+4 4 3+4 4 5+5 4 3+4 4 3+4 4 3+4 4 | | |
| | 5 2 3+3 4 3+3 4 3+4 4 5+5 4 3+4 4 3+4 4 3+4 4 | | |
| | 5 2 4+4 4 4+4 4 3+4 4 6+6 4 3+4 4 3+4 4 3+4 4 | | |
| | 6 2 3+3 4 4+4 4 3+4 4 6+6 4 3+4 4 3+4 4 3+4 4 | | |
| | 6 2 4+4 4 3+3 4 3+4 4 5+5 4 3+4 4 3+4 4 3+4 4 | | |
| | 5 2 4+4 4 3+3 4 3+4 4 6+6 4 3+4 4 3+4 4 3+4 4 | | |
| | 5 2 3+3 4 4+4 4 3+4 4 5+5 4 3+4 4 3+4 4 3+4 4 | | |
| | 6 2 4+4 4 4+4 4 3+4 4 6+6 4 3+4 4 3+4 4 3+4 4 | | |
| | 6 2 3+3 4 3+3 4 3+4 4 5+5 4 3+4 4 3+4 4 3+4 4 | | |
| | 5 2 4+4 4 3+3 4 3+4 4 6+6 4 3+4 4 3+4 4 3+4 4 | | |
| | 5 2 3+3 4 4+4 4 3+4 4 6+6 4 3+4 4 3+4 4 3+4 4 | | |
| | 6 2 3+3 4 4+4 4 3+4 4 5+5 4 3+4 4 3+4 4 3+4 4 | | |
| | 6 2 4+4 4 3+3 4 3+4 4 5+5 4 3+4 4 3+4 4 3+4 4 | | |
| | 6 2 3+3 4 4+4 4 3+4 4 5+5 4 3+4 4 3+4 4 3+4 4 | | |
| | 5 2 4+4 4 3+3 4 3+4 4 6+6 4 3+4 4 3+4 4 3+4 4 | | |
| | 5 2 4+4 4 4+4 4 3+4 4 6+6 4 3+4 4 3+4 4 3+4 4 | | |
| | 6 2 3+3 4 3+3 4 3+4 4 5+5 4 3+4 4 3+4 4 3+4 4 | | |
| | 5 2 3+3 4 4+4 4 3+4 4 6+6 4 3+4 4 3+4 4 3+4 4 | | |
| | 6 2 4+4 4 3+3 4 3+4 4 5+5 4 3+4 4 3+4 4 3+4 4 | | |
| | 6 2 4+4 4 4+4 4 3+4 4 6+6 4 3+4 4 3+4 4 3+4 4 | | |
| | 5 2 3+3 4 3+3 4 3+4 4 5+5 4 3+4 4 3+4 4 3+4 4 | | |
| | 6 2 4+4 4 3+3 4 3+4 4 6+6 4 3+4 4 3+4 4 3+4 4 | | |
| | 6 2 3+3 4 4+4 4 3+4 4 5+5 4 3+4 4 3+4 4 3+4 4 | | |
| | 5 2 4+4 4 4+4 4 3+4 4 5+5 4 3+4 4 3+4 4 3+4 4 | | |
| | 5 2 3+3 4 3+3 4 3+4 4 5+5 4 3+4 4 3+4 4 3+4 4 | | |
| | 5 2 3+3 4 4+4 4 3+4 4 6+6 4 3+4 4 3+4 4 3+4 4 | | |
| | 6 2 4+4 4 3+3 4 3+4 4 6+6 4 3+4 4 3+4 4 3+4 4 | | |
| | 5 2 3+3 4 3+3 4 3+4 4 5+5 4 3+4 4 3+4 4 3+4 4 | | |
| | 5 2 4+4 4 4+4 4 3+4 4 6+6 4 3+4 4 3+4 4 3+4 4 | | |
| | 5 2 4+4 4 3+3 4 3+4 4 5+5 4 3+4 4 3+4 4 3+4 4 | | |
| | 6 2 3+3 4 4+4 4 3+4 4 6+6 4 3+4 4 3+4 4 3+4 4 | | |
| | 6 2 4+4 4 4+4 4 3+4 4 6+6 4 3+4 4 3+4 4 3+4 4 | | |
| | 6 2 3+3 4 3+3 4 3+4 4 5+5 4 3+4 4 3+4 4 3+4 4 | | |
| | 6 2 4+4 4 3+3 4 3+4 4 5+5 4 3+4 4 3+4 4 3+4 4 | | |
| | 5 2 3+3 4 4+4 4 3+4 4 6+6 4 3+4 4 3+4 4 3+4 4 | | |
| | 6 2 3+3 4 3+3 4 3+4 4 6+6 4 3+4 4 3+4 4 3+4 4 | | |

Continued on next column

**TABLE XXXVII– continued**

| TQC | ICSD | NoE | SF | | | | | | |
|---|---|---|---|---|---|---|---|---|---|
| | 5 2 | 4+4 4 | 4+4 4 | 3+4 4 | 5+5 4 | 3+4 4 | 3+4 4 | 3+4 4 | |
| | 6 2 | | | | | | | | |
| | 5 2 | | | | | | | | |
| | 5 2 | | | | | | | | |
| | 5 2 | | | | | | | | |
| | 6 2 | | | | | | | | |
| | 6 2 | | | | | | | | |
| | 5 2 | | | | | | | | |
| | 5 2 | | | | | | | | |
| | 6 2 | | | | | | | | |
| | 5 2 | | | | | | | | |
| | 6 2 | | | | | | | | |
| | 5 2 | | | | | | | | |
| | 5 2 | | | | | | | | |
| | 6 2 | | | | | | | | |
| | 5 2 | | | | | | | | |
| | 5 2 | | | | | | | | |
| | 6 2 | | | | | | | | |
| | 6 2 | | | | | | | | |
| | 6 2 | | | | | | | | |
| | 5 2 | | | | | | | | |
| | 5 2 | | | | | | | | |
| | 6 2 | | | | | | | | |
| | 5 2 | | | | | | | | |
| | 5 2 | | | | | | | | |
| | 6 2 | | | | | | | | |
| | 5 2 | | | | | | | | |
| | 6 2 | | | | | | | | |
| | 5 2 | | | | | | | | |
| | 6 2 | | | | | | | | |
| | 5 2 | | | | | | | | |
| | 6 2 | | | | | | | | |
| | 6 2 | | | | | | | | |
| | 6 2 | | | | | | | | |
| | 6 2 | | | | | | | | |
| | 5 2 | | | | | | | | |
| | 6 2 | | | | | | | | |
| | 5 2 | | | | | | | | |
| | 5 2 | | | | | | | | |
| | 5 2 | | | | | | | | |
| | 5 2 | | | | | | | | |
| | 5 2 | | | | | | | | |
| | 6 2 | | | | | | | | |
| | 6 2 | | | | | | | | |
| | 6 2 | | | | | | | | |
| | 5 2 | | | | | | | | |
| | 5 2 | | | | | | | | |
| | 6 2 | | | | | | | | |
| | 5 2 | | | | | | | | |
| 62.2.194 | 150105 | 312 | Ce1Cu6 | | | | | | |
| | | -GM | -R | -S | -T | -U | -X | -Y | -Z |
| | 5 2 | 3+3 4 | 3+3 4 | 3+4 4 | 5+5 4 | 3+4 4 | 3+4 4 | 3+4 4 | |
| | 5 2 | 4+4 4 | 4+4 4 | 3+4 4 | 6+6 4 | 3+4 4 | 3+4 4 | 3+4 4 | |
| | 6 2 | 3+3 4 | 3+3 4 | 3+4 4 | 5+5 4 | 3+4 4 | 3+4 4 | 3+4 4 | |
| | 6 2 | 4+4 4 | 4+4 4 | 3+4 4 | 6+6 4 | 3+4 4 | 3+4 4 | 3+4 4 | |
| | 5 2 | 4+4 4 | 4+4 4 | 3+4 4 | 5+5 4 | 3+4 4 | 3+4 4 | 3+4 4 | |
| | 5 2 | 3+3 4 | 3+3 4 | 3+4 4 | 6+6 4 | 3+4 4 | 3+4 4 | 3+4 4 | |
| | 6 2 | 3+3 4 | 3+3 4 | 3+4 4 | 5+5 4 | 3+4 4 | 3+4 4 | 3+4 4 | |
| | 6 2 | 4+4 4 | 4+4 4 | 3+4 4 | 6+6 4 | 3+4 4 | 3+4 4 | 3+4 4 | |
| | 5 2 | 3+3 4 | 4+4 4 | 3+4 4 | 6+6 4 | 3+4 4 | 3+4 4 | 3+4 4 | |
| | 5 2 | 4+4 4 | 3+3 4 | 3+4 4 | 5+5 4 | 3+4 4 | 3+4 4 | 3+4 4 | |

Continued on next column

**TABLE XXXVII– continued**

| TQC | ICSD | NoE | SF | | | | | |
|---|---|---|---|---|---|---|---|---|
| | 6 2 | 4+4 4 | 3+3 4 | 3+4 4 | 5+5 4 | 3+4 4 | 3+4 4 | 3+4 4 |
| | 5 2 | 3+3 4 | 4+4 4 | 3+4 4 | 6+6 4 | 3+4 4 | 3+4 4 | 3+4 4 |
| | 6 2 | 3+3 4 | 3+3 4 | 3+4 4 | 5+5 4 | 3+4 4 | 3+4 4 | 3+4 4 |
| | 5 2 | 4+4 4 | 4+4 4 | 3+4 4 | 6+6 4 | 3+4 4 | 3+4 4 | 3+4 4 |
| | 6 2 | 4+4 4 | 3+3 4 | 3+4 4 | 6+6 4 | 3+4 4 | 3+4 4 | 3+4 4 |
| | 6 2 | 3+3 4 | 4+4 4 | 3+4 4 | 5+5 4 | 3+4 4 | 3+4 4 | 3+4 4 |
| | 5 2 | 3+3 4 | 4+4 4 | 3+4 4 | 5+5 4 | 3+4 4 | 3+4 4 | 3+4 4 |
| | 6 2 | 4+4 4 | 3+3 4 | 3+4 4 | 6+6 4 | 3+4 4 | 3+4 4 | 3+4 4 |
| | 5 2 | 4+4 4 | 4+4 4 | 3+4 4 | 6+6 4 | 3+4 4 | 3+4 4 | 3+4 4 |
| | 6 2 | 3+3 4 | 3+3 4 | 3+4 4 | 5+5 4 | 3+4 4 | 3+4 4 | 3+4 4 |
| | 5 2 | 3+3 4 | 4+4 4 | 3+4 4 | 6+6 4 | 3+4 4 | 3+4 4 | 3+4 4 |
| | 6 2 | 4+4 4 | 3+3 4 | 3+4 4 | 6+6 4 | 3+4 4 | 3+4 4 | 3+4 4 |
| | 6 2 | 4+4 4 | 4+4 4 | 3+4 4 | 5+5 4 | 3+4 4 | 3+4 4 | 3+4 4 |
| | 5 2 | 3+3 4 | 3+3 4 | 3+4 4 | 6+6 4 | 3+4 4 | 3+4 4 | 3+4 4 |
| | 5 2 | 4+4 4 | 3+3 4 | 3+4 4 | 5+5 4 | 3+4 4 | 3+4 4 | 3+4 4 |
| | 6 2 | 3+3 4 | 4+4 4 | 3+4 4 | 6+6 4 | 3+4 4 | 3+4 4 | 3+4 4 |
| | 5 2 | 3+3 4 | 3+3 4 | 3+4 4 | 5+5 4 | 3+4 4 | 3+4 4 | 3+4 4 |
| | 6 2 | 4+4 4 | 4+4 4 | 3+4 4 | 5+5 4 | 3+4 4 | 3+4 4 | 3+4 4 |
| | 6 2 | 3+3 4 | 3+3 4 | 3+4 4 | 5+5 4 | 3+4 4 | 3+4 4 | 3+4 4 |
| | 6 2 | 4+4 4 | 4+4 4 | 3+4 4 | 6+6 4 | 3+4 4 | 3+4 4 | 3+4 4 |
| | 6 2 | 3+3 4 | 4+4 4 | 3+4 4 | 5+5 4 | 3+4 4 | 3+4 4 | 3+4 4 |
| | 5 2 | 4+4 4 | 3+3 4 | 3+4 4 | 6+6 4 | 3+4 4 | 3+4 4 | 3+4 4 |
| | 5 2 | 4+4 4 | 3+3 4 | 3+4 4 | 5+5 4 | 3+4 4 | 3+4 4 | 3+4 4 |
| | 6 2 | 3+3 4 | 4+4 4 | 3+4 4 | 6+6 4 | 3+4 4 | 3+4 4 | 3+4 4 |
| | 5 2 | 3+3 4 | 4+4 4 | 3+4 4 | 5+5 4 | 3+4 4 | 3+4 4 | 3+4 4 |
| | 5 2 | 4+4 4 | 3+3 4 | 3+4 4 | 6+6 4 | 3+4 4 | 3+4 4 | 3+4 4 |
| | 5 2 | 3+3 4 | 3+3 4 | 3+4 4 | 5+5 4 | 3+4 4 | 3+4 4 | 3+4 4 |
| | 6 2 | 4+4 4 | 4+4 4 | 3+4 4 | 6+6 4 | 3+4 4 | 3+4 4 | 3+4 4 |
| | 5 2 | 4+4 4 | 4+4 4 | 3+4 4 | 6+6 4 | 3+4 4 | 3+4 4 | 3+4 4 |
| | 6 2 | 3+3 4 | 3+3 4 | 3+4 4 | 5+5 4 | 3+4 4 | 3+4 4 | 3+4 4 |
| | 5 2 | 3+3 4 | 4+4 4 | 3+4 4 | 6+6 4 | 3+4 4 | 3+4 4 | 3+4 4 |
| | 5 2 | 4+4 4 | 3+3 4 | 3+4 4 | 5+5 4 | 3+4 4 | 3+4 4 | 3+4 4 |
| | 5 2 | 4+4 4 | 4+4 4 | 3+4 4 | 6+6 4 | 3+4 4 | 3+4 4 | 3+4 4 |
| | 6 2 | 3+3 4 | 3+3 4 | 3+4 4 | 5+5 4 | 3+4 4 | 3+4 4 | 3+4 4 |
| | 5 2 | 3+3 4 | 4+4 4 | 3+4 4 | 6+6 4 | 3+4 4 | 3+4 4 | 3+4 4 |
| | 5 2 | 4+4 4 | 3+3 4 | 3+4 4 | 5+5 4 | 3+4 4 | 3+4 4 | 3+4 4 |
| | 5 2 | 3+3 4 | 3+3 4 | 3+4 4 | 6+6 4 | 3+4 4 | 3+4 4 | 3+4 4 |
| | 6 2 | 4+4 4 | 4+4 4 | 3+4 4 | 5+5 4 | 3+4 4 | 3+4 4 | 3+4 4 |
| | 5 2 | 3+3 4 | 4+4 4 | 3+4 4 | 5+5 4 | 3+4 4 | 3+4 4 | 3+4 4 |
| | 6 2 | 4+4 4 | 3+3 4 | 3+4 4 | 6+6 4 | 3+4 4 | 3+4 4 | 3+4 4 |
| | 6 2 | 3+3 4 | 3+3 4 | 3+4 4 | 6+6 4 | 3+4 4 | 3+4 4 | 3+4 4 |
| | 6 2 | 4+4 4 | 4+4 4 | 3+4 4 | 5+5 4 | 3+4 4 | 3+4 4 | 3+4 4 |
| | 5 2 | 4+4 4 | 3+3 4 | 3+4 4 | 6+6 4 | 3+4 4 | 3+4 4 | 3+4 4 |
| | 5 2 | 3+3 4 | 4+4 4 | 3+4 4 | 5+5 4 | 3+4 4 | 3+4 4 | 3+4 4 |
| | 6 2 | 4+4 4 | 3+3 4 | 3+4 4 | 5+5 4 | 3+4 4 | 3+4 4 | 3+4 4 |
| | 5 2 | 3+3 4 | 4+4 4 | 3+4 4 | 6+6 4 | 3+4 4 | 3+4 4 | 3+4 4 |
| | 6 2 | 3+3 4 | 4+4 4 | 3+4 4 | 5+5 4 | 3+4 4 | 3+4 4 | 3+4 4 |
| | 5 2 | 4+4 4 | 3+3 4 | 3+4 4 | 6+6 4 | 3+4 4 | 3+4 4 | 3+4 4 |
| | 6 2 | 4+4 4 | 4+4 4 | 3+4 4 | 5+5 4 | 3+4 4 | 3+4 4 | 3+4 4 |
| | 6 2 | 3+3 4 | 3+3 4 | 3+4 4 | 6+6 4 | 3+4 4 | 3+4 4 | 3+4 4 |
| | 5 2 | 3+3 4 | 3+3 4 | 3+4 4 | 5+5 4 | 3+4 4 | 3+4 4 | 3+4 4 |
| | 6 2 | 4+4 4 | 4+4 4 | 3+4 4 | 6+6 4 | 3+4 4 | 3+4 4 | 3+4 4 |
| | 5 2 | 4+4 4 | 3+3 4 | 3+4 4 | 6+6 4 | 3+4 4 | 3+4 4 | 3+4 4 |
| | 6 2 | 3+3 4 | 4+4 4 | 3+4 4 | 6+6 4 | 3+4 4 | 3+4 4 | 3+4 4 |
| | 6 2 | 4+4 4 | 4+4 4 | 3+4 4 | 5+5 4 | 3+4 4 | 3+4 4 | 3+4 4 |
| | 5 2 | 3+3 4 | 3+3 4 | 3+4 4 | 5+5 4 | 3+4 4 | 3+4 4 | 3+4 4 |
| | 6 2 | 3+3 4 | 4+4 4 | 3+4 4 | 5+5 4 | 3+4 4 | 3+4 4 | 3+4 4 |
| | 5 2 | 4+4 4 | 3+3 4 | 3+4 4 | 6+6 4 | 3+4 4 | 3+4 4 | 3+4 4 |
| | 6 2 | 3+3 4 | 3+3 4 | 3+4 4 | 5+5 4 | 3+4 4 | 3+4 4 | 3+4 4 |
| | 5 2 | 4+4 4 | 4+4 4 | 3+4 4 | 6+6 4 | 3+4 4 | 3+4 4 | 3+4 4 |
| | 5 2 | 3+3 4 | 3+3 4 | 3+4 4 | 5+5 4 | 3+4 4 | 3+4 4 | 3+4 4 |

Continued on next column

**TABLE XXXVII– continued**

| TQC | ICSD | NoE | SF | | | | | |
|---|---|---|---|---|---|---|---|---|
| | 6 2 | 4+4 4 | 4+4 4 | 3+4 4 | 6+6 4 | 3+4 4 | 3+4 4 | 3+4 4 |
| | 6 2 | 3+3 4 | 3+3 4 | 3+4 4 | 6+6 4 | 3+4 4 | 3+4 4 | 3+4 4 |
| | 5 2 | 4+4 4 | 4+4 4 | 3+4 4 | 5+5 4 | 3+4 4 | 3+4 4 | 3+4 4 |
| | 6 2 | 3+3 4 | 4+4 4 | 3+4 4 | 5+5 4 | 3+4 4 | 3+4 4 | 3+4 4 |
| | 6 2 | 4+4 4 | 3+3 4 | 3+4 4 | 6+6 4 | 3+4 4 | 3+4 4 | 3+4 4 |
| | 5 2 | 3+3 4 | 3+3 4 | 3+4 4 | 6+6 4 | 3+4 4 | 3+4 4 | 3+4 4 |
| | 5 2 | 4+4 4 | 4+4 4 | 3+4 4 | 5+5 4 | 3+4 4 | 3+4 4 | 3+4 4 |
| | 6 2 | | | | | | | |
| | 6 2 | | | | | | | |
| | 5 2 | | | | | | | |
| | 5 2 | | | | | | | |
| | 6 2 | | | | | | | |
| | 6 2 | | | | | | | |
| | 5 2 | | | | | | | |
| | 6 2 | | | | | | | |
| | 5 2 | | | | | | | |
| | 6 2 | | | | | | | |
| | 6 2 | | | | | | | |
| | 6 2 | | | | | | | |
| | 5 2 | | | | | | | |
| | 5 2 | | | | | | | |
| | 6 2 | | | | | | | |
| | 5 2 | | | | | | | |
| | 5 2 | | | | | | | |
| | 6 2 | | | | | | | |
| | 6 2 | | | | | | | |
| | 5 2 | | | | | | | |
| | 5 2 | | | | | | | |
| | 6 2 | | | | | | | |
| | 5 2 | | | | | | | |
| | 6 2 | | | | | | | |
| | 5 2 | | | | | | | |
| | 5 2 | | | | | | | |
| | 6 2 | | | | | | | |
| | 5 2 | | | | | | | |
| | 5 2 | | | | | | | |
| | 6 2 | | | | | | | |
| | 5 2 | | | | | | | |
| | 6 2 | | | | | | | |
| | 5 2 | | | | | | | |
| | 6 2 | | | | | | | |
| | 6 2 | | | | | | | |
| | 5 2 | | | | | | | |
| | 5 2 | | | | | | | |
| | 6 2 | | | | | | | |
| | 5 2 | | | | | | | |
| | 5 2 | | | | | | | |
| | 6 2 | | | | | | | |
| | 6 2 | | | | | | | |
| | 5 2 | | | | | | | |
| | 6 2 | | | | | | | |
| | 5 2 | | | | | | | |
| | 6 2 | | | | | | | |
| | 6 2 | | | | | | | |
| | 5 2 | | | | | | | |
| | 6 2 | | | | | | | |
| | 5 2 | | | | | | | |
| | 6 2 | | | | | | | |
| | 5 2 | | | | | | | |
| | 5 2 | | | | | | | |
| | 6 2 | | | | | | | |
| | 6 2 | | | | | | | |
| | 5 2 | | | | | | | |
| | 6 2 | | | | | | | |
| | 5 2 | | | | | | | |
| | 6 2 | | | | | | | |
| | 6 2 | | | | | | | |
| | 5 2 | | | | | | | |
| | 6 2 | | | | | | | |
| | 6 2 | | | | | | | |
| | 5 2 | | | | | | | |
| | 6 2 | | | | | | | |
| | 5 2 | | | | | | | |
| | 5 2 | | | | | | | |
| | 5 2 | | | | | | | |
| | 6 2 | | | | | | | |
| | 5 2 | | | | | | | |
| | 6 2 | | | | | | | |
| | 5 2 | | | | | | | |
| | 5 2 | | | | | | | |
| | 5 2 | | | | | | | |
| | 5 2 | | | | | | | |
| | 6 2 | | | | | | | |
| | 6 2 | | | | | | | |
| | 5 2 | | | | | | | |
| 62.2.195 | 237728 | 128 | Cd1Pt1Sr1 | | | | | |
| | -GM | -R | -S | -T | -U | -X | -Y | -Z |
| | 5 2 | 3+3 4 | 3+3 4 | 3+4 4 | 6+6 4 | 3+4 4 | 3+4 4 | 3+4 4 |
| | 5 2 | 4+4 4 | 4+4 4 | 3+4 4 | 5+5 4 | 3+4 4 | 3+4 4 | 3+4 4 |
| | 6 2 | 4+4 4 | 3+3 4 | 3+4 4 | 6+6 4 | 3+4 4 | 3+4 4 | 3+4 4 |
| | 6 2 | 3+3 4 | 4+4 4 | 3+4 4 | 5+5 4 | 3+4 4 | 3+4 4 | 3+4 4 |
| | 5 2 | 3+3 4 | 3+3 4 | 3+4 4 | 6+6 4 | 3+4 4 | 3+4 4 | 3+4 4 |
| | 5 2 | 4+4 4 | 4+4 4 | 3+4 4 | 5+5 4 | 3+4 4 | 3+4 4 | 3+4 4 |
| | 6 2 | 4+4 4 | 4+4 4 | 3+4 4 | 6+6 4 | 3+4 4 | 3+4 4 | 3+4 4 |
| | 6 2 | 3+3 4 | 3+3 4 | 3+4 4 | 5+5 4 | 3+4 4 | 3+4 4 | 3+4 4 |
| | 5 2 | 4+4 4 | 3+3 4 | 3+4 4 | 6+6 4 | 3+4 4 | 3+4 4 | 3+4 4 |
| | 5 2 | 3+3 4 | 4+4 4 | 3+4 4 | 5+5 4 | 3+4 4 | 3+4 4 | 3+4 4 |
| | 6 2 | 3+3 4 | 3+3 4 | 3+4 4 | 5+5 4 | 3+4 4 | 3+4 4 | 3+4 4 |
| | 5 2 | 4+4 4 | 4+4 4 | 3+4 4 | 6+6 4 | 3+4 4 | 3+4 4 | 3+4 4 |
| | 6 2 | 3+3 4 | 4+4 4 | 3+4 4 | 6+6 4 | 3+4 4 | 3+4 4 | 3+4 4 |
| | 6 2 | 4+4 4 | 3+3 4 | 3+4 4 | 5+5 4 | 3+4 4 | 3+4 4 | 3+4 4 |
| | 5 2 | 3+3 4 | 3+3 4 | 3+4 4 | 5+5 4 | 3+4 4 | 3+4 4 | 3+4 4 |
| | 6 2 | 4+4 4 | 4+4 4 | 3+4 4 | 6+6 4 | 3+4 4 | 3+4 4 | 3+4 4 |
| | 6 2 | 3+3 4 | 3+3 4 | 3+4 4 | 5+5 4 | 3+4 4 | 3+4 4 | 3+4 4 |
| | 6 2 | 4+4 4 | 4+4 4 | 3+4 4 | 6+6 4 | 3+4 4 | 3+4 4 | 3+4 4 |
| | 5 2 | 4+4 4 | 4+4 4 | 3+4 4 | 5+5 4 | 3+4 4 | 3+4 4 | 3+4 4 |
| | 5 2 | 3+3 4 | 3+3 4 | 3+4 4 | 6+6 4 | 3+4 4 | 3+4 4 | 3+4 4 |
| | 5 2 | 3+3 4 | 4+4 4 | 3+4 4 | 6+6 4 | 3+4 4 | 3+4 4 | 3+4 4 |
| | 6 2 | 4+4 4 | 3+3 4 | 3+4 4 | 5+5 4 | 3+4 4 | 3+4 4 | 3+4 4 |
| | 5 2 | 3+3 4 | 3+3 4 | 3+4 4 | 5+5 4 | 3+4 4 | 3+4 4 | 3+4 4 |
| | 6 2 | 4+4 4 | 4+4 4 | 3+4 4 | 6+6 4 | 3+4 4 | 3+4 4 | 3+4 4 |
| | 6 2 | 4+4 4 | 3+3 4 | 3+4 4 | 5+5 4 | 3+4 4 | 3+4 4 | 3+4 4 |
| | 6 2 | 3+3 4 | 4+4 4 | 3+4 4 | 6+6 4 | 3+4 4 | 3+4 4 | 3+4 4 |
| | 5 2 | 3+3 4 | 3+3 4 | 3+4 4 | 5+5 4 | 3+4 4 | 3+4 4 | 3+4 4 |
| | 5 2 | 4+4 4 | 4+4 4 | 3+4 4 | 6+6 4 | 3+4 4 | 3+4 4 | 3+4 4 |
| | 5 2 | 4+4 4 | 4+4 4 | 3+4 4 | 5+5 4 | 3+4 4 | 3+4 4 | 3+4 4 |
| | 5 2 | 3+3 4 | 3+3 4 | 3+4 4 | 6+6 4 | 3+4 4 | 3+4 4 | 3+4 4 |
| | 6 2 | 4+4 4 | 4+4 4 | 3+4 4 | 5+5 4 | 3+4 4 | 3+4 4 | 3+4 4 |
| | 6 2 | 3+3 4 | 3+3 4 | 3+4 4 | 6+6 4 | 3+4 4 | 3+4 4 | 3+4 4 |
| | 5 2 | | | | | | | |
| | 5 2 | | | | | | | |
| | 6 2 | | | | | | | |

Continued on next column

Continued on next column

**TABLE XXXVII– continued**

| TQC | ICSD | NoE | SF | | | | | |
|---|---|---|---|---|---|---|---|---|
| | 6 2 | | | | | | | |
| | 5 2 | | | | | | | |
| | 5 2 | | | | | | | |
| | 6 2 | | | | | | | |
| | 6 2 | | | | | | | |
| | 6 2 | | | | | | | |
| | 5 2 | | | | | | | |
| | 6 2 | | | | | | | |
| | 5 2 | | | | | | | |
| | 5 2 | | | | | | | |
| | 6 2 | | | | | | | |
| | 5 2 | | | | | | | |
| | 6 2 | | | | | | | |
| | 6 2 | | | | | | | |
| | 5 2 | | | | | | | |
| | 5 2 | | | | | | | |
| | 6 2 | | | | | | | |
| | 5 2 | | | | | | | |
| | 5 2 | | | | | | | |
| | 6 2 | | | | | | | |
| | 5 2 | | | | | | | |
| | 6 2 | | | | | | | |
| | 5 2 | | | | | | | |
| | 6 2 | | | | | | | |
| | 6 2 | | | | | | | |
| | 5 2 | | | | | | | |
| | 6 2 | | | | | | | |
| | 5 2 | | | | | | | |
| | 5 2 | | | | | | | |
| 62.2.196 | 12159 | 72 | Cd1Na1Sb1 | | | | | |
| | -GM | -R | -S | -T | -U | -X | -Y | -Z |
| | 5 2 | 4+4 4 | 4+4 4 | 3+4 4 | 5+5 4 | 3+4 4 | 3+4 4 | 3+4 4 |
| | 6 2 | 3+3 4 | 3+3 4 | 3+4 4 | 6+6 4 | 3+4 4 | 3+4 4 | 3+4 4 |
| | 6 2 | 3+3 4 | 3+3 4 | 3+4 4 | 5+5 4 | 3+4 4 | 3+4 4 | 3+4 4 |
| | 5 2 | 4+4 4 | 4+4 4 | 3+4 4 | 6+6 4 | 3+4 4 | 3+4 4 | 3+4 4 |
| | 6 2 | 3+3 4 | 3+3 4 | 3+4 4 | 6+6 4 | 3+4 4 | 3+4 4 | 3+4 4 |
| | 5 2 | 4+4 4 | 4+4 4 | 3+4 4 | 5+5 4 | 3+4 4 | 3+4 4 | 3+4 4 |
| | 6 2 | 4+4 4 | 4+4 4 | 3+4 4 | 6+6 4 | 3+4 4 | 3+4 4 | 3+4 4 |
| | 5 2 | 3+3 4 | 3+3 4 | 3+4 4 | 5+5 4 | 3+4 4 | 3+4 4 | 3+4 4 |
| | 5 2 | 3+3 4 | 4+4 4 | 3+4 4 | 6+6 4 | 3+4 4 | 3+4 4 | 3+4 4 |
| | 6 2 | 4+4 4 | 3+3 4 | 3+4 4 | 5+5 4 | 3+4 4 | 3+4 4 | 3+4 4 |
| | 5 2 | 4+4 4 | 4+4 4 | 3+4 4 | 6+6 4 | 3+4 4 | 3+4 4 | 3+4 4 |
| | 6 2 | 3+3 4 | 3+3 4 | 3+4 4 | 5+5 4 | 3+4 4 | 3+4 4 | 3+4 4 |
| | 5 2 | 3+3 4 | 4+4 4 | 3+4 4 | 6+6 4 | 3+4 4 | 3+4 4 | 3+4 4 |
| | 6 2 | 4+4 4 | 3+3 4 | 3+4 4 | 5+5 4 | 3+4 4 | 3+4 4 | 3+4 4 |
| | 5 2 | 4+4 4 | 3+3 4 | 3+4 4 | 6+6 4 | 3+4 4 | 3+4 4 | 3+4 4 |
| | 5 2 | 3+3 4 | 4+4 4 | 3+4 4 | 6+6 4 | 3+4 4 | 3+4 4 | 3+4 4 |
| | 5 2 | 4+4 4 | 4+4 4 | 3+4 4 | 5+5 4 | 3+4 4 | 3+4 4 | 3+4 4 |
| | 6 2 | 3+3 4 | 3+3 4 | 3+4 4 | 5+5 4 | 3+4 4 | 3+4 4 | 3+4 4 |
| | 6 2 | | | | | | | |
| | 5 2 | | | | | | | |
| | 6 2 | | | | | | | |
| | 5 2 | | | | | | | |
| | 6 2 | | | | | | | |
| | 6 2 | | | | | | | |
| | 5 2 | | | | | | | |
| | 6 2 | | | | | | | |
| | 5 2 | | | | | | | |
| | 5 2 | | | | | | | |
| | 6 2 | | | | | | | |
| | 5 2 | | | | | | | |

Continued on next column

**TABLE XXXVII– continued**

| TQC | ICSD | NoE | SF | | | | | |
|---|---|---|---|---|---|---|---|---|
| | 5 2 | | | | | | | |
| | 6 2 | | | | | | | |
| | 5 2 | | | | | | | |
| | 6 2 | | | | | | | |
| | 6 2 | | | | | | | |
| | 5 2 | | | | | | | |
| 62.2.197 | 72642 | 64 | Ca1Pt1Si1 | | | | | |
| | -GM | -R | -S | -T | -U | -X | -Y | -Z |
| | 5 2 | 4+4 4 | 4+4 4 | 3+4 4 | 5+5 4 | 3+4 4 | 3+4 4 | 3+4 4 |
| | 5 2 | 3+3 4 | 3+3 4 | 3+4 4 | 6+6 4 | 3+4 4 | 3+4 4 | 3+4 4 |
| | 6 2 | 3+3 4 | 4+4 4 | 3+4 4 | 6+6 4 | 3+4 4 | 3+4 4 | 3+4 4 |
| | 6 2 | 4+4 4 | 3+3 4 | 3+4 4 | 5+5 4 | 3+4 4 | 3+4 4 | 3+4 4 |
| | 6 2 | 4+4 4 | 3+3 4 | 3+4 4 | 5+5 4 | 3+4 4 | 3+4 4 | 3+4 4 |
| | 5 2 | 3+3 4 | 4+4 4 | 3+4 4 | 6+6 4 | 3+4 4 | 3+4 4 | 3+4 4 |
| | 6 2 | 4+4 4 | 4+4 4 | 3+4 4 | 5+5 4 | 3+4 4 | 3+4 4 | 3+4 4 |
| | 5 2 | 3+3 4 | 3+3 4 | 3+4 4 | 6+6 4 | 3+4 4 | 3+4 4 | 3+4 4 |
| | 6 2 | 3+3 4 | 4+4 4 | 3+4 4 | 5+5 4 | 3+4 4 | 3+4 4 | 3+4 4 |
| | 5 2 | 4+4 4 | 3+3 4 | 3+4 4 | 6+6 4 | 3+4 4 | 3+4 4 | 3+4 4 |
| | 5 2 | 4+4 4 | 3+3 4 | 3+4 4 | 5+5 4 | 3+4 4 | 3+4 4 | 3+4 4 |
| | 5 2 | 3+3 4 | 4+4 4 | 3+4 4 | 6+6 4 | 3+4 4 | 3+4 4 | 3+4 4 |
| | 6 2 | 3+3 4 | 3+3 4 | 3+4 4 | 5+5 4 | 3+4 4 | 3+4 4 | 3+4 4 |
| | 5 2 | 4+4 4 | 4+4 4 | 3+4 4 | 6+6 4 | 3+4 4 | 3+4 4 | 3+4 4 |
| | 6 2 | 3+3 4 | 3+3 4 | 3+4 4 | 5+5 4 | 3+4 4 | 3+4 4 | 3+4 4 |
| | 5 2 | 4+4 4 | 4+4 4 | 3+4 4 | 5+5 4 | 3+4 4 | 3+4 4 | 3+4 4 |
| | 6 2 | | | | | | | |
| | 5 2 | | | | | | | |
| | 6 2 | | | | | | | |
| | 5 2 | | | | | | | |
| | 6 2 | | | | | | | |
| | 5 2 | | | | | | | |
| | 6 2 | | | | | | | |
| | 6 2 | | | | | | | |
| | 5 2 | | | | | | | |
| | 6 2 | | | | | | | |
| | 5 2 | | | | | | | |
| | 5 2 | | | | | | | |
| | 5 2 | | | | | | | |
| | 6 2 | | | | | | | |
| | 6 2 | | | | | | | |
| | 5 2 | | | | | | | |
| 62.2.198 | 426616 | 56 | Ca1Mg1Pd1 | | | | | |
| | -GM | -R | -S | -T | -U | -X | -Y | -Z |
| | 5 2 | 3+3 4 | 4+4 4 | 3+4 4 | 5+5 4 | 3+4 4 | 3+4 4 | 3+4 4 |
| | 5 2 | 4+4 4 | 3+3 4 | 3+4 4 | 6+6 4 | 3+4 4 | 3+4 4 | 3+4 4 |
| | 6 2 | 4+4 4 | 4+4 4 | 3+4 4 | 6+6 4 | 3+4 4 | 3+4 4 | 3+4 4 |
| | 6 2 | 3+3 4 | 3+3 4 | 3+4 4 | 5+5 4 | 3+4 4 | 3+4 4 | 3+4 4 |
| | 6 2 | 4+4 4 | 3+3 4 | 3+4 4 | 6+6 4 | 3+4 4 | 3+4 4 | 3+4 4 |
| | 5 2 | 3+3 4 | 4+4 4 | 3+4 4 | 5+5 4 | 3+4 4 | 3+4 4 | 3+4 4 |
| | 6 2 | 3+3 4 | 3+3 4 | 3+4 4 | 5+5 4 | 3+4 4 | 3+4 4 | 3+4 4 |
| | 5 2 | 4+4 4 | 4+4 4 | 3+4 4 | 6+6 4 | 3+4 4 | 3+4 4 | 3+4 4 |
| | 5 2 | 3+3 4 | 4+4 4 | 3+4 4 | 5+5 4 | 3+4 4 | 3+4 4 | 3+4 4 |
| | 6 2 | 4+4 4 | 3+3 4 | 3+4 4 | 6+6 4 | 3+4 4 | 3+4 4 | 3+4 4 |
| | 5 2 | 3+3 4 | 4+4 4 | 3+4 4 | 5+5 4 | 3+4 4 | 3+4 4 | 3+4 4 |
| | 5 2 | 4+4 4 | 3+3 4 | 3+4 4 | 6+6 4 | 3+4 4 | 3+4 4 | 3+4 4 |
| | 6 2 | 4+4 4 | 3+3 4 | 3+4 4 | 5+5 4 | 3+4 4 | 3+4 4 | 3+4 4 |
| | 5 2 | 3+3 4 | 4+4 4 | 3+4 4 | 6+6 4 | 3+4 4 | 3+4 4 | 3+4 4 |
| | 5 2 | | | | | | | |
| | 6 2 | | | | | | | |
| | 5 2 | | | | | | | |
| | 6 2 | | | | | | | |
| | 5 2 | | | | | | | |

Continued on next column

**TABLE XXXVII– continued**

| TQC | ICSD | NoE | SF | | | | | |
|---|---|---|---|---|---|---|---|---|
| | 6 2 | | | | | | | |
| | 5 2 | | | | | | | |
| | 6 2 | | | | | | | |
| | 6 2 | | | | | | | |
| | 5 2 | | | | | | | |
| | 6 2 | | | | | | | |
| | 6 2 | | | | | | | |
| | 5 2 | | | | | | | |
| | 5 2 | | | | | | | |
| 62.2.199 | 410841 | 56 | Ca1In1Rh1 | | | | | |
| | -GM | -R | -S | -T | -U | -X | -Y | -Z |
| | 5 2 | 4+4 4 | 4+4 4 | 3+4 4 | 5+5 4 | 3+4 4 | 3+4 4 | 3+4 4 |
| | 5 2 | 3+3 4 | 3+3 4 | 3+4 4 | 6+6 4 | 3+4 4 | 3+4 4 | 3+4 4 |
| | 6 2 | 4+4 4 | 4+4 4 | 3+4 4 | 6+6 4 | 3+4 4 | 3+4 4 | 3+4 4 |
| | 6 2 | 3+3 4 | 3+3 4 | 3+4 4 | 5+5 4 | 3+4 4 | 3+4 4 | 3+4 4 |
| | 6 2 | 4+4 4 | 3+3 4 | 3+4 4 | 5+5 4 | 3+4 4 | 3+4 4 | 3+4 4 |
| | 6 2 | 3+3 4 | 4+4 4 | 3+4 4 | 5+5 4 | 3+4 4 | 3+4 4 | 3+4 4 |
| | 5 2 | 4+4 4 | 3+3 4 | 3+4 4 | 6+6 4 | 3+4 4 | 3+4 4 | 3+4 4 |
| | 5 2 | 3+3 4 | 4+4 4 | 3+4 4 | 6+6 4 | 3+4 4 | 3+4 4 | 3+4 4 |
| | 5 2 | 3+3 4 | 4+4 4 | 3+4 4 | 5+5 4 | 3+4 4 | 3+4 4 | 3+4 4 |
| | 5 2 | 4+4 4 | 3+3 4 | 3+4 4 | 6+6 4 | 3+4 4 | 3+4 4 | 3+4 4 |
| | 5 2 | 3+3 4 | 4+4 4 | 3+4 4 | 5+5 4 | 3+4 4 | 3+4 4 | 3+4 4 |
| | 6 2 | 4+4 4 | 3+3 4 | 3+4 4 | 6+6 4 | 3+4 4 | 3+4 4 | 3+4 4 |
| | 5 2 | 4+4 4 | 4+4 4 | 3+4 4 | 5+5 4 | 3+4 4 | 3+4 4 | 3+4 4 |
| | 6 2 | 3+3 4 | 3+3 4 | 3+4 4 | 6+6 4 | 3+4 4 | 3+4 4 | 3+4 4 |
| | 5 2 | | | | | | | |
| | 5 2 | | | | | | | |
| | 6 2 | | | | | | | |
| | 5 2 | | | | | | | |
| | 6 2 | | | | | | | |
| | 6 2 | | | | | | | |
| | 6 2 | | | | | | | |
| | 5 2 | | | | | | | |
| | 6 2 | | | | | | | |
| | 5 2 | | | | | | | |
| | 6 2 | | | | | | | |
| | 6 2 | | | | | | | |
| | 5 2 | | | | | | | |
| | 5 2 | | | | | | | |
| 62.2.200 | 106343 | 64 | Ca1Ge1Pt1 | | | | | |
| | -GM | -R | -S | -T | -U | -X | -Y | -Z |
| | 5 2 | 3+3 4 | 4+4 4 | 3+4 4 | 5+5 4 | 3+4 4 | 3+4 4 | 3+4 4 |
| | 5 2 | 4+4 4 | 3+3 4 | 3+4 4 | 6+6 4 | 3+4 4 | 3+4 4 | 3+4 4 |
| | 6 2 | 4+4 4 | 3+3 4 | 3+4 4 | 6+6 4 | 3+4 4 | 3+4 4 | 3+4 4 |
| | 6 2 | 3+3 4 | 4+4 4 | 3+4 4 | 5+5 4 | 3+4 4 | 3+4 4 | 3+4 4 |
| | 6 2 | 3+3 4 | 4+4 4 | 3+4 4 | 5+5 4 | 3+4 4 | 3+4 4 | 3+4 4 |
| | 5 2 | 4+4 4 | 3+3 4 | 3+4 4 | 6+6 4 | 3+4 4 | 3+4 4 | 3+4 4 |
| | 6 2 | 3+3 4 | 4+4 4 | 3+4 4 | 5+5 4 | 3+4 4 | 3+4 4 | 3+4 4 |
| | 5 2 | 4+4 4 | 3+3 4 | 3+4 4 | 6+6 4 | 3+4 4 | 3+4 4 | 3+4 4 |
| | 6 2 | 4+4 4 | 4+4 4 | 3+4 4 | 5+5 4 | 3+4 4 | 3+4 4 | 3+4 4 |
| | 5 2 | 3+3 4 | 3+3 4 | 3+4 4 | 6+6 4 | 3+4 4 | 3+4 4 | 3+4 4 |
| | 5 2 | 4+4 4 | 3+3 4 | 3+4 4 | 5+5 4 | 3+4 4 | 3+4 4 | 3+4 4 |
| | 5 2 | 3+3 4 | 4+4 4 | 3+4 4 | 6+6 4 | 3+4 4 | 3+4 4 | 3+4 4 |
| | 6 2 | 3+3 4 | 3+3 4 | 3+4 4 | 5+5 4 | 3+4 4 | 3+4 4 | 3+4 4 |
| | 5 2 | 4+4 4 | 4+4 4 | 3+4 4 | 6+6 4 | 3+4 4 | 3+4 4 | 3+4 4 |
| | 6 2 | 4+4 4 | 3+3 4 | 3+4 4 | 5+5 4 | 3+4 4 | 3+4 4 | 3+4 4 |
| | 5 2 | 3+3 4 | 4+4 4 | 3+4 4 | 5+5 4 | 3+4 4 | 3+4 4 | 3+4 4 |
| | 6 2 | | | | | | | |
| | 5 2 | | | | | | | |
| | 6 2 | | | | | | | |
| | 6 2 | | | | | | | |

Continued on next column

**TABLE XXXVII– continued**

| TQC | ICSD | NoE | SF | | | | | |
|---|---|---|---|---|---|---|---|---|
| | 5 2 | | | | | | | |
| | 5 2 | | | | | | | |
| | 6 2 | | | | | | | |
| | 6 2 | | | | | | | |
| | 5 2 | | | | | | | |
| | 5 2 | | | | | | | |
| | 6 2 | | | | | | | |
| | 5 2 | | | | | | | |
| | 5 2 | | | | | | | |
| | 6 2 | | | | | | | |
| | 6 2 | | | | | | | |
| | 5 2 | | | | | | | |
| 62.2.201 | 291004 | 200 | Ca1Fe2O4 | | | | | |
| | -GM | -R | -S | -T | -U | -X | -Y | -Z |
| | 5 2 | 3+3 4 | 3+3 4 | 3+4 4 | 5+5 4 | 3+4 4 | 3+4 4 | 3+4 4 |
| | 5 2 | 4+4 4 | 4+4 4 | 3+4 4 | 6+6 4 | 3+4 4 | 3+4 4 | 3+4 4 |
| | 6 2 | 3+3 4 | 3+3 4 | 3+4 4 | 6+6 4 | 3+4 4 | 3+4 4 | 3+4 4 |
| | 6 2 | 4+4 4 | 4+4 4 | 3+4 4 | 5+5 4 | 3+4 4 | 3+4 4 | 3+4 4 |
| | 5 2 | 3+3 4 | 3+3 4 | 3+4 4 | 6+6 4 | 3+4 4 | 3+4 4 | 3+4 4 |
| | 6 2 | 4+4 4 | 4+4 4 | 3+4 4 | 5+5 4 | 3+4 4 | 3+4 4 | 3+4 4 |
| | 5 2 | 3+3 4 | 3+3 4 | 3+4 4 | 5+5 4 | 3+4 4 | 3+4 4 | 3+4 4 |
| | 6 2 | 4+4 4 | 4+4 4 | 3+4 4 | 6+6 4 | 3+4 4 | 3+4 4 | 3+4 4 |
| | 5 2 | 3+3 4 | 3+3 4 | 3+4 4 | 6+6 4 | 3+4 4 | 3+4 4 | 3+4 4 |
| | 6 2 | 4+4 4 | 4+4 4 | 3+4 4 | 5+5 4 | 3+4 4 | 3+4 4 | 3+4 4 |
| | 5 2 | 4+4 4 | 3+3 4 | 3+4 4 | 5+5 4 | 3+4 4 | 3+4 4 | 3+4 4 |
| | 6 2 | 3+3 4 | 4+4 4 | 3+4 4 | 6+6 4 | 3+4 4 | 3+4 4 | 3+4 4 |
| | 5 2 | 3+3 4 | 3+3 4 | 3+4 4 | 6+6 4 | 3+4 4 | 3+4 4 | 3+4 4 |
| | 5 2 | 4+4 4 | 4+4 4 | 3+4 4 | 5+5 4 | 3+4 4 | 3+4 4 | 3+4 4 |
| | 6 2 | 3+3 4 | 3+3 4 | 3+4 4 | 6+6 4 | 3+4 4 | 3+4 4 | 3+4 4 |
| | 6 2 | 4+4 4 | 4+4 4 | 3+4 4 | 5+5 4 | 3+4 4 | 3+4 4 | 3+4 4 |
| | 5 2 | 3+3 4 | 3+3 4 | 3+4 4 | 6+6 4 | 3+4 4 | 3+4 4 | 3+4 4 |
| | 6 2 | 4+4 4 | 4+4 4 | 3+4 4 | 5+5 4 | 3+4 4 | 3+4 4 | 3+4 4 |
| | 5 2 | 3+3 4 | 3+3 4 | 3+4 4 | 5+5 4 | 3+4 4 | 3+4 4 | 3+4 4 |
| | 5 2 | 4+4 4 | 4+4 4 | 3+4 4 | 6+6 4 | 3+4 4 | 3+4 4 | 3+4 4 |
| | 6 2 | 3+3 4 | 4+4 4 | 3+4 4 | 6+6 4 | 3+4 4 | 3+4 4 | 3+4 4 |
| | 5 2 | 4+4 4 | 3+3 4 | 3+4 4 | 5+5 4 | 3+4 4 | 3+4 4 | 3+4 4 |
| | 6 2 | 4+4 4 | 3+3 4 | 3+4 4 | 5+5 4 | 3+4 4 | 3+4 4 | 3+4 4 |
| | 5 2 | 3+3 4 | 4+4 4 | 3+4 4 | 6+6 4 | 3+4 4 | 3+4 4 | 3+4 4 |
| | 6 2 | 3+3 4 | 3+3 4 | 3+4 4 | 5+5 4 | 3+4 4 | 3+4 4 | 3+4 4 |
| | 5 2 | 4+4 4 | 4+4 4 | 3+4 4 | 6+6 4 | 3+4 4 | 3+4 4 | 3+4 4 |
| | 5 2 | 3+3 4 | 3+3 4 | 3+4 4 | 5+5 4 | 3+4 4 | 3+4 4 | 3+4 4 |
| | 6 2 | 4+4 4 | 4+4 4 | 3+4 4 | 6+6 4 | 3+4 4 | 3+4 4 | 3+4 4 |
| | 6 2 | 3+3 4 | 3+3 4 | 3+4 4 | 5+5 4 | 3+4 4 | 3+4 4 | 3+4 4 |
| | 5 2 | 4+4 4 | 4+4 4 | 3+4 4 | 6+6 4 | 3+4 4 | 3+4 4 | 3+4 4 |
| | 6 2 | 4+4 4 | 4+4 4 | 3+4 4 | 6+6 4 | 3+4 4 | 3+4 4 | 3+4 4 |
| | 6 2 | 3+3 4 | 3+3 4 | 3+4 4 | 6+6 4 | 3+4 4 | 3+4 4 | 3+4 4 |
| | 6 2 | 3+3 4 | 3+3 4 | 3+4 4 | 5+5 4 | 3+4 4 | 3+4 4 | 3+4 4 |
| | 6 2 | 4+4 4 | 4+4 4 | 3+4 4 | 5+5 4 | 3+4 4 | 3+4 4 | 3+4 4 |
| | 5 2 | 3+3 4 | 4+4 4 | 3+4 4 | 6+6 4 | 3+4 4 | 3+4 4 | 3+4 4 |
| | 5 2 | 4+4 4 | 3+3 4 | 3+4 4 | 5+5 4 | 3+4 4 | 3+4 4 | 3+4 4 |
| | 5 2 | 3+3 4 | 4+4 4 | 3+4 4 | 6+6 4 | 3+4 4 | 3+4 4 | 3+4 4 |
| | 6 2 | 4+4 4 | 3+3 4 | 3+4 4 | 5+5 4 | 3+4 4 | 3+4 4 | 3+4 4 |
| | 5 2 | 3+3 4 | 3+3 4 | 3+4 4 | 6+6 4 | 3+4 4 | 3+4 4 | 3+4 4 |
| | 6 2 | 4+4 4 | 4+4 4 | 3+4 4 | 5+5 4 | 3+4 4 | 3+4 4 | 3+4 4 |
| | 5 2 | 3+3 4 | 3+3 4 | 3+4 4 | 5+5 4 | 3+4 4 | 3+4 4 | 3+4 4 |
| | 5 2 | 4+4 4 | 4+4 4 | 3+4 4 | 6+6 4 | 3+4 4 | 3+4 4 | 3+4 4 |
| | 5 2 | 4+4 4 | 4+4 4 | 3+4 4 | 5+5 4 | 3+4 4 | 3+4 4 | 3+4 4 |
| | 6 2 | 3+3 4 | 3+3 4 | 3+4 4 | 6+6 4 | 3+4 4 | 3+4 4 | 3+4 4 |
| | 6 2 | 4+4 4 | 3+3 4 | 3+4 4 | 6+6 4 | 3+4 4 | 3+4 4 | 3+4 4 |
| | 5 2 | 3+3 4 | 4+4 4 | 3+4 4 | 5+5 4 | 3+4 4 | 3+4 4 | 3+4 4 |
| | 5 2 | 4+4 4 | 4+4 4 | 3+4 4 | 5+5 4 | 3+4 4 | 3+4 4 | 3+4 4 |

Continued on next column

**TABLE XXXVII– continued**

| TQC | ICSD | NoE | SF | | | | | |
|---|---|---|---|---|---|---|---|---|
| | 5 2 | 3+3 4 | 3+3 4 | 3+4 4 | 6+6 4 | 3+4 4 | 3+4 4 | 3+4 4 |
| | 5 2 | 3+3 4 | 4+4 4 | 3+4 4 | 5+5 4 | 3+4 4 | 3+4 4 | 3+4 4 |
| | 6 2 | 4+4 4 | 3+3 4 | 3+4 4 | 6+6 4 | 3+4 4 | 3+4 4 | 3+4 4 |
| | 6 2 | | | | | | | |
| | 5 2 | | | | | | | |
| | 6 2 | | | | | | | |
| | 5 2 | | | | | | | |
| | 6 2 | | | | | | | |
| | 5 2 | | | | | | | |
| | 6 2 | | | | | | | |
| | 6 2 | | | | | | | |
| | 6 2 | | | | | | | |
| | 5 2 | | | | | | | |
| | 5 2 | | | | | | | |
| | 5 2 | | | | | | | |
| | 6 2 | | | | | | | |
| | 5 2 | | | | | | | |
| | 5 2 | | | | | | | |
| | 5 2 | | | | | | | |
| | 5 2 | | | | | | | |
| | 6 2 | | | | | | | |
| | 6 2 | | | | | | | |
| | 5 2 | | | | | | | |
| | 6 2 | | | | | | | |
| | 5 2 | | | | | | | |
| | 6 2 | | | | | | | |
| | 5 2 | | | | | | | |
| | 6 2 | | | | | | | |
| | 6 2 | | | | | | | |
| | 6 2 | | | | | | | |
| | 6 2 | | | | | | | |
| | 6 2 | | | | | | | |
| | 6 2 | | | | | | | |
| | 5 2 | | | | | | | |
| | 6 2 | | | | | | | |
| | 5 2 | | | | | | | |
| | 6 2 | | | | | | | |
| | 6 2 | | | | | | | |
| | 6 2 | | | | | | | |
| | 5 2 | | | | | | | |
| | 5 2 | | | | | | | |
| | 5 2 | | | | | | | |
| | 5 2 | | | | | | | |
| | 6 2 | | | | | | | |
| | 6 2 | | | | | | | |
| | 6 2 | | | | | | | |
| | 5 2 | | | | | | | |
| | 5 2 | | | | | | | |
| | 6 2 | | | | | | | |
| | 5 2 | | | | | | | |
| | 6 2 | | | | | | | |
| | 6 2 | | | | | | | |
| | 6 2 | | | | | | | |
| 62.2.202 | 426610 | 96 | Ca1Cd1Pt1 | | | | | |
| | -GM | -R | -S | -T | -U | -X | -Y | -Z |
| | 6 2 | 4+4 4 | 4+4 4 | 3+4 4 | 5+5 4 | 3+4 4 | 3+4 4 | 3+4 4 |
| | 5 2 | 3+3 4 | 3+3 4 | 3+4 4 | 6+6 4 | 3+4 4 | 3+4 4 | 3+4 4 |
| | 6 2 | 3+3 4 | 3+3 4 | 3+4 4 | 5+5 4 | 3+4 4 | 3+4 4 | 3+4 4 |
| | 5 2 | 4+4 4 | 4+4 4 | 3+4 4 | 6+6 4 | 3+4 4 | 3+4 4 | 3+4 4 |
| | 5 2 | 3+3 4 | 3+3 4 | 3+4 4 | 5+5 4 | 3+4 4 | 3+4 4 | 3+4 4 |
| | 6 2 | 4+4 4 | 4+4 4 | 3+4 4 | 6+6 4 | 3+4 4 | 3+4 4 | 3+4 4 |

Continued on next column

**TABLE XXXVII– continued**

| TQC | ICSD | NoE | SF | | | | | |
|---|---|---|---|---|---|---|---|---|
| | 6 2 | 3+3 4 | 4+4 4 | 3+4 4 | 6+6 4 | 3+4 4 | 3+4 4 | 3+4 4 |
| | 5 2 | 4+4 4 | 3+3 4 | 3+4 4 | 5+5 4 | 3+4 4 | 3+4 4 | 3+4 4 |
| | 6 2 | 4+4 4 | 4+4 4 | 3+4 4 | 6+6 4 | 3+4 4 | 3+4 4 | 3+4 4 |
| | 6 2 | 3+3 4 | 3+3 4 | 3+4 4 | 5+5 4 | 3+4 4 | 3+4 4 | 3+4 4 |
| | 5 2 | 4+4 4 | 3+3 4 | 3+4 4 | 6+6 4 | 3+4 4 | 3+4 4 | 3+4 4 |
| | 5 2 | 3+3 4 | 4+4 4 | 3+4 4 | 5+5 4 | 3+4 4 | 3+4 4 | 3+4 4 |
| | 5 2 | 3+3 4 | 3+3 4 | 3+4 4 | 5+5 4 | 3+4 4 | 3+4 4 | 3+4 4 |
| | 5 2 | 4+4 4 | 4+4 4 | 3+4 4 | 6+6 4 | 3+4 4 | 3+4 4 | 3+4 4 |
| | 6 2 | 3+3 4 | 4+4 4 | 3+4 4 | 6+6 4 | 3+4 4 | 3+4 4 | 3+4 4 |
| | 6 2 | 4+4 4 | 3+3 4 | 3+4 4 | 5+5 4 | 3+4 4 | 3+4 4 | 3+4 4 |
| | 5 2 | 4+4 4 | 4+4 4 | 3+4 4 | 6+6 4 | 3+4 4 | 3+4 4 | 3+4 4 |
| | 5 2 | 3+3 4 | 3+3 4 | 3+4 4 | 5+5 4 | 3+4 4 | 3+4 4 | 3+4 4 |
| | 6 2 | 3+3 4 | 3+3 4 | 3+4 4 | 6+6 4 | 3+4 4 | 3+4 4 | 3+4 4 |
| | 6 2 | 4+4 4 | 4+4 4 | 3+4 4 | 5+5 4 | 3+4 4 | 3+4 4 | 3+4 4 |
| | 5 2 | 4+4 4 | 3+3 4 | 3+4 4 | 6+6 4 | 3+4 4 | 3+4 4 | 3+4 4 |
| | 5 2 | 3+3 4 | 4+4 4 | 3+4 4 | 5+5 4 | 3+4 4 | 3+4 4 | 3+4 4 |
| | 6 2 | 3+3 4 | 3+3 4 | 3+4 4 | 6+6 4 | 3+4 4 | 3+4 4 | 3+4 4 |
| | 6 2 | 4+4 4 | 4+4 4 | 3+4 4 | 5+5 4 | 3+4 4 | 3+4 4 | 3+4 4 |
| | 6 2 | | | | | | | |
| | 5 2 | | | | | | | |
| | 6 2 | | | | | | | |
| | 5 2 | | | | | | | |
| | 5 2 | | | | | | | |
| | 5 2 | | | | | | | |
| | 6 2 | | | | | | | |
| | 6 2 | | | | | | | |
| | 5 2 | | | | | | | |
| | 6 2 | | | | | | | |
| | 5 2 | | | | | | | |
| | 6 2 | | | | | | | |
| | 5 2 | | | | | | | |
| | 5 2 | | | | | | | |
| | 5 2 | | | | | | | |
| | 6 2 | | | | | | | |
| | 6 2 | | | | | | | |
| | 5 2 | | | | | | | |
| | 5 2 | | | | | | | |
| | 6 2 | | | | | | | |
| | 6 2 | | | | | | | |
| | 6 2 | | | | | | | |
| | 5 2 | | | | | | | |
| | 5 2 | | | | | | | |
| 62.2.203 | 426611 | 96 | Ca1Cd1Pd1 | | | | | |
| | -GM | -R | -S | -T | -U | -X | -Y | -Z |
| | 6 2 | 3+3 4 | 3+3 4 | 3+4 4 | 5+5 4 | 3+4 4 | 3+4 4 | 3+4 4 |
| | 5 2 | 4+4 4 | 4+4 4 | 3+4 4 | 6+6 4 | 3+4 4 | 3+4 4 | 3+4 4 |
| | 5 2 | 3+3 4 | 3+3 4 | 3+4 4 | 6+6 4 | 3+4 4 | 3+4 4 | 3+4 4 |
| | 6 2 | 4+4 4 | 4+4 4 | 3+4 4 | 5+5 4 | 3+4 4 | 3+4 4 | 3+4 4 |
| | 5 2 | 4+4 4 | 4+4 4 | 3+4 4 | 6+6 4 | 3+4 4 | 3+4 4 | 3+4 4 |
| | 6 2 | 3+3 4 | 3+3 4 | 3+4 4 | 5+5 4 | 3+4 4 | 3+4 4 | 3+4 4 |
| | 6 2 | 4+4 4 | 4+4 4 | 3+4 4 | 5+5 4 | 3+4 4 | 3+4 4 | 3+4 4 |
| | 5 2 | 3+3 4 | 3+3 4 | 3+4 4 | 6+6 4 | 3+4 4 | 3+4 4 | 3+4 4 |
| | 6 2 | 3+3 4 | 3+3 4 | 3+4 4 | 5+5 4 | 3+4 4 | 3+4 4 | 3+4 4 |
| | 6 2 | 4+4 4 | 4+4 4 | 3+4 4 | 6+6 4 | 3+4 4 | 3+4 4 | 3+4 4 |
| | 5 2 | 3+3 4 | 4+4 4 | 3+4 4 | 5+5 4 | 3+4 4 | 3+4 4 | 3+4 4 |
| | 5 2 | 4+4 4 | 3+3 4 | 3+4 4 | 6+6 4 | 3+4 4 | 3+4 4 | 3+4 4 |
| | 5 2 | 4+4 4 | 4+4 4 | 3+4 4 | 6+6 4 | 3+4 4 | 3+4 4 | 3+4 4 |
| | 5 2 | 3+3 4 | 3+3 4 | 3+4 4 | 5+5 4 | 3+4 4 | 3+4 4 | 3+4 4 |
| | 6 2 | 4+4 4 | 3+3 4 | 3+4 4 | 5+5 4 | 3+4 4 | 3+4 4 | 3+4 4 |
| | 6 2 | 3+3 4 | 4+4 4 | 3+4 4 | 6+6 4 | 3+4 4 | 3+4 4 | 3+4 4 |
| | 5 2 | 3+3 4 | 3+3 4 | 3+4 4 | 5+5 4 | 3+4 4 | 3+4 4 | 3+4 4 |

Continued on next column

**TABLE XXXVII– continued**

| TQC | ICSD | NoE | SF | | | | | |
|---|---|---|---|---|---|---|---|---|
| | 5 2 | 4+4 4 | 4+4 4 | 3+4 4 | 6+6 4 | 3+4 4 | 3+4 4 | 3+4 4 |
| | 6 2 | 4+4 4 | 4+4 4 | 3+4 4 | 5+5 4 | 3+4 4 | 3+4 4 | 3+4 4 |
| | 6 2 | 3+3 4 | 3+3 4 | 3+4 4 | 6+6 4 | 3+4 4 | 3+4 4 | 3+4 4 |
| | 5 2 | 3+3 4 | 4+4 4 | 3+4 4 | 5+5 4 | 3+4 4 | 3+4 4 | 3+4 4 |
| | 5 2 | 4+4 4 | 3+3 4 | 3+4 4 | 6+6 4 | 3+4 4 | 3+4 4 | 3+4 4 |
| | 6 2 | 4+4 4 | 4+4 4 | 3+4 4 | 5+5 4 | 3+4 4 | 3+4 4 | 3+4 4 |
| | 6 2 | 3+3 4 | 3+3 4 | 3+4 4 | 6+6 4 | 3+4 4 | 3+4 4 | 3+4 4 |
| | 6 2 | | | | | | | |
| | 5 2 | | | | | | | |
| | 6 2 | | | | | | | |
| | 5 2 | | | | | | | |
| | 5 2 | | | | | | | |
| | 5 2 | | | | | | | |
| | 6 2 | | | | | | | |
| | 5 2 | | | | | | | |
| | 6 2 | | | | | | | |
| | 5 2 | | | | | | | |
| | 6 2 | | | | | | | |
| | 5 2 | | | | | | | |
| | 6 2 | | | | | | | |
| | 5 2 | | | | | | | |
| | 5 2 | | | | | | | |
| | 6 2 | | | | | | | |
| | 6 2 | | | | | | | |
| | 5 2 | | | | | | | |
| | 6 2 | | | | | | | |
| | 5 2 | | | | | | | |
| | 6 2 | | | | | | | |
| | 6 2 | | | | | | | |
| | 5 2 | | | | | | | |
| | 5 2 | | | | | | | |
| 62.2.204 | 68116 | 112 | C2U1W1 | | | | | |
| | -GM | -R | -S | -T | -U | -X | -Y | -Z |
| | 5 2 | 3+3 4 | 3+3 4 | 3+4 4 | 5+5 4 | 3+4 4 | 3+4 4 | 3+4 4 |
| | 6 2 | 4+4 4 | 4+4 4 | 3+4 4 | 6+6 4 | 3+4 4 | 3+4 4 | 3+4 4 |
| | 5 2 | 3+3 4 | 3+3 4 | 3+4 4 | 5+5 4 | 3+4 4 | 3+4 4 | 3+4 4 |
| | 6 2 | 4+4 4 | 4+4 4 | 3+4 4 | 6+6 4 | 3+4 4 | 3+4 4 | 3+4 4 |
| | 5 2 | 4+4 4 | 3+3 4 | 3+4 4 | 5+5 4 | 3+4 4 | 3+4 4 | 3+4 4 |
| | 6 2 | 3+3 4 | 4+4 4 | 3+4 4 | 6+6 4 | 3+4 4 | 3+4 4 | 3+4 4 |
| | 5 2 | 3+3 4 | 4+4 4 | 3+4 4 | 5+5 4 | 3+4 4 | 3+4 4 | 3+4 4 |
| | 6 2 | 4+4 4 | 3+3 4 | 3+4 4 | 6+6 4 | 3+4 4 | 3+4 4 | 3+4 4 |
| | 5 2 | 3+3 4 | 4+4 4 | 3+4 4 | 6+6 4 | 3+4 4 | 3+4 4 | 3+4 4 |
| | 6 2 | 4+4 4 | 3+3 4 | 3+4 4 | 5+5 4 | 3+4 4 | 3+4 4 | 3+4 4 |
| | 5 2 | 4+4 4 | 3+3 4 | 3+4 4 | 6+6 4 | 3+4 4 | 3+4 4 | 3+4 4 |
| | 6 2 | 3+3 4 | 4+4 4 | 3+4 4 | 5+5 4 | 3+4 4 | 3+4 4 | 3+4 4 |
| | 5 2 | 4+4 4 | 4+4 4 | 3+4 4 | 5+5 4 | 3+4 4 | 3+4 4 | 3+4 4 |
| | 5 2 | 3+3 4 | 3+3 4 | 3+4 4 | 6+6 4 | 3+4 4 | 3+4 4 | 3+4 4 |
| | 6 2 | 3+3 4 | 3+3 4 | 3+4 4 | 5+5 4 | 3+4 4 | 3+4 4 | 3+4 4 |
| | 6 2 | 4+4 4 | 4+4 4 | 3+4 4 | 6+6 4 | 3+4 4 | 3+4 4 | 3+4 4 |
| | 5 2 | 3+3 4 | 4+4 4 | 3+4 4 | 6+6 4 | 3+4 4 | 3+4 4 | 3+4 4 |
| | 6 2 | 4+4 4 | 3+3 4 | 3+4 4 | 6+6 4 | 3+4 4 | 3+4 4 | 3+4 4 |
| | 5 2 | 4+4 4 | 4+4 4 | 3+4 4 | 5+5 4 | 3+4 4 | 3+4 4 | 3+4 4 |
| | 6 2 | 3+3 4 | 3+3 4 | 3+4 4 | 5+5 4 | 3+4 4 | 3+4 4 | 3+4 4 |
| | 6 2 | 3+3 4 | 3+3 4 | 3+4 4 | 6+6 4 | 3+4 4 | 3+4 4 | 3+4 4 |
| | 5 2 | 4+4 4 | 4+4 4 | 3+4 4 | 5+5 4 | 3+4 4 | 3+4 4 | 3+4 4 |
| | 5 2 | 4+4 4 | 3+3 4 | 3+4 4 | 5+5 4 | 3+4 4 | 3+4 4 | 3+4 4 |
| | 6 2 | 3+3 4 | 4+4 4 | 3+4 4 | 6+6 4 | 3+4 4 | 3+4 4 | 3+4 4 |
| | 6 2 | 4+4 4 | 3+3 4 | 3+4 4 | 6+6 4 | 3+4 4 | 3+4 4 | 3+4 4 |
| | 5 2 | 3+3 4 | 4+4 4 | 3+4 4 | 5+5 4 | 3+4 4 | 3+4 4 | 3+4 4 |
| | 6 2 | 4+4 4 | 4+4 4 | 3+4 4 | 6+6 4 | 3+4 4 | 3+4 4 | 3+4 4 |
| | 6 2 | 3+3 4 | 3+3 4 | 3+4 4 | 5+5 4 | 3+4 4 | 3+4 4 | 3+4 4 |

Continued on next column

**TABLE XXXVII– continued**

| TQC | ICSD | NoE | SF | | | | | |
|---|---|---|---|---|---|---|---|---|
| | 5 2 | | | | | | | |
| | 5 2 | | | | | | | |
| | 5 2 | | | | | | | |
| | 6 2 | | | | | | | |
| | 6 2 | | | | | | | |
| | 6 2 | | | | | | | |
| | 5 2 | | | | | | | |
| | 5 2 | | | | | | | |
| | 5 2 | | | | | | | |
| | 6 2 | | | | | | | |
| | 6 2 | | | | | | | |
| | 5 2 | | | | | | | |
| | 5 2 | | | | | | | |
| | 5 2 | | | | | | | |
| | 6 2 | | | | | | | |
| | 6 2 | | | | | | | |
| | 6 2 | | | | | | | |
| | 6 2 | | | | | | | |
| | 5 2 | | | | | | | |
| | 6 2 | | | | | | | |
| | 5 2 | | | | | | | |
| | 5 2 | | | | | | | |
| | 6 2 | | | | | | | |
| | 5 2 | | | | | | | |
| | 5 2 | | | | | | | |
| | 5 2 | | | | | | | |
| | 5 2 | | | | | | | |
| | 6 2 | | | | | | | |
| 62.2.205 | 16549 | 112 | C2Mo1U1 | | | | | |
| | -GM | -R | -S | -T | -U | -X | -Y | -Z |
| | 5 2 | 3+3 4 | 3+3 4 | 3+4 4 | 6+6 4 | 3+4 4 | 3+4 4 | 3+4 4 |
| | 6 2 | 4+4 4 | 4+4 4 | 3+4 4 | 5+5 4 | 3+4 4 | 3+4 4 | 3+4 4 |
| | 5 2 | 4+4 4 | 4+4 4 | 3+4 4 | 6+6 4 | 3+4 4 | 3+4 4 | 3+4 4 |
| | 6 2 | 3+3 4 | 3+3 4 | 3+4 4 | 5+5 4 | 3+4 4 | 3+4 4 | 3+4 4 |
| | 5 2 | 3+3 4 | 4+4 4 | 3+4 4 | 6+6 4 | 3+4 4 | 3+4 4 | 3+4 4 |
| | 6 2 | 4+4 4 | 3+3 4 | 3+4 4 | 5+5 4 | 3+4 4 | 3+4 4 | 3+4 4 |
| | 5 2 | 4+4 4 | 3+3 4 | 3+4 4 | 6+6 4 | 3+4 4 | 3+4 4 | 3+4 4 |
| | 6 2 | 3+3 4 | 4+4 4 | 3+4 4 | 5+5 4 | 3+4 4 | 3+4 4 | 3+4 4 |
| | 5 2 | 4+4 4 | 3+3 4 | 3+4 4 | 6+6 4 | 3+4 4 | 3+4 4 | 3+4 4 |
| | 6 2 | 3+3 4 | 4+4 4 | 3+4 4 | 5+5 4 | 3+4 4 | 3+4 4 | 3+4 4 |
| | 5 2 | 3+3 4 | 4+4 4 | 3+4 4 | 5+5 4 | 3+4 4 | 3+4 4 | 3+4 4 |
| | 6 2 | 4+4 4 | 3+3 4 | 3+4 4 | 6+6 4 | 3+4 4 | 3+4 4 | 3+4 4 |
| | 5 2 | 3+3 4 | 3+3 4 | 3+4 4 | 6+6 4 | 3+4 4 | 3+4 4 | 3+4 4 |
| | 5 2 | 4+4 4 | 4+4 4 | 3+4 4 | 5+5 4 | 3+4 4 | 3+4 4 | 3+4 4 |
| | 6 2 | 3+3 4 | 4+4 4 | 3+4 4 | 6+6 4 | 3+4 4 | 3+4 4 | 3+4 4 |
| | 6 2 | 4+4 4 | 3+3 4 | 3+4 4 | 5+5 4 | 3+4 4 | 3+4 4 | 3+4 4 |
| | 5 2 | 4+4 4 | 3+3 4 | 3+4 4 | 5+5 4 | 3+4 4 | 3+4 4 | 3+4 4 |
| | 6 2 | 3+3 4 | 4+4 4 | 3+4 4 | 5+5 4 | 3+4 4 | 3+4 4 | 3+4 4 |
| | 6 2 | 3+3 4 | 3+3 4 | 3+4 4 | 6+6 4 | 3+4 4 | 3+4 4 | 3+4 4 |
| | 5 2 | 4+4 4 | 4+4 4 | 3+4 4 | 6+6 4 | 3+4 4 | 3+4 4 | 3+4 4 |
| | 6 2 | 4+4 4 | 4+4 4 | 3+4 4 | 6+6 4 | 3+4 4 | 3+4 4 | 3+4 4 |
| | 5 2 | 3+3 4 | 3+3 4 | 3+4 4 | 5+5 4 | 3+4 4 | 3+4 4 | 3+4 4 |
| | 5 2 | 3+3 4 | 4+4 4 | 3+4 4 | 6+6 4 | 3+4 4 | 3+4 4 | 3+4 4 |
| | 6 2 | 4+4 4 | 3+3 4 | 3+4 4 | 5+5 4 | 3+4 4 | 3+4 4 | 3+4 4 |
| | 5 2 | 3+3 4 | 4+4 4 | 3+4 4 | 5+5 4 | 3+4 4 | 3+4 4 | 3+4 4 |
| | 6 2 | 4+4 4 | 3+3 4 | 3+4 4 | 6+6 4 | 3+4 4 | 3+4 4 | 3+4 4 |
| | 6 2 | 3+3 4 | 3+3 4 | 3+4 4 | 5+5 4 | 3+4 4 | 3+4 4 | 3+4 4 |
| | 6 2 | 4+4 4 | 4+4 4 | 3+4 4 | 6+6 4 | 3+4 4 | 3+4 4 | 3+4 4 |
| | 5 2 | | | | | | | |
| | 5 2 | | | | | | | |
| | 6 2 | | | | | | | |

Continued on next column

**TABLE XXXVII– continued**

| TQC | ICSD | NoE | SF | | | | | |
|---|---|---|---|---|---|---|---|---|
| | 5 2 | | | | | | | |
| | 6 2 | | | | | | | |
| | 5 2 | | | | | | | |
| | 5 2 | | | | | | | |
| | 6 2 | | | | | | | |
| | 6 2 | | | | | | | |
| | 5 2 | | | | | | | |
| | 5 2 | | | | | | | |
| | 5 2 | | | | | | | |
| | 6 2 | | | | | | | |
| | 5 2 | | | | | | | |
| | 6 2 | | | | | | | |
| | 6 2 | | | | | | | |
| | 6 2 | | | | | | | |
| | 5 2 | | | | | | | |
| | 6 2 | | | | | | | |
| | 6 2 | | | | | | | |
| | 5 2 | | | | | | | |
| | 5 2 | | | | | | | |
| | 6 2 | | | | | | | |
| | 5 2 | | | | | | | |
| | 5 2 | | | | | | | |
| | 5 2 | | | | | | | |
| | 5 2 | | | | | | | |
| | 6 2 | | | | | | | |
| 62.2.206 | 151477 | 104 | C2Cr3 | | | | | |
| | -GM | -R | -S | -T | -U | -X | -Y | -Z |
| | 5 2 | 3+3 4 | 3+3 4 | 3+4 4 | 5+5 4 | 3+4 4 | 3+4 4 | 3+4 4 |
| | 6 2 | 4+4 4 | 4+4 4 | 3+4 4 | 6+6 4 | 3+4 4 | 3+4 4 | 3+4 4 |
| | 5 2 | 3+3 4 | 3+3 4 | 3+4 4 | 6+6 4 | 3+4 4 | 3+4 4 | 3+4 4 |
| | 5 2 | 4+4 4 | 4+4 4 | 3+4 4 | 5+5 4 | 3+4 4 | 3+4 4 | 3+4 4 |
| | 6 2 | 3+3 4 | 3+3 4 | 3+4 4 | 5+5 4 | 3+4 4 | 3+4 4 | 3+4 4 |
| | 6 2 | 4+4 4 | 4+4 4 | 3+4 4 | 6+6 4 | 3+4 4 | 3+4 4 | 3+4 4 |
| | 6 2 | 4+4 4 | 3+3 4 | 3+4 4 | 6+6 4 | 3+4 4 | 3+4 4 | 3+4 4 |
| | 5 2 | 3+3 4 | 4+4 4 | 3+4 4 | 5+5 4 | 3+4 4 | 3+4 4 | 3+4 4 |
| | 6 2 | 3+3 4 | 3+3 4 | 3+4 4 | 6+6 4 | 3+4 4 | 3+4 4 | 3+4 4 |
| | 5 2 | 4+4 4 | 4+4 4 | 3+4 4 | 5+5 4 | 3+4 4 | 3+4 4 | 3+4 4 |
| | 5 2 | 4+4 4 | 3+3 4 | 3+4 4 | 6+6 4 | 3+4 4 | 3+4 4 | 3+4 4 |
| | 6 2 | 3+3 4 | 4+4 4 | 3+4 4 | 5+5 4 | 3+4 4 | 3+4 4 | 3+4 4 |
| | 6 2 | 3+3 4 | 4+4 4 | 3+4 4 | 5+5 4 | 3+4 4 | 3+4 4 | 3+4 4 |
| | 5 2 | 4+4 4 | 3+3 4 | 3+4 4 | 6+6 4 | 3+4 4 | 3+4 4 | 3+4 4 |
| | 6 2 | 4+4 4 | 3+3 4 | 3+4 4 | 5+5 4 | 3+4 4 | 3+4 4 | 3+4 4 |
| | 5 2 | 3+3 4 | 4+4 4 | 3+4 4 | 5+5 4 | 3+4 4 | 3+4 4 | 3+4 4 |
| | 5 2 | 3+3 4 | 3+3 4 | 3+4 4 | 6+6 4 | 3+4 4 | 3+4 4 | 3+4 4 |
| | 6 2 | 4+4 4 | 4+4 4 | 3+4 4 | 6+6 4 | 3+4 4 | 3+4 4 | 3+4 4 |
| | 6 2 | 3+3 4 | 3+3 4 | 3+4 4 | 5+5 4 | 3+4 4 | 3+4 4 | 3+4 4 |
| | 6 2 | 4+4 4 | 4+4 4 | 3+4 4 | 6+6 4 | 3+4 4 | 3+4 4 | 3+4 4 |
| | 5 2 | 4+4 4 | 3+3 4 | 3+4 4 | 5+5 4 | 3+4 4 | 3+4 4 | 3+4 4 |
| | 5 2 | 3+3 4 | 4+4 4 | 3+4 4 | 6+6 4 | 3+4 4 | 3+4 4 | 3+4 4 |
| | 6 2 | 3+3 4 | 4+4 4 | 3+4 4 | 5+5 4 | 3+4 4 | 3+4 4 | 3+4 4 |
| | 5 2 | 4+4 4 | 3+3 4 | 3+4 4 | 6+6 4 | 3+4 4 | 3+4 4 | 3+4 4 |
| | 6 2 | 3+3 4 | 3+3 4 | 3+4 4 | 6+6 4 | 3+4 4 | 3+4 4 | 3+4 4 |
| | 6 2 | 4+4 4 | 4+4 4 | 3+4 4 | 5+5 4 | 3+4 4 | 3+4 4 | 3+4 4 |
| | 5 2 | | | | | | | |
| | 5 2 | | | | | | | |
| | 6 2 | | | | | | | |
| | 6 2 | | | | | | | |
| | 6 2 | | | | | | | |
| | 5 2 | | | | | | | |
| | 5 2 | | | | | | | |
| | 5 2 | | | | | | | |

Continued on next column

**TABLE XXXVII– continued**

| TQC | ICSD | NoE | SF | | | | | |
|---|---|---|---|---|---|---|---|---|
| | 6 2 | | | | | | | |
| | 6 2 | | | | | | | |
| | 5 2 | | | | | | | |
| | 5 2 | | | | | | | |
| | 6 2 | | | | | | | |
| | 5 2 | | | | | | | |
| | 6 2 | | | | | | | |
| | 5 2 | | | | | | | |
| | 5 2 | | | | | | | |
| | 5 2 | | | | | | | |
| | 6 2 | | | | | | | |
| | 5 2 | | | | | | | |
| | 6 2 | | | | | | | |
| | 5 2 | | | | | | | |
| | 5 2 | | | | | | | |
| | 5 2 | | | | | | | |
| | 5 2 | | | | | | | |
| | 6 2 | | | | | | | |
| 62.2.207 | 15086 | 104 | C2Cr3 | | | | | |
| | -GM | -R | -S | -T | -U | -X | -Y | -Z |
| | 5 2 | 3+3 4 | 3+3 4 | 3+4 4 | 5+5 4 | 3+4 4 | 3+4 4 | 3+4 4 |
| | 6 2 | 4+4 4 | 4+4 4 | 3+4 4 | 6+6 4 | 3+4 4 | 3+4 4 | 3+4 4 |
| | 5 2 | 4+4 4 | 3+3 4 | 3+4 4 | 6+6 4 | 3+4 4 | 3+4 4 | 3+4 4 |
| | 5 2 | 3+3 4 | 4+4 4 | 3+4 4 | 5+5 4 | 3+4 4 | 3+4 4 | 3+4 4 |
| | 6 2 | 3+3 4 | 3+3 4 | 3+4 4 | 5+5 4 | 3+4 4 | 3+4 4 | 3+4 4 |
| | 6 2 | 4+4 4 | 4+4 4 | 3+4 4 | 6+6 4 | 3+4 4 | 3+4 4 | 3+4 4 |
| | 6 2 | 3+3 4 | 3+3 4 | 3+4 4 | 6+6 4 | 3+4 4 | 3+4 4 | 3+4 4 |
| | 5 2 | 4+4 4 | 4+4 4 | 3+4 4 | 5+5 4 | 3+4 4 | 3+4 4 | 3+4 4 |
| | 6 2 | 4+4 4 | 3+3 4 | 3+4 4 | 6+6 4 | 3+4 4 | 3+4 4 | 3+4 4 |
| | 5 2 | 3+3 4 | 4+4 4 | 3+4 4 | 5+5 4 | 3+4 4 | 3+4 4 | 3+4 4 |
| | 5 2 | 3+3 4 | 3+3 4 | 3+4 4 | 6+6 4 | 3+4 4 | 3+4 4 | 3+4 4 |
| | 6 2 | 4+4 4 | 4+4 4 | 3+4 4 | 5+5 4 | 3+4 4 | 3+4 4 | 3+4 4 |
| | 6 2 | 4+4 4 | 3+3 4 | 3+4 4 | 5+5 4 | 3+4 4 | 3+4 4 | 3+4 4 |
| | 5 2 | 3+3 4 | 4+4 4 | 3+4 4 | 6+6 4 | 3+4 4 | 3+4 4 | 3+4 4 |
| | 6 2 | 3+3 4 | 4+4 4 | 3+4 4 | 5+5 4 | 3+4 4 | 3+4 4 | 3+4 4 |
| | 5 2 | 4+4 4 | 3+3 4 | 3+4 4 | 5+5 4 | 3+4 4 | 3+4 4 | 3+4 4 |
| | 5 2 | 3+3 4 | 3+3 4 | 3+4 4 | 6+6 4 | 3+4 4 | 3+4 4 | 3+4 4 |
| | 6 2 | 4+4 4 | 4+4 4 | 3+4 4 | 6+6 4 | 3+4 4 | 3+4 4 | 3+4 4 |
| | 6 2 | 4+4 4 | 3+3 4 | 3+4 4 | 5+5 4 | 3+4 4 | 3+4 4 | 3+4 4 |
| | 6 2 | 3+3 4 | 4+4 4 | 3+4 4 | 6+6 4 | 3+4 4 | 3+4 4 | 3+4 4 |
| | 5 2 | 3+3 4 | 3+3 4 | 3+4 4 | 5+5 4 | 3+4 4 | 3+4 4 | 3+4 4 |
| | 5 2 | 4+4 4 | 4+4 4 | 3+4 4 | 6+6 4 | 3+4 4 | 3+4 4 | 3+4 4 |
| | 5 2 | 4+4 4 | 3+3 4 | 3+4 4 | 5+5 4 | 3+4 4 | 3+4 4 | 3+4 4 |
| | 6 2 | 3+3 4 | 4+4 4 | 3+4 4 | 6+6 4 | 3+4 4 | 3+4 4 | 3+4 4 |
| | 6 2 | 4+4 4 | 4+4 4 | 3+4 4 | 6+6 4 | 3+4 4 | 3+4 4 | 3+4 4 |
| | 6 2 | 3+3 4 | 3+3 4 | 3+4 4 | 5+5 4 | 3+4 4 | 3+4 4 | 3+4 4 |
| | 5 2 | | | | | | | |
| | 5 2 | | | | | | | |
| | 6 2 | | | | | | | |
| | 6 2 | | | | | | | |
| | 6 2 | | | | | | | |
| | 5 2 | | | | | | | |
| | 5 2 | | | | | | | |
| | 5 2 | | | | | | | |
| | 6 2 | | | | | | | |
| | 6 2 | | | | | | | |
| | 5 2 | | | | | | | |
| | 5 2 | | | | | | | |
| | 6 2 | | | | | | | |
| | 5 2 | | | | | | | |
| | 6 2 | | | | | | | |

Continued on next column

**TABLE XXXVII– continued**

| TQC | ICSD | NoE | SF | | | | | |
|---|---|---|---|---|---|---|---|---|
| | 5 2 | | | | | | | |
| | 5 2 | | | | | | | |
| | 5 2 | | | | | | | |
| | 5 2 | | | | | | | |
| | 6 2 | | | | | | | |
| | 6 2 | | | | | | | |
| | 5 2 | | | | | | | |
| | 5 2 | | | | | | | |
| | 5 2 | | | | | | | |
| | 5 2 | | | | | | | |
| | 6 2 | | | | | | | |
| 62.2.208 | 247128 | 136 | C1H2Cl1Cu1N2 | | | | | |
| | -GM | -R | -S | -T | -U | -X | -Y | -Z |
| | 5 2 | 3+3 4 | 3+3 4 | 3+4 4 | 5+5 4 | 3+4 4 | 3+4 4 | 3+4 4 |
| | 5 2 | 4+4 4 | 4+4 4 | 3+4 4 | 6+6 4 | 3+4 4 | 3+4 4 | 3+4 4 |
| | 6 2 | 3+3 4 | 3+3 4 | 3+4 4 | 6+6 4 | 3+4 4 | 3+4 4 | 3+4 4 |
| | 6 2 | 4+4 4 | 4+4 4 | 3+4 4 | 5+5 4 | 3+4 4 | 3+4 4 | 3+4 4 |
| | 6 2 | 3+3 4 | 3+3 4 | 3+4 4 | 6+6 4 | 3+4 4 | 3+4 4 | 3+4 4 |
| | 5 2 | 4+4 4 | 4+4 4 | 3+4 4 | 5+5 4 | 3+4 4 | 3+4 4 | 3+4 4 |
| | 6 2 | 3+3 4 | 3+3 4 | 3+4 4 | 6+6 4 | 3+4 4 | 3+4 4 | 3+4 4 |
| | 5 2 | 4+4 4 | 4+4 4 | 3+4 4 | 5+5 4 | 3+4 4 | 3+4 4 | 3+4 4 |
| | 5 2 | 3+3 4 | 3+3 4 | 3+4 4 | 5+5 4 | 3+4 4 | 3+4 4 | 3+4 4 |
| | 5 2 | 4+4 4 | 4+4 4 | 3+4 4 | 6+6 4 | 3+4 4 | 3+4 4 | 3+4 4 |
| | 6 2 | 3+3 4 | 3+3 4 | 3+4 4 | 6+6 4 | 3+4 4 | 3+4 4 | 3+4 4 |
| | 6 2 | 4+4 4 | 4+4 4 | 3+4 4 | 5+5 4 | 3+4 4 | 3+4 4 | 3+4 4 |
| | 5 2 | 3+3 4 | 3+3 4 | 3+4 4 | 6+6 4 | 3+4 4 | 3+4 4 | 3+4 4 |
| | 6 2 | 4+4 4 | 4+4 4 | 3+4 4 | 5+5 4 | 3+4 4 | 3+4 4 | 3+4 4 |
| | 5 2 | 3+3 4 | 3+3 4 | 3+4 4 | 5+5 4 | 3+4 4 | 3+4 4 | 3+4 4 |
| | 6 2 | 4+4 4 | 4+4 4 | 3+4 4 | 6+6 4 | 3+4 4 | 3+4 4 | 3+4 4 |
| | 5 2 | 4+4 4 | 4+4 4 | 3+4 4 | 5+5 4 | 3+4 4 | 3+4 4 | 3+4 4 |
| | 5 2 | 3+3 4 | 3+3 4 | 3+4 4 | 6+6 4 | 3+4 4 | 3+4 4 | 3+4 4 |
| | 6 2 | 3+3 4 | 3+3 4 | 3+4 4 | 6+6 4 | 3+4 4 | 3+4 4 | 3+4 4 |
| | 6 2 | 4+4 4 | 4+4 4 | 3+4 4 | 5+5 4 | 3+4 4 | 3+4 4 | 3+4 4 |
| | 5 2 | 3+3 4 | 3+3 4 | 3+4 4 | 6+6 4 | 3+4 4 | 3+4 4 | 3+4 4 |
| | 5 2 | 4+4 4 | 4+4 4 | 3+4 4 | 5+5 4 | 3+4 4 | 3+4 4 | 3+4 4 |
| | 5 2 | 3+3 4 | 3+3 4 | 3+4 4 | 6+6 4 | 3+4 4 | 3+4 4 | 3+4 4 |
| | 5 2 | 4+4 4 | 4+4 4 | 3+4 4 | 5+5 4 | 3+4 4 | 3+4 4 | 3+4 4 |
| | 6 2 | 4+4 4 | 3+3 4 | 3+4 4 | 5+5 4 | 3+4 4 | 3+4 4 | 3+4 4 |
| | 6 2 | 3+3 4 | 4+4 4 | 3+4 4 | 6+6 4 | 3+4 4 | 3+4 4 | 3+4 4 |
| | 6 2 | 3+3 4 | 4+4 4 | 3+4 4 | 6+6 4 | 3+4 4 | 3+4 4 | 3+4 4 |
| | 6 2 | 4+4 4 | 3+3 4 | 3+4 4 | 5+5 4 | 3+4 4 | 3+4 4 | 3+4 4 |
| | 5 2 | 3+3 4 | 4+4 4 | 3+4 4 | 5+5 4 | 3+4 4 | 3+4 4 | 3+4 4 |
| | 6 2 | 4+4 4 | 3+3 4 | 3+4 4 | 6+6 4 | 3+4 4 | 3+4 4 | 3+4 4 |
| | 6 2 | 3+3 4 | 3+3 4 | 3+4 4 | 6+6 4 | 3+4 4 | 3+4 4 | 3+4 4 |
| | 5 2 | 4+4 4 | 4+4 4 | 3+4 4 | 6+6 4 | 3+4 4 | 3+4 4 | 3+4 4 |
| | 6 2 | 3+3 4 | 3+3 4 | 3+4 4 | 5+5 4 | 3+4 4 | 3+4 4 | 3+4 4 |
| | 6 2 | 4+4 4 | 4+4 4 | 3+4 4 | 5+5 4 | 3+4 4 | 3+4 4 | 3+4 4 |
| | 5 2 | | | | | | | |
| | 5 2 | | | | | | | |
| | 5 2 | | | | | | | |
| | 5 2 | | | | | | | |
| | 6 2 | | | | | | | |
| | 6 2 | | | | | | | |
| | 5 2 | | | | | | | |
| | 5 2 | | | | | | | |
| | 6 2 | | | | | | | |
| | 6 2 | | | | | | | |
| | 5 2 | | | | | | | |
| | 5 2 | | | | | | | |
| | 6 2 | | | | | | | |
| | 6 2 | | | | | | | |

Continued on next column

**TABLE XXXVII– continued**

| TQC | ICSD | NoE | SF | | | | | |
|---|---|---|---|---|---|---|---|---|
| | 5 2 | | | | | | | |
| | 6 2 | | | | | | | |
| | 5 2 | | | | | | | |
| | 6 2 | | | | | | | |
| | 5 2 | | | | | | | |
| | 5 2 | | | | | | | |
| | 6 2 | | | | | | | |
| | 6 2 | | | | | | | |
| | 6 2 | | | | | | | |
| | 6 2 | | | | | | | |
| | 5 2 | | | | | | | |
| | 5 2 | | | | | | | |
| | 5 2 | | | | | | | |
| | 5 2 | | | | | | | |
| | 5 2 | | | | | | | |
| | 6 2 | | | | | | | |
| | 6 2 | | | | | | | |
| | 5 2 | | | | | | | |
| | 5 2 | | | | | | | |
| | 6 2 | | | | | | | |
| 62.2.209 | 236383 | 96 | Bi3Co1 | | | | | |
| | -GM | -R | -S | -T | -U | -X | -Y | -Z |
| | 5 2 | 3+3 4 | 3+3 4 | 3+4 4 | 5+5 4 | 3+4 4 | 3+4 4 | 3+4 4 |
| | 5 2 | 4+4 4 | 4+4 4 | 3+4 4 | 6+6 4 | 3+4 4 | 3+4 4 | 3+4 4 |
| | 6 2 | 3+3 4 | 4+4 4 | 3+4 4 | 6+6 4 | 3+4 4 | 3+4 4 | 3+4 4 |
| | 6 2 | 4+4 4 | 3+3 4 | 3+4 4 | 5+5 4 | 3+4 4 | 3+4 4 | 3+4 4 |
| | 5 2 | 3+3 4 | 3+3 4 | 3+4 4 | 6+6 4 | 3+4 4 | 3+4 4 | 3+4 4 |
| | 6 2 | 4+4 4 | 4+4 4 | 3+4 4 | 5+5 4 | 3+4 4 | 3+4 4 | 3+4 4 |
| | 6 2 | 4+4 4 | 3+3 4 | 3+4 4 | 5+5 4 | 3+4 4 | 3+4 4 | 3+4 4 |
| | 5 2 | 3+3 4 | 4+4 4 | 3+4 4 | 6+6 4 | 3+4 4 | 3+4 4 | 3+4 4 |
| | 5 2 | 4+4 4 | 3+3 4 | 3+4 4 | 5+5 4 | 3+4 4 | 3+4 4 | 3+4 4 |
| | 6 2 | 3+3 4 | 4+4 4 | 3+4 4 | 6+6 4 | 3+4 4 | 3+4 4 | 3+4 4 |
| | 5 2 | 3+3 4 | 4+4 4 | 3+4 4 | 6+6 4 | 3+4 4 | 3+4 4 | 3+4 4 |
| | 6 2 | 4+4 4 | 3+3 4 | 3+4 4 | 6+6 4 | 3+4 4 | 3+4 4 | 3+4 4 |
| | 5 2 | 3+3 4 | 4+4 4 | 3+4 4 | 5+5 4 | 3+4 4 | 3+4 4 | 3+4 4 |
| | 6 2 | 4+4 4 | 3+3 4 | 3+4 4 | 6+6 4 | 3+4 4 | 3+4 4 | 3+4 4 |
| | 5 2 | 4+4 4 | 3+3 4 | 3+4 4 | 5+5 4 | 3+4 4 | 3+4 4 | 3+4 4 |
| | 6 2 | 3+3 4 | 4+4 4 | 3+4 4 | 5+5 4 | 3+4 4 | 3+4 4 | 3+4 4 |
| | 6 2 | 3+3 4 | 3+3 4 | 3+4 4 | 6+6 4 | 3+4 4 | 3+4 4 | 3+4 4 |
| | 5 2 | 4+4 4 | 4+4 4 | 3+4 4 | 5+5 4 | 3+4 4 | 3+4 4 | 3+4 4 |
| | 5 2 | 3+3 4 | 3+3 4 | 3+4 4 | 5+5 4 | 3+4 4 | 3+4 4 | 3+4 4 |
| | 6 2 | 4+4 4 | 4+4 4 | 3+4 4 | 6+6 4 | 3+4 4 | 3+4 4 | 3+4 4 |
| | 6 2 | 4+4 4 | 4+4 4 | 3+4 4 | 5+5 4 | 3+4 4 | 3+4 4 | 3+4 4 |
| | 5 2 | 3+3 4 | 3+3 4 | 3+4 4 | 6+6 4 | 3+4 4 | 3+4 4 | 3+4 4 |
| | 6 2 | 4+4 4 | 4+4 4 | 3+4 4 | 5+5 4 | 3+4 4 | 3+4 4 | 3+4 4 |
| | 6 2 | 3+3 4 | 3+3 4 | 3+4 4 | 6+6 4 | 3+4 4 | 3+4 4 | 3+4 4 |
| | 5 2 | | | | | | | |
| | 5 2 | | | | | | | |
| | 6 2 | | | | | | | |
| | 5 2 | | | | | | | |
| | 5 2 | | | | | | | |
| | 6 2 | | | | | | | |
| | 5 2 | | | | | | | |
| | 5 2 | | | | | | | |
| | 6 2 | | | | | | | |
| | 5 2 | | | | | | | |
| | 5 2 | | | | | | | |
| | 5 2 | | | | | | | |
| | 5 2 | | | | | | | |
| | 6 2 | | | | | | | |
| | 6 2 | | | | | | | |

Continued on next column

**TABLE XXXVII– continued**

| TQC | ICSD | NoE | SF | | | | | |
|---|---|---|---|---|---|---|---|---|
| | 6 2 | | | | | | | |
| | 5 2 | | | | | | | |
| | 6 2 | | | | | | | |
| | 5 2 | | | | | | | |
| | 6 2 | | | | | | | |
| | 5 2 | | | | | | | |
| | 5 2 | | | | | | | |
| | 6 2 | | | | | | | |
| | 6 2 | | | | | | | |
| 62.2.210 | 193272 | 96 | Bi3Co1 | | | | | |
| | -GM | -R | -S | -T | -U | -X | -Y | -Z |
| | 5 2 | 3+3 4 | 3+3 4 | 3+4 4 | 6+6 4 | 3+4 4 | 3+4 4 | 3+4 4 |
| | 5 2 | 4+4 4 | 4+4 4 | 3+4 4 | 5+5 4 | 3+4 4 | 3+4 4 | 3+4 4 |
| | 6 2 | 3+3 4 | 3+3 4 | 3+4 4 | 5+5 4 | 3+4 4 | 3+4 4 | 3+4 4 |
| | 6 2 | 4+4 4 | 4+4 4 | 3+4 4 | 6+6 4 | 3+4 4 | 3+4 4 | 3+4 4 |
| | 5 2 | 4+4 4 | 3+3 4 | 3+4 4 | 5+5 4 | 3+4 4 | 3+4 4 | 3+4 4 |
| | 6 2 | 3+3 4 | 4+4 4 | 3+4 4 | 6+6 4 | 3+4 4 | 3+4 4 | 3+4 4 |
| | 6 2 | 3+3 4 | 4+4 4 | 3+4 4 | 6+6 4 | 3+4 4 | 3+4 4 | 3+4 4 |
| | 5 2 | 4+4 4 | 3+3 4 | 3+4 4 | 5+5 4 | 3+4 4 | 3+4 4 | 3+4 4 |
| | 5 2 | 3+3 4 | 4+4 4 | 3+4 4 | 6+6 4 | 3+4 4 | 3+4 4 | 3+4 4 |
| | 6 2 | 4+4 4 | 3+3 4 | 3+4 4 | 5+5 4 | 3+4 4 | 3+4 4 | 3+4 4 |
| | 5 2 | 4+4 4 | 3+3 4 | 3+4 4 | 5+5 4 | 3+4 4 | 3+4 4 | 3+4 4 |
| | 6 2 | 3+3 4 | 4+4 4 | 3+4 4 | 5+5 4 | 3+4 4 | 3+4 4 | 3+4 4 |
| | 5 2 | 4+4 4 | 3+3 4 | 3+4 4 | 6+6 4 | 3+4 4 | 3+4 4 | 3+4 4 |
| | 6 2 | 3+3 4 | 4+4 4 | 3+4 4 | 5+5 4 | 3+4 4 | 3+4 4 | 3+4 4 |
| | 5 2 | 3+3 4 | 4+4 4 | 3+4 4 | 6+6 4 | 3+4 4 | 3+4 4 | 3+4 4 |
| | 6 2 | 4+4 4 | 3+3 4 | 3+4 4 | 6+6 4 | 3+4 4 | 3+4 4 | 3+4 4 |
| | 6 2 | 4+4 4 | 4+4 4 | 3+4 4 | 5+5 4 | 3+4 4 | 3+4 4 | 3+4 4 |
| | 5 2 | 3+3 4 | 3+3 4 | 3+4 4 | 6+6 4 | 3+4 4 | 3+4 4 | 3+4 4 |
| | 5 2 | 4+4 4 | 4+4 4 | 3+4 4 | 6+6 4 | 3+4 4 | 3+4 4 | 3+4 4 |
| | 6 2 | 3+3 4 | 3+3 4 | 3+4 4 | 5+5 4 | 3+4 4 | 3+4 4 | 3+4 4 |
| | 5 2 | 3+3 4 | 3+3 4 | 3+4 4 | 6+6 4 | 3+4 4 | 3+4 4 | 3+4 4 |
| | 6 2 | 4+4 4 | 4+4 4 | 3+4 4 | 5+5 4 | 3+4 4 | 3+4 4 | 3+4 4 |
| | 6 2 | 3+3 4 | 3+3 4 | 3+4 4 | 6+6 4 | 3+4 4 | 3+4 4 | 3+4 4 |
| | 6 2 | 4+4 4 | 4+4 4 | 3+4 4 | 5+5 4 | 3+4 4 | 3+4 4 | 3+4 4 |
| | 5 2 | | | | | | | |
| | 5 2 | | | | | | | |
| | 6 2 | | | | | | | |
| | 5 2 | | | | | | | |
| | 5 2 | | | | | | | |
| | 6 2 | | | | | | | |
| | 5 2 | | | | | | | |
| | 5 2 | | | | | | | |
| | 6 2 | | | | | | | |
| | 5 2 | | | | | | | |
| | 5 2 | | | | | | | |
| | 5 2 | | | | | | | |
| | 5 2 | | | | | | | |
| | 6 2 | | | | | | | |
| | 6 2 | | | | | | | |
| | 6 2 | | | | | | | |
| | 5 2 | | | | | | | |
| | 6 2 | | | | | | | |
| | 5 2 | | | | | | | |
| | 6 2 | | | | | | | |
| | 5 2 | | | | | | | |
| | 5 2 | | | | | | | |
| | 6 2 | | | | | | | |
| | 6 2 | | | | | | | |
| 62.2.211 | 428110 | 120 | Bi1Li1Yb1 | | | | | |
| | -GM | -R | -S | -T | -U | -X | -Y | -Z |

Continued on next column

**TABLE XXXVII– continued**

| TQC | ICSD | NoE | SF | | | | | |
|---|---|---|---|---|---|---|---|---|
| | 5 2 | 3+3 4 | 3+3 4 | 3+4 4 | 5+5 4 | 3+4 4 | 3+4 4 | 3+4 4 |
| | 6 2 | 4+4 4 | 4+4 4 | 3+4 4 | 6+6 4 | 3+4 4 | 3+4 4 | 3+4 4 |
| | 5 2 | 3+3 4 | 3+3 4 | 3+4 4 | 6+6 4 | 3+4 4 | 3+4 4 | 3+4 4 |
| | 6 2 | 4+4 4 | 4+4 4 | 3+4 4 | 5+5 4 | 3+4 4 | 3+4 4 | 3+4 4 |
| | 5 2 | 3+3 4 | 3+3 4 | 3+4 4 | 6+6 4 | 3+4 4 | 3+4 4 | 3+4 4 |
| | 6 2 | 4+4 4 | 4+4 4 | 3+4 4 | 5+5 4 | 3+4 4 | 3+4 4 | 3+4 4 |
| | 5 2 | 4+4 4 | 3+3 4 | 3+4 4 | 6+6 4 | 3+4 4 | 3+4 4 | 3+4 4 |
| | 6 2 | 3+3 4 | 4+4 4 | 3+4 4 | 5+5 4 | 3+4 4 | 3+4 4 | 3+4 4 |
| | 5 2 | 4+4 4 | 3+3 4 | 3+4 4 | 5+5 4 | 3+4 4 | 3+4 4 | 3+4 4 |
| | 6 2 | 3+3 4 | 4+4 4 | 3+4 4 | 6+6 4 | 3+4 4 | 3+4 4 | 3+4 4 |
| | 5 2 | 3+3 4 | 4+4 4 | 3+4 4 | 5+5 4 | 3+4 4 | 3+4 4 | 3+4 4 |
| | 5 2 | 4+4 4 | 3+3 4 | 3+4 4 | 6+6 4 | 3+4 4 | 3+4 4 | 3+4 4 |
| | 6 2 | 4+4 4 | 4+4 4 | 3+4 4 | 5+5 4 | 3+4 4 | 3+4 4 | 3+4 4 |
| | 6 2 | 3+3 4 | 3+3 4 | 3+4 4 | 6+6 4 | 3+4 4 | 3+4 4 | 3+4 4 |
| | 5 2 | 3+3 4 | 3+3 4 | 3+4 4 | 5+5 4 | 3+4 4 | 3+4 4 | 3+4 4 |
| | 6 2 | 4+4 4 | 4+4 4 | 3+4 4 | 6+6 4 | 3+4 4 | 3+4 4 | 3+4 4 |
| | 5 2 | 3+3 4 | 3+3 4 | 3+4 4 | 5+5 4 | 3+4 4 | 3+4 4 | 3+4 4 |
| | 6 2 | 4+4 4 | 4+4 4 | 3+4 4 | 6+6 4 | 3+4 4 | 3+4 4 | 3+4 4 |
| | 5 2 | 3+3 4 | 3+3 4 | 3+4 4 | 5+5 4 | 3+4 4 | 3+4 4 | 3+4 4 |
| | 6 2 | 4+4 4 | 4+4 4 | 3+4 4 | 6+6 4 | 3+4 4 | 3+4 4 | 3+4 4 |
| | 5 2 | 3+3 4 | 3+3 4 | 3+4 4 | 5+5 4 | 3+4 4 | 3+4 4 | 3+4 4 |
| | 6 2 | 4+4 4 | 4+4 4 | 3+4 4 | 6+6 4 | 3+4 4 | 3+4 4 | 3+4 4 |
| | 5 2 | 3+3 4 | 3+3 4 | 3+4 4 | 5+5 4 | 3+4 4 | 3+4 4 | 3+4 4 |
| | 6 2 | 4+4 4 | 4+4 4 | 3+4 4 | 6+6 4 | 3+4 4 | 3+4 4 | 3+4 4 |
| | 5 2 | 3+3 4 | 4+4 4 | 3+4 4 | 5+5 4 | 3+4 4 | 3+4 4 | 3+4 4 |
| | 5 2 | 4+4 4 | 3+3 4 | 3+4 4 | 5+5 4 | 3+4 4 | 3+4 4 | 3+4 4 |
| | 6 2 | 3+3 4 | 3+3 4 | 3+4 4 | 6+6 4 | 3+4 4 | 3+4 4 | 3+4 4 |
| | 5 2 | 4+4 4 | 4+4 4 | 3+4 4 | 6+6 4 | 3+4 4 | 3+4 4 | 3+4 4 |
| | 6 2 | 4+4 4 | 3+3 4 | 3+4 4 | 5+5 4 | 3+4 4 | 3+4 4 | 3+4 4 |
| | 5 2 | 3+3 4 | 4+4 4 | 3+4 4 | 6+6 4 | 3+4 4 | 3+4 4 | 3+4 4 |
| | 6 2 | | | | | | | |
| | 6 2 | | | | | | | |
| | 5 2 | | | | | | | |
| | 5 2 | | | | | | | |
| | 6 2 | | | | | | | |
| | 6 2 | | | | | | | |
| | 6 2 | | | | | | | |
| | 5 2 | | | | | | | |
| | 6 2 | | | | | | | |
| | 5 2 | | | | | | | |
| | 5 2 | | | | | | | |
| | 5 2 | | | | | | | |
| | 6 2 | | | | | | | |
| | 6 2 | | | | | | | |
| | 6 2 | | | | | | | |
| | 5 2 | | | | | | | |
| | 6 2 | | | | | | | |
| | 5 2 | | | | | | | |
| | 6 2 | | | | | | | |
| | 5 2 | | | | | | | |
| | 6 2 | | | | | | | |
| | 5 2 | | | | | | | |
| | 6 2 | | | | | | | |
| | 5 2 | | | | | | | |
| | 6 2 | | | | | | | |
| | 5 2 | | | | | | | |
| | 5 2 | | | | | | | |
| | 5 2 | | | | | | | |
| | 6 2 | | | | | | | |
| | 5 2 | | | | | | | |
| 62.2.212 | 52698 | 96 | Ba2Si1 | | | | | |

Continued on next column

**TABLE XXXVII– continued**

| TQC | ICSD | NoE | SF | | | | | |
|---|---|---|---|---|---|---|---|---|
| | -GM | -R | -S | -T | -U | -X | -Y | -Z |
| | 5 2 | 3+3 4 | 3+3 4 | 3+4 4 | 5+5 4 | 3+4 4 | 3+4 4 | 3+4 4 |
| | 6 2 | 4+4 4 | 4+4 4 | 3+4 4 | 6+6 4 | 3+4 4 | 3+4 4 | 3+4 4 |
| | 5 2 | 3+3 4 | 3+3 4 | 3+4 4 | 6+6 4 | 3+4 4 | 3+4 4 | 3+4 4 |
| | 6 2 | 4+4 4 | 4+4 4 | 3+4 4 | 5+5 4 | 3+4 4 | 3+4 4 | 3+4 4 |
| | 5 2 | 4+4 4 | 4+4 4 | 3+4 4 | 5+5 4 | 3+4 4 | 3+4 4 | 3+4 4 |
| | 5 2 | 3+3 4 | 3+3 4 | 3+4 4 | 6+6 4 | 3+4 4 | 3+4 4 | 3+4 4 |
| | 6 2 | 4+4 4 | 3+3 4 | 3+4 4 | 6+6 4 | 3+4 4 | 3+4 4 | 3+4 4 |
| | 6 2 | 3+3 4 | 4+4 4 | 3+4 4 | 5+5 4 | 3+4 4 | 3+4 4 | 3+4 4 |
| | 5 2 | 3+3 4 | 3+3 4 | 3+4 4 | 5+5 4 | 3+4 4 | 3+4 4 | 3+4 4 |
| | 5 2 | 4+4 4 | 4+4 4 | 3+4 4 | 6+6 4 | 3+4 4 | 3+4 4 | 3+4 4 |
| | 6 2 | 3+3 4 | 3+3 4 | 3+4 4 | 5+5 4 | 3+4 4 | 3+4 4 | 3+4 4 |
| | 6 2 | 4+4 4 | 4+4 4 | 3+4 4 | 6+6 4 | 3+4 4 | 3+4 4 | 3+4 4 |
| | 5 2 | 4+4 4 | 4+4 4 | 3+4 4 | 6+6 4 | 3+4 4 | 3+4 4 | 3+4 4 |
| | 5 2 | 3+3 4 | 3+3 4 | 3+4 4 | 5+5 4 | 3+4 4 | 3+4 4 | 3+4 4 |
| | 6 2 | 4+4 4 | 4+4 4 | 3+4 4 | 6+6 4 | 3+4 4 | 3+4 4 | 3+4 4 |
| | 6 2 | 3+3 4 | 3+3 4 | 3+4 4 | 5+5 4 | 3+4 4 | 3+4 4 | 3+4 4 |
| | 5 2 | 4+4 4 | 3+3 4 | 3+4 4 | 6+6 4 | 3+4 4 | 3+4 4 | 3+4 4 |
| | 6 2 | 3+3 4 | 4+4 4 | 3+4 4 | 5+5 4 | 3+4 4 | 3+4 4 | 3+4 4 |
| | 5 2 | 4+4 4 | 3+3 4 | 3+4 4 | 6+6 4 | 3+4 4 | 3+4 4 | 3+4 4 |
| | 6 2 | 3+3 4 | 4+4 4 | 3+4 4 | 5+5 4 | 3+4 4 | 3+4 4 | 3+4 4 |
| | 6 2 | 3+3 4 | 4+4 4 | 3+4 4 | 5+5 4 | 3+4 4 | 3+4 4 | 3+4 4 |
| | 5 2 | 4+4 4 | 3+3 4 | 3+4 4 | 5+5 4 | 3+4 4 | 3+4 4 | 3+4 4 |
| | 5 2 | 3+3 4 | 4+4 4 | 3+4 4 | 6+6 4 | 3+4 4 | 3+4 4 | 3+4 4 |
| | 6 2 | 4+4 4 | 3+3 4 | 3+4 4 | 6+6 4 | 3+4 4 | 3+4 4 | 3+4 4 |
| | 5 2 | | | | | | | |
| | 5 2 | | | | | | | |
| | 5 2 | | | | | | | |
| | 6 2 | | | | | | | |
| | 5 2 | | | | | | | |
| | 6 2 | | | | | | | |
| | 6 2 | | | | | | | |
| | 6 2 | | | | | | | |
| | 5 2 | | | | | | | |
| | 6 2 | | | | | | | |
| | 6 2 | | | | | | | |
| | 5 2 | | | | | | | |
| | 6 2 | | | | | | | |
| | 5 2 | | | | | | | |
| | 6 2 | | | | | | | |
| | 5 2 | | | | | | | |
| | 5 2 | | | | | | | |
| | 5 2 | | | | | | | |
| | 6 2 | | | | | | | |
| | 6 2 | | | | | | | |
| | 6 2 | | | | | | | |
| | 6 2 | | | | | | | |
| | 5 2 | | | | | | | |
| | 6 2 | | | | | | | |
| 62.2.213 | 416561 | 96 | Ba2Pb1 | | | | | |
| | -GM | -R | -S | -T | -U | -X | -Y | -Z |
| | 5 2 | 3+3 4 | 3+3 4 | 3+4 4 | 6+6 4 | 3+4 4 | 3+4 4 | 3+4 4 |
| | 6 2 | 4+4 4 | 4+4 4 | 3+4 4 | 5+5 4 | 3+4 4 | 3+4 4 | 3+4 4 |
| | 5 2 | 3+3 4 | 3+3 4 | 3+4 4 | 5+5 4 | 3+4 4 | 3+4 4 | 3+4 4 |
| | 6 2 | 4+4 4 | 4+4 4 | 3+4 4 | 6+6 4 | 3+4 4 | 3+4 4 | 3+4 4 |
| | 5 2 | 4+4 4 | 3+3 4 | 3+4 4 | 6+6 4 | 3+4 4 | 3+4 4 | 3+4 4 |
| | 5 2 | 3+3 4 | 4+4 4 | 3+4 4 | 5+5 4 | 3+4 4 | 3+4 4 | 3+4 4 |
| | 6 2 | 4+4 4 | 4+4 4 | 3+4 4 | 5+5 4 | 3+4 4 | 3+4 4 | 3+4 4 |
| | 6 2 | 3+3 4 | 3+3 4 | 3+4 4 | 6+6 4 | 3+4 4 | 3+4 4 | 3+4 4 |
| | 5 2 | 3+3 4 | 4+4 4 | 3+4 4 | 6+6 4 | 3+4 4 | 3+4 4 | 3+4 4 |
| | 6 2 | 4+4 4 | 3+3 4 | 3+4 4 | 5+5 4 | 3+4 4 | 3+4 4 | 3+4 4 |

Continued on next column

**TABLE XXXVII– continued**

| TQC | ICSD | NoE | SF | | | | | |
|---|---|---|---|---|---|---|---|---|
| | 5 2 | 3+3 4 | 4+4 4 | 3+4 4 | 6+6 4 | 3+4 4 | 3+4 4 | 3+4 4 |
| | 6 2 | 4+4 4 | 3+3 4 | 3+4 4 | 5+5 4 | 3+4 4 | 3+4 4 | 3+4 4 |
| | 5 2 | 4+4 4 | 3+3 4 | 3+4 4 | 5+5 4 | 3+4 4 | 3+4 4 | 3+4 4 |
| | 5 2 | 3+3 4 | 4+4 4 | 3+4 4 | 6+6 4 | 3+4 4 | 3+4 4 | 3+4 4 |
| | 6 2 | 4+4 4 | 3+3 4 | 3+4 4 | 5+5 4 | 3+4 4 | 3+4 4 | 3+4 4 |
| | 6 2 | 3+3 4 | 4+4 4 | 3+4 4 | 6+6 4 | 3+4 4 | 3+4 4 | 3+4 4 |
| | 5 2 | 4+4 4 | 4+4 4 | 3+4 4 | 5+5 4 | 3+4 4 | 3+4 4 | 3+4 4 |
| | 6 2 | 3+3 4 | 3+3 4 | 3+4 4 | 6+6 4 | 3+4 4 | 3+4 4 | 3+4 4 |
| | 5 2 | 3+3 4 | 3+3 4 | 3+4 4 | 5+5 4 | 3+4 4 | 3+4 4 | 3+4 4 |
| | 6 2 | 4+4 4 | 4+4 4 | 3+4 4 | 6+6 4 | 3+4 4 | 3+4 4 | 3+4 4 |
| | 6 2 | 4+4 4 | 4+4 4 | 3+4 4 | 6+6 4 | 3+4 4 | 3+4 4 | 3+4 4 |
| | 5 2 | 3+3 4 | 3+3 4 | 3+4 4 | 5+5 4 | 3+4 4 | 3+4 4 | 3+4 4 |
| | 5 2 | 4+4 4 | 4+4 4 | 3+4 4 | 6+6 4 | 3+4 4 | 3+4 4 | 3+4 4 |
| | 6 2 | 3+3 4 | 3+3 4 | 3+4 4 | 5+5 4 | 3+4 4 | 3+4 4 | 3+4 4 |
| | 5 2 | | | | | | | |
| | 5 2 | | | | | | | |
| | 6 2 | | | | | | | |
| | 5 2 | | | | | | | |
| | 6 2 | | | | | | | |
| | 5 2 | | | | | | | |
| | 6 2 | | | | | | | |
| | 6 2 | | | | | | | |
| | 5 2 | | | | | | | |
| | 6 2 | | | | | | | |
| | 6 2 | | | | | | | |
| | 5 2 | | | | | | | |
| | 6 2 | | | | | | | |
| | 5 2 | | | | | | | |
| | 6 2 | | | | | | | |
| | 5 2 | | | | | | | |
| | 5 2 | | | | | | | |
| | 6 2 | | | | | | | |
| | 5 2 | | | | | | | |
| | 6 2 | | | | | | | |
| | 6 2 | | | | | | | |
| | 5 2 | | | | | | | |
| | 6 2 | | | | | | | |
| | 6 2 | | | | | | | |
| 62.2.214 | 90336 | 304 | Ba1Ho2Te4 | | | | | |
| | -GM | -R | -S | -T | -U | -X | -Y | -Z |
| | 5 2 | 3+3 4 | 3+3 4 | 3+4 4 | 5+5 4 | 3+4 4 | 3+4 4 | 3+4 4 |
| | 5 2 | 4+4 4 | 4+4 4 | 3+4 4 | 6+6 4 | 3+4 4 | 3+4 4 | 3+4 4 |
| | 6 2 | 3+3 4 | 3+3 4 | 3+4 4 | 5+5 4 | 3+4 4 | 3+4 4 | 3+4 4 |
| | 6 2 | 4+4 4 | 4+4 4 | 3+4 4 | 6+6 4 | 3+4 4 | 3+4 4 | 3+4 4 |
| | 5 2 | 3+3 4 | 3+3 4 | 3+4 4 | 5+5 4 | 3+4 4 | 3+4 4 | 3+4 4 |
| | 5 2 | 4+4 4 | 4+4 4 | 3+4 4 | 6+6 4 | 3+4 4 | 3+4 4 | 3+4 4 |
| | 6 2 | 3+3 4 | 3+3 4 | 3+4 4 | 5+5 4 | 3+4 4 | 3+4 4 | 3+4 4 |
| | 6 2 | 4+4 4 | 4+4 4 | 3+4 4 | 6+6 4 | 3+4 4 | 3+4 4 | 3+4 4 |
| | 5 2 | 3+3 4 | 3+3 4 | 3+4 4 | 6+6 4 | 3+4 4 | 3+4 4 | 3+4 4 |
| | 5 2 | 4+4 4 | 4+4 4 | 3+4 4 | 5+5 4 | 3+4 4 | 3+4 4 | 3+4 4 |
| | 6 2 | 3+3 4 | 3+3 4 | 3+4 4 | 5+5 4 | 3+4 4 | 3+4 4 | 3+4 4 |
| | 6 2 | 4+4 4 | 4+4 4 | 3+4 4 | 6+6 4 | 3+4 4 | 3+4 4 | 3+4 4 |
| | 5 2 | 3+3 4 | 3+3 4 | 3+4 4 | 5+5 4 | 3+4 4 | 3+4 4 | 3+4 4 |
| | 5 2 | 4+4 4 | 4+4 4 | 3+4 4 | 6+6 4 | 3+4 4 | 3+4 4 | 3+4 4 |
| | 6 2 | 3+3 4 | 3+3 4 | 3+4 4 | 6+6 4 | 3+4 4 | 3+4 4 | 3+4 4 |
| | 6 2 | 4+4 4 | 4+4 4 | 3+4 4 | 5+5 4 | 3+4 4 | 3+4 4 | 3+4 4 |
| | 5 2 | 3+3 4 | 3+3 4 | 3+4 4 | 6+6 4 | 3+4 4 | 3+4 4 | 3+4 4 |
| | 5 2 | 4+4 4 | 4+4 4 | 3+4 4 | 5+5 4 | 3+4 4 | 3+4 4 | 3+4 4 |
| | 6 2 | 3+3 4 | 3+3 4 | 3+4 4 | 5+5 4 | 3+4 4 | 3+4 4 | 3+4 4 |
| | 6 2 | 4+4 4 | 4+4 4 | 3+4 4 | 6+6 4 | 3+4 4 | 3+4 4 | 3+4 4 |
| | 5 2 | 3+3 4 | 3+3 4 | 3+4 4 | 5+5 4 | 3+4 4 | 3+4 4 | 3+4 4 |

Continued on next column

**TABLE XXXVII– continued**

| TQC | ICSD | NoE | SF |
|---|---|---|---|
| | 5 2 4+4 4 4+4 4 3+4 4 6+6 4 3+4 4 3+4 4 3+4 4 | | |
| | 6 2 3+3 4 3+3 4 3+4 4 5+5 4 3+4 4 3+4 4 3+4 4 | | |
| | 6 2 4+4 4 4+4 4 3+4 4 6+6 4 3+4 4 3+4 4 3+4 4 | | |
| | 5 2 3+3 4 4+4 4 3+4 4 6+6 4 3+4 4 3+4 4 3+4 4 | | |
| | 5 2 4+4 4 3+3 4 3+4 4 5+5 4 3+4 4 3+4 4 3+4 4 | | |
| | 5 2 4+4 4 3+3 4 3+4 4 5+5 4 3+4 4 3+4 4 3+4 4 | | |
| | 5 2 3+3 4 4+4 4 3+4 4 6+6 4 3+4 4 3+4 4 3+4 4 | | |
| | 6 2 3+3 4 4+4 4 3+4 4 5+5 4 3+4 4 3+4 4 3+4 4 | | |
| | 6 2 4+4 4 3+3 4 3+4 4 6+6 4 3+4 4 3+4 4 3+4 4 | | |
| | 6 2 3+3 4 3+3 4 3+4 4 5+5 4 3+4 4 3+4 4 3+4 4 | | |
| | 6 2 4+4 4 4+4 4 3+4 4 6+6 4 3+4 4 3+4 4 3+4 4 | | |
| | 5 2 3+3 4 4+4 4 3+4 4 6+6 4 3+4 4 3+4 4 3+4 4 | | |
| | 5 2 4+4 4 3+3 4 3+4 4 5+5 4 3+4 4 3+4 4 3+4 4 | | |
| | 6 2 3+3 4 4+4 4 3+4 4 5+5 4 3+4 4 3+4 4 3+4 4 | | |
| | 6 2 4+4 4 3+3 4 3+4 4 6+6 4 3+4 4 3+4 4 3+4 4 | | |
| | 5 2 4+4 4 3+3 4 3+4 4 6+6 4 3+4 4 3+4 4 3+4 4 | | |
| | 6 2 3+3 4 4+4 4 3+4 4 5+5 4 3+4 4 3+4 4 3+4 4 | | |
| | 5 2 4+4 4 3+3 4 3+4 4 5+5 4 3+4 4 3+4 4 3+4 4 | | |
| | 6 2 3+3 4 4+4 4 3+4 4 6+6 4 3+4 4 3+4 4 3+4 4 | | |
| | 5 2 3+3 4 4+4 4 3+4 4 5+5 4 3+4 4 3+4 4 3+4 4 | | |
| | 6 2 4+4 4 3+3 4 3+4 4 6+6 4 3+4 4 3+4 4 3+4 4 | | |
| | 5 2 4+4 4 4+4 4 3+4 4 5+5 4 3+4 4 3+4 4 3+4 4 | | |
| | 6 2 3+3 4 3+3 4 3+4 4 6+6 4 3+4 4 3+4 4 3+4 4 | | |
| | 5 2 3+3 4 4+4 4 3+4 4 5+5 4 3+4 4 3+4 4 3+4 4 | | |
| | 6 2 4+4 4 3+3 4 3+4 4 6+6 4 3+4 4 3+4 4 3+4 4 | | |
| | 5 2 4+4 4 4+4 4 3+4 4 5+5 4 3+4 4 3+4 4 3+4 4 | | |
| | 6 2 3+3 4 3+3 4 3+4 4 6+6 4 3+4 4 3+4 4 3+4 4 | | |
| | 5 2 3+3 4 3+3 4 3+4 4 5+5 4 3+4 4 3+4 4 3+4 4 | | |
| | 5 2 4+4 4 4+4 4 3+4 4 6+6 4 3+4 4 3+4 4 3+4 4 | | |
| | 6 2 4+4 4 3+3 4 3+4 4 6+6 4 3+4 4 3+4 4 3+4 4 | | |
| | 5 2 3+3 4 4+4 4 3+4 4 5+5 4 3+4 4 3+4 4 3+4 4 | | |
| | 6 2 3+3 4 4+4 4 3+4 4 6+6 4 3+4 4 3+4 4 3+4 4 | | |
| | 5 2 4+4 4 3+3 4 3+4 4 5+5 4 3+4 4 3+4 4 3+4 4 | | |
| | 6 2 3+3 4 3+3 4 3+4 4 5+5 4 3+4 4 3+4 4 3+4 4 | | |
| | 6 2 4+4 4 4+4 4 3+4 4 6+6 4 3+4 4 3+4 4 3+4 4 | | |
| | 5 2 3+3 4 3+3 4 3+4 4 5+5 4 3+4 4 3+4 4 3+4 4 | | |
| | 6 2 4+4 4 4+4 4 3+4 4 6+6 4 3+4 4 3+4 4 3+4 4 | | |
| | 5 2 3+3 4 3+3 4 3+4 4 5+5 4 3+4 4 3+4 4 3+4 4 | | |
| | 6 2 4+4 4 4+4 4 3+4 4 6+6 4 3+4 4 3+4 4 3+4 4 | | |
| | 5 2 4+4 4 4+4 4 3+4 4 5+5 4 3+4 4 3+4 4 3+4 4 | | |
| | 6 2 3+3 4 3+3 4 3+4 4 6+6 4 3+4 4 3+4 4 3+4 4 | | |
| | 5 2 3+3 4 3+3 4 3+4 4 6+6 4 3+4 4 3+4 4 3+4 4 | | |
| | 6 2 4+4 4 4+4 4 3+4 4 5+5 4 3+4 4 3+4 4 3+4 4 | | |
| | 6 2 3+3 4 4+4 4 3+4 4 6+6 4 3+4 4 3+4 4 3+4 4 | | |
| | 6 2 4+4 4 3+3 4 3+4 4 5+5 4 3+4 4 3+4 4 3+4 4 | | |
| | 5 2 4+4 4 4+4 4 3+4 4 6+6 4 3+4 4 3+4 4 3+4 4 | | |
| | 5 2 3+3 4 3+3 4 3+4 4 5+5 4 3+4 4 3+4 4 3+4 4 | | |
| | 5 2 3+3 4 3+3 4 3+4 4 5+5 4 3+4 4 3+4 4 3+4 4 | | |
| | 6 2 4+4 4 4+4 4 3+4 4 6+6 4 3+4 4 3+4 4 3+4 4 | | |
| | 6 2 3+3 4 4+4 4 3+4 4 5+5 4 3+4 4 3+4 4 3+4 4 | | |
| | 5 2 4+4 4 3+3 4 3+4 4 6+6 4 3+4 4 3+4 4 3+4 4 | | |
| | 5 2 3+3 4 4+4 4 3+4 4 5+5 4 3+4 4 3+4 4 3+4 4 | | |
| | 5 2 4+4 4 3+3 4 3+4 4 6+6 4 3+4 4 3+4 4 3+4 4 | | |
| | 6 2 4+4 4 3+3 4 3+4 4 6+6 4 3+4 4 3+4 4 3+4 4 | | |
| | 5 2 3+3 4 4+4 4 3+4 4 5+5 4 3+4 4 3+4 4 3+4 4 | | |
| | 6 2 | | |
| | 5 2 | | |
| | 6 2 | | |
| | 6 2 | | |
| | 5 2 | | |
| | 5 2 | | |

Continued on next column

| TQC | ICSD | NoE | SF |
|---|---|---|---|
| | 6 2 | | |
| | 5 2 | | |
| | 5 2 | | |
| | 6 2 | | |
| | 5 2 | | |
| | 6 2 | | |
| | 5 2 | | |
| | 5 2 | | |
| | 6 2 | | |
| | 5 2 | | |
| | 6 2 | | |
| | 5 2 | | |
| | 6 2 | | |
| | 5 2 | | |
| | 6 2 | | |
| | 6 2 | | |
| | 5 2 | | |
| | 5 2 | | |
| | 6 2 | | |
| | 6 2 | | |
| | 6 2 | | |
| | 5 2 | | |
| | 6 2 | | |
| | 5 2 | | |
| | 6 2 | | |
| | 6 2 | | |
| | 5 2 | | |
| | 5 2 | | |
| | 5 2 | | |
| | 6 2 | | |
| | 5 2 | | |
| | 5 2 | | |
| | 6 2 | | |
| | 5 2 | | |
| | 6 2 | | |
| | 6 2 | | |
| | 6 2 | | |
| | 5 2 | | |
| | 5 2 | | |
| | 6 2 | | |
| | 6 2 | | |
| | 5 2 | | |
| | 5 2 | | |
| | 6 2 | | |
| | 5 2 | | |
| | 6 2 | | |
| | 5 2 | | |
| | 5 2 | | |
| | 6 2 | | |
| | 6 2 | | |
| | 5 2 | | |
| | 6 2 | | |
| | 6 2 | | |
| | 6 2 | | |
| | 5 2 | | |
| | 5 2 | | |
| | 5 2 | | |
| | 5 2 | | |
| | 5 2 | | |
| | 5 2 | | |
| | 6 2 | | |

Continued on next column

**TABLE XXXVII– continued**

| TQC | ICSD | NoE | SF | | | | | |
|---|---|---|---|---|---|---|---|---|
| | 6 2 | | | | | | | |
| | 5 2 | | | | | | | |
| | 6 2 | | | | | | | |
| | 6 2 | | | | | | | |
| | 6 2 | | | | | | | |
| | 5 2 | | | | | | | |
| | 5 2 | | | | | | | |
| | 6 2 | | | | | | | |
| | 5 2 | | | | | | | |
| 62.2.215 | 424857 | 328 | Ba1Hg6 | | | | | |
| | -GM | -R | -S | -T | -U | -X | -Y | -Z |
| | 5 2 | 3+3 4 | 3+3 4 | 3+4 4 | 5+5 4 | 3+4 4 | 3+4 4 | 3+4 4 |
| | 5 2 | 4+4 4 | 4+4 4 | 3+4 4 | 6+6 4 | 3+4 4 | 3+4 4 | 3+4 4 |
| | 6 2 | 3+3 4 | 3+3 4 | 3+4 4 | 5+5 4 | 3+4 4 | 3+4 4 | 3+4 4 |
| | 6 2 | 4+4 4 | 4+4 4 | 3+4 4 | 6+6 4 | 3+4 4 | 3+4 4 | 3+4 4 |
| | 5 2 | 4+4 4 | 3+3 4 | 3+4 4 | 5+5 4 | 3+4 4 | 3+4 4 | 3+4 4 |
| | 6 2 | 3+3 4 | 4+4 4 | 3+4 4 | 6+6 4 | 3+4 4 | 3+4 4 | 3+4 4 |
| | 5 2 | 3+3 4 | 4+4 4 | 3+4 4 | 6+6 4 | 3+4 4 | 3+4 4 | 3+4 4 |
| | 6 2 | 4+4 4 | 3+3 4 | 3+4 4 | 5+5 4 | 3+4 4 | 3+4 4 | 3+4 4 |
| | 5 2 | 3+3 4 | 4+4 4 | 3+4 4 | 6+6 4 | 3+4 4 | 3+4 4 | 3+4 4 |
| | 6 2 | 4+4 4 | 3+3 4 | 3+4 4 | 5+5 4 | 3+4 4 | 3+4 4 | 3+4 4 |
| | 5 2 | 4+4 4 | 3+3 4 | 3+4 4 | 6+6 4 | 3+4 4 | 3+4 4 | 3+4 4 |
| | 6 2 | 3+3 4 | 4+4 4 | 3+4 4 | 5+5 4 | 3+4 4 | 3+4 4 | 3+4 4 |
| | 5 2 | 3+3 4 | 4+4 4 | 3+4 4 | 5+5 4 | 3+4 4 | 3+4 4 | 3+4 4 |
| | 6 2 | 4+4 4 | 3+3 4 | 3+4 4 | 6+6 4 | 3+4 4 | 3+4 4 | 3+4 4 |
| | 6 2 | 4+4 4 | 3+3 4 | 3+4 4 | 6+6 4 | 3+4 4 | 3+4 4 | 3+4 4 |
| | 5 2 | 3+3 4 | 4+4 4 | 3+4 4 | 5+5 4 | 3+4 4 | 3+4 4 | 3+4 4 |
| | 5 2 | 3+3 4 | 3+3 4 | 3+4 4 | 6+6 4 | 3+4 4 | 3+4 4 | 3+4 4 |
| | 6 2 | 4+4 4 | 4+4 4 | 3+4 4 | 5+5 4 | 3+4 4 | 3+4 4 | 3+4 4 |
| | 6 2 | 3+3 4 | 4+4 4 | 3+4 4 | 6+6 4 | 3+4 4 | 3+4 4 | 3+4 4 |
| | 5 2 | 4+4 4 | 3+3 4 | 3+4 4 | 5+5 4 | 3+4 4 | 3+4 4 | 3+4 4 |
| | 5 2 | 4+4 4 | 3+3 4 | 3+4 4 | 6+6 4 | 3+4 4 | 3+4 4 | 3+4 4 |
| | 5 2 | 3+3 4 | 4+4 4 | 3+4 4 | 5+5 4 | 3+4 4 | 3+4 4 | 3+4 4 |
| | 6 2 | 3+3 4 | 4+4 4 | 3+4 4 | 6+6 4 | 3+4 4 | 3+4 4 | 3+4 4 |
| | 5 2 | 4+4 4 | 3+3 4 | 3+4 4 | 5+5 4 | 3+4 4 | 3+4 4 | 3+4 4 |
| | 6 2 | 3+3 4 | 3+3 4 | 3+4 4 | 5+5 4 | 3+4 4 | 3+4 4 | 3+4 4 |
| | 5 2 | 4+4 4 | 4+4 4 | 3+4 4 | 6+6 4 | 3+4 4 | 3+4 4 | 3+4 4 |
| | 6 2 | 4+4 4 | 4+4 4 | 3+4 4 | 5+5 4 | 3+4 4 | 3+4 4 | 3+4 4 |
| | 5 2 | 3+3 4 | 3+3 4 | 3+4 4 | 6+6 4 | 3+4 4 | 3+4 4 | 3+4 4 |
| | 6 2 | 3+3 4 | 3+3 4 | 3+4 4 | 5+5 4 | 3+4 4 | 3+4 4 | 3+4 4 |
| | 5 2 | 4+4 4 | 4+4 4 | 3+4 4 | 6+6 4 | 3+4 4 | 3+4 4 | 3+4 4 |
| | 6 2 | 4+4 4 | 3+3 4 | 3+4 4 | 5+5 4 | 3+4 4 | 3+4 4 | 3+4 4 |
| | 5 2 | 3+3 4 | 4+4 4 | 3+4 4 | 6+6 4 | 3+4 4 | 3+4 4 | 3+4 4 |
| | 6 2 | 3+3 4 | 3+3 4 | 3+4 4 | 6+6 4 | 3+4 4 | 3+4 4 | 3+4 4 |
| | 5 2 | 4+4 4 | 4+4 4 | 3+4 4 | 5+5 4 | 3+4 4 | 3+4 4 | 3+4 4 |
| | 6 2 | 4+4 4 | 3+3 4 | 3+4 4 | 6+6 4 | 3+4 4 | 3+4 4 | 3+4 4 |
| | 6 2 | 3+3 4 | 4+4 4 | 3+4 4 | 5+5 4 | 3+4 4 | 3+4 4 | 3+4 4 |
| | 5 2 | 4+4 4 | 3+3 4 | 3+4 4 | 6+6 4 | 3+4 4 | 3+4 4 | 3+4 4 |
| | 6 2 | 3+3 4 | 4+4 4 | 3+4 4 | 5+5 4 | 3+4 4 | 3+4 4 | 3+4 4 |
| | 5 2 | 3+3 4 | 4+4 4 | 3+4 4 | 5+5 4 | 3+4 4 | 3+4 4 | 3+4 4 |
| | 5 2 | 4+4 4 | 3+3 4 | 3+4 4 | 6+6 4 | 3+4 4 | 3+4 4 | 3+4 4 |
| | 5 2 | 4+4 4 | 4+4 4 | 3+4 4 | 5+5 4 | 3+4 4 | 3+4 4 | 3+4 4 |
| | 6 2 | 3+3 4 | 3+3 4 | 3+4 4 | 6+6 4 | 3+4 4 | 3+4 4 | 3+4 4 |
| | 6 2 | 4+4 4 | 3+3 4 | 3+4 4 | 5+5 4 | 3+4 4 | 3+4 4 | 3+4 4 |
| | 5 2 | 3+3 4 | 4+4 4 | 3+4 4 | 6+6 4 | 3+4 4 | 3+4 4 | 3+4 4 |
| | 6 2 | 3+3 4 | 4+4 4 | 3+4 4 | 5+5 4 | 3+4 4 | 3+4 4 | 3+4 4 |
| | 5 2 | 4+4 4 | 3+3 4 | 3+4 4 | 6+6 4 | 3+4 4 | 3+4 4 | 3+4 4 |
| | 5 2 | 4+4 4 | 4+4 4 | 3+4 4 | 6+6 4 | 3+4 4 | 3+4 4 | 3+4 4 |
| | 6 2 | 3+3 4 | 3+3 4 | 3+4 4 | 5+5 4 | 3+4 4 | 3+4 4 | 3+4 4 |
| | 6 2 | 4+4 4 | 4+4 4 | 3+4 4 | 6+6 4 | 3+4 4 | 3+4 4 | 3+4 4 |
| | 5 2 | 3+3 4 | 3+3 4 | 3+4 4 | 5+5 4 | 3+4 4 | 3+4 4 | 3+4 4 |
| | 6 2 | 3+3 4 | 3+3 4 | 3+4 4 | 6+6 4 | 3+4 4 | 3+4 4 | 3+4 4 |
| | 5 2 | 4+4 4 | 4+4 4 | 3+4 4 | 5+5 4 | 3+4 4 | 3+4 4 | 3+4 4 |
| | 5 2 | 4+4 4 | 3+3 4 | 3+4 4 | 5+5 4 | 3+4 4 | 3+4 4 | 3+4 4 |
| | 6 2 | 3+3 4 | 4+4 4 | 3+4 4 | 6+6 4 | 3+4 4 | 3+4 4 | 3+4 4 |
| | 6 2 | 3+3 4 | 3+3 4 | 3+4 4 | 5+5 4 | 3+4 4 | 3+4 4 | 3+4 4 |
| | 6 2 | 4+4 4 | 4+4 4 | 3+4 4 | 6+6 4 | 3+4 4 | 3+4 4 | 3+4 4 |
| | 5 2 | 3+3 4 | 3+3 4 | 3+4 4 | 5+5 4 | 3+4 4 | 3+4 4 | 3+4 4 |
| | 5 2 | 4+4 4 | 4+4 4 | 3+4 4 | 6+6 4 | 3+4 4 | 3+4 4 | 3+4 4 |
| | 5 2 | 3+3 4 | 4+4 4 | 3+4 4 | 5+5 4 | 3+4 4 | 3+4 4 | 3+4 4 |
| | 6 2 | 4+4 4 | 3+3 4 | 3+4 4 | 5+5 4 | 3+4 4 | 3+4 4 | 3+4 4 |
| | 6 2 | 3+3 4 | 3+3 4 | 3+4 4 | 6+6 4 | 3+4 4 | 3+4 4 | 3+4 4 |
| | 6 2 | 4+4 4 | 4+4 4 | 3+4 4 | 5+5 4 | 3+4 4 | 3+4 4 | 3+4 4 |
| | 5 2 | 3+3 4 | 4+4 4 | 3+4 4 | 6+6 4 | 3+4 4 | 3+4 4 | 3+4 4 |
| | 5 2 | 4+4 4 | 3+3 4 | 3+4 4 | 6+6 4 | 3+4 4 | 3+4 4 | 3+4 4 |
| | 6 2 | 3+3 4 | 4+4 4 | 3+4 4 | 5+5 4 | 3+4 4 | 3+4 4 | 3+4 4 |
| | 5 2 | 4+4 4 | 3+3 4 | 3+4 4 | 6+6 4 | 3+4 4 | 3+4 4 | 3+4 4 |
| | 6 2 | 4+4 4 | 3+3 4 | 3+4 4 | 5+5 4 | 3+4 4 | 3+4 4 | 3+4 4 |
| | 5 2 | 3+3 4 | 4+4 4 | 3+4 4 | 6+6 4 | 3+4 4 | 3+4 4 | 3+4 4 |
| | 6 2 | 3+3 4 | 4+4 4 | 3+4 4 | 5+5 4 | 3+4 4 | 3+4 4 | 3+4 4 |
| | 5 2 | 4+4 4 | 3+3 4 | 3+4 4 | 6+6 4 | 3+4 4 | 3+4 4 | 3+4 4 |
| | 6 2 | 3+3 4 | 4+4 4 | 3+4 4 | 6+6 4 | 3+4 4 | 3+4 4 | 3+4 4 |
| | 6 2 | 4+4 4 | 3+3 4 | 3+4 4 | 5+5 4 | 3+4 4 | 3+4 4 | 3+4 4 |
| | 5 2 | 4+4 4 | 3+3 4 | 3+4 4 | 5+5 4 | 3+4 4 | 3+4 4 | 3+4 4 |
| | 5 2 | 3+3 4 | 4+4 4 | 3+4 4 | 6+6 4 | 3+4 4 | 3+4 4 | 3+4 4 |
| | 6 2 | 3+3 4 | 3+3 4 | 3+4 4 | 5+5 4 | 3+4 4 | 3+4 4 | 3+4 4 |
| | 6 2 | 4+4 4 | 4+4 4 | 3+4 4 | 6+6 4 | 3+4 4 | 3+4 4 | 3+4 4 |
| | 6 2 | 4+4 4 | 3+3 4 | 3+4 4 | 5+5 4 | 3+4 4 | 3+4 4 | 3+4 4 |
| | 6 2 | 3+3 4 | 4+4 4 | 3+4 4 | 6+6 4 | 3+4 4 | 3+4 4 | 3+4 4 |
| | 5 2 | 3+3 4 | 3+3 4 | 3+4 4 | 6+6 4 | 3+4 4 | 3+4 4 | 3+4 4 |
| | 5 2 | 4+4 4 | 4+4 4 | 3+4 4 | 5+5 4 | 3+4 4 | 3+4 4 | 3+4 4 |
| | 6 2 | 4+4 4 | 4+4 4 | 3+4 4 | 5+5 4 | 3+4 4 | 3+4 4 | 3+4 4 |
| | 5 2 | 3+3 4 | 3+3 4 | 3+4 4 | 6+6 4 | 3+4 4 | 3+4 4 | 3+4 4 |
| | 5 2 | | | | | | | |
| | 5 2 | | | | | | | |
| | 6 2 | | | | | | | |
| | 5 2 | | | | | | | |
| | 6 2 | | | | | | | |
| | 6 2 | | | | | | | |
| | 5 2 | | | | | | | |
| | 5 2 | | | | | | | |
| | 6 2 | | | | | | | |
| | 6 2 | | | | | | | |
| | 5 2 | | | | | | | |
| | 5 2 | | | | | | | |
| | 6 2 | | | | | | | |
| | 6 2 | | | | | | | |
| | 5 2 | | | | | | | |
| | 5 2 | | | | | | | |
| | 6 2 | | | | | | | |
| | 5 2 | | | | | | | |
| | 6 2 | | | | | | | |
| | 6 2 | | | | | | | |
| | 6 2 | | | | | | | |
| | 5 2 | | | | | | | |
| | 6 2 | | | | | | | |
| | 5 2 | | | | | | | |
| | 5 2 | | | | | | | |
| | 5 2 | | | | | | | |
| | 6 2 | | | | | | | |
| | 5 2 | | | | | | | |
| | 6 2 | | | | | | | |

Continued on next column

Continued on next column

**TABLE XXXVII– continued**

| TQC | ICSD | NoE | SF | | | | |
|---|---|---|---|---|---|---|---|
| | 6 2 | | | | | | |
| | 5 2 | | | | | | |
| | 5 2 | | | | | | |
| | 6 2 | | | | | | |
| | 6 2 | | | | | | |
| | 6 2 | | | | | | |
| | 6 2 | | | | | | |
| | 5 2 | | | | | | |
| | 5 2 | | | | | | |
| | 6 2 | | | | | | |
| | 5 2 | | | | | | |
| | 5 2 | | | | | | |
| | 6 2 | | | | | | |
| | 5 2 | | | | | | |
| | 6 2 | | | | | | |
| | 5 2 | | | | | | |
| | 6 2 | | | | | | |
| | 5 2 | | | | | | |
| | 6 2 | | | | | | |
| | 6 2 | | | | | | |
| | 5 2 | | | | | | |
| | 6 2 | | | | | | |
| | 5 2 | | | | | | |
| | 5 2 | | | | | | |
| | 6 2 | | | | | | |
| | 6 2 | | | | | | |
| | 5 2 | | | | | | |
| | 5 2 | | | | | | |
| | 6 2 | | | | | | |
| | 5 2 | | | | | | |
| | 5 2 | | | | | | |
| | 6 2 | | | | | | |
| | 6 2 | | | | | | |
| | 5 2 | | | | | | |
| | 6 2 | | | | | | |
| | 5 2 | | | | | | |
| | 5 2 | | | | | | |
| | 6 2 | | | | | | |
| | 5 2 | | | | | | |
| | 6 2 | | | | | | |
| | 6 2 | | | | | | |
| | 5 2 | | | | | | |
| | 6 2 | | | | | | |
| | 6 2 | | | | | | |
| | 6 2 | | | | | | |
| | 5 2 | | | | | | |
| | 5 2 | | | | | | |
| | 5 2 | | | | | | |
| | 5 2 | | | | | | |
| | 6 2 | | | | | | |
| | 5 2 | | | | | | |
| | 5 2 | | | | | | |
| | 6 2 | | | | | | |
| 62.2.216 | 237304 | 96 | Ba1Ge1Ni1 | | | | |
| -GM | -R | -S | -T | -U | -X | -Y | -Z |
| 5 2 | 3+3 4 | 3+3 4 | 3+4 4 | 5+5 4 | 3+4 4 | 3+4 4 | 3+4 4 |
| 6 2 | 4+4 4 | 4+4 4 | 3+4 4 | 6+6 4 | 3+4 4 | 3+4 4 | 3+4 4 |
| 5 2 | 3+3 4 | 3+3 4 | 3+4 4 | 5+5 4 | 3+4 4 | 3+4 4 | 3+4 4 |
| 6 2 | 4+4 4 | 4+4 4 | 3+4 4 | 6+6 4 | 3+4 4 | 3+4 4 | 3+4 4 |
| 5 2 | 4+4 4 | 3+3 4 | 3+4 4 | 5+5 4 | 3+4 4 | 3+4 4 | 3+4 4 |
| 6 2 | 3+3 4 | 4+4 4 | 3+4 4 | 6+6 4 | 3+4 4 | 3+4 4 | 3+4 4 |

Continued on next column

**TABLE XXXVII– continued**

| TQC | ICSD | NoE | SF | | | | |
|---|---|---|---|---|---|---|---|
| 5 2 | 3+3 4 | 4+4 4 | 3+4 4 | 5+5 4 | 3+4 4 | 3+4 4 | 3+4 4 |
| 6 2 | 4+4 4 | 3+3 4 | 3+4 4 | 6+6 4 | 3+4 4 | 3+4 4 | 3+4 4 |
| 5 2 | 3+3 4 | 3+3 4 | 3+4 4 | 6+6 4 | 3+4 4 | 3+4 4 | 3+4 4 |
| 6 2 | 4+4 4 | 4+4 4 | 3+4 4 | 5+5 4 | 3+4 4 | 3+4 4 | 3+4 4 |
| 5 2 | 4+4 4 | 4+4 4 | 3+4 4 | 5+5 4 | 3+4 4 | 3+4 4 | 3+4 4 |
| 5 2 | 3+3 4 | 3+3 4 | 3+4 4 | 6+6 4 | 3+4 4 | 3+4 4 | 3+4 4 |
| 6 2 | 4+4 4 | 4+4 4 | 3+4 4 | 5+5 4 | 3+4 4 | 3+4 4 | 3+4 4 |
| 5 2 | 3+3 4 | 3+3 4 | 3+4 4 | 6+6 4 | 3+4 4 | 3+4 4 | 3+4 4 |
| 6 2 | 3+3 4 | 4+4 4 | 3+4 4 | 6+6 4 | 3+4 4 | 3+4 4 | 3+4 4 |
| 6 2 | 4+4 4 | 3+3 4 | 3+4 4 | 5+5 4 | 3+4 4 | 3+4 4 | 3+4 4 |
| 5 2 | 3+3 4 | 3+3 4 | 3+4 4 | 5+5 4 | 3+4 4 | 3+4 4 | 3+4 4 |
| 5 2 | 4+4 4 | 4+4 4 | 3+4 4 | 6+6 4 | 3+4 4 | 3+4 4 | 3+4 4 |
| 6 2 | 3+3 4 | 3+3 4 | 3+4 4 | 5+5 4 | 3+4 4 | 3+4 4 | 3+4 4 |
| 6 2 | 4+4 4 | 4+4 4 | 3+4 4 | 6+6 4 | 3+4 4 | 3+4 4 | 3+4 4 |
| 6 2 | 4+4 4 | 3+3 4 | 3+4 4 | 5+5 4 | 3+4 4 | 3+4 4 | 3+4 4 |
| 5 2 | 3+3 4 | 4+4 4 | 3+4 4 | 6+6 4 | 3+4 4 | 3+4 4 | 3+4 4 |
| 6 2 | 3+3 4 | 4+4 4 | 3+4 4 | 6+6 4 | 3+4 4 | 3+4 4 | 3+4 4 |
| 5 2 | 4+4 4 | 3+3 4 | 3+4 4 | 6+6 4 | 3+4 4 | 3+4 4 | 3+4 4 |
| 6 2 | | | | | | | |
| 6 2 | | | | | | | |
| 5 2 | | | | | | | |
| 5 2 | | | | | | | |
| 6 2 | | | | | | | |
| 6 2 | | | | | | | |
| 5 2 | | | | | | | |
| 5 2 | | | | | | | |
| 6 2 | | | | | | | |
| 6 2 | | | | | | | |
| 5 2 | | | | | | | |
| 5 2 | | | | | | | |
| 5 2 | | | | | | | |
| 5 2 | | | | | | | |
| 6 2 | | | | | | | |
| 6 2 | | | | | | | |
| 6 2 | | | | | | | |
| 5 2 | | | | | | | |
| 5 2 | | | | | | | |
| 6 2 | | | | | | | |
| 5 2 | | | | | | | |
| 5 2 | | | | | | | |
| 5 2 | | | | | | | |
| 6 2 | | | | | | | |
| 62.2.217 | 424315 | 176 | Ba1Fe2Se3 | | | | |
| -GM | -R | -S | -T | -U | -X | -Y | -Z |
| 5 2 | 3+3 4 | 3+3 4 | 3+4 4 | 6+6 4 | 3+4 4 | 3+4 4 | 3+4 4 |
| 5 2 | 4+4 4 | 4+4 4 | 3+4 4 | 5+5 4 | 3+4 4 | 3+4 4 | 3+4 4 |
| 6 2 | 3+3 4 | 4+4 4 | 3+4 4 | 5+5 4 | 3+4 4 | 3+4 4 | 3+4 4 |
| 6 2 | 4+4 4 | 3+3 4 | 3+4 4 | 6+6 4 | 3+4 4 | 3+4 4 | 3+4 4 |
| 5 2 | 4+4 4 | 3+3 4 | 3+4 4 | 5+5 4 | 3+4 4 | 3+4 4 | 3+4 4 |
| 6 2 | 3+3 4 | 4+4 4 | 3+4 4 | 6+6 4 | 3+4 4 | 3+4 4 | 3+4 4 |
| 5 2 | 4+4 4 | 3+3 4 | 3+4 4 | 6+6 4 | 3+4 4 | 3+4 4 | 3+4 4 |
| 6 2 | 3+3 4 | 4+4 4 | 3+4 4 | 5+5 4 | 3+4 4 | 3+4 4 | 3+4 4 |
| 5 2 | 3+3 4 | 4+4 4 | 3+4 4 | 6+6 4 | 3+4 4 | 3+4 4 | 3+4 4 |
| 5 2 | 4+4 4 | 3+3 4 | 3+4 4 | 5+5 4 | 3+4 4 | 3+4 4 | 3+4 4 |
| 6 2 | 4+4 4 | 3+3 4 | 3+4 4 | 5+5 4 | 3+4 4 | 3+4 4 | 3+4 4 |
| 6 2 | 3+3 4 | 4+4 4 | 3+4 4 | 6+6 4 | 3+4 4 | 3+4 4 | 3+4 4 |
| 6 2 | 3+3 4 | 4+4 4 | 3+4 4 | 5+5 4 | 3+4 4 | 3+4 4 | 3+4 4 |
| 5 2 | 4+4 4 | 3+3 4 | 3+4 4 | 6+6 4 | 3+4 4 | 3+4 4 | 3+4 4 |
| 5 2 | 3+3 4 | 4+4 4 | 3+4 4 | 5+5 4 | 3+4 4 | 3+4 4 | 3+4 4 |
| 6 2 | 4+4 4 | 3+3 4 | 3+4 4 | 6+6 4 | 3+4 4 | 3+4 4 | 3+4 4 |
| 5 2 | 3+3 4 | 3+3 4 | 3+4 4 | 5+5 4 | 3+4 4 | 3+4 4 | 3+4 4 |

Continued on next column

**TABLE XXXVII– continued**

| TQC | ICSD | NoE | SF |
|---|---|---|---|
| | 6 2 4+4 4 4+4 4 3+4 4 6+6 4 3+4 4 3+4 4 3+4 4 | | |
| | 5 2 4+4 4 4+4 4 3+4 4 5+5 4 3+4 4 3+4 4 3+4 4 | | |
| | 6 2 3+3 4 3+3 4 3+4 4 6+6 4 3+4 4 3+4 4 3+4 4 | | |
| | 5 2 4+4 4 3+3 4 3+4 4 5+5 4 3+4 4 3+4 4 3+4 4 | | |
| | 6 2 3+3 4 4+4 4 3+4 4 6+6 4 3+4 4 3+4 4 3+4 4 | | |
| | 6 2 3+3 4 3+3 4 3+4 4 6+6 4 3+4 4 3+4 4 3+4 4 | | |
| | 5 2 4+4 4 4+4 4 3+4 4 6+6 4 3+4 4 3+4 4 3+4 4 | | |
| | 6 2 4+4 4 4+4 4 3+4 4 5+5 4 3+4 4 3+4 4 3+4 4 | | |
| | 5 2 3+3 4 3+3 4 3+4 4 6+6 4 3+4 4 3+4 4 3+4 4 | | |
| | 6 2 3+3 4 4+4 4 3+4 4 5+5 4 3+4 4 3+4 4 3+4 4 | | |
| | 5 2 4+4 4 3+3 4 3+4 4 6+6 4 3+4 4 3+4 4 3+4 4 | | |
| | 5 2 3+3 4 4+4 4 3+4 4 5+5 4 3+4 4 3+4 4 3+4 4 | | |
| | 6 2 4+4 4 3+3 4 3+4 4 5+5 4 3+4 4 3+4 4 3+4 4 | | |
| | 5 2 4+4 4 3+3 4 3+4 4 5+5 4 3+4 4 3+4 4 3+4 4 | | |
| | 5 2 3+3 4 4+4 4 3+4 4 6+6 4 3+4 4 3+4 4 3+4 4 | | |
| | 6 2 4+4 4 3+3 4 3+4 4 5+5 4 3+4 4 3+4 4 3+4 4 | | |
| | 5 2 3+3 4 4+4 4 3+4 4 6+6 4 3+4 4 3+4 4 3+4 4 | | |
| | 6 2 3+3 4 4+4 4 3+4 4 5+5 4 3+4 4 3+4 4 3+4 4 | | |
| | 6 2 4+4 4 3+3 4 3+4 4 6+6 4 3+4 4 3+4 4 3+4 4 | | |
| | 5 2 3+3 4 3+3 4 3+4 4 5+5 4 3+4 4 3+4 4 3+4 4 | | |
| | 6 2 4+4 4 4+4 4 3+4 4 5+5 4 3+4 4 3+4 4 3+4 4 | | |
| | 5 2 3+3 4 4+4 4 3+4 4 6+6 4 3+4 4 3+4 4 3+4 4 | | |
| | 6 2 4+4 4 3+3 4 3+4 4 6+6 4 3+4 4 3+4 4 3+4 4 | | |
| | 6 2 4+4 4 4+4 4 3+4 4 6+6 4 3+4 4 3+4 4 3+4 4 | | |
| | 6 2 3+3 4 3+3 4 3+4 4 5+5 4 3+4 4 3+4 4 3+4 4 | | |
| | 5 2 4+4 4 4+4 4 3+4 4 6+6 4 3+4 4 3+4 4 3+4 4 | | |
| | 5 2 3+3 4 3+3 4 3+4 4 5+5 4 3+4 4 3+4 4 3+4 4 | | |
| | 6 2 | | |
| | 6 2 | | |
| | 5 2 | | |
| | 5 2 | | |
| | 6 2 | | |
| | 6 2 | | |
| | 6 2 | | |
| | 5 2 | | |
| | 5 2 | | |
| | 6 2 | | |
| | 5 2 | | |
| | 5 2 | | |
| | 6 2 | | |
| | 5 2 | | |
| | 5 2 | | |
| | 6 2 | | |
| | 6 2 | | |
| | 6 2 | | |
| | 5 2 | | |
| | 5 2 | | |
| | 5 2 | | |
| | 5 2 | | |
| | 6 2 | | |
| | 6 2 | | |
| | 6 2 | | |
| | 5 2 | | |
| | 5 2 | | |
| | 5 2 | | |
| | 6 2 | | |
| | 5 2 | | |
| | 5 2 | | |
| | 5 2 | | |
| | 6 2 | | |
| | 6 2 | | |

Continued on next column

**TABLE XXXVII– continued**

| TQC | ICSD | NoE | SF |
|---|---|---|---|
| | 6 2 | | |
| | 6 2 | | |
| | 6 2 | | |
| | 6 2 | | |
| | 5 2 | | |
| | 5 2 | | |
| | 5 2 | | |
| | 6 2 | | |
| | 5 2 | | |
| | 5 2 | | |
| 62.2.218 | 424314 | 176 | Ba1Fe2Se3 |
| | -GM -R -S -T -U -X -Y -Z | | |
| | 5 2 3+3 4 3+3 4 3+4 4 5+5 4 3+4 4 3+4 4 3+4 4 | | |
| | 5 2 4+4 4 4+4 4 3+4 4 6+6 4 3+4 4 3+4 4 3+4 4 | | |
| | 6 2 3+3 4 3+3 4 3+4 4 6+6 4 3+4 4 3+4 4 3+4 4 | | |
| | 6 2 4+4 4 4+4 4 3+4 4 5+5 4 3+4 4 3+4 4 3+4 4 | | |
| | 5 2 4+4 4 4+4 4 3+4 4 6+6 4 3+4 4 3+4 4 3+4 4 | | |
| | 6 2 3+3 4 3+3 4 3+4 4 5+5 4 3+4 4 3+4 4 3+4 4 | | |
| | 5 2 4+4 4 4+4 4 3+4 4 5+5 4 3+4 4 3+4 4 3+4 4 | | |
| | 6 2 3+3 4 3+3 4 3+4 4 6+6 4 3+4 4 3+4 4 3+4 4 | | |
| | 5 2 3+3 4 3+3 4 3+4 4 5+5 4 3+4 4 3+4 4 3+4 4 | | |
| | 5 2 4+4 4 4+4 4 3+4 4 6+6 4 3+4 4 3+4 4 3+4 4 | | |
| | 6 2 4+4 4 4+4 4 3+4 4 6+6 4 3+4 4 3+4 4 3+4 4 | | |
| | 6 2 3+3 4 3+3 4 3+4 4 5+5 4 3+4 4 3+4 4 3+4 4 | | |
| | 6 2 3+3 4 3+3 4 3+4 4 6+6 4 3+4 4 3+4 4 3+4 4 | | |
| | 5 2 4+4 4 4+4 4 3+4 4 5+5 4 3+4 4 3+4 4 3+4 4 | | |
| | 5 2 3+3 4 3+3 4 3+4 4 6+6 4 3+4 4 3+4 4 3+4 4 | | |
| | 6 2 4+4 4 4+4 4 3+4 4 5+5 4 3+4 4 3+4 4 3+4 4 | | |
| | 5 2 3+3 4 4+4 4 3+4 4 6+6 4 3+4 4 3+4 4 3+4 4 | | |
| | 6 2 4+4 4 3+3 4 3+4 4 5+5 4 3+4 4 3+4 4 3+4 4 | | |
| | 5 2 4+4 4 3+3 4 3+4 4 6+6 4 3+4 4 3+4 4 3+4 4 | | |
| | 6 2 3+3 4 4+4 4 3+4 4 5+5 4 3+4 4 3+4 4 3+4 4 | | |
| | 5 2 4+4 4 4+4 4 3+4 4 5+5 4 3+4 4 3+4 4 3+4 4 | | |
| | 6 2 3+3 4 3+3 4 3+4 4 6+6 4 3+4 4 3+4 4 3+4 4 | | |
| | 6 2 3+3 4 4+4 4 3+4 4 5+5 4 3+4 4 3+4 4 3+4 4 | | |
| | 5 2 4+4 4 3+3 4 3+4 4 5+5 4 3+4 4 3+4 4 3+4 4 | | |
| | 6 2 4+4 4 3+3 4 3+4 4 6+6 4 3+4 4 3+4 4 3+4 4 | | |
| | 5 2 3+3 4 4+4 4 3+4 4 5+5 4 3+4 4 3+4 4 3+4 4 | | |
| | 6 2 3+3 4 3+3 4 3+4 4 6+6 4 3+4 4 3+4 4 3+4 4 | | |
| | 5 2 4+4 4 4+4 4 3+4 4 5+5 4 3+4 4 3+4 4 3+4 4 | | |
| | 5 2 3+3 4 3+3 4 3+4 4 6+6 4 3+4 4 3+4 4 3+4 4 | | |
| | 6 2 4+4 4 4+4 4 3+4 4 6+6 4 3+4 4 3+4 4 3+4 4 | | |
| | 5 2 4+4 4 4+4 4 3+4 4 6+6 4 3+4 4 3+4 4 3+4 4 | | |
| | 5 2 3+3 4 3+3 4 3+4 4 5+5 4 3+4 4 3+4 4 3+4 4 | | |
| | 6 2 4+4 4 4+4 4 3+4 4 6+6 4 3+4 4 3+4 4 3+4 4 | | |
| | 5 2 3+3 4 3+3 4 3+4 4 5+5 4 3+4 4 3+4 4 3+4 4 | | |
| | 6 2 3+3 4 3+3 4 3+4 4 6+6 4 3+4 4 3+4 4 3+4 4 | | |
| | 6 2 4+4 4 4+4 4 3+4 4 5+5 4 3+4 4 3+4 4 3+4 4 | | |
| | 5 2 3+3 4 3+3 4 3+4 4 6+6 4 3+4 4 3+4 4 3+4 4 | | |
| | 6 2 4+4 4 4+4 4 3+4 4 6+6 4 3+4 4 3+4 4 3+4 4 | | |
| | 6 2 3+3 4 3+3 4 3+4 4 5+5 4 3+4 4 3+4 4 3+4 4 | | |
| | 6 2 4+4 4 4+4 4 3+4 4 5+5 4 3+4 4 3+4 4 3+4 4 | | |
| | 5 2 4+4 4 3+3 4 3+4 4 5+5 4 3+4 4 3+4 4 3+4 4 | | |
| | 6 2 3+3 4 4+4 4 3+4 4 6+6 4 3+4 4 3+4 4 3+4 4 | | |
| | 5 2 4+4 4 3+3 4 3+4 4 5+5 4 3+4 4 3+4 4 3+4 4 | | |
| | 5 2 3+3 4 4+4 4 3+4 4 6+6 4 3+4 4 3+4 4 3+4 4 | | |
| | 6 2 | | |
| | 6 2 | | |
| | 6 2 | | |
| | 5 2 | | |
| | 5 2 | | |

Continued on next column

**TABLE XXXVII– continued**

| TQC | ICSD | NoE | SF | | | | | |
|---|---|---|---|---|---|---|---|---|
| | 6 2 | | | | | | | |
| | 6 2 | | | | | | | |
| | 5 2 | | | | | | | |
| | 5 2 | | | | | | | |
| | 6 2 | | | | | | | |
| | 5 2 | | | | | | | |
| | 5 2 | | | | | | | |
| | 5 2 | | | | | | | |
| | 6 2 | | | | | | | |
| | 5 2 | | | | | | | |
| | 6 2 | | | | | | | |
| | 6 2 | | | | | | | |
| | 6 2 | | | | | | | |
| | 5 2 | | | | | | | |
| | 5 2 | | | | | | | |
| | 5 2 | | | | | | | |
| | 5 2 | | | | | | | |
| | 6 2 | | | | | | | |
| | 6 2 | | | | | | | |
| | 6 2 | | | | | | | |
| | 5 2 | | | | | | | |
| | 5 2 | | | | | | | |
| | 5 2 | | | | | | | |
| | 6 2 | | | | | | | |
| | 5 2 | | | | | | | |
| | 5 2 | | | | | | | |
| | 5 2 | | | | | | | |
| | 6 2 | | | | | | | |
| | 6 2 | | | | | | | |
| | 6 2 | | | | | | | |
| | 6 2 | | | | | | | |
| | 6 2 | | | | | | | |
| | 6 2 | | | | | | | |
| | 5 2 | | | | | | | |
| | 5 2 | | | | | | | |
| | 5 2 | | | | | | | |
| | 6 2 | | | | | | | |
| | 5 2 | | | | | | | |
| | 5 2 | | | | | | | |
| 62.2.219 | 290595 | 176 | Ba1Fe2Se3 | | | | | |
| | -GM | -R | -S | -T | -U | -X | -Y | -Z |
| | 5 2 | 3+3 4 | 3+3 4 | 3+4 4 | 5+5 4 | 3+4 4 | 3+4 4 | 3+4 4 |
| | 5 2 | 4+4 4 | 4+4 4 | 3+4 4 | 6+6 4 | 3+4 4 | 3+4 4 | 3+4 4 |
| | 6 2 | 4+4 4 | 4+4 4 | 3+4 4 | 5+5 4 | 3+4 4 | 3+4 4 | 3+4 4 |
| | 6 2 | 3+3 4 | 3+3 4 | 3+4 4 | 6+6 4 | 3+4 4 | 3+4 4 | 3+4 4 |
| | 5 2 | 3+3 4 | 3+3 4 | 3+4 4 | 6+6 4 | 3+4 4 | 3+4 4 | 3+4 4 |
| | 6 2 | 4+4 4 | 4+4 4 | 3+4 4 | 5+5 4 | 3+4 4 | 3+4 4 | 3+4 4 |
| | 5 2 | 3+3 4 | 3+3 4 | 3+4 4 | 5+5 4 | 3+4 4 | 3+4 4 | 3+4 4 |
| | 6 2 | 4+4 4 | 4+4 4 | 3+4 4 | 6+6 4 | 3+4 4 | 3+4 4 | 3+4 4 |
| | 5 2 | 4+4 4 | 4+4 4 | 3+4 4 | 5+5 4 | 3+4 4 | 3+4 4 | 3+4 4 |
| | 5 2 | 3+3 4 | 3+3 4 | 3+4 4 | 6+6 4 | 3+4 4 | 3+4 4 | 3+4 4 |
| | 6 2 | 4+4 4 | 4+4 4 | 3+4 4 | 6+6 4 | 3+4 4 | 3+4 4 | 3+4 4 |
| | 6 2 | 3+3 4 | 3+3 4 | 3+4 4 | 5+5 4 | 3+4 4 | 3+4 4 | 3+4 4 |
| | 6 2 | 3+3 4 | 3+3 4 | 3+4 4 | 6+6 4 | 3+4 4 | 3+4 4 | 3+4 4 |
| | 5 2 | 4+4 4 | 4+4 4 | 3+4 4 | 5+5 4 | 3+4 4 | 3+4 4 | 3+4 4 |
| | 5 2 | 4+4 4 | 4+4 4 | 3+4 4 | 5+5 4 | 3+4 4 | 3+4 4 | 3+4 4 |
| | 6 2 | 3+3 4 | 3+3 4 | 3+4 4 | 6+6 4 | 3+4 4 | 3+4 4 | 3+4 4 |
| | 5 2 | 4+4 4 | 3+3 4 | 3+4 4 | 6+6 4 | 3+4 4 | 3+4 4 | 3+4 4 |
| | 6 2 | 3+3 4 | 4+4 4 | 3+4 4 | 5+5 4 | 3+4 4 | 3+4 4 | 3+4 4 |
| | 5 2 | 3+3 4 | 4+4 4 | 3+4 4 | 6+6 4 | 3+4 4 | 3+4 4 | 3+4 4 |
| | 6 2 | 4+4 4 | 3+3 4 | 3+4 4 | 5+5 4 | 3+4 4 | 3+4 4 | 3+4 4 |

Continued on next column

**TABLE XXXVII– continued**

| TQC | ICSD | NoE | SF | | | | | |
|---|---|---|---|---|---|---|---|---|
| | 6 2 | 3+3 4 | 3+3 4 | 3+4 4 | 5+5 4 | 3+4 4 | 3+4 4 | 3+4 4 |
| | 5 2 | 4+4 4 | 4+4 4 | 3+4 4 | 6+6 4 | 3+4 4 | 3+4 4 | 3+4 4 |
| | 5 2 | 3+3 4 | 3+3 4 | 3+4 4 | 5+5 4 | 3+4 4 | 3+4 4 | 3+4 4 |
| | 6 2 | 4+4 4 | 4+4 4 | 3+4 4 | 5+5 4 | 3+4 4 | 3+4 4 | 3+4 4 |
| | 6 2 | 4+4 4 | 3+3 4 | 3+4 4 | 6+6 4 | 3+4 4 | 3+4 4 | 3+4 4 |
| | 5 2 | 3+3 4 | 4+4 4 | 3+4 4 | 6+6 4 | 3+4 4 | 3+4 4 | 3+4 4 |
| | 6 2 | 3+3 4 | 4+4 4 | 3+4 4 | 5+5 4 | 3+4 4 | 3+4 4 | 3+4 4 |
| | 5 2 | 4+4 4 | 3+3 4 | 3+4 4 | 5+5 4 | 3+4 4 | 3+4 4 | 3+4 4 |
| | 5 2 | 4+4 4 | 4+4 4 | 3+4 4 | 6+6 4 | 3+4 4 | 3+4 4 | 3+4 4 |
| | 6 2 | 3+3 4 | 3+3 4 | 3+4 4 | 6+6 4 | 3+4 4 | 3+4 4 | 3+4 4 |
| | 5 2 | 3+3 4 | 3+3 4 | 3+4 4 | 6+6 4 | 3+4 4 | 3+4 4 | 3+4 4 |
| | 5 2 | 4+4 4 | 4+4 4 | 3+4 4 | 5+5 4 | 3+4 4 | 3+4 4 | 3+4 4 |
| | 6 2 | 3+3 4 | 3+3 4 | 3+4 4 | 6+6 4 | 3+4 4 | 3+4 4 | 3+4 4 |
| | 5 2 | 4+4 4 | 4+4 4 | 3+4 4 | 5+5 4 | 3+4 4 | 3+4 4 | 3+4 4 |
| | 6 2 | 4+4 4 | 4+4 4 | 3+4 4 | 6+6 4 | 3+4 4 | 3+4 4 | 3+4 4 |
| | 5 2 | 3+3 4 | 3+3 4 | 3+4 4 | 5+5 4 | 3+4 4 | 3+4 4 | 3+4 4 |
| | 6 2 | 3+3 4 | 3+3 4 | 3+4 4 | 6+6 4 | 3+4 4 | 3+4 4 | 3+4 4 |
| | 6 2 | 4+4 4 | 4+4 4 | 3+4 4 | 6+6 4 | 3+4 4 | 3+4 4 | 3+4 4 |
| | 5 2 | 3+3 4 | 3+3 4 | 3+4 4 | 5+5 4 | 3+4 4 | 3+4 4 | 3+4 4 |
| | 6 2 | 4+4 4 | 4+4 4 | 3+4 4 | 5+5 4 | 3+4 4 | 3+4 4 | 3+4 4 |
| | 6 2 | 3+3 4 | 3+3 4 | 3+4 4 | 5+5 4 | 3+4 4 | 3+4 4 | 3+4 4 |
| | 6 2 | 4+4 4 | 4+4 4 | 3+4 4 | 6+6 4 | 3+4 4 | 3+4 4 | 3+4 4 |
| | 5 2 | 4+4 4 | 4+4 4 | 3+4 4 | 6+6 4 | 3+4 4 | 3+4 4 | 3+4 4 |
| | 5 2 | 3+3 4 | 3+3 4 | 3+4 4 | 5+5 4 | 3+4 4 | 3+4 4 | 3+4 4 |
| | 6 2 | | | | | | | |
| | 6 2 | | | | | | | |
| | 5 2 | | | | | | | |
| | 5 2 | | | | | | | |
| | 6 2 | | | | | | | |
| | 6 2 | | | | | | | |
| | 5 2 | | | | | | | |
| | 6 2 | | | | | | | |
| | 5 2 | | | | | | | |
| | 6 2 | | | | | | | |
| | 5 2 | | | | | | | |
| | 6 2 | | | | | | | |
| | 5 2 | | | | | | | |
| | 5 2 | | | | | | | |
| | 5 2 | | | | | | | |
| | 6 2 | | | | | | | |
| | 6 2 | | | | | | | |
| | 6 2 | | | | | | | |
| | 5 2 | | | | | | | |
| | 5 2 | | | | | | | |
| | 5 2 | | | | | | | |
| | 5 2 | | | | | | | |
| | 6 2 | | | | | | | |
| | 6 2 | | | | | | | |
| | 6 2 | | | | | | | |
| | 5 2 | | | | | | | |
| | 5 2 | | | | | | | |
| | 6 2 | | | | | | | |
| | 5 2 | | | | | | | |
| | 5 2 | | | | | | | |
| | 5 2 | | | | | | | |
| | 5 2 | | | | | | | |
| | 6 2 | | | | | | | |
| | 6 2 | | | | | | | |
| | 6 2 | | | | | | | |
| | 6 2 | | | | | | | |
| | 6 2 | | | | | | | |

Continued on next column

**TABLE XXXVII– continued**

| TQC | ICSD | NoE | SF | | | | | |
|---|---|---|---|---|---|---|---|---|
| | 6 2 | | | | | | | |
| | 5 2 | | | | | | | |
| | 5 2 | | | | | | | |
| | 5 2 | | | | | | | |
| | 5 2 | | | | | | | |
| | 5 2 | | | | | | | |
| | 6 2 | | | | | | | |
| 62.2.220 | 185512 | 176 | Ba1Fe2Se3 | | | | | |
| | -GM | -R | -S | -T | -U | -X | -Y | -Z |
| | 5 2 | 3+3 4 | 3+3 4 | 3+4 4 | 6+6 4 | 3+4 4 | 3+4 4 | 3+4 4 |
| | 5 2 | 4+4 4 | 4+4 4 | 3+4 4 | 5+5 4 | 3+4 4 | 3+4 4 | 3+4 4 |
| | 6 2 | 4+4 4 | 4+4 4 | 3+4 4 | 5+5 4 | 3+4 4 | 3+4 4 | 3+4 4 |
| | 6 2 | 3+3 4 | 3+3 4 | 3+4 4 | 6+6 4 | 3+4 4 | 3+4 4 | 3+4 4 |
| | 5 2 | 3+3 4 | 3+3 4 | 3+4 4 | 5+5 4 | 3+4 4 | 3+4 4 | 3+4 4 |
| | 6 2 | 4+4 4 | 4+4 4 | 3+4 4 | 6+6 4 | 3+4 4 | 3+4 4 | 3+4 4 |
| | 5 2 | 3+3 4 | 3+3 4 | 3+4 4 | 6+6 4 | 3+4 4 | 3+4 4 | 3+4 4 |
| | 6 2 | 4+4 4 | 4+4 4 | 3+4 4 | 5+5 4 | 3+4 4 | 3+4 4 | 3+4 4 |
| | 5 2 | 4+4 4 | 4+4 4 | 3+4 4 | 6+6 4 | 3+4 4 | 3+4 4 | 3+4 4 |
| | 5 2 | 3+3 4 | 3+3 4 | 3+4 4 | 5+5 4 | 3+4 4 | 3+4 4 | 3+4 4 |
| | 6 2 | 3+3 4 | 3+3 4 | 3+4 4 | 5+5 4 | 3+4 4 | 3+4 4 | 3+4 4 |
| | 6 2 | 4+4 4 | 4+4 4 | 3+4 4 | 6+6 4 | 3+4 4 | 3+4 4 | 3+4 4 |
| | 6 2 | 4+4 4 | 4+4 4 | 3+4 4 | 5+5 4 | 3+4 4 | 3+4 4 | 3+4 4 |
| | 5 2 | 3+3 4 | 3+3 4 | 3+4 4 | 6+6 4 | 3+4 4 | 3+4 4 | 3+4 4 |
| | 5 2 | 4+4 4 | 4+4 4 | 3+4 4 | 5+5 4 | 3+4 4 | 3+4 4 | 3+4 4 |
| | 6 2 | 3+3 4 | 3+3 4 | 3+4 4 | 6+6 4 | 3+4 4 | 3+4 4 | 3+4 4 |
| | 5 2 | 4+4 4 | 3+3 4 | 3+4 4 | 5+5 4 | 3+4 4 | 3+4 4 | 3+4 4 |
| | 6 2 | 3+3 4 | 4+4 4 | 3+4 4 | 6+6 4 | 3+4 4 | 3+4 4 | 3+4 4 |
| | 5 2 | 3+3 4 | 4+4 4 | 3+4 4 | 5+5 4 | 3+4 4 | 3+4 4 | 3+4 4 |
| | 6 2 | 4+4 4 | 3+3 4 | 3+4 4 | 6+6 4 | 3+4 4 | 3+4 4 | 3+4 4 |
| | 6 2 | 3+3 4 | 3+3 4 | 3+4 4 | 5+5 4 | 3+4 4 | 3+4 4 | 3+4 4 |
| | 5 2 | 4+4 4 | 4+4 4 | 3+4 4 | 6+6 4 | 3+4 4 | 3+4 4 | 3+4 4 |
| | 6 2 | 4+4 4 | 3+3 4 | 3+4 4 | 6+6 4 | 3+4 4 | 3+4 4 | 3+4 4 |
| | 5 2 | 3+3 4 | 4+4 4 | 3+4 4 | 6+6 4 | 3+4 4 | 3+4 4 | 3+4 4 |
| | 6 2 | 3+3 4 | 4+4 4 | 3+4 4 | 5+5 4 | 3+4 4 | 3+4 4 | 3+4 4 |
| | 5 2 | 4+4 4 | 3+3 4 | 3+4 4 | 6+6 4 | 3+4 4 | 3+4 4 | 3+4 4 |
| | 6 2 | 4+4 4 | 4+4 4 | 3+4 4 | 5+5 4 | 3+4 4 | 3+4 4 | 3+4 4 |
| | 5 2 | 3+3 4 | 3+3 4 | 3+4 4 | 6+6 4 | 3+4 4 | 3+4 4 | 3+4 4 |
| | 5 2 | 3+3 4 | 4+4 4 | 3+4 4 | 5+5 4 | 3+4 4 | 3+4 4 | 3+4 4 |
| | 6 2 | 4+4 4 | 3+3 4 | 3+4 4 | 5+5 4 | 3+4 4 | 3+4 4 | 3+4 4 |
| | 5 2 | 3+3 4 | 3+3 4 | 3+4 4 | 5+5 4 | 3+4 4 | 3+4 4 | 3+4 4 |
| | 5 2 | 4+4 4 | 4+4 4 | 3+4 4 | 6+6 4 | 3+4 4 | 3+4 4 | 3+4 4 |
| | 5 2 | 3+3 4 | 3+3 4 | 3+4 4 | 5+5 4 | 3+4 4 | 3+4 4 | 3+4 4 |
| | 6 2 | 4+4 4 | 4+4 4 | 3+4 4 | 6+6 4 | 3+4 4 | 3+4 4 | 3+4 4 |
| | 6 2 | 4+4 4 | 4+4 4 | 3+4 4 | 5+5 4 | 3+4 4 | 3+4 4 | 3+4 4 |
| | 6 2 | 3+3 4 | 3+3 4 | 3+4 4 | 6+6 4 | 3+4 4 | 3+4 4 | 3+4 4 |
| | 5 2 | 4+4 4 | 3+3 4 | 3+4 4 | 5+5 4 | 3+4 4 | 3+4 4 | 3+4 4 |
| | 6 2 | 3+3 4 | 4+4 4 | 3+4 4 | 5+5 4 | 3+4 4 | 3+4 4 | 3+4 4 |
| | 5 2 | 3+3 4 | 4+4 4 | 3+4 4 | 6+6 4 | 3+4 4 | 3+4 4 | 3+4 4 |
| | 6 2 | 4+4 4 | 3+3 4 | 3+4 4 | 6+6 4 | 3+4 4 | 3+4 4 | 3+4 4 |
| | 6 2 | 3+3 4 | 4+4 4 | 3+4 4 | 6+6 4 | 3+4 4 | 3+4 4 | 3+4 4 |
| | 6 2 | 4+4 4 | 3+3 4 | 3+4 4 | 5+5 4 | 3+4 4 | 3+4 4 | 3+4 4 |
| | 5 2 | 3+3 4 | 4+4 4 | 3+4 4 | 6+6 4 | 3+4 4 | 3+4 4 | 3+4 4 |
| | 5 2 | 4+4 4 | 3+3 4 | 3+4 4 | 5+5 4 | 3+4 4 | 3+4 4 | 3+4 4 |
| | 6 2 | | | | | | | |
| | 6 2 | | | | | | | |
| | 5 2 | | | | | | | |
| | 5 2 | | | | | | | |
| | 6 2 | | | | | | | |
| | 6 2 | | | | | | | |
| | 6 2 | | | | | | | |
| | 5 2 | | | | | | | |

Continued on next column

**TABLE XXXVII– continued**

| TQC | ICSD | NoE | SF | | | | | |
|---|---|---|---|---|---|---|---|---|
| | 6 2 | | | | | | | |
| | 5 2 | | | | | | | |
| | 5 2 | | | | | | | |
| | 6 2 | | | | | | | |
| | 5 2 | | | | | | | |
| | 5 2 | | | | | | | |
| | 6 2 | | | | | | | |
| | 5 2 | | | | | | | |
| | 6 2 | | | | | | | |
| | 6 2 | | | | | | | |
| | 5 2 | | | | | | | |
| | 5 2 | | | | | | | |
| | 5 2 | | | | | | | |
| | 5 2 | | | | | | | |
| | 6 2 | | | | | | | |
| | 6 2 | | | | | | | |
| | 6 2 | | | | | | | |
| | 5 2 | | | | | | | |
| | 5 2 | | | | | | | |
| | 5 2 | | | | | | | |
| | 6 2 | | | | | | | |
| | 5 2 | | | | | | | |
| | 5 2 | | | | | | | |
| | 5 2 | | | | | | | |
| | 6 2 | | | | | | | |
| | 6 2 | | | | | | | |
| | 6 2 | | | | | | | |
| | 6 2 | | | | | | | |
| | 6 2 | | | | | | | |
| | 6 2 | | | | | | | |
| | 5 2 | | | | | | | |
| | 5 2 | | | | | | | |
| | 5 2 | | | | | | | |
| | 6 2 | | | | | | | |
| | 5 2 | | | | | | | |
| | 5 2 | | | | | | | |
| 62.2.221 | 41893 | 112 | B2Co1Nb1 | | | | | |
| | -GM | -R | -S | -T | -U | -X | -Y | -Z |
| | 5 2 | 3+3 4 | 3+3 4 | 3+4 4 | 6+6 4 | 3+4 4 | 3+4 4 | 3+4 4 |
| | 5 2 | 4+4 4 | 4+4 4 | 3+4 4 | 5+5 4 | 3+4 4 | 3+4 4 | 3+4 4 |
| | 6 2 | 3+3 4 | 3+3 4 | 3+4 4 | 6+6 4 | 3+4 4 | 3+4 4 | 3+4 4 |
| | 6 2 | 4+4 4 | 4+4 4 | 3+4 4 | 5+5 4 | 3+4 4 | 3+4 4 | 3+4 4 |
| | 5 2 | 4+4 4 | 4+4 4 | 3+4 4 | 6+6 4 | 3+4 4 | 3+4 4 | 3+4 4 |
| | 5 2 | 3+3 4 | 3+3 4 | 3+4 4 | 5+5 4 | 3+4 4 | 3+4 4 | 3+4 4 |
| | 6 2 | 3+3 4 | 3+3 4 | 3+4 4 | 6+6 4 | 3+4 4 | 3+4 4 | 3+4 4 |
| | 6 2 | 4+4 4 | 4+4 4 | 3+4 4 | 5+5 4 | 3+4 4 | 3+4 4 | 3+4 4 |
| | 5 2 | 3+3 4 | 4+4 4 | 3+4 4 | 6+6 4 | 3+4 4 | 3+4 4 | 3+4 4 |
| | 5 2 | 4+4 4 | 3+3 4 | 3+4 4 | 5+5 4 | 3+4 4 | 3+4 4 | 3+4 4 |
| | 6 2 | 4+4 4 | 3+3 4 | 3+4 4 | 5+5 4 | 3+4 4 | 3+4 4 | 3+4 4 |
| | 6 2 | 3+3 4 | 4+4 4 | 3+4 4 | 6+6 4 | 3+4 4 | 3+4 4 | 3+4 4 |
| | 5 2 | 3+3 4 | 3+3 4 | 3+4 4 | 6+6 4 | 3+4 4 | 3+4 4 | 3+4 4 |
| | 5 2 | 4+4 4 | 4+4 4 | 3+4 4 | 5+5 4 | 3+4 4 | 3+4 4 | 3+4 4 |
| | 6 2 | 4+4 4 | 3+3 4 | 3+4 4 | 6+6 4 | 3+4 4 | 3+4 4 | 3+4 4 |
| | 6 2 | 3+3 4 | 4+4 4 | 3+4 4 | 5+5 4 | 3+4 4 | 3+4 4 | 3+4 4 |
| | 5 2 | 3+3 4 | 4+4 4 | 3+4 4 | 5+5 4 | 3+4 4 | 3+4 4 | 3+4 4 |
| | 6 2 | 4+4 4 | 3+3 4 | 3+4 4 | 6+6 4 | 3+4 4 | 3+4 4 | 3+4 4 |
| | 6 2 | 4+4 4 | 4+4 4 | 3+4 4 | 6+6 4 | 3+4 4 | 3+4 4 | 3+4 4 |
| | 5 2 | 3+3 4 | 3+3 4 | 3+4 4 | 5+5 4 | 3+4 4 | 3+4 4 | 3+4 4 |
| | 5 2 | 4+4 4 | 3+3 4 | 3+4 4 | 5+5 4 | 3+4 4 | 3+4 4 | 3+4 4 |
| | 5 2 | 3+3 4 | 4+4 4 | 3+4 4 | 5+5 4 | 3+4 4 | 3+4 4 | 3+4 4 |
| | 6 2 | 4+4 4 | 3+3 4 | 3+4 4 | 6+6 4 | 3+4 4 | 3+4 4 | 3+4 4 |

Continued on next column

**TABLE XXXVII– continued**

| TQC | ICSD | NoE | SF | | | | | |
|---|---|---|---|---|---|---|---|---|
| | 6 2 | 3+3 4 | 4+4 4 | 3+4 4 | 6+6 4 | 3+4 4 | 3+4 4 | 3+4 4 |
| | 5 2 | 3+3 4 | 3+3 4 | 3+4 4 | 5+5 4 | 3+4 4 | 3+4 4 | 3+4 4 |
| | 6 2 | 4+4 4 | 4+4 4 | 3+4 4 | 6+6 4 | 3+4 4 | 3+4 4 | 3+4 4 |
| | 6 2 | 4+4 4 | 4+4 4 | 3+4 4 | 5+5 4 | 3+4 4 | 3+4 4 | 3+4 4 |
| | 6 2 | 3+3 4 | 3+3 4 | 3+4 4 | 6+6 4 | 3+4 4 | 3+4 4 | 3+4 4 |
| | 5 2 | | | | | | | |
| | 5 2 | | | | | | | |
| | 6 2 | | | | | | | |
| | 6 2 | | | | | | | |
| | 6 2 | | | | | | | |
| | 5 2 | | | | | | | |
| | 5 2 | | | | | | | |
| | 5 2 | | | | | | | |
| | 6 2 | | | | | | | |
| | 6 2 | | | | | | | |
| | 5 2 | | | | | | | |
| | 5 2 | | | | | | | |
| | 5 2 | | | | | | | |
| | 5 2 | | | | | | | |
| | 5 2 | | | | | | | |
| | 6 2 | | | | | | | |
| | 6 2 | | | | | | | |
| | 6 2 | | | | | | | |
| | 5 2 | | | | | | | |
| | 6 2 | | | | | | | |
| | 5 2 | | | | | | | |
| | 6 2 | | | | | | | |
| | 5 2 | | | | | | | |
| | 6 2 | | | | | | | |
| | 5 2 | | | | | | | |
| | 5 2 | | | | | | | |
| | 5 2 | | | | | | | |
| | 5 2 | | | | | | | |
| 62.2.222 | 421745 | 136 | B1O3Pr1 | | | | | |
| | -GM | -R | -S | -T | -U | -X | -Y | -Z |
| | 5 2 | 3+3 4 | 3+3 4 | 3+4 4 | 6+6 4 | 3+4 4 | 3+4 4 | 3+4 4 |
| | 5 2 | 4+4 4 | 4+4 4 | 3+4 4 | 5+5 4 | 3+4 4 | 3+4 4 | 3+4 4 |
| | 6 2 | 4+4 4 | 3+3 4 | 3+4 4 | 5+5 4 | 3+4 4 | 3+4 4 | 3+4 4 |
| | 6 2 | 3+3 4 | 4+4 4 | 3+4 4 | 6+6 4 | 3+4 4 | 3+4 4 | 3+4 4 |
| | 6 2 | 4+4 4 | 4+4 4 | 3+4 4 | 6+6 4 | 3+4 4 | 3+4 4 | 3+4 4 |
| | 5 2 | 3+3 4 | 3+3 4 | 3+4 4 | 5+5 4 | 3+4 4 | 3+4 4 | 3+4 4 |
| | 5 2 | 3+3 4 | 4+4 4 | 3+4 4 | 6+6 4 | 3+4 4 | 3+4 4 | 3+4 4 |
| | 6 2 | 4+4 4 | 3+3 4 | 3+4 4 | 5+5 4 | 3+4 4 | 3+4 4 | 3+4 4 |
| | 5 2 | 3+3 4 | 3+3 4 | 3+4 4 | 6+6 4 | 3+4 4 | 3+4 4 | 3+4 4 |
| | 6 2 | 4+4 4 | 4+4 4 | 3+4 4 | 5+5 4 | 3+4 4 | 3+4 4 | 3+4 4 |
| | 5 2 | 4+4 4 | 3+3 4 | 3+4 4 | 6+6 4 | 3+4 4 | 3+4 4 | 3+4 4 |
| | 5 2 | 3+3 4 | 4+4 4 | 3+4 4 | 5+5 4 | 3+4 4 | 3+4 4 | 3+4 4 |
| | 6 2 | 4+4 4 | 4+4 4 | 3+4 4 | 6+6 4 | 3+4 4 | 3+4 4 | 3+4 4 |
| | 5 2 | 3+3 4 | 3+3 4 | 3+4 4 | 5+5 4 | 3+4 4 | 3+4 4 | 3+4 4 |
| | 6 2 | 3+3 4 | 3+3 4 | 3+4 4 | 5+5 4 | 3+4 4 | 3+4 4 | 3+4 4 |
| | 6 2 | 4+4 4 | 4+4 4 | 3+4 4 | 6+6 4 | 3+4 4 | 3+4 4 | 3+4 4 |
| | 5 2 | 4+4 4 | 3+3 4 | 3+4 4 | 6+6 4 | 3+4 4 | 3+4 4 | 3+4 4 |
| | 6 2 | 3+3 4 | 4+4 4 | 3+4 4 | 5+5 4 | 3+4 4 | 3+4 4 | 3+4 4 |
| | 6 2 | 4+4 4 | 4+4 4 | 3+4 4 | 6+6 4 | 3+4 4 | 3+4 4 | 3+4 4 |
| | 5 2 | 3+3 4 | 3+3 4 | 3+4 4 | 5+5 4 | 3+4 4 | 3+4 4 | 3+4 4 |
| | 5 2 | 3+3 4 | 3+3 4 | 3+4 4 | 5+5 4 | 3+4 4 | 3+4 4 | 3+4 4 |
| | 6 2 | 4+4 4 | 4+4 4 | 3+4 4 | 6+6 4 | 3+4 4 | 3+4 4 | 3+4 4 |
| | 5 2 | 3+3 4 | 3+3 4 | 3+4 4 | 6+6 4 | 3+4 4 | 3+4 4 | 3+4 4 |
| | 6 2 | 4+4 4 | 4+4 4 | 3+4 4 | 6+6 4 | 3+4 4 | 3+4 4 | 3+4 4 |
| | 6 2 | 3+3 4 | 4+4 4 | 3+4 4 | 5+5 4 | 3+4 4 | 3+4 4 | 3+4 4 |
| | 6 2 | 4+4 4 | 3+3 4 | 3+4 4 | 5+5 4 | 3+4 4 | 3+4 4 | 3+4 4 |

Continued on next column

**TABLE XXXVII– continued**

| TQC | ICSD | NoE | SF | | | | | |
|---|---|---|---|---|---|---|---|---|
| | 5 2 | 4+4 4 | 4+4 4 | 3+4 4 | 5+5 4 | 3+4 4 | 3+4 4 | 3+4 4 |
| | 5 2 | 3+3 4 | 3+3 4 | 3+4 4 | 5+5 4 | 3+4 4 | 3+4 4 | 3+4 4 |
| | 5 2 | 3+3 4 | 4+4 4 | 3+4 4 | 6+6 4 | 3+4 4 | 3+4 4 | 3+4 4 |
| | 6 2 | 4+4 4 | 3+3 4 | 3+4 4 | 6+6 4 | 3+4 4 | 3+4 4 | 3+4 4 |
| | 6 2 | 3+3 4 | 3+3 4 | 3+4 4 | 5+5 4 | 3+4 4 | 3+4 4 | 3+4 4 |
| | 5 2 | 4+4 4 | 4+4 4 | 3+4 4 | 6+6 4 | 3+4 4 | 3+4 4 | 3+4 4 |
| | 6 2 | 3+3 4 | 3+3 4 | 3+4 4 | 6+6 4 | 3+4 4 | 3+4 4 | 3+4 4 |
| | 5 2 | 4+4 4 | 4+4 4 | 3+4 4 | 5+5 4 | 3+4 4 | 3+4 4 | 3+4 4 |
| | 5 2 | | | | | | | |
| | 6 2 | | | | | | | |
| | 5 2 | | | | | | | |
| | 6 2 | | | | | | | |
| | 5 2 | | | | | | | |
| | 6 2 | | | | | | | |
| | 5 2 | | | | | | | |
| | 6 2 | | | | | | | |
| | 5 2 | | | | | | | |
| | 5 2 | | | | | | | |
| | 5 2 | | | | | | | |
| | 6 2 | | | | | | | |
| | 6 2 | | | | | | | |
| | 6 2 | | | | | | | |
| | 6 2 | | | | | | | |
| | 5 2 | | | | | | | |
| | 5 2 | | | | | | | |
| | 5 2 | | | | | | | |
| | 5 2 | | | | | | | |
| | 5 2 | | | | | | | |
| | 6 2 | | | | | | | |
| | 6 2 | | | | | | | |
| | 6 2 | | | | | | | |
| | 5 2 | | | | | | | |
| | 5 2 | | | | | | | |
| | 6 2 | | | | | | | |
| | 5 2 | | | | | | | |
| | 6 2 | | | | | | | |
| | 6 2 | | | | | | | |
| | 6 2 | | | | | | | |
| | 5 2 | | | | | | | |
| | 5 2 | | | | | | | |
| | 6 2 | | | | | | | |
| | 5 2 | | | | | | | |
| 62.2.223 | 42894 | 72 | B1Co1Mo1 | | | | | |
| | -GM | -R | -S | -T | -U | -X | -Y | -Z |
| | 5 2 | 4+4 4 | 3+3 4 | 3+4 4 | 6+6 4 | 3+4 4 | 3+4 4 | 3+4 4 |
| | 6 2 | 3+3 4 | 4+4 4 | 3+4 4 | 5+5 4 | 3+4 4 | 3+4 4 | 3+4 4 |
| | 5 2 | 3+3 4 | 4+4 4 | 3+4 4 | 6+6 4 | 3+4 4 | 3+4 4 | 3+4 4 |
| | 6 2 | 4+4 4 | 3+3 4 | 3+4 4 | 5+5 4 | 3+4 4 | 3+4 4 | 3+4 4 |
| | 5 2 | 3+3 4 | 4+4 4 | 3+4 4 | 5+5 4 | 3+4 4 | 3+4 4 | 3+4 4 |
| | 6 2 | 4+4 4 | 3+3 4 | 3+4 4 | 5+5 4 | 3+4 4 | 3+4 4 | 3+4 4 |
| | 6 2 | 4+4 4 | 3+3 4 | 3+4 4 | 6+6 4 | 3+4 4 | 3+4 4 | 3+4 4 |
| | 5 2 | 3+3 4 | 4+4 4 | 3+4 4 | 6+6 4 | 3+4 4 | 3+4 4 | 3+4 4 |
| | 5 2 | 3+3 4 | 4+4 4 | 3+4 4 | 6+6 4 | 3+4 4 | 3+4 4 | 3+4 4 |
| | 6 2 | 4+4 4 | 3+3 4 | 3+4 4 | 5+5 4 | 3+4 4 | 3+4 4 | 3+4 4 |
| | 5 2 | 3+3 4 | 3+3 4 | 3+4 4 | 6+6 4 | 3+4 4 | 3+4 4 | 3+4 4 |
| | 6 2 | 4+4 4 | 4+4 4 | 3+4 4 | 5+5 4 | 3+4 4 | 3+4 4 | 3+4 4 |
| | 6 2 | 4+4 4 | 3+3 4 | 3+4 4 | 6+6 4 | 3+4 4 | 3+4 4 | 3+4 4 |
| | 6 2 | 3+3 4 | 4+4 4 | 3+4 4 | 5+5 4 | 3+4 4 | 3+4 4 | 3+4 4 |
| | 6 2 | 4+4 4 | 4+4 4 | 3+4 4 | 6+6 4 | 3+4 4 | 3+4 4 | 3+4 4 |
| | 5 2 | 3+3 4 | 3+3 4 | 3+4 4 | 6+6 4 | 3+4 4 | 3+4 4 | 3+4 4 |
| | 6 2 | 3+3 4 | 3+3 4 | 3+4 4 | 5+5 4 | 3+4 4 | 3+4 4 | 3+4 4 |

Continued on next column

**TABLE XXXVII– continued**

| TQC | ICSD | NoE | SF | | | | | |
|---|---|---|---|---|---|---|---|---|
| | 5 2 | 4+4 4 | 4+4 4 | 3+4 4 | 5+5 4 | 3+4 4 | 3+4 4 | 3+4 4 |
| | 6 2 | | | | | | | |
| | 5 2 | | | | | | | |
| | 5 2 | | | | | | | |
| | 6 2 | | | | | | | |
| | 5 2 | | | | | | | |
| | 5 2 | | | | | | | |
| | 5 2 | | | | | | | |
| | 6 2 | | | | | | | |
| | 5 2 | | | | | | | |
| | 5 2 | | | | | | | |
| | 6 2 | | | | | | | |
| | 5 2 | | | | | | | |
| | 5 2 | | | | | | | |
| | 6 2 | | | | | | | |
| | 5 2 | | | | | | | |
| | 6 2 | | | | | | | |
| | 5 2 | | | | | | | |
| | 6 2 | | | | | | | |
| 62.2.224 | 50842 | 96 | B1Ca2F1N2 | | | | | |
| | -GM | -R | -S | -T | -U | -X | -Y | -Z |
| | 5 2 | 3+3 4 | 3+3 4 | 3+4 4 | 5+5 4 | 3+4 4 | 3+4 4 | 3+4 4 |
| | 5 2 | 4+4 4 | 4+4 4 | 3+4 4 | 6+6 4 | 3+4 4 | 3+4 4 | 3+4 4 |
| | 6 2 | 3+3 4 | 3+3 4 | 3+4 4 | 6+6 4 | 3+4 4 | 3+4 4 | 3+4 4 |
| | 6 2 | 4+4 4 | 4+4 4 | 3+4 4 | 5+5 4 | 3+4 4 | 3+4 4 | 3+4 4 |
| | 5 2 | 3+3 4 | 3+3 4 | 3+4 4 | 5+5 4 | 3+4 4 | 3+4 4 | 3+4 4 |
| | 5 2 | 4+4 4 | 4+4 4 | 3+4 4 | 6+6 4 | 3+4 4 | 3+4 4 | 3+4 4 |
| | 6 2 | 4+4 4 | 3+3 4 | 3+4 4 | 5+5 4 | 3+4 4 | 3+4 4 | 3+4 4 |
| | 6 2 | 3+3 4 | 4+4 4 | 3+4 4 | 6+6 4 | 3+4 4 | 3+4 4 | 3+4 4 |
| | 5 2 | 3+3 4 | 3+3 4 | 3+4 4 | 5+5 4 | 3+4 4 | 3+4 4 | 3+4 4 |
| | 6 2 | 4+4 4 | 4+4 4 | 3+4 4 | 5+5 4 | 3+4 4 | 3+4 4 | 3+4 4 |
| | 6 2 | 3+3 4 | 3+3 4 | 3+4 4 | 6+6 4 | 3+4 4 | 3+4 4 | 3+4 4 |
| | 5 2 | 4+4 4 | 4+4 4 | 3+4 4 | 5+5 4 | 3+4 4 | 3+4 4 | 3+4 4 |
| | 6 2 | 3+3 4 | 4+4 4 | 3+4 4 | 6+6 4 | 3+4 4 | 3+4 4 | 3+4 4 |
| | 5 2 | 4+4 4 | 3+3 4 | 3+4 4 | 6+6 4 | 3+4 4 | 3+4 4 | 3+4 4 |
| | 5 2 | 4+4 4 | 3+3 4 | 3+4 4 | 6+6 4 | 3+4 4 | 3+4 4 | 3+4 4 |
| | 6 2 | 3+3 4 | 4+4 4 | 3+4 4 | 5+5 4 | 3+4 4 | 3+4 4 | 3+4 4 |
| | 5 2 | 4+4 4 | 3+3 4 | 3+4 4 | 5+5 4 | 3+4 4 | 3+4 4 | 3+4 4 |
| | 5 2 | 3+3 4 | 4+4 4 | 3+4 4 | 6+6 4 | 3+4 4 | 3+4 4 | 3+4 4 |
| | 5 2 | 3+3 4 | 3+3 4 | 3+4 4 | 5+5 4 | 3+4 4 | 3+4 4 | 3+4 4 |
| | 5 2 | 4+4 4 | 4+4 4 | 3+4 4 | 6+6 4 | 3+4 4 | 3+4 4 | 3+4 4 |
| | 6 2 | 3+3 4 | 3+3 4 | 3+4 4 | 6+6 4 | 3+4 4 | 3+4 4 | 3+4 4 |
| | 5 2 | 4+4 4 | 4+4 4 | 3+4 4 | 5+5 4 | 3+4 4 | 3+4 4 | 3+4 4 |
| | 6 2 | 3+3 4 | 3+3 4 | 3+4 4 | 6+6 4 | 3+4 4 | 3+4 4 | 3+4 4 |
| | 5 2 | 4+4 4 | 4+4 4 | 3+4 4 | 5+5 4 | 3+4 4 | 3+4 4 | 3+4 4 |
| | 6 2 | | | | | | | |
| | 6 2 | | | | | | | |
| | 6 2 | | | | | | | |
| | 6 2 | | | | | | | |
| | 5 2 | | | | | | | |
| | 6 2 | | | | | | | |
| | 5 2 | | | | | | | |
| | 6 2 | | | | | | | |
| | 5 2 | | | | | | | |
| | 5 2 | | | | | | | |
| | 5 2 | | | | | | | |
| | 6 2 | | | | | | | |
| | 6 2 | | | | | | | |
| | 6 2 | | | | | | | |
| | 5 2 | | | | | | | |
| | 6 2 | | | | | | | |

Continued on next column

**TABLE XXXVII– continued**

| TQC | ICSD | NoE | SF | | | | | |
|---|---|---|---|---|---|---|---|---|
| | 5 2 | | | | | | | |
| | 6 2 | | | | | | | |
| | 5 2 | | | | | | | |
| | 5 2 | | | | | | | |
| | 6 2 | | | | | | | |
| | 6 2 | | | | | | | |
| | 5 2 | | | | | | | |
| | 5 2 | | | | | | | |
| 62.2.225 | 263057 | 192 | Au3Ga2Rb1 | | | | | |
| | -GM | -R | -S | -T | -U | -X | -Y | -Z |
| | 5 2 | 3+3 4 | 3+3 4 | 3+4 4 | 5+5 4 | 3+4 4 | 3+4 4 | 3+4 4 |
| | 5 2 | 4+4 4 | 4+4 4 | 3+4 4 | 6+6 4 | 3+4 4 | 3+4 4 | 3+4 4 |
| | 6 2 | 3+3 4 | 3+3 4 | 3+4 4 | 5+5 4 | 3+4 4 | 3+4 4 | 3+4 4 |
| | 6 2 | 4+4 4 | 4+4 4 | 3+4 4 | 6+6 4 | 3+4 4 | 3+4 4 | 3+4 4 |
| | 5 2 | 3+3 4 | 4+4 4 | 3+4 4 | 5+5 4 | 3+4 4 | 3+4 4 | 3+4 4 |
| | 5 2 | 4+4 4 | 3+3 4 | 3+4 4 | 6+6 4 | 3+4 4 | 3+4 4 | 3+4 4 |
| | 6 2 | 4+4 4 | 3+3 4 | 3+4 4 | 5+5 4 | 3+4 4 | 3+4 4 | 3+4 4 |
| | 6 2 | 3+3 4 | 4+4 4 | 3+4 4 | 6+6 4 | 3+4 4 | 3+4 4 | 3+4 4 |
| | 5 2 | 4+4 4 | 3+3 4 | 3+4 4 | 6+6 4 | 3+4 4 | 3+4 4 | 3+4 4 |
| | 5 2 | 3+3 4 | 4+4 4 | 3+4 4 | 5+5 4 | 3+4 4 | 3+4 4 | 3+4 4 |
| | 6 2 | 3+3 4 | 4+4 4 | 3+4 4 | 5+5 4 | 3+4 4 | 3+4 4 | 3+4 4 |
| | 5 2 | 4+4 4 | 3+3 4 | 3+4 4 | 6+6 4 | 3+4 4 | 3+4 4 | 3+4 4 |
| | 5 2 | 4+4 4 | 4+4 4 | 3+4 4 | 5+5 4 | 3+4 4 | 3+4 4 | 3+4 4 |
| | 6 2 | 3+3 4 | 3+3 4 | 3+4 4 | 6+6 4 | 3+4 4 | 3+4 4 | 3+4 4 |
| | 6 2 | 4+4 4 | 3+3 4 | 3+4 4 | 5+5 4 | 3+4 4 | 3+4 4 | 3+4 4 |
| | 6 2 | 3+3 4 | 4+4 4 | 3+4 4 | 6+6 4 | 3+4 4 | 3+4 4 | 3+4 4 |
| | 5 2 | 3+3 4 | 3+3 4 | 3+4 4 | 5+5 4 | 3+4 4 | 3+4 4 | 3+4 4 |
| | 6 2 | 4+4 4 | 4+4 4 | 3+4 4 | 6+6 4 | 3+4 4 | 3+4 4 | 3+4 4 |
| | 5 2 | 4+4 4 | 4+4 4 | 3+4 4 | 6+6 4 | 3+4 4 | 3+4 4 | 3+4 4 |
| | 6 2 | 3+3 4 | 3+3 4 | 3+4 4 | 5+5 4 | 3+4 4 | 3+4 4 | 3+4 4 |
| | 6 2 | 3+3 4 | 4+4 4 | 3+4 4 | 5+5 4 | 3+4 4 | 3+4 4 | 3+4 4 |
| | 5 2 | 4+4 4 | 3+3 4 | 3+4 4 | 6+6 4 | 3+4 4 | 3+4 4 | 3+4 4 |
| | 6 2 | 4+4 4 | 4+4 4 | 3+4 4 | 5+5 4 | 3+4 4 | 3+4 4 | 3+4 4 |
| | 6 2 | 3+3 4 | 3+3 4 | 3+4 4 | 6+6 4 | 3+4 4 | 3+4 4 | 3+4 4 |
| | 5 2 | 4+4 4 | 3+3 4 | 3+4 4 | 6+6 4 | 3+4 4 | 3+4 4 | 3+4 4 |
| | 5 2 | 3+3 4 | 4+4 4 | 3+4 4 | 5+5 4 | 3+4 4 | 3+4 4 | 3+4 4 |
| | 5 2 | 3+3 4 | 3+3 4 | 3+4 4 | 6+6 4 | 3+4 4 | 3+4 4 | 3+4 4 |
| | 6 2 | 4+4 4 | 4+4 4 | 3+4 4 | 5+5 4 | 3+4 4 | 3+4 4 | 3+4 4 |
| | 6 2 | 4+4 4 | 3+3 4 | 3+4 4 | 6+6 4 | 3+4 4 | 3+4 4 | 3+4 4 |
| | 5 2 | 3+3 4 | 4+4 4 | 3+4 4 | 5+5 4 | 3+4 4 | 3+4 4 | 3+4 4 |
| | 5 2 | 3+3 4 | 3+3 4 | 3+4 4 | 6+6 4 | 3+4 4 | 3+4 4 | 3+4 4 |
| | 6 2 | 4+4 4 | 4+4 4 | 3+4 4 | 5+5 4 | 3+4 4 | 3+4 4 | 3+4 4 |
| | 5 2 | 3+3 4 | 4+4 4 | 3+4 4 | 6+6 4 | 3+4 4 | 3+4 4 | 3+4 4 |
| | 6 2 | 4+4 4 | 3+3 4 | 3+4 4 | 6+6 4 | 3+4 4 | 3+4 4 | 3+4 4 |
| | 5 2 | 4+4 4 | 4+4 4 | 3+4 4 | 5+5 4 | 3+4 4 | 3+4 4 | 3+4 4 |
| | 6 2 | 3+3 4 | 3+3 4 | 3+4 4 | 5+5 4 | 3+4 4 | 3+4 4 | 3+4 4 |
| | 6 2 | 3+3 4 | 4+4 4 | 3+4 4 | 5+5 4 | 3+4 4 | 3+4 4 | 3+4 4 |
| | 5 2 | 4+4 4 | 3+3 4 | 3+4 4 | 6+6 4 | 3+4 4 | 3+4 4 | 3+4 4 |
| | 5 2 | 3+3 4 | 3+3 4 | 3+4 4 | 6+6 4 | 3+4 4 | 3+4 4 | 3+4 4 |
| | 6 2 | 4+4 4 | 4+4 4 | 3+4 4 | 5+5 4 | 3+4 4 | 3+4 4 | 3+4 4 |
| | 5 2 | 4+4 4 | 4+4 4 | 3+4 4 | 6+6 4 | 3+4 4 | 3+4 4 | 3+4 4 |
| | 6 2 | 3+3 4 | 3+3 4 | 3+4 4 | 5+5 4 | 3+4 4 | 3+4 4 | 3+4 4 |
| | 6 2 | 3+3 4 | 3+3 4 | 3+4 4 | 6+6 4 | 3+4 4 | 3+4 4 | 3+4 4 |
| | 5 2 | 4+4 4 | 4+4 4 | 3+4 4 | 6+6 4 | 3+4 4 | 3+4 4 | 3+4 4 |
| | 6 2 | 4+4 4 | 3+3 4 | 3+4 4 | 5+5 4 | 3+4 4 | 3+4 4 | 3+4 4 |
| | 5 2 | 3+3 4 | 4+4 4 | 3+4 4 | 5+5 4 | 3+4 4 | 3+4 4 | 3+4 4 |
| | 6 2 | 3+3 4 | 4+4 4 | 3+4 4 | 6+6 4 | 3+4 4 | 3+4 4 | 3+4 4 |
| | 5 2 | 4+4 4 | 3+3 4 | 3+4 4 | 5+5 4 | 3+4 4 | 3+4 4 | 3+4 4 |
| | 5 2 | | | | | | | |
| | 6 2 | | | | | | | |
| | 5 2 | | | | | | | |

Continued on next column

**TABLE XXXVII– continued**

| TQC | ICSD | NoE | SF | | | | | |
|---|---|---|---|---|---|---|---|---|
| | 6 2 | | | | | | | |
| | 6 2 | | | | | | | |
| | 5 2 | | | | | | | |
| | 6 2 | | | | | | | |
| | 5 2 | | | | | | | |
| | 5 2 | | | | | | | |
| | 6 2 | | | | | | | |
| | 6 2 | | | | | | | |
| | 5 2 | | | | | | | |
| | 6 2 | | | | | | | |
| | 5 2 | | | | | | | |
| | 5 2 | | | | | | | |
| | 6 2 | | | | | | | |
| | 5 2 | | | | | | | |
| | 6 2 | | | | | | | |
| | 6 2 | | | | | | | |
| | 5 2 | | | | | | | |
| | 5 2 | | | | | | | |
| | 5 2 | | | | | | | |
| | 6 2 | | | | | | | |
| | 6 2 | | | | | | | |
| | 5 2 | | | | | | | |
| | 6 2 | | | | | | | |
| | 5 2 | | | | | | | |
| | 5 2 | | | | | | | |
| | 6 2 | | | | | | | |
| | 5 2 | | | | | | | |
| | 6 2 | | | | | | | |
| | 6 2 | | | | | | | |
| | 5 2 | | | | | | | |
| | 5 2 | | | | | | | |
| | 5 2 | | | | | | | |
| | 6 2 | | | | | | | |
| | 5 2 | | | | | | | |
| | 6 2 | | | | | | | |
| | 6 2 | | | | | | | |
| | 5 2 | | | | | | | |
| | 6 2 | | | | | | | |
| | 6 2 | | | | | | | |
| | 5 2 | | | | | | | |
| | 5 2 | | | | | | | |
| | 6 2 | | | | | | | |
| | 6 2 | | | | | | | |
| | 5 2 | | | | | | | |
| | 5 2 | | | | | | | |
| 62.2.226 | 263058 | 192 | Au3Cs1Ga2 | | | | | |
| | -GM | -R | -S | -T | -U | -X | -Y | -Z |
| | 5 2 | 3+3 4 | 3+3 4 | 3+4 4 | 6+6 4 | 3+4 4 | 3+4 4 | 3+4 4 |
| | 5 2 | 4+4 4 | 4+4 4 | 3+4 4 | 5+5 4 | 3+4 4 | 3+4 4 | 3+4 4 |
| | 6 2 | 4+4 4 | 4+4 4 | 3+4 4 | 6+6 4 | 3+4 4 | 3+4 4 | 3+4 4 |
| | 6 2 | 3+3 4 | 3+3 4 | 3+4 4 | 5+5 4 | 3+4 4 | 3+4 4 | 3+4 4 |
| | 5 2 | 3+3 4 | 3+3 4 | 3+4 4 | 5+5 4 | 3+4 4 | 3+4 4 | 3+4 4 |
| | 5 2 | 4+4 4 | 4+4 4 | 3+4 4 | 6+6 4 | 3+4 4 | 3+4 4 | 3+4 4 |
| | 6 2 | 4+4 4 | 4+4 4 | 3+4 4 | 6+6 4 | 3+4 4 | 3+4 4 | 3+4 4 |
| | 5 2 | 3+3 4 | 3+3 4 | 3+4 4 | 5+5 4 | 3+4 4 | 3+4 4 | 3+4 4 |
| | 6 2 | 3+3 4 | 3+3 4 | 3+4 4 | 6+6 4 | 3+4 4 | 3+4 4 | 3+4 4 |
| | 6 2 | 4+4 4 | 4+4 4 | 3+4 4 | 5+5 4 | 3+4 4 | 3+4 4 | 3+4 4 |
| | 5 2 | 3+3 4 | 3+3 4 | 3+4 4 | 6+6 4 | 3+4 4 | 3+4 4 | 3+4 4 |
| | 5 2 | 4+4 4 | 4+4 4 | 3+4 4 | 5+5 4 | 3+4 4 | 3+4 4 | 3+4 4 |
| | 5 2 | 4+4 4 | 3+3 4 | 3+4 4 | 6+6 4 | 3+4 4 | 3+4 4 | 3+4 4 |
| | 5 2 | 3+3 4 | 4+4 4 | 3+4 4 | 5+5 4 | 3+4 4 | 3+4 4 | 3+4 4 |
| | 6 2 | 4+4 4 | 4+4 4 | 3+4 4 | 6+6 4 | 3+4 4 | 3+4 4 | 3+4 4 |
| | 6 2 | 3+3 4 | 3+3 4 | 3+4 4 | 5+5 4 | 3+4 4 | 3+4 4 | 3+4 4 |
| | 5 2 | 3+3 4 | 4+4 4 | 3+4 4 | 5+5 4 | 3+4 4 | 3+4 4 | 3+4 4 |
| | 6 2 | 4+4 4 | 3+3 4 | 3+4 4 | 6+6 4 | 3+4 4 | 3+4 4 | 3+4 4 |
| | 6 2 | 4+4 4 | 3+3 4 | 3+4 4 | 5+5 4 | 3+4 4 | 3+4 4 | 3+4 4 |
| | 6 2 | 3+3 4 | 4+4 4 | 3+4 4 | 6+6 4 | 3+4 4 | 3+4 4 | 3+4 4 |
| | 6 2 | 3+3 4 | 3+3 4 | 3+4 4 | 6+6 4 | 3+4 4 | 3+4 4 | 3+4 4 |
| | 5 2 | 4+4 4 | 4+4 4 | 3+4 4 | 5+5 4 | 3+4 4 | 3+4 4 | 3+4 4 |
| | 6 2 | 4+4 4 | 3+3 4 | 3+4 4 | 6+6 4 | 3+4 4 | 3+4 4 | 3+4 4 |
| | 6 2 | 3+3 4 | 4+4 4 | 3+4 4 | 5+5 4 | 3+4 4 | 3+4 4 | 3+4 4 |
| | 5 2 | 4+4 4 | 4+4 4 | 3+4 4 | 5+5 4 | 3+4 4 | 3+4 4 | 3+4 4 |
| | 5 2 | 3+3 4 | 3+3 4 | 3+4 4 | 6+6 4 | 3+4 4 | 3+4 4 | 3+4 4 |
| | 5 2 | 3+3 4 | 4+4 4 | 3+4 4 | 5+5 4 | 3+4 4 | 3+4 4 | 3+4 4 |
| | 6 2 | 4+4 4 | 3+3 4 | 3+4 4 | 6+6 4 | 3+4 4 | 3+4 4 | 3+4 4 |
| | 6 2 | 4+4 4 | 4+4 4 | 3+4 4 | 5+5 4 | 3+4 4 | 3+4 4 | 3+4 4 |
| | 5 2 | 3+3 4 | 3+3 4 | 3+4 4 | 6+6 4 | 3+4 4 | 3+4 4 | 3+4 4 |
| | 5 2 | 3+3 4 | 4+4 4 | 3+4 4 | 5+5 4 | 3+4 4 | 3+4 4 | 3+4 4 |
| | 6 2 | 4+4 4 | 3+3 4 | 3+4 4 | 6+6 4 | 3+4 4 | 3+4 4 | 3+4 4 |
| | 5 2 | 3+3 4 | 3+3 4 | 3+4 4 | 5+5 4 | 3+4 4 | 3+4 4 | 3+4 4 |
| | 6 2 | 4+4 4 | 4+4 4 | 3+4 4 | 6+6 4 | 3+4 4 | 3+4 4 | 3+4 4 |
| | 5 2 | 4+4 4 | 3+3 4 | 3+4 4 | 5+5 4 | 3+4 4 | 3+4 4 | 3+4 4 |
| | 6 2 | 3+3 4 | 4+4 4 | 3+4 4 | 6+6 4 | 3+4 4 | 3+4 4 | 3+4 4 |
| | 5 2 | 3+3 4 | 3+3 4 | 3+4 4 | 6+6 4 | 3+4 4 | 3+4 4 | 3+4 4 |
| | 6 2 | 4+4 4 | 4+4 4 | 3+4 4 | 5+5 4 | 3+4 4 | 3+4 4 | 3+4 4 |
| | 5 2 | 3+3 4 | 4+4 4 | 3+4 4 | 5+5 4 | 3+4 4 | 3+4 4 | 3+4 4 |
| | 6 2 | 4+4 4 | 3+3 4 | 3+4 4 | 6+6 4 | 3+4 4 | 3+4 4 | 3+4 4 |
| | 5 2 | 4+4 4 | 3+3 4 | 3+4 4 | 5+5 4 | 3+4 4 | 3+4 4 | 3+4 4 |
| | 6 2 | 3+3 4 | 4+4 4 | 3+4 4 | 6+6 4 | 3+4 4 | 3+4 4 | 3+4 4 |
| | 6 2 | 3+3 4 | 4+4 4 | 3+4 4 | 5+5 4 | 3+4 4 | 3+4 4 | 3+4 4 |
| | 5 2 | 4+4 4 | 3+3 4 | 3+4 4 | 5+5 4 | 3+4 4 | 3+4 4 | 3+4 4 |
| | 6 2 | 4+4 4 | 4+4 4 | 3+4 4 | 6+6 4 | 3+4 4 | 3+4 4 | 3+4 4 |
| | 5 2 | 3+3 4 | 3+3 4 | 3+4 4 | 6+6 4 | 3+4 4 | 3+4 4 | 3+4 4 |
| | 6 2 | 3+3 4 | 4+4 4 | 3+4 4 | 5+5 4 | 3+4 4 | 3+4 4 | 3+4 4 |
| | 5 2 | 4+4 4 | 3+3 4 | 3+4 4 | 6+6 4 | 3+4 4 | 3+4 4 | 3+4 4 |
| | 5 2 | | | | | | | |
| | 6 2 | | | | | | | |
| | 5 2 | | | | | | | |
| | 6 2 | | | | | | | |
| | 6 2 | | | | | | | |
| | 5 2 | | | | | | | |
| | 6 2 | | | | | | | |
| | 5 2 | | | | | | | |
| | 5 2 | | | | | | | |
| | 6 2 | | | | | | | |
| | 6 2 | | | | | | | |
| | 5 2 | | | | | | | |
| | 5 2 | | | | | | | |
| | 6 2 | | | | | | | |
| | 5 2 | | | | | | | |
| | 6 2 | | | | | | | |
| | 5 2 | | | | | | | |
| | 6 2 | | | | | | | |
| | 5 2 | | | | | | | |
| | 6 2 | | | | | | | |
| | 5 2 | | | | | | | |
| | 6 2 | | | | | | | |
| | 5 2 | | | | | | | |
| | 6 2 | | | | | | | |
| | 5 2 | | | | | | | |
| | 6 2 | | | | | | | |
| | 5 2 | | | | | | | |

Continued on next column

Continued on next column

**TABLE XXXVII– continued**

| TQC | ICSD | NoE | SF | | | | | |
|---|---|---|---|---|---|---|---|---|
| | 6 2 | | | | | | | |
| | 5 2 | | | | | | | |
| | 5 2 | | | | | | | |
| | 6 2 | | | | | | | |
| | 6 2 | | | | | | | |
| | 5 2 | | | | | | | |
| | 5 2 | | | | | | | |
| | 5 2 | | | | | | | |
| | 6 2 | | | | | | | |
| | 5 2 | | | | | | | |
| | 6 2 | | | | | | | |
| | 5 2 | | | | | | | |
| | 6 2 | | | | | | | |
| | 6 2 | | | | | | | |
| | 6 2 | | | | | | | |
| | 5 2 | | | | | | | |
| | 5 2 | | | | | | | |
| | 6 2 | | | | | | | |
| | 6 2 | | | | | | | |
| | 5 2 | | | | | | | |
| | 5 2 | | | | | | | |
| 62.2.227 | 418239 | 264 | Au3Ca3In1 | | | | | |
| | -GM | -R | -S | -T | -U | -X | -Y | -Z |
| | 5 2 | 3+3 4 | 3+3 4 | 3+4 4 | 5+5 4 | 3+4 4 | 3+4 4 | 3+4 4 |
| | 5 2 | 4+4 4 | 4+4 4 | 3+4 4 | 6+6 4 | 3+4 4 | 3+4 4 | 3+4 4 |
| | 6 2 | 3+3 4 | 3+3 4 | 3+4 4 | 5+5 4 | 3+4 4 | 3+4 4 | 3+4 4 |
| | 6 2 | 4+4 4 | 4+4 4 | 3+4 4 | 6+6 4 | 3+4 4 | 3+4 4 | 3+4 4 |
| | 5 2 | 3+3 4 | 3+3 4 | 3+4 4 | 5+5 4 | 3+4 4 | 3+4 4 | 3+4 4 |
| | 5 2 | 4+4 4 | 4+4 4 | 3+4 4 | 6+6 4 | 3+4 4 | 3+4 4 | 3+4 4 |
| | 6 2 | 3+3 4 | 3+3 4 | 3+4 4 | 5+5 4 | 3+4 4 | 3+4 4 | 3+4 4 |
| | 6 2 | 4+4 4 | 4+4 4 | 3+4 4 | 6+6 4 | 3+4 4 | 3+4 4 | 3+4 4 |
| | 5 2 | 4+4 4 | 3+3 4 | 3+4 4 | 5+5 4 | 3+4 4 | 3+4 4 | 3+4 4 |
| | 5 2 | 3+3 4 | 4+4 4 | 3+4 4 | 6+6 4 | 3+4 4 | 3+4 4 | 3+4 4 |
| | 6 2 | 3+3 4 | 3+3 4 | 3+4 4 | 5+5 4 | 3+4 4 | 3+4 4 | 3+4 4 |
| | 6 2 | 4+4 4 | 4+4 4 | 3+4 4 | 6+6 4 | 3+4 4 | 3+4 4 | 3+4 4 |
| | 5 2 | 3+3 4 | 3+3 4 | 3+4 4 | 6+6 4 | 3+4 4 | 3+4 4 | 3+4 4 |
| | 5 2 | 4+4 4 | 4+4 4 | 3+4 4 | 5+5 4 | 3+4 4 | 3+4 4 | 3+4 4 |
| | 6 2 | 3+3 4 | 4+4 4 | 3+4 4 | 6+6 4 | 3+4 4 | 3+4 4 | 3+4 4 |
| | 6 2 | 4+4 4 | 3+3 4 | 3+4 4 | 5+5 4 | 3+4 4 | 3+4 4 | 3+4 4 |
| | 5 2 | 3+3 4 | 3+3 4 | 3+4 4 | 5+5 4 | 3+4 4 | 3+4 4 | 3+4 4 |
| | 6 2 | 4+4 4 | 4+4 4 | 3+4 4 | 6+6 4 | 3+4 4 | 3+4 4 | 3+4 4 |
| | 5 2 | 3+3 4 | 3+3 4 | 3+4 4 | 6+6 4 | 3+4 4 | 3+4 4 | 3+4 4 |
| | 6 2 | 4+4 4 | 4+4 4 | 3+4 4 | 5+5 4 | 3+4 4 | 3+4 4 | 3+4 4 |
| | 5 2 | 3+3 4 | 3+3 4 | 3+4 4 | 5+5 4 | 3+4 4 | 3+4 4 | 3+4 4 |
| | 5 2 | 4+4 4 | 4+4 4 | 3+4 4 | 6+6 4 | 3+4 4 | 3+4 4 | 3+4 4 |
| | 6 2 | 3+3 4 | 3+3 4 | 3+4 4 | 5+5 4 | 3+4 4 | 3+4 4 | 3+4 4 |
| | 6 2 | 4+4 4 | 4+4 4 | 3+4 4 | 6+6 4 | 3+4 4 | 3+4 4 | 3+4 4 |
| | 6 2 | 4+4 4 | 3+3 4 | 3+4 4 | 6+6 4 | 3+4 4 | 3+4 4 | 3+4 4 |
| | 6 2 | 3+3 4 | 4+4 4 | 3+4 4 | 5+5 4 | 3+4 4 | 3+4 4 | 3+4 4 |
| | 5 2 | 4+4 4 | 3+3 4 | 3+4 4 | 5+5 4 | 3+4 4 | 3+4 4 | 3+4 4 |
| | 5 2 | 3+3 4 | 4+4 4 | 3+4 4 | 6+6 4 | 3+4 4 | 3+4 4 | 3+4 4 |
| | 6 2 | 3+3 4 | 3+3 4 | 3+4 4 | 5+5 4 | 3+4 4 | 3+4 4 | 3+4 4 |
| | 6 2 | 4+4 4 | 4+4 4 | 3+4 4 | 6+6 4 | 3+4 4 | 3+4 4 | 3+4 4 |
| | 5 2 | 4+4 4 | 3+3 4 | 3+4 4 | 6+6 4 | 3+4 4 | 3+4 4 | 3+4 4 |
| | 5 2 | 3+3 4 | 4+4 4 | 3+4 4 | 5+5 4 | 3+4 4 | 3+4 4 | 3+4 4 |
| | 6 2 | 3+3 4 | 3+3 4 | 3+4 4 | 6+6 4 | 3+4 4 | 3+4 4 | 3+4 4 |
| | 6 2 | 4+4 4 | 4+4 4 | 3+4 4 | 5+5 4 | 3+4 4 | 3+4 4 | 3+4 4 |
| | 5 2 | 4+4 4 | 4+4 4 | 3+4 4 | 5+5 4 | 3+4 4 | 3+4 4 | 3+4 4 |
| | 5 2 | 3+3 4 | 3+3 4 | 3+4 4 | 6+6 4 | 3+4 4 | 3+4 4 | 3+4 4 |
| | 5 2 | 3+3 4 | 4+4 4 | 3+4 4 | 5+5 4 | 3+4 4 | 3+4 4 | 3+4 4 |
| | 5 2 | 4+4 4 | 3+3 4 | 3+4 4 | 6+6 4 | 3+4 4 | 3+4 4 | 3+4 4 |

Continued on next column

**TABLE XXXVII– continued**

| TQC | ICSD | NoE | SF | | | | | |
|---|---|---|---|---|---|---|---|---|
| | 6 2 | 3+3 4 | 4+4 4 | 3+4 4 | 5+5 4 | 3+4 4 | 3+4 4 | 3+4 4 |
| | 6 2 | 4+4 4 | 3+3 4 | 3+4 4 | 6+6 4 | 3+4 4 | 3+4 4 | 3+4 4 |
| | 5 2 | 3+3 4 | 3+3 4 | 3+4 4 | 5+5 4 | 3+4 4 | 3+4 4 | 3+4 4 |
| | 5 2 | 4+4 4 | 4+4 4 | 3+4 4 | 6+6 4 | 3+4 4 | 3+4 4 | 3+4 4 |
| | 6 2 | 3+3 4 | 4+4 4 | 3+4 4 | 6+6 4 | 3+4 4 | 3+4 4 | 3+4 4 |
| | 6 2 | 4+4 4 | 3+3 4 | 3+4 4 | 5+5 4 | 3+4 4 | 3+4 4 | 3+4 4 |
| | 5 2 | 4+4 4 | 4+4 4 | 3+4 4 | 5+5 4 | 3+4 4 | 3+4 4 | 3+4 4 |
| | 5 2 | 3+3 4 | 3+3 4 | 3+4 4 | 6+6 4 | 3+4 4 | 3+4 4 | 3+4 4 |
| | 6 2 | 4+4 4 | 3+3 4 | 3+4 4 | 5+5 4 | 3+4 4 | 3+4 4 | 3+4 4 |
| | 6 2 | 3+3 4 | 4+4 4 | 3+4 4 | 6+6 4 | 3+4 4 | 3+4 4 | 3+4 4 |
| | 5 2 | 4+4 4 | 3+3 4 | 3+4 4 | 5+5 4 | 3+4 4 | 3+4 4 | 3+4 4 |
| | 6 2 | 3+3 4 | 4+4 4 | 3+4 4 | 6+6 4 | 3+4 4 | 3+4 4 | 3+4 4 |
| | 5 2 | 4+4 4 | 3+3 4 | 3+4 4 | 5+5 4 | 3+4 4 | 3+4 4 | 3+4 4 |
| | 6 2 | 3+3 4 | 4+4 4 | 3+4 4 | 6+6 4 | 3+4 4 | 3+4 4 | 3+4 4 |
| | 5 2 | 4+4 4 | 3+3 4 | 3+4 4 | 6+6 4 | 3+4 4 | 3+4 4 | 3+4 4 |
| | 5 2 | 3+3 4 | 4+4 4 | 3+4 4 | 5+5 4 | 3+4 4 | 3+4 4 | 3+4 4 |
| | 5 2 | 3+3 4 | 4+4 4 | 3+4 4 | 5+5 4 | 3+4 4 | 3+4 4 | 3+4 4 |
| | 5 2 | 4+4 4 | 3+3 4 | 3+4 4 | 6+6 4 | 3+4 4 | 3+4 4 | 3+4 4 |
| | 6 2 | 4+4 4 | 3+3 4 | 3+4 4 | 5+5 4 | 3+4 4 | 3+4 4 | 3+4 4 |
| | 6 2 | 3+3 4 | 4+4 4 | 3+4 4 | 6+6 4 | 3+4 4 | 3+4 4 | 3+4 4 |
| | 5 2 | 3+3 4 | 3+3 4 | 3+4 4 | 5+5 4 | 3+4 4 | 3+4 4 | 3+4 4 |
| | 6 2 | 4+4 4 | 4+4 4 | 3+4 4 | 6+6 4 | 3+4 4 | 3+4 4 | 3+4 4 |
| | 5 2 | 3+3 4 | 3+3 4 | 3+4 4 | 6+6 4 | 3+4 4 | 3+4 4 | 3+4 4 |
| | 5 2 | 4+4 4 | 4+4 4 | 3+4 4 | 5+5 4 | 3+4 4 | 3+4 4 | 3+4 4 |
| | 6 2 | 4+4 4 | 3+3 4 | 3+4 4 | 6+6 4 | 3+4 4 | 3+4 4 | 3+4 4 |
| | 5 2 | 3+3 4 | 4+4 4 | 3+4 4 | 5+5 4 | 3+4 4 | 3+4 4 | 3+4 4 |
| | 6 2 | 4+4 4 | 3+3 4 | 3+4 4 | 6+6 4 | 3+4 4 | 3+4 4 | 3+4 4 |
| | 6 2 | 3+3 4 | 4+4 4 | 3+4 4 | 5+5 4 | 3+4 4 | 3+4 4 | 3+4 4 |
| | 6 2 | | | | | | | |
| | 6 2 | | | | | | | |
| | 5 2 | | | | | | | |
| | 5 2 | | | | | | | |
| | 6 2 | | | | | | | |
| | 5 2 | | | | | | | |
| | 5 2 | | | | | | | |
| | 6 2 | | | | | | | |
| | 6 2 | | | | | | | |
| | 5 2 | | | | | | | |
| | 6 2 | | | | | | | |
| | 5 2 | | | | | | | |
| | 5 2 | | | | | | | |
| | 5 2 | | | | | | | |
| | 6 2 | | | | | | | |
| | 6 2 | | | | | | | |
| | 6 2 | | | | | | | |
| | 6 2 | | | | | | | |
| | 6 2 | | | | | | | |
| | 5 2 | | | | | | | |
| | 5 2 | | | | | | | |
| | 6 2 | | | | | | | |
| | 5 2 | | | | | | | |
| | 6 2 | | | | | | | |
| | 5 2 | | | | | | | |
| | 5 2 | | | | | | | |
| | 6 2 | | | | | | | |
| | 6 2 | | | | | | | |
| | 6 2 | | | | | | | |
| | 5 2 | | | | | | | |
| | 5 2 | | | | | | | |
| | 5 2 | | | | | | | |
| | 6 2 | | | | | | | |

Continued on next column

**TABLE XXXVII– continued**

| TQC | ICSD NoE SF |
|---|---|
| | 6 2 |
| | 5 2 |
| | 6 2 |
| | 5 2 |
| | 5 2 |
| | 5 2 |
| | 5 2 |
| | 6 2 |
| | 6 2 |
| | 5 2 |
| | 5 2 |
| | 6 2 |
| | 6 2 |
| | 6 2 |
| | 6 2 |
| | 5 2 |
| | 5 2 |
| | 5 2 |
| | 6 2 |
| | 6 2 |
| | 5 2 |
| | 6 2 |
| | 6 2 |
| | 6 2 |
| | 5 2 |
| | 5 2 |
| | 5 2 |
| | 5 2 |
| | 6 2 |
| | 6 2 |
| | 6 2 |
| | 5 2 |
| | 5 2 |
| 62.2.228 | 249563 152 Au2In2Sr1 |
| | -GM -R -S -T -U -X -Y -Z |
| | 5 2 3+3 4 3+3 4 3+4 4 5+5 4 3+4 4 3+4 4 3+4 4 |
| | 5 2 4+4 4 4+4 4 3+4 4 6+6 4 3+4 4 3+4 4 3+4 4 |
| | 6 2 3+3 4 3+3 4 3+4 4 5+5 4 3+4 4 3+4 4 3+4 4 |
| | 6 2 4+4 4 4+4 4 3+4 4 6+6 4 3+4 4 3+4 4 3+4 4 |
| | 5 2 3+3 4 4+4 4 3+4 4 5+5 4 3+4 4 3+4 4 3+4 4 |
| | 6 2 4+4 4 3+3 4 3+4 4 6+6 4 3+4 4 3+4 4 3+4 4 |
| | 5 2 3+3 4 3+3 4 3+4 4 5+5 4 3+4 4 3+4 4 3+4 4 |
| | 6 2 4+4 4 4+4 4 3+4 4 6+6 4 3+4 4 3+4 4 3+4 4 |
| | 5 2 3+3 4 4+4 4 3+4 4 5+5 4 3+4 4 3+4 4 3+4 4 |
| | 6 2 4+4 4 3+3 4 3+4 4 6+6 4 3+4 4 3+4 4 3+4 4 |
| | 5 2 4+4 4 3+3 4 3+4 4 6+6 4 3+4 4 3+4 4 3+4 4 |
| | 6 2 3+3 4 4+4 4 3+4 4 5+5 4 3+4 4 3+4 4 3+4 4 |
| | 5 2 3+3 4 4+4 4 3+4 4 5+5 4 3+4 4 3+4 4 3+4 4 |
| | 6 2 4+4 4 3+3 4 3+4 4 6+6 4 3+4 4 3+4 4 3+4 4 |
| | 5 2 3+3 4 3+3 4 3+4 4 5+5 4 3+4 4 3+4 4 3+4 4 |
| | 6 2 4+4 4 4+4 4 3+4 4 6+6 4 3+4 4 3+4 4 3+4 4 |
| | 5 2 4+4 4 4+4 4 3+4 4 6+6 4 3+4 4 3+4 4 3+4 4 |
| | 5 2 3+3 4 3+3 4 3+4 4 5+5 4 3+4 4 3+4 4 3+4 4 |
| | 6 2 4+4 4 3+3 4 3+4 4 5+5 4 3+4 4 3+4 4 3+4 4 |
| | 5 2 3+3 4 4+4 4 3+4 4 6+6 4 3+4 4 3+4 4 3+4 4 |
| | 6 2 3+3 4 4+4 4 3+4 4 6+6 4 3+4 4 3+4 4 3+4 4 |
| | 5 2 4+4 4 3+3 4 3+4 4 5+5 4 3+4 4 3+4 4 3+4 4 |
| | 6 2 3+3 4 3+3 4 3+4 4 6+6 4 3+4 4 3+4 4 3+4 4 |
| | 6 2 4+4 4 4+4 4 3+4 4 5+5 4 3+4 4 3+4 4 3+4 4 |
| | 5 2 3+3 4 4+4 4 3+4 4 5+5 4 3+4 4 3+4 4 3+4 4 |
| | 5 2 4+4 4 3+3 4 3+4 4 6+6 4 3+4 4 3+4 4 3+4 4 |
| | 5 2 4+4 4 3+3 4 3+4 4 5+5 4 3+4 4 3+4 4 3+4 4 |
| | 6 2 3+3 4 4+4 4 3+4 4 6+6 4 3+4 4 3+4 4 3+4 4 |
| | 6 2 3+3 4 4+4 4 3+4 4 6+6 4 3+4 4 3+4 4 3+4 4 |
| | 5 2 4+4 4 3+3 4 3+4 4 5+5 4 3+4 4 3+4 4 3+4 4 |
| | 6 2 4+4 4 3+3 4 3+4 4 6+6 4 3+4 4 3+4 4 3+4 4 |
| | 5 2 3+3 4 4+4 4 3+4 4 5+5 4 3+4 4 3+4 4 3+4 4 |
| | 6 2 3+3 4 3+3 4 3+4 4 6+6 4 3+4 4 3+4 4 3+4 4 |
| | 5 2 4+4 4 4+4 4 3+4 4 5+5 4 3+4 4 3+4 4 3+4 4 |
| | 6 2 4+4 4 4+4 4 3+4 4 5+5 4 3+4 4 3+4 4 3+4 4 |
| | 5 2 3+3 4 3+3 4 3+4 4 6+6 4 3+4 4 3+4 4 3+4 4 |
| | 6 2 4+4 4 3+3 4 3+4 4 5+5 4 3+4 4 3+4 4 3+4 4 |
| | 5 2 3+3 4 4+4 4 3+4 4 6+6 4 3+4 4 3+4 4 3+4 4 |
| | 6 2 |
| | 6 2 |
| | 5 2 |
| | 6 2 |
| | 6 2 |
| | 5 2 |
| | 6 2 |
| | 5 2 |
| | 5 2 |
| | 6 2 |
| | 5 2 |
| | 5 2 |
| | 6 2 |
| | 5 2 |
| | 5 2 |
| | 6 2 |
| | 6 2 |
| | 5 2 |
| | 6 2 |
| | 5 2 |
| | 6 2 |
| | 6 2 |
| | 6 2 |
| | 5 2 |
| | 5 2 |
| | 6 2 |
| | 5 2 |
| | 5 2 |
| | 6 2 |
| | 5 2 |
| | 6 2 |
| | 5 2 |
| | 5 2 |
| | 6 2 |
| | 5 2 |
| | 5 2 |
| | 5 2 |
| | 6 2 |
| 62.2.229 | 249562 152 Au2Ba1In2 |
| | -GM -R -S -T -U -X -Y -Z |
| | 5 2 3+3 4 3+3 4 3+4 4 5+5 4 3+4 4 3+4 4 3+4 4 |
| | 6 2 4+4 4 4+4 4 3+4 4 6+6 4 3+4 4 3+4 4 3+4 4 |
| | 5 2 3+3 4 3+3 4 3+4 4 5+5 4 3+4 4 3+4 4 3+4 4 |
| | 6 2 4+4 4 4+4 4 3+4 4 6+6 4 3+4 4 3+4 4 3+4 4 |
| | 5 2 3+3 4 4+4 4 3+4 4 5+5 4 3+4 4 3+4 4 3+4 4 |
| | 6 2 4+4 4 3+3 4 3+4 4 6+6 4 3+4 4 3+4 4 3+4 4 |
| | 5 2 4+4 4 3+3 4 3+4 4 6+6 4 3+4 4 3+4 4 3+4 4 |
| | 6 2 3+3 4 4+4 4 3+4 4 5+5 4 3+4 4 3+4 4 3+4 4 |
| | 5 2 3+3 4 4+4 4 3+4 4 5+5 4 3+4 4 3+4 4 3+4 4 |

Continued on next column

**TABLE XXXVII– continued**

| TQC | ICSD | NoE | SF | | | | | |
|---|---|---|---|---|---|---|---|---|
| | 6 2 | 4+4 4 | 3+3 4 | 3+4 4 | 6+6 4 | 3+4 4 | 3+4 4 | 3+4 4 |
| | 5 2 | 4+4 4 | 3+3 4 | 3+4 4 | 6+6 4 | 3+4 4 | 3+4 4 | 3+4 4 |
| | 6 2 | 3+3 4 | 4+4 4 | 3+4 4 | 5+5 4 | 3+4 4 | 3+4 4 | 3+4 4 |
| | 5 2 | 3+3 4 | 4+4 4 | 3+4 4 | 5+5 4 | 3+4 4 | 3+4 4 | 3+4 4 |
| | 6 2 | 4+4 4 | 3+3 4 | 3+4 4 | 6+6 4 | 3+4 4 | 3+4 4 | 3+4 4 |
| | 5 2 | 3+3 4 | 3+3 4 | 3+4 4 | 6+6 4 | 3+4 4 | 3+4 4 | 3+4 4 |
| | 6 2 | 4+4 4 | 4+4 4 | 3+4 4 | 5+5 4 | 3+4 4 | 3+4 4 | 3+4 4 |
| | 5 2 | 4+4 4 | 4+4 4 | 3+4 4 | 6+6 4 | 3+4 4 | 3+4 4 | 3+4 4 |
| | 5 2 | 3+3 4 | 3+3 4 | 3+4 4 | 5+5 4 | 3+4 4 | 3+4 4 | 3+4 4 |
| | 6 2 | 4+4 4 | 3+3 4 | 3+4 4 | 5+5 4 | 3+4 4 | 3+4 4 | 3+4 4 |
| | 5 2 | 3+3 4 | 4+4 4 | 3+4 4 | 6+6 4 | 3+4 4 | 3+4 4 | 3+4 4 |
| | 5 2 | 4+4 4 | 4+4 4 | 3+4 4 | 6+6 4 | 3+4 4 | 3+4 4 | 3+4 4 |
| | 6 2 | 3+3 4 | 3+3 4 | 3+4 4 | 5+5 4 | 3+4 4 | 3+4 4 | 3+4 4 |
| | 6 2 | 3+3 4 | 4+4 4 | 3+4 4 | 6+6 4 | 3+4 4 | 3+4 4 | 3+4 4 |
| | 6 2 | 4+4 4 | 3+3 4 | 3+4 4 | 5+5 4 | 3+4 4 | 3+4 4 | 3+4 4 |
| | 5 2 | 3+3 4 | 3+3 4 | 3+4 4 | 5+5 4 | 3+4 4 | 3+4 4 | 3+4 4 |
| | 5 2 | 4+4 4 | 4+4 4 | 3+4 4 | 6+6 4 | 3+4 4 | 3+4 4 | 3+4 4 |
| | 6 2 | 4+4 4 | 3+3 4 | 3+4 4 | 6+6 4 | 3+4 4 | 3+4 4 | 3+4 4 |
| | 6 2 | 3+3 4 | 4+4 4 | 3+4 4 | 5+5 4 | 3+4 4 | 3+4 4 | 3+4 4 |
| | 5 2 | 3+3 4 | 4+4 4 | 3+4 4 | 6+6 4 | 3+4 4 | 3+4 4 | 3+4 4 |
| | 5 2 | 4+4 4 | 3+3 4 | 3+4 4 | 5+5 4 | 3+4 4 | 3+4 4 | 3+4 4 |
| | 6 2 | 4+4 4 | 3+3 4 | 3+4 4 | 6+6 4 | 3+4 4 | 3+4 4 | 3+4 4 |
| | 5 2 | 3+3 4 | 4+4 4 | 3+4 4 | 5+5 4 | 3+4 4 | 3+4 4 | 3+4 4 |
| | 6 2 | 3+3 4 | 3+3 4 | 3+4 4 | 6+6 4 | 3+4 4 | 3+4 4 | 3+4 4 |
| | 5 2 | 4+4 4 | 4+4 4 | 3+4 4 | 5+5 4 | 3+4 4 | 3+4 4 | 3+4 4 |
| | 6 2 | 4+4 4 | 4+4 4 | 3+4 4 | 5+5 4 | 3+4 4 | 3+4 4 | 3+4 4 |
| | 5 2 | 3+3 4 | 3+3 4 | 3+4 4 | 5+5 4 | 3+4 4 | 3+4 4 | 3+4 4 |
| | 6 2 | 4+4 4 | 3+3 4 | 3+4 4 | 6+6 4 | 3+4 4 | 3+4 4 | 3+4 4 |
| | 5 2 | 3+3 4 | 4+4 4 | 3+4 4 | 6+6 4 | 3+4 4 | 3+4 4 | 3+4 4 |
| | 6 2 | | | | | | | |
| | 6 2 | | | | | | | |
| | 5 2 | | | | | | | |
| | 6 2 | | | | | | | |
| | 5 2 | | | | | | | |
| | 6 2 | | | | | | | |
| | 5 2 | | | | | | | |
| | 5 2 | | | | | | | |
| | 6 2 | | | | | | | |
| | 6 2 | | | | | | | |
| | 5 2 | | | | | | | |
| | 5 2 | | | | | | | |
| | 6 2 | | | | | | | |
| | 6 2 | | | | | | | |
| | 5 2 | | | | | | | |
| | 6 2 | | | | | | | |
| | 5 2 | | | | | | | |
| | 6 2 | | | | | | | |
| | 5 2 | | | | | | | |
| | 5 2 | | | | | | | |
| | 6 2 | | | | | | | |
| | 6 2 | | | | | | | |
| | 5 2 | | | | | | | |
| | 6 2 | | | | | | | |
| | 5 2 | | | | | | | |
| | 6 2 | | | | | | | |
| | 5 2 | | | | | | | |
| | 5 2 | | | | | | | |
| | 6 2 | | | | | | | |
| | 5 2 | | | | | | | |
| | 6 2 | | | | | | | |
| | 5 2 | | | | | | | |

Continued on next column

**TABLE XXXVII– continued**

| TQC | ICSD | NoE | SF | | | | | |
|---|---|---|---|---|---|---|---|---|
| | 5 2 | | | | | | | |
| | 5 2 | | | | | | | |
| | 6 2 | | | | | | | |
| | 5 2 | | | | | | | |
| | 5 2 | | | | | | | |
| | 5 2 | | | | | | | |
| 62.2.230 | 391422 | 96 | Au1In1Sr1 | | | | | |
| | -GM | -R | -S | -T | -U | -X | -Y | -Z |
| | 5 2 | 3+3 4 | 3+3 4 | 3+4 4 | 5+5 4 | 3+4 4 | 3+4 4 | 3+4 4 |
| | 5 2 | 4+4 4 | 4+4 4 | 3+4 4 | 6+6 4 | 3+4 4 | 3+4 4 | 3+4 4 |
| | 6 2 | 3+3 4 | 3+3 4 | 3+4 4 | 5+5 4 | 3+4 4 | 3+4 4 | 3+4 4 |
| | 6 2 | 4+4 4 | 4+4 4 | 3+4 4 | 6+6 4 | 3+4 4 | 3+4 4 | 3+4 4 |
| | 5 2 | 4+4 4 | 3+3 4 | 3+4 4 | 5+5 4 | 3+4 4 | 3+4 4 | 3+4 4 |
| | 5 2 | 3+3 4 | 4+4 4 | 3+4 4 | 6+6 4 | 3+4 4 | 3+4 4 | 3+4 4 |
| | 6 2 | 3+3 4 | 4+4 4 | 3+4 4 | 5+5 4 | 3+4 4 | 3+4 4 | 3+4 4 |
| | 6 2 | 4+4 4 | 3+3 4 | 3+4 4 | 6+6 4 | 3+4 4 | 3+4 4 | 3+4 4 |
| | 5 2 | 3+3 4 | 4+4 4 | 3+4 4 | 6+6 4 | 3+4 4 | 3+4 4 | 3+4 4 |
| | 5 2 | 4+4 4 | 3+3 4 | 3+4 4 | 5+5 4 | 3+4 4 | 3+4 4 | 3+4 4 |
| | 6 2 | 3+3 4 | 4+4 4 | 3+4 4 | 6+6 4 | 3+4 4 | 3+4 4 | 3+4 4 |
| | 5 2 | 4+4 4 | 3+3 4 | 3+4 4 | 5+5 4 | 3+4 4 | 3+4 4 | 3+4 4 |
| | 6 2 | 4+4 4 | 3+3 4 | 3+4 4 | 6+6 4 | 3+4 4 | 3+4 4 | 3+4 4 |
| | 6 2 | 3+3 4 | 4+4 4 | 3+4 4 | 5+5 4 | 3+4 4 | 3+4 4 | 3+4 4 |
| | 5 2 | 4+4 4 | 3+3 4 | 3+4 4 | 5+5 4 | 3+4 4 | 3+4 4 | 3+4 4 |
| | 6 2 | 3+3 4 | 4+4 4 | 3+4 4 | 6+6 4 | 3+4 4 | 3+4 4 | 3+4 4 |
| | 5 2 | 3+3 4 | 3+3 4 | 3+4 4 | 6+6 4 | 3+4 4 | 3+4 4 | 3+4 4 |
| | 5 2 | 4+4 4 | 4+4 4 | 3+4 4 | 5+5 4 | 3+4 4 | 3+4 4 | 3+4 4 |
| | 6 2 | 3+3 4 | 4+4 4 | 3+4 4 | 6+6 4 | 3+4 4 | 3+4 4 | 3+4 4 |
| | 6 2 | 4+4 4 | 3+3 4 | 3+4 4 | 5+5 4 | 3+4 4 | 3+4 4 | 3+4 4 |
| | 5 2 | 3+3 4 | 4+4 4 | 3+4 4 | 6+6 4 | 3+4 4 | 3+4 4 | 3+4 4 |
| | 6 2 | 4+4 4 | 3+3 4 | 3+4 4 | 5+5 4 | 3+4 4 | 3+4 4 | 3+4 4 |
| | 5 2 | 3+3 4 | 4+4 4 | 3+4 4 | 6+6 4 | 3+4 4 | 3+4 4 | 3+4 4 |
| | 6 2 | 4+4 4 | 3+3 4 | 3+4 4 | 6+6 4 | 3+4 4 | 3+4 4 | 3+4 4 |
| | 5 2 | | | | | | | |
| | 5 2 | | | | | | | |
| | 6 2 | | | | | | | |
| | 6 2 | | | | | | | |
| | 6 2 | | | | | | | |
| | 5 2 | | | | | | | |
| | 6 2 | | | | | | | |
| | 5 2 | | | | | | | |
| | 5 2 | | | | | | | |
| | 6 2 | | | | | | | |
| | 5 2 | | | | | | | |
| | 5 2 | | | | | | | |
| | 6 2 | | | | | | | |
| | 5 2 | | | | | | | |
| | 6 2 | | | | | | | |
| | 6 2 | | | | | | | |
| | 6 2 | | | | | | | |
| | 5 2 | | | | | | | |
| | 5 2 | | | | | | | |
| | 5 2 | | | | | | | |
| | 6 2 | | | | | | | |
| | 5 2 | | | | | | | |
| | 6 2 | | | | | | | |
| | 5 2 | | | | | | | |
| 62.2.231 | 420674 | 160 | Au1Eu1Zn1 | | | | | |
| | -GM | -R | -S | -T | -U | -X | -Y | -Z |
| | 5 2 | 3+3 4 | 3+3 4 | 3+4 4 | 6+6 4 | 3+4 4 | 3+4 4 | 3+4 4 |
| | 6 2 | 4+4 4 | 4+4 4 | 3+4 4 | 5+5 4 | 3+4 4 | 3+4 4 | 3+4 4 |
| | 5 2 | 3+3 4 | 3+3 4 | 3+4 4 | 6+6 4 | 3+4 4 | 3+4 4 | 3+4 4 |

Continued on next column

**TABLE XXXVII– continued**

| TQC | ICSD NoE SF |
|---|---|
| | 6 2 4+4 4 4+4 4 3+4 4 5+5 4 3+4 4 3+4 4 3+4 4 |
| | 5 2 4+4 4 3+3 4 3+4 4 6+6 4 3+4 4 3+4 4 3+4 4 |
| | 6 2 3+3 4 4+4 4 3+4 4 5+5 4 3+4 4 3+4 4 3+4 4 |
| | 5 2 3+3 4 4+4 4 3+4 4 6+6 4 3+4 4 3+4 4 3+4 4 |
| | 6 2 4+4 4 3+3 4 3+4 4 5+5 4 3+4 4 3+4 4 3+4 4 |
| | 5 2 3+3 4 4+4 4 3+4 4 5+5 4 3+4 4 3+4 4 3+4 4 |
| | 6 2 4+4 4 3+3 4 3+4 4 6+6 4 3+4 4 3+4 4 3+4 4 |
| | 5 2 4+4 4 4+4 4 3+4 4 5+5 4 3+4 4 3+4 4 3+4 4 |
| | 5 2 3+3 4 3+3 4 3+4 4 6+6 4 3+4 4 3+4 4 3+4 4 |
| | 6 2 3+3 4 3+3 4 3+4 4 5+5 4 3+4 4 3+4 4 3+4 4 |
| | 6 2 4+4 4 4+4 4 3+4 4 6+6 4 3+4 4 3+4 4 3+4 4 |
| | 5 2 4+4 4 4+4 4 3+4 4 5+5 4 3+4 4 3+4 4 3+4 4 |
| | 6 2 3+3 4 3+3 4 3+4 4 5+5 4 3+4 4 3+4 4 3+4 4 |
| | 5 2 4+4 4 3+3 4 3+4 4 6+6 4 3+4 4 3+4 4 3+4 4 |
| | 6 2 3+3 4 4+4 4 3+4 4 6+6 4 3+4 4 3+4 4 3+4 4 |
| | 5 2 3+3 4 4+4 4 3+4 4 5+5 4 3+4 4 3+4 4 3+4 4 |
| | 6 2 4+4 4 3+3 4 3+4 4 6+6 4 3+4 4 3+4 4 3+4 4 |
| | 5 2 3+3 4 4+4 4 3+4 4 5+5 4 3+4 4 3+4 4 3+4 4 |
| | 5 2 4+4 4 3+3 4 3+4 4 6+6 4 3+4 4 3+4 4 3+4 4 |
| | 6 2 4+4 4 3+3 4 3+4 4 5+5 4 3+4 4 3+4 4 3+4 4 |
| | 5 2 3+3 4 4+4 4 3+4 4 6+6 4 3+4 4 3+4 4 3+4 4 |
| | 6 2 4+4 4 3+3 4 3+4 4 5+5 4 3+4 4 3+4 4 3+4 4 |
| | 6 2 3+3 4 4+4 4 3+4 4 6+6 4 3+4 4 3+4 4 3+4 4 |
| | 5 2 3+3 4 3+3 4 3+4 4 6+6 4 3+4 4 3+4 4 3+4 4 |
| | 5 2 4+4 4 4+4 4 3+4 4 5+5 4 3+4 4 3+4 4 3+4 4 |
| | 5 2 3+3 4 4+4 4 3+4 4 5+5 4 3+4 4 3+4 4 3+4 4 |
| | 5 2 4+4 4 3+3 4 3+4 4 6+6 4 3+4 4 3+4 4 3+4 4 |
| | 6 2 4+4 4 4+4 4 3+4 4 5+5 4 3+4 4 3+4 4 3+4 4 |
| | 6 2 3+3 4 3+3 4 3+4 4 6+6 4 3+4 4 3+4 4 3+4 4 |
| | 6 2 3+3 4 4+4 4 3+4 4 5+5 4 3+4 4 3+4 4 3+4 4 |
| | 5 2 4+4 4 3+3 4 3+4 4 5+5 4 3+4 4 3+4 4 3+4 4 |
| | 5 2 3+3 4 4+4 4 3+4 4 6+6 4 3+4 4 3+4 4 3+4 4 |
| | 6 2 4+4 4 3+3 4 3+4 4 6+6 4 3+4 4 3+4 4 3+4 4 |
| | 6 2 3+3 4 3+3 4 3+4 4 5+5 4 3+4 4 3+4 4 3+4 4 |
| | 5 2 4+4 4 4+4 4 3+4 4 6+6 4 3+4 4 3+4 4 3+4 4 |
| | 6 2 3+3 4 4+4 4 3+4 4 5+5 4 3+4 4 3+4 4 3+4 4 |
| | 6 2 4+4 4 3+3 4 3+4 4 5+5 4 3+4 4 3+4 4 3+4 4 |
| | 5 2 |
| | 6 2 |
| | 5 2 |
| | 6 2 |
| | 5 2 |
| | 5 2 |
| | 6 2 |
| | 6 2 |
| | 6 2 |
| | 5 2 |
| | 6 2 |
| | 5 2 |
| | 5 2 |
| | 6 2 |
| | 6 2 |
| | 5 2 |
| | 5 2 |
| | 5 2 |
| | 6 2 |
| | 6 2 |
| | 6 2 |
| | 5 2 |
| | 5 2 |
| | 5 2 |

Continued on next column

**TABLE XXXVII– continued**

| TQC | ICSD NoE SF |
|---|---|
| | 6 2 |
| | 5 2 |
| | 6 2 |
| | 6 2 |
| | 5 2 |
| | 6 2 |
| | 6 2 |
| | 5 2 |
| | 5 2 |
| | 6 2 |
| | 6 2 |
| | 5 2 |
| | 5 2 |
| | 6 2 |
| | 5 2 |
| | 5 2 |
| 62.2.232 | 411544 160 Au1Cd1Eu1 |
| | -GM -R -S -T -U -X -Y -Z |
| | 5 2 3+3 4 3+3 4 3+4 4 6+6 4 3+4 4 3+4 4 3+4 4 |
| | 5 2 4+4 4 4+4 4 3+4 4 5+5 4 3+4 4 3+4 4 3+4 4 |
| | 6 2 3+3 4 3+3 4 3+4 4 6+6 4 3+4 4 3+4 4 3+4 4 |
| | 6 2 4+4 4 4+4 4 3+4 4 5+5 4 3+4 4 3+4 4 3+4 4 |
| | 5 2 4+4 4 3+3 4 3+4 4 6+6 4 3+4 4 3+4 4 3+4 4 |
| | 6 2 3+3 4 4+4 4 3+4 4 5+5 4 3+4 4 3+4 4 3+4 4 |
| | 5 2 3+3 4 4+4 4 3+4 4 6+6 4 3+4 4 3+4 4 3+4 4 |
| | 6 2 4+4 4 3+3 4 3+4 4 5+5 4 3+4 4 3+4 4 3+4 4 |
| | 5 2 3+3 4 3+3 4 3+4 4 6+6 4 3+4 4 3+4 4 3+4 4 |
| | 6 2 4+4 4 4+4 4 3+4 4 5+5 4 3+4 4 3+4 4 3+4 4 |
| | 5 2 3+3 4 4+4 4 3+4 4 6+6 4 3+4 4 3+4 4 3+4 4 |
| | 5 2 4+4 4 3+3 4 3+4 4 5+5 4 3+4 4 3+4 4 3+4 4 |
| | 6 2 4+4 4 4+4 4 3+4 4 6+6 4 3+4 4 3+4 4 3+4 4 |
| | 6 2 3+3 4 3+3 4 3+4 4 5+5 4 3+4 4 3+4 4 3+4 4 |
| | 5 2 3+3 4 3+3 4 3+4 4 5+5 4 3+4 4 3+4 4 3+4 4 |
| | 6 2 4+4 4 4+4 4 3+4 4 6+6 4 3+4 4 3+4 4 3+4 4 |
| | 6 2 3+3 4 3+3 4 3+4 4 5+5 4 3+4 4 3+4 4 3+4 4 |
| | 5 2 4+4 4 4+4 4 3+4 4 6+6 4 3+4 4 3+4 4 3+4 4 |
| | 6 2 3+3 4 4+4 4 3+4 4 5+5 4 3+4 4 3+4 4 3+4 4 |
| | 5 2 4+4 4 3+3 4 3+4 4 6+6 4 3+4 4 3+4 4 3+4 4 |
| | 6 2 4+4 4 4+4 4 3+4 4 5+5 4 3+4 4 3+4 4 3+4 4 |
| | 5 2 3+3 4 3+3 4 3+4 4 6+6 4 3+4 4 3+4 4 3+4 4 |
| | 6 2 4+4 4 3+3 4 3+4 4 5+5 4 3+4 4 3+4 4 3+4 4 |
| | 5 2 3+3 4 4+4 4 3+4 4 6+6 4 3+4 4 3+4 4 3+4 4 |
| | 6 2 3+3 4 3+3 4 3+4 4 5+5 4 3+4 4 3+4 4 3+4 4 |
| | 6 2 4+4 4 4+4 4 3+4 4 6+6 4 3+4 4 3+4 4 3+4 4 |
| | 5 2 4+4 4 3+3 4 3+4 4 5+5 4 3+4 4 3+4 4 3+4 4 |
| | 5 2 3+3 4 4+4 4 3+4 4 6+6 4 3+4 4 3+4 4 3+4 4 |
| | 5 2 3+3 4 4+4 4 3+4 4 5+5 4 3+4 4 3+4 4 3+4 4 |
| | 5 2 4+4 4 3+3 4 3+4 4 6+6 4 3+4 4 3+4 4 3+4 4 |
| | 6 2 4+4 4 4+4 4 3+4 4 5+5 4 3+4 4 3+4 4 3+4 4 |
| | 6 2 3+3 4 3+3 4 3+4 4 6+6 4 3+4 4 3+4 4 3+4 4 |
| | 5 2 3+3 4 4+4 4 3+4 4 5+5 4 3+4 4 3+4 4 3+4 4 |
| | 5 2 4+4 4 3+3 4 3+4 4 6+6 4 3+4 4 3+4 4 3+4 4 |
| | 6 2 4+4 4 3+3 4 3+4 4 5+5 4 3+4 4 3+4 4 3+4 4 |
| | 6 2 3+3 4 4+4 4 3+4 4 6+6 4 3+4 4 3+4 4 3+4 4 |
| | 5 2 3+3 4 3+3 4 3+4 4 5+5 4 3+4 4 3+4 4 3+4 4 |
| | 5 2 4+4 4 4+4 4 3+4 4 6+6 4 3+4 4 3+4 4 3+4 4 |
| | 6 2 3+3 4 3+3 4 3+4 4 5+5 4 3+4 4 3+4 4 3+4 4 |
| | 6 2 4+4 4 4+4 4 3+4 4 6+6 4 3+4 4 3+4 4 3+4 4 |
| | 5 2 |
| | 6 2 |
| | 5 2 |

Continued on next column

**TABLE XXXVII– continued**

| TQC | ICSD | NoE | SF | | | | | |
|---|---|---|---|---|---|---|---|---|
| | 5 2 | | | | | | | |
| | 6 2 | | | | | | | |
| | 5 2 | | | | | | | |
| | 6 2 | | | | | | | |
| | 6 2 | | | | | | | |
| | 6 2 | | | | | | | |
| | 5 2 | | | | | | | |
| | 5 2 | | | | | | | |
| | 6 2 | | | | | | | |
| | 5 2 | | | | | | | |
| | 6 2 | | | | | | | |
| | 5 2 | | | | | | | |
| | 6 2 | | | | | | | |
| | 5 2 | | | | | | | |
| | 5 2 | | | | | | | |
| | 6 2 | | | | | | | |
| | 6 2 | | | | | | | |
| | 5 2 | | | | | | | |
| | 6 2 | | | | | | | |
| | 5 2 | | | | | | | |
| | 5 2 | | | | | | | |
| | 6 2 | | | | | | | |
| | 5 2 | | | | | | | |
| | 6 2 | | | | | | | |
| | 5 2 | | | | | | | |
| | 6 2 | | | | | | | |
| | 5 2 | | | | | | | |
| | 6 2 | | | | | | | |
| | 5 2 | | | | | | | |
| | 6 2 | | | | | | | |
| | 5 2 | | | | | | | |
| | 6 2 | | | | | | | |
| | 5 2 | | | | | | | |
| | 6 2 | | | | | | | |
| | 5 2 | | | | | | | |
| | 6 2 | | | | | | | |
| | 5 2 | | | | | | | |
| 62.2.233 | 408579 | 64 | Au1Ca1In1 | | | | | |
| | -GM | -R | -S | -T | -U | -X | -Y | -Z |
| | 5 2 | 3+3 4 | 4+4 4 | 3+4 4 | 6+6 4 | 3+4 4 | 3+4 4 | 3+4 4 |
| | 5 2 | 4+4 4 | 3+3 4 | 3+4 4 | 5+5 4 | 3+4 4 | 3+4 4 | 3+4 4 |
| | 6 2 | 3+3 4 | 4+4 4 | 3+4 4 | 6+6 4 | 3+4 4 | 3+4 4 | 3+4 4 |
| | 6 2 | 4+4 4 | 3+3 4 | 3+4 4 | 5+5 4 | 3+4 4 | 3+4 4 | 3+4 4 |
| | 5 2 | 4+4 4 | 3+3 4 | 3+4 4 | 6+6 4 | 3+4 4 | 3+4 4 | 3+4 4 |
| | 6 2 | 3+3 4 | 4+4 4 | 3+4 4 | 5+5 4 | 3+4 4 | 3+4 4 | 3+4 4 |
| | 5 2 | 4+4 4 | 3+3 4 | 3+4 4 | 5+5 4 | 3+4 4 | 3+4 4 | 3+4 4 |
| | 6 2 | 3+3 4 | 4+4 4 | 3+4 4 | 6+6 4 | 3+4 4 | 3+4 4 | 3+4 4 |
| | 5 2 | 3+3 4 | 3+3 4 | 3+4 4 | 6+6 4 | 3+4 4 | 3+4 4 | 3+4 4 |
| | 5 2 | 4+4 4 | 4+4 4 | 3+4 4 | 5+5 4 | 3+4 4 | 3+4 4 | 3+4 4 |
| | 6 2 | 3+3 4 | 4+4 4 | 3+4 4 | 6+6 4 | 3+4 4 | 3+4 4 | 3+4 4 |
| | 6 2 | 4+4 4 | 3+3 4 | 3+4 4 | 5+5 4 | 3+4 4 | 3+4 4 | 3+4 4 |
| | 6 2 | 3+3 4 | 4+4 4 | 3+4 4 | 6+6 4 | 3+4 4 | 3+4 4 | 3+4 4 |
| | 5 2 | 4+4 4 | 3+3 4 | 3+4 4 | 5+5 4 | 3+4 4 | 3+4 4 | 3+4 4 |
| | 6 2 | 3+3 4 | 4+4 4 | 3+4 4 | 6+6 4 | 3+4 4 | 3+4 4 | 3+4 4 |
| | 5 2 | 4+4 4 | 3+3 4 | 3+4 4 | 6+6 4 | 3+4 4 | 3+4 4 | 3+4 4 |
| | 5 2 | | | | | | | |
| | 5 2 | | | | | | | |
| | 6 2 | | | | | | | |
| | 5 2 | | | | | | | |
| | 6 2 | | | | | | | |
| | 5 2 | | | | | | | |

Continued on next column

**TABLE XXXVII– continued**

| TQC | ICSD | NoE | SF | | | | | |
|---|---|---|---|---|---|---|---|---|
| | 6 2 | | | | | | | |
| | 6 2 | | | | | | | |
| | 6 2 | | | | | | | |
| | 5 2 | | | | | | | |
| | 5 2 | | | | | | | |
| | 5 2 | | | | | | | |
| | 6 2 | | | | | | | |
| | 5 2 | | | | | | | |
| | 6 2 | | | | | | | |
| | 5 2 | | | | | | | |
| 62.2.234 | 106273 | 64 | Au1Ca1Ga1 | | | | | |
| | -GM | -R | -S | -T | -U | -X | -Y | -Z |
| | 5 2 | 3+3 4 | 4+4 4 | 3+4 4 | 5+5 4 | 3+4 4 | 3+4 4 | 3+4 4 |
| | 5 2 | 4+4 4 | 3+3 4 | 3+4 4 | 6+6 4 | 3+4 4 | 3+4 4 | 3+4 4 |
| | 6 2 | 3+3 4 | 4+4 4 | 3+4 4 | 5+5 4 | 3+4 4 | 3+4 4 | 3+4 4 |
| | 6 2 | 4+4 4 | 3+3 4 | 3+4 4 | 6+6 4 | 3+4 4 | 3+4 4 | 3+4 4 |
| | 5 2 | 4+4 4 | 3+3 4 | 3+4 4 | 5+5 4 | 3+4 4 | 3+4 4 | 3+4 4 |
| | 6 2 | 3+3 4 | 4+4 4 | 3+4 4 | 6+6 4 | 3+4 4 | 3+4 4 | 3+4 4 |
| | 5 2 | 4+4 4 | 3+3 4 | 3+4 4 | 5+5 4 | 3+4 4 | 3+4 4 | 3+4 4 |
| | 6 2 | 3+3 4 | 4+4 4 | 3+4 4 | 6+6 4 | 3+4 4 | 3+4 4 | 3+4 4 |
| | 5 2 | 3+3 4 | 3+3 4 | 3+4 4 | 5+5 4 | 3+4 4 | 3+4 4 | 3+4 4 |
| | 5 2 | 4+4 4 | 4+4 4 | 3+4 4 | 6+6 4 | 3+4 4 | 3+4 4 | 3+4 4 |
| | 6 2 | 3+3 4 | 4+4 4 | 3+4 4 | 5+5 4 | 3+4 4 | 3+4 4 | 3+4 4 |
| | 6 2 | 4+4 4 | 3+3 4 | 3+4 4 | 6+6 4 | 3+4 4 | 3+4 4 | 3+4 4 |
| | 6 2 | 4+4 4 | 4+4 4 | 3+4 4 | 5+5 4 | 3+4 4 | 3+4 4 | 3+4 4 |
| | 5 2 | 3+3 4 | 3+3 4 | 3+4 4 | 6+6 4 | 3+4 4 | 3+4 4 | 3+4 4 |
| | 6 2 | 3+3 4 | 4+4 4 | 3+4 4 | 5+5 4 | 3+4 4 | 3+4 4 | 3+4 4 |
| | 5 2 | 4+4 4 | 3+3 4 | 3+4 4 | 5+5 4 | 3+4 4 | 3+4 4 | 3+4 4 |
| | 5 2 | | | | | | | |
| | 6 2 | | | | | | | |
| | 5 2 | | | | | | | |
| | 5 2 | | | | | | | |
| | 6 2 | | | | | | | |
| | 5 2 | | | | | | | |
| | 6 2 | | | | | | | |
| | 6 2 | | | | | | | |
| | 6 2 | | | | | | | |
| | 5 2 | | | | | | | |
| | 5 2 | | | | | | | |
| | 5 2 | | | | | | | |
| | 6 2 | | | | | | | |
| | 5 2 | | | | | | | |
| | 6 2 | | | | | | | |
| | 5 2 | | | | | | | |
| 62.2.235 | 168665 | 56 | As2Zr1 | | | | | |
| | -GM | -R | -S | -T | -U | -X | -Y | -Z |
| | 5 2 | 4+4 4 | 3+3 4 | 3+4 4 | 5+5 4 | 3+4 4 | 3+4 4 | 3+4 4 |
| | 5 2 | 3+3 4 | 4+4 4 | 3+4 4 | 6+6 4 | 3+4 4 | 3+4 4 | 3+4 4 |
| | 6 2 | 3+3 4 | 3+3 4 | 3+4 4 | 5+5 4 | 3+4 4 | 3+4 4 | 3+4 4 |
| | 6 2 | 4+4 4 | 4+4 4 | 3+4 4 | 6+6 4 | 3+4 4 | 3+4 4 | 3+4 4 |
| | 5 2 | 3+3 4 | 4+4 4 | 3+4 4 | 6+6 4 | 3+4 4 | 3+4 4 | 3+4 4 |
| | 5 2 | 4+4 4 | 3+3 4 | 3+4 4 | 5+5 4 | 3+4 4 | 3+4 4 | 3+4 4 |
| | 6 2 | 4+4 4 | 4+4 4 | 3+4 4 | 5+5 4 | 3+4 4 | 3+4 4 | 3+4 4 |
| | 6 2 | 3+3 4 | 3+3 4 | 3+4 4 | 6+6 4 | 3+4 4 | 3+4 4 | 3+4 4 |
| | 5 2 | 3+3 4 | 3+3 4 | 3+4 4 | 5+5 4 | 3+4 4 | 3+4 4 | 3+4 4 |
| | 6 2 | 4+4 4 | 4+4 4 | 3+4 4 | 6+6 4 | 3+4 4 | 3+4 4 | 3+4 4 |
| | 6 2 | 3+3 4 | 3+3 4 | 3+4 4 | 6+6 4 | 3+4 4 | 3+4 4 | 3+4 4 |
| | 5 2 | 4+4 4 | 4+4 4 | 3+4 4 | 5+5 4 | 3+4 4 | 3+4 4 | 3+4 4 |
| | 5 2 | 4+4 4 | 4+4 4 | 3+4 4 | 5+5 4 | 3+4 4 | 3+4 4 | 3+4 4 |
| | 5 2 | 3+3 4 | 3+3 4 | 3+4 4 | 6+6 4 | 3+4 4 | 3+4 4 | 3+4 4 |
| | 5 2 | | | | | | | |

Continued on next column

**TABLE XXXVII– continued**

| TQC | ICSD | NoE | SF | | | | | |
|---|---|---|---|---|---|---|---|---|
| | 5 2 | | | | | | | |
| | 6 2 | | | | | | | |
| | 5 2 | | | | | | | |
| | 5 2 | | | | | | | |
| | 6 2 | | | | | | | |
| | 5 2 | | | | | | | |
| | 6 2 | | | | | | | |
| | 6 2 | | | | | | | |
| | 5 2 | | | | | | | |
| | 6 2 | | | | | | | |
| | 5 2 | | | | | | | |
| | 6 2 | | | | | | | |
| | 6 2 | | | | | | | |
| 62.2.236 | 42446 | 40 | As1V1 | | | | | |
| | -GM | -R | -S | -T | -U | -X | -Y | -Z |
| | 5 2 | 4+4 4 | 4+4 4 | 3+4 4 | 6+6 4 | 3+4 4 | 3+4 4 | 3+4 4 |
| | 5 2 | 3+3 4 | 3+3 4 | 3+4 4 | 5+5 4 | 3+4 4 | 3+4 4 | 3+4 4 |
| | 6 2 | 3+3 4 | 3+3 4 | 3+4 4 | 6+6 4 | 3+4 4 | 3+4 4 | 3+4 4 |
| | 6 2 | 4+4 4 | 4+4 4 | 3+4 4 | 5+5 4 | 3+4 4 | 3+4 4 | 3+4 4 |
| | 5 2 | 3+3 4 | 3+3 4 | 3+4 4 | 6+6 4 | 3+4 4 | 3+4 4 | 3+4 4 |
| | 6 2 | 4+4 4 | 4+4 4 | 3+4 4 | 6+6 4 | 3+4 4 | 3+4 4 | 3+4 4 |
| | 5 2 | 4+4 4 | 3+3 4 | 3+4 4 | 5+5 4 | 3+4 4 | 3+4 4 | 3+4 4 |
| | 5 2 | 3+3 4 | 4+4 4 | 3+4 4 | 5+5 4 | 3+4 4 | 3+4 4 | 3+4 4 |
| | 6 2 | 4+4 4 | 3+3 4 | 3+4 4 | 6+6 4 | 3+4 4 | 3+4 4 | 3+4 4 |
| | 5 2 | 3+3 4 | 4+4 4 | 3+4 4 | 5+5 4 | 3+4 4 | 3+4 4 | 3+4 4 |
| | 6 2 | | | | | | | |
| | 5 2 | | | | | | | |
| | 6 2 | | | | | | | |
| | 5 2 | | | | | | | |
| | 6 2 | | | | | | | |
| | 5 2 | | | | | | | |
| | 6 2 | | | | | | | |
| | 5 2 | | | | | | | |
| | 5 2 | | | | | | | |
| | 5 2 | | | | | | | |
| 62.2.237 | 51201 | 160 | As1Li1Ni1O4 | | | | | |
| | -GM | -R | -S | -T | -U | -X | -Y | -Z |
| | 5 2 | 3+3 4 | 3+3 4 | 3+4 4 | 6+6 4 | 3+4 4 | 3+4 4 | 3+4 4 |
| | 5 2 | 4+4 4 | 4+4 4 | 3+4 4 | 5+5 4 | 3+4 4 | 3+4 4 | 3+4 4 |
| | 6 2 | 3+3 4 | 4+4 4 | 3+4 4 | 5+5 4 | 3+4 4 | 3+4 4 | 3+4 4 |
| | 6 2 | 4+4 4 | 3+3 4 | 3+4 4 | 6+6 4 | 3+4 4 | 3+4 4 | 3+4 4 |
| | 6 2 | 4+4 4 | 4+4 4 | 3+4 4 | 5+5 4 | 3+4 4 | 3+4 4 | 3+4 4 |
| | 6 2 | 3+3 4 | 3+3 4 | 3+4 4 | 6+6 4 | 3+4 4 | 3+4 4 | 3+4 4 |
| | 5 2 | 3+3 4 | 3+3 4 | 3+4 4 | 5+5 4 | 3+4 4 | 3+4 4 | 3+4 4 |
| | 6 2 | 4+4 4 | 4+4 4 | 3+4 4 | 6+6 4 | 3+4 4 | 3+4 4 | 3+4 4 |
| | 5 2 | 3+3 4 | 3+3 4 | 3+4 4 | 6+6 4 | 3+4 4 | 3+4 4 | 3+4 4 |
| | 5 2 | 4+4 4 | 4+4 4 | 3+4 4 | 5+5 4 | 3+4 4 | 3+4 4 | 3+4 4 |
| | 6 2 | 3+3 4 | 4+4 4 | 3+4 4 | 6+6 4 | 3+4 4 | 3+4 4 | 3+4 4 |
| | 6 2 | 4+4 4 | 3+3 4 | 3+4 4 | 6+6 4 | 3+4 4 | 3+4 4 | 3+4 4 |
| | 5 2 | 4+4 4 | 3+3 4 | 3+4 4 | 5+5 4 | 3+4 4 | 3+4 4 | 3+4 4 |
| | 6 2 | 3+3 4 | 4+4 4 | 3+4 4 | 6+6 4 | 3+4 4 | 3+4 4 | 3+4 4 |
| | 5 2 | 4+4 4 | 3+3 4 | 3+4 4 | 5+5 4 | 3+4 4 | 3+4 4 | 3+4 4 |
| | 5 2 | 3+3 4 | 4+4 4 | 3+4 4 | 5+5 4 | 3+4 4 | 3+4 4 | 3+4 4 |
| | 5 2 | 4+4 4 | 3+3 4 | 3+4 4 | 5+5 4 | 3+4 4 | 3+4 4 | 3+4 4 |
| | 6 2 | 3+3 4 | 4+4 4 | 3+4 4 | 6+6 4 | 3+4 4 | 3+4 4 | 3+4 4 |
| | 5 2 | 4+4 4 | 4+4 4 | 3+4 4 | 5+5 4 | 3+4 4 | 3+4 4 | 3+4 4 |
| | 6 2 | 3+3 4 | 3+3 4 | 3+4 4 | 5+5 4 | 3+4 4 | 3+4 4 | 3+4 4 |
| | 5 2 | 4+4 4 | 4+4 4 | 3+4 4 | 6+6 4 | 3+4 4 | 3+4 4 | 3+4 4 |
| | 5 2 | 3+3 4 | 3+3 4 | 3+4 4 | 6+6 4 | 3+4 4 | 3+4 4 | 3+4 4 |
| | 6 2 | 4+4 4 | 3+3 4 | 3+4 4 | 6+6 4 | 3+4 4 | 3+4 4 | 3+4 4 |
| | 6 2 | 3+3 4 | 4+4 4 | 3+4 4 | 6+6 4 | 3+4 4 | 3+4 4 | 3+4 4 |
| | 5 2 | 3+3 4 | 4+4 4 | 3+4 4 | 5+5 4 | 3+4 4 | 3+4 4 | 3+4 4 |
| | 6 2 | 4+4 4 | 3+3 4 | 3+4 4 | 5+5 4 | 3+4 4 | 3+4 4 | 3+4 4 |
| | 5 2 | 4+4 4 | 4+4 4 | 3+4 4 | 6+6 4 | 3+4 4 | 3+4 4 | 3+4 4 |
| | 5 2 | 3+3 4 | 3+3 4 | 3+4 4 | 6+6 4 | 3+4 4 | 3+4 4 | 3+4 4 |
| | 5 2 | 3+3 4 | 3+3 4 | 3+4 4 | 5+5 4 | 3+4 4 | 3+4 4 | 3+4 4 |
| | 6 2 | 4+4 4 | 4+4 4 | 3+4 4 | 5+5 4 | 3+4 4 | 3+4 4 | 3+4 4 |
| | 6 2 | 3+3 4 | 3+3 4 | 3+4 4 | 5+5 4 | 3+4 4 | 3+4 4 | 3+4 4 |
| | 6 2 | 4+4 4 | 4+4 4 | 3+4 4 | 6+6 4 | 3+4 4 | 3+4 4 | 3+4 4 |
| | 6 2 | 3+3 4 | 4+4 4 | 3+4 4 | 6+6 4 | 3+4 4 | 3+4 4 | 3+4 4 |
| | 6 2 | 4+4 4 | 3+3 4 | 3+4 4 | 5+5 4 | 3+4 4 | 3+4 4 | 3+4 4 |
| | 5 2 | 4+4 4 | 3+3 4 | 3+4 4 | 6+6 4 | 3+4 4 | 3+4 4 | 3+4 4 |
| | 6 2 | 3+3 4 | 4+4 4 | 3+4 4 | 5+5 4 | 3+4 4 | 3+4 4 | 3+4 4 |
| | 5 2 | 4+4 4 | 3+3 4 | 3+4 4 | 6+6 4 | 3+4 4 | 3+4 4 | 3+4 4 |
| | 6 2 | 3+3 4 | 4+4 4 | 3+4 4 | 5+5 4 | 3+4 4 | 3+4 4 | 3+4 4 |
| | 6 2 | 3+3 4 | 4+4 4 | 3+4 4 | 5+5 4 | 3+4 4 | 3+4 4 | 3+4 4 |
| | 5 2 | 4+4 4 | 3+3 4 | 3+4 4 | 6+6 4 | 3+4 4 | 3+4 4 | 3+4 4 |
| | 5 2 | | | | | | | |
| | 5 2 | | | | | | | |
| | 6 2 | | | | | | | |
| | 6 2 | | | | | | | |
| | 5 2 | | | | | | | |
| | 6 2 | | | | | | | |
| | 5 2 | | | | | | | |
| | 5 2 | | | | | | | |
| | 6 2 | | | | | | | |
| | 5 2 | | | | | | | |
| | 5 2 | | | | | | | |
| | 6 2 | | | | | | | |
| | 5 2 | | | | | | | |
| | 5 2 | | | | | | | |
| | 6 2 | | | | | | | |
| | 6 2 | | | | | | | |
| | 6 2 | | | | | | | |
| | 5 2 | | | | | | | |
| | 6 2 | | | | | | | |
| | 5 2 | | | | | | | |
| | 6 2 | | | | | | | |
| | 6 2 | | | | | | | |
| | 5 2 | | | | | | | |
| | 5 2 | | | | | | | |
| | 5 2 | | | | | | | |
| | 5 2 | | | | | | | |
| | 6 2 | | | | | | | |
| | 6 2 | | | | | | | |
| | 5 2 | | | | | | | |
| | 6 2 | | | | | | | |
| | 5 2 | | | | | | | |
| | 6 2 | | | | | | | |
| | 5 2 | | | | | | | |
| | 6 2 | | | | | | | |
| | 5 2 | | | | | | | |
| | 6 2 | | | | | | | |
| | 6 2 | | | | | | | |
| | 5 2 | | | | | | | |
| | 6 2 | | | | | | | |
| | 6 2 | | | | | | | |
| 62.2.238 | 245184 | 160 | As1Li1Ni1O4 | | | | | |
| | -GM | -R | -S | -T | -U | -X | -Y | -Z |
| | 5 2 | 3+3 4 | 3+3 4 | 3+4 4 | 6+6 4 | 3+4 4 | 3+4 4 | 3+4 4 |
| | 5 2 | 4+4 4 | 4+4 4 | 3+4 4 | 5+5 4 | 3+4 4 | 3+4 4 | 3+4 4 |
| | 6 2 | 3+3 4 | 4+4 4 | 3+4 4 | 5+5 4 | 3+4 4 | 3+4 4 | 3+4 4 |

Continued on next column

**TABLE XXXVII– continued**

| TQC | ICSD NoE SF |
|---|---|
| | 6 2 4+4 4 3+3 4 3+4 4 6+6 4 3+4 4 3+4 4 3+4 4 |
| | 6 2 3+3 4 3+3 4 3+4 4 6+6 4 3+4 4 3+4 4 3+4 4 |
| | 5 2 4+4 4 4+4 4 3+4 4 5+5 4 3+4 4 3+4 4 3+4 4 |
| | 6 2 3+3 4 3+3 4 3+4 4 5+5 4 3+4 4 3+4 4 3+4 4 |
| | 6 2 4+4 4 4+4 4 3+4 4 6+6 4 3+4 4 3+4 4 3+4 4 |
| | 5 2 3+3 4 3+3 4 3+4 4 6+6 4 3+4 4 3+4 4 3+4 4 |
| | 6 2 4+4 4 4+4 4 3+4 4 5+5 4 3+4 4 3+4 4 3+4 4 |
| | 5 2 3+3 4 3+3 4 3+4 4 6+6 4 3+4 4 3+4 4 3+4 4 |
| | 5 2 4+4 4 4+4 4 3+4 4 6+6 4 3+4 4 3+4 4 3+4 4 |
| | 6 2 4+4 4 3+3 4 3+4 4 5+5 4 3+4 4 3+4 4 3+4 4 |
| | 6 2 3+3 4 4+4 4 3+4 4 6+6 4 3+4 4 3+4 4 3+4 4 |
| | 5 2 4+4 4 4+4 4 3+4 4 5+5 4 3+4 4 3+4 4 3+4 4 |
| | 5 2 3+3 4 3+3 4 3+4 4 5+5 4 3+4 4 3+4 4 3+4 4 |
| | 5 2 4+4 4 4+4 4 3+4 4 5+5 4 3+4 4 3+4 4 3+4 4 |
| | 5 2 3+3 4 3+3 4 3+4 4 6+6 4 3+4 4 3+4 4 3+4 4 |
| | 6 2 4+4 4 4+4 4 3+4 4 5+5 4 3+4 4 3+4 4 3+4 4 |
| | 6 2 3+3 4 3+3 4 3+4 4 5+5 4 3+4 4 3+4 4 3+4 4 |
| | 5 2 4+4 4 3+3 4 3+4 4 6+6 4 3+4 4 3+4 4 3+4 4 |
| | 5 2 3+3 4 4+4 4 3+4 4 6+6 4 3+4 4 3+4 4 3+4 4 |
| | 6 2 4+4 4 3+3 4 3+4 4 5+5 4 3+4 4 3+4 4 3+4 4 |
| | 6 2 3+3 4 4+4 4 3+4 4 5+5 4 3+4 4 3+4 4 3+4 4 |
| | 5 2 3+3 4 4+4 4 3+4 4 6+6 4 3+4 4 3+4 4 3+4 4 |
| | 6 2 4+4 4 3+3 4 3+4 4 6+6 4 3+4 4 3+4 4 3+4 4 |
| | 5 2 4+4 4 3+3 4 3+4 4 6+6 4 3+4 4 3+4 4 3+4 4 |
| | 5 2 3+3 4 4+4 4 3+4 4 6+6 4 3+4 4 3+4 4 3+4 4 |
| | 5 2 3+3 4 3+3 4 3+4 4 5+5 4 3+4 4 3+4 4 3+4 4 |
| | 6 2 4+4 4 4+4 4 3+4 4 5+5 4 3+4 4 3+4 4 3+4 4 |
| | 6 2 3+3 4 3+3 4 3+4 4 5+5 4 3+4 4 3+4 4 3+4 4 |
| | 6 2 4+4 4 4+4 4 3+4 4 6+6 4 3+4 4 3+4 4 3+4 4 |
| | 6 2 4+4 4 4+4 4 3+4 4 6+6 4 3+4 4 3+4 4 3+4 4 |
| | 5 2 3+3 4 3+3 4 3+4 4 5+5 4 3+4 4 3+4 4 3+4 4 |
| | 6 2 3+3 4 3+3 4 3+4 4 6+6 4 3+4 4 3+4 4 3+4 4 |
| | 6 2 4+4 4 4+4 4 3+4 4 5+5 4 3+4 4 3+4 4 3+4 4 |
| | 5 2 4+4 4 3+3 4 3+4 4 6+6 4 3+4 4 3+4 4 3+4 4 |
| | 6 2 3+3 4 4+4 4 3+4 4 5+5 4 3+4 4 3+4 4 3+4 4 |
| | 6 2 3+3 4 4+4 4 3+4 4 5+5 4 3+4 4 3+4 4 3+4 4 |
| | 5 2 4+4 4 3+3 4 3+4 4 6+6 4 3+4 4 3+4 4 3+4 4 |
| | 6 2 |
| | 6 2 |
| | 6 2 |
| | 5 2 |
| | 5 2 |
| | 5 2 |
| | 5 2 |
| | 5 2 |
| | 5 2 |
| | 6 2 |
| | 6 2 |
| | 5 2 |
| | 5 2 |
| | 5 2 |
| | 6 2 |
| | 6 2 |
| | 5 2 |
| | 6 2 |
| | 5 2 |
| | 6 2 |
| | 6 2 |
| | 6 2 |
| | 5 2 |
| | 5 2 |

Continued on next column

**TABLE XXXVII– continued**

| TQC | ICSD NoE SF |
|---|---|
| | 5 2 |
| | 6 2 |
| | 5 2 |
| | 6 2 |
| | 5 2 |
| | 5 2 |
| | 6 2 |
| | 6 2 |
| | 5 2 |
| | 6 2 |
| | 5 2 |
| | 6 2 |
| | 6 2 |
| | 5 2 |
| | 6 2 |
| | 6 2 |
| 62.2.239 | 412296 144 As1Cu1Mg1 |
| | -GM -R -S -T -U -X -Y -Z |
| | 5 2 3+3 4 3+3 4 3+4 4 5+5 4 3+4 4 3+4 4 3+4 4 |
| | 6 2 4+4 4 4+4 4 3+4 4 6+6 4 3+4 4 3+4 4 3+4 4 |
| | 5 2 3+3 4 3+3 4 3+4 4 5+5 4 3+4 4 3+4 4 3+4 4 |
| | 5 2 4+4 4 4+4 4 3+4 4 6+6 4 3+4 4 3+4 4 3+4 4 |
| | 6 2 3+3 4 4+4 4 3+4 4 6+6 4 3+4 4 3+4 4 3+4 4 |
| | 5 2 4+4 4 3+3 4 3+4 4 5+5 4 3+4 4 3+4 4 3+4 4 |
| | 6 2 3+3 4 4+4 4 3+4 4 6+6 4 3+4 4 3+4 4 3+4 4 |
| | 6 2 4+4 4 3+3 4 3+4 4 5+5 4 3+4 4 3+4 4 3+4 4 |
| | 6 2 4+4 4 3+3 4 3+4 4 5+5 4 3+4 4 3+4 4 3+4 4 |
| | 5 2 3+3 4 4+4 4 3+4 4 6+6 4 3+4 4 3+4 4 3+4 4 |
| | 5 2 3+3 4 3+3 4 3+4 4 6+6 4 3+4 4 3+4 4 3+4 4 |
| | 5 2 4+4 4 4+4 4 3+4 4 6+6 4 3+4 4 3+4 4 3+4 4 |
| | 6 2 4+4 4 4+4 4 3+4 4 5+5 4 3+4 4 3+4 4 3+4 4 |
| | 6 2 3+3 4 3+3 4 3+4 4 5+5 4 3+4 4 3+4 4 3+4 4 |
| | 5 2 3+3 4 4+4 4 3+4 4 6+6 4 3+4 4 3+4 4 3+4 4 |
| | 6 2 4+4 4 3+3 4 3+4 4 5+5 4 3+4 4 3+4 4 3+4 4 |
| | 5 2 3+3 4 3+3 4 3+4 4 5+5 4 3+4 4 3+4 4 3+4 4 |
| | 6 2 4+4 4 4+4 4 3+4 4 6+6 4 3+4 4 3+4 4 3+4 4 |
| | 6 2 4+4 4 3+3 4 3+4 4 5+5 4 3+4 4 3+4 4 3+4 4 |
| | 6 2 3+3 4 4+4 4 3+4 4 6+6 4 3+4 4 3+4 4 3+4 4 |
| | 5 2 3+3 4 4+4 4 3+4 4 5+5 4 3+4 4 3+4 4 3+4 4 |
| | 6 2 4+4 4 3+3 4 3+4 4 6+6 4 3+4 4 3+4 4 3+4 4 |
| | 6 2 4+4 4 3+3 4 3+4 4 5+5 4 3+4 4 3+4 4 3+4 4 |
| | 5 2 3+3 4 4+4 4 3+4 4 6+6 4 3+4 4 3+4 4 3+4 4 |
| | 6 2 3+3 4 4+4 4 3+4 4 5+5 4 3+4 4 3+4 4 3+4 4 |
| | 5 2 4+4 4 3+3 4 3+4 4 5+5 4 3+4 4 3+4 4 3+4 4 |
| | 6 2 3+3 4 4+4 4 3+4 4 6+6 4 3+4 4 3+4 4 3+4 4 |
| | 5 2 4+4 4 3+3 4 3+4 4 6+6 4 3+4 4 3+4 4 3+4 4 |
| | 5 2 4+4 4 4+4 4 3+4 4 5+5 4 3+4 4 3+4 4 3+4 4 |
| | 5 2 3+3 4 3+3 4 3+4 4 6+6 4 3+4 4 3+4 4 3+4 4 |
| | 6 2 3+3 4 3+3 4 3+4 4 5+5 4 3+4 4 3+4 4 3+4 4 |
| | 5 2 4+4 4 4+4 4 3+4 4 5+5 4 3+4 4 3+4 4 3+4 4 |
| | 5 2 4+4 4 3+3 4 3+4 4 6+6 4 3+4 4 3+4 4 3+4 4 |
| | 6 2 3+3 4 4+4 4 3+4 4 6+6 4 3+4 4 3+4 4 3+4 4 |
| | 5 2 3+3 4 4+4 4 3+4 4 5+5 4 3+4 4 3+4 4 3+4 4 |
| | 6 2 4+4 4 3+3 4 3+4 4 6+6 4 3+4 4 3+4 4 3+4 4 |
| | 5 2 |
| | 6 2 |
| | 5 2 |
| | 6 2 |
| | 5 2 |
| | 6 2 |
| | 5 2 |

Continued on next column

**TABLE XXXVII– continued**

| TQC | ICSD | NoE | SF | | | | | |
|---|---|---|---|---|---|---|---|---|
| | 5 2 | | | | | | | |
| | 6 2 | | | | | | | |
| | 5 2 | | | | | | | |
| | 5 2 | | | | | | | |
| | 6 2 | | | | | | | |
| | 5 2 | | | | | | | |
| | 6 2 | | | | | | | |
| | 6 2 | | | | | | | |
| | 5 2 | | | | | | | |
| | 5 2 | | | | | | | |
| | 6 2 | | | | | | | |
| | 5 2 | | | | | | | |
| | 6 2 | | | | | | | |
| | 5 2 | | | | | | | |
| | 6 2 | | | | | | | |
| | 5 2 | | | | | | | |
| | 5 2 | | | | | | | |
| | 6 2 | | | | | | | |
| | 5 2 | | | | | | | |
| | 6 2 | | | | | | | |
| | 6 2 | | | | | | | |
| | 5 2 | | | | | | | |
| | 6 2 | | | | | | | |
| | 6 2 | | | | | | | |
| | 5 2 | | | | | | | |
| | 5 2 | | | | | | | |
| | 6 2 | | | | | | | |
| | 5 2 | | | | | | | |
| | 6 2 | | | | | | | |
| 62.2.240 | 406953 | 72 | As1Co1Hf1 | | | | | |
| | -GM | -R | -S | -T | -U | -X | -Y | -Z |
| | 5 2 | 3+3 4 | 3+3 4 | 3+4 4 | 5+5 4 | 3+4 4 | 3+4 4 | 3+4 4 |
| | 6 2 | 4+4 4 | 4+4 4 | 3+4 4 | 6+6 4 | 3+4 4 | 3+4 4 | 3+4 4 |
| | 5 2 | 4+4 4 | 4+4 4 | 3+4 4 | 5+5 4 | 3+4 4 | 3+4 4 | 3+4 4 |
| | 6 2 | 3+3 4 | 3+3 4 | 3+4 4 | 6+6 4 | 3+4 4 | 3+4 4 | 3+4 4 |
| | 5 2 | 3+3 4 | 4+4 4 | 3+4 4 | 6+6 4 | 3+4 4 | 3+4 4 | 3+4 4 |
| | 6 2 | 4+4 4 | 3+3 4 | 3+4 4 | 6+6 4 | 3+4 4 | 3+4 4 | 3+4 4 |
| | 5 2 | 4+4 4 | 3+3 4 | 3+4 4 | 5+5 4 | 3+4 4 | 3+4 4 | 3+4 4 |
| | 6 2 | 3+3 4 | 4+4 4 | 3+4 4 | 5+5 4 | 3+4 4 | 3+4 4 | 3+4 4 |
| | 5 2 | 4+4 4 | 3+3 4 | 3+4 4 | 5+5 4 | 3+4 4 | 3+4 4 | 3+4 4 |
| | 5 2 | 3+3 4 | 4+4 4 | 3+4 4 | 6+6 4 | 3+4 4 | 3+4 4 | 3+4 4 |
| | 6 2 | 3+3 4 | 3+3 4 | 3+4 4 | 5+5 4 | 3+4 4 | 3+4 4 | 3+4 4 |
| | 6 2 | 4+4 4 | 4+4 4 | 3+4 4 | 6+6 4 | 3+4 4 | 3+4 4 | 3+4 4 |
| | 6 2 | 4+4 4 | 4+4 4 | 3+4 4 | 5+5 4 | 3+4 4 | 3+4 4 | 3+4 4 |
| | 6 2 | 3+3 4 | 3+3 4 | 3+4 4 | 6+6 4 | 3+4 4 | 3+4 4 | 3+4 4 |
| | 5 2 | 3+3 4 | 3+3 4 | 3+4 4 | 5+5 4 | 3+4 4 | 3+4 4 | 3+4 4 |
| | 6 2 | 4+4 4 | 4+4 4 | 3+4 4 | 5+5 4 | 3+4 4 | 3+4 4 | 3+4 4 |
| | 5 2 | 3+3 4 | 3+3 4 | 3+4 4 | 6+6 4 | 3+4 4 | 3+4 4 | 3+4 4 |
| | 6 2 | 4+4 4 | 4+4 4 | 3+4 4 | 6+6 4 | 3+4 4 | 3+4 4 | 3+4 4 |
| | 5 2 | | | | | | | |
| | 5 2 | | | | | | | |
| | 6 2 | | | | | | | |
| | 5 2 | | | | | | | |
| | 5 2 | | | | | | | |
| | 5 2 | | | | | | | |
| | 6 2 | | | | | | | |
| | 6 2 | | | | | | | |
| | 5 2 | | | | | | | |
| | 5 2 | | | | | | | |
| | 5 2 | | | | | | | |
| | 5 2 | | | | | | | |

Continued on next column

**TABLE XXXVII– continued**

| TQC | ICSD | NoE | SF | | | | | |
|---|---|---|---|---|---|---|---|---|
| | 6 2 | | | | | | | |
| | 6 2 | | | | | | | |
| | 5 2 | | | | | | | |
| | 5 2 | | | | | | | |
| | 6 2 | | | | | | | |
| | 6 2 | | | | | | | |
| 62.2.241 | 48030 | 56 | As1Co1 | | | | | |
| | -GM | -R | -S | -T | -U | -X | -Y | -Z |
| | 5 2 | 3+3 4 | 4+4 4 | 3+4 4 | 6+6 4 | 3+4 4 | 3+4 4 | 3+4 4 |
| | 5 2 | 4+4 4 | 3+3 4 | 3+4 4 | 5+5 4 | 3+4 4 | 3+4 4 | 3+4 4 |
| | 6 2 | 3+3 4 | 4+4 4 | 3+4 4 | 6+6 4 | 3+4 4 | 3+4 4 | 3+4 4 |
| | 6 2 | 4+4 4 | 3+3 4 | 3+4 4 | 5+5 4 | 3+4 4 | 3+4 4 | 3+4 4 |
| | 5 2 | 4+4 4 | 3+3 4 | 3+4 4 | 6+6 4 | 3+4 4 | 3+4 4 | 3+4 4 |
| | 6 2 | 3+3 4 | 4+4 4 | 3+4 4 | 6+6 4 | 3+4 4 | 3+4 4 | 3+4 4 |
| | 5 2 | 4+4 4 | 4+4 4 | 3+4 4 | 5+5 4 | 3+4 4 | 3+4 4 | 3+4 4 |
| | 6 2 | 3+3 4 | 3+3 4 | 3+4 4 | 5+5 4 | 3+4 4 | 3+4 4 | 3+4 4 |
| | 5 2 | 3+3 4 | 3+3 4 | 3+4 4 | 6+6 4 | 3+4 4 | 3+4 4 | 3+4 4 |
| | 5 2 | 4+4 4 | 4+4 4 | 3+4 4 | 5+5 4 | 3+4 4 | 3+4 4 | 3+4 4 |
| | 6 2 | 4+4 4 | 3+3 4 | 3+4 4 | 5+5 4 | 3+4 4 | 3+4 4 | 3+4 4 |
| | 5 2 | 3+3 4 | 4+4 4 | 3+4 4 | 6+6 4 | 3+4 4 | 3+4 4 | 3+4 4 |
| | 6 2 | 3+3 4 | 4+4 4 | 3+4 4 | 5+5 4 | 3+4 4 | 3+4 4 | 3+4 4 |
| | 5 2 | 4+4 4 | 3+3 4 | 3+4 4 | 6+6 4 | 3+4 4 | 3+4 4 | 3+4 4 |
| | 5 2 | | | | | | | |
| | 5 2 | | | | | | | |
| | 6 2 | | | | | | | |
| | 5 2 | | | | | | | |
| | 6 2 | | | | | | | |
| | 5 2 | | | | | | | |
| | 5 2 | | | | | | | |
| | 6 2 | | | | | | | |
| | 6 2 | | | | | | | |
| | 6 2 | | | | | | | |
| | 6 2 | | | | | | | |
| | 5 2 | | | | | | | |
| | 5 2 | | | | | | | |
| | 5 2 | | | | | | | |
| 62.2.242 | 48029 | 56 | As1Co1 | | | | | |
| | -GM | -R | -S | -T | -U | -X | -Y | -Z |
| | 5 2 | 3+3 4 | 4+4 4 | 3+4 4 | 6+6 4 | 3+4 4 | 3+4 4 | 3+4 4 |
| | 5 2 | 4+4 4 | 3+3 4 | 3+4 4 | 5+5 4 | 3+4 4 | 3+4 4 | 3+4 4 |
| | 6 2 | 3+3 4 | 4+4 4 | 3+4 4 | 6+6 4 | 3+4 4 | 3+4 4 | 3+4 4 |
| | 6 2 | 4+4 4 | 3+3 4 | 3+4 4 | 5+5 4 | 3+4 4 | 3+4 4 | 3+4 4 |
| | 5 2 | 4+4 4 | 3+3 4 | 3+4 4 | 6+6 4 | 3+4 4 | 3+4 4 | 3+4 4 |
| | 6 2 | 3+3 4 | 4+4 4 | 3+4 4 | 6+6 4 | 3+4 4 | 3+4 4 | 3+4 4 |
| | 5 2 | 4+4 4 | 4+4 4 | 3+4 4 | 5+5 4 | 3+4 4 | 3+4 4 | 3+4 4 |
| | 6 2 | 3+3 4 | 3+3 4 | 3+4 4 | 5+5 4 | 3+4 4 | 3+4 4 | 3+4 4 |
| | 5 2 | 3+3 4 | 3+3 4 | 3+4 4 | 6+6 4 | 3+4 4 | 3+4 4 | 3+4 4 |
| | 5 2 | 4+4 4 | 4+4 4 | 3+4 4 | 5+5 4 | 3+4 4 | 3+4 4 | 3+4 4 |
| | 6 2 | 4+4 4 | 3+3 4 | 3+4 4 | 5+5 4 | 3+4 4 | 3+4 4 | 3+4 4 |
| | 5 2 | 3+3 4 | 4+4 4 | 3+4 4 | 6+6 4 | 3+4 4 | 3+4 4 | 3+4 4 |
| | 6 2 | 3+3 4 | 4+4 4 | 3+4 4 | 5+5 4 | 3+4 4 | 3+4 4 | 3+4 4 |
| | 5 2 | 4+4 4 | 3+3 4 | 3+4 4 | 6+6 4 | 3+4 4 | 3+4 4 | 3+4 4 |
| | 5 2 | | | | | | | |
| | 5 2 | | | | | | | |
| | 6 2 | | | | | | | |
| | 5 2 | | | | | | | |
| | 6 2 | | | | | | | |
| | 5 2 | | | | | | | |
| | 5 2 | | | | | | | |
| | 6 2 | | | | | | | |
| | 6 2 | | | | | | | |

Continued on next column

**TABLE XXXVII– continued**

| TQC | ICSD | NoE | SF | | | | | |
|---|---|---|---|---|---|---|---|---|
| | 6 2 | | | | | | | |
| | 6 2 | | | | | | | |
| | 5 2 | | | | | | | |
| | 5 2 | | | | | | | |
| | 5 2 | | | | | | | |
| 62.2.243 | 48028 | 56 | As1Co1 | | | | | |
| | -GM | -R | -S | -T | -U | -X | -Y | -Z |
| | 5 2 | 3+3 4 | 4+4 4 | 3+4 4 | 6+6 4 | 3+4 4 | 3+4 4 | 3+4 4 |
| | 5 2 | 4+4 4 | 3+3 4 | 3+4 4 | 5+5 4 | 3+4 4 | 3+4 4 | 3+4 4 |
| | 6 2 | 3+3 4 | 4+4 4 | 3+4 4 | 6+6 4 | 3+4 4 | 3+4 4 | 3+4 4 |
| | 6 2 | 4+4 4 | 3+3 4 | 3+4 4 | 5+5 4 | 3+4 4 | 3+4 4 | 3+4 4 |
| | 5 2 | 4+4 4 | 3+3 4 | 3+4 4 | 6+6 4 | 3+4 4 | 3+4 4 | 3+4 4 |
| | 6 2 | 3+3 4 | 4+4 4 | 3+4 4 | 6+6 4 | 3+4 4 | 3+4 4 | 3+4 4 |
| | 5 2 | 4+4 4 | 4+4 4 | 3+4 4 | 5+5 4 | 3+4 4 | 3+4 4 | 3+4 4 |
| | 6 2 | 3+3 4 | 3+3 4 | 3+4 4 | 5+5 4 | 3+4 4 | 3+4 4 | 3+4 4 |
| | 5 2 | 3+3 4 | 3+3 4 | 3+4 4 | 6+6 4 | 3+4 4 | 3+4 4 | 3+4 4 |
| | 5 2 | 4+4 4 | 4+4 4 | 3+4 4 | 5+5 4 | 3+4 4 | 3+4 4 | 3+4 4 |
| | 6 2 | 4+4 4 | 3+3 4 | 3+4 4 | 5+5 4 | 3+4 4 | 3+4 4 | 3+4 4 |
| | 5 2 | 3+3 4 | 4+4 4 | 3+4 4 | 6+6 4 | 3+4 4 | 3+4 4 | 3+4 4 |
| | 6 2 | 3+3 4 | 4+4 4 | 3+4 4 | 5+5 4 | 3+4 4 | 3+4 4 | 3+4 4 |
| | 5 2 | 4+4 4 | 3+3 4 | 3+4 4 | 6+6 4 | 3+4 4 | 3+4 4 | 3+4 4 |
| | 5 2 | | | | | | | |
| | 5 2 | | | | | | | |
| | 6 2 | | | | | | | |
| | 5 2 | | | | | | | |
| | 6 2 | | | | | | | |
| | 5 2 | | | | | | | |
| | 5 2 | | | | | | | |
| | 6 2 | | | | | | | |
| | 6 2 | | | | | | | |
| | 6 2 | | | | | | | |
| | 6 2 | | | | | | | |
| | 5 2 | | | | | | | |
| | 5 2 | | | | | | | |
| | 5 2 | | | | | | | |
| 62.2.244 | 48026 | 56 | As1Co1 | | | | | |
| | -GM | -R | -S | -T | -U | -X | -Y | -Z |
| | 5 2 | 3+3 4 | 4+4 4 | 3+4 4 | 6+6 4 | 3+4 4 | 3+4 4 | 3+4 4 |
| | 5 2 | 4+4 4 | 3+3 4 | 3+4 4 | 5+5 4 | 3+4 4 | 3+4 4 | 3+4 4 |
| | 6 2 | 3+3 4 | 4+4 4 | 3+4 4 | 6+6 4 | 3+4 4 | 3+4 4 | 3+4 4 |
| | 6 2 | 4+4 4 | 3+3 4 | 3+4 4 | 5+5 4 | 3+4 4 | 3+4 4 | 3+4 4 |
| | 5 2 | 4+4 4 | 3+3 4 | 3+4 4 | 6+6 4 | 3+4 4 | 3+4 4 | 3+4 4 |
| | 6 2 | 3+3 4 | 4+4 4 | 3+4 4 | 6+6 4 | 3+4 4 | 3+4 4 | 3+4 4 |
| | 5 2 | 4+4 4 | 4+4 4 | 3+4 4 | 5+5 4 | 3+4 4 | 3+4 4 | 3+4 4 |
| | 6 2 | 3+3 4 | 3+3 4 | 3+4 4 | 5+5 4 | 3+4 4 | 3+4 4 | 3+4 4 |
| | 5 2 | 3+3 4 | 3+3 4 | 3+4 4 | 6+6 4 | 3+4 4 | 3+4 4 | 3+4 4 |
| | 5 2 | 4+4 4 | 4+4 4 | 3+4 4 | 5+5 4 | 3+4 4 | 3+4 4 | 3+4 4 |
| | 6 2 | 4+4 4 | 3+3 4 | 3+4 4 | 5+5 4 | 3+4 4 | 3+4 4 | 3+4 4 |
| | 5 2 | 3+3 4 | 4+4 4 | 3+4 4 | 6+6 4 | 3+4 4 | 3+4 4 | 3+4 4 |
| | 6 2 | 3+3 4 | 4+4 4 | 3+4 4 | 5+5 4 | 3+4 4 | 3+4 4 | 3+4 4 |
| | 5 2 | 4+4 4 | 3+3 4 | 3+4 4 | 6+6 4 | 3+4 4 | 3+4 4 | 3+4 4 |
| | 5 2 | | | | | | | |
| | 5 2 | | | | | | | |
| | 5 2 | | | | | | | |
| | 6 2 | | | | | | | |
| | 6 2 | | | | | | | |
| | 5 2 | | | | | | | |
| | 5 2 | | | | | | | |
| | 6 2 | | | | | | | |
| | 6 2 | | | | | | | |
| | 6 2 | | | | | | | |

Continued on next column

**TABLE XXXVII– continued**

| TQC | ICSD | NoE | SF | | | | | |
|---|---|---|---|---|---|---|---|---|
| | 6 2 | | | | | | | |
| | 5 2 | | | | | | | |
| | 5 2 | | | | | | | |
| | 5 2 | | | | | | | |
| 62.2.245 | 9571 | 72 | As1Cd1Na1 | | | | | |
| | -GM | -R | -S | -T | -U | -X | -Y | -Z |
| | 5 2 | 3+3 4 | 3+3 4 | 3+4 4 | 6+6 4 | 3+4 4 | 3+4 4 | 3+4 4 |
| | 6 2 | 4+4 4 | 4+4 4 | 3+4 4 | 5+5 4 | 3+4 4 | 3+4 4 | 3+4 4 |
| | 5 2 | 4+4 4 | 4+4 4 | 3+4 4 | 5+5 4 | 3+4 4 | 3+4 4 | 3+4 4 |
| | 6 2 | 3+3 4 | 3+3 4 | 3+4 4 | 6+6 4 | 3+4 4 | 3+4 4 | 3+4 4 |
| | 6 2 | 4+4 4 | 4+4 4 | 3+4 4 | 6+6 4 | 3+4 4 | 3+4 4 | 3+4 4 |
| | 5 2 | 3+3 4 | 3+3 4 | 3+4 4 | 5+5 4 | 3+4 4 | 3+4 4 | 3+4 4 |
| | 6 2 | 3+3 4 | 3+3 4 | 3+4 4 | 6+6 4 | 3+4 4 | 3+4 4 | 3+4 4 |
| | 5 2 | 4+4 4 | 4+4 4 | 3+4 4 | 5+5 4 | 3+4 4 | 3+4 4 | 3+4 4 |
| | 5 2 | 3+3 4 | 3+3 4 | 3+4 4 | 6+6 4 | 3+4 4 | 3+4 4 | 3+4 4 |
| | 5 2 | 4+4 4 | 4+4 4 | 3+4 4 | 5+5 4 | 3+4 4 | 3+4 4 | 3+4 4 |
| | 6 2 | 3+3 4 | 3+3 4 | 3+4 4 | 6+6 4 | 3+4 4 | 3+4 4 | 3+4 4 |
| | 6 2 | 4+4 4 | 4+4 4 | 3+4 4 | 5+5 4 | 3+4 4 | 3+4 4 | 3+4 4 |
| | 6 2 | 3+3 4 | 3+3 4 | 3+4 4 | 6+6 4 | 3+4 4 | 3+4 4 | 3+4 4 |
| | 5 2 | 4+4 4 | 4+4 4 | 3+4 4 | 5+5 4 | 3+4 4 | 3+4 4 | 3+4 4 |
| | 5 2 | 3+3 4 | 4+4 4 | 3+4 4 | 6+6 4 | 3+4 4 | 3+4 4 | 3+4 4 |
| | 5 2 | 4+4 4 | 3+3 4 | 3+4 4 | 6+6 4 | 3+4 4 | 3+4 4 | 3+4 4 |
| | 6 2 | 3+3 4 | 3+3 4 | 3+4 4 | 5+5 4 | 3+4 4 | 3+4 4 | 3+4 4 |
| | 5 2 | 4+4 4 | 4+4 4 | 3+4 4 | 5+5 4 | 3+4 4 | 3+4 4 | 3+4 4 |
| | 6 2 | | | | | | | |
| | 5 2 | | | | | | | |
| | 6 2 | | | | | | | |
| | 5 2 | | | | | | | |
| | 6 2 | | | | | | | |
| | 6 2 | | | | | | | |
| | 5 2 | | | | | | | |
| | 6 2 | | | | | | | |
| | 5 2 | | | | | | | |
| | 5 2 | | | | | | | |
| | 6 2 | | | | | | | |
| | 5 2 | | | | | | | |
| | 5 2 | | | | | | | |
| | 6 2 | | | | | | | |
| | 5 2 | | | | | | | |
| | 5 2 | | | | | | | |
| | 6 2 | | | | | | | |
| | 6 2 | | | | | | | |
| 62.2.246 | 249802 | 184 | As1.01Ce1Se0.99 | | | | | |
| | -GM | -R | -S | -T | -U | -X | -Y | -Z |
| | 5 2 | 3+3 4 | 3+3 4 | 3+4 4 | 5+5 4 | 3+4 4 | 3+4 4 | 3+4 4 |
| | 6 2 | 4+4 4 | 4+4 4 | 3+4 4 | 6+6 4 | 3+4 4 | 3+4 4 | 3+4 4 |
| | 5 2 | 3+3 4 | 3+3 4 | 3+4 4 | 5+5 4 | 3+4 4 | 3+4 4 | 3+4 4 |
| | 6 2 | 4+4 4 | 4+4 4 | 3+4 4 | 6+6 4 | 3+4 4 | 3+4 4 | 3+4 4 |
| | 5 2 | 3+3 4 | 3+3 4 | 3+4 4 | 6+6 4 | 3+4 4 | 3+4 4 | 3+4 4 |
| | 6 2 | 4+4 4 | 4+4 4 | 3+4 4 | 5+5 4 | 3+4 4 | 3+4 4 | 3+4 4 |
| | 5 2 | 3+3 4 | 3+3 4 | 3+4 4 | 5+5 4 | 3+4 4 | 3+4 4 | 3+4 4 |
| | 6 2 | 4+4 4 | 4+4 4 | 3+4 4 | 6+6 4 | 3+4 4 | 3+4 4 | 3+4 4 |
| | 5 2 | 3+3 4 | 4+4 4 | 3+4 4 | 6+6 4 | 3+4 4 | 3+4 4 | 3+4 4 |
| | 5 2 | 4+4 4 | 3+3 4 | 3+4 4 | 5+5 4 | 3+4 4 | 3+4 4 | 3+4 4 |
| | 6 2 | 4+4 4 | 3+3 4 | 3+4 4 | 6+6 4 | 3+4 4 | 3+4 4 | 3+4 4 |
| | 6 2 | 3+3 4 | 4+4 4 | 3+4 4 | 5+5 4 | 3+4 4 | 3+4 4 | 3+4 4 |
| | 5 2 | 3+3 4 | 3+3 4 | 3+4 4 | 6+6 4 | 3+4 4 | 3+4 4 | 3+4 4 |
| | 6 2 | 4+4 4 | 4+4 4 | 3+4 4 | 5+5 4 | 3+4 4 | 3+4 4 | 3+4 4 |
| | 5 2 | 4+4 4 | 3+3 4 | 3+4 4 | 5+5 4 | 3+4 4 | 3+4 4 | 3+4 4 |
| | 6 2 | 3+3 4 | 4+4 4 | 3+4 4 | 6+6 4 | 3+4 4 | 3+4 4 | 3+4 4 |
| | 5 2 | 3+3 4 | 4+4 4 | 3+4 4 | 6+6 4 | 3+4 4 | 3+4 4 | 3+4 4 |

Continued on next column

**TABLE XXXVII– continued**

| TQC | ICSD | NoE | SF |
|---|---|---|---|
| | 6 2 4+4 4 3+3 4 3+4 4 5+5 4 3+4 4 3+4 4 3+4 4 | | |
| | 5 2 3+3 4 3+3 4 3+4 4 6+6 4 3+4 4 3+4 4 3+4 4 | | |
| | 6 2 4+4 4 4+4 4 3+4 4 5+5 4 3+4 4 3+4 4 3+4 4 | | |
| | 5 2 3+3 4 3+3 4 3+4 4 6+6 4 3+4 4 3+4 4 3+4 4 | | |
| | 6 2 4+4 4 4+4 4 3+4 4 5+5 4 3+4 4 3+4 4 3+4 4 | | |
| | 5 2 3+3 4 3+3 4 3+4 4 6+6 4 3+4 4 3+4 4 3+4 4 | | |
| | 6 2 4+4 4 4+4 4 3+4 4 5+5 4 3+4 4 3+4 4 3+4 4 | | |
| | 5 2 3+3 4 3+3 4 3+4 4 5+5 4 3+4 4 3+4 4 3+4 4 | | |
| | 6 2 4+4 4 4+4 4 3+4 4 6+6 4 3+4 4 3+4 4 3+4 4 | | |
| | 5 2 4+4 4 3+3 4 3+4 4 5+5 4 3+4 4 3+4 4 3+4 4 | | |
| | 6 2 3+3 4 4+4 4 3+4 4 5+5 4 3+4 4 3+4 4 3+4 4 | | |
| | 5 2 3+3 4 3+3 4 3+4 4 6+6 4 3+4 4 3+4 4 3+4 4 | | |
| | 6 2 4+4 4 4+4 4 3+4 4 6+6 4 3+4 4 3+4 4 3+4 4 | | |
| | 5 2 4+4 4 4+4 4 3+4 4 5+5 4 3+4 4 3+4 4 3+4 4 | | |
| | 6 2 3+3 4 3+3 4 3+4 4 6+6 4 3+4 4 3+4 4 3+4 4 | | |
| | 5 2 3+3 4 3+3 4 3+4 4 6+6 4 3+4 4 3+4 4 3+4 4 | | |
| | 5 2 4+4 4 4+4 4 3+4 4 5+5 4 3+4 4 3+4 4 3+4 4 | | |
| | 6 2 3+3 4 4+4 4 3+4 4 5+5 4 3+4 4 3+4 4 3+4 4 | | |
| | 6 2 4+4 4 3+3 4 3+4 4 6+6 4 3+4 4 3+4 4 3+4 4 | | |
| | 5 2 3+3 4 4+4 4 3+4 4 6+6 4 3+4 4 3+4 4 3+4 4 | | |
| | 5 2 4+4 4 3+3 4 3+4 4 5+5 4 3+4 4 3+4 4 3+4 4 | | |
| | 6 2 4+4 4 3+3 4 3+4 4 6+6 4 3+4 4 3+4 4 3+4 4 | | |
| | 6 2 3+3 4 4+4 4 3+4 4 5+5 4 3+4 4 3+4 4 3+4 4 | | |
| | 5 2 3+3 4 3+3 4 3+4 4 5+5 4 3+4 4 3+4 4 3+4 4 | | |
| | 5 2 4+4 4 4+4 4 3+4 4 6+6 4 3+4 4 3+4 4 3+4 4 | | |
| | 6 2 3+3 4 4+4 4 3+4 4 6+6 4 3+4 4 3+4 4 3+4 4 | | |
| | 6 2 4+4 4 3+3 4 3+4 4 6+6 4 3+4 4 3+4 4 3+4 4 | | |
| | 5 2 3+3 4 3+3 4 3+4 4 5+5 4 3+4 4 3+4 4 3+4 4 | | |
| | 5 2 4+4 4 4+4 4 3+4 4 5+5 4 3+4 4 3+4 4 3+4 4 | | |
| | 6 2 | | |
| | 6 2 | | |
| | 5 2 | | |
| | 6 2 | | |
| | 5 2 | | |
| | 6 2 | | |
| | 6 2 | | |
| | 6 2 | | |
| | 5 2 | | |
| | 6 2 | | |
| | 5 2 | | |
| | 6 2 | | |
| | 5 2 | | |
| | 5 2 | | |
| | 5 2 | | |
| | 6 2 | | |
| | 5 2 | | |
| | 5 2 | | |
| | 5 2 | | |
| | 6 2 | | |
| | 5 2 | | |
| | 6 2 | | |
| | 6 2 | | |
| | 5 2 | | |
| | 6 2 | | |
| | 5 2 | | |
| | 5 2 | | |
| | 6 2 | | |
| | 5 2 | | |
| | 6 2 | | |
| | 5 2 | | |
| | 6 2 | | |

Continued on next column

**TABLE XXXVII– continued**

| TQC | ICSD | NoE | SF |
|---|---|---|---|
| | 5 2 | | |
| | 6 2 | | |
| | 5 2 | | |
| | 6 2 | | |
| | 5 2 | | |
| | 6 2 | | |
| | 5 2 | | |
| | 5 2 | | |
| | 5 2 | | |
| | 6 2 | | |
| | 6 2 | | |
| | 5 2 | | |
| | 6 2 | | |
| | 6 2 | | |
| 62.2.247 | 150173 | 104 | Al1Pr1Pt1 |
| | -GM -R -S -T -U -X -Y -Z | | |
| | 5 2 3+3 4 3+3 4 3+4 4 6+6 4 3+4 4 3+4 4 3+4 4 | | |
| | 5 2 4+4 4 4+4 4 3+4 4 5+5 4 3+4 4 3+4 4 3+4 4 | | |
| | 6 2 4+4 4 4+4 4 3+4 4 6+6 4 3+4 4 3+4 4 3+4 4 | | |
| | 6 2 3+3 4 3+3 4 3+4 4 5+5 4 3+4 4 3+4 4 3+4 4 | | |
| | 5 2 3+3 4 3+3 4 3+4 4 6+6 4 3+4 4 3+4 4 3+4 4 | | |
| | 5 2 4+4 4 4+4 4 3+4 4 5+5 4 3+4 4 3+4 4 3+4 4 | | |
| | 6 2 4+4 4 3+3 4 3+4 4 6+6 4 3+4 4 3+4 4 3+4 4 | | |
| | 6 2 3+3 4 4+4 4 3+4 4 5+5 4 3+4 4 3+4 4 3+4 4 | | |
| | 5 2 4+4 4 3+3 4 3+4 4 5+5 4 3+4 4 3+4 4 3+4 4 | | |
| | 5 2 3+3 4 4+4 4 3+4 4 6+6 4 3+4 4 3+4 4 3+4 4 | | |
| | 6 2 4+4 4 4+4 4 3+4 4 5+5 4 3+4 4 3+4 4 3+4 4 | | |
| | 5 2 3+3 4 3+3 4 3+4 4 6+6 4 3+4 4 3+4 4 3+4 4 | | |
| | 6 2 3+3 4 4+4 4 3+4 4 5+5 4 3+4 4 3+4 4 3+4 4 | | |
| | 6 2 4+4 4 3+3 4 3+4 4 6+6 4 3+4 4 3+4 4 3+4 4 | | |
| | 5 2 3+3 4 4+4 4 3+4 4 5+5 4 3+4 4 3+4 4 3+4 4 | | |
| | 6 2 4+4 4 3+3 4 3+4 4 6+6 4 3+4 4 3+4 4 3+4 4 | | |
| | 5 2 4+4 4 4+4 4 3+4 4 5+5 4 3+4 4 3+4 4 3+4 4 | | |
| | 5 2 3+3 4 3+3 4 3+4 4 6+6 4 3+4 4 3+4 4 3+4 4 | | |
| | 6 2 4+4 4 3+3 4 3+4 4 5+5 4 3+4 4 3+4 4 3+4 4 | | |
| | 6 2 3+3 4 4+4 4 3+4 4 6+6 4 3+4 4 3+4 4 3+4 4 | | |
| | 6 2 4+4 4 3+3 4 3+4 4 5+5 4 3+4 4 3+4 4 3+4 4 | | |
| | 5 2 3+3 4 4+4 4 3+4 4 6+6 4 3+4 4 3+4 4 3+4 4 | | |
| | 5 2 4+4 4 3+3 4 3+4 4 5+5 4 3+4 4 3+4 4 3+4 4 | | |
| | 6 2 3+3 4 4+4 4 3+4 4 5+5 4 3+4 4 3+4 4 3+4 4 | | |
| | 6 2 3+3 4 3+3 4 3+4 4 6+6 4 3+4 4 3+4 4 3+4 4 | | |
| | 5 2 4+4 4 4+4 4 3+4 4 5+5 4 3+4 4 3+4 4 3+4 4 | | |
| | 5 2 | | |
| | 6 2 | | |
| | 5 2 | | |
| | 5 2 | | |
| | 6 2 | | |
| | 5 2 | | |
| | 6 2 | | |
| | 6 2 | | |
| | 5 2 | | |
| | 5 2 | | |
| | 6 2 | | |
| | 5 2 | | |
| | 6 2 | | |
| | 6 2 | | |
| | 6 2 | | |
| | 5 2 | | |
| | 5 2 | | |
| | 5 2 | | |
| | 6 2 | | |

Continued on next column

**TABLE XXXVII– continued**

| TQC | ICSD | NoE | SF | | | | | |
|---|---|---|---|---|---|---|---|---|
| | 5 2 | | | | | | | |
| | 6 2 | | | | | | | |
| | 5 2 | | | | | | | |
| | 6 2 | | | | | | | |
| | 6 2 | | | | | | | |
| | 6 2 | | | | | | | |
| | 5 2 | | | | | | | |
| 62.2.248 | 160051 | 184 | Al1Ce1Ru1 | | | | | |
| | -GM | -R | -S | -T | -U | -X | -Y | -Z |
| | 5 2 | 3+3 4 | 3+3 4 | 3+4 4 | 6+6 4 | 3+4 4 | 3+4 4 | 3+4 4 |
| | 5 2 | 4+4 4 | 4+4 4 | 3+4 4 | 5+5 4 | 3+4 4 | 3+4 4 | 3+4 4 |
| | 6 2 | 3+3 4 | 3+3 4 | 3+4 4 | 6+6 4 | 3+4 4 | 3+4 4 | 3+4 4 |
| | 5 2 | 4+4 4 | 4+4 4 | 3+4 4 | 5+5 4 | 3+4 4 | 3+4 4 | 3+4 4 |
| | 6 2 | 3+3 4 | 4+4 4 | 3+4 4 | 6+6 4 | 3+4 4 | 3+4 4 | 3+4 4 |
| | 6 2 | 4+4 4 | 3+3 4 | 3+4 4 | 5+5 4 | 3+4 4 | 3+4 4 | 3+4 4 |
| | 5 2 | 3+3 4 | 3+3 4 | 3+4 4 | 6+6 4 | 3+4 4 | 3+4 4 | 3+4 4 |
| | 6 2 | 4+4 4 | 4+4 4 | 3+4 4 | 5+5 4 | 3+4 4 | 3+4 4 | 3+4 4 |
| | 5 2 | 4+4 4 | 3+3 4 | 3+4 4 | 6+6 4 | 3+4 4 | 3+4 4 | 3+4 4 |
| | 5 2 | 3+3 4 | 4+4 4 | 3+4 4 | 5+5 4 | 3+4 4 | 3+4 4 | 3+4 4 |
| | 6 2 | 3+3 4 | 4+4 4 | 3+4 4 | 6+6 4 | 3+4 4 | 3+4 4 | 3+4 4 |
| | 5 2 | 4+4 4 | 3+3 4 | 3+4 4 | 5+5 4 | 3+4 4 | 3+4 4 | 3+4 4 |
| | 6 2 | 3+3 4 | 4+4 4 | 3+4 4 | 6+6 4 | 3+4 4 | 3+4 4 | 3+4 4 |
| | 6 2 | 4+4 4 | 3+3 4 | 3+4 4 | 5+5 4 | 3+4 4 | 3+4 4 | 3+4 4 |
| | 5 2 | 4+4 4 | 3+3 4 | 3+4 4 | 6+6 4 | 3+4 4 | 3+4 4 | 3+4 4 |
| | 6 2 | 3+3 4 | 4+4 4 | 3+4 4 | 5+5 4 | 3+4 4 | 3+4 4 | 3+4 4 |
| | 5 2 | 3+3 4 | 3+3 4 | 3+4 4 | 5+5 4 | 3+4 4 | 3+4 4 | 3+4 4 |
| | 5 2 | 4+4 4 | 4+4 4 | 3+4 4 | 6+6 4 | 3+4 4 | 3+4 4 | 3+4 4 |
| | 6 2 | 3+3 4 | 3+3 4 | 3+4 4 | 5+5 4 | 3+4 4 | 3+4 4 | 3+4 4 |
| | 5 2 | 4+4 4 | 4+4 4 | 3+4 4 | 6+6 4 | 3+4 4 | 3+4 4 | 3+4 4 |
| | 6 2 | 3+3 4 | 4+4 4 | 3+4 4 | 5+5 4 | 3+4 4 | 3+4 4 | 3+4 4 |
| | 6 2 | 4+4 4 | 3+3 4 | 3+4 4 | 6+6 4 | 3+4 4 | 3+4 4 | 3+4 4 |
| | 5 2 | 4+4 4 | 4+4 4 | 3+4 4 | 5+5 4 | 3+4 4 | 3+4 4 | 3+4 4 |
| | 5 2 | 3+3 4 | 3+3 4 | 3+4 4 | 6+6 4 | 3+4 4 | 3+4 4 | 3+4 4 |
| | 6 2 | 4+4 4 | 3+3 4 | 3+4 4 | 5+5 4 | 3+4 4 | 3+4 4 | 3+4 4 |
| | 5 2 | 3+3 4 | 4+4 4 | 3+4 4 | 5+5 4 | 3+4 4 | 3+4 4 | 3+4 4 |
| | 6 2 | 4+4 4 | 3+3 4 | 3+4 4 | 6+6 4 | 3+4 4 | 3+4 4 | 3+4 4 |
| | 6 2 | 3+3 4 | 4+4 4 | 3+4 4 | 6+6 4 | 3+4 4 | 3+4 4 | 3+4 4 |
| | 5 2 | 3+3 4 | 3+3 4 | 3+4 4 | 5+5 4 | 3+4 4 | 3+4 4 | 3+4 4 |
| | 6 2 | 4+4 4 | 4+4 4 | 3+4 4 | 5+5 4 | 3+4 4 | 3+4 4 | 3+4 4 |
| | 5 2 | 4+4 4 | 3+3 4 | 3+4 4 | 6+6 4 | 3+4 4 | 3+4 4 | 3+4 4 |
| | 6 2 | 3+3 4 | 4+4 4 | 3+4 4 | 6+6 4 | 3+4 4 | 3+4 4 | 3+4 4 |
| | 5 2 | 3+3 4 | 4+4 4 | 3+4 4 | 5+5 4 | 3+4 4 | 3+4 4 | 3+4 4 |
| | 5 2 | 4+4 4 | 3+3 4 | 3+4 4 | 5+5 4 | 3+4 4 | 3+4 4 | 3+4 4 |
| | 6 2 | 3+3 4 | 3+3 4 | 3+4 4 | 6+6 4 | 3+4 4 | 3+4 4 | 3+4 4 |
| | 6 2 | 4+4 4 | 4+4 4 | 3+4 4 | 6+6 4 | 3+4 4 | 3+4 4 | 3+4 4 |
| | 5 2 | 4+4 4 | 4+4 4 | 3+4 4 | 5+5 4 | 3+4 4 | 3+4 4 | 3+4 4 |
| | 5 2 | 3+3 4 | 3+3 4 | 3+4 4 | 6+6 4 | 3+4 4 | 3+4 4 | 3+4 4 |
| | 6 2 | 4+4 4 | 4+4 4 | 3+4 4 | 5+5 4 | 3+4 4 | 3+4 4 | 3+4 4 |
| | 6 2 | 3+3 4 | 3+3 4 | 3+4 4 | 6+6 4 | 3+4 4 | 3+4 4 | 3+4 4 |
| | 6 2 | 4+4 4 | 4+4 4 | 3+4 4 | 5+5 4 | 3+4 4 | 3+4 4 | 3+4 4 |
| | 5 2 | 3+3 4 | 3+3 4 | 3+4 4 | 6+6 4 | 3+4 4 | 3+4 4 | 3+4 4 |
| | 6 2 | 3+3 4 | 3+3 4 | 3+4 4 | 5+5 4 | 3+4 4 | 3+4 4 | 3+4 4 |
| | 5 2 | 4+4 4 | 4+4 4 | 3+4 4 | 6+6 4 | 3+4 4 | 3+4 4 | 3+4 4 |
| | 6 2 | 3+3 4 | 3+3 4 | 3+4 4 | 5+5 4 | 3+4 4 | 3+4 4 | 3+4 4 |
| | 5 2 | 4+4 4 | 4+4 4 | 3+4 4 | 5+5 4 | 3+4 4 | 3+4 4 | 3+4 4 |
| | 5 2 | | | | | | | |
| | 5 2 | | | | | | | |
| | 5 2 | | | | | | | |
| | 5 2 | | | | | | | |
| | 6 2 | | | | | | | |
| | 5 2 | | | | | | | |

Continued on next column

| TQC | ICSD | NoE | SF | | | | | |
|---|---|---|---|---|---|---|---|---|
| | 6 2 | | | | | | | |
| | 6 2 | | | | | | | |
| | 5 2 | | | | | | | |
| | 6 2 | | | | | | | |
| | 6 2 | | | | | | | |
| | 5 2 | | | | | | | |
| | 6 2 | | | | | | | |
| | 5 2 | | | | | | | |
| | 5 2 | | | | | | | |
| | 5 2 | | | | | | | |
| | 6 2 | | | | | | | |
| | 6 2 | | | | | | | |
| | 5 2 | | | | | | | |
| | 5 2 | | | | | | | |
| | 6 2 | | | | | | | |
| | 6 2 | | | | | | | |
| | 5 2 | | | | | | | |
| | 6 2 | | | | | | | |
| | 6 2 | | | | | | | |
| | 6 2 | | | | | | | |
| | 5 2 | | | | | | | |
| | 6 2 | | | | | | | |
| | 6 2 | | | | | | | |
| | 5 2 | | | | | | | |
| | 6 2 | | | | | | | |
| | 5 2 | | | | | | | |
| | 5 2 | | | | | | | |
| | 6 2 | | | | | | | |
| | 5 2 | | | | | | | |
| | 6 2 | | | | | | | |
| | 5 2 | | | | | | | |
| | 5 2 | | | | | | | |
| | 5 2 | | | | | | | |
| | 6 2 | | | | | | | |
| | 6 2 | | | | | | | |
| | 6 2 | | | | | | | |
| | 5 2 | | | | | | | |
| | 6 2 | | | | | | | |
| | 5 2 | | | | | | | |
| | 5 2 | | | | | | | |
| 62.2.249 | 55623 | 192 | Al1Ce1Rh1 | | | | | |
| | -GM | -R | -S | -T | -U | -X | -Y | -Z |
| | 5 2 | 3+3 4 | 3+3 4 | 3+4 4 | 5+5 4 | 3+4 4 | 3+4 4 | 3+4 4 |
| | 5 2 | 4+4 4 | 4+4 4 | 3+4 4 | 6+6 4 | 3+4 4 | 3+4 4 | 3+4 4 |
| | 6 2 | 3+3 4 | 3+3 4 | 3+4 4 | 5+5 4 | 3+4 4 | 3+4 4 | 3+4 4 |
| | 6 2 | 4+4 4 | 4+4 4 | 3+4 4 | 6+6 4 | 3+4 4 | 3+4 4 | 3+4 4 |
| | 5 2 | 4+4 4 | 3+3 4 | 3+4 4 | 5+5 4 | 3+4 4 | 3+4 4 | 3+4 4 |
| | 6 2 | 3+3 4 | 4+4 4 | 3+4 4 | 6+6 4 | 3+4 4 | 3+4 4 | 3+4 4 |
| | 5 2 | 3+3 4 | 4+4 4 | 3+4 4 | 5+5 4 | 3+4 4 | 3+4 4 | 3+4 4 |
| | 6 2 | 4+4 4 | 3+3 4 | 3+4 4 | 6+6 4 | 3+4 4 | 3+4 4 | 3+4 4 |
| | 5 2 | 3+3 4 | 4+4 4 | 3+4 4 | 5+5 4 | 3+4 4 | 3+4 4 | 3+4 4 |
| | 5 2 | 4+4 4 | 3+3 4 | 3+4 4 | 6+6 4 | 3+4 4 | 3+4 4 | 3+4 4 |
| | 6 2 | 4+4 4 | 3+3 4 | 3+4 4 | 5+5 4 | 3+4 4 | 3+4 4 | 3+4 4 |
| | 5 2 | 3+3 4 | 4+4 4 | 3+4 4 | 6+6 4 | 3+4 4 | 3+4 4 | 3+4 4 |
| | 6 2 | 4+4 4 | 3+3 4 | 3+4 4 | 5+5 4 | 3+4 4 | 3+4 4 | 3+4 4 |
| | 6 2 | 3+3 4 | 4+4 4 | 3+4 4 | 6+6 4 | 3+4 4 | 3+4 4 | 3+4 4 |
| | 5 2 | 3+3 4 | 4+4 4 | 3+4 4 | 5+5 4 | 3+4 4 | 3+4 4 | 3+4 4 |
| | 6 2 | 4+4 4 | 3+3 4 | 3+4 4 | 6+6 4 | 3+4 4 | 3+4 4 | 3+4 4 |
| | 5 2 | 3+3 4 | 3+3 4 | 3+4 4 | 6+6 4 | 3+4 4 | 3+4 4 | 3+4 4 |
| | 5 2 | 4+4 4 | 4+4 4 | 3+4 4 | 5+5 4 | 3+4 4 | 3+4 4 | 3+4 4 |
| | 6 2 | 4+4 4 | 3+3 4 | 3+4 4 | 6+6 4 | 3+4 4 | 3+4 4 | 3+4 4 |

Continued on next column

**TABLE XXXVII– continued**

| TQC | ICSD | NoE | SF |
|---|---|---|---|
| | 5 2 3+3 4 4+4 4 3+4 4 5+5 4 3+4 4 3+4 4 3+4 4 | | |
| | 6 2 4+4 4 3+3 4 3+4 4 6+6 4 3+4 4 3+4 4 3+4 4 | | |
| | 5 2 3+3 4 4+4 4 3+4 4 5+5 4 3+4 4 3+4 4 3+4 4 | | |
| | 6 2 3+3 4 4+4 4 3+4 4 6+6 4 3+4 4 3+4 4 3+4 4 | | |
| | 5 2 4+4 4 3+3 4 3+4 4 6+6 4 3+4 4 3+4 4 3+4 4 | | |
| | 6 2 3+3 4 4+4 4 3+4 4 5+5 4 3+4 4 3+4 4 3+4 4 | | |
| | 5 2 4+4 4 3+3 4 3+4 4 6+6 4 3+4 4 3+4 4 3+4 4 | | |
| | 6 2 3+3 4 4+4 4 3+4 4 5+5 4 3+4 4 3+4 4 3+4 4 | | |
| | 6 2 4+4 4 3+3 4 3+4 4 6+6 4 3+4 4 3+4 4 3+4 4 | | |
| | 5 2 4+4 4 3+3 4 3+4 4 5+5 4 3+4 4 3+4 4 3+4 4 | | |
| | 6 2 3+3 4 4+4 4 3+4 4 5+5 4 3+4 4 3+4 4 3+4 4 | | |
| | 5 2 3+3 4 4+4 4 3+4 4 6+6 4 3+4 4 3+4 4 3+4 4 | | |
| | 6 2 4+4 4 3+3 4 3+4 4 5+5 4 3+4 4 3+4 4 3+4 4 | | |
| | 5 2 3+3 4 3+3 4 3+4 4 6+6 4 3+4 4 3+4 4 3+4 4 | | |
| | 5 2 4+4 4 4+4 4 3+4 4 5+5 4 3+4 4 3+4 4 3+4 4 | | |
| | 6 2 4+4 4 3+3 4 3+4 4 6+6 4 3+4 4 3+4 4 3+4 4 | | |
| | 6 2 3+3 4 4+4 4 3+4 4 5+5 4 3+4 4 3+4 4 3+4 4 | | |
| | 5 2 3+3 4 3+3 4 3+4 4 6+6 4 3+4 4 3+4 4 3+4 4 | | |
| | 5 2 4+4 4 4+4 4 3+4 4 5+5 4 3+4 4 3+4 4 3+4 4 | | |
| | 6 2 3+3 4 4+4 4 3+4 4 6+6 4 3+4 4 3+4 4 3+4 4 | | |
| | 6 2 4+4 4 3+3 4 3+4 4 5+5 4 3+4 4 3+4 4 3+4 4 | | |
| | 6 2 3+3 4 4+4 4 3+4 4 6+6 4 3+4 4 3+4 4 3+4 4 | | |
| | 5 2 4+4 4 3+3 4 3+4 4 5+5 4 3+4 4 3+4 4 3+4 4 | | |
| | 6 2 4+4 4 4+4 4 3+4 4 6+6 4 3+4 4 3+4 4 3+4 4 | | |
| | 5 2 3+3 4 3+3 4 3+4 4 6+6 4 3+4 4 3+4 4 3+4 4 | | |
| | 6 2 4+4 4 4+4 4 3+4 4 5+5 4 3+4 4 3+4 4 3+4 4 | | |
| | 5 2 3+3 4 3+3 4 3+4 4 6+6 4 3+4 4 3+4 4 3+4 4 | | |
| | 5 2 4+4 4 4+4 4 3+4 4 5+5 4 3+4 4 3+4 4 3+4 4 | | |
| | 5 2 3+3 4 3+3 4 3+4 4 6+6 4 3+4 4 3+4 4 3+4 4 | | |
| | 5 2 | | |
| | 5 2 | | |
| | 6 2 | | |
| | 5 2 | | |
| | 6 2 | | |
| | 6 2 | | |
| | 5 2 | | |
| | 6 2 | | |
| | 6 2 | | |
| | 5 2 | | |
| | 6 2 | | |
| | 5 2 | | |
| | 6 2 | | |
| | 5 2 | | |
| | 6 2 | | |
| | 5 2 | | |
| | 5 2 | | |
| | 6 2 | | |
| | 6 2 | | |
| | 5 2 | | |
| | 6 2 | | |
| | 5 2 | | |
| | 6 2 | | |
| | 6 2 | | |
| | 6 2 | | |
| | 5 2 | | |
| | 5 2 | | |
| | 6 2 | | |
| | 6 2 | | |
| | 5 2 | | |
| | 5 2 | | |
| | 6 2 | | |

Continued on next column

**TABLE XXXVII– continued**

| TQC | ICSD | NoE | SF |
|---|---|---|---|
| | 6 2 | | |
| | 5 2 | | |
| | 5 2 | | |
| | 5 2 | | |
| | 5 2 | | |
| | 6 2 | | |
| | 6 2 | | |
| | 6 2 | | |
| | 5 2 | | |
| | 5 2 | | |
| | 6 2 | | |
| | 5 2 | | |
| | 6 2 | | |
| | 6 2 | | |
| | 6 2 | | |
| | 5 2 | | |
| 62.2.250 | 160052 | 192 | Al1Ce1Rh1 |
| | -GM -R -S -T -U -X -Y -Z | | |
| | 5 2 3+3 4 3+3 4 3+4 4 6+6 4 3+4 4 3+4 4 3+4 4 | | |
| | 5 2 4+4 4 4+4 4 3+4 4 5+5 4 3+4 4 3+4 4 3+4 4 | | |
| | 6 2 3+3 4 3+3 4 3+4 4 6+6 4 3+4 4 3+4 4 3+4 4 | | |
| | 6 2 4+4 4 4+4 4 3+4 4 5+5 4 3+4 4 3+4 4 3+4 4 | | |
| | 5 2 3+3 4 4+4 4 3+4 4 6+6 4 3+4 4 3+4 4 3+4 4 | | |
| | 6 2 4+4 4 3+3 4 3+4 4 5+5 4 3+4 4 3+4 4 3+4 4 | | |
| | 5 2 4+4 4 3+3 4 3+4 4 6+6 4 3+4 4 3+4 4 3+4 4 | | |
| | 6 2 3+3 4 4+4 4 3+4 4 5+5 4 3+4 4 3+4 4 3+4 4 | | |
| | 5 2 4+4 4 3+3 4 3+4 4 6+6 4 3+4 4 3+4 4 3+4 4 | | |
| | 5 2 3+3 4 4+4 4 3+4 4 5+5 4 3+4 4 3+4 4 3+4 4 | | |
| | 6 2 3+3 4 4+4 4 3+4 4 6+6 4 3+4 4 3+4 4 3+4 4 | | |
| | 5 2 4+4 4 3+3 4 3+4 4 5+5 4 3+4 4 3+4 4 3+4 4 | | |
| | 6 2 3+3 4 4+4 4 3+4 4 6+6 4 3+4 4 3+4 4 3+4 4 | | |
| | 6 2 4+4 4 3+3 4 3+4 4 5+5 4 3+4 4 3+4 4 3+4 4 | | |
| | 5 2 4+4 4 3+3 4 3+4 4 6+6 4 3+4 4 3+4 4 3+4 4 | | |
| | 6 2 3+3 4 4+4 4 3+4 4 5+5 4 3+4 4 3+4 4 3+4 4 | | |
| | 5 2 3+3 4 3+3 4 3+4 4 5+5 4 3+4 4 3+4 4 3+4 4 | | |
| | 5 2 4+4 4 4+4 4 3+4 4 6+6 4 3+4 4 3+4 4 3+4 4 | | |
| | 6 2 3+3 4 3+3 4 3+4 4 5+5 4 3+4 4 3+4 4 3+4 4 | | |
| | 5 2 4+4 4 4+4 4 3+4 4 6+6 4 3+4 4 3+4 4 3+4 4 | | |
| | 6 2 3+3 4 4+4 4 3+4 4 5+5 4 3+4 4 3+4 4 3+4 4 | | |
| | 5 2 4+4 4 3+3 4 3+4 4 6+6 4 3+4 4 3+4 4 3+4 4 | | |
| | 6 2 4+4 4 3+3 4 3+4 4 5+5 4 3+4 4 3+4 4 3+4 4 | | |
| | 5 2 3+3 4 4+4 4 3+4 4 5+5 4 3+4 4 3+4 4 3+4 4 | | |
| | 6 2 4+4 4 3+3 4 3+4 4 6+6 4 3+4 4 3+4 4 3+4 4 | | |
| | 5 2 3+3 4 4+4 4 3+4 4 5+5 4 3+4 4 3+4 4 3+4 4 | | |
| | 6 2 4+4 4 3+3 4 3+4 4 6+6 4 3+4 4 3+4 4 3+4 4 | | |
| | 6 2 3+3 4 4+4 4 3+4 4 5+5 4 3+4 4 3+4 4 3+4 4 | | |
| | 5 2 3+3 4 3+3 4 3+4 4 6+6 4 3+4 4 3+4 4 3+4 4 | | |
| | 6 2 4+4 4 4+4 4 3+4 4 6+6 4 3+4 4 3+4 4 3+4 4 | | |
| | 5 2 3+3 4 3+3 4 3+4 4 5+5 4 3+4 4 3+4 4 3+4 4 | | |
| | 6 2 4+4 4 4+4 4 3+4 4 6+6 4 3+4 4 3+4 4 3+4 4 | | |
| | 5 2 4+4 4 4+4 4 3+4 4 5+5 4 3+4 4 3+4 4 3+4 4 | | |
| | 5 2 3+3 4 3+3 4 3+4 4 6+6 4 3+4 4 3+4 4 3+4 4 | | |
| | 6 2 3+3 4 4+4 4 3+4 4 5+5 4 3+4 4 3+4 4 3+4 4 | | |
| | 6 2 4+4 4 3+3 4 3+4 4 6+6 4 3+4 4 3+4 4 3+4 4 | | |
| | 5 2 4+4 4 4+4 4 3+4 4 5+5 4 3+4 4 3+4 4 3+4 4 | | |
| | 5 2 3+3 4 3+3 4 3+4 4 6+6 4 3+4 4 3+4 4 3+4 4 | | |
| | 6 2 4+4 4 3+3 4 3+4 4 5+5 4 3+4 4 3+4 4 3+4 4 | | |
| | 6 2 3+3 4 4+4 4 3+4 4 6+6 4 3+4 4 3+4 4 3+4 4 | | |
| | 6 2 4+4 4 3+3 4 3+4 4 5+5 4 3+4 4 3+4 4 3+4 4 | | |
| | 5 2 3+3 4 4+4 4 3+4 4 6+6 4 3+4 4 3+4 4 3+4 4 | | |
| | 6 2 3+3 4 3+3 4 3+4 4 5+5 4 3+4 4 3+4 4 3+4 4 | | |

Continued on next column

**TABLE XXXVII– continued**

| TQC | ICSD | NoE | SF | | | | | |
|---|---|---|---|---|---|---|---|---|
| | 5 2 | 4+4 4 | 4+4 4 | 3+4 4 | 5+5 4 | 3+4 4 | 3+4 4 | 3+4 4 |
| | 6 2 | 3+3 4 | 3+3 4 | 3+4 4 | 6+6 4 | 3+4 4 | 3+4 4 | 3+4 4 |
| | 5 2 | 4+4 4 | 4+4 4 | 3+4 4 | 5+5 4 | 3+4 4 | 3+4 4 | 3+4 4 |
| | 5 2 | 3+3 4 | 3+3 4 | 3+4 4 | 6+6 4 | 3+4 4 | 3+4 4 | 3+4 4 |
| | 5 2 | 4+4 4 | 4+4 4 | 3+4 4 | 5+5 4 | 3+4 4 | 3+4 4 | 3+4 4 |
| | 5 2 | | | | | | | |
| | 5 2 | | | | | | | |
| | 6 2 | | | | | | | |
| | 5 2 | | | | | | | |
| | 6 2 | | | | | | | |
| | 6 2 | | | | | | | |
| | 5 2 | | | | | | | |
| | 6 2 | | | | | | | |
| | 6 2 | | | | | | | |
| | 5 2 | | | | | | | |
| | 6 2 | | | | | | | |
| | 5 2 | | | | | | | |
| | 6 2 | | | | | | | |
| | 5 2 | | | | | | | |
| | 6 2 | | | | | | | |
| | 5 2 | | | | | | | |
| | 5 2 | | | | | | | |
| | 6 2 | | | | | | | |
| | 6 2 | | | | | | | |
| | 5 2 | | | | | | | |
| | 6 2 | | | | | | | |
| | 5 2 | | | | | | | |
| | 6 2 | | | | | | | |
| | 6 2 | | | | | | | |
| | 6 2 | | | | | | | |
| | 5 2 | | | | | | | |
| | 5 2 | | | | | | | |
| | 6 2 | | | | | | | |
| | 6 2 | | | | | | | |
| | 5 2 | | | | | | | |
| | 5 2 | | | | | | | |
| | 6 2 | | | | | | | |
| | 6 2 | | | | | | | |
| | 5 2 | | | | | | | |
| | 5 2 | | | | | | | |
| | 5 2 | | | | | | | |
| | 5 2 | | | | | | | |
| | 6 2 | | | | | | | |
| | 6 2 | | | | | | | |
| | 6 2 | | | | | | | |
| | 5 2 | | | | | | | |
| | 5 2 | | | | | | | |
| | 6 2 | | | | | | | |
| | 5 2 | | | | | | | |
| | 6 2 | | | | | | | |
| | 6 2 | | | | | | | |
| | 6 2 | | | | | | | |
| | 5 2 | | | | | | | |
| 62.2.251 | 370027 | 152 | Al1Au1Yb1 | | | | | |
| | -GM | -R | -S | -T | -U | -X | -Y | -Z |
| | 5 2 | 3+3 4 | 3+3 4 | 3+4 4 | 6+6 4 | 3+4 4 | 3+4 4 | 3+4 4 |
| | 5 2 | 4+4 4 | 4+4 4 | 3+4 4 | 5+5 4 | 3+4 4 | 3+4 4 | 3+4 4 |
| | 6 2 | 3+3 4 | 3+3 4 | 3+4 4 | 6+6 4 | 3+4 4 | 3+4 4 | 3+4 4 |
| | 6 2 | 4+4 4 | 4+4 4 | 3+4 4 | 5+5 4 | 3+4 4 | 3+4 4 | 3+4 4 |
| | 5 2 | 4+4 4 | 3+3 4 | 3+4 4 | 6+6 4 | 3+4 4 | 3+4 4 | 3+4 4 |
| | 5 2 | 3+3 4 | 4+4 4 | 3+4 4 | 5+5 4 | 3+4 4 | 3+4 4 | 3+4 4 |

Continued on next column

**TABLE XXXVII– continued**

| TQC | ICSD | NoE | SF | | | | | |
|---|---|---|---|---|---|---|---|---|
| | 6 2 | 3+3 4 | 4+4 4 | 3+4 4 | 6+6 4 | 3+4 4 | 3+4 4 | 3+4 4 |
| | 6 2 | 4+4 4 | 3+3 4 | 3+4 4 | 5+5 4 | 3+4 4 | 3+4 4 | 3+4 4 |
| | 5 2 | 3+3 4 | 4+4 4 | 3+4 4 | 5+5 4 | 3+4 4 | 3+4 4 | 3+4 4 |
| | 5 2 | 4+4 4 | 3+3 4 | 3+4 4 | 6+6 4 | 3+4 4 | 3+4 4 | 3+4 4 |
| | 6 2 | 3+3 4 | 3+3 4 | 3+4 4 | 5+5 4 | 3+4 4 | 3+4 4 | 3+4 4 |
| | 5 2 | 4+4 4 | 4+4 4 | 3+4 4 | 6+6 4 | 3+4 4 | 3+4 4 | 3+4 4 |
| | 6 2 | 4+4 4 | 3+3 4 | 3+4 4 | 5+5 4 | 3+4 4 | 3+4 4 | 3+4 4 |
| | 6 2 | 3+3 4 | 4+4 4 | 3+4 4 | 6+6 4 | 3+4 4 | 3+4 4 | 3+4 4 |
| | 5 2 | 4+4 4 | 3+3 4 | 3+4 4 | 5+5 4 | 3+4 4 | 3+4 4 | 3+4 4 |
| | 6 2 | 3+3 4 | 4+4 4 | 3+4 4 | 6+6 4 | 3+4 4 | 3+4 4 | 3+4 4 |
| | 5 2 | 3+3 4 | 3+3 4 | 3+4 4 | 5+5 4 | 3+4 4 | 3+4 4 | 3+4 4 |
| | 5 2 | 4+4 4 | 4+4 4 | 3+4 4 | 6+6 4 | 3+4 4 | 3+4 4 | 3+4 4 |
| | 6 2 | 3+3 4 | 4+4 4 | 3+4 4 | 5+5 4 | 3+4 4 | 3+4 4 | 3+4 4 |
| | 6 2 | 4+4 4 | 3+3 4 | 3+4 4 | 6+6 4 | 3+4 4 | 3+4 4 | 3+4 4 |
| | 5 2 | 4+4 4 | 4+4 4 | 3+4 4 | 5+5 4 | 3+4 4 | 3+4 4 | 3+4 4 |
| | 5 2 | 3+3 4 | 3+3 4 | 3+4 4 | 6+6 4 | 3+4 4 | 3+4 4 | 3+4 4 |
| | 6 2 | 3+3 4 | 4+4 4 | 3+4 4 | 5+5 4 | 3+4 4 | 3+4 4 | 3+4 4 |
| | 6 2 | 4+4 4 | 3+3 4 | 3+4 4 | 6+6 4 | 3+4 4 | 3+4 4 | 3+4 4 |
| | 5 2 | 4+4 4 | 4+4 4 | 3+4 4 | 5+5 4 | 3+4 4 | 3+4 4 | 3+4 4 |
| | 5 2 | 3+3 4 | 3+3 4 | 3+4 4 | 6+6 4 | 3+4 4 | 3+4 4 | 3+4 4 |
| | 6 2 | 3+3 4 | 3+3 4 | 3+4 4 | 5+5 4 | 3+4 4 | 3+4 4 | 3+4 4 |
| | 6 2 | 4+4 4 | 4+4 4 | 3+4 4 | 6+6 4 | 3+4 4 | 3+4 4 | 3+4 4 |
| | 6 2 | 3+3 4 | 4+4 4 | 3+4 4 | 5+5 4 | 3+4 4 | 3+4 4 | 3+4 4 |
| | 5 2 | 4+4 4 | 3+3 4 | 3+4 4 | 5+5 4 | 3+4 4 | 3+4 4 | 3+4 4 |
| | 6 2 | 4+4 4 | 3+3 4 | 3+4 4 | 6+6 4 | 3+4 4 | 3+4 4 | 3+4 4 |
| | 5 2 | 3+3 4 | 4+4 4 | 3+4 4 | 6+6 4 | 3+4 4 | 3+4 4 | 3+4 4 |
| | 5 2 | 4+4 4 | 3+3 4 | 3+4 4 | 5+5 4 | 3+4 4 | 3+4 4 | 3+4 4 |
| | 6 2 | 3+3 4 | 4+4 4 | 3+4 4 | 6+6 4 | 3+4 4 | 3+4 4 | 3+4 4 |
| | 5 2 | 4+4 4 | 4+4 4 | 3+4 4 | 5+5 4 | 3+4 4 | 3+4 4 | 3+4 4 |
| | 5 2 | 3+3 4 | 3+3 4 | 3+4 4 | 6+6 4 | 3+4 4 | 3+4 4 | 3+4 4 |
| | 5 2 | 3+3 4 | 3+3 4 | 3+4 4 | 5+5 4 | 3+4 4 | 3+4 4 | 3+4 4 |
| | 6 2 | 4+4 4 | 4+4 4 | 3+4 4 | 5+5 4 | 3+4 4 | 3+4 4 | 3+4 4 |
| | 6 2 | | | | | | | |
| | 6 2 | | | | | | | |
| | 5 2 | | | | | | | |
| | 6 2 | | | | | | | |
| | 5 2 | | | | | | | |
| | 5 2 | | | | | | | |
| | 5 2 | | | | | | | |
| | 6 2 | | | | | | | |
| | 6 2 | | | | | | | |
| | 5 2 | | | | | | | |
| | 6 2 | | | | | | | |
| | 5 2 | | | | | | | |
| | 6 2 | | | | | | | |
| | 5 2 | | | | | | | |
| | 6 2 | | | | | | | |
| | 6 2 | | | | | | | |
| | 5 2 | | | | | | | |
| | 5 2 | | | | | | | |
| | 6 2 | | | | | | | |
| | 5 2 | | | | | | | |
| | 6 2 | | | | | | | |
| | 5 2 | | | | | | | |
| | 6 2 | | | | | | | |
| | 6 2 | | | | | | | |
| | 5 2 | | | | | | | |
| | 6 2 | | | | | | | |
| | 5 2 | | | | | | | |
| | 6 2 | | | | | | | |
| | 5 2 | | | | | | | |

Continued on next column

**TABLE XXXVII– continued**

| TQC | ICSD | NoE | SF | | | | | |
|---|---|---|---|---|---|---|---|---|
| | 5 2 | | | | | | | |
| | 6 2 | | | | | | | |
| | 6 2 | | | | | | | |
| | 6 2 | | | | | | | |
| | 5 2 | | | | | | | |
| | 6 2 | | | | | | | |
| | 5 2 | | | | | | | |
| | 5 2 | | | | | | | |
| | 5 2 | | | | | | | |
| 62.2.252 | 370015 | 64 | Al1Au1Ca1 | | | | | |
| | -GM | -R | -S | -T | -U | -X | -Y | -Z |
| | 5 2 | 3+3 4 | 4+4 4 | 3+4 4 | 5+5 4 | 3+4 4 | 3+4 4 | 3+4 4 |
| | 5 2 | 4+4 4 | 3+3 4 | 3+4 4 | 6+6 4 | 3+4 4 | 3+4 4 | 3+4 4 |
| | 6 2 | 3+3 4 | 4+4 4 | 3+4 4 | 5+5 4 | 3+4 4 | 3+4 4 | 3+4 4 |
| | 6 2 | 4+4 4 | 3+3 4 | 3+4 4 | 6+6 4 | 3+4 4 | 3+4 4 | 3+4 4 |
| | 5 2 | 4+4 4 | 3+3 4 | 3+4 4 | 5+5 4 | 3+4 4 | 3+4 4 | 3+4 4 |
| | 5 2 | 3+3 4 | 4+4 4 | 3+4 4 | 6+6 4 | 3+4 4 | 3+4 4 | 3+4 4 |
| | 6 2 | 4+4 4 | 3+3 4 | 3+4 4 | 5+5 4 | 3+4 4 | 3+4 4 | 3+4 4 |
| | 6 2 | 3+3 4 | 4+4 4 | 3+4 4 | 6+6 4 | 3+4 4 | 3+4 4 | 3+4 4 |
| | 5 2 | 3+3 4 | 3+3 4 | 3+4 4 | 5+5 4 | 3+4 4 | 3+4 4 | 3+4 4 |
| | 5 2 | 4+4 4 | 4+4 4 | 3+4 4 | 6+6 4 | 3+4 4 | 3+4 4 | 3+4 4 |
| | 6 2 | 3+3 4 | 4+4 4 | 3+4 4 | 5+5 4 | 3+4 4 | 3+4 4 | 3+4 4 |
| | 6 2 | 4+4 4 | 3+3 4 | 3+4 4 | 6+6 4 | 3+4 4 | 3+4 4 | 3+4 4 |
| | 6 2 | 3+3 4 | 4+4 4 | 3+4 4 | 5+5 4 | 3+4 4 | 3+4 4 | 3+4 4 |
| | 5 2 | 4+4 4 | 3+3 4 | 3+4 4 | 6+6 4 | 3+4 4 | 3+4 4 | 3+4 4 |
| | 5 2 | 3+3 4 | 4+4 4 | 3+4 4 | 5+5 4 | 3+4 4 | 3+4 4 | 3+4 4 |
| | 6 2 | 4+4 4 | 3+3 4 | 3+4 4 | 5+5 4 | 3+4 4 | 3+4 4 | 3+4 4 |
| | 5 2 | | | | | | | |
| | 5 2 | | | | | | | |
| | 6 2 | | | | | | | |
| | 5 2 | | | | | | | |
| | 6 2 | | | | | | | |
| | 5 2 | | | | | | | |
| | 6 2 | | | | | | | |
| | 6 2 | | | | | | | |
| | 6 2 | | | | | | | |
| | 5 2 | | | | | | | |
| | 5 2 | | | | | | | |
| | 5 2 | | | | | | | |
| | 6 2 | | | | | | | |
| | 5 2 | | | | | | | |
| | 6 2 | | | | | | | |
| | 5 2 | | | | | | | |
| 62.2.253 | 83983 | 160 | Ag1Sb1Yb1 | | | | | |
| | -GM | -R | -S | -T | -U | -X | -Y | -Z |
| | 5 2 | 3+3 4 | 3+3 4 | 3+4 4 | 5+5 4 | 3+4 4 | 3+4 4 | 3+4 4 |
| | 6 2 | 4+4 4 | 4+4 4 | 3+4 4 | 6+6 4 | 3+4 4 | 3+4 4 | 3+4 4 |
| | 5 2 | 3+3 4 | 3+3 4 | 3+4 4 | 5+5 4 | 3+4 4 | 3+4 4 | 3+4 4 |
| | 6 2 | 4+4 4 | 4+4 4 | 3+4 4 | 6+6 4 | 3+4 4 | 3+4 4 | 3+4 4 |
| | 5 2 | 3+3 4 | 3+3 4 | 3+4 4 | 5+5 4 | 3+4 4 | 3+4 4 | 3+4 4 |
| | 6 2 | 4+4 4 | 4+4 4 | 3+4 4 | 6+6 4 | 3+4 4 | 3+4 4 | 3+4 4 |
| | 5 2 | 4+4 4 | 3+3 4 | 3+4 4 | 5+5 4 | 3+4 4 | 3+4 4 | 3+4 4 |
| | 6 2 | 3+3 4 | 4+4 4 | 3+4 4 | 6+6 4 | 3+4 4 | 3+4 4 | 3+4 4 |
| | 5 2 | 4+4 4 | 4+4 4 | 3+4 4 | 6+6 4 | 3+4 4 | 3+4 4 | 3+4 4 |
| | 6 2 | 3+3 4 | 3+3 4 | 3+4 4 | 5+5 4 | 3+4 4 | 3+4 4 | 3+4 4 |
| | 5 2 | 4+4 4 | 3+3 4 | 3+4 4 | 6+6 4 | 3+4 4 | 3+4 4 | 3+4 4 |
| | 5 2 | 3+3 4 | 4+4 4 | 3+4 4 | 5+5 4 | 3+4 4 | 3+4 4 | 3+4 4 |
| | 6 2 | 3+3 4 | 3+3 4 | 3+4 4 | 5+5 4 | 3+4 4 | 3+4 4 | 3+4 4 |
| | 5 2 | 4+4 4 | 4+4 4 | 3+4 4 | 5+5 4 | 3+4 4 | 3+4 4 | 3+4 4 |
| | 6 2 | 3+3 4 | 3+3 4 | 3+4 4 | 6+6 4 | 3+4 4 | 3+4 4 | 3+4 4 |
| | 6 2 | 4+4 4 | 4+4 4 | 3+4 4 | 6+6 4 | 3+4 4 | 3+4 4 | 3+4 4 |

Continued on next column

**TABLE XXXVII– continued**

| TQC | ICSD | NoE | SF | | | | | |
|---|---|---|---|---|---|---|---|---|
| | 5 2 | 4+4 4 | 4+4 4 | 3+4 4 | 6+6 4 | 3+4 4 | 3+4 4 | 3+4 4 |
| | 6 2 | 3+3 4 | 3+3 4 | 3+4 4 | 5+5 4 | 3+4 4 | 3+4 4 | 3+4 4 |
| | 6 2 | 4+4 4 | 4+4 4 | 3+4 4 | 6+6 4 | 3+4 4 | 3+4 4 | 3+4 4 |
| | 5 2 | 3+3 4 | 3+3 4 | 3+4 4 | 5+5 4 | 3+4 4 | 3+4 4 | 3+4 4 |
| | 6 2 | 3+3 4 | 3+3 4 | 3+4 4 | 6+6 4 | 3+4 4 | 3+4 4 | 3+4 4 |
| | 5 2 | 4+4 4 | 4+4 4 | 3+4 4 | 5+5 4 | 3+4 4 | 3+4 4 | 3+4 4 |
| | 6 2 | 4+4 4 | 3+3 4 | 3+4 4 | 6+6 4 | 3+4 4 | 3+4 4 | 3+4 4 |
| | 5 2 | 3+3 4 | 4+4 4 | 3+4 4 | 6+6 4 | 3+4 4 | 3+4 4 | 3+4 4 |
| | 6 2 | 3+3 4 | 4+4 4 | 3+4 4 | 5+5 4 | 3+4 4 | 3+4 4 | 3+4 4 |
| | 6 2 | 4+4 4 | 3+3 4 | 3+4 4 | 5+5 4 | 3+4 4 | 3+4 4 | 3+4 4 |
| | 5 2 | 3+3 4 | 3+3 4 | 3+4 4 | 6+6 4 | 3+4 4 | 3+4 4 | 3+4 4 |
| | 5 2 | 4+4 4 | 4+4 4 | 3+4 4 | 5+5 4 | 3+4 4 | 3+4 4 | 3+4 4 |
| | 5 2 | 3+3 4 | 3+3 4 | 3+4 4 | 6+6 4 | 3+4 4 | 3+4 4 | 3+4 4 |
| | 6 2 | 4+4 4 | 4+4 4 | 3+4 4 | 6+6 4 | 3+4 4 | 3+4 4 | 3+4 4 |
| | 6 2 | 4+4 4 | 4+4 4 | 3+4 4 | 5+5 4 | 3+4 4 | 3+4 4 | 3+4 4 |
| | 5 2 | 3+3 4 | 3+3 4 | 3+4 4 | 5+5 4 | 3+4 4 | 3+4 4 | 3+4 4 |
| | 5 2 | 3+3 4 | 3+3 4 | 3+4 4 | 6+6 4 | 3+4 4 | 3+4 4 | 3+4 4 |
| | 6 2 | 4+4 4 | 4+4 4 | 3+4 4 | 5+5 4 | 3+4 4 | 3+4 4 | 3+4 4 |
| | 5 2 | 3+3 4 | 3+3 4 | 3+4 4 | 6+6 4 | 3+4 4 | 3+4 4 | 3+4 4 |
| | 5 2 | 4+4 4 | 4+4 4 | 3+4 4 | 5+5 4 | 3+4 4 | 3+4 4 | 3+4 4 |
| | 6 2 | 3+3 4 | 3+3 4 | 3+4 4 | 6+6 4 | 3+4 4 | 3+4 4 | 3+4 4 |
| | 5 2 | 4+4 4 | 4+4 4 | 3+4 4 | 5+5 4 | 3+4 4 | 3+4 4 | 3+4 4 |
| | 6 2 | 4+4 4 | 4+4 4 | 3+4 4 | 6+6 4 | 3+4 4 | 3+4 4 | 3+4 4 |
| | 6 2 | 3+3 4 | 3+3 4 | 3+4 4 | 5+5 4 | 3+4 4 | 3+4 4 | 3+4 4 |
| | 5 2 | | | | | | | |
| | 6 2 | | | | | | | |
| | 5 2 | | | | | | | |
| | 5 2 | | | | | | | |
| | 6 2 | | | | | | | |
| | 5 2 | | | | | | | |
| | 6 2 | | | | | | | |
| | 5 2 | | | | | | | |
| | 5 2 | | | | | | | |
| | 6 2 | | | | | | | |
| | 6 2 | | | | | | | |
| | 6 2 | | | | | | | |
| | 5 2 | | | | | | | |
| | 6 2 | | | | | | | |
| | 5 2 | | | | | | | |
| | 5 2 | | | | | | | |
| | 6 2 | | | | | | | |
| | 6 2 | | | | | | | |
| | 5 2 | | | | | | | |
| | 5 2 | | | | | | | |
| | 5 2 | | | | | | | |
| | 6 2 | | | | | | | |
| | 6 2 | | | | | | | |
| | 6 2 | | | | | | | |
| | 6 2 | | | | | | | |
| | 5 2 | | | | | | | |
| | 6 2 | | | | | | | |
| | 5 2 | | | | | | | |
| | 5 2 | | | | | | | |
| | 6 2 | | | | | | | |
| | 5 2 | | | | | | | |
| | 6 2 | | | | | | | |
| | 6 2 | | | | | | | |
| | 5 2 | | | | | | | |
| | 5 2 | | | | | | | |
| | 5 2 | | | | | | | |
| | 6 2 | | | | | | | |

Continued on next column

**TABLE XXXVII– continued**

| TQC | ICSD | NoE | SF | | | | | |
|---|---|---|---|---|---|---|---|---|
| | 5 2 | | | | | | | |
| | 6 2 | | | | | | | |
| | 5 2 | | | | | | | |
| 62.2.254 | 420615 | 160 | Ag1As2Tb1 | | | | | |
| | -GM | -R | -S | -T | -U | -X | -Y | -Z |
| | 5 2 | 3+3 4 | 3+3 4 | 3+4 4 | 5+5 4 | 3+4 4 | 3+4 4 | 3+4 4 |
| | 5 2 | 4+4 4 | 4+4 4 | 3+4 4 | 6+6 4 | 3+4 4 | 3+4 4 | 3+4 4 |
| | 6 2 | 3+3 4 | 3+3 4 | 3+4 4 | 5+5 4 | 3+4 4 | 3+4 4 | 3+4 4 |
| | 6 2 | 4+4 4 | 4+4 4 | 3+4 4 | 6+6 4 | 3+4 4 | 3+4 4 | 3+4 4 |
| | 5 2 | 3+3 4 | 3+3 4 | 3+4 4 | 6+6 4 | 3+4 4 | 3+4 4 | 3+4 4 |
| | 5 2 | 4+4 4 | 4+4 4 | 3+4 4 | 5+5 4 | 3+4 4 | 3+4 4 | 3+4 4 |
| | 6 2 | 3+3 4 | 3+3 4 | 3+4 4 | 5+5 4 | 3+4 4 | 3+4 4 | 3+4 4 |
| | 6 2 | 4+4 4 | 4+4 4 | 3+4 4 | 6+6 4 | 3+4 4 | 3+4 4 | 3+4 4 |
| | 5 2 | 3+3 4 | 3+3 4 | 3+4 4 | 6+6 4 | 3+4 4 | 3+4 4 | 3+4 4 |
| | 5 2 | 4+4 4 | 4+4 4 | 3+4 4 | 5+5 4 | 3+4 4 | 3+4 4 | 3+4 4 |
| | 6 2 | 3+3 4 | 3+3 4 | 3+4 4 | 6+6 4 | 3+4 4 | 3+4 4 | 3+4 4 |
| | 6 2 | 4+4 4 | 4+4 4 | 3+4 4 | 5+5 4 | 3+4 4 | 3+4 4 | 3+4 4 |
| | 5 2 | 3+3 4 | 3+3 4 | 3+4 4 | 5+5 4 | 3+4 4 | 3+4 4 | 3+4 4 |
| | 5 2 | 4+4 4 | 4+4 4 | 3+4 4 | 6+6 4 | 3+4 4 | 3+4 4 | 3+4 4 |
| | 6 2 | 4+4 4 | 3+3 4 | 3+4 4 | 5+5 4 | 3+4 4 | 3+4 4 | 3+4 4 |
| | 6 2 | 3+3 4 | 4+4 4 | 3+4 4 | 6+6 4 | 3+4 4 | 3+4 4 | 3+4 4 |
| | 5 2 | 3+3 4 | 3+3 4 | 3+4 4 | 5+5 4 | 3+4 4 | 3+4 4 | 3+4 4 |
| | 5 2 | 4+4 4 | 4+4 4 | 3+4 4 | 6+6 4 | 3+4 4 | 3+4 4 | 3+4 4 |
| | 6 2 | 4+4 4 | 3+3 4 | 3+4 4 | 5+5 4 | 3+4 4 | 3+4 4 | 3+4 4 |
| | 5 2 | 3+3 4 | 4+4 4 | 3+4 4 | 6+6 4 | 3+4 4 | 3+4 4 | 3+4 4 |
| | 6 2 | 3+3 4 | 4+4 4 | 3+4 4 | 5+5 4 | 3+4 4 | 3+4 4 | 3+4 4 |
| | 5 2 | 4+4 4 | 3+3 4 | 3+4 4 | 6+6 4 | 3+4 4 | 3+4 4 | 3+4 4 |
| | 6 2 | 3+3 4 | 3+3 4 | 3+4 4 | 5+5 4 | 3+4 4 | 3+4 4 | 3+4 4 |
| | 6 2 | 4+4 4 | 4+4 4 | 3+4 4 | 6+6 4 | 3+4 4 | 3+4 4 | 3+4 4 |
| | 6 2 | 3+3 4 | 3+3 4 | 3+4 4 | 5+5 4 | 3+4 4 | 3+4 4 | 3+4 4 |
| | 6 2 | 4+4 4 | 4+4 4 | 3+4 4 | 6+6 4 | 3+4 4 | 3+4 4 | 3+4 4 |
| | 5 2 | 4+4 4 | 3+3 4 | 3+4 4 | 6+6 4 | 3+4 4 | 3+4 4 | 3+4 4 |
| | 5 2 | 3+3 4 | 4+4 4 | 3+4 4 | 5+5 4 | 3+4 4 | 3+4 4 | 3+4 4 |
| | 5 2 | 3+3 4 | 3+3 4 | 3+4 4 | 5+5 4 | 3+4 4 | 3+4 4 | 3+4 4 |
| | 6 2 | 4+4 4 | 4+4 4 | 3+4 4 | 6+6 4 | 3+4 4 | 3+4 4 | 3+4 4 |
| | 5 2 | 3+3 4 | 4+4 4 | 3+4 4 | 6+6 4 | 3+4 4 | 3+4 4 | 3+4 4 |
| | 6 2 | 4+4 4 | 3+3 4 | 3+4 4 | 5+5 4 | 3+4 4 | 3+4 4 | 3+4 4 |
| | 5 2 | 3+3 4 | 3+3 4 | 3+4 4 | 5+5 4 | 3+4 4 | 3+4 4 | 3+4 4 |
| | 6 2 | 4+4 4 | 4+4 4 | 3+4 4 | 6+6 4 | 3+4 4 | 3+4 4 | 3+4 4 |
| | 6 2 | 3+3 4 | 4+4 4 | 3+4 4 | 6+6 4 | 3+4 4 | 3+4 4 | 3+4 4 |
| | 5 2 | 4+4 4 | 3+3 4 | 3+4 4 | 5+5 4 | 3+4 4 | 3+4 4 | 3+4 4 |
| | 5 2 | 3+3 4 | 3+3 4 | 3+4 4 | 5+5 4 | 3+4 4 | 3+4 4 | 3+4 4 |
| | 6 2 | 4+4 4 | 4+4 4 | 3+4 4 | 6+6 4 | 3+4 4 | 3+4 4 | 3+4 4 |
| | 6 2 | 3+3 4 | 3+3 4 | 3+4 4 | 6+6 4 | 3+4 4 | 3+4 4 | 3+4 4 |
| | 5 2 | 4+4 4 | 4+4 4 | 3+4 4 | 5+5 4 | 3+4 4 | 3+4 4 | 3+4 4 |
| | 6 2 | | | | | | | |
| | 6 2 | | | | | | | |
| | 5 2 | | | | | | | |
| | 5 2 | | | | | | | |
| | 5 2 | | | | | | | |
| | 5 2 | | | | | | | |
| | 6 2 | | | | | | | |
| | 5 2 | | | | | | | |
| | 5 2 | | | | | | | |
| | 6 2 | | | | | | | |
| | 5 2 | | | | | | | |
| | 6 2 | | | | | | | |
| | 5 2 | | | | | | | |
| | 6 2 | | | | | | | |
| | 6 2 | | | | | | | |
| | 5 2 | | | | | | | |

Continued on next column

**TABLE XXXVII– continued**

| TQC | ICSD | NoE | SF | | | | | |
|---|---|---|---|---|---|---|---|---|
| | 5 2 | | | | | | | |
| | 6 2 | | | | | | | |
| | 5 2 | | | | | | | |
| | 6 2 | | | | | | | |
| | 5 2 | | | | | | | |
| | 5 2 | | | | | | | |
| | 6 2 | | | | | | | |
| | 6 2 | | | | | | | |
| | 6 2 | | | | | | | |
| | 6 2 | | | | | | | |
| | 5 2 | | | | | | | |
| | 5 2 | | | | | | | |
| | 5 2 | | | | | | | |
| | 5 2 | | | | | | | |
| | 6 2 | | | | | | | |
| | 5 2 | | | | | | | |
| | 6 2 | | | | | | | |
| | 5 2 | | | | | | | |
| | 6 2 | | | | | | | |
| | 5 2 | | | | | | | |
| | 5 2 | | | | | | | |
| | 6 2 | | | | | | | |
| | 6 2 | | | | | | | |
| | 5 2 | | | | | | | |
| 62.2.255 | 98736 | 136 | Ag1As2Pr1 | | | | | |
| | -GM | -R | -S | -T | -U | -X | -Y | -Z |
| | 5 2 | 3+3 4 | 3+3 4 | 3+4 4 | 5+5 4 | 3+4 4 | 3+4 4 | 3+4 4 |
| | 5 2 | 4+4 4 | 4+4 4 | 3+4 4 | 6+6 4 | 3+4 4 | 3+4 4 | 3+4 4 |
| | 6 2 | 3+3 4 | 3+3 4 | 3+4 4 | 6+6 4 | 3+4 4 | 3+4 4 | 3+4 4 |
| | 6 2 | 4+4 4 | 4+4 4 | 3+4 4 | 5+5 4 | 3+4 4 | 3+4 4 | 3+4 4 |
| | 5 2 | 3+3 4 | 3+3 4 | 3+4 4 | 6+6 4 | 3+4 4 | 3+4 4 | 3+4 4 |
| | 5 2 | 4+4 4 | 4+4 4 | 3+4 4 | 5+5 4 | 3+4 4 | 3+4 4 | 3+4 4 |
| | 6 2 | 3+3 4 | 3+3 4 | 3+4 4 | 6+6 4 | 3+4 4 | 3+4 4 | 3+4 4 |
| | 6 2 | 4+4 4 | 4+4 4 | 3+4 4 | 5+5 4 | 3+4 4 | 3+4 4 | 3+4 4 |
| | 5 2 | 3+3 4 | 3+3 4 | 3+4 4 | 6+6 4 | 3+4 4 | 3+4 4 | 3+4 4 |
| | 5 2 | 4+4 4 | 4+4 4 | 3+4 4 | 5+5 4 | 3+4 4 | 3+4 4 | 3+4 4 |
| | 6 2 | 3+3 4 | 3+3 4 | 3+4 4 | 6+6 4 | 3+4 4 | 3+4 4 | 3+4 4 |
| | 6 2 | 4+4 4 | 4+4 4 | 3+4 4 | 5+5 4 | 3+4 4 | 3+4 4 | 3+4 4 |
| | 5 2 | 3+3 4 | 3+3 4 | 3+4 4 | 5+5 4 | 3+4 4 | 3+4 4 | 3+4 4 |
| | 5 2 | 4+4 4 | 4+4 4 | 3+4 4 | 6+6 4 | 3+4 4 | 3+4 4 | 3+4 4 |
| | 6 2 | 3+3 4 | 3+3 4 | 3+4 4 | 5+5 4 | 3+4 4 | 3+4 4 | 3+4 4 |
| | 6 2 | 4+4 4 | 4+4 4 | 3+4 4 | 6+6 4 | 3+4 4 | 3+4 4 | 3+4 4 |
| | 5 2 | 4+4 4 | 3+3 4 | 3+4 4 | 5+5 4 | 3+4 4 | 3+4 4 | 3+4 4 |
| | 5 2 | 3+3 4 | 4+4 4 | 3+4 4 | 6+6 4 | 3+4 4 | 3+4 4 | 3+4 4 |
| | 6 2 | 3+3 4 | 3+3 4 | 3+4 4 | 5+5 4 | 3+4 4 | 3+4 4 | 3+4 4 |
| | 5 2 | 4+4 4 | 4+4 4 | 3+4 4 | 6+6 4 | 3+4 4 | 3+4 4 | 3+4 4 |
| | 6 2 | 3+3 4 | 3+3 4 | 3+4 4 | 5+5 4 | 3+4 4 | 3+4 4 | 3+4 4 |
| | 5 2 | 4+4 4 | 4+4 4 | 3+4 4 | 6+6 4 | 3+4 4 | 3+4 4 | 3+4 4 |
| | 6 2 | 3+3 4 | 3+3 4 | 3+4 4 | 5+5 4 | 3+4 4 | 3+4 4 | 3+4 4 |
| | 6 2 | 4+4 4 | 4+4 4 | 3+4 4 | 6+6 4 | 3+4 4 | 3+4 4 | 3+4 4 |
| | 6 2 | 4+4 4 | 3+3 4 | 3+4 4 | 5+5 4 | 3+4 4 | 3+4 4 | 3+4 4 |
| | 5 2 | 3+3 4 | 4+4 4 | 3+4 4 | 6+6 4 | 3+4 4 | 3+4 4 | 3+4 4 |
| | 6 2 | 3+3 4 | 3+3 4 | 3+4 4 | 6+6 4 | 3+4 4 | 3+4 4 | 3+4 4 |
| | 5 2 | 4+4 4 | 4+4 4 | 3+4 4 | 5+5 4 | 3+4 4 | 3+4 4 | 3+4 4 |
| | 5 2 | 4+4 4 | 4+4 4 | 3+4 4 | 5+5 4 | 3+4 4 | 3+4 4 | 3+4 4 |
| | 6 2 | 3+3 4 | 3+3 4 | 3+4 4 | 6+6 4 | 3+4 4 | 3+4 4 | 3+4 4 |
| | 5 2 | 4+4 4 | 3+3 4 | 3+4 4 | 6+6 4 | 3+4 4 | 3+4 4 | 3+4 4 |
| | 6 2 | 3+3 4 | 4+4 4 | 3+4 4 | 5+5 4 | 3+4 4 | 3+4 4 | 3+4 4 |
| | 5 2 | 3+3 4 | 3+3 4 | 3+4 4 | 6+6 4 | 3+4 4 | 3+4 4 | 3+4 4 |
| | 6 2 | 4+4 4 | 4+4 4 | 3+4 4 | 5+5 4 | 3+4 4 | 3+4 4 | 3+4 4 |
| | 6 2 | | | | | | | |

Continued on next column

**TABLE XXXVII– continued**

| TQC | ICSD | NoE | SF |
|---|---|---|---|
| | 5 2 | | |
| | 5 2 | | |
| | 6 2 | | |
| | 5 2 | | |
| | 6 2 | | |
| | 6 2 | | |
| | 5 2 | | |
| | 5 2 | | |
| | 6 2 | | |
| | 6 2 | | |
| | 5 2 | | |
| | 5 2 | | |
| | 5 2 | | |
| | 6 2 | | |
| | 5 2 | | |
| | 5 2 | | |
| | 6 2 | | |
| | 5 2 | | |
| | 6 2 | | |
| | 5 2 | | |
| | 6 2 | | |
| | 6 2 | | |
| | 5 2 | | |
| | 5 2 | | |
| | 5 2 | | |
| | 6 2 | | |
| | 5 2 | | |
| | 6 2 | | |
| | 5 2 | | |
| | 6 2 | | |
| | 5 2 | | |
| | 6 2 | | |
| | 6 2 | | |
| 62.2.256 | 420611 | 136 | Ag1As2Pr1 |
| | -GM -R -S -T -U -X -Y -Z | | |
| | 5 2 3+3 4 3+3 4 3+4 4 5+5 4 3+4 4 3+4 4 3+4 4 | | |
| | 5 2 4+4 4 4+4 4 3+4 4 6+6 4 3+4 4 3+4 4 3+4 4 | | |
| | 6 2 3+3 4 3+3 4 3+4 4 5+5 4 3+4 4 3+4 4 3+4 4 | | |
| | 6 2 4+4 4 4+4 4 3+4 4 6+6 4 3+4 4 3+4 4 3+4 4 | | |
| | 5 2 3+3 4 3+3 4 3+4 4 5+5 4 3+4 4 3+4 4 3+4 4 | | |
| | 5 2 4+4 4 4+4 4 3+4 4 6+6 4 3+4 4 3+4 4 3+4 4 | | |
| | 6 2 3+3 4 3+3 4 3+4 4 5+5 4 3+4 4 3+4 4 3+4 4 | | |
| | 6 2 4+4 4 4+4 4 3+4 4 6+6 4 3+4 4 3+4 4 3+4 4 | | |
| | 5 2 3+3 4 3+3 4 3+4 4 5+5 4 3+4 4 3+4 4 3+4 4 | | |
| | 5 2 4+4 4 4+4 4 3+4 4 6+6 4 3+4 4 3+4 4 3+4 4 | | |
| | 6 2 3+3 4 3+3 4 3+4 4 5+5 4 3+4 4 3+4 4 3+4 4 | | |
| | 6 2 4+4 4 4+4 4 3+4 4 6+6 4 3+4 4 3+4 4 3+4 4 | | |
| | 5 2 3+3 4 3+3 4 3+4 4 6+6 4 3+4 4 3+4 4 3+4 4 | | |
| | 5 2 4+4 4 4+4 4 3+4 4 5+5 4 3+4 4 3+4 4 3+4 4 | | |
| | 6 2 3+3 4 3+3 4 3+4 4 6+6 4 3+4 4 3+4 4 3+4 4 | | |
| | 6 2 4+4 4 4+4 4 3+4 4 5+5 4 3+4 4 3+4 4 3+4 4 | | |
| | 5 2 4+4 4 3+3 4 3+4 4 5+5 4 3+4 4 3+4 4 3+4 4 | | |
| | 5 2 3+3 4 4+4 4 3+4 4 6+6 4 3+4 4 3+4 4 3+4 4 | | |
| | 6 2 3+3 4 3+3 4 3+4 4 6+6 4 3+4 4 3+4 4 3+4 4 | | |
| | 5 2 4+4 4 4+4 4 3+4 4 5+5 4 3+4 4 3+4 4 3+4 4 | | |
| | 6 2 3+3 4 3+3 4 3+4 4 6+6 4 3+4 4 3+4 4 3+4 4 | | |
| | 5 2 4+4 4 4+4 4 3+4 4 5+5 4 3+4 4 3+4 4 3+4 4 | | |
| | 6 2 3+3 4 3+3 4 3+4 4 6+6 4 3+4 4 3+4 4 3+4 4 | | |
| | 6 2 4+4 4 4+4 4 3+4 4 5+5 4 3+4 4 3+4 4 3+4 4 | | |
| | 6 2 4+4 4 3+3 4 3+4 4 6+6 4 3+4 4 3+4 4 3+4 4 | | |
| | 5 2 3+3 4 4+4 4 3+4 4 5+5 4 3+4 4 3+4 4 3+4 4 | | |

Continued on next column

**TABLE XXXVII– continued**

| TQC | ICSD | NoE | SF |
|---|---|---|---|
| | 6 2 3+3 4 3+3 4 3+4 4 5+5 4 3+4 4 3+4 4 3+4 4 | | |
| | 5 2 4+4 4 4+4 4 3+4 4 6+6 4 3+4 4 3+4 4 3+4 4 | | |
| | 5 2 4+4 4 4+4 4 3+4 4 6+6 4 3+4 4 3+4 4 3+4 4 | | |
| | 6 2 3+3 4 3+3 4 3+4 4 5+5 4 3+4 4 3+4 4 3+4 4 | | |
| | 5 2 4+4 4 3+3 4 3+4 4 5+5 4 3+4 4 3+4 4 3+4 4 | | |
| | 6 2 3+3 4 4+4 4 3+4 4 6+6 4 3+4 4 3+4 4 3+4 4 | | |
| | 5 2 3+3 4 3+3 4 3+4 4 5+5 4 3+4 4 3+4 4 3+4 4 | | |
| | 6 2 4+4 4 4+4 4 3+4 4 6+6 4 3+4 4 3+4 4 3+4 4 | | |
| | 6 2 | | |
| | 5 2 | | |
| | 5 2 | | |
| | 6 2 | | |
| | 5 2 | | |
| | 6 2 | | |
| | 6 2 | | |
| | 5 2 | | |
| | 5 2 | | |
| | 6 2 | | |
| | 6 2 | | |
| | 5 2 | | |
| | 5 2 | | |
| | 5 2 | | |
| | 6 2 | | |
| | 5 2 | | |
| | 5 2 | | |
| | 6 2 | | |
| | 5 2 | | |
| | 6 2 | | |
| | 5 2 | | |
| | 6 2 | | |
| | 6 2 | | |
| | 5 2 | | |
| | 5 2 | | |
| | 5 2 | | |
| | 6 2 | | |
| | 5 2 | | |
| | 6 2 | | |
| | 5 2 | | |
| | 6 2 | | |
| | 5 2 | | |
| | 6 2 | | |
| | 6 2 | | |
| 62.2.257 | 412816 | 136 | Ag1As2Pr1 |
| | -GM -R -S -T -U -X -Y -Z | | |
| | 5 2 3+3 4 3+3 4 3+4 4 5+5 4 3+4 4 3+4 4 3+4 4 | | |
| | 5 2 4+4 4 4+4 4 3+4 4 6+6 4 3+4 4 3+4 4 3+4 4 | | |
| | 6 2 3+3 4 3+3 4 3+4 4 6+6 4 3+4 4 3+4 4 3+4 4 | | |
| | 6 2 4+4 4 4+4 4 3+4 4 5+5 4 3+4 4 3+4 4 3+4 4 | | |
| | 5 2 3+3 4 3+3 4 3+4 4 6+6 4 3+4 4 3+4 4 3+4 4 | | |
| | 5 2 4+4 4 4+4 4 3+4 4 5+5 4 3+4 4 3+4 4 3+4 4 | | |
| | 6 2 3+3 4 3+3 4 3+4 4 6+6 4 3+4 4 3+4 4 3+4 4 | | |
| | 6 2 4+4 4 4+4 4 3+4 4 5+5 4 3+4 4 3+4 4 3+4 4 | | |
| | 5 2 3+3 4 3+3 4 3+4 4 6+6 4 3+4 4 3+4 4 3+4 4 | | |
| | 5 2 4+4 4 4+4 4 3+4 4 5+5 4 3+4 4 3+4 4 3+4 4 | | |
| | 6 2 3+3 4 3+3 4 3+4 4 6+6 4 3+4 4 3+4 4 3+4 4 | | |
| | 6 2 4+4 4 4+4 4 3+4 4 5+5 4 3+4 4 3+4 4 3+4 4 | | |
| | 5 2 3+3 4 3+3 4 3+4 4 5+5 4 3+4 4 3+4 4 3+4 4 | | |
| | 5 2 4+4 4 4+4 4 3+4 4 6+6 4 3+4 4 3+4 4 3+4 4 | | |
| | 6 2 3+3 4 3+3 4 3+4 4 5+5 4 3+4 4 3+4 4 3+4 4 | | |
| | 6 2 4+4 4 4+4 4 3+4 4 6+6 4 3+4 4 3+4 4 3+4 4 | | |
| | 5 2 4+4 4 3+3 4 3+4 4 5+5 4 3+4 4 3+4 4 3+4 4 | | |

Continued on next column

**TABLE XXXVII– continued**

| TQC | ICSD | NoE | SF | | | | | |
|---|---|---|---|---|---|---|---|---|
| | 5 2 | 3+3 4 | 4+4 4 | 3+4 4 | 6+6 4 | 3+4 4 | 3+4 4 | 3+4 4 |
| | 6 2 | 3+3 4 | 3+3 4 | 3+4 4 | 5+5 4 | 3+4 4 | 3+4 4 | 3+4 4 |
| | 5 2 | 4+4 4 | 4+4 4 | 3+4 4 | 6+6 4 | 3+4 4 | 3+4 4 | 3+4 4 |
| | 6 2 | 3+3 4 | 3+3 4 | 3+4 4 | 5+5 4 | 3+4 4 | 3+4 4 | 3+4 4 |
| | 5 2 | 4+4 4 | 4+4 4 | 3+4 4 | 6+6 4 | 3+4 4 | 3+4 4 | 3+4 4 |
| | 6 2 | 3+3 4 | 3+3 4 | 3+4 4 | 5+5 4 | 3+4 4 | 3+4 4 | 3+4 4 |
| | 6 2 | 4+4 4 | 4+4 4 | 3+4 4 | 6+6 4 | 3+4 4 | 3+4 4 | 3+4 4 |
| | 6 2 | 4+4 4 | 3+3 4 | 3+4 4 | 5+5 4 | 3+4 4 | 3+4 4 | 3+4 4 |
| | 5 2 | 3+3 4 | 4+4 4 | 3+4 4 | 6+6 4 | 3+4 4 | 3+4 4 | 3+4 4 |
| | 6 2 | 3+3 4 | 3+3 4 | 3+4 4 | 6+6 4 | 3+4 4 | 3+4 4 | 3+4 4 |
| | 5 2 | 4+4 4 | 4+4 4 | 3+4 4 | 5+5 4 | 3+4 4 | 3+4 4 | 3+4 4 |
| | 5 2 | 4+4 4 | 4+4 4 | 3+4 4 | 5+5 4 | 3+4 4 | 3+4 4 | 3+4 4 |
| | 6 2 | 3+3 4 | 3+3 4 | 3+4 4 | 6+6 4 | 3+4 4 | 3+4 4 | 3+4 4 |
| | 5 2 | 4+4 4 | 3+3 4 | 3+4 4 | 6+6 4 | 3+4 4 | 3+4 4 | 3+4 4 |
| | 6 2 | 3+3 4 | 4+4 4 | 3+4 4 | 5+5 4 | 3+4 4 | 3+4 4 | 3+4 4 |
| | 5 2 | 3+3 4 | 3+3 4 | 3+4 4 | 6+6 4 | 3+4 4 | 3+4 4 | 3+4 4 |
| | 6 2 | 4+4 4 | 4+4 4 | 3+4 4 | 5+5 4 | 3+4 4 | 3+4 4 | 3+4 4 |
| | 6 2 | | | | | | | |
| | 5 2 | | | | | | | |
| | 5 2 | | | | | | | |
| | 6 2 | | | | | | | |
| | 5 2 | | | | | | | |
| | 6 2 | | | | | | | |
| | 6 2 | | | | | | | |
| | 5 2 | | | | | | | |
| | 5 2 | | | | | | | |
| | 6 2 | | | | | | | |
| | 6 2 | | | | | | | |
| | 5 2 | | | | | | | |
| | 5 2 | | | | | | | |
| | 5 2 | | | | | | | |
| | 6 2 | | | | | | | |
| | 5 2 | | | | | | | |
| | 5 2 | | | | | | | |
| | 6 2 | | | | | | | |
| | 5 2 | | | | | | | |
| | 6 2 | | | | | | | |
| | 5 2 | | | | | | | |
| | 6 2 | | | | | | | |
| | 6 2 | | | | | | | |
| | 5 2 | | | | | | | |
| | 5 2 | | | | | | | |
| | 5 2 | | | | | | | |
| | 6 2 | | | | | | | |
| | 5 2 | | | | | | | |
| | 6 2 | | | | | | | |
| | 5 2 | | | | | | | |
| | 6 2 | | | | | | | |
| | 5 2 | | | | | | | |
| | 6 2 | | | | | | | |
| | 6 2 | | | | | | | |

The End

TABLE XXXVIII: The computed irreducible representations for the valence bands for the compounds of SEBR and NLC in space group 63. See the list of tables.

| TQC | ICSD | NoE | SF | | | |
|---|---|---|---|---|---|---|
| 63.1.1 | 68416 | 64 | Te3U1 | | | |
| | -GM | -T | -Y | -Z | -R | -S |

Continued on next column

**TABLE XXXVIII– continued**

| TQC | ICSD | NoE | SF | | | |
|---|---|---|---|---|---|---|
| | 5 2 | 3+4 4 | 5 2 | 3+4 4 | 2+2 4 | 3+4 2 |
| | 6 2 | 3+4 4 | 6 2 | 3+4 4 | 2+2 4 | 5+6 2 |
| | 5 2 | 3+4 4 | 6 2 | 3+4 4 | 2+2 4 | 3+4 2 |
| | 6 2 | 3+4 4 | 5 2 | 3+4 4 | 2+2 4 | 5+6 2 |
| | 5 2 | 3+4 4 | 6 2 | 3+4 4 | 2+2 4 | 3+4 2 |
| | 6 2 | 3+4 4 | 5 2 | 3+4 4 | 2+2 4 | 5+6 2 |
| | 5 2 | 3+4 4 | 6 2 | 3+4 4 | 2+2 4 | 3+4 2 |
| | 6 2 | 3+4 4 | 5 2 | 3+4 4 | 2+2 4 | 5+6 2 |
| | 5 2 | 3+4 4 | 5 2 | 3+4 4 | 2+2 4 | 3+4 2 |
| | 6 2 | 3+4 4 | 6 2 | 3+4 4 | 2+2 4 | 5+6 2 |
| | 5 2 | 3+4 4 | 6 2 | 3+4 4 | 2+2 4 | 5+6 2 |
| | 6 2 | 3+4 4 | 5 2 | 3+4 4 | 2+2 4 | 3+4 2 |
| | 5 2 | 3+4 4 | 5 2 | 3+4 4 | 2+2 4 | 3+4 2 |
| | 6 2 | 3+4 4 | 6 2 | 3+4 4 | 2+2 4 | 5+6 2 |
| | 5 2 | 3+4 4 | 5 2 | 3+4 4 | 2+2 4 | 3+4 2 |
| | 5 2 | 3+4 4 | 6 2 | 3+4 4 | 2+2 4 | 5+6 2 |
| | 6 2 | | 5 2 | | | 5+6 2 |
| | 5 2 | | 6 2 | | | 3+4 2 |
| | 6 2 | | 5 2 | | | 5+6 2 |
| | 6 2 | | 6 2 | | | 3+4 2 |
| | 5 2 | | 6 2 | | | 3+4 2 |
| | 5 2 | | 6 2 | | | 5+6 2 |
| | 5 2 | | 6 2 | | | 5+6 2 |
| | 6 2 | | 5 2 | | | 3+4 2 |
| | 6 2 | | 5 2 | | | 3+4 2 |
| | 5 2 | | 6 2 | | | 5+6 2 |
| | 6 2 | | 5 2 | | | 5+6 2 |
| | 6 2 | | 5 2 | | | 3+4 2 |
| | 6 2 | | 5 2 | | | 3+4 2 |
| | 5 2 | | 6 2 | | | 5+6 2 |
| | 5 2 | | 6 2 | | | 3+4 2 |
| | 5 2 | | 6 2 | | | 5+6 2 |
| 63.1.2 | 106074 | 28 | Sn1Sr1 | | | |
| | -GM | -T | -Y | -Z | -R | -S |
| | 5 2 | 3+4 4 | 5 2 | 3+4 4 | 2+2 4 | 3+4 2 |
| | 6 2 | 3+4 4 | 6 2 | 3+4 4 | 2+2 4 | 5+6 2 |
| | 5 2 | 3+4 4 | 5 2 | 3+4 4 | 2+2 4 | 3+4 2 |
| | 6 2 | 3+4 4 | 6 2 | 3+4 4 | 2+2 4 | 5+6 2 |
| | 5 2 | 3+4 4 | 5 2 | 3+4 4 | 2+2 4 | 3+4 2 |
| | 5 2 | 3+4 4 | 6 2 | 3+4 4 | 2+2 4 | 5+6 2 |
| | 6 2 | 3+4 4 | 5 2 | 3+4 4 | 2+2 4 | 3+4 2 |
| | 6 2 | | 6 2 | | | 5+6 2 |
| | 5 2 | | 5 2 | | | 5+6 2 |
| | 6 2 | | 6 2 | | | 3+4 2 |
| | 5 2 | | 6 2 | | | 3+4 2 |
| | 6 2 | | 5 2 | | | 5+6 2 |
| | 5 2 | | 5 2 | | | 5+6 2 |
| | 6 2 | | 6 2 | | | 5+6 2 |
| 63.1.3 | 416971 | 68 | Si6Sr1 | | | |
| | -GM | -T | -Y | -Z | -R | -S |
| | 5 2 | 3+4 4 | 5 2 | 3+4 4 | 2+2 4 | 3+4 2 |
| | 6 2 | 3+4 4 | 6 2 | 3+4 4 | 2+2 4 | 5+6 2 |
| | 5 2 | 3+4 4 | 6 2 | 3+4 4 | 2+2 4 | 3+4 2 |
| | 6 2 | 3+4 4 | 5 2 | 3+4 4 | 2+2 4 | 5+6 2 |
| | 5 2 | 3+4 4 | 6 2 | 3+4 4 | 2+2 4 | 3+4 2 |
| | 6 2 | 3+4 4 | 5 2 | 3+4 4 | 2+2 4 | 5+6 2 |
| | 5 2 | 3+4 4 | 6 2 | 3+4 4 | 2+2 4 | 3+4 2 |
| | 6 2 | 3+4 4 | 5 2 | 3+4 4 | 2+2 4 | 5+6 2 |
| | 5 2 | 3+4 4 | 5 2 | 3+4 4 | 2+2 4 | 3+4 2 |
| | 5 2 | 3+4 4 | 6 2 | 3+4 4 | 2+2 4 | 3+4 2 |
| | 6 2 | 3+4 4 | 5 2 | 3+4 4 | 2+2 4 | 5+6 2 |

Continued on next column

**TABLE XXXVIII– continued**

| TQC | ICSD | | NoE | | SF | |
|---|---|---|---|---|---|---|
| | 6 2 | 3+4 4 | 6 2 | 3+4 4 | 2+2 4 | 5+6 2 |
| | 5 2 | 3+4 4 | 6 2 | 3+4 4 | 2+2 4 | 3+4 2 |
| | 5 2 | 3+4 4 | 6 2 | 3+4 4 | 2+2 4 | 5+6 2 |
| | 5 2 | 3+4 4 | 5 2 | 3+4 4 | 2+2 4 | 3+4 2 |
| | 6 2 | 3+4 4 | 5 2 | 3+4 4 | 2+2 4 | 5+6 2 |
| | 6 2 | 3+4 4 | 5 2 | 3+4 4 | 2+2 4 | 3+4 2 |
| | 6 2 | | 6 2 | | | 5+6 2 |
| | 5 2 | | 6 2 | | | 5+6 2 |
| | 6 2 | | 6 2 | | | 3+4 2 |
| | 6 2 | | 5 2 | | | 5+6 2 |
| | 5 2 | | 5 2 | | | 3+4 2 |
| | 5 2 | | 5 2 | | | 5+6 2 |
| | 5 2 | | 6 2 | | | 3+4 2 |
| | 6 2 | | 6 2 | | | 5+6 2 |
| | 6 2 | | 5 2 | | | 3+4 2 |
| | 5 2 | | 6 2 | | | 3+4 2 |
| | 5 2 | | 5 2 | | | 5+6 2 |
| | 5 2 | | 5 2 | | | 5+6 2 |
| | 5 2 | | 6 2 | | | 3+4 2 |
| | 6 2 | | 6 2 | | | 3+4 2 |
| | 5 2 | | 6 2 | | | 3+4 2 |
| | 5 2 | | 5 2 | | | 3+4 2 |
| | 6 2 | | 5 2 | | | 5+6 2 |
| 63.1.4 | 240779 | | 68 | | Si6Sr1 | |
| | -GM | -T | -Y | -Z | -R | -S |
| | 5 2 | 3+4 4 | 5 2 | 3+4 4 | 2+2 4 | 3+4 2 |
| | 6 2 | 3+4 4 | 6 2 | 3+4 4 | 2+2 4 | 5+6 2 |
| | 5 2 | 3+4 4 | 6 2 | 3+4 4 | 2+2 4 | 3+4 2 |
| | 6 2 | 3+4 4 | 5 2 | 3+4 4 | 2+2 4 | 5+6 2 |
| | 5 2 | 3+4 4 | 6 2 | 3+4 4 | 2+2 4 | 3+4 2 |
| | 6 2 | 3+4 4 | 5 2 | 3+4 4 | 2+2 4 | 5+6 2 |
| | 5 2 | 3+4 4 | 6 2 | 3+4 4 | 2+2 4 | 3+4 2 |
| | 6 2 | 3+4 4 | 5 2 | 3+4 4 | 2+2 4 | 5+6 2 |
| | 5 2 | 3+4 4 | 5 2 | 3+4 4 | 2+2 4 | 3+4 2 |
| | 5 2 | 3+4 4 | 6 2 | 3+4 4 | 2+2 4 | 3+4 2 |
| | 6 2 | 3+4 4 | 5 2 | 3+4 4 | 2+2 4 | 5+6 2 |
| | 6 2 | 3+4 4 | 6 2 | 3+4 4 | 2+2 4 | 5+6 2 |
| | 5 2 | 3+4 4 | 6 2 | 3+4 4 | 2+2 4 | 3+4 2 |
| | 5 2 | 3+4 4 | 6 2 | 3+4 4 | 2+2 4 | 5+6 2 |
| | 5 2 | 3+4 4 | 5 2 | 3+4 4 | 2+2 4 | 3+4 2 |
| | 6 2 | 3+4 4 | 5 2 | 3+4 4 | 2+2 4 | 5+6 2 |
| | 6 2 | 3+4 4 | 5 2 | 3+4 4 | 2+2 4 | 3+4 2 |
| | 6 2 | | 6 2 | | | 5+6 2 |
| | 5 2 | | 6 2 | | | 5+6 2 |
| | 6 2 | | 6 2 | | | 3+4 2 |
| | 6 2 | | 5 2 | | | 5+6 2 |
| | 5 2 | | 5 2 | | | 3+4 2 |
| | 5 2 | | 5 2 | | | 5+6 2 |
| | 5 2 | | 6 2 | | | 3+4 2 |
| | 6 2 | | 6 2 | | | 5+6 2 |
| | 6 2 | | 5 2 | | | 3+4 2 |
| | 5 2 | | 6 2 | | | 3+4 2 |
| | 5 2 | | 5 2 | | | 5+6 2 |
| | 5 2 | | 5 2 | | | 5+6 2 |
| | 5 2 | | 6 2 | | | 3+4 2 |
| | 6 2 | | 6 2 | | | 3+4 2 |
| | 5 2 | | 6 2 | | | 3+4 2 |
| | 5 2 | | 5 2 | | | 3+4 2 |
| | 6 2 | | 5 2 | | | 5+6 2 |
| 63.1.5 | 78997 | | 28 | | Si1Sr1 | |
| | -GM | -T | -Y | -Z | -R | -S |

Continued on next column

**TABLE XXXVIII– continued**

| TQC | ICSD | | NoE | | SF | |
|---|---|---|---|---|---|---|
| | 5 2 | 3+4 4 | 5 2 | 3+4 4 | 2+2 4 | 3+4 2 |
| | 6 2 | 3+4 4 | 6 2 | 3+4 4 | 2+2 4 | 5+6 2 |
| | 5 2 | 3+4 4 | 5 2 | 3+4 4 | 2+2 4 | 3+4 2 |
| | 6 2 | 3+4 4 | 6 2 | 3+4 4 | 2+2 4 | 5+6 2 |
| | 5 2 | 3+4 4 | 5 2 | 3+4 4 | 2+2 4 | 3+4 2 |
| | 5 2 | 3+4 4 | 6 2 | 3+4 4 | 2+2 4 | 5+6 2 |
| | 6 2 | 3+4 4 | 6 2 | 3+4 4 | 2+2 4 | 3+4 2 |
| | 6 2 | | 5 2 | | | 5+6 2 |
| | 5 2 | | 5 2 | | | 5+6 2 |
| | 6 2 | | 6 2 | | | 3+4 2 |
| | 5 2 | | 6 2 | | | 3+4 2 |
| | 6 2 | | 5 2 | | | 5+6 2 |
| | 5 2 | | 5 2 | | | 5+6 2 |
| | 6 2 | | 6 2 | | | 5+6 2 |
| 63.1.6 | 87915 | | 84 | | Pd3Te2 | |
| | -GM | -T | -Y | -Z | -R | -S |
| | 5 2 | 3+4 4 | 6 2 | 3+4 4 | 2+2 4 | 3+4 2 |
| | 6 2 | 3+4 4 | 5 2 | 3+4 4 | 2+2 4 | 5+6 2 |
| | 5 2 | 3+4 4 | 5 2 | 3+4 4 | 2+2 4 | 3+4 2 |
| | 5 2 | 3+4 4 | 5 2 | 3+4 4 | 2+2 4 | 3+4 2 |
| | 6 2 | 3+4 4 | 6 2 | 3+4 4 | 2+2 4 | 5+6 2 |
| | 6 2 | 3+4 4 | 6 2 | 3+4 4 | 2+2 4 | 3+4 2 |
| | 5 2 | 3+4 4 | 5 2 | 3+4 4 | 2+2 4 | 5+6 2 |
| | 6 2 | 3+4 4 | 5 2 | 3+4 4 | 2+2 4 | 5+6 2 |
| | 6 2 | 3+4 4 | 6 2 | 3+4 4 | 2+2 4 | 5+6 2 |
| | 5 2 | 3+4 4 | 6 2 | 3+4 4 | 2+2 4 | 3+4 2 |
| | 6 2 | 3+4 4 | 5 2 | 3+4 4 | 2+2 4 | 5+6 2 |
| | 6 2 | 3+4 4 | 6 2 | 3+4 4 | 2+2 4 | 3+4 2 |
| | 5 2 | 3+4 4 | 5 2 | 3+4 4 | 2+2 4 | 5+6 2 |
| | 6 2 | 3+4 4 | 5 2 | 3+4 4 | 2+2 4 | 5+6 2 |
| | 5 2 | 3+4 4 | 6 2 | 3+4 4 | 2+2 4 | 5+6 2 |
| | 6 2 | 3+4 4 | 6 2 | 3+4 4 | 2+2 4 | 3+4 2 |
| | 5 2 | 3+4 4 | 6 2 | 3+4 4 | 2+2 4 | 5+6 2 |
| | 6 2 | 3+4 4 | 6 2 | 3+4 4 | 2+2 4 | 3+4 2 |
| | 5 2 | 3+4 4 | 5 2 | 3+4 4 | 2+2 4 | 5+6 2 |
| | 5 2 | 3+4 4 | 5 2 | 3+4 4 | 2+2 4 | 3+4 2 |
| | 6 2 | 3+4 4 | 6 2 | 3+4 4 | 2+2 4 | 5+6 2 |
| | 5 2 | | 6 2 | | | 3+4 2 |
| | 6 2 | | 6 2 | | | 5+6 2 |
| | 5 2 | | 5 2 | | | 3+4 2 |
| | 5 2 | | 5 2 | | | 5+6 2 |
| | 6 2 | | 5 2 | | | 3+4 2 |
| | 5 2 | | 6 2 | | | 5+6 2 |
| | 5 2 | | 5 2 | | | 3+4 2 |
| | 6 2 | | 6 2 | | | 3+4 2 |
| | 6 2 | | 5 2 | | | 5+6 2 |
| | 5 2 | | 6 2 | | | 3+4 2 |
| | 5 2 | | 6 2 | | | 5+6 2 |
| | 6 2 | | 6 2 | | | 3+4 2 |
| | 5 2 | | 6 2 | | | 5+6 2 |
| | 6 2 | | 5 2 | | | 3+4 2 |
| | 5 2 | | 6 2 | | | 5+6 2 |
| | 6 2 | | 5 2 | | | 3+4 2 |
| | 5 2 | | 5 2 | | | 3+4 2 |
| | 6 2 | | 6 2 | | | 5+6 2 |
| | 6 2 | | 5 2 | | | 5+6 2 |
| | 6 2 | | 6 2 | | | 3+4 2 |
| | 5 2 | | 6 2 | | | 3+4 2 |
| 63.1.7 | 77902 | | 84 | | Pd3Te2 | |
| | -GM | -T | -Y | -Z | -R | -S |
| | 5 2 | 3+4 4 | 6 2 | 3+4 4 | 2+2 4 | 3+4 2 |

Continued on next column

**TABLE XXXVIII– continued**

| TQC | ICSD | | NoE | | SF | |
|---|---|---|---|---|---|---|
| | 6 2 | 3+4 4 | 5 2 | 3+4 4 | 2+2 4 | 5+6 2 |
| | 5 2 | 3+4 4 | 5 2 | 3+4 4 | 2+2 4 | 3+4 2 |
| | 5 2 | 3+4 4 | 5 2 | 3+4 4 | 2+2 4 | 3+4 2 |
| | 6 2 | 3+4 4 | 6 2 | 3+4 4 | 2+2 4 | 5+6 2 |
| | 6 2 | 3+4 4 | 6 2 | 3+4 4 | 2+2 4 | 3+4 2 |
| | 5 2 | 3+4 4 | 5 2 | 3+4 4 | 2+2 4 | 5+6 2 |
| | 6 2 | 3+4 4 | 5 2 | 3+4 4 | 2+2 4 | 5+6 2 |
| | 6 2 | 3+4 4 | 6 2 | 3+4 4 | 2+2 4 | 5+6 2 |
| | 5 2 | 3+4 4 | 6 2 | 3+4 4 | 2+2 4 | 3+4 2 |
| | 6 2 | 3+4 4 | 5 2 | 3+4 4 | 2+2 4 | 5+6 2 |
| | 6 2 | 3+4 4 | 6 2 | 3+4 4 | 2+2 4 | 3+4 2 |
| | 5 2 | 3+4 4 | 5 2 | 3+4 4 | 2+2 4 | 5+6 2 |
| | 6 2 | 3+4 4 | 5 2 | 3+4 4 | 2+2 4 | 5+6 2 |
| | 5 2 | 3+4 4 | 6 2 | 3+4 4 | 2+2 4 | 5+6 2 |
| | 6 2 | 3+4 4 | 6 2 | 3+4 4 | 2+2 4 | 3+4 2 |
| | 5 2 | 3+4 4 | 6 2 | 3+4 4 | 2+2 4 | 5+6 2 |
| | 6 2 | 3+4 4 | 6 2 | 3+4 4 | 2+2 4 | 3+4 2 |
| | 5 2 | 3+4 4 | 5 2 | 3+4 4 | 2+2 4 | 5+6 2 |
| | 5 2 | 3+4 4 | 5 2 | 3+4 4 | 2+2 4 | 3+4 2 |
| | 6 2 | 3+4 4 | 6 2 | 3+4 4 | 2+2 4 | 5+6 2 |
| | 5 2 | | 6 2 | | | 3+4 2 |
| | 6 2 | | 6 2 | | | 5+6 2 |
| | 5 2 | | 5 2 | | | 3+4 2 |
| | 5 2 | | 5 2 | | | 5+6 2 |
| | 6 2 | | 5 2 | | | 3+4 2 |
| | 5 2 | | 6 2 | | | 5+6 2 |
| | 5 2 | | 5 2 | | | 3+4 2 |
| | 6 2 | | 6 2 | | | 3+4 2 |
| | 6 2 | | 6 2 | | | 5+6 2 |
| | 5 2 | | 5 2 | | | 3+4 2 |
| | 5 2 | | 6 2 | | | 5+6 2 |
| | 6 2 | | 6 2 | | | 3+4 2 |
| | 5 2 | | 6 2 | | | 5+6 2 |
| | 6 2 | | 5 2 | | | 3+4 2 |
| | 5 2 | | 6 2 | | | 5+6 2 |
| | 6 2 | | 5 2 | | | 3+4 2 |
| | 5 2 | | 5 2 | | | 3+4 2 |
| | 6 2 | | 6 2 | | | 5+6 2 |
| | 6 2 | | 5 2 | | | 5+6 2 |
| | 6 2 | | 6 2 | | | 3+4 2 |
| | 5 2 | | 6 2 | | | 3+4 2 |
| 63.1.8 | 411798 | | 84 | | Pd1Sn2Yb1 | |
| | -GM | -T | -Y | -Z | -R | -S |
| | 5 2 | 3+4 4 | 5 2 | 3+4 4 | 2+2 4 | 3+4 2 |
| | 6 2 | 3+4 4 | 6 2 | 3+4 4 | 2+2 4 | 5+6 2 |
| | 5 2 | 3+4 4 | 5 2 | 3+4 4 | 2+2 4 | 5+6 2 |
| | 6 2 | 3+4 4 | 6 2 | 3+4 4 | 2+2 4 | 3+4 2 |
| | 5 2 | 3+4 4 | 5 2 | 3+4 4 | 2+2 4 | 5+6 2 |
| | 6 2 | 3+4 4 | 6 2 | 3+4 4 | 2+2 4 | 3+4 2 |
| | 5 2 | 3+4 4 | 6 2 | 3+4 4 | 2+2 4 | 5+6 2 |
| | 6 2 | 3+4 4 | 5 2 | 3+4 4 | 2+2 4 | 3+4 2 |
| | 5 2 | 3+4 4 | 6 2 | 3+4 4 | 2+2 4 | 3+4 2 |
| | 5 2 | 3+4 4 | 5 2 | 3+4 4 | 2+2 4 | 5+6 2 |
| | 6 2 | 3+4 4 | 6 2 | 3+4 4 | 2+2 4 | 3+4 2 |
| | 6 2 | 3+4 4 | 5 2 | 3+4 4 | 2+2 4 | 5+6 2 |
| | 6 2 | 3+4 4 | 5 2 | 3+4 4 | 2+2 4 | 5+6 2 |
| | 5 2 | 3+4 4 | 6 2 | 3+4 4 | 2+2 4 | 3+4 2 |
| | 5 2 | 3+4 4 | 6 2 | 3+4 4 | 2+2 4 | 3+4 2 |
| | 5 2 | 3+4 4 | 5 2 | 3+4 4 | 2+2 4 | 5+6 2 |
| | 5 2 | 3+4 4 | 6 2 | 3+4 4 | 2+2 4 | 3+4 2 |
| | 6 2 | 3+4 4 | 5 2 | 3+4 4 | 2+2 4 | 5+6 2 |

Continued on next column

**TABLE XXXVIII– continued**

| TQC | ICSD | | NoE | | SF | |
|---|---|---|---|---|---|---|
| | 6 2 | 3+4 4 | 5 2 | 3+4 4 | 2+2 4 | 3+4 2 |
| | 5 2 | 3+4 4 | 6 2 | 3+4 4 | 2+2 4 | 5+6 2 |
| | 6 2 | 3+4 4 | 6 2 | 3+4 4 | 2+2 4 | 3+4 2 |
| | 6 2 | | 5 2 | | | 5+6 2 |
| | 5 2 | | 6 2 | | | 3+4 2 |
| | 5 2 | | 5 2 | | | 5+6 2 |
| | 5 2 | | 6 2 | | | 5+6 2 |
| | 6 2 | | 5 2 | | | 3+4 2 |
| | 6 2 | | 6 2 | | | 3+4 2 |
| | 6 2 | | 5 2 | | | 5+6 2 |
| | 5 2 | | 6 2 | | | 3+4 2 |
| | 5 2 | | 5 2 | | | 5+6 2 |
| | 6 2 | | 6 2 | | | 5+6 2 |
| | 5 2 | | 5 2 | | | 3+4 2 |
| | 5 2 | | 6 2 | | | 3+4 2 |
| | 6 2 | | 5 2 | | | 5+6 2 |
| | 6 2 | | 5 2 | | | 3+4 2 |
| | 5 2 | | 6 2 | | | 5+6 2 |
| | 6 2 | | 5 2 | | | 3+4 2 |
| | 5 2 | | 6 2 | | | 5+6 2 |
| | 6 2 | | 6 2 | | | 3+4 2 |
| | 5 2 | | 6 2 | | | 5+6 2 |
| | 5 2 | | 5 2 | | | 3+4 2 |
| | 6 2 | | 6 2 | | | 3+4 2 |
| 63.1.9 | 99253 | | 212 | | O7Re1Tb3 | |
| | -GM | -T | -Y | -Z | -R | -S |
| | 5 2 | 3+4 4 | 5 2 | 3+4 4 | 2+2 4 | 5+6 2 |
| | 5 2 | 3+4 4 | 5 2 | 3+4 4 | 2+2 4 | 5+6 2 |
| | 5 2 | 3+4 4 | 6 2 | 3+4 4 | 2+2 4 | 3+4 2 |
| | 5 2 | 3+4 4 | 6 2 | 3+4 4 | 2+2 4 | 5+6 2 |
| | 6 2 | 3+4 4 | 5 2 | 3+4 4 | 2+2 4 | 3+4 2 |
| | 6 2 | 3+4 4 | 5 2 | 3+4 4 | 2+2 4 | 5+6 2 |
| | 6 2 | 3+4 4 | 6 2 | 3+4 4 | 2+2 4 | 3+4 2 |
| | 5 2 | 3+4 4 | 6 2 | 3+4 4 | 2+2 4 | 3+4 2 |
| | 6 2 | 3+4 4 | 6 2 | 3+4 4 | 2+2 4 | 5+6 2 |
| | 5 2 | 3+4 4 | 5 2 | 3+4 4 | 2+2 4 | 3+4 2 |
| | 6 2 | 3+4 4 | 6 2 | 3+4 4 | 2+2 4 | 5+6 2 |
| | 6 2 | 3+4 4 | 5 2 | 3+4 4 | 2+2 4 | 3+4 2 |
| | 5 2 | 3+4 4 | 5 2 | 3+4 4 | 2+2 4 | 3+4 2 |
| | 5 2 | 3+4 4 | 6 2 | 3+4 4 | 2+2 4 | 3+4 2 |
| | 6 2 | 3+4 4 | 6 2 | 3+4 4 | 2+2 4 | 5+6 2 |
| | 5 2 | 3+4 4 | 6 2 | 3+4 4 | 2+2 4 | 3+4 2 |
| | 6 2 | 3+4 4 | 5 2 | 3+4 4 | 2+2 4 | 3+4 2 |
| | 6 2 | 3+4 4 | 6 2 | 3+4 4 | 2+2 4 | 5+6 2 |
| | 5 2 | 3+4 4 | 5 2 | 3+4 4 | 2+2 4 | 5+6 2 |
| | 6 2 | 3+4 4 | 6 2 | 3+4 4 | 2+2 4 | 3+4 2 |
| | 6 2 | 3+4 4 | 5 2 | 3+4 4 | 2+2 4 | 3+4 2 |
| | 5 2 | 3+4 4 | 6 2 | 3+4 4 | 2+2 4 | 5+6 2 |
| | 6 2 | 3+4 4 | 6 2 | 3+4 4 | 2+2 4 | 3+4 2 |
| | 6 2 | 3+4 4 | 6 2 | 3+4 4 | 2+2 4 | 5+6 2 |
| | 6 2 | 3+4 4 | 5 2 | 3+4 4 | 2+2 4 | 3+4 2 |
| | 6 2 | 3+4 4 | 6 2 | 3+4 4 | 2+2 4 | 3+4 2 |
| | 5 2 | 3+4 4 | 5 2 | 3+4 4 | 2+2 4 | 3+4 2 |
| | 5 2 | 3+4 4 | 5 2 | 3+4 4 | 2+2 4 | 5+6 2 |
| | 5 2 | 3+4 4 | 6 2 | 3+4 4 | 2+2 4 | 3+4 2 |
| | 6 2 | 3+4 4 | 6 2 | 3+4 4 | 2+2 4 | 5+6 2 |
| | 6 2 | 3+4 4 | 5 2 | 3+4 4 | 2+2 4 | 5+6 2 |
| | 5 2 | 3+4 4 | 5 2 | 3+4 4 | 2+2 4 | 3+4 2 |
| | 6 2 | 3+4 4 | 5 2 | 3+4 4 | 2+2 4 | 5+6 2 |
| | 6 2 | 3+4 4 | 5 2 | 3+4 4 | 2+2 4 | 5+6 2 |
| | 5 2 | 3+4 4 | 6 2 | 3+4 4 | 2+2 4 | 3+4 2 |

Continued on next column

**TABLE XXXVIII– continued**

| TQC | ICSD | NoE | SF |
|---|---|---|---|
| | 5 2 3+4 4 | 6 2 3+4 4 2+2 4 | 5+6 2 |
| | 6 2 3+4 4 | 6 2 3+4 4 2+2 4 | 3+4 2 |
| | 6 2 3+4 4 | 6 2 3+4 4 2+2 4 | 3+4 2 |
| | 5 2 3+4 4 | 5 2 3+4 4 2+2 4 | 3+4 2 |
| | 5 2 3+4 4 | 5 2 3+4 4 2+2 4 | 3+4 2 |
| | 5 2 3+4 4 | 5 2 3+4 4 2+2 4 | 3+4 2 |
| | 5 2 3+4 4 | 5 2 3+4 4 2+2 4 | 3+4 2 |
| | 5 2 3+4 4 | 6 2 3+4 4 2+2 4 | 3+4 2 |
| | 5 2 3+4 4 | 5 2 3+4 4 2+2 4 | 5+6 2 |
| | 5 2 3+4 4 | 6 2 3+4 4 2+2 4 | 5+6 2 |
| | 6 2 3+4 4 | 6 2 3+4 4 2+2 4 | 3+4 2 |
| | 5 2 3+4 4 | 5 2 3+4 4 2+2 4 | 5+6 2 |
| | 6 2 3+4 4 | 5 2 3+4 4 2+2 4 | 3+4 2 |
| | 6 2 3+4 4 | 5 2 3+4 4 2+2 4 | 3+4 2 |
| | 5 2 3+4 4 | 5 2 3+4 4 2+2 4 | 5+6 2 |
| | 5 2 3+4 4 | 6 2 3+4 4 2+2 4 | 3+4 2 |
| | 6 2 3+4 4 | 6 2 3+4 4 2+2 4 | 5+6 2 |
| | 5 2 3+4 4 | 6 2 3+4 4 2+2 4 | 5+6 2 |
| | 6 2 | 5 2 | 5+6 2 |
| | 5 2 | 6 2 | 3+4 2 |
| | 5 2 | 5 2 | 5+6 2 |
| | 5 2 | 5 2 | 5+6 2 |
| | 6 2 | 6 2 | 5+6 2 |
| | 6 2 | 5 2 | 3+4 2 |
| | 5 2 | 6 2 | 5+6 2 |
| | 6 2 | 5 2 | 3+4 2 |
| | 5 2 | 6 2 | 5+6 2 |
| | 6 2 | 5 2 | 5+6 2 |
| | 6 2 | 6 2 | 3+4 2 |
| | 5 2 | 6 2 | 5+6 2 |
| | 6 2 | 5 2 | 3+4 2 |
| | 6 2 | 5 2 | 3+4 2 |
| | 5 2 | 5 2 | 5+6 2 |
| | 6 2 | 6 2 | 3+4 2 |
| | 5 2 | 6 2 | 5+6 2 |
| | 6 2 | 6 2 | 3+4 2 |
| | 6 2 | 5 2 | 5+6 2 |
| | 6 2 | 6 2 | 5+6 2 |
| | 6 2 | 6 2 | 3+4 2 |
| | 6 2 | 5 2 | 3+4 2 |
| | 5 2 | 6 2 | 3+4 2 |
| | 6 2 | 6 2 | 5+6 2 |
| | 6 2 | 6 2 | 3+4 2 |
| | 5 2 | 6 2 | 5+6 2 |
| | 6 2 | 5 2 | 5+6 2 |
| | 5 2 | 6 2 | 3+4 2 |
| | 6 2 | 6 2 | 3+4 2 |
| | 5 2 | 6 2 | 3+4 2 |
| | 6 2 | 5 2 | 3+4 2 |
| | 6 2 | 6 2 | 3+4 2 |
| | 6 2 | 5 2 | 3+4 2 |
| | 5 2 | 6 2 | 3+4 2 |
| | 6 2 | 5 2 | 5+6 2 |
| | 6 2 | 6 2 | 3+4 2 |
| | 5 2 | 5 2 | 5+6 2 |
| | 6 2 | 6 2 | 3+4 2 |
| | 5 2 | 6 2 | 5+6 2 |
| | 6 2 | 5 2 | 3+4 2 |
| | 6 2 | 6 2 | 5+6 2 |
| | 6 2 | 5 2 | 3+4 2 |
| | 5 2 | 6 2 | 5+6 2 |

Continued on next column

**TABLE XXXVIII– continued**

| TQC | ICSD | NoE | SF |
|---|---|---|---|
| | 5 2 | 6 2 | 5+6 2 |
| | 6 2 | 6 2 | 3+4 2 |
| | 5 2 | 5 2 | 3+4 2 |
| | 6 2 | 5 2 | 3+4 2 |
| | 6 2 | 6 2 | 3+4 2 |
| | 5 2 | 5 2 | 3+4 2 |
| | 6 2 | 6 2 | 3+4 2 |
| | 5 2 | 6 2 | 3+4 2 |
| | 6 2 | 5 2 | 3+4 2 |
| | 6 2 | 6 2 | 5+6 2 |
| 63.1.10 | 96410 | 176 | O7Pr3Re1 |
| | -GM -T | -Y -Z -R | -S |
| | 5 2 3+4 4 | 5 2 3+4 4 2+2 4 | 5+6 2 |
| | 5 2 3+4 4 | 5 2 3+4 4 2+2 4 | 5+6 2 |
| | 5 2 3+4 4 | 6 2 3+4 4 2+2 4 | 3+4 2 |
| | 6 2 3+4 4 | 5 2 3+4 4 2+2 4 | 5+6 2 |
| | 5 2 3+4 4 | 6 2 3+4 4 2+2 4 | 3+4 2 |
| | 6 2 3+4 4 | 5 2 3+4 4 2+2 4 | 5+6 2 |
| | 6 2 3+4 4 | 6 2 3+4 4 2+2 4 | 3+4 2 |
| | 5 2 3+4 4 | 5 2 3+4 4 2+2 4 | 3+4 2 |
| | 6 2 3+4 4 | 5 2 3+4 4 2+2 4 | 5+6 2 |
| | 5 2 3+4 4 | 6 2 3+4 4 2+2 4 | 3+4 2 |
| | 5 2 3+4 4 | 6 2 3+4 4 2+2 4 | 3+4 2 |
| | 6 2 3+4 4 | 6 2 3+4 4 2+2 4 | 5+6 2 |
| | 5 2 3+4 4 | 5 2 3+4 4 2+2 4 | 3+4 2 |
| | 6 2 3+4 4 | 6 2 3+4 4 2+2 4 | 5+6 2 |
| | 6 2 3+4 4 | 5 2 3+4 4 2+2 4 | 5+6 2 |
| | 5 2 3+4 4 | 5 2 3+4 4 2+2 4 | 3+4 2 |
| | 6 2 3+4 4 | 6 2 3+4 4 2+2 4 | 3+4 2 |
| | 5 2 3+4 4 | 5 2 3+4 4 2+2 4 | 5+6 2 |
| | 5 2 3+4 4 | 6 2 3+4 4 2+2 4 | 3+4 2 |
| | 6 2 3+4 4 | 5 2 3+4 4 2+2 4 | 3+4 2 |
| | 6 2 3+4 4 | 5 2 3+4 4 2+2 4 | 5+6 2 |
| | 5 2 3+4 4 | 6 2 3+4 4 2+2 4 | 5+6 2 |
| | 6 2 3+4 4 | 6 2 3+4 4 2+2 4 | 3+4 2 |
| | 6 2 3+4 4 | 6 2 3+4 4 2+2 4 | 3+4 2 |
| | 6 2 3+4 4 | 6 2 3+4 4 2+2 4 | 5+6 2 |
| | 5 2 3+4 4 | 5 2 3+4 4 2+2 4 | 5+6 2 |
| | 5 2 3+4 4 | 6 2 3+4 4 2+2 4 | 3+4 2 |
| | 6 2 3+4 4 | 6 2 3+4 4 2+2 4 | 3+4 2 |
| | 6 2 3+4 4 | 5 2 3+4 4 2+2 4 | 3+4 2 |
| | 6 2 3+4 4 | 6 2 3+4 4 2+2 4 | 3+4 2 |
| | 6 2 3+4 4 | 6 2 3+4 4 2+2 4 | 5+6 2 |
| | 5 2 3+4 4 | 5 2 3+4 4 2+2 4 | 3+4 2 |
| | 6 2 3+4 4 | 5 2 3+4 4 2+2 4 | 5+6 2 |
| | 5 2 3+4 4 | 6 2 3+4 4 2+2 4 | 3+4 2 |
| | 6 2 3+4 4 | 6 2 3+4 4 2+2 4 | 5+6 2 |
| | 6 2 3+4 4 | 6 2 3+4 4 2+2 4 | 5+6 2 |
| | 5 2 3+4 4 | 5 2 3+4 4 2+2 4 | 3+4 2 |
| | 6 2 3+4 4 | 6 2 3+4 4 2+2 4 | 3+4 2 |
| | 5 2 3+4 4 | 5 2 3+4 4 2+2 4 | 3+4 2 |
| | 5 2 3+4 4 | 5 2 3+4 4 2+2 4 | 3+4 2 |
| | 5 2 3+4 4 | 5 2 3+4 4 2+2 4 | 3+4 2 |
| | 5 2 3+4 4 | 5 2 3+4 4 2+2 4 | 3+4 2 |
| | 5 2 3+4 4 | 5 2 3+4 4 2+2 4 | 3+4 2 |
| | 5 2 3+4 4 | 5 2 3+4 4 2+2 4 | 3+4 2 |
| | 5 2 | 5 2 | 3+4 2 |
| | 5 2 | 5 2 | 3+4 2 |
| | 5 2 | 5 2 | 3+4 2 |
| | 6 2 | 6 2 | 5+6 2 |
| | 5 2 | 6 2 | 5+6 2 |

Continued on next column

**TABLE XXXVIII– continued**

| TQC | ICSD | | NoE | | SF | |
|---|---|---|---|---|---|---|
| | 5 2 | | 6 2 | | | 5+6 2 |
| | 6 2 | | 5 2 | | | 5+6 2 |
| | 5 2 | | 6 2 | | | 5+6 2 |
| | 6 2 | | 5 2 | | | 5+6 2 |
| | 6 2 | | 5 2 | | | 5+6 2 |
| | 5 2 | | 5 2 | | | 3+4 2 |
| | 6 2 | | 6 2 | | | 5+6 2 |
| | 5 2 | | 5 2 | | | 3+4 2 |
| | 6 2 | | 6 2 | | | 5+6 2 |
| | 5 2 | | 5 2 | | | 3+4 2 |
| | 6 2 | | 6 2 | | | 5+6 2 |
| | 6 2 | | 6 2 | | | 3+4 2 |
| | 6 2 | | 5 2 | | | 5+6 2 |
| | 6 2 | | 6 2 | | | 5+6 2 |
| | 5 2 | | 6 2 | | | 5+6 2 |
| | 6 2 | | 6 2 | | | 3+4 2 |
| | 5 2 | | 6 2 | | | 5+6 2 |
| | 6 2 | | 6 2 | | | 3+4 2 |
| | 6 2 | | 6 2 | | | 5+6 2 |
| | 5 2 | | 6 2 | | | 3+4 2 |
| | 5 2 | | 5 2 | | | 5+6 2 |
| | 6 2 | | 6 2 | | | 3+4 2 |
| | 6 2 | | 5 2 | | | 5+6 2 |
| | 6 2 | | 6 2 | | | 3+4 2 |
| | 6 2 | | 5 2 | | | 5+6 2 |
| | 6 2 | | 6 2 | | | 3+4 2 |
| | 5 2 | | 5 2 | | | 5+6 2 |
| | 6 2 | | 6 2 | | | 5+6 2 |
| | 6 2 | | 6 2 | | | 3+4 2 |
| | 5 2 | | 6 2 | | | 5+6 2 |
| | 6 2 | | 5 2 | | | 3+4 2 |
| | 5 2 | | 5 2 | | | 3+4 2 |
| | 5 2 | | 5 2 | | | 3+4 2 |
| | 5 2 | | 5 2 | | | 3+4 2 |
| | 6 2 | | 6 2 | | | 3+4 2 |
| | 5 2 | | 5 2 | | | 3+4 2 |
| | 6 2 | | 6 2 | | | 5+6 2 |
| | 6 2 | | 6 2 | | | 3+4 2 |
| | 6 2 | | 6 2 | | | 3+4 2 |
| 63.1.11 | 96149 | | 76 | | Np1Si1 | |
| | -GM | -T | -Y | -Z | -R | -S |
| | 5 2 | 3+4 4 | 6 2 | 3+4 4 | 2+2 4 | 5+6 2 |
| | 6 2 | 3+4 4 | 5 2 | 3+4 4 | 2+2 4 | 3+4 2 |
| | 5 2 | 3+4 4 | 5 2 | 3+4 4 | 2+2 4 | 5+6 2 |
| | 6 2 | 3+4 4 | 6 2 | 3+4 4 | 2+2 4 | 3+4 2 |
| | 5 2 | 3+4 4 | 5 2 | 3+4 4 | 2+2 4 | 5+6 2 |
| | 6 2 | 3+4 4 | 6 2 | 3+4 4 | 2+2 4 | 3+4 2 |
| | 5 2 | 3+4 4 | 5 2 | 3+4 4 | 2+2 4 | 5+6 2 |
| | 6 2 | 3+4 4 | 6 2 | 3+4 4 | 2+2 4 | 3+4 2 |
| | 5 2 | 3+4 4 | 5 2 | 3+4 4 | 2+2 4 | 5+6 2 |
| | 6 2 | 3+4 4 | 6 2 | 3+4 4 | 2+2 4 | 3+4 2 |
| | 5 2 | 3+4 4 | 6 2 | 3+4 4 | 2+2 4 | 5+6 2 |
| | 5 2 | 3+4 4 | 5 2 | 3+4 4 | 2+2 4 | 3+4 2 |
| | 6 2 | 3+4 4 | 6 2 | 3+4 4 | 2+2 4 | 5+6 2 |
| | 6 2 | 3+4 4 | 5 2 | 3+4 4 | 2+2 4 | 3+4 2 |
| | 5 2 | 3+4 4 | 5 2 | 3+4 4 | 2+2 4 | 5+6 2 |
| | 6 2 | 3+4 4 | 6 2 | 3+4 4 | 2+2 4 | 3+4 2 |
| | 5 2 | 3+4 4 | 5 2 | 3+4 4 | 2+2 4 | 3+4 2 |
| | 6 2 | 3+4 4 | 6 2 | 3+4 4 | 2+2 4 | 3+4 2 |
| | 5 2 | 3+4 4 | 5 2 | 3+4 4 | 2+2 4 | 5+6 2 |
| | 6 2 | | 6 2 | | | 5+6 2 |

Continued on next column

**TABLE XXXVIII– continued**

| TQC | ICSD | | NoE | | SF | |
|---|---|---|---|---|---|---|
| | 5 2 | | 5 2 | | | 5+6 2 |
| | 6 2 | | 6 2 | | | 3+4 2 |
| | 5 2 | | 6 2 | | | 3+4 2 |
| | 6 2 | | 5 2 | | | 5+6 2 |
| | 6 2 | | 6 2 | | | 3+4 2 |
| | 5 2 | | 5 2 | | | 3+4 2 |
| | 5 2 | | 5 2 | | | 3+4 2 |
| | 5 2 | | 6 2 | | | 5+6 2 |
| | 6 2 | | 5 2 | | | 3+4 2 |
| | 5 2 | | 5 2 | | | 5+6 2 |
| | 6 2 | | 6 2 | | | 5+6 2 |
| | 6 2 | | 5 2 | | | 3+4 2 |
| | 6 2 | | 6 2 | | | 5+6 2 |
| | 5 2 | | 6 2 | | | 3+4 2 |
| | 6 2 | | 6 2 | | | 3+4 2 |
| | 6 2 | | 5 2 | | | 5+6 2 |
| | 5 2 | | 5 2 | | | 3+4 2 |
| | 6 2 | | 5 2 | | | 5+6 2 |
| 63.1.12 | 86102 | | 84 | | Mg1O5V2 | |
| | -GM | -T | -Y | -Z | -R | -S |
| | 5 2 | 3+4 4 | 6 2 | 3+4 4 | 2+2 4 | 3+4 2 |
| | 5 2 | 3+4 4 | 6 2 | 3+4 4 | 2+2 4 | 5+6 2 |
| | 6 2 | 3+4 4 | 5 2 | 3+4 4 | 2+2 4 | 3+4 2 |
| | 6 2 | 3+4 4 | 5 2 | 3+4 4 | 2+2 4 | 5+6 2 |
| | 5 2 | 3+4 4 | 5 2 | 3+4 4 | 2+2 4 | 5+6 2 |
| | 6 2 | 3+4 4 | 6 2 | 3+4 4 | 2+2 4 | 3+4 2 |
| | 5 2 | 3+4 4 | 5 2 | 3+4 4 | 2+2 4 | 3+4 2 |
| | 6 2 | 3+4 4 | 6 2 | 3+4 4 | 2+2 4 | 5+6 2 |
| | 5 2 | 3+4 4 | 5 2 | 3+4 4 | 2+2 4 | 3+4 2 |
| | 6 2 | 3+4 4 | 6 2 | 3+4 4 | 2+2 4 | 5+6 2 |
| | 5 2 | 3+4 4 | 5 2 | 3+4 4 | 2+2 4 | 3+4 2 |
| | 5 2 | 3+4 4 | 6 2 | 3+4 4 | 2+2 4 | 5+6 2 |
| | 5 2 | 3+4 4 | 6 2 | 3+4 4 | 2+2 4 | 3+4 2 |
| | 5 2 | 3+4 4 | 5 2 | 3+4 4 | 2+2 4 | 5+6 2 |
| | 6 2 | 3+4 4 | 6 2 | 3+4 4 | 2+2 4 | 5+6 2 |
| | 6 2 | 3+4 4 | 6 2 | 3+4 4 | 2+2 4 | 3+4 2 |
| | 5 2 | 3+4 4 | 6 2 | 3+4 4 | 2+2 4 | 3+4 2 |
| | 6 2 | 3+4 4 | 5 2 | 3+4 4 | 2+2 4 | 3+4 2 |
| | 6 2 | 3+4 4 | 6 2 | 3+4 4 | 2+2 4 | 5+6 2 |
| | 5 2 | 3+4 4 | 5 2 | 3+4 4 | 2+2 4 | 3+4 2 |
| | 5 2 | 3+4 4 | 5 2 | 3+4 4 | 2+2 4 | 5+6 2 |
| | 6 2 | | 5 2 | | | 5+6 2 |
| | 6 2 | | 5 2 | | | 3+4 2 |
| | 5 2 | | 6 2 | | | 5+6 2 |
| | 6 2 | | 6 2 | | | 5+6 2 |
| | 6 2 | | 6 2 | | | 3+4 2 |
| | 5 2 | | 5 2 | | | 5+6 2 |
| | 5 2 | | 5 2 | | | 3+4 2 |
| | 6 2 | | 6 2 | | | 5+6 2 |
| | 6 2 | | 5 2 | | | 3+4 2 |
| | 5 2 | | 6 2 | | | 3+4 2 |
| | 6 2 | | 6 2 | | | 5+6 2 |
| | 5 2 | | 6 2 | | | 3+4 2 |
| | 5 2 | | 5 2 | | | 5+6 2 |
| | 6 2 | | 5 2 | | | 5+6 2 |
| | 5 2 | | 6 2 | | | 3+4 2 |
| | 5 2 | | 5 2 | | | 5+6 2 |
| | 6 2 | | 5 2 | | | 3+4 2 |
| | 6 2 | | 6 2 | | | 3+4 2 |
| | 6 2 | | 5 2 | | | 5+6 2 |
| | 5 2 | | 6 2 | | | 3+4 2 |

Continued on next column

**TABLE XXXVIII– continued**

| TQC | ICSD | | NoE | | SF | |
|---|---|---|---|---|---|---|
| | 5 2 | | 6 2 | | | 5+6 2 |
| 63.1.13 | 50981 | | 84 | | Mg1O5V2 | |
| | -GM | -T | -Y | -Z | -R | -S |
| | 5 2 | 3+4 4 | 6 2 | 3+4 4 | 2+2 4 | 3+4 2 |
| | 5 2 | 3+4 4 | 6 2 | 3+4 4 | 2+2 4 | 5+6 2 |
| | 6 2 | 3+4 4 | 5 2 | 3+4 4 | 2+2 4 | 3+4 2 |
| | 6 2 | 3+4 4 | 5 2 | 3+4 4 | 2+2 4 | 5+6 2 |
| | 5 2 | 3+4 4 | 5 2 | 3+4 4 | 2+2 4 | 5+6 2 |
| | 6 2 | 3+4 4 | 6 2 | 3+4 4 | 2+2 4 | 3+4 2 |
| | 5 2 | 3+4 4 | 5 2 | 3+4 4 | 2+2 4 | 3+4 2 |
| | 6 2 | 3+4 4 | 6 2 | 3+4 4 | 2+2 4 | 5+6 2 |
| | 5 2 | 3+4 4 | 5 2 | 3+4 4 | 2+2 4 | 3+4 2 |
| | 6 2 | 3+4 4 | 6 2 | 3+4 4 | 2+2 4 | 5+6 2 |
| | 5 2 | 3+4 4 | 5 2 | 3+4 4 | 2+2 4 | 3+4 2 |
| | 5 2 | 3+4 4 | 6 2 | 3+4 4 | 2+2 4 | 5+6 2 |
| | 5 2 | 3+4 4 | 6 2 | 3+4 4 | 2+2 4 | 3+4 2 |
| | 5 2 | 3+4 4 | 5 2 | 3+4 4 | 2+2 4 | 5+6 2 |
| | 6 2 | 3+4 4 | 6 2 | 3+4 4 | 2+2 4 | 5+6 2 |
| | 6 2 | 3+4 4 | 6 2 | 3+4 4 | 2+2 4 | 3+4 2 |
| | 6 2 | 3+4 4 | 6 2 | 3+4 4 | 2+2 4 | 3+4 2 |
| | 5 2 | 3+4 4 | 5 2 | 3+4 4 | 2+2 4 | 3+4 2 |
| | 6 2 | 3+4 4 | 6 2 | 3+4 4 | 2+2 4 | 5+6 2 |
| | 5 2 | 3+4 4 | 5 2 | 3+4 4 | 2+2 4 | 3+4 2 |
| | 6 2 | 3+4 4 | 5 2 | 3+4 4 | 2+2 4 | 5+6 2 |
| | 5 2 | | 5 2 | | | 5+6 2 |
| | 6 2 | | 5 2 | | | 3+4 2 |
| | 5 2 | | 6 2 | | | 5+6 2 |
| | 6 2 | | 6 2 | | | 5+6 2 |
| | 6 2 | | 6 2 | | | 3+4 2 |
| | 5 2 | | 5 2 | | | 5+6 2 |
| | 5 2 | | 5 2 | | | 3+4 2 |
| | 6 2 | | 6 2 | | | 5+6 2 |
| | 6 2 | | 5 2 | | | 3+4 2 |
| | 5 2 | | 6 2 | | | 3+4 2 |
| | 6 2 | | 6 2 | | | 5+6 2 |
| | 5 2 | | 6 2 | | | 3+4 2 |
| | 5 2 | | 5 2 | | | 5+6 2 |
| | 6 2 | | 5 2 | | | 5+6 2 |
| | 5 2 | | 6 2 | | | 3+4 2 |
| | 5 2 | | 5 2 | | | 5+6 2 |
| | 6 2 | | 5 2 | | | 3+4 2 |
| | 6 2 | | 6 2 | | | 3+4 2 |
| | 6 2 | | 5 2 | | | 5+6 2 |
| | 5 2 | | 6 2 | | | 3+4 2 |
| | 5 2 | | 6 2 | | | 5+6 2 |
| 63.1.14 | 50979 | | 84 | | Mg1O5V2 | |
| | -GM | -T | -Y | -Z | -R | -S |
| | 5 2 | 3+4 4 | 6 2 | 3+4 4 | 2+2 4 | 3+4 2 |
| | 5 2 | 3+4 4 | 6 2 | 3+4 4 | 2+2 4 | 5+6 2 |
| | 6 2 | 3+4 4 | 5 2 | 3+4 4 | 2+2 4 | 3+4 2 |
| | 6 2 | 3+4 4 | 5 2 | 3+4 4 | 2+2 4 | 5+6 2 |
| | 5 2 | 3+4 4 | 5 2 | 3+4 4 | 2+2 4 | 5+6 2 |
| | 6 2 | 3+4 4 | 6 2 | 3+4 4 | 2+2 4 | 3+4 2 |
| | 5 2 | 3+4 4 | 5 2 | 3+4 4 | 2+2 4 | 3+4 2 |
| | 6 2 | 3+4 4 | 6 2 | 3+4 4 | 2+2 4 | 5+6 2 |
| | 5 2 | 3+4 4 | 5 2 | 3+4 4 | 2+2 4 | 3+4 2 |
| | 6 2 | 3+4 4 | 6 2 | 3+4 4 | 2+2 4 | 5+6 2 |
| | 5 2 | 3+4 4 | 5 2 | 3+4 4 | 2+2 4 | 3+4 2 |
| | 5 2 | 3+4 4 | 6 2 | 3+4 4 | 2+2 4 | 5+6 2 |
| | 5 2 | 3+4 4 | 6 2 | 3+4 4 | 2+2 4 | 3+4 2 |
| | 5 2 | 3+4 4 | 5 2 | 3+4 4 | 2+2 4 | 5+6 2 |

Continued on next column

**TABLE XXXVIII– continued**

| TQC | ICSD | | NoE | | SF | |
|---|---|---|---|---|---|---|
| | 6 2 | 3+4 4 | 6 2 | 3+4 4 | 2+2 4 | 5+6 2 |
| | 6 2 | 3+4 4 | 6 2 | 3+4 4 | 2+2 4 | 3+4 2 |
| | 6 2 | 3+4 4 | 6 2 | 3+4 4 | 2+2 4 | 3+4 2 |
| | 5 2 | 3+4 4 | 5 2 | 3+4 4 | 2+2 4 | 3+4 2 |
| | 6 2 | 3+4 4 | 6 2 | 3+4 4 | 2+2 4 | 5+6 2 |
| | 5 2 | 3+4 4 | 5 2 | 3+4 4 | 2+2 4 | 3+4 2 |
| | 6 2 | 3+4 4 | 5 2 | 3+4 4 | 2+2 4 | 5+6 2 |
| | 5 2 | | 5 2 | | | 5+6 2 |
| | 6 2 | | 5 2 | | | 3+4 2 |
| | 5 2 | | 6 2 | | | 5+6 2 |
| | 6 2 | | 6 2 | | | 5+6 2 |
| | 6 2 | | 6 2 | | | 3+4 2 |
| | 5 2 | | 5 2 | | | 5+6 2 |
| | 5 2 | | 5 2 | | | 3+4 2 |
| | 6 2 | | 6 2 | | | 5+6 2 |
| | 6 2 | | 5 2 | | | 3+4 2 |
| | 5 2 | | 6 2 | | | 3+4 2 |
| | 6 2 | | 6 2 | | | 5+6 2 |
| | 5 2 | | 6 2 | | | 3+4 2 |
| | 5 2 | | 5 2 | | | 5+6 2 |
| | 6 2 | | 5 2 | | | 5+6 2 |
| | 5 2 | | 6 2 | | | 3+4 2 |
| | 5 2 | | 5 2 | | | 5+6 2 |
| | 6 2 | | 5 2 | | | 3+4 2 |
| | 6 2 | | 6 2 | | | 3+4 2 |
| | 6 2 | | 5 2 | | | 5+6 2 |
| | 5 2 | | 6 2 | | | 3+4 2 |
| | 5 2 | | 6 2 | | | 5+6 2 |
| 63.1.15 | 250752 | | 172 | | La2Sn1Zn5 | |
| | -GM | -T | -Y | -Z | -R | -S |
| | 5 2 | 3+4 4 | 5 2 | 3+4 4 | 2+2 4 | 3+4 2 |
| | 5 2 | 3+4 4 | 5 2 | 3+4 4 | 2+2 4 | 3+4 2 |
| | 5 2 | 3+4 4 | 6 2 | 3+4 4 | 2+2 4 | 3+4 2 |
| | 6 2 | 3+4 4 | 5 2 | 3+4 4 | 2+2 4 | 5+6 2 |
| | 6 2 | 3+4 4 | 6 2 | 3+4 4 | 2+2 4 | 5+6 2 |
| | 6 2 | 3+4 4 | 6 2 | 3+4 4 | 2+2 4 | 5+6 2 |
| | 5 2 | 3+4 4 | 6 2 | 3+4 4 | 2+2 4 | 3+4 2 |
| | 6 2 | 3+4 4 | 5 2 | 3+4 4 | 2+2 4 | 5+6 2 |
| | 6 2 | 3+4 4 | 6 2 | 3+4 4 | 2+2 4 | 5+6 2 |
| | 6 2 | 3+4 4 | 6 2 | 3+4 4 | 2+2 4 | 5+6 2 |
| | 6 2 | 3+4 4 | 6 2 | 3+4 4 | 2+2 4 | 5+6 2 |
| | 6 2 | 3+4 4 | 6 2 | 3+4 4 | 2+2 4 | 5+6 2 |
| | 5 2 | 3+4 4 | 5 2 | 3+4 4 | 2+2 4 | 3+4 2 |
| | 6 2 | 3+4 4 | 6 2 | 3+4 4 | 2+2 4 | 5+6 2 |
| | 5 2 | 3+4 4 | 6 2 | 3+4 4 | 2+2 4 | 3+4 2 |
| | 6 2 | 3+4 4 | 5 2 | 3+4 4 | 2+2 4 | 5+6 2 |
| | 5 2 | 3+4 4 | 5 2 | 3+4 4 | 2+2 4 | 3+4 2 |
| | 6 2 | 3+4 4 | 6 2 | 3+4 4 | 2+2 4 | 5+6 2 |
| | 5 2 | 3+4 4 | 5 2 | 3+4 4 | 2+2 4 | 3+4 2 |
| | 6 2 | 3+4 4 | 6 2 | 3+4 4 | 2+2 4 | 5+6 2 |
| | 5 2 | 3+4 4 | 6 2 | 3+4 4 | 2+2 4 | 3+4 2 |
| | 6 2 | 3+4 4 | 5 2 | 3+4 4 | 2+2 4 | 5+6 2 |
| | 5 2 | 3+4 4 | 6 2 | 3+4 4 | 2+2 4 | 3+4 2 |
| | 5 2 | 3+4 4 | 6 2 | 3+4 4 | 2+2 4 | 5+6 2 |
| | 5 2 | 3+4 4 | 5 2 | 3+4 4 | 2+2 4 | 3+4 2 |
| | 6 2 | 3+4 4 | 6 2 | 3+4 4 | 2+2 4 | 5+6 2 |
| | 5 2 | 3+4 4 | 5 2 | 3+4 4 | 2+2 4 | 5+6 2 |
| | 6 2 | 3+4 4 | 6 2 | 3+4 4 | 2+2 4 | 3+4 2 |
| | 6 2 | 3+4 4 | 6 2 | 3+4 4 | 2+2 4 | 3+4 2 |
| | 5 2 | 3+4 4 | 5 2 | 3+4 4 | 2+2 4 | 5+6 2 |
| | 6 2 | 3+4 4 | 5 2 | 3+4 4 | 2+2 4 | 3+4 2 |

Continued on next column

**TABLE XXXVIII– continued**

| TQC | ICSD | | NoE | | SF | |
|---|---|---|---|---|---|---|
| | 5 2 | 3+4 4 | 5 2 | 3+4 4 | 2+2 4 | 5+6 2 |
| | 6 2 | 3+4 4 | 6 2 | 3+4 4 | 2+2 4 | 3+4 2 |
| | 5 2 | 3+4 4 | 6 2 | 3+4 4 | 2+2 4 | 5+6 2 |
| | 5 2 | 3+4 4 | 5 2 | 3+4 4 | 2+2 4 | 3+4 2 |
| | 6 2 | 3+4 4 | 6 2 | 3+4 4 | 2+2 4 | 3+4 2 |
| | 6 2 | 3+4 4 | 5 2 | 3+4 4 | 2+2 4 | 5+6 2 |
| | 6 2 | 3+4 4 | 6 2 | 3+4 4 | 2+2 4 | 5+6 2 |
| | 5 2 | 3+4 4 | 5 2 | 3+4 4 | 2+2 4 | 3+4 2 |
| | 6 2 | 3+4 4 | 5 2 | 3+4 4 | 2+2 4 | 5+6 2 |
| | 5 2 | 3+4 4 | 6 2 | 3+4 4 | 2+2 4 | 3+4 2 |
| | 5 2 | 3+4 4 | 6 2 | 3+4 4 | 2+2 4 | 5+6 2 |
| | 6 2 | 3+4 4 | 5 2 | 3+4 4 | 2+2 4 | 3+4 2 |
| | 5 2 | | 5 2 | | | 5+6 2 |
| | 6 2 | | 5 2 | | | 3+4 2 |
| | 6 2 | | 6 2 | | | 5+6 2 |
| | 5 2 | | 5 2 | | | 3+4 2 |
| | 5 2 | | 6 2 | | | 5+6 2 |
| | 6 2 | | 5 2 | | | 5+6 2 |
| | 6 2 | | 6 2 | | | 3+4 2 |
| | 5 2 | | 5 2 | | | 5+6 2 |
| | 5 2 | | 6 2 | | | 3+4 2 |
| | 6 2 | | 6 2 | | | 3+4 2 |
| | 6 2 | | 5 2 | | | 5+6 2 |
| | 5 2 | | 5 2 | | | 3+4 2 |
| | 6 2 | | 5 2 | | | 5+6 2 |
| | 5 2 | | 6 2 | | | 3+4 2 |
| | 6 2 | | 6 2 | | | 5+6 2 |
| | 6 2 | | 5 2 | | | 5+6 2 |
| | 5 2 | | 5 2 | | | 3+4 2 |
| | 6 2 | | 6 2 | | | 3+4 2 |
| | 5 2 | | 5 2 | | | 5+6 2 |
| | 6 2 | | 6 2 | | | 5+6 2 |
| | 5 2 | | 5 2 | | | 3+4 2 |
| | 6 2 | | 6 2 | | | 3+4 2 |
| | 5 2 | | 5 2 | | | 5+6 2 |
| | 5 2 | | 6 2 | | | 5+6 2 |
| | 6 2 | | 6 2 | | | 3+4 2 |
| | 5 2 | | 6 2 | | | 3+4 2 |
| | 6 2 | | 5 2 | | | 5+6 2 |
| | 5 2 | | 6 2 | | | 3+4 2 |
| | 5 2 | | 5 2 | | | 3+4 2 |
| | 6 2 | | 5 2 | | | 5+6 2 |
| | 6 2 | | 6 2 | | | 5+6 2 |
| | 5 2 | | 6 2 | | | 3+4 2 |
| | 5 2 | | 5 2 | | | 5+6 2 |
| | 5 2 | | 5 2 | | | 5+6 2 |
| | 6 2 | | 6 2 | | | 3+4 2 |
| | 5 2 | | 5 2 | | | 3+4 2 |
| | 6 2 | | 6 2 | | | 3+4 2 |
| | 5 2 | | 6 2 | | | 5+6 2 |
| | 5 2 | | 5 2 | | | 3+4 2 |
| | 6 2 | | 6 2 | | | 3+4 2 |
| | 5 2 | | 5 2 | | | 5+6 2 |
| | 5 2 | | 5 2 | | | 3+4 2 |
| | 6 2 | | 6 2 | | | 5+6 2 |
| 63.1.16 | 250756 | | 172 | | La2Pb1Zn5 | |
| | -GM | -T | -Y | -Z | -R | -S |
| | 5 2 | 3+4 4 | 5 2 | 3+4 4 | 2+2 4 | 3+4 2 |
| | 5 2 | 3+4 4 | 5 2 | 3+4 4 | 2+2 4 | 3+4 2 |
| | 5 2 | 3+4 4 | 6 2 | 3+4 4 | 2+2 4 | 3+4 2 |
| | 6 2 | 3+4 4 | 5 2 | 3+4 4 | 2+2 4 | 5+6 2 |

Continued on next column

**TABLE XXXVIII– continued**

| TQC | ICSD | | NoE | | SF | |
|---|---|---|---|---|---|---|
| | 6 2 | 3+4 4 | 6 2 | 3+4 4 | 2+2 4 | 5+6 2 |
| | 6 2 | 3+4 4 | 6 2 | 3+4 4 | 2+2 4 | 5+6 2 |
| | 5 2 | 3+4 4 | 6 2 | 3+4 4 | 2+2 4 | 3+4 2 |
| | 6 2 | 3+4 4 | 5 2 | 3+4 4 | 2+2 4 | 5+6 2 |
| | 6 2 | 3+4 4 | 6 2 | 3+4 4 | 2+2 4 | 5+6 2 |
| | 6 2 | 3+4 4 | 6 2 | 3+4 4 | 2+2 4 | 5+6 2 |
| | 6 2 | 3+4 4 | 6 2 | 3+4 4 | 2+2 4 | 5+6 2 |
| | 6 2 | 3+4 4 | 6 2 | 3+4 4 | 2+2 4 | 5+6 2 |
| | 5 2 | 3+4 4 | 5 2 | 3+4 4 | 2+2 4 | 3+4 2 |
| | 6 2 | 3+4 4 | 6 2 | 3+4 4 | 2+2 4 | 5+6 2 |
| | 5 2 | 3+4 4 | 6 2 | 3+4 4 | 2+2 4 | 3+4 2 |
| | 6 2 | 3+4 4 | 5 2 | 3+4 4 | 2+2 4 | 5+6 2 |
| | 5 2 | 3+4 4 | 5 2 | 3+4 4 | 2+2 4 | 3+4 2 |
| | 6 2 | 3+4 4 | 6 2 | 3+4 4 | 2+2 4 | 5+6 2 |
| | 5 2 | 3+4 4 | 6 2 | 3+4 4 | 2+2 4 | 3+4 2 |
| | 6 2 | 3+4 4 | 5 2 | 3+4 4 | 2+2 4 | 5+6 2 |
| | 6 2 | 3+4 4 | 6 2 | 3+4 4 | 2+2 4 | 3+4 2 |
| | 5 2 | 3+4 4 | 5 2 | 3+4 4 | 2+2 4 | 5+6 2 |
| | 5 2 | 3+4 4 | 6 2 | 3+4 4 | 2+2 4 | 3+4 2 |
| | 5 2 | 3+4 4 | 6 2 | 3+4 4 | 2+2 4 | 5+6 2 |
| | 5 2 | 3+4 4 | 5 2 | 3+4 4 | 2+2 4 | 3+4 2 |
| | 6 2 | 3+4 4 | 6 2 | 3+4 4 | 2+2 4 | 5+6 2 |
| | 6 2 | 3+4 4 | 5 2 | 3+4 4 | 2+2 4 | 3+4 2 |
| | 5 2 | 3+4 4 | 6 2 | 3+4 4 | 2+2 4 | 5+6 2 |
| | 6 2 | 3+4 4 | 6 2 | 3+4 4 | 2+2 4 | 3+4 2 |
| | 5 2 | 3+4 4 | 5 2 | 3+4 4 | 2+2 4 | 5+6 2 |
| | 6 2 | 3+4 4 | 5 2 | 3+4 4 | 2+2 4 | 3+4 2 |
| | 5 2 | 3+4 4 | 5 2 | 3+4 4 | 2+2 4 | 5+6 2 |
| | 6 2 | 3+4 4 | 6 2 | 3+4 4 | 2+2 4 | 3+4 2 |
| | 5 2 | 3+4 4 | 6 2 | 3+4 4 | 2+2 4 | 5+6 2 |
| | 5 2 | 3+4 4 | 5 2 | 3+4 4 | 2+2 4 | 3+4 2 |
| | 6 2 | 3+4 4 | 6 2 | 3+4 4 | 2+2 4 | 5+6 2 |
| | 6 2 | 3+4 4 | 5 2 | 3+4 4 | 2+2 4 | 3+4 2 |
| | 6 2 | 3+4 4 | 5 2 | 3+4 4 | 2+2 4 | 5+6 2 |
| | 5 2 | 3+4 4 | 6 2 | 3+4 4 | 2+2 4 | 3+4 2 |
| | 6 2 | 3+4 4 | 5 2 | 3+4 4 | 2+2 4 | 5+6 2 |
| | 5 2 | 3+4 4 | 6 2 | 3+4 4 | 2+2 4 | 3+4 2 |
| | 5 2 | 3+4 4 | 6 2 | 3+4 4 | 2+2 4 | 5+6 2 |
| | 6 2 | 3+4 4 | 5 2 | 3+4 4 | 2+2 4 | 3+4 2 |
| | 5 2 | | 5 2 | | | 5+6 2 |
| | 6 2 | | 6 2 | | | 3+4 2 |
| | 6 2 | | 5 2 | | | 5+6 2 |
| | 5 2 | | 5 2 | | | 5+6 2 |
| | 5 2 | | 6 2 | | | 3+4 2 |
| | 6 2 | | 6 2 | | | 5+6 2 |
| | 6 2 | | 5 2 | | | 3+4 2 |
| | 5 2 | | 6 2 | | | 5+6 2 |
| | 5 2 | | 5 2 | | | 3+4 2 |
| | 6 2 | | 6 2 | | | 5+6 2 |
| | 6 2 | | 5 2 | | | 3+4 2 |
| | 5 2 | | 5 2 | | | 5+6 2 |
| | 6 2 | | 5 2 | | | 3+4 2 |
| | 6 2 | | 6 2 | | | 5+6 2 |
| | 5 2 | | 5 2 | | | 3+4 2 |
| | 6 2 | | 6 2 | | | 5+6 2 |
| | 5 2 | | 5 2 | | | 3+4 2 |
| | 6 2 | | 6 2 | | | 5+6 2 |
| | 5 2 | | 5 2 | | | 3+4 2 |
| | 6 2 | | 6 2 | | | 3+4 2 |
| | 5 2 | | 5 2 | | | 5+6 2 |
| | 6 2 | | 6 2 | | | 3+4 2 |

Continued on next column

**TABLE XXXVIII– continued**

| TQC | ICSD | | NoE | SF | | |
|---|---|---|---|---|---|---|
| | 6 2 | | 5 2 | | | 5+6 2 |
| | 5 2 | | 6 2 | | | 3+4 2 |
| | 5 2 | | 6 2 | | | 5+6 2 |
| | 5 2 | | 6 2 | | | 3+4 2 |
| | 6 2 | | 5 2 | | | 5+6 2 |
| | 5 2 | | 6 2 | | | 3+4 2 |
| | 5 2 | | 5 2 | | | 3+4 2 |
| | 6 2 | | 5 2 | | | 5+6 2 |
| | 6 2 | | 6 2 | | | 5+6 2 |
| | 5 2 | | 6 2 | | | 3+4 2 |
| | 5 2 | | 5 2 | | | 5+6 2 |
| | 6 2 | | 5 2 | | | 3+4 2 |
| | 5 2 | | 6 2 | | | 5+6 2 |
| | 5 2 | | 5 2 | | | 3+4 2 |
| | 6 2 | | 6 2 | | | 3+4 2 |
| | 5 2 | | 6 2 | | | 5+6 2 |
| | 5 2 | | 5 2 | | | 3+4 2 |
| | 6 2 | | 6 2 | | | 3+4 2 |
| | 5 2 | | 5 2 | | | 5+6 2 |
| | 5 2 | | 6 2 | | | 3+4 2 |
| | 6 2 | | 5 2 | | | 5+6 2 |
| 63.1.17 | 423611 | | 72 | La2Li2Si3 | | |
| | -GM | -T | -Y | -Z | -R | -S |
| | 5 2 | 3+4 4 | 5 2 | 3+4 4 | 2+2 4 | 3+4 2 |
| | 6 2 | 3+4 4 | 6 2 | 3+4 4 | 2+2 4 | 5+6 2 |
| | 5 2 | 3+4 4 | 5 2 | 3+4 4 | 2+2 4 | 3+4 2 |
| | 6 2 | 3+4 4 | 6 2 | 3+4 4 | 2+2 4 | 5+6 2 |
| | 6 2 | 3+4 4 | 6 2 | 3+4 4 | 2+2 4 | 5+6 2 |
| | 5 2 | 3+4 4 | 5 2 | 3+4 4 | 2+2 4 | 3+4 2 |
| | 5 2 | 3+4 4 | 5 2 | 3+4 4 | 2+2 4 | 3+4 2 |
| | 6 2 | 3+4 4 | 6 2 | 3+4 4 | 2+2 4 | 5+6 2 |
| | 5 2 | 3+4 4 | 5 2 | 3+4 4 | 2+2 4 | 3+4 2 |
| | 6 2 | 3+4 4 | 6 2 | 3+4 4 | 2+2 4 | 5+6 2 |
| | 5 2 | 3+4 4 | 5 2 | 3+4 4 | 2+2 4 | 3+4 2 |
| | 6 2 | 3+4 4 | 6 2 | 3+4 4 | 2+2 4 | 5+6 2 |
| | 5 2 | 3+4 4 | 5 2 | 3+4 4 | 2+2 4 | 5+6 2 |
| | 6 2 | 3+4 4 | 6 2 | 3+4 4 | 2+2 4 | 3+4 2 |
| | 5 2 | 3+4 4 | 5 2 | 3+4 4 | 2+2 4 | 3+4 2 |
| | 6 2 | 3+4 4 | 6 2 | 3+4 4 | 2+2 4 | 5+6 2 |
| | 5 2 | 3+4 4 | 5 2 | 3+4 4 | 2+2 4 | 5+6 2 |
| | 6 2 | 3+4 4 | 6 2 | 3+4 4 | 2+2 4 | 5+6 2 |
| | 5 2 | | 5 2 | | | 3+4 2 |
| | 5 2 | | 6 2 | | | 3+4 2 |
| | 6 2 | | 5 2 | | | 5+6 2 |
| | 6 2 | | 6 2 | | | 3+4 2 |
| | 6 2 | | 5 2 | | | 3+4 2 |
| | 5 2 | | 6 2 | | | 5+6 2 |
| | 5 2 | | 6 2 | | | 3+4 2 |
| | 6 2 | | 6 2 | | | 5+6 2 |
| | 6 2 | | 5 2 | | | 5+6 2 |
| | 5 2 | | 6 2 | | | 5+6 2 |
| | 6 2 | | 5 2 | | | 3+4 2 |
| | 5 2 | | 6 2 | | | 3+4 2 |
| | 5 2 | | 5 2 | | | 5+6 2 |
| | 5 2 | | 6 2 | | | 3+4 2 |
| | 5 2 | | 6 2 | | | 5+6 2 |
| | 5 2 | | 5 2 | | | 5+6 2 |
| | 6 2 | | 5 2 | | | 3+4 2 |
| | 5 2 | | 5 2 | | | 5+6 2 |
| 63.1.18 | 410732 | | 112 | La1Rh1Sn2 | | |
| | -GM | -T | -Y | -Z | -R | -S |

Continued on next column

**TABLE XXXVIII– continued**

| TQC | ICSD | | NoE | SF | | |
|---|---|---|---|---|---|---|
| | 5 2 | 3+4 4 | 5 2 | 3+4 4 | 2+2 4 | 5+6 2 |
| | 5 2 | 3+4 4 | 5 2 | 3+4 4 | 2+2 4 | 5+6 2 |
| | 5 2 | 3+4 4 | 6 2 | 3+4 4 | 2+2 4 | 3+4 2 |
| | 6 2 | 3+4 4 | 5 2 | 3+4 4 | 2+2 4 | 5+6 2 |
| | 6 2 | 3+4 4 | 6 2 | 3+4 4 | 2+2 4 | 3+4 2 |
| | 6 2 | 3+4 4 | 6 2 | 3+4 4 | 2+2 4 | 3+4 2 |
| | 5 2 | 3+4 4 | 6 2 | 3+4 4 | 2+2 4 | 3+4 2 |
| | 6 2 | 3+4 4 | 5 2 | 3+4 4 | 2+2 4 | 5+6 2 |
| | 6 2 | 3+4 4 | 6 2 | 3+4 4 | 2+2 4 | 3+4 2 |
| | 6 2 | 3+4 4 | 6 2 | 3+4 4 | 2+2 4 | 3+4 2 |
| | 6 2 | 3+4 4 | 6 2 | 3+4 4 | 2+2 4 | 3+4 2 |
| | 6 2 | 3+4 4 | 6 2 | 3+4 4 | 2+2 4 | 3+4 2 |
| | 5 2 | 3+4 4 | 6 2 | 3+4 4 | 2+2 4 | 5+6 2 |
| | 6 2 | 3+4 4 | 5 2 | 3+4 4 | 2+2 4 | 3+4 2 |
| | 5 2 | 3+4 4 | 6 2 | 3+4 4 | 2+2 4 | 5+6 2 |
| | 6 2 | 3+4 4 | 5 2 | 3+4 4 | 2+2 4 | 3+4 2 |
| | 5 2 | 3+4 4 | 6 2 | 3+4 4 | 2+2 4 | 5+6 2 |
| | 6 2 | 3+4 4 | 5 2 | 3+4 4 | 2+2 4 | 3+4 2 |
| | 5 2 | 3+4 4 | 6 2 | 3+4 4 | 2+2 4 | 3+4 2 |
| | 5 2 | 3+4 4 | 5 2 | 3+4 4 | 2+2 4 | 5+6 2 |
| | 6 2 | 3+4 4 | 6 2 | 3+4 4 | 2+2 4 | 5+6 2 |
| | 5 2 | 3+4 4 | 5 2 | 3+4 4 | 2+2 4 | 5+6 2 |
| | 6 2 | 3+4 4 | 6 2 | 3+4 4 | 2+2 4 | 3+4 2 |
| | 6 2 | 3+4 4 | 5 2 | 3+4 4 | 2+2 4 | 3+4 2 |
| | 5 2 | 3+4 4 | 5 2 | 3+4 4 | 2+2 4 | 3+4 2 |
| | 6 2 | 3+4 4 | 5 2 | 3+4 4 | 2+2 4 | 5+6 2 |
| | 5 2 | 3+4 4 | 6 2 | 3+4 4 | 2+2 4 | 5+6 2 |
| | 5 2 | 3+4 4 | 5 2 | 3+4 4 | 2+2 4 | 3+4 2 |
| | 5 2 | | 6 2 | | | 3+4 2 |
| | 5 2 | | 6 2 | | | 5+6 2 |
| | 5 2 | | 5 2 | | | 5+6 2 |
| | 6 2 | | 6 2 | | | 3+4 2 |
| | 6 2 | | 6 2 | | | 5+6 2 |
| | 6 2 | | 6 2 | | | 5+6 2 |
| | 6 2 | | 5 2 | | | 3+4 2 |
| | 5 2 | | 5 2 | | | 3+4 2 |
| | 5 2 | | 6 2 | | | 3+4 2 |
| | 5 2 | | 5 2 | | | 5+6 2 |
| | 5 2 | | 6 2 | | | 3+4 2 |
| | 6 2 | | 5 2 | | | 5+6 2 |
| | 6 2 | | 6 2 | | | 5+6 2 |
| | 5 2 | | 6 2 | | | 5+6 2 |
| | 6 2 | | 5 2 | | | 3+4 2 |
| | 5 2 | | 5 2 | | | 5+6 2 |
| | 6 2 | | 6 2 | | | 5+6 2 |
| | 6 2 | | 5 2 | | | 3+4 2 |
| | 6 2 | | 5 2 | | | 3+4 2 |
| | 6 2 | | 6 2 | | | 3+4 2 |
| | 5 2 | | 6 2 | | | 5+6 2 |
| | 6 2 | | 6 2 | | | 5+6 2 |
| | 5 2 | | 5 2 | | | 5+6 2 |
| | 5 2 | | 5 2 | | | 3+4 2 |
| | 5 2 | | 6 2 | | | 3+4 2 |
| | 5 2 | | 5 2 | | | 5+6 2 |
| | 6 2 | | 6 2 | | | 3+4 2 |
| | 5 2 | | 5 2 | | | 5+6 2 |
| 63.1.19 | 38311 | | 152 | La1Ni5P3 | | |
| | -GM | -T | -Y | -Z | -R | -S |
| | 5 2 | 3+4 4 | 5 2 | 3+4 4 | 2+2 4 | 3+4 2 |
| | 6 2 | 3+4 4 | 6 2 | 3+4 4 | 2+2 4 | 5+6 2 |
| | 5 2 | 3+4 4 | 6 2 | 3+4 4 | 2+2 4 | 3+4 2 |

Continued on next column

**TABLE XXXVIII– continued**

| TQC | ICSD | | NoE | | SF | |
|---|---|---|---|---|---|---|
| | 6 2 | 3+4 4 | 5 2 | 3+4 4 | 2+2 4 | 5+6 2 |
| | 5 2 | 3+4 4 | 5 2 | 3+4 4 | 2+2 4 | 3+4 2 |
| | 6 2 | 3+4 4 | 6 2 | 3+4 4 | 2+2 4 | 5+6 2 |
| | 5 2 | 3+4 4 | 5 2 | 3+4 4 | 2+2 4 | 5+6 2 |
| | 6 2 | 3+4 4 | 6 2 | 3+4 4 | 2+2 4 | 3+4 2 |
| | 5 2 | 3+4 4 | 5 2 | 3+4 4 | 2+2 4 | 3+4 2 |
| | 5 2 | 3+4 4 | 5 2 | 3+4 4 | 2+2 4 | 3+4 2 |
| | 6 2 | 3+4 4 | 6 2 | 3+4 4 | 2+2 4 | 5+6 2 |
| | 5 2 | 3+4 4 | 6 2 | 3+4 4 | 2+2 4 | 3+4 2 |
| | 6 2 | 3+4 4 | 5 2 | 3+4 4 | 2+2 4 | 5+6 2 |
| | 6 2 | 3+4 4 | 6 2 | 3+4 4 | 2+2 4 | 5+6 2 |
| | 5 2 | 3+4 4 | 6 2 | 3+4 4 | 2+2 4 | 5+6 2 |
| | 5 2 | 3+4 4 | 5 2 | 3+4 4 | 2+2 4 | 5+6 2 |
| | 6 2 | 3+4 4 | 5 2 | 3+4 4 | 2+2 4 | 5+6 2 |
| | 5 2 | 3+4 4 | 6 2 | 3+4 4 | 2+2 4 | 3+4 2 |
| | 6 2 | 3+4 4 | 5 2 | 3+4 4 | 2+2 4 | 3+4 2 |
| | 5 2 | 3+4 4 | 5 2 | 3+4 4 | 2+2 4 | 3+4 2 |
| | 5 2 | 3+4 4 | 6 2 | 3+4 4 | 2+2 4 | 5+6 2 |
| | 5 2 | 3+4 4 | 5 2 | 3+4 4 | 2+2 4 | 3+4 2 |
| | 6 2 | 3+4 4 | 5 2 | 3+4 4 | 2+2 4 | 5+6 2 |
| | 5 2 | 3+4 4 | 5 2 | 3+4 4 | 2+2 4 | 3+4 2 |
| | 6 2 | 3+4 4 | 5 2 | 3+4 4 | 2+2 4 | 5+6 2 |
| | 5 2 | 3+4 4 | 6 2 | 3+4 4 | 2+2 4 | 3+4 2 |
| | 6 2 | 3+4 4 | 5 2 | 3+4 4 | 2+2 4 | 3+4 2 |
| | 6 2 | 3+4 4 | 6 2 | 3+4 4 | 2+2 4 | 5+6 2 |
| | 5 2 | 3+4 4 | 6 2 | 3+4 4 | 2+2 4 | 3+4 2 |
| | 6 2 | 3+4 4 | 5 2 | 3+4 4 | 2+2 4 | 5+6 2 |
| | 6 2 | 3+4 4 | 6 2 | 3+4 4 | 2+2 4 | 5+6 2 |
| | 6 2 | 3+4 4 | 6 2 | 3+4 4 | 2+2 4 | 3+4 2 |
| | 5 2 | 3+4 4 | 6 2 | 3+4 4 | 2+2 4 | 5+6 2 |
| | 5 2 | 3+4 4 | 5 2 | 3+4 4 | 2+2 4 | 5+6 2 |
| | 5 2 | 3+4 4 | 5 2 | 3+4 4 | 2+2 4 | 3+4 2 |
| | 5 2 | 3+4 4 | 5 2 | 3+4 4 | 2+2 4 | 5+6 2 |
| | 6 2 | 3+4 4 | 6 2 | 3+4 4 | 2+2 4 | 5+6 2 |
| | 5 2 | 3+4 4 | 6 2 | 3+4 4 | 2+2 4 | 5+6 2 |
| | 5 2 | | 6 2 | | | 3+4 2 |
| | 5 2 | | 5 2 | | | 3+4 2 |
| | 5 2 | | 5 2 | | | 3+4 2 |
| | 6 2 | | 6 2 | | | 5+6 2 |
| | 6 2 | | 5 2 | | | 5+6 2 |
| | 5 2 | | 5 2 | | | 3+4 2 |
| | 6 2 | | 5 2 | | | 5+6 2 |
| | 5 2 | | 6 2 | | | 5+6 2 |
| | 6 2 | | 5 2 | | | 3+4 2 |
| | 6 2 | | 5 2 | | | 5+6 2 |
| | 5 2 | | 6 2 | | | 3+4 2 |
| | 6 2 | | 5 2 | | | 3+4 2 |
| | 5 2 | | 6 2 | | | 5+6 2 |
| | 6 2 | | 5 2 | | | 5+6 2 |
| | 6 2 | | 5 2 | | | 3+4 2 |
| | 5 2 | | 6 2 | | | 3+4 2 |
| | 6 2 | | 6 2 | | | 5+6 2 |
| | 5 2 | | 5 2 | | | 3+4 2 |
| | 5 2 | | 6 2 | | | 5+6 2 |
| | 5 2 | | 6 2 | | | 3+4 2 |
| | 5 2 | | 5 2 | | | 5+6 2 |
| | 5 2 | | 6 2 | | | 3+4 2 |
| | 6 2 | | 5 2 | | | 5+6 2 |
| | 5 2 | | 6 2 | | | 5+6 2 |
| | 6 2 | | 6 2 | | | 5+6 2 |
| | 5 2 | | 5 2 | | | 5+6 2 |

Continued on next column

**TABLE XXXVIII– continued**

| TQC | ICSD | | NoE | | SF | |
|---|---|---|---|---|---|---|
| | 5 2 | | 6 2 | | | 5+6 2 |
| | 6 2 | | 5 2 | | | 3+4 2 |
| | 6 2 | | 6 2 | | | 3+4 2 |
| | 5 2 | | 5 2 | | | 5+6 2 |
| | 6 2 | | 5 2 | | | 3+4 2 |
| | 5 2 | | 6 2 | | | 5+6 2 |
| | 5 2 | | 5 2 | | | 5+6 2 |
| | 6 2 | | 5 2 | | | 5+6 2 |
| | 5 2 | | 6 2 | | | 3+4 2 |
| | 5 2 | | 5 2 | | | 5+6 2 |
| | 6 2 | | 6 2 | | | 3+4 2 |
| | 6 2 | | 6 2 | | | 3+4 2 |
| 63.1.20 | 165363 | | 96 | | In5Y3 | |
| | -GM | -T | -Y | -Z | -R | -S |
| | 5 2 | 3+4 4 | 5 2 | 3+4 4 | 2+2 4 | 3+4 2 |
| | 5 2 | 3+4 4 | 5 2 | 3+4 4 | 2+2 4 | 3+4 2 |
| | 6 2 | 3+4 4 | 6 2 | 3+4 4 | 2+2 4 | 5+6 2 |
| | 6 2 | 3+4 4 | 6 2 | 3+4 4 | 2+2 4 | 5+6 2 |
| | 5 2 | 3+4 4 | 5 2 | 3+4 4 | 2+2 4 | 3+4 2 |
| | 6 2 | 3+4 4 | 6 2 | 3+4 4 | 2+2 4 | 5+6 2 |
| | 5 2 | 3+4 4 | 6 2 | 3+4 4 | 2+2 4 | 3+4 2 |
| | 5 2 | 3+4 4 | 6 2 | 3+4 4 | 2+2 4 | 3+4 2 |
| | 6 2 | 3+4 4 | 5 2 | 3+4 4 | 2+2 4 | 5+6 2 |
| | 6 2 | 3+4 4 | 5 2 | 3+4 4 | 2+2 4 | 5+6 2 |
| | 5 2 | 3+4 4 | 5 2 | 3+4 4 | 2+2 4 | 5+6 2 |
| | 6 2 | 3+4 4 | 6 2 | 3+4 4 | 2+2 4 | 3+4 2 |
| | 5 2 | 3+4 4 | 6 2 | 3+4 4 | 2+2 4 | 3+4 2 |
| | 5 2 | 3+4 4 | 6 2 | 3+4 4 | 2+2 4 | 5+6 2 |
| | 6 2 | 3+4 4 | 5 2 | 3+4 4 | 2+2 4 | 3+4 2 |
| | 6 2 | 3+4 4 | 5 2 | 3+4 4 | 2+2 4 | 5+6 2 |
| | 5 2 | 3+4 4 | 5 2 | 3+4 4 | 2+2 4 | 3+4 2 |
| | 5 2 | 3+4 4 | 6 2 | 3+4 4 | 2+2 4 | 3+4 2 |
| | 6 2 | 3+4 4 | 5 2 | 3+4 4 | 2+2 4 | 5+6 2 |
| | 6 2 | 3+4 4 | 6 2 | 3+4 4 | 2+2 4 | 5+6 2 |
| | 5 2 | 3+4 4 | 5 2 | 3+4 4 | 2+2 4 | 3+4 2 |
| | 6 2 | 3+4 4 | 6 2 | 3+4 4 | 2+2 4 | 5+6 2 |
| | 5 2 | 3+4 4 | 5 2 | 3+4 4 | 2+2 4 | 5+6 2 |
| | 6 2 | 3+4 4 | 6 2 | 3+4 4 | 2+2 4 | 3+4 2 |
| | 5 2 | | 5 2 | | | 3+4 2 |
| | 6 2 | | 6 2 | | | 5+6 2 |
| | 5 2 | | 6 2 | | | 5+6 2 |
| | 6 2 | | 5 2 | | | 3+4 2 |
| | 5 2 | | 5 2 | | | 5+6 2 |
| | 5 2 | | 5 2 | | | 3+4 2 |
| | 6 2 | | 6 2 | | | 5+6 2 |
| | 6 2 | | 5 2 | | | 3+4 2 |
| | 6 2 | | 6 2 | | | 5+6 2 |
| | 5 2 | | 6 2 | | | 3+4 2 |
| | 6 2 | | 5 2 | | | 3+4 2 |
| | 5 2 | | 6 2 | | | 5+6 2 |
| | 5 2 | | 6 2 | | | 3+4 2 |
| | 5 2 | | 5 2 | | | 5+6 2 |
| | 5 2 | | 5 2 | | | 5+6 2 |
| | 6 2 | | 6 2 | | | 5+6 2 |
| | 6 2 | | 5 2 | | | 5+6 2 |
| | 6 2 | | 6 2 | | | 3+4 2 |
| | 5 2 | | 5 2 | | | 5+6 2 |
| | 6 2 | | 5 2 | | | 3+4 2 |
| | 5 2 | | 6 2 | | | 3+4 2 |
| | 6 2 | | 6 2 | | | 3+4 2 |
| | 5 2 | | 5 2 | | | 5+6 2 |

Continued on next column

**TABLE XXXVIII– continued**

| TQC | ICSD | | NoE | SF | | |
|---|---|---|---|---|---|---|
| | 6 2 | | 6 2 | | | 5+6 2 |
| 63.1.21 | 165362 | | 96 | In5La3 | | |
| | -GM | -T | -Y | -Z | -R | -S |
| | 5 2 | 3+4 4 | 6 2 | 3+4 4 | 2+2 4 | 3+4 2 |
| | 5 2 | 3+4 4 | 6 2 | 3+4 4 | 2+2 4 | 3+4 2 |
| | 6 2 | 3+4 4 | 5 2 | 3+4 4 | 2+2 4 | 5+6 2 |
| | 6 2 | 3+4 4 | 5 2 | 3+4 4 | 2+2 4 | 5+6 2 |
| | 6 2 | 3+4 4 | 5 2 | 3+4 4 | 2+2 4 | 3+4 2 |
| | 5 2 | 3+4 4 | 6 2 | 3+4 4 | 2+2 4 | 5+6 2 |
| | 5 2 | 3+4 4 | 6 2 | 3+4 4 | 2+2 4 | 3+4 2 |
| | 5 2 | 3+4 4 | 6 2 | 3+4 4 | 2+2 4 | 3+4 2 |
| | 6 2 | 3+4 4 | 5 2 | 3+4 4 | 2+2 4 | 5+6 2 |
| | 6 2 | 3+4 4 | 5 2 | 3+4 4 | 2+2 4 | 5+6 2 |
| | 5 2 | 3+4 4 | 5 2 | 3+4 4 | 2+2 4 | 5+6 2 |
| | 6 2 | 3+4 4 | 6 2 | 3+4 4 | 2+2 4 | 3+4 2 |
| | 5 2 | 3+4 4 | 5 2 | 3+4 4 | 2+2 4 | 3+4 2 |
| | 5 2 | 3+4 4 | 6 2 | 3+4 4 | 2+2 4 | 5+6 2 |
| | 6 2 | 3+4 4 | 6 2 | 3+4 4 | 2+2 4 | 3+4 2 |
| | 5 2 | 3+4 4 | 5 2 | 3+4 4 | 2+2 4 | 5+6 2 |
| | 6 2 | 3+4 4 | 5 2 | 3+4 4 | 2+2 4 | 5+6 2 |
| | 5 2 | 3+4 4 | 6 2 | 3+4 4 | 2+2 4 | 3+4 2 |
| | 6 2 | 3+4 4 | 6 2 | 3+4 4 | 2+2 4 | 3+4 2 |
| | 5 2 | 3+4 4 | 5 2 | 3+4 4 | 2+2 4 | 5+6 2 |
| | 5 2 | 3+4 4 | 5 2 | 3+4 4 | 2+2 4 | 3+4 2 |
| | 6 2 | 3+4 4 | 6 2 | 3+4 4 | 2+2 4 | 5+6 2 |
| | 6 2 | 3+4 4 | 6 2 | 3+4 4 | 2+2 4 | 3+4 2 |
| | 6 2 | 3+4 4 | 5 2 | 3+4 4 | 2+2 4 | 5+6 2 |
| | 5 2 | | 5 2 | | | 3+4 2 |
| | 6 2 | | 6 2 | | | 5+6 2 |
| | 5 2 | | 6 2 | | | 3+4 2 |
| | 6 2 | | 5 2 | | | 5+6 2 |
| | 6 2 | | 5 2 | | | 5+6 2 |
| | 5 2 | | 5 2 | | | 3+4 2 |
| | 5 2 | | 6 2 | | | 5+6 2 |
| | 6 2 | | 6 2 | | | 3+4 2 |
| | 6 2 | | 5 2 | | | 5+6 2 |
| | 5 2 | | 6 2 | | | 3+4 2 |
| | 6 2 | | 6 2 | | | 5+6 2 |
| | 5 2 | | 5 2 | | | 3+4 2 |
| | 5 2 | | 6 2 | | | 3+4 2 |
| | 5 2 | | 5 2 | | | 5+6 2 |
| | 6 2 | | 5 2 | | | 5+6 2 |
| | 5 2 | | 5 2 | | | 5+6 2 |
| | 5 2 | | 6 2 | | | 3+4 2 |
| | 6 2 | | 5 2 | | | 5+6 2 |
| | 6 2 | | 6 2 | | | 3+4 2 |
| | 6 2 | | 5 2 | | | 5+6 2 |
| | 6 2 | | 6 2 | | | 5+6 2 |
| | 5 2 | | 6 2 | | | 3+4 2 |
| | 6 2 | | 5 2 | | | 3+4 2 |
| | 6 2 | | 5 2 | | | 5+6 2 |
| 63.1.22 | 418178 | | 64 | In4Ni1Sr1 | | |
| | -GM | -T | -Y | -Z | -R | -S |
| | 5 2 | 3+4 4 | 5 2 | 3+4 4 | 2+2 4 | 3+4 2 |
| | 6 2 | 3+4 4 | 6 2 | 3+4 4 | 2+2 4 | 5+6 2 |
| | 5 2 | 3+4 4 | 5 2 | 3+4 4 | 2+2 4 | 3+4 2 |
| | 6 2 | 3+4 4 | 6 2 | 3+4 4 | 2+2 4 | 5+6 2 |
| | 5 2 | 3+4 4 | 5 2 | 3+4 4 | 2+2 4 | 3+4 2 |
| | 6 2 | 3+4 4 | 6 2 | 3+4 4 | 2+2 4 | 5+6 2 |
| | 5 2 | 3+4 4 | 6 2 | 3+4 4 | 2+2 4 | 3+4 2 |
| | 6 2 | 3+4 4 | 5 2 | 3+4 4 | 2+2 4 | 5+6 2 |

Continued on next column

**TABLE XXXVIII– continued**

| TQC | ICSD | | NoE | SF | | |
|---|---|---|---|---|---|---|
| | 5 2 | 3+4 4 | 6 2 | 3+4 4 | 2+2 4 | 5+6 2 |
| | 6 2 | 3+4 4 | 5 2 | 3+4 4 | 2+2 4 | 3+4 2 |
| | 5 2 | 3+4 4 | 6 2 | 3+4 4 | 2+2 4 | 3+4 2 |
| | 5 2 | 3+4 4 | 6 2 | 3+4 4 | 2+2 4 | 5+6 2 |
| | 5 2 | 3+4 4 | 5 2 | 3+4 4 | 2+2 4 | 5+6 2 |
| | 6 2 | 3+4 4 | 5 2 | 3+4 4 | 2+2 4 | 5+6 2 |
| | 6 2 | 3+4 4 | 5 2 | 3+4 4 | 2+2 4 | 3+4 2 |
| | 6 2 | 3+4 4 | 6 2 | 3+4 4 | 2+2 4 | 3+4 2 |
| | 5 2 | | 5 2 | | | 5+6 2 |
| | 5 2 | | 5 2 | | | 5+6 2 |
| | 5 2 | | 6 2 | | | 3+4 2 |
| | 6 2 | | 6 2 | | | 3+4 2 |
| | 6 2 | | 6 2 | | | 5+6 2 |
| | 5 2 | | 5 2 | | | 3+4 2 |
| | 6 2 | | 6 2 | | | 5+6 2 |
| | 5 2 | | 5 2 | | | 3+4 2 |
| | 6 2 | | 5 2 | | | 5+6 2 |
| | 5 2 | | 6 2 | | | 3+4 2 |
| | 6 2 | | 6 2 | | | 3+4 2 |
| | 5 2 | | 6 2 | | | 5+6 2 |
| | 6 2 | | 6 2 | | | 3+4 2 |
| | 5 2 | | 5 2 | | | 3+4 2 |
| | 6 2 | | 5 2 | | | 5+6 2 |
| | 5 2 | | 6 2 | | | 3+4 2 |
| 63.1.23 | 424100 | | 80 | In2Pt1Yb1 | | |
| | -GM | -T | -Y | -Z | -R | -S |
| | 5 2 | 3+4 4 | 5 2 | 3+4 4 | 2+2 4 | 3+4 2 |
| | 6 2 | 3+4 4 | 6 2 | 3+4 4 | 2+2 4 | 5+6 2 |
| | 5 2 | 3+4 4 | 5 2 | 3+4 4 | 2+2 4 | 5+6 2 |
| | 6 2 | 3+4 4 | 6 2 | 3+4 4 | 2+2 4 | 3+4 2 |
| | 5 2 | 3+4 4 | 5 2 | 3+4 4 | 2+2 4 | 5+6 2 |
| | 6 2 | 3+4 4 | 6 2 | 3+4 4 | 2+2 4 | 3+4 2 |
| | 5 2 | 3+4 4 | 6 2 | 3+4 4 | 2+2 4 | 3+4 2 |
| | 6 2 | 3+4 4 | 5 2 | 3+4 4 | 2+2 4 | 5+6 2 |
| | 5 2 | 3+4 4 | 6 2 | 3+4 4 | 2+2 4 | 3+4 2 |
| | 5 2 | 3+4 4 | 5 2 | 3+4 4 | 2+2 4 | 5+6 2 |
| | 6 2 | 3+4 4 | 6 2 | 3+4 4 | 2+2 4 | 3+4 2 |
| | 6 2 | 3+4 4 | 5 2 | 3+4 4 | 2+2 4 | 5+6 2 |
| | 5 2 | 3+4 4 | 5 2 | 3+4 4 | 2+2 4 | 5+6 2 |
| | 5 2 | 3+4 4 | 6 2 | 3+4 4 | 2+2 4 | 3+4 2 |
| | 6 2 | 3+4 4 | 6 2 | 3+4 4 | 2+2 4 | 3+4 2 |
| | 5 2 | 3+4 4 | 5 2 | 3+4 4 | 2+2 4 | 5+6 2 |
| | 5 2 | 3+4 4 | 6 2 | 3+4 4 | 2+2 4 | 3+4 2 |
| | 6 2 | 3+4 4 | 5 2 | 3+4 4 | 2+2 4 | 3+4 2 |
| | 6 2 | 3+4 4 | 5 2 | 3+4 4 | 2+2 4 | 5+6 2 |
| | 5 2 | 3+4 4 | 6 2 | 3+4 4 | 2+2 4 | 5+6 2 |
| | 6 2 | | 6 2 | | | 3+4 2 |
| | 5 2 | | 5 2 | | | 5+6 2 |
| | 6 2 | | 6 2 | | | 3+4 2 |
| | 5 2 | | 5 2 | | | 5+6 2 |
| | 6 2 | | 6 2 | | | 3+4 2 |
| | 5 2 | | 5 2 | | | 5+6 2 |
| | 6 2 | | 6 2 | | | 3+4 2 |
| | 6 2 | | 5 2 | | | 5+6 2 |
| | 5 2 | | 6 2 | | | 3+4 2 |
| | 5 2 | | 5 2 | | | 5+6 2 |
| | 5 2 | | 6 2 | | | 5+6 2 |
| | 6 2 | | 5 2 | | | 3+4 2 |
| | 6 2 | | 6 2 | | | 3+4 2 |
| | 5 2 | | 5 2 | | | 5+6 2 |
| | 6 2 | | 6 2 | | | 3+4 2 |

Continued on next column

**TABLE XXXVIII– continued**

| TQC | ICSD | | NoE | | SF | |
|---|---|---|---|---|---|---|
| | 6 2 | | 5 2 | | | 5+6 2 |
| | 5 2 | | 6 2 | | | 3+4 2 |
| | 5 2 | | 5 2 | | | 5+6 2 |
| | 5 2 | | 6 2 | | | 5+6 2 |
| | 5 2 | | 5 2 | | | 3+4 2 |
| 63.1.24 | 410437 | | 80 | | In2Pd1Yb1 | |
| | -GM | -T | -Y | -Z | -R | -S |
| | 5 2 | 3+4 4 | 5 2 | 3+4 4 | 2+2 4 | 3+4 2 |
| | 6 2 | 3+4 4 | 6 2 | 3+4 4 | 2+2 4 | 5+6 2 |
| | 5 2 | 3+4 4 | 5 2 | 3+4 4 | 2+2 4 | 3+4 2 |
| | 6 2 | 3+4 4 | 6 2 | 3+4 4 | 2+2 4 | 5+6 2 |
| | 5 2 | 3+4 4 | 5 2 | 3+4 4 | 2+2 4 | 3+4 2 |
| | 6 2 | 3+4 4 | 6 2 | 3+4 4 | 2+2 4 | 5+6 2 |
| | 6 2 | 3+4 4 | 6 2 | 3+4 4 | 2+2 4 | 3+4 2 |
| | 5 2 | 3+4 4 | 5 2 | 3+4 4 | 2+2 4 | 5+6 2 |
| | 5 2 | 3+4 4 | 6 2 | 3+4 4 | 2+2 4 | 5+6 2 |
| | 5 2 | 3+4 4 | 5 2 | 3+4 4 | 2+2 4 | 3+4 2 |
| | 6 2 | 3+4 4 | 6 2 | 3+4 4 | 2+2 4 | 5+6 2 |
| | 6 2 | 3+4 4 | 5 2 | 3+4 4 | 2+2 4 | 3+4 2 |
| | 5 2 | 3+4 4 | 5 2 | 3+4 4 | 2+2 4 | 3+4 2 |
| | 5 2 | 3+4 4 | 6 2 | 3+4 4 | 2+2 4 | 5+6 2 |
| | 6 2 | 3+4 4 | 6 2 | 3+4 4 | 2+2 4 | 5+6 2 |
| | 5 2 | 3+4 4 | 5 2 | 3+4 4 | 2+2 4 | 3+4 2 |
| | 5 2 | 3+4 4 | 6 2 | 3+4 4 | 2+2 4 | 5+6 2 |
| | 6 2 | 3+4 4 | 5 2 | 3+4 4 | 2+2 4 | 3+4 2 |
| | 6 2 | 3+4 4 | 6 2 | 3+4 4 | 2+2 4 | 3+4 2 |
| | 5 2 | 3+4 4 | 5 2 | 3+4 4 | 2+2 4 | 5+6 2 |
| | 6 2 | | 6 2 | | | 5+6 2 |
| | 6 2 | | 5 2 | | | 5+6 2 |
| | 5 2 | | 6 2 | | | 3+4 2 |
| | 5 2 | | 5 2 | | | 3+4 2 |
| | 6 2 | | 6 2 | | | 5+6 2 |
| | 5 2 | | 5 2 | | | 5+6 2 |
| | 6 2 | | 6 2 | | | 3+4 2 |
| | 6 2 | | 5 2 | | | 3+4 2 |
| | 5 2 | | 6 2 | | | 5+6 2 |
| | 5 2 | | 5 2 | | | 3+4 2 |
| | 5 2 | | 6 2 | | | 3+4 2 |
| | 6 2 | | 5 2 | | | 5+6 2 |
| | 5 2 | | 6 2 | | | 5+6 2 |
| | 6 2 | | 5 2 | | | 3+4 2 |
| | 6 2 | | 6 2 | | | 5+6 2 |
| | 6 2 | | 5 2 | | | 3+4 2 |
| | 5 2 | | 5 2 | | | 3+4 2 |
| | 5 2 | | 6 2 | | | 5+6 2 |
| | 5 2 | | 6 2 | | | 3+4 2 |
| | 5 2 | | 5 2 | | | 5+6 2 |
| 63.1.25 | 414077 | | 80 | | In2Ni1Yb1 | |
| | -GM | -T | -Y | -Z | -R | -S |
| | 5 2 | 3+4 4 | 5 2 | 3+4 4 | 2+2 4 | 3+4 2 |
| | 6 2 | 3+4 4 | 6 2 | 3+4 4 | 2+2 4 | 5+6 2 |
| | 5 2 | 3+4 4 | 5 2 | 3+4 4 | 2+2 4 | 3+4 2 |
| | 6 2 | 3+4 4 | 6 2 | 3+4 4 | 2+2 4 | 5+6 2 |
| | 5 2 | 3+4 4 | 5 2 | 3+4 4 | 2+2 4 | 3+4 2 |
| | 6 2 | 3+4 4 | 6 2 | 3+4 4 | 2+2 4 | 5+6 2 |
| | 6 2 | 3+4 4 | 6 2 | 3+4 4 | 2+2 4 | 3+4 2 |
| | 5 2 | 3+4 4 | 5 2 | 3+4 4 | 2+2 4 | 5+6 2 |
| | 5 2 | 3+4 4 | 6 2 | 3+4 4 | 2+2 4 | 5+6 2 |
| | 5 2 | 3+4 4 | 5 2 | 3+4 4 | 2+2 4 | 3+4 2 |
| | 6 2 | 3+4 4 | 6 2 | 3+4 4 | 2+2 4 | 5+6 2 |
| | 5 2 | 3+4 4 | 5 2 | 3+4 4 | 2+2 4 | 3+4 2 |

Continued on next column

**TABLE XXXVIII– continued**

| TQC | ICSD | | NoE | | SF | |
|---|---|---|---|---|---|---|
| | 6 2 | 3+4 4 | 5 2 | 3+4 4 | 2+2 4 | 3+4 2 |
| | 6 2 | 3+4 4 | 6 2 | 3+4 4 | 2+2 4 | 5+6 2 |
| | 5 2 | 3+4 4 | 6 2 | 3+4 4 | 2+2 4 | 5+6 2 |
| | 5 2 | 3+4 4 | 5 2 | 3+4 4 | 2+2 4 | 5+6 2 |
| | 5 2 | 3+4 4 | 6 2 | 3+4 4 | 2+2 4 | 3+4 2 |
| | 6 2 | 3+4 4 | 5 2 | 3+4 4 | 2+2 4 | 3+4 2 |
| | 5 2 | 3+4 4 | 6 2 | 3+4 4 | 2+2 4 | 3+4 2 |
| | 6 2 | 3+4 4 | 6 2 | 3+4 4 | 2+2 4 | 5+6 2 |
| | 5 2 | | 6 2 | | | 5+6 2 |
| | 6 2 | | 5 2 | | | 3+4 2 |
| | 5 2 | | 5 2 | | | 5+6 2 |
| | 6 2 | | 6 2 | | | 3+4 2 |
| | 6 2 | | 5 2 | | | 5+6 2 |
| | 5 2 | | 5 2 | | | 3+4 2 |
| | 6 2 | | 6 2 | | | 3+4 2 |
| | 6 2 | | 5 2 | | | 5+6 2 |
| | 5 2 | | 6 2 | | | 5+6 2 |
| | 5 2 | | 5 2 | | | 3+4 2 |
| | 5 2 | | 6 2 | | | 3+4 2 |
| | 6 2 | | 5 2 | | | 5+6 2 |
| | 5 2 | | 6 2 | | | 5+6 2 |
| | 6 2 | | 5 2 | | | 3+4 2 |
| | 6 2 | | 6 2 | | | 5+6 2 |
| | 6 2 | | 5 2 | | | 3+4 2 |
| | 5 2 | | 6 2 | | | 3+4 2 |
| | 5 2 | | 5 2 | | | 5+6 2 |
| | 5 2 | | 6 2 | | | 3+4 2 |
| | 5 2 | | 5 2 | | | 5+6 2 |
| 63.1.26 | 412431 | | 52 | | In2La1Rh1 | |
| | -GM | -T | -Y | -Z | -R | -S |
| | 5 2 | 3+4 4 | 5 2 | 3+4 4 | 2+2 4 | 3+4 2 |
| | 6 2 | 3+4 4 | 6 2 | 3+4 4 | 2+2 4 | 5+6 2 |
| | 5 2 | 3+4 4 | 5 2 | 3+4 4 | 2+2 4 | 3+4 2 |
| | 6 2 | 3+4 4 | 6 2 | 3+4 4 | 2+2 4 | 5+6 2 |
| | 5 2 | 3+4 4 | 5 2 | 3+4 4 | 2+2 4 | 3+4 2 |
| | 6 2 | 3+4 4 | 6 2 | 3+4 4 | 2+2 4 | 5+6 2 |
| | 5 2 | 3+4 4 | 5 2 | 3+4 4 | 2+2 4 | 3+4 2 |
| | 6 2 | 3+4 4 | 6 2 | 3+4 4 | 2+2 4 | 5+6 2 |
| | 5 2 | 3+4 4 | 6 2 | 3+4 4 | 2+2 4 | 5+6 2 |
| | 5 2 | 3+4 4 | 5 2 | 3+4 4 | 2+2 4 | 3+4 2 |
| | 6 2 | 3+4 4 | 6 2 | 3+4 4 | 2+2 4 | 5+6 2 |
| | 6 2 | 3+4 4 | 5 2 | 3+4 4 | 2+2 4 | 3+4 2 |
| | 5 2 | 3+4 4 | 6 2 | 3+4 4 | 2+2 4 | 5+6 2 |
| | 5 2 | | 5 2 | | | 3+4 2 |
| | 5 2 | | 6 2 | | | 5+6 2 |
| | 6 2 | | 5 2 | | | 5+6 2 |
| | 5 2 | | 6 2 | | | 3+4 2 |
| | 6 2 | | 6 2 | | | 3+4 2 |
| | 6 2 | | 5 2 | | | 5+6 2 |
| | 5 2 | | 6 2 | | | 3+4 2 |
| | 6 2 | | 5 2 | | | 5+6 2 |
| | 5 2 | | 5 2 | | | 5+6 2 |
| | 5 2 | | 6 2 | | | 3+4 2 |
| | 6 2 | | 5 2 | | | 3+4 2 |
| | 5 2 | | 6 2 | | | 5+6 2 |
| | 5 2 | | 5 2 | | | 3+4 2 |
| 63.1.27 | 414479 | | 52 | | In2Ir1La1 | |
| | -GM | -T | -Y | -Z | -R | -S |
| | 5 2 | 3+4 4 | 5 2 | 3+4 4 | 2+2 4 | 3+4 2 |
| | 6 2 | 3+4 4 | 6 2 | 3+4 4 | 2+2 4 | 5+6 2 |
| | 5 2 | 3+4 4 | 5 2 | 3+4 4 | 2+2 4 | 3+4 2 |

Continued on next column

**TABLE XXXVIII– continued**

| TQC | ICSD | | NoE | SF | | |
|---|---|---|---|---|---|---|
| | 6 2 | 3+4 4 | 6 2 | 3+4 4 | 2+2 4 | 5+6 2 |
| | 5 2 | 3+4 4 | 5 2 | 3+4 4 | 2+2 4 | 3+4 2 |
| | 6 2 | 3+4 4 | 6 2 | 3+4 4 | 2+2 4 | 5+6 2 |
| | 5 2 | 3+4 4 | 5 2 | 3+4 4 | 2+2 4 | 3+4 2 |
| | 6 2 | 3+4 4 | 6 2 | 3+4 4 | 2+2 4 | 5+6 2 |
| | 5 2 | 3+4 4 | 6 2 | 3+4 4 | 2+2 4 | 5+6 2 |
| | 5 2 | 3+4 4 | 5 2 | 3+4 4 | 2+2 4 | 3+4 2 |
| | 6 2 | 3+4 4 | 6 2 | 3+4 4 | 2+2 4 | 5+6 2 |
| | 6 2 | 3+4 4 | 5 2 | 3+4 4 | 2+2 4 | 3+4 2 |
| | 5 2 | 3+4 4 | 5 2 | 3+4 4 | 2+2 4 | 5+6 2 |
| | 5 2 | | 6 2 | | | 3+4 2 |
| | 6 2 | | 6 2 | | | 5+6 2 |
| | 5 2 | | 5 2 | | | 3+4 2 |
| | 5 2 | | 6 2 | | | 5+6 2 |
| | 6 2 | | 5 2 | | | 3+4 2 |
| | 6 2 | | 6 2 | | | 5+6 2 |
| | 5 2 | | 5 2 | | | 3+4 2 |
| | 6 2 | | 6 2 | | | 5+6 2 |
| | 5 2 | | 5 2 | | | 3+4 2 |
| | 5 2 | | 6 2 | | | 5+6 2 |
| | 6 2 | | 5 2 | | | 3+4 2 |
| | 5 2 | | 6 2 | | | 5+6 2 |
| | 5 2 | | 5 2 | | | 3+4 2 |
| 63.1.28 | 200067 | | 88 | Hf3Ni2Si3 | | |
| | -GM | -T | -Y | -Z | -R | -S |
| | 5 2 | 3+4 4 | 5 2 | 3+4 4 | 2+2 4 | 3+4 2 |
| | 5 2 | 3+4 4 | 5 2 | 3+4 4 | 2+2 4 | 3+4 2 |
| | 6 2 | 3+4 4 | 6 2 | 3+4 4 | 2+2 4 | 5+6 2 |
| | 5 2 | 3+4 4 | 6 2 | 3+4 4 | 2+2 4 | 3+4 2 |
| | 6 2 | 3+4 4 | 5 2 | 3+4 4 | 2+2 4 | 5+6 2 |
| | 6 2 | 3+4 4 | 6 2 | 3+4 4 | 2+2 4 | 5+6 2 |
| | 5 2 | 3+4 4 | 6 2 | 3+4 4 | 2+2 4 | 5+6 2 |
| | 5 2 | 3+4 4 | 6 2 | 3+4 4 | 2+2 4 | 5+6 2 |
| | 6 2 | 3+4 4 | 5 2 | 3+4 4 | 2+2 4 | 3+4 2 |
| | 6 2 | 3+4 4 | 5 2 | 3+4 4 | 2+2 4 | 3+4 2 |
| | 6 2 | 3+4 4 | 5 2 | 3+4 4 | 2+2 4 | 3+4 2 |
| | 6 2 | 3+4 4 | 6 2 | 3+4 4 | 2+2 4 | 5+6 2 |
| | 6 2 | 3+4 4 | 6 2 | 3+4 4 | 2+2 4 | 3+4 2 |
| | 5 2 | 3+4 4 | 5 2 | 3+4 4 | 2+2 4 | 5+6 2 |
| | 5 2 | 3+4 4 | 5 2 | 3+4 4 | 2+2 4 | 3+4 2 |
| | 6 2 | 3+4 4 | 5 2 | 3+4 4 | 2+2 4 | 5+6 2 |
| | 5 2 | 3+4 4 | 5 2 | 3+4 4 | 2+2 4 | 3+4 2 |
| | 5 2 | 3+4 4 | 6 2 | 3+4 4 | 2+2 4 | 5+6 2 |
| | 6 2 | 3+4 4 | 5 2 | 3+4 4 | 2+2 4 | 3+4 2 |
| | 5 2 | 3+4 4 | 6 2 | 3+4 4 | 2+2 4 | 5+6 2 |
| | 5 2 | 3+4 4 | 6 2 | 3+4 4 | 2+2 4 | 3+4 2 |
| | 6 2 | 3+4 4 | 6 2 | 3+4 4 | 2+2 4 | 5+6 2 |
| | 5 2 | | 6 2 | | | 3+4 2 |
| | 6 2 | | 6 2 | | | 5+6 2 |
| | 5 2 | | 5 2 | | | 3+4 2 |
| | 5 2 | | 5 2 | | | 3+4 2 |
| | 5 2 | | 6 2 | | | 5+6 2 |
| | 6 2 | | 5 2 | | | 3+4 2 |
| | 6 2 | | 6 2 | | | 5+6 2 |
| | 5 2 | | 5 2 | | | 5+6 2 |
| | 5 2 | | 6 2 | | | 3+4 2 |
| | 6 2 | | 6 2 | | | 3+4 2 |
| | 5 2 | | 5 2 | | | 3+4 2 |
| | 5 2 | | 6 2 | | | 5+6 2 |
| | 6 2 | | 6 2 | | | 5+6 2 |
| | 5 2 | | 6 2 | | | 5+6 2 |

Continued on next column

**TABLE XXXVIII– continued**

| TQC | ICSD | | NoE | SF | | |
|---|---|---|---|---|---|---|
| | 6 2 | | 5 2 | | | 3+4 2 |
| | 5 2 | | 5 2 | | | 3+4 2 |
| | 6 2 | | 6 2 | | | 5+6 2 |
| | 5 2 | | 5 2 | | | 5+6 2 |
| | 6 2 | | 5 2 | | | 3+4 2 |
| | 5 2 | | 5 2 | | | 3+4 2 |
| | 6 2 | | 6 2 | | | 5+6 2 |
| | 5 2 | | 5 2 | | | 5+6 2 |
| 63.1.29 | 26225 | | 36 | H5Li4Rh1 | | |
| | -GM | -T | -Y | -Z | -R | -S |
| | 5 2 | 3+4 4 | 6 2 | 3+4 4 | 2+2 4 | 3+4 2 |
| | 6 2 | 3+4 4 | 5 2 | 3+4 4 | 2+2 4 | 5+6 2 |
| | 5 2 | 3+4 4 | 5 2 | 3+4 4 | 2+2 4 | 3+4 2 |
| | 5 2 | 3+4 4 | 5 2 | 3+4 4 | 2+2 4 | 5+6 2 |
| | 6 2 | 3+4 4 | 6 2 | 3+4 4 | 2+2 4 | 5+6 2 |
| | 6 2 | 3+4 4 | 6 2 | 3+4 4 | 2+2 4 | 3+4 2 |
| | 5 2 | 3+4 4 | 5 2 | 3+4 4 | 2+2 4 | 3+4 2 |
| | 6 2 | 3+4 4 | 6 2 | 3+4 4 | 2+2 4 | 3+4 2 |
| | 5 2 | 3+4 4 | 6 2 | 3+4 4 | 2+2 4 | 5+6 2 |
| | 6 2 | | 5 2 | | | 5+6 2 |
| | 5 2 | | 6 2 | | | 3+4 2 |
| | 6 2 | | 6 2 | | | 5+6 2 |
| | 5 2 | | 5 2 | | | 3+4 2 |
| | 5 2 | | 5 2 | | | 5+6 2 |
| | 6 2 | | 6 2 | | | 3+4 2 |
| | 6 2 | | 5 2 | | | 5+6 2 |
| | 5 2 | | 5 2 | | | 3+4 2 |
| | 5 2 | | 5 2 | | | 5+6 2 |
| 63.1.30 | 262431 | | 72 | Ge3La2Li2 | | |
| | -GM | -T | -Y | -Z | -R | -S |
| | 5 2 | 3+4 4 | 5 2 | 3+4 4 | 2+2 4 | 5+6 2 |
| | 6 2 | 3+4 4 | 6 2 | 3+4 4 | 2+2 4 | 3+4 2 |
| | 5 2 | 3+4 4 | 5 2 | 3+4 4 | 2+2 4 | 5+6 2 |
| | 6 2 | 3+4 4 | 6 2 | 3+4 4 | 2+2 4 | 3+4 2 |
| | 6 2 | 3+4 4 | 6 2 | 3+4 4 | 2+2 4 | 3+4 2 |
| | 5 2 | 3+4 4 | 5 2 | 3+4 4 | 2+2 4 | 5+6 2 |
| | 5 2 | 3+4 4 | 5 2 | 3+4 4 | 2+2 4 | 5+6 2 |
| | 6 2 | 3+4 4 | 6 2 | 3+4 4 | 2+2 4 | 3+4 2 |
| | 5 2 | 3+4 4 | 5 2 | 3+4 4 | 2+2 4 | 5+6 2 |
| | 6 2 | 3+4 4 | 6 2 | 3+4 4 | 2+2 4 | 3+4 2 |
| | 5 2 | 3+4 4 | 5 2 | 3+4 4 | 2+2 4 | 5+6 2 |
| | 6 2 | 3+4 4 | 6 2 | 3+4 4 | 2+2 4 | 3+4 2 |
| | 5 2 | 3+4 4 | 6 2 | 3+4 4 | 2+2 4 | 3+4 2 |
| | 6 2 | 3+4 4 | 5 2 | 3+4 4 | 2+2 4 | 5+6 2 |
| | 5 2 | 3+4 4 | 5 2 | 3+4 4 | 2+2 4 | 5+6 2 |
| | 6 2 | 3+4 4 | 6 2 | 3+4 4 | 2+2 4 | 3+4 2 |
| | 5 2 | 3+4 4 | 5 2 | 3+4 4 | 2+2 4 | 3+4 2 |
| | 6 2 | 3+4 4 | 6 2 | 3+4 4 | 2+2 4 | 3+4 2 |
| | 5 2 | | 5 2 | | | 5+6 2 |
| | 5 2 | | 6 2 | | | 5+6 2 |
| | 6 2 | | 5 2 | | | 3+4 2 |
| | 6 2 | | 6 2 | | | 5+6 2 |
| | 6 2 | | 6 2 | | | 5+6 2 |
| | 5 2 | | 5 2 | | | 3+4 2 |
| | 5 2 | | 6 2 | | | 5+6 2 |
| | 6 2 | | 5 2 | | | 3+4 2 |
| | 6 2 | | 6 2 | | | 3+4 2 |
| | 5 2 | | 6 2 | | | 3+4 2 |
| | 6 2 | | 5 2 | | | 5+6 2 |
| | 5 2 | | 6 2 | | | 5+6 2 |
| | 5 2 | | 5 2 | | | 3+4 2 |

Continued on next column

**TABLE XXXVIII– continued**

| TQC | ICSD | | NoE | | SF | |
|---|---|---|---|---|---|---|
| | 5 2 | | 6 2 | | | 5+6 2 |
| | 5 2 | | 6 2 | | | 3+4 2 |
| | 5 2 | | 5 2 | | | 3+4 2 |
| | 6 2 | | 5 2 | | | 5+6 2 |
| | 5 2 | | 5 2 | | | 3+4 2 |
| 63.1.31 | 42375 | | 36 | | Ge1Nd1 | |
| | -GM | -T | -Y | -Z | -R | -S |
| | 5 2 | 3+4 4 | 5 2 | 3+4 4 | 2+2 4 | 3+4 2 |
| | 6 2 | 3+4 4 | 6 2 | 3+4 4 | 2+2 4 | 5+6 2 |
| | 5 2 | 3+4 4 | 6 2 | 3+4 4 | 2+2 4 | 3+4 2 |
| | 6 2 | 3+4 4 | 5 2 | 3+4 4 | 2+2 4 | 5+6 2 |
| | 5 2 | 3+4 4 | 5 2 | 3+4 4 | 2+2 4 | 3+4 2 |
| | 5 2 | 3+4 4 | 6 2 | 3+4 4 | 2+2 4 | 5+6 2 |
| | 6 2 | 3+4 4 | 6 2 | 3+4 4 | 2+2 4 | 3+4 2 |
| | 6 2 | 3+4 4 | 5 2 | 3+4 4 | 2+2 4 | 5+6 2 |
| | 5 2 | 3+4 4 | 5 2 | 3+4 4 | 2+2 4 | 5+6 2 |
| | 6 2 | | 6 2 | | | 3+4 2 |
| | 5 2 | | 5 2 | | | 3+4 2 |
| | 6 2 | | 5 2 | | | 5+6 2 |
| | 5 2 | | 6 2 | | | 5+6 2 |
| | 6 2 | | 6 2 | | | 5+6 2 |
| | 5 2 | | 5 2 | | | 5+6 2 |
| | 6 2 | | 5 2 | | | 3+4 2 |
| | 6 2 | | 6 2 | | | 3+4 2 |
| | 6 2 | | 5 2 | | | 5+6 2 |
| 63.1.32 | 42374 | | 36 | | Ge1Nd1 | |
| | -GM | -T | -Y | -Z | -R | -S |
| | 5 2 | 3+4 4 | 5 2 | 3+4 4 | 2+2 4 | 3+4 2 |
| | 6 2 | 3+4 4 | 6 2 | 3+4 4 | 2+2 4 | 5+6 2 |
| | 5 2 | 3+4 4 | 6 2 | 3+4 4 | 2+2 4 | 3+4 2 |
| | 6 2 | 3+4 4 | 5 2 | 3+4 4 | 2+2 4 | 5+6 2 |
| | 5 2 | 3+4 4 | 5 2 | 3+4 4 | 2+2 4 | 3+4 2 |
| | 5 2 | 3+4 4 | 6 2 | 3+4 4 | 2+2 4 | 5+6 2 |
| | 6 2 | 3+4 4 | 6 2 | 3+4 4 | 2+2 4 | 3+4 2 |
| | 6 2 | 3+4 4 | 5 2 | 3+4 4 | 2+2 4 | 5+6 2 |
| | 5 2 | 3+4 4 | 5 2 | 3+4 4 | 2+2 4 | 5+6 2 |
| | 6 2 | | 6 2 | | | 3+4 2 |
| | 5 2 | | 5 2 | | | 3+4 2 |
| | 6 2 | | 6 2 | | | 5+6 2 |
| | 5 2 | | 5 2 | | | 5+6 2 |
| | 6 2 | | 6 2 | | | 5+6 2 |
| | 5 2 | | 5 2 | | | 5+6 2 |
| | 6 2 | | 5 2 | | | 3+4 2 |
| | 6 2 | | 6 2 | | | 3+4 2 |
| | 6 2 | | 5 2 | | | 5+6 2 |
| 63.1.33 | 170875 | | 208 | | Gd3O7Os1 | |
| | -GM | -T | -Y | -Z | -R | -S |
| | 5 2 | 3+4 4 | 5 2 | 3+4 4 | 2+2 4 | 3+4 2 |
| | 5 2 | 3+4 4 | 5 2 | 3+4 4 | 2+2 4 | 3+4 2 |
| | 5 2 | 3+4 4 | 6 2 | 3+4 4 | 2+2 4 | 5+6 2 |
| | 6 2 | 3+4 4 | 5 2 | 3+4 4 | 2+2 4 | 3+4 2 |
| | 5 2 | 3+4 4 | 6 2 | 3+4 4 | 2+2 4 | 3+4 2 |
| | 6 2 | 3+4 4 | 5 2 | 3+4 4 | 2+2 4 | 5+6 2 |
| | 6 2 | 3+4 4 | 6 2 | 3+4 4 | 2+2 4 | 5+6 2 |
| | 6 2 | 3+4 4 | 6 2 | 3+4 4 | 2+2 4 | 5+6 2 |
| | 5 2 | 3+4 4 | 6 2 | 3+4 4 | 2+2 4 | 3+4 2 |
| | 6 2 | 3+4 4 | 5 2 | 3+4 4 | 2+2 4 | 5+6 2 |
| | 5 2 | 3+4 4 | 6 2 | 3+4 4 | 2+2 4 | 3+4 2 |
| | 6 2 | 3+4 4 | 5 2 | 3+4 4 | 2+2 4 | 5+6 2 |
| | 5 2 | 3+4 4 | 6 2 | 3+4 4 | 2+2 4 | 5+6 2 |
| | 6 2 | 3+4 4 | 5 2 | 3+4 4 | 2+2 4 | 5+6 2 |

Continued on next column

**TABLE XXXVIII– continued**

| TQC | ICSD | | NoE | | SF | |
|---|---|---|---|---|---|---|
| | 5 2 | 3+4 4 | 5 2 | 3+4 4 | 2+2 4 | 3+4 2 |
| | 6 2 | 3+4 4 | 6 2 | 3+4 4 | 2+2 4 | 5+6 2 |
| | 5 2 | 3+4 4 | 6 2 | 3+4 4 | 2+2 4 | 5+6 2 |
| | 6 2 | 3+4 4 | 6 2 | 3+4 4 | 2+2 4 | 3+4 2 |
| | 5 2 | 3+4 4 | 5 2 | 3+4 4 | 2+2 4 | 3+4 2 |
| | 6 2 | 3+4 4 | 6 2 | 3+4 4 | 2+2 4 | 5+6 2 |
| | 6 2 | 3+4 4 | 6 2 | 3+4 4 | 2+2 4 | 5+6 2 |
| | 6 2 | 3+4 4 | 5 2 | 3+4 4 | 2+2 4 | 3+4 2 |
| | 5 2 | 3+4 4 | 6 2 | 3+4 4 | 2+2 4 | 5+6 2 |
| | 6 2 | 3+4 4 | 5 2 | 3+4 4 | 2+2 4 | 5+6 2 |
| | 6 2 | 3+4 4 | 6 2 | 3+4 4 | 2+2 4 | 5+6 2 |
| | 5 2 | 3+4 4 | 5 2 | 3+4 4 | 2+2 4 | 3+4 2 |
| | 6 2 | 3+4 4 | 6 2 | 3+4 4 | 2+2 4 | 5+6 2 |
| | 5 2 | 3+4 4 | 5 2 | 3+4 4 | 2+2 4 | 5+6 2 |
| | 5 2 | 3+4 4 | 6 2 | 3+4 4 | 2+2 4 | 3+4 2 |
| | 6 2 | 3+4 4 | 5 2 | 3+4 4 | 2+2 4 | 3+4 2 |
| | 6 2 | 3+4 4 | 6 2 | 3+4 4 | 2+2 4 | 5+6 2 |
| | 5 2 | 3+4 4 | 5 2 | 3+4 4 | 2+2 4 | 3+4 2 |
| | 5 2 | 3+4 4 | 5 2 | 3+4 4 | 2+2 4 | 5+6 2 |
| | 6 2 | 3+4 4 | 6 2 | 3+4 4 | 2+2 4 | 3+4 2 |
| | 6 2 | 3+4 4 | 6 2 | 3+4 4 | 2+2 4 | 3+4 2 |
| | 6 2 | 3+4 4 | 6 2 | 3+4 4 | 2+2 4 | 3+4 2 |
| | 5 2 | 3+4 4 | 5 2 | 3+4 4 | 2+2 4 | 5+6 2 |
| | 6 2 | 3+4 4 | 6 2 | 3+4 4 | 2+2 4 | 5+6 2 |
| | 5 2 | 3+4 4 | 5 2 | 3+4 4 | 2+2 4 | 5+6 2 |
| | 5 2 | 3+4 4 | 5 2 | 3+4 4 | 2+2 4 | 5+6 2 |
| | 5 2 | 3+4 4 | 5 2 | 3+4 4 | 2+2 4 | 5+6 2 |
| | 5 2 | 3+4 4 | 5 2 | 3+4 4 | 2+2 4 | 5+6 2 |
| | 5 2 | 3+4 4 | 5 2 | 3+4 4 | 2+2 4 | 5+6 2 |
| | 5 2 | 3+4 4 | 5 2 | 3+4 4 | 2+2 4 | 5+6 2 |
| | 5 2 | 3+4 4 | 5 2 | 3+4 4 | 2+2 4 | 5+6 2 |
| | 5 2 | 3+4 4 | 5 2 | 3+4 4 | 2+2 4 | 5+6 2 |
| | 5 2 | 3+4 4 | 6 2 | 3+4 4 | 2+2 4 | 5+6 2 |
| | 6 2 | 3+4 4 | 5 2 | 3+4 4 | 2+2 4 | 3+4 2 |
| | 6 2 | 3+4 4 | 6 2 | 3+4 4 | 2+2 4 | 3+4 2 |
| | 5 2 | 3+4 4 | 6 2 | 3+4 4 | 2+2 4 | 3+4 2 |
| | 5 2 | 3+4 4 | 5 2 | 3+4 4 | 2+2 4 | 3+4 2 |
| | 6 2 | 3+4 4 | 6 2 | 3+4 4 | 2+2 4 | 5+6 2 |
| | 6 2 | | 5 2 | | | 3+4 2 |
| | 5 2 | | 6 2 | | | 3+4 2 |
| | 5 2 | | 5 2 | | | 5+6 2 |
| | 6 2 | | 6 2 | | | 3+4 2 |
| | 5 2 | | 5 2 | | | 3+4 2 |
| | 6 2 | | 6 2 | | | 3+4 2 |
| | 6 2 | | 5 2 | | | 3+4 2 |
| | 5 2 | | 5 2 | | | 5+6 2 |
| | 6 2 | | 6 2 | | | 3+4 2 |
| | 6 2 | | 5 2 | | | 5+6 2 |
| | 6 2 | | 6 2 | | | 3+4 2 |
| | 5 2 | | 6 2 | | | 3+4 2 |
| | 6 2 | | 6 2 | | | 5+6 2 |
| | 5 2 | | 6 2 | | | 3+4 2 |
| | 6 2 | | 6 2 | | | 3+4 2 |
| | 6 2 | | 5 2 | | | 5+6 2 |
| | 5 2 | | 6 2 | | | 5+6 2 |
| | 6 2 | | 6 2 | | | 5+6 2 |
| | 6 2 | | 5 2 | | | 3+4 2 |
| | 5 2 | | 5 2 | | | 3+4 2 |
| | 6 2 | | 6 2 | | | 5+6 2 |
| | 6 2 | | 6 2 | | | 3+4 2 |
| | 5 2 | | 6 2 | | | 5+6 2 |

Continued on next column

**TABLE XXXVIII– continued**

| TQC | ICSD | NoE | SF | | | |
|---|---|---|---|---|---|---|
| | 6 2 | | 6 2 | | | 3+4 2 |
| | 6 2 | | 5 2 | | | 3+4 2 |
| | 6 2 | | 6 2 | | | 5+6 2 |
| | 6 2 | | 6 2 | | | 3+4 2 |
| | 5 2 | | 5 2 | | | 5+6 2 |
| | 5 2 | | 6 2 | | | 5+6 2 |
| | 6 2 | | 6 2 | | | 5+6 2 |
| | 6 2 | | 5 2 | | | 5+6 2 |
| | 5 2 | | 6 2 | | | 5+6 2 |
| | 6 2 | | 6 2 | | | 3+4 2 |
| | 6 2 | | 5 2 | | | 5+6 2 |
| | 5 2 | | 6 2 | | | 5+6 2 |
| | 6 2 | | 6 2 | | | 3+4 2 |
| | 6 2 | | 5 2 | | | 5+6 2 |
| | 5 2 | | 5 2 | | | 5+6 2 |
| | 6 2 | | 6 2 | | | 3+4 2 |
| | 5 2 | | 5 2 | | | 5+6 2 |
| | 6 2 | | 6 2 | | | 3+4 2 |
| | 5 2 | | 6 2 | | | 5+6 2 |
| | 6 2 | | 6 2 | | | 5+6 2 |
| | 6 2 | | 5 2 | | | 3+4 2 |
| | 5 2 | | 6 2 | | | 5+6 2 |
| | 6 2 | | 6 2 | | | 3+4 2 |
| | 5 2 | | 5 2 | | | 5+6 2 |
| | 6 2 | | 5 2 | | | 5+6 2 |
| | 5 2 | | 5 2 | | | 5+6 2 |
| | 5 2 | | 6 2 | | | 5+6 2 |
| | 6 2 | | 6 2 | | | 3+4 2 |
| | 6 2 | | 5 2 | | | 5+6 2 |
| 63.1.34 | 423067 | 80 | Ga4Gd1Ni1 | | | |
| | -GM | -T | -Y | -Z | -R | -S |
| | 5 2 | 3+4 4 | 5 2 | 3+4 4 | 2+2 4 | 3+4 2 |
| | 6 2 | 3+4 4 | 6 2 | 3+4 4 | 2+2 4 | 5+6 2 |
| | 5 2 | 3+4 4 | 5 2 | 3+4 4 | 2+2 4 | 3+4 2 |
| | 6 2 | 3+4 4 | 6 2 | 3+4 4 | 2+2 4 | 5+6 2 |
| | 5 2 | 3+4 4 | 5 2 | 3+4 4 | 2+2 4 | 3+4 2 |
| | 6 2 | 3+4 4 | 6 2 | 3+4 4 | 2+2 4 | 5+6 2 |
| | 5 2 | 3+4 4 | 6 2 | 3+4 4 | 2+2 4 | 3+4 2 |
| | 6 2 | 3+4 4 | 5 2 | 3+4 4 | 2+2 4 | 5+6 2 |
| | 5 2 | 3+4 4 | 6 2 | 3+4 4 | 2+2 4 | 5+6 2 |
| | 6 2 | 3+4 4 | 5 2 | 3+4 4 | 2+2 4 | 3+4 2 |
| | 5 2 | 3+4 4 | 6 2 | 3+4 4 | 2+2 4 | 3+4 2 |
| | 5 2 | 3+4 4 | 6 2 | 3+4 4 | 2+2 4 | 5+6 2 |
| | 5 2 | 3+4 4 | 5 2 | 3+4 4 | 2+2 4 | 5+6 2 |
| | 6 2 | 3+4 4 | 5 2 | 3+4 4 | 2+2 4 | 5+6 2 |
| | 6 2 | 3+4 4 | 5 2 | 3+4 4 | 2+2 4 | 3+4 2 |
| | 6 2 | 3+4 4 | 6 2 | 3+4 4 | 2+2 4 | 3+4 2 |
| | 5 2 | 3+4 4 | 5 2 | 3+4 4 | 2+2 4 | 5+6 2 |
| | 5 2 | 3+4 4 | 6 2 | 3+4 4 | 2+2 4 | 5+6 2 |
| | 5 2 | 3+4 4 | 5 2 | 3+4 4 | 2+2 4 | 3+4 2 |
| | 6 2 | 3+4 4 | 6 2 | 3+4 4 | 2+2 4 | 3+4 2 |
| | 6 2 | | 6 2 | | | 5+6 2 |
| | 5 2 | | 5 2 | | | 5+6 2 |
| | 5 2 | | 6 2 | | | 3+4 2 |
| | 5 2 | | 5 2 | | | 3+4 2 |
| | 6 2 | | 5 2 | | | 5+6 2 |
| | 6 2 | | 6 2 | | | 3+4 2 |
| | 5 2 | | 6 2 | | | 3+4 2 |
| | 6 2 | | 6 2 | | | 5+6 2 |
| | 5 2 | | 5 2 | | | 3+4 2 |
| | 6 2 | | 6 2 | | | 3+4 2 |

Continued on next column

**TABLE XXXVIII– continued**

| TQC | ICSD | NoE | SF | | | |
|---|---|---|---|---|---|---|
| | 6 2 | | 5 2 | | | 5+6 2 |
| | 5 2 | | 5 2 | | | 5+6 2 |
| | 6 2 | | 6 2 | | | 3+4 2 |
| | 5 2 | | 6 2 | | | 3+4 2 |
| | 6 2 | | 5 2 | | | 5+6 2 |
| | 5 2 | | 6 2 | | | 5+6 2 |
| | 6 2 | | 6 2 | | | 3+4 2 |
| | 5 2 | | 5 2 | | | 3+4 2 |
| | 6 2 | | 5 2 | | | 3+4 2 |
| | 6 2 | | 5 2 | | | 5+6 2 |
| 63.1.35 | 189439 | 68 | Fe2O3 | | | |
| | -GM | -T | -Y | -Z | -R | -S |
| | 5 2 | 3+4 4 | 6 2 | 3+4 4 | 2+2 4 | 3+4 2 |
| | 5 2 | 3+4 4 | 5 2 | 3+4 4 | 2+2 4 | 5+6 2 |
| | 6 2 | 3+4 4 | 6 2 | 3+4 4 | 2+2 4 | 3+4 2 |
| | 5 2 | 3+4 4 | 5 2 | 3+4 4 | 2+2 4 | 5+6 2 |
| | 6 2 | 3+4 4 | 5 2 | 3+4 4 | 2+2 4 | 3+4 2 |
| | 6 2 | 3+4 4 | 6 2 | 3+4 4 | 2+2 4 | 5+6 2 |
| | 5 2 | 3+4 4 | 5 2 | 3+4 4 | 2+2 4 | 3+4 2 |
| | 5 2 | 3+4 4 | 5 2 | 3+4 4 | 2+2 4 | 3+4 2 |
| | 5 2 | 3+4 4 | 6 2 | 3+4 4 | 2+2 4 | 3+4 2 |
| | 5 2 | 3+4 4 | 6 2 | 3+4 4 | 2+2 4 | 5+6 2 |
| | 6 2 | 3+4 4 | 6 2 | 3+4 4 | 2+2 4 | 5+6 2 |
| | 6 2 | 3+4 4 | 5 2 | 3+4 4 | 2+2 4 | 5+6 2 |
| | 5 2 | 3+4 4 | 5 2 | 3+4 4 | 2+2 4 | 3+4 2 |
| | 5 2 | 3+4 4 | 5 2 | 3+4 4 | 2+2 4 | 3+4 2 |
| | 5 2 | 3+4 4 | 5 2 | 3+4 4 | 2+2 4 | 3+4 2 |
| | 6 2 | 3+4 4 | 6 2 | 3+4 4 | 2+2 4 | 3+4 2 |
| | 6 2 | 3+4 4 | 6 2 | 3+4 4 | 2+2 4 | 5+6 2 |
| | 5 2 | | 6 2 | | | 5+6 2 |
| | 5 2 | | 5 2 | | | 5+6 2 |
| | 6 2 | | 6 2 | | | 3+4 2 |
| | 6 2 | | 5 2 | | | 5+6 2 |
| | 6 2 | | 6 2 | | | 3+4 2 |
| | 6 2 | | 5 2 | | | 5+6 2 |
| | 5 2 | | 6 2 | | | 5+6 2 |
| | 6 2 | | 5 2 | | | 3+4 2 |
| | 5 2 | | 5 2 | | | 3+4 2 |
| | 6 2 | | 6 2 | | | 3+4 2 |
| | 5 2 | | 5 2 | | | 3+4 2 |
| | 5 2 | | 5 2 | | | 5+6 2 |
| | 5 2 | | 5 2 | | | 5+6 2 |
| | 6 2 | | 5 2 | | | 3+4 2 |
| | 5 2 | | 5 2 | | | 3+4 2 |
| | 5 2 | | 5 2 | | | 3+4 2 |
| | 6 2 | | 5 2 | | | 3+4 2 |
| 63.1.36 | 261037 | 80 | Fe1Se3U1 | | | |
| | -GM | -T | -Y | -Z | -R | -S |
| | 5 2 | 3+4 4 | 5 2 | 3+4 4 | 2+2 4 | 3+4 2 |
| | 6 2 | 3+4 4 | 6 2 | 3+4 4 | 2+2 4 | 5+6 2 |
| | 5 2 | 3+4 4 | 6 2 | 3+4 4 | 2+2 4 | 3+4 2 |
| | 6 2 | 3+4 4 | 5 2 | 3+4 4 | 2+2 4 | 5+6 2 |
| | 5 2 | 3+4 4 | 6 2 | 3+4 4 | 2+2 4 | 3+4 2 |
| | 6 2 | 3+4 4 | 5 2 | 3+4 4 | 2+2 4 | 5+6 2 |
| | 5 2 | 3+4 4 | 6 2 | 3+4 4 | 2+2 4 | 3+4 2 |
| | 6 2 | 3+4 4 | 5 2 | 3+4 4 | 2+2 4 | 5+6 2 |
| | 5 2 | 3+4 4 | 6 2 | 3+4 4 | 2+2 4 | 5+6 2 |
| | 6 2 | 3+4 4 | 5 2 | 3+4 4 | 2+2 4 | 3+4 2 |
| | 5 2 | 3+4 4 | 5 2 | 3+4 4 | 2+2 4 | 5+6 2 |
| | 6 2 | 3+4 4 | 5 2 | 3+4 4 | 2+2 4 | 3+4 2 |
| | 6 2 | 3+4 4 | 6 2 | 3+4 4 | 2+2 4 | 5+6 2 |

Continued on next column

**TABLE XXXVIII– continued**

| TQC | ICSD | | NoE | | SF | |
|---|---|---|---|---|---|---|
| | 5 2 | 3+4 4 | 6 2 | 3+4 4 | 2+2 4 | 3+4 2 |
| | 5 2 | 3+4 4 | 5 2 | 3+4 4 | 2+2 4 | 3+4 2 |
| | 5 2 | 3+4 4 | 5 2 | 3+4 4 | 2+2 4 | 3+4 2 |
| | 6 2 | 3+4 4 | 6 2 | 3+4 4 | 2+2 4 | 5+6 2 |
| | 5 2 | 3+4 4 | 6 2 | 3+4 4 | 2+2 4 | 5+6 2 |
| | 6 2 | 3+4 4 | 5 2 | 3+4 4 | 2+2 4 | 5+6 2 |
| | 5 2 | 3+4 4 | 6 2 | 3+4 4 | 2+2 4 | 3+4 2 |
| | 5 2 | | 6 2 | | | 3+4 2 |
| | 6 2 | | 5 2 | | | 5+6 2 |
| | 6 2 | | 5 2 | | | 3+4 2 |
| | 5 2 | | 6 2 | | | 3+4 2 |
| | 5 2 | | 5 2 | | | 5+6 2 |
| | 5 2 | | 6 2 | | | 5+6 2 |
| | 5 2 | | 6 2 | | | 3+4 2 |
| | 6 2 | | 5 2 | | | 5+6 2 |
| | 6 2 | | 6 2 | | | 3+4 2 |
| | 6 2 | | 5 2 | | | 3+4 2 |
| | 5 2 | | 5 2 | | | 5+6 2 |
| | 6 2 | | 6 2 | | | 5+6 2 |
| | 5 2 | | 5 2 | | | 3+4 2 |
| | 5 2 | | 5 2 | | | 3+4 2 |
| | 5 2 | | 5 2 | | | 3+4 2 |
| | 5 2 | | 5 2 | | | 3+4 2 |
| | 5 2 | | 5 2 | | | 3+4 2 |
| | 5 2 | | 5 2 | | | 3+4 2 |
| | 6 2 | | 5 2 | | | 3+4 2 |
| | 5 2 | | 5 2 | | | 5+6 2 |
| 63.1.37 | 245105 | | 200 | | Eu3O7Re1 | |
| | -GM | -T | -Y | -Z | -R | -S |
| | 5 2 | 3+4 4 | 5 2 | 3+4 4 | 2+2 4 | 3+4 2 |
| | 5 2 | 3+4 4 | 6 2 | 3+4 4 | 2+2 4 | 5+6 2 |
| | 6 2 | 3+4 4 | 5 2 | 3+4 4 | 2+2 4 | 3+4 2 |
| | 5 2 | 3+4 4 | 6 2 | 3+4 4 | 2+2 4 | 3+4 2 |
| | 6 2 | 3+4 4 | 5 2 | 3+4 4 | 2+2 4 | 5+6 2 |
| | 5 2 | 3+4 4 | 5 2 | 3+4 4 | 2+2 4 | 3+4 2 |
| | 5 2 | 3+4 4 | 6 2 | 3+4 4 | 2+2 4 | 5+6 2 |
| | 5 2 | 3+4 4 | 6 2 | 3+4 4 | 2+2 4 | 3+4 2 |
| | 6 2 | 3+4 4 | 5 2 | 3+4 4 | 2+2 4 | 5+6 2 |
| | 6 2 | 3+4 4 | 5 2 | 3+4 4 | 2+2 4 | 3+4 2 |
| | 6 2 | 3+4 4 | 6 2 | 3+4 4 | 2+2 4 | 5+6 2 |
| | 5 2 | 3+4 4 | 6 2 | 3+4 4 | 2+2 4 | 5+6 2 |
| | 6 2 | 3+4 4 | 5 2 | 3+4 4 | 2+2 4 | 5+6 2 |
| | 5 2 | 3+4 4 | 6 2 | 3+4 4 | 2+2 4 | 3+4 2 |
| | 6 2 | 3+4 4 | 5 2 | 3+4 4 | 2+2 4 | 3+4 2 |
| | 6 2 | 3+4 4 | 6 2 | 3+4 4 | 2+2 4 | 5+6 2 |
| | 5 2 | 3+4 4 | 5 2 | 3+4 4 | 2+2 4 | 5+6 2 |
| | 6 2 | 3+4 4 | 6 2 | 3+4 4 | 2+2 4 | 3+4 2 |
| | 5 2 | 3+4 4 | 5 2 | 3+4 4 | 2+2 4 | 5+6 2 |
| | 6 2 | 3+4 4 | 6 2 | 3+4 4 | 2+2 4 | 3+4 2 |
| | 6 2 | 3+4 4 | 6 2 | 3+4 4 | 2+2 4 | 5+6 2 |
| | 5 2 | 3+4 4 | 5 2 | 3+4 4 | 2+2 4 | 3+4 2 |
| | 6 2 | 3+4 4 | 6 2 | 3+4 4 | 2+2 4 | 5+6 2 |
| | 6 2 | 3+4 4 | 6 2 | 3+4 4 | 2+2 4 | 5+6 2 |
| | 5 2 | 3+4 4 | 6 2 | 3+4 4 | 2+2 4 | 5+6 2 |
| | 6 2 | 3+4 4 | 5 2 | 3+4 4 | 2+2 4 | 3+4 2 |
| | 5 2 | 3+4 4 | 6 2 | 3+4 4 | 2+2 4 | 5+6 2 |
| | 6 2 | 3+4 4 | 5 2 | 3+4 4 | 2+2 4 | 3+4 2 |
| | 5 2 | 3+4 4 | 6 2 | 3+4 4 | 2+2 4 | 5+6 2 |
| | 6 2 | 3+4 4 | 6 2 | 3+4 4 | 2+2 4 | 5+6 2 |
| | 6 2 | 3+4 4 | 5 2 | 3+4 4 | 2+2 4 | 3+4 2 |
| | 6 2 | 3+4 4 | 5 2 | 3+4 4 | 2+2 4 | 3+4 2 |

Continued on next column

**TABLE XXXVIII– continued**

| TQC | ICSD | | NoE | | SF | |
|---|---|---|---|---|---|---|
| | 5 2 | 3+4 4 | 6 2 | 3+4 4 | 2+2 4 | 5+6 2 |
| | 6 2 | 3+4 4 | 5 2 | 3+4 4 | 2+2 4 | 3+4 2 |
| | 6 2 | 3+4 4 | 5 2 | 3+4 4 | 2+2 4 | 5+6 2 |
| | 5 2 | 3+4 4 | 6 2 | 3+4 4 | 2+2 4 | 5+6 2 |
| | 5 2 | 3+4 4 | 6 2 | 3+4 4 | 2+2 4 | 3+4 2 |
| | 6 2 | 3+4 4 | 6 2 | 3+4 4 | 2+2 4 | 5+6 2 |
| | 5 2 | 3+4 4 | 5 2 | 3+4 4 | 2+2 4 | 5+6 2 |
| | 5 2 | 3+4 4 | 5 2 | 3+4 4 | 2+2 4 | 5+6 2 |
| | 5 2 | 3+4 4 | 5 2 | 3+4 4 | 2+2 4 | 5+6 2 |
| | 5 2 | 3+4 4 | 5 2 | 3+4 4 | 2+2 4 | 3+4 2 |
| | 5 2 | 3+4 4 | 6 2 | 3+4 4 | 2+2 4 | 5+6 2 |
| | 5 2 | 3+4 4 | 6 2 | 3+4 4 | 2+2 4 | 5+6 2 |
| | 6 2 | 3+4 4 | 6 2 | 3+4 4 | 2+2 4 | 3+4 2 |
| | 5 2 | 3+4 4 | 5 2 | 3+4 4 | 2+2 4 | 5+6 2 |
| | 6 2 | 3+4 4 | 5 2 | 3+4 4 | 2+2 4 | 3+4 2 |
| | 5 2 | 3+4 4 | 5 2 | 3+4 4 | 2+2 4 | 5+6 2 |
| | 5 2 | 3+4 4 | 5 2 | 3+4 4 | 2+2 4 | 3+4 2 |
| | 6 2 | 3+4 4 | 6 2 | 3+4 4 | 2+2 4 | 5+6 2 |
| | 6 2 | | 5 2 | | | 5+6 2 |
| | 5 2 | | 6 2 | | | 5+6 2 |
| | 5 2 | | 6 2 | | | 3+4 2 |
| | 5 2 | | 6 2 | | | 3+4 2 |
| | 6 2 | | 5 2 | | | 3+4 2 |
| | 6 2 | | 6 2 | | | 3+4 2 |
| | 5 2 | | 5 2 | | | 3+4 2 |
| | 6 2 | | 6 2 | | | 3+4 2 |
| | 6 2 | | 5 2 | | | 3+4 2 |
| | 6 2 | | 6 2 | | | 5+6 2 |
| | 5 2 | | 5 2 | | | 5+6 2 |
| | 6 2 | | 6 2 | | | 3+4 2 |
| | 5 2 | | 5 2 | | | 3+4 2 |
| | 6 2 | | 5 2 | | | 5+6 2 |
| | 5 2 | | 6 2 | | | 5+6 2 |
| | 6 2 | | 6 2 | | | 3+4 2 |
| | 6 2 | | 6 2 | | | 5+6 2 |
| | 5 2 | | 5 2 | | | 3+4 2 |
| | 5 2 | | 6 2 | | | 5+6 2 |
| | 6 2 | | 5 2 | | | 3+4 2 |
| | 6 2 | | 5 2 | | | 3+4 2 |
| | 5 2 | | 5 2 | | | 5+6 2 |
| | 6 2 | | 6 2 | | | 3+4 2 |
| | 5 2 | | 6 2 | | | 3+4 2 |
| | 6 2 | | 5 2 | | | 5+6 2 |
| | 6 2 | | 6 2 | | | 3+4 2 |
| | 6 2 | | 6 2 | | | 5+6 2 |
| | 6 2 | | 6 2 | | | 3+4 2 |
| | 6 2 | | 6 2 | | | 5+6 2 |
| | 5 2 | | 5 2 | | | 5+6 2 |
| | 5 2 | | 5 2 | | | 5+6 2 |
| | 5 2 | | 5 2 | | | 5+6 2 |
| | 5 2 | | 5 2 | | | 5+6 2 |
| | 6 2 | | 6 2 | | | 3+4 2 |
| | 6 2 | | 5 2 | | | 5+6 2 |
| | 6 2 | | 6 2 | | | 5+6 2 |
| | 5 2 | | 6 2 | | | 3+4 2 |
| | 6 2 | | 5 2 | | | 5+6 2 |
| | 5 2 | | 6 2 | | | 3+4 2 |
| | 5 2 | | 5 2 | | | 5+6 2 |
| | 6 2 | | 6 2 | | | 5+6 2 |
| | 6 2 | | 6 2 | | | 5+6 2 |
| | 6 2 | | 6 2 | | | 3+4 2 |

Continued on next column

**TABLE XXXVIII– continued**

| TQC | ICSD | | NoE | SF | | |
|---|---|---|---|---|---|---|
| | 5 2 | | 6 2 | | | 5+6 2 |
| | 5 2 | | 5 2 | | | 3+4 2 |
| | 6 2 | | 6 2 | | | 5+6 2 |
| | 6 2 | | 6 2 | | | 5+6 2 |
| | 6 2 | | 5 2 | | | 3+4 2 |
| | 5 2 | | 6 2 | | | 5+6 2 |
| | 6 2 | | 5 2 | | | 5+6 2 |
| 63.1.38 | 400536 | | 172 | Eu2O7Ta2 | | |
| | -GM | -T | -Y | -Z | -R | -S |
| | 5 2 | 3+4 4 | 5 2 | 3+4 4 | 2+2 4 | 5+6 2 |
| | 6 2 | 3+4 4 | 6 2 | 3+4 4 | 2+2 4 | 3+4 2 |
| | 5 2 | 3+4 4 | 6 2 | 3+4 4 | 2+2 4 | 3+4 2 |
| | 6 2 | 3+4 4 | 5 2 | 3+4 4 | 2+2 4 | 5+6 2 |
| | 5 2 | 3+4 4 | 6 2 | 3+4 4 | 2+2 4 | 3+4 2 |
| | 5 2 | 3+4 4 | 5 2 | 3+4 4 | 2+2 4 | 5+6 2 |
| | 6 2 | 3+4 4 | 6 2 | 3+4 4 | 2+2 4 | 3+4 2 |
| | 6 2 | 3+4 4 | 5 2 | 3+4 4 | 2+2 4 | 5+6 2 |
| | 5 2 | 3+4 4 | 5 2 | 3+4 4 | 2+2 4 | 3+4 2 |
| | 6 2 | 3+4 4 | 6 2 | 3+4 4 | 2+2 4 | 5+6 2 |
| | 5 2 | 3+4 4 | 5 2 | 3+4 4 | 2+2 4 | 3+4 2 |
| | 5 2 | 3+4 4 | 6 2 | 3+4 4 | 2+2 4 | 5+6 2 |
| | 6 2 | 3+4 4 | 6 2 | 3+4 4 | 2+2 4 | 3+4 2 |
| | 6 2 | 3+4 4 | 5 2 | 3+4 4 | 2+2 4 | 5+6 2 |
| | 5 2 | 3+4 4 | 6 2 | 3+4 4 | 2+2 4 | 5+6 2 |
| | 5 2 | 3+4 4 | 5 2 | 3+4 4 | 2+2 4 | 3+4 2 |
| | 6 2 | 3+4 4 | 5 2 | 3+4 4 | 2+2 4 | 3+4 2 |
| | 5 2 | 3+4 4 | 6 2 | 3+4 4 | 2+2 4 | 5+6 2 |
| | 6 2 | 3+4 4 | 5 2 | 3+4 4 | 2+2 4 | 3+4 2 |
| | 5 2 | 3+4 4 | 5 2 | 3+4 4 | 2+2 4 | 5+6 2 |
| | 6 2 | 3+4 4 | 6 2 | 3+4 4 | 2+2 4 | 3+4 2 |
| | 5 2 | 3+4 4 | 5 2 | 3+4 4 | 2+2 4 | 3+4 2 |
| | 5 2 | 3+4 4 | 6 2 | 3+4 4 | 2+2 4 | 5+6 2 |
| | 6 2 | 3+4 4 | 6 2 | 3+4 4 | 2+2 4 | 3+4 2 |
| | 6 2 | 3+4 4 | 5 2 | 3+4 4 | 2+2 4 | 5+6 2 |
| | 6 2 | 3+4 4 | 6 2 | 3+4 4 | 2+2 4 | 3+4 2 |
| | 5 2 | 3+4 4 | 5 2 | 3+4 4 | 2+2 4 | 5+6 2 |
| | 6 2 | 3+4 4 | 5 2 | 3+4 4 | 2+2 4 | 3+4 2 |
| | 5 2 | 3+4 4 | 5 2 | 3+4 4 | 2+2 4 | 3+4 2 |
| | 5 2 | 3+4 4 | 6 2 | 3+4 4 | 2+2 4 | 5+6 2 |
| | 6 2 | 3+4 4 | 6 2 | 3+4 4 | 2+2 4 | 5+6 2 |
| | 6 2 | 3+4 4 | 6 2 | 3+4 4 | 2+2 4 | 5+6 2 |
| | 6 2 | 3+4 4 | 6 2 | 3+4 4 | 2+2 4 | 3+4 2 |
| | 5 2 | 3+4 4 | 5 2 | 3+4 4 | 2+2 4 | 3+4 2 |
| | 5 2 | 3+4 4 | 5 2 | 3+4 4 | 2+2 4 | 5+6 2 |
| | 6 2 | 3+4 4 | 5 2 | 3+4 4 | 2+2 4 | 5+6 2 |
| | 5 2 | 3+4 4 | 6 2 | 3+4 4 | 2+2 4 | 3+4 2 |
| | 6 2 | 3+4 4 | 6 2 | 3+4 4 | 2+2 4 | 5+6 2 |
| | 5 2 | 3+4 4 | 5 2 | 3+4 4 | 2+2 4 | 5+6 2 |
| | 6 2 | 3+4 4 | 6 2 | 3+4 4 | 2+2 4 | 3+4 2 |
| | 5 2 | 3+4 4 | 6 2 | 3+4 4 | 2+2 4 | 3+4 2 |
| | 6 2 | 3+4 4 | 6 2 | 3+4 4 | 2+2 4 | 5+6 2 |
| | 5 2 | 3+4 4 | 6 2 | 3+4 4 | 2+2 4 | 3+4 2 |
| | 6 2 | | 5 2 | | | 5+6 2 |
| | 5 2 | | 6 2 | | | 3+4 2 |
| | 6 2 | | 5 2 | | | 5+6 2 |
| | 6 2 | | 6 2 | | | 3+4 2 |
| | 5 2 | | 6 2 | | | 5+6 2 |
| | 6 2 | | 5 2 | | | 5+6 2 |
| | 5 2 | | 5 2 | | | 3+4 2 |
| | 5 2 | | 5 2 | | | 5+6 2 |
| | 6 2 | | 6 2 | | | 3+4 2 |

Continued on next column

**TABLE XXXVIII– continued**

| TQC | ICSD | | NoE | SF | | |
|---|---|---|---|---|---|---|
| | 5 2 | | 5 2 | | | 5+6 2 |
| | 5 2 | | 5 2 | | | 5+6 2 |
| | 6 2 | | 6 2 | | | 3+4 2 |
| | 6 2 | | 5 2 | | | 5+6 2 |
| | 5 2 | | 5 2 | | | 3+4 2 |
| | 5 2 | | 6 2 | | | 5+6 2 |
| | 6 2 | | 5 2 | | | 3+4 2 |
| | 5 2 | | 5 2 | | | 3+4 2 |
| | 6 2 | | 6 2 | | | 3+4 2 |
| | 5 2 | | 6 2 | | | 5+6 2 |
| | 6 2 | | 6 2 | | | 5+6 2 |
| | 5 2 | | 5 2 | | | 3+4 2 |
| | 6 2 | | 6 2 | | | 5+6 2 |
| | 6 2 | | 6 2 | | | 3+4 2 |
| | 6 2 | | 5 2 | | | 5+6 2 |
| | 5 2 | | 6 2 | | | 5+6 2 |
| | 6 2 | | 5 2 | | | 3+4 2 |
| | 6 2 | | 6 2 | | | 5+6 2 |
| | 6 2 | | 6 2 | | | 5+6 2 |
| | 6 2 | | 6 2 | | | 5+6 2 |
| | 5 2 | | 6 2 | | | 3+4 2 |
| | 5 2 | | 5 2 | | | 5+6 2 |
| | 5 2 | | 5 2 | | | 3+4 2 |
| | 6 2 | | 6 2 | | | 3+4 2 |
| | 5 2 | | 5 2 | | | 5+6 2 |
| | 6 2 | | 6 2 | | | 5+6 2 |
| | 5 2 | | 5 2 | | | 3+4 2 |
| | 6 2 | | 5 2 | | | 5+6 2 |
| | 5 2 | | 6 2 | | | 3+4 2 |
| | 6 2 | | 6 2 | | | 5+6 2 |
| | 6 2 | | 5 2 | | | 3+4 2 |
| | 6 2 | | 6 2 | | | 5+6 2 |
| | 5 2 | | 5 2 | | | 3+4 2 |
| | 5 2 | | 5 2 | | | 5+6 2 |
| 63.1.39 | 40812 | | 164 | Eu1Ni5P3 | | |
| | -GM | -T | -Y | -Z | -R | -S |
| | 5 2 | 3+4 4 | 5 2 | 3+4 4 | 2+2 4 | 3+4 2 |
| | 6 2 | 3+4 4 | 6 2 | 3+4 4 | 2+2 4 | 5+6 2 |
| | 5 2 | 3+4 4 | 5 2 | 3+4 4 | 2+2 4 | 3+4 2 |
| | 6 2 | 3+4 4 | 6 2 | 3+4 4 | 2+2 4 | 5+6 2 |
| | 5 2 | 3+4 4 | 6 2 | 3+4 4 | 2+2 4 | 3+4 2 |
| | 6 2 | 3+4 4 | 5 2 | 3+4 4 | 2+2 4 | 5+6 2 |
| | 5 2 | 3+4 4 | 5 2 | 3+4 4 | 2+2 4 | 3+4 2 |
| | 6 2 | 3+4 4 | 6 2 | 3+4 4 | 2+2 4 | 5+6 2 |
| | 5 2 | 3+4 4 | 5 2 | 3+4 4 | 2+2 4 | 5+6 2 |
| | 5 2 | 3+4 4 | 5 2 | 3+4 4 | 2+2 4 | 5+6 2 |
| | 6 2 | 3+4 4 | 6 2 | 3+4 4 | 2+2 4 | 3+4 2 |
| | 5 2 | 3+4 4 | 6 2 | 3+4 4 | 2+2 4 | 5+6 2 |
| | 6 2 | 3+4 4 | 5 2 | 3+4 4 | 2+2 4 | 3+4 2 |
| | 6 2 | 3+4 4 | 6 2 | 3+4 4 | 2+2 4 | 3+4 2 |
| | 5 2 | 3+4 4 | 6 2 | 3+4 4 | 2+2 4 | 3+4 2 |
| | 5 2 | 3+4 4 | 5 2 | 3+4 4 | 2+2 4 | 3+4 2 |
| | 6 2 | 3+4 4 | 5 2 | 3+4 4 | 2+2 4 | 3+4 2 |
| | 5 2 | 3+4 4 | 6 2 | 3+4 4 | 2+2 4 | 5+6 2 |
| | 6 2 | 3+4 4 | 5 2 | 3+4 4 | 2+2 4 | 5+6 2 |
| | 5 2 | 3+4 4 | 5 2 | 3+4 4 | 2+2 4 | 5+6 2 |
| | 5 2 | 3+4 4 | 6 2 | 3+4 4 | 2+2 4 | 3+4 2 |
| | 5 2 | 3+4 4 | 5 2 | 3+4 4 | 2+2 4 | 5+6 2 |
| | 6 2 | 3+4 4 | 5 2 | 3+4 4 | 2+2 4 | 3+4 2 |
| | 5 2 | 3+4 4 | 5 2 | 3+4 4 | 2+2 4 | 5+6 2 |
| | 6 2 | 3+4 4 | 5 2 | 3+4 4 | 2+2 4 | 3+4 2 |

Continued on next column

**TABLE XXXVIII– continued**

| TQC | ICSD | | NoE | SF | | |
|---|---|---|---|---|---|---|
| | 5 2 | 3+4 4 | 6 2 | 3+4 4 | 2+2 4 | 5+6 2 |
| | 6 2 | 3+4 4 | 5 2 | 3+4 4 | 2+2 4 | 5+6 2 |
| | 5 2 | 3+4 4 | 6 2 | 3+4 4 | 2+2 4 | 3+4 2 |
| | 6 2 | 3+4 4 | 6 2 | 3+4 4 | 2+2 4 | 5+6 2 |
| | 6 2 | 3+4 4 | 5 2 | 3+4 4 | 2+2 4 | 3+4 2 |
| | 6 2 | 3+4 4 | 6 2 | 3+4 4 | 2+2 4 | 3+4 2 |
| | 6 2 | 3+4 4 | 6 2 | 3+4 4 | 2+2 4 | 5+6 2 |
| | 5 2 | 3+4 4 | 5 2 | 3+4 4 | 2+2 4 | 3+4 2 |
| | 5 2 | 3+4 4 | 6 2 | 3+4 4 | 2+2 4 | 3+4 2 |
| | 5 2 | 3+4 4 | 5 2 | 3+4 4 | 2+2 4 | 5+6 2 |
| | 5 2 | 3+4 4 | 5 2 | 3+4 4 | 2+2 4 | 3+4 2 |
| | 6 2 | 3+4 4 | 6 2 | 3+4 4 | 2+2 4 | 3+4 2 |
| | 5 2 | 3+4 4 | 6 2 | 3+4 4 | 2+2 4 | 5+6 2 |
| | 5 2 | 3+4 4 | 6 2 | 3+4 4 | 2+2 4 | 3+4 2 |
| | 5 2 | 3+4 4 | 5 2 | 3+4 4 | 2+2 4 | 5+6 2 |
| | 5 2 | 3+4 4 | 5 2 | 3+4 4 | 2+2 4 | 5+6 2 |
| | 6 2 | | 6 2 | | | 3+4 2 |
| | 6 2 | | 5 2 | | | 3+4 2 |
| | 6 2 | | 5 2 | | | 5+6 2 |
| | 5 2 | | 5 2 | | | 3+4 2 |
| | 6 2 | | 6 2 | | | 3+4 2 |
| | 5 2 | | 5 2 | | | 5+6 2 |
| | 6 2 | | 5 2 | | | 3+4 2 |
| | 5 2 | | 6 2 | | | 5+6 2 |
| | 6 2 | | 5 2 | | | 5+6 2 |
| | 5 2 | | 6 2 | | | 3+4 2 |
| | 6 2 | | 5 2 | | | 3+4 2 |
| | 6 2 | | 5 2 | | | 5+6 2 |
| | 5 2 | | 6 2 | | | 3+4 2 |
| | 6 2 | | 5 2 | | | 5+6 2 |
| | 5 2 | | 6 2 | | | 5+6 2 |
| | 5 2 | | 6 2 | | | 3+4 2 |
| | 5 2 | | 6 2 | | | 5+6 2 |
| | 5 2 | | 5 2 | | | 3+4 2 |
| | 5 2 | | 6 2 | | | 5+6 2 |
| | 5 2 | | 5 2 | | | 3+4 2 |
| | 6 2 | | 6 2 | | | 3+4 2 |
| | 5 2 | | 6 2 | | | 3+4 2 |
| | 6 2 | | 5 2 | | | 3+4 2 |
| | 5 2 | | 6 2 | | | 3+4 2 |
| | 6 2 | | 5 2 | | | 5+6 2 |
| | 5 2 | | 5 2 | | | 5+6 2 |
| | 6 2 | | 5 2 | | | 3+4 2 |
| | 6 2 | | 6 2 | | | 3+4 2 |
| | 5 2 | | 5 2 | | | 5+6 2 |
| | 5 2 | | 5 2 | | | 3+4 2 |
| | 5 2 | | 6 2 | | | 3+4 2 |
| | 6 2 | | 6 2 | | | 5+6 2 |
| | 5 2 | | 6 2 | | | 5+6 2 |
| | 6 2 | | 5 2 | | | 3+4 2 |
| | 6 2 | | 5 2 | | | 3+4 2 |
| | 5 2 | | 5 2 | | | 5+6 2 |
| | 6 2 | | 6 2 | | | 3+4 2 |
| | 5 2 | | 6 2 | | | 3+4 2 |
| | 5 2 | | 6 2 | | | 5+6 2 |
| | 5 2 | | 5 2 | | | 5+6 2 |
| | 6 2 | | 5 2 | | | 5+6 2 |
| 63.1.40 | 237971 | | 76 | Eu1In4Ir1 | | |
| | -GM | -T | -Y | -Z | -R | -S |
| | 5 2 | 3+4 4 | 5 2 | 3+4 4 | 2+2 4 | 3+4 2 |
| | 6 2 | 3+4 4 | 6 2 | 3+4 4 | 2+2 4 | 5+6 2 |

Continued on next column

**TABLE XXXVIII– continued**

| TQC | ICSD | | NoE | SF | | |
|---|---|---|---|---|---|---|
| | 5 2 | 3+4 4 | 5 2 | 3+4 4 | 2+2 4 | 3+4 2 |
| | 6 2 | 3+4 4 | 6 2 | 3+4 4 | 2+2 4 | 5+6 2 |
| | 5 2 | 3+4 4 | 5 2 | 3+4 4 | 2+2 4 | 5+6 2 |
| | 6 2 | 3+4 4 | 6 2 | 3+4 4 | 2+2 4 | 3+4 2 |
| | 5 2 | 3+4 4 | 6 2 | 3+4 4 | 2+2 4 | 5+6 2 |
| | 6 2 | 3+4 4 | 5 2 | 3+4 4 | 2+2 4 | 3+4 2 |
| | 5 2 | 3+4 4 | 6 2 | 3+4 4 | 2+2 4 | 3+4 2 |
| | 6 2 | 3+4 4 | 5 2 | 3+4 4 | 2+2 4 | 5+6 2 |
| | 5 2 | 3+4 4 | 6 2 | 3+4 4 | 2+2 4 | 5+6 2 |
| | 5 2 | 3+4 4 | 5 2 | 3+4 4 | 2+2 4 | 3+4 2 |
| | 5 2 | 3+4 4 | 6 2 | 3+4 4 | 2+2 4 | 3+4 2 |
| | 6 2 | 3+4 4 | 5 2 | 3+4 4 | 2+2 4 | 5+6 2 |
| | 6 2 | 3+4 4 | 5 2 | 3+4 4 | 2+2 4 | 3+4 2 |
| | 6 2 | 3+4 4 | 6 2 | 3+4 4 | 2+2 4 | 5+6 2 |
| | 5 2 | 3+4 4 | 5 2 | 3+4 4 | 2+2 4 | 3+4 2 |
| | 5 2 | 3+4 4 | 6 2 | 3+4 4 | 2+2 4 | 3+4 2 |
| | 5 2 | 3+4 4 | 5 2 | 3+4 4 | 2+2 4 | 5+6 2 |
| | 6 2 | | 6 2 | | | 5+6 2 |
| | 6 2 | | 6 2 | | | 3+4 2 |
| | 6 2 | | 5 2 | | | 5+6 2 |
| | 5 2 | | 6 2 | | | 3+4 2 |
| | 5 2 | | 5 2 | | | 5+6 2 |
| | 5 2 | | 5 2 | | | 3+4 2 |
| | 6 2 | | 6 2 | | | 5+6 2 |
| | 5 2 | | 6 2 | | | 5+6 2 |
| | 6 2 | | 6 2 | | | 3+4 2 |
| | 6 2 | | 5 2 | | | 5+6 2 |
| | 5 2 | | 6 2 | | | 5+6 2 |
| | 6 2 | | 5 2 | | | 3+4 2 |
| | 5 2 | | 5 2 | | | 3+4 2 |
| | 6 2 | | 6 2 | | | 5+6 2 |
| | 6 2 | | 6 2 | | | 5+6 2 |
| | 5 2 | | 6 2 | | | 3+4 2 |
| | 5 2 | | 5 2 | | | 5+6 2 |
| | 6 2 | | 5 2 | | | 3+4 2 |
| | 5 2 | | 6 2 | | | 5+6 2 |
| 63.1.41 | 80501 | | 164 | Er3Ge4 | | |
| | -GM | -T | -Y | -Z | -R | -S |
| | 5 2 | 3+4 4 | 5 2 | 3+4 4 | 2+2 4 | 3+4 2 |
| | 6 2 | 3+4 4 | 6 2 | 3+4 4 | 2+2 4 | 5+6 2 |
| | 5 2 | 3+4 4 | 6 2 | 3+4 4 | 2+2 4 | 3+4 2 |
| | 5 2 | 3+4 4 | 6 2 | 3+4 4 | 2+2 4 | 3+4 2 |
| | 6 2 | 3+4 4 | 5 2 | 3+4 4 | 2+2 4 | 5+6 2 |
| | 6 2 | 3+4 4 | 5 2 | 3+4 4 | 2+2 4 | 5+6 2 |
| | 5 2 | 3+4 4 | 6 2 | 3+4 4 | 2+2 4 | 3+4 2 |
| | 6 2 | 3+4 4 | 5 2 | 3+4 4 | 2+2 4 | 5+6 2 |
| | 5 2 | 3+4 4 | 6 2 | 3+4 4 | 2+2 4 | 5+6 2 |
| | 6 2 | 3+4 4 | 6 2 | 3+4 4 | 2+2 4 | 3+4 2 |
| | 5 2 | 3+4 4 | 5 2 | 3+4 4 | 2+2 4 | 5+6 2 |
| | 6 2 | 3+4 4 | 5 2 | 3+4 4 | 2+2 4 | 3+4 2 |
| | 5 2 | 3+4 4 | 6 2 | 3+4 4 | 2+2 4 | 5+6 2 |
| | 6 2 | 3+4 4 | 5 2 | 3+4 4 | 2+2 4 | 3+4 2 |
| | 5 2 | 3+4 4 | 6 2 | 3+4 4 | 2+2 4 | 3+4 2 |
| | 5 2 | 3+4 4 | 5 2 | 3+4 4 | 2+2 4 | 5+6 2 |
| | 6 2 | 3+4 4 | 6 2 | 3+4 4 | 2+2 4 | 5+6 2 |
| | 6 2 | 3+4 4 | 5 2 | 3+4 4 | 2+2 4 | 3+4 2 |
| | 5 2 | 3+4 4 | 6 2 | 3+4 4 | 2+2 4 | 3+4 2 |
| | 5 2 | 3+4 4 | 5 2 | 3+4 4 | 2+2 4 | 5+6 2 |
| | 6 2 | 3+4 4 | 6 2 | 3+4 4 | 2+2 4 | 3+4 2 |
| | 5 2 | 3+4 4 | 6 2 | 3+4 4 | 2+2 4 | 5+6 2 |
| | 6 2 | 3+4 4 | 5 2 | 3+4 4 | 2+2 4 | 5+6 2 |

Continued on next column

**TABLE XXXVIII– continued**

| TQC | ICSD | | NoE | | SF | |
|---|---|---|---|---|---|---|
| | 6 2 | 3+4 4 | 5 2 | 3+4 4 | 2+2 4 | 3+4 2 |
| | 5 2 | 3+4 4 | 5 2 | 3+4 4 | 2+2 4 | 5+6 2 |
| | 5 2 | 3+4 4 | 6 2 | 3+4 4 | 2+2 4 | 3+4 2 |
| | 6 2 | 3+4 4 | 5 2 | 3+4 4 | 2+2 4 | 5+6 2 |
| | 5 2 | 3+4 4 | 5 2 | 3+4 4 | 2+2 4 | 3+4 2 |
| | 6 2 | 3+4 4 | 6 2 | 3+4 4 | 2+2 4 | 3+4 2 |
| | 6 2 | 3+4 4 | 6 2 | 3+4 4 | 2+2 4 | 3+4 2 |
| | 5 2 | 3+4 4 | 5 2 | 3+4 4 | 2+2 4 | 5+6 2 |
| | 5 2 | 3+4 4 | 5 2 | 3+4 4 | 2+2 4 | 3+4 2 |
| | 6 2 | 3+4 4 | 6 2 | 3+4 4 | 2+2 4 | 5+6 2 |
| | 6 2 | 3+4 4 | 6 2 | 3+4 4 | 2+2 4 | 5+6 2 |
| | 5 2 | 3+4 4 | 5 2 | 3+4 4 | 2+2 4 | 3+4 2 |
| | 6 2 | 3+4 4 | 6 2 | 3+4 4 | 2+2 4 | 5+6 2 |
| | 5 2 | 3+4 4 | 6 2 | 3+4 4 | 2+2 4 | 5+6 2 |
| | 6 2 | 3+4 4 | 5 2 | 3+4 4 | 2+2 4 | 5+6 2 |
| | 6 2 | 3+4 4 | 6 2 | 3+4 4 | 2+2 4 | 3+4 2 |
| | 6 2 | 3+4 4 | 6 2 | 3+4 4 | 2+2 4 | 3+4 2 |
| | 6 2 | 3+4 4 | 5 2 | 3+4 4 | 2+2 4 | 3+4 2 |
| | 5 2 | | 5 2 | | | 5+6 2 |
| | 5 2 | | 5 2 | | | 3+4 2 |
| | 6 2 | | 6 2 | | | 5+6 2 |
| | 6 2 | | 5 2 | | | 5+6 2 |
| | 5 2 | | 5 2 | | | 5+6 2 |
| | 6 2 | | 6 2 | | | 3+4 2 |
| | 5 2 | | 6 2 | | | 5+6 2 |
| | 5 2 | | 5 2 | | | 3+4 2 |
| | 6 2 | | 5 2 | | | 5+6 2 |
| | 5 2 | | 6 2 | | | 3+4 2 |
| | 6 2 | | 6 2 | | | 3+4 2 |
| | 5 2 | | 6 2 | | | 5+6 2 |
| | 6 2 | | 5 2 | | | 5+6 2 |
| | 5 2 | | 6 2 | | | 3+4 2 |
| | 5 2 | | 5 2 | | | 5+6 2 |
| | 6 2 | | 6 2 | | | 3+4 2 |
| | 5 2 | | 6 2 | | | 5+6 2 |
| | 6 2 | | 5 2 | | | 3+4 2 |
| | 6 2 | | 5 2 | | | 3+4 2 |
| | 5 2 | | 5 2 | | | 5+6 2 |
| | 5 2 | | 6 2 | | | 3+4 2 |
| | 5 2 | | 5 2 | | | 5+6 2 |
| | 6 2 | | 6 2 | | | 5+6 2 |
| | 6 2 | | 6 2 | | | 3+4 2 |
| | 5 2 | | 5 2 | | | 3+4 2 |
| | 6 2 | | 5 2 | | | 5+6 2 |
| | 5 2 | | 6 2 | | | 5+6 2 |
| | 6 2 | | 5 2 | | | 3+4 2 |
| | 5 2 | | 5 2 | | | 5+6 2 |
| | 6 2 | | 6 2 | | | 3+4 2 |
| | 6 2 | | 5 2 | | | 5+6 2 |
| | 5 2 | | 6 2 | | | 3+4 2 |
| | 5 2 | | 5 2 | | | 5+6 2 |
| | 6 2 | | 6 2 | | | 3+4 2 |
| | 5 2 | | 5 2 | | | 3+4 2 |
| | 6 2 | | 6 2 | | | 5+6 2 |
| | 5 2 | | 6 2 | | | 3+4 2 |
| | 6 2 | | 5 2 | | | 5+6 2 |
| | 5 2 | | 6 2 | | | 3+4 2 |
| | 6 2 | | 5 2 | | | 5+6 2 |
| | 5 2 | | 6 2 | | | 5+6 2 |
| 63.1.42 | 171141 | | 80 | | Er1Ni1Si2 | |
| | -GM | -T | -Y | -Z | -R | -S |
| | 5 2 | 3+4 4 | 5 2 | 3+4 4 | 2+2 4 | 3+4 2 |
| | 6 2 | 3+4 4 | 6 2 | 3+4 4 | 2+2 4 | 5+6 2 |
| | 5 2 | 3+4 4 | 5 2 | 3+4 4 | 2+2 4 | 3+4 2 |
| | 6 2 | 3+4 4 | 6 2 | 3+4 4 | 2+2 4 | 5+6 2 |
| | 5 2 | 3+4 4 | 5 2 | 3+4 4 | 2+2 4 | 3+4 2 |
| | 6 2 | 3+4 4 | 6 2 | 3+4 4 | 2+2 4 | 5+6 2 |
| | 5 2 | 3+4 4 | 5 2 | 3+4 4 | 2+2 4 | 3+4 2 |
| | 6 2 | 3+4 4 | 6 2 | 3+4 4 | 2+2 4 | 5+6 2 |
| | 5 2 | 3+4 4 | 6 2 | 3+4 4 | 2+2 4 | 5+6 2 |
| | 5 2 | 3+4 4 | 5 2 | 3+4 4 | 2+2 4 | 3+4 2 |
| | 6 2 | 3+4 4 | 6 2 | 3+4 4 | 2+2 4 | 5+6 2 |
| | 6 2 | 3+4 4 | 5 2 | 3+4 4 | 2+2 4 | 3+4 2 |
| | 6 2 | 3+4 4 | 5 2 | 3+4 4 | 2+2 4 | 3+4 2 |
| | 5 2 | 3+4 4 | 6 2 | 3+4 4 | 2+2 4 | 5+6 2 |
| | 5 2 | 3+4 4 | 6 2 | 3+4 4 | 2+2 4 | 5+6 2 |
| | 5 2 | 3+4 4 | 6 2 | 3+4 4 | 2+2 4 | 3+4 2 |
| | 5 2 | 3+4 4 | 5 2 | 3+4 4 | 2+2 4 | 3+4 2 |
| | 5 2 | 3+4 4 | 5 2 | 3+4 4 | 2+2 4 | 5+6 2 |
| | 6 2 | 3+4 4 | 6 2 | 3+4 4 | 2+2 4 | 5+6 2 |
| | 5 2 | 3+4 4 | 6 2 | 3+4 4 | 2+2 4 | 5+6 2 |
| | 6 2 | | 5 2 | | | 3+4 2 |
| | 5 2 | | 5 2 | | | 5+6 2 |
| | 6 2 | | 5 2 | | | 3+4 2 |
| | 6 2 | | 6 2 | | | 5+6 2 |
| | 6 2 | | 6 2 | | | 3+4 2 |
| | 5 2 | | 6 2 | | | 3+4 2 |
| | 6 2 | | 5 2 | | | 5+6 2 |
| | 5 2 | | 5 2 | | | 5+6 2 |
| | 5 2 | | 5 2 | | | 3+4 2 |
| | 6 2 | | 6 2 | | | 3+4 2 |
| | 5 2 | | 6 2 | | | 5+6 2 |
| | 6 2 | | 6 2 | | | 3+4 2 |
| | 5 2 | | 6 2 | | | 5+6 2 |
| | 6 2 | | 5 2 | | | 3+4 2 |
| | 6 2 | | 5 2 | | | 5+6 2 |
| | 5 2 | | 6 2 | | | 3+4 2 |
| | 6 2 | | 5 2 | | | 5+6 2 |
| | 5 2 | | 5 2 | | | 5+6 2 |
| | 5 2 | | 6 2 | | | 3+4 2 |
| | 6 2 | | 6 2 | | | 3+4 2 |
| 63.1.43 | 423066 | | 88 | | Er1Ga4Ni1 | |
| | -GM | -T | -Y | -Z | -R | -S |
| | 5 2 | 3+4 4 | 5 2 | 3+4 4 | 2+2 4 | 3+4 2 |
| | 6 2 | 3+4 4 | 6 2 | 3+4 4 | 2+2 4 | 5+6 2 |
| | 5 2 | 3+4 4 | 5 2 | 3+4 4 | 2+2 4 | 5+6 2 |
| | 6 2 | 3+4 4 | 6 2 | 3+4 4 | 2+2 4 | 3+4 2 |
| | 5 2 | 3+4 4 | 5 2 | 3+4 4 | 2+2 4 | 5+6 2 |
| | 6 2 | 3+4 4 | 6 2 | 3+4 4 | 2+2 4 | 3+4 2 |
| | 5 2 | 3+4 4 | 6 2 | 3+4 4 | 2+2 4 | 5+6 2 |
| | 6 2 | 3+4 4 | 5 2 | 3+4 4 | 2+2 4 | 3+4 2 |
| | 5 2 | 3+4 4 | 6 2 | 3+4 4 | 2+2 4 | 3+4 2 |
| | 6 2 | 3+4 4 | 5 2 | 3+4 4 | 2+2 4 | 5+6 2 |
| | 5 2 | 3+4 4 | 6 2 | 3+4 4 | 2+2 4 | 5+6 2 |
| | 5 2 | 3+4 4 | 6 2 | 3+4 4 | 2+2 4 | 3+4 2 |
| | 5 2 | 3+4 4 | 5 2 | 3+4 4 | 2+2 4 | 3+4 2 |
| | 6 2 | 3+4 4 | 5 2 | 3+4 4 | 2+2 4 | 3+4 2 |
| | 6 2 | 3+4 4 | 5 2 | 3+4 4 | 2+2 4 | 5+6 2 |
| | 6 2 | 3+4 4 | 6 2 | 3+4 4 | 2+2 4 | 5+6 2 |
| | 5 2 | 3+4 4 | 5 2 | 3+4 4 | 2+2 4 | 3+4 2 |
| | 5 2 | 3+4 4 | 6 2 | 3+4 4 | 2+2 4 | 3+4 2 |
| | 5 2 | 3+4 4 | 5 2 | 3+4 4 | 2+2 4 | 5+6 2 |

Continued on next column

**TABLE XXXVIII– continued**

| TQC | ICSD | | NoE | | SF | |
|---|---|---|---|---|---|---|
| | 6 2 | 3+4 4 | 6 2 | 3+4 4 | 2+2 4 | 5+6 2 |
| | 6 2 | 3+4 4 | 6 2 | 3+4 4 | 2+2 4 | 3+4 2 |
| | 5 2 | 3+4 4 | 5 2 | 3+4 4 | 2+2 4 | 5+6 2 |
| | 5 2 | | 6 2 | | | 3+4 2 |
| | 5 2 | | 5 2 | | | 5+6 2 |
| | 6 2 | | 5 2 | | | 3+4 2 |
| | 6 2 | | 6 2 | | | 5+6 2 |
| | 6 2 | | 6 2 | | | 3+4 2 |
| | 5 2 | | 6 2 | | | 5+6 2 |
| | 6 2 | | 5 2 | | | 5+6 2 |
| | 5 2 | | 5 2 | | | 5+6 2 |
| | 6 2 | | 5 2 | | | 3+4 2 |
| | 5 2 | | 6 2 | | | 3+4 2 |
| | 6 2 | | 6 2 | | | 5+6 2 |
| | 5 2 | | 6 2 | | | 3+4 2 |
| | 6 2 | | 6 2 | | | 5+6 2 |
| | 5 2 | | 5 2 | | | 5+6 2 |
| | 6 2 | | 5 2 | | | 3+4 2 |
| | 5 2 | | 6 2 | | | 5+6 2 |
| | 6 2 | | 5 2 | | | 3+4 2 |
| | 6 2 | | 5 2 | | | 5+6 2 |
| | 5 2 | | 6 2 | | | 3+4 2 |
| | 5 2 | | 6 2 | | | 3+4 2 |
| | 5 2 | | 6 2 | | | 5+6 2 |
| | 6 2 | | 5 2 | | | 5+6 2 |
| 63.1.44 | 152349 | | 56 | | Dy1Sn2 | |
| | -GM | -T | -Y | -Z | -R | -S |
| | 5 2 | 3+4 4 | 5 2 | 3+4 4 | 2+2 4 | 3+4 2 |
| | 6 2 | 3+4 4 | 6 2 | 3+4 4 | 2+2 4 | 5+6 2 |
| | 5 2 | 3+4 4 | 5 2 | 3+4 4 | 2+2 4 | 3+4 2 |
| | 6 2 | 3+4 4 | 6 2 | 3+4 4 | 2+2 4 | 5+6 2 |
| | 5 2 | 3+4 4 | 5 2 | 3+4 4 | 2+2 4 | 3+4 2 |
| | 5 2 | 3+4 4 | 6 2 | 3+4 4 | 2+2 4 | 5+6 2 |
| | 6 2 | 3+4 4 | 5 2 | 3+4 4 | 2+2 4 | 3+4 2 |
| | 6 2 | 3+4 4 | 6 2 | 3+4 4 | 2+2 4 | 5+6 2 |
| | 5 2 | 3+4 4 | 6 2 | 3+4 4 | 2+2 4 | 5+6 2 |
| | 5 2 | 3+4 4 | 5 2 | 3+4 4 | 2+2 4 | 3+4 2 |
| | 6 2 | 3+4 4 | 6 2 | 3+4 4 | 2+2 4 | 5+6 2 |
| | 6 2 | 3+4 4 | 5 2 | 3+4 4 | 2+2 4 | 3+4 2 |
| | 5 2 | 3+4 4 | 5 2 | 3+4 4 | 2+2 4 | 3+4 2 |
| | 5 2 | 3+4 4 | 6 2 | 3+4 4 | 2+2 4 | 5+6 2 |
| | 6 2 | | 6 2 | | | 5+6 2 |
| | 5 2 | | 5 2 | | | 5+6 2 |
| | 6 2 | | 6 2 | | | 3+4 2 |
| | 5 2 | | 6 2 | | | 5+6 2 |
| | 5 2 | | 5 2 | | | 3+4 2 |
| | 6 2 | | 5 2 | | | 5+6 2 |
| | 6 2 | | 6 2 | | | 3+4 2 |
| | 5 2 | | 5 2 | | | 5+6 2 |
| | 6 2 | | 6 2 | | | 3+4 2 |
| | 5 2 | | 6 2 | | | 5+6 2 |
| | 5 2 | | 5 2 | | | 3+4 2 |
| | 6 2 | | 6 2 | | | 3+4 2 |
| | 5 2 | | 5 2 | | | 5+6 2 |
| | 6 2 | | 6 2 | | | 5+6 2 |
| 63.1.45 | 180214 | | 48 | | Dy1Si1 | |
| | -GM | -T | -Y | -Z | -R | -S |
| | 5 2 | 3+4 4 | 6 2 | 3+4 4 | 2+2 4 | 3+4 2 |
| | 6 2 | 3+4 4 | 5 2 | 3+4 4 | 2+2 4 | 5+6 2 |
| | 5 2 | 3+4 4 | 6 2 | 3+4 4 | 2+2 4 | 3+4 2 |
| | 6 2 | 3+4 4 | 5 2 | 3+4 4 | 2+2 4 | 5+6 2 |

Continued on next column

**TABLE XXXVIII– continued**

| TQC | ICSD | | NoE | | SF | |
|---|---|---|---|---|---|---|
| | 5 2 | 3+4 4 | 6 2 | 3+4 4 | 2+2 4 | 3+4 2 |
| | 5 2 | 3+4 4 | 5 2 | 3+4 4 | 2+2 4 | 5+6 2 |
| | 6 2 | 3+4 4 | 6 2 | 3+4 4 | 2+2 4 | 3+4 2 |
| | 6 2 | 3+4 4 | 5 2 | 3+4 4 | 2+2 4 | 5+6 2 |
| | 5 2 | 3+4 4 | 5 2 | 3+4 4 | 2+2 4 | 5+6 2 |
| | 6 2 | 3+4 4 | 6 2 | 3+4 4 | 2+2 4 | 3+4 2 |
| | 5 2 | 3+4 4 | 6 2 | 3+4 4 | 2+2 4 | 3+4 2 |
| | 6 2 | 3+4 4 | 5 2 | 3+4 4 | 2+2 4 | 5+6 2 |
| | 5 2 | | 5 2 | | | 5+6 2 |
| | 6 2 | | 6 2 | | | 5+6 2 |
| | 6 2 | | 5 2 | | | 3+4 2 |
| | 5 2 | | 6 2 | | | 5+6 2 |
| | 6 2 | | 5 2 | | | 3+4 2 |
| | 5 2 | | 6 2 | | | 5+6 2 |
| | 6 2 | | 5 2 | | | 3+4 2 |
| | 5 2 | | 5 2 | | | 5+6 2 |
| | 6 2 | | 6 2 | | | 3+4 2 |
| | 5 2 | | 5 2 | | | 5+6 2 |
| | 6 2 | | 6 2 | | | 3+4 2 |
| | 6 2 | | 5 2 | | | 5+6 2 |
| 63.1.46 | 154699 | | 48 | | Dy1Si1 | |
| | -GM | -T | -Y | -Z | -R | -S |
| | 5 2 | 3+4 4 | 6 2 | 3+4 4 | 2+2 4 | 3+4 2 |
| | 6 2 | 3+4 4 | 5 2 | 3+4 4 | 2+2 4 | 5+6 2 |
| | 5 2 | 3+4 4 | 6 2 | 3+4 4 | 2+2 4 | 3+4 2 |
| | 6 2 | 3+4 4 | 5 2 | 3+4 4 | 2+2 4 | 5+6 2 |
| | 5 2 | 3+4 4 | 6 2 | 3+4 4 | 2+2 4 | 3+4 2 |
| | 5 2 | 3+4 4 | 5 2 | 3+4 4 | 2+2 4 | 5+6 2 |
| | 6 2 | 3+4 4 | 6 2 | 3+4 4 | 2+2 4 | 3+4 2 |
| | 6 2 | 3+4 4 | 5 2 | 3+4 4 | 2+2 4 | 5+6 2 |
| | 5 2 | 3+4 4 | 5 2 | 3+4 4 | 2+2 4 | 5+6 2 |
| | 6 2 | 3+4 4 | 6 2 | 3+4 4 | 2+2 4 | 3+4 2 |
| | 5 2 | 3+4 4 | 5 2 | 3+4 4 | 2+2 4 | 3+4 2 |
| | 6 2 | 3+4 4 | 6 2 | 3+4 4 | 2+2 4 | 5+6 2 |
| | 5 2 | | 5 2 | | | 5+6 2 |
| | 6 2 | | 6 2 | | | 5+6 2 |
| | 6 2 | | 5 2 | | | 3+4 2 |
| | 5 2 | | 6 2 | | | 5+6 2 |
| | 5 2 | | 5 2 | | | 3+4 2 |
| | 6 2 | | 6 2 | | | 5+6 2 |
| | 5 2 | | 5 2 | | | 3+4 2 |
| | 6 2 | | 5 2 | | | 5+6 2 |
| | 6 2 | | 6 2 | | | 3+4 2 |
| | 5 2 | | 5 2 | | | 5+6 2 |
| | 6 2 | | 6 2 | | | 3+4 2 |
| | 6 2 | | 5 2 | | | 5+6 2 |
| 63.1.47 | 57225 | | 64 | | Dy1Ge3 | |
| | -GM | -T | -Y | -Z | -R | -S |
| | 5 2 | 3+4 4 | 5 2 | 3+4 4 | 2+2 4 | 5+6 2 |
| | 6 2 | 3+4 4 | 6 2 | 3+4 4 | 2+2 4 | 3+4 2 |
| | 5 2 | 3+4 4 | 5 2 | 3+4 4 | 2+2 4 | 5+6 2 |
| | 6 2 | 3+4 4 | 6 2 | 3+4 4 | 2+2 4 | 3+4 2 |
| | 5 2 | 3+4 4 | 5 2 | 3+4 4 | 2+2 4 | 5+6 2 |
| | 6 2 | 3+4 4 | 6 2 | 3+4 4 | 2+2 4 | 3+4 2 |
| | 5 2 | 3+4 4 | 5 2 | 3+4 4 | 2+2 4 | 5+6 2 |
| | 6 2 | 3+4 4 | 6 2 | 3+4 4 | 2+2 4 | 3+4 2 |
| | 5 2 | 3+4 4 | 6 2 | 3+4 4 | 2+2 4 | 3+4 2 |
| | 5 2 | 3+4 4 | 5 2 | 3+4 4 | 2+2 4 | 5+6 2 |
| | 6 2 | 3+4 4 | 5 2 | 3+4 4 | 2+2 4 | 3+4 2 |
| | 6 2 | 3+4 4 | 6 2 | 3+4 4 | 2+2 4 | 5+6 2 |
| | 6 2 | 3+4 4 | 5 2 | 3+4 4 | 2+2 4 | 3+4 2 |

Continued on next column

**TABLE XXXVIII– continued**

| TQC | ICSD | | NoE | | SF | |
|---|---|---|---|---|---|---|
| | 5 2 | 3+4 4 | 6 2 | 3+4 4 | 2+2 4 | 5+6 2 |
| | 5 2 | 3+4 4 | 6 2 | 3+4 4 | 2+2 4 | 5+6 2 |
| | 5 2 | 3+4 4 | 6 2 | 3+4 4 | 2+2 4 | 3+4 2 |
| | 5 2 | | 6 2 | | | 3+4 2 |
| | 5 2 | | 5 2 | | | 5+6 2 |
| | 6 2 | | 5 2 | | | 3+4 2 |
| | 6 2 | | 6 2 | | | 3+4 2 |
| | 6 2 | | 5 2 | | | 5+6 2 |
| | 6 2 | | 5 2 | | | 3+4 2 |
| | 5 2 | | 6 2 | | | 5+6 2 |
| | 5 2 | | 5 2 | | | 3+4 2 |
| | 5 2 | | 6 2 | | | 5+6 2 |
| | 6 2 | | 6 2 | | | 3+4 2 |
| | 5 2 | | 5 2 | | | 5+6 2 |
| | 6 2 | | 5 2 | | | 3+4 2 |
| | 6 2 | | 5 2 | | | 5+6 2 |
| | 6 2 | | 6 2 | | | 3+4 2 |
| | 6 2 | | 5 2 | | | 5+6 2 |
| | 5 2 | | 6 2 | | | 3+4 2 |
| 63.1.48 | 59858 | | 228 | | Cu3Ho2Rb1S5 | |
| | -GM | -T | -Y | -Z | -R | -S |
| | 5 2 | 3+4 4 | 6 2 | 3+4 4 | 2+2 4 | 3+4 2 |
| | 5 2 | 3+4 4 | 6 2 | 3+4 4 | 2+2 4 | 3+4 2 |
| | 6 2 | 3+4 4 | 5 2 | 3+4 4 | 2+2 4 | 5+6 2 |
| | 6 2 | 3+4 4 | 5 2 | 3+4 4 | 2+2 4 | 5+6 2 |
| | 5 2 | 3+4 4 | 5 2 | 3+4 4 | 2+2 4 | 3+4 2 |
| | 6 2 | 3+4 4 | 6 2 | 3+4 4 | 2+2 4 | 5+6 2 |
| | 5 2 | 3+4 4 | 6 2 | 3+4 4 | 2+2 4 | 3+4 2 |
| | 5 2 | 3+4 4 | 6 2 | 3+4 4 | 2+2 4 | 5+6 2 |
| | 6 2 | 3+4 4 | 5 2 | 3+4 4 | 2+2 4 | 3+4 2 |
| | 6 2 | 3+4 4 | 5 2 | 3+4 4 | 2+2 4 | 5+6 2 |
| | 5 2 | 3+4 4 | 6 2 | 3+4 4 | 2+2 4 | 3+4 2 |
| | 5 2 | 3+4 4 | 6 2 | 3+4 4 | 2+2 4 | 5+6 2 |
| | 6 2 | 3+4 4 | 5 2 | 3+4 4 | 2+2 4 | 3+4 2 |
| | 6 2 | 3+4 4 | 5 2 | 3+4 4 | 2+2 4 | 5+6 2 |
| | 5 2 | 3+4 4 | 6 2 | 3+4 4 | 2+2 4 | 3+4 2 |
| | 5 2 | 3+4 4 | 6 2 | 3+4 4 | 2+2 4 | 5+6 2 |
| | 6 2 | 3+4 4 | 5 2 | 3+4 4 | 2+2 4 | 3+4 2 |
| | 6 2 | 3+4 4 | 5 2 | 3+4 4 | 2+2 4 | 5+6 2 |
| | 5 2 | 3+4 4 | 6 2 | 3+4 4 | 2+2 4 | 5+6 2 |
| | 5 2 | 3+4 4 | 5 2 | 3+4 4 | 2+2 4 | 3+4 2 |
| | 6 2 | 3+4 4 | 6 2 | 3+4 4 | 2+2 4 | 3+4 2 |
| | 6 2 | 3+4 4 | 5 2 | 3+4 4 | 2+2 4 | 3+4 2 |
| | 6 2 | 3+4 4 | 5 2 | 3+4 4 | 2+2 4 | 5+6 2 |
| | 5 2 | 3+4 4 | 5 2 | 3+4 4 | 2+2 4 | 5+6 2 |
| | 6 2 | 3+4 4 | 6 2 | 3+4 4 | 2+2 4 | 3+4 2 |
| | 6 2 | 3+4 4 | 6 2 | 3+4 4 | 2+2 4 | 5+6 2 |
| | 5 2 | 3+4 4 | 5 2 | 3+4 4 | 2+2 4 | 5+6 2 |
| | 5 2 | 3+4 4 | 6 2 | 3+4 4 | 2+2 4 | 3+4 2 |
| | 5 2 | 3+4 4 | 5 2 | 3+4 4 | 2+2 4 | 3+4 2 |
| | 6 2 | 3+4 4 | 6 2 | 3+4 4 | 2+2 4 | 5+6 2 |
| | 5 2 | 3+4 4 | 5 2 | 3+4 4 | 2+2 4 | 3+4 2 |
| | 6 2 | 3+4 4 | 6 2 | 3+4 4 | 2+2 4 | 5+6 2 |
| | 5 2 | 3+4 4 | 5 2 | 3+4 4 | 2+2 4 | 3+4 2 |
| | 6 2 | 3+4 4 | 6 2 | 3+4 4 | 2+2 4 | 5+6 2 |
| | 5 2 | 3+4 4 | 5 2 | 3+4 4 | 2+2 4 | 3+4 2 |
| | 5 2 | 3+4 4 | 5 2 | 3+4 4 | 2+2 4 | 3+4 2 |
| | 6 2 | 3+4 4 | 6 2 | 3+4 4 | 2+2 4 | 5+6 2 |
| | 6 2 | 3+4 4 | 5 2 | 3+4 4 | 2+2 4 | 5+6 2 |
| | 6 2 | 3+4 4 | 6 2 | 3+4 4 | 2+2 4 | 5+6 2 |
| | 5 2 | 3+4 4 | 5 2 | 3+4 4 | 2+2 4 | 3+4 2 |
| | 5 2 | 3+4 4 | 6 2 | 3+4 4 | 2+2 4 | 3+4 2 |
| | 6 2 | 3+4 4 | 6 2 | 3+4 4 | 2+2 4 | 3+4 2 |
| | 5 2 | 3+4 4 | 6 2 | 3+4 4 | 2+2 4 | 5+6 2 |
| | 5 2 | 3+4 4 | 6 2 | 3+4 4 | 2+2 4 | 5+6 2 |
| | 5 2 | 3+4 4 | 5 2 | 3+4 4 | 2+2 4 | 5+6 2 |
| | 6 2 | 3+4 4 | 6 2 | 3+4 4 | 2+2 4 | 5+6 2 |
| | 6 2 | 3+4 4 | 5 2 | 3+4 4 | 2+2 4 | 3+4 2 |
| | 6 2 | 3+4 4 | 6 2 | 3+4 4 | 2+2 4 | 3+4 2 |
| | 6 2 | 3+4 4 | 6 2 | 3+4 4 | 2+2 4 | 5+6 2 |
| | 5 2 | 3+4 4 | 5 2 | 3+4 4 | 2+2 4 | 5+6 2 |
| | 6 2 | 3+4 4 | 6 2 | 3+4 4 | 2+2 4 | 5+6 2 |
| | 5 2 | 3+4 4 | 5 2 | 3+4 4 | 2+2 4 | 3+4 2 |
| | 5 2 | 3+4 4 | 6 2 | 3+4 4 | 2+2 4 | 5+6 2 |
| | 5 2 | 3+4 4 | 5 2 | 3+4 4 | 2+2 4 | 3+4 2 |
| | 6 2 | 3+4 4 | 5 2 | 3+4 4 | 2+2 4 | 5+6 2 |
| | 6 2 | 3+4 4 | 6 2 | 3+4 4 | 2+2 4 | 3+4 2 |
| | 5 2 | 3+4 4 | 6 2 | 3+4 4 | 2+2 4 | 5+6 2 |
| | 5 2 | | 5 2 | | | 3+4 2 |
| | 5 2 | | 5 2 | | | 3+4 2 |
| | 5 2 | | 5 2 | | | 5+6 2 |
| | 6 2 | | 5 2 | | | 3+4 2 |
| | 6 2 | | 6 2 | | | 5+6 2 |
| | 6 2 | | 5 2 | | | 3+4 2 |
| | 5 2 | | 5 2 | | | 3+4 2 |
| | 5 2 | | 5 2 | | | 3+4 2 |
| | 5 2 | | 6 2 | | | 3+4 2 |
| | 6 2 | | 5 2 | | | 5+6 2 |
| | 6 2 | | 6 2 | | | 5+6 2 |
| | 6 2 | | 6 2 | | | 3+4 2 |
| | 6 2 | | 6 2 | | | 5+6 2 |
| | 5 2 | | 5 2 | | | 3+4 2 |
| | 6 2 | | 6 2 | | | 5+6 2 |
| | 6 2 | | 6 2 | | | 5+6 2 |
| | 6 2 | | 5 2 | | | 3+4 2 |
| | 5 2 | | 6 2 | | | 3+4 2 |
| | 6 2 | | 5 2 | | | 5+6 2 |
| | 5 2 | | 5 2 | | | 3+4 2 |
| | 6 2 | | 5 2 | | | 5+6 2 |
| | 5 2 | | 6 2 | | | 3+4 2 |
| | 5 2 | | 5 2 | | | 5+6 2 |
| | 5 2 | | 6 2 | | | 3+4 2 |
| | 5 2 | | 5 2 | | | 3+4 2 |
| | 6 2 | | 5 2 | | | 3+4 2 |
| | 6 2 | | 6 2 | | | 5+6 2 |
| | 5 2 | | 6 2 | | | 3+4 2 |
| | 6 2 | | 6 2 | | | 5+6 2 |
| | 5 2 | | 6 2 | | | 5+6 2 |
| | 6 2 | | 6 2 | | | 3+4 2 |
| | 5 2 | | 5 2 | | | 3+4 2 |
| | 5 2 | | 6 2 | | | 5+6 2 |
| | 6 2 | | 6 2 | | | 3+4 2 |
| | 5 2 | | 5 2 | | | 5+6 2 |
| | 6 2 | | 6 2 | | | 5+6 2 |
| | 5 2 | | 5 2 | | | 3+4 2 |
| | 6 2 | | 5 2 | | | 5+6 2 |
| | 5 2 | | 5 2 | | | 3+4 2 |
| | 6 2 | | 6 2 | | | 3+4 2 |
| | 5 2 | | 6 2 | | | 5+6 2 |
| | 6 2 | | 5 2 | | | 5+6 2 |
| | 6 2 | | 5 2 | | | 3+4 2 |
| | 5 2 | | 6 2 | | | 5+6 2 |

Continued on next column

**TABLE XXXVIII– continued**

| TQC | ICSD | | NoE | | SF | |
|---|---|---|---|---|---|---|
| | 5 2 | | 5 2 | | | 3+4 2 |
| | 6 2 | | 5 2 | | | 5+6 2 |
| | 5 2 | | 6 2 | | | 3+4 2 |
| | 6 2 | | 5 2 | | | 5+6 2 |
| | 6 2 | | 6 2 | | | 5+6 2 |
| | 5 2 | | 6 2 | | | 3+4 2 |
| | 5 2 | | 6 2 | | | 5+6 2 |
| | 5 2 | | 5 2 | | | 3+4 2 |
| | 6 2 | | 5 2 | | | 5+6 2 |
| | 6 2 | | 6 2 | | | 3+4 2 |
| | 5 2 | | 6 2 | | | 5+6 2 |
| | 6 2 | | 5 2 | | | 3+4 2 |
| | 5 2 | | 6 2 | | | 5+6 2 |
| 63.1.49 | 174371 | | 228 | Cu3Ho2K1Se5 | | |
| | -GM | -T | -Y | -Z | -R | -S |
| | 5 2 | 3+4 4 | 6 2 | 3+4 4 | 2+2 4 | 3+4 2 |
| | 5 2 | 3+4 4 | 6 2 | 3+4 4 | 2+2 4 | 3+4 2 |
| | 6 2 | 3+4 4 | 5 2 | 3+4 4 | 2+2 4 | 5+6 2 |
| | 6 2 | 3+4 4 | 5 2 | 3+4 4 | 2+2 4 | 5+6 2 |
| | 5 2 | 3+4 4 | 5 2 | 3+4 4 | 2+2 4 | 3+4 2 |
| | 6 2 | 3+4 4 | 6 2 | 3+4 4 | 2+2 4 | 5+6 2 |
| | 5 2 | 3+4 4 | 6 2 | 3+4 4 | 2+2 4 | 3+4 2 |
| | 5 2 | 3+4 4 | 6 2 | 3+4 4 | 2+2 4 | 5+6 2 |
| | 6 2 | 3+4 4 | 5 2 | 3+4 4 | 2+2 4 | 3+4 2 |
| | 6 2 | 3+4 4 | 5 2 | 3+4 4 | 2+2 4 | 5+6 2 |
| | 5 2 | 3+4 4 | 6 2 | 3+4 4 | 2+2 4 | 3+4 2 |
| | 5 2 | 3+4 4 | 6 2 | 3+4 4 | 2+2 4 | 5+6 2 |
| | 6 2 | 3+4 4 | 5 2 | 3+4 4 | 2+2 4 | 3+4 2 |
| | 6 2 | 3+4 4 | 5 2 | 3+4 4 | 2+2 4 | 5+6 2 |
| | 5 2 | 3+4 4 | 6 2 | 3+4 4 | 2+2 4 | 3+4 2 |
| | 5 2 | 3+4 4 | 6 2 | 3+4 4 | 2+2 4 | 5+6 2 |
| | 6 2 | 3+4 4 | 5 2 | 3+4 4 | 2+2 4 | 3+4 2 |
| | 6 2 | 3+4 4 | 5 2 | 3+4 4 | 2+2 4 | 5+6 2 |
| | 5 2 | 3+4 4 | 6 2 | 3+4 4 | 2+2 4 | 3+4 2 |
| | 6 2 | 3+4 4 | 5 2 | 3+4 4 | 2+2 4 | 5+6 2 |
| | 5 2 | 3+4 4 | 6 2 | 3+4 4 | 2+2 4 | 3+4 2 |
| | 6 2 | 3+4 4 | 5 2 | 3+4 4 | 2+2 4 | 5+6 2 |
| | 5 2 | 3+4 4 | 5 2 | 3+4 4 | 2+2 4 | 3+4 2 |
| | 6 2 | 3+4 4 | 6 2 | 3+4 4 | 2+2 4 | 5+6 2 |
| | 5 2 | 3+4 4 | 6 2 | 3+4 4 | 2+2 4 | 5+6 2 |
| | 5 2 | 3+4 4 | 6 2 | 3+4 4 | 2+2 4 | 3+4 2 |
| | 6 2 | 3+4 4 | 5 2 | 3+4 4 | 2+2 4 | 5+6 2 |
| | 6 2 | 3+4 4 | 5 2 | 3+4 4 | 2+2 4 | 3+4 2 |
| | 6 2 | 3+4 4 | 5 2 | 3+4 4 | 2+2 4 | 5+6 2 |
| | 5 2 | 3+4 4 | 5 2 | 3+4 4 | 2+2 4 | 3+4 2 |
| | 6 2 | 3+4 4 | 6 2 | 3+4 4 | 2+2 4 | 5+6 2 |
| | 6 2 | 3+4 4 | 5 2 | 3+4 4 | 2+2 4 | 3+4 2 |
| | 5 2 | 3+4 4 | 6 2 | 3+4 4 | 2+2 4 | 5+6 2 |
| | 5 2 | 3+4 4 | 6 2 | 3+4 4 | 2+2 4 | 3+4 2 |
| | 5 2 | 3+4 4 | 5 2 | 3+4 4 | 2+2 4 | 5+6 2 |
| | 5 2 | 3+4 4 | 5 2 | 3+4 4 | 2+2 4 | 5+6 2 |
| | 6 2 | 3+4 4 | 6 2 | 3+4 4 | 2+2 4 | 3+4 2 |
| | 6 2 | 3+4 4 | 5 2 | 3+4 4 | 2+2 4 | 3+4 2 |
| | 6 2 | 3+4 4 | 6 2 | 3+4 4 | 2+2 4 | 3+4 2 |
| | 5 2 | 3+4 4 | 5 2 | 3+4 4 | 2+2 4 | 5+6 2 |
| | 5 2 | 3+4 4 | 6 2 | 3+4 4 | 2+2 4 | 5+6 2 |
| | 5 2 | 3+4 4 | 6 2 | 3+4 4 | 2+2 4 | 5+6 2 |
| | 6 2 | 3+4 4 | 6 2 | 3+4 4 | 2+2 4 | 3+4 2 |
| | 6 2 | 3+4 4 | 6 2 | 3+4 4 | 2+2 4 | 3+4 2 |
| | 5 2 | 3+4 4 | 6 2 | 3+4 4 | 2+2 4 | 3+4 2 |
| | 6 2 | 3+4 4 | 6 2 | 3+4 4 | 2+2 4 | 3+4 2 |

Continued on next column

**TABLE XXXVIII– continued**

| TQC | ICSD | | NoE | | SF | |
|---|---|---|---|---|---|---|
| | 5 2 | 3+4 4 | 6 2 | 3+4 4 | 2+2 4 | 5+6 2 |
| | 6 2 | 3+4 4 | 5 2 | 3+4 4 | 2+2 4 | 5+6 2 |
| | 6 2 | 3+4 4 | 5 2 | 3+4 4 | 2+2 4 | 3+4 2 |
| | 5 2 | 3+4 4 | 5 2 | 3+4 4 | 2+2 4 | 3+4 2 |
| | 5 2 | 3+4 4 | 6 2 | 3+4 4 | 2+2 4 | 3+4 2 |
| | 6 2 | 3+4 4 | 6 2 | 3+4 4 | 2+2 4 | 5+6 2 |
| | 5 2 | 3+4 4 | 5 2 | 3+4 4 | 2+2 4 | 3+4 2 |
| | 6 2 | 3+4 4 | 5 2 | 3+4 4 | 2+2 4 | 5+6 2 |
| | 5 2 | 3+4 4 | 5 2 | 3+4 4 | 2+2 4 | 3+4 2 |
| | 6 2 | 3+4 4 | 6 2 | 3+4 4 | 2+2 4 | 5+6 2 |
| | 5 2 | 3+4 4 | 6 2 | 3+4 4 | 2+2 4 | 3+4 2 |
| | 5 2 | | 5 2 | | | 5+6 2 |
| | 5 2 | | 5 2 | | | 3+4 2 |
| | 5 2 | | 5 2 | | | 5+6 2 |
| | 6 2 | | 5 2 | | | 5+6 2 |
| | 5 2 | | 5 2 | | | 3+4 2 |
| | 5 2 | | 6 2 | | | 5+6 2 |
| | 6 2 | | 5 2 | | | 5+6 2 |
| | 6 2 | | 5 2 | | | 5+6 2 |
| | 5 2 | | 6 2 | | | 5+6 2 |
| | 6 2 | | 5 2 | | | 3+4 2 |
| | 6 2 | | 6 2 | | | 3+4 2 |
| | 6 2 | | 6 2 | | | 5+6 2 |
| | 6 2 | | 6 2 | | | 3+4 2 |
| | 5 2 | | 5 2 | | | 3+4 2 |
| | 6 2 | | 5 2 | | | 5+6 2 |
| | 6 2 | | 6 2 | | | 5+6 2 |
| | 6 2 | | 6 2 | | | 3+4 2 |
| | 5 2 | | 6 2 | | | 3+4 2 |
| | 6 2 | | 5 2 | | | 5+6 2 |
| | 6 2 | | 5 2 | | | 5+6 2 |
| | 5 2 | | 6 2 | | | 3+4 2 |
| | 5 2 | | 6 2 | | | 5+6 2 |
| | 5 2 | | 5 2 | | | 3+4 2 |
| | 5 2 | | 5 2 | | | 5+6 2 |
| | 5 2 | | 5 2 | | | 5+6 2 |
| | 6 2 | | 6 2 | | | 5+6 2 |
| | 6 2 | | 5 2 | | | 3+4 2 |
| | 6 2 | | 6 2 | | | 5+6 2 |
| | 5 2 | | 6 2 | | | 3+4 2 |
| | 5 2 | | 6 2 | | | 5+6 2 |
| | 6 2 | | 6 2 | | | 3+4 2 |
| | 5 2 | | 5 2 | | | 5+6 2 |
| | 5 2 | | 6 2 | | | 3+4 2 |
| | 6 2 | | 5 2 | | | 3+4 2 |
| | 6 2 | | 6 2 | | | 5+6 2 |
| | 5 2 | | 5 2 | | | 3+4 2 |
| | 5 2 | | 5 2 | | | 5+6 2 |
| | 6 2 | | 6 2 | | | 3+4 2 |
| | 5 2 | | 5 2 | | | 5+6 2 |
| | 6 2 | | 6 2 | | | 3+4 2 |
| | 5 2 | | 6 2 | | | 5+6 2 |
| | 6 2 | | 5 2 | | | 3+4 2 |
| | 6 2 | | 6 2 | | | 5+6 2 |
| | 5 2 | | 5 2 | | | 3+4 2 |
| | 5 2 | | 5 2 | | | 5+6 2 |
| | 6 2 | | 6 2 | | | 3+4 2 |
| | 5 2 | | 5 2 | | | 5+6 2 |
| | 6 2 | | 6 2 | | | 3+4 2 |
| | 6 2 | | 5 2 | | | 3+4 2 |
| | 5 2 | | 6 2 | | | 3+4 2 |

Continued on next column

**TABLE XXXVIII– continued**

| TQC | ICSD | NoE | SF | | | |
|---|---|---|---|---|---|---|
| | 5 2 | | 6 2 | | | 5+6 2 |
| | 5 2 | | 5 2 | | | 3+4 2 |
| | 6 2 | | 5 2 | | | 5+6 2 |
| | 6 2 | | 6 2 | | | 3+4 2 |
| | 5 2 | | 6 2 | | | 5+6 2 |
| | 6 2 | | 6 2 | | | 5+6 2 |
| | 5 2 | | 5 2 | | | 3+4 2 |
| 63.1.50 | 423716 | 220 | Cu3Er2S5Tl1 | | | |
| | -GM | -T | -Y | -Z | -R | -S |
| | 5 2 | 3+4 4 | 6 2 | 3+4 4 | 2+2 4 | 3+4 2 |
| | 5 2 | 3+4 4 | 6 2 | 3+4 4 | 2+2 4 | 3+4 2 |
| | 6 2 | 3+4 4 | 5 2 | 3+4 4 | 2+2 4 | 5+6 2 |
| | 6 2 | 3+4 4 | 5 2 | 3+4 4 | 2+2 4 | 5+6 2 |
| | 5 2 | 3+4 4 | 6 2 | 3+4 4 | 2+2 4 | 3+4 2 |
| | 5 2 | 3+4 4 | 6 2 | 3+4 4 | 2+2 4 | 5+6 2 |
| | 6 2 | 3+4 4 | 5 2 | 3+4 4 | 2+2 4 | 3+4 2 |
| | 6 2 | 3+4 4 | 5 2 | 3+4 4 | 2+2 4 | 5+6 2 |
| | 5 2 | 3+4 4 | 6 2 | 3+4 4 | 2+2 4 | 3+4 2 |
| | 5 2 | 3+4 4 | 6 2 | 3+4 4 | 2+2 4 | 5+6 2 |
| | 6 2 | 3+4 4 | 5 2 | 3+4 4 | 2+2 4 | 3+4 2 |
| | 6 2 | 3+4 4 | 5 2 | 3+4 4 | 2+2 4 | 5+6 2 |
| | 5 2 | 3+4 4 | 6 2 | 3+4 4 | 2+2 4 | 3+4 2 |
| | 5 2 | 3+4 4 | 6 2 | 3+4 4 | 2+2 4 | 5+6 2 |
| | 6 2 | 3+4 4 | 5 2 | 3+4 4 | 2+2 4 | 3+4 2 |
| | 6 2 | 3+4 4 | 5 2 | 3+4 4 | 2+2 4 | 5+6 2 |
| | 5 2 | 3+4 4 | 6 2 | 3+4 4 | 2+2 4 | 5+6 2 |
| | 5 2 | 3+4 4 | 6 2 | 3+4 4 | 2+2 4 | 3+4 2 |
| | 6 2 | 3+4 4 | 5 2 | 3+4 4 | 2+2 4 | 5+6 2 |
| | 6 2 | 3+4 4 | 5 2 | 3+4 4 | 2+2 4 | 3+4 2 |
| | 6 2 | 3+4 4 | 5 2 | 3+4 4 | 2+2 4 | 5+6 2 |
| | 5 2 | 3+4 4 | 5 2 | 3+4 4 | 2+2 4 | 5+6 2 |
| | 6 2 | 3+4 4 | 6 2 | 3+4 4 | 2+2 4 | 3+4 2 |
| | 6 2 | 3+4 4 | 6 2 | 3+4 4 | 2+2 4 | 3+4 2 |
| | 5 2 | 3+4 4 | 5 2 | 3+4 4 | 2+2 4 | 5+6 2 |
| | 5 2 | 3+4 4 | 6 2 | 3+4 4 | 2+2 4 | 3+4 2 |
| | 5 2 | 3+4 4 | 5 2 | 3+4 4 | 2+2 4 | 5+6 2 |
| | 6 2 | 3+4 4 | 6 2 | 3+4 4 | 2+2 4 | 3+4 2 |
| | 5 2 | 3+4 4 | 5 2 | 3+4 4 | 2+2 4 | 5+6 2 |
| | 5 2 | 3+4 4 | 5 2 | 3+4 4 | 2+2 4 | 5+6 2 |
| | 6 2 | 3+4 4 | 5 2 | 3+4 4 | 2+2 4 | 3+4 2 |
| | 6 2 | 3+4 4 | 6 2 | 3+4 4 | 2+2 4 | 3+4 2 |
| | 6 2 | 3+4 4 | 6 2 | 3+4 4 | 2+2 4 | 3+4 2 |
| | 5 2 | 3+4 4 | 5 2 | 3+4 4 | 2+2 4 | 5+6 2 |
| | 5 2 | 3+4 4 | 6 2 | 3+4 4 | 2+2 4 | 5+6 2 |
| | 6 2 | 3+4 4 | 6 2 | 3+4 4 | 2+2 4 | 3+4 2 |
| | 6 2 | 3+4 4 | 6 2 | 3+4 4 | 2+2 4 | 3+4 2 |
| | 5 2 | 3+4 4 | 6 2 | 3+4 4 | 2+2 4 | 3+4 2 |
| | 5 2 | 3+4 4 | 6 2 | 3+4 4 | 2+2 4 | 5+6 2 |
| | 6 2 | 3+4 4 | 5 2 | 3+4 4 | 2+2 4 | 5+6 2 |
| | 6 2 | 3+4 4 | 6 2 | 3+4 4 | 2+2 4 | 3+4 2 |
| | 5 2 | 3+4 4 | 5 2 | 3+4 4 | 2+2 4 | 3+4 2 |
| | 5 2 | 3+4 4 | 6 2 | 3+4 4 | 2+2 4 | 3+4 2 |
| | 6 2 | 3+4 4 | 5 2 | 3+4 4 | 2+2 4 | 5+6 2 |
| | 5 2 | 3+4 4 | 6 2 | 3+4 4 | 2+2 4 | 3+4 2 |
| | 6 2 | 3+4 4 | 5 2 | 3+4 4 | 2+2 4 | 5+6 2 |
| | 5 2 | 3+4 4 | 5 2 | 3+4 4 | 2+2 4 | 3+4 2 |
| | 6 2 | 3+4 4 | 6 2 | 3+4 4 | 2+2 4 | 5+6 2 |
| | 5 2 | 3+4 4 | 6 2 | 3+4 4 | 2+2 4 | 3+4 2 |
| | 5 2 | 3+4 4 | 5 2 | 3+4 4 | 2+2 4 | 5+6 2 |
| | 6 2 | 3+4 4 | 5 2 | 3+4 4 | 2+2 4 | 5+6 2 |
| | 5 2 | 3+4 4 | 5 2 | 3+4 4 | 2+2 4 | 5+6 2 |

Continued on next column

**TABLE XXXVIII– continued**

| TQC | ICSD | NoE | SF | | | |
|---|---|---|---|---|---|---|
| | 5 2 | 3+4 4 | 5 2 | 3+4 4 | 2+2 4 | 3+4 2 |
| | 6 2 | 3+4 4 | 6 2 | 3+4 4 | 2+2 4 | 3+4 2 |
| | 5 2 | 3+4 4 | 6 2 | 3+4 4 | 2+2 4 | 5+6 2 |
| | 6 2 | | 5 2 | | | 5+6 2 |
| | 5 2 | | 5 2 | | | 5+6 2 |
| | 5 2 | | 5 2 | | | 3+4 2 |
| | 6 2 | | 6 2 | | | 5+6 2 |
| | 6 2 | | 5 2 | | | 5+6 2 |
| | 5 2 | | 5 2 | | | 3+4 2 |
| | 6 2 | | 6 2 | | | 3+4 2 |
| | 6 2 | | 6 2 | | | 5+6 2 |
| | 6 2 | | 6 2 | | | 3+4 2 |
| | 5 2 | | 5 2 | | | 3+4 2 |
| | 6 2 | | 6 2 | | | 5+6 2 |
| | 6 2 | | 6 2 | | | 3+4 2 |
| | 6 2 | | 5 2 | | | 5+6 2 |
| | 5 2 | | 5 2 | | | 3+4 2 |
| | 5 2 | | 6 2 | | | 5+6 2 |
| | 6 2 | | 5 2 | | | 3+4 2 |
| | 5 2 | | 6 2 | | | 5+6 2 |
| | 6 2 | | 5 2 | | | 5+6 2 |
| | 5 2 | | 5 2 | | | 3+4 2 |
| | 5 2 | | 6 2 | | | 5+6 2 |
| | 6 2 | | 5 2 | | | 5+6 2 |
| | 5 2 | | 5 2 | | | 5+6 2 |
| | 6 2 | | 6 2 | | | 3+4 2 |
| | 5 2 | | 6 2 | | | 5+6 2 |
| | 6 2 | | 6 2 | | | 3+4 2 |
| | 5 2 | | 5 2 | | | 5+6 2 |
| | 6 2 | | 6 2 | | | 3+4 2 |
| | 5 2 | | 5 2 | | | 5+6 2 |
| | 6 2 | | 6 2 | | | 3+4 2 |
| | 5 2 | | 6 2 | | | 3+4 2 |
| | 6 2 | | 5 2 | | | 5+6 2 |
| | 5 2 | | 6 2 | | | 3+4 2 |
| | 5 2 | | 6 2 | | | 5+6 2 |
| | 6 2 | | 5 2 | | | 3+4 2 |
| | 5 2 | | 5 2 | | | 5+6 2 |
| | 6 2 | | 6 2 | | | 5+6 2 |
| | 5 2 | | 6 2 | | | 3+4 2 |
| | 6 2 | | 5 2 | | | 3+4 2 |
| | 5 2 | | 5 2 | | | 5+6 2 |
| | 6 2 | | 6 2 | | | 3+4 2 |
| | 5 2 | | 5 2 | | | 5+6 2 |
| | 6 2 | | 5 2 | | | 3+4 2 |
| | 5 2 | | 6 2 | | | 5+6 2 |
| | 6 2 | | 6 2 | | | 3+4 2 |
| | 6 2 | | 5 2 | | | 3+4 2 |
| | 5 2 | | 6 2 | | | 5+6 2 |
| | 5 2 | | 6 2 | | | 3+4 2 |
| | 5 2 | | 5 2 | | | 3+4 2 |
| | 6 2 | | 5 2 | | | 5+6 2 |
| | 6 2 | | 6 2 | | | 5+6 2 |
| | 5 2 | | 6 2 | | | 3+4 2 |
| | 5 2 | | 5 2 | | | 5+6 2 |
| | 5 2 | | 6 2 | | | 3+4 2 |
| | 6 2 | | 5 2 | | | 5+6 2 |
| | 6 2 | | 6 2 | | | 5+6 2 |
| 63.1.51 | 174372 | 232 | Cu3Er2K1Se5 | | | |
| | -GM | -T | -Y | -Z | -R | -S |
| | 5 2 | 3+4 4 | 5 2 | 3+4 4 | 2+2 4 | 3+4 2 |

Continued on next column

**TABLE XXXVIII– continued**

| TQC | ICSD | NoE | SF |
|---|---|---|---|
| | 5 2 3+4 4 | 5 2 3+4 4 2+2 4 | 3+4 2 |
| | 6 2 3+4 4 | 6 2 3+4 4 2+2 4 | 5+6 2 |
| | 6 2 3+4 4 | 6 2 3+4 4 2+2 4 | 5+6 2 |
| | 5 2 3+4 4 | 5 2 3+4 4 2+2 4 | 3+4 2 |
| | 6 2 3+4 4 | 6 2 3+4 4 2+2 4 | 5+6 2 |
| | 5 2 3+4 4 | 6 2 3+4 4 2+2 4 | 3+4 2 |
| | 5 2 3+4 4 | 6 2 3+4 4 2+2 4 | 3+4 2 |
| | 6 2 3+4 4 | 5 2 3+4 4 2+2 4 | 5+6 2 |
| | 6 2 3+4 4 | 5 2 3+4 4 2+2 4 | 5+6 2 |
| | 5 2 3+4 4 | 6 2 3+4 4 2+2 4 | 3+4 2 |
| | 5 2 3+4 4 | 6 2 3+4 4 2+2 4 | 3+4 2 |
| | 6 2 3+4 4 | 5 2 3+4 4 2+2 4 | 5+6 2 |
| | 6 2 3+4 4 | 5 2 3+4 4 2+2 4 | 5+6 2 |
| | 5 2 3+4 4 | 6 2 3+4 4 2+2 4 | 3+4 2 |
| | 5 2 3+4 4 | 6 2 3+4 4 2+2 4 | 3+4 2 |
| | 6 2 3+4 4 | 5 2 3+4 4 2+2 4 | 5+6 2 |
| | 6 2 3+4 4 | 5 2 3+4 4 2+2 4 | 5+6 2 |
| | 5 2 3+4 4 | 5 2 3+4 4 2+2 4 | 3+4 2 |
| | 6 2 3+4 4 | 6 2 3+4 4 2+2 4 | 5+6 2 |
| | 5 2 3+4 4 | 5 2 3+4 4 2+2 4 | 3+4 2 |
| | 6 2 3+4 4 | 6 2 3+4 4 2+2 4 | 5+6 2 |
| | 5 2 3+4 4 | 5 2 3+4 4 2+2 4 | 3+4 2 |
| | 6 2 3+4 4 | 6 2 3+4 4 2+2 4 | 5+6 2 |
| | 5 2 3+4 4 | 6 2 3+4 4 2+2 4 | 3+4 2 |
| | 5 2 3+4 4 | 6 2 3+4 4 2+2 4 | 3+4 2 |
| | 6 2 3+4 4 | 5 2 3+4 4 2+2 4 | 5+6 2 |
| | 6 2 3+4 4 | 5 2 3+4 4 2+2 4 | 5+6 2 |
| | 6 2 3+4 4 | 5 2 3+4 4 2+2 4 | 3+4 2 |
| | 5 2 3+4 4 | 6 2 3+4 4 2+2 4 | 5+6 2 |
| | 6 2 3+4 4 | 5 2 3+4 4 2+2 4 | 5+6 2 |
| | 6 2 3+4 4 | 5 2 3+4 4 2+2 4 | 5+6 2 |
| | 5 2 3+4 4 | 6 2 3+4 4 2+2 4 | 3+4 2 |
| | 5 2 3+4 4 | 6 2 3+4 4 2+2 4 | 3+4 2 |
| | 5 2 3+4 4 | 5 2 3+4 4 2+2 4 | 5+6 2 |
| | 5 2 3+4 4 | 5 2 3+4 4 2+2 4 | 5+6 2 |
| | 6 2 3+4 4 | 6 2 3+4 4 2+2 4 | 3+4 2 |
| | 6 2 3+4 4 | 5 2 3+4 4 2+2 4 | 3+4 2 |
| | 6 2 3+4 4 | 6 2 3+4 4 2+2 4 | 3+4 2 |
| | 5 2 3+4 4 | 5 2 3+4 4 2+2 4 | 5+6 2 |
| | 5 2 3+4 4 | 6 2 3+4 4 2+2 4 | 5+6 2 |
| | 5 2 3+4 4 | 6 2 3+4 4 2+2 4 | 5+6 2 |
| | 5 2 3+4 4 | 6 2 3+4 4 2+2 4 | 3+4 2 |
| | 5 2 3+4 4 | 6 2 3+4 4 2+2 4 | 3+4 2 |
| | 6 2 3+4 4 | 6 2 3+4 4 2+2 4 | 5+6 2 |
| | 6 2 3+4 4 | 5 2 3+4 4 2+2 4 | 3+4 2 |
| | 6 2 3+4 4 | 5 2 3+4 4 2+2 4 | 5+6 2 |
| | 5 2 3+4 4 | 6 2 3+4 4 2+2 4 | 3+4 2 |
| | 6 2 3+4 4 | 5 2 3+4 4 2+2 4 | 5+6 2 |
| | 5 2 3+4 4 | 6 2 3+4 4 2+2 4 | 3+4 2 |
| | 6 2 3+4 4 | 5 2 3+4 4 2+2 4 | 5+6 2 |
| | 6 2 3+4 4 | 6 2 3+4 4 2+2 4 | 3+4 2 |
| | 5 2 3+4 4 | 6 2 3+4 4 2+2 4 | 3+4 2 |
| | 5 2 3+4 4 | 6 2 3+4 4 2+2 4 | 3+4 2 |
| | 6 2 3+4 4 | 5 2 3+4 4 2+2 4 | 5+6 2 |
| | 6 2 3+4 4 | 5 2 3+4 4 2+2 4 | 3+4 2 |
| | 5 2 3+4 4 | 6 2 3+4 4 2+2 4 | 3+4 2 |
| | 5 2 3+4 4 | 5 2 3+4 4 2+2 4 | 3+4 2 |
| | 5 2 | 5 2 | 5+6 2 |
| | 5 2 | 5 2 | 5+6 2 |
| | 6 2 | 5 2 | 5+6 2 |
| | 5 2 | 5 2 | 3+4 2 |

Continued on next column

**TABLE XXXVIII– continued**

| TQC | ICSD | NoE | SF |
|---|---|---|---|
| | 5 2 | 5 2 | 5+6 2 |
| | 6 2 | 5 2 | 5+6 2 |
| | 6 2 | 6 2 | 5+6 2 |
| | 5 2 | 6 2 | 5+6 2 |
| | 6 2 | 5 2 | 3+4 2 |
| | 6 2 | 6 2 | 3+4 2 |
| | 6 2 | 6 2 | 3+4 2 |
| | 6 2 | 5 2 | 5+6 2 |
| | 5 2 | 6 2 | 3+4 2 |
| | 6 2 | 6 2 | 5+6 2 |
| | 6 2 | 5 2 | 3+4 2 |
| | 6 2 | 6 2 | 5+6 2 |
| | 6 2 | 5 2 | 3+4 2 |
| | 5 2 | 6 2 | 5+6 2 |
| | 5 2 | 5 2 | 5+6 2 |
| | 5 2 | 6 2 | 3+4 2 |
| | 6 2 | 5 2 | 5+6 2 |
| | 5 2 | 6 2 | 3+4 2 |
| | 6 2 | 5 2 | 5+6 2 |
| | 6 2 | 5 2 | 3+4 2 |
| | 6 2 | 6 2 | 5+6 2 |
| | 5 2 | 6 2 | 3+4 2 |
| | 5 2 | 5 2 | 3+4 2 |
| | 5 2 | 6 2 | 5+6 2 |
| | 5 2 | 5 2 | 3+4 2 |
| | 6 2 | 6 2 | 5+6 2 |
| | 6 2 | 5 2 | 5+6 2 |
| | 6 2 | 6 2 | 3+4 2 |
| | 5 2 | 6 2 | 3+4 2 |
| | 5 2 | 5 2 | 5+6 2 |
| | 5 2 | 5 2 | 5+6 2 |
| | 5 2 | 6 2 | 3+4 2 |
| | 6 2 | 5 2 | 5+6 2 |
| | 6 2 | 6 2 | 5+6 2 |
| | 5 2 | 6 2 | 5+6 2 |
| | 6 2 | 6 2 | 3+4 2 |
| | 5 2 | 6 2 | 3+4 2 |
| | 6 2 | 5 2 | 5+6 2 |
| | 6 2 | 5 2 | 3+4 2 |
| | 6 2 | 5 2 | 5+6 2 |
| | 5 2 | 5 2 | 3+4 2 |
| | 5 2 | 6 2 | 5+6 2 |
| | 5 2 | 6 2 | 3+4 2 |
| | 6 2 | 5 2 | 3+4 2 |
| | 5 2 | 6 2 | 3+4 2 |
| | 5 2 | 5 2 | 5+6 2 |
| | 5 2 | 6 2 | 3+4 2 |
| | 6 2 | 5 2 | 5+6 2 |
| | 6 2 | 6 2 | 5+6 2 |
| | 6 2 | 6 2 | 3+4 2 |
| | 5 2 | 5 2 | 5+6 2 |
| | 6 2 | 6 2 | 3+4 2 |
| | 5 2 | 5 2 | 5+6 2 |
| | 5 2 | 6 2 | 5+6 2 |
| 63.1.52 | 95028 | 172 | Cu1Ho2K1Se4 |
| | -GM -T | -Y -Z | -R -S |
| | 5 2 3+4 4 | 5 2 3+4 4 2+2 4 | 3+4 2 |
| | 5 2 3+4 4 | 5 2 3+4 4 2+2 4 | 3+4 2 |
| | 6 2 3+4 4 | 6 2 3+4 4 2+2 4 | 5+6 2 |
| | 6 2 3+4 4 | 6 2 3+4 4 2+2 4 | 5+6 2 |
| | 5 2 3+4 4 | 5 2 3+4 4 2+2 4 | 3+4 2 |

Continued on next column

**TABLE XXXVIII– continued**

| TQC | ICSD | | NoE | | SF | |
|---|---|---|---|---|---|---|
| | 6 2 | 3+4 4 | 6 2 | 3+4 4 | 2+2 4 | 5+6 2 |
| | 5 2 | 3+4 4 | 6 2 | 3+4 4 | 2+2 4 | 5+6 2 |
| | 5 2 | 3+4 4 | 6 2 | 3+4 4 | 2+2 4 | 5+6 2 |
| | 6 2 | 3+4 4 | 5 2 | 3+4 4 | 2+2 4 | 3+4 2 |
| | 6 2 | 3+4 4 | 5 2 | 3+4 4 | 2+2 4 | 3+4 2 |
| | 5 2 | 3+4 4 | 6 2 | 3+4 4 | 2+2 4 | 3+4 2 |
| | 5 2 | 3+4 4 | 5 2 | 3+4 4 | 2+2 4 | 3+4 2 |
| | 6 2 | 3+4 4 | 6 2 | 3+4 4 | 2+2 4 | 5+6 2 |
| | 6 2 | 3+4 4 | 5 2 | 3+4 4 | 2+2 4 | 5+6 2 |
| | 5 2 | 3+4 4 | 6 2 | 3+4 4 | 2+2 4 | 5+6 2 |
| | 5 2 | 3+4 4 | 6 2 | 3+4 4 | 2+2 4 | 5+6 2 |
| | 6 2 | 3+4 4 | 5 2 | 3+4 4 | 2+2 4 | 3+4 2 |
| | 6 2 | 3+4 4 | 5 2 | 3+4 4 | 2+2 4 | 3+4 2 |
| | 5 2 | 3+4 4 | 6 2 | 3+4 4 | 2+2 4 | 5+6 2 |
| | 6 2 | 3+4 4 | 5 2 | 3+4 4 | 2+2 4 | 3+4 2 |
| | 5 2 | 3+4 4 | 6 2 | 3+4 4 | 2+2 4 | 5+6 2 |
| | 6 2 | 3+4 4 | 5 2 | 3+4 4 | 2+2 4 | 3+4 2 |
| | 5 2 | 3+4 4 | 6 2 | 3+4 4 | 2+2 4 | 3+4 2 |
| | 6 2 | 3+4 4 | 5 2 | 3+4 4 | 2+2 4 | 5+6 2 |
| | 5 2 | 3+4 4 | 6 2 | 3+4 4 | 2+2 4 | 3+4 2 |
| | 6 2 | 3+4 4 | 5 2 | 3+4 4 | 2+2 4 | 5+6 2 |
| | 5 2 | 3+4 4 | 6 2 | 3+4 4 | 2+2 4 | 5+6 2 |
| | 5 2 | 3+4 4 | 6 2 | 3+4 4 | 2+2 4 | 3+4 2 |
| | 6 2 | 3+4 4 | 5 2 | 3+4 4 | 2+2 4 | 5+6 2 |
| | 6 2 | 3+4 4 | 5 2 | 3+4 4 | 2+2 4 | 3+4 2 |
| | 5 2 | 3+4 4 | 5 2 | 3+4 4 | 2+2 4 | 3+4 2 |
| | 5 2 | 3+4 4 | 5 2 | 3+4 4 | 2+2 4 | 3+4 2 |
| | 6 2 | 3+4 4 | 6 2 | 3+4 4 | 2+2 4 | 5+6 2 |
| | 5 2 | 3+4 4 | 5 2 | 3+4 4 | 2+2 4 | 3+4 2 |
| | 6 2 | 3+4 4 | 6 2 | 3+4 4 | 2+2 4 | 5+6 2 |
| | 5 2 | 3+4 4 | 5 2 | 3+4 4 | 2+2 4 | 5+6 2 |
| | 5 2 | 3+4 4 | 6 2 | 3+4 4 | 2+2 4 | 3+4 2 |
| | 6 2 | 3+4 4 | 6 2 | 3+4 4 | 2+2 4 | 3+4 2 |
| | 5 2 | 3+4 4 | 5 2 | 3+4 4 | 2+2 4 | 5+6 2 |
| | 6 2 | 3+4 4 | 6 2 | 3+4 4 | 2+2 4 | 3+4 2 |
| | 6 2 | 3+4 4 | 5 2 | 3+4 4 | 2+2 4 | 5+6 2 |
| | 5 2 | 3+4 4 | 6 2 | 3+4 4 | 2+2 4 | 3+4 2 |
| | 6 2 | 3+4 4 | 5 2 | 3+4 4 | 2+2 4 | 5+6 2 |
| | 5 2 | | 5 2 | | | 5+6 2 |
| | 6 2 | | 6 2 | | | 3+4 2 |
| | 6 2 | | 6 2 | | | 5+6 2 |
| | 5 2 | | 5 2 | | | 5+6 2 |
| | 5 2 | | 6 2 | | | 3+4 2 |
| | 6 2 | | 6 2 | | | 3+4 2 |
| | 6 2 | | 5 2 | | | 5+6 2 |
| | 5 2 | | 6 2 | | | 5+6 2 |
| | 6 2 | | 6 2 | | | 5+6 2 |
| | 6 2 | | 5 2 | | | 3+4 2 |
| | 5 2 | | 6 2 | | | 5+6 2 |
| | 6 2 | | 6 2 | | | 3+4 2 |
| | 5 2 | | 5 2 | | | 5+6 2 |
| | 6 2 | | 6 2 | | | 5+6 2 |
| | 5 2 | | 6 2 | | | 3+4 2 |
| | 6 2 | | 6 2 | | | 5+6 2 |
| | 6 2 | | 5 2 | | | 3+4 2 |
| | 6 2 | | 6 2 | | | 5+6 2 |
| | 6 2 | | 5 2 | | | 5+6 2 |
| | 5 2 | | 5 2 | | | 3+4 2 |
| | 5 2 | | 6 2 | | | 5+6 2 |
| | 5 2 | | 5 2 | | | 3+4 2 |
| | 6 2 | | 6 2 | | | 5+6 2 |
| | 5 2 | | 6 2 | | | 3+4 2 |
| | 6 2 | | 5 2 | | | 3+4 2 |
| | 5 2 | | 5 2 | | | 5+6 2 |
| | 5 2 | | 6 2 | | | 3+4 2 |
| | 6 2 | | 5 2 | | | 5+6 2 |
| | 6 2 | | 6 2 | | | 3+4 2 |
| | 5 2 | | 6 2 | | | 5+6 2 |
| | 6 2 | | 5 2 | | | 3+4 2 |
| | 5 2 | | 6 2 | | | 5+6 2 |
| | 6 2 | | 5 2 | | | 5+6 2 |
| | 6 2 | | 6 2 | | | 3+4 2 |
| | 6 2 | | 5 2 | | | 5+6 2 |
| | 5 2 | | 6 2 | | | 3+4 2 |
| | 5 2 | | 5 2 | | | 3+4 2 |
| | 5 2 | | 6 2 | | | 5+6 2 |
| | 5 2 | | 5 2 | | | 3+4 2 |
| | 6 2 | | 5 2 | | | 5+6 2 |
| | 5 2 | | 5 2 | | | 5+6 2 |
| | 5 2 | | 6 2 | | | 3+4 2 |
| | 6 2 | | 6 2 | | | 3+4 2 |
| 63.1.53 | 97561 | | 172 | | Cu1Ho2K1S4 | |
| | -GM | -T | -Y | -Z | -R | -S |
| | 5 2 | 3+4 4 | 5 2 | 3+4 4 | 2+2 4 | 3+4 2 |
| | 5 2 | 3+4 4 | 6 2 | 3+4 4 | 2+2 4 | 5+6 2 |
| | 6 2 | 3+4 4 | 6 2 | 3+4 4 | 2+2 4 | 5+6 2 |
| | 6 2 | 3+4 4 | 5 2 | 3+4 4 | 2+2 4 | 3+4 2 |
| | 5 2 | 3+4 4 | 5 2 | 3+4 4 | 2+2 4 | 3+4 2 |
| | 6 2 | 3+4 4 | 6 2 | 3+4 4 | 2+2 4 | 5+6 2 |
| | 5 2 | 3+4 4 | 6 2 | 3+4 4 | 2+2 4 | 5+6 2 |
| | 5 2 | 3+4 4 | 6 2 | 3+4 4 | 2+2 4 | 5+6 2 |
| | 6 2 | 3+4 4 | 5 2 | 3+4 4 | 2+2 4 | 3+4 2 |
| | 6 2 | 3+4 4 | 5 2 | 3+4 4 | 2+2 4 | 3+4 2 |
| | 5 2 | 3+4 4 | 6 2 | 3+4 4 | 2+2 4 | 3+4 2 |
| | 5 2 | 3+4 4 | 5 2 | 3+4 4 | 2+2 4 | 3+4 2 |
| | 6 2 | 3+4 4 | 6 2 | 3+4 4 | 2+2 4 | 5+6 2 |
| | 6 2 | 3+4 4 | 5 2 | 3+4 4 | 2+2 4 | 5+6 2 |
| | 5 2 | 3+4 4 | 6 2 | 3+4 4 | 2+2 4 | 5+6 2 |
| | 5 2 | 3+4 4 | 6 2 | 3+4 4 | 2+2 4 | 5+6 2 |
| | 6 2 | 3+4 4 | 5 2 | 3+4 4 | 2+2 4 | 3+4 2 |
| | 6 2 | 3+4 4 | 5 2 | 3+4 4 | 2+2 4 | 3+4 2 |
| | 6 2 | 3+4 4 | 6 2 | 3+4 4 | 2+2 4 | 5+6 2 |
| | 5 2 | 3+4 4 | 5 2 | 3+4 4 | 2+2 4 | 3+4 2 |
| | 5 2 | 3+4 4 | 6 2 | 3+4 4 | 2+2 4 | 5+6 2 |
| | 6 2 | 3+4 4 | 5 2 | 3+4 4 | 2+2 4 | 3+4 2 |
| | 5 2 | 3+4 4 | 6 2 | 3+4 4 | 2+2 4 | 3+4 2 |
| | 5 2 | 3+4 4 | 5 2 | 3+4 4 | 2+2 4 | 5+6 2 |
| | 6 2 | 3+4 4 | 6 2 | 3+4 4 | 2+2 4 | 3+4 2 |
| | 6 2 | 3+4 4 | 5 2 | 3+4 4 | 2+2 4 | 5+6 2 |
| | 5 2 | 3+4 4 | 6 2 | 3+4 4 | 2+2 4 | 5+6 2 |
| | 5 2 | 3+4 4 | 5 2 | 3+4 4 | 2+2 4 | 3+4 2 |
| | 6 2 | 3+4 4 | 6 2 | 3+4 4 | 2+2 4 | 5+6 2 |
| | 6 2 | 3+4 4 | 5 2 | 3+4 4 | 2+2 4 | 3+4 2 |
| | 5 2 | 3+4 4 | 5 2 | 3+4 4 | 2+2 4 | 3+4 2 |
| | 5 2 | 3+4 4 | 5 2 | 3+4 4 | 2+2 4 | 3+4 2 |
| | 6 2 | 3+4 4 | 6 2 | 3+4 4 | 2+2 4 | 5+6 2 |
| | 5 2 | 3+4 4 | 5 2 | 3+4 4 | 2+2 4 | 3+4 2 |
| | 6 2 | 3+4 4 | 6 2 | 3+4 4 | 2+2 4 | 5+6 2 |
| | 5 2 | 3+4 4 | 5 2 | 3+4 4 | 2+2 4 | 3+4 2 |
| | 6 2 | 3+4 4 | 6 2 | 3+4 4 | 2+2 4 | 5+6 2 |
| | 5 2 | 3+4 4 | 6 2 | 3+4 4 | 2+2 4 | 3+4 2 |
| | 6 2 | 3+4 4 | 5 2 | 3+4 4 | 2+2 4 | 5+6 2 |

Continued on next column

**TABLE XXXVIII– continued**

| TQC | ICSD | | NoE | | SF | |
|---|---|---|---|---|---|---|
| | 5 2 | 3+4 4 | 5 2 | 3+4 4 | 2+2 4 | 3+4 2 |
| | 5 2 | 3+4 4 | 6 2 | 3+4 4 | 2+2 4 | 5+6 2 |
| | 6 2 | 3+4 4 | 6 2 | 3+4 4 | 2+2 4 | 3+4 2 |
| | 6 2 | 3+4 4 | 5 2 | 3+4 4 | 2+2 4 | 5+6 2 |
| | 5 2 | | 5 2 | | | 5+6 2 |
| | 6 2 | | 6 2 | | | 3+4 2 |
| | 6 2 | | 6 2 | | | 5+6 2 |
| | 5 2 | | 5 2 | | | 5+6 2 |
| | 5 2 | | 6 2 | | | 3+4 2 |
| | 6 2 | | 6 2 | | | 3+4 2 |
| | 5 2 | | 5 2 | | | 5+6 2 |
| | 6 2 | | 5 2 | | | 5+6 2 |
| | 6 2 | | 6 2 | | | 5+6 2 |
| | 6 2 | | 6 2 | | | 3+4 2 |
| | 6 2 | | 6 2 | | | 5+6 2 |
| | 5 2 | | 6 2 | | | 3+4 2 |
| | 6 2 | | 5 2 | | | 5+6 2 |
| | 5 2 | | 6 2 | | | 5+6 2 |
| | 6 2 | | 6 2 | | | 5+6 2 |
| | 5 2 | | 6 2 | | | 3+4 2 |
| | 6 2 | | 5 2 | | | 3+4 2 |
| | 6 2 | | 6 2 | | | 5+6 2 |
| | 6 2 | | 5 2 | | | 5+6 2 |
| | 5 2 | | 5 2 | | | 3+4 2 |
| | 5 2 | | 6 2 | | | 5+6 2 |
| | 6 2 | | 5 2 | | | 3+4 2 |
| | 5 2 | | 6 2 | | | 5+6 2 |
| | 5 2 | | 5 2 | | | 3+4 2 |
| | 6 2 | | 6 2 | | | 3+4 2 |
| | 5 2 | | 6 2 | | | 5+6 2 |
| | 5 2 | | 5 2 | | | 3+4 2 |
| | 6 2 | | 5 2 | | | 5+6 2 |
| | 6 2 | | 6 2 | | | 3+4 2 |
| | 5 2 | | 6 2 | | | 5+6 2 |
| | 6 2 | | 6 2 | | | 3+4 2 |
| | 5 2 | | 5 2 | | | 5+6 2 |
| | 6 2 | | 6 2 | | | 5+6 2 |
| | 6 2 | | 5 2 | | | 3+4 2 |
| | 6 2 | | 5 2 | | | 5+6 2 |
| | 5 2 | | 6 2 | | | 3+4 2 |
| | 5 2 | | 6 2 | | | 3+4 2 |
| | 5 2 | | 5 2 | | | 5+6 2 |
| | 5 2 | | 5 2 | | | 3+4 2 |
| | 6 2 | | 5 2 | | | 5+6 2 |
| | 5 2 | | 5 2 | | | 5+6 2 |
| | 5 2 | | 6 2 | | | 3+4 2 |
| | 6 2 | | 6 2 | | | 3+4 2 |
| 63.1.54 | 106440 | | 80 | | Cu1Eu1In4 | |
| | -GM | -T | -Y | -Z | -R | -S |
| | 5 2 | 3+4 4 | 5 2 | 3+4 4 | 2+2 4 | 3+4 2 |
| | 6 2 | 3+4 4 | 6 2 | 3+4 4 | 2+2 4 | 5+6 2 |
| | 5 2 | 3+4 4 | 5 2 | 3+4 4 | 2+2 4 | 5+6 2 |
| | 6 2 | 3+4 4 | 6 2 | 3+4 4 | 2+2 4 | 3+4 2 |
| | 5 2 | 3+4 4 | 5 2 | 3+4 4 | 2+2 4 | 5+6 2 |
| | 6 2 | 3+4 4 | 6 2 | 3+4 4 | 2+2 4 | 3+4 2 |
| | 5 2 | 3+4 4 | 6 2 | 3+4 4 | 2+2 4 | 5+6 2 |
| | 6 2 | 3+4 4 | 5 2 | 3+4 4 | 2+2 4 | 3+4 2 |
| | 5 2 | 3+4 4 | 6 2 | 3+4 4 | 2+2 4 | 3+4 2 |
| | 6 2 | 3+4 4 | 5 2 | 3+4 4 | 2+2 4 | 5+6 2 |
| | 5 2 | 3+4 4 | 6 2 | 3+4 4 | 2+2 4 | 5+6 2 |
| | 5 2 | 3+4 4 | 5 2 | 3+4 4 | 2+2 4 | 3+4 2 |

Continued on next column

**TABLE XXXVIII– continued**

| TQC | ICSD | | NoE | | SF | |
|---|---|---|---|---|---|---|
| | 5 2 | 3+4 4 | 6 2 | 3+4 4 | 2+2 4 | 3+4 2 |
| | 6 2 | 3+4 4 | 5 2 | 3+4 4 | 2+2 4 | 5+6 2 |
| | 6 2 | 3+4 4 | 5 2 | 3+4 4 | 2+2 4 | 3+4 2 |
| | 6 2 | 3+4 4 | 6 2 | 3+4 4 | 2+2 4 | 5+6 2 |
| | 5 2 | 3+4 4 | 5 2 | 3+4 4 | 2+2 4 | 3+4 2 |
| | 5 2 | 3+4 4 | 6 2 | 3+4 4 | 2+2 4 | 3+4 2 |
| | 5 2 | 3+4 4 | 6 2 | 3+4 4 | 2+2 4 | 5+6 2 |
| | 6 2 | 3+4 4 | 5 2 | 3+4 4 | 2+2 4 | 3+4 2 |
| | 6 2 | | 5 2 | | | 5+6 2 |
| | 5 2 | | 6 2 | | | 5+6 2 |
| | 5 2 | | 6 2 | | | 3+4 2 |
| | 6 2 | | 5 2 | | | 3+4 2 |
| | 6 2 | | 5 2 | | | 5+6 2 |
| | 5 2 | | 6 2 | | | 5+6 2 |
| | 6 2 | | 6 2 | | | 3+4 2 |
| | 5 2 | | 6 2 | | | 3+4 2 |
| | 6 2 | | 5 2 | | | 5+6 2 |
| | 5 2 | | 6 2 | | | 5+6 2 |
| | 6 2 | | 5 2 | | | 5+6 2 |
| | 5 2 | | 5 2 | | | 3+4 2 |
| | 6 2 | | 6 2 | | | 5+6 2 |
| | 5 2 | | 6 2 | | | 5+6 2 |
| | 5 2 | | 5 2 | | | 3+4 2 |
| | 6 2 | | 6 2 | | | 3+4 2 |
| | 6 2 | | 5 2 | | | 5+6 2 |
| | 5 2 | | 6 2 | | | 5+6 2 |
| | 6 2 | | 5 2 | | | 3+4 2 |
| | 5 2 | | 6 2 | | | 5+6 2 |
| 63.1.55 | 429089 | | 112 | | Cs1Sc1Se5U1 | |
| | -GM | -T | -Y | -Z | -R | -S |
| | 5 2 | 3+4 4 | 5 2 | 3+4 4 | 2+2 4 | 3+4 2 |
| | 6 2 | 3+4 4 | 6 2 | 3+4 4 | 2+2 4 | 5+6 2 |
| | 5 2 | 3+4 4 | 5 2 | 3+4 4 | 2+2 4 | 3+4 2 |
| | 6 2 | 3+4 4 | 6 2 | 3+4 4 | 2+2 4 | 5+6 2 |
| | 5 2 | 3+4 4 | 5 2 | 3+4 4 | 2+2 4 | 3+4 2 |
| | 6 2 | 3+4 4 | 6 2 | 3+4 4 | 2+2 4 | 5+6 2 |
| | 5 2 | 3+4 4 | 5 2 | 3+4 4 | 2+2 4 | 3+4 2 |
| | 6 2 | 3+4 4 | 6 2 | 3+4 4 | 2+2 4 | 5+6 2 |
| | 5 2 | 3+4 4 | 5 2 | 3+4 4 | 2+2 4 | 3+4 2 |
| | 6 2 | 3+4 4 | 6 2 | 3+4 4 | 2+2 4 | 5+6 2 |
| | 5 2 | 3+4 4 | 6 2 | 3+4 4 | 2+2 4 | 3+4 2 |
| | 6 2 | 3+4 4 | 5 2 | 3+4 4 | 2+2 4 | 5+6 2 |
| | 5 2 | 3+4 4 | 5 2 | 3+4 4 | 2+2 4 | 3+4 2 |
| | 6 2 | 3+4 4 | 6 2 | 3+4 4 | 2+2 4 | 5+6 2 |
| | 5 2 | 3+4 4 | 5 2 | 3+4 4 | 2+2 4 | 5+6 2 |
| | 5 2 | 3+4 4 | 5 2 | 3+4 4 | 2+2 4 | 3+4 2 |
| | 6 2 | 3+4 4 | 6 2 | 3+4 4 | 2+2 4 | 5+6 2 |
| | 6 2 | 3+4 4 | 6 2 | 3+4 4 | 2+2 4 | 3+4 2 |
| | 5 2 | 3+4 4 | 6 2 | 3+4 4 | 2+2 4 | 5+6 2 |
| | 6 2 | 3+4 4 | 5 2 | 3+4 4 | 2+2 4 | 3+4 2 |
| | 6 2 | 3+4 4 | 5 2 | 3+4 4 | 2+2 4 | 3+4 2 |
| | 5 2 | 3+4 4 | 6 2 | 3+4 4 | 2+2 4 | 5+6 2 |
| | 6 2 | 3+4 4 | 5 2 | 3+4 4 | 2+2 4 | 3+4 2 |
| | 5 2 | 3+4 4 | 6 2 | 3+4 4 | 2+2 4 | 5+6 2 |
| | 6 2 | 3+4 4 | 5 2 | 3+4 4 | 2+2 4 | 3+4 2 |
| | 5 2 | 3+4 4 | 6 2 | 3+4 4 | 2+2 4 | 5+6 2 |
| | 5 2 | 3+4 4 | 6 2 | 3+4 4 | 2+2 4 | 3+4 2 |
| | 5 2 | 3+4 4 | 5 2 | 3+4 4 | 2+2 4 | 5+6 2 |
| | 6 2 | | 5 2 | | | 3+4 2 |
| | 6 2 | | 6 2 | | | 3+4 2 |
| | 5 2 | | 5 2 | | | 5+6 2 |

Continued on next column

**TABLE XXXVIII– continued**

| TQC | ICSD | | NoE | | SF | |
|---|---|---|---|---|---|---|
| | 6 2 | | 6 2 | | | 3+4 2 |
| | 5 2 | | 6 2 | | | 5+6 2 |
| | 6 2 | | 5 2 | | | 5+6 2 |
| | 5 2 | | 5 2 | | | 5+6 2 |
| | 6 2 | | 6 2 | | | 3+4 2 |
| | 5 2 | | 5 2 | | | 5+6 2 |
| | 5 2 | | 5 2 | | | 3+4 2 |
| | 6 2 | | 5 2 | | | 5+6 2 |
| | 5 2 | | 6 2 | | | 3+4 2 |
| | 5 2 | | 5 2 | | | 3+4 2 |
| | 5 2 | | 6 2 | | | 3+4 2 |
| | 6 2 | | 5 2 | | | 5+6 2 |
| | 5 2 | | 6 2 | | | 3+4 2 |
| | 6 2 | | 5 2 | | | 5+6 2 |
| | 5 2 | | 6 2 | | | 3+4 2 |
| | 6 2 | | 5 2 | | | 5+6 2 |
| | 5 2 | | 5 2 | | | 5+6 2 |
| | 5 2 | | 6 2 | | | 3+4 2 |
| | 6 2 | | 5 2 | | | 5+6 2 |
| | 6 2 | | 6 2 | | | 3+4 2 |
| | 6 2 | | 6 2 | | | 5+6 2 |
| | 6 2 | | 6 2 | | | 3+4 2 |
| | 6 2 | | 6 2 | | | 5+6 2 |
| | 5 2 | | 5 2 | | | 3+4 2 |
| | 5 2 | | 5 2 | | | 5+6 2 |
| 63.1.56 | 413682 | | 240 | | Cs1Cu3Se5Yb2 | |
| | -GM | -T | -Y | -Z | -R | -S |
| | 5 2 | 3+4 4 | 5 2 | 3+4 4 | 2+2 4 | 3+4 2 |
| | 5 2 | 3+4 4 | 5 2 | 3+4 4 | 2+2 4 | 3+4 2 |
| | 6 2 | 3+4 4 | 6 2 | 3+4 4 | 2+2 4 | 5+6 2 |
| | 6 2 | 3+4 4 | 6 2 | 3+4 4 | 2+2 4 | 5+6 2 |
| | 5 2 | 3+4 4 | 6 2 | 3+4 4 | 2+2 4 | 3+4 2 |
| | 5 2 | 3+4 4 | 6 2 | 3+4 4 | 2+2 4 | 3+4 2 |
| | 6 2 | 3+4 4 | 5 2 | 3+4 4 | 2+2 4 | 5+6 2 |
| | 6 2 | 3+4 4 | 5 2 | 3+4 4 | 2+2 4 | 5+6 2 |
| | 5 2 | 3+4 4 | 6 2 | 3+4 4 | 2+2 4 | 3+4 2 |
| | 5 2 | 3+4 4 | 6 2 | 3+4 4 | 2+2 4 | 5+6 2 |
| | 6 2 | 3+4 4 | 5 2 | 3+4 4 | 2+2 4 | 3+4 2 |
| | 6 2 | 3+4 4 | 5 2 | 3+4 4 | 2+2 4 | 5+6 2 |
| | 5 2 | 3+4 4 | 6 2 | 3+4 4 | 2+2 4 | 3+4 2 |
| | 5 2 | 3+4 4 | 6 2 | 3+4 4 | 2+2 4 | 5+6 2 |
| | 6 2 | 3+4 4 | 5 2 | 3+4 4 | 2+2 4 | 3+4 2 |
| | 6 2 | 3+4 4 | 5 2 | 3+4 4 | 2+2 4 | 5+6 2 |
| | 5 2 | 3+4 4 | 5 2 | 3+4 4 | 2+2 4 | 3+4 2 |
| | 6 2 | 3+4 4 | 6 2 | 3+4 4 | 2+2 4 | 5+6 2 |
| | 5 2 | 3+4 4 | 6 2 | 3+4 4 | 2+2 4 | 3+4 2 |
| | 5 2 | 3+4 4 | 5 2 | 3+4 4 | 2+2 4 | 5+6 2 |
| | 6 2 | 3+4 4 | 6 2 | 3+4 4 | 2+2 4 | 5+6 2 |
| | 6 2 | 3+4 4 | 5 2 | 3+4 4 | 2+2 4 | 5+6 2 |
| | 5 2 | 3+4 4 | 5 2 | 3+4 4 | 2+2 4 | 3+4 2 |
| | 6 2 | 3+4 4 | 6 2 | 3+4 4 | 2+2 4 | 3+4 2 |
| | 6 2 | 3+4 4 | 5 2 | 3+4 4 | 2+2 4 | 5+6 2 |
| | 6 2 | 3+4 4 | 6 2 | 3+4 4 | 2+2 4 | 3+4 2 |
| | 5 2 | 3+4 4 | 6 2 | 3+4 4 | 2+2 4 | 5+6 2 |
| | 5 2 | 3+4 4 | 5 2 | 3+4 4 | 2+2 4 | 3+4 2 |
| | 5 2 | 3+4 4 | 5 2 | 3+4 4 | 2+2 4 | 3+4 2 |
| | 6 2 | 3+4 4 | 6 2 | 3+4 4 | 2+2 4 | 5+6 2 |
| | 5 2 | 3+4 4 | 5 2 | 3+4 4 | 2+2 4 | 3+4 2 |
| | 6 2 | 3+4 4 | 6 2 | 3+4 4 | 2+2 4 | 5+6 2 |
| | 5 2 | 3+4 4 | 5 2 | 3+4 4 | 2+2 4 | 5+6 2 |
| | 6 2 | 3+4 4 | 6 2 | 3+4 4 | 2+2 4 | 3+4 2 |

Continued on next column

**TABLE XXXVIII– continued**

| TQC | ICSD | | NoE | | SF | |
|---|---|---|---|---|---|---|
| | 5 2 | 3+4 4 | 5 2 | 3+4 4 | 2+2 4 | 5+6 2 |
| | 5 2 | 3+4 4 | 5 2 | 3+4 4 | 2+2 4 | 5+6 2 |
| | 6 2 | 3+4 4 | 6 2 | 3+4 4 | 2+2 4 | 3+4 2 |
| | 6 2 | 3+4 4 | 5 2 | 3+4 4 | 2+2 4 | 3+4 2 |
| | 6 2 | 3+4 4 | 6 2 | 3+4 4 | 2+2 4 | 3+4 2 |
| | 5 2 | 3+4 4 | 5 2 | 3+4 4 | 2+2 4 | 5+6 2 |
| | 6 2 | 3+4 4 | 6 2 | 3+4 4 | 2+2 4 | 5+6 2 |
| | 5 2 | 3+4 4 | 6 2 | 3+4 4 | 2+2 4 | 5+6 2 |
| | 5 2 | 3+4 4 | 6 2 | 3+4 4 | 2+2 4 | 3+4 2 |
| | 6 2 | 3+4 4 | 5 2 | 3+4 4 | 2+2 4 | 3+4 2 |
| | 5 2 | 3+4 4 | 6 2 | 3+4 4 | 2+2 4 | 3+4 2 |
| | 6 2 | 3+4 4 | 6 2 | 3+4 4 | 2+2 4 | 3+4 2 |
| | 5 2 | 3+4 4 | 5 2 | 3+4 4 | 2+2 4 | 5+6 2 |
| | 5 2 | 3+4 4 | 6 2 | 3+4 4 | 2+2 4 | 5+6 2 |
| | 6 2 | 3+4 4 | 6 2 | 3+4 4 | 2+2 4 | 3+4 2 |
| | 6 2 | 3+4 4 | 6 2 | 3+4 4 | 2+2 4 | 3+4 2 |
| | 6 2 | 3+4 4 | 5 2 | 3+4 4 | 2+2 4 | 3+4 2 |
| | 5 2 | 3+4 4 | 5 2 | 3+4 4 | 2+2 4 | 5+6 2 |
| | 5 2 | 3+4 4 | 6 2 | 3+4 4 | 2+2 4 | 5+6 2 |
| | 5 2 | 3+4 4 | 5 2 | 3+4 4 | 2+2 4 | 5+6 2 |
| | 6 2 | 3+4 4 | 6 2 | 3+4 4 | 2+2 4 | 3+4 2 |
| | 6 2 | 3+4 4 | 5 2 | 3+4 4 | 2+2 4 | 3+4 2 |
| | 5 2 | 3+4 4 | 6 2 | 3+4 4 | 2+2 4 | 3+4 2 |
| | 5 2 | 3+4 4 | 5 2 | 3+4 4 | 2+2 4 | 5+6 2 |
| | 5 2 | 3+4 4 | 5 2 | 3+4 4 | 2+2 4 | 5+6 2 |
| | 5 2 | 3+4 4 | 5 2 | 3+4 4 | 2+2 4 | 3+4 2 |
| | 6 2 | | 5 2 | | | 3+4 2 |
| | 5 2 | | 5 2 | | | 5+6 2 |
| | 6 2 | | 6 2 | | | 5+6 2 |
| | 5 2 | | 5 2 | | | 5+6 2 |
| | 5 2 | | 5 2 | | | 5+6 2 |
| | 6 2 | | 5 2 | | | 3+4 2 |
| | 6 2 | | 6 2 | | | 5+6 2 |
| | 6 2 | | 6 2 | | | 3+4 2 |
| | 6 2 | | 6 2 | | | 5+6 2 |
| | 5 2 | | 5 2 | | | 3+4 2 |
| | 6 2 | | 6 2 | | | 3+4 2 |
| | 6 2 | | 6 2 | | | 5+6 2 |
| | 6 2 | | 6 2 | | | 5+6 2 |
| | 6 2 | | 5 2 | | | 5+6 2 |
| | 5 2 | | 5 2 | | | 3+4 2 |
| | 6 2 | | 6 2 | | | 3+4 2 |
| | 5 2 | | 5 2 | | | 5+6 2 |
| | 5 2 | | 5 2 | | | 5+6 2 |
| | 6 2 | | 6 2 | | | 3+4 2 |
| | 6 2 | | 6 2 | | | 5+6 2 |
| | 5 2 | | 5 2 | | | 3+4 2 |
| | 5 2 | | 6 2 | | | 5+6 2 |
| | 5 2 | | 5 2 | | | 3+4 2 |
| | 6 2 | | 5 2 | | | 3+4 2 |
| | 6 2 | | 6 2 | | | 5+6 2 |
| | 5 2 | | 6 2 | | | 3+4 2 |
| | 6 2 | | 5 2 | | | 5+6 2 |
| | 6 2 | | 6 2 | | | 3+4 2 |
| | 5 2 | | 6 2 | | | 5+6 2 |
| | 6 2 | | 5 2 | | | 5+6 2 |
| | 5 2 | | 6 2 | | | 3+4 2 |
| | 5 2 | | 5 2 | | | 5+6 2 |
| | 5 2 | | 5 2 | | | 3+4 2 |
| | 5 2 | | 6 2 | | | 5+6 2 |
| | 6 2 | | 6 2 | | | 5+6 2 |

Continued on next column

**TABLE XXXVIII– continued**

| TQC | ICSD | | NoE | | SF | |
|---|---|---|---|---|---|---|
| | 5 2 | | 6 2 | | | 5+6 2 |
| | 5 2 | | 5 2 | | | 3+4 2 |
| | 6 2 | | 6 2 | | | 3+4 2 |
| | 6 2 | | 5 2 | | | 3+4 2 |
| | 5 2 | | 5 2 | | | 5+6 2 |
| | 6 2 | | 5 2 | | | 5+6 2 |
| | 5 2 | | 6 2 | | | 3+4 2 |
| | 6 2 | | 5 2 | | | 5+6 2 |
| | 6 2 | | 6 2 | | | 3+4 2 |
| | 5 2 | | 5 2 | | | 3+4 2 |
| | 6 2 | | 6 2 | | | 3+4 2 |
| | 5 2 | | 6 2 | | | 3+4 2 |
| | 5 2 | | 5 2 | | | 5+6 2 |
| | 5 2 | | 6 2 | | | 3+4 2 |
| | 6 2 | | 5 2 | | | 5+6 2 |
| | 6 2 | | 6 2 | | | 5+6 2 |
| | 5 2 | | 6 2 | | | 3+4 2 |
| | 6 2 | | 5 2 | | | 5+6 2 |
| | 5 2 | | 5 2 | | | 3+4 2 |
| | 5 2 | | 6 2 | | | 5+6 2 |
| | 5 2 | | 6 2 | | | 5+6 2 |
| | 5 2 | | 6 2 | | | 3+4 2 |
| | 6 2 | | 5 2 | | | 5+6 2 |
| | 6 2 | | 5 2 | | | 3+4 2 |
| | 6 2 | | 6 2 | | | 5+6 2 |
| 63.1.57 | 413677 | | 228 | | Cs1Cu3Ho2Se5 | |
| | -GM | -T | -Y | -Z | -R | -S |
| | 5 2 | 3+4 4 | 6 2 | 3+4 4 | 2+2 4 | 3+4 2 |
| | 5 2 | 3+4 4 | 6 2 | 3+4 4 | 2+2 4 | 3+4 2 |
| | 6 2 | 3+4 4 | 5 2 | 3+4 4 | 2+2 4 | 5+6 2 |
| | 6 2 | 3+4 4 | 5 2 | 3+4 4 | 2+2 4 | 5+6 2 |
| | 5 2 | 3+4 4 | 6 2 | 3+4 4 | 2+2 4 | 3+4 2 |
| | 5 2 | 3+4 4 | 6 2 | 3+4 4 | 2+2 4 | 3+4 2 |
| | 6 2 | 3+4 4 | 5 2 | 3+4 4 | 2+2 4 | 5+6 2 |
| | 6 2 | 3+4 4 | 5 2 | 3+4 4 | 2+2 4 | 5+6 2 |
| | 5 2 | 3+4 4 | 5 2 | 3+4 4 | 2+2 4 | 3+4 2 |
| | 6 2 | 3+4 4 | 6 2 | 3+4 4 | 2+2 4 | 5+6 2 |
| | 5 2 | 3+4 4 | 6 2 | 3+4 4 | 2+2 4 | 3+4 2 |
| | 5 2 | 3+4 4 | 6 2 | 3+4 4 | 2+2 4 | 5+6 2 |
| | 6 2 | 3+4 4 | 5 2 | 3+4 4 | 2+2 4 | 3+4 2 |
| | 6 2 | 3+4 4 | 5 2 | 3+4 4 | 2+2 4 | 5+6 2 |
| | 5 2 | 3+4 4 | 6 2 | 3+4 4 | 2+2 4 | 3+4 2 |
| | 5 2 | 3+4 4 | 6 2 | 3+4 4 | 2+2 4 | 5+6 2 |
| | 6 2 | 3+4 4 | 5 2 | 3+4 4 | 2+2 4 | 3+4 2 |
| | 6 2 | 3+4 4 | 5 2 | 3+4 4 | 2+2 4 | 5+6 2 |
| | 5 2 | 3+4 4 | 6 2 | 3+4 4 | 2+2 4 | 3+4 2 |
| | 5 2 | 3+4 4 | 5 2 | 3+4 4 | 2+2 4 | 5+6 2 |
| | 6 2 | 3+4 4 | 6 2 | 3+4 4 | 2+2 4 | 5+6 2 |
| | 6 2 | 3+4 4 | 5 2 | 3+4 4 | 2+2 4 | 5+6 2 |
| | 5 2 | 3+4 4 | 5 2 | 3+4 4 | 2+2 4 | 3+4 2 |
| | 6 2 | 3+4 4 | 6 2 | 3+4 4 | 2+2 4 | 3+4 2 |
| | 6 2 | 3+4 4 | 5 2 | 3+4 4 | 2+2 4 | 5+6 2 |
| | 6 2 | 3+4 4 | 5 2 | 3+4 4 | 2+2 4 | 3+4 2 |
| | 5 2 | 3+4 4 | 6 2 | 3+4 4 | 2+2 4 | 5+6 2 |
| | 5 2 | 3+4 4 | 6 2 | 3+4 4 | 2+2 4 | 3+4 2 |
| | 5 2 | 3+4 4 | 5 2 | 3+4 4 | 2+2 4 | 3+4 2 |
| | 6 2 | 3+4 4 | 6 2 | 3+4 4 | 2+2 4 | 5+6 2 |
| | 5 2 | 3+4 4 | 5 2 | 3+4 4 | 2+2 4 | 3+4 2 |
| | 6 2 | 3+4 4 | 6 2 | 3+4 4 | 2+2 4 | 5+6 2 |
| | 5 2 | 3+4 4 | 5 2 | 3+4 4 | 2+2 4 | 5+6 2 |
| | 6 2 | 3+4 4 | 6 2 | 3+4 4 | 2+2 4 | 3+4 2 |

Continued on next column

| TQC | ICSD | | NoE | | SF | |
|---|---|---|---|---|---|---|
| | 5 2 | 3+4 4 | 5 2 | 3+4 4 | 2+2 4 | 5+6 2 |
| | 5 2 | 3+4 4 | 5 2 | 3+4 4 | 2+2 4 | 5+6 2 |
| | 6 2 | 3+4 4 | 6 2 | 3+4 4 | 2+2 4 | 3+4 2 |
| | 6 2 | 3+4 4 | 5 2 | 3+4 4 | 2+2 4 | 3+4 2 |
| | 6 2 | 3+4 4 | 6 2 | 3+4 4 | 2+2 4 | 3+4 2 |
| | 5 2 | 3+4 4 | 5 2 | 3+4 4 | 2+2 4 | 5+6 2 |
| | 5 2 | 3+4 4 | 6 2 | 3+4 4 | 2+2 4 | 5+6 2 |
| | 6 2 | 3+4 4 | 6 2 | 3+4 4 | 2+2 4 | 5+6 2 |
| | 5 2 | 3+4 4 | 6 2 | 3+4 4 | 2+2 4 | 3+4 2 |
| | 6 2 | 3+4 4 | 5 2 | 3+4 4 | 2+2 4 | 3+4 2 |
| | 5 2 | 3+4 4 | 6 2 | 3+4 4 | 2+2 4 | 3+4 2 |
| | 6 2 | 3+4 4 | 6 2 | 3+4 4 | 2+2 4 | 3+4 2 |
| | 5 2 | 3+4 4 | 6 2 | 3+4 4 | 2+2 4 | 5+6 2 |
| | 5 2 | 3+4 4 | 5 2 | 3+4 4 | 2+2 4 | 5+6 2 |
| | 6 2 | 3+4 4 | 6 2 | 3+4 4 | 2+2 4 | 3+4 2 |
| | 6 2 | 3+4 4 | 6 2 | 3+4 4 | 2+2 4 | 3+4 2 |
| | 6 2 | 3+4 4 | 5 2 | 3+4 4 | 2+2 4 | 3+4 2 |
| | 5 2 | 3+4 4 | 5 2 | 3+4 4 | 2+2 4 | 5+6 2 |
| | 5 2 | 3+4 4 | 6 2 | 3+4 4 | 2+2 4 | 5+6 2 |
| | 5 2 | 3+4 4 | 5 2 | 3+4 4 | 2+2 4 | 3+4 2 |
| | 6 2 | 3+4 4 | 6 2 | 3+4 4 | 2+2 4 | 5+6 2 |
| | 6 2 | 3+4 4 | 5 2 | 3+4 4 | 2+2 4 | 3+4 2 |
| | 5 2 | 3+4 4 | 6 2 | 3+4 4 | 2+2 4 | 3+4 2 |
| | 5 2 | | 5 2 | | | 5+6 2 |
| | 5 2 | | 5 2 | | | 5+6 2 |
| | 5 2 | | 5 2 | | | 3+4 2 |
| | 6 2 | | 5 2 | | | 5+6 2 |
| | 5 2 | | 5 2 | | | 3+4 2 |
| | 6 2 | | 6 2 | | | 5+6 2 |
| | 5 2 | | 5 2 | | | 5+6 2 |
| | 6 2 | | 5 2 | | | 5+6 2 |
| | 5 2 | | 5 2 | | | 5+6 2 |
| | 6 2 | | 6 2 | | | 3+4 2 |
| | 6 2 | | 6 2 | | | 3+4 2 |
| | 6 2 | | 6 2 | | | 5+6 2 |
| | 6 2 | | 6 2 | | | 3+4 2 |
| | 5 2 | | 5 2 | | | 3+4 2 |
| | 6 2 | | 6 2 | | | 5+6 2 |
| | 6 2 | | 6 2 | | | 5+6 2 |
| | 6 2 | | 5 2 | | | 5+6 2 |
| | 5 2 | | 6 2 | | | 3+4 2 |
| | 6 2 | | 5 2 | | | 3+4 2 |
| | 6 2 | | 6 2 | | | 5+6 2 |
| | 5 2 | | 5 2 | | | 5+6 2 |
| | 5 2 | | 6 2 | | | 3+4 2 |
| | 5 2 | | 5 2 | | | 5+6 2 |
| | 5 2 | | 5 2 | | | 3+4 2 |
| | 5 2 | | 5 2 | | | 5+6 2 |
| | 6 2 | | 5 2 | | | 5+6 2 |
| | 6 2 | | 6 2 | | | 5+6 2 |
| | 6 2 | | 6 2 | | | 3+4 2 |
| | 5 2 | | 6 2 | | | 3+4 2 |
| | 5 2 | | 6 2 | | | 3+4 2 |
| | 6 2 | | 6 2 | | | 5+6 2 |
| | 5 2 | | 5 2 | | | 5+6 2 |
| | 5 2 | | 6 2 | | | 3+4 2 |
| | 6 2 | | 5 2 | | | 5+6 2 |
| | 5 2 | | 6 2 | | | 3+4 2 |
| | 6 2 | | 5 2 | | | 3+4 2 |
| | 5 2 | | 5 2 | | | 5+6 2 |
| | 6 2 | | 6 2 | | | 3+4 2 |

Continued on next column

**TABLE XXXVIII– continued**

| TQC | ICSD | | NoE | SF | | |
|---|---|---|---|---|---|---|
| | 5 2 | | 5 2 | | | 5+6 2 |
| | 5 2 | | 6 2 | | | 5+6 2 |
| | 6 2 | | 6 2 | | | 3+4 2 |
| | 6 2 | | 5 2 | | | 5+6 2 |
| | 5 2 | | 5 2 | | | 3+4 2 |
| | 6 2 | | 6 2 | | | 3+4 2 |
| | 5 2 | | 5 2 | | | 5+6 2 |
| | 5 2 | | 6 2 | | | 3+4 2 |
| | 6 2 | | 5 2 | | | 5+6 2 |
| | 6 2 | | 5 2 | | | 3+4 2 |
| | 6 2 | | 6 2 | | | 3+4 2 |
| | 5 2 | | 6 2 | | | 3+4 2 |
| | 5 2 | | 6 2 | | | 5+6 2 |
| | 5 2 | | 5 2 | | | 3+4 2 |
| | 6 2 | | 5 2 | | | 5+6 2 |
| | 6 2 | | 6 2 | | | 5+6 2 |
| | 5 2 | | 6 2 | | | 3+4 2 |
| | 5 2 | | 6 2 | | | 5+6 2 |
| | 6 2 | | 5 2 | | | 3+4 2 |
| 63.1.58 | 412074 | | 232 | Cs1Cu3Er2S5 | | |
| | -GM | -T | -Y | -Z | -R | -S |
| | 5 2 | 3+4 4 | 6 2 | 3+4 4 | 2+2 4 | 3+4 2 |
| | 5 2 | 3+4 4 | 6 2 | 3+4 4 | 2+2 4 | 3+4 2 |
| | 6 2 | 3+4 4 | 5 2 | 3+4 4 | 2+2 4 | 5+6 2 |
| | 6 2 | 3+4 4 | 5 2 | 3+4 4 | 2+2 4 | 5+6 2 |
| | 5 2 | 3+4 4 | 6 2 | 3+4 4 | 2+2 4 | 3+4 2 |
| | 5 2 | 3+4 4 | 6 2 | 3+4 4 | 2+2 4 | 3+4 2 |
| | 6 2 | 3+4 4 | 5 2 | 3+4 4 | 2+2 4 | 5+6 2 |
| | 6 2 | 3+4 4 | 5 2 | 3+4 4 | 2+2 4 | 5+6 2 |
| | 5 2 | 3+4 4 | 6 2 | 3+4 4 | 2+2 4 | 3+4 2 |
| | 6 2 | 3+4 4 | 5 2 | 3+4 4 | 2+2 4 | 5+6 2 |
| | 5 2 | 3+4 4 | 6 2 | 3+4 4 | 2+2 4 | 3+4 2 |
| | 5 2 | 3+4 4 | 6 2 | 3+4 4 | 2+2 4 | 5+6 2 |
| | 6 2 | 3+4 4 | 5 2 | 3+4 4 | 2+2 4 | 3+4 2 |
| | 6 2 | 3+4 4 | 5 2 | 3+4 4 | 2+2 4 | 5+6 2 |
| | 5 2 | 3+4 4 | 6 2 | 3+4 4 | 2+2 4 | 3+4 2 |
| | 5 2 | 3+4 4 | 6 2 | 3+4 4 | 2+2 4 | 5+6 2 |
| | 6 2 | 3+4 4 | 5 2 | 3+4 4 | 2+2 4 | 3+4 2 |
| | 6 2 | 3+4 4 | 5 2 | 3+4 4 | 2+2 4 | 5+6 2 |
| | 5 2 | 3+4 4 | 6 2 | 3+4 4 | 2+2 4 | 3+4 2 |
| | 5 2 | 3+4 4 | 5 2 | 3+4 4 | 2+2 4 | 5+6 2 |
| | 6 2 | 3+4 4 | 6 2 | 3+4 4 | 2+2 4 | 5+6 2 |
| | 6 2 | 3+4 4 | 5 2 | 3+4 4 | 2+2 4 | 5+6 2 |
| | 5 2 | 3+4 4 | 5 2 | 3+4 4 | 2+2 4 | 3+4 2 |
| | 6 2 | 3+4 4 | 6 2 | 3+4 4 | 2+2 4 | 3+4 2 |
| | 6 2 | 3+4 4 | 5 2 | 3+4 4 | 2+2 4 | 5+6 2 |
| | 6 2 | 3+4 4 | 5 2 | 3+4 4 | 2+2 4 | 3+4 2 |
| | 5 2 | 3+4 4 | 6 2 | 3+4 4 | 2+2 4 | 5+6 2 |
| | 5 2 | 3+4 4 | 6 2 | 3+4 4 | 2+2 4 | 3+4 2 |
| | 5 2 | 3+4 4 | 5 2 | 3+4 4 | 2+2 4 | 3+4 2 |
| | 6 2 | 3+4 4 | 6 2 | 3+4 4 | 2+2 4 | 5+6 2 |
| | 5 2 | 3+4 4 | 5 2 | 3+4 4 | 2+2 4 | 3+4 2 |
| | 6 2 | 3+4 4 | 6 2 | 3+4 4 | 2+2 4 | 5+6 2 |
| | 5 2 | 3+4 4 | 5 2 | 3+4 4 | 2+2 4 | 5+6 2 |
| | 6 2 | 3+4 4 | 6 2 | 3+4 4 | 2+2 4 | 3+4 2 |
| | 5 2 | 3+4 4 | 5 2 | 3+4 4 | 2+2 4 | 5+6 2 |
| | 5 2 | 3+4 4 | 5 2 | 3+4 4 | 2+2 4 | 5+6 2 |
| | 6 2 | 3+4 4 | 6 2 | 3+4 4 | 2+2 4 | 3+4 2 |
| | 6 2 | 3+4 4 | 5 2 | 3+4 4 | 2+2 4 | 3+4 2 |
| | 6 2 | 3+4 4 | 6 2 | 3+4 4 | 2+2 4 | 3+4 2 |
| | 5 2 | 3+4 4 | 5 2 | 3+4 4 | 2+2 4 | 5+6 2 |

Continued on next column

**TABLE XXXVIII– continued**

| TQC | ICSD | | NoE | SF | | |
|---|---|---|---|---|---|---|
| | 6 2 | 3+4 4 | 6 2 | 3+4 4 | 2+2 4 | 5+6 2 |
| | 6 2 | 3+4 4 | 5 2 | 3+4 4 | 2+2 4 | 5+6 2 |
| | 5 2 | 3+4 4 | 6 2 | 3+4 4 | 2+2 4 | 3+4 2 |
| | 5 2 | 3+4 4 | 6 2 | 3+4 4 | 2+2 4 | 3+4 2 |
| | 5 2 | 3+4 4 | 6 2 | 3+4 4 | 2+2 4 | 3+4 2 |
| | 5 2 | 3+4 4 | 6 2 | 3+4 4 | 2+2 4 | 3+4 2 |
| | 5 2 | 3+4 4 | 6 2 | 3+4 4 | 2+2 4 | 5+6 2 |
| | 6 2 | 3+4 4 | 5 2 | 3+4 4 | 2+2 4 | 5+6 2 |
| | 6 2 | 3+4 4 | 6 2 | 3+4 4 | 2+2 4 | 3+4 2 |
| | 6 2 | 3+4 4 | 5 2 | 3+4 4 | 2+2 4 | 3+4 2 |
| | 6 2 | 3+4 4 | 5 2 | 3+4 4 | 2+2 4 | 3+4 2 |
| | 5 2 | 3+4 4 | 6 2 | 3+4 4 | 2+2 4 | 5+6 2 |
| | 5 2 | 3+4 4 | 6 2 | 3+4 4 | 2+2 4 | 5+6 2 |
| | 5 2 | 3+4 4 | 5 2 | 3+4 4 | 2+2 4 | 5+6 2 |
| | 6 2 | 3+4 4 | 6 2 | 3+4 4 | 2+2 4 | 3+4 2 |
| | 6 2 | 3+4 4 | 6 2 | 3+4 4 | 2+2 4 | 3+4 2 |
| | 5 2 | 3+4 4 | 5 2 | 3+4 4 | 2+2 4 | 3+4 2 |
| | 5 2 | 3+4 4 | 5 2 | 3+4 4 | 2+2 4 | 5+6 2 |
| | 5 2 | | 5 2 | | | 5+6 2 |
| | 5 2 | | 5 2 | | | 3+4 2 |
| | 6 2 | | 5 2 | | | 5+6 2 |
| | 6 2 | | 6 2 | | | 3+4 2 |
| | 6 2 | | 5 2 | | | 5+6 2 |
| | 5 2 | | 5 2 | | | 5+6 2 |
| | 5 2 | | 5 2 | | | 5+6 2 |
| | 5 2 | | 6 2 | | | 5+6 2 |
| | 6 2 | | 5 2 | | | 3+4 2 |
| | 6 2 | | 6 2 | | | 5+6 2 |
| | 6 2 | | 6 2 | | | 3+4 2 |
| | 6 2 | | 6 2 | | | 3+4 2 |
| | 5 2 | | 5 2 | | | 5+6 2 |
| | 6 2 | | 6 2 | | | 3+4 2 |
| | 6 2 | | 6 2 | | | 5+6 2 |
| | 6 2 | | 5 2 | | | 5+6 2 |
| | 5 2 | | 6 2 | | | 3+4 2 |
| | 6 2 | | 5 2 | | | 3+4 2 |
| | 5 2 | | 5 2 | | | 5+6 2 |
| | 6 2 | | 6 2 | | | 3+4 2 |
| | 5 2 | | 5 2 | | | 5+6 2 |
| | 5 2 | | 5 2 | | | 5+6 2 |
| | 5 2 | | 6 2 | | | 3+4 2 |
| | 6 2 | | 5 2 | | | 5+6 2 |
| | 5 2 | | 5 2 | | | 3+4 2 |
| | 6 2 | | 6 2 | | | 3+4 2 |
| | 5 2 | | 6 2 | | | 5+6 2 |
| | 6 2 | | 6 2 | | | 5+6 2 |
| | 5 2 | | 6 2 | | | 3+4 2 |
| | 5 2 | | 5 2 | | | 5+6 2 |
| | 6 2 | | 5 2 | | | 5+6 2 |
| | 6 2 | | 6 2 | | | 3+4 2 |
| | 5 2 | | 5 2 | | | 5+6 2 |
| | 5 2 | | 6 2 | | | 3+4 2 |
| | 6 2 | | 6 2 | | | 3+4 2 |
| | 5 2 | | 5 2 | | | 5+6 2 |
| | 6 2 | | 6 2 | | | 3+4 2 |
| | 5 2 | | 5 2 | | | 5+6 2 |
| | 5 2 | | 6 2 | | | 5+6 2 |
| | 6 2 | | 6 2 | | | 3+4 2 |
| | 6 2 | | 5 2 | | | 5+6 2 |
| | 5 2 | | 5 2 | | | 3+4 2 |
| | 6 2 | | 5 2 | | | 5+6 2 |

Continued on next column

**TABLE XXXVIII– continued**

| TQC | ICSD | | NoE | | SF | |
|---|---|---|---|---|---|---|
| | 5 2 | | 6 2 | | | 3+4 2 |
| | 5 2 | | 5 2 | | | 3+4 2 |
| | 6 2 | | 6 2 | | | 5+6 2 |
| | 6 2 | | 5 2 | | | 3+4 2 |
| | 6 2 | | 6 2 | | | 3+4 2 |
| | 5 2 | | 6 2 | | | 3+4 2 |
| | 5 2 | | 6 2 | | | 5+6 2 |
| | 5 2 | | 5 2 | | | 3+4 2 |
| | 6 2 | | 5 2 | | | 5+6 2 |
| | 6 2 | | 6 2 | | | 5+6 2 |
| | 5 2 | | 6 2 | | | 3+4 2 |
| | 5 2 | | 5 2 | | | 5+6 2 |
| | 5 2 | | 6 2 | | | 3+4 2 |
| | 6 2 | | 5 2 | | | 5+6 2 |
| | 6 2 | | 6 2 | | | 5+6 2 |
| 63.1.59 | 413674 | | 224 | | Cs1Cu3Dy2Se5 | |
| | -GM | -T | -Y | -Z | -R | -S |
| | 5 2 | 3+4 4 | 6 2 | 3+4 4 | 2+2 4 | 3+4 2 |
| | 5 2 | 3+4 4 | 6 2 | 3+4 4 | 2+2 4 | 3+4 2 |
| | 6 2 | 3+4 4 | 5 2 | 3+4 4 | 2+2 4 | 5+6 2 |
| | 6 2 | 3+4 4 | 5 2 | 3+4 4 | 2+2 4 | 5+6 2 |
| | 5 2 | 3+4 4 | 6 2 | 3+4 4 | 2+2 4 | 3+4 2 |
| | 5 2 | 3+4 4 | 6 2 | 3+4 4 | 2+2 4 | 3+4 2 |
| | 6 2 | 3+4 4 | 5 2 | 3+4 4 | 2+2 4 | 5+6 2 |
| | 6 2 | 3+4 4 | 5 2 | 3+4 4 | 2+2 4 | 5+6 2 |
| | 5 2 | 3+4 4 | 5 2 | 3+4 4 | 2+2 4 | 3+4 2 |
| | 6 2 | 3+4 4 | 6 2 | 3+4 4 | 2+2 4 | 5+6 2 |
| | 5 2 | 3+4 4 | 6 2 | 3+4 4 | 2+2 4 | 3+4 2 |
| | 5 2 | 3+4 4 | 6 2 | 3+4 4 | 2+2 4 | 5+6 2 |
| | 6 2 | 3+4 4 | 5 2 | 3+4 4 | 2+2 4 | 3+4 2 |
| | 6 2 | 3+4 4 | 5 2 | 3+4 4 | 2+2 4 | 5+6 2 |
| | 5 2 | 3+4 4 | 6 2 | 3+4 4 | 2+2 4 | 3+4 2 |
| | 5 2 | 3+4 4 | 6 2 | 3+4 4 | 2+2 4 | 5+6 2 |
| | 6 2 | 3+4 4 | 5 2 | 3+4 4 | 2+2 4 | 3+4 2 |
| | 6 2 | 3+4 4 | 5 2 | 3+4 4 | 2+2 4 | 5+6 2 |
| | 5 2 | 3+4 4 | 6 2 | 3+4 4 | 2+2 4 | 5+6 2 |
| | 5 2 | 3+4 4 | 5 2 | 3+4 4 | 2+2 4 | 3+4 2 |
| | 6 2 | 3+4 4 | 6 2 | 3+4 4 | 2+2 4 | 3+4 2 |
| | 6 2 | 3+4 4 | 5 2 | 3+4 4 | 2+2 4 | 3+4 2 |
| | 5 2 | 3+4 4 | 5 2 | 3+4 4 | 2+2 4 | 5+6 2 |
| | 6 2 | 3+4 4 | 6 2 | 3+4 4 | 2+2 4 | 5+6 2 |
| | 6 2 | 3+4 4 | 5 2 | 3+4 4 | 2+2 4 | 3+4 2 |
| | 6 2 | 3+4 4 | 5 2 | 3+4 4 | 2+2 4 | 5+6 2 |
| | 5 2 | 3+4 4 | 6 2 | 3+4 4 | 2+2 4 | 3+4 2 |
| | 5 2 | 3+4 4 | 6 2 | 3+4 4 | 2+2 4 | 5+6 2 |
| | 5 2 | 3+4 4 | 5 2 | 3+4 4 | 2+2 4 | 3+4 2 |
| | 6 2 | 3+4 4 | 6 2 | 3+4 4 | 2+2 4 | 5+6 2 |
| | 5 2 | 3+4 4 | 5 2 | 3+4 4 | 2+2 4 | 3+4 2 |
| | 6 2 | 3+4 4 | 6 2 | 3+4 4 | 2+2 4 | 5+6 2 |
| | 5 2 | 3+4 4 | 5 2 | 3+4 4 | 2+2 4 | 3+4 2 |
| | 6 2 | 3+4 4 | 6 2 | 3+4 4 | 2+2 4 | 5+6 2 |
| | 5 2 | 3+4 4 | 5 2 | 3+4 4 | 2+2 4 | 3+4 2 |
| | 5 2 | 3+4 4 | 5 2 | 3+4 4 | 2+2 4 | 3+4 2 |
| | 6 2 | 3+4 4 | 6 2 | 3+4 4 | 2+2 4 | 5+6 2 |
| | 6 2 | 3+4 4 | 5 2 | 3+4 4 | 2+2 4 | 5+6 2 |
| | 6 2 | 3+4 4 | 6 2 | 3+4 4 | 2+2 4 | 5+6 2 |
| | 5 2 | 3+4 4 | 5 2 | 3+4 4 | 2+2 4 | 3+4 2 |
| | 5 2 | 3+4 4 | 6 2 | 3+4 4 | 2+2 4 | 3+4 2 |
| | 6 2 | 3+4 4 | 6 2 | 3+4 4 | 2+2 4 | 3+4 2 |
| | 5 2 | 3+4 4 | 6 2 | 3+4 4 | 2+2 4 | 5+6 2 |
| | 6 2 | 3+4 4 | 5 2 | 3+4 4 | 2+2 4 | 5+6 2 |

Continued on next column

**TABLE XXXVIII– continued**

| TQC | ICSD | | NoE | | SF | |
|---|---|---|---|---|---|---|
| | 5 2 | 3+4 4 | 6 2 | 3+4 4 | 2+2 4 | 5+6 2 |
| | 5 2 | 3+4 4 | 6 2 | 3+4 4 | 2+2 4 | 5+6 2 |
| | 6 2 | 3+4 4 | 6 2 | 3+4 4 | 2+2 4 | 3+4 2 |
| | 5 2 | 3+4 4 | 5 2 | 3+4 4 | 2+2 4 | 3+4 2 |
| | 6 2 | 3+4 4 | 6 2 | 3+4 4 | 2+2 4 | 5+6 2 |
| | 6 2 | 3+4 4 | 6 2 | 3+4 4 | 2+2 4 | 5+6 2 |
| | 6 2 | 3+4 4 | 5 2 | 3+4 4 | 2+2 4 | 5+6 2 |
| | 5 2 | 3+4 4 | 5 2 | 3+4 4 | 2+2 4 | 3+4 2 |
| | 5 2 | 3+4 4 | 6 2 | 3+4 4 | 2+2 4 | 3+4 2 |
| | 5 2 | 3+4 4 | 5 2 | 3+4 4 | 2+2 4 | 5+6 2 |
| | 6 2 | 3+4 4 | 6 2 | 3+4 4 | 2+2 4 | 3+4 2 |
| | 6 2 | 3+4 4 | 6 2 | 3+4 4 | 2+2 4 | 5+6 2 |
| | 5 2 | | 5 2 | | | 5+6 2 |
| | 5 2 | | 5 2 | | | 3+4 2 |
| | 5 2 | | 5 2 | | | 3+4 2 |
| | 5 2 | | 5 2 | | | 5+6 2 |
| | 6 2 | | 5 2 | | | 3+4 2 |
| | 5 2 | | 5 2 | | | 5+6 2 |
| | 6 2 | | 6 2 | | | 3+4 2 |
| | 5 2 | | 5 2 | | | 3+4 2 |
| | 6 2 | | 5 2 | | | 3+4 2 |
| | 5 2 | | 5 2 | | | 3+4 2 |
| | 6 2 | | 6 2 | | | 5+6 2 |
| | 6 2 | | 6 2 | | | 5+6 2 |
| | 6 2 | | 6 2 | | | 3+4 2 |
| | 6 2 | | 6 2 | | | 5+6 2 |
| | 5 2 | | 5 2 | | | 5+6 2 |
| | 6 2 | | 6 2 | | | 3+4 2 |
| | 6 2 | | 6 2 | | | 3+4 2 |
| | 6 2 | | 5 2 | | | 3+4 2 |
| | 5 2 | | 6 2 | | | 5+6 2 |
| | 6 2 | | 5 2 | | | 5+6 2 |
| | 6 2 | | 6 2 | | | 3+4 2 |
| | 5 2 | | 5 2 | | | 3+4 2 |
| | 5 2 | | 6 2 | | | 5+6 2 |
| | 5 2 | | 5 2 | | | 3+4 2 |
| | 5 2 | | 5 2 | | | 5+6 2 |
| | 5 2 | | 5 2 | | | 3+4 2 |
| | 6 2 | | 6 2 | | | 3+4 2 |
| | 6 2 | | 6 2 | | | 3+4 2 |
| | 5 2 | | 5 2 | | | 5+6 2 |
| | 5 2 | | 6 2 | | | 5+6 2 |
| | 6 2 | | 6 2 | | | 5+6 2 |
| | 6 2 | | 6 2 | | | 3+4 2 |
| | 5 2 | | 5 2 | | | 3+4 2 |
| | 5 2 | | 6 2 | | | 3+4 2 |
| | 6 2 | | 5 2 | | | 5+6 2 |
| | 5 2 | | 5 2 | | | 5+6 2 |
| | 6 2 | | 6 2 | | | 5+6 2 |
| | 5 2 | | 5 2 | | | 3+4 2 |
| | 6 2 | | 6 2 | | | 5+6 2 |
| | 5 2 | | 5 2 | | | 3+4 2 |
| | 5 2 | | 6 2 | | | 5+6 2 |
| | 6 2 | | 6 2 | | | 3+4 2 |
| | 6 2 | | 5 2 | | | 5+6 2 |
| | 5 2 | | 6 2 | | | 3+4 2 |
| | 6 2 | | 5 2 | | | 5+6 2 |
| | 5 2 | | 5 2 | | | 5+6 2 |
| | 5 2 | | 6 2 | | | 3+4 2 |
| | 6 2 | | 5 2 | | | 3+4 2 |
| | 6 2 | | 5 2 | | | 5+6 2 |

Continued on next column

**TABLE XXXVIII– continued**

| TQC | ICSD | | NoE | | SF | |
|---|---|---|---|---|---|---|
| | 6 2 | | 6 2 | | | 5+6 2 |
| | 5 2 | | 6 2 | | | 3+4 2 |
| | 5 2 | | 6 2 | | | 5+6 2 |
| | 5 2 | | 5 2 | | | 3+4 2 |
| | 6 2 | | 5 2 | | | 5+6 2 |
| | 6 2 | | 6 2 | | | 3+4 2 |
| | 5 2 | | 6 2 | | | 5+6 2 |
| 63.1.60 | 236871 | | 76 | | Cr2Mg1O4 | |
| | -GM | -T | -Y | -Z | -R | -S |
| | 5 2 | 3+4 4 | 6 2 | 3+4 4 | 2+2 4 | 3+4 2 |
| | 5 2 | 3+4 4 | 5 2 | 3+4 4 | 2+2 4 | 3+4 2 |
| | 6 2 | 3+4 4 | 5 2 | 3+4 4 | 2+2 4 | 5+6 2 |
| | 5 2 | 3+4 4 | 6 2 | 3+4 4 | 2+2 4 | 3+4 2 |
| | 5 2 | 3+4 4 | 5 2 | 3+4 4 | 2+2 4 | 5+6 2 |
| | 6 2 | 3+4 4 | 5 2 | 3+4 4 | 2+2 4 | 3+4 2 |
| | 6 2 | 3+4 4 | 6 2 | 3+4 4 | 2+2 4 | 3+4 2 |
| | 5 2 | 3+4 4 | 5 2 | 3+4 4 | 2+2 4 | 5+6 2 |
| | 6 2 | 3+4 4 | 6 2 | 3+4 4 | 2+2 4 | 5+6 2 |
| | 6 2 | 3+4 4 | 6 2 | 3+4 4 | 2+2 4 | 5+6 2 |
| | 5 2 | 3+4 4 | 6 2 | 3+4 4 | 2+2 4 | 3+4 2 |
| | 6 2 | 3+4 4 | 5 2 | 3+4 4 | 2+2 4 | 5+6 2 |
| | 6 2 | 3+4 4 | 5 2 | 3+4 4 | 2+2 4 | 5+6 2 |
| | 5 2 | 3+4 4 | 6 2 | 3+4 4 | 2+2 4 | 5+6 2 |
| | 5 2 | 3+4 4 | 6 2 | 3+4 4 | 2+2 4 | 3+4 2 |
| | 6 2 | 3+4 4 | 5 2 | 3+4 4 | 2+2 4 | 3+4 2 |
| | 6 2 | 3+4 4 | 6 2 | 3+4 4 | 2+2 4 | 5+6 2 |
| | 5 2 | 3+4 4 | 6 2 | 3+4 4 | 2+2 4 | 3+4 2 |
| | 5 2 | 3+4 4 | 6 2 | 3+4 4 | 2+2 4 | 5+6 2 |
| | 6 2 | | 5 2 | | | 5+6 2 |
| | 5 2 | | 6 2 | | | 3+4 2 |
| | 6 2 | | 5 2 | | | 5+6 2 |
| | 5 2 | | 6 2 | | | 3+4 2 |
| | 5 2 | | 5 2 | | | 3+4 2 |
| | 6 2 | | 6 2 | | | 5+6 2 |
| | 6 2 | | 6 2 | | | 5+6 2 |
| | 6 2 | | 6 2 | | | 5+6 2 |
| | 6 2 | | 5 2 | | | 5+6 2 |
| | 5 2 | | 5 2 | | | 3+4 2 |
| | 6 2 | | 6 2 | | | 3+4 2 |
| | 6 2 | | 6 2 | | | 5+6 2 |
| | 6 2 | | 5 2 | | | 5+6 2 |
| | 5 2 | | 5 2 | | | 3+4 2 |
| | 5 2 | | 5 2 | | | 3+4 2 |
| | 6 2 | | 6 2 | | | 3+4 2 |
| | 5 2 | | 6 2 | | | 5+6 2 |
| | 6 2 | | 6 2 | | | 5+6 2 |
| | 6 2 | | 5 2 | | | 3+4 2 |
| 63.1.61 | 20952 | | 192 | | Co3Ga9Ho2 | |
| | -GM | -T | -Y | -Z | -R | -S |
| | 5 2 | 3+4 4 | 5 2 | 3+4 4 | 2+2 4 | 3+4 2 |
| | 6 2 | 3+4 4 | 6 2 | 3+4 4 | 2+2 4 | 5+6 2 |
| | 5 2 | 3+4 4 | 5 2 | 3+4 4 | 2+2 4 | 3+4 2 |
| | 6 2 | 3+4 4 | 6 2 | 3+4 4 | 2+2 4 | 5+6 2 |
| | 5 2 | 3+4 4 | 6 2 | 3+4 4 | 2+2 4 | 3+4 2 |
| | 6 2 | 3+4 4 | 5 2 | 3+4 4 | 2+2 4 | 5+6 2 |
| | 5 2 | 3+4 4 | 6 2 | 3+4 4 | 2+2 4 | 3+4 2 |
| | 6 2 | 3+4 4 | 5 2 | 3+4 4 | 2+2 4 | 5+6 2 |
| | 5 2 | 3+4 4 | 6 2 | 3+4 4 | 2+2 4 | 3+4 2 |
| | 6 2 | 3+4 4 | 5 2 | 3+4 4 | 2+2 4 | 5+6 2 |
| | 5 2 | 3+4 4 | 5 2 | 3+4 4 | 2+2 4 | 3+4 2 |
| | 6 2 | 3+4 4 | 6 2 | 3+4 4 | 2+2 4 | 5+6 2 |

Continued on next column

**TABLE XXXVIII– continued**

| TQC | ICSD | | NoE | | SF | |
|---|---|---|---|---|---|---|
| | 5 2 | 3+4 4 | 5 2 | 3+4 4 | 2+2 4 | 3+4 2 |
| | 6 2 | 3+4 4 | 6 2 | 3+4 4 | 2+2 4 | 5+6 2 |
| | 5 2 | 3+4 4 | 6 2 | 3+4 4 | 2+2 4 | 3+4 2 |
| | 6 2 | 3+4 4 | 5 2 | 3+4 4 | 2+2 4 | 5+6 2 |
| | 5 2 | 3+4 4 | 5 2 | 3+4 4 | 2+2 4 | 5+6 2 |
| | 6 2 | 3+4 4 | 6 2 | 3+4 4 | 2+2 4 | 3+4 2 |
| | 5 2 | 3+4 4 | 6 2 | 3+4 4 | 2+2 4 | 5+6 2 |
| | 5 2 | 3+4 4 | 5 2 | 3+4 4 | 2+2 4 | 3+4 2 |
| | 6 2 | 3+4 4 | 6 2 | 3+4 4 | 2+2 4 | 5+6 2 |
| | 5 2 | 3+4 4 | 5 2 | 3+4 4 | 2+2 4 | 3+4 2 |
| | 6 2 | 3+4 4 | 5 2 | 3+4 4 | 2+2 4 | 3+4 2 |
| | 5 2 | 3+4 4 | 6 2 | 3+4 4 | 2+2 4 | 5+6 2 |
| | 6 2 | 3+4 4 | 6 2 | 3+4 4 | 2+2 4 | 5+6 2 |
| | 5 2 | 3+4 4 | 5 2 | 3+4 4 | 2+2 4 | 3+4 2 |
| | 5 2 | 3+4 4 | 5 2 | 3+4 4 | 2+2 4 | 5+6 2 |
| | 6 2 | 3+4 4 | 6 2 | 3+4 4 | 2+2 4 | 3+4 2 |
| | 6 2 | 3+4 4 | 6 2 | 3+4 4 | 2+2 4 | 5+6 2 |
| | 5 2 | 3+4 4 | 5 2 | 3+4 4 | 2+2 4 | 3+4 2 |
| | 6 2 | 3+4 4 | 5 2 | 3+4 4 | 2+2 4 | 3+4 2 |
| | 5 2 | 3+4 4 | 6 2 | 3+4 4 | 2+2 4 | 5+6 2 |
| | 6 2 | 3+4 4 | 6 2 | 3+4 4 | 2+2 4 | 3+4 2 |
| | 5 2 | 3+4 4 | 5 2 | 3+4 4 | 2+2 4 | 5+6 2 |
| | 6 2 | 3+4 4 | 5 2 | 3+4 4 | 2+2 4 | 3+4 2 |
| | 6 2 | 3+4 4 | 6 2 | 3+4 4 | 2+2 4 | 3+4 2 |
| | 5 2 | 3+4 4 | 6 2 | 3+4 4 | 2+2 4 | 3+4 2 |
| | 6 2 | 3+4 4 | 5 2 | 3+4 4 | 2+2 4 | 3+4 2 |
| | 5 2 | 3+4 4 | 6 2 | 3+4 4 | 2+2 4 | 5+6 2 |
| | 5 2 | 3+4 4 | 5 2 | 3+4 4 | 2+2 4 | 5+6 2 |
| | 5 2 | 3+4 4 | 5 2 | 3+4 4 | 2+2 4 | 5+6 2 |
| | 5 2 | 3+4 4 | 5 2 | 3+4 4 | 2+2 4 | 3+4 2 |
| | 5 2 | 3+4 4 | 6 2 | 3+4 4 | 2+2 4 | 3+4 2 |
| | 5 2 | 3+4 4 | 6 2 | 3+4 4 | 2+2 4 | 5+6 2 |
| | 5 2 | 3+4 4 | 5 2 | 3+4 4 | 2+2 4 | 3+4 2 |
| | 6 2 | 3+4 4 | 6 2 | 3+4 4 | 2+2 4 | 5+6 2 |
| | 6 2 | 3+4 4 | 5 2 | 3+4 4 | 2+2 4 | 5+6 2 |
| | 6 2 | 3+4 4 | 6 2 | 3+4 4 | 2+2 4 | 5+6 2 |
| | 5 2 | | 6 2 | | | 3+4 2 |
| | 5 2 | | 5 2 | | | 3+4 2 |
| | 6 2 | | 5 2 | | | 5+6 2 |
| | 5 2 | | 5 2 | | | 3+4 2 |
| | 6 2 | | 5 2 | | | 5+6 2 |
| | 5 2 | | 6 2 | | | 5+6 2 |
| | 5 2 | | 5 2 | | | 5+6 2 |
| | 5 2 | | 6 2 | | | 3+4 2 |
| | 6 2 | | 5 2 | | | 3+4 2 |
| | 5 2 | | 5 2 | | | 3+4 2 |
| | 5 2 | | 6 2 | | | 5+6 2 |
| | 6 2 | | 5 2 | | | 3+4 2 |
| | 6 2 | | 6 2 | | | 5+6 2 |
| | 5 2 | | 5 2 | | | 3+4 2 |
| | 5 2 | | 6 2 | | | 5+6 2 |
| | 6 2 | | 5 2 | | | 3+4 2 |
| | 5 2 | | 6 2 | | | 5+6 2 |
| | 6 2 | | 5 2 | | | 3+4 2 |
| | 6 2 | | 5 2 | | | 3+4 2 |
| | 5 2 | | 6 2 | | | 5+6 2 |
| | 5 2 | | 5 2 | | | 3+4 2 |
| | 6 2 | | 6 2 | | | 3+4 2 |
| | 5 2 | | 5 2 | | | 5+6 2 |
| | 6 2 | | 6 2 | | | 3+4 2 |
| | 5 2 | | 5 2 | | | 3+4 2 |

Continued on next column

**TABLE XXXVIII– continued**

| TQC | ICSD | | NoE | | SF | |
|---|---|---|---|---|---|---|
| | 5 2 | | 6 2 | | | 5+6 2 |
| | 6 2 | | 5 2 | | | 5+6 2 |
| | 5 2 | | 6 2 | | | 3+4 2 |
| | 5 2 | | 5 2 | | | 5+6 2 |
| | 6 2 | | 5 2 | | | 3+4 2 |
| | 6 2 | | 6 2 | | | 5+6 2 |
| | 6 2 | | 5 2 | | | 3+4 2 |
| | 5 2 | | 6 2 | | | 5+6 2 |
| | 6 2 | | 5 2 | | | 3+4 2 |
| | 5 2 | | 6 2 | | | 3+4 2 |
| | 5 2 | | 6 2 | | | 3+4 2 |
| | 6 2 | | 5 2 | | | 3+4 2 |
| | 5 2 | | 6 2 | | | 5+6 2 |
| | 6 2 | | 5 2 | | | 3+4 2 |
| | 5 2 | | 6 2 | | | 5+6 2 |
| | 6 2 | | 5 2 | | | 3+4 2 |
| | 6 2 | | 5 2 | | | 5+6 2 |
| | 5 2 | | 5 2 | | | 5+6 2 |
| | 6 2 | | 6 2 | | | 3+4 2 |
| | 5 2 | | 6 2 | | | 3+4 2 |
| | 6 2 | | 5 2 | | | 5+6 2 |
| | 5 2 | | 6 2 | | | 3+4 2 |
| | 6 2 | | 6 2 | | | 5+6 2 |
| 63.1.62 | 427218 | | 124 | | Ce3Ge3Mn2 | |
| | -GM | -T | -Y | -Z | -R | -S |
| | 5 2 | 3+4 4 | 5 2 | 3+4 4 | 2+2 4 | 3+4 2 |
| | 6 2 | 3+4 4 | 6 2 | 3+4 4 | 2+2 4 | 5+6 2 |
| | 5 2 | 3+4 4 | 5 2 | 3+4 4 | 2+2 4 | 3+4 2 |
| | 5 2 | 3+4 4 | 5 2 | 3+4 4 | 2+2 4 | 3+4 2 |
| | 6 2 | 3+4 4 | 6 2 | 3+4 4 | 2+2 4 | 5+6 2 |
| | 6 2 | 3+4 4 | 6 2 | 3+4 4 | 2+2 4 | 5+6 2 |
| | 5 2 | 3+4 4 | 5 2 | 3+4 4 | 2+2 4 | 3+4 2 |
| | 6 2 | 3+4 4 | 5 2 | 3+4 4 | 2+2 4 | 3+4 2 |
| | 5 2 | 3+4 4 | 6 2 | 3+4 4 | 2+2 4 | 5+6 2 |
| | 6 2 | 3+4 4 | 6 2 | 3+4 4 | 2+2 4 | 5+6 2 |
| | 5 2 | 3+4 4 | 5 2 | 3+4 4 | 2+2 4 | 3+4 2 |
| | 6 2 | 3+4 4 | 6 2 | 3+4 4 | 2+2 4 | 5+6 2 |
| | 5 2 | 3+4 4 | 5 2 | 3+4 4 | 2+2 4 | 3+4 2 |
| | 6 2 | 3+4 4 | 6 2 | 3+4 4 | 2+2 4 | 5+6 2 |
| | 5 2 | 3+4 4 | 6 2 | 3+4 4 | 2+2 4 | 3+4 2 |
| | 6 2 | 3+4 4 | 5 2 | 3+4 4 | 2+2 4 | 5+6 2 |
| | 5 2 | 3+4 4 | 5 2 | 3+4 4 | 2+2 4 | 3+4 2 |
| | 5 2 | 3+4 4 | 5 2 | 3+4 4 | 2+2 4 | 5+6 2 |
| | 6 2 | 3+4 4 | 6 2 | 3+4 4 | 2+2 4 | 3+4 2 |
| | 6 2 | 3+4 4 | 6 2 | 3+4 4 | 2+2 4 | 5+6 2 |
| | 5 2 | 3+4 4 | 5 2 | 3+4 4 | 2+2 4 | 3+4 2 |
| | 6 2 | 3+4 4 | 6 2 | 3+4 4 | 2+2 4 | 5+6 2 |
| | 5 2 | 3+4 4 | 6 2 | 3+4 4 | 2+2 4 | 3+4 2 |
| | 6 2 | 3+4 4 | 5 2 | 3+4 4 | 2+2 4 | 5+6 2 |
| | 5 2 | 3+4 4 | 5 2 | 3+4 4 | 2+2 4 | 5+6 2 |
| | 5 2 | 3+4 4 | 5 2 | 3+4 4 | 2+2 4 | 5+6 2 |
| | 6 2 | 3+4 4 | 6 2 | 3+4 4 | 2+2 4 | 3+4 2 |
| | 5 2 | 3+4 4 | 6 2 | 3+4 4 | 2+2 4 | 5+6 2 |
| | 6 2 | 3+4 4 | 5 2 | 3+4 4 | 2+2 4 | 3+4 2 |
| | 6 2 | 3+4 4 | 6 2 | 3+4 4 | 2+2 4 | 3+4 2 |
| | 5 2 | 3+4 4 | 6 2 | 3+4 4 | 2+2 4 | 3+4 2 |
| | 5 2 | | 5 2 | | | 5+6 2 |
| | 6 2 | | 6 2 | | | 3+4 2 |
| | 6 2 | | 5 2 | | | 5+6 2 |
| | 6 2 | | 5 2 | | | 5+6 2 |
| | 6 2 | | 6 2 | | | 3+4 2 |

Continued on next column

**TABLE XXXVIII– continued**

| TQC | ICSD | | NoE | | SF | |
|---|---|---|---|---|---|---|
| | 5 2 | | 6 2 | | | 5+6 2 |
| | 6 2 | | 6 2 | | | 5+6 2 |
| | 5 2 | | 5 2 | | | 3+4 2 |
| | 5 2 | | 5 2 | | | 5+6 2 |
| | 5 2 | | 5 2 | | | 3+4 2 |
| | 6 2 | | 5 2 | | | 3+4 2 |
| | 5 2 | | 6 2 | | | 5+6 2 |
| | 6 2 | | 5 2 | | | 5+6 2 |
| | 6 2 | | 6 2 | | | 5+6 2 |
| | 6 2 | | 5 2 | | | 3+4 2 |
| | 5 2 | | 6 2 | | | 5+6 2 |
| | 5 2 | | 5 2 | | | 3+4 2 |
| | 5 2 | | 6 2 | | | 3+4 2 |
| | 5 2 | | 6 2 | | | 5+6 2 |
| | 5 2 | | 6 2 | | | 3+4 2 |
| | 6 2 | | 6 2 | | | 5+6 2 |
| | 6 2 | | 5 2 | | | 3+4 2 |
| | 5 2 | | 5 2 | | | 3+4 2 |
| | 5 2 | | 5 2 | | | 5+6 2 |
| | 5 2 | | 6 2 | | | 5+6 2 |
| | 6 2 | | 6 2 | | | 3+4 2 |
| | 6 2 | | 6 2 | | | 5+6 2 |
| | 5 2 | | 5 2 | | | 5+6 2 |
| | 6 2 | | 5 2 | | | 5+6 2 |
| | 5 2 | | 6 2 | | | 3+4 2 |
| | 5 2 | | 5 2 | | | 3+4 2 |
| 63.1.63 | 250753 | | 176 | | Ce2Sn1Zn5 | |
| | -GM | -T | -Y | -Z | -R | -S |
| | 5 2 | 3+4 4 | 5 2 | 3+4 4 | 2+2 4 | 3+4 2 |
| | 5 2 | 3+4 4 | 5 2 | 3+4 4 | 2+2 4 | 3+4 2 |
| | 5 2 | 3+4 4 | 5 2 | 3+4 4 | 2+2 4 | 3+4 2 |
| | 6 2 | 3+4 4 | 6 2 | 3+4 4 | 2+2 4 | 5+6 2 |
| | 6 2 | 3+4 4 | 6 2 | 3+4 4 | 2+2 4 | 5+6 2 |
| | 6 2 | 3+4 4 | 6 2 | 3+4 4 | 2+2 4 | 5+6 2 |
| | 5 2 | 3+4 4 | 6 2 | 3+4 4 | 2+2 4 | 3+4 2 |
| | 6 2 | 3+4 4 | 5 2 | 3+4 4 | 2+2 4 | 5+6 2 |
| | 6 2 | 3+4 4 | 6 2 | 3+4 4 | 2+2 4 | 5+6 2 |
| | 6 2 | 3+4 4 | 6 2 | 3+4 4 | 2+2 4 | 5+6 2 |
| | 6 2 | 3+4 4 | 6 2 | 3+4 4 | 2+2 4 | 5+6 2 |
| | 6 2 | 3+4 4 | 6 2 | 3+4 4 | 2+2 4 | 5+6 2 |
| | 5 2 | 3+4 4 | 5 2 | 3+4 4 | 2+2 4 | 3+4 2 |
| | 5 2 | 3+4 4 | 6 2 | 3+4 4 | 2+2 4 | 5+6 2 |
| | 6 2 | 3+4 4 | 6 2 | 3+4 4 | 2+2 4 | 3+4 2 |
| | 6 2 | 3+4 4 | 5 2 | 3+4 4 | 2+2 4 | 5+6 2 |
| | 5 2 | 3+4 4 | 5 2 | 3+4 4 | 2+2 4 | 3+4 2 |
| | 6 2 | 3+4 4 | 6 2 | 3+4 4 | 2+2 4 | 5+6 2 |
| | 5 2 | 3+4 4 | 6 2 | 3+4 4 | 2+2 4 | 3+4 2 |
| | 6 2 | 3+4 4 | 5 2 | 3+4 4 | 2+2 4 | 5+6 2 |
| | 6 2 | 3+4 4 | 6 2 | 3+4 4 | 2+2 4 | 3+4 2 |
| | 5 2 | 3+4 4 | 5 2 | 3+4 4 | 2+2 4 | 5+6 2 |
| | 5 2 | 3+4 4 | 6 2 | 3+4 4 | 2+2 4 | 3+4 2 |
| | 5 2 | 3+4 4 | 6 2 | 3+4 4 | 2+2 4 | 5+6 2 |
| | 5 2 | 3+4 4 | 5 2 | 3+4 4 | 2+2 4 | 3+4 2 |
| | 6 2 | 3+4 4 | 6 2 | 3+4 4 | 2+2 4 | 5+6 2 |
| | 6 2 | 3+4 4 | 5 2 | 3+4 4 | 2+2 4 | 5+6 2 |
| | 5 2 | 3+4 4 | 6 2 | 3+4 4 | 2+2 4 | 3+4 2 |
| | 6 2 | 3+4 4 | 5 2 | 3+4 4 | 2+2 4 | 3+4 2 |
| | 5 2 | 3+4 4 | 6 2 | 3+4 4 | 2+2 4 | 5+6 2 |
| | 6 2 | 3+4 4 | 5 2 | 3+4 4 | 2+2 4 | 3+4 2 |
| | 5 2 | 3+4 4 | 5 2 | 3+4 4 | 2+2 4 | 5+6 2 |
| | 6 2 | 3+4 4 | 6 2 | 3+4 4 | 2+2 4 | 3+4 2 |

Continued on next column

**TABLE XXXVIII– continued**

| TQC | ICSD | | NoE | | SF | | |
|---|---|---|---|---|---|---|---|
| | 5 2 | 3+4 4 | 6 2 | 3+4 4 | 2+2 4 | 5+6 2 |
| | 5 2 | 3+4 4 | 5 2 | 3+4 4 | 2+2 4 | 3+4 2 |
| | 6 2 | 3+4 4 | 6 2 | 3+4 4 | 2+2 4 | 3+4 2 |
| | 6 2 | 3+4 4 | 5 2 | 3+4 4 | 2+2 4 | 5+6 2 |
| | 6 2 | 3+4 4 | 6 2 | 3+4 4 | 2+2 4 | 5+6 2 |
| | 5 2 | 3+4 4 | 5 2 | 3+4 4 | 2+2 4 | 3+4 2 |
| | 6 2 | 3+4 4 | 5 2 | 3+4 4 | 2+2 4 | 5+6 2 |
| | 5 2 | 3+4 4 | 6 2 | 3+4 4 | 2+2 4 | 3+4 2 |
| | 5 2 | 3+4 4 | 6 2 | 3+4 4 | 2+2 4 | 5+6 2 |
| | 6 2 | 3+4 4 | 5 2 | 3+4 4 | 2+2 4 | 3+4 2 |
| | 5 2 | 3+4 4 | 5 2 | 3+4 4 | 2+2 4 | 3+4 2 |
| | 6 2 | | 6 2 | | | 5+6 2 |
| | 6 2 | | 5 2 | | | 5+6 2 |
| | 5 2 | | 5 2 | | | 3+4 2 |
| | 5 2 | | 6 2 | | | 5+6 2 |
| | 6 2 | | 6 2 | | | 5+6 2 |
| | 5 2 | | 5 2 | | | 3+4 2 |
| | 6 2 | | 6 2 | | | 5+6 2 |
| | 5 2 | | 5 2 | | | 3+4 2 |
| | 6 2 | | 6 2 | | | 5+6 2 |
| | 6 2 | | 5 2 | | | 3+4 2 |
| | 5 2 | | 5 2 | | | 3+4 2 |
| | 6 2 | | 5 2 | | | 5+6 2 |
| | 5 2 | | 6 2 | | | 5+6 2 |
| | 6 2 | | 5 2 | | | 3+4 2 |
| | 6 2 | | 6 2 | | | 5+6 2 |
| | 5 2 | | 5 2 | | | 3+4 2 |
| | 6 2 | | 6 2 | | | 5+6 2 |
| | 5 2 | | 5 2 | | | 3+4 2 |
| | 6 2 | | 5 2 | | | 5+6 2 |
| | 5 2 | | 6 2 | | | 3+4 2 |
| | 6 2 | | 6 2 | | | 3+4 2 |
| | 5 2 | | 5 2 | | | 5+6 2 |
| | 5 2 | | 6 2 | | | 5+6 2 |
| | 6 2 | | 6 2 | | | 3+4 2 |
| | 5 2 | | 6 2 | | | 3+4 2 |
| | 6 2 | | 5 2 | | | 5+6 2 |
| | 5 2 | | 6 2 | | | 3+4 2 |
| | 5 2 | | 5 2 | | | 3+4 2 |
| | 6 2 | | 5 2 | | | 5+6 2 |
| | 6 2 | | 6 2 | | | 5+6 2 |
| | 5 2 | | 6 2 | | | 3+4 2 |
| | 5 2 | | 5 2 | | | 5+6 2 |
| | 6 2 | | 5 2 | | | 5+6 2 |
| | 5 2 | | 6 2 | | | 3+4 2 |
| | 5 2 | | 5 2 | | | 3+4 2 |
| | 6 2 | | 6 2 | | | 3+4 2 |
| | 5 2 | | 6 2 | | | 5+6 2 |
| | 6 2 | | 5 2 | | | 3+4 2 |
| | 5 2 | | 6 2 | | | 5+6 2 |
| | 5 2 | | 5 2 | | | 3+4 2 |
| | 5 2 | | 6 2 | | | 3+4 2 |
| | 6 2 | | 5 2 | | | 5+6 2 |
| | 5 2 | | 6 2 | | | 5+6 2 |
| | 6 2 | | 5 2 | | | 5+6 2 |
| 63.1.64 | 250757 | | 176 | | Ce2Pb1Zn5 | |
| | -GM | -T | -Y | -Z | -R | -S |
| | 5 2 | 3+4 4 | 5 2 | 3+4 4 | 2+2 4 | 3+4 2 |
| | 5 2 | 3+4 4 | 5 2 | 3+4 4 | 2+2 4 | 3+4 2 |
| | 5 2 | 3+4 4 | 5 2 | 3+4 4 | 2+2 4 | 3+4 2 |
| | 6 2 | 3+4 4 | 6 2 | 3+4 4 | 2+2 4 | 5+6 2 |
| | 6 2 | 3+4 4 | 6 2 | 3+4 4 | 2+2 4 | 5+6 2 |
| | 6 2 | 3+4 4 | 6 2 | 3+4 4 | 2+2 4 | 5+6 2 |
| | 5 2 | 3+4 4 | 6 2 | 3+4 4 | 2+2 4 | 3+4 2 |
| | 6 2 | 3+4 4 | 5 2 | 3+4 4 | 2+2 4 | 5+6 2 |
| | 6 2 | 3+4 4 | 6 2 | 3+4 4 | 2+2 4 | 5+6 2 |
| | 6 2 | 3+4 4 | 6 2 | 3+4 4 | 2+2 4 | 5+6 2 |
| | 6 2 | 3+4 4 | 6 2 | 3+4 4 | 2+2 4 | 5+6 2 |
| | 6 2 | 3+4 4 | 6 2 | 3+4 4 | 2+2 4 | 5+6 2 |
| | 5 2 | 3+4 4 | 5 2 | 3+4 4 | 2+2 4 | 3+4 2 |
| | 6 2 | 3+4 4 | 6 2 | 3+4 4 | 2+2 4 | 5+6 2 |
| | 5 2 | 3+4 4 | 6 2 | 3+4 4 | 2+2 4 | 3+4 2 |
| | 6 2 | 3+4 4 | 5 2 | 3+4 4 | 2+2 4 | 5+6 2 |
| | 5 2 | 3+4 4 | 5 2 | 3+4 4 | 2+2 4 | 3+4 2 |
| | 6 2 | 3+4 4 | 6 2 | 3+4 4 | 2+2 4 | 5+6 2 |
| | 5 2 | 3+4 4 | 6 2 | 3+4 4 | 2+2 4 | 3+4 2 |
| | 6 2 | 3+4 4 | 5 2 | 3+4 4 | 2+2 4 | 5+6 2 |
| | 6 2 | 3+4 4 | 6 2 | 3+4 4 | 2+2 4 | 3+4 2 |
| | 5 2 | 3+4 4 | 5 2 | 3+4 4 | 2+2 4 | 5+6 2 |
| | 5 2 | 3+4 4 | 6 2 | 3+4 4 | 2+2 4 | 3+4 2 |
| | 5 2 | 3+4 4 | 6 2 | 3+4 4 | 2+2 4 | 5+6 2 |
| | 5 2 | 3+4 4 | 5 2 | 3+4 4 | 2+2 4 | 3+4 2 |
| | 6 2 | 3+4 4 | 6 2 | 3+4 4 | 2+2 4 | 5+6 2 |
| | 6 2 | 3+4 4 | 5 2 | 3+4 4 | 2+2 4 | 3+4 2 |
| | 6 2 | 3+4 4 | 6 2 | 3+4 4 | 2+2 4 | 5+6 2 |
| | 5 2 | 3+4 4 | 5 2 | 3+4 4 | 2+2 4 | 3+4 2 |
| | 5 2 | 3+4 4 | 6 2 | 3+4 4 | 2+2 4 | 5+6 2 |
| | 6 2 | 3+4 4 | 5 2 | 3+4 4 | 2+2 4 | 3+4 2 |
| | 6 2 | 3+4 4 | 5 2 | 3+4 4 | 2+2 4 | 5+6 2 |
| | 5 2 | 3+4 4 | 6 2 | 3+4 4 | 2+2 4 | 3+4 2 |
| | 5 2 | 3+4 4 | 6 2 | 3+4 4 | 2+2 4 | 5+6 2 |
| | 5 2 | 3+4 4 | 5 2 | 3+4 4 | 2+2 4 | 3+4 2 |
| | 6 2 | 3+4 4 | 6 2 | 3+4 4 | 2+2 4 | 5+6 2 |
| | 6 2 | 3+4 4 | 5 2 | 3+4 4 | 2+2 4 | 3+4 2 |
| | 6 2 | 3+4 4 | 6 2 | 3+4 4 | 2+2 4 | 5+6 2 |
| | 5 2 | 3+4 4 | 5 2 | 3+4 4 | 2+2 4 | 3+4 2 |
| | 6 2 | 3+4 4 | 5 2 | 3+4 4 | 2+2 4 | 5+6 2 |
| | 5 2 | 3+4 4 | 6 2 | 3+4 4 | 2+2 4 | 3+4 2 |
| | 5 2 | 3+4 4 | 6 2 | 3+4 4 | 2+2 4 | 5+6 2 |
| | 6 2 | 3+4 4 | 5 2 | 3+4 4 | 2+2 4 | 3+4 2 |
| | 5 2 | 3+4 4 | 5 2 | 3+4 4 | 2+2 4 | 3+4 2 |
| | 6 2 | | 6 2 | | | 5+6 2 |
| | 6 2 | | 5 2 | | | 5+6 2 |
| | 5 2 | | 6 2 | | | 5+6 2 |
| | 5 2 | | 5 2 | | | 3+4 2 |
| | 6 2 | | 6 2 | | | 5+6 2 |
| | 5 2 | | 5 2 | | | 3+4 2 |
| | 6 2 | | 6 2 | | | 5+6 2 |
| | 5 2 | | 5 2 | | | 3+4 2 |
| | 6 2 | | 6 2 | | | 5+6 2 |
| | 6 2 | | 5 2 | | | 3+4 2 |
| | 6 2 | | 5 2 | | | 5+6 2 |
| | 6 2 | | 5 2 | | | 3+4 2 |
| | 5 2 | | 6 2 | | | 5+6 2 |
| | 5 2 | | 5 2 | | | 3+4 2 |
| | 6 2 | | 6 2 | | | 5+6 2 |
| | 5 2 | | 5 2 | | | 3+4 2 |
| | 6 2 | | 6 2 | | | 3+4 2 |
| | 6 2 | | 5 2 | | | 5+6 2 |
| | 5 2 | | 6 2 | | | 3+4 2 |
| | 6 2 | | 5 2 | | | 5+6 2 |
| | 5 2 | | 6 2 | | | 3+4 2 |

Continued on next column

Continued on next column

**TABLE XXXVIII– continued**

| TQC | ICSD | | NoE | | SF | |
|---|---|---|---|---|---|---|
| | 6 2 | | 5 2 | | | 5+6 2 |
| | 5 2 | | 6 2 | | | 3+4 2 |
| | 5 2 | | 6 2 | | | 5+6 2 |
| | 5 2 | | 6 2 | | | 3+4 2 |
| | 6 2 | | 5 2 | | | 5+6 2 |
| | 5 2 | | 6 2 | | | 3+4 2 |
| | 5 2 | | 5 2 | | | 3+4 2 |
| | 6 2 | | 5 2 | | | 5+6 2 |
| | 6 2 | | 6 2 | | | 5+6 2 |
| | 5 2 | | 6 2 | | | 3+4 2 |
| | 5 2 | | 5 2 | | | 5+6 2 |
| | 6 2 | | 5 2 | | | 3+4 2 |
| | 5 2 | | 6 2 | | | 5+6 2 |
| | 5 2 | | 5 2 | | | 3+4 2 |
| | 6 2 | | 6 2 | | | 3+4 2 |
| | 5 2 | | 6 2 | | | 5+6 2 |
| | 6 2 | | 5 2 | | | 3+4 2 |
| | 5 2 | | 6 2 | | | 3+4 2 |
| | 5 2 | | 6 2 | | | 5+6 2 |
| | 5 2 | | 5 2 | | | 3+4 2 |
| | 6 2 | | 5 2 | | | 5+6 2 |
| | 5 2 | | 6 2 | | | 5+6 2 |
| | 6 2 | | 5 2 | | | 5+6 2 |
| 63.1.65 | 63036 | | 76 | | Ce2Ge3Li2 | |
| | -GM | -T | -Y | -Z | -R | -S |
| | 5 2 | 3+4 4 | 5 2 | 3+4 4 | 2+2 4 | 3+4 2 |
| | 6 2 | 3+4 4 | 6 2 | 3+4 4 | 2+2 4 | 5+6 2 |
| | 5 2 | 3+4 4 | 5 2 | 3+4 4 | 2+2 4 | 3+4 2 |
| | 6 2 | 3+4 4 | 6 2 | 3+4 4 | 2+2 4 | 5+6 2 |
| | 6 2 | 3+4 4 | 6 2 | 3+4 4 | 2+2 4 | 5+6 2 |
| | 5 2 | 3+4 4 | 5 2 | 3+4 4 | 2+2 4 | 3+4 2 |
| | 5 2 | 3+4 4 | 5 2 | 3+4 4 | 2+2 4 | 3+4 2 |
| | 6 2 | 3+4 4 | 6 2 | 3+4 4 | 2+2 4 | 5+6 2 |
| | 5 2 | 3+4 4 | 5 2 | 3+4 4 | 2+2 4 | 3+4 2 |
| | 6 2 | 3+4 4 | 6 2 | 3+4 4 | 2+2 4 | 5+6 2 |
| | 5 2 | 3+4 4 | 5 2 | 3+4 4 | 2+2 4 | 3+4 2 |
| | 6 2 | 3+4 4 | 6 2 | 3+4 4 | 2+2 4 | 5+6 2 |
| | 5 2 | 3+4 4 | 5 2 | 3+4 4 | 2+2 4 | 5+6 2 |
| | 6 2 | 3+4 4 | 6 2 | 3+4 4 | 2+2 4 | 3+4 2 |
| | 5 2 | 3+4 4 | 5 2 | 3+4 4 | 2+2 4 | 3+4 2 |
| | 6 2 | 3+4 4 | 6 2 | 3+4 4 | 2+2 4 | 5+6 2 |
| | 5 2 | 3+4 4 | 5 2 | 3+4 4 | 2+2 4 | 5+6 2 |
| | 6 2 | 3+4 4 | 6 2 | 3+4 4 | 2+2 4 | 5+6 2 |
| | 5 2 | 3+4 4 | 5 2 | 3+4 4 | 2+2 4 | 3+4 2 |
| | 5 2 | | 6 2 | | | 3+4 2 |
| | 6 2 | | 5 2 | | | 5+6 2 |
| | 6 2 | | 6 2 | | | 3+4 2 |
| | 6 2 | | 5 2 | | | 3+4 2 |
| | 5 2 | | 6 2 | | | 5+6 2 |
| | 5 2 | | 6 2 | | | 3+4 2 |
| | 6 2 | | 5 2 | | | 5+6 2 |
| | 5 2 | | 6 2 | | | 5+6 2 |
| | 6 2 | | 6 2 | | | 5+6 2 |
| | 6 2 | | 5 2 | | | 3+4 2 |
| | 5 2 | | 6 2 | | | 3+4 2 |
| | 5 2 | | 5 2 | | | 5+6 2 |
| | 5 2 | | 6 2 | | | 3+4 2 |
| | 5 2 | | 6 2 | | | 5+6 2 |
| | 5 2 | | 5 2 | | | 5+6 2 |
| | 6 2 | | 5 2 | | | 3+4 2 |
| | 5 2 | | 5 2 | | | 5+6 2 |

Continued on next column

**TABLE XXXVIII– continued**

| TQC | ICSD | | NoE | | SF | |
|---|---|---|---|---|---|---|
| | 6 2 | | 5 2 | | | 3+4 2 |
| | 6 2 | | 5 2 | | | 3+4 2 |
| 63.1.66 | 262430 | | 76 | | Ce2Ge3Li2 | |
| | -GM | -T | -Y | -Z | -R | -S |
| | 5 2 | 3+4 4 | 5 2 | 3+4 4 | 2+2 4 | 5+6 2 |
| | 6 2 | 3+4 4 | 6 2 | 3+4 4 | 2+2 4 | 3+4 2 |
| | 5 2 | 3+4 4 | 5 2 | 3+4 4 | 2+2 4 | 5+6 2 |
| | 6 2 | 3+4 4 | 6 2 | 3+4 4 | 2+2 4 | 3+4 2 |
| | 6 2 | 3+4 4 | 6 2 | 3+4 4 | 2+2 4 | 3+4 2 |
| | 5 2 | 3+4 4 | 5 2 | 3+4 4 | 2+2 4 | 5+6 2 |
| | 5 2 | 3+4 4 | 5 2 | 3+4 4 | 2+2 4 | 5+6 2 |
| | 6 2 | 3+4 4 | 6 2 | 3+4 4 | 2+2 4 | 3+4 2 |
| | 5 2 | 3+4 4 | 5 2 | 3+4 4 | 2+2 4 | 5+6 2 |
| | 6 2 | 3+4 4 | 6 2 | 3+4 4 | 2+2 4 | 3+4 2 |
| | 5 2 | 3+4 4 | 5 2 | 3+4 4 | 2+2 4 | 5+6 2 |
| | 6 2 | 3+4 4 | 6 2 | 3+4 4 | 2+2 4 | 3+4 2 |
| | 5 2 | 3+4 4 | 5 2 | 3+4 4 | 2+2 4 | 3+4 2 |
| | 6 2 | 3+4 4 | 6 2 | 3+4 4 | 2+2 4 | 5+6 2 |
| | 5 2 | 3+4 4 | 5 2 | 3+4 4 | 2+2 4 | 5+6 2 |
| | 6 2 | 3+4 4 | 6 2 | 3+4 4 | 2+2 4 | 3+4 2 |
| | 5 2 | 3+4 4 | 5 2 | 3+4 4 | 2+2 4 | 3+4 2 |
| | 6 2 | 3+4 4 | 6 2 | 3+4 4 | 2+2 4 | 3+4 2 |
| | 5 2 | 3+4 4 | 5 2 | 3+4 4 | 2+2 4 | 5+6 2 |
| | 5 2 | | 6 2 | | | 5+6 2 |
| | 6 2 | | 5 2 | | | 3+4 2 |
| | 6 2 | | 6 2 | | | 5+6 2 |
| | 6 2 | | 6 2 | | | 5+6 2 |
| | 5 2 | | 5 2 | | | 3+4 2 |
| | 5 2 | | 6 2 | | | 5+6 2 |
| | 6 2 | | 5 2 | | | 3+4 2 |
| | 5 2 | | 6 2 | | | 3+4 2 |
| | 6 2 | | 6 2 | | | 3+4 2 |
| | 6 2 | | 5 2 | | | 5+6 2 |
| | 5 2 | | 6 2 | | | 5+6 2 |
| | 5 2 | | 5 2 | | | 3+4 2 |
| | 5 2 | | 6 2 | | | 5+6 2 |
| | 5 2 | | 6 2 | | | 3+4 2 |
| | 6 2 | | 5 2 | | | 3+4 2 |
| | 5 2 | | 5 2 | | | 5+6 2 |
| | 5 2 | | 5 2 | | | 3+4 2 |
| | 6 2 | | 5 2 | | | 5+6 2 |
| | 6 2 | | 5 2 | | | 5+6 2 |
| 63.1.67 | 99133 | | 68 | | Ce1Rh2Si1 | |
| | -GM | -T | -Y | -Z | -R | -S |
| | 5 2 | 3+4 4 | 5 2 | 3+4 4 | 2+2 4 | 5+6 2 |
| | 6 2 | 3+4 4 | 6 2 | 3+4 4 | 2+2 4 | 3+4 2 |
| | 5 2 | 3+4 4 | 5 2 | 3+4 4 | 2+2 4 | 5+6 2 |
| | 6 2 | 3+4 4 | 6 2 | 3+4 4 | 2+2 4 | 3+4 2 |
| | 5 2 | 3+4 4 | 5 2 | 3+4 4 | 2+2 4 | 5+6 2 |
| | 6 2 | 3+4 4 | 6 2 | 3+4 4 | 2+2 4 | 3+4 2 |
| | 5 2 | 3+4 4 | 5 2 | 3+4 4 | 2+2 4 | 5+6 2 |
| | 6 2 | 3+4 4 | 6 2 | 3+4 4 | 2+2 4 | 3+4 2 |
| | 5 2 | 3+4 4 | 6 2 | 3+4 4 | 2+2 4 | 5+6 2 |
| | 6 2 | 3+4 4 | 5 2 | 3+4 4 | 2+2 4 | 3+4 2 |
| | 5 2 | 3+4 4 | 5 2 | 3+4 4 | 2+2 4 | 5+6 2 |
| | 6 2 | 3+4 4 | 6 2 | 3+4 4 | 2+2 4 | 3+4 2 |
| | 6 2 | 3+4 4 | 5 2 | 3+4 4 | 2+2 4 | 5+6 2 |
| | 5 2 | 3+4 4 | 6 2 | 3+4 4 | 2+2 4 | 3+4 2 |
| | 6 2 | 3+4 4 | 5 2 | 3+4 4 | 2+2 4 | 3+4 2 |
| | 6 2 | 3+4 4 | 6 2 | 3+4 4 | 2+2 4 | 5+6 2 |
| | 5 2 | 3+4 4 | 5 2 | 3+4 4 | 2+2 4 | 3+4 2 |

Continued on next column

**TABLE XXXVIII– continued**

| TQC | ICSD | NoE | SF | | | |
|---|---|---|---|---|---|---|
| | 5 2 | | 5 2 | | | 5+6 2 |
| | 6 2 | | 6 2 | | | 3+4 2 |
| | 5 2 | | 6 2 | | | 5+6 2 |
| | 6 2 | | 5 2 | | | 3+4 2 |
| | 5 2 | | 6 2 | | | 5+6 2 |
| | 6 2 | | 6 2 | | | 5+6 2 |
| | 5 2 | | 6 2 | | | 3+4 2 |
| | 6 2 | | 5 2 | | | 5+6 2 |
| | 5 2 | | 5 2 | | | 3+4 2 |
| | 5 2 | | 5 2 | | | 5+6 2 |
| | 5 2 | | 6 2 | | | 3+4 2 |
| | 6 2 | | 6 2 | | | 5+6 2 |
| | 5 2 | | 6 2 | | | 3+4 2 |
| | 5 2 | | 5 2 | | | 3+4 2 |
| | 5 2 | | 6 2 | | | 5+6 2 |
| | 6 2 | | 6 2 | | | 5+6 2 |
| | 6 2 | | 5 2 | | | 3+4 2 |
| 63.1.68 | 421810 | 212 | Cd2F1Sb5Sr5 | | | |
| | -GM | -T | -Y | -Z | -R | -S |
| | 5 2 | 3+4 4 | 5 2 | 3+4 4 | 2+2 4 | 3+4 2 |
| | 6 2 | 3+4 4 | 6 2 | 3+4 4 | 2+2 4 | 5+6 2 |
| | 5 2 | 3+4 4 | 5 2 | 3+4 4 | 2+2 4 | 3+4 2 |
| | 6 2 | 3+4 4 | 6 2 | 3+4 4 | 2+2 4 | 5+6 2 |
| | 5 2 | 3+4 4 | 6 2 | 3+4 4 | 2+2 4 | 3+4 2 |
| | 6 2 | 3+4 4 | 5 2 | 3+4 4 | 2+2 4 | 5+6 2 |
| | 5 2 | 3+4 4 | 6 2 | 3+4 4 | 2+2 4 | 3+4 2 |
| | 6 2 | 3+4 4 | 5 2 | 3+4 4 | 2+2 4 | 5+6 2 |
| | 5 2 | 3+4 4 | 5 2 | 3+4 4 | 2+2 4 | 3+4 2 |
| | 5 2 | 3+4 4 | 5 2 | 3+4 4 | 2+2 4 | 3+4 2 |
| | 5 2 | 3+4 4 | 5 2 | 3+4 4 | 2+2 4 | 3+4 2 |
| | 6 2 | 3+4 4 | 6 2 | 3+4 4 | 2+2 4 | 5+6 2 |
| | 5 2 | 3+4 4 | 5 2 | 3+4 4 | 2+2 4 | 3+4 2 |
| | 6 2 | 3+4 4 | 6 2 | 3+4 4 | 2+2 4 | 5+6 2 |
| | 6 2 | 3+4 4 | 6 2 | 3+4 4 | 2+2 4 | 3+4 2 |
| | 5 2 | 3+4 4 | 5 2 | 3+4 4 | 2+2 4 | 5+6 2 |
| | 5 2 | 3+4 4 | 6 2 | 3+4 4 | 2+2 4 | 3+4 2 |
| | 6 2 | 3+4 4 | 5 2 | 3+4 4 | 2+2 4 | 5+6 2 |
| | 5 2 | 3+4 4 | 6 2 | 3+4 4 | 2+2 4 | 3+4 2 |
| | 6 2 | 3+4 4 | 6 2 | 3+4 4 | 2+2 4 | 5+6 2 |
| | 6 2 | 3+4 4 | 5 2 | 3+4 4 | 2+2 4 | 5+6 2 |
| | 6 2 | 3+4 4 | 6 2 | 3+4 4 | 2+2 4 | 5+6 2 |
| | 5 2 | 3+4 4 | 5 2 | 3+4 4 | 2+2 4 | 3+4 2 |
| | 6 2 | 3+4 4 | 6 2 | 3+4 4 | 2+2 4 | 5+6 2 |
| | 5 2 | 3+4 4 | 5 2 | 3+4 4 | 2+2 4 | 3+4 2 |
| | 6 2 | 3+4 4 | 6 2 | 3+4 4 | 2+2 4 | 5+6 2 |
| | 6 2 | 3+4 4 | 6 2 | 3+4 4 | 2+2 4 | 5+6 2 |
| | 5 2 | 3+4 4 | 5 2 | 3+4 4 | 2+2 4 | 3+4 2 |
| | 5 2 | 3+4 4 | 6 2 | 3+4 4 | 2+2 4 | 5+6 2 |
| | 6 2 | 3+4 4 | 5 2 | 3+4 4 | 2+2 4 | 3+4 2 |
| | 5 2 | 3+4 4 | 6 2 | 3+4 4 | 2+2 4 | 5+6 2 |
| | 6 2 | 3+4 4 | 5 2 | 3+4 4 | 2+2 4 | 3+4 2 |
| | 5 2 | 3+4 4 | 6 2 | 3+4 4 | 2+2 4 | 3+4 2 |
| | 6 2 | 3+4 4 | 5 2 | 3+4 4 | 2+2 4 | 5+6 2 |
| | 5 2 | 3+4 4 | 6 2 | 3+4 4 | 2+2 4 | 5+6 2 |
| | 6 2 | 3+4 4 | 6 2 | 3+4 4 | 2+2 4 | 5+6 2 |
| | 6 2 | 3+4 4 | 6 2 | 3+4 4 | 2+2 4 | 3+4 2 |
| | 5 2 | 3+4 4 | 6 2 | 3+4 4 | 2+2 4 | 5+6 2 |
| | 6 2 | 3+4 4 | 5 2 | 3+4 4 | 2+2 4 | 3+4 2 |
| | 6 2 | 3+4 4 | 6 2 | 3+4 4 | 2+2 4 | 5+6 2 |
| | 6 2 | 3+4 4 | 6 2 | 3+4 4 | 2+2 4 | 5+6 2 |
| | 6 2 | 3+4 4 | 5 2 | 3+4 4 | 2+2 4 | 5+6 2 |

Continued on next column

**TABLE XXXVIII– continued**

| TQC | ICSD | NoE | SF | | | |
|---|---|---|---|---|---|---|
| | 5 2 | 3+4 4 | 5 2 | 3+4 4 | 2+2 4 | 3+4 2 |
| | 6 2 | 3+4 4 | 6 2 | 3+4 4 | 2+2 4 | 5+6 2 |
| | 5 2 | 3+4 4 | 5 2 | 3+4 4 | 2+2 4 | 3+4 2 |
| | 6 2 | 3+4 4 | 6 2 | 3+4 4 | 2+2 4 | 5+6 2 |
| | 5 2 | 3+4 4 | 6 2 | 3+4 4 | 2+2 4 | 3+4 2 |
| | 6 2 | 3+4 4 | 5 2 | 3+4 4 | 2+2 4 | 5+6 2 |
| | 5 2 | 3+4 4 | 6 2 | 3+4 4 | 2+2 4 | 3+4 2 |
| | 6 2 | 3+4 4 | 5 2 | 3+4 4 | 2+2 4 | 5+6 2 |
| | 5 2 | 3+4 4 | 5 2 | 3+4 4 | 2+2 4 | 3+4 2 |
| | 6 2 | 3+4 4 | 6 2 | 3+4 4 | 2+2 4 | 5+6 2 |
| | 5 2 | 3+4 4 | 6 2 | 3+4 4 | 2+2 4 | 3+4 2 |
| | 6 2 | | 5 2 | | | 5+6 2 |
| | 5 2 | | 6 2 | | | 3+4 2 |
| | 6 2 | | 5 2 | | | 5+6 2 |
| | 5 2 | | 5 2 | | | 3+4 2 |
| | 6 2 | | 6 2 | | | 5+6 2 |
| | 5 2 | | 5 2 | | | 3+4 2 |
| | 6 2 | | 6 2 | | | 5+6 2 |
| | 6 2 | | 5 2 | | | 3+4 2 |
| | 5 2 | | 6 2 | | | 5+6 2 |
| | 5 2 | | 5 2 | | | 3+4 2 |
| | 6 2 | | 6 2 | | | 5+6 2 |
| | 5 2 | | 5 2 | | | 3+4 2 |
| | 5 2 | | 6 2 | | | 5+6 2 |
| | 6 2 | | 6 2 | | | 3+4 2 |
| | 6 2 | | 5 2 | | | 5+6 2 |
| | 6 2 | | 5 2 | | | 3+4 2 |
| | 5 2 | | 6 2 | | | 5+6 2 |
| | 5 2 | | 5 2 | | | 5+6 2 |
| | 6 2 | | 6 2 | | | 3+4 2 |
| | 5 2 | | 6 2 | | | 3+4 2 |
| | 6 2 | | 5 2 | | | 5+6 2 |
| | 5 2 | | 6 2 | | | 3+4 2 |
| | 6 2 | | 5 2 | | | 5+6 2 |
| | 6 2 | | 5 2 | | | 3+4 2 |
| | 5 2 | | 6 2 | | | 5+6 2 |
| | 5 2 | | 5 2 | | | 3+4 2 |
| | 6 2 | | 5 2 | | | 3+4 2 |
| | 6 2 | | 6 2 | | | 5+6 2 |
| | 5 2 | | 5 2 | | | 5+6 2 |
| | 5 2 | | 6 2 | | | 3+4 2 |
| | 6 2 | | 6 2 | | | 5+6 2 |
| | 5 2 | | 6 2 | | | 3+4 2 |
| | 6 2 | | 6 2 | | | 5+6 2 |
| | 5 2 | | 5 2 | | | 5+6 2 |
| | 6 2 | | 5 2 | | | 3+4 2 |
| | 5 2 | | 6 2 | | | 3+4 2 |
| | 6 2 | | 5 2 | | | 3+4 2 |
| | 6 2 | | 5 2 | | | 5+6 2 |
| | 5 2 | | 6 2 | | | 5+6 2 |
| | 6 2 | | 5 2 | | | 5+6 2 |
| | 6 2 | | 5 2 | | | 3+4 2 |
| | 5 2 | | 5 2 | | | 3+4 2 |
| | 5 2 | | 6 2 | | | 3+4 2 |
| | 5 2 | | 5 2 | | | 5+6 2 |
| | 5 2 | | 6 2 | | | 3+4 2 |
| | 5 2 | | 6 2 | | | 3+4 2 |
| | 5 2 | | 6 2 | | | 5+6 2 |
| | 6 2 | | 6 2 | | | 3+4 2 |
| | 5 2 | | 5 2 | | | 5+6 2 |
| | 6 2 | | 5 2 | | | 5+6 2 |

Continued on next column

**TABLE XXXVIII– continued**

| TQC | ICSD | NoE | SF | | | |
|---|---|---|---|---|---|---|
| | 6 2 | | 5 2 | | | 5+6 2 |
| | 5 2 | | 6 2 | | | 3+4 2 |
| | 6 2 | | 5 2 | | | 5+6 2 |
| 63.1.69 | 424107 | 136 | Ca2Sn6Zn3 | | | |
| | -GM | -T | -Y | -Z | -R | -S |
| | 5 2 | 3+4 4 | 6 2 | 3+4 4 | 2+2 4 | 5+6 2 |
| | 5 2 | 3+4 4 | 5 2 | 3+4 4 | 2+2 4 | 3+4 2 |
| | 6 2 | 3+4 4 | 5 2 | 3+4 4 | 2+2 4 | 3+4 2 |
| | 6 2 | 3+4 4 | 6 2 | 3+4 4 | 2+2 4 | 5+6 2 |
| | 5 2 | 3+4 4 | 6 2 | 3+4 4 | 2+2 4 | 5+6 2 |
| | 6 2 | 3+4 4 | 5 2 | 3+4 4 | 2+2 4 | 3+4 2 |
| | 5 2 | 3+4 4 | 6 2 | 3+4 4 | 2+2 4 | 3+4 2 |
| | 5 2 | 3+4 4 | 6 2 | 3+4 4 | 2+2 4 | 5+6 2 |
| | 6 2 | 3+4 4 | 5 2 | 3+4 4 | 2+2 4 | 3+4 2 |
| | 6 2 | 3+4 4 | 6 2 | 3+4 4 | 2+2 4 | 5+6 2 |
| | 5 2 | 3+4 4 | 5 2 | 3+4 4 | 2+2 4 | 5+6 2 |
| | 5 2 | 3+4 4 | 6 2 | 3+4 4 | 2+2 4 | 3+4 2 |
| | 6 2 | 3+4 4 | 5 2 | 3+4 4 | 2+2 4 | 5+6 2 |
| | 5 2 | 3+4 4 | 5 2 | 3+4 4 | 2+2 4 | 3+4 2 |
| | 5 2 | 3+4 4 | 6 2 | 3+4 4 | 2+2 4 | 3+4 2 |
| | 5 2 | 3+4 4 | 5 2 | 3+4 4 | 2+2 4 | 5+6 2 |
| | 6 2 | 3+4 4 | 6 2 | 3+4 4 | 2+2 4 | 5+6 2 |
| | 6 2 | 3+4 4 | 6 2 | 3+4 4 | 2+2 4 | 3+4 2 |
| | 6 2 | 3+4 4 | 5 2 | 3+4 4 | 2+2 4 | 3+4 2 |
| | 5 2 | 3+4 4 | 6 2 | 3+4 4 | 2+2 4 | 5+6 2 |
| | 6 2 | 3+4 4 | 5 2 | 3+4 4 | 2+2 4 | 5+6 2 |
| | 6 2 | 3+4 4 | 5 2 | 3+4 4 | 2+2 4 | 3+4 2 |
| | 6 2 | 3+4 4 | 6 2 | 3+4 4 | 2+2 4 | 3+4 2 |
| | 5 2 | 3+4 4 | 6 2 | 3+4 4 | 2+2 4 | 5+6 2 |
| | 5 2 | 3+4 4 | 6 2 | 3+4 4 | 2+2 4 | 3+4 2 |
| | 6 2 | 3+4 4 | 5 2 | 3+4 4 | 2+2 4 | 5+6 2 |
| | 5 2 | 3+4 4 | 5 2 | 3+4 4 | 2+2 4 | 3+4 2 |
| | 6 2 | 3+4 4 | 6 2 | 3+4 4 | 2+2 4 | 5+6 2 |
| | 5 2 | 3+4 4 | 5 2 | 3+4 4 | 2+2 4 | 5+6 2 |
| | 5 2 | 3+4 4 | 6 2 | 3+4 4 | 2+2 4 | 5+6 2 |
| | 5 2 | 3+4 4 | 5 2 | 3+4 4 | 2+2 4 | 3+4 2 |
| | 6 2 | 3+4 4 | 6 2 | 3+4 4 | 2+2 4 | 3+4 2 |
| | 6 2 | 3+4 4 | 6 2 | 3+4 4 | 2+2 4 | 5+6 2 |
| | 5 2 | 3+4 4 | 5 2 | 3+4 4 | 2+2 4 | 3+4 2 |
| | 5 2 | | 5 2 | | | 5+6 2 |
| | 6 2 | | 5 2 | | | 3+4 2 |
| | 5 2 | | 6 2 | | | 3+4 2 |
| | 6 2 | | 6 2 | | | 5+6 2 |
| | 5 2 | | 5 2 | | | 3+4 2 |
| | 6 2 | | 6 2 | | | 5+6 2 |
| | 5 2 | | 5 2 | | | 3+4 2 |
| | 6 2 | | 5 2 | | | 5+6 2 |
| | 6 2 | | 5 2 | | | 5+6 2 |
| | 6 2 | | 5 2 | | | 3+4 2 |
| | 5 2 | | 6 2 | | | 3+4 2 |
| | 6 2 | | 6 2 | | | 5+6 2 |
| | 5 2 | | 5 2 | | | 3+4 2 |
| | 5 2 | | 6 2 | | | 5+6 2 |
| | 6 2 | | 6 2 | | | 5+6 2 |
| | 6 2 | | 5 2 | | | 3+4 2 |
| | 5 2 | | 6 2 | | | 5+6 2 |
| | 6 2 | | 5 2 | | | 3+4 2 |
| | 5 2 | | 6 2 | | | 5+6 2 |
| | 5 2 | | 6 2 | | | 3+4 2 |
| | 6 2 | | 5 2 | | | 3+4 2 |
| | 5 2 | | 5 2 | | | 5+6 2 |

Continued on next column

**TABLE XXXVIII– continued**

| TQC | ICSD | NoE | SF | | | |
|---|---|---|---|---|---|---|
| | 6 2 | | 5 2 | | | 3+4 2 |
| | 6 2 | | 5 2 | | | 5+6 2 |
| | 5 2 | | 6 2 | | | 5+6 2 |
| | 5 2 | | 5 2 | | | 3+4 2 |
| | 5 2 | | 5 2 | | | 5+6 2 |
| | 6 2 | | 6 2 | | | 3+4 2 |
| | 5 2 | | 6 2 | | | 3+4 2 |
| | 6 2 | | 5 2 | | | 3+4 2 |
| | 5 2 | | 6 2 | | | 5+6 2 |
| | 5 2 | | 6 2 | | | 5+6 2 |
| | 6 2 | | 6 2 | | | 5+6 2 |
| | 5 2 | | 5 2 | | | 3+4 2 |
| 63.1.70 | 96228 | 24 | Ca2In1N1 | | | |
| | -GM | -T | -Y | -Z | -R | -S |
| | 5 2 | 3+4 4 | 6 2 | 3+4 4 | 2+2 4 | 3+4 2 |
| | 6 2 | 3+4 4 | 5 2 | 3+4 4 | 2+2 4 | 5+6 2 |
| | 5 2 | 3+4 4 | 5 2 | 3+4 4 | 2+2 4 | 3+4 2 |
| | 6 2 | 3+4 4 | 6 2 | 3+4 4 | 2+2 4 | 5+6 2 |
| | 5 2 | 3+4 4 | 5 2 | 3+4 4 | 2+2 4 | 5+6 2 |
| | 6 2 | 3+4 4 | 6 2 | 3+4 4 | 2+2 4 | 3+4 2 |
| | 5 2 | | 6 2 | | | 5+6 2 |
| | 5 2 | | 5 2 | | | 3+4 2 |
| | 6 2 | | 5 2 | | | 3+4 2 |
| | 5 2 | | 5 2 | | | 5+6 2 |
| | 6 2 | | 6 2 | | | 5+6 2 |
| | 5 2 | | 5 2 | | | 3+4 2 |
| 63.1.71 | 165193 | 12 | Ca1Sn1 | | | |
| | -GM | -T | -Y | -Z | -R | -S |
| | 5 2 | 3+4 4 | 5 2 | 3+4 4 | 2+2 4 | 5+6 2 |
| | 6 2 | 3+4 4 | 6 2 | 3+4 4 | 2+2 4 | 3+4 2 |
| | 5 2 | 3+4 4 | 6 2 | 3+4 4 | 2+2 4 | 3+4 2 |
| | 6 2 | | 5 2 | | | 5+6 2 |
| | 5 2 | | 5 2 | | | 5+6 2 |
| | 6 2 | | 6 2 | | | 5+6 2 |
| 63.1.72 | 410774 | 40 | Ca1Pd1Sn2 | | | |
| | -GM | -T | -Y | -Z | -R | -S |
| | 5 2 | 3+4 4 | 6 2 | 3+4 4 | 2+2 4 | 5+6 2 |
| | 5 2 | 3+4 4 | 5 2 | 3+4 4 | 2+2 4 | 3+4 2 |
| | 6 2 | 3+4 4 | 6 2 | 3+4 4 | 2+2 4 | 5+6 2 |
| | 6 2 | 3+4 4 | 5 2 | 3+4 4 | 2+2 4 | 3+4 2 |
| | 5 2 | 3+4 4 | 5 2 | 3+4 4 | 2+2 4 | 3+4 2 |
| | 6 2 | 3+4 4 | 6 2 | 3+4 4 | 2+2 4 | 5+6 2 |
| | 5 2 | 3+4 4 | 6 2 | 3+4 4 | 2+2 4 | 5+6 2 |
| | 5 2 | 3+4 4 | 5 2 | 3+4 4 | 2+2 4 | 3+4 2 |
| | 5 2 | 3+4 4 | 6 2 | 3+4 4 | 2+2 4 | 5+6 2 |
| | 6 2 | 3+4 4 | 5 2 | 3+4 4 | 2+2 4 | 3+4 2 |
| | 6 2 | | 5 2 | | | 5+6 2 |
| | 5 2 | | 6 2 | | | 3+4 2 |
| | 6 2 | | 6 2 | | | 5+6 2 |
| | 6 2 | | 5 2 | | | 3+4 2 |
| | 5 2 | | 5 2 | | | 5+6 2 |
| | 5 2 | | 6 2 | | | 3+4 2 |
| | 6 2 | | 6 2 | | | 3+4 2 |
| | 5 2 | | 6 2 | | | 5+6 2 |
| | 5 2 | | 5 2 | | | 5+6 2 |
| | 6 2 | | 6 2 | | | 3+4 2 |
| 63.1.73 | 408880 | 36 | Ca1In2Pt1 | | | |
| | -GM | -T | -Y | -Z | -R | -S |
| | 5 2 | 3+4 4 | 6 2 | 3+4 4 | 2+2 4 | 5+6 2 |
| | 5 2 | 3+4 4 | 5 2 | 3+4 4 | 2+2 4 | 3+4 2 |
| | 6 2 | 3+4 4 | 6 2 | 3+4 4 | 2+2 4 | 5+6 2 |

Continued on next column

**TABLE XXXVIII– continued**

| TQC | ICSD | | NoE | | SF | |
|---|---|---|---|---|---|---|
| | 6 2 | 3+4 4 | 5 2 | 3+4 4 | 2+2 4 | 3+4 2 |
| | 5 2 | 3+4 4 | 5 2 | 3+4 4 | 2+2 4 | 3+4 2 |
| | 5 2 | 3+4 4 | 6 2 | 3+4 4 | 2+2 4 | 5+6 2 |
| | 6 2 | 3+4 4 | 6 2 | 3+4 4 | 2+2 4 | 5+6 2 |
| | 5 2 | 3+4 4 | 5 2 | 3+4 4 | 2+2 4 | 3+4 2 |
| | 5 2 | 3+4 4 | 6 2 | 3+4 4 | 2+2 4 | 5+6 2 |
| | 6 2 | | 5 2 | | | 3+4 2 |
| | 6 2 | | 5 2 | | | 5+6 2 |
| | 5 2 | | 6 2 | | | 3+4 2 |
| | 6 2 | | 6 2 | | | 5+6 2 |
| | 6 2 | | 5 2 | | | 3+4 2 |
| | 5 2 | | 6 2 | | | 5+6 2 |
| | 5 2 | | 5 2 | | | 3+4 2 |
| | 5 2 | | 6 2 | | | 3+4 2 |
| | 5 2 | | 5 2 | | | 5+6 2 |
| 63.1.74 | 408881 | | 36 | | Ca1In2Pd1 | |
| | -GM | -T | -Y | -Z | -R | -S |
| | 5 2 | 3+4 4 | 6 2 | 3+4 4 | 2+2 4 | 5+6 2 |
| | 5 2 | 3+4 4 | 5 2 | 3+4 4 | 2+2 4 | 3+4 2 |
| | 6 2 | 3+4 4 | 6 2 | 3+4 4 | 2+2 4 | 5+6 2 |
| | 6 2 | 3+4 4 | 5 2 | 3+4 4 | 2+2 4 | 3+4 2 |
| | 5 2 | 3+4 4 | 5 2 | 3+4 4 | 2+2 4 | 3+4 2 |
| | 5 2 | 3+4 4 | 6 2 | 3+4 4 | 2+2 4 | 5+6 2 |
| | 5 2 | 3+4 4 | 6 2 | 3+4 4 | 2+2 4 | 5+6 2 |
| | 6 2 | 3+4 4 | 5 2 | 3+4 4 | 2+2 4 | 3+4 2 |
| | 5 2 | 3+4 4 | 6 2 | 3+4 4 | 2+2 4 | 5+6 2 |
| | 6 2 | | 5 2 | | | 3+4 2 |
| | 6 2 | | 6 2 | | | 3+4 2 |
| | 5 2 | | 5 2 | | | 5+6 2 |
| | 6 2 | | 6 2 | | | 5+6 2 |
| | 6 2 | | 5 2 | | | 3+4 2 |
| | 5 2 | | 5 2 | | | 5+6 2 |
| | 5 2 | | 6 2 | | | 3+4 2 |
| | 5 2 | | 6 2 | | | 5+6 2 |
| | 6 2 | | 6 2 | | | 3+4 2 |
| 63.1.75 | 16356 | | 184 | | Ca1Fe5O7 | |
| | -GM | -T | -Y | -Z | -R | -S |
| | 5 2 | 3+4 4 | 5 2 | 3+4 4 | 2+2 4 | 3+4 2 |
| | 6 2 | 3+4 4 | 6 2 | 3+4 4 | 2+2 4 | 5+6 2 |
| | 5 2 | 3+4 4 | 5 2 | 3+4 4 | 2+2 4 | 3+4 2 |
| | 6 2 | 3+4 4 | 6 2 | 3+4 4 | 2+2 4 | 5+6 2 |
| | 5 2 | 3+4 4 | 5 2 | 3+4 4 | 2+2 4 | 3+4 2 |
| | 6 2 | 3+4 4 | 6 2 | 3+4 4 | 2+2 4 | 5+6 2 |
| | 5 2 | 3+4 4 | 5 2 | 3+4 4 | 2+2 4 | 3+4 2 |
| | 6 2 | 3+4 4 | 6 2 | 3+4 4 | 2+2 4 | 5+6 2 |
| | 5 2 | 3+4 4 | 6 2 | 3+4 4 | 2+2 4 | 3+4 2 |
| | 5 2 | 3+4 4 | 6 2 | 3+4 4 | 2+2 4 | 3+4 2 |
| | 6 2 | 3+4 4 | 5 2 | 3+4 4 | 2+2 4 | 5+6 2 |
| | 6 2 | 3+4 4 | 5 2 | 3+4 4 | 2+2 4 | 5+6 2 |
| | 5 2 | 3+4 4 | 6 2 | 3+4 4 | 2+2 4 | 3+4 2 |
| | 5 2 | 3+4 4 | 6 2 | 3+4 4 | 2+2 4 | 5+6 2 |
| | 6 2 | 3+4 4 | 5 2 | 3+4 4 | 2+2 4 | 3+4 2 |
| | 6 2 | 3+4 4 | 5 2 | 3+4 4 | 2+2 4 | 5+6 2 |
| | 5 2 | 3+4 4 | 5 2 | 3+4 4 | 2+2 4 | 5+6 2 |
| | 6 2 | 3+4 4 | 6 2 | 3+4 4 | 2+2 4 | 3+4 2 |
| | 5 2 | 3+4 4 | 5 2 | 3+4 4 | 2+2 4 | 3+4 2 |
| | 6 2 | 3+4 4 | 6 2 | 3+4 4 | 2+2 4 | 5+6 2 |
| | 5 2 | 3+4 4 | 5 2 | 3+4 4 | 2+2 4 | 3+4 2 |
| | 6 2 | 3+4 4 | 6 2 | 3+4 4 | 2+2 4 | 5+6 2 |
| | 5 2 | 3+4 4 | 5 2 | 3+4 4 | 2+2 4 | 3+4 2 |
| | 5 2 | 3+4 4 | 5 2 | 3+4 4 | 2+2 4 | 3+4 2 |

Continued on next column

**TABLE XXXVIII– continued**

| TQC | ICSD | | NoE | | SF | |
|---|---|---|---|---|---|---|
| | 6 2 | 3+4 4 | 6 2 | 3+4 4 | 2+2 4 | 5+6 2 |
| | 6 2 | 3+4 4 | 6 2 | 3+4 4 | 2+2 4 | 5+6 2 |
| | 5 2 | 3+4 4 | 6 2 | 3+4 4 | 2+2 4 | 3+4 2 |
| | 6 2 | 3+4 4 | 6 2 | 3+4 4 | 2+2 4 | 5+6 2 |
| | 5 2 | 3+4 4 | 5 2 | 3+4 4 | 2+2 4 | 5+6 2 |
| | 6 2 | 3+4 4 | 5 2 | 3+4 4 | 2+2 4 | 3+4 2 |
| | 6 2 | 3+4 4 | 6 2 | 3+4 4 | 2+2 4 | 5+6 2 |
| | 5 2 | 3+4 4 | 5 2 | 3+4 4 | 2+2 4 | 5+6 2 |
| | 5 2 | 3+4 4 | 6 2 | 3+4 4 | 2+2 4 | 3+4 2 |
| | 6 2 | 3+4 4 | 5 2 | 3+4 4 | 2+2 4 | 3+4 2 |
| | 5 2 | 3+4 4 | 5 2 | 3+4 4 | 2+2 4 | 5+6 2 |
| | 5 2 | 3+4 4 | 6 2 | 3+4 4 | 2+2 4 | 3+4 2 |
| | 5 2 | 3+4 4 | 5 2 | 3+4 4 | 2+2 4 | 5+6 2 |
| | 5 2 | 3+4 4 | 5 2 | 3+4 4 | 2+2 4 | 3+4 2 |
| | 5 2 | 3+4 4 | 6 2 | 3+4 4 | 2+2 4 | 3+4 2 |
| | 6 2 | 3+4 4 | 5 2 | 3+4 4 | 2+2 4 | 3+4 2 |
| | 6 2 | 3+4 4 | 6 2 | 3+4 4 | 2+2 4 | 5+6 2 |
| | 5 2 | 3+4 4 | 6 2 | 3+4 4 | 2+2 4 | 5+6 2 |
| | 5 2 | 3+4 4 | 5 2 | 3+4 4 | 2+2 4 | 3+4 2 |
| | 5 2 | 3+4 4 | 5 2 | 3+4 4 | 2+2 4 | 3+4 2 |
| | 6 2 | 3+4 4 | 6 2 | 3+4 4 | 2+2 4 | 3+4 2 |
| | 6 2 | 3+4 4 | 6 2 | 3+4 4 | 2+2 4 | 5+6 2 |
| | 6 2 | | 6 2 | | | 5+6 2 |
| | 6 2 | | 5 2 | | | 3+4 2 |
| | 6 2 | | 5 2 | | | 3+4 2 |
| | 5 2 | | 6 2 | | | 5+6 2 |
| | 5 2 | | 5 2 | | | 3+4 2 |
| | 6 2 | | 5 2 | | | 3+4 2 |
| | 5 2 | | 6 2 | | | 5+6 2 |
| | 5 2 | | 5 2 | | | 3+4 2 |
| | 5 2 | | 6 2 | | | 5+6 2 |
| | 5 2 | | 6 2 | | | 3+4 2 |
| | 6 2 | | 5 2 | | | 5+6 2 |
| | 6 2 | | 6 2 | | | 5+6 2 |
| | 5 2 | | 6 2 | | | 5+6 2 |
| | 6 2 | | 6 2 | | | 5+6 2 |
| | 6 2 | | 5 2 | | | 3+4 2 |
| | 6 2 | | 5 2 | | | 5+6 2 |
| | 6 2 | | 6 2 | | | 3+4 2 |
| | 6 2 | | 5 2 | | | 5+6 2 |
| | 5 2 | | 5 2 | | | 3+4 2 |
| | 5 2 | | 5 2 | | | 3+4 2 |
| | 6 2 | | 6 2 | | | 3+4 2 |
| | 6 2 | | 6 2 | | | 3+4 2 |
| | 5 2 | | 5 2 | | | 5+6 2 |
| | 5 2 | | 5 2 | | | 5+6 2 |
| | 5 2 | | 5 2 | | | 5+6 2 |
| | 5 2 | | 6 2 | | | 3+4 2 |
| | 6 2 | | 6 2 | | | 5+6 2 |
| | 6 2 | | 5 2 | | | 5+6 2 |
| | 5 2 | | 5 2 | | | 3+4 2 |
| | 5 2 | | 6 2 | | | 3+4 2 |
| | 6 2 | | 5 2 | | | 3+4 2 |
| | 5 2 | | 5 2 | | | 3+4 2 |
| | 5 2 | | 5 2 | | | 5+6 2 |
| | 6 2 | | 6 2 | | | 5+6 2 |
| | 5 2 | | 6 2 | | | 5+6 2 |
| | 6 2 | | 5 2 | | | 3+4 2 |
| | 6 2 | | 6 2 | | | 3+4 2 |
| | 5 2 | | 5 2 | | | 5+6 2 |
| | 5 2 | | 6 2 | | | 3+4 2 |

Continued on next column

**TABLE XXXVIII– continued**

| TQC | ICSD | NoE | SF | | | |
|---|---|---|---|---|---|---|
| | 6 2 | | 5 2 | | | 3+4 2 |
| | 5 2 | | 6 2 | | | 3+4 2 |
| | 5 2 | | 5 2 | | | 5+6 2 |
| | 5 2 | | 6 2 | | | 3+4 2 |
| | 5 2 | | 5 2 | | | 5+6 2 |
| | 6 2 | | 5 2 | | | 3+4 2 |
| | 6 2 | | 5 2 | | | 5+6 2 |
| 63.1.76 | 75133 | 72 | C1Fe2Si1Th1 | | | |
| | -GM | -T | -Y | -Z | -R | -S |
| | 5 2 | 3+4 4 | 5 2 | 3+4 4 | 2+2 4 | 3+4 2 |
| | 6 2 | 3+4 4 | 6 2 | 3+4 4 | 2+2 4 | 5+6 2 |
| | 5 2 | 3+4 4 | 5 2 | 3+4 4 | 2+2 4 | 3+4 2 |
| | 6 2 | 3+4 4 | 6 2 | 3+4 4 | 2+2 4 | 5+6 2 |
| | 5 2 | 3+4 4 | 5 2 | 3+4 4 | 2+2 4 | 5+6 2 |
| | 6 2 | 3+4 4 | 6 2 | 3+4 4 | 2+2 4 | 3+4 2 |
| | 5 2 | 3+4 4 | 5 2 | 3+4 4 | 2+2 4 | 3+4 2 |
| | 6 2 | 3+4 4 | 6 2 | 3+4 4 | 2+2 4 | 5+6 2 |
| | 5 2 | 3+4 4 | 5 2 | 3+4 4 | 2+2 4 | 5+6 2 |
| | 5 2 | 3+4 4 | 5 2 | 3+4 4 | 2+2 4 | 5+6 2 |
| | 5 2 | 3+4 4 | 6 2 | 3+4 4 | 2+2 4 | 3+4 2 |
| | 6 2 | 3+4 4 | 5 2 | 3+4 4 | 2+2 4 | 5+6 2 |
| | 6 2 | 3+4 4 | 6 2 | 3+4 4 | 2+2 4 | 3+4 2 |
| | 5 2 | 3+4 4 | 6 2 | 3+4 4 | 2+2 4 | 3+4 2 |
| | 6 2 | 3+4 4 | 5 2 | 3+4 4 | 2+2 4 | 5+6 2 |
| | 6 2 | 3+4 4 | 6 2 | 3+4 4 | 2+2 4 | 3+4 2 |
| | 6 2 | 3+4 4 | 6 2 | 3+4 4 | 2+2 4 | 3+4 2 |
| | 6 2 | 3+4 4 | 6 2 | 3+4 4 | 2+2 4 | 3+4 2 |
| | 6 2 | | 6 2 | | | 5+6 2 |
| | 6 2 | | 5 2 | | | 3+4 2 |
| | 5 2 | | 5 2 | | | 3+4 2 |
| | 6 2 | | 5 2 | | | 3+4 2 |
| | 5 2 | | 6 2 | | | 5+6 2 |
| | 6 2 | | 6 2 | | | 3+4 2 |
| | 5 2 | | 6 2 | | | 5+6 2 |
| | 5 2 | | 6 2 | | | 5+6 2 |
| | 6 2 | | 5 2 | | | 5+6 2 |
| | 5 2 | | 5 2 | | | 3+4 2 |
| | 6 2 | | 5 2 | | | 5+6 2 |
| | 5 2 | | 6 2 | | | 3+4 2 |
| | 5 2 | | 5 2 | | | 5+6 2 |
| | 6 2 | | 6 2 | | | 5+6 2 |
| | 5 2 | | 6 2 | | | 3+4 2 |
| | 5 2 | | 5 2 | | | 5+6 2 |
| | 5 2 | | 6 2 | | | 3+4 2 |
| | 5 2 | | 5 2 | | | 5+6 2 |
| 63.1.77 | 421809 | 212 | Ba5Cd2F1Sb5 | | | |
| | -GM | -T | -Y | -Z | -R | -S |
| | 5 2 | 3+4 4 | 5 2 | 3+4 4 | 2+2 4 | 3+4 2 |
| | 6 2 | 3+4 4 | 6 2 | 3+4 4 | 2+2 4 | 5+6 2 |
| | 5 2 | 3+4 4 | 6 2 | 3+4 4 | 2+2 4 | 3+4 2 |
| | 6 2 | 3+4 4 | 5 2 | 3+4 4 | 2+2 4 | 5+6 2 |
| | 5 2 | 3+4 4 | 6 2 | 3+4 4 | 2+2 4 | 3+4 2 |
| | 6 2 | 3+4 4 | 5 2 | 3+4 4 | 2+2 4 | 5+6 2 |
| | 5 2 | 3+4 4 | 5 2 | 3+4 4 | 2+2 4 | 3+4 2 |
| | 5 2 | 3+4 4 | 5 2 | 3+4 4 | 2+2 4 | 3+4 2 |
| | 5 2 | 3+4 4 | 5 2 | 3+4 4 | 2+2 4 | 3+4 2 |
| | 6 2 | 3+4 4 | 6 2 | 3+4 4 | 2+2 4 | 5+6 2 |
| | 5 2 | 3+4 4 | 5 2 | 3+4 4 | 2+2 4 | 3+4 2 |
| | 6 2 | 3+4 4 | 6 2 | 3+4 4 | 2+2 4 | 5+6 2 |
| | 6 2 | 3+4 4 | 6 2 | 3+4 4 | 2+2 4 | 5+6 2 |
| | 5 2 | 3+4 4 | 5 2 | 3+4 4 | 2+2 4 | 3+4 2 |

Continued on next column

**TABLE XXXVIII– continued**

| TQC | ICSD | NoE | SF | | | |
|---|---|---|---|---|---|---|
| | 5 2 | 3+4 4 | 6 2 | 3+4 4 | 2+2 4 | 3+4 2 |
| | 6 2 | 3+4 4 | 5 2 | 3+4 4 | 2+2 4 | 5+6 2 |
| | 6 2 | 3+4 4 | 6 2 | 3+4 4 | 2+2 4 | 5+6 2 |
| | 6 2 | 3+4 4 | 6 2 | 3+4 4 | 2+2 4 | 5+6 2 |
| | 5 2 | 3+4 4 | 6 2 | 3+4 4 | 2+2 4 | 3+4 2 |
| | 6 2 | 3+4 4 | 5 2 | 3+4 4 | 2+2 4 | 5+6 2 |
| | 5 2 | 3+4 4 | 5 2 | 3+4 4 | 2+2 4 | 5+6 2 |
| | 6 2 | 3+4 4 | 6 2 | 3+4 4 | 2+2 4 | 3+4 2 |
| | 6 2 | 3+4 4 | 6 2 | 3+4 4 | 2+2 4 | 5+6 2 |
| | 5 2 | 3+4 4 | 5 2 | 3+4 4 | 2+2 4 | 3+4 2 |
| | 6 2 | 3+4 4 | 6 2 | 3+4 4 | 2+2 4 | 5+6 2 |
| | 5 2 | 3+4 4 | 5 2 | 3+4 4 | 2+2 4 | 3+4 2 |
| | 5 2 | 3+4 4 | 6 2 | 3+4 4 | 2+2 4 | 5+6 2 |
| | 6 2 | 3+4 4 | 5 2 | 3+4 4 | 2+2 4 | 3+4 2 |
| | 6 2 | 3+4 4 | 6 2 | 3+4 4 | 2+2 4 | 5+6 2 |
| | 6 2 | 3+4 4 | 6 2 | 3+4 4 | 2+2 4 | 5+6 2 |
| | 5 2 | 3+4 4 | 5 2 | 3+4 4 | 2+2 4 | 5+6 2 |
| | 5 2 | 3+4 4 | 6 2 | 3+4 4 | 2+2 4 | 3+4 2 |
| | 6 2 | 3+4 4 | 6 2 | 3+4 4 | 2+2 4 | 3+4 2 |
| | 6 2 | 3+4 4 | 6 2 | 3+4 4 | 2+2 4 | 5+6 2 |
| | 5 2 | 3+4 4 | 6 2 | 3+4 4 | 2+2 4 | 5+6 2 |
| | 6 2 | 3+4 4 | 6 2 | 3+4 4 | 2+2 4 | 3+4 2 |
| | 6 2 | 3+4 4 | 5 2 | 3+4 4 | 2+2 4 | 5+6 2 |
| | 5 2 | 3+4 4 | 5 2 | 3+4 4 | 2+2 4 | 5+6 2 |
| | 6 2 | 3+4 4 | 5 2 | 3+4 4 | 2+2 4 | 3+4 2 |
| | 5 2 | 3+4 4 | 6 2 | 3+4 4 | 2+2 4 | 5+6 2 |
| | 6 2 | 3+4 4 | 5 2 | 3+4 4 | 2+2 4 | 3+4 2 |
| | 6 2 | 3+4 4 | 6 2 | 3+4 4 | 2+2 4 | 5+6 2 |
| | 5 2 | 3+4 4 | 5 2 | 3+4 4 | 2+2 4 | 3+4 2 |
| | 6 2 | 3+4 4 | 6 2 | 3+4 4 | 2+2 4 | 5+6 2 |
| | 5 2 | 3+4 4 | 5 2 | 3+4 4 | 2+2 4 | 3+4 2 |
| | 6 2 | 3+4 4 | 6 2 | 3+4 4 | 2+2 4 | 3+4 2 |
| | 5 2 | 3+4 4 | 5 2 | 3+4 4 | 2+2 4 | 5+6 2 |
| | 6 2 | 3+4 4 | 6 2 | 3+4 4 | 2+2 4 | 5+6 2 |
| | 5 2 | 3+4 4 | 6 2 | 3+4 4 | 2+2 4 | 3+4 2 |
| | 6 2 | 3+4 4 | 5 2 | 3+4 4 | 2+2 4 | 5+6 2 |
| | 5 2 | 3+4 4 | 5 2 | 3+4 4 | 2+2 4 | 3+4 2 |
| | 6 2 | 3+4 4 | 6 2 | 3+4 4 | 2+2 4 | 5+6 2 |
| | 5 2 | 3+4 4 | 6 2 | 3+4 4 | 2+2 4 | 3+4 2 |
| | 6 2 | | 5 2 | | | 5+6 2 |
| | 5 2 | | 5 2 | | | 3+4 2 |
| | 6 2 | | 6 2 | | | 5+6 2 |
| | 5 2 | | 6 2 | | | 3+4 2 |
| | 6 2 | | 5 2 | | | 5+6 2 |
| | 5 2 | | 5 2 | | | 3+4 2 |
| | 6 2 | | 6 2 | | | 5+6 2 |
| | 6 2 | | 5 2 | | | 3+4 2 |
| | 5 2 | | 6 2 | | | 5+6 2 |
| | 6 2 | | 5 2 | | | 3+4 2 |
| | 5 2 | | 6 2 | | | 5+6 2 |
| | 5 2 | | 6 2 | | | 3+4 2 |
| | 5 2 | | 5 2 | | | 5+6 2 |
| | 6 2 | | 6 2 | | | 3+4 2 |
| | 6 2 | | 5 2 | | | 3+4 2 |
| | 6 2 | | 5 2 | | | 5+6 2 |
| | 5 2 | | 6 2 | | | 3+4 2 |
| | 5 2 | | 5 2 | | | 5+6 2 |
| | 6 2 | | 6 2 | | | 5+6 2 |
| | 5 2 | | 6 2 | | | 3+4 2 |
| | 6 2 | | 5 2 | | | 5+6 2 |
| | 5 2 | | 6 2 | | | 3+4 2 |

Continued on next column

**TABLE XXXVIII– continued**

| TQC | ICSD | | NoE | SF | | |
|---|---|---|---|---|---|---|
| | 6 2 | | 5 2 | | | 5+6 2 |
| | 5 2 | | 5 2 | | | 3+4 2 |
| | 6 2 | | 6 2 | | | 5+6 2 |
| | 5 2 | | 5 2 | | | 3+4 2 |
| | 6 2 | | 6 2 | | | 5+6 2 |
| | 6 2 | | 5 2 | | | 3+4 2 |
| | 5 2 | | 5 2 | | | 5+6 2 |
| | 5 2 | | 6 2 | | | 3+4 2 |
| | 6 2 | | 6 2 | | | 5+6 2 |
| | 5 2 | | 6 2 | | | 3+4 2 |
| | 6 2 | | 5 2 | | | 5+6 2 |
| | 5 2 | | 6 2 | | | 3+4 2 |
| | 5 2 | | 5 2 | | | 5+6 2 |
| | 6 2 | | 5 2 | | | 3+4 2 |
| | 6 2 | | 5 2 | | | 3+4 2 |
| | 6 2 | | 6 2 | | | 5+6 2 |
| | 5 2 | | 5 2 | | | 3+4 2 |
| | 6 2 | | 5 2 | | | 5+6 2 |
| | 6 2 | | 6 2 | | | 3+4 2 |
| | 5 2 | | 5 2 | | | 5+6 2 |
| | 5 2 | | 6 2 | | | 3+4 2 |
| | 5 2 | | 5 2 | | | 5+6 2 |
| | 5 2 | | 6 2 | | | 3+4 2 |
| | 5 2 | | 6 2 | | | 3+4 2 |
| | 5 2 | | 6 2 | | | 5+6 2 |
| | 6 2 | | 5 2 | | | 3+4 2 |
| | 5 2 | | 6 2 | | | 5+6 2 |
| | 6 2 | | 5 2 | | | 3+4 2 |
| | 6 2 | | 5 2 | | | 5+6 2 |
| | 5 2 | | 6 2 | | | 5+6 2 |
| | 6 2 | | 5 2 | | | 5+6 2 |
| 63.1.78 | 411979 | | 100 | Ba3Pb5 | | |
| | -GM | -T | -Y | -Z | -R | -S |
| | 5 2 | 3+4 4 | 5 2 | 3+4 4 | 2+2 4 | 3+4 2 |
| | 6 2 | 3+4 4 | 6 2 | 3+4 4 | 2+2 4 | 5+6 2 |
| | 5 2 | 3+4 4 | 6 2 | 3+4 4 | 2+2 4 | 5+6 2 |
| | 5 2 | 3+4 4 | 6 2 | 3+4 4 | 2+2 4 | 5+6 2 |
| | 6 2 | 3+4 4 | 5 2 | 3+4 4 | 2+2 4 | 3+4 2 |
| | 6 2 | 3+4 4 | 5 2 | 3+4 4 | 2+2 4 | 3+4 2 |
| | 6 2 | 3+4 4 | 5 2 | 3+4 4 | 2+2 4 | 5+6 2 |
| | 5 2 | 3+4 4 | 6 2 | 3+4 4 | 2+2 4 | 3+4 2 |
| | 5 2 | 3+4 4 | 6 2 | 3+4 4 | 2+2 4 | 5+6 2 |
| | 5 2 | 3+4 4 | 5 2 | 3+4 4 | 2+2 4 | 3+4 2 |
| | 6 2 | 3+4 4 | 6 2 | 3+4 4 | 2+2 4 | 3+4 2 |
| | 6 2 | 3+4 4 | 5 2 | 3+4 4 | 2+2 4 | 5+6 2 |
| | 5 2 | 3+4 4 | 5 2 | 3+4 4 | 2+2 4 | 3+4 2 |
| | 6 2 | 3+4 4 | 6 2 | 3+4 4 | 2+2 4 | 3+4 2 |
| | 5 2 | 3+4 4 | 6 2 | 3+4 4 | 2+2 4 | 5+6 2 |
| | 6 2 | 3+4 4 | 5 2 | 3+4 4 | 2+2 4 | 5+6 2 |
| | 5 2 | 3+4 4 | 5 2 | 3+4 4 | 2+2 4 | 3+4 2 |
| | 6 2 | 3+4 4 | 6 2 | 3+4 4 | 2+2 4 | 5+6 2 |
| | 5 2 | 3+4 4 | 6 2 | 3+4 4 | 2+2 4 | 5+6 2 |
| | 5 2 | 3+4 4 | 5 2 | 3+4 4 | 2+2 4 | 5+6 2 |
| | 5 2 | 3+4 4 | 6 2 | 3+4 4 | 2+2 4 | 3+4 2 |
| | 6 2 | 3+4 4 | 5 2 | 3+4 4 | 2+2 4 | 3+4 2 |
| | 6 2 | 3+4 4 | 6 2 | 3+4 4 | 2+2 4 | 5+6 2 |
| | 6 2 | 3+4 4 | 5 2 | 3+4 4 | 2+2 4 | 3+4 2 |
| | 5 2 | 3+4 4 | 5 2 | 3+4 4 | 2+2 4 | 3+4 2 |
| | 6 2 | | 6 2 | | | 5+6 2 |
| | 5 2 | | 6 2 | | | 5+6 2 |
| | 6 2 | | 5 2 | | | 3+4 2 |

Continued on next column

**TABLE XXXVIII– continued**

| TQC | ICSD | | NoE | SF | | |
|---|---|---|---|---|---|---|
| | 6 2 | | 5 2 | | | 3+4 2 |
| | 5 2 | | 6 2 | | | 5+6 2 |
| | 5 2 | | 5 2 | | | 3+4 2 |
| | 6 2 | | 6 2 | | | 5+6 2 |
| | 5 2 | | 5 2 | | | 3+4 2 |
| | 6 2 | | 6 2 | | | 5+6 2 |
| | 5 2 | | 6 2 | | | 5+6 2 |
| | 6 2 | | 5 2 | | | 3+4 2 |
| | 5 2 | | 6 2 | | | 5+6 2 |
| | 5 2 | | 5 2 | | | 3+4 2 |
| | 6 2 | | 5 2 | | | 3+4 2 |
| | 6 2 | | 6 2 | | | 5+6 2 |
| | 5 2 | | 5 2 | | | 3+4 2 |
| | 6 2 | | 5 2 | | | 5+6 2 |
| | 5 2 | | 6 2 | | | 3+4 2 |
| | 6 2 | | 5 2 | | | 3+4 2 |
| | 5 2 | | 6 2 | | | 5+6 2 |
| | 6 2 | | 5 2 | | | 3+4 2 |
| | 6 2 | | 5 2 | | | 5+6 2 |
| | 5 2 | | 6 2 | | | 5+6 2 |
| | 5 2 | | 6 2 | | | 3+4 2 |
| | 6 2 | | 6 2 | | | 3+4 2 |
| 63.1.79 | 165184 | | 100 | Ba3Pb5 | | |
| | -GM | -T | -Y | -Z | -R | -S |
| | 5 2 | 3+4 4 | 5 2 | 3+4 4 | 2+2 4 | 5+6 2 |
| | 6 2 | 3+4 4 | 6 2 | 3+4 4 | 2+2 4 | 3+4 2 |
| | 5 2 | 3+4 4 | 6 2 | 3+4 4 | 2+2 4 | 3+4 2 |
| | 5 2 | 3+4 4 | 6 2 | 3+4 4 | 2+2 4 | 3+4 2 |
| | 6 2 | 3+4 4 | 5 2 | 3+4 4 | 2+2 4 | 5+6 2 |
| | 6 2 | 3+4 4 | 5 2 | 3+4 4 | 2+2 4 | 5+6 2 |
| | 6 2 | 3+4 4 | 5 2 | 3+4 4 | 2+2 4 | 3+4 2 |
| | 5 2 | 3+4 4 | 6 2 | 3+4 4 | 2+2 4 | 5+6 2 |
| | 5 2 | 3+4 4 | 6 2 | 3+4 4 | 2+2 4 | 3+4 2 |
| | 5 2 | 3+4 4 | 5 2 | 3+4 4 | 2+2 4 | 5+6 2 |
| | 6 2 | 3+4 4 | 6 2 | 3+4 4 | 2+2 4 | 5+6 2 |
| | 6 2 | 3+4 4 | 5 2 | 3+4 4 | 2+2 4 | 3+4 2 |
| | 5 2 | 3+4 4 | 5 2 | 3+4 4 | 2+2 4 | 5+6 2 |
| | 6 2 | 3+4 4 | 6 2 | 3+4 4 | 2+2 4 | 3+4 2 |
| | 6 2 | 3+4 4 | 6 2 | 3+4 4 | 2+2 4 | 5+6 2 |
| | 5 2 | 3+4 4 | 5 2 | 3+4 4 | 2+2 4 | 3+4 2 |
| | 5 2 | 3+4 4 | 5 2 | 3+4 4 | 2+2 4 | 5+6 2 |
| | 6 2 | 3+4 4 | 6 2 | 3+4 4 | 2+2 4 | 3+4 2 |
| | 5 2 | 3+4 4 | 6 2 | 3+4 4 | 2+2 4 | 3+4 2 |
| | 5 2 | 3+4 4 | 5 2 | 3+4 4 | 2+2 4 | 3+4 2 |
| | 5 2 | 3+4 4 | 6 2 | 3+4 4 | 2+2 4 | 5+6 2 |
| | 6 2 | 3+4 4 | 5 2 | 3+4 4 | 2+2 4 | 5+6 2 |
| | 6 2 | 3+4 4 | 6 2 | 3+4 4 | 2+2 4 | 3+4 2 |
| | 6 2 | 3+4 4 | 5 2 | 3+4 4 | 2+2 4 | 5+6 2 |
| | 5 2 | 3+4 4 | 5 2 | 3+4 4 | 2+2 4 | 3+4 2 |
| | 6 2 | | 6 2 | | | 5+6 2 |
| | 5 2 | | 6 2 | | | 3+4 2 |
| | 6 2 | | 5 2 | | | 5+6 2 |
| | 6 2 | | 5 2 | | | 5+6 2 |
| | 5 2 | | 6 2 | | | 3+4 2 |
| | 5 2 | | 5 2 | | | 5+6 2 |
| | 6 2 | | 6 2 | | | 3+4 2 |
| | 5 2 | | 5 2 | | | 5+6 2 |
| | 6 2 | | 6 2 | | | 3+4 2 |
| | 5 2 | | 6 2 | | | 3+4 2 |
| | 6 2 | | 5 2 | | | 5+6 2 |
| | 5 2 | | 6 2 | | | 3+4 2 |

Continued on next column

**TABLE XXXVIII– continued**

| TQC | ICSD | | NoE | | SF | |
|---|---|---|---|---|---|---|
| | 5 2 | | 5 2 | | | 5+6 2 |
| | 6 2 | | 5 2 | | | 5+6 2 |
| | 6 2 | | 6 2 | | | 3+4 2 |
| | 5 2 | | 5 2 | | | 5+6 2 |
| | 6 2 | | 5 2 | | | 3+4 2 |
| | 5 2 | | 6 2 | | | 5+6 2 |
| | 6 2 | | 5 2 | | | 5+6 2 |
| | 5 2 | | 6 2 | | | 3+4 2 |
| | 6 2 | | 5 2 | | | 5+6 2 |
| | 6 2 | | 5 2 | | | 3+4 2 |
| | 5 2 | | 6 2 | | | 3+4 2 |
| | 5 2 | | 6 2 | | | 5+6 2 |
| | 6 2 | | 6 2 | | | 5+6 2 |
| 63.1.80 | 710014 | | 68 | | Ba1Si6 | |
| | -GM | -T | -Y | -Z | -R | -S |
| | 5 2 | 3+4 4 | 5 2 | 3+4 4 | 2+2 4 | 3+4 2 |
| | 6 2 | 3+4 4 | 6 2 | 3+4 4 | 2+2 4 | 5+6 2 |
| | 5 2 | 3+4 4 | 6 2 | 3+4 4 | 2+2 4 | 3+4 2 |
| | 6 2 | 3+4 4 | 5 2 | 3+4 4 | 2+2 4 | 5+6 2 |
| | 5 2 | 3+4 4 | 6 2 | 3+4 4 | 2+2 4 | 3+4 2 |
| | 6 2 | 3+4 4 | 5 2 | 3+4 4 | 2+2 4 | 5+6 2 |
| | 5 2 | 3+4 4 | 6 2 | 3+4 4 | 2+2 4 | 5+6 2 |
| | 6 2 | 3+4 4 | 5 2 | 3+4 4 | 2+2 4 | 3+4 2 |
| | 5 2 | 3+4 4 | 6 2 | 3+4 4 | 2+2 4 | 3+4 2 |
| | 5 2 | 3+4 4 | 5 2 | 3+4 4 | 2+2 4 | 3+4 2 |
| | 6 2 | 3+4 4 | 5 2 | 3+4 4 | 2+2 4 | 5+6 2 |
| | 5 2 | 3+4 4 | 6 2 | 3+4 4 | 2+2 4 | 5+6 2 |
| | 6 2 | 3+4 4 | 6 2 | 3+4 4 | 2+2 4 | 3+4 2 |
| | 5 2 | 3+4 4 | 6 2 | 3+4 4 | 2+2 4 | 5+6 2 |
| | 5 2 | 3+4 4 | 5 2 | 3+4 4 | 2+2 4 | 3+4 2 |
| | 6 2 | 3+4 4 | 5 2 | 3+4 4 | 2+2 4 | 5+6 2 |
| | 6 2 | 3+4 4 | 5 2 | 3+4 4 | 2+2 4 | 5+6 2 |
| | 6 2 | | 6 2 | | | 3+4 2 |
| | 5 2 | | 6 2 | | | 5+6 2 |
| | 6 2 | | 6 2 | | | 3+4 2 |
| | 6 2 | | 5 2 | | | 5+6 2 |
| | 5 2 | | 5 2 | | | 3+4 2 |
| | 5 2 | | 5 2 | | | 5+6 2 |
| | 6 2 | | 6 2 | | | 3+4 2 |
| | 5 2 | | 6 2 | | | 3+4 2 |
| | 6 2 | | 5 2 | | | 5+6 2 |
| | 5 2 | | 6 2 | | | 5+6 2 |
| | 5 2 | | 5 2 | | | 3+4 2 |
| | 5 2 | | 5 2 | | | 3+4 2 |
| | 5 2 | | 6 2 | | | 5+6 2 |
| | 6 2 | | 6 2 | | | 3+4 2 |
| | 5 2 | | 6 2 | | | 3+4 2 |
| | 5 2 | | 5 2 | | | 3+4 2 |
| | 6 2 | | 5 2 | | | 5+6 2 |
| 63.1.81 | 245295 | | 68 | | Ba1Si6 | |
| | -GM | -T | -Y | -Z | -R | -S |
| | 5 2 | 3+4 4 | 5 2 | 3+4 4 | 2+2 4 | 3+4 2 |
| | 6 2 | 3+4 4 | 6 2 | 3+4 4 | 2+2 4 | 5+6 2 |
| | 5 2 | 3+4 4 | 6 2 | 3+4 4 | 2+2 4 | 3+4 2 |
| | 6 2 | 3+4 4 | 5 2 | 3+4 4 | 2+2 4 | 5+6 2 |
| | 5 2 | 3+4 4 | 6 2 | 3+4 4 | 2+2 4 | 5+6 2 |
| | 6 2 | 3+4 4 | 5 2 | 3+4 4 | 2+2 4 | 3+4 2 |
| | 5 2 | 3+4 4 | 6 2 | 3+4 4 | 2+2 4 | 3+4 2 |
| | 6 2 | 3+4 4 | 5 2 | 3+4 4 | 2+2 4 | 5+6 2 |
| | 5 2 | 3+4 4 | 6 2 | 3+4 4 | 2+2 4 | 5+6 2 |
| | 5 2 | 3+4 4 | 5 2 | 3+4 4 | 2+2 4 | 5+6 2 |

Continued on next column

**TABLE XXXVIII– continued**

| TQC | ICSD | | NoE | | SF | |
|---|---|---|---|---|---|---|
| | 6 2 | 3+4 4 | 5 2 | 3+4 4 | 2+2 4 | 3+4 2 |
| | 5 2 | 3+4 4 | 6 2 | 3+4 4 | 2+2 4 | 3+4 2 |
| | 6 2 | 3+4 4 | 6 2 | 3+4 4 | 2+2 4 | 5+6 2 |
| | 5 2 | 3+4 4 | 6 2 | 3+4 4 | 2+2 4 | 3+4 2 |
| | 5 2 | 3+4 4 | 5 2 | 3+4 4 | 2+2 4 | 5+6 2 |
| | 6 2 | 3+4 4 | 5 2 | 3+4 4 | 2+2 4 | 3+4 2 |
| | 6 2 | 3+4 4 | 5 2 | 3+4 4 | 2+2 4 | 3+4 2 |
| | 6 2 | | 6 2 | | | 5+6 2 |
| | 5 2 | | 6 2 | | | 3+4 2 |
| | 6 2 | | 6 2 | | | 5+6 2 |
| | 6 2 | | 5 2 | | | 3+4 2 |
| | 5 2 | | 5 2 | | | 5+6 2 |
| | 6 2 | | 5 2 | | | 3+4 2 |
| | 5 2 | | 6 2 | | | 5+6 2 |
| | 5 2 | | 6 2 | | | 3+4 2 |
| | 6 2 | | 5 2 | | | 5+6 2 |
| | 5 2 | | 6 2 | | | 3+4 2 |
| | 5 2 | | 5 2 | | | 5+6 2 |
| | 5 2 | | 5 2 | | | 5+6 2 |
| | 5 2 | | 6 2 | | | 3+4 2 |
| | 6 2 | | 6 2 | | | 5+6 2 |
| | 5 2 | | 5 2 | | | 5+6 2 |
| | 5 2 | | 6 2 | | | 5+6 2 |
| | 6 2 | | 5 2 | | | 3+4 2 |
| 63.1.82 | 78998 | | 28 | | Ba1Si1 | |
| | -GM | -T | -Y | -Z | -R | -S |
| | 5 2 | 3+4 4 | 5 2 | 3+4 4 | 2+2 4 | 3+4 2 |
| | 6 2 | 3+4 4 | 6 2 | 3+4 4 | 2+2 4 | 5+6 2 |
| | 5 2 | 3+4 4 | 6 2 | 3+4 4 | 2+2 4 | 3+4 2 |
| | 6 2 | 3+4 4 | 5 2 | 3+4 4 | 2+2 4 | 5+6 2 |
| | 5 2 | 3+4 4 | 5 2 | 3+4 4 | 2+2 4 | 3+4 2 |
| | 5 2 | 3+4 4 | 6 2 | 3+4 4 | 2+2 4 | 5+6 2 |
| | 6 2 | 3+4 4 | 6 2 | 3+4 4 | 2+2 4 | 3+4 2 |
| | 6 2 | | 5 2 | | | 5+6 2 |
| | 5 2 | | 5 2 | | | 5+6 2 |
| | 6 2 | | 6 2 | | | 3+4 2 |
| | 6 2 | | 6 2 | | | 3+4 2 |
| | 5 2 | | 5 2 | | | 5+6 2 |
| | 5 2 | | 5 2 | | | 5+6 2 |
| | 6 2 | | 6 2 | | | 5+6 2 |
| 63.1.83 | 427286 | | 116 | | Ba1Ru3Sn6 | |
| | -GM | -T | -Y | -Z | -R | -S |
| | 5 2 | 3+4 4 | 5 2 | 3+4 4 | 2+2 4 | 5+6 2 |
| | 6 2 | 3+4 4 | 6 2 | 3+4 4 | 2+2 4 | 3+4 2 |
| | 5 2 | 3+4 4 | 5 2 | 3+4 4 | 2+2 4 | 5+6 2 |
| | 6 2 | 3+4 4 | 6 2 | 3+4 4 | 2+2 4 | 3+4 2 |
| | 5 2 | 3+4 4 | 5 2 | 3+4 4 | 2+2 4 | 5+6 2 |
| | 6 2 | 3+4 4 | 6 2 | 3+4 4 | 2+2 4 | 3+4 2 |
| | 5 2 | 3+4 4 | 5 2 | 3+4 4 | 2+2 4 | 5+6 2 |
| | 6 2 | 3+4 4 | 6 2 | 3+4 4 | 2+2 4 | 3+4 2 |
| | 5 2 | 3+4 4 | 6 2 | 3+4 4 | 2+2 4 | 3+4 2 |
| | 5 2 | 3+4 4 | 5 2 | 3+4 4 | 2+2 4 | 5+6 2 |
| | 6 2 | 3+4 4 | 5 2 | 3+4 4 | 2+2 4 | 3+4 2 |
| | 6 2 | 3+4 4 | 6 2 | 3+4 4 | 2+2 4 | 5+6 2 |
| | 5 2 | 3+4 4 | 5 2 | 3+4 4 | 2+2 4 | 3+4 2 |
| | 6 2 | 3+4 4 | 6 2 | 3+4 4 | 2+2 4 | 5+6 2 |
| | 6 2 | 3+4 4 | 6 2 | 3+4 4 | 2+2 4 | 5+6 2 |
| | 5 2 | 3+4 4 | 5 2 | 3+4 4 | 2+2 4 | 3+4 2 |
| | 6 2 | 3+4 4 | 5 2 | 3+4 4 | 2+2 4 | 5+6 2 |
| | 6 2 | 3+4 4 | 6 2 | 3+4 4 | 2+2 4 | 3+4 2 |
| | 5 2 | 3+4 4 | 6 2 | 3+4 4 | 2+2 4 | 5+6 2 |

Continued on next column

**TABLE XXXVIII– continued**

| TQC | ICSD | | NoE | | SF | |
|---|---|---|---|---|---|---|
| | 5 2 | 3+4 4 | 5 2 | 3+4 4 | 2+2 4 | 3+4 2 |
| | 6 2 | 3+4 4 | 5 2 | 3+4 4 | 2+2 4 | 5+6 2 |
| | 5 2 | 3+4 4 | 5 2 | 3+4 4 | 2+2 4 | 3+4 2 |
| | 6 2 | 3+4 4 | 6 2 | 3+4 4 | 2+2 4 | 5+6 2 |
| | 6 2 | 3+4 4 | 6 2 | 3+4 4 | 2+2 4 | 3+4 2 |
| | 5 2 | 3+4 4 | 5 2 | 3+4 4 | 2+2 4 | 3+4 2 |
| | 6 2 | 3+4 4 | 5 2 | 3+4 4 | 2+2 4 | 5+6 2 |
| | 6 2 | 3+4 4 | 6 2 | 3+4 4 | 2+2 4 | 5+6 2 |
| | 5 2 | 3+4 4 | 6 2 | 3+4 4 | 2+2 4 | 5+6 2 |
| | 6 2 | 3+4 4 | 6 2 | 3+4 4 | 2+2 4 | 3+4 2 |
| | 5 2 | | 5 2 | | | 3+4 2 |
| | 5 2 | | 5 2 | | | 3+4 2 |
| | 6 2 | | 6 2 | | | 5+6 2 |
| | 6 2 | | 5 2 | | | 3+4 2 |
| | 5 2 | | 5 2 | | | 5+6 2 |
| | 5 2 | | 6 2 | | | 5+6 2 |
| | 6 2 | | 6 2 | | | 3+4 2 |
| | 5 2 | | 6 2 | | | 5+6 2 |
| | 6 2 | | 5 2 | | | 3+4 2 |
| | 5 2 | | 6 2 | | | 5+6 2 |
| | 5 2 | | 5 2 | | | 3+4 2 |
| | 6 2 | | 5 2 | | | 5+6 2 |
| | 5 2 | | 5 2 | | | 3+4 2 |
| | 5 2 | | 6 2 | | | 3+4 2 |
| | 5 2 | | 6 2 | | | 3+4 2 |
| | 6 2 | | 6 2 | | | 5+6 2 |
| | 5 2 | | 6 2 | | | 5+6 2 |
| | 6 2 | | 5 2 | | | 3+4 2 |
| | 5 2 | | 6 2 | | | 5+6 2 |
| | 5 2 | | 5 2 | | | 3+4 2 |
| | 5 2 | | 5 2 | | | 3+4 2 |
| | 6 2 | | 6 2 | | | 5+6 2 |
| | 5 2 | | 5 2 | | | 5+6 2 |
| | 5 2 | | 6 2 | | | 3+4 2 |
| | 6 2 | | 6 2 | | | 3+4 2 |
| | 6 2 | | 5 2 | | | 5+6 2 |
| | 5 2 | | 5 2 | | | 3+4 2 |
| | 6 2 | | 5 2 | | | 5+6 2 |
| | 5 2 | | 6 2 | | | 5+6 2 |
| 63.1.84 | 58664 | | 28 | | Ba1Pb1 | |
| | -GM | -T | -Y | -Z | -R | -S |
| | 5 2 | 3+4 4 | 5 2 | 3+4 4 | 2+2 4 | 3+4 2 |
| | 6 2 | 3+4 4 | 6 2 | 3+4 4 | 2+2 4 | 5+6 2 |
| | 5 2 | 3+4 4 | 5 2 | 3+4 4 | 2+2 4 | 3+4 2 |
| | 6 2 | 3+4 4 | 6 2 | 3+4 4 | 2+2 4 | 5+6 2 |
| | 5 2 | 3+4 4 | 5 2 | 3+4 4 | 2+2 4 | 3+4 2 |
| | 5 2 | 3+4 4 | 6 2 | 3+4 4 | 2+2 4 | 5+6 2 |
| | 6 2 | 3+4 4 | 5 2 | 3+4 4 | 2+2 4 | 3+4 2 |
| | 6 2 | | 6 2 | | | 5+6 2 |
| | 5 2 | | 5 2 | | | 5+6 2 |
| | 6 2 | | 6 2 | | | 3+4 2 |
| | 5 2 | | 5 2 | | | 3+4 2 |
| | 6 2 | | 6 2 | | | 5+6 2 |
| | 5 2 | | 6 2 | | | 5+6 2 |
| | 6 2 | | 5 2 | | | 5+6 2 |
| 63.1.85 | 251583 | | 68 | | Ba1Ge6 | |
| | -GM | -T | -Y | -Z | -R | -S |
| | 5 2 | 3+4 4 | 5 2 | 3+4 4 | 2+2 4 | 3+4 2 |
| | 6 2 | 3+4 4 | 6 2 | 3+4 4 | 2+2 4 | 5+6 2 |
| | 5 2 | 3+4 4 | 6 2 | 3+4 4 | 2+2 4 | 3+4 2 |
| | 6 2 | 3+4 4 | 5 2 | 3+4 4 | 2+2 4 | 5+6 2 |

Continued on next column

**TABLE XXXVIII– continued**

| TQC | ICSD | | NoE | | SF | |
|---|---|---|---|---|---|---|
| | 5 2 | 3+4 4 | 6 2 | 3+4 4 | 2+2 4 | 3+4 2 |
| | 6 2 | 3+4 4 | 5 2 | 3+4 4 | 2+2 4 | 5+6 2 |
| | 5 2 | 3+4 4 | 6 2 | 3+4 4 | 2+2 4 | 3+4 2 |
| | 6 2 | 3+4 4 | 5 2 | 3+4 4 | 2+2 4 | 5+6 2 |
| | 5 2 | 3+4 4 | 6 2 | 3+4 4 | 2+2 4 | 3+4 2 |
| | 5 2 | 3+4 4 | 5 2 | 3+4 4 | 2+2 4 | 3+4 2 |
| | 6 2 | 3+4 4 | 5 2 | 3+4 4 | 2+2 4 | 5+6 2 |
| | 6 2 | 3+4 4 | 6 2 | 3+4 4 | 2+2 4 | 5+6 2 |
| | 5 2 | 3+4 4 | 6 2 | 3+4 4 | 2+2 4 | 3+4 2 |
| | 5 2 | 3+4 4 | 6 2 | 3+4 4 | 2+2 4 | 5+6 2 |
| | 5 2 | 3+4 4 | 5 2 | 3+4 4 | 2+2 4 | 3+4 2 |
| | 6 2 | 3+4 4 | 5 2 | 3+4 4 | 2+2 4 | 5+6 2 |
| | 6 2 | 3+4 4 | 5 2 | 3+4 4 | 2+2 4 | 3+4 2 |
| | 5 2 | | 6 2 | | | 5+6 2 |
| | 6 2 | | 6 2 | | | 5+6 2 |
| | 6 2 | | 6 2 | | | 3+4 2 |
| | 6 2 | | 6 2 | | | 5+6 2 |
| | 6 2 | | 5 2 | | | 3+4 2 |
| | 5 2 | | 5 2 | | | 5+6 2 |
| | 5 2 | | 5 2 | | | 3+4 2 |
| | 5 2 | | 5 2 | | | 5+6 2 |
| | 6 2 | | 6 2 | | | 3+4 2 |
| | 5 2 | | 5 2 | | | 5+6 2 |
| | 5 2 | | 6 2 | | | 3+4 2 |
| | 5 2 | | 5 2 | | | 3+4 2 |
| | 5 2 | | 6 2 | | | 5+6 2 |
| | 6 2 | | 5 2 | | | 3+4 2 |
| | 6 2 | | 6 2 | | | 3+4 2 |
| | 5 2 | | 6 2 | | | 3+4 2 |
| | 5 2 | | 5 2 | | | 5+6 2 |
| 63.1.86 | 290596 | | 88 | | Ba1Fe2Se3 | |
| | -GM | -T | -Y | -Z | -R | -S |
| | 5 2 | 3+4 4 | 5 2 | 3+4 4 | 2+2 4 | 5+6 2 |
| | 6 2 | 3+4 4 | 6 2 | 3+4 4 | 2+2 4 | 3+4 2 |
| | 5 2 | 3+4 4 | 6 2 | 3+4 4 | 2+2 4 | 3+4 2 |
| | 6 2 | 3+4 4 | 5 2 | 3+4 4 | 2+2 4 | 5+6 2 |
| | 5 2 | 3+4 4 | 5 2 | 3+4 4 | 2+2 4 | 3+4 2 |
| | 6 2 | 3+4 4 | 6 2 | 3+4 4 | 2+2 4 | 5+6 2 |
| | 5 2 | 3+4 4 | 6 2 | 3+4 4 | 2+2 4 | 5+6 2 |
| | 6 2 | 3+4 4 | 5 2 | 3+4 4 | 2+2 4 | 3+4 2 |
| | 5 2 | 3+4 4 | 5 2 | 3+4 4 | 2+2 4 | 5+6 2 |
| | 6 2 | 3+4 4 | 6 2 | 3+4 4 | 2+2 4 | 3+4 2 |
| | 5 2 | 3+4 4 | 6 2 | 3+4 4 | 2+2 4 | 5+6 2 |
| | 6 2 | 3+4 4 | 5 2 | 3+4 4 | 2+2 4 | 3+4 2 |
| | 6 2 | 3+4 4 | 5 2 | 3+4 4 | 2+2 4 | 3+4 2 |
| | 5 2 | 3+4 4 | 6 2 | 3+4 4 | 2+2 4 | 5+6 2 |
| | 5 2 | 3+4 4 | 5 2 | 3+4 4 | 2+2 4 | 5+6 2 |
| | 6 2 | 3+4 4 | 6 2 | 3+4 4 | 2+2 4 | 3+4 2 |
| | 5 2 | 3+4 4 | 5 2 | 3+4 4 | 2+2 4 | 3+4 2 |
| | 6 2 | 3+4 4 | 6 2 | 3+4 4 | 2+2 4 | 5+6 2 |
| | 5 2 | 3+4 4 | 5 2 | 3+4 4 | 2+2 4 | 3+4 2 |
| | 6 2 | 3+4 4 | 6 2 | 3+4 4 | 2+2 4 | 5+6 2 |
| | 5 2 | 3+4 4 | 6 2 | 3+4 4 | 2+2 4 | 3+4 2 |
| | 6 2 | 3+4 4 | 6 2 | 3+4 4 | 2+2 4 | 5+6 2 |
| | 6 2 | | 5 2 | | | 5+6 2 |
| | 5 2 | | 6 2 | | | 5+6 2 |
| | 6 2 | | 5 2 | | | 3+4 2 |
| | 5 2 | | 6 2 | | | 3+4 2 |
| | 5 2 | | 6 2 | | | 5+6 2 |
| | 6 2 | | 5 2 | | | 5+6 2 |
| | 5 2 | | 5 2 | | | 3+4 2 |

Continued on next column

**TABLE XXXVIII– continued**

| TQC | ICSD | | NoE | SF | | |
|---|---|---|---|---|---|---|
| | 6 2 | | 5 2 | | | 3+4 2 |
| | 6 2 | | 6 2 | | | 3+4 2 |
| | 5 2 | | 5 2 | | | 5+6 2 |
| | 5 2 | | 5 2 | | | 3+4 2 |
| | 6 2 | | 6 2 | | | 5+6 2 |
| | 5 2 | | 6 2 | | | 3+4 2 |
| | 6 2 | | 5 2 | | | 5+6 2 |
| | 5 2 | | 5 2 | | | 3+4 2 |
| | 6 2 | | 5 2 | | | 3+4 2 |
| | 5 2 | | 6 2 | | | 5+6 2 |
| | 6 2 | | 6 2 | | | 5+6 2 |
| | 6 2 | | 6 2 | | | 5+6 2 |
| | 5 2 | | 5 2 | | | 3+4 2 |
| | 5 2 | | 5 2 | | | 3+4 2 |
| | 5 2 | | 5 2 | | | 5+6 2 |
| 63.1.87 | 290594 | | 88 | Ba1Fe2Se3 | | |
| | -GM | -T | -Y | -Z | -R | -S |
| | 5 2 | 3+4 4 | 5 2 | 3+4 4 | 2+2 4 | 5+6 2 |
| | 6 2 | 3+4 4 | 6 2 | 3+4 4 | 2+2 4 | 3+4 2 |
| | 5 2 | 3+4 4 | 6 2 | 3+4 4 | 2+2 4 | 3+4 2 |
| | 6 2 | 3+4 4 | 5 2 | 3+4 4 | 2+2 4 | 5+6 2 |
| | 5 2 | 3+4 4 | 6 2 | 3+4 4 | 2+2 4 | 3+4 2 |
| | 6 2 | 3+4 4 | 5 2 | 3+4 4 | 2+2 4 | 5+6 2 |
| | 5 2 | 3+4 4 | 6 2 | 3+4 4 | 2+2 4 | 5+6 2 |
| | 6 2 | 3+4 4 | 5 2 | 3+4 4 | 2+2 4 | 3+4 2 |
| | 5 2 | 3+4 4 | 6 2 | 3+4 4 | 2+2 4 | 3+4 2 |
| | 6 2 | 3+4 4 | 5 2 | 3+4 4 | 2+2 4 | 5+6 2 |
| | 5 2 | 3+4 4 | 5 2 | 3+4 4 | 2+2 4 | 3+4 2 |
| | 6 2 | 3+4 4 | 6 2 | 3+4 4 | 2+2 4 | 5+6 2 |
| | 6 2 | 3+4 4 | 5 2 | 3+4 4 | 2+2 4 | 5+6 2 |
| | 5 2 | 3+4 4 | 6 2 | 3+4 4 | 2+2 4 | 3+4 2 |
| | 5 2 | 3+4 4 | 5 2 | 3+4 4 | 2+2 4 | 5+6 2 |
| | 6 2 | 3+4 4 | 5 2 | 3+4 4 | 2+2 4 | 3+4 2 |
| | 5 2 | 3+4 4 | 6 2 | 3+4 4 | 2+2 4 | 3+4 2 |
| | 6 2 | 3+4 4 | 6 2 | 3+4 4 | 2+2 4 | 5+6 2 |
| | 5 2 | 3+4 4 | 5 2 | 3+4 4 | 2+2 4 | 3+4 2 |
| | 5 2 | 3+4 4 | 6 2 | 3+4 4 | 2+2 4 | 5+6 2 |
| | 6 2 | 3+4 4 | 6 2 | 3+4 4 | 2+2 4 | 5+6 2 |
| | 6 2 | 3+4 4 | 6 2 | 3+4 4 | 2+2 4 | 5+6 2 |
| | 5 2 | | 5 2 | | | 3+4 2 |
| | 6 2 | | 6 2 | | | 5+6 2 |
| | 6 2 | | 5 2 | | | 3+4 2 |
| | 5 2 | | 6 2 | | | 5+6 2 |
| | 6 2 | | 6 2 | | | 3+4 2 |
| | 6 2 | | 5 2 | | | 5+6 2 |
| | 5 2 | | 5 2 | | | 3+4 2 |
| | 5 2 | | 5 2 | | | 3+4 2 |
| | 6 2 | | 6 2 | | | 3+4 2 |
| | 5 2 | | 5 2 | | | 5+6 2 |
| | 5 2 | | 5 2 | | | 3+4 2 |
| | 6 2 | | 6 2 | | | 5+6 2 |
| | 5 2 | | 6 2 | | | 3+4 2 |
| | 6 2 | | 5 2 | | | 5+6 2 |
| | 5 2 | | 5 2 | | | 3+4 2 |
| | 6 2 | | 5 2 | | | 5+6 2 |
| | 5 2 | | 6 2 | | | 3+4 2 |
| | 6 2 | | 6 2 | | | 5+6 2 |
| | 6 2 | | 6 2 | | | 5+6 2 |
| | 5 2 | | 5 2 | | | 3+4 2 |
| | 5 2 | | 5 2 | | | 3+4 2 |
| | 5 2 | | 5 2 | | | 5+6 2 |

Continued on next column

**TABLE XXXVIII– continued**

| TQC | ICSD | | NoE | SF | | |
|---|---|---|---|---|---|---|
| 63.1.88 | 16307 | | 88 | Ba1Fe2S3 | | |
| | -GM | -T | -Y | -Z | -R | -S |
| | 5 2 | 3+4 4 | 5 2 | 3+4 4 | 2+2 4 | 5+6 2 |
| | 6 2 | 3+4 4 | 6 2 | 3+4 4 | 2+2 4 | 3+4 2 |
| | 5 2 | 3+4 4 | 5 2 | 3+4 4 | 2+2 4 | 3+4 2 |
| | 6 2 | 3+4 4 | 6 2 | 3+4 4 | 2+2 4 | 5+6 2 |
| | 5 2 | 3+4 4 | 5 2 | 3+4 4 | 2+2 4 | 3+4 2 |
| | 6 2 | 3+4 4 | 6 2 | 3+4 4 | 2+2 4 | 5+6 2 |
| | 5 2 | 3+4 4 | 6 2 | 3+4 4 | 2+2 4 | 5+6 2 |
| | 6 2 | 3+4 4 | 5 2 | 3+4 4 | 2+2 4 | 3+4 2 |
| | 5 2 | 3+4 4 | 6 2 | 3+4 4 | 2+2 4 | 5+6 2 |
| | 6 2 | 3+4 4 | 5 2 | 3+4 4 | 2+2 4 | 3+4 2 |
| | 5 2 | 3+4 4 | 6 2 | 3+4 4 | 2+2 4 | 3+4 2 |
| | 6 2 | 3+4 4 | 5 2 | 3+4 4 | 2+2 4 | 5+6 2 |
| | 6 2 | 3+4 4 | 5 2 | 3+4 4 | 2+2 4 | 5+6 2 |
| | 5 2 | 3+4 4 | 6 2 | 3+4 4 | 2+2 4 | 3+4 2 |
| | 5 2 | 3+4 4 | 5 2 | 3+4 4 | 2+2 4 | 5+6 2 |
| | 6 2 | 3+4 4 | 6 2 | 3+4 4 | 2+2 4 | 3+4 2 |
| | 5 2 | 3+4 4 | 6 2 | 3+4 4 | 2+2 4 | 3+4 2 |
| | 6 2 | 3+4 4 | 5 2 | 3+4 4 | 2+2 4 | 5+6 2 |
| | 6 2 | 3+4 4 | 5 2 | 3+4 4 | 2+2 4 | 5+6 2 |
| | 5 2 | 3+4 4 | 6 2 | 3+4 4 | 2+2 4 | 3+4 2 |
| | 5 2 | 3+4 4 | 5 2 | 3+4 4 | 2+2 4 | 5+6 2 |
| | 6 2 | 3+4 4 | 6 2 | 3+4 4 | 2+2 4 | 3+4 2 |
| | 6 2 | | 6 2 | | | 5+6 2 |
| | 6 2 | | 6 2 | | | 5+6 2 |
| | 5 2 | | 5 2 | | | 3+4 2 |
| | 5 2 | | 6 2 | | | 3+4 2 |
| | 5 2 | | 6 2 | | | 5+6 2 |
| | 6 2 | | 5 2 | | | 5+6 2 |
| | 5 2 | | 5 2 | | | 3+4 2 |
| | 6 2 | | 5 2 | | | 3+4 2 |
| | 6 2 | | 6 2 | | | 3+4 2 |
| | 5 2 | | 5 2 | | | 3+4 2 |
| | 5 2 | | 5 2 | | | 5+6 2 |
| | 6 2 | | 6 2 | | | 5+6 2 |
| | 5 2 | | 6 2 | | | 3+4 2 |
| | 6 2 | | 5 2 | | | 5+6 2 |
| | 5 2 | | 5 2 | | | 3+4 2 |
| | 5 2 | | 5 2 | | | 3+4 2 |
| | 6 2 | | 6 2 | | | 5+6 2 |
| | 6 2 | | 6 2 | | | 5+6 2 |
| | 6 2 | | 6 2 | | | 5+6 2 |
| | 5 2 | | 5 2 | | | 3+4 2 |
| | 5 2 | | 5 2 | | | 3+4 2 |
| | 5 2 | | 5 2 | | | 5+6 2 |
| 63.1.89 | 64596 | | 124 | B7Fe1Y3 | | |
| | -GM | -T | -Y | -Z | -R | -S |
| | 5 2 | 3+4 4 | 6 2 | 3+4 4 | 2+2 4 | 3+4 2 |
| | 6 2 | 3+4 4 | 5 2 | 3+4 4 | 2+2 4 | 5+6 2 |
| | 5 2 | 3+4 4 | 5 2 | 3+4 4 | 2+2 4 | 5+6 2 |
| | 5 2 | 3+4 4 | 5 2 | 3+4 4 | 2+2 4 | 5+6 2 |
| | 6 2 | 3+4 4 | 6 2 | 3+4 4 | 2+2 4 | 3+4 2 |
| | 6 2 | 3+4 4 | 6 2 | 3+4 4 | 2+2 4 | 3+4 2 |
| | 6 2 | 3+4 4 | 5 2 | 3+4 4 | 2+2 4 | 5+6 2 |
| | 5 2 | 3+4 4 | 6 2 | 3+4 4 | 2+2 4 | 3+4 2 |
| | 5 2 | 3+4 4 | 5 2 | 3+4 4 | 2+2 4 | 5+6 2 |
| | 5 2 | 3+4 4 | 6 2 | 3+4 4 | 2+2 4 | 3+4 2 |
| | 6 2 | 3+4 4 | 6 2 | 3+4 4 | 2+2 4 | 5+6 2 |
| | 6 2 | 3+4 4 | 5 2 | 3+4 4 | 2+2 4 | 3+4 2 |
| | 6 2 | 3+4 4 | 5 2 | 3+4 4 | 2+2 4 | 3+4 2 |

Continued on next column

**TABLE XXXVIII– continued**

| TQC | ICSD | | NoE | SF | | |
|---|---|---|---|---|---|---|
| | 5 2 | 3+4 4 | 5 2 | 3+4 4 | 2+2 4 | 5+6 2 |
| | 5 2 | 3+4 4 | 6 2 | 3+4 4 | 2+2 4 | 5+6 2 |
| | 5 2 | 3+4 4 | 6 2 | 3+4 4 | 2+2 4 | 3+4 2 |
| | 6 2 | 3+4 4 | 6 2 | 3+4 4 | 2+2 4 | 5+6 2 |
| | 6 2 | 3+4 4 | 5 2 | 3+4 4 | 2+2 4 | 3+4 2 |
| | 6 2 | 3+4 4 | 5 2 | 3+4 4 | 2+2 4 | 5+6 2 |
| | 5 2 | 3+4 4 | 6 2 | 3+4 4 | 2+2 4 | 5+6 2 |
| | 5 2 | 3+4 4 | 5 2 | 3+4 4 | 2+2 4 | 3+4 2 |
| | 6 2 | 3+4 4 | 5 2 | 3+4 4 | 2+2 4 | 3+4 2 |
| | 5 2 | 3+4 4 | 6 2 | 3+4 4 | 2+2 4 | 5+6 2 |
| | 6 2 | 3+4 4 | 6 2 | 3+4 4 | 2+2 4 | 3+4 2 |
| | 5 2 | 3+4 4 | 5 2 | 3+4 4 | 2+2 4 | 3+4 2 |
| | 6 2 | 3+4 4 | 6 2 | 3+4 4 | 2+2 4 | 5+6 2 |
| | 5 2 | 3+4 4 | 6 2 | 3+4 4 | 2+2 4 | 3+4 2 |
| | 5 2 | 3+4 4 | 5 2 | 3+4 4 | 2+2 4 | 3+4 2 |
| | 6 2 | 3+4 4 | 6 2 | 3+4 4 | 2+2 4 | 5+6 2 |
| | 5 2 | 3+4 4 | 5 2 | 3+4 4 | 2+2 4 | 5+6 2 |
| | 6 2 | 3+4 4 | 6 2 | 3+4 4 | 2+2 4 | 3+4 2 |
| | 5 2 | | 6 2 | | | 5+6 2 |
| | 6 2 | | 5 2 | | | 3+4 2 |
| | 6 2 | | 5 2 | | | 5+6 2 |
| | 5 2 | | 5 2 | | | 3+4 2 |
| | 5 2 | | 6 2 | | | 5+6 2 |
| | 6 2 | | 6 2 | | | 3+4 2 |
| | 5 2 | | 5 2 | | | 3+4 2 |
| | 6 2 | | 5 2 | | | 5+6 2 |
| | 5 2 | | 6 2 | | | 5+6 2 |
| | 5 2 | | 6 2 | | | 3+4 2 |
| | 6 2 | | 6 2 | | | 5+6 2 |
| | 6 2 | | 5 2 | | | 5+6 2 |
| | 6 2 | | 6 2 | | | 3+4 2 |
| | 5 2 | | 6 2 | | | 3+4 2 |
| | 5 2 | | 5 2 | | | 5+6 2 |
| | 5 2 | | 5 2 | | | 5+6 2 |
| | 6 2 | | 5 2 | | | 3+4 2 |
| | 5 2 | | 6 2 | | | 3+4 2 |
| | 6 2 | | 6 2 | | | 5+6 2 |
| | 6 2 | | 5 2 | | | 3+4 2 |
| | 5 2 | | 5 2 | | | 5+6 2 |
| | 5 2 | | 5 2 | | | 3+4 2 |
| | 5 2 | | 6 2 | | | 5+6 2 |
| | 6 2 | | 6 2 | | | 3+4 2 |
| | 6 2 | | 5 2 | | | 3+4 2 |
| | 5 2 | | 6 2 | | | 5+6 2 |
| | 5 2 | | 6 2 | | | 3+4 2 |
| | 6 2 | | 5 2 | | | 5+6 2 |
| | 5 2 | | 5 2 | | | 5+6 2 |
| | 6 2 | | 6 2 | | | 3+4 2 |
| | 5 2 | | 5 2 | | | 3+4 2 |
| 63.1.90 | 91243 | | 76 | B4Mg1Os3 | | |
| | -GM | -T | -Y | -Z | -R | -S |
| | 5 2 | 3+4 4 | 6 2 | 3+4 4 | 2+2 4 | 3+4 2 |
| | 6 2 | 3+4 4 | 5 2 | 3+4 4 | 2+2 4 | 5+6 2 |
| | 5 2 | 3+4 4 | 6 2 | 3+4 4 | 2+2 4 | 5+6 2 |
| | 5 2 | 3+4 4 | 5 2 | 3+4 4 | 2+2 4 | 3+4 2 |
| | 6 2 | 3+4 4 | 6 2 | 3+4 4 | 2+2 4 | 3+4 2 |
| | 6 2 | 3+4 4 | 5 2 | 3+4 4 | 2+2 4 | 5+6 2 |
| | 5 2 | 3+4 4 | 5 2 | 3+4 4 | 2+2 4 | 3+4 2 |
| | 6 2 | 3+4 4 | 5 2 | 3+4 4 | 2+2 4 | 5+6 2 |
| | 6 2 | 3+4 4 | 6 2 | 3+4 4 | 2+2 4 | 5+6 2 |
| | 5 2 | 3+4 4 | 5 2 | 3+4 4 | 2+2 4 | 5+6 2 |

Continued on next column

| TQC | ICSD | | NoE | SF | | |
|---|---|---|---|---|---|---|
| | 6 2 | 3+4 4 | 6 2 | 3+4 4 | 2+2 4 | 3+4 2 |
| | 5 2 | 3+4 4 | 5 2 | 3+4 4 | 2+2 4 | 3+4 2 |
| | 6 2 | 3+4 4 | 6 2 | 3+4 4 | 2+2 4 | 3+4 2 |
| | 5 2 | 3+4 4 | 5 2 | 3+4 4 | 2+2 4 | 5+6 2 |
| | 5 2 | 3+4 4 | 6 2 | 3+4 4 | 2+2 4 | 3+4 2 |
| | 6 2 | 3+4 4 | 5 2 | 3+4 4 | 2+2 4 | 5+6 2 |
| | 5 2 | 3+4 4 | 6 2 | 3+4 4 | 2+2 4 | 3+4 2 |
| | 6 2 | 3+4 4 | 6 2 | 3+4 4 | 2+2 4 | 5+6 2 |
| | 5 2 | 3+4 4 | 5 2 | 3+4 4 | 2+2 4 | 3+4 2 |
| | 6 2 | | 6 2 | | | 3+4 2 |
| | 5 2 | | 6 2 | | | 5+6 2 |
| | 6 2 | | 5 2 | | | 5+6 2 |
| | 5 2 | | 5 2 | | | 5+6 2 |
| | 5 2 | | 5 2 | | | 3+4 2 |
| | 6 2 | | 6 2 | | | 5+6 2 |
| | 6 2 | | 5 2 | | | 3+4 2 |
| | 5 2 | | 6 2 | | | 3+4 2 |
| | 6 2 | | 6 2 | | | 3+4 2 |
| | 5 2 | | 6 2 | | | 5+6 2 |
| | 5 2 | | 6 2 | | | 5+6 2 |
| | 6 2 | | 5 2 | | | 3+4 2 |
| | 6 2 | | 5 2 | | | 5+6 2 |
| | 5 2 | | 6 2 | | | 3+4 2 |
| | 5 2 | | 6 2 | | | 5+6 2 |
| | 6 2 | | 6 2 | | | 3+4 2 |
| | 5 2 | | 5 2 | | | 5+6 2 |
| | 5 2 | | 5 2 | | | 5+6 2 |
| | 6 2 | | 6 2 | | | 3+4 2 |
| 63.1.91 | 420401 | | 124 | B4Fe1O7 | | |
| | -GM | -T | -Y | -Z | -R | -S |
| | 5 2 | 3+4 4 | 6 2 | 3+4 4 | 2+2 4 | 5+6 2 |
| | 6 2 | 3+4 4 | 5 2 | 3+4 4 | 2+2 4 | 3+4 2 |
| | 6 2 | 3+4 4 | 6 2 | 3+4 4 | 2+2 4 | 3+4 2 |
| | 5 2 | 3+4 4 | 5 2 | 3+4 4 | 2+2 4 | 5+6 2 |
| | 6 2 | 3+4 4 | 5 2 | 3+4 4 | 2+2 4 | 3+4 2 |
| | 5 2 | 3+4 4 | 6 2 | 3+4 4 | 2+2 4 | 5+6 2 |
| | 5 2 | 3+4 4 | 5 2 | 3+4 4 | 2+2 4 | 5+6 2 |
| | 6 2 | 3+4 4 | 6 2 | 3+4 4 | 2+2 4 | 3+4 2 |
| | 5 2 | 3+4 4 | 5 2 | 3+4 4 | 2+2 4 | 5+6 2 |
| | 6 2 | 3+4 4 | 6 2 | 3+4 4 | 2+2 4 | 3+4 2 |
| | 5 2 | 3+4 4 | 6 2 | 3+4 4 | 2+2 4 | 3+4 2 |
| | 5 2 | 3+4 4 | 5 2 | 3+4 4 | 2+2 4 | 3+4 2 |
| | 6 2 | 3+4 4 | 6 2 | 3+4 4 | 2+2 4 | 5+6 2 |
| | 6 2 | 3+4 4 | 5 2 | 3+4 4 | 2+2 4 | 5+6 2 |
| | 6 2 | 3+4 4 | 6 2 | 3+4 4 | 2+2 4 | 3+4 2 |
| | 5 2 | 3+4 4 | 5 2 | 3+4 4 | 2+2 4 | 5+6 2 |
| | 5 2 | 3+4 4 | 5 2 | 3+4 4 | 2+2 4 | 5+6 2 |
| | 6 2 | 3+4 4 | 5 2 | 3+4 4 | 2+2 4 | 3+4 2 |
| | 5 2 | 3+4 4 | 6 2 | 3+4 4 | 2+2 4 | 5+6 2 |
| | 5 2 | 3+4 4 | 6 2 | 3+4 4 | 2+2 4 | 3+4 2 |
| | 6 2 | 3+4 4 | 6 2 | 3+4 4 | 2+2 4 | 5+6 2 |
| | 6 2 | 3+4 4 | 5 2 | 3+4 4 | 2+2 4 | 3+4 2 |
| | 6 2 | 3+4 4 | 6 2 | 3+4 4 | 2+2 4 | 5+6 2 |
| | 5 2 | 3+4 4 | 5 2 | 3+4 4 | 2+2 4 | 3+4 2 |
| | 6 2 | 3+4 4 | 5 2 | 3+4 4 | 2+2 4 | 3+4 2 |
| | 6 2 | 3+4 4 | 6 2 | 3+4 4 | 2+2 4 | 3+4 2 |
| | 5 2 | 3+4 4 | 6 2 | 3+4 4 | 2+2 4 | 5+6 2 |
| | 5 2 | 3+4 4 | 5 2 | 3+4 4 | 2+2 4 | 5+6 2 |
| | 6 2 | 3+4 4 | 5 2 | 3+4 4 | 2+2 4 | 3+4 2 |
| | 5 2 | 3+4 4 | 5 2 | 3+4 4 | 2+2 4 | 5+6 2 |
| | 5 2 | 3+4 4 | 6 2 | 3+4 4 | 2+2 4 | 3+4 2 |

Continued on next column

**TABLE XXXVIII– continued**

| TQC | ICSD | NoE | SF | | | |
|---|---|---|---|---|---|---|
| | 6 2 | | 6 2 | | | 5+6 2 |
| | 5 2 | | 6 2 | | | 3+4 2 |
| | 6 2 | | 5 2 | | | 5+6 2 |
| | 5 2 | | 5 2 | | | 5+6 2 |
| | 5 2 | | 6 2 | | | 3+4 2 |
| | 6 2 | | 6 2 | | | 5+6 2 |
| | 5 2 | | 5 2 | | | 3+4 2 |
| | 5 2 | | 6 2 | | | 5+6 2 |
| | 5 2 | | 6 2 | | | 3+4 2 |
| | 6 2 | | 5 2 | | | 3+4 2 |
| | 6 2 | | 6 2 | | | 5+6 2 |
| | 5 2 | | 6 2 | | | 3+4 2 |
| | 5 2 | | 5 2 | | | 5+6 2 |
| | 6 2 | | 5 2 | | | 3+4 2 |
| | 6 2 | | 5 2 | | | 3+4 2 |
| | 6 2 | | 5 2 | | | 5+6 2 |
| | 5 2 | | 6 2 | | | 5+6 2 |
| | 5 2 | | 5 2 | | | 5+6 2 |
| | 6 2 | | 6 2 | | | 3+4 2 |
| | 6 2 | | 5 2 | | | 3+4 2 |
| | 5 2 | | 6 2 | | | 3+4 2 |
| | 6 2 | | 5 2 | | | 5+6 2 |
| | 6 2 | | 6 2 | | | 3+4 2 |
| | 6 2 | | 6 2 | | | 5+6 2 |
| | 5 2 | | 5 2 | | | 5+6 2 |
| | 5 2 | | 5 2 | | | 5+6 2 |
| | 6 2 | | 6 2 | | | 3+4 2 |
| | 5 2 | | 5 2 | | | 3+4 2 |
| | 6 2 | | 6 2 | | | 5+6 2 |
| | 6 2 | | 6 2 | | | 5+6 2 |
| | 6 2 | | 6 2 | | | 3+4 2 |
| 63.1.92 | 418162 | 108 | B2Cr2Ir1 | | | |
| | -GM | -T | -Y | -Z | -R | -S |
| | 5 2 | 3+4 4 | 6 2 | 3+4 4 | 2+2 4 | 5+6 2 |
| | 6 2 | 3+4 4 | 5 2 | 3+4 4 | 2+2 4 | 3+4 2 |
| | 5 2 | 3+4 4 | 6 2 | 3+4 4 | 2+2 4 | 5+6 2 |
| | 6 2 | 3+4 4 | 5 2 | 3+4 4 | 2+2 4 | 3+4 2 |
| | 5 2 | 3+4 4 | 6 2 | 3+4 4 | 2+2 4 | 3+4 2 |
| | 6 2 | 3+4 4 | 5 2 | 3+4 4 | 2+2 4 | 5+6 2 |
| | 5 2 | 3+4 4 | 5 2 | 3+4 4 | 2+2 4 | 5+6 2 |
| | 5 2 | 3+4 4 | 6 2 | 3+4 4 | 2+2 4 | 3+4 2 |
| | 6 2 | 3+4 4 | 5 2 | 3+4 4 | 2+2 4 | 5+6 2 |
| | 6 2 | 3+4 4 | 6 2 | 3+4 4 | 2+2 4 | 3+4 2 |
| | 5 2 | 3+4 4 | 6 2 | 3+4 4 | 2+2 4 | 5+6 2 |
| | 6 2 | 3+4 4 | 5 2 | 3+4 4 | 2+2 4 | 3+4 2 |
| | 6 2 | 3+4 4 | 5 2 | 3+4 4 | 2+2 4 | 5+6 2 |
| | 5 2 | 3+4 4 | 6 2 | 3+4 4 | 2+2 4 | 3+4 2 |
| | 5 2 | 3+4 4 | 5 2 | 3+4 4 | 2+2 4 | 3+4 2 |
| | 5 2 | 3+4 4 | 5 2 | 3+4 4 | 2+2 4 | 5+6 2 |
| | 6 2 | 3+4 4 | 6 2 | 3+4 4 | 2+2 4 | 3+4 2 |
| | 6 2 | 3+4 4 | 6 2 | 3+4 4 | 2+2 4 | 5+6 2 |
| | 5 2 | 3+4 4 | 5 2 | 3+4 4 | 2+2 4 | 3+4 2 |
| | 6 2 | 3+4 4 | 6 2 | 3+4 4 | 2+2 4 | 3+4 2 |
| | 6 2 | 3+4 4 | 6 2 | 3+4 4 | 2+2 4 | 5+6 2 |
| | 5 2 | 3+4 4 | 5 2 | 3+4 4 | 2+2 4 | 5+6 2 |
| | 5 2 | 3+4 4 | 6 2 | 3+4 4 | 2+2 4 | 5+6 2 |
| | 6 2 | 3+4 4 | 5 2 | 3+4 4 | 2+2 4 | 3+4 2 |
| | 6 2 | 3+4 4 | 6 2 | 3+4 4 | 2+2 4 | 3+4 2 |
| | 6 2 | 3+4 4 | 5 2 | 3+4 4 | 2+2 4 | 5+6 2 |
| | 6 2 | 3+4 4 | 6 2 | 3+4 4 | 2+2 4 | 3+4 2 |
| | 5 2 | | 5 2 | | | 5+6 2 |

Continued on next column

**TABLE XXXVIII– continued**

| TQC | ICSD | NoE | SF | | | |
|---|---|---|---|---|---|---|
| | 6 2 | | 6 2 | | | 3+4 2 |
| | 5 2 | | 6 2 | | | 5+6 2 |
| | 6 2 | | 5 2 | | | 3+4 2 |
| | 5 2 | | 6 2 | | | 3+4 2 |
| | 5 2 | | 5 2 | | | 5+6 2 |
| | 6 2 | | 5 2 | | | 3+4 2 |
| | 5 2 | | 5 2 | | | 5+6 2 |
| | 5 2 | | 6 2 | | | 5+6 2 |
| | 5 2 | | 5 2 | | | 3+4 2 |
| | 6 2 | | 6 2 | | | 3+4 2 |
| | 5 2 | | 5 2 | | | 5+6 2 |
| | 5 2 | | 5 2 | | | 3+4 2 |
| | 6 2 | | 5 2 | | | 5+6 2 |
| | 6 2 | | 6 2 | | | 5+6 2 |
| | 5 2 | | 5 2 | | | 3+4 2 |
| | 5 2 | | 6 2 | | | 5+6 2 |
| | 6 2 | | 6 2 | | | 5+6 2 |
| | 5 2 | | 5 2 | | | 5+6 2 |
| | 6 2 | | 5 2 | | | 3+4 2 |
| | 5 2 | | 6 2 | | | 3+4 2 |
| | 5 2 | | 6 2 | | | 5+6 2 |
| | 5 2 | | 6 2 | | | 3+4 2 |
| | 6 2 | | 5 2 | | | 5+6 2 |
| | 5 2 | | 6 2 | | | 3+4 2 |
| | 6 2 | | 6 2 | | | 5+6 2 |
| | 5 2 | | 5 2 | | | 5+6 2 |
| 63.1.93 | 237221 | 80 | Au1Eu1In4 | | | |
| | -GM | -T | -Y | -Z | -R | -S |
| | 5 2 | 3+4 4 | 5 2 | 3+4 4 | 2+2 4 | 3+4 2 |
| | 6 2 | 3+4 4 | 6 2 | 3+4 4 | 2+2 4 | 5+6 2 |
| | 5 2 | 3+4 4 | 5 2 | 3+4 4 | 2+2 4 | 3+4 2 |
| | 6 2 | 3+4 4 | 6 2 | 3+4 4 | 2+2 4 | 5+6 2 |
| | 5 2 | 3+4 4 | 5 2 | 3+4 4 | 2+2 4 | 5+6 2 |
| | 6 2 | 3+4 4 | 6 2 | 3+4 4 | 2+2 4 | 3+4 2 |
| | 5 2 | 3+4 4 | 6 2 | 3+4 4 | 2+2 4 | 5+6 2 |
| | 6 2 | 3+4 4 | 5 2 | 3+4 4 | 2+2 4 | 3+4 2 |
| | 5 2 | 3+4 4 | 6 2 | 3+4 4 | 2+2 4 | 3+4 2 |
| | 6 2 | 3+4 4 | 5 2 | 3+4 4 | 2+2 4 | 5+6 2 |
| | 5 2 | 3+4 4 | 5 2 | 3+4 4 | 2+2 4 | 5+6 2 |
| | 5 2 | 3+4 4 | 6 2 | 3+4 4 | 2+2 4 | 3+4 2 |
| | 6 2 | 3+4 4 | 5 2 | 3+4 4 | 2+2 4 | 3+4 2 |
| | 5 2 | 3+4 4 | 6 2 | 3+4 4 | 2+2 4 | 5+6 2 |
| | 6 2 | 3+4 4 | 6 2 | 3+4 4 | 2+2 4 | 3+4 2 |
| | 5 2 | 3+4 4 | 5 2 | 3+4 4 | 2+2 4 | 5+6 2 |
| | 6 2 | 3+4 4 | 5 2 | 3+4 4 | 2+2 4 | 3+4 2 |
| | 5 2 | 3+4 4 | 6 2 | 3+4 4 | 2+2 4 | 5+6 2 |
| | 6 2 | 3+4 4 | 6 2 | 3+4 4 | 2+2 4 | 3+4 2 |
| | 5 2 | 3+4 4 | 5 2 | 3+4 4 | 2+2 4 | 5+6 2 |
| | 6 2 | | 5 2 | | | 3+4 2 |
| | 5 2 | | 6 2 | | | 3+4 2 |
| | 6 2 | | 6 2 | | | 5+6 2 |
| | 5 2 | | 5 2 | | | 3+4 2 |
| | 6 2 | | 5 2 | | | 5+6 2 |
| | 5 2 | | 6 2 | | | 5+6 2 |
| | 6 2 | | 6 2 | | | 3+4 2 |
| | 5 2 | | 6 2 | | | 3+4 2 |
| | 5 2 | | 5 2 | | | 5+6 2 |
| | 6 2 | | 6 2 | | | 5+6 2 |
| | 6 2 | | 5 2 | | | 5+6 2 |
| | 5 2 | | 5 2 | | | 3+4 2 |
| | 6 2 | | 6 2 | | | 5+6 2 |

Continued on next column

**TABLE XXXVIII– continued**

| TQC | ICSD | NoE | SF | | | |
|---|---|---|---|---|---|---|
| | 5 2 | | 5 2 | | | 5+6 2 |
| | 6 2 | | 6 2 | | | 3+4 2 |
| | 5 2 | | 6 2 | | | 3+4 2 |
| | 6 2 | | 5 2 | | | 5+6 2 |
| | 5 2 | | 6 2 | | | 5+6 2 |
| | 6 2 | | 5 2 | | | 5+6 2 |
| | 6 2 | | 5 2 | | | 3+4 2 |
| 63.1.94 | 33916 | 164 | As3Eu1Ni5 | | | |
| | -GM | -T | -Y | -Z | -R | -S |
| | 5 2 | 3+4 4 | 5 2 | 3+4 4 | 2+2 4 | 3+4 2 |
| | 6 2 | 3+4 4 | 6 2 | 3+4 4 | 2+2 4 | 5+6 2 |
| | 5 2 | 3+4 4 | 5 2 | 3+4 4 | 2+2 4 | 3+4 2 |
| | 6 2 | 3+4 4 | 6 2 | 3+4 4 | 2+2 4 | 5+6 2 |
| | 5 2 | 3+4 4 | 6 2 | 3+4 4 | 2+2 4 | 3+4 2 |
| | 6 2 | 3+4 4 | 5 2 | 3+4 4 | 2+2 4 | 5+6 2 |
| | 5 2 | 3+4 4 | 5 2 | 3+4 4 | 2+2 4 | 3+4 2 |
| | 6 2 | 3+4 4 | 6 2 | 3+4 4 | 2+2 4 | 5+6 2 |
| | 5 2 | 3+4 4 | 5 2 | 3+4 4 | 2+2 4 | 3+4 2 |
| | 5 2 | 3+4 4 | 5 2 | 3+4 4 | 2+2 4 | 3+4 2 |
| | 6 2 | 3+4 4 | 6 2 | 3+4 4 | 2+2 4 | 5+6 2 |
| | 5 2 | 3+4 4 | 6 2 | 3+4 4 | 2+2 4 | 3+4 2 |
| | 6 2 | 3+4 4 | 5 2 | 3+4 4 | 2+2 4 | 5+6 2 |
| | 6 2 | 3+4 4 | 6 2 | 3+4 4 | 2+2 4 | 5+6 2 |
| | 5 2 | 3+4 4 | 6 2 | 3+4 4 | 2+2 4 | 5+6 2 |
| | 5 2 | 3+4 4 | 5 2 | 3+4 4 | 2+2 4 | 5+6 2 |
| | 6 2 | 3+4 4 | 5 2 | 3+4 4 | 2+2 4 | 5+6 2 |
| | 5 2 | 3+4 4 | 6 2 | 3+4 4 | 2+2 4 | 3+4 2 |
| | 6 2 | 3+4 4 | 5 2 | 3+4 4 | 2+2 4 | 3+4 2 |
| | 5 2 | 3+4 4 | 5 2 | 3+4 4 | 2+2 4 | 3+4 2 |
| | 6 2 | 3+4 4 | 6 2 | 3+4 4 | 2+2 4 | 5+6 2 |
| | 5 2 | 3+4 4 | 5 2 | 3+4 4 | 2+2 4 | 3+4 2 |
| | 5 2 | 3+4 4 | 5 2 | 3+4 4 | 2+2 4 | 5+6 2 |
| | 5 2 | 3+4 4 | 5 2 | 3+4 4 | 2+2 4 | 3+4 2 |
| | 5 2 | 3+4 4 | 5 2 | 3+4 4 | 2+2 4 | 5+6 2 |
| | 6 2 | 3+4 4 | 6 2 | 3+4 4 | 2+2 4 | 3+4 2 |
| | 6 2 | 3+4 4 | 5 2 | 3+4 4 | 2+2 4 | 3+4 2 |
| | 5 2 | 3+4 4 | 6 2 | 3+4 4 | 2+2 4 | 5+6 2 |
| | 6 2 | 3+4 4 | 6 2 | 3+4 4 | 2+2 4 | 3+4 2 |
| | 6 2 | 3+4 4 | 5 2 | 3+4 4 | 2+2 4 | 5+6 2 |
| | 6 2 | 3+4 4 | 6 2 | 3+4 4 | 2+2 4 | 5+6 2 |
| | 6 2 | 3+4 4 | 6 2 | 3+4 4 | 2+2 4 | 3+4 2 |
| | 5 2 | 3+4 4 | 6 2 | 3+4 4 | 2+2 4 | 5+6 2 |
| | 5 2 | 3+4 4 | 5 2 | 3+4 4 | 2+2 4 | 5+6 2 |
| | 5 2 | 3+4 4 | 5 2 | 3+4 4 | 2+2 4 | 5+6 2 |
| | 5 2 | 3+4 4 | 6 2 | 3+4 4 | 2+2 4 | 5+6 2 |
| | 5 2 | 3+4 4 | 5 2 | 3+4 4 | 2+2 4 | 3+4 2 |
| | 6 2 | 3+4 4 | 6 2 | 3+4 4 | 2+2 4 | 3+4 2 |
| | 5 2 | 3+4 4 | 6 2 | 3+4 4 | 2+2 4 | 5+6 2 |
| | 5 2 | 3+4 4 | 5 2 | 3+4 4 | 2+2 4 | 3+4 2 |
| | 5 2 | 3+4 4 | 5 2 | 3+4 4 | 2+2 4 | 3+4 2 |
| | 6 2 | | 6 2 | | | 5+6 2 |
| | 6 2 | | 5 2 | | | 5+6 2 |
| | 6 2 | | 5 2 | | | 3+4 2 |
| | 5 2 | | 6 2 | | | 5+6 2 |
| | 6 2 | | 5 2 | | | 5+6 2 |
| | 5 2 | | 5 2 | | | 3+4 2 |
| | 6 2 | | 6 2 | | | 3+4 2 |
| | 5 2 | | 5 2 | | | 5+6 2 |
| | 5 2 | | 5 2 | | | 3+4 2 |
| | 6 2 | | 6 2 | | | 5+6 2 |
| | 6 2 | | 5 2 | | | 5+6 2 |

Continued on next column

**TABLE XXXVIII– continued**

| TQC | ICSD | NoE | SF | | | |
|---|---|---|---|---|---|---|
| | 5 2 | | 6 2 | | | 3+4 2 |
| | 6 2 | | 5 2 | | | 3+4 2 |
| | 5 2 | | 6 2 | | | 5+6 2 |
| | 6 2 | | 5 2 | | | 3+4 2 |
| | 5 2 | | 6 2 | | | 5+6 2 |
| | 5 2 | | 6 2 | | | 3+4 2 |
| | 5 2 | | 5 2 | | | 5+6 2 |
| | 5 2 | | 6 2 | | | 3+4 2 |
| | 5 2 | | 6 2 | | | 5+6 2 |
| | 6 2 | | 5 2 | | | 5+6 2 |
| | 5 2 | | 6 2 | | | 5+6 2 |
| | 6 2 | | 5 2 | | | 5+6 2 |
| | 5 2 | | 6 2 | | | 5+6 2 |
| | 6 2 | | 5 2 | | | 3+4 2 |
| | 6 2 | | 5 2 | | | 3+4 2 |
| | 5 2 | | 6 2 | | | 5+6 2 |
| | 6 2 | | 5 2 | | | 5+6 2 |
| | 5 2 | | 5 2 | | | 3+4 2 |
| | 5 2 | | 5 2 | | | 5+6 2 |
| | 5 2 | | 6 2 | | | 5+6 2 |
| | 6 2 | | 6 2 | | | 3+4 2 |
| | 5 2 | | 6 2 | | | 3+4 2 |
| | 6 2 | | 5 2 | | | 5+6 2 |
| | 6 2 | | 5 2 | | | 3+4 2 |
| | 5 2 | | 5 2 | | | 5+6 2 |
| | 5 2 | | 6 2 | | | 5+6 2 |
| | 6 2 | | 6 2 | | | 3+4 2 |
| | 5 2 | | 5 2 | | | 5+6 2 |
| | 5 2 | | 6 2 | | | 3+4 2 |
| | 6 2 | | 5 2 | | | 3+4 2 |
| 63.1.95 | 104653 | 152 | Al9Co3Y2 | | | |
| | -GM | -T | -Y | -Z | -R | -S |
| | 5 2 | 3+4 4 | 5 2 | 3+4 4 | 2+2 4 | 3+4 2 |
| | 6 2 | 3+4 4 | 5 2 | 3+4 4 | 2+2 4 | 5+6 2 |
| | 5 2 | 3+4 4 | 6 2 | 3+4 4 | 2+2 4 | 3+4 2 |
| | 6 2 | 3+4 4 | 6 2 | 3+4 4 | 2+2 4 | 5+6 2 |
| | 5 2 | 3+4 4 | 5 2 | 3+4 4 | 2+2 4 | 3+4 2 |
| | 6 2 | 3+4 4 | 6 2 | 3+4 4 | 2+2 4 | 5+6 2 |
| | 5 2 | 3+4 4 | 6 2 | 3+4 4 | 2+2 4 | 3+4 2 |
| | 6 2 | 3+4 4 | 5 2 | 3+4 4 | 2+2 4 | 5+6 2 |
| | 5 2 | 3+4 4 | 5 2 | 3+4 4 | 2+2 4 | 3+4 2 |
| | 6 2 | 3+4 4 | 6 2 | 3+4 4 | 2+2 4 | 5+6 2 |
| | 5 2 | 3+4 4 | 6 2 | 3+4 4 | 2+2 4 | 3+4 2 |
| | 6 2 | 3+4 4 | 5 2 | 3+4 4 | 2+2 4 | 5+6 2 |
| | 5 2 | 3+4 4 | 5 2 | 3+4 4 | 2+2 4 | 3+4 2 |
| | 6 2 | 3+4 4 | 6 2 | 3+4 4 | 2+2 4 | 5+6 2 |
| | 5 2 | 3+4 4 | 5 2 | 3+4 4 | 2+2 4 | 3+4 2 |
| | 6 2 | 3+4 4 | 6 2 | 3+4 4 | 2+2 4 | 5+6 2 |
| | 5 2 | 3+4 4 | 5 2 | 3+4 4 | 2+2 4 | 5+6 2 |
| | 5 2 | 3+4 4 | 6 2 | 3+4 4 | 2+2 4 | 3+4 2 |
| | 6 2 | 3+4 4 | 6 2 | 3+4 4 | 2+2 4 | 5+6 2 |
| | 5 2 | 3+4 4 | 5 2 | 3+4 4 | 2+2 4 | 3+4 2 |
| | 6 2 | 3+4 4 | 6 2 | 3+4 4 | 2+2 4 | 3+4 2 |
| | 5 2 | 3+4 4 | 5 2 | 3+4 4 | 2+2 4 | 5+6 2 |
| | 6 2 | 3+4 4 | 5 2 | 3+4 4 | 2+2 4 | 3+4 2 |
| | 5 2 | 3+4 4 | 6 2 | 3+4 4 | 2+2 4 | 5+6 2 |
| | 6 2 | 3+4 4 | 6 2 | 3+4 4 | 2+2 4 | 3+4 2 |
| | 5 2 | 3+4 4 | 5 2 | 3+4 4 | 2+2 4 | 5+6 2 |
| | 6 2 | 3+4 4 | 5 2 | 3+4 4 | 2+2 4 | 3+4 2 |
| | 5 2 | 3+4 4 | 6 2 | 3+4 4 | 2+2 4 | 5+6 2 |
| | 6 2 | 3+4 4 | 6 2 | 3+4 4 | 2+2 4 | 5+6 2 |

Continued on next column

**TABLE XXXVIII– continued**

| TQC | ICSD | | NoE | | SF | |
|---|---|---|---|---|---|---|
| | 5 2 | 3+4 4 | 6 2 | 3+4 4 | 2+2 4 | 3+4 2 |
| | 5 2 | 3+4 4 | 5 2 | 3+4 4 | 2+2 4 | 3+4 2 |
| | 6 2 | 3+4 4 | 5 2 | 3+4 4 | 2+2 4 | 5+6 2 |
| | 6 2 | 3+4 4 | 5 2 | 3+4 4 | 2+2 4 | 5+6 2 |
| | 5 2 | 3+4 4 | 6 2 | 3+4 4 | 2+2 4 | 3+4 2 |
| | 5 2 | 3+4 4 | 6 2 | 3+4 4 | 2+2 4 | 3+4 2 |
| | 6 2 | 3+4 4 | 5 2 | 3+4 4 | 2+2 4 | 3+4 2 |
| | 6 2 | 3+4 4 | 6 2 | 3+4 4 | 2+2 4 | 3+4 2 |
| | 5 2 | 3+4 4 | 5 2 | 3+4 4 | 2+2 4 | 3+4 2 |
| | 6 2 | | 6 2 | | | 5+6 2 |
| | 5 2 | | 5 2 | | | 5+6 2 |
| | 5 2 | | 5 2 | | | 5+6 2 |
| | 5 2 | | 6 2 | | | 3+4 2 |
| | 5 2 | | 5 2 | | | 3+4 2 |
| | 5 2 | | 6 2 | | | 5+6 2 |
| | 6 2 | | 5 2 | | | 3+4 2 |
| | 5 2 | | 6 2 | | | 5+6 2 |
| | 6 2 | | 5 2 | | | 5+6 2 |
| | 5 2 | | 6 2 | | | 3+4 2 |
| | 6 2 | | 5 2 | | | 5+6 2 |
| | 5 2 | | 6 2 | | | 3+4 2 |
| | 5 2 | | 5 2 | | | 3+4 2 |
| | 5 2 | | 5 2 | | | 5+6 2 |
| | 6 2 | | 5 2 | | | 5+6 2 |
| | 5 2 | | 6 2 | | | 5+6 2 |
| | 5 2 | | 5 2 | | | 3+4 2 |
| | 5 2 | | 6 2 | | | 3+4 2 |
| | 6 2 | | 5 2 | | | 5+6 2 |
| | 6 2 | | 5 2 | | | 3+4 2 |
| | 5 2 | | 6 2 | | | 3+4 2 |
| | 5 2 | | 5 2 | | | 5+6 2 |
| | 6 2 | | 5 2 | | | 3+4 2 |
| | 5 2 | | 5 2 | | | 3+4 2 |
| | 6 2 | | 5 2 | | | 3+4 2 |
| | 5 2 | | 6 2 | | | 3+4 2 |
| | 5 2 | | 6 2 | | | 5+6 2 |
| | 6 2 | | 5 2 | | | 5+6 2 |
| | 6 2 | | 6 2 | | | 3+4 2 |
| | 6 2 | | 5 2 | | | 5+6 2 |
| | 6 2 | | 5 2 | | | 5+6 2 |
| | 5 2 | | 6 2 | | | 3+4 2 |
| | 5 2 | | 5 2 | | | 3+4 2 |
| | 6 2 | | 6 2 | | | 3+4 2 |
| | 5 2 | | 6 2 | | | 3+4 2 |
| | 6 2 | | 6 2 | | | 5+6 2 |
| | 5 2 | | 5 2 | | | 3+4 2 |
| | 5 2 | | 5 2 | | | 5+6 2 |
| 63.1.96 | 58157 | | 52 | | Al6Ru1 | |
| | -GM | -T | -Y | -Z | -R | -S |
| | 5 2 | 3+4 4 | 5 2 | 3+4 4 | 2+2 4 | 3+4 2 |
| | 5 2 | 3+4 4 | 6 2 | 3+4 4 | 2+2 4 | 5+6 2 |
| | 6 2 | 3+4 4 | 5 2 | 3+4 4 | 2+2 4 | 3+4 2 |
| | 5 2 | 3+4 4 | 6 2 | 3+4 4 | 2+2 4 | 5+6 2 |
| | 5 2 | 3+4 4 | 6 2 | 3+4 4 | 2+2 4 | 5+6 2 |
| | 6 2 | 3+4 4 | 6 2 | 3+4 4 | 2+2 4 | 3+4 2 |
| | 6 2 | 3+4 4 | 5 2 | 3+4 4 | 2+2 4 | 5+6 2 |
| | 6 2 | 3+4 4 | 5 2 | 3+4 4 | 2+2 4 | 5+6 2 |
| | 6 2 | 3+4 4 | 6 2 | 3+4 4 | 2+2 4 | 3+4 2 |
| | 6 2 | 3+4 4 | 5 2 | 3+4 4 | 2+2 4 | 3+4 2 |
| | 5 2 | 3+4 4 | 5 2 | 3+4 4 | 2+2 4 | 5+6 2 |
| | 5 2 | 3+4 4 | 6 2 | 3+4 4 | 2+2 4 | 3+4 2 |

Continued on next column

| TQC | ICSD | | NoE | | SF | |
|---|---|---|---|---|---|---|
| | 5 2 | 3+4 4 | 5 2 | 3+4 4 | 2+2 4 | 3+4 2 |
| | 5 2 | | 5 2 | | | 5+6 2 |
| | 6 2 | | 5 2 | | | 3+4 2 |
| | 6 2 | | 5 2 | | | 5+6 2 |
| | 5 2 | | 6 2 | | | 5+6 2 |
| | 5 2 | | 5 2 | | | 3+4 2 |
| | 5 2 | | 6 2 | | | 3+4 2 |
| | 5 2 | | 6 2 | | | 5+6 2 |
| | 6 2 | | 6 2 | | | 3+4 2 |
| | 6 2 | | 6 2 | | | 3+4 2 |
| | 6 2 | | 5 2 | | | 5+6 2 |
| | 5 2 | | 6 2 | | | 5+6 2 |
| | 5 2 | | 6 2 | | | 5+6 2 |
| | 6 2 | | 6 2 | | | 3+4 2 |
| 63.1.97 | 370014 | | 80 | | Al2Ni1Yb1 | |
| | -GM | -T | -Y | -Z | -R | -S |
| | 5 2 | 3+4 4 | 5 2 | 3+4 4 | 2+2 4 | 5+6 2 |
| | 6 2 | 3+4 4 | 6 2 | 3+4 4 | 2+2 4 | 3+4 2 |
| | 5 2 | 3+4 4 | 5 2 | 3+4 4 | 2+2 4 | 5+6 2 |
| | 6 2 | 3+4 4 | 6 2 | 3+4 4 | 2+2 4 | 3+4 2 |
| | 5 2 | 3+4 4 | 5 2 | 3+4 4 | 2+2 4 | 5+6 2 |
| | 6 2 | 3+4 4 | 6 2 | 3+4 4 | 2+2 4 | 3+4 2 |
| | 5 2 | 3+4 4 | 5 2 | 3+4 4 | 2+2 4 | 5+6 2 |
| | 6 2 | 3+4 4 | 6 2 | 3+4 4 | 2+2 4 | 3+4 2 |
| | 5 2 | 3+4 4 | 6 2 | 3+4 4 | 2+2 4 | 3+4 2 |
| | 6 2 | 3+4 4 | 5 2 | 3+4 4 | 2+2 4 | 5+6 2 |
| | 5 2 | 3+4 4 | 6 2 | 3+4 4 | 2+2 4 | 3+4 2 |
| | 5 2 | 3+4 4 | 5 2 | 3+4 4 | 2+2 4 | 5+6 2 |
| | 6 2 | 3+4 4 | 5 2 | 3+4 4 | 2+2 4 | 3+4 2 |
| | 6 2 | 3+4 4 | 6 2 | 3+4 4 | 2+2 4 | 5+6 2 |
| | 5 2 | 3+4 4 | 6 2 | 3+4 4 | 2+2 4 | 3+4 2 |
| | 5 2 | 3+4 4 | 5 2 | 3+4 4 | 2+2 4 | 3+4 2 |
| | 5 2 | 3+4 4 | 6 2 | 3+4 4 | 2+2 4 | 5+6 2 |
| | 6 2 | 3+4 4 | 5 2 | 3+4 4 | 2+2 4 | 5+6 2 |
| | 6 2 | 3+4 4 | 6 2 | 3+4 4 | 2+2 4 | 3+4 2 |
| | 5 2 | 3+4 4 | 6 2 | 3+4 4 | 2+2 4 | 5+6 2 |
| | 6 2 | | 5 2 | | | 5+6 2 |
| | 5 2 | | 6 2 | | | 3+4 2 |
| | 5 2 | | 5 2 | | | 3+4 2 |
| | 6 2 | | 6 2 | | | 3+4 2 |
| | 5 2 | | 5 2 | | | 5+6 2 |
| | 6 2 | | 5 2 | | | 3+4 2 |
| | 6 2 | | 6 2 | | | 5+6 2 |
| | 6 2 | | 5 2 | | | 5+6 2 |
| | 5 2 | | 6 2 | | | 3+4 2 |
| | 5 2 | | 5 2 | | | 5+6 2 |
| | 5 2 | | 6 2 | | | 5+6 2 |
| | 6 2 | | 5 2 | | | 3+4 2 |
| | 5 2 | | 6 2 | | | 3+4 2 |
| | 6 2 | | 5 2 | | | 5+6 2 |
| | 6 2 | | 6 2 | | | 3+4 2 |
| | 6 2 | | 5 2 | | | 5+6 2 |
| | 5 2 | | 6 2 | | | 3+4 2 |
| | 5 2 | | 5 2 | | | 5+6 2 |
| | 5 2 | | 6 2 | | | 5+6 2 |
| | 5 2 | | 5 2 | | | 3+4 2 |
| 63.1.98 | 423614 | | 64 | | Al2Ge1La2 | |
| | -GM | -T | -Y | -Z | -R | -S |
| | 5 2 | 3+4 4 | 6 2 | 3+4 4 | 2+2 4 | 3+4 2 |
| | 6 2 | 3+4 4 | 5 2 | 3+4 4 | 2+2 4 | 5+6 2 |
| | 5 2 | 3+4 4 | 5 2 | 3+4 4 | 2+2 4 | 5+6 2 |

Continued on next column

**TABLE XXXVIII– continued**

| TQC | ICSD | | NoE | SF | | |
|---|---|---|---|---|---|---|
| | 6 2 | 3+4 4 | 6 2 | 3+4 4 | 2+2 4 | 3+4 2 |
| | 5 2 | 3+4 4 | 6 2 | 3+4 4 | 2+2 4 | 3+4 2 |
| | 6 2 | 3+4 4 | 5 2 | 3+4 4 | 2+2 4 | 5+6 2 |
| | 5 2 | 3+4 4 | 5 2 | 3+4 4 | 2+2 4 | 5+6 2 |
| | 6 2 | 3+4 4 | 6 2 | 3+4 4 | 2+2 4 | 3+4 2 |
| | 5 2 | 3+4 4 | 6 2 | 3+4 4 | 2+2 4 | 3+4 2 |
| | 5 2 | 3+4 4 | 6 2 | 3+4 4 | 2+2 4 | 5+6 2 |
| | 6 2 | 3+4 4 | 5 2 | 3+4 4 | 2+2 4 | 3+4 2 |
| | 6 2 | 3+4 4 | 5 2 | 3+4 4 | 2+2 4 | 5+6 2 |
| | 5 2 | 3+4 4 | 5 2 | 3+4 4 | 2+2 4 | 3+4 2 |
| | 6 2 | 3+4 4 | 6 2 | 3+4 4 | 2+2 4 | 5+6 2 |
| | 5 2 | 3+4 4 | 5 2 | 3+4 4 | 2+2 4 | 5+6 2 |
| | 6 2 | 3+4 4 | 6 2 | 3+4 4 | 2+2 4 | 3+4 2 |
| | 5 2 | | 6 2 | | | 5+6 2 |
| | 6 2 | | 5 2 | | | 3+4 2 |
| | 5 2 | | 5 2 | | | 3+4 2 |
| | 6 2 | | 6 2 | | | 5+6 2 |
| | 5 2 | | 5 2 | | | 3+4 2 |
| | 6 2 | | 6 2 | | | 5+6 2 |
| | 5 2 | | 5 2 | | | 5+6 2 |
| | 6 2 | | 5 2 | | | 3+4 2 |
| | 5 2 | | 5 2 | | | 3+4 2 |
| | 6 2 | | 6 2 | | | 3+4 2 |
| | 5 2 | | 6 2 | | | 5+6 2 |
| | 6 2 | | 5 2 | | | 3+4 2 |
| | 5 2 | | 6 2 | | | 5+6 2 |
| | 6 2 | | 5 2 | | | 5+6 2 |
| | 5 2 | | 5 2 | | | 5+6 2 |
| | 5 2 | | 6 2 | | | 3+4 2 |
| 63.1.99 | 99138 | | 36 | Al1Si1Y1 | | |
| | -GM | -T | -Y | -Z | -R | -S |
| | 5 2 | 3+4 4 | 6 2 | 3+4 4 | 2+2 4 | 3+4 2 |
| | 6 2 | 3+4 4 | 5 2 | 3+4 4 | 2+2 4 | 5+6 2 |
| | 5 2 | 3+4 4 | 6 2 | 3+4 4 | 2+2 4 | 5+6 2 |
| | 6 2 | 3+4 4 | 5 2 | 3+4 4 | 2+2 4 | 3+4 2 |
| | 5 2 | 3+4 4 | 6 2 | 3+4 4 | 2+2 4 | 5+6 2 |
| | 5 2 | 3+4 4 | 5 2 | 3+4 4 | 2+2 4 | 3+4 2 |
| | 6 2 | 3+4 4 | 6 2 | 3+4 4 | 2+2 4 | 5+6 2 |
| | 6 2 | 3+4 4 | 5 2 | 3+4 4 | 2+2 4 | 3+4 2 |
| | 5 2 | 3+4 4 | 5 2 | 3+4 4 | 2+2 4 | 5+6 2 |
| | 6 2 | | 6 2 | | | 3+4 2 |
| | 5 2 | | 5 2 | | | 3+4 2 |
| | 5 2 | | 6 2 | | | 5+6 2 |
| | 6 2 | | 5 2 | | | 5+6 2 |
| | 6 2 | | 6 2 | | | 3+4 2 |
| | 6 2 | | 6 2 | | | 5+6 2 |
| | 6 2 | | 5 2 | | | 5+6 2 |
| | 5 2 | | 5 2 | | | 3+4 2 |
| | 5 2 | | 6 2 | | | 5+6 2 |
| 63.1.100 | 415373 | | 64 | Al1Lu1Si1 | | |
| | -GM | -T | -Y | -Z | -R | -S |
| | 5 2 | 3+4 4 | 6 2 | 3+4 4 | 2+2 4 | 3+4 2 |
| | 6 2 | 3+4 4 | 5 2 | 3+4 4 | 2+2 4 | 5+6 2 |
| | 5 2 | 3+4 4 | 6 2 | 3+4 4 | 2+2 4 | 3+4 2 |
| | 6 2 | 3+4 4 | 5 2 | 3+4 4 | 2+2 4 | 5+6 2 |
| | 5 2 | 3+4 4 | 6 2 | 3+4 4 | 2+2 4 | 3+4 2 |
| | 5 2 | 3+4 4 | 6 2 | 3+4 4 | 2+2 4 | 5+6 2 |
| | 6 2 | 3+4 4 | 5 2 | 3+4 4 | 2+2 4 | 3+4 2 |
| | 6 2 | 3+4 4 | 5 2 | 3+4 4 | 2+2 4 | 5+6 2 |
| | 5 2 | 3+4 4 | 5 2 | 3+4 4 | 2+2 4 | 3+4 2 |
| | 6 2 | 3+4 4 | 6 2 | 3+4 4 | 2+2 4 | 5+6 2 |

Continued on next column

**TABLE XXXVIII– continued**

| TQC | ICSD | | NoE | SF | | |
|---|---|---|---|---|---|---|
| | 5 2 | 3+4 4 | 6 2 | 3+4 4 | 2+2 4 | 5+6 2 |
| | 5 2 | 3+4 4 | 5 2 | 3+4 4 | 2+2 4 | 3+4 2 |
| | 6 2 | 3+4 4 | 5 2 | 3+4 4 | 2+2 4 | 5+6 2 |
| | 6 2 | 3+4 4 | 6 2 | 3+4 4 | 2+2 4 | 3+4 2 |
| | 6 2 | 3+4 4 | 5 2 | 3+4 4 | 2+2 4 | 3+4 2 |
| | 5 2 | 3+4 4 | 6 2 | 3+4 4 | 2+2 4 | 5+6 2 |
| | 5 2 | | 5 2 | | | 5+6 2 |
| | 5 2 | | 6 2 | | | 3+4 2 |
| | 6 2 | | 5 2 | | | 3+4 2 |
| | 6 2 | | 6 2 | | | 5+6 2 |
| | 6 2 | | 5 2 | | | 3+4 2 |
| | 5 2 | | 5 2 | | | 5+6 2 |
| | 6 2 | | 5 2 | | | 3+4 2 |
| | 5 2 | | 6 2 | | | 5+6 2 |
| | 5 2 | | 6 2 | | | 5+6 2 |
| | 5 2 | | 5 2 | | | 3+4 2 |
| | 6 2 | | 6 2 | | | 3+4 2 |
| | 6 2 | | 6 2 | | | 5+6 2 |
| | 6 2 | | 6 2 | | | 3+4 2 |
| | 6 2 | | 5 2 | | | 3+4 2 |
| | 5 2 | | 5 2 | | | 5+6 2 |
| | 5 2 | | 6 2 | | | 3+4 2 |
| 63.1.101 | 76295 | | 36 | Al1Ge1Y1 | | |
| | -GM | -T | -Y | -Z | -R | -S |
| | 5 2 | 3+4 4 | 6 2 | 3+4 4 | 2+2 4 | 3+4 2 |
| | 6 2 | 3+4 4 | 5 2 | 3+4 4 | 2+2 4 | 5+6 2 |
| | 5 2 | 3+4 4 | 6 2 | 3+4 4 | 2+2 4 | 5+6 2 |
| | 6 2 | 3+4 4 | 5 2 | 3+4 4 | 2+2 4 | 3+4 2 |
| | 5 2 | 3+4 4 | 6 2 | 3+4 4 | 2+2 4 | 5+6 2 |
| | 5 2 | 3+4 4 | 5 2 | 3+4 4 | 2+2 4 | 3+4 2 |
| | 6 2 | 3+4 4 | 6 2 | 3+4 4 | 2+2 4 | 5+6 2 |
| | 6 2 | 3+4 4 | 5 2 | 3+4 4 | 2+2 4 | 3+4 2 |
| | 5 2 | 3+4 4 | 5 2 | 3+4 4 | 2+2 4 | 5+6 2 |
| | 6 2 | | 6 2 | | | 3+4 2 |
| | 5 2 | | 5 2 | | | 3+4 2 |
| | 5 2 | | 5 2 | | | 5+6 2 |
| | 6 2 | | 6 2 | | | 5+6 2 |
| | 6 2 | | 6 2 | | | 3+4 2 |
| | 6 2 | | 6 2 | | | 5+6 2 |
| | 5 2 | | 5 2 | | | 5+6 2 |
| | 6 2 | | 5 2 | | | 3+4 2 |
| | 5 2 | | 6 2 | | | 5+6 2 |
| 63.1.102 | 426322 | | 92 | Al1Fe1.5Si2 | | |
| | -GM | -T | -Y | -Z | -R | -S |
| | 5 2 | 3+4 4 | 6 2 | 3+4 4 | 2+2 4 | 5+6 2 |
| | 5 2 | 3+4 4 | 6 2 | 3+4 4 | 2+2 4 | 3+4 2 |
| | 6 2 | 3+4 4 | 5 2 | 3+4 4 | 2+2 4 | 3+4 2 |
| | 5 2 | 3+4 4 | 5 2 | 3+4 4 | 2+2 4 | 5+6 2 |
| | 6 2 | 3+4 4 | 6 2 | 3+4 4 | 2+2 4 | 3+4 2 |
| | 6 2 | 3+4 4 | 5 2 | 3+4 4 | 2+2 4 | 5+6 2 |
| | 5 2 | 3+4 4 | 6 2 | 3+4 4 | 2+2 4 | 5+6 2 |
| | 6 2 | 3+4 4 | 5 2 | 3+4 4 | 2+2 4 | 3+4 2 |
| | 5 2 | 3+4 4 | 5 2 | 3+4 4 | 2+2 4 | 3+4 2 |
| | 5 2 | 3+4 4 | 6 2 | 3+4 4 | 2+2 4 | 5+6 2 |
| | 6 2 | 3+4 4 | 5 2 | 3+4 4 | 2+2 4 | 3+4 2 |
| | 5 2 | 3+4 4 | 6 2 | 3+4 4 | 2+2 4 | 5+6 2 |
| | 6 2 | 3+4 4 | 6 2 | 3+4 4 | 2+2 4 | 5+6 2 |
| | 5 2 | 3+4 4 | 5 2 | 3+4 4 | 2+2 4 | 3+4 2 |
| | 6 2 | 3+4 4 | 6 2 | 3+4 4 | 2+2 4 | 3+4 2 |
| | 5 2 | 3+4 4 | 6 2 | 3+4 4 | 2+2 4 | 3+4 2 |
| | 5 2 | 3+4 4 | 5 2 | 3+4 4 | 2+2 4 | 5+6 2 |

Continued on next column

**TABLE XXXVIII– continued**

| TQC | ICSD | NoE | SF | | | |
|---|---|---|---|---|---|---|
| | 6 2 | 3+4 4 | 6 2 | 3+4 4 | 2+2 4 | 5+6 2 |
| | 5 2 | 3+4 4 | 5 2 | 3+4 4 | 2+2 4 | 5+6 2 |
| | 6 2 | 3+4 4 | 6 2 | 3+4 4 | 2+2 4 | 3+4 2 |
| | 5 2 | 3+4 4 | 6 2 | 3+4 4 | 2+2 4 | 3+4 2 |
| | 6 2 | 3+4 4 | 5 2 | 3+4 4 | 2+2 4 | 3+4 2 |
| | 6 2 | 3+4 4 | 5 2 | 3+4 4 | 2+2 4 | 3+4 2 |
| | 5 2 | | 5 2 | | | 5+6 2 |
| | 5 2 | | 6 2 | | | 5+6 2 |
| | 6 2 | | 6 2 | | | 3+4 2 |
| | 6 2 | | 6 2 | | | 5+6 2 |
| | 5 2 | | 6 2 | | | 3+4 2 |
| | 6 2 | | 5 2 | | | 3+4 2 |
| | 6 2 | | 6 2 | | | 3+4 2 |
| | 5 2 | | 5 2 | | | 5+6 2 |
| | 5 2 | | 5 2 | | | 5+6 2 |
| | 5 2 | | 5 2 | | | 5+6 2 |
| | 6 2 | | 6 2 | | | 3+4 2 |
| | 6 2 | | 5 2 | | | 5+6 2 |
| | 5 2 | | 5 2 | | | 5+6 2 |
| | 5 2 | | 6 2 | | | 3+4 2 |
| | 6 2 | | 6 2 | | | 5+6 2 |
| | 5 2 | | 5 2 | | | 3+4 2 |
| | 6 2 | | 5 2 | | | 5+6 2 |
| | 6 2 | | 6 2 | | | 3+4 2 |
| | 5 2 | | 5 2 | | | 5+6 2 |
| | 6 2 | | 6 2 | | | 3+4 2 |
| | 5 2 | | 5 2 | | | 5+6 2 |
| | 6 2 | | 6 2 | | | 3+4 2 |
| | 6 2 | | 5 2 | | | 3+4 2 |
| 63.1.103 | 93684 | | 220 | Ag3Cs1Se5Tb2 | | |
| | -GM | -T | -Y | -Z | -R | -S |
| | 5 2 | 3+4 4 | 6 2 | 3+4 4 | 2+2 4 | 3+4 2 |
| | 5 2 | 3+4 4 | 6 2 | 3+4 4 | 2+2 4 | 3+4 2 |
| | 6 2 | 3+4 4 | 5 2 | 3+4 4 | 2+2 4 | 5+6 2 |
| | 6 2 | 3+4 4 | 5 2 | 3+4 4 | 2+2 4 | 5+6 2 |
| | 5 2 | 3+4 4 | 6 2 | 3+4 4 | 2+2 4 | 3+4 2 |
| | 5 2 | 3+4 4 | 6 2 | 3+4 4 | 2+2 4 | 5+6 2 |
| | 6 2 | 3+4 4 | 5 2 | 3+4 4 | 2+2 4 | 3+4 2 |
| | 6 2 | 3+4 4 | 5 2 | 3+4 4 | 2+2 4 | 5+6 2 |
| | 5 2 | 3+4 4 | 5 2 | 3+4 4 | 2+2 4 | 3+4 2 |
| | 6 2 | 3+4 4 | 6 2 | 3+4 4 | 2+2 4 | 5+6 2 |
| | 5 2 | 3+4 4 | 6 2 | 3+4 4 | 2+2 4 | 3+4 2 |
| | 5 2 | 3+4 4 | 6 2 | 3+4 4 | 2+2 4 | 5+6 2 |
| | 6 2 | 3+4 4 | 5 2 | 3+4 4 | 2+2 4 | 3+4 2 |
| | 6 2 | 3+4 4 | 5 2 | 3+4 4 | 2+2 4 | 5+6 2 |
| | 5 2 | 3+4 4 | 6 2 | 3+4 4 | 2+2 4 | 3+4 2 |
| | 5 2 | 3+4 4 | 6 2 | 3+4 4 | 2+2 4 | 5+6 2 |
| | 6 2 | 3+4 4 | 5 2 | 3+4 4 | 2+2 4 | 3+4 2 |
| | 6 2 | 3+4 4 | 5 2 | 3+4 4 | 2+2 4 | 5+6 2 |
| | 5 2 | 3+4 4 | 6 2 | 3+4 4 | 2+2 4 | 3+4 2 |
| | 5 2 | 3+4 4 | 6 2 | 3+4 4 | 2+2 4 | 5+6 2 |
| | 6 2 | 3+4 4 | 5 2 | 3+4 4 | 2+2 4 | 3+4 2 |
| | 6 2 | 3+4 4 | 5 2 | 3+4 4 | 2+2 4 | 5+6 2 |
| | 5 2 | 3+4 4 | 5 2 | 3+4 4 | 2+2 4 | 3+4 2 |
| | 6 2 | 3+4 4 | 5 2 | 3+4 4 | 2+2 4 | 5+6 2 |
| | 6 2 | 3+4 4 | 6 2 | 3+4 4 | 2+2 4 | 3+4 2 |
| | 5 2 | 3+4 4 | 6 2 | 3+4 4 | 2+2 4 | 5+6 2 |
| | 6 2 | 3+4 4 | 5 2 | 3+4 4 | 2+2 4 | 5+6 2 |
| | 5 2 | 3+4 4 | 6 2 | 3+4 4 | 2+2 4 | 3+4 2 |
| | 5 2 | 3+4 4 | 6 2 | 3+4 4 | 2+2 4 | 5+6 2 |
| | 6 2 | 3+4 4 | 5 2 | 3+4 4 | 2+2 4 | 3+4 2 |

Continued on next column

**TABLE XXXVIII– continued**

| TQC | ICSD | NoE | SF | | | |
|---|---|---|---|---|---|---|
| | 5 2 | 3+4 4 | 6 2 | 3+4 4 | 2+2 4 | 5+6 2 |
| | 6 2 | 3+4 4 | 5 2 | 3+4 4 | 2+2 4 | 3+4 2 |
| | 5 2 | 3+4 4 | 6 2 | 3+4 4 | 2+2 4 | 3+4 2 |
| | 6 2 | 3+4 4 | 5 2 | 3+4 4 | 2+2 4 | 5+6 2 |
| | 6 2 | 3+4 4 | 6 2 | 3+4 4 | 2+2 4 | 3+4 2 |
| | 5 2 | 3+4 4 | 5 2 | 3+4 4 | 2+2 4 | 3+4 2 |
| | 6 2 | 3+4 4 | 6 2 | 3+4 4 | 2+2 4 | 5+6 2 |
| | 5 2 | 3+4 4 | 5 2 | 3+4 4 | 2+2 4 | 5+6 2 |
| | 6 2 | 3+4 4 | 6 2 | 3+4 4 | 2+2 4 | 5+6 2 |
| | 6 2 | 3+4 4 | 5 2 | 3+4 4 | 2+2 4 | 3+4 2 |
| | 5 2 | 3+4 4 | 6 2 | 3+4 4 | 2+2 4 | 3+4 2 |
| | 6 2 | 3+4 4 | 5 2 | 3+4 4 | 2+2 4 | 5+6 2 |
| | 5 2 | 3+4 4 | 6 2 | 3+4 4 | 2+2 4 | 3+4 2 |
| | 6 2 | 3+4 4 | 6 2 | 3+4 4 | 2+2 4 | 5+6 2 |
| | 5 2 | 3+4 4 | 6 2 | 3+4 4 | 2+2 4 | 3+4 2 |
| | 6 2 | 3+4 4 | 5 2 | 3+4 4 | 2+2 4 | 5+6 2 |
| | 5 2 | 3+4 4 | 5 2 | 3+4 4 | 2+2 4 | 5+6 2 |
| | 5 2 | 3+4 4 | 6 2 | 3+4 4 | 2+2 4 | 3+4 2 |
| | 5 2 | 3+4 4 | 5 2 | 3+4 4 | 2+2 4 | 5+6 2 |
| | 6 2 | 3+4 4 | 5 2 | 3+4 4 | 2+2 4 | 3+4 2 |
| | 5 2 | 3+4 4 | 5 2 | 3+4 4 | 2+2 4 | 5+6 2 |
| | 5 2 | 3+4 4 | 5 2 | 3+4 4 | 2+2 4 | 5+6 2 |
| | 6 2 | 3+4 4 | 6 2 | 3+4 4 | 2+2 4 | 3+4 2 |
| | 6 2 | 3+4 4 | 6 2 | 3+4 4 | 2+2 4 | 3+4 2 |
| | 6 2 | 3+4 4 | 5 2 | 3+4 4 | 2+2 4 | 5+6 2 |
| | 5 2 | | 6 2 | | | 3+4 2 |
| | 5 2 | | 5 2 | | | 5+6 2 |
| | 6 2 | | 6 2 | | | 5+6 2 |
| | 5 2 | | 5 2 | | | 3+4 2 |
| | 5 2 | | 5 2 | | | 3+4 2 |
| | 5 2 | | 6 2 | | | 5+6 2 |
| | 6 2 | | 5 2 | | | 3+4 2 |
| | 6 2 | | 6 2 | | | 3+4 2 |
| | 6 2 | | 5 2 | | | 5+6 2 |
| | 5 2 | | 6 2 | | | 3+4 2 |
| | 5 2 | | 5 2 | | | 5+6 2 |
| | 5 2 | | 6 2 | | | 5+6 2 |
| | 6 2 | | 6 2 | | | 5+6 2 |
| | 6 2 | | 5 2 | | | 3+4 2 |
| | 5 2 | | 5 2 | | | 5+6 2 |
| | 6 2 | | 6 2 | | | 3+4 2 |
| | 6 2 | | 5 2 | | | 5+6 2 |
| | 5 2 | | 6 2 | | | 3+4 2 |
| | 6 2 | | 5 2 | | | 3+4 2 |
| | 5 2 | | 6 2 | | | 3+4 2 |
| | 6 2 | | 6 2 | | | 5+6 2 |
| | 5 2 | | 6 2 | | | 3+4 2 |
| | 6 2 | | 6 2 | | | 3+4 2 |
| | 5 2 | | 6 2 | | | 5+6 2 |
| | 6 2 | | 5 2 | | | 3+4 2 |
| | 5 2 | | 6 2 | | | 5+6 2 |
| | 5 2 | | 6 2 | | | 3+4 2 |
| | 5 2 | | 5 2 | | | 3+4 2 |
| | 6 2 | | 6 2 | | | 5+6 2 |
| | 5 2 | | 5 2 | | | 3+4 2 |
| | 6 2 | | 5 2 | | | 5+6 2 |
| | 5 2 | | 5 2 | | | 5+6 2 |
| | 6 2 | | 5 2 | | | 3+4 2 |
| | 5 2 | | 5 2 | | | 3+4 2 |
| | 6 2 | | 6 2 | | | 5+6 2 |
| | 6 2 | | 5 2 | | | 3+4 2 |

Continued on next column

**TABLE XXXVIII– continued**

| TQC | ICSD | | NoE | | SF | |
|---|---|---|---|---|---|---|
| | 6 2 | | 6 2 | | | 5+6 2 |
| | 5 2 | | 5 2 | | | 5+6 2 |
| | 5 2 | | 6 2 | | | 5+6 2 |
| | 5 2 | | 6 2 | | | 3+4 2 |
| | 6 2 | | 5 2 | | | 5+6 2 |
| | 6 2 | | 6 2 | | | 3+4 2 |
| | 5 2 | | 5 2 | | | 5+6 2 |
| | 6 2 | | 6 2 | | | 3+4 2 |
| | 5 2 | | 6 2 | | | 5+6 2 |
| | 6 2 | | 5 2 | | | 3+4 2 |
| | 5 2 | | 5 2 | | | 5+6 2 |
| | 5 2 | | 6 2 | | | 3+4 2 |
| | 6 2 | | 5 2 | | | 5+6 2 |
| | 6 2 | | 6 2 | | | 3+4 2 |
| | 6 2 | | 5 2 | | | 5+6 2 |
| | 5 2 | | 6 2 | | | 5+6 2 |
| | 5 2 | | 6 2 | | | 3+4 2 |
| | 5 2 | | 5 2 | | | 3+4 2 |
| | 6 2 | | 6 2 | | | 5+6 2 |
| 63.2.1 | 43419 | | 28 | | U1 | |
| | -GM | -T | -Y | -Z | -R | -S |
| | 5 2 | 3+4 4 | 5 2 | 3+4 4 | 2+2 4 | 3+4 2 |
| | 6 2 | 3+4 4 | 6 2 | 3+4 4 | 2+2 4 | 5+6 2 |
| | 5 2 | 3+4 4 | 5 2 | 3+4 4 | 2+2 4 | 3+4 2 |
| | 6 2 | 3+4 4 | 6 2 | 3+4 4 | 2+2 4 | 5+6 2 |
| | 5 2 | 3+4 4 | 5 2 | 3+4 4 | 2+2 4 | 3+4 2 |
| | 5 2 | 3+4 4 | 6 2 | 3+4 4 | 2+2 4 | 5+6 2 |
| | 6 2 | 3+4 4 | 6 2 | 3+4 4 | 2+2 4 | 3+4 2 |
| | 6 2 | | 5 2 | | | 5+6 2 |
| | 5 2 | | 5 2 | | | 5+6 2 |
| | 6 2 | | 5 2 | | | 3+4 2 |
| | 5 2 | | 6 2 | | | 5+6 2 |
| | 6 2 | | 5 2 | | | 3+4 2 |
| | 6 2 | | 6 2 | | | 5+6 2 |
| | 5 2 | | 5 2 | | | 5+6 2 |
| 63.2.2 | 43339 | | 28 | | U1 | |
| | -GM | -T | -Y | -Z | -R | -S |
| | 5 2 | 3+4 4 | 5 2 | 3+4 4 | 2+2 4 | 3+4 2 |
| | 6 2 | 3+4 4 | 6 2 | 3+4 4 | 2+2 4 | 5+6 2 |
| | 5 2 | 3+4 4 | 5 2 | 3+4 4 | 2+2 4 | 3+4 2 |
| | 6 2 | 3+4 4 | 6 2 | 3+4 4 | 2+2 4 | 5+6 2 |
| | 5 2 | 3+4 4 | 5 2 | 3+4 4 | 2+2 4 | 3+4 2 |
| | 5 2 | 3+4 4 | 6 2 | 3+4 4 | 2+2 4 | 5+6 2 |
| | 6 2 | 3+4 4 | 6 2 | 3+4 4 | 2+2 4 | 3+4 2 |
| | 6 2 | | 5 2 | | | 5+6 2 |
| | 5 2 | | 5 2 | | | 5+6 2 |
| | 6 2 | | 5 2 | | | 3+4 2 |
| | 5 2 | | 6 2 | | | 5+6 2 |
| | 6 2 | | 5 2 | | | 3+4 2 |
| | 6 2 | | 6 2 | | | 5+6 2 |
| | 5 2 | | 5 2 | | | 5+6 2 |
| 63.2.3 | 411981 | | 100 | | Sn5Sr3 | |
| | -GM | -T | -Y | -Z | -R | -S |
| | 5 2 | 3+4 4 | 5 2 | 3+4 4 | 2+2 4 | 5+6 2 |
| | 5 2 | 3+4 4 | 5 2 | 3+4 4 | 2+2 4 | 5+6 2 |
| | 6 2 | 3+4 4 | 6 2 | 3+4 4 | 2+2 4 | 3+4 2 |
| | 6 2 | 3+4 4 | 6 2 | 3+4 4 | 2+2 4 | 3+4 2 |
| | 5 2 | 3+4 4 | 5 2 | 3+4 4 | 2+2 4 | 3+4 2 |
| | 6 2 | 3+4 4 | 6 2 | 3+4 4 | 2+2 4 | 5+6 2 |
| | 5 2 | 3+4 4 | 6 2 | 3+4 4 | 2+2 4 | 5+6 2 |
| | 5 2 | 3+4 4 | 6 2 | 3+4 4 | 2+2 4 | 5+6 2 |

Continued on next column

**TABLE XXXVIII– continued**

| TQC | ICSD | | NoE | | SF | |
|---|---|---|---|---|---|---|
| | 6 2 | 3+4 4 | 5 2 | 3+4 4 | 2+2 4 | 3+4 2 |
| | 6 2 | 3+4 4 | 5 2 | 3+4 4 | 2+2 4 | 3+4 2 |
| | 5 2 | 3+4 4 | 5 2 | 3+4 4 | 2+2 4 | 3+4 2 |
| | 6 2 | 3+4 4 | 6 2 | 3+4 4 | 2+2 4 | 5+6 2 |
| | 5 2 | 3+4 4 | 6 2 | 3+4 4 | 2+2 4 | 5+6 2 |
| | 5 2 | 3+4 4 | 5 2 | 3+4 4 | 2+2 4 | 5+6 2 |
| | 5 2 | 3+4 4 | 6 2 | 3+4 4 | 2+2 4 | 3+4 2 |
| | 6 2 | 3+4 4 | 5 2 | 3+4 4 | 2+2 4 | 3+4 2 |
| | 5 2 | 3+4 4 | 6 2 | 3+4 4 | 2+2 4 | 5+6 2 |
| | 6 2 | 3+4 4 | 5 2 | 3+4 4 | 2+2 4 | 5+6 2 |
| | 6 2 | 3+4 4 | 6 2 | 3+4 4 | 2+2 4 | 3+4 2 |
| | 6 2 | 3+4 4 | 5 2 | 3+4 4 | 2+2 4 | 3+4 2 |
| | 5 2 | 3+4 4 | 5 2 | 3+4 4 | 2+2 4 | 3+4 2 |
| | 6 2 | 3+4 4 | 6 2 | 3+4 4 | 2+2 4 | 5+6 2 |
| | 5 2 | 3+4 4 | 5 2 | 3+4 4 | 2+2 4 | 3+4 2 |
| | 6 2 | 3+4 4 | 6 2 | 3+4 4 | 2+2 4 | 5+6 2 |
| | 5 2 | 3+4 4 | 5 2 | 3+4 4 | 2+2 4 | 3+4 2 |
| | 6 2 | | 6 2 | | | 5+6 2 |
| | 5 2 | | 6 2 | | | 3+4 2 |
| | 6 2 | | 5 2 | | | 3+4 2 |
| | 5 2 | | 5 2 | | | 5+6 2 |
| | 6 2 | | 5 2 | | | 5+6 2 |
| | 5 2 | | 6 2 | | | 3+4 2 |
| | 6 2 | | 6 2 | | | 5+6 2 |
| | 6 2 | | 5 2 | | | 3+4 2 |
| | 5 2 | | 6 2 | | | 5+6 2 |
| | 5 2 | | 6 2 | | | 5+6 2 |
| | 6 2 | | 5 2 | | | 3+4 2 |
| | 5 2 | | 6 2 | | | 5+6 2 |
| | 5 2 | | 5 2 | | | 3+4 2 |
| | 5 2 | | 5 2 | | | 3+4 2 |
| | 6 2 | | 5 2 | | | 3+4 2 |
| | 6 2 | | 5 2 | | | 5+6 2 |
| | 6 2 | | 6 2 | | | 5+6 2 |
| | 5 2 | | 5 2 | | | 3+4 2 |
| | 6 2 | | 6 2 | | | 5+6 2 |
| | 6 2 | | 6 2 | | | 3+4 2 |
| | 5 2 | | 5 2 | | | 5+6 2 |
| | 6 2 | | 5 2 | | | 3+4 2 |
| | 5 2 | | 6 2 | | | 5+6 2 |
| | 5 2 | | 6 2 | | | 3+4 2 |
| | 5 2 | | 6 2 | | | 3+4 2 |
| 63.2.4 | 165183 | | 100 | | Sn5Sr3 | |
| | -GM | -T | -Y | -Z | -R | -S |
| | 5 2 | 3+4 4 | 5 2 | 3+4 4 | 2+2 4 | 5+6 2 |
| | 5 2 | 3+4 4 | 5 2 | 3+4 4 | 2+2 4 | 5+6 2 |
| | 6 2 | 3+4 4 | 6 2 | 3+4 4 | 2+2 4 | 3+4 2 |
| | 6 2 | 3+4 4 | 6 2 | 3+4 4 | 2+2 4 | 3+4 2 |
| | 6 2 | 3+4 4 | 6 2 | 3+4 4 | 2+2 4 | 3+4 2 |
| | 5 2 | 3+4 4 | 5 2 | 3+4 4 | 2+2 4 | 5+6 2 |
| | 5 2 | 3+4 4 | 5 2 | 3+4 4 | 2+2 4 | 5+6 2 |
| | 5 2 | 3+4 4 | 5 2 | 3+4 4 | 2+2 4 | 5+6 2 |
| | 6 2 | 3+4 4 | 6 2 | 3+4 4 | 2+2 4 | 3+4 2 |
| | 6 2 | 3+4 4 | 6 2 | 3+4 4 | 2+2 4 | 3+4 2 |
| | 5 2 | 3+4 4 | 5 2 | 3+4 4 | 2+2 4 | 5+6 2 |
| | 6 2 | 3+4 4 | 6 2 | 3+4 4 | 2+2 4 | 3+4 2 |
| | 5 2 | 3+4 4 | 5 2 | 3+4 4 | 2+2 4 | 5+6 2 |
| | 5 2 | 3+4 4 | 5 2 | 3+4 4 | 2+2 4 | 5+6 2 |
| | 6 2 | 3+4 4 | 6 2 | 3+4 4 | 2+2 4 | 3+4 2 |
| | 6 2 | 3+4 4 | 6 2 | 3+4 4 | 2+2 4 | 3+4 2 |
| | 5 2 | 3+4 4 | 6 2 | 3+4 4 | 2+2 4 | 3+4 2 |

Continued on next column

**TABLE XXXVIII– continued**

| TQC | ICSD | | NoE | SF | | |
|---|---|---|---|---|---|---|
| | 6 2 | 3+4 4 | 5 2 | 3+4 4 | 2+2 4 | 5+6 2 |
| | 6 2 | 3+4 4 | 6 2 | 3+4 4 | 2+2 4 | 3+4 2 |
| | 5 2 | 3+4 4 | 5 2 | 3+4 4 | 2+2 4 | 5+6 2 |
| | 5 2 | 3+4 4 | 5 2 | 3+4 4 | 2+2 4 | 5+6 2 |
| | 6 2 | 3+4 4 | 6 2 | 3+4 4 | 2+2 4 | 3+4 2 |
| | 5 2 | 3+4 4 | 5 2 | 3+4 4 | 2+2 4 | 5+6 2 |
| | 6 2 | 3+4 4 | 6 2 | 3+4 4 | 2+2 4 | 3+4 2 |
| | 5 2 | 3+4 4 | 5 2 | 3+4 4 | 2+2 4 | 5+6 2 |
| | 6 2 | | 6 2 | | | 3+4 2 |
| | 5 2 | | 6 2 | | | 5+6 2 |
| | 6 2 | | 5 2 | | | 3+4 2 |
| | 6 2 | | 5 2 | | | 5+6 2 |
| | 5 2 | | 6 2 | | | 3+4 2 |
| | 5 2 | | 5 2 | | | 5+6 2 |
| | 6 2 | | 6 2 | | | 3+4 2 |
| | 5 2 | | 5 2 | | | 5+6 2 |
| | 6 2 | | 6 2 | | | 3+4 2 |
| | 5 2 | | 6 2 | | | 3+4 2 |
| | 6 2 | | 6 2 | | | 5+6 2 |
| | 5 2 | | 5 2 | | | 3+4 2 |
| | 5 2 | | 5 2 | | | 5+6 2 |
| | 6 2 | | 5 2 | | | 5+6 2 |
| | 6 2 | | 5 2 | | | 5+6 2 |
| | 5 2 | | 5 2 | | | 3+4 2 |
| | 6 2 | | 5 2 | | | 5+6 2 |
| | 5 2 | | 6 2 | | | 3+4 2 |
| | 6 2 | | 6 2 | | | 3+4 2 |
| | 6 2 | | 5 2 | | | 3+4 2 |
| | 5 2 | | 6 2 | | | 5+6 2 |
| | 5 2 | | 6 2 | | | 5+6 2 |
| | 6 2 | | 6 2 | | | 3+4 2 |
| | 5 2 | | 5 2 | | | 5+6 2 |
| | 5 2 | | 6 2 | | | 3+4 2 |
| 63.2.5 | 54756 | | 52 | Sn4Sr1 | | |
| | -GM | -T | -Y | -Z | -R | -S |
| | 5 2 | 3+4 4 | 5 2 | 3+4 4 | 2+2 4 | 3+4 2 |
| | 6 2 | 3+4 4 | 6 2 | 3+4 4 | 2+2 4 | 5+6 2 |
| | 5 2 | 3+4 4 | 5 2 | 3+4 4 | 2+2 4 | 5+6 2 |
| | 6 2 | 3+4 4 | 6 2 | 3+4 4 | 2+2 4 | 3+4 2 |
| | 5 2 | 3+4 4 | 5 2 | 3+4 4 | 2+2 4 | 3+4 2 |
| | 6 2 | 3+4 4 | 6 2 | 3+4 4 | 2+2 4 | 5+6 2 |
| | 5 2 | 3+4 4 | 6 2 | 3+4 4 | 2+2 4 | 5+6 2 |
| | 6 2 | 3+4 4 | 5 2 | 3+4 4 | 2+2 4 | 3+4 2 |
| | 5 2 | 3+4 4 | 6 2 | 3+4 4 | 2+2 4 | 3+4 2 |
| | 6 2 | 3+4 4 | 5 2 | 3+4 4 | 2+2 4 | 5+6 2 |
| | 5 2 | 3+4 4 | 6 2 | 3+4 4 | 2+2 4 | 5+6 2 |
| | 5 2 | 3+4 4 | 6 2 | 3+4 4 | 2+2 4 | 3+4 2 |
| | 5 2 | 3+4 4 | 5 2 | 3+4 4 | 2+2 4 | 3+4 2 |
| | 6 2 | | 5 2 | | | 3+4 2 |
| | 5 2 | | 5 2 | | | 5+6 2 |
| | 6 2 | | 5 2 | | | 3+4 2 |
| | 6 2 | | 6 2 | | | 5+6 2 |
| | 5 2 | | 5 2 | | | 3+4 2 |
| | 6 2 | | 6 2 | | | 5+6 2 |
| | 6 2 | | 6 2 | | | 5+6 2 |
| | 5 2 | | 6 2 | | | 5+6 2 |
| | 5 2 | | 6 2 | | | 5+6 2 |
| | 5 2 | | 6 2 | | | 3+4 2 |
| | 6 2 | | 5 2 | | | 5+6 2 |
| | 6 2 | | 6 2 | | | 3+4 2 |
| | 6 2 | | 5 2 | | | 3+4 2 |

Continued on next column

**TABLE XXXVIII– continued**

| TQC | ICSD | | NoE | SF | | |
|---|---|---|---|---|---|---|
| 63.2.6 | 281669 | | 52 | Sn4Sr1 | | |
| | -GM | -T | -Y | -Z | -R | -S |
| | 5 2 | 3+4 4 | 5 2 | 3+4 4 | 2+2 4 | 3+4 2 |
| | 6 2 | 3+4 4 | 6 2 | 3+4 4 | 2+2 4 | 5+6 2 |
| | 5 2 | 3+4 4 | 5 2 | 3+4 4 | 2+2 4 | 5+6 2 |
| | 6 2 | 3+4 4 | 6 2 | 3+4 4 | 2+2 4 | 3+4 2 |
| | 5 2 | 3+4 4 | 5 2 | 3+4 4 | 2+2 4 | 3+4 2 |
| | 6 2 | 3+4 4 | 6 2 | 3+4 4 | 2+2 4 | 5+6 2 |
| | 5 2 | 3+4 4 | 6 2 | 3+4 4 | 2+2 4 | 5+6 2 |
| | 6 2 | 3+4 4 | 5 2 | 3+4 4 | 2+2 4 | 3+4 2 |
| | 5 2 | 3+4 4 | 6 2 | 3+4 4 | 2+2 4 | 3+4 2 |
| | 6 2 | 3+4 4 | 5 2 | 3+4 4 | 2+2 4 | 5+6 2 |
| | 5 2 | 3+4 4 | 6 2 | 3+4 4 | 2+2 4 | 5+6 2 |
| | 5 2 | 3+4 4 | 6 2 | 3+4 4 | 2+2 4 | 3+4 2 |
| | 5 2 | 3+4 4 | 5 2 | 3+4 4 | 2+2 4 | 3+4 2 |
| | 6 2 | | 5 2 | | | 3+4 2 |
| | 5 2 | | 5 2 | | | 5+6 2 |
| | 6 2 | | 5 2 | | | 3+4 2 |
| | 6 2 | | 6 2 | | | 5+6 2 |
| | 5 2 | | 5 2 | | | 3+4 2 |
| | 6 2 | | 6 2 | | | 5+6 2 |
| | 6 2 | | 6 2 | | | 5+6 2 |
| | 5 2 | | 6 2 | | | 5+6 2 |
| | 5 2 | | 6 2 | | | 5+6 2 |
| | 5 2 | | 6 2 | | | 3+4 2 |
| | 6 2 | | 5 2 | | | 5+6 2 |
| | 6 2 | | 6 2 | | | 3+4 2 |
| | 6 2 | | 5 2 | | | 3+4 2 |
| 63.2.7 | 73619 | | 100 | Si3Tb2 | | |
| | -GM | -T | -Y | -Z | -R | -S |
| | 5 2 | 3+4 4 | 6 2 | 3+4 4 | 2+2 4 | 3+4 2 |
| | 6 2 | 3+4 4 | 5 2 | 3+4 4 | 2+2 4 | 5+6 2 |
| | 5 2 | 3+4 4 | 5 2 | 3+4 4 | 2+2 4 | 5+6 2 |
| | 6 2 | 3+4 4 | 6 2 | 3+4 4 | 2+2 4 | 3+4 2 |
| | 5 2 | 3+4 4 | 6 2 | 3+4 4 | 2+2 4 | 5+6 2 |
| | 5 2 | 3+4 4 | 5 2 | 3+4 4 | 2+2 4 | 3+4 2 |
| | 6 2 | 3+4 4 | 6 2 | 3+4 4 | 2+2 4 | 5+6 2 |
| | 6 2 | 3+4 4 | 5 2 | 3+4 4 | 2+2 4 | 3+4 2 |
| | 5 2 | 3+4 4 | 6 2 | 3+4 4 | 2+2 4 | 5+6 2 |
| | 5 2 | 3+4 4 | 5 2 | 3+4 4 | 2+2 4 | 3+4 2 |
| | 6 2 | 3+4 4 | 6 2 | 3+4 4 | 2+2 4 | 3+4 2 |
| | 5 2 | 3+4 4 | 6 2 | 3+4 4 | 2+2 4 | 5+6 2 |
| | 6 2 | 3+4 4 | 5 2 | 3+4 4 | 2+2 4 | 5+6 2 |
| | 5 2 | 3+4 4 | 5 2 | 3+4 4 | 2+2 4 | 3+4 2 |
| | 6 2 | 3+4 4 | 6 2 | 3+4 4 | 2+2 4 | 5+6 2 |
| | 6 2 | 3+4 4 | 5 2 | 3+4 4 | 2+2 4 | 3+4 2 |
| | 5 2 | 3+4 4 | 5 2 | 3+4 4 | 2+2 4 | 3+4 2 |
| | 6 2 | 3+4 4 | 6 2 | 3+4 4 | 2+2 4 | 5+6 2 |
| | 5 2 | 3+4 4 | 5 2 | 3+4 4 | 2+2 4 | 3+4 2 |
| | 6 2 | 3+4 4 | 6 2 | 3+4 4 | 2+2 4 | 5+6 2 |
| | 5 2 | 3+4 4 | 5 2 | 3+4 4 | 2+2 4 | 5+6 2 |
| | 5 2 | 3+4 4 | 5 2 | 3+4 4 | 2+2 4 | 3+4 2 |
| | 6 2 | 3+4 4 | 6 2 | 3+4 4 | 2+2 4 | 3+4 2 |
| | 6 2 | 3+4 4 | 6 2 | 3+4 4 | 2+2 4 | 3+4 2 |
| | 5 2 | 3+4 4 | 5 2 | 3+4 4 | 2+2 4 | 5+6 2 |
| | 6 2 | | 5 2 | | | 5+6 2 |
| | 6 2 | | 6 2 | | | 5+6 2 |
| | 5 2 | | 6 2 | | | 3+4 2 |
| | 5 2 | | 5 2 | | | 3+4 2 |
| | 6 2 | | 6 2 | | | 3+4 2 |
| | 5 2 | | 5 2 | | | 5+6 2 |

Continued on next column

**TABLE XXXVIII– continued**

| TQC | ICSD | NoE | SF | | | |
|---|---|---|---|---|---|---|
| | 6 2 | | 5 2 | | | 5+6 2 |
| | 5 2 | | 6 2 | | | 3+4 2 |
| | 6 2 | | 5 2 | | | 3+4 2 |
| | 5 2 | | 6 2 | | | 5+6 2 |
| | 6 2 | | 5 2 | | | 3+4 2 |
| | 6 2 | | 6 2 | | | 5+6 2 |
| | 5 2 | | 6 2 | | | 3+4 2 |
| | 5 2 | | 5 2 | | | 5+6 2 |
| | 6 2 | | 5 2 | | | 3+4 2 |
| | 6 2 | | 6 2 | | | 5+6 2 |
| | 5 2 | | 5 2 | | | 5+6 2 |
| | 6 2 | | 5 2 | | | 3+4 2 |
| | 5 2 | | 6 2 | | | 5+6 2 |
| | 5 2 | | 6 2 | | | 3+4 2 |
| | 6 2 | | 6 2 | | | 3+4 2 |
| | 6 2 | | 5 2 | | | 5+6 2 |
| | 5 2 | | 5 2 | | | 3+4 2 |
| | 6 2 | | 6 2 | | | 5+6 2 |
| | 5 2 | | 5 2 | | | 3+4 2 |
| 63.2.8 | 95179 | 24 | Si2Zr1 | | | |
| | -GM | -T | -Y | -Z | -R | -S |
| | 5 2 | 3+4 4 | 5 2 | 3+4 4 | 2+2 4 | 5+6 2 |
| | 5 2 | 3+4 4 | 6 2 | 3+4 4 | 2+2 4 | 3+4 2 |
| | 6 2 | 3+4 4 | 6 2 | 3+4 4 | 2+2 4 | 5+6 2 |
| | 6 2 | 3+4 4 | 5 2 | 3+4 4 | 2+2 4 | 3+4 2 |
| | 5 2 | 3+4 4 | 5 2 | 3+4 4 | 2+2 4 | 3+4 2 |
| | 5 2 | 3+4 4 | 6 2 | 3+4 4 | 2+2 4 | 5+6 2 |
| | 5 2 | | 5 2 | | | 5+6 2 |
| | 6 2 | | 6 2 | | | 5+6 2 |
| | 6 2 | | 6 2 | | | 3+4 2 |
| | 5 2 | | 6 2 | | | 3+4 2 |
| | 5 2 | | 5 2 | | | 3+4 2 |
| | 6 2 | | 5 2 | | | 3+4 2 |
| 63.2.9 | 20284 | 16 | Si1Zr1 | | | |
| | -GM | -T | -Y | -Z | -R | -S |
| | 5 2 | 3+4 4 | 5 2 | 3+4 4 | 2+2 4 | 3+4 2 |
| | 6 2 | 3+4 4 | 6 2 | 3+4 4 | 2+2 4 | 5+6 2 |
| | 5 2 | 3+4 4 | 5 2 | 3+4 4 | 2+2 4 | 5+6 2 |
| | 6 2 | 3+4 4 | 5 2 | 3+4 4 | 2+2 4 | 3+4 2 |
| | 6 2 | | 5 2 | | | 3+4 2 |
| | 5 2 | | 6 2 | | | 3+4 2 |
| | 5 2 | | 6 2 | | | 5+6 2 |
| | 5 2 | | 6 2 | | | 5+6 2 |
| 63.2.10 | 22235 | 68 | Sb2Yb1 | | | |
| | -GM | -T | -Y | -Z | -R | -S |
| | 5 2 | 3+4 4 | 5 2 | 3+4 4 | 2+2 4 | 3+4 2 |
| | 6 2 | 3+4 4 | 6 2 | 3+4 4 | 2+2 4 | 5+6 2 |
| | 5 2 | 3+4 4 | 5 2 | 3+4 4 | 2+2 4 | 3+4 2 |
| | 6 2 | 3+4 4 | 6 2 | 3+4 4 | 2+2 4 | 5+6 2 |
| | 5 2 | 3+4 4 | 5 2 | 3+4 4 | 2+2 4 | 3+4 2 |
| | 6 2 | 3+4 4 | 6 2 | 3+4 4 | 2+2 4 | 5+6 2 |
| | 5 2 | 3+4 4 | 5 2 | 3+4 4 | 2+2 4 | 3+4 2 |
| | 6 2 | 3+4 4 | 6 2 | 3+4 4 | 2+2 4 | 5+6 2 |
| | 5 2 | 3+4 4 | 6 2 | 3+4 4 | 2+2 4 | 5+6 2 |
| | 5 2 | 3+4 4 | 5 2 | 3+4 4 | 2+2 4 | 3+4 2 |
| | 6 2 | 3+4 4 | 6 2 | 3+4 4 | 2+2 4 | 5+6 2 |
| | 6 2 | 3+4 4 | 5 2 | 3+4 4 | 2+2 4 | 3+4 2 |
| | 5 2 | 3+4 4 | 5 2 | 3+4 4 | 2+2 4 | 3+4 2 |
| | 6 2 | 3+4 4 | 6 2 | 3+4 4 | 2+2 4 | 5+6 2 |
| | 5 2 | 3+4 4 | 6 2 | 3+4 4 | 2+2 4 | 5+6 2 |
| | 5 2 | 3+4 4 | 5 2 | 3+4 4 | 2+2 4 | 5+6 2 |

Continued on next column

**TABLE XXXVIII– continued**

| TQC | ICSD | NoE | SF | | | |
|---|---|---|---|---|---|---|
| | 6 2 | 3+4 4 | 6 2 | 3+4 4 | 2+2 4 | 3+4 2 |
| | 5 2 | | 5 2 | | | 5+6 2 |
| | 5 2 | | 6 2 | | | 3+4 2 |
| | 6 2 | | 6 2 | | | 5+6 2 |
| | 5 2 | | 5 2 | | | 3+4 2 |
| | 6 2 | | 6 2 | | | 3+4 2 |
| | 6 2 | | 5 2 | | | 5+6 2 |
| | 5 2 | | 6 2 | | | 3+4 2 |
| | 5 2 | | 5 2 | | | 5+6 2 |
| | 5 2 | | 5 2 | | | 5+6 2 |
| | 6 2 | | 6 2 | | | 3+4 2 |
| | 6 2 | | 6 2 | | | 5+6 2 |
| | 5 2 | | 5 2 | | | 3+4 2 |
| | 6 2 | | 6 2 | | | 3+4 2 |
| | 5 2 | | 5 2 | | | 5+6 2 |
| | 6 2 | | 5 2 | | | 3+4 2 |
| | 5 2 | | 6 2 | | | 5+6 2 |
| | 6 2 | | 5 2 | | | 5+6 2 |
| 63.2.11 | 163454 | 52 | Rh1Zr1 | | | |
| | -GM | -T | -Y | -Z | -R | -S |
| | 5 2 | 3+4 4 | 5 2 | 3+4 4 | 2+2 4 | 3+4 2 |
| | 5 2 | 3+4 4 | 6 2 | 3+4 4 | 2+2 4 | 5+6 2 |
| | 6 2 | 3+4 4 | 5 2 | 3+4 4 | 2+2 4 | 3+4 2 |
| | 6 2 | 3+4 4 | 6 2 | 3+4 4 | 2+2 4 | 5+6 2 |
| | 5 2 | 3+4 4 | 6 2 | 3+4 4 | 2+2 4 | 3+4 2 |
| | 5 2 | 3+4 4 | 5 2 | 3+4 4 | 2+2 4 | 3+4 2 |
| | 6 2 | 3+4 4 | 5 2 | 3+4 4 | 2+2 4 | 5+6 2 |
| | 6 2 | 3+4 4 | 5 2 | 3+4 4 | 2+2 4 | 5+6 2 |
| | 5 2 | 3+4 4 | 6 2 | 3+4 4 | 2+2 4 | 5+6 2 |
| | 6 2 | 3+4 4 | 6 2 | 3+4 4 | 2+2 4 | 3+4 2 |
| | 6 2 | 3+4 4 | 6 2 | 3+4 4 | 2+2 4 | 5+6 2 |
| | 5 2 | 3+4 4 | 6 2 | 3+4 4 | 2+2 4 | 5+6 2 |
| | 6 2 | 3+4 4 | 5 2 | 3+4 4 | 2+2 4 | 3+4 2 |
| | 5 2 | | 6 2 | | | 5+6 2 |
| | 5 2 | | 6 2 | | | 3+4 2 |
| | 5 2 | | 5 2 | | | 3+4 2 |
| | 6 2 | | 5 2 | | | 5+6 2 |
| | 5 2 | | 5 2 | | | 3+4 2 |
| | 6 2 | | 5 2 | | | 5+6 2 |
| | 5 2 | | 6 2 | | | 3+4 2 |
| | 6 2 | | 5 2 | | | 3+4 2 |
| | 5 2 | | 5 2 | | | 3+4 2 |
| | 6 2 | | 6 2 | | | 3+4 2 |
| | 6 2 | | 6 2 | | | 5+6 2 |
| | 5 2 | | 5 2 | | | 5+6 2 |
| | 6 2 | | 5 2 | | | 3+4 2 |
| 63.2.12 | 55546 | 28 | Pd1Zr1 | | | |
| | -GM | -T | -Y | -Z | -R | -S |
| | 5 2 | 3+4 4 | 5 2 | 3+4 4 | 2+2 4 | 3+4 2 |
| | 6 2 | 3+4 4 | 6 2 | 3+4 4 | 2+2 4 | 5+6 2 |
| | 5 2 | 3+4 4 | 5 2 | 3+4 4 | 2+2 4 | 3+4 2 |
| | 5 2 | 3+4 4 | 6 2 | 3+4 4 | 2+2 4 | 3+4 2 |
| | 6 2 | 3+4 4 | 5 2 | 3+4 4 | 2+2 4 | 5+6 2 |
| | 6 2 | 3+4 4 | 6 2 | 3+4 4 | 2+2 4 | 5+6 2 |
| | 5 2 | 3+4 4 | 6 2 | 3+4 4 | 2+2 4 | 3+4 2 |
| | 6 2 | | 6 2 | | | 5+6 2 |
| | 5 2 | | 5 2 | | | 3+4 2 |
| | 5 2 | | 6 2 | | | 5+6 2 |
| | 6 2 | | 5 2 | | | 3+4 2 |
| | 6 2 | | 5 2 | | | 5+6 2 |
| | 5 2 | | 5 2 | | | 5+6 2 |

Continued on next column

**TABLE XXXVIII– continued**

| TQC | ICSD | | NoE | SF | | |
|---|---|---|---|---|---|---|
| | 6 2 | | 5 2 | | | 5+6 2 |
| 63.2.13 | 165577 | | 52 | Pd1Sr1Tl2 | | |
| | -GM | -T | -Y | -Z | -R | -S |
| | 5 2 | 3+4 4 | 5 2 | 3+4 4 | 2+2 4 | 5+6 2 |
| | 6 2 | 3+4 4 | 6 2 | 3+4 4 | 2+2 4 | 3+4 2 |
| | 5 2 | 3+4 4 | 5 2 | 3+4 4 | 2+2 4 | 5+6 2 |
| | 6 2 | 3+4 4 | 6 2 | 3+4 4 | 2+2 4 | 3+4 2 |
| | 5 2 | 3+4 4 | 5 2 | 3+4 4 | 2+2 4 | 5+6 2 |
| | 6 2 | 3+4 4 | 6 2 | 3+4 4 | 2+2 4 | 3+4 2 |
| | 6 2 | 3+4 4 | 6 2 | 3+4 4 | 2+2 4 | 5+6 2 |
| | 5 2 | 3+4 4 | 5 2 | 3+4 4 | 2+2 4 | 3+4 2 |
| | 5 2 | 3+4 4 | 6 2 | 3+4 4 | 2+2 4 | 3+4 2 |
| | 5 2 | 3+4 4 | 5 2 | 3+4 4 | 2+2 4 | 5+6 2 |
| | 6 2 | 3+4 4 | 6 2 | 3+4 4 | 2+2 4 | 3+4 2 |
| | 6 2 | 3+4 4 | 5 2 | 3+4 4 | 2+2 4 | 5+6 2 |
| | 5 2 | 3+4 4 | 5 2 | 3+4 4 | 2+2 4 | 5+6 2 |
| | 5 2 | | 6 2 | | | 3+4 2 |
| | 6 2 | | 6 2 | | | 3+4 2 |
| | 5 2 | | 5 2 | | | 5+6 2 |
| | 5 2 | | 6 2 | | | 3+4 2 |
| | 6 2 | | 5 2 | | | 5+6 2 |
| | 6 2 | | 6 2 | | | 5+6 2 |
| | 5 2 | | 5 2 | | | 3+4 2 |
| | 6 2 | | 6 2 | | | 3+4 2 |
| | 6 2 | | 5 2 | | | 3+4 2 |
| | 5 2 | | 5 2 | | | 5+6 2 |
| | 6 2 | | 6 2 | | | 5+6 2 |
| | 5 2 | | 6 2 | | | 3+4 2 |
| | 5 2 | | 5 2 | | | 5+6 2 |
| 63.2.14 | 78010 | | 172 | O7Pr3Sb1 | | |
| | -GM | -T | -Y | -Z | -R | -S |
| | 5 2 | 3+4 4 | 5 2 | 3+4 4 | 2+2 4 | 5+6 2 |
| | 5 2 | 3+4 4 | 5 2 | 3+4 4 | 2+2 4 | 5+6 2 |
| | 5 2 | 3+4 4 | 6 2 | 3+4 4 | 2+2 4 | 3+4 2 |
| | 6 2 | 3+4 4 | 5 2 | 3+4 4 | 2+2 4 | 5+6 2 |
| | 5 2 | 3+4 4 | 6 2 | 3+4 4 | 2+2 4 | 3+4 2 |
| | 6 2 | 3+4 4 | 5 2 | 3+4 4 | 2+2 4 | 5+6 2 |
| | 6 2 | 3+4 4 | 5 2 | 3+4 4 | 2+2 4 | 3+4 2 |
| | 5 2 | 3+4 4 | 6 2 | 3+4 4 | 2+2 4 | 3+4 2 |
| | 5 2 | 3+4 4 | 5 2 | 3+4 4 | 2+2 4 | 3+4 2 |
| | 5 2 | 3+4 4 | 6 2 | 3+4 4 | 2+2 4 | 5+6 2 |
| | 6 2 | 3+4 4 | 6 2 | 3+4 4 | 2+2 4 | 3+4 2 |
| | 6 2 | 3+4 4 | 6 2 | 3+4 4 | 2+2 4 | 5+6 2 |
| | 6 2 | 3+4 4 | 6 2 | 3+4 4 | 2+2 4 | 5+6 2 |
| | 5 2 | 3+4 4 | 5 2 | 3+4 4 | 2+2 4 | 5+6 2 |
| | 6 2 | 3+4 4 | 5 2 | 3+4 4 | 2+2 4 | 3+4 2 |
| | 6 2 | 3+4 4 | 6 2 | 3+4 4 | 2+2 4 | 5+6 2 |
| | 6 2 | 3+4 4 | 5 2 | 3+4 4 | 2+2 4 | 3+4 2 |
| | 5 2 | 3+4 4 | 6 2 | 3+4 4 | 2+2 4 | 5+6 2 |
| | 6 2 | 3+4 4 | 6 2 | 3+4 4 | 2+2 4 | 5+6 2 |
| | 5 2 | 3+4 4 | 6 2 | 3+4 4 | 2+2 4 | 3+4 2 |
| | 6 2 | 3+4 4 | 5 2 | 3+4 4 | 2+2 4 | 3+4 2 |
| | 5 2 | 3+4 4 | 5 2 | 3+4 4 | 2+2 4 | 3+4 2 |
| | 6 2 | 3+4 4 | 6 2 | 3+4 4 | 2+2 4 | 3+4 2 |
| | 5 2 | 3+4 4 | 5 2 | 3+4 4 | 2+2 4 | 5+6 2 |
| | 6 2 | 3+4 4 | 6 2 | 3+4 4 | 2+2 4 | 3+4 2 |
| | 5 2 | 3+4 4 | 5 2 | 3+4 4 | 2+2 4 | 5+6 2 |
| | 6 2 | 3+4 4 | 6 2 | 3+4 4 | 2+2 4 | 3+4 2 |
| | 5 2 | 3+4 4 | 6 2 | 3+4 4 | 2+2 4 | 3+4 2 |
| | 6 2 | 3+4 4 | 5 2 | 3+4 4 | 2+2 4 | 3+4 2 |
| | 6 2 | 3+4 4 | 6 2 | 3+4 4 | 2+2 4 | 3+4 2 |
| | 6 2 | 3+4 4 | 6 2 | 3+4 4 | 2+2 4 | 5+6 2 |
| | 5 2 | 3+4 4 | 5 2 | 3+4 4 | 2+2 4 | 5+6 2 |
| | 6 2 | 3+4 4 | 5 2 | 3+4 4 | 2+2 4 | 3+4 2 |
| | 6 2 | 3+4 4 | 6 2 | 3+4 4 | 2+2 4 | 3+4 2 |
| | 5 2 | 3+4 4 | 6 2 | 3+4 4 | 2+2 4 | 5+6 2 |
| | 6 2 | 3+4 4 | 6 2 | 3+4 4 | 2+2 4 | 5+6 2 |
| | 5 2 | 3+4 4 | 5 2 | 3+4 4 | 2+2 4 | 3+4 2 |
| | 6 2 | 3+4 4 | 6 2 | 3+4 4 | 2+2 4 | 3+4 2 |
| | 5 2 | 3+4 4 | 5 2 | 3+4 4 | 2+2 4 | 3+4 2 |
| | 5 2 | 3+4 4 | 6 2 | 3+4 4 | 2+2 4 | 3+4 2 |
| | 6 2 | 3+4 4 | 5 2 | 3+4 4 | 2+2 4 | 5+6 2 |
| | 6 2 | 3+4 4 | 6 2 | 3+4 4 | 2+2 4 | 5+6 2 |
| | 6 2 | 3+4 4 | 6 2 | 3+4 4 | 2+2 4 | 5+6 2 |
| | 6 2 | | 6 2 | | | 5+6 2 |
| | 6 2 | | 6 2 | | | 5+6 2 |
| | 5 2 | | 5 2 | | | 3+4 2 |
| | 5 2 | | 5 2 | | | 5+6 2 |
| | 5 2 | | 5 2 | | | 5+6 2 |
| | 5 2 | | 6 2 | | | 3+4 2 |
| | 6 2 | | 5 2 | | | 5+6 2 |
| | 5 2 | | 5 2 | | | 5+6 2 |
| | 5 2 | | 5 2 | | | 3+4 2 |
| | 5 2 | | 5 2 | | | 3+4 2 |
| | 6 2 | | 6 2 | | | 5+6 2 |
| | 5 2 | | 6 2 | | | 3+4 2 |
| | 5 2 | | 5 2 | | | 5+6 2 |
| | 5 2 | | 5 2 | | | 5+6 2 |
| | 6 2 | | 6 2 | | | 3+4 2 |
| | 5 2 | | 6 2 | | | 3+4 2 |
| | 6 2 | | 5 2 | | | 5+6 2 |
| | 6 2 | | 6 2 | | | 3+4 2 |
| | 5 2 | | 6 2 | | | 5+6 2 |
| | 6 2 | | 5 2 | | | 3+4 2 |
| | 6 2 | | 6 2 | | | 5+6 2 |
| | 6 2 | | 5 2 | | | 3+4 2 |
| | 5 2 | | 6 2 | | | 5+6 2 |
| | 5 2 | | 6 2 | | | 5+6 2 |
| | 6 2 | | 6 2 | | | 5+6 2 |
| | 5 2 | | 5 2 | | | 3+4 2 |
| | 5 2 | | 6 2 | | | 3+4 2 |
| | 6 2 | | 6 2 | | | 3+4 2 |
| | 6 2 | | 6 2 | | | 5+6 2 |
| | 6 2 | | 5 2 | | | 3+4 2 |
| | 5 2 | | 6 2 | | | 3+4 2 |
| | 6 2 | | 5 2 | | | 5+6 2 |
| | 6 2 | | 5 2 | | | 3+4 2 |
| | 5 2 | | 5 2 | | | 3+4 2 |
| | 6 2 | | 6 2 | | | 5+6 2 |
| | 6 2 | | 5 2 | | | 3+4 2 |
| | 5 2 | | 5 2 | | | 3+4 2 |
| | 5 2 | | 5 2 | | | 3+4 2 |
| | 6 2 | | 6 2 | | | 5+6 2 |
| | 6 2 | | 5 2 | | | 3+4 2 |
| | 5 2 | | 6 2 | | | 3+4 2 |
| | 6 2 | | 6 2 | | | 3+4 2 |
| | 5 2 | | 6 2 | | | 5+6 2 |
| 63.2.15 | 96322 | | 28 | Ni1Zr1 | | |
| | -GM | -T | -Y | -Z | -R | -S |
| | 5 2 | 3+4 4 | 5 2 | 3+4 4 | 2+2 4 | 3+4 2 |
| | 6 2 | 3+4 4 | 6 2 | 3+4 4 | 2+2 4 | 5+6 2 |
| | 6 2 | 3+4 4 | 5 2 | 3+4 4 | 2+2 4 | 3+4 2 |

Continued on next column

**TABLE XXXVIII– continued**

| TQC | ICSD | | NoE | | SF | |
|---|---|---|---|---|---|---|
| | 5 2 | 3+4 4 | 6 2 | 3+4 4 | 2+2 4 | 5+6 2 |
| | 5 2 | 3+4 4 | 5 2 | 3+4 4 | 2+2 4 | 5+6 2 |
| | 6 2 | 3+4 4 | 6 2 | 3+4 4 | 2+2 4 | 3+4 2 |
| | 5 2 | 3+4 4 | 6 2 | 3+4 4 | 2+2 4 | 5+6 2 |
| | 6 2 | | 6 2 | | | 5+6 2 |
| | 5 2 | | 5 2 | | | 3+4 2 |
| | 6 2 | | 6 2 | | | 3+4 2 |
| | 5 2 | | 5 2 | | | 5+6 2 |
| | 6 2 | | 5 2 | | | 3+4 2 |
| | 5 2 | | 5 2 | | | 3+4 2 |
| | 6 2 | | 5 2 | | | 3+4 2 |
| 63.2.16 | 105478 | | 28 | | Ni1Zr1 | |
| | -GM | -T | -Y | -Z | -R | -S |
| | 5 2 | 3+4 4 | 5 2 | 3+4 4 | 2+2 4 | 5+6 2 |
| | 6 2 | 3+4 4 | 6 2 | 3+4 4 | 2+2 4 | 3+4 2 |
| | 6 2 | 3+4 4 | 5 2 | 3+4 4 | 2+2 4 | 5+6 2 |
| | 5 2 | 3+4 4 | 6 2 | 3+4 4 | 2+2 4 | 3+4 2 |
| | 5 2 | 3+4 4 | 5 2 | 3+4 4 | 2+2 4 | 3+4 2 |
| | 6 2 | 3+4 4 | 6 2 | 3+4 4 | 2+2 4 | 5+6 2 |
| | 6 2 | 3+4 4 | 6 2 | 3+4 4 | 2+2 4 | 3+4 2 |
| | 5 2 | | 6 2 | | | 3+4 2 |
| | 5 2 | | 5 2 | | | 5+6 2 |
| | 6 2 | | 6 2 | | | 5+6 2 |
| | 5 2 | | 5 2 | | | 3+4 2 |
| | 6 2 | | 5 2 | | | 5+6 2 |
| | 5 2 | | 5 2 | | | 5+6 2 |
| | 6 2 | | 5 2 | | | 5+6 2 |
| 63.2.17 | 71964 | | 56 | | Ni1O1Zr3 | |
| | -GM | -T | -Y | -Z | -R | -S |
| | 5 2 | 3+4 4 | 5 2 | 3+4 4 | 2+2 4 | 3+4 2 |
| | 5 2 | 3+4 4 | 5 2 | 3+4 4 | 2+2 4 | 3+4 2 |
| | 6 2 | 3+4 4 | 6 2 | 3+4 4 | 2+2 4 | 5+6 2 |
| | 6 2 | 3+4 4 | 6 2 | 3+4 4 | 2+2 4 | 5+6 2 |
| | 6 2 | 3+4 4 | 6 2 | 3+4 4 | 2+2 4 | 5+6 2 |
| | 6 2 | 3+4 4 | 6 2 | 3+4 4 | 2+2 4 | 5+6 2 |
| | 6 2 | 3+4 4 | 6 2 | 3+4 4 | 2+2 4 | 5+6 2 |
| | 6 2 | 3+4 4 | 6 2 | 3+4 4 | 2+2 4 | 5+6 2 |
| | 5 2 | 3+4 4 | 6 2 | 3+4 4 | 2+2 4 | 3+4 2 |
| | 6 2 | 3+4 4 | 5 2 | 3+4 4 | 2+2 4 | 5+6 2 |
| | 6 2 | 3+4 4 | 5 2 | 3+4 4 | 2+2 4 | 3+4 2 |
| | 5 2 | 3+4 4 | 5 2 | 3+4 4 | 2+2 4 | 3+4 2 |
| | 5 2 | 3+4 4 | 6 2 | 3+4 4 | 2+2 4 | 5+6 2 |
| | 6 2 | 3+4 4 | 5 2 | 3+4 4 | 2+2 4 | 3+4 2 |
| | 5 2 | | 6 2 | | | 5+6 2 |
| | 6 2 | | 6 2 | | | 3+4 2 |
| | 5 2 | | 5 2 | | | 5+6 2 |
| | 5 2 | | 6 2 | | | 5+6 2 |
| | 6 2 | | 5 2 | | | 5+6 2 |
| | 5 2 | | 5 2 | | | 3+4 2 |
| | 6 2 | | 6 2 | | | 3+4 2 |
| | 5 2 | | 6 2 | | | 3+4 2 |
| | 6 2 | | 6 2 | | | 5+6 2 |
| | 6 2 | | 5 2 | | | 3+4 2 |
| | 6 2 | | 5 2 | | | 5+6 2 |
| | 5 2 | | 5 2 | | | 3+4 2 |
| | 5 2 | | 6 2 | | | 5+6 2 |
| | 6 2 | | 5 2 | | | 3+4 2 |
| 63.2.18 | 63084 | | 184 | | Nd3O7Ru1 | |
| | -GM | -T | -Y | -Z | -R | -S |
| | 5 2 | 3+4 4 | 5 2 | 3+4 4 | 2+2 4 | 3+4 2 |
| | 5 2 | 3+4 4 | 5 2 | 3+4 4 | 2+2 4 | 3+4 2 |

Continued on next column

**TABLE XXXVIII– continued**

| TQC | ICSD | | NoE | | SF | |
|---|---|---|---|---|---|---|
| | 5 2 | 3+4 4 | 6 2 | 3+4 4 | 2+2 4 | 5+6 2 |
| | 6 2 | 3+4 4 | 5 2 | 3+4 4 | 2+2 4 | 3+4 2 |
| | 5 2 | 3+4 4 | 6 2 | 3+4 4 | 2+2 4 | 3+4 2 |
| | 6 2 | 3+4 4 | 5 2 | 3+4 4 | 2+2 4 | 5+6 2 |
| | 6 2 | 3+4 4 | 6 2 | 3+4 4 | 2+2 4 | 5+6 2 |
| | 5 2 | 3+4 4 | 5 2 | 3+4 4 | 2+2 4 | 5+6 2 |
| | 6 2 | 3+4 4 | 6 2 | 3+4 4 | 2+2 4 | 3+4 2 |
| | 5 2 | 3+4 4 | 6 2 | 3+4 4 | 2+2 4 | 5+6 2 |
| | 6 2 | 3+4 4 | 6 2 | 3+4 4 | 2+2 4 | 3+4 2 |
| | 6 2 | 3+4 4 | 5 2 | 3+4 4 | 2+2 4 | 5+6 2 |
| | 5 2 | 3+4 4 | 5 2 | 3+4 4 | 2+2 4 | 3+4 2 |
| | 5 2 | 3+4 4 | 6 2 | 3+4 4 | 2+2 4 | 5+6 2 |
| | 6 2 | 3+4 4 | 5 2 | 3+4 4 | 2+2 4 | 5+6 2 |
| | 6 2 | 3+4 4 | 6 2 | 3+4 4 | 2+2 4 | 3+4 2 |
| | 5 2 | 3+4 4 | 6 2 | 3+4 4 | 2+2 4 | 5+6 2 |
| | 5 2 | 3+4 4 | 5 2 | 3+4 4 | 2+2 4 | 5+6 2 |
| | 6 2 | 3+4 4 | 6 2 | 3+4 4 | 2+2 4 | 3+4 2 |
| | 5 2 | 3+4 4 | 5 2 | 3+4 4 | 2+2 4 | 5+6 2 |
| | 6 2 | 3+4 4 | 5 2 | 3+4 4 | 2+2 4 | 5+6 2 |
| | 5 2 | 3+4 4 | 6 2 | 3+4 4 | 2+2 4 | 3+4 2 |
| | 6 2 | 3+4 4 | 6 2 | 3+4 4 | 2+2 4 | 3+4 2 |
| | 6 2 | 3+4 4 | 5 2 | 3+4 4 | 2+2 4 | 5+6 2 |
| | 5 2 | 3+4 4 | 6 2 | 3+4 4 | 2+2 4 | 3+4 2 |
| | 6 2 | 3+4 4 | 5 2 | 3+4 4 | 2+2 4 | 3+4 2 |
| | 6 2 | 3+4 4 | 6 2 | 3+4 4 | 2+2 4 | 5+6 2 |
| | 5 2 | 3+4 4 | 6 2 | 3+4 4 | 2+2 4 | 5+6 2 |
| | 6 2 | 3+4 4 | 5 2 | 3+4 4 | 2+2 4 | 5+6 2 |
| | 6 2 | 3+4 4 | 6 2 | 3+4 4 | 2+2 4 | 5+6 2 |
| | 6 2 | 3+4 4 | 6 2 | 3+4 4 | 2+2 4 | 3+4 2 |
| | 6 2 | 3+4 4 | 5 2 | 3+4 4 | 2+2 4 | 5+6 2 |
| | 5 2 | 3+4 4 | 5 2 | 3+4 4 | 2+2 4 | 3+4 2 |
| | 5 2 | 3+4 4 | 6 2 | 3+4 4 | 2+2 4 | 5+6 2 |
| | 6 2 | 3+4 4 | 6 2 | 3+4 4 | 2+2 4 | 3+4 2 |
| | 6 2 | 3+4 4 | 6 2 | 3+4 4 | 2+2 4 | 3+4 2 |
| | 5 2 | 3+4 4 | 5 2 | 3+4 4 | 2+2 4 | 5+6 2 |
| | 6 2 | 3+4 4 | 5 2 | 3+4 4 | 2+2 4 | 5+6 2 |
| | 5 2 | 3+4 4 | 6 2 | 3+4 4 | 2+2 4 | 5+6 2 |
| | 5 2 | 3+4 4 | 5 2 | 3+4 4 | 2+2 4 | 5+6 2 |
| | 5 2 | 3+4 4 | 5 2 | 3+4 4 | 2+2 4 | 5+6 2 |
| | 5 2 | 3+4 4 | 5 2 | 3+4 4 | 2+2 4 | 5+6 2 |
| | 5 2 | 3+4 4 | 5 2 | 3+4 4 | 2+2 4 | 5+6 2 |
| | 5 2 | 3+4 4 | 5 2 | 3+4 4 | 2+2 4 | 5+6 2 |
| | 5 2 | 3+4 4 | 5 2 | 3+4 4 | 2+2 4 | 5+6 2 |
| | 5 2 | 3+4 4 | 5 2 | 3+4 4 | 2+2 4 | 5+6 2 |
| | 5 2 | | 6 2 | | | 3+4 2 |
| | 5 2 | | 5 2 | | | 5+6 2 |
| | 5 2 | | 6 2 | | | 3+4 2 |
| | 6 2 | | 5 2 | | | 3+4 2 |
| | 5 2 | | 5 2 | | | 3+4 2 |
| | 6 2 | | 6 2 | | | 3+4 2 |
| | 6 2 | | 6 2 | | | 3+4 2 |
| | 6 2 | | 5 2 | | | 3+4 2 |
| | 5 2 | | 5 2 | | | 5+6 2 |
| | 6 2 | | 5 2 | | | 5+6 2 |
| | 5 2 | | 6 2 | | | 5+6 2 |
| | 6 2 | | 6 2 | | | 3+4 2 |
| | 6 2 | | 5 2 | | | 3+4 2 |
| | 5 2 | | 6 2 | | | 3+4 2 |
| | 5 2 | | 6 2 | | | 5+6 2 |
| | 6 2 | | 5 2 | | | 3+4 2 |
| | 5 2 | | 6 2 | | | 3+4 2 |

Continued on next column

**TABLE XXXVIII– continued**

| TQC | ICSD | | NoE | | SF | |
|---|---|---|---|---|---|---|
| | 6 2 | | 6 2 | | | 5+6 2 |
| | 6 2 | | 6 2 | | | 5+6 2 |
| | 6 2 | | 6 2 | | | 3+4 2 |
| | 6 2 | | 5 2 | | | 5+6 2 |
| | 6 2 | | 6 2 | | | 5+6 2 |
| | 6 2 | | 6 2 | | | 3+4 2 |
| | 5 2 | | 6 2 | | | 3+4 2 |
| | 5 2 | | 6 2 | | | 3+4 2 |
| | 6 2 | | 6 2 | | | 3+4 2 |
| | 6 2 | | 5 2 | | | 5+6 2 |
| | 6 2 | | 5 2 | | | 3+4 2 |
| | 5 2 | | 6 2 | | | 5+6 2 |
| | 6 2 | | 5 2 | | | 3+4 2 |
| | 5 2 | | 6 2 | | | 5+6 2 |
| | 6 2 | | 6 2 | | | 3+4 2 |
| | 6 2 | | 6 2 | | | 5+6 2 |
| | 6 2 | | 5 2 | | | 3+4 2 |
| | 5 2 | | 5 2 | | | 5+6 2 |
| | 5 2 | | 5 2 | | | 5+6 2 |
| | 5 2 | | 5 2 | | | 5+6 2 |
| | 5 2 | | 5 2 | | | 5+6 2 |
| | 5 2 | | 5 2 | | | 5+6 2 |
| | 6 2 | | 6 2 | | | 5+6 2 |
| | 6 2 | | 6 2 | | | 5+6 2 |
| | 6 2 | | 6 2 | | | 3+4 2 |
| | 6 2 | | 5 2 | | | 5+6 2 |
| | 5 2 | | 6 2 | | | 3+4 2 |
| | 6 2 | | 6 2 | | | 5+6 2 |
| | 5 2 | | 6 2 | | | 5+6 2 |
| 63.2.19 | 190762 | | 184 | | Nd3O7Ru1 | |
| | -GM | -T | -Y | -Z | -R | -S |
| | 5 2 | 3+4 4 | 5 2 | 3+4 4 | 2+2 4 | 3+4 2 |
| | 5 2 | 3+4 4 | 5 2 | 3+4 4 | 2+2 4 | 3+4 2 |
| | 5 2 | 3+4 4 | 6 2 | 3+4 4 | 2+2 4 | 5+6 2 |
| | 6 2 | 3+4 4 | 5 2 | 3+4 4 | 2+2 4 | 3+4 2 |
| | 5 2 | 3+4 4 | 6 2 | 3+4 4 | 2+2 4 | 3+4 2 |
| | 6 2 | 3+4 4 | 5 2 | 3+4 4 | 2+2 4 | 5+6 2 |
| | 6 2 | 3+4 4 | 6 2 | 3+4 4 | 2+2 4 | 5+6 2 |
| | 5 2 | 3+4 4 | 5 2 | 3+4 4 | 2+2 4 | 5+6 2 |
| | 6 2 | 3+4 4 | 6 2 | 3+4 4 | 2+2 4 | 3+4 2 |
| | 5 2 | 3+4 4 | 6 2 | 3+4 4 | 2+2 4 | 5+6 2 |
| | 6 2 | 3+4 4 | 5 2 | 3+4 4 | 2+2 4 | 3+4 2 |
| | 6 2 | 3+4 4 | 6 2 | 3+4 4 | 2+2 4 | 5+6 2 |
| | 5 2 | 3+4 4 | 5 2 | 3+4 4 | 2+2 4 | 5+6 2 |
| | 5 2 | 3+4 4 | 5 2 | 3+4 4 | 2+2 4 | 3+4 2 |
| | 6 2 | 3+4 4 | 6 2 | 3+4 4 | 2+2 4 | 5+6 2 |
| | 6 2 | 3+4 4 | 6 2 | 3+4 4 | 2+2 4 | 3+4 2 |
| | 5 2 | 3+4 4 | 6 2 | 3+4 4 | 2+2 4 | 5+6 2 |
| | 6 2 | 3+4 4 | 5 2 | 3+4 4 | 2+2 4 | 5+6 2 |
| | 5 2 | 3+4 4 | 5 2 | 3+4 4 | 2+2 4 | 3+4 2 |
| | 5 2 | 3+4 4 | 6 2 | 3+4 4 | 2+2 4 | 5+6 2 |
| | 6 2 | 3+4 4 | 5 2 | 3+4 4 | 2+2 4 | 5+6 2 |
| | 5 2 | 3+4 4 | 6 2 | 3+4 4 | 2+2 4 | 3+4 2 |
| | 6 2 | 3+4 4 | 5 2 | 3+4 4 | 2+2 4 | 5+6 2 |
| | 6 2 | 3+4 4 | 6 2 | 3+4 4 | 2+2 4 | 3+4 2 |
| | 5 2 | 3+4 4 | 6 2 | 3+4 4 | 2+2 4 | 3+4 2 |
| | 6 2 | 3+4 4 | 5 2 | 3+4 4 | 2+2 4 | 3+4 2 |
| | 5 2 | 3+4 4 | 6 2 | 3+4 4 | 2+2 4 | 5+6 2 |
| | 6 2 | 3+4 4 | 6 2 | 3+4 4 | 2+2 4 | 5+6 2 |
| | 6 2 | 3+4 4 | 5 2 | 3+4 4 | 2+2 4 | 5+6 2 |
| | 6 2 | 3+4 4 | 6 2 | 3+4 4 | 2+2 4 | 5+6 2 |

Continued on next column

**TABLE XXXVIII– continued**

| TQC | ICSD | | NoE | | SF | |
|---|---|---|---|---|---|---|
| | 6 2 | 3+4 4 | 6 2 | 3+4 4 | 2+2 4 | 3+4 2 |
| | 6 2 | 3+4 4 | 5 2 | 3+4 4 | 2+2 4 | 5+6 2 |
| | 5 2 | 3+4 4 | 5 2 | 3+4 4 | 2+2 4 | 3+4 2 |
| | 5 2 | 3+4 4 | 6 2 | 3+4 4 | 2+2 4 | 5+6 2 |
| | 6 2 | 3+4 4 | 6 2 | 3+4 4 | 2+2 4 | 3+4 2 |
| | 6 2 | 3+4 4 | 6 2 | 3+4 4 | 2+2 4 | 3+4 2 |
| | 5 2 | 3+4 4 | 5 2 | 3+4 4 | 2+2 4 | 5+6 2 |
| | 6 2 | 3+4 4 | 6 2 | 3+4 4 | 2+2 4 | 5+6 2 |
| | 5 2 | 3+4 4 | 5 2 | 3+4 4 | 2+2 4 | 5+6 2 |
| | 5 2 | 3+4 4 | 5 2 | 3+4 4 | 2+2 4 | 5+6 2 |
| | 5 2 | 3+4 4 | 5 2 | 3+4 4 | 2+2 4 | 5+6 2 |
| | 5 2 | 3+4 4 | 5 2 | 3+4 4 | 2+2 4 | 5+6 2 |
| | 5 2 | 3+4 4 | 5 2 | 3+4 4 | 2+2 4 | 5+6 2 |
| | 5 2 | 3+4 4 | 5 2 | 3+4 4 | 2+2 4 | 5+6 2 |
| | 5 2 | 3+4 4 | 5 2 | 3+4 4 | 2+2 4 | 5+6 2 |
| | 5 2 | 3+4 4 | 5 2 | 3+4 4 | 2+2 4 | 5+6 2 |
| | 5 2 | | 5 2 | | | 5+6 2 |
| | 5 2 | | 6 2 | | | 3+4 2 |
| | 5 2 | | 6 2 | | | 3+4 2 |
| | 6 2 | | 5 2 | | | 3+4 2 |
| | 5 2 | | 5 2 | | | 3+4 2 |
| | 6 2 | | 6 2 | | | 3+4 2 |
| | 6 2 | | 6 2 | | | 3+4 2 |
| | 6 2 | | 5 2 | | | 3+4 2 |
| | 5 2 | | 5 2 | | | 5+6 2 |
| | 5 2 | | 5 2 | | | 5+6 2 |
| | 6 2 | | 6 2 | | | 5+6 2 |
| | 6 2 | | 6 2 | | | 3+4 2 |
| | 6 2 | | 5 2 | | | 5+6 2 |
| | 5 2 | | 6 2 | | | 3+4 2 |
| | 6 2 | | 5 2 | | | 3+4 2 |
| | 5 2 | | 6 2 | | | 3+4 2 |
| | 5 2 | | 6 2 | | | 3+4 2 |
| | 6 2 | | 6 2 | | | 5+6 2 |
| | 6 2 | | 6 2 | | | 5+6 2 |
| | 6 2 | | 6 2 | | | 3+4 2 |
| | 6 2 | | 5 2 | | | 5+6 2 |
| | 6 2 | | 6 2 | | | 5+6 2 |
| | 6 2 | | 6 2 | | | 3+4 2 |
| | 5 2 | | 6 2 | | | 3+4 2 |
| | 5 2 | | 6 2 | | | 3+4 2 |
| | 6 2 | | 6 2 | | | 3+4 2 |
| | 6 2 | | 6 2 | | | 5+6 2 |
| | 6 2 | | 5 2 | | | 3+4 2 |
| | 6 2 | | 5 2 | | | 5+6 2 |
| | 5 2 | | 5 2 | | | 3+4 2 |
| | 6 2 | | 6 2 | | | 3+4 2 |
| | 5 2 | | 6 2 | | | 5+6 2 |
| | 6 2 | | 6 2 | | | 3+4 2 |
| | 6 2 | | 5 2 | | | 5+6 2 |
| | 5 2 | | 5 2 | | | 5+6 2 |
| | 5 2 | | 5 2 | | | 5+6 2 |
| | 5 2 | | 5 2 | | | 5+6 2 |
| | 5 2 | | 5 2 | | | 5+6 2 |
| | 5 2 | | 5 2 | | | 5+6 2 |
| | 6 2 | | 6 2 | | | 5+6 2 |
| | 6 2 | | 6 2 | | | 5+6 2 |
| | 6 2 | | 6 2 | | | 3+4 2 |
| | 6 2 | | 6 2 | | | 5+6 2 |
| | 5 2 | | 5 2 | | | 3+4 2 |
| | 6 2 | | 6 2 | | | 5+6 2 |

Continued on next column

**TABLE XXXVIII– continued**

| TQC | ICSD | NoE | SF | | | |
|---|---|---|---|---|---|---|
| | 5 2 | | 6 2 | | | 5+6 2 |
| 63.2.20 | 71692 | 64 | Nd1Ni1Si2 | | | |
| | -GM | -T | -Y | -Z | -R | -S |
| | 5 2 | 3+4 4 | 5 2 | 3+4 4 | 2+2 4 | 3+4 2 |
| | 6 2 | 3+4 4 | 6 2 | 3+4 4 | 2+2 4 | 5+6 2 |
| | 5 2 | 3+4 4 | 5 2 | 3+4 4 | 2+2 4 | 3+4 2 |
| | 6 2 | 3+4 4 | 6 2 | 3+4 4 | 2+2 4 | 5+6 2 |
| | 5 2 | 3+4 4 | 5 2 | 3+4 4 | 2+2 4 | 3+4 2 |
| | 6 2 | 3+4 4 | 6 2 | 3+4 4 | 2+2 4 | 5+6 2 |
| | 5 2 | 3+4 4 | 5 2 | 3+4 4 | 2+2 4 | 3+4 2 |
| | 6 2 | 3+4 4 | 6 2 | 3+4 4 | 2+2 4 | 5+6 2 |
| | 5 2 | 3+4 4 | 6 2 | 3+4 4 | 2+2 4 | 5+6 2 |
| | 5 2 | 3+4 4 | 5 2 | 3+4 4 | 2+2 4 | 3+4 2 |
| | 6 2 | 3+4 4 | 6 2 | 3+4 4 | 2+2 4 | 5+6 2 |
| | 6 2 | 3+4 4 | 5 2 | 3+4 4 | 2+2 4 | 3+4 2 |
| | 5 2 | 3+4 4 | 5 2 | 3+4 4 | 2+2 4 | 3+4 2 |
| | 5 2 | 3+4 4 | 6 2 | 3+4 4 | 2+2 4 | 5+6 2 |
| | 6 2 | 3+4 4 | 6 2 | 3+4 4 | 2+2 4 | 5+6 2 |
| | 5 2 | 3+4 4 | 6 2 | 3+4 4 | 2+2 4 | 3+4 2 |
| | 5 2 | | 5 2 | | | 3+4 2 |
| | 5 2 | | 5 2 | | | 5+6 2 |
| | 5 2 | | 6 2 | | | 5+6 2 |
| | 6 2 | | 5 2 | | | 5+6 2 |
| | 6 2 | | 6 2 | | | 3+4 2 |
| | 6 2 | | 5 2 | | | 5+6 2 |
| | 5 2 | | 5 2 | | | 3+4 2 |
| | 6 2 | | 6 2 | | | 5+6 2 |
| | 5 2 | | 6 2 | | | 3+4 2 |
| | 6 2 | | 6 2 | | | 3+4 2 |
| | 5 2 | | 6 2 | | | 5+6 2 |
| | 6 2 | | 5 2 | | | 3+4 2 |
| | 5 2 | | 5 2 | | | 5+6 2 |
| | 6 2 | | 5 2 | | | 5+6 2 |
| | 6 2 | | 6 2 | | | 3+4 2 |
| | 5 2 | | 5 2 | | | 5+6 2 |
| 63.2.21 | 71691 | 64 | Nd1Ni1Si2 | | | |
| | -GM | -T | -Y | -Z | -R | -S |
| | 5 2 | 3+4 4 | 5 2 | 3+4 4 | 2+2 4 | 5+6 2 |
| | 6 2 | 3+4 4 | 6 2 | 3+4 4 | 2+2 4 | 3+4 2 |
| | 5 2 | 3+4 4 | 5 2 | 3+4 4 | 2+2 4 | 5+6 2 |
| | 6 2 | 3+4 4 | 6 2 | 3+4 4 | 2+2 4 | 3+4 2 |
| | 5 2 | 3+4 4 | 5 2 | 3+4 4 | 2+2 4 | 5+6 2 |
| | 6 2 | 3+4 4 | 6 2 | 3+4 4 | 2+2 4 | 3+4 2 |
| | 5 2 | 3+4 4 | 5 2 | 3+4 4 | 2+2 4 | 5+6 2 |
| | 6 2 | 3+4 4 | 6 2 | 3+4 4 | 2+2 4 | 3+4 2 |
| | 5 2 | 3+4 4 | 6 2 | 3+4 4 | 2+2 4 | 3+4 2 |
| | 5 2 | 3+4 4 | 5 2 | 3+4 4 | 2+2 4 | 5+6 2 |
| | 6 2 | 3+4 4 | 6 2 | 3+4 4 | 2+2 4 | 3+4 2 |
| | 6 2 | 3+4 4 | 5 2 | 3+4 4 | 2+2 4 | 5+6 2 |
| | 5 2 | 3+4 4 | 5 2 | 3+4 4 | 2+2 4 | 5+6 2 |
| | 5 2 | 3+4 4 | 6 2 | 3+4 4 | 2+2 4 | 3+4 2 |
| | 6 2 | 3+4 4 | 6 2 | 3+4 4 | 2+2 4 | 3+4 2 |
| | 5 2 | 3+4 4 | 6 2 | 3+4 4 | 2+2 4 | 5+6 2 |
| | 5 2 | | 5 2 | | | 5+6 2 |
| | 5 2 | | 5 2 | | | 3+4 2 |
| | 6 2 | | 6 2 | | | 3+4 2 |
| | 5 2 | | 5 2 | | | 3+4 2 |
| | 6 2 | | 6 2 | | | 5+6 2 |
| | 6 2 | | 5 2 | | | 3+4 2 |
| | 5 2 | | 5 2 | | | 5+6 2 |
| | 6 2 | | 6 2 | | | 3+4 2 |

Continued on next column

**TABLE XXXVIII– continued**

| TQC | ICSD | NoE | SF | | | |
|---|---|---|---|---|---|---|
| | 5 2 | | 6 2 | | | 5+6 2 |
| | 6 2 | | 6 2 | | | 5+6 2 |
| | 5 2 | | 6 2 | | | 3+4 2 |
| | 6 2 | | 5 2 | | | 5+6 2 |
| | 5 2 | | 5 2 | | | 3+4 2 |
| | 6 2 | | 5 2 | | | 3+4 2 |
| | 6 2 | | 6 2 | | | 5+6 2 |
| | 5 2 | | 5 2 | | | 3+4 2 |
| 63.2.22 | 425493 | 48 | Mg2Pd1Sr1 | | | |
| | -GM | -T | -Y | -Z | -R | -S |
| | 5 2 | 3+4 4 | 5 2 | 3+4 4 | 2+2 4 | 3+4 2 |
| | 6 2 | 3+4 4 | 6 2 | 3+4 4 | 2+2 4 | 5+6 2 |
| | 5 2 | 3+4 4 | 5 2 | 3+4 4 | 2+2 4 | 3+4 2 |
| | 6 2 | 3+4 4 | 6 2 | 3+4 4 | 2+2 4 | 5+6 2 |
| | 5 2 | 3+4 4 | 5 2 | 3+4 4 | 2+2 4 | 3+4 2 |
| | 6 2 | 3+4 4 | 6 2 | 3+4 4 | 2+2 4 | 5+6 2 |
| | 6 2 | 3+4 4 | 6 2 | 3+4 4 | 2+2 4 | 3+4 2 |
| | 5 2 | 3+4 4 | 5 2 | 3+4 4 | 2+2 4 | 5+6 2 |
| | 5 2 | 3+4 4 | 6 2 | 3+4 4 | 2+2 4 | 3+4 2 |
| | 6 2 | 3+4 4 | 5 2 | 3+4 4 | 2+2 4 | 5+6 2 |
| | 5 2 | 3+4 4 | 6 2 | 3+4 4 | 2+2 4 | 5+6 2 |
| | 5 2 | 3+4 4 | 5 2 | 3+4 4 | 2+2 4 | 3+4 2 |
| | 6 2 | | 6 2 | | | 5+6 2 |
| | 5 2 | | 5 2 | | | 3+4 2 |
| | 5 2 | | 6 2 | | | 5+6 2 |
| | 6 2 | | 6 2 | | | 3+4 2 |
| | 6 2 | | 5 2 | | | 5+6 2 |
| | 5 2 | | 5 2 | | | 3+4 2 |
| | 6 2 | | 6 2 | | | 3+4 2 |
| | 5 2 | | 5 2 | | | 5+6 2 |
| | 6 2 | | 6 2 | | | 5+6 2 |
| | 6 2 | | 5 2 | | | 5+6 2 |
| | 5 2 | | 6 2 | | | 3+4 2 |
| | 5 2 | | 5 2 | | | 3+4 2 |
| 63.2.23 | 163453 | 52 | Ir1Zr1 | | | |
| | -GM | -T | -Y | -Z | -R | -S |
| | 5 2 | 3+4 4 | 5 2 | 3+4 4 | 2+2 4 | 3+4 2 |
| | 6 2 | 3+4 4 | 6 2 | 3+4 4 | 2+2 4 | 5+6 2 |
| | 5 2 | 3+4 4 | 6 2 | 3+4 4 | 2+2 4 | 3+4 2 |
| | 6 2 | 3+4 4 | 5 2 | 3+4 4 | 2+2 4 | 5+6 2 |
| | 5 2 | 3+4 4 | 6 2 | 3+4 4 | 2+2 4 | 3+4 2 |
| | 6 2 | 3+4 4 | 5 2 | 3+4 4 | 2+2 4 | 5+6 2 |
| | 5 2 | 3+4 4 | 5 2 | 3+4 4 | 2+2 4 | 3+4 2 |
| | 6 2 | 3+4 4 | 5 2 | 3+4 4 | 2+2 4 | 5+6 2 |
| | 5 2 | 3+4 4 | 6 2 | 3+4 4 | 2+2 4 | 5+6 2 |
| | 6 2 | 3+4 4 | 6 2 | 3+4 4 | 2+2 4 | 3+4 2 |
| | 6 2 | 3+4 4 | 6 2 | 3+4 4 | 2+2 4 | 5+6 2 |
| | 6 2 | 3+4 4 | 6 2 | 3+4 4 | 2+2 4 | 3+4 2 |
| | 5 2 | 3+4 4 | 5 2 | 3+4 4 | 2+2 4 | 5+6 2 |
| | 5 2 | | 6 2 | | | 3+4 2 |
| | 5 2 | | 6 2 | | | 5+6 2 |
| | 5 2 | | 5 2 | | | 3+4 2 |
| | 6 2 | | 5 2 | | | 5+6 2 |
| | 5 2 | | 5 2 | | | 3+4 2 |
| | 6 2 | | 5 2 | | | 3+4 2 |
| | 5 2 | | 5 2 | | | 5+6 2 |
| | 6 2 | | 6 2 | | | 3+4 2 |
| | 5 2 | | 6 2 | | | 3+4 2 |
| | 6 2 | | 5 2 | | | 5+6 2 |
| | 6 2 | | 6 2 | | | 3+4 2 |
| | 5 2 | | 5 2 | | | 5+6 2 |

Continued on next column

**TABLE XXXVIII– continued**

| TQC | ICSD | | NoE | | SF | |
|---|---|---|---|---|---|---|
| | 6 2 | | 5 2 | | | 3+4 2 |
| 63.2.24 | 412908 | | 92 | | In4Pt1Yb1 | |
| | -GM | -T | -Y | -Z | -R | -S |
| | 5 2 | 3+4 4 | 5 2 | 3+4 4 | 2+2 4 | 3+4 2 |
| | 6 2 | 3+4 4 | 6 2 | 3+4 4 | 2+2 4 | 5+6 2 |
| | 5 2 | 3+4 4 | 5 2 | 3+4 4 | 2+2 4 | 3+4 2 |
| | 6 2 | 3+4 4 | 6 2 | 3+4 4 | 2+2 4 | 5+6 2 |
| | 5 2 | 3+4 4 | 5 2 | 3+4 4 | 2+2 4 | 3+4 2 |
| | 6 2 | 3+4 4 | 6 2 | 3+4 4 | 2+2 4 | 5+6 2 |
| | 5 2 | 3+4 4 | 6 2 | 3+4 4 | 2+2 4 | 3+4 2 |
| | 6 2 | 3+4 4 | 5 2 | 3+4 4 | 2+2 4 | 5+6 2 |
| | 5 2 | 3+4 4 | 6 2 | 3+4 4 | 2+2 4 | 3+4 2 |
| | 6 2 | 3+4 4 | 5 2 | 3+4 4 | 2+2 4 | 5+6 2 |
| | 5 2 | 3+4 4 | 6 2 | 3+4 4 | 2+2 4 | 5+6 2 |
| | 5 2 | 3+4 4 | 5 2 | 3+4 4 | 2+2 4 | 3+4 2 |
| | 6 2 | 3+4 4 | 5 2 | 3+4 4 | 2+2 4 | 3+4 2 |
| | 5 2 | 3+4 4 | 6 2 | 3+4 4 | 2+2 4 | 5+6 2 |
| | 6 2 | 3+4 4 | 5 2 | 3+4 4 | 2+2 4 | 3+4 2 |
| | 6 2 | 3+4 4 | 6 2 | 3+4 4 | 2+2 4 | 5+6 2 |
| | 5 2 | 3+4 4 | 6 2 | 3+4 4 | 2+2 4 | 3+4 2 |
| | 5 2 | 3+4 4 | 5 2 | 3+4 4 | 2+2 4 | 3+4 2 |
| | 5 2 | 3+4 4 | 5 2 | 3+4 4 | 2+2 4 | 5+6 2 |
| | 6 2 | 3+4 4 | 6 2 | 3+4 4 | 2+2 4 | 3+4 2 |
| | 6 2 | 3+4 4 | 6 2 | 3+4 4 | 2+2 4 | 5+6 2 |
| | 5 2 | 3+4 4 | 5 2 | 3+4 4 | 2+2 4 | 5+6 2 |
| | 6 2 | 3+4 4 | 6 2 | 3+4 4 | 2+2 4 | 3+4 2 |
| | 5 2 | | 5 2 | | | 3+4 2 |
| | 6 2 | | 5 2 | | | 5+6 2 |
| | 5 2 | | 6 2 | | | 5+6 2 |
| | 6 2 | | 6 2 | | | 3+4 2 |
| | 5 2 | | 5 2 | | | 5+6 2 |
| | 6 2 | | 6 2 | | | 5+6 2 |
| | 5 2 | | 5 2 | | | 5+6 2 |
| | 6 2 | | 5 2 | | | 3+4 2 |
| | 5 2 | | 6 2 | | | 3+4 2 |
| | 6 2 | | 6 2 | | | 5+6 2 |
| | 5 2 | | 6 2 | | | 3+4 2 |
| | 6 2 | | 6 2 | | | 5+6 2 |
| | 5 2 | | 5 2 | | | 3+4 2 |
| | 6 2 | | 5 2 | | | 5+6 2 |
| | 5 2 | | 5 2 | | | 3+4 2 |
| | 6 2 | | 5 2 | | | 5+6 2 |
| | 6 2 | | 6 2 | | | 5+6 2 |
| | 5 2 | | 6 2 | | | 3+4 2 |
| | 5 2 | | 6 2 | | | 5+6 2 |
| | 6 2 | | 6 2 | | | 3+4 2 |
| | 5 2 | | 5 2 | | | 5+6 2 |
| | 6 2 | | 5 2 | | | 5+6 2 |
| | 6 2 | | 6 2 | | | 3+4 2 |
| 63.2.25 | 418180 | | 64 | | In4Pt1Sr1 | |
| | -GM | -T | -Y | -Z | -R | -S |
| | 5 2 | 3+4 4 | 5 2 | 3+4 4 | 2+2 4 | 3+4 2 |
| | 6 2 | 3+4 4 | 6 2 | 3+4 4 | 2+2 4 | 5+6 2 |
| | 5 2 | 3+4 4 | 6 2 | 3+4 4 | 2+2 4 | 3+4 2 |
| | 6 2 | 3+4 4 | 5 2 | 3+4 4 | 2+2 4 | 5+6 2 |
| | 5 2 | 3+4 4 | 6 2 | 3+4 4 | 2+2 4 | 3+4 2 |
| | 6 2 | 3+4 4 | 5 2 | 3+4 4 | 2+2 4 | 5+6 2 |
| | 5 2 | 3+4 4 | 6 2 | 3+4 4 | 2+2 4 | 3+4 2 |
| | 6 2 | 3+4 4 | 5 2 | 3+4 4 | 2+2 4 | 5+6 2 |
| | 5 2 | 3+4 4 | 6 2 | 3+4 4 | 2+2 4 | 5+6 2 |
| | 6 2 | 3+4 4 | 5 2 | 3+4 4 | 2+2 4 | 3+4 2 |

Continued on next column

**TABLE XXXVIII– continued**

| TQC | ICSD | | NoE | | SF | |
|---|---|---|---|---|---|---|
| | 5 2 | 3+4 4 | 5 2 | 3+4 4 | 2+2 4 | 3+4 2 |
| | 5 2 | 3+4 4 | 6 2 | 3+4 4 | 2+2 4 | 5+6 2 |
| | 6 2 | 3+4 4 | 6 2 | 3+4 4 | 2+2 4 | 5+6 2 |
| | 5 2 | 3+4 4 | 5 2 | 3+4 4 | 2+2 4 | 3+4 2 |
| | 6 2 | 3+4 4 | 5 2 | 3+4 4 | 2+2 4 | 5+6 2 |
| | 5 2 | 3+4 4 | 6 2 | 3+4 4 | 2+2 4 | 3+4 2 |
| | 6 2 | | 5 2 | | | 5+6 2 |
| | 5 2 | | 6 2 | | | 5+6 2 |
| | 5 2 | | 6 2 | | | 3+4 2 |
| | 6 2 | | 5 2 | | | 5+6 2 |
| | 6 2 | | 5 2 | | | 3+4 2 |
| | 5 2 | | 6 2 | | | 3+4 2 |
| | 6 2 | | 6 2 | | | 5+6 2 |
| | 5 2 | | 5 2 | | | 5+6 2 |
| | 6 2 | | 5 2 | | | 3+4 2 |
| | 5 2 | | 6 2 | | | 3+4 2 |
| | 6 2 | | 6 2 | | | 5+6 2 |
| | 5 2 | | 6 2 | | | 3+4 2 |
| | 6 2 | | 6 2 | | | 3+4 2 |
| | 5 2 | | 5 2 | | | 5+6 2 |
| | 6 2 | | 5 2 | | | 3+4 2 |
| | 6 2 | | 6 2 | | | 5+6 2 |
| 63.2.26 | 411398 | | 92 | | In4Pd1Yb1 | |
| | -GM | -T | -Y | -Z | -R | -S |
| | 5 2 | 3+4 4 | 5 2 | 3+4 4 | 2+2 4 | 3+4 2 |
| | 6 2 | 3+4 4 | 6 2 | 3+4 4 | 2+2 4 | 5+6 2 |
| | 5 2 | 3+4 4 | 5 2 | 3+4 4 | 2+2 4 | 3+4 2 |
| | 6 2 | 3+4 4 | 6 2 | 3+4 4 | 2+2 4 | 5+6 2 |
| | 5 2 | 3+4 4 | 5 2 | 3+4 4 | 2+2 4 | 3+4 2 |
| | 6 2 | 3+4 4 | 6 2 | 3+4 4 | 2+2 4 | 5+6 2 |
| | 5 2 | 3+4 4 | 5 2 | 3+4 4 | 2+2 4 | 3+4 2 |
| | 6 2 | 3+4 4 | 6 2 | 3+4 4 | 2+2 4 | 5+6 2 |
| | 5 2 | 3+4 4 | 6 2 | 3+4 4 | 2+2 4 | 5+6 2 |
| | 6 2 | 3+4 4 | 5 2 | 3+4 4 | 2+2 4 | 3+4 2 |
| | 5 2 | 3+4 4 | 6 2 | 3+4 4 | 2+2 4 | 3+4 2 |
| | 5 2 | 3+4 4 | 5 2 | 3+4 4 | 2+2 4 | 5+6 2 |
| | 5 2 | 3+4 4 | 6 2 | 3+4 4 | 2+2 4 | 5+6 2 |
| | 6 2 | 3+4 4 | 5 2 | 3+4 4 | 2+2 4 | 3+4 2 |
| | 6 2 | 3+4 4 | 5 2 | 3+4 4 | 2+2 4 | 5+6 2 |
| | 6 2 | 3+4 4 | 6 2 | 3+4 4 | 2+2 4 | 3+4 2 |
| | 5 2 | 3+4 4 | 6 2 | 3+4 4 | 2+2 4 | 5+6 2 |
| | 5 2 | 3+4 4 | 5 2 | 3+4 4 | 2+2 4 | 5+6 2 |
| | 5 2 | 3+4 4 | 6 2 | 3+4 4 | 2+2 4 | 3+4 2 |
| | 6 2 | 3+4 4 | 5 2 | 3+4 4 | 2+2 4 | 5+6 2 |
| | 6 2 | 3+4 4 | 6 2 | 3+4 4 | 2+2 4 | 3+4 2 |
| | 5 2 | 3+4 4 | 5 2 | 3+4 4 | 2+2 4 | 5+6 2 |
| | 6 2 | 3+4 4 | 6 2 | 3+4 4 | 2+2 4 | 3+4 2 |
| | 5 2 | | 5 2 | | | 5+6 2 |
| | 6 2 | | 5 2 | | | 3+4 2 |
| | 5 2 | | 6 2 | | | 3+4 2 |
| | 6 2 | | 6 2 | | | 3+4 2 |
| | 5 2 | | 6 2 | | | 5+6 2 |
| | 6 2 | | 5 2 | | | 3+4 2 |
| | 5 2 | | 5 2 | | | 5+6 2 |
| | 6 2 | | 5 2 | | | 3+4 2 |
| | 5 2 | | 6 2 | | | 5+6 2 |
| | 6 2 | | 6 2 | | | 3+4 2 |
| | 5 2 | | 6 2 | | | 5+6 2 |
| | 6 2 | | 6 2 | | | 3+4 2 |
| | 5 2 | | 5 2 | | | 5+6 2 |
| | 6 2 | | 5 2 | | | 3+4 2 |

Continued on next column

**TABLE XXXVIII– continued**

| TQC | ICSD | NoE | SF | | | |
|---|---|---|---|---|---|---|
| | 5 2 | | 5 2 | | | 5+6 2 |
| | 6 2 | | 5 2 | | | 3+4 2 |
| | 6 2 | | 6 2 | | | 3+4 2 |
| | 5 2 | | 6 2 | | | 5+6 2 |
| | 5 2 | | 6 2 | | | 3+4 2 |
| | 6 2 | | 6 2 | | | 3+4 2 |
| | 5 2 | | 5 2 | | | 5+6 2 |
| | 6 2 | | 5 2 | | | 3+4 2 |
| | 6 2 | | 6 2 | | | 5+6 2 |
| 63.2.27 | 418179 | 64 | In4Pd1Sr1 | | | |
| | -GM | -T | -Y | -Z | -R | -S |
| | 5 2 | 3+4 4 | 5 2 | 3+4 4 | 2+2 4 | 3+4 2 |
| | 6 2 | 3+4 4 | 6 2 | 3+4 4 | 2+2 4 | 5+6 2 |
| | 5 2 | 3+4 4 | 5 2 | 3+4 4 | 2+2 4 | 3+4 2 |
| | 6 2 | 3+4 4 | 6 2 | 3+4 4 | 2+2 4 | 5+6 2 |
| | 5 2 | 3+4 4 | 5 2 | 3+4 4 | 2+2 4 | 3+4 2 |
| | 6 2 | 3+4 4 | 6 2 | 3+4 4 | 2+2 4 | 5+6 2 |
| | 5 2 | 3+4 4 | 6 2 | 3+4 4 | 2+2 4 | 3+4 2 |
| | 6 2 | 3+4 4 | 5 2 | 3+4 4 | 2+2 4 | 5+6 2 |
| | 5 2 | 3+4 4 | 6 2 | 3+4 4 | 2+2 4 | 5+6 2 |
| | 6 2 | 3+4 4 | 5 2 | 3+4 4 | 2+2 4 | 3+4 2 |
| | 5 2 | 3+4 4 | 6 2 | 3+4 4 | 2+2 4 | 3+4 2 |
| | 5 2 | 3+4 4 | 5 2 | 3+4 4 | 2+2 4 | 5+6 2 |
| | 5 2 | 3+4 4 | 6 2 | 3+4 4 | 2+2 4 | 5+6 2 |
| | 6 2 | 3+4 4 | 5 2 | 3+4 4 | 2+2 4 | 3+4 2 |
| | 6 2 | 3+4 4 | 5 2 | 3+4 4 | 2+2 4 | 5+6 2 |
| | 6 2 | 3+4 4 | 6 2 | 3+4 4 | 2+2 4 | 3+4 2 |
| | 5 2 | | 5 2 | | | 5+6 2 |
| | 5 2 | | 6 2 | | | 5+6 2 |
| | 5 2 | | 6 2 | | | 3+4 2 |
| | 6 2 | | 5 2 | | | 5+6 2 |
| | 5 2 | | 6 2 | | | 3+4 2 |
| | 6 2 | | 5 2 | | | 5+6 2 |
| | 5 2 | | 6 2 | | | 3+4 2 |
| | 6 2 | | 5 2 | | | 5+6 2 |
| | 6 2 | | 5 2 | | | 3+4 2 |
| | 5 2 | | 6 2 | | | 3+4 2 |
| | 6 2 | | 6 2 | | | 5+6 2 |
| | 5 2 | | 6 2 | | | 3+4 2 |
| | 6 2 | | 5 2 | | | 3+4 2 |
| | 5 2 | | 6 2 | | | 5+6 2 |
| | 6 2 | | 5 2 | | | 3+4 2 |
| | 6 2 | | 6 2 | | | 5+6 2 |
| 63.2.28 | 415937 | 92 | In4Pb1Sr3 | | | |
| | -GM | -T | -Y | -Z | -R | -S |
| | 5 2 | 3+4 4 | 5 2 | 3+4 4 | 2+2 4 | 3+4 2 |
| | 6 2 | 3+4 4 | 6 2 | 3+4 4 | 2+2 4 | 5+6 2 |
| | 5 2 | 3+4 4 | 5 2 | 3+4 4 | 2+2 4 | 3+4 2 |
| | 5 2 | 3+4 4 | 5 2 | 3+4 4 | 2+2 4 | 3+4 2 |
| | 6 2 | 3+4 4 | 6 2 | 3+4 4 | 2+2 4 | 5+6 2 |
| | 6 2 | 3+4 4 | 6 2 | 3+4 4 | 2+2 4 | 5+6 2 |
| | 5 2 | 3+4 4 | 5 2 | 3+4 4 | 2+2 4 | 3+4 2 |
| | 6 2 | 3+4 4 | 6 2 | 3+4 4 | 2+2 4 | 5+6 2 |
| | 5 2 | 3+4 4 | 5 2 | 3+4 4 | 2+2 4 | 3+4 2 |
| | 6 2 | 3+4 4 | 5 2 | 3+4 4 | 2+2 4 | 3+4 2 |
| | 5 2 | 3+4 4 | 6 2 | 3+4 4 | 2+2 4 | 5+6 2 |
| | 6 2 | 3+4 4 | 6 2 | 3+4 4 | 2+2 4 | 5+6 2 |
| | 5 2 | 3+4 4 | 5 2 | 3+4 4 | 2+2 4 | 3+4 2 |
| | 6 2 | 3+4 4 | 6 2 | 3+4 4 | 2+2 4 | 5+6 2 |
| | 5 2 | 3+4 4 | 5 2 | 3+4 4 | 2+2 4 | 3+4 2 |
| | 6 2 | 3+4 4 | 6 2 | 3+4 4 | 2+2 4 | 5+6 2 |

Continued on next column

**TABLE XXXVIII– continued**

| TQC | ICSD | NoE | SF | | | |
|---|---|---|---|---|---|---|
| | 5 2 | 3+4 4 | 5 2 | 3+4 4 | 2+2 4 | 3+4 2 |
| | 6 2 | 3+4 4 | 6 2 | 3+4 4 | 2+2 4 | 5+6 2 |
| | 5 2 | 3+4 4 | 5 2 | 3+4 4 | 2+2 4 | 3+4 2 |
| | 6 2 | 3+4 4 | 6 2 | 3+4 4 | 2+2 4 | 5+6 2 |
| | 5 2 | 3+4 4 | 5 2 | 3+4 4 | 2+2 4 | 3+4 2 |
| | 5 2 | 3+4 4 | 6 2 | 3+4 4 | 2+2 4 | 3+4 2 |
| | 6 2 | 3+4 4 | 5 2 | 3+4 4 | 2+2 4 | 5+6 2 |
| | 6 2 | | 6 2 | | | 5+6 2 |
| | 5 2 | | 6 2 | | | 5+6 2 |
| | 6 2 | | 5 2 | | | 3+4 2 |
| | 5 2 | | 6 2 | | | 5+6 2 |
| | 5 2 | | 6 2 | | | 5+6 2 |
| | 6 2 | | 5 2 | | | 3+4 2 |
| | 6 2 | | 5 2 | | | 3+4 2 |
| | 6 2 | | 5 2 | | | 5+6 2 |
| | 5 2 | | 5 2 | | | 5+6 2 |
| | 5 2 | | 6 2 | | | 3+4 2 |
| | 6 2 | | 6 2 | | | 3+4 2 |
| | 5 2 | | 6 2 | | | 5+6 2 |
| | 5 2 | | 5 2 | | | 3+4 2 |
| | 6 2 | | 6 2 | | | 3+4 2 |
| | 5 2 | | 6 2 | | | 5+6 2 |
| | 5 2 | | 6 2 | | | 3+4 2 |
| | 5 2 | | 5 2 | | | 5+6 2 |
| | 6 2 | | 5 2 | | | 5+6 2 |
| | 5 2 | | 5 2 | | | 3+4 2 |
| | 6 2 | | 6 2 | | | 5+6 2 |
| | 5 2 | | 5 2 | | | 5+6 2 |
| | 6 2 | | 5 2 | | | 3+4 2 |
| | 6 2 | | 6 2 | | | 5+6 2 |
| 63.2.29 | 410400 | 52 | In2Pt1Sr1 | | | |
| | -GM | -T | -Y | -Z | -R | -S |
| | 5 2 | 3+4 4 | 5 2 | 3+4 4 | 2+2 4 | 3+4 2 |
| | 6 2 | 3+4 4 | 6 2 | 3+4 4 | 2+2 4 | 5+6 2 |
| | 5 2 | 3+4 4 | 5 2 | 3+4 4 | 2+2 4 | 3+4 2 |
| | 6 2 | 3+4 4 | 6 2 | 3+4 4 | 2+2 4 | 5+6 2 |
| | 5 2 | 3+4 4 | 5 2 | 3+4 4 | 2+2 4 | 3+4 2 |
| | 6 2 | 3+4 4 | 6 2 | 3+4 4 | 2+2 4 | 5+6 2 |
| | 6 2 | 3+4 4 | 6 2 | 3+4 4 | 2+2 4 | 3+4 2 |
| | 5 2 | 3+4 4 | 5 2 | 3+4 4 | 2+2 4 | 5+6 2 |
| | 5 2 | 3+4 4 | 6 2 | 3+4 4 | 2+2 4 | 5+6 2 |
| | 6 2 | 3+4 4 | 5 2 | 3+4 4 | 2+2 4 | 3+4 2 |
| | 5 2 | 3+4 4 | 6 2 | 3+4 4 | 2+2 4 | 5+6 2 |
| | 6 2 | 3+4 4 | 5 2 | 3+4 4 | 2+2 4 | 3+4 2 |
| | 5 2 | 3+4 4 | 5 2 | 3+4 4 | 2+2 4 | 5+6 2 |
| | 5 2 | | 6 2 | | | 3+4 2 |
| | 6 2 | | 6 2 | | | 5+6 2 |
| | 5 2 | | 5 2 | | | 3+4 2 |
| | 6 2 | | 6 2 | | | 5+6 2 |
| | 5 2 | | 5 2 | | | 3+4 2 |
| | 6 2 | | 5 2 | | | 5+6 2 |
| | 5 2 | | 6 2 | | | 3+4 2 |
| | 6 2 | | 6 2 | | | 5+6 2 |
| | 6 2 | | 5 2 | | | 3+4 2 |
| | 5 2 | | 6 2 | | | 5+6 2 |
| | 5 2 | | 5 2 | | | 3+4 2 |
| | 5 2 | | 6 2 | | | 5+6 2 |
| | 6 2 | | 5 2 | | | 3+4 2 |
| 63.2.30 | 410402 | 52 | In2Pd1Sr1 | | | |
| | -GM | -T | -Y | -Z | -R | -S |
| | 5 2 | 3+4 4 | 5 2 | 3+4 4 | 2+2 4 | 3+4 2 |

Continued on next column

**TABLE XXXVIII– continued**

| TQC | ICSD | | NoE | SF | | |
|---|---|---|---|---|---|---|
| | 6 2 | 3+4 4 | 6 2 | 3+4 4 | 2+2 4 | 5+6 2 |
| | 5 2 | 3+4 4 | 5 2 | 3+4 4 | 2+2 4 | 3+4 2 |
| | 6 2 | 3+4 4 | 6 2 | 3+4 4 | 2+2 4 | 5+6 2 |
| | 5 2 | 3+4 4 | 5 2 | 3+4 4 | 2+2 4 | 3+4 2 |
| | 6 2 | 3+4 4 | 6 2 | 3+4 4 | 2+2 4 | 5+6 2 |
| | 6 2 | 3+4 4 | 6 2 | 3+4 4 | 2+2 4 | 3+4 2 |
| | 5 2 | 3+4 4 | 5 2 | 3+4 4 | 2+2 4 | 5+6 2 |
| | 5 2 | 3+4 4 | 6 2 | 3+4 4 | 2+2 4 | 5+6 2 |
| | 5 2 | 3+4 4 | 5 2 | 3+4 4 | 2+2 4 | 3+4 2 |
| | 6 2 | 3+4 4 | 6 2 | 3+4 4 | 2+2 4 | 5+6 2 |
| | 6 2 | 3+4 4 | 5 2 | 3+4 4 | 2+2 4 | 3+4 2 |
| | 5 2 | 3+4 4 | 5 2 | 3+4 4 | 2+2 4 | 3+4 2 |
| | 5 2 | | 6 2 | | | 5+6 2 |
| | 5 2 | | 6 2 | | | 5+6 2 |
| | 6 2 | | 5 2 | | | 3+4 2 |
| | 5 2 | | 6 2 | | | 5+6 2 |
| | 6 2 | | 5 2 | | | 3+4 2 |
| | 6 2 | | 6 2 | | | 3+4 2 |
| | 5 2 | | 6 2 | | | 5+6 2 |
| | 6 2 | | 5 2 | | | 5+6 2 |
| | 6 2 | | 5 2 | | | 5+6 2 |
| | 5 2 | | 5 2 | | | 3+4 2 |
| | 5 2 | | 6 2 | | | 3+4 2 |
| | 5 2 | | 6 2 | | | 5+6 2 |
| | 6 2 | | 5 2 | | | 3+4 2 |
| 63.2.31 | 190810 | | 76 | Hf2Pt3 | | |
| | -GM | -T | -Y | -Z | -R | -S |
| | 5 2 | 3+4 4 | 6 2 | 3+4 4 | 2+2 4 | 5+6 2 |
| | 6 2 | 3+4 4 | 5 2 | 3+4 4 | 2+2 4 | 3+4 2 |
| | 5 2 | 3+4 4 | 6 2 | 3+4 4 | 2+2 4 | 5+6 2 |
| | 6 2 | 3+4 4 | 5 2 | 3+4 4 | 2+2 4 | 3+4 2 |
| | 6 2 | 3+4 4 | 5 2 | 3+4 4 | 2+2 4 | 3+4 2 |
| | 5 2 | 3+4 4 | 5 2 | 3+4 4 | 2+2 4 | 5+6 2 |
| | 6 2 | 3+4 4 | 6 2 | 3+4 4 | 2+2 4 | 5+6 2 |
| | 5 2 | 3+4 4 | 6 2 | 3+4 4 | 2+2 4 | 3+4 2 |
| | 5 2 | 3+4 4 | 5 2 | 3+4 4 | 2+2 4 | 3+4 2 |
| | 6 2 | 3+4 4 | 5 2 | 3+4 4 | 2+2 4 | 5+6 2 |
| | 6 2 | 3+4 4 | 5 2 | 3+4 4 | 2+2 4 | 3+4 2 |
| | 5 2 | 3+4 4 | 6 2 | 3+4 4 | 2+2 4 | 5+6 2 |
| | 6 2 | 3+4 4 | 6 2 | 3+4 4 | 2+2 4 | 3+4 2 |
| | 5 2 | 3+4 4 | 6 2 | 3+4 4 | 2+2 4 | 5+6 2 |
| | 5 2 | 3+4 4 | 6 2 | 3+4 4 | 2+2 4 | 5+6 2 |
| | 6 2 | 3+4 4 | 6 2 | 3+4 4 | 2+2 4 | 3+4 2 |
| | 5 2 | 3+4 4 | 5 2 | 3+4 4 | 2+2 4 | 3+4 2 |
| | 6 2 | 3+4 4 | 6 2 | 3+4 4 | 2+2 4 | 5+6 2 |
| | 5 2 | 3+4 4 | 5 2 | 3+4 4 | 2+2 4 | 3+4 2 |
| | 5 2 | | 6 2 | | | 5+6 2 |
| | 5 2 | | 5 2 | | | 3+4 2 |
| | 6 2 | | 5 2 | | | 5+6 2 |
| | 5 2 | | 6 2 | | | 3+4 2 |
| | 6 2 | | 6 2 | | | 5+6 2 |
| | 5 2 | | 5 2 | | | 3+4 2 |
| | 6 2 | | 5 2 | | | 5+6 2 |
| | 5 2 | | 5 2 | | | 3+4 2 |
| | 6 2 | | 6 2 | | | 5+6 2 |
| | 6 2 | | 6 2 | | | 3+4 2 |
| | 6 2 | | 5 2 | | | 5+6 2 |
| | 5 2 | | 5 2 | | | 3+4 2 |
| | 5 2 | | 5 2 | | | 5+6 2 |
| | 6 2 | | 6 2 | | | 3+4 2 |
| | 5 2 | | 5 2 | | | 5+6 2 |
| | 5 2 | | 5 2 | | | 5+6 2 |
| | 6 2 | | 6 2 | | | 3+4 2 |
| | 5 2 | | 6 2 | | | 5+6 2 |
| | 6 2 | | 6 2 | | | 3+4 2 |
| 63.2.32 | 15051 | | 68 | Hf1Te5 | | |
| | -GM | -T | -Y | -Z | -R | -S |
| | 5 2 | 3+4 4 | 6 2 | 3+4 4 | 2+2 4 | 3+4 2 |
| | 5 2 | 3+4 4 | 6 2 | 3+4 4 | 2+2 4 | 5+6 2 |
| | 6 2 | 3+4 4 | 5 2 | 3+4 4 | 2+2 4 | 5+6 2 |
| | 5 2 | 3+4 4 | 6 2 | 3+4 4 | 2+2 4 | 3+4 2 |
| | 6 2 | 3+4 4 | 5 2 | 3+4 4 | 2+2 4 | 3+4 2 |
| | 5 2 | 3+4 4 | 6 2 | 3+4 4 | 2+2 4 | 5+6 2 |
| | 6 2 | 3+4 4 | 5 2 | 3+4 4 | 2+2 4 | 3+4 2 |
| | 6 2 | 3+4 4 | 5 2 | 3+4 4 | 2+2 4 | 5+6 2 |
| | 5 2 | 3+4 4 | 6 2 | 3+4 4 | 2+2 4 | 3+4 2 |
| | 6 2 | 3+4 4 | 5 2 | 3+4 4 | 2+2 4 | 5+6 2 |
| | 6 2 | 3+4 4 | 5 2 | 3+4 4 | 2+2 4 | 3+4 2 |
| | 5 2 | 3+4 4 | 6 2 | 3+4 4 | 2+2 4 | 5+6 2 |
| | 5 2 | 3+4 4 | 6 2 | 3+4 4 | 2+2 4 | 3+4 2 |
| | 6 2 | 3+4 4 | 5 2 | 3+4 4 | 2+2 4 | 5+6 2 |
| | 6 2 | 3+4 4 | 5 2 | 3+4 4 | 2+2 4 | 3+4 2 |
| | 5 2 | 3+4 4 | 6 2 | 3+4 4 | 2+2 4 | 5+6 2 |
| | 5 2 | 3+4 4 | 6 2 | 3+4 4 | 2+2 4 | 3+4 2 |
| | 5 2 | | 6 2 | | | 5+6 2 |
| | 5 2 | | 5 2 | | | 5+6 2 |
| | 6 2 | | 6 2 | | | 3+4 2 |
| | 5 2 | | 6 2 | | | 5+6 2 |
| | 6 2 | | 5 2 | | | 3+4 2 |
| | 5 2 | | 6 2 | | | 5+6 2 |
| | 6 2 | | 5 2 | | | 3+4 2 |
| | 5 2 | | 5 2 | | | 5+6 2 |
| | 6 2 | | 6 2 | | | 3+4 2 |
| | 5 2 | | 5 2 | | | 5+6 2 |
| | 6 2 | | 6 2 | | | 3+4 2 |
| | 6 2 | | 5 2 | | | 5+6 2 |
| | 5 2 | | 6 2 | | | 3+4 2 |
| | 6 2 | | 5 2 | | | 3+4 2 |
| | 5 2 | | 6 2 | | | 3+4 2 |
| | 5 2 | | 6 2 | | | 5+6 2 |
| | 5 2 | | 5 2 | | | 5+6 2 |
| 63.2.33 | 171235 | | 172 | Ge6Pr2Zn3 | | |
| | -GM | -T | -Y | -Z | -R | -S |
| | 5 2 | 3+4 4 | 5 2 | 3+4 4 | 2+2 4 | 3+4 2 |
| | 6 2 | 3+4 4 | 6 2 | 3+4 4 | 2+2 4 | 5+6 2 |
| | 5 2 | 3+4 4 | 5 2 | 3+4 4 | 2+2 4 | 3+4 2 |
| | 6 2 | 3+4 4 | 6 2 | 3+4 4 | 2+2 4 | 5+6 2 |
| | 5 2 | 3+4 4 | 5 2 | 3+4 4 | 2+2 4 | 3+4 2 |
| | 6 2 | 3+4 4 | 6 2 | 3+4 4 | 2+2 4 | 5+6 2 |
| | 5 2 | 3+4 4 | 5 2 | 3+4 4 | 2+2 4 | 3+4 2 |
| | 6 2 | 3+4 4 | 6 2 | 3+4 4 | 2+2 4 | 5+6 2 |
| | 5 2 | 3+4 4 | 5 2 | 3+4 4 | 2+2 4 | 3+4 2 |
| | 6 2 | 3+4 4 | 6 2 | 3+4 4 | 2+2 4 | 5+6 2 |
| | 5 2 | 3+4 4 | 5 2 | 3+4 4 | 2+2 4 | 5+6 2 |
| | 6 2 | 3+4 4 | 6 2 | 3+4 4 | 2+2 4 | 3+4 2 |
| | 5 2 | 3+4 4 | 5 2 | 3+4 4 | 2+2 4 | 3+4 2 |
| | 6 2 | 3+4 4 | 6 2 | 3+4 4 | 2+2 4 | 5+6 2 |
| | 5 2 | 3+4 4 | 5 2 | 3+4 4 | 2+2 4 | 3+4 2 |
| | 6 2 | 3+4 4 | 6 2 | 3+4 4 | 2+2 4 | 5+6 2 |
| | 5 2 | 3+4 4 | 6 2 | 3+4 4 | 2+2 4 | 3+4 2 |
| | 5 2 | 3+4 4 | 5 2 | 3+4 4 | 2+2 4 | 5+6 2 |
| | 6 2 | 3+4 4 | 5 2 | 3+4 4 | 2+2 4 | 5+6 2 |

Continued on next column

**TABLE XXXVIII– continued**

| TQC | -GM | -T | -Y | -Z | -R | -S |
|---|---|---|---|---|---|---|
| | ICSD | | NoE | | SF | |
| | 5 2 | 3+4 4 | 6 2 | 3+4 4 | 2+2 4 | 3+4 2 |
| | 5 2 | 3+4 4 | 6 2 | 3+4 4 | 2+2 4 | 3+4 2 |
| | 6 2 | 3+4 4 | 6 2 | 3+4 4 | 2+2 4 | 5+6 2 |
| | 5 2 | 3+4 4 | 5 2 | 3+4 4 | 2+2 4 | 3+4 2 |
| | 6 2 | 3+4 4 | 6 2 | 3+4 4 | 2+2 4 | 5+6 2 |
| | 5 2 | 3+4 4 | 6 2 | 3+4 4 | 2+2 4 | 3+4 2 |
| | 5 2 | 3+4 4 | 5 2 | 3+4 4 | 2+2 4 | 5+6 2 |
| | 6 2 | 3+4 4 | 5 2 | 3+4 4 | 2+2 4 | 3+4 2 |
| | 5 2 | 3+4 4 | 5 2 | 3+4 4 | 2+2 4 | 5+6 2 |
| | 6 2 | 3+4 4 | 5 2 | 3+4 4 | 2+2 4 | 3+4 2 |
| | 6 2 | 3+4 4 | 6 2 | 3+4 4 | 2+2 4 | 5+6 2 |
| | 5 2 | 3+4 4 | 6 2 | 3+4 4 | 2+2 4 | 5+6 2 |
| | 6 2 | 3+4 4 | 6 2 | 3+4 4 | 2+2 4 | 3+4 2 |
| | 5 2 | 3+4 4 | 5 2 | 3+4 4 | 2+2 4 | 5+6 2 |
| | 6 2 | 3+4 4 | 6 2 | 3+4 4 | 2+2 4 | 3+4 2 |
| | 6 2 | 3+4 4 | 5 2 | 3+4 4 | 2+2 4 | 3+4 2 |
| | 6 2 | 3+4 4 | 5 2 | 3+4 4 | 2+2 4 | 5+6 2 |
| | 5 2 | 3+4 4 | 6 2 | 3+4 4 | 2+2 4 | 3+4 2 |
| | 5 2 | 3+4 4 | 6 2 | 3+4 4 | 2+2 4 | 5+6 2 |
| | 6 2 | 3+4 4 | 6 2 | 3+4 4 | 2+2 4 | 3+4 2 |
| | 6 2 | 3+4 4 | 5 2 | 3+4 4 | 2+2 4 | 5+6 2 |
| | 5 2 | 3+4 4 | 5 2 | 3+4 4 | 2+2 4 | 5+6 2 |
| | 6 2 | 3+4 4 | 5 2 | 3+4 4 | 2+2 4 | 3+4 2 |
| | 5 2 | 3+4 4 | 6 2 | 3+4 4 | 2+2 4 | 5+6 2 |
| | 5 2 | | 5 2 | | | 3+4 2 |
| | 5 2 | | 6 2 | | | 3+4 2 |
| | 5 2 | | 5 2 | | | 5+6 2 |
| | 6 2 | | 6 2 | | | 3+4 2 |
| | 6 2 | | 6 2 | | | 5+6 2 |
| | 6 2 | | 5 2 | | | 3+4 2 |
| | 5 2 | | 6 2 | | | 5+6 2 |
| | 5 2 | | 5 2 | | | 3+4 2 |
| | 6 2 | | 5 2 | | | 5+6 2 |
| | 5 2 | | 6 2 | | | 3+4 2 |
| | 6 2 | | 6 2 | | | 5+6 2 |
| | 6 2 | | 5 2 | | | 5+6 2 |
| | 6 2 | | 5 2 | | | 3+4 2 |
| | 6 2 | | 5 2 | | | 3+4 2 |
| | 5 2 | | 6 2 | | | 5+6 2 |
| | 5 2 | | 6 2 | | | 5+6 2 |
| | 5 2 | | 5 2 | | | 3+4 2 |
| | 6 2 | | 6 2 | | | 3+4 2 |
| | 5 2 | | 5 2 | | | 5+6 2 |
| | 6 2 | | 5 2 | | | 5+6 2 |
| | 6 2 | | 6 2 | | | 3+4 2 |
| | 6 2 | | 5 2 | | | 3+4 2 |
| | 6 2 | | 5 2 | | | 5+6 2 |
| | 5 2 | | 5 2 | | | 3+4 2 |
| | 6 2 | | 6 2 | | | 5+6 2 |
| | 5 2 | | 5 2 | | | 3+4 2 |
| | 5 2 | | 6 2 | | | 5+6 2 |
| | 6 2 | | 5 2 | | | 5+6 2 |
| | 5 2 | | 6 2 | | | 3+4 2 |
| | 5 2 | | 6 2 | | | 5+6 2 |
| | 5 2 | | 5 2 | | | 3+4 2 |
| | 5 2 | | 6 2 | | | 5+6 2 |
| | 5 2 | | 5 2 | | | 3+4 2 |
| | 6 2 | | 6 2 | | | 5+6 2 |
| | 5 2 | | 6 2 | | | 3+4 2 |
| | 5 2 | | 6 2 | | | 3+4 2 |
| | 5 2 | | 6 2 | | | 5+6 2 |

Continued on next column

**TABLE XXXVIII– continued**

| TQC | ICSD | | NoE | | SF | |
|---|---|---|---|---|---|---|
| | 6 2 | | 6 2 | | | 3+4 2 |
| | 6 2 | | 5 2 | | | 5+6 2 |
| | 6 2 | | 6 2 | | | 3+4 2 |
| | 5 2 | | 6 2 | | | 5+6 2 |
| | 6 2 | | 5 2 | | | 5+6 2 |
| | 5 2 | | 5 2 | | | 3+4 2 |
| 63.2.34 | 171234 | | 176 | Ge6Nd2Zn3 | | |
| | -GM | -T | -Y | -Z | -R | -S |
| | 5 2 | 3+4 4 | 5 2 | 3+4 4 | 2+2 4 | 3+4 2 |
| | 6 2 | 3+4 4 | 6 2 | 3+4 4 | 2+2 4 | 5+6 2 |
| | 5 2 | 3+4 4 | 5 2 | 3+4 4 | 2+2 4 | 3+4 2 |
| | 6 2 | 3+4 4 | 6 2 | 3+4 4 | 2+2 4 | 5+6 2 |
| | 5 2 | 3+4 4 | 5 2 | 3+4 4 | 2+2 4 | 3+4 2 |
| | 6 2 | 3+4 4 | 6 2 | 3+4 4 | 2+2 4 | 5+6 2 |
| | 5 2 | 3+4 4 | 5 2 | 3+4 4 | 2+2 4 | 3+4 2 |
| | 6 2 | 3+4 4 | 6 2 | 3+4 4 | 2+2 4 | 5+6 2 |
| | 5 2 | 3+4 4 | 5 2 | 3+4 4 | 2+2 4 | 5+6 2 |
| | 6 2 | 3+4 4 | 6 2 | 3+4 4 | 2+2 4 | 3+4 2 |
| | 5 2 | 3+4 4 | 5 2 | 3+4 4 | 2+2 4 | 3+4 2 |
| | 6 2 | 3+4 4 | 6 2 | 3+4 4 | 2+2 4 | 5+6 2 |
| | 5 2 | 3+4 4 | 5 2 | 3+4 4 | 2+2 4 | 3+4 2 |
| | 6 2 | 3+4 4 | 6 2 | 3+4 4 | 2+2 4 | 5+6 2 |
| | 5 2 | 3+4 4 | 5 2 | 3+4 4 | 2+2 4 | 3+4 2 |
| | 6 2 | 3+4 4 | 6 2 | 3+4 4 | 2+2 4 | 5+6 2 |
| | 5 2 | 3+4 4 | 6 2 | 3+4 4 | 2+2 4 | 3+4 2 |
| | 5 2 | 3+4 4 | 5 2 | 3+4 4 | 2+2 4 | 5+6 2 |
| | 6 2 | 3+4 4 | 5 2 | 3+4 4 | 2+2 4 | 3+4 2 |
| | 5 2 | 3+4 4 | 6 2 | 3+4 4 | 2+2 4 | 5+6 2 |
| | 5 2 | 3+4 4 | 6 2 | 3+4 4 | 2+2 4 | 5+6 2 |
| | 6 2 | 3+4 4 | 6 2 | 3+4 4 | 2+2 4 | 3+4 2 |
| | 5 2 | 3+4 4 | 5 2 | 3+4 4 | 2+2 4 | 5+6 2 |
| | 6 2 | 3+4 4 | 6 2 | 3+4 4 | 2+2 4 | 3+4 2 |
| | 5 2 | 3+4 4 | 6 2 | 3+4 4 | 2+2 4 | 5+6 2 |
| | 6 2 | 3+4 4 | 5 2 | 3+4 4 | 2+2 4 | 3+4 2 |
| | 6 2 | 3+4 4 | 5 2 | 3+4 4 | 2+2 4 | 5+6 2 |
| | 5 2 | 3+4 4 | 5 2 | 3+4 4 | 2+2 4 | 3+4 2 |
| | 5 2 | 3+4 4 | 5 2 | 3+4 4 | 2+2 4 | 5+6 2 |
| | 5 2 | 3+4 4 | 6 2 | 3+4 4 | 2+2 4 | 3+4 2 |
| | 6 2 | 3+4 4 | 6 2 | 3+4 4 | 2+2 4 | 3+4 2 |
| | 6 2 | 3+4 4 | 6 2 | 3+4 4 | 2+2 4 | 5+6 2 |
| | 5 2 | 3+4 4 | 5 2 | 3+4 4 | 2+2 4 | 3+4 2 |
| | 6 2 | 3+4 4 | 6 2 | 3+4 4 | 2+2 4 | 5+6 2 |
| | 6 2 | 3+4 4 | 5 2 | 3+4 4 | 2+2 4 | 5+6 2 |
| | 6 2 | 3+4 4 | 5 2 | 3+4 4 | 2+2 4 | 3+4 2 |
| | 5 2 | 3+4 4 | 6 2 | 3+4 4 | 2+2 4 | 5+6 2 |
| | 5 2 | 3+4 4 | 6 2 | 3+4 4 | 2+2 4 | 3+4 2 |
| | 6 2 | 3+4 4 | 5 2 | 3+4 4 | 2+2 4 | 5+6 2 |
| | 6 2 | 3+4 4 | 6 2 | 3+4 4 | 2+2 4 | 3+4 2 |
| | 5 2 | 3+4 4 | 5 2 | 3+4 4 | 2+2 4 | 5+6 2 |
| | 5 2 | 3+4 4 | 5 2 | 3+4 4 | 2+2 4 | 3+4 2 |
| | 6 2 | 3+4 4 | 6 2 | 3+4 4 | 2+2 4 | 3+4 2 |
| | 5 2 | 3+4 4 | 6 2 | 3+4 4 | 2+2 4 | 5+6 2 |
| | 5 2 | | 6 2 | | | 5+6 2 |
| | 5 2 | | 5 2 | | | 3+4 2 |
| | 6 2 | | 5 2 | | | 5+6 2 |
| | 6 2 | | 6 2 | | | 3+4 2 |
| | 6 2 | | 5 2 | | | 3+4 2 |
| | 5 2 | | 6 2 | | | 5+6 2 |
| | 5 2 | | 5 2 | | | 5+6 2 |
| | 6 2 | | 5 2 | | | 3+4 2 |
| | 5 2 | | 6 2 | | | 5+6 2 |

Continued on next column

**TABLE XXXVIII– continued**

| TQC | ICSD | | NoE | | SF | |
|---|---|---|---|---|---|---|
| | 6 2 | | 6 2 | | | 3+4 2 |
| | 6 2 | | 5 2 | | | 3+4 2 |
| | 6 2 | | 5 2 | | | 5+6 2 |
| | 6 2 | | 5 2 | | | 3+4 2 |
| | 5 2 | | 6 2 | | | 5+6 2 |
| | 5 2 | | 6 2 | | | 3+4 2 |
| | 5 2 | | 5 2 | | | 5+6 2 |
| | 6 2 | | 6 2 | | | 5+6 2 |
| | 5 2 | | 5 2 | | | 3+4 2 |
| | 6 2 | | 5 2 | | | 3+4 2 |
| | 6 2 | | 6 2 | | | 5+6 2 |
| | 6 2 | | 5 2 | | | 5+6 2 |
| | 6 2 | | 5 2 | | | 3+4 2 |
| | 5 2 | | 6 2 | | | 5+6 2 |
| | 6 2 | | 5 2 | | | 3+4 2 |
| | 5 2 | | 5 2 | | | 5+6 2 |
| | 5 2 | | 6 2 | | | 3+4 2 |
| | 6 2 | | 5 2 | | | 3+4 2 |
| | 5 2 | | 6 2 | | | 5+6 2 |
| | 5 2 | | 6 2 | | | 3+4 2 |
| | 5 2 | | 6 2 | | | 5+6 2 |
| | 5 2 | | 5 2 | | | 3+4 2 |
| | 5 2 | | 5 2 | | | 3+4 2 |
| | 6 2 | | 6 2 | | | 5+6 2 |
| | 5 2 | | 6 2 | | | 5+6 2 |
| | 5 2 | | 6 2 | | | 5+6 2 |
| | 5 2 | | 6 2 | | | 3+4 2 |
| | 6 2 | | 6 2 | | | 5+6 2 |
| | 6 2 | | 6 2 | | | 3+4 2 |
| | 6 2 | | 5 2 | | | 5+6 2 |
| | 5 2 | | 6 2 | | | 3+4 2 |
| | 6 2 | | 5 2 | | | 3+4 2 |
| | 5 2 | | 5 2 | | | 5+6 2 |
| | 5 2 | | 5 2 | | | 5+6 2 |
| | 6 2 | | 6 2 | | | 3+4 2 |
| 63.2.35 | 171232 | | 164 | Ge6La2Zn3 | | |
| | -GM | -T | -Y | -Z | -R | -S |
| | 5 2 | 3+4 4 | 5 2 | 3+4 4 | 2+2 4 | 3+4 2 |
| | 6 2 | 3+4 4 | 6 2 | 3+4 4 | 2+2 4 | 5+6 2 |
| | 5 2 | 3+4 4 | 5 2 | 3+4 4 | 2+2 4 | 3+4 2 |
| | 6 2 | 3+4 4 | 6 2 | 3+4 4 | 2+2 4 | 5+6 2 |
| | 5 2 | 3+4 4 | 5 2 | 3+4 4 | 2+2 4 | 3+4 2 |
| | 6 2 | 3+4 4 | 6 2 | 3+4 4 | 2+2 4 | 5+6 2 |
| | 5 2 | 3+4 4 | 5 2 | 3+4 4 | 2+2 4 | 5+6 2 |
| | 6 2 | 3+4 4 | 6 2 | 3+4 4 | 2+2 4 | 3+4 2 |
| | 5 2 | 3+4 4 | 5 2 | 3+4 4 | 2+2 4 | 3+4 2 |
| | 6 2 | 3+4 4 | 6 2 | 3+4 4 | 2+2 4 | 5+6 2 |
| | 5 2 | 3+4 4 | 5 2 | 3+4 4 | 2+2 4 | 5+6 2 |
| | 6 2 | 3+4 4 | 6 2 | 3+4 4 | 2+2 4 | 3+4 2 |
| | 5 2 | 3+4 4 | 5 2 | 3+4 4 | 2+2 4 | 3+4 2 |
| | 6 2 | 3+4 4 | 6 2 | 3+4 4 | 2+2 4 | 5+6 2 |
| | 5 2 | 3+4 4 | 5 2 | 3+4 4 | 2+2 4 | 3+4 2 |
| | 6 2 | 3+4 4 | 6 2 | 3+4 4 | 2+2 4 | 5+6 2 |
| | 5 2 | 3+4 4 | 6 2 | 3+4 4 | 2+2 4 | 3+4 2 |
| | 5 2 | 3+4 4 | 5 2 | 3+4 4 | 2+2 4 | 5+6 2 |
| | 6 2 | 3+4 4 | 5 2 | 3+4 4 | 2+2 4 | 5+6 2 |
| | 5 2 | 3+4 4 | 6 2 | 3+4 4 | 2+2 4 | 3+4 2 |
| | 5 2 | 3+4 4 | 6 2 | 3+4 4 | 2+2 4 | 3+4 2 |
| | 6 2 | 3+4 4 | 6 2 | 3+4 4 | 2+2 4 | 5+6 2 |
| | 5 2 | 3+4 4 | 5 2 | 3+4 4 | 2+2 4 | 3+4 2 |
| | 6 2 | 3+4 4 | 6 2 | 3+4 4 | 2+2 4 | 5+6 2 |

Continued on next column

**TABLE XXXVIII– continued**

| TQC | ICSD | | NoE | | SF | |
|---|---|---|---|---|---|---|
| | 5 2 | 3+4 4 | 6 2 | 3+4 4 | 2+2 4 | 3+4 2 |
| | 6 2 | 3+4 4 | 5 2 | 3+4 4 | 2+2 4 | 5+6 2 |
| | 5 2 | 3+4 4 | 5 2 | 3+4 4 | 2+2 4 | 3+4 2 |
| | 6 2 | 3+4 4 | 5 2 | 3+4 4 | 2+2 4 | 5+6 2 |
| | 5 2 | 3+4 4 | 5 2 | 3+4 4 | 2+2 4 | 3+4 2 |
| | 6 2 | 3+4 4 | 6 2 | 3+4 4 | 2+2 4 | 5+6 2 |
| | 5 2 | 3+4 4 | 6 2 | 3+4 4 | 2+2 4 | 5+6 2 |
| | 6 2 | 3+4 4 | 6 2 | 3+4 4 | 2+2 4 | 3+4 2 |
| | 5 2 | 3+4 4 | 5 2 | 3+4 4 | 2+2 4 | 5+6 2 |
| | 6 2 | 3+4 4 | 6 2 | 3+4 4 | 2+2 4 | 3+4 2 |
| | 6 2 | 3+4 4 | 5 2 | 3+4 4 | 2+2 4 | 3+4 2 |
| | 6 2 | 3+4 4 | 5 2 | 3+4 4 | 2+2 4 | 5+6 2 |
| | 5 2 | 3+4 4 | 6 2 | 3+4 4 | 2+2 4 | 3+4 2 |
| | 5 2 | 3+4 4 | 6 2 | 3+4 4 | 2+2 4 | 5+6 2 |
| | 6 2 | 3+4 4 | 5 2 | 3+4 4 | 2+2 4 | 3+4 2 |
| | 6 2 | 3+4 4 | 6 2 | 3+4 4 | 2+2 4 | 5+6 2 |
| | 5 2 | 3+4 4 | 5 2 | 3+4 4 | 2+2 4 | 3+4 2 |
| | 6 2 | | 5 2 | | | 5+6 2 |
| | 5 2 | | 6 2 | | | 5+6 2 |
| | 5 2 | | 5 2 | | | 3+4 2 |
| | 5 2 | | 6 2 | | | 3+4 2 |
| | 5 2 | | 5 2 | | | 5+6 2 |
| | 6 2 | | 6 2 | | | 3+4 2 |
| | 6 2 | | 6 2 | | | 5+6 2 |
| | 5 2 | | 5 2 | | | 3+4 2 |
| | 6 2 | | 6 2 | | | 5+6 2 |
| | 5 2 | | 5 2 | | | 3+4 2 |
| | 6 2 | | 5 2 | | | 5+6 2 |
| | 5 2 | | 6 2 | | | 3+4 2 |
| | 6 2 | | 6 2 | | | 5+6 2 |
| | 6 2 | | 5 2 | | | 5+6 2 |
| | 6 2 | | 5 2 | | | 3+4 2 |
| | 6 2 | | 5 2 | | | 3+4 2 |
| | 5 2 | | 6 2 | | | 5+6 2 |
| | 5 2 | | 6 2 | | | 5+6 2 |
| | 5 2 | | 5 2 | | | 3+4 2 |
| | 6 2 | | 6 2 | | | 3+4 2 |
| | 5 2 | | 5 2 | | | 5+6 2 |
| | 6 2 | | 5 2 | | | 5+6 2 |
| | 6 2 | | 6 2 | | | 3+4 2 |
| | 6 2 | | 5 2 | | | 3+4 2 |
| | 6 2 | | 5 2 | | | 5+6 2 |
| | 5 2 | | 5 2 | | | 3+4 2 |
| | 6 2 | | 5 2 | | | 5+6 2 |
| | 5 2 | | 6 2 | | | 3+4 2 |
| | 5 2 | | 6 2 | | | 5+6 2 |
| | 5 2 | | 5 2 | | | 5+6 2 |
| | 6 2 | | 6 2 | | | 3+4 2 |
| | 5 2 | | 6 2 | | | 5+6 2 |
| | 5 2 | | 5 2 | | | 3+4 2 |
| | 5 2 | | 5 2 | | | 5+6 2 |
| | 6 2 | | 6 2 | | | 3+4 2 |
| | 5 2 | | 6 2 | | | 5+6 2 |
| | 5 2 | | 6 2 | | | 3+4 2 |
| | 5 2 | | 6 2 | | | 3+4 2 |
| | 5 2 | | 6 2 | | | 5+6 2 |
| | 6 2 | | 5 2 | | | 3+4 2 |
| | 6 2 | | 6 2 | | | 5+6 2 |
| 63.2.36 | 56047 | | 40 | Ge2Th1 | | |
| | -GM | -T | -Y | -Z | -R | -S |
| | 5 2 | 3+4 4 | 5 2 | 3+4 4 | 2+2 4 | 3+4 2 |

Continued on next column

**TABLE XXXVIII– continued**

| TQC | ICSD | | NoE | | SF | |
|---|---|---|---|---|---|---|
| | 6 2 | 3+4 4 | 6 2 | 3+4 4 | 2+2 4 | 5+6 2 |
| | 5 2 | 3+4 4 | 5 2 | 3+4 4 | 2+2 4 | 3+4 2 |
| | 6 2 | 3+4 4 | 6 2 | 3+4 4 | 2+2 4 | 5+6 2 |
| | 5 2 | 3+4 4 | 5 2 | 3+4 4 | 2+2 4 | 5+6 2 |
| | 6 2 | 3+4 4 | 6 2 | 3+4 4 | 2+2 4 | 3+4 2 |
| | 5 2 | 3+4 4 | 5 2 | 3+4 4 | 2+2 4 | 3+4 2 |
| | 6 2 | 3+4 4 | 6 2 | 3+4 4 | 2+2 4 | 5+6 2 |
| | 5 2 | 3+4 4 | 5 2 | 3+4 4 | 2+2 4 | 5+6 2 |
| | 5 2 | 3+4 4 | 6 2 | 3+4 4 | 2+2 4 | 3+4 2 |
| | 6 2 | | 6 2 | | | 5+6 2 |
| | 6 2 | | 5 2 | | | 3+4 2 |
| | 6 2 | | 5 2 | | | 3+4 2 |
| | 5 2 | | 6 2 | | | 5+6 2 |
| | 5 2 | | 5 2 | | | 5+6 2 |
| | 6 2 | | 6 2 | | | 5+6 2 |
| | 5 2 | | 6 2 | | | 3+4 2 |
| | 5 2 | | 6 2 | | | 3+4 2 |
| | 5 2 | | 5 2 | | | 5+6 2 |
| | 6 2 | | 5 2 | | | 3+4 2 |
| 63.2.37 | 240343 | | 44 | | Ge2Ni1Sr1 | |
| | -GM | -T | -Y | -Z | -R | -S |
| | 5 2 | 3+4 4 | 6 2 | 3+4 4 | 2+2 4 | 5+6 2 |
| | 5 2 | 3+4 4 | 5 2 | 3+4 4 | 2+2 4 | 3+4 2 |
| | 6 2 | 3+4 4 | 6 2 | 3+4 4 | 2+2 4 | 5+6 2 |
| | 5 2 | 3+4 4 | 5 2 | 3+4 4 | 2+2 4 | 3+4 2 |
| | 6 2 | 3+4 4 | 5 2 | 3+4 4 | 2+2 4 | 5+6 2 |
| | 6 2 | 3+4 4 | 6 2 | 3+4 4 | 2+2 4 | 3+4 2 |
| | 5 2 | 3+4 4 | 6 2 | 3+4 4 | 2+2 4 | 3+4 2 |
| | 5 2 | 3+4 4 | 6 2 | 3+4 4 | 2+2 4 | 5+6 2 |
| | 6 2 | 3+4 4 | 5 2 | 3+4 4 | 2+2 4 | 5+6 2 |
| | 6 2 | 3+4 4 | 5 2 | 3+4 4 | 2+2 4 | 3+4 2 |
| | 5 2 | 3+4 4 | 6 2 | 3+4 4 | 2+2 4 | 5+6 2 |
| | 5 2 | | 6 2 | | | 3+4 2 |
| | 5 2 | | 5 2 | | | 5+6 2 |
| | 6 2 | | 5 2 | | | 3+4 2 |
| | 5 2 | | 6 2 | | | 5+6 2 |
| | 5 2 | | 5 2 | | | 3+4 2 |
| | 6 2 | | 6 2 | | | 5+6 2 |
| | 6 2 | | 5 2 | | | 3+4 2 |
| | 6 2 | | 6 2 | | | 5+6 2 |
| | 5 2 | | 5 2 | | | 3+4 2 |
| | 6 2 | | 5 2 | | | 3+4 2 |
| | 5 2 | | 6 2 | | | 5+6 2 |
| 63.2.38 | 240728 | | 36 | | Ge1Na2Zn1 | |
| | -GM | -T | -Y | -Z | -R | -S |
| | 5 2 | 3+4 4 | 5 2 | 3+4 4 | 2+2 4 | 5+6 2 |
| | 5 2 | 3+4 4 | 5 2 | 3+4 4 | 2+2 4 | 3+4 2 |
| | 6 2 | 3+4 4 | 6 2 | 3+4 4 | 2+2 4 | 5+6 2 |
| | 5 2 | 3+4 4 | 5 2 | 3+4 4 | 2+2 4 | 5+6 2 |
| | 5 2 | 3+4 4 | 5 2 | 3+4 4 | 2+2 4 | 5+6 2 |
| | 5 2 | 3+4 4 | 5 2 | 3+4 4 | 2+2 4 | 5+6 2 |
| | 5 2 | 3+4 4 | 5 2 | 3+4 4 | 2+2 4 | 5+6 2 |
| | 5 2 | 3+4 4 | 5 2 | 3+4 4 | 2+2 4 | 5+6 2 |
| | 5 2 | 3+4 4 | 5 2 | 3+4 4 | 2+2 4 | 5+6 2 |
| | 5 2 | | 5 2 | | | 5+6 2 |
| | 5 2 | | 5 2 | | | 5+6 2 |
| | 5 2 | | 5 2 | | | 5+6 2 |
| | 5 2 | | 5 2 | | | 5+6 2 |
| | 5 2 | | 6 2 | | | 3+4 2 |
| | 6 2 | | 5 2 | | | 5+6 2 |
| | 6 2 | | 6 2 | | | 3+4 2 |

Continued on next column

**TABLE XXXVIII– continued**

| TQC | ICSD | | NoE | | SF | |
|---|---|---|---|---|---|---|
| | 5 2 | | 6 2 | | | 3+4 2 |
| | 6 2 | | 6 2 | | | 5+6 2 |
| 63.2.39 | 163917 | | 140 | | Gd3Ge4 | |
| | -GM | -T | -Y | -Z | -R | -S |
| | 5 2 | 3+4 4 | 5 2 | 3+4 4 | 2+2 4 | 3+4 2 |
| | 6 2 | 3+4 4 | 6 2 | 3+4 4 | 2+2 4 | 5+6 2 |
| | 5 2 | 3+4 4 | 6 2 | 3+4 4 | 2+2 4 | 3+4 2 |
| | 5 2 | 3+4 4 | 6 2 | 3+4 4 | 2+2 4 | 3+4 2 |
| | 6 2 | 3+4 4 | 5 2 | 3+4 4 | 2+2 4 | 5+6 2 |
| | 6 2 | 3+4 4 | 5 2 | 3+4 4 | 2+2 4 | 5+6 2 |
| | 5 2 | 3+4 4 | 6 2 | 3+4 4 | 2+2 4 | 3+4 2 |
| | 6 2 | 3+4 4 | 5 2 | 3+4 4 | 2+2 4 | 5+6 2 |
| | 5 2 | 3+4 4 | 6 2 | 3+4 4 | 2+2 4 | 5+6 2 |
| | 6 2 | 3+4 4 | 6 2 | 3+4 4 | 2+2 4 | 3+4 2 |
| | 5 2 | 3+4 4 | 5 2 | 3+4 4 | 2+2 4 | 5+6 2 |
| | 6 2 | 3+4 4 | 5 2 | 3+4 4 | 2+2 4 | 3+4 2 |
| | 5 2 | 3+4 4 | 6 2 | 3+4 4 | 2+2 4 | 5+6 2 |
| | 6 2 | 3+4 4 | 5 2 | 3+4 4 | 2+2 4 | 3+4 2 |
| | 5 2 | 3+4 4 | 6 2 | 3+4 4 | 2+2 4 | 3+4 2 |
| | 5 2 | 3+4 4 | 5 2 | 3+4 4 | 2+2 4 | 5+6 2 |
| | 6 2 | 3+4 4 | 6 2 | 3+4 4 | 2+2 4 | 5+6 2 |
| | 6 2 | 3+4 4 | 5 2 | 3+4 4 | 2+2 4 | 3+4 2 |
| | 5 2 | 3+4 4 | 6 2 | 3+4 4 | 2+2 4 | 3+4 2 |
| | 5 2 | 3+4 4 | 5 2 | 3+4 4 | 2+2 4 | 5+6 2 |
| | 6 2 | 3+4 4 | 6 2 | 3+4 4 | 2+2 4 | 3+4 2 |
| | 5 2 | 3+4 4 | 6 2 | 3+4 4 | 2+2 4 | 5+6 2 |
| | 6 2 | 3+4 4 | 5 2 | 3+4 4 | 2+2 4 | 5+6 2 |
| | 6 2 | 3+4 4 | 5 2 | 3+4 4 | 2+2 4 | 3+4 2 |
| | 5 2 | 3+4 4 | 5 2 | 3+4 4 | 2+2 4 | 5+6 2 |
| | 5 2 | 3+4 4 | 5 2 | 3+4 4 | 2+2 4 | 3+4 2 |
| | 6 2 | 3+4 4 | 6 2 | 3+4 4 | 2+2 4 | 5+6 2 |
| | 5 2 | 3+4 4 | 5 2 | 3+4 4 | 2+2 4 | 3+4 2 |
| | 6 2 | 3+4 4 | 6 2 | 3+4 4 | 2+2 4 | 3+4 2 |
| | 6 2 | 3+4 4 | 6 2 | 3+4 4 | 2+2 4 | 3+4 2 |
| | 5 2 | 3+4 4 | 5 2 | 3+4 4 | 2+2 4 | 5+6 2 |
| | 5 2 | 3+4 4 | 5 2 | 3+4 4 | 2+2 4 | 3+4 2 |
| | 6 2 | 3+4 4 | 6 2 | 3+4 4 | 2+2 4 | 5+6 2 |
| | 6 2 | 3+4 4 | 6 2 | 3+4 4 | 2+2 4 | 5+6 2 |
| | 5 2 | 3+4 4 | 5 2 | 3+4 4 | 2+2 4 | 5+6 2 |
| | 6 2 | | 6 2 | | | 5+6 2 |
| | 5 2 | | 6 2 | | | 3+4 2 |
| | 6 2 | | 5 2 | | | 5+6 2 |
| | 6 2 | | 6 2 | | | 3+4 2 |
| | 6 2 | | 6 2 | | | 3+4 2 |
| | 6 2 | | 5 2 | | | 3+4 2 |
| | 5 2 | | 5 2 | | | 5+6 2 |
| | 5 2 | | 5 2 | | | 5+6 2 |
| | 6 2 | | 5 2 | | | 3+4 2 |
| | 5 2 | | 6 2 | | | 5+6 2 |
| | 6 2 | | 6 2 | | | 5+6 2 |
| | 5 2 | | 6 2 | | | 3+4 2 |
| | 6 2 | | 5 2 | | | 5+6 2 |
| | 5 2 | | 5 2 | | | 3+4 2 |
| | 5 2 | | 6 2 | | | 5+6 2 |
| | 6 2 | | 5 2 | | | 3+4 2 |
| | 5 2 | | 6 2 | | | 5+6 2 |
| | 6 2 | | 5 2 | | | 3+4 2 |
| | 5 2 | | 5 2 | | | 5+6 2 |
| | 6 2 | | 6 2 | | | 3+4 2 |
| | 5 2 | | 6 2 | | | 5+6 2 |
| | 6 2 | | 5 2 | | | 3+4 2 |

Continued on next column

**TABLE XXXVIII– continued**

| TQC | ICSD | | NoE | SF | | |
|---|---|---|---|---|---|---|
| | 6 2 | | 6 2 | | | 5+6 2 |
| | 5 2 | | 6 2 | | | 3+4 2 |
| | 6 2 | | 5 2 | | | 5+6 2 |
| | 5 2 | | 5 2 | | | 3+4 2 |
| | 5 2 | | 6 2 | | | 5+6 2 |
| | 5 2 | | 5 2 | | | 3+4 2 |
| | 6 2 | | 6 2 | | | 5+6 2 |
| | 6 2 | | 5 2 | | | 3+4 2 |
| | 5 2 | | 6 2 | | | 5+6 2 |
| | 5 2 | | 5 2 | | | 3+4 2 |
| | 6 2 | | 6 2 | | | 5+6 2 |
| | 5 2 | | 5 2 | | | 3+4 2 |
| | 6 2 | | 5 2 | | | 5+6 2 |
| 63.2.40 | 104087 | | 56 | Gd1Ni1 | | |
| | -GM | -T | -Y | -Z | -R | -S |
| | 5 2 | 3+4 4 | 6 2 | 3+4 4 | 2+2 4 | 3+4 2 |
| | 6 2 | 3+4 4 | 5 2 | 3+4 4 | 2+2 4 | 5+6 2 |
| | 5 2 | 3+4 4 | 6 2 | 3+4 4 | 2+2 4 | 3+4 2 |
| | 6 2 | 3+4 4 | 5 2 | 3+4 4 | 2+2 4 | 5+6 2 |
| | 5 2 | 3+4 4 | 6 2 | 3+4 4 | 2+2 4 | 3+4 2 |
| | 5 2 | 3+4 4 | 5 2 | 3+4 4 | 2+2 4 | 5+6 2 |
| | 6 2 | 3+4 4 | 6 2 | 3+4 4 | 2+2 4 | 3+4 2 |
| | 6 2 | 3+4 4 | 5 2 | 3+4 4 | 2+2 4 | 5+6 2 |
| | 5 2 | 3+4 4 | 5 2 | 3+4 4 | 2+2 4 | 5+6 2 |
| | 6 2 | 3+4 4 | 6 2 | 3+4 4 | 2+2 4 | 3+4 2 |
| | 6 2 | 3+4 4 | 5 2 | 3+4 4 | 2+2 4 | 5+6 2 |
| | 5 2 | 3+4 4 | 6 2 | 3+4 4 | 2+2 4 | 3+4 2 |
| | 5 2 | 3+4 4 | 6 2 | 3+4 4 | 2+2 4 | 5+6 2 |
| | 6 2 | 3+4 4 | 5 2 | 3+4 4 | 2+2 4 | 3+4 2 |
| | 6 2 | | 6 2 | | | 3+4 2 |
| | 5 2 | | 6 2 | | | 3+4 2 |
| | 5 2 | | 5 2 | | | 5+6 2 |
| | 6 2 | | 5 2 | | | 5+6 2 |
| | 6 2 | | 6 2 | | | 3+4 2 |
| | 5 2 | | 5 2 | | | 5+6 2 |
| | 6 2 | | 5 2 | | | 5+6 2 |
| | 5 2 | | 5 2 | | | 3+4 2 |
| | 5 2 | | 6 2 | | | 5+6 2 |
| | 5 2 | | 6 2 | | | 3+4 2 |
| | 6 2 | | 5 2 | | | 3+4 2 |
| | 6 2 | | 6 2 | | | 5+6 2 |
| | 5 2 | | 5 2 | | | 5+6 2 |
| | 6 2 | | 5 2 | | | 3+4 2 |
| 63.2.41 | 423070 | | 92 | Ga4Ni1Yb1 | | |
| | -GM | -T | -Y | -Z | -R | -S |
| | 5 2 | 3+4 4 | 5 2 | 3+4 4 | 2+2 4 | 3+4 2 |
| | 6 2 | 3+4 4 | 6 2 | 3+4 4 | 2+2 4 | 5+6 2 |
| | 5 2 | 3+4 4 | 5 2 | 3+4 4 | 2+2 4 | 5+6 2 |
| | 6 2 | 3+4 4 | 6 2 | 3+4 4 | 2+2 4 | 3+4 2 |
| | 5 2 | 3+4 4 | 5 2 | 3+4 4 | 2+2 4 | 5+6 2 |
| | 6 2 | 3+4 4 | 6 2 | 3+4 4 | 2+2 4 | 3+4 2 |
| | 5 2 | 3+4 4 | 6 2 | 3+4 4 | 2+2 4 | 5+6 2 |
| | 6 2 | 3+4 4 | 5 2 | 3+4 4 | 2+2 4 | 3+4 2 |
| | 5 2 | 3+4 4 | 6 2 | 3+4 4 | 2+2 4 | 3+4 2 |
| | 6 2 | 3+4 4 | 5 2 | 3+4 4 | 2+2 4 | 5+6 2 |
| | 5 2 | 3+4 4 | 6 2 | 3+4 4 | 2+2 4 | 5+6 2 |
| | 5 2 | 3+4 4 | 6 2 | 3+4 4 | 2+2 4 | 3+4 2 |
| | 5 2 | 3+4 4 | 5 2 | 3+4 4 | 2+2 4 | 3+4 2 |
| | 6 2 | 3+4 4 | 5 2 | 3+4 4 | 2+2 4 | 3+4 2 |
| | 6 2 | 3+4 4 | 5 2 | 3+4 4 | 2+2 4 | 5+6 2 |
| | 6 2 | 3+4 4 | 6 2 | 3+4 4 | 2+2 4 | 5+6 2 |

Continued on next column

**TABLE XXXVIII– continued**

| TQC | ICSD | | NoE | SF | | |
|---|---|---|---|---|---|---|
| | 5 2 | 3+4 4 | 5 2 | 3+4 4 | 2+2 4 | 3+4 2 |
| | 5 2 | 3+4 4 | 5 2 | 3+4 4 | 2+2 4 | 3+4 2 |
| | 5 2 | 3+4 4 | 6 2 | 3+4 4 | 2+2 4 | 5+6 2 |
| | 6 2 | 3+4 4 | 6 2 | 3+4 4 | 2+2 4 | 5+6 2 |
| | 6 2 | 3+4 4 | 6 2 | 3+4 4 | 2+2 4 | 3+4 2 |
| | 5 2 | 3+4 4 | 5 2 | 3+4 4 | 2+2 4 | 5+6 2 |
| | 5 2 | 3+4 4 | 6 2 | 3+4 4 | 2+2 4 | 3+4 2 |
| | 6 2 | | 5 2 | | | 5+6 2 |
| | 5 2 | | 5 2 | | | 3+4 2 |
| | 6 2 | | 6 2 | | | 5+6 2 |
| | 5 2 | | 6 2 | | | 5+6 2 |
| | 6 2 | | 5 2 | | | 3+4 2 |
| | 6 2 | | 5 2 | | | 5+6 2 |
| | 5 2 | | 6 2 | | | 5+6 2 |
| | 5 2 | | 5 2 | | | 3+4 2 |
| | 6 2 | | 6 2 | | | 3+4 2 |
| | 6 2 | | 6 2 | | | 5+6 2 |
| | 5 2 | | 6 2 | | | 3+4 2 |
| | 6 2 | | 6 2 | | | 5+6 2 |
| | 5 2 | | 5 2 | | | 5+6 2 |
| | 6 2 | | 5 2 | | | 3+4 2 |
| | 5 2 | | 6 2 | | | 3+4 2 |
| | 6 2 | | 5 2 | | | 5+6 2 |
| | 6 2 | | 5 2 | | | 5+6 2 |
| | 5 2 | | 6 2 | | | 3+4 2 |
| | 5 2 | | 6 2 | | | 5+6 2 |
| | 5 2 | | 6 2 | | | 3+4 2 |
| | 6 2 | | 5 2 | | | 5+6 2 |
| | 6 2 | | 6 2 | | | 3+4 2 |
| | 6 2 | | 5 2 | | | 5+6 2 |
| 63.2.42 | 154676 | | 92 | Ga4Ni1Yb1 | | |
| | -GM | -T | -Y | -Z | -R | -S |
| | 5 2 | 3+4 4 | 5 2 | 3+4 4 | 2+2 4 | 3+4 2 |
| | 6 2 | 3+4 4 | 6 2 | 3+4 4 | 2+2 4 | 5+6 2 |
| | 5 2 | 3+4 4 | 5 2 | 3+4 4 | 2+2 4 | 5+6 2 |
| | 6 2 | 3+4 4 | 6 2 | 3+4 4 | 2+2 4 | 3+4 2 |
| | 5 2 | 3+4 4 | 5 2 | 3+4 4 | 2+2 4 | 5+6 2 |
| | 6 2 | 3+4 4 | 6 2 | 3+4 4 | 2+2 4 | 3+4 2 |
| | 5 2 | 3+4 4 | 6 2 | 3+4 4 | 2+2 4 | 5+6 2 |
| | 6 2 | 3+4 4 | 5 2 | 3+4 4 | 2+2 4 | 3+4 2 |
| | 5 2 | 3+4 4 | 6 2 | 3+4 4 | 2+2 4 | 3+4 2 |
| | 6 2 | 3+4 4 | 5 2 | 3+4 4 | 2+2 4 | 5+6 2 |
| | 5 2 | 3+4 4 | 6 2 | 3+4 4 | 2+2 4 | 5+6 2 |
| | 5 2 | 3+4 4 | 6 2 | 3+4 4 | 2+2 4 | 3+4 2 |
| | 5 2 | 3+4 4 | 5 2 | 3+4 4 | 2+2 4 | 3+4 2 |
| | 6 2 | 3+4 4 | 5 2 | 3+4 4 | 2+2 4 | 3+4 2 |
| | 6 2 | 3+4 4 | 5 2 | 3+4 4 | 2+2 4 | 5+6 2 |
| | 6 2 | 3+4 4 | 6 2 | 3+4 4 | 2+2 4 | 5+6 2 |
| | 5 2 | 3+4 4 | 5 2 | 3+4 4 | 2+2 4 | 3+4 2 |
| | 5 2 | 3+4 4 | 5 2 | 3+4 4 | 2+2 4 | 3+4 2 |
| | 5 2 | 3+4 4 | 6 2 | 3+4 4 | 2+2 4 | 5+6 2 |
| | 6 2 | 3+4 4 | 6 2 | 3+4 4 | 2+2 4 | 5+6 2 |
| | 6 2 | 3+4 4 | 6 2 | 3+4 4 | 2+2 4 | 3+4 2 |
| | 5 2 | 3+4 4 | 5 2 | 3+4 4 | 2+2 4 | 5+6 2 |
| | 5 2 | 3+4 4 | 6 2 | 3+4 4 | 2+2 4 | 3+4 2 |
| | 6 2 | | 5 2 | | | 5+6 2 |
| | 5 2 | | 5 2 | | | 3+4 2 |
| | 6 2 | | 6 2 | | | 5+6 2 |
| | 5 2 | | 6 2 | | | 5+6 2 |
| | 6 2 | | 5 2 | | | 3+4 2 |
| | 6 2 | | 5 2 | | | 5+6 2 |

Continued on next column

**TABLE XXXVIII– continued**

| TQC | ICSD | | NoE | | SF | |
|---|---|---|---|---|---|---|
| | 5 2 | | 6 2 | | | 5+6 2 |
| | 5 2 | | 5 2 | | | 3+4 2 |
| | 6 2 | | 6 2 | | | 3+4 2 |
| | 6 2 | | 6 2 | | | 5+6 2 |
| | 5 2 | | 6 2 | | | 3+4 2 |
| | 6 2 | | 6 2 | | | 5+6 2 |
| | 5 2 | | 5 2 | | | 5+6 2 |
| | 6 2 | | 5 2 | | | 3+4 2 |
| | 5 2 | | 5 2 | | | 3+4 2 |
| | 6 2 | | 6 2 | | | 5+6 2 |
| | 6 2 | | 5 2 | | | 5+6 2 |
| | 5 2 | | 6 2 | | | 3+4 2 |
| | 5 2 | | 6 2 | | | 5+6 2 |
| | 5 2 | | 6 2 | | | 3+4 2 |
| | 6 2 | | 5 2 | | | 5+6 2 |
| | 6 2 | | 6 2 | | | 3+4 2 |
| | 6 2 | | 5 2 | | | 5+6 2 |
| 63.2.43 | 262740 | | 80 | | Ga2Ni1Yb1 | |
| | -GM | -T | -Y | -Z | -R | -S |
| | 5 2 | 3+4 4 | 5 2 | 3+4 4 | 2+2 4 | 3+4 2 |
| | 6 2 | 3+4 4 | 6 2 | 3+4 4 | 2+2 4 | 5+6 2 |
| | 5 2 | 3+4 4 | 5 2 | 3+4 4 | 2+2 4 | 3+4 2 |
| | 6 2 | 3+4 4 | 6 2 | 3+4 4 | 2+2 4 | 5+6 2 |
| | 5 2 | 3+4 4 | 5 2 | 3+4 4 | 2+2 4 | 3+4 2 |
| | 6 2 | 3+4 4 | 6 2 | 3+4 4 | 2+2 4 | 5+6 2 |
| | 5 2 | 3+4 4 | 5 2 | 3+4 4 | 2+2 4 | 3+4 2 |
| | 6 2 | 3+4 4 | 6 2 | 3+4 4 | 2+2 4 | 5+6 2 |
| | 5 2 | 3+4 4 | 6 2 | 3+4 4 | 2+2 4 | 5+6 2 |
| | 5 2 | 3+4 4 | 5 2 | 3+4 4 | 2+2 4 | 3+4 2 |
| | 6 2 | 3+4 4 | 6 2 | 3+4 4 | 2+2 4 | 5+6 2 |
| | 6 2 | 3+4 4 | 5 2 | 3+4 4 | 2+2 4 | 3+4 2 |
| | 5 2 | 3+4 4 | 5 2 | 3+4 4 | 2+2 4 | 3+4 2 |
| | 6 2 | 3+4 4 | 6 2 | 3+4 4 | 2+2 4 | 5+6 2 |
| | 5 2 | 3+4 4 | 6 2 | 3+4 4 | 2+2 4 | 5+6 2 |
| | 5 2 | 3+4 4 | 5 2 | 3+4 4 | 2+2 4 | 5+6 2 |
| | 5 2 | 3+4 4 | 5 2 | 3+4 4 | 2+2 4 | 3+4 2 |
| | 6 2 | 3+4 4 | 6 2 | 3+4 4 | 2+2 4 | 3+4 2 |
| | 6 2 | 3+4 4 | 6 2 | 3+4 4 | 2+2 4 | 5+6 2 |
| | 5 2 | 3+4 4 | 6 2 | 3+4 4 | 2+2 4 | 3+4 2 |
| | 6 2 | | 5 2 | | | 5+6 2 |
| | 5 2 | | 6 2 | | | 5+6 2 |
| | 6 2 | | 6 2 | | | 3+4 2 |
| | 5 2 | | 5 2 | | | 3+4 2 |
| | 5 2 | | 5 2 | | | 5+6 2 |
| | 6 2 | | 6 2 | | | 3+4 2 |
| | 6 2 | | 5 2 | | | 5+6 2 |
| | 6 2 | | 5 2 | | | 3+4 2 |
| | 5 2 | | 6 2 | | | 5+6 2 |
| | 5 2 | | 5 2 | | | 3+4 2 |
| | 5 2 | | 6 2 | | | 3+4 2 |
| | 6 2 | | 5 2 | | | 5+6 2 |
| | 6 2 | | 6 2 | | | 5+6 2 |
| | 5 2 | | 5 2 | | | 3+4 2 |
| | 6 2 | | 6 2 | | | 5+6 2 |
| | 6 2 | | 5 2 | | | 3+4 2 |
| | 5 2 | | 5 2 | | | 3+4 2 |
| | 5 2 | | 6 2 | | | 5+6 2 |
| | 5 2 | | 6 2 | | | 3+4 2 |
| | 5 2 | | 6 2 | | | 5+6 2 |
| 63.2.44 | 262739 | | 80 | | Ga2Ni1Yb1 | |
| | -GM | -T | -Y | -Z | -R | -S |

Continued on next column

**TABLE XXXVIII– continued**

| TQC | ICSD | | NoE | | SF | |
|---|---|---|---|---|---|---|
| | 5 2 | 3+4 4 | 5 2 | 3+4 4 | 2+2 4 | 5+6 2 |
| | 6 2 | 3+4 4 | 6 2 | 3+4 4 | 2+2 4 | 3+4 2 |
| | 5 2 | 3+4 4 | 5 2 | 3+4 4 | 2+2 4 | 5+6 2 |
| | 6 2 | 3+4 4 | 6 2 | 3+4 4 | 2+2 4 | 3+4 2 |
| | 5 2 | 3+4 4 | 5 2 | 3+4 4 | 2+2 4 | 5+6 2 |
| | 6 2 | 3+4 4 | 6 2 | 3+4 4 | 2+2 4 | 3+4 2 |
| | 5 2 | 3+4 4 | 5 2 | 3+4 4 | 2+2 4 | 5+6 2 |
| | 6 2 | 3+4 4 | 6 2 | 3+4 4 | 2+2 4 | 3+4 2 |
| | 5 2 | 3+4 4 | 6 2 | 3+4 4 | 2+2 4 | 3+4 2 |
| | 5 2 | 3+4 4 | 5 2 | 3+4 4 | 2+2 4 | 5+6 2 |
| | 6 2 | 3+4 4 | 6 2 | 3+4 4 | 2+2 4 | 3+4 2 |
| | 6 2 | 3+4 4 | 5 2 | 3+4 4 | 2+2 4 | 5+6 2 |
| | 5 2 | 3+4 4 | 5 2 | 3+4 4 | 2+2 4 | 5+6 2 |
| | 6 2 | 3+4 4 | 6 2 | 3+4 4 | 2+2 4 | 3+4 2 |
| | 5 2 | 3+4 4 | 6 2 | 3+4 4 | 2+2 4 | 3+4 2 |
| | 5 2 | 3+4 4 | 5 2 | 3+4 4 | 2+2 4 | 3+4 2 |
| | 5 2 | 3+4 4 | 5 2 | 3+4 4 | 2+2 4 | 5+6 2 |
| | 6 2 | 3+4 4 | 6 2 | 3+4 4 | 2+2 4 | 5+6 2 |
| | 6 2 | 3+4 4 | 6 2 | 3+4 4 | 2+2 4 | 5+6 2 |
| | 5 2 | 3+4 4 | 6 2 | 3+4 4 | 2+2 4 | 3+4 2 |
| | 6 2 | | 5 2 | | | 3+4 2 |
| | 5 2 | | 6 2 | | | 3+4 2 |
| | 5 2 | | 6 2 | | | 5+6 2 |
| | 6 2 | | 5 2 | | | 5+6 2 |
| | 5 2 | | 5 2 | | | 3+4 2 |
| | 6 2 | | 6 2 | | | 3+4 2 |
| | 6 2 | | 5 2 | | | 5+6 2 |
| | 6 2 | | 5 2 | | | 5+6 2 |
| | 5 2 | | 6 2 | | | 3+4 2 |
| | 5 2 | | 5 2 | | | 5+6 2 |
| | 5 2 | | 6 2 | | | 5+6 2 |
| | 6 2 | | 5 2 | | | 3+4 2 |
| | 6 2 | | 6 2 | | | 3+4 2 |
| | 5 2 | | 5 2 | | | 5+6 2 |
| | 6 2 | | 6 2 | | | 3+4 2 |
| | 6 2 | | 5 2 | | | 5+6 2 |
| | 5 2 | | 5 2 | | | 3+4 2 |
| | 5 2 | | 6 2 | | | 5+6 2 |
| | 5 2 | | 6 2 | | | 5+6 2 |
| | 5 2 | | 6 2 | | | 3+4 2 |
| 63.2.45 | 262738 | | 80 | | Ga2Ni1Yb1 | |
| | -GM | -T | -Y | -Z | -R | -S |
| | 5 2 | 3+4 4 | 5 2 | 3+4 4 | 2+2 4 | 5+6 2 |
| | 6 2 | 3+4 4 | 6 2 | 3+4 4 | 2+2 4 | 3+4 2 |
| | 5 2 | 3+4 4 | 5 2 | 3+4 4 | 2+2 4 | 5+6 2 |
| | 6 2 | 3+4 4 | 6 2 | 3+4 4 | 2+2 4 | 3+4 2 |
| | 5 2 | 3+4 4 | 5 2 | 3+4 4 | 2+2 4 | 5+6 2 |
| | 6 2 | 3+4 4 | 6 2 | 3+4 4 | 2+2 4 | 3+4 2 |
| | 5 2 | 3+4 4 | 5 2 | 3+4 4 | 2+2 4 | 5+6 2 |
| | 6 2 | 3+4 4 | 6 2 | 3+4 4 | 2+2 4 | 3+4 2 |
| | 5 2 | 3+4 4 | 6 2 | 3+4 4 | 2+2 4 | 3+4 2 |
| | 6 2 | 3+4 4 | 5 2 | 3+4 4 | 2+2 4 | 5+6 2 |
| | 5 2 | 3+4 4 | 6 2 | 3+4 4 | 2+2 4 | 3+4 2 |
| | 6 2 | 3+4 4 | 5 2 | 3+4 4 | 2+2 4 | 5+6 2 |
| | 5 2 | 3+4 4 | 5 2 | 3+4 4 | 2+2 4 | 5+6 2 |
| | 6 2 | 3+4 4 | 6 2 | 3+4 4 | 2+2 4 | 3+4 2 |
| | 5 2 | 3+4 4 | 6 2 | 3+4 4 | 2+2 4 | 3+4 2 |
| | 5 2 | 3+4 4 | 5 2 | 3+4 4 | 2+2 4 | 3+4 2 |
| | 5 2 | 3+4 4 | 5 2 | 3+4 4 | 2+2 4 | 5+6 2 |
| | 6 2 | 3+4 4 | 6 2 | 3+4 4 | 2+2 4 | 5+6 2 |
| | 6 2 | 3+4 4 | 6 2 | 3+4 4 | 2+2 4 | 5+6 2 |

Continued on next column

**TABLE XXXVIII– continued**

| TQC | ICSD | | NoE | SF | | |
|---|---|---|---|---|---|---|
| | 5 2 | 3+4 4 | 6 2 | 3+4 4 | 2+2 4 | 3+4 2 |
| | 6 2 | | 5 2 | | | 3+4 2 |
| | 5 2 | | 6 2 | | | 3+4 2 |
| | 5 2 | | 6 2 | | | 5+6 2 |
| | 6 2 | | 5 2 | | | 5+6 2 |
| | 5 2 | | 5 2 | | | 3+4 2 |
| | 6 2 | | 6 2 | | | 3+4 2 |
| | 6 2 | | 5 2 | | | 5+6 2 |
| | 6 2 | | 5 2 | | | 5+6 2 |
| | 5 2 | | 6 2 | | | 3+4 2 |
| | 5 2 | | 5 2 | | | 5+6 2 |
| | 5 2 | | 6 2 | | | 5+6 2 |
| | 6 2 | | 5 2 | | | 3+4 2 |
| | 6 2 | | 6 2 | | | 3+4 2 |
| | 5 2 | | 5 2 | | | 5+6 2 |
| | 6 2 | | 6 2 | | | 3+4 2 |
| | 6 2 | | 5 2 | | | 5+6 2 |
| | 5 2 | | 5 2 | | | 3+4 2 |
| | 5 2 | | 6 2 | | | 5+6 2 |
| | 5 2 | | 6 2 | | | 5+6 2 |
| | 5 2 | | 6 2 | | | 3+4 2 |
| 63.2.46 | 262737 | | 80 | Ga2Ni1Yb1 | | |
| | -GM | -T | -Y | -Z | -R | -S |
| | 5 2 | 3+4 4 | 5 2 | 3+4 4 | 2+2 4 | 3+4 2 |
| | 6 2 | 3+4 4 | 6 2 | 3+4 4 | 2+2 4 | 5+6 2 |
| | 5 2 | 3+4 4 | 5 2 | 3+4 4 | 2+2 4 | 3+4 2 |
| | 6 2 | 3+4 4 | 6 2 | 3+4 4 | 2+2 4 | 5+6 2 |
| | 5 2 | 3+4 4 | 5 2 | 3+4 4 | 2+2 4 | 3+4 2 |
| | 6 2 | 3+4 4 | 6 2 | 3+4 4 | 2+2 4 | 5+6 2 |
| | 5 2 | 3+4 4 | 5 2 | 3+4 4 | 2+2 4 | 3+4 2 |
| | 6 2 | 3+4 4 | 6 2 | 3+4 4 | 2+2 4 | 5+6 2 |
| | 5 2 | 3+4 4 | 6 2 | 3+4 4 | 2+2 4 | 5+6 2 |
| | 6 2 | 3+4 4 | 5 2 | 3+4 4 | 2+2 4 | 3+4 2 |
| | 5 2 | 3+4 4 | 6 2 | 3+4 4 | 2+2 4 | 5+6 2 |
| | 6 2 | 3+4 4 | 5 2 | 3+4 4 | 2+2 4 | 3+4 2 |
| | 5 2 | 3+4 4 | 5 2 | 3+4 4 | 2+2 4 | 3+4 2 |
| | 6 2 | 3+4 4 | 6 2 | 3+4 4 | 2+2 4 | 5+6 2 |
| | 5 2 | 3+4 4 | 6 2 | 3+4 4 | 2+2 4 | 5+6 2 |
| | 5 2 | 3+4 4 | 5 2 | 3+4 4 | 2+2 4 | 5+6 2 |
| | 5 2 | 3+4 4 | 5 2 | 3+4 4 | 2+2 4 | 3+4 2 |
| | 6 2 | 3+4 4 | 6 2 | 3+4 4 | 2+2 4 | 3+4 2 |
| | 6 2 | 3+4 4 | 6 2 | 3+4 4 | 2+2 4 | 5+6 2 |
| | 5 2 | 3+4 4 | 6 2 | 3+4 4 | 2+2 4 | 3+4 2 |
| | 6 2 | | 5 2 | | | 5+6 2 |
| | 5 2 | | 6 2 | | | 5+6 2 |
| | 6 2 | | 6 2 | | | 3+4 2 |
| | 5 2 | | 5 2 | | | 3+4 2 |
| | 5 2 | | 5 2 | | | 5+6 2 |
| | 6 2 | | 6 2 | | | 5+6 2 |
| | 6 2 | | 5 2 | | | 3+4 2 |
| | 6 2 | | 5 2 | | | 3+4 2 |
| | 5 2 | | 6 2 | | | 5+6 2 |
| | 5 2 | | 5 2 | | | 3+4 2 |
| | 5 2 | | 6 2 | | | 3+4 2 |
| | 6 2 | | 5 2 | | | 5+6 2 |
| | 6 2 | | 6 2 | | | 5+6 2 |
| | 5 2 | | 5 2 | | | 3+4 2 |
| | 6 2 | | 6 2 | | | 5+6 2 |
| | 6 2 | | 5 2 | | | 3+4 2 |
| | 5 2 | | 5 2 | | | 3+4 2 |
| | 5 2 | | 6 2 | | | 5+6 2 |

Continued on next column

**TABLE XXXVIII– continued**

| TQC | ICSD | | NoE | SF | | |
|---|---|---|---|---|---|---|
| | 5 2 | | 6 2 | | | 3+4 2 |
| | 5 2 | | 6 2 | | | 5+6 2 |
| 63.2.47 | 423623 | | 28 | Ga1La1 | | |
| | -GM | -T | -Y | -Z | -R | -S |
| | 5 2 | 3+4 4 | 6 2 | 3+4 4 | 2+2 4 | 5+6 2 |
| | 6 2 | 3+4 4 | 5 2 | 3+4 4 | 2+2 4 | 3+4 2 |
| | 5 2 | 3+4 4 | 6 2 | 3+4 4 | 2+2 4 | 5+6 2 |
| | 6 2 | 3+4 4 | 5 2 | 3+4 4 | 2+2 4 | 3+4 2 |
| | 5 2 | 3+4 4 | 6 2 | 3+4 4 | 2+2 4 | 5+6 2 |
| | 5 2 | 3+4 4 | 5 2 | 3+4 4 | 2+2 4 | 3+4 2 |
| | 6 2 | 3+4 4 | 6 2 | 3+4 4 | 2+2 4 | 5+6 2 |
| | 6 2 | | 5 2 | | | 3+4 2 |
| | 5 2 | | 5 2 | | | 3+4 2 |
| | 6 2 | | 6 2 | | | 5+6 2 |
| | 5 2 | | 5 2 | | | 5+6 2 |
| | 6 2 | | 5 2 | | | 3+4 2 |
| | 6 2 | | 6 2 | | | 3+4 2 |
| | 5 2 | | 5 2 | | | 3+4 2 |
| 63.2.48 | 2795 | | 60 | Ga1 | | |
| | -GM | -T | -Y | -Z | -R | -S |
| | 5 2 | 3+4 4 | 5 2 | 3+4 4 | 2+2 4 | 3+4 2 |
| | 6 2 | 3+4 4 | 6 2 | 3+4 4 | 2+2 4 | 5+6 2 |
| | 6 2 | 3+4 4 | 5 2 | 3+4 4 | 2+2 4 | 3+4 2 |
| | 5 2 | 3+4 4 | 6 2 | 3+4 4 | 2+2 4 | 5+6 2 |
| | 5 2 | 3+4 4 | 6 2 | 3+4 4 | 2+2 4 | 3+4 2 |
| | 6 2 | 3+4 4 | 5 2 | 3+4 4 | 2+2 4 | 5+6 2 |
| | 5 2 | 3+4 4 | 5 2 | 3+4 4 | 2+2 4 | 5+6 2 |
| | 5 2 | 3+4 4 | 6 2 | 3+4 4 | 2+2 4 | 5+6 2 |
| | 6 2 | 3+4 4 | 5 2 | 3+4 4 | 2+2 4 | 3+4 2 |
| | 6 2 | 3+4 4 | 6 2 | 3+4 4 | 2+2 4 | 3+4 2 |
| | 5 2 | 3+4 4 | 6 2 | 3+4 4 | 2+2 4 | 5+6 2 |
| | 5 2 | 3+4 4 | 5 2 | 3+4 4 | 2+2 4 | 3+4 2 |
| | 5 2 | 3+4 4 | 5 2 | 3+4 4 | 2+2 4 | 3+4 2 |
| | 6 2 | 3+4 4 | 6 2 | 3+4 4 | 2+2 4 | 5+6 2 |
| | 6 2 | 3+4 4 | 5 2 | 3+4 4 | 2+2 4 | 5+6 2 |
| | 6 2 | | 5 2 | | | 3+4 2 |
| | 5 2 | | 6 2 | | | 5+6 2 |
| | 6 2 | | 6 2 | | | 3+4 2 |
| | 6 2 | | 5 2 | | | 5+6 2 |
| | 5 2 | | 5 2 | | | 3+4 2 |
| | 6 2 | | 6 2 | | | 3+4 2 |
| | 5 2 | | 6 2 | | | 5+6 2 |
| | 5 2 | | 5 2 | | | 5+6 2 |
| | 6 2 | | 6 2 | | | 3+4 2 |
| | 5 2 | | 5 2 | | | 3+4 2 |
| | 5 2 | | 6 2 | | | 5+6 2 |
| | 6 2 | | 5 2 | | | 3+4 2 |
| | 6 2 | | 6 2 | | | 5+6 2 |
| | 5 2 | | 6 2 | | | 3+4 2 |
| | 5 2 | | 5 2 | | | 5+6 2 |
| 63.2.49 | 290249 | | 124 | Fe4O5 | | |
| | -GM | -T | -Y | -Z | -R | -S |
| | 5 2 | 3+4 4 | 6 2 | 3+4 4 | 2+2 4 | 3+4 2 |
| | 5 2 | 3+4 4 | 5 2 | 3+4 4 | 2+2 4 | 5+6 2 |
| | 6 2 | 3+4 4 | 6 2 | 3+4 4 | 2+2 4 | 3+4 2 |
| | 5 2 | 3+4 4 | 5 2 | 3+4 4 | 2+2 4 | 5+6 2 |
| | 6 2 | 3+4 4 | 6 2 | 3+4 4 | 2+2 4 | 3+4 2 |
| | 6 2 | 3+4 4 | 5 2 | 3+4 4 | 2+2 4 | 5+6 2 |
| | 5 2 | 3+4 4 | 6 2 | 3+4 4 | 2+2 4 | 3+4 2 |
| | 5 2 | 3+4 4 | 5 2 | 3+4 4 | 2+2 4 | 5+6 2 |
| | 6 2 | 3+4 4 | 5 2 | 3+4 4 | 2+2 4 | 3+4 2 |

Continued on next column

**TABLE XXXVIII– continued**

| TQC | ICSD | NoE | SF |
|---|---|---|---|
| | 6 2 3+4 4 | 6 2 3+4 4 2+2 4 | 5+6 2 |
| | 5 2 3+4 4 | 5 2 3+4 4 2+2 4 | 3+4 2 |
| | 6 2 3+4 4 | 5 2 3+4 4 2+2 4 | 3+4 2 |
| | 5 2 3+4 4 | 6 2 3+4 4 2+2 4 | 5+6 2 |
| | 5 2 3+4 4 | 6 2 3+4 4 2+2 4 | 5+6 2 |
| | 6 2 3+4 4 | 5 2 3+4 4 2+2 4 | 3+4 2 |
| | 6 2 3+4 4 | 6 2 3+4 4 2+2 4 | 5+6 2 |
| | 5 2 3+4 4 | 5 2 3+4 4 2+2 4 | 5+6 2 |
| | 6 2 3+4 4 | 6 2 3+4 4 2+2 4 | 5+6 2 |
| | 5 2 3+4 4 | 5 2 3+4 4 2+2 4 | 3+4 2 |
| | 5 2 3+4 4 | 6 2 3+4 4 2+2 4 | 3+4 2 |
| | 5 2 3+4 4 | 6 2 3+4 4 2+2 4 | 3+4 2 |
| | 5 2 3+4 4 | 5 2 3+4 4 2+2 4 | 3+4 2 |
| | 6 2 3+4 4 | 5 2 3+4 4 2+2 4 | 5+6 2 |
| | 5 2 3+4 4 | 5 2 3+4 4 2+2 4 | 5+6 2 |
| | 6 2 3+4 4 | 5 2 3+4 4 2+2 4 | 3+4 2 |
| | 6 2 3+4 4 | 6 2 3+4 4 2+2 4 | 3+4 2 |
| | 5 2 3+4 4 | 5 2 3+4 4 2+2 4 | 3+4 2 |
| | 5 2 3+4 4 | 6 2 3+4 4 2+2 4 | 5+6 2 |
| | 5 2 3+4 4 | 6 2 3+4 4 2+2 4 | 3+4 2 |
| | 5 2 3+4 4 | 6 2 3+4 4 2+2 4 | 5+6 2 |
| | 6 2 3+4 4 | 6 2 3+4 4 2+2 4 | 3+4 2 |
| | 5 2 | 6 2 | 5+6 2 |
| | 6 2 | 5 2 | 5+6 2 |
| | 6 2 | 5 2 | 3+4 2 |
| | 5 2 | 5 2 | 5+6 2 |
| | 6 2 | 6 2 | 5+6 2 |
| | 6 2 | 5 2 | 3+4 2 |
| | 6 2 | 6 2 | 3+4 2 |
| | 6 2 | 6 2 | 5+6 2 |
| | 6 2 | 5 2 | 5+6 2 |
| | 5 2 | 5 2 | 3+4 2 |
| | 5 2 | 5 2 | 3+4 2 |
| | 6 2 | 6 2 | 5+6 2 |
| | 5 2 | 5 2 | 3+4 2 |
| | 6 2 | 6 2 | 3+4 2 |
| | 5 2 | 6 2 | 3+4 2 |
| | 5 2 | 5 2 | 5+6 2 |
| | 5 2 | 6 2 | 5+6 2 |
| | 6 2 | 6 2 | 3+4 2 |
| | 5 2 | 5 2 | 5+6 2 |
| | 5 2 | 5 2 | 3+4 2 |
| | 6 2 | 5 2 | 5+6 2 |
| | 6 2 | 5 2 | 5+6 2 |
| | 5 2 | 5 2 | 3+4 2 |
| | 5 2 | 5 2 | 3+4 2 |
| | 5 2 | 5 2 | 5+6 2 |
| | 6 2 | 5 2 | 3+4 2 |
| | 5 2 | 6 2 | 5+6 2 |
| | 6 2 | 5 2 | 3+4 2 |
| | 6 2 | 6 2 | 3+4 2 |
| | 5 2 | 5 2 | 3+4 2 |
| | 6 2 | 6 2 | 3+4 2 |
| 63.2.50 | 98 | 80 | Fe1S3U1 |
| | -GM -T | -Y -Z | -R -S |
| | 5 2 3+4 4 | 6 2 3+4 4 2+2 4 | 3+4 2 |
| | 6 2 3+4 4 | 5 2 3+4 4 2+2 4 | 5+6 2 |
| | 5 2 3+4 4 | 6 2 3+4 4 2+2 4 | 3+4 2 |
| | 6 2 3+4 4 | 5 2 3+4 4 2+2 4 | 5+6 2 |
| | 5 2 3+4 4 | 6 2 3+4 4 2+2 4 | 3+4 2 |
| | 6 2 3+4 4 | 5 2 3+4 4 2+2 4 | 5+6 2 |

Continued on next column

**TABLE XXXVIII– continued**

| TQC | ICSD | NoE | SF |
|---|---|---|---|
| | 5 2 3+4 4 | 6 2 3+4 4 2+2 4 | 3+4 2 |
| | 6 2 3+4 4 | 5 2 3+4 4 2+2 4 | 5+6 2 |
| | 5 2 3+4 4 | 6 2 3+4 4 2+2 4 | 5+6 2 |
| | 6 2 3+4 4 | 5 2 3+4 4 2+2 4 | 3+4 2 |
| | 5 2 3+4 4 | 5 2 3+4 4 2+2 4 | 5+6 2 |
| | 6 2 3+4 4 | 5 2 3+4 4 2+2 4 | 3+4 2 |
| | 6 2 3+4 4 | 6 2 3+4 4 2+2 4 | 5+6 2 |
| | 5 2 3+4 4 | 6 2 3+4 4 2+2 4 | 3+4 2 |
| | 5 2 3+4 4 | 5 2 3+4 4 2+2 4 | 3+4 2 |
| | 5 2 3+4 4 | 5 2 3+4 4 2+2 4 | 3+4 2 |
| | 6 2 3+4 4 | 6 2 3+4 4 2+2 4 | 5+6 2 |
| | 5 2 3+4 4 | 6 2 3+4 4 2+2 4 | 5+6 2 |
| | 6 2 3+4 4 | 5 2 3+4 4 2+2 4 | 5+6 2 |
| | 5 2 3+4 4 | 6 2 3+4 4 2+2 4 | 3+4 2 |
| | 5 2 | 6 2 | 3+4 2 |
| | 6 2 | 5 2 | 3+4 2 |
| | 6 2 | 5 2 | 5+6 2 |
| | 5 2 | 5 2 | 3+4 2 |
| | 5 2 | 6 2 | 5+6 2 |
| | 5 2 | 6 2 | 5+6 2 |
| | 5 2 | 6 2 | 5+6 2 |
| | 6 2 | 5 2 | 3+4 2 |
| | 6 2 | 6 2 | 3+4 2 |
| | 6 2 | 5 2 | 3+4 2 |
| | 6 2 | 6 2 | 5+6 2 |
| | 5 2 | 5 2 | 5+6 2 |
| | 5 2 | 5 2 | 3+4 2 |
| | 5 2 | 5 2 | 3+4 2 |
| | 6 2 | 5 2 | 3+4 2 |
| | 5 2 | 5 2 | 3+4 2 |
| | 5 2 | 5 2 | 3+4 2 |
| | 5 2 | 5 2 | 3+4 2 |
| | 5 2 | 5 2 | 3+4 2 |
| | 5 2 | 6 2 | 5+6 2 |
| 63.2.51 | 261040 | 80 | Fe1S3U1 |
| | -GM -T | -Y -Z | -R -S |
| | 5 2 3+4 4 | 6 2 3+4 4 2+2 4 | 3+4 2 |
| | 6 2 3+4 4 | 5 2 3+4 4 2+2 4 | 5+6 2 |
| | 5 2 3+4 4 | 6 2 3+4 4 2+2 4 | 3+4 2 |
| | 6 2 3+4 4 | 5 2 3+4 4 2+2 4 | 5+6 2 |
| | 5 2 3+4 4 | 6 2 3+4 4 2+2 4 | 3+4 2 |
| | 6 2 3+4 4 | 5 2 3+4 4 2+2 4 | 5+6 2 |
| | 5 2 3+4 4 | 6 2 3+4 4 2+2 4 | 3+4 2 |
| | 6 2 3+4 4 | 5 2 3+4 4 2+2 4 | 5+6 2 |
| | 5 2 3+4 4 | 6 2 3+4 4 2+2 4 | 5+6 2 |
| | 6 2 3+4 4 | 5 2 3+4 4 2+2 4 | 3+4 2 |
| | 5 2 3+4 4 | 5 2 3+4 4 2+2 4 | 5+6 2 |
| | 6 2 3+4 4 | 5 2 3+4 4 2+2 4 | 3+4 2 |
| | 6 2 3+4 4 | 6 2 3+4 4 2+2 4 | 5+6 2 |
| | 5 2 3+4 4 | 6 2 3+4 4 2+2 4 | 3+4 2 |
| | 5 2 3+4 4 | 5 2 3+4 4 2+2 4 | 3+4 2 |
| | 5 2 3+4 4 | 5 2 3+4 4 2+2 4 | 3+4 2 |
| | 6 2 3+4 4 | 6 2 3+4 4 2+2 4 | 5+6 2 |
| | 5 2 3+4 4 | 6 2 3+4 4 2+2 4 | 5+6 2 |
| | 6 2 3+4 4 | 5 2 3+4 4 2+2 4 | 5+6 2 |
| | 5 2 3+4 4 | 6 2 3+4 4 2+2 4 | 3+4 2 |
| | 5 2 | 6 2 | 3+4 2 |
| | 6 2 | 5 2 | 3+4 2 |
| | 6 2 | 5 2 | 5+6 2 |
| | 5 2 | 5 2 | 3+4 2 |
| | 5 2 | 6 2 | 5+6 2 |

Continued on next column

**TABLE XXXVIII– continued**

| TQC | ICSD | | NoE | | SF | |
|---|---|---|---|---|---|---|
| | 5 2 | | 6 2 | | | 5+6 2 |
| | 5 2 | | 6 2 | | | 5+6 2 |
| | 6 2 | | 5 2 | | | 3+4 2 |
| | 6 2 | | 6 2 | | | 3+4 2 |
| | 6 2 | | 5 2 | | | 3+4 2 |
| | 6 2 | | 6 2 | | | 5+6 2 |
| | 5 2 | | 5 2 | | | 5+6 2 |
| | 5 2 | | 5 2 | | | 3+4 2 |
| | 5 2 | | 5 2 | | | 3+4 2 |
| | 6 2 | | 5 2 | | | 3+4 2 |
| | 5 2 | | 5 2 | | | 3+4 2 |
| | 5 2 | | 5 2 | | | 3+4 2 |
| | 5 2 | | 5 2 | | | 3+4 2 |
| | 5 2 | | 5 2 | | | 3+4 2 |
| | 5 2 | | 6 2 | | | 5+6 2 |
| 63.2.52 | 261039 | | 80 | | Fe1S3U1 | |
| | -GM | -T | -Y | -Z | -R | -S |
| | 5 2 | 3+4 4 | 6 2 | 3+4 4 | 2+2 4 | 3+4 2 |
| | 6 2 | 3+4 4 | 5 2 | 3+4 4 | 2+2 4 | 5+6 2 |
| | 5 2 | 3+4 4 | 6 2 | 3+4 4 | 2+2 4 | 3+4 2 |
| | 6 2 | 3+4 4 | 5 2 | 3+4 4 | 2+2 4 | 5+6 2 |
| | 5 2 | 3+4 4 | 6 2 | 3+4 4 | 2+2 4 | 3+4 2 |
| | 6 2 | 3+4 4 | 5 2 | 3+4 4 | 2+2 4 | 5+6 2 |
| | 5 2 | 3+4 4 | 6 2 | 3+4 4 | 2+2 4 | 3+4 2 |
| | 6 2 | 3+4 4 | 5 2 | 3+4 4 | 2+2 4 | 5+6 2 |
| | 5 2 | 3+4 4 | 6 2 | 3+4 4 | 2+2 4 | 5+6 2 |
| | 6 2 | 3+4 4 | 5 2 | 3+4 4 | 2+2 4 | 3+4 2 |
| | 5 2 | 3+4 4 | 5 2 | 3+4 4 | 2+2 4 | 5+6 2 |
| | 6 2 | 3+4 4 | 5 2 | 3+4 4 | 2+2 4 | 3+4 2 |
| | 6 2 | 3+4 4 | 6 2 | 3+4 4 | 2+2 4 | 5+6 2 |
| | 5 2 | 3+4 4 | 6 2 | 3+4 4 | 2+2 4 | 3+4 2 |
| | 5 2 | 3+4 4 | 5 2 | 3+4 4 | 2+2 4 | 3+4 2 |
| | 5 2 | 3+4 4 | 5 2 | 3+4 4 | 2+2 4 | 3+4 2 |
| | 6 2 | 3+4 4 | 6 2 | 3+4 4 | 2+2 4 | 5+6 2 |
| | 5 2 | 3+4 4 | 6 2 | 3+4 4 | 2+2 4 | 5+6 2 |
| | 6 2 | 3+4 4 | 5 2 | 3+4 4 | 2+2 4 | 5+6 2 |
| | 5 2 | 3+4 4 | 6 2 | 3+4 4 | 2+2 4 | 3+4 2 |
| | 5 2 | | 6 2 | | | 3+4 2 |
| | 6 2 | | 5 2 | | | 3+4 2 |
| | 6 2 | | 5 2 | | | 5+6 2 |
| | 5 2 | | 5 2 | | | 3+4 2 |
| | 5 2 | | 6 2 | | | 5+6 2 |
| | 5 2 | | 6 2 | | | 5+6 2 |
| | 5 2 | | 6 2 | | | 3+4 2 |
| | 6 2 | | 5 2 | | | 5+6 2 |
| | 6 2 | | 6 2 | | | 3+4 2 |
| | 6 2 | | 5 2 | | | 3+4 2 |
| | 6 2 | | 6 2 | | | 5+6 2 |
| | 5 2 | | 5 2 | | | 5+6 2 |
| | 5 2 | | 5 2 | | | 3+4 2 |
| | 5 2 | | 5 2 | | | 3+4 2 |
| | 6 2 | | 5 2 | | | 3+4 2 |
| | 5 2 | | 5 2 | | | 3+4 2 |
| | 5 2 | | 5 2 | | | 3+4 2 |
| | 5 2 | | 5 2 | | | 3+4 2 |
| | 5 2 | | 5 2 | | | 3+4 2 |
| | 5 2 | | 6 2 | | | 5+6 2 |
| 63.2.53 | 261038 | | 80 | | Fe1S3U1 | |
| | -GM | -T | -Y | -Z | -R | -S |
| | 5 2 | 3+4 4 | 6 2 | 3+4 4 | 2+2 4 | 3+4 2 |
| | 6 2 | 3+4 4 | 5 2 | 3+4 4 | 2+2 4 | 5+6 2 |
| | 5 2 | 3+4 4 | 6 2 | 3+4 4 | 2+2 4 | 3+4 2 |
| | 6 2 | 3+4 4 | 5 2 | 3+4 4 | 2+2 4 | 5+6 2 |
| | 5 2 | 3+4 4 | 6 2 | 3+4 4 | 2+2 4 | 3+4 2 |
| | 6 2 | 3+4 4 | 5 2 | 3+4 4 | 2+2 4 | 5+6 2 |
| | 5 2 | 3+4 4 | 6 2 | 3+4 4 | 2+2 4 | 3+4 2 |
| | 6 2 | 3+4 4 | 5 2 | 3+4 4 | 2+2 4 | 5+6 2 |
| | 5 2 | 3+4 4 | 6 2 | 3+4 4 | 2+2 4 | 5+6 2 |
| | 6 2 | 3+4 4 | 5 2 | 3+4 4 | 2+2 4 | 3+4 2 |
| | 5 2 | 3+4 4 | 5 2 | 3+4 4 | 2+2 4 | 5+6 2 |
| | 6 2 | 3+4 4 | 5 2 | 3+4 4 | 2+2 4 | 3+4 2 |
| | 6 2 | 3+4 4 | 6 2 | 3+4 4 | 2+2 4 | 5+6 2 |
| | 5 2 | 3+4 4 | 6 2 | 3+4 4 | 2+2 4 | 3+4 2 |
| | 5 2 | 3+4 4 | 5 2 | 3+4 4 | 2+2 4 | 3+4 2 |
| | 5 2 | 3+4 4 | 5 2 | 3+4 4 | 2+2 4 | 3+4 2 |
| | 6 2 | 3+4 4 | 6 2 | 3+4 4 | 2+2 4 | 5+6 2 |
| | 5 2 | 3+4 4 | 6 2 | 3+4 4 | 2+2 4 | 5+6 2 |
| | 6 2 | 3+4 4 | 5 2 | 3+4 4 | 2+2 4 | 5+6 2 |
| | 5 2 | 3+4 4 | 6 2 | 3+4 4 | 2+2 4 | 3+4 2 |
| | 5 2 | | 6 2 | | | 3+4 2 |
| | 6 2 | | 5 2 | | | 3+4 2 |
| | 6 2 | | 5 2 | | | 5+6 2 |
| | 5 2 | | 5 2 | | | 3+4 2 |
| | 5 2 | | 6 2 | | | 5+6 2 |
| | 5 2 | | 6 2 | | | 5+6 2 |
| | 5 2 | | 6 2 | | | 3+4 2 |
| | 6 2 | | 5 2 | | | 5+6 2 |
| | 6 2 | | 6 2 | | | 3+4 2 |
| | 6 2 | | 5 2 | | | 3+4 2 |
| | 6 2 | | 6 2 | | | 5+6 2 |
| | 5 2 | | 5 2 | | | 5+6 2 |
| | 5 2 | | 5 2 | | | 3+4 2 |
| | 5 2 | | 5 2 | | | 3+4 2 |
| | 6 2 | | 5 2 | | | 3+4 2 |
| | 5 2 | | 5 2 | | | 3+4 2 |
| | 5 2 | | 5 2 | | | 3+4 2 |
| | 5 2 | | 5 2 | | | 3+4 2 |
| | 5 2 | | 5 2 | | | 3+4 2 |
| | 5 2 | | 6 2 | | | 5+6 2 |
| 63.2.54 | 410700 | | 68 | | Eu1Ir1Sn2 | |
| | -GM | -T | -Y | -Z | -R | -S |
| | 5 2 | 3+4 4 | 5 2 | 3+4 4 | 2+2 4 | 3+4 2 |
| | 6 2 | 3+4 4 | 6 2 | 3+4 4 | 2+2 4 | 5+6 2 |
| | 5 2 | 3+4 4 | 5 2 | 3+4 4 | 2+2 4 | 3+4 2 |
| | 6 2 | 3+4 4 | 6 2 | 3+4 4 | 2+2 4 | 5+6 2 |
| | 5 2 | 3+4 4 | 5 2 | 3+4 4 | 2+2 4 | 3+4 2 |
| | 6 2 | 3+4 4 | 6 2 | 3+4 4 | 2+2 4 | 5+6 2 |
| | 6 2 | 3+4 4 | 6 2 | 3+4 4 | 2+2 4 | 3+4 2 |
| | 5 2 | 3+4 4 | 5 2 | 3+4 4 | 2+2 4 | 5+6 2 |
| | 5 2 | 3+4 4 | 6 2 | 3+4 4 | 2+2 4 | 5+6 2 |
| | 5 2 | 3+4 4 | 5 2 | 3+4 4 | 2+2 4 | 3+4 2 |
| | 6 2 | 3+4 4 | 6 2 | 3+4 4 | 2+2 4 | 5+6 2 |
| | 6 2 | 3+4 4 | 5 2 | 3+4 4 | 2+2 4 | 3+4 2 |
| | 6 2 | 3+4 4 | 5 2 | 3+4 4 | 2+2 4 | 5+6 2 |
| | 5 2 | 3+4 4 | 6 2 | 3+4 4 | 2+2 4 | 3+4 2 |
| | 5 2 | 3+4 4 | 6 2 | 3+4 4 | 2+2 4 | 5+6 2 |
| | 5 2 | 3+4 4 | 5 2 | 3+4 4 | 2+2 4 | 3+4 2 |
| | 5 2 | 3+4 4 | 6 2 | 3+4 4 | 2+2 4 | 5+6 2 |
| | 6 2 | | 5 2 | | | 3+4 2 |
| | 6 2 | | 5 2 | | | 5+6 2 |
| | 6 2 | | 6 2 | | | 3+4 2 |
| | 5 2 | | 6 2 | | | 5+6 2 |

Continued on next column

Continued on next column

**TABLE XXXVIII– continued**

| TQC | ICSD | | NoE | | SF | |
|---|---|---|---|---|---|---|
| | 6 2 | | 5 2 | | | 3+4 2 |
| | 5 2 | | 6 2 | | | 5+6 2 |
| | 5 2 | | 6 2 | | | 3+4 2 |
| | 5 2 | | 5 2 | | | 3+4 2 |
| | 6 2 | | 5 2 | | | 5+6 2 |
| | 6 2 | | 5 2 | | | 5+6 2 |
| | 5 2 | | 6 2 | | | 5+6 2 |
| | 6 2 | | 6 2 | | | 3+4 2 |
| | 6 2 | | 5 2 | | | 3+4 2 |
| | 5 2 | | 6 2 | | | 5+6 2 |
| | 5 2 | | 5 2 | | | 3+4 2 |
| | 5 2 | | 6 2 | | | 5+6 2 |
| | 6 2 | | 5 2 | | | 5+6 2 |
| 63.2.55 | 411268 | | 64 | | Eu1In2Rh1 | |
| | -GM | -T | -Y | -Z | -R | -S |
| | 5 2 | 3+4 4 | 5 2 | 3+4 4 | 2+2 4 | 3+4 2 |
| | 6 2 | 3+4 4 | 6 2 | 3+4 4 | 2+2 4 | 5+6 2 |
| | 5 2 | 3+4 4 | 5 2 | 3+4 4 | 2+2 4 | 3+4 2 |
| | 6 2 | 3+4 4 | 6 2 | 3+4 4 | 2+2 4 | 5+6 2 |
| | 5 2 | 3+4 4 | 5 2 | 3+4 4 | 2+2 4 | 3+4 2 |
| | 6 2 | 3+4 4 | 6 2 | 3+4 4 | 2+2 4 | 5+6 2 |
| | 6 2 | 3+4 4 | 6 2 | 3+4 4 | 2+2 4 | 3+4 2 |
| | 5 2 | 3+4 4 | 5 2 | 3+4 4 | 2+2 4 | 5+6 2 |
| | 5 2 | 3+4 4 | 6 2 | 3+4 4 | 2+2 4 | 5+6 2 |
| | 5 2 | 3+4 4 | 5 2 | 3+4 4 | 2+2 4 | 3+4 2 |
| | 6 2 | 3+4 4 | 6 2 | 3+4 4 | 2+2 4 | 5+6 2 |
| | 6 2 | 3+4 4 | 5 2 | 3+4 4 | 2+2 4 | 3+4 2 |
| | 5 2 | 3+4 4 | 5 2 | 3+4 4 | 2+2 4 | 3+4 2 |
| | 5 2 | 3+4 4 | 6 2 | 3+4 4 | 2+2 4 | 5+6 2 |
| | 6 2 | 3+4 4 | 6 2 | 3+4 4 | 2+2 4 | 5+6 2 |
| | 5 2 | 3+4 4 | 5 2 | 3+4 4 | 2+2 4 | 3+4 2 |
| | 5 2 | | 6 2 | | | 5+6 2 |
| | 6 2 | | 5 2 | | | 3+4 2 |
| | 6 2 | | 6 2 | | | 5+6 2 |
| | 5 2 | | 6 2 | | | 3+4 2 |
| | 6 2 | | 5 2 | | | 5+6 2 |
| | 6 2 | | 5 2 | | | 5+6 2 |
| | 5 2 | | 6 2 | | | 3+4 2 |
| | 5 2 | | 5 2 | | | 3+4 2 |
| | 6 2 | | 6 2 | | | 5+6 2 |
| | 5 2 | | 5 2 | | | 5+6 2 |
| | 6 2 | | 6 2 | | | 3+4 2 |
| | 6 2 | | 5 2 | | | 3+4 2 |
| | 5 2 | | 6 2 | | | 5+6 2 |
| | 5 2 | | 5 2 | | | 3+4 2 |
| | 5 2 | | 6 2 | | | 3+4 2 |
| | 6 2 | | 5 2 | | | 5+6 2 |
| 63.2.56 | 404599 | | 68 | | Eu1Ge2Ir1 | |
| | -GM | -T | -Y | -Z | -R | -S |
| | 5 2 | 3+4 4 | 5 2 | 3+4 4 | 2+2 4 | 5+6 2 |
| | 6 2 | 3+4 4 | 6 2 | 3+4 4 | 2+2 4 | 3+4 2 |
| | 5 2 | 3+4 4 | 5 2 | 3+4 4 | 2+2 4 | 5+6 2 |
| | 6 2 | 3+4 4 | 6 2 | 3+4 4 | 2+2 4 | 3+4 2 |
| | 5 2 | 3+4 4 | 5 2 | 3+4 4 | 2+2 4 | 5+6 2 |
| | 6 2 | 3+4 4 | 6 2 | 3+4 4 | 2+2 4 | 3+4 2 |
| | 5 2 | 3+4 4 | 5 2 | 3+4 4 | 2+2 4 | 5+6 2 |
| | 6 2 | 3+4 4 | 6 2 | 3+4 4 | 2+2 4 | 3+4 2 |
| | 5 2 | 3+4 4 | 6 2 | 3+4 4 | 2+2 4 | 3+4 2 |
| | 5 2 | 3+4 4 | 5 2 | 3+4 4 | 2+2 4 | 5+6 2 |
| | 6 2 | 3+4 4 | 5 2 | 3+4 4 | 2+2 4 | 3+4 2 |
| | 6 2 | 3+4 4 | 6 2 | 3+4 4 | 2+2 4 | 5+6 2 |

Continued on next column

| TQC | ICSD | | NoE | | SF | |
|---|---|---|---|---|---|---|
| | 6 2 | 3+4 4 | 5 2 | 3+4 4 | 2+2 4 | 5+6 2 |
| | 5 2 | 3+4 4 | 6 2 | 3+4 4 | 2+2 4 | 3+4 2 |
| | 5 2 | 3+4 4 | 6 2 | 3+4 4 | 2+2 4 | 5+6 2 |
| | 5 2 | 3+4 4 | 6 2 | 3+4 4 | 2+2 4 | 3+4 2 |
| | 5 2 | 3+4 4 | 5 2 | 3+4 4 | 2+2 4 | 5+6 2 |
| | 6 2 | | 5 2 | | | 3+4 2 |
| | 6 2 | | 5 2 | | | 3+4 2 |
| | 5 2 | | 6 2 | | | 5+6 2 |
| | 5 2 | | 6 2 | | | 3+4 2 |
| | 6 2 | | 5 2 | | | 3+4 2 |
| | 5 2 | | 6 2 | | | 5+6 2 |
| | 6 2 | | 5 2 | | | 3+4 2 |
| | 6 2 | | 6 2 | | | 5+6 2 |
| | 5 2 | | 6 2 | | | 5+6 2 |
| | 6 2 | | 6 2 | | | 3+4 2 |
| | 6 2 | | 5 2 | | | 5+6 2 |
| | 5 2 | | 5 2 | | | 3+4 2 |
| | 5 2 | | 5 2 | | | 5+6 2 |
| | 6 2 | | 6 2 | | | 3+4 2 |
| | 5 2 | | 6 2 | | | 5+6 2 |
| | 6 2 | | 5 2 | | | 3+4 2 |
| | 5 2 | | 5 2 | | | 3+4 2 |
| 63.2.57 | 425774 | | 188 | | Er3Ge3Ru2 | |
| | -GM | -T | -Y | -Z | -R | -S |
| | 5 2 | 3+4 4 | 5 2 | 3+4 4 | 2+2 4 | 3+4 2 |
| | 5 2 | 3+4 4 | 5 2 | 3+4 4 | 2+2 4 | 3+4 2 |
| | 6 2 | 3+4 4 | 6 2 | 3+4 4 | 2+2 4 | 5+6 2 |
| | 6 2 | 3+4 4 | 6 2 | 3+4 4 | 2+2 4 | 5+6 2 |
| | 5 2 | 3+4 4 | 5 2 | 3+4 4 | 2+2 4 | 3+4 2 |
| | 6 2 | 3+4 4 | 6 2 | 3+4 4 | 2+2 4 | 5+6 2 |
| | 5 2 | 3+4 4 | 5 2 | 3+4 4 | 2+2 4 | 3+4 2 |
| | 5 2 | 3+4 4 | 5 2 | 3+4 4 | 2+2 4 | 3+4 2 |
| | 6 2 | 3+4 4 | 6 2 | 3+4 4 | 2+2 4 | 5+6 2 |
| | 6 2 | 3+4 4 | 6 2 | 3+4 4 | 2+2 4 | 5+6 2 |
| | 5 2 | 3+4 4 | 5 2 | 3+4 4 | 2+2 4 | 3+4 2 |
| | 6 2 | 3+4 4 | 6 2 | 3+4 4 | 2+2 4 | 5+6 2 |
| | 5 2 | 3+4 4 | 5 2 | 3+4 4 | 2+2 4 | 3+4 2 |
| | 5 2 | 3+4 4 | 5 2 | 3+4 4 | 2+2 4 | 5+6 2 |
| | 6 2 | 3+4 4 | 6 2 | 3+4 4 | 2+2 4 | 3+4 2 |
| | 5 2 | 3+4 4 | 6 2 | 3+4 4 | 2+2 4 | 5+6 2 |
| | 6 2 | 3+4 4 | 5 2 | 3+4 4 | 2+2 4 | 3+4 2 |
| | 6 2 | 3+4 4 | 5 2 | 3+4 4 | 2+2 4 | 5+6 2 |
| | 6 2 | 3+4 4 | 6 2 | 3+4 4 | 2+2 4 | 3+4 2 |
| | 5 2 | 3+4 4 | 6 2 | 3+4 4 | 2+2 4 | 5+6 2 |
| | 5 2 | 3+4 4 | 5 2 | 3+4 4 | 2+2 4 | 3+4 2 |
| | 6 2 | 3+4 4 | 6 2 | 3+4 4 | 2+2 4 | 5+6 2 |
| | 5 2 | 3+4 4 | 5 2 | 3+4 4 | 2+2 4 | 3+4 2 |
| | 6 2 | 3+4 4 | 6 2 | 3+4 4 | 2+2 4 | 5+6 2 |
| | 5 2 | 3+4 4 | 5 2 | 3+4 4 | 2+2 4 | 5+6 2 |
| | 5 2 | 3+4 4 | 5 2 | 3+4 4 | 2+2 4 | 5+6 2 |
| | 6 2 | 3+4 4 | 6 2 | 3+4 4 | 2+2 4 | 3+4 2 |
| | 5 2 | 3+4 4 | 6 2 | 3+4 4 | 2+2 4 | 3+4 2 |
| | 6 2 | 3+4 4 | 5 2 | 3+4 4 | 2+2 4 | 5+6 2 |
| | 6 2 | 3+4 4 | 6 2 | 3+4 4 | 2+2 4 | 3+4 2 |
| | 5 2 | 3+4 4 | 5 2 | 3+4 4 | 2+2 4 | 3+4 2 |
| | 5 2 | 3+4 4 | 5 2 | 3+4 4 | 2+2 4 | 3+4 2 |
| | 6 2 | 3+4 4 | 6 2 | 3+4 4 | 2+2 4 | 5+6 2 |
| | 6 2 | 3+4 4 | 6 2 | 3+4 4 | 2+2 4 | 5+6 2 |
| | 6 2 | 3+4 4 | 5 2 | 3+4 4 | 2+2 4 | 3+4 2 |
| | 6 2 | 3+4 4 | 6 2 | 3+4 4 | 2+2 4 | 5+6 2 |
| | 5 2 | 3+4 4 | 6 2 | 3+4 4 | 2+2 4 | 5+6 2 |

Continued on next column

**TABLE XXXVIII– continued**

| TQC | ICSD | | NoE | | SF | |
|---|---|---|---|---|---|---|
| | 5 2 | 3+4 4 | 5 2 | 3+4 4 | 2+2 4 | 3+4 2 |
| | 6 2 | 3+4 4 | 5 2 | 3+4 4 | 2+2 4 | 5+6 2 |
| | 5 2 | 3+4 4 | 5 2 | 3+4 4 | 2+2 4 | 5+6 2 |
| | 6 2 | 3+4 4 | 6 2 | 3+4 4 | 2+2 4 | 3+4 2 |
| | 5 2 | 3+4 4 | 5 2 | 3+4 4 | 2+2 4 | 3+4 2 |
| | 6 2 | 3+4 4 | 6 2 | 3+4 4 | 2+2 4 | 5+6 2 |
| | 5 2 | 3+4 4 | 5 2 | 3+4 4 | 2+2 4 | 3+4 2 |
| | 5 2 | 3+4 4 | 6 2 | 3+4 4 | 2+2 4 | 3+4 2 |
| | 5 2 | 3+4 4 | 6 2 | 3+4 4 | 2+2 4 | 5+6 2 |
| | 5 2 | 3+4 4 | 6 2 | 3+4 4 | 2+2 4 | 5+6 2 |
| | 6 2 | | 6 2 | | | 3+4 2 |
| | 6 2 | | 5 2 | | | 5+6 2 |
| | 5 2 | | 5 2 | | | 5+6 2 |
| | 5 2 | | 6 2 | | | 3+4 2 |
| | 6 2 | | 6 2 | | | 5+6 2 |
| | 6 2 | | 5 2 | | | 5+6 2 |
| | 5 2 | | 5 2 | | | 3+4 2 |
| | 6 2 | | 6 2 | | | 5+6 2 |
| | 5 2 | | 5 2 | | | 3+4 2 |
| | 6 2 | | 6 2 | | | 3+4 2 |
| | 6 2 | | 5 2 | | | 5+6 2 |
| | 5 2 | | 6 2 | | | 3+4 2 |
| | 6 2 | | 5 2 | | | 3+4 2 |
| | 5 2 | | 5 2 | | | 5+6 2 |
| | 5 2 | | 6 2 | | | 3+4 2 |
| | 6 2 | | 6 2 | | | 5+6 2 |
| | 5 2 | | 5 2 | | | 3+4 2 |
| | 6 2 | | 6 2 | | | 5+6 2 |
| | 5 2 | | 6 2 | | | 3+4 2 |
| | 6 2 | | 5 2 | | | 5+6 2 |
| | 5 2 | | 6 2 | | | 3+4 2 |
| | 6 2 | | 5 2 | | | 5+6 2 |
| | 5 2 | | 5 2 | | | 5+6 2 |
| | 6 2 | | 5 2 | | | 3+4 2 |
| | 6 2 | | 6 2 | | | 5+6 2 |
| | 5 2 | | 6 2 | | | 5+6 2 |
| | 5 2 | | 6 2 | | | 3+4 2 |
| | 5 2 | | 5 2 | | | 3+4 2 |
| | 6 2 | | 6 2 | | | 5+6 2 |
| | 5 2 | | 6 2 | | | 3+4 2 |
| | 6 2 | | 5 2 | | | 3+4 2 |
| | 6 2 | | 5 2 | | | 5+6 2 |
| | 6 2 | | 6 2 | | | 5+6 2 |
| | 5 2 | | 6 2 | | | 3+4 2 |
| | 5 2 | | 5 2 | | | 3+4 2 |
| | 6 2 | | 5 2 | | | 5+6 2 |
| | 6 2 | | 5 2 | | | 5+6 2 |
| | 5 2 | | 6 2 | | | 3+4 2 |
| | 6 2 | | 5 2 | | | 3+4 2 |
| | 5 2 | | 5 2 | | | 5+6 2 |
| | 5 2 | | 6 2 | | | 3+4 2 |
| | 6 2 | | 6 2 | | | 5+6 2 |
| | 6 2 | | 5 2 | | | 3+4 2 |
| | 5 2 | | 6 2 | | | 5+6 2 |
| | 6 2 | | 5 2 | | | 3+4 2 |
| | 5 2 | | 5 2 | | | 5+6 2 |
| | 5 2 | | 6 2 | | | 3+4 2 |
| 63.2.58 | 55524 | | 68 | | Er1Ge3 | |
| | -GM | -T | -Y | -Z | -R | -S |
| | 5 2 | 3+4 4 | 5 2 | 3+4 4 | 2+2 4 | 3+4 2 |
| | 6 2 | 3+4 4 | 6 2 | 3+4 4 | 2+2 4 | 5+6 2 |

Continued on next column

**TABLE XXXVIII– continued**

| TQC | ICSD | | NoE | | SF | |
|---|---|---|---|---|---|---|
| | 5 2 | 3+4 4 | 5 2 | 3+4 4 | 2+2 4 | 3+4 2 |
| | 6 2 | 3+4 4 | 6 2 | 3+4 4 | 2+2 4 | 5+6 2 |
| | 5 2 | 3+4 4 | 5 2 | 3+4 4 | 2+2 4 | 3+4 2 |
| | 6 2 | 3+4 4 | 6 2 | 3+4 4 | 2+2 4 | 5+6 2 |
| | 5 2 | 3+4 4 | 5 2 | 3+4 4 | 2+2 4 | 3+4 2 |
| | 6 2 | 3+4 4 | 6 2 | 3+4 4 | 2+2 4 | 5+6 2 |
| | 5 2 | 3+4 4 | 6 2 | 3+4 4 | 2+2 4 | 5+6 2 |
| | 5 2 | 3+4 4 | 5 2 | 3+4 4 | 2+2 4 | 3+4 2 |
| | 6 2 | 3+4 4 | 5 2 | 3+4 4 | 2+2 4 | 5+6 2 |
| | 6 2 | 3+4 4 | 6 2 | 3+4 4 | 2+2 4 | 3+4 2 |
| | 6 2 | 3+4 4 | 5 2 | 3+4 4 | 2+2 4 | 5+6 2 |
| | 5 2 | 3+4 4 | 6 2 | 3+4 4 | 2+2 4 | 3+4 2 |
| | 5 2 | 3+4 4 | 6 2 | 3+4 4 | 2+2 4 | 3+4 2 |
| | 5 2 | 3+4 4 | 6 2 | 3+4 4 | 2+2 4 | 5+6 2 |
| | 5 2 | 3+4 4 | 6 2 | 3+4 4 | 2+2 4 | 5+6 2 |
| | 5 2 | | 5 2 | | | 3+4 2 |
| | 6 2 | | 5 2 | | | 5+6 2 |
| | 6 2 | | 6 2 | | | 5+6 2 |
| | 6 2 | | 5 2 | | | 3+4 2 |
| | 5 2 | | 5 2 | | | 3+4 2 |
| | 5 2 | | 6 2 | | | 5+6 2 |
| | 6 2 | | 6 2 | | | 3+4 2 |
| | 5 2 | | 5 2 | | | 5+6 2 |
| | 6 2 | | 5 2 | | | 5+6 2 |
| | 5 2 | | 6 2 | | | 3+4 2 |
| | 6 2 | | 5 2 | | | 5+6 2 |
| | 6 2 | | 5 2 | | | 3+4 2 |
| | 6 2 | | 6 2 | | | 5+6 2 |
| | 5 2 | | 5 2 | | | 3+4 2 |
| | 6 2 | | 6 2 | | | 5+6 2 |
| | 5 2 | | 5 2 | | | 3+4 2 |
| | 5 2 | | 6 2 | | | 3+4 2 |
| 63.2.59 | 55523 | | 68 | | Er1Ge3 | |
| | -GM | -T | -Y | -Z | -R | -S |
| | 5 2 | 3+4 4 | 5 2 | 3+4 4 | 2+2 4 | 5+6 2 |
| | 6 2 | 3+4 4 | 6 2 | 3+4 4 | 2+2 4 | 3+4 2 |
| | 5 2 | 3+4 4 | 5 2 | 3+4 4 | 2+2 4 | 5+6 2 |
| | 6 2 | 3+4 4 | 6 2 | 3+4 4 | 2+2 4 | 3+4 2 |
| | 5 2 | 3+4 4 | 5 2 | 3+4 4 | 2+2 4 | 5+6 2 |
| | 6 2 | 3+4 4 | 6 2 | 3+4 4 | 2+2 4 | 3+4 2 |
| | 5 2 | 3+4 4 | 5 2 | 3+4 4 | 2+2 4 | 5+6 2 |
| | 6 2 | 3+4 4 | 6 2 | 3+4 4 | 2+2 4 | 3+4 2 |
| | 5 2 | 3+4 4 | 6 2 | 3+4 4 | 2+2 4 | 3+4 2 |
| | 5 2 | 3+4 4 | 5 2 | 3+4 4 | 2+2 4 | 5+6 2 |
| | 6 2 | 3+4 4 | 5 2 | 3+4 4 | 2+2 4 | 3+4 2 |
| | 6 2 | 3+4 4 | 6 2 | 3+4 4 | 2+2 4 | 5+6 2 |
| | 6 2 | 3+4 4 | 5 2 | 3+4 4 | 2+2 4 | 3+4 2 |
| | 5 2 | 3+4 4 | 6 2 | 3+4 4 | 2+2 4 | 5+6 2 |
| | 5 2 | 3+4 4 | 6 2 | 3+4 4 | 2+2 4 | 5+6 2 |
| | 5 2 | 3+4 4 | 6 2 | 3+4 4 | 2+2 4 | 3+4 2 |
| | 5 2 | 3+4 4 | 6 2 | 3+4 4 | 2+2 4 | 3+4 2 |
| | 5 2 | | 5 2 | | | 5+6 2 |
| | 6 2 | | 5 2 | | | 3+4 2 |
| | 6 2 | | 6 2 | | | 3+4 2 |
| | 6 2 | | 5 2 | | | 5+6 2 |
| | 5 2 | | 5 2 | | | 3+4 2 |
| | 5 2 | | 6 2 | | | 5+6 2 |
| | 6 2 | | 6 2 | | | 5+6 2 |
| | 5 2 | | 5 2 | | | 3+4 2 |
| | 6 2 | | 5 2 | | | 3+4 2 |
| | 6 2 | | 6 2 | | | 5+6 2 |

Continued on next column

**TABLE XXXVIII– continued**

| TQC | ICSD | | NoE | | SF | |
|---|---|---|---|---|---|---|
| | 5 2 | | 5 2 | | | 3+4 2 |
| | 6 2 | | 5 2 | | | 5+6 2 |
| | 6 2 | | 6 2 | | | 3+4 2 |
| | 5 2 | | 5 2 | | | 5+6 2 |
| | 6 2 | | 6 2 | | | 3+4 2 |
| | 5 2 | | 5 2 | | | 5+6 2 |
| | 5 2 | | 6 2 | | | 5+6 2 |
| 63.2.60 | 59896 | | 160 | Cu1Gd2Rb1S4 | | |
| | -GM | -T | -Y | -Z | -R | -S |
| | 5 2 | 3+4 4 | 6 2 | 3+4 4 | 2+2 4 | 5+6 2 |
| | 5 2 | 3+4 4 | 6 2 | 3+4 4 | 2+2 4 | 5+6 2 |
| | 6 2 | 3+4 4 | 5 2 | 3+4 4 | 2+2 4 | 3+4 2 |
| | 6 2 | 3+4 4 | 5 2 | 3+4 4 | 2+2 4 | 3+4 2 |
| | 5 2 | 3+4 4 | 5 2 | 3+4 4 | 2+2 4 | 3+4 2 |
| | 6 2 | 3+4 4 | 6 2 | 3+4 4 | 2+2 4 | 5+6 2 |
| | 5 2 | 3+4 4 | 6 2 | 3+4 4 | 2+2 4 | 5+6 2 |
| | 5 2 | 3+4 4 | 6 2 | 3+4 4 | 2+2 4 | 5+6 2 |
| | 6 2 | 3+4 4 | 5 2 | 3+4 4 | 2+2 4 | 3+4 2 |
| | 6 2 | 3+4 4 | 5 2 | 3+4 4 | 2+2 4 | 3+4 2 |
| | 5 2 | 3+4 4 | 6 2 | 3+4 4 | 2+2 4 | 3+4 2 |
| | 5 2 | 3+4 4 | 5 2 | 3+4 4 | 2+2 4 | 3+4 2 |
| | 6 2 | 3+4 4 | 6 2 | 3+4 4 | 2+2 4 | 5+6 2 |
| | 6 2 | 3+4 4 | 5 2 | 3+4 4 | 2+2 4 | 5+6 2 |
| | 5 2 | 3+4 4 | 6 2 | 3+4 4 | 2+2 4 | 5+6 2 |
| | 5 2 | 3+4 4 | 6 2 | 3+4 4 | 2+2 4 | 5+6 2 |
| | 6 2 | 3+4 4 | 5 2 | 3+4 4 | 2+2 4 | 3+4 2 |
| | 6 2 | 3+4 4 | 5 2 | 3+4 4 | 2+2 4 | 3+4 2 |
| | 5 2 | 3+4 4 | 6 2 | 3+4 4 | 2+2 4 | 5+6 2 |
| | 6 2 | 3+4 4 | 5 2 | 3+4 4 | 2+2 4 | 3+4 2 |
| | 5 2 | 3+4 4 | 6 2 | 3+4 4 | 2+2 4 | 3+4 2 |
| | 5 2 | 3+4 4 | 5 2 | 3+4 4 | 2+2 4 | 3+4 2 |
| | 6 2 | 3+4 4 | 5 2 | 3+4 4 | 2+2 4 | 5+6 2 |
| | 6 2 | 3+4 4 | 6 2 | 3+4 4 | 2+2 4 | 5+6 2 |
| | 5 2 | 3+4 4 | 5 2 | 3+4 4 | 2+2 4 | 3+4 2 |
| | 5 2 | 3+4 4 | 5 2 | 3+4 4 | 2+2 4 | 3+4 2 |
| | 6 2 | 3+4 4 | 6 2 | 3+4 4 | 2+2 4 | 5+6 2 |
| | 5 2 | 3+4 4 | 5 2 | 3+4 4 | 2+2 4 | 3+4 2 |
| | 6 2 | 3+4 4 | 6 2 | 3+4 4 | 2+2 4 | 5+6 2 |
| | 5 2 | 3+4 4 | 5 2 | 3+4 4 | 2+2 4 | 3+4 2 |
| | 6 2 | 3+4 4 | 6 2 | 3+4 4 | 2+2 4 | 5+6 2 |
| | 5 2 | 3+4 4 | 5 2 | 3+4 4 | 2+2 4 | 3+4 2 |
| | 6 2 | 3+4 4 | 6 2 | 3+4 4 | 2+2 4 | 5+6 2 |
| | 5 2 | 3+4 4 | 5 2 | 3+4 4 | 2+2 4 | 3+4 2 |
| | 6 2 | 3+4 4 | 6 2 | 3+4 4 | 2+2 4 | 5+6 2 |
| | 5 2 | 3+4 4 | 6 2 | 3+4 4 | 2+2 4 | 3+4 2 |
| | 5 2 | 3+4 4 | 5 2 | 3+4 4 | 2+2 4 | 5+6 2 |
| | 6 2 | 3+4 4 | 6 2 | 3+4 4 | 2+2 4 | 3+4 2 |
| | 6 2 | 3+4 4 | 5 2 | 3+4 4 | 2+2 4 | 5+6 2 |
| | 5 2 | 3+4 4 | 5 2 | 3+4 4 | 2+2 4 | 3+4 2 |
| | 5 2 | | 6 2 | | | 5+6 2 |
| | 6 2 | | 6 2 | | | 3+4 2 |
| | 6 2 | | 5 2 | | | 5+6 2 |
| | 5 2 | | 6 2 | | | 5+6 2 |
| | 6 2 | | 5 2 | | | 5+6 2 |
| | 6 2 | | 6 2 | | | 3+4 2 |
| | 5 2 | | 5 2 | | | 3+4 2 |
| | 6 2 | | 6 2 | | | 5+6 2 |
| | 5 2 | | 6 2 | | | 3+4 2 |
| | 5 2 | | 5 2 | | | 5+6 2 |
| | 6 2 | | 5 2 | | | 5+6 2 |
| | 6 2 | | 6 2 | | | 5+6 2 |
| | 6 2 | | 6 2 | | | 3+4 2 |
| | 6 2 | | 6 2 | | | 5+6 2 |
| | 5 2 | | 6 2 | | | 5+6 2 |
| | 6 2 | | 5 2 | | | 3+4 2 |
| | 5 2 | | 6 2 | | | 5+6 2 |
| | 6 2 | | 6 2 | | | 5+6 2 |
| | 5 2 | | 6 2 | | | 3+4 2 |
| | 6 2 | | 5 2 | | | 3+4 2 |
| | 6 2 | | 6 2 | | | 5+6 2 |
| | 6 2 | | 5 2 | | | 5+6 2 |
| | 5 2 | | 6 2 | | | 3+4 2 |
| | 6 2 | | 5 2 | | | 5+6 2 |
| | 5 2 | | 6 2 | | | 3+4 2 |
| | 5 2 | | 5 2 | | | 5+6 2 |
| | 5 2 | | 6 2 | | | 3+4 2 |
| | 6 2 | | 6 2 | | | 3+4 2 |
| | 5 2 | | 5 2 | | | 5+6 2 |
| | 6 2 | | 5 2 | | | 3+4 2 |
| | 6 2 | | 5 2 | | | 5+6 2 |
| | 5 2 | | 6 2 | | | 3+4 2 |
| | 5 2 | | 5 2 | | | 5+6 2 |
| | 5 2 | | 6 2 | | | 3+4 2 |
| | 6 2 | | 5 2 | | | 5+6 2 |
| | 6 2 | | 6 2 | | | 5+6 2 |
| | 6 2 | | 6 2 | | | 3+4 2 |
| | 6 2 | | 5 2 | | | 5+6 2 |
| | 5 2 | | 5 2 | | | 3+4 2 |
| | 5 2 | | 6 2 | | | 3+4 2 |
| 63.2.61 | 416795 | | 72 | Cu1Eu1Sn2 | | |
| | -GM | -T | -Y | -Z | -R | -S |
| | 5 2 | 3+4 4 | 5 2 | 3+4 4 | 2+2 4 | 3+4 2 |
| | 6 2 | 3+4 4 | 6 2 | 3+4 4 | 2+2 4 | 5+6 2 |
| | 5 2 | 3+4 4 | 5 2 | 3+4 4 | 2+2 4 | 5+6 2 |
| | 6 2 | 3+4 4 | 6 2 | 3+4 4 | 2+2 4 | 3+4 2 |
| | 5 2 | 3+4 4 | 5 2 | 3+4 4 | 2+2 4 | 5+6 2 |
| | 6 2 | 3+4 4 | 6 2 | 3+4 4 | 2+2 4 | 3+4 2 |
| | 5 2 | 3+4 4 | 5 2 | 3+4 4 | 2+2 4 | 5+6 2 |
| | 6 2 | 3+4 4 | 6 2 | 3+4 4 | 2+2 4 | 3+4 2 |
| | 5 2 | 3+4 4 | 6 2 | 3+4 4 | 2+2 4 | 3+4 2 |
| | 5 2 | 3+4 4 | 5 2 | 3+4 4 | 2+2 4 | 5+6 2 |
| | 6 2 | 3+4 4 | 6 2 | 3+4 4 | 2+2 4 | 3+4 2 |
| | 6 2 | 3+4 4 | 5 2 | 3+4 4 | 2+2 4 | 5+6 2 |
| | 6 2 | 3+4 4 | 5 2 | 3+4 4 | 2+2 4 | 5+6 2 |
| | 5 2 | 3+4 4 | 6 2 | 3+4 4 | 2+2 4 | 3+4 2 |
| | 5 2 | 3+4 4 | 6 2 | 3+4 4 | 2+2 4 | 5+6 2 |
| | 5 2 | 3+4 4 | 6 2 | 3+4 4 | 2+2 4 | 3+4 2 |
| | 5 2 | 3+4 4 | 5 2 | 3+4 4 | 2+2 4 | 5+6 2 |
| | 6 2 | 3+4 4 | 5 2 | 3+4 4 | 2+2 4 | 3+4 2 |
| | 6 2 | | 5 2 | | | 3+4 2 |
| | 5 2 | | 6 2 | | | 5+6 2 |
| | 6 2 | | 6 2 | | | 3+4 2 |
| | 5 2 | | 5 2 | | | 5+6 2 |
| | 6 2 | | 6 2 | | | 3+4 2 |
| | 5 2 | | 5 2 | | | 5+6 2 |
| | 5 2 | | 5 2 | | | 3+4 2 |
| | 6 2 | | 6 2 | | | 5+6 2 |
| | 5 2 | | 6 2 | | | 3+4 2 |
| | 6 2 | | 6 2 | | | 5+6 2 |
| | 6 2 | | 5 2 | | | 3+4 2 |
| | 6 2 | | 5 2 | | | 5+6 2 |
| | 5 2 | | 6 2 | | | 3+4 2 |

Continued on next column

**TABLE XXXVIII– continued**

| TQC | ICSD | NoE | SF | | | |
|---|---|---|---|---|---|---|
| | 5 2 | | 5 2 | | | 5+6 2 |
| | 6 2 | | 6 2 | | | 3+4 2 |
| | 5 2 | | 5 2 | | | 5+6 2 |
| | 6 2 | | 6 2 | | | 3+4 2 |
| | 5 2 | | 5 2 | | | 3+4 2 |
| 63.2.62 | 413681 | | 236 | Cs1Cu3Se5Tm2 | | |
| | -GM | -T | -Y | -Z | -R | -S |
| | 5 2 | 3+4 4 | 5 2 | 3+4 4 | 2+2 4 | 3+4 2 |
| | 5 2 | 3+4 4 | 5 2 | 3+4 4 | 2+2 4 | 3+4 2 |
| | 6 2 | 3+4 4 | 6 2 | 3+4 4 | 2+2 4 | 5+6 2 |
| | 6 2 | 3+4 4 | 6 2 | 3+4 4 | 2+2 4 | 5+6 2 |
| | 5 2 | 3+4 4 | 6 2 | 3+4 4 | 2+2 4 | 3+4 2 |
| | 5 2 | 3+4 4 | 6 2 | 3+4 4 | 2+2 4 | 3+4 2 |
| | 6 2 | 3+4 4 | 5 2 | 3+4 4 | 2+2 4 | 5+6 2 |
| | 6 2 | 3+4 4 | 5 2 | 3+4 4 | 2+2 4 | 5+6 2 |
| | 5 2 | 3+4 4 | 6 2 | 3+4 4 | 2+2 4 | 3+4 2 |
| | 5 2 | 3+4 4 | 6 2 | 3+4 4 | 2+2 4 | 5+6 2 |
| | 6 2 | 3+4 4 | 5 2 | 3+4 4 | 2+2 4 | 3+4 2 |
| | 6 2 | 3+4 4 | 5 2 | 3+4 4 | 2+2 4 | 5+6 2 |
| | 5 2 | 3+4 4 | 6 2 | 3+4 4 | 2+2 4 | 3+4 2 |
| | 5 2 | 3+4 4 | 6 2 | 3+4 4 | 2+2 4 | 5+6 2 |
| | 6 2 | 3+4 4 | 5 2 | 3+4 4 | 2+2 4 | 3+4 2 |
| | 6 2 | 3+4 4 | 5 2 | 3+4 4 | 2+2 4 | 5+6 2 |
| | 5 2 | 3+4 4 | 5 2 | 3+4 4 | 2+2 4 | 3+4 2 |
| | 6 2 | 3+4 4 | 6 2 | 3+4 4 | 2+2 4 | 5+6 2 |
| | 5 2 | 3+4 4 | 6 2 | 3+4 4 | 2+2 4 | 3+4 2 |
| | 5 2 | 3+4 4 | 5 2 | 3+4 4 | 2+2 4 | 5+6 2 |
| | 6 2 | 3+4 4 | 6 2 | 3+4 4 | 2+2 4 | 5+6 2 |
| | 6 2 | 3+4 4 | 5 2 | 3+4 4 | 2+2 4 | 5+6 2 |
| | 5 2 | 3+4 4 | 5 2 | 3+4 4 | 2+2 4 | 3+4 2 |
| | 6 2 | 3+4 4 | 6 2 | 3+4 4 | 2+2 4 | 3+4 2 |
| | 6 2 | 3+4 4 | 5 2 | 3+4 4 | 2+2 4 | 5+6 2 |
| | 6 2 | 3+4 4 | 6 2 | 3+4 4 | 2+2 4 | 3+4 2 |
| | 5 2 | 3+4 4 | 5 2 | 3+4 4 | 2+2 4 | 5+6 2 |
| | 5 2 | 3+4 4 | 6 2 | 3+4 4 | 2+2 4 | 3+4 2 |
| | 5 2 | 3+4 4 | 5 2 | 3+4 4 | 2+2 4 | 3+4 2 |
| | 6 2 | 3+4 4 | 6 2 | 3+4 4 | 2+2 4 | 5+6 2 |
| | 5 2 | 3+4 4 | 5 2 | 3+4 4 | 2+2 4 | 3+4 2 |
| | 6 2 | 3+4 4 | 6 2 | 3+4 4 | 2+2 4 | 5+6 2 |
| | 5 2 | 3+4 4 | 5 2 | 3+4 4 | 2+2 4 | 5+6 2 |
| | 6 2 | 3+4 4 | 6 2 | 3+4 4 | 2+2 4 | 3+4 2 |
| | 5 2 | 3+4 4 | 5 2 | 3+4 4 | 2+2 4 | 5+6 2 |
| | 5 2 | 3+4 4 | 5 2 | 3+4 4 | 2+2 4 | 5+6 2 |
| | 6 2 | 3+4 4 | 6 2 | 3+4 4 | 2+2 4 | 3+4 2 |
| | 6 2 | 3+4 4 | 5 2 | 3+4 4 | 2+2 4 | 3+4 2 |
| | 6 2 | 3+4 4 | 6 2 | 3+4 4 | 2+2 4 | 3+4 2 |
| | 5 2 | 3+4 4 | 5 2 | 3+4 4 | 2+2 4 | 5+6 2 |
| | 5 2 | 3+4 4 | 6 2 | 3+4 4 | 2+2 4 | 5+6 2 |
| | 6 2 | 3+4 4 | 6 2 | 3+4 4 | 2+2 4 | 5+6 2 |
| | 5 2 | 3+4 4 | 6 2 | 3+4 4 | 2+2 4 | 3+4 2 |
| | 6 2 | 3+4 4 | 6 2 | 3+4 4 | 2+2 4 | 3+4 2 |
| | 5 2 | 3+4 4 | 5 2 | 3+4 4 | 2+2 4 | 3+4 2 |
| | 6 2 | 3+4 4 | 6 2 | 3+4 4 | 2+2 4 | 3+4 2 |
| | 5 2 | 3+4 4 | 6 2 | 3+4 4 | 2+2 4 | 5+6 2 |
| | 5 2 | 3+4 4 | 5 2 | 3+4 4 | 2+2 4 | 5+6 2 |
| | 6 2 | 3+4 4 | 6 2 | 3+4 4 | 2+2 4 | 3+4 2 |
| | 6 2 | 3+4 4 | 6 2 | 3+4 4 | 2+2 4 | 3+4 2 |
| | 6 2 | 3+4 4 | 5 2 | 3+4 4 | 2+2 4 | 3+4 2 |
| | 5 2 | 3+4 4 | 5 2 | 3+4 4 | 2+2 4 | 5+6 2 |
| | 5 2 | 3+4 4 | 6 2 | 3+4 4 | 2+2 4 | 5+6 2 |
| | 5 2 | 3+4 4 | 5 2 | 3+4 4 | 2+2 4 | 3+4 2 |

Continued on next column

**TABLE XXXVIII– continued**

| TQC | ICSD | NoE | SF | | | |
|---|---|---|---|---|---|---|
| | 6 2 | 3+4 4 | 6 2 | 3+4 4 | 2+2 4 | 5+6 2 |
| | 6 2 | 3+4 4 | 5 2 | 3+4 4 | 2+2 4 | 3+4 2 |
| | 5 2 | 3+4 4 | 6 2 | 3+4 4 | 2+2 4 | 3+4 2 |
| | 5 2 | 3+4 4 | 5 2 | 3+4 4 | 2+2 4 | 5+6 2 |
| | 5 2 | 3+4 4 | 5 2 | 3+4 4 | 2+2 4 | 5+6 2 |
| | 5 2 | | 5 2 | | | 3+4 2 |
| | 6 2 | | 5 2 | | | 3+4 2 |
| | 5 2 | | 5 2 | | | 5+6 2 |
| | 6 2 | | 6 2 | | | 5+6 2 |
| | 5 2 | | 5 2 | | | 5+6 2 |
| | 6 2 | | 5 2 | | | 5+6 2 |
| | 5 2 | | 5 2 | | | 5+6 2 |
| | 6 2 | | 6 2 | | | 3+4 2 |
| | 6 2 | | 6 2 | | | 3+4 2 |
| | 6 2 | | 6 2 | | | 5+6 2 |
| | 6 2 | | 6 2 | | | 3+4 2 |
| | 5 2 | | 5 2 | | | 3+4 2 |
| | 6 2 | | 6 2 | | | 5+6 2 |
| | 6 2 | | 6 2 | | | 5+6 2 |
| | 6 2 | | 5 2 | | | 5+6 2 |
| | 5 2 | | 6 2 | | | 3+4 2 |
| | 6 2 | | 5 2 | | | 3+4 2 |
| | 6 2 | | 5 2 | | | 5+6 2 |
| | 5 2 | | 6 2 | | | 3+4 2 |
| | 5 2 | | 5 2 | | | 5+6 2 |
| | 5 2 | | 5 2 | | | 5+6 2 |
| | 6 2 | | 6 2 | | | 3+4 2 |
| | 5 2 | | 6 2 | | | 5+6 2 |
| | 6 2 | | 6 2 | | | 3+4 2 |
| | 5 2 | | 5 2 | | | 3+4 2 |
| | 6 2 | | 6 2 | | | 5+6 2 |
| | 5 2 | | 6 2 | | | 3+4 2 |
| | 5 2 | | 5 2 | | | 5+6 2 |
| | 6 2 | | 5 2 | | | 5+6 2 |
| | 6 2 | | 5 2 | | | 3+4 2 |
| | 5 2 | | 6 2 | | | 5+6 2 |
| | 6 2 | | 5 2 | | | 3+4 2 |
| | 5 2 | | 6 2 | | | 5+6 2 |
| | 5 2 | | 5 2 | | | 5+6 2 |
| | 5 2 | | 6 2 | | | 3+4 2 |
| | 6 2 | | 6 2 | | | 5+6 2 |
| | 5 2 | | 6 2 | | | 5+6 2 |
| | 5 2 | | 5 2 | | | 3+4 2 |
| | 6 2 | | 6 2 | | | 3+4 2 |
| | 6 2 | | 5 2 | | | 5+6 2 |
| | 5 2 | | 5 2 | | | 3+4 2 |
| | 6 2 | | 6 2 | | | 5+6 2 |
| | 5 2 | | 5 2 | | | 3+4 2 |
| | 5 2 | | 5 2 | | | 5+6 2 |
| | 6 2 | | 6 2 | | | 3+4 2 |
| | 6 2 | | 5 2 | | | 3+4 2 |
| | 6 2 | | 6 2 | | | 3+4 2 |
| | 5 2 | | 6 2 | | | 3+4 2 |
| | 5 2 | | 6 2 | | | 5+6 2 |
| | 5 2 | | 5 2 | | | 3+4 2 |
| | 6 2 | | 5 2 | | | 5+6 2 |
| | 6 2 | | 6 2 | | | 5+6 2 |
| | 5 2 | | 6 2 | | | 3+4 2 |
| | 5 2 | | 5 2 | | | 5+6 2 |
| | 6 2 | | 6 2 | | | 3+4 2 |
| | 5 2 | | 5 2 | | | 5+6 2 |

Continued on next column

**TABLE XXXVIII– continued**

| TQC | ICSD | | NoE | SF | | |
|---|---|---|---|---|---|---|
| | 5 2 | | 6 2 | | | 5+6 2 |
| | 5 2 | | 6 2 | | | 3+4 2 |
| | 6 2 | | 5 2 | | | 5+6 2 |
| 63.2.63 | 96226 | | 220 | Cs1Cu3Se5Tb2 | | |
| | -GM | -T | -Y | -Z | -R | -S |
| | 5 2 | 3+4 4 | 6 2 | 3+4 4 | 2+2 4 | 3+4 2 |
| | 5 2 | 3+4 4 | 6 2 | 3+4 4 | 2+2 4 | 3+4 2 |
| | 6 2 | 3+4 4 | 5 2 | 3+4 4 | 2+2 4 | 5+6 2 |
| | 6 2 | 3+4 4 | 5 2 | 3+4 4 | 2+2 4 | 5+6 2 |
| | 5 2 | 3+4 4 | 6 2 | 3+4 4 | 2+2 4 | 3+4 2 |
| | 5 2 | 3+4 4 | 6 2 | 3+4 4 | 2+2 4 | 5+6 2 |
| | 6 2 | 3+4 4 | 5 2 | 3+4 4 | 2+2 4 | 3+4 2 |
| | 6 2 | 3+4 4 | 5 2 | 3+4 4 | 2+2 4 | 5+6 2 |
| | 5 2 | 3+4 4 | 5 2 | 3+4 4 | 2+2 4 | 3+4 2 |
| | 6 2 | 3+4 4 | 6 2 | 3+4 4 | 2+2 4 | 5+6 2 |
| | 5 2 | 3+4 4 | 6 2 | 3+4 4 | 2+2 4 | 3+4 2 |
| | 5 2 | 3+4 4 | 6 2 | 3+4 4 | 2+2 4 | 5+6 2 |
| | 6 2 | 3+4 4 | 5 2 | 3+4 4 | 2+2 4 | 3+4 2 |
| | 6 2 | 3+4 4 | 5 2 | 3+4 4 | 2+2 4 | 5+6 2 |
| | 5 2 | 3+4 4 | 6 2 | 3+4 4 | 2+2 4 | 3+4 2 |
| | 5 2 | 3+4 4 | 6 2 | 3+4 4 | 2+2 4 | 5+6 2 |
| | 6 2 | 3+4 4 | 5 2 | 3+4 4 | 2+2 4 | 3+4 2 |
| | 6 2 | 3+4 4 | 5 2 | 3+4 4 | 2+2 4 | 5+6 2 |
| | 5 2 | 3+4 4 | 6 2 | 3+4 4 | 2+2 4 | 5+6 2 |
| | 5 2 | 3+4 4 | 5 2 | 3+4 4 | 2+2 4 | 3+4 2 |
| | 6 2 | 3+4 4 | 6 2 | 3+4 4 | 2+2 4 | 3+4 2 |
| | 6 2 | 3+4 4 | 5 2 | 3+4 4 | 2+2 4 | 3+4 2 |
| | 6 2 | 3+4 4 | 5 2 | 3+4 4 | 2+2 4 | 5+6 2 |
| | 5 2 | 3+4 4 | 5 2 | 3+4 4 | 2+2 4 | 5+6 2 |
| | 6 2 | 3+4 4 | 6 2 | 3+4 4 | 2+2 4 | 3+4 2 |
| | 6 2 | 3+4 4 | 5 2 | 3+4 4 | 2+2 4 | 5+6 2 |
| | 5 2 | 3+4 4 | 6 2 | 3+4 4 | 2+2 4 | 3+4 2 |
| | 5 2 | 3+4 4 | 6 2 | 3+4 4 | 2+2 4 | 5+6 2 |
| | 5 2 | 3+4 4 | 5 2 | 3+4 4 | 2+2 4 | 3+4 2 |
| | 6 2 | 3+4 4 | 6 2 | 3+4 4 | 2+2 4 | 5+6 2 |
| | 5 2 | 3+4 4 | 5 2 | 3+4 4 | 2+2 4 | 3+4 2 |
| | 6 2 | 3+4 4 | 6 2 | 3+4 4 | 2+2 4 | 5+6 2 |
| | 5 2 | 3+4 4 | 5 2 | 3+4 4 | 2+2 4 | 3+4 2 |
| | 6 2 | 3+4 4 | 6 2 | 3+4 4 | 2+2 4 | 5+6 2 |
| | 5 2 | 3+4 4 | 5 2 | 3+4 4 | 2+2 4 | 3+4 2 |
| | 5 2 | 3+4 4 | 5 2 | 3+4 4 | 2+2 4 | 3+4 2 |
| | 6 2 | 3+4 4 | 6 2 | 3+4 4 | 2+2 4 | 5+6 2 |
| | 6 2 | 3+4 4 | 5 2 | 3+4 4 | 2+2 4 | 5+6 2 |
| | 6 2 | 3+4 4 | 6 2 | 3+4 4 | 2+2 4 | 5+6 2 |
| | 5 2 | 3+4 4 | 5 2 | 3+4 4 | 2+2 4 | 3+4 2 |
| | 5 2 | 3+4 4 | 6 2 | 3+4 4 | 2+2 4 | 3+4 2 |
| | 6 2 | 3+4 4 | 6 2 | 3+4 4 | 2+2 4 | 3+4 2 |
| | 5 2 | 3+4 4 | 6 2 | 3+4 4 | 2+2 4 | 5+6 2 |
| | 6 2 | 3+4 4 | 5 2 | 3+4 4 | 2+2 4 | 5+6 2 |
| | 5 2 | 3+4 4 | 6 2 | 3+4 4 | 2+2 4 | 5+6 2 |
| | 5 2 | 3+4 4 | 6 2 | 3+4 4 | 2+2 4 | 5+6 2 |
| | 5 2 | 3+4 4 | 6 2 | 3+4 4 | 2+2 4 | 3+4 2 |
| | 6 2 | 3+4 4 | 5 2 | 3+4 4 | 2+2 4 | 3+4 2 |
| | 6 2 | 3+4 4 | 6 2 | 3+4 4 | 2+2 4 | 5+6 2 |
| | 6 2 | 3+4 4 | 5 2 | 3+4 4 | 2+2 4 | 5+6 2 |
| | 6 2 | 3+4 4 | 6 2 | 3+4 4 | 2+2 4 | 5+6 2 |
| | 5 2 | 3+4 4 | 5 2 | 3+4 4 | 2+2 4 | 3+4 2 |
| | 5 2 | 3+4 4 | 6 2 | 3+4 4 | 2+2 4 | 5+6 2 |
| | 5 2 | 3+4 4 | 5 2 | 3+4 4 | 2+2 4 | 3+4 2 |
| | 6 2 | 3+4 4 | 6 2 | 3+4 4 | 2+2 4 | 3+4 2 |
| | 6 2 | | 6 2 | | | 5+6 2 |

Continued on next column

**TABLE XXXVIII– continued**

| TQC | ICSD | | NoE | SF | | |
|---|---|---|---|---|---|---|
| | 5 2 | | 5 2 | | | 5+6 2 |
| | 5 2 | | 5 2 | | | 3+4 2 |
| | 5 2 | | 5 2 | | | 3+4 2 |
| | 5 2 | | 5 2 | | | 5+6 2 |
| | 6 2 | | 5 2 | | | 3+4 2 |
| | 5 2 | | 5 2 | | | 5+6 2 |
| | 6 2 | | 6 2 | | | 3+4 2 |
| | 5 2 | | 5 2 | | | 3+4 2 |
| | 6 2 | | 5 2 | | | 3+4 2 |
| | 5 2 | | 5 2 | | | 3+4 2 |
| | 6 2 | | 6 2 | | | 5+6 2 |
| | 6 2 | | 6 2 | | | 5+6 2 |
| | 6 2 | | 6 2 | | | 3+4 2 |
| | 6 2 | | 6 2 | | | 5+6 2 |
| | 5 2 | | 5 2 | | | 5+6 2 |
| | 6 2 | | 6 2 | | | 3+4 2 |
| | 6 2 | | 6 2 | | | 3+4 2 |
| | 6 2 | | 5 2 | | | 5+6 2 |
| | 5 2 | | 6 2 | | | 3+4 2 |
| | 6 2 | | 5 2 | | | 5+6 2 |
| | 6 2 | | 6 2 | | | 3+4 2 |
| | 5 2 | | 5 2 | | | 3+4 2 |
| | 5 2 | | 6 2 | | | 5+6 2 |
| | 5 2 | | 5 2 | | | 3+4 2 |
| | 5 2 | | 5 2 | | | 5+6 2 |
| | 5 2 | | 5 2 | | | 3+4 2 |
| | 6 2 | | 6 2 | | | 3+4 2 |
| | 6 2 | | 6 2 | | | 3+4 2 |
| | 5 2 | | 6 2 | | | 5+6 2 |
| | 5 2 | | 6 2 | | | 5+6 2 |
| | 6 2 | | 5 2 | | | 5+6 2 |
| | 6 2 | | 5 2 | | | 3+4 2 |
| | 6 2 | | 6 2 | | | 3+4 2 |
| | 5 2 | | 5 2 | | | 3+4 2 |
| | 5 2 | | 5 2 | | | 5+6 2 |
| | 5 2 | | 6 2 | | | 5+6 2 |
| | 6 2 | | 5 2 | | | 5+6 2 |
| | 5 2 | | 6 2 | | | 5+6 2 |
| | 6 2 | | 6 2 | | | 3+4 2 |
| | 5 2 | | 5 2 | | | 5+6 2 |
| | 5 2 | | 5 2 | | | 3+4 2 |
| | 6 2 | | 6 2 | | | 5+6 2 |
| | 6 2 | | 6 2 | | | 3+4 2 |
| | 5 2 | | 6 2 | | | 5+6 2 |
| | 6 2 | | 5 2 | | | 3+4 2 |
| | 5 2 | | 5 2 | | | 5+6 2 |
| | 5 2 | | 6 2 | | | 3+4 2 |
| | 6 2 | | 5 2 | | | 3+4 2 |
| | 6 2 | | 6 2 | | | 5+6 2 |
| | 6 2 | | 5 2 | | | 5+6 2 |
| | 5 2 | | 6 2 | | | 3+4 2 |
| | 5 2 | | 6 2 | | | 5+6 2 |
| | 6 2 | | 5 2 | | | 3+4 2 |
| | 5 2 | | 5 2 | | | 5+6 2 |
| 63.2.64 | 413680 | | 220 | Cs1Cu3Se5Tb2 | | |
| | -GM | -T | -Y | -Z | -R | -S |
| | 5 2 | 3+4 4 | 6 2 | 3+4 4 | 2+2 4 | 3+4 2 |
| | 5 2 | 3+4 4 | 6 2 | 3+4 4 | 2+2 4 | 3+4 2 |
| | 6 2 | 3+4 4 | 5 2 | 3+4 4 | 2+2 4 | 5+6 2 |
| | 6 2 | 3+4 4 | 5 2 | 3+4 4 | 2+2 4 | 5+6 2 |
| | 5 2 | 3+4 4 | 6 2 | 3+4 4 | 2+2 4 | 3+4 2 |

Continued on next column

**TABLE XXXVIII– continued**

| TQC | ICSD | NoE | SF | | | |
|---|---|---|---|---|---|---|
| | 5 2 | 3+4 4 | 6 2 | 3+4 4 | 2+2 4 | 5+6 2 |
| | 6 2 | 3+4 4 | 5 2 | 3+4 4 | 2+2 4 | 3+4 2 |
| | 6 2 | 3+4 4 | 5 2 | 3+4 4 | 2+2 4 | 5+6 2 |
| | 5 2 | 3+4 4 | 5 2 | 3+4 4 | 2+2 4 | 3+4 2 |
| | 6 2 | 3+4 4 | 6 2 | 3+4 4 | 2+2 4 | 5+6 2 |
| | 5 2 | 3+4 4 | 6 2 | 3+4 4 | 2+2 4 | 3+4 2 |
| | 5 2 | 3+4 4 | 6 2 | 3+4 4 | 2+2 4 | 5+6 2 |
| | 6 2 | 3+4 4 | 5 2 | 3+4 4 | 2+2 4 | 3+4 2 |
| | 6 2 | 3+4 4 | 5 2 | 3+4 4 | 2+2 4 | 5+6 2 |
| | 5 2 | 3+4 4 | 6 2 | 3+4 4 | 2+2 4 | 3+4 2 |
| | 5 2 | 3+4 4 | 6 2 | 3+4 4 | 2+2 4 | 5+6 2 |
| | 6 2 | 3+4 4 | 5 2 | 3+4 4 | 2+2 4 | 3+4 2 |
| | 6 2 | 3+4 4 | 5 2 | 3+4 4 | 2+2 4 | 5+6 2 |
| | 5 2 | 3+4 4 | 6 2 | 3+4 4 | 2+2 4 | 3+4 2 |
| | 5 2 | 3+4 4 | 5 2 | 3+4 4 | 2+2 4 | 5+6 2 |
| | 6 2 | 3+4 4 | 6 2 | 3+4 4 | 2+2 4 | 5+6 2 |
| | 6 2 | 3+4 4 | 5 2 | 3+4 4 | 2+2 4 | 5+6 2 |
| | 6 2 | 3+4 4 | 5 2 | 3+4 4 | 2+2 4 | 3+4 2 |
| | 5 2 | 3+4 4 | 6 2 | 3+4 4 | 2+2 4 | 3+4 2 |
| | 6 2 | 3+4 4 | 5 2 | 3+4 4 | 2+2 4 | 5+6 2 |
| | 6 2 | 3+4 4 | 5 2 | 3+4 4 | 2+2 4 | 5+6 2 |
| | 5 2 | 3+4 4 | 6 2 | 3+4 4 | 2+2 4 | 3+4 2 |
| | 5 2 | 3+4 4 | 6 2 | 3+4 4 | 2+2 4 | 3+4 2 |
| | 5 2 | 3+4 4 | 5 2 | 3+4 4 | 2+2 4 | 3+4 2 |
| | 6 2 | 3+4 4 | 6 2 | 3+4 4 | 2+2 4 | 5+6 2 |
| | 5 2 | 3+4 4 | 5 2 | 3+4 4 | 2+2 4 | 3+4 2 |
| | 6 2 | 3+4 4 | 6 2 | 3+4 4 | 2+2 4 | 5+6 2 |
| | 5 2 | 3+4 4 | 5 2 | 3+4 4 | 2+2 4 | 5+6 2 |
| | 6 2 | 3+4 4 | 6 2 | 3+4 4 | 2+2 4 | 3+4 2 |
| | 5 2 | 3+4 4 | 5 2 | 3+4 4 | 2+2 4 | 5+6 2 |
| | 5 2 | 3+4 4 | 5 2 | 3+4 4 | 2+2 4 | 5+6 2 |
| | 6 2 | 3+4 4 | 6 2 | 3+4 4 | 2+2 4 | 3+4 2 |
| | 6 2 | 3+4 4 | 5 2 | 3+4 4 | 2+2 4 | 3+4 2 |
| | 6 2 | 3+4 4 | 6 2 | 3+4 4 | 2+2 4 | 3+4 2 |
| | 5 2 | 3+4 4 | 5 2 | 3+4 4 | 2+2 4 | 5+6 2 |
| | 5 2 | 3+4 4 | 6 2 | 3+4 4 | 2+2 4 | 5+6 2 |
| | 6 2 | 3+4 4 | 6 2 | 3+4 4 | 2+2 4 | 5+6 2 |
| | 5 2 | 3+4 4 | 5 2 | 3+4 4 | 2+2 4 | 3+4 2 |
| | 6 2 | 3+4 4 | 6 2 | 3+4 4 | 2+2 4 | 3+4 2 |
| | 5 2 | 3+4 4 | 6 2 | 3+4 4 | 2+2 4 | 3+4 2 |
| | 5 2 | 3+4 4 | 6 2 | 3+4 4 | 2+2 4 | 3+4 2 |
| | 6 2 | 3+4 4 | 6 2 | 3+4 4 | 2+2 4 | 5+6 2 |
| | 5 2 | 3+4 4 | 5 2 | 3+4 4 | 2+2 4 | 5+6 2 |
| | 6 2 | 3+4 4 | 6 2 | 3+4 4 | 2+2 4 | 3+4 2 |
| | 6 2 | 3+4 4 | 6 2 | 3+4 4 | 2+2 4 | 3+4 2 |
| | 6 2 | 3+4 4 | 5 2 | 3+4 4 | 2+2 4 | 3+4 2 |
| | 5 2 | 3+4 4 | 5 2 | 3+4 4 | 2+2 4 | 5+6 2 |
| | 5 2 | 3+4 4 | 6 2 | 3+4 4 | 2+2 4 | 3+4 2 |
| | 5 2 | 3+4 4 | 5 2 | 3+4 4 | 2+2 4 | 5+6 2 |
| | 6 2 | 3+4 4 | 6 2 | 3+4 4 | 2+2 4 | 5+6 2 |
| | 6 2 | | 6 2 | | | 3+4 2 |
| | 5 2 | | 5 2 | | | 3+4 2 |
| | 5 2 | | 5 2 | | | 5+6 2 |
| | 5 2 | | 5 2 | | | 5+6 2 |
| | 5 2 | | 5 2 | | | 3+4 2 |
| | 6 2 | | 5 2 | | | 5+6 2 |
| | 5 2 | | 5 2 | | | 3+4 2 |
| | 6 2 | | 6 2 | | | 5+6 2 |
| | 5 2 | | 5 2 | | | 5+6 2 |
| | 6 2 | | 5 2 | | | 5+6 2 |
| | 5 2 | | 5 2 | | | 5+6 2 |
| | 6 2 | | 6 2 | | | 3+4 2 |
| | 6 2 | | 6 2 | | | 3+4 2 |
| | 6 2 | | 6 2 | | | 5+6 2 |
| | 6 2 | | 6 2 | | | 3+4 2 |
| | 5 2 | | 5 2 | | | 3+4 2 |
| | 6 2 | | 6 2 | | | 5+6 2 |
| | 6 2 | | 6 2 | | | 5+6 2 |
| | 6 2 | | 5 2 | | | 3+4 2 |
| | 5 2 | | 6 2 | | | 5+6 2 |
| | 6 2 | | 5 2 | | | 3+4 2 |
| | 6 2 | | 6 2 | | | 5+6 2 |
| | 5 2 | | 5 2 | | | 5+6 2 |
| | 5 2 | | 6 2 | | | 3+4 2 |
| | 5 2 | | 5 2 | | | 5+6 2 |
| | 5 2 | | 5 2 | | | 3+4 2 |
| | 5 2 | | 5 2 | | | 5+6 2 |
| | 6 2 | | 6 2 | | | 5+6 2 |
| | 6 2 | | 6 2 | | | 5+6 2 |
| | 5 2 | | 6 2 | | | 3+4 2 |
| | 5 2 | | 6 2 | | | 3+4 2 |
| | 6 2 | | 5 2 | | | 3+4 2 |
| | 6 2 | | 5 2 | | | 5+6 2 |
| | 6 2 | | 6 2 | | | 5+6 2 |
| | 5 2 | | 5 2 | | | 5+6 2 |
| | 5 2 | | 5 2 | | | 3+4 2 |
| | 5 2 | | 6 2 | | | 3+4 2 |
| | 6 2 | | 5 2 | | | 3+4 2 |
| | 5 2 | | 6 2 | | | 3+4 2 |
| | 6 2 | | 6 2 | | | 5+6 2 |
| | 5 2 | | 5 2 | | | 3+4 2 |
| | 5 2 | | 5 2 | | | 5+6 2 |
| | 6 2 | | 6 2 | | | 3+4 2 |
| | 6 2 | | 6 2 | | | 5+6 2 |
| | 5 2 | | 6 2 | | | 3+4 2 |
| | 6 2 | | 5 2 | | | 5+6 2 |
| | 5 2 | | 5 2 | | | 3+4 2 |
| | 5 2 | | 6 2 | | | 5+6 2 |
| | 6 2 | | 5 2 | | | 5+6 2 |
| | 6 2 | | 6 2 | | | 3+4 2 |
| | 6 2 | | 5 2 | | | 3+4 2 |
| | 5 2 | | 6 2 | | | 3+4 2 |
| | 5 2 | | 6 2 | | | 5+6 2 |
| | 6 2 | | 5 2 | | | 3+4 2 |
| | 5 2 | | 5 2 | | | 5+6 2 |
| 63.2.65 | 193548 | 76 | Cr2Mg1O4 | | | |
| | -GM | -T | -Y | -Z | -R | -S |
| | 5 2 | 3+4 4 | 6 2 | 3+4 4 | 2+2 4 | 3+4 2 |
| | 5 2 | 3+4 4 | 5 2 | 3+4 4 | 2+2 4 | 5+6 2 |
| | 6 2 | 3+4 4 | 6 2 | 3+4 4 | 2+2 4 | 3+4 2 |
| | 5 2 | 3+4 4 | 5 2 | 3+4 4 | 2+2 4 | 3+4 2 |
| | 6 2 | 3+4 4 | 5 2 | 3+4 4 | 2+2 4 | 5+6 2 |
| | 5 2 | 3+4 4 | 5 2 | 3+4 4 | 2+2 4 | 3+4 2 |
| | 6 2 | 3+4 4 | 6 2 | 3+4 4 | 2+2 4 | 3+4 2 |
| | 5 2 | 3+4 4 | 5 2 | 3+4 4 | 2+2 4 | 5+6 2 |
| | 6 2 | 3+4 4 | 6 2 | 3+4 4 | 2+2 4 | 5+6 2 |
| | 6 2 | 3+4 4 | 6 2 | 3+4 4 | 2+2 4 | 5+6 2 |
| | 5 2 | 3+4 4 | 6 2 | 3+4 4 | 2+2 4 | 3+4 2 |
| | 6 2 | 3+4 4 | 5 2 | 3+4 4 | 2+2 4 | 5+6 2 |
| | 5 2 | 3+4 4 | 5 2 | 3+4 4 | 2+2 4 | 5+6 2 |
| | 6 2 | 3+4 4 | 6 2 | 3+4 4 | 2+2 4 | 3+4 2 |
| | 5 2 | 3+4 4 | 5 2 | 3+4 4 | 2+2 4 | 5+6 2 |

Continued on next column

**TABLE XXXVIII– continued**

| TQC | ICSD | | NoE | | SF | |
|---|---|---|---|---|---|---|
| | 6 2 | 3+4 4 | 6 2 | 3+4 4 | 2+2 4 | 3+4 2 |
| | 6 2 | 3+4 4 | 6 2 | 3+4 4 | 2+2 4 | 5+6 2 |
| | 5 2 | 3+4 4 | 6 2 | 3+4 4 | 2+2 4 | 3+4 2 |
| | 5 2 | 3+4 4 | 6 2 | 3+4 4 | 2+2 4 | 5+6 2 |
| | 5 2 | | 5 2 | | | 5+6 2 |
| | 6 2 | | 5 2 | | | 5+6 2 |
| | 6 2 | | 6 2 | | | 3+4 2 |
| | 5 2 | | 5 2 | | | 3+4 2 |
| | 5 2 | | 6 2 | | | 3+4 2 |
| | 6 2 | | 6 2 | | | 5+6 2 |
| | 6 2 | | 6 2 | | | 5+6 2 |
| | 6 2 | | 6 2 | | | 5+6 2 |
| | 6 2 | | 5 2 | | | 5+6 2 |
| | 5 2 | | 5 2 | | | 3+4 2 |
| | 6 2 | | 6 2 | | | 3+4 2 |
| | 6 2 | | 6 2 | | | 5+6 2 |
| | 6 2 | | 5 2 | | | 5+6 2 |
| | 5 2 | | 5 2 | | | 3+4 2 |
| | 5 2 | | 5 2 | | | 3+4 2 |
| | 6 2 | | 6 2 | | | 3+4 2 |
| | 5 2 | | 6 2 | | | 5+6 2 |
| | 6 2 | | 6 2 | | | 5+6 2 |
| | 5 2 | | 5 2 | | | 3+4 2 |
| 63.2.66 | 192389 | | 88 | | Cr2Fe1O4 | |
| | -GM | -T | -Y | -Z | -R | -S |
| | 5 2 | 3+4 4 | 6 2 | 3+4 4 | 2+2 4 | 5+6 2 |
| | 5 2 | 3+4 4 | 5 2 | 3+4 4 | 2+2 4 | 3+4 2 |
| | 6 2 | 3+4 4 | 6 2 | 3+4 4 | 2+2 4 | 3+4 2 |
| | 5 2 | 3+4 4 | 5 2 | 3+4 4 | 2+2 4 | 3+4 2 |
| | 6 2 | 3+4 4 | 6 2 | 3+4 4 | 2+2 4 | 3+4 2 |
| | 5 2 | 3+4 4 | 5 2 | 3+4 4 | 2+2 4 | 5+6 2 |
| | 6 2 | 3+4 4 | 5 2 | 3+4 4 | 2+2 4 | 3+4 2 |
| | 5 2 | 3+4 4 | 5 2 | 3+4 4 | 2+2 4 | 5+6 2 |
| | 6 2 | 3+4 4 | 6 2 | 3+4 4 | 2+2 4 | 5+6 2 |
| | 6 2 | 3+4 4 | 6 2 | 3+4 4 | 2+2 4 | 5+6 2 |
| | 5 2 | 3+4 4 | 5 2 | 3+4 4 | 2+2 4 | 3+4 2 |
| | 6 2 | 3+4 4 | 6 2 | 3+4 4 | 2+2 4 | 5+6 2 |
| | 5 2 | 3+4 4 | 5 2 | 3+4 4 | 2+2 4 | 3+4 2 |
| | 5 2 | 3+4 4 | 6 2 | 3+4 4 | 2+2 4 | 5+6 2 |
| | 6 2 | 3+4 4 | 5 2 | 3+4 4 | 2+2 4 | 5+6 2 |
| | 6 2 | 3+4 4 | 6 2 | 3+4 4 | 2+2 4 | 3+4 2 |
| | 6 2 | 3+4 4 | 6 2 | 3+4 4 | 2+2 4 | 3+4 2 |
| | 5 2 | 3+4 4 | 6 2 | 3+4 4 | 2+2 4 | 5+6 2 |
| | 5 2 | 3+4 4 | 6 2 | 3+4 4 | 2+2 4 | 3+4 2 |
| | 5 2 | 3+4 4 | 5 2 | 3+4 4 | 2+2 4 | 5+6 2 |
| | 5 2 | 3+4 4 | 5 2 | 3+4 4 | 2+2 4 | 5+6 2 |
| | 6 2 | 3+4 4 | 6 2 | 3+4 4 | 2+2 4 | 5+6 2 |
| | 5 2 | | 6 2 | | | 5+6 2 |
| | 6 2 | | 5 2 | | | 3+4 2 |
| | 6 2 | | 6 2 | | | 3+4 2 |
| | 6 2 | | 6 2 | | | 5+6 2 |
| | 5 2 | | 5 2 | | | 5+6 2 |
| | 6 2 | | 6 2 | | | 3+4 2 |
| | 6 2 | | 6 2 | | | 5+6 2 |
| | 6 2 | | 5 2 | | | 3+4 2 |
| | 6 2 | | 5 2 | | | 5+6 2 |
| | 6 2 | | 6 2 | | | 5+6 2 |
| | 5 2 | | 5 2 | | | 3+4 2 |
| | 5 2 | | 5 2 | | | 5+6 2 |
| | 6 2 | | 6 2 | | | 3+4 2 |
| | 5 2 | | 5 2 | | | 3+4 2 |
| | 5 2 | | 5 2 | | | 5+6 2 |
| | 5 2 | | 6 2 | | | 3+4 2 |
| | 5 2 | | 6 2 | | | 3+4 2 |
| | 5 2 | | 6 2 | | | 3+4 2 |
| | 6 2 | | 5 2 | | | 5+6 2 |
| | 5 2 | | 5 2 | | | 3+4 2 |
| | 6 2 | | 5 2 | | | 5+6 2 |
| | 6 2 | | 6 2 | | | 3+4 2 |
| 63.2.67 | 20951 | | 152 | | Co3Ga9Y2 | |
| | -GM | -T | -Y | -Z | -R | -S |
| | 5 2 | 3+4 4 | 5 2 | 3+4 4 | 2+2 4 | 3+4 2 |
| | 6 2 | 3+4 4 | 5 2 | 3+4 4 | 2+2 4 | 3+4 2 |
| | 5 2 | 3+4 4 | 6 2 | 3+4 4 | 2+2 4 | 5+6 2 |
| | 6 2 | 3+4 4 | 6 2 | 3+4 4 | 2+2 4 | 5+6 2 |
| | 5 2 | 3+4 4 | 5 2 | 3+4 4 | 2+2 4 | 3+4 2 |
| | 6 2 | 3+4 4 | 6 2 | 3+4 4 | 2+2 4 | 5+6 2 |
| | 5 2 | 3+4 4 | 6 2 | 3+4 4 | 2+2 4 | 5+6 2 |
| | 6 2 | 3+4 4 | 5 2 | 3+4 4 | 2+2 4 | 3+4 2 |
| | 5 2 | 3+4 4 | 5 2 | 3+4 4 | 2+2 4 | 3+4 2 |
| | 5 2 | 3+4 4 | 6 2 | 3+4 4 | 2+2 4 | 5+6 2 |
| | 6 2 | 3+4 4 | 6 2 | 3+4 4 | 2+2 4 | 3+4 2 |
| | 5 2 | 3+4 4 | 5 2 | 3+4 4 | 2+2 4 | 5+6 2 |
| | 6 2 | 3+4 4 | 5 2 | 3+4 4 | 2+2 4 | 3+4 2 |
| | 6 2 | 3+4 4 | 6 2 | 3+4 4 | 2+2 4 | 5+6 2 |
| | 5 2 | 3+4 4 | 6 2 | 3+4 4 | 2+2 4 | 3+4 2 |
| | 6 2 | 3+4 4 | 5 2 | 3+4 4 | 2+2 4 | 5+6 2 |
| | 5 2 | 3+4 4 | 5 2 | 3+4 4 | 2+2 4 | 3+4 2 |
| | 6 2 | 3+4 4 | 6 2 | 3+4 4 | 2+2 4 | 5+6 2 |
| | 5 2 | 3+4 4 | 6 2 | 3+4 4 | 2+2 4 | 3+4 2 |
| | 5 2 | 3+4 4 | 5 2 | 3+4 4 | 2+2 4 | 5+6 2 |
| | 6 2 | 3+4 4 | 6 2 | 3+4 4 | 2+2 4 | 5+6 2 |
| | 5 2 | 3+4 4 | 5 2 | 3+4 4 | 2+2 4 | 3+4 2 |
| | 6 2 | 3+4 4 | 6 2 | 3+4 4 | 2+2 4 | 5+6 2 |
| | 5 2 | 3+4 4 | 5 2 | 3+4 4 | 2+2 4 | 3+4 2 |
| | 6 2 | 3+4 4 | 6 2 | 3+4 4 | 2+2 4 | 3+4 2 |
| | 5 2 | 3+4 4 | 5 2 | 3+4 4 | 2+2 4 | 5+6 2 |
| | 5 2 | 3+4 4 | 5 2 | 3+4 4 | 2+2 4 | 3+4 2 |
| | 6 2 | 3+4 4 | 6 2 | 3+4 4 | 2+2 4 | 5+6 2 |
| | 6 2 | 3+4 4 | 6 2 | 3+4 4 | 2+2 4 | 3+4 2 |
| | 5 2 | 3+4 4 | 5 2 | 3+4 4 | 2+2 4 | 5+6 2 |
| | 6 2 | 3+4 4 | 5 2 | 3+4 4 | 2+2 4 | 5+6 2 |
| | 5 2 | 3+4 4 | 6 2 | 3+4 4 | 2+2 4 | 3+4 2 |
| | 5 2 | 3+4 4 | 6 2 | 3+4 4 | 2+2 4 | 5+6 2 |
| | 6 2 | 3+4 4 | 5 2 | 3+4 4 | 2+2 4 | 3+4 2 |
| | 6 2 | 3+4 4 | 5 2 | 3+4 4 | 2+2 4 | 5+6 2 |
| | 6 2 | 3+4 4 | 6 2 | 3+4 4 | 2+2 4 | 5+6 2 |
| | 5 2 | 3+4 4 | 6 2 | 3+4 4 | 2+2 4 | 5+6 2 |
| | 6 2 | 3+4 4 | 5 2 | 3+4 4 | 2+2 4 | 5+6 2 |
| | 5 2 | | 6 2 | | | 3+4 2 |
| | 5 2 | | 5 2 | | | 3+4 2 |
| | 5 2 | | 5 2 | | | 3+4 2 |
| | 5 2 | | 5 2 | | | 5+6 2 |
| | 5 2 | | 6 2 | | | 5+6 2 |
| | 5 2 | | 6 2 | | | 3+4 2 |
| | 5 2 | | 5 2 | | | 3+4 2 |
| | 6 2 | | 6 2 | | | 5+6 2 |
| | 6 2 | | 5 2 | | | 3+4 2 |
| | 6 2 | | 6 2 | | | 3+4 2 |
| | 5 2 | | 6 2 | | | 5+6 2 |
| | 6 2 | | 5 2 | | | 5+6 2 |
| | 5 2 | | 5 2 | | | 3+4 2 |

Continued on next column

**TABLE XXXVIII– continued**

| TQC | ICSD | | NoE | | SF | |
|---|---|---|---|---|---|---|
| | 5 2 | | 5 2 | | | 3+4 2 |
| | 6 2 | | 5 2 | | | 5+6 2 |
| | 5 2 | | 6 2 | | | 3+4 2 |
| | 5 2 | | 5 2 | | | 5+6 2 |
| | 5 2 | | 5 2 | | | 3+4 2 |
| | 5 2 | | 6 2 | | | 5+6 2 |
| | 6 2 | | 5 2 | | | 3+4 2 |
| | 5 2 | | 6 2 | | | 5+6 2 |
| | 6 2 | | 5 2 | | | 3+4 2 |
| | 5 2 | | 6 2 | | | 5+6 2 |
| | 5 2 | | 5 2 | | | 5+6 2 |
| | 6 2 | | 5 2 | | | 5+6 2 |
| | 5 2 | | 6 2 | | | 5+6 2 |
| | 5 2 | | 5 2 | | | 5+6 2 |
| | 6 2 | | 5 2 | | | 3+4 2 |
| | 6 2 | | 5 2 | | | 3+4 2 |
| | 6 2 | | 6 2 | | | 5+6 2 |
| | 5 2 | | 6 2 | | | 3+4 2 |
| | 6 2 | | 5 2 | | | 5+6 2 |
| | 5 2 | | 6 2 | | | 5+6 2 |
| | 5 2 | | 6 2 | | | 5+6 2 |
| | 6 2 | | 5 2 | | | 5+6 2 |
| | 6 2 | | 6 2 | | | 3+4 2 |
| | 5 2 | | 5 2 | | | 5+6 2 |
| | 6 2 | | 5 2 | | | 3+4 2 |
| 63.2.68 | 102378 | | 88 | | Co2Eu1Ga3 | |
| | -GM | -T | -Y | -Z | -R | -S |
| | 5 2 | 3+4 4 | 5 2 | 3+4 4 | 2+2 4 | 5+6 2 |
| | 6 2 | 3+4 4 | 6 2 | 3+4 4 | 2+2 4 | 3+4 2 |
| | 5 2 | 3+4 4 | 5 2 | 3+4 4 | 2+2 4 | 5+6 2 |
| | 6 2 | 3+4 4 | 6 2 | 3+4 4 | 2+2 4 | 3+4 2 |
| | 5 2 | 3+4 4 | 5 2 | 3+4 4 | 2+2 4 | 5+6 2 |
| | 5 2 | 3+4 4 | 5 2 | 3+4 4 | 2+2 4 | 5+6 2 |
| | 6 2 | 3+4 4 | 6 2 | 3+4 4 | 2+2 4 | 3+4 2 |
| | 6 2 | 3+4 4 | 6 2 | 3+4 4 | 2+2 4 | 3+4 2 |
| | 5 2 | 3+4 4 | 5 2 | 3+4 4 | 2+2 4 | 3+4 2 |
| | 6 2 | 3+4 4 | 6 2 | 3+4 4 | 2+2 4 | 5+6 2 |
| | 5 2 | 3+4 4 | 6 2 | 3+4 4 | 2+2 4 | 3+4 2 |
| | 5 2 | 3+4 4 | 5 2 | 3+4 4 | 2+2 4 | 5+6 2 |
| | 6 2 | 3+4 4 | 5 2 | 3+4 4 | 2+2 4 | 3+4 2 |
| | 6 2 | 3+4 4 | 6 2 | 3+4 4 | 2+2 4 | 5+6 2 |
| | 5 2 | 3+4 4 | 5 2 | 3+4 4 | 2+2 4 | 3+4 2 |
| | 6 2 | 3+4 4 | 6 2 | 3+4 4 | 2+2 4 | 3+4 2 |
| | 5 2 | 3+4 4 | 5 2 | 3+4 4 | 2+2 4 | 5+6 2 |
| | 5 2 | 3+4 4 | 6 2 | 3+4 4 | 2+2 4 | 5+6 2 |
| | 5 2 | 3+4 4 | 5 2 | 3+4 4 | 2+2 4 | 5+6 2 |
| | 6 2 | 3+4 4 | 5 2 | 3+4 4 | 2+2 4 | 3+4 2 |
| | 5 2 | 3+4 4 | 6 2 | 3+4 4 | 2+2 4 | 3+4 2 |
| | 6 2 | 3+4 4 | 6 2 | 3+4 4 | 2+2 4 | 3+4 2 |
| | 5 2 | | 6 2 | | | 5+6 2 |
| | 5 2 | | 6 2 | | | 5+6 2 |
| | 6 2 | | 5 2 | | | 3+4 2 |
| | 5 2 | | 5 2 | | | 5+6 2 |
| | 6 2 | | 6 2 | | | 3+4 2 |
| | 6 2 | | 5 2 | | | 5+6 2 |
| | 6 2 | | 6 2 | | | 3+4 2 |
| | 5 2 | | 6 2 | | | 3+4 2 |
| | 6 2 | | 5 2 | | | 5+6 2 |
| | 5 2 | | 6 2 | | | 3+4 2 |
| | 6 2 | | 6 2 | | | 5+6 2 |
| | 5 2 | | 5 2 | | | 3+4 2 |

Continued on next column

**TABLE XXXVIII– continued**

| TQC | ICSD | | NoE | | SF | |
|---|---|---|---|---|---|---|
| | 6 2 | | 5 2 | | | 5+6 2 |
| | 6 2 | | 6 2 | | | 3+4 2 |
| | 5 2 | | 5 2 | | | 5+6 2 |
| | 6 2 | | 6 2 | | | 3+4 2 |
| | 5 2 | | 5 2 | | | 5+6 2 |
| | 5 2 | | 6 2 | | | 3+4 2 |
| | 6 2 | | 6 2 | | | 5+6 2 |
| | 5 2 | | 5 2 | | | 5+6 2 |
| | 6 2 | | 6 2 | | | 3+4 2 |
| | 6 2 | | 5 2 | | | 5+6 2 |
| 63.2.69 | 246169 | | 60 | | Ce1Pt1Si2 | |
| | -GM | -T | -Y | -Z | -R | -S |
| | 5 2 | 3+4 4 | 5 2 | 3+4 4 | 2+2 4 | 5+6 2 |
| | 6 2 | 3+4 4 | 6 2 | 3+4 4 | 2+2 4 | 3+4 2 |
| | 5 2 | 3+4 4 | 5 2 | 3+4 4 | 2+2 4 | 5+6 2 |
| | 6 2 | 3+4 4 | 6 2 | 3+4 4 | 2+2 4 | 3+4 2 |
| | 5 2 | 3+4 4 | 5 2 | 3+4 4 | 2+2 4 | 5+6 2 |
| | 6 2 | 3+4 4 | 6 2 | 3+4 4 | 2+2 4 | 3+4 2 |
| | 5 2 | 3+4 4 | 5 2 | 3+4 4 | 2+2 4 | 5+6 2 |
| | 6 2 | 3+4 4 | 6 2 | 3+4 4 | 2+2 4 | 3+4 2 |
| | 5 2 | 3+4 4 | 6 2 | 3+4 4 | 2+2 4 | 5+6 2 |
| | 5 2 | 3+4 4 | 5 2 | 3+4 4 | 2+2 4 | 3+4 2 |
| | 6 2 | 3+4 4 | 5 2 | 3+4 4 | 2+2 4 | 3+4 2 |
| | 6 2 | 3+4 4 | 6 2 | 3+4 4 | 2+2 4 | 5+6 2 |
| | 5 2 | 3+4 4 | 6 2 | 3+4 4 | 2+2 4 | 3+4 2 |
| | 5 2 | 3+4 4 | 5 2 | 3+4 4 | 2+2 4 | 5+6 2 |
| | 6 2 | 3+4 4 | 6 2 | 3+4 4 | 2+2 4 | 3+4 2 |
| | 5 2 | | 6 2 | | | 5+6 2 |
| | 5 2 | | 5 2 | | | 3+4 2 |
| | 6 2 | | 6 2 | | | 5+6 2 |
| | 5 2 | | 5 2 | | | 5+6 2 |
| | 6 2 | | 5 2 | | | 3+4 2 |
| | 6 2 | | 5 2 | | | 3+4 2 |
| | 5 2 | | 6 2 | | | 5+6 2 |
| | 5 2 | | 5 2 | | | 3+4 2 |
| | 6 2 | | 6 2 | | | 3+4 2 |
| | 6 2 | | 6 2 | | | 5+6 2 |
| | 5 2 | | 6 2 | | | 5+6 2 |
| | 6 2 | | 6 2 | | | 3+4 2 |
| | 5 2 | | 5 2 | | | 5+6 2 |
| | 5 2 | | 5 2 | | | 5+6 2 |
| | 6 2 | | 5 2 | | | 3+4 2 |
| 63.2.70 | 42711 | | 60 | | Ce1Ni1Si2 | |
| | -GM | -T | -Y | -Z | -R | -S |
| | 5 2 | 3+4 4 | 5 2 | 3+4 4 | 2+2 4 | 5+6 2 |
| | 6 2 | 3+4 4 | 6 2 | 3+4 4 | 2+2 4 | 3+4 2 |
| | 5 2 | 3+4 4 | 5 2 | 3+4 4 | 2+2 4 | 5+6 2 |
| | 6 2 | 3+4 4 | 6 2 | 3+4 4 | 2+2 4 | 3+4 2 |
| | 5 2 | 3+4 4 | 5 2 | 3+4 4 | 2+2 4 | 5+6 2 |
| | 6 2 | 3+4 4 | 6 2 | 3+4 4 | 2+2 4 | 3+4 2 |
| | 5 2 | 3+4 4 | 5 2 | 3+4 4 | 2+2 4 | 5+6 2 |
| | 6 2 | 3+4 4 | 6 2 | 3+4 4 | 2+2 4 | 3+4 2 |
| | 5 2 | 3+4 4 | 6 2 | 3+4 4 | 2+2 4 | 3+4 2 |
| | 5 2 | 3+4 4 | 5 2 | 3+4 4 | 2+2 4 | 5+6 2 |
| | 6 2 | 3+4 4 | 6 2 | 3+4 4 | 2+2 4 | 3+4 2 |
| | 6 2 | 3+4 4 | 5 2 | 3+4 4 | 2+2 4 | 5+6 2 |
| | 5 2 | 3+4 4 | 5 2 | 3+4 4 | 2+2 4 | 5+6 2 |
| | 5 2 | 3+4 4 | 6 2 | 3+4 4 | 2+2 4 | 3+4 2 |
| | 6 2 | 3+4 4 | 6 2 | 3+4 4 | 2+2 4 | 5+6 2 |
| | 5 2 | | 6 2 | | | 3+4 2 |
| | 5 2 | | 5 2 | | | 5+6 2 |

Continued on next column

**TABLE XXXVIII– continued**

| TQC | ICSD | | NoE | | SF | |
|---|---|---|---|---|---|---|
| | 5 2 | | 5 2 | | | 3+4 2 |
| | 6 2 | | 6 2 | | | 3+4 2 |
| | 6 2 | | 5 2 | | | 3+4 2 |
| | 5 2 | | 6 2 | | | 5+6 2 |
| | 6 2 | | 5 2 | | | 3+4 2 |
| | 5 2 | | 5 2 | | | 5+6 2 |
| | 6 2 | | 6 2 | | | 3+4 2 |
| | 6 2 | | 6 2 | | | 5+6 2 |
| | 5 2 | | 6 2 | | | 5+6 2 |
| | 5 2 | | 6 2 | | | 3+4 2 |
| | 6 2 | | 5 2 | | | 5+6 2 |
| | 5 2 | | 5 2 | | | 3+4 2 |
| | 6 2 | | 5 2 | | | 3+4 2 |
| 63.2.71 | 154440 | | 44 | Ce1Ni1 | | |
| | -GM | -T | -Y | -Z | -R | -S |
| | 5 2 | 3+4 4 | 6 2 | 3+4 4 | 2+2 4 | 3+4 2 |
| | 6 2 | 3+4 4 | 5 2 | 3+4 4 | 2+2 4 | 5+6 2 |
| | 5 2 | 3+4 4 | 6 2 | 3+4 4 | 2+2 4 | 3+4 2 |
| | 6 2 | 3+4 4 | 5 2 | 3+4 4 | 2+2 4 | 5+6 2 |
| | 5 2 | 3+4 4 | 6 2 | 3+4 4 | 2+2 4 | 3+4 2 |
| | 5 2 | 3+4 4 | 5 2 | 3+4 4 | 2+2 4 | 5+6 2 |
| | 6 2 | 3+4 4 | 6 2 | 3+4 4 | 2+2 4 | 3+4 2 |
| | 6 2 | 3+4 4 | 5 2 | 3+4 4 | 2+2 4 | 5+6 2 |
| | 5 2 | 3+4 4 | 5 2 | 3+4 4 | 2+2 4 | 5+6 2 |
| | 6 2 | 3+4 4 | 6 2 | 3+4 4 | 2+2 4 | 3+4 2 |
| | 5 2 | 3+4 4 | 5 2 | 3+4 4 | 2+2 4 | 5+6 2 |
| | 6 2 | | 6 2 | | | 3+4 2 |
| | 5 2 | | 6 2 | | | 3+4 2 |
| | 6 2 | | 5 2 | | | 5+6 2 |
| | 6 2 | | 6 2 | | | 3+4 2 |
| | 5 2 | | 6 2 | | | 5+6 2 |
| | 6 2 | | 5 2 | | | 3+4 2 |
| | 5 2 | | 5 2 | | | 5+6 2 |
| | 6 2 | | 5 2 | | | 5+6 2 |
| | 5 2 | | 6 2 | | | 3+4 2 |
| | 6 2 | | 5 2 | | | 5+6 2 |
| | 5 2 | | 5 2 | | | 5+6 2 |
| 63.2.72 | 157363 | | 68 | Ce1In4Pd1 | | |
| | -GM | -T | -Y | -Z | -R | -S |
| | 5 2 | 3+4 4 | 5 2 | 3+4 4 | 2+2 4 | 3+4 2 |
| | 6 2 | 3+4 4 | 6 2 | 3+4 4 | 2+2 4 | 5+6 2 |
| | 5 2 | 3+4 4 | 5 2 | 3+4 4 | 2+2 4 | 5+6 2 |
| | 6 2 | 3+4 4 | 6 2 | 3+4 4 | 2+2 4 | 3+4 2 |
| | 5 2 | 3+4 4 | 5 2 | 3+4 4 | 2+2 4 | 5+6 2 |
| | 6 2 | 3+4 4 | 6 2 | 3+4 4 | 2+2 4 | 3+4 2 |
| | 5 2 | 3+4 4 | 6 2 | 3+4 4 | 2+2 4 | 5+6 2 |
| | 6 2 | 3+4 4 | 5 2 | 3+4 4 | 2+2 4 | 3+4 2 |
| | 5 2 | 3+4 4 | 6 2 | 3+4 4 | 2+2 4 | 3+4 2 |
| | 6 2 | 3+4 4 | 5 2 | 3+4 4 | 2+2 4 | 5+6 2 |
| | 5 2 | 3+4 4 | 6 2 | 3+4 4 | 2+2 4 | 5+6 2 |
| | 5 2 | 3+4 4 | 6 2 | 3+4 4 | 2+2 4 | 3+4 2 |
| | 5 2 | 3+4 4 | 5 2 | 3+4 4 | 2+2 4 | 3+4 2 |
| | 6 2 | 3+4 4 | 5 2 | 3+4 4 | 2+2 4 | 5+6 2 |
| | 6 2 | 3+4 4 | 5 2 | 3+4 4 | 2+2 4 | 3+4 2 |
| | 6 2 | 3+4 4 | 6 2 | 3+4 4 | 2+2 4 | 5+6 2 |
| | 5 2 | 3+4 4 | 6 2 | 3+4 4 | 2+2 4 | 3+4 2 |
| | 5 2 | | 5 2 | | | 3+4 2 |
| | 5 2 | | 6 2 | | | 5+6 2 |
| | 6 2 | | 5 2 | | | 3+4 2 |
| | 5 2 | | 6 2 | | | 5+6 2 |
| | 6 2 | | 5 2 | | | 3+4 2 |

Continued on next column

**TABLE XXXVIII– continued**

| TQC | ICSD | | NoE | | SF | |
|---|---|---|---|---|---|---|
| | 5 2 | | 6 2 | | | 5+6 2 |
| | 6 2 | | 5 2 | | | 3+4 2 |
| | 6 2 | | 5 2 | | | 5+6 2 |
| | 5 2 | | 6 2 | | | 5+6 2 |
| | 6 2 | | 6 2 | | | 5+6 2 |
| | 5 2 | | 6 2 | | | 5+6 2 |
| | 5 2 | | 5 2 | | | 3+4 2 |
| | 6 2 | | 6 2 | | | 3+4 2 |
| | 6 2 | | 5 2 | | | 5+6 2 |
| | 6 2 | | 6 2 | | | 5+6 2 |
| | 5 2 | | 5 2 | | | 3+4 2 |
| | 6 2 | | 6 2 | | | 5+6 2 |
| 63.2.73 | 150132 | | 56 | Ce1In2Pd1 | | |
| | -GM | -T | -Y | -Z | -R | -S |
| | 5 2 | 3+4 4 | 5 2 | 3+4 4 | 2+2 4 | 5+6 2 |
| | 6 2 | 3+4 4 | 6 2 | 3+4 4 | 2+2 4 | 3+4 2 |
| | 5 2 | 3+4 4 | 5 2 | 3+4 4 | 2+2 4 | 5+6 2 |
| | 6 2 | 3+4 4 | 6 2 | 3+4 4 | 2+2 4 | 3+4 2 |
| | 5 2 | 3+4 4 | 5 2 | 3+4 4 | 2+2 4 | 5+6 2 |
| | 6 2 | 3+4 4 | 6 2 | 3+4 4 | 2+2 4 | 3+4 2 |
| | 5 2 | 3+4 4 | 5 2 | 3+4 4 | 2+2 4 | 5+6 2 |
| | 6 2 | 3+4 4 | 6 2 | 3+4 4 | 2+2 4 | 3+4 2 |
| | 5 2 | 3+4 4 | 6 2 | 3+4 4 | 2+2 4 | 3+4 2 |
| | 6 2 | 3+4 4 | 5 2 | 3+4 4 | 2+2 4 | 5+6 2 |
| | 5 2 | 3+4 4 | 6 2 | 3+4 4 | 2+2 4 | 3+4 2 |
| | 6 2 | 3+4 4 | 5 2 | 3+4 4 | 2+2 4 | 5+6 2 |
| | 5 2 | 3+4 4 | 6 2 | 3+4 4 | 2+2 4 | 5+6 2 |
| | 5 2 | 3+4 4 | 5 2 | 3+4 4 | 2+2 4 | 3+4 2 |
| | 6 2 | | 6 2 | | | 3+4 2 |
| | 5 2 | | 5 2 | | | 5+6 2 |
| | 5 2 | | 6 2 | | | 3+4 2 |
| | 6 2 | | 6 2 | | | 5+6 2 |
| | 6 2 | | 5 2 | | | 3+4 2 |
| | 5 2 | | 5 2 | | | 5+6 2 |
| | 6 2 | | 6 2 | | | 3+4 2 |
| | 5 2 | | 5 2 | | | 3+4 2 |
| | 6 2 | | 5 2 | | | 5+6 2 |
| | 5 2 | | 6 2 | | | 5+6 2 |
| | 5 2 | | 6 2 | | | 3+4 2 |
| | 5 2 | | 5 2 | | | 3+4 2 |
| | 6 2 | | 5 2 | | | 5+6 2 |
| | 6 2 | | 6 2 | | | 3+4 2 |
| 63.2.74 | 106394 | | 60 | Ce1Ge2Ni1 | | |
| | -GM | -T | -Y | -Z | -R | -S |
| | 5 2 | 3+4 4 | 5 2 | 3+4 4 | 2+2 4 | 3+4 2 |
| | 6 2 | 3+4 4 | 6 2 | 3+4 4 | 2+2 4 | 5+6 2 |
| | 5 2 | 3+4 4 | 5 2 | 3+4 4 | 2+2 4 | 3+4 2 |
| | 6 2 | 3+4 4 | 6 2 | 3+4 4 | 2+2 4 | 5+6 2 |
| | 5 2 | 3+4 4 | 5 2 | 3+4 4 | 2+2 4 | 3+4 2 |
| | 6 2 | 3+4 4 | 6 2 | 3+4 4 | 2+2 4 | 5+6 2 |
| | 5 2 | 3+4 4 | 5 2 | 3+4 4 | 2+2 4 | 3+4 2 |
| | 6 2 | 3+4 4 | 6 2 | 3+4 4 | 2+2 4 | 5+6 2 |
| | 5 2 | 3+4 4 | 6 2 | 3+4 4 | 2+2 4 | 5+6 2 |
| | 5 2 | 3+4 4 | 5 2 | 3+4 4 | 2+2 4 | 3+4 2 |
| | 6 2 | 3+4 4 | 6 2 | 3+4 4 | 2+2 4 | 5+6 2 |
| | 6 2 | 3+4 4 | 5 2 | 3+4 4 | 2+2 4 | 3+4 2 |
| | 5 2 | 3+4 4 | 5 2 | 3+4 4 | 2+2 4 | 3+4 2 |
| | 5 2 | 3+4 4 | 6 2 | 3+4 4 | 2+2 4 | 5+6 2 |
| | 6 2 | 3+4 4 | 6 2 | 3+4 4 | 2+2 4 | 5+6 2 |
| | 5 2 | | 6 2 | | | 3+4 2 |
| | 5 2 | | 5 2 | | | 5+6 2 |

Continued on next column

**TABLE XXXVIII– continued**

| TQC | ICSD | | NoE | SF | | |
|---|---|---|---|---|---|---|
| | 5 2 | | 5 2 | | | 3+4 2 |
| | 6 2 | | 6 2 | | | 5+6 2 |
| | 5 2 | | 5 2 | | | 5+6 2 |
| | 6 2 | | 6 2 | | | 3+4 2 |
| | 6 2 | | 5 2 | | | 5+6 2 |
| | 5 2 | | 5 2 | | | 3+4 2 |
| | 6 2 | | 6 2 | | | 5+6 2 |
| | 6 2 | | 6 2 | | | 3+4 2 |
| | 5 2 | | 6 2 | | | 3+4 2 |
| | 5 2 | | 6 2 | | | 5+6 2 |
| | 6 2 | | 5 2 | | | 3+4 2 |
| | 6 2 | | 5 2 | | | 5+6 2 |
| | 5 2 | | 5 2 | | | 5+6 2 |
| 63.2.75 | 425494 | | 88 | Cd2Pd1Sr1 | | |
| | -GM | -T | -Y | -Z | -R | -S |
| | 5 2 | 3+4 4 | 5 2 | 3+4 4 | 2+2 4 | 5+6 2 |
| | 6 2 | 3+4 4 | 6 2 | 3+4 4 | 2+2 4 | 3+4 2 |
| | 5 2 | 3+4 4 | 5 2 | 3+4 4 | 2+2 4 | 5+6 2 |
| | 6 2 | 3+4 4 | 6 2 | 3+4 4 | 2+2 4 | 3+4 2 |
| | 5 2 | 3+4 4 | 5 2 | 3+4 4 | 2+2 4 | 5+6 2 |
| | 6 2 | 3+4 4 | 6 2 | 3+4 4 | 2+2 4 | 3+4 2 |
| | 6 2 | 3+4 4 | 5 2 | 3+4 4 | 2+2 4 | 5+6 2 |
| | 5 2 | 3+4 4 | 6 2 | 3+4 4 | 2+2 4 | 3+4 2 |
| | 6 2 | 3+4 4 | 5 2 | 3+4 4 | 2+2 4 | 3+4 2 |
| | 5 2 | 3+4 4 | 6 2 | 3+4 4 | 2+2 4 | 5+6 2 |
| | 5 2 | 3+4 4 | 6 2 | 3+4 4 | 2+2 4 | 5+6 2 |
| | 6 2 | 3+4 4 | 5 2 | 3+4 4 | 2+2 4 | 3+4 2 |
| | 5 2 | 3+4 4 | 5 2 | 3+4 4 | 2+2 4 | 5+6 2 |
| | 6 2 | 3+4 4 | 6 2 | 3+4 4 | 2+2 4 | 3+4 2 |
| | 5 2 | 3+4 4 | 5 2 | 3+4 4 | 2+2 4 | 3+4 2 |
| | 6 2 | 3+4 4 | 6 2 | 3+4 4 | 2+2 4 | 5+6 2 |
| | 6 2 | 3+4 4 | 5 2 | 3+4 4 | 2+2 4 | 3+4 2 |
| | 5 2 | 3+4 4 | 6 2 | 3+4 4 | 2+2 4 | 5+6 2 |
| | 6 2 | 3+4 4 | 5 2 | 3+4 4 | 2+2 4 | 3+4 2 |
| | 5 2 | 3+4 4 | 5 2 | 3+4 4 | 2+2 4 | 5+6 2 |
| | 5 2 | 3+4 4 | 6 2 | 3+4 4 | 2+2 4 | 3+4 2 |
| | 6 2 | 3+4 4 | 6 2 | 3+4 4 | 2+2 4 | 5+6 2 |
| | 5 2 | | 6 2 | | | 3+4 2 |
| | 5 2 | | 6 2 | | | 5+6 2 |
| | 6 2 | | 5 2 | | | 3+4 2 |
| | 5 2 | | 5 2 | | | 5+6 2 |
| | 6 2 | | 6 2 | | | 5+6 2 |
| | 6 2 | | 5 2 | | | 3+4 2 |
| | 5 2 | | 6 2 | | | 3+4 2 |
| | 6 2 | | 5 2 | | | 5+6 2 |
| | 5 2 | | 6 2 | | | 3+4 2 |
| | 5 2 | | 5 2 | | | 5+6 2 |
| | 6 2 | | 5 2 | | | 5+6 2 |
| | 5 2 | | 6 2 | | | 3+4 2 |
| | 6 2 | | 6 2 | | | 3+4 2 |
| | 5 2 | | 5 2 | | | 5+6 2 |
| | 5 2 | | 6 2 | | | 3+4 2 |
| | 6 2 | | 5 2 | | | 5+6 2 |
| | 6 2 | | 6 2 | | | 5+6 2 |
| | 5 2 | | 5 2 | | | 3+4 2 |
| | 6 2 | | 6 2 | | | 3+4 2 |
| | 5 2 | | 5 2 | | | 3+4 2 |
| | 6 2 | | 6 2 | | | 5+6 2 |
| | 5 2 | | 5 2 | | | 5+6 2 |
| 63.2.76 | 425064 | | 96 | Cd2Nd1Pd1 | | |
| | -GM | -T | -Y | -Z | -R | -S |

Continued on next column

**TABLE XXXVIII– continued**

| TQC | ICSD | | NoE | SF | | |
|---|---|---|---|---|---|---|
| | 5 2 | 3+4 4 | 5 2 | 3+4 4 | 2+2 4 | 3+4 2 |
| | 6 2 | 3+4 4 | 6 2 | 3+4 4 | 2+2 4 | 5+6 2 |
| | 5 2 | 3+4 4 | 5 2 | 3+4 4 | 2+2 4 | 3+4 2 |
| | 6 2 | 3+4 4 | 6 2 | 3+4 4 | 2+2 4 | 5+6 2 |
| | 5 2 | 3+4 4 | 5 2 | 3+4 4 | 2+2 4 | 5+6 2 |
| | 6 2 | 3+4 4 | 6 2 | 3+4 4 | 2+2 4 | 3+4 2 |
| | 5 2 | 3+4 4 | 5 2 | 3+4 4 | 2+2 4 | 3+4 2 |
| | 6 2 | 3+4 4 | 6 2 | 3+4 4 | 2+2 4 | 5+6 2 |
| | 6 2 | 3+4 4 | 5 2 | 3+4 4 | 2+2 4 | 5+6 2 |
| | 5 2 | 3+4 4 | 5 2 | 3+4 4 | 2+2 4 | 3+4 2 |
| | 6 2 | 3+4 4 | 6 2 | 3+4 4 | 2+2 4 | 5+6 2 |
| | 5 2 | 3+4 4 | 6 2 | 3+4 4 | 2+2 4 | 3+4 2 |
| | 5 2 | 3+4 4 | 5 2 | 3+4 4 | 2+2 4 | 3+4 2 |
| | 6 2 | 3+4 4 | 6 2 | 3+4 4 | 2+2 4 | 5+6 2 |
| | 5 2 | 3+4 4 | 5 2 | 3+4 4 | 2+2 4 | 5+6 2 |
| | 6 2 | 3+4 4 | 6 2 | 3+4 4 | 2+2 4 | 3+4 2 |
| | 6 2 | 3+4 4 | 5 2 | 3+4 4 | 2+2 4 | 5+6 2 |
| | 5 2 | 3+4 4 | 6 2 | 3+4 4 | 2+2 4 | 3+4 2 |
| | 5 2 | 3+4 4 | 5 2 | 3+4 4 | 2+2 4 | 5+6 2 |
| | 6 2 | 3+4 4 | 6 2 | 3+4 4 | 2+2 4 | 3+4 2 |
| | 6 2 | 3+4 4 | 5 2 | 3+4 4 | 2+2 4 | 3+4 2 |
| | 5 2 | 3+4 4 | 6 2 | 3+4 4 | 2+2 4 | 5+6 2 |
| | 5 2 | 3+4 4 | 6 2 | 3+4 4 | 2+2 4 | 5+6 2 |
| | 6 2 | 3+4 4 | 6 2 | 3+4 4 | 2+2 4 | 3+4 2 |
| | 5 2 | | 5 2 | | | 5+6 2 |
| | 5 2 | | 5 2 | | | 3+4 2 |
| | 6 2 | | 6 2 | | | 3+4 2 |
| | 6 2 | | 5 2 | | | 5+6 2 |
| | 5 2 | | 6 2 | | | 5+6 2 |
| | 5 2 | | 5 2 | | | 5+6 2 |
| | 6 2 | | 6 2 | | | 3+4 2 |
| | 5 2 | | 5 2 | | | 3+4 2 |
| | 6 2 | | 6 2 | | | 3+4 2 |
| | 5 2 | | 5 2 | | | 5+6 2 |
| | 6 2 | | 5 2 | | | 5+6 2 |
| | 5 2 | | 6 2 | | | 3+4 2 |
| | 5 2 | | 6 2 | | | 5+6 2 |
| | 6 2 | | 5 2 | | | 3+4 2 |
| | 5 2 | | 6 2 | | | 3+4 2 |
| | 6 2 | | 6 2 | | | 5+6 2 |
| | 6 2 | | 5 2 | | | 5+6 2 |
| | 5 2 | | 5 2 | | | 5+6 2 |
| | 5 2 | | 5 2 | | | 3+4 2 |
| | 6 2 | | 6 2 | | | 3+4 2 |
| | 6 2 | | 6 2 | | | 5+6 2 |
| | 5 2 | | 5 2 | | | 5+6 2 |
| | 6 2 | | 5 2 | | | 3+4 2 |
| | 5 2 | | 6 2 | | | 5+6 2 |
| 63.2.77 | 421432 | | 120 | Ca4Ge3Ni4 | | |
| | -GM | -T | -Y | -Z | -R | -S |
| | 5 2 | 3+4 4 | 6 2 | 3+4 4 | 2+2 4 | 3+4 2 |
| | 5 2 | 3+4 4 | 6 2 | 3+4 4 | 2+2 4 | 5+6 2 |
| | 6 2 | 3+4 4 | 5 2 | 3+4 4 | 2+2 4 | 3+4 2 |
| | 6 2 | 3+4 4 | 5 2 | 3+4 4 | 2+2 4 | 5+6 2 |
| | 5 2 | 3+4 4 | 6 2 | 3+4 4 | 2+2 4 | 5+6 2 |
| | 6 2 | 3+4 4 | 5 2 | 3+4 4 | 2+2 4 | 3+4 2 |
| | 5 2 | 3+4 4 | 6 2 | 3+4 4 | 2+2 4 | 3+4 2 |
| | 5 2 | 3+4 4 | 6 2 | 3+4 4 | 2+2 4 | 3+4 2 |
| | 6 2 | 3+4 4 | 5 2 | 3+4 4 | 2+2 4 | 5+6 2 |
| | 6 2 | 3+4 4 | 5 2 | 3+4 4 | 2+2 4 | 5+6 2 |
| | 6 2 | 3+4 4 | 5 2 | 3+4 4 | 2+2 4 | 5+6 2 |

Continued on next column

**TABLE XXXVIII– continued**

| TQC | ICSD | | NoE | SF | | |
|---|---|---|---|---|---|---|
| | 6 2 | 3+4 4 | 5 2 | 3+4 4 | 2+2 4 | 3+4 2 |
| | 5 2 | 3+4 4 | 6 2 | 3+4 4 | 2+2 4 | 3+4 2 |
| | 5 2 | 3+4 4 | 5 2 | 3+4 4 | 2+2 4 | 3+4 2 |
| | 5 2 | 3+4 4 | 5 2 | 3+4 4 | 2+2 4 | 5+6 2 |
| | 5 2 | 3+4 4 | 6 2 | 3+4 4 | 2+2 4 | 5+6 2 |
| | 6 2 | 3+4 4 | 6 2 | 3+4 4 | 2+2 4 | 3+4 2 |
| | 6 2 | 3+4 4 | 6 2 | 3+4 4 | 2+2 4 | 5+6 2 |
| | 5 2 | 3+4 4 | 5 2 | 3+4 4 | 2+2 4 | 3+4 2 |
| | 5 2 | 3+4 4 | 6 2 | 3+4 4 | 2+2 4 | 5+6 2 |
| | 6 2 | 3+4 4 | 5 2 | 3+4 4 | 2+2 4 | 3+4 2 |
| | 5 2 | 3+4 4 | 6 2 | 3+4 4 | 2+2 4 | 5+6 2 |
| | 6 2 | 3+4 4 | 5 2 | 3+4 4 | 2+2 4 | 5+6 2 |
| | 5 2 | 3+4 4 | 6 2 | 3+4 4 | 2+2 4 | 3+4 2 |
| | 5 2 | 3+4 4 | 6 2 | 3+4 4 | 2+2 4 | 3+4 2 |
| | 6 2 | 3+4 4 | 5 2 | 3+4 4 | 2+2 4 | 3+4 2 |
| | 5 2 | 3+4 4 | 6 2 | 3+4 4 | 2+2 4 | 5+6 2 |
| | 6 2 | 3+4 4 | 6 2 | 3+4 4 | 2+2 4 | 5+6 2 |
| | 5 2 | 3+4 4 | 5 2 | 3+4 4 | 2+2 4 | 5+6 2 |
| | 6 2 | 3+4 4 | 5 2 | 3+4 4 | 2+2 4 | 5+6 2 |
| | 6 2 | | 5 2 | | | 3+4 2 |
| | 6 2 | | 5 2 | | | 3+4 2 |
| | 6 2 | | 5 2 | | | 3+4 2 |
| | 5 2 | | 5 2 | | | 5+6 2 |
| | 6 2 | | 6 2 | | | 3+4 2 |
| | 5 2 | | 6 2 | | | 5+6 2 |
| | 6 2 | | 6 2 | | | 3+4 2 |
| | 5 2 | | 6 2 | | | 5+6 2 |
| | 6 2 | | 5 2 | | | 3+4 2 |
| | 5 2 | | 5 2 | | | 5+6 2 |
| | 6 2 | | 6 2 | | | 3+4 2 |
| | 5 2 | | 6 2 | | | 5+6 2 |
| | 6 2 | | 5 2 | | | 3+4 2 |
| | 5 2 | | 5 2 | | | 5+6 2 |
| | 6 2 | | 5 2 | | | 3+4 2 |
| | 5 2 | | 6 2 | | | 3+4 2 |
| | 6 2 | | 5 2 | | | 5+6 2 |
| | 5 2 | | 6 2 | | | 5+6 2 |
| | 6 2 | | 6 2 | | | 5+6 2 |
| | 5 2 | | 6 2 | | | 3+4 2 |
| | 6 2 | | 6 2 | | | 3+4 2 |
| | 6 2 | | 5 2 | | | 5+6 2 |
| | 5 2 | | 5 2 | | | 3+4 2 |
| | 5 2 | | 6 2 | | | 5+6 2 |
| | 5 2 | | 6 2 | | | 3+4 2 |
| | 5 2 | | 6 2 | | | 5+6 2 |
| | 5 2 | | 6 2 | | | 3+4 2 |
| | 6 2 | | 5 2 | | | 5+6 2 |
| | 5 2 | | 5 2 | | | 3+4 2 |
| | 6 2 | | 5 2 | | | 5+6 2 |
| 63.2.78 | 58948 | | 36 | Ca3Zn1 | | |
| | -GM | -T | -Y | -Z | -R | -S |
| | 5 2 | 3+4 4 | 5 2 | 3+4 4 | 2+2 4 | 3+4 2 |
| | 6 2 | 3+4 4 | 6 2 | 3+4 4 | 2+2 4 | 5+6 2 |
| | 5 2 | 3+4 4 | 5 2 | 3+4 4 | 2+2 4 | 3+4 2 |
| | 6 2 | 3+4 4 | 6 2 | 3+4 4 | 2+2 4 | 5+6 2 |
| | 5 2 | 3+4 4 | 5 2 | 3+4 4 | 2+2 4 | 3+4 2 |
| | 6 2 | 3+4 4 | 6 2 | 3+4 4 | 2+2 4 | 5+6 2 |
| | 5 2 | 3+4 4 | 5 2 | 3+4 4 | 2+2 4 | 3+4 2 |
| | 6 2 | 3+4 4 | 6 2 | 3+4 4 | 2+2 4 | 5+6 2 |
| | 5 2 | 3+4 4 | 5 2 | 3+4 4 | 2+2 4 | 3+4 2 |
| | 6 2 | | 6 2 | | | 5+6 2 |

Continued on next column

**TABLE XXXVIII– continued**

| TQC | ICSD | | NoE | SF | | |
|---|---|---|---|---|---|---|
| | 5 2 | | 6 2 | | | 3+4 2 |
| | 6 2 | | 5 2 | | | 5+6 2 |
| | 5 2 | | 5 2 | | | 3+4 2 |
| | 5 2 | | 6 2 | | | 5+6 2 |
| | 6 2 | | 6 2 | | | 3+4 2 |
| | 6 2 | | 5 2 | | | 5+6 2 |
| | 5 2 | | 5 2 | | | 5+6 2 |
| | 6 2 | | 5 2 | | | 3+4 2 |
| 63.2.79 | 78996 | | 12 | Ca1Si1 | | |
| | -GM | -T | -Y | -Z | -R | -S |
| | 5 2 | 3+4 4 | 5 2 | 3+4 4 | 2+2 4 | 5+6 2 |
| | 6 2 | 3+4 4 | 6 2 | 3+4 4 | 2+2 4 | 3+4 2 |
| | 5 2 | 3+4 4 | 6 2 | 3+4 4 | 2+2 4 | 3+4 2 |
| | 6 2 | | 5 2 | | | 5+6 2 |
| | 5 2 | | 5 2 | | | 5+6 2 |
| | 6 2 | | 5 2 | | | 5+6 2 |
| 63.2.80 | 190659 | | 12 | Ca1Si1 | | |
| | -GM | -T | -Y | -Z | -R | -S |
| | 5 2 | 3+4 4 | 5 2 | 3+4 4 | 2+2 4 | 5+6 2 |
| | 6 2 | 3+4 4 | 6 2 | 3+4 4 | 2+2 4 | 3+4 2 |
| | 5 2 | 3+4 4 | 6 2 | 3+4 4 | 2+2 4 | 3+4 2 |
| | 6 2 | | 5 2 | | | 5+6 2 |
| | 5 2 | | 5 2 | | | 5+6 2 |
| | 6 2 | | 5 2 | | | 5+6 2 |
| 63.2.81 | 410890 | | 48 | Ca1In4Pd1 | | |
| | -GM | -T | -Y | -Z | -R | -S |
| | 5 2 | 3+4 4 | 6 2 | 3+4 4 | 2+2 4 | 3+4 2 |
| | 6 2 | 3+4 4 | 5 2 | 3+4 4 | 2+2 4 | 5+6 2 |
| | 5 2 | 3+4 4 | 6 2 | 3+4 4 | 2+2 4 | 5+6 2 |
| | 5 2 | 3+4 4 | 5 2 | 3+4 4 | 2+2 4 | 3+4 2 |
| | 6 2 | 3+4 4 | 6 2 | 3+4 4 | 2+2 4 | 3+4 2 |
| | 5 2 | 3+4 4 | 5 2 | 3+4 4 | 2+2 4 | 5+6 2 |
| | 6 2 | 3+4 4 | 5 2 | 3+4 4 | 2+2 4 | 3+4 2 |
| | 6 2 | 3+4 4 | 6 2 | 3+4 4 | 2+2 4 | 5+6 2 |
| | 5 2 | 3+4 4 | 6 2 | 3+4 4 | 2+2 4 | 3+4 2 |
| | 5 2 | 3+4 4 | 5 2 | 3+4 4 | 2+2 4 | 3+4 2 |
| | 5 2 | 3+4 4 | 6 2 | 3+4 4 | 2+2 4 | 5+6 2 |
| | 6 2 | 3+4 4 | 5 2 | 3+4 4 | 2+2 4 | 3+4 2 |
| | 6 2 | | 6 2 | | | 5+6 2 |
| | 5 2 | | 5 2 | | | 3+4 2 |
| | 6 2 | | 6 2 | | | 5+6 2 |
| | 5 2 | | 5 2 | | | 3+4 2 |
| | 6 2 | | 5 2 | | | 5+6 2 |
| | 5 2 | | 6 2 | | | 5+6 2 |
| | 6 2 | | 6 2 | | | 5+6 2 |
| | 5 2 | | 6 2 | | | 3+4 2 |
| | 6 2 | | 6 2 | | | 5+6 2 |
| | 5 2 | | 5 2 | | | 3+4 2 |
| | 6 2 | | 5 2 | | | 5+6 2 |
| | 6 2 | | 6 2 | | | 3+4 2 |
| 63.2.82 | 106347 | | 48 | Ca1In4Ni1 | | |
| | -GM | -T | -Y | -Z | -R | -S |
| | 5 2 | 3+4 4 | 6 2 | 3+4 4 | 2+2 4 | 5+6 2 |
| | 6 2 | 3+4 4 | 5 2 | 3+4 4 | 2+2 4 | 3+4 2 |
| | 5 2 | 3+4 4 | 6 2 | 3+4 4 | 2+2 4 | 3+4 2 |
| | 5 2 | 3+4 4 | 5 2 | 3+4 4 | 2+2 4 | 5+6 2 |
| | 5 2 | 3+4 4 | 6 2 | 3+4 4 | 2+2 4 | 5+6 2 |
| | 6 2 | 3+4 4 | 5 2 | 3+4 4 | 2+2 4 | 5+6 2 |
| | 6 2 | 3+4 4 | 5 2 | 3+4 4 | 2+2 4 | 3+4 2 |
| | 6 2 | 3+4 4 | 6 2 | 3+4 4 | 2+2 4 | 3+4 2 |
| | 5 2 | 3+4 4 | 5 2 | 3+4 4 | 2+2 4 | 5+6 2 |

Continued on next column

**TABLE XXXVIII– continued**

| TQC | ICSD | | NoE | SF | | |
|---|---|---|---|---|---|---|
| | 5 2 | 3+4 4 | 5 2 | 3+4 4 | 2+2 4 | 3+4 2 |
| | 5 2 | 3+4 4 | 6 2 | 3+4 4 | 2+2 4 | 5+6 2 |
| | 6 2 | 3+4 4 | 6 2 | 3+4 4 | 2+2 4 | 3+4 2 |
| | 6 2 | | 6 2 | | | 3+4 2 |
| | 5 2 | | 5 2 | | | 5+6 2 |
| | 6 2 | | 6 2 | | | 5+6 2 |
| | 5 2 | | 5 2 | | | 3+4 2 |
| | 6 2 | | 5 2 | | | 3+4 2 |
| | 5 2 | | 6 2 | | | 5+6 2 |
| | 6 2 | | 6 2 | | | 3+4 2 |
| | 5 2 | | 6 2 | | | 5+6 2 |
| | 6 2 | | 6 2 | | | 3+4 2 |
| | 5 2 | | 5 2 | | | 3+4 2 |
| | 6 2 | | 5 2 | | | 5+6 2 |
| | 6 2 | | 6 2 | | | 5+6 2 |
| 63.2.83 | 240342 | | 44 | Ca1Ge2Ni1 | | |
| | -GM | -T | -Y | -Z | -R | -S |
| | 5 2 | 3+4 4 | 6 2 | 3+4 4 | 2+2 4 | 5+6 2 |
| | 5 2 | 3+4 4 | 5 2 | 3+4 4 | 2+2 4 | 3+4 2 |
| | 6 2 | 3+4 4 | 6 2 | 3+4 4 | 2+2 4 | 5+6 2 |
| | 5 2 | 3+4 4 | 5 2 | 3+4 4 | 2+2 4 | 3+4 2 |
| | 6 2 | 3+4 4 | 5 2 | 3+4 4 | 2+2 4 | 5+6 2 |
| | 6 2 | 3+4 4 | 6 2 | 3+4 4 | 2+2 4 | 3+4 2 |
| | 5 2 | 3+4 4 | 6 2 | 3+4 4 | 2+2 4 | 3+4 2 |
| | 5 2 | 3+4 4 | 6 2 | 3+4 4 | 2+2 4 | 5+6 2 |
| | 6 2 | 3+4 4 | 5 2 | 3+4 4 | 2+2 4 | 5+6 2 |
| | 6 2 | 3+4 4 | 5 2 | 3+4 4 | 2+2 4 | 3+4 2 |
| | 5 2 | 3+4 4 | 6 2 | 3+4 4 | 2+2 4 | 5+6 2 |
| | 5 2 | | 5 2 | | | 3+4 2 |
| | 5 2 | | 6 2 | | | 5+6 2 |
| | 6 2 | | 5 2 | | | 3+4 2 |
| | 5 2 | | 6 2 | | | 5+6 2 |
| | 5 2 | | 5 2 | | | 3+4 2 |
| | 6 2 | | 6 2 | | | 3+4 2 |
| | 6 2 | | 5 2 | | | 5+6 2 |
| | 6 2 | | 6 2 | | | 5+6 2 |
| | 5 2 | | 5 2 | | | 3+4 2 |
| | 6 2 | | 5 2 | | | 5+6 2 |
| | 5 2 | | 6 2 | | | 3+4 2 |
| 63.2.84 | 710075 | | 36 | Ca1Ga2Ni1 | | |
| | -GM | -T | -Y | -Z | -R | -S |
| | 5 2 | 3+4 4 | 6 2 | 3+4 4 | 2+2 4 | 3+4 2 |
| | 5 2 | 3+4 4 | 5 2 | 3+4 4 | 2+2 4 | 5+6 2 |
| | 6 2 | 3+4 4 | 6 2 | 3+4 4 | 2+2 4 | 3+4 2 |
| | 6 2 | 3+4 4 | 5 2 | 3+4 4 | 2+2 4 | 5+6 2 |
| | 5 2 | 3+4 4 | 5 2 | 3+4 4 | 2+2 4 | 5+6 2 |
| | 6 2 | 3+4 4 | 6 2 | 3+4 4 | 2+2 4 | 3+4 2 |
| | 5 2 | 3+4 4 | 6 2 | 3+4 4 | 2+2 4 | 3+4 2 |
| | 5 2 | 3+4 4 | 5 2 | 3+4 4 | 2+2 4 | 3+4 2 |
| | 5 2 | 3+4 4 | 6 2 | 3+4 4 | 2+2 4 | 5+6 2 |
| | 6 2 | | 5 2 | | | 5+6 2 |
| | 6 2 | | 5 2 | | | 5+6 2 |
| | 6 2 | | 6 2 | | | 3+4 2 |
| | 5 2 | | 6 2 | | | 5+6 2 |
| | 6 2 | | 5 2 | | | 3+4 2 |
| | 5 2 | | 6 2 | | | 3+4 2 |
| | 5 2 | | 5 2 | | | 5+6 2 |
| | 5 2 | | 6 2 | | | 5+6 2 |
| | 6 2 | | 5 2 | | | 3+4 2 |
| 63.2.85 | 16355 | | 140 | Ca1Fe4O6 | | |
| | -GM | -T | -Y | -Z | -R | -S |

Continued on next column

**TABLE XXXVIII– continued**

| TQC | ICSD | | NoE | SF | | |
|---|---|---|---|---|---|---|
| | 5 2 | 3+4 4 | 5 2 | 3+4 4 | 2+2 4 | 3+4 2 |
| | 5 2 | 3+4 4 | 6 2 | 3+4 4 | 2+2 4 | 3+4 2 |
| | 6 2 | 3+4 4 | 5 2 | 3+4 4 | 2+2 4 | 5+6 2 |
| | 6 2 | 3+4 4 | 6 2 | 3+4 4 | 2+2 4 | 5+6 2 |
| | 5 2 | 3+4 4 | 6 2 | 3+4 4 | 2+2 4 | 5+6 2 |
| | 5 2 | 3+4 4 | 5 2 | 3+4 4 | 2+2 4 | 5+6 2 |
| | 5 2 | 3+4 4 | 6 2 | 3+4 4 | 2+2 4 | 5+6 2 |
| | 6 2 | 3+4 4 | 5 2 | 3+4 4 | 2+2 4 | 5+6 2 |
| | 6 2 | 3+4 4 | 5 2 | 3+4 4 | 2+2 4 | 3+4 2 |
| | 5 2 | 3+4 4 | 5 2 | 3+4 4 | 2+2 4 | 3+4 2 |
| | 6 2 | 3+4 4 | 6 2 | 3+4 4 | 2+2 4 | 3+4 2 |
| | 5 2 | 3+4 4 | 5 2 | 3+4 4 | 2+2 4 | 5+6 2 |
| | 6 2 | 3+4 4 | 6 2 | 3+4 4 | 2+2 4 | 3+4 2 |
| | 6 2 | 3+4 4 | 6 2 | 3+4 4 | 2+2 4 | 3+4 2 |
| | 5 2 | 3+4 4 | 5 2 | 3+4 4 | 2+2 4 | 5+6 2 |
| | 5 2 | 3+4 4 | 5 2 | 3+4 4 | 2+2 4 | 3+4 2 |
| | 6 2 | 3+4 4 | 6 2 | 3+4 4 | 2+2 4 | 3+4 2 |
| | 6 2 | 3+4 4 | 6 2 | 3+4 4 | 2+2 4 | 5+6 2 |
| | 5 2 | 3+4 4 | 5 2 | 3+4 4 | 2+2 4 | 3+4 2 |
| | 5 2 | 3+4 4 | 6 2 | 3+4 4 | 2+2 4 | 3+4 2 |
| | 6 2 | 3+4 4 | 6 2 | 3+4 4 | 2+2 4 | 5+6 2 |
| | 6 2 | 3+4 4 | 5 2 | 3+4 4 | 2+2 4 | 5+6 2 |
| | 5 2 | 3+4 4 | 5 2 | 3+4 4 | 2+2 4 | 3+4 2 |
| | 5 2 | 3+4 4 | 6 2 | 3+4 4 | 2+2 4 | 5+6 2 |
| | 6 2 | 3+4 4 | 6 2 | 3+4 4 | 2+2 4 | 5+6 2 |
| | 6 2 | 3+4 4 | 5 2 | 3+4 4 | 2+2 4 | 5+6 2 |
| | 6 2 | 3+4 4 | 6 2 | 3+4 4 | 2+2 4 | 3+4 2 |
| | 5 2 | 3+4 4 | 6 2 | 3+4 4 | 2+2 4 | 5+6 2 |
| | 5 2 | 3+4 4 | 6 2 | 3+4 4 | 2+2 4 | 3+4 2 |
| | 6 2 | 3+4 4 | 5 2 | 3+4 4 | 2+2 4 | 3+4 2 |
| | 6 2 | 3+4 4 | 5 2 | 3+4 4 | 2+2 4 | 3+4 2 |
| | 6 2 | 3+4 4 | 6 2 | 3+4 4 | 2+2 4 | 3+4 2 |
| | 6 2 | 3+4 4 | 6 2 | 3+4 4 | 2+2 4 | 5+6 2 |
| | 5 2 | 3+4 4 | 6 2 | 3+4 4 | 2+2 4 | 5+6 2 |
| | 5 2 | 3+4 4 | 5 2 | 3+4 4 | 2+2 4 | 3+4 2 |
| | 5 2 | | 5 2 | | | 5+6 2 |
| | 6 2 | | 5 2 | | | 3+4 2 |
| | 5 2 | | 6 2 | | | 5+6 2 |
| | 5 2 | | 6 2 | | | 3+4 2 |
| | 5 2 | | 6 2 | | | 5+6 2 |
| | 6 2 | | 5 2 | | | 5+6 2 |
| | 6 2 | | 6 2 | | | 3+4 2 |
| | 5 2 | | 5 2 | | | 3+4 2 |
| | 6 2 | | 5 2 | | | 3+4 2 |
| | 6 2 | | 6 2 | | | 3+4 2 |
| | 6 2 | | 5 2 | | | 3+4 2 |
| | 6 2 | | 6 2 | | | 3+4 2 |
| | 6 2 | | 6 2 | | | 5+6 2 |
| | 5 2 | | 5 2 | | | 5+6 2 |
| | 5 2 | | 5 2 | | | 5+6 2 |
| | 6 2 | | 6 2 | | | 3+4 2 |
| | 6 2 | | 6 2 | | | 3+4 2 |
| | 5 2 | | 5 2 | | | 5+6 2 |
| | 5 2 | | 5 2 | | | 3+4 2 |
| | 5 2 | | 6 2 | | | 5+6 2 |
| | 5 2 | | 6 2 | | | 3+4 2 |
| | 6 2 | | 5 2 | | | 5+6 2 |
| | 6 2 | | 6 2 | | | 3+4 2 |
| | 5 2 | | 5 2 | | | 5+6 2 |
| | 6 2 | | 6 2 | | | 5+6 2 |
| | 6 2 | | 6 2 | | | 3+4 2 |

Continued on next column

**TABLE XXXVIII– continued**

| TQC | ICSD | | NoE | SF | | |
|---|---|---|---|---|---|---|
| | 5 2 | | 5 2 | | | 3+4 2 |
| | 6 2 | | 6 2 | | | 5+6 2 |
| | 5 2 | | 5 2 | | | 5+6 2 |
| | 6 2 | | 5 2 | | | 3+4 2 |
| | 5 2 | | 6 2 | | | 3+4 2 |
| | 5 2 | | 5 2 | | | 5+6 2 |
| | 6 2 | | 6 2 | | | 3+4 2 |
| | 5 2 | | 5 2 | | | 3+4 2 |
| | 6 2 | | 6 2 | | | 5+6 2 |
| 63.2.86 | 260482 | | 88 | C1P1Ru2Tb1 | | |
| | -GM | -T | -Y | -Z | -R | -S |
| | 5 2 | 3+4 4 | 5 2 | 3+4 4 | 2+2 4 | 5+6 2 |
| | 6 2 | 3+4 4 | 6 2 | 3+4 4 | 2+2 4 | 3+4 2 |
| | 5 2 | 3+4 4 | 5 2 | 3+4 4 | 2+2 4 | 5+6 2 |
| | 6 2 | 3+4 4 | 6 2 | 3+4 4 | 2+2 4 | 3+4 2 |
| | 5 2 | 3+4 4 | 5 2 | 3+4 4 | 2+2 4 | 5+6 2 |
| | 6 2 | 3+4 4 | 6 2 | 3+4 4 | 2+2 4 | 3+4 2 |
| | 5 2 | 3+4 4 | 5 2 | 3+4 4 | 2+2 4 | 5+6 2 |
| | 6 2 | 3+4 4 | 6 2 | 3+4 4 | 2+2 4 | 3+4 2 |
| | 5 2 | 3+4 4 | 6 2 | 3+4 4 | 2+2 4 | 5+6 2 |
| | 6 2 | 3+4 4 | 5 2 | 3+4 4 | 2+2 4 | 3+4 2 |
| | 5 2 | 3+4 4 | 5 2 | 3+4 4 | 2+2 4 | 3+4 2 |
| | 5 2 | 3+4 4 | 5 2 | 3+4 4 | 2+2 4 | 3+4 2 |
| | 6 2 | 3+4 4 | 6 2 | 3+4 4 | 2+2 4 | 5+6 2 |
| | 5 2 | 3+4 4 | 5 2 | 3+4 4 | 2+2 4 | 3+4 2 |
| | 6 2 | 3+4 4 | 6 2 | 3+4 4 | 2+2 4 | 5+6 2 |
| | 6 2 | 3+4 4 | 5 2 | 3+4 4 | 2+2 4 | 5+6 2 |
| | 6 2 | 3+4 4 | 5 2 | 3+4 4 | 2+2 4 | 5+6 2 |
| | 6 2 | 3+4 4 | 6 2 | 3+4 4 | 2+2 4 | 3+4 2 |
| | 6 2 | 3+4 4 | 6 2 | 3+4 4 | 2+2 4 | 5+6 2 |
| | 5 2 | 3+4 4 | 6 2 | 3+4 4 | 2+2 4 | 5+6 2 |
| | 5 2 | 3+4 4 | 6 2 | 3+4 4 | 2+2 4 | 5+6 2 |
| | 6 2 | 3+4 4 | 6 2 | 3+4 4 | 2+2 4 | 3+4 2 |
| | 5 2 | | 5 2 | | | 5+6 2 |
| | 6 2 | | 6 2 | | | 5+6 2 |
| | 5 2 | | 6 2 | | | 3+4 2 |
| | 6 2 | | 5 2 | | | 3+4 2 |
| | 6 2 | | 5 2 | | | 5+6 2 |
| | 5 2 | | 6 2 | | | 3+4 2 |
| | 5 2 | | 5 2 | | | 3+4 2 |
| | 6 2 | | 6 2 | | | 3+4 2 |
| | 6 2 | | 6 2 | | | 5+6 2 |
| | 5 2 | | 5 2 | | | 3+4 2 |
| | 5 2 | | 5 2 | | | 3+4 2 |
| | 5 2 | | 6 2 | | | 5+6 2 |
| | 6 2 | | 6 2 | | | 5+6 2 |
| | 5 2 | | 5 2 | | | 3+4 2 |
| | 5 2 | | 6 2 | | | 5+6 2 |
| | 6 2 | | 5 2 | | | 5+6 2 |
| | 5 2 | | 5 2 | | | 3+4 2 |
| | 6 2 | | 6 2 | | | 3+4 2 |
| | 5 2 | | 6 2 | | | 5+6 2 |
| | 6 2 | | 5 2 | | | 3+4 2 |
| | 5 2 | | 6 2 | | | 3+4 2 |
| | 5 2 | | 6 2 | | | 5+6 2 |
| 63.2.87 | 159578 | | 84 | C1Gd1Ru2Si1 | | |
| | -GM | -T | -Y | -Z | -R | -S |
| | 5 2 | 3+4 4 | 5 2 | 3+4 4 | 2+2 4 | 3+4 2 |
| | 6 2 | 3+4 4 | 6 2 | 3+4 4 | 2+2 4 | 5+6 2 |
| | 5 2 | 3+4 4 | 5 2 | 3+4 4 | 2+2 4 | 3+4 2 |
| | 6 2 | 3+4 4 | 6 2 | 3+4 4 | 2+2 4 | 5+6 2 |

Continued on next column

**TABLE XXXVIII– continued**

| TQC | ICSD | | NoE | SF | | |
|---|---|---|---|---|---|---|
| | 5 2 | 3+4 4 | 5 2 | 3+4 4 | 2+2 4 | 3+4 2 |
| | 6 2 | 3+4 4 | 6 2 | 3+4 4 | 2+2 4 | 5+6 2 |
| | 5 2 | 3+4 4 | 5 2 | 3+4 4 | 2+2 4 | 3+4 2 |
| | 6 2 | 3+4 4 | 6 2 | 3+4 4 | 2+2 4 | 5+6 2 |
| | 5 2 | 3+4 4 | 5 2 | 3+4 4 | 2+2 4 | 5+6 2 |
| | 5 2 | 3+4 4 | 5 2 | 3+4 4 | 2+2 4 | 5+6 2 |
| | 5 2 | 3+4 4 | 6 2 | 3+4 4 | 2+2 4 | 3+4 2 |
| | 6 2 | 3+4 4 | 5 2 | 3+4 4 | 2+2 4 | 5+6 2 |
| | 6 2 | 3+4 4 | 6 2 | 3+4 4 | 2+2 4 | 3+4 2 |
| | 6 2 | 3+4 4 | 6 2 | 3+4 4 | 2+2 4 | 3+4 2 |
| | 6 2 | 3+4 4 | 5 2 | 3+4 4 | 2+2 4 | 5+6 2 |
| | 5 2 | 3+4 4 | 6 2 | 3+4 4 | 2+2 4 | 3+4 2 |
| | 6 2 | 3+4 4 | 6 2 | 3+4 4 | 2+2 4 | 3+4 2 |
| | 6 2 | 3+4 4 | 6 2 | 3+4 4 | 2+2 4 | 3+4 2 |
| | 6 2 | 3+4 4 | 5 2 | 3+4 4 | 2+2 4 | 3+4 2 |
| | 6 2 | 3+4 4 | 5 2 | 3+4 4 | 2+2 4 | 3+4 2 |
| | 5 2 | 3+4 4 | 6 2 | 3+4 4 | 2+2 4 | 5+6 2 |
| | 5 2 | | 5 2 | | | 5+6 2 |
| | 6 2 | | 6 2 | | | 3+4 2 |
| | 6 2 | | 6 2 | | | 5+6 2 |
| | 5 2 | | 6 2 | | | 3+4 2 |
| | 5 2 | | 5 2 | | | 5+6 2 |
| | 6 2 | | 5 2 | | | 3+4 2 |
| | 5 2 | | 5 2 | | | 5+6 2 |
| | 6 2 | | 6 2 | | | 3+4 2 |
| | 5 2 | | 6 2 | | | 5+6 2 |
| | 6 2 | | 6 2 | | | 5+6 2 |
| | 5 2 | | 5 2 | | | 5+6 2 |
| | 5 2 | | 5 2 | | | 5+6 2 |
| | 5 2 | | 6 2 | | | 3+4 2 |
| | 6 2 | | 6 2 | | | 3+4 2 |
| | 5 2 | | 5 2 | | | 3+4 2 |
| | 5 2 | | 5 2 | | | 5+6 2 |
| | 6 2 | | 5 2 | | | 3+4 2 |
| | 6 2 | | 6 2 | | | 5+6 2 |
| | 5 2 | | 6 2 | | | 5+6 2 |
| | 5 2 | | 6 2 | | | 3+4 2 |
| | 6 2 | | 5 2 | | | 5+6 2 |
| 63.2.88 | 50633 | | 88 | C1Dy1Ru2Si1 | | |
| | -GM | -T | -Y | -Z | -R | -S |
| | 5 2 | 3+4 4 | 5 2 | 3+4 4 | 2+2 4 | 3+4 2 |
| | 6 2 | 3+4 4 | 6 2 | 3+4 4 | 2+2 4 | 5+6 2 |
| | 5 2 | 3+4 4 | 5 2 | 3+4 4 | 2+2 4 | 3+4 2 |
| | 6 2 | 3+4 4 | 6 2 | 3+4 4 | 2+2 4 | 5+6 2 |
| | 5 2 | 3+4 4 | 5 2 | 3+4 4 | 2+2 4 | 3+4 2 |
| | 6 2 | 3+4 4 | 6 2 | 3+4 4 | 2+2 4 | 5+6 2 |
| | 5 2 | 3+4 4 | 5 2 | 3+4 4 | 2+2 4 | 3+4 2 |
| | 6 2 | 3+4 4 | 6 2 | 3+4 4 | 2+2 4 | 5+6 2 |
| | 5 2 | 3+4 4 | 5 2 | 3+4 4 | 2+2 4 | 5+6 2 |
| | 5 2 | 3+4 4 | 5 2 | 3+4 4 | 2+2 4 | 5+6 2 |
| | 5 2 | 3+4 4 | 6 2 | 3+4 4 | 2+2 4 | 3+4 2 |
| | 6 2 | 3+4 4 | 5 2 | 3+4 4 | 2+2 4 | 5+6 2 |
| | 6 2 | 3+4 4 | 6 2 | 3+4 4 | 2+2 4 | 3+4 2 |
| | 6 2 | 3+4 4 | 6 2 | 3+4 4 | 2+2 4 | 3+4 2 |
| | 6 2 | 3+4 4 | 5 2 | 3+4 4 | 2+2 4 | 5+6 2 |
| | 5 2 | 3+4 4 | 6 2 | 3+4 4 | 2+2 4 | 3+4 2 |
| | 6 2 | 3+4 4 | 6 2 | 3+4 4 | 2+2 4 | 3+4 2 |
| | 6 2 | 3+4 4 | 6 2 | 3+4 4 | 2+2 4 | 3+4 2 |
| | 6 2 | 3+4 4 | 5 2 | 3+4 4 | 2+2 4 | 3+4 2 |
| | 6 2 | 3+4 4 | 5 2 | 3+4 4 | 2+2 4 | 3+4 2 |
| | 5 2 | 3+4 4 | 6 2 | 3+4 4 | 2+2 4 | 5+6 2 |

Continued on next column

**TABLE XXXVIII– continued**

| TQC | ICSD | NoE | SF | | | |
|---|---|---|---|---|---|---|
| | 5 2 | 3+4 4 | 5 2 | 3+4 4 | 2+2 4 | 5+6 2 |
| | 6 2 | | 6 2 | | | 3+4 2 |
| | 6 2 | | 6 2 | | | 5+6 2 |
| | 5 2 | | 6 2 | | | 3+4 2 |
| | 5 2 | | 5 2 | | | 5+6 2 |
| | 6 2 | | 5 2 | | | 3+4 2 |
| | 5 2 | | 5 2 | | | 5+6 2 |
| | 5 2 | | 6 2 | | | 3+4 2 |
| | 6 2 | | 6 2 | | | 5+6 2 |
| | 6 2 | | 6 2 | | | 5+6 2 |
| | 5 2 | | 5 2 | | | 5+6 2 |
| | 5 2 | | 5 2 | | | 3+4 2 |
| | 5 2 | | 6 2 | | | 5+6 2 |
| | 6 2 | | 6 2 | | | 3+4 2 |
| | 5 2 | | 5 2 | | | 3+4 2 |
| | 6 2 | | 5 2 | | | 5+6 2 |
| | 5 2 | | 5 2 | | | 5+6 2 |
| | 6 2 | | 6 2 | | | 5+6 2 |
| | 5 2 | | 6 2 | | | 3+4 2 |
| | 5 2 | | 6 2 | | | 3+4 2 |
| | 6 2 | | 5 2 | | | 5+6 2 |
| | 5 2 | | 6 2 | | | 3+4 2 |
| | 6 2 | | 5 2 | | | 3+4 2 |
| 63.2.89 | 40773 | | 88 | C1Dy1Fe2Si1 | | |
| | -GM | -T | -Y | -Z | -R | -S |
| | 5 2 | 3+4 4 | 5 2 | 3+4 4 | 2+2 4 | 3+4 2 |
| | 6 2 | 3+4 4 | 6 2 | 3+4 4 | 2+2 4 | 5+6 2 |
| | 5 2 | 3+4 4 | 5 2 | 3+4 4 | 2+2 4 | 3+4 2 |
| | 6 2 | 3+4 4 | 6 2 | 3+4 4 | 2+2 4 | 5+6 2 |
| | 5 2 | 3+4 4 | 5 2 | 3+4 4 | 2+2 4 | 3+4 2 |
| | 6 2 | 3+4 4 | 6 2 | 3+4 4 | 2+2 4 | 5+6 2 |
| | 5 2 | 3+4 4 | 5 2 | 3+4 4 | 2+2 4 | 3+4 2 |
| | 6 2 | 3+4 4 | 6 2 | 3+4 4 | 2+2 4 | 5+6 2 |
| | 5 2 | 3+4 4 | 5 2 | 3+4 4 | 2+2 4 | 5+6 2 |
| | 5 2 | 3+4 4 | 5 2 | 3+4 4 | 2+2 4 | 5+6 2 |
| | 5 2 | 3+4 4 | 6 2 | 3+4 4 | 2+2 4 | 3+4 2 |
| | 6 2 | 3+4 4 | 5 2 | 3+4 4 | 2+2 4 | 5+6 2 |
| | 6 2 | 3+4 4 | 6 2 | 3+4 4 | 2+2 4 | 3+4 2 |
| | 5 2 | 3+4 4 | 6 2 | 3+4 4 | 2+2 4 | 3+4 2 |
| | 6 2 | 3+4 4 | 5 2 | 3+4 4 | 2+2 4 | 5+6 2 |
| | 6 2 | 3+4 4 | 6 2 | 3+4 4 | 2+2 4 | 3+4 2 |
| | 6 2 | 3+4 4 | 6 2 | 3+4 4 | 2+2 4 | 3+4 2 |
| | 6 2 | 3+4 4 | 6 2 | 3+4 4 | 2+2 4 | 3+4 2 |
| | 6 2 | 3+4 4 | 6 2 | 3+4 4 | 2+2 4 | 5+6 2 |
| | 6 2 | 3+4 4 | 5 2 | 3+4 4 | 2+2 4 | 3+4 2 |
| | 5 2 | 3+4 4 | 5 2 | 3+4 4 | 2+2 4 | 3+4 2 |
| | 5 2 | 3+4 4 | 6 2 | 3+4 4 | 2+2 4 | 3+4 2 |
| | 6 2 | | 5 2 | | | 5+6 2 |
| | 6 2 | | 6 2 | | | 3+4 2 |
| | 6 2 | | 6 2 | | | 5+6 2 |
| | 5 2 | | 5 2 | | | 5+6 2 |
| | 5 2 | | 5 2 | | | 3+4 2 |
| | 5 2 | | 6 2 | | | 5+6 2 |
| | 6 2 | | 5 2 | | | 5+6 2 |
| | 5 2 | | 6 2 | | | 5+6 2 |
| | 5 2 | | 6 2 | | | 3+4 2 |
| | 6 2 | | 6 2 | | | 5+6 2 |
| | 5 2 | | 5 2 | | | 3+4 2 |
| | 6 2 | | 5 2 | | | 3+4 2 |
| | 5 2 | | 6 2 | | | 3+4 2 |
| | 6 2 | | 5 2 | | | 5+6 2 |

Continued on next column

**TABLE XXXVIII– continued**

| TQC | ICSD | NoE | SF | | | |
|---|---|---|---|---|---|---|
| | 5 2 | | 5 2 | | | 5+6 2 |
| | 5 2 | | 6 2 | | | 5+6 2 |
| | 6 2 | | 5 2 | | | 3+4 2 |
| | 5 2 | | 6 2 | | | 5+6 2 |
| | 5 2 | | 6 2 | | | 5+6 2 |
| | 6 2 | | 6 2 | | | 3+4 2 |
| | 5 2 | | 5 2 | | | 3+4 2 |
| | 6 2 | | 5 2 | | | 3+4 2 |
| 63.2.90 | 411622 | | 104 | C1B3Nb3 | | |
| | -GM | -T | -Y | -Z | -R | -S |
| | 5 2 | 3+4 4 | 5 2 | 3+4 4 | 2+2 4 | 3+4 2 |
| | 5 2 | 3+4 4 | 6 2 | 3+4 4 | 2+2 4 | 5+6 2 |
| | 6 2 | 3+4 4 | 6 2 | 3+4 4 | 2+2 4 | 3+4 2 |
| | 6 2 | 3+4 4 | 5 2 | 3+4 4 | 2+2 4 | 5+6 2 |
| | 5 2 | 3+4 4 | 5 2 | 3+4 4 | 2+2 4 | 3+4 2 |
| | 6 2 | 3+4 4 | 6 2 | 3+4 4 | 2+2 4 | 5+6 2 |
| | 5 2 | 3+4 4 | 5 2 | 3+4 4 | 2+2 4 | 3+4 2 |
| | 5 2 | 3+4 4 | 6 2 | 3+4 4 | 2+2 4 | 5+6 2 |
| | 6 2 | 3+4 4 | 6 2 | 3+4 4 | 2+2 4 | 3+4 2 |
| | 6 2 | 3+4 4 | 5 2 | 3+4 4 | 2+2 4 | 5+6 2 |
| | 5 2 | 3+4 4 | 6 2 | 3+4 4 | 2+2 4 | 3+4 2 |
| | 6 2 | 3+4 4 | 5 2 | 3+4 4 | 2+2 4 | 5+6 2 |
| | 5 2 | 3+4 4 | 5 2 | 3+4 4 | 2+2 4 | 3+4 2 |
| | 5 2 | 3+4 4 | 6 2 | 3+4 4 | 2+2 4 | 5+6 2 |
| | 6 2 | 3+4 4 | 6 2 | 3+4 4 | 2+2 4 | 3+4 2 |
| | 5 2 | 3+4 4 | 6 2 | 3+4 4 | 2+2 4 | 5+6 2 |
| | 5 2 | 3+4 4 | 5 2 | 3+4 4 | 2+2 4 | 3+4 2 |
| | 6 2 | 3+4 4 | 6 2 | 3+4 4 | 2+2 4 | 5+6 2 |
| | 6 2 | 3+4 4 | 5 2 | 3+4 4 | 2+2 4 | 3+4 2 |
| | 6 2 | 3+4 4 | 5 2 | 3+4 4 | 2+2 4 | 5+6 2 |
| | 5 2 | 3+4 4 | 6 2 | 3+4 4 | 2+2 4 | 3+4 2 |
| | 6 2 | 3+4 4 | 5 2 | 3+4 4 | 2+2 4 | 5+6 2 |
| | 5 2 | 3+4 4 | 6 2 | 3+4 4 | 2+2 4 | 3+4 2 |
| | 6 2 | 3+4 4 | 5 2 | 3+4 4 | 2+2 4 | 5+6 2 |
| | 5 2 | 3+4 4 | 5 2 | 3+4 4 | 2+2 4 | 5+6 2 |
| | 5 2 | 3+4 4 | 6 2 | 3+4 4 | 2+2 4 | 3+4 2 |
| | 6 2 | | 6 2 | | | 5+6 2 |
| | 6 2 | | 5 2 | | | 3+4 2 |
| | 5 2 | | 5 2 | | | 5+6 2 |
| | 6 2 | | 6 2 | | | 3+4 2 |
| | 5 2 | | 5 2 | | | 5+6 2 |
| | 6 2 | | 6 2 | | | 3+4 2 |
| | 5 2 | | 6 2 | | | 3+4 2 |
| | 6 2 | | 5 2 | | | 5+6 2 |
| | 5 2 | | 6 2 | | | 5+6 2 |
| | 6 2 | | 5 2 | | | 3+4 2 |
| | 5 2 | | 5 2 | | | 5+6 2 |
| | 6 2 | | 5 2 | | | 5+6 2 |
| | 5 2 | | 6 2 | | | 3+4 2 |
| | 5 2 | | 5 2 | | | 5+6 2 |
| | 6 2 | | 5 2 | | | 3+4 2 |
| | 6 2 | | 6 2 | | | 3+4 2 |
| | 5 2 | | 5 2 | | | 5+6 2 |
| | 5 2 | | 6 2 | | | 5+6 2 |
| | 6 2 | | 6 2 | | | 3+4 2 |
| | 6 2 | | 5 2 | | | 3+4 2 |
| | 5 2 | | 6 2 | | | 5+6 2 |
| | 5 2 | | 6 2 | | | 5+6 2 |
| | 6 2 | | 5 2 | | | 3+4 2 |
| | 5 2 | | 6 2 | | | 3+4 2 |
| | 6 2 | | 5 2 | | | 3+4 2 |

Continued on next column

**TABLE XXXVIII– continued**

| TQC | ICSD | NoE | SF | | | |
|---|---|---|---|---|---|---|
| | 5 2 | | 6 2 | | | 3+4 2 |
| 63.2.91 | 107610 | 64 | Be1Pd3 | | | |
| | -GM | -T | -Y | -Z | -R | -S |
| | 5 2 | 3+4 4 | 6 2 | 3+4 4 | 2+2 4 | 5+6 2 |
| | 6 2 | 3+4 4 | 5 2 | 3+4 4 | 2+2 4 | 3+4 2 |
| | 5 2 | 3+4 4 | 5 2 | 3+4 4 | 2+2 4 | 3+4 2 |
| | 6 2 | 3+4 4 | 6 2 | 3+4 4 | 2+2 4 | 5+6 2 |
| | 5 2 | 3+4 4 | 6 2 | 3+4 4 | 2+2 4 | 5+6 2 |
| | 6 2 | 3+4 4 | 5 2 | 3+4 4 | 2+2 4 | 3+4 2 |
| | 6 2 | 3+4 4 | 5 2 | 3+4 4 | 2+2 4 | 5+6 2 |
| | 5 2 | 3+4 4 | 5 2 | 3+4 4 | 2+2 4 | 5+6 2 |
| | 6 2 | 3+4 4 | 6 2 | 3+4 4 | 2+2 4 | 3+4 2 |
| | 6 2 | 3+4 4 | 6 2 | 3+4 4 | 2+2 4 | 5+6 2 |
| | 5 2 | 3+4 4 | 5 2 | 3+4 4 | 2+2 4 | 3+4 2 |
| | 5 2 | 3+4 4 | 6 2 | 3+4 4 | 2+2 4 | 5+6 2 |
| | 5 2 | 3+4 4 | 6 2 | 3+4 4 | 2+2 4 | 3+4 2 |
| | 6 2 | 3+4 4 | 6 2 | 3+4 4 | 2+2 4 | 3+4 2 |
| | 5 2 | 3+4 4 | 5 2 | 3+4 4 | 2+2 4 | 5+6 2 |
| | 6 2 | 3+4 4 | 6 2 | 3+4 4 | 2+2 4 | 3+4 2 |
| | 5 2 | | 5 2 | | | 3+4 2 |
| | 5 2 | | 6 2 | | | 3+4 2 |
| | 6 2 | | 5 2 | | | 3+4 2 |
| | 6 2 | | 6 2 | | | 5+6 2 |
| | 5 2 | | 5 2 | | | 5+6 2 |
| | 5 2 | | 5 2 | | | 5+6 2 |
| | 6 2 | | 5 2 | | | 3+4 2 |
| | 5 2 | | 6 2 | | | 5+6 2 |
| | 5 2 | | 5 2 | | | 3+4 2 |
| | 5 2 | | 6 2 | | | 5+6 2 |
| | 6 2 | | 5 2 | | | 3+4 2 |
| | 6 2 | | 6 2 | | | 5+6 2 |
| | 6 2 | | 6 2 | | | 3+4 2 |
| | 5 2 | | 6 2 | | | 5+6 2 |
| | 6 2 | | 5 2 | | | 5+6 2 |
| | 5 2 | | 5 2 | | | 3+4 2 |
| 63.2.92 | 411980 | 100 | Ba3Sn5 | | | |
| | -GM | -T | -Y | -Z | -R | -S |
| | 5 2 | 3+4 4 | 6 2 | 3+4 4 | 2+2 4 | 5+6 2 |
| | 5 2 | 3+4 4 | 6 2 | 3+4 4 | 2+2 4 | 5+6 2 |
| | 6 2 | 3+4 4 | 5 2 | 3+4 4 | 2+2 4 | 3+4 2 |
| | 6 2 | 3+4 4 | 5 2 | 3+4 4 | 2+2 4 | 3+4 2 |
| | 6 2 | 3+4 4 | 5 2 | 3+4 4 | 2+2 4 | 3+4 2 |
| | 5 2 | 3+4 4 | 6 2 | 3+4 4 | 2+2 4 | 5+6 2 |
| | 5 2 | 3+4 4 | 6 2 | 3+4 4 | 2+2 4 | 5+6 2 |
| | 5 2 | 3+4 4 | 5 2 | 3+4 4 | 2+2 4 | 5+6 2 |
| | 6 2 | 3+4 4 | 6 2 | 3+4 4 | 2+2 4 | 3+4 2 |
| | 6 2 | 3+4 4 | 5 2 | 3+4 4 | 2+2 4 | 3+4 2 |
| | 5 2 | 3+4 4 | 5 2 | 3+4 4 | 2+2 4 | 3+4 2 |
| | 6 2 | 3+4 4 | 6 2 | 3+4 4 | 2+2 4 | 5+6 2 |
| | 5 2 | 3+4 4 | 5 2 | 3+4 4 | 2+2 4 | 3+4 2 |
| | 5 2 | 3+4 4 | 6 2 | 3+4 4 | 2+2 4 | 5+6 2 |
| | 6 2 | 3+4 4 | 6 2 | 3+4 4 | 2+2 4 | 5+6 2 |
| | 6 2 | 3+4 4 | 5 2 | 3+4 4 | 2+2 4 | 3+4 2 |
| | 5 2 | 3+4 4 | 5 2 | 3+4 4 | 2+2 4 | 3+4 2 |
| | 5 2 | 3+4 4 | 6 2 | 3+4 4 | 2+2 4 | 5+6 2 |
| | 6 2 | 3+4 4 | 6 2 | 3+4 4 | 2+2 4 | 5+6 2 |
| | 5 2 | 3+4 4 | 5 2 | 3+4 4 | 2+2 4 | 3+4 2 |
| | 6 2 | 3+4 4 | 5 2 | 3+4 4 | 2+2 4 | 5+6 2 |
| | 5 2 | 3+4 4 | 6 2 | 3+4 4 | 2+2 4 | 3+4 2 |
| | 6 2 | 3+4 4 | 6 2 | 3+4 4 | 2+2 4 | 5+6 2 |
| | 6 2 | 3+4 4 | 5 2 | 3+4 4 | 2+2 4 | 3+4 2 |

Continued on next column

**TABLE XXXVIII– continued**

| TQC | ICSD | NoE | SF | | | |
|---|---|---|---|---|---|---|
| | 5 2 | 3+4 4 | 5 2 | 3+4 4 | 2+2 4 | 5+6 2 |
| | 6 2 | | 6 2 | | | 3+4 2 |
| | 5 2 | | 6 2 | | | 3+4 2 |
| | 6 2 | | 5 2 | | | 5+6 2 |
| | 6 2 | | 5 2 | | | 3+4 2 |
| | 5 2 | | 6 2 | | | 5+6 2 |
| | 5 2 | | 5 2 | | | 3+4 2 |
| | 6 2 | | 6 2 | | | 5+6 2 |
| | 5 2 | | 5 2 | | | 3+4 2 |
| | 6 2 | | 6 2 | | | 5+6 2 |
| | 5 2 | | 6 2 | | | 5+6 2 |
| | 6 2 | | 5 2 | | | 3+4 2 |
| | 5 2 | | 6 2 | | | 5+6 2 |
| | 6 2 | | 5 2 | | | 3+4 2 |
| | 5 2 | | 5 2 | | | 3+4 2 |
| | 6 2 | | 5 2 | | | 3+4 2 |
| | 5 2 | | 5 2 | | | 5+6 2 |
| | 5 2 | | 6 2 | | | 3+4 2 |
| | 6 2 | | 6 2 | | | 5+6 2 |
| | 6 2 | | 5 2 | | | 5+6 2 |
| | 6 2 | | 6 2 | | | 3+4 2 |
| | 6 2 | | 5 2 | | | 5+6 2 |
| | 5 2 | | 6 2 | | | 3+4 2 |
| | 5 2 | | 5 2 | | | 5+6 2 |
| | 5 2 | | 6 2 | | | 3+4 2 |
| | 5 2 | | 6 2 | | | 3+4 2 |
| 63.2.93 | 167166 | 100 | Ba3Sn5 | | | |
| | -GM | -T | -Y | -Z | -R | -S |
| | 5 2 | 3+4 4 | 6 2 | 3+4 4 | 2+2 4 | 3+4 2 |
| | 5 2 | 3+4 4 | 6 2 | 3+4 4 | 2+2 4 | 3+4 2 |
| | 6 2 | 3+4 4 | 5 2 | 3+4 4 | 2+2 4 | 5+6 2 |
| | 6 2 | 3+4 4 | 5 2 | 3+4 4 | 2+2 4 | 5+6 2 |
| | 6 2 | 3+4 4 | 5 2 | 3+4 4 | 2+2 4 | 3+4 2 |
| | 5 2 | 3+4 4 | 6 2 | 3+4 4 | 2+2 4 | 5+6 2 |
| | 5 2 | 3+4 4 | 6 2 | 3+4 4 | 2+2 4 | 3+4 2 |
| | 5 2 | 3+4 4 | 6 2 | 3+4 4 | 2+2 4 | 3+4 2 |
| | 6 2 | 3+4 4 | 5 2 | 3+4 4 | 2+2 4 | 5+6 2 |
| | 6 2 | 3+4 4 | 5 2 | 3+4 4 | 2+2 4 | 5+6 2 |
| | 5 2 | 3+4 4 | 5 2 | 3+4 4 | 2+2 4 | 5+6 2 |
| | 6 2 | 3+4 4 | 6 2 | 3+4 4 | 2+2 4 | 3+4 2 |
| | 5 2 | 3+4 4 | 5 2 | 3+4 4 | 2+2 4 | 5+6 2 |
| | 5 2 | 3+4 4 | 6 2 | 3+4 4 | 2+2 4 | 3+4 2 |
| | 6 2 | 3+4 4 | 6 2 | 3+4 4 | 2+2 4 | 3+4 2 |
| | 6 2 | 3+4 4 | 5 2 | 3+4 4 | 2+2 4 | 5+6 2 |
| | 5 2 | 3+4 4 | 5 2 | 3+4 4 | 2+2 4 | 5+6 2 |
| | 5 2 | 3+4 4 | 6 2 | 3+4 4 | 2+2 4 | 3+4 2 |
| | 6 2 | 3+4 4 | 6 2 | 3+4 4 | 2+2 4 | 3+4 2 |
| | 5 2 | 3+4 4 | 5 2 | 3+4 4 | 2+2 4 | 5+6 2 |
| | 6 2 | 3+4 4 | 5 2 | 3+4 4 | 2+2 4 | 3+4 2 |
| | 5 2 | 3+4 4 | 6 2 | 3+4 4 | 2+2 4 | 5+6 2 |
| | 6 2 | 3+4 4 | 6 2 | 3+4 4 | 2+2 4 | 3+4 2 |
| | 6 2 | 3+4 4 | 5 2 | 3+4 4 | 2+2 4 | 5+6 2 |
| | 5 2 | 3+4 4 | 5 2 | 3+4 4 | 2+2 4 | 3+4 2 |
| | 6 2 | | 6 2 | | | 5+6 2 |
| | 5 2 | | 6 2 | | | 3+4 2 |
| | 6 2 | | 5 2 | | | 5+6 2 |
| | 5 2 | | 5 2 | | | 5+6 2 |
| | 6 2 | | 6 2 | | | 3+4 2 |
| | 5 2 | | 5 2 | | | 5+6 2 |
| | 6 2 | | 6 2 | | | 3+4 2 |
| | 5 2 | | 5 2 | | | 5+6 2 |

Continued on next column

**TABLE XXXVIII– continued**

| TQC | ICSD | | NoE | SF | | |
|---|---|---|---|---|---|---|
| | 6 2 | | 6 2 | | | 3+4 2 |
| | 5 2 | | 6 2 | | | 3+4 2 |
| | 6 2 | | 5 2 | | | 5+6 2 |
| | 5 2 | | 6 2 | | | 3+4 2 |
| | 6 2 | | 5 2 | | | 5+6 2 |
| | 5 2 | | 5 2 | | | 5+6 2 |
| | 6 2 | | 5 2 | | | 5+6 2 |
| | 5 2 | | 5 2 | | | 3+4 2 |
| | 5 2 | | 6 2 | | | 5+6 2 |
| | 6 2 | | 6 2 | | | 3+4 2 |
| | 6 2 | | 5 2 | | | 3+4 2 |
| | 6 2 | | 6 2 | | | 5+6 2 |
| | 5 2 | | 5 2 | | | 3+4 2 |
| | 6 2 | | 6 2 | | | 5+6 2 |
| | 5 2 | | 5 2 | | | 3+4 2 |
| | 5 2 | | 6 2 | | | 5+6 2 |
| | 5 2 | | 6 2 | | | 5+6 2 |
| 63.2.94 | 167163 | | 100 | Ba3Ge5 | | |
| | -GM | -T | -Y | -Z | -R | -S |
| | 5 2 | 3+4 4 | 5 2 | 3+4 4 | 2+2 4 | 3+4 2 |
| | 6 2 | 3+4 4 | 6 2 | 3+4 4 | 2+2 4 | 5+6 2 |
| | 5 2 | 3+4 4 | 6 2 | 3+4 4 | 2+2 4 | 5+6 2 |
| | 5 2 | 3+4 4 | 6 2 | 3+4 4 | 2+2 4 | 5+6 2 |
| | 6 2 | 3+4 4 | 5 2 | 3+4 4 | 2+2 4 | 3+4 2 |
| | 6 2 | 3+4 4 | 5 2 | 3+4 4 | 2+2 4 | 3+4 2 |
| | 6 2 | 3+4 4 | 5 2 | 3+4 4 | 2+2 4 | 5+6 2 |
| | 5 2 | 3+4 4 | 6 2 | 3+4 4 | 2+2 4 | 3+4 2 |
| | 5 2 | 3+4 4 | 6 2 | 3+4 4 | 2+2 4 | 3+4 2 |
| | 5 2 | 3+4 4 | 5 2 | 3+4 4 | 2+2 4 | 5+6 2 |
| | 6 2 | 3+4 4 | 6 2 | 3+4 4 | 2+2 4 | 3+4 2 |
| | 6 2 | 3+4 4 | 5 2 | 3+4 4 | 2+2 4 | 5+6 2 |
| | 5 2 | 3+4 4 | 5 2 | 3+4 4 | 2+2 4 | 3+4 2 |
| | 6 2 | 3+4 4 | 6 2 | 3+4 4 | 2+2 4 | 5+6 2 |
| | 6 2 | 3+4 4 | 6 2 | 3+4 4 | 2+2 4 | 3+4 2 |
| | 5 2 | 3+4 4 | 5 2 | 3+4 4 | 2+2 4 | 5+6 2 |
| | 5 2 | 3+4 4 | 5 2 | 3+4 4 | 2+2 4 | 3+4 2 |
| | 6 2 | 3+4 4 | 6 2 | 3+4 4 | 2+2 4 | 5+6 2 |
| | 5 2 | 3+4 4 | 6 2 | 3+4 4 | 2+2 4 | 5+6 2 |
| | 5 2 | 3+4 4 | 5 2 | 3+4 4 | 2+2 4 | 3+4 2 |
| | 5 2 | 3+4 4 | 6 2 | 3+4 4 | 2+2 4 | 5+6 2 |
| | 6 2 | 3+4 4 | 5 2 | 3+4 4 | 2+2 4 | 3+4 2 |
| | 6 2 | 3+4 4 | 6 2 | 3+4 4 | 2+2 4 | 5+6 2 |
| | 6 2 | 3+4 4 | 5 2 | 3+4 4 | 2+2 4 | 3+4 2 |
| | 5 2 | 3+4 4 | 5 2 | 3+4 4 | 2+2 4 | 5+6 2 |
| | 6 2 | | 6 2 | | | 3+4 2 |
| | 5 2 | | 6 2 | | | 3+4 2 |
| | 6 2 | | 5 2 | | | 5+6 2 |
| | 6 2 | | 5 2 | | | 3+4 2 |
| | 5 2 | | 6 2 | | | 5+6 2 |
| | 6 2 | | 5 2 | | | 3+4 2 |
| | 5 2 | | 6 2 | | | 5+6 2 |
| | 5 2 | | 5 2 | | | 3+4 2 |
| | 6 2 | | 6 2 | | | 5+6 2 |
| | 5 2 | | 6 2 | | | 5+6 2 |
| | 6 2 | | 5 2 | | | 3+4 2 |
| | 6 2 | | 6 2 | | | 3+4 2 |
| | 6 2 | | 5 2 | | | 5+6 2 |
| | 5 2 | | 5 2 | | | 3+4 2 |
| | 5 2 | | 6 2 | | | 3+4 2 |
| | 5 2 | | 5 2 | | | 5+6 2 |
| | 5 2 | | 6 2 | | | 3+4 2 |

Continued on next column

**TABLE XXXVIII– continued**

| TQC | ICSD | | NoE | SF | | |
|---|---|---|---|---|---|---|
| | 6 2 | | 5 2 | | | 5+6 2 |
| | 6 2 | | 6 2 | | | 5+6 2 |
| | 6 2 | | 5 2 | | | 5+6 2 |
| | 5 2 | | 5 2 | | | 3+4 2 |
| | 6 2 | | 6 2 | | | 3+4 2 |
| | 5 2 | | 5 2 | | | 5+6 2 |
| | 5 2 | | 6 2 | | | 3+4 2 |
| | 5 2 | | 6 2 | | | 3+4 2 |
| 63.2.95 | 429723 | | 148 | Ba2Ge4Ir4Mg1 | | |
| | -GM | -T | -Y | -Z | -R | -S |
| | 5 2 | 3+4 4 | 6 2 | 3+4 4 | 2+2 4 | 3+4 2 |
| | 5 2 | 3+4 4 | 6 2 | 3+4 4 | 2+2 4 | 3+4 2 |
| | 6 2 | 3+4 4 | 5 2 | 3+4 4 | 2+2 4 | 5+6 2 |
| | 6 2 | 3+4 4 | 5 2 | 3+4 4 | 2+2 4 | 5+6 2 |
| | 5 2 | 3+4 4 | 6 2 | 3+4 4 | 2+2 4 | 3+4 2 |
| | 5 2 | 3+4 4 | 6 2 | 3+4 4 | 2+2 4 | 3+4 2 |
| | 6 2 | 3+4 4 | 5 2 | 3+4 4 | 2+2 4 | 5+6 2 |
| | 6 2 | 3+4 4 | 5 2 | 3+4 4 | 2+2 4 | 5+6 2 |
| | 5 2 | 3+4 4 | 6 2 | 3+4 4 | 2+2 4 | 5+6 2 |
| | 5 2 | 3+4 4 | 6 2 | 3+4 4 | 2+2 4 | 5+6 2 |
| | 6 2 | 3+4 4 | 5 2 | 3+4 4 | 2+2 4 | 3+4 2 |
| | 6 2 | 3+4 4 | 5 2 | 3+4 4 | 2+2 4 | 3+4 2 |
| | 5 2 | 3+4 4 | 6 2 | 3+4 4 | 2+2 4 | 3+4 2 |
| | 5 2 | 3+4 4 | 6 2 | 3+4 4 | 2+2 4 | 3+4 2 |
| | 6 2 | 3+4 4 | 5 2 | 3+4 4 | 2+2 4 | 5+6 2 |
| | 6 2 | 3+4 4 | 5 2 | 3+4 4 | 2+2 4 | 5+6 2 |
| | 5 2 | 3+4 4 | 5 2 | 3+4 4 | 2+2 4 | 5+6 2 |
| | 5 2 | 3+4 4 | 5 2 | 3+4 4 | 2+2 4 | 5+6 2 |
| | 6 2 | 3+4 4 | 6 2 | 3+4 4 | 2+2 4 | 3+4 2 |
| | 5 2 | 3+4 4 | 6 2 | 3+4 4 | 2+2 4 | 3+4 2 |
| | 6 2 | 3+4 4 | 5 2 | 3+4 4 | 2+2 4 | 5+6 2 |
| | 6 2 | 3+4 4 | 6 2 | 3+4 4 | 2+2 4 | 5+6 2 |
| | 5 2 | 3+4 4 | 5 2 | 3+4 4 | 2+2 4 | 3+4 2 |
| | 6 2 | 3+4 4 | 6 2 | 3+4 4 | 2+2 4 | 3+4 2 |
| | 5 2 | 3+4 4 | 6 2 | 3+4 4 | 2+2 4 | 5+6 2 |
| | 6 2 | 3+4 4 | 5 2 | 3+4 4 | 2+2 4 | 3+4 2 |
| | 6 2 | 3+4 4 | 5 2 | 3+4 4 | 2+2 4 | 5+6 2 |
| | 5 2 | 3+4 4 | 6 2 | 3+4 4 | 2+2 4 | 3+4 2 |
| | 5 2 | 3+4 4 | 5 2 | 3+4 4 | 2+2 4 | 3+4 2 |
| | 6 2 | 3+4 4 | 6 2 | 3+4 4 | 2+2 4 | 5+6 2 |
| | 5 2 | 3+4 4 | 5 2 | 3+4 4 | 2+2 4 | 3+4 2 |
| | 6 2 | 3+4 4 | 6 2 | 3+4 4 | 2+2 4 | 5+6 2 |
| | 6 2 | 3+4 4 | 5 2 | 3+4 4 | 2+2 4 | 5+6 2 |
| | 5 2 | 3+4 4 | 5 2 | 3+4 4 | 2+2 4 | 3+4 2 |
| | 6 2 | 3+4 4 | 6 2 | 3+4 4 | 2+2 4 | 5+6 2 |
| | 5 2 | 3+4 4 | 5 2 | 3+4 4 | 2+2 4 | 3+4 2 |
| | 6 2 | 3+4 4 | 6 2 | 3+4 4 | 2+2 4 | 3+4 2 |
| | 5 2 | | 5 2 | | | 5+6 2 |
| | 5 2 | | 5 2 | | | 5+6 2 |
| | 6 2 | | 6 2 | | | 3+4 2 |
| | 5 2 | | 6 2 | | | 5+6 2 |
| | 6 2 | | 5 2 | | | 3+4 2 |
| | 5 2 | | 6 2 | | | 5+6 2 |
| | 6 2 | | 6 2 | | | 5+6 2 |
| | 6 2 | | 5 2 | | | 3+4 2 |
| | 5 2 | | 6 2 | | | 3+4 2 |
| | 5 2 | | 5 2 | | | 5+6 2 |
| | 6 2 | | 5 2 | | | 3+4 2 |
| | 6 2 | | 6 2 | | | 3+4 2 |
| | 5 2 | | 6 2 | | | 5+6 2 |
| | 6 2 | | 5 2 | | | 5+6 2 |

Continued on next column

**TABLE XXXVIII– continued**

| TQC | ICSD | | NoE | | SF | |
|---|---|---|---|---|---|---|
| | 5 2 | | 5 2 | | | 3+4 2 |
| | 5 2 | | 6 2 | | | 5+6 2 |
| | 6 2 | | 6 2 | | | 3+4 2 |
| | 6 2 | | 5 2 | | | 5+6 2 |
| | 5 2 | | 6 2 | | | 5+6 2 |
| | 5 2 | | 5 2 | | | 3+4 2 |
| | 5 2 | | 5 2 | | | 3+4 2 |
| | 6 2 | | 6 2 | | | 5+6 2 |
| | 6 2 | | 5 2 | | | 3+4 2 |
| | 5 2 | | 6 2 | | | 5+6 2 |
| | 5 2 | | 6 2 | | | 5+6 2 |
| | 6 2 | | 5 2 | | | 5+6 2 |
| | 6 2 | | 6 2 | | | 3+4 2 |
| | 5 2 | | 5 2 | | | 3+4 2 |
| | 5 2 | | 6 2 | | | 5+6 2 |
| | 6 2 | | 6 2 | | | 3+4 2 |
| | 6 2 | | 5 2 | | | 5+6 2 |
| | 6 2 | | 6 2 | | | 3+4 2 |
| | 5 2 | | 5 2 | | | 3+4 2 |
| | 6 2 | | 5 2 | | | 5+6 2 |
| | 5 2 | | 5 2 | | | 3+4 2 |
| | 6 2 | | 6 2 | | | 5+6 2 |
| | 5 2 | | 5 2 | | | 3+4 2 |
| 63.2.96 | 418610 | | 140 | | Ba1Zn5 | |
| | -GM | -T | -Y | -Z | -R | -S |
| | 5 2 | 3+4 4 | 5 2 | 3+4 4 | 2+2 4 | 5+6 2 |
| | 6 2 | 3+4 4 | 6 2 | 3+4 4 | 2+2 4 | 3+4 2 |
| | 5 2 | 3+4 4 | 5 2 | 3+4 4 | 2+2 4 | 5+6 2 |
| | 6 2 | 3+4 4 | 6 2 | 3+4 4 | 2+2 4 | 3+4 2 |
| | 5 2 | 3+4 4 | 5 2 | 3+4 4 | 2+2 4 | 5+6 2 |
| | 5 2 | 3+4 4 | 6 2 | 3+4 4 | 2+2 4 | 5+6 2 |
| | 6 2 | 3+4 4 | 5 2 | 3+4 4 | 2+2 4 | 3+4 2 |
| | 6 2 | 3+4 4 | 6 2 | 3+4 4 | 2+2 4 | 3+4 2 |
| | 5 2 | 3+4 4 | 6 2 | 3+4 4 | 2+2 4 | 3+4 2 |
| | 6 2 | 3+4 4 | 5 2 | 3+4 4 | 2+2 4 | 5+6 2 |
| | 6 2 | 3+4 4 | 6 2 | 3+4 4 | 2+2 4 | 5+6 2 |
| | 6 2 | 3+4 4 | 5 2 | 3+4 4 | 2+2 4 | 3+4 2 |
| | 5 2 | 3+4 4 | 6 2 | 3+4 4 | 2+2 4 | 5+6 2 |
| | 6 2 | 3+4 4 | 5 2 | 3+4 4 | 2+2 4 | 5+6 2 |
| | 5 2 | 3+4 4 | 6 2 | 3+4 4 | 2+2 4 | 3+4 2 |
| | 5 2 | 3+4 4 | 5 2 | 3+4 4 | 2+2 4 | 3+4 2 |
| | 6 2 | 3+4 4 | 6 2 | 3+4 4 | 2+2 4 | 3+4 2 |
| | 5 2 | 3+4 4 | 5 2 | 3+4 4 | 2+2 4 | 5+6 2 |
| | 5 2 | 3+4 4 | 5 2 | 3+4 4 | 2+2 4 | 3+4 2 |
| | 5 2 | 3+4 4 | 6 2 | 3+4 4 | 2+2 4 | 5+6 2 |
| | 6 2 | 3+4 4 | 6 2 | 3+4 4 | 2+2 4 | 3+4 2 |
| | 5 2 | 3+4 4 | 5 2 | 3+4 4 | 2+2 4 | 5+6 2 |
| | 6 2 | 3+4 4 | 6 2 | 3+4 4 | 2+2 4 | 3+4 2 |
| | 6 2 | 3+4 4 | 5 2 | 3+4 4 | 2+2 4 | 5+6 2 |
| | 5 2 | 3+4 4 | 6 2 | 3+4 4 | 2+2 4 | 5+6 2 |
| | 6 2 | 3+4 4 | 5 2 | 3+4 4 | 2+2 4 | 3+4 2 |
| | 5 2 | 3+4 4 | 6 2 | 3+4 4 | 2+2 4 | 5+6 2 |
| | 6 2 | 3+4 4 | 5 2 | 3+4 4 | 2+2 4 | 3+4 2 |
| | 5 2 | 3+4 4 | 5 2 | 3+4 4 | 2+2 4 | 3+4 2 |
| | 6 2 | 3+4 4 | 6 2 | 3+4 4 | 2+2 4 | 5+6 2 |
| | 5 2 | 3+4 4 | 6 2 | 3+4 4 | 2+2 4 | 3+4 2 |
| | 6 2 | 3+4 4 | 5 2 | 3+4 4 | 2+2 4 | 3+4 2 |
| | 5 2 | 3+4 4 | 5 2 | 3+4 4 | 2+2 4 | 5+6 2 |
| | 6 2 | 3+4 4 | 6 2 | 3+4 4 | 2+2 4 | 3+4 2 |
| | 5 2 | 3+4 4 | 5 2 | 3+4 4 | 2+2 4 | 5+6 2 |
| | 5 2 | | 6 2 | | | 5+6 2 |

Continued on next column

**TABLE XXXVIII– continued**

| TQC | ICSD | | NoE | | SF | |
|---|---|---|---|---|---|---|
| | 6 2 | | 6 2 | | | 5+6 2 |
| | 5 2 | | 5 2 | | | 3+4 2 |
| | 6 2 | | 6 2 | | | 5+6 2 |
| | 5 2 | | 5 2 | | | 3+4 2 |
| | 6 2 | | 5 2 | | | 5+6 2 |
| | 5 2 | | 5 2 | | | 3+4 2 |
| | 6 2 | | 6 2 | | | 5+6 2 |
| | 5 2 | | 6 2 | | | 3+4 2 |
| | 6 2 | | 6 2 | | | 3+4 2 |
| | 5 2 | | 5 2 | | | 3+4 2 |
| | 6 2 | | 6 2 | | | 5+6 2 |
| | 5 2 | | 5 2 | | | 3+4 2 |
| | 5 2 | | 5 2 | | | 5+6 2 |
| | 5 2 | | 6 2 | | | 5+6 2 |
| | 6 2 | | 6 2 | | | 3+4 2 |
| | 6 2 | | 5 2 | | | 3+4 2 |
| | 5 2 | | 5 2 | | | 5+6 2 |
| | 6 2 | | 6 2 | | | 5+6 2 |
| | 5 2 | | 5 2 | | | 3+4 2 |
| | 6 2 | | 6 2 | | | 5+6 2 |
| | 6 2 | | 5 2 | | | 3+4 2 |
| | 5 2 | | 6 2 | | | 5+6 2 |
| | 6 2 | | 6 2 | | | 3+4 2 |
| | 6 2 | | 5 2 | | | 5+6 2 |
| | 5 2 | | 6 2 | | | 3+4 2 |
| | 6 2 | | 5 2 | | | 5+6 2 |
| | 5 2 | | 5 2 | | | 3+4 2 |
| | 5 2 | | 6 2 | | | 5+6 2 |
| | 6 2 | | 6 2 | | | 3+4 2 |
| | 6 2 | | 5 2 | | | 3+4 2 |
| | 5 2 | | 6 2 | | | 5+6 2 |
| | 5 2 | | 6 2 | | | 5+6 2 |
| | 5 2 | | 5 2 | | | 3+4 2 |
| | 6 2 | | 5 2 | | | 3+4 2 |
| 63.2.97 | 405111 | | 60 | | Ba1Pd1Sb2 | |
| | -GM | -T | -Y | -Z | -R | -S |
| | 5 2 | 3+4 4 | 5 2 | 3+4 4 | 2+2 4 | 5+6 2 |
| | 6 2 | 3+4 4 | 6 2 | 3+4 4 | 2+2 4 | 3+4 2 |
| | 5 2 | 3+4 4 | 5 2 | 3+4 4 | 2+2 4 | 3+4 2 |
| | 6 2 | 3+4 4 | 6 2 | 3+4 4 | 2+2 4 | 5+6 2 |
| | 5 2 | 3+4 4 | 5 2 | 3+4 4 | 2+2 4 | 3+4 2 |
| | 6 2 | 3+4 4 | 6 2 | 3+4 4 | 2+2 4 | 5+6 2 |
| | 5 2 | 3+4 4 | 5 2 | 3+4 4 | 2+2 4 | 5+6 2 |
| | 6 2 | 3+4 4 | 6 2 | 3+4 4 | 2+2 4 | 3+4 2 |
| | 5 2 | 3+4 4 | 6 2 | 3+4 4 | 2+2 4 | 3+4 2 |
| | 5 2 | 3+4 4 | 5 2 | 3+4 4 | 2+2 4 | 5+6 2 |
| | 6 2 | 3+4 4 | 5 2 | 3+4 4 | 2+2 4 | 3+4 2 |
| | 6 2 | 3+4 4 | 6 2 | 3+4 4 | 2+2 4 | 5+6 2 |
| | 5 2 | 3+4 4 | 6 2 | 3+4 4 | 2+2 4 | 3+4 2 |
| | 5 2 | 3+4 4 | 6 2 | 3+4 4 | 2+2 4 | 5+6 2 |
| | 6 2 | 3+4 4 | 5 2 | 3+4 4 | 2+2 4 | 5+6 2 |
| | 5 2 | | 6 2 | | | 3+4 2 |
| | 5 2 | | 5 2 | | | 5+6 2 |
| | 6 2 | | 5 2 | | | 3+4 2 |
| | 5 2 | | 5 2 | | | 3+4 2 |
| | 6 2 | | 6 2 | | | 5+6 2 |
| | 5 2 | | 5 2 | | | 3+4 2 |
| | 6 2 | | 6 2 | | | 3+4 2 |
| | 6 2 | | 5 2 | | | 5+6 2 |
| | 6 2 | | 6 2 | | | 3+4 2 |
| | 5 2 | | 6 2 | | | 5+6 2 |

Continued on next column

**TABLE XXXVIII– continued**

| TQC | ICSD | | NoE | SF | | |
|---|---|---|---|---|---|---|
| | 5 2 | | 6 2 | | | 5+6 2 |
| | 6 2 | | 5 2 | | | 3+4 2 |
| | 5 2 | | 6 2 | | | 5+6 2 |
| | 6 2 | | 5 2 | | | 5+6 2 |
| | 5 2 | | 5 2 | | | 3+4 2 |
| 63.2.98 | 14110 | | 76 | Ba1Ni2Si2 | | |
| | -GM | -T | -Y | -Z | -R | -S |
| | 5 2 | 3+4 4 | 5 2 | 3+4 4 | 2+2 4 | 3+4 2 |
| | 6 2 | 3+4 4 | 6 2 | 3+4 4 | 2+2 4 | 5+6 2 |
| | 5 2 | 3+4 4 | 5 2 | 3+4 4 | 2+2 4 | 5+6 2 |
| | 6 2 | 3+4 4 | 6 2 | 3+4 4 | 2+2 4 | 3+4 2 |
| | 5 2 | 3+4 4 | 6 2 | 3+4 4 | 2+2 4 | 5+6 2 |
| | 6 2 | 3+4 4 | 5 2 | 3+4 4 | 2+2 4 | 3+4 2 |
| | 5 2 | 3+4 4 | 5 2 | 3+4 4 | 2+2 4 | 3+4 2 |
| | 6 2 | 3+4 4 | 6 2 | 3+4 4 | 2+2 4 | 5+6 2 |
| | 5 2 | 3+4 4 | 5 2 | 3+4 4 | 2+2 4 | 5+6 2 |
| | 5 2 | 3+4 4 | 5 2 | 3+4 4 | 2+2 4 | 5+6 2 |
| | 6 2 | 3+4 4 | 6 2 | 3+4 4 | 2+2 4 | 3+4 2 |
| | 6 2 | 3+4 4 | 6 2 | 3+4 4 | 2+2 4 | 3+4 2 |
| | 6 2 | 3+4 4 | 6 2 | 3+4 4 | 2+2 4 | 3+4 2 |
| | 5 2 | 3+4 4 | 6 2 | 3+4 4 | 2+2 4 | 5+6 2 |
| | 5 2 | 3+4 4 | 5 2 | 3+4 4 | 2+2 4 | 5+6 2 |
| | 6 2 | 3+4 4 | 5 2 | 3+4 4 | 2+2 4 | 3+4 2 |
| | 5 2 | 3+4 4 | 6 2 | 3+4 4 | 2+2 4 | 5+6 2 |
| | 6 2 | 3+4 4 | 5 2 | 3+4 4 | 2+2 4 | 5+6 2 |
| | 6 2 | 3+4 4 | 5 2 | 3+4 4 | 2+2 4 | 3+4 2 |
| | 5 2 | | 5 2 | | | 3+4 2 |
| | 5 2 | | 6 2 | | | 3+4 2 |
| | 5 2 | | 5 2 | | | 5+6 2 |
| | 6 2 | | 6 2 | | | 3+4 2 |
| | 5 2 | | 6 2 | | | 5+6 2 |
| | 5 2 | | 6 2 | | | 5+6 2 |
| | 6 2 | | 6 2 | | | 5+6 2 |
| | 6 2 | | 5 2 | | | 3+4 2 |
| | 6 2 | | 5 2 | | | 5+6 2 |
| | 5 2 | | 5 2 | | | 3+4 2 |
| | 5 2 | | 6 2 | | | 5+6 2 |
| | 6 2 | | 6 2 | | | 3+4 2 |
| | 6 2 | | 5 2 | | | 3+4 2 |
| | 5 2 | | 5 2 | | | 5+6 2 |
| | 5 2 | | 6 2 | | | 3+4 2 |
| | 5 2 | | 6 2 | | | 3+4 2 |
| | 5 2 | | 6 2 | | | 3+4 2 |
| | 5 2 | | 6 2 | | | 5+6 2 |
| | 6 2 | | 5 2 | | | 3+4 2 |
| 63.2.99 | 411228 | | 52 | Ba1In2Pt1 | | |
| | -GM | -T | -Y | -Z | -R | -S |
| | 5 2 | 3+4 4 | 5 2 | 3+4 4 | 2+2 4 | 3+4 2 |
| | 6 2 | 3+4 4 | 6 2 | 3+4 4 | 2+2 4 | 5+6 2 |
| | 5 2 | 3+4 4 | 5 2 | 3+4 4 | 2+2 4 | 3+4 2 |
| | 6 2 | 3+4 4 | 6 2 | 3+4 4 | 2+2 4 | 5+6 2 |
| | 5 2 | 3+4 4 | 5 2 | 3+4 4 | 2+2 4 | 3+4 2 |
| | 6 2 | 3+4 4 | 6 2 | 3+4 4 | 2+2 4 | 5+6 2 |
| | 5 2 | 3+4 4 | 5 2 | 3+4 4 | 2+2 4 | 3+4 2 |
| | 6 2 | 3+4 4 | 6 2 | 3+4 4 | 2+2 4 | 5+6 2 |
| | 5 2 | 3+4 4 | 6 2 | 3+4 4 | 2+2 4 | 3+4 2 |
| | 6 2 | 3+4 4 | 5 2 | 3+4 4 | 2+2 4 | 5+6 2 |
| | 5 2 | 3+4 4 | 6 2 | 3+4 4 | 2+2 4 | 5+6 2 |
| | 6 2 | 3+4 4 | 5 2 | 3+4 4 | 2+2 4 | 3+4 2 |
| | 5 2 | 3+4 4 | 6 2 | 3+4 4 | 2+2 4 | 5+6 2 |
| | 5 2 | | 5 2 | | | 3+4 2 |
| | 6 2 | | 6 2 | | | 5+6 2 |
| | 5 2 | | 5 2 | | | 3+4 2 |
| | 6 2 | | 6 2 | | | 5+6 2 |
| | 6 2 | | 5 2 | | | 3+4 2 |
| | 5 2 | | 5 2 | | | 5+6 2 |
| | 5 2 | | 6 2 | | | 3+4 2 |
| | 6 2 | | 6 2 | | | 5+6 2 |
| | 6 2 | | 5 2 | | | 3+4 2 |
| | 5 2 | | 6 2 | | | 5+6 2 |
| | 5 2 | | 5 2 | | | 3+4 2 |
| | 5 2 | | 6 2 | | | 5+6 2 |
| | 6 2 | | 5 2 | | | 3+4 2 |
| 63.2.100 | 414228 | | 52 | Ba1In1 | | |
| | -GM | -T | -Y | -Z | -R | -S |
| | 5 2 | 3+4 4 | 5 2 | 3+4 4 | 2+2 4 | 3+4 2 |
| | 6 2 | 3+4 4 | 6 2 | 3+4 4 | 2+2 4 | 5+6 2 |
| | 6 2 | 3+4 4 | 5 2 | 3+4 4 | 2+2 4 | 3+4 2 |
| | 5 2 | 3+4 4 | 6 2 | 3+4 4 | 2+2 4 | 5+6 2 |
| | 5 2 | 3+4 4 | 5 2 | 3+4 4 | 2+2 4 | 3+4 2 |
| | 6 2 | 3+4 4 | 6 2 | 3+4 4 | 2+2 4 | 5+6 2 |
| | 5 2 | 3+4 4 | 6 2 | 3+4 4 | 2+2 4 | 5+6 2 |
| | 6 2 | 3+4 4 | 5 2 | 3+4 4 | 2+2 4 | 3+4 2 |
| | 5 2 | 3+4 4 | 5 2 | 3+4 4 | 2+2 4 | 3+4 2 |
| | 6 2 | 3+4 4 | 6 2 | 3+4 4 | 2+2 4 | 5+6 2 |
| | 6 2 | 3+4 4 | 5 2 | 3+4 4 | 2+2 4 | 3+4 2 |
| | 5 2 | 3+4 4 | 6 2 | 3+4 4 | 2+2 4 | 5+6 2 |
| | 5 2 | 3+4 4 | 5 2 | 3+4 4 | 2+2 4 | 5+6 2 |
| | 5 2 | | 6 2 | | | 3+4 2 |
| | 6 2 | | 5 2 | | | 3+4 2 |
| | 6 2 | | 6 2 | | | 5+6 2 |
| | 5 2 | | 5 2 | | | 5+6 2 |
| | 6 2 | | 6 2 | | | 3+4 2 |
| | 6 2 | | 6 2 | | | 3+4 2 |
| | 5 2 | | 5 2 | | | 5+6 2 |
| | 5 2 | | 5 2 | | | 5+6 2 |
| | 5 2 | | 6 2 | | | 5+6 2 |
| | 6 2 | | 6 2 | | | 3+4 2 |
| | 5 2 | | 5 2 | | | 5+6 2 |
| | 6 2 | | 6 2 | | | 3+4 2 |
| | 5 2 | | 5 2 | | | 3+4 2 |
| 63.2.101 | 429115 | | 88 | Ba1Cd2Pt1 | | |
| | -GM | -T | -Y | -Z | -R | -S |
| | 5 2 | 3+4 4 | 5 2 | 3+4 4 | 2+2 4 | 5+6 2 |
| | 6 2 | 3+4 4 | 6 2 | 3+4 4 | 2+2 4 | 3+4 2 |
| | 5 2 | 3+4 4 | 5 2 | 3+4 4 | 2+2 4 | 5+6 2 |
| | 6 2 | 3+4 4 | 6 2 | 3+4 4 | 2+2 4 | 3+4 2 |
| | 5 2 | 3+4 4 | 5 2 | 3+4 4 | 2+2 4 | 5+6 2 |
| | 6 2 | 3+4 4 | 6 2 | 3+4 4 | 2+2 4 | 3+4 2 |
| | 5 2 | 3+4 4 | 5 2 | 3+4 4 | 2+2 4 | 5+6 2 |
| | 6 2 | 3+4 4 | 6 2 | 3+4 4 | 2+2 4 | 3+4 2 |
| | 6 2 | 3+4 4 | 5 2 | 3+4 4 | 2+2 4 | 3+4 2 |
| | 5 2 | 3+4 4 | 6 2 | 3+4 4 | 2+2 4 | 5+6 2 |
| | 5 2 | 3+4 4 | 6 2 | 3+4 4 | 2+2 4 | 3+4 2 |
| | 6 2 | 3+4 4 | 5 2 | 3+4 4 | 2+2 4 | 5+6 2 |
| | 5 2 | 3+4 4 | 6 2 | 3+4 4 | 2+2 4 | 3+4 2 |
| | 6 2 | 3+4 4 | 5 2 | 3+4 4 | 2+2 4 | 5+6 2 |
| | 5 2 | 3+4 4 | 5 2 | 3+4 4 | 2+2 4 | 3+4 2 |
| | 6 2 | 3+4 4 | 6 2 | 3+4 4 | 2+2 4 | 5+6 2 |
| | 6 2 | 3+4 4 | 5 2 | 3+4 4 | 2+2 4 | 3+4 2 |
| | 5 2 | 3+4 4 | 6 2 | 3+4 4 | 2+2 4 | 5+6 2 |
| | 6 2 | 3+4 4 | 5 2 | 3+4 4 | 2+2 4 | 3+4 2 |

Continued on next column

**TABLE XXXVIII– continued**

| TQC | ICSD | | NoE | | SF | |
|---|---|---|---|---|---|---|
| | 5 2 | 3+4 4 | 5 2 | 3+4 4 | 2+2 4 | 5+6 2 |
| | 6 2 | 3+4 4 | 6 2 | 3+4 4 | 2+2 4 | 3+4 2 |
| | 5 2 | 3+4 4 | 6 2 | 3+4 4 | 2+2 4 | 5+6 2 |
| | 5 2 | | 6 2 | | | 3+4 2 |
| | 5 2 | | 6 2 | | | 5+6 2 |
| | 6 2 | | 5 2 | | | 3+4 2 |
| | 5 2 | | 5 2 | | | 5+6 2 |
| | 6 2 | | 6 2 | | | 5+6 2 |
| | 6 2 | | 5 2 | | | 3+4 2 |
| | 5 2 | | 6 2 | | | 5+6 2 |
| | 6 2 | | 5 2 | | | 3+4 2 |
| | 5 2 | | 6 2 | | | 3+4 2 |
| | 5 2 | | 6 2 | | | 5+6 2 |
| | 6 2 | | 5 2 | | | 3+4 2 |
| | 5 2 | | 5 2 | | | 5+6 2 |
| | 6 2 | | 6 2 | | | 3+4 2 |
| | 5 2 | | 5 2 | | | 5+6 2 |
| | 6 2 | | 6 2 | | | 3+4 2 |
| | 5 2 | | 5 2 | | | 5+6 2 |
| | 6 2 | | 5 2 | | | 3+4 2 |
| | 5 2 | | 6 2 | | | 5+6 2 |
| | 6 2 | | 6 2 | | | 3+4 2 |
| | 5 2 | | 5 2 | | | 3+4 2 |
| | 6 2 | | 6 2 | | | 5+6 2 |
| | 5 2 | | 5 2 | | | 5+6 2 |
| 63.2.102 | 39663 | | 120 | | B7W1Y3 | |
| | -GM | -T | -Y | -Z | -R | -S |
| | 5 2 | 3+4 4 | 6 2 | 3+4 4 | 2+2 4 | 3+4 2 |
| | 6 2 | 3+4 4 | 5 2 | 3+4 4 | 2+2 4 | 5+6 2 |
| | 5 2 | 3+4 4 | 5 2 | 3+4 4 | 2+2 4 | 3+4 2 |
| | 5 2 | 3+4 4 | 5 2 | 3+4 4 | 2+2 4 | 3+4 2 |
| | 6 2 | 3+4 4 | 6 2 | 3+4 4 | 2+2 4 | 5+6 2 |
| | 6 2 | 3+4 4 | 6 2 | 3+4 4 | 2+2 4 | 5+6 2 |
| | 6 2 | 3+4 4 | 5 2 | 3+4 4 | 2+2 4 | 3+4 2 |
| | 5 2 | 3+4 4 | 6 2 | 3+4 4 | 2+2 4 | 5+6 2 |
| | 5 2 | 3+4 4 | 5 2 | 3+4 4 | 2+2 4 | 3+4 2 |
| | 5 2 | 3+4 4 | 6 2 | 3+4 4 | 2+2 4 | 3+4 2 |
| | 6 2 | 3+4 4 | 6 2 | 3+4 4 | 2+2 4 | 5+6 2 |
| | 6 2 | 3+4 4 | 5 2 | 3+4 4 | 2+2 4 | 5+6 2 |
| | 6 2 | 3+4 4 | 5 2 | 3+4 4 | 2+2 4 | 3+4 2 |
| | 5 2 | 3+4 4 | 6 2 | 3+4 4 | 2+2 4 | 5+6 2 |
| | 5 2 | 3+4 4 | 5 2 | 3+4 4 | 2+2 4 | 3+4 2 |
| | 5 2 | 3+4 4 | 6 2 | 3+4 4 | 2+2 4 | 5+6 2 |
| | 6 2 | 3+4 4 | 6 2 | 3+4 4 | 2+2 4 | 3+4 2 |
| | 6 2 | 3+4 4 | 5 2 | 3+4 4 | 2+2 4 | 5+6 2 |
| | 5 2 | 3+4 4 | 5 2 | 3+4 4 | 2+2 4 | 3+4 2 |
| | 6 2 | 3+4 4 | 6 2 | 3+4 4 | 2+2 4 | 3+4 2 |
| | 5 2 | 3+4 4 | 5 2 | 3+4 4 | 2+2 4 | 5+6 2 |
| | 6 2 | 3+4 4 | 5 2 | 3+4 4 | 2+2 4 | 5+6 2 |
| | 5 2 | 3+4 4 | 6 2 | 3+4 4 | 2+2 4 | 3+4 2 |
| | 6 2 | 3+4 4 | 6 2 | 3+4 4 | 2+2 4 | 5+6 2 |
| | 5 2 | 3+4 4 | 5 2 | 3+4 4 | 2+2 4 | 5+6 2 |
| | 6 2 | 3+4 4 | 6 2 | 3+4 4 | 2+2 4 | 3+4 2 |
| | 5 2 | 3+4 4 | 6 2 | 3+4 4 | 2+2 4 | 5+6 2 |
| | 5 2 | 3+4 4 | 5 2 | 3+4 4 | 2+2 4 | 5+6 2 |
| | 6 2 | 3+4 4 | 6 2 | 3+4 4 | 2+2 4 | 3+4 2 |
| | 5 2 | 3+4 4 | 5 2 | 3+4 4 | 2+2 4 | 3+4 2 |
| | 5 2 | | 6 2 | | | 5+6 2 |
| | 6 2 | | 6 2 | | | 3+4 2 |
| | 5 2 | | 5 2 | | | 3+4 2 |
| | 6 2 | | 5 2 | | | 5+6 2 |

Continued on next column

**TABLE XXXVIII– continued**

| TQC | ICSD | | NoE | | SF | |
|---|---|---|---|---|---|---|
| | 6 2 | | 5 2 | | | 3+4 2 |
| | 5 2 | | 6 2 | | | 5+6 2 |
| | 5 2 | | 6 2 | | | 5+6 2 |
| | 6 2 | | 6 2 | | | 5+6 2 |
| | 6 2 | | 5 2 | | | 3+4 2 |
| | 6 2 | | 6 2 | | | 3+4 2 |
| | 6 2 | | 5 2 | | | 5+6 2 |
| | 5 2 | | 6 2 | | | 3+4 2 |
| | 5 2 | | 5 2 | | | 5+6 2 |
| | 5 2 | | 5 2 | | | 3+4 2 |
| | 6 2 | | 5 2 | | | 5+6 2 |
| | 6 2 | | 6 2 | | | 3+4 2 |
| | 5 2 | | 6 2 | | | 3+4 2 |
| | 5 2 | | 5 2 | | | 5+6 2 |
| | 5 2 | | 6 2 | | | 5+6 2 |
| | 6 2 | | 5 2 | | | 3+4 2 |
| | 6 2 | | 6 2 | | | 5+6 2 |
| | 5 2 | | 5 2 | | | 3+4 2 |
| | 5 2 | | 5 2 | | | 5+6 2 |
| | 5 2 | | 6 2 | | | 3+4 2 |
| | 6 2 | | 6 2 | | | 5+6 2 |
| | 5 2 | | 5 2 | | | 5+6 2 |
| | 5 2 | | 6 2 | | | 3+4 2 |
| | 6 2 | | 6 2 | | | 5+6 2 |
| | 5 2 | | 5 2 | | | 3+4 2 |
| | 6 2 | | 6 2 | | | 5+6 2 |
| 63.2.103 | 62282 | | 192 | | B7Co4Nb3 | |
| | -GM | -T | -Y | -Z | -R | -S |
| | 5 2 | 3+4 4 | 5 2 | 3+4 4 | 2+2 4 | 3+4 2 |
| | 6 2 | 3+4 4 | 6 2 | 3+4 4 | 2+2 4 | 5+6 2 |
| | 5 2 | 3+4 4 | 5 2 | 3+4 4 | 2+2 4 | 3+4 2 |
| | 5 2 | 3+4 4 | 5 2 | 3+4 4 | 2+2 4 | 3+4 2 |
| | 6 2 | 3+4 4 | 6 2 | 3+4 4 | 2+2 4 | 5+6 2 |
| | 6 2 | 3+4 4 | 6 2 | 3+4 4 | 2+2 4 | 5+6 2 |
| | 5 2 | 3+4 4 | 5 2 | 3+4 4 | 2+2 4 | 3+4 2 |
| | 6 2 | 3+4 4 | 6 2 | 3+4 4 | 2+2 4 | 5+6 2 |
| | 5 2 | 3+4 4 | 5 2 | 3+4 4 | 2+2 4 | 3+4 2 |
| | 5 2 | 3+4 4 | 5 2 | 3+4 4 | 2+2 4 | 3+4 2 |
| | 6 2 | 3+4 4 | 6 2 | 3+4 4 | 2+2 4 | 5+6 2 |
| | 6 2 | 3+4 4 | 6 2 | 3+4 4 | 2+2 4 | 5+6 2 |
| | 5 2 | 3+4 4 | 5 2 | 3+4 4 | 2+2 4 | 5+6 2 |
| | 6 2 | 3+4 4 | 6 2 | 3+4 4 | 2+2 4 | 3+4 2 |
| | 5 2 | 3+4 4 | 5 2 | 3+4 4 | 2+2 4 | 5+6 2 |
| | 5 2 | 3+4 4 | 5 2 | 3+4 4 | 2+2 4 | 3+4 2 |
| | 6 2 | 3+4 4 | 6 2 | 3+4 4 | 2+2 4 | 3+4 2 |
| | 6 2 | 3+4 4 | 6 2 | 3+4 4 | 2+2 4 | 5+6 2 |
| | 5 2 | 3+4 4 | 5 2 | 3+4 4 | 2+2 4 | 3+4 2 |
| | 6 2 | 3+4 4 | 6 2 | 3+4 4 | 2+2 4 | 5+6 2 |
| | 5 2 | 3+4 4 | 5 2 | 3+4 4 | 2+2 4 | 3+4 2 |
| | 5 2 | 3+4 4 | 5 2 | 3+4 4 | 2+2 4 | 3+4 2 |
| | 6 2 | 3+4 4 | 6 2 | 3+4 4 | 2+2 4 | 5+6 2 |
| | 6 2 | 3+4 4 | 6 2 | 3+4 4 | 2+2 4 | 5+6 2 |
| | 5 2 | 3+4 4 | 5 2 | 3+4 4 | 2+2 4 | 5+6 2 |
| | 6 2 | 3+4 4 | 6 2 | 3+4 4 | 2+2 4 | 3+4 2 |
| | 5 2 | 3+4 4 | 5 2 | 3+4 4 | 2+2 4 | 5+6 2 |
| | 5 2 | 3+4 4 | 6 2 | 3+4 4 | 2+2 4 | 5+6 2 |
| | 6 2 | 3+4 4 | 5 2 | 3+4 4 | 2+2 4 | 3+4 2 |
| | 6 2 | 3+4 4 | 6 2 | 3+4 4 | 2+2 4 | 5+6 2 |
| | 5 2 | 3+4 4 | 5 2 | 3+4 4 | 2+2 4 | 3+4 2 |
| | 6 2 | 3+4 4 | 6 2 | 3+4 4 | 2+2 4 | 3+4 2 |
| | 5 2 | 3+4 4 | 6 2 | 3+4 4 | 2+2 4 | 5+6 2 |

Continued on next column

**TABLE XXXVIII– continued**

| TQC | ICSD | NoE | SF | | | |
|---|---|---|---|---|---|---|
| | 5 2 | 3+4 4 | 5 2 | 3+4 4 | 2+2 4 | 3+4 2 |
| | 6 2 | 3+4 4 | 6 2 | 3+4 4 | 2+2 4 | 5+6 2 |
| | 6 2 | 3+4 4 | 6 2 | 3+4 4 | 2+2 4 | 3+4 2 |
| | 5 2 | 3+4 4 | 5 2 | 3+4 4 | 2+2 4 | 5+6 2 |
| | 6 2 | 3+4 4 | 5 2 | 3+4 4 | 2+2 4 | 3+4 2 |
| | 5 2 | 3+4 4 | 5 2 | 3+4 4 | 2+2 4 | 5+6 2 |
| | 5 2 | 3+4 4 | 6 2 | 3+4 4 | 2+2 4 | 5+6 2 |
| | 6 2 | 3+4 4 | 5 2 | 3+4 4 | 2+2 4 | 3+4 2 |
| | 5 2 | 3+4 4 | 5 2 | 3+4 4 | 2+2 4 | 3+4 2 |
| | 6 2 | 3+4 4 | 6 2 | 3+4 4 | 2+2 4 | 3+4 2 |
| | 6 2 | 3+4 4 | 6 2 | 3+4 4 | 2+2 4 | 5+6 2 |
| | 6 2 | 3+4 4 | 6 2 | 3+4 4 | 2+2 4 | 5+6 2 |
| | 6 2 | 3+4 4 | 5 2 | 3+4 4 | 2+2 4 | 5+6 2 |
| | 6 2 | 3+4 4 | 6 2 | 3+4 4 | 2+2 4 | 3+4 2 |
| | 5 2 | 3+4 4 | 6 2 | 3+4 4 | 2+2 4 | 3+4 2 |
| | 5 2 | | 5 2 | | | 3+4 2 |
| | 5 2 | | 6 2 | | | 5+6 2 |
| | 5 2 | | 5 2 | | | 5+6 2 |
| | 6 2 | | 5 2 | | | 5+6 2 |
| | 6 2 | | 5 2 | | | 3+4 2 |
| | 6 2 | | 5 2 | | | 3+4 2 |
| | 5 2 | | 6 2 | | | 5+6 2 |
| | 6 2 | | 5 2 | | | 3+4 2 |
| | 6 2 | | 5 2 | | | 5+6 2 |
| | 5 2 | | 5 2 | | | 5+6 2 |
| | 5 2 | | 6 2 | | | 5+6 2 |
| | 6 2 | | 5 2 | | | 5+6 2 |
| | 5 2 | | 5 2 | | | 3+4 2 |
| | 5 2 | | 6 2 | | | 3+4 2 |
| | 5 2 | | 5 2 | | | 5+6 2 |
| | 5 2 | | 6 2 | | | 3+4 2 |
| | 6 2 | | 6 2 | | | 3+4 2 |
| | 5 2 | | 6 2 | | | 5+6 2 |
| | 6 2 | | 6 2 | | | 3+4 2 |
| | 5 2 | | 5 2 | | | 3+4 2 |
| | 6 2 | | 6 2 | | | 5+6 2 |
| | 6 2 | | 5 2 | | | 5+6 2 |
| | 5 2 | | 6 2 | | | 3+4 2 |
| | 5 2 | | 5 2 | | | 5+6 2 |
| | 6 2 | | 6 2 | | | 3+4 2 |
| | 5 2 | | 6 2 | | | 3+4 2 |
| | 5 2 | | 5 2 | | | 3+4 2 |
| | 6 2 | | 5 2 | | | 5+6 2 |
| | 6 2 | | 6 2 | | | 3+4 2 |
| | 6 2 | | 6 2 | | | 5+6 2 |
| | 5 2 | | 5 2 | | | 3+4 2 |
| | 5 2 | | 5 2 | | | 5+6 2 |
| | 6 2 | | 6 2 | | | 3+4 2 |
| | 5 2 | | 6 2 | | | 5+6 2 |
| | 6 2 | | 6 2 | | | 5+6 2 |
| | 5 2 | | 5 2 | | | 3+4 2 |
| | 6 2 | | 6 2 | | | 5+6 2 |
| | 5 2 | | 5 2 | | | 3+4 2 |
| | 5 2 | | 5 2 | | | 5+6 2 |
| | 6 2 | | 5 2 | | | 3+4 2 |
| | 6 2 | | 5 2 | | | 3+4 2 |
| | 5 2 | | 6 2 | | | 3+4 2 |
| | 5 2 | | 6 2 | | | 5+6 2 |
| | 6 2 | | 5 2 | | | 5+6 2 |
| | 6 2 | | 6 2 | | | 3+4 2 |
| | 5 2 | | 6 2 | | | 5+6 2 |
| | 5 2 | | 6 2 | | | 3+4 2 |
| | 5 2 | | 6 2 | | | 3+4 2 |
| 63.2.104 | 43662 | | 48 | | B1Re3 | |
| | -GM | -T | -Y | -Z | -R | -S |
| | 5 2 | 3+4 4 | 6 2 | 3+4 4 | 2+2 4 | 5+6 2 |
| | 6 2 | 3+4 4 | 5 2 | 3+4 4 | 2+2 4 | 3+4 2 |
| | 5 2 | 3+4 4 | 5 2 | 3+4 4 | 2+2 4 | 3+4 2 |
| | 6 2 | 3+4 4 | 6 2 | 3+4 4 | 2+2 4 | 5+6 2 |
| | 5 2 | 3+4 4 | 6 2 | 3+4 4 | 2+2 4 | 3+4 2 |
| | 6 2 | 3+4 4 | 5 2 | 3+4 4 | 2+2 4 | 5+6 2 |
| | 6 2 | 3+4 4 | 5 2 | 3+4 4 | 2+2 4 | 5+6 2 |
| | 5 2 | 3+4 4 | 5 2 | 3+4 4 | 2+2 4 | 3+4 2 |
| | 6 2 | 3+4 4 | 6 2 | 3+4 4 | 2+2 4 | 5+6 2 |
| | 6 2 | 3+4 4 | 5 2 | 3+4 4 | 2+2 4 | 5+6 2 |
| | 6 2 | 3+4 4 | 6 2 | 3+4 4 | 2+2 4 | 3+4 2 |
| | 5 2 | 3+4 4 | 6 2 | 3+4 4 | 2+2 4 | 3+4 2 |
| | 5 2 | | 5 2 | | | 5+6 2 |
| | 5 2 | | 5 2 | | | 5+6 2 |
| | 5 2 | | 6 2 | | | 3+4 2 |
| | 5 2 | | 6 2 | | | 3+4 2 |
| | 6 2 | | 6 2 | | | 5+6 2 |
| | 5 2 | | 5 2 | | | 5+6 2 |
| | 6 2 | | 5 2 | | | 3+4 2 |
| | 6 2 | | 6 2 | | | 3+4 2 |
| | 5 2 | | 6 2 | | | 3+4 2 |
| | 5 2 | | 5 2 | | | 3+4 2 |
| | 6 2 | | 6 2 | | | 5+6 2 |
| | 5 2 | | 5 2 | | | 5+6 2 |
| 63.2.105 | 107505 | | 36 | | Au1In2Na1 | |
| | -GM | -T | -Y | -Z | -R | -S |
| | 5 2 | 3+4 4 | 6 2 | 3+4 4 | 2+2 4 | 5+6 2 |
| | 6 2 | 3+4 4 | 5 2 | 3+4 4 | 2+2 4 | 3+4 2 |
| | 5 2 | 3+4 4 | 5 2 | 3+4 4 | 2+2 4 | 5+6 2 |
| | 5 2 | 3+4 4 | 6 2 | 3+4 4 | 2+2 4 | 3+4 2 |
| | 6 2 | 3+4 4 | 6 2 | 3+4 4 | 2+2 4 | 3+4 2 |
| | 5 2 | 3+4 4 | 5 2 | 3+4 4 | 2+2 4 | 5+6 2 |
| | 6 2 | 3+4 4 | 6 2 | 3+4 4 | 2+2 4 | 5+6 2 |
| | 5 2 | 3+4 4 | 5 2 | 3+4 4 | 2+2 4 | 3+4 2 |
| | 6 2 | 3+4 4 | 6 2 | 3+4 4 | 2+2 4 | 5+6 2 |
| | 5 2 | | 5 2 | | | 3+4 2 |
| | 6 2 | | 6 2 | | | 3+4 2 |
| | 5 2 | | 5 2 | | | 5+6 2 |
| | 6 2 | | 6 2 | | | 5+6 2 |
| | 6 2 | | 5 2 | | | 3+4 2 |
| | 5 2 | | 5 2 | | | 5+6 2 |
| | 5 2 | | 6 2 | | | 3+4 2 |
| | 5 2 | | 6 2 | | | 5+6 2 |
| | 6 2 | | 5 2 | | | 3+4 2 |
| 63.2.106 | 411390 | | 68 | | Au1Eu1In2 | |
| | -GM | -T | -Y | -Z | -R | -S |
| | 5 2 | 3+4 4 | 5 2 | 3+4 4 | 2+2 4 | 3+4 2 |
| | 6 2 | 3+4 4 | 6 2 | 3+4 4 | 2+2 4 | 5+6 2 |
| | 5 2 | 3+4 4 | 5 2 | 3+4 4 | 2+2 4 | 3+4 2 |
| | 6 2 | 3+4 4 | 6 2 | 3+4 4 | 2+2 4 | 5+6 2 |
| | 5 2 | 3+4 4 | 5 2 | 3+4 4 | 2+2 4 | 3+4 2 |
| | 5 2 | 3+4 4 | 6 2 | 3+4 4 | 2+2 4 | 5+6 2 |
| | 6 2 | 3+4 4 | 5 2 | 3+4 4 | 2+2 4 | 3+4 2 |
| | 6 2 | 3+4 4 | 6 2 | 3+4 4 | 2+2 4 | 5+6 2 |
| | 5 2 | 3+4 4 | 6 2 | 3+4 4 | 2+2 4 | 5+6 2 |
| | 6 2 | 3+4 4 | 5 2 | 3+4 4 | 2+2 4 | 3+4 2 |
| | 5 2 | 3+4 4 | 5 2 | 3+4 4 | 2+2 4 | 5+6 2 |

Continued on next column

**TABLE XXXVIII– continued**

| TQC | ICSD | | NoE | | SF | |
|---|---|---|---|---|---|---|
| | 5 2 | 3+4 4 | 6 2 | 3+4 4 | 2+2 4 | 3+4 2 |
| | 6 2 | 3+4 4 | 6 2 | 3+4 4 | 2+2 4 | 3+4 2 |
| | 5 2 | 3+4 4 | 5 2 | 3+4 4 | 2+2 4 | 5+6 2 |
| | 6 2 | 3+4 4 | 6 2 | 3+4 4 | 2+2 4 | 5+6 2 |
| | 5 2 | 3+4 4 | 5 2 | 3+4 4 | 2+2 4 | 3+4 2 |
| | 6 2 | 3+4 4 | 6 2 | 3+4 4 | 2+2 4 | 5+6 2 |
| | 5 2 | | 5 2 | | | 3+4 2 |
| | 6 2 | | 5 2 | | | 5+6 2 |
| | 5 2 | | 6 2 | | | 3+4 2 |
| | 6 2 | | 6 2 | | | 5+6 2 |
| | 5 2 | | 5 2 | | | 3+4 2 |
| | 6 2 | | 5 2 | | | 5+6 2 |
| | 5 2 | | 6 2 | | | 3+4 2 |
| | 5 2 | | 6 2 | | | 3+4 2 |
| | 6 2 | | 5 2 | | | 5+6 2 |
| | 5 2 | | 6 2 | | | 5+6 2 |
| | 6 2 | | 5 2 | | | 3+4 2 |
| | 6 2 | | 6 2 | | | 5+6 2 |
| | 5 2 | | 5 2 | | | 3+4 2 |
| | 5 2 | | 6 2 | | | 3+4 2 |
| | 5 2 | | 5 2 | | | 5+6 2 |
| | 6 2 | | 5 2 | | | 5+6 2 |
| | 6 2 | | 6 2 | | | 5+6 2 |
| 63.2.107 | 107515 | | 108 | | Au1Co1Na4O5 | |
| | -GM | -T | -Y | -Z | -R | -S |
| | 5 2 | 3+4 4 | 5 2 | 3+4 4 | 2+2 4 | 3+4 2 |
| | 5 2 | 3+4 4 | 5 2 | 3+4 4 | 2+2 4 | 3+4 2 |
| | 6 2 | 3+4 4 | 6 2 | 3+4 4 | 2+2 4 | 5+6 2 |
| | 5 2 | 3+4 4 | 5 2 | 3+4 4 | 2+2 4 | 3+4 2 |
| | 5 2 | 3+4 4 | 5 2 | 3+4 4 | 2+2 4 | 3+4 2 |
| | 6 2 | 3+4 4 | 6 2 | 3+4 4 | 2+2 4 | 5+6 2 |
| | 6 2 | 3+4 4 | 6 2 | 3+4 4 | 2+2 4 | 5+6 2 |
| | 6 2 | 3+4 4 | 6 2 | 3+4 4 | 2+2 4 | 5+6 2 |
| | 5 2 | 3+4 4 | 5 2 | 3+4 4 | 2+2 4 | 3+4 2 |
| | 6 2 | 3+4 4 | 6 2 | 3+4 4 | 2+2 4 | 5+6 2 |
| | 5 2 | 3+4 4 | 5 2 | 3+4 4 | 2+2 4 | 3+4 2 |
| | 5 2 | 3+4 4 | 5 2 | 3+4 4 | 2+2 4 | 3+4 2 |
| | 5 2 | 3+4 4 | 5 2 | 3+4 4 | 2+2 4 | 3+4 2 |
| | 5 2 | 3+4 4 | 5 2 | 3+4 4 | 2+2 4 | 3+4 2 |
| | 5 2 | 3+4 4 | 5 2 | 3+4 4 | 2+2 4 | 3+4 2 |
| | 5 2 | 3+4 4 | 5 2 | 3+4 4 | 2+2 4 | 3+4 2 |
| | 5 2 | 3+4 4 | 5 2 | 3+4 4 | 2+2 4 | 3+4 2 |
| | 5 2 | 3+4 4 | 5 2 | 3+4 4 | 2+2 4 | 3+4 2 |
| | 5 2 | 3+4 4 | 5 2 | 3+4 4 | 2+2 4 | 3+4 2 |
| | 5 2 | 3+4 4 | 5 2 | 3+4 4 | 2+2 4 | 3+4 2 |
| | 6 2 | 3+4 4 | 6 2 | 3+4 4 | 2+2 4 | 5+6 2 |
| | 6 2 | 3+4 4 | 6 2 | 3+4 4 | 2+2 4 | 5+6 2 |
| | 6 2 | 3+4 4 | 6 2 | 3+4 4 | 2+2 4 | 5+6 2 |
| | 6 2 | 3+4 4 | 6 2 | 3+4 4 | 2+2 4 | 5+6 2 |
| | 6 2 | 3+4 4 | 6 2 | 3+4 4 | 2+2 4 | 5+6 2 |
| | 5 2 | 3+4 4 | 6 2 | 3+4 4 | 2+2 4 | 3+4 2 |
| | 5 2 | 3+4 4 | 5 2 | 3+4 4 | 2+2 4 | 5+6 2 |
| | 6 2 | | 5 2 | | | 3+4 2 |
| | 6 2 | | 6 2 | | | 5+6 2 |
| | 6 2 | | 6 2 | | | 5+6 2 |
| | 5 2 | | 6 2 | | | 3+4 2 |
| | 5 2 | | 5 2 | | | 5+6 2 |
| | 6 2 | | 5 2 | | | 3+4 2 |
| | 5 2 | | 5 2 | | | 3+4 2 |
| | 5 2 | | 6 2 | | | 5+6 2 |
| | 5 2 | | 5 2 | | | 3+4 2 |

Continued on next column

**TABLE XXXVIII– continued**

| TQC | ICSD | | NoE | | SF | |
|---|---|---|---|---|---|---|
| | 6 2 | | 6 2 | | | 5+6 2 |
| | 6 2 | | 6 2 | | | 3+4 2 |
| | 6 2 | | 5 2 | | | 5+6 2 |
| | 5 2 | | 6 2 | | | 3+4 2 |
| | 6 2 | | 5 2 | | | 5+6 2 |
| | 6 2 | | 6 2 | | | 3+4 2 |
| | 6 2 | | 6 2 | | | 5+6 2 |
| | 5 2 | | 5 2 | | | 5+6 2 |
| | 5 2 | | 5 2 | | | 3+4 2 |
| | 5 2 | | 5 2 | | | 3+4 2 |
| | 5 2 | | 5 2 | | | 3+4 2 |
| | 5 2 | | 5 2 | | | 3+4 2 |
| | 5 2 | | 5 2 | | | 3+4 2 |
| | 6 2 | | 6 2 | | | 5+6 2 |
| | 5 2 | | 6 2 | | | 5+6 2 |
| | 6 2 | | 6 2 | | | 5+6 2 |
| | 5 2 | | 5 2 | | | 3+4 2 |
| | 6 2 | | 6 2 | | | 3+4 2 |
| 63.2.108 | 290262 | | 196 | | As4Cd5Rb2 | |
| | -GM | -T | -Y | -Z | -R | -S |
| | 5 2 | 3+4 4 | 6 2 | 3+4 4 | 2+2 4 | 3+4 2 |
| | 5 2 | 3+4 4 | 6 2 | 3+4 4 | 2+2 4 | 3+4 2 |
| | 6 2 | 3+4 4 | 5 2 | 3+4 4 | 2+2 4 | 5+6 2 |
| | 6 2 | 3+4 4 | 5 2 | 3+4 4 | 2+2 4 | 5+6 2 |
| | 5 2 | 3+4 4 | 6 2 | 3+4 4 | 2+2 4 | 5+6 2 |
| | 5 2 | 3+4 4 | 6 2 | 3+4 4 | 2+2 4 | 3+4 2 |
| | 6 2 | 3+4 4 | 5 2 | 3+4 4 | 2+2 4 | 5+6 2 |
| | 6 2 | 3+4 4 | 5 2 | 3+4 4 | 2+2 4 | 3+4 2 |
| | 6 2 | 3+4 4 | 6 2 | 3+4 4 | 2+2 4 | 5+6 2 |
| | 5 2 | 3+4 4 | 5 2 | 3+4 4 | 2+2 4 | 5+6 2 |
| | 5 2 | 3+4 4 | 5 2 | 3+4 4 | 2+2 4 | 3+4 2 |
| | 5 2 | 3+4 4 | 6 2 | 3+4 4 | 2+2 4 | 3+4 2 |
| | 6 2 | 3+4 4 | 6 2 | 3+4 4 | 2+2 4 | 3+4 2 |
| | 5 2 | 3+4 4 | 6 2 | 3+4 4 | 2+2 4 | 3+4 2 |
| | 5 2 | 3+4 4 | 5 2 | 3+4 4 | 2+2 4 | 5+6 2 |
| | 6 2 | 3+4 4 | 5 2 | 3+4 4 | 2+2 4 | 5+6 2 |
| | 6 2 | 3+4 4 | 5 2 | 3+4 4 | 2+2 4 | 5+6 2 |
| | 6 2 | 3+4 4 | 6 2 | 3+4 4 | 2+2 4 | 3+4 2 |
| | 6 2 | 3+4 4 | 6 2 | 3+4 4 | 2+2 4 | 3+4 2 |
| | 5 2 | 3+4 4 | 6 2 | 3+4 4 | 2+2 4 | 5+6 2 |
| | 5 2 | 3+4 4 | 5 2 | 3+4 4 | 2+2 4 | 5+6 2 |
| | 6 2 | 3+4 4 | 5 2 | 3+4 4 | 2+2 4 | 3+4 2 |
| | 6 2 | 3+4 4 | 6 2 | 3+4 4 | 2+2 4 | 5+6 2 |
| | 5 2 | 3+4 4 | 5 2 | 3+4 4 | 2+2 4 | 3+4 2 |
| | 5 2 | 3+4 4 | 6 2 | 3+4 4 | 2+2 4 | 5+6 2 |
| | 5 2 | 3+4 4 | 5 2 | 3+4 4 | 2+2 4 | 3+4 2 |
| | 6 2 | 3+4 4 | 5 2 | 3+4 4 | 2+2 4 | 5+6 2 |
| | 6 2 | 3+4 4 | 6 2 | 3+4 4 | 2+2 4 | 3+4 2 |
| | 6 2 | 3+4 4 | 5 2 | 3+4 4 | 2+2 4 | 5+6 2 |
| | 5 2 | 3+4 4 | 6 2 | 3+4 4 | 2+2 4 | 3+4 2 |
| | 6 2 | 3+4 4 | 5 2 | 3+4 4 | 2+2 4 | 5+6 2 |
| | 5 2 | 3+4 4 | 6 2 | 3+4 4 | 2+2 4 | 3+4 2 |
| | 5 2 | 3+4 4 | 5 2 | 3+4 4 | 2+2 4 | 5+6 2 |
| | 6 2 | 3+4 4 | 6 2 | 3+4 4 | 2+2 4 | 3+4 2 |
| | 6 2 | 3+4 4 | 5 2 | 3+4 4 | 2+2 4 | 3+4 2 |
| | 5 2 | 3+4 4 | 5 2 | 3+4 4 | 2+2 4 | 5+6 2 |
| | 5 2 | 3+4 4 | 6 2 | 3+4 4 | 2+2 4 | 5+6 2 |
| | 6 2 | 3+4 4 | 6 2 | 3+4 4 | 2+2 4 | 3+4 2 |
| | 5 2 | 3+4 4 | 5 2 | 3+4 4 | 2+2 4 | 5+6 2 |
| | 6 2 | 3+4 4 | 6 2 | 3+4 4 | 2+2 4 | 3+4 2 |
| | 6 2 | 3+4 4 | 6 2 | 3+4 4 | 2+2 4 | 5+6 2 |

Continued on next column

**TABLE XXXVIII– continued**

| TQC | ICSD | NoE | SF | | | |
|---|---|---|---|---|---|---|
| | 5 2 | 3+4 4 | 5 2 | 3+4 4 | 2+2 4 | 3+4 2 |
| | 5 2 | 3+4 4 | 6 2 | 3+4 4 | 2+2 4 | 5+6 2 |
| | 6 2 | 3+4 4 | 6 2 | 3+4 4 | 2+2 4 | 3+4 2 |
| | 5 2 | 3+4 4 | 5 2 | 3+4 4 | 2+2 4 | 3+4 2 |
| | 6 2 | 3+4 4 | 5 2 | 3+4 4 | 2+2 4 | 5+6 2 |
| | 6 2 | 3+4 4 | 5 2 | 3+4 4 | 2+2 4 | 3+4 2 |
| | 5 2 | 3+4 4 | 6 2 | 3+4 4 | 2+2 4 | 5+6 2 |
| | 6 2 | 3+4 4 | 5 2 | 3+4 4 | 2+2 4 | 5+6 2 |
| | 5 2 | | 6 2 | | | 3+4 2 |
| | 5 2 | | 5 2 | | | 5+6 2 |
| | 5 2 | | 6 2 | | | 3+4 2 |
| | 6 2 | | 5 2 | | | 5+6 2 |
| | 5 2 | | 5 2 | | | 3+4 2 |
| | 5 2 | | 6 2 | | | 3+4 2 |
| | 6 2 | | 6 2 | | | 5+6 2 |
| | 6 2 | | 5 2 | | | 3+4 2 |
| | 6 2 | | 5 2 | | | 5+6 2 |
| | 5 2 | | 6 2 | | | 5+6 2 |
| | 6 2 | | 6 2 | | | 3+4 2 |
| | 5 2 | | 5 2 | | | 5+6 2 |
| | 6 2 | | 6 2 | | | 3+4 2 |
| | 5 2 | | 5 2 | | | 5+6 2 |
| | 6 2 | | 6 2 | | | 3+4 2 |
| | 5 2 | | 5 2 | | | 5+6 2 |
| | 6 2 | | 6 2 | | | 3+4 2 |
| | 5 2 | | 5 2 | | | 5+6 2 |
| | 6 2 | | 6 2 | | | 3+4 2 |
| | 5 2 | | 6 2 | | | 3+4 2 |
| | 6 2 | | 5 2 | | | 5+6 2 |
| | 5 2 | | 6 2 | | | 5+6 2 |
| | 6 2 | | 5 2 | | | 3+4 2 |
| | 5 2 | | 5 2 | | | 5+6 2 |
| | 6 2 | | 6 2 | | | 3+4 2 |
| | 5 2 | | 5 2 | | | 3+4 2 |
| | 6 2 | | 6 2 | | | 5+6 2 |
| | 5 2 | | 5 2 | | | 5+6 2 |
| | 6 2 | | 6 2 | | | 3+4 2 |
| | 6 2 | | 6 2 | | | 5+6 2 |
| | 5 2 | | 5 2 | | | 3+4 2 |
| | 6 2 | | 5 2 | | | 5+6 2 |
| | 5 2 | | 6 2 | | | 3+4 2 |
| | 6 2 | | 5 2 | | | 5+6 2 |
| | 5 2 | | 6 2 | | | 3+4 2 |
| | 6 2 | | 5 2 | | | 3+4 2 |
| | 5 2 | | 6 2 | | | 5+6 2 |
| | 5 2 | | 5 2 | | | 5+6 2 |
| | 6 2 | | 6 2 | | | 3+4 2 |
| | 5 2 | | 5 2 | | | 5+6 2 |
| | 5 2 | | 5 2 | | | 5+6 2 |
| | 6 2 | | 6 2 | | | 3+4 2 |
| | 5 2 | | 6 2 | | | 5+6 2 |
| | 6 2 | | 6 2 | | | 3+4 2 |
| | 5 2 | | 5 2 | | | 5+6 2 |
| | 5 2 | | 5 2 | | | 3+4 2 |
| | 6 2 | | 5 2 | | | 5+6 2 |
| | 5 2 | | 6 2 | | | 3+4 2 |
| | 6 2 | | 6 2 | | | 3+4 2 |
| 63.2.109 | 15032 | 268 | As3Nb4 | | | |
| | -GM | -T | -Y | -Z | -R | -S |
| | 5 2 | 3+4 4 | 5 2 | 3+4 4 | 2+2 4 | 3+4 2 |
| | 6 2 | 3+4 4 | 6 2 | 3+4 4 | 2+2 4 | 5+6 2 |

Continued on next column

**TABLE XXXVIII– continued**

| TQC | ICSD | NoE | SF | | | |
|---|---|---|---|---|---|---|
| | 5 2 | 3+4 4 | 6 2 | 3+4 4 | 2+2 4 | 3+4 2 |
| | 6 2 | 3+4 4 | 5 2 | 3+4 4 | 2+2 4 | 5+6 2 |
| | 5 2 | 3+4 4 | 6 2 | 3+4 4 | 2+2 4 | 3+4 2 |
| | 5 2 | 3+4 4 | 6 2 | 3+4 4 | 2+2 4 | 3+4 2 |
| | 6 2 | 3+4 4 | 5 2 | 3+4 4 | 2+2 4 | 5+6 2 |
| | 6 2 | 3+4 4 | 5 2 | 3+4 4 | 2+2 4 | 5+6 2 |
| | 5 2 | 3+4 4 | 5 2 | 3+4 4 | 2+2 4 | 3+4 2 |
| | 6 2 | 3+4 4 | 6 2 | 3+4 4 | 2+2 4 | 5+6 2 |
| | 5 2 | 3+4 4 | 5 2 | 3+4 4 | 2+2 4 | 3+4 2 |
| | 5 2 | 3+4 4 | 5 2 | 3+4 4 | 2+2 4 | 3+4 2 |
| | 6 2 | 3+4 4 | 6 2 | 3+4 4 | 2+2 4 | 5+6 2 |
| | 6 2 | 3+4 4 | 6 2 | 3+4 4 | 2+2 4 | 5+6 2 |
| | 5 2 | 3+4 4 | 5 2 | 3+4 4 | 2+2 4 | 3+4 2 |
| | 5 2 | 3+4 4 | 5 2 | 3+4 4 | 2+2 4 | 3+4 2 |
| | 6 2 | 3+4 4 | 6 2 | 3+4 4 | 2+2 4 | 3+4 2 |
| | 5 2 | 3+4 4 | 5 2 | 3+4 4 | 2+2 4 | 5+6 2 |
| | 5 2 | 3+4 4 | 6 2 | 3+4 4 | 2+2 4 | 3+4 2 |
| | 6 2 | 3+4 4 | 5 2 | 3+4 4 | 2+2 4 | 5+6 2 |
| | 5 2 | 3+4 4 | 6 2 | 3+4 4 | 2+2 4 | 5+6 2 |
| | 5 2 | 3+4 4 | 6 2 | 3+4 4 | 2+2 4 | 5+6 2 |
| | 6 2 | 3+4 4 | 5 2 | 3+4 4 | 2+2 4 | 3+4 2 |
| | 6 2 | 3+4 4 | 5 2 | 3+4 4 | 2+2 4 | 3+4 2 |
| | 6 2 | 3+4 4 | 6 2 | 3+4 4 | 2+2 4 | 3+4 2 |
| | 5 2 | 3+4 4 | 5 2 | 3+4 4 | 2+2 4 | 5+6 2 |
| | 6 2 | 3+4 4 | 6 2 | 3+4 4 | 2+2 4 | 5+6 2 |
| | 6 2 | 3+4 4 | 6 2 | 3+4 4 | 2+2 4 | 5+6 2 |
| | 5 2 | 3+4 4 | 5 2 | 3+4 4 | 2+2 4 | 3+4 2 |
| | 5 2 | 3+4 4 | 5 2 | 3+4 4 | 2+2 4 | 3+4 2 |
| | 6 2 | 3+4 4 | 6 2 | 3+4 4 | 2+2 4 | 5+6 2 |
| | 6 2 | 3+4 4 | 6 2 | 3+4 4 | 2+2 4 | 5+6 2 |
| | 6 2 | 3+4 4 | 5 2 | 3+4 4 | 2+2 4 | 5+6 2 |
| | 5 2 | 3+4 4 | 6 2 | 3+4 4 | 2+2 4 | 5+6 2 |
| | 5 2 | 3+4 4 | 5 2 | 3+4 4 | 2+2 4 | 3+4 2 |
| | 6 2 | 3+4 4 | 6 2 | 3+4 4 | 2+2 4 | 5+6 2 |
| | 5 2 | 3+4 4 | 6 2 | 3+4 4 | 2+2 4 | 3+4 2 |
| | 5 2 | 3+4 4 | 6 2 | 3+4 4 | 2+2 4 | 3+4 2 |
| | 6 2 | 3+4 4 | 6 2 | 3+4 4 | 2+2 4 | 5+6 2 |
| | 6 2 | 3+4 4 | 5 2 | 3+4 4 | 2+2 4 | 3+4 2 |
| | 5 2 | 3+4 4 | 6 2 | 3+4 4 | 2+2 4 | 5+6 2 |
| | 5 2 | 3+4 4 | 5 2 | 3+4 4 | 2+2 4 | 3+4 2 |
| | 6 2 | 3+4 4 | 6 2 | 3+4 4 | 2+2 4 | 5+6 2 |
| | 6 2 | 3+4 4 | 6 2 | 3+4 4 | 2+2 4 | 3+4 2 |
| | 6 2 | 3+4 4 | 5 2 | 3+4 4 | 2+2 4 | 5+6 2 |
| | 5 2 | 3+4 4 | 6 2 | 3+4 4 | 2+2 4 | 5+6 2 |
| | 5 2 | 3+4 4 | 5 2 | 3+4 4 | 2+2 4 | 3+4 2 |
| | 6 2 | 3+4 4 | 6 2 | 3+4 4 | 2+2 4 | 3+4 2 |
| | 6 2 | 3+4 4 | 5 2 | 3+4 4 | 2+2 4 | 5+6 2 |
| | 5 2 | 3+4 4 | 6 2 | 3+4 4 | 2+2 4 | 5+6 2 |
| | 6 2 | 3+4 4 | 5 2 | 3+4 4 | 2+2 4 | 3+4 2 |
| | 6 2 | 3+4 4 | 6 2 | 3+4 4 | 2+2 4 | 5+6 2 |
| | 5 2 | 3+4 4 | 5 2 | 3+4 4 | 2+2 4 | 3+4 2 |
| | 6 2 | 3+4 4 | 5 2 | 3+4 4 | 2+2 4 | 3+4 2 |
| | 5 2 | 3+4 4 | 6 2 | 3+4 4 | 2+2 4 | 5+6 2 |
| | 5 2 | 3+4 4 | 5 2 | 3+4 4 | 2+2 4 | 5+6 2 |
| | 6 2 | 3+4 4 | 5 2 | 3+4 4 | 2+2 4 | 3+4 2 |
| | 6 2 | 3+4 4 | 6 2 | 3+4 4 | 2+2 4 | 5+6 2 |
| | 5 2 | 3+4 4 | 5 2 | 3+4 4 | 2+2 4 | 3+4 2 |
| | 5 2 | 3+4 4 | 5 2 | 3+4 4 | 2+2 4 | 3+4 2 |
| | 6 2 | 3+4 4 | 6 2 | 3+4 4 | 2+2 4 | 5+6 2 |
| | 6 2 | 3+4 4 | 6 2 | 3+4 4 | 2+2 4 | 5+6 2 |
| | 6 2 | 3+4 4 | 6 2 | 3+4 4 | 2+2 4 | 5+6 2 |

Continued on next column

**TABLE XXXVIII– continued**

| TQC | ICSD | | NoE | | SF | |
|---|---|---|---|---|---|---|
| | 6 2 | 3+4 4 | 6 2 | 3+4 4 | 2+2 4 | 5+6 2 |
| | 5 2 | 3+4 4 | 6 2 | 3+4 4 | 2+2 4 | 3+4 2 |
| | 5 2 | 3+4 4 | 5 2 | 3+4 4 | 2+2 4 | 3+4 2 |
| | 6 2 | 3+4 4 | 5 2 | 3+4 4 | 2+2 4 | 5+6 2 |
| | 6 2 | | 5 2 | | | 5+6 2 |
| | 6 2 | | 6 2 | | | 3+4 2 |
| | 5 2 | | 6 2 | | | 5+6 2 |
| | 6 2 | | 6 2 | | | 3+4 2 |
| | 5 2 | | 5 2 | | | 5+6 2 |
| | 5 2 | | 5 2 | | | 3+4 2 |
| | 6 2 | | 6 2 | | | 5+6 2 |
| | 5 2 | | 6 2 | | | 3+4 2 |
| | 6 2 | | 5 2 | | | 5+6 2 |
| | 5 2 | | 5 2 | | | 5+6 2 |
| | 5 2 | | 6 2 | | | 3+4 2 |
| | 6 2 | | 5 2 | | | 5+6 2 |
| | 5 2 | | 5 2 | | | 3+4 2 |
| | 6 2 | | 5 2 | | | 3+4 2 |
| | 5 2 | | 6 2 | | | 3+4 2 |
| | 5 2 | | 5 2 | | | 5+6 2 |
| | 6 2 | | 5 2 | | | 3+4 2 |
| | 6 2 | | 6 2 | | | 5+6 2 |
| | 6 2 | | 6 2 | | | 3+4 2 |
| | 5 2 | | 6 2 | | | 5+6 2 |
| | 6 2 | | 6 2 | | | 3+4 2 |
| | 5 2 | | 6 2 | | | 3+4 2 |
| | 5 2 | | 5 2 | | | 3+4 2 |
| | 6 2 | | 5 2 | | | 5+6 2 |
| | 6 2 | | 6 2 | | | 3+4 2 |
| | 6 2 | | 5 2 | | | 5+6 2 |
| | 6 2 | | 6 2 | | | 3+4 2 |
| | 6 2 | | 5 2 | | | 5+6 2 |
| | 5 2 | | 6 2 | | | 5+6 2 |
| | 6 2 | | 6 2 | | | 5+6 2 |
| | 5 2 | | 5 2 | | | 5+6 2 |
| | 5 2 | | 5 2 | | | 3+4 2 |
| | 5 2 | | 5 2 | | | 3+4 2 |
| | 5 2 | | 5 2 | | | 3+4 2 |
| | 5 2 | | 6 2 | | | 5+6 2 |
| | 5 2 | | 5 2 | | | 5+6 2 |
| | 6 2 | | 5 2 | | | 5+6 2 |
| | 5 2 | | 5 2 | | | 3+4 2 |
| | 6 2 | | 6 2 | | | 3+4 2 |
| | 5 2 | | 5 2 | | | 5+6 2 |
| | 6 2 | | 6 2 | | | 5+6 2 |
| | 5 2 | | 5 2 | | | 3+4 2 |
| | 5 2 | | 5 2 | | | 3+4 2 |
| | 6 2 | | 6 2 | | | 5+6 2 |
| | 6 2 | | 6 2 | | | 5+6 2 |
| | 6 2 | | 5 2 | | | 3+4 2 |
| | 5 2 | | 6 2 | | | 3+4 2 |
| | 5 2 | | 5 2 | | | 3+4 2 |
| | 6 2 | | 6 2 | | | 3+4 2 |
| | 5 2 | | 6 2 | | | 5+6 2 |
| | 6 2 | | 6 2 | | | 5+6 2 |
| | 5 2 | | 5 2 | | | 3+4 2 |
| | 6 2 | | 6 2 | | | 5+6 2 |
| | 6 2 | | 6 2 | | | 3+4 2 |
| | 5 2 | | 5 2 | | | 3+4 2 |
| | 5 2 | | 5 2 | | | 5+6 2 |
| | 5 2 | | 6 2 | | | 3+4 2 |
| | 6 2 | | 6 2 | | | 5+6 2 |
| | 5 2 | | 6 2 | | | 3+4 2 |
| | 5 2 | | 5 2 | | | 5+6 2 |
| | 6 2 | | 5 2 | | | 5+6 2 |
| | 6 2 | | 6 2 | | | 3+4 2 |
| | 6 2 | | 5 2 | | | 3+4 2 |
| | 6 2 | | 6 2 | | | 5+6 2 |
| | 5 2 | | 5 2 | | | 5+6 2 |
| | 5 2 | | 5 2 | | | 5+6 2 |
| | 6 2 | | 5 2 | | | 3+4 2 |
| 63.2.110 | 405110 | | 60 | | As2Ba1Pd1 | |
| | -GM | -T | -Y | -Z | -R | -S |
| | 5 2 | 3+4 4 | 5 2 | 3+4 4 | 2+2 4 | 5+6 2 |
| | 6 2 | 3+4 4 | 6 2 | 3+4 4 | 2+2 4 | 3+4 2 |
| | 5 2 | 3+4 4 | 5 2 | 3+4 4 | 2+2 4 | 3+4 2 |
| | 6 2 | 3+4 4 | 6 2 | 3+4 4 | 2+2 4 | 5+6 2 |
| | 5 2 | 3+4 4 | 6 2 | 3+4 4 | 2+2 4 | 3+4 2 |
| | 5 2 | 3+4 4 | 5 2 | 3+4 4 | 2+2 4 | 5+6 2 |
| | 6 2 | 3+4 4 | 6 2 | 3+4 4 | 2+2 4 | 5+6 2 |
| | 5 2 | 3+4 4 | 5 2 | 3+4 4 | 2+2 4 | 3+4 2 |
| | 6 2 | 3+4 4 | 5 2 | 3+4 4 | 2+2 4 | 3+4 2 |
| | 5 2 | 3+4 4 | 6 2 | 3+4 4 | 2+2 4 | 5+6 2 |
| | 6 2 | 3+4 4 | 5 2 | 3+4 4 | 2+2 4 | 3+4 2 |
| | 6 2 | 3+4 4 | 6 2 | 3+4 4 | 2+2 4 | 5+6 2 |
| | 5 2 | 3+4 4 | 6 2 | 3+4 4 | 2+2 4 | 3+4 2 |
| | 5 2 | 3+4 4 | 6 2 | 3+4 4 | 2+2 4 | 5+6 2 |
| | 6 2 | 3+4 4 | 5 2 | 3+4 4 | 2+2 4 | 5+6 2 |
| | 5 2 | | 6 2 | | | 3+4 2 |
| | 5 2 | | 5 2 | | | 3+4 2 |
| | 5 2 | | 5 2 | | | 5+6 2 |
| | 6 2 | | 6 2 | | | 3+4 2 |
| | 5 2 | | 5 2 | | | 5+6 2 |
| | 6 2 | | 5 2 | | | 3+4 2 |
| | 6 2 | | 6 2 | | | 3+4 2 |
| | 6 2 | | 5 2 | | | 5+6 2 |
| | 6 2 | | 6 2 | | | 3+4 2 |
| | 5 2 | | 6 2 | | | 5+6 2 |
| | 5 2 | | 5 2 | | | 5+6 2 |
| | 6 2 | | 6 2 | | | 3+4 2 |
| | 5 2 | | 6 2 | | | 5+6 2 |
| | 6 2 | | 5 2 | | | 5+6 2 |
| | 5 2 | | 5 2 | | | 3+4 2 |
| 63.2.111 | 153384 | | 48 | | Al2Ba1Si2 | |
| | -GM | -T | -Y | -Z | -R | -S |
| | 5 2 | 3+4 4 | 5 2 | 3+4 4 | 2+2 4 | 3+4 2 |
| | 6 2 | 3+4 4 | 6 2 | 3+4 4 | 2+2 4 | 5+6 2 |
| | 5 2 | 3+4 4 | 5 2 | 3+4 4 | 2+2 4 | 3+4 2 |
| | 6 2 | 3+4 4 | 6 2 | 3+4 4 | 2+2 4 | 5+6 2 |
| | 5 2 | 3+4 4 | 6 2 | 3+4 4 | 2+2 4 | 5+6 2 |
| | 6 2 | 3+4 4 | 5 2 | 3+4 4 | 2+2 4 | 3+4 2 |
| | 5 2 | 3+4 4 | 5 2 | 3+4 4 | 2+2 4 | 3+4 2 |
| | 6 2 | 3+4 4 | 6 2 | 3+4 4 | 2+2 4 | 5+6 2 |
| | 5 2 | 3+4 4 | 6 2 | 3+4 4 | 2+2 4 | 3+4 2 |
| | 5 2 | 3+4 4 | 6 2 | 3+4 4 | 2+2 4 | 5+6 2 |
| | 6 2 | 3+4 4 | 5 2 | 3+4 4 | 2+2 4 | 5+6 2 |
| | 6 2 | 3+4 4 | 5 2 | 3+4 4 | 2+2 4 | 3+4 2 |
| | 5 2 | | 5 2 | | | 5+6 2 |
| | 6 2 | | 5 2 | | | 5+6 2 |
| | 5 2 | | 6 2 | | | 3+4 2 |
| | 5 2 | | 6 2 | | | 3+4 2 |
| | 6 2 | | 6 2 | | | 3+4 2 |

Continued on next column

Continued on next column

**TABLE XXXVIII– continued**

| TQC | ICSD | NoE | SF | | | |
|---|---|---|---|---|---|---|
| | 5 2 | | 6 2 | | | 5+6 2 |
| | 5 2 | | 6 2 | | | 3+4 2 |
| | 5 2 | | 5 2 | | | 3+4 2 |
| | 6 2 | | 5 2 | | | 5+6 2 |
| | 5 2 | | 5 2 | | | 5+6 2 |
| | 6 2 | | 6 2 | | | 5+6 2 |
| | 6 2 | | 5 2 | | | 5+6 2 |
| 63.2.112 | 415372 | 60 | Al1Si1Tm1 | | | |
| | -GM | -T | -Y | -Z | -R | -S |
| | 5 2 | 3+4 4 | 6 2 | 3+4 4 | 2+2 4 | 3+4 2 |
| | 6 2 | 3+4 4 | 5 2 | 3+4 4 | 2+2 4 | 5+6 2 |
| | 5 2 | 3+4 4 | 6 2 | 3+4 4 | 2+2 4 | 5+6 2 |
| | 6 2 | 3+4 4 | 5 2 | 3+4 4 | 2+2 4 | 3+4 2 |
| | 5 2 | 3+4 4 | 6 2 | 3+4 4 | 2+2 4 | 5+6 2 |
| | 5 2 | 3+4 4 | 5 2 | 3+4 4 | 2+2 4 | 3+4 2 |
| | 6 2 | 3+4 4 | 6 2 | 3+4 4 | 2+2 4 | 5+6 2 |
| | 6 2 | 3+4 4 | 5 2 | 3+4 4 | 2+2 4 | 3+4 2 |
| | 5 2 | 3+4 4 | 5 2 | 3+4 4 | 2+2 4 | 5+6 2 |
| | 6 2 | 3+4 4 | 6 2 | 3+4 4 | 2+2 4 | 3+4 2 |
| | 5 2 | 3+4 4 | 5 2 | 3+4 4 | 2+2 4 | 3+4 2 |
| | 5 2 | 3+4 4 | 6 2 | 3+4 4 | 2+2 4 | 5+6 2 |
| | 6 2 | 3+4 4 | 5 2 | 3+4 4 | 2+2 4 | 5+6 2 |
| | 6 2 | 3+4 4 | 6 2 | 3+4 4 | 2+2 4 | 3+4 2 |
| | 6 2 | 3+4 4 | 5 2 | 3+4 4 | 2+2 4 | 5+6 2 |
| | 5 2 | | 6 2 | | | 5+6 2 |
| | 6 2 | | 5 2 | | | 3+4 2 |
| | 5 2 | | 5 2 | | | 3+4 2 |
| | 6 2 | | 6 2 | | | 5+6 2 |
| | 5 2 | | 6 2 | | | 5+6 2 |
| | 6 2 | | 6 2 | | | 3+4 2 |
| | 6 2 | | 5 2 | | | 5+6 2 |
| | 5 2 | | 5 2 | | | 3+4 2 |
| | 5 2 | | 5 2 | | | 5+6 2 |
| | 6 2 | | 6 2 | | | 3+4 2 |
| | 5 2 | | 5 2 | | | 5+6 2 |
| | 6 2 | | 6 2 | | | 3+4 2 |
| | 5 2 | | 5 2 | | | 3+4 2 |
| | 6 2 | | 6 2 | | | 5+6 2 |
| | 6 2 | | 6 2 | | | 5+6 2 |
| 63.2.113 | 54470 | 56 | Al1La1 | | | |
| | -GM | -T | -Y | -Z | -R | -S |
| | 5 2 | 3+4 4 | 6 2 | 3+4 4 | 2+2 4 | 5+6 2 |
| | 5 2 | 3+4 4 | 6 2 | 3+4 4 | 2+2 4 | 3+4 2 |
| | 6 2 | 3+4 4 | 5 2 | 3+4 4 | 2+2 4 | 3+4 2 |
| | 6 2 | 3+4 4 | 5 2 | 3+4 4 | 2+2 4 | 5+6 2 |
| | 5 2 | 3+4 4 | 6 2 | 3+4 4 | 2+2 4 | 3+4 2 |
| | 5 2 | 3+4 4 | 6 2 | 3+4 4 | 2+2 4 | 5+6 2 |
| | 6 2 | 3+4 4 | 5 2 | 3+4 4 | 2+2 4 | 5+6 2 |
| | 6 2 | 3+4 4 | 5 2 | 3+4 4 | 2+2 4 | 3+4 2 |
| | 5 2 | 3+4 4 | 6 2 | 3+4 4 | 2+2 4 | 5+6 2 |
| | 5 2 | 3+4 4 | 6 2 | 3+4 4 | 2+2 4 | 3+4 2 |
| | 6 2 | 3+4 4 | 5 2 | 3+4 4 | 2+2 4 | 3+4 2 |
| | 5 2 | 3+4 4 | 5 2 | 3+4 4 | 2+2 4 | 5+6 2 |
| | 6 2 | 3+4 4 | 6 2 | 3+4 4 | 2+2 4 | 3+4 2 |
| | 5 2 | 3+4 4 | 6 2 | 3+4 4 | 2+2 4 | 5+6 2 |
| | 6 2 | | 5 2 | | | 3+4 2 |
| | 6 2 | | 5 2 | | | 5+6 2 |
| | 5 2 | | 5 2 | | | 3+4 2 |
| | 6 2 | | 6 2 | | | 5+6 2 |
| | 5 2 | | 5 2 | | | 3+4 2 |
| | 5 2 | | 5 2 | | | 3+4 2 |

Continued on next column

**TABLE XXXVIII– continued**

| TQC | ICSD | NoE | SF | | | |
|---|---|---|---|---|---|---|
| | 6 2 | | 6 2 | | | 5+6 2 |
| | 6 2 | | 5 2 | | | 5+6 2 |
| | 6 2 | | 6 2 | | | 5+6 2 |
| | 6 2 | | 5 2 | | | 3+4 2 |
| | 5 2 | | 6 2 | | | 5+6 2 |
| | 6 2 | | 5 2 | | | 5+6 2 |
| | 5 2 | | 6 2 | | | 5+6 2 |
| | 6 2 | | 6 2 | | | 3+4 2 |
| 63.2.114 | 251809 | 24 | Al1B1W1 | | | |
| | -GM | -T | -Y | -Z | -R | -S |
| | 5 2 | 3+4 4 | 5 2 | 3+4 4 | 2+2 4 | 5+6 2 |
| | 6 2 | 3+4 4 | 6 2 | 3+4 4 | 2+2 4 | 3+4 2 |
| | 5 2 | 3+4 4 | 6 2 | 3+4 4 | 2+2 4 | 5+6 2 |
| | 5 2 | 3+4 4 | 5 2 | 3+4 4 | 2+2 4 | 3+4 2 |
| | 6 2 | 3+4 4 | 6 2 | 3+4 4 | 2+2 4 | 5+6 2 |
| | 6 2 | 3+4 4 | 5 2 | 3+4 4 | 2+2 4 | 5+6 2 |
| | 5 2 | | 6 2 | | | 3+4 2 |
| | 5 2 | | 5 2 | | | 5+6 2 |
| | 6 2 | | 5 2 | | | 3+4 2 |
| | 5 2 | | 5 2 | | | 3+4 2 |
| | 5 2 | | 6 2 | | | 3+4 2 |
| | 6 2 | | 6 2 | | | 3+4 2 |
| 63.2.115 | 251808 | 24 | Al1B1Mo1 | | | |
| | -GM | -T | -Y | -Z | -R | -S |
| | 5 2 | 3+4 4 | 5 2 | 3+4 4 | 2+2 4 | 3+4 2 |
| | 5 2 | 3+4 4 | 6 2 | 3+4 4 | 2+2 4 | 5+6 2 |
| | 6 2 | 3+4 4 | 6 2 | 3+4 4 | 2+2 4 | 3+4 2 |
| | 5 2 | 3+4 4 | 5 2 | 3+4 4 | 2+2 4 | 5+6 2 |
| | 6 2 | 3+4 4 | 6 2 | 3+4 4 | 2+2 4 | 3+4 2 |
| | 5 2 | 3+4 4 | 5 2 | 3+4 4 | 2+2 4 | 3+4 2 |
| | 6 2 | | 6 2 | | | 5+6 2 |
| | 5 2 | | 5 2 | | | 3+4 2 |
| | 6 2 | | 5 2 | | | 5+6 2 |
| | 5 2 | | 5 2 | | | 5+6 2 |
| | 6 2 | | 6 2 | | | 5+6 2 |
| | 5 2 | | 6 2 | | | 5+6 2 |
| 63.2.116 | 182114 | 120 | Al10Ru2U1 | | | |
| | -GM | -T | -Y | -Z | -R | -S |
| | 5 2 | 3+4 4 | 5 2 | 3+4 4 | 2+2 4 | 3+4 2 |
| | 6 2 | 3+4 4 | 6 2 | 3+4 4 | 2+2 4 | 5+6 2 |
| | 5 2 | 3+4 4 | 5 2 | 3+4 4 | 2+2 4 | 5+6 2 |
| | 6 2 | 3+4 4 | 6 2 | 3+4 4 | 2+2 4 | 3+4 2 |
| | 5 2 | 3+4 4 | 5 2 | 3+4 4 | 2+2 4 | 5+6 2 |
| | 6 2 | 3+4 4 | 6 2 | 3+4 4 | 2+2 4 | 3+4 2 |
| | 5 2 | 3+4 4 | 5 2 | 3+4 4 | 2+2 4 | 5+6 2 |
| | 6 2 | 3+4 4 | 6 2 | 3+4 4 | 2+2 4 | 3+4 2 |
| | 5 2 | 3+4 4 | 5 2 | 3+4 4 | 2+2 4 | 3+4 2 |
| | 6 2 | 3+4 4 | 6 2 | 3+4 4 | 2+2 4 | 5+6 2 |
| | 5 2 | 3+4 4 | 6 2 | 3+4 4 | 2+2 4 | 3+4 2 |
| | 6 2 | 3+4 4 | 5 2 | 3+4 4 | 2+2 4 | 3+4 2 |
| | 5 2 | 3+4 4 | 5 2 | 3+4 4 | 2+2 4 | 5+6 2 |
| | 5 2 | 3+4 4 | 6 2 | 3+4 4 | 2+2 4 | 5+6 2 |
| | 6 2 | 3+4 4 | 6 2 | 3+4 4 | 2+2 4 | 5+6 2 |
| | 5 2 | 3+4 4 | 6 2 | 3+4 4 | 2+2 4 | 3+4 2 |
| | 6 2 | 3+4 4 | 5 2 | 3+4 4 | 2+2 4 | 3+4 2 |
| | 5 2 | 3+4 4 | 5 2 | 3+4 4 | 2+2 4 | 5+6 2 |
| | 5 2 | 3+4 4 | 5 2 | 3+4 4 | 2+2 4 | 5+6 2 |
| | 6 2 | 3+4 4 | 6 2 | 3+4 4 | 2+2 4 | 3+4 2 |
| | 6 2 | 3+4 4 | 6 2 | 3+4 4 | 2+2 4 | 3+4 2 |
| | 5 2 | 3+4 4 | 6 2 | 3+4 4 | 2+2 4 | 5+6 2 |
| | 6 2 | 3+4 4 | 6 2 | 3+4 4 | 2+2 4 | 3+4 2 |

Continued on next column

**TABLE XXXVIII– continued**

| TQC | ICSD | | NoE | | SF | |
|---|---|---|---|---|---|---|
| | 5 2 | 3+4 4 | 6 2 | 3+4 4 | 2+2 4 | 5+6 2 |
| | 5 2 | 3+4 4 | 6 2 | 3+4 4 | 2+2 4 | 3+4 2 |
| | 5 2 | 3+4 4 | 6 2 | 3+4 4 | 2+2 4 | 5+6 2 |
| | 6 2 | 3+4 4 | 6 2 | 3+4 4 | 2+2 4 | 3+4 2 |
| | 5 2 | 3+4 4 | 6 2 | 3+4 4 | 2+2 4 | 5+6 2 |
| | 5 2 | 3+4 4 | 6 2 | 3+4 4 | 2+2 4 | 5+6 2 |
| | 5 2 | 3+4 4 | 6 2 | 3+4 4 | 2+2 4 | 3+4 2 |
| | 6 2 | | 5 2 | | | 5+6 2 |
| | 5 2 | | 5 2 | | | 3+4 2 |
| | 5 2 | | 6 2 | | | 5+6 2 |
| | 5 2 | | 5 2 | | | 3+4 2 |
| | 6 2 | | 6 2 | | | 5+6 2 |
| | 5 2 | | 5 2 | | | 3+4 2 |
| | 6 2 | | 6 2 | | | 3+4 2 |
| | 5 2 | | 6 2 | | | 5+6 2 |
| | 5 2 | | 5 2 | | | 3+4 2 |
| | 5 2 | | 5 2 | | | 5+6 2 |
| | 6 2 | | 5 2 | | | 5+6 2 |
| | 6 2 | | 6 2 | | | 5+6 2 |
| | 5 2 | | 6 2 | | | 3+4 2 |
| | 5 2 | | 5 2 | | | 3+4 2 |
| | 5 2 | | 6 2 | | | 5+6 2 |
| | 5 2 | | 6 2 | | | 5+6 2 |
| | 5 2 | | 6 2 | | | 3+4 2 |
| | 6 2 | | 5 2 | | | 3+4 2 |
| | 5 2 | | 5 2 | | | 5+6 2 |
| | 5 2 | | 5 2 | | | 3+4 2 |
| | 5 2 | | 6 2 | | | 3+4 2 |
| | 5 2 | | 6 2 | | | 5+6 2 |
| | 5 2 | | 5 2 | | | 3+4 2 |
| | 6 2 | | 6 2 | | | 5+6 2 |
| | 6 2 | | 5 2 | | | 3+4 2 |
| | 5 2 | | 6 2 | | | 3+4 2 |
| | 5 2 | | 5 2 | | | 5+6 2 |
| | 6 2 | | 6 2 | | | 5+6 2 |
| | 6 2 | | 6 2 | | | 3+4 2 |
| | 6 2 | | 6 2 | | | 5+6 2 |
| 63.2.117 | 99179 | | 120 | | Al10Fe2U1 | |
| | -GM | -T | -Y | -Z | -R | -S |
| | 5 2 | 3+4 4 | 5 2 | 3+4 4 | 2+2 4 | 3+4 2 |
| | 6 2 | 3+4 4 | 6 2 | 3+4 4 | 2+2 4 | 5+6 2 |
| | 5 2 | 3+4 4 | 5 2 | 3+4 4 | 2+2 4 | 5+6 2 |
| | 6 2 | 3+4 4 | 6 2 | 3+4 4 | 2+2 4 | 3+4 2 |
| | 5 2 | 3+4 4 | 5 2 | 3+4 4 | 2+2 4 | 5+6 2 |
| | 6 2 | 3+4 4 | 6 2 | 3+4 4 | 2+2 4 | 3+4 2 |
| | 5 2 | 3+4 4 | 5 2 | 3+4 4 | 2+2 4 | 5+6 2 |
| | 6 2 | 3+4 4 | 6 2 | 3+4 4 | 2+2 4 | 3+4 2 |
| | 5 2 | 3+4 4 | 5 2 | 3+4 4 | 2+2 4 | 3+4 2 |
| | 6 2 | 3+4 4 | 6 2 | 3+4 4 | 2+2 4 | 5+6 2 |
| | 5 2 | 3+4 4 | 6 2 | 3+4 4 | 2+2 4 | 3+4 2 |
| | 6 2 | 3+4 4 | 5 2 | 3+4 4 | 2+2 4 | 3+4 2 |
| | 5 2 | 3+4 4 | 5 2 | 3+4 4 | 2+2 4 | 5+6 2 |
| | 5 2 | 3+4 4 | 6 2 | 3+4 4 | 2+2 4 | 5+6 2 |
| | 6 2 | 3+4 4 | 6 2 | 3+4 4 | 2+2 4 | 5+6 2 |
| | 6 2 | 3+4 4 | 6 2 | 3+4 4 | 2+2 4 | 3+4 2 |
| | 5 2 | 3+4 4 | 5 2 | 3+4 4 | 2+2 4 | 3+4 2 |
| | 6 2 | 3+4 4 | 5 2 | 3+4 4 | 2+2 4 | 5+6 2 |
| | 5 2 | 3+4 4 | 5 2 | 3+4 4 | 2+2 4 | 5+6 2 |
| | 6 2 | 3+4 4 | 6 2 | 3+4 4 | 2+2 4 | 3+4 2 |
| | 5 2 | 3+4 4 | 6 2 | 3+4 4 | 2+2 4 | 3+4 2 |
| | 6 2 | 3+4 4 | 6 2 | 3+4 4 | 2+2 4 | 5+6 2 |

Continued on next column

**TABLE XXXVIII– continued**

| TQC | ICSD | | NoE | | SF | |
|---|---|---|---|---|---|---|
| | 5 2 | 3+4 4 | 6 2 | 3+4 4 | 2+2 4 | 3+4 2 |
| | 5 2 | 3+4 4 | 6 2 | 3+4 4 | 2+2 4 | 5+6 2 |
| | 5 2 | 3+4 4 | 6 2 | 3+4 4 | 2+2 4 | 5+6 2 |
| | 5 2 | 3+4 4 | 5 2 | 3+4 4 | 2+2 4 | 3+4 2 |
| | 6 2 | 3+4 4 | 6 2 | 3+4 4 | 2+2 4 | 5+6 2 |
| | 6 2 | 3+4 4 | 5 2 | 3+4 4 | 2+2 4 | 3+4 2 |
| | 5 2 | 3+4 4 | 5 2 | 3+4 4 | 2+2 4 | 5+6 2 |
| | 6 2 | 3+4 4 | 5 2 | 3+4 4 | 2+2 4 | 3+4 2 |
| | 6 2 | | 6 2 | | | 5+6 2 |
| | 5 2 | | 6 2 | | | 3+4 2 |
| | 6 2 | | 5 2 | | | 5+6 2 |
| | 6 2 | | 5 2 | | | 3+4 2 |
| | 5 2 | | 5 2 | | | 3+4 2 |
| | 5 2 | | 6 2 | | | 5+6 2 |
| | 5 2 | | 6 2 | | | 3+4 2 |
| | 5 2 | | 5 2 | | | 5+6 2 |
| | 5 2 | | 5 2 | | | 5+6 2 |
| | 5 2 | | 5 2 | | | 3+4 2 |
| | 6 2 | | 6 2 | | | 5+6 2 |
| | 5 2 | | 6 2 | | | 3+4 2 |
| | 5 2 | | 5 2 | | | 5+6 2 |
| | 5 2 | | 6 2 | | | 3+4 2 |
| | 5 2 | | 6 2 | | | 5+6 2 |
| | 5 2 | | 6 2 | | | 3+4 2 |
| | 5 2 | | 6 2 | | | 5+6 2 |
| | 5 2 | | 6 2 | | | 3+4 2 |
| | 5 2 | | 6 2 | | | 3+4 2 |
| | 6 2 | | 6 2 | | | 5+6 2 |
| | 5 2 | | 6 2 | | | 3+4 2 |
| | 5 2 | | 5 2 | | | 5+6 2 |
| | 5 2 | | 6 2 | | | 3+4 2 |
| | 6 2 | | 5 2 | | | 3+4 2 |
| | 5 2 | | 6 2 | | | 5+6 2 |
| | 5 2 | | 5 2 | | | 3+4 2 |
| | 6 2 | | 6 2 | | | 5+6 2 |
| | 5 2 | | 6 2 | | | 5+6 2 |
| | 6 2 | | 6 2 | | | 3+4 2 |
| | 6 2 | | 6 2 | | | 5+6 2 |
| 63.2.118 | 152354 | | 116 | | Al10Fe2Th1 | |
| | -GM | -T | -Y | -Z | -R | -S |
| | 5 2 | 3+4 4 | 5 2 | 3+4 4 | 2+2 4 | 3+4 2 |
| | 6 2 | 3+4 4 | 6 2 | 3+4 4 | 2+2 4 | 5+6 2 |
| | 5 2 | 3+4 4 | 5 2 | 3+4 4 | 2+2 4 | 3+4 2 |
| | 6 2 | 3+4 4 | 6 2 | 3+4 4 | 2+2 4 | 5+6 2 |
| | 5 2 | 3+4 4 | 6 2 | 3+4 4 | 2+2 4 | 5+6 2 |
| | 5 2 | 3+4 4 | 5 2 | 3+4 4 | 2+2 4 | 3+4 2 |
| | 6 2 | 3+4 4 | 5 2 | 3+4 4 | 2+2 4 | 3+4 2 |
| | 6 2 | 3+4 4 | 6 2 | 3+4 4 | 2+2 4 | 5+6 2 |
| | 5 2 | 3+4 4 | 5 2 | 3+4 4 | 2+2 4 | 5+6 2 |
| | 5 2 | 3+4 4 | 6 2 | 3+4 4 | 2+2 4 | 3+4 2 |
| | 6 2 | 3+4 4 | 5 2 | 3+4 4 | 2+2 4 | 5+6 2 |
| | 6 2 | 3+4 4 | 6 2 | 3+4 4 | 2+2 4 | 3+4 2 |
| | 5 2 | 3+4 4 | 5 2 | 3+4 4 | 2+2 4 | 5+6 2 |
| | 5 2 | 3+4 4 | 6 2 | 3+4 4 | 2+2 4 | 3+4 2 |
| | 6 2 | 3+4 4 | 6 2 | 3+4 4 | 2+2 4 | 3+4 2 |
| | 6 2 | 3+4 4 | 6 2 | 3+4 4 | 2+2 4 | 5+6 2 |
| | 6 2 | 3+4 4 | 5 2 | 3+4 4 | 2+2 4 | 3+4 2 |
| | 5 2 | 3+4 4 | 5 2 | 3+4 4 | 2+2 4 | 5+6 2 |
| | 5 2 | 3+4 4 | 6 2 | 3+4 4 | 2+2 4 | 3+4 2 |
| | 6 2 | 3+4 4 | 5 2 | 3+4 4 | 2+2 4 | 5+6 2 |
| | 6 2 | 3+4 4 | 6 2 | 3+4 4 | 2+2 4 | 5+6 2 |

Continued on next column

**TABLE XXXVIII– continued**

| TQC | ICSD | | NoE | | SF | |
|---|---|---|---|---|---|---|
| | 5 2 | 3+4 4 | 6 2 | 3+4 4 | 2+2 4 | 3+4 2 |
| | 5 2 | 3+4 4 | 6 2 | 3+4 4 | 2+2 4 | 5+6 2 |
| | 5 2 | 3+4 4 | 5 2 | 3+4 4 | 2+2 4 | 3+4 2 |
| | 6 2 | 3+4 4 | 5 2 | 3+4 4 | 2+2 4 | 3+4 2 |
| | 5 2 | 3+4 4 | 6 2 | 3+4 4 | 2+2 4 | 5+6 2 |
| | 5 2 | 3+4 4 | 6 2 | 3+4 4 | 2+2 4 | 3+4 2 |
| | 6 2 | 3+4 4 | 5 2 | 3+4 4 | 2+2 4 | 5+6 2 |
| | 5 2 | 3+4 4 | 6 2 | 3+4 4 | 2+2 4 | 3+4 2 |
| | 6 2 | | 6 2 | | | 5+6 2 |
| | 5 2 | | 5 2 | | | 3+4 2 |
| | 5 2 | | 5 2 | | | 5+6 2 |
| | 6 2 | | 6 2 | | | 5+6 2 |
| | 5 2 | | 5 2 | | | 5+6 2 |
| | 6 2 | | 6 2 | | | 3+4 2 |
| | 6 2 | | 5 2 | | | 5+6 2 |
| | 5 2 | | 6 2 | | | 3+4 2 |
| | 5 2 | | 5 2 | | | 5+6 2 |
| | 5 2 | | 5 2 | | | 3+4 2 |
| | 5 2 | | 5 2 | | | 3+4 2 |
| | 6 2 | | 6 2 | | | 3+4 2 |
| | 5 2 | | 5 2 | | | 5+6 2 |
| | 5 2 | | 6 2 | | | 3+4 2 |
| | 5 2 | | 6 2 | | | 5+6 2 |
| | 5 2 | | 6 2 | | | 3+4 2 |
| | 5 2 | | 6 2 | | | 5+6 2 |
| | 5 2 | | 6 2 | | | 5+6 2 |
| | 6 2 | | 6 2 | | | 5+6 2 |
| | 5 2 | | 6 2 | | | 3+4 2 |
| | 5 2 | | 6 2 | | | 3+4 2 |
| | 5 2 | | 6 2 | | | 5+6 2 |
| | 6 2 | | 5 2 | | | 5+6 2 |
| | 5 2 | | 6 2 | | | 3+4 2 |
| | 5 2 | | 5 2 | | | 5+6 2 |
| | 6 2 | | 6 2 | | | 3+4 2 |
| | 5 2 | | 6 2 | | | 3+4 2 |
| | 5 2 | | 5 2 | | | 5+6 2 |
| | 5 2 | | 6 2 | | | 3+4 2 |
| 63.2.119 | 59912 | | 116 | Al10Ce1Ru2 | | |
| | -GM | -T | -Y | -Z | -R | -S |
| | 5 2 | 3+4 4 | 5 2 | 3+4 4 | 2+2 4 | 3+4 2 |
| | 6 2 | 3+4 4 | 6 2 | 3+4 4 | 2+2 4 | 5+6 2 |
| | 5 2 | 3+4 4 | 5 2 | 3+4 4 | 2+2 4 | 3+4 2 |
| | 6 2 | 3+4 4 | 6 2 | 3+4 4 | 2+2 4 | 5+6 2 |
| | 5 2 | 3+4 4 | 6 2 | 3+4 4 | 2+2 4 | 3+4 2 |
| | 5 2 | 3+4 4 | 5 2 | 3+4 4 | 2+2 4 | 5+6 2 |
| | 6 2 | 3+4 4 | 5 2 | 3+4 4 | 2+2 4 | 3+4 2 |
| | 6 2 | 3+4 4 | 6 2 | 3+4 4 | 2+2 4 | 5+6 2 |
| | 5 2 | 3+4 4 | 5 2 | 3+4 4 | 2+2 4 | 5+6 2 |
| | 6 2 | 3+4 4 | 6 2 | 3+4 4 | 2+2 4 | 3+4 2 |
| | 5 2 | 3+4 4 | 6 2 | 3+4 4 | 2+2 4 | 5+6 2 |
| | 6 2 | 3+4 4 | 5 2 | 3+4 4 | 2+2 4 | 5+6 2 |
| | 5 2 | 3+4 4 | 5 2 | 3+4 4 | 2+2 4 | 3+4 2 |
| | 5 2 | 3+4 4 | 6 2 | 3+4 4 | 2+2 4 | 3+4 2 |
| | 6 2 | 3+4 4 | 6 2 | 3+4 4 | 2+2 4 | 3+4 2 |
| | 5 2 | 3+4 4 | 6 2 | 3+4 4 | 2+2 4 | 5+6 2 |
| | 6 2 | 3+4 4 | 5 2 | 3+4 4 | 2+2 4 | 5+6 2 |
| | 5 2 | 3+4 4 | 5 2 | 3+4 4 | 2+2 4 | 3+4 2 |
| | 5 2 | 3+4 4 | 5 2 | 3+4 4 | 2+2 4 | 3+4 2 |
| | 6 2 | 3+4 4 | 6 2 | 3+4 4 | 2+2 4 | 5+6 2 |
| | 6 2 | 3+4 4 | 6 2 | 3+4 4 | 2+2 4 | 5+6 2 |
| | 5 2 | 3+4 4 | 6 2 | 3+4 4 | 2+2 4 | 3+4 2 |

Continued on next column

**TABLE XXXVIII– continued**

| TQC | ICSD | | NoE | | SF | |
|---|---|---|---|---|---|---|
| | 6 2 | 3+4 4 | 6 2 | 3+4 4 | 2+2 4 | 5+6 2 |
| | 5 2 | 3+4 4 | 6 2 | 3+4 4 | 2+2 4 | 3+4 2 |
| | 5 2 | 3+4 4 | 6 2 | 3+4 4 | 2+2 4 | 3+4 2 |
| | 5 2 | 3+4 4 | 6 2 | 3+4 4 | 2+2 4 | 5+6 2 |
| | 6 2 | 3+4 4 | 6 2 | 3+4 4 | 2+2 4 | 5+6 2 |
| | 5 2 | 3+4 4 | 6 2 | 3+4 4 | 2+2 4 | 3+4 2 |
| | 5 2 | 3+4 4 | 6 2 | 3+4 4 | 2+2 4 | 3+4 2 |
| | 5 2 | | 6 2 | | | 5+6 2 |
| | 5 2 | | 5 2 | | | 3+4 2 |
| | 5 2 | | 6 2 | | | 5+6 2 |
| | 6 2 | | 5 2 | | | 3+4 2 |
| | 5 2 | | 5 2 | | | 5+6 2 |
| | 6 2 | | 6 2 | | | 3+4 2 |
| | 6 2 | | 5 2 | | | 5+6 2 |
| | 5 2 | | 6 2 | | | 5+6 2 |
| | 5 2 | | 6 2 | | | 3+4 2 |
| | 5 2 | | 5 2 | | | 5+6 2 |
| | 6 2 | | 5 2 | | | 3+4 2 |
| | 5 2 | | 5 2 | | | 3+4 2 |
| | 6 2 | | 6 2 | | | 3+4 2 |
| | 5 2 | | 5 2 | | | 5+6 2 |
| | 5 2 | | 6 2 | | | 5+6 2 |
| | 5 2 | | 6 2 | | | 3+4 2 |
| | 5 2 | | 6 2 | | | 3+4 2 |
| | 5 2 | | 6 2 | | | 5+6 2 |
| | 6 2 | | 5 2 | | | 5+6 2 |
| | 5 2 | | 5 2 | | | 5+6 2 |
| | 5 2 | | 5 2 | | | 3+4 2 |
| | 5 2 | | 6 2 | | | 5+6 2 |
| | 5 2 | | 6 2 | | | 3+4 2 |
| | 5 2 | | 5 2 | | | 5+6 2 |
| | 6 2 | | 6 2 | | | 3+4 2 |
| | 6 2 | | 5 2 | | | 5+6 2 |
| | 5 2 | | 6 2 | | | 5+6 2 |
| | 5 2 | | 5 2 | | | 3+4 2 |
| | 6 2 | | 6 2 | | | 3+4 2 |
| 63.2.120 | 238167 | | 232 | Ag1.2Cs0.4Er0.8Te2 | | |
| | -GM | -T | -Y | -Z | -R | -S |
| | 5 2 | 3+4 4 | 5 2 | 3+4 4 | 2+2 4 | 3+4 2 |
| | 5 2 | 3+4 4 | 5 2 | 3+4 4 | 2+2 4 | 3+4 2 |
| | 6 2 | 3+4 4 | 6 2 | 3+4 4 | 2+2 4 | 5+6 2 |
| | 6 2 | 3+4 4 | 6 2 | 3+4 4 | 2+2 4 | 5+6 2 |
| | 5 2 | 3+4 4 | 6 2 | 3+4 4 | 2+2 4 | 3+4 2 |
| | 5 2 | 3+4 4 | 6 2 | 3+4 4 | 2+2 4 | 3+4 2 |
| | 6 2 | 3+4 4 | 5 2 | 3+4 4 | 2+2 4 | 5+6 2 |
| | 6 2 | 3+4 4 | 5 2 | 3+4 4 | 2+2 4 | 5+6 2 |
| | 5 2 | 3+4 4 | 6 2 | 3+4 4 | 2+2 4 | 3+4 2 |
| | 5 2 | 3+4 4 | 6 2 | 3+4 4 | 2+2 4 | 3+4 2 |
| | 6 2 | 3+4 4 | 5 2 | 3+4 4 | 2+2 4 | 5+6 2 |
| | 6 2 | 3+4 4 | 5 2 | 3+4 4 | 2+2 4 | 5+6 2 |
| | 5 2 | 3+4 4 | 6 2 | 3+4 4 | 2+2 4 | 3+4 2 |
| | 6 2 | 3+4 4 | 5 2 | 3+4 4 | 2+2 4 | 5+6 2 |
| | 5 2 | 3+4 4 | 6 2 | 3+4 4 | 2+2 4 | 3+4 2 |
| | 5 2 | 3+4 4 | 6 2 | 3+4 4 | 2+2 4 | 5+6 2 |
| | 6 2 | 3+4 4 | 5 2 | 3+4 4 | 2+2 4 | 3+4 2 |
| | 6 2 | 3+4 4 | 5 2 | 3+4 4 | 2+2 4 | 5+6 2 |
| | 5 2 | 3+4 4 | 6 2 | 3+4 4 | 2+2 4 | 3+4 2 |
| | 5 2 | 3+4 4 | 5 2 | 3+4 4 | 2+2 4 | 5+6 2 |
| | 6 2 | 3+4 4 | 6 2 | 3+4 4 | 2+2 4 | 3+4 2 |
| | 6 2 | 3+4 4 | 5 2 | 3+4 4 | 2+2 4 | 3+4 2 |
| | 5 2 | 3+4 4 | 5 2 | 3+4 4 | 2+2 4 | 5+6 2 |

Continued on next column

**TABLE XXXVIII– continued**

| TQC | ICSD | NoE | SF |
|---|---|---|---|
| | 6 2 3+4 4 | 6 2 3+4 4 | 2+2 4 5+6 2 |
| | 5 2 3+4 4 | 6 2 3+4 4 | 2+2 4 3+4 2 |
| | 6 2 3+4 4 | 5 2 3+4 4 | 2+2 4 5+6 2 |
| | 5 2 3+4 4 | 6 2 3+4 4 | 2+2 4 3+4 2 |
| | 6 2 3+4 4 | 5 2 3+4 4 | 2+2 4 5+6 2 |
| | 5 2 3+4 4 | 5 2 3+4 4 | 2+2 4 3+4 2 |
| | 6 2 3+4 4 | 6 2 3+4 4 | 2+2 4 5+6 2 |
| | 5 2 3+4 4 | 6 2 3+4 4 | 2+2 4 5+6 2 |
| | 6 2 3+4 4 | 5 2 3+4 4 | 2+2 4 3+4 2 |
| | 5 2 3+4 4 | 5 2 3+4 4 | 2+2 4 3+4 2 |
| | 6 2 3+4 4 | 6 2 3+4 4 | 2+2 4 5+6 2 |
| | 6 2 3+4 4 | 6 2 3+4 4 | 2+2 4 3+4 2 |
| | 5 2 3+4 4 | 5 2 3+4 4 | 2+2 4 3+4 2 |
| | 6 2 3+4 4 | 6 2 3+4 4 | 2+2 4 5+6 2 |
| | 5 2 3+4 4 | 5 2 3+4 4 | 2+2 4 5+6 2 |
| | 6 2 3+4 4 | 5 2 3+4 4 | 2+2 4 5+6 2 |
| | 5 2 3+4 4 | 6 2 3+4 4 | 2+2 4 3+4 2 |
| | 6 2 3+4 4 | 6 2 3+4 4 | 2+2 4 3+4 2 |
| | 6 2 3+4 4 | 5 2 3+4 4 | 2+2 4 3+4 2 |
| | 5 2 3+4 4 | 6 2 3+4 4 | 2+2 4 5+6 2 |
| | 6 2 3+4 4 | 6 2 3+4 4 | 2+2 4 5+6 2 |
| | 5 2 3+4 4 | 6 2 3+4 4 | 2+2 4 3+4 2 |
| | 6 2 3+4 4 | 5 2 3+4 4 | 2+2 4 5+6 2 |
| | 5 2 3+4 4 | 6 2 3+4 4 | 2+2 4 5+6 2 |
| | 5 2 3+4 4 | 5 2 3+4 4 | 2+2 4 3+4 2 |
| | 5 2 3+4 4 | 5 2 3+4 4 | 2+2 4 5+6 2 |
| | 6 2 3+4 4 | 5 2 3+4 4 | 2+2 4 3+4 2 |
| | 5 2 3+4 4 | 5 2 3+4 4 | 2+2 4 5+6 2 |
| | 5 2 3+4 4 | 6 2 3+4 4 | 2+2 4 3+4 2 |
| | 6 2 3+4 4 | 5 2 3+4 4 | 2+2 4 5+6 2 |
| | 6 2 3+4 4 | 5 2 3+4 4 | 2+2 4 3+4 2 |
| | 6 2 3+4 4 | 6 2 3+4 4 | 2+2 4 5+6 2 |
| | 5 2 3+4 4 | 6 2 3+4 4 | 2+2 4 3+4 2 |
| | 5 2 3+4 4 | 5 2 3+4 4 | 2+2 4 5+6 2 |
| | 5 2 3+4 4 | 6 2 3+4 4 | 2+2 4 5+6 2 |
| | 6 2 | 5 2 | 3+4 2 |
| | 5 2 | 5 2 | 3+4 2 |
| | 5 2 | 6 2 | 5+6 2 |
| | 6 2 | 6 2 | 3+4 2 |
| | 6 2 | 5 2 | 5+6 2 |
| | 6 2 | 5 2 | 3+4 2 |
| | 5 2 | 6 2 | 3+4 2 |
| | 5 2 | 5 2 | 5+6 2 |
| | 5 2 | 6 2 | 5+6 2 |
| | 6 2 | 5 2 | 5+6 2 |
| | 6 2 | 6 2 | 3+4 2 |
| | 6 2 | 5 2 | 3+4 2 |
| | 5 2 | 6 2 | 5+6 2 |
| | 6 2 | 6 2 | 5+6 2 |
| | 5 2 | 5 2 | 3+4 2 |
| | 5 2 | 6 2 | 3+4 2 |
| | 6 2 | 5 2 | 3+4 2 |
| | 6 2 | 6 2 | 5+6 2 |
| | 5 2 | 6 2 | 3+4 2 |
| | 6 2 | 6 2 | 3+4 2 |
| | 5 2 | 6 2 | 5+6 2 |
| | 6 2 | 6 2 | 3+4 2 |
| | 5 2 | 5 2 | 5+6 2 |
| | 5 2 | 5 2 | 3+4 2 |
| | 6 2 | 5 2 | 5+6 2 |
| | 6 2 | 5 2 | 3+4 2 |

Continued on next column

**TABLE XXXVIII– continued**

| TQC | ICSD | NoE | SF |
|---|---|---|---|
| | 5 2 | 6 2 | 5+6 2 |
| | 6 2 | 6 2 | 3+4 2 |
| | 5 2 | 5 2 | 3+4 2 |
| | 5 2 | 6 2 | 3+4 2 |
| | 6 2 | 5 2 | 5+6 2 |
| | 6 2 | 6 2 | 5+6 2 |
| | 5 2 | 5 2 | 3+4 2 |
| | 5 2 | 6 2 | 5+6 2 |
| | 6 2 | 5 2 | 5+6 2 |
| | 5 2 | 6 2 | 3+4 2 |
| | 6 2 | 5 2 | 3+4 2 |
| | 5 2 | 6 2 | 5+6 2 |
| | 5 2 | 6 2 | 3+4 2 |
| | 6 2 | 5 2 | 5+6 2 |
| | 6 2 | 5 2 | 5+6 2 |
| | 5 2 | 6 2 | 3+4 2 |
| | 6 2 | 5 2 | 5+6 2 |
| | 5 2 | 5 2 | 3+4 2 |
| | 6 2 | 5 2 | 5+6 2 |
| | 5 2 | 6 2 | 3+4 2 |
| | 6 2 | 6 2 | 5+6 2 |
| | 6 2 | 5 2 | 5+6 2 |
| | 5 2 | 6 2 | 3+4 2 |
| | 5 2 | 5 2 | 5+6 2 |
| | 6 2 | 6 2 | 3+4 2 |
| | 5 2 | 5 2 | 5+6 2 |
| | 6 2 | 6 2 | 3+4 2 |
| | 5 2 | 6 2 | 5+6 2 |
| | 6 2 | 6 2 | 3+4 2 |
| | 5 2 | 5 2 | 5+6 2 |
| | 5 2 | 5 2 | 3+4 2 |
| | 5 2 | 6 2 | 5+6 2 |

The End

TABLE XXXIX: The computed irreducible representations for the valence bands for the compounds of SEBR and NLC in space group 64. See the list of tables.

| TQC | ICSD | NoE | SF | | | |
|---|---|---|---|---|---|---|
| 64.1.1 | 411192 | 80 | Sn3Ti2 | | | |
| | -GM | -T | -Y | -Z | -R | -S |
| | 5 2 | 3+4 4 | 5 2 | 3+4 4 | 3+3 2 | 2+2 4 |
| | 5 2 | 3+4 4 | 6 2 | 3+4 4 | 4+4 2 | 2+2 4 |
| | 6 2 | 3+4 4 | 6 2 | 3+4 4 | 6+6 2 | 2+2 4 |
| | 6 2 | 3+4 4 | 6 2 | 3+4 4 | 5+5 2 | 2+2 4 |
| | 5 2 | 3+4 4 | 5 2 | 3+4 4 | 5+5 2 | 2+2 4 |
| | 6 2 | 3+4 4 | 5 2 | 3+4 4 | 6+6 2 | 2+2 4 |
| | 5 2 | 3+4 4 | 6 2 | 3+4 4 | 3+3 2 | 2+2 4 |
| | 6 2 | 3+4 4 | 6 2 | 3+4 4 | 4+4 2 | 2+2 4 |
| | 6 2 | 3+4 4 | 6 2 | 3+4 4 | 6+6 2 | 2+2 4 |
| | 6 2 | 3+4 4 | 5 2 | 3+4 4 | 5+5 2 | 2+2 4 |
| | 5 2 | 3+4 4 | 5 2 | 3+4 4 | 3+3 2 | 2+2 4 |
| | 5 2 | 3+4 4 | 5 2 | 3+4 4 | 4+4 2 | 2+2 4 |
| | 5 2 | 3+4 4 | 5 2 | 3+4 4 | 4+4 2 | 2+2 4 |
| | 5 2 | 3+4 4 | 5 2 | 3+4 4 | 3+3 2 | 2+2 4 |
| | 6 2 | 3+4 4 | 6 2 | 3+4 4 | 5+5 2 | 2+2 4 |
| | 5 2 | 3+4 4 | 6 2 | 3+4 4 | 6+6 2 | 2+2 4 |
| | 6 2 | 3+4 4 | 5 2 | 3+4 4 | 6+6 2 | 2+2 4 |
| | 5 2 | 3+4 4 | 5 2 | 3+4 4 | 5+5 2 | 2+2 4 |
| | 6 2 | 3+4 4 | 6 2 | 3+4 4 | 3+3 2 | 2+2 4 |

Continued on next column

**TABLE XXXIX– continued**

| TQC | ICSD | | NoE | | SF | |
|---|---|---|---|---|---|---|
| | 6 2 | 3+4 4 | 5 2 | 3+4 4 | 4+4 2 | 2+2 4 |
| | 6 2 | | 6 2 | | 3+3 2 | |
| | 6 2 | | 6 2 | | 4+4 2 | |
| | 5 2 | | 6 2 | | 6+6 2 | |
| | 5 2 | | 6 2 | | 5+5 2 | |
| | 5 2 | | 5 2 | | 5+5 2 | |
| | 5 2 | | 6 2 | | 6+6 2 | |
| | 6 2 | | 6 2 | | 3+3 2 | |
| | 6 2 | | 6 2 | | 4+4 2 | |
| | 5 2 | | 5 2 | | 5+5 2 | |
| | 6 2 | | 5 2 | | 6+6 2 | |
| | 5 2 | | 6 2 | | 3+3 2 | |
| | 5 2 | | 5 2 | | 4+4 2 | |
| | 5 2 | | 5 2 | | 5+5 2 | |
| | 6 2 | | 5 2 | | 3+3 2 | |
| | 5 2 | | 6 2 | | 4+4 2 | |
| | 5 2 | | 6 2 | | 6+6 2 | |
| | 6 2 | | 6 2 | | 6+6 2 | |
| | 6 2 | | 6 2 | | 5+5 2 | |
| | 5 2 | | 5 2 | | 4+4 2 | |
| | 5 2 | | 5 2 | | 3+3 2 | |
| 64.1.2 | 410902 | | 80 | | Sn3Ti2 | |
| | -GM | -T | -Y | -Z | -R | -S |
| | 5 2 | 3+4 4 | 5 2 | 3+4 4 | 5+5 2 | 2+2 4 |
| | 5 2 | 3+4 4 | 6 2 | 3+4 4 | 6+6 2 | 2+2 4 |
| | 6 2 | 3+4 4 | 6 2 | 3+4 4 | 4+4 2 | 2+2 4 |
| | 6 2 | 3+4 4 | 6 2 | 3+4 4 | 3+3 2 | 2+2 4 |
| | 5 2 | 3+4 4 | 5 2 | 3+4 4 | 3+3 2 | 2+2 4 |
| | 6 2 | 3+4 4 | 5 2 | 3+4 4 | 4+4 2 | 2+2 4 |
| | 5 2 | 3+4 4 | 6 2 | 3+4 4 | 5+5 2 | 2+2 4 |
| | 6 2 | 3+4 4 | 6 2 | 3+4 4 | 6+6 2 | 2+2 4 |
| | 6 2 | 3+4 4 | 6 2 | 3+4 4 | 4+4 2 | 2+2 4 |
| | 6 2 | 3+4 4 | 5 2 | 3+4 4 | 3+3 2 | 2+2 4 |
| | 5 2 | 3+4 4 | 5 2 | 3+4 4 | 5+5 2 | 2+2 4 |
| | 5 2 | 3+4 4 | 5 2 | 3+4 4 | 6+6 2 | 2+2 4 |
| | 5 2 | 3+4 4 | 5 2 | 3+4 4 | 6+6 2 | 2+2 4 |
| | 5 2 | 3+4 4 | 5 2 | 3+4 4 | 5+5 2 | 2+2 4 |
| | 6 2 | 3+4 4 | 6 2 | 3+4 4 | 3+3 2 | 2+2 4 |
| | 5 2 | 3+4 4 | 6 2 | 3+4 4 | 4+4 2 | 2+2 4 |
| | 6 2 | 3+4 4 | 5 2 | 3+4 4 | 4+4 2 | 2+2 4 |
| | 6 2 | 3+4 4 | 5 2 | 3+4 4 | 3+3 2 | 2+2 4 |
| | 5 2 | 3+4 4 | 6 2 | 3+4 4 | 5+5 2 | 2+2 4 |
| | 6 2 | 3+4 4 | 5 2 | 3+4 4 | 6+6 2 | 2+2 4 |
| | 6 2 | | 6 2 | | 5+5 2 | |
| | 6 2 | | 6 2 | | 6+6 2 | |
| | 5 2 | | 6 2 | | 4+4 2 | |
| | 5 2 | | 6 2 | | 3+3 2 | |
| | 5 2 | | 5 2 | | 3+3 2 | |
| | 5 2 | | 6 2 | | 4+4 2 | |
| | 6 2 | | 6 2 | | 5+5 2 | |
| | 6 2 | | 6 2 | | 6+6 2 | |
| | 5 2 | | 5 2 | | 3+3 2 | |
| | 6 2 | | 5 2 | | 4+4 2 | |
| | 5 2 | | 6 2 | | 5+5 2 | |
| | 5 2 | | 5 2 | | 6+6 2 | |
| | 5 2 | | 5 2 | | 3+3 2 | |
| | 6 2 | | 5 2 | | 5+5 2 | |
| | 5 2 | | 6 2 | | 6+6 2 | |
| | 5 2 | | 6 2 | | 4+4 2 | |
| | 6 2 | | 6 2 | | 4+4 2 | |
| | 6 2 | | 6 2 | | 3+3 2 | |

Continued on next column

**TABLE XXXIX– continued**

| TQC | ICSD | | NoE | | SF | |
|---|---|---|---|---|---|---|
| | 5 2 | | 5 2 | | 6+6 2 | |
| | 5 2 | | 5 2 | | 5+5 2 | |
| 64.1.3 | 413279 | | 88 | | Pd8Sn24 | |
| | -GM | -T | -Y | -Z | -R | -S |
| | 5 2 | 3+4 4 | 5 2 | 3+4 4 | 5+5 2 | 2+2 4 |
| | 6 2 | 3+4 4 | 6 2 | 3+4 4 | 6+6 2 | 2+2 4 |
| | 5 2 | 3+4 4 | 6 2 | 3+4 4 | 3+3 2 | 2+2 4 |
| | 6 2 | 3+4 4 | 5 2 | 3+4 4 | 4+4 2 | 2+2 4 |
| | 5 2 | 3+4 4 | 6 2 | 3+4 4 | 4+4 2 | 2+2 4 |
| | 6 2 | 3+4 4 | 5 2 | 3+4 4 | 3+3 2 | 2+2 4 |
| | 5 2 | 3+4 4 | 5 2 | 3+4 4 | 6+6 2 | 2+2 4 |
| | 6 2 | 3+4 4 | 6 2 | 3+4 4 | 5+5 2 | 2+2 4 |
| | 6 2 | 3+4 4 | 5 2 | 3+4 4 | 6+6 2 | 2+2 4 |
| | 5 2 | 3+4 4 | 6 2 | 3+4 4 | 5+5 2 | 2+2 4 |
| | 5 2 | 3+4 4 | 5 2 | 3+4 4 | 4+4 2 | 2+2 4 |
| | 6 2 | 3+4 4 | 6 2 | 3+4 4 | 3+3 2 | 2+2 4 |
| | 5 2 | 3+4 4 | 6 2 | 3+4 4 | 5+5 2 | 2+2 4 |
| | 6 2 | 3+4 4 | 5 2 | 3+4 4 | 6+6 2 | 2+2 4 |
| | 6 2 | 3+4 4 | 6 2 | 3+4 4 | 4+4 2 | 2+2 4 |
| | 6 2 | 3+4 4 | 6 2 | 3+4 4 | 3+3 2 | 2+2 4 |
| | 5 2 | 3+4 4 | 5 2 | 3+4 4 | 3+3 2 | 2+2 4 |
| | 5 2 | 3+4 4 | 6 2 | 3+4 4 | 4+4 2 | 2+2 4 |
| | 6 2 | 3+4 4 | 5 2 | 3+4 4 | 6+6 2 | 2+2 4 |
| | 5 2 | 3+4 4 | 6 2 | 3+4 4 | 5+5 2 | 2+2 4 |
| | 5 2 | 3+4 4 | 5 2 | 3+4 4 | 5+5 2 | 2+2 4 |
| | 5 2 | 3+4 4 | 5 2 | 3+4 4 | 6+6 2 | 2+2 4 |
| | 6 2 | | 5 2 | | 3+3 2 | |
| | 5 2 | | 6 2 | | 4+4 2 | |
| | 5 2 | | 5 2 | | 4+4 2 | |
| | 6 2 | | 5 2 | | 3+3 2 | |
| | 6 2 | | 6 2 | | 6+6 2 | |
| | 6 2 | | 6 2 | | 5+5 2 | |
| | 5 2 | | 6 2 | | 3+3 2 | |
| | 6 2 | | 6 2 | | 4+4 2 | |
| | 5 2 | | 5 2 | | 6+6 2 | |
| | 6 2 | | 5 2 | | 5+5 2 | |
| | 6 2 | | 5 2 | | 4+4 2 | |
| | 5 2 | | 6 2 | | 3+3 2 | |
| | 6 2 | | 5 2 | | 5+5 2 | |
| | 5 2 | | 6 2 | | 6+6 2 | |
| | 6 2 | | 5 2 | | 6+6 2 | |
| | 5 2 | | 5 2 | | 5+5 2 | |
| | 5 2 | | 6 2 | | 5+5 2 | |
| | 5 2 | | 5 2 | | 3+3 2 | |
| | 6 2 | | 6 2 | | 6+6 2 | |
| | 5 2 | | 5 2 | | 4+4 2 | |
| | 5 2 | | 5 2 | | 4+4 2 | |
| | 6 2 | | 5 2 | | 3+3 2 | |
| 64.1.4 | 417180 | | 20 | | P1 | |
| | -GM | -T | -Y | -Z | -R | -S |
| | 5 2 | 3+4 4 | 5 2 | 3+4 4 | 3+3 2 | 2+2 4 |
| | 6 2 | 3+4 4 | 6 2 | 3+4 4 | 4+4 2 | 2+2 4 |
| | 5 2 | 3+4 4 | 5 2 | 3+4 4 | 5+5 2 | 2+2 4 |
| | 5 2 | 3+4 4 | 5 2 | 3+4 4 | 6+6 2 | 2+2 4 |
| | 6 2 | 3+4 4 | 6 2 | 3+4 4 | 4+4 2 | 2+2 4 |
| | 6 2 | | 6 2 | | 3+3 2 | |
| | 6 2 | | 6 2 | | 6+6 2 | |
| | 5 2 | | 5 2 | | 5+5 2 | |
| | 5 2 | | 5 2 | | 3+3 2 | |
| | 5 2 | | 6 2 | | 4+4 2 | |
| 64.1.5 | 23836 | | 20 | | P1 | |

Continued on next column

**TABLE XXXIX– continued**

| TQC | ICSD | | NoE | | SF | |
|---|---|---|---|---|---|---|
| | -GM | -T | -Y | -Z | -R | -S |
| | 5 2 | 3+4 4 | 5 2 | 3+4 4 | 3+3 2 | 2+2 4 |
| | 6 2 | 3+4 4 | 6 2 | 3+4 4 | 4+4 2 | 2+2 4 |
| | 5 2 | 3+4 4 | 5 2 | 3+4 4 | 5+5 2 | 2+2 4 |
| | 5 2 | 3+4 4 | 5 2 | 3+4 4 | 6+6 2 | 2+2 4 |
| | 6 2 | 3+4 4 | 6 2 | 3+4 4 | 4+4 2 | 2+2 4 |
| | 6 2 | | 6 2 | | 3+3 2 | |
| | 6 2 | | 6 2 | | 6+6 2 | |
| | 5 2 | | 5 2 | | 5+5 2 | |
| | 5 2 | | 5 2 | | 3+3 2 | |
| | 5 2 | | 6 2 | | 4+4 2 | |
| 64.1.6 | 150873 | | 20 | | P1 | |
| | -GM | -T | -Y | -Z | -R | -S |
| | 5 2 | 3+4 4 | 5 2 | 3+4 4 | 3+3 2 | 2+2 4 |
| | 6 2 | 3+4 4 | 6 2 | 3+4 4 | 4+4 2 | 2+2 4 |
| | 5 2 | 3+4 4 | 5 2 | 3+4 4 | 5+5 2 | 2+2 4 |
| | 5 2 | 3+4 4 | 5 2 | 3+4 4 | 6+6 2 | 2+2 4 |
| | 6 2 | 3+4 4 | 6 2 | 3+4 4 | 4+4 2 | 2+2 4 |
| | 6 2 | | 6 2 | | 3+3 2 | |
| | 6 2 | | 6 2 | | 6+6 2 | |
| | 5 2 | | 5 2 | | 5+5 2 | |
| | 5 2 | | 5 2 | | 3+3 2 | |
| | 5 2 | | 6 2 | | 4+4 2 | |
| 64.1.7 | 77831 | | 68 | | La1S1 | |
| | -GM | -T | -Y | -Z | -R | -S |
| | 5 2 | 3+4 4 | 5 2 | 3+4 4 | 3+3 2 | 2+2 4 |
| | 6 2 | 3+4 4 | 6 2 | 3+4 4 | 4+4 2 | 2+2 4 |
| | 5 2 | 3+4 4 | 5 2 | 3+4 4 | 5+5 2 | 2+2 4 |
| | 6 2 | 3+4 4 | 6 2 | 3+4 4 | 6+6 2 | 2+2 4 |
| | 5 2 | 3+4 4 | 5 2 | 3+4 4 | 3+3 2 | 2+2 4 |
| | 6 2 | 3+4 4 | 6 2 | 3+4 4 | 4+4 2 | 2+2 4 |
| | 5 2 | 3+4 4 | 5 2 | 3+4 4 | 5+5 2 | 2+2 4 |
| | 6 2 | 3+4 4 | 6 2 | 3+4 4 | 6+6 2 | 2+2 4 |
| | 5 2 | 3+4 4 | 5 2 | 3+4 4 | 3+3 2 | 2+2 4 |
| | 6 2 | 3+4 4 | 6 2 | 3+4 4 | 4+4 2 | 2+2 4 |
| | 5 2 | 3+4 4 | 5 2 | 3+4 4 | 5+5 2 | 2+2 4 |
| | 6 2 | 3+4 4 | 6 2 | 3+4 4 | 6+6 2 | 2+2 4 |
| | 5 2 | 3+4 4 | 5 2 | 3+4 4 | 3+3 2 | 2+2 4 |
| | 6 2 | 3+4 4 | 6 2 | 3+4 4 | 4+4 2 | 2+2 4 |
| | 5 2 | 3+4 4 | 5 2 | 3+4 4 | 5+5 2 | 2+2 4 |
| | 6 2 | 3+4 4 | 6 2 | 3+4 4 | 6+6 2 | 2+2 4 |
| | 5 2 | 3+4 4 | 5 2 | 3+4 4 | 3+3 2 | 2+2 4 |
| | 6 2 | | 6 2 | | 4+4 2 | |
| | 5 2 | | 5 2 | | 5+5 2 | |
| | 6 2 | | 6 2 | | 6+6 2 | |
| | 5 2 | | 5 2 | | 3+3 2 | |
| | 6 2 | | 5 2 | | 4+4 2 | |
| | 5 2 | | 6 2 | | 5+5 2 | |
| | 6 2 | | 6 2 | | 6+6 2 | |
| | 5 2 | | 5 2 | | 3+3 2 | |
| | 6 2 | | 6 2 | | 4+4 2 | |
| | 5 2 | | 5 2 | | 5+5 2 | |
| | 6 2 | | 6 2 | | 6+6 2 | |
| | 5 2 | | 5 2 | | 3+3 2 | |
| | 6 2 | | 6 2 | | 4+4 2 | |
| | 5 2 | | 5 2 | | 5+5 2 | |
| | 6 2 | | 6 2 | | 6+6 2 | |
| | 5 2 | | 6 2 | | 5+5 2 | |
| | 6 2 | | 5 2 | | 6+6 2 | |
| 64.1.8 | 79878 | | 132 | | K1Na2Ni1O2 | |
| | -GM | -T | -Y | -Z | -R | -S |
| | 5 2 | 3+4 4 | 5 2 | 3+4 4 | 3+3 2 | 2+2 4 |
| | 6 2 | 3+4 4 | 6 2 | 3+4 4 | 4+4 2 | 2+2 4 |
| | 5 2 | 3+4 4 | 5 2 | 3+4 4 | 5+5 2 | 2+2 4 |
| | 6 2 | 3+4 4 | 6 2 | 3+4 4 | 6+6 2 | 2+2 4 |
| | 5 2 | 3+4 4 | 5 2 | 3+4 4 | 5+5 2 | 2+2 4 |
| | 6 2 | 3+4 4 | 5 2 | 3+4 4 | 6+6 2 | 2+2 4 |
| | 5 2 | 3+4 4 | 6 2 | 3+4 4 | 3+3 2 | 2+2 4 |
| | 6 2 | 3+4 4 | 6 2 | 3+4 4 | 4+4 2 | 2+2 4 |
| | 5 2 | 3+4 4 | 5 2 | 3+4 4 | 5+5 2 | 2+2 4 |
| | 6 2 | 3+4 4 | 6 2 | 3+4 4 | 6+6 2 | 2+2 4 |
| | 5 2 | 3+4 4 | 5 2 | 3+4 4 | 3+3 2 | 2+2 4 |
| | 6 2 | 3+4 4 | 6 2 | 3+4 4 | 4+4 2 | 2+2 4 |
| | 5 2 | 3+4 4 | 6 2 | 3+4 4 | 3+3 2 | 2+2 4 |
| | 6 2 | 3+4 4 | 5 2 | 3+4 4 | 6+6 2 | 2+2 4 |
| | 5 2 | 3+4 4 | 6 2 | 3+4 4 | 4+4 2 | 2+2 4 |
| | 6 2 | 3+4 4 | 5 2 | 3+4 4 | 5+5 2 | 2+2 4 |
| | 5 2 | 3+4 4 | 6 2 | 3+4 4 | 4+4 2 | 2+2 4 |
| | 6 2 | 3+4 4 | 6 2 | 3+4 4 | 5+5 2 | 2+2 4 |
| | 5 2 | 3+4 4 | 5 2 | 3+4 4 | 3+3 2 | 2+2 4 |
| | 6 2 | 3+4 4 | 5 2 | 3+4 4 | 6+6 2 | 2+2 4 |
| | 5 2 | 3+4 4 | 6 2 | 3+4 4 | 6+6 2 | 2+2 4 |
| | 5 2 | 3+4 4 | 5 2 | 3+4 4 | 3+3 2 | 2+2 4 |
| | 6 2 | 3+4 4 | 6 2 | 3+4 4 | 5+5 2 | 2+2 4 |
| | 6 2 | 3+4 4 | 5 2 | 3+4 4 | 4+4 2 | 2+2 4 |
| | 5 2 | 3+4 4 | 6 2 | 3+4 4 | 5+5 2 | 2+2 4 |
| | 6 2 | 3+4 4 | 5 2 | 3+4 4 | 6+6 2 | 2+2 4 |
| | 5 2 | 3+4 4 | 5 2 | 3+4 4 | 3+3 2 | 2+2 4 |
| | 6 2 | 3+4 4 | 6 2 | 3+4 4 | 4+4 2 | 2+2 4 |
| | 6 2 | 3+4 4 | 5 2 | 3+4 4 | 5+5 2 | 2+2 4 |
| | 5 2 | 3+4 4 | 5 2 | 3+4 4 | 6+6 2 | 2+2 4 |
| | 6 2 | 3+4 4 | 6 2 | 3+4 4 | 3+3 2 | 2+2 4 |
| | 5 2 | 3+4 4 | 6 2 | 3+4 4 | 4+4 2 | 2+2 4 |
| | 5 2 | 3+4 4 | 5 2 | 3+4 4 | 5+5 2 | 2+2 4 |
| | 5 2 | | 6 2 | | 6+6 2 | |
| | 6 2 | | 5 2 | | 3+3 2 | |
| | 5 2 | | 6 2 | | 4+4 2 | |
| | 6 2 | | 6 2 | | 3+3 2 | |
| | 5 2 | | 5 2 | | 4+4 2 | |
| | 6 2 | | 5 2 | | 5+5 2 | |
| | 5 2 | | 6 2 | | 6+6 2 | |
| | 6 2 | | 6 2 | | 5+5 2 | |
| | 6 2 | | 5 2 | | 6+6 2 | |
| | 5 2 | | 5 2 | | 3+3 2 | |
| | 5 2 | | 5 2 | | 4+4 2 | |
| | 6 2 | | 6 2 | | 5+5 2 | |
| | 5 2 | | 5 2 | | 6+6 2 | |
| | 6 2 | | 6 2 | | 4+4 2 | |
| | 6 2 | | 6 2 | | 3+3 2 | |
| | 5 2 | | 6 2 | | 3+3 2 | |
| | 5 2 | | 6 2 | | 4+4 2 | |
| | 6 2 | | 5 2 | | 6+6 2 | |
| | 5 2 | | 6 2 | | 5+5 2 | |
| | 6 2 | | 5 2 | | 3+3 2 | |
| | 5 2 | | 6 2 | | 4+4 2 | |
| | 6 2 | | 5 2 | | 5+5 2 | |
| | 6 2 | | 5 2 | | 6+6 2 | |
| | 5 2 | | 6 2 | | 3+3 2 | |
| | 5 2 | | 6 2 | | 4+4 2 | |
| | 6 2 | | 5 2 | | 5+5 2 | |
| | 6 2 | | 5 2 | | 6+6 2 | |
| | 6 2 | | 5 2 | | 6+6 2 | |

Continued on next column

**TABLE XXXIX– continued**

| TQC | ICSD | | NoE | | SF | |
|---|---|---|---|---|---|---|
| | 5 2 | | 5 2 | | 3+3 2 | |
| | 6 2 | | 6 2 | | 4+4 2 | |
| | 5 2 | | 6 2 | | 5+5 2 | |
| | 5 2 | | 5 2 | | 5+5 2 | |
| | 5 2 | | 6 2 | | 6+6 2 | |
| 64.1.9 | 40442 | | 108 | | Hf2Ni3Si4 | |
| | -GM | -T | -Y | -Z | -R | -S |
| | 5 2 | 3+4 4 | 5 2 | 3+4 4 | 6+6 2 | 2+2 4 |
| | 6 2 | 3+4 4 | 6 2 | 3+4 4 | 5+5 2 | 2+2 4 |
| | 5 2 | 3+4 4 | 6 2 | 3+4 4 | 4+4 2 | 2+2 4 |
| | 5 2 | 3+4 4 | 5 2 | 3+4 4 | 3+3 2 | 2+2 4 |
| | 6 2 | 3+4 4 | 6 2 | 3+4 4 | 3+3 2 | 2+2 4 |
| | 6 2 | 3+4 4 | 5 2 | 3+4 4 | 4+4 2 | 2+2 4 |
| | 5 2 | 3+4 4 | 5 2 | 3+4 4 | 5+5 2 | 2+2 4 |
| | 6 2 | 3+4 4 | 6 2 | 3+4 4 | 6+6 2 | 2+2 4 |
| | 5 2 | 3+4 4 | 5 2 | 3+4 4 | 5+5 2 | 2+2 4 |
| | 6 2 | 3+4 4 | 5 2 | 3+4 4 | 6+6 2 | 2+2 4 |
| | 5 2 | 3+4 4 | 5 2 | 3+4 4 | 3+3 2 | 2+2 4 |
| | 5 2 | 3+4 4 | 6 2 | 3+4 4 | 4+4 2 | 2+2 4 |
| | 6 2 | 3+4 4 | 6 2 | 3+4 4 | 5+5 2 | 2+2 4 |
| | 5 2 | 3+4 4 | 6 2 | 3+4 4 | 6+6 2 | 2+2 4 |
| | 6 2 | 3+4 4 | 5 2 | 3+4 4 | 6+6 2 | 2+2 4 |
| | 5 2 | 3+4 4 | 6 2 | 3+4 4 | 5+5 2 | 2+2 4 |
| | 5 2 | 3+4 4 | 5 2 | 3+4 4 | 4+4 2 | 2+2 4 |
| | 6 2 | 3+4 4 | 6 2 | 3+4 4 | 3+3 2 | 2+2 4 |
| | 5 2 | 3+4 4 | 6 2 | 3+4 4 | 5+5 2 | 2+2 4 |
| | 6 2 | 3+4 4 | 5 2 | 3+4 4 | 6+6 2 | 2+2 4 |
| | 5 2 | 3+4 4 | 5 2 | 3+4 4 | 3+3 2 | 2+2 4 |
| | 5 2 | 3+4 4 | 6 2 | 3+4 4 | 4+4 2 | 2+2 4 |
| | 6 2 | 3+4 4 | 5 2 | 3+4 4 | 5+5 2 | 2+2 4 |
| | 5 2 | 3+4 4 | 5 2 | 3+4 4 | 6+6 2 | 2+2 4 |
| | 5 2 | 3+4 4 | 6 2 | 3+4 4 | 6+6 2 | 2+2 4 |
| | 6 2 | 3+4 4 | 5 2 | 3+4 4 | 5+5 2 | 2+2 4 |
| | 5 2 | 3+4 4 | 5 2 | 3+4 4 | 3+3 2 | 2+2 4 |
| | 5 2 | | 5 2 | | 4+4 2 | |
| | 5 2 | | 6 2 | | 6+6 2 | |
| | 6 2 | | 6 2 | | 5+5 2 | |
| | 6 2 | | 5 2 | | 6+6 2 | |
| | 5 2 | | 5 2 | | 5+5 2 | |
| | 5 2 | | 6 2 | | 4+4 2 | |
| | 6 2 | | 5 2 | | 3+3 2 | |
| | 5 2 | | 5 2 | | 6+6 2 | |
| | 5 2 | | 6 2 | | 5+5 2 | |
| | 6 2 | | 5 2 | | 5+5 2 | |
| | 5 2 | | 6 2 | | 4+4 2 | |
| | 5 2 | | 5 2 | | 3+3 2 | |
| | 6 2 | | 5 2 | | 6+6 2 | |
| | 5 2 | | 6 2 | | 4+4 2 | |
| | 6 2 | | 6 2 | | 3+3 2 | |
| | 6 2 | | 6 2 | | 5+5 2 | |
| | 5 2 | | 5 2 | | 6+6 2 | |
| | 5 2 | | 6 2 | | 4+4 2 | |
| | 6 2 | | 5 2 | | 3+3 2 | |
| | 5 2 | | 6 2 | | 5+5 2 | |
| | 6 2 | | 5 2 | | 6+6 2 | |
| | 5 2 | | 5 2 | | 4+4 2 | |
| | 5 2 | | 6 2 | | 3+3 2 | |
| | 6 2 | | 5 2 | | 5+5 2 | |
| | 5 2 | | 5 2 | | 6+6 2 | |
| | 6 2 | | 5 2 | | 3+3 2 | |
| | 6 2 | | 6 2 | | 4+4 2 | |

Continued on next column

**TABLE XXXIX– continued**

| TQC | ICSD | | NoE | | SF | |
|---|---|---|---|---|---|---|
| 64.1.10 | 90341 | | 72 | | Ge2Ni1 | |
| | -GM | -T | -Y | -Z | -R | -S |
| | 5 2 | 3+4 4 | 5 2 | 3+4 4 | 5+5 2 | 2+2 4 |
| | 6 2 | 3+4 4 | 6 2 | 3+4 4 | 6+6 2 | 2+2 4 |
| | 5 2 | 3+4 4 | 6 2 | 3+4 4 | 3+3 2 | 2+2 4 |
| | 5 2 | 3+4 4 | 5 2 | 3+4 4 | 4+4 2 | 2+2 4 |
| | 5 2 | 3+4 4 | 5 2 | 3+4 4 | 5+5 2 | 2+2 4 |
| | 6 2 | 3+4 4 | 6 2 | 3+4 4 | 6+6 2 | 2+2 4 |
| | 6 2 | 3+4 4 | 6 2 | 3+4 4 | 4+4 2 | 2+2 4 |
| | 6 2 | 3+4 4 | 5 2 | 3+4 4 | 3+3 2 | 2+2 4 |
| | 5 2 | 3+4 4 | 5 2 | 3+4 4 | 5+5 2 | 2+2 4 |
| | 6 2 | 3+4 4 | 6 2 | 3+4 4 | 6+6 2 | 2+2 4 |
| | 6 2 | 3+4 4 | 6 2 | 3+4 4 | 3+3 2 | 2+2 4 |
| | 5 2 | 3+4 4 | 6 2 | 3+4 4 | 4+4 2 | 2+2 4 |
| | 6 2 | 3+4 4 | 6 2 | 3+4 4 | 6+6 2 | 2+2 4 |
| | 5 2 | 3+4 4 | 6 2 | 3+4 4 | 5+5 2 | 2+2 4 |
| | 6 2 | 3+4 4 | 6 2 | 3+4 4 | 4+4 2 | 2+2 4 |
| | 5 2 | 3+4 4 | 5 2 | 3+4 4 | 3+3 2 | 2+2 4 |
| | 5 2 | 3+4 4 | 5 2 | 3+4 4 | 4+4 2 | 2+2 4 |
| | 5 2 | 3+4 4 | 5 2 | 3+4 4 | 6+6 2 | 2+2 4 |
| | 5 2 | | 6 2 | | 3+3 2 | |
| | 6 2 | | 6 2 | | 5+5 2 | |
| | 5 2 | | 5 2 | | 3+3 2 | |
| | 5 2 | | 5 2 | | 4+4 2 | |
| | 6 2 | | 6 2 | | 5+5 2 | |
| | 5 2 | | 5 2 | | 6+6 2 | |
| | 6 2 | | 5 2 | | 3+3 2 | |
| | 6 2 | | 6 2 | | 6+6 2 | |
| | 5 2 | | 6 2 | | 4+4 2 | |
| | 6 2 | | 5 2 | | 5+5 2 | |
| | 5 2 | | 5 2 | | 3+3 2 | |
| | 5 2 | | 5 2 | | 4+4 2 | |
| | 6 2 | | 5 2 | | 5+5 2 | |
| | 6 2 | | 6 2 | | 6+6 2 | |
| | 5 2 | | 5 2 | | 5+5 2 | |
| | 6 2 | | 5 2 | | 6+6 2 | |
| | 6 2 | | 6 2 | | 3+3 2 | |
| | 5 2 | | 5 2 | | 4+4 2 | |
| 64.1.11 | 150569 | | 84 | | Ga1Sn2V2 | |
| | -GM | -T | -Y | -Z | -R | -S |
| | 5 2 | 3+4 4 | 6 2 | 3+4 4 | 3+3 2 | 2+2 4 |
| | 5 2 | 3+4 4 | 5 2 | 3+4 4 | 4+4 2 | 2+2 4 |
| | 6 2 | 3+4 4 | 6 2 | 3+4 4 | 5+5 2 | 2+2 4 |
| | 5 2 | 3+4 4 | 5 2 | 3+4 4 | 6+6 2 | 2+2 4 |
| | 6 2 | 3+4 4 | 6 2 | 3+4 4 | 4+4 2 | 2+2 4 |
| | 5 2 | 3+4 4 | 5 2 | 3+4 4 | 3+3 2 | 2+2 4 |
| | 6 2 | 3+4 4 | 6 2 | 3+4 4 | 6+6 2 | 2+2 4 |
| | 6 2 | 3+4 4 | 6 2 | 3+4 4 | 5+5 2 | 2+2 4 |
| | 6 2 | 3+4 4 | 5 2 | 3+4 4 | 5+5 2 | 2+2 4 |
| | 6 2 | 3+4 4 | 6 2 | 3+4 4 | 6+6 2 | 2+2 4 |
| | 5 2 | 3+4 4 | 5 2 | 3+4 4 | 4+4 2 | 2+2 4 |
| | 5 2 | 3+4 4 | 5 2 | 3+4 4 | 3+3 2 | 2+2 4 |
| | 6 2 | 3+4 4 | 6 2 | 3+4 4 | 3+3 2 | 2+2 4 |
| | 5 2 | 3+4 4 | 5 2 | 3+4 4 | 4+4 2 | 2+2 4 |
| | 5 2 | 3+4 4 | 5 2 | 3+4 4 | 5+5 2 | 2+2 4 |
| | 5 2 | 3+4 4 | 5 2 | 3+4 4 | 6+6 2 | 2+2 4 |
| | 6 2 | 3+4 4 | 6 2 | 3+4 4 | 4+4 2 | 2+2 4 |
| | 5 2 | 3+4 4 | 5 2 | 3+4 4 | 6+6 2 | 2+2 4 |
| | 6 2 | 3+4 4 | 5 2 | 3+4 4 | 3+3 2 | 2+2 4 |
| | 6 2 | 3+4 4 | 6 2 | 3+4 4 | 5+5 2 | 2+2 4 |
| | 6 2 | 3+4 4 | 6 2 | 3+4 4 | 5+5 2 | 2+2 4 |

Continued on next column

**TABLE XXXIX– continued**

| TQC | ICSD | | NoE | | SF | |
|---|---|---|---|---|---|---|
| | 5 2 | | 6 2 | | 6+6 2 | |
| | 5 2 | | 6 2 | | 3+3 2 | |
| | 6 2 | | 6 2 | | 4+4 2 | |
| | 5 2 | | 6 2 | | 5+5 2 | |
| | 5 2 | | 6 2 | | 6+6 2 | |
| | 6 2 | | 6 2 | | 4+4 2 | |
| | 6 2 | | 5 2 | | 3+3 2 | |
| | 6 2 | | 5 2 | | 3+3 2 | |
| | 5 2 | | 5 2 | | 4+4 2 | |
| | 5 2 | | 6 2 | | 6+6 2 | |
| | 5 2 | | 5 2 | | 5+5 2 | |
| | 5 2 | | 6 2 | | 3+3 2 | |
| | 6 2 | | 5 2 | | 4+4 2 | |
| | 6 2 | | 6 2 | | 6+6 2 | |
| | 5 2 | | 5 2 | | 5+5 2 | |
| | 6 2 | | 6 2 | | 5+5 2 | |
| | 5 2 | | 5 2 | | 6+6 2 | |
| | 5 2 | | 6 2 | | 3+3 2 | |
| | 5 2 | | 5 2 | | 4+4 2 | |
| | 6 2 | | 5 2 | | 4+4 2 | |
| | 6 2 | | 5 2 | | 3+3 2 | |
| 64.1.12 | 106752 | | 84 | | Ga1Sn2V2 | |
| | -GM | -T | -Y | -Z | -R | -S |
| | 5 2 | 3+4 4 | 6 2 | 3+4 4 | 5+5 2 | 2+2 4 |
| | 5 2 | 3+4 4 | 5 2 | 3+4 4 | 6+6 2 | 2+2 4 |
| | 6 2 | 3+4 4 | 6 2 | 3+4 4 | 3+3 2 | 2+2 4 |
| | 5 2 | 3+4 4 | 5 2 | 3+4 4 | 4+4 2 | 2+2 4 |
| | 6 2 | 3+4 4 | 6 2 | 3+4 4 | 6+6 2 | 2+2 4 |
| | 5 2 | 3+4 4 | 5 2 | 3+4 4 | 5+5 2 | 2+2 4 |
| | 6 2 | 3+4 4 | 6 2 | 3+4 4 | 4+4 2 | 2+2 4 |
| | 6 2 | 3+4 4 | 6 2 | 3+4 4 | 3+3 2 | 2+2 4 |
| | 6 2 | 3+4 4 | 5 2 | 3+4 4 | 3+3 2 | 2+2 4 |
| | 6 2 | 3+4 4 | 6 2 | 3+4 4 | 4+4 2 | 2+2 4 |
| | 5 2 | 3+4 4 | 5 2 | 3+4 4 | 6+6 2 | 2+2 4 |
| | 5 2 | 3+4 4 | 5 2 | 3+4 4 | 5+5 2 | 2+2 4 |
| | 6 2 | 3+4 4 | 6 2 | 3+4 4 | 5+5 2 | 2+2 4 |
| | 5 2 | 3+4 4 | 5 2 | 3+4 4 | 6+6 2 | 2+2 4 |
| | 5 2 | 3+4 4 | 5 2 | 3+4 4 | 3+3 2 | 2+2 4 |
| | 5 2 | 3+4 4 | 5 2 | 3+4 4 | 4+4 2 | 2+2 4 |
| | 6 2 | 3+4 4 | 6 2 | 3+4 4 | 6+6 2 | 2+2 4 |
| | 5 2 | 3+4 4 | 5 2 | 3+4 4 | 4+4 2 | 2+2 4 |
| | 6 2 | 3+4 4 | 5 2 | 3+4 4 | 3+3 2 | 2+2 4 |
| | 6 2 | 3+4 4 | 6 2 | 3+4 4 | 5+5 2 | 2+2 4 |
| | 6 2 | 3+4 4 | 6 2 | 3+4 4 | 3+3 2 | 2+2 4 |
| | 5 2 | | 6 2 | | 4+4 2 | |
| | 5 2 | | 6 2 | | 5+5 2 | |
| | 6 2 | | 6 2 | | 6+6 2 | |
| | 5 2 | | 6 2 | | 3+3 2 | |
| | 5 2 | | 6 2 | | 4+4 2 | |
| | 6 2 | | 6 2 | | 6+6 2 | |
| | 6 2 | | 5 2 | | 5+5 2 | |
| | 6 2 | | 5 2 | | 5+5 2 | |
| | 5 2 | | 5 2 | | 6+6 2 | |
| | 5 2 | | 6 2 | | 4+4 2 | |
| | 5 2 | | 5 2 | | 3+3 2 | |
| | 5 2 | | 6 2 | | 5+5 2 | |
| | 6 2 | | 5 2 | | 6+6 2 | |
| | 6 2 | | 6 2 | | 4+4 2 | |
| | 5 2 | | 5 2 | | 3+3 2 | |
| | 6 2 | | 6 2 | | 3+3 2 | |
| | 5 2 | | 5 2 | | 4+4 2 | |

Continued on next column

**TABLE XXXIX– continued**

| TQC | ICSD | | NoE | | SF | |
|---|---|---|---|---|---|---|
| | 5 2 | | 6 2 | | 5+5 2 | |
| | 5 2 | | 5 2 | | 6+6 2 | |
| | 6 2 | | 5 2 | | 6+6 2 | |
| | 6 2 | | 5 2 | | 5+5 2 | |
| 64.1.13 | 71196 | | 136 | | Co2Nb2Te4 | |
| | -GM | -T | -Y | -Z | -R | -S |
| | 5 2 | 3+4 4 | 5 2 | 3+4 4 | 3+3 2 | 2+2 4 |
| | 6 2 | 3+4 4 | 6 2 | 3+4 4 | 4+4 2 | 2+2 4 |
| | 5 2 | 3+4 4 | 5 2 | 3+4 4 | 5+5 2 | 2+2 4 |
| | 6 2 | 3+4 4 | 6 2 | 3+4 4 | 6+6 2 | 2+2 4 |
| | 5 2 | 3+4 4 | 5 2 | 3+4 4 | 3+3 2 | 2+2 4 |
| | 6 2 | 3+4 4 | 6 2 | 3+4 4 | 4+4 2 | 2+2 4 |
| | 5 2 | 3+4 4 | 5 2 | 3+4 4 | 5+5 2 | 2+2 4 |
| | 6 2 | 3+4 4 | 6 2 | 3+4 4 | 6+6 2 | 2+2 4 |
| | 5 2 | 3+4 4 | 5 2 | 3+4 4 | 3+3 2 | 2+2 4 |
| | 6 2 | 3+4 4 | 6 2 | 3+4 4 | 5+5 2 | 2+2 4 |
| | 5 2 | 3+4 4 | 5 2 | 3+4 4 | 4+4 2 | 2+2 4 |
| | 5 2 | 3+4 4 | 5 2 | 3+4 4 | 6+6 2 | 2+2 4 |
| | 6 2 | 3+4 4 | 6 2 | 3+4 4 | 4+4 2 | 2+2 4 |
| | 6 2 | 3+4 4 | 6 2 | 3+4 4 | 3+3 2 | 2+2 4 |
| | 5 2 | 3+4 4 | 5 2 | 3+4 4 | 6+6 2 | 2+2 4 |
| | 6 2 | 3+4 4 | 6 2 | 3+4 4 | 5+5 2 | 2+2 4 |
| | 5 2 | 3+4 4 | 5 2 | 3+4 4 | 6+6 2 | 2+2 4 |
| | 6 2 | 3+4 4 | 6 2 | 3+4 4 | 5+5 2 | 2+2 4 |
| | 6 2 | 3+4 4 | 5 2 | 3+4 4 | 3+3 2 | 2+2 4 |
| | 5 2 | 3+4 4 | 6 2 | 3+4 4 | 4+4 2 | 2+2 4 |
| | 6 2 | 3+4 4 | 6 2 | 3+4 4 | 3+3 2 | 2+2 4 |
| | 6 2 | 3+4 4 | 5 2 | 3+4 4 | 4+4 2 | 2+2 4 |
| | 5 2 | 3+4 4 | 6 2 | 3+4 4 | 5+5 2 | 2+2 4 |
| | 5 2 | 3+4 4 | 5 2 | 3+4 4 | 6+6 2 | 2+2 4 |
| | 5 2 | 3+4 4 | 5 2 | 3+4 4 | 6+6 2 | 2+2 4 |
| | 5 2 | 3+4 4 | 6 2 | 3+4 4 | 5+5 2 | 2+2 4 |
| | 6 2 | 3+4 4 | 6 2 | 3+4 4 | 3+3 2 | 2+2 4 |
| | 6 2 | 3+4 4 | 5 2 | 3+4 4 | 4+4 2 | 2+2 4 |
| | 5 2 | 3+4 4 | 6 2 | 3+4 4 | 6+6 2 | 2+2 4 |
| | 6 2 | 3+4 4 | 5 2 | 3+4 4 | 5+5 2 | 2+2 4 |
| | 5 2 | 3+4 4 | 6 2 | 3+4 4 | 3+3 2 | 2+2 4 |
| | 6 2 | 3+4 4 | 6 2 | 3+4 4 | 4+4 2 | 2+2 4 |
| | 6 2 | 3+4 4 | 5 2 | 3+4 4 | 5+5 2 | 2+2 4 |
| | 5 2 | 3+4 4 | 6 2 | 3+4 4 | 6+6 2 | 2+2 4 |
| | 5 2 | | 5 2 | | 4+4 2 | |
| | 6 2 | | 6 2 | | 3+3 2 | |
| | 6 2 | | 5 2 | | 4+4 2 | |
| | 5 2 | | 5 2 | | 3+3 2 | |
| | 5 2 | | 5 2 | | 5+5 2 | |
| | 6 2 | | 6 2 | | 6+6 2 | |
| | 6 2 | | 6 2 | | 4+4 2 | |
| | 5 2 | | 6 2 | | 3+3 2 | |
| | 6 2 | | 5 2 | | 5+5 2 | |
| | 5 2 | | 5 2 | | 6+6 2 | |
| | 5 2 | | 5 2 | | 3+3 2 | |
| | 6 2 | | 5 2 | | 4+4 2 | |
| | 5 2 | | 6 2 | | 6+6 2 | |
| | 6 2 | | 6 2 | | 5+5 2 | |
| | 5 2 | | 6 2 | | 4+4 2 | |
| | 6 2 | | 6 2 | | 3+3 2 | |
| | 6 2 | | 5 2 | | 6+6 2 | |
| | 5 2 | | 6 2 | | 5+5 2 | |
| | 6 2 | | 6 2 | | 3+3 2 | |
| | 6 2 | | 5 2 | | 4+4 2 | |
| | 5 2 | | 5 2 | | 6+6 2 | |

Continued on next column

**TABLE XXXIX– continued**

| TQC | ICSD | | NoE | | SF | |
|---|---|---|---|---|---|---|
| | 5 2 | | 5 2 | | 5+5 2 | |
| | 5 2 | | 5 2 | | 6+6 2 | |
| | 5 2 | | 6 2 | | 5+5 2 | |
| | 5 2 | | 5 2 | | 4+4 2 | |
| | 6 2 | | 5 2 | | 6+6 2 | |
| | 5 2 | | 5 2 | | 3+3 2 | |
| | 5 2 | | 5 2 | | 5+5 2 | |
| | 5 2 | | 6 2 | | 6+6 2 | |
| | 6 2 | | 5 2 | | 5+5 2 | |
| | 6 2 | | 6 2 | | 4+4 2 | |
| | 5 2 | | 6 2 | | 3+3 2 | |
| | 5 2 | | 6 2 | | 3+3 2 | |
| | 6 2 | | 6 2 | | 4+4 2 | |
| 64.1.14 | 71519 | | 136 | | Co1Nb1Te2 | |
| | -GM | -T | -Y | -Z | -R | -S |
| | 5 2 | 3+4 4 | 5 2 | 3+4 4 | 3+3 2 | 2+2 4 |
| | 6 2 | 3+4 4 | 6 2 | 3+4 4 | 4+4 2 | 2+2 4 |
| | 5 2 | 3+4 4 | 5 2 | 3+4 4 | 5+5 2 | 2+2 4 |
| | 6 2 | 3+4 4 | 6 2 | 3+4 4 | 6+6 2 | 2+2 4 |
| | 5 2 | 3+4 4 | 5 2 | 3+4 4 | 3+3 2 | 2+2 4 |
| | 6 2 | 3+4 4 | 6 2 | 3+4 4 | 4+4 2 | 2+2 4 |
| | 5 2 | 3+4 4 | 5 2 | 3+4 4 | 5+5 2 | 2+2 4 |
| | 6 2 | 3+4 4 | 6 2 | 3+4 4 | 6+6 2 | 2+2 4 |
| | 5 2 | 3+4 4 | 5 2 | 3+4 4 | 3+3 2 | 2+2 4 |
| | 6 2 | 3+4 4 | 6 2 | 3+4 4 | 5+5 2 | 2+2 4 |
| | 5 2 | 3+4 4 | 5 2 | 3+4 4 | 4+4 2 | 2+2 4 |
| | 5 2 | 3+4 4 | 5 2 | 3+4 4 | 6+6 2 | 2+2 4 |
| | 6 2 | 3+4 4 | 6 2 | 3+4 4 | 4+4 2 | 2+2 4 |
| | 6 2 | 3+4 4 | 6 2 | 3+4 4 | 3+3 2 | 2+2 4 |
| | 5 2 | 3+4 4 | 5 2 | 3+4 4 | 6+6 2 | 2+2 4 |
| | 6 2 | 3+4 4 | 6 2 | 3+4 4 | 5+5 2 | 2+2 4 |
| | 5 2 | 3+4 4 | 5 2 | 3+4 4 | 6+6 2 | 2+2 4 |
| | 6 2 | 3+4 4 | 6 2 | 3+4 4 | 5+5 2 | 2+2 4 |
| | 6 2 | 3+4 4 | 5 2 | 3+4 4 | 3+3 2 | 2+2 4 |
| | 5 2 | 3+4 4 | 6 2 | 3+4 4 | 4+4 2 | 2+2 4 |
| | 6 2 | 3+4 4 | 6 2 | 3+4 4 | 3+3 2 | 2+2 4 |
| | 6 2 | 3+4 4 | 5 2 | 3+4 4 | 4+4 2 | 2+2 4 |
| | 5 2 | 3+4 4 | 6 2 | 3+4 4 | 5+5 2 | 2+2 4 |
| | 5 2 | 3+4 4 | 5 2 | 3+4 4 | 6+6 2 | 2+2 4 |
| | 5 2 | 3+4 4 | 5 2 | 3+4 4 | 6+6 2 | 2+2 4 |
| | 5 2 | 3+4 4 | 6 2 | 3+4 4 | 5+5 2 | 2+2 4 |
| | 6 2 | 3+4 4 | 6 2 | 3+4 4 | 3+3 2 | 2+2 4 |
| | 6 2 | 3+4 4 | 5 2 | 3+4 4 | 4+4 2 | 2+2 4 |
| | 5 2 | 3+4 4 | 6 2 | 3+4 4 | 6+6 2 | 2+2 4 |
| | 6 2 | 3+4 4 | 5 2 | 3+4 4 | 5+5 2 | 2+2 4 |
| | 6 2 | 3+4 4 | 6 2 | 3+4 4 | 3+3 2 | 2+2 4 |
| | 5 2 | 3+4 4 | 6 2 | 3+4 4 | 4+4 2 | 2+2 4 |
| | 6 2 | 3+4 4 | 5 2 | 3+4 4 | 5+5 2 | 2+2 4 |
| | 5 2 | 3+4 4 | 6 2 | 3+4 4 | 6+6 2 | 2+2 4 |
| | 5 2 | | 5 2 | | 4+4 2 | |
| | 6 2 | | 6 2 | | 3+3 2 | |
| | 6 2 | | 5 2 | | 4+4 2 | |
| | 5 2 | | 5 2 | | 3+3 2 | |
| | 6 2 | | 5 2 | | 5+5 2 | |
| | 5 2 | | 6 2 | | 6+6 2 | |
| | 6 2 | | 6 2 | | 4+4 2 | |
| | 5 2 | | 6 2 | | 3+3 2 | |
| | 6 2 | | 5 2 | | 5+5 2 | |
| | 5 2 | | 5 2 | | 6+6 2 | |
| | 5 2 | | 5 2 | | 3+3 2 | |
| | 6 2 | | 5 2 | | 4+4 2 | |

Continued on next column

**TABLE XXXIX– continued**

| TQC | ICSD | | NoE | | SF | |
|---|---|---|---|---|---|---|
| | 5 2 | | 6 2 | | 6+6 2 | |
| | 6 2 | | 6 2 | | 5+5 2 | |
| | 5 2 | | 6 2 | | 4+4 2 | |
| | 6 2 | | 6 2 | | 3+3 2 | |
| | 6 2 | | 5 2 | | 6+6 2 | |
| | 5 2 | | 6 2 | | 5+5 2 | |
| | 6 2 | | 6 2 | | 3+3 2 | |
| | 6 2 | | 5 2 | | 4+4 2 | |
| | 5 2 | | 5 2 | | 6+6 2 | |
| | 5 2 | | 5 2 | | 5+5 2 | |
| | 5 2 | | 5 2 | | 6+6 2 | |
| | 5 2 | | 6 2 | | 5+5 2 | |
| | 6 2 | | 5 2 | | 4+4 2 | |
| | 5 2 | | 5 2 | | 6+6 2 | |
| | 5 2 | | 5 2 | | 3+3 2 | |
| | 5 2 | | 5 2 | | 5+5 2 | |
| | 5 2 | | 6 2 | | 6+6 2 | |
| | 6 2 | | 5 2 | | 5+5 2 | |
| | 6 2 | | 6 2 | | 4+4 2 | |
| | 5 2 | | 6 2 | | 3+3 2 | |
| | 5 2 | | 6 2 | | 3+3 2 | |
| | 6 2 | | 6 2 | | 4+4 2 | |
| 64.1.15 | 424937 | | 120 | | Co1Cs1O2 | |
| | -GM | -T | -Y | -Z | -R | -S |
| | 5 2 | 3+4 4 | 6 2 | 3+4 4 | 5+5 2 | 2+2 4 |
| | 5 2 | 3+4 4 | 6 2 | 3+4 4 | 6+6 2 | 2+2 4 |
| | 6 2 | 3+4 4 | 5 2 | 3+4 4 | 3+3 2 | 2+2 4 |
| | 6 2 | 3+4 4 | 5 2 | 3+4 4 | 4+4 2 | 2+2 4 |
| | 5 2 | 3+4 4 | 5 2 | 3+4 4 | 3+3 2 | 2+2 4 |
| | 6 2 | 3+4 4 | 6 2 | 3+4 4 | 4+4 2 | 2+2 4 |
| | 6 2 | 3+4 4 | 6 2 | 3+4 4 | 6+6 2 | 2+2 4 |
| | 5 2 | 3+4 4 | 5 2 | 3+4 4 | 5+5 2 | 2+2 4 |
| | 5 2 | 3+4 4 | 5 2 | 3+4 4 | 3+3 2 | 2+2 4 |
| | 5 2 | 3+4 4 | 5 2 | 3+4 4 | 4+4 2 | 2+2 4 |
| | 6 2 | 3+4 4 | 6 2 | 3+4 4 | 5+5 2 | 2+2 4 |
| | 6 2 | 3+4 4 | 6 2 | 3+4 4 | 6+6 2 | 2+2 4 |
| | 5 2 | 3+4 4 | 6 2 | 3+4 4 | 5+5 2 | 2+2 4 |
| | 5 2 | 3+4 4 | 6 2 | 3+4 4 | 6+6 2 | 2+2 4 |
| | 6 2 | 3+4 4 | 5 2 | 3+4 4 | 4+4 2 | 2+2 4 |
| | 6 2 | 3+4 4 | 5 2 | 3+4 4 | 3+3 2 | 2+2 4 |
| | 5 2 | 3+4 4 | 6 2 | 3+4 4 | 6+6 2 | 2+2 4 |
| | 5 2 | 3+4 4 | 6 2 | 3+4 4 | 5+5 2 | 2+2 4 |
| | 6 2 | 3+4 4 | 5 2 | 3+4 4 | 3+3 2 | 2+2 4 |
| | 5 2 | 3+4 4 | 5 2 | 3+4 4 | 4+4 2 | 2+2 4 |
| | 5 2 | 3+4 4 | 6 2 | 3+4 4 | 5+5 2 | 2+2 4 |
| | 6 2 | 3+4 4 | 6 2 | 3+4 4 | 6+6 2 | 2+2 4 |
| | 6 2 | 3+4 4 | 5 2 | 3+4 4 | 3+3 2 | 2+2 4 |
| | 6 2 | 3+4 4 | 5 2 | 3+4 4 | 4+4 2 | 2+2 4 |
| | 5 2 | 3+4 4 | 5 2 | 3+4 4 | 3+3 2 | 2+2 4 |
| | 6 2 | 3+4 4 | 6 2 | 3+4 4 | 4+4 2 | 2+2 4 |
| | 6 2 | 3+4 4 | 6 2 | 3+4 4 | 5+5 2 | 2+2 4 |
| | 5 2 | 3+4 4 | 5 2 | 3+4 4 | 6+6 2 | 2+2 4 |
| | 5 2 | 3+4 4 | 5 2 | 3+4 4 | 3+3 2 | 2+2 4 |
| | 5 2 | 3+4 4 | 5 2 | 3+4 4 | 4+4 2 | 2+2 4 |
| | 6 2 | | 6 2 | | 5+5 2 | |
| | 6 2 | | 6 2 | | 6+6 2 | |
| | 6 2 | | 6 2 | | 4+4 2 | |
| | 5 2 | | 5 2 | | 3+3 2 | |
| | 5 2 | | 5 2 | | 6+6 2 | |
| | 6 2 | | 6 2 | | 5+5 2 | |
| | 5 2 | | 5 2 | | 3+3 2 | |

Continued on next column

**TABLE XXXIX– continued**

| TQC | ICSD | | NoE | | SF | |
|---|---|---|---|---|---|---|
| | 6 2 | | 5 2 | | 4+4 2 | |
| | 5 2 | | 5 2 | | 6+6 2 | |
| | 5 2 | | 6 2 | | 5+5 2 | |
| | 6 2 | | 6 2 | | 5+5 2 | |
| | 6 2 | | 6 2 | | 6+6 2 | |
| | 6 2 | | 5 2 | | 4+4 2 | |
| | 5 2 | | 6 2 | | 3+3 2 | |
| | 6 2 | | 6 2 | | 5+5 2 | |
| | 5 2 | | 5 2 | | 6+6 2 | |
| | 6 2 | | 6 2 | | 4+4 2 | |
| | 5 2 | | 5 2 | | 3+3 2 | |
| | 5 2 | | 5 2 | | 4+4 2 | |
| | 5 2 | | 5 2 | | 3+3 2 | |
| | 6 2 | | 6 2 | | 5+5 2 | |
| | 6 2 | | 6 2 | | 6+6 2 | |
| | 5 2 | | 5 2 | | 4+4 2 | |
| | 5 2 | | 5 2 | | 3+3 2 | |
| | 6 2 | | 6 2 | | 6+6 2 | |
| | 5 2 | | 5 2 | | 5+5 2 | |
| | 5 2 | | 5 2 | | 3+3 2 | |
| | 6 2 | | 6 2 | | 4+4 2 | |
| | 6 2 | | 6 2 | | 5+5 2 | |
| | 5 2 | | 5 2 | | 6+6 2 | |
| 64.1.16 | 40372 | | 72 | | Ca5Ga2N4 | |
| | -GM | -T | -Y | -Z | -R | -S |
| | 5 2 | 3+4 4 | 5 2 | 3+4 4 | 5+5 2 | 2+2 4 |
| | 5 2 | 3+4 4 | 6 2 | 3+4 4 | 6+6 2 | 2+2 4 |
| | 6 2 | 3+4 4 | 5 2 | 3+4 4 | 3+3 2 | 2+2 4 |
| | 6 2 | 3+4 4 | 6 2 | 3+4 4 | 4+4 2 | 2+2 4 |
| | 5 2 | 3+4 4 | 5 2 | 3+4 4 | 5+5 2 | 2+2 4 |
| | 6 2 | 3+4 4 | 6 2 | 3+4 4 | 6+6 2 | 2+2 4 |
| | 5 2 | 3+4 4 | 5 2 | 3+4 4 | 3+3 2 | 2+2 4 |
| | 6 2 | 3+4 4 | 6 2 | 3+4 4 | 4+4 2 | 2+2 4 |
| | 5 2 | 3+4 4 | 5 2 | 3+4 4 | 5+5 2 | 2+2 4 |
| | 6 2 | 3+4 4 | 6 2 | 3+4 4 | 6+6 2 | 2+2 4 |
| | 5 2 | 3+4 4 | 5 2 | 3+4 4 | 3+3 2 | 2+2 4 |
| | 6 2 | 3+4 4 | 6 2 | 3+4 4 | 4+4 2 | 2+2 4 |
| | 5 2 | 3+4 4 | 6 2 | 3+4 4 | 3+3 2 | 2+2 4 |
| | 6 2 | 3+4 4 | 5 2 | 3+4 4 | 5+5 2 | 2+2 4 |
| | 5 2 | 3+4 4 | 5 2 | 3+4 4 | 6+6 2 | 2+2 4 |
| | 5 2 | 3+4 4 | 5 2 | 3+4 4 | 4+4 2 | 2+2 4 |
| | 6 2 | 3+4 4 | 6 2 | 3+4 4 | 4+4 2 | 2+2 4 |
| | 6 2 | 3+4 4 | 5 2 | 3+4 4 | 3+3 2 | 2+2 4 |
| | 5 2 | | 6 2 | | 6+6 2 | |
| | 6 2 | | 5 2 | | 5+5 2 | |
| | 5 2 | | 6 2 | | 5+5 2 | |
| | 5 2 | | 6 2 | | 6+6 2 | |
| | 5 2 | | 5 2 | | 3+3 2 | |
| | 6 2 | | 6 2 | | 5+5 2 | |
| | 5 2 | | 5 2 | | 4+4 2 | |
| | 6 2 | | 6 2 | | 6+6 2 | |
| | 5 2 | | 6 2 | | 4+4 2 | |
| | 5 2 | | 5 2 | | 3+3 2 | |
| | 6 2 | | 5 2 | | 5+5 2 | |
| | 6 2 | | 5 2 | | 6+6 2 | |
| | 5 2 | | 5 2 | | 3+3 2 | |
| | 6 2 | | 6 2 | | 4+4 2 | |
| | 6 2 | | 6 2 | | 5+5 2 | |
| | 6 2 | | 6 2 | | 6+6 2 | |
| | 5 2 | | 5 2 | | 4+4 2 | |
| | 5 2 | | 5 2 | | 3+3 2 | |
| 64.1.17 | 103352 | | 192 | | Ca2Mn1O3.5 | |
| | -GM | -T | -Y | -Z | -R | -S |
| | 5 2 | 3+4 4 | 6 2 | 3+4 4 | 3+3 2 | 2+2 4 |
| | 5 2 | 3+4 4 | 6 2 | 3+4 4 | 4+4 2 | 2+2 4 |
| | 6 2 | 3+4 4 | 5 2 | 3+4 4 | 5+5 2 | 2+2 4 |
| | 6 2 | 3+4 4 | 5 2 | 3+4 4 | 6+6 2 | 2+2 4 |
| | 6 2 | 3+4 4 | 5 2 | 3+4 4 | 5+5 2 | 2+2 4 |
| | 5 2 | 3+4 4 | 6 2 | 3+4 4 | 6+6 2 | 2+2 4 |
| | 5 2 | 3+4 4 | 6 2 | 3+4 4 | 3+3 2 | 2+2 4 |
| | 6 2 | 3+4 4 | 5 2 | 3+4 4 | 4+4 2 | 2+2 4 |
| | 5 2 | 3+4 4 | 6 2 | 3+4 4 | 4+4 2 | 2+2 4 |
| | 5 2 | 3+4 4 | 5 2 | 3+4 4 | 3+3 2 | 2+2 4 |
| | 6 2 | 3+4 4 | 5 2 | 3+4 4 | 6+6 2 | 2+2 4 |
| | 6 2 | 3+4 4 | 5 2 | 3+4 4 | 5+5 2 | 2+2 4 |
| | 5 2 | 3+4 4 | 6 2 | 3+4 4 | 3+3 2 | 2+2 4 |
| | 5 2 | 3+4 4 | 5 2 | 3+4 4 | 4+4 2 | 2+2 4 |
| | 6 2 | 3+4 4 | 6 2 | 3+4 4 | 6+6 2 | 2+2 4 |
| | 6 2 | 3+4 4 | 6 2 | 3+4 4 | 5+5 2 | 2+2 4 |
| | 6 2 | 3+4 4 | 6 2 | 3+4 4 | 5+5 2 | 2+2 4 |
| | 5 2 | 3+4 4 | 5 2 | 3+4 4 | 6+6 2 | 2+2 4 |
| | 6 2 | 3+4 4 | 6 2 | 3+4 4 | 4+4 2 | 2+2 4 |
| | 5 2 | 3+4 4 | 5 2 | 3+4 4 | 3+3 2 | 2+2 4 |
| | 5 2 | 3+4 4 | 6 2 | 3+4 4 | 4+4 2 | 2+2 4 |
| | 6 2 | 3+4 4 | 6 2 | 3+4 4 | 3+3 2 | 2+2 4 |
| | 5 2 | 3+4 4 | 5 2 | 3+4 4 | 6+6 2 | 2+2 4 |
| | 5 2 | 3+4 4 | 6 2 | 3+4 4 | 5+5 2 | 2+2 4 |
| | 6 2 | 3+4 4 | 5 2 | 3+4 4 | 6+6 2 | 2+2 4 |
| | 5 2 | 3+4 4 | 5 2 | 3+4 4 | 5+5 2 | 2+2 4 |
| | 6 2 | 3+4 4 | 6 2 | 3+4 4 | 3+3 2 | 2+2 4 |
| | 5 2 | 3+4 4 | 5 2 | 3+4 4 | 4+4 2 | 2+2 4 |
| | 6 2 | 3+4 4 | 6 2 | 3+4 4 | 3+3 2 | 2+2 4 |
| | 6 2 | 3+4 4 | 5 2 | 3+4 4 | 4+4 2 | 2+2 4 |
| | 5 2 | 3+4 4 | 6 2 | 3+4 4 | 5+5 2 | 2+2 4 |
| | 6 2 | 3+4 4 | 5 2 | 3+4 4 | 6+6 2 | 2+2 4 |
| | 5 2 | 3+4 4 | 5 2 | 3+4 4 | 3+3 2 | 2+2 4 |
| | 6 2 | 3+4 4 | 6 2 | 3+4 4 | 4+4 2 | 2+2 4 |
| | 5 2 | 3+4 4 | 5 2 | 3+4 4 | 5+5 2 | 2+2 4 |
| | 6 2 | 3+4 4 | 6 2 | 3+4 4 | 6+6 2 | 2+2 4 |
| | 5 2 | 3+4 4 | 5 2 | 3+4 4 | 3+3 2 | 2+2 4 |
| | 5 2 | 3+4 4 | 5 2 | 3+4 4 | 4+4 2 | 2+2 4 |
| | 6 2 | 3+4 4 | 6 2 | 3+4 4 | 5+5 2 | 2+2 4 |
| | 6 2 | 3+4 4 | 5 2 | 3+4 4 | 6+6 2 | 2+2 4 |
| | 5 2 | 3+4 4 | 6 2 | 3+4 4 | 3+3 2 | 2+2 4 |
| | 5 2 | 3+4 4 | 5 2 | 3+4 4 | 4+4 2 | 2+2 4 |
| | 5 2 | 3+4 4 | 5 2 | 3+4 4 | 5+5 2 | 2+2 4 |
| | 6 2 | 3+4 4 | 6 2 | 3+4 4 | 6+6 2 | 2+2 4 |
| | 6 2 | 3+4 4 | 6 2 | 3+4 4 | 3+3 2 | 2+2 4 |
| | 5 2 | 3+4 4 | 5 2 | 3+4 4 | 4+4 2 | 2+2 4 |
| | 6 2 | 3+4 4 | 6 2 | 3+4 4 | 5+5 2 | 2+2 4 |
| | 5 2 | 3+4 4 | 6 2 | 3+4 4 | 6+6 2 | 2+2 4 |
| | 6 2 | | 5 2 | | 3+3 2 | |
| | 5 2 | | 6 2 | | 4+4 2 | |
| | 6 2 | | 5 2 | | 5+5 2 | |
| | 6 2 | | 5 2 | | 6+6 2 | |
| | 5 2 | | 6 2 | | 3+3 2 | |
| | 5 2 | | 6 2 | | 4+4 2 | |
| | 6 2 | | 6 2 | | 5+5 2 | |
| | 6 2 | | 5 2 | | 6+6 2 | |
| | 5 2 | | 5 2 | | 5+5 2 | |
| | 6 2 | | 6 2 | | 6+6 2 | |
| | 5 2 | | 6 2 | | 3+3 2 | |

Continued on next column

Continued on next column

**TABLE XXXIX– continued**

| TQC | ICSD | | NoE | | SF | |
|---|---|---|---|---|---|---|
| | 6 2 | | 6 2 | | 4+4 2 | |
| | 6 2 | | 6 2 | | 5+5 2 | |
| | 6 2 | | 5 2 | | 6+6 2 | |
| | 5 2 | | 6 2 | | 3+3 2 | |
| | 6 2 | | 5 2 | | 4+4 2 | |
| | 6 2 | | 6 2 | | 5+5 2 | |
| | 5 2 | | 5 2 | | 6+6 2 | |
| | 6 2 | | 5 2 | | 3+3 2 | |
| | 6 2 | | 6 2 | | 4+4 2 | |
| | 5 2 | | 6 2 | | 5+5 2 | |
| | 6 2 | | 6 2 | | 6+6 2 | |
| | 5 2 | | 5 2 | | 3+3 2 | |
| | 5 2 | | 6 2 | | 4+4 2 | |
| | 6 2 | | 5 2 | | 5+5 2 | |
| | 5 2 | | 5 2 | | 6+6 2 | |
| | 5 2 | | 5 2 | | 5+5 2 | |
| | 5 2 | | 5 2 | | 6+6 2 | |
| | 6 2 | | 6 2 | | 3+3 2 | |
| | 6 2 | | 5 2 | | 4+4 2 | |
| | 5 2 | | 6 2 | | 4+4 2 | |
| | 5 2 | | 6 2 | | 3+3 2 | |
| | 5 2 | | 6 2 | | 5+5 2 | |
| | 6 2 | | 5 2 | | 6+6 2 | |
| | 5 2 | | 6 2 | | 5+5 2 | |
| | 6 2 | | 6 2 | | 6+6 2 | |
| | 6 2 | | 5 2 | | 3+3 2 | |
| | 6 2 | | 6 2 | | 4+4 2 | |
| | 6 2 | | 5 2 | | 5+5 2 | |
| | 6 2 | | 6 2 | | 6+6 2 | |
| | 5 2 | | 5 2 | | 3+3 2 | |
| | 5 2 | | 5 2 | | 4+4 2 | |
| | 5 2 | | 6 2 | | 3+3 2 | |
| | 6 2 | | 6 2 | | 4+4 2 | |
| | 5 2 | | 5 2 | | 5+5 2 | |
| | 6 2 | | 6 2 | | 6+6 2 | |
| | 6 2 | | 5 2 | | 3+3 2 | |
| | 5 2 | | 5 2 | | 4+4 2 | |
| 64.1.18 | 186936 | | 172 | | Ba2Ge4Pd5 | |
| | -GM | -T | -Y | -Z | -R | -S |
| | 5 2 | 3+4 4 | 5 2 | 3+4 4 | 5+5 2 | 2+2 4 |
| | 6 2 | 3+4 4 | 6 2 | 3+4 4 | 6+6 2 | 2+2 4 |
| | 5 2 | 3+4 4 | 5 2 | 3+4 4 | 3+3 2 | 2+2 4 |
| | 6 2 | 3+4 4 | 6 2 | 3+4 4 | 4+4 2 | 2+2 4 |
| | 5 2 | 3+4 4 | 5 2 | 3+4 4 | 6+6 2 | 2+2 4 |
| | 6 2 | 3+4 4 | 6 2 | 3+4 4 | 5+5 2 | 2+2 4 |
| | 5 2 | 3+4 4 | 5 2 | 3+4 4 | 3+3 2 | 2+2 4 |
| | 6 2 | 3+4 4 | 6 2 | 3+4 4 | 4+4 2 | 2+2 4 |
| | 5 2 | 3+4 4 | 5 2 | 3+4 4 | 5+5 2 | 2+2 4 |
| | 5 2 | 3+4 4 | 6 2 | 3+4 4 | 6+6 2 | 2+2 4 |
| | 6 2 | 3+4 4 | 5 2 | 3+4 4 | 3+3 2 | 2+2 4 |
| | 6 2 | 3+4 4 | 6 2 | 3+4 4 | 4+4 2 | 2+2 4 |
| | 5 2 | 3+4 4 | 5 2 | 3+4 4 | 6+6 2 | 2+2 4 |
| | 6 2 | 3+4 4 | 6 2 | 3+4 4 | 5+5 2 | 2+2 4 |
| | 5 2 | 3+4 4 | 5 2 | 3+4 4 | 4+4 2 | 2+2 4 |
| | 6 2 | 3+4 4 | 6 2 | 3+4 4 | 3+3 2 | 2+2 4 |
| | 5 2 | 3+4 4 | 5 2 | 3+4 4 | 3+3 2 | 2+2 4 |
| | 6 2 | 3+4 4 | 6 2 | 3+4 4 | 4+4 2 | 2+2 4 |
| | 5 2 | 3+4 4 | 6 2 | 3+4 4 | 6+6 2 | 2+2 4 |
| | 6 2 | 3+4 4 | 5 2 | 3+4 4 | 5+5 2 | 2+2 4 |
| | 5 2 | 3+4 4 | 5 2 | 3+4 4 | 3+3 2 | 2+2 4 |
| | 6 2 | 3+4 4 | 5 2 | 3+4 4 | 4+4 2 | 2+2 4 |

Continued on next column

**TABLE XXXIX– continued**

| TQC | ICSD | | NoE | | SF | |
|---|---|---|---|---|---|---|
| | 5 2 | 3+4 4 | 6 2 | 3+4 4 | 5+5 2 | 2+2 4 |
| | 6 2 | 3+4 4 | 6 2 | 3+4 4 | 6+6 2 | 2+2 4 |
| | 5 2 | 3+4 4 | 6 2 | 3+4 4 | 3+3 2 | 2+2 4 |
| | 6 2 | 3+4 4 | 5 2 | 3+4 4 | 4+4 2 | 2+2 4 |
| | 6 2 | 3+4 4 | 5 2 | 3+4 4 | 3+3 2 | 2+2 4 |
| | 5 2 | 3+4 4 | 5 2 | 3+4 4 | 4+4 2 | 2+2 4 |
| | 5 2 | 3+4 4 | 6 2 | 3+4 4 | 5+5 2 | 2+2 4 |
| | 6 2 | 3+4 4 | 5 2 | 3+4 4 | 6+6 2 | 2+2 4 |
| | 5 2 | 3+4 4 | 5 2 | 3+4 4 | 5+5 2 | 2+2 4 |
| | 5 2 | 3+4 4 | 6 2 | 3+4 4 | 6+6 2 | 2+2 4 |
| | 6 2 | 3+4 4 | 6 2 | 3+4 4 | 3+3 2 | 2+2 4 |
| | 5 2 | 3+4 4 | 5 2 | 3+4 4 | 4+4 2 | 2+2 4 |
| | 6 2 | 3+4 4 | 6 2 | 3+4 4 | 4+4 2 | 2+2 4 |
| | 5 2 | 3+4 4 | 5 2 | 3+4 4 | 3+3 2 | 2+2 4 |
| | 6 2 | 3+4 4 | 6 2 | 3+4 4 | 6+6 2 | 2+2 4 |
| | 5 2 | 3+4 4 | 5 2 | 3+4 4 | 5+5 2 | 2+2 4 |
| | 6 2 | 3+4 4 | 6 2 | 3+4 4 | 4+4 2 | 2+2 4 |
| | 6 2 | 3+4 4 | 5 2 | 3+4 4 | 3+3 2 | 2+2 4 |
| | 5 2 | 3+4 4 | 5 2 | 3+4 4 | 5+5 2 | 2+2 4 |
| | 5 2 | 3+4 4 | 6 2 | 3+4 4 | 6+6 2 | 2+2 4 |
| | 6 2 | 3+4 4 | 5 2 | 3+4 4 | 4+4 2 | 2+2 4 |
| | 5 2 | | 5 2 | | 3+3 2 | |
| | 6 2 | | 6 2 | | 6+6 2 | |
| | 5 2 | | 5 2 | | 5+5 2 | |
| | 6 2 | | 5 2 | | 4+4 2 | |
| | 5 2 | | 6 2 | | 3+3 2 | |
| | 5 2 | | 5 2 | | 6+6 2 | |
| | 5 2 | | 6 2 | | 5+5 2 | |
| | 5 2 | | 5 2 | | 3+3 2 | |
| | 5 2 | | 6 2 | | 4+4 2 | |
| | 6 2 | | 6 2 | | 5+5 2 | |
| | 5 2 | | 5 2 | | 6+6 2 | |
| | 5 2 | | 6 2 | | 3+3 2 | |
| | 5 2 | | 5 2 | | 5+5 2 | |
| | 6 2 | | 5 2 | | 4+4 2 | |
| | 6 2 | | 6 2 | | 6+6 2 | |
| | 6 2 | | 6 2 | | 4+4 2 | |
| | 6 2 | | 5 2 | | 3+3 2 | |
| | 5 2 | | 5 2 | | 3+3 2 | |
| | 6 2 | | 5 2 | | 4+4 2 | |
| | 5 2 | | 6 2 | | 6+6 2 | |
| | 6 2 | | 6 2 | | 5+5 2 | |
| | 5 2 | | 5 2 | | 4+4 2 | |
| | 5 2 | | 5 2 | | 3+3 2 | |
| | 5 2 | | 6 2 | | 4+4 2 | |
| | 5 2 | | 5 2 | | 3+3 2 | |
| | 5 2 | | 5 2 | | 5+5 2 | |
| | 6 2 | | 6 2 | | 3+3 2 | |
| | 5 2 | | 5 2 | | 6+6 2 | |
| | 5 2 | | 6 2 | | 4+4 2 | |
| | 6 2 | | 5 2 | | 4+4 2 | |
| | 5 2 | | 6 2 | | 3+3 2 | |
| | 6 2 | | 6 2 | | 5+5 2 | |
| | 5 2 | | 5 2 | | 6+6 2 | |
| | 5 2 | | 5 2 | | 6+6 2 | |
| | 5 2 | | 5 2 | | 5+5 2 | |
| | 6 2 | | 6 2 | | 3+3 2 | |
| | 5 2 | | 6 2 | | 4+4 2 | |
| | 6 2 | | 5 2 | | 5+5 2 | |
| | 6 2 | | 6 2 | | 6+6 2 | |
| | 5 2 | | 5 2 | | 4+4 2 | |

Continued on next column

**TABLE XXXIX– continued**

| TQC | ICSD | | NoE | | SF | |
|---|---|---|---|---|---|---|
| | 5 2 | | 5 2 | | 3+3 2 | |
| | 6 2 | | 6 2 | | 4+4 2 | |
| | 5 2 | | 5 2 | | 3+3 2 | |
| 64.1.19 | 58636 | | 152 | | Ba2Bi4Cd3 | |
| | -GM | -T | -Y | -Z | -R | -S |
| | 5 2 | 3+4 4 | 5 2 | 3+4 4 | 5+5 2 | 2+2 4 |
| | 5 2 | 3+4 4 | 5 2 | 3+4 4 | 6+6 2 | 2+2 4 |
| | 6 2 | 3+4 4 | 6 2 | 3+4 4 | 3+3 2 | 2+2 4 |
| | 6 2 | 3+4 4 | 6 2 | 3+4 4 | 4+4 2 | 2+2 4 |
| | 5 2 | 3+4 4 | 5 2 | 3+4 4 | 6+6 2 | 2+2 4 |
| | 5 2 | 3+4 4 | 6 2 | 3+4 4 | 5+5 2 | 2+2 4 |
| | 6 2 | 3+4 4 | 5 2 | 3+4 4 | 3+3 2 | 2+2 4 |
| | 6 2 | 3+4 4 | 6 2 | 3+4 4 | 4+4 2 | 2+2 4 |
| | 5 2 | 3+4 4 | 5 2 | 3+4 4 | 5+5 2 | 2+2 4 |
| | 6 2 | 3+4 4 | 6 2 | 3+4 4 | 3+3 2 | 2+2 4 |
| | 5 2 | 3+4 4 | 5 2 | 3+4 4 | 6+6 2 | 2+2 4 |
| | 5 2 | 3+4 4 | 6 2 | 3+4 4 | 4+4 2 | 2+2 4 |
| | 6 2 | 3+4 4 | 5 2 | 3+4 4 | 6+6 2 | 2+2 4 |
| | 6 2 | 3+4 4 | 6 2 | 3+4 4 | 4+4 2 | 2+2 4 |
| | 5 2 | 3+4 4 | 6 2 | 3+4 4 | 5+5 2 | 2+2 4 |
| | 6 2 | 3+4 4 | 5 2 | 3+4 4 | 3+3 2 | 2+2 4 |
| | 5 2 | 3+4 4 | 5 2 | 3+4 4 | 4+4 2 | 2+2 4 |
| | 6 2 | 3+4 4 | 6 2 | 3+4 4 | 3+3 2 | 2+2 4 |
| | 5 2 | 3+4 4 | 6 2 | 3+4 4 | 6+6 2 | 2+2 4 |
| | 5 2 | 3+4 4 | 6 2 | 3+4 4 | 5+5 2 | 2+2 4 |
| | 6 2 | 3+4 4 | 5 2 | 3+4 4 | 3+3 2 | 2+2 4 |
| | 5 2 | 3+4 4 | 5 2 | 3+4 4 | 4+4 2 | 2+2 4 |
| | 6 2 | 3+4 4 | 5 2 | 3+4 4 | 6+6 2 | 2+2 4 |
| | 6 2 | 3+4 4 | 6 2 | 3+4 4 | 5+5 2 | 2+2 4 |
| | 5 2 | 3+4 4 | 5 2 | 3+4 4 | 3+3 2 | 2+2 4 |
| | 5 2 | 3+4 4 | 5 2 | 3+4 4 | 4+4 2 | 2+2 4 |
| | 5 2 | 3+4 4 | 5 2 | 3+4 4 | 3+3 2 | 2+2 4 |
| | 5 2 | 3+4 4 | 5 2 | 3+4 4 | 4+4 2 | 2+2 4 |
| | 6 2 | 3+4 4 | 6 2 | 3+4 4 | 5+5 2 | 2+2 4 |
| | 5 2 | 3+4 4 | 5 2 | 3+4 4 | 4+4 2 | 2+2 4 |
| | 5 2 | 3+4 4 | 6 2 | 3+4 4 | 6+6 2 | 2+2 4 |
| | 6 2 | 3+4 4 | 6 2 | 3+4 4 | 3+3 2 | 2+2 4 |
| | 5 2 | 3+4 4 | 5 2 | 3+4 4 | 5+5 2 | 2+2 4 |
| | 5 2 | 3+4 4 | 6 2 | 3+4 4 | 4+4 2 | 2+2 4 |
| | 6 2 | 3+4 4 | 5 2 | 3+4 4 | 6+6 2 | 2+2 4 |
| | 6 2 | 3+4 4 | 5 2 | 3+4 4 | 3+3 2 | 2+2 4 |
| | 5 2 | 3+4 4 | 5 2 | 3+4 4 | 4+4 2 | 2+2 4 |
| | 5 2 | 3+4 4 | 5 2 | 3+4 4 | 3+3 2 | 2+2 4 |
| | 5 2 | | 5 2 | | 4+4 2 | |
| | 5 2 | | 5 2 | | 3+3 2 | |
| | 5 2 | | 6 2 | | 4+4 2 | |
| | 5 2 | | 5 2 | | 3+3 2 | |
| | 6 2 | | 5 2 | | 6+6 2 | |
| | 5 2 | | 5 2 | | 5+5 2 | |
| | 5 2 | | 6 2 | | 3+3 2 | |
| | 6 2 | | 6 2 | | 4+4 2 | |
| | 5 2 | | 5 2 | | 6+6 2 | |
| | 6 2 | | 6 2 | | 5+5 2 | |
| | 6 2 | | 5 2 | | 3+3 2 | |
| | 5 2 | | 6 2 | | 4+4 2 | |
| | 6 2 | | 6 2 | | 6+6 2 | |
| | 5 2 | | 5 2 | | 5+5 2 | |
| | 6 2 | | 5 2 | | 3+3 2 | |
| | 5 2 | | 5 2 | | 4+4 2 | |
| | 5 2 | | 5 2 | | 3+3 2 | |
| | 6 2 | | 5 2 | | 4+4 2 | |

Continued on next column

**TABLE XXXIX– continued**

| TQC | ICSD | | NoE | | SF | |
|---|---|---|---|---|---|---|
| | 5 2 | | 6 2 | | 6+6 2 | |
| | 5 2 | | 6 2 | | 5+5 2 | |
| | 6 2 | | 6 2 | | 3+3 2 | |
| | 5 2 | | 5 2 | | 4+4 2 | |
| | 6 2 | | 6 2 | | 5+5 2 | |
| | 5 2 | | 5 2 | | 6+6 2 | |
| | 6 2 | | 6 2 | | 6+6 2 | |
| | 6 2 | | 6 2 | | 4+4 2 | |
| | 5 2 | | 5 2 | | 5+5 2 | |
| | 6 2 | | 6 2 | | 3+3 2 | |
| | 5 2 | | 6 2 | | 3+3 2 | |
| | 6 2 | | 5 2 | | 5+5 2 | |
| | 6 2 | | 6 2 | | 6+6 2 | |
| | 5 2 | | 5 2 | | 4+4 2 | |
| | 6 2 | | 6 2 | | 5+5 2 | |
| | 5 2 | | 6 2 | | 6+6 2 | |
| | 5 2 | | 5 2 | | 4+4 2 | |
| | 5 2 | | 5 2 | | 3+3 2 | |
| | 6 2 | | 5 2 | | 5+5 2 | |
| | 5 2 | | 5 2 | | 6+6 2 | |
| 64.1.20 | 416447 | | 152 | | Ba2Bi4Cd3 | |
| | -GM | -T | -Y | -Z | -R | -S |
| | 5 2 | 3+4 4 | 5 2 | 3+4 4 | 5+5 2 | 2+2 4 |
| | 5 2 | 3+4 4 | 5 2 | 3+4 4 | 6+6 2 | 2+2 4 |
| | 6 2 | 3+4 4 | 6 2 | 3+4 4 | 3+3 2 | 2+2 4 |
| | 6 2 | 3+4 4 | 6 2 | 3+4 4 | 4+4 2 | 2+2 4 |
| | 5 2 | 3+4 4 | 5 2 | 3+4 4 | 5+5 2 | 2+2 4 |
| | 5 2 | 3+4 4 | 6 2 | 3+4 4 | 6+6 2 | 2+2 4 |
| | 6 2 | 3+4 4 | 5 2 | 3+4 4 | 3+3 2 | 2+2 4 |
| | 6 2 | 3+4 4 | 6 2 | 3+4 4 | 4+4 2 | 2+2 4 |
| | 5 2 | 3+4 4 | 5 2 | 3+4 4 | 6+6 2 | 2+2 4 |
| | 6 2 | 3+4 4 | 6 2 | 3+4 4 | 4+4 2 | 2+2 4 |
| | 5 2 | 3+4 4 | 5 2 | 3+4 4 | 5+5 2 | 2+2 4 |
| | 5 2 | 3+4 4 | 6 2 | 3+4 4 | 3+3 2 | 2+2 4 |
| | 6 2 | 3+4 4 | 5 2 | 3+4 4 | 5+5 2 | 2+2 4 |
| | 6 2 | 3+4 4 | 6 2 | 3+4 4 | 3+3 2 | 2+2 4 |
| | 5 2 | 3+4 4 | 6 2 | 3+4 4 | 6+6 2 | 2+2 4 |
| | 6 2 | 3+4 4 | 5 2 | 3+4 4 | 4+4 2 | 2+2 4 |
| | 5 2 | 3+4 4 | 5 2 | 3+4 4 | 3+3 2 | 2+2 4 |
| | 6 2 | 3+4 4 | 6 2 | 3+4 4 | 4+4 2 | 2+2 4 |
| | 5 2 | 3+4 4 | 6 2 | 3+4 4 | 5+5 2 | 2+2 4 |
| | 5 2 | 3+4 4 | 6 2 | 3+4 4 | 6+6 2 | 2+2 4 |
| | 6 2 | 3+4 4 | 5 2 | 3+4 4 | 4+4 2 | 2+2 4 |
| | 5 2 | 3+4 4 | 5 2 | 3+4 4 | 3+3 2 | 2+2 4 |
| | 6 2 | 3+4 4 | 5 2 | 3+4 4 | 5+5 2 | 2+2 4 |
| | 6 2 | 3+4 4 | 6 2 | 3+4 4 | 6+6 2 | 2+2 4 |
| | 5 2 | 3+4 4 | 5 2 | 3+4 4 | 4+4 2 | 2+2 4 |
| | 5 2 | 3+4 4 | 5 2 | 3+4 4 | 3+3 2 | 2+2 4 |
| | 5 2 | 3+4 4 | 5 2 | 3+4 4 | 4+4 2 | 2+2 4 |
| | 5 2 | 3+4 4 | 5 2 | 3+4 4 | 3+3 2 | 2+2 4 |
| | 6 2 | 3+4 4 | 6 2 | 3+4 4 | 6+6 2 | 2+2 4 |
| | 5 2 | 3+4 4 | 5 2 | 3+4 4 | 3+3 2 | 2+2 4 |
| | 5 2 | 3+4 4 | 6 2 | 3+4 4 | 4+4 2 | 2+2 4 |
| | 6 2 | 3+4 4 | 6 2 | 3+4 4 | 5+5 2 | 2+2 4 |
| | 5 2 | 3+4 4 | 5 2 | 3+4 4 | 6+6 2 | 2+2 4 |
| | 5 2 | 3+4 4 | 6 2 | 3+4 4 | 3+3 2 | 2+2 4 |
| | 6 2 | 3+4 4 | 5 2 | 3+4 4 | 5+5 2 | 2+2 4 |
| | 6 2 | 3+4 4 | 5 2 | 3+4 4 | 4+4 2 | 2+2 4 |
| | 5 2 | 3+4 4 | 5 2 | 3+4 4 | 3+3 2 | 2+2 4 |
| | 5 2 | 3+4 4 | 5 2 | 3+4 4 | 4+4 2 | 2+2 4 |
| | 5 2 | | 5 2 | | 3+3 2 | |

Continued on next column

**TABLE XXXIX– continued**

| TQC | ICSD | NoE | SF |
|---|---|---|---|
| | 5 2 | 5 2 | 4+4 2 |
| | 5 2 | 5 2 | 3+3 2 |
| | 5 2 | 6 2 | 4+4 2 |
| | 6 2 | 5 2 | 5+5 2 |
| | 5 2 | 5 2 | 6+6 2 |
| | 5 2 | 6 2 | 3+3 2 |
| | 6 2 | 6 2 | 4+4 2 |
| | 5 2 | 5 2 | 5+5 2 |
| | 6 2 | 6 2 | 6+6 2 |
| | 6 2 | 5 2 | 4+4 2 |
| | 5 2 | 6 2 | 3+3 2 |
| | 6 2 | 6 2 | 5+5 2 |
| | 5 2 | 5 2 | 6+6 2 |
| | 6 2 | 5 2 | 4+4 2 |
| | 5 2 | 5 2 | 3+3 2 |
| | 5 2 | 5 2 | 4+4 2 |
| | 6 2 | 5 2 | 3+3 2 |
| | 5 2 | 6 2 | 5+5 2 |
| | 5 2 | 6 2 | 6+6 2 |
| | 6 2 | 6 2 | 4+4 2 |
| | 5 2 | 5 2 | 3+3 2 |
| | 6 2 | 6 2 | 6+6 2 |
| | 5 2 | 5 2 | 5+5 2 |
| | 6 2 | 6 2 | 5+5 2 |
| | 6 2 | 6 2 | 3+3 2 |
| | 5 2 | 5 2 | 6+6 2 |
| | 6 2 | 6 2 | 4+4 2 |
| | 5 2 | 6 2 | 6+6 2 |
| | 6 2 | 5 2 | 4+4 2 |
| | 6 2 | 6 2 | 5+5 2 |
| | 5 2 | 5 2 | 3+3 2 |
| | 6 2 | 6 2 | 6+6 2 |
| | 5 2 | 6 2 | 5+5 2 |
| | 5 2 | 5 2 | 3+3 2 |
| | 5 2 | 5 2 | 4+4 2 |
| | 6 2 | 5 2 | 6+6 2 |
| | 5 2 | 5 2 | 5+5 2 |
| 64.1.21 | 416446 | 152 | Ba2Bi4Cd3 |
| | -GM -T | -Y -Z | -R -S |
| | 5 2 3+4 4 | 5 2 3+4 4 | 3+3 2 2+2 4 |
| | 5 2 3+4 4 | 5 2 3+4 4 | 4+4 2 2+2 4 |
| | 6 2 3+4 4 | 6 2 3+4 4 | 5+5 2 2+2 4 |
| | 6 2 3+4 4 | 6 2 3+4 4 | 6+6 2 2+2 4 |
| | 5 2 3+4 4 | 5 2 3+4 4 | 4+4 2 2+2 4 |
| | 5 2 3+4 4 | 6 2 3+4 4 | 3+3 2 2+2 4 |
| | 6 2 3+4 4 | 5 2 3+4 4 | 5+5 2 2+2 4 |
| | 6 2 3+4 4 | 6 2 3+4 4 | 6+6 2 2+2 4 |
| | 5 2 3+4 4 | 5 2 3+4 4 | 3+3 2 2+2 4 |
| | 6 2 3+4 4 | 6 2 3+4 4 | 5+5 2 2+2 4 |
| | 5 2 3+4 4 | 5 2 3+4 4 | 4+4 2 2+2 4 |
| | 5 2 3+4 4 | 6 2 3+4 4 | 6+6 2 2+2 4 |
| | 6 2 3+4 4 | 5 2 3+4 4 | 4+4 2 2+2 4 |
| | 6 2 3+4 4 | 6 2 3+4 4 | 6+6 2 2+2 4 |
| | 5 2 3+4 4 | 6 2 3+4 4 | 3+3 2 2+2 4 |
| | 6 2 3+4 4 | 5 2 3+4 4 | 5+5 2 2+2 4 |
| | 5 2 3+4 4 | 5 2 3+4 4 | 6+6 2 2+2 4 |
| | 6 2 3+4 4 | 6 2 3+4 4 | 5+5 2 2+2 4 |
| | 5 2 3+4 4 | 6 2 3+4 4 | 4+4 2 2+2 4 |
| | 5 2 3+4 4 | 6 2 3+4 4 | 3+3 2 2+2 4 |
| | 6 2 3+4 4 | 5 2 3+4 4 | 5+5 2 2+2 4 |
| | 5 2 3+4 4 | 5 2 3+4 4 | 6+6 2 2+2 4 |
| | 6 2 3+4 4 | 5 2 3+4 4 | 4+4 2 2+2 4 |
| | 6 2 3+4 4 | 6 2 3+4 4 | 3+3 2 2+2 4 |
| | 5 2 3+4 4 | 5 2 3+4 4 | 5+5 2 2+2 4 |
| | 5 2 3+4 4 | 5 2 3+4 4 | 6+6 2 2+2 4 |
| | 5 2 3+4 4 | 5 2 3+4 4 | 5+5 2 2+2 4 |
| | 5 2 3+4 4 | 5 2 3+4 4 | 6+6 2 2+2 4 |
| | 6 2 3+4 4 | 6 2 3+4 4 | 3+3 2 2+2 4 |
| | 5 2 3+4 4 | 5 2 3+4 4 | 6+6 2 2+2 4 |
| | 5 2 3+4 4 | 6 2 3+4 4 | 4+4 2 2+2 4 |
| | 6 2 3+4 4 | 6 2 3+4 4 | 5+5 2 2+2 4 |
| | 5 2 3+4 4 | 5 2 3+4 4 | 3+3 2 2+2 4 |
| | 5 2 3+4 4 | 6 2 3+4 4 | 6+6 2 2+2 4 |
| | 6 2 3+4 4 | 5 2 3+4 4 | 4+4 2 2+2 4 |
| | 6 2 3+4 4 | 5 2 3+4 4 | 5+5 2 2+2 4 |
| | 5 2 3+4 4 | 5 2 3+4 4 | 6+6 2 2+2 4 |
| | 5 2 3+4 4 | 5 2 3+4 4 | 5+5 2 2+2 4 |
| | 5 2 | 5 2 | 6+6 2 |
| | 5 2 | 5 2 | 5+5 2 |
| | 5 2 | 5 2 | 6+6 2 |
| | 5 2 | 6 2 | 5+5 2 |
| | 6 2 | 5 2 | 4+4 2 |
| | 5 2 | 5 2 | 3+3 2 |
| | 5 2 | 6 2 | 5+5 2 |
| | 6 2 | 6 2 | 6+6 2 |
| | 5 2 | 5 2 | 4+4 2 |
| | 6 2 | 6 2 | 3+3 2 |
| | 6 2 | 5 2 | 5+5 2 |
| | 5 2 | 6 2 | 6+6 2 |
| | 6 2 | 6 2 | 4+4 2 |
| | 5 2 | 5 2 | 3+3 2 |
| | 6 2 | 5 2 | 5+5 2 |
| | 5 2 | 5 2 | 6+6 2 |
| | 5 2 | 5 2 | 5+5 2 |
| | 6 2 | 5 2 | 6+6 2 |
| | 5 2 | 6 2 | 4+4 2 |
| | 5 2 | 6 2 | 3+3 2 |
| | 6 2 | 6 2 | 5+5 2 |
| | 5 2 | 5 2 | 6+6 2 |
| | 6 2 | 6 2 | 3+3 2 |
| | 5 2 | 5 2 | 4+4 2 |
| | 6 2 | 6 2 | 4+4 2 |
| | 6 2 | 6 2 | 6+6 2 |
| | 5 2 | 5 2 | 3+3 2 |
| | 6 2 | 6 2 | 5+5 2 |
| | 5 2 | 6 2 | 3+3 2 |
| | 6 2 | 5 2 | 5+5 2 |
| | 6 2 | 6 2 | 4+4 2 |
| | 5 2 | 5 2 | 6+6 2 |
| | 6 2 | 6 2 | 3+3 2 |
| | 5 2 | 6 2 | 4+4 2 |
| | 5 2 | 5 2 | 6+6 2 |
| | 5 2 | 5 2 | 5+5 2 |
| | 6 2 | 5 2 | 3+3 2 |
| | 5 2 | 5 2 | 4+4 2 |
| 64.1.22 | 416445 | 152 | Ba2Bi4Cd3 |
| | -GM -T | -Y -Z | -R -S |
| | 5 2 3+4 4 | 5 2 3+4 4 | 5+5 2 2+2 4 |
| | 5 2 3+4 4 | 5 2 3+4 4 | 6+6 2 2+2 4 |
| | 6 2 3+4 4 | 6 2 3+4 4 | 3+3 2 2+2 4 |
| | 6 2 3+4 4 | 6 2 3+4 4 | 4+4 2 2+2 4 |
| | 5 2 3+4 4 | 5 2 3+4 4 | 5+5 2 2+2 4 |

Continued on next column

**TABLE XXXIX– continued**

| TQC | ICSD | NoE | SF |
|---|---|---|---|
| | 5 2 3+4 4 | 6 2 3+4 4 | 6+6 2 2+2 4 |
| | 6 2 3+4 4 | 5 2 3+4 4 | 3+3 2 2+2 4 |
| | 6 2 3+4 4 | 6 2 3+4 4 | 4+4 2 2+2 4 |
| | 5 2 3+4 4 | 5 2 3+4 4 | 6+6 2 2+2 4 |
| | 6 2 3+4 4 | 6 2 3+4 4 | 4+4 2 2+2 4 |
| | 5 2 3+4 4 | 5 2 3+4 4 | 5+5 2 2+2 4 |
| | 5 2 3+4 4 | 6 2 3+4 4 | 3+3 2 2+2 4 |
| | 6 2 3+4 4 | 6 2 3+4 4 | 5+5 2 2+2 4 |
| | 6 2 3+4 4 | 5 2 3+4 4 | 3+3 2 2+2 4 |
| | 5 2 3+4 4 | 6 2 3+4 4 | 6+6 2 2+2 4 |
| | 6 2 3+4 4 | 5 2 3+4 4 | 4+4 2 2+2 4 |
| | 5 2 3+4 4 | 5 2 3+4 4 | 3+3 2 2+2 4 |
| | 6 2 3+4 4 | 6 2 3+4 4 | 4+4 2 2+2 4 |
| | 5 2 3+4 4 | 6 2 3+4 4 | 5+5 2 2+2 4 |
| | 5 2 3+4 4 | 6 2 3+4 4 | 6+6 2 2+2 4 |
| | 6 2 3+4 4 | 5 2 3+4 4 | 4+4 2 2+2 4 |
| | 5 2 3+4 4 | 5 2 3+4 4 | 3+3 2 2+2 4 |
| | 6 2 3+4 4 | 5 2 3+4 4 | 5+5 2 2+2 4 |
| | 6 2 3+4 4 | 6 2 3+4 4 | 6+6 2 2+2 4 |
| | 5 2 3+4 4 | 5 2 3+4 4 | 4+4 2 2+2 4 |
| | 5 2 3+4 4 | 5 2 3+4 4 | 3+3 2 2+2 4 |
| | 5 2 3+4 4 | 5 2 3+4 4 | 4+4 2 2+2 4 |
| | 5 2 3+4 4 | 5 2 3+4 4 | 3+3 2 2+2 4 |
| | 6 2 3+4 4 | 6 2 3+4 4 | 6+6 2 2+2 4 |
| | 5 2 3+4 4 | 5 2 3+4 4 | 3+3 2 2+2 4 |
| | 5 2 3+4 4 | 6 2 3+4 4 | 4+4 2 2+2 4 |
| | 6 2 3+4 4 | 6 2 3+4 4 | 5+5 2 2+2 4 |
| | 5 2 3+4 4 | 5 2 3+4 4 | 6+6 2 2+2 4 |
| | 5 2 3+4 4 | 6 2 3+4 4 | 3+3 2 2+2 4 |
| | 6 2 3+4 4 | 5 2 3+4 4 | 5+5 2 2+2 4 |
| | 6 2 3+4 4 | 5 2 3+4 4 | 4+4 2 2+2 4 |
| | 5 2 3+4 4 | 5 2 3+4 4 | 3+3 2 2+2 4 |
| | 5 2 3+4 4 | 5 2 3+4 4 | 4+4 2 2+2 4 |
| | 5 2 | 5 2 | 3+3 2 |
| | 5 2 | 5 2 | 4+4 2 |
| | 5 2 | 6 2 | 3+3 2 |
| | 5 2 | 5 2 | 4+4 2 |
| | 6 2 | 5 2 | 5+5 2 |
| | 5 2 | 5 2 | 6+6 2 |
| | 5 2 | 6 2 | 3+3 2 |
| | 6 2 | 6 2 | 4+4 2 |
| | 5 2 | 5 2 | 5+5 2 |
| | 6 2 | 6 2 | 6+6 2 |
| | 6 2 | 5 2 | 4+4 2 |
| | 5 2 | 6 2 | 3+3 2 |
| | 6 2 | 6 2 | 5+5 2 |
| | 5 2 | 5 2 | 6+6 2 |
| | 6 2 | 5 2 | 4+4 2 |
| | 5 2 | 5 2 | 3+3 2 |
| | 5 2 | 5 2 | 4+4 2 |
| | 6 2 | 5 2 | 3+3 2 |
| | 5 2 | 6 2 | 5+5 2 |
| | 5 2 | 6 2 | 6+6 2 |
| | 6 2 | 6 2 | 4+4 2 |
| | 5 2 | 5 2 | 3+3 2 |
| | 6 2 | 6 2 | 6+6 2 |
| | 5 2 | 5 2 | 5+5 2 |
| | 6 2 | 6 2 | 5+5 2 |
| | 6 2 | 6 2 | 3+3 2 |
| | 5 2 | 5 2 | 4+4 2 |
| | 6 2 | 6 2 | 6+6 2 |
| | 5 2 | 6 2 | 6+6 2 |
| | 6 2 | 5 2 | 4+4 2 |
| | 6 2 | 6 2 | 5+5 2 |
| | 5 2 | 5 2 | 3+3 2 |
| | 6 2 | 6 2 | 6+6 2 |
| | 5 2 | 6 2 | 5+5 2 |
| | 5 2 | 5 2 | 3+3 2 |
| | 5 2 | 5 2 | 4+4 2 |
| | 6 2 | 5 2 | 6+6 2 |
| | 5 2 | 5 2 | 5+5 2 |
| 64.1.23 | 416444 | 152 | Ba2Bi4Cd3 |
| | -GM -T | -Y -Z | -R -S |
| | 5 2 3+4 4 | 5 2 3+4 4 | 3+3 2 2+2 4 |
| | 5 2 3+4 4 | 5 2 3+4 4 | 4+4 2 2+2 4 |
| | 6 2 3+4 4 | 6 2 3+4 4 | 5+5 2 2+2 4 |
| | 6 2 3+4 4 | 6 2 3+4 4 | 6+6 2 2+2 4 |
| | 5 2 3+4 4 | 5 2 3+4 4 | 3+3 2 2+2 4 |
| | 5 2 3+4 4 | 6 2 3+4 4 | 4+4 2 2+2 4 |
| | 6 2 3+4 4 | 5 2 3+4 4 | 5+5 2 2+2 4 |
| | 6 2 3+4 4 | 6 2 3+4 4 | 6+6 2 2+2 4 |
| | 5 2 3+4 4 | 5 2 3+4 4 | 4+4 2 2+2 4 |
| | 6 2 3+4 4 | 6 2 3+4 4 | 6+6 2 2+2 4 |
| | 5 2 3+4 4 | 5 2 3+4 4 | 3+3 2 2+2 4 |
| | 5 2 3+4 4 | 6 2 3+4 4 | 5+5 2 2+2 4 |
| | 6 2 3+4 4 | 5 2 3+4 4 | 3+3 2 2+2 4 |
| | 6 2 3+4 4 | 6 2 3+4 4 | 5+5 2 2+2 4 |
| | 5 2 3+4 4 | 6 2 3+4 4 | 4+4 2 2+2 4 |
| | 6 2 3+4 4 | 5 2 3+4 4 | 6+6 2 2+2 4 |
| | 5 2 3+4 4 | 5 2 3+4 4 | 5+5 2 2+2 4 |
| | 6 2 3+4 4 | 6 2 3+4 4 | 6+6 2 2+2 4 |
| | 5 2 3+4 4 | 6 2 3+4 4 | 3+3 2 2+2 4 |
| | 5 2 3+4 4 | 6 2 3+4 4 | 4+4 2 2+2 4 |
| | 6 2 3+4 4 | 5 2 3+4 4 | 6+6 2 2+2 4 |
| | 5 2 3+4 4 | 5 2 3+4 4 | 5+5 2 2+2 4 |
| | 6 2 3+4 4 | 5 2 3+4 4 | 3+3 2 2+2 4 |
| | 6 2 3+4 4 | 6 2 3+4 4 | 4+4 2 2+2 4 |
| | 5 2 3+4 4 | 5 2 3+4 4 | 6+6 2 2+2 4 |
| | 5 2 3+4 4 | 5 2 3+4 4 | 5+5 2 2+2 4 |
| | 5 2 3+4 4 | 5 2 3+4 4 | 6+6 2 2+2 4 |
| | 5 2 3+4 4 | 5 2 3+4 4 | 5+5 2 2+2 4 |
| | 6 2 3+4 4 | 6 2 3+4 4 | 4+4 2 2+2 4 |
| | 5 2 3+4 4 | 5 2 3+4 4 | 5+5 2 2+2 4 |
| | 5 2 3+4 4 | 6 2 3+4 4 | 3+3 2 2+2 4 |
| | 6 2 3+4 4 | 6 2 3+4 4 | 6+6 2 2+2 4 |
| | 5 2 3+4 4 | 5 2 3+4 4 | 4+4 2 2+2 4 |
| | 5 2 3+4 4 | 6 2 3+4 4 | 5+5 2 2+2 4 |
| | 6 2 3+4 4 | 5 2 3+4 4 | 3+3 2 2+2 4 |
| | 6 2 3+4 4 | 5 2 3+4 4 | 6+6 2 2+2 4 |
| | 5 2 3+4 4 | 5 2 3+4 4 | 5+5 2 2+2 4 |
| | 5 2 3+4 4 | 5 2 3+4 4 | 6+6 2 2+2 4 |
| | 5 2 | 5 2 | 5+5 2 |
| | 5 2 | 5 2 | 6+6 2 |
| | 5 2 | 6 2 | 5+5 2 |
| | 5 2 | 5 2 | 6+6 2 |
| | 6 2 | 5 2 | 3+3 2 |
| | 5 2 | 5 2 | 4+4 2 |
| | 5 2 | 6 2 | 5+5 2 |
| | 6 2 | 6 2 | 6+6 2 |
| | 5 2 | 5 2 | 3+3 2 |
| | 6 2 | 6 2 | 4+4 2 |
| | 6 2 | 5 2 | 6+6 2 |

Continued on next column

**TABLE XXXIX– continued**

| TQC | ICSD | | NoE | | SF | |
|---|---|---|---|---|---|---|
| | 5 2 | | 6 2 | | 5+5 2 | |
| | 6 2 | | 6 2 | | 3+3 2 | |
| | 5 2 | | 5 2 | | 4+4 2 | |
| | 6 2 | | 5 2 | | 6+6 2 | |
| | 5 2 | | 5 2 | | 5+5 2 | |
| | 5 2 | | 5 2 | | 6+6 2 | |
| | 6 2 | | 5 2 | | 5+5 2 | |
| | 5 2 | | 6 2 | | 3+3 2 | |
| | 5 2 | | 6 2 | | 4+4 2 | |
| | 6 2 | | 6 2 | | 6+6 2 | |
| | 5 2 | | 5 2 | | 5+5 2 | |
| | 6 2 | | 6 2 | | 4+4 2 | |
| | 5 2 | | 5 2 | | 3+3 2 | |
| | 6 2 | | 6 2 | | 3+3 2 | |
| | 6 2 | | 6 2 | | 5+5 2 | |
| | 5 2 | | 5 2 | | 4+4 2 | |
| | 6 2 | | 6 2 | | 6+6 2 | |
| | 5 2 | | 6 2 | | 4+4 2 | |
| | 6 2 | | 5 2 | | 6+6 2 | |
| | 6 2 | | 6 2 | | 3+3 2 | |
| | 5 2 | | 5 2 | | 5+5 2 | |
| | 6 2 | | 6 2 | | 4+4 2 | |
| | 5 2 | | 6 2 | | 3+3 2 | |
| | 5 2 | | 5 2 | | 5+5 2 | |
| | 5 2 | | 5 2 | | 6+6 2 | |
| | 6 2 | | 5 2 | | 4+4 2 | |
| | 5 2 | | 5 2 | | 3+3 2 | |
| 64.1.24 | 412877 | | 180 | | Ba1Fe1K1O3 | |
| | -GM | -T | -Y | -Z | -R | -S |
| | 5 2 | 3+4 4 | 5 2 | 3+4 4 | 3+3 2 | 2+2 4 |
| | 5 2 | 3+4 4 | 5 2 | 3+4 4 | 4+4 2 | 2+2 4 |
| | 6 2 | 3+4 4 | 6 2 | 3+4 4 | 5+5 2 | 2+2 4 |
| | 6 2 | 3+4 4 | 6 2 | 3+4 4 | 6+6 2 | 2+2 4 |
| | 5 2 | 3+4 4 | 5 2 | 3+4 4 | 5+5 2 | 2+2 4 |
| | 5 2 | 3+4 4 | 5 2 | 3+4 4 | 6+6 2 | 2+2 4 |
| | 6 2 | 3+4 4 | 6 2 | 3+4 4 | 3+3 2 | 2+2 4 |
| | 6 2 | 3+4 4 | 6 2 | 3+4 4 | 4+4 2 | 2+2 4 |
| | 6 2 | 3+4 4 | 5 2 | 3+4 4 | 5+5 2 | 2+2 4 |
| | 5 2 | 3+4 4 | 6 2 | 3+4 4 | 3+3 2 | 2+2 4 |
| | 5 2 | 3+4 4 | 5 2 | 3+4 4 | 4+4 2 | 2+2 4 |
| | 6 2 | 3+4 4 | 6 2 | 3+4 4 | 6+6 2 | 2+2 4 |
| | 6 2 | 3+4 4 | 6 2 | 3+4 4 | 3+3 2 | 2+2 4 |
| | 6 2 | 3+4 4 | 6 2 | 3+4 4 | 4+4 2 | 2+2 4 |
| | 5 2 | 3+4 4 | 5 2 | 3+4 4 | 6+6 2 | 2+2 4 |
| | 5 2 | 3+4 4 | 5 2 | 3+4 4 | 5+5 2 | 2+2 4 |
| | 5 2 | 3+4 4 | 5 2 | 3+4 4 | 5+5 2 | 2+2 4 |
| | 5 2 | 3+4 4 | 5 2 | 3+4 4 | 6+6 2 | 2+2 4 |
| | 6 2 | 3+4 4 | 6 2 | 3+4 4 | 3+3 2 | 2+2 4 |
| | 6 2 | 3+4 4 | 6 2 | 3+4 4 | 4+4 2 | 2+2 4 |
| | 5 2 | 3+4 4 | 5 2 | 3+4 4 | 4+4 2 | 2+2 4 |
| | 5 2 | 3+4 4 | 6 2 | 3+4 4 | 3+3 2 | 2+2 4 |
| | 6 2 | 3+4 4 | 5 2 | 3+4 4 | 5+5 2 | 2+2 4 |
| | 6 2 | 3+4 4 | 6 2 | 3+4 4 | 6+6 2 | 2+2 4 |
| | 5 2 | 3+4 4 | 5 2 | 3+4 4 | 3+3 2 | 2+2 4 |
| | 5 2 | 3+4 4 | 5 2 | 3+4 4 | 4+4 2 | 2+2 4 |
| | 5 2 | 3+4 4 | 6 2 | 3+4 4 | 6+6 2 | 2+2 4 |
| | 6 2 | 3+4 4 | 6 2 | 3+4 4 | 5+5 2 | 2+2 4 |
| | 6 2 | 3+4 4 | 5 2 | 3+4 4 | 3+3 2 | 2+2 4 |
| | 6 2 | 3+4 4 | 6 2 | 3+4 4 | 4+4 2 | 2+2 4 |
| | 5 2 | 3+4 4 | 5 2 | 3+4 4 | 5+5 2 | 2+2 4 |
| | 6 2 | 3+4 4 | 6 2 | 3+4 4 | 6+6 2 | 2+2 4 |
| | 5 2 | 3+4 4 | 5 2 | 3+4 4 | 6+6 2 | 2+2 4 |
| | 5 2 | 3+4 4 | 6 2 | 3+4 4 | 5+5 2 | 2+2 4 |
| | 6 2 | 3+4 4 | 5 2 | 3+4 4 | 3+3 2 | 2+2 4 |
| | 6 2 | 3+4 4 | 6 2 | 3+4 4 | 4+4 2 | 2+2 4 |
| | 5 2 | 3+4 4 | 5 2 | 3+4 4 | 5+5 2 | 2+2 4 |
| | 5 2 | 3+4 4 | 6 2 | 3+4 4 | 6+6 2 | 2+2 4 |
| | 5 2 | 3+4 4 | 6 2 | 3+4 4 | 4+4 2 | 2+2 4 |
| | 6 2 | 3+4 4 | 5 2 | 3+4 4 | 5+5 2 | 2+2 4 |
| | 5 2 | 3+4 4 | 5 2 | 3+4 4 | 6+6 2 | 2+2 4 |
| | 6 2 | 3+4 4 | 5 2 | 3+4 4 | 3+3 2 | 2+2 4 |
| | 6 2 | 3+4 4 | 6 2 | 3+4 4 | 3+3 2 | 2+2 4 |
| | 6 2 | 3+4 4 | 6 2 | 3+4 4 | 4+4 2 | 2+2 4 |
| | 6 2 | 3+4 4 | 5 2 | 3+4 4 | 5+5 2 | 2+2 4 |
| | 5 2 | | 6 2 | | 4+4 2 | |
| | 5 2 | | 6 2 | | 6+6 2 | |
| | 6 2 | | 5 2 | | 3+3 2 | |
| | 5 2 | | 5 2 | | 3+3 2 | |
| | 6 2 | | 5 2 | | 4+4 2 | |
| | 5 2 | | 6 2 | | 6+6 2 | |
| | 6 2 | | 6 2 | | 5+5 2 | |
| | 5 2 | | 6 2 | | 6+6 2 | |
| | 6 2 | | 5 2 | | 5+5 2 | |
| | 5 2 | | 5 2 | | 3+3 2 | |
| | 5 2 | | 5 2 | | 4+4 2 | |
| | 6 2 | | 6 2 | | 6+6 2 | |
| | 6 2 | | 6 2 | | 5+5 2 | |
| | 5 2 | | 6 2 | | 4+4 2 | |
| | 6 2 | | 6 2 | | 3+3 2 | |
| | 6 2 | | 5 2 | | 3+3 2 | |
| | 5 2 | | 5 2 | | 4+4 2 | |
| | 6 2 | | 5 2 | | 5+5 2 | |
| | 5 2 | | 6 2 | | 6+6 2 | |
| | 6 2 | | 6 2 | | 4+4 2 | |
| | 5 2 | | 6 2 | | 3+3 2 | |
| | 6 2 | | 5 2 | | 5+5 2 | |
| | 5 2 | | 5 2 | | 6+6 2 | |
| | 5 2 | | 6 2 | | 4+4 2 | |
| | 5 2 | | 5 2 | | 3+3 2 | |
| | 6 2 | | 6 2 | | 5+5 2 | |
| | 6 2 | | 5 2 | | 6+6 2 | |
| | 5 2 | | 6 2 | | 3+3 2 | |
| | 6 2 | | 6 2 | | 4+4 2 | |
| | 5 2 | | 5 2 | | 5+5 2 | |
| | 5 2 | | 5 2 | | 6+6 2 | |
| | 6 2 | | 6 2 | | 6+6 2 | |
| | 6 2 | | 5 2 | | 5+5 2 | |
| | 6 2 | | 5 2 | | 3+3 2 | |
| | 5 2 | | 6 2 | | 4+4 2 | |
| | 5 2 | | 6 2 | | 4+4 2 | |
| | 6 2 | | 5 2 | | 5+5 2 | |
| | 6 2 | | 6 2 | | 3+3 2 | |
| | 5 2 | | 6 2 | | 6+6 2 | |
| | 6 2 | | 5 2 | | 5+5 2 | |
| | 5 2 | | 5 2 | | 6+6 2 | |
| | 6 2 | | 5 2 | | 4+4 2 | |
| | 5 2 | | 5 2 | | 3+3 2 | |
| | 6 2 | | 5 2 | | 6+6 2 | |
| | 6 2 | | 6 2 | | 5+5 2 | |
| 64.1.25 | 63708 | | 176 | | As2Ni3O8 | |
| | -GM | -T | -Y | -Z | -R | -S |
| | 5 2 | 3+4 4 | 5 2 | 3+4 4 | 5+5 2 | 2+2 4 |

Continued on next column

**TABLE XXXIX– continued**

| TQC | ICSD | NoE | SF |
|---|---|---|---|
| | 5 2 3+4 4 | 6 2 3+4 4 | 6+6 2 2+2 4 |
| | 6 2 3+4 4 | 5 2 3+4 4 | 3+3 2 2+2 4 |
| | 6 2 3+4 4 | 6 2 3+4 4 | 4+4 2 2+2 4 |
| | 6 2 3+4 4 | 6 2 3+4 4 | 3+3 2 2+2 4 |
| | 5 2 3+4 4 | 5 2 3+4 4 | 4+4 2 2+2 4 |
| | 5 2 3+4 4 | 5 2 3+4 4 | 5+5 2 2+2 4 |
| | 6 2 3+4 4 | 6 2 3+4 4 | 6+6 2 2+2 4 |
| | 6 2 3+4 4 | 6 2 3+4 4 | 4+4 2 2+2 4 |
| | 5 2 3+4 4 | 6 2 3+4 4 | 5+5 2 2+2 4 |
| | 6 2 3+4 4 | 5 2 3+4 4 | 6+6 2 2+2 4 |
| | 6 2 3+4 4 | 5 2 3+4 4 | 3+3 2 2+2 4 |
| | 6 2 3+4 4 | 6 2 3+4 4 | 3+3 2 2+2 4 |
| | 5 2 3+4 4 | 6 2 3+4 4 | 4+4 2 2+2 4 |
| | 5 2 3+4 4 | 5 2 3+4 4 | 6+6 2 2+2 4 |
| | 5 2 3+4 4 | 5 2 3+4 4 | 5+5 2 2+2 4 |
| | 5 2 3+4 4 | 5 2 3+4 4 | 5+5 2 2+2 4 |
| | 6 2 3+4 4 | 6 2 3+4 4 | 6+6 2 2+2 4 |
| | 6 2 3+4 4 | 5 2 3+4 4 | 3+3 2 2+2 4 |
| | 5 2 3+4 4 | 6 2 3+4 4 | 4+4 2 2+2 4 |
| | 5 2 3+4 4 | 5 2 3+4 4 | 3+3 2 2+2 4 |
| | 6 2 3+4 4 | 5 2 3+4 4 | 4+4 2 2+2 4 |
| | 5 2 3+4 4 | 5 2 3+4 4 | 5+5 2 2+2 4 |
| | 5 2 3+4 4 | 6 2 3+4 4 | 6+6 2 2+2 4 |
| | 6 2 3+4 4 | 5 2 3+4 4 | 6+6 2 2+2 4 |
| | 5 2 3+4 4 | 5 2 3+4 4 | 5+5 2 2+2 4 |
| | 6 2 3+4 4 | 6 2 3+4 4 | 4+4 2 2+2 4 |
| | 5 2 3+4 4 | 6 2 3+4 4 | 3+3 2 2+2 4 |
| | 6 2 3+4 4 | 6 2 3+4 4 | 6+6 2 2+2 4 |
| | 5 2 3+4 4 | 5 2 3+4 4 | 5+5 2 2+2 4 |
| | 6 2 3+4 4 | 6 2 3+4 4 | 4+4 2 2+2 4 |
| | 6 2 3+4 4 | 6 2 3+4 4 | 3+3 2 2+2 4 |
| | 6 2 3+4 4 | 6 2 3+4 4 | 3+3 2 2+2 4 |
| | 5 2 3+4 4 | 5 2 3+4 4 | 4+4 2 2+2 4 |
| | 6 2 3+4 4 | 6 2 3+4 4 | 3+3 2 2+2 4 |
| | 5 2 3+4 4 | 6 2 3+4 4 | 4+4 2 2+2 4 |
| | 6 2 3+4 4 | 5 2 3+4 4 | 5+5 2 2+2 4 |
| | 5 2 3+4 4 | 5 2 3+4 4 | 6+6 2 2+2 4 |
| | 5 2 3+4 4 | 5 2 3+4 4 | 4+4 2 2+2 4 |
| | 5 2 3+4 4 | 6 2 3+4 4 | 3+3 2 2+2 4 |
| | 5 2 3+4 4 | 5 2 3+4 4 | 5+5 2 2+2 4 |
| | 6 2 3+4 4 | 6 2 3+4 4 | 6+6 2 2+2 4 |
| | 5 2 3+4 4 | 6 2 3+4 4 | 5+5 2 2+2 4 |
| | 5 2 3+4 4 | 5 2 3+4 4 | 6+6 2 2+2 4 |
| | 6 2 | 6 2 | 6+6 2 |
| | 6 2 | 5 2 | 5+5 2 |
| | 5 2 | 6 2 | 4+4 2 |
| | 5 2 | 5 2 | 3+3 2 |
| | 5 2 | 5 2 | 3+3 2 |
| | 6 2 | 6 2 | 4+4 2 |
| | 6 2 | 5 2 | 6+6 2 |
| | 5 2 | 5 2 | 5+5 2 |
| | 6 2 | 6 2 | 4+4 2 |
| | 6 2 | 5 2 | 3+3 2 |
| | 5 2 | 6 2 | 3+3 2 |
| | 6 2 | 5 2 | 5+5 2 |
| | 5 2 | 6 2 | 6+6 2 |
| | 5 2 | 5 2 | 4+4 2 |
| | 6 2 | 6 2 | 5+5 2 |
| | 6 2 | 6 2 | 6+6 2 |
| | 6 2 | 6 2 | 4+4 2 |
| | 6 2 | 6 2 | 3+3 2 |

Continued on next column

**TABLE XXXIX– continued**

| TQC | ICSD | NoE | SF |
|---|---|---|---|
| | 5 2 | 5 2 | 5+5 2 |
| | 6 2 | 5 2 | 6+6 2 |
| | 5 2 | 5 2 | 3+3 2 |
| | 6 2 | 6 2 | 4+4 2 |
| | 5 2 | 5 2 | 4+4 2 |
| | 5 2 | 5 2 | 3+3 2 |
| | 5 2 | 5 2 | 6+6 2 |
| | 6 2 | 5 2 | 5+5 2 |
| | 5 2 | 6 2 | 3+3 2 |
| | 6 2 | 6 2 | 6+6 2 |
| | 5 2 | 5 2 | 4+4 2 |
| | 6 2 | 5 2 | 5+5 2 |
| | 5 2 | 6 2 | 5+5 2 |
| | 5 2 | 5 2 | 3+3 2 |
| | 5 2 | 6 2 | 6+6 2 |
| | 6 2 | 5 2 | 4+4 2 |
| | 5 2 | 5 2 | 4+4 2 |
| | 6 2 | 6 2 | 3+3 2 |
| | 5 2 | 5 2 | 3+3 2 |
| | 5 2 | 5 2 | 4+4 2 |
| | 6 2 | 6 2 | 4+4 2 |
| | 6 2 | 5 2 | 3+3 2 |
| | 5 2 | 5 2 | 4+4 2 |
| | 5 2 | 6 2 | 3+3 2 |
| | 5 2 | 6 2 | 5+5 2 |
| | 5 2 | 5 2 | 6+6 2 |
| 64.1.26 | 68347 | 180 | Ag1Fe2S3 |
| | -GM -T | -Y -Z | -R -S |
| | 5 2 3+4 4 | 6 2 3+4 4 | 3+3 2 2+2 4 |
| | 5 2 3+4 4 | 6 2 3+4 4 | 4+4 2 2+2 4 |
| | 6 2 3+4 4 | 5 2 3+4 4 | 5+5 2 2+2 4 |
| | 6 2 3+4 4 | 6 2 3+4 4 | 6+6 2 2+2 4 |
| | 5 2 3+4 4 | 6 2 3+4 4 | 5+5 2 2+2 4 |
| | 6 2 3+4 4 | 5 2 3+4 4 | 6+6 2 2+2 4 |
| | 5 2 3+4 4 | 5 2 3+4 4 | 3+3 2 2+2 4 |
| | 5 2 3+4 4 | 6 2 3+4 4 | 4+4 2 2+2 4 |
| | 6 2 3+4 4 | 5 2 3+4 4 | 5+5 2 2+2 4 |
| | 6 2 3+4 4 | 5 2 3+4 4 | 6+6 2 2+2 4 |
| | 6 2 3+4 4 | 6 2 3+4 4 | 3+3 2 2+2 4 |
| | 5 2 3+4 4 | 5 2 3+4 4 | 4+4 2 2+2 4 |
| | 5 2 3+4 4 | 5 2 3+4 4 | 6+6 2 2+2 4 |
| | 5 2 3+4 4 | 6 2 3+4 4 | 5+5 2 2+2 4 |
| | 6 2 3+4 4 | 5 2 3+4 4 | 3+3 2 2+2 4 |
| | 6 2 3+4 4 | 6 2 3+4 4 | 4+4 2 2+2 4 |
| | 5 2 3+4 4 | 6 2 3+4 4 | 4+4 2 2+2 4 |
| | 5 2 3+4 4 | 5 2 3+4 4 | 3+3 2 2+2 4 |
| | 5 2 3+4 4 | 6 2 3+4 4 | 5+5 2 2+2 4 |
| | 6 2 3+4 4 | 6 2 3+4 4 | 6+6 2 2+2 4 |
| | 6 2 3+4 4 | 5 2 3+4 4 | 4+4 2 2+2 4 |
| | 5 2 3+4 4 | 5 2 3+4 4 | 6+6 2 2+2 4 |
| | 6 2 3+4 4 | 6 2 3+4 4 | 3+3 2 2+2 4 |
| | 5 2 3+4 4 | 5 2 3+4 4 | 5+5 2 2+2 4 |
| | 6 2 3+4 4 | 5 2 3+4 4 | 6+6 2 2+2 4 |
| | 5 2 3+4 4 | 5 2 3+4 4 | 5+5 2 2+2 4 |
| | 6 2 3+4 4 | 6 2 3+4 4 | 3+3 2 2+2 4 |
| | 6 2 3+4 4 | 6 2 3+4 4 | 4+4 2 2+2 4 |
| | 5 2 3+4 4 | 5 2 3+4 4 | 5+5 2 2+2 4 |
| | 6 2 3+4 4 | 6 2 3+4 4 | 6+6 2 2+2 4 |
| | 5 2 3+4 4 | 5 2 3+4 4 | 4+4 2 2+2 4 |
| | 5 2 3+4 4 | 6 2 3+4 4 | 3+3 2 2+2 4 |
| | 6 2 3+4 4 | 6 2 3+4 4 | 6+6 2 2+2 4 |

Continued on next column

**TABLE XXXIX– continued**

| TQC | ICSD | NoE | SF |
|---|---|---|---|
| | 6 2 3+4 4 | 5 2 3+4 4 | 5+5 2 2+2 4 |
| | 5 2 3+4 4 | 6 2 3+4 4 | 3+3 2 2+2 4 |
| | 6 2 3+4 4 | 5 2 3+4 4 | 4+4 2 2+2 4 |
| | 5 2 3+4 4 | 5 2 3+4 4 | 5+5 2 2+2 4 |
| | 6 2 3+4 4 | 6 2 3+4 4 | 6+6 2 2+2 4 |
| | 6 2 3+4 4 | 5 2 3+4 4 | 4+4 2 2+2 4 |
| | 5 2 3+4 4 | 6 2 3+4 4 | 3+3 2 2+2 4 |
| | 6 2 3+4 4 | 5 2 3+4 4 | 6+6 2 2+2 4 |
| | 6 2 3+4 4 | 5 2 3+4 4 | 3+3 2 2+2 4 |
| | 5 2 3+4 4 | 6 2 3+4 4 | 4+4 2 2+2 4 |
| | 6 2 3+4 4 | 6 2 3+4 4 | 5+5 2 2+2 4 |
| | 5 2 3+4 4 | 6 2 3+4 4 | 6+6 2 2+2 4 |
| | 6 2 | 5 2 | 5+5 2 |
| | 5 2 | 5 2 | 3+3 2 |
| | 6 2 | 5 2 | 4+4 2 |
| | 5 2 | 6 2 | 4+4 2 |
| | 5 2 | 5 2 | 3+3 2 |
| | 6 2 | 5 2 | 6+6 2 |
| | 6 2 | 6 2 | 5+5 2 |
| | 5 2 | 6 2 | 4+4 2 |
| | 6 2 | 6 2 | 3+3 2 |
| | 5 2 | 6 2 | 5+5 2 |
| | 5 2 | 5 2 | 6+6 2 |
| | 6 2 | 5 2 | 3+3 2 |
| | 5 2 | 6 2 | 4+4 2 |
| | 5 2 | 5 2 | 6+6 2 |
| | 6 2 | 5 2 | 5+5 2 |
| | 6 2 | 6 2 | 3+3 2 |
| | 5 2 | 5 2 | 4+4 2 |
| | 5 2 | 6 2 | 5+5 2 |
| | 6 2 | 6 2 | 6+6 2 |
| | 5 2 | 6 2 | 4+4 2 |
| | 6 2 | 5 2 | 3+3 2 |
| | 5 2 | 5 2 | 5+5 2 |
| | 6 2 | 6 2 | 6+6 2 |
| | 5 2 | 6 2 | 3+3 2 |
| | 5 2 | 6 2 | 5+5 2 |
| | 6 2 | 5 2 | 6+6 2 |
| | 6 2 | 5 2 | 4+4 2 |
| | 5 2 | 6 2 | 4+4 2 |
| | 5 2 | 6 2 | 3+3 2 |
| | 6 2 | 6 2 | 5+5 2 |
| | 6 2 | 5 2 | 6+6 2 |
| | 5 2 | 6 2 | 3+3 2 |
| | 5 2 | 5 2 | 4+4 2 |
| | 5 2 | 6 2 | 5+5 2 |
| | 5 2 | 6 2 | 6+6 2 |
| | 5 2 | 5 2 | 6+6 2 |
| | 6 2 | 5 2 | 5+5 2 |
| | 5 2 | 6 2 | 3+3 2 |
| | 6 2 | 5 2 | 4+4 2 |
| | 6 2 | 6 2 | 5+5 2 |
| | 6 2 | 5 2 | 6+6 2 |
| | 5 2 | 6 2 | 6+6 2 |
| | 6 2 | 5 2 | 5+5 2 |
| | 5 2 | 6 2 | 3+3 2 |
| | 6 2 | 6 2 | 4+4 2 |

The End

TABLE XL: The computed irreducible representations for the valence bands for the compounds of SEBR and NLC in space group 65. See the list of tables.

| TQC | ICSD | | NoE | | SF | |
|---|---|---|---|---|---|---|
| 65.1.1 | 87827 | | 66 | | Ni1Si3Y1 | |
| | -GM | -T | -Y | -Z | -R | -S |
| | 5 2 | 5 2 | 6 2 | 6 2 | 3+4 2 | 3+4 2 |
| | 6 2 | 6 2 | 5 2 | 5 2 | 5+6 2 | 5+6 2 |
| | 5 2 | 5 2 | 6 2 | 6 2 | 3+4 2 | 3+4 2 |
| | 6 2 | 6 2 | 5 2 | 5 2 | 5+6 2 | 5+6 2 |
| | 5 2 | 5 2 | 6 2 | 6 2 | 3+4 2 | 3+4 2 |
| | 6 2 | 6 2 | 5 2 | 5 2 | 5+6 2 | 5+6 2 |
| | 5 2 | 5 2 | 6 2 | 6 2 | 3+4 2 | 3+4 2 |
| | 6 2 | 6 2 | 5 2 | 5 2 | 5+6 2 | 5+6 2 |
| | 5 2 | 5 2 | 5 2 | 5 2 | 5+6 2 | 3+4 2 |
| | 5 2 | 6 2 | 6 2 | 5 2 | 5+6 2 | 5+6 2 |
| | 6 2 | 6 2 | 6 2 | 6 2 | 3+4 2 | 5+6 2 |
| | 6 2 | 6 2 | 5 2 | 6 2 | 5+6 2 | 3+4 2 |
| | 5 2 | 5 2 | 5 2 | 5 2 | 3+4 2 | 5+6 2 |
| | 5 2 | 5 2 | 5 2 | 6 2 | 3+4 2 | 3+4 2 |
| | 6 2 | 5 2 | 6 2 | 6 2 | 3+4 2 | 3+4 2 |
| | 6 2 | 5 2 | 6 2 | 5 2 | 3+4 2 | 5+6 2 |
| | 6 2 | 6 2 | 6 2 | 5 2 | 5+6 2 | 3+4 2 |
| | 5 2 | 6 2 | 5 2 | 6 2 | 5+6 2 | 5+6 2 |
| | 5 2 | 6 2 | 5 2 | 5 2 | 3+4 2 | 3+4 2 |
| | 5 2 | 5 2 | 6 2 | 5 2 | 5+6 2 | 3+4 2 |
| | 5 2 | 5 2 | 5 2 | 5 2 | 3+4 2 | 5+6 2 |
| | 6 2 | 6 2 | 6 2 | 6 2 | 5+6 2 | 5+6 2 |
| | 5 2 | 6 2 | 6 2 | 6 2 | 5+6 2 | 5+6 2 |
| | 6 2 | 6 2 | 5 2 | 6 2 | 5+6 2 | 3+4 2 |
| | 5 2 | 5 2 | 6 2 | 5 2 | 3+4 2 | 5+6 2 |
| | 6 2 | 6 2 | 5 2 | 5 2 | 5+6 2 | 3+4 2 |
| | 6 2 | 5 2 | 6 2 | 6 2 | 3+4 2 | 5+6 2 |
| | 6 2 | 6 2 | 5 2 | 6 2 | 5+6 2 | 5+6 2 |
| | 5 2 | 5 2 | 6 2 | 6 2 | 3+4 2 | 3+4 2 |
| | 6 2 | 6 2 | 5 2 | 5 2 | 3+4 2 | 3+4 2 |
| | 5 2 | 5 2 | 5 2 | 6 2 | 5+6 2 | 5+6 2 |
| | 5 2 | 5 2 | 5 2 | 5 2 | 3+4 2 | 3+4 2 |
| | 5 2 | 6 2 | 6 2 | 5 2 | 5+6 2 | 3+4 2 |
| 65.1.2 | 78760 | | 42 | | Mn2S4Sn1 | |
| | -GM | -T | -Y | -Z | -R | -S |
| | 5 2 | 6 2 | 5 2 | 5 2 | 3+4 2 | 3+4 2 |
| | 6 2 | 6 2 | 6 2 | 6 2 | 5+6 2 | 5+6 2 |
| | 6 2 | 5 2 | 6 2 | 6 2 | 5+6 2 | 5+6 2 |
| | 5 2 | 5 2 | 5 2 | 5 2 | 3+4 2 | 3+4 2 |
| | 5 2 | 5 2 | 5 2 | 5 2 | 3+4 2 | 3+4 2 |
| | 5 2 | 5 2 | 6 2 | 6 2 | 5+6 2 | 5+6 2 |
| | 6 2 | 6 2 | 6 2 | 6 2 | 5+6 2 | 3+4 2 |
| | 5 2 | 6 2 | 6 2 | 6 2 | 3+4 2 | 5+6 2 |
| | 6 2 | 6 2 | 6 2 | 5 2 | 5+6 2 | 5+6 2 |
| | 5 2 | 5 2 | 5 2 | 6 2 | 5+6 2 | 5+6 2 |
| | 6 2 | 5 2 | 5 2 | 5 2 | 3+4 2 | 3+4 2 |
| | 5 2 | 6 2 | 6 2 | 6 2 | 3+4 2 | 3+4 2 |
| | 5 2 | 6 2 | 6 2 | 6 2 | 5+6 2 | 5+6 2 |
| | 6 2 | 6 2 | 5 2 | 5 2 | 5+6 2 | 3+4 2 |
| | 6 2 | 5 2 | 5 2 | 5 2 | 3+4 2 | 5+6 2 |
| | 6 2 | 5 2 | 5 2 | 5 2 | 3+4 2 | 3+4 2 |
| | 5 2 | 5 2 | 6 2 | 6 2 | 3+4 2 | 3+4 2 |
| | 5 2 | 5 2 | 6 2 | 6 2 | 5+6 2 | 5+6 2 |
| | 5 2 | 5 2 | 6 2 | 6 2 | 3+4 2 | 3+4 2 |
| | 5 2 | 5 2 | 6 2 | 6 2 | 5+6 2 | 5+6 2 |
| | 5 2 | 5 2 | 6 2 | 6 2 | 5+6 2 | 3+4 2 |

Continued on next column

**TABLE XL– continued**

| TQC | ICSD | NoE | SF | | | |
|---|---|---|---|---|---|---|
| 65.1.3 | 240761 | 64 | Mg3Ni2Tb2 | | | |
| | -GM | -T | -Y | -Z | -R | -S |
| | 5 2 | 6 2 | 6 2 | 5 2 | 3+4 2 | 3+4 2 |
| | 6 2 | 5 2 | 5 2 | 6 2 | 5+6 2 | 5+6 2 |
| | 5 2 | 6 2 | 6 2 | 5 2 | 3+4 2 | 3+4 2 |
| | 6 2 | 5 2 | 5 2 | 6 2 | 5+6 2 | 5+6 2 |
| | 5 2 | 6 2 | 6 2 | 5 2 | 3+4 2 | 3+4 2 |
| | 6 2 | 5 2 | 5 2 | 6 2 | 5+6 2 | 5+6 2 |
| | 5 2 | 6 2 | 6 2 | 5 2 | 3+4 2 | 3+4 2 |
| | 6 2 | 5 2 | 5 2 | 6 2 | 5+6 2 | 5+6 2 |
| | 5 2 | 5 2 | 5 2 | 6 2 | 3+4 2 | 5+6 2 |
| | 5 2 | 6 2 | 6 2 | 6 2 | 5+6 2 | 3+4 2 |
| | 6 2 | 5 2 | 5 2 | 5 2 | 3+4 2 | 5+6 2 |
| | 5 2 | 6 2 | 6 2 | 5 2 | 5+6 2 | 3+4 2 |
| | 6 2 | 6 2 | 5 2 | 6 2 | 3+4 2 | 3+4 2 |
| | 6 2 | 5 2 | 6 2 | 6 2 | 5+6 2 | 5+6 2 |
| | 5 2 | 6 2 | 6 2 | 5 2 | 5+6 2 | 3+4 2 |
| | 6 2 | 5 2 | 5 2 | 5 2 | 3+4 2 | 5+6 2 |
| | 6 2 | 6 2 | 5 2 | 5 2 | 5+6 2 | 3+4 2 |
| | 5 2 | 5 2 | 5 2 | 5 2 | 3+4 2 | 5+6 2 |
| | 5 2 | 6 2 | 6 2 | 6 2 | 5+6 2 | 3+4 2 |
| | 5 2 | 5 2 | 6 2 | 5 2 | 3+4 2 | 5+6 2 |
| | 6 2 | 5 2 | 5 2 | 6 2 | 5+6 2 | 3+4 2 |
| | 6 2 | 6 2 | 6 2 | 6 2 | 5+6 2 | 5+6 2 |
| | 5 2 | 6 2 | 6 2 | 5 2 | 3+4 2 | 3+4 2 |
| | 6 2 | 5 2 | 5 2 | 6 2 | 5+6 2 | 5+6 2 |
| | 5 2 | 5 2 | 6 2 | 5 2 | 3+4 2 | 3+4 2 |
| | 5 2 | 6 2 | 6 2 | 6 2 | 3+4 2 | 5+6 2 |
| | 6 2 | 6 2 | 5 2 | 5 2 | 5+6 2 | 3+4 2 |
| | 6 2 | 5 2 | 5 2 | 6 2 | 5+6 2 | 5+6 2 |
| | 5 2 | 5 2 | 6 2 | 5 2 | 3+4 2 | 3+4 2 |
| | 5 2 | 6 2 | 5 2 | 6 2 | 3+4 2 | 5+6 2 |
| | 6 2 | 5 2 | 6 2 | 6 2 | 5+6 2 | 5+6 2 |
| | 5 2 | 6 2 | 6 2 | 5 2 | 3+4 2 | 3+4 2 |
| 65.1.4 | 261871 | 84 | La2Sn5 | | | |
| | -GM | -T | -Y | -Z | -R | -S |
| | 5 2 | 5 2 | 5 2 | 5 2 | 5+6 2 | 5+6 2 |
| | 6 2 | 6 2 | 6 2 | 6 2 | 3+4 2 | 3+4 2 |
| | 5 2 | 6 2 | 6 2 | 5 2 | 3+4 2 | 3+4 2 |
| | 6 2 | 5 2 | 5 2 | 6 2 | 5+6 2 | 5+6 2 |
| | 5 2 | 5 2 | 5 2 | 5 2 | 5+6 2 | 5+6 2 |
| | 6 2 | 6 2 | 6 2 | 6 2 | 3+4 2 | 3+4 2 |
| | 5 2 | 6 2 | 6 2 | 5 2 | 3+4 2 | 3+4 2 |
| | 6 2 | 5 2 | 5 2 | 6 2 | 5+6 2 | 5+6 2 |
| | 5 2 | 5 2 | 5 2 | 5 2 | 3+4 2 | 5+6 2 |
| | 5 2 | 6 2 | 6 2 | 6 2 | 5+6 2 | 3+4 2 |
| | 6 2 | 6 2 | 6 2 | 5 2 | 5+6 2 | 3+4 2 |
| | 6 2 | 5 2 | 6 2 | 6 2 | 3+4 2 | 5+6 2 |
| | 5 2 | 5 2 | 5 2 | 5 2 | 5+6 2 | 5+6 2 |
| | 6 2 | 6 2 | 5 2 | 6 2 | 3+4 2 | 3+4 2 |
| | 5 2 | 6 2 | 6 2 | 5 2 | 3+4 2 | 3+4 2 |
| | 6 2 | 5 2 | 5 2 | 6 2 | 5+6 2 | 5+6 2 |
| | 5 2 | 6 2 | 5 2 | 6 2 | 5+6 2 | 3+4 2 |
| | 6 2 | 5 2 | 6 2 | 5 2 | 3+4 2 | 5+6 2 |
| | 5 2 | 6 2 | 5 2 | 6 2 | 5+6 2 | 3+4 2 |
| | 5 2 | 5 2 | 6 2 | 5 2 | 3+4 2 | 5+6 2 |
| | 6 2 | 5 2 | 6 2 | 6 2 | 3+4 2 | 3+4 2 |
| | 5 2 | 6 2 | 5 2 | 6 2 | 3+4 2 | 5+6 2 |
| | 6 2 | 5 2 | 5 2 | 5 2 | 5+6 2 | 3+4 2 |
| | 5 2 | 5 2 | 5 2 | 5 2 | 3+4 2 | 5+6 2 |
| | 6 2 | 6 2 | 6 2 | 6 2 | 5+6 2 | 5+6 2 |

Continued on next column

**TABLE XL– continued**

| TQC | ICSD | NoE | SF | | | |
|---|---|---|---|---|---|---|
| | 5 2 | 6 2 | 5 2 | 5 2 | 3+4 2 | 3+4 2 |
| | 6 2 | 6 2 | 6 2 | 6 2 | 5+6 2 | 3+4 2 |
| | 5 2 | 5 2 | 6 2 | 5 2 | 5+6 2 | 5+6 2 |
| | 6 2 | 5 2 | 5 2 | 5 2 | 3+4 2 | 5+6 2 |
| | 6 2 | 6 2 | 6 2 | 6 2 | 3+4 2 | 3+4 2 |
| | 6 2 | 5 2 | 6 2 | 5 2 | 5+6 2 | 5+6 2 |
| | 5 2 | 5 2 | 6 2 | 5 2 | 5+6 2 | 3+4 2 |
| | 5 2 | 5 2 | 5 2 | 6 2 | 5+6 2 | 5+6 2 |
| | 6 2 | 6 2 | 6 2 | 6 2 | 5+6 2 | 5+6 2 |
| | 6 2 | 6 2 | 5 2 | 6 2 | 3+4 2 | 3+4 2 |
| | 5 2 | 6 2 | 6 2 | 6 2 | 3+4 2 | 3+4 2 |
| | 6 2 | 5 2 | 6 2 | 5 2 | 3+4 2 | 3+4 2 |
| | 5 2 | 5 2 | 5 2 | 5 2 | 5+6 2 | 5+6 2 |
| | 6 2 | 6 2 | 5 2 | 5 2 | 5+6 2 | 5+6 2 |
| | 5 2 | 6 2 | 6 2 | 6 2 | 3+4 2 | 3+4 2 |
| | 6 2 | 6 2 | 5 2 | 6 2 | 5+6 2 | 3+4 2 |
| | 5 2 | 5 2 | 6 2 | 6 2 | 5+6 2 | 3+4 2 |
| 65.1.5 | 167637 | 188 | In9Ni4Y11 | | | |
| | -GM | -T | -Y | -Z | -R | -S |
| | 5 2 | 5 2 | 5 2 | 5 2 | 3+4 2 | 3+4 2 |
| | 6 2 | 6 2 | 6 2 | 6 2 | 5+6 2 | 5+6 2 |
| | 5 2 | 5 2 | 5 2 | 5 2 | 3+4 2 | 3+4 2 |
| | 6 2 | 6 2 | 6 2 | 6 2 | 5+6 2 | 5+6 2 |
| | 5 2 | 5 2 | 5 2 | 5 2 | 3+4 2 | 3+4 2 |
| | 5 2 | 5 2 | 5 2 | 5 2 | 3+4 2 | 3+4 2 |
| | 6 2 | 6 2 | 6 2 | 6 2 | 5+6 2 | 5+6 2 |
| | 5 2 | 6 2 | 6 2 | 5 2 | 3+4 2 | 3+4 2 |
| | 5 2 | 6 2 | 6 2 | 5 2 | 5+6 2 | 5+6 2 |
| | 6 2 | 5 2 | 5 2 | 6 2 | 3+4 2 | 3+4 2 |
| | 6 2 | 5 2 | 5 2 | 6 2 | 5+6 2 | 5+6 2 |
| | 6 2 | 6 2 | 6 2 | 6 2 | 5+6 2 | 5+6 2 |
| | 5 2 | 5 2 | 5 2 | 5 2 | 3+4 2 | 3+4 2 |
| | 6 2 | 5 2 | 5 2 | 5 2 | 3+4 2 | 3+4 2 |
| | 5 2 | 6 2 | 6 2 | 6 2 | 5+6 2 | 5+6 2 |
| | 6 2 | 6 2 | 6 2 | 6 2 | 5+6 2 | 5+6 2 |
| | 5 2 | 5 2 | 5 2 | 5 2 | 5+6 2 | 3+4 2 |
| | 6 2 | 6 2 | 6 2 | 6 2 | 3+4 2 | 5+6 2 |
| | 5 2 | 6 2 | 6 2 | 5 2 | 3+4 2 | 3+4 2 |
| | 5 2 | 6 2 | 5 2 | 5 2 | 5+6 2 | 5+6 2 |
| | 6 2 | 5 2 | 6 2 | 6 2 | 3+4 2 | 3+4 2 |
| | 6 2 | 5 2 | 5 2 | 6 2 | 5+6 2 | 5+6 2 |
| | 6 2 | 6 2 | 6 2 | 6 2 | 5+6 2 | 5+6 2 |
| | 5 2 | 5 2 | 5 2 | 5 2 | 3+4 2 | 3+4 2 |
| | 6 2 | 6 2 | 6 2 | 6 2 | 5+6 2 | 5+6 2 |
| | 6 2 | 5 2 | 6 2 | 5 2 | 3+4 2 | 3+4 2 |
| | 5 2 | 5 2 | 5 2 | 5 2 | 5+6 2 | 5+6 2 |
| | 5 2 | 6 2 | 5 2 | 6 2 | 3+4 2 | 3+4 2 |
| | 5 2 | 5 2 | 5 2 | 5 2 | 5+6 2 | 5+6 2 |
| | 6 2 | 6 2 | 6 2 | 6 2 | 3+4 2 | 3+4 2 |
| | 6 2 | 6 2 | 6 2 | 6 2 | 5+6 2 | 5+6 2 |
| | 5 2 | 6 2 | 6 2 | 6 2 | 3+4 2 | 3+4 2 |
| | 6 2 | 5 2 | 6 2 | 5 2 | 5+6 2 | 5+6 2 |
| | 5 2 | 6 2 | 5 2 | 5 2 | 3+4 2 | 5+6 2 |
| | 6 2 | 5 2 | 5 2 | 6 2 | 5+6 2 | 3+4 2 |
| | 5 2 | 6 2 | 6 2 | 6 2 | 5+6 2 | 5+6 2 |
| | 6 2 | 5 2 | 5 2 | 5 2 | 3+4 2 | 3+4 2 |
| | 6 2 | 6 2 | 6 2 | 6 2 | 5+6 2 | 5+6 2 |
| | 5 2 | 5 2 | 5 2 | 5 2 | 3+4 2 | 3+4 2 |
| | 6 2 | 6 2 | 6 2 | 6 2 | 5+6 2 | 5+6 2 |
| | 5 2 | 6 2 | 5 2 | 6 2 | 5+6 2 | 3+4 2 |
| | 6 2 | 5 2 | 6 2 | 5 2 | 3+4 2 | 5+6 2 |

Continued on next column

**TABLE XL– continued**

| TQC | ICSD | NoE | SF |
|---|---|---|---|
| | 6 2 5 2 | 6 2 5 2 | 3+4 2 5+6 2 |
| | 5 2 6 2 | 5 2 6 2 | 5+6 2 3+4 2 |
| | 5 2 5 2 | 6 2 6 2 | 3+4 2 3+4 2 |
| | 6 2 6 2 | 5 2 5 2 | 5+6 2 5+6 2 |
| | 5 2 5 2 | 5 2 5 2 | 3+4 2 3+4 2 |
| | 6 2 6 2 | 6 2 6 2 | 5+6 2 5+6 2 |
| | 5 2 6 2 | 5 2 6 2 | 3+4 2 5+6 2 |
| | 5 2 5 2 | 5 2 5 2 | 3+4 2 5+6 2 |
| | 6 2 6 2 | 6 2 6 2 | 5+6 2 3+4 2 |
| | 5 2 5 2 | 6 2 6 2 | 3+4 2 3+4 2 |
| | 6 2 6 2 | 5 2 5 2 | 5+6 2 5+6 2 |
| | 5 2 5 2 | 5 2 5 2 | 3+4 2 3+4 2 |
| | 6 2 6 2 | 6 2 6 2 | 5+6 2 3+4 2 |
| | 6 2 5 2 | 6 2 5 2 | 3+4 2 5+6 2 |
| | 5 2 5 2 | 5 2 5 2 | 5+6 2 5+6 2 |
| | 6 2 6 2 | 6 2 6 2 | 5+6 2 3+4 2 |
| | 5 2 5 2 | 5 2 6 2 | 3+4 2 5+6 2 |
| | 5 2 6 2 | 5 2 5 2 | 5+6 2 3+4 2 |
| | 6 2 5 2 | 6 2 5 2 | 5+6 2 3+4 2 |
| | 6 2 5 2 | 6 2 6 2 | 3+4 2 5+6 2 |
| | 6 2 6 2 | 6 2 5 2 | 3+4 2 3+4 2 |
| | 5 2 6 2 | 6 2 6 2 | 3+4 2 5+6 2 |
| | 5 2 5 2 | 5 2 6 2 | 5+6 2 3+4 2 |
| | 5 2 5 2 | 5 2 5 2 | 5+6 2 5+6 2 |
| | 6 2 5 2 | 5 2 6 2 | 3+4 2 5+6 2 |
| | 6 2 5 2 | 5 2 5 2 | 5+6 2 3+4 2 |
| | 5 2 6 2 | 6 2 6 2 | 5+6 2 5+6 2 |
| | 5 2 5 2 | 6 2 5 2 | 3+4 2 3+4 2 |
| | 6 2 5 2 | 6 2 5 2 | 5+6 2 3+4 2 |
| | 6 2 6 2 | 5 2 6 2 | 3+4 2 3+4 2 |
| | 5 2 5 2 | 5 2 5 2 | 3+4 2 5+6 2 |
| | 6 2 6 2 | 6 2 6 2 | 5+6 2 5+6 2 |
| | 5 2 6 2 | 6 2 5 2 | 3+4 2 3+4 2 |
| | 6 2 6 2 | 5 2 5 2 | 3+4 2 3+4 2 |
| | 5 2 6 2 | 5 2 6 2 | 5+6 2 5+6 2 |
| | 6 2 5 2 | 5 2 6 2 | 3+4 2 5+6 2 |
| | 5 2 6 2 | 6 2 5 2 | 5+6 2 5+6 2 |
| | 5 2 6 2 | 6 2 6 2 | 3+4 2 3+4 2 |
| | 6 2 5 2 | 5 2 6 2 | 5+6 2 5+6 2 |
| | 5 2 6 2 | 6 2 5 2 | 3+4 2 3+4 2 |
| | 6 2 5 2 | 5 2 5 2 | 5+6 2 3+4 2 |
| | 6 2 5 2 | 6 2 6 2 | 5+6 2 5+6 2 |
| | 5 2 6 2 | 6 2 5 2 | 3+4 2 3+4 2 |
| | 6 2 5 2 | 5 2 6 2 | 5+6 2 5+6 2 |
| | 5 2 6 2 | 5 2 6 2 | 3+4 2 3+4 2 |
| | 6 2 5 2 | 6 2 5 2 | 5+6 2 5+6 2 |
| | 5 2 5 2 | 6 2 6 2 | 5+6 2 3+4 2 |
| | 5 2 5 2 | 5 2 5 2 | 3+4 2 5+6 2 |
| | 6 2 6 2 | 5 2 5 2 | 3+4 2 3+4 2 |
| | 5 2 6 2 | 5 2 5 2 | 3+4 2 5+6 2 |
| | 5 2 5 2 | 6 2 6 2 | 5+6 2 5+6 2 |
| | 6 2 5 2 | 6 2 5 2 | 5+6 2 3+4 2 |
| 65.1.6 | 167624 | 188 | In9La11Ni4 |
| | -GM -T | -Y -Z | -R -S |
| | 5 2 5 2 | 5 2 5 2 | 3+4 2 3+4 2 |
| | 6 2 6 2 | 6 2 6 2 | 5+6 2 5+6 2 |
| | 5 2 6 2 | 6 2 5 2 | 3+4 2 3+4 2 |
| | 5 2 6 2 | 6 2 5 2 | 5+6 2 5+6 2 |
| | 6 2 5 2 | 5 2 6 2 | 3+4 2 3+4 2 |
| | 6 2 5 2 | 5 2 6 2 | 5+6 2 5+6 2 |
| | 5 2 5 2 | 5 2 5 2 | 3+4 2 3+4 2 |

Continued on next column

**TABLE XL– continued**

| TQC | ICSD | NoE | SF |
|---|---|---|---|
| | 5 2 5 2 | 5 2 5 2 | 5+6 2 5+6 2 |
| | 6 2 6 2 | 6 2 6 2 | 3+4 2 3+4 2 |
| | 6 2 6 2 | 6 2 6 2 | 3+4 2 3+4 2 |
| | 5 2 5 2 | 5 2 5 2 | 5+6 2 5+6 2 |
| | 6 2 6 2 | 6 2 6 2 | 5+6 2 5+6 2 |
| | 5 2 5 2 | 5 2 5 2 | 3+4 2 3+4 2 |
| | 5 2 6 2 | 6 2 5 2 | 5+6 2 5+6 2 |
| | 5 2 6 2 | 6 2 5 2 | 3+4 2 3+4 2 |
| | 6 2 5 2 | 5 2 6 2 | 5+6 2 5+6 2 |
| | 6 2 5 2 | 6 2 6 2 | 3+4 2 3+4 2 |
| | 6 2 6 2 | 5 2 6 2 | 5+6 2 5+6 2 |
| | 6 2 5 2 | 6 2 5 2 | 3+4 2 3+4 2 |
| | 5 2 6 2 | 5 2 6 2 | 5+6 2 5+6 2 |
| | 5 2 5 2 | 5 2 5 2 | 5+6 2 5+6 2 |
| | 6 2 6 2 | 6 2 6 2 | 3+4 2 3+4 2 |
| | 6 2 6 2 | 6 2 6 2 | 5+6 2 5+6 2 |
| | 5 2 5 2 | 5 2 5 2 | 3+4 2 3+4 2 |
| | 5 2 6 2 | 6 2 6 2 | 5+6 2 5+6 2 |
| | 5 2 5 2 | 6 2 5 2 | 3+4 2 3+4 2 |
| | 6 2 6 2 | 6 2 5 2 | 5+6 2 5+6 2 |
| | 6 2 6 2 | 5 2 5 2 | 3+4 2 3+4 2 |
| | 5 2 5 2 | 5 2 6 2 | 5+6 2 5+6 2 |
| | 6 2 6 2 | 6 2 6 2 | 3+4 2 3+4 2 |
| | 6 2 5 2 | 5 2 6 2 | 5+6 2 3+4 2 |
| | 5 2 6 2 | 6 2 5 2 | 5+6 2 5+6 2 |
| | 6 2 5 2 | 5 2 5 2 | 3+4 2 5+6 2 |
| | 5 2 6 2 | 6 2 6 2 | 5+6 2 5+6 2 |
| | 6 2 5 2 | 5 2 5 2 | 5+6 2 3+4 2 |
| | 6 2 6 2 | 6 2 6 2 | 3+4 2 5+6 2 |
| | 5 2 5 2 | 5 2 5 2 | 3+4 2 3+4 2 |
| | 6 2 6 2 | 6 2 6 2 | 5+6 2 5+6 2 |
| | 5 2 6 2 | 5 2 6 2 | 3+4 2 5+6 2 |
| | 6 2 5 2 | 6 2 6 2 | 5+6 2 3+4 2 |
| | 6 2 6 2 | 6 2 6 2 | 3+4 2 3+4 2 |
| | 5 2 5 2 | 5 2 5 2 | 5+6 2 5+6 2 |
| | 6 2 5 2 | 6 2 5 2 | 5+6 2 3+4 2 |
| | 5 2 6 2 | 5 2 6 2 | 3+4 2 5+6 2 |
| | 5 2 5 2 | 5 2 5 2 | 3+4 2 3+4 2 |
| | 5 2 6 2 | 6 2 6 2 | 5+6 2 5+6 2 |
| | 6 2 6 2 | 5 2 6 2 | 3+4 2 3+4 2 |
| | 6 2 5 2 | 6 2 5 2 | 5+6 2 5+6 2 |
| | 5 2 6 2 | 5 2 6 2 | 3+4 2 5+6 2 |
| | 5 2 6 2 | 6 2 5 2 | 3+4 2 5+6 2 |
| | 6 2 5 2 | 5 2 5 2 | 5+6 2 3+4 2 |
| | 5 2 6 2 | 6 2 6 2 | 3+4 2 3+4 2 |
| | 6 2 5 2 | 5 2 6 2 | 5+6 2 5+6 2 |
| | 5 2 5 2 | 5 2 5 2 | 5+6 2 3+4 2 |
| | 6 2 5 2 | 6 2 6 2 | 3+4 2 3+4 2 |
| | 6 2 5 2 | 6 2 5 2 | 3+4 2 5+6 2 |
| | 5 2 6 2 | 5 2 5 2 | 5+6 2 3+4 2 |
| | 5 2 5 2 | 6 2 5 2 | 5+6 2 5+6 2 |
| | 6 2 6 2 | 5 2 5 2 | 3+4 2 3+4 2 |
| | 6 2 5 2 | 6 2 6 2 | 5+6 2 5+6 2 |
| | 5 2 5 2 | 5 2 6 2 | 3+4 2 3+4 2 |
| | 6 2 5 2 | 6 2 5 2 | 5+6 2 5+6 2 |
| | 5 2 5 2 | 6 2 6 2 | 3+4 2 3+4 2 |
| | 6 2 6 2 | 5 2 5 2 | 5+6 2 3+4 2 |
| | 6 2 5 2 | 5 2 5 2 | 3+4 2 5+6 2 |
| | 6 2 6 2 | 6 2 6 2 | 5+6 2 3+4 2 |
| | 5 2 6 2 | 5 2 6 2 | 3+4 2 5+6 2 |
| | 5 2 5 2 | 6 2 6 2 | 5+6 2 5+6 2 |

Continued on next column

**TABLE XL– continued**

| TQC | ICSD | NoE | SF | | | |
|---|---|---|---|---|---|---|
| | 6 2 | 6 2 | 5 2 | 6 2 | 3+4 2 | 3+4 2 |
| | 5 2 | 5 2 | 6 2 | 5 2 | 3+4 2 | 5+6 2 |
| | 5 2 | 6 2 | 5 2 | 6 2 | 3+4 2 | 5+6 2 |
| | 6 2 | 5 2 | 6 2 | 5 2 | 5+6 2 | 3+4 2 |
| | 5 2 | 6 2 | 6 2 | 5 2 | 5+6 2 | 3+4 2 |
| | 6 2 | 5 2 | 5 2 | 6 2 | 3+4 2 | 3+4 2 |
| | 6 2 | 6 2 | 5 2 | 5 2 | 5+6 2 | 5+6 2 |
| | 5 2 | 5 2 | 6 2 | 6 2 | 3+4 2 | 5+6 2 |
| | 5 2 | 6 2 | 5 2 | 5 2 | 5+6 2 | 3+4 2 |
| | 5 2 | 6 2 | 5 2 | 6 2 | 3+4 2 | 5+6 2 |
| | 6 2 | 5 2 | 6 2 | 5 2 | 5+6 2 | 3+4 2 |
| | 5 2 | 5 2 | 6 2 | 6 2 | 5+6 2 | 5+6 2 |
| | 6 2 | 6 2 | 5 2 | 5 2 | 5+6 2 | 3+4 2 |
| | 5 2 | 5 2 | 6 2 | 6 2 | 3+4 2 | 5+6 2 |
| | 6 2 | 6 2 | 5 2 | 6 2 | 3+4 2 | 3+4 2 |
| | 6 2 | 5 2 | 5 2 | 5 2 | 5+6 2 | 5+6 2 |
| | 5 2 | 6 2 | 6 2 | 6 2 | 3+4 2 | 5+6 2 |
| | 6 2 | 6 2 | 5 2 | 5 2 | 5+6 2 | 3+4 2 |
| | 5 2 | 6 2 | 6 2 | 6 2 | 3+4 2 | 3+4 2 |
| | 5 2 | 5 2 | 5 2 | 5 2 | 5+6 2 | 5+6 2 |
| | 5 2 | 6 2 | 6 2 | 6 2 | 3+4 2 | 3+4 2 |
| | 6 2 | 5 2 | 5 2 | 5 2 | 5+6 2 | 5+6 2 |
| | 5 2 | 5 2 | 6 2 | 5 2 | 5+6 2 | 3+4 2 |
| | 6 2 | 6 2 | 6 2 | 5 2 | 5+6 2 | 3+4 2 |
| | 5 2 | 5 2 | 5 2 | 5 2 | 3+4 2 | 5+6 2 |
| | 5 2 | 6 2 | 6 2 | 6 2 | 3+4 2 | 5+6 2 |
| 65.1.7 | 415513 | 156 | In16Pt7Tb2 | | | |
| | -GM | -T | -Y | -Z | -R | -S |
| | 5 2 | 5 2 | 5 2 | 5 2 | 3+4 2 | 3+4 2 |
| | 6 2 | 6 2 | 6 2 | 6 2 | 5+6 2 | 5+6 2 |
| | 5 2 | 6 2 | 5 2 | 6 2 | 5+6 2 | 3+4 2 |
| | 6 2 | 5 2 | 6 2 | 5 2 | 3+4 2 | 5+6 2 |
| | 5 2 | 6 2 | 5 2 | 6 2 | 5+6 2 | 3+4 2 |
| | 6 2 | 5 2 | 6 2 | 5 2 | 3+4 2 | 5+6 2 |
| | 6 2 | 5 2 | 6 2 | 5 2 | 3+4 2 | 5+6 2 |
| | 5 2 | 6 2 | 5 2 | 6 2 | 5+6 2 | 3+4 2 |
| | 5 2 | 5 2 | 6 2 | 6 2 | 3+4 2 | 5+6 2 |
| | 6 2 | 6 2 | 5 2 | 5 2 | 5+6 2 | 3+4 2 |
| | 5 2 | 6 2 | 6 2 | 5 2 | 3+4 2 | 5+6 2 |
| | 5 2 | 6 2 | 5 2 | 5 2 | 5+6 2 | 3+4 2 |
| | 6 2 | 5 2 | 5 2 | 6 2 | 3+4 2 | 5+6 2 |
| | 6 2 | 6 2 | 5 2 | 6 2 | 5+6 2 | 3+4 2 |
| | 5 2 | 5 2 | 6 2 | 5 2 | 5+6 2 | 5+6 2 |
| | 5 2 | 5 2 | 6 2 | 5 2 | 3+4 2 | 3+4 2 |
| | 6 2 | 5 2 | 6 2 | 6 2 | 3+4 2 | 3+4 2 |
| | 6 2 | 6 2 | 5 2 | 6 2 | 5+6 2 | 5+6 2 |
| | 5 2 | 6 2 | 5 2 | 5 2 | 3+4 2 | 5+6 2 |
| | 5 2 | 5 2 | 6 2 | 5 2 | 3+4 2 | 3+4 2 |
| | 6 2 | 5 2 | 6 2 | 6 2 | 5+6 2 | 3+4 2 |
| | 5 2 | 5 2 | 5 2 | 6 2 | 3+4 2 | 5+6 2 |
| | 5 2 | 6 2 | 6 2 | 6 2 | 5+6 2 | 3+4 2 |
| | 6 2 | 6 2 | 5 2 | 5 2 | 5+6 2 | 5+6 2 |
| | 5 2 | 6 2 | 5 2 | 6 2 | 3+4 2 | 5+6 2 |
| | 6 2 | 5 2 | 6 2 | 6 2 | 3+4 2 | 5+6 2 |
| | 5 2 | 6 2 | 5 2 | 5 2 | 5+6 2 | 3+4 2 |
| | 5 2 | 5 2 | 5 2 | 6 2 | 3+4 2 | 5+6 2 |
| | 6 2 | 6 2 | 6 2 | 6 2 | 5+6 2 | 3+4 2 |
| | 6 2 | 6 2 | 5 2 | 5 2 | 3+4 2 | 5+6 2 |
| | 5 2 | 5 2 | 6 2 | 5 2 | 5+6 2 | 3+4 2 |
| | 5 2 | 5 2 | 5 2 | 5 2 | 3+4 2 | 5+6 2 |
| | 6 2 | 6 2 | 5 2 | 5 2 | 5+6 2 | 5+6 2 |

Continued on next column

**TABLE XL– continued**

| TQC | ICSD | NoE | SF | | | |
|---|---|---|---|---|---|---|
| | 6 2 | 6 2 | 6 2 | 6 2 | 3+4 2 | 3+4 2 |
| | 6 2 | 5 2 | 6 2 | 6 2 | 3+4 2 | 5+6 2 |
| | 5 2 | 5 2 | 5 2 | 6 2 | 5+6 2 | 3+4 2 |
| | 5 2 | 6 2 | 5 2 | 6 2 | 5+6 2 | 5+6 2 |
| | 6 2 | 6 2 | 6 2 | 5 2 | 3+4 2 | 3+4 2 |
| | 5 2 | 6 2 | 5 2 | 6 2 | 3+4 2 | 5+6 2 |
| | 5 2 | 6 2 | 5 2 | 5 2 | 5+6 2 | 3+4 2 |
| | 6 2 | 5 2 | 6 2 | 5 2 | 3+4 2 | 5+6 2 |
| | 5 2 | 5 2 | 6 2 | 6 2 | 5+6 2 | 3+4 2 |
| | 6 2 | 6 2 | 5 2 | 6 2 | 3+4 2 | 5+6 2 |
| | 5 2 | 5 2 | 5 2 | 5 2 | 5+6 2 | 5+6 2 |
| | 6 2 | 6 2 | 6 2 | 6 2 | 3+4 2 | 3+4 2 |
| | 5 2 | 5 2 | 6 2 | 5 2 | 3+4 2 | 5+6 2 |
| | 5 2 | 6 2 | 5 2 | 5 2 | 5+6 2 | 3+4 2 |
| | 6 2 | 6 2 | 6 2 | 6 2 | 3+4 2 | 3+4 2 |
| | 6 2 | 5 2 | 5 2 | 6 2 | 5+6 2 | 5+6 2 |
| | 5 2 | 5 2 | 6 2 | 5 2 | 5+6 2 | 5+6 2 |
| | 5 2 | 6 2 | 5 2 | 5 2 | 3+4 2 | 3+4 2 |
| | 5 2 | 6 2 | 5 2 | 6 2 | 3+4 2 | 5+6 2 |
| | 6 2 | 5 2 | 5 2 | 5 2 | 5+6 2 | 5+6 2 |
| | 5 2 | 6 2 | 5 2 | 6 2 | 3+4 2 | 3+4 2 |
| | 6 2 | 6 2 | 6 2 | 6 2 | 5+6 2 | 5+6 2 |
| | 5 2 | 5 2 | 5 2 | 6 2 | 3+4 2 | 5+6 2 |
| | 6 2 | 5 2 | 6 2 | 6 2 | 3+4 2 | 3+4 2 |
| | 5 2 | 6 2 | 5 2 | 5 2 | 3+4 2 | 5+6 2 |
| | 6 2 | 6 2 | 5 2 | 5 2 | 5+6 2 | 3+4 2 |
| | 5 2 | 5 2 | 5 2 | 6 2 | 5+6 2 | 3+4 2 |
| | 6 2 | 6 2 | 6 2 | 5 2 | 5+6 2 | 3+4 2 |
| | 6 2 | 6 2 | 6 2 | 6 2 | 3+4 2 | 5+6 2 |
| | 5 2 | 5 2 | 5 2 | 5 2 | 3+4 2 | 5+6 2 |
| | 6 2 | 6 2 | 6 2 | 6 2 | 5+6 2 | 3+4 2 |
| | 6 2 | 6 2 | 5 2 | 5 2 | 5+6 2 | 3+4 2 |
| | 5 2 | 5 2 | 6 2 | 5 2 | 3+4 2 | 5+6 2 |
| | 5 2 | 6 2 | 5 2 | 5 2 | 3+4 2 | 5+6 2 |
| | 6 2 | 5 2 | 6 2 | 6 2 | 5+6 2 | 3+4 2 |
| | 5 2 | 6 2 | 6 2 | 5 2 | 3+4 2 | 3+4 2 |
| | 6 2 | 5 2 | 5 2 | 6 2 | 3+4 2 | 5+6 2 |
| | 6 2 | 5 2 | 5 2 | 6 2 | 5+6 2 | 3+4 2 |
| | 5 2 | 5 2 | 6 2 | 5 2 | 5+6 2 | 5+6 2 |
| | 5 2 | 6 2 | 6 2 | 5 2 | 3+4 2 | 3+4 2 |
| | 6 2 | 6 2 | 6 2 | 6 2 | 5+6 2 | 5+6 2 |
| | 5 2 | 5 2 | 5 2 | 6 2 | 3+4 2 | 5+6 2 |
| | 5 2 | 6 2 | 5 2 | 5 2 | 5+6 2 | 3+4 2 |
| | 6 2 | 5 2 | 6 2 | 6 2 | 3+4 2 | 3+4 2 |
| | 5 2 | 5 2 | 6 2 | 5 2 | 5+6 2 | 5+6 2 |
| 65.1.8 | 54612 | 66 | Ho2Ni2Pb1 | | | |
| | -GM | -T | -Y | -Z | -R | -S |
| | 5 2 | 5 2 | 6 2 | 6 2 | 3+4 2 | 5+6 2 |
| | 6 2 | 6 2 | 5 2 | 5 2 | 5+6 2 | 3+4 2 |
| | 5 2 | 5 2 | 6 2 | 6 2 | 3+4 2 | 5+6 2 |
| | 6 2 | 6 2 | 5 2 | 5 2 | 5+6 2 | 3+4 2 |
| | 5 2 | 5 2 | 6 2 | 6 2 | 5+6 2 | 5+6 2 |
| | 6 2 | 6 2 | 5 2 | 5 2 | 3+4 2 | 3+4 2 |
| | 5 2 | 5 2 | 6 2 | 6 2 | 3+4 2 | 5+6 2 |
| | 6 2 | 6 2 | 5 2 | 5 2 | 5+6 2 | 3+4 2 |
| | 5 2 | 5 2 | 5 2 | 5 2 | 3+4 2 | 3+4 2 |
| | 5 2 | 6 2 | 6 2 | 6 2 | 5+6 2 | 5+6 2 |
| | 6 2 | 6 2 | 5 2 | 5 2 | 5+6 2 | 3+4 2 |
| | 6 2 | 5 2 | 6 2 | 6 2 | 3+4 2 | 5+6 2 |
| | 5 2 | 5 2 | 5 2 | 6 2 | 3+4 2 | 5+6 2 |
| | 5 2 | 6 2 | 6 2 | 6 2 | 5+6 2 | 3+4 2 |

Continued on next column

**TABLE XL– continued**

| TQC | ICSD | NoE | SF |
|---|---|---|---|
| | 6 2 5 2 | 5 2 5 2 | 5+6 2 5+6 2 |
| | 6 2 6 2 | 6 2 5 2 | 5+6 2 3+4 2 |
| | 6 2 5 2 | 5 2 6 2 | 3+4 2 5+6 2 |
| | 5 2 5 2 | 6 2 6 2 | 3+4 2 3+4 2 |
| | 6 2 6 2 | 6 2 5 2 | 5+6 2 5+6 2 |
| | 5 2 6 2 | 5 2 5 2 | 3+4 2 3+4 2 |
| | 5 2 5 2 | 6 2 6 2 | 5+6 2 5+6 2 |
| | 6 2 6 2 | 5 2 6 2 | 5+6 2 3+4 2 |
| | 5 2 6 2 | 6 2 5 2 | 3+4 2 3+4 2 |
| | 6 2 5 2 | 6 2 6 2 | 5+6 2 5+6 2 |
| | 5 2 5 2 | 5 2 5 2 | 3+4 2 3+4 2 |
| | 6 2 6 2 | 5 2 5 2 | 3+4 2 5+6 2 |
| | 5 2 5 2 | 5 2 6 2 | 5+6 2 3+4 2 |
| | 5 2 6 2 | 6 2 5 2 | 5+6 2 5+6 2 |
| | 6 2 6 2 | 6 2 6 2 | 3+4 2 5+6 2 |
| | 6 2 5 2 | 5 2 5 2 | 5+6 2 3+4 2 |
| | 5 2 6 2 | 6 2 5 2 | 3+4 2 3+4 2 |
| | 5 2 5 2 | 6 2 6 2 | 5+6 2 5+6 2 |
| | 5 2 6 2 | 5 2 6 2 | 3+4 2 5+6 2 |
| 65.1.9 | 60646 | 66 | Ge3Ni1Y1 |
| | -GM -T | -Y -Z | -R -S |
| | 5 2 5 2 | 6 2 6 2 | 3+4 2 3+4 2 |
| | 6 2 6 2 | 5 2 5 2 | 5+6 2 5+6 2 |
| | 5 2 5 2 | 6 2 6 2 | 3+4 2 3+4 2 |
| | 6 2 6 2 | 5 2 5 2 | 5+6 2 5+6 2 |
| | 5 2 5 2 | 6 2 6 2 | 3+4 2 3+4 2 |
| | 6 2 6 2 | 5 2 5 2 | 5+6 2 5+6 2 |
| | 5 2 5 2 | 6 2 6 2 | 3+4 2 3+4 2 |
| | 6 2 6 2 | 5 2 5 2 | 5+6 2 5+6 2 |
| | 5 2 6 2 | 5 2 5 2 | 3+4 2 5+6 2 |
| | 5 2 5 2 | 6 2 5 2 | 3+4 2 3+4 2 |
| | 6 2 6 2 | 6 2 6 2 | 5+6 2 3+4 2 |
| | 6 2 6 2 | 5 2 6 2 | 3+4 2 5+6 2 |
| | 5 2 5 2 | 5 2 5 2 | 5+6 2 3+4 2 |
| | 5 2 5 2 | 6 2 6 2 | 5+6 2 5+6 2 |
| | 6 2 5 2 | 5 2 6 2 | 5+6 2 5+6 2 |
| | 6 2 5 2 | 6 2 5 2 | 5+6 2 3+4 2 |
| | 6 2 6 2 | 6 2 5 2 | 3+4 2 5+6 2 |
| | 5 2 6 2 | 5 2 5 2 | 3+4 2 3+4 2 |
| | 5 2 6 2 | 5 2 6 2 | 5+6 2 5+6 2 |
| | 5 2 5 2 | 6 2 5 2 | 3+4 2 3+4 2 |
| | 5 2 5 2 | 5 2 5 2 | 5+6 2 5+6 2 |
| | 6 2 6 2 | 6 2 6 2 | 3+4 2 5+6 2 |
| | 5 2 6 2 | 6 2 6 2 | 3+4 2 3+4 2 |
| | 6 2 6 2 | 6 2 6 2 | 3+4 2 3+4 2 |
| | 5 2 5 2 | 5 2 5 2 | 5+6 2 5+6 2 |
| | 6 2 6 2 | 5 2 5 2 | 3+4 2 3+4 2 |
| | 6 2 5 2 | 6 2 6 2 | 5+6 2 3+4 2 |
| | 6 2 6 2 | 5 2 6 2 | 3+4 2 3+4 2 |
| | 5 2 5 2 | 6 2 6 2 | 5+6 2 5+6 2 |
| | 6 2 6 2 | 5 2 5 2 | 5+6 2 5+6 2 |
| | 5 2 5 2 | 5 2 6 2 | 3+4 2 3+4 2 |
| | 5 2 5 2 | 5 2 5 2 | 5+6 2 5+6 2 |
| | 6 2 5 2 | 6 2 5 2 | 3+4 2 5+6 2 |
| 65.1.10 | 421933 | 62 | Gd2Mg3Ni2 |
| | -GM -T | -Y -Z | -R -S |
| | 5 2 6 2 | 6 2 5 2 | 3+4 2 3+4 2 |
| | 6 2 5 2 | 5 2 6 2 | 5+6 2 5+6 2 |
| | 5 2 6 2 | 6 2 5 2 | 3+4 2 3+4 2 |
| | 6 2 5 2 | 5 2 6 2 | 5+6 2 5+6 2 |
| | 5 2 6 2 | 6 2 5 2 | 3+4 2 3+4 2 |

Continued on next column

**TABLE XL– continued**

| TQC | ICSD | NoE | SF |
|---|---|---|---|
| | 6 2 5 2 | 5 2 6 2 | 5+6 2 5+6 2 |
| | 5 2 6 2 | 6 2 5 2 | 3+4 2 3+4 2 |
| | 6 2 5 2 | 5 2 6 2 | 5+6 2 5+6 2 |
| | 5 2 5 2 | 6 2 6 2 | 3+4 2 5+6 2 |
| | 5 2 6 2 | 5 2 6 2 | 5+6 2 3+4 2 |
| | 6 2 5 2 | 5 2 5 2 | 5+6 2 5+6 2 |
| | 5 2 6 2 | 6 2 5 2 | 3+4 2 3+4 2 |
| | 6 2 5 2 | 5 2 6 2 | 3+4 2 3+4 2 |
| | 6 2 6 2 | 6 2 5 2 | 5+6 2 5+6 2 |
| | 5 2 6 2 | 6 2 6 2 | 5+6 2 3+4 2 |
| | 6 2 5 2 | 5 2 5 2 | 3+4 2 5+6 2 |
| | 6 2 6 2 | 5 2 5 2 | 5+6 2 3+4 2 |
| | 5 2 5 2 | 5 2 5 2 | 3+4 2 5+6 2 |
| | 5 2 6 2 | 6 2 5 2 | 5+6 2 3+4 2 |
| | 5 2 5 2 | 6 2 6 2 | 3+4 2 5+6 2 |
| | 6 2 5 2 | 5 2 6 2 | 5+6 2 3+4 2 |
| | 6 2 6 2 | 6 2 6 2 | 5+6 2 5+6 2 |
| | 5 2 6 2 | 6 2 5 2 | 3+4 2 3+4 2 |
| | 6 2 5 2 | 5 2 6 2 | 3+4 2 5+6 2 |
| | 5 2 5 2 | 6 2 5 2 | 5+6 2 3+4 2 |
| | 5 2 6 2 | 6 2 6 2 | 3+4 2 5+6 2 |
| | 6 2 6 2 | 5 2 5 2 | 5+6 2 3+4 2 |
| | 6 2 5 2 | 5 2 6 2 | 5+6 2 5+6 2 |
| | 5 2 5 2 | 6 2 5 2 | 3+4 2 3+4 2 |
| | 5 2 6 2 | 6 2 6 2 | 3+4 2 5+6 2 |
| | 6 2 5 2 | 5 2 6 2 | 5+6 2 5+6 2 |
| 65.1.11 | 20337 | 86 | Fe2Si7U3 |
| | -GM -T | -Y -Z | -R -S |
| | 5 2 6 2 | 6 2 5 2 | 3+4 2 3+4 2 |
| | 6 2 5 2 | 5 2 6 2 | 5+6 2 5+6 2 |
| | 5 2 6 2 | 5 2 6 2 | 5+6 2 3+4 2 |
| | 5 2 6 2 | 6 2 5 2 | 3+4 2 3+4 2 |
| | 6 2 5 2 | 6 2 5 2 | 3+4 2 5+6 2 |
| | 6 2 5 2 | 5 2 6 2 | 5+6 2 5+6 2 |
| | 6 2 5 2 | 6 2 5 2 | 3+4 2 5+6 2 |
| | 5 2 5 2 | 6 2 6 2 | 3+4 2 5+6 2 |
| | 6 2 6 2 | 5 2 5 2 | 5+6 2 3+4 2 |
| | 6 2 5 2 | 6 2 5 2 | 3+4 2 5+6 2 |
| | 5 2 6 2 | 6 2 5 2 | 3+4 2 3+4 2 |
| | 6 2 5 2 | 5 2 6 2 | 5+6 2 5+6 2 |
| | 5 2 5 2 | 5 2 6 2 | 5+6 2 3+4 2 |
| | 5 2 6 2 | 6 2 6 2 | 5+6 2 3+4 2 |
| | 6 2 6 2 | 6 2 6 2 | 3+4 2 5+6 2 |
| | 5 2 5 2 | 5 2 5 2 | 5+6 2 5+6 2 |
| | 6 2 5 2 | 5 2 5 2 | 5+6 2 3+4 2 |
| | 5 2 5 2 | 5 2 5 2 | 3+4 2 5+6 2 |
| | 6 2 6 2 | 6 2 6 2 | 3+4 2 5+6 2 |
| | 5 2 6 2 | 5 2 5 2 | 5+6 2 3+4 2 |
| | 5 2 6 2 | 5 2 5 2 | 3+4 2 3+4 2 |
| | 6 2 5 2 | 6 2 6 2 | 3+4 2 5+6 2 |
| | 6 2 6 2 | 6 2 6 2 | 5+6 2 3+4 2 |
| | 6 2 5 2 | 6 2 6 2 | 5+6 2 3+4 2 |
| | 5 2 6 2 | 5 2 6 2 | 3+4 2 3+4 2 |
| | 5 2 6 2 | 5 2 6 2 | 3+4 2 5+6 2 |
| | 5 2 5 2 | 6 2 5 2 | 5+6 2 5+6 2 |
| | 5 2 5 2 | 6 2 5 2 | 5+6 2 5+6 2 |
| | 6 2 5 2 | 5 2 5 2 | 3+4 2 3+4 2 |
| | 6 2 5 2 | 6 2 6 2 | 3+4 2 3+4 2 |
| | 5 2 6 2 | 5 2 5 2 | 5+6 2 5+6 2 |
| | 6 2 6 2 | 6 2 6 2 | 3+4 2 5+6 2 |
| | 6 2 5 2 | 5 2 6 2 | 5+6 2 3+4 2 |

Continued on next column

**TABLE XL– continued**

| TQC | ICSD | NoE | SF |
|---|---|---|---|
| | 6 2 6 2 | 6 2 6 2 | 5+6 2 5+6 2 |
| | 6 2 6 2 | 6 2 5 2 | 3+4 2 3+4 2 |
| | 6 2 6 2 | 6 2 5 2 | 3+4 2 3+4 2 |
| | 5 2 5 2 | 5 2 5 2 | 3+4 2 5+6 2 |
| | 5 2 5 2 | 5 2 6 2 | 5+6 2 5+6 2 |
| | 5 2 5 2 | 6 2 5 2 | 3+4 2 3+4 2 |
| | 6 2 6 2 | 6 2 6 2 | 5+6 2 3+4 2 |
| | 6 2 5 2 | 6 2 6 2 | 5+6 2 5+6 2 |
| | 6 2 5 2 | 5 2 5 2 | 3+4 2 5+6 2 |
| | 5 2 6 2 | 6 2 5 2 | 3+4 2 3+4 2 |
| 65.1.12 | 261512 | 76 | Fe2O5Sr3 |
| | -GM -T | -Y -Z | -R -S |
| | 5 2 5 2 | 6 2 6 2 | 3+4 2 3+4 2 |
| | 6 2 6 2 | 5 2 5 2 | 5+6 2 5+6 2 |
| | 5 2 6 2 | 5 2 6 2 | 3+4 2 5+6 2 |
| | 6 2 6 2 | 6 2 5 2 | 5+6 2 3+4 2 |
| | 5 2 5 2 | 5 2 6 2 | 3+4 2 5+6 2 |
| | 5 2 5 2 | 6 2 5 2 | 3+4 2 3+4 2 |
| | 6 2 6 2 | 5 2 5 2 | 5+6 2 3+4 2 |
| | 5 2 5 2 | 5 2 6 2 | 3+4 2 5+6 2 |
| | 5 2 5 2 | 6 2 6 2 | 5+6 2 3+4 2 |
| | 6 2 5 2 | 6 2 6 2 | 3+4 2 5+6 2 |
| | 6 2 6 2 | 5 2 5 2 | 5+6 2 3+4 2 |
| | 5 2 5 2 | 6 2 5 2 | 3+4 2 3+4 2 |
| | 6 2 5 2 | 5 2 6 2 | 5+6 2 5+6 2 |
| | 6 2 6 2 | 6 2 5 2 | 5+6 2 3+4 2 |
| | 5 2 5 2 | 6 2 5 2 | 5+6 2 5+6 2 |
| | 6 2 6 2 | 6 2 5 2 | 3+4 2 3+4 2 |
| | 6 2 5 2 | 5 2 5 2 | 5+6 2 3+4 2 |
| | 6 2 6 2 | 6 2 5 2 | 5+6 2 5+6 2 |
| | 5 2 6 2 | 6 2 6 2 | 3+4 2 3+4 2 |
| | 5 2 5 2 | 5 2 5 2 | 5+6 2 5+6 2 |
| | 6 2 5 2 | 5 2 6 2 | 3+4 2 3+4 2 |
| | 6 2 6 2 | 6 2 5 2 | 3+4 2 5+6 2 |
| | 5 2 6 2 | 6 2 6 2 | 3+4 2 3+4 2 |
| | 5 2 6 2 | 5 2 6 2 | 5+6 2 5+6 2 |
| | 6 2 6 2 | 5 2 6 2 | 5+6 2 5+6 2 |
| | 6 2 5 2 | 6 2 6 2 | 3+4 2 3+4 2 |
| | 6 2 6 2 | 6 2 6 2 | 5+6 2 5+6 2 |
| | 5 2 5 2 | 5 2 5 2 | 5+6 2 5+6 2 |
| | 5 2 5 2 | 6 2 5 2 | 5+6 2 3+4 2 |
| | 6 2 6 2 | 5 2 5 2 | 3+4 2 5+6 2 |
| | 6 2 5 2 | 6 2 6 2 | 5+6 2 3+4 2 |
| | 6 2 6 2 | 6 2 6 2 | 5+6 2 5+6 2 |
| | 5 2 6 2 | 5 2 6 2 | 3+4 2 5+6 2 |
| | 6 2 5 2 | 6 2 5 2 | 5+6 2 3+4 2 |
| | 5 2 5 2 | 5 2 5 2 | 3+4 2 3+4 2 |
| | 6 2 5 2 | 6 2 5 2 | 3+4 2 3+4 2 |
| | 5 2 6 2 | 5 2 6 2 | 5+6 2 5+6 2 |
| | 5 2 6 2 | 5 2 6 2 | 5+6 2 5+6 2 |
| 65.1.13 | 238922 | 86 | Fe2Ge7U3 |
| | -GM -T | -Y -Z | -R -S |
| | 5 2 5 2 | 6 2 6 2 | 3+4 2 3+4 2 |
| | 6 2 6 2 | 5 2 5 2 | 5+6 2 5+6 2 |
| | 5 2 5 2 | 5 2 5 2 | 3+4 2 3+4 2 |
| | 5 2 5 2 | 6 2 6 2 | 3+4 2 3+4 2 |
| | 6 2 6 2 | 6 2 6 2 | 5+6 2 5+6 2 |
| | 6 2 6 2 | 5 2 5 2 | 5+6 2 5+6 2 |
| | 6 2 6 2 | 6 2 6 2 | 5+6 2 5+6 2 |
| | 5 2 6 2 | 6 2 5 2 | 5+6 2 5+6 2 |
| | 6 2 5 2 | 6 2 6 2 | 3+4 2 3+4 2 |
| | 6 2 6 2 | 5 2 6 2 | 5+6 2 5+6 2 |
| | 5 2 5 2 | 6 2 6 2 | 3+4 2 3+4 2 |
| | 6 2 6 2 | 5 2 5 2 | 5+6 2 5+6 2 |
| | 5 2 6 2 | 5 2 5 2 | 3+4 2 3+4 2 |
| | 5 2 5 2 | 6 2 5 2 | 3+4 2 3+4 2 |
| | 6 2 5 2 | 6 2 6 2 | 5+6 2 5+6 2 |
| | 5 2 6 2 | 5 2 5 2 | 3+4 2 5+6 2 |
| | 6 2 6 2 | 5 2 6 2 | 3+4 2 5+6 2 |
| | 5 2 6 2 | 5 2 6 2 | 5+6 2 3+4 2 |
| | 6 2 5 2 | 6 2 6 2 | 5+6 2 5+6 2 |
| | 5 2 5 2 | 6 2 5 2 | 3+4 2 3+4 2 |
| | 5 2 5 2 | 5 2 6 2 | 5+6 2 3+4 2 |
| | 6 2 6 2 | 5 2 5 2 | 5+6 2 5+6 2 |
| | 6 2 6 2 | 6 2 5 2 | 3+4 2 3+4 2 |
| | 6 2 5 2 | 6 2 5 2 | 3+4 2 3+4 2 |
| | 5 2 5 2 | 5 2 5 2 | 5+6 2 3+4 2 |
| | 5 2 5 2 | 5 2 5 2 | 5+6 2 5+6 2 |
| | 5 2 6 2 | 6 2 6 2 | 3+4 2 5+6 2 |
| | 5 2 6 2 | 6 2 6 2 | 3+4 2 3+4 2 |
| | 6 2 6 2 | 5 2 6 2 | 5+6 2 5+6 2 |
| | 6 2 6 2 | 6 2 5 2 | 5+6 2 5+6 2 |
| | 5 2 5 2 | 5 2 5 2 | 3+4 2 3+4 2 |
| | 6 2 6 2 | 6 2 6 2 | 5+6 2 5+6 2 |
| | 6 2 5 2 | 5 2 5 2 | 3+4 2 5+6 2 |
| | 6 2 5 2 | 6 2 5 2 | 3+4 2 3+4 2 |
| | 6 2 5 2 | 6 2 6 2 | 5+6 2 3+4 2 |
| | 5 2 5 2 | 5 2 6 2 | 5+6 2 3+4 2 |
| | 6 2 6 2 | 6 2 6 2 | 5+6 2 5+6 2 |
| | 5 2 6 2 | 5 2 5 2 | 5+6 2 5+6 2 |
| | 6 2 6 2 | 6 2 6 2 | 3+4 2 3+4 2 |
| | 5 2 5 2 | 6 2 5 2 | 3+4 2 3+4 2 |
| | 6 2 6 2 | 6 2 5 2 | 3+4 2 5+6 2 |
| | 6 2 6 2 | 5 2 6 2 | 5+6 2 5+6 2 |
| | 5 2 5 2 | 6 2 6 2 | 5+6 2 3+4 2 |
| 65.1.14 | 174439 | 94 | Er3Sn7 |
| | -GM -T | -Y -Z | -R -S |
| | 5 2 6 2 | 6 2 5 2 | 3+4 2 3+4 2 |
| | 6 2 5 2 | 5 2 6 2 | 5+6 2 5+6 2 |
| | 5 2 6 2 | 5 2 6 2 | 5+6 2 3+4 2 |
| | 5 2 6 2 | 6 2 5 2 | 3+4 2 3+4 2 |
| | 6 2 5 2 | 5 2 6 2 | 5+6 2 5+6 2 |
| | 6 2 5 2 | 6 2 5 2 | 3+4 2 5+6 2 |
| | 5 2 6 2 | 6 2 5 2 | 3+4 2 3+4 2 |
| | 6 2 5 2 | 5 2 6 2 | 5+6 2 5+6 2 |
| | 5 2 5 2 | 6 2 5 2 | 3+4 2 5+6 2 |
| | 6 2 6 2 | 6 2 5 2 | 3+4 2 3+4 2 |
| | 6 2 5 2 | 6 2 5 2 | 5+6 2 5+6 2 |
| | 6 2 5 2 | 5 2 6 2 | 3+4 2 5+6 2 |
| | 5 2 5 2 | 5 2 5 2 | 5+6 2 5+6 2 |
| | 6 2 5 2 | 6 2 6 2 | 5+6 2 3+4 2 |
| | 5 2 6 2 | 6 2 6 2 | 3+4 2 3+4 2 |
| | 5 2 6 2 | 5 2 6 2 | 3+4 2 5+6 2 |
| | 6 2 5 2 | 5 2 5 2 | 3+4 2 3+4 2 |
| | 5 2 5 2 | 5 2 5 2 | 5+6 2 5+6 2 |
| | 6 2 6 2 | 6 2 5 2 | 5+6 2 5+6 2 |
| | 6 2 6 2 | 6 2 6 2 | 3+4 2 3+4 2 |
| | 6 2 6 2 | 6 2 5 2 | 5+6 2 3+4 2 |
| | 5 2 5 2 | 5 2 6 2 | 3+4 2 5+6 2 |
| | 6 2 5 2 | 6 2 6 2 | 5+6 2 3+4 2 |
| | 5 2 5 2 | 5 2 6 2 | 3+4 2 3+4 2 |
| | 5 2 6 2 | 6 2 6 2 | 3+4 2 5+6 2 |

Continued on next column

**TABLE XL– continued**

| TQC | ICSD | NoE | SF |
|---|---|---|---|
| | 5 2 6 2 | 5 2 5 2 | 5+6 2 5+6 2 |
| | 6 2 5 2 | 5 2 6 2 | 5+6 2 5+6 2 |
| | 6 2 5 2 | 6 2 5 2 | 3+4 2 3+4 2 |
| | 5 2 5 2 | 5 2 5 2 | 3+4 2 5+6 2 |
| | 6 2 6 2 | 6 2 5 2 | 5+6 2 5+6 2 |
| | 6 2 5 2 | 6 2 6 2 | 5+6 2 3+4 2 |
| | 6 2 6 2 | 6 2 5 2 | 3+4 2 5+6 2 |
| | 5 2 5 2 | 6 2 5 2 | 3+4 2 3+4 2 |
| | 6 2 5 2 | 5 2 6 2 | 3+4 2 5+6 2 |
| | 5 2 6 2 | 6 2 6 2 | 5+6 2 3+4 2 |
| | 6 2 6 2 | 5 2 5 2 | 3+4 2 5+6 2 |
| | 5 2 5 2 | 5 2 5 2 | 3+4 2 5+6 2 |
| | 6 2 5 2 | 6 2 6 2 | 3+4 2 3+4 2 |
| | 6 2 5 2 | 6 2 5 2 | 5+6 2 5+6 2 |
| | 5 2 6 2 | 6 2 5 2 | 5+6 2 5+6 2 |
| | 6 2 5 2 | 5 2 6 2 | 3+4 2 3+4 2 |
| | 6 2 6 2 | 6 2 5 2 | 5+6 2 5+6 2 |
| | 5 2 5 2 | 6 2 5 2 | 3+4 2 3+4 2 |
| | 6 2 5 2 | 5 2 5 2 | 3+4 2 5+6 2 |
| | 5 2 6 2 | 6 2 6 2 | 3+4 2 5+6 2 |
| | 6 2 5 2 | 5 2 5 2 | 3+4 2 3+4 2 |
| | 6 2 5 2 | 6 2 6 2 | 5+6 2 5+6 2 |
| 65.1.15 | 165783 | 74 | Dy2Ni1Sn6 |
| | -GM -T | -Y -Z | -R -S |
| | 5 2 5 2 | 6 2 6 2 | 3+4 2 3+4 2 |
| | 6 2 6 2 | 5 2 5 2 | 5+6 2 5+6 2 |
| | 5 2 5 2 | 6 2 6 2 | 3+4 2 3+4 2 |
| | 6 2 6 2 | 5 2 5 2 | 5+6 2 5+6 2 |
| | 5 2 5 2 | 6 2 6 2 | 3+4 2 3+4 2 |
| | 6 2 6 2 | 5 2 5 2 | 5+6 2 5+6 2 |
| | 5 2 5 2 | 6 2 6 2 | 3+4 2 3+4 2 |
| | 6 2 6 2 | 5 2 5 2 | 5+6 2 5+6 2 |
| | 5 2 5 2 | 5 2 5 2 | 5+6 2 3+4 2 |
| | 6 2 6 2 | 6 2 6 2 | 5+6 2 5+6 2 |
| | 5 2 6 2 | 6 2 5 2 | 3+4 2 5+6 2 |
| | 6 2 5 2 | 5 2 6 2 | 5+6 2 3+4 2 |
| | 5 2 6 2 | 5 2 5 2 | 3+4 2 3+4 2 |
| | 5 2 5 2 | 5 2 6 2 | 3+4 2 5+6 2 |
| | 5 2 5 2 | 5 2 6 2 | 3+4 2 3+4 2 |
| | 5 2 6 2 | 5 2 5 2 | 3+4 2 5+6 2 |
| | 6 2 5 2 | 6 2 5 2 | 5+6 2 3+4 2 |
| | 6 2 5 2 | 6 2 5 2 | 5+6 2 3+4 2 |
| | 6 2 5 2 | 6 2 5 2 | 5+6 2 3+4 2 |
| | 5 2 5 2 | 6 2 5 2 | 3+4 2 3+4 2 |
| | 5 2 6 2 | 5 2 5 2 | 3+4 2 3+4 2 |
| | 5 2 5 2 | 5 2 5 2 | 5+6 2 5+6 2 |
| | 5 2 5 2 | 5 2 5 2 | 3+4 2 3+4 2 |
| | 6 2 6 2 | 6 2 6 2 | 3+4 2 5+6 2 |
| | 6 2 6 2 | 5 2 5 2 | 3+4 2 3+4 2 |
| | 5 2 5 2 | 5 2 6 2 | 3+4 2 5+6 2 |
| | 6 2 6 2 | 5 2 5 2 | 5+6 2 3+4 2 |
| | 5 2 5 2 | 6 2 5 2 | 3+4 2 5+6 2 |
| | 5 2 6 2 | 6 2 6 2 | 5+6 2 3+4 2 |
| | 6 2 5 2 | 6 2 6 2 | 3+4 2 5+6 2 |
| | 5 2 6 2 | 5 2 6 2 | 5+6 2 3+4 2 |
| | 6 2 6 2 | 5 2 5 2 | 3+4 2 5+6 2 |
| | 5 2 5 2 | 5 2 5 2 | 5+6 2 3+4 2 |
| | 5 2 6 2 | 5 2 6 2 | 3+4 2 5+6 2 |
| | 6 2 6 2 | 6 2 6 2 | 5+6 2 3+4 2 |
| | 5 2 5 2 | 6 2 5 2 | 3+4 2 5+6 2 |
| | 6 2 5 2 | 5 2 5 2 | 5+6 2 5+6 2 |

Continued on next column

**TABLE XL– continued**

| TQC | ICSD | NoE | SF |
|---|---|---|---|
| 65.1.16 | 418215 | 152 | Cu1Mg4Tb1 |
| | -GM -T | -Y -Z | -R -S |
| | 5 2 5 2 | 5 2 5 2 | 5+6 2 5+6 2 |
| | 6 2 6 2 | 6 2 6 2 | 3+4 2 3+4 2 |
| | 6 2 6 2 | 6 2 6 2 | 3+4 2 3+4 2 |
| | 5 2 5 2 | 5 2 5 2 | 5+6 2 5+6 2 |
| | 6 2 6 2 | 6 2 6 2 | 3+4 2 3+4 2 |
| | 5 2 5 2 | 5 2 5 2 | 5+6 2 5+6 2 |
| | 5 2 5 2 | 5 2 5 2 | 5+6 2 5+6 2 |
| | 6 2 6 2 | 6 2 6 2 | 3+4 2 3+4 2 |
| | 6 2 5 2 | 6 2 5 2 | 5+6 2 3+4 2 |
| | 5 2 5 2 | 5 2 5 2 | 5+6 2 5+6 2 |
| | 5 2 6 2 | 5 2 6 2 | 3+4 2 5+6 2 |
| | 6 2 6 2 | 6 2 6 2 | 3+4 2 3+4 2 |
| | 6 2 6 2 | 6 2 6 2 | 3+4 2 3+4 2 |
| | 5 2 5 2 | 5 2 5 2 | 5+6 2 3+4 2 |
| | 6 2 5 2 | 6 2 5 2 | 5+6 2 5+6 2 |
| | 5 2 6 2 | 5 2 6 2 | 3+4 2 5+6 2 |
| | 5 2 5 2 | 6 2 6 2 | 5+6 2 3+4 2 |
| | 5 2 5 2 | 5 2 5 2 | 3+4 2 5+6 2 |
| | 6 2 6 2 | 6 2 6 2 | 3+4 2 3+4 2 |
| | 6 2 6 2 | 5 2 5 2 | 5+6 2 5+6 2 |
| | 5 2 5 2 | 5 2 5 2 | 3+4 2 3+4 2 |
| | 6 2 6 2 | 6 2 5 2 | 5+6 2 5+6 2 |
| | 5 2 6 2 | 5 2 6 2 | 3+4 2 3+4 2 |
| | 5 2 6 2 | 6 2 6 2 | 5+6 2 5+6 2 |
| | 6 2 5 2 | 5 2 5 2 | 3+4 2 5+6 2 |
| | 6 2 5 2 | 6 2 6 2 | 5+6 2 3+4 2 |
| | 6 2 6 2 | 5 2 6 2 | 3+4 2 5+6 2 |
| | 5 2 6 2 | 6 2 5 2 | 5+6 2 3+4 2 |
| | 5 2 5 2 | 6 2 6 2 | 3+4 2 3+4 2 |
| | 5 2 6 2 | 6 2 5 2 | 3+4 2 5+6 2 |
| | 6 2 5 2 | 5 2 6 2 | 5+6 2 5+6 2 |
| | 5 2 5 2 | 5 2 5 2 | 5+6 2 3+4 2 |
| | 5 2 6 2 | 5 2 5 2 | 5+6 2 5+6 2 |
| | 6 2 5 2 | 6 2 6 2 | 3+4 2 3+4 2 |
| | 6 2 6 2 | 6 2 6 2 | 3+4 2 5+6 2 |
| | 5 2 5 2 | 5 2 5 2 | 5+6 2 3+4 2 |
| | 6 2 6 2 | 6 2 5 2 | 3+4 2 3+4 2 |
| | 5 2 5 2 | 6 2 6 2 | 5+6 2 5+6 2 |
| | 5 2 6 2 | 5 2 6 2 | 5+6 2 5+6 2 |
| | 6 2 5 2 | 5 2 5 2 | 3+4 2 5+6 2 |
| | 6 2 6 2 | 5 2 5 2 | 5+6 2 3+4 2 |
| | 5 2 5 2 | 6 2 6 2 | 3+4 2 3+4 2 |
| | 5 2 6 2 | 6 2 6 2 | 5+6 2 3+4 2 |
| | 6 2 5 2 | 5 2 5 2 | 5+6 2 5+6 2 |
| | 6 2 6 2 | 5 2 5 2 | 3+4 2 5+6 2 |
| | 6 2 5 2 | 6 2 5 2 | 3+4 2 3+4 2 |
| | 6 2 6 2 | 6 2 6 2 | 3+4 2 3+4 2 |
| | 5 2 5 2 | 5 2 6 2 | 5+6 2 5+6 2 |
| | 5 2 6 2 | 6 2 6 2 | 3+4 2 5+6 2 |
| | 5 2 6 2 | 5 2 5 2 | 5+6 2 3+4 2 |
| | 6 2 5 2 | 5 2 5 2 | 3+4 2 3+4 2 |
| | 6 2 5 2 | 5 2 6 2 | 3+4 2 5+6 2 |
| | 5 2 5 2 | 6 2 6 2 | 5+6 2 5+6 2 |
| | 5 2 5 2 | 6 2 5 2 | 5+6 2 3+4 2 |
| | 6 2 6 2 | 5 2 6 2 | 3+4 2 5+6 2 |
| | 6 2 6 2 | 6 2 5 2 | 3+4 2 3+4 2 |
| | 5 2 5 2 | 6 2 6 2 | 5+6 2 5+6 2 |
| | 5 2 6 2 | 5 2 5 2 | 5+6 2 3+4 2 |
| | 6 2 5 2 | 6 2 5 2 | 3+4 2 5+6 2 |

Continued on next column

**TABLE XL– continued**

| TQC | ICSD | NoE | SF |
|---|---|---|---|
| | 5 2 6 2 | 5 2 6 2 | 5+6 2 3+4 2 |
| | 6 2 5 2 | 5 2 5 2 | 3+4 2 5+6 2 |
| | 5 2 6 2 | 6 2 6 2 | 5+6 2 3+4 2 |
| | 6 2 5 2 | 5 2 5 2 | 3+4 2 5+6 2 |
| | 5 2 5 2 | 5 2 5 2 | 5+6 2 3+4 2 |
| | 6 2 6 2 | 6 2 6 2 | 3+4 2 5+6 2 |
| | 5 2 6 2 | 6 2 6 2 | 5+6 2 3+4 2 |
| | 6 2 6 2 | 6 2 6 2 | 3+4 2 5+6 2 |
| | 6 2 5 2 | 5 2 5 2 | 3+4 2 3+4 2 |
| | 5 2 6 2 | 6 2 6 2 | 5+6 2 5+6 2 |
| | 5 2 5 2 | 5 2 6 2 | 5+6 2 3+4 2 |
| | 6 2 5 2 | 6 2 5 2 | 5+6 2 5+6 2 |
| | 5 2 5 2 | 5 2 6 2 | 5+6 2 3+4 2 |
| | 6 2 6 2 | 5 2 5 2 | 3+4 2 5+6 2 |
| | 5 2 6 2 | 6 2 5 2 | 5+6 2 3+4 2 |
| | 6 2 6 2 | 5 2 5 2 | 3+4 2 5+6 2 |
| | 6 2 5 2 | 6 2 6 2 | 5+6 2 3+4 2 |
| 65.1.17 | 424317 | 132 | Ce7Ge10Li8 |
| | -GM -T | -Y -Z | -R -S |
| | 5 2 5 2 | 5 2 5 2 | 5+6 2 5+6 2 |
| | 5 2 5 2 | 6 2 6 2 | 3+4 2 3+4 2 |
| | 6 2 6 2 | 5 2 5 2 | 3+4 2 3+4 2 |
| | 6 2 6 2 | 6 2 6 2 | 5+6 2 5+6 2 |
| | 5 2 5 2 | 5 2 5 2 | 3+4 2 3+4 2 |
| | 5 2 5 2 | 6 2 6 2 | 3+4 2 3+4 2 |
| | 6 2 6 2 | 5 2 5 2 | 5+6 2 5+6 2 |
| | 6 2 6 2 | 5 2 5 2 | 5+6 2 5+6 2 |
| | 5 2 5 2 | 5 2 6 2 | 3+4 2 3+4 2 |
| | 5 2 5 2 | 6 2 5 2 | 5+6 2 5+6 2 |
| | 6 2 6 2 | 6 2 6 2 | 3+4 2 3+4 2 |
| | 5 2 5 2 | 6 2 6 2 | 3+4 2 3+4 2 |
| | 6 2 6 2 | 6 2 5 2 | 5+6 2 5+6 2 |
| | 6 2 6 2 | 5 2 6 2 | 5+6 2 5+6 2 |
| | 5 2 5 2 | 6 2 6 2 | 3+4 2 5+6 2 |
| | 6 2 6 2 | 5 2 6 2 | 5+6 2 3+4 2 |
| | 6 2 6 2 | 6 2 5 2 | 5+6 2 5+6 2 |
| | 5 2 5 2 | 5 2 5 2 | 3+4 2 5+6 2 |
| | 5 2 5 2 | 6 2 6 2 | 5+6 2 3+4 2 |
| | 6 2 6 2 | 5 2 5 2 | 3+4 2 5+6 2 |
| | 5 2 6 2 | 5 2 5 2 | 5+6 2 3+4 2 |
| | 5 2 6 2 | 6 2 6 2 | 3+4 2 3+4 2 |
| | 6 2 5 2 | 6 2 5 2 | 5+6 2 5+6 2 |
| | 6 2 6 2 | 5 2 6 2 | 5+6 2 3+4 2 |
| | 6 2 5 2 | 6 2 6 2 | 3+4 2 5+6 2 |
| | 5 2 5 2 | 6 2 6 2 | 5+6 2 5+6 2 |
| | 6 2 6 2 | 6 2 6 2 | 3+4 2 3+4 2 |
| | 6 2 6 2 | 5 2 5 2 | 5+6 2 5+6 2 |
| | 5 2 6 2 | 6 2 5 2 | 3+4 2 3+4 2 |
| | 5 2 6 2 | 5 2 6 2 | 5+6 2 5+6 2 |
| | 6 2 6 2 | 5 2 5 2 | 3+4 2 5+6 2 |
| | 5 2 5 2 | 6 2 6 2 | 3+4 2 3+4 2 |
| | 5 2 5 2 | 6 2 5 2 | 5+6 2 5+6 2 |
| | 6 2 5 2 | 5 2 6 2 | 3+4 2 3+4 2 |
| | 5 2 6 2 | 5 2 5 2 | 5+6 2 5+6 2 |
| | 6 2 6 2 | 6 2 6 2 | 3+4 2 3+4 2 |
| | 6 2 5 2 | 6 2 5 2 | 5+6 2 5+6 2 |
| | 6 2 5 2 | 5 2 6 2 | 5+6 2 3+4 2 |
| | 6 2 5 2 | 5 2 6 2 | 5+6 2 3+4 2 |
| | 5 2 6 2 | 6 2 5 2 | 3+4 2 3+4 2 |
| | 5 2 5 2 | 5 2 5 2 | 3+4 2 5+6 2 |
| | 6 2 5 2 | 6 2 6 2 | 5+6 2 5+6 2 |
| | 5 2 6 2 | 6 2 5 2 | 5+6 2 3+4 2 |
| | 6 2 5 2 | 5 2 5 2 | 3+4 2 3+4 2 |
| | 5 2 6 2 | 5 2 6 2 | 3+4 2 5+6 2 |
| | 6 2 6 2 | 6 2 5 2 | 5+6 2 5+6 2 |
| | 6 2 6 2 | 5 2 5 2 | 5+6 2 3+4 2 |
| | 5 2 5 2 | 5 2 6 2 | 3+4 2 5+6 2 |
| | 6 2 6 2 | 6 2 6 2 | 3+4 2 3+4 2 |
| | 6 2 5 2 | 6 2 6 2 | 3+4 2 3+4 2 |
| | 5 2 5 2 | 6 2 5 2 | 3+4 2 5+6 2 |
| | 5 2 5 2 | 5 2 5 2 | 5+6 2 3+4 2 |
| | 5 2 6 2 | 6 2 6 2 | 5+6 2 5+6 2 |
| | 6 2 6 2 | 5 2 5 2 | 3+4 2 5+6 2 |
| | 5 2 6 2 | 5 2 6 2 | 5+6 2 3+4 2 |
| | 5 2 5 2 | 6 2 5 2 | 3+4 2 5+6 2 |
| | 6 2 6 2 | 5 2 5 2 | 5+6 2 3+4 2 |
| | 5 2 5 2 | 5 2 5 2 | 3+4 2 5+6 2 |
| | 5 2 6 2 | 6 2 6 2 | 5+6 2 3+4 2 |
| | 5 2 5 2 | 5 2 6 2 | 3+4 2 3+4 2 |
| | 5 2 6 2 | 6 2 5 2 | 5+6 2 5+6 2 |
| | 6 2 6 2 | 6 2 6 2 | 3+4 2 3+4 2 |
| | 5 2 6 2 | 5 2 6 2 | 5+6 2 5+6 2 |
| | 6 2 5 2 | 6 2 6 2 | 3+4 2 5+6 2 |
| | 5 2 5 2 | 6 2 5 2 | 5+6 2 3+4 2 |
| | 6 2 5 2 | 5 2 6 2 | 3+4 2 5+6 2 |
| 65.1.18 | 63667 | 64 | Ce3Sn7 |
| | -GM -T | -Y -Z | -R -S |
| | 5 2 6 2 | 6 2 5 2 | 3+4 2 3+4 2 |
| | 6 2 5 2 | 5 2 6 2 | 5+6 2 5+6 2 |
| | 5 2 5 2 | 5 2 5 2 | 3+4 2 3+4 2 |
| | 5 2 6 2 | 6 2 5 2 | 3+4 2 3+4 2 |
| | 6 2 5 2 | 5 2 6 2 | 5+6 2 5+6 2 |
| | 6 2 6 2 | 6 2 6 2 | 5+6 2 5+6 2 |
| | 5 2 6 2 | 6 2 5 2 | 5+6 2 5+6 2 |
| | 6 2 5 2 | 6 2 6 2 | 3+4 2 3+4 2 |
| | 6 2 6 2 | 5 2 6 2 | 5+6 2 5+6 2 |
| | 5 2 6 2 | 6 2 5 2 | 5+6 2 5+6 2 |
| | 6 2 6 2 | 6 2 6 2 | 3+4 2 3+4 2 |
| | 6 2 5 2 | 5 2 6 2 | 5+6 2 5+6 2 |
| | 5 2 6 2 | 5 2 6 2 | 3+4 2 5+6 2 |
| | 6 2 5 2 | 6 2 6 2 | 5+6 2 3+4 2 |
| | 5 2 5 2 | 6 2 5 2 | 3+4 2 3+4 2 |
| | 5 2 5 2 | 5 2 5 2 | 3+4 2 5+6 2 |
| | 6 2 6 2 | 5 2 6 2 | 3+4 2 3+4 2 |
| | 5 2 6 2 | 5 2 6 2 | 5+6 2 5+6 2 |
| | 6 2 6 2 | 6 2 5 2 | 5+6 2 5+6 2 |
| | 5 2 5 2 | 6 2 5 2 | 5+6 2 3+4 2 |
| | 6 2 6 2 | 5 2 5 2 | 3+4 2 3+4 2 |
| | 5 2 5 2 | 6 2 6 2 | 5+6 2 5+6 2 |
| | 6 2 5 2 | 5 2 5 2 | 3+4 2 3+4 2 |
| | 6 2 5 2 | 6 2 6 2 | 5+6 2 3+4 2 |
| | 5 2 5 2 | 5 2 5 2 | 5+6 2 5+6 2 |
| | 5 2 5 2 | 6 2 6 2 | 5+6 2 3+4 2 |
| | 5 2 6 2 | 5 2 5 2 | 3+4 2 3+4 2 |
| | 6 2 6 2 | 5 2 5 2 | 3+4 2 5+6 2 |
| | 6 2 5 2 | 6 2 6 2 | 5+6 2 3+4 2 |
| | 6 2 6 2 | 6 2 6 2 | 3+4 2 5+6 2 |
| | 6 2 6 2 | 6 2 5 2 | 3+4 2 5+6 2 |
| | 5 2 5 2 | 6 2 6 2 | 5+6 2 5+6 2 |
| 65.1.19 | 261436 | 62 | Cd1Cu2Nd2 |
| | -GM -T | -Y -Z | -R -S |
| | 5 2 5 2 | 5 2 5 2 | 5+6 2 5+6 2 |

Continued on next column

Continued on next column

**TABLE XL– continued**

| TQC | ICSD NoE SF |
|---|---|
| | 6 2 6 2 6 2 6 2 3+4 2 3+4 2 |
| | 5 2 5 2 5 2 5 2 5+6 2 5+6 2 |
| | 6 2 6 2 6 2 6 2 3+4 2 3+4 2 |
| | 5 2 5 2 5 2 5 2 5+6 2 5+6 2 |
| | 6 2 6 2 6 2 6 2 3+4 2 3+4 2 |
| | 5 2 6 2 6 2 5 2 5+6 2 5+6 2 |
| | 6 2 5 2 5 2 6 2 3+4 2 3+4 2 |
| | 5 2 6 2 5 2 6 2 5+6 2 3+4 2 |
| | 5 2 6 2 5 2 6 2 5+6 2 3+4 2 |
| | 5 2 6 2 5 2 6 2 5+6 2 3+4 2 |
| | 5 2 6 2 5 2 6 2 5+6 2 3+4 2 |
| | 5 2 6 2 5 2 6 2 5+6 2 3+4 2 |
| | 5 2 5 2 6 2 6 2 5+6 2 3+4 2 |
| | 5 2 6 2 5 2 5 2 3+4 2 5+6 2 |
| | 6 2 6 2 6 2 5 2 5+6 2 3+4 2 |
| | 6 2 5 2 5 2 6 2 3+4 2 5+6 2 |
| | 5 2 6 2 6 2 6 2 5+6 2 3+4 2 |
| | 6 2 5 2 5 2 5 2 3+4 2 5+6 2 |
| | 5 2 6 2 5 2 6 2 5+6 2 3+4 2 |
| | 6 2 5 2 5 2 5 2 3+4 2 5+6 2 |
| | 5 2 6 2 6 2 6 2 5+6 2 3+4 2 |
| | 6 2 5 2 6 2 5 2 3+4 2 5+6 2 |
| | 5 2 6 2 6 2 5 2 5+6 2 3+4 2 |
| | 6 2 5 2 5 2 6 2 3+4 2 5+6 2 |
| | 5 2 6 2 5 2 6 2 5+6 2 3+4 2 |
| | 5 2 5 2 6 2 5 2 3+4 2 5+6 2 |
| | 6 2 5 2 6 2 6 2 3+4 2 5+6 2 |
| | 6 2 5 2 5 2 6 2 5+6 2 3+4 2 |
| | 5 2 6 2 6 2 5 2 3+4 2 3+4 2 |
| | 5 2 5 2 5 2 5 2 3+4 2 5+6 2 |
| 65.1.20 | 419476 56 C3B2Tb2 |
| | -GM -T -Y -Z -R -S |
| | 5 2 6 2 6 2 5 2 3+4 2 3+4 2 |
| | 6 2 5 2 5 2 6 2 5+6 2 5+6 2 |
| | 5 2 6 2 6 2 5 2 3+4 2 3+4 2 |
| | 6 2 5 2 5 2 6 2 5+6 2 5+6 2 |
| | 5 2 6 2 6 2 5 2 5+6 2 5+6 2 |
| | 6 2 5 2 5 2 6 2 3+4 2 3+4 2 |
| | 5 2 6 2 6 2 5 2 3+4 2 3+4 2 |
| | 6 2 5 2 5 2 6 2 5+6 2 5+6 2 |
| | 5 2 6 2 5 2 6 2 3+4 2 5+6 2 |
| | 6 2 5 2 6 2 5 2 5+6 2 3+4 2 |
| | 5 2 5 2 5 2 6 2 3+4 2 5+6 2 |
| | 6 2 6 2 6 2 5 2 5+6 2 3+4 2 |
| | 5 2 5 2 6 2 5 2 5+6 2 3+4 2 |
| | 6 2 6 2 6 2 6 2 3+4 2 5+6 2 |
| | 6 2 5 2 6 2 5 2 5+6 2 3+4 2 |
| | 5 2 5 2 5 2 6 2 5+6 2 3+4 2 |
| | 5 2 6 2 6 2 6 2 3+4 2 5+6 2 |
| | 6 2 5 2 5 2 5 2 5+6 2 5+6 2 |
| | 6 2 5 2 5 2 6 2 3+4 2 3+4 2 |
| | 5 2 6 2 6 2 5 2 3+4 2 5+6 2 |
| | 6 2 6 2 5 2 6 2 5+6 2 3+4 2 |
| | 5 2 5 2 6 2 5 2 5+6 2 5+6 2 |
| | 6 2 6 2 6 2 6 2 3+4 2 3+4 2 |
| | 5 2 5 2 5 2 5 2 5+6 2 5+6 2 |
| | 5 2 6 2 5 2 6 2 3+4 2 3+4 2 |
| | 6 2 5 2 5 2 5 2 3+4 2 3+4 2 |
| | 5 2 6 2 6 2 5 2 5+6 2 5+6 2 |
| | 6 2 5 2 6 2 6 2 5+6 2 5+6 2 |
| 65.1.21 | 413212 154 Br2Ce4Ga5 |
| | -GM -T -Y -Z -R -S |
| | 5 2 5 2 6 2 6 2 3+4 2 3+4 2 |
| | 5 2 5 2 6 2 6 2 5+6 2 5+6 2 |
| | 6 2 6 2 5 2 5 2 3+4 2 3+4 2 |
| | 6 2 6 2 5 2 5 2 5+6 2 5+6 2 |
| | 5 2 5 2 5 2 5 2 3+4 2 3+4 2 |
| | 6 2 6 2 6 2 6 2 5+6 2 5+6 2 |
| | 5 2 5 2 5 2 5 2 3+4 2 3+4 2 |
| | 6 2 6 2 6 2 6 2 5+6 2 5+6 2 |
| | 5 2 5 2 6 2 6 2 3+4 2 3+4 2 |
| | 6 2 6 2 5 2 5 2 5+6 2 5+6 2 |
| | 5 2 5 2 6 2 6 2 5+6 2 5+6 2 |
| | 6 2 6 2 5 2 5 2 3+4 2 3+4 2 |
| | 5 2 5 2 6 2 6 2 3+4 2 5+6 2 |
| | 6 2 6 2 5 2 5 2 5+6 2 3+4 2 |
| | 5 2 5 2 5 2 5 2 3+4 2 3+4 2 |
| | 6 2 6 2 6 2 6 2 5+6 2 5+6 2 |
| | 6 2 6 2 6 2 6 2 3+4 2 3+4 2 |
| | 5 2 5 2 5 2 5 2 5+6 2 5+6 2 |
| | 5 2 5 2 5 2 5 2 5+6 2 5+6 2 |
| | 6 2 6 2 6 2 6 2 3+4 2 3+4 2 |
| | 6 2 5 2 6 2 6 2 5+6 2 3+4 2 |
| | 5 2 6 2 5 2 5 2 3+4 2 5+6 2 |
| | 5 2 5 2 6 2 6 2 3+4 2 5+6 2 |
| | 6 2 6 2 5 2 5 2 5+6 2 3+4 2 |
| | 6 2 5 2 5 2 6 2 3+4 2 3+4 2 |
| | 5 2 6 2 6 2 5 2 5+6 2 5+6 2 |
| | 5 2 5 2 5 2 6 2 3+4 2 3+4 2 |
| | 6 2 6 2 6 2 5 2 5+6 2 5+6 2 |
| | 6 2 6 2 6 2 6 2 5+6 2 3+4 2 |
| | 5 2 5 2 5 2 5 2 3+4 2 5+6 2 |
| | 6 2 5 2 5 2 5 2 3+4 2 5+6 2 |
| | 5 2 6 2 6 2 6 2 5+6 2 3+4 2 |
| | 5 2 6 2 6 2 6 2 3+4 2 3+4 2 |
| | 6 2 5 2 5 2 5 2 5+6 2 5+6 2 |
| | 5 2 6 2 5 2 6 2 3+4 2 5+6 2 |
| | 6 2 5 2 6 2 5 2 5+6 2 3+4 2 |
| | 5 2 6 2 5 2 6 2 5+6 2 3+4 2 |
| | 5 2 6 2 6 2 5 2 3+4 2 5+6 2 |
| | 6 2 5 2 6 2 5 2 3+4 2 3+4 2 |
| | 6 2 6 2 5 2 5 2 5+6 2 5+6 2 |
| | 5 2 5 2 5 2 6 2 5+6 2 3+4 2 |
| | 6 2 6 2 6 2 5 2 3+4 2 5+6 2 |
| | 5 2 5 2 5 2 6 2 3+4 2 3+4 2 |
| | 5 2 6 2 5 2 5 2 5+6 2 5+6 2 |
| | 6 2 5 2 6 2 6 2 5+6 2 3+4 2 |
| | 6 2 6 2 6 2 5 2 3+4 2 5+6 2 |
| | 6 2 5 2 5 2 6 2 3+4 2 3+4 2 |
| | 5 2 5 2 6 2 6 2 5+6 2 5+6 2 |
| | 5 2 6 2 6 2 5 2 5+6 2 5+6 2 |
| | 6 2 6 2 5 2 6 2 3+4 2 3+4 2 |
| | 5 2 6 2 6 2 5 2 5+6 2 5+6 2 |
| | 6 2 5 2 5 2 6 2 3+4 2 3+4 2 |
| | 5 2 5 2 6 2 6 2 3+4 2 3+4 2 |
| | 5 2 6 2 5 2 6 2 5+6 2 5+6 2 |
| | 6 2 6 2 6 2 6 2 3+4 2 5+6 2 |
| | 5 2 5 2 6 2 5 2 5+6 2 3+4 2 |
| | 6 2 5 2 5 2 5 2 5+6 2 3+4 2 |
| | 5 2 5 2 5 2 5 2 3+4 2 5+6 2 |
| | 6 2 6 2 5 2 5 2 3+4 2 3+4 2 |
| | 6 2 5 2 6 2 6 2 5+6 2 5+6 2 |

Continued on next column

**TABLE XL– continued**

| TQC | ICSD | NoE | SF |
|---|---|---|---|
| | 6 2 6 2 | 5 2 5 2 | 3+4 2 3+4 2 |
| | 6 2 5 2 | 5 2 5 2 | 5+6 2 3+4 2 |
| | 5 2 6 2 | 6 2 6 2 | 3+4 2 5+6 2 |
| | 5 2 6 2 | 5 2 5 2 | 5+6 2 5+6 2 |
| | 5 2 5 2 | 5 2 6 2 | 5+6 2 3+4 2 |
| | 5 2 5 2 | 6 2 5 2 | 5+6 2 5+6 2 |
| | 6 2 5 2 | 6 2 6 2 | 3+4 2 3+4 2 |
| | 5 2 6 2 | 6 2 5 2 | 5+6 2 5+6 2 |
| | 5 2 6 2 | 6 2 6 2 | 3+4 2 3+4 2 |
| | 6 2 6 2 | 6 2 5 2 | 3+4 2 3+4 2 |
| | 5 2 5 2 | 6 2 5 2 | 3+4 2 5+6 2 |
| | 6 2 6 2 | 5 2 6 2 | 5+6 2 5+6 2 |
| | 6 2 6 2 | 5 2 6 2 | 5+6 2 5+6 2 |
| | 5 2 5 2 | 6 2 6 2 | 3+4 2 3+4 2 |
| | 6 2 5 2 | 6 2 6 2 | 3+4 2 5+6 2 |
| | 6 2 5 2 | 5 2 5 2 | 5+6 2 3+4 2 |
| | 5 2 6 2 | 5 2 5 2 | 5+6 2 5+6 2 |
| 65.1.22 | 78621 | 136 | Ba2Cu4O8Yb1 |
| | -GM -T | -Y -Z | -R -S |
| | 5 2 6 2 | 5 2 6 2 | 3+4 2 5+6 2 |
| | 6 2 5 2 | 6 2 5 2 | 5+6 2 3+4 2 |
| | 5 2 5 2 | 5 2 5 2 | 3+4 2 3+4 2 |
| | 6 2 6 2 | 6 2 6 2 | 5+6 2 5+6 2 |
| | 6 2 5 2 | 6 2 5 2 | 5+6 2 3+4 2 |
| | 6 2 5 2 | 6 2 5 2 | 5+6 2 3+4 2 |
| | 5 2 6 2 | 6 2 5 2 | 3+4 2 5+6 2 |
| | 5 2 5 2 | 5 2 6 2 | 5+6 2 3+4 2 |
| | 6 2 6 2 | 5 2 6 2 | 3+4 2 5+6 2 |
| | 6 2 6 2 | 6 2 6 2 | 3+4 2 5+6 2 |
| | 5 2 5 2 | 5 2 5 2 | 5+6 2 3+4 2 |
| | 6 2 5 2 | 6 2 6 2 | 3+4 2 5+6 2 |
| | 6 2 6 2 | 5 2 5 2 | 5+6 2 3+4 2 |
| | 5 2 5 2 | 6 2 5 2 | 5+6 2 3+4 2 |
| | 5 2 6 2 | 5 2 5 2 | 3+4 2 5+6 2 |
| | 6 2 5 2 | 6 2 6 2 | 5+6 2 3+4 2 |
| | 5 2 6 2 | 5 2 5 2 | 3+4 2 5+6 2 |
| | 6 2 5 2 | 6 2 6 2 | 5+6 2 3+4 2 |
| | 5 2 6 2 | 5 2 5 2 | 3+4 2 3+4 2 |
| | 6 2 5 2 | 6 2 6 2 | 5+6 2 5+6 2 |
| | 5 2 5 2 | 6 2 5 2 | 3+4 2 3+4 2 |
| | 5 2 6 2 | 6 2 6 2 | 5+6 2 5+6 2 |
| | 6 2 6 2 | 5 2 5 2 | 3+4 2 3+4 2 |
| | 5 2 6 2 | 5 2 5 2 | 5+6 2 5+6 2 |
| | 5 2 5 2 | 5 2 6 2 | 3+4 2 5+6 2 |
| | 5 2 6 2 | 5 2 6 2 | 3+4 2 3+4 2 |
| | 6 2 6 2 | 6 2 6 2 | 3+4 2 5+6 2 |
| | 6 2 5 2 | 6 2 6 2 | 5+6 2 5+6 2 |
| | 5 2 6 2 | 6 2 5 2 | 5+6 2 3+4 2 |
| | 6 2 6 2 | 5 2 6 2 | 5+6 2 5+6 2 |
| | 6 2 5 2 | 5 2 5 2 | 3+4 2 5+6 2 |
| | 5 2 5 2 | 5 2 6 2 | 5+6 2 3+4 2 |
| | 5 2 5 2 | 6 2 5 2 | 5+6 2 5+6 2 |
| | 5 2 5 2 | 6 2 5 2 | 3+4 2 3+4 2 |
| | 6 2 6 2 | 5 2 5 2 | 5+6 2 3+4 2 |
| | 5 2 6 2 | 6 2 5 2 | 5+6 2 3+4 2 |
| | 6 2 5 2 | 5 2 6 2 | 3+4 2 3+4 2 |
| | 6 2 6 2 | 6 2 5 2 | 3+4 2 5+6 2 |
| | 5 2 5 2 | 5 2 6 2 | 3+4 2 3+4 2 |
| | 6 2 5 2 | 6 2 6 2 | 5+6 2 5+6 2 |
| | 6 2 6 2 | 5 2 6 2 | 3+4 2 5+6 2 |
| | 6 2 5 2 | 6 2 5 2 | 5+6 2 3+4 2 |

Continued on next column

**TABLE XL– continued**

| TQC | ICSD | NoE | SF |
|---|---|---|---|
| | 5 2 6 2 | 6 2 6 2 | 3+4 2 5+6 2 |
| | 6 2 5 2 | 5 2 6 2 | 5+6 2 3+4 2 |
| | 5 2 5 2 | 5 2 5 2 | 5+6 2 5+6 2 |
| | 6 2 6 2 | 6 2 5 2 | 3+4 2 3+4 2 |
| | 6 2 6 2 | 5 2 6 2 | 5+6 2 5+6 2 |
| | 5 2 5 2 | 5 2 5 2 | 5+6 2 3+4 2 |
| | 6 2 6 2 | 6 2 5 2 | 3+4 2 3+4 2 |
| | 5 2 5 2 | 6 2 6 2 | 5+6 2 5+6 2 |
| | 6 2 5 2 | 6 2 5 2 | 3+4 2 3+4 2 |
| | 5 2 5 2 | 5 2 5 2 | 3+4 2 3+4 2 |
| | 6 2 5 2 | 5 2 6 2 | 5+6 2 5+6 2 |
| | 5 2 6 2 | 6 2 5 2 | 5+6 2 3+4 2 |
| | 6 2 5 2 | 6 2 5 2 | 5+6 2 3+4 2 |
| | 5 2 6 2 | 5 2 6 2 | 3+4 2 5+6 2 |
| | 5 2 5 2 | 6 2 5 2 | 5+6 2 5+6 2 |
| | 6 2 6 2 | 5 2 6 2 | 5+6 2 3+4 2 |
| | 6 2 5 2 | 6 2 6 2 | 3+4 2 5+6 2 |
| | 6 2 6 2 | 6 2 6 2 | 5+6 2 3+4 2 |
| | 6 2 6 2 | 6 2 5 2 | 3+4 2 3+4 2 |
| | 5 2 5 2 | 5 2 5 2 | 5+6 2 3+4 2 |
| | 6 2 5 2 | 6 2 5 2 | 3+4 2 5+6 2 |
| | 5 2 5 2 | 5 2 6 2 | 5+6 2 3+4 2 |
| | 6 2 5 2 | 6 2 5 2 | 5+6 2 5+6 2 |
| | 6 2 6 2 | 5 2 5 2 | 5+6 2 3+4 2 |
| | 5 2 5 2 | 6 2 5 2 | 5+6 2 3+4 2 |
| | 6 2 5 2 | 6 2 5 2 | 3+4 2 3+4 2 |
| 65.1.23 | 154010 | 158 | Ba1Mn2O6Tb1 |
| | -GM -T | -Y -Z | -R -S |
| | 5 2 6 2 | 5 2 6 2 | 5+6 2 3+4 2 |
| | 6 2 5 2 | 6 2 5 2 | 3+4 2 5+6 2 |
| | 5 2 5 2 | 5 2 5 2 | 5+6 2 5+6 2 |
| | 6 2 6 2 | 6 2 6 2 | 3+4 2 3+4 2 |
| | 5 2 6 2 | 5 2 6 2 | 5+6 2 3+4 2 |
| | 6 2 5 2 | 6 2 5 2 | 3+4 2 5+6 2 |
| | 6 2 5 2 | 6 2 6 2 | 3+4 2 5+6 2 |
| | 5 2 6 2 | 5 2 5 2 | 5+6 2 3+4 2 |
| | 5 2 6 2 | 5 2 6 2 | 5+6 2 3+4 2 |
| | 6 2 5 2 | 6 2 5 2 | 3+4 2 5+6 2 |
| | 5 2 6 2 | 5 2 6 2 | 3+4 2 3+4 2 |
| | 6 2 5 2 | 6 2 5 2 | 5+6 2 5+6 2 |
| | 6 2 5 2 | 5 2 6 2 | 5+6 2 3+4 2 |
| | 5 2 6 2 | 6 2 5 2 | 3+4 2 5+6 2 |
| | 6 2 6 2 | 6 2 5 2 | 3+4 2 3+4 2 |
| | 6 2 6 2 | 5 2 6 2 | 3+4 2 5+6 2 |
| | 5 2 5 2 | 6 2 5 2 | 5+6 2 5+6 2 |
| | 6 2 5 2 | 5 2 6 2 | 5+6 2 3+4 2 |
| | 5 2 5 2 | 6 2 5 2 | 5+6 2 5+6 2 |
| | 5 2 6 2 | 5 2 6 2 | 3+4 2 3+4 2 |
| | 6 2 6 2 | 5 2 5 2 | 5+6 2 3+4 2 |
| | 5 2 5 2 | 6 2 6 2 | 3+4 2 5+6 2 |
| | 5 2 5 2 | 5 2 5 2 | 5+6 2 5+6 2 |
| | 6 2 6 2 | 6 2 6 2 | 3+4 2 3+4 2 |
| | 5 2 5 2 | 5 2 5 2 | 5+6 2 5+6 2 |
| | 5 2 6 2 | 6 2 5 2 | 3+4 2 3+4 2 |
| | 6 2 5 2 | 5 2 6 2 | 5+6 2 5+6 2 |
| | 6 2 6 2 | 6 2 6 2 | 3+4 2 3+4 2 |
| | 5 2 5 2 | 6 2 6 2 | 3+4 2 3+4 2 |
| | 6 2 6 2 | 5 2 5 2 | 5+6 2 5+6 2 |
| | 6 2 6 2 | 6 2 5 2 | 3+4 2 3+4 2 |
| | 6 2 5 2 | 5 2 5 2 | 5+6 2 5+6 2 |
| | 5 2 5 2 | 5 2 6 2 | 5+6 2 3+4 2 |

Continued on next column

**TABLE XL– continued**

| TQC | ICSD | NoE | SF |
|---|---|---|---|
| | 6 2 6 2 | 6 2 6 2 | 3+4 2 5+6 2 |
| | 5 2 5 2 | 6 2 5 2 | 5+6 2 5+6 2 |
| | 6 2 6 2 | 5 2 6 2 | 3+4 2 3+4 2 |
| | 5 2 5 2 | 5 2 5 2 | 5+6 2 5+6 2 |
| | 6 2 6 2 | 6 2 6 2 | 3+4 2 3+4 2 |
| | 5 2 6 2 | 5 2 6 2 | 5+6 2 3+4 2 |
| | 5 2 5 2 | 6 2 5 2 | 3+4 2 5+6 2 |
| | 5 2 6 2 | 6 2 5 2 | 3+4 2 5+6 2 |
| | 6 2 5 2 | 5 2 6 2 | 5+6 2 3+4 2 |
| | 5 2 5 2 | 6 2 6 2 | 3+4 2 3+4 2 |
| | 6 2 6 2 | 5 2 6 2 | 3+4 2 5+6 2 |
| | 6 2 5 2 | 6 2 5 2 | 3+4 2 5+6 2 |
| | 5 2 6 2 | 5 2 5 2 | 5+6 2 3+4 2 |
| | 5 2 6 2 | 5 2 6 2 | 3+4 2 5+6 2 |
| | 5 2 5 2 | 5 2 5 2 | 5+6 2 3+4 2 |
| | 6 2 5 2 | 6 2 6 2 | 5+6 2 5+6 2 |
| | 6 2 6 2 | 5 2 5 2 | 5+6 2 3+4 2 |
| | 6 2 6 2 | 5 2 6 2 | 3+4 2 3+4 2 |
| | 6 2 5 2 | 6 2 6 2 | 5+6 2 5+6 2 |
| | 6 2 5 2 | 6 2 5 2 | 3+4 2 3+4 2 |
| | 5 2 6 2 | 6 2 5 2 | 5+6 2 3+4 2 |
| | 5 2 5 2 | 6 2 6 2 | 5+6 2 5+6 2 |
| | 5 2 6 2 | 6 2 6 2 | 3+4 2 5+6 2 |
| | 5 2 6 2 | 5 2 6 2 | 5+6 2 5+6 2 |
| | 6 2 5 2 | 5 2 5 2 | 5+6 2 3+4 2 |
| | 5 2 6 2 | 5 2 5 2 | 3+4 2 5+6 2 |
| | 6 2 5 2 | 6 2 6 2 | 3+4 2 3+4 2 |
| | 5 2 6 2 | 5 2 5 2 | 3+4 2 3+4 2 |
| | 6 2 5 2 | 6 2 5 2 | 3+4 2 5+6 2 |
| | 5 2 6 2 | 5 2 6 2 | 5+6 2 5+6 2 |
| | 6 2 5 2 | 6 2 5 2 | 5+6 2 3+4 2 |
| | 6 2 6 2 | 5 2 5 2 | 3+4 2 5+6 2 |
| | 5 2 5 2 | 6 2 6 2 | 5+6 2 3+4 2 |
| | 5 2 5 2 | 5 2 5 2 | 5+6 2 5+6 2 |
| | 5 2 6 2 | 6 2 6 2 | 3+4 2 3+4 2 |
| | 5 2 5 2 | 6 2 5 2 | 5+6 2 3+4 2 |
| | 5 2 6 2 | 5 2 6 2 | 3+4 2 5+6 2 |
| | 6 2 6 2 | 5 2 5 2 | 3+4 2 5+6 2 |
| | 6 2 5 2 | 6 2 6 2 | 5+6 2 3+4 2 |
| | 6 2 6 2 | 6 2 5 2 | 3+4 2 3+4 2 |
| | 5 2 5 2 | 5 2 5 2 | 5+6 2 5+6 2 |
| | 6 2 6 2 | 6 2 6 2 | 3+4 2 5+6 2 |
| | 5 2 5 2 | 5 2 6 2 | 5+6 2 3+4 2 |
| | 6 2 6 2 | 5 2 6 2 | 3+4 2 5+6 2 |
| | 6 2 5 2 | 6 2 6 2 | 5+6 2 3+4 2 |
| | 5 2 6 2 | 5 2 5 2 | 3+4 2 3+4 2 |
| 65.1.24 | 106291 | 104 | Au7Rb3 |
| | -GM -T | -Y -Z | -R -S |
| | 5 2 5 2 | 6 2 6 2 | 5+6 2 3+4 2 |
| | 6 2 6 2 | 5 2 5 2 | 3+4 2 5+6 2 |
| | 5 2 5 2 | 5 2 5 2 | 3+4 2 3+4 2 |
| | 5 2 5 2 | 6 2 6 2 | 5+6 2 3+4 2 |
| | 6 2 6 2 | 5 2 5 2 | 3+4 2 5+6 2 |
| | 6 2 6 2 | 6 2 6 2 | 5+6 2 5+6 2 |
| | 5 2 5 2 | 6 2 6 2 | 5+6 2 3+4 2 |
| | 5 2 6 2 | 5 2 6 2 | 3+4 2 5+6 2 |
| | 6 2 5 2 | 6 2 5 2 | 5+6 2 3+4 2 |
| | 6 2 6 2 | 5 2 5 2 | 3+4 2 5+6 2 |
| | 6 2 6 2 | 6 2 6 2 | 5+6 2 5+6 2 |
| | 6 2 6 2 | 6 2 6 2 | 5+6 2 5+6 2 |
| | 5 2 6 2 | 6 2 5 2 | 3+4 2 5+6 2 |

Continued on next column

**TABLE XL– continued**

| TQC | ICSD | NoE | SF |
|---|---|---|---|
| | 5 2 6 2 | 5 2 6 2 | 5+6 2 3+4 2 |
| | 5 2 6 2 | 6 2 6 2 | 3+4 2 3+4 2 |
| | 5 2 6 2 | 6 2 5 2 | 3+4 2 5+6 2 |
| | 5 2 5 2 | 6 2 5 2 | 5+6 2 5+6 2 |
| | 5 2 6 2 | 5 2 5 2 | 5+6 2 5+6 2 |
| | 5 2 6 2 | 5 2 5 2 | 3+4 2 5+6 2 |
| | 6 2 6 2 | 5 2 6 2 | 3+4 2 3+4 2 |
| | 5 2 5 2 | 6 2 5 2 | 5+6 2 5+6 2 |
| | 5 2 5 2 | 5 2 5 2 | 5+6 2 3+4 2 |
| | 6 2 6 2 | 6 2 6 2 | 3+4 2 5+6 2 |
| | 5 2 6 2 | 5 2 5 2 | 3+4 2 3+4 2 |
| | 6 2 6 2 | 6 2 5 2 | 3+4 2 3+4 2 |
| | 5 2 5 2 | 6 2 6 2 | 5+6 2 5+6 2 |
| | 6 2 6 2 | 6 2 6 2 | 3+4 2 5+6 2 |
| | 5 2 5 2 | 5 2 6 2 | 5+6 2 3+4 2 |
| | 6 2 6 2 | 6 2 5 2 | 3+4 2 5+6 2 |
| | 5 2 6 2 | 5 2 6 2 | 5+6 2 3+4 2 |
| | 5 2 5 2 | 6 2 5 2 | 3+4 2 5+6 2 |
| | 5 2 6 2 | 5 2 6 2 | 3+4 2 5+6 2 |
| | 6 2 6 2 | 6 2 5 2 | 3+4 2 3+4 2 |
| | 6 2 6 2 | 5 2 6 2 | 5+6 2 3+4 2 |
| | 5 2 5 2 | 6 2 5 2 | 3+4 2 5+6 2 |
| | 6 2 6 2 | 6 2 5 2 | 5+6 2 3+4 2 |
| | 5 2 5 2 | 5 2 6 2 | 5+6 2 5+6 2 |
| | 6 2 5 2 | 5 2 6 2 | 3+4 2 5+6 2 |
| | 5 2 6 2 | 5 2 5 2 | 3+4 2 3+4 2 |
| | 5 2 6 2 | 6 2 6 2 | 5+6 2 3+4 2 |
| | 5 2 5 2 | 6 2 5 2 | 5+6 2 5+6 2 |
| | 6 2 6 2 | 5 2 6 2 | 3+4 2 5+6 2 |
| | 5 2 5 2 | 6 2 5 2 | 5+6 2 3+4 2 |
| | 5 2 6 2 | 5 2 6 2 | 3+4 2 5+6 2 |
| | 5 2 6 2 | 6 2 5 2 | 5+6 2 3+4 2 |
| | 5 2 5 2 | 5 2 5 2 | 3+4 2 5+6 2 |
| | 6 2 6 2 | 6 2 5 2 | 5+6 2 5+6 2 |
| | 5 2 6 2 | 6 2 6 2 | 3+4 2 3+4 2 |
| | 6 2 6 2 | 5 2 6 2 | 3+4 2 5+6 2 |
| | 5 2 6 2 | 6 2 5 2 | 3+4 2 3+4 2 |
| | 5 2 6 2 | 6 2 5 2 | 5+6 2 5+6 2 |
| | 5 2 5 2 | 5 2 5 2 | 5+6 2 3+4 2 |
| 65.1.25 | 106290 | 104 | Au7Rb3 |
| | -GM -T | -Y -Z | -R -S |
| | 5 2 5 2 | 6 2 6 2 | 3+4 2 5+6 2 |
| | 6 2 6 2 | 5 2 5 2 | 5+6 2 3+4 2 |
| | 5 2 5 2 | 5 2 5 2 | 5+6 2 5+6 2 |
| | 5 2 5 2 | 6 2 6 2 | 3+4 2 5+6 2 |
| | 6 2 6 2 | 5 2 5 2 | 5+6 2 3+4 2 |
| | 6 2 6 2 | 6 2 6 2 | 3+4 2 3+4 2 |
| | 5 2 5 2 | 6 2 6 2 | 3+4 2 5+6 2 |
| | 5 2 6 2 | 5 2 6 2 | 5+6 2 3+4 2 |
| | 6 2 5 2 | 6 2 5 2 | 3+4 2 5+6 2 |
| | 6 2 6 2 | 5 2 5 2 | 5+6 2 3+4 2 |
| | 6 2 6 2 | 6 2 6 2 | 3+4 2 3+4 2 |
| | 6 2 6 2 | 6 2 6 2 | 3+4 2 3+4 2 |
| | 5 2 6 2 | 6 2 5 2 | 5+6 2 3+4 2 |
| | 5 2 6 2 | 5 2 6 2 | 3+4 2 5+6 2 |
| | 5 2 6 2 | 6 2 6 2 | 5+6 2 5+6 2 |
| | 5 2 6 2 | 6 2 5 2 | 5+6 2 3+4 2 |
| | 5 2 5 2 | 6 2 5 2 | 3+4 2 3+4 2 |
| | 5 2 6 2 | 5 2 5 2 | 3+4 2 3+4 2 |
| | 5 2 6 2 | 5 2 5 2 | 5+6 2 3+4 2 |
| | 6 2 6 2 | 5 2 6 2 | 5+6 2 5+6 2 |

Continued on next column

**TABLE XL– continued**

| TQC | ICSD | NoE | SF |
|---|---|---|---|
| | 5 2 5 2 | 6 2 5 2 | 3+4 2 3+4 2 |
| | 5 2 5 2 | 5 2 5 2 | 3+4 2 5+6 2 |
| | 6 2 6 2 | 6 2 6 2 | 5+6 2 3+4 2 |
| | 5 2 6 2 | 5 2 5 2 | 5+6 2 5+6 2 |
| | 6 2 6 2 | 6 2 5 2 | 5+6 2 5+6 2 |
| | 5 2 5 2 | 6 2 6 2 | 3+4 2 3+4 2 |
| | 6 2 6 2 | 6 2 6 2 | 5+6 2 3+4 2 |
| | 5 2 5 2 | 5 2 6 2 | 3+4 2 5+6 2 |
| | 6 2 6 2 | 6 2 5 2 | 5+6 2 3+4 2 |
| | 5 2 6 2 | 5 2 6 2 | 3+4 2 5+6 2 |
| | 5 2 5 2 | 6 2 5 2 | 5+6 2 3+4 2 |
| | 5 2 6 2 | 5 2 6 2 | 5+6 2 3+4 2 |
| | 6 2 6 2 | 6 2 5 2 | 5+6 2 5+6 2 |
| | 6 2 6 2 | 5 2 6 2 | 3+4 2 3+4 2 |
| | 5 2 5 2 | 6 2 5 2 | 5+6 2 5+6 2 |
| | 6 2 6 2 | 6 2 5 2 | 3+4 2 5+6 2 |
| | 5 2 5 2 | 5 2 6 2 | 3+4 2 3+4 2 |
| | 6 2 5 2 | 5 2 6 2 | 5+6 2 3+4 2 |
| | 5 2 6 2 | 5 2 5 2 | 5+6 2 5+6 2 |
| | 5 2 6 2 | 6 2 6 2 | 3+4 2 5+6 2 |
| | 5 2 5 2 | 6 2 5 2 | 3+4 2 3+4 2 |
| | 6 2 6 2 | 5 2 6 2 | 5+6 2 3+4 2 |
| | 5 2 5 2 | 6 2 5 2 | 3+4 2 5+6 2 |
| | 5 2 6 2 | 5 2 6 2 | 5+6 2 3+4 2 |
| | 5 2 6 2 | 6 2 5 2 | 3+4 2 5+6 2 |
| | 5 2 5 2 | 5 2 5 2 | 5+6 2 3+4 2 |
| | 6 2 6 2 | 6 2 5 2 | 3+4 2 3+4 2 |
| | 5 2 6 2 | 6 2 6 2 | 5+6 2 5+6 2 |
| | 6 2 6 2 | 5 2 6 2 | 5+6 2 3+4 2 |
| | 5 2 6 2 | 6 2 5 2 | 5+6 2 5+6 2 |
| | 5 2 6 2 | 6 2 5 2 | 3+4 2 3+4 2 |
| | 5 2 5 2 | 5 2 5 2 | 3+4 2 5+6 2 |
| 65.1.26 | 418529 | 64 | Au4Ce1Si2 |
| | -GM -T | -Y -Z | -R -S |
| | 5 2 5 2 | 5 2 5 2 | 3+4 2 3+4 2 |
| | 6 2 6 2 | 6 2 6 2 | 5+6 2 5+6 2 |
| | 6 2 6 2 | 6 2 6 2 | 5+6 2 5+6 2 |
| | 6 2 6 2 | 6 2 6 2 | 5+6 2 5+6 2 |
| | 5 2 6 2 | 5 2 6 2 | 3+4 2 5+6 2 |
| | 6 2 5 2 | 6 2 5 2 | 5+6 2 3+4 2 |
| | 5 2 6 2 | 5 2 5 2 | 3+4 2 3+4 2 |
| | 5 2 5 2 | 5 2 6 2 | 5+6 2 5+6 2 |
| | 5 2 6 2 | 6 2 6 2 | 3+4 2 3+4 2 |
| | 6 2 5 2 | 5 2 5 2 | 5+6 2 5+6 2 |
| | 6 2 5 2 | 6 2 6 2 | 3+4 2 3+4 2 |
| | 6 2 6 2 | 5 2 5 2 | 5+6 2 5+6 2 |
| | 5 2 6 2 | 6 2 5 2 | 3+4 2 3+4 2 |
| | 6 2 5 2 | 6 2 5 2 | 5+6 2 5+6 2 |
| | 5 2 6 2 | 5 2 6 2 | 3+4 2 3+4 2 |
| | 5 2 5 2 | 6 2 6 2 | 5+6 2 5+6 2 |
| | 6 2 6 2 | 6 2 5 2 | 3+4 2 3+4 2 |
| | 5 2 5 2 | 5 2 6 2 | 5+6 2 5+6 2 |
| | 5 2 6 2 | 6 2 5 2 | 3+4 2 3+4 2 |
| | 6 2 6 2 | 5 2 6 2 | 3+4 2 5+6 2 |
| | 5 2 5 2 | 6 2 5 2 | 5+6 2 3+4 2 |
| | 6 2 5 2 | 5 2 6 2 | 5+6 2 5+6 2 |
| | 5 2 5 2 | 6 2 5 2 | 3+4 2 3+4 2 |
| | 6 2 6 2 | 5 2 6 2 | 5+6 2 5+6 2 |
| | 6 2 6 2 | 6 2 5 2 | 3+4 2 3+4 2 |
| | 5 2 5 2 | 5 2 6 2 | 5+6 2 5+6 2 |
| | 6 2 6 2 | 5 2 6 2 | 3+4 2 5+6 2 |
| | 5 2 5 2 | 6 2 5 2 | 5+6 2 3+4 2 |
| | 6 2 6 2 | 6 2 6 2 | 3+4 2 5+6 2 |
| | 5 2 5 2 | 5 2 5 2 | 3+4 2 3+4 2 |
| | 6 2 5 2 | 5 2 5 2 | 5+6 2 3+4 2 |
| | 6 2 6 2 | 6 2 6 2 | 3+4 2 5+6 2 |
| 65.2.1 | 15927 | 50 | Mg1O3V1 |
| | -GM -T | -Y -Z | -R -S |
| | 5 2 5 2 | 6 2 5 2 | 5+6 2 5+6 2 |
| | 5 2 6 2 | 5 2 6 2 | 5+6 2 3+4 2 |
| | 5 2 5 2 | 6 2 5 2 | 3+4 2 5+6 2 |
| | 6 2 5 2 | 5 2 6 2 | 3+4 2 3+4 2 |
| | 6 2 6 2 | 5 2 5 2 | 5+6 2 5+6 2 |
| | 5 2 6 2 | 5 2 6 2 | 5+6 2 3+4 2 |
| | 6 2 6 2 | 6 2 6 2 | 3+4 2 3+4 2 |
| | 5 2 6 2 | 5 2 5 2 | 3+4 2 3+4 2 |
| | 6 2 5 2 | 6 2 6 2 | 3+4 2 5+6 2 |
| | 5 2 6 2 | 6 2 6 2 | 5+6 2 3+4 2 |
| | 5 2 6 2 | 6 2 6 2 | 3+4 2 3+4 2 |
| | 6 2 6 2 | 5 2 5 2 | 5+6 2 5+6 2 |
| | 6 2 5 2 | 6 2 5 2 | 3+4 2 5+6 2 |
| | 6 2 5 2 | 6 2 6 2 | 5+6 2 3+4 2 |
| | 5 2 5 2 | 6 2 5 2 | 5+6 2 3+4 2 |
| | 5 2 6 2 | 5 2 6 2 | 3+4 2 5+6 2 |
| | 6 2 6 2 | 5 2 6 2 | 3+4 2 3+4 2 |
| | 6 2 6 2 | 6 2 5 2 | 5+6 2 5+6 2 |
| | 6 2 5 2 | 5 2 6 2 | 3+4 2 3+4 2 |
| | 6 2 5 2 | 5 2 6 2 | 3+4 2 5+6 2 |
| | 5 2 6 2 | 6 2 5 2 | 5+6 2 3+4 2 |
| | 6 2 5 2 | 6 2 5 2 | 3+4 2 5+6 2 |
| | 5 2 5 2 | 6 2 5 2 | 5+6 2 5+6 2 |
| | 6 2 5 2 | 6 2 5 2 | 3+4 2 5+6 2 |
| | 6 2 5 2 | 6 2 5 2 | 3+4 2 5+6 2 |
| 65.2.2 | 104809 | 84 | Lu2Ni1Sn6 |
| | -GM -T | -Y -Z | -R -S |
| | 5 2 5 2 | 5 2 5 2 | 3+4 2 3+4 2 |
| | 6 2 6 2 | 6 2 6 2 | 5+6 2 5+6 2 |
| | 5 2 6 2 | 6 2 5 2 | 3+4 2 3+4 2 |
| | 6 2 5 2 | 5 2 6 2 | 5+6 2 5+6 2 |
| | 5 2 6 2 | 6 2 5 2 | 3+4 2 3+4 2 |
| | 5 2 5 2 | 5 2 6 2 | 5+6 2 5+6 2 |
| | 6 2 6 2 | 6 2 5 2 | 3+4 2 3+4 2 |
| | 6 2 5 2 | 5 2 6 2 | 5+6 2 5+6 2 |
| | 5 2 6 2 | 5 2 6 2 | 5+6 2 5+6 2 |
| | 6 2 5 2 | 6 2 5 2 | 5+6 2 3+4 2 |
| | 5 2 5 2 | 6 2 6 2 | 3+4 2 3+4 2 |
| | 6 2 6 2 | 5 2 5 2 | 5+6 2 5+6 2 |
| | 5 2 5 2 | 5 2 6 2 | 3+4 2 5+6 2 |
| | 6 2 6 2 | 6 2 5 2 | 3+4 2 3+4 2 |
| | 5 2 5 2 | 5 2 5 2 | 3+4 2 3+4 2 |
| | 6 2 5 2 | 6 2 6 2 | 5+6 2 5+6 2 |
| | 5 2 6 2 | 5 2 6 2 | 3+4 2 3+4 2 |
| | 6 2 5 2 | 6 2 5 2 | 5+6 2 5+6 2 |
| | 5 2 6 2 | 5 2 6 2 | 5+6 2 3+4 2 |
| | 5 2 6 2 | 6 2 5 2 | 3+4 2 5+6 2 |
| | 5 2 5 2 | 5 2 5 2 | 3+4 2 3+4 2 |
| | 6 2 6 2 | 6 2 6 2 | 3+4 2 5+6 2 |
| | 6 2 5 2 | 5 2 5 2 | 5+6 2 3+4 2 |
| | 6 2 5 2 | 5 2 6 2 | 3+4 2 5+6 2 |
| | 5 2 6 2 | 6 2 5 2 | 5+6 2 3+4 2 |
| | 6 2 5 2 | 6 2 6 2 | 5+6 2 5+6 2 |
| | 5 2 6 2 | 5 2 5 2 | 3+4 2 3+4 2 |

Continued on next column

**TABLE XL– continued**

| TQC | ICSD NoE SF |
|---|---|
| | 5 2 6 2 5 2 6 2 5+6 2 5+6 2 |
| | 5 2 6 2 5 2 5 2 3+4 2 5+6 2 |
| | 5 2 5 2 5 2 6 2 3+4 2 3+4 2 |
| | 6 2 6 2 6 2 6 2 5+6 2 5+6 2 |
| | 6 2 6 2 6 2 6 2 5+6 2 5+6 2 |
| | 6 2 6 2 6 2 6 2 5+6 2 5+6 2 |
| | 5 2 6 2 6 2 6 2 5+6 2 5+6 2 |
| | 5 2 5 2 5 2 6 2 3+4 2 5+6 2 |
| | 5 2 6 2 5 2 6 2 3+4 2 3+4 2 |
| | 5 2 6 2 5 2 5 2 3+4 2 5+6 2 |
| | 6 2 5 2 6 2 6 2 3+4 2 3+4 2 |
| | 5 2 5 2 5 2 5 2 3+4 2 5+6 2 |
| | 6 2 5 2 5 2 5 2 3+4 2 3+4 2 |
| | 5 2 6 2 5 2 5 2 5+6 2 3+4 2 |
| | 5 2 6 2 6 2 6 2 3+4 2 5+6 2 |
| 65.2.3 | 106923 102 Lu1Ni1Sn4 |
| | -GM -T -Y -Z -R -S |
| | 5 2 6 2 6 2 5 2 3+4 2 3+4 2 |
| | 6 2 5 2 5 2 6 2 5+6 2 5+6 2 |
| | 5 2 6 2 6 2 5 2 3+4 2 3+4 2 |
| | 6 2 5 2 5 2 6 2 5+6 2 5+6 2 |
| | 5 2 6 2 6 2 5 2 3+4 2 3+4 2 |
| | 5 2 5 2 5 2 6 2 5+6 2 5+6 2 |
| | 6 2 6 2 6 2 5 2 3+4 2 3+4 2 |
| | 6 2 5 2 5 2 6 2 5+6 2 5+6 2 |
| | 5 2 6 2 5 2 6 2 3+4 2 3+4 2 |
| | 6 2 5 2 6 2 5 2 3+4 2 5+6 2 |
| | 5 2 5 2 6 2 6 2 5+6 2 5+6 2 |
| | 5 2 6 2 5 2 5 2 3+4 2 3+4 2 |
| | 6 2 5 2 5 2 6 2 5+6 2 3+4 2 |
| | 5 2 6 2 5 2 5 2 5+6 2 5+6 2 |
| | 6 2 5 2 6 2 6 2 5+6 2 5+6 2 |
| | 5 2 6 2 5 2 5 2 3+4 2 3+4 2 |
| | 6 2 5 2 6 2 5 2 3+4 2 3+4 2 |
| | 5 2 5 2 6 2 6 2 3+4 2 5+6 2 |
| | 6 2 6 2 5 2 6 2 5+6 2 3+4 2 |
| | 5 2 5 2 5 2 5 2 5+6 2 5+6 2 |
| | 6 2 6 2 6 2 6 2 3+4 2 3+4 2 |
| | 5 2 6 2 6 2 5 2 5+6 2 5+6 2 |
| | 6 2 6 2 5 2 5 2 3+4 2 3+4 2 |
| | 6 2 5 2 6 2 6 2 5+6 2 5+6 2 |
| | 5 2 5 2 5 2 6 2 3+4 2 3+4 2 |
| | 5 2 6 2 5 2 5 2 5+6 2 5+6 2 |
| | 5 2 5 2 6 2 6 2 3+4 2 3+4 2 |
| | 6 2 6 2 5 2 5 2 5+6 2 5+6 2 |
| | 6 2 5 2 6 2 6 2 3+4 2 3+4 2 |
| | 6 2 6 2 6 2 5 2 5+6 2 5+6 2 |
| | 6 2 6 2 6 2 5 2 3+4 2 5+6 2 |
| | 6 2 5 2 6 2 6 2 5+6 2 3+4 2 |
| | 5 2 5 2 6 2 6 2 3+4 2 3+4 2 |
| | 6 2 6 2 5 2 5 2 3+4 2 5+6 2 |
| | 5 2 5 2 5 2 5 2 5+6 2 5+6 2 |
| | 5 2 6 2 6 2 6 2 5+6 2 3+4 2 |
| | 6 2 6 2 5 2 6 2 5+6 2 3+4 2 |
| | 5 2 6 2 5 2 6 2 5+6 2 3+4 2 |
| | 5 2 6 2 6 2 5 2 3+4 2 3+4 2 |
| | 6 2 5 2 5 2 5 2 3+4 2 5+6 2 |
| | 5 2 5 2 6 2 6 2 5+6 2 5+6 2 |
| | 6 2 5 2 6 2 5 2 5+6 2 5+6 2 |
| | 6 2 6 2 5 2 6 2 3+4 2 3+4 2 |
| | 5 2 5 2 6 2 6 2 5+6 2 5+6 2 |

Continued on next column

**TABLE XL– continued**

| TQC | ICSD NoE SF |
|---|---|
| | 6 2 5 2 5 2 6 2 3+4 2 5+6 2 |
| | 5 2 5 2 5 2 5 2 5+6 2 3+4 2 |
| | 5 2 6 2 5 2 5 2 3+4 2 5+6 2 |
| | 5 2 6 2 6 2 5 2 5+6 2 3+4 2 |
| | 5 2 5 2 5 2 5 2 3+4 2 5+6 2 |
| | 6 2 6 2 6 2 5 2 3+4 2 3+4 2 |
| | 6 2 5 2 6 2 6 2 5+6 2 5+6 2 |
| 65.2.4 | 83885 44 Ge2U1 |
| | -GM -T -Y -Z -R -S |
| | 5 2 5 2 6 2 6 2 5+6 2 3+4 2 |
| | 6 2 6 2 5 2 5 2 3+4 2 5+6 2 |
| | 5 2 5 2 6 2 6 2 5+6 2 3+4 2 |
| | 6 2 6 2 5 2 5 2 3+4 2 5+6 2 |
| | 5 2 6 2 6 2 6 2 3+4 2 3+4 2 |
| | 6 2 5 2 5 2 5 2 5+6 2 5+6 2 |
| | 5 2 5 2 6 2 6 2 5+6 2 3+4 2 |
| | 6 2 6 2 5 2 5 2 3+4 2 5+6 2 |
| | 5 2 6 2 6 2 5 2 5+6 2 3+4 2 |
| | 5 2 5 2 5 2 6 2 3+4 2 5+6 2 |
| | 6 2 6 2 5 2 5 2 3+4 2 5+6 2 |
| | 5 2 5 2 5 2 6 2 3+4 2 3+4 2 |
| | 6 2 5 2 6 2 5 2 5+6 2 5+6 2 |
| | 6 2 6 2 6 2 6 2 5+6 2 3+4 2 |
| | 6 2 6 2 6 2 5 2 3+4 2 3+4 2 |
| | 5 2 6 2 6 2 5 2 5+6 2 3+4 2 |
| | 5 2 5 2 5 2 6 2 3+4 2 5+6 2 |
| | 6 2 6 2 6 2 6 2 5+6 2 5+6 2 |
| | 6 2 5 2 5 2 5 2 5+6 2 5+6 2 |
| | 5 2 6 2 6 2 6 2 5+6 2 3+4 2 |
| | 5 2 5 2 6 2 5 2 3+4 2 3+4 2 |
| | 6 2 6 2 5 2 5 2 3+4 2 5+6 2 |
| 65.2.5 | 82702 44 Ge2U1 |
| | -GM -T -Y -Z -R -S |
| | 5 2 6 2 6 2 5 2 3+4 2 3+4 2 |
| | 6 2 5 2 5 2 6 2 5+6 2 5+6 2 |
| | 5 2 6 2 6 2 5 2 3+4 2 3+4 2 |
| | 6 2 5 2 5 2 6 2 5+6 2 5+6 2 |
| | 5 2 5 2 6 2 5 2 5+6 2 3+4 2 |
| | 6 2 6 2 5 2 6 2 3+4 2 5+6 2 |
| | 5 2 6 2 6 2 5 2 3+4 2 3+4 2 |
| | 6 2 5 2 5 2 6 2 5+6 2 5+6 2 |
| | 5 2 5 2 6 2 6 2 3+4 2 3+4 2 |
| | 5 2 6 2 5 2 5 2 5+6 2 5+6 2 |
| | 6 2 5 2 5 2 6 2 5+6 2 5+6 2 |
| | 5 2 6 2 5 2 5 2 5+6 2 3+4 2 |
| | 6 2 6 2 6 2 6 2 3+4 2 5+6 2 |
| | 6 2 5 2 6 2 5 2 3+4 2 3+4 2 |
| | 6 2 5 2 6 2 6 2 5+6 2 3+4 2 |
| | 5 2 5 2 6 2 6 2 3+4 2 3+4 2 |
| | 5 2 6 2 5 2 5 2 5+6 2 5+6 2 |
| | 6 2 5 2 6 2 5 2 3+4 2 5+6 2 |
| | 6 2 6 2 5 2 6 2 3+4 2 5+6 2 |
| | 5 2 5 2 6 2 5 2 3+4 2 3+4 2 |
| | 5 2 5 2 6 2 6 2 5+6 2 3+4 2 |
| | 6 2 6 2 5 2 6 2 5+6 2 5+6 2 |
| 65.2.6 | 236995 44 Ge2U1 |
| | -GM -T -Y -Z -R -S |
| | 5 2 5 2 6 2 6 2 5+6 2 3+4 2 |
| | 6 2 6 2 5 2 5 2 3+4 2 5+6 2 |
| | 5 2 5 2 6 2 6 2 5+6 2 3+4 2 |
| | 6 2 6 2 5 2 5 2 3+4 2 5+6 2 |

Continued on next column

**TABLE XL– continued**

| TQC | ICSD | NoE | SF |
|---|---|---|---|
| | 5 2 6 2 | 6 2 6 2 | 3+4 2 3+4 2 |
| | 6 2 5 2 | 5 2 5 2 | 5+6 2 5+6 2 |
| | 5 2 5 2 | 6 2 6 2 | 5+6 2 3+4 2 |
| | 6 2 6 2 | 5 2 5 2 | 3+4 2 5+6 2 |
| | 5 2 6 2 | 6 2 5 2 | 5+6 2 3+4 2 |
| | 5 2 5 2 | 5 2 6 2 | 3+4 2 5+6 2 |
| | 6 2 6 2 | 5 2 5 2 | 3+4 2 5+6 2 |
| | 5 2 5 2 | 5 2 6 2 | 3+4 2 3+4 2 |
| | 6 2 5 2 | 6 2 5 2 | 5+6 2 5+6 2 |
| | 6 2 6 2 | 6 2 6 2 | 5+6 2 3+4 2 |
| | 6 2 6 2 | 6 2 5 2 | 3+4 2 3+4 2 |
| | 5 2 6 2 | 6 2 5 2 | 5+6 2 3+4 2 |
| | 5 2 5 2 | 5 2 6 2 | 3+4 2 5+6 2 |
| | 6 2 6 2 | 6 2 6 2 | 5+6 2 5+6 2 |
| | 6 2 5 2 | 5 2 5 2 | 5+6 2 5+6 2 |
| | 5 2 6 2 | 6 2 6 2 | 5+6 2 3+4 2 |
| | 5 2 5 2 | 6 2 5 2 | 3+4 2 3+4 2 |
| | 6 2 6 2 | 5 2 5 2 | 3+4 2 5+6 2 |
| 65.2.7 | 56046 | 40 | Ge2Th1 |
| | -GM -T | -Y -Z | -R -S |
| | 5 2 6 2 | 6 2 5 2 | 3+4 2 3+4 2 |
| | 6 2 5 2 | 5 2 6 2 | 5+6 2 5+6 2 |
| | 5 2 6 2 | 6 2 5 2 | 3+4 2 3+4 2 |
| | 6 2 5 2 | 5 2 6 2 | 5+6 2 5+6 2 |
| | 5 2 5 2 | 6 2 5 2 | 5+6 2 3+4 2 |
| | 6 2 6 2 | 5 2 6 2 | 3+4 2 5+6 2 |
| | 5 2 6 2 | 6 2 5 2 | 3+4 2 3+4 2 |
| | 6 2 5 2 | 5 2 6 2 | 5+6 2 5+6 2 |
| | 5 2 5 2 | 6 2 6 2 | 5+6 2 3+4 2 |
| | 5 2 6 2 | 5 2 5 2 | 3+4 2 5+6 2 |
| | 6 2 5 2 | 5 2 6 2 | 5+6 2 5+6 2 |
| | 5 2 6 2 | 5 2 5 2 | 5+6 2 3+4 2 |
| | 6 2 6 2 | 6 2 5 2 | 3+4 2 5+6 2 |
| | 5 2 5 2 | 6 2 6 2 | 3+4 2 3+4 2 |
| | 6 2 5 2 | 6 2 6 2 | 5+6 2 3+4 2 |
| | 5 2 5 2 | 5 2 5 2 | 3+4 2 3+4 2 |
| | 6 2 6 2 | 6 2 6 2 | 3+4 2 5+6 2 |
| | 6 2 6 2 | 6 2 5 2 | 5+6 2 5+6 2 |
| | 6 2 6 2 | 5 2 6 2 | 3+4 2 3+4 2 |
| | 5 2 5 2 | 6 2 5 2 | 3+4 2 5+6 2 |
| 65.2.8 | 193219 | 32 | Ge2Pb1Sr2 |
| | -GM -T | -Y -Z | -R -S |
| | 5 2 5 2 | 5 2 6 2 | 3+4 2 5+6 2 |
| | 6 2 6 2 | 6 2 5 2 | 5+6 2 3+4 2 |
| | 5 2 6 2 | 6 2 6 2 | 3+4 2 5+6 2 |
| | 6 2 5 2 | 5 2 5 2 | 5+6 2 3+4 2 |
| | 5 2 6 2 | 6 2 6 2 | 5+6 2 5+6 2 |
| | 6 2 5 2 | 5 2 5 2 | 3+4 2 3+4 2 |
| | 5 2 5 2 | 6 2 6 2 | 3+4 2 5+6 2 |
| | 6 2 6 2 | 5 2 5 2 | 5+6 2 3+4 2 |
| | 5 2 6 2 | 6 2 5 2 | 3+4 2 3+4 2 |
| | 5 2 5 2 | 5 2 5 2 | 5+6 2 5+6 2 |
| | 6 2 5 2 | 5 2 6 2 | 3+4 2 3+4 2 |
| | 6 2 5 2 | 6 2 6 2 | 5+6 2 5+6 2 |
| | 5 2 6 2 | 5 2 5 2 | 5+6 2 3+4 2 |
| | 6 2 6 2 | 6 2 6 2 | 3+4 2 5+6 2 |
| | 5 2 6 2 | 6 2 5 2 | 3+4 2 5+6 2 |
| | 6 2 6 2 | 5 2 6 2 | 5+6 2 3+4 2 |
| 65.2.9 | 40487 | 240 | Ge2Lu1Mn1 |
| | -GM -T | -Y -Z | -R -S |
| | 5 2 5 2 | 5 2 5 2 | 3+4 2 3+4 2 |

Continued on next column

**TABLE XL– continued**

| TQC | ICSD | NoE | SF |
|---|---|---|---|
| | 6 2 6 2 | 6 2 6 2 | 5+6 2 5+6 2 |
| | 5 2 5 2 | 5 2 5 2 | 3+4 2 3+4 2 |
| | 6 2 6 2 | 6 2 6 2 | 5+6 2 5+6 2 |
| | 5 2 5 2 | 5 2 5 2 | 3+4 2 3+4 2 |
| | 6 2 6 2 | 6 2 6 2 | 5+6 2 5+6 2 |
| | 6 2 6 2 | 6 2 6 2 | 3+4 2 5+6 2 |
| | 5 2 5 2 | 5 2 5 2 | 5+6 2 3+4 2 |
| | 5 2 5 2 | 5 2 5 2 | 3+4 2 3+4 2 |
| | 6 2 6 2 | 6 2 6 2 | 5+6 2 5+6 2 |
| | 5 2 5 2 | 5 2 6 2 | 3+4 2 3+4 2 |
| | 6 2 6 2 | 6 2 5 2 | 5+6 2 5+6 2 |
| | 6 2 5 2 | 6 2 5 2 | 3+4 2 5+6 2 |
| | 5 2 6 2 | 5 2 6 2 | 5+6 2 3+4 2 |
| | 5 2 5 2 | 5 2 5 2 | 5+6 2 3+4 2 |
| | 6 2 6 2 | 6 2 6 2 | 3+4 2 5+6 2 |
| | 5 2 6 2 | 5 2 6 2 | 3+4 2 5+6 2 |
| | 6 2 5 2 | 6 2 5 2 | 5+6 2 3+4 2 |
| | 6 2 5 2 | 6 2 5 2 | 5+6 2 3+4 2 |
| | 5 2 6 2 | 5 2 6 2 | 3+4 2 5+6 2 |
| | 5 2 5 2 | 6 2 6 2 | 3+4 2 5+6 2 |
| | 6 2 6 2 | 5 2 5 2 | 5+6 2 3+4 2 |
| | 5 2 5 2 | 6 2 5 2 | 3+4 2 5+6 2 |
| | 6 2 6 2 | 5 2 6 2 | 5+6 2 3+4 2 |
| | 5 2 6 2 | 5 2 5 2 | 5+6 2 5+6 2 |
| | 5 2 5 2 | 6 2 6 2 | 3+4 2 3+4 2 |
| | 5 2 6 2 | 5 2 5 2 | 3+4 2 3+4 2 |
| | 6 2 5 2 | 6 2 6 2 | 5+6 2 5+6 2 |
| | 5 2 5 2 | 5 2 6 2 | 3+4 2 3+4 2 |
| | 6 2 6 2 | 6 2 5 2 | 5+6 2 5+6 2 |
| | 6 2 6 2 | 5 2 6 2 | 5+6 2 3+4 2 |
| | 5 2 5 2 | 5 2 5 2 | 3+4 2 5+6 2 |
| | 6 2 6 2 | 6 2 5 2 | 3+4 2 3+4 2 |
| | 5 2 5 2 | 5 2 6 2 | 5+6 2 5+6 2 |
| | 6 2 6 2 | 6 2 6 2 | 3+4 2 3+4 2 |
| | 6 2 5 2 | 6 2 5 2 | 5+6 2 5+6 2 |
| | 5 2 5 2 | 5 2 6 2 | 3+4 2 5+6 2 |
| | 6 2 6 2 | 6 2 5 2 | 5+6 2 3+4 2 |
| | 5 2 5 2 | 5 2 5 2 | 3+4 2 3+4 2 |
| | 5 2 6 2 | 5 2 6 2 | 5+6 2 3+4 2 |
| | 6 2 5 2 | 6 2 6 2 | 3+4 2 5+6 2 |
| | 5 2 6 2 | 5 2 5 2 | 5+6 2 3+4 2 |
| | 6 2 5 2 | 6 2 5 2 | 3+4 2 5+6 2 |
| | 5 2 6 2 | 6 2 6 2 | 5+6 2 5+6 2 |
| | 6 2 5 2 | 5 2 5 2 | 3+4 2 3+4 2 |
| | 5 2 6 2 | 5 2 6 2 | 5+6 2 3+4 2 |
| | 6 2 6 2 | 6 2 6 2 | 3+4 2 5+6 2 |
| | 6 2 5 2 | 6 2 5 2 | 5+6 2 5+6 2 |
| | 5 2 5 2 | 5 2 6 2 | 3+4 2 3+4 2 |
| | 6 2 6 2 | 5 2 6 2 | 5+6 2 5+6 2 |
| | 5 2 6 2 | 6 2 5 2 | 3+4 2 3+4 2 |
| | 5 2 5 2 | 6 2 6 2 | 5+6 2 5+6 2 |
| | 6 2 6 2 | 5 2 5 2 | 3+4 2 3+4 2 |
| | 6 2 5 2 | 6 2 5 2 | 5+6 2 5+6 2 |
| | 6 2 5 2 | 5 2 5 2 | 5+6 2 5+6 2 |
| | 5 2 6 2 | 6 2 6 2 | 3+4 2 3+4 2 |
| | 5 2 6 2 | 6 2 5 2 | 5+6 2 3+4 2 |
| | 6 2 6 2 | 5 2 6 2 | 3+4 2 5+6 2 |
| | 6 2 5 2 | 6 2 5 2 | 3+4 2 5+6 2 |
| | 5 2 5 2 | 5 2 5 2 | 5+6 2 3+4 2 |
| | 6 2 6 2 | 6 2 6 2 | 3+4 2 3+4 2 |
| | 5 2 6 2 | 6 2 6 2 | 5+6 2 5+6 2 |

Continued on next column

**TABLE XL– continued**

| TQC | ICSD | NoE | SF |
|---|---|---|---|
| | 5 2 5 2 | 5 2 5 2 | 5+6 2 3+4 2 |
| | 6 2 5 2 | 5 2 6 2 | 3+4 2 3+4 2 |
| | 5 2 5 2 | 5 2 6 2 | 3+4 2 5+6 2 |
| | 5 2 6 2 | 5 2 6 2 | 5+6 2 3+4 2 |
| | 6 2 6 2 | 6 2 5 2 | 3+4 2 5+6 2 |
| | 5 2 5 2 | 5 2 5 2 | 5+6 2 3+4 2 |
| | 6 2 5 2 | 6 2 5 2 | 3+4 2 5+6 2 |
| | 6 2 6 2 | 6 2 6 2 | 5+6 2 3+4 2 |
| | 6 2 6 2 | 5 2 5 2 | 5+6 2 3+4 2 |
| | 5 2 5 2 | 6 2 5 2 | 3+4 2 5+6 2 |
| | 6 2 5 2 | 5 2 6 2 | 5+6 2 5+6 2 |
| | 5 2 6 2 | 6 2 6 2 | 3+4 2 5+6 2 |
| | 6 2 6 2 | 6 2 5 2 | 3+4 2 5+6 2 |
| | 5 2 5 2 | 5 2 5 2 | 5+6 2 3+4 2 |
| | 5 2 5 2 | 6 2 6 2 | 3+4 2 3+4 2 |
| | 6 2 6 2 | 5 2 6 2 | 3+4 2 3+4 2 |
| | 5 2 5 2 | 6 2 5 2 | 5+6 2 5+6 2 |
| | 5 2 5 2 | 5 2 6 2 | 5+6 2 5+6 2 |
| | 6 2 6 2 | 5 2 5 2 | 3+4 2 3+4 2 |
| | 6 2 5 2 | 6 2 5 2 | 5+6 2 5+6 2 |
| | 5 2 6 2 | 6 2 6 2 | 3+4 2 5+6 2 |
| | 6 2 6 2 | 5 2 5 2 | 3+4 2 3+4 2 |
| | 6 2 5 2 | 6 2 6 2 | 5+6 2 3+4 2 |
| | 5 2 6 2 | 5 2 6 2 | 5+6 2 5+6 2 |
| | 5 2 6 2 | 5 2 5 2 | 5+6 2 3+4 2 |
| | 5 2 5 2 | 6 2 6 2 | 5+6 2 5+6 2 |
| | 6 2 5 2 | 6 2 6 2 | 3+4 2 5+6 2 |
| | 5 2 6 2 | 6 2 5 2 | 3+4 2 5+6 2 |
| | 6 2 5 2 | 5 2 6 2 | 3+4 2 3+4 2 |
| | 5 2 5 2 | 6 2 5 2 | 3+4 2 5+6 2 |
| | 6 2 5 2 | 5 2 6 2 | 5+6 2 3+4 2 |
| | 5 2 6 2 | 6 2 6 2 | 5+6 2 3+4 2 |
| | 6 2 6 2 | 5 2 5 2 | 5+6 2 3+4 2 |
| | 5 2 5 2 | 5 2 6 2 | 3+4 2 5+6 2 |
| | 6 2 6 2 | 6 2 5 2 | 5+6 2 5+6 2 |
| | 5 2 5 2 | 5 2 5 2 | 3+4 2 5+6 2 |
| | 5 2 6 2 | 6 2 6 2 | 5+6 2 3+4 2 |
| | 5 2 6 2 | 6 2 6 2 | 3+4 2 3+4 2 |
| | 6 2 5 2 | 5 2 5 2 | 5+6 2 5+6 2 |
| | 5 2 6 2 | 5 2 5 2 | 5+6 2 3+4 2 |
| | 6 2 5 2 | 5 2 6 2 | 3+4 2 5+6 2 |
| | 6 2 6 2 | 6 2 6 2 | 3+4 2 5+6 2 |
| | 5 2 6 2 | 6 2 5 2 | 3+4 2 3+4 2 |
| | 5 2 5 2 | 5 2 5 2 | 5+6 2 5+6 2 |
| | 6 2 6 2 | 6 2 5 2 | 3+4 2 3+4 2 |
| | 5 2 6 2 | 5 2 5 2 | 5+6 2 5+6 2 |
| | 5 2 5 2 | 6 2 6 2 | 5+6 2 3+4 2 |
| | 5 2 6 2 | 6 2 5 2 | 5+6 2 5+6 2 |
| | 5 2 5 2 | 5 2 6 2 | 3+4 2 3+4 2 |
| | 6 2 5 2 | 5 2 5 2 | 3+4 2 5+6 2 |
| | 5 2 5 2 | 6 2 6 2 | 5+6 2 3+4 2 |
| | 6 2 6 2 | 5 2 6 2 | 5+6 2 3+4 2 |
| | 6 2 6 2 | 6 2 5 2 | 3+4 2 3+4 2 |
| | 6 2 5 2 | 5 2 6 2 | 5+6 2 5+6 2 |
| | 5 2 6 2 | 6 2 6 2 | 3+4 2 3+4 2 |
| | 5 2 6 2 | 5 2 5 2 | 3+4 2 5+6 2 |
| | 6 2 5 2 | 5 2 5 2 | 3+4 2 5+6 2 |
| | 6 2 5 2 | 5 2 6 2 | 5+6 2 3+4 2 |
| 65.2.10 | 104134 | 82 | Gd3Sn7 |
| | -GM -T | -Y -Z | -R -S |
| | 5 2 6 2 | 6 2 5 2 | 3+4 2 3+4 2 |
| | 6 2 5 2 | 5 2 6 2 | 5+6 2 5+6 2 |
| | 5 2 6 2 | 5 2 6 2 | 5+6 2 3+4 2 |
| | 5 2 6 2 | 6 2 5 2 | 3+4 2 3+4 2 |
| | 6 2 5 2 | 6 2 5 2 | 3+4 2 5+6 2 |
| | 6 2 5 2 | 5 2 6 2 | 5+6 2 5+6 2 |
| | 5 2 5 2 | 6 2 5 2 | 3+4 2 5+6 2 |
| | 6 2 6 2 | 6 2 5 2 | 3+4 2 3+4 2 |
| | 6 2 5 2 | 6 2 6 2 | 5+6 2 5+6 2 |
| | 6 2 6 2 | 5 2 5 2 | 3+4 2 5+6 2 |
| | 5 2 5 2 | 6 2 5 2 | 3+4 2 3+4 2 |
| | 6 2 5 2 | 5 2 6 2 | 5+6 2 5+6 2 |
| | 5 2 5 2 | 5 2 5 2 | 5+6 2 5+6 2 |
| | 6 2 6 2 | 6 2 6 2 | 3+4 2 3+4 2 |
| | 5 2 6 2 | 6 2 6 2 | 5+6 2 3+4 2 |
| | 5 2 5 2 | 5 2 5 2 | 3+4 2 5+6 2 |
| | 6 2 5 2 | 5 2 6 2 | 5+6 2 3+4 2 |
| | 5 2 5 2 | 5 2 5 2 | 3+4 2 5+6 2 |
| | 6 2 6 2 | 6 2 5 2 | 5+6 2 5+6 2 |
| | 6 2 6 2 | 6 2 6 2 | 3+4 2 3+4 2 |
| | 6 2 6 2 | 6 2 5 2 | 5+6 2 3+4 2 |
| | 5 2 5 2 | 5 2 6 2 | 3+4 2 5+6 2 |
| | 5 2 5 2 | 6 2 6 2 | 5+6 2 3+4 2 |
| | 6 2 6 2 | 5 2 6 2 | 3+4 2 5+6 2 |
| | 5 2 5 2 | 5 2 6 2 | 3+4 2 3+4 2 |
| | 5 2 6 2 | 6 2 5 2 | 5+6 2 3+4 2 |
| | 5 2 6 2 | 5 2 6 2 | 5+6 2 5+6 2 |
| | 6 2 5 2 | 5 2 5 2 | 3+4 2 5+6 2 |
| | 6 2 5 2 | 6 2 5 2 | 3+4 2 5+6 2 |
| | 6 2 5 2 | 6 2 6 2 | 5+6 2 3+4 2 |
| | 6 2 6 2 | 6 2 5 2 | 5+6 2 5+6 2 |
| | 5 2 5 2 | 6 2 5 2 | 3+4 2 3+4 2 |
| | 6 2 6 2 | 6 2 5 2 | 3+4 2 5+6 2 |
| | 6 2 5 2 | 5 2 5 2 | 5+6 2 5+6 2 |
| | 5 2 5 2 | 5 2 6 2 | 3+4 2 3+4 2 |
| | 5 2 6 2 | 6 2 6 2 | 3+4 2 5+6 2 |
| | 6 2 5 2 | 6 2 5 2 | 5+6 2 5+6 2 |
| | 6 2 5 2 | 5 2 5 2 | 3+4 2 3+4 2 |
| | 6 2 6 2 | 6 2 6 2 | 3+4 2 5+6 2 |
| | 5 2 5 2 | 6 2 5 2 | 5+6 2 3+4 2 |
| | 6 2 6 2 | 5 2 6 2 | 3+4 2 5+6 2 |
| 65.2.11 | 91064 | 138 | Fe4O11Sr4 |
| | -GM -T | -Y -Z | -R -S |
| | 5 2 6 2 | 5 2 6 2 | 3+4 2 5+6 2 |
| | 5 2 6 2 | 6 2 5 2 | 3+4 2 3+4 2 |
| | 6 2 5 2 | 5 2 6 2 | 5+6 2 5+6 2 |
| | 5 2 6 2 | 5 2 6 2 | 5+6 2 3+4 2 |
| | 5 2 5 2 | 6 2 5 2 | 5+6 2 5+6 2 |
| | 6 2 5 2 | 5 2 6 2 | 5+6 2 3+4 2 |
| | 5 2 6 2 | 6 2 5 2 | 3+4 2 5+6 2 |
| | 5 2 5 2 | 5 2 5 2 | 3+4 2 3+4 2 |
| | 6 2 5 2 | 6 2 6 2 | 5+6 2 5+6 2 |
| | 6 2 6 2 | 5 2 5 2 | 3+4 2 5+6 2 |
| | 6 2 6 2 | 5 2 5 2 | 5+6 2 3+4 2 |
| | 5 2 6 2 | 5 2 6 2 | 3+4 2 3+4 2 |
| | 5 2 5 2 | 6 2 5 2 | 5+6 2 5+6 2 |
| | 6 2 5 2 | 5 2 6 2 | 3+4 2 3+4 2 |
| | 5 2 6 2 | 6 2 6 2 | 5+6 2 5+6 2 |
| | 6 2 5 2 | 6 2 5 2 | 5+6 2 3+4 2 |
| | 6 2 6 2 | 6 2 6 2 | 3+4 2 3+4 2 |
| | 5 2 5 2 | 5 2 5 2 | 5+6 2 5+6 2 |
| | 6 2 5 2 | 6 2 5 2 | 3+4 2 5+6 2 |

Continued on next column

**TABLE XL– continued**

| TQC | ICSD | NoE | SF |
|---|---|---|---|
| | 6 2 5 2 | 6 2 5 2 | 5+6 2 3+4 2 |
| | 6 2 5 2 | 6 2 5 2 | 5+6 2 3+4 2 |
| | 5 2 6 2 | 6 2 6 2 | 3+4 2 3+4 2 |
| | 6 2 5 2 | 6 2 5 2 | 5+6 2 5+6 2 |
| | 6 2 6 2 | 5 2 6 2 | 3+4 2 3+4 2 |
| | 5 2 5 2 | 5 2 5 2 | 5+6 2 5+6 2 |
| | 6 2 5 2 | 6 2 5 2 | 3+4 2 5+6 2 |
| | 6 2 5 2 | 6 2 5 2 | 3+4 2 5+6 2 |
| | 5 2 5 2 | 6 2 6 2 | 5+6 2 3+4 2 |
| | 5 2 5 2 | 5 2 6 2 | 3+4 2 5+6 2 |
| | 6 2 6 2 | 6 2 5 2 | 3+4 2 3+4 2 |
| | 5 2 5 2 | 6 2 6 2 | 5+6 2 3+4 2 |
| | 5 2 6 2 | 6 2 6 2 | 3+4 2 5+6 2 |
| | 6 2 6 2 | 6 2 6 2 | 3+4 2 5+6 2 |
| | 6 2 5 2 | 5 2 6 2 | 3+4 2 3+4 2 |
| | 5 2 5 2 | 6 2 6 2 | 5+6 2 3+4 2 |
| | 5 2 6 2 | 6 2 5 2 | 3+4 2 5+6 2 |
| | 6 2 5 2 | 5 2 6 2 | 5+6 2 3+4 2 |
| | 5 2 6 2 | 6 2 5 2 | 5+6 2 5+6 2 |
| | 6 2 6 2 | 5 2 6 2 | 3+4 2 3+4 2 |
| | 6 2 6 2 | 5 2 5 2 | 5+6 2 5+6 2 |
| | 6 2 6 2 | 6 2 6 2 | 3+4 2 3+4 2 |
| | 6 2 6 2 | 5 2 5 2 | 5+6 2 3+4 2 |
| | 5 2 5 2 | 5 2 6 2 | 5+6 2 5+6 2 |
| | 6 2 5 2 | 6 2 5 2 | 3+4 2 3+4 2 |
| | 5 2 5 2 | 6 2 6 2 | 3+4 2 5+6 2 |
| | 5 2 5 2 | 6 2 6 2 | 3+4 2 3+4 2 |
| | 5 2 5 2 | 6 2 6 2 | 5+6 2 5+6 2 |
| | 5 2 6 2 | 5 2 5 2 | 5+6 2 5+6 2 |
| | 6 2 6 2 | 5 2 5 2 | 3+4 2 5+6 2 |
| | 6 2 6 2 | 6 2 6 2 | 3+4 2 3+4 2 |
| | 6 2 5 2 | 5 2 5 2 | 5+6 2 3+4 2 |
| | 6 2 5 2 | 5 2 5 2 | 5+6 2 3+4 2 |
| | 6 2 6 2 | 5 2 5 2 | 3+4 2 5+6 2 |
| | 6 2 6 2 | 6 2 5 2 | 3+4 2 3+4 2 |
| | 5 2 6 2 | 5 2 5 2 | 5+6 2 5+6 2 |
| | 5 2 5 2 | 5 2 5 2 | 5+6 2 3+4 2 |
| | 6 2 6 2 | 6 2 6 2 | 3+4 2 3+4 2 |
| | 6 2 6 2 | 6 2 6 2 | 3+4 2 5+6 2 |
| | 5 2 5 2 | 5 2 6 2 | 3+4 2 5+6 2 |
| | 6 2 6 2 | 6 2 5 2 | 5+6 2 3+4 2 |
| | 5 2 5 2 | 6 2 6 2 | 5+6 2 5+6 2 |
| | 5 2 5 2 | 6 2 6 2 | 5+6 2 3+4 2 |
| | 5 2 5 2 | 6 2 6 2 | 3+4 2 5+6 2 |
| | 5 2 5 2 | 5 2 5 2 | 5+6 2 5+6 2 |
| | 5 2 5 2 | 6 2 5 2 | 3+4 2 3+4 2 |
| | 5 2 6 2 | 5 2 5 2 | 5+6 2 5+6 2 |
| | 6 2 5 2 | 5 2 5 2 | 5+6 2 3+4 2 |
| | 5 2 6 2 | 5 2 6 2 | 3+4 2 5+6 2 |
| | 5 2 5 2 | 6 2 6 2 | 5+6 2 3+4 2 |
| 65.2.12 | 171189 | 88 | Er1Ni1Si3 |
| | -GM -T | -Y -Z | -R -S |
| | 5 2 6 2 | 6 2 5 2 | 3+4 2 3+4 2 |
| | 6 2 5 2 | 5 2 6 2 | 5+6 2 5+6 2 |
| | 5 2 6 2 | 6 2 5 2 | 3+4 2 3+4 2 |
| | 6 2 5 2 | 5 2 6 2 | 5+6 2 5+6 2 |
| | 5 2 6 2 | 6 2 5 2 | 3+4 2 3+4 2 |
| | 6 2 5 2 | 5 2 6 2 | 5+6 2 5+6 2 |
| | 5 2 6 2 | 6 2 5 2 | 3+4 2 3+4 2 |
| | 6 2 5 2 | 5 2 6 2 | 5+6 2 5+6 2 |
| | 5 2 6 2 | 5 2 6 2 | 3+4 2 3+4 2 |

Continued on next column

**TABLE XL– continued**

| TQC | ICSD | NoE | SF |
|---|---|---|---|
| | 5 2 5 2 | 6 2 5 2 | 3+4 2 5+6 2 |
| | 6 2 5 2 | 6 2 6 2 | 5+6 2 5+6 2 |
| | 6 2 5 2 | 5 2 5 2 | 3+4 2 3+4 2 |
| | 5 2 6 2 | 5 2 6 2 | 5+6 2 5+6 2 |
| | 5 2 6 2 | 5 2 5 2 | 5+6 2 3+4 2 |
| | 6 2 6 2 | 6 2 5 2 | 5+6 2 3+4 2 |
| | 6 2 6 2 | 6 2 6 2 | 5+6 2 5+6 2 |
| | 6 2 5 2 | 6 2 5 2 | 3+4 2 3+4 2 |
| | 5 2 5 2 | 5 2 6 2 | 3+4 2 5+6 2 |
| | 5 2 5 2 | 5 2 6 2 | 5+6 2 3+4 2 |
| | 5 2 6 2 | 6 2 6 2 | 3+4 2 3+4 2 |
| | 5 2 6 2 | 5 2 6 2 | 3+4 2 5+6 2 |
| | 6 2 5 2 | 6 2 5 2 | 5+6 2 5+6 2 |
| | 5 2 5 2 | 6 2 5 2 | 3+4 2 5+6 2 |
| | 6 2 6 2 | 5 2 5 2 | 5+6 2 3+4 2 |
| | 5 2 5 2 | 6 2 6 2 | 3+4 2 5+6 2 |
| | 6 2 5 2 | 5 2 6 2 | 3+4 2 3+4 2 |
| | 6 2 6 2 | 6 2 5 2 | 5+6 2 5+6 2 |
| | 6 2 5 2 | 5 2 5 2 | 3+4 2 5+6 2 |
| | 5 2 6 2 | 6 2 6 2 | 5+6 2 3+4 2 |
| | 6 2 5 2 | 5 2 6 2 | 5+6 2 3+4 2 |
| | 5 2 6 2 | 5 2 5 2 | 3+4 2 5+6 2 |
| | 5 2 6 2 | 6 2 6 2 | 5+6 2 3+4 2 |
| | 6 2 5 2 | 6 2 5 2 | 3+4 2 5+6 2 |
| | 6 2 6 2 | 5 2 5 2 | 3+4 2 5+6 2 |
| | 5 2 5 2 | 6 2 5 2 | 5+6 2 3+4 2 |
| | 5 2 5 2 | 5 2 6 2 | 3+4 2 3+4 2 |
| | 6 2 6 2 | 5 2 6 2 | 5+6 2 5+6 2 |
| | 5 2 5 2 | 5 2 5 2 | 5+6 2 3+4 2 |
| | 6 2 6 2 | 6 2 6 2 | 3+4 2 3+4 2 |
| | 5 2 5 2 | 5 2 5 2 | 3+4 2 5+6 2 |
| | 5 2 6 2 | 6 2 6 2 | 5+6 2 5+6 2 |
| | 6 2 6 2 | 6 2 5 2 | 3+4 2 3+4 2 |
| | 5 2 5 2 | 5 2 6 2 | 5+6 2 5+6 2 |
| | 6 2 6 2 | 6 2 5 2 | 3+4 2 3+4 2 |
| 65.2.13 | 154835 | 120 | Dy1Fe4Ge2 |
| | -GM -T | -Y -Z | -R -S |
| | 5 2 6 2 | 5 2 6 2 | 5+6 2 3+4 2 |
| | 5 2 5 2 | 5 2 5 2 | 5+6 2 5+6 2 |
| | 6 2 5 2 | 6 2 5 2 | 3+4 2 5+6 2 |
| | 6 2 6 2 | 6 2 6 2 | 3+4 2 3+4 2 |
| | 6 2 5 2 | 6 2 5 2 | 3+4 2 5+6 2 |
| | 6 2 6 2 | 6 2 6 2 | 3+4 2 3+4 2 |
| | 6 2 5 2 | 6 2 5 2 | 3+4 2 5+6 2 |
| | 6 2 6 2 | 6 2 6 2 | 3+4 2 3+4 2 |
| | 5 2 6 2 | 5 2 5 2 | 5+6 2 5+6 2 |
| | 6 2 5 2 | 5 2 6 2 | 5+6 2 3+4 2 |
| | 6 2 5 2 | 6 2 5 2 | 3+4 2 3+4 2 |
| | 5 2 6 2 | 6 2 6 2 | 3+4 2 5+6 2 |
| | 5 2 5 2 | 6 2 5 2 | 5+6 2 5+6 2 |
| | 5 2 6 2 | 6 2 6 2 | 5+6 2 3+4 2 |
| | 6 2 6 2 | 5 2 5 2 | 3+4 2 3+4 2 |
| | 6 2 5 2 | 5 2 6 2 | 3+4 2 5+6 2 |
| | 5 2 6 2 | 5 2 5 2 | 5+6 2 3+4 2 |
| | 5 2 5 2 | 5 2 6 2 | 5+6 2 5+6 2 |
| | 6 2 6 2 | 5 2 5 2 | 3+4 2 3+4 2 |
| | 6 2 5 2 | 5 2 6 2 | 3+4 2 5+6 2 |
| | 5 2 6 2 | 6 2 5 2 | 5+6 2 5+6 2 |
| | 5 2 5 2 | 6 2 6 2 | 5+6 2 3+4 2 |
| | 5 2 6 2 | 6 2 6 2 | 3+4 2 5+6 2 |
| | 6 2 5 2 | 6 2 5 2 | 3+4 2 3+4 2 |

Continued on next column

**TABLE XL– continued**

| TQC | ICSD | NoE | SF |
|---|---|---|---|
| | 6 2 5 2 | 5 2 6 2 | 5+6 2 5+6 2 |
| | 6 2 6 2 | 5 2 5 2 | 5+6 2 3+4 2 |
| | 5 2 5 2 | 6 2 5 2 | 3+4 2 5+6 2 |
| | 6 2 6 2 | 6 2 6 2 | 3+4 2 3+4 2 |
| | 5 2 5 2 | 5 2 6 2 | 5+6 2 5+6 2 |
| | 6 2 6 2 | 5 2 5 2 | 5+6 2 3+4 2 |
| | 6 2 5 2 | 5 2 5 2 | 5+6 2 3+4 2 |
| | 5 2 6 2 | 5 2 6 2 | 3+4 2 5+6 2 |
| | 6 2 6 2 | 6 2 6 2 | 3+4 2 5+6 2 |
| | 6 2 5 2 | 6 2 5 2 | 5+6 2 3+4 2 |
| | 5 2 5 2 | 5 2 6 2 | 3+4 2 3+4 2 |
| | 5 2 6 2 | 6 2 5 2 | 3+4 2 5+6 2 |
| | 6 2 6 2 | 6 2 5 2 | 5+6 2 3+4 2 |
| | 5 2 5 2 | 6 2 6 2 | 5+6 2 5+6 2 |
| | 5 2 6 2 | 5 2 6 2 | 3+4 2 3+4 2 |
| | 6 2 5 2 | 6 2 6 2 | 3+4 2 5+6 2 |
| | 6 2 5 2 | 5 2 5 2 | 3+4 2 3+4 2 |
| | 6 2 6 2 | 6 2 5 2 | 3+4 2 5+6 2 |
| | 6 2 5 2 | 5 2 5 2 | 5+6 2 3+4 2 |
| | 6 2 6 2 | 6 2 6 2 | 3+4 2 5+6 2 |
| | 6 2 5 2 | 6 2 6 2 | 3+4 2 5+6 2 |
| | 5 2 6 2 | 6 2 5 2 | 3+4 2 3+4 2 |
| | 6 2 5 2 | 6 2 5 2 | 3+4 2 5+6 2 |
| | 6 2 6 2 | 6 2 6 2 | 5+6 2 3+4 2 |
| | 5 2 6 2 | 6 2 5 2 | 5+6 2 3+4 2 |
| | 5 2 6 2 | 6 2 6 2 | 3+4 2 5+6 2 |
| | 6 2 5 2 | 6 2 5 2 | 3+4 2 3+4 2 |
| | 6 2 5 2 | 6 2 6 2 | 5+6 2 5+6 2 |
| | 5 2 5 2 | 6 2 5 2 | 3+4 2 5+6 2 |
| | 5 2 6 2 | 6 2 6 2 | 3+4 2 3+4 2 |
| | 6 2 5 2 | 6 2 6 2 | 5+6 2 3+4 2 |
| | 5 2 6 2 | 5 2 5 2 | 5+6 2 5+6 2 |
| | 5 2 5 2 | 6 2 6 2 | 3+4 2 3+4 2 |
| | 6 2 6 2 | 6 2 5 2 | 3+4 2 5+6 2 |
| | 5 2 5 2 | 6 2 6 2 | 3+4 2 3+4 2 |
| | 6 2 6 2 | 6 2 5 2 | 3+4 2 5+6 2 |
| 65.2.14 | 419475 | 120 | Cu1Mg4Y1 |
| | -GM -T | -Y -Z | -R -S |
| | 5 2 5 2 | 5 2 5 2 | 3+4 2 3+4 2 |
| | 6 2 6 2 | 6 2 6 2 | 5+6 2 5+6 2 |
| | 5 2 5 2 | 5 2 5 2 | 3+4 2 3+4 2 |
| | 6 2 6 2 | 6 2 6 2 | 5+6 2 5+6 2 |
| | 6 2 6 2 | 6 2 6 2 | 3+4 2 3+4 2 |
| | 5 2 5 2 | 5 2 5 2 | 5+6 2 5+6 2 |
| | 5 2 5 2 | 5 2 5 2 | 5+6 2 5+6 2 |
| | 6 2 6 2 | 6 2 6 2 | 3+4 2 3+4 2 |
| | 6 2 5 2 | 6 2 5 2 | 5+6 2 3+4 2 |
| | 5 2 6 2 | 5 2 6 2 | 3+4 2 5+6 2 |
| | 5 2 6 2 | 5 2 6 2 | 3+4 2 5+6 2 |
| | 6 2 5 2 | 6 2 5 2 | 5+6 2 3+4 2 |
| | 5 2 6 2 | 5 2 6 2 | 3+4 2 3+4 2 |
| | 6 2 5 2 | 6 2 5 2 | 5+6 2 5+6 2 |
| | 6 2 5 2 | 6 2 5 2 | 5+6 2 3+4 2 |
| | 5 2 6 2 | 5 2 6 2 | 3+4 2 5+6 2 |
| | 5 2 5 2 | 6 2 6 2 | 5+6 2 3+4 2 |
| | 5 2 5 2 | 5 2 5 2 | 3+4 2 5+6 2 |
| | 6 2 6 2 | 6 2 6 2 | 3+4 2 3+4 2 |
| | 6 2 6 2 | 5 2 5 2 | 5+6 2 5+6 2 |
| | 5 2 5 2 | 5 2 5 2 | 3+4 2 3+4 2 |
| | 6 2 6 2 | 6 2 5 2 | 5+6 2 5+6 2 |
| | 5 2 6 2 | 5 2 6 2 | 3+4 2 3+4 2 |

Continued on next column

**TABLE XL– continued**

| TQC | ICSD | NoE | SF |
|---|---|---|---|
| | 5 2 6 2 | 6 2 6 2 | 5+6 2 5+6 2 |
| | 6 2 5 2 | 5 2 5 2 | 3+4 2 5+6 2 |
| | 6 2 6 2 | 6 2 6 2 | 5+6 2 3+4 2 |
| | 5 2 5 2 | 5 2 6 2 | 3+4 2 5+6 2 |
| | 6 2 6 2 | 6 2 5 2 | 5+6 2 3+4 2 |
| | 5 2 5 2 | 6 2 6 2 | 3+4 2 5+6 2 |
| | 5 2 6 2 | 6 2 5 2 | 3+4 2 3+4 2 |
| | 5 2 5 2 | 5 2 6 2 | 5+6 2 5+6 2 |
| | 6 2 5 2 | 5 2 5 2 | 5+6 2 3+4 2 |
| | 5 2 6 2 | 5 2 5 2 | 5+6 2 5+6 2 |
| | 6 2 5 2 | 6 2 6 2 | 3+4 2 3+4 2 |
| | 6 2 5 2 | 6 2 6 2 | 3+4 2 5+6 2 |
| | 5 2 6 2 | 5 2 5 2 | 5+6 2 3+4 2 |
| | 6 2 6 2 | 6 2 6 2 | 3+4 2 3+4 2 |
| | 5 2 5 2 | 5 2 5 2 | 5+6 2 5+6 2 |
| | 5 2 6 2 | 6 2 6 2 | 5+6 2 5+6 2 |
| | 6 2 5 2 | 5 2 5 2 | 3+4 2 5+6 2 |
| | 6 2 6 2 | 5 2 5 2 | 5+6 2 3+4 2 |
| | 5 2 5 2 | 6 2 6 2 | 3+4 2 3+4 2 |
| | 5 2 5 2 | 5 2 6 2 | 5+6 2 3+4 2 |
| | 6 2 6 2 | 6 2 5 2 | 5+6 2 5+6 2 |
| | 6 2 6 2 | 5 2 5 2 | 3+4 2 5+6 2 |
| | 6 2 5 2 | 6 2 6 2 | 3+4 2 3+4 2 |
| | 6 2 6 2 | 6 2 5 2 | 3+4 2 3+4 2 |
| | 5 2 5 2 | 5 2 6 2 | 5+6 2 5+6 2 |
| | 5 2 6 2 | 6 2 6 2 | 3+4 2 5+6 2 |
| | 6 2 5 2 | 5 2 5 2 | 5+6 2 3+4 2 |
| | 5 2 5 2 | 5 2 5 2 | 3+4 2 5+6 2 |
| | 6 2 6 2 | 5 2 6 2 | 3+4 2 3+4 2 |
| | 5 2 5 2 | 6 2 6 2 | 5+6 2 5+6 2 |
| | 5 2 5 2 | 6 2 5 2 | 5+6 2 3+4 2 |
| | 6 2 6 2 | 6 2 6 2 | 3+4 2 5+6 2 |
| | 5 2 6 2 | 5 2 5 2 | 3+4 2 3+4 2 |
| | 6 2 6 2 | 6 2 6 2 | 5+6 2 5+6 2 |
| | 5 2 5 2 | 5 2 6 2 | 5+6 2 3+4 2 |
| | 6 2 5 2 | 5 2 6 2 | 5+6 2 5+6 2 |
| | 5 2 5 2 | 6 2 5 2 | 5+6 2 3+4 2 |
| 65.2.15 | 418093 | 88 | Co2Ge7U3 |
| | -GM -T | -Y -Z | -R -S |
| | 5 2 6 2 | 6 2 5 2 | 3+4 2 3+4 2 |
| | 6 2 5 2 | 5 2 6 2 | 5+6 2 5+6 2 |
| | 5 2 6 2 | 5 2 6 2 | 3+4 2 5+6 2 |
| | 5 2 6 2 | 6 2 5 2 | 5+6 2 3+4 2 |
| | 6 2 5 2 | 6 2 5 2 | 3+4 2 3+4 2 |
| | 6 2 5 2 | 5 2 6 2 | 5+6 2 5+6 2 |
| | 6 2 5 2 | 6 2 5 2 | 5+6 2 3+4 2 |
| | 5 2 5 2 | 6 2 6 2 | 5+6 2 3+4 2 |
| | 6 2 6 2 | 6 2 5 2 | 3+4 2 5+6 2 |
| | 6 2 5 2 | 5 2 5 2 | 5+6 2 3+4 2 |
| | 5 2 6 2 | 6 2 5 2 | 3+4 2 5+6 2 |
| | 6 2 5 2 | 5 2 6 2 | 5+6 2 3+4 2 |
| | 5 2 5 2 | 5 2 6 2 | 3+4 2 5+6 2 |
| | 6 2 6 2 | 6 2 6 2 | 3+4 2 5+6 2 |
| | 5 2 6 2 | 6 2 5 2 | 5+6 2 3+4 2 |
| | 5 2 5 2 | 5 2 6 2 | 3+4 2 3+4 2 |
| | 6 2 5 2 | 5 2 5 2 | 3+4 2 3+4 2 |
| | 5 2 5 2 | 5 2 5 2 | 5+6 2 5+6 2 |
| | 6 2 6 2 | 6 2 5 2 | 5+6 2 3+4 2 |
| | 5 2 6 2 | 6 2 6 2 | 3+4 2 5+6 2 |
| | 5 2 6 2 | 5 2 5 2 | 5+6 2 5+6 2 |
| | 6 2 5 2 | 5 2 6 2 | 5+6 2 3+4 2 |

Continued on next column

**TABLE XL– continued**

| TQC | ICSD | NoE | SF |
|---|---|---|---|
| | 6 2 5 2 | 6 2 6 2 | 3+4 2 5+6 2 |
| | 6 2 6 2 | 6 2 6 2 | 3+4 2 5+6 2 |
| | 5 2 6 2 | 5 2 6 2 | 5+6 2 5+6 2 |
| | 5 2 6 2 | 5 2 6 2 | 5+6 2 3+4 2 |
| | 5 2 5 2 | 6 2 5 2 | 3+4 2 3+4 2 |
| | 5 2 5 2 | 6 2 5 2 | 3+4 2 3+4 2 |
| | 6 2 5 2 | 5 2 5 2 | 5+6 2 5+6 2 |
| | 6 2 5 2 | 6 2 6 2 | 5+6 2 3+4 2 |
| | 5 2 6 2 | 5 2 6 2 | 3+4 2 5+6 2 |
| | 6 2 5 2 | 6 2 5 2 | 5+6 2 3+4 2 |
| | 6 2 6 2 | 6 2 6 2 | 3+4 2 3+4 2 |
| | 6 2 6 2 | 5 2 6 2 | 3+4 2 5+6 2 |
| | 5 2 6 2 | 6 2 5 2 | 5+6 2 5+6 2 |
| | 6 2 6 2 | 5 2 5 2 | 5+6 2 5+6 2 |
| | 6 2 5 2 | 5 2 5 2 | 5+6 2 3+4 2 |
| | 5 2 5 2 | 6 2 6 2 | 3+4 2 3+4 2 |
| | 6 2 5 2 | 6 2 5 2 | 5+6 2 5+6 2 |
| | 6 2 6 2 | 6 2 6 2 | 3+4 2 3+4 2 |
| | 5 2 5 2 | 6 2 6 2 | 3+4 2 5+6 2 |
| | 5 2 5 2 | 5 2 5 2 | 5+6 2 3+4 2 |
| | 6 2 5 2 | 6 2 5 2 | 5+6 2 3+4 2 |
| | 6 2 6 2 | 5 2 5 2 | 5+6 2 5+6 2 |
| 65.2.16 | 72872 | 40 | Co1O4Re1 |
| | -GM -T | -Y -Z | -R -S |
| | 5 2 6 2 | 5 2 6 2 | 3+4 2 5+6 2 |
| | 5 2 6 2 | 5 2 6 2 | 3+4 2 5+6 2 |
| | 6 2 5 2 | 6 2 5 2 | 5+6 2 3+4 2 |
| | 6 2 5 2 | 6 2 5 2 | 5+6 2 3+4 2 |
| | 5 2 6 2 | 5 2 6 2 | 3+4 2 5+6 2 |
| | 5 2 6 2 | 5 2 6 2 | 3+4 2 5+6 2 |
| | 5 2 6 2 | 5 2 6 2 | 3+4 2 5+6 2 |
| | 5 2 5 2 | 5 2 6 2 | 3+4 2 3+4 2 |
| | 5 2 5 2 | 5 2 5 2 | 5+6 2 5+6 2 |
| | 6 2 6 2 | 6 2 5 2 | 5+6 2 3+4 2 |
| | 6 2 5 2 | 6 2 6 2 | 3+4 2 3+4 2 |
| | 6 2 5 2 | 6 2 5 2 | 5+6 2 3+4 2 |
| | 6 2 6 2 | 5 2 5 2 | 5+6 2 5+6 2 |
| | 6 2 5 2 | 6 2 5 2 | 3+4 2 3+4 2 |
| | 6 2 5 2 | 6 2 5 2 | 5+6 2 3+4 2 |
| | 5 2 6 2 | 6 2 6 2 | 5+6 2 5+6 2 |
| | 5 2 5 2 | 5 2 5 2 | 3+4 2 3+4 2 |
| | 5 2 5 2 | 5 2 5 2 | 3+4 2 5+6 2 |
| | 5 2 5 2 | 5 2 5 2 | 3+4 2 3+4 2 |
| | 5 2 6 2 | 5 2 5 2 | 3+4 2 3+4 2 |
| 65.2.17 | 16497 | 88 | Ce3Ni2Si8 |
| | -GM -T | -Y -Z | -R -S |
| | 5 2 5 2 | 5 2 5 2 | 3+4 2 3+4 2 |
| | 5 2 5 2 | 6 2 6 2 | 3+4 2 3+4 2 |
| | 6 2 6 2 | 5 2 5 2 | 5+6 2 5+6 2 |
| | 6 2 6 2 | 6 2 6 2 | 5+6 2 5+6 2 |
| | 5 2 5 2 | 6 2 6 2 | 3+4 2 3+4 2 |
| | 6 2 6 2 | 5 2 5 2 | 5+6 2 5+6 2 |
| | 6 2 6 2 | 6 2 6 2 | 5+6 2 5+6 2 |
| | 5 2 6 2 | 6 2 6 2 | 5+6 2 5+6 2 |
| | 6 2 5 2 | 6 2 5 2 | 3+4 2 3+4 2 |
| | 6 2 6 2 | 5 2 6 2 | 5+6 2 5+6 2 |
| | 5 2 5 2 | 6 2 6 2 | 3+4 2 3+4 2 |
| | 6 2 6 2 | 5 2 5 2 | 5+6 2 5+6 2 |
| | 5 2 6 2 | 5 2 6 2 | 3+4 2 5+6 2 |
| | 6 2 5 2 | 6 2 5 2 | 5+6 2 3+4 2 |
| | 5 2 5 2 | 6 2 5 2 | 3+4 2 3+4 2 |

Continued on next column

**TABLE XL– continued**

| TQC | ICSD | NoE | SF |
|---|---|---|---|
| | 5 2 6 2 | 5 2 6 2 | 3+4 2 5+6 2 |
| | 6 2 6 2 | 5 2 5 2 | 5+6 2 5+6 2 |
| | 5 2 6 2 | 5 2 6 2 | 5+6 2 3+4 2 |
| | 6 2 5 2 | 6 2 6 2 | 3+4 2 5+6 2 |
| | 5 2 5 2 | 6 2 5 2 | 5+6 2 3+4 2 |
| | 6 2 5 2 | 5 2 5 2 | 3+4 2 3+4 2 |
| | 5 2 5 2 | 6 2 5 2 | 5+6 2 3+4 2 |
| | 6 2 6 2 | 5 2 6 2 | 3+4 2 5+6 2 |
| | 6 2 6 2 | 6 2 6 2 | 3+4 2 5+6 2 |
| | 5 2 5 2 | 5 2 6 2 | 5+6 2 5+6 2 |
| | 5 2 6 2 | 5 2 5 2 | 3+4 2 3+4 2 |
| | 5 2 5 2 | 6 2 5 2 | 5+6 2 3+4 2 |
| | 5 2 6 2 | 6 2 6 2 | 3+4 2 5+6 2 |
| | 6 2 6 2 | 6 2 5 2 | 5+6 2 3+4 2 |
| | 5 2 5 2 | 5 2 5 2 | 5+6 2 5+6 2 |
| | 6 2 6 2 | 6 2 6 2 | 3+4 2 3+4 2 |
| | 6 2 6 2 | 6 2 5 2 | 5+6 2 5+6 2 |
| | 6 2 5 2 | 5 2 5 2 | 3+4 2 3+4 2 |
| | 6 2 6 2 | 5 2 6 2 | 5+6 2 5+6 2 |
| | 5 2 5 2 | 6 2 5 2 | 3+4 2 3+4 2 |
| | 5 2 5 2 | 5 2 6 2 | 5+6 2 3+4 2 |
| | 6 2 6 2 | 5 2 6 2 | 3+4 2 5+6 2 |
| | 5 2 5 2 | 6 2 5 2 | 3+4 2 5+6 2 |
| | 6 2 5 2 | 5 2 6 2 | 5+6 2 3+4 2 |
| | 6 2 6 2 | 6 2 6 2 | 5+6 2 5+6 2 |
| | 6 2 6 2 | 6 2 6 2 | 3+4 2 5+6 2 |
| | 6 2 6 2 | 5 2 5 2 | 5+6 2 3+4 2 |
| | 5 2 6 2 | 6 2 5 2 | 5+6 2 5+6 2 |
| | 5 2 5 2 | 5 2 6 2 | 3+4 2 3+4 2 |
| 65.2.18 | 174528 | 52 | Ce2Si7 |
| | -GM -T | -Y -Z | -R -S |
| | 5 2 6 2 | 6 2 5 2 | 5+6 2 5+6 2 |
| | 6 2 5 2 | 5 2 6 2 | 3+4 2 3+4 2 |
| | 5 2 6 2 | 6 2 5 2 | 5+6 2 5+6 2 |
| | 6 2 5 2 | 5 2 6 2 | 3+4 2 3+4 2 |
| | 5 2 6 2 | 6 2 5 2 | 3+4 2 5+6 2 |
| | 6 2 5 2 | 5 2 6 2 | 5+6 2 3+4 2 |
| | 5 2 6 2 | 6 2 5 2 | 5+6 2 5+6 2 |
| | 6 2 5 2 | 5 2 6 2 | 3+4 2 3+4 2 |
| | 5 2 6 2 | 5 2 6 2 | 5+6 2 3+4 2 |
| | 6 2 5 2 | 6 2 5 2 | 3+4 2 5+6 2 |
| | 5 2 5 2 | 6 2 5 2 | 3+4 2 5+6 2 |
| | 5 2 5 2 | 5 2 6 2 | 3+4 2 3+4 2 |
| | 6 2 6 2 | 6 2 6 2 | 5+6 2 5+6 2 |
| | 5 2 6 2 | 5 2 6 2 | 5+6 2 3+4 2 |
| | 6 2 6 2 | 6 2 5 2 | 3+4 2 5+6 2 |
| | 6 2 5 2 | 5 2 6 2 | 5+6 2 3+4 2 |
| | 5 2 5 2 | 5 2 5 2 | 5+6 2 5+6 2 |
| | 6 2 6 2 | 6 2 6 2 | 3+4 2 3+4 2 |
| | 5 2 5 2 | 6 2 5 2 | 3+4 2 3+4 2 |
| | 6 2 5 2 | 5 2 5 2 | 5+6 2 3+4 2 |
| | 5 2 6 2 | 6 2 5 2 | 3+4 2 5+6 2 |
| | 6 2 6 2 | 5 2 5 2 | 5+6 2 3+4 2 |
| | 5 2 5 2 | 5 2 6 2 | 3+4 2 5+6 2 |
| | 6 2 5 2 | 6 2 6 2 | 5+6 2 5+6 2 |
| | 6 2 5 2 | 6 2 5 2 | 3+4 2 3+4 2 |
| | 5 2 6 2 | 6 2 5 2 | 5+6 2 3+4 2 |
| 65.2.19 | 39221 | 234 | C4Er4Ni13 |
| | -GM -T | -Y -Z | -R -S |
| | 5 2 6 2 | 5 2 6 2 | 5+6 2 3+4 2 |
| | 6 2 5 2 | 6 2 5 2 | 3+4 2 5+6 2 |

Continued on next column

**TABLE XL– continued**

| TQC | ICSD | NoE | SF |
|---|---|---|---|
| | 5 2 5 2 | 5 2 5 2 | 3+4 2 3+4 2 |
| | 6 2 6 2 | 6 2 6 2 | 5+6 2 5+6 2 |
| | 5 2 6 2 | 5 2 6 2 | 5+6 2 3+4 2 |
| | 6 2 5 2 | 6 2 5 2 | 3+4 2 5+6 2 |
| | 5 2 5 2 | 5 2 6 2 | 3+4 2 5+6 2 |
| | 6 2 6 2 | 6 2 5 2 | 5+6 2 3+4 2 |
| | 5 2 5 2 | 5 2 5 2 | 3+4 2 3+4 2 |
| | 6 2 6 2 | 6 2 6 2 | 5+6 2 5+6 2 |
| | 6 2 6 2 | 6 2 6 2 | 5+6 2 5+6 2 |
| | 5 2 5 2 | 5 2 5 2 | 3+4 2 3+4 2 |
| | 5 2 6 2 | 5 2 6 2 | 3+4 2 5+6 2 |
| | 6 2 5 2 | 6 2 5 2 | 5+6 2 3+4 2 |
| | 5 2 6 2 | 5 2 6 2 | 3+4 2 5+6 2 |
| | 6 2 5 2 | 6 2 5 2 | 5+6 2 3+4 2 |
| | 5 2 5 2 | 5 2 5 2 | 3+4 2 3+4 2 |
| | 5 2 6 2 | 6 2 5 2 | 5+6 2 5+6 2 |
| | 5 2 6 2 | 6 2 5 2 | 3+4 2 3+4 2 |
| | 6 2 6 2 | 6 2 6 2 | 5+6 2 5+6 2 |
| | 6 2 5 2 | 5 2 6 2 | 3+4 2 5+6 2 |
| | 5 2 5 2 | 5 2 6 2 | 5+6 2 3+4 2 |
| | 6 2 5 2 | 6 2 6 2 | 3+4 2 5+6 2 |
| | 5 2 5 2 | 5 2 6 2 | 5+6 2 3+4 2 |
| | 6 2 6 2 | 5 2 6 2 | 5+6 2 5+6 2 |
| | 6 2 5 2 | 6 2 6 2 | 3+4 2 5+6 2 |
| | 6 2 5 2 | 5 2 5 2 | 5+6 2 5+6 2 |
| | 6 2 6 2 | 6 2 6 2 | 5+6 2 3+4 2 |
| | 6 2 5 2 | 6 2 6 2 | 3+4 2 3+4 2 |
| | 6 2 6 2 | 5 2 5 2 | 3+4 2 3+4 2 |
| | 5 2 5 2 | 5 2 6 2 | 5+6 2 5+6 2 |
| | 6 2 5 2 | 5 2 5 2 | 5+6 2 5+6 2 |
| | 6 2 6 2 | 5 2 5 2 | 5+6 2 3+4 2 |
| | 5 2 5 2 | 6 2 5 2 | 3+4 2 5+6 2 |
| | 5 2 6 2 | 5 2 5 2 | 5+6 2 5+6 2 |
| | 5 2 5 2 | 6 2 5 2 | 5+6 2 3+4 2 |
| | 5 2 5 2 | 5 2 6 2 | 3+4 2 3+4 2 |
| | 5 2 6 2 | 6 2 6 2 | 3+4 2 5+6 2 |
| | 5 2 5 2 | 6 2 5 2 | 3+4 2 5+6 2 |
| | 5 2 5 2 | 6 2 5 2 | 3+4 2 3+4 2 |
| | 6 2 6 2 | 6 2 6 2 | 5+6 2 5+6 2 |
| | 5 2 6 2 | 5 2 6 2 | 3+4 2 3+4 2 |
| | 5 2 5 2 | 5 2 5 2 | 5+6 2 3+4 2 |
| | 6 2 6 2 | 6 2 5 2 | 5+6 2 5+6 2 |
| | 5 2 5 2 | 5 2 6 2 | 3+4 2 3+4 2 |
| | 5 2 6 2 | 6 2 6 2 | 5+6 2 5+6 2 |
| | 5 2 5 2 | 6 2 6 2 | 5+6 2 5+6 2 |
| | 5 2 5 2 | 6 2 5 2 | 3+4 2 5+6 2 |
| | 6 2 5 2 | 5 2 6 2 | 5+6 2 3+4 2 |
| | 5 2 6 2 | 5 2 5 2 | 3+4 2 5+6 2 |
| | 5 2 6 2 | 5 2 6 2 | 3+4 2 3+4 2 |
| | 6 2 6 2 | 6 2 5 2 | 3+4 2 5+6 2 |
| | 6 2 5 2 | 6 2 6 2 | 5+6 2 3+4 2 |
| | 6 2 5 2 | 6 2 6 2 | 3+4 2 3+4 2 |
| | 6 2 5 2 | 5 2 6 2 | 5+6 2 5+6 2 |
| | 6 2 6 2 | 5 2 5 2 | 3+4 2 5+6 2 |
| | 6 2 5 2 | 5 2 6 2 | 5+6 2 5+6 2 |
| | 5 2 5 2 | 5 2 5 2 | 5+6 2 3+4 2 |
| | 6 2 6 2 | 5 2 6 2 | 3+4 2 3+4 2 |
| | 5 2 5 2 | 5 2 6 2 | 5+6 2 3+4 2 |
| | 6 2 5 2 | 5 2 5 2 | 3+4 2 5+6 2 |
| | 5 2 5 2 | 6 2 5 2 | 5+6 2 3+4 2 |
| | 6 2 6 2 | 5 2 5 2 | 3+4 2 3+4 2 |

Continued on next column

**TABLE XL– continued**

| TQC | ICSD | NoE | SF |
|---|---|---|---|
| | 5 2 5 2 | 6 2 6 2 | 3+4 2 5+6 2 |
| | 6 2 6 2 | 6 2 5 2 | 5+6 2 5+6 2 |
| | 5 2 5 2 | 6 2 6 2 | 5+6 2 5+6 2 |
| | 5 2 5 2 | 5 2 6 2 | 3+4 2 3+4 2 |
| | 5 2 6 2 | 6 2 5 2 | 5+6 2 3+4 2 |
| | 5 2 6 2 | 6 2 6 2 | 3+4 2 3+4 2 |
| | 6 2 6 2 | 6 2 5 2 | 5+6 2 3+4 2 |
| | 5 2 5 2 | 6 2 6 2 | 5+6 2 5+6 2 |
| | 5 2 6 2 | 6 2 6 2 | 3+4 2 3+4 2 |
| | 6 2 5 2 | 5 2 5 2 | 5+6 2 5+6 2 |
| | 6 2 5 2 | 6 2 6 2 | 3+4 2 5+6 2 |
| | 5 2 5 2 | 5 2 5 2 | 3+4 2 3+4 2 |
| | 5 2 6 2 | 5 2 5 2 | 5+6 2 5+6 2 |
| | 5 2 6 2 | 6 2 6 2 | 5+6 2 3+4 2 |
| | 6 2 5 2 | 6 2 5 2 | 5+6 2 5+6 2 |
| | 5 2 6 2 | 5 2 6 2 | 5+6 2 5+6 2 |
| | 5 2 5 2 | 6 2 5 2 | 3+4 2 3+4 2 |
| | 6 2 5 2 | 6 2 6 2 | 5+6 2 3+4 2 |
| | 5 2 6 2 | 6 2 5 2 | 3+4 2 5+6 2 |
| | 6 2 6 2 | 5 2 5 2 | 3+4 2 3+4 2 |
| | 5 2 5 2 | 6 2 5 2 | 5+6 2 5+6 2 |
| | 6 2 6 2 | 5 2 6 2 | 3+4 2 3+4 2 |
| | 5 2 5 2 | 5 2 6 2 | 3+4 2 5+6 2 |
| | 5 2 5 2 | 6 2 5 2 | 5+6 2 3+4 2 |
| | 6 2 6 2 | 6 2 6 2 | 3+4 2 5+6 2 |
| | 5 2 6 2 | 6 2 6 2 | 5+6 2 3+4 2 |
| | 6 2 6 2 | 5 2 5 2 | 3+4 2 5+6 2 |
| | 5 2 5 2 | 5 2 5 2 | 5+6 2 3+4 2 |
| | 5 2 6 2 | 6 2 6 2 | 3+4 2 5+6 2 |
| | 6 2 5 2 | 5 2 6 2 | 5+6 2 3+4 2 |
| | 5 2 5 2 | 5 2 5 2 | 5+6 2 5+6 2 |
| | 5 2 5 2 | 6 2 5 2 | 3+4 2 5+6 2 |
| | 6 2 6 2 | 6 2 6 2 | 3+4 2 5+6 2 |
| | 5 2 5 2 | 5 2 6 2 | 5+6 2 5+6 2 |
| | 6 2 5 2 | 5 2 6 2 | 3+4 2 3+4 2 |
| | 5 2 6 2 | 6 2 6 2 | 5+6 2 5+6 2 |
| | 5 2 6 2 | 6 2 5 2 | 3+4 2 3+4 2 |
| | 6 2 6 2 | 6 2 6 2 | 5+6 2 5+6 2 |
| | 5 2 5 2 | 6 2 5 2 | 5+6 2 3+4 2 |
| | 6 2 5 2 | 5 2 5 2 | 5+6 2 3+4 2 |
| | 5 2 5 2 | 5 2 6 2 | 3+4 2 5+6 2 |
| | 6 2 5 2 | 6 2 5 2 | 5+6 2 5+6 2 |
| | 5 2 6 2 | 6 2 5 2 | 3+4 2 3+4 2 |
| | 5 2 6 2 | 5 2 5 2 | 3+4 2 5+6 2 |
| | 6 2 5 2 | 5 2 6 2 | 5+6 2 3+4 2 |
| | 5 2 6 2 | 6 2 6 2 | 3+4 2 3+4 2 |
| | 5 2 6 2 | 5 2 6 2 | 5+6 2 5+6 2 |
| | 6 2 5 2 | 6 2 5 2 | 3+4 2 3+4 2 |
| | 5 2 6 2 | 6 2 5 2 | 5+6 2 5+6 2 |
| | 6 2 5 2 | 5 2 5 2 | 5+6 2 3+4 2 |
| | 5 2 5 2 | 5 2 6 2 | 3+4 2 5+6 2 |
| | 6 2 5 2 | 6 2 6 2 | 5+6 2 3+4 2 |
| | 6 2 6 2 | 6 2 5 2 | 3+4 2 5+6 2 |
| | 6 2 5 2 | 5 2 6 2 | 5+6 2 5+6 2 |
| 65.2.20 | 185392 | 148 | Br2Cd2Hg2S2 |
| | -GM -T | -Y -Z | -R -S |
| | 5 2 6 2 | 6 2 5 2 | 3+4 2 3+4 2 |
| | 6 2 5 2 | 5 2 6 2 | 5+6 2 5+6 2 |
| | 5 2 5 2 | 5 2 5 2 | 5+6 2 5+6 2 |
| | 5 2 6 2 | 5 2 6 2 | 5+6 2 3+4 2 |
| | 6 2 5 2 | 6 2 5 2 | 3+4 2 5+6 2 |

Continued on next column

**TABLE XL– continued**

| TQC | ICSD | NoE | SF |
|---|---|---|---|
| | 6 2 5 2 | 6 2 5 2 | 3+4 2 5+6 2 |
| | 5 2 5 2 | 5 2 5 2 | 3+4 2 3+4 2 |
| | 5 2 6 2 | 5 2 6 2 | 5+6 2 3+4 2 |
| | 6 2 5 2 | 6 2 5 2 | 3+4 2 5+6 2 |
| | 5 2 6 2 | 5 2 6 2 | 5+6 2 3+4 2 |
| | 6 2 6 2 | 6 2 6 2 | 5+6 2 5+6 2 |
| | 5 2 5 2 | 5 2 5 2 | 3+4 2 3+4 2 |
| | 5 2 6 2 | 5 2 6 2 | 5+6 2 3+4 2 |
| | 6 2 5 2 | 6 2 5 2 | 3+4 2 5+6 2 |
| | 5 2 5 2 | 6 2 5 2 | 3+4 2 3+4 2 |
| | 6 2 6 2 | 5 2 6 2 | 5+6 2 5+6 2 |
| | 5 2 5 2 | 5 2 5 2 | 5+6 2 3+4 2 |
| | 6 2 6 2 | 6 2 6 2 | 3+4 2 5+6 2 |
| | 6 2 6 2 | 5 2 6 2 | 3+4 2 3+4 2 |
| | 5 2 5 2 | 6 2 5 2 | 5+6 2 5+6 2 |
| | 6 2 6 2 | 5 2 6 2 | 5+6 2 3+4 2 |
| | 5 2 5 2 | 6 2 5 2 | 3+4 2 5+6 2 |
| | 6 2 5 2 | 6 2 5 2 | 3+4 2 5+6 2 |
| | 5 2 6 2 | 5 2 6 2 | 5+6 2 3+4 2 |
| | 6 2 6 2 | 6 2 6 2 | 3+4 2 5+6 2 |
| | 5 2 5 2 | 5 2 5 2 | 5+6 2 3+4 2 |
| | 6 2 6 2 | 6 2 5 2 | 3+4 2 3+4 2 |
| | 5 2 5 2 | 5 2 6 2 | 5+6 2 5+6 2 |
| | 5 2 6 2 | 6 2 5 2 | 3+4 2 3+4 2 |
| | 6 2 5 2 | 5 2 6 2 | 5+6 2 3+4 2 |
| | 6 2 6 2 | 5 2 6 2 | 3+4 2 5+6 2 |
| | 5 2 5 2 | 6 2 5 2 | 3+4 2 5+6 2 |
| | 6 2 5 2 | 6 2 5 2 | 5+6 2 5+6 2 |
| | 5 2 6 2 | 6 2 6 2 | 5+6 2 3+4 2 |
| | 6 2 6 2 | 5 2 6 2 | 5+6 2 5+6 2 |
| | 5 2 5 2 | 5 2 5 2 | 3+4 2 3+4 2 |
| | 5 2 6 2 | 5 2 6 2 | 5+6 2 3+4 2 |
| | 6 2 5 2 | 6 2 5 2 | 3+4 2 5+6 2 |
| | 6 2 6 2 | 6 2 6 2 | 5+6 2 3+4 2 |
| | 6 2 6 2 | 5 2 6 2 | 3+4 2 5+6 2 |
| | 5 2 5 2 | 6 2 5 2 | 3+4 2 3+4 2 |
| | 5 2 5 2 | 5 2 5 2 | 5+6 2 5+6 2 |
| | 6 2 5 2 | 6 2 5 2 | 3+4 2 5+6 2 |
| | 5 2 6 2 | 5 2 6 2 | 5+6 2 3+4 2 |
| | 6 2 6 2 | 6 2 6 2 | 5+6 2 5+6 2 |
| | 5 2 5 2 | 6 2 6 2 | 3+4 2 3+4 2 |
| | 6 2 6 2 | 5 2 5 2 | 3+4 2 3+4 2 |
| | 5 2 5 2 | 5 2 5 2 | 5+6 2 5+6 2 |
| | 5 2 5 2 | 6 2 6 2 | 3+4 2 3+4 2 |
| | 6 2 6 2 | 6 2 6 2 | 5+6 2 5+6 2 |
| | 6 2 6 2 | 5 2 6 2 | 3+4 2 3+4 2 |
| | 6 2 6 2 | 5 2 5 2 | 5+6 2 5+6 2 |
| | 5 2 5 2 | 6 2 6 2 | 3+4 2 5+6 2 |
| | 5 2 6 2 | 5 2 5 2 | 5+6 2 5+6 2 |
| | 6 2 6 2 | 6 2 5 2 | 3+4 2 3+4 2 |
| | 6 2 5 2 | 6 2 5 2 | 5+6 2 3+4 2 |
| | 5 2 6 2 | 5 2 5 2 | 3+4 2 5+6 2 |
| | 5 2 5 2 | 5 2 6 2 | 3+4 2 3+4 2 |
| | 6 2 6 2 | 6 2 6 2 | 5+6 2 3+4 2 |
| | 6 2 5 2 | 5 2 5 2 | 3+4 2 5+6 2 |
| | 5 2 6 2 | 6 2 6 2 | 5+6 2 3+4 2 |
| | 6 2 5 2 | 6 2 6 2 | 3+4 2 5+6 2 |
| | 5 2 5 2 | 5 2 5 2 | 5+6 2 3+4 2 |
| | 5 2 6 2 | 6 2 6 2 | 5+6 2 5+6 2 |
| | 6 2 6 2 | 6 2 6 2 | 5+6 2 3+4 2 |
| | 6 2 5 2 | 6 2 5 2 | 3+4 2 3+4 2 |
| | 5 2 5 2 | 5 2 6 2 | 5+6 2 5+6 2 |
| | 6 2 6 2 | 6 2 5 2 | 3+4 2 3+4 2 |
| | 6 2 6 2 | 5 2 6 2 | 3+4 2 5+6 2 |
| | 6 2 6 2 | 6 2 6 2 | 5+6 2 5+6 2 |
| | 6 2 6 2 | 6 2 6 2 | 3+4 2 3+4 2 |
| | 6 2 6 2 | 6 2 6 2 | 5+6 2 5+6 2 |
| | 5 2 5 2 | 5 2 5 2 | 5+6 2 5+6 2 |
| | 5 2 5 2 | 6 2 5 2 | 3+4 2 5+6 2 |
| 65.2.21 | 391060 | 184 | Ba3Ge4 |
| | -GM -T | -Y -Z | -R -S |
| | 5 2 6 2 | 5 2 6 2 | 5+6 2 3+4 2 |
| | 6 2 5 2 | 6 2 5 2 | 3+4 2 5+6 2 |
| | 5 2 5 2 | 5 2 5 2 | 5+6 2 5+6 2 |
| | 6 2 6 2 | 6 2 5 2 | 3+4 2 3+4 2 |
| | 5 2 5 2 | 5 2 6 2 | 3+4 2 3+4 2 |
| | 6 2 6 2 | 5 2 6 2 | 5+6 2 5+6 2 |
| | 5 2 5 2 | 6 2 5 2 | 5+6 2 5+6 2 |
| | 6 2 6 2 | 6 2 6 2 | 3+4 2 3+4 2 |
| | 5 2 5 2 | 5 2 5 2 | 3+4 2 3+4 2 |
| | 6 2 6 2 | 6 2 6 2 | 5+6 2 5+6 2 |
| | 5 2 6 2 | 5 2 6 2 | 3+4 2 5+6 2 |
| | 6 2 5 2 | 6 2 5 2 | 5+6 2 3+4 2 |
| | 5 2 6 2 | 5 2 6 2 | 5+6 2 3+4 2 |
| | 6 2 5 2 | 6 2 5 2 | 3+4 2 5+6 2 |
| | 5 2 5 2 | 5 2 5 2 | 5+6 2 3+4 2 |
| | 5 2 6 2 | 6 2 6 2 | 3+4 2 5+6 2 |
| | 6 2 6 2 | 6 2 6 2 | 5+6 2 3+4 2 |
| | 5 2 5 2 | 5 2 5 2 | 3+4 2 5+6 2 |
| | 6 2 5 2 | 5 2 5 2 | 5+6 2 5+6 2 |
| | 6 2 6 2 | 6 2 6 2 | 3+4 2 3+4 2 |
| | 5 2 5 2 | 5 2 5 2 | 3+4 2 3+4 2 |
| | 6 2 6 2 | 6 2 6 2 | 5+6 2 5+6 2 |
| | 5 2 6 2 | 5 2 6 2 | 3+4 2 5+6 2 |
| | 6 2 5 2 | 6 2 5 2 | 5+6 2 3+4 2 |
| | 5 2 5 2 | 6 2 6 2 | 3+4 2 3+4 2 |
| | 6 2 6 2 | 5 2 5 2 | 5+6 2 5+6 2 |
| | 5 2 6 2 | 6 2 5 2 | 5+6 2 3+4 2 |
| | 5 2 5 2 | 5 2 5 2 | 5+6 2 5+6 2 |
| | 6 2 5 2 | 6 2 6 2 | 3+4 2 3+4 2 |
| | 6 2 6 2 | 5 2 6 2 | 3+4 2 5+6 2 |
| | 5 2 6 2 | 5 2 6 2 | 5+6 2 3+4 2 |
| | 6 2 5 2 | 5 2 6 2 | 3+4 2 5+6 2 |
| | 5 2 6 2 | 6 2 5 2 | 5+6 2 3+4 2 |
| | 5 2 5 2 | 5 2 5 2 | 3+4 2 5+6 2 |
| | 6 2 6 2 | 6 2 5 2 | 5+6 2 5+6 2 |
| | 5 2 5 2 | 5 2 6 2 | 3+4 2 3+4 2 |
| | 6 2 5 2 | 6 2 5 2 | 5+6 2 3+4 2 |
| | 5 2 5 2 | 5 2 6 2 | 3+4 2 5+6 2 |
| | 6 2 6 2 | 6 2 5 2 | 5+6 2 3+4 2 |
| | 5 2 6 2 | 6 2 5 2 | 3+4 2 5+6 2 |
| | 6 2 5 2 | 5 2 6 2 | 3+4 2 5+6 2 |
| | 5 2 6 2 | 5 2 6 2 | 5+6 2 3+4 2 |
| | 6 2 5 2 | 6 2 5 2 | 5+6 2 5+6 2 |
| | 6 2 5 2 | 6 2 6 2 | 3+4 2 3+4 2 |
| | 5 2 6 2 | 5 2 5 2 | 3+4 2 3+4 2 |
| | 5 2 6 2 | 5 2 6 2 | 3+4 2 5+6 2 |
| | 6 2 6 2 | 6 2 6 2 | 5+6 2 5+6 2 |
| | 6 2 5 2 | 6 2 5 2 | 5+6 2 3+4 2 |
| | 5 2 6 2 | 5 2 6 2 | 3+4 2 5+6 2 |
| | 6 2 5 2 | 6 2 5 2 | 5+6 2 3+4 2 |
| | 5 2 6 2 | 6 2 5 2 | 5+6 2 5+6 2 |

Continued on next column

**TABLE XL– continued**

| TQC | ICSD | NoE | SF |
|---|---|---|---|
| | 6 2 5 2 | 5 2 6 2 | 3+4 2 3+4 2 |
| | 5 2 6 2 | 6 2 5 2 | 3+4 2 3+4 2 |
| | 6 2 6 2 | 5 2 5 2 | 5+6 2 5+6 2 |
| | 6 2 5 2 | 6 2 6 2 | 5+6 2 3+4 2 |
| | 5 2 5 2 | 5 2 6 2 | 3+4 2 5+6 2 |
| | 5 2 6 2 | 6 2 5 2 | 5+6 2 5+6 2 |
| | 5 2 5 2 | 6 2 6 2 | 3+4 2 3+4 2 |
| | 6 2 5 2 | 5 2 6 2 | 3+4 2 5+6 2 |
| | 6 2 6 2 | 5 2 5 2 | 5+6 2 3+4 2 |
| | 5 2 6 2 | 6 2 5 2 | 5+6 2 3+4 2 |
| | 5 2 5 2 | 6 2 6 2 | 3+4 2 5+6 2 |
| | 6 2 6 2 | 5 2 5 2 | 5+6 2 5+6 2 |
| | 6 2 5 2 | 5 2 6 2 | 3+4 2 3+4 2 |
| | 5 2 6 2 | 6 2 5 2 | 3+4 2 5+6 2 |
| | 5 2 5 2 | 6 2 6 2 | 3+4 2 3+4 2 |
| | 6 2 6 2 | 5 2 6 2 | 5+6 2 5+6 2 |
| | 5 2 5 2 | 5 2 6 2 | 3+4 2 3+4 2 |
| | 6 2 5 2 | 5 2 5 2 | 5+6 2 5+6 2 |
| | 6 2 6 2 | 6 2 6 2 | 5+6 2 3+4 2 |
| | 5 2 5 2 | 5 2 5 2 | 3+4 2 5+6 2 |
| | 5 2 6 2 | 6 2 5 2 | 5+6 2 3+4 2 |
| | 6 2 5 2 | 6 2 6 2 | 5+6 2 5+6 2 |
| | 6 2 6 2 | 6 2 5 2 | 5+6 2 3+4 2 |
| | 6 2 5 2 | 5 2 5 2 | 3+4 2 3+4 2 |
| | 6 2 6 2 | 6 2 6 2 | 3+4 2 5+6 2 |
| | 5 2 5 2 | 6 2 6 2 | 5+6 2 5+6 2 |
| | 6 2 6 2 | 5 2 5 2 | 3+4 2 3+4 2 |
| | 5 2 5 2 | 5 2 6 2 | 3+4 2 3+4 2 |
| | 5 2 5 2 | 6 2 5 2 | 5+6 2 5+6 2 |
| | 5 2 6 2 | 6 2 5 2 | 3+4 2 3+4 2 |
| | 5 2 6 2 | 5 2 6 2 | 5+6 2 5+6 2 |
| | 6 2 6 2 | 5 2 5 2 | 5+6 2 3+4 2 |
| | 5 2 5 2 | 6 2 6 2 | 3+4 2 5+6 2 |
| | 6 2 6 2 | 5 2 5 2 | 5+6 2 3+4 2 |
| | 5 2 5 2 | 6 2 6 2 | 5+6 2 5+6 2 |
| | 6 2 6 2 | 5 2 5 2 | 3+4 2 5+6 2 |
| | 6 2 5 2 | 5 2 6 2 | 3+4 2 3+4 2 |
| | 5 2 6 2 | 6 2 5 2 | 3+4 2 5+6 2 |
| | 5 2 5 2 | 5 2 6 2 | 3+4 2 5+6 2 |
| | 6 2 6 2 | 6 2 6 2 | 5+6 2 3+4 2 |
| | 5 2 6 2 | 6 2 5 2 | 5+6 2 3+4 2 |
| 65.2.22 | 75687 | 134 | Ba2Cu4Er1O8 |
| | -GM -T | -Y -Z | -R -S |
| | 5 2 6 2 | 5 2 6 2 | 3+4 2 5+6 2 |
| | 6 2 5 2 | 6 2 5 2 | 5+6 2 3+4 2 |
| | 5 2 6 2 | 5 2 6 2 | 3+4 2 5+6 2 |
| | 6 2 5 2 | 6 2 5 2 | 5+6 2 3+4 2 |
| | 6 2 5 2 | 6 2 5 2 | 5+6 2 3+4 2 |
| | 6 2 5 2 | 6 2 5 2 | 5+6 2 3+4 2 |
| | 5 2 6 2 | 5 2 5 2 | 3+4 2 5+6 2 |
| | 5 2 6 2 | 6 2 6 2 | 5+6 2 5+6 2 |
| | 6 2 6 2 | 5 2 6 2 | 3+4 2 3+4 2 |
| | 6 2 5 2 | 6 2 6 2 | 3+4 2 5+6 2 |
| | 5 2 5 2 | 5 2 5 2 | 5+6 2 3+4 2 |
| | 6 2 5 2 | 6 2 5 2 | 5+6 2 3+4 2 |
| | 6 2 6 2 | 5 2 6 2 | 3+4 2 5+6 2 |
| | 5 2 5 2 | 6 2 5 2 | 5+6 2 3+4 2 |
| | 5 2 6 2 | 5 2 5 2 | 3+4 2 5+6 2 |
| | 6 2 5 2 | 6 2 6 2 | 5+6 2 3+4 2 |
| | 5 2 6 2 | 5 2 5 2 | 3+4 2 5+6 2 |
| | 6 2 5 2 | 6 2 6 2 | 5+6 2 3+4 2 |

Continued on next column

**TABLE XL– continued**

| TQC | ICSD | NoE | SF |
|---|---|---|---|
| | 5 2 6 2 | 5 2 5 2 | 3+4 2 5+6 2 |
| | 6 2 5 2 | 6 2 6 2 | 5+6 2 3+4 2 |
| | 5 2 5 2 | 6 2 5 2 | 3+4 2 3+4 2 |
| | 5 2 6 2 | 6 2 6 2 | 5+6 2 5+6 2 |
| | 5 2 6 2 | 5 2 5 2 | 3+4 2 3+4 2 |
| | 5 2 5 2 | 5 2 6 2 | 5+6 2 5+6 2 |
| | 6 2 6 2 | 5 2 5 2 | 3+4 2 5+6 2 |
| | 5 2 6 2 | 5 2 6 2 | 3+4 2 3+4 2 |
| | 6 2 6 2 | 6 2 6 2 | 5+6 2 5+6 2 |
| | 6 2 6 2 | 6 2 6 2 | 3+4 2 5+6 2 |
| | 5 2 6 2 | 6 2 5 2 | 5+6 2 5+6 2 |
| | 5 2 5 2 | 5 2 6 2 | 5+6 2 3+4 2 |
| | 6 2 5 2 | 5 2 5 2 | 5+6 2 3+4 2 |
| | 6 2 5 2 | 5 2 5 2 | 3+4 2 5+6 2 |
| | 5 2 5 2 | 6 2 5 2 | 5+6 2 3+4 2 |
| | 5 2 5 2 | 6 2 6 2 | 5+6 2 5+6 2 |
| | 6 2 5 2 | 6 2 5 2 | 3+4 2 3+4 2 |
| | 6 2 6 2 | 6 2 5 2 | 5+6 2 3+4 2 |
| | 5 2 6 2 | 5 2 6 2 | 3+4 2 3+4 2 |
| | 6 2 6 2 | 5 2 5 2 | 3+4 2 5+6 2 |
| | 5 2 5 2 | 6 2 6 2 | 3+4 2 5+6 2 |
| | 6 2 6 2 | 6 2 6 2 | 5+6 2 3+4 2 |
| | 5 2 5 2 | 5 2 5 2 | 3+4 2 5+6 2 |
| | 6 2 5 2 | 5 2 6 2 | 5+6 2 3+4 2 |
| | 6 2 5 2 | 6 2 6 2 | 5+6 2 3+4 2 |
| | 5 2 6 2 | 5 2 5 2 | 3+4 2 5+6 2 |
| | 6 2 5 2 | 6 2 6 2 | 5+6 2 3+4 2 |
| | 6 2 6 2 | 5 2 5 2 | 3+4 2 5+6 2 |
| | 5 2 5 2 | 6 2 6 2 | 3+4 2 5+6 2 |
| | 6 2 6 2 | 5 2 5 2 | 5+6 2 3+4 2 |
| | 6 2 6 2 | 5 2 5 2 | 3+4 2 5+6 2 |
| | 5 2 5 2 | 5 2 6 2 | 5+6 2 3+4 2 |
| | 5 2 5 2 | 6 2 5 2 | 5+6 2 5+6 2 |
| | 6 2 6 2 | 5 2 6 2 | 5+6 2 3+4 2 |
| | 6 2 6 2 | 6 2 6 2 | 3+4 2 5+6 2 |
| | 5 2 5 2 | 5 2 5 2 | 5+6 2 3+4 2 |
| | 5 2 6 2 | 6 2 6 2 | 3+4 2 3+4 2 |
| | 6 2 5 2 | 6 2 5 2 | 5+6 2 5+6 2 |
| | 5 2 5 2 | 5 2 6 2 | 5+6 2 3+4 2 |
| | 6 2 5 2 | 6 2 5 2 | 3+4 2 3+4 2 |
| | 6 2 6 2 | 6 2 5 2 | 5+6 2 5+6 2 |
| | 6 2 5 2 | 6 2 6 2 | 5+6 2 3+4 2 |
| | 6 2 5 2 | 6 2 5 2 | 3+4 2 3+4 2 |
| | 5 2 6 2 | 5 2 5 2 | 3+4 2 3+4 2 |
| | 6 2 5 2 | 6 2 5 2 | 5+6 2 5+6 2 |
| | 5 2 5 2 | 5 2 6 2 | 5+6 2 3+4 2 |
| | 6 2 5 2 | 6 2 5 2 | 5+6 2 5+6 2 |
| | 6 2 6 2 | 5 2 5 2 | 5+6 2 3+4 2 |
| | 5 2 5 2 | 6 2 6 2 | 5+6 2 3+4 2 |
| 65.2.23 | 74127 | 134 | Ba2Cu4Er1O8 |
| | -GM -T | -Y -Z | -R -S |
| | 5 2 6 2 | 5 2 6 2 | 3+4 2 5+6 2 |
| | 6 2 5 2 | 6 2 5 2 | 5+6 2 3+4 2 |
| | 5 2 6 2 | 5 2 6 2 | 3+4 2 5+6 2 |
| | 6 2 5 2 | 6 2 5 2 | 5+6 2 3+4 2 |
| | 6 2 5 2 | 6 2 5 2 | 5+6 2 3+4 2 |
| | 6 2 5 2 | 6 2 5 2 | 5+6 2 3+4 2 |
| | 5 2 6 2 | 5 2 5 2 | 3+4 2 5+6 2 |
| | 5 2 6 2 | 6 2 6 2 | 5+6 2 5+6 2 |
| | 6 2 6 2 | 6 2 6 2 | 3+4 2 3+4 2 |
| | 6 2 5 2 | 5 2 6 2 | 3+4 2 5+6 2 |

Continued on next column

**TABLE XL– continued**

| TQC | ICSD | NoE | SF |
|---|---|---|---|
| | 5 2 5 2 | 5 2 5 2 | 5+6 2 3+4 2 |
| | 6 2 5 2 | 6 2 5 2 | 5+6 2 3+4 2 |
| | 6 2 6 2 | 5 2 6 2 | 3+4 2 5+6 2 |
| | 5 2 5 2 | 6 2 5 2 | 5+6 2 3+4 2 |
| | 5 2 6 2 | 5 2 5 2 | 3+4 2 5+6 2 |
| | 6 2 5 2 | 6 2 6 2 | 5+6 2 3+4 2 |
| | 5 2 6 2 | 5 2 6 2 | 3+4 2 5+6 2 |
| | 6 2 5 2 | 6 2 5 2 | 5+6 2 3+4 2 |
| | 5 2 6 2 | 5 2 5 2 | 3+4 2 5+6 2 |
| | 6 2 5 2 | 6 2 6 2 | 5+6 2 3+4 2 |
| | 5 2 5 2 | 6 2 5 2 | 3+4 2 3+4 2 |
| | 5 2 6 2 | 6 2 6 2 | 5+6 2 5+6 2 |
| | 5 2 6 2 | 5 2 5 2 | 3+4 2 3+4 2 |
| | 5 2 5 2 | 5 2 6 2 | 5+6 2 5+6 2 |
| | 6 2 6 2 | 5 2 5 2 | 3+4 2 5+6 2 |
| | 5 2 6 2 | 5 2 6 2 | 3+4 2 3+4 2 |
| | 6 2 6 2 | 6 2 6 2 | 5+6 2 5+6 2 |
| | 6 2 6 2 | 6 2 6 2 | 3+4 2 5+6 2 |
| | 5 2 6 2 | 6 2 5 2 | 5+6 2 5+6 2 |
| | 5 2 5 2 | 5 2 6 2 | 5+6 2 3+4 2 |
| | 6 2 5 2 | 5 2 5 2 | 5+6 2 3+4 2 |
| | 6 2 5 2 | 5 2 5 2 | 3+4 2 5+6 2 |
| | 5 2 5 2 | 6 2 5 2 | 5+6 2 3+4 2 |
| | 5 2 5 2 | 6 2 6 2 | 5+6 2 5+6 2 |
| | 6 2 5 2 | 6 2 5 2 | 3+4 2 3+4 2 |
| | 6 2 6 2 | 6 2 5 2 | 5+6 2 3+4 2 |
| | 5 2 6 2 | 5 2 6 2 | 3+4 2 3+4 2 |
| | 6 2 6 2 | 5 2 5 2 | 3+4 2 5+6 2 |
| | 6 2 5 2 | 6 2 6 2 | 3+4 2 5+6 2 |
| | 5 2 6 2 | 6 2 6 2 | 5+6 2 3+4 2 |
| | 5 2 5 2 | 5 2 5 2 | 3+4 2 5+6 2 |
| | 6 2 5 2 | 5 2 6 2 | 5+6 2 3+4 2 |
| | 6 2 5 2 | 6 2 6 2 | 5+6 2 3+4 2 |
| | 5 2 6 2 | 5 2 5 2 | 3+4 2 5+6 2 |
| | 6 2 5 2 | 5 2 6 2 | 5+6 2 3+4 2 |
| | 6 2 6 2 | 6 2 5 2 | 3+4 2 5+6 2 |
| | 5 2 5 2 | 6 2 6 2 | 3+4 2 5+6 2 |
| | 6 2 6 2 | 5 2 5 2 | 5+6 2 3+4 2 |
| | 6 2 6 2 | 5 2 5 2 | 3+4 2 5+6 2 |
| | 5 2 5 2 | 5 2 6 2 | 5+6 2 3+4 2 |
| | 5 2 5 2 | 6 2 5 2 | 5+6 2 5+6 2 |
| | 6 2 6 2 | 5 2 6 2 | 5+6 2 3+4 2 |
| | 6 2 6 2 | 6 2 6 2 | 3+4 2 5+6 2 |
| | 5 2 5 2 | 5 2 5 2 | 5+6 2 3+4 2 |
| | 5 2 6 2 | 6 2 6 2 | 3+4 2 3+4 2 |
| | 6 2 5 2 | 6 2 5 2 | 5+6 2 5+6 2 |
| | 5 2 5 2 | 5 2 5 2 | 5+6 2 3+4 2 |
| | 6 2 5 2 | 6 2 6 2 | 3+4 2 3+4 2 |
| | 6 2 6 2 | 6 2 5 2 | 5+6 2 5+6 2 |
| | 6 2 5 2 | 6 2 6 2 | 5+6 2 3+4 2 |
| | 6 2 5 2 | 6 2 5 2 | 3+4 2 3+4 2 |
| | 5 2 6 2 | 5 2 5 2 | 3+4 2 3+4 2 |
| | 6 2 5 2 | 6 2 5 2 | 5+6 2 3+4 2 |
| | 5 2 5 2 | 5 2 6 2 | 5+6 2 5+6 2 |
| | 6 2 5 2 | 6 2 5 2 | 5+6 2 5+6 2 |
| | 6 2 6 2 | 5 2 5 2 | 5+6 2 3+4 2 |
| | 5 2 5 2 | 6 2 6 2 | 5+6 2 3+4 2 |
| 65.2.24 | 67642 | 134 | Ba2Cu4Er1O8 |
| | -GM -T | -Y -Z | -R -S |
| | 5 2 6 2 | 5 2 6 2 | 3+4 2 5+6 2 |
| | 6 2 5 2 | 6 2 5 2 | 5+6 2 3+4 2 |
| | 5 2 6 2 | 5 2 6 2 | 3+4 2 5+6 2 |
| | 6 2 5 2 | 6 2 5 2 | 5+6 2 3+4 2 |
| | 6 2 5 2 | 6 2 5 2 | 5+6 2 3+4 2 |
| | 6 2 5 2 | 6 2 5 2 | 5+6 2 3+4 2 |
| | 5 2 6 2 | 5 2 5 2 | 3+4 2 5+6 2 |
| | 5 2 6 2 | 6 2 6 2 | 5+6 2 3+4 2 |
| | 6 2 6 2 | 6 2 6 2 | 3+4 2 5+6 2 |
| | 6 2 5 2 | 5 2 6 2 | 3+4 2 5+6 2 |
| | 5 2 5 2 | 5 2 5 2 | 5+6 2 3+4 2 |
| | 6 2 5 2 | 6 2 5 2 | 5+6 2 3+4 2 |
| | 6 2 6 2 | 5 2 6 2 | 3+4 2 5+6 2 |
| | 5 2 5 2 | 6 2 5 2 | 5+6 2 3+4 2 |
| | 5 2 6 2 | 5 2 5 2 | 3+4 2 5+6 2 |
| | 6 2 5 2 | 6 2 6 2 | 5+6 2 3+4 2 |
| | 5 2 6 2 | 5 2 5 2 | 3+4 2 5+6 2 |
| | 6 2 5 2 | 6 2 6 2 | 5+6 2 3+4 2 |
| | 5 2 6 2 | 5 2 5 2 | 3+4 2 5+6 2 |
| | 6 2 5 2 | 6 2 6 2 | 5+6 2 3+4 2 |
| | 5 2 5 2 | 6 2 5 2 | 3+4 2 3+4 2 |
| | 5 2 6 2 | 6 2 6 2 | 5+6 2 5+6 2 |
| | 5 2 6 2 | 5 2 5 2 | 3+4 2 3+4 2 |
| | 6 2 6 2 | 5 2 5 2 | 5+6 2 5+6 2 |
| | 5 2 5 2 | 5 2 6 2 | 3+4 2 5+6 2 |
| | 5 2 6 2 | 5 2 6 2 | 3+4 2 3+4 2 |
| | 6 2 6 2 | 6 2 6 2 | 5+6 2 5+6 2 |
| | 6 2 6 2 | 6 2 6 2 | 3+4 2 5+6 2 |
| | 5 2 5 2 | 6 2 6 2 | 5+6 2 5+6 2 |
| | 5 2 6 2 | 5 2 5 2 | 5+6 2 3+4 2 |
| | 6 2 5 2 | 5 2 5 2 | 3+4 2 3+4 2 |
| | 6 2 5 2 | 5 2 5 2 | 5+6 2 5+6 2 |
| | 5 2 5 2 | 6 2 5 2 | 5+6 2 3+4 2 |
| | 5 2 5 2 | 6 2 6 2 | 5+6 2 5+6 2 |
| | 6 2 6 2 | 6 2 5 2 | 3+4 2 3+4 2 |
| | 6 2 6 2 | 5 2 5 2 | 5+6 2 3+4 2 |
| | 5 2 5 2 | 6 2 6 2 | 3+4 2 3+4 2 |
| | 6 2 6 2 | 5 2 5 2 | 3+4 2 5+6 2 |
| | 5 2 5 2 | 6 2 6 2 | 3+4 2 5+6 2 |
| | 6 2 6 2 | 6 2 6 2 | 5+6 2 3+4 2 |
| | 6 2 5 2 | 5 2 5 2 | 3+4 2 5+6 2 |
| | 5 2 5 2 | 6 2 6 2 | 5+6 2 3+4 2 |
| | 6 2 5 2 | 5 2 6 2 | 3+4 2 3+4 2 |
| | 6 2 6 2 | 5 2 5 2 | 5+6 2 5+6 2 |
| | 5 2 5 2 | 6 2 6 2 | 5+6 2 3+4 2 |
| | 6 2 6 2 | 5 2 5 2 | 3+4 2 5+6 2 |
| | 5 2 5 2 | 6 2 6 2 | 3+4 2 5+6 2 |
| | 6 2 6 2 | 5 2 5 2 | 5+6 2 3+4 2 |
| | 6 2 6 2 | 5 2 5 2 | 3+4 2 3+4 2 |
| | 5 2 5 2 | 5 2 6 2 | 5+6 2 5+6 2 |
| | 5 2 5 2 | 6 2 5 2 | 5+6 2 3+4 2 |
| | 6 2 6 2 | 5 2 6 2 | 3+4 2 5+6 2 |
| | 6 2 6 2 | 6 2 6 2 | 5+6 2 5+6 2 |
| | 5 2 5 2 | 5 2 5 2 | 5+6 2 3+4 2 |
| | 5 2 6 2 | 6 2 6 2 | 5+6 2 3+4 2 |
| | 6 2 5 2 | 5 2 5 2 | 3+4 2 5+6 2 |
| | 6 2 5 2 | 6 2 5 2 | 5+6 2 3+4 2 |
| | 5 2 5 2 | 6 2 6 2 | 3+4 2 3+4 2 |
| | 6 2 6 2 | 6 2 5 2 | 5+6 2 5+6 2 |
| | 6 2 5 2 | 6 2 6 2 | 5+6 2 3+4 2 |
| | 6 2 5 2 | 6 2 5 2 | 3+4 2 3+4 2 |
| | 5 2 6 2 | 5 2 5 2 | 3+4 2 3+4 2 |
| | 6 2 5 2 | 6 2 5 2 | 5+6 2 5+6 2 |

Continued on next column

**TABLE XL– continued**

| TQC | ICSD | NoE | SF | | | |
|---|---|---|---|---|---|---|
| | 5 2 | 5 2 | 5 2 | 6 2 | 5+6 2 | 3+4 2 |
| | 6 2 | 5 2 | 6 2 | 5 2 | 5+6 2 | 5+6 2 |
| | 6 2 | 6 2 | 5 2 | 5 2 | 5+6 2 | 3+4 2 |
| | 5 2 | 5 2 | 6 2 | 6 2 | 5+6 2 | 3+4 2 |
| 65.2.25 | 39394 | 96 | B6Ni3Yb2 | | | |
| | -GM | -T | -Y | -Z | -R | -S |
| | 5 2 | 6 2 | 6 2 | 5 2 | 3+4 2 | 3+4 2 |
| | 6 2 | 5 2 | 5 2 | 6 2 | 5+6 2 | 5+6 2 |
| | 5 2 | 6 2 | 6 2 | 5 2 | 5+6 2 | 5+6 2 |
| | 6 2 | 5 2 | 5 2 | 6 2 | 3+4 2 | 3+4 2 |
| | 5 2 | 6 2 | 6 2 | 6 2 | 3+4 2 | 5+6 2 |
| | 6 2 | 5 2 | 5 2 | 5 2 | 5+6 2 | 3+4 2 |
| | 5 2 | 6 2 | 5 2 | 5 2 | 5+6 2 | 5+6 2 |
| | 6 2 | 5 2 | 6 2 | 6 2 | 3+4 2 | 3+4 2 |
| | 5 2 | 5 2 | 5 2 | 6 2 | 3+4 2 | 5+6 2 |
| | 6 2 | 6 2 | 6 2 | 5 2 | 5+6 2 | 3+4 2 |
| | 5 2 | 5 2 | 6 2 | 6 2 | 3+4 2 | 3+4 2 |
| | 5 2 | 6 2 | 5 2 | 6 2 | 5+6 2 | 5+6 2 |
| | 5 2 | 6 2 | 6 2 | 5 2 | 5+6 2 | 5+6 2 |
| | 6 2 | 5 2 | 5 2 | 5 2 | 3+4 2 | 3+4 2 |
| | 6 2 | 6 2 | 5 2 | 6 2 | 3+4 2 | 3+4 2 |
| | 5 2 | 5 2 | 5 2 | 6 2 | 5+6 2 | 5+6 2 |
| | 6 2 | 6 2 | 6 2 | 5 2 | 5+6 2 | 3+4 2 |
| | 6 2 | 6 2 | 5 2 | 5 2 | 3+4 2 | 3+4 2 |
| | 5 2 | 5 2 | 5 2 | 6 2 | 5+6 2 | 3+4 2 |
| | 5 2 | 5 2 | 6 2 | 6 2 | 5+6 2 | 5+6 2 |
| | 5 2 | 6 2 | 5 2 | 5 2 | 5+6 2 | 3+4 2 |
| | 6 2 | 6 2 | 6 2 | 6 2 | 3+4 2 | 5+6 2 |
| | 6 2 | 5 2 | 5 2 | 5 2 | 3+4 2 | 5+6 2 |
| | 5 2 | 6 2 | 6 2 | 6 2 | 3+4 2 | 3+4 2 |
| | 5 2 | 5 2 | 5 2 | 5 2 | 5+6 2 | 5+6 2 |
| | 5 2 | 6 2 | 6 2 | 6 2 | 5+6 2 | 3+4 2 |
| | 6 2 | 6 2 | 6 2 | 5 2 | 5+6 2 | 3+4 2 |
| | 5 2 | 5 2 | 5 2 | 6 2 | 3+4 2 | 5+6 2 |
| | 6 2 | 5 2 | 5 2 | 6 2 | 3+4 2 | 3+4 2 |
| | 6 2 | 5 2 | 5 2 | 6 2 | 5+6 2 | 3+4 2 |
| | 5 2 | 6 2 | 6 2 | 5 2 | 3+4 2 | 5+6 2 |
| | 5 2 | 6 2 | 6 2 | 5 2 | 5+6 2 | 5+6 2 |
| | 6 2 | 6 2 | 5 2 | 6 2 | 5+6 2 | 3+4 2 |
| | 5 2 | 5 2 | 6 2 | 5 2 | 3+4 2 | 5+6 2 |
| | 6 2 | 5 2 | 5 2 | 6 2 | 5+6 2 | 3+4 2 |
| | 5 2 | 6 2 | 5 2 | 6 2 | 3+4 2 | 3+4 2 |
| | 5 2 | 5 2 | 6 2 | 6 2 | 3+4 2 | 3+4 2 |
| | 6 2 | 6 2 | 6 2 | 6 2 | 5+6 2 | 5+6 2 |
| | 5 2 | 6 2 | 5 2 | 5 2 | 5+6 2 | 5+6 2 |
| | 5 2 | 6 2 | 6 2 | 5 2 | 5+6 2 | 3+4 2 |
| | 6 2 | 5 2 | 5 2 | 5 2 | 3+4 2 | 5+6 2 |
| | 5 2 | 6 2 | 5 2 | 6 2 | 5+6 2 | 3+4 2 |
| | 6 2 | 5 2 | 6 2 | 5 2 | 3+4 2 | 5+6 2 |
| | 5 2 | 5 2 | 6 2 | 6 2 | 3+4 2 | 3+4 2 |
| | 6 2 | 6 2 | 5 2 | 5 2 | 5+6 2 | 5+6 2 |
| | 5 2 | 6 2 | 5 2 | 6 2 | 5+6 2 | 3+4 2 |
| | 5 2 | 5 2 | 5 2 | 5 2 | 3+4 2 | 5+6 2 |
| | 6 2 | 6 2 | 6 2 | 6 2 | 5+6 2 | 3+4 2 |
| 65.2.26 | 68002 | 98 | B6Lu2Ni3 | | | |
| | -GM | -T | -Y | -Z | -R | -S |
| | 5 2 | 5 2 | 5 2 | 5 2 | 3+4 2 | 3+4 2 |
| | 6 2 | 6 2 | 6 2 | 6 2 | 5+6 2 | 5+6 2 |
| | 5 2 | 6 2 | 6 2 | 5 2 | 5+6 2 | 5+6 2 |
| | 6 2 | 5 2 | 5 2 | 6 2 | 3+4 2 | 3+4 2 |
| | 5 2 | 6 2 | 6 2 | 5 2 | 3+4 2 | 5+6 2 |
| | 6 2 | 5 2 | 5 2 | 6 2 | 5+6 2 | 3+4 2 |
| | 5 2 | 6 2 | 5 2 | 5 2 | 5+6 2 | 5+6 2 |
| | 6 2 | 5 2 | 6 2 | 6 2 | 3+4 2 | 3+4 2 |
| | 5 2 | 5 2 | 5 2 | 6 2 | 3+4 2 | 5+6 2 |
| | 6 2 | 6 2 | 6 2 | 6 2 | 5+6 2 | 3+4 2 |
| | 5 2 | 5 2 | 6 2 | 5 2 | 3+4 2 | 3+4 2 |
| | 5 2 | 6 2 | 5 2 | 5 2 | 5+6 2 | 5+6 2 |
| | 5 2 | 6 2 | 6 2 | 6 2 | 5+6 2 | 5+6 2 |
| | 6 2 | 5 2 | 5 2 | 5 2 | 3+4 2 | 3+4 2 |
| | 6 2 | 6 2 | 5 2 | 6 2 | 3+4 2 | 3+4 2 |
| | 6 2 | 5 2 | 6 2 | 6 2 | 5+6 2 | 5+6 2 |
| | 5 2 | 5 2 | 5 2 | 5 2 | 5+6 2 | 3+4 2 |
| | 5 2 | 6 2 | 6 2 | 6 2 | 3+4 2 | 5+6 2 |
| | 6 2 | 5 2 | 5 2 | 5 2 | 3+4 2 | 5+6 2 |
| | 5 2 | 6 2 | 6 2 | 5 2 | 5+6 2 | 3+4 2 |
| | 6 2 | 5 2 | 5 2 | 6 2 | 5+6 2 | 3+4 2 |
| | 5 2 | 6 2 | 5 2 | 6 2 | 3+4 2 | 5+6 2 |
| | 6 2 | 6 2 | 5 2 | 5 2 | 5+6 2 | 3+4 2 |
| | 6 2 | 6 2 | 6 2 | 6 2 | 3+4 2 | 3+4 2 |
| | 6 2 | 5 2 | 6 2 | 5 2 | 5+6 2 | 5+6 2 |
| | 5 2 | 6 2 | 5 2 | 6 2 | 3+4 2 | 3+4 2 |
| | 5 2 | 5 2 | 5 2 | 5 2 | 5+6 2 | 3+4 2 |
| | 5 2 | 5 2 | 6 2 | 6 2 | 3+4 2 | 5+6 2 |
| | 6 2 | 6 2 | 6 2 | 6 2 | 5+6 2 | 5+6 2 |
| | 5 2 | 6 2 | 5 2 | 5 2 | 3+4 2 | 3+4 2 |
| | 6 2 | 5 2 | 6 2 | 5 2 | 5+6 2 | 5+6 2 |
| | 6 2 | 6 2 | 5 2 | 5 2 | 3+4 2 | 3+4 2 |
| | 5 2 | 6 2 | 5 2 | 6 2 | 5+6 2 | 3+4 2 |
| | 5 2 | 5 2 | 5 2 | 5 2 | 5+6 2 | 5+6 2 |
| | 5 2 | 5 2 | 6 2 | 6 2 | 3+4 2 | 3+4 2 |
| | 6 2 | 5 2 | 5 2 | 6 2 | 5+6 2 | 5+6 2 |
| | 6 2 | 6 2 | 6 2 | 5 2 | 3+4 2 | 5+6 2 |
| | 5 2 | 6 2 | 6 2 | 6 2 | 5+6 2 | 3+4 2 |
| | 5 2 | 5 2 | 5 2 | 5 2 | 3+4 2 | 3+4 2 |
| | 5 2 | 6 2 | 6 2 | 5 2 | 5+6 2 | 5+6 2 |
| | 6 2 | 5 2 | 6 2 | 6 2 | 3+4 2 | 3+4 2 |
| | 5 2 | 5 2 | 5 2 | 6 2 | 5+6 2 | 5+6 2 |
| | 6 2 | 6 2 | 5 2 | 6 2 | 3+4 2 | 3+4 2 |
| | 5 2 | 5 2 | 5 2 | 6 2 | 5+6 2 | 3+4 2 |
| | 6 2 | 6 2 | 6 2 | 6 2 | 3+4 2 | 5+6 2 |
| | 5 2 | 6 2 | 6 2 | 5 2 | 5+6 2 | 3+4 2 |
| | 5 2 | 6 2 | 5 2 | 6 2 | 5+6 2 | 5+6 2 |
| | 5 2 | 6 2 | 5 2 | 5 2 | 5+6 2 | 3+4 2 |
| | 6 2 | 5 2 | 6 2 | 5 2 | 3+4 2 | 3+4 2 |
| 65.2.27 | 417990 | 144 | B4Rh8Zn5 | | | |
| | -GM | -T | -Y | -Z | -R | -S |
| | 5 2 | 6 2 | 6 2 | 5 2 | 5+6 2 | 5+6 2 |
| | 6 2 | 6 2 | 5 2 | 5 2 | 3+4 2 | 3+4 2 |
| | 5 2 | 5 2 | 6 2 | 6 2 | 3+4 2 | 3+4 2 |
| | 6 2 | 5 2 | 5 2 | 6 2 | 5+6 2 | 5+6 2 |
| | 5 2 | 6 2 | 5 2 | 6 2 | 5+6 2 | 5+6 2 |
| | 5 2 | 6 2 | 6 2 | 5 2 | 3+4 2 | 3+4 2 |
| | 5 2 | 5 2 | 5 2 | 6 2 | 3+4 2 | 3+4 2 |
| | 6 2 | 6 2 | 5 2 | 5 2 | 5+6 2 | 5+6 2 |
| | 5 2 | 5 2 | 6 2 | 6 2 | 5+6 2 | 3+4 2 |
| | 6 2 | 6 2 | 6 2 | 5 2 | 5+6 2 | 3+4 2 |
| | 5 2 | 6 2 | 5 2 | 6 2 | 5+6 2 | 5+6 2 |
| | 5 2 | 5 2 | 6 2 | 5 2 | 3+4 2 | 3+4 2 |
| | 6 2 | 6 2 | 6 2 | 6 2 | 5+6 2 | 3+4 2 |
| | 5 2 | 5 2 | 5 2 | 5 2 | 3+4 2 | 5+6 2 |
| | 6 2 | 6 2 | 5 2 | 6 2 | 5+6 2 | 5+6 2 |

Continued on next column

**TABLE XL– continued**

| TQC | ICSD | NoE | SF |
|---|---|---|---|
| | 5 2 5 2 | 6 2 5 2 | 5+6 2 3+4 2 |
| | 6 2 6 2 | 5 2 6 2 | 3+4 2 3+4 2 |
| | 6 2 6 2 | 6 2 5 2 | 5+6 2 3+4 2 |
| | 5 2 6 2 | 5 2 6 2 | 3+4 2 5+6 2 |
| | 5 2 6 2 | 5 2 5 2 | 5+6 2 3+4 2 |
| | 5 2 5 2 | 5 2 6 2 | 3+4 2 5+6 2 |
| | 5 2 6 2 | 6 2 6 2 | 5+6 2 3+4 2 |
| | 6 2 5 2 | 5 2 6 2 | 3+4 2 3+4 2 |
| | 5 2 6 2 | 6 2 6 2 | 5+6 2 5+6 2 |
| | 5 2 6 2 | 5 2 5 2 | 3+4 2 3+4 2 |
| | 6 2 5 2 | 5 2 6 2 | 3+4 2 5+6 2 |
| | 5 2 6 2 | 6 2 6 2 | 3+4 2 3+4 2 |
| | 6 2 6 2 | 5 2 5 2 | 5+6 2 5+6 2 |
| | 6 2 5 2 | 5 2 5 2 | 5+6 2 3+4 2 |
| | 5 2 5 2 | 5 2 6 2 | 5+6 2 5+6 2 |
| | 5 2 5 2 | 6 2 5 2 | 5+6 2 3+4 2 |
| | 6 2 6 2 | 6 2 5 2 | 5+6 2 5+6 2 |
| | 5 2 6 2 | 5 2 5 2 | 3+4 2 3+4 2 |
| | 6 2 6 2 | 6 2 6 2 | 3+4 2 5+6 2 |
| | 5 2 6 2 | 5 2 5 2 | 3+4 2 3+4 2 |
| | 5 2 5 2 | 6 2 6 2 | 5+6 2 5+6 2 |
| | 5 2 5 2 | 6 2 6 2 | 5+6 2 3+4 2 |
| | 6 2 6 2 | 5 2 6 2 | 3+4 2 3+4 2 |
| | 5 2 6 2 | 6 2 5 2 | 5+6 2 3+4 2 |
| | 5 2 5 2 | 5 2 5 2 | 5+6 2 5+6 2 |
| | 6 2 6 2 | 6 2 6 2 | 3+4 2 5+6 2 |
| | 5 2 5 2 | 6 2 6 2 | 3+4 2 5+6 2 |
| | 5 2 5 2 | 5 2 5 2 | 5+6 2 5+6 2 |
| | 6 2 6 2 | 6 2 5 2 | 3+4 2 3+4 2 |
| | 5 2 5 2 | 6 2 5 2 | 5+6 2 3+4 2 |
| | 6 2 5 2 | 6 2 6 2 | 3+4 2 5+6 2 |
| | 6 2 6 2 | 5 2 5 2 | 5+6 2 3+4 2 |
| | 5 2 6 2 | 5 2 5 2 | 3+4 2 5+6 2 |
| | 5 2 6 2 | 6 2 6 2 | 5+6 2 3+4 2 |
| | 5 2 5 2 | 5 2 5 2 | 5+6 2 3+4 2 |
| | 6 2 6 2 | 6 2 6 2 | 3+4 2 5+6 2 |
| | 6 2 6 2 | 5 2 5 2 | 5+6 2 5+6 2 |
| | 6 2 5 2 | 5 2 5 2 | 3+4 2 3+4 2 |
| | 5 2 6 2 | 6 2 5 2 | 5+6 2 3+4 2 |
| | 6 2 5 2 | 5 2 6 2 | 3+4 2 3+4 2 |
| | 5 2 5 2 | 5 2 5 2 | 3+4 2 5+6 2 |
| | 6 2 5 2 | 6 2 6 2 | 5+6 2 5+6 2 |
| | 5 2 6 2 | 6 2 5 2 | 3+4 2 5+6 2 |
| | 5 2 6 2 | 6 2 6 2 | 5+6 2 5+6 2 |
| | 5 2 6 2 | 6 2 6 2 | 5+6 2 3+4 2 |
| | 6 2 6 2 | 6 2 5 2 | 3+4 2 3+4 2 |
| | 6 2 5 2 | 6 2 6 2 | 5+6 2 5+6 2 |
| | 5 2 6 2 | 5 2 6 2 | 3+4 2 5+6 2 |
| | 5 2 6 2 | 5 2 5 2 | 5+6 2 3+4 2 |
| | 6 2 5 2 | 6 2 5 2 | 3+4 2 5+6 2 |
| | 5 2 5 2 | 6 2 6 2 | 3+4 2 3+4 2 |
| | 6 2 6 2 | 5 2 5 2 | 5+6 2 3+4 2 |
| | 5 2 6 2 | 6 2 6 2 | 3+4 2 5+6 2 |
| | 6 2 6 2 | 5 2 5 2 | 5+6 2 5+6 2 |
| | 5 2 5 2 | 6 2 5 2 | 5+6 2 3+4 2 |
| | 5 2 5 2 | 6 2 5 2 | 3+4 2 3+4 2 |
| | 6 2 6 2 | 5 2 5 2 | 3+4 2 3+4 2 |
| 65.2.28 | 413191 | 146 | Al5Br2La4 |
| | -GM -T | -Y -Z | -R -S |
| | 5 2 5 2 | 6 2 6 2 | 3+4 2 3+4 2 |
| | 5 2 5 2 | 6 2 6 2 | 5+6 2 5+6 2 |
| | 6 2 6 2 | 5 2 5 2 | 3+4 2 3+4 2 |
| | 6 2 6 2 | 5 2 5 2 | 5+6 2 5+6 2 |
| | 5 2 5 2 | 5 2 5 2 | 3+4 2 3+4 2 |
| | 6 2 6 2 | 6 2 6 2 | 5+6 2 5+6 2 |
| | 5 2 5 2 | 5 2 5 2 | 3+4 2 3+4 2 |
| | 6 2 6 2 | 6 2 6 2 | 5+6 2 5+6 2 |
| | 5 2 6 2 | 6 2 6 2 | 5+6 2 5+6 2 |
| | 6 2 5 2 | 5 2 5 2 | 3+4 2 3+4 2 |
| | 5 2 5 2 | 5 2 5 2 | 3+4 2 3+4 2 |
| | 6 2 6 2 | 6 2 6 2 | 5+6 2 5+6 2 |
| | 5 2 6 2 | 6 2 6 2 | 5+6 2 3+4 2 |
| | 6 2 5 2 | 5 2 5 2 | 3+4 2 5+6 2 |
| | 5 2 5 2 | 5 2 5 2 | 5+6 2 5+6 2 |
| | 6 2 6 2 | 6 2 6 2 | 3+4 2 3+4 2 |
| | 6 2 6 2 | 6 2 6 2 | 5+6 2 5+6 2 |
| | 5 2 5 2 | 5 2 5 2 | 3+4 2 3+4 2 |
| | 5 2 5 2 | 5 2 5 2 | 3+4 2 3+4 2 |
| | 6 2 6 2 | 6 2 6 2 | 5+6 2 5+6 2 |
| | 6 2 5 2 | 6 2 6 2 | 3+4 2 5+6 2 |
| | 5 2 6 2 | 5 2 5 2 | 5+6 2 3+4 2 |
| | 5 2 5 2 | 6 2 6 2 | 5+6 2 3+4 2 |
| | 6 2 6 2 | 5 2 5 2 | 3+4 2 5+6 2 |
| | 5 2 5 2 | 5 2 6 2 | 3+4 2 3+4 2 |
| | 6 2 6 2 | 6 2 5 2 | 5+6 2 5+6 2 |
| | 5 2 6 2 | 6 2 6 2 | 5+6 2 5+6 2 |
| | 6 2 5 2 | 5 2 5 2 | 3+4 2 3+4 2 |
| | 5 2 6 2 | 6 2 5 2 | 3+4 2 3+4 2 |
| | 6 2 5 2 | 5 2 6 2 | 5+6 2 5+6 2 |
| | 5 2 6 2 | 6 2 6 2 | 5+6 2 5+6 2 |
| | 6 2 5 2 | 5 2 5 2 | 3+4 2 3+4 2 |
| | 5 2 6 2 | 5 2 5 2 | 5+6 2 5+6 2 |
| | 6 2 5 2 | 6 2 6 2 | 3+4 2 3+4 2 |
| | 5 2 5 2 | 5 2 5 2 | 3+4 2 3+4 2 |
| | 6 2 6 2 | 6 2 6 2 | 5+6 2 5+6 2 |
| | 5 2 6 2 | 5 2 6 2 | 3+4 2 5+6 2 |
| | 5 2 5 2 | 6 2 5 2 | 5+6 2 3+4 2 |
| | 6 2 6 2 | 6 2 5 2 | 3+4 2 5+6 2 |
| | 6 2 5 2 | 5 2 6 2 | 5+6 2 3+4 2 |
| | 5 2 6 2 | 5 2 5 2 | 3+4 2 5+6 2 |
| | 6 2 6 2 | 6 2 5 2 | 5+6 2 3+4 2 |
| | 5 2 5 2 | 5 2 6 2 | 3+4 2 5+6 2 |
| | 6 2 6 2 | 6 2 5 2 | 5+6 2 3+4 2 |
| | 5 2 5 2 | 5 2 6 2 | 3+4 2 5+6 2 |
| | 6 2 6 2 | 6 2 5 2 | 5+6 2 3+4 2 |
| | 5 2 5 2 | 5 2 6 2 | 3+4 2 5+6 2 |
| | 5 2 6 2 | 5 2 5 2 | 5+6 2 3+4 2 |
| | 6 2 6 2 | 6 2 6 2 | 3+4 2 5+6 2 |
| | 5 2 6 2 | 6 2 5 2 | 5+6 2 3+4 2 |
| | 6 2 5 2 | 5 2 6 2 | 5+6 2 3+4 2 |
| | 6 2 6 2 | 6 2 6 2 | 5+6 2 5+6 2 |
| | 5 2 5 2 | 5 2 5 2 | 3+4 2 5+6 2 |
| | 6 2 5 2 | 6 2 5 2 | 3+4 2 3+4 2 |
| | 5 2 5 2 | 6 2 6 2 | 3+4 2 3+4 2 |
| | 5 2 6 2 | 5 2 6 2 | 5+6 2 5+6 2 |
| | 6 2 5 2 | 6 2 6 2 | 3+4 2 5+6 2 |
| | 5 2 5 2 | 5 2 5 2 | 5+6 2 3+4 2 |
| | 6 2 6 2 | 5 2 5 2 | 5+6 2 5+6 2 |
| | 6 2 5 2 | 6 2 6 2 | 3+4 2 3+4 2 |
| | 6 2 6 2 | 5 2 5 2 | 5+6 2 5+6 2 |
| | 6 2 5 2 | 5 2 6 2 | 3+4 2 3+4 2 |
| | 5 2 6 2 | 6 2 5 2 | 3+4 2 5+6 2 |

Continued on next column

**TABLE XL– continued**

| TQC | ICSD | NoE | SF |
|---|---|---|---|
| | 5 2 6 2 | 5 2 5 2 | 5+6 2 3+4 2 |
| | 5 2 5 2 | 5 2 6 2 | 3+4 2 5+6 2 |
| | 5 2 5 2 | 6 2 5 2 | 3+4 2 3+4 2 |
| | 5 2 6 2 | 6 2 5 2 | 5+6 2 3+4 2 |
| | 6 2 6 2 | 6 2 6 2 | 5+6 2 5+6 2 |
| | 5 2 5 2 | 6 2 6 2 | 3+4 2 5+6 2 |
| | 6 2 5 2 | 6 2 5 2 | 5+6 2 5+6 2 |
| | 5 2 6 2 | 5 2 6 2 | 5+6 2 3+4 2 |
| | 5 2 5 2 | 6 2 5 2 | 3+4 2 3+4 2 |
| | 6 2 5 2 | 6 2 6 2 | 3+4 2 5+6 2 |
| 65.2.29 | 260798 | 78 | Al1B4Yb1 |
| | -GM -T | -Y -Z | -R -S |
| | 5 2 5 2 | 5 2 5 2 | 5+6 2 5+6 2 |
| | 6 2 6 2 | 6 2 6 2 | 3+4 2 3+4 2 |
| | 5 2 6 2 | 6 2 5 2 | 5+6 2 5+6 2 |
| | 6 2 5 2 | 5 2 6 2 | 3+4 2 3+4 2 |
| | 5 2 5 2 | 6 2 6 2 | 3+4 2 5+6 2 |
| | 6 2 6 2 | 5 2 5 2 | 5+6 2 3+4 2 |
| | 5 2 6 2 | 6 2 5 2 | 5+6 2 5+6 2 |
| | 6 2 5 2 | 5 2 6 2 | 3+4 2 3+4 2 |
| | 5 2 6 2 | 5 2 6 2 | 5+6 2 3+4 2 |
| | 6 2 5 2 | 6 2 5 2 | 3+4 2 5+6 2 |
| | 5 2 5 2 | 6 2 5 2 | 3+4 2 5+6 2 |
| | 5 2 6 2 | 5 2 6 2 | 3+4 2 3+4 2 |
| | 6 2 5 2 | 5 2 6 2 | 5+6 2 3+4 2 |
| | 5 2 6 2 | 6 2 5 2 | 5+6 2 5+6 2 |
| | 6 2 6 2 | 6 2 6 2 | 3+4 2 5+6 2 |
| | 6 2 5 2 | 5 2 6 2 | 5+6 2 3+4 2 |
| | 5 2 6 2 | 6 2 5 2 | 3+4 2 5+6 2 |
| | 6 2 5 2 | 6 2 5 2 | 5+6 2 5+6 2 |
| | 5 2 5 2 | 5 2 5 2 | 5+6 2 3+4 2 |
| | 5 2 6 2 | 5 2 6 2 | 3+4 2 3+4 2 |
| | 5 2 6 2 | 5 2 5 2 | 3+4 2 5+6 2 |
| | 6 2 5 2 | 6 2 6 2 | 5+6 2 3+4 2 |
| | 6 2 6 2 | 5 2 6 2 | 3+4 2 5+6 2 |
| | 6 2 5 2 | 6 2 6 2 | 5+6 2 3+4 2 |
| | 5 2 5 2 | 5 2 5 2 | 3+4 2 3+4 2 |
| | 5 2 6 2 | 6 2 6 2 | 5+6 2 5+6 2 |
| | 6 2 6 2 | 5 2 5 2 | 3+4 2 5+6 2 |
| | 6 2 5 2 | 6 2 5 2 | 5+6 2 3+4 2 |
| | 5 2 6 2 | 5 2 6 2 | 5+6 2 3+4 2 |
| | 5 2 5 2 | 6 2 5 2 | 3+4 2 5+6 2 |
| | 5 2 5 2 | 6 2 6 2 | 5+6 2 3+4 2 |
| | 5 2 6 2 | 5 2 6 2 | 3+4 2 5+6 2 |
| | 6 2 6 2 | 5 2 6 2 | 3+4 2 3+4 2 |
| | 6 2 5 2 | 6 2 5 2 | 5+6 2 3+4 2 |
| | 6 2 5 2 | 6 2 5 2 | 3+4 2 5+6 2 |
| | 5 2 6 2 | 5 2 5 2 | 5+6 2 5+6 2 |
| | 5 2 6 2 | 6 2 6 2 | 5+6 2 3+4 2 |
| | 6 2 5 2 | 5 2 5 2 | 5+6 2 5+6 2 |
| | 5 2 6 2 | 5 2 6 2 | 3+4 2 5+6 2 |
| 65.2.30 | 167533 | 78 | Al1B4Yb1 |
| | -GM -T | -Y -Z | -R -S |
| | 5 2 5 2 | 5 2 5 2 | 5+6 2 5+6 2 |
| | 6 2 6 2 | 6 2 6 2 | 3+4 2 3+4 2 |
| | 5 2 6 2 | 6 2 5 2 | 5+6 2 5+6 2 |
| | 6 2 5 2 | 5 2 6 2 | 3+4 2 3+4 2 |
| | 5 2 5 2 | 6 2 6 2 | 3+4 2 5+6 2 |
| | 6 2 6 2 | 5 2 5 2 | 5+6 2 3+4 2 |
| | 5 2 6 2 | 6 2 5 2 | 5+6 2 5+6 2 |
| | 6 2 5 2 | 5 2 6 2 | 3+4 2 3+4 2 |

Continued on next column

**TABLE XL– continued**

| TQC | ICSD | NoE | SF |
|---|---|---|---|
| | 5 2 6 2 | 5 2 6 2 | 5+6 2 3+4 2 |
| | 6 2 5 2 | 6 2 5 2 | 3+4 2 5+6 2 |
| | 5 2 5 2 | 6 2 5 2 | 3+4 2 5+6 2 |
| | 5 2 6 2 | 5 2 6 2 | 3+4 2 3+4 2 |
| | 6 2 5 2 | 5 2 6 2 | 5+6 2 3+4 2 |
| | 5 2 6 2 | 6 2 5 2 | 5+6 2 5+6 2 |
| | 6 2 6 2 | 6 2 6 2 | 3+4 2 5+6 2 |
| | 6 2 5 2 | 5 2 6 2 | 5+6 2 3+4 2 |
| | 5 2 6 2 | 6 2 5 2 | 3+4 2 5+6 2 |
| | 6 2 5 2 | 6 2 5 2 | 5+6 2 5+6 2 |
| | 5 2 5 2 | 5 2 5 2 | 5+6 2 3+4 2 |
| | 5 2 6 2 | 5 2 6 2 | 3+4 2 3+4 2 |
| | 5 2 6 2 | 5 2 5 2 | 3+4 2 5+6 2 |
| | 6 2 5 2 | 6 2 6 2 | 5+6 2 3+4 2 |
| | 6 2 6 2 | 5 2 6 2 | 3+4 2 5+6 2 |
| | 6 2 5 2 | 6 2 6 2 | 5+6 2 3+4 2 |
| | 5 2 5 2 | 5 2 5 2 | 3+4 2 3+4 2 |
| | 5 2 6 2 | 6 2 6 2 | 5+6 2 5+6 2 |
| | 6 2 6 2 | 5 2 5 2 | 3+4 2 5+6 2 |
| | 6 2 5 2 | 6 2 5 2 | 5+6 2 3+4 2 |
| | 5 2 6 2 | 5 2 6 2 | 5+6 2 3+4 2 |
| | 5 2 5 2 | 6 2 5 2 | 3+4 2 5+6 2 |
| | 5 2 5 2 | 6 2 6 2 | 5+6 2 3+4 2 |
| | 5 2 6 2 | 5 2 6 2 | 3+4 2 5+6 2 |
| | 6 2 6 2 | 5 2 6 2 | 3+4 2 3+4 2 |
| | 6 2 5 2 | 6 2 5 2 | 5+6 2 3+4 2 |
| | 6 2 5 2 | 6 2 5 2 | 3+4 2 5+6 2 |
| | 5 2 6 2 | 5 2 5 2 | 5+6 2 5+6 2 |
| | 5 2 6 2 | 6 2 6 2 | 5+6 2 3+4 2 |
| | 6 2 5 2 | 5 2 5 2 | 5+6 2 5+6 2 |
| | 5 2 6 2 | 5 2 6 2 | 3+4 2 5+6 2 |
| 65.2.31 | 167535 | 80 | Al1B4Lu1 |
| | -GM -T | -Y -Z | -R -S |
| | 5 2 5 2 | 5 2 5 2 | 5+6 2 5+6 2 |
| | 6 2 6 2 | 6 2 6 2 | 3+4 2 3+4 2 |
| | 5 2 6 2 | 6 2 5 2 | 5+6 2 5+6 2 |
| | 6 2 5 2 | 5 2 6 2 | 3+4 2 3+4 2 |
| | 5 2 5 2 | 6 2 5 2 | 5+6 2 5+6 2 |
| | 6 2 6 2 | 5 2 6 2 | 3+4 2 3+4 2 |
| | 5 2 6 2 | 6 2 5 2 | 5+6 2 5+6 2 |
| | 6 2 5 2 | 5 2 6 2 | 3+4 2 3+4 2 |
| | 5 2 6 2 | 5 2 6 2 | 5+6 2 3+4 2 |
| | 6 2 5 2 | 6 2 5 2 | 3+4 2 5+6 2 |
| | 5 2 5 2 | 6 2 5 2 | 3+4 2 5+6 2 |
| | 5 2 6 2 | 5 2 6 2 | 3+4 2 3+4 2 |
| | 6 2 5 2 | 5 2 6 2 | 5+6 2 3+4 2 |
| | 5 2 6 2 | 6 2 5 2 | 5+6 2 5+6 2 |
| | 6 2 6 2 | 6 2 6 2 | 3+4 2 5+6 2 |
| | 6 2 5 2 | 5 2 6 2 | 5+6 2 3+4 2 |
| | 6 2 6 2 | 5 2 6 2 | 3+4 2 3+4 2 |
| | 5 2 5 2 | 5 2 5 2 | 3+4 2 3+4 2 |
| | 6 2 5 2 | 6 2 6 2 | 5+6 2 5+6 2 |
| | 5 2 6 2 | 6 2 5 2 | 3+4 2 5+6 2 |
| | 5 2 5 2 | 6 2 6 2 | 5+6 2 5+6 2 |
| | 6 2 6 2 | 5 2 5 2 | 5+6 2 3+4 2 |
| | 5 2 5 2 | 5 2 5 2 | 3+4 2 5+6 2 |
| | 6 2 6 2 | 6 2 5 2 | 5+6 2 3+4 2 |
| | 5 2 5 2 | 6 2 5 2 | 5+6 2 5+6 2 |
| | 5 2 6 2 | 5 2 6 2 | 3+4 2 3+4 2 |
| | 6 2 5 2 | 6 2 6 2 | 3+4 2 5+6 2 |
| | 5 2 6 2 | 5 2 5 2 | 3+4 2 3+4 2 |

Continued on next column

**TABLE XL– continued**

| TQC | ICSD | NoE | SF | | | |
|---|---|---|---|---|---|---|
| | 6 2 | 5 2 | 6 2 | 6 2 | 5+6 2 | 5+6 2 |
| | 5 2 | 5 2 | 6 2 | 6 2 | 3+4 2 | 3+4 2 |
| | 6 2 | 6 2 | 5 2 | 5 2 | 5+6 2 | 5+6 2 |
| | 6 2 | 6 2 | 6 2 | 5 2 | 5+6 2 | 5+6 2 |
| | 5 2 | 5 2 | 5 2 | 5 2 | 3+4 2 | 3+4 2 |
| | 6 2 | 6 2 | 5 2 | 6 2 | 5+6 2 | 3+4 2 |
| | 6 2 | 6 2 | 5 2 | 5 2 | 3+4 2 | 5+6 2 |
| | 5 2 | 6 2 | 6 2 | 6 2 | 5+6 2 | 3+4 2 |
| | 5 2 | 5 2 | 5 2 | 5 2 | 3+4 2 | 3+4 2 |
| | 5 2 | 5 2 | 6 2 | 6 2 | 5+6 2 | 5+6 2 |
| | 5 2 | 5 2 | 6 2 | 6 2 | 5+6 2 | 5+6 2 |
| | 5 2 | 6 2 | 5 2 | 6 2 | 3+4 2 | 3+4 2 |

The End

TABLE XLI: The computed irreducible representations for the valence bands for the compounds of SEBR and NLC in space group 67. See the list of tables.

| TQC | ICSD | NoE | SF | | | |
|---|---|---|---|---|---|---|
| 67.2.1 | 167666 | 52 | Al2Cu1Ir1 | | | |
| | -GM | -T | -Y | -Z | -R | -S |
| | 5 2 | 6 2 | 6 2 | 6 2 | 2+2 4 | 2+2 4 |
| | 6 2 | 5 2 | 5 2 | 5 2 | 2+2 4 | 2+2 4 |
| | 5 2 | 6 2 | 5 2 | 5 2 | 2+2 4 | 2+2 4 |
| | 6 2 | 5 2 | 5 2 | 5 2 | 2+2 4 | 2+2 4 |
| | 6 2 | 6 2 | 6 2 | 6 2 | 2+2 4 | 2+2 4 |
| | 5 2 | 6 2 | 5 2 | 6 2 | 2+2 4 | 2+2 4 |
| | 5 2 | 5 2 | 6 2 | 5 2 | 2+2 4 | 2+2 4 |
| | 6 2 | 6 2 | 6 2 | 6 2 | 2+2 4 | 2+2 4 |
| | 5 2 | 6 2 | 6 2 | 6 2 | 2+2 4 | 2+2 4 |
| | 6 2 | 6 2 | 5 2 | 5 2 | 2+2 4 | 2+2 4 |
| | 5 2 | 5 2 | 6 2 | 5 2 | 2+2 4 | 2+2 4 |
| | 6 2 | 5 2 | 6 2 | 6 2 | 2+2 4 | 2+2 4 |
| | 5 2 | 5 2 | 5 2 | 5 2 | 2+2 4 | 2+2 4 |
| | 6 2 | 5 2 | 6 2 | 5 2 | | |
| | 6 2 | 6 2 | 5 2 | 6 2 | | |
| | 5 2 | 6 2 | 5 2 | 6 2 | | |
| | 5 2 | 5 2 | 6 2 | 6 2 | | |
| | 5 2 | 5 2 | 5 2 | 5 2 | | |
| | 5 2 | 5 2 | 5 2 | 5 2 | | |
| | 5 2 | 6 2 | 6 2 | 5 2 | | |
| | 5 2 | 5 2 | 5 2 | 6 2 | | |
| | 6 2 | 6 2 | 6 2 | 6 2 | | |
| | 6 2 | 5 2 | 6 2 | 6 2 | | |
| | 5 2 | 6 2 | 6 2 | 5 2 | | |
| | 6 2 | 5 2 | 5 2 | 6 2 | | |
| | 6 2 | 5 2 | 6 2 | 5 2 | | |

The End

TABLE XLII: The computed irreducible representations for the valence bands for the compounds of SEBR and NLC in space group 69. See the list of tables.

| TQC | ICSD | NoE | SF | | |
|---|---|---|---|---|---|
| 69.1.1 | 262559 | 196 | Pt3Rb2Se6U1 | | |
| | -GM | -Y | -Z | -T | -L |
| | 5 2 | 5 2 | 5 2 | 5 2 | 2+2 2 |
| | 6 2 | 6 2 | 6 2 | 6 2 | 3+3 2 |
| | 5 2 | 5 2 | 5 2 | 5 2 | 2+2 2 |
| | 6 2 | 6 2 | 6 2 | 6 2 | 3+3 2 |

Continued on next column

**TABLE XLII– continued**

| TQC | ICSD | NoE | SF | | |
|---|---|---|---|---|---|
| | 5 2 | 5 2 | 6 2 | 6 2 | 2+2 2 |
| | 6 2 | 6 2 | 5 2 | 5 2 | 3+3 2 |
| | 5 2 | 5 2 | 5 2 | 5 2 | 2+2 2 |
| | 6 2 | 6 2 | 6 2 | 6 2 | 3+3 2 |
| | 6 2 | 5 2 | 6 2 | 5 2 | 2+2 2 |
| | 5 2 | 6 2 | 5 2 | 6 2 | 3+3 2 |
| | 6 2 | 5 2 | 6 2 | 5 2 | 3+3 2 |
| | 5 2 | 6 2 | 5 2 | 6 2 | 2+2 2 |
| | 5 2 | 6 2 | 5 2 | 6 2 | 3+3 2 |
| | 6 2 | 5 2 | 6 2 | 5 2 | 2+2 2 |
| | 6 2 | 5 2 | 6 2 | 6 2 | 3+3 2 |
| | 5 2 | 6 2 | 5 2 | 5 2 | 2+2 2 |
| | 5 2 | 6 2 | 6 2 | 6 2 | 2+2 2 |
| | 6 2 | 5 2 | 5 2 | 5 2 | 3+3 2 |
| | 5 2 | 5 2 | 6 2 | 6 2 | 2+2 2 |
| | 6 2 | 6 2 | 6 2 | 5 2 | 3+3 2 |
| | 6 2 | 5 2 | 5 2 | 5 2 | 2+2 2 |
| | 6 2 | 6 2 | 6 2 | 6 2 | 3+3 2 |
| | 5 2 | 5 2 | 5 2 | 5 2 | 2+2 2 |
| | 5 2 | 6 2 | 5 2 | 6 2 | 3+3 2 |
| | 5 2 | 6 2 | 6 2 | 5 2 | 3+3 2 |
| | 6 2 | 5 2 | 5 2 | 6 2 | 2+2 2 |
| | 6 2 | 6 2 | 6 2 | 6 2 | 2+2 2 |
| | 5 2 | 5 2 | 5 2 | 5 2 | 3+3 2 |
| | 5 2 | 6 2 | 5 2 | 5 2 | 2+2 2 |
| | 6 2 | 5 2 | 6 2 | 6 2 | 3+3 2 |
| | 5 2 | 6 2 | 6 2 | 6 2 | 3+3 2 |
| | 6 2 | 5 2 | 5 2 | 5 2 | 2+2 2 |
| | 6 2 | 5 2 | 6 2 | 5 2 | 3+3 2 |
| | 5 2 | 6 2 | 5 2 | 6 2 | 2+2 2 |
| | 5 2 | 5 2 | 6 2 | 6 2 | 2+2 2 |
| | 6 2 | 6 2 | 5 2 | 5 2 | 3+3 2 |
| | 5 2 | 6 2 | 5 2 | 5 2 | 3+3 2 |
| | 6 2 | 5 2 | 6 2 | 6 2 | 2+2 2 |
| | 6 2 | 6 2 | 6 2 | 6 2 | 3+3 2 |
| | 5 2 | 5 2 | 5 2 | 5 2 | 2+2 2 |
| | 6 2 | 6 2 | 6 2 | 6 2 | 3+3 2 |
| | 5 2 | 5 2 | 5 2 | 5 2 | 2+2 2 |
| | 5 2 | 5 2 | 5 2 | 5 2 | 2+2 2 |
| | 6 2 | 6 2 | 6 2 | 6 2 | 3+3 2 |
| | 5 2 | 5 2 | 5 2 | 5 2 | 2+2 2 |
| | 6 2 | 6 2 | 6 2 | 6 2 | 3+3 2 |
| | 6 2 | 6 2 | 6 2 | 6 2 | 3+3 2 |
| | 6 2 | 6 2 | 6 2 | 6 2 | 3+3 2 |
| | 5 2 | 5 2 | 5 2 | 5 2 | 2+2 2 |
| | 5 2 | 5 2 | 5 2 | 5 2 | 2+2 2 |
| | 5 2 | 5 2 | 5 2 | 5 2 | 2+2 2 |
| | 5 2 | 6 2 | 5 2 | 6 2 | 3+3 2 |
| | 6 2 | 5 2 | 6 2 | 5 2 | 2+2 2 |
| | 6 2 | 6 2 | 6 2 | 6 2 | 3+3 2 |
| | 6 2 | 6 2 | 6 2 | 5 2 | 2+2 2 |
| | 5 2 | 5 2 | 5 2 | 6 2 | 3+3 2 |
| | 5 2 | 5 2 | 5 2 | 5 2 | 2+2 2 |
| | 6 2 | 5 2 | 5 2 | 5 2 | 2+2 2 |
| | 5 2 | 5 2 | 6 2 | 5 2 | 2+2 2 |
| | 6 2 | 6 2 | 6 2 | 6 2 | 3+3 2 |
| | 6 2 | 5 2 | 5 2 | 6 2 | 2+2 2 |
| | 5 2 | 6 2 | 6 2 | 5 2 | 3+3 2 |
| | 5 2 | 6 2 | 5 2 | 6 2 | 3+3 2 |
| | 5 2 | 6 2 | 5 2 | 6 2 | 3+3 2 |
| | 6 2 | 5 2 | 6 2 | 5 2 | 2+2 2 |

Continued on next column

**TABLE XLII– continued**

| TQC | ICSD | NoE | SF |
|---|---|---|---|
| | 6 2 6 2 | 6 2 6 2 | 3+3 2 |
| | 6 2 6 2 | 6 2 6 2 | 3+3 2 |
| | 5 2 5 2 | 6 2 5 2 | 2+2 2 |
| | 6 2 6 2 | 5 2 6 2 | 3+3 2 |
| | 5 2 6 2 | 5 2 6 2 | 3+3 2 |
| | 5 2 5 2 | 6 2 6 2 | 2+2 2 |
| | 6 2 6 2 | 6 2 5 2 | 3+3 2 |
| | 6 2 5 2 | 6 2 6 2 | 2+2 2 |
| | 5 2 6 2 | 5 2 5 2 | 3+3 2 |
| | 6 2 6 2 | 6 2 5 2 | 3+3 2 |
| | 5 2 5 2 | 5 2 6 2 | 2+2 2 |
| | 5 2 5 2 | 5 2 5 2 | 2+2 2 |
| | 5 2 6 2 | 5 2 5 2 | 3+3 2 |
| | 6 2 5 2 | 5 2 6 2 | 2+2 2 |
| | 5 2 6 2 | 5 2 6 2 | 3+3 2 |
| | 5 2 5 2 | 6 2 5 2 | 2+2 2 |
| | 6 2 5 2 | 5 2 6 2 | 2+2 2 |
| | 6 2 5 2 | 6 2 5 2 | 3+3 2 |
| | 6 2 6 2 | 6 2 5 2 | 2+2 2 |
| | 6 2 5 2 | 5 2 6 2 | 3+3 2 |
| | 5 2 6 2 | 6 2 5 2 | 2+2 2 |
| | 5 2 6 2 | 5 2 5 2 | 3+3 2 |
| | 5 2 5 2 | 5 2 6 2 | 2+2 2 |
| | 6 2 6 2 | 6 2 5 2 | 2+2 2 |
| | 6 2 5 2 | 6 2 6 2 | 3+3 2 |
| | 5 2 5 2 | 5 2 5 2 | 2+2 2 |
| | 6 2 5 2 | 6 2 5 2 | 3+3 2 |
| | 5 2 6 2 | 5 2 6 2 | 2+2 2 |
| | 6 2 6 2 | 6 2 6 2 | 3+3 2 |
| | 6 2 5 2 | 6 2 5 2 | 2+2 2 |
| | 5 2 6 2 | 5 2 6 2 | 3+3 2 |
| | 5 2 5 2 | 5 2 5 2 | 2+2 2 |
| | 5 2 6 2 | 5 2 6 2 | 3+3 2 |
| 69.1.2 | 262561 | 196 | Pd3Rb2Se6U1 |
| | -GM -Y | -Z -T | -L |
| | 5 2 5 2 | 5 2 5 2 | 2+2 2 |
| | 6 2 6 2 | 6 2 6 2 | 3+3 2 |
| | 5 2 5 2 | 5 2 5 2 | 2+2 2 |
| | 6 2 6 2 | 6 2 6 2 | 3+3 2 |
| | 5 2 5 2 | 6 2 6 2 | 2+2 2 |
| | 6 2 6 2 | 5 2 5 2 | 3+3 2 |
| | 5 2 5 2 | 5 2 5 2 | 2+2 2 |
| | 6 2 6 2 | 6 2 6 2 | 3+3 2 |
| | 6 2 5 2 | 6 2 5 2 | 2+2 2 |
| | 5 2 6 2 | 5 2 6 2 | 3+3 2 |
| | 6 2 5 2 | 6 2 5 2 | 2+2 2 |
| | 5 2 6 2 | 5 2 6 2 | 3+3 2 |
| | 5 2 6 2 | 5 2 6 2 | 3+3 2 |
| | 6 2 5 2 | 6 2 5 2 | 2+2 2 |
| | 6 2 5 2 | 6 2 6 2 | 3+3 2 |
| | 5 2 6 2 | 5 2 5 2 | 2+2 2 |
| | 5 2 6 2 | 6 2 6 2 | 3+3 2 |
| | 6 2 5 2 | 5 2 5 2 | 2+2 2 |
| | 5 2 5 2 | 6 2 5 2 | 3+3 2 |
| | 6 2 6 2 | 5 2 6 2 | 2+2 2 |
| | 6 2 5 2 | 6 2 5 2 | 2+2 2 |
| | 6 2 6 2 | 6 2 6 2 | 3+3 2 |
| | 5 2 5 2 | 5 2 5 2 | 2+2 2 |
| | 5 2 6 2 | 5 2 6 2 | 3+3 2 |
| | 5 2 6 2 | 6 2 5 2 | 3+3 2 |
| | 6 2 5 2 | 5 2 6 2 | 2+2 2 |

Continued on next column

**TABLE XLII– continued**

| TQC | ICSD | NoE | SF |
|---|---|---|---|
| | 6 2 6 2 | 6 2 5 2 | 2+2 2 |
| | 5 2 5 2 | 5 2 6 2 | 3+3 2 |
| | 5 2 6 2 | 6 2 5 2 | 2+2 2 |
| | 6 2 5 2 | 5 2 6 2 | 3+3 2 |
| | 5 2 6 2 | 6 2 6 2 | 3+3 2 |
| | 6 2 5 2 | 5 2 5 2 | 2+2 2 |
| | 6 2 5 2 | 6 2 5 2 | 3+3 2 |
| | 5 2 6 2 | 5 2 6 2 | 2+2 2 |
| | 5 2 5 2 | 6 2 6 2 | 2+2 2 |
| | 6 2 6 2 | 5 2 5 2 | 3+3 2 |
| | 5 2 6 2 | 5 2 5 2 | 2+2 2 |
| | 6 2 5 2 | 6 2 6 2 | 3+3 2 |
| | 6 2 6 2 | 6 2 6 2 | 3+3 2 |
| | 6 2 6 2 | 6 2 5 2 | 2+2 2 |
| | 5 2 5 2 | 5 2 6 2 | 3+3 2 |
| | 6 2 5 2 | 5 2 6 2 | 2+2 2 |
| | 5 2 6 2 | 6 2 5 2 | 3+3 2 |
| | 5 2 6 2 | 5 2 5 2 | 2+2 2 |
| | 5 2 5 2 | 5 2 6 2 | 3+3 2 |
| | 6 2 5 2 | 6 2 5 2 | 2+2 2 |
| | 6 2 6 2 | 6 2 6 2 | 3+3 2 |
| | 5 2 6 2 | 5 2 6 2 | 3+3 2 |
| | 5 2 5 2 | 5 2 5 2 | 2+2 2 |
| | 6 2 5 2 | 6 2 5 2 | 2+2 2 |
| | 5 2 5 2 | 5 2 5 2 | 2+2 2 |
| | 5 2 6 2 | 5 2 6 2 | 3+3 2 |
| | 6 2 5 2 | 6 2 5 2 | 2+2 2 |
| | 6 2 6 2 | 6 2 5 2 | 2+2 2 |
| | 6 2 5 2 | 6 2 6 2 | 3+3 2 |
| | 5 2 6 2 | 5 2 6 2 | 3+3 2 |
| | 5 2 5 2 | 5 2 5 2 | 2+2 2 |
| | 6 2 5 2 | 6 2 5 2 | 2+2 2 |
| | 5 2 5 2 | 5 2 5 2 | 2+2 2 |
| | 6 2 5 2 | 6 2 6 2 | 3+3 2 |
| | 6 2 6 2 | 5 2 5 2 | 2+2 2 |
| | 5 2 6 2 | 5 2 6 2 | 3+3 2 |
| | 5 2 6 2 | 6 2 6 2 | 3+3 2 |
| | 5 2 6 2 | 6 2 6 2 | 3+3 2 |
| | 6 2 5 2 | 5 2 5 2 | 2+2 2 |
| | 6 2 6 2 | 6 2 6 2 | 3+3 2 |
| | 6 2 6 2 | 6 2 6 2 | 3+3 2 |
| | 6 2 5 2 | 6 2 5 2 | 2+2 2 |
| | 5 2 6 2 | 6 2 6 2 | 3+3 2 |
| | 6 2 5 2 | 5 2 6 2 | 3+3 2 |
| | 5 2 6 2 | 6 2 6 2 | 2+2 2 |
| | 5 2 6 2 | 5 2 6 2 | 3+3 2 |
| | 5 2 5 2 | 5 2 5 2 | 3+3 2 |
| | 6 2 6 2 | 6 2 5 2 | 2+2 2 |
| | 5 2 6 2 | 5 2 5 2 | 3+3 2 |
| | 5 2 5 2 | 6 2 5 2 | 2+2 2 |
| | 6 2 6 2 | 5 2 6 2 | 2+2 2 |
| | 5 2 5 2 | 5 2 5 2 | 3+3 2 |
| | 5 2 6 2 | 5 2 6 2 | 3+3 2 |
| | 6 2 5 2 | 5 2 6 2 | 2+2 2 |
| | 6 2 5 2 | 6 2 5 2 | 2+2 2 |
| | 5 2 5 2 | 5 2 6 2 | 2+2 2 |
| | 5 2 6 2 | 6 2 5 2 | 3+3 2 |
| | 6 2 5 2 | 5 2 5 2 | 2+2 2 |
| | 5 2 5 2 | 6 2 6 2 | 2+2 2 |
| | 6 2 6 2 | 5 2 5 2 | 3+3 2 |
| | 6 2 5 2 | 6 2 5 2 | 2+2 2 |

Continued on next column

**TABLE XLII– continued**

| TQC | ICSD | NoE | SF |
|---|---|---|---|
| | 5 2 6 2 | 5 2 5 2 | 3+3 2 |
| | 6 2 5 2 | 6 2 6 2 | 2+2 2 |
| | 6 2 6 2 | 6 2 6 2 | 2+2 2 |
| | 6 2 5 2 | 6 2 5 2 | 3+3 2 |
| | 6 2 6 2 | 6 2 5 2 | 3+3 2 |
| | 5 2 5 2 | 6 2 6 2 | 2+2 2 |
| | 6 2 6 2 | 5 2 6 2 | 3+3 2 |
| | 5 2 5 2 | 5 2 5 2 | 2+2 2 |
| | 5 2 6 2 | 5 2 6 2 | 3+3 2 |
| | 5 2 5 2 | 5 2 5 2 | 2+2 2 |
| | 5 2 6 2 | 5 2 6 2 | 3+3 2 |
| 69.1.3 | 262562 | 196 | Pd3Rb2S6U1 |
| | -GM -Y | -Z -T | -L |
| | 5 2 5 2 | 5 2 5 2 | 2+2 2 |
| | 6 2 6 2 | 6 2 6 2 | 3+3 2 |
| | 5 2 5 2 | 5 2 5 2 | 2+2 2 |
| | 6 2 6 2 | 6 2 6 2 | 3+3 2 |
| | 5 2 5 2 | 6 2 6 2 | 2+2 2 |
| | 6 2 6 2 | 5 2 5 2 | 3+3 2 |
| | 5 2 5 2 | 5 2 5 2 | 2+2 2 |
| | 6 2 6 2 | 6 2 6 2 | 3+3 2 |
| | 6 2 5 2 | 6 2 5 2 | 2+2 2 |
| | 5 2 6 2 | 5 2 6 2 | 3+3 2 |
| | 6 2 5 2 | 6 2 5 2 | 2+2 2 |
| | 5 2 6 2 | 5 2 6 2 | 3+3 2 |
| | 5 2 6 2 | 5 2 6 2 | 3+3 2 |
| | 6 2 5 2 | 6 2 5 2 | 2+2 2 |
| | 6 2 5 2 | 6 2 5 2 | 3+3 2 |
| | 6 2 6 2 | 5 2 6 2 | 2+2 2 |
| | 5 2 6 2 | 6 2 6 2 | 2+2 2 |
| | 5 2 5 2 | 5 2 6 2 | 3+3 2 |
| | 6 2 5 2 | 6 2 5 2 | 2+2 2 |
| | 5 2 6 2 | 6 2 5 2 | 3+3 2 |
| | 6 2 5 2 | 5 2 5 2 | 2+2 2 |
| | 6 2 6 2 | 6 2 6 2 | 3+3 2 |
| | 5 2 5 2 | 5 2 5 2 | 2+2 2 |
| | 5 2 6 2 | 5 2 6 2 | 3+3 2 |
| | 5 2 6 2 | 6 2 5 2 | 3+3 2 |
| | 6 2 5 2 | 5 2 6 2 | 2+2 2 |
| | 5 2 6 2 | 6 2 6 2 | 2+2 2 |
| | 6 2 5 2 | 5 2 5 2 | 3+3 2 |
| | 5 2 6 2 | 6 2 5 2 | 3+3 2 |
| | 5 2 5 2 | 5 2 6 2 | 2+2 2 |
| | 6 2 6 2 | 6 2 5 2 | 3+3 2 |
| | 6 2 5 2 | 5 2 6 2 | 2+2 2 |
| | 5 2 5 2 | 6 2 6 2 | 3+3 2 |
| | 6 2 6 2 | 6 2 5 2 | 2+2 2 |
| | 5 2 5 2 | 5 2 6 2 | 2+2 2 |
| | 6 2 6 2 | 5 2 5 2 | 3+3 2 |
| | 5 2 5 2 | 5 2 5 2 | 2+2 2 |
| | 6 2 6 2 | 6 2 6 2 | 3+3 2 |
| | 6 2 6 2 | 6 2 6 2 | 3+3 2 |
| | 6 2 5 2 | 6 2 6 2 | 2+2 2 |
| | 6 2 6 2 | 5 2 5 2 | 3+3 2 |
| | 5 2 5 2 | 6 2 5 2 | 2+2 2 |
| | 5 2 6 2 | 5 2 5 2 | 3+3 2 |
| | 5 2 5 2 | 5 2 6 2 | 2+2 2 |
| | 6 2 6 2 | 5 2 6 2 | 3+3 2 |
| | 5 2 5 2 | 6 2 6 2 | 2+2 2 |
| | 6 2 6 2 | 6 2 5 2 | 3+3 2 |
| | 6 2 6 2 | 6 2 6 2 | 3+3 2 |

Continued on next column

**TABLE XLII– continued**

| TQC | ICSD | NoE | SF |
|---|---|---|---|
| | 5 2 5 2 | 5 2 5 2 | 2+2 2 |
| | 5 2 5 2 | 5 2 5 2 | 2+2 2 |
| | 5 2 5 2 | 5 2 5 2 | 2+2 2 |
| | 5 2 6 2 | 5 2 6 2 | 3+3 2 |
| | 6 2 5 2 | 6 2 6 2 | 2+2 2 |
| | 6 2 6 2 | 6 2 5 2 | 3+3 2 |
| | 5 2 5 2 | 6 2 5 2 | 2+2 2 |
| | 6 2 6 2 | 5 2 6 2 | 3+3 2 |
| | 5 2 5 2 | 5 2 5 2 | 2+2 2 |
| | 6 2 5 2 | 6 2 5 2 | 2+2 2 |
| | 5 2 5 2 | 5 2 6 2 | 2+2 2 |
| | 6 2 5 2 | 5 2 5 2 | 2+2 2 |
| | 5 2 6 2 | 6 2 5 2 | 3+3 2 |
| | 6 2 6 2 | 5 2 6 2 | 3+3 2 |
| | 5 2 6 2 | 6 2 6 2 | 3+3 2 |
| | 5 2 6 2 | 6 2 6 2 | 3+3 2 |
| | 6 2 5 2 | 5 2 5 2 | 2+2 2 |
| | 6 2 6 2 | 6 2 6 2 | 3+3 2 |
| | 6 2 6 2 | 6 2 6 2 | 3+3 2 |
| | 6 2 5 2 | 6 2 5 2 | 2+2 2 |
| | 5 2 6 2 | 5 2 6 2 | 3+3 2 |
| | 5 2 5 2 | 5 2 6 2 | 3+3 2 |
| | 6 2 6 2 | 6 2 6 2 | 2+2 2 |
| | 6 2 6 2 | 6 2 6 2 | 3+3 2 |
| | 6 2 6 2 | 6 2 5 2 | 3+3 2 |
| | 5 2 5 2 | 6 2 5 2 | 2+2 2 |
| | 5 2 6 2 | 5 2 5 2 | 3+3 2 |
| | 5 2 6 2 | 5 2 5 2 | 2+2 2 |
| | 5 2 5 2 | 5 2 6 2 | 3+3 2 |
| | 5 2 5 2 | 5 2 5 2 | 2+2 2 |
| | 5 2 6 2 | 5 2 6 2 | 3+3 2 |
| | 6 2 5 2 | 5 2 6 2 | 2+2 2 |
| | 5 2 5 2 | 6 2 5 2 | 2+2 2 |
| | 6 2 5 2 | 5 2 5 2 | 2+2 2 |
| | 5 2 6 2 | 6 2 6 2 | 3+3 2 |
| | 6 2 5 2 | 5 2 5 2 | 2+2 2 |
| | 5 2 5 2 | 5 2 5 2 | 2+2 2 |
| | 6 2 6 2 | 6 2 6 2 | 3+3 2 |
| | 6 2 5 2 | 6 2 5 2 | 2+2 2 |
| | 5 2 6 2 | 5 2 5 2 | 3+3 2 |
| | 6 2 5 2 | 6 2 6 2 | 2+2 2 |
| | 6 2 5 2 | 6 2 6 2 | 2+2 2 |
| | 6 2 6 2 | 6 2 5 2 | 3+3 2 |
| | 5 2 5 2 | 5 2 5 2 | 2+2 2 |
| | 6 2 6 2 | 6 2 6 2 | 3+3 2 |
| | 5 2 6 2 | 6 2 6 2 | 3+3 2 |
| | 6 2 5 2 | 5 2 5 2 | 2+2 2 |
| | 5 2 6 2 | 5 2 6 2 | 3+3 2 |
| | 5 2 5 2 | 5 2 5 2 | 2+2 2 |
| | 5 2 6 2 | 5 2 6 2 | 3+3 2 |
| 69.1.4 | 262563 | 196 | K2Pd3S6U1 |
| | -GM -Y | -Z -T | -L |
| | 5 2 5 2 | 5 2 5 2 | 2+2 2 |
| | 6 2 6 2 | 6 2 6 2 | 3+3 2 |
| | 5 2 5 2 | 5 2 5 2 | 2+2 2 |
| | 6 2 6 2 | 6 2 6 2 | 3+3 2 |
| | 5 2 5 2 | 5 2 5 2 | 2+2 2 |
| | 6 2 6 2 | 6 2 6 2 | 3+3 2 |
| | 5 2 5 2 | 5 2 5 2 | 2+2 2 |
| | 6 2 6 2 | 6 2 6 2 | 3+3 2 |
| | 6 2 5 2 | 6 2 5 2 | 2+2 2 |

Continued on next column

**TABLE XLII– continued**

| TQC | ICSD | NoE | SF |
|---|---|---|---|
| | 5 2 6 2 | 5 2 6 2 | 3+3 2 |
| | 6 2 5 2 | 6 2 5 2 | 2+2 2 |
| | 5 2 6 2 | 5 2 6 2 | 3+3 2 |
| | 5 2 5 2 | 5 2 6 2 | 3+3 2 |
| | 6 2 6 2 | 6 2 5 2 | 2+2 2 |
| | 5 2 6 2 | 5 2 6 2 | 3+3 2 |
| | 6 2 5 2 | 6 2 5 2 | 2+2 2 |
| | 5 2 6 2 | 6 2 5 2 | 2+2 2 |
| | 6 2 5 2 | 5 2 6 2 | 3+3 2 |
| | 5 2 6 2 | 5 2 5 2 | 3+3 2 |
| | 5 2 5 2 | 6 2 6 2 | 2+2 2 |
| | 6 2 6 2 | 5 2 5 2 | 3+3 2 |
| | 6 2 5 2 | 6 2 6 2 | 2+2 2 |
| | 5 2 5 2 | 6 2 6 2 | 3+3 2 |
| | 5 2 6 2 | 5 2 6 2 | 2+2 2 |
| | 6 2 5 2 | 5 2 5 2 | 3+3 2 |
| | 6 2 6 2 | 6 2 5 2 | 2+2 2 |
| | 6 2 6 2 | 6 2 5 2 | 3+3 2 |
| | 6 2 5 2 | 5 2 6 2 | 2+2 2 |
| | 6 2 5 2 | 6 2 5 2 | 2+2 2 |
| | 5 2 6 2 | 5 2 5 2 | 3+3 2 |
| | 6 2 6 2 | 6 2 6 2 | 2+2 2 |
| | 5 2 5 2 | 6 2 6 2 | 3+3 2 |
| | 5 2 5 2 | 5 2 5 2 | 2+2 2 |
| | 5 2 6 2 | 6 2 6 2 | 3+3 2 |
| | 5 2 5 2 | 5 2 6 2 | 2+2 2 |
| | 6 2 6 2 | 5 2 5 2 | 3+3 2 |
| | 5 2 6 2 | 5 2 5 2 | 2+2 2 |
| | 6 2 5 2 | 6 2 6 2 | 3+3 2 |
| | 6 2 6 2 | 6 2 6 2 | 3+3 2 |
| | 6 2 5 2 | 6 2 5 2 | 2+2 2 |
| | 6 2 6 2 | 5 2 6 2 | 3+3 2 |
| | 5 2 5 2 | 6 2 5 2 | 2+2 2 |
| | 5 2 6 2 | 5 2 6 2 | 2+2 2 |
| | 5 2 5 2 | 5 2 5 2 | 3+3 2 |
| | 6 2 6 2 | 5 2 6 2 | 3+3 2 |
| | 5 2 5 2 | 6 2 6 2 | 2+2 2 |
| | 6 2 6 2 | 6 2 5 2 | 3+3 2 |
| | 6 2 6 2 | 6 2 6 2 | 3+3 2 |
| | 5 2 5 2 | 5 2 5 2 | 2+2 2 |
| | 5 2 5 2 | 5 2 5 2 | 2+2 2 |
| | 5 2 5 2 | 5 2 5 2 | 2+2 2 |
| | 5 2 6 2 | 5 2 6 2 | 3+3 2 |
| | 6 2 5 2 | 6 2 5 2 | 2+2 2 |
| | 6 2 6 2 | 6 2 6 2 | 3+3 2 |
| | 5 2 5 2 | 5 2 5 2 | 2+2 2 |
| | 5 2 6 2 | 6 2 6 2 | 3+3 2 |
| | 6 2 5 2 | 5 2 5 2 | 2+2 2 |
| | 6 2 5 2 | 6 2 5 2 | 2+2 2 |
| | 5 2 5 2 | 5 2 6 2 | 2+2 2 |
| | 6 2 5 2 | 5 2 5 2 | 2+2 2 |
| | 5 2 6 2 | 6 2 5 2 | 3+3 2 |
| | 5 2 6 2 | 5 2 6 2 | 3+3 2 |
| | 6 2 6 2 | 6 2 6 2 | 3+3 2 |
| | 5 2 6 2 | 6 2 6 2 | 3+3 2 |
| | 6 2 5 2 | 5 2 5 2 | 2+2 2 |
| | 6 2 6 2 | 6 2 6 2 | 3+3 2 |
| | 6 2 6 2 | 6 2 6 2 | 3+3 2 |
| | 6 2 5 2 | 6 2 5 2 | 2+2 2 |
| | 5 2 6 2 | 6 2 6 2 | 3+3 2 |
| | 5 2 5 2 | 5 2 6 2 | 2+2 2 |

Continued on next column

**TABLE XLII– continued**

| TQC | ICSD | NoE | SF |
|---|---|---|---|
| | 6 2 6 2 | 5 2 6 2 | 3+3 2 |
| | 6 2 6 2 | 6 2 6 2 | 3+3 2 |
| | 6 2 5 2 | 6 2 5 2 | 3+3 2 |
| | 5 2 6 2 | 6 2 5 2 | 2+2 2 |
| | 5 2 6 2 | 5 2 5 2 | 3+3 2 |
| | 5 2 6 2 | 5 2 5 2 | 2+2 2 |
| | 5 2 5 2 | 5 2 6 2 | 3+3 2 |
| | 5 2 5 2 | 5 2 5 2 | 2+2 2 |
| | 6 2 6 2 | 5 2 6 2 | 3+3 2 |
| | 5 2 5 2 | 5 2 6 2 | 2+2 2 |
| | 5 2 5 2 | 6 2 5 2 | 2+2 2 |
| | 6 2 5 2 | 5 2 6 2 | 2+2 2 |
| | 5 2 6 2 | 6 2 5 2 | 3+3 2 |
| | 6 2 5 2 | 5 2 5 2 | 2+2 2 |
| | 5 2 5 2 | 5 2 5 2 | 2+2 2 |
| | 6 2 6 2 | 6 2 6 2 | 3+3 2 |
| | 6 2 5 2 | 6 2 5 2 | 2+2 2 |
| | 5 2 6 2 | 5 2 5 2 | 3+3 2 |
| | 6 2 5 2 | 6 2 6 2 | 2+2 2 |
| | 6 2 5 2 | 6 2 6 2 | 3+3 2 |
| | 6 2 6 2 | 6 2 5 2 | 2+2 2 |
| | 5 2 5 2 | 5 2 5 2 | 3+3 2 |
| | 6 2 6 2 | 6 2 6 2 | 2+2 2 |
| | 5 2 6 2 | 6 2 6 2 | 3+3 2 |
| | 5 2 5 2 | 5 2 6 2 | 2+2 2 |
| | 6 2 6 2 | 5 2 5 2 | 3+3 2 |
| | 5 2 5 2 | 5 2 5 2 | 2+2 2 |
| | 5 2 6 2 | 5 2 6 2 | 3+3 2 |
| 69.1.5 | 414574 | 66 | Ge6Li2Sr4 |
| | -GM -Y | -Z -T | -L |
| | 5 2 5 2 | 5 2 6 2 | 2+2 2 |
| | 6 2 6 2 | 6 2 5 2 | 3+3 2 |
| | 5 2 5 2 | 5 2 5 2 | 2+2 2 |
| | 6 2 6 2 | 6 2 6 2 | 3+3 2 |
| | 5 2 5 2 | 5 2 6 2 | 2+2 2 |
| | 6 2 6 2 | 6 2 5 2 | 3+3 2 |
| | 5 2 5 2 | 5 2 5 2 | 3+3 2 |
| | 6 2 6 2 | 6 2 6 2 | 2+2 2 |
| | 5 2 6 2 | 5 2 6 2 | 2+2 2 |
| | 6 2 5 2 | 6 2 5 2 | 3+3 2 |
| | 5 2 5 2 | 6 2 6 2 | 2+2 2 |
| | 6 2 6 2 | 5 2 5 2 | 3+3 2 |
| | 5 2 5 2 | 5 2 5 2 | 3+3 2 |
| | 6 2 6 2 | 6 2 6 2 | 2+2 2 |
| | 5 2 6 2 | 6 2 5 2 | 2+2 2 |
| | 6 2 5 2 | 5 2 6 2 | 3+3 2 |
| | 5 2 5 2 | 5 2 5 2 | 3+3 2 |
| | 6 2 6 2 | 6 2 6 2 | 2+2 2 |
| | 6 2 6 2 | 6 2 6 2 | 2+2 2 |
| | 5 2 5 2 | 5 2 5 2 | 3+3 2 |
| | 5 2 5 2 | 5 2 5 2 | 3+3 2 |
| | 6 2 6 2 | 6 2 6 2 | 2+2 2 |
| | 5 2 6 2 | 6 2 6 2 | 2+2 2 |
| | 6 2 6 2 | 5 2 6 2 | 2+2 2 |
| | 5 2 5 2 | 6 2 5 2 | 3+3 2 |
| | 6 2 5 2 | 5 2 5 2 | 3+3 2 |
| | 6 2 6 2 | 6 2 5 2 | 2+2 2 |
| | 5 2 5 2 | 6 2 6 2 | 3+3 2 |
| | 6 2 6 2 | 5 2 6 2 | 3+3 2 |
| | 6 2 5 2 | 6 2 5 2 | 2+2 2 |
| | 5 2 5 2 | 5 2 5 2 | 3+3 2 |

Continued on next column

**TABLE XLII– continued**

| TQC | ICSD | NoE | SF |
|---|---|---|---|
| | 5 2 6 2 | 5 2 6 2 | 2+2 2 |
| | 6 2 5 2 | 5 2 6 2 | 2+2 2 |
| 69.1.6 | 68676 | 60 | Cu2O3Sr2 |
| | -GM -Y | -Z -T | -L |
| | 5 2 5 2 | 5 2 5 2 | 3+3 2 |
| | 6 2 6 2 | 6 2 6 2 | 2+2 2 |
| | 5 2 5 2 | 5 2 5 2 | 2+2 2 |
| | 6 2 6 2 | 6 2 6 2 | 3+3 2 |
| | 5 2 6 2 | 6 2 5 2 | 2+2 2 |
| | 5 2 5 2 | 5 2 5 2 | 3+3 2 |
| | 6 2 5 2 | 5 2 6 2 | 2+2 2 |
| | 5 2 6 2 | 5 2 6 2 | 3+3 2 |
| | 6 2 5 2 | 6 2 5 2 | 3+3 2 |
| | 5 2 6 2 | 6 2 6 2 | 2+2 2 |
| | 6 2 5 2 | 5 2 5 2 | 2+2 2 |
| | 6 2 6 2 | 6 2 6 2 | 3+3 2 |
| | 5 2 6 2 | 6 2 5 2 | 2+2 2 |
| | 5 2 6 2 | 6 2 5 2 | 3+3 2 |
| | 6 2 6 2 | 6 2 6 2 | 3+3 2 |
| | 6 2 6 2 | 6 2 6 2 | 2+2 2 |
| | 6 2 5 2 | 5 2 6 2 | 3+3 2 |
| | 5 2 5 2 | 5 2 5 2 | 2+2 2 |
| | 5 2 5 2 | 5 2 6 2 | 3+3 2 |
| | 6 2 6 2 | 6 2 5 2 | 3+3 2 |
| | 6 2 5 2 | 5 2 6 2 | 2+2 2 |
| | 5 2 6 2 | 6 2 5 2 | 3+3 2 |
| | 6 2 5 2 | 5 2 6 2 | 2+2 2 |
| | 5 2 6 2 | 6 2 5 2 | 3+3 2 |
| | 6 2 5 2 | 5 2 6 2 | 2+2 2 |
| | 5 2 5 2 | 5 2 5 2 | 3+3 2 |
| | 5 2 6 2 | 6 2 5 2 | 2+2 2 |
| | 6 2 6 2 | 6 2 6 2 | 3+3 2 |
| | 6 2 5 2 | 5 2 6 2 | 2+2 2 |
| | 5 2 6 2 | 6 2 6 2 | 3+3 2 |
| 69.1.7 | 262555 | 196 | Cs2Pt3Se6U1 |
| | -GM -Y | -Z -T | -L |
| | 5 2 5 2 | 5 2 5 2 | 2+2 2 |
| | 6 2 6 2 | 6 2 6 2 | 3+3 2 |
| | 5 2 5 2 | 5 2 5 2 | 2+2 2 |
| | 6 2 6 2 | 6 2 6 2 | 3+3 2 |
| | 5 2 6 2 | 6 2 5 2 | 2+2 2 |
| | 6 2 5 2 | 5 2 6 2 | 3+3 2 |
| | 5 2 5 2 | 6 2 6 2 | 2+2 2 |
| | 6 2 6 2 | 5 2 5 2 | 3+3 2 |
| | 6 2 5 2 | 6 2 5 2 | 2+2 2 |
| | 5 2 6 2 | 5 2 6 2 | 3+3 2 |
| | 6 2 5 2 | 6 2 5 2 | 3+3 2 |
| | 5 2 6 2 | 5 2 6 2 | 2+2 2 |
| | 5 2 6 2 | 5 2 6 2 | 3+3 2 |
| | 6 2 5 2 | 6 2 5 2 | 2+2 2 |
| | 6 2 5 2 | 6 2 6 2 | 3+3 2 |
| | 5 2 6 2 | 5 2 5 2 | 2+2 2 |
| | 6 2 6 2 | 6 2 6 2 | 2+2 2 |
| | 5 2 5 2 | 5 2 5 2 | 3+3 2 |
| | 5 2 5 2 | 6 2 6 2 | 2+2 2 |
| | 6 2 6 2 | 6 2 5 2 | 3+3 2 |
| | 6 2 5 2 | 5 2 5 2 | 2+2 2 |
| | 6 2 6 2 | 6 2 6 2 | 3+3 2 |
| | 5 2 5 2 | 5 2 5 2 | 2+2 2 |
| | 5 2 6 2 | 5 2 6 2 | 3+3 2 |
| | 6 2 6 2 | 6 2 5 2 | 2+2 2 |

Continued on next column

**TABLE XLII– continued**

| TQC | ICSD | NoE | SF |
|---|---|---|---|
| | 5 2 5 2 | 6 2 6 2 | 3+3 2 |
| | 5 2 5 2 | 5 2 6 2 | 3+3 2 |
| | 6 2 6 2 | 5 2 5 2 | 2+2 2 |
| | 5 2 5 2 | 6 2 6 2 | 3+3 2 |
| | 6 2 5 2 | 6 2 5 2 | 2+2 2 |
| | 5 2 6 2 | 5 2 6 2 | 2+2 2 |
| | 6 2 6 2 | 5 2 5 2 | 3+3 2 |
| | 5 2 6 2 | 6 2 5 2 | 3+3 2 |
| | 5 2 5 2 | 6 2 6 2 | 2+2 2 |
| | 6 2 6 2 | 5 2 5 2 | 2+2 2 |
| | 6 2 5 2 | 5 2 6 2 | 3+3 2 |
| | 5 2 6 2 | 5 2 5 2 | 3+3 2 |
| | 6 2 5 2 | 6 2 6 2 | 2+2 2 |
| | 6 2 5 2 | 5 2 6 2 | 3+3 2 |
| | 5 2 6 2 | 6 2 5 2 | 2+2 2 |
| | 6 2 6 2 | 5 2 6 2 | 3+3 2 |
| | 5 2 5 2 | 6 2 5 2 | 2+2 2 |
| | 5 2 5 2 | 5 2 5 2 | 2+2 2 |
| | 5 2 6 2 | 5 2 6 2 | 3+3 2 |
| | 6 2 5 2 | 6 2 5 2 | 2+2 2 |
| | 6 2 6 2 | 6 2 6 2 | 3+3 2 |
| | 6 2 6 2 | 6 2 6 2 | 3+3 2 |
| | 6 2 6 2 | 5 2 5 2 | 3+3 2 |
| | 5 2 5 2 | 6 2 6 2 | 2+2 2 |
| | 5 2 5 2 | 5 2 5 2 | 2+2 2 |
| | 5 2 5 2 | 5 2 5 2 | 2+2 2 |
| | 5 2 5 2 | 5 2 6 2 | 2+2 2 |
| | 6 2 6 2 | 6 2 5 2 | 3+3 2 |
| | 6 2 6 2 | 6 2 6 2 | 3+3 2 |
| | 6 2 6 2 | 6 2 5 2 | 2+2 2 |
| | 5 2 5 2 | 5 2 6 2 | 3+3 2 |
| | 5 2 5 2 | 5 2 5 2 | 2+2 2 |
| | 6 2 5 2 | 5 2 5 2 | 2+2 2 |
| | 5 2 5 2 | 6 2 5 2 | 2+2 2 |
| | 6 2 6 2 | 6 2 6 2 | 3+3 2 |
| | 6 2 6 2 | 5 2 6 2 | 2+2 2 |
| | 5 2 5 2 | 6 2 5 2 | 3+3 2 |
| | 5 2 6 2 | 5 2 6 2 | 3+3 2 |
| | 5 2 6 2 | 5 2 6 2 | 3+3 2 |
| | 6 2 5 2 | 6 2 5 2 | 2+2 2 |
| | 6 2 6 2 | 6 2 6 2 | 3+3 2 |
| | 6 2 6 2 | 6 2 6 2 | 3+3 2 |
| | 5 2 5 2 | 6 2 5 2 | 2+2 2 |
| | 6 2 6 2 | 5 2 6 2 | 3+3 2 |
| | 5 2 6 2 | 5 2 6 2 | 3+3 2 |
| | 5 2 5 2 | 6 2 6 2 | 2+2 2 |
| | 6 2 6 2 | 6 2 5 2 | 3+3 2 |
| | 6 2 6 2 | 6 2 6 2 | 2+2 2 |
| | 5 2 5 2 | 5 2 5 2 | 3+3 2 |
| | 6 2 6 2 | 6 2 5 2 | 3+3 2 |
| | 5 2 5 2 | 5 2 6 2 | 2+2 2 |
| | 5 2 5 2 | 5 2 5 2 | 3+3 2 |
| | 5 2 6 2 | 5 2 5 2 | 2+2 2 |
| | 6 2 5 2 | 5 2 6 2 | 2+2 2 |
| | 5 2 6 2 | 5 2 6 2 | 3+3 2 |
| | 5 2 5 2 | 5 2 5 2 | 2+2 2 |
| | 6 2 5 2 | 6 2 5 2 | 2+2 2 |
| | 6 2 6 2 | 6 2 6 2 | 3+3 2 |
| | 6 2 5 2 | 6 2 5 2 | 2+2 2 |
| | 6 2 5 2 | 5 2 6 2 | 2+2 2 |
| | 5 2 6 2 | 6 2 5 2 | 3+3 2 |

Continued on next column

**TABLE XLII– continued**

| TQC | ICSD | NoE | SF |
|---|---|---|---|
| | 5 2 6 2 | 5 2 5 2 | 3+3 2 |
| | 5 2 5 2 | 5 2 6 2 | 2+2 2 |
| | 6 2 5 2 | 6 2 5 2 | 2+2 2 |
| | 5 2 6 2 | 5 2 6 2 | 3+3 2 |
| | 6 2 5 2 | 6 2 5 2 | 2+2 2 |
| | 6 2 5 2 | 6 2 5 2 | 3+3 2 |
| | 5 2 6 2 | 5 2 6 2 | 2+2 2 |
| | 6 2 6 2 | 6 2 6 2 | 3+3 2 |
| | 6 2 5 2 | 5 2 5 2 | 2+2 2 |
| | 5 2 6 2 | 6 2 6 2 | 3+3 2 |
| | 5 2 5 2 | 5 2 5 2 | 2+2 2 |
| | 5 2 6 2 | 5 2 6 2 | 3+3 2 |
| 69.1.8 | 262556 | 196 | Cs2Pt3S6U1 |
| | -GM -Y | -Z -T | -L |
| | 5 2 5 2 | 5 2 5 2 | 2+2 2 |
| | 6 2 6 2 | 6 2 6 2 | 3+3 2 |
| | 5 2 5 2 | 5 2 5 2 | 2+2 2 |
| | 6 2 6 2 | 6 2 6 2 | 3+3 2 |
| | 5 2 6 2 | 6 2 5 2 | 2+2 2 |
| | 6 2 5 2 | 5 2 6 2 | 3+3 2 |
| | 5 2 5 2 | 6 2 6 2 | 2+2 2 |
| | 6 2 6 2 | 5 2 5 2 | 3+3 2 |
| | 6 2 5 2 | 6 2 5 2 | 2+2 2 |
| | 5 2 6 2 | 5 2 6 2 | 3+3 2 |
| | 6 2 5 2 | 6 2 5 2 | 3+3 2 |
| | 5 2 6 2 | 5 2 6 2 | 2+2 2 |
| | 5 2 6 2 | 5 2 6 2 | 3+3 2 |
| | 6 2 5 2 | 6 2 5 2 | 2+2 2 |
| | 6 2 5 2 | 6 2 6 2 | 3+3 2 |
| | 5 2 6 2 | 5 2 5 2 | 2+2 2 |
| | 6 2 5 2 | 6 2 5 2 | 2+2 2 |
| | 6 2 6 2 | 6 2 6 2 | 3+3 2 |
| | 6 2 6 2 | 6 2 6 2 | 2+2 2 |
| | 5 2 5 2 | 5 2 5 2 | 3+3 2 |
| | 5 2 5 2 | 5 2 5 2 | 2+2 2 |
| | 5 2 6 2 | 5 2 6 2 | 3+3 2 |
| | 6 2 5 2 | 5 2 5 2 | 2+2 2 |
| | 5 2 6 2 | 6 2 6 2 | 3+3 2 |
| | 5 2 6 2 | 6 2 5 2 | 3+3 2 |
| | 6 2 5 2 | 5 2 6 2 | 2+2 2 |
| | 5 2 5 2 | 6 2 6 2 | 3+3 2 |
| | 6 2 6 2 | 5 2 5 2 | 2+2 2 |
| | 6 2 5 2 | 6 2 6 2 | 3+3 2 |
| | 5 2 6 2 | 6 2 5 2 | 2+2 2 |
| | 5 2 6 2 | 5 2 6 2 | 3+3 2 |
| | 5 2 6 2 | 5 2 5 2 | 2+2 2 |
| | 6 2 5 2 | 6 2 6 2 | 3+3 2 |
| | 5 2 5 2 | 6 2 5 2 | 2+2 2 |
| | 6 2 5 2 | 5 2 6 2 | 2+2 2 |
| | 6 2 6 2 | 5 2 5 2 | 3+3 2 |
| | 5 2 6 2 | 5 2 5 2 | 3+3 2 |
| | 6 2 5 2 | 6 2 6 2 | 2+2 2 |
| | 6 2 6 2 | 6 2 6 2 | 3+3 2 |
| | 5 2 5 2 | 5 2 6 2 | 2+2 2 |
| | 6 2 6 2 | 6 2 5 2 | 3+3 2 |
| | 5 2 5 2 | 5 2 5 2 | 2+2 2 |
| | 5 2 5 2 | 5 2 6 2 | 2+2 2 |
| | 5 2 6 2 | 5 2 5 2 | 3+3 2 |
| | 6 2 5 2 | 6 2 5 2 | 2+2 2 |
| | 6 2 6 2 | 6 2 6 2 | 3+3 2 |
| | 6 2 6 2 | 6 2 6 2 | 3+3 2 |

Continued on next column

**TABLE XLII– continued**

| TQC | ICSD | NoE | SF |
|---|---|---|---|
| | 6 2 6 2 | 5 2 5 2 | 2+2 2 |
| | 5 2 5 2 | 6 2 6 2 | 3+3 2 |
| | 5 2 5 2 | 5 2 5 2 | 2+2 2 |
| | 5 2 5 2 | 5 2 5 2 | 2+2 2 |
| | 5 2 6 2 | 5 2 6 2 | 3+3 2 |
| | 6 2 6 2 | 6 2 5 2 | 3+3 2 |
| | 6 2 5 2 | 6 2 6 2 | 2+2 2 |
| | 6 2 6 2 | 6 2 6 2 | 3+3 2 |
| | 5 2 5 2 | 5 2 5 2 | 2+2 2 |
| | 5 2 5 2 | 5 2 5 2 | 2+2 2 |
| | 5 2 5 2 | 5 2 5 2 | 2+2 2 |
| | 6 2 5 2 | 6 2 5 2 | 2+2 2 |
| | 6 2 6 2 | 6 2 6 2 | 3+3 2 |
| | 6 2 6 2 | 5 2 5 2 | 2+2 2 |
| | 5 2 5 2 | 5 2 6 2 | 3+3 2 |
| | 5 2 6 2 | 6 2 6 2 | 3+3 2 |
| | 5 2 6 2 | 5 2 6 2 | 3+3 2 |
| | 6 2 5 2 | 6 2 5 2 | 2+2 2 |
| | 6 2 6 2 | 6 2 6 2 | 3+3 2 |
| | 6 2 6 2 | 6 2 6 2 | 3+3 2 |
| | 6 2 5 2 | 6 2 5 2 | 2+2 2 |
| | 5 2 6 2 | 5 2 6 2 | 3+3 2 |
| | 5 2 6 2 | 5 2 6 2 | 3+3 2 |
| | 5 2 5 2 | 6 2 6 2 | 2+2 2 |
| | 6 2 6 2 | 5 2 5 2 | 3+3 2 |
| | 6 2 5 2 | 6 2 5 2 | 2+2 2 |
| | 6 2 6 2 | 6 2 6 2 | 3+3 2 |
| | 5 2 6 2 | 6 2 5 2 | 3+3 2 |
| | 5 2 6 2 | 5 2 6 2 | 2+2 2 |
| | 5 2 5 2 | 5 2 5 2 | 3+3 2 |
| | 5 2 5 2 | 5 2 6 2 | 2+2 2 |
| | 6 2 5 2 | 5 2 5 2 | 2+2 2 |
| | 5 2 6 2 | 5 2 6 2 | 3+3 2 |
| | 5 2 5 2 | 5 2 5 2 | 2+2 2 |
| | 6 2 5 2 | 6 2 5 2 | 2+2 2 |
| | 6 2 6 2 | 6 2 6 2 | 3+3 2 |
| | 6 2 5 2 | 6 2 5 2 | 2+2 2 |
| | 5 2 5 2 | 5 2 5 2 | 2+2 2 |
| | 5 2 6 2 | 6 2 6 2 | 3+3 2 |
| | 6 2 5 2 | 5 2 5 2 | 2+2 2 |
| | 5 2 6 2 | 5 2 6 2 | 3+3 2 |
| | 5 2 5 2 | 5 2 5 2 | 2+2 2 |
| | 6 2 6 2 | 6 2 6 2 | 2+2 2 |
| | 5 2 5 2 | 6 2 5 2 | 3+3 2 |
| | 6 2 5 2 | 5 2 5 2 | 3+3 2 |
| | 6 2 6 2 | 6 2 6 2 | 2+2 2 |
| | 6 2 6 2 | 6 2 6 2 | 3+3 2 |
| | 5 2 5 2 | 5 2 5 2 | 2+2 2 |
| | 6 2 6 2 | 6 2 6 2 | 3+3 2 |
| | 5 2 5 2 | 5 2 6 2 | 3+3 2 |
| | 6 2 6 2 | 5 2 5 2 | 2+2 2 |
| 69.1.9 | 262557 | 196 | Cs2Pd3Se6U1 |
| | -GM -Y | -Z -T | -L |
| | 5 2 5 2 | 5 2 5 2 | 2+2 2 |
| | 6 2 6 2 | 6 2 6 2 | 3+3 2 |
| | 5 2 5 2 | 5 2 5 2 | 2+2 2 |
| | 6 2 6 2 | 6 2 6 2 | 3+3 2 |
| | 5 2 6 2 | 6 2 5 2 | 2+2 2 |
| | 6 2 5 2 | 5 2 6 2 | 3+3 2 |
| | 5 2 5 2 | 6 2 6 2 | 2+2 2 |
| | 6 2 6 2 | 5 2 5 2 | 3+3 2 |

Continued on next column

**TABLE XLII– continued**

| TQC | ICSD | NoE | SF |
|---|---|---|---|
| | 6 2 5 2 | 6 2 5 2 | 2+2 2 |
| | 5 2 6 2 | 5 2 6 2 | 3+3 2 |
| | 6 2 5 2 | 6 2 5 2 | 2+2 2 |
| | 5 2 6 2 | 5 2 6 2 | 3+3 2 |
| | 5 2 6 2 | 5 2 6 2 | 3+3 2 |
| | 6 2 5 2 | 6 2 5 2 | 2+2 2 |
| | 6 2 5 2 | 6 2 5 2 | 3+3 2 |
| | 6 2 6 2 | 6 2 6 2 | 2+2 2 |
| | 5 2 6 2 | 5 2 6 2 | 2+2 2 |
| | 5 2 5 2 | 5 2 5 2 | 3+3 2 |
| | 5 2 5 2 | 5 2 5 2 | 3+3 2 |
| | 6 2 6 2 | 6 2 6 2 | 2+2 2 |
| | 6 2 5 2 | 6 2 5 2 | 2+2 2 |
| | 6 2 6 2 | 6 2 6 2 | 3+3 2 |
| | 5 2 5 2 | 5 2 5 2 | 2+2 2 |
| | 5 2 6 2 | 5 2 6 2 | 3+3 2 |
| | 5 2 6 2 | 6 2 5 2 | 3+3 2 |
| | 6 2 5 2 | 5 2 6 2 | 2+2 2 |
| | 5 2 5 2 | 6 2 6 2 | 2+2 2 |
| | 6 2 6 2 | 5 2 5 2 | 3+3 2 |
| | 6 2 6 2 | 6 2 5 2 | 3+3 2 |
| | 5 2 6 2 | 5 2 5 2 | 2+2 2 |
| | 5 2 5 2 | 6 2 6 2 | 3+3 2 |
| | 6 2 5 2 | 5 2 6 2 | 2+2 2 |
| | 5 2 5 2 | 6 2 6 2 | 3+3 2 |
| | 5 2 5 2 | 6 2 5 2 | 2+2 2 |
| | 6 2 6 2 | 5 2 6 2 | 2+2 2 |
| | 6 2 6 2 | 5 2 5 2 | 3+3 2 |
| | 5 2 5 2 | 5 2 5 2 | 2+2 2 |
| | 6 2 6 2 | 6 2 6 2 | 3+3 2 |
| | 6 2 6 2 | 6 2 6 2 | 3+3 2 |
| | 6 2 5 2 | 5 2 6 2 | 2+2 2 |
| | 5 2 6 2 | 6 2 5 2 | 3+3 2 |
| | 6 2 5 2 | 5 2 5 2 | 2+2 2 |
| | 5 2 6 2 | 6 2 6 2 | 3+3 2 |
| | 5 2 6 2 | 5 2 6 2 | 2+2 2 |
| | 6 2 5 2 | 5 2 5 2 | 3+3 2 |
| | 5 2 5 2 | 6 2 6 2 | 2+2 2 |
| | 6 2 6 2 | 6 2 5 2 | 3+3 2 |
| | 5 2 6 2 | 5 2 5 2 | 3+3 2 |
| | 6 2 5 2 | 5 2 6 2 | 2+2 2 |
| | 5 2 5 2 | 6 2 5 2 | 2+2 2 |
| | 5 2 5 2 | 5 2 5 2 | 2+2 2 |
| | 5 2 6 2 | 5 2 5 2 | 2+2 2 |
| | 6 2 5 2 | 6 2 6 2 | 3+3 2 |
| | 6 2 6 2 | 6 2 6 2 | 3+3 2 |
| | 6 2 6 2 | 6 2 5 2 | 2+2 2 |
| | 6 2 5 2 | 5 2 6 2 | 3+3 2 |
| | 5 2 5 2 | 5 2 5 2 | 2+2 2 |
| | 5 2 5 2 | 6 2 5 2 | 2+2 2 |
| | 5 2 5 2 | 5 2 5 2 | 2+2 2 |
| | 6 2 6 2 | 6 2 6 2 | 3+3 2 |
| | 5 2 5 2 | 5 2 5 2 | 2+2 2 |
| | 6 2 6 2 | 5 2 6 2 | 3+3 2 |
| | 5 2 6 2 | 6 2 6 2 | 3+3 2 |
| | 5 2 6 2 | 6 2 6 2 | 3+3 2 |
| | 6 2 5 2 | 5 2 5 2 | 2+2 2 |
| | 6 2 6 2 | 6 2 6 2 | 3+3 2 |
| | 6 2 6 2 | 6 2 6 2 | 3+3 2 |
| | 6 2 5 2 | 6 2 5 2 | 2+2 2 |
| | 6 2 6 2 | 6 2 6 2 | 3+3 2 |

Continued on next column

**TABLE XLII– continued**

| TQC | ICSD | NoE | SF |
|---|---|---|---|
| | 5 2 6 2 | 5 2 6 2 | 3+3 2 |
| | 5 2 5 2 | 5 2 6 2 | 2+2 2 |
| | 5 2 6 2 | 6 2 5 2 | 3+3 2 |
| | 5 2 5 2 | 5 2 6 2 | 3+3 2 |
| | 6 2 6 2 | 6 2 5 2 | 2+2 2 |
| | 5 2 6 2 | 5 2 5 2 | 3+3 2 |
| | 6 2 5 2 | 6 2 5 2 | 2+2 2 |
| | 5 2 6 2 | 5 2 5 2 | 2+2 2 |
| | 5 2 5 2 | 5 2 6 2 | 3+3 2 |
| | 5 2 6 2 | 5 2 6 2 | 3+3 2 |
| | 6 2 5 2 | 6 2 6 2 | 2+2 2 |
| | 6 2 5 2 | 5 2 5 2 | 2+2 2 |
| | 5 2 5 2 | 5 2 5 2 | 2+2 2 |
| | 6 2 6 2 | 6 2 6 2 | 3+3 2 |
| | 5 2 5 2 | 5 2 5 2 | 2+2 2 |
| | 5 2 5 2 | 6 2 5 2 | 2+2 2 |
| | 6 2 5 2 | 5 2 6 2 | 3+3 2 |
| | 6 2 6 2 | 6 2 5 2 | 2+2 2 |
| | 5 2 6 2 | 5 2 5 2 | 3+3 2 |
| | 6 2 5 2 | 6 2 6 2 | 2+2 2 |
| | 6 2 5 2 | 6 2 6 2 | 2+2 2 |
| | 6 2 6 2 | 6 2 5 2 | 3+3 2 |
| | 6 2 6 2 | 6 2 6 2 | 3+3 2 |
| | 6 2 5 2 | 6 2 5 2 | 2+2 2 |
| | 5 2 6 2 | 5 2 6 2 | 3+3 2 |
| | 5 2 5 2 | 5 2 5 2 | 2+2 2 |
| | 5 2 6 2 | 5 2 6 2 | 3+3 2 |
| | 5 2 5 2 | 5 2 5 2 | 2+2 2 |
| | 5 2 6 2 | 5 2 6 2 | 3+3 2 |
| 69.1.10 | 261242 | 196 | Cs2Pd3Se6U1 |
| | -GM -Y | -Z -T | -L |
| | 5 2 5 2 | 5 2 5 2 | 2+2 2 |
| | 6 2 6 2 | 6 2 6 2 | 3+3 2 |
| | 5 2 5 2 | 5 2 5 2 | 2+2 2 |
| | 6 2 6 2 | 6 2 6 2 | 3+3 2 |
| | 5 2 6 2 | 6 2 5 2 | 2+2 2 |
| | 6 2 5 2 | 5 2 6 2 | 3+3 2 |
| | 5 2 5 2 | 6 2 6 2 | 2+2 2 |
| | 6 2 6 2 | 5 2 5 2 | 3+3 2 |
| | 6 2 5 2 | 6 2 5 2 | 2+2 2 |
| | 5 2 6 2 | 5 2 6 2 | 3+3 2 |
| | 6 2 5 2 | 6 2 5 2 | 2+2 2 |
| | 5 2 6 2 | 5 2 6 2 | 3+3 2 |
| | 5 2 6 2 | 5 2 6 2 | 3+3 2 |
| | 6 2 5 2 | 6 2 5 2 | 2+2 2 |
| | 6 2 5 2 | 6 2 5 2 | 3+3 2 |
| | 6 2 6 2 | 6 2 6 2 | 2+2 2 |
| | 5 2 6 2 | 5 2 6 2 | 2+2 2 |
| | 5 2 5 2 | 5 2 5 2 | 3+3 2 |
| | 5 2 5 2 | 5 2 5 2 | 3+3 2 |
| | 6 2 6 2 | 6 2 6 2 | 2+2 2 |
| | 6 2 5 2 | 6 2 5 2 | 2+2 2 |
| | 6 2 6 2 | 6 2 6 2 | 3+3 2 |
| | 5 2 5 2 | 5 2 5 2 | 2+2 2 |
| | 5 2 6 2 | 5 2 6 2 | 3+3 2 |
| | 5 2 6 2 | 6 2 5 2 | 3+3 2 |
| | 6 2 5 2 | 5 2 6 2 | 2+2 2 |
| | 5 2 5 2 | 6 2 6 2 | 2+2 2 |
| | 6 2 6 2 | 5 2 5 2 | 3+3 2 |
| | 6 2 6 2 | 6 2 5 2 | 3+3 2 |
| | 5 2 6 2 | 5 2 5 2 | 2+2 2 |

Continued on next column

**TABLE XLII– continued**

| TQC | ICSD | NoE | SF |
|---|---|---|---|
| | 5 2 5 2 | 6 2 6 2 | 3+3 2 |
| | 6 2 5 2 | 5 2 6 2 | 2+2 2 |
| | 5 2 5 2 | 6 2 6 2 | 3+3 2 |
| | 5 2 5 2 | 6 2 5 2 | 2+2 2 |
| | 6 2 6 2 | 5 2 6 2 | 2+2 2 |
| | 6 2 6 2 | 5 2 5 2 | 3+3 2 |
| | 5 2 5 2 | 5 2 5 2 | 2+2 2 |
| | 6 2 6 2 | 6 2 6 2 | 3+3 2 |
| | 6 2 6 2 | 6 2 6 2 | 3+3 2 |
| | 6 2 5 2 | 5 2 6 2 | 2+2 2 |
| | 5 2 6 2 | 6 2 5 2 | 3+3 2 |
| | 6 2 5 2 | 5 2 5 2 | 2+2 2 |
| | 5 2 6 2 | 6 2 6 2 | 3+3 2 |
| | 5 2 6 2 | 5 2 6 2 | 2+2 2 |
| | 6 2 5 2 | 5 2 5 2 | 3+3 2 |
| | 5 2 5 2 | 6 2 6 2 | 2+2 2 |
| | 6 2 6 2 | 6 2 5 2 | 3+3 2 |
| | 5 2 6 2 | 5 2 5 2 | 3+3 2 |
| | 6 2 5 2 | 5 2 6 2 | 2+2 2 |
| | 5 2 5 2 | 6 2 5 2 | 2+2 2 |
| | 5 2 5 2 | 5 2 5 2 | 2+2 2 |
| | 5 2 6 2 | 5 2 5 2 | 2+2 2 |
| | 6 2 5 2 | 6 2 6 2 | 3+3 2 |
| | 6 2 6 2 | 6 2 6 2 | 3+3 2 |
| | 6 2 6 2 | 6 2 5 2 | 2+2 2 |
| | 6 2 5 2 | 5 2 6 2 | 3+3 2 |
| | 5 2 5 2 | 5 2 5 2 | 2+2 2 |
| | 5 2 5 2 | 6 2 5 2 | 2+2 2 |
| | 5 2 5 2 | 5 2 5 2 | 2+2 2 |
| | 6 2 6 2 | 6 2 6 2 | 3+3 2 |
| | 5 2 5 2 | 5 2 5 2 | 2+2 2 |
| | 6 2 6 2 | 5 2 6 2 | 3+3 2 |
| | 5 2 6 2 | 6 2 6 2 | 3+3 2 |
| | 5 2 6 2 | 6 2 6 2 | 3+3 2 |
| | 6 2 5 2 | 5 2 5 2 | 2+2 2 |
| | 6 2 6 2 | 6 2 6 2 | 3+3 2 |
| | 6 2 6 2 | 6 2 6 2 | 3+3 2 |
| | 6 2 5 2 | 6 2 5 2 | 2+2 2 |
| | 6 2 6 2 | 6 2 6 2 | 3+3 2 |
| | 5 2 6 2 | 5 2 6 2 | 3+3 2 |
| | 5 2 5 2 | 5 2 6 2 | 2+2 2 |
| | 5 2 6 2 | 6 2 5 2 | 3+3 2 |
| | 5 2 5 2 | 5 2 6 2 | 3+3 2 |
| | 6 2 6 2 | 6 2 5 2 | 2+2 2 |
| | 5 2 6 2 | 5 2 5 2 | 3+3 2 |
| | 6 2 5 2 | 6 2 5 2 | 2+2 2 |
| | 5 2 6 2 | 5 2 5 2 | 2+2 2 |
| | 5 2 5 2 | 5 2 6 2 | 3+3 2 |
| | 5 2 6 2 | 5 2 6 2 | 3+3 2 |
| | 6 2 5 2 | 6 2 6 2 | 2+2 2 |
| | 6 2 5 2 | 5 2 5 2 | 2+2 2 |
| | 5 2 5 2 | 5 2 5 2 | 2+2 2 |
| | 6 2 6 2 | 6 2 6 2 | 3+3 2 |
| | 5 2 5 2 | 5 2 5 2 | 2+2 2 |
| | 5 2 5 2 | 6 2 5 2 | 2+2 2 |
| | 6 2 5 2 | 5 2 6 2 | 3+3 2 |
| | 6 2 6 2 | 6 2 5 2 | 2+2 2 |
| | 5 2 6 2 | 5 2 5 2 | 3+3 2 |
| | 6 2 5 2 | 6 2 6 2 | 2+2 2 |
| | 6 2 5 2 | 6 2 6 2 | 2+2 2 |
| | 6 2 6 2 | 6 2 5 2 | 3+3 2 |

Continued on next column

**TABLE XLII– continued**

| TQC | ICSD | NoE | SF |
|---|---|---|---|
| | 6 2 6 2 | 6 2 6 2 | 3+3 2 |
| | 6 2 5 2 | 6 2 5 2 | 2+2 2 |
| | 5 2 6 2 | 5 2 6 2 | 3+3 2 |
| | 5 2 5 2 | 5 2 5 2 | 2+2 2 |
| | 5 2 6 2 | 5 2 6 2 | 3+3 2 |
| | 5 2 5 2 | 5 2 5 2 | 2+2 2 |
| | 5 2 6 2 | 5 2 6 2 | 3+3 2 |
| 69.1.11 | 262558 | 196 | Cs2Pd3S6U1 |
| | -GM -Y | -Z -T | -L |
| | 5 2 5 2 | 5 2 5 2 | 2+2 2 |
| | 6 2 6 2 | 6 2 6 2 | 3+3 2 |
| | 5 2 5 2 | 5 2 5 2 | 2+2 2 |
| | 6 2 6 2 | 6 2 6 2 | 3+3 2 |
| | 5 2 6 2 | 6 2 5 2 | 2+2 2 |
| | 6 2 5 2 | 5 2 6 2 | 3+3 2 |
| | 5 2 5 2 | 6 2 6 2 | 2+2 2 |
| | 6 2 6 2 | 5 2 5 2 | 3+3 2 |
| | 6 2 5 2 | 6 2 5 2 | 2+2 2 |
| | 5 2 6 2 | 5 2 6 2 | 3+3 2 |
| | 6 2 5 2 | 5 2 5 2 | 2+2 2 |
| | 5 2 6 2 | 6 2 6 2 | 3+3 2 |
| | 5 2 6 2 | 5 2 6 2 | 3+3 2 |
| | 6 2 5 2 | 6 2 5 2 | 2+2 2 |
| | 6 2 5 2 | 6 2 5 2 | 3+3 2 |
| | 6 2 6 2 | 6 2 6 2 | 2+2 2 |
| | 5 2 5 2 | 5 2 5 2 | 2+2 2 |
| | 5 2 6 2 | 5 2 6 2 | 3+3 2 |
| | 6 2 5 2 | 6 2 6 2 | 2+2 2 |
| | 5 2 6 2 | 6 2 5 2 | 3+3 2 |
| | 6 2 5 2 | 5 2 5 2 | 2+2 2 |
| | 6 2 6 2 | 5 2 6 2 | 3+3 2 |
| | 5 2 5 2 | 6 2 5 2 | 2+2 2 |
| | 5 2 6 2 | 5 2 6 2 | 3+3 2 |
| | 5 2 6 2 | 6 2 5 2 | 3+3 2 |
| | 6 2 5 2 | 5 2 6 2 | 2+2 2 |
| | 5 2 5 2 | 6 2 6 2 | 3+3 2 |
| | 6 2 6 2 | 5 2 5 2 | 2+2 2 |
| | 6 2 6 2 | 6 2 5 2 | 3+3 2 |
| | 5 2 6 2 | 5 2 5 2 | 2+2 2 |
| | 5 2 5 2 | 6 2 6 2 | 3+3 2 |
| | 6 2 5 2 | 5 2 6 2 | 2+2 2 |
| | 5 2 5 2 | 6 2 6 2 | 3+3 2 |
| | 5 2 5 2 | 6 2 5 2 | 2+2 2 |
| | 6 2 6 2 | 5 2 6 2 | 2+2 2 |
| | 6 2 6 2 | 5 2 5 2 | 3+3 2 |
| | 5 2 5 2 | 5 2 5 2 | 2+2 2 |
| | 6 2 6 2 | 6 2 6 2 | 3+3 2 |
| | 6 2 6 2 | 6 2 6 2 | 3+3 2 |
| | 6 2 5 2 | 5 2 6 2 | 2+2 2 |
| | 5 2 6 2 | 6 2 5 2 | 3+3 2 |
| | 6 2 5 2 | 5 2 5 2 | 2+2 2 |
| | 5 2 6 2 | 5 2 6 2 | 3+3 2 |
| | 5 2 5 2 | 6 2 5 2 | 2+2 2 |
| | 6 2 6 2 | 5 2 6 2 | 3+3 2 |
| | 6 2 5 2 | 6 2 6 2 | 2+2 2 |
| | 5 2 6 2 | 6 2 5 2 | 3+3 2 |
| | 6 2 6 2 | 6 2 5 2 | 3+3 2 |
| | 5 2 5 2 | 5 2 6 2 | 2+2 2 |
| | 5 2 5 2 | 5 2 5 2 | 2+2 2 |
| | 5 2 5 2 | 5 2 5 2 | 2+2 2 |
| | 5 2 6 2 | 5 2 6 2 | 3+3 2 |

Continued on next column

**TABLE XLII– continued**

| TQC | ICSD | NoE | SF |
|---|---|---|---|
| | 6 2 6 2 | 6 2 6 2 | 3+3 2 |
| | 6 2 5 2 | 6 2 5 2 | 2+2 2 |
| | 6 2 6 2 | 6 2 5 2 | 3+3 2 |
| | 5 2 5 2 | 5 2 6 2 | 2+2 2 |
| | 5 2 5 2 | 5 2 5 2 | 2+2 2 |
| | 6 2 5 2 | 6 2 5 2 | 2+2 2 |
| | 5 2 5 2 | 5 2 6 2 | 2+2 2 |
| | 5 2 6 2 | 5 2 5 2 | 3+3 2 |
| | 6 2 5 2 | 6 2 5 2 | 2+2 2 |
| | 6 2 6 2 | 5 2 6 2 | 3+3 2 |
| | 5 2 6 2 | 6 2 6 2 | 3+3 2 |
| | 5 2 6 2 | 6 2 6 2 | 3+3 2 |
| | 6 2 5 2 | 5 2 5 2 | 2+2 2 |
| | 6 2 6 2 | 6 2 6 2 | 3+3 2 |
| | 6 2 6 2 | 6 2 6 2 | 3+3 2 |
| | 6 2 5 2 | 6 2 5 2 | 2+2 2 |
| | 5 2 6 2 | 5 2 6 2 | 3+3 2 |
| | 6 2 5 2 | 5 2 6 2 | 3+3 2 |
| | 5 2 6 2 | 6 2 6 2 | 2+2 2 |
| | 6 2 6 2 | 6 2 5 2 | 3+3 2 |
| | 6 2 5 2 | 6 2 6 2 | 3+3 2 |
| | 5 2 6 2 | 6 2 5 2 | 2+2 2 |
| | 5 2 6 2 | 5 2 5 2 | 3+3 2 |
| | 5 2 6 2 | 5 2 5 2 | 2+2 2 |
| | 5 2 5 2 | 5 2 6 2 | 3+3 2 |
| | 5 2 5 2 | 5 2 5 2 | 2+2 2 |
| | 5 2 6 2 | 5 2 6 2 | 3+3 2 |
| | 6 2 5 2 | 5 2 6 2 | 2+2 2 |
| | 6 2 5 2 | 5 2 5 2 | 2+2 2 |
| | 5 2 5 2 | 6 2 5 2 | 2+2 2 |
| | 5 2 6 2 | 6 2 6 2 | 3+3 2 |
| | 6 2 5 2 | 5 2 5 2 | 2+2 2 |
| | 5 2 5 2 | 5 2 5 2 | 2+2 2 |
| | 6 2 6 2 | 6 2 6 2 | 3+3 2 |
| | 6 2 5 2 | 6 2 5 2 | 2+2 2 |
| | 5 2 6 2 | 5 2 5 2 | 3+3 2 |
| | 6 2 5 2 | 6 2 6 2 | 2+2 2 |
| | 6 2 5 2 | 6 2 6 2 | 2+2 2 |
| | 6 2 5 2 | 6 2 5 2 | 3+3 2 |
| | 5 2 6 2 | 5 2 5 2 | 2+2 2 |
| | 6 2 6 2 | 6 2 6 2 | 3+3 2 |
| | 5 2 6 2 | 6 2 6 2 | 3+3 2 |
| | 6 2 5 2 | 5 2 5 2 | 2+2 2 |
| | 5 2 6 2 | 5 2 6 2 | 3+3 2 |
| | 5 2 5 2 | 5 2 6 2 | 2+2 2 |
| | 5 2 6 2 | 5 2 5 2 | 3+3 2 |
| 69.1.12 | 73160 | 174 | Ba2Na1Ni3O6 |
| | -GM -Y | -Z -T | -L |
| | 5 2 6 2 | 6 2 5 2 | 2+2 2 |
| | 6 2 5 2 | 5 2 6 2 | 3+3 2 |
| | 5 2 5 2 | 6 2 6 2 | 2+2 2 |
| | 6 2 6 2 | 5 2 5 2 | 3+3 2 |
| | 5 2 5 2 | 5 2 5 2 | 2+2 2 |
| | 6 2 6 2 | 6 2 6 2 | 3+3 2 |
| | 6 2 6 2 | 6 2 6 2 | 3+3 2 |
| | 5 2 5 2 | 5 2 5 2 | 2+2 2 |
| | 5 2 5 2 | 5 2 6 2 | 2+2 2 |
| | 5 2 6 2 | 5 2 5 2 | 3+3 2 |
| | 6 2 5 2 | 6 2 5 2 | 2+2 2 |
| | 5 2 6 2 | 6 2 6 2 | 3+3 2 |
| | 6 2 6 2 | 6 2 5 2 | 2+2 2 |

Continued on next column

**TABLE XLII– continued**

| TQC | ICSD | NoE | SF |
|---|---|---|---|
| | 6 2 5 2 | 5 2 6 2 | 3+3 2 |
| | 6 2 5 2 | 5 2 6 2 | 3+3 2 |
| | 5 2 6 2 | 6 2 5 2 | 2+2 2 |
| | 5 2 6 2 | 6 2 5 2 | 2+2 2 |
| | 6 2 5 2 | 5 2 6 2 | 3+3 2 |
| | 5 2 5 2 | 6 2 6 2 | 2+2 2 |
| | 6 2 6 2 | 5 2 5 2 | 3+3 2 |
| | 5 2 6 2 | 6 2 5 2 | 2+2 2 |
| | 6 2 6 2 | 6 2 5 2 | 3+3 2 |
| | 5 2 5 2 | 5 2 6 2 | 3+3 2 |
| | 6 2 5 2 | 5 2 6 2 | 2+2 2 |
| | 5 2 5 2 | 6 2 6 2 | 2+2 2 |
| | 5 2 5 2 | 6 2 5 2 | 3+3 2 |
| | 6 2 6 2 | 5 2 6 2 | 2+2 2 |
| | 6 2 6 2 | 5 2 5 2 | 3+3 2 |
| | 5 2 6 2 | 5 2 5 2 | 2+2 2 |
| | 6 2 5 2 | 6 2 6 2 | 3+3 2 |
| | 5 2 6 2 | 5 2 6 2 | 3+3 2 |
| | 6 2 6 2 | 6 2 5 2 | 3+3 2 |
| | 6 2 5 2 | 5 2 5 2 | 2+2 2 |
| | 6 2 5 2 | 6 2 6 2 | 2+2 2 |
| | 5 2 5 2 | 6 2 5 2 | 3+3 2 |
| | 5 2 6 2 | 5 2 6 2 | 2+2 2 |
| | 6 2 6 2 | 5 2 6 2 | 3+3 2 |
| | 5 2 5 2 | 5 2 6 2 | 2+2 2 |
| | 5 2 5 2 | 5 2 5 2 | 3+3 2 |
| | 6 2 6 2 | 6 2 5 2 | 2+2 2 |
| | 6 2 5 2 | 6 2 6 2 | 2+2 2 |
| | 6 2 6 2 | 5 2 5 2 | 3+3 2 |
| | 5 2 5 2 | 6 2 5 2 | 2+2 2 |
| | 6 2 6 2 | 6 2 6 2 | 3+3 2 |
| | 5 2 6 2 | 5 2 6 2 | 2+2 2 |
| | 5 2 5 2 | 6 2 5 2 | 3+3 2 |
| | 5 2 5 2 | 5 2 6 2 | 3+3 2 |
| | 6 2 6 2 | 6 2 5 2 | 2+2 2 |
| | 6 2 5 2 | 5 2 5 2 | 2+2 2 |
| | 5 2 5 2 | 6 2 5 2 | 2+2 2 |
| | 6 2 5 2 | 5 2 5 2 | 2+2 2 |
| | 5 2 6 2 | 6 2 6 2 | 3+3 2 |
| | 5 2 6 2 | 6 2 6 2 | 3+3 2 |
| | 6 2 5 2 | 5 2 6 2 | 3+3 2 |
| | 5 2 6 2 | 5 2 5 2 | 2+2 2 |
| | 6 2 6 2 | 6 2 6 2 | 3+3 2 |
| | 6 2 6 2 | 6 2 6 2 | 3+3 2 |
| | 5 2 5 2 | 6 2 5 2 | 2+2 2 |
| | 6 2 6 2 | 6 2 6 2 | 3+3 2 |
| | 5 2 5 2 | 5 2 5 2 | 2+2 2 |
| | 6 2 6 2 | 5 2 6 2 | 3+3 2 |
| | 5 2 5 2 | 5 2 5 2 | 2+2 2 |
| | 5 2 6 2 | 6 2 5 2 | 2+2 2 |
| | 6 2 6 2 | 5 2 6 2 | 3+3 2 |
| | 6 2 5 2 | 5 2 6 2 | 3+3 2 |
| | 5 2 6 2 | 6 2 6 2 | 3+3 2 |
| | 6 2 5 2 | 5 2 6 2 | 2+2 2 |
| | 5 2 5 2 | 6 2 5 2 | 2+2 2 |
| | 5 2 6 2 | 5 2 5 2 | 3+3 2 |
| | 5 2 6 2 | 5 2 5 2 | 2+2 2 |
| | 5 2 5 2 | 5 2 5 2 | 3+3 2 |
| | 6 2 6 2 | 6 2 6 2 | 2+2 2 |
| | 5 2 5 2 | 5 2 6 2 | 3+3 2 |
| | 5 2 5 2 | 5 2 5 2 | 2+2 2 |

Continued on next column

**TABLE XLII– continued**

| TQC | ICSD | NoE | SF |
|---|---|---|---|
| | 6 2 6 2 | 6 2 5 2 | 2+2 2 |
| | 6 2 5 2 | 6 2 5 2 | 3+3 2 |
| | 6 2 5 2 | 6 2 6 2 | 2+2 2 |
| | 6 2 5 2 | 6 2 5 2 | 3+3 2 |
| | 6 2 6 2 | 6 2 6 2 | 2+2 2 |
| | 5 2 5 2 | 5 2 5 2 | 3+3 2 |
| | 6 2 6 2 | 6 2 6 2 | 2+2 2 |
| | 5 2 6 2 | 5 2 6 2 | 3+3 2 |
| | 5 2 6 2 | 5 2 6 2 | 2+2 2 |
| | 6 2 5 2 | 6 2 5 2 | 3+3 2 |
| | 5 2 5 2 | 6 2 5 2 | 2+2 2 |
| | 6 2 6 2 | 5 2 6 2 | 3+3 2 |
| | 5 2 6 2 | 6 2 6 2 | 3+3 2 |
| 69.1.13 | 59588 | 174 | Ba2Na1Ni3O6 |
| | -GM -Y | -Z -T | -L |
| | 5 2 6 2 | 6 2 5 2 | 2+2 2 |
| | 6 2 5 2 | 5 2 6 2 | 3+3 2 |
| | 5 2 5 2 | 6 2 6 2 | 2+2 2 |
| | 6 2 6 2 | 5 2 5 2 | 3+3 2 |
| | 5 2 5 2 | 5 2 5 2 | 3+3 2 |
| | 6 2 6 2 | 6 2 6 2 | 2+2 2 |
| | 6 2 6 2 | 6 2 6 2 | 2+2 2 |
| | 5 2 5 2 | 5 2 5 2 | 3+3 2 |
| | 5 2 5 2 | 5 2 5 2 | 3+3 2 |
| | 6 2 6 2 | 5 2 6 2 | 2+2 2 |
| | 5 2 6 2 | 6 2 5 2 | 3+3 2 |
| | 5 2 5 2 | 6 2 6 2 | 2+2 2 |
| | 6 2 6 2 | 6 2 5 2 | 3+3 2 |
| | 6 2 5 2 | 5 2 6 2 | 2+2 2 |
| | 5 2 5 2 | 5 2 6 2 | 2+2 2 |
| | 6 2 6 2 | 6 2 5 2 | 3+3 2 |
| | 5 2 6 2 | 6 2 5 2 | 2+2 2 |
| | 6 2 5 2 | 5 2 6 2 | 3+3 2 |
| | 5 2 5 2 | 6 2 6 2 | 2+2 2 |
| | 6 2 6 2 | 5 2 5 2 | 3+3 2 |
| | 5 2 6 2 | 6 2 5 2 | 2+2 2 |
| | 6 2 6 2 | 6 2 5 2 | 3+3 2 |
| | 5 2 5 2 | 5 2 6 2 | 2+2 2 |
| | 6 2 5 2 | 6 2 6 2 | 3+3 2 |
| | 5 2 5 2 | 5 2 6 2 | 2+2 2 |
| | 5 2 5 2 | 6 2 5 2 | 3+3 2 |
| | 6 2 6 2 | 5 2 6 2 | 3+3 2 |
| | 6 2 6 2 | 5 2 5 2 | 2+2 2 |
| | 5 2 6 2 | 5 2 5 2 | 3+3 2 |
| | 6 2 5 2 | 6 2 6 2 | 2+2 2 |
| | 5 2 6 2 | 5 2 6 2 | 2+2 2 |
| | 6 2 6 2 | 6 2 5 2 | 2+2 2 |
| | 6 2 5 2 | 5 2 5 2 | 3+3 2 |
| | 6 2 5 2 | 6 2 6 2 | 3+3 2 |
| | 5 2 5 2 | 6 2 5 2 | 2+2 2 |
| | 5 2 6 2 | 5 2 6 2 | 3+3 2 |
| | 6 2 6 2 | 5 2 6 2 | 2+2 2 |
| | 5 2 5 2 | 5 2 6 2 | 3+3 2 |
| | 5 2 5 2 | 5 2 5 2 | 2+2 2 |
| | 6 2 6 2 | 6 2 5 2 | 3+3 2 |
| | 6 2 5 2 | 5 2 6 2 | 3+3 2 |
| | 6 2 6 2 | 6 2 5 2 | 2+2 2 |
| | 5 2 5 2 | 6 2 5 2 | 3+3 2 |
| | 6 2 6 2 | 6 2 6 2 | 2+2 2 |
| | 5 2 6 2 | 5 2 6 2 | 3+3 2 |
| | 5 2 5 2 | 6 2 5 2 | 2+2 2 |

Continued on next column

**TABLE XLII– continued**

| TQC | ICSD | NoE | SF |
|---|---|---|---|
| | 5 2 5 2 | 5 2 6 2 | 2+2 2 |
| | 6 2 6 2 | 6 2 5 2 | 3+3 2 |
| | 6 2 5 2 | 5 2 5 2 | 3+3 2 |
| | 5 2 5 2 | 6 2 5 2 | 3+3 2 |
| | 5 2 5 2 | 5 2 6 2 | 2+2 2 |
| | 6 2 6 2 | 6 2 5 2 | 3+3 2 |
| | 5 2 6 2 | 6 2 6 2 | 2+2 2 |
| | 6 2 5 2 | 5 2 6 2 | 2+2 2 |
| | 5 2 6 2 | 5 2 5 2 | 3+3 2 |
| | 6 2 6 2 | 6 2 6 2 | 2+2 2 |
| | 6 2 6 2 | 6 2 6 2 | 2+2 2 |
| | 5 2 5 2 | 6 2 5 2 | 3+3 2 |
| | 6 2 6 2 | 6 2 6 2 | 2+2 2 |
| | 5 2 5 2 | 5 2 5 2 | 3+3 2 |
| | 6 2 6 2 | 5 2 6 2 | 2+2 2 |
| | 5 2 5 2 | 6 2 5 2 | 3+3 2 |
| | 5 2 6 2 | 5 2 5 2 | 3+3 2 |
| | 6 2 6 2 | 5 2 6 2 | 2+2 2 |
| | 6 2 5 2 | 5 2 6 2 | 2+2 2 |
| | 5 2 6 2 | 6 2 6 2 | 2+2 2 |
| | 5 2 5 2 | 5 2 6 2 | 3+3 2 |
| | 6 2 5 2 | 6 2 5 2 | 3+3 2 |
| | 5 2 6 2 | 5 2 5 2 | 2+2 2 |
| | 5 2 6 2 | 5 2 5 2 | 2+2 2 |
| | 5 2 5 2 | 6 2 5 2 | 3+3 2 |
| | 6 2 6 2 | 5 2 6 2 | 2+2 2 |
| | 5 2 5 2 | 5 2 6 2 | 3+3 2 |
| | 5 2 5 2 | 5 2 5 2 | 3+3 2 |
| | 6 2 6 2 | 6 2 5 2 | 3+3 2 |
| | 6 2 5 2 | 6 2 5 2 | 2+2 2 |
| | 6 2 5 2 | 6 2 6 2 | 3+3 2 |
| | 6 2 5 2 | 6 2 6 2 | 2+2 2 |
| | 6 2 6 2 | 6 2 5 2 | 2+2 2 |
| | 5 2 5 2 | 5 2 5 2 | 3+3 2 |
| | 6 2 6 2 | 5 2 6 2 | 3+3 2 |
| | 5 2 6 2 | 6 2 6 2 | 2+2 2 |
| | 5 2 6 2 | 5 2 6 2 | 3+3 2 |
| | 6 2 5 2 | 6 2 5 2 | 2+2 2 |
| | 5 2 5 2 | 6 2 5 2 | 3+3 2 |
| | 6 2 6 2 | 5 2 6 2 | 2+2 2 |
| | 5 2 6 2 | 6 2 6 2 | 2+2 2 |
| 69.1.14 | 72328 | 180 | Ba2Cu3Na1O6 |
| | -GM -Y | -Z -T | -L |
| | 5 2 6 2 | 6 2 5 2 | 2+2 2 |
| | 6 2 5 2 | 5 2 6 2 | 3+3 2 |
| | 5 2 5 2 | 6 2 6 2 | 2+2 2 |
| | 6 2 6 2 | 5 2 5 2 | 3+3 2 |
| | 5 2 5 2 | 5 2 5 2 | 3+3 2 |
| | 6 2 6 2 | 6 2 6 2 | 2+2 2 |
| | 6 2 6 2 | 6 2 6 2 | 2+2 2 |
| | 5 2 5 2 | 5 2 5 2 | 3+3 2 |
| | 5 2 5 2 | 5 2 5 2 | 3+3 2 |
| | 6 2 6 2 | 6 2 5 2 | 3+3 2 |
| | 5 2 6 2 | 5 2 6 2 | 2+2 2 |
| | 6 2 5 2 | 6 2 5 2 | 2+2 2 |
| | 5 2 5 2 | 5 2 6 2 | 3+3 2 |
| | 5 2 6 2 | 6 2 6 2 | 2+2 2 |
| | 6 2 6 2 | 5 2 6 2 | 2+2 2 |
| | 6 2 5 2 | 6 2 5 2 | 3+3 2 |
| | 5 2 6 2 | 6 2 5 2 | 3+3 2 |
| | 6 2 5 2 | 5 2 6 2 | 2+2 2 |

Continued on next column

**TABLE XLII– continued**

| TQC | ICSD | NoE | SF |
|---|---|---|---|
| | 5 2 5 2 | 6 2 6 2 | 2+2 2 |
| | 6 2 6 2 | 5 2 5 2 | 3+3 2 |
| | 5 2 6 2 | 6 2 5 2 | 2+2 2 |
| | 6 2 6 2 | 5 2 5 2 | 3+3 2 |
| | 5 2 5 2 | 6 2 6 2 | 3+3 2 |
| | 6 2 5 2 | 5 2 6 2 | 2+2 2 |
| | 5 2 5 2 | 6 2 6 2 | 2+2 2 |
| | 5 2 5 2 | 6 2 5 2 | 3+3 2 |
| | 6 2 6 2 | 5 2 6 2 | 3+3 2 |
| | 6 2 6 2 | 5 2 5 2 | 2+2 2 |
| | 5 2 6 2 | 5 2 5 2 | 3+3 2 |
| | 6 2 5 2 | 6 2 6 2 | 2+2 2 |
| | 6 2 6 2 | 6 2 6 2 | 2+2 2 |
| | 5 2 5 2 | 5 2 5 2 | 3+3 2 |
| | 6 2 6 2 | 5 2 6 2 | 2+2 2 |
| | 5 2 5 2 | 6 2 5 2 | 3+3 2 |
| | 6 2 6 2 | 5 2 6 2 | 2+2 2 |
| | 5 2 5 2 | 6 2 5 2 | 3+3 2 |
| | 6 2 6 2 | 6 2 6 2 | 2+2 2 |
| | 5 2 6 2 | 5 2 6 2 | 3+3 2 |
| | 6 2 5 2 | 5 2 5 2 | 2+2 2 |
| | 5 2 5 2 | 6 2 5 2 | 3+3 2 |
| | 6 2 6 2 | 5 2 6 2 | 2+2 2 |
| | 6 2 5 2 | 6 2 5 2 | 3+3 2 |
| | 5 2 6 2 | 5 2 6 2 | 2+2 2 |
| | 6 2 5 2 | 5 2 6 2 | 3+3 2 |
| | 5 2 5 2 | 6 2 5 2 | 2+2 2 |
| | 6 2 6 2 | 6 2 6 2 | 3+3 2 |
| | 5 2 6 2 | 6 2 5 2 | 2+2 2 |
| | 5 2 5 2 | 5 2 5 2 | 3+3 2 |
| | 5 2 5 2 | 5 2 5 2 | 3+3 2 |
| | 6 2 5 2 | 5 2 5 2 | 3+3 2 |
| | 5 2 5 2 | 5 2 5 2 | 3+3 2 |
| | 5 2 5 2 | 5 2 5 2 | 3+3 2 |
| | 5 2 5 2 | 6 2 5 2 | 2+2 2 |
| | 5 2 6 2 | 5 2 6 2 | 2+2 2 |
| | 6 2 6 2 | 6 2 6 2 | 2+2 2 |
| | 6 2 6 2 | 6 2 6 2 | 2+2 2 |
| | 6 2 6 2 | 6 2 6 2 | 3+3 2 |
| | 6 2 6 2 | 6 2 6 2 | 2+2 2 |
| | 6 2 6 2 | 6 2 6 2 | 2+2 2 |
| | 5 2 5 2 | 5 2 6 2 | 2+2 2 |
| | 6 2 6 2 | 5 2 5 2 | 3+3 2 |
| | 5 2 6 2 | 6 2 6 2 | 2+2 2 |
| | 5 2 6 2 | 6 2 5 2 | 2+2 2 |
| | 6 2 5 2 | 5 2 6 2 | 3+3 2 |
| | 5 2 5 2 | 6 2 5 2 | 2+2 2 |
| | 5 2 6 2 | 6 2 6 2 | 3+3 2 |
| | 6 2 5 2 | 5 2 5 2 | 2+2 2 |
| | 5 2 6 2 | 6 2 6 2 | 3+3 2 |
| | 5 2 6 2 | 5 2 5 2 | 3+3 2 |
| | 6 2 6 2 | 5 2 5 2 | 3+3 2 |
| | 5 2 5 2 | 6 2 6 2 | 2+2 2 |
| | 6 2 5 2 | 5 2 5 2 | 3+3 2 |
| | 6 2 5 2 | 5 2 6 2 | 2+2 2 |
| | 5 2 6 2 | 5 2 5 2 | 3+3 2 |
| | 6 2 6 2 | 6 2 6 2 | 2+2 2 |
| | 6 2 5 2 | 5 2 5 2 | 3+3 2 |
| | 6 2 5 2 | 6 2 6 2 | 3+3 2 |
| | 5 2 5 2 | 6 2 5 2 | 3+3 2 |
| | 5 2 5 2 | 5 2 5 2 | 2+2 2 |

Continued on next column

**TABLE XLII– continued**

| TQC | ICSD | NoE | SF |
|---|---|---|---|
| | 6 2 6 2 | 6 2 6 2 | 2+2 2 |
| | 5 2 5 2 | 5 2 5 2 | 3+3 2 |
| | 6 2 5 2 | 6 2 5 2 | 2+2 2 |
| | 6 2 6 2 | 6 2 6 2 | 3+3 2 |
| | 6 2 5 2 | 5 2 6 2 | 2+2 2 |
| | 5 2 6 2 | 6 2 5 2 | 3+3 2 |
| | 6 2 5 2 | 5 2 6 2 | 2+2 2 |
| | 5 2 6 2 | 5 2 5 2 | 3+3 2 |
| | 5 2 6 2 | 6 2 6 2 | 2+2 2 |
| | 6 2 5 2 | 6 2 5 2 | 3+3 2 |
| | 5 2 5 2 | 5 2 6 2 | 2+2 2 |
| 69.1.15 | 59589 | 180 | Ba2Cu3Na1O6 |
| | -GM -Y | -Z -T | -L |
| | 5 2 6 2 | 6 2 5 2 | 2+2 2 |
| | 6 2 5 2 | 5 2 6 2 | 3+3 2 |
| | 5 2 5 2 | 6 2 6 2 | 2+2 2 |
| | 6 2 6 2 | 5 2 5 2 | 3+3 2 |
| | 5 2 5 2 | 5 2 5 2 | 3+3 2 |
| | 6 2 6 2 | 6 2 6 2 | 2+2 2 |
| | 6 2 6 2 | 6 2 6 2 | 2+2 2 |
| | 5 2 5 2 | 5 2 5 2 | 3+3 2 |
| | 5 2 5 2 | 5 2 5 2 | 3+3 2 |
| | 6 2 6 2 | 6 2 6 2 | 2+2 2 |
| | 6 2 6 2 | 5 2 5 2 | 3+3 2 |
| | 5 2 5 2 | 6 2 6 2 | 2+2 2 |
| | 5 2 5 2 | 5 2 5 2 | 3+3 2 |
| | 6 2 6 2 | 6 2 6 2 | 2+2 2 |
| | 6 2 5 2 | 5 2 6 2 | 2+2 2 |
| | 5 2 6 2 | 6 2 5 2 | 3+3 2 |
| | 5 2 6 2 | 6 2 5 2 | 3+3 2 |
| | 6 2 5 2 | 5 2 6 2 | 2+2 2 |
| | 5 2 5 2 | 6 2 6 2 | 2+2 2 |
| | 6 2 6 2 | 5 2 5 2 | 3+3 2 |
| | 5 2 6 2 | 6 2 5 2 | 3+3 2 |
| | 6 2 6 2 | 5 2 5 2 | 2+2 2 |
| | 5 2 5 2 | 6 2 6 2 | 2+2 2 |
| | 6 2 5 2 | 5 2 6 2 | 3+3 2 |
| | 5 2 5 2 | 6 2 6 2 | 2+2 2 |
| | 5 2 5 2 | 6 2 5 2 | 3+3 2 |
| | 6 2 6 2 | 5 2 6 2 | 3+3 2 |
| | 6 2 6 2 | 5 2 5 2 | 2+2 2 |
| | 5 2 6 2 | 5 2 5 2 | 3+3 2 |
| | 6 2 5 2 | 6 2 6 2 | 2+2 2 |
| | 5 2 6 2 | 6 2 6 2 | 2+2 2 |
| | 6 2 5 2 | 5 2 5 2 | 2+2 2 |
| | 6 2 6 2 | 5 2 6 2 | 3+3 2 |
| | 6 2 5 2 | 6 2 5 2 | 3+3 2 |
| | 5 2 5 2 | 6 2 5 2 | 3+3 2 |
| | 5 2 6 2 | 5 2 6 2 | 2+2 2 |
| | 6 2 6 2 | 6 2 5 2 | 3+3 2 |
| | 5 2 5 2 | 5 2 6 2 | 2+2 2 |
| | 5 2 6 2 | 5 2 6 2 | 2+2 2 |
| | 6 2 5 2 | 5 2 6 2 | 3+3 2 |
| | 6 2 6 2 | 6 2 5 2 | 3+3 2 |
| | 5 2 5 2 | 5 2 5 2 | 2+2 2 |
| | 6 2 5 2 | 6 2 6 2 | 3+3 2 |
| | 6 2 6 2 | 6 2 5 2 | 2+2 2 |
| | 5 2 6 2 | 6 2 6 2 | 2+2 2 |
| | 6 2 5 2 | 5 2 6 2 | 3+3 2 |
| | 5 2 6 2 | 6 2 5 2 | 2+2 2 |
| | 5 2 5 2 | 5 2 5 2 | 3+3 2 |

Continued on next column

**TABLE XLII– continued**

| TQC | ICSD | NoE | SF |
|---|---|---|---|
| | 5 2 5 2 | 5 2 5 2 | 3+3 2 |
| | 5 2 5 2 | 5 2 5 2 | 3+3 2 |
| | 6 2 5 2 | 5 2 5 2 | 3+3 2 |
| | 5 2 5 2 | 5 2 5 2 | 3+3 2 |
| | 5 2 6 2 | 6 2 6 2 | 2+2 2 |
| | 5 2 6 2 | 5 2 5 2 | 2+2 2 |
| | 6 2 6 2 | 6 2 6 2 | 2+2 2 |
| | 6 2 5 2 | 6 2 6 2 | 2+2 2 |
| | 6 2 6 2 | 6 2 6 2 | 3+3 2 |
| | 6 2 6 2 | 6 2 6 2 | 2+2 2 |
| | 6 2 6 2 | 6 2 6 2 | 2+2 2 |
| | 6 2 5 2 | 6 2 6 2 | 2+2 2 |
| | 5 2 6 2 | 5 2 6 2 | 3+3 2 |
| | 5 2 6 2 | 5 2 5 2 | 2+2 2 |
| | 5 2 6 2 | 5 2 5 2 | 2+2 2 |
| | 6 2 5 2 | 6 2 6 2 | 3+3 2 |
| | 5 2 5 2 | 6 2 5 2 | 3+3 2 |
| | 6 2 5 2 | 6 2 6 2 | 2+2 2 |
| | 5 2 6 2 | 5 2 5 2 | 2+2 2 |
| | 6 2 6 2 | 6 2 6 2 | 3+3 2 |
| | 5 2 6 2 | 5 2 5 2 | 3+3 2 |
| | 5 2 6 2 | 5 2 6 2 | 2+2 2 |
| | 5 2 5 2 | 6 2 5 2 | 3+3 2 |
| | 6 2 5 2 | 5 2 5 2 | 3+3 2 |
| | 5 2 5 2 | 5 2 5 2 | 2+2 2 |
| | 6 2 6 2 | 6 2 6 2 | 2+2 2 |
| | 6 2 5 2 | 5 2 5 2 | 3+3 2 |
| | 6 2 6 2 | 5 2 6 2 | 3+3 2 |
| | 6 2 5 2 | 6 2 5 2 | 3+3 2 |
| | 5 2 5 2 | 6 2 5 2 | 3+3 2 |
| | 5 2 5 2 | 5 2 6 2 | 2+2 2 |
| | 6 2 5 2 | 6 2 5 2 | 2+2 2 |
| | 6 2 6 2 | 6 2 6 2 | 3+3 2 |
| | 5 2 6 2 | 5 2 6 2 | 3+3 2 |
| | 6 2 5 2 | 6 2 5 2 | 2+2 2 |
| | 5 2 5 2 | 5 2 6 2 | 2+2 2 |
| | 6 2 6 2 | 6 2 5 2 | 3+3 2 |
| | 5 2 5 2 | 5 2 6 2 | 2+2 2 |
| | 6 2 6 2 | 5 2 5 2 | 3+3 2 |
| | 5 2 6 2 | 6 2 6 2 | 2+2 2 |
| | 6 2 5 2 | 6 2 5 2 | 3+3 2 |
| | 5 2 5 2 | 5 2 6 2 | 2+2 2 |
| 69.1.16 | 401239 | 180 | Ba2Cu3Li1O6 |
| | -GM -Y | -Z -T | -L |
| | 5 2 6 2 | 6 2 5 2 | 2+2 2 |
| | 6 2 5 2 | 5 2 6 2 | 3+3 2 |
| | 5 2 5 2 | 6 2 6 2 | 2+2 2 |
| | 6 2 6 2 | 5 2 5 2 | 3+3 2 |
| | 5 2 5 2 | 5 2 5 2 | 3+3 2 |
| | 6 2 6 2 | 6 2 6 2 | 2+2 2 |
| | 6 2 6 2 | 6 2 6 2 | 2+2 2 |
| | 5 2 5 2 | 5 2 5 2 | 3+3 2 |
| | 5 2 5 2 | 5 2 5 2 | 3+3 2 |
| | 6 2 6 2 | 6 2 5 2 | 2+2 2 |
| | 6 2 6 2 | 5 2 6 2 | 3+3 2 |
| | 5 2 5 2 | 6 2 6 2 | 2+2 2 |
| | 5 2 5 2 | 5 2 6 2 | 3+3 2 |
| | 5 2 6 2 | 5 2 5 2 | 2+2 2 |
| | 6 2 6 2 | 6 2 6 2 | 2+2 2 |
| | 6 2 5 2 | 6 2 5 2 | 3+3 2 |
| | 5 2 6 2 | 6 2 5 2 | 3+3 2 |

Continued on next column

**TABLE XLII– continued**

| TQC | ICSD | NoE | SF |
|---|---|---|---|
| | 6 2 5 2 | 5 2 6 2 | 2+2 2 |
| | 5 2 5 2 | 6 2 6 2 | 2+2 2 |
| | 6 2 6 2 | 5 2 5 2 | 3+3 2 |
| | 5 2 6 2 | 6 2 5 2 | 2+2 2 |
| | 5 2 6 2 | 6 2 5 2 | 3+3 2 |
| | 6 2 5 2 | 5 2 6 2 | 3+3 2 |
| | 6 2 5 2 | 5 2 6 2 | 2+2 2 |
| | 5 2 5 2 | 6 2 6 2 | 2+2 2 |
| | 5 2 5 2 | 6 2 5 2 | 3+3 2 |
| | 6 2 6 2 | 5 2 6 2 | 3+3 2 |
| | 6 2 6 2 | 5 2 5 2 | 2+2 2 |
| | 5 2 5 2 | 5 2 5 2 | 3+3 2 |
| | 6 2 6 2 | 6 2 6 2 | 2+2 2 |
| | 6 2 6 2 | 5 2 6 2 | 2+2 2 |
| | 5 2 5 2 | 6 2 5 2 | 3+3 2 |
| | 5 2 5 2 | 5 2 6 2 | 2+2 2 |
| | 6 2 6 2 | 6 2 5 2 | 3+3 2 |
| | 6 2 6 2 | 5 2 6 2 | 2+2 2 |
| | 5 2 5 2 | 6 2 5 2 | 3+3 2 |
| | 6 2 6 2 | 6 2 6 2 | 3+3 2 |
| | 5 2 5 2 | 5 2 6 2 | 2+2 2 |
| | 5 2 6 2 | 5 2 5 2 | 2+2 2 |
| | 6 2 5 2 | 5 2 6 2 | 3+3 2 |
| | 6 2 6 2 | 6 2 5 2 | 3+3 2 |
| | 6 2 5 2 | 5 2 5 2 | 2+2 2 |
| | 5 2 5 2 | 6 2 6 2 | 3+3 2 |
| | 6 2 6 2 | 6 2 5 2 | 2+2 2 |
| | 6 2 6 2 | 6 2 6 2 | 2+2 2 |
| | 5 2 5 2 | 6 2 6 2 | 3+3 2 |
| | 5 2 5 2 | 5 2 5 2 | 2+2 2 |
| | 5 2 6 2 | 5 2 5 2 | 3+3 2 |
| | 5 2 5 2 | 5 2 5 2 | 3+3 2 |
| | 5 2 5 2 | 5 2 5 2 | 3+3 2 |
| | 6 2 5 2 | 5 2 5 2 | 3+3 2 |
| | 5 2 6 2 | 5 2 5 2 | 3+3 2 |
| | 5 2 6 2 | 6 2 6 2 | 2+2 2 |
| | 5 2 5 2 | 5 2 5 2 | 2+2 2 |
| | 6 2 5 2 | 6 2 6 2 | 2+2 2 |
| | 6 2 6 2 | 6 2 6 2 | 2+2 2 |
| | 6 2 6 2 | 6 2 6 2 | 3+3 2 |
| | 6 2 6 2 | 6 2 6 2 | 2+2 2 |
| | 6 2 5 2 | 6 2 6 2 | 2+2 2 |
| | 5 2 6 2 | 5 2 6 2 | 2+2 2 |
| | 6 2 6 2 | 5 2 5 2 | 3+3 2 |
| | 5 2 6 2 | 6 2 6 2 | 2+2 2 |
| | 5 2 6 2 | 5 2 6 2 | 2+2 2 |
| | 6 2 5 2 | 6 2 5 2 | 3+3 2 |
| | 5 2 5 2 | 6 2 5 2 | 2+2 2 |
| | 5 2 6 2 | 5 2 6 2 | 3+3 2 |
| | 6 2 5 2 | 6 2 6 2 | 2+2 2 |
| | 5 2 6 2 | 6 2 5 2 | 3+3 2 |
| | 6 2 6 2 | 5 2 5 2 | 3+3 2 |
| | 5 2 5 2 | 5 2 5 2 | 2+2 2 |
| | 5 2 6 2 | 5 2 6 2 | 3+3 2 |
| | 6 2 5 2 | 5 2 5 2 | 3+3 2 |
| | 5 2 5 2 | 6 2 6 2 | 2+2 2 |
| | 5 2 6 2 | 5 2 5 2 | 3+3 2 |
| | 6 2 6 2 | 5 2 6 2 | 2+2 2 |
| | 6 2 5 2 | 6 2 5 2 | 3+3 2 |
| | 6 2 5 2 | 6 2 5 2 | 3+3 2 |
| | 5 2 5 2 | 5 2 6 2 | 2+2 2 |

Continued on next column

**TABLE XLII– continued**

| TQC | ICSD NoE SF |
|---|---|
| | 6 2 5 2 6 2 6 2 3+3 2 |
| | 6 2 6 2 6 2 5 2 2+2 2 |
| | 6 2 5 2 6 2 5 2 3+3 2 |
| | 5 2 5 2 5 2 6 2 2+2 2 |
| | 6 2 6 2 6 2 6 2 3+3 2 |
| | 6 2 5 2 6 2 5 2 2+2 2 |
| | 5 2 6 2 5 2 5 2 3+3 2 |
| | 6 2 5 2 5 2 6 2 2+2 2 |
| | 5 2 6 2 5 2 5 2 3+3 2 |
| | 5 2 6 2 6 2 6 2 2+2 2 |
| | 6 2 5 2 6 2 6 2 3+3 2 |
| | 5 2 5 2 5 2 5 2 2+2 2 |
| 69.1.17 | 174521 114 Au1Eu2Ge3 |
| | -GM -Y -Z -T -L |
| | 5 2 5 2 5 2 6 2 2+2 2 |
| | 6 2 6 2 6 2 5 2 3+3 2 |
| | 5 2 5 2 5 2 5 2 2+2 2 |
| | 6 2 6 2 6 2 6 2 3+3 2 |
| | 5 2 5 2 5 2 6 2 2+2 2 |
| | 6 2 6 2 6 2 5 2 3+3 2 |
| | 5 2 6 2 6 2 5 2 2+2 2 |
| | 6 2 5 2 5 2 6 2 3+3 2 |
| | 5 2 6 2 5 2 6 2 2+2 2 |
| | 5 2 5 2 6 2 5 2 3+3 2 |
| | 6 2 5 2 5 2 5 2 3+3 2 |
| | 5 2 6 2 6 2 6 2 2+2 2 |
| | 6 2 5 2 5 2 6 2 3+3 2 |
| | 6 2 6 2 6 2 5 2 2+2 2 |
| | 5 2 6 2 6 2 5 2 2+2 2 |
| | 6 2 5 2 5 2 6 2 3+3 2 |
| | 5 2 5 2 5 2 5 2 3+3 2 |
| | 6 2 6 2 6 2 6 2 2+2 2 |
| | 6 2 6 2 6 2 6 2 2+2 2 |
| | 5 2 5 2 5 2 5 2 3+3 2 |
| | 5 2 5 2 5 2 5 2 3+3 2 |
| | 6 2 6 2 6 2 6 2 2+2 2 |
| | 5 2 5 2 5 2 5 2 3+3 2 |
| | 5 2 6 2 5 2 6 2 2+2 2 |
| | 6 2 6 2 6 2 5 2 2+2 2 |
| | 6 2 5 2 6 2 6 2 3+3 2 |
| | 5 2 6 2 5 2 6 2 2+2 2 |
| | 5 2 5 2 5 2 5 2 3+3 2 |
| | 6 2 6 2 6 2 6 2 2+2 2 |
| | 5 2 5 2 5 2 5 2 3+3 2 |
| | 6 2 5 2 6 2 6 2 2+2 2 |
| | 5 2 6 2 6 2 5 2 3+3 2 |
| | 6 2 6 2 6 2 5 2 2+2 2 |
| | 6 2 5 2 5 2 6 2 3+3 2 |
| | 6 2 6 2 6 2 6 2 2+2 2 |
| | 6 2 5 2 6 2 5 2 3+3 2 |
| | 5 2 6 2 5 2 6 2 3+3 2 |
| | 5 2 5 2 5 2 5 2 2+2 2 |
| | 6 2 5 2 6 2 5 2 3+3 2 |
| | 6 2 6 2 6 2 5 2 3+3 2 |
| | 5 2 5 2 5 2 6 2 2+2 2 |
| | 6 2 6 2 5 2 6 2 2+2 2 |
| | 6 2 5 2 5 2 6 2 2+2 2 |
| | 5 2 6 2 6 2 5 2 3+3 2 |
| | 5 2 5 2 5 2 6 2 2+2 2 |
| | 5 2 5 2 5 2 5 2 3+3 2 |
| | 5 2 6 2 6 2 6 2 3+3 2 |

Continued on next column

**TABLE XLII– continued**

| TQC | ICSD NoE SF |
|---|---|
| | 6 2 5 2 5 2 5 2 2+2 2 |
| | 5 2 6 2 6 2 5 2 3+3 2 |
| | 6 2 5 2 6 2 5 2 2+2 2 |
| | 5 2 6 2 6 2 6 2 3+3 2 |
| | 6 2 6 2 5 2 6 2 2+2 2 |
| | 5 2 5 2 6 2 6 2 3+3 2 |
| | 6 2 6 2 5 2 5 2 2+2 2 |
| | 5 2 5 2 5 2 6 2 2+2 2 |
| | 6 2 6 2 5 2 5 2 3+3 2 |
| | 5 2 5 2 5 2 5 2 3+3 2 |
| 69.1.18 | 410522 54 Ag2Ca4Si6 |
| | -GM -Y -Z -T -L |
| | 5 2 5 2 5 2 5 2 3+3 2 |
| | 6 2 6 2 6 2 6 2 2+2 2 |
| | 6 2 6 2 6 2 6 2 2+2 2 |
| | 5 2 5 2 5 2 5 2 3+3 2 |
| | 5 2 5 2 5 2 5 2 3+3 2 |
| | 6 2 6 2 6 2 6 2 2+2 2 |
| | 5 2 6 2 5 2 5 2 3+3 2 |
| | 5 2 5 2 5 2 6 2 2+2 2 |
| | 6 2 6 2 6 2 6 2 2+2 2 |
| | 5 2 5 2 5 2 5 2 3+3 2 |
| | 6 2 6 2 6 2 5 2 3+3 2 |
| | 5 2 5 2 5 2 6 2 2+2 2 |
| | 6 2 6 2 6 2 6 2 2+2 2 |
| | 5 2 5 2 5 2 5 2 3+3 2 |
| | 6 2 5 2 6 2 5 2 3+3 2 |
| | 5 2 6 2 6 2 6 2 2+2 2 |
| | 6 2 6 2 6 2 6 2 2+2 2 |
| | 6 2 6 2 5 2 6 2 3+3 2 |
| | 5 2 5 2 6 2 5 2 2+2 2 |
| | 6 2 5 2 5 2 5 2 3+3 2 |
| | 5 2 5 2 6 2 5 2 2+2 2 |
| | 6 2 6 2 5 2 6 2 3+3 2 |
| | 6 2 6 2 6 2 5 2 3+3 2 |
| | 6 2 5 2 6 2 6 2 2+2 2 |
| | 5 2 5 2 5 2 5 2 3+3 2 |
| | 5 2 6 2 5 2 6 2 2+2 2 |
| | 6 2 5 2 5 2 6 2 2+2 2 |

The End

TABLE XLIII: The computed irreducible representations for the valence bands for the compounds of SEBR and NLC in space group 70. See the list of tables.

| TQC | ICSD | NoE | SF | |
|---|---|---|---|---|
| 70.1.1 | 1089 | 24 | Si2Ti1 | |
| | -GM | -Y | -Z | -T -L |
| | 5 2 | 3+4 4 | 3+4 4 | 3+4 4 2+2 2 |
| | 6 2 | 3+4 4 | 3+4 4 | 3+4 4 3+3 2 |
| | 5 2 | 3+4 4 | 3+4 4 | 3+4 4 2+2 2 |
| | 5 2 | 3+4 4 | 3+4 4 | 3+4 4 3+3 2 |
| | 6 2 | 3+4 4 | 3+4 4 | 3+4 4 3+3 2 |
| | 6 2 | 3+4 4 | 3+4 4 | 3+4 4 2+2 2 |
| | 6 2 | | | 3+3 2 |
| | 5 2 | | | 2+2 2 |
| | 5 2 | | | 2+2 2 |
| | 5 2 | | | 2+2 2 |
| | 6 2 | | | 3+3 2 |
| | 5 2 | | | 3+3 2 |

Continued on next column

**TABLE XLIII– continued**

| TQC | ICSD | NoE | SF | | |
|---|---|---|---|---|---|
| 70.1.2 | 97545 | 52 | Sb1Sn1Ti1 | | |
| | -GM | -Y | -Z | -T | -L |
| | 5 2 | 3+4 4 | 3+4 4 | 3+4 4 | 3+3 2 |
| | 6 2 | 3+4 4 | 3+4 4 | 3+4 4 | 2+2 2 |
| | 5 2 | 3+4 4 | 3+4 4 | 3+4 4 | 3+3 2 |
| | 6 2 | 3+4 4 | 3+4 4 | 3+4 4 | 2+2 2 |
| | 6 2 | 3+4 4 | 3+4 4 | 3+4 4 | 3+3 2 |
| | 5 2 | 3+4 4 | 3+4 4 | 3+4 4 | 2+2 2 |
| | 6 2 | 3+4 4 | 3+4 4 | 3+4 4 | 2+2 2 |
| | 5 2 | 3+4 4 | 3+4 4 | 3+4 4 | 3+3 2 |
| | 5 2 | 3+4 4 | 3+4 4 | 3+4 4 | 3+3 2 |
| | 5 2 | 3+4 4 | 3+4 4 | 3+4 4 | 2+2 2 |
| | 6 2 | 3+4 4 | 3+4 4 | 3+4 4 | 2+2 2 |
| | 5 2 | 3+4 4 | 3+4 4 | 3+4 4 | 3+3 2 |
| | 5 2 | 3+4 4 | 3+4 4 | 3+4 4 | 3+3 2 |
| | 6 2 | | | | 2+2 2 |
| | 6 2 | | | | 3+3 2 |
| | 6 2 | | | | 3+3 2 |
| | 5 2 | | | | 2+2 2 |
| | 6 2 | | | | 2+2 2 |
| | 5 2 | | | | 3+3 2 |
| | 5 2 | | | | 3+3 2 |
| | 6 2 | | | | 2+2 2 |
| | 6 2 | | | | 3+3 2 |
| | 6 2 | | | | 2+2 2 |
| | 5 2 | | | | 2+2 2 |
| | 5 2 | | | | 3+3 2 |
| | 5 2 | | | | 2+2 2 |
| 70.1.3 | 88369 | 96 | O4Tb1V1 | | |
| | -GM | -Y | -Z | -T | -L |
| | 5 2 | 3+4 4 | 3+4 4 | 3+4 4 | 3+3 2 |
| | 6 2 | 3+4 4 | 3+4 4 | 3+4 4 | 2+2 2 |
| | 5 2 | 3+4 4 | 3+4 4 | 3+4 4 | 3+3 2 |
| | 6 2 | 3+4 4 | 3+4 4 | 3+4 4 | 2+2 2 |
| | 5 2 | 3+4 4 | 3+4 4 | 3+4 4 | 3+3 2 |
| | 5 2 | 3+4 4 | 3+4 4 | 3+4 4 | 2+2 2 |
| | 6 2 | 3+4 4 | 3+4 4 | 3+4 4 | 3+3 2 |
| | 6 2 | 3+4 4 | 3+4 4 | 3+4 4 | 2+2 2 |
| | 5 2 | 3+4 4 | 3+4 4 | 3+4 4 | 3+3 2 |
| | 6 2 | 3+4 4 | 3+4 4 | 3+4 4 | 2+2 2 |
| | 6 2 | 3+4 4 | 3+4 4 | 3+4 4 | 3+3 2 |
| | 5 2 | 3+4 4 | 3+4 4 | 3+4 4 | 2+2 2 |
| | 6 2 | 3+4 4 | 3+4 4 | 3+4 4 | 2+2 2 |
| | 5 2 | 3+4 4 | 3+4 4 | 3+4 4 | 3+3 2 |
| | 6 2 | 3+4 4 | 3+4 4 | 3+4 4 | 2+2 2 |
| | 5 2 | 3+4 4 | 3+4 4 | 3+4 4 | 3+3 2 |
| | 6 2 | 3+4 4 | 3+4 4 | 3+4 4 | 3+3 2 |
| | 5 2 | 3+4 4 | 3+4 4 | 3+4 4 | 2+2 2 |
| | 5 2 | 3+4 4 | 3+4 4 | 3+4 4 | 3+3 2 |
| | 6 2 | 3+4 4 | 3+4 4 | 3+4 4 | 2+2 2 |
| | 5 2 | 3+4 4 | 3+4 4 | 3+4 4 | 2+2 2 |
| | 5 2 | 3+4 4 | 3+4 4 | 3+4 4 | 3+3 2 |
| | 5 2 | 3+4 4 | 3+4 4 | 3+4 4 | 3+3 2 |
| | 5 2 | 3+4 4 | 3+4 4 | 3+4 4 | 2+2 2 |
| | 5 2 | | | | 3+3 2 |
| | 6 2 | | | | 3+3 2 |
| | 6 2 | | | | 2+2 2 |
| | 6 2 | | | | 2+2 2 |
| | 5 2 | | | | 3+3 2 |
| | 6 2 | | | | 2+2 2 |
| | 5 2 | | | | 3+3 2 |

Continued on next column

**TABLE XLIII– continued**

| TQC | ICSD | NoE | SF | | |
|---|---|---|---|---|---|
| | 6 2 | | | | 2+2 2 |
| | 6 2 | | | | 2+2 2 |
| | 6 2 | | | | 3+3 2 |
| | 6 2 | | | | 3+3 2 |
| | 5 2 | | | | 3+3 2 |
| | 5 2 | | | | 2+2 2 |
| | 5 2 | | | | 2+2 2 |
| | 6 2 | | | | 3+3 2 |
| | 6 2 | | | | 2+2 2 |
| | 5 2 | | | | 3+3 2 |
| | 6 2 | | | | 2+2 2 |
| | 5 2 | | | | 3+3 2 |
| | 6 2 | | | | 2+2 2 |
| | 5 2 | | | | 3+3 2 |
| | 6 2 | | | | 2+2 2 |
| | 5 2 | | | | 3+3 2 |
| | 5 2 | | | | 2+2 2 |
| 70.1.4 | 106552 | 84 | Nb1Sn2 | | |
| | -GM | -Y | -Z | -T | -L |
| | 5 2 | 3+4 4 | 3+4 4 | 3+4 4 | 2+2 2 |
| | 5 2 | 3+4 4 | 3+4 4 | 3+4 4 | 2+2 2 |
| | 6 2 | 3+4 4 | 3+4 4 | 3+4 4 | 3+3 2 |
| | 6 2 | 3+4 4 | 3+4 4 | 3+4 4 | 3+3 2 |
| | 5 2 | 3+4 4 | 3+4 4 | 3+4 4 | 2+2 2 |
| | 5 2 | 3+4 4 | 3+4 4 | 3+4 4 | 2+2 2 |
| | 6 2 | 3+4 4 | 3+4 4 | 3+4 4 | 3+3 2 |
| | 6 2 | 3+4 4 | 3+4 4 | 3+4 4 | 3+3 2 |
| | 5 2 | 3+4 4 | 3+4 4 | 3+4 4 | 2+2 2 |
| | 5 2 | 3+4 4 | 3+4 4 | 3+4 4 | 3+3 2 |
| | 6 2 | 3+4 4 | 3+4 4 | 3+4 4 | 2+2 2 |
| | 6 2 | 3+4 4 | 3+4 4 | 3+4 4 | 3+3 2 |
| | 5 2 | 3+4 4 | 3+4 4 | 3+4 4 | 2+2 2 |
| | 5 2 | 3+4 4 | 3+4 4 | 3+4 4 | 3+3 2 |
| | 6 2 | 3+4 4 | 3+4 4 | 3+4 4 | 2+2 2 |
| | 6 2 | 3+4 4 | 3+4 4 | 3+4 4 | 3+3 2 |
| | 5 2 | 3+4 4 | 3+4 4 | 3+4 4 | 3+3 2 |
| | 6 2 | 3+4 4 | 3+4 4 | 3+4 4 | 2+2 2 |
| | 5 2 | 3+4 4 | 3+4 4 | 3+4 4 | 3+3 2 |
| | 5 2 | 3+4 4 | 3+4 4 | 3+4 4 | 2+2 2 |
| | 6 2 | 3+4 4 | 3+4 4 | 3+4 4 | 3+3 2 |
| | 6 2 | | | | 2+2 2 |
| | 6 2 | | | | 3+3 2 |
| | 5 2 | | | | 2+2 2 |
| | 5 2 | | | | 3+3 2 |
| | 6 2 | | | | 2+2 2 |
| | 5 2 | | | | 2+2 2 |
| | 6 2 | | | | 3+3 2 |
| | 6 2 | | | | 3+3 2 |
| | 5 2 | | | | 3+3 2 |
| | 5 2 | | | | 2+2 2 |
| | 6 2 | | | | 3+3 2 |
| | 5 2 | | | | 2+2 2 |
| | 6 2 | | | | 3+3 2 |
| | 6 2 | | | | 2+2 2 |
| | 5 2 | | | | 2+2 2 |
| | 5 2 | | | | 3+3 2 |
| | 5 2 | | | | 2+2 2 |
| | 6 2 | | | | 3+3 2 |
| | 5 2 | | | | 2+2 2 |
| | 6 2 | | | | 3+3 2 |
| | 5 2 | | | | 3+3 2 |

Continued on next column

**TABLE XLIII– continued**

| TQC | ICSD | | NoE | SF | |
|---|---|---|---|---|---|
| 70.1.5 | 429511 | | 68 | Ga2Rh2Sr1 | |
| | -GM | -Y | -Z | -T | -L |
| | 5 2 | 3+4 4 | 3+4 4 | 3+4 4 | 2+2 2 |
| | 6 2 | 3+4 4 | 3+4 4 | 3+4 4 | 3+3 2 |
| | 5 2 | 3+4 4 | 3+4 4 | 3+4 4 | 2+2 2 |
| | 6 2 | 3+4 4 | 3+4 4 | 3+4 4 | 3+3 2 |
| | 5 2 | 3+4 4 | 3+4 4 | 3+4 4 | 2+2 2 |
| | 6 2 | 3+4 4 | 3+4 4 | 3+4 4 | 3+3 2 |
| | 5 2 | 3+4 4 | 3+4 4 | 3+4 4 | 2+2 2 |
| | 6 2 | 3+4 4 | 3+4 4 | 3+4 4 | 3+3 2 |
| | 5 2 | 3+4 4 | 3+4 4 | 3+4 4 | 3+3 2 |
| | 6 2 | 3+4 4 | 3+4 4 | 3+4 4 | 2+2 2 |
| | 6 2 | 3+4 4 | 3+4 4 | 3+4 4 | 3+3 2 |
| | 5 2 | 3+4 4 | 3+4 4 | 3+4 4 | 2+2 2 |
| | 5 2 | 3+4 4 | 3+4 4 | 3+4 4 | 2+2 2 |
| | 6 2 | 3+4 4 | 3+4 4 | 3+4 4 | 3+3 2 |
| | 6 2 | 3+4 4 | 3+4 4 | 3+4 4 | 2+2 2 |
| | 6 2 | 3+4 4 | 3+4 4 | 3+4 4 | 2+2 2 |
| | 5 2 | 3+4 4 | 3+4 4 | 3+4 4 | 3+3 2 |
| | 5 2 | | | | 3+3 2 |
| | 5 2 | | | | 2+2 2 |
| | 6 2 | | | | 3+3 2 |
| | 5 2 | | | | 2+2 2 |
| | 6 2 | | | | 3+3 2 |
| | 6 2 | | | | 2+2 2 |
| | 5 2 | | | | 3+3 2 |
| | 6 2 | | | | 3+3 2 |
| | 5 2 | | | | 2+2 2 |
| | 6 2 | | | | 2+2 2 |
| | 5 2 | | | | 3+3 2 |
| | 6 2 | | | | 2+2 2 |
| | 5 2 | | | | 3+3 2 |
| | 5 2 | | | | 2+2 2 |
| | 5 2 | | | | 3+3 2 |
| | 5 2 | | | | 3+3 2 |
| | 6 2 | | | | 3+3 2 |
| 70.1.6 | 89074 | | 80 | Cr2O4Si1 | |
| | -GM | -Y | -Z | -T | -L |
| | 5 2 | 3+4 4 | 3+4 4 | 3+4 4 | 2+2 2 |
| | 6 2 | 3+4 4 | 3+4 4 | 3+4 4 | 3+3 2 |
| | 5 2 | 3+4 4 | 3+4 4 | 3+4 4 | 3+3 2 |
| | 6 2 | 3+4 4 | 3+4 4 | 3+4 4 | 3+3 2 |
| | 5 2 | 3+4 4 | 3+4 4 | 3+4 4 | 2+2 2 |
| | 6 2 | 3+4 4 | 3+4 4 | 3+4 4 | 2+2 2 |
| | 5 2 | 3+4 4 | 3+4 4 | 3+4 4 | 2+2 2 |
| | 6 2 | 3+4 4 | 3+4 4 | 3+4 4 | 3+3 2 |
| | 5 2 | 3+4 4 | 3+4 4 | 3+4 4 | 3+3 2 |
| | 6 2 | 3+4 4 | 3+4 4 | 3+4 4 | 2+2 2 |
| | 5 2 | 3+4 4 | 3+4 4 | 3+4 4 | 3+3 2 |
| | 5 2 | 3+4 4 | 3+4 4 | 3+4 4 | 2+2 2 |
| | 5 2 | 3+4 4 | 3+4 4 | 3+4 4 | 3+3 2 |
| | 6 2 | 3+4 4 | 3+4 4 | 3+4 4 | 2+2 2 |
| | 6 2 | 3+4 4 | 3+4 4 | 3+4 4 | 2+2 2 |
| | 6 2 | 3+4 4 | 3+4 4 | 3+4 4 | 3+3 2 |
| | 6 2 | 3+4 4 | 3+4 4 | 3+4 4 | 3+3 2 |
| | 5 2 | 3+4 4 | 3+4 4 | 3+4 4 | 3+3 2 |
| | 5 2 | 3+4 4 | 3+4 4 | 3+4 4 | 2+2 2 |
| | 6 2 | 3+4 4 | 3+4 4 | 3+4 4 | 3+3 2 |
| | 5 2 | | | | 2+2 2 |
| | 6 2 | | | | 2+2 2 |
| | 5 2 | | | | 3+3 2 |
| | 5 2 | | | | 2+2 2 |
| | 5 2 | | | | 2+2 2 |
| | 6 2 | | | | 3+3 2 |
| | 6 2 | | | | 3+3 2 |
| | 6 2 | | | | 2+2 2 |
| | 5 2 | | | | 3+3 2 |
| | 6 2 | | | | 2+2 2 |
| | 5 2 | | | | 3+3 2 |
| | 6 2 | | | | 2+2 2 |
| | 5 2 | | | | 3+3 2 |
| | 6 2 | | | | 2+2 2 |
| | 5 2 | | | | 3+3 2 |
| | 6 2 | | | | 2+2 2 |
| | 6 2 | | | | 3+3 2 |
| | 5 2 | | | | 3+3 2 |
| | 5 2 | | | | 2+2 2 |
| | 5 2 | | | | 2+2 2 |
| 70.1.7 | 89073 | | 80 | Cr2O4Si1 | |
| | -GM | -Y | -Z | -T | -L |
| | 5 2 | 3+4 4 | 3+4 4 | 3+4 4 | 3+3 2 |
| | 6 2 | 3+4 4 | 3+4 4 | 3+4 4 | 2+2 2 |
| | 5 2 | 3+4 4 | 3+4 4 | 3+4 4 | 2+2 2 |
| | 6 2 | 3+4 4 | 3+4 4 | 3+4 4 | 2+2 2 |
| | 5 2 | 3+4 4 | 3+4 4 | 3+4 4 | 3+3 2 |
| | 6 2 | 3+4 4 | 3+4 4 | 3+4 4 | 3+3 2 |
| | 5 2 | 3+4 4 | 3+4 4 | 3+4 4 | 3+3 2 |
| | 6 2 | 3+4 4 | 3+4 4 | 3+4 4 | 2+2 2 |
| | 5 2 | 3+4 4 | 3+4 4 | 3+4 4 | 2+2 2 |
| | 6 2 | 3+4 4 | 3+4 4 | 3+4 4 | 3+3 2 |
| | 5 2 | 3+4 4 | 3+4 4 | 3+4 4 | 2+2 2 |
| | 5 2 | 3+4 4 | 3+4 4 | 3+4 4 | 3+3 2 |
| | 5 2 | 3+4 4 | 3+4 4 | 3+4 4 | 2+2 2 |
| | 6 2 | 3+4 4 | 3+4 4 | 3+4 4 | 3+3 2 |
| | 6 2 | 3+4 4 | 3+4 4 | 3+4 4 | 3+3 2 |
| | 6 2 | 3+4 4 | 3+4 4 | 3+4 4 | 2+2 2 |
| | 6 2 | 3+4 4 | 3+4 4 | 3+4 4 | 2+2 2 |
| | 5 2 | 3+4 4 | 3+4 4 | 3+4 4 | 2+2 2 |
| | 5 2 | 3+4 4 | 3+4 4 | 3+4 4 | 3+3 2 |
| | 6 2 | 3+4 4 | 3+4 4 | 3+4 4 | 2+2 2 |
| | 5 2 | | | | 3+3 2 |
| | 6 2 | | | | 3+3 2 |
| | 5 2 | | | | 2+2 2 |
| | 5 2 | | | | 3+3 2 |
| | 5 2 | | | | 3+3 2 |
| | 6 2 | | | | 2+2 2 |
| | 6 2 | | | | 2+2 2 |
| | 6 2 | | | | 3+3 2 |
| | 5 2 | | | | 2+2 2 |
| | 6 2 | | | | 3+3 2 |
| | 5 2 | | | | 2+2 2 |
| | 6 2 | | | | 3+3 2 |
| | 5 2 | | | | 2+2 2 |
| | 6 2 | | | | 3+3 2 |
| | 5 2 | | | | 2+2 2 |
| | 6 2 | | | | 3+3 2 |
| | 5 2 | | | | 2+2 2 |
| | 6 2 | | | | 2+2 2 |
| | 5 2 | | | | 3+3 2 |
| | 5 2 | | | | 3+3 2 |
| 70.1.8 | 89072 | | 80 | Cr2O4Si1 | |
| | -GM | -Y | -Z | -T | -L |

Continued on next column

Continued on next column

**TABLE XLIII– continued**

| TQC | ICSD | | NoE | SF | |
|---|---|---|---|---|---|
| | 5 2 | 3+4 4 | 3+4 4 | 3+4 4 | 2+2 2 |
| | 6 2 | 3+4 4 | 3+4 4 | 3+4 4 | 3+3 2 |
| | 5 2 | 3+4 4 | 3+4 4 | 3+4 4 | 3+3 2 |
| | 6 2 | 3+4 4 | 3+4 4 | 3+4 4 | 3+3 2 |
| | 5 2 | 3+4 4 | 3+4 4 | 3+4 4 | 2+2 2 |
| | 6 2 | 3+4 4 | 3+4 4 | 3+4 4 | 2+2 2 |
| | 5 2 | 3+4 4 | 3+4 4 | 3+4 4 | 2+2 2 |
| | 6 2 | 3+4 4 | 3+4 4 | 3+4 4 | 3+3 2 |
| | 5 2 | 3+4 4 | 3+4 4 | 3+4 4 | 3+3 2 |
| | 6 2 | 3+4 4 | 3+4 4 | 3+4 4 | 2+2 2 |
| | 5 2 | 3+4 4 | 3+4 4 | 3+4 4 | 3+3 2 |
| | 5 2 | 3+4 4 | 3+4 4 | 3+4 4 | 2+2 2 |
| | 5 2 | 3+4 4 | 3+4 4 | 3+4 4 | 3+3 2 |
| | 6 2 | 3+4 4 | 3+4 4 | 3+4 4 | 2+2 2 |
| | 6 2 | 3+4 4 | 3+4 4 | 3+4 4 | 2+2 2 |
| | 6 2 | 3+4 4 | 3+4 4 | 3+4 4 | 3+3 2 |
| | 6 2 | 3+4 4 | 3+4 4 | 3+4 4 | 3+3 2 |
| | 5 2 | 3+4 4 | 3+4 4 | 3+4 4 | 3+3 2 |
| | 5 2 | 3+4 4 | 3+4 4 | 3+4 4 | 2+2 2 |
| | 6 2 | 3+4 4 | 3+4 4 | 3+4 4 | 3+3 2 |
| | 5 2 | | | | 2+2 2 |
| | 6 2 | | | | 2+2 2 |
| | 5 2 | | | | 3+3 2 |
| | 5 2 | | | | 2+2 2 |
| | 5 2 | | | | 2+2 2 |
| | 6 2 | | | | 3+3 2 |
| | 6 2 | | | | 3+3 2 |
| | 6 2 | | | | 2+2 2 |
| | 5 2 | | | | 3+3 2 |
| | 6 2 | | | | 2+2 2 |
| | 5 2 | | | | 3+3 2 |
| | 6 2 | | | | 2+2 2 |
| | 5 2 | | | | 3+3 2 |
| | 6 2 | | | | 2+2 2 |
| | 5 2 | | | | 3+3 2 |
| | 6 2 | | | | 2+2 2 |
| | 5 2 | | | | 3+3 2 |
| | 6 2 | | | | 2+2 2 |
| | 5 2 | | | | 3+3 2 |
| | 5 2 | | | | 2+2 2 |
| 70.1.9 | 89071 | | 80 | Cr2O4Si1 | |
| | -GM | -Y | -Z | -T | -L |
| | 5 2 | 3+4 4 | 3+4 4 | 3+4 4 | 2+2 2 |
| | 6 2 | 3+4 4 | 3+4 4 | 3+4 4 | 3+3 2 |
| | 5 2 | 3+4 4 | 3+4 4 | 3+4 4 | 3+3 2 |
| | 6 2 | 3+4 4 | 3+4 4 | 3+4 4 | 3+3 2 |
| | 5 2 | 3+4 4 | 3+4 4 | 3+4 4 | 2+2 2 |
| | 6 2 | 3+4 4 | 3+4 4 | 3+4 4 | 2+2 2 |
| | 5 2 | 3+4 4 | 3+4 4 | 3+4 4 | 2+2 2 |
| | 6 2 | 3+4 4 | 3+4 4 | 3+4 4 | 3+3 2 |
| | 5 2 | 3+4 4 | 3+4 4 | 3+4 4 | 3+3 2 |
| | 6 2 | 3+4 4 | 3+4 4 | 3+4 4 | 2+2 2 |
| | 5 2 | 3+4 4 | 3+4 4 | 3+4 4 | 3+3 2 |
| | 5 2 | 3+4 4 | 3+4 4 | 3+4 4 | 2+2 2 |
| | 5 2 | 3+4 4 | 3+4 4 | 3+4 4 | 3+3 2 |
| | 6 2 | 3+4 4 | 3+4 4 | 3+4 4 | 2+2 2 |
| | 6 2 | 3+4 4 | 3+4 4 | 3+4 4 | 2+2 2 |
| | 6 2 | 3+4 4 | 3+4 4 | 3+4 4 | 3+3 2 |
| | 6 2 | 3+4 4 | 3+4 4 | 3+4 4 | 3+3 2 |
| | 5 2 | 3+4 4 | 3+4 4 | 3+4 4 | 3+3 2 |
| | 6 2 | 3+4 4 | 3+4 4 | 3+4 4 | 2+2 2 |

Continued on next column

**TABLE XLIII– continued**

| TQC | ICSD | | NoE | SF | |
|---|---|---|---|---|---|
| | 5 2 | 3+4 4 | 3+4 4 | 3+4 4 | 3+3 2 |
| | 5 2 | | | | 2+2 2 |
| | 6 2 | | | | 2+2 2 |
| | 5 2 | | | | 3+3 2 |
| | 5 2 | | | | 2+2 2 |
| | 5 2 | | | | 2+2 2 |
| | 6 2 | | | | 3+3 2 |
| | 6 2 | | | | 3+3 2 |
| | 6 2 | | | | 2+2 2 |
| | 5 2 | | | | 3+3 2 |
| | 6 2 | | | | 2+2 2 |
| | 5 2 | | | | 3+3 2 |
| | 6 2 | | | | 2+2 2 |
| | 5 2 | | | | 3+3 2 |
| | 6 2 | | | | 2+2 2 |
| | 5 2 | | | | 3+3 2 |
| | 6 2 | | | | 2+2 2 |
| | 5 2 | | | | 3+3 2 |
| | 5 2 | | | | 2+2 2 |
| | 6 2 | | | | 3+3 2 |
| | 5 2 | | | | 2+2 2 |
| 70.1.10 | 89070 | | 80 | Cr2O4Si1 | |
| | -GM | -Y | -Z | -T | -L |
| | 5 2 | 3+4 4 | 3+4 4 | 3+4 4 | 3+3 2 |
| | 6 2 | 3+4 4 | 3+4 4 | 3+4 4 | 2+2 2 |
| | 5 2 | 3+4 4 | 3+4 4 | 3+4 4 | 2+2 2 |
| | 6 2 | 3+4 4 | 3+4 4 | 3+4 4 | 2+2 2 |
| | 5 2 | 3+4 4 | 3+4 4 | 3+4 4 | 3+3 2 |
| | 6 2 | 3+4 4 | 3+4 4 | 3+4 4 | 3+3 2 |
| | 5 2 | 3+4 4 | 3+4 4 | 3+4 4 | 3+3 2 |
| | 6 2 | 3+4 4 | 3+4 4 | 3+4 4 | 2+2 2 |
| | 5 2 | 3+4 4 | 3+4 4 | 3+4 4 | 2+2 2 |
| | 6 2 | 3+4 4 | 3+4 4 | 3+4 4 | 3+3 2 |
| | 5 2 | 3+4 4 | 3+4 4 | 3+4 4 | 2+2 2 |
| | 5 2 | 3+4 4 | 3+4 4 | 3+4 4 | 3+3 2 |
| | 5 2 | 3+4 4 | 3+4 4 | 3+4 4 | 2+2 2 |
| | 6 2 | 3+4 4 | 3+4 4 | 3+4 4 | 3+3 2 |
| | 6 2 | 3+4 4 | 3+4 4 | 3+4 4 | 3+3 2 |
| | 6 2 | 3+4 4 | 3+4 4 | 3+4 4 | 2+2 2 |
| | 6 2 | 3+4 4 | 3+4 4 | 3+4 4 | 2+2 2 |
| | 5 2 | 3+4 4 | 3+4 4 | 3+4 4 | 3+3 2 |
| | 6 2 | 3+4 4 | 3+4 4 | 3+4 4 | 2+2 2 |
| | 5 2 | 3+4 4 | 3+4 4 | 3+4 4 | 2+2 2 |
| | 5 2 | | | | 3+3 2 |
| | 6 2 | | | | 3+3 2 |
| | 5 2 | | | | 2+2 2 |
| | 5 2 | | | | 3+3 2 |
| | 5 2 | | | | 3+3 2 |
| | 6 2 | | | | 2+2 2 |
| | 6 2 | | | | 3+3 2 |
| | 6 2 | | | | 2+2 2 |
| | 5 2 | | | | 2+2 2 |
| | 6 2 | | | | 3+3 2 |
| | 5 2 | | | | 2+2 2 |
| | 6 2 | | | | 3+3 2 |
| | 5 2 | | | | 2+2 2 |
| | 6 2 | | | | 3+3 2 |
| | 5 2 | | | | 2+2 2 |
| | 6 2 | | | | 3+3 2 |
| | 5 2 | | | | 2+2 2 |
| | 5 2 | | | | 3+3 2 |

Continued on next column

**TABLE XLIII– continued**

| TQC | ICSD | NoE | SF | | |
|---|---|---|---|---|---|
| | 6 2 | | | | 2+2 2 |
| | 5 2 | | | | 3+3 2 |
| 70.1.11 | 89069 | 80 | Cr2O4Si1 | | |
| | -GM | -Y | -Z | -T | -L |
| | 5 2 | 3+4 4 | 3+4 4 | 3+4 4 | 3+3 2 |
| | 6 2 | 3+4 4 | 3+4 4 | 3+4 4 | 2+2 2 |
| | 5 2 | 3+4 4 | 3+4 4 | 3+4 4 | 2+2 2 |
| | 6 2 | 3+4 4 | 3+4 4 | 3+4 4 | 2+2 2 |
| | 5 2 | 3+4 4 | 3+4 4 | 3+4 4 | 3+3 2 |
| | 6 2 | 3+4 4 | 3+4 4 | 3+4 4 | 3+3 2 |
| | 5 2 | 3+4 4 | 3+4 4 | 3+4 4 | 3+3 2 |
| | 6 2 | 3+4 4 | 3+4 4 | 3+4 4 | 2+2 2 |
| | 5 2 | 3+4 4 | 3+4 4 | 3+4 4 | 2+2 2 |
| | 6 2 | 3+4 4 | 3+4 4 | 3+4 4 | 3+3 2 |
| | 5 2 | 3+4 4 | 3+4 4 | 3+4 4 | 2+2 2 |
| | 5 2 | 3+4 4 | 3+4 4 | 3+4 4 | 3+3 2 |
| | 5 2 | 3+4 4 | 3+4 4 | 3+4 4 | 2+2 2 |
| | 6 2 | 3+4 4 | 3+4 4 | 3+4 4 | 3+3 2 |
| | 6 2 | 3+4 4 | 3+4 4 | 3+4 4 | 3+3 2 |
| | 6 2 | 3+4 4 | 3+4 4 | 3+4 4 | 2+2 2 |
| | 6 2 | 3+4 4 | 3+4 4 | 3+4 4 | 2+2 2 |
| | 5 2 | 3+4 4 | 3+4 4 | 3+4 4 | 3+3 2 |
| | 6 2 | 3+4 4 | 3+4 4 | 3+4 4 | 2+2 2 |
| | 5 2 | 3+4 4 | 3+4 4 | 3+4 4 | 2+2 2 |
| | 5 2 | | | | 3+3 2 |
| | 6 2 | | | | 3+3 2 |
| | 5 2 | | | | 2+2 2 |
| | 5 2 | | | | 3+3 2 |
| | 5 2 | | | | 3+3 2 |
| | 6 2 | | | | 2+2 2 |
| | 6 2 | | | | 2+2 2 |
| | 6 2 | | | | 3+3 2 |
| | 5 2 | | | | 3+3 2 |
| | 6 2 | | | | 2+2 2 |
| | 5 2 | | | | 2+2 2 |
| | 6 2 | | | | 3+3 2 |
| | 5 2 | | | | 2+2 2 |
| | 6 2 | | | | 3+3 2 |
| | 5 2 | | | | 2+2 2 |
| | 6 2 | | | | 3+3 2 |
| | 5 2 | | | | 2+2 2 |
| | 5 2 | | | | 3+3 2 |
| | 6 2 | | | | 2+2 2 |
| | 5 2 | | | | 3+3 2 |
| 70.1.12 | 75639 | 80 | Cr2O4Si1 | | |
| | -GM | -Y | -Z | -T | -L |
| | 5 2 | 3+4 4 | 3+4 4 | 3+4 4 | 3+3 2 |
| | 6 2 | 3+4 4 | 3+4 4 | 3+4 4 | 2+2 2 |
| | 5 2 | 3+4 4 | 3+4 4 | 3+4 4 | 2+2 2 |
| | 6 2 | 3+4 4 | 3+4 4 | 3+4 4 | 2+2 2 |
| | 5 2 | 3+4 4 | 3+4 4 | 3+4 4 | 3+3 2 |
| | 6 2 | 3+4 4 | 3+4 4 | 3+4 4 | 3+3 2 |
| | 5 2 | 3+4 4 | 3+4 4 | 3+4 4 | 3+3 2 |
| | 6 2 | 3+4 4 | 3+4 4 | 3+4 4 | 2+2 2 |
| | 5 2 | 3+4 4 | 3+4 4 | 3+4 4 | 2+2 2 |
| | 6 2 | 3+4 4 | 3+4 4 | 3+4 4 | 3+3 2 |
| | 5 2 | 3+4 4 | 3+4 4 | 3+4 4 | 2+2 2 |
| | 5 2 | 3+4 4 | 3+4 4 | 3+4 4 | 3+3 2 |
| | 5 2 | 3+4 4 | 3+4 4 | 3+4 4 | 2+2 2 |
| | 6 2 | 3+4 4 | 3+4 4 | 3+4 4 | 3+3 2 |
| | 6 2 | 3+4 4 | 3+4 4 | 3+4 4 | 3+3 2 |

Continued on next column

**TABLE XLIII– continued**

| TQC | ICSD | NoE | SF | | |
|---|---|---|---|---|---|
| | 6 2 | 3+4 4 | 3+4 4 | 3+4 4 | 2+2 2 |
| | 6 2 | 3+4 4 | 3+4 4 | 3+4 4 | 2+2 2 |
| | 5 2 | 3+4 4 | 3+4 4 | 3+4 4 | 3+3 2 |
| | 6 2 | 3+4 4 | 3+4 4 | 3+4 4 | 2+2 2 |
| | 5 2 | 3+4 4 | 3+4 4 | 3+4 4 | 2+2 2 |
| | 5 2 | | | | 3+3 2 |
| | 6 2 | | | | 3+3 2 |
| | 5 2 | | | | 2+2 2 |
| | 5 2 | | | | 3+3 2 |
| | 5 2 | | | | 3+3 2 |
| | 6 2 | | | | 2+2 2 |
| | 6 2 | | | | 3+3 2 |
| | 6 2 | | | | 2+2 2 |
| | 5 2 | | | | 2+2 2 |
| | 6 2 | | | | 3+3 2 |
| | 5 2 | | | | 2+2 2 |
| | 6 2 | | | | 3+3 2 |
| | 5 2 | | | | 2+2 2 |
| | 6 2 | | | | 3+3 2 |
| | 5 2 | | | | 2+2 2 |
| | 6 2 | | | | 3+3 2 |
| | 5 2 | | | | 2+2 2 |
| | 5 2 | | | | 3+3 2 |
| | 6 2 | | | | 2+2 2 |
| | 5 2 | | | | 3+3 2 |
| 70.1.13 | 96610 | 84 | Cl2Cu2O1 | | |
| | -GM | -Y | -Z | -T | -L |
| | 5 2 | 3+4 4 | 3+4 4 | 3+4 4 | 3+3 2 |
| | 6 2 | 3+4 4 | 3+4 4 | 3+4 4 | 2+2 2 |
| | 5 2 | 3+4 4 | 3+4 4 | 3+4 4 | 2+2 2 |
| | 5 2 | 3+4 4 | 3+4 4 | 3+4 4 | 3+3 2 |
| | 6 2 | 3+4 4 | 3+4 4 | 3+4 4 | 2+2 2 |
| | 6 2 | 3+4 4 | 3+4 4 | 3+4 4 | 3+3 2 |
| | 5 2 | 3+4 4 | 3+4 4 | 3+4 4 | 3+3 2 |
| | 5 2 | 3+4 4 | 3+4 4 | 3+4 4 | 3+3 2 |
| | 5 2 | 3+4 4 | 3+4 4 | 3+4 4 | 2+2 2 |
| | 6 2 | 3+4 4 | 3+4 4 | 3+4 4 | 2+2 2 |
| | 5 2 | 3+4 4 | 3+4 4 | 3+4 4 | 3+3 2 |
| | 6 2 | 3+4 4 | 3+4 4 | 3+4 4 | 2+2 2 |
| | 5 2 | 3+4 4 | 3+4 4 | 3+4 4 | 3+3 2 |
| | 6 2 | 3+4 4 | 3+4 4 | 3+4 4 | 3+3 2 |
| | 5 2 | 3+4 4 | 3+4 4 | 3+4 4 | 2+2 2 |
| | 5 2 | 3+4 4 | 3+4 4 | 3+4 4 | 3+3 2 |
| | 5 2 | 3+4 4 | 3+4 4 | 3+4 4 | 3+3 2 |
| | 6 2 | 3+4 4 | 3+4 4 | 3+4 4 | 2+2 2 |
| | 5 2 | 3+4 4 | 3+4 4 | 3+4 4 | 2+2 2 |
| | 5 2 | 3+4 4 | 3+4 4 | 3+4 4 | 3+3 2 |
| | 6 2 | 3+4 4 | 3+4 4 | 3+4 4 | 2+2 2 |
| | 6 2 | | | | 3+3 2 |
| | 5 2 | | | | 2+2 2 |
| | 5 2 | | | | 3+3 2 |
| | 5 2 | | | | 2+2 2 |
| | 6 2 | | | | 3+3 2 |
| | 5 2 | | | | 3+3 2 |
| | 5 2 | | | | 3+3 2 |
| | 5 2 | | | | 3+3 2 |
| | 5 2 | | | | 3+3 2 |
| | 5 2 | | | | 2+2 2 |
| | 6 2 | | | | 3+3 2 |
| | 5 2 | | | | 2+2 2 |
| | 5 2 | | | | 3+3 2 |

Continued on next column

**TABLE XLIII– continued**

| TQC | ICSD | | NoE | SF | |
|---|---|---|---|---|---|
| | 5 2 | | | | 2+2 2 |
| | 5 2 | | | | 3+3 2 |
| | 6 2 | | | | 2+2 2 |
| | 5 2 | | | | 3+3 2 |
| | 5 2 | | | | 3+3 2 |
| | 5 2 | | | | 3+3 2 |
| | 5 2 | | | | 3+3 2 |
| | 5 2 | | | | 2+2 2 |
| 70.1.14 | 77143 | | 68 | Ba1Ge2Ru2 | |
| | -GM | -Y | -Z | -T | -L |
| | 5 2 | 3+4 4 | 3+4 4 | 3+4 4 | 2+2 2 |
| | 6 2 | 3+4 4 | 3+4 4 | 3+4 4 | 3+3 2 |
| | 5 2 | 3+4 4 | 3+4 4 | 3+4 4 | 2+2 2 |
| | 6 2 | 3+4 4 | 3+4 4 | 3+4 4 | 3+3 2 |
| | 5 2 | 3+4 4 | 3+4 4 | 3+4 4 | 2+2 2 |
| | 5 2 | 3+4 4 | 3+4 4 | 3+4 4 | 3+3 2 |
| | 6 2 | 3+4 4 | 3+4 4 | 3+4 4 | 2+2 2 |
| | 6 2 | 3+4 4 | 3+4 4 | 3+4 4 | 3+3 2 |
| | 5 2 | 3+4 4 | 3+4 4 | 3+4 4 | 3+3 2 |
| | 6 2 | 3+4 4 | 3+4 4 | 3+4 4 | 2+2 2 |
| | 6 2 | 3+4 4 | 3+4 4 | 3+4 4 | 3+3 2 |
| | 5 2 | 3+4 4 | 3+4 4 | 3+4 4 | 2+2 2 |
| | 5 2 | 3+4 4 | 3+4 4 | 3+4 4 | 3+3 2 |
| | 6 2 | 3+4 4 | 3+4 4 | 3+4 4 | 2+2 2 |
| | 6 2 | 3+4 4 | 3+4 4 | 3+4 4 | 2+2 2 |
| | 6 2 | 3+4 4 | 3+4 4 | 3+4 4 | 3+3 2 |
| | 5 2 | 3+4 4 | 3+4 4 | 3+4 4 | 2+2 2 |
| | 5 2 | | | | 2+2 2 |
| | 6 2 | | | | 3+3 2 |
| | 5 2 | | | | 2+2 2 |
| | 6 2 | | | | 3+3 2 |
| | 5 2 | | | | 3+3 2 |
| | 5 2 | | | | 3+3 2 |
| | 6 2 | | | | 2+2 2 |
| | 5 2 | | | | 3+3 2 |
| | 6 2 | | | | 2+2 2 |
| | 6 2 | | | | 3+3 2 |
| | 5 2 | | | | 2+2 2 |
| | 6 2 | | | | 2+2 2 |
| | 5 2 | | | | 3+3 2 |
| | 5 2 | | | | 2+2 2 |
| | 5 2 | | | | 3+3 2 |
| | 5 2 | | | | 3+3 2 |
| | 6 2 | | | | 3+3 2 |
| 70.1.15 | 40620 | | 84 | B2Gd1Ir2 | |
| | -GM | -Y | -Z | -T | -L |
| | 5 2 | 3+4 4 | 3+4 4 | 3+4 4 | 2+2 2 |
| | 6 2 | 3+4 4 | 3+4 4 | 3+4 4 | 3+3 2 |
| | 5 2 | 3+4 4 | 3+4 4 | 3+4 4 | 2+2 2 |
| | 6 2 | 3+4 4 | 3+4 4 | 3+4 4 | 3+3 2 |
| | 5 2 | 3+4 4 | 3+4 4 | 3+4 4 | 2+2 2 |
| | 5 2 | 3+4 4 | 3+4 4 | 3+4 4 | 3+3 2 |
| | 6 2 | 3+4 4 | 3+4 4 | 3+4 4 | 2+2 2 |
| | 6 2 | 3+4 4 | 3+4 4 | 3+4 4 | 3+3 2 |
| | 5 2 | 3+4 4 | 3+4 4 | 3+4 4 | 3+3 2 |
| | 6 2 | 3+4 4 | 3+4 4 | 3+4 4 | 2+2 2 |
| | 6 2 | 3+4 4 | 3+4 4 | 3+4 4 | 3+3 2 |
| | 5 2 | 3+4 4 | 3+4 4 | 3+4 4 | 2+2 2 |
| | 5 2 | 3+4 4 | 3+4 4 | 3+4 4 | 3+3 2 |
| | 6 2 | 3+4 4 | 3+4 4 | 3+4 4 | 2+2 2 |
| | 6 2 | 3+4 4 | 3+4 4 | 3+4 4 | 2+2 2 |

Continued on next column

**TABLE XLIII– continued**

| TQC | ICSD | | NoE | SF | |
|---|---|---|---|---|---|
| | 6 2 | 3+4 4 | 3+4 4 | 3+4 4 | 2+2 2 |
| | 5 2 | 3+4 4 | 3+4 4 | 3+4 4 | 3+3 2 |
| | 5 2 | 3+4 4 | 3+4 4 | 3+4 4 | 2+2 2 |
| | 5 2 | 3+4 4 | 3+4 4 | 3+4 4 | 3+3 2 |
| | 6 2 | 3+4 4 | 3+4 4 | 3+4 4 | 2+2 2 |
| | 5 2 | 3+4 4 | 3+4 4 | 3+4 4 | 3+3 2 |
| | 6 2 | | | | 3+3 2 |
| | 5 2 | | | | 3+3 2 |
| | 6 2 | | | | 3+3 2 |
| | 6 2 | | | | 2+2 2 |
| | 5 2 | | | | 2+2 2 |
| | 6 2 | | | | 2+2 2 |
| | 5 2 | | | | 3+3 2 |
| | 6 2 | | | | 2+2 2 |
| | 5 2 | | | | 3+3 2 |
| | 5 2 | | | | 3+3 2 |
| | 5 2 | | | | 2+2 2 |
| | 5 2 | | | | 3+3 2 |
| | 6 2 | | | | 3+3 2 |
| | 6 2 | | | | 2+2 2 |
| | 5 2 | | | | 3+3 2 |
| | 6 2 | | | | 2+2 2 |
| | 5 2 | | | | 3+3 2 |
| | 5 2 | | | | 2+2 2 |
| | 6 2 | | | | 3+3 2 |
| | 6 2 | | | | 2+2 2 |
| | 5 2 | | | | 3+3 2 |
| 70.1.16 | 40799 | | 68 | B2Ce1Ru2 | |
| | -GM | -Y | -Z | -T | -L |
| | 5 2 | 3+4 4 | 3+4 4 | 3+4 4 | 2+2 2 |
| | 6 2 | 3+4 4 | 3+4 4 | 3+4 4 | 3+3 2 |
| | 5 2 | 3+4 4 | 3+4 4 | 3+4 4 | 2+2 2 |
| | 6 2 | 3+4 4 | 3+4 4 | 3+4 4 | 3+3 2 |
| | 5 2 | 3+4 4 | 3+4 4 | 3+4 4 | 2+2 2 |
| | 5 2 | 3+4 4 | 3+4 4 | 3+4 4 | 3+3 2 |
| | 6 2 | 3+4 4 | 3+4 4 | 3+4 4 | 2+2 2 |
| | 6 2 | 3+4 4 | 3+4 4 | 3+4 4 | 3+3 2 |
| | 5 2 | 3+4 4 | 3+4 4 | 3+4 4 | 3+3 2 |
| | 6 2 | 3+4 4 | 3+4 4 | 3+4 4 | 2+2 2 |
| | 6 2 | 3+4 4 | 3+4 4 | 3+4 4 | 3+3 2 |
| | 5 2 | 3+4 4 | 3+4 4 | 3+4 4 | 2+2 2 |
| | 5 2 | 3+4 4 | 3+4 4 | 3+4 4 | 2+2 2 |
| | 6 2 | 3+4 4 | 3+4 4 | 3+4 4 | 3+3 2 |
| | 6 2 | 3+4 4 | 3+4 4 | 3+4 4 | 2+2 2 |
| | 6 2 | 3+4 4 | 3+4 4 | 3+4 4 | 2+2 2 |
| | 5 2 | 3+4 4 | 3+4 4 | 3+4 4 | 3+3 2 |
| | 5 2 | | | | 3+3 2 |
| | 6 2 | | | | 2+2 2 |
| | 5 2 | | | | 3+3 2 |
| | 6 2 | | | | 2+2 2 |
| | 6 2 | | | | 3+3 2 |
| | 5 2 | | | | 3+3 2 |
| | 5 2 | | | | 2+2 2 |
| | 6 2 | | | | 3+3 2 |
| | 6 2 | | | | 2+2 2 |
| | 5 2 | | | | 2+2 2 |
| | 5 2 | | | | 3+3 2 |
| | 6 2 | | | | 3+3 2 |
| | 5 2 | | | | 2+2 2 |
| | 5 2 | | | | 3+3 2 |
| | 5 2 | | | | 2+2 2 |

Continued on next column

**TABLE XLIII– continued**

| TQC | ICSD | | NoE | SF | |
|---|---|---|---|---|---|
| | 5 2 | | | | 3+3 2 |
| | 6 2 | | | | 3+3 2 |
| 70.1.17 | 40618 | | 72 | B2Ce1Ir2 | |
| | -GM | -Y | -Z | -T | -L |
| | 5 2 | 3+4 4 | 3+4 4 | 3+4 4 | 3+3 2 |
| | 6 2 | 3+4 4 | 3+4 4 | 3+4 4 | 2+2 2 |
| | 5 2 | 3+4 4 | 3+4 4 | 3+4 4 | 3+3 2 |
| | 6 2 | 3+4 4 | 3+4 4 | 3+4 4 | 2+2 2 |
| | 5 2 | 3+4 4 | 3+4 4 | 3+4 4 | 3+3 2 |
| | 5 2 | 3+4 4 | 3+4 4 | 3+4 4 | 2+2 2 |
| | 6 2 | 3+4 4 | 3+4 4 | 3+4 4 | 3+3 2 |
| | 6 2 | 3+4 4 | 3+4 4 | 3+4 4 | 2+2 2 |
| | 5 2 | 3+4 4 | 3+4 4 | 3+4 4 | 2+2 2 |
| | 6 2 | 3+4 4 | 3+4 4 | 3+4 4 | 3+3 2 |
| | 6 2 | 3+4 4 | 3+4 4 | 3+4 4 | 2+2 2 |
| | 5 2 | 3+4 4 | 3+4 4 | 3+4 4 | 3+3 2 |
| | 5 2 | 3+4 4 | 3+4 4 | 3+4 4 | 2+2 2 |
| | 6 2 | 3+4 4 | 3+4 4 | 3+4 4 | 3+3 2 |
| | 6 2 | 3+4 4 | 3+4 4 | 3+4 4 | 3+3 2 |
| | 6 2 | 3+4 4 | 3+4 4 | 3+4 4 | 3+3 2 |
| | 5 2 | 3+4 4 | 3+4 4 | 3+4 4 | 2+2 2 |
| | 5 2 | 3+4 4 | 3+4 4 | 3+4 4 | 3+3 2 |
| | 5 2 | | | | 2+2 2 |
| | 6 2 | | | | 3+3 2 |
| | 6 2 | | | | 2+2 2 |
| | 5 2 | | | | 2+2 2 |
| | 5 2 | | | | 2+2 2 |
| | 6 2 | | | | 2+2 2 |
| | 6 2 | | | | 3+3 2 |
| | 5 2 | | | | 3+3 2 |
| | 6 2 | | | | 2+2 2 |
| | 5 2 | | | | 3+3 2 |
| | 6 2 | | | | 3+3 2 |
| | 5 2 | | | | 2+2 2 |
| | 5 2 | | | | 2+2 2 |
| | 5 2 | | | | 3+3 2 |
| | 5 2 | | | | 2+2 2 |
| | 6 2 | | | | 2+2 2 |
| | 6 2 | | | | 3+3 2 |
| | 5 2 | | | | 2+2 2 |
| 70.1.18 | 49737 | | 52 | B2Ca1Rh2 | |
| | -GM | -Y | -Z | -T | -L |
| | 5 2 | 3+4 4 | 3+4 4 | 3+4 4 | 3+3 2 |
| | 6 2 | 3+4 4 | 3+4 4 | 3+4 4 | 2+2 2 |
| | 6 2 | 3+4 4 | 3+4 4 | 3+4 4 | 3+3 2 |
| | 5 2 | 3+4 4 | 3+4 4 | 3+4 4 | 2+2 2 |
| | 5 2 | 3+4 4 | 3+4 4 | 3+4 4 | 3+3 2 |
| | 6 2 | 3+4 4 | 3+4 4 | 3+4 4 | 2+2 2 |
| | 6 2 | 3+4 4 | 3+4 4 | 3+4 4 | 2+2 2 |
| | 6 2 | 3+4 4 | 3+4 4 | 3+4 4 | 2+2 2 |
| | 5 2 | 3+4 4 | 3+4 4 | 3+4 4 | 3+3 2 |
| | 5 2 | 3+4 4 | 3+4 4 | 3+4 4 | 2+2 2 |
| | 5 2 | 3+4 4 | 3+4 4 | 3+4 4 | 3+3 2 |
| | 6 2 | 3+4 4 | 3+4 4 | 3+4 4 | 2+2 2 |
| | 5 2 | 3+4 4 | 3+4 4 | 3+4 4 | 3+3 2 |
| | 5 2 | | | | 3+3 2 |
| | 6 2 | | | | 3+3 2 |
| | 6 2 | | | | 2+2 2 |
| | 6 2 | | | | 3+3 2 |
| | 5 2 | | | | 2+2 2 |
| | 6 2 | | | | 2+2 2 |

Continued on next column

**TABLE XLIII– continued**

| TQC | ICSD | | NoE | SF | |
|---|---|---|---|---|---|
| | 5 2 | | | | 3+3 2 |
| | 6 2 | | | | 2+2 2 |
| | 5 2 | | | | 3+3 2 |
| | 5 2 | | | | 3+3 2 |
| | 5 2 | | | | 2+2 2 |
| | 5 2 | | | | 3+3 2 |
| | 6 2 | | | | 3+3 2 |
| 70.1.19 | 49738 | | 52 | B2Ca1Ir2 | |
| | -GM | -Y | -Z | -T | -L |
| | 5 2 | 3+4 4 | 3+4 4 | 3+4 4 | 3+3 2 |
| | 6 2 | 3+4 4 | 3+4 4 | 3+4 4 | 2+2 2 |
| | 5 2 | 3+4 4 | 3+4 4 | 3+4 4 | 3+3 2 |
| | 6 2 | 3+4 4 | 3+4 4 | 3+4 4 | 2+2 2 |
| | 5 2 | 3+4 4 | 3+4 4 | 3+4 4 | 3+3 2 |
| | 6 2 | 3+4 4 | 3+4 4 | 3+4 4 | 2+2 2 |
| | 6 2 | 3+4 4 | 3+4 4 | 3+4 4 | 2+2 2 |
| | 6 2 | 3+4 4 | 3+4 4 | 3+4 4 | 2+2 2 |
| | 5 2 | 3+4 4 | 3+4 4 | 3+4 4 | 3+3 2 |
| | 5 2 | 3+4 4 | 3+4 4 | 3+4 4 | 2+2 2 |
| | 5 2 | 3+4 4 | 3+4 4 | 3+4 4 | 3+3 2 |
| | 6 2 | 3+4 4 | 3+4 4 | 3+4 4 | 2+2 2 |
| | 5 2 | 3+4 4 | 3+4 4 | 3+4 4 | 3+3 2 |
| | 5 2 | | | | 3+3 2 |
| | 6 2 | | | | 3+3 2 |
| | 6 2 | | | | 2+2 2 |
| | 6 2 | | | | 3+3 2 |
| | 5 2 | | | | 2+2 2 |
| | 6 2 | | | | 2+2 2 |
| | 5 2 | | | | 3+3 2 |
| | 6 2 | | | | 2+2 2 |
| | 5 2 | | | | 3+3 2 |
| | 5 2 | | | | 3+3 2 |
| | 5 2 | | | | 2+2 2 |
| | 5 2 | | | | 3+3 2 |
| | 6 2 | | | | 3+3 2 |
| 70.1.20 | 75028 | | 52 | B1Ir1Li1 | |
| | -GM | -Y | -Z | -T | -L |
| | 5 2 | 3+4 4 | 3+4 4 | 3+4 4 | 2+2 2 |
| | 6 2 | 3+4 4 | 3+4 4 | 3+4 4 | 3+3 2 |
| | 6 2 | 3+4 4 | 3+4 4 | 3+4 4 | 2+2 2 |
| | 5 2 | 3+4 4 | 3+4 4 | 3+4 4 | 3+3 2 |
| | 5 2 | 3+4 4 | 3+4 4 | 3+4 4 | 2+2 2 |
| | 6 2 | 3+4 4 | 3+4 4 | 3+4 4 | 3+3 2 |
| | 6 2 | 3+4 4 | 3+4 4 | 3+4 4 | 3+3 2 |
| | 6 2 | 3+4 4 | 3+4 4 | 3+4 4 | 3+3 2 |
| | 5 2 | 3+4 4 | 3+4 4 | 3+4 4 | 2+2 2 |
| | 5 2 | 3+4 4 | 3+4 4 | 3+4 4 | 2+2 2 |
| | 6 2 | 3+4 4 | 3+4 4 | 3+4 4 | 3+3 2 |
| | 5 2 | 3+4 4 | 3+4 4 | 3+4 4 | 3+3 2 |
| | 5 2 | 3+4 4 | 3+4 4 | 3+4 4 | 2+2 2 |
| | 6 2 | | | | 2+2 2 |
| | 5 2 | | | | 2+2 2 |
| | 6 2 | | | | 2+2 2 |
| | 6 2 | | | | 3+3 2 |
| | 6 2 | | | | 3+3 2 |
| | 5 2 | | | | 3+3 2 |
| | 5 2 | | | | 2+2 2 |
| | 6 2 | | | | 2+2 2 |
| | 5 2 | | | | 3+3 2 |
| | 5 2 | | | | 2+2 2 |
| | 5 2 | | | | 3+3 2 |

Continued on next column

**TABLE XLIII– continued**

| TQC | ICSD | NoE | SF | | |
|---|---|---|---|---|---|
| | 6 2 | | | | 2+2 2 |
| | 5 2 | | | | 2+2 2 |
| 70.1.21 | 75027 | 52 | B1Ir1Li1 | | |
| | -GM | -Y | -Z | -T | -L |
| | 5 2 | 3+4 4 | 3+4 4 | 3+4 4 | 2+2 2 |
| | 6 2 | 3+4 4 | 3+4 4 | 3+4 4 | 3+3 2 |
| | 6 2 | 3+4 4 | 3+4 4 | 3+4 4 | 2+2 2 |
| | 5 2 | 3+4 4 | 3+4 4 | 3+4 4 | 3+3 2 |
| | 5 2 | 3+4 4 | 3+4 4 | 3+4 4 | 2+2 2 |
| | 6 2 | 3+4 4 | 3+4 4 | 3+4 4 | 3+3 2 |
| | 6 2 | 3+4 4 | 3+4 4 | 3+4 4 | 3+3 2 |
| | 6 2 | 3+4 4 | 3+4 4 | 3+4 4 | 3+3 2 |
| | 5 2 | 3+4 4 | 3+4 4 | 3+4 4 | 2+2 2 |
| | 5 2 | 3+4 4 | 3+4 4 | 3+4 4 | 2+2 2 |
| | 6 2 | 3+4 4 | 3+4 4 | 3+4 4 | 3+3 2 |
| | 5 2 | 3+4 4 | 3+4 4 | 3+4 4 | 3+3 2 |
| | 5 2 | 3+4 4 | 3+4 4 | 3+4 4 | 2+2 2 |
| | 6 2 | | | | 2+2 2 |
| | 5 2 | | | | 2+2 2 |
| | 6 2 | | | | 2+2 2 |
| | 6 2 | | | | 3+3 2 |
| | 6 2 | | | | 3+3 2 |
| | 5 2 | | | | 3+3 2 |
| | 5 2 | | | | 2+2 2 |
| | 6 2 | | | | 2+2 2 |
| | 5 2 | | | | 3+3 2 |
| | 5 2 | | | | 2+2 2 |
| | 5 2 | | | | 3+3 2 |
| | 6 2 | | | | 2+2 2 |
| | 5 2 | | | | 2+2 2 |
| 70.1.22 | 418163 | 60 | B1Cr2 | | |
| | -GM | -Y | -Z | -T | -L |
| | 5 2 | 3+4 4 | 3+4 4 | 3+4 4 | 3+3 2 |
| | 5 2 | 3+4 4 | 3+4 4 | 3+4 4 | 2+2 2 |
| | 6 2 | 3+4 4 | 3+4 4 | 3+4 4 | 3+3 2 |
| | 6 2 | 3+4 4 | 3+4 4 | 3+4 4 | 2+2 2 |
| | 6 2 | 3+4 4 | 3+4 4 | 3+4 4 | 3+3 2 |
| | 5 2 | 3+4 4 | 3+4 4 | 3+4 4 | 2+2 2 |
| | 5 2 | 3+4 4 | 3+4 4 | 3+4 4 | 2+2 2 |
| | 5 2 | 3+4 4 | 3+4 4 | 3+4 4 | 3+3 2 |
| | 6 2 | 3+4 4 | 3+4 4 | 3+4 4 | 2+2 2 |
| | 6 2 | 3+4 4 | 3+4 4 | 3+4 4 | 3+3 2 |
| | 5 2 | 3+4 4 | 3+4 4 | 3+4 4 | 3+3 2 |
| | 6 2 | 3+4 4 | 3+4 4 | 3+4 4 | 2+2 2 |
| | 6 2 | 3+4 4 | 3+4 4 | 3+4 4 | 3+3 2 |
| | 5 2 | 3+4 4 | 3+4 4 | 3+4 4 | 3+3 2 |
| | 6 2 | 3+4 4 | 3+4 4 | 3+4 4 | 2+2 2 |
| | 5 2 | | | | 3+3 2 |
| | 6 2 | | | | 3+3 2 |
| | 5 2 | | | | 2+2 2 |
| | 5 2 | | | | 3+3 2 |
| | 5 2 | | | | 3+3 2 |
| | 6 2 | | | | 2+2 2 |
| | 5 2 | | | | 2+2 2 |
| | 5 2 | | | | 3+3 2 |
| | 6 2 | | | | 2+2 2 |
| | 6 2 | | | | 2+2 2 |
| | 5 2 | | | | 3+3 2 |
| | 5 2 | | | | 2+2 2 |
| | 6 2 | | | | 2+2 2 |
| | 6 2 | | | | 2+2 2 |

Continued on next column

**TABLE XLIII– continued**

| TQC | ICSD | NoE | SF | | |
|---|---|---|---|---|---|
| | 5 2 | | | | 3+3 2 |
| 70.1.23 | 410521 | 228 | Ag2Eu4Si6 | | |
| | -GM | -Y | -Z | -T | -L |
| | 5 2 | 3+4 4 | 3+4 4 | 3+4 4 | 2+2 2 |
| | 5 2 | 3+4 4 | 3+4 4 | 3+4 4 | 3+3 2 |
| | 6 2 | 3+4 4 | 3+4 4 | 3+4 4 | 3+3 2 |
| | 6 2 | 3+4 4 | 3+4 4 | 3+4 4 | 2+2 2 |
| | 5 2 | 3+4 4 | 3+4 4 | 3+4 4 | 2+2 2 |
| | 5 2 | 3+4 4 | 3+4 4 | 3+4 4 | 2+2 2 |
| | 6 2 | 3+4 4 | 3+4 4 | 3+4 4 | 3+3 2 |
| | 6 2 | 3+4 4 | 3+4 4 | 3+4 4 | 3+3 2 |
| | 5 2 | 3+4 4 | 3+4 4 | 3+4 4 | 3+3 2 |
| | 5 2 | 3+4 4 | 3+4 4 | 3+4 4 | 2+2 2 |
| | 6 2 | 3+4 4 | 3+4 4 | 3+4 4 | 3+3 2 |
| | 6 2 | 3+4 4 | 3+4 4 | 3+4 4 | 2+2 2 |
| | 5 2 | 3+4 4 | 3+4 4 | 3+4 4 | 2+2 2 |
| | 5 2 | 3+4 4 | 3+4 4 | 3+4 4 | 3+3 2 |
| | 6 2 | 3+4 4 | 3+4 4 | 3+4 4 | 2+2 2 |
| | 6 2 | 3+4 4 | 3+4 4 | 3+4 4 | 3+3 2 |
| | 5 2 | 3+4 4 | 3+4 4 | 3+4 4 | 3+3 2 |
| | 5 2 | 3+4 4 | 3+4 4 | 3+4 4 | 2+2 2 |
| | 6 2 | 3+4 4 | 3+4 4 | 3+4 4 | 3+3 2 |
| | 6 2 | 3+4 4 | 3+4 4 | 3+4 4 | 2+2 2 |
| | 5 2 | 3+4 4 | 3+4 4 | 3+4 4 | 3+3 2 |
| | 5 2 | 3+4 4 | 3+4 4 | 3+4 4 | 2+2 2 |
| | 6 2 | 3+4 4 | 3+4 4 | 3+4 4 | 3+3 2 |
| | 6 2 | 3+4 4 | 3+4 4 | 3+4 4 | 2+2 2 |
| | 5 2 | 3+4 4 | 3+4 4 | 3+4 4 | 3+3 2 |
| | 6 2 | 3+4 4 | 3+4 4 | 3+4 4 | 2+2 2 |
| | 5 2 | 3+4 4 | 3+4 4 | 3+4 4 | 2+2 2 |
| | 5 2 | 3+4 4 | 3+4 4 | 3+4 4 | 3+3 2 |
| | 6 2 | 3+4 4 | 3+4 4 | 3+4 4 | 2+2 2 |
| | 5 2 | 3+4 4 | 3+4 4 | 3+4 4 | 2+2 2 |
| | 6 2 | 3+4 4 | 3+4 4 | 3+4 4 | 3+3 2 |
| | 6 2 | 3+4 4 | 3+4 4 | 3+4 4 | 3+3 2 |
| | 5 2 | 3+4 4 | 3+4 4 | 3+4 4 | 2+2 2 |
| | 6 2 | 3+4 4 | 3+4 4 | 3+4 4 | 3+3 2 |
| | 6 2 | 3+4 4 | 3+4 4 | 3+4 4 | 3+3 2 |
| | 5 2 | 3+4 4 | 3+4 4 | 3+4 4 | 2+2 2 |
| | 6 2 | 3+4 4 | 3+4 4 | 3+4 4 | 3+3 2 |
| | 5 2 | 3+4 4 | 3+4 4 | 3+4 4 | 2+2 2 |
| | 6 2 | 3+4 4 | 3+4 4 | 3+4 4 | 2+2 2 |
| | 5 2 | 3+4 4 | 3+4 4 | 3+4 4 | 3+3 2 |
| | 6 2 | 3+4 4 | 3+4 4 | 3+4 4 | 3+3 2 |
| | 5 2 | 3+4 4 | 3+4 4 | 3+4 4 | 2+2 2 |
| | 6 2 | 3+4 4 | 3+4 4 | 3+4 4 | 3+3 2 |
| | 5 2 | 3+4 4 | 3+4 4 | 3+4 4 | 2+2 2 |
| | 6 2 | 3+4 4 | 3+4 4 | 3+4 4 | 2+2 2 |
| | 5 2 | 3+4 4 | 3+4 4 | 3+4 4 | 2+2 2 |
| | 6 2 | 3+4 4 | 3+4 4 | 3+4 4 | 3+3 2 |
| | 5 2 | 3+4 4 | 3+4 4 | 3+4 4 | 3+3 2 |
| | 5 2 | 3+4 4 | 3+4 4 | 3+4 4 | 2+2 2 |
| | 6 2 | 3+4 4 | 3+4 4 | 3+4 4 | 3+3 2 |
| | 6 2 | 3+4 4 | 3+4 4 | 3+4 4 | 3+3 2 |
| | 5 2 | 3+4 4 | 3+4 4 | 3+4 4 | 2+2 2 |
| | 6 2 | 3+4 4 | 3+4 4 | 3+4 4 | 2+2 2 |
| | 5 2 | 3+4 4 | 3+4 4 | 3+4 4 | 3+3 2 |
| | 6 2 | 3+4 4 | 3+4 4 | 3+4 4 | 2+2 2 |
| | 5 2 | 3+4 4 | 3+4 4 | 3+4 4 | 3+3 2 |
| | 5 2 | 3+4 4 | 3+4 4 | 3+4 4 | 3+3 2 |
| | 6 2 | | | | 2+2 2 |

Continued on next column

**TABLE XLIII– continued**

| TQC | ICSD | NoE | SF |
|---|---|---|---|
| | 5 2 | | 3+3 2 |
| | 6 2 | | 2+2 2 |
| | 6 2 | | 3+3 2 |
| | 5 2 | | 2+2 2 |
| | 5 2 | | 3+3 2 |
| | 6 2 | | 2+2 2 |
| | 5 2 | | 2+2 2 |
| | 5 2 | | 3+3 2 |
| | 6 2 | | 2+2 2 |
| | 6 2 | | 3+3 2 |
| | 5 2 | | 3+3 2 |
| | 6 2 | | 2+2 2 |
| | 5 2 | | 3+3 2 |
| | 6 2 | | 3+3 2 |
| | 6 2 | | 2+2 2 |
| | 5 2 | | 2+2 2 |
| | 6 2 | | 3+3 2 |
| | 5 2 | | 2+2 2 |
| | 6 2 | | 2+2 2 |
| | 5 2 | | 3+3 2 |
| | 6 2 | | 2+2 2 |
| | 5 2 | | 3+3 2 |
| | 5 2 | | 3+3 2 |
| | 6 2 | | 2+2 2 |
| | 6 2 | | 3+3 2 |
| | 5 2 | | 2+2 2 |
| | 6 2 | | 3+3 2 |
| | 5 2 | | 2+2 2 |
| | 6 2 | | 3+3 2 |
| | 6 2 | | 2+2 2 |
| | 5 2 | | 3+3 2 |
| | 6 2 | | 2+2 2 |
| | 6 2 | | 3+3 2 |
| | 5 2 | | 3+3 2 |
| | 5 2 | | 2+2 2 |
| | 6 2 | | 2+2 2 |
| | 5 2 | | 2+2 2 |
| | 6 2 | | 3+3 2 |
| | 5 2 | | 3+3 2 |
| | 6 2 | | 2+2 2 |
| | 5 2 | | 3+3 2 |
| | 5 2 | | 2+2 2 |
| | 6 2 | | 3+3 2 |
| | 5 2 | | 2+2 2 |
| | 5 2 | | 3+3 2 |
| | 6 2 | | 2+2 2 |
| | 5 2 | | 2+2 2 |
| | 5 2 | | 3+3 2 |
| | 6 2 | | 3+3 2 |
| | 6 2 | | 2+2 2 |
| | 6 2 | | 3+3 2 |
| | 5 2 | | 2+2 2 |
| | 5 2 | | 2+2 2 |
| | 6 2 | | 2+2 2 |
| | 6 2 | | 3+3 2 |
| | 6 2 | | 3+3 2 |

The End

TABLE XLIV: The computed irreducible representations for the valence bands for the compounds of SEBR and NLC in space group 71. See the list of tables.

| TQC | ICSD | | NoE | | SF | |
|---|---|---|---|---|---|---|
| 71.1.1 | 82655 | | 52 | | Te2U1 | |
| | -GM | -X | -T | -R | -S | -W |
| | 5 2 | 5 2 | 3+4 2 | 3+4 2 | 3+4 2 | 5 2 |
| | 6 2 | 6 2 | 5+6 2 | 5+6 2 | 5+6 2 | 5 2 |
| | 5 2 | 5 2 | 3+4 2 | 3+4 2 | 3+4 2 | 5 2 |
| | 6 2 | 6 2 | 5+6 2 | 5+6 2 | 5+6 2 | 5 2 |
| | 5 2 | 5 2 | 5+6 2 | 3+4 2 | 5+6 2 | 5 2 |
| | 6 2 | 6 2 | 3+4 2 | 5+6 2 | 3+4 2 | 5 2 |
| | 5 2 | 6 2 | 3+4 2 | 3+4 2 | 3+4 2 | 5 2 |
| | 6 2 | 5 2 | 5+6 2 | 5+6 2 | 5+6 2 | 5 2 |
| | 5 2 | 5 2 | 5+6 2 | 3+4 2 | 5+6 2 | 5 2 |
| | 5 2 | 6 2 | 3+4 2 | 5+6 2 | 3+4 2 | 5 2 |
| | 6 2 | 5 2 | 5+6 2 | 5+6 2 | 3+4 2 | 5 2 |
| | 6 2 | 6 2 | 3+4 2 | 3+4 2 | 5+6 2 | 5 2 |
| | 5 2 | 5 2 | 5+6 2 | 5+6 2 | 3+4 2 | 5 2 |
| | 6 2 | 6 2 | 3+4 2 | 3+4 2 | 5+6 2 | 5 2 |
| | 5 2 | 6 2 | 3+4 2 | 3+4 2 | 5+6 2 | 5 2 |
| | 6 2 | 6 2 | 5+6 2 | 5+6 2 | 5+6 2 | 5 2 |
| | 5 2 | 5 2 | 5+6 2 | 5+6 2 | 3+4 2 | 5 2 |
| | 5 2 | 6 2 | 5+6 2 | 3+4 2 | 3+4 2 | 5 2 |
| | 5 2 | 6 2 | 3+4 2 | 5+6 2 | 3+4 2 | 5 2 |
| | 6 2 | 5 2 | 3+4 2 | 3+4 2 | 5+6 2 | 5 2 |
| | 5 2 | 5 2 | 5+6 2 | 5+6 2 | 3+4 2 | 5 2 |
| | 6 2 | 5 2 | 5+6 2 | 3+4 2 | 3+4 2 | 5 2 |
| | 6 2 | 5 2 | 3+4 2 | 5+6 2 | 5+6 2 | 5 2 |
| | 5 2 | 5 2 | 3+4 2 | 3+4 2 | 5+6 2 | 5 2 |
| | 6 2 | 6 2 | 5+6 2 | 3+4 2 | 3+4 2 | 5 2 |
| | 5 2 | 6 2 | 5+6 2 | 5+6 2 | 5+6 2 | 5 2 |
| 71.1.2 | 82654 | | 52 | | Te2U1 | |
| | -GM | -X | -T | -R | -S | -W |
| | 5 2 | 5 2 | 3+4 2 | 3+4 2 | 3+4 2 | 5 2 |
| | 6 2 | 6 2 | 5+6 2 | 5+6 2 | 5+6 2 | 5 2 |
| | 5 2 | 5 2 | 3+4 2 | 3+4 2 | 3+4 2 | 5 2 |
| | 6 2 | 6 2 | 5+6 2 | 5+6 2 | 5+6 2 | 5 2 |
| | 5 2 | 5 2 | 5+6 2 | 3+4 2 | 5+6 2 | 5 2 |
| | 6 2 | 6 2 | 3+4 2 | 5+6 2 | 3+4 2 | 5 2 |
| | 5 2 | 6 2 | 3+4 2 | 3+4 2 | 3+4 2 | 5 2 |
| | 6 2 | 5 2 | 5+6 2 | 5+6 2 | 5+6 2 | 5 2 |
| | 5 2 | 5 2 | 5+6 2 | 3+4 2 | 5+6 2 | 5 2 |
| | 5 2 | 6 2 | 3+4 2 | 5+6 2 | 3+4 2 | 5 2 |
| | 6 2 | 5 2 | 5+6 2 | 5+6 2 | 3+4 2 | 5 2 |
| | 6 2 | 6 2 | 3+4 2 | 3+4 2 | 5+6 2 | 5 2 |
| | 5 2 | 5 2 | 5+6 2 | 5+6 2 | 3+4 2 | 5 2 |
| | 6 2 | 6 2 | 3+4 2 | 3+4 2 | 5+6 2 | 5 2 |
| | 5 2 | 6 2 | 3+4 2 | 3+4 2 | 5+6 2 | 5 2 |
| | 6 2 | 6 2 | 5+6 2 | 5+6 2 | 5+6 2 | 5 2 |
| | 5 2 | 5 2 | 5+6 2 | 5+6 2 | 3+4 2 | 5 2 |
| | 5 2 | 6 2 | 5+6 2 | 3+4 2 | 3+4 2 | 5 2 |
| | 5 2 | 6 2 | 3+4 2 | 5+6 2 | 3+4 2 | 5 2 |
| | 6 2 | 5 2 | 3+4 2 | 3+4 2 | 5+6 2 | 5 2 |
| | 5 2 | 5 2 | 5+6 2 | 5+6 2 | 3+4 2 | 5 2 |
| | 6 2 | 5 2 | 5+6 2 | 3+4 2 | 3+4 2 | 5 2 |
| | 6 2 | 5 2 | 3+4 2 | 5+6 2 | 5+6 2 | 5 2 |
| | 5 2 | 5 2 | 3+4 2 | 3+4 2 | 5+6 2 | 5 2 |
| | 6 2 | 6 2 | 5+6 2 | 3+4 2 | 3+4 2 | 5 2 |
| | 5 2 | 6 2 | 5+6 2 | 5+6 2 | 5+6 2 | 5 2 |
| 71.1.3 | 82653 | | 52 | | Te2U1 | |
| | -GM | -X | -T | -R | -S | -W |

Continued on next column

**TABLE XLIV– continued**

| TQC | ICSD NoE SF |
|---|---|
| | 5 2 5 2 5+6 2 3+4 2 5+6 2 5 2 |
| | 6 2 6 2 3+4 2 5+6 2 3+4 2 5 2 |
| | 5 2 5 2 5+6 2 3+4 2 5+6 2 5 2 |
| | 6 2 6 2 3+4 2 5+6 2 3+4 2 5 2 |
| | 5 2 5 2 3+4 2 3+4 2 3+4 2 5 2 |
| | 6 2 6 2 5+6 2 5+6 2 5+6 2 5 2 |
| | 5 2 6 2 5+6 2 3+4 2 5+6 2 5 2 |
| | 6 2 5 2 3+4 2 5+6 2 3+4 2 5 2 |
| | 5 2 5 2 3+4 2 3+4 2 3+4 2 5 2 |
| | 5 2 6 2 5+6 2 5+6 2 5+6 2 5 2 |
| | 6 2 5 2 3+4 2 5+6 2 5+6 2 5 2 |
| | 6 2 6 2 5+6 2 3+4 2 3+4 2 5 2 |
| | 5 2 5 2 3+4 2 5+6 2 5+6 2 5 2 |
| | 6 2 6 2 5+6 2 3+4 2 3+4 2 5 2 |
| | 5 2 6 2 5+6 2 3+4 2 3+4 2 5 2 |
| | 6 2 6 2 3+4 2 5+6 2 3+4 2 5 2 |
| | 5 2 5 2 3+4 2 5+6 2 5+6 2 5 2 |
| | 5 2 6 2 3+4 2 3+4 2 5+6 2 5 2 |
| | 5 2 6 2 5+6 2 5+6 2 5+6 2 5 2 |
| | 6 2 5 2 5+6 2 3+4 2 3+4 2 5 2 |
| | 5 2 5 2 3+4 2 5+6 2 5+6 2 5 2 |
| | 6 2 5 2 3+4 2 3+4 2 5+6 2 5 2 |
| | 6 2 5 2 5+6 2 5+6 2 3+4 2 5 2 |
| | 5 2 5 2 5+6 2 3+4 2 3+4 2 5 2 |
| | 6 2 6 2 3+4 2 3+4 2 5+6 2 5 2 |
| | 5 2 6 2 3+4 2 5+6 2 3+4 2 5 2 |
| 71.1.4 | 82652 52 Te2U1 |
| | -GM -X -T -R -S -W |
| | 5 2 5 2 3+4 2 3+4 2 3+4 2 5 2 |
| | 6 2 6 2 5+6 2 5+6 2 5+6 2 5 2 |
| | 5 2 5 2 3+4 2 3+4 2 3+4 2 5 2 |
| | 6 2 6 2 5+6 2 5+6 2 5+6 2 5 2 |
| | 5 2 5 2 5+6 2 3+4 2 5+6 2 5 2 |
| | 6 2 6 2 3+4 2 5+6 2 3+4 2 5 2 |
| | 5 2 6 2 3+4 2 3+4 2 3+4 2 5 2 |
| | 6 2 5 2 5+6 2 5+6 2 5+6 2 5 2 |
| | 5 2 5 2 5+6 2 3+4 2 5+6 2 5 2 |
| | 5 2 6 2 3+4 2 5+6 2 3+4 2 5 2 |
| | 6 2 5 2 5+6 2 5+6 2 3+4 2 5 2 |
| | 6 2 6 2 3+4 2 3+4 2 5+6 2 5 2 |
| | 5 2 5 2 5+6 2 5+6 2 3+4 2 5 2 |
| | 6 2 6 2 3+4 2 3+4 2 5+6 2 5 2 |
| | 5 2 6 2 3+4 2 3+4 2 5+6 2 5 2 |
| | 6 2 6 2 5+6 2 5+6 2 5+6 2 5 2 |
| | 5 2 5 2 5+6 2 5+6 2 3+4 2 5 2 |
| | 5 2 6 2 5+6 2 3+4 2 3+4 2 5 2 |
| | 5 2 6 2 3+4 2 5+6 2 3+4 2 5 2 |
| | 6 2 5 2 3+4 2 3+4 2 5+6 2 5 2 |
| | 5 2 5 2 5+6 2 5+6 2 3+4 2 5 2 |
| | 6 2 5 2 5+6 2 3+4 2 3+4 2 5 2 |
| | 6 2 5 2 3+4 2 5+6 2 5+6 2 5 2 |
| | 5 2 5 2 3+4 2 3+4 2 5+6 2 5 2 |
| | 6 2 6 2 5+6 2 3+4 2 3+4 2 5 2 |
| | 5 2 6 2 5+6 2 5+6 2 5+6 2 5 2 |
| 71.1.5 | 82651 52 Te2U1 |
| | -GM -X -T -R -S -W |
| | 5 2 5 2 3+4 2 3+4 2 3+4 2 5 2 |
| | 6 2 6 2 5+6 2 5+6 2 5+6 2 5 2 |
| | 5 2 5 2 3+4 2 3+4 2 3+4 2 5 2 |
| | 6 2 6 2 5+6 2 5+6 2 5+6 2 5 2 |
| | 5 2 5 2 5+6 2 3+4 2 5+6 2 5 2 |

Continued on next column

**TABLE XLIV– continued**

| TQC | ICSD NoE SF |
|---|---|
| | 6 2 6 2 3+4 2 5+6 2 3+4 2 5 2 |
| | 5 2 6 2 3+4 2 3+4 2 3+4 2 5 2 |
| | 6 2 5 2 5+6 2 5+6 2 5+6 2 5 2 |
| | 5 2 5 2 5+6 2 3+4 2 5+6 2 5 2 |
| | 5 2 6 2 3+4 2 5+6 2 3+4 2 5 2 |
| | 6 2 5 2 5+6 2 5+6 2 3+4 2 5 2 |
| | 6 2 6 2 3+4 2 3+4 2 5+6 2 5 2 |
| | 5 2 5 2 5+6 2 5+6 2 3+4 2 5 2 |
| | 6 2 6 2 3+4 2 3+4 2 5+6 2 5 2 |
| | 5 2 6 2 3+4 2 3+4 2 5+6 2 5 2 |
| | 6 2 6 2 5+6 2 5+6 2 5+6 2 5 2 |
| | 5 2 5 2 5+6 2 5+6 2 3+4 2 5 2 |
| | 5 2 6 2 5+6 2 3+4 2 3+4 2 5 2 |
| | 5 2 6 2 3+4 2 5+6 2 3+4 2 5 2 |
| | 6 2 5 2 3+4 2 3+4 2 5+6 2 5 2 |
| | 5 2 5 2 5+6 2 5+6 2 3+4 2 5 2 |
| | 6 2 5 2 5+6 2 3+4 2 3+4 2 5 2 |
| | 6 2 5 2 3+4 2 5+6 2 5+6 2 5 2 |
| | 5 2 5 2 3+4 2 3+4 2 5+6 2 5 2 |
| | 6 2 6 2 5+6 2 3+4 2 3+4 2 5 2 |
| | 5 2 6 2 5+6 2 5+6 2 5+6 2 5 2 |
| 71.1.6 | 82650 52 Te2U1 |
| | -GM -X -T -R -S -W |
| | 5 2 5 2 3+4 2 3+4 2 3+4 2 5 2 |
| | 6 2 6 2 5+6 2 5+6 2 5+6 2 5 2 |
| | 5 2 5 2 3+4 2 3+4 2 3+4 2 5 2 |
| | 6 2 6 2 5+6 2 5+6 2 5+6 2 5 2 |
| | 5 2 5 2 5+6 2 3+4 2 5+6 2 5 2 |
| | 6 2 6 2 3+4 2 5+6 2 3+4 2 5 2 |
| | 5 2 6 2 3+4 2 3+4 2 3+4 2 5 2 |
| | 6 2 5 2 5+6 2 5+6 2 5+6 2 5 2 |
| | 5 2 5 2 5+6 2 3+4 2 5+6 2 5 2 |
| | 5 2 6 2 3+4 2 5+6 2 3+4 2 5 2 |
| | 6 2 5 2 5+6 2 5+6 2 3+4 2 5 2 |
| | 6 2 6 2 3+4 2 3+4 2 5+6 2 5 2 |
| | 5 2 5 2 5+6 2 5+6 2 3+4 2 5 2 |
| | 6 2 6 2 3+4 2 3+4 2 5+6 2 5 2 |
| | 5 2 6 2 3+4 2 3+4 2 5+6 2 5 2 |
| | 6 2 6 2 5+6 2 5+6 2 5+6 2 5 2 |
| | 5 2 5 2 5+6 2 5+6 2 3+4 2 5 2 |
| | 5 2 6 2 5+6 2 3+4 2 3+4 2 5 2 |
| | 5 2 6 2 3+4 2 5+6 2 3+4 2 5 2 |
| | 6 2 5 2 3+4 2 3+4 2 5+6 2 5 2 |
| | 5 2 5 2 5+6 2 5+6 2 3+4 2 5 2 |
| | 6 2 5 2 5+6 2 3+4 2 3+4 2 5 2 |
| | 6 2 5 2 3+4 2 5+6 2 5+6 2 5 2 |
| | 5 2 5 2 3+4 2 3+4 2 5+6 2 5 2 |
| | 6 2 6 2 5+6 2 3+4 2 3+4 2 5 2 |
| | 5 2 6 2 5+6 2 5+6 2 5+6 2 5 2 |
| 71.1.7 | 82649 52 Te2U1 |
| | -GM -X -T -R -S -W |
| | 5 2 5 2 3+4 2 3+4 2 3+4 2 5 2 |
| | 6 2 6 2 5+6 2 5+6 2 5+6 2 5 2 |
| | 5 2 5 2 3+4 2 3+4 2 3+4 2 5 2 |
| | 6 2 6 2 5+6 2 5+6 2 5+6 2 5 2 |
| | 5 2 5 2 5+6 2 3+4 2 5+6 2 5 2 |
| | 6 2 6 2 3+4 2 5+6 2 3+4 2 5 2 |
| | 5 2 6 2 3+4 2 3+4 2 3+4 2 5 2 |
| | 6 2 5 2 5+6 2 5+6 2 5+6 2 5 2 |
| | 5 2 5 2 5+6 2 3+4 2 5+6 2 5 2 |
| | 5 2 6 2 3+4 2 5+6 2 3+4 2 5 2 |

Continued on next column

**TABLE XLIV– continued**

| TQC | ICSD | NoE | SF |
|---|---|---|---|
| | 6 2 5 2 5+6 2 5+6 2 3+4 2 5 2 | | |
| | 6 2 6 2 3+4 2 3+4 2 5+6 2 5 2 | | |
| | 5 2 5 2 5+6 2 5+6 2 3+4 2 5 2 | | |
| | 6 2 6 2 3+4 2 3+4 2 5+6 2 5 2 | | |
| | 5 2 6 2 3+4 2 3+4 2 5+6 2 5 2 | | |
| | 6 2 6 2 5+6 2 5+6 2 5+6 2 5 2 | | |
| | 5 2 5 2 5+6 2 5+6 2 3+4 2 5 2 | | |
| | 5 2 6 2 5+6 2 3+4 2 3+4 2 5 2 | | |
| | 5 2 6 2 3+4 2 5+6 2 3+4 2 5 2 | | |
| | 6 2 5 2 3+4 2 3+4 2 5+6 2 5 2 | | |
| | 5 2 5 2 5+6 2 5+6 2 3+4 2 5 2 | | |
| | 6 2 5 2 5+6 2 3+4 2 3+4 2 5 2 | | |
| | 6 2 5 2 3+4 2 5+6 2 5+6 2 5 2 | | |
| | 5 2 5 2 3+4 2 3+4 2 5+6 2 5 2 | | |
| | 6 2 6 2 5+6 2 3+4 2 3+4 2 5 2 | | |
| | 5 2 6 2 5+6 2 5+6 2 5+6 2 5 2 | | |
| 71.1.8 | 82648 | 52 | Te2U1 |
| | -GM -X -T -R -S -W | | |
| | 5 2 5 2 5+6 2 3+4 2 5+6 2 5 2 | | |
| | 6 2 6 2 3+4 2 5+6 2 3+4 2 5 2 | | |
| | 5 2 5 2 5+6 2 3+4 2 5+6 2 5 2 | | |
| | 6 2 6 2 3+4 2 5+6 2 3+4 2 5 2 | | |
| | 5 2 5 2 3+4 2 3+4 2 3+4 2 5 2 | | |
| | 6 2 6 2 5+6 2 5+6 2 5+6 2 5 2 | | |
| | 5 2 6 2 5+6 2 3+4 2 5+6 2 5 2 | | |
| | 6 2 5 2 3+4 2 5+6 2 3+4 2 5 2 | | |
| | 5 2 5 2 3+4 2 3+4 2 3+4 2 5 2 | | |
| | 5 2 6 2 5+6 2 5+6 2 5+6 2 5 2 | | |
| | 6 2 5 2 3+4 2 5+6 2 5+6 2 5 2 | | |
| | 6 2 6 2 5+6 2 3+4 2 3+4 2 5 2 | | |
| | 5 2 5 2 3+4 2 5+6 2 5+6 2 5 2 | | |
| | 6 2 6 2 5+6 2 3+4 2 3+4 2 5 2 | | |
| | 5 2 6 2 5+6 2 3+4 2 3+4 2 5 2 | | |
| | 6 2 6 2 3+4 2 5+6 2 3+4 2 5 2 | | |
| | 5 2 5 2 3+4 2 5+6 2 5+6 2 5 2 | | |
| | 5 2 6 2 3+4 2 3+4 2 5+6 2 5 2 | | |
| | 5 2 6 2 5+6 2 5+6 2 5+6 2 5 2 | | |
| | 6 2 5 2 5+6 2 3+4 2 3+4 2 5 2 | | |
| | 5 2 5 2 3+4 2 5+6 2 5+6 2 5 2 | | |
| | 6 2 5 2 3+4 2 3+4 2 5+6 2 5 2 | | |
| | 6 2 5 2 5+6 2 5+6 2 3+4 2 5 2 | | |
| | 5 2 5 2 5+6 2 3+4 2 3+4 2 5 2 | | |
| | 6 2 6 2 3+4 2 3+4 2 5+6 2 5 2 | | |
| | 5 2 6 2 3+4 2 5+6 2 3+4 2 5 2 | | |
| 71.1.9 | 82647 | 52 | Te2U1 |
| | -GM -X -T -R -S -W | | |
| | 5 2 5 2 3+4 2 5+6 2 5+6 2 5 2 | | |
| | 6 2 6 2 5+6 2 3+4 2 3+4 2 5 2 | | |
| | 5 2 5 2 3+4 2 5+6 2 5+6 2 5 2 | | |
| | 6 2 6 2 5+6 2 3+4 2 3+4 2 5 2 | | |
| | 5 2 5 2 5+6 2 5+6 2 3+4 2 5 2 | | |
| | 6 2 6 2 3+4 2 3+4 2 5+6 2 5 2 | | |
| | 5 2 6 2 3+4 2 5+6 2 5+6 2 5 2 | | |
| | 6 2 5 2 5+6 2 3+4 2 3+4 2 5 2 | | |
| | 5 2 5 2 5+6 2 5+6 2 3+4 2 5 2 | | |
| | 5 2 6 2 3+4 2 3+4 2 5+6 2 5 2 | | |
| | 6 2 5 2 5+6 2 3+4 2 5+6 2 5 2 | | |
| | 6 2 6 2 3+4 2 5+6 2 3+4 2 5 2 | | |
| | 5 2 5 2 5+6 2 3+4 2 5+6 2 5 2 | | |
| | 6 2 6 2 3+4 2 5+6 2 3+4 2 5 2 | | |
| | 5 2 6 2 3+4 2 5+6 2 3+4 2 5 2 | | |

Continued on next column

**TABLE XLIV– continued**

| TQC | ICSD | NoE | SF |
|---|---|---|---|
| | 6 2 6 2 5+6 2 3+4 2 3+4 2 5 2 | | |
| | 5 2 5 2 5+6 2 3+4 2 5+6 2 5 2 | | |
| | 5 2 6 2 5+6 2 5+6 2 5+6 2 5 2 | | |
| | 5 2 6 2 3+4 2 3+4 2 5+6 2 5 2 | | |
| | 6 2 5 2 3+4 2 5+6 2 3+4 2 5 2 | | |
| | 5 2 5 2 5+6 2 3+4 2 5+6 2 5 2 | | |
| | 6 2 5 2 5+6 2 5+6 2 5+6 2 5 2 | | |
| | 6 2 5 2 3+4 2 3+4 2 3+4 2 5 2 | | |
| | 5 2 5 2 3+4 2 5+6 2 3+4 2 5 2 | | |
| | 6 2 6 2 5+6 2 5+6 2 5+6 2 5 2 | | |
| | 5 2 6 2 5+6 2 3+4 2 3+4 2 5 2 | | |
| 71.1.10 | 82646 | 52 | Te2U1 |
| | -GM -X -T -R -S -W | | |
| | 5 2 5 2 5+6 2 5+6 2 3+4 2 5 2 | | |
| | 6 2 6 2 3+4 2 3+4 2 5+6 2 5 2 | | |
| | 5 2 5 2 5+6 2 5+6 2 3+4 2 5 2 | | |
| | 6 2 6 2 3+4 2 3+4 2 5+6 2 5 2 | | |
| | 5 2 5 2 3+4 2 5+6 2 5+6 2 5 2 | | |
| | 6 2 6 2 5+6 2 3+4 2 3+4 2 5 2 | | |
| | 5 2 6 2 5+6 2 5+6 2 3+4 2 5 2 | | |
| | 6 2 5 2 3+4 2 3+4 2 5+6 2 5 2 | | |
| | 5 2 5 2 3+4 2 5+6 2 5+6 2 5 2 | | |
| | 5 2 6 2 5+6 2 3+4 2 3+4 2 5 2 | | |
| | 6 2 5 2 3+4 2 3+4 2 3+4 2 5 2 | | |
| | 6 2 6 2 5+6 2 5+6 2 5+6 2 5 2 | | |
| | 5 2 5 2 3+4 2 3+4 2 3+4 2 5 2 | | |
| | 6 2 6 2 5+6 2 5+6 2 5+6 2 5 2 | | |
| | 5 2 6 2 5+6 2 5+6 2 5+6 2 5 2 | | |
| | 6 2 6 2 3+4 2 3+4 2 5+6 2 5 2 | | |
| | 5 2 5 2 3+4 2 3+4 2 3+4 2 5 2 | | |
| | 5 2 6 2 3+4 2 5+6 2 3+4 2 5 2 | | |
| | 5 2 6 2 5+6 2 3+4 2 3+4 2 5 2 | | |
| | 6 2 5 2 5+6 2 5+6 2 5+6 2 5 2 | | |
| | 5 2 5 2 3+4 2 3+4 2 3+4 2 5 2 | | |
| | 6 2 5 2 3+4 2 5+6 2 3+4 2 5 2 | | |
| | 6 2 5 2 5+6 2 3+4 2 5+6 2 5 2 | | |
| | 5 2 5 2 5+6 2 5+6 2 5+6 2 5 2 | | |
| | 6 2 6 2 3+4 2 5+6 2 3+4 2 5 2 | | |
| | 5 2 6 2 3+4 2 3+4 2 5+6 2 5 2 | | |
| 71.1.11 | 82645 | 52 | Te2U1 |
| | -GM -X -T -R -S -W | | |
| | 5 2 5 2 3+4 2 3+4 2 3+4 2 5 2 | | |
| | 6 2 6 2 5+6 2 5+6 2 5+6 2 5 2 | | |
| | 5 2 5 2 3+4 2 3+4 2 3+4 2 5 2 | | |
| | 6 2 6 2 5+6 2 5+6 2 5+6 2 5 2 | | |
| | 5 2 5 2 5+6 2 3+4 2 5+6 2 5 2 | | |
| | 6 2 6 2 3+4 2 5+6 2 3+4 2 5 2 | | |
| | 5 2 6 2 3+4 2 3+4 2 3+4 2 5 2 | | |
| | 6 2 5 2 5+6 2 5+6 2 5+6 2 5 2 | | |
| | 5 2 5 2 5+6 2 3+4 2 5+6 2 5 2 | | |
| | 5 2 6 2 3+4 2 5+6 2 3+4 2 5 2 | | |
| | 6 2 5 2 5+6 2 5+6 2 3+4 2 5 2 | | |
| | 6 2 6 2 3+4 2 3+4 2 5+6 2 5 2 | | |
| | 5 2 5 2 5+6 2 5+6 2 3+4 2 5 2 | | |
| | 6 2 6 2 3+4 2 3+4 2 5+6 2 5 2 | | |
| | 5 2 6 2 3+4 2 3+4 2 5+6 2 5 2 | | |
| | 6 2 6 2 5+6 2 5+6 2 5+6 2 5 2 | | |
| | 5 2 5 2 5+6 2 5+6 2 3+4 2 5 2 | | |
| | 5 2 6 2 5+6 2 3+4 2 3+4 2 5 2 | | |
| | 5 2 6 2 3+4 2 5+6 2 3+4 2 5 2 | | |
| | 6 2 5 2 3+4 2 3+4 2 5+6 2 5 2 | | |

Continued on next column

**TABLE XLIV– continued**

| TQC | ICSD | NoE | SF |
|---|---|---|---|
| | 5 2 5 2 5+6 2 5+6 2 3+4 2 5 2 | | |
| | 6 2 5 2 5+6 2 3+4 2 3+4 2 5 2 | | |
| | 6 2 5 2 3+4 2 5+6 2 5+6 2 5 2 | | |
| | 5 2 5 2 3+4 2 3+4 2 5+6 2 5 2 | | |
| | 6 2 6 2 5+6 2 3+4 2 3+4 2 5 2 | | |
| | 5 2 6 2 5+6 2 5+6 2 5+6 2 5 2 | | |
| 71.1.12 | 82644 | 52 | Te2U1 |
| | -GM -X -T -R -S -W | | |
| | 5 2 5 2 3+4 2 3+4 2 3+4 2 5 2 | | |
| | 6 2 6 2 5+6 2 5+6 2 5+6 2 5 2 | | |
| | 5 2 5 2 3+4 2 3+4 2 3+4 2 5 2 | | |
| | 6 2 6 2 5+6 2 5+6 2 5+6 2 5 2 | | |
| | 5 2 5 2 5+6 2 3+4 2 5+6 2 5 2 | | |
| | 6 2 6 2 3+4 2 5+6 2 3+4 2 5 2 | | |
| | 5 2 6 2 3+4 2 3+4 2 3+4 2 5 2 | | |
| | 6 2 5 2 5+6 2 5+6 2 5+6 2 5 2 | | |
| | 5 2 5 2 5+6 2 3+4 2 5+6 2 5 2 | | |
| | 5 2 6 2 3+4 2 5+6 2 3+4 2 5 2 | | |
| | 6 2 5 2 5+6 2 5+6 2 3+4 2 5 2 | | |
| | 6 2 6 2 3+4 2 3+4 2 5+6 2 5 2 | | |
| | 5 2 5 2 5+6 2 5+6 2 3+4 2 5 2 | | |
| | 6 2 6 2 3+4 2 3+4 2 5+6 2 5 2 | | |
| | 5 2 6 2 3+4 2 3+4 2 5+6 2 5 2 | | |
| | 6 2 6 2 5+6 2 5+6 2 5+6 2 5 2 | | |
| | 5 2 5 2 5+6 2 5+6 2 3+4 2 5 2 | | |
| | 5 2 6 2 5+6 2 3+4 2 3+4 2 5 2 | | |
| | 5 2 6 2 3+4 2 5+6 2 3+4 2 5 2 | | |
| | 6 2 5 2 3+4 2 3+4 2 5+6 2 5 2 | | |
| | 5 2 5 2 5+6 2 5+6 2 3+4 2 5 2 | | |
| | 6 2 5 2 5+6 2 3+4 2 3+4 2 5 2 | | |
| | 6 2 5 2 3+4 2 5+6 2 5+6 2 5 2 | | |
| | 5 2 5 2 3+4 2 3+4 2 5+6 2 5 2 | | |
| | 6 2 6 2 5+6 2 3+4 2 3+4 2 5 2 | | |
| | 5 2 6 2 5+6 2 5+6 2 5+6 2 5 2 | | |
| 71.1.13 | 82643 | 52 | Te2U1 |
| | -GM -X -T -R -S -W | | |
| | 5 2 5 2 3+4 2 3+4 2 3+4 2 5 2 | | |
| | 6 2 6 2 5+6 2 5+6 2 5+6 2 5 2 | | |
| | 5 2 5 2 3+4 2 3+4 2 3+4 2 5 2 | | |
| | 6 2 6 2 5+6 2 5+6 2 5+6 2 5 2 | | |
| | 5 2 5 2 5+6 2 3+4 2 5+6 2 5 2 | | |
| | 6 2 6 2 3+4 2 5+6 2 3+4 2 5 2 | | |
| | 5 2 6 2 3+4 2 3+4 2 3+4 2 5 2 | | |
| | 6 2 5 2 5+6 2 5+6 2 5+6 2 5 2 | | |
| | 5 2 5 2 5+6 2 3+4 2 5+6 2 5 2 | | |
| | 5 2 6 2 3+4 2 5+6 2 3+4 2 5 2 | | |
| | 6 2 5 2 5+6 2 5+6 2 3+4 2 5 2 | | |
| | 6 2 6 2 3+4 2 3+4 2 5+6 2 5 2 | | |
| | 5 2 5 2 5+6 2 5+6 2 3+4 2 5 2 | | |
| | 6 2 6 2 3+4 2 3+4 2 5+6 2 5 2 | | |
| | 5 2 6 2 3+4 2 3+4 2 5+6 2 5 2 | | |
| | 6 2 6 2 5+6 2 5+6 2 5+6 2 5 2 | | |
| | 5 2 5 2 5+6 2 5+6 2 3+4 2 5 2 | | |
| | 5 2 6 2 5+6 2 3+4 2 3+4 2 5 2 | | |
| | 5 2 6 2 3+4 2 5+6 2 3+4 2 5 2 | | |
| | 6 2 5 2 3+4 2 3+4 2 5+6 2 5 2 | | |
| | 5 2 5 2 5+6 2 5+6 2 3+4 2 5 2 | | |
| | 6 2 5 2 5+6 2 3+4 2 3+4 2 5 2 | | |
| | 6 2 5 2 3+4 2 5+6 2 5+6 2 5 2 | | |
| | 5 2 5 2 3+4 2 3+4 2 5+6 2 5 2 | | |
| | 6 2 6 2 5+6 2 3+4 2 3+4 2 5 2 | | |

Continued on next column

**TABLE XLIV– continued**

| TQC | ICSD | NoE | SF |
|---|---|---|---|
| | 5 2 6 2 5+6 2 5+6 2 5+6 2 5 2 | | |
| 71.1.14 | 82642 | 52 | Te2U1 |
| | -GM -X -T -R -S -W | | |
| | 5 2 5 2 3+4 2 3+4 2 3+4 2 5 2 | | |
| | 6 2 6 2 5+6 2 5+6 2 5+6 2 5 2 | | |
| | 5 2 5 2 3+4 2 3+4 2 3+4 2 5 2 | | |
| | 6 2 6 2 5+6 2 5+6 2 5+6 2 5 2 | | |
| | 5 2 5 2 5+6 2 3+4 2 5+6 2 5 2 | | |
| | 6 2 6 2 3+4 2 5+6 2 3+4 2 5 2 | | |
| | 5 2 6 2 3+4 2 3+4 2 3+4 2 5 2 | | |
| | 6 2 5 2 5+6 2 5+6 2 5+6 2 5 2 | | |
| | 5 2 5 2 5+6 2 3+4 2 5+6 2 5 2 | | |
| | 5 2 6 2 3+4 2 5+6 2 3+4 2 5 2 | | |
| | 6 2 5 2 5+6 2 5+6 2 3+4 2 5 2 | | |
| | 6 2 6 2 3+4 2 3+4 2 5+6 2 5 2 | | |
| | 5 2 5 2 5+6 2 5+6 2 3+4 2 5 2 | | |
| | 6 2 6 2 3+4 2 3+4 2 5+6 2 5 2 | | |
| | 5 2 6 2 3+4 2 3+4 2 5+6 2 5 2 | | |
| | 6 2 6 2 5+6 2 5+6 2 5+6 2 5 2 | | |
| | 5 2 5 2 5+6 2 5+6 2 3+4 2 5 2 | | |
| | 5 2 6 2 5+6 2 3+4 2 3+4 2 5 2 | | |
| | 5 2 6 2 3+4 2 5+6 2 3+4 2 5 2 | | |
| | 6 2 5 2 3+4 2 3+4 2 5+6 2 5 2 | | |
| | 5 2 5 2 5+6 2 5+6 2 3+4 2 5 2 | | |
| | 6 2 5 2 5+6 2 3+4 2 3+4 2 5 2 | | |
| | 6 2 5 2 3+4 2 5+6 2 5+6 2 5 2 | | |
| | 5 2 5 2 3+4 2 3+4 2 5+6 2 5 2 | | |
| | 6 2 6 2 5+6 2 3+4 2 3+4 2 5 2 | | |
| | 5 2 6 2 5+6 2 5+6 2 5+6 2 5 2 | | |
| 71.1.15 | 63170 | 52 | Te2U1 |
| | -GM -X -T -R -S -W | | |
| | 5 2 5 2 3+4 2 3+4 2 3+4 2 5 2 | | |
| | 6 2 6 2 5+6 2 5+6 2 5+6 2 5 2 | | |
| | 5 2 5 2 3+4 2 3+4 2 3+4 2 5 2 | | |
| | 6 2 6 2 5+6 2 5+6 2 5+6 2 5 2 | | |
| | 5 2 5 2 5+6 2 3+4 2 5+6 2 5 2 | | |
| | 6 2 6 2 3+4 2 5+6 2 3+4 2 5 2 | | |
| | 5 2 6 2 3+4 2 3+4 2 3+4 2 5 2 | | |
| | 6 2 5 2 5+6 2 5+6 2 5+6 2 5 2 | | |
| | 5 2 5 2 5+6 2 3+4 2 5+6 2 5 2 | | |
| | 5 2 6 2 3+4 2 5+6 2 3+4 2 5 2 | | |
| | 6 2 5 2 5+6 2 5+6 2 3+4 2 5 2 | | |
| | 6 2 6 2 3+4 2 3+4 2 5+6 2 5 2 | | |
| | 5 2 5 2 5+6 2 5+6 2 3+4 2 5 2 | | |
| | 6 2 6 2 3+4 2 3+4 2 5+6 2 5 2 | | |
| | 5 2 6 2 3+4 2 3+4 2 5+6 2 5 2 | | |
| | 6 2 6 2 5+6 2 5+6 2 5+6 2 5 2 | | |
| | 5 2 5 2 5+6 2 5+6 2 3+4 2 5 2 | | |
| | 5 2 6 2 5+6 2 3+4 2 3+4 2 5 2 | | |
| | 5 2 6 2 3+4 2 5+6 2 3+4 2 5 2 | | |
| | 6 2 5 2 3+4 2 3+4 2 5+6 2 5 2 | | |
| | 5 2 5 2 5+6 2 5+6 2 3+4 2 5 2 | | |
| | 6 2 5 2 5+6 2 3+4 2 3+4 2 5 2 | | |
| | 6 2 5 2 3+4 2 5+6 2 5+6 2 5 2 | | |
| | 5 2 5 2 3+4 2 3+4 2 5+6 2 5 2 | | |
| | 6 2 6 2 5+6 2 3+4 2 3+4 2 5 2 | | |
| | 5 2 6 2 5+6 2 5+6 2 5+6 2 5 2 | | |
| 71.1.16 | 40694 | 52 | Te2U1 |
| | -GM -X -T -R -S -W | | |
| | 5 2 5 2 3+4 2 3+4 2 3+4 2 5 2 | | |
| | 6 2 6 2 5+6 2 5+6 2 5+6 2 5 2 | | |

Continued on next column

**TABLE XLIV– continued**

| TQC | ICSD | NoE | SF |
|---|---|---|---|
| | 5 2 5 2 3+4 2 3+4 2 3+4 2 5 2 | | |
| | 6 2 6 2 5+6 2 5+6 2 5+6 2 5 2 | | |
| | 5 2 5 2 5+6 2 3+4 2 5+6 2 5 2 | | |
| | 6 2 6 2 3+4 2 5+6 2 3+4 2 5 2 | | |
| | 5 2 6 2 3+4 2 3+4 2 3+4 2 5 2 | | |
| | 6 2 5 2 5+6 2 5+6 2 5+6 2 5 2 | | |
| | 5 2 5 2 5+6 2 3+4 2 3+4 2 5 2 | | |
| | 5 2 6 2 3+4 2 5+6 2 5+6 2 5 2 | | |
| | 6 2 5 2 5+6 2 5+6 2 3+4 2 5 2 | | |
| | 6 2 6 2 3+4 2 3+4 2 5+6 2 5 2 | | |
| | 5 2 5 2 5+6 2 5+6 2 3+4 2 5 2 | | |
| | 6 2 6 2 3+4 2 3+4 2 5+6 2 5 2 | | |
| | 5 2 6 2 3+4 2 3+4 2 5+6 2 5 2 | | |
| | 6 2 6 2 5+6 2 5+6 2 5+6 2 5 2 | | |
| | 5 2 5 2 5+6 2 5+6 2 3+4 2 5 2 | | |
| | 5 2 6 2 5+6 2 3+4 2 3+4 2 5 2 | | |
| | 5 2 6 2 3+4 2 5+6 2 3+4 2 5 2 | | |
| | 6 2 5 2 3+4 2 3+4 2 3+4 2 5 2 | | |
| | 5 2 5 2 5+6 2 5+6 2 5+6 2 5 2 | | |
| | 6 2 5 2 5+6 2 3+4 2 3+4 2 5 2 | | |
| | 6 2 5 2 3+4 2 5+6 2 5+6 2 5 2 | | |
| | 5 2 5 2 3+4 2 3+4 2 5+6 2 5 2 | | |
| | 6 2 6 2 5+6 2 3+4 2 3+4 2 5 2 | | |
| | 5 2 6 2 5+6 2 5+6 2 5+6 2 5 2 | | |
| 71.1.17 | 416820 | 70 | Ni2Tb2Zn1 |
| | -GM -X -T -R -S -W | | |
| | 5 2 6 2 3+4 2 5+6 2 5+6 2 5 2 | | |
| | 6 2 5 2 5+6 2 3+4 2 3+4 2 5 2 | | |
| | 5 2 6 2 3+4 2 5+6 2 5+6 2 5 2 | | |
| | 6 2 5 2 5+6 2 3+4 2 3+4 2 5 2 | | |
| | 5 2 6 2 3+4 2 5+6 2 5+6 2 5 2 | | |
| | 5 2 5 2 5+6 2 3+4 2 3+4 2 5 2 | | |
| | 6 2 6 2 3+4 2 5+6 2 5+6 2 5 2 | | |
| | 6 2 5 2 5+6 2 3+4 2 3+4 2 5 2 | | |
| | 5 2 5 2 3+4 2 3+4 2 3+4 2 5 2 | | |
| | 5 2 5 2 3+4 2 3+4 2 3+4 2 5 2 | | |
| | 5 2 5 2 3+4 2 3+4 2 3+4 2 5 2 | | |
| | 5 2 5 2 3+4 2 3+4 2 3+4 2 5 2 | | |
| | 5 2 5 2 3+4 2 3+4 2 3+4 2 5 2 | | |
| | 5 2 5 2 3+4 2 3+4 2 3+4 2 5 2 | | |
| | 6 2 6 2 5+6 2 5+6 2 5+6 2 5 2 | | |
| | 6 2 5 2 5+6 2 5+6 2 3+4 2 5 2 | | |
| | 5 2 6 2 3+4 2 3+4 2 5+6 2 5 2 | | |
| | 6 2 6 2 3+4 2 5+6 2 3+4 2 5 2 | | |
| | 5 2 6 2 3+4 2 3+4 2 5+6 2 5 2 | | |
| | 6 2 5 2 5+6 2 5+6 2 3+4 2 5 2 | | |
| | 5 2 6 2 5+6 2 3+4 2 5+6 2 5 2 | | |
| | 6 2 5 2 3+4 2 3+4 2 5+6 2 5 2 | | |
| | 5 2 5 2 3+4 2 5+6 2 5+6 2 5 2 | | |
| | 5 2 6 2 5+6 2 3+4 2 3+4 2 5 2 | | |
| | 6 2 5 2 5+6 2 5+6 2 3+4 2 5 2 | | |
| | 5 2 5 2 5+6 2 5+6 2 5+6 2 5 2 | | |
| | 6 2 6 2 5+6 2 5+6 2 3+4 2 5 2 | | |
| | 5 2 5 2 3+4 2 3+4 2 5+6 2 5 2 | | |
| | 5 2 5 2 3+4 2 5+6 2 5+6 2 5 2 | | |
| | 6 2 6 2 5+6 2 3+4 2 3+4 2 5 2 | | |
| | 6 2 6 2 3+4 2 3+4 2 3+4 2 5 2 | | |
| | 6 2 5 2 5+6 2 3+4 2 3+4 2 5 2 | | |
| | 5 2 6 2 5+6 2 5+6 2 5+6 2 5 2 | | |
| | 6 2 5 2 3+4 2 5+6 2 3+4 2 5 2 | | |
| | 5 2 6 2 5+6 2 3+4 2 5+6 2 5 2 | | |

Continued on next column

**TABLE XLIV– continued**

| TQC | ICSD | NoE | SF |
|---|---|---|---|
| 71.1.18 | 85406 | 120 | Ni2P2Sr1 |
| | -GM -X -T -R -S -W | | |
| | 5 2 5 2 3+4 2 3+4 2 3+4 2 5 2 | | |
| | 5 2 5 2 5+6 2 5+6 2 3+4 2 5 2 | | |
| | 6 2 6 2 3+4 2 3+4 2 5+6 2 5 2 | | |
| | 6 2 6 2 5+6 2 5+6 2 5+6 2 5 2 | | |
| | 5 2 5 2 5+6 2 5+6 2 5+6 2 5 2 | | |
| | 6 2 6 2 3+4 2 3+4 2 3+4 2 5 2 | | |
| | 6 2 6 2 5+6 2 5+6 2 5+6 2 5 2 | | |
| | 6 2 6 2 5+6 2 5+6 2 5+6 2 5 2 | | |
| | 5 2 5 2 3+4 2 5+6 2 3+4 2 5 2 | | |
| | 6 2 6 2 5+6 2 3+4 2 5+6 2 5 2 | | |
| | 5 2 5 2 5+6 2 3+4 2 3+4 2 5 2 | | |
| | 6 2 6 2 3+4 2 5+6 2 5+6 2 5 2 | | |
| | 5 2 5 2 3+4 2 5+6 2 5+6 2 5 2 | | |
| | 5 2 6 2 5+6 2 3+4 2 5+6 2 5 2 | | |
| | 5 2 6 2 3+4 2 5+6 2 3+4 2 5 2 | | |
| | 6 2 5 2 3+4 2 3+4 2 3+4 2 5 2 | | |
| | 6 2 5 2 5+6 2 5+6 2 5+6 2 5 2 | | |
| | 6 2 6 2 5+6 2 3+4 2 3+4 2 5 2 | | |
| | 5 2 5 2 3+4 2 5+6 2 5+6 2 5 2 | | |
| | 6 2 5 2 3+4 2 3+4 2 3+4 2 5 2 | | |
| | 5 2 6 2 5+6 2 3+4 2 3+4 2 5 2 | | |
| | 6 2 6 2 5+6 2 5+6 2 5+6 2 5 2 | | |
| | 5 2 6 2 5+6 2 5+6 2 3+4 2 5 2 | | |
| | 5 2 6 2 3+4 2 3+4 2 5+6 2 5 2 | | |
| | 6 2 5 2 3+4 2 3+4 2 5+6 2 5 2 | | |
| | 6 2 5 2 5+6 2 5+6 2 3+4 2 5 2 | | |
| | 5 2 5 2 5+6 2 5+6 2 3+4 2 5 2 | | |
| | 5 2 5 2 3+4 2 3+4 2 5+6 2 5 2 | | |
| | 6 2 5 2 3+4 2 3+4 2 3+4 2 5 2 | | |
| | 6 2 6 2 5+6 2 5+6 2 5+6 2 5 2 | | |
| | 6 2 6 2 5+6 2 3+4 2 3+4 2 5 2 | | |
| | 6 2 6 2 3+4 2 5+6 2 5+6 2 5 2 | | |
| | 5 2 6 2 3+4 2 5+6 2 5+6 2 5 2 | | |
| | 5 2 5 2 3+4 2 3+4 2 3+4 2 5 2 | | |
| | 5 2 5 2 5+6 2 5+6 2 5+6 2 5 2 | | |
| | 5 2 6 2 3+4 2 3+4 2 3+4 2 5 2 | | |
| | 6 2 6 2 5+6 2 5+6 2 5+6 2 5 2 | | |
| | 6 2 5 2 3+4 2 3+4 2 3+4 2 5 2 | | |
| | 5 2 6 2 5+6 2 3+4 2 5+6 2 5 2 | | |
| | 5 2 5 2 5+6 2 5+6 2 3+4 2 5 2 | | |
| | 6 2 6 2 3+4 2 3+4 2 3+4 2 5 2 | | |
| | 5 2 5 2 5+6 2 5+6 2 5+6 2 5 2 | | |
| | 6 2 6 2 3+4 2 5+6 2 3+4 2 5 2 | | |
| | 5 2 5 2 5+6 2 3+4 2 5+6 2 5 2 | | |
| | 6 2 6 2 3+4 2 5+6 2 5+6 2 5 2 | | |
| | 6 2 5 2 3+4 2 3+4 2 3+4 2 5 2 | | |
| | 5 2 5 2 5+6 2 5+6 2 5+6 2 5 2 | | |
| | 6 2 6 2 5+6 2 3+4 2 3+4 2 5 2 | | |
| | 5 2 6 2 3+4 2 5+6 2 5+6 2 5 2 | | |
| | 5 2 6 2 3+4 2 3+4 2 3+4 2 5 2 | | |
| | 5 2 5 2 5+6 2 5+6 2 5+6 2 5 2 | | |
| | 5 2 6 2 3+4 2 3+4 2 3+4 2 5 2 | | |
| | 6 2 6 2 5+6 2 5+6 2 5+6 2 5 2 | | |
| | 5 2 5 2 3+4 2 3+4 2 3+4 2 5 2 | | |
| | 6 2 6 2 5+6 2 3+4 2 3+4 2 5 2 | | |
| | 6 2 5 2 5+6 2 5+6 2 5+6 2 5 2 | | |
| | 5 2 5 2 3+4 2 3+4 2 3+4 2 5 2 | | |
| | 5 2 5 2 5+6 2 3+4 2 5+6 2 5 2 | | |
| | 6 2 6 2 3+4 2 5+6 2 3+4 2 5 2 | | |

Continued on next column

**TABLE XLIV– continued**

| TQC | ICSD NoE SF |
|---|---|
| | 5 2 6 2 5+6 2 5+6 2 5+6 2 5 2 |
| 71.1.19 | 25307 54 Li4Sb4Sr3 |
| | -GM -X -T -R -S -W |
| | 5 2 5 2 5+6 2 5+6 2 3+4 2 5 2 |
| | 6 2 6 2 3+4 2 3+4 2 5+6 2 5 2 |
| | 5 2 5 2 5+6 2 3+4 2 5+6 2 5 2 |
| | 5 2 5 2 5+6 2 5+6 2 3+4 2 5 2 |
| | 6 2 6 2 3+4 2 3+4 2 5+6 2 5 2 |
| | 6 2 6 2 3+4 2 5+6 2 3+4 2 5 2 |
| | 5 2 5 2 5+6 2 5+6 2 3+4 2 5 2 |
| | 6 2 6 2 3+4 2 3+4 2 5+6 2 5 2 |
| | 5 2 6 2 3+4 2 5+6 2 3+4 2 5 2 |
| | 6 2 5 2 5+6 2 3+4 2 5+6 2 5 2 |
| | 6 2 6 2 3+4 2 5+6 2 3+4 2 5 2 |
| | 6 2 6 2 3+4 2 5+6 2 3+4 2 5 2 |
| | 5 2 5 2 3+4 2 5+6 2 5+6 2 5 2 |
| | 5 2 6 2 5+6 2 3+4 2 5+6 2 5 2 |
| | 6 2 5 2 3+4 2 5+6 2 3+4 2 5 2 |
| | 6 2 6 2 5+6 2 3+4 2 3+4 2 5 2 |
| | 5 2 5 2 5+6 2 3+4 2 3+4 2 5 2 |
| | 6 2 6 2 3+4 2 5+6 2 5+6 2 5 2 |
| | 5 2 6 2 5+6 2 5+6 2 5+6 2 5 2 |
| | 6 2 5 2 3+4 2 3+4 2 3+4 2 5 2 |
| | 5 2 6 2 5+6 2 3+4 2 5+6 2 5 2 |
| | 5 2 6 2 3+4 2 5+6 2 3+4 2 5 2 |
| | 5 2 6 2 5+6 2 3+4 2 5+6 2 5 2 |
| | 5 2 5 2 3+4 2 5+6 2 3+4 2 5 2 |
| | 6 2 5 2 5+6 2 3+4 2 5+6 2 5 2 |
| | 6 2 5 2 3+4 2 5+6 2 5+6 2 5 2 |
| | 6 2 5 2 3+4 2 3+4 2 3+4 2 5 2 |
| 71.1.20 | 416817 54 La2Ni2Zn1 |
| | -GM -X -T -R -S -W |
| | 5 2 6 2 3+4 2 5+6 2 5+6 2 5 2 |
| | 6 2 5 2 5+6 2 3+4 2 3+4 2 5 2 |
| | 5 2 6 2 3+4 2 5+6 2 5+6 2 5 2 |
| | 6 2 5 2 5+6 2 3+4 2 3+4 2 5 2 |
| | 5 2 6 2 3+4 2 5+6 2 5+6 2 5 2 |
| | 5 2 5 2 5+6 2 3+4 2 3+4 2 5 2 |
| | 6 2 6 2 3+4 2 5+6 2 5+6 2 5 2 |
| | 6 2 5 2 5+6 2 3+4 2 3+4 2 5 2 |
| | 5 2 5 2 3+4 2 3+4 2 3+4 2 5 2 |
| | 5 2 5 2 3+4 2 3+4 2 3+4 2 5 2 |
| | 5 2 5 2 3+4 2 3+4 2 3+4 2 5 2 |
| | 5 2 5 2 3+4 2 3+4 2 3+4 2 5 2 |
| | 5 2 5 2 3+4 2 3+4 2 3+4 2 5 2 |
| | 5 2 5 2 3+4 2 3+4 2 3+4 2 5 2 |
| | 6 2 5 2 5+6 2 5+6 2 5+6 2 5 2 |
| | 6 2 6 2 5+6 2 5+6 2 3+4 2 5 2 |
| | 5 2 6 2 3+4 2 3+4 2 5+6 2 5 2 |
| | 6 2 6 2 3+4 2 5+6 2 5+6 2 5 2 |
| | 5 2 6 2 5+6 2 3+4 2 3+4 2 5 2 |
| | 6 2 5 2 5+6 2 5+6 2 3+4 2 5 2 |
| | 6 2 6 2 3+4 2 3+4 2 5+6 2 5 2 |
| | 5 2 5 2 3+4 2 3+4 2 5+6 2 5 2 |
| | 5 2 5 2 3+4 2 5+6 2 3+4 2 5 2 |
| | 5 2 6 2 5+6 2 3+4 2 5+6 2 5 2 |
| | 6 2 5 2 5+6 2 5+6 2 3+4 2 5 2 |
| | 6 2 5 2 5+6 2 3+4 2 3+4 2 5 2 |
| | 5 2 5 2 5+6 2 5+6 2 5+6 2 5 2 |
| 71.1.21 | 56747 48 H1Li1Nd2O3 |
| | -GM -X -T -R -S -W |

Continued on next column

**TABLE XLIV– continued**

| TQC | ICSD NoE SF |
|---|---|
| | 5 2 6 2 5+6 2 5+6 2 3+4 2 5 2 |
| | 6 2 5 2 3+4 2 3+4 2 5+6 2 5 2 |
| | 5 2 5 2 3+4 2 5+6 2 5+6 2 5 2 |
| | 6 2 6 2 5+6 2 3+4 2 3+4 2 5 2 |
| | 5 2 5 2 3+4 2 5+6 2 5+6 2 5 2 |
| | 5 2 6 2 5+6 2 3+4 2 3+4 2 5 2 |
| | 6 2 5 2 3+4 2 5+6 2 5+6 2 5 2 |
| | 5 2 6 2 5+6 2 3+4 2 3+4 2 5 2 |
| | 6 2 5 2 3+4 2 5+6 2 3+4 2 5 2 |
| | 6 2 6 2 5+6 2 3+4 2 5+6 2 5 2 |
| | 5 2 5 2 3+4 2 5+6 2 5+6 2 5 2 |
| | 6 2 5 2 5+6 2 3+4 2 3+4 2 5 2 |
| | 5 2 6 2 5+6 2 3+4 2 5+6 2 5 2 |
| | 6 2 6 2 3+4 2 5+6 2 3+4 2 5 2 |
| | 6 2 6 2 5+6 2 3+4 2 5+6 2 5 2 |
| | 5 2 6 2 3+4 2 5+6 2 3+4 2 5 2 |
| | 5 2 6 2 5+6 2 5+6 2 5+6 2 5 2 |
| | 5 2 6 2 3+4 2 3+4 2 3+4 2 5 2 |
| | 6 2 5 2 5+6 2 3+4 2 5+6 2 5 2 |
| | 6 2 5 2 5+6 2 3+4 2 3+4 2 5 2 |
| | 6 2 5 2 5+6 2 3+4 2 3+4 2 5 2 |
| | 6 2 5 2 3+4 2 5+6 2 5+6 2 5 2 |
| | 5 2 6 2 3+4 2 5+6 2 3+4 2 5 2 |
| | 5 2 6 2 5+6 2 5+6 2 5+6 2 5 2 |
| 71.1.22 | 56746 48 H1Li1Nd2O3 |
| | -GM -X -T -R -S -W |
| | 5 2 6 2 5+6 2 5+6 2 3+4 2 5 2 |
| | 6 2 5 2 3+4 2 3+4 2 5+6 2 5 2 |
| | 5 2 5 2 3+4 2 5+6 2 5+6 2 5 2 |
| | 6 2 6 2 5+6 2 3+4 2 3+4 2 5 2 |
| | 5 2 5 2 3+4 2 5+6 2 5+6 2 5 2 |
| | 5 2 6 2 5+6 2 3+4 2 3+4 2 5 2 |
| | 6 2 5 2 3+4 2 5+6 2 5+6 2 5 2 |
| | 5 2 6 2 5+6 2 3+4 2 3+4 2 5 2 |
| | 6 2 5 2 3+4 2 5+6 2 3+4 2 5 2 |
| | 6 2 6 2 5+6 2 3+4 2 5+6 2 5 2 |
| | 5 2 5 2 3+4 2 5+6 2 5+6 2 5 2 |
| | 6 2 5 2 5+6 2 3+4 2 3+4 2 5 2 |
| | 5 2 6 2 5+6 2 3+4 2 5+6 2 5 2 |
| | 6 2 6 2 3+4 2 5+6 2 3+4 2 5 2 |
| | 6 2 6 2 5+6 2 3+4 2 5+6 2 5 2 |
| | 5 2 6 2 3+4 2 5+6 2 3+4 2 5 2 |
| | 5 2 6 2 5+6 2 5+6 2 5+6 2 5 2 |
| | 5 2 6 2 3+4 2 3+4 2 3+4 2 5 2 |
| | 6 2 5 2 5+6 2 3+4 2 5+6 2 5 2 |
| | 6 2 5 2 5+6 2 3+4 2 3+4 2 5 2 |
| | 6 2 5 2 5+6 2 3+4 2 3+4 2 5 2 |
| | 6 2 5 2 3+4 2 5+6 2 5+6 2 5 2 |
| | 5 2 6 2 3+4 2 5+6 2 3+4 2 5 2 |
| | 5 2 6 2 5+6 2 5+6 2 5+6 2 5 2 |
| 71.1.23 | 56745 44 H1Ce2Li1O3 |
| | -GM -X -T -R -S -W |
| | 5 2 6 2 5+6 2 5+6 2 3+4 2 5 2 |
| | 6 2 5 2 3+4 2 3+4 2 5+6 2 5 2 |
| | 5 2 5 2 3+4 2 5+6 2 5+6 2 5 2 |
| | 6 2 6 2 5+6 2 3+4 2 3+4 2 5 2 |
| | 5 2 5 2 3+4 2 5+6 2 5+6 2 5 2 |
| | 6 2 6 2 5+6 2 3+4 2 3+4 2 5 2 |
| | 5 2 5 2 3+4 2 5+6 2 5+6 2 5 2 |
| | 5 2 6 2 5+6 2 3+4 2 3+4 2 5 2 |
| | 6 2 5 2 3+4 2 5+6 2 3+4 2 5 2 |

Continued on next column

**TABLE XLIV– continued**

| TQC | ICSD NoE SF |
|---|---|
| | 6 2 6 2 5+6 2 3+4 2 5+6 2 5 2 |
| | 5 2 5 2 3+4 2 5+6 2 5+6 2 5 2 |
| | 6 2 5 2 5+6 2 3+4 2 3+4 2 5 2 |
| | 5 2 6 2 5+6 2 3+4 2 5+6 2 5 2 |
| | 6 2 6 2 3+4 2 5+6 2 3+4 2 5 2 |
| | 6 2 6 2 5+6 2 3+4 2 5+6 2 5 2 |
| | 5 2 6 2 3+4 2 5+6 2 3+4 2 5 2 |
| | 5 2 6 2 5+6 2 5+6 2 5+6 2 5 2 |
| | 5 2 6 2 3+4 2 3+4 2 3+4 2 5 2 |
| | 6 2 5 2 5+6 2 3+4 2 5+6 2 5 2 |
| | 6 2 5 2 5+6 2 3+4 2 3+4 2 5 2 |
| | 6 2 5 2 5+6 2 3+4 2 3+4 2 5 2 |
| | 6 2 5 2 3+4 2 5+6 2 5+6 2 5 2 |
| 71.1.24 | 710017 34 Ge5Nd1 |
| | -GM -X -T -R -S -W |
| | 5 2 5 2 3+4 2 3+4 2 3+4 2 5 2 |
| | 6 2 6 2 5+6 2 5+6 2 5+6 2 5 2 |
| | 6 2 6 2 5+6 2 5+6 2 5+6 2 5 2 |
| | 6 2 6 2 5+6 2 5+6 2 5+6 2 5 2 |
| | 5 2 6 2 5+6 2 5+6 2 3+4 2 5 2 |
| | 5 2 5 2 3+4 2 3+4 2 5+6 2 5 2 |
| | 6 2 6 2 3+4 2 5+6 2 5+6 2 5 2 |
| | 5 2 5 2 3+4 2 3+4 2 3+4 2 5 2 |
| | 6 2 5 2 5+6 2 5+6 2 5+6 2 5 2 |
| | 6 2 5 2 5+6 2 3+4 2 3+4 2 5 2 |
| | 5 2 6 2 3+4 2 3+4 2 3+4 2 5 2 |
| | 5 2 6 2 3+4 2 3+4 2 3+4 2 5 2 |
| | 6 2 6 2 5+6 2 5+6 2 5+6 2 5 2 |
| | 5 2 6 2 3+4 2 5+6 2 5+6 2 5 2 |
| | 5 2 6 2 5+6 2 3+4 2 3+4 2 5 2 |
| | 6 2 6 2 5+6 2 5+6 2 5+6 2 5 2 |
| | 6 2 6 2 5+6 2 5+6 2 5+6 2 5 2 |
| 71.1.25 | 246814 34 Ge5Nd1 |
| | -GM -X -T -R -S -W |
| | 5 2 5 2 3+4 2 3+4 2 3+4 2 5 2 |
| | 6 2 6 2 5+6 2 5+6 2 5+6 2 5 2 |
| | 6 2 6 2 5+6 2 5+6 2 5+6 2 5 2 |
| | 6 2 6 2 5+6 2 5+6 2 5+6 2 5 2 |
| | 5 2 6 2 5+6 2 5+6 2 3+4 2 5 2 |
| | 5 2 5 2 3+4 2 3+4 2 5+6 2 5 2 |
| | 6 2 6 2 3+4 2 5+6 2 5+6 2 5 2 |
| | 5 2 5 2 3+4 2 3+4 2 3+4 2 5 2 |
| | 6 2 5 2 5+6 2 5+6 2 5+6 2 5 2 |
| | 6 2 5 2 5+6 2 3+4 2 3+4 2 5 2 |
| | 5 2 6 2 3+4 2 3+4 2 3+4 2 5 2 |
| | 5 2 6 2 3+4 2 3+4 2 3+4 2 5 2 |
| | 6 2 6 2 5+6 2 5+6 2 5+6 2 5 2 |
| | 5 2 6 2 3+4 2 5+6 2 5+6 2 5 2 |
| | 5 2 6 2 5+6 2 3+4 2 3+4 2 5 2 |
| | 6 2 6 2 5+6 2 5+6 2 5+6 2 5 2 |
| | 6 2 6 2 5+6 2 5+6 2 5+6 2 5 2 |
| 71.1.26 | 411182 128 Ge4Pd4Yb3 |
| | -GM -X -T -R -S -W |
| | 5 2 5 2 3+4 2 3+4 2 3+4 2 5 2 |
| | 5 2 5 2 3+4 2 5+6 2 5+6 2 5 2 |
| | 6 2 6 2 5+6 2 3+4 2 3+4 2 5 2 |
| | 6 2 6 2 5+6 2 5+6 2 5+6 2 5 2 |
| | 5 2 5 2 3+4 2 5+6 2 5+6 2 5 2 |
| | 6 2 6 2 5+6 2 3+4 2 3+4 2 5 2 |
| | 6 2 6 2 5+6 2 5+6 2 5+6 2 5 2 |
| | 6 2 6 2 5+6 2 5+6 2 5+6 2 5 2 |
| | 5 2 5 2 3+4 2 5+6 2 5+6 2 5 2 |
| | 6 2 6 2 5+6 2 3+4 2 3+4 2 5 2 |
| | 5 2 5 2 3+4 2 5+6 2 5+6 2 5 2 |
| | 6 2 6 2 5+6 2 3+4 2 3+4 2 5 2 |
| | 5 2 6 2 5+6 2 5+6 2 3+4 2 5 2 |
| | 5 2 5 2 3+4 2 3+4 2 3+4 2 5 2 |
| | 6 2 5 2 5+6 2 5+6 2 5+6 2 5 2 |
| | 6 2 6 2 3+4 2 3+4 2 5+6 2 5 2 |
| | 5 2 6 2 3+4 2 3+4 2 3+4 2 5 2 |
| | 6 2 5 2 5+6 2 5+6 2 5+6 2 5 2 |
| | 5 2 5 2 3+4 2 3+4 2 3+4 2 5 2 |
| | 6 2 5 2 5+6 2 5+6 2 5+6 2 5 2 |
| | 6 2 6 2 3+4 2 3+4 2 3+4 2 5 2 |
| | 6 2 5 2 5+6 2 5+6 2 5+6 2 5 2 |
| | 5 2 6 2 3+4 2 3+4 2 5+6 2 5 2 |
| | 5 2 6 2 3+4 2 5+6 2 3+4 2 5 2 |
| | 5 2 5 2 5+6 2 3+4 2 5+6 2 5 2 |
| | 5 2 5 2 5+6 2 3+4 2 3+4 2 5 2 |
| | 6 2 6 2 5+6 2 5+6 2 5+6 2 5 2 |
| | 5 2 6 2 3+4 2 5+6 2 3+4 2 5 2 |
| | 6 2 5 2 3+4 2 3+4 2 3+4 2 5 2 |
| | 6 2 6 2 5+6 2 3+4 2 5+6 2 5 2 |
| | 5 2 6 2 3+4 2 5+6 2 3+4 2 5 2 |
| | 5 2 6 2 5+6 2 5+6 2 5+6 2 5 2 |
| | 6 2 5 2 5+6 2 5+6 2 5+6 2 5 2 |
| | 5 2 5 2 5+6 2 3+4 2 3+4 2 5 2 |
| | 6 2 5 2 3+4 2 3+4 2 5+6 2 5 2 |
| | 5 2 6 2 3+4 2 5+6 2 3+4 2 5 2 |
| | 6 2 5 2 5+6 2 5+6 2 5+6 2 5 2 |
| | 6 2 6 2 3+4 2 3+4 2 3+4 2 5 2 |
| | 5 2 6 2 5+6 2 5+6 2 3+4 2 5 2 |
| | 6 2 6 2 3+4 2 3+4 2 3+4 2 5 2 |
| | 5 2 5 2 3+4 2 5+6 2 5+6 2 5 2 |
| | 6 2 6 2 5+6 2 3+4 2 3+4 2 5 2 |
| | 5 2 6 2 5+6 2 5+6 2 5+6 2 5 2 |
| | 6 2 5 2 3+4 2 5+6 2 5+6 2 5 2 |
| | 6 2 6 2 5+6 2 5+6 2 5+6 2 5 2 |
| | 6 2 6 2 5+6 2 3+4 2 3+4 2 5 2 |
| | 5 2 5 2 3+4 2 5+6 2 5+6 2 5 2 |
| | 6 2 6 2 5+6 2 5+6 2 5+6 2 5 2 |
| | 6 2 6 2 5+6 2 3+4 2 3+4 2 5 2 |
| | 5 2 5 2 3+4 2 3+4 2 5+6 2 5 2 |
| | 5 2 6 2 3+4 2 5+6 2 3+4 2 5 2 |
| | 5 2 5 2 5+6 2 5+6 2 5+6 2 5 2 |
| | 6 2 5 2 5+6 2 3+4 2 5+6 2 5 2 |
| | 6 2 6 2 5+6 2 5+6 2 3+4 2 5 2 |
| | 6 2 6 2 5+6 2 5+6 2 5+6 2 5 2 |
| | 5 2 6 2 3+4 2 5+6 2 3+4 2 5 2 |
| | 6 2 5 2 3+4 2 3+4 2 5+6 2 5 2 |
| | 6 2 6 2 5+6 2 5+6 2 5+6 2 5 2 |
| | 6 2 6 2 5+6 2 5+6 2 3+4 2 5 2 |
| | 6 2 6 2 3+4 2 3+4 2 5+6 2 5 2 |
| | 5 2 5 2 5+6 2 5+6 2 5+6 2 5 2 |
| | 6 2 6 2 5+6 2 3+4 2 3+4 2 5 2 |
| | 5 2 5 2 3+4 2 5+6 2 5+6 2 5 2 |
| | 5 2 6 2 5+6 2 5+6 2 5+6 2 5 2 |
| 71.1.27 | 262435 62 Ge4Li4Nd3 |
| | -GM -X -T -R -S -W |
| | 5 2 5 2 3+4 2 5+6 2 5+6 2 5 2 |
| | 6 2 6 2 5+6 2 3+4 2 3+4 2 5 2 |
| | 5 2 5 2 3+4 2 3+4 2 3+4 2 5 2 |

Continued on next column

**TABLE XLIV– continued**

| TQC | ICSD NoE SF |
|---|---|
| | 5 2 5 2 3+4 2 5+6 2 5+6 2 5 2 |
| | 6 2 6 2 5+6 2 3+4 2 3+4 2 5 2 |
| | 6 2 6 2 5+6 2 5+6 2 5+6 2 5 2 |
| | 6 2 6 2 5+6 2 5+6 2 5+6 2 5 2 |
| | 5 2 5 2 3+4 2 5+6 2 3+4 2 5 2 |
| | 5 2 5 2 5+6 2 3+4 2 5+6 2 5 2 |
| | 6 2 6 2 3+4 2 3+4 2 5+6 2 5 2 |
| | 6 2 6 2 5+6 2 5+6 2 3+4 2 5 2 |
| | 6 2 6 2 5+6 2 5+6 2 5+6 2 5 2 |
| | 5 2 5 2 5+6 2 5+6 2 3+4 2 5 2 |
| | 5 2 6 2 3+4 2 3+4 2 3+4 2 5 2 |
| | 6 2 5 2 5+6 2 5+6 2 5+6 2 5 2 |
| | 6 2 6 2 3+4 2 3+4 2 5+6 2 5 2 |
| | 5 2 5 2 3+4 2 3+4 2 5+6 2 5 2 |
| | 6 2 6 2 3+4 2 5+6 2 3+4 2 5 2 |
| | 5 2 6 2 5+6 2 3+4 2 3+4 2 5 2 |
| | 6 2 5 2 5+6 2 5+6 2 3+4 2 5 2 |
| | 5 2 6 2 3+4 2 3+4 2 3+4 2 5 2 |
| | 5 2 6 2 3+4 2 3+4 2 5+6 2 5 2 |
| | 5 2 5 2 5+6 2 5+6 2 5+6 2 5 2 |
| | 5 2 6 2 3+4 2 5+6 2 3+4 2 5 2 |
| | 6 2 5 2 5+6 2 3+4 2 5+6 2 5 2 |
| | 6 2 5 2 5+6 2 5+6 2 3+4 2 5 2 |
| | 6 2 5 2 5+6 2 3+4 2 5+6 2 5 2 |
| | 5 2 6 2 3+4 2 5+6 2 5+6 2 5 2 |
| | 6 2 6 2 5+6 2 3+4 2 3+4 2 5 2 |
| | 5 2 6 2 5+6 2 5+6 2 5+6 2 5 2 |
| | 6 2 5 2 3+4 2 5+6 2 5+6 2 5 2 |
| 71.1.28 | 170342 168 Ge2Ni1Yb1 |
| | -GM -X -T -R -S -W |
| | 5 2 5 2 3+4 2 3+4 2 3+4 2 5 2 |
| | 6 2 6 2 5+6 2 5+6 2 5+6 2 5 2 |
| | 5 2 5 2 3+4 2 3+4 2 3+4 2 5 2 |
| | 6 2 6 2 5+6 2 5+6 2 5+6 2 5 2 |
| | 5 2 5 2 3+4 2 5+6 2 5+6 2 5 2 |
| | 6 2 6 2 5+6 2 3+4 2 3+4 2 5 2 |
| | 5 2 6 2 3+4 2 3+4 2 3+4 2 5 2 |
| | 6 2 5 2 5+6 2 5+6 2 5+6 2 5 2 |
| | 5 2 5 2 3+4 2 5+6 2 5+6 2 5 2 |
| | 6 2 6 2 5+6 2 3+4 2 3+4 2 5 2 |
| | 5 2 5 2 3+4 2 5+6 2 5+6 2 5 2 |
| | 6 2 6 2 5+6 2 3+4 2 3+4 2 5 2 |
| | 5 2 6 2 5+6 2 5+6 2 3+4 2 5 2 |
| | 6 2 5 2 3+4 2 3+4 2 5+6 2 5 2 |
| | 5 2 6 2 3+4 2 3+4 2 5+6 2 5 2 |
| | 6 2 5 2 5+6 2 5+6 2 3+4 2 5 2 |
| | 5 2 6 2 5+6 2 3+4 2 5+6 2 5 2 |
| | 6 2 5 2 3+4 2 5+6 2 3+4 2 5 2 |
| | 5 2 5 2 5+6 2 5+6 2 5+6 2 5 2 |
| | 5 2 6 2 5+6 2 3+4 2 3+4 2 5 2 |
| | 5 2 5 2 3+4 2 5+6 2 3+4 2 5 2 |
| | 6 2 6 2 5+6 2 3+4 2 5+6 2 5 2 |
| | 6 2 5 2 3+4 2 3+4 2 3+4 2 5 2 |
| | 6 2 6 2 3+4 2 5+6 2 5+6 2 5 2 |
| | 5 2 5 2 5+6 2 3+4 2 3+4 2 5 2 |
| | 6 2 6 2 3+4 2 5+6 2 5+6 2 5 2 |
| | 6 2 5 2 3+4 2 5+6 2 5+6 2 5 2 |
| | 5 2 6 2 5+6 2 3+4 2 5+6 2 5 2 |
| | 5 2 6 2 3+4 2 5+6 2 3+4 2 5 2 |
| | 5 2 6 2 5+6 2 3+4 2 3+4 2 5 2 |
| | 5 2 5 2 5+6 2 3+4 2 3+4 2 5 2 |

Continued on next column

**TABLE XLIV– continued**

| TQC | ICSD NoE SF |
|---|---|
| | 6 2 6 2 3+4 2 3+4 2 5+6 2 5 2 |
| | 5 2 6 2 3+4 2 5+6 2 3+4 2 5 2 |
| | 6 2 5 2 3+4 2 5+6 2 5+6 2 5 2 |
| | 5 2 6 2 5+6 2 5+6 2 5+6 2 5 2 |
| | 5 2 5 2 5+6 2 3+4 2 3+4 2 5 2 |
| | 6 2 5 2 5+6 2 5+6 2 5+6 2 5 2 |
| | 5 2 6 2 3+4 2 3+4 2 3+4 2 5 2 |
| | 6 2 5 2 5+6 2 3+4 2 3+4 2 5 2 |
| | 5 2 6 2 3+4 2 5+6 2 5+6 2 5 2 |
| | 6 2 6 2 5+6 2 5+6 2 3+4 2 5 2 |
| | 6 2 5 2 5+6 2 3+4 2 5+6 2 5 2 |
| | 6 2 6 2 3+4 2 5+6 2 5+6 2 5 2 |
| | 5 2 5 2 5+6 2 3+4 2 3+4 2 5 2 |
| | 6 2 5 2 3+4 2 3+4 2 3+4 2 5 2 |
| | 5 2 5 2 3+4 2 5+6 2 5+6 2 5 2 |
| | 6 2 6 2 5+6 2 5+6 2 5+6 2 5 2 |
| | 5 2 5 2 5+6 2 3+4 2 3+4 2 5 2 |
| | 6 2 5 2 3+4 2 3+4 2 3+4 2 5 2 |
| | 5 2 6 2 5+6 2 5+6 2 5+6 2 5 2 |
| | 6 2 6 2 3+4 2 3+4 2 5+6 2 5 2 |
| | 6 2 5 2 3+4 2 5+6 2 3+4 2 5 2 |
| | 5 2 6 2 5+6 2 3+4 2 5+6 2 5 2 |
| | 5 2 5 2 3+4 2 3+4 2 3+4 2 5 2 |
| | 5 2 6 2 5+6 2 5+6 2 3+4 2 5 2 |
| | 6 2 5 2 3+4 2 5+6 2 5+6 2 5 2 |
| | 5 2 6 2 5+6 2 5+6 2 3+4 2 5 2 |
| | 6 2 5 2 3+4 2 3+4 2 5+6 2 5 2 |
| | 5 2 6 2 5+6 2 5+6 2 5+6 2 5 2 |
| | 6 2 5 2 3+4 2 3+4 2 3+4 2 5 2 |
| | 6 2 6 2 5+6 2 3+4 2 3+4 2 5 2 |
| | 5 2 5 2 5+6 2 5+6 2 5+6 2 5 2 |
| | 6 2 5 2 3+4 2 3+4 2 5+6 2 5 2 |
| | 5 2 6 2 3+4 2 5+6 2 3+4 2 5 2 |
| | 5 2 6 2 3+4 2 3+4 2 5+6 2 5 2 |
| | 6 2 6 2 3+4 2 5+6 2 3+4 2 5 2 |
| | 5 2 5 2 3+4 2 5+6 2 3+4 2 5 2 |
| | 6 2 6 2 5+6 2 3+4 2 5+6 2 5 2 |
| | 6 2 5 2 5+6 2 5+6 2 3+4 2 5 2 |
| | 5 2 6 2 5+6 2 3+4 2 5+6 2 5 2 |
| | 6 2 5 2 5+6 2 3+4 2 3+4 2 5 2 |
| | 5 2 5 2 5+6 2 5+6 2 5+6 2 5 2 |
| | 6 2 6 2 3+4 2 5+6 2 3+4 2 5 2 |
| | 5 2 5 2 5+6 2 3+4 2 5+6 2 5 2 |
| | 5 2 6 2 3+4 2 5+6 2 3+4 2 5 2 |
| | 5 2 5 2 3+4 2 3+4 2 5+6 2 5 2 |
| | 6 2 6 2 5+6 2 3+4 2 3+4 2 5 2 |
| | 6 2 6 2 3+4 2 5+6 2 5+6 2 5 2 |
| | 6 2 5 2 3+4 2 3+4 2 3+4 2 5 2 |
| | 5 2 6 2 5+6 2 5+6 2 3+4 2 5 2 |
| | 5 2 6 2 3+4 2 3+4 2 5+6 2 5 2 |
| | 6 2 5 2 5+6 2 5+6 2 3+4 2 5 2 |
| | 5 2 5 2 3+4 2 5+6 2 5+6 2 5 2 |
| | 6 2 6 2 5+6 2 3+4 2 5+6 2 5 2 |
| 71.1.29 | 170343 172 Ge2Lu1Ni1 |
| | -GM -X -T -R -S -W |
| | 5 2 5 2 3+4 2 3+4 2 3+4 2 5 2 |
| | 6 2 6 2 5+6 2 5+6 2 5+6 2 5 2 |
| | 5 2 5 2 3+4 2 3+4 2 3+4 2 5 2 |
| | 6 2 6 2 5+6 2 5+6 2 5+6 2 5 2 |
| | 5 2 5 2 5+6 2 3+4 2 5+6 2 5 2 |
| | 6 2 6 2 3+4 2 5+6 2 3+4 2 5 2 |

Continued on next column

**TABLE XLIV– continued**

| TQC | ICSD NoE SF |
|---|---|
| | 5 2 6 2 3+4 2 3+4 2 3+4 2 5 2 |
| | 6 2 5 2 5+6 2 5+6 2 5+6 2 5 2 |
| | 5 2 5 2 5+6 2 3+4 2 5+6 2 5 2 |
| | 6 2 6 2 3+4 2 5+6 2 3+4 2 5 2 |
| | 6 2 6 2 5+6 2 5+6 2 5+6 2 5 2 |
| | 5 2 5 2 3+4 2 3+4 2 3+4 2 5 2 |
| | 5 2 6 2 3+4 2 3+4 2 3+4 2 5 2 |
| | 6 2 5 2 5+6 2 5+6 2 5+6 2 5 2 |
| | 5 2 6 2 5+6 2 3+4 2 5+6 2 5 2 |
| | 6 2 5 2 3+4 2 5+6 2 3+4 2 5 2 |
| | 5 2 6 2 3+4 2 5+6 2 3+4 2 5 2 |
| | 6 2 5 2 5+6 2 3+4 2 5+6 2 5 2 |
| | 5 2 5 2 3+4 2 3+4 2 5+6 2 5 2 |
| | 5 2 6 2 3+4 2 5+6 2 3+4 2 5 2 |
| | 5 2 5 2 5+6 2 3+4 2 3+4 2 5 2 |
| | 6 2 6 2 3+4 2 5+6 2 5+6 2 5 2 |
| | 6 2 5 2 5+6 2 5+6 2 3+4 2 5 2 |
| | 6 2 6 2 5+6 2 3+4 2 5+6 2 5 2 |
| | 6 2 6 2 3+4 2 5+6 2 3+4 2 5 2 |
| | 5 2 5 2 5+6 2 3+4 2 5+6 2 5 2 |
| | 5 2 5 2 3+4 2 3+4 2 3+4 2 5 2 |
| | 6 2 6 2 3+4 2 5+6 2 5+6 2 5 2 |
| | 5 2 5 2 5+6 2 3+4 2 3+4 2 5 2 |
| | 6 2 6 2 5+6 2 5+6 2 5+6 2 5 2 |
| | 6 2 5 2 3+4 2 3+4 2 3+4 2 5 2 |
| | 5 2 6 2 5+6 2 5+6 2 5+6 2 5 2 |
| | 5 2 5 2 3+4 2 3+4 2 5+6 2 5 2 |
| | 5 2 6 2 5+6 2 5+6 2 3+4 2 5 2 |
| | 6 2 5 2 3+4 2 5+6 2 3+4 2 5 2 |
| | 6 2 6 2 5+6 2 3+4 2 5+6 2 5 2 |
| | 5 2 5 2 3+4 2 5+6 2 3+4 2 5 2 |
| | 6 2 6 2 5+6 2 3+4 2 5+6 2 5 2 |
| | 6 2 6 2 5+6 2 3+4 2 5+6 2 5 2 |
| | 5 2 5 2 5+6 2 5+6 2 5+6 2 5 2 |
| | 6 2 6 2 3+4 2 3+4 2 3+4 2 5 2 |
| | 5 2 5 2 3+4 2 5+6 2 3+4 2 5 2 |
| | 6 2 5 2 5+6 2 5+6 2 5+6 2 5 2 |
| | 5 2 5 2 3+4 2 3+4 2 5+6 2 5 2 |
| | 5 2 6 2 3+4 2 5+6 2 3+4 2 5 2 |
| | 6 2 6 2 3+4 2 3+4 2 3+4 2 5 2 |
| | 6 2 5 2 5+6 2 3+4 2 5+6 2 5 2 |
| | 5 2 6 2 5+6 2 5+6 2 5+6 2 5 2 |
| | 5 2 6 2 5+6 2 3+4 2 3+4 2 5 2 |
| | 5 2 5 2 3+4 2 5+6 2 3+4 2 5 2 |
| | 5 2 6 2 3+4 2 5+6 2 5+6 2 5 2 |
| | 6 2 5 2 5+6 2 3+4 2 3+4 2 5 2 |
| | 6 2 6 2 5+6 2 5+6 2 5+6 2 5 2 |
| | 6 2 5 2 3+4 2 3+4 2 3+4 2 5 2 |
| | 6 2 5 2 5+6 2 3+4 2 5+6 2 5 2 |
| | 5 2 6 2 3+4 2 5+6 2 5+6 2 5 2 |
| | 5 2 6 2 5+6 2 3+4 2 3+4 2 5 2 |
| | 5 2 6 2 3+4 2 5+6 2 3+4 2 5 2 |
| | 5 2 6 2 3+4 2 5+6 2 3+4 2 5 2 |
| | 6 2 6 2 5+6 2 5+6 2 3+4 2 5 2 |
| | 5 2 5 2 5+6 2 3+4 2 5+6 2 5 2 |
| | 6 2 5 2 3+4 2 3+4 2 3+4 2 5 2 |
| | 5 2 6 2 5+6 2 3+4 2 5+6 2 5 2 |
| | 5 2 5 2 3+4 2 5+6 2 5+6 2 5 2 |
| | 6 2 5 2 3+4 2 3+4 2 5+6 2 5 2 |
| | 6 2 6 2 5+6 2 5+6 2 3+4 2 5 2 |
| | 5 2 5 2 3+4 2 5+6 2 3+4 2 5 2 |

Continued on next column

**TABLE XLIV– continued**

| TQC | ICSD NoE SF |
|---|---|
| | 5 2 6 2 3+4 2 3+4 2 5+6 2 5 2 |
| | 6 2 6 2 5+6 2 3+4 2 3+4 2 5 2 |
| | 6 2 5 2 5+6 2 5+6 2 5+6 2 5 2 |
| | 5 2 6 2 3+4 2 3+4 2 5+6 2 5 2 |
| | 6 2 5 2 3+4 2 5+6 2 3+4 2 5 2 |
| | 5 2 5 2 5+6 2 5+6 2 3+4 2 5 2 |
| | 6 2 5 2 5+6 2 3+4 2 5+6 2 5 2 |
| | 6 2 5 2 3+4 2 3+4 2 5+6 2 5 2 |
| | 5 2 6 2 5+6 2 5+6 2 3+4 2 5 2 |
| | 5 2 6 2 3+4 2 5+6 2 3+4 2 5 2 |
| | 6 2 5 2 3+4 2 5+6 2 5+6 2 5 2 |
| | 6 2 6 2 5+6 2 3+4 2 5+6 2 5 2 |
| | 5 2 6 2 5+6 2 5+6 2 5+6 2 5 2 |
| | 5 2 5 2 5+6 2 3+4 2 3+4 2 5 2 |
| | 6 2 6 2 5+6 2 3+4 2 3+4 2 5 2 |
| | 5 2 5 2 5+6 2 3+4 2 3+4 2 5 2 |
| | 6 2 6 2 3+4 2 5+6 2 5+6 2 5 2 |
| | 5 2 5 2 3+4 2 3+4 2 3+4 2 5 2 |
| | 6 2 6 2 3+4 2 3+4 2 3+4 2 5 2 |
| 71.1.30 | 76509 112 Ge2Ir1Y1 |
| | -GM -X -T -R -S -W |
| | 5 2 5 2 3+4 2 3+4 2 3+4 2 5 2 |
| | 6 2 6 2 5+6 2 5+6 2 5+6 2 5 2 |
| | 5 2 5 2 3+4 2 3+4 2 3+4 2 5 2 |
| | 6 2 6 2 5+6 2 5+6 2 5+6 2 5 2 |
| | 5 2 5 2 5+6 2 5+6 2 3+4 2 5 2 |
| | 6 2 6 2 3+4 2 3+4 2 5+6 2 5 2 |
| | 5 2 6 2 3+4 2 3+4 2 3+4 2 5 2 |
| | 6 2 5 2 5+6 2 5+6 2 5+6 2 5 2 |
| | 5 2 5 2 5+6 2 5+6 2 3+4 2 5 2 |
| | 6 2 6 2 3+4 2 3+4 2 5+6 2 5 2 |
| | 6 2 6 2 3+4 2 3+4 2 3+4 2 5 2 |
| | 5 2 5 2 5+6 2 5+6 2 5+6 2 5 2 |
| | 5 2 6 2 3+4 2 3+4 2 3+4 2 5 2 |
| | 6 2 5 2 5+6 2 5+6 2 5+6 2 5 2 |
| | 5 2 6 2 3+4 2 3+4 2 3+4 2 5 2 |
| | 6 2 5 2 5+6 2 5+6 2 5+6 2 5 2 |
| | 5 2 6 2 3+4 2 3+4 2 3+4 2 5 2 |
| | 6 2 5 2 5+6 2 5+6 2 5+6 2 5 2 |
| | 5 2 5 2 3+4 2 5+6 2 3+4 2 5 2 |
| | 5 2 6 2 3+4 2 3+4 2 5+6 2 5 2 |
| | 5 2 5 2 5+6 2 5+6 2 5+6 2 5 2 |
| | 6 2 5 2 3+4 2 3+4 2 3+4 2 5 2 |
| | 6 2 6 2 5+6 2 3+4 2 3+4 2 5 2 |
| | 6 2 6 2 5+6 2 5+6 2 5+6 2 5 2 |
| | 5 2 5 2 3+4 2 5+6 2 5+6 2 5 2 |
| | 6 2 6 2 5+6 2 3+4 2 3+4 2 5 2 |
| | 6 2 5 2 5+6 2 3+4 2 3+4 2 5 2 |
| | 5 2 6 2 3+4 2 5+6 2 5+6 2 5 2 |
| | 5 2 6 2 3+4 2 3+4 2 5+6 2 5 2 |
| | 5 2 6 2 5+6 2 3+4 2 3+4 2 5 2 |
| | 6 2 5 2 5+6 2 5+6 2 3+4 2 5 2 |
| | 6 2 6 2 3+4 2 5+6 2 5+6 2 5 2 |
| | 5 2 6 2 5+6 2 5+6 2 3+4 2 5 2 |
| | 5 2 5 2 3+4 2 3+4 2 5+6 2 5 2 |
| | 5 2 6 2 3+4 2 3+4 2 3+4 2 5 2 |
| | 6 2 5 2 5+6 2 5+6 2 5+6 2 5 2 |
| | 5 2 5 2 3+4 2 5+6 2 3+4 2 5 2 |
| | 5 2 6 2 5+6 2 3+4 2 5+6 2 5 2 |
| | 6 2 5 2 3+4 2 3+4 2 5+6 2 5 2 |
| | 5 2 6 2 5+6 2 5+6 2 3+4 2 5 2 |

Continued on next column

**TABLE XLIV– continued**

| TQC | ICSD NoE SF |
|---|---|
| | 6 2 6 2 3+4 2 3+4 2 5+6 2 5 2 |
| | 6 2 5 2 5+6 2 5+6 2 3+4 2 5 2 |
| | 5 2 5 2 3+4 2 5+6 2 5+6 2 5 2 |
| | 6 2 6 2 3+4 2 3+4 2 3+4 2 5 2 |
| | 6 2 5 2 5+6 2 3+4 2 5+6 2 5 2 |
| | 5 2 5 2 3+4 2 5+6 2 3+4 2 5 2 |
| | 6 2 5 2 5+6 2 5+6 2 3+4 2 5 2 |
| | 5 2 6 2 3+4 2 3+4 2 5+6 2 5 2 |
| | 6 2 5 2 5+6 2 3+4 2 5+6 2 5 2 |
| | 5 2 6 2 3+4 2 5+6 2 3+4 2 5 2 |
| | 6 2 6 2 5+6 2 3+4 2 3+4 2 5 2 |
| | 5 2 6 2 5+6 2 3+4 2 5+6 2 5 2 |
| | 5 2 5 2 5+6 2 5+6 2 3+4 2 5 2 |
| | 6 2 6 2 5+6 2 5+6 2 5+6 2 5 2 |
| | 6 2 5 2 3+4 2 5+6 2 5+6 2 5 2 |
| | 5 2 6 2 5+6 2 3+4 2 3+4 2 5 2 |
| 71.1.31 | 88227 156 Ge2Ho1Pt1 |
| | -GM -X -T -R -S -W |
| | 5 2 5 2 3+4 2 3+4 2 3+4 2 5 2 |
| | 6 2 6 2 5+6 2 5+6 2 5+6 2 5 2 |
| | 5 2 5 2 3+4 2 3+4 2 3+4 2 5 2 |
| | 6 2 6 2 5+6 2 5+6 2 5+6 2 5 2 |
| | 5 2 5 2 3+4 2 5+6 2 5+6 2 5 2 |
| | 6 2 6 2 5+6 2 3+4 2 3+4 2 5 2 |
| | 5 2 6 2 3+4 2 3+4 2 3+4 2 5 2 |
| | 6 2 5 2 5+6 2 5+6 2 5+6 2 5 2 |
| | 5 2 5 2 3+4 2 5+6 2 5+6 2 5 2 |
| | 6 2 6 2 5+6 2 3+4 2 3+4 2 5 2 |
| | 5 2 5 2 3+4 2 5+6 2 5+6 2 5 2 |
| | 6 2 6 2 5+6 2 3+4 2 3+4 2 5 2 |
| | 5 2 6 2 5+6 2 3+4 2 3+4 2 5 2 |
| | 6 2 5 2 3+4 2 5+6 2 5+6 2 5 2 |
| | 5 2 6 2 3+4 2 5+6 2 5+6 2 5 2 |
| | 6 2 5 2 5+6 2 3+4 2 3+4 2 5 2 |
| | 5 2 6 2 5+6 2 3+4 2 5+6 2 5 2 |
| | 5 2 5 2 3+4 2 5+6 2 3+4 2 5 2 |
| | 5 2 6 2 5+6 2 5+6 2 5+6 2 5 2 |
| | 6 2 5 2 5+6 2 3+4 2 3+4 2 5 2 |
| | 5 2 5 2 3+4 2 5+6 2 3+4 2 5 2 |
| | 6 2 6 2 5+6 2 3+4 2 5+6 2 5 2 |
| | 6 2 5 2 3+4 2 3+4 2 5+6 2 5 2 |
| | 6 2 6 2 3+4 2 5+6 2 3+4 2 5 2 |
| | 5 2 5 2 5+6 2 5+6 2 3+4 2 5 2 |
| | 6 2 6 2 3+4 2 3+4 2 5+6 2 5 2 |
| | 6 2 5 2 3+4 2 3+4 2 5+6 2 5 2 |
| | 5 2 6 2 5+6 2 5+6 2 3+4 2 5 2 |
| | 5 2 6 2 5+6 2 3+4 2 3+4 2 5 2 |
| | 5 2 6 2 3+4 2 5+6 2 5+6 2 5 2 |
| | 6 2 5 2 3+4 2 3+4 2 5+6 2 5 2 |
| | 5 2 6 2 5+6 2 5+6 2 3+4 2 5 2 |
| | 6 2 5 2 3+4 2 5+6 2 5+6 2 5 2 |
| | 5 2 6 2 5+6 2 3+4 2 3+4 2 5 2 |
| | 5 2 5 2 5+6 2 3+4 2 5+6 2 5 2 |
| | 6 2 6 2 5+6 2 5+6 2 3+4 2 5 2 |
| | 6 2 5 2 3+4 2 3+4 2 5+6 2 5 2 |
| | 5 2 6 2 3+4 2 5+6 2 3+4 2 5 2 |
| | 6 2 5 2 5+6 2 5+6 2 5+6 2 5 2 |
| | 5 2 5 2 3+4 2 3+4 2 3+4 2 5 2 |
| | 5 2 6 2 3+4 2 5+6 2 5+6 2 5 2 |
| | 6 2 6 2 5+6 2 3+4 2 3+4 2 5 2 |
| | 5 2 5 2 5+6 2 5+6 2 5+6 2 5 2 |

Continued on next column

**TABLE XLIV– continued**

| TQC | ICSD NoE SF |
|---|---|
| | 6 2 5 2 3+4 2 3+4 2 3+4 2 5 2 |
| | 6 2 6 2 5+6 2 3+4 2 5+6 2 5 2 |
| | 5 2 5 2 5+6 2 5+6 2 3+4 2 5 2 |
| | 5 2 5 2 3+4 2 5+6 2 3+4 2 5 2 |
| | 6 2 6 2 5+6 2 3+4 2 3+4 2 5 2 |
| | 6 2 6 2 3+4 2 3+4 2 5+6 2 5 2 |
| | 6 2 5 2 5+6 2 3+4 2 5+6 2 5 2 |
| | 5 2 5 2 3+4 2 5+6 2 5+6 2 5 2 |
| | 6 2 6 2 3+4 2 5+6 2 5+6 2 5 2 |
| | 5 2 6 2 3+4 2 3+4 2 3+4 2 5 2 |
| | 5 2 5 2 3+4 2 5+6 2 3+4 2 5 2 |
| | 6 2 5 2 3+4 2 3+4 2 3+4 2 5 2 |
| | 5 2 6 2 5+6 2 5+6 2 5+6 2 5 2 |
| | 6 2 5 2 5+6 2 3+4 2 3+4 2 5 2 |
| | 5 2 6 2 5+6 2 5+6 2 3+4 2 5 2 |
| | 6 2 5 2 3+4 2 3+4 2 5+6 2 5 2 |
| | 5 2 6 2 5+6 2 5+6 2 5+6 2 5 2 |
| | 6 2 5 2 3+4 2 5+6 2 3+4 2 5 2 |
| | 5 2 6 2 3+4 2 3+4 2 5+6 2 5 2 |
| | 6 2 6 2 5+6 2 3+4 2 3+4 2 5 2 |
| | 5 2 5 2 3+4 2 5+6 2 5+6 2 5 2 |
| | 6 2 6 2 5+6 2 3+4 2 5+6 2 5 2 |
| | 5 2 6 2 3+4 2 5+6 2 3+4 2 5 2 |
| | 6 2 5 2 5+6 2 5+6 2 3+4 2 5 2 |
| | 5 2 6 2 3+4 2 3+4 2 5+6 2 5 2 |
| | 6 2 5 2 5+6 2 5+6 2 3+4 2 5 2 |
| | 6 2 5 2 5+6 2 3+4 2 5+6 2 5 2 |
| | 5 2 6 2 5+6 2 3+4 2 3+4 2 5 2 |
| | 6 2 6 2 3+4 2 5+6 2 5+6 2 5 2 |
| | 5 2 5 2 5+6 2 3+4 2 3+4 2 5 2 |
| | 5 2 6 2 5+6 2 3+4 2 5+6 2 5 2 |
| | 6 2 5 2 3+4 2 5+6 2 3+4 2 5 2 |
| | 5 2 5 2 3+4 2 5+6 2 5+6 2 5 2 |
| | 6 2 6 2 3+4 2 5+6 2 5+6 2 5 2 |
| | 5 2 6 2 5+6 2 3+4 2 3+4 2 5 2 |
| 71.1.32 | 300260 156 Ge2Ho1Pt1 |
| | -GM -X -T -R -S -W |
| | 5 2 5 2 3+4 2 3+4 2 3+4 2 5 2 |
| | 6 2 6 2 5+6 2 5+6 2 5+6 2 5 2 |
| | 5 2 5 2 3+4 2 3+4 2 3+4 2 5 2 |
| | 6 2 6 2 5+6 2 5+6 2 5+6 2 5 2 |
| | 5 2 5 2 3+4 2 3+4 2 3+4 2 5 2 |
| | 6 2 6 2 5+6 2 5+6 2 5+6 2 5 2 |
| | 5 2 6 2 3+4 2 3+4 2 3+4 2 5 2 |
| | 6 2 5 2 5+6 2 5+6 2 5+6 2 5 2 |
| | 5 2 5 2 5+6 2 3+4 2 5+6 2 5 2 |
| | 6 2 6 2 3+4 2 5+6 2 3+4 2 5 2 |
| | 5 2 5 2 3+4 2 3+4 2 3+4 2 5 2 |
| | 6 2 6 2 5+6 2 5+6 2 5+6 2 5 2 |
| | 5 2 6 2 5+6 2 3+4 2 5+6 2 5 2 |
| | 6 2 5 2 3+4 2 5+6 2 3+4 2 5 2 |
| | 5 2 6 2 3+4 2 3+4 2 3+4 2 5 2 |
| | 6 2 5 2 5+6 2 5+6 2 5+6 2 5 2 |
| | 5 2 6 2 5+6 2 5+6 2 3+4 2 5 2 |
| | 6 2 5 2 3+4 2 3+4 2 5+6 2 5 2 |
| | 5 2 5 2 5+6 2 3+4 2 3+4 2 5 2 |
| | 5 2 6 2 5+6 2 5+6 2 5+6 2 5 2 |
| | 5 2 5 2 3+4 2 3+4 2 5+6 2 5 2 |
| | 6 2 5 2 5+6 2 5+6 2 3+4 2 5 2 |
| | 6 2 6 2 3+4 2 5+6 2 3+4 2 5 2 |
| | 6 2 6 2 3+4 2 3+4 2 5+6 2 5 2 |

Continued on next column

**TABLE XLIV– continued**

| TQC | ICSD | NoE | SF |
|---|---|---|---|
| | 5 2 5 2 5+6 2 3+4 2 5+6 2 5 2 | | |
| | 6 2 6 2 3+4 2 5+6 2 3+4 2 5 2 | | |
| | 6 2 5 2 3+4 2 5+6 2 3+4 2 5 2 | | |
| | 5 2 6 2 5+6 2 3+4 2 5+6 2 5 2 | | |
| | 5 2 6 2 5+6 2 5+6 2 5+6 2 5 2 | | |
| | 5 2 6 2 3+4 2 3+4 2 3+4 2 5 2 | | |
| | 6 2 5 2 3+4 2 5+6 2 3+4 2 5 2 | | |
| | 6 2 6 2 5+6 2 3+4 2 5+6 2 5 2 | | |
| | 5 2 5 2 3+4 2 3+4 2 3+4 2 5 2 | | |
| | 5 2 6 2 5+6 2 5+6 2 5+6 2 5 2 | | |
| | 5 2 5 2 5+6 2 5+6 2 3+4 2 5 2 | | |
| | 6 2 6 2 5+6 2 3+4 2 5+6 2 5 2 | | |
| | 5 2 5 2 3+4 2 5+6 2 3+4 2 5 2 | | |
| | 6 2 6 2 3+4 2 3+4 2 5+6 2 5 2 | | |
| | 5 2 5 2 5+6 2 3+4 2 3+4 2 5 2 | | |
| | 6 2 6 2 3+4 2 5+6 2 5+6 2 5 2 | | |
| | 5 2 6 2 3+4 2 3+4 2 5+6 2 5 2 | | |
| | 6 2 5 2 5+6 2 5+6 2 3+4 2 5 2 | | |
| | 5 2 5 2 5+6 2 3+4 2 5+6 2 5 2 | | |
| | 6 2 6 2 3+4 2 5+6 2 3+4 2 5 2 | | |
| | 5 2 5 2 5+6 2 5+6 2 3+4 2 5 2 | | |
| | 6 2 5 2 3+4 2 3+4 2 5+6 2 5 2 | | |
| | 5 2 5 2 5+6 2 3+4 2 5+6 2 5 2 | | |
| | 6 2 6 2 5+6 2 5+6 2 5+6 2 5 2 | | |
| | 6 2 6 2 3+4 2 5+6 2 3+4 2 5 2 | | |
| | 5 2 5 2 5+6 2 5+6 2 3+4 2 5 2 | | |
| | 6 2 5 2 3+4 2 3+4 2 3+4 2 5 2 | | |
| | 5 2 6 2 3+4 2 3+4 2 3+4 2 5 2 | | |
| | 6 2 6 2 3+4 2 5+6 2 5+6 2 5 2 | | |
| | 5 2 5 2 3+4 2 3+4 2 5+6 2 5 2 | | |
| | 5 2 5 2 3+4 2 5+6 2 5+6 2 5 2 | | |
| | 6 2 6 2 5+6 2 3+4 2 3+4 2 5 2 | | |
| | 6 2 6 2 5+6 2 5+6 2 5+6 2 5 2 | | |
| | 5 2 5 2 5+6 2 3+4 2 5+6 2 5 2 | | |
| | 6 2 5 2 5+6 2 5+6 2 3+4 2 5 2 | | |
| | 5 2 6 2 3+4 2 3+4 2 3+4 2 5 2 | | |
| | 6 2 5 2 3+4 2 3+4 2 5+6 2 5 2 | | |
| | 5 2 6 2 3+4 2 5+6 2 3+4 2 5 2 | | |
| | 6 2 6 2 5+6 2 5+6 2 5+6 2 5 2 | | |
| | 5 2 5 2 3+4 2 3+4 2 3+4 2 5 2 | | |
| | 6 2 6 2 5+6 2 5+6 2 5+6 2 5 2 | | |
| | 5 2 6 2 3+4 2 3+4 2 3+4 2 5 2 | | |
| | 6 2 5 2 5+6 2 3+4 2 5+6 2 5 2 | | |
| | 5 2 5 2 3+4 2 5+6 2 3+4 2 5 2 | | |
| | 6 2 6 2 5+6 2 3+4 2 5+6 2 5 2 | | |
| | 6 2 5 2 5+6 2 5+6 2 5+6 2 5 2 | | |
| | 5 2 6 2 5+6 2 5+6 2 3+4 2 5 2 | | |
| | 5 2 5 2 5+6 2 3+4 2 3+4 2 5 2 | | |
| | 6 2 6 2 3+4 2 5+6 2 3+4 2 5 2 | | |
| | 5 2 5 2 5+6 2 3+4 2 5+6 2 5 2 | | |
| | 6 2 5 2 3+4 2 3+4 2 5+6 2 5 2 | | |
| | 5 2 6 2 3+4 2 5+6 2 3+4 2 5 2 | | |
| | 6 2 5 2 3+4 2 5+6 2 5+6 2 5 2 | | |
| | 5 2 6 2 5+6 2 3+4 2 3+4 2 5 2 | | |
| 71.1.33 | 174530 | 38 | Gd1Si5 |
| | -GM -X -T -R -S -W | | |
| | 5 2 5 2 3+4 2 3+4 2 3+4 2 5 2 | | |
| | 6 2 6 2 5+6 2 5+6 2 5+6 2 5 2 | | |
| | 6 2 6 2 5+6 2 5+6 2 5+6 2 5 2 | | |
| | 6 2 6 2 5+6 2 5+6 2 5+6 2 5 2 | | |
| | 5 2 6 2 5+6 2 5+6 2 3+4 2 5 2 | | |

Continued on next column

**TABLE XLIV– continued**

| TQC | ICSD | NoE | SF |
|---|---|---|---|
| | 5 2 5 2 3+4 2 3+4 2 5+6 2 5 2 | | |
| | 6 2 6 2 3+4 2 5+6 2 5+6 2 5 2 | | |
| | 5 2 5 2 3+4 2 3+4 2 3+4 2 5 2 | | |
| | 6 2 5 2 5+6 2 3+4 2 5+6 2 5 2 | | |
| | 6 2 6 2 5+6 2 5+6 2 3+4 2 5 2 | | |
| | 5 2 5 2 3+4 2 3+4 2 3+4 2 5 2 | | |
| | 5 2 6 2 3+4 2 3+4 2 3+4 2 5 2 | | |
| | 6 2 6 2 5+6 2 5+6 2 5+6 2 5 2 | | |
| | 5 2 6 2 5+6 2 5+6 2 5+6 2 5 2 | | |
| | 5 2 6 2 3+4 2 3+4 2 3+4 2 5 2 | | |
| | 6 2 6 2 5+6 2 5+6 2 5+6 2 5 2 | | |
| | 6 2 6 2 5+6 2 5+6 2 5+6 2 5 2 | | |
| | 6 2 6 2 5+6 2 5+6 2 5+6 2 5 2 | | |
| | 6 2 6 2 5+6 2 5+6 2 5+6 2 5 2 | | |
| 71.1.34 | 710016 | 38 | Gd1Ge5 |
| | -GM -X -T -R -S -W | | |
| | 5 2 5 2 3+4 2 3+4 2 3+4 2 5 2 | | |
| | 6 2 6 2 5+6 2 5+6 2 5+6 2 5 2 | | |
| | 6 2 6 2 5+6 2 5+6 2 5+6 2 5 2 | | |
| | 6 2 6 2 5+6 2 5+6 2 5+6 2 5 2 | | |
| | 5 2 6 2 5+6 2 5+6 2 3+4 2 5 2 | | |
| | 5 2 5 2 3+4 2 3+4 2 5+6 2 5 2 | | |
| | 6 2 6 2 3+4 2 5+6 2 5+6 2 5 2 | | |
| | 5 2 5 2 3+4 2 3+4 2 3+4 2 5 2 | | |
| | 6 2 5 2 5+6 2 5+6 2 5+6 2 5 2 | | |
| | 6 2 5 2 5+6 2 3+4 2 3+4 2 5 2 | | |
| | 5 2 6 2 3+4 2 3+4 2 3+4 2 5 2 | | |
| | 5 2 6 2 3+4 2 3+4 2 3+4 2 5 2 | | |
| | 6 2 6 2 5+6 2 5+6 2 5+6 2 5 2 | | |
| | 5 2 6 2 5+6 2 5+6 2 5+6 2 5 2 | | |
| | 5 2 6 2 3+4 2 3+4 2 3+4 2 5 2 | | |
| | 6 2 6 2 5+6 2 5+6 2 5+6 2 5 2 | | |
| | 6 2 6 2 5+6 2 5+6 2 5+6 2 5 2 | | |
| | 6 2 6 2 5+6 2 5+6 2 5+6 2 5 2 | | |
| | 6 2 6 2 5+6 2 5+6 2 5+6 2 5 2 | | |
| 71.1.35 | 154471 | 80 | Fe6Ga6U1 |
| | -GM -X -T -R -S -W | | |
| | 5 2 5 2 3+4 2 3+4 2 3+4 2 5 2 | | |
| | 6 2 6 2 5+6 2 5+6 2 5+6 2 5 2 | | |
| | 6 2 6 2 5+6 2 5+6 2 5+6 2 5 2 | | |
| | 6 2 6 2 5+6 2 5+6 2 5+6 2 5 2 | | |
| | 5 2 5 2 5+6 2 5+6 2 3+4 2 5 2 | | |
| | 6 2 6 2 3+4 2 3+4 2 5+6 2 5 2 | | |
| | 5 2 5 2 3+4 2 3+4 2 5+6 2 5 2 | | |
| | 6 2 6 2 3+4 2 5+6 2 3+4 2 5 2 | | |
| | 5 2 5 2 5+6 2 3+4 2 5+6 2 5 2 | | |
| | 5 2 6 2 5+6 2 5+6 2 3+4 2 5 2 | | |
| | 5 2 6 2 5+6 2 5+6 2 3+4 2 5 2 | | |
| | 5 2 6 2 3+4 2 3+4 2 5+6 2 5 2 | | |
| | 5 2 6 2 3+4 2 3+4 2 3+4 2 5 2 | | |
| | 6 2 6 2 5+6 2 3+4 2 5+6 2 5 2 | | |
| | 6 2 5 2 3+4 2 5+6 2 5+6 2 5 2 | | |
| | 5 2 6 2 3+4 2 5+6 2 3+4 2 5 2 | | |
| | 5 2 6 2 5+6 2 3+4 2 5+6 2 5 2 | | |
| | 6 2 6 2 3+4 2 5+6 2 5+6 2 5 2 | | |
| | 5 2 5 2 5+6 2 3+4 2 3+4 2 5 2 | | |
| | 5 2 5 2 5+6 2 3+4 2 3+4 2 5 2 | | |
| | 5 2 6 2 3+4 2 5+6 2 5+6 2 5 2 | | |
| | 5 2 6 2 5+6 2 3+4 2 3+4 2 5 2 | | |
| | 5 2 6 2 5+6 2 5+6 2 3+4 2 5 2 | | |
| | 5 2 5 2 3+4 2 5+6 2 5+6 2 5 2 | | |

Continued on next column

**TABLE XLIV– continued**

| TQC | ICSD | NoE | SF |
|---|---|---|---|
| | 5 2 6 2 3+4 2 3+4 2 5+6 2 5 2 | | |
| | 6 2 6 2 3+4 2 3+4 2 3+4 2 5 2 | | |
| | 6 2 6 2 5+6 2 5+6 2 3+4 2 5 2 | | |
| | 6 2 5 2 5+6 2 3+4 2 5+6 2 5 2 | | |
| | 5 2 5 2 3+4 2 5+6 2 3+4 2 5 2 | | |
| | 5 2 5 2 3+4 2 5+6 2 5+6 2 5 2 | | |
| | 6 2 6 2 5+6 2 3+4 2 3+4 2 5 2 | | |
| | 5 2 5 2 5+6 2 5+6 2 5+6 2 5 2 | | |
| | 6 2 5 2 5+6 2 5+6 2 5+6 2 5 2 | | |
| | 5 2 6 2 3+4 2 5+6 2 5+6 2 5 2 | | |
| | 6 2 6 2 5+6 2 3+4 2 3+4 2 5 2 | | |
| | 5 2 5 2 3+4 2 3+4 2 3+4 2 5 2 | | |
| | 5 2 6 2 5+6 2 5+6 2 3+4 2 5 2 | | |
| | 6 2 6 2 5+6 2 3+4 2 5+6 2 5 2 | | |
| | 5 2 6 2 3+4 2 5+6 2 5+6 2 5 2 | | |
| | 5 2 6 2 3+4 2 3+4 2 3+4 2 5 2 | | |
| 71.1.36 | 187334 | 90 | Fe4Ge4U3 |
| | -GM -X -T -R -S -W | | |
| | 5 2 5 2 3+4 2 3+4 2 3+4 2 5 2 | | |
| | 5 2 5 2 3+4 2 5+6 2 5+6 2 5 2 | | |
| | 6 2 6 2 5+6 2 3+4 2 3+4 2 5 2 | | |
| | 6 2 6 2 5+6 2 5+6 2 5+6 2 5 2 | | |
| | 5 2 5 2 3+4 2 5+6 2 5+6 2 5 2 | | |
| | 6 2 6 2 5+6 2 3+4 2 3+4 2 5 2 | | |
| | 6 2 6 2 5+6 2 5+6 2 5+6 2 5 2 | | |
| | 5 2 5 2 5+6 2 5+6 2 5+6 2 5 2 | | |
| | 6 2 6 2 3+4 2 3+4 2 3+4 2 5 2 | | |
| | 6 2 6 2 5+6 2 5+6 2 5+6 2 5 2 | | |
| | 5 2 5 2 3+4 2 3+4 2 5+6 2 5 2 | | |
| | 6 2 6 2 5+6 2 5+6 2 3+4 2 5 2 | | |
| | 5 2 5 2 5+6 2 5+6 2 3+4 2 5 2 | | |
| | 5 2 6 2 3+4 2 3+4 2 3+4 2 5 2 | | |
| | 6 2 5 2 5+6 2 5+6 2 5+6 2 5 2 | | |
| | 6 2 6 2 3+4 2 3+4 2 5+6 2 5 2 | | |
| | 6 2 6 2 3+4 2 3+4 2 3+4 2 5 2 | | |
| | 5 2 5 2 5+6 2 5+6 2 5+6 2 5 2 | | |
| | 5 2 6 2 3+4 2 3+4 2 3+4 2 5 2 | | |
| | 6 2 5 2 5+6 2 3+4 2 3+4 2 5 2 | | |
| | 5 2 6 2 3+4 2 5+6 2 3+4 2 5 2 | | |
| | 6 2 5 2 3+4 2 5+6 2 5+6 2 5 2 | | |
| | 5 2 5 2 5+6 2 5+6 2 5+6 2 5 2 | | |
| | 6 2 5 2 5+6 2 3+4 2 5+6 2 5 2 | | |
| | 5 2 5 2 5+6 2 3+4 2 3+4 2 5 2 | | |
| | 5 2 6 2 3+4 2 5+6 2 5+6 2 5 2 | | |
| | 6 2 5 2 3+4 2 3+4 2 3+4 2 5 2 | | |
| | 5 2 6 2 5+6 2 3+4 2 3+4 2 5 2 | | |
| | 5 2 6 2 3+4 2 5+6 2 3+4 2 5 2 | | |
| | 6 2 6 2 3+4 2 5+6 2 5+6 2 5 2 | | |
| | 5 2 5 2 5+6 2 3+4 2 5+6 2 5 2 | | |
| | 6 2 6 2 5+6 2 5+6 2 3+4 2 5 2 | | |
| | 6 2 6 2 3+4 2 5+6 2 3+4 2 5 2 | | |
| | 5 2 6 2 5+6 2 3+4 2 5+6 2 5 2 | | |
| | 6 2 5 2 5+6 2 3+4 2 5+6 2 5 2 | | |
| | 5 2 5 2 3+4 2 5+6 2 3+4 2 5 2 | | |
| | 5 2 6 2 5+6 2 3+4 2 3+4 2 5 2 | | |
| | 6 2 6 2 3+4 2 5+6 2 5+6 2 5 2 | | |
| | 5 2 5 2 5+6 2 3+4 2 3+4 2 5 2 | | |
| | 6 2 6 2 3+4 2 5+6 2 5+6 2 5 2 | | |
| | 6 2 5 2 5+6 2 3+4 2 3+4 2 5 2 | | |
| | 5 2 5 2 3+4 2 3+4 2 5+6 2 5 2 | | |
| | 6 2 6 2 5+6 2 5+6 2 5+6 2 5 2 | | |

Continued on next column

**TABLE XLIV– continued**

| TQC | ICSD | NoE | SF |
|---|---|---|---|
| | 5 2 6 2 3+4 2 5+6 2 3+4 2 5 2 | | |
| | 5 2 6 2 5+6 2 5+6 2 5+6 2 5 2 | | |
| 71.1.37 | 83190 | 102 | Cu8Ge8Nd6 |
| | -GM -X -T -R -S -W | | |
| | 5 2 5 2 3+4 2 3+4 2 3+4 2 5 2 | | |
| | 5 2 5 2 3+4 2 5+6 2 5+6 2 5 2 | | |
| | 6 2 6 2 5+6 2 3+4 2 3+4 2 5 2 | | |
| | 6 2 6 2 5+6 2 5+6 2 5+6 2 5 2 | | |
| | 5 2 5 2 3+4 2 5+6 2 5+6 2 5 2 | | |
| | 6 2 6 2 5+6 2 3+4 2 3+4 2 5 2 | | |
| | 6 2 6 2 5+6 2 5+6 2 5+6 2 5 2 | | |
| | 5 2 5 2 3+4 2 5+6 2 5+6 2 5 2 | | |
| | 6 2 6 2 5+6 2 3+4 2 3+4 2 5 2 | | |
| | 5 2 6 2 5+6 2 5+6 2 5+6 2 5 2 | | |
| | 6 2 5 2 3+4 2 3+4 2 5+6 2 5 2 | | |
| | 6 2 6 2 5+6 2 5+6 2 3+4 2 5 2 | | |
| | 5 2 5 2 5+6 2 5+6 2 3+4 2 5 2 | | |
| | 5 2 6 2 3+4 2 3+4 2 3+4 2 5 2 | | |
| | 6 2 5 2 5+6 2 5+6 2 5+6 2 5 2 | | |
| | 6 2 6 2 3+4 2 3+4 2 5+6 2 5 2 | | |
| | 6 2 6 2 3+4 2 3+4 2 3+4 2 5 2 | | |
| | 5 2 5 2 5+6 2 5+6 2 5+6 2 5 2 | | |
| | 5 2 6 2 5+6 2 3+4 2 3+4 2 5 2 | | |
| | 6 2 5 2 3+4 2 5+6 2 5+6 2 5 2 | | |
| | 5 2 5 2 3+4 2 3+4 2 3+4 2 5 2 | | |
| | 6 2 6 2 5+6 2 5+6 2 5+6 2 5 2 | | |
| | 5 2 5 2 3+4 2 5+6 2 5+6 2 5 2 | | |
| | 6 2 5 2 3+4 2 3+4 2 3+4 2 5 2 | | |
| | 5 2 6 2 5+6 2 3+4 2 5+6 2 5 2 | | |
| | 5 2 5 2 3+4 2 3+4 2 3+4 2 5 2 | | |
| | 6 2 6 2 5+6 2 5+6 2 3+4 2 5 2 | | |
| | 5 2 6 2 5+6 2 5+6 2 5+6 2 5 2 | | |
| | 5 2 5 2 3+4 2 3+4 2 3+4 2 5 2 | | |
| | 6 2 6 2 5+6 2 5+6 2 5+6 2 5 2 | | |
| | 5 2 5 2 5+6 2 3+4 2 3+4 2 5 2 | | |
| | 6 2 6 2 3+4 2 5+6 2 5+6 2 5 2 | | |
| | 5 2 6 2 5+6 2 3+4 2 3+4 2 5 2 | | |
| | 6 2 5 2 3+4 2 5+6 2 5+6 2 5 2 | | |
| | 5 2 5 2 5+6 2 3+4 2 3+4 2 5 2 | | |
| | 6 2 6 2 3+4 2 5+6 2 5+6 2 5 2 | | |
| | 6 2 6 2 5+6 2 5+6 2 3+4 2 5 2 | | |
| | 6 2 5 2 3+4 2 3+4 2 5+6 2 5 2 | | |
| | 5 2 6 2 5+6 2 3+4 2 3+4 2 5 2 | | |
| | 6 2 6 2 3+4 2 5+6 2 3+4 2 5 2 | | |
| | 5 2 5 2 3+4 2 5+6 2 5+6 2 5 2 | | |
| | 5 2 5 2 3+4 2 3+4 2 3+4 2 5 2 | | |
| | 5 2 6 2 5+6 2 5+6 2 5+6 2 5 2 | | |
| | 5 2 6 2 3+4 2 3+4 2 5+6 2 5 2 | | |
| | 6 2 5 2 5+6 2 3+4 2 3+4 2 5 2 | | |
| | 6 2 5 2 5+6 2 5+6 2 3+4 2 5 2 | | |
| | 5 2 5 2 5+6 2 3+4 2 5+6 2 5 2 | | |
| | 6 2 6 2 3+4 2 5+6 2 5+6 2 5 2 | | |
| | 6 2 6 2 5+6 2 5+6 2 3+4 2 5 2 | | |
| | 6 2 6 2 5+6 2 3+4 2 5+6 2 5 2 | | |
| | 5 2 6 2 3+4 2 5+6 2 5+6 2 5 2 | | |
| 71.1.38 | 102956 | 114 | Cu8Gd6Sn8 |
| | -GM -X -T -R -S -W | | |
| | 5 2 5 2 3+4 2 5+6 2 5+6 2 5 2 | | |
| | 6 2 6 2 5+6 2 3+4 2 3+4 2 5 2 | | |
| | 5 2 5 2 3+4 2 3+4 2 3+4 2 5 2 | | |
| | 6 2 6 2 5+6 2 5+6 2 5+6 2 5 2 | | |

Continued on next column

**TABLE XLIV– continued**

| TQC | ICSD NoE SF |
|---|---|
| | 5 2 5 2 3+4 2 5+6 2 5+6 2 5 2 |
| | 6 2 6 2 5+6 2 3+4 2 3+4 2 5 2 |
| | 6 2 6 2 5+6 2 5+6 2 5+6 2 5 2 |
| | 5 2 5 2 3+4 2 5+6 2 5+6 2 5 2 |
| | 6 2 6 2 5+6 2 3+4 2 3+4 2 5 2 |
| | 5 2 6 2 5+6 2 5+6 2 5+6 2 5 2 |
| | 6 2 5 2 3+4 2 3+4 2 5+6 2 5 2 |
| | 6 2 6 2 5+6 2 5+6 2 3+4 2 5 2 |
| | 5 2 6 2 5+6 2 5+6 2 3+4 2 5 2 |
| | 5 2 5 2 3+4 2 3+4 2 3+4 2 5 2 |
| | 6 2 5 2 5+6 2 5+6 2 5+6 2 5 2 |
| | 6 2 6 2 3+4 2 3+4 2 5+6 2 5 2 |
| | 5 2 6 2 3+4 2 3+4 2 3+4 2 5 2 |
| | 6 2 5 2 5+6 2 5+6 2 5+6 2 5 2 |
| | 5 2 5 2 3+4 2 3+4 2 3+4 2 5 2 |
| | 6 2 5 2 5+6 2 5+6 2 5+6 2 5 2 |
| | 5 2 6 2 3+4 2 3+4 2 5+6 2 5 2 |
| | 6 2 6 2 5+6 2 5+6 2 3+4 2 5 2 |
| | 6 2 5 2 3+4 2 5+6 2 5+6 2 5 2 |
| | 5 2 5 2 5+6 2 3+4 2 3+4 2 5 2 |
| | 5 2 6 2 3+4 2 3+4 2 5+6 2 5 2 |
| | 6 2 6 2 3+4 2 3+4 2 5+6 2 5 2 |
| | 5 2 5 2 5+6 2 5+6 2 3+4 2 5 2 |
| | 5 2 6 2 5+6 2 5+6 2 3+4 2 5 2 |
| | 6 2 5 2 5+6 2 3+4 2 5+6 2 5 2 |
| | 5 2 6 2 3+4 2 5+6 2 3+4 2 5 2 |
| | 6 2 5 2 5+6 2 3+4 2 3+4 2 5 2 |
| | 5 2 6 2 3+4 2 5+6 2 5+6 2 5 2 |
| | 5 2 6 2 5+6 2 5+6 2 3+4 2 5 2 |
| | 6 2 5 2 5+6 2 3+4 2 5+6 2 5 2 |
| | 5 2 5 2 3+4 2 3+4 2 5+6 2 5 2 |
| | 6 2 6 2 3+4 2 5+6 2 3+4 2 5 2 |
| | 6 2 6 2 5+6 2 5+6 2 5+6 2 5 2 |
| | 6 2 6 2 3+4 2 3+4 2 3+4 2 5 2 |
| | 5 2 5 2 5+6 2 3+4 2 3+4 2 5 2 |
| | 5 2 5 2 3+4 2 5+6 2 5+6 2 5 2 |
| | 6 2 6 2 3+4 2 3+4 2 3+4 2 5 2 |
| | 5 2 5 2 3+4 2 5+6 2 5+6 2 5 2 |
| | 5 2 6 2 5+6 2 5+6 2 3+4 2 5 2 |
| | 5 2 6 2 5+6 2 3+4 2 5+6 2 5 2 |
| | 6 2 5 2 3+4 2 3+4 2 3+4 2 5 2 |
| | 6 2 5 2 5+6 2 5+6 2 3+4 2 5 2 |
| | 6 2 6 2 5+6 2 5+6 2 5+6 2 5 2 |
| | 6 2 6 2 3+4 2 5+6 2 5+6 2 5 2 |
| | 6 2 6 2 5+6 2 5+6 2 3+4 2 5 2 |
| | 5 2 6 2 5+6 2 3+4 2 5+6 2 5 2 |
| | 6 2 5 2 3+4 2 5+6 2 5+6 2 5 2 |
| | 5 2 6 2 5+6 2 3+4 2 5+6 2 5 2 |
| | 5 2 6 2 5+6 2 5+6 2 3+4 2 5 2 |
| | 6 2 5 2 3+4 2 3+4 2 5+6 2 5 2 |
| | 6 2 5 2 5+6 2 3+4 2 3+4 2 5 2 |
| | 5 2 6 2 5+6 2 5+6 2 5+6 2 5 2 |
| | 6 2 6 2 3+4 2 5+6 2 5+6 2 5 2 |
| 71.1.39 | 413485 158 Cu6Sn5Yb3 |
| | -GM -X -T -R -S -W |
| | 5 2 5 2 5+6 2 5+6 2 3+4 2 5 2 |
| | 5 2 5 2 3+4 2 3+4 2 3+4 2 5 2 |
| | 6 2 6 2 5+6 2 5+6 2 5+6 2 5 2 |
| | 6 2 6 2 3+4 2 3+4 2 5+6 2 5 2 |
| | 5 2 5 2 3+4 2 3+4 2 3+4 2 5 2 |
| | 6 2 6 2 5+6 2 5+6 2 5+6 2 5 2 |

Continued on next column

**TABLE XLIV– continued**

| TQC | ICSD NoE SF |
|---|---|
| | 6 2 6 2 3+4 2 3+4 2 5+6 2 5 2 |
| | 6 2 6 2 3+4 2 3+4 2 5+6 2 5 2 |
| | 5 2 5 2 3+4 2 3+4 2 3+4 2 5 2 |
| | 6 2 6 2 5+6 2 5+6 2 5+6 2 5 2 |
| | 5 2 5 2 3+4 2 3+4 2 3+4 2 5 2 |
| | 6 2 6 2 5+6 2 5+6 2 5+6 2 5 2 |
| | 5 2 6 2 3+4 2 3+4 2 5+6 2 5 2 |
| | 5 2 5 2 5+6 2 5+6 2 3+4 2 5 2 |
| | 5 2 6 2 5+6 2 3+4 2 5+6 2 5 2 |
| | 6 2 5 2 5+6 2 5+6 2 3+4 2 5 2 |
| | 6 2 5 2 3+4 2 3+4 2 5+6 2 5 2 |
| | 6 2 5 2 3+4 2 5+6 2 3+4 2 5 2 |
| | 6 2 6 2 3+4 2 3+4 2 3+4 2 5 2 |
| | 5 2 6 2 5+6 2 5+6 2 3+4 2 5 2 |
| | 5 2 6 2 3+4 2 5+6 2 5+6 2 5 2 |
| | 6 2 5 2 5+6 2 5+6 2 5+6 2 5 2 |
| | 6 2 6 2 5+6 2 3+4 2 5+6 2 5 2 |
| | 5 2 6 2 3+4 2 5+6 2 3+4 2 5 2 |
| | 5 2 5 2 3+4 2 3+4 2 5+6 2 5 2 |
| | 6 2 6 2 5+6 2 3+4 2 3+4 2 5 2 |
| | 5 2 5 2 3+4 2 5+6 2 5+6 2 5 2 |
| | 5 2 6 2 5+6 2 5+6 2 3+4 2 5 2 |
| | 6 2 5 2 3+4 2 3+4 2 3+4 2 5 2 |
| | 5 2 5 2 5+6 2 3+4 2 3+4 2 5 2 |
| | 6 2 6 2 5+6 2 5+6 2 3+4 2 5 2 |
| | 6 2 6 2 3+4 2 3+4 2 5+6 2 5 2 |
| | 5 2 5 2 3+4 2 5+6 2 5+6 2 5 2 |
| | 6 2 6 2 5+6 2 3+4 2 3+4 2 5 2 |
| | 6 2 5 2 3+4 2 3+4 2 5+6 2 5 2 |
| | 5 2 6 2 5+6 2 5+6 2 3+4 2 5 2 |
| | 5 2 5 2 5+6 2 5+6 2 5+6 2 5 2 |
| | 5 2 6 2 3+4 2 3+4 2 5+6 2 5 2 |
| | 5 2 5 2 3+4 2 3+4 2 3+4 2 5 2 |
| | 5 2 5 2 5+6 2 5+6 2 5+6 2 5 2 |
| | 6 2 6 2 5+6 2 3+4 2 5+6 2 5 2 |
| | 5 2 6 2 3+4 2 5+6 2 3+4 2 5 2 |
| | 6 2 6 2 5+6 2 5+6 2 5+6 2 5 2 |
| | 6 2 6 2 3+4 2 3+4 2 3+4 2 5 2 |
| | 6 2 5 2 3+4 2 5+6 2 3+4 2 5 2 |
| | 5 2 5 2 5+6 2 3+4 2 5+6 2 5 2 |
| | 6 2 5 2 3+4 2 5+6 2 3+4 2 5 2 |
| | 6 2 6 2 3+4 2 3+4 2 5+6 2 5 2 |
| | 5 2 5 2 5+6 2 3+4 2 5+6 2 5 2 |
| | 6 2 5 2 5+6 2 5+6 2 3+4 2 5 2 |
| | 5 2 5 2 3+4 2 3+4 2 5+6 2 5 2 |
| | 6 2 6 2 5+6 2 5+6 2 5+6 2 5 2 |
| | 6 2 6 2 5+6 2 5+6 2 3+4 2 5 2 |
| | 5 2 5 2 3+4 2 3+4 2 3+4 2 5 2 |
| | 6 2 6 2 3+4 2 3+4 2 3+4 2 5 2 |
| | 5 2 6 2 5+6 2 5+6 2 3+4 2 5 2 |
| | 5 2 5 2 3+4 2 3+4 2 5+6 2 5 2 |
| | 6 2 6 2 5+6 2 3+4 2 3+4 2 5 2 |
| | 6 2 6 2 3+4 2 3+4 2 5+6 2 5 2 |
| | 6 2 6 2 5+6 2 5+6 2 5+6 2 5 2 |
| | 5 2 5 2 3+4 2 3+4 2 5+6 2 5 2 |
| | 6 2 6 2 3+4 2 5+6 2 3+4 2 5 2 |
| | 5 2 6 2 3+4 2 3+4 2 5+6 2 5 2 |
| | 6 2 5 2 3+4 2 5+6 2 5+6 2 5 2 |
| | 5 2 6 2 5+6 2 3+4 2 3+4 2 5 2 |
| | 6 2 5 2 3+4 2 5+6 2 3+4 2 5 2 |
| | 6 2 6 2 5+6 2 5+6 2 5+6 2 5 2 |

Continued on next column

**TABLE XLIV– continued**

| TQC | ICSD NoE SF |
|---|---|
| | 5 2 5 2 3+4 2 3+4 2 5+6 2 5 2 |
| | 6 2 6 2 3+4 2 3+4 2 5+6 2 5 2 |
| | 5 2 6 2 5+6 2 5+6 2 3+4 2 5 2 |
| | 6 2 6 2 3+4 2 3+4 2 5+6 2 5 2 |
| | 6 2 5 2 5+6 2 3+4 2 3+4 2 5 2 |
| | 6 2 5 2 3+4 2 3+4 2 3+4 2 5 2 |
| | 6 2 6 2 3+4 2 3+4 2 5+6 2 5 2 |
| | 5 2 6 2 3+4 2 5+6 2 5+6 2 5 2 |
| | 5 2 6 2 5+6 2 3+4 2 5+6 2 5 2 |
| | 5 2 6 2 5+6 2 5+6 2 3+4 2 5 2 |
| | 6 2 5 2 3+4 2 3+4 2 5+6 2 5 2 |
| | 5 2 6 2 3+4 2 5+6 2 5+6 2 5 2 |
| 71.1.40 | 426132 132 Cu4Ge4Yb3 |
| | -GM -X -T -R -S -W |
| | 5 2 5 2 5+6 2 3+4 2 5+6 2 5 2 |
| | 5 2 5 2 5+6 2 5+6 2 3+4 2 5 2 |
| | 6 2 6 2 3+4 2 3+4 2 5+6 2 5 2 |
| | 6 2 6 2 3+4 2 5+6 2 3+4 2 5 2 |
| | 5 2 5 2 5+6 2 5+6 2 3+4 2 5 2 |
| | 6 2 6 2 3+4 2 3+4 2 5+6 2 5 2 |
| | 6 2 6 2 3+4 2 5+6 2 3+4 2 5 2 |
| | 5 2 5 2 3+4 2 5+6 2 3+4 2 5 2 |
| | 6 2 6 2 5+6 2 5+6 2 5+6 2 5 2 |
| | 5 2 5 2 3+4 2 3+4 2 3+4 2 5 2 |
| | 6 2 6 2 5+6 2 3+4 2 3+4 2 5 2 |
| | 6 2 6 2 3+4 2 5+6 2 5+6 2 5 2 |
| | 5 2 6 2 3+4 2 5+6 2 5+6 2 5 2 |
| | 5 2 5 2 5+6 2 3+4 2 5+6 2 5 2 |
| | 6 2 5 2 3+4 2 5+6 2 3+4 2 5 2 |
| | 6 2 6 2 5+6 2 3+4 2 3+4 2 5 2 |
| | 6 2 6 2 5+6 2 3+4 2 5+6 2 5 2 |
| | 5 2 5 2 3+4 2 5+6 2 3+4 2 5 2 |
| | 5 2 5 2 5+6 2 3+4 2 5+6 2 5 2 |
| | 6 2 6 2 3+4 2 5+6 2 3+4 2 5 2 |
| | 5 2 5 2 5+6 2 3+4 2 3+4 2 5 2 |
| | 6 2 6 2 3+4 2 5+6 2 3+4 2 5 2 |
| | 6 2 5 2 5+6 2 5+6 2 5+6 2 5 2 |
| | 5 2 5 2 3+4 2 3+4 2 5+6 2 5 2 |
| | 6 2 6 2 5+6 2 3+4 2 3+4 2 5 2 |
| | 5 2 5 2 5+6 2 5+6 2 3+4 2 5 2 |
| | 5 2 6 2 3+4 2 3+4 2 5+6 2 5 2 |
| | 5 2 6 2 3+4 2 5+6 2 5+6 2 5 2 |
| | 5 2 5 2 3+4 2 3+4 2 5+6 2 5 2 |
| | 6 2 6 2 3+4 2 5+6 2 3+4 2 5 2 |
| | 6 2 5 2 5+6 2 3+4 2 5+6 2 5 2 |
| | 5 2 6 2 5+6 2 5+6 2 3+4 2 5 2 |
| | 5 2 6 2 3+4 2 5+6 2 3+4 2 5 2 |
| | 6 2 5 2 3+4 2 3+4 2 5+6 2 5 2 |
| | 5 2 5 2 5+6 2 5+6 2 3+4 2 5 2 |
| | 6 2 6 2 5+6 2 3+4 2 5+6 2 5 2 |
| | 6 2 6 2 3+4 2 5+6 2 3+4 2 5 2 |
| | 6 2 5 2 5+6 2 3+4 2 5+6 2 5 2 |
| | 5 2 6 2 3+4 2 5+6 2 5+6 2 5 2 |
| | 6 2 6 2 5+6 2 3+4 2 3+4 2 5 2 |
| | 5 2 5 2 5+6 2 5+6 2 5+6 2 5 2 |
| | 5 2 6 2 5+6 2 3+4 2 3+4 2 5 2 |
| | 6 2 5 2 3+4 2 3+4 2 5+6 2 5 2 |
| | 6 2 6 2 3+4 2 5+6 2 3+4 2 5 2 |
| | 6 2 6 2 3+4 2 5+6 2 3+4 2 5 2 |
| | 5 2 6 2 3+4 2 5+6 2 5+6 2 5 2 |
| | 6 2 5 2 5+6 2 5+6 2 3+4 2 5 2 |

Continued on next column

**TABLE XLIV– continued**

| TQC | ICSD NoE SF |
|---|---|
| | 6 2 6 2 3+4 2 5+6 2 3+4 2 5 2 |
| | 5 2 6 2 5+6 2 3+4 2 3+4 2 5 2 |
| | 6 2 5 2 3+4 2 3+4 2 5+6 2 5 2 |
| | 5 2 5 2 5+6 2 5+6 2 3+4 2 5 2 |
| | 5 2 6 2 3+4 2 5+6 2 5+6 2 5 2 |
| | 5 2 6 2 3+4 2 3+4 2 3+4 2 5 2 |
| | 6 2 5 2 3+4 2 3+4 2 5+6 2 5 2 |
| | 6 2 6 2 5+6 2 5+6 2 5+6 2 5 2 |
| | 6 2 5 2 3+4 2 5+6 2 3+4 2 5 2 |
| | 6 2 6 2 3+4 2 5+6 2 3+4 2 5 2 |
| | 6 2 6 2 5+6 2 3+4 2 5+6 2 5 2 |
| | 6 2 5 2 3+4 2 5+6 2 3+4 2 5 2 |
| | 5 2 6 2 5+6 2 5+6 2 3+4 2 5 2 |
| | 6 2 6 2 3+4 2 3+4 2 5+6 2 5 2 |
| | 5 2 5 2 3+4 2 5+6 2 3+4 2 5 2 |
| | 5 2 6 2 5+6 2 5+6 2 3+4 2 5 2 |
| | 6 2 6 2 3+4 2 3+4 2 5+6 2 5 2 |
| | 5 2 6 2 5+6 2 5+6 2 3+4 2 5 2 |
| | 5 2 5 2 3+4 2 3+4 2 3+4 2 5 2 |
| 71.1.41 | 82565 102 Cu4Ge4U3 |
| | -GM -X -T -R -S -W |
| | 5 2 5 2 3+4 2 3+4 2 3+4 2 5 2 |
| | 6 2 6 2 5+6 2 5+6 2 5+6 2 5 2 |
| | 5 2 5 2 3+4 2 5+6 2 5+6 2 5 2 |
| | 5 2 5 2 3+4 2 3+4 2 3+4 2 5 2 |
| | 6 2 6 2 5+6 2 5+6 2 5+6 2 5 2 |
| | 6 2 6 2 5+6 2 3+4 2 3+4 2 5 2 |
| | 6 2 6 2 5+6 2 3+4 2 3+4 2 5 2 |
| | 5 2 5 2 3+4 2 3+4 2 5+6 2 5 2 |
| | 5 2 6 2 5+6 2 5+6 2 3+4 2 5 2 |
| | 6 2 5 2 3+4 2 5+6 2 3+4 2 5 2 |
| | 6 2 6 2 5+6 2 3+4 2 5+6 2 5 2 |
| | 6 2 6 2 5+6 2 3+4 2 3+4 2 5 2 |
| | 5 2 5 2 5+6 2 3+4 2 5+6 2 5 2 |
| | 5 2 6 2 3+4 2 5+6 2 5+6 2 5 2 |
| | 6 2 5 2 5+6 2 3+4 2 3+4 2 5 2 |
| | 6 2 6 2 3+4 2 5+6 2 3+4 2 5 2 |
| | 6 2 6 2 3+4 2 5+6 2 5+6 2 5 2 |
| | 5 2 5 2 5+6 2 3+4 2 3+4 2 5 2 |
| | 5 2 5 2 3+4 2 5+6 2 5+6 2 5 2 |
| | 6 2 6 2 5+6 2 5+6 2 3+4 2 5 2 |
| | 5 2 6 2 3+4 2 3+4 2 5+6 2 5 2 |
| | 6 2 5 2 3+4 2 3+4 2 3+4 2 5 2 |
| | 5 2 5 2 5+6 2 3+4 2 3+4 2 5 2 |
| | 6 2 6 2 3+4 2 5+6 2 5+6 2 5 2 |
| | 5 2 5 2 5+6 2 5+6 2 3+4 2 5 2 |
| | 5 2 5 2 5+6 2 5+6 2 5+6 2 5 2 |
| | 6 2 6 2 3+4 2 3+4 2 5+6 2 5 2 |
| | 5 2 5 2 5+6 2 3+4 2 3+4 2 5 2 |
| | 5 2 6 2 3+4 2 5+6 2 5+6 2 5 2 |
| | 6 2 6 2 5+6 2 5+6 2 3+4 2 5 2 |
| | 5 2 5 2 5+6 2 3+4 2 5+6 2 5 2 |
| | 6 2 6 2 3+4 2 3+4 2 3+4 2 5 2 |
| | 5 2 6 2 5+6 2 5+6 2 5+6 2 5 2 |
| | 6 2 5 2 3+4 2 3+4 2 5+6 2 5 2 |
| | 5 2 5 2 5+6 2 5+6 2 3+4 2 5 2 |
| | 6 2 6 2 3+4 2 3+4 2 3+4 2 5 2 |
| | 6 2 6 2 5+6 2 3+4 2 5+6 2 5 2 |
| | 6 2 5 2 3+4 2 5+6 2 3+4 2 5 2 |
| | 5 2 6 2 5+6 2 5+6 2 5+6 2 5 2 |
| | 6 2 5 2 3+4 2 3+4 2 5+6 2 5 2 |

Continued on next column

**TABLE XLIV– continued**

| TQC | ICSD NoE SF |
|---|---|
| | 5 2 6 2 3+4 2 3+4 2 5+6 2 5 2 |
| | 5 2 5 2 5+6 2 5+6 2 3+4 2 5 2 |
| | 5 2 6 2 3+4 2 5+6 2 3+4 2 5 2 |
| | 5 2 5 2 5+6 2 3+4 2 5+6 2 5 2 |
| | 5 2 6 2 3+4 2 5+6 2 3+4 2 5 2 |
| | 6 2 5 2 5+6 2 5+6 2 5+6 2 5 2 |
| | 6 2 5 2 5+6 2 3+4 2 3+4 2 5 2 |
| | 6 2 6 2 3+4 2 3+4 2 5+6 2 5 2 |
| | 5 2 6 2 5+6 2 3+4 2 3+4 2 5 2 |
| | 6 2 5 2 5+6 2 5+6 2 3+4 2 5 2 |
| | 6 2 6 2 5+6 2 3+4 2 3+4 2 5 2 |
| 71.1.42 | 81142 102 Cu4Ge4U3 |
| | -GM -X -T -R -S -W |
| | 5 2 5 2 3+4 2 5+6 2 5+6 2 5 2 |
| | 6 2 6 2 5+6 2 3+4 2 3+4 2 5 2 |
| | 5 2 5 2 3+4 2 3+4 2 3+4 2 5 2 |
| | 5 2 5 2 3+4 2 5+6 2 5+6 2 5 2 |
| | 6 2 6 2 5+6 2 3+4 2 3+4 2 5 2 |
| | 6 2 6 2 5+6 2 5+6 2 5+6 2 5 2 |
| | 6 2 6 2 5+6 2 5+6 2 5+6 2 5 2 |
| | 5 2 5 2 3+4 2 5+6 2 3+4 2 5 2 |
| | 5 2 6 2 5+6 2 3+4 2 5+6 2 5 2 |
| | 6 2 5 2 3+4 2 3+4 2 5+6 2 5 2 |
| | 6 2 6 2 5+6 2 5+6 2 3+4 2 5 2 |
| | 6 2 6 2 5+6 2 5+6 2 5+6 2 5 2 |
| | 5 2 5 2 5+6 2 5+6 2 3+4 2 5 2 |
| | 5 2 6 2 3+4 2 3+4 2 3+4 2 5 2 |
| | 6 2 5 2 5+6 2 5+6 2 5+6 2 5 2 |
| | 6 2 6 2 3+4 2 3+4 2 5+6 2 5 2 |
| | 6 2 6 2 3+4 2 3+4 2 3+4 2 5 2 |
| | 5 2 5 2 5+6 2 5+6 2 5+6 2 5 2 |
| | 5 2 5 2 3+4 2 3+4 2 3+4 2 5 2 |
| | 6 2 6 2 5+6 2 3+4 2 5+6 2 5 2 |
| | 5 2 6 2 3+4 2 5+6 2 3+4 2 5 2 |
| | 6 2 5 2 3+4 2 5+6 2 5+6 2 5 2 |
| | 5 2 5 2 5+6 2 5+6 2 5+6 2 5 2 |
| | 6 2 6 2 3+4 2 3+4 2 3+4 2 5 2 |
| | 5 2 5 2 5+6 2 3+4 2 5+6 2 5 2 |
| | 5 2 5 2 5+6 2 3+4 2 3+4 2 5 2 |
| | 6 2 6 2 3+4 2 5+6 2 3+4 2 5 2 |
| | 5 2 5 2 5+6 2 5+6 2 5+6 2 5 2 |
| | 5 2 6 2 3+4 2 3+4 2 3+4 2 5 2 |
| | 6 2 6 2 5+6 2 3+4 2 5+6 2 5 2 |
| | 5 2 5 2 5+6 2 5+6 2 3+4 2 5 2 |
| | 6 2 6 2 3+4 2 5+6 2 5+6 2 5 2 |
| | 5 2 6 2 5+6 2 3+4 2 3+4 2 5 2 |
| | 6 2 5 2 3+4 2 5+6 2 3+4 2 5 2 |
| | 5 2 5 2 5+6 2 3+4 2 5+6 2 5 2 |
| | 6 2 6 2 3+4 2 5+6 2 5+6 2 5 2 |
| | 6 2 6 2 5+6 2 5+6 2 3+4 2 5 2 |
| | 6 2 5 2 3+4 2 3+4 2 5+6 2 5 2 |
| | 5 2 6 2 5+6 2 3+4 2 3+4 2 5 2 |
| | 6 2 5 2 3+4 2 5+6 2 3+4 2 5 2 |
| | 5 2 6 2 3+4 2 5+6 2 3+4 2 5 2 |
| | 5 2 5 2 5+6 2 3+4 2 5+6 2 5 2 |
| | 5 2 6 2 3+4 2 3+4 2 5+6 2 5 2 |
| | 5 2 5 2 5+6 2 5+6 2 3+4 2 5 2 |
| | 5 2 6 2 3+4 2 3+4 2 5+6 2 5 2 |
| | 6 2 5 2 5+6 2 3+4 2 3+4 2 5 2 |
| | 6 2 5 2 5+6 2 5+6 2 5+6 2 5 2 |
| | 6 2 6 2 3+4 2 5+6 2 3+4 2 5 2 |

Continued on next column

**TABLE XLIV– continued**

| TQC | ICSD NoE SF |
|---|---|
| | 5 2 6 2 5+6 2 5+6 2 5+6 2 5 2 |
| | 6 2 5 2 5+6 2 3+4 2 5+6 2 5 2 |
| | 6 2 6 2 5+6 2 5+6 2 5+6 2 5 2 |
| 71.1.43 | 42717 114 Cu4Gd3Ge4 |
| | -GM -X -T -R -S -W |
| | 5 2 5 2 3+4 2 5+6 2 5+6 2 5 2 |
| | 5 2 5 2 3+4 2 3+4 2 3+4 2 5 2 |
| | 6 2 6 2 5+6 2 5+6 2 5+6 2 5 2 |
| | 6 2 6 2 5+6 2 3+4 2 3+4 2 5 2 |
| | 5 2 5 2 3+4 2 3+4 2 3+4 2 5 2 |
| | 6 2 6 2 5+6 2 5+6 2 5+6 2 5 2 |
| | 6 2 6 2 5+6 2 3+4 2 3+4 2 5 2 |
| | 5 2 5 2 5+6 2 3+4 2 3+4 2 5 2 |
| | 6 2 6 2 3+4 2 5+6 2 5+6 2 5 2 |
| | 5 2 5 2 5+6 2 3+4 2 3+4 2 5 2 |
| | 6 2 6 2 3+4 2 5+6 2 3+4 2 5 2 |
| | 6 2 6 2 5+6 2 3+4 2 5+6 2 5 2 |
| | 5 2 5 2 5+6 2 3+4 2 5+6 2 5 2 |
| | 5 2 6 2 3+4 2 5+6 2 5+6 2 5 2 |
| | 6 2 5 2 5+6 2 3+4 2 3+4 2 5 2 |
| | 6 2 6 2 3+4 2 5+6 2 3+4 2 5 2 |
| | 6 2 6 2 3+4 2 5+6 2 5+6 2 5 2 |
| | 5 2 5 2 5+6 2 3+4 2 3+4 2 5 2 |
| | 5 2 5 2 3+4 2 5+6 2 5+6 2 5 2 |
| | 6 2 6 2 5+6 2 3+4 2 3+4 2 5 2 |
| | 5 2 5 2 3+4 2 5+6 2 3+4 2 5 2 |
| | 6 2 6 2 5+6 2 3+4 2 5+6 2 5 2 |
| | 6 2 5 2 3+4 2 3+4 2 3+4 2 5 2 |
| | 5 2 5 2 3+4 2 5+6 2 5+6 2 5 2 |
| | 5 2 6 2 5+6 2 5+6 2 3+4 2 5 2 |
| | 6 2 5 2 3+4 2 5+6 2 3+4 2 5 2 |
| | 5 2 6 2 5+6 2 3+4 2 5+6 2 5 2 |
| | 5 2 6 2 5+6 2 3+4 2 5+6 2 5 2 |
| | 5 2 5 2 5+6 2 5+6 2 5+6 2 5 2 |
| | 6 2 6 2 3+4 2 3+4 2 3+4 2 5 2 |
| | 6 2 5 2 5+6 2 5+6 2 5+6 2 5 2 |
| | 5 2 6 2 3+4 2 3+4 2 3+4 2 5 2 |
| | 5 2 6 2 5+6 2 5+6 2 5+6 2 5 2 |
| | 6 2 5 2 3+4 2 3+4 2 3+4 2 5 2 |
| | 5 2 5 2 5+6 2 5+6 2 5+6 2 5 2 |
| | 6 2 6 2 3+4 2 3+4 2 3+4 2 5 2 |
| | 6 2 6 2 5+6 2 3+4 2 3+4 2 5 2 |
| | 6 2 5 2 3+4 2 5+6 2 5+6 2 5 2 |
| | 5 2 6 2 5+6 2 3+4 2 5+6 2 5 2 |
| | 6 2 6 2 3+4 2 5+6 2 5+6 2 5 2 |
| | 5 2 5 2 3+4 2 5+6 2 3+4 2 5 2 |
| | 5 2 5 2 3+4 2 3+4 2 3+4 2 5 2 |
| | 5 2 6 2 5+6 2 3+4 2 5+6 2 5 2 |
| | 6 2 6 2 3+4 2 5+6 2 3+4 2 5 2 |
| | 5 2 5 2 5+6 2 5+6 2 5+6 2 5 2 |
| | 6 2 5 2 5+6 2 3+4 2 5+6 2 5 2 |
| | 6 2 6 2 5+6 2 3+4 2 3+4 2 5 2 |
| | 6 2 6 2 3+4 2 3+4 2 3+4 2 5 2 |
| | 6 2 6 2 5+6 2 3+4 2 5+6 2 5 2 |
| | 5 2 6 2 5+6 2 5+6 2 3+4 2 5 2 |
| | 6 2 5 2 3+4 2 3+4 2 3+4 2 5 2 |
| | 5 2 6 2 5+6 2 5+6 2 3+4 2 5 2 |
| | 5 2 6 2 3+4 2 3+4 2 5+6 2 5 2 |
| | 6 2 5 2 5+6 2 5+6 2 3+4 2 5 2 |
| | 6 2 5 2 5+6 2 3+4 2 5+6 2 5 2 |
| | 6 2 6 2 5+6 2 5+6 2 3+4 2 5 2 |

Continued on next column

**TABLE XLIV– continued**

| TQC | ICSD NoE SF |
|---|---|
| | 5 2 6 2 3+4 2 3+4 2 3+4 2 5 2 |
| 71.1.44 | 55650 126 Cu4Er3Si4 |
| | -GM -X -T -R -S -W |
| | 5 2 5 2 3+4 2 5+6 2 5+6 2 5 2 |
| | 5 2 5 2 3+4 2 3+4 2 3+4 2 5 2 |
| | 6 2 6 2 5+6 2 5+6 2 5+6 2 5 2 |
| | 6 2 6 2 5+6 2 3+4 2 3+4 2 5 2 |
| | 5 2 5 2 3+4 2 3+4 2 3+4 2 5 2 |
| | 6 2 6 2 5+6 2 5+6 2 5+6 2 5 2 |
| | 6 2 6 2 5+6 2 3+4 2 3+4 2 5 2 |
| | 5 2 5 2 5+6 2 3+4 2 3+4 2 5 2 |
| | 6 2 6 2 3+4 2 3+4 2 5+6 2 5 2 |
| | 5 2 5 2 5+6 2 5+6 2 3+4 2 5 2 |
| | 6 2 6 2 3+4 2 5+6 2 3+4 2 5 2 |
| | 6 2 6 2 5+6 2 3+4 2 5+6 2 5 2 |
| | 5 2 6 2 5+6 2 3+4 2 5+6 2 5 2 |
| | 5 2 5 2 3+4 2 5+6 2 5+6 2 5 2 |
| | 6 2 5 2 5+6 2 3+4 2 3+4 2 5 2 |
| | 6 2 6 2 3+4 2 5+6 2 3+4 2 5 2 |
| | 6 2 6 2 3+4 2 5+6 2 5+6 2 5 2 |
| | 5 2 5 2 5+6 2 3+4 2 3+4 2 5 2 |
| | 5 2 5 2 3+4 2 5+6 2 5+6 2 5 2 |
| | 6 2 6 2 5+6 2 3+4 2 3+4 2 5 2 |
| | 5 2 5 2 3+4 2 5+6 2 3+4 2 5 2 |
| | 6 2 6 2 5+6 2 3+4 2 5+6 2 5 2 |
| | 6 2 5 2 3+4 2 3+4 2 3+4 2 5 2 |
| | 5 2 5 2 5+6 2 5+6 2 5+6 2 5 2 |
| | 5 2 6 2 3+4 2 5+6 2 3+4 2 5 2 |
| | 6 2 5 2 3+4 2 5+6 2 5+6 2 5 2 |
| | 5 2 6 2 5+6 2 3+4 2 3+4 2 5 2 |
| | 5 2 6 2 5+6 2 3+4 2 5+6 2 5 2 |
| | 5 2 5 2 5+6 2 5+6 2 5+6 2 5 2 |
| | 6 2 6 2 5+6 2 3+4 2 3+4 2 5 2 |
| | 6 2 6 2 3+4 2 5+6 2 5+6 2 5 2 |
| | 5 2 5 2 3+4 2 3+4 2 3+4 2 5 2 |
| | 5 2 6 2 5+6 2 5+6 2 5+6 2 5 2 |
| | 6 2 5 2 3+4 2 3+4 2 3+4 2 5 2 |
| | 5 2 5 2 5+6 2 5+6 2 3+4 2 5 2 |
| | 6 2 6 2 3+4 2 3+4 2 5+6 2 5 2 |
| | 6 2 6 2 5+6 2 3+4 2 3+4 2 5 2 |
| | 6 2 5 2 3+4 2 5+6 2 5+6 2 5 2 |
| | 5 2 6 2 5+6 2 3+4 2 5+6 2 5 2 |
| | 6 2 6 2 3+4 2 5+6 2 5+6 2 5 2 |
| | 5 2 5 2 3+4 2 5+6 2 3+4 2 5 2 |
| | 5 2 6 2 3+4 2 3+4 2 3+4 2 5 2 |
| | 6 2 5 2 5+6 2 3+4 2 5+6 2 5 2 |
| | 6 2 6 2 5+6 2 5+6 2 3+4 2 5 2 |
| | 5 2 5 2 5+6 2 3+4 2 5+6 2 5 2 |
| | 6 2 6 2 3+4 2 3+4 2 3+4 2 5 2 |
| | 6 2 5 2 5+6 2 3+4 2 3+4 2 5 2 |
| | 6 2 6 2 3+4 2 5+6 2 5+6 2 5 2 |
| | 5 2 6 2 5+6 2 3+4 2 3+4 2 5 2 |
| | 6 2 6 2 5+6 2 5+6 2 3+4 2 5 2 |
| | 5 2 5 2 3+4 2 3+4 2 5+6 2 5 2 |
| | 5 2 6 2 5+6 2 3+4 2 3+4 2 5 2 |
| | 5 2 6 2 3+4 2 5+6 2 3+4 2 5 2 |
| | 6 2 5 2 5+6 2 5+6 2 5+6 2 5 2 |
| | 6 2 5 2 5+6 2 3+4 2 5+6 2 5 2 |
| | 6 2 6 2 5+6 2 3+4 2 3+4 2 5 2 |
| | 5 2 6 2 3+4 2 5+6 2 3+4 2 5 2 |
| | 6 2 5 2 5+6 2 3+4 2 5+6 2 5 2 |

Continued on next column

**TABLE XLIV– continued**

| TQC | ICSD NoE SF |
|---|---|
| | 6 2 6 2 3+4 2 3+4 2 3+4 2 5 2 |
| | 6 2 6 2 5+6 2 3+4 2 3+4 2 5 2 |
| | 6 2 5 2 5+6 2 5+6 2 3+4 2 5 2 |
| | 5 2 6 2 3+4 2 3+4 2 5+6 2 5 2 |
| | 5 2 6 2 5+6 2 3+4 2 3+4 2 5 2 |
| 71.1.45 | 201112 58 Cs1Fe1S2 |
| | -GM -X -T -R -S -W |
| | 5 2 5 2 3+4 2 3+4 2 3+4 2 5 2 |
| | 6 2 6 2 5+6 2 5+6 2 5+6 2 5 2 |
| | 5 2 5 2 5+6 2 5+6 2 5+6 2 5 2 |
| | 6 2 6 2 3+4 2 3+4 2 3+4 2 5 2 |
| | 5 2 5 2 5+6 2 3+4 2 5+6 2 5 2 |
| | 6 2 6 2 3+4 2 5+6 2 3+4 2 5 2 |
| | 5 2 6 2 3+4 2 3+4 2 3+4 2 5 2 |
| | 6 2 5 2 5+6 2 5+6 2 5+6 2 5 2 |
| | 5 2 6 2 3+4 2 3+4 2 3+4 2 5 2 |
| | 5 2 5 2 5+6 2 5+6 2 5+6 2 5 2 |
| | 6 2 6 2 3+4 2 3+4 2 3+4 2 5 2 |
| | 6 2 5 2 5+6 2 5+6 2 5+6 2 5 2 |
| | 5 2 5 2 5+6 2 5+6 2 5+6 2 5 2 |
| | 6 2 6 2 3+4 2 3+4 2 3+4 2 5 2 |
| | 6 2 6 2 3+4 2 3+4 2 5+6 2 5 2 |
| | 6 2 5 2 5+6 2 3+4 2 3+4 2 5 2 |
| | 5 2 6 2 5+6 2 5+6 2 5+6 2 5 2 |
| | 5 2 6 2 3+4 2 3+4 2 3+4 2 5 2 |
| | 6 2 5 2 5+6 2 5+6 2 5+6 2 5 2 |
| | 5 2 5 2 3+4 2 5+6 2 3+4 2 5 2 |
| | 5 2 5 2 5+6 2 5+6 2 5+6 2 5 2 |
| | 6 2 6 2 3+4 2 3+4 2 3+4 2 5 2 |
| | 6 2 6 2 5+6 2 3+4 2 3+4 2 5 2 |
| | 5 2 5 2 3+4 2 5+6 2 5+6 2 5 2 |
| | 5 2 5 2 5+6 2 5+6 2 5+6 2 5 2 |
| | 6 2 5 2 3+4 2 5+6 2 3+4 2 5 2 |
| | 5 2 6 2 5+6 2 3+4 2 5+6 2 5 2 |
| | 5 2 5 2 3+4 2 5+6 2 3+4 2 5 2 |
| | 5 2 5 2 3+4 2 5+6 2 3+4 2 5 2 |
| 71.1.46 | 247288 88 Ce3Rh4Si4 |
| | -GM -X -T -R -S -W |
| | 5 2 5 2 3+4 2 3+4 2 3+4 2 5 2 |
| | 6 2 6 2 5+6 2 5+6 2 5+6 2 5 2 |
| | 5 2 5 2 3+4 2 3+4 2 3+4 2 5 2 |
| | 6 2 6 2 5+6 2 5+6 2 3+4 2 5 2 |
| | 5 2 5 2 3+4 2 3+4 2 5+6 2 5 2 |
| | 6 2 6 2 5+6 2 5+6 2 5+6 2 5 2 |
| | 6 2 6 2 5+6 2 5+6 2 5+6 2 5 2 |
| | 5 2 5 2 3+4 2 3+4 2 5+6 2 5 2 |
| | 6 2 5 2 5+6 2 5+6 2 3+4 2 5 2 |
| | 5 2 6 2 5+6 2 3+4 2 3+4 2 5 2 |
| | 6 2 6 2 3+4 2 5+6 2 5+6 2 5 2 |
| | 6 2 6 2 5+6 2 5+6 2 5+6 2 5 2 |
| | 5 2 5 2 5+6 2 5+6 2 3+4 2 5 2 |
| | 5 2 6 2 3+4 2 3+4 2 3+4 2 5 2 |
| | 6 2 5 2 5+6 2 5+6 2 5+6 2 5 2 |
| | 6 2 6 2 3+4 2 3+4 2 5+6 2 5 2 |
| | 5 2 6 2 3+4 2 3+4 2 3+4 2 5 2 |
| | 6 2 5 2 5+6 2 5+6 2 5+6 2 5 2 |
| | 6 2 5 2 3+4 2 3+4 2 3+4 2 5 2 |
| | 6 2 5 2 3+4 2 5+6 2 5+6 2 5 2 |
| | 5 2 6 2 5+6 2 3+4 2 3+4 2 5 2 |
| | 5 2 5 2 5+6 2 5+6 2 3+4 2 5 2 |
| | 6 2 6 2 3+4 2 3+4 2 5+6 2 5 2 |

Continued on next column

**TABLE XLIV– continued**

| TQC | ICSD | NoE | SF |
|---|---|---|---|
| | 5 2 5 2 5+6 2 3+4 2 3+4 2 5 2 | | |
| | 5 2 5 2 3+4 2 5+6 2 5+6 2 5 2 | | |
| | 6 2 6 2 5+6 2 5+6 2 5+6 2 5 2 | | |
| | 5 2 6 2 3+4 2 3+4 2 3+4 2 5 2 | | |
| | 5 2 5 2 5+6 2 3+4 2 5+6 2 5 2 | | |
| | 6 2 6 2 3+4 2 5+6 2 3+4 2 5 2 | | |
| | 6 2 5 2 5+6 2 3+4 2 3+4 2 5 2 | | |
| | 5 2 6 2 3+4 2 5+6 2 5+6 2 5 2 | | |
| | 5 2 6 2 5+6 2 3+4 2 3+4 2 5 2 | | |
| | 6 2 5 2 3+4 2 5+6 2 5+6 2 5 2 | | |
| | 5 2 6 2 5+6 2 3+4 2 5+6 2 5 2 | | |
| | 6 2 5 2 3+4 2 5+6 2 3+4 2 5 2 | | |
| | 5 2 6 2 5+6 2 5+6 2 5+6 2 5 2 | | |
| | 6 2 6 2 3+4 2 5+6 2 5+6 2 5 2 | | |
| | 5 2 6 2 3+4 2 3+4 2 3+4 2 5 2 | | |
| | 5 2 6 2 5+6 2 5+6 2 3+4 2 5 2 | | |
| | 6 2 6 2 3+4 2 3+4 2 3+4 2 5 2 | | |
| | 6 2 6 2 5+6 2 3+4 2 5+6 2 5 2 | | |
| | 5 2 5 2 3+4 2 5+6 2 5+6 2 5 2 | | |
| | 5 2 5 2 5+6 2 3+4 2 5+6 2 5 2 | | |
| | 6 2 5 2 5+6 2 5+6 2 3+4 2 5 2 | | |
| 71.1.47 | 174076 | 92 | Ce3Pd4Si4 |
| | -GM -X | -T | -R -S -W |
| | 5 2 5 2 3+4 2 3+4 2 3+4 2 5 2 | | |
| | 6 2 6 2 5+6 2 5+6 2 5+6 2 5 2 | | |
| | 5 2 5 2 3+4 2 3+4 2 3+4 2 5 2 | | |
| | 6 2 6 2 5+6 2 3+4 2 3+4 2 5 2 | | |
| | 5 2 5 2 3+4 2 5+6 2 5+6 2 5 2 | | |
| | 6 2 6 2 5+6 2 5+6 2 5+6 2 5 2 | | |
| | 6 2 6 2 5+6 2 5+6 2 5+6 2 5 2 | | |
| | 5 2 5 2 3+4 2 3+4 2 3+4 2 5 2 | | |
| | 6 2 5 2 5+6 2 5+6 2 5+6 2 5 2 | | |
| | 5 2 6 2 5+6 2 3+4 2 3+4 2 5 2 | | |
| | 6 2 6 2 3+4 2 5+6 2 5+6 2 5 2 | | |
| | 6 2 6 2 5+6 2 5+6 2 5+6 2 5 2 | | |
| | 5 2 6 2 5+6 2 5+6 2 3+4 2 5 2 | | |
| | 5 2 5 2 3+4 2 3+4 2 3+4 2 5 2 | | |
| | 6 2 5 2 5+6 2 5+6 2 5+6 2 5 2 | | |
| | 6 2 6 2 3+4 2 3+4 2 5+6 2 5 2 | | |
| | 5 2 6 2 3+4 2 3+4 2 3+4 2 5 2 | | |
| | 6 2 5 2 5+6 2 5+6 2 5+6 2 5 2 | | |
| | 6 2 5 2 3+4 2 3+4 2 3+4 2 5 2 | | |
| | 5 2 6 2 5+6 2 5+6 2 5+6 2 5 2 | | |
| | 6 2 5 2 3+4 2 3+4 2 3+4 2 5 2 | | |
| | 5 2 5 2 5+6 2 5+6 2 3+4 2 5 2 | | |
| | 5 2 6 2 3+4 2 3+4 2 5+6 2 5 2 | | |
| | 6 2 5 2 5+6 2 3+4 2 5+6 2 5 2 | | |
| | 5 2 5 2 3+4 2 5+6 2 3+4 2 5 2 | | |
| | 6 2 6 2 5+6 2 5+6 2 5+6 2 5 2 | | |
| | 5 2 6 2 3+4 2 3+4 2 3+4 2 5 2 | | |
| | 5 2 5 2 5+6 2 3+4 2 5+6 2 5 2 | | |
| | 6 2 6 2 3+4 2 5+6 2 3+4 2 5 2 | | |
| | 5 2 5 2 5+6 2 3+4 2 3+4 2 5 2 | | |
| | 6 2 6 2 3+4 2 5+6 2 5+6 2 5 2 | | |
| | 5 2 6 2 5+6 2 3+4 2 3+4 2 5 2 | | |
| | 6 2 5 2 3+4 2 5+6 2 5+6 2 5 2 | | |
| | 5 2 6 2 5+6 2 3+4 2 3+4 2 5 2 | | |
| | 6 2 6 2 5+6 2 5+6 2 5+6 2 5 2 | | |
| | 5 2 5 2 3+4 2 5+6 2 5+6 2 5 2 | | |
| | 6 2 6 2 3+4 2 3+4 2 3+4 2 5 2 | | |
| | 5 2 5 2 5+6 2 5+6 2 5+6 2 5 2 | | |

Continued on next column

**TABLE XLIV– continued**

| TQC | ICSD | NoE | SF |
|---|---|---|---|
| | 5 2 6 2 3+4 2 5+6 2 3+4 2 5 2 | | |
| | 5 2 6 2 5+6 2 3+4 2 3+4 2 5 2 | | |
| | 6 2 6 2 3+4 2 5+6 2 5+6 2 5 2 | | |
| | 6 2 6 2 3+4 2 3+4 2 5+6 2 5 2 | | |
| | 6 2 5 2 5+6 2 3+4 2 3+4 2 5 2 | | |
| | 5 2 5 2 5+6 2 5+6 2 5+6 2 5 2 | | |
| | 5 2 5 2 5+6 2 5+6 2 3+4 2 5 2 | | |
| | 5 2 6 2 3+4 2 3+4 2 5+6 2 5 2 | | |
| 71.1.48 | 262434 | 56 | Ce3Ge4Li4 |
| | -GM -X | -T | -R -S -W |
| | 5 2 5 2 3+4 2 5+6 2 5+6 2 5 2 | | |
| | 6 2 6 2 5+6 2 3+4 2 3+4 2 5 2 | | |
| | 5 2 5 2 3+4 2 3+4 2 3+4 2 5 2 | | |
| | 5 2 5 2 3+4 2 5+6 2 5+6 2 5 2 | | |
| | 6 2 6 2 5+6 2 3+4 2 3+4 2 5 2 | | |
| | 6 2 6 2 5+6 2 5+6 2 5+6 2 5 2 | | |
| | 6 2 6 2 5+6 2 5+6 2 5+6 2 5 2 | | |
| | 5 2 5 2 3+4 2 5+6 2 3+4 2 5 2 | | |
| | 5 2 5 2 5+6 2 3+4 2 5+6 2 5 2 | | |
| | 6 2 6 2 3+4 2 3+4 2 5+6 2 5 2 | | |
| | 6 2 6 2 5+6 2 5+6 2 3+4 2 5 2 | | |
| | 6 2 6 2 5+6 2 5+6 2 5+6 2 5 2 | | |
| | 5 2 5 2 5+6 2 5+6 2 3+4 2 5 2 | | |
| | 5 2 6 2 3+4 2 3+4 2 3+4 2 5 2 | | |
| | 6 2 5 2 5+6 2 5+6 2 5+6 2 5 2 | | |
| | 6 2 6 2 3+4 2 3+4 2 5+6 2 5 2 | | |
| | 5 2 5 2 3+4 2 3+4 2 5+6 2 5 2 | | |
| | 6 2 6 2 3+4 2 3+4 2 3+4 2 5 2 | | |
| | 5 2 6 2 5+6 2 5+6 2 3+4 2 5 2 | | |
| | 6 2 5 2 5+6 2 5+6 2 3+4 2 5 2 | | |
| | 5 2 6 2 3+4 2 3+4 2 3+4 2 5 2 | | |
| | 5 2 6 2 3+4 2 3+4 2 5+6 2 5 2 | | |
| | 5 2 5 2 5+6 2 5+6 2 5+6 2 5 2 | | |
| | 5 2 5 2 3+4 2 5+6 2 3+4 2 5 2 | | |
| | 5 2 6 2 5+6 2 3+4 2 5+6 2 5 2 | | |
| | 6 2 5 2 5+6 2 5+6 2 3+4 2 5 2 | | |
| | 6 2 5 2 5+6 2 3+4 2 5+6 2 5 2 | | |
| | 6 2 5 2 3+4 2 5+6 2 5+6 2 5 2 | | |
| 71.1.49 | 151160 | 96 | Ce3Cu4Sn4 |
| | -GM -X | -T | -R -S -W |
| | 5 2 5 2 5+6 2 3+4 2 5+6 2 5 2 | | |
| | 6 2 6 2 3+4 2 5+6 2 3+4 2 5 2 | | |
| | 5 2 5 2 5+6 2 5+6 2 3+4 2 5 2 | | |
| | 6 2 6 2 3+4 2 3+4 2 5+6 2 5 2 | | |
| | 5 2 5 2 5+6 2 5+6 2 3+4 2 5 2 | | |
| | 6 2 6 2 3+4 2 3+4 2 5+6 2 5 2 | | |
| | 6 2 6 2 3+4 2 3+4 2 5+6 2 5 2 | | |
| | 5 2 5 2 5+6 2 3+4 2 5+6 2 5 2 | | |
| | 5 2 6 2 3+4 2 5+6 2 3+4 2 5 2 | | |
| | 6 2 6 2 3+4 2 5+6 2 5+6 2 5 2 | | |
| | 6 2 5 2 5+6 2 3+4 2 3+4 2 5 2 | | |
| | 6 2 6 2 3+4 2 3+4 2 5+6 2 5 2 | | |
| | 5 2 5 2 3+4 2 3+4 2 3+4 2 5 2 | | |
| | 5 2 6 2 5+6 2 5+6 2 3+4 2 5 2 | | |
| | 6 2 5 2 3+4 2 3+4 2 5+6 2 5 2 | | |
| | 6 2 6 2 5+6 2 5+6 2 5+6 2 5 2 | | |
| | 6 2 6 2 5+6 2 5+6 2 3+4 2 5 2 | | |
| | 5 2 5 2 3+4 2 3+4 2 5+6 2 5 2 | | |
| | 5 2 5 2 3+4 2 5+6 2 3+4 2 5 2 | | |
| | 6 2 6 2 5+6 2 3+4 2 5+6 2 5 2 | | |
| | 5 2 6 2 5+6 2 5+6 2 3+4 2 5 2 | | |

Continued on next column

**TABLE XLIV– continued**

| TQC | ICSD NoE SF |
|---|---|
| | 6 2 5 2 3+4 2 3+4 2 5+6 2 5 2 |
| | 5 2 5 2 5+6 2 3+4 2 3+4 2 5 2 |
| | 6 2 6 2 5+6 2 5+6 2 5+6 2 5 2 |
| | 5 2 5 2 3+4 2 5+6 2 5+6 2 5 2 |
| | 6 2 5 2 5+6 2 5+6 2 3+4 2 5 2 |
| | 5 2 6 2 3+4 2 3+4 2 5+6 2 5 2 |
| | 5 2 6 2 3+4 2 3+4 2 3+4 2 5 2 |
| | 5 2 5 2 5+6 2 5+6 2 5+6 2 5 2 |
| | 6 2 6 2 3+4 2 3+4 2 3+4 2 5 2 |
| | 5 2 5 2 3+4 2 5+6 2 3+4 2 5 2 |
| | 6 2 6 2 5+6 2 3+4 2 5+6 2 5 2 |
| | 5 2 6 2 3+4 2 5+6 2 3+4 2 5 2 |
| | 6 2 5 2 3+4 2 3+4 2 5+6 2 5 2 |
| | 5 2 5 2 5+6 2 5+6 2 3+4 2 5 2 |
| | 6 2 6 2 5+6 2 3+4 2 5+6 2 5 2 |
| | 6 2 6 2 3+4 2 3+4 2 3+4 2 5 2 |
| | 6 2 6 2 5+6 2 5+6 2 5+6 2 5 2 |
| | 5 2 5 2 3+4 2 5+6 2 3+4 2 5 2 |
| | 5 2 5 2 5+6 2 3+4 2 3+4 2 5 2 |
| | 6 2 5 2 5+6 2 3+4 2 5+6 2 5 2 |
| | 5 2 6 2 5+6 2 5+6 2 3+4 2 5 2 |
| | 5 2 6 2 3+4 2 5+6 2 5+6 2 5 2 |
| | 5 2 6 2 5+6 2 3+4 2 5+6 2 5 2 |
| | 5 2 5 2 3+4 2 5+6 2 3+4 2 5 2 |
| | 6 2 5 2 3+4 2 5+6 2 3+4 2 5 2 |
| | 6 2 5 2 3+4 2 3+4 2 5+6 2 5 2 |
| | 6 2 6 2 5+6 2 3+4 2 5+6 2 5 2 |
| 71.1.50 | 416816 56 Ce2Ni2Zn1 |
| | -GM -X -T -R -S -W |
| | 5 2 6 2 3+4 2 5+6 2 5+6 2 5 2 |
| | 6 2 5 2 5+6 2 3+4 2 3+4 2 5 2 |
| | 5 2 6 2 3+4 2 5+6 2 5+6 2 5 2 |
| | 6 2 5 2 5+6 2 3+4 2 3+4 2 5 2 |
| | 5 2 6 2 3+4 2 5+6 2 5+6 2 5 2 |
| | 5 2 5 2 5+6 2 3+4 2 3+4 2 5 2 |
| | 6 2 6 2 3+4 2 5+6 2 5+6 2 5 2 |
| | 6 2 5 2 5+6 2 3+4 2 3+4 2 5 2 |
| | 5 2 5 2 3+4 2 3+4 2 3+4 2 5 2 |
| | 5 2 5 2 3+4 2 3+4 2 3+4 2 5 2 |
| | 5 2 5 2 3+4 2 3+4 2 3+4 2 5 2 |
| | 5 2 5 2 3+4 2 3+4 2 3+4 2 5 2 |
| | 5 2 5 2 3+4 2 3+4 2 3+4 2 5 2 |
| | 5 2 5 2 3+4 2 3+4 2 3+4 2 5 2 |
| | 6 2 5 2 5+6 2 5+6 2 5+6 2 5 2 |
| | 6 2 6 2 5+6 2 5+6 2 3+4 2 5 2 |
| | 5 2 6 2 3+4 2 3+4 2 5+6 2 5 2 |
| | 6 2 6 2 3+4 2 5+6 2 3+4 2 5 2 |
| | 5 2 6 2 5+6 2 3+4 2 5+6 2 5 2 |
| | 6 2 5 2 5+6 2 5+6 2 3+4 2 5 2 |
| | 6 2 5 2 3+4 2 3+4 2 5+6 2 5 2 |
| | 5 2 6 2 3+4 2 3+4 2 5+6 2 5 2 |
| | 5 2 5 2 3+4 2 5+6 2 3+4 2 5 2 |
| | 6 2 6 2 5+6 2 3+4 2 5+6 2 5 2 |
| | 5 2 5 2 5+6 2 5+6 2 3+4 2 5 2 |
| | 6 2 5 2 5+6 2 3+4 2 3+4 2 5 2 |
| | 5 2 6 2 5+6 2 5+6 2 5+6 2 5 2 |
| | 5 2 5 2 5+6 2 5+6 2 3+4 2 5 2 |
| 71.1.51 | 174529 32 Ce1Si5 |
| | -GM -X -T -R -S -W |
| | 5 2 5 2 5+6 2 5+6 2 3+4 2 5 2 |
| | 6 2 6 2 3+4 2 3+4 2 5+6 2 5 2 |

Continued on next column

**TABLE XLIV– continued**

| TQC | ICSD NoE SF |
|---|---|
| | 6 2 6 2 3+4 2 3+4 2 5+6 2 5 2 |
| | 6 2 6 2 3+4 2 3+4 2 5+6 2 5 2 |
| | 5 2 5 2 3+4 2 3+4 2 3+4 2 5 2 |
| | 5 2 6 2 5+6 2 5+6 2 5+6 2 5 2 |
| | 6 2 6 2 5+6 2 3+4 2 5+6 2 5 2 |
| | 5 2 5 2 5+6 2 5+6 2 3+4 2 5 2 |
| | 6 2 5 2 3+4 2 5+6 2 5+6 2 5 2 |
| | 5 2 6 2 3+4 2 5+6 2 3+4 2 5 2 |
| | 6 2 5 2 5+6 2 3+4 2 3+4 2 5 2 |
| | 5 2 6 2 5+6 2 5+6 2 3+4 2 5 2 |
| | 6 2 6 2 3+4 2 3+4 2 5+6 2 5 2 |
| | 5 2 6 2 5+6 2 3+4 2 3+4 2 5 2 |
| | 5 2 6 2 3+4 2 5+6 2 5+6 2 5 2 |
| | 6 2 6 2 3+4 2 3+4 2 5+6 2 5 2 |
| 71.1.52 | 246812 32 Ce1Ge5 |
| | -GM -X -T -R -S -W |
| | 5 2 5 2 3+4 2 3+4 2 3+4 2 5 2 |
| | 6 2 6 2 5+6 2 5+6 2 5+6 2 5 2 |
| | 6 2 6 2 5+6 2 5+6 2 5+6 2 5 2 |
| | 6 2 6 2 5+6 2 5+6 2 5+6 2 5 2 |
| | 5 2 6 2 5+6 2 5+6 2 3+4 2 5 2 |
| | 5 2 5 2 3+4 2 3+4 2 5+6 2 5 2 |
| | 6 2 6 2 3+4 2 5+6 2 5+6 2 5 2 |
| | 5 2 5 2 3+4 2 3+4 2 3+4 2 5 2 |
| | 6 2 5 2 5+6 2 5+6 2 5+6 2 5 2 |
| | 6 2 5 2 5+6 2 3+4 2 3+4 2 5 2 |
| | 5 2 6 2 3+4 2 3+4 2 3+4 2 5 2 |
| | 5 2 6 2 3+4 2 3+4 2 3+4 2 5 2 |
| | 6 2 6 2 5+6 2 5+6 2 5+6 2 5 2 |
| | 5 2 6 2 3+4 2 5+6 2 5+6 2 5 2 |
| | 5 2 6 2 5+6 2 3+4 2 3+4 2 5 2 |
| | 6 2 6 2 5+6 2 5+6 2 5+6 2 5 2 |
| 71.1.53 | 429175 26 Ca2Mg1Pt2 |
| | -GM -X -T -R -S -W |
| | 5 2 5 2 5+6 2 3+4 2 3+4 2 5 2 |
| | 5 2 5 2 3+4 2 5+6 2 5+6 2 5 2 |
| | 6 2 6 2 5+6 2 5+6 2 3+4 2 5 2 |
| | 6 2 6 2 3+4 2 3+4 2 5+6 2 5 2 |
| | 5 2 6 2 5+6 2 5+6 2 5+6 2 5 2 |
| | 6 2 5 2 3+4 2 3+4 2 3+4 2 5 2 |
| | 6 2 6 2 3+4 2 5+6 2 3+4 2 5 2 |
| | 5 2 6 2 5+6 2 3+4 2 5+6 2 5 2 |
| | 6 2 5 2 3+4 2 3+4 2 5+6 2 5 2 |
| | 5 2 6 2 3+4 2 5+6 2 3+4 2 5 2 |
| | 5 2 5 2 5+6 2 3+4 2 3+4 2 5 2 |
| | 5 2 5 2 5+6 2 5+6 2 5+6 2 5 2 |
| | 6 2 5 2 5+6 2 3+4 2 3+4 2 5 2 |
| 71.1.54 | 429174 26 Ca2Mg1Pd2 |
| | -GM -X -T -R -S -W |
| | 5 2 5 2 5+6 2 3+4 2 3+4 2 5 2 |
| | 5 2 5 2 3+4 2 5+6 2 5+6 2 5 2 |
| | 6 2 6 2 5+6 2 5+6 2 5+6 2 5 2 |
| | 6 2 6 2 3+4 2 3+4 2 3+4 2 5 2 |
| | 5 2 6 2 3+4 2 3+4 2 3+4 2 5 2 |
| | 6 2 5 2 5+6 2 5+6 2 5+6 2 5 2 |
| | 5 2 6 2 3+4 2 5+6 2 5+6 2 5 2 |
| | 6 2 6 2 3+4 2 3+4 2 3+4 2 5 2 |
| | 6 2 5 2 5+6 2 3+4 2 5+6 2 5 2 |
| | 5 2 6 2 3+4 2 5+6 2 3+4 2 5 2 |
| | 5 2 5 2 5+6 2 3+4 2 5+6 2 5 2 |
| | 5 2 5 2 5+6 2 5+6 2 3+4 2 5 2 |

Continued on next column

**TABLE XLIV– continued**

| TQC | ICSD | NoE | SF |
|---|---|---|---|
| | 6 2 5 2 5+6 2 3+4 2 3+4 2 5 2 | | |
| 71.1.55 | 425471 | 36 | Ca2Cd1Pd2 |
| | -GM -X -T -R -S -W | | |
| | 5 2 5 2 3+4 2 3+4 2 3+4 2 5 2 | | |
| | 5 2 5 2 3+4 2 3+4 2 3+4 2 5 2 | | |
| | 5 2 5 2 3+4 2 3+4 2 3+4 2 5 2 | | |
| | 5 2 5 2 3+4 2 3+4 2 3+4 2 5 2 | | |
| | 5 2 5 2 3+4 2 3+4 2 3+4 2 5 2 | | |
| | 5 2 5 2 3+4 2 3+4 2 3+4 2 5 2 | | |
| | 6 2 6 2 5+6 2 5+6 2 5+6 2 5 2 | | |
| | 5 2 5 2 5+6 2 5+6 2 5+6 2 5 2 | | |
| | 6 2 6 2 3+4 2 5+6 2 3+4 2 5 2 | | |
| | 6 2 6 2 3+4 2 3+4 2 5+6 2 5 2 | | |
| | 5 2 5 2 5+6 2 3+4 2 3+4 2 5 2 | | |
| | 5 2 6 2 3+4 2 5+6 2 5+6 2 5 2 | | |
| | 6 2 6 2 5+6 2 5+6 2 5+6 2 5 2 | | |
| | 6 2 5 2 3+4 2 3+4 2 3+4 2 5 2 | | |
| | 5 2 6 2 5+6 2 3+4 2 5+6 2 5 2 | | |
| | 5 2 5 2 5+6 2 5+6 2 3+4 2 5 2 | | |
| | 5 2 5 2 3+4 2 3+4 2 3+4 2 5 2 | | |
| | 6 2 5 2 5+6 2 3+4 2 3+4 2 5 2 | | |
| 71.1.56 | 40950 | 34 | C4Co1Sc3 |
| | -GM -X -T -R -S -W | | |
| | 5 2 5 2 5+6 2 5+6 2 5+6 2 5 2 | | |
| | 6 2 6 2 3+4 2 3+4 2 3+4 2 5 2 | | |
| | 5 2 5 2 3+4 2 3+4 2 5+6 2 5 2 | | |
| | 6 2 5 2 5+6 2 5+6 2 3+4 2 5 2 | | |
| | 5 2 6 2 3+4 2 5+6 2 5+6 2 5 2 | | |
| | 5 2 6 2 5+6 2 3+4 2 3+4 2 5 2 | | |
| | 6 2 5 2 5+6 2 5+6 2 5+6 2 5 2 | | |
| | 5 2 5 2 3+4 2 3+4 2 3+4 2 5 2 | | |
| | 6 2 6 2 5+6 2 3+4 2 5+6 2 5 2 | | |
| | 6 2 5 2 3+4 2 5+6 2 3+4 2 5 2 | | |
| | 5 2 6 2 5+6 2 3+4 2 5+6 2 5 2 | | |
| | 5 2 5 2 5+6 2 3+4 2 5+6 2 5 2 | | |
| | 5 2 5 2 3+4 2 3+4 2 5+6 2 5 2 | | |
| | 5 2 6 2 5+6 2 3+4 2 3+4 2 5 2 | | |
| | 5 2 5 2 3+4 2 5+6 2 5+6 2 5 2 | | |
| | 5 2 5 2 3+4 2 3+4 2 5+6 2 5 2 | | |
| | 5 2 6 2 5+6 2 5+6 2 3+4 2 5 2 | | |
| 71.1.57 | 236393 | 34 | C4Co1Sc3 |
| | -GM -X -T -R -S -W | | |
| | 5 2 5 2 5+6 2 5+6 2 5+6 2 5 2 | | |
| | 6 2 6 2 3+4 2 3+4 2 3+4 2 5 2 | | |
| | 5 2 5 2 3+4 2 3+4 2 5+6 2 5 2 | | |
| | 6 2 5 2 5+6 2 5+6 2 3+4 2 5 2 | | |
| | 5 2 6 2 3+4 2 5+6 2 5+6 2 5 2 | | |
| | 5 2 6 2 5+6 2 3+4 2 3+4 2 5 2 | | |
| | 6 2 5 2 5+6 2 5+6 2 5+6 2 5 2 | | |
| | 5 2 5 2 3+4 2 3+4 2 3+4 2 5 2 | | |
| | 6 2 6 2 5+6 2 3+4 2 5+6 2 5 2 | | |
| | 6 2 6 2 3+4 2 5+6 2 3+4 2 5 2 | | |
| | 5 2 5 2 5+6 2 3+4 2 5+6 2 5 2 | | |
| | 5 2 5 2 5+6 2 3+4 2 5+6 2 5 2 | | |
| | 5 2 5 2 3+4 2 3+4 2 5+6 2 5 2 | | |
| | 5 2 6 2 5+6 2 3+4 2 3+4 2 5 2 | | |
| | 5 2 5 2 3+4 2 5+6 2 5+6 2 5 2 | | |
| | 5 2 5 2 3+4 2 3+4 2 5+6 2 5 2 | | |
| | 6 2 6 2 5+6 2 5+6 2 3+4 2 5 2 | | |
| 71.1.58 | 173167 | 34 | C4Co1Sc3 |
| | -GM -X -T -R -S -W | | |
| | 5 2 5 2 5+6 2 5+6 2 5+6 2 5 2 | | |
| | 6 2 6 2 3+4 2 3+4 2 3+4 2 5 2 | | |
| | 5 2 5 2 3+4 2 3+4 2 5+6 2 5 2 | | |
| | 6 2 5 2 5+6 2 5+6 2 3+4 2 5 2 | | |
| | 5 2 6 2 3+4 2 5+6 2 5+6 2 5 2 | | |
| | 5 2 6 2 5+6 2 3+4 2 3+4 2 5 2 | | |
| | 6 2 5 2 5+6 2 5+6 2 5+6 2 5 2 | | |
| | 5 2 5 2 3+4 2 3+4 2 3+4 2 5 2 | | |
| | 6 2 6 2 5+6 2 3+4 2 5+6 2 5 2 | | |
| | 6 2 6 2 3+4 2 5+6 2 3+4 2 5 2 | | |
| | 5 2 5 2 5+6 2 3+4 2 5+6 2 5 2 | | |
| | 5 2 5 2 5+6 2 3+4 2 5+6 2 5 2 | | |
| | 5 2 5 2 3+4 2 3+4 2 5+6 2 5 2 | | |
| | 5 2 6 2 5+6 2 3+4 2 3+4 2 5 2 | | |
| | 5 2 5 2 3+4 2 5+6 2 5+6 2 5 2 | | |
| | 5 2 5 2 3+4 2 3+4 2 5+6 2 5 2 | | |
| | 6 2 6 2 5+6 2 5+6 2 3+4 2 5 2 | | |
| 71.1.59 | 94741 | 68 | C2Bi2Ca1O8 |
| | -GM -X -T -R -S -W | | |
| | 5 2 5 2 3+4 2 5+6 2 5+6 2 5 2 | | |
| | 6 2 6 2 5+6 2 3+4 2 3+4 2 5 2 | | |
| | 5 2 6 2 3+4 2 3+4 2 3+4 2 5 2 | | |
| | 6 2 5 2 5+6 2 5+6 2 5+6 2 5 2 | | |
| | 6 2 6 2 5+6 2 5+6 2 5+6 2 5 2 | | |
| | 5 2 5 2 3+4 2 3+4 2 3+4 2 5 2 | | |
| | 5 2 6 2 3+4 2 3+4 2 3+4 2 5 2 | | |
| | 6 2 5 2 5+6 2 5+6 2 5+6 2 5 2 | | |
| | 5 2 5 2 3+4 2 3+4 2 3+4 2 5 2 | | |
| | 6 2 6 2 5+6 2 5+6 2 5+6 2 5 2 | | |
| | 6 2 5 2 5+6 2 3+4 2 3+4 2 5 2 | | |
| | 5 2 6 2 3+4 2 5+6 2 5+6 2 5 2 | | |
| | 5 2 5 2 5+6 2 5+6 2 5+6 2 5 2 | | |
| | 6 2 5 2 3+4 2 3+4 2 3+4 2 5 2 | | |
| | 5 2 6 2 3+4 2 3+4 2 5+6 2 5 2 | | |
| | 6 2 6 2 5+6 2 5+6 2 3+4 2 5 2 | | |
| | 5 2 6 2 5+6 2 3+4 2 3+4 2 5 2 | | |
| | 5 2 6 2 3+4 2 3+4 2 5+6 2 5 2 | | |
| | 6 2 5 2 3+4 2 5+6 2 5+6 2 5 2 | | |
| | 5 2 5 2 5+6 2 5+6 2 3+4 2 5 2 | | |
| | 5 2 6 2 5+6 2 3+4 2 3+4 2 5 2 | | |
| | 6 2 5 2 3+4 2 5+6 2 5+6 2 5 2 | | |
| | 6 2 6 2 5+6 2 5+6 2 5+6 2 5 2 | | |
| | 5 2 6 2 3+4 2 3+4 2 3+4 2 5 2 | | |
| | 5 2 6 2 3+4 2 5+6 2 5+6 2 5 2 | | |
| | 6 2 5 2 5+6 2 3+4 2 3+4 2 5 2 | | |
| | 5 2 5 2 5+6 2 3+4 2 5+6 2 5 2 | | |
| | 6 2 6 2 3+4 2 3+4 2 3+4 2 5 2 | | |
| | 6 2 5 2 5+6 2 5+6 2 3+4 2 5 2 | | |
| | 6 2 5 2 5+6 2 5+6 2 5+6 2 5 2 | | |
| | 6 2 6 2 3+4 2 5+6 2 5+6 2 5 2 | | |
| | 5 2 5 2 3+4 2 3+4 2 3+4 2 5 2 | | |
| | 5 2 5 2 5+6 2 3+4 2 5+6 2 5 2 | | |
| | 6 2 6 2 5+6 2 5+6 2 3+4 2 5 2 | | |
| 71.1.60 | 238028 | 52 | C1Co2Pr2Si1 |
| | -GM -X -T -R -S -W | | |
| | 5 2 6 2 3+4 2 5+6 2 5+6 2 5 2 | | |
| | 6 2 5 2 5+6 2 3+4 2 3+4 2 5 2 | | |
| | 5 2 6 2 3+4 2 5+6 2 5+6 2 5 2 | | |
| | 6 2 5 2 5+6 2 3+4 2 3+4 2 5 2 | | |
| | 5 2 5 2 3+4 2 5+6 2 5+6 2 5 2 | | |
| | 5 2 6 2 5+6 2 3+4 2 3+4 2 5 2 | | |

Continued on next column

**TABLE XLIV– continued**

| TQC | ICSD | NoE | SF |
|---|---|---|---|
| | 6 2 6 2 3+4 2 5+6 2 5+6 2 5 2 | | |
| | 6 2 5 2 5+6 2 3+4 2 3+4 2 5 2 | | |
| | 5 2 5 2 5+6 2 5+6 2 3+4 2 5 2 | | |
| | 5 2 5 2 3+4 2 3+4 2 3+4 2 5 2 | | |
| | 6 2 6 2 3+4 2 3+4 2 5+6 2 5 2 | | |
| | 6 2 6 2 5+6 2 5+6 2 5+6 2 5 2 | | |
| | 6 2 6 2 3+4 2 3+4 2 5+6 2 5 2 | | |
| | 6 2 6 2 3+4 2 3+4 2 5+6 2 5 2 | | |
| | 6 2 6 2 5+6 2 5+6 2 5+6 2 5 2 | | |
| | 5 2 5 2 3+4 2 5+6 2 3+4 2 5 2 | | |
| | 6 2 5 2 5+6 2 5+6 2 5+6 2 5 2 | | |
| | 5 2 6 2 5+6 2 3+4 2 3+4 2 5 2 | | |
| | 6 2 5 2 5+6 2 3+4 2 3+4 2 5 2 | | |
| | 5 2 6 2 5+6 2 5+6 2 3+4 2 5 2 | | |
| | 5 2 5 2 3+4 2 3+4 2 5+6 2 5 2 | | |
| | 6 2 6 2 5+6 2 5+6 2 3+4 2 5 2 | | |
| | 5 2 5 2 3+4 2 3+4 2 5+6 2 5 2 | | |
| | 6 2 6 2 3+4 2 5+6 2 3+4 2 5 2 | | |
| | 5 2 5 2 3+4 2 3+4 2 5+6 2 5 2 | | |
| | 6 2 5 2 5+6 2 5+6 2 5+6 2 5 2 | | |
| 71.1.61 | 418403 | 56 | C1B4Dy2 |
| | -GM -X -T -R -S -W | | |
| | 5 2 6 2 3+4 2 3+4 2 3+4 2 5 2 | | |
| | 6 2 5 2 5+6 2 5+6 2 5+6 2 5 2 | | |
| | 5 2 6 2 3+4 2 3+4 2 3+4 2 5 2 | | |
| | 6 2 5 2 5+6 2 5+6 2 5+6 2 5 2 | | |
| | 5 2 5 2 5+6 2 3+4 2 5+6 2 5 2 | | |
| | 6 2 6 2 3+4 2 5+6 2 3+4 2 5 2 | | |
| | 5 2 6 2 3+4 2 5+6 2 3+4 2 5 2 | | |
| | 6 2 5 2 5+6 2 3+4 2 5+6 2 5 2 | | |
| | 5 2 5 2 5+6 2 3+4 2 5+6 2 5 2 | | |
| | 6 2 5 2 3+4 2 5+6 2 3+4 2 5 2 | | |
| | 5 2 6 2 3+4 2 5+6 2 5+6 2 5 2 | | |
| | 6 2 6 2 5+6 2 3+4 2 3+4 2 5 2 | | |
| | 6 2 6 2 5+6 2 5+6 2 3+4 2 5 2 | | |
| | 6 2 6 2 3+4 2 5+6 2 5+6 2 5 2 | | |
| | 5 2 5 2 3+4 2 3+4 2 3+4 2 5 2 | | |
| | 6 2 5 2 3+4 2 3+4 2 5+6 2 5 2 | | |
| | 5 2 6 2 5+6 2 5+6 2 5+6 2 5 2 | | |
| | 6 2 5 2 5+6 2 5+6 2 3+4 2 5 2 | | |
| | 5 2 6 2 3+4 2 3+4 2 3+4 2 5 2 | | |
| | 5 2 5 2 3+4 2 5+6 2 5+6 2 5 2 | | |
| | 6 2 6 2 5+6 2 3+4 2 3+4 2 5 2 | | |
| | 5 2 5 2 3+4 2 5+6 2 5+6 2 5 2 | | |
| | 5 2 6 2 5+6 2 3+4 2 5+6 2 5 2 | | |
| | 6 2 5 2 5+6 2 5+6 2 3+4 2 5 2 | | |
| | 5 2 6 2 3+4 2 3+4 2 3+4 2 5 2 | | |
| | 6 2 5 2 3+4 2 5+6 2 5+6 2 5 2 | | |
| | 6 2 5 2 5+6 2 3+4 2 3+4 2 5 2 | | |
| | 5 2 5 2 3+4 2 5+6 2 5+6 2 5 2 | | |
| 71.1.62 | 428114 | 54 | Ba3Li4Sb4 |
| | -GM -X -T -R -S -W | | |
| | 5 2 5 2 3+4 2 5+6 2 5+6 2 5 2 | | |
| | 6 2 6 2 5+6 2 3+4 2 3+4 2 5 2 | | |
| | 5 2 5 2 3+4 2 3+4 2 3+4 2 5 2 | | |
| | 5 2 6 2 5+6 2 5+6 2 5+6 2 5 2 | | |
| | 6 2 5 2 3+4 2 3+4 2 3+4 2 5 2 | | |
| | 6 2 6 2 5+6 2 5+6 2 5+6 2 5 2 | | |
| | 5 2 6 2 5+6 2 5+6 2 5+6 2 5 2 | | |
| | 6 2 5 2 5+6 2 5+6 2 3+4 2 5 2 | | |
| | 6 2 6 2 3+4 2 3+4 2 5+6 2 5 2 | | |

Continued on next column

**TABLE XLIV– continued**

| TQC | ICSD | NoE | SF |
|---|---|---|---|
| | 5 2 6 2 3+4 2 3+4 2 5+6 2 5 2 | | |
| | 6 2 5 2 5+6 2 5+6 2 3+4 2 5 2 | | |
| | 6 2 6 2 5+6 2 5+6 2 5+6 2 5 2 | | |
| | 5 2 5 2 5+6 2 5+6 2 3+4 2 5 2 | | |
| | 5 2 6 2 3+4 2 3+4 2 3+4 2 5 2 | | |
| | 6 2 5 2 5+6 2 5+6 2 5+6 2 5 2 | | |
| | 6 2 6 2 3+4 2 3+4 2 5+6 2 5 2 | | |
| | 5 2 5 2 3+4 2 3+4 2 5+6 2 5 2 | | |
| | 6 2 6 2 5+6 2 5+6 2 3+4 2 5 2 | | |
| | 5 2 6 2 3+4 2 3+4 2 3+4 2 5 2 | | |
| | 6 2 5 2 5+6 2 5+6 2 5+6 2 5 2 | | |
| | 5 2 6 2 3+4 2 3+4 2 3+4 2 5 2 | | |
| | 5 2 6 2 5+6 2 5+6 2 5+6 2 5 2 | | |
| | 5 2 5 2 3+4 2 3+4 2 3+4 2 5 2 | | |
| | 5 2 6 2 3+4 2 5+6 2 5+6 2 5 2 | | |
| | 6 2 5 2 5+6 2 3+4 2 3+4 2 5 2 | | |
| | 6 2 5 2 5+6 2 5+6 2 3+4 2 5 2 | | |
| | 6 2 5 2 5+6 2 3+4 2 5+6 2 5 2 | | |
| 71.1.63 | 290306 | 162 | Ba3Cd1Hg10 |
| | -GM -X -T -R -S -W | | |
| | 5 2 5 2 3+4 2 3+4 2 3+4 2 5 2 | | |
| | 5 2 5 2 3+4 2 3+4 2 3+4 2 5 2 | | |
| | 6 2 6 2 5+6 2 5+6 2 5+6 2 5 2 | | |
| | 6 2 6 2 5+6 2 5+6 2 5+6 2 5 2 | | |
| | 5 2 6 2 5+6 2 5+6 2 3+4 2 5 2 | | |
| | 6 2 5 2 3+4 2 3+4 2 5+6 2 5 2 | | |
| | 6 2 6 2 5+6 2 5+6 2 5+6 2 5 2 | | |
| | 6 2 6 2 5+6 2 5+6 2 5+6 2 5 2 | | |
| | 5 2 6 2 3+4 2 3+4 2 3+4 2 5 2 | | |
| | 6 2 5 2 5+6 2 5+6 2 5+6 2 5 2 | | |
| | 5 2 5 2 3+4 2 3+4 2 3+4 2 5 2 | | |
| | 6 2 6 2 5+6 2 5+6 2 5+6 2 5 2 | | |
| | 5 2 5 2 3+4 2 5+6 2 5+6 2 5 2 | | |
| | 5 2 5 2 3+4 2 5+6 2 5+6 2 5 2 | | |
| | 5 2 5 2 3+4 2 5+6 2 5+6 2 5 2 | | |
| | 5 2 5 2 3+4 2 5+6 2 5+6 2 5 2 | | |
| | 5 2 5 2 5+6 2 5+6 2 5+6 2 5 2 | | |
| | 5 2 5 2 3+4 2 3+4 2 3+4 2 5 2 | | |
| | 6 2 6 2 3+4 2 5+6 2 5+6 2 5 2 | | |
| | 6 2 6 2 5+6 2 3+4 2 3+4 2 5 2 | | |
| | 5 2 6 2 3+4 2 5+6 2 5+6 2 5 2 | | |
| | 5 2 5 2 3+4 2 3+4 2 3+4 2 5 2 | | |
| | 5 2 6 2 5+6 2 5+6 2 3+4 2 5 2 | | |
| | 6 2 5 2 5+6 2 3+4 2 5+6 2 5 2 | | |
| | 6 2 6 2 3+4 2 3+4 2 3+4 2 5 2 | | |
| | 5 2 5 2 5+6 2 5+6 2 5+6 2 5 2 | | |
| | 5 2 6 2 3+4 2 3+4 2 3+4 2 5 2 | | |
| | 6 2 5 2 3+4 2 5+6 2 5+6 2 5 2 | | |
| | 5 2 6 2 5+6 2 3+4 2 5+6 2 5 2 | | |
| | 6 2 5 2 3+4 2 5+6 2 3+4 2 5 2 | | |
| | 6 2 6 2 5+6 2 5+6 2 3+4 2 5 2 | | |
| | 6 2 6 2 5+6 2 3+4 2 5+6 2 5 2 | | |
| | 6 2 5 2 5+6 2 3+4 2 3+4 2 5 2 | | |
| | 6 2 6 2 5+6 2 3+4 2 3+4 2 5 2 | | |
| | 5 2 5 2 3+4 2 5+6 2 5+6 2 5 2 | | |
| | 5 2 5 2 3+4 2 5+6 2 5+6 2 5 2 | | |
| | 5 2 5 2 3+4 2 5+6 2 5+6 2 5 2 | | |
| | 5 2 6 2 3+4 2 5+6 2 3+4 2 5 2 | | |
| | 6 2 5 2 5+6 2 3+4 2 5+6 2 5 2 | | |
| | 5 2 6 2 3+4 2 3+4 2 3+4 2 5 2 | | |
| | 6 2 5 2 5+6 2 5+6 2 5+6 2 5 2 | | |

Continued on next column

**TABLE XLIV– continued**

| TQC | ICSD NoE SF |
|---|---|
| | 5 2 6 2 3+4 2 5+6 2 3+4 2 5 2 |
| | 6 2 6 2 5+6 2 3+4 2 5+6 2 5 2 |
| | 5 2 5 2 3+4 2 3+4 2 3+4 2 5 2 |
| | 5 2 5 2 5+6 2 5+6 2 5+6 2 5 2 |
| | 6 2 6 2 3+4 2 3+4 2 3+4 2 5 2 |
| | 6 2 5 2 5+6 2 5+6 2 3+4 2 5 2 |
| | 5 2 6 2 3+4 2 5+6 2 5+6 2 5 2 |
| | 6 2 5 2 5+6 2 3+4 2 5+6 2 5 2 |
| | 6 2 6 2 5+6 2 5+6 2 5+6 2 5 2 |
| | 6 2 5 2 3+4 2 3+4 2 3+4 2 5 2 |
| | 5 2 6 2 5+6 2 3+4 2 5+6 2 5 2 |
| | 5 2 5 2 3+4 2 5+6 2 3+4 2 5 2 |
| | 6 2 6 2 5+6 2 3+4 2 3+4 2 5 2 |
| | 5 2 5 2 5+6 2 5+6 2 3+4 2 5 2 |
| | 6 2 6 2 3+4 2 3+4 2 5+6 2 5 2 |
| | 5 2 5 2 3+4 2 5+6 2 5+6 2 5 2 |
| | 6 2 6 2 5+6 2 3+4 2 3+4 2 5 2 |
| | 5 2 5 2 3+4 2 3+4 2 5+6 2 5 2 |
| | 5 2 5 2 5+6 2 5+6 2 3+4 2 5 2 |
| | 6 2 6 2 3+4 2 3+4 2 5+6 2 5 2 |
| | 6 2 5 2 5+6 2 5+6 2 3+4 2 5 2 |
| | 5 2 6 2 3+4 2 5+6 2 5+6 2 5 2 |
| | 5 2 5 2 5+6 2 5+6 2 3+4 2 5 2 |
| | 6 2 6 2 3+4 2 3+4 2 3+4 2 5 2 |
| | 5 2 6 2 3+4 2 3+4 2 5+6 2 5 2 |
| | 6 2 5 2 5+6 2 5+6 2 5+6 2 5 2 |
| | 5 2 6 2 5+6 2 5+6 2 3+4 2 5 2 |
| | 6 2 5 2 3+4 2 3+4 2 5+6 2 5 2 |
| | 5 2 6 2 3+4 2 3+4 2 5+6 2 5 2 |
| | 5 2 5 2 3+4 2 5+6 2 3+4 2 5 2 |
| | 6 2 5 2 5+6 2 3+4 2 3+4 2 5 2 |
| | 6 2 6 2 5+6 2 5+6 2 5+6 2 5 2 |
| | 5 2 6 2 3+4 2 5+6 2 3+4 2 5 2 |
| | 6 2 5 2 5+6 2 3+4 2 5+6 2 5 2 |
| | 5 2 5 2 5+6 2 5+6 2 5+6 2 5 2 |
| | 6 2 5 2 3+4 2 3+4 2 3+4 2 5 2 |
| | 6 2 6 2 3+4 2 3+4 2 5+6 2 5 2 |
| | 5 2 6 2 3+4 2 5+6 2 3+4 2 5 2 |
| | 5 2 6 2 5+6 2 3+4 2 3+4 2 5 2 |
| | 5 2 6 2 5+6 2 3+4 2 3+4 2 5 2 |
| 71.1.64 | 428115 54 Ba3Bi4Li4 |
| | -GM -X -T -R -S -W |
| | 5 2 5 2 3+4 2 5+6 2 5+6 2 5 2 |
| | 6 2 6 2 5+6 2 3+4 2 3+4 2 5 2 |
| | 5 2 5 2 3+4 2 3+4 2 3+4 2 5 2 |
| | 5 2 5 2 3+4 2 5+6 2 5+6 2 5 2 |
| | 6 2 6 2 5+6 2 3+4 2 3+4 2 5 2 |
| | 6 2 6 2 5+6 2 5+6 2 5+6 2 5 2 |
| | 5 2 5 2 5+6 2 5+6 2 3+4 2 5 2 |
| | 6 2 6 2 3+4 2 3+4 2 5+6 2 5 2 |
| | 5 2 6 2 5+6 2 5+6 2 5+6 2 5 2 |
| | 6 2 5 2 3+4 2 3+4 2 5+6 2 5 2 |
| | 6 2 6 2 5+6 2 5+6 2 3+4 2 5 2 |
| | 6 2 6 2 5+6 2 5+6 2 5+6 2 5 2 |
| | 5 2 5 2 5+6 2 5+6 2 3+4 2 5 2 |
| | 5 2 6 2 3+4 2 3+4 2 3+4 2 5 2 |
| | 6 2 5 2 5+6 2 5+6 2 5+6 2 5 2 |
| | 6 2 6 2 3+4 2 3+4 2 5+6 2 5 2 |
| | 5 2 5 2 3+4 2 3+4 2 3+4 2 5 2 |
| | 6 2 6 2 5+6 2 5+6 2 5+6 2 5 2 |
| | 5 2 6 2 3+4 2 5+6 2 3+4 2 5 2 |

Continued on next column

**TABLE XLIV– continued**

| TQC | ICSD NoE SF |
|---|---|
| | 5 2 5 2 5+6 2 3+4 2 5+6 2 5 2 |
| | 6 2 6 2 3+4 2 3+4 2 3+4 2 5 2 |
| | 5 2 5 2 5+6 2 5+6 2 3+4 2 5 2 |
| | 6 2 6 2 3+4 2 3+4 2 5+6 2 5 2 |
| | 5 2 6 2 5+6 2 5+6 2 5+6 2 5 2 |
| | 5 2 5 2 3+4 2 3+4 2 3+4 2 5 2 |
| | 6 2 5 2 5+6 2 5+6 2 3+4 2 5 2 |
| | 6 2 5 2 5+6 2 3+4 2 5+6 2 5 2 |
| 71.1.65 | 72631 96 Ba1Ni1O5Tm2 |
| | -GM -X -T -R -S -W |
| | 5 2 6 2 3+4 2 3+4 2 3+4 2 5 2 |
| | 6 2 5 2 5+6 2 5+6 2 5+6 2 5 2 |
| | 5 2 5 2 5+6 2 5+6 2 3+4 2 5 2 |
| | 5 2 6 2 3+4 2 3+4 2 5+6 2 5 2 |
| | 6 2 5 2 5+6 2 5+6 2 3+4 2 5 2 |
| | 5 2 6 2 5+6 2 3+4 2 5+6 2 5 2 |
| | 6 2 5 2 3+4 2 5+6 2 3+4 2 5 2 |
| | 5 2 6 2 5+6 2 3+4 2 5+6 2 5 2 |
| | 6 2 5 2 3+4 2 5+6 2 3+4 2 5 2 |
| | 5 2 5 2 5+6 2 3+4 2 5+6 2 5 2 |
| | 5 2 6 2 3+4 2 3+4 2 5+6 2 5 2 |
| | 6 2 5 2 5+6 2 5+6 2 3+4 2 5 2 |
| | 6 2 5 2 3+4 2 3+4 2 3+4 2 5 2 |
| | 5 2 6 2 5+6 2 5+6 2 5+6 2 5 2 |
| | 6 2 6 2 3+4 2 3+4 2 5+6 2 5 2 |
| | 6 2 6 2 3+4 2 3+4 2 5+6 2 5 2 |
| | 6 2 6 2 3+4 2 3+4 2 5+6 2 5 2 |
| | 6 2 5 2 3+4 2 5+6 2 3+4 2 5 2 |
| | 5 2 6 2 3+4 2 5+6 2 3+4 2 5 2 |
| | 6 2 6 2 5+6 2 3+4 2 3+4 2 5 2 |
| | 5 2 6 2 3+4 2 5+6 2 3+4 2 5 2 |
| | 5 2 5 2 3+4 2 3+4 2 5+6 2 5 2 |
| | 5 2 6 2 5+6 2 5+6 2 3+4 2 5 2 |
| | 6 2 6 2 3+4 2 5+6 2 5+6 2 5 2 |
| | 6 2 5 2 5+6 2 3+4 2 3+4 2 5 2 |
| | 5 2 6 2 3+4 2 3+4 2 3+4 2 5 2 |
| | 6 2 5 2 5+6 2 5+6 2 5+6 2 5 2 |
| | 6 2 5 2 3+4 2 3+4 2 5+6 2 5 2 |
| | 6 2 6 2 5+6 2 5+6 2 3+4 2 5 2 |
| | 6 2 6 2 3+4 2 3+4 2 3+4 2 5 2 |
| | 5 2 5 2 5+6 2 5+6 2 5+6 2 5 2 |
| | 6 2 6 2 3+4 2 5+6 2 5+6 2 5 2 |
| | 5 2 5 2 3+4 2 3+4 2 3+4 2 5 2 |
| | 5 2 5 2 5+6 2 3+4 2 5+6 2 5 2 |
| | 5 2 6 2 3+4 2 3+4 2 3+4 2 5 2 |
| | 5 2 5 2 5+6 2 5+6 2 5+6 2 5 2 |
| | 5 2 5 2 3+4 2 3+4 2 3+4 2 5 2 |
| | 6 2 6 2 5+6 2 5+6 2 5+6 2 5 2 |
| | 5 2 5 2 3+4 2 5+6 2 3+4 2 5 2 |
| | 6 2 6 2 3+4 2 3+4 2 3+4 2 5 2 |
| | 6 2 5 2 3+4 2 3+4 2 3+4 2 5 2 |
| | 5 2 5 2 3+4 2 3+4 2 3+4 2 5 2 |
| | 5 2 5 2 5+6 2 3+4 2 5+6 2 5 2 |
| | 5 2 6 2 5+6 2 5+6 2 3+4 2 5 2 |
| | 5 2 6 2 3+4 2 3+4 2 5+6 2 5 2 |
| | 5 2 5 2 5+6 2 5+6 2 5+6 2 5 2 |
| | 6 2 6 2 3+4 2 3+4 2 3+4 2 5 2 |
| | 6 2 5 2 5+6 2 5+6 2 5+6 2 5 2 |
| 71.1.66 | 71415 96 Ba1Ni1O5Tm2 |
| | -GM -X -T -R -S -W |
| | 5 2 6 2 3+4 2 3+4 2 3+4 2 5 2 |

Continued on next column

**TABLE XLIV– continued**

| TQC | ICSD NoE SF |
|---|---|
| | 6 2 5 2 5+6 2 5+6 2 5+6 2 5 2 |
| | 5 2 5 2 3+4 2 5+6 2 5+6 2 5 2 |
| | 5 2 6 2 5+6 2 3+4 2 3+4 2 5 2 |
| | 6 2 5 2 3+4 2 5+6 2 5+6 2 5 2 |
| | 5 2 6 2 3+4 2 3+4 2 3+4 2 5 2 |
| | 6 2 5 2 5+6 2 5+6 2 5+6 2 5 2 |
| | 5 2 6 2 3+4 2 3+4 2 3+4 2 5 2 |
| | 6 2 5 2 5+6 2 5+6 2 5+6 2 5 2 |
| | 5 2 5 2 3+4 2 3+4 2 3+4 2 5 2 |
| | 5 2 6 2 5+6 2 3+4 2 3+4 2 5 2 |
| | 6 2 5 2 3+4 2 5+6 2 5+6 2 5 2 |
| | 6 2 5 2 5+6 2 3+4 2 5+6 2 5 2 |
| | 5 2 6 2 3+4 2 5+6 2 3+4 2 5 2 |
| | 6 2 6 2 5+6 2 3+4 2 3+4 2 5 2 |
| | 6 2 6 2 5+6 2 3+4 2 3+4 2 5 2 |
| | 6 2 6 2 5+6 2 3+4 2 3+4 2 5 2 |
| | 6 2 5 2 5+6 2 5+6 2 5+6 2 5 2 |
| | 5 2 6 2 5+6 2 5+6 2 5+6 2 5 2 |
| | 6 2 6 2 3+4 2 3+4 2 5+6 2 5 2 |
| | 5 2 6 2 5+6 2 5+6 2 5+6 2 5 2 |
| | 5 2 5 2 5+6 2 3+4 2 3+4 2 5 2 |
| | 5 2 6 2 3+4 2 5+6 2 5+6 2 5 2 |
| | 6 2 6 2 5+6 2 5+6 2 3+4 2 5 2 |
| | 6 2 6 2 3+4 2 3+4 2 5+6 2 5 2 |
| | 5 2 5 2 5+6 2 3+4 2 5+6 2 5 2 |
| | 6 2 5 2 3+4 2 5+6 2 3+4 2 5 2 |
| | 6 2 5 2 5+6 2 3+4 2 3+4 2 5 2 |
| | 6 2 6 2 3+4 2 5+6 2 5+6 2 5 2 |
| | 6 2 6 2 5+6 2 3+4 2 5+6 2 5 2 |
| | 5 2 5 2 3+4 2 5+6 2 3+4 2 5 2 |
| | 6 2 6 2 5+6 2 5+6 2 3+4 2 5 2 |
| | 5 2 5 2 5+6 2 3+4 2 5+6 2 5 2 |
| | 5 2 5 2 3+4 2 3+4 2 3+4 2 5 2 |
| | 5 2 6 2 5+6 2 3+4 2 5+6 2 5 2 |
| | 5 2 5 2 3+4 2 5+6 2 3+4 2 5 2 |
| | 5 2 5 2 5+6 2 3+4 2 5+6 2 5 2 |
| | 6 2 6 2 3+4 2 5+6 2 3+4 2 5 2 |
| | 5 2 5 2 5+6 2 5+6 2 5+6 2 5 2 |
| | 6 2 6 2 5+6 2 3+4 2 5+6 2 5 2 |
| | 6 2 5 2 5+6 2 3+4 2 5+6 2 5 2 |
| | 5 2 5 2 5+6 2 3+4 2 5+6 2 5 2 |
| | 5 2 5 2 3+4 2 3+4 2 3+4 2 5 2 |
| | 5 2 6 2 3+4 2 5+6 2 5+6 2 5 2 |
| | 5 2 6 2 5+6 2 3+4 2 3+4 2 5 2 |
| | 5 2 5 2 3+4 2 5+6 2 3+4 2 5 2 |
| | 6 2 6 2 5+6 2 3+4 2 3+4 2 5 2 |
| | 6 2 5 2 3+4 2 5+6 2 5+6 2 5 2 |
| 71.1.67 | 67831 96 Ba1Ni1O5Tm2 |
| | -GM -X -T -R -S -W |
| | 5 2 6 2 3+4 2 3+4 2 3+4 2 5 2 |
| | 6 2 5 2 5+6 2 5+6 2 5+6 2 5 2 |
| | 5 2 5 2 5+6 2 5+6 2 3+4 2 5 2 |
| | 5 2 6 2 3+4 2 3+4 2 5+6 2 5 2 |
| | 6 2 5 2 5+6 2 5+6 2 3+4 2 5 2 |
| | 5 2 6 2 5+6 2 3+4 2 5+6 2 5 2 |
| | 6 2 5 2 3+4 2 5+6 2 3+4 2 5 2 |
| | 5 2 6 2 5+6 2 3+4 2 5+6 2 5 2 |
| | 6 2 5 2 3+4 2 5+6 2 3+4 2 5 2 |
| | 5 2 5 2 5+6 2 3+4 2 5+6 2 5 2 |
| | 5 2 6 2 3+4 2 3+4 2 5+6 2 5 2 |
| | 6 2 5 2 5+6 2 5+6 2 3+4 2 5 2 |

Continued on next column

**TABLE XLIV– continued**

| TQC | ICSD NoE SF |
|---|---|
| | 6 2 5 2 3+4 2 3+4 2 3+4 2 5 2 |
| | 5 2 6 2 5+6 2 5+6 2 5+6 2 5 2 |
| | 6 2 6 2 3+4 2 3+4 2 5+6 2 5 2 |
| | 6 2 6 2 3+4 2 3+4 2 5+6 2 5 2 |
| | 6 2 6 2 3+4 2 3+4 2 5+6 2 5 2 |
| | 6 2 5 2 3+4 2 5+6 2 3+4 2 5 2 |
| | 5 2 6 2 3+4 2 5+6 2 3+4 2 5 2 |
| | 6 2 6 2 5+6 2 3+4 2 3+4 2 5 2 |
| | 5 2 6 2 3+4 2 5+6 2 3+4 2 5 2 |
| | 5 2 5 2 3+4 2 3+4 2 5+6 2 5 2 |
| | 5 2 6 2 5+6 2 5+6 2 3+4 2 5 2 |
| | 6 2 6 2 3+4 2 5+6 2 5+6 2 5 2 |
| | 6 2 5 2 5+6 2 3+4 2 3+4 2 5 2 |
| | 5 2 6 2 3+4 2 3+4 2 3+4 2 5 2 |
| | 6 2 5 2 5+6 2 5+6 2 5+6 2 5 2 |
| | 6 2 5 2 3+4 2 3+4 2 5+6 2 5 2 |
| | 6 2 6 2 5+6 2 5+6 2 3+4 2 5 2 |
| | 6 2 6 2 3+4 2 3+4 2 3+4 2 5 2 |
| | 5 2 5 2 5+6 2 5+6 2 5+6 2 5 2 |
| | 6 2 6 2 3+4 2 5+6 2 5+6 2 5 2 |
| | 5 2 5 2 3+4 2 3+4 2 3+4 2 5 2 |
| | 5 2 5 2 5+6 2 3+4 2 5+6 2 5 2 |
| | 5 2 6 2 3+4 2 3+4 2 3+4 2 5 2 |
| | 5 2 5 2 5+6 2 5+6 2 5+6 2 5 2 |
| | 5 2 5 2 3+4 2 3+4 2 3+4 2 5 2 |
| | 6 2 6 2 5+6 2 5+6 2 5+6 2 5 2 |
| | 5 2 5 2 3+4 2 5+6 2 3+4 2 5 2 |
| | 6 2 6 2 3+4 2 3+4 2 3+4 2 5 2 |
| | 6 2 5 2 3+4 2 3+4 2 3+4 2 5 2 |
| | 5 2 5 2 3+4 2 3+4 2 3+4 2 5 2 |
| | 5 2 5 2 5+6 2 3+4 2 5+6 2 5 2 |
| | 5 2 6 2 5+6 2 5+6 2 3+4 2 5 2 |
| | 5 2 6 2 3+4 2 3+4 2 5+6 2 5 2 |
| | 5 2 5 2 5+6 2 5+6 2 5+6 2 5 2 |
| | 6 2 6 2 3+4 2 3+4 2 3+4 2 5 2 |
| | 6 2 5 2 5+6 2 5+6 2 5+6 2 5 2 |
| 71.1.68 | 62761 96 Ba1Ni1O5Tm2 |
| | -GM -X -T -R -S -W |
| | 5 2 6 2 3+4 2 3+4 2 3+4 2 5 2 |
| | 6 2 5 2 5+6 2 5+6 2 5+6 2 5 2 |
| | 5 2 5 2 5+6 2 3+4 2 5+6 2 5 2 |
| | 5 2 6 2 3+4 2 5+6 2 3+4 2 5 2 |
| | 6 2 5 2 5+6 2 3+4 2 5+6 2 5 2 |
| | 5 2 6 2 5+6 2 5+6 2 3+4 2 5 2 |
| | 6 2 5 2 3+4 2 3+4 2 5+6 2 5 2 |
| | 5 2 6 2 5+6 2 5+6 2 3+4 2 5 2 |
| | 6 2 5 2 3+4 2 3+4 2 5+6 2 5 2 |
| | 5 2 5 2 5+6 2 5+6 2 3+4 2 5 2 |
| | 5 2 6 2 3+4 2 5+6 2 3+4 2 5 2 |
| | 6 2 5 2 5+6 2 3+4 2 5+6 2 5 2 |
| | 6 2 5 2 3+4 2 5+6 2 5+6 2 5 2 |
| | 5 2 6 2 5+6 2 3+4 2 3+4 2 5 2 |
| | 6 2 6 2 3+4 2 5+6 2 3+4 2 5 2 |
| | 6 2 6 2 3+4 2 5+6 2 3+4 2 5 2 |
| | 6 2 6 2 3+4 2 5+6 2 3+4 2 5 2 |
| | 6 2 5 2 3+4 2 3+4 2 5+6 2 5 2 |
| | 5 2 6 2 3+4 2 3+4 2 5+6 2 5 2 |
| | 6 2 6 2 5+6 2 5+6 2 5+6 2 5 2 |
| | 5 2 6 2 3+4 2 3+4 2 5+6 2 5 2 |
| | 5 2 5 2 3+4 2 5+6 2 3+4 2 5 2 |
| | 5 2 6 2 5+6 2 3+4 2 5+6 2 5 2 |

Continued on next column

**TABLE XLIV– continued**

| TQC | ICSD | NoE | SF |
|---|---|---|---|
| | 6 2 6 2 3+4 2 3+4 2 3+4 2 5 2 | | |
| | 6 2 5 2 5+6 2 5+6 2 5+6 2 5 2 | | |
| | 5 2 6 2 3+4 2 5+6 2 5+6 2 5 2 | | |
| | 6 2 5 2 5+6 2 3+4 2 3+4 2 5 2 | | |
| | 6 2 5 2 3+4 2 5+6 2 3+4 2 5 2 | | |
| | 6 2 6 2 5+6 2 3+4 2 5+6 2 5 2 | | |
| | 6 2 6 2 3+4 2 5+6 2 5+6 2 5 2 | | |
| | 5 2 5 2 5+6 2 3+4 2 3+4 2 5 2 | | |
| | 6 2 6 2 3+4 2 3+4 2 3+4 2 5 2 | | |
| | 5 2 5 2 3+4 2 5+6 2 5+6 2 5 2 | | |
| | 5 2 5 2 5+6 2 5+6 2 3+4 2 5 2 | | |
| | 5 2 6 2 3+4 2 5+6 2 5+6 2 5 2 | | |
| | 5 2 5 2 5+6 2 3+4 2 3+4 2 5 2 | | |
| | 5 2 5 2 3+4 2 5+6 2 5+6 2 5 2 | | |
| | 6 2 6 2 5+6 2 3+4 2 3+4 2 5 2 | | |
| | 5 2 5 2 3+4 2 3+4 2 5+6 2 5 2 | | |
| | 6 2 6 2 3+4 2 5+6 2 5+6 2 5 2 | | |
| | 6 2 5 2 3+4 2 5+6 2 5+6 2 5 2 | | |
| | 5 2 5 2 3+4 2 5+6 2 5+6 2 5 2 | | |
| | 5 2 5 2 5+6 2 5+6 2 3+4 2 5 2 | | |
| | 5 2 6 2 5+6 2 3+4 2 5+6 2 5 2 | | |
| | 5 2 6 2 3+4 2 5+6 2 3+4 2 5 2 | | |
| | 5 2 5 2 5+6 2 3+4 2 3+4 2 5 2 | | |
| | 6 2 6 2 3+4 2 5+6 2 5+6 2 5 2 | | |
| | 6 2 5 2 5+6 2 3+4 2 3+4 2 5 2 | | |
| 71.1.69 | 62426 | 96 | Ba1Ni1O5Tm2 |
| | -GM -X -T -R -S -W | | |
| | 5 2 6 2 3+4 2 3+4 2 3+4 2 5 2 | | |
| | 6 2 5 2 5+6 2 5+6 2 5+6 2 5 2 | | |
| | 5 2 5 2 5+6 2 3+4 2 5+6 2 5 2 | | |
| | 5 2 6 2 3+4 2 5+6 2 3+4 2 5 2 | | |
| | 6 2 5 2 5+6 2 3+4 2 5+6 2 5 2 | | |
| | 5 2 6 2 5+6 2 5+6 2 3+4 2 5 2 | | |
| | 6 2 5 2 3+4 2 3+4 2 5+6 2 5 2 | | |
| | 5 2 6 2 5+6 2 5+6 2 3+4 2 5 2 | | |
| | 6 2 5 2 3+4 2 3+4 2 5+6 2 5 2 | | |
| | 5 2 5 2 5+6 2 5+6 2 3+4 2 5 2 | | |
| | 5 2 6 2 3+4 2 5+6 2 3+4 2 5 2 | | |
| | 6 2 5 2 5+6 2 3+4 2 5+6 2 5 2 | | |
| | 6 2 5 2 3+4 2 5+6 2 5+6 2 5 2 | | |
| | 5 2 6 2 5+6 2 3+4 2 3+4 2 5 2 | | |
| | 6 2 6 2 3+4 2 5+6 2 3+4 2 5 2 | | |
| | 6 2 6 2 3+4 2 5+6 2 3+4 2 5 2 | | |
| | 6 2 6 2 3+4 2 5+6 2 3+4 2 5 2 | | |
| | 6 2 5 2 3+4 2 3+4 2 5+6 2 5 2 | | |
| | 5 2 6 2 3+4 2 3+4 2 5+6 2 5 2 | | |
| | 6 2 6 2 5+6 2 5+6 2 5+6 2 5 2 | | |
| | 5 2 6 2 3+4 2 3+4 2 5+6 2 5 2 | | |
| | 5 2 5 2 3+4 2 5+6 2 3+4 2 5 2 | | |
| | 5 2 6 2 5+6 2 3+4 2 5+6 2 5 2 | | |
| | 6 2 6 2 3+4 2 3+4 2 3+4 2 5 2 | | |
| | 6 2 5 2 5+6 2 5+6 2 5+6 2 5 2 | | |
| | 5 2 6 2 3+4 2 5+6 2 5+6 2 5 2 | | |
| | 6 2 5 2 5+6 2 3+4 2 3+4 2 5 2 | | |
| | 6 2 5 2 3+4 2 5+6 2 3+4 2 5 2 | | |
| | 6 2 6 2 5+6 2 3+4 2 5+6 2 5 2 | | |
| | 6 2 6 2 3+4 2 5+6 2 5+6 2 5 2 | | |
| | 5 2 5 2 5+6 2 3+4 2 3+4 2 5 2 | | |
| | 6 2 6 2 3+4 2 3+4 2 3+4 2 5 2 | | |
| | 5 2 5 2 3+4 2 5+6 2 5+6 2 5 2 | | |
| | 5 2 5 2 5+6 2 5+6 2 3+4 2 5 2 | | |

Continued on next column

**TABLE XLIV– continued**

| TQC | ICSD | NoE | SF |
|---|---|---|---|
| | 5 2 6 2 3+4 2 5+6 2 5+6 2 5 2 | | |
| | 5 2 5 2 5+6 2 3+4 2 3+4 2 5 2 | | |
| | 5 2 5 2 3+4 2 5+6 2 5+6 2 5 2 | | |
| | 6 2 6 2 5+6 2 3+4 2 3+4 2 5 2 | | |
| | 5 2 5 2 3+4 2 3+4 2 5+6 2 5 2 | | |
| | 6 2 6 2 3+4 2 5+6 2 5+6 2 5 2 | | |
| | 6 2 5 2 3+4 2 5+6 2 5+6 2 5 2 | | |
| | 5 2 5 2 3+4 2 5+6 2 5+6 2 5 2 | | |
| | 5 2 5 2 5+6 2 5+6 2 3+4 2 5 2 | | |
| | 5 2 6 2 5+6 2 3+4 2 5+6 2 5 2 | | |
| | 5 2 6 2 3+4 2 5+6 2 3+4 2 5 2 | | |
| | 5 2 5 2 5+6 2 3+4 2 3+4 2 5 2 | | |
| | 6 2 6 2 3+4 2 5+6 2 5+6 2 5 2 | | |
| | 6 2 5 2 5+6 2 3+4 2 3+4 2 5 2 | | |
| 71.1.70 | 66078 | 88 | Ba1Ni1O5Tb2 |
| | -GM -X -T -R -S -W | | |
| | 5 2 6 2 3+4 2 3+4 2 3+4 2 5 2 | | |
| | 6 2 5 2 5+6 2 5+6 2 5+6 2 5 2 | | |
| | 5 2 5 2 3+4 2 3+4 2 3+4 2 5 2 | | |
| | 5 2 6 2 3+4 2 5+6 2 5+6 2 5 2 | | |
| | 6 2 5 2 5+6 2 3+4 2 3+4 2 5 2 | | |
| | 5 2 6 2 3+4 2 5+6 2 5+6 2 5 2 | | |
| | 6 2 5 2 5+6 2 3+4 2 3+4 2 5 2 | | |
| | 5 2 6 2 3+4 2 5+6 2 5+6 2 5 2 | | |
| | 5 2 5 2 5+6 2 3+4 2 3+4 2 5 2 | | |
| | 6 2 5 2 3+4 2 5+6 2 5+6 2 5 2 | | |
| | 6 2 6 2 5+6 2 5+6 2 3+4 2 5 2 | | |
| | 5 2 5 2 3+4 2 3+4 2 5+6 2 5 2 | | |
| | 6 2 5 2 5+6 2 3+4 2 3+4 2 5 2 | | |
| | 5 2 6 2 3+4 2 5+6 2 5+6 2 5 2 | | |
| | 6 2 6 2 5+6 2 5+6 2 5+6 2 5 2 | | |
| | 6 2 6 2 5+6 2 5+6 2 5+6 2 5 2 | | |
| | 6 2 6 2 5+6 2 5+6 2 5+6 2 5 2 | | |
| | 6 2 5 2 5+6 2 3+4 2 3+4 2 5 2 | | |
| | 5 2 6 2 5+6 2 3+4 2 3+4 2 5 2 | | |
| | 6 2 6 2 3+4 2 5+6 2 3+4 2 5 2 | | |
| | 5 2 6 2 5+6 2 3+4 2 3+4 2 5 2 | | |
| | 5 2 6 2 5+6 2 3+4 2 5+6 2 5 2 | | |
| | 5 2 5 2 3+4 2 5+6 2 3+4 2 5 2 | | |
| | 6 2 6 2 3+4 2 3+4 2 5+6 2 5 2 | | |
| | 6 2 6 2 5+6 2 5+6 2 3+4 2 5 2 | | |
| | 5 2 6 2 3+4 2 3+4 2 5+6 2 5 2 | | |
| | 6 2 5 2 5+6 2 5+6 2 5+6 2 5 2 | | |
| | 6 2 5 2 5+6 2 3+4 2 3+4 2 5 2 | | |
| | 6 2 6 2 5+6 2 5+6 2 3+4 2 5 2 | | |
| | 5 2 5 2 3+4 2 5+6 2 3+4 2 5 2 | | |
| | 6 2 5 2 3+4 2 3+4 2 5+6 2 5 2 | | |
| | 6 2 6 2 5+6 2 3+4 2 5+6 2 5 2 | | |
| | 5 2 5 2 5+6 2 5+6 2 3+4 2 5 2 | | |
| | 5 2 5 2 3+4 2 5+6 2 3+4 2 5 2 | | |
| | 5 2 5 2 5+6 2 5+6 2 5+6 2 5 2 | | |
| | 5 2 6 2 3+4 2 3+4 2 3+4 2 5 2 | | |
| | 5 2 5 2 5+6 2 5+6 2 5+6 2 5 2 | | |
| | 6 2 6 2 5+6 2 5+6 2 3+4 2 5 2 | | |
| | 5 2 5 2 3+4 2 3+4 2 5+6 2 5 2 | | |
| | 6 2 5 2 5+6 2 5+6 2 3+4 2 5 2 | | |
| | 5 2 6 2 5+6 2 3+4 2 3+4 2 5 2 | | |
| | 6 2 5 2 5+6 2 5+6 2 3+4 2 5 2 | | |
| | 5 2 5 2 3+4 2 5+6 2 3+4 2 5 2 | | |
| | 5 2 6 2 3+4 2 3+4 2 5+6 2 5 2 | | |
| 71.1.71 | 75035 | 76 | Ba1Ni1O5Pr2 |

Continued on next column

**TABLE XLIV– continued**

| TQC | ICSD | NoE | SF | | | |
|---|---|---|---|---|---|---|
| | -GM | -X | -T | -R | -S | -W |
| | 5 2 | 6 2 | 3+4 2 | 5+6 2 | 5+6 2 | 5 2 |
| | 6 2 | 5 2 | 5+6 2 | 3+4 2 | 3+4 2 | 5 2 |
| | 5 2 | 5 2 | 3+4 2 | 3+4 2 | 3+4 2 | 5 2 |
| | 5 2 | 5 2 | 3+4 2 | 5+6 2 | 5+6 2 | 5 2 |
| | 6 2 | 6 2 | 5+6 2 | 3+4 2 | 3+4 2 | 5 2 |
| | 5 2 | 6 2 | 3+4 2 | 5+6 2 | 5+6 2 | 5 2 |
| | 5 2 | 5 2 | 5+6 2 | 3+4 2 | 3+4 2 | 5 2 |
| | 6 2 | 5 2 | 3+4 2 | 5+6 2 | 5+6 2 | 5 2 |
| | 6 2 | 6 2 | 5+6 2 | 5+6 2 | 3+4 2 | 5 2 |
| | 5 2 | 5 2 | 3+4 2 | 3+4 2 | 5+6 2 | 5 2 |
| | 6 2 | 6 2 | 5+6 2 | 3+4 2 | 3+4 2 | 5 2 |
| | 5 2 | 5 2 | 3+4 2 | 5+6 2 | 5+6 2 | 5 2 |
| | 6 2 | 6 2 | 5+6 2 | 3+4 2 | 3+4 2 | 5 2 |
| | 5 2 | 5 2 | 3+4 2 | 5+6 2 | 5+6 2 | 5 2 |
| | 6 2 | 6 2 | 5+6 2 | 5+6 2 | 5+6 2 | 5 2 |
| | 6 2 | 6 2 | 5+6 2 | 5+6 2 | 5+6 2 | 5 2 |
| | 6 2 | 6 2 | 5+6 2 | 5+6 2 | 5+6 2 | 5 2 |
| | 6 2 | 5 2 | 5+6 2 | 3+4 2 | 3+4 2 | 5 2 |
| | 5 2 | 6 2 | 5+6 2 | 3+4 2 | 3+4 2 | 5 2 |
| | 6 2 | 6 2 | 3+4 2 | 5+6 2 | 3+4 2 | 5 2 |
| | 5 2 | 6 2 | 5+6 2 | 3+4 2 | 3+4 2 | 5 2 |
| | 5 2 | 5 2 | 3+4 2 | 3+4 2 | 3+4 2 | 5 2 |
| | 5 2 | 6 2 | 5+6 2 | 3+4 2 | 5+6 2 | 5 2 |
| | 6 2 | 5 2 | 3+4 2 | 5+6 2 | 5+6 2 | 5 2 |
| | 6 2 | 5 2 | 5+6 2 | 5+6 2 | 3+4 2 | 5 2 |
| | 6 2 | 6 2 | 3+4 2 | 5+6 2 | 3+4 2 | 5 2 |
| | 5 2 | 5 2 | 5+6 2 | 3+4 2 | 3+4 2 | 5 2 |
| | 6 2 | 6 2 | 5+6 2 | 5+6 2 | 5+6 2 | 5 2 |
| | 6 2 | 5 2 | 3+4 2 | 5+6 2 | 5+6 2 | 5 2 |
| | 5 2 | 6 2 | 5+6 2 | 3+4 2 | 3+4 2 | 5 2 |
| | 6 2 | 6 2 | 3+4 2 | 3+4 2 | 5+6 2 | 5 2 |
| | 6 2 | 6 2 | 5+6 2 | 3+4 2 | 5+6 2 | 5 2 |
| | 5 2 | 5 2 | 5+6 2 | 5+6 2 | 3+4 2 | 5 2 |
| | 5 2 | 5 2 | 5+6 2 | 5+6 2 | 3+4 2 | 5 2 |
| | 5 2 | 5 2 | 5+6 2 | 5+6 2 | 3+4 2 | 5 2 |
| | 5 2 | 5 2 | 5+6 2 | 5+6 2 | 3+4 2 | 5 2 |
| | 5 2 | 5 2 | 3+4 2 | 5+6 2 | 3+4 2 | 5 2 |
| | 5 2 | 5 2 | 5+6 2 | 3+4 2 | 5+6 2 | 5 2 |
| 71.1.72 | 72626 | 78 | Ba1Nd2Ni1O5 | | | |
| | -GM | -X | -T | -R | -S | -W |
| | 5 2 | 6 2 | 5+6 2 | 3+4 2 | 5+6 2 | 5 2 |
| | 6 2 | 5 2 | 3+4 2 | 5+6 2 | 3+4 2 | 5 2 |
| | 5 2 | 5 2 | 5+6 2 | 5+6 2 | 3+4 2 | 5 2 |
| | 5 2 | 5 2 | 5+6 2 | 3+4 2 | 5+6 2 | 5 2 |
| | 6 2 | 6 2 | 3+4 2 | 5+6 2 | 3+4 2 | 5 2 |
| | 5 2 | 5 2 | 5+6 2 | 3+4 2 | 5+6 2 | 5 2 |
| | 6 2 | 6 2 | 3+4 2 | 5+6 2 | 3+4 2 | 5 2 |
| | 5 2 | 5 2 | 5+6 2 | 3+4 2 | 5+6 2 | 5 2 |
| | 5 2 | 6 2 | 3+4 2 | 3+4 2 | 5+6 2 | 5 2 |
| | 6 2 | 5 2 | 5+6 2 | 5+6 2 | 3+4 2 | 5 2 |
| | 6 2 | 6 2 | 3+4 2 | 5+6 2 | 3+4 2 | 5 2 |
| | 5 2 | 5 2 | 5+6 2 | 3+4 2 | 5+6 2 | 5 2 |
| | 6 2 | 5 2 | 3+4 2 | 5+6 2 | 3+4 2 | 5 2 |
| | 5 2 | 6 2 | 5+6 2 | 3+4 2 | 5+6 2 | 5 2 |
| | 6 2 | 6 2 | 3+4 2 | 3+4 2 | 5+6 2 | 5 2 |
| | 6 2 | 6 2 | 3+4 2 | 3+4 2 | 5+6 2 | 5 2 |
| | 6 2 | 6 2 | 3+4 2 | 3+4 2 | 5+6 2 | 5 2 |
| | 6 2 | 5 2 | 3+4 2 | 5+6 2 | 3+4 2 | 5 2 |
| | 5 2 | 6 2 | 3+4 2 | 5+6 2 | 3+4 2 | 5 2 |
| | 6 2 | 6 2 | 5+6 2 | 3+4 2 | 3+4 2 | 5 2 |

Continued on next column

**TABLE XLIV– continued**

| TQC | ICSD | NoE | SF | | | |
|---|---|---|---|---|---|---|
| | 5 2 | 6 2 | 3+4 2 | 5+6 2 | 3+4 2 | 5 2 |
| | 5 2 | 5 2 | 3+4 2 | 5+6 2 | 3+4 2 | 5 2 |
| | 5 2 | 6 2 | 5+6 2 | 3+4 2 | 5+6 2 | 5 2 |
| | 6 2 | 6 2 | 5+6 2 | 5+6 2 | 5+6 2 | 5 2 |
| | 6 2 | 5 2 | 3+4 2 | 3+4 2 | 3+4 2 | 5 2 |
| | 6 2 | 5 2 | 5+6 2 | 3+4 2 | 3+4 2 | 5 2 |
| | 5 2 | 5 2 | 3+4 2 | 5+6 2 | 3+4 2 | 5 2 |
| | 6 2 | 6 2 | 3+4 2 | 3+4 2 | 5+6 2 | 5 2 |
| | 6 2 | 6 2 | 5+6 2 | 5+6 2 | 5+6 2 | 5 2 |
| | 5 2 | 5 2 | 3+4 2 | 3+4 2 | 3+4 2 | 5 2 |
| | 6 2 | 6 2 | 5+6 2 | 5+6 2 | 5+6 2 | 5 2 |
| | 6 2 | 6 2 | 3+4 2 | 5+6 2 | 5+6 2 | 5 2 |
| | 5 2 | 5 2 | 3+4 2 | 3+4 2 | 3+4 2 | 5 2 |
| | 5 2 | 5 2 | 3+4 2 | 3+4 2 | 3+4 2 | 5 2 |
| | 5 2 | 5 2 | 3+4 2 | 3+4 2 | 3+4 2 | 5 2 |
| | 5 2 | 5 2 | 3+4 2 | 3+4 2 | 3+4 2 | 5 2 |
| | 5 2 | 5 2 | 5+6 2 | 3+4 2 | 3+4 2 | 5 2 |
| | 5 2 | 6 2 | 3+4 2 | 5+6 2 | 5+6 2 | 5 2 |
| | 5 2 | 5 2 | 5+6 2 | 3+4 2 | 5+6 2 | 5 2 |
| 71.1.73 | 62627 | 78 | Ba1Nd2Ni1O5 | | | |
| | -GM | -X | -T | -R | -S | -W |
| | 5 2 | 6 2 | 3+4 2 | 5+6 2 | 5+6 2 | 5 2 |
| | 6 2 | 5 2 | 5+6 2 | 3+4 2 | 3+4 2 | 5 2 |
| | 5 2 | 5 2 | 3+4 2 | 3+4 2 | 3+4 2 | 5 2 |
| | 5 2 | 5 2 | 3+4 2 | 5+6 2 | 5+6 2 | 5 2 |
| | 6 2 | 6 2 | 5+6 2 | 3+4 2 | 3+4 2 | 5 2 |
| | 5 2 | 5 2 | 3+4 2 | 5+6 2 | 5+6 2 | 5 2 |
| | 6 2 | 6 2 | 5+6 2 | 3+4 2 | 3+4 2 | 5 2 |
| | 5 2 | 5 2 | 3+4 2 | 5+6 2 | 5+6 2 | 5 2 |
| | 5 2 | 6 2 | 5+6 2 | 5+6 2 | 5+6 2 | 5 2 |
| | 6 2 | 5 2 | 3+4 2 | 3+4 2 | 3+4 2 | 5 2 |
| | 6 2 | 6 2 | 5+6 2 | 3+4 2 | 3+4 2 | 5 2 |
| | 5 2 | 5 2 | 3+4 2 | 5+6 2 | 5+6 2 | 5 2 |
| | 6 2 | 5 2 | 5+6 2 | 3+4 2 | 3+4 2 | 5 2 |
| | 5 2 | 6 2 | 3+4 2 | 5+6 2 | 5+6 2 | 5 2 |
| | 6 2 | 6 2 | 5+6 2 | 5+6 2 | 5+6 2 | 5 2 |
| | 6 2 | 6 2 | 5+6 2 | 5+6 2 | 5+6 2 | 5 2 |
| | 6 2 | 6 2 | 5+6 2 | 5+6 2 | 5+6 2 | 5 2 |
| | 6 2 | 5 2 | 5+6 2 | 3+4 2 | 3+4 2 | 5 2 |
| | 6 2 | 6 2 | 5+6 2 | 3+4 2 | 3+4 2 | 5 2 |
| | 5 2 | 6 2 | 3+4 2 | 5+6 2 | 3+4 2 | 5 2 |
| | 5 2 | 6 2 | 5+6 2 | 3+4 2 | 3+4 2 | 5 2 |
| | 5 2 | 5 2 | 3+4 2 | 3+4 2 | 3+4 2 | 5 2 |
| | 5 2 | 6 2 | 5+6 2 | 3+4 2 | 5+6 2 | 5 2 |
| | 6 2 | 6 2 | 3+4 2 | 5+6 2 | 5+6 2 | 5 2 |
| | 6 2 | 5 2 | 5+6 2 | 5+6 2 | 3+4 2 | 5 2 |
| | 6 2 | 5 2 | 3+4 2 | 5+6 2 | 3+4 2 | 5 2 |
| | 5 2 | 5 2 | 5+6 2 | 3+4 2 | 5+6 2 | 5 2 |
| | 6 2 | 6 2 | 5+6 2 | 5+6 2 | 3+4 2 | 5 2 |
| | 6 2 | 6 2 | 3+4 2 | 3+4 2 | 5+6 2 | 5 2 |
| | 5 2 | 5 2 | 5+6 2 | 5+6 2 | 3+4 2 | 5 2 |
| | 6 2 | 6 2 | 3+4 2 | 3+4 2 | 5+6 2 | 5 2 |
| | 6 2 | 6 2 | 5+6 2 | 3+4 2 | 5+6 2 | 5 2 |
| | 5 2 | 5 2 | 5+6 2 | 5+6 2 | 3+4 2 | 5 2 |
| | 5 2 | 5 2 | 5+6 2 | 5+6 2 | 3+4 2 | 5 2 |
| | 5 2 | 5 2 | 5+6 2 | 5+6 2 | 3+4 2 | 5 2 |
| | 5 2 | 5 2 | 5+6 2 | 5+6 2 | 3+4 2 | 5 2 |
| | 5 2 | 5 2 | 3+4 2 | 5+6 2 | 5+6 2 | 5 2 |
| | 5 2 | 6 2 | 5+6 2 | 3+4 2 | 3+4 2 | 5 2 |
| | 5 2 | 5 2 | 3+4 2 | 5+6 2 | 5+6 2 | 5 2 |
| 71.1.74 | 68425 | 86 | Ba1Gd2Ni1O5 | | | |

Continued on next column

**TABLE XLIV– continued**

| TQC | ICSD | | NoE | | SF | | | | | | | |
|---|---|---|---|---|---|---|---|---|---|---|---|---|
| | -GM | | -X | | -T | | -R | | -S | | -W | |
| | 5 | 2 | 6 | 2 | 3+4 | 2 | 3+4 | 2 | 3+4 | 2 | 5 | 2 |
| | 6 | 2 | 5 | 2 | 5+6 | 2 | 5+6 | 2 | 5+6 | 2 | 5 | 2 |
| | 5 | 2 | 5 | 2 | 5+6 | 2 | 3+4 | 2 | 5+6 | 2 | 5 | 2 |
| | 5 | 2 | 6 | 2 | 5+6 | 2 | 5+6 | 2 | 3+4 | 2 | 5 | 2 |
| | 6 | 2 | 5 | 2 | 3+4 | 2 | 3+4 | 2 | 5+6 | 2 | 5 | 2 |
| | 5 | 2 | 5 | 2 | 5+6 | 2 | 5+6 | 2 | 3+4 | 2 | 5 | 2 |
| | 6 | 2 | 6 | 2 | 3+4 | 2 | 3+4 | 2 | 5+6 | 2 | 5 | 2 |
| | 5 | 2 | 6 | 2 | 5+6 | 2 | 5+6 | 2 | 3+4 | 2 | 5 | 2 |
| | 6 | 2 | 5 | 2 | 3+4 | 2 | 3+4 | 2 | 5+6 | 2 | 5 | 2 |
| | 5 | 2 | 5 | 2 | 5+6 | 2 | 5+6 | 2 | 3+4 | 2 | 5 | 2 |
| | 6 | 2 | 6 | 2 | 3+4 | 2 | 5+6 | 2 | 5+6 | 2 | 5 | 2 |
| | 5 | 2 | 5 | 2 | 5+6 | 2 | 3+4 | 2 | 3+4 | 2 | 5 | 2 |
| | 6 | 2 | 5 | 2 | 3+4 | 2 | 3+4 | 2 | 5+6 | 2 | 5 | 2 |
| | 5 | 2 | 6 | 2 | 5+6 | 2 | 5+6 | 2 | 3+4 | 2 | 5 | 2 |
| | 6 | 2 | 6 | 2 | 3+4 | 2 | 5+6 | 2 | 3+4 | 2 | 5 | 2 |
| | 6 | 2 | 6 | 2 | 3+4 | 2 | 5+6 | 2 | 3+4 | 2 | 5 | 2 |
| | 6 | 2 | 6 | 2 | 3+4 | 2 | 5+6 | 2 | 3+4 | 2 | 5 | 2 |
| | 6 | 2 | 5 | 2 | 3+4 | 2 | 3+4 | 2 | 5+6 | 2 | 5 | 2 |
| | 6 | 2 | 6 | 2 | 3+4 | 2 | 3+4 | 2 | 5+6 | 2 | 5 | 2 |
| | 5 | 2 | 6 | 2 | 5+6 | 2 | 5+6 | 2 | 5+6 | 2 | 5 | 2 |
| | 5 | 2 | 6 | 2 | 3+4 | 2 | 3+4 | 2 | 5+6 | 2 | 5 | 2 |
| | 5 | 2 | 5 | 2 | 3+4 | 2 | 5+6 | 2 | 5+6 | 2 | 5 | 2 |
| | 5 | 2 | 6 | 2 | 5+6 | 2 | 3+4 | 2 | 3+4 | 2 | 5 | 2 |
| | 6 | 2 | 6 | 2 | 5+6 | 2 | 3+4 | 2 | 3+4 | 2 | 5 | 2 |
| | 6 | 2 | 5 | 2 | 3+4 | 2 | 5+6 | 2 | 5+6 | 2 | 5 | 2 |
| | 5 | 2 | 5 | 2 | 3+4 | 2 | 5+6 | 2 | 5+6 | 2 | 5 | 2 |
| | 6 | 2 | 6 | 2 | 5+6 | 2 | 3+4 | 2 | 3+4 | 2 | 5 | 2 |
| | 6 | 2 | 5 | 2 | 3+4 | 2 | 5+6 | 2 | 5+6 | 2 | 5 | 2 |
| | 6 | 2 | 6 | 2 | 5+6 | 2 | 3+4 | 2 | 3+4 | 2 | 5 | 2 |
| | 6 | 2 | 6 | 2 | 3+4 | 2 | 5+6 | 2 | 5+6 | 2 | 5 | 2 |
| | 5 | 2 | 5 | 2 | 5+6 | 2 | 3+4 | 2 | 3+4 | 2 | 5 | 2 |
| | 6 | 2 | 6 | 2 | 3+4 | 2 | 3+4 | 2 | 3+4 | 2 | 5 | 2 |
| | 5 | 2 | 5 | 2 | 3+4 | 2 | 5+6 | 2 | 5+6 | 2 | 5 | 2 |
| | 5 | 2 | 5 | 2 | 3+4 | 2 | 5+6 | 2 | 5+6 | 2 | 5 | 2 |
| | 5 | 2 | 5 | 2 | 5+6 | 2 | 5+6 | 2 | 3+4 | 2 | 5 | 2 |
| | 5 | 2 | 5 | 2 | 3+4 | 2 | 5+6 | 2 | 5+6 | 2 | 5 | 2 |
| | 5 | 2 | 6 | 2 | 5+6 | 2 | 3+4 | 2 | 3+4 | 2 | 5 | 2 |
| | 5 | 2 | 5 | 2 | 3+4 | 2 | 5+6 | 2 | 5+6 | 2 | 5 | 2 |
| | 5 | 2 | 6 | 2 | 5+6 | 2 | 3+4 | 2 | 3+4 | 2 | 5 | 2 |
| | 6 | 2 | 6 | 2 | 3+4 | 2 | 5+6 | 2 | 5+6 | 2 | 5 | 2 |
| | 6 | 2 | 5 | 2 | 3+4 | 2 | 3+4 | 2 | 5+6 | 2 | 5 | 2 |
| | 6 | 2 | 5 | 2 | 3+4 | 2 | 5+6 | 2 | 5+6 | 2 | 5 | 2 |
| | 5 | 2 | 5 | 2 | 5+6 | 2 | 5+6 | 2 | 5+6 | 2 | 5 | 2 |
| 71.1.75 | 62758 | | 86 | | Ba1Gd2Ni1O5 | | | | | | | |
| | -GM | | -X | | -T | | -R | | -S | | -W | |
| | 5 | 2 | 6 | 2 | 3+4 | 2 | 3+4 | 2 | 3+4 | 2 | 5 | 2 |
| | 6 | 2 | 5 | 2 | 5+6 | 2 | 5+6 | 2 | 5+6 | 2 | 5 | 2 |
| | 5 | 2 | 5 | 2 | 3+4 | 2 | 3+4 | 2 | 3+4 | 2 | 5 | 2 |
| | 5 | 2 | 6 | 2 | 3+4 | 2 | 5+6 | 2 | 5+6 | 2 | 5 | 2 |
| | 6 | 2 | 5 | 2 | 5+6 | 2 | 3+4 | 2 | 3+4 | 2 | 5 | 2 |
| | 5 | 2 | 5 | 2 | 3+4 | 2 | 5+6 | 2 | 5+6 | 2 | 5 | 2 |
| | 6 | 2 | 6 | 2 | 5+6 | 2 | 3+4 | 2 | 3+4 | 2 | 5 | 2 |
| | 5 | 2 | 6 | 2 | 3+4 | 2 | 5+6 | 2 | 5+6 | 2 | 5 | 2 |
| | 6 | 2 | 5 | 2 | 5+6 | 2 | 3+4 | 2 | 3+4 | 2 | 5 | 2 |
| | 5 | 2 | 5 | 2 | 3+4 | 2 | 5+6 | 2 | 5+6 | 2 | 5 | 2 |
| | 6 | 2 | 6 | 2 | 5+6 | 2 | 5+6 | 2 | 3+4 | 2 | 5 | 2 |
| | 5 | 2 | 5 | 2 | 3+4 | 2 | 3+4 | 2 | 5+6 | 2 | 5 | 2 |
| | 6 | 2 | 5 | 2 | 5+6 | 2 | 3+4 | 2 | 3+4 | 2 | 5 | 2 |
| | 5 | 2 | 6 | 2 | 3+4 | 2 | 5+6 | 2 | 5+6 | 2 | 5 | 2 |
| | 6 | 2 | 6 | 2 | 5+6 | 2 | 5+6 | 2 | 5+6 | 2 | 5 | 2 |

Continued on next column

**TABLE XLIV– continued**

| TQC | ICSD | | NoE | | SF | | | | | | | |
|---|---|---|---|---|---|---|---|---|---|---|---|---|
| | 6 | 2 | 6 | 2 | 5+6 | 2 | 5+6 | 2 | 5+6 | 2 | 5 | 2 |
| | 6 | 2 | 6 | 2 | 5+6 | 2 | 5+6 | 2 | 5+6 | 2 | 5 | 2 |
| | 6 | 2 | 5 | 2 | 5+6 | 2 | 3+4 | 2 | 3+4 | 2 | 5 | 2 |
| | 6 | 2 | 6 | 2 | 5+6 | 2 | 3+4 | 2 | 3+4 | 2 | 5 | 2 |
| | 5 | 2 | 6 | 2 | 3+4 | 2 | 5+6 | 2 | 3+4 | 2 | 5 | 2 |
| | 5 | 2 | 6 | 2 | 5+6 | 2 | 3+4 | 2 | 3+4 | 2 | 5 | 2 |
| | 5 | 2 | 5 | 2 | 5+6 | 2 | 5+6 | 2 | 3+4 | 2 | 5 | 2 |
| | 5 | 2 | 6 | 2 | 3+4 | 2 | 3+4 | 2 | 5+6 | 2 | 5 | 2 |
| | 6 | 2 | 6 | 2 | 3+4 | 2 | 3+4 | 2 | 5+6 | 2 | 5 | 2 |
| | 6 | 2 | 5 | 2 | 5+6 | 2 | 5+6 | 2 | 3+4 | 2 | 5 | 2 |
| | 5 | 2 | 6 | 2 | 5+6 | 2 | 5+6 | 2 | 3+4 | 2 | 5 | 2 |
| | 6 | 2 | 5 | 2 | 3+4 | 2 | 3+4 | 2 | 5+6 | 2 | 5 | 2 |
| | 6 | 2 | 5 | 2 | 5+6 | 2 | 5+6 | 2 | 3+4 | 2 | 5 | 2 |
| | 6 | 2 | 6 | 2 | 5+6 | 2 | 3+4 | 2 | 5+6 | 2 | 5 | 2 |
| | 6 | 2 | 6 | 2 | 3+4 | 2 | 5+6 | 2 | 3+4 | 2 | 5 | 2 |
| | 5 | 2 | 5 | 2 | 3+4 | 2 | 3+4 | 2 | 5+6 | 2 | 5 | 2 |
| | 6 | 2 | 6 | 2 | 5+6 | 2 | 3+4 | 2 | 5+6 | 2 | 5 | 2 |
| | 5 | 2 | 5 | 2 | 5+6 | 2 | 5+6 | 2 | 3+4 | 2 | 5 | 2 |
| | 5 | 2 | 5 | 2 | 5+6 | 2 | 5+6 | 2 | 3+4 | 2 | 5 | 2 |
| | 5 | 2 | 5 | 2 | 3+4 | 2 | 5+6 | 2 | 5+6 | 2 | 5 | 2 |
| | 5 | 2 | 5 | 2 | 5+6 | 2 | 5+6 | 2 | 3+4 | 2 | 5 | 2 |
| | 5 | 2 | 6 | 2 | 3+4 | 2 | 3+4 | 2 | 5+6 | 2 | 5 | 2 |
| | 5 | 2 | 5 | 2 | 5+6 | 2 | 5+6 | 2 | 3+4 | 2 | 5 | 2 |
| | 5 | 2 | 6 | 2 | 3+4 | 2 | 3+4 | 2 | 5+6 | 2 | 5 | 2 |
| | 6 | 2 | 6 | 2 | 5+6 | 2 | 5+6 | 2 | 3+4 | 2 | 5 | 2 |
| | 6 | 2 | 5 | 2 | 5+6 | 2 | 3+4 | 2 | 3+4 | 2 | 5 | 2 |
| | 6 | 2 | 5 | 2 | 5+6 | 2 | 5+6 | 2 | 3+4 | 2 | 5 | 2 |
| | 5 | 2 | 5 | 2 | 3+4 | 2 | 5+6 | 2 | 3+4 | 2 | 5 | 2 |
| 71.1.76 | 62423 | | 86 | | Ba1Gd2Ni1O5 | | | | | | | |
| | -GM | | -X | | -T | | -R | | -S | | -W | |
| | 5 | 2 | 6 | 2 | 3+4 | 2 | 3+4 | 2 | 3+4 | 2 | 5 | 2 |
| | 6 | 2 | 5 | 2 | 5+6 | 2 | 5+6 | 2 | 5+6 | 2 | 5 | 2 |
| | 5 | 2 | 5 | 2 | 3+4 | 2 | 3+4 | 2 | 3+4 | 2 | 5 | 2 |
| | 5 | 2 | 6 | 2 | 3+4 | 2 | 5+6 | 2 | 5+6 | 2 | 5 | 2 |
| | 6 | 2 | 5 | 2 | 5+6 | 2 | 3+4 | 2 | 3+4 | 2 | 5 | 2 |
| | 5 | 2 | 5 | 2 | 3+4 | 2 | 5+6 | 2 | 5+6 | 2 | 5 | 2 |
| | 6 | 2 | 6 | 2 | 5+6 | 2 | 3+4 | 2 | 3+4 | 2 | 5 | 2 |
| | 5 | 2 | 6 | 2 | 3+4 | 2 | 5+6 | 2 | 5+6 | 2 | 5 | 2 |
| | 6 | 2 | 5 | 2 | 5+6 | 2 | 3+4 | 2 | 3+4 | 2 | 5 | 2 |
| | 5 | 2 | 5 | 2 | 3+4 | 2 | 5+6 | 2 | 5+6 | 2 | 5 | 2 |
| | 6 | 2 | 6 | 2 | 5+6 | 2 | 5+6 | 2 | 3+4 | 2 | 5 | 2 |
| | 5 | 2 | 5 | 2 | 3+4 | 2 | 3+4 | 2 | 5+6 | 2 | 5 | 2 |
| | 6 | 2 | 5 | 2 | 5+6 | 2 | 3+4 | 2 | 3+4 | 2 | 5 | 2 |
| | 5 | 2 | 6 | 2 | 3+4 | 2 | 5+6 | 2 | 5+6 | 2 | 5 | 2 |
| | 6 | 2 | 6 | 2 | 5+6 | 2 | 5+6 | 2 | 5+6 | 2 | 5 | 2 |
| | 6 | 2 | 6 | 2 | 5+6 | 2 | 5+6 | 2 | 5+6 | 2 | 5 | 2 |
| | 6 | 2 | 6 | 2 | 5+6 | 2 | 5+6 | 2 | 5+6 | 2 | 5 | 2 |
| | 6 | 2 | 5 | 2 | 5+6 | 2 | 3+4 | 2 | 3+4 | 2 | 5 | 2 |
| | 6 | 2 | 6 | 2 | 5+6 | 2 | 3+4 | 2 | 3+4 | 2 | 5 | 2 |
| | 5 | 2 | 6 | 2 | 3+4 | 2 | 5+6 | 2 | 3+4 | 2 | 5 | 2 |
| | 5 | 2 | 6 | 2 | 5+6 | 2 | 3+4 | 2 | 3+4 | 2 | 5 | 2 |
| | 5 | 2 | 5 | 2 | 5+6 | 2 | 5+6 | 2 | 3+4 | 2 | 5 | 2 |
| | 5 | 2 | 6 | 2 | 3+4 | 2 | 3+4 | 2 | 5+6 | 2 | 5 | 2 |
| | 6 | 2 | 6 | 2 | 3+4 | 2 | 3+4 | 2 | 5+6 | 2 | 5 | 2 |
| | 6 | 2 | 5 | 2 | 5+6 | 2 | 5+6 | 2 | 3+4 | 2 | 5 | 2 |
| | 5 | 2 | 6 | 2 | 5+6 | 2 | 5+6 | 2 | 3+4 | 2 | 5 | 2 |
| | 6 | 2 | 5 | 2 | 3+4 | 2 | 3+4 | 2 | 5+6 | 2 | 5 | 2 |
| | 6 | 2 | 5 | 2 | 5+6 | 2 | 5+6 | 2 | 3+4 | 2 | 5 | 2 |
| | 6 | 2 | 6 | 2 | 5+6 | 2 | 3+4 | 2 | 5+6 | 2 | 5 | 2 |
| | 6 | 2 | 6 | 2 | 3+4 | 2 | 5+6 | 2 | 3+4 | 2 | 5 | 2 |
| | 5 | 2 | 5 | 2 | 3+4 | 2 | 3+4 | 2 | 5+6 | 2 | 5 | 2 |

Continued on next column

**TABLE XLIV– continued**

| TQC | ICSD | NoE | SF |
|---|---|---|---|
| | 6 2 6 2 5+6 2 3+4 2 5+6 2 5 2 | | |
| | 5 2 5 2 5+6 2 5+6 2 3+4 2 5 2 | | |
| | 5 2 5 2 5+6 2 5+6 2 3+4 2 5 2 | | |
| | 5 2 5 2 3+4 2 5+6 2 5+6 2 5 2 | | |
| | 5 2 5 2 5+6 2 5+6 2 3+4 2 5 2 | | |
| | 5 2 6 2 3+4 2 3+4 2 5+6 2 5 2 | | |
| | 5 2 5 2 5+6 2 5+6 2 3+4 2 5 2 | | |
| | 5 2 6 2 3+4 2 3+4 2 5+6 2 5 2 | | |
| | 6 2 6 2 5+6 2 5+6 2 3+4 2 5 2 | | |
| | 6 2 5 2 5+6 2 3+4 2 3+4 2 5 2 | | |
| | 6 2 5 2 5+6 2 5+6 2 3+4 2 5 2 | | |
| | 5 2 5 2 3+4 2 5+6 2 3+4 2 5 2 | | |
| 71.1.77 | 72630 | 94 | Ba1Er2Ni1O5 |
| | -GM -X -T -R -S -W | | |
| | 5 2 6 2 3+4 2 3+4 2 3+4 2 5 2 | | |
| | 6 2 5 2 5+6 2 5+6 2 5+6 2 5 2 | | |
| | 5 2 5 2 5+6 2 5+6 2 3+4 2 5 2 | | |
| | 5 2 6 2 5+6 2 3+4 2 5+6 2 5 2 | | |
| | 6 2 5 2 3+4 2 5+6 2 3+4 2 5 2 | | |
| | 5 2 6 2 5+6 2 3+4 2 5+6 2 5 2 | | |
| | 6 2 5 2 3+4 2 5+6 2 3+4 2 5 2 | | |
| | 5 2 6 2 5+6 2 3+4 2 5+6 2 5 2 | | |
| | 6 2 5 2 3+4 2 5+6 2 3+4 2 5 2 | | |
| | 5 2 5 2 5+6 2 3+4 2 5+6 2 5 2 | | |
| | 5 2 6 2 3+4 2 3+4 2 3+4 2 5 2 | | |
| | 6 2 5 2 5+6 2 5+6 2 5+6 2 5 2 | | |
| | 6 2 5 2 3+4 2 5+6 2 3+4 2 5 2 | | |
| | 5 2 6 2 5+6 2 3+4 2 5+6 2 5 2 | | |
| | 6 2 6 2 3+4 2 3+4 2 5+6 2 5 2 | | |
| | 6 2 6 2 3+4 2 3+4 2 5+6 2 5 2 | | |
| | 6 2 6 2 3+4 2 3+4 2 5+6 2 5 2 | | |
| | 6 2 5 2 3+4 2 5+6 2 3+4 2 5 2 | | |
| | 5 2 6 2 3+4 2 5+6 2 3+4 2 5 2 | | |
| | 6 2 6 2 5+6 2 3+4 2 3+4 2 5 2 | | |
| | 5 2 6 2 3+4 2 5+6 2 3+4 2 5 2 | | |
| | 5 2 5 2 3+4 2 3+4 2 5+6 2 5 2 | | |
| | 5 2 6 2 5+6 2 5+6 2 3+4 2 5 2 | | |
| | 6 2 6 2 3+4 2 5+6 2 5+6 2 5 2 | | |
| | 6 2 5 2 5+6 2 3+4 2 3+4 2 5 2 | | |
| | 5 2 6 2 3+4 2 3+4 2 3+4 2 5 2 | | |
| | 6 2 5 2 5+6 2 5+6 2 5+6 2 5 2 | | |
| | 6 2 5 2 3+4 2 3+4 2 5+6 2 5 2 | | |
| | 6 2 6 2 5+6 2 5+6 2 3+4 2 5 2 | | |
| | 6 2 6 2 3+4 2 3+4 2 3+4 2 5 2 | | |
| | 5 2 5 2 5+6 2 5+6 2 5+6 2 5 2 | | |
| | 6 2 6 2 3+4 2 5+6 2 5+6 2 5 2 | | |
| | 5 2 5 2 3+4 2 3+4 2 3+4 2 5 2 | | |
| | 5 2 5 2 5+6 2 3+4 2 3+4 2 5 2 | | |
| | 5 2 5 2 3+4 2 3+4 2 5+6 2 5 2 | | |
| | 5 2 6 2 5+6 2 5+6 2 5+6 2 5 2 | | |
| | 5 2 5 2 3+4 2 3+4 2 3+4 2 5 2 | | |
| | 6 2 6 2 5+6 2 5+6 2 5+6 2 5 2 | | |
| | 5 2 5 2 3+4 2 3+4 2 3+4 2 5 2 | | |
| | 6 2 6 2 3+4 2 5+6 2 3+4 2 5 2 | | |
| | 6 2 5 2 3+4 2 3+4 2 3+4 2 5 2 | | |
| | 5 2 5 2 3+4 2 3+4 2 3+4 2 5 2 | | |
| | 5 2 5 2 5+6 2 3+4 2 5+6 2 5 2 | | |
| | 5 2 6 2 5+6 2 5+6 2 3+4 2 5 2 | | |
| | 5 2 6 2 3+4 2 3+4 2 5+6 2 5 2 | | |
| | 5 2 5 2 5+6 2 5+6 2 5+6 2 5 2 | | |
| | 6 2 6 2 3+4 2 3+4 2 3+4 2 5 2 | | |

Continued on next column

| TQC | ICSD | NoE | SF |
|---|---|---|---|
| 71.1.78 | 69324 | 94 | Ba1Er2Ni1O5 |
| | -GM -X -T -R -S -W | | |
| | 5 2 6 2 3+4 2 3+4 2 3+4 2 5 2 | | |
| | 6 2 5 2 5+6 2 5+6 2 5+6 2 5 2 | | |
| | 5 2 5 2 5+6 2 5+6 2 3+4 2 5 2 | | |
| | 5 2 6 2 5+6 2 3+4 2 5+6 2 5 2 | | |
| | 6 2 5 2 3+4 2 5+6 2 3+4 2 5 2 | | |
| | 5 2 6 2 5+6 2 3+4 2 5+6 2 5 2 | | |
| | 6 2 5 2 3+4 2 5+6 2 3+4 2 5 2 | | |
| | 5 2 6 2 5+6 2 3+4 2 5+6 2 5 2 | | |
| | 6 2 5 2 3+4 2 5+6 2 3+4 2 5 2 | | |
| | 5 2 5 2 5+6 2 3+4 2 5+6 2 5 2 | | |
| | 5 2 6 2 3+4 2 3+4 2 3+4 2 5 2 | | |
| | 6 2 5 2 5+6 2 5+6 2 5+6 2 5 2 | | |
| | 6 2 5 2 3+4 2 5+6 2 3+4 2 5 2 | | |
| | 5 2 6 2 5+6 2 3+4 2 5+6 2 5 2 | | |
| | 6 2 6 2 3+4 2 3+4 2 5+6 2 5 2 | | |
| | 6 2 6 2 3+4 2 3+4 2 5+6 2 5 2 | | |
| | 6 2 6 2 3+4 2 3+4 2 5+6 2 5 2 | | |
| | 6 2 5 2 3+4 2 5+6 2 3+4 2 5 2 | | |
| | 5 2 6 2 3+4 2 5+6 2 3+4 2 5 2 | | |
| | 6 2 6 2 5+6 2 3+4 2 3+4 2 5 2 | | |
| | 5 2 6 2 3+4 2 5+6 2 3+4 2 5 2 | | |
| | 5 2 5 2 3+4 2 3+4 2 5+6 2 5 2 | | |
| | 5 2 6 2 5+6 2 5+6 2 3+4 2 5 2 | | |
| | 6 2 6 2 3+4 2 5+6 2 5+6 2 5 2 | | |
| | 6 2 5 2 5+6 2 3+4 2 3+4 2 5 2 | | |
| | 5 2 6 2 3+4 2 3+4 2 3+4 2 5 2 | | |
| | 6 2 5 2 5+6 2 5+6 2 5+6 2 5 2 | | |
| | 6 2 5 2 3+4 2 3+4 2 5+6 2 5 2 | | |
| | 6 2 6 2 5+6 2 5+6 2 3+4 2 5 2 | | |
| | 6 2 6 2 3+4 2 3+4 2 3+4 2 5 2 | | |
| | 5 2 5 2 5+6 2 5+6 2 5+6 2 5 2 | | |
| | 6 2 6 2 3+4 2 5+6 2 5+6 2 5 2 | | |
| | 5 2 5 2 3+4 2 3+4 2 3+4 2 5 2 | | |
| | 5 2 5 2 5+6 2 3+4 2 3+4 2 5 2 | | |
| | 5 2 5 2 3+4 2 3+4 2 5+6 2 5 2 | | |
| | 5 2 6 2 5+6 2 5+6 2 5+6 2 5 2 | | |
| | 5 2 5 2 3+4 2 3+4 2 3+4 2 5 2 | | |
| | 6 2 6 2 5+6 2 3+4 2 5+6 2 5 2 | | |
| | 5 2 5 2 3+4 2 5+6 2 3+4 2 5 2 | | |
| | 6 2 6 2 3+4 2 5+6 2 3+4 2 5 2 | | |
| | 6 2 5 2 3+4 2 3+4 2 3+4 2 5 2 | | |
| | 5 2 5 2 3+4 2 3+4 2 3+4 2 5 2 | | |
| | 5 2 5 2 5+6 2 3+4 2 5+6 2 5 2 | | |
| | 5 2 6 2 5+6 2 5+6 2 3+4 2 5 2 | | |
| | 5 2 6 2 3+4 2 3+4 2 5+6 2 5 2 | | |
| | 5 2 5 2 5+6 2 5+6 2 5+6 2 5 2 | | |
| | 6 2 6 2 3+4 2 3+4 2 3+4 2 5 2 | | |
| 71.1.79 | 69323 | 94 | Ba1Er2Ni1O5 |
| | -GM -X -T -R -S -W | | |
| | 5 2 6 2 3+4 2 3+4 2 3+4 2 5 2 | | |
| | 6 2 5 2 5+6 2 5+6 2 5+6 2 5 2 | | |
| | 5 2 5 2 3+4 2 5+6 2 5+6 2 5 2 | | |
| | 5 2 6 2 3+4 2 3+4 2 3+4 2 5 2 | | |
| | 6 2 5 2 5+6 2 5+6 2 5+6 2 5 2 | | |
| | 5 2 6 2 3+4 2 3+4 2 3+4 2 5 2 | | |
| | 6 2 5 2 5+6 2 5+6 2 5+6 2 5 2 | | |
| | 5 2 6 2 3+4 2 3+4 2 3+4 2 5 2 | | |
| | 6 2 5 2 5+6 2 5+6 2 5+6 2 5 2 | | |
| | 5 2 5 2 3+4 2 3+4 2 3+4 2 5 2 | | |

Continued on next column

**TABLE XLIV– continued**

| TQC | ICSD | NoE | SF |
|---|---|---|---|
| | 5 2 6 2 5+6 2 3+4 2 5+6 2 5 2 | | |
| | 6 2 5 2 3+4 2 5+6 2 3+4 2 5 2 | | |
| | 6 2 5 2 5+6 2 5+6 2 5+6 2 5 2 | | |
| | 5 2 6 2 3+4 2 3+4 2 3+4 2 5 2 | | |
| | 6 2 6 2 5+6 2 3+4 2 3+4 2 5 2 | | |
| | 6 2 6 2 5+6 2 3+4 2 3+4 2 5 2 | | |
| | 6 2 6 2 5+6 2 3+4 2 3+4 2 5 2 | | |
| | 6 2 5 2 5+6 2 5+6 2 5+6 2 5 2 | | |
| | 5 2 6 2 5+6 2 5+6 2 5+6 2 5 2 | | |
| | 6 2 6 2 3+4 2 3+4 2 5+6 2 5 2 | | |
| | 5 2 6 2 5+6 2 5+6 2 5+6 2 5 2 | | |
| | 5 2 5 2 5+6 2 3+4 2 3+4 2 5 2 | | |
| | 5 2 6 2 3+4 2 5+6 2 5+6 2 5 2 | | |
| | 6 2 6 2 5+6 2 5+6 2 3+4 2 5 2 | | |
| | 6 2 5 2 3+4 2 3+4 2 5+6 2 5 2 | | |
| | 5 2 6 2 5+6 2 3+4 2 5+6 2 5 2 | | |
| | 6 2 5 2 3+4 2 5+6 2 3+4 2 5 2 | | |
| | 6 2 5 2 5+6 2 3+4 2 3+4 2 5 2 | | |
| | 6 2 6 2 3+4 2 5+6 2 5+6 2 5 2 | | |
| | 6 2 6 2 5+6 2 3+4 2 5+6 2 5 2 | | |
| | 5 2 5 2 3+4 2 5+6 2 3+4 2 5 2 | | |
| | 6 2 6 2 5+6 2 5+6 2 3+4 2 5 2 | | |
| | 5 2 5 2 5+6 2 3+4 2 5+6 2 5 2 | | |
| | 5 2 5 2 3+4 2 3+4 2 5+6 2 5 2 | | |
| | 5 2 5 2 5+6 2 3+4 2 3+4 2 5 2 | | |
| | 5 2 6 2 3+4 2 5+6 2 3+4 2 5 2 | | |
| | 5 2 5 2 5+6 2 3+4 2 5+6 2 5 2 | | |
| | 6 2 6 2 3+4 2 3+4 2 3+4 2 5 2 | | |
| | 5 2 5 2 5+6 2 5+6 2 5+6 2 5 2 | | |
| | 6 2 6 2 5+6 2 5+6 2 5+6 2 5 2 | | |
| | 6 2 5 2 5+6 2 3+4 2 5+6 2 5 2 | | |
| | 5 2 5 2 5+6 2 3+4 2 5+6 2 5 2 | | |
| | 5 2 5 2 3+4 2 3+4 2 3+4 2 5 2 | | |
| | 5 2 6 2 3+4 2 5+6 2 5+6 2 5 2 | | |
| | 5 2 6 2 5+6 2 3+4 2 3+4 2 5 2 | | |
| | 5 2 5 2 3+4 2 5+6 2 3+4 2 5 2 | | |
| | 6 2 6 2 5+6 2 3+4 2 5+6 2 5 2 | | |
| 71.1.80 | 69315 | 94 | Ba1Er2Ni1O5 |
| | -GM -X -T -R -S -W | | |
| | 5 2 6 2 3+4 2 3+4 2 3+4 2 5 2 | | |
| | 6 2 5 2 5+6 2 5+6 2 5+6 2 5 2 | | |
| | 5 2 5 2 3+4 2 5+6 2 5+6 2 5 2 | | |
| | 5 2 6 2 3+4 2 3+4 2 3+4 2 5 2 | | |
| | 6 2 5 2 5+6 2 5+6 2 5+6 2 5 2 | | |
| | 5 2 6 2 3+4 2 3+4 2 3+4 2 5 2 | | |
| | 6 2 5 2 5+6 2 5+6 2 5+6 2 5 2 | | |
| | 5 2 6 2 3+4 2 3+4 2 3+4 2 5 2 | | |
| | 6 2 5 2 5+6 2 5+6 2 5+6 2 5 2 | | |
| | 5 2 5 2 3+4 2 3+4 2 3+4 2 5 2 | | |
| | 5 2 6 2 5+6 2 3+4 2 5+6 2 5 2 | | |
| | 6 2 5 2 3+4 2 5+6 2 3+4 2 5 2 | | |
| | 6 2 5 2 5+6 2 5+6 2 5+6 2 5 2 | | |
| | 5 2 6 2 3+4 2 3+4 2 3+4 2 5 2 | | |
| | 6 2 6 2 5+6 2 3+4 2 3+4 2 5 2 | | |
| | 6 2 6 2 5+6 2 3+4 2 3+4 2 5 2 | | |
| | 6 2 6 2 5+6 2 3+4 2 3+4 2 5 2 | | |
| | 6 2 5 2 5+6 2 5+6 2 5+6 2 5 2 | | |
| | 5 2 6 2 5+6 2 5+6 2 5+6 2 5 2 | | |
| | 6 2 6 2 3+4 2 3+4 2 5+6 2 5 2 | | |
| | 5 2 6 2 5+6 2 5+6 2 5+6 2 5 2 | | |
| | 5 2 5 2 5+6 2 3+4 2 3+4 2 5 2 | | |

Continued on next column

**TABLE XLIV– continued**

| TQC | ICSD | NoE | SF |
|---|---|---|---|
| | 5 2 6 2 3+4 2 5+6 2 5+6 2 5 2 | | |
| | 6 2 6 2 5+6 2 5+6 2 3+4 2 5 2 | | |
| | 6 2 6 2 3+4 2 3+4 2 5+6 2 5 2 | | |
| | 5 2 5 2 5+6 2 3+4 2 5+6 2 5 2 | | |
| | 6 2 5 2 3+4 2 5+6 2 3+4 2 5 2 | | |
| | 6 2 5 2 5+6 2 3+4 2 5+6 2 5 2 | | |
| | 6 2 6 2 3+4 2 5+6 2 3+4 2 5 2 | | |
| | 6 2 6 2 5+6 2 3+4 2 5+6 2 5 2 | | |
| | 5 2 5 2 3+4 2 5+6 2 3+4 2 5 2 | | |
| | 6 2 6 2 5+6 2 5+6 2 3+4 2 5 2 | | |
| | 5 2 5 2 5+6 2 3+4 2 5+6 2 5 2 | | |
| | 5 2 5 2 3+4 2 3+4 2 5+6 2 5 2 | | |
| | 5 2 5 2 5+6 2 3+4 2 3+4 2 5 2 | | |
| | 5 2 6 2 3+4 2 5+6 2 3+4 2 5 2 | | |
| | 5 2 5 2 5+6 2 3+4 2 5+6 2 5 2 | | |
| | 6 2 6 2 3+4 2 3+4 2 3+4 2 5 2 | | |
| | 5 2 5 2 5+6 2 5+6 2 5+6 2 5 2 | | |
| | 6 2 6 2 5+6 2 5+6 2 5+6 2 5 2 | | |
| | 6 2 5 2 5+6 2 3+4 2 5+6 2 5 2 | | |
| | 5 2 5 2 5+6 2 3+4 2 5+6 2 5 2 | | |
| | 5 2 5 2 3+4 2 3+4 2 3+4 2 5 2 | | |
| | 5 2 6 2 3+4 2 5+6 2 5+6 2 5 2 | | |
| | 5 2 6 2 5+6 2 3+4 2 3+4 2 5 2 | | |
| | 5 2 5 2 3+4 2 5+6 2 3+4 2 5 2 | | |
| | 6 2 6 2 5+6 2 3+4 2 5+6 2 5 2 | | |
| 71.1.81 | 62760 | 94 | Ba1Er2Ni1O5 |
| | -GM -X -T -R -S -W | | |
| | 5 2 6 2 3+4 2 3+4 2 3+4 2 5 2 | | |
| | 6 2 5 2 5+6 2 5+6 2 5+6 2 5 2 | | |
| | 5 2 5 2 3+4 2 3+4 2 3+4 2 5 2 | | |
| | 5 2 6 2 3+4 2 5+6 2 5+6 2 5 2 | | |
| | 6 2 5 2 5+6 2 3+4 2 3+4 2 5 2 | | |
| | 5 2 6 2 3+4 2 5+6 2 5+6 2 5 2 | | |
| | 6 2 5 2 5+6 2 3+4 2 3+4 2 5 2 | | |
| | 5 2 6 2 3+4 2 5+6 2 5+6 2 5 2 | | |
| | 6 2 5 2 5+6 2 3+4 2 3+4 2 5 2 | | |
| | 5 2 5 2 3+4 2 5+6 2 5+6 2 5 2 | | |
| | 5 2 6 2 5+6 2 5+6 2 3+4 2 5 2 | | |
| | 6 2 5 2 3+4 2 3+4 2 5+6 2 5 2 | | |
| | 6 2 5 2 5+6 2 3+4 2 3+4 2 5 2 | | |
| | 5 2 6 2 3+4 2 5+6 2 5+6 2 5 2 | | |
| | 6 2 6 2 5+6 2 5+6 2 5+6 2 5 2 | | |
| | 6 2 6 2 5+6 2 5+6 2 5+6 2 5 2 | | |
| | 6 2 6 2 5+6 2 5+6 2 5+6 2 5 2 | | |
| | 6 2 5 2 5+6 2 3+4 2 3+4 2 5 2 | | |
| | 5 2 6 2 5+6 2 3+4 2 3+4 2 5 2 | | |
| | 6 2 6 2 3+4 2 5+6 2 3+4 2 5 2 | | |
| | 5 2 6 2 5+6 2 3+4 2 3+4 2 5 2 | | |
| | 5 2 5 2 5+6 2 5+6 2 5+6 2 5 2 | | |
| | 5 2 6 2 3+4 2 3+4 2 3+4 2 5 2 | | |
| | 6 2 6 2 5+6 2 3+4 2 5+6 2 5 2 | | |
| | 6 2 5 2 3+4 2 5+6 2 3+4 2 5 2 | | |
| | 5 2 6 2 5+6 2 5+6 2 3+4 2 5 2 | | |
| | 6 2 5 2 3+4 2 3+4 2 5+6 2 5 2 | | |
| | 6 2 5 2 5+6 2 5+6 2 5+6 2 5 2 | | |
| | 6 2 6 2 3+4 2 3+4 2 3+4 2 5 2 | | |
| | 6 2 6 2 5+6 2 5+6 2 3+4 2 5 2 | | |
| | 5 2 5 2 3+4 2 3+4 2 5+6 2 5 2 | | |
| | 6 2 6 2 5+6 2 3+4 2 5+6 2 5 2 | | |
| | 5 2 5 2 5+6 2 5+6 2 3+4 2 5 2 | | |
| | 5 2 5 2 3+4 2 5+6 2 3+4 2 5 2 | | |

Continued on next column

**TABLE XLIV– continued**

| TQC | ICSD | NoE | SF |
|---|---|---|---|
| | 5 2 5 2 5+6 2 5+6 2 5+6 2 5 2 | | |
| | 5 2 6 2 3+4 2 3+4 2 5+6 2 5 2 | | |
| | 5 2 5 2 5+6 2 5+6 2 3+4 2 5 2 | | |
| | 6 2 6 2 3+4 2 3+4 2 5+6 2 5 2 | | |
| | 5 2 5 2 5+6 2 5+6 2 3+4 2 5 2 | | |
| | 6 2 6 2 5+6 2 3+4 2 3+4 2 5 2 | | |
| | 6 2 5 2 5+6 2 5+6 2 3+4 2 5 2 | | |
| | 5 2 5 2 5+6 2 5+6 2 3+4 2 5 2 | | |
| | 5 2 5 2 3+4 2 5+6 2 5+6 2 5 2 | | |
| | 5 2 6 2 3+4 2 3+4 2 3+4 2 5 2 | | |
| | 5 2 6 2 5+6 2 5+6 2 5+6 2 5 2 | | |
| | 5 2 5 2 3+4 2 3+4 2 5+6 2 5 2 | | |
| | 6 2 6 2 5+6 2 5+6 2 3+4 2 5 2 | | |
| 71.1.82 | 62425 | 94 | Ba1Er2Ni1O5 |
| | -GM -X -T -R -S -W | | |
| | 5 2 6 2 3+4 2 3+4 2 3+4 2 5 2 | | |
| | 6 2 5 2 5+6 2 5+6 2 5+6 2 5 2 | | |
| | 5 2 5 2 3+4 2 3+4 2 3+4 2 5 2 | | |
| | 5 2 6 2 3+4 2 5+6 2 5+6 2 5 2 | | |
| | 6 2 5 2 5+6 2 3+4 2 3+4 2 5 2 | | |
| | 5 2 6 2 3+4 2 5+6 2 5+6 2 5 2 | | |
| | 6 2 5 2 5+6 2 3+4 2 3+4 2 5 2 | | |
| | 5 2 6 2 3+4 2 5+6 2 5+6 2 5 2 | | |
| | 6 2 5 2 5+6 2 3+4 2 3+4 2 5 2 | | |
| | 5 2 5 2 3+4 2 5+6 2 5+6 2 5 2 | | |
| | 5 2 6 2 5+6 2 5+6 2 3+4 2 5 2 | | |
| | 6 2 5 2 3+4 2 3+4 2 5+6 2 5 2 | | |
| | 6 2 5 2 5+6 2 3+4 2 3+4 2 5 2 | | |
| | 5 2 6 2 3+4 2 5+6 2 5+6 2 5 2 | | |
| | 6 2 6 2 5+6 2 5+6 2 5+6 2 5 2 | | |
| | 6 2 6 2 5+6 2 5+6 2 5+6 2 5 2 | | |
| | 6 2 6 2 5+6 2 5+6 2 5+6 2 5 2 | | |
| | 6 2 5 2 5+6 2 3+4 2 3+4 2 5 2 | | |
| | 5 2 6 2 5+6 2 3+4 2 3+4 2 5 2 | | |
| | 6 2 6 2 3+4 2 5+6 2 3+4 2 5 2 | | |
| | 5 2 6 2 5+6 2 3+4 2 3+4 2 5 2 | | |
| | 5 2 5 2 5+6 2 5+6 2 5+6 2 5 2 | | |
| | 5 2 6 2 3+4 2 3+4 2 3+4 2 5 2 | | |
| | 6 2 6 2 5+6 2 3+4 2 5+6 2 5 2 | | |
| | 6 2 5 2 3+4 2 5+6 2 3+4 2 5 2 | | |
| | 5 2 6 2 5+6 2 5+6 2 3+4 2 5 2 | | |
| | 6 2 5 2 3+4 2 3+4 2 5+6 2 5 2 | | |
| | 6 2 5 2 5+6 2 5+6 2 5+6 2 5 2 | | |
| | 6 2 6 2 3+4 2 3+4 2 3+4 2 5 2 | | |
| | 6 2 6 2 5+6 2 5+6 2 3+4 2 5 2 | | |
| | 5 2 5 2 3+4 2 3+4 2 5+6 2 5 2 | | |
| | 6 2 6 2 5+6 2 3+4 2 5+6 2 5 2 | | |
| | 5 2 5 2 5+6 2 5+6 2 3+4 2 5 2 | | |
| | 5 2 5 2 3+4 2 5+6 2 3+4 2 5 2 | | |
| | 5 2 5 2 5+6 2 5+6 2 5+6 2 5 2 | | |
| | 5 2 6 2 3+4 2 3+4 2 5+6 2 5 2 | | |
| | 5 2 5 2 5+6 2 5+6 2 3+4 2 5 2 | | |
| | 6 2 6 2 3+4 2 3+4 2 5+6 2 5 2 | | |
| | 5 2 5 2 5+6 2 5+6 2 3+4 2 5 2 | | |
| | 6 2 6 2 5+6 2 3+4 2 3+4 2 5 2 | | |
| | 6 2 5 2 5+6 2 5+6 2 3+4 2 5 2 | | |
| | 5 2 5 2 5+6 2 5+6 2 3+4 2 5 2 | | |
| | 5 2 5 2 3+4 2 5+6 2 5+6 2 5 2 | | |
| | 5 2 6 2 3+4 2 3+4 2 3+4 2 5 2 | | |
| | 5 2 6 2 5+6 2 5+6 2 5+6 2 5 2 | | |
| | 5 2 5 2 3+4 2 3+4 2 5+6 2 5 2 | | |

Continued on next column

**TABLE XLIV– continued**

| TQC | ICSD | NoE | SF |
|---|---|---|---|
| | 6 2 6 2 5+6 2 5+6 2 3+4 2 5 2 | | |
| 71.1.83 | 85046 | 90 | Ba1Dy2Ni1O5 |
| | -GM -X -T -R -S -W | | |
| | 5 2 6 2 3+4 2 3+4 2 3+4 2 5 2 | | |
| | 6 2 5 2 5+6 2 5+6 2 5+6 2 5 2 | | |
| | 5 2 5 2 3+4 2 3+4 2 3+4 2 5 2 | | |
| | 5 2 6 2 3+4 2 5+6 2 5+6 2 5 2 | | |
| | 6 2 5 2 5+6 2 3+4 2 3+4 2 5 2 | | |
| | 5 2 6 2 3+4 2 5+6 2 5+6 2 5 2 | | |
| | 6 2 5 2 5+6 2 3+4 2 3+4 2 5 2 | | |
| | 5 2 6 2 3+4 2 5+6 2 5+6 2 5 2 | | |
| | 6 2 5 2 5+6 2 3+4 2 3+4 2 5 2 | | |
| | 5 2 5 2 3+4 2 5+6 2 5+6 2 5 2 | | |
| | 6 2 6 2 5+6 2 5+6 2 3+4 2 5 2 | | |
| | 5 2 5 2 3+4 2 3+4 2 5+6 2 5 2 | | |
| | 6 2 5 2 5+6 2 3+4 2 3+4 2 5 2 | | |
| | 5 2 6 2 3+4 2 5+6 2 5+6 2 5 2 | | |
| | 6 2 6 2 5+6 2 5+6 2 5+6 2 5 2 | | |
| | 6 2 6 2 5+6 2 5+6 2 5+6 2 5 2 | | |
| | 6 2 6 2 5+6 2 5+6 2 5+6 2 5 2 | | |
| | 6 2 5 2 5+6 2 3+4 2 3+4 2 5 2 | | |
| | 6 2 6 2 5+6 2 3+4 2 3+4 2 5 2 | | |
| | 5 2 6 2 3+4 2 5+6 2 3+4 2 5 2 | | |
| | 5 2 6 2 5+6 2 3+4 2 3+4 2 5 2 | | |
| | 5 2 5 2 5+6 2 5+6 2 5+6 2 5 2 | | |
| | 5 2 6 2 3+4 2 3+4 2 3+4 2 5 2 | | |
| | 6 2 6 2 5+6 2 3+4 2 5+6 2 5 2 | | |
| | 6 2 5 2 3+4 2 5+6 2 3+4 2 5 2 | | |
| | 5 2 5 2 5+6 2 5+6 2 3+4 2 5 2 | | |
| | 6 2 6 2 3+4 2 3+4 2 5+6 2 5 2 | | |
| | 6 2 5 2 5+6 2 5+6 2 5+6 2 5 2 | | |
| | 6 2 6 2 3+4 2 3+4 2 3+4 2 5 2 | | |
| | 6 2 6 2 5+6 2 5+6 2 3+4 2 5 2 | | |
| | 5 2 5 2 3+4 2 3+4 2 5+6 2 5 2 | | |
| | 6 2 6 2 5+6 2 3+4 2 5+6 2 5 2 | | |
| | 5 2 5 2 5+6 2 5+6 2 3+4 2 5 2 | | |
| | 5 2 5 2 5+6 2 5+6 2 3+4 2 5 2 | | |
| | 5 2 5 2 3+4 2 5+6 2 5+6 2 5 2 | | |
| | 5 2 6 2 3+4 2 5+6 2 3+4 2 5 2 | | |
| | 5 2 5 2 5+6 2 3+4 2 5+6 2 5 2 | | |
| | 5 2 5 2 5+6 2 5+6 2 5+6 2 5 2 | | |
| | 6 2 6 2 3+4 2 3+4 2 3+4 2 5 2 | | |
| | 5 2 6 2 5+6 2 3+4 2 3+4 2 5 2 | | |
| | 6 2 5 2 5+6 2 5+6 2 3+4 2 5 2 | | |
| | 6 2 5 2 5+6 2 5+6 2 3+4 2 5 2 | | |
| | 5 2 5 2 3+4 2 5+6 2 3+4 2 5 2 | | |
| | 5 2 6 2 3+4 2 3+4 2 5+6 2 5 2 | | |
| | 5 2 6 2 5+6 2 5+6 2 5+6 2 5 2 | | |
| 71.1.84 | 72627 | 90 | Ba1Dy2Ni1O5 |
| | -GM -X -T -R -S -W | | |
| | 5 2 6 2 3+4 2 3+4 2 3+4 2 5 2 | | |
| | 6 2 5 2 5+6 2 5+6 2 5+6 2 5 2 | | |
| | 5 2 5 2 5+6 2 3+4 2 5+6 2 5 2 | | |
| | 5 2 6 2 5+6 2 5+6 2 3+4 2 5 2 | | |
| | 6 2 5 2 3+4 2 3+4 2 5+6 2 5 2 | | |
| | 5 2 6 2 5+6 2 5+6 2 3+4 2 5 2 | | |
| | 6 2 5 2 3+4 2 3+4 2 5+6 2 5 2 | | |
| | 5 2 6 2 5+6 2 5+6 2 3+4 2 5 2 | | |
| | 6 2 5 2 3+4 2 3+4 2 5+6 2 5 2 | | |
| | 5 2 5 2 5+6 2 5+6 2 3+4 2 5 2 | | |
| | 6 2 6 2 3+4 2 5+6 2 5+6 2 5 2 | | |

Continued on next column

**TABLE XLIV– continued**

| TQC | ICSD NoE SF |
|---|---|
| | 5 2 5 2 5+6 2 3+4 2 3+4 2 5 2 |
| | 6 2 5 2 3+4 2 3+4 2 5+6 2 5 2 |
| | 5 2 6 2 5+6 2 5+6 2 3+4 2 5 2 |
| | 6 2 6 2 3+4 2 5+6 2 3+4 2 5 2 |
| | 6 2 6 2 3+4 2 5+6 2 3+4 2 5 2 |
| | 6 2 6 2 3+4 2 5+6 2 3+4 2 5 2 |
| | 6 2 5 2 3+4 2 3+4 2 5+6 2 5 2 |
| | 6 2 6 2 3+4 2 3+4 2 5+6 2 5 2 |
| | 5 2 6 2 5+6 2 5+6 2 5+6 2 5 2 |
| | 5 2 6 2 3+4 2 3+4 2 5+6 2 5 2 |
| | 5 2 5 2 3+4 2 5+6 2 3+4 2 5 2 |
| | 5 2 6 2 5+6 2 3+4 2 5+6 2 5 2 |
| | 6 2 6 2 3+4 2 3+4 2 3+4 2 5 2 |
| | 6 2 5 2 5+6 2 5+6 2 5+6 2 5 2 |
| | 5 2 6 2 3+4 2 5+6 2 5+6 2 5 2 |
| | 6 2 5 2 5+6 2 3+4 2 3+4 2 5 2 |
| | 6 2 5 2 3+4 2 5+6 2 3+4 2 5 2 |
| | 6 2 6 2 5+6 2 3+4 2 5+6 2 5 2 |
| | 6 2 6 2 3+4 2 5+6 2 5+6 2 5 2 |
| | 5 2 5 2 5+6 2 3+4 2 3+4 2 5 2 |
| | 6 2 6 2 3+4 2 3+4 2 3+4 2 5 2 |
| | 5 2 5 2 3+4 2 5+6 2 5+6 2 5 2 |
| | 5 2 5 2 3+4 2 5+6 2 5+6 2 5 2 |
| | 5 2 5 2 5+6 2 5+6 2 3+4 2 5 2 |
| | 5 2 6 2 5+6 2 3+4 2 3+4 2 5 2 |
| | 5 2 5 2 3+4 2 5+6 2 5+6 2 5 2 |
| | 6 2 6 2 3+4 2 5+6 2 3+4 2 5 2 |
| | 5 2 5 2 5+6 2 3+4 2 5+6 2 5 2 |
| | 5 2 6 2 3+4 2 3+4 2 5+6 2 5 2 |
| | 6 2 5 2 3+4 2 5+6 2 5+6 2 5 2 |
| | 6 2 5 2 3+4 2 5+6 2 5+6 2 5 2 |
| | 5 2 5 2 5+6 2 5+6 2 3+4 2 5 2 |
| | 5 2 6 2 5+6 2 3+4 2 5+6 2 5 2 |
| | 5 2 6 2 3+4 2 5+6 2 3+4 2 5 2 |
| 71.1.85 | 167153 48 B6Fe2U1 |
| | -GM -X -T -R -S -W |
| | 5 2 5 2 3+4 2 5+6 2 5+6 2 5 2 |
| | 6 2 6 2 5+6 2 3+4 2 3+4 2 5 2 |
| | 6 2 6 2 5+6 2 3+4 2 3+4 2 5 2 |
| | 6 2 6 2 5+6 2 3+4 2 3+4 2 5 2 |
| | 5 2 5 2 5+6 2 3+4 2 5+6 2 5 2 |
| | 5 2 6 2 3+4 2 5+6 2 5+6 2 5 2 |
| | 6 2 6 2 3+4 2 5+6 2 3+4 2 5 2 |
| | 5 2 5 2 5+6 2 3+4 2 3+4 2 5 2 |
| | 6 2 5 2 5+6 2 5+6 2 5+6 2 5 2 |
| | 5 2 6 2 3+4 2 3+4 2 3+4 2 5 2 |
| | 5 2 6 2 3+4 2 5+6 2 5+6 2 5 2 |
| | 6 2 6 2 3+4 2 3+4 2 5+6 2 5 2 |
| | 5 2 5 2 5+6 2 5+6 2 3+4 2 5 2 |
| | 6 2 6 2 3+4 2 3+4 2 3+4 2 5 2 |
| | 5 2 5 2 5+6 2 5+6 2 5+6 2 5 2 |
| | 5 2 6 2 5+6 2 3+4 2 5+6 2 5 2 |
| | 6 2 5 2 3+4 2 3+4 2 5+6 2 5 2 |
| | 5 2 5 2 5+6 2 5+6 2 3+4 2 5 2 |
| | 6 2 5 2 3+4 2 5+6 2 3+4 2 5 2 |
| | 6 2 6 2 5+6 2 3+4 2 3+4 2 5 2 |
| | 6 2 6 2 5+6 2 3+4 2 3+4 2 5 2 |
| | 5 2 6 2 3+4 2 5+6 2 3+4 2 5 2 |
| | 6 2 5 2 3+4 2 5+6 2 5+6 2 5 2 |
| | 6 2 5 2 5+6 2 3+4 2 3+4 2 5 2 |
| 71.1.86 | 81543 42 B6Cr2Th1 |
| | -GM -X -T -R -S -W |
| | 5 2 5 2 5+6 2 3+4 2 5+6 2 5 2 |
| | 6 2 6 2 3+4 2 5+6 2 3+4 2 5 2 |
| | 6 2 6 2 3+4 2 5+6 2 3+4 2 5 2 |
| | 6 2 6 2 3+4 2 5+6 2 3+4 2 5 2 |
| | 5 2 5 2 3+4 2 3+4 2 5+6 2 5 2 |
| | 5 2 6 2 5+6 2 5+6 2 3+4 2 5 2 |
| | 5 2 6 2 5+6 2 3+4 2 5+6 2 5 2 |
| | 6 2 5 2 3+4 2 5+6 2 3+4 2 5 2 |
| | 6 2 5 2 3+4 2 3+4 2 5+6 2 5 2 |
| | 5 2 6 2 5+6 2 5+6 2 3+4 2 5 2 |
| | 5 2 6 2 5+6 2 3+4 2 5+6 2 5 2 |
| | 6 2 6 2 5+6 2 5+6 2 5+6 2 5 2 |
| | 5 2 5 2 3+4 2 3+4 2 3+4 2 5 2 |
| | 6 2 5 2 3+4 2 5+6 2 3+4 2 5 2 |
| | 6 2 6 2 5+6 2 3+4 2 5+6 2 5 2 |
| | 5 2 5 2 3+4 2 5+6 2 5+6 2 5 2 |
| | 5 2 6 2 5+6 2 5+6 2 5+6 2 5 2 |
| | 5 2 5 2 3+4 2 3+4 2 3+4 2 5 2 |
| | 6 2 5 2 5+6 2 3+4 2 3+4 2 5 2 |
| | 6 2 6 2 3+4 2 5+6 2 3+4 2 5 2 |
| | 6 2 6 2 3+4 2 5+6 2 3+4 2 5 2 |
| 71.1.87 | 81544 42 B6Ce1Cr2 |
| | -GM -X -T -R -S -W |
| | 5 2 5 2 3+4 2 3+4 2 3+4 2 5 2 |
| | 6 2 6 2 5+6 2 5+6 2 5+6 2 5 2 |
| | 6 2 6 2 5+6 2 5+6 2 5+6 2 5 2 |
| | 6 2 6 2 5+6 2 5+6 2 5+6 2 5 2 |
| | 5 2 5 2 5+6 2 5+6 2 3+4 2 5 2 |
| | 5 2 6 2 3+4 2 3+4 2 5+6 2 5 2 |
| | 5 2 6 2 3+4 2 5+6 2 3+4 2 5 2 |
| | 6 2 5 2 5+6 2 3+4 2 5+6 2 5 2 |
| | 6 2 6 2 5+6 2 3+4 2 3+4 2 5 2 |
| | 5 2 5 2 3+4 2 5+6 2 5+6 2 5 2 |
| | 5 2 6 2 3+4 2 3+4 2 3+4 2 5 2 |
| | 6 2 6 2 5+6 2 5+6 2 3+4 2 5 2 |
| | 6 2 5 2 3+4 2 3+4 2 5+6 2 5 2 |
| | 5 2 5 2 5+6 2 5+6 2 5+6 2 5 2 |
| | 6 2 6 2 5+6 2 3+4 2 3+4 2 5 2 |
| | 5 2 5 2 3+4 2 5+6 2 3+4 2 5 2 |
| | 5 2 6 2 3+4 2 5+6 2 3+4 2 5 2 |
| | 6 2 5 2 5+6 2 3+4 2 5+6 2 5 2 |
| | 5 2 5 2 3+4 2 3+4 2 5+6 2 5 2 |
| | 6 2 6 2 5+6 2 5+6 2 5+6 2 5 2 |
| | 6 2 6 2 5+6 2 5+6 2 5+6 2 5 2 |
| 71.1.88 | 40791 30 B4Cr3 |
| | -GM -X -T -R -S -W |
| | 5 2 5 2 3+4 2 5+6 2 5+6 2 5 2 |
| | 6 2 6 2 5+6 2 3+4 2 3+4 2 5 2 |
| | 5 2 5 2 5+6 2 3+4 2 5+6 2 5 2 |
| | 6 2 6 2 3+4 2 5+6 2 3+4 2 5 2 |
| | 5 2 5 2 3+4 2 5+6 2 5+6 2 5 2 |
| | 6 2 5 2 5+6 2 3+4 2 3+4 2 5 2 |
| | 5 2 5 2 5+6 2 3+4 2 5+6 2 5 2 |
| | 5 2 5 2 5+6 2 3+4 2 5+6 2 5 2 |
| | 6 2 6 2 3+4 2 5+6 2 3+4 2 5 2 |
| | 6 2 5 2 5+6 2 5+6 2 5+6 2 5 2 |
| | 5 2 6 2 3+4 2 5+6 2 3+4 2 5 2 |
| | 5 2 6 2 5+6 2 3+4 2 3+4 2 5 2 |
| | 5 2 6 2 3+4 2 3+4 2 5+6 2 5 2 |
| | 5 2 5 2 3+4 2 3+4 2 5+6 2 5 2 |

Continued on next column

Continued on next column

**TABLE XLIV– continued**

| TQC | ICSD NoE SF |
|---|---|
| | 5 2 6 2 5+6 2 3+4 2 5+6 2 5 2 |
| 71.1.89 | 423380 18 B4Cr1 |
| | -GM -X -T -R -S -W |
| | 5 2 5 2 5+6 2 3+4 2 5+6 2 5 2 |
| | 5 2 6 2 3+4 2 5+6 2 3+4 2 5 2 |
| | 6 2 6 2 3+4 2 5+6 2 5+6 2 5 2 |
| | 6 2 5 2 5+6 2 3+4 2 3+4 2 5 2 |
| | 5 2 6 2 5+6 2 5+6 2 5+6 2 5 2 |
| | 5 2 5 2 3+4 2 5+6 2 3+4 2 5 2 |
| | 5 2 5 2 3+4 2 3+4 2 5+6 2 5 2 |
| | 5 2 5 2 5+6 2 5+6 2 5+6 2 5 2 |
| | 5 2 6 2 3+4 2 5+6 2 5+6 2 5 2 |
| 71.1.90 | 407576 68 B2N4Nd3 |
| | -GM -X -T -R -S -W |
| | 5 2 5 2 3+4 2 3+4 2 3+4 2 5 2 |
| | 5 2 6 2 3+4 2 5+6 2 5+6 2 5 2 |
| | 6 2 5 2 5+6 2 3+4 2 3+4 2 5 2 |
| | 6 2 6 2 5+6 2 5+6 2 5+6 2 5 2 |
| | 5 2 6 2 3+4 2 5+6 2 3+4 2 5 2 |
| | 6 2 5 2 5+6 2 3+4 2 5+6 2 5 2 |
| | 6 2 6 2 5+6 2 5+6 2 5+6 2 5 2 |
| | 5 2 5 2 3+4 2 3+4 2 3+4 2 5 2 |
| | 6 2 6 2 5+6 2 5+6 2 5+6 2 5 2 |
| | 5 2 6 2 5+6 2 5+6 2 5+6 2 5 2 |
| | 6 2 6 2 3+4 2 3+4 2 3+4 2 5 2 |
| | 6 2 5 2 5+6 2 5+6 2 5+6 2 5 2 |
| | 5 2 5 2 3+4 2 3+4 2 3+4 2 5 2 |
| | 6 2 6 2 5+6 2 5+6 2 5+6 2 5 2 |
| | 6 2 5 2 5+6 2 3+4 2 3+4 2 5 2 |
| | 5 2 6 2 3+4 2 5+6 2 5+6 2 5 2 |
| | 5 2 5 2 5+6 2 5+6 2 3+4 2 5 2 |
| | 5 2 6 2 3+4 2 3+4 2 5+6 2 5 2 |
| | 6 2 5 2 3+4 2 5+6 2 3+4 2 5 2 |
| | 6 2 6 2 3+4 2 3+4 2 5+6 2 5 2 |
| | 5 2 5 2 5+6 2 3+4 2 3+4 2 5 2 |
| | 5 2 5 2 5+6 2 5+6 2 3+4 2 5 2 |
| | 6 2 6 2 3+4 2 3+4 2 5+6 2 5 2 |
| | 6 2 6 2 5+6 2 5+6 2 3+4 2 5 2 |
| | 5 2 6 2 5+6 2 3+4 2 5+6 2 5 2 |
| | 6 2 5 2 3+4 2 3+4 2 3+4 2 5 2 |
| | 5 2 6 2 3+4 2 5+6 2 5+6 2 5 2 |
| | 6 2 5 2 5+6 2 5+6 2 5+6 2 5 2 |
| | 5 2 5 2 3+4 2 5+6 2 5+6 2 5 2 |
| | 5 2 5 2 5+6 2 3+4 2 3+4 2 5 2 |
| | 6 2 6 2 5+6 2 5+6 2 3+4 2 5 2 |
| | 6 2 5 2 5+6 2 5+6 2 5+6 2 5 2 |
| | 6 2 6 2 3+4 2 3+4 2 5+6 2 5 2 |
| | 6 2 6 2 5+6 2 5+6 2 5+6 2 5 2 |
| 71.1.91 | 68610 62 B2Ce3N4 |
| | -GM -X -T -R -S -W |
| | 5 2 5 2 3+4 2 3+4 2 3+4 2 5 2 |
| | 5 2 6 2 3+4 2 5+6 2 5+6 2 5 2 |
| | 6 2 5 2 5+6 2 3+4 2 3+4 2 5 2 |
| | 6 2 6 2 5+6 2 5+6 2 5+6 2 5 2 |
| | 5 2 6 2 3+4 2 5+6 2 3+4 2 5 2 |
| | 6 2 5 2 5+6 2 3+4 2 5+6 2 5 2 |
| | 6 2 6 2 5+6 2 5+6 2 5+6 2 5 2 |
| | 5 2 5 2 3+4 2 3+4 2 3+4 2 5 2 |
| | 6 2 6 2 5+6 2 5+6 2 5+6 2 5 2 |
| | 5 2 6 2 3+4 2 3+4 2 5+6 2 5 2 |
| | 5 2 6 2 5+6 2 5+6 2 3+4 2 5 2 |

Continued on next column

**TABLE XLIV– continued**

| TQC | ICSD NoE SF |
|---|---|
| | 6 2 5 2 5+6 2 5+6 2 5+6 2 5 2 |
| | 6 2 5 2 3+4 2 3+4 2 3+4 2 5 2 |
| | 6 2 5 2 5+6 2 5+6 2 5+6 2 5 2 |
| | 6 2 6 2 5+6 2 3+4 2 3+4 2 5 2 |
| | 5 2 6 2 3+4 2 5+6 2 5+6 2 5 2 |
| | 5 2 5 2 5+6 2 5+6 2 3+4 2 5 2 |
| | 5 2 6 2 3+4 2 3+4 2 5+6 2 5 2 |
| | 6 2 5 2 3+4 2 5+6 2 3+4 2 5 2 |
| | 6 2 6 2 3+4 2 3+4 2 5+6 2 5 2 |
| | 5 2 5 2 5+6 2 3+4 2 3+4 2 5 2 |
| | 5 2 5 2 5+6 2 3+4 2 3+4 2 5 2 |
| | 6 2 6 2 3+4 2 5+6 2 5+6 2 5 2 |
| | 6 2 6 2 5+6 2 5+6 2 3+4 2 5 2 |
| | 5 2 6 2 3+4 2 3+4 2 5+6 2 5 2 |
| | 6 2 5 2 5+6 2 3+4 2 3+4 2 5 2 |
| | 6 2 6 2 3+4 2 5+6 2 5+6 2 5 2 |
| | 5 2 5 2 5+6 2 5+6 2 5+6 2 5 2 |
| | 5 2 5 2 3+4 2 5+6 2 5+6 2 5 2 |
| | 5 2 5 2 5+6 2 3+4 2 3+4 2 5 2 |
| | 6 2 5 2 5+6 2 5+6 2 3+4 2 5 2 |
| 71.1.92 | 281420 30 As1Te1Ti1 |
| | -GM -X -T -R -S -W |
| | 5 2 6 2 3+4 2 3+4 2 5+6 2 5 2 |
| | 5 2 5 2 5+6 2 5+6 2 3+4 2 5 2 |
| | 6 2 5 2 3+4 2 5+6 2 3+4 2 5 2 |
| | 6 2 6 2 5+6 2 3+4 2 5+6 2 5 2 |
| | 5 2 5 2 5+6 2 3+4 2 5+6 2 5 2 |
| | 6 2 6 2 3+4 2 5+6 2 3+4 2 5 2 |
| | 5 2 6 2 5+6 2 5+6 2 5+6 2 5 2 |
| | 6 2 5 2 3+4 2 3+4 2 3+4 2 5 2 |
| | 5 2 6 2 3+4 2 3+4 2 3+4 2 5 2 |
| | 5 2 6 2 5+6 2 3+4 2 5+6 2 5 2 |
| | 6 2 5 2 3+4 2 5+6 2 5+6 2 5 2 |
| | 5 2 6 2 3+4 2 3+4 2 3+4 2 5 2 |
| | 5 2 5 2 5+6 2 5+6 2 3+4 2 5 2 |
| | 6 2 5 2 5+6 2 5+6 2 3+4 2 5 2 |
| | 5 2 5 2 3+4 2 3+4 2 5+6 2 5 2 |
| 71.1.93 | 58046 48 Al5Ni2Pr1 |
| | -GM -X -T -R -S -W |
| | 5 2 5 2 3+4 2 3+4 2 3+4 2 5 2 |
| | 6 2 6 2 5+6 2 5+6 2 5+6 2 5 2 |
| | 6 2 6 2 5+6 2 5+6 2 5+6 2 5 2 |
| | 6 2 6 2 5+6 2 5+6 2 5+6 2 5 2 |
| | 5 2 5 2 5+6 2 5+6 2 3+4 2 5 2 |
| | 6 2 6 2 3+4 2 3+4 2 5+6 2 5 2 |
| | 5 2 6 2 3+4 2 3+4 2 3+4 2 5 2 |
| | 5 2 5 2 3+4 2 5+6 2 5+6 2 5 2 |
| | 6 2 5 2 5+6 2 3+4 2 5+6 2 5 2 |
| | 5 2 5 2 5+6 2 5+6 2 3+4 2 5 2 |
| | 5 2 6 2 3+4 2 3+4 2 3+4 2 5 2 |
| | 6 2 6 2 5+6 2 5+6 2 3+4 2 5 2 |
| | 6 2 6 2 3+4 2 3+4 2 5+6 2 5 2 |
| | 5 2 6 2 5+6 2 5+6 2 5+6 2 5 2 |
| | 5 2 6 2 3+4 2 3+4 2 5+6 2 5 2 |
| | 5 2 5 2 5+6 2 5+6 2 3+4 2 5 2 |
| | 6 2 5 2 3+4 2 3+4 2 5+6 2 5 2 |
| | 6 2 5 2 3+4 2 5+6 2 3+4 2 5 2 |
| | 6 2 6 2 3+4 2 3+4 2 3+4 2 5 2 |
| | 6 2 6 2 5+6 2 5+6 2 5+6 2 5 2 |
| | 5 2 6 2 5+6 2 3+4 2 3+4 2 5 2 |
| | 5 2 6 2 5+6 2 3+4 2 3+4 2 5 2 |

Continued on next column

**TABLE XLIV– continued**

| TQC | ICSD NoE SF |
|---|---|
| | 6 2 6 2 3+4 2 5+6 2 5+6 2 5 2 |
| | 6 2 6 2 5+6 2 5+6 2 5+6 2 5 2 |
| 71.1.94 | 418387 142 Al4O1Si6Sr10 |
| | -GM -X -T -R -S -W |
| | 5 2 5 2 3+4 2 3+4 2 3+4 2 5 2 |
| | 6 2 6 2 5+6 2 5+6 2 5+6 2 5 2 |
| | 5 2 6 2 3+4 2 3+4 2 3+4 2 5 2 |
| | 5 2 6 2 5+6 2 5+6 2 5+6 2 5 2 |
| | 6 2 5 2 3+4 2 3+4 2 3+4 2 5 2 |
| | 6 2 5 2 5+6 2 5+6 2 5+6 2 5 2 |
| | 5 2 5 2 3+4 2 5+6 2 5+6 2 5 2 |
| | 5 2 5 2 3+4 2 3+4 2 3+4 2 5 2 |
| | 6 2 6 2 5+6 2 5+6 2 5+6 2 5 2 |
| | 5 2 5 2 5+6 2 3+4 2 5+6 2 5 2 |
| | 5 2 5 2 3+4 2 3+4 2 3+4 2 5 2 |
| | 6 2 6 2 5+6 2 5+6 2 5+6 2 5 2 |
| | 5 2 5 2 3+4 2 3+4 2 3+4 2 5 2 |
| | 5 2 6 2 3+4 2 3+4 2 3+4 2 5 2 |
| | 5 2 6 2 5+6 2 5+6 2 5+6 2 5 2 |
| | 6 2 5 2 5+6 2 3+4 2 3+4 2 5 2 |
| | 6 2 5 2 3+4 2 5+6 2 5+6 2 5 2 |
| | 6 2 6 2 5+6 2 3+4 2 3+4 2 5 2 |
| | 6 2 6 2 5+6 2 5+6 2 5+6 2 5 2 |
| | 5 2 5 2 3+4 2 3+4 2 3+4 2 5 2 |
| | 6 2 6 2 5+6 2 5+6 2 5+6 2 5 2 |
| | 6 2 6 2 3+4 2 5+6 2 3+4 2 5 2 |
| | 5 2 5 2 5+6 2 3+4 2 5+6 2 5 2 |
| | 6 2 6 2 3+4 2 5+6 2 3+4 2 5 2 |
| | 5 2 5 2 3+4 2 3+4 2 3+4 2 5 2 |
| | 5 2 6 2 5+6 2 3+4 2 3+4 2 5 2 |
| | 5 2 6 2 3+4 2 5+6 2 5+6 2 5 2 |
| | 6 2 5 2 5+6 2 3+4 2 3+4 2 5 2 |
| | 6 2 6 2 3+4 2 5+6 2 5+6 2 5 2 |
| | 5 2 6 2 5+6 2 3+4 2 3+4 2 5 2 |
| | 6 2 5 2 5+6 2 5+6 2 3+4 2 5 2 |
| | 5 2 6 2 3+4 2 3+4 2 5+6 2 5 2 |
| | 6 2 5 2 3+4 2 3+4 2 3+4 2 5 2 |
| | 6 2 6 2 5+6 2 5+6 2 5+6 2 5 2 |
| | 6 2 5 2 5+6 2 3+4 2 3+4 2 5 2 |
| | 6 2 6 2 3+4 2 5+6 2 3+4 2 5 2 |
| | 5 2 5 2 5+6 2 3+4 2 5+6 2 5 2 |
| | 6 2 6 2 5+6 2 5+6 2 5+6 2 5 2 |
| | 6 2 6 2 3+4 2 5+6 2 3+4 2 5 2 |
| | 6 2 6 2 3+4 2 5+6 2 3+4 2 5 2 |
| | 5 2 5 2 3+4 2 3+4 2 3+4 2 5 2 |
| | 5 2 5 2 5+6 2 5+6 2 3+4 2 5 2 |
| | 6 2 6 2 3+4 2 3+4 2 5+6 2 5 2 |
| | 6 2 6 2 3+4 2 3+4 2 5+6 2 5 2 |
| | 5 2 5 2 5+6 2 5+6 2 3+4 2 5 2 |
| | 5 2 5 2 5+6 2 5+6 2 5+6 2 5 2 |
| | 6 2 6 2 3+4 2 3+4 2 3+4 2 5 2 |
| | 6 2 6 2 3+4 2 3+4 2 5+6 2 5 2 |
| | 5 2 5 2 5+6 2 5+6 2 3+4 2 5 2 |
| | 5 2 5 2 5+6 2 5+6 2 5+6 2 5 2 |
| | 6 2 6 2 5+6 2 5+6 2 3+4 2 5 2 |
| | 6 2 6 2 5+6 2 5+6 2 5+6 2 5 2 |
| | 6 2 5 2 5+6 2 3+4 2 5+6 2 5 2 |
| | 5 2 6 2 3+4 2 3+4 2 5+6 2 5 2 |
| | 6 2 6 2 3+4 2 5+6 2 3+4 2 5 2 |
| | 6 2 6 2 5+6 2 5+6 2 5+6 2 5 2 |
| | 5 2 5 2 5+6 2 3+4 2 3+4 2 5 2 |

Continued on next column

**TABLE XLIV– continued**

| TQC | ICSD NoE SF |
|---|---|
| | 6 2 6 2 3+4 2 5+6 2 5+6 2 5 2 |
| | 5 2 5 2 5+6 2 3+4 2 3+4 2 5 2 |
| | 6 2 6 2 3+4 2 5+6 2 5+6 2 5 2 |
| | 5 2 5 2 3+4 2 3+4 2 5+6 2 5 2 |
| | 6 2 5 2 5+6 2 5+6 2 5+6 2 5 2 |
| | 5 2 6 2 3+4 2 5+6 2 5+6 2 5 2 |
| | 6 2 6 2 5+6 2 3+4 2 3+4 2 5 2 |
| | 5 2 6 2 3+4 2 3+4 2 3+4 2 5 2 |
| | 5 2 5 2 5+6 2 3+4 2 3+4 2 5 2 |
| | 5 2 5 2 3+4 2 5+6 2 5+6 2 5 2 |
| | 6 2 5 2 5+6 2 5+6 2 5+6 2 5 2 |
| | 6 2 6 2 3+4 2 3+4 2 3+4 2 5 2 |
| | 6 2 6 2 5+6 2 5+6 2 5+6 2 5 2 |
| | 5 2 5 2 5+6 2 5+6 2 3+4 2 5 2 |
| 71.1.95 | 9564 44 Al2Sn2Sr3 |
| | -GM -X -T -R -S -W |
| | 5 2 6 2 3+4 2 3+4 2 3+4 2 5 2 |
| | 6 2 5 2 5+6 2 5+6 2 5+6 2 5 2 |
| | 5 2 5 2 3+4 2 3+4 2 3+4 2 5 2 |
| | 5 2 6 2 3+4 2 3+4 2 3+4 2 5 2 |
| | 6 2 5 2 5+6 2 5+6 2 5+6 2 5 2 |
| | 6 2 6 2 5+6 2 5+6 2 5+6 2 5 2 |
| | 5 2 6 2 5+6 2 3+4 2 5+6 2 5 2 |
| | 5 2 6 2 3+4 2 5+6 2 3+4 2 5 2 |
| | 6 2 5 2 5+6 2 5+6 2 5+6 2 5 2 |
| | 6 2 5 2 3+4 2 3+4 2 3+4 2 5 2 |
| | 6 2 6 2 5+6 2 5+6 2 5+6 2 5 2 |
| | 6 2 6 2 5+6 2 5+6 2 5+6 2 5 2 |
| | 5 2 5 2 3+4 2 5+6 2 5+6 2 5 2 |
| | 6 2 6 2 5+6 2 3+4 2 3+4 2 5 2 |
| | 5 2 5 2 5+6 2 5+6 2 3+4 2 5 2 |
| | 6 2 6 2 3+4 2 3+4 2 5+6 2 5 2 |
| | 5 2 5 2 5+6 2 5+6 2 3+4 2 5 2 |
| | 6 2 6 2 3+4 2 3+4 2 5+6 2 5 2 |
| | 6 2 6 2 5+6 2 3+4 2 3+4 2 5 2 |
| | 5 2 5 2 3+4 2 3+4 2 3+4 2 5 2 |
| | 5 2 5 2 5+6 2 5+6 2 5+6 2 5 2 |
| | 6 2 5 2 3+4 2 5+6 2 5+6 2 5 2 |
| 71.1.96 | 100127 44 Al2Si2Sr3 |
| | -GM -X -T -R -S -W |
| | 5 2 6 2 3+4 2 3+4 2 3+4 2 5 2 |
| | 6 2 5 2 5+6 2 5+6 2 5+6 2 5 2 |
| | 5 2 5 2 5+6 2 3+4 2 5+6 2 5 2 |
| | 5 2 6 2 3+4 2 5+6 2 3+4 2 5 2 |
| | 6 2 5 2 5+6 2 3+4 2 5+6 2 5 2 |
| | 6 2 6 2 3+4 2 5+6 2 3+4 2 5 2 |
| | 5 2 6 2 3+4 2 3+4 2 3+4 2 5 2 |
| | 5 2 6 2 5+6 2 5+6 2 5+6 2 5 2 |
| | 6 2 5 2 3+4 2 5+6 2 3+4 2 5 2 |
| | 6 2 5 2 5+6 2 3+4 2 5+6 2 5 2 |
| | 6 2 6 2 3+4 2 5+6 2 3+4 2 5 2 |
| | 6 2 6 2 3+4 2 5+6 2 3+4 2 5 2 |
| | 5 2 5 2 5+6 2 5+6 2 3+4 2 5 2 |
| | 6 2 6 2 3+4 2 3+4 2 5+6 2 5 2 |
| | 5 2 5 2 3+4 2 5+6 2 5+6 2 5 2 |
| | 6 2 6 2 5+6 2 3+4 2 3+4 2 5 2 |
| | 5 2 5 2 3+4 2 5+6 2 5+6 2 5 2 |
| | 6 2 6 2 5+6 2 3+4 2 3+4 2 5 2 |
| | 5 2 6 2 3+4 2 3+4 2 5+6 2 5 2 |
| | 6 2 5 2 5+6 2 3+4 2 5+6 2 5 2 |
| | 5 2 5 2 3+4 2 5+6 2 3+4 2 5 2 |

Continued on next column

**TABLE XLIV– continued**

| TQC | ICSD | NoE | SF |
|---|---|---|---|
| | 6 2 5 2 5+6 2 5+6 2 3+4 2 5 2 | | |
| 71.1.97 | 52617 | 44 | Al2Ge2Sr3 |
| | -GM -X -T -R -S -W | | |
| | 5 2 6 2 3+4 2 3+4 2 3+4 2 5 2 | | |
| | 6 2 5 2 5+6 2 5+6 2 5+6 2 5 2 | | |
| | 5 2 5 2 3+4 2 3+4 2 3+4 2 5 2 | | |
| | 5 2 6 2 3+4 2 5+6 2 3+4 2 5 2 | | |
| | 6 2 5 2 5+6 2 3+4 2 5+6 2 5 2 | | |
| | 6 2 6 2 5+6 2 5+6 2 5+6 2 5 2 | | |
| | 5 2 6 2 3+4 2 3+4 2 3+4 2 5 2 | | |
| | 5 2 6 2 5+6 2 5+6 2 5+6 2 5 2 | | |
| | 6 2 5 2 5+6 2 5+6 2 5+6 2 5 2 | | |
| | 6 2 5 2 3+4 2 3+4 2 3+4 2 5 2 | | |
| | 6 2 6 2 5+6 2 5+6 2 5+6 2 5 2 | | |
| | 6 2 6 2 5+6 2 5+6 2 5+6 2 5 2 | | |
| | 5 2 5 2 3+4 2 5+6 2 5+6 2 5 2 | | |
| | 6 2 6 2 5+6 2 3+4 2 3+4 2 5 2 | | |
| | 5 2 5 2 5+6 2 5+6 2 3+4 2 5 2 | | |
| | 6 2 6 2 3+4 2 3+4 2 5+6 2 5 2 | | |
| | 5 2 5 2 5+6 2 5+6 2 3+4 2 5 2 | | |
| | 6 2 6 2 3+4 2 3+4 2 5+6 2 5 2 | | |
| | 5 2 6 2 5+6 2 3+4 2 3+4 2 5 2 | | |
| | 6 2 5 2 3+4 2 3+4 2 3+4 2 5 2 | | |
| | 5 2 5 2 5+6 2 5+6 2 5+6 2 5 2 | | |
| | 6 2 5 2 3+4 2 5+6 2 5+6 2 5 2 | | |
| 71.1.98 | 100126 | 20 | Al2Ca3Si2 |
| | -GM -X -T -R -S -W | | |
| | 5 2 5 2 3+4 2 5+6 2 5+6 2 5 2 | | |
| | 6 2 6 2 5+6 2 3+4 2 3+4 2 5 2 | | |
| | 5 2 5 2 5+6 2 5+6 2 3+4 2 5 2 | | |
| | 6 2 6 2 3+4 2 3+4 2 5+6 2 5 2 | | |
| | 5 2 6 2 5+6 2 5+6 2 3+4 2 5 2 | | |
| | 6 2 5 2 3+4 2 3+4 2 5+6 2 5 2 | | |
| | 5 2 6 2 5+6 2 3+4 2 3+4 2 5 2 | | |
| | 6 2 5 2 3+4 2 3+4 2 3+4 2 5 2 | | |
| | 5 2 5 2 5+6 2 5+6 2 5+6 2 5 2 | | |
| | 6 2 5 2 3+4 2 5+6 2 5+6 2 5 2 | | |
| 71.1.99 | 52616 | 20 | Al2Ca3Ge2 |
| | -GM -X -T -R -S -W | | |
| | 5 2 5 2 5+6 2 5+6 2 3+4 2 5 2 | | |
| | 6 2 6 2 3+4 2 3+4 2 5+6 2 5 2 | | |
| | 5 2 5 2 3+4 2 5+6 2 5+6 2 5 2 | | |
| | 6 2 6 2 5+6 2 3+4 2 3+4 2 5 2 | | |
| | 5 2 6 2 3+4 2 5+6 2 5+6 2 5 2 | | |
| | 6 2 5 2 5+6 2 3+4 2 3+4 2 5 2 | | |
| | 5 2 6 2 3+4 2 3+4 2 5+6 2 5 2 | | |
| | 6 2 5 2 5+6 2 3+4 2 5+6 2 5 2 | | |
| | 5 2 5 2 3+4 2 5+6 2 3+4 2 5 2 | | |
| | 5 2 5 2 5+6 2 5+6 2 3+4 2 5 2 | | |
| 71.1.100 | 9565 | 44 | Al2Ba3Sn2 |
| | -GM -X -T -R -S -W | | |
| | 5 2 6 2 3+4 2 3+4 2 3+4 2 5 2 | | |
| | 6 2 5 2 5+6 2 5+6 2 5+6 2 5 2 | | |
| | 5 2 5 2 5+6 2 3+4 2 5+6 2 5 2 | | |
| | 5 2 6 2 3+4 2 5+6 2 3+4 2 5 2 | | |
| | 6 2 5 2 5+6 2 3+4 2 5+6 2 5 2 | | |
| | 6 2 6 2 3+4 2 5+6 2 3+4 2 5 2 | | |
| | 5 2 6 2 3+4 2 5+6 2 3+4 2 5 2 | | |
| | 5 2 6 2 5+6 2 3+4 2 5+6 2 5 2 | | |
| | 6 2 5 2 3+4 2 5+6 2 3+4 2 5 2 | | |
| | 6 2 5 2 5+6 2 3+4 2 5+6 2 5 2 | | |

Continued on next column

**TABLE XLIV– continued**

| TQC | ICSD | NoE | SF |
|---|---|---|---|
| | 6 2 6 2 3+4 2 5+6 2 3+4 2 5 2 | | |
| | 6 2 6 2 3+4 2 5+6 2 3+4 2 5 2 | | |
| | 5 2 5 2 5+6 2 5+6 2 5+6 2 5 2 | | |
| | 6 2 6 2 3+4 2 3+4 2 3+4 2 5 2 | | |
| | 5 2 5 2 3+4 2 5+6 2 5+6 2 5 2 | | |
| | 6 2 6 2 5+6 2 3+4 2 3+4 2 5 2 | | |
| | 5 2 5 2 3+4 2 3+4 2 5+6 2 5 2 | | |
| | 6 2 6 2 5+6 2 5+6 2 3+4 2 5 2 | | |
| | 5 2 6 2 5+6 2 3+4 2 5+6 2 5 2 | | |
| | 6 2 5 2 3+4 2 3+4 2 5+6 2 5 2 | | |
| | 5 2 5 2 3+4 2 5+6 2 3+4 2 5 2 | | |
| | 6 2 5 2 5+6 2 5+6 2 3+4 2 5 2 | | |
| 71.1.101 | 100128 | 44 | Al2Ba3Si2 |
| | -GM -X -T -R -S -W | | |
| | 5 2 6 2 3+4 2 3+4 2 3+4 2 5 2 | | |
| | 6 2 5 2 5+6 2 5+6 2 5+6 2 5 2 | | |
| | 5 2 5 2 5+6 2 3+4 2 5+6 2 5 2 | | |
| | 5 2 6 2 3+4 2 5+6 2 3+4 2 5 2 | | |
| | 6 2 5 2 5+6 2 3+4 2 5+6 2 5 2 | | |
| | 6 2 6 2 3+4 2 5+6 2 3+4 2 5 2 | | |
| | 5 2 6 2 3+4 2 3+4 2 3+4 2 5 2 | | |
| | 6 2 6 2 5+6 2 5+6 2 5+6 2 5 2 | | |
| | 5 2 5 2 3+4 2 5+6 2 3+4 2 5 2 | | |
| | 6 2 5 2 5+6 2 3+4 2 5+6 2 5 2 | | |
| | 6 2 6 2 3+4 2 5+6 2 3+4 2 5 2 | | |
| | 6 2 6 2 3+4 2 5+6 2 3+4 2 5 2 | | |
| | 5 2 5 2 5+6 2 5+6 2 5+6 2 5 2 | | |
| | 6 2 6 2 3+4 2 3+4 2 3+4 2 5 2 | | |
| | 5 2 5 2 3+4 2 5+6 2 5+6 2 5 2 | | |
| | 6 2 6 2 5+6 2 3+4 2 3+4 2 5 2 | | |
| | 5 2 5 2 3+4 2 5+6 2 5+6 2 5 2 | | |
| | 6 2 6 2 5+6 2 3+4 2 3+4 2 5 2 | | |
| | 5 2 6 2 5+6 2 3+4 2 5+6 2 5 2 | | |
| | 6 2 5 2 3+4 2 3+4 2 5+6 2 5 2 | | |
| | 5 2 5 2 5+6 2 5+6 2 3+4 2 5 2 | | |
| | 6 2 5 2 3+4 2 5+6 2 5+6 2 5 2 | | |
| 71.1.102 | 52612 | 44 | Al2Ba3Ge2 |
| | -GM -X -T -R -S -W | | |
| | 5 2 6 2 3+4 2 3+4 2 3+4 2 5 2 | | |
| | 6 2 5 2 5+6 2 5+6 2 5+6 2 5 2 | | |
| | 5 2 5 2 5+6 2 3+4 2 5+6 2 5 2 | | |
| | 5 2 6 2 3+4 2 5+6 2 3+4 2 5 2 | | |
| | 6 2 5 2 5+6 2 3+4 2 5+6 2 5 2 | | |
| | 6 2 6 2 3+4 2 5+6 2 3+4 2 5 2 | | |
| | 5 2 6 2 3+4 2 3+4 2 3+4 2 5 2 | | |
| | 5 2 6 2 5+6 2 5+6 2 5+6 2 5 2 | | |
| | 6 2 5 2 3+4 2 5+6 2 3+4 2 5 2 | | |
| | 6 2 5 2 5+6 2 3+4 2 5+6 2 5 2 | | |
| | 6 2 6 2 3+4 2 5+6 2 3+4 2 5 2 | | |
| | 6 2 6 2 3+4 2 5+6 2 3+4 2 5 2 | | |
| | 5 2 5 2 5+6 2 3+4 2 5+6 2 5 2 | | |
| | 6 2 6 2 3+4 2 5+6 2 3+4 2 5 2 | | |
| | 5 2 5 2 3+4 2 5+6 2 5+6 2 5 2 | | |
| | 6 2 6 2 5+6 2 3+4 2 3+4 2 5 2 | | |
| | 5 2 5 2 3+4 2 3+4 2 5+6 2 5 2 | | |
| | 5 2 6 2 5+6 2 5+6 2 3+4 2 5 2 | | |
| | 6 2 5 2 5+6 2 3+4 2 5+6 2 5 2 | | |
| | 6 2 6 2 3+4 2 5+6 2 5+6 2 5 2 | | |
| | 5 2 5 2 3+4 2 3+4 2 3+4 2 5 2 | | |
| | 6 2 5 2 5+6 2 5+6 2 5+6 2 5 2 | | |
| 71.1.103 | 163921 | 162 | Al16Ce2Pt9 |

Continued on next column

**TABLE XLIV– continued**

| TQC | ICSD | NoE | SF | | | |
|---|---|---|---|---|---|---|
| | -GM | -X | -T | -R | -S | -W |
| | 5 2 | 5 2 | 3+4 2 | 3+4 2 | 3+4 2 | 5 2 |
| | 6 2 | 6 2 | 5+6 2 | 5+6 2 | 5+6 2 | 5 2 |
| | 5 2 | 5 2 | 3+4 2 | 3+4 2 | 3+4 2 | 5 2 |
| | 6 2 | 6 2 | 5+6 2 | 5+6 2 | 5+6 2 | 5 2 |
| | 5 2 | 6 2 | 3+4 2 | 3+4 2 | 3+4 2 | 5 2 |
| | 6 2 | 5 2 | 5+6 2 | 5+6 2 | 5+6 2 | 5 2 |
| | 5 2 | 6 2 | 3+4 2 | 3+4 2 | 3+4 2 | 5 2 |
| | 6 2 | 5 2 | 5+6 2 | 5+6 2 | 5+6 2 | 5 2 |
| | 5 2 | 5 2 | 5+6 2 | 5+6 2 | 5+6 2 | 5 2 |
| | 6 2 | 6 2 | 3+4 2 | 3+4 2 | 3+4 2 | 5 2 |
| | 5 2 | 6 2 | 5+6 2 | 3+4 2 | 3+4 2 | 5 2 |
| | 6 2 | 5 2 | 5+6 2 | 5+6 2 | 5+6 2 | 5 2 |
| | 5 2 | 5 2 | 3+4 2 | 3+4 2 | 5+6 2 | 5 2 |
| | 5 2 | 6 2 | 3+4 2 | 5+6 2 | 3+4 2 | 5 2 |
| | 6 2 | 6 2 | 5+6 2 | 3+4 2 | 5+6 2 | 5 2 |
| | 5 2 | 5 2 | 5+6 2 | 5+6 2 | 5+6 2 | 5 2 |
| | 6 2 | 5 2 | 3+4 2 | 3+4 2 | 3+4 2 | 5 2 |
| | 5 2 | 5 2 | 3+4 2 | 5+6 2 | 3+4 2 | 5 2 |
| | 6 2 | 6 2 | 5+6 2 | 3+4 2 | 5+6 2 | 5 2 |
| | 5 2 | 6 2 | 3+4 2 | 5+6 2 | 3+4 2 | 5 2 |
| | 6 2 | 5 2 | 5+6 2 | 3+4 2 | 5+6 2 | 5 2 |
| | 5 2 | 5 2 | 3+4 2 | 5+6 2 | 3+4 2 | 5 2 |
| | 5 2 | 6 2 | 5+6 2 | 5+6 2 | 5+6 2 | 5 2 |
| | 6 2 | 5 2 | 3+4 2 | 3+4 2 | 3+4 2 | 5 2 |
| | 5 2 | 5 2 | 5+6 2 | 3+4 2 | 5+6 2 | 5 2 |
| | 6 2 | 6 2 | 3+4 2 | 5+6 2 | 3+4 2 | 5 2 |
| | 5 2 | 6 2 | 3+4 2 | 5+6 2 | 5+6 2 | 5 2 |
| | 6 2 | 5 2 | 5+6 2 | 3+4 2 | 3+4 2 | 5 2 |
| | 5 2 | 6 2 | 5+6 2 | 3+4 2 | 5+6 2 | 5 2 |
| | 5 2 | 5 2 | 3+4 2 | 5+6 2 | 5+6 2 | 5 2 |
| | 6 2 | 6 2 | 5+6 2 | 3+4 2 | 3+4 2 | 5 2 |
| | 6 2 | 5 2 | 3+4 2 | 3+4 2 | 5+6 2 | 5 2 |
| | 5 2 | 6 2 | 5+6 2 | 5+6 2 | 3+4 2 | 5 2 |
| | 5 2 | 5 2 | 5+6 2 | 3+4 2 | 5+6 2 | 5 2 |
| | 6 2 | 6 2 | 3+4 2 | 5+6 2 | 3+4 2 | 5 2 |
| | 6 2 | 5 2 | 5+6 2 | 3+4 2 | 5+6 2 | 5 2 |
| | 5 2 | 6 2 | 5+6 2 | 5+6 2 | 3+4 2 | 5 2 |
| | 5 2 | 6 2 | 3+4 2 | 3+4 2 | 5+6 2 | 5 2 |
| | 6 2 | 5 2 | 5+6 2 | 3+4 2 | 3+4 2 | 5 2 |
| | 5 2 | 5 2 | 3+4 2 | 5+6 2 | 5+6 2 | 5 2 |
| | 6 2 | 6 2 | 5+6 2 | 3+4 2 | 3+4 2 | 5 2 |
| | 6 2 | 5 2 | 3+4 2 | 5+6 2 | 5+6 2 | 5 2 |
| | 5 2 | 6 2 | 5+6 2 | 3+4 2 | 3+4 2 | 5 2 |
| | 5 2 | 5 2 | 3+4 2 | 5+6 2 | 5+6 2 | 5 2 |
| | 5 2 | 6 2 | 5+6 2 | 3+4 2 | 3+4 2 | 5 2 |
| | 6 2 | 6 2 | 3+4 2 | 5+6 2 | 5+6 2 | 5 2 |
| | 5 2 | 5 2 | 5+6 2 | 3+4 2 | 5+6 2 | 5 2 |
| | 6 2 | 5 2 | 5+6 2 | 3+4 2 | 3+4 2 | 5 2 |
| | 6 2 | 6 2 | 3+4 2 | 5+6 2 | 3+4 2 | 5 2 |
| | 5 2 | 5 2 | 5+6 2 | 3+4 2 | 5+6 2 | 5 2 |
| | 5 2 | 5 2 | 5+6 2 | 5+6 2 | 5+6 2 | 5 2 |
| | 6 2 | 6 2 | 3+4 2 | 5+6 2 | 3+4 2 | 5 2 |
| | 5 2 | 6 2 | 5+6 2 | 3+4 2 | 5+6 2 | 5 2 |
| | 5 2 | 5 2 | 3+4 2 | 3+4 2 | 3+4 2 | 5 2 |
| | 5 2 | 6 2 | 5+6 2 | 3+4 2 | 5+6 2 | 5 2 |
| | 5 2 | 5 2 | 3+4 2 | 5+6 2 | 3+4 2 | 5 2 |
| | 6 2 | 5 2 | 3+4 2 | 3+4 2 | 5+6 2 | 5 2 |
| | 6 2 | 5 2 | 5+6 2 | 5+6 2 | 3+4 2 | 5 2 |
| | 6 2 | 6 2 | 5+6 2 | 3+4 2 | 5+6 2 | 5 2 |
| | 5 2 | 5 2 | 3+4 2 | 5+6 2 | 3+4 2 | 5 2 |

Continued on next column

**TABLE XLIV– continued**

| TQC | ICSD | NoE | SF | | | |
|---|---|---|---|---|---|---|
| | 5 2 | 6 2 | 5+6 2 | 3+4 2 | 5+6 2 | 5 2 |
| | 6 2 | 5 2 | 3+4 2 | 5+6 2 | 5+6 2 | 5 2 |
| | 6 2 | 5 2 | 5+6 2 | 3+4 2 | 3+4 2 | 5 2 |
| | 5 2 | 5 2 | 3+4 2 | 5+6 2 | 5+6 2 | 5 2 |
| | 5 2 | 6 2 | 5+6 2 | 3+4 2 | 3+4 2 | 5 2 |
| | 6 2 | 6 2 | 3+4 2 | 5+6 2 | 5+6 2 | 5 2 |
| | 6 2 | 5 2 | 5+6 2 | 3+4 2 | 3+4 2 | 5 2 |
| | 5 2 | 6 2 | 5+6 2 | 5+6 2 | 5+6 2 | 5 2 |
| | 5 2 | 6 2 | 3+4 2 | 3+4 2 | 5+6 2 | 5 2 |
| | 6 2 | 5 2 | 3+4 2 | 5+6 2 | 3+4 2 | 5 2 |
| | 6 2 | 6 2 | 5+6 2 | 3+4 2 | 3+4 2 | 5 2 |
| | 6 2 | 5 2 | 3+4 2 | 3+4 2 | 5+6 2 | 5 2 |
| | 5 2 | 6 2 | 5+6 2 | 5+6 2 | 5+6 2 | 5 2 |
| | 5 2 | 5 2 | 3+4 2 | 5+6 2 | 3+4 2 | 5 2 |
| | 5 2 | 6 2 | 5+6 2 | 3+4 2 | 3+4 2 | 5 2 |
| | 6 2 | 6 2 | 5+6 2 | 5+6 2 | 5+6 2 | 5 2 |
| | 5 2 | 5 2 | 3+4 2 | 3+4 2 | 3+4 2 | 5 2 |
| | 6 2 | 6 2 | 3+4 2 | 5+6 2 | 3+4 2 | 5 2 |
| | 6 2 | 6 2 | 5+6 2 | 3+4 2 | 5+6 2 | 5 2 |
| | 5 2 | 5 2 | 3+4 2 | 5+6 2 | 5+6 2 | 5 2 |
| | 5 2 | 5 2 | 5+6 2 | 3+4 2 | 5+6 2 | 5 2 |
| 71.1.104 | 157372 | 114 | Ag4Gd3Sn4 | | | |
| | -GM | -X | -T | -R | -S | -W |
| | 5 2 | 5 2 | 5+6 2 | 5+6 2 | 3+4 2 | 5 2 |
| | 5 2 | 5 2 | 5+6 2 | 3+4 2 | 5+6 2 | 5 2 |
| | 6 2 | 6 2 | 3+4 2 | 5+6 2 | 3+4 2 | 5 2 |
| | 6 2 | 6 2 | 3+4 2 | 3+4 2 | 5+6 2 | 5 2 |
| | 5 2 | 5 2 | 5+6 2 | 3+4 2 | 5+6 2 | 5 2 |
| | 6 2 | 6 2 | 3+4 2 | 5+6 2 | 3+4 2 | 5 2 |
| | 6 2 | 6 2 | 3+4 2 | 3+4 2 | 5+6 2 | 5 2 |
| | 5 2 | 5 2 | 3+4 2 | 3+4 2 | 5+6 2 | 5 2 |
| | 6 2 | 6 2 | 5+6 2 | 3+4 2 | 3+4 2 | 5 2 |
| | 5 2 | 5 2 | 3+4 2 | 5+6 2 | 5+6 2 | 5 2 |
| | 6 2 | 6 2 | 5+6 2 | 3+4 2 | 5+6 2 | 5 2 |
| | 6 2 | 6 2 | 3+4 2 | 5+6 2 | 3+4 2 | 5 2 |
| | 5 2 | 6 2 | 3+4 2 | 3+4 2 | 3+4 2 | 5 2 |
| | 5 2 | 5 2 | 5+6 2 | 5+6 2 | 3+4 2 | 5 2 |
| | 6 2 | 5 2 | 3+4 2 | 3+4 2 | 5+6 2 | 5 2 |
| | 6 2 | 6 2 | 5+6 2 | 5+6 2 | 5+6 2 | 5 2 |
| | 6 2 | 5 2 | 5+6 2 | 3+4 2 | 3+4 2 | 5 2 |
| | 6 2 | 5 2 | 3+4 2 | 5+6 2 | 5+6 2 | 5 2 |
| | 5 2 | 6 2 | 5+6 2 | 3+4 2 | 3+4 2 | 5 2 |
| | 5 2 | 6 2 | 3+4 2 | 5+6 2 | 3+4 2 | 5 2 |
| | 6 2 | 5 2 | 5+6 2 | 5+6 2 | 5+6 2 | 5 2 |
| | 5 2 | 6 2 | 3+4 2 | 3+4 2 | 5+6 2 | 5 2 |
| | 6 2 | 5 2 | 5+6 2 | 3+4 2 | 3+4 2 | 5 2 |
| | 5 2 | 6 2 | 3+4 2 | 5+6 2 | 5+6 2 | 5 2 |
| | 5 2 | 6 2 | 5+6 2 | 3+4 2 | 3+4 2 | 5 2 |
| | 5 2 | 5 2 | 3+4 2 | 5+6 2 | 5+6 2 | 5 2 |
| | 6 2 | 6 2 | 3+4 2 | 5+6 2 | 5+6 2 | 5 2 |
| | 5 2 | 5 2 | 5+6 2 | 3+4 2 | 3+4 2 | 5 2 |
| | 6 2 | 6 2 | 5+6 2 | 5+6 2 | 5+6 2 | 5 2 |
| | 5 2 | 5 2 | 3+4 2 | 3+4 2 | 3+4 2 | 5 2 |
| | 5 2 | 6 2 | 5+6 2 | 5+6 2 | 3+4 2 | 5 2 |
| | 6 2 | 5 2 | 3+4 2 | 3+4 2 | 5+6 2 | 5 2 |
| | 5 2 | 6 2 | 3+4 2 | 5+6 2 | 3+4 2 | 5 2 |
| | 6 2 | 5 2 | 5+6 2 | 3+4 2 | 5+6 2 | 5 2 |
| | 6 2 | 5 2 | 3+4 2 | 5+6 2 | 5+6 2 | 5 2 |
| | 6 2 | 6 2 | 5+6 2 | 3+4 2 | 3+4 2 | 5 2 |
| | 6 2 | 6 2 | 5+6 2 | 5+6 2 | 3+4 2 | 5 2 |
| | 5 2 | 5 2 | 3+4 2 | 3+4 2 | 5+6 2 | 5 2 |

Continued on next column

**TABLE XLIV– continued**

| TQC | ICSD | NoE | SF |
|---|---|---|---|
| | 5 2 6 2 3+4 2 3+4 2 5+6 2 5 2 | | |
| | 6 2 5 2 5+6 2 5+6 2 3+4 2 5 2 | | |
| | 5 2 6 2 5+6 2 5+6 2 5+6 2 5 2 | | |
| | 5 2 5 2 5+6 2 3+4 2 3+4 2 5 2 | | |
| | 5 2 6 2 3+4 2 3+4 2 3+4 2 5 2 | | |
| | 6 2 6 2 3+4 2 5+6 2 5+6 2 5 2 | | |
| | 6 2 5 2 3+4 2 3+4 2 3+4 2 5 2 | | |
| | 5 2 5 2 5+6 2 5+6 2 3+4 2 5 2 | | |
| | 6 2 6 2 3+4 2 3+4 2 5+6 2 5 2 | | |
| | 6 2 6 2 5+6 2 3+4 2 5+6 2 5 2 | | |
| | 6 2 6 2 3+4 2 3+4 2 3+4 2 5 2 | | |
| | 5 2 6 2 3+4 2 5+6 2 5+6 2 5 2 | | |
| | 6 2 5 2 5+6 2 3+4 2 5+6 2 5 2 | | |
| | 5 2 6 2 3+4 2 5+6 2 5+6 2 5 2 | | |
| | 5 2 6 2 3+4 2 3+4 2 3+4 2 5 2 | | |
| | 6 2 5 2 5+6 2 5+6 2 5+6 2 5 2 | | |
| | 6 2 5 2 3+4 2 3+4 2 3+4 2 5 2 | | |
| | 6 2 6 2 3+4 2 5+6 2 5+6 2 5 2 | | |
| | 5 2 6 2 5+6 2 3+4 2 5+6 2 5 2 | | |
| 71.1.105 | 417491 | 96 | Ag4Ce3Ge4 |
| | -GM -X -T -R -S -W | | |
| | 5 2 5 2 3+4 2 5+6 2 5+6 2 5 2 | | |
| | 5 2 5 2 3+4 2 3+4 2 3+4 2 5 2 | | |
| | 6 2 6 2 5+6 2 5+6 2 5+6 2 5 2 | | |
| | 6 2 6 2 5+6 2 3+4 2 3+4 2 5 2 | | |
| | 5 2 5 2 3+4 2 3+4 2 3+4 2 5 2 | | |
| | 6 2 6 2 5+6 2 5+6 2 5+6 2 5 2 | | |
| | 6 2 6 2 5+6 2 3+4 2 3+4 2 5 2 | | |
| | 5 2 5 2 3+4 2 3+4 2 5+6 2 5 2 | | |
| | 6 2 6 2 5+6 2 5+6 2 3+4 2 5 2 | | |
| | 5 2 5 2 3+4 2 3+4 2 3+4 2 5 2 | | |
| | 6 2 6 2 5+6 2 5+6 2 3+4 2 5 2 | | |
| | 6 2 6 2 5+6 2 3+4 2 5+6 2 5 2 | | |
| | 5 2 5 2 5+6 2 3+4 2 5+6 2 5 2 | | |
| | 5 2 6 2 3+4 2 5+6 2 5+6 2 5 2 | | |
| | 6 2 5 2 5+6 2 3+4 2 3+4 2 5 2 | | |
| | 6 2 6 2 3+4 2 5+6 2 3+4 2 5 2 | | |
| | 6 2 5 2 3+4 2 3+4 2 5+6 2 5 2 | | |
| | 5 2 5 2 5+6 2 5+6 2 3+4 2 5 2 | | |
| | 6 2 6 2 3+4 2 5+6 2 5+6 2 5 2 | | |
| | 5 2 6 2 5+6 2 3+4 2 5+6 2 5 2 | | |
| | 6 2 5 2 3+4 2 5+6 2 3+4 2 5 2 | | |
| | 5 2 6 2 5+6 2 3+4 2 3+4 2 5 2 | | |
| | 5 2 6 2 3+4 2 3+4 2 5+6 2 5 2 | | |
| | 6 2 5 2 5+6 2 5+6 2 3+4 2 5 2 | | |
| | 5 2 6 2 3+4 2 3+4 2 5+6 2 5 2 | | |
| | 5 2 5 2 5+6 2 5+6 2 3+4 2 5 2 | | |
| | 6 2 6 2 5+6 2 5+6 2 3+4 2 5 2 | | |
| | 5 2 5 2 3+4 2 3+4 2 5+6 2 5 2 | | |
| | 6 2 6 2 3+4 2 5+6 2 5+6 2 5 2 | | |
| | 5 2 5 2 5+6 2 3+4 2 3+4 2 5 2 | | |
| | 5 2 6 2 3+4 2 5+6 2 5+6 2 5 2 | | |
| | 6 2 5 2 5+6 2 3+4 2 3+4 2 5 2 | | |
| | 5 2 6 2 5+6 2 5+6 2 5+6 2 5 2 | | |
| | 6 2 5 2 3+4 2 3+4 2 3+4 2 5 2 | | |
| | 6 2 5 2 5+6 2 5+6 2 5+6 2 5 2 | | |
| | 6 2 6 2 3+4 2 3+4 2 3+4 2 5 2 | | |
| | 6 2 6 2 3+4 2 5+6 2 5+6 2 5 2 | | |
| | 5 2 5 2 5+6 2 3+4 2 3+4 2 5 2 | | |
| | 5 2 6 2 5+6 2 3+4 2 5+6 2 5 2 | | |
| | 6 2 5 2 3+4 2 5+6 2 5+6 2 5 2 | | |

Continued on next column

**TABLE XLIV– continued**

| TQC | ICSD | NoE | SF |
|---|---|---|---|
| | 5 2 5 2 3+4 2 3+4 2 3+4 2 5 2 | | |
| | 5 2 6 2 5+6 2 5+6 2 5+6 2 5 2 | | |
| | 5 2 6 2 3+4 2 3+4 2 3+4 2 5 2 | | |
| | 5 2 6 2 3+4 2 5+6 2 3+4 2 5 2 | | |
| | 5 2 5 2 5+6 2 5+6 2 5+6 2 5 2 | | |
| | 6 2 5 2 5+6 2 5+6 2 5+6 2 5 2 | | |
| | 6 2 5 2 5+6 2 3+4 2 3+4 2 5 2 | | |
| | 6 2 6 2 3+4 2 3+4 2 3+4 2 5 2 | | |

The End

TABLE XLV: The computed irreducible representations for the valence bands for the compounds of SEBR and NLC in space group 72. See the list of tables.

| TQC | ICSD | NoE | SF |
|---|---|---|---|
| 72.1.1 | 84206 | 144 | Ni3Si5Y2 |
| | -GM -X -S -R -T -W | | |
| | 5 2 6 2 2+2 4 2+2 4 3+4 2 5+5 4 | | |
| | 5 2 6 2 2+2 4 2+2 4 5+6 2 5+5 4 | | |
| | 6 2 5 2 2+2 4 2+2 4 3+4 2 5+5 4 | | |
| | 6 2 5 2 2+2 4 2+2 4 5+6 2 5+5 4 | | |
| | 5 2 6 2 2+2 4 2+2 4 3+4 2 5+5 4 | | |
| | 5 2 6 2 2+2 4 2+2 4 5+6 2 5+5 4 | | |
| | 6 2 5 2 2+2 4 2+2 4 3+4 2 5+5 4 | | |
| | 6 2 5 2 2+2 4 2+2 4 5+6 2 5+5 4 | | |
| | 5 2 6 2 2+2 4 2+2 4 3+4 2 5+5 4 | | |
| | 5 2 6 2 2+2 4 2+2 4 5+6 2 5+5 4 | | |
| | 6 2 5 2 2+2 4 2+2 4 3+4 2 5+5 4 | | |
| | 5 2 5 2 2+2 4 2+2 4 5+6 2 5+5 4 | | |
| | 6 2 6 2 2+2 4 2+2 4 3+4 2 5+5 4 | | |
| | 5 2 6 2 2+2 4 2+2 4 5+6 2 5+5 4 | | |
| | 6 2 5 2 2+2 4 2+2 4 3+4 2 5+5 4 | | |
| | 6 2 5 2 2+2 4 2+2 4 5+6 2 5+5 4 | | |
| | 5 2 5 2 2+2 4 2+2 4 5+6 2 5+5 4 | | |
| | 6 2 6 2 2+2 4 2+2 4 3+4 2 5+5 4 | | |
| | 5 2 5 2 2+2 4 2+2 4 3+4 2 5+5 4 | | |
| | 6 2 6 2 2+2 4 2+2 4 3+4 2 5+5 4 | | |
| | 5 2 5 2 2+2 4 2+2 4 5+6 2 5+5 4 | | |
| | 6 2 6 2 2+2 4 2+2 4 5+6 2 5+5 4 | | |
| | 6 2 5 2 2+2 4 2+2 4 5+6 2 5+5 4 | | |
| | 5 2 5 2 2+2 4 2+2 4 3+4 2 5+5 4 | | |
| | 5 2 6 2 2+2 4 2+2 4 3+4 2 5+5 4 | | |
| | 6 2 6 2 2+2 4 2+2 4 5+6 2 5+5 4 | | |
| | 5 2 6 2 2+2 4 2+2 4 5+6 2 5+5 4 | | |
| | 6 2 5 2 2+2 4 2+2 4 3+4 2 5+5 4 | | |
| | 5 2 6 2 2+2 4 2+2 4 3+4 2 5+5 4 | | |
| | 5 2 5 2 2+2 4 2+2 4 5+6 2 5+5 4 | | |
| | 6 2 5 2 2+2 4 2+2 4 3+4 2 5+5 4 | | |
| | 6 2 6 2 2+2 4 2+2 4 5+6 2 5+5 4 | | |
| | 5 2 6 2 2+2 4 2+2 4 5+6 2 5+5 4 | | |
| | 5 2 6 2 2+2 4 2+2 4 3+4 2 5+5 4 | | |
| | 5 2 5 2 2+2 4 2+2 4 3+4 2 5+5 4 | | |
| | 6 2 5 2 2+2 4 2+2 4 5+6 2 5+5 4 | | |
| | 6 2 5 2 3+4 2 | | |
| | 5 2 5 2 5+6 2 | | |
| | 6 2 6 2 3+4 2 | | |
| | 5 2 6 2 5+6 2 | | |
| | 6 2 5 2 3+4 2 | | |
| | 5 2 6 2 3+4 2 | | |
| | 6 2 5 2 5+6 2 | | |

Continued on next column

**TABLE XLV– continued**

| TQC | ICSD | NoE | SF |
|---|---|---|---|
| | 6 2 5 2 | | 3+4 2 |
| | 5 2 6 2 | | 5+6 2 |
| | 6 2 5 2 | | 3+4 2 |
| | 5 2 6 2 | | 5+6 2 |
| | 6 2 5 2 | | 3+4 2 |
| | 6 2 6 2 | | 5+6 2 |
| | 5 2 6 2 | | 5+6 2 |
| | 6 2 5 2 | | 3+4 2 |
| | 5 2 6 2 | | 5+6 2 |
| | 6 2 6 2 | | 3+4 2 |
| | 5 2 5 2 | | 3+4 2 |
| | 5 2 5 2 | | 5+6 2 |
| | 6 2 6 2 | | 5+6 2 |
| | 5 2 5 2 | | 5+6 2 |
| | 6 2 6 2 | | 3+4 2 |
| | 5 2 5 2 | | 5+6 2 |
| | 6 2 6 2 | | 3+4 2 |
| | 5 2 5 2 | | 5+6 2 |
| | 6 2 5 2 | | 5+6 2 |
| | 5 2 6 2 | | 3+4 2 |
| | 5 2 5 2 | | 3+4 2 |
| | 5 2 5 2 | | 5+6 2 |
| | 6 2 6 2 | | 3+4 2 |
| | 5 2 5 2 | | 5+6 2 |
| | 6 2 5 2 | | 3+4 2 |
| | 5 2 6 2 | | 3+4 2 |
| | 6 2 5 2 | | 5+6 2 |
| | 5 2 6 2 | | 5+6 2 |
| | 6 2 6 2 | | 3+4 2 |
| 72.1.2 | 428478 | 168 | Eu2Pt3Si5 |
| | -GM -X | -S | -R -T -W |
| | 5 2 6 2 | 2+2 4 | 2+2 4 3+4 2 5+5 4 |
| | 5 2 5 2 | 2+2 4 | 2+2 4 5+6 2 5+5 4 |
| | 6 2 6 2 | 2+2 4 | 2+2 4 3+4 2 5+5 4 |
| | 6 2 5 2 | 2+2 4 | 2+2 4 5+6 2 5+5 4 |
| | 5 2 6 2 | 2+2 4 | 2+2 4 3+4 2 5+5 4 |
| | 5 2 6 2 | 2+2 4 | 2+2 4 5+6 2 5+5 4 |
| | 6 2 5 2 | 2+2 4 | 2+2 4 3+4 2 5+5 4 |
| | 6 2 5 2 | 2+2 4 | 2+2 4 5+6 2 5+5 4 |
| | 5 2 6 2 | 2+2 4 | 2+2 4 3+4 2 5+5 4 |
| | 6 2 5 2 | 2+2 4 | 2+2 4 5+6 2 5+5 4 |
| | 5 2 6 2 | 2+2 4 | 2+2 4 3+4 2 5+5 4 |
| | 5 2 5 2 | 2+2 4 | 2+2 4 5+6 2 5+5 4 |
| | 6 2 6 2 | 2+2 4 | 2+2 4 3+4 2 5+5 4 |
| | 5 2 6 2 | 2+2 4 | 2+2 4 5+6 2 5+5 4 |
| | 6 2 5 2 | 2+2 4 | 2+2 4 3+4 2 5+5 4 |
| | 6 2 5 2 | 2+2 4 | 2+2 4 5+6 2 5+5 4 |
| | 5 2 5 2 | 2+2 4 | 2+2 4 5+6 2 5+5 4 |
| | 6 2 6 2 | 2+2 4 | 2+2 4 3+4 2 5+5 4 |
| | 5 2 5 2 | 2+2 4 | 2+2 4 3+4 2 5+5 4 |
| | 6 2 6 2 | 2+2 4 | 2+2 4 5+6 2 5+5 4 |
| | 5 2 5 2 | 2+2 4 | 2+2 4 3+4 2 5+5 4 |
| | 6 2 6 2 | 2+2 4 | 2+2 4 5+6 2 5+5 4 |
| | 6 2 5 2 | 2+2 4 | 2+2 4 3+4 2 5+5 4 |
| | 6 2 6 2 | 2+2 4 | 2+2 4 5+6 2 5+5 4 |
| | 5 2 5 2 | 2+2 4 | 2+2 4 5+6 2 5+5 4 |
| | 5 2 6 2 | 2+2 4 | 2+2 4 3+4 2 5+5 4 |
| | 5 2 6 2 | 2+2 4 | 2+2 4 5+6 2 5+5 4 |
| | 5 2 5 2 | 2+2 4 | 2+2 4 3+4 2 5+5 4 |
| | 6 2 6 2 | 2+2 4 | 2+2 4 3+4 2 5+5 4 |
| | 5 2 5 2 | 2+2 4 | 2+2 4 5+6 2 5+5 4 |

Continued on next column

**TABLE XLV– continued**

| TQC | ICSD | NoE | SF |
|---|---|---|---|
| | 6 2 6 2 | 2+2 4 | 2+2 4 3+4 2 5+5 4 |
| | 5 2 6 2 | 2+2 4 | 2+2 4 5+6 2 5+5 4 |
| | 6 2 5 2 | 2+2 4 | 2+2 4 3+4 2 5+5 4 |
| | 5 2 5 2 | 2+2 4 | 2+2 4 5+6 2 5+5 4 |
| | 6 2 6 2 | 2+2 4 | 2+2 4 3+4 2 5+5 4 |
| | 6 2 5 2 | 2+2 4 | 2+2 4 3+4 2 5+5 4 |
| | 5 2 6 2 | 2+2 4 | 2+2 4 5+6 2 5+5 4 |
| | 5 2 5 2 | 2+2 4 | 2+2 4 3+4 2 5+5 4 |
| | 6 2 5 2 | 2+2 4 | 2+2 4 5+6 2 5+5 4 |
| | 6 2 6 2 | 2+2 4 | 2+2 4 3+4 2 5+5 4 |
| | 6 2 6 2 | 2+2 4 | 2+2 4 5+6 2 5+5 4 |
| | 5 2 6 2 | 2+2 4 | 2+2 4 3+4 2 5+5 4 |
| | 5 2 5 2 | | 5+6 2 |
| | 6 2 5 2 | | 3+4 2 |
| | 5 2 5 2 | | 5+6 2 |
| | 5 2 6 2 | | 5+6 2 |
| | 6 2 6 2 | | 3+4 2 |
| | 5 2 5 2 | | 3+4 2 |
| | 6 2 5 2 | | 5+6 2 |
| | 6 2 5 2 | | 5+6 2 |
| | 6 2 6 2 | | 3+4 2 |
| | 5 2 6 2 | | 5+6 2 |
| | 5 2 5 2 | | 3+4 2 |
| | 6 2 6 2 | | 5+6 2 |
| | 6 2 5 2 | | 5+6 2 |
| | 5 2 6 2 | | 3+4 2 |
| | 5 2 5 2 | | 5+6 2 |
| | 6 2 5 2 | | 5+6 2 |
| | 5 2 6 2 | | 3+4 2 |
| | 5 2 5 2 | | 5+6 2 |
| | 6 2 5 2 | | 3+4 2 |
| | 5 2 6 2 | | 5+6 2 |
| | 6 2 5 2 | | 3+4 2 |
| | 5 2 6 2 | | 5+6 2 |
| | 5 2 5 2 | | 3+4 2 |
| | 6 2 6 2 | | 3+4 2 |
| | 6 2 5 2 | | 5+6 2 |
| | 5 2 6 2 | | 3+4 2 |
| | 6 2 5 2 | | 5+6 2 |
| | 5 2 6 2 | | 3+4 2 |
| | 5 2 5 2 | | 5+6 2 |
| | 6 2 6 2 | | 3+4 2 |
| | 5 2 6 2 | | 5+6 2 |
| | 6 2 5 2 | | 5+6 2 |
| | 6 2 6 2 | | 3+4 2 |
| | 5 2 5 2 | | 3+4 2 |
| | 6 2 6 2 | | 5+6 2 |
| | 5 2 5 2 | | 3+4 2 |
| | 5 2 6 2 | | 5+6 2 |
| | 6 2 5 2 | | 3+4 2 |
| | 6 2 6 2 | | 5+6 2 |
| | 5 2 5 2 | | 3+4 2 |
| | 5 2 5 2 | | 5+6 2 |
| | 6 2 6 2 | | 3+4 2 |
| 72.2.1 | 170063 | 76 | Pb5Sc6 |
| | -GM -X | -S | -R -T -W |
| | 5 2 5 2 | 2+2 4 | 2+2 4 3+4 2 5+5 4 |
| | 5 2 6 2 | 2+2 4 | 2+2 4 5+6 2 5+5 4 |
| | 5 2 5 2 | 2+2 4 | 2+2 4 3+4 2 5+5 4 |
| | 5 2 6 2 | 2+2 4 | 2+2 4 5+6 2 5+5 4 |
| | 6 2 6 2 | 2+2 4 | 2+2 4 3+4 2 5+5 4 |

Continued on next column

**TABLE XLV– continued**

| TQC | ICSD | | NoE | | SF | |
|---|---|---|---|---|---|---|
| | 6 2 | 5 2 | 2+2 4 | 2+2 4 | 5+6 2 | 5+5 4 |
| | 5 2 | 5 2 | 2+2 4 | 2+2 4 | 3+4 2 | 5+5 4 |
| | 6 2 | 5 2 | 2+2 4 | 2+2 4 | 5+6 2 | 5+5 4 |
| | 6 2 | 6 2 | 2+2 4 | 2+2 4 | 3+4 2 | 5+5 4 |
| | 6 2 | 6 2 | 2+2 4 | 2+2 4 | 5+6 2 | 5+5 4 |
| | 6 2 | 5 2 | 2+2 4 | 2+2 4 | 5+6 2 | 5+5 4 |
| | 5 2 | 6 2 | 2+2 4 | 2+2 4 | 3+4 2 | 5+5 4 |
| | 6 2 | 5 2 | 2+2 4 | 2+2 4 | 5+6 2 | 5+5 4 |
| | 5 2 | 6 2 | 2+2 4 | 2+2 4 | 3+4 2 | 5+5 4 |
| | 6 2 | 6 2 | 2+2 4 | 2+2 4 | 5+6 2 | 5+5 4 |
| | 6 2 | 5 2 | 2+2 4 | 2+2 4 | 3+4 2 | 5+5 4 |
| | 5 2 | 5 2 | 2+2 4 | 2+2 4 | 5+6 2 | 5+5 4 |
| | 5 2 | 6 2 | 2+2 4 | 2+2 4 | 3+4 2 | 5+5 4 |
| | 5 2 | 6 2 | 2+2 4 | 2+2 4 | 3+4 2 | 5+5 4 |
| | 5 2 | 5 2 | | | 5+6 2 | |
| | 6 2 | 6 2 | | | 3+4 2 | |
| | 5 2 | 6 2 | | | 5+6 2 | |
| | 5 2 | 6 2 | | | 5+6 2 | |
| | 6 2 | 5 2 | | | 3+4 2 | |
| | 5 2 | 5 2 | | | 5+6 2 | |
| | 5 2 | 6 2 | | | 3+4 2 | |
| | 5 2 | 5 2 | | | 3+4 2 | |
| | 6 2 | 5 2 | | | 5+6 2 | |
| | 6 2 | 6 2 | | | 5+6 2 | |
| | 5 2 | 5 2 | | | 3+4 2 | |
| | 5 2 | 5 2 | | | 5+6 2 | |
| | 6 2 | 6 2 | | | 3+4 2 | |
| | 5 2 | 5 2 | | | 5+6 2 | |
| | 6 2 | 5 2 | | | 3+4 2 | |
| | 6 2 | 5 2 | | | 5+6 2 | |
| | 5 2 | 6 2 | | | 3+4 2 | |
| | 5 2 | 6 2 | | | 5+6 2 | |
| | 5 2 | 6 2 | | | 3+4 2 | |
| 72.2.2 | 416512 | | 84 | | Ni1Pd2Te2 | |
| | -GM | -X | -S | -R | -T | -W |
| | 5 2 | 6 2 | 2+2 4 | 2+2 4 | 3+4 2 | 5+5 4 |
| | 5 2 | 5 2 | 2+2 4 | 2+2 4 | 5+6 2 | 5+5 4 |
| | 6 2 | 6 2 | 2+2 4 | 2+2 4 | 5+6 2 | 5+5 4 |
| | 6 2 | 5 2 | 2+2 4 | 2+2 4 | 3+4 2 | 5+5 4 |
| | 5 2 | 5 2 | 2+2 4 | 2+2 4 | 3+4 2 | 5+5 4 |
| | 6 2 | 6 2 | 2+2 4 | 2+2 4 | 5+6 2 | 5+5 4 |
| | 5 2 | 6 2 | 2+2 4 | 2+2 4 | 3+4 2 | 5+5 4 |
| | 5 2 | 6 2 | 2+2 4 | 2+2 4 | 5+6 2 | 5+5 4 |
| | 5 2 | 5 2 | 2+2 4 | 2+2 4 | 3+4 2 | 5+5 4 |
| | 6 2 | 6 2 | 2+2 4 | 2+2 4 | 5+6 2 | 5+5 4 |
| | 5 2 | 6 2 | 2+2 4 | 2+2 4 | 3+4 2 | 5+5 4 |
| | 6 2 | 6 2 | 2+2 4 | 2+2 4 | 5+6 2 | 5+5 4 |
| | 6 2 | 5 2 | 2+2 4 | 2+2 4 | 3+4 2 | 5+5 4 |
| | 5 2 | 5 2 | 2+2 4 | 2+2 4 | 5+6 2 | 5+5 4 |
| | 6 2 | 6 2 | 2+2 4 | 2+2 4 | 5+6 2 | 5+5 4 |
| | 6 2 | 5 2 | 2+2 4 | 2+2 4 | 3+4 2 | 5+5 4 |
| | 6 2 | 5 2 | 2+2 4 | 2+2 4 | 3+4 2 | 5+5 4 |
| | 5 2 | 6 2 | 2+2 4 | 2+2 4 | 5+6 2 | 5+5 4 |
| | 5 2 | 5 2 | 2+2 4 | 2+2 4 | 5+6 2 | 5+5 4 |
| | 6 2 | 5 2 | 2+2 4 | 2+2 4 | 3+4 2 | 5+5 4 |
| | 5 2 | 6 2 | 2+2 4 | 2+2 4 | 5+6 2 | 5+5 4 |
| | 6 2 | 6 2 | | | 5+6 2 | |
| | 5 2 | 5 2 | | | 3+4 2 | |
| | 6 2 | 6 2 | | | 5+6 2 | |
| | 5 2 | 5 2 | | | 3+4 2 | |
| | 5 2 | 5 2 | | | 3+4 2 | |

Continued on next column

**TABLE XLV– continued**

| TQC | ICSD | | NoE | | SF | |
|---|---|---|---|---|---|---|
| | 5 2 | 5 2 | | | 5+6 2 | |
| | 5 2 | 6 2 | | | 3+4 2 | |
| | 6 2 | 5 2 | | | 3+4 2 | |
| | 6 2 | 6 2 | | | 5+6 2 | |
| | 6 2 | 5 2 | | | 3+4 2 | |
| | 6 2 | 6 2 | | | 5+6 2 | |
| | 6 2 | 6 2 | | | 5+6 2 | |
| | 6 2 | 5 2 | | | 3+4 2 | |
| | 5 2 | 5 2 | | | 5+6 2 | |
| | 5 2 | 5 2 | | | 3+4 2 | |
| | 6 2 | 6 2 | | | 3+4 2 | |
| | 5 2 | 5 2 | | | 3+4 2 | |
| | 5 2 | 5 2 | | | 5+6 2 | |
| | 5 2 | 6 2 | | | 5+6 2 | |
| | 6 2 | 6 2 | | | 5+6 2 | |
| | 6 2 | 5 2 | | | 3+4 2 | |
| 72.2.3 | 416515 | | 80 | | Fe1Pd2Se2 | |
| | -GM | -X | -S | -R | -T | -W |
| | 5 2 | 6 2 | 2+2 4 | 2+2 4 | 3+4 2 | 5+5 4 |
| | 5 2 | 5 2 | 2+2 4 | 2+2 4 | 5+6 2 | 5+5 4 |
| | 6 2 | 6 2 | 2+2 4 | 2+2 4 | 5+6 2 | 5+5 4 |
| | 6 2 | 5 2 | 2+2 4 | 2+2 4 | 3+4 2 | 5+5 4 |
| | 5 2 | 5 2 | 2+2 4 | 2+2 4 | 3+4 2 | 5+5 4 |
| | 6 2 | 6 2 | 2+2 4 | 2+2 4 | 3+4 2 | 5+5 4 |
| | 5 2 | 6 2 | 2+2 4 | 2+2 4 | 5+6 2 | 5+5 4 |
| | 5 2 | 5 2 | 2+2 4 | 2+2 4 | 5+6 2 | 5+5 4 |
| | 5 2 | 6 2 | 2+2 4 | 2+2 4 | 5+6 2 | 5+5 4 |
| | 6 2 | 6 2 | 2+2 4 | 2+2 4 | 3+4 2 | 5+5 4 |
| | 6 2 | 6 2 | 2+2 4 | 2+2 4 | 3+4 2 | 5+5 4 |
| | 5 2 | 6 2 | 2+2 4 | 2+2 4 | 5+6 2 | 5+5 4 |
| | 5 2 | 5 2 | 2+2 4 | 2+2 4 | 3+4 2 | 5+5 4 |
| | 6 2 | 5 2 | 2+2 4 | 2+2 4 | 5+6 2 | 5+5 4 |
| | 6 2 | 6 2 | 2+2 4 | 2+2 4 | 5+6 2 | 5+5 4 |
| | 6 2 | 5 2 | 2+2 4 | 2+2 4 | 3+4 2 | 5+5 4 |
| | 6 2 | 5 2 | 2+2 4 | 2+2 4 | 3+4 2 | 5+5 4 |
| | 5 2 | 5 2 | 2+2 4 | 2+2 4 | 5+6 2 | 5+5 4 |
| | 5 2 | 6 2 | 2+2 4 | 2+2 4 | 5+6 2 | 5+5 4 |
| | 6 2 | 5 2 | 2+2 4 | 2+2 4 | 3+4 2 | 5+5 4 |
| | 6 2 | 6 2 | | | 5+6 2 | |
| | 5 2 | 5 2 | | | 5+6 2 | |
| | 6 2 | 6 2 | | | 3+4 2 | |
| | 5 2 | 5 2 | | | 3+4 2 | |
| | 5 2 | 6 2 | | | 5+6 2 | |
| | 5 2 | 6 2 | | | 3+4 2 | |
| | 6 2 | 5 2 | | | 5+6 2 | |
| | 5 2 | 6 2 | | | 3+4 2 | |
| | 5 2 | 5 2 | | | 5+6 2 | |
| | 6 2 | 5 2 | | | 3+4 2 | |
| | 6 2 | 5 2 | | | 5+6 2 | |
| | 5 2 | 5 2 | | | 3+4 2 | |
| | 6 2 | 6 2 | | | 3+4 2 | |
| | 6 2 | 5 2 | | | 5+6 2 | |
| | 6 2 | 6 2 | | | 3+4 2 | |
| | 6 2 | 5 2 | | | 3+4 2 | |
| | 5 2 | 6 2 | | | 5+6 2 | |
| | 5 2 | 5 2 | | | 3+4 2 | |
| | 6 2 | 6 2 | | | 5+6 2 | |
| | 5 2 | 5 2 | | | 5+6 2 | |
| 72.2.4 | 68792 | | 176 | | Er2Ru3Si5 | |
| | -GM | -X | -S | -R | -T | -W |
| | 5 2 | 6 2 | 2+2 4 | 2+2 4 | 3+4 2 | 5+5 4 |

Continued on next column

**TABLE XLV– continued**

| TQC | ICSD | NoE | SF | | | |
|---|---|---|---|---|---|---|
| | 5 2 | 6 2 | 2+2 4 | 2+2 4 | 5+6 2 | 5+5 4 |
| | 6 2 | 5 2 | 2+2 4 | 2+2 4 | 3+4 2 | 5+5 4 |
| | 6 2 | 5 2 | 2+2 4 | 2+2 4 | 5+6 2 | 5+5 4 |
| | 5 2 | 6 2 | 2+2 4 | 2+2 4 | 3+4 2 | 5+5 4 |
| | 5 2 | 6 2 | 2+2 4 | 2+2 4 | 5+6 2 | 5+5 4 |
| | 6 2 | 5 2 | 2+2 4 | 2+2 4 | 3+4 2 | 5+5 4 |
| | 6 2 | 5 2 | 2+2 4 | 2+2 4 | 5+6 2 | 5+5 4 |
| | 5 2 | 6 2 | 2+2 4 | 2+2 4 | 5+6 2 | 5+5 4 |
| | 5 2 | 6 2 | 2+2 4 | 2+2 4 | 3+4 2 | 5+5 4 |
| | 6 2 | 5 2 | 2+2 4 | 2+2 4 | 3+4 2 | 5+5 4 |
| | 5 2 | 5 2 | 2+2 4 | 2+2 4 | 5+6 2 | 5+5 4 |
| | 5 2 | 6 2 | 2+2 4 | 2+2 4 | 5+6 2 | 5+5 4 |
| | 6 2 | 6 2 | 2+2 4 | 2+2 4 | 3+4 2 | 5+5 4 |
| | 6 2 | 5 2 | 2+2 4 | 2+2 4 | 3+4 2 | 5+5 4 |
| | 6 2 | 5 2 | 2+2 4 | 2+2 4 | 5+6 2 | 5+5 4 |
| | 5 2 | 5 2 | 2+2 4 | 2+2 4 | 3+4 2 | 5+5 4 |
| | 6 2 | 6 2 | 2+2 4 | 2+2 4 | 5+6 2 | 5+5 4 |
| | 5 2 | 5 2 | 2+2 4 | 2+2 4 | 5+6 2 | 5+5 4 |
| | 6 2 | 6 2 | 2+2 4 | 2+2 4 | 5+6 2 | 5+5 4 |
| | 5 2 | 5 2 | 2+2 4 | 2+2 4 | 3+4 2 | 5+5 4 |
| | 6 2 | 5 2 | 2+2 4 | 2+2 4 | 3+4 2 | 5+5 4 |
| | 6 2 | 6 2 | 2+2 4 | 2+2 4 | 5+6 2 | 5+5 4 |
| | 5 2 | 5 2 | 2+2 4 | 2+2 4 | 3+4 2 | 5+5 4 |
| | 6 2 | 6 2 | 2+2 4 | 2+2 4 | 5+6 2 | 5+5 4 |
| | 5 2 | 6 2 | 2+2 4 | 2+2 4 | 3+4 2 | 5+5 4 |
| | 6 2 | 6 2 | 2+2 4 | 2+2 4 | 3+4 2 | 5+5 4 |
| | 5 2 | 5 2 | 2+2 4 | 2+2 4 | 5+6 2 | 5+5 4 |
| | 5 2 | 6 2 | 2+2 4 | 2+2 4 | 5+6 2 | 5+5 4 |
| | 6 2 | 5 2 | 2+2 4 | 2+2 4 | 3+4 2 | 5+5 4 |
| | 5 2 | 6 2 | 2+2 4 | 2+2 4 | 3+4 2 | 5+5 4 |
| | 6 2 | 5 2 | 2+2 4 | 2+2 4 | 5+6 2 | 5+5 4 |
| | 5 2 | 5 2 | 2+2 4 | 2+2 4 | 3+4 2 | 5+5 4 |
| | 5 2 | 6 2 | 2+2 4 | 2+2 4 | 5+6 2 | 5+5 4 |
| | 6 2 | 6 2 | 2+2 4 | 2+2 4 | 5+6 2 | 5+5 4 |
| | 6 2 | 6 2 | 2+2 4 | 2+2 4 | 5+6 2 | 5+5 4 |
| | 5 2 | 5 2 | 2+2 4 | 2+2 4 | 3+4 2 | 5+5 4 |
| | 6 2 | 5 2 | 2+2 4 | 2+2 4 | 5+6 2 | 5+5 4 |
| | 6 2 | 6 2 | 2+2 4 | 2+2 4 | 5+6 2 | 5+5 4 |
| | 5 2 | 6 2 | 2+2 4 | 2+2 4 | 3+4 2 | 5+5 4 |
| | 5 2 | 5 2 | 2+2 4 | 2+2 4 | 5+6 2 | 5+5 4 |
| | 5 2 | 5 2 | 2+2 4 | 2+2 4 | 5+6 2 | 5+5 4 |
| | 5 2 | 5 2 | 2+2 4 | 2+2 4 | 3+4 2 | 5+5 4 |
| | 6 2 | 5 2 | 2+2 4 | 2+2 4 | 3+4 2 | 5+5 4 |
| | 5 2 | 6 2 | | | 3+4 2 | |
| | 6 2 | 6 2 | | | 3+4 2 | |
| | 6 2 | 5 2 | | | 5+6 2 | |
| | 5 2 | 6 2 | | | 5+6 2 | |
| | 6 2 | 5 2 | | | 3+4 2 | |
| | 6 2 | 6 2 | | | 3+4 2 | |
| | 5 2 | 5 2 | | | 5+6 2 | |
| | 6 2 | 6 2 | | | 5+6 2 | |
| | 5 2 | 5 2 | | | 5+6 2 | |
| | 6 2 | 6 2 | | | 3+4 2 | |
| | 6 2 | 5 2 | | | 3+4 2 | |
| | 5 2 | 6 2 | | | 3+4 2 | |
| | 6 2 | 5 2 | | | 5+6 2 | |
| | 6 2 | 6 2 | | | 5+6 2 | |
| | 5 2 | 6 2 | | | 3+4 2 | |
| | 6 2 | 5 2 | | | 3+4 2 | |
| | 5 2 | 5 2 | | | 5+6 2 | |
| | 5 2 | 5 2 | | | 3+4 2 | |
| | 5 2 | 6 2 | | | 3+4 2 | |
| | 6 2 | 6 2 | | | 5+6 2 | |
| | 6 2 | 6 2 | | | 5+6 2 | |
| | 5 2 | 5 2 | | | 5+6 2 | |
| | 5 2 | 6 2 | | | 3+4 2 | |
| | 6 2 | 5 2 | | | 3+4 2 | |
| | 6 2 | 6 2 | | | 5+6 2 | |
| | 5 2 | 5 2 | | | 3+4 2 | |
| | 6 2 | 6 2 | | | 3+4 2 | |
| | 5 2 | 5 2 | | | 5+6 2 | |
| | 6 2 | 5 2 | | | 5+6 2 | |
| | 5 2 | 5 2 | | | 3+4 2 | |
| | 5 2 | 6 2 | | | 3+4 2 | |
| | 6 2 | 5 2 | | | 3+4 2 | |
| | 5 2 | 6 2 | | | 5+6 2 | |
| | 5 2 | 5 2 | | | 5+6 2 | |
| | 5 2 | 5 2 | | | 3+4 2 | |
| | 6 2 | 5 2 | | | 5+6 2 | |
| | 6 2 | 6 2 | | | 3+4 2 | |
| | 5 2 | 6 2 | | | 3+4 2 | |
| | 6 2 | 6 2 | | | 5+6 2 | |
| | 5 2 | 6 2 | | | 3+4 2 | |
| | 5 2 | 5 2 | | | 5+6 2 | |
| | 6 2 | 5 2 | | | 5+6 2 | |
| | 5 2 | 6 2 | | | 3+4 2 | |
| | 6 2 | 6 2 | | | 3+4 2 | |
| 72.2.5 | 54107 | 100 | Cr4Sc2Si5 | | | |
| | -GM | -X | -S | -R | -T | -W |
| | 5 2 | 5 2 | 2+2 4 | 2+2 4 | 5+6 2 | 5+5 4 |
| | 5 2 | 6 2 | 2+2 4 | 2+2 4 | 3+4 2 | 5+5 4 |
| | 5 2 | 5 2 | 2+2 4 | 2+2 4 | 3+4 2 | 5+5 4 |
| | 5 2 | 6 2 | 2+2 4 | 2+2 4 | 5+6 2 | 5+5 4 |
| | 6 2 | 6 2 | 2+2 4 | 2+2 4 | 3+4 2 | 5+5 4 |
| | 6 2 | 5 2 | 2+2 4 | 2+2 4 | 5+6 2 | 5+5 4 |
| | 5 2 | 6 2 | 2+2 4 | 2+2 4 | 3+4 2 | 5+5 4 |
| | 6 2 | 5 2 | 2+2 4 | 2+2 4 | 5+6 2 | 5+5 4 |
| | 6 2 | 5 2 | 2+2 4 | 2+2 4 | 3+4 2 | 5+5 4 |
| | 6 2 | 5 2 | 2+2 4 | 2+2 4 | 3+4 2 | 5+5 4 |
| | 6 2 | 6 2 | 2+2 4 | 2+2 4 | 5+6 2 | 5+5 4 |
| | 6 2 | 6 2 | 2+2 4 | 2+2 4 | 5+6 2 | 5+5 4 |
| | 5 2 | 5 2 | 2+2 4 | 2+2 4 | 3+4 2 | 5+5 4 |
| | 5 2 | 6 2 | 2+2 4 | 2+2 4 | 5+6 2 | 5+5 4 |
| | 6 2 | 6 2 | 2+2 4 | 2+2 4 | 5+6 2 | 5+5 4 |
| | 5 2 | 6 2 | 2+2 4 | 2+2 4 | 5+6 2 | 5+5 4 |
| | 6 2 | 5 2 | 2+2 4 | 2+2 4 | 3+4 2 | 5+5 4 |
| | 5 2 | 5 2 | 2+2 4 | 2+2 4 | 3+4 2 | 5+5 4 |
| | 5 2 | 6 2 | 2+2 4 | 2+2 4 | 3+4 2 | 5+5 4 |
| | 5 2 | 5 2 | 2+2 4 | 2+2 4 | 3+4 2 | 5+5 4 |
| | 5 2 | 6 2 | 2+2 4 | 2+2 4 | 5+6 2 | 5+5 4 |
| | 5 2 | 5 2 | 2+2 4 | 2+2 4 | 5+6 2 | 5+5 4 |
| | 6 2 | 6 2 | 2+2 4 | 2+2 4 | 5+6 2 | 5+5 4 |
| | 5 2 | 5 2 | 2+2 4 | 2+2 4 | 5+6 2 | 5+5 4 |
| | 6 2 | 6 2 | 2+2 4 | 2+2 4 | 3+4 2 | 5+5 4 |
| | 6 2 | 5 2 | | | 5+6 2 | |
| | 5 2 | 6 2 | | | 3+4 2 | |
| | 5 2 | 6 2 | | | 5+6 2 | |
| | 5 2 | 5 2 | | | 3+4 2 | |
| | 5 2 | 5 2 | | | 5+6 2 | |
| | 6 2 | 6 2 | | | 3+4 2 | |
| | 5 2 | 6 2 | | | 5+6 2 | |
| | 6 2 | 5 2 | | | 3+4 2 | |

Continued on next column

Continued on next column

**TABLE XLV– continued**

| TQC | ICSD | NoE | SF | | | |
|---|---|---|---|---|---|---|
| | 6 2 | 5 2 | | | 5+6 2 | |
| | 6 2 | 5 2 | | | 3+4 2 | |
| | 5 2 | 6 2 | | | 5+6 2 | |
| | 6 2 | 5 2 | | | 3+4 2 | |
| | 5 2 | 6 2 | | | 3+4 2 | |
| | 5 2 | 5 2 | | | 5+6 2 | |
| | 6 2 | 6 2 | | | 3+4 2 | |
| | 5 2 | 5 2 | | | 3+4 2 | |
| | 6 2 | 5 2 | | | 5+6 2 | |
| | 5 2 | 6 2 | | | 3+4 2 | |
| | 6 2 | 6 2 | | | 5+6 2 | |
| | 5 2 | 5 2 | | | 3+4 2 | |
| | 6 2 | 6 2 | | | 5+6 2 | |
| | 6 2 | 5 2 | | | 5+6 2 | |
| | 6 2 | 6 2 | | | 3+4 2 | |
| | 5 2 | 6 2 | | | 5+6 2 | |
| | 6 2 | 5 2 | | | 3+4 2 | |
| 72.2.6 | 40758 | 140 | C4Fe1Tm2 | | | |
| | -GM | -X | -S | -R | -T | -W |
| | 5 2 | 6 2 | 2+2 4 | 2+2 4 | 5+6 2 | 5+5 4 |
| | 5 2 | 5 2 | 2+2 4 | 2+2 4 | 3+4 2 | 5+5 4 |
| | 6 2 | 6 2 | 2+2 4 | 2+2 4 | 3+4 2 | 5+5 4 |
| | 6 2 | 5 2 | 2+2 4 | 2+2 4 | 5+6 2 | 5+5 4 |
| | 5 2 | 6 2 | 2+2 4 | 2+2 4 | 3+4 2 | 5+5 4 |
| | 5 2 | 5 2 | 2+2 4 | 2+2 4 | 5+6 2 | 5+5 4 |
| | 6 2 | 6 2 | 2+2 4 | 2+2 4 | 5+6 2 | 5+5 4 |
| | 6 2 | 5 2 | 2+2 4 | 2+2 4 | 3+4 2 | 5+5 4 |
| | 5 2 | 6 2 | 2+2 4 | 2+2 4 | 5+6 2 | 5+5 4 |
| | 5 2 | 5 2 | 2+2 4 | 2+2 4 | 3+4 2 | 5+5 4 |
| | 6 2 | 5 2 | 2+2 4 | 2+2 4 | 5+6 2 | 5+5 4 |
| | 5 2 | 6 2 | 2+2 4 | 2+2 4 | 3+4 2 | 5+5 4 |
| | 6 2 | 6 2 | 2+2 4 | 2+2 4 | 5+6 2 | 5+5 4 |
| | 5 2 | 5 2 | 2+2 4 | 2+2 4 | 3+4 2 | 5+5 4 |
| | 6 2 | 6 2 | 2+2 4 | 2+2 4 | 5+6 2 | 5+5 4 |
| | 6 2 | 5 2 | 2+2 4 | 2+2 4 | 3+4 2 | 5+5 4 |
| | 5 2 | 6 2 | 2+2 4 | 2+2 4 | 5+6 2 | 5+5 4 |
| | 6 2 | 5 2 | 2+2 4 | 2+2 4 | 3+4 2 | 5+5 4 |
| | 6 2 | 5 2 | 2+2 4 | 2+2 4 | 3+4 2 | 5+5 4 |
| | 5 2 | 6 2 | 2+2 4 | 2+2 4 | 5+6 2 | 5+5 4 |
| | 5 2 | 5 2 | 2+2 4 | 2+2 4 | 3+4 2 | 5+5 4 |
| | 6 2 | 6 2 | 2+2 4 | 2+2 4 | 5+6 2 | 5+5 4 |
| | 6 2 | 6 2 | 2+2 4 | 2+2 4 | 5+6 2 | 5+5 4 |
| | 5 2 | 5 2 | 2+2 4 | 2+2 4 | 3+4 2 | 5+5 4 |
| | 6 2 | 5 2 | 2+2 4 | 2+2 4 | 5+6 2 | 5+5 4 |
| | 5 2 | 6 2 | 2+2 4 | 2+2 4 | 3+4 2 | 5+5 4 |
| | 6 2 | 6 2 | 2+2 4 | 2+2 4 | 5+6 2 | 5+5 4 |
| | 5 2 | 5 2 | 2+2 4 | 2+2 4 | 5+6 2 | 5+5 4 |
| | 5 2 | 5 2 | 2+2 4 | 2+2 4 | 3+4 2 | 5+5 4 |
| | 5 2 | 6 2 | 2+2 4 | 2+2 4 | 3+4 2 | 5+5 4 |
| | 6 2 | 5 2 | 2+2 4 | 2+2 4 | 5+6 2 | 5+5 4 |
| | 6 2 | 6 2 | 2+2 4 | 2+2 4 | 3+4 2 | 5+5 4 |
| | 6 2 | 6 2 | 2+2 4 | 2+2 4 | 5+6 2 | 5+5 4 |
| | 5 2 | 5 2 | 2+2 4 | 2+2 4 | 3+4 2 | 5+5 4 |
| | 5 2 | 6 2 | 2+2 4 | 2+2 4 | 5+6 2 | 5+5 4 |
| | 6 2 | 5 2 | | | 3+4 2 | |
| | 5 2 | 5 2 | | | 3+4 2 | |
| | 6 2 | 5 2 | | | 5+6 2 | |
| | 6 2 | 6 2 | | | 5+6 2 | |
| | 5 2 | 6 2 | | | 3+4 2 | |
| | 6 2 | 6 2 | | | 5+6 2 | |
| | 6 2 | 5 2 | | | 3+4 2 | |

Continued on next column

**TABLE XLV– continued**

| TQC | ICSD | NoE | SF | | | |
|---|---|---|---|---|---|---|
| | 5 2 | 5 2 | | | 5+6 2 | |
| | 5 2 | 6 2 | | | 3+4 2 | |
| | 6 2 | 6 2 | | | 5+6 2 | |
| | 5 2 | 5 2 | | | 3+4 2 | |
| | 5 2 | 5 2 | | | 5+6 2 | |
| | 6 2 | 6 2 | | | 3+4 2 | |
| | 5 2 | 5 2 | | | 5+6 2 | |
| | 6 2 | 5 2 | | | 3+4 2 | |
| | 6 2 | 6 2 | | | 5+6 2 | |
| | 5 2 | 6 2 | | | 3+4 2 | |
| | 5 2 | 5 2 | | | 3+4 2 | |
| | 6 2 | 6 2 | | | 5+6 2 | |
| | 6 2 | 6 2 | | | 3+4 2 | |
| | 6 2 | 5 2 | | | 5+6 2 | |
| | 5 2 | 6 2 | | | 3+4 2 | |
| | 5 2 | 5 2 | | | 5+6 2 | |
| | 5 2 | 5 2 | | | 5+6 2 | |
| | 6 2 | 6 2 | | | 3+4 2 | |
| | 5 2 | 5 2 | | | 5+6 2 | |
| | 6 2 | 6 2 | | | 3+4 2 | |
| | 5 2 | 5 2 | | | 5+6 2 | |
| | 6 2 | 5 2 | | | 3+4 2 | |
| | 5 2 | 6 2 | | | 5+6 2 | |
| | 5 2 | 6 2 | | | 3+4 2 | |
| | 5 2 | 5 2 | | | 5+6 2 | |
| | 6 2 | 5 2 | | | 3+4 2 | |
| | 5 2 | 6 2 | | | 3+4 2 | |
| | 6 2 | 6 2 | | | 3+4 2 | |
| 72.2.7 | 28232 | 196 | Br2Hg5O4 | | | |
| | -GM | -X | -S | -R | -T | -W |
| | 5 2 | 6 2 | 2+2 4 | 2+2 4 | 3+4 2 | 5+5 4 |
| | 6 2 | 5 2 | 2+2 4 | 2+2 4 | 5+6 2 | 5+5 4 |
| | 5 2 | 6 2 | 2+2 4 | 2+2 4 | 3+4 2 | 5+5 4 |
| | 6 2 | 6 2 | 2+2 4 | 2+2 4 | 5+6 2 | 5+5 4 |
| | 5 2 | 5 2 | 2+2 4 | 2+2 4 | 5+6 2 | 5+5 4 |
| | 6 2 | 5 2 | 2+2 4 | 2+2 4 | 3+4 2 | 5+5 4 |
| | 5 2 | 6 2 | 2+2 4 | 2+2 4 | 5+6 2 | 5+5 4 |
| | 6 2 | 5 2 | 2+2 4 | 2+2 4 | 3+4 2 | 5+5 4 |
| | 5 2 | 5 2 | 2+2 4 | 2+2 4 | 5+6 2 | 5+5 4 |
| | 5 2 | 5 2 | 2+2 4 | 2+2 4 | 5+6 2 | 5+5 4 |
| | 6 2 | 6 2 | 2+2 4 | 2+2 4 | 3+4 2 | 5+5 4 |
| | 6 2 | 6 2 | 2+2 4 | 2+2 4 | 3+4 2 | 5+5 4 |
| | 5 2 | 6 2 | 2+2 4 | 2+2 4 | 3+4 2 | 5+5 4 |
| | 6 2 | 5 2 | 2+2 4 | 2+2 4 | 5+6 2 | 5+5 4 |
| | 5 2 | 5 2 | 2+2 4 | 2+2 4 | 3+4 2 | 5+5 4 |
| | 5 2 | 6 2 | 2+2 4 | 2+2 4 | 3+4 2 | 5+5 4 |
| | 6 2 | 5 2 | 2+2 4 | 2+2 4 | 5+6 2 | 5+5 4 |
| | 6 2 | 6 2 | 2+2 4 | 2+2 4 | 5+6 2 | 5+5 4 |
| | 5 2 | 5 2 | 2+2 4 | 2+2 4 | 3+4 2 | 5+5 4 |
| | 6 2 | 6 2 | 2+2 4 | 2+2 4 | 5+6 2 | 5+5 4 |
| | 5 2 | 5 2 | 2+2 4 | 2+2 4 | 5+6 2 | 5+5 4 |
| | 6 2 | 6 2 | 2+2 4 | 2+2 4 | 3+4 2 | 5+5 4 |
| | 5 2 | 6 2 | 2+2 4 | 2+2 4 | 5+6 2 | 5+5 4 |
| | 6 2 | 6 2 | 2+2 4 | 2+2 4 | 3+4 2 | 5+5 4 |
| | 5 2 | 5 2 | 2+2 4 | 2+2 4 | 3+4 2 | 5+5 4 |
| | 6 2 | 6 2 | 2+2 4 | 2+2 4 | 5+6 2 | 5+5 4 |
| | 5 2 | 5 2 | 2+2 4 | 2+2 4 | 5+6 2 | 5+5 4 |
| | 6 2 | 5 2 | 2+2 4 | 2+2 4 | 3+4 2 | 5+5 4 |
| | 6 2 | 6 2 | 2+2 4 | 2+2 4 | 5+6 2 | 5+5 4 |
| | 5 2 | 5 2 | 2+2 4 | 2+2 4 | 3+4 2 | 5+5 4 |
| | 6 2 | 6 2 | 2+2 4 | 2+2 4 | 5+6 2 | 5+5 4 |

Continued on next column

**TABLE XLV– continued**

| TQC | ICSD | NoE | SF | | |
|---|---|---|---|---|---|
| | 5 2 5 2 | 2+2 4 | 2+2 4 | 3+4 2 | 5+5 4 |
| | 5 2 6 2 | 2+2 4 | 2+2 4 | 5+6 2 | 5+5 4 |
| | 5 2 5 2 | 2+2 4 | 2+2 4 | 3+4 2 | 5+5 4 |
| | 6 2 5 2 | 2+2 4 | 2+2 4 | 5+6 2 | 5+5 4 |
| | 6 2 6 2 | 2+2 4 | 2+2 4 | 3+4 2 | 5+5 4 |
| | 5 2 6 2 | 2+2 4 | 2+2 4 | 3+4 2 | 5+5 4 |
| | 6 2 5 2 | 2+2 4 | 2+2 4 | 5+6 2 | 5+5 4 |
| | 5 2 6 2 | 2+2 4 | 2+2 4 | 5+6 2 | 5+5 4 |
| | 6 2 5 2 | 2+2 4 | 2+2 4 | 3+4 2 | 5+5 4 |
| | 5 2 6 2 | 2+2 4 | 2+2 4 | 3+4 2 | 5+5 4 |
| | 6 2 6 2 | 2+2 4 | 2+2 4 | 5+6 2 | 5+5 4 |
| | 5 2 5 2 | 2+2 4 | 2+2 4 | 3+4 2 | 5+5 4 |
| | 6 2 6 2 | 2+2 4 | 2+2 4 | 5+6 2 | 5+5 4 |
| | 6 2 5 2 | 2+2 4 | 2+2 4 | 5+6 2 | 5+5 4 |
| | 5 2 5 2 | 2+2 4 | 2+2 4 | 3+4 2 | 5+5 4 |
| | 6 2 6 2 | 2+2 4 | 2+2 4 | 5+6 2 | 5+5 4 |
| | 5 2 6 2 | 2+2 4 | 2+2 4 | 5+6 2 | 5+5 4 |
| | 5 2 5 2 | 2+2 4 | 2+2 4 | 3+4 2 | 5+5 4 |
| | 6 2 6 2 | | | 3+4 2 | |
| | 6 2 5 2 | | | 5+6 2 | |
| | 5 2 6 2 | | | 5+6 2 | |
| | 6 2 5 2 | | | 3+4 2 | |
| | 5 2 5 2 | | | 3+4 2 | |
| | 5 2 6 2 | | | 5+6 2 | |
| | 6 2 5 2 | | | 3+4 2 | |
| | 5 2 5 2 | | | 3+4 2 | |
| | 5 2 6 2 | | | 5+6 2 | |
| | 6 2 6 2 | | | 3+4 2 | |
| | 5 2 6 2 | | | 5+6 2 | |
| | 6 2 5 2 | | | 3+4 2 | |
| | 5 2 5 2 | | | 5+6 2 | |
| | 6 2 6 2 | | | 5+6 2 | |
| | 6 2 5 2 | | | 3+4 2 | |
| | 5 2 6 2 | | | 3+4 2 | |
| | 6 2 6 2 | | | 5+6 2 | |
| | 6 2 5 2 | | | 5+6 2 | |
| | 5 2 5 2 | | | 3+4 2 | |
| | 5 2 6 2 | | | 3+4 2 | |
| | 5 2 5 2 | | | 3+4 2 | |
| | 6 2 5 2 | | | 5+6 2 | |
| | 5 2 5 2 | | | 5+6 2 | |
| | 6 2 6 2 | | | 3+4 2 | |
| | 6 2 6 2 | | | 5+6 2 | |
| | 5 2 6 2 | | | 5+6 2 | |
| | 5 2 5 2 | | | 5+6 2 | |
| | 6 2 5 2 | | | 3+4 2 | |
| | 5 2 5 2 | | | 3+4 2 | |
| | 6 2 6 2 | | | 3+4 2 | |
| | 5 2 6 2 | | | 5+6 2 | |
| | 6 2 5 2 | | | 3+4 2 | |
| | 6 2 6 2 | | | 5+6 2 | |
| | 5 2 5 2 | | | 3+4 2 | |
| | 6 2 6 2 | | | 3+4 2 | |
| | 5 2 5 2 | | | 5+6 2 | |
| | 5 2 6 2 | | | 5+6 2 | |
| | 5 2 6 2 | | | 3+4 2 | |
| | 6 2 5 2 | | | 5+6 2 | |
| | 6 2 6 2 | | | 3+4 2 | |
| | 6 2 5 2 | | | 5+6 2 | |
| | 5 2 6 2 | | | 5+6 2 | |
| | 6 2 6 2 | | | 3+4 2 | |

Continued on next column

**TABLE XLV– continued**

| TQC | ICSD | NoE | SF | | |
|---|---|---|---|---|---|
| | 5 2 5 2 | | | 3+4 2 | |
| | 5 2 5 2 | | | 5+6 2 | |
| | 6 2 5 2 | | | 3+4 2 | |
| | 6 2 6 2 | | | 5+6 2 | |
| | 5 2 5 2 | | | 5+6 2 | |
| | 5 2 6 2 | | | 3+4 2 | |

The End

TABLE XLVI: The computed irreducible representations for the valence bands for the compounds of SEBR and NLC in space group 74. See the list of tables.

| TQC | ICSD | NoE | SF | | | |
|---|---|---|---|---|---|---|
| 74.2.1 | 106234 | 96 | Yb1Zn2 | | | |
| | -GM | -X | -S | -R | -T | -W |
| | 5 2 | 5 2 | 3+4 2 | 3+4 2 | 2+2 4 | 2+4 2 |
| | 6 2 | 6 2 | 5+6 2 | 5+6 2 | 2+2 4 | 3+5 2 |
| | 5 2 | 6 2 | 3+4 2 | 3+4 2 | 2+2 4 | 2+4 2 |
| | 6 2 | 5 2 | 5+6 2 | 5+6 2 | 2+2 4 | 3+5 2 |
| | 5 2 | 5 2 | 3+4 2 | 3+4 2 | 2+2 4 | 3+5 2 |
| | 5 2 | 6 2 | 5+6 2 | 5+6 2 | 2+2 4 | 2+4 2 |
| | 6 2 | 6 2 | 3+4 2 | 3+4 2 | 2+2 4 | 2+4 2 |
| | 6 2 | 5 2 | 5+6 2 | 5+6 2 | 2+2 4 | 3+5 2 |
| | 5 2 | 6 2 | 3+4 2 | 5+6 2 | 2+2 4 | 2+4 2 |
| | 5 2 | 5 2 | 5+6 2 | 3+4 2 | 2+2 4 | 3+5 2 |
| | 5 2 | 5 2 | 3+4 2 | 5+6 2 | 2+2 4 | 2+4 2 |
| | 6 2 | 6 2 | 5+6 2 | 5+6 2 | 2+2 4 | 3+5 2 |
| | 5 2 | 5 2 | 5+6 2 | 3+4 2 | 2+2 4 | 3+5 2 |
| | 6 2 | 6 2 | 3+4 2 | 5+6 2 | 2+2 4 | 2+4 2 |
| | 5 2 | 5 2 | 5+6 2 | 3+4 2 | 2+2 4 | 3+5 2 |
| | 6 2 | 6 2 | 3+4 2 | 3+4 2 | 2+2 4 | 2+4 2 |
| | 6 2 | 5 2 | 5+6 2 | 5+6 2 | 2+2 4 | 2+4 2 |
| | 5 2 | 6 2 | 3+4 2 | 3+4 2 | 2+2 4 | 3+5 2 |
| | 6 2 | 5 2 | 5+6 2 | 5+6 2 | 2+2 4 | 2+4 2 |
| | 5 2 | 6 2 | 3+4 2 | 3+4 2 | 2+2 4 | 3+5 2 |
| | 6 2 | 6 2 | 5+6 2 | 5+6 2 | 2+2 4 | 3+5 2 |
| | 6 2 | 5 2 | 3+4 2 | 3+4 2 | 2+2 4 | 2+4 2 |
| | 6 2 | 6 2 | 5+6 2 | 3+4 2 | 2+2 4 | 3+5 2 |
| | 5 2 | 5 2 | 3+4 2 | 5+6 2 | 2+2 4 | 2+4 2 |
| | 6 2 | 6 2 | 5+6 2 | 3+4 2 | | 2+4 2 |
| | 5 2 | 5 2 | 3+4 2 | 3+4 2 | | 3+5 2 |
| | 6 2 | 6 2 | 3+4 2 | 5+6 2 | | 2+4 2 |
| | 5 2 | 5 2 | 5+6 2 | 5+6 2 | | 3+5 2 |
| | 5 2 | 5 2 | 5+6 2 | 5+6 2 | | 2+4 2 |
| | 6 2 | 6 2 | 3+4 2 | 3+4 2 | | 3+5 2 |
| | 5 2 | 6 2 | 5+6 2 | 5+6 2 | | 3+5 2 |
| | 5 2 | 5 2 | 3+4 2 | 3+4 2 | | 2+4 2 |
| | 6 2 | 5 2 | 3+4 2 | 5+6 2 | | 2+4 2 |
| | 5 2 | 6 2 | 5+6 2 | 3+4 2 | | 3+5 2 |
| | 6 2 | 6 2 | 5+6 2 | 5+6 2 | | 2+4 2 |
| | 6 2 | 5 2 | 3+4 2 | 3+4 2 | | 3+5 2 |
| | 5 2 | 5 2 | 5+6 2 | 5+6 2 | | 2+4 2 |
| | 6 2 | 6 2 | 3+4 2 | 3+4 2 | | 3+5 2 |
| | 5 2 | 5 2 | 3+4 2 | 5+6 2 | | 3+5 2 |
| | 6 2 | 6 2 | 5+6 2 | 5+6 2 | | 2+4 2 |
| | 5 2 | 5 2 | 5+6 2 | 3+4 2 | | 3+5 2 |
| | 6 2 | 6 2 | 5+6 2 | 3+4 2 | | 2+4 2 |
| | 5 2 | 5 2 | 3+4 2 | 5+6 2 | | 2+4 2 |
| | 6 2 | 6 2 | 5+6 2 | 3+4 2 | | 3+5 2 |
| | 6 2 | 5 2 | 3+4 2 | 5+6 2 | | 3+5 2 |

Continued on next column

**TABLE XLVI– continued**

| TQC | ICSD | NoE | SF | | | |
|---|---|---|---|---|---|---|
| | 6 2 | 6 2 | 5+6 2 | 3+4 2 | | 2+4 2 |
| | 5 2 | 6 2 | 3+4 2 | 5+6 2 | | 2+4 2 |
| | 5 2 | 5 2 | 3+4 2 | 5+6 2 | | 3+5 2 |
| 74.2.2 | 429139 | 88 | Ga2La1Rh3 | | | |
| | -GM | -X | -S | -R | -T | -W |
| | 5 2 | 6 2 | 5+6 2 | 3+4 2 | 2+2 4 | 2+4 2 |
| | 6 2 | 5 2 | 3+4 2 | 5+6 2 | 2+2 4 | 3+5 2 |
| | 5 2 | 6 2 | 5+6 2 | 3+4 2 | 2+2 4 | 2+4 2 |
| | 6 2 | 5 2 | 3+4 2 | 5+6 2 | 2+2 4 | 3+5 2 |
| | 5 2 | 6 2 | 5+6 2 | 3+4 2 | 2+2 4 | 3+5 2 |
| | 6 2 | 5 2 | 3+4 2 | 5+6 2 | 2+2 4 | 2+4 2 |
| | 5 2 | 6 2 | 5+6 2 | 5+6 2 | 2+2 4 | 2+4 2 |
| | 6 2 | 5 2 | 3+4 2 | 3+4 2 | 2+2 4 | 3+5 2 |
| | 5 2 | 5 2 | 3+4 2 | 3+4 2 | 2+2 4 | 2+4 2 |
| | 6 2 | 6 2 | 5+6 2 | 5+6 2 | 2+2 4 | 3+5 2 |
| | 5 2 | 5 2 | 3+4 2 | 3+4 2 | 2+2 4 | 3+5 2 |
| | 6 2 | 6 2 | 5+6 2 | 5+6 2 | 2+2 4 | 2+4 2 |
| | 5 2 | 6 2 | 3+4 2 | 3+4 2 | 2+2 4 | 2+4 2 |
| | 5 2 | 5 2 | 3+4 2 | 5+6 2 | 2+2 4 | 3+5 2 |
| | 6 2 | 5 2 | 3+4 2 | 3+4 2 | 2+2 4 | 3+5 2 |
| | 5 2 | 5 2 | 3+4 2 | 5+6 2 | 2+2 4 | 2+4 2 |
| | 5 2 | 6 2 | 3+4 2 | 3+4 2 | 2+2 4 | 2+4 2 |
| | 6 2 | 6 2 | 5+6 2 | 5+6 2 | 2+2 4 | 3+5 2 |
| | 5 2 | 6 2 | 3+4 2 | 5+6 2 | 2+2 4 | 2+4 2 |
| | 5 2 | 6 2 | 3+4 2 | 3+4 2 | 2+2 4 | 3+5 2 |
| | 6 2 | 6 2 | 5+6 2 | 5+6 2 | 2+2 4 | 3+5 2 |
| | 5 2 | 5 2 | 3+4 2 | 3+4 2 | 2+2 4 | 2+4 2 |
| | 5 2 | 6 2 | 5+6 2 | 3+4 2 | | 2+4 2 |
| | 6 2 | 6 2 | 5+6 2 | 3+4 2 | | 3+5 2 |
| | 5 2 | 6 2 | 5+6 2 | 5+6 2 | | 2+4 2 |
| | 5 2 | 6 2 | 5+6 2 | 5+6 2 | | 3+5 2 |
| | 5 2 | 5 2 | 3+4 2 | 5+6 2 | | 2+4 2 |
| | 6 2 | 6 2 | 3+4 2 | 3+4 2 | | 3+5 2 |
| | 6 2 | 5 2 | 3+4 2 | 5+6 2 | | 2+4 2 |
| | 5 2 | 5 2 | 5+6 2 | 5+6 2 | | 3+5 2 |
| | 5 2 | 6 2 | 3+4 2 | 3+4 2 | | 2+4 2 |
| | 6 2 | 5 2 | 5+6 2 | 3+4 2 | | 3+5 2 |
| | 5 2 | 5 2 | 5+6 2 | 5+6 2 | | 3+5 2 |
| | 6 2 | 5 2 | 3+4 2 | 3+4 2 | | 2+4 2 |
| | 5 2 | 6 2 | 3+4 2 | 3+4 2 | | 3+5 2 |
| | 6 2 | 6 2 | 5+6 2 | 5+6 2 | | 2+4 2 |
| | 5 2 | 6 2 | 3+4 2 | 3+4 2 | | 3+5 2 |
| | 5 2 | 6 2 | 3+4 2 | 5+6 2 | | 2+4 2 |
| | 5 2 | 6 2 | 5+6 2 | 5+6 2 | | 2+4 2 |
| | 6 2 | 5 2 | 5+6 2 | 3+4 2 | | 3+5 2 |
| | 5 2 | 6 2 | 3+4 2 | 3+4 2 | | 2+4 2 |
| | 5 2 | 5 2 | 3+4 2 | 5+6 2 | | 3+5 2 |
| | 6 2 | 6 2 | 3+4 2 | 3+4 2 | | 2+4 2 |
| | 5 2 | 6 2 | 3+4 2 | 5+6 2 | | 3+5 2 |
| 74.2.3 | 238760 | 88 | Fe2O4Si1 | | | |
| | -GM | -X | -S | -R | -T | -W |
| | 5 2 | 6 2 | 5+6 2 | 5+6 2 | 2+2 4 | 2+4 2 |
| | 6 2 | 5 2 | 3+4 2 | 3+4 2 | 2+2 4 | 3+5 2 |
| | 5 2 | 6 2 | 3+4 2 | 5+6 2 | 2+2 4 | 2+4 2 |
| | 6 2 | 5 2 | 5+6 2 | 3+4 2 | 2+2 4 | 3+5 2 |
| | 5 2 | 5 2 | 3+4 2 | 5+6 2 | 2+2 4 | 2+4 2 |
| | 6 2 | 5 2 | 3+4 2 | 3+4 2 | 2+2 4 | 3+5 2 |
| | 6 2 | 6 2 | 5+6 2 | 5+6 2 | 2+2 4 | 2+4 2 |
| | 5 2 | 6 2 | 5+6 2 | 3+4 2 | 2+2 4 | 3+5 2 |
| | 5 2 | 6 2 | 5+6 2 | 5+6 2 | 2+2 4 | 2+4 2 |
| | 5 2 | 6 2 | 5+6 2 | 3+4 2 | 2+2 4 | 3+5 2 |

Continued on next column

**TABLE XLVI– continued**

| TQC | ICSD | NoE | SF | | | |
|---|---|---|---|---|---|---|
| | 6 2 | 5 2 | 3+4 2 | 5+6 2 | 2+2 4 | 3+5 2 |
| | 6 2 | 5 2 | 3+4 2 | 3+4 2 | 2+2 4 | 2+4 2 |
| | 5 2 | 5 2 | 3+4 2 | 5+6 2 | 2+2 4 | 2+4 2 |
| | 5 2 | 5 2 | 3+4 2 | 3+4 2 | 2+2 4 | 3+5 2 |
| | 6 2 | 6 2 | 5+6 2 | 3+4 2 | 2+2 4 | 2+4 2 |
| | 6 2 | 5 2 | 3+4 2 | 5+6 2 | 2+2 4 | 3+5 2 |
| | 6 2 | 6 2 | 3+4 2 | 5+6 2 | 2+2 4 | 2+4 2 |
| | 5 2 | 6 2 | 5+6 2 | 5+6 2 | 2+2 4 | 3+5 2 |
| | 5 2 | 5 2 | 3+4 2 | 3+4 2 | 2+2 4 | 3+5 2 |
| | 5 2 | 5 2 | 5+6 2 | 3+4 2 | 2+2 4 | 2+4 2 |
| | 5 2 | 6 2 | 5+6 2 | 3+4 2 | 2+2 4 | 3+5 2 |
| | 5 2 | 5 2 | 3+4 2 | 5+6 2 | 2+2 4 | 2+4 2 |
| | 5 2 | 5 2 | 5+6 2 | 5+6 2 | | 2+4 2 |
| | 6 2 | 6 2 | 5+6 2 | 5+6 2 | | 3+5 2 |
| | 5 2 | 6 2 | 3+4 2 | 5+6 2 | | 2+4 2 |
| | 5 2 | 6 2 | 3+4 2 | 5+6 2 | | 3+5 2 |
| | 6 2 | 6 2 | 5+6 2 | 3+4 2 | | 2+4 2 |
| | 6 2 | 5 2 | 5+6 2 | 5+6 2 | | 3+5 2 |
| | 6 2 | 5 2 | 3+4 2 | 3+4 2 | | 2+4 2 |
| | 6 2 | 5 2 | 5+6 2 | 3+4 2 | | 3+5 2 |
| | 6 2 | 6 2 | 3+4 2 | 3+4 2 | | 2+4 2 |
| | 6 2 | 6 2 | 5+6 2 | 3+4 2 | | 3+5 2 |
| | 5 2 | 5 2 | 3+4 2 | 5+6 2 | | 2+4 2 |
| | 5 2 | 6 2 | 3+4 2 | 5+6 2 | | 3+5 2 |
| | 5 2 | 5 2 | 3+4 2 | 5+6 2 | | 3+5 2 |
| | 5 2 | 5 2 | 5+6 2 | 5+6 2 | | 2+4 2 |
| | 5 2 | 6 2 | 5+6 2 | 3+4 2 | | 3+5 2 |
| | 5 2 | 5 2 | 5+6 2 | 3+4 2 | | 2+4 2 |
| | 5 2 | 6 2 | 3+4 2 | 5+6 2 | | 3+5 2 |
| | 5 2 | 5 2 | 3+4 2 | 3+4 2 | | 2+4 2 |
| | 5 2 | 6 2 | 3+4 2 | 5+6 2 | | 3+5 2 |
| | 5 2 | 6 2 | 3+4 2 | 5+6 2 | | 2+4 2 |
| | 5 2 | 6 2 | 3+4 2 | 5+6 2 | | 3+5 2 |
| | 5 2 | 6 2 | 3+4 2 | 5+6 2 | | 2+4 2 |
| 74.2.4 | 238759 | 88 | Fe2O4Si1 | | | |
| | -GM | -X | -S | -R | -T | -W |
| | 5 2 | 6 2 | 5+6 2 | 5+6 2 | 2+2 4 | 2+4 2 |
| | 6 2 | 5 2 | 3+4 2 | 3+4 2 | 2+2 4 | 3+5 2 |
| | 5 2 | 6 2 | 3+4 2 | 5+6 2 | 2+2 4 | 2+4 2 |
| | 6 2 | 5 2 | 5+6 2 | 5+6 2 | 2+2 4 | 3+5 2 |
| | 5 2 | 5 2 | 3+4 2 | 3+4 2 | 2+2 4 | 2+4 2 |
| | 6 2 | 5 2 | 3+4 2 | 3+4 2 | 2+2 4 | 3+5 2 |
| | 6 2 | 6 2 | 5+6 2 | 5+6 2 | 2+2 4 | 2+4 2 |
| | 5 2 | 6 2 | 5+6 2 | 3+4 2 | 2+2 4 | 3+5 2 |
| | 5 2 | 6 2 | 5+6 2 | 5+6 2 | 2+2 4 | 2+4 2 |
| | 5 2 | 6 2 | 5+6 2 | 3+4 2 | 2+2 4 | 3+5 2 |
| | 6 2 | 5 2 | 3+4 2 | 5+6 2 | 2+2 4 | 3+5 2 |
| | 6 2 | 5 2 | 3+4 2 | 3+4 2 | 2+2 4 | 2+4 2 |
| | 5 2 | 5 2 | 3+4 2 | 5+6 2 | 2+2 4 | 2+4 2 |
| | 5 2 | 5 2 | 3+4 2 | 3+4 2 | 2+2 4 | 3+5 2 |
| | 6 2 | 6 2 | 5+6 2 | 3+4 2 | 2+2 4 | 2+4 2 |
| | 6 2 | 5 2 | 3+4 2 | 5+6 2 | 2+2 4 | 3+5 2 |
| | 6 2 | 6 2 | 3+4 2 | 5+6 2 | 2+2 4 | 2+4 2 |
| | 5 2 | 6 2 | 5+6 2 | 5+6 2 | 2+2 4 | 3+5 2 |
| | 5 2 | 5 2 | 3+4 2 | 3+4 2 | 2+2 4 | 3+5 2 |
| | 5 2 | 5 2 | 5+6 2 | 3+4 2 | 2+2 4 | 2+4 2 |
| | 5 2 | 6 2 | 5+6 2 | 3+4 2 | 2+2 4 | 3+5 2 |
| | 5 2 | 5 2 | 3+4 2 | 5+6 2 | 2+2 4 | 2+4 2 |
| | 5 2 | 5 2 | 5+6 2 | 5+6 2 | | 2+4 2 |
| | 6 2 | 6 2 | 5+6 2 | 5+6 2 | | 3+5 2 |
| | 5 2 | 6 2 | 3+4 2 | 5+6 2 | | 2+4 2 |

Continued on next column

**TABLE XLVI– continued**

| TQC | ICSD | NoE | SF | | | |
|---|---|---|---|---|---|---|
| | 5 2 | 6 2 | 3+4 2 | 5+6 2 | | 3+5 2 |
| | 6 2 | 6 2 | 5+6 2 | 3+4 2 | | 2+4 2 |
| | 6 2 | 5 2 | 5+6 2 | 5+6 2 | | 3+5 2 |
| | 6 2 | 5 2 | 3+4 2 | 3+4 2 | | 2+4 2 |
| | 6 2 | 5 2 | 5+6 2 | 3+4 2 | | 3+5 2 |
| | 6 2 | 6 2 | 3+4 2 | 3+4 2 | | 2+4 2 |
| | 6 2 | 6 2 | 5+6 2 | 3+4 2 | | 3+5 2 |
| | 5 2 | 5 2 | 3+4 2 | 5+6 2 | | 2+4 2 |
| | 5 2 | 6 2 | 3+4 2 | 5+6 2 | | 3+5 2 |
| | 5 2 | 5 2 | 3+4 2 | 5+6 2 | | 3+5 2 |
| | 5 2 | 5 2 | 5+6 2 | 5+6 2 | | 2+4 2 |
| | 5 2 | 6 2 | 5+6 2 | 3+4 2 | | 3+5 2 |
| | 5 2 | 5 2 | 5+6 2 | 3+4 2 | | 2+4 2 |
| | 5 2 | 5 2 | 3+4 2 | 5+6 2 | | 3+5 2 |
| | 5 2 | 6 2 | 3+4 2 | 3+4 2 | | 2+4 2 |
| | 5 2 | 6 2 | 3+4 2 | 5+6 2 | | 3+5 2 |
| | 5 2 | 6 2 | 3+4 2 | 5+6 2 | | 2+4 2 |
| | 5 2 | 6 2 | 3+4 2 | 5+6 2 | | 3+5 2 |
| | 5 2 | 6 2 | 3+4 2 | 5+6 2 | | 2+4 2 |
| 74.2.5 | 238758 | 88 | Fe2O4Si1 | | | |
| | -GM | -X | -S | -R | -T | -W |
| | 5 2 | 6 2 | 5+6 2 | 5+6 2 | 2+2 4 | 2+4 2 |
| | 6 2 | 5 2 | 3+4 2 | 3+4 2 | 2+2 4 | 3+5 2 |
| | 5 2 | 6 2 | 5+6 2 | 5+6 2 | 2+2 4 | 2+4 2 |
| | 6 2 | 5 2 | 3+4 2 | 3+4 2 | 2+2 4 | 3+5 2 |
| | 5 2 | 5 2 | 3+4 2 | 5+6 2 | 2+2 4 | 2+4 2 |
| | 6 2 | 5 2 | 3+4 2 | 3+4 2 | 2+2 4 | 3+5 2 |
| | 6 2 | 6 2 | 5+6 2 | 5+6 2 | 2+2 4 | 2+4 2 |
| | 5 2 | 6 2 | 5+6 2 | 3+4 2 | 2+2 4 | 3+5 2 |
| | 5 2 | 6 2 | 5+6 2 | 5+6 2 | 2+2 4 | 2+4 2 |
| | 5 2 | 6 2 | 5+6 2 | 3+4 2 | 2+2 4 | 3+5 2 |
| | 6 2 | 5 2 | 3+4 2 | 5+6 2 | 2+2 4 | 3+5 2 |
| | 6 2 | 5 2 | 3+4 2 | 3+4 2 | 2+2 4 | 2+4 2 |
| | 5 2 | 5 2 | 3+4 2 | 5+6 2 | 2+2 4 | 2+4 2 |
| | 5 2 | 5 2 | 3+4 2 | 3+4 2 | 2+2 4 | 3+5 2 |
| | 6 2 | 6 2 | 5+6 2 | 3+4 2 | 2+2 4 | 2+4 2 |
| | 6 2 | 5 2 | 3+4 2 | 5+6 2 | 2+2 4 | 3+5 2 |
| | 6 2 | 6 2 | 3+4 2 | 5+6 2 | 2+2 4 | 2+4 2 |
| | 5 2 | 6 2 | 5+6 2 | 5+6 2 | 2+2 4 | 3+5 2 |
| | 5 2 | 5 2 | 3+4 2 | 3+4 2 | 2+2 4 | 3+5 2 |
| | 5 2 | 6 2 | 5+6 2 | 3+4 2 | 2+2 4 | 2+4 2 |
| | 5 2 | 5 2 | 5+6 2 | 3+4 2 | 2+2 4 | 2+4 2 |
| | 6 2 | 5 2 | 3+4 2 | 5+6 2 | 2+2 4 | 3+5 2 |
| | 5 2 | 5 2 | 5+6 2 | 5+6 2 | | 2+4 2 |
| | 5 2 | 6 2 | 5+6 2 | 5+6 2 | | 3+5 2 |
| | 5 2 | 6 2 | 5+6 2 | 5+6 2 | | 2+4 2 |
| | 5 2 | 6 2 | 3+4 2 | 5+6 2 | | 3+5 2 |
| | 6 2 | 6 2 | 5+6 2 | 5+6 2 | | 2+4 2 |
| | 6 2 | 5 2 | 3+4 2 | 3+4 2 | | 3+5 2 |
| | 6 2 | 5 2 | 3+4 2 | 3+4 2 | | 2+4 2 |
| | 6 2 | 5 2 | 5+6 2 | 3+4 2 | | 3+5 2 |
| | 6 2 | 6 2 | 3+4 2 | 3+4 2 | | 2+4 2 |
| | 6 2 | 6 2 | 5+6 2 | 3+4 2 | | 3+5 2 |
| | 5 2 | 6 2 | 3+4 2 | 5+6 2 | | 2+4 2 |
| | 5 2 | 5 2 | 3+4 2 | 5+6 2 | | 3+5 2 |
| | 5 2 | 5 2 | 3+4 2 | 5+6 2 | | 3+5 2 |
| | 5 2 | 5 2 | 5+6 2 | 5+6 2 | | 2+4 2 |
| | 5 2 | 6 2 | 5+6 2 | 3+4 2 | | 3+5 2 |
| | 5 2 | 5 2 | 5+6 2 | 3+4 2 | | 2+4 2 |
| | 5 2 | 6 2 | 3+4 2 | 5+6 2 | | 3+5 2 |
| | 5 2 | 5 2 | 3+4 2 | 3+4 2 | | 2+4 2 |

Continued on next column

**TABLE XLVI– continued**

| TQC | ICSD | NoE | SF | | | |
|---|---|---|---|---|---|---|
| | 5 2 | 6 2 | 3+4 2 | 5+6 2 | | 3+5 2 |
| | 5 2 | 6 2 | 3+4 2 | 5+6 2 | | 2+4 2 |
| | 5 2 | 6 2 | 3+4 2 | 5+6 2 | | 3+5 2 |
| | 5 2 | 6 2 | 3+4 2 | 5+6 2 | | 2+4 2 |
| 74.2.6 | 93229 | 164 | Fe2Mo4O7 | | | |
| | -GM | -X | -S | -R | -T | -W |
| | 5 2 | 5 2 | 5+6 2 | 3+4 2 | 2+2 4 | 2+4 2 |
| | 6 2 | 6 2 | 3+4 2 | 3+4 2 | 2+2 4 | 3+5 2 |
| | 5 2 | 5 2 | 3+4 2 | 5+6 2 | 2+2 4 | 2+4 2 |
| | 6 2 | 6 2 | 5+6 2 | 5+6 2 | 2+2 4 | 3+5 2 |
| | 5 2 | 6 2 | 3+4 2 | 5+6 2 | 2+2 4 | 2+4 2 |
| | 6 2 | 6 2 | 5+6 2 | 3+4 2 | 2+2 4 | 3+5 2 |
| | 6 2 | 5 2 | 3+4 2 | 5+6 2 | 2+2 4 | 3+5 2 |
| | 5 2 | 6 2 | 5+6 2 | 3+4 2 | 2+2 4 | 2+4 2 |
| | 5 2 | 5 2 | 3+4 2 | 5+6 2 | 2+2 4 | 2+4 2 |
| | 5 2 | 5 2 | 5+6 2 | 5+6 2 | 2+2 4 | 3+5 2 |
| | 6 2 | 5 2 | 3+4 2 | 3+4 2 | 2+2 4 | 2+4 2 |
| | 5 2 | 5 2 | 5+6 2 | 5+6 2 | 2+2 4 | 3+5 2 |
| | 6 2 | 5 2 | 5+6 2 | 3+4 2 | 2+2 4 | 2+4 2 |
| | 5 2 | 6 2 | 3+4 2 | 5+6 2 | 2+2 4 | 3+5 2 |
| | 6 2 | 6 2 | 3+4 2 | 5+6 2 | 2+2 4 | 2+4 2 |
| | 5 2 | 6 2 | 5+6 2 | 5+6 2 | 2+2 4 | 3+5 2 |
| | 6 2 | 5 2 | 3+4 2 | 5+6 2 | 2+2 4 | 3+5 2 |
| | 6 2 | 6 2 | 3+4 2 | 3+4 2 | 2+2 4 | 2+4 2 |
| | 5 2 | 6 2 | 5+6 2 | 5+6 2 | 2+2 4 | 2+4 2 |
| | 6 2 | 5 2 | 5+6 2 | 3+4 2 | 2+2 4 | 3+5 2 |
| | 5 2 | 5 2 | 3+4 2 | 3+4 2 | 2+2 4 | 2+4 2 |
| | 5 2 | 6 2 | 3+4 2 | 3+4 2 | 2+2 4 | 3+5 2 |
| | 5 2 | 5 2 | 5+6 2 | 5+6 2 | 2+2 4 | 2+4 2 |
| | 6 2 | 5 2 | 5+6 2 | 3+4 2 | 2+2 4 | 3+5 2 |
| | 5 2 | 6 2 | 3+4 2 | 3+4 2 | 2+2 4 | 3+5 2 |
| | 6 2 | 6 2 | 5+6 2 | 3+4 2 | 2+2 4 | 2+4 2 |
| | 6 2 | 6 2 | 3+4 2 | 3+4 2 | 2+2 4 | 2+4 2 |
| | 6 2 | 5 2 | 5+6 2 | 3+4 2 | 2+2 4 | 3+5 2 |
| | 6 2 | 6 2 | 3+4 2 | 3+4 2 | 2+2 4 | 2+4 2 |
| | 5 2 | 6 2 | 3+4 2 | 5+6 2 | 2+2 4 | 3+5 2 |
| | 6 2 | 6 2 | 5+6 2 | 5+6 2 | 2+2 4 | 2+4 2 |
| | 5 2 | 5 2 | 3+4 2 | 3+4 2 | 2+2 4 | 3+5 2 |
| | 6 2 | 5 2 | 5+6 2 | 5+6 2 | 2+2 4 | 2+4 2 |
| | 5 2 | 5 2 | 3+4 2 | 3+4 2 | 2+2 4 | 3+5 2 |
| | 6 2 | 5 2 | 5+6 2 | 5+6 2 | 2+2 4 | 2+4 2 |
| | 5 2 | 6 2 | 3+4 2 | 3+4 2 | 2+2 4 | 3+5 2 |
| | 5 2 | 6 2 | 5+6 2 | 3+4 2 | 2+2 4 | 2+4 2 |
| | 5 2 | 6 2 | 3+4 2 | 3+4 2 | 2+2 4 | 3+5 2 |
| | 6 2 | 6 2 | 5+6 2 | 5+6 2 | 2+2 4 | 2+4 2 |
| | 5 2 | 5 2 | 3+4 2 | 5+6 2 | 2+2 4 | 3+5 2 |
| | 5 2 | 5 2 | 5+6 2 | 3+4 2 | 2+2 4 | 3+5 2 |
| | 5 2 | 6 2 | 3+4 2 | 5+6 2 | | 2+4 2 |
| | 6 2 | 5 2 | 5+6 2 | 3+4 2 | | 3+5 2 |
| | 5 2 | 6 2 | 3+4 2 | 5+6 2 | | 2+4 2 |
| | 6 2 | 6 2 | 5+6 2 | 3+4 2 | | 2+4 2 |
| | 6 2 | 6 2 | 3+4 2 | 5+6 2 | | 3+5 2 |
| | 6 2 | 6 2 | 3+4 2 | 3+4 2 | | 2+4 2 |
| | 6 2 | 5 2 | 5+6 2 | 3+4 2 | | 3+5 2 |
| | 5 2 | 5 2 | 3+4 2 | 5+6 2 | | 3+5 2 |
| | 6 2 | 5 2 | 5+6 2 | 3+4 2 | | 2+4 2 |
| | 6 2 | 6 2 | 5+6 2 | 5+6 2 | | 2+4 2 |
| | 5 2 | 5 2 | 3+4 2 | 5+6 2 | | 3+5 2 |
| | 5 2 | 6 2 | 5+6 2 | 3+4 2 | | 3+5 2 |
| | 6 2 | 5 2 | 3+4 2 | 5+6 2 | | 2+4 2 |
| | 6 2 | 6 2 | 5+6 2 | 3+4 2 | | 2+4 2 |

Continued on next column

**TABLE XLVI– continued**

| TQC | ICSD | NoE | SF | | | |
|---|---|---|---|---|---|---|
| | 6 2 | 6 2 | 5+6 2 | 5+6 2 | | 3+5 2 |
| | 5 2 | 5 2 | 3+4 2 | 3+4 2 | | 2+4 2 |
| | 6 2 | 5 2 | 5+6 2 | 3+4 2 | | 3+5 2 |
| | 5 2 | 5 2 | 3+4 2 | 5+6 2 | | 2+4 2 |
| | 5 2 | 5 2 | 5+6 2 | 3+4 2 | | 3+5 2 |
| | 5 2 | 5 2 | 3+4 2 | 3+4 2 | | 2+4 2 |
| | 6 2 | 6 2 | 5+6 2 | 5+6 2 | | 3+5 2 |
| | 6 2 | 6 2 | 3+4 2 | 3+4 2 | | 3+5 2 |
| | 6 2 | 6 2 | 5+6 2 | 3+4 2 | | 2+4 2 |
| | 6 2 | 6 2 | 3+4 2 | 5+6 2 | | 3+5 2 |
| | 5 2 | 6 2 | 5+6 2 | 5+6 2 | | 2+4 2 |
| | 5 2 | 6 2 | 3+4 2 | 5+6 2 | | 2+4 2 |
| | 5 2 | 5 2 | 5+6 2 | 5+6 2 | | 3+5 2 |
| | 6 2 | 5 2 | 3+4 2 | 3+4 2 | | 2+4 2 |
| | 6 2 | 5 2 | 3+4 2 | 3+4 2 | | 3+5 2 |
| | 5 2 | 6 2 | 3+4 2 | 3+4 2 | | 2+4 2 |
| | 5 2 | 6 2 | 5+6 2 | 5+6 2 | | 3+5 2 |
| | 5 2 | 6 2 | 3+4 2 | 3+4 2 | | 3+5 2 |
| | 6 2 | 6 2 | 3+4 2 | 5+6 2 | | 2+4 2 |
| | 5 2 | 6 2 | 3+4 2 | 5+6 2 | | 2+4 2 |
| | 5 2 | 6 2 | 5+6 2 | 3+4 2 | | 3+5 2 |
| | 5 2 | 5 2 | 5+6 2 | 3+4 2 | | 2+4 2 |
| | 5 2 | 6 2 | 3+4 2 | 5+6 2 | | 3+5 2 |
| | 6 2 | 5 2 | 3+4 2 | 5+6 2 | | 3+5 2 |
| | 6 2 | 6 2 | 5+6 2 | 3+4 2 | | 2+4 2 |
| | 5 2 | 6 2 | 3+4 2 | 3+4 2 | | 3+5 2 |
| | 5 2 | 6 2 | 5+6 2 | 5+6 2 | | 2+4 2 |
| 74.2.7 | 74288 | 128 | Cs1F5Ni1Pd1 | | | |
| | -GM | -X | -S | -R | -T | -W |
| | 5 2 | 6 2 | 3+4 2 | 3+4 2 | 2+2 4 | 2+4 2 |
| | 5 2 | 6 2 | 3+4 2 | 5+6 2 | 2+2 4 | 3+5 2 |
| | 5 2 | 5 2 | 5+6 2 | 3+4 2 | 2+2 4 | 2+4 2 |
| | 6 2 | 5 2 | 3+4 2 | 5+6 2 | 2+2 4 | 3+5 2 |
| | 6 2 | 6 2 | 3+4 2 | 3+4 2 | 2+2 4 | 2+4 2 |
| | 5 2 | 6 2 | 5+6 2 | 5+6 2 | 2+2 4 | 3+5 2 |
| | 6 2 | 5 2 | 5+6 2 | 5+6 2 | 2+2 4 | 2+4 2 |
| | 6 2 | 5 2 | 5+6 2 | 3+4 2 | 2+2 4 | 3+5 2 |
| | 5 2 | 5 2 | 3+4 2 | 3+4 2 | 2+2 4 | 2+4 2 |
| | 6 2 | 6 2 | 5+6 2 | 5+6 2 | 2+2 4 | 3+5 2 |
| | 5 2 | 6 2 | 5+6 2 | 3+4 2 | 2+2 4 | 2+4 2 |
| | 6 2 | 5 2 | 3+4 2 | 5+6 2 | 2+2 4 | 3+5 2 |
| | 5 2 | 6 2 | 5+6 2 | 5+6 2 | 2+2 4 | 2+4 2 |
| | 6 2 | 5 2 | 3+4 2 | 3+4 2 | 2+2 4 | 3+5 2 |
| | 5 2 | 6 2 | 3+4 2 | 3+4 2 | 2+2 4 | 3+5 2 |
| | 5 2 | 6 2 | 3+4 2 | 5+6 2 | 2+2 4 | 2+4 2 |
| | 5 2 | 5 2 | 5+6 2 | 3+4 2 | 2+2 4 | 2+4 2 |
| | 5 2 | 6 2 | 3+4 2 | 5+6 2 | 2+2 4 | 3+5 2 |
| | 5 2 | 5 2 | 5+6 2 | 3+4 2 | 2+2 4 | 2+4 2 |
| | 6 2 | 5 2 | 3+4 2 | 5+6 2 | 2+2 4 | 3+5 2 |
| | 6 2 | 6 2 | 3+4 2 | 3+4 2 | 2+2 4 | 2+4 2 |
| | 5 2 | 6 2 | 5+6 2 | 5+6 2 | 2+2 4 | 3+5 2 |
| | 5 2 | 6 2 | 3+4 2 | 3+4 2 | 2+2 4 | 2+4 2 |
| | 6 2 | 5 2 | 3+4 2 | 3+4 2 | 2+2 4 | 3+5 2 |
| | 5 2 | 6 2 | 5+6 2 | 3+4 2 | 2+2 4 | 3+5 2 |
| | 6 2 | 5 2 | 3+4 2 | 3+4 2 | 2+2 4 | 2+4 2 |
| | 5 2 | 5 2 | 5+6 2 | 5+6 2 | 2+2 4 | 3+5 2 |
| | 6 2 | 5 2 | 5+6 2 | 5+6 2 | 2+2 4 | 2+4 2 |
| | 6 2 | 5 2 | 3+4 2 | 5+6 2 | 2+2 4 | 3+5 2 |
| | 5 2 | 6 2 | 3+4 2 | 3+4 2 | 2+2 4 | 2+4 2 |
| | 6 2 | 6 2 | 5+6 2 | 3+4 2 | 2+2 4 | 2+4 2 |
| | 5 2 | 6 2 | 3+4 2 | 3+4 2 | 2+2 4 | 3+5 2 |

Continued on next column

**TABLE XLVI– continued**

| TQC | ICSD | NoE | SF | | | |
|---|---|---|---|---|---|---|
| | 5 2 | 5 2 | 5+6 2 | 5+6 2 | | 2+4 2 |
| | 6 2 | 6 2 | 5+6 2 | 5+6 2 | | 3+5 2 |
| | 5 2 | 5 2 | 3+4 2 | 5+6 2 | | 2+4 2 |
| | 6 2 | 6 2 | 5+6 2 | 5+6 2 | | 3+5 2 |
| | 6 2 | 6 2 | 3+4 2 | 3+4 2 | | 3+5 2 |
| | 6 2 | 5 2 | 5+6 2 | 3+4 2 | | 2+4 2 |
| | 6 2 | 5 2 | 5+6 2 | 5+6 2 | | 3+5 2 |
| | 6 2 | 5 2 | 5+6 2 | 5+6 2 | | 2+4 2 |
| | 5 2 | 5 2 | 5+6 2 | 3+4 2 | | 2+4 2 |
| | 6 2 | 5 2 | 3+4 2 | 3+4 2 | | 3+5 2 |
| | 5 2 | 6 2 | 5+6 2 | 3+4 2 | | 2+4 2 |
| | 5 2 | 5 2 | 3+4 2 | 5+6 2 | | 3+5 2 |
| | 5 2 | 6 2 | 3+4 2 | 5+6 2 | | 2+4 2 |
| | 6 2 | 6 2 | 5+6 2 | 5+6 2 | | 3+5 2 |
| | 6 2 | 6 2 | 3+4 2 | 3+4 2 | | 2+4 2 |
| | 6 2 | 5 2 | 5+6 2 | 5+6 2 | | 3+5 2 |
| | 5 2 | 6 2 | 3+4 2 | 3+4 2 | | 3+5 2 |
| | 5 2 | 6 2 | 3+4 2 | 3+4 2 | | 2+4 2 |
| | 5 2 | 6 2 | 3+4 2 | 5+6 2 | | 3+5 2 |
| | 5 2 | 6 2 | 3+4 2 | 3+4 2 | | 2+4 2 |
| | 5 2 | 6 2 | 3+4 2 | 5+6 2 | | 2+4 2 |
| | 5 2 | 6 2 | 3+4 2 | 3+4 2 | | 3+5 2 |
| | 5 2 | 5 2 | 3+4 2 | 5+6 2 | | 2+4 2 |
| | 5 2 | 5 2 | 3+4 2 | 3+4 2 | | 3+5 2 |
| | 5 2 | 6 2 | 5+6 2 | 5+6 2 | | 3+5 2 |
| | 5 2 | 5 2 | 5+6 2 | 3+4 2 | | 2+4 2 |
| | 5 2 | 5 2 | 3+4 2 | 3+4 2 | | 2+4 2 |
| | 5 2 | 5 2 | 5+6 2 | 3+4 2 | | 3+5 2 |
| | 5 2 | 5 2 | 3+4 2 | 3+4 2 | | 3+5 2 |
| | 5 2 | 5 2 | 3+4 2 | 3+4 2 | | 2+4 2 |
| | 5 2 | 6 2 | 3+4 2 | 3+4 2 | | 3+5 2 |
| | 5 2 | 5 2 | 3+4 2 | 3+4 2 | | 2+4 2 |
| 74.2.8 | 10156 | 184 | Co2O4Si1 | | | |
| | -GM | -X | -S | -R | -T | -W |
| | 5 2 | 5 2 | 5+6 2 | 5+6 2 | 2+2 4 | 2+4 2 |
| | 6 2 | 6 2 | 3+4 2 | 3+4 2 | 2+2 4 | 3+5 2 |
| | 5 2 | 6 2 | 5+6 2 | 3+4 2 | 2+2 4 | 2+4 2 |
| | 6 2 | 5 2 | 3+4 2 | 5+6 2 | 2+2 4 | 3+5 2 |
| | 5 2 | 6 2 | 3+4 2 | 3+4 2 | 2+2 4 | 2+4 2 |
| | 6 2 | 5 2 | 5+6 2 | 5+6 2 | 2+2 4 | 3+5 2 |
| | 6 2 | 5 2 | 3+4 2 | 3+4 2 | 2+2 4 | 2+4 2 |
| | 6 2 | 5 2 | 3+4 2 | 5+6 2 | 2+2 4 | 3+5 2 |
| | 5 2 | 5 2 | 5+6 2 | 5+6 2 | 2+2 4 | 2+4 2 |
| | 6 2 | 6 2 | 5+6 2 | 3+4 2 | 2+2 4 | 3+5 2 |
| | 5 2 | 6 2 | 3+4 2 | 5+6 2 | 2+2 4 | 2+4 2 |
| | 5 2 | 6 2 | 3+4 2 | 3+4 2 | 2+2 4 | 3+5 2 |
| | 6 2 | 5 2 | 5+6 2 | 5+6 2 | 2+2 4 | 2+4 2 |
| | 5 2 | 6 2 | 5+6 2 | 3+4 2 | 2+2 4 | 3+5 2 |
| | 5 2 | 5 2 | 3+4 2 | 3+4 2 | 2+2 4 | 2+4 2 |
| | 6 2 | 6 2 | 5+6 2 | 5+6 2 | 2+2 4 | 3+5 2 |
| | 6 2 | 6 2 | 3+4 2 | 3+4 2 | 2+2 4 | 2+4 2 |
| | 5 2 | 5 2 | 5+6 2 | 5+6 2 | 2+2 4 | 3+5 2 |
| | 5 2 | 5 2 | 5+6 2 | 5+6 2 | 2+2 4 | 2+4 2 |
| | 6 2 | 6 2 | 3+4 2 | 3+4 2 | 2+2 4 | 3+5 2 |
| | 5 2 | 6 2 | 3+4 2 | 3+4 2 | 2+2 4 | 2+4 2 |
| | 6 2 | 6 2 | 5+6 2 | 5+6 2 | 2+2 4 | 3+5 2 |
| | 5 2 | 5 2 | 3+4 2 | 3+4 2 | 2+2 4 | 2+4 2 |
| | 5 2 | 6 2 | 5+6 2 | 5+6 2 | 2+2 4 | 3+5 2 |
| | 5 2 | 5 2 | 5+6 2 | 3+4 2 | 2+2 4 | 2+4 2 |
| | 6 2 | 5 2 | 3+4 2 | 5+6 2 | 2+2 4 | 3+5 2 |
| | 6 2 | 6 2 | 3+4 2 | 5+6 2 | 2+2 4 | 3+5 2 |

Continued on next column

**TABLE XLVI– continued**

| TQC | ICSD | NoE | SF | | | |
|---|---|---|---|---|---|---|
| | 5 2 | 6 2 | 5+6 2 | 3+4 2 | 2+2 4 | 2+4 2 |
| | 6 2 | 6 2 | 3+4 2 | 3+4 2 | 2+2 4 | 2+4 2 |
| | 6 2 | 5 2 | 5+6 2 | 3+4 2 | 2+2 4 | 3+5 2 |
| | 5 2 | 5 2 | 3+4 2 | 5+6 2 | 2+2 4 | 2+4 2 |
| | 6 2 | 5 2 | 5+6 2 | 5+6 2 | 2+2 4 | 3+5 2 |
| | 5 2 | 5 2 | 3+4 2 | 5+6 2 | 2+2 4 | 2+4 2 |
| | 6 2 | 6 2 | 5+6 2 | 3+4 2 | 2+2 4 | 3+5 2 |
| | 6 2 | 5 2 | 3+4 2 | 5+6 2 | 2+2 4 | 2+4 2 |
| | 6 2 | 5 2 | 3+4 2 | 3+4 2 | 2+2 4 | 3+5 2 |
| | 5 2 | 6 2 | 5+6 2 | 5+6 2 | 2+2 4 | 2+4 2 |
| | 6 2 | 6 2 | 3+4 2 | 3+4 2 | 2+2 4 | 3+5 2 |
| | 5 2 | 6 2 | 3+4 2 | 5+6 2 | 2+2 4 | 2+4 2 |
| | 5 2 | 5 2 | 5+6 2 | 3+4 2 | 2+2 4 | 3+5 2 |
| | 5 2 | 6 2 | 5+6 2 | 5+6 2 | 2+2 4 | 3+5 2 |
| | 5 2 | 5 2 | 5+6 2 | 3+4 2 | 2+2 4 | 2+4 2 |
| | 5 2 | 6 2 | 3+4 2 | 3+4 2 | 2+2 4 | 3+5 2 |
| | 5 2 | 5 2 | 3+4 2 | 5+6 2 | 2+2 4 | 2+4 2 |
| | 6 2 | 5 2 | 5+6 2 | 3+4 2 | 2+2 4 | 2+4 2 |
| | 6 2 | 5 2 | 5+6 2 | 5+6 2 | 2+2 4 | 3+5 2 |
| | 6 2 | 6 2 | 5+6 2 | 3+4 2 | | 2+4 2 |
| | 5 2 | 5 2 | 5+6 2 | 5+6 2 | | 3+5 2 |
| | 6 2 | 6 2 | 5+6 2 | 3+4 2 | | 3+5 2 |
| | 5 2 | 6 2 | 3+4 2 | 5+6 2 | | 2+4 2 |
| | 6 2 | 6 2 | 3+4 2 | 5+6 2 | | 2+4 2 |
| | 6 2 | 6 2 | 3+4 2 | 3+4 2 | | 3+5 2 |
| | 5 2 | 5 2 | 3+4 2 | 3+4 2 | | 2+4 2 |
| | 5 2 | 6 2 | 3+4 2 | 5+6 2 | | 3+5 2 |
| | 5 2 | 6 2 | 3+4 2 | 5+6 2 | | 3+5 2 |
| | 5 2 | 5 2 | 3+4 2 | 5+6 2 | | 2+4 2 |
| | 6 2 | 5 2 | 5+6 2 | 3+4 2 | | 2+4 2 |
| | 6 2 | 6 2 | 5+6 2 | 3+4 2 | | 3+5 2 |
| | 5 2 | 5 2 | 3+4 2 | 3+4 2 | | 3+5 2 |
| | 5 2 | 5 2 | 3+4 2 | 3+4 2 | | 2+4 2 |
| | 6 2 | 6 2 | 5+6 2 | 5+6 2 | | 2+4 2 |
| | 6 2 | 5 2 | 5+6 2 | 5+6 2 | | 3+5 2 |
| | 6 2 | 5 2 | 5+6 2 | 3+4 2 | | 2+4 2 |
| | 6 2 | 6 2 | 5+6 2 | 5+6 2 | | 3+5 2 |
| | 5 2 | 5 2 | 3+4 2 | 3+4 2 | | 2+4 2 |
| | 5 2 | 6 2 | 3+4 2 | 5+6 2 | | 3+5 2 |
| | 5 2 | 6 2 | 3+4 2 | 5+6 2 | | 2+4 2 |
| | 5 2 | 6 2 | 3+4 2 | 3+4 2 | | 3+5 2 |
| | 6 2 | 6 2 | 5+6 2 | 3+4 2 | | 3+5 2 |
| | 5 2 | 6 2 | 3+4 2 | 5+6 2 | | 2+4 2 |
| | 6 2 | 6 2 | 5+6 2 | 3+4 2 | | 3+5 2 |
| | 5 2 | 5 2 | 3+4 2 | 5+6 2 | | 2+4 2 |
| | 5 2 | 5 2 | 5+6 2 | 3+4 2 | | 3+5 2 |
| | 6 2 | 6 2 | 3+4 2 | 3+4 2 | | 2+4 2 |
| | 5 2 | 5 2 | 3+4 2 | 5+6 2 | | 2+4 2 |
| | 5 2 | 5 2 | 3+4 2 | 5+6 2 | | 3+5 2 |
| | 6 2 | 6 2 | 5+6 2 | 3+4 2 | | 2+4 2 |
| | 6 2 | 5 2 | 5+6 2 | 5+6 2 | | 3+5 2 |
| | 6 2 | 6 2 | 5+6 2 | 5+6 2 | | 2+4 2 |
| | 6 2 | 5 2 | 5+6 2 | 3+4 2 | | 3+5 2 |
| | 6 2 | 6 2 | 5+6 2 | 3+4 2 | | 3+5 2 |
| | 5 2 | 5 2 | 3+4 2 | 5+6 2 | | 2+4 2 |
| | 5 2 | 6 2 | 3+4 2 | 5+6 2 | | 3+5 2 |
| | 5 2 | 6 2 | 5+6 2 | 3+4 2 | | 2+4 2 |
| | 5 2 | 6 2 | 3+4 2 | 5+6 2 | | 3+5 2 |
| | 5 2 | 6 2 | 3+4 2 | 3+4 2 | | 2+4 2 |
| | 5 2 | 5 2 | 3+4 2 | 3+4 2 | | 2+4 2 |
| | 6 2 | 5 2 | 3+4 2 | 3+4 2 | | 3+5 2 |

Continued on next column

**TABLE XLVI– continued**

| TQC | ICSD | NoE | SF | | | |
|---|---|---|---|---|---|---|
| | 5 2 | 6 2 | 5+6 2 | 5+6 2 | | 2+4 2 |
| | 5 2 | 5 2 | 3+4 2 | 3+4 2 | | 3+5 2 |
| | 6 2 | 6 2 | 3+4 2 | 5+6 2 | | 3+5 2 |
| | 6 2 | 6 2 | 3+4 2 | 3+4 2 | | 2+4 2 |
| 74.2.9 | 246171 | 180 | Ce3Pt5Si1 | | | |
| | -GM | -X | -S | -R | -T | -W |
| | 5 2 | 5 2 | 3+4 2 | 5+6 2 | 2+2 4 | 2+4 2 |
| | 6 2 | 5 2 | 5+6 2 | 3+4 2 | 2+2 4 | 3+5 2 |
| | 5 2 | 6 2 | 3+4 2 | 5+6 2 | 2+2 4 | 2+4 2 |
| | 5 2 | 6 2 | 5+6 2 | 5+6 2 | 2+2 4 | 3+5 2 |
| | 6 2 | 5 2 | 5+6 2 | 3+4 2 | 2+2 4 | 2+4 2 |
| | 6 2 | 6 2 | 3+4 2 | 3+4 2 | 2+2 4 | 3+5 2 |
| | 5 2 | 5 2 | 5+6 2 | 3+4 2 | 2+2 4 | 2+4 2 |
| | 5 2 | 6 2 | 3+4 2 | 5+6 2 | 2+2 4 | 3+5 2 |
| | 6 2 | 5 2 | 3+4 2 | 5+6 2 | 2+2 4 | 3+5 2 |
| | 5 2 | 6 2 | 5+6 2 | 5+6 2 | 2+2 4 | 2+4 2 |
| | 6 2 | 5 2 | 5+6 2 | 3+4 2 | 2+2 4 | 3+5 2 |
| | 6 2 | 6 2 | 3+4 2 | 3+4 2 | 2+2 4 | 2+4 2 |
| | 5 2 | 5 2 | 3+4 2 | 5+6 2 | 2+2 4 | 2+4 2 |
| | 6 2 | 6 2 | 5+6 2 | 3+4 2 | 2+2 4 | 3+5 2 |
| | 5 2 | 5 2 | 3+4 2 | 5+6 2 | 2+2 4 | 2+4 2 |
| | 6 2 | 6 2 | 5+6 2 | 5+6 2 | 2+2 4 | 3+5 2 |
| | 5 2 | 5 2 | 5+6 2 | 3+4 2 | 2+2 4 | 3+5 2 |
| | 5 2 | 6 2 | 3+4 2 | 3+4 2 | 2+2 4 | 2+4 2 |
| | 5 2 | 5 2 | 3+4 2 | 5+6 2 | 2+2 4 | 3+5 2 |
| | 6 2 | 6 2 | 5+6 2 | 5+6 2 | 2+2 4 | 2+4 2 |
| | 6 2 | 5 2 | 3+4 2 | 3+4 2 | 2+2 4 | 3+5 2 |
| | 6 2 | 5 2 | 5+6 2 | 5+6 2 | 2+2 4 | 2+4 2 |
| | 5 2 | 6 2 | 3+4 2 | 3+4 2 | 2+2 4 | 2+4 2 |
| | 6 2 | 6 2 | 5+6 2 | 3+4 2 | 2+2 4 | 3+5 2 |
| | 5 2 | 6 2 | 5+6 2 | 3+4 2 | 2+2 4 | 2+4 2 |
| | 6 2 | 5 2 | 3+4 2 | 5+6 2 | 2+2 4 | 3+5 2 |
| | 5 2 | 6 2 | 5+6 2 | 5+6 2 | 2+2 4 | 3+5 2 |
| | 5 2 | 5 2 | 5+6 2 | 3+4 2 | 2+2 4 | 2+4 2 |
| | 5 2 | 6 2 | 3+4 2 | 3+4 2 | 2+2 4 | 2+4 2 |
| | 6 2 | 6 2 | 5+6 2 | 5+6 2 | 2+2 4 | 3+5 2 |
| | 6 2 | 6 2 | 5+6 2 | 3+4 2 | 2+2 4 | 3+5 2 |
| | 5 2 | 5 2 | 3+4 2 | 5+6 2 | 2+2 4 | 2+4 2 |
| | 5 2 | 6 2 | 5+6 2 | 3+4 2 | 2+2 4 | 3+5 2 |
| | 5 2 | 6 2 | 3+4 2 | 3+4 2 | 2+2 4 | 2+4 2 |
| | 5 2 | 5 2 | 3+4 2 | 5+6 2 | 2+2 4 | 3+5 2 |
| | 5 2 | 6 2 | 5+6 2 | 3+4 2 | 2+2 4 | 2+4 2 |
| | 6 2 | 5 2 | 3+4 2 | 5+6 2 | 2+2 4 | 2+4 2 |
| | 5 2 | 5 2 | 5+6 2 | 5+6 2 | 2+2 4 | 3+5 2 |
| | 5 2 | 6 2 | 5+6 2 | 3+4 2 | 2+2 4 | 3+5 2 |
| | 6 2 | 5 2 | 3+4 2 | 3+4 2 | 2+2 4 | 2+4 2 |
| | 6 2 | 5 2 | 5+6 2 | 5+6 2 | 2+2 4 | 2+4 2 |
| | 5 2 | 6 2 | 5+6 2 | 3+4 2 | 2+2 4 | 3+5 2 |
| | 6 2 | 6 2 | 3+4 2 | 5+6 2 | 2+2 4 | 3+5 2 |
| | 5 2 | 5 2 | 5+6 2 | 3+4 2 | 2+2 4 | 2+4 2 |
| | 6 2 | 6 2 | 3+4 2 | 5+6 2 | 2+2 4 | 2+4 2 |
| | 6 2 | 5 2 | 5+6 2 | 3+4 2 | | 3+5 2 |
| | 5 2 | 6 2 | 3+4 2 | 5+6 2 | | 3+5 2 |
| | 5 2 | 5 2 | 5+6 2 | 3+4 2 | | 2+4 2 |
| | 6 2 | 6 2 | 5+6 2 | 5+6 2 | | 3+5 2 |
| | 5 2 | 6 2 | 3+4 2 | 5+6 2 | | 2+4 2 |
| | 6 2 | 6 2 | 5+6 2 | 3+4 2 | | 3+5 2 |
| | 5 2 | 5 2 | 3+4 2 | 5+6 2 | | 2+4 2 |
| | 6 2 | 6 2 | 5+6 2 | 3+4 2 | | 2+4 2 |
| | 5 2 | 5 2 | 5+6 2 | 5+6 2 | | 3+5 2 |
| | 6 2 | 6 2 | 3+4 2 | 3+4 2 | | 2+4 2 |

Continued on next column

**TABLE XLVI– continued**

| TQC | ICSD | NoE | SF | | | |
|---|---|---|---|---|---|---|
| | 5 2 | 6 2 | 5+6 2 | 5+6 2 | | 3+5 2 |
| | 6 2 | 5 2 | 3+4 2 | 5+6 2 | | 2+4 2 |
| | 5 2 | 6 2 | 5+6 2 | 3+4 2 | | 3+5 2 |
| | 6 2 | 6 2 | 3+4 2 | 5+6 2 | | 2+4 2 |
| | 5 2 | 5 2 | 5+6 2 | 5+6 2 | | 3+5 2 |
| | 6 2 | 6 2 | 3+4 2 | 3+4 2 | | 2+4 2 |
| | 5 2 | 5 2 | 5+6 2 | 3+4 2 | | 3+5 2 |
| | 6 2 | 6 2 | 5+6 2 | 3+4 2 | | 3+5 2 |
| | 6 2 | 5 2 | 3+4 2 | 5+6 2 | | 2+4 2 |
| | 5 2 | 6 2 | 3+4 2 | 3+4 2 | | 2+4 2 |
| | 5 2 | 5 2 | 5+6 2 | 5+6 2 | | 3+5 2 |
| | 5 2 | 6 2 | 3+4 2 | 3+4 2 | | 2+4 2 |
| | 6 2 | 6 2 | 3+4 2 | 5+6 2 | | 3+5 2 |
| | 5 2 | 5 2 | 5+6 2 | 5+6 2 | | 2+4 2 |
| | 6 2 | 6 2 | 5+6 2 | 3+4 2 | | 3+5 2 |
| | 5 2 | 5 2 | 5+6 2 | 5+6 2 | | 3+5 2 |
| | 5 2 | 6 2 | 3+4 2 | 3+4 2 | | 2+4 2 |
| | 6 2 | 6 2 | 3+4 2 | 5+6 2 | | 3+5 2 |
| | 5 2 | 5 2 | 5+6 2 | 3+4 2 | | 3+5 2 |
| | 6 2 | 6 2 | 5+6 2 | 5+6 2 | | 2+4 2 |
| | 5 2 | 5 2 | 3+4 2 | 3+4 2 | | 2+4 2 |
| | 6 2 | 5 2 | 5+6 2 | 5+6 2 | | 2+4 2 |
| | 6 2 | 6 2 | 5+6 2 | 3+4 2 | | 3+5 2 |
| | 5 2 | 6 2 | 5+6 2 | 5+6 2 | | 2+4 2 |
| | 6 2 | 5 2 | 5+6 2 | 3+4 2 | | 3+5 2 |
| | 5 2 | 6 2 | 3+4 2 | 3+4 2 | | 2+4 2 |
| | 6 2 | 5 2 | 5+6 2 | 5+6 2 | | 3+5 2 |
| | 5 2 | 6 2 | 3+4 2 | 5+6 2 | | 3+5 2 |
| | 5 2 | 5 2 | 5+6 2 | 3+4 2 | | 2+4 2 |
| | 5 2 | 6 2 | 3+4 2 | 3+4 2 | | 3+5 2 |
| | 5 2 | 6 2 | 5+6 2 | 5+6 2 | | 2+4 2 |
| | 6 2 | 6 2 | 3+4 2 | 3+4 2 | | 2+4 2 |
| | 6 2 | 5 2 | 3+4 2 | 5+6 2 | | 3+5 2 |
| | 6 2 | 5 2 | 5+6 2 | 3+4 2 | | 3+5 2 |
| | 6 2 | 5 2 | 5+6 2 | 5+6 2 | | 2+4 2 |
| 74.2.10 | 93244 | 180 | Ce3Pd5Si1 | | | |
| | -GM | -X | -S | -R | -T | -W |
| | 5 2 | 6 2 | 5+6 2 | 5+6 2 | 2+2 4 | 2+4 2 |
| | 6 2 | 5 2 | 3+4 2 | 3+4 2 | 2+2 4 | 3+5 2 |
| | 5 2 | 5 2 | 5+6 2 | 5+6 2 | 2+2 4 | 2+4 2 |
| | 6 2 | 6 2 | 3+4 2 | 5+6 2 | 2+2 4 | 3+5 2 |
| | 5 2 | 5 2 | 3+4 2 | 3+4 2 | 2+2 4 | 2+4 2 |
| | 6 2 | 6 2 | 5+6 2 | 3+4 2 | 2+2 4 | 3+5 2 |
| | 5 2 | 6 2 | 5+6 2 | 3+4 2 | 2+2 4 | 2+4 2 |
| | 5 2 | 5 2 | 3+4 2 | 5+6 2 | 2+2 4 | 3+5 2 |
| | 6 2 | 5 2 | 5+6 2 | 5+6 2 | 2+2 4 | 3+5 2 |
| | 6 2 | 6 2 | 3+4 2 | 5+6 2 | 2+2 4 | 2+4 2 |
| | 5 2 | 5 2 | 3+4 2 | 3+4 2 | 2+2 4 | 3+5 2 |
| | 6 2 | 6 2 | 5+6 2 | 3+4 2 | 2+2 4 | 2+4 2 |
| | 5 2 | 5 2 | 5+6 2 | 3+4 2 | 2+2 4 | 2+4 2 |
| | 6 2 | 6 2 | 3+4 2 | 5+6 2 | 2+2 4 | 3+5 2 |
| | 5 2 | 5 2 | 5+6 2 | 5+6 2 | 2+2 4 | 2+4 2 |
| | 6 2 | 6 2 | 3+4 2 | 3+4 2 | 2+2 4 | 3+5 2 |
| | 5 2 | 6 2 | 3+4 2 | 5+6 2 | 2+2 4 | 3+5 2 |
| | 5 2 | 5 2 | 5+6 2 | 3+4 2 | 2+2 4 | 2+4 2 |
| | 5 2 | 5 2 | 5+6 2 | 5+6 2 | 2+2 4 | 2+4 2 |
| | 6 2 | 6 2 | 3+4 2 | 5+6 2 | 2+2 4 | 3+5 2 |
| | 6 2 | 5 2 | 5+6 2 | 3+4 2 | 2+2 4 | 3+5 2 |
| | 6 2 | 6 2 | 3+4 2 | 3+4 2 | 2+2 4 | 2+4 2 |
| | 5 2 | 5 2 | 5+6 2 | 5+6 2 | 2+2 4 | 2+4 2 |
| | 6 2 | 6 2 | 3+4 2 | 3+4 2 | 2+2 4 | 3+5 2 |

Continued on next column

**TABLE XLVI– continued**

| TQC | ICSD | NoE | SF | | | |
|---|---|---|---|---|---|---|
| | 5 2 | 6 2 | 3+4 2 | 3+4 2 | 2+2 4 | 2+4 2 |
| | 6 2 | 5 2 | 5+6 2 | 5+6 2 | 2+2 4 | 3+5 2 |
| | 5 2 | 6 2 | 3+4 2 | 5+6 2 | 2+2 4 | 3+5 2 |
| | 5 2 | 5 2 | 3+4 2 | 3+4 2 | 2+2 4 | 2+4 2 |
| | 5 2 | 6 2 | 3+4 2 | 3+4 2 | 2+2 4 | 2+4 2 |
| | 6 2 | 6 2 | 5+6 2 | 5+6 2 | 2+2 4 | 3+5 2 |
| | 5 2 | 6 2 | 5+6 2 | 3+4 2 | 2+2 4 | 3+5 2 |
| | 6 2 | 5 2 | 3+4 2 | 5+6 2 | 2+2 4 | 2+4 2 |
| | 5 2 | 6 2 | 5+6 2 | 3+4 2 | 2+2 4 | 2+4 2 |
| | 5 2 | 5 2 | 3+4 2 | 3+4 2 | 2+2 4 | 3+5 2 |
| | 5 2 | 5 2 | 5+6 2 | 3+4 2 | 2+2 4 | 2+4 2 |
| | 5 2 | 5 2 | 3+4 2 | 5+6 2 | 2+2 4 | 3+5 2 |
| | 5 2 | 6 2 | 5+6 2 | 3+4 2 | 2+2 4 | 3+5 2 |
| | 6 2 | 5 2 | 3+4 2 | 5+6 2 | 2+2 4 | 2+4 2 |
| | 6 2 | 6 2 | 3+4 2 | 3+4 2 | 2+2 4 | 3+5 2 |
| | 5 2 | 6 2 | 5+6 2 | 5+6 2 | 2+2 4 | 2+4 2 |
| | 5 2 | 5 2 | 3+4 2 | 5+6 2 | 2+2 4 | 3+5 2 |
| | 5 2 | 5 2 | 5+6 2 | 3+4 2 | 2+2 4 | 2+4 2 |
| | 6 2 | 6 2 | 3+4 2 | 3+4 2 | 2+2 4 | 3+5 2 |
| | 6 2 | 6 2 | 3+4 2 | 5+6 2 | 2+2 4 | 2+4 2 |
| | 5 2 | 6 2 | 5+6 2 | 5+6 2 | 2+2 4 | 2+4 2 |
| | 6 2 | 5 2 | 3+4 2 | 3+4 2 | | 3+5 2 |
| | 6 2 | 5 2 | 5+6 2 | 3+4 2 | | 2+4 2 |
| | 5 2 | 6 2 | 3+4 2 | 5+6 2 | | 3+5 2 |
| | 6 2 | 6 2 | 3+4 2 | 3+4 2 | | 3+5 2 |
| | 5 2 | 6 2 | 5+6 2 | 5+6 2 | | 2+4 2 |
| | 5 2 | 6 2 | 3+4 2 | 5+6 2 | | 3+5 2 |
| | 6 2 | 5 2 | 3+4 2 | 5+6 2 | | 2+4 2 |
| | 6 2 | 6 2 | 3+4 2 | 3+4 2 | | 3+5 2 |
| | 5 2 | 5 2 | 5+6 2 | 5+6 2 | | 2+4 2 |
| | 6 2 | 6 2 | 3+4 2 | 3+4 2 | | 2+4 2 |
| | 5 2 | 5 2 | 5+6 2 | 5+6 2 | | 3+5 2 |
| | 6 2 | 6 2 | 5+6 2 | 5+6 2 | | 3+5 2 |
| | 6 2 | 6 2 | 5+6 2 | 3+4 2 | | 2+4 2 |
| | 5 2 | 6 2 | 3+4 2 | 5+6 2 | | 2+4 2 |
| | 5 2 | 5 2 | 3+4 2 | 5+6 2 | | 3+5 2 |
| | 6 2 | 6 2 | 3+4 2 | 3+4 2 | | 3+5 2 |
| | 6 2 | 5 2 | 5+6 2 | 3+4 2 | | 2+4 2 |
| | 5 2 | 6 2 | 5+6 2 | 3+4 2 | | 3+5 2 |
| | 6 2 | 5 2 | 3+4 2 | 5+6 2 | | 2+4 2 |
| | 5 2 | 6 2 | 5+6 2 | 3+4 2 | | 2+4 2 |
| | 5 2 | 6 2 | 3+4 2 | 5+6 2 | | 3+5 2 |
| | 5 2 | 5 2 | 5+6 2 | 5+6 2 | | 3+5 2 |
| | 5 2 | 6 2 | 5+6 2 | 3+4 2 | | 2+4 2 |
| | 6 2 | 5 2 | 3+4 2 | 5+6 2 | | 3+5 2 |
| | 6 2 | 5 2 | 3+4 2 | 3+4 2 | | 2+4 2 |
| | 5 2 | 6 2 | 3+4 2 | 5+6 2 | | 2+4 2 |
| | 5 2 | 5 2 | 5+6 2 | 3+4 2 | | 3+5 2 |
| | 6 2 | 6 2 | 5+6 2 | 5+6 2 | | 3+5 2 |
| | 5 2 | 6 2 | 3+4 2 | 3+4 2 | | 2+4 2 |
| | 5 2 | 6 2 | 3+4 2 | 3+4 2 | | 3+5 2 |
| | 5 2 | 5 2 | 5+6 2 | 5+6 2 | | 2+4 2 |
| | 6 2 | 6 2 | 3+4 2 | 5+6 2 | | 2+4 2 |
| | 6 2 | 6 2 | 3+4 2 | 3+4 2 | | 3+5 2 |
| | 6 2 | 6 2 | 3+4 2 | 3+4 2 | | 3+5 2 |
| | 5 2 | 5 2 | 3+4 2 | 5+6 2 | | 2+4 2 |
| | 5 2 | 5 2 | 5+6 2 | 3+4 2 | | 2+4 2 |
| | 6 2 | 5 2 | 3+4 2 | 5+6 2 | | 3+5 2 |
| | 5 2 | 6 2 | 3+4 2 | 5+6 2 | | 3+5 2 |
| | 6 2 | 5 2 | 5+6 2 | 5+6 2 | | 2+4 2 |
| | 5 2 | 6 2 | 5+6 2 | 3+4 2 | | 3+5 2 |

Continued on next column

**TABLE XLVI– continued**

| TQC | ICSD | NoE | SF | | | |
|---|---|---|---|---|---|---|
| | 6 2 | 6 2 | 3+4 2 | 3+4 2 | | 2+4 2 |
| | 5 2 | 5 2 | 5+6 2 | 3+4 2 | | 2+4 2 |
| | 6 2 | 6 2 | 5+6 2 | 5+6 2 | | 3+5 2 |
| | 6 2 | 5 2 | 3+4 2 | 5+6 2 | | 3+5 2 |
| | 6 2 | 5 2 | 3+4 2 | 3+4 2 | | 2+4 2 |
| 74.2.11 | 260381 | 68 | Cd2Sr1 | | | |
| | -GM | -X | -S | -R | -T | -W |
| | 5 2 | 5 2 | 5+6 2 | 5+6 2 | 2+2 4 | 2+4 2 |
| | 6 2 | 6 2 | 3+4 2 | 3+4 2 | 2+2 4 | 3+5 2 |
| | 5 2 | 5 2 | 5+6 2 | 5+6 2 | 2+2 4 | 2+4 2 |
| | 6 2 | 6 2 | 3+4 2 | 3+4 2 | 2+2 4 | 3+5 2 |
| | 5 2 | 5 2 | 5+6 2 | 5+6 2 | 2+2 4 | 3+5 2 |
| | 5 2 | 6 2 | 3+4 2 | 3+4 2 | 2+2 4 | 2+4 2 |
| | 6 2 | 5 2 | 5+6 2 | 5+6 2 | 2+2 4 | 2+4 2 |
| | 6 2 | 6 2 | 3+4 2 | 3+4 2 | 2+2 4 | 3+5 2 |
| | 5 2 | 6 2 | 3+4 2 | 5+6 2 | 2+2 4 | 2+4 2 |
| | 6 2 | 5 2 | 5+6 2 | 3+4 2 | 2+2 4 | 3+5 2 |
| | 5 2 | 6 2 | 3+4 2 | 5+6 2 | 2+2 4 | 2+4 2 |
| | 5 2 | 6 2 | 5+6 2 | 3+4 2 | 2+2 4 | 3+5 2 |
| | 5 2 | 5 2 | 3+4 2 | 3+4 2 | 2+2 4 | 2+4 2 |
| | 6 2 | 5 2 | 3+4 2 | 5+6 2 | 2+2 4 | 3+5 2 |
| | 6 2 | 6 2 | 5+6 2 | 3+4 2 | 2+2 4 | 3+5 2 |
| | 6 2 | 5 2 | 5+6 2 | 5+6 2 | 2+2 4 | 2+4 2 |
| | 5 2 | 6 2 | 3+4 2 | 3+4 2 | 2+2 4 | 2+4 2 |
| | 6 2 | 5 2 | 5+6 2 | 5+6 2 | | 3+5 2 |
| | 5 2 | 6 2 | 3+4 2 | 3+4 2 | | 2+4 2 |
| | 6 2 | 5 2 | 5+6 2 | 5+6 2 | | 3+5 2 |
| | 5 2 | 5 2 | 5+6 2 | 5+6 2 | | 3+5 2 |
| | 6 2 | 6 2 | 5+6 2 | 3+4 2 | | 2+4 2 |
| | 6 2 | 5 2 | 3+4 2 | 3+4 2 | | 3+5 2 |
| | 6 2 | 5 2 | 3+4 2 | 5+6 2 | | 2+4 2 |
| | 5 2 | 6 2 | 5+6 2 | 3+4 2 | | 2+4 2 |
| | 5 2 | 6 2 | 5+6 2 | 5+6 2 | | 3+5 2 |
| | 6 2 | 5 2 | 3+4 2 | 5+6 2 | | 2+4 2 |
| | 5 2 | 6 2 | 3+4 2 | 3+4 2 | | 3+5 2 |
| | 5 2 | 6 2 | 3+4 2 | 3+4 2 | | 2+4 2 |
| | 6 2 | 5 2 | 5+6 2 | 5+6 2 | | 3+5 2 |
| | 5 2 | 5 2 | 3+4 2 | 3+4 2 | | 3+5 2 |
| | 5 2 | 6 2 | 5+6 2 | 5+6 2 | | 2+4 2 |
| | 6 2 | 6 2 | 3+4 2 | 5+6 2 | | 2+4 2 |
| | 6 2 | 5 2 | 3+4 2 | 3+4 2 | | 3+5 2 |
| 74.2.12 | 97094 | 124 | Ca2O7Os2 | | | |
| | -GM | -X | -S | -R | -T | -W |
| | 5 2 | 6 2 | 3+4 2 | 5+6 2 | 2+2 4 | 2+4 2 |
| | 5 2 | 5 2 | 3+4 2 | 3+4 2 | 2+2 4 | 3+5 2 |
| | 5 2 | 5 2 | 3+4 2 | 5+6 2 | 2+2 4 | 3+5 2 |
| | 5 2 | 6 2 | 5+6 2 | 5+6 2 | 2+2 4 | 2+4 2 |
| | 5 2 | 6 2 | 3+4 2 | 3+4 2 | 2+2 4 | 3+5 2 |
| | 6 2 | 6 2 | 5+6 2 | 5+6 2 | 2+2 4 | 2+4 2 |
| | 6 2 | 5 2 | 3+4 2 | 3+4 2 | 2+2 4 | 2+4 2 |
| | 6 2 | 5 2 | 5+6 2 | 5+6 2 | 2+2 4 | 3+5 2 |
| | 6 2 | 6 2 | 5+6 2 | 3+4 2 | 2+2 4 | 3+5 2 |
| | 6 2 | 5 2 | 5+6 2 | 3+4 2 | 2+2 4 | 2+4 2 |
| | 5 2 | 5 2 | 5+6 2 | 5+6 2 | 2+2 4 | 2+4 2 |
| | 5 2 | 5 2 | 3+4 2 | 5+6 2 | 2+2 4 | 3+5 2 |
| | 6 2 | 6 2 | 5+6 2 | 3+4 2 | 2+2 4 | 2+4 2 |
| | 6 2 | 6 2 | 3+4 2 | 3+4 2 | 2+2 4 | 3+5 2 |
| | 5 2 | 6 2 | 3+4 2 | 5+6 2 | 2+2 4 | 3+5 2 |
| | 5 2 | 5 2 | 3+4 2 | 5+6 2 | 2+2 4 | 2+4 2 |
| | 5 2 | 6 2 | 3+4 2 | 5+6 2 | 2+2 4 | 2+4 2 |
| | 5 2 | 6 2 | 3+4 2 | 3+4 2 | 2+2 4 | 3+5 2 |

Continued on next column

**TABLE XLVI– continued**

| TQC | ICSD | NoE | SF | | | |
|---|---|---|---|---|---|---|
| | 5 2 | 5 2 | 3+4 2 | 5+6 2 | 2+2 4 | 3+5 2 |
| | 5 2 | 6 2 | 5+6 2 | 3+4 2 | 2+2 4 | 2+4 2 |
| | 5 2 | 6 2 | 3+4 2 | 5+6 2 | 2+2 4 | 3+5 2 |
| | 5 2 | 5 2 | 5+6 2 | 3+4 2 | 2+2 4 | 2+4 2 |
| | 5 2 | 5 2 | 3+4 2 | 5+6 2 | 2+2 4 | 3+5 2 |
| | 5 2 | 6 2 | 3+4 2 | 5+6 2 | 2+2 4 | 2+4 2 |
| | 5 2 | 5 2 | 5+6 2 | 3+4 2 | 2+2 4 | 2+4 2 |
| | 5 2 | 6 2 | 3+4 2 | 5+6 2 | 2+2 4 | 3+5 2 |
| | 6 2 | 5 2 | 5+6 2 | 3+4 2 | 2+2 4 | 2+4 2 |
| | 5 2 | 5 2 | 5+6 2 | 5+6 2 | 2+2 4 | 3+5 2 |
| | 6 2 | 5 2 | 3+4 2 | 3+4 2 | 2+2 4 | 2+4 2 |
| | 6 2 | 6 2 | 5+6 2 | 5+6 2 | 2+2 4 | 3+5 2 |
| | 5 2 | 6 2 | 5+6 2 | 5+6 2 | 2+2 4 | 2+4 2 |
| | 6 2 | 5 2 | 3+4 2 | 5+6 2 | | 3+5 2 |
| | 6 2 | 5 2 | 5+6 2 | 5+6 2 | | 3+5 2 |
| | 5 2 | 5 2 | 3+4 2 | 3+4 2 | | 2+4 2 |
| | 6 2 | 5 2 | 3+4 2 | 3+4 2 | | 2+4 2 |
| | 5 2 | 6 2 | 5+6 2 | 5+6 2 | | 3+5 2 |
| | 6 2 | 6 2 | 5+6 2 | 3+4 2 | | 3+5 2 |
| | 5 2 | 5 2 | 3+4 2 | 5+6 2 | | 2+4 2 |
| | 5 2 | 6 2 | 3+4 2 | 3+4 2 | | 3+5 2 |
| | 6 2 | 5 2 | 5+6 2 | 3+4 2 | | 2+4 2 |
| | 5 2 | 5 2 | 5+6 2 | 5+6 2 | | 2+4 2 |
| | 6 2 | 6 2 | 5+6 2 | 3+4 2 | | 3+5 2 |
| | 6 2 | 6 2 | 3+4 2 | 3+4 2 | | 3+5 2 |
| | 6 2 | 6 2 | 5+6 2 | 3+4 2 | | 2+4 2 |
| | 6 2 | 5 2 | 5+6 2 | 5+6 2 | | 3+5 2 |
| | 6 2 | 6 2 | 3+4 2 | 3+4 2 | | 2+4 2 |
| | 6 2 | 6 2 | 5+6 2 | 5+6 2 | | 3+5 2 |
| | 6 2 | 6 2 | 5+6 2 | 3+4 2 | | 2+4 2 |
| | 6 2 | 5 2 | 5+6 2 | 3+4 2 | | 3+5 2 |
| | 6 2 | 6 2 | 3+4 2 | 3+4 2 | | 2+4 2 |
| | 5 2 | 6 2 | 5+6 2 | 3+4 2 | | 3+5 2 |
| | 6 2 | 5 2 | 3+4 2 | 3+4 2 | | 2+4 2 |
| | 5 2 | 5 2 | 5+6 2 | 5+6 2 | | 3+5 2 |
| | 6 2 | 5 2 | 3+4 2 | 5+6 2 | | 2+4 2 |
| | 6 2 | 5 2 | 5+6 2 | 3+4 2 | | 2+4 2 |
| | 6 2 | 6 2 | 3+4 2 | 5+6 2 | | 3+5 2 |
| | 5 2 | 6 2 | 3+4 2 | 5+6 2 | | 3+5 2 |
| | 5 2 | 6 2 | 3+4 2 | 5+6 2 | | 2+4 2 |
| | 5 2 | 5 2 | 3+4 2 | 5+6 2 | | 3+5 2 |
| | 5 2 | 5 2 | 3+4 2 | 3+4 2 | | 2+4 2 |
| | 5 2 | 6 2 | 5+6 2 | 5+6 2 | | 2+4 2 |
| | 5 2 | 5 2 | 3+4 2 | 3+4 2 | | 3+5 2 |
| 74.2.13 | 97093 | 124 | Ca2O7Os2 | | | |
| | -GM | -X | -S | -R | -T | -W |
| | 5 2 | 6 2 | 3+4 2 | 5+6 2 | 2+2 4 | 2+4 2 |
| | 5 2 | 5 2 | 3+4 2 | 3+4 2 | 2+2 4 | 3+5 2 |
| | 5 2 | 6 2 | 3+4 2 | 5+6 2 | 2+2 4 | 3+5 2 |
| | 5 2 | 5 2 | 5+6 2 | 5+6 2 | 2+2 4 | 2+4 2 |
| | 5 2 | 6 2 | 3+4 2 | 3+4 2 | 2+2 4 | 3+5 2 |
| | 6 2 | 6 2 | 5+6 2 | 5+6 2 | 2+2 4 | 2+4 2 |
| | 6 2 | 5 2 | 3+4 2 | 3+4 2 | 2+2 4 | 2+4 2 |
| | 6 2 | 6 2 | 5+6 2 | 5+6 2 | 2+2 4 | 3+5 2 |
| | 6 2 | 5 2 | 5+6 2 | 3+4 2 | 2+2 4 | 3+5 2 |
| | 6 2 | 5 2 | 5+6 2 | 3+4 2 | 2+2 4 | 2+4 2 |
| | 5 2 | 5 2 | 5+6 2 | 5+6 2 | 2+2 4 | 2+4 2 |
| | 5 2 | 5 2 | 3+4 2 | 5+6 2 | 2+2 4 | 3+5 2 |
| | 6 2 | 6 2 | 5+6 2 | 3+4 2 | 2+2 4 | 2+4 2 |
| | 6 2 | 6 2 | 3+4 2 | 3+4 2 | 2+2 4 | 3+5 2 |
| | 5 2 | 6 2 | 3+4 2 | 5+6 2 | 2+2 4 | 3+5 2 |

Continued on next column

**TABLE XLVI– continued**

| TQC | ICSD | NoE | SF | | | |
|---|---|---|---|---|---|---|
| | 5 2 | 6 2 | 3+4 2 | 5+6 2 | 2+2 4 | 2+4 2 |
| | 5 2 | 5 2 | 3+4 2 | 5+6 2 | 2+2 4 | 2+4 2 |
| | 5 2 | 6 2 | 3+4 2 | 3+4 2 | 2+2 4 | 3+5 2 |
| | 5 2 | 5 2 | 3+4 2 | 5+6 2 | 2+2 4 | 3+5 2 |
| | 5 2 | 6 2 | 5+6 2 | 3+4 2 | 2+2 4 | 2+4 2 |
| | 5 2 | 6 2 | 3+4 2 | 5+6 2 | 2+2 4 | 3+5 2 |
| | 5 2 | 5 2 | 5+6 2 | 3+4 2 | 2+2 4 | 2+4 2 |
| | 5 2 | 5 2 | 3+4 2 | 5+6 2 | 2+2 4 | 3+5 2 |
| | 5 2 | 6 2 | 3+4 2 | 5+6 2 | 2+2 4 | 2+4 2 |
| | 5 2 | 5 2 | 5+6 2 | 3+4 2 | 2+2 4 | 2+4 2 |
| | 5 2 | 6 2 | 3+4 2 | 5+6 2 | 2+2 4 | 3+5 2 |
| | 6 2 | 5 2 | 5+6 2 | 3+4 2 | 2+2 4 | 2+4 2 |
| | 5 2 | 5 2 | 5+6 2 | 5+6 2 | 2+2 4 | 3+5 2 |
| | 6 2 | 5 2 | 3+4 2 | 3+4 2 | 2+2 4 | 2+4 2 |
| | 6 2 | 6 2 | 5+6 2 | 5+6 2 | 2+2 4 | 3+5 2 |
| | 5 2 | 6 2 | 5+6 2 | 5+6 2 | 2+2 4 | 2+4 2 |
| | 6 2 | 5 2 | 3+4 2 | 5+6 2 | | 3+5 2 |
| | 5 2 | 5 2 | 5+6 2 | 5+6 2 | | 3+5 2 |
| | 6 2 | 5 2 | 3+4 2 | 3+4 2 | | 2+4 2 |
| | 6 2 | 5 2 | 3+4 2 | 3+4 2 | | 2+4 2 |
| | 5 2 | 6 2 | 5+6 2 | 5+6 2 | | 3+5 2 |
| | 6 2 | 6 2 | 5+6 2 | 3+4 2 | | 3+5 2 |
| | 5 2 | 5 2 | 3+4 2 | 5+6 2 | | 2+4 2 |
| | 5 2 | 6 2 | 3+4 2 | 3+4 2 | | 3+5 2 |
| | 6 2 | 5 2 | 5+6 2 | 3+4 2 | | 2+4 2 |
| | 5 2 | 5 2 | 5+6 2 | 5+6 2 | | 2+4 2 |
| | 6 2 | 6 2 | 5+6 2 | 3+4 2 | | 3+5 2 |
| | 6 2 | 6 2 | 3+4 2 | 3+4 2 | | 3+5 2 |
| | 6 2 | 6 2 | 5+6 2 | 3+4 2 | | 2+4 2 |
| | 6 2 | 5 2 | 5+6 2 | 5+6 2 | | 3+5 2 |
| | 6 2 | 6 2 | 3+4 2 | 3+4 2 | | 2+4 2 |
| | 6 2 | 6 2 | 5+6 2 | 5+6 2 | | 3+5 2 |
| | 6 2 | 6 2 | 5+6 2 | 3+4 2 | | 2+4 2 |
| | 6 2 | 5 2 | 5+6 2 | 3+4 2 | | 3+5 2 |
| | 6 2 | 6 2 | 3+4 2 | 3+4 2 | | 2+4 2 |
| | 5 2 | 6 2 | 5+6 2 | 3+4 2 | | 3+5 2 |
| | 6 2 | 5 2 | 3+4 2 | 3+4 2 | | 2+4 2 |
| | 5 2 | 5 2 | 5+6 2 | 5+6 2 | | 2+4 2 |
| | 6 2 | 5 2 | 3+4 2 | 5+6 2 | | 3+5 2 |
| | 6 2 | 5 2 | 5+6 2 | 3+4 2 | | 2+4 2 |
| | 6 2 | 6 2 | 3+4 2 | 5+6 2 | | 3+5 2 |
| | 5 2 | 6 2 | 3+4 2 | 5+6 2 | | 3+5 2 |
| | 5 2 | 6 2 | 3+4 2 | 5+6 2 | | 2+4 2 |
| | 5 2 | 5 2 | 3+4 2 | 5+6 2 | | 3+5 2 |
| | 5 2 | 5 2 | 3+4 2 | 3+4 2 | | 2+4 2 |
| | 5 2 | 5 2 | 5+6 2 | 5+6 2 | | 2+4 2 |
| | 5 2 | 6 2 | 3+4 2 | 3+4 2 | | 3+5 2 |
| 74.2.14 | 97092 | 124 | Ca2O7Os2 | | | |
| | -GM | -X | -S | -R | -T | -W |
| | 5 2 | 6 2 | 3+4 2 | 5+6 2 | 2+2 4 | 2+4 2 |
| | 5 2 | 5 2 | 3+4 2 | 3+4 2 | 2+2 4 | 3+5 2 |
| | 5 2 | 5 2 | 3+4 2 | 5+6 2 | 2+2 4 | 3+5 2 |
| | 5 2 | 6 2 | 5+6 2 | 5+6 2 | 2+2 4 | 2+4 2 |
| | 5 2 | 6 2 | 3+4 2 | 3+4 2 | 2+2 4 | 3+5 2 |
| | 6 2 | 6 2 | 5+6 2 | 5+6 2 | 2+2 4 | 2+4 2 |
| | 6 2 | 5 2 | 3+4 2 | 3+4 2 | 2+2 4 | 2+4 2 |
| | 6 2 | 5 2 | 5+6 2 | 5+6 2 | 2+2 4 | 3+5 2 |
| | 6 2 | 6 2 | 5+6 2 | 3+4 2 | 2+2 4 | 3+5 2 |
| | 5 2 | 5 2 | 5+6 2 | 3+4 2 | 2+2 4 | 2+4 2 |
| | 6 2 | 5 2 | 5+6 2 | 5+6 2 | 2+2 4 | 2+4 2 |
| | 5 2 | 5 2 | 3+4 2 | 5+6 2 | 2+2 4 | 3+5 2 |

Continued on next column

**TABLE XLVI– continued**

| TQC | ICSD | NoE | SF | | | |
|---|---|---|---|---|---|---|
| | 6 2 | 6 2 | 5+6 2 | 3+4 2 | 2+2 4 | 2+4 2 |
| | 6 2 | 6 2 | 3+4 2 | 3+4 2 | 2+2 4 | 3+5 2 |
| | 5 2 | 6 2 | 3+4 2 | 5+6 2 | 2+2 4 | 3+5 2 |
| | 5 2 | 6 2 | 3+4 2 | 5+6 2 | 2+2 4 | 2+4 2 |
| | 5 2 | 5 2 | 3+4 2 | 5+6 2 | 2+2 4 | 2+4 2 |
| | 5 2 | 6 2 | 3+4 2 | 3+4 2 | 2+2 4 | 3+5 2 |
| | 5 2 | 5 2 | 3+4 2 | 5+6 2 | 2+2 4 | 3+5 2 |
| | 5 2 | 6 2 | 5+6 2 | 3+4 2 | 2+2 4 | 2+4 2 |
| | 5 2 | 6 2 | 3+4 2 | 5+6 2 | 2+2 4 | 3+5 2 |
| | 5 2 | 5 2 | 5+6 2 | 3+4 2 | 2+2 4 | 2+4 2 |
| | 5 2 | 5 2 | 3+4 2 | 5+6 2 | 2+2 4 | 3+5 2 |
| | 5 2 | 6 2 | 3+4 2 | 5+6 2 | 2+2 4 | 2+4 2 |
| | 5 2 | 5 2 | 5+6 2 | 3+4 2 | 2+2 4 | 2+4 2 |
| | 5 2 | 6 2 | 3+4 2 | 5+6 2 | 2+2 4 | 3+5 2 |
| | 6 2 | 5 2 | 5+6 2 | 3+4 2 | 2+2 4 | 2+4 2 |
| | 5 2 | 5 2 | 5+6 2 | 5+6 2 | 2+2 4 | 3+5 2 |
| | 6 2 | 5 2 | 3+4 2 | 3+4 2 | 2+2 4 | 2+4 2 |
| | 6 2 | 6 2 | 5+6 2 | 5+6 2 | 2+2 4 | 3+5 2 |
| | 5 2 | 6 2 | 5+6 2 | 5+6 2 | 2+2 4 | 2+4 2 |
| | 6 2 | 5 2 | 3+4 2 | 5+6 2 | | 3+5 2 |
| | 6 2 | 5 2 | 5+6 2 | 5+6 2 | | 3+5 2 |
| | 5 2 | 5 2 | 3+4 2 | 3+4 2 | | 2+4 2 |
| | 6 2 | 5 2 | 5+6 2 | 3+4 2 | | 2+4 2 |
| | 5 2 | 6 2 | 3+4 2 | 5+6 2 | | 3+5 2 |
| | 6 2 | 6 2 | 5+6 2 | 3+4 2 | | 3+5 2 |
| | 5 2 | 5 2 | 3+4 2 | 5+6 2 | | 2+4 2 |
| | 5 2 | 6 2 | 3+4 2 | 3+4 2 | | 3+5 2 |
| | 6 2 | 5 2 | 5+6 2 | 3+4 2 | | 2+4 2 |
| | 5 2 | 5 2 | 5+6 2 | 5+6 2 | | 2+4 2 |
| | 6 2 | 6 2 | 5+6 2 | 3+4 2 | | 3+5 2 |
| | 6 2 | 6 2 | 3+4 2 | 3+4 2 | | 3+5 2 |
| | 6 2 | 6 2 | 5+6 2 | 3+4 2 | | 2+4 2 |
| | 6 2 | 5 2 | 5+6 2 | 5+6 2 | | 3+5 2 |
| | 6 2 | 6 2 | 3+4 2 | 3+4 2 | | 2+4 2 |
| | 6 2 | 6 2 | 5+6 2 | 5+6 2 | | 3+5 2 |
| | 6 2 | 6 2 | 5+6 2 | 3+4 2 | | 2+4 2 |
| | 6 2 | 5 2 | 5+6 2 | 3+4 2 | | 3+5 2 |
| | 6 2 | 6 2 | 3+4 2 | 3+4 2 | | 2+4 2 |
| | 5 2 | 6 2 | 3+4 2 | 3+4 2 | | 3+5 2 |
| | 6 2 | 5 2 | 5+6 2 | 3+4 2 | | 2+4 2 |
| | 5 2 | 5 2 | 5+6 2 | 5+6 2 | | 2+4 2 |
| | 6 2 | 5 2 | 3+4 2 | 5+6 2 | | 3+5 2 |
| | 6 2 | 5 2 | 5+6 2 | 3+4 2 | | 2+4 2 |
| | 6 2 | 6 2 | 3+4 2 | 5+6 2 | | 3+5 2 |
| | 5 2 | 6 2 | 3+4 2 | 5+6 2 | | 3+5 2 |
| | 5 2 | 6 2 | 3+4 2 | 5+6 2 | | 2+4 2 |
| | 5 2 | 5 2 | 3+4 2 | 5+6 2 | | 3+5 2 |
| | 5 2 | 5 2 | 3+4 2 | 3+4 2 | | 2+4 2 |
| | 5 2 | 5 2 | 5+6 2 | 5+6 2 | | 3+5 2 |
| | 5 2 | 6 2 | 3+4 2 | 3+4 2 | | 2+4 2 |
| 74.2.15 | 97091 | 124 | Ca2O7Os2 | | | |
| | -GM | -X | -S | -R | -T | -W |
| | 5 2 | 6 2 | 3+4 2 | 5+6 2 | 2+2 4 | 2+4 2 |
| | 5 2 | 5 2 | 3+4 2 | 3+4 2 | 2+2 4 | 3+5 2 |
| | 5 2 | 5 2 | 3+4 2 | 5+6 2 | 2+2 4 | 3+5 2 |
| | 5 2 | 6 2 | 5+6 2 | 5+6 2 | 2+2 4 | 2+4 2 |
| | 5 2 | 6 2 | 3+4 2 | 3+4 2 | 2+2 4 | 3+5 2 |
| | 6 2 | 6 2 | 5+6 2 | 5+6 2 | 2+2 4 | 2+4 2 |
| | 6 2 | 5 2 | 3+4 2 | 3+4 2 | 2+2 4 | 2+4 2 |
| | 6 2 | 5 2 | 5+6 2 | 5+6 2 | 2+2 4 | 3+5 2 |
| | 6 2 | 6 2 | 5+6 2 | 3+4 2 | 2+2 4 | 3+5 2 |

Continued on next column

**TABLE XLVI– continued**

| TQC | ICSD | NoE | SF | | | |
|---|---|---|---|---|---|---|
| | 5 2 | 5 2 | 5+6 2 | 3+4 2 | 2+2 4 | 2+4 2 |
| | 6 2 | 5 2 | 5+6 2 | 5+6 2 | 2+2 4 | 2+4 2 |
| | 5 2 | 5 2 | 3+4 2 | 5+6 2 | 2+2 4 | 3+5 2 |
| | 6 2 | 6 2 | 5+6 2 | 3+4 2 | 2+2 4 | 2+4 2 |
| | 6 2 | 6 2 | 3+4 2 | 3+4 2 | 2+2 4 | 3+5 2 |
| | 5 2 | 6 2 | 3+4 2 | 5+6 2 | 2+2 4 | 3+5 2 |
| | 5 2 | 6 2 | 3+4 2 | 5+6 2 | 2+2 4 | 2+4 2 |
| | 5 2 | 5 2 | 3+4 2 | 5+6 2 | 2+2 4 | 2+4 2 |
| | 5 2 | 6 2 | 3+4 2 | 3+4 2 | 2+2 4 | 3+5 2 |
| | 5 2 | 5 2 | 3+4 2 | 5+6 2 | 2+2 4 | 3+5 2 |
| | 5 2 | 6 2 | 5+6 2 | 3+4 2 | 2+2 4 | 2+4 2 |
| | 5 2 | 6 2 | 3+4 2 | 5+6 2 | 2+2 4 | 3+5 2 |
| | 5 2 | 5 2 | 5+6 2 | 3+4 2 | 2+2 4 | 2+4 2 |
| | 5 2 | 5 2 | 3+4 2 | 5+6 2 | 2+2 4 | 3+5 2 |
| | 5 2 | 6 2 | 3+4 2 | 5+6 2 | 2+2 4 | 2+4 2 |
| | 5 2 | 5 2 | 5+6 2 | 3+4 2 | 2+2 4 | 2+4 2 |
| | 5 2 | 6 2 | 3+4 2 | 5+6 2 | 2+2 4 | 3+5 2 |
| | 6 2 | 5 2 | 5+6 2 | 3+4 2 | 2+2 4 | 2+4 2 |
| | 5 2 | 5 2 | 5+6 2 | 5+6 2 | 2+2 4 | 3+5 2 |
| | 6 2 | 5 2 | 3+4 2 | 3+4 2 | 2+2 4 | 2+4 2 |
| | 6 2 | 6 2 | 5+6 2 | 5+6 2 | 2+2 4 | 3+5 2 |
| | 5 2 | 6 2 | 5+6 2 | 5+6 2 | 2+2 4 | 2+4 2 |
| | 6 2 | 5 2 | 3+4 2 | 5+6 2 | | 3+5 2 |
| | 6 2 | 5 2 | 5+6 2 | 5+6 2 | | 3+5 2 |
| | 5 2 | 5 2 | 3+4 2 | 3+4 2 | | 2+4 2 |
| | 6 2 | 5 2 | 5+6 2 | 3+4 2 | | 2+4 2 |
| | 5 2 | 6 2 | 3+4 2 | 5+6 2 | | 3+5 2 |
| | 6 2 | 6 2 | 5+6 2 | 3+4 2 | | 3+5 2 |
| | 5 2 | 5 2 | 3+4 2 | 5+6 2 | | 2+4 2 |
| | 5 2 | 6 2 | 3+4 2 | 3+4 2 | | 3+5 2 |
| | 6 2 | 5 2 | 5+6 2 | 3+4 2 | | 2+4 2 |
| | 5 2 | 5 2 | 5+6 2 | 5+6 2 | | 2+4 2 |
| | 6 2 | 6 2 | 5+6 2 | 3+4 2 | | 3+5 2 |
| | 6 2 | 6 2 | 3+4 2 | 3+4 2 | | 3+5 2 |
| | 6 2 | 6 2 | 5+6 2 | 3+4 2 | | 2+4 2 |
| | 6 2 | 5 2 | 5+6 2 | 5+6 2 | | 3+5 2 |
| | 6 2 | 6 2 | 3+4 2 | 3+4 2 | | 2+4 2 |
| | 6 2 | 6 2 | 5+6 2 | 5+6 2 | | 3+5 2 |
| | 6 2 | 6 2 | 5+6 2 | 3+4 2 | | 2+4 2 |
| | 6 2 | 5 2 | 5+6 2 | 3+4 2 | | 3+5 2 |
| | 6 2 | 6 2 | 3+4 2 | 3+4 2 | | 2+4 2 |
| | 5 2 | 6 2 | 3+4 2 | 3+4 2 | | 3+5 2 |
| | 6 2 | 5 2 | 5+6 2 | 3+4 2 | | 2+4 2 |
| | 5 2 | 5 2 | 5+6 2 | 5+6 2 | | 2+4 2 |
| | 6 2 | 5 2 | 3+4 2 | 5+6 2 | | 3+5 2 |
| | 6 2 | 5 2 | 5+6 2 | 3+4 2 | | 2+4 2 |
| | 6 2 | 6 2 | 3+4 2 | 5+6 2 | | 3+5 2 |
| | 5 2 | 6 2 | 3+4 2 | 5+6 2 | | 3+5 2 |
| | 5 2 | 6 2 | 3+4 2 | 5+6 2 | | 2+4 2 |
| | 5 2 | 5 2 | 3+4 2 | 5+6 2 | | 3+5 2 |
| | 5 2 | 5 2 | 3+4 2 | 3+4 2 | | 2+4 2 |
| | 5 2 | 5 2 | 5+6 2 | 5+6 2 | | 3+5 2 |
| | 5 2 | 6 2 | 3+4 2 | 3+4 2 | | 2+4 2 |
| 74.2.16 | 418971 | 52 | Ca1Zn2 | | | |
| | -GM | -X | -S | -R | -T | -W |
| | 5 2 | 5 2 | 3+4 2 | 5+6 2 | 2+2 4 | 2+4 2 |
| | 5 2 | 6 2 | 5+6 2 | 3+4 2 | 2+2 4 | 3+5 2 |
| | 6 2 | 5 2 | 3+4 2 | 5+6 2 | 2+2 4 | 2+4 2 |
| | 5 2 | 6 2 | 5+6 2 | 3+4 2 | 2+2 4 | 3+5 2 |
| | 5 2 | 6 2 | 3+4 2 | 3+4 2 | 2+2 4 | 3+5 2 |
| | 6 2 | 5 2 | 3+4 2 | 5+6 2 | 2+2 4 | 2+4 2 |

Continued on next column

**TABLE XLVI– continued**

| TQC | ICSD | NoE | SF | | | |
|---|---|---|---|---|---|---|
| | 5 2 | 6 2 | 5+6 2 | 3+4 2 | 2+2 4 | 3+5 2 |
| | 6 2 | 5 2 | 5+6 2 | 5+6 2 | 2+2 4 | 2+4 2 |
| | 6 2 | 6 2 | 3+4 2 | 3+4 2 | 2+2 4 | 2+4 2 |
| | 5 2 | 5 2 | 5+6 2 | 5+6 2 | 2+2 4 | 3+5 2 |
| | 6 2 | 6 2 | 3+4 2 | 3+4 2 | 2+2 4 | 2+4 2 |
| | 5 2 | 5 2 | 5+6 2 | 5+6 2 | 2+2 4 | 3+5 2 |
| | 6 2 | 5 2 | 3+4 2 | 3+4 2 | 2+2 4 | 3+5 2 |
| | 6 2 | 6 2 | 5+6 2 | 5+6 2 | | 2+4 2 |
| | 6 2 | 5 2 | 5+6 2 | 3+4 2 | | 3+5 2 |
| | 5 2 | 5 2 | 3+4 2 | 5+6 2 | | 2+4 2 |
| | 6 2 | 6 2 | 5+6 2 | 3+4 2 | | 2+4 2 |
| | 5 2 | 6 2 | 5+6 2 | 5+6 2 | | 3+5 2 |
| | 6 2 | 5 2 | 3+4 2 | 5+6 2 | | 3+5 2 |
| | 5 2 | 6 2 | 3+4 2 | 3+4 2 | | 2+4 2 |
| | 5 2 | 6 2 | 3+4 2 | 3+4 2 | | 2+4 2 |
| | 6 2 | 5 2 | 5+6 2 | 5+6 2 | | 3+5 2 |
| | 5 2 | 5 2 | 3+4 2 | 3+4 2 | | 2+4 2 |
| | 5 2 | 6 2 | 5+6 2 | 5+6 2 | | 3+5 2 |
| | 6 2 | 6 2 | 3+4 2 | 5+6 2 | | 2+4 2 |
| | 6 2 | 5 2 | 3+4 2 | 3+4 2 | | 3+5 2 |
| 74.2.17 | 420576 | 52 | Ca1Cd2 | | | |
| | -GM | -X | -S | -R | -T | -W |
| | 5 2 | 6 2 | 5+6 2 | 5+6 2 | 2+2 4 | 3+5 2 |
| | 6 2 | 5 2 | 3+4 2 | 3+4 2 | 2+2 4 | 2+4 2 |
| | 5 2 | 6 2 | 5+6 2 | 5+6 2 | 2+2 4 | 3+5 2 |
| | 5 2 | 6 2 | 3+4 2 | 3+4 2 | 2+2 4 | 2+4 2 |
| | 5 2 | 5 2 | 5+6 2 | 3+4 2 | 2+2 4 | 2+4 2 |
| | 6 2 | 5 2 | 5+6 2 | 5+6 2 | 2+2 4 | 3+5 2 |
| | 6 2 | 6 2 | 3+4 2 | 3+4 2 | 2+2 4 | 2+4 2 |
| | 6 2 | 5 2 | 3+4 2 | 5+6 2 | 2+2 4 | 3+5 2 |
| | 6 2 | 6 2 | 5+6 2 | 3+4 2 | 2+2 4 | 3+5 2 |
| | 5 2 | 5 2 | 3+4 2 | 5+6 2 | 2+2 4 | 2+4 2 |
| | 5 2 | 6 2 | 5+6 2 | 3+4 2 | 2+2 4 | 3+5 2 |
| | 6 2 | 5 2 | 3+4 2 | 5+6 2 | 2+2 4 | 2+4 2 |
| | 6 2 | 5 2 | 3+4 2 | 5+6 2 | 2+2 4 | 2+4 2 |
| | 5 2 | 6 2 | 3+4 2 | 3+4 2 | | 3+5 2 |
| | 6 2 | 5 2 | 5+6 2 | 3+4 2 | | 2+4 2 |
| | 6 2 | 5 2 | 5+6 2 | 5+6 2 | | 3+5 2 |
| | 5 2 | 6 2 | 3+4 2 | 3+4 2 | | 3+5 2 |
| | 5 2 | 6 2 | 3+4 2 | 5+6 2 | | 2+4 2 |
| | 6 2 | 5 2 | 5+6 2 | 5+6 2 | | 2+4 2 |
| | 5 2 | 6 2 | 5+6 2 | 3+4 2 | | 3+5 2 |
| | 5 2 | 6 2 | 5+6 2 | 3+4 2 | | 2+4 2 |
| | 6 2 | 5 2 | 3+4 2 | 5+6 2 | | 3+5 2 |
| | 5 2 | 5 2 | 3+4 2 | 3+4 2 | | 3+5 2 |
| | 5 2 | 6 2 | 5+6 2 | 5+6 2 | | 2+4 2 |
| | 6 2 | 6 2 | 5+6 2 | 5+6 2 | | 3+5 2 |
| | 6 2 | 5 2 | 5+6 2 | 3+4 2 | | 2+4 2 |
| 74.2.18 | 414229 | 32 | Ba1In2 | | | |
| | -GM | -X | -S | -R | -T | -W |
| | 5 2 | 5 2 | 3+4 2 | 3+4 2 | 2+2 4 | 2+4 2 |
| | 6 2 | 6 2 | 5+6 2 | 5+6 2 | 2+2 4 | 3+5 2 |
| | 5 2 | 5 2 | 3+4 2 | 3+4 2 | 2+2 4 | 2+4 2 |
| | 6 2 | 6 2 | 5+6 2 | 5+6 2 | 2+2 4 | 3+5 2 |
| | 5 2 | 5 2 | 3+4 2 | 3+4 2 | 2+2 4 | 3+5 2 |
| | 5 2 | 6 2 | 5+6 2 | 5+6 2 | 2+2 4 | 2+4 2 |
| | 6 2 | 5 2 | 3+4 2 | 3+4 2 | 2+2 4 | 2+4 2 |
| | 6 2 | 6 2 | 5+6 2 | 5+6 2 | 2+2 4 | 3+5 2 |
| | 5 2 | 5 2 | 5+6 2 | 5+6 2 | | 2+4 2 |
| | 6 2 | 6 2 | 3+4 2 | 3+4 2 | | 3+5 2 |
| | 5 2 | 6 2 | 5+6 2 | 3+4 2 | | 3+5 2 |

Continued on next column

**TABLE XLVI– continued**

| TQC | ICSD | NoE | SF | | | |
|---|---|---|---|---|---|---|
| | 6 2 | 5 2 | 3+4 2 | 5+6 2 | | 2+4 2 |
| | 5 2 | 5 2 | 5+6 2 | 5+6 2 | | 2+4 2 |
| | 6 2 | 6 2 | 3+4 2 | 5+6 2 | | 3+5 2 |
| | 5 2 | 5 2 | 5+6 2 | 3+4 2 | | 3+5 2 |
| | 5 2 | 6 2 | 5+6 2 | 3+4 2 | | 2+4 2 |
| 74.2.19 | 251586 | 60 | Ba1Ge5 | | | |
| | -GM | -X | -S | -R | -T | -W |
| | 5 2 | 6 2 | 5+6 2 | 3+4 2 | 2+2 4 | 2+4 2 |
| | 6 2 | 5 2 | 3+4 2 | 5+6 2 | 2+2 4 | 3+5 2 |
| | 5 2 | 6 2 | 5+6 2 | 3+4 2 | 2+2 4 | 2+4 2 |
| | 6 2 | 5 2 | 3+4 2 | 5+6 2 | 2+2 4 | 3+5 2 |
| | 5 2 | 5 2 | 3+4 2 | 3+4 2 | 2+2 4 | 2+4 2 |
| | 6 2 | 6 2 | 5+6 2 | 5+6 2 | 2+2 4 | 3+5 2 |
| | 5 2 | 6 2 | 5+6 2 | 3+4 2 | 2+2 4 | 3+5 2 |
| | 6 2 | 5 2 | 3+4 2 | 5+6 2 | 2+2 4 | 2+4 2 |
| | 5 2 | 5 2 | 5+6 2 | 5+6 2 | 2+2 4 | 3+5 2 |
| | 6 2 | 6 2 | 3+4 2 | 3+4 2 | 2+2 4 | 2+4 2 |
| | 5 2 | 5 2 | 3+4 2 | 3+4 2 | 2+2 4 | 2+4 2 |
| | 6 2 | 6 2 | 5+6 2 | 5+6 2 | 2+2 4 | 3+5 2 |
| | 5 2 | 5 2 | 3+4 2 | 3+4 2 | 2+2 4 | 2+4 2 |
| | 6 2 | 6 2 | 5+6 2 | 3+4 2 | 2+2 4 | 3+5 2 |
| | 5 2 | 5 2 | 3+4 2 | 5+6 2 | 2+2 4 | 3+5 2 |
| | 6 2 | 6 2 | 5+6 2 | 5+6 2 | | 2+4 2 |
| | 5 2 | 5 2 | 3+4 2 | 3+4 2 | | 3+5 2 |
| | 5 2 | 6 2 | 5+6 2 | 5+6 2 | | 2+4 2 |
| | 6 2 | 5 2 | 3+4 2 | 5+6 2 | | 2+4 2 |
| | 6 2 | 6 2 | 5+6 2 | 3+4 2 | | 3+5 2 |
| | 5 2 | 6 2 | 3+4 2 | 5+6 2 | | 2+4 2 |
| | 6 2 | 6 2 | 5+6 2 | 3+4 2 | | 3+5 2 |
| | 6 2 | 5 2 | 3+4 2 | 3+4 2 | | 2+4 2 |
| | 5 2 | 5 2 | 5+6 2 | 5+6 2 | | 3+5 2 |
| | 5 2 | 6 2 | 3+4 2 | 3+4 2 | | 3+5 2 |
| | 6 2 | 5 2 | 3+4 2 | 3+4 2 | | 2+4 2 |
| | 5 2 | 5 2 | 5+6 2 | 5+6 2 | | 3+5 2 |
| | 5 2 | 5 2 | 5+6 2 | 3+4 2 | | 2+4 2 |
| | 6 2 | 6 2 | 5+6 2 | 5+6 2 | | 3+5 2 |
| | 5 2 | 5 2 | 3+4 2 | 5+6 2 | | 2+4 2 |
| 74.2.20 | 194880 | 60 | Ba1Ge5 | | | |
| | -GM | -X | -S | -R | -T | -W |
| | 5 2 | 6 2 | 5+6 2 | 3+4 2 | 2+2 4 | 2+4 2 |
| | 6 2 | 5 2 | 3+4 2 | 5+6 2 | 2+2 4 | 3+5 2 |
| | 5 2 | 6 2 | 5+6 2 | 3+4 2 | 2+2 4 | 2+4 2 |
| | 6 2 | 5 2 | 3+4 2 | 5+6 2 | 2+2 4 | 3+5 2 |
| | 5 2 | 5 2 | 3+4 2 | 3+4 2 | 2+2 4 | 2+4 2 |
| | 6 2 | 6 2 | 5+6 2 | 5+6 2 | 2+2 4 | 3+5 2 |
| | 5 2 | 6 2 | 5+6 2 | 3+4 2 | 2+2 4 | 3+5 2 |
| | 6 2 | 5 2 | 3+4 2 | 5+6 2 | 2+2 4 | 2+4 2 |
| | 5 2 | 5 2 | 5+6 2 | 5+6 2 | 2+2 4 | 3+5 2 |
| | 6 2 | 6 2 | 3+4 2 | 3+4 2 | 2+2 4 | 2+4 2 |
| | 5 2 | 5 2 | 3+4 2 | 3+4 2 | 2+2 4 | 2+4 2 |
| | 6 2 | 6 2 | 5+6 2 | 5+6 2 | 2+2 4 | 3+5 2 |
| | 5 2 | 5 2 | 3+4 2 | 3+4 2 | 2+2 4 | 2+4 2 |
| | 6 2 | 6 2 | 5+6 2 | 3+4 2 | 2+2 4 | 3+5 2 |
| | 5 2 | 5 2 | 3+4 2 | 5+6 2 | 2+2 4 | 3+5 2 |
| | 6 2 | 6 2 | 5+6 2 | 5+6 2 | | 2+4 2 |
| | 5 2 | 5 2 | 3+4 2 | 3+4 2 | | 3+5 2 |
| | 5 2 | 6 2 | 5+6 2 | 5+6 2 | | 2+4 2 |
| | 6 2 | 5 2 | 3+4 2 | 5+6 2 | | 2+4 2 |
| | 6 2 | 6 2 | 5+6 2 | 3+4 2 | | 3+5 2 |
| | 5 2 | 6 2 | 3+4 2 | 5+6 2 | | 2+4 2 |
| | 6 2 | 6 2 | 5+6 2 | 3+4 2 | | 3+5 2 |

Continued on next column

**TABLE XLVI– continued**

| TQC | ICSD | NoE | SF | | | |
|---|---|---|---|---|---|---|
| | 6 2 | 5 2 | 3+4 2 | 3+4 2 | | 2+4 2 |
| | 5 2 | 5 2 | 5+6 2 | 5+6 2 | | 3+5 2 |
| | 5 2 | 6 2 | 3+4 2 | 3+4 2 | | 3+5 2 |
| | 6 2 | 5 2 | 3+4 2 | 3+4 2 | | 2+4 2 |
| | 5 2 | 5 2 | 5+6 2 | 5+6 2 | | 3+5 2 |
| | 5 2 | 5 2 | 5+6 2 | 3+4 2 | | 2+4 2 |
| | 6 2 | 6 2 | 5+6 2 | 5+6 2 | | 3+5 2 |
| | 5 2 | 5 2 | 3+4 2 | 5+6 2 | | 2+4 2 |
| 74.2.21 | 260668 | 68 | Ba1Cd2 | | | |
| | -GM | -X | -S | -R | -T | -W |
| | 5 2 | 5 2 | 5+6 2 | 5+6 2 | 2+2 4 | 2+4 2 |
| | 6 2 | 6 2 | 3+4 2 | 3+4 2 | 2+2 4 | 3+5 2 |
| | 5 2 | 5 2 | 5+6 2 | 5+6 2 | 2+2 4 | 2+4 2 |
| | 6 2 | 6 2 | 3+4 2 | 3+4 2 | 2+2 4 | 3+5 2 |
| | 5 2 | 5 2 | 5+6 2 | 5+6 2 | 2+2 4 | 3+5 2 |
| | 5 2 | 6 2 | 3+4 2 | 3+4 2 | 2+2 4 | 2+4 2 |
| | 6 2 | 5 2 | 5+6 2 | 5+6 2 | 2+2 4 | 2+4 2 |
| | 6 2 | 6 2 | 3+4 2 | 3+4 2 | 2+2 4 | 3+5 2 |
| | 5 2 | 6 2 | 3+4 2 | 5+6 2 | 2+2 4 | 3+5 2 |
| | 6 2 | 5 2 | 3+4 2 | 3+4 2 | 2+2 4 | 2+4 2 |
| | 5 2 | 6 2 | 5+6 2 | 3+4 2 | 2+2 4 | 2+4 2 |
| | 5 2 | 6 2 | 5+6 2 | 5+6 2 | 2+2 4 | 3+5 2 |
| | 6 2 | 5 2 | 3+4 2 | 3+4 2 | 2+2 4 | 2+4 2 |
| | 5 2 | 5 2 | 3+4 2 | 5+6 2 | 2+2 4 | 3+5 2 |
| | 6 2 | 6 2 | 5+6 2 | 3+4 2 | 2+2 4 | 3+5 2 |
| | 6 2 | 5 2 | 5+6 2 | 5+6 2 | 2+2 4 | 2+4 2 |
| | 5 2 | 6 2 | 3+4 2 | 3+4 2 | 2+2 4 | 2+4 2 |
| | 6 2 | 5 2 | 5+6 2 | 5+6 2 | | 3+5 2 |
| | 5 2 | 6 2 | 3+4 2 | 3+4 2 | | 2+4 2 |
| | 6 2 | 5 2 | 5+6 2 | 5+6 2 | | 3+5 2 |
| | 5 2 | 5 2 | 5+6 2 | 5+6 2 | | 2+4 2 |
| | 6 2 | 6 2 | 5+6 2 | 3+4 2 | | 3+5 2 |
| | 6 2 | 5 2 | 3+4 2 | 3+4 2 | | 3+5 2 |
| | 6 2 | 5 2 | 3+4 2 | 5+6 2 | | 2+4 2 |
| | 5 2 | 6 2 | 5+6 2 | 3+4 2 | | 2+4 2 |
| | 5 2 | 6 2 | 5+6 2 | 5+6 2 | | 3+5 2 |
| | 6 2 | 5 2 | 3+4 2 | 5+6 2 | | 2+4 2 |
| | 5 2 | 6 2 | 3+4 2 | 3+4 2 | | 3+5 2 |
| | 5 2 | 6 2 | 3+4 2 | 3+4 2 | | 2+4 2 |
| | 6 2 | 5 2 | 5+6 2 | 5+6 2 | | 3+5 2 |
| | 5 2 | 5 2 | 3+4 2 | 3+4 2 | | 3+5 2 |
| | 6 2 | 6 2 | 5+6 2 | 5+6 2 | | 2+4 2 |
| | 5 2 | 6 2 | 3+4 2 | 3+4 2 | | 2+4 2 |
| | 6 2 | 5 2 | 3+4 2 | 5+6 2 | | 3+5 2 |
| 74.2.22 | 55542 | 48 | Au2Ca1 | | | |
| | -GM | -X | -S | -R | -T | -W |
| | 5 2 | 5 2 | 3+4 2 | 3+4 2 | 2+2 4 | 2+4 2 |
| | 6 2 | 6 2 | 5+6 2 | 5+6 2 | 2+2 4 | 3+5 2 |
| | 5 2 | 5 2 | 5+6 2 | 5+6 2 | 2+2 4 | 2+4 2 |
| | 5 2 | 6 2 | 3+4 2 | 3+4 2 | 2+2 4 | 3+5 2 |
| | 6 2 | 6 2 | 3+4 2 | 3+4 2 | 2+2 4 | 3+5 2 |
| | 5 2 | 6 2 | 5+6 2 | 5+6 2 | 2+2 4 | 2+4 2 |
| | 5 2 | 6 2 | 3+4 2 | 3+4 2 | 2+2 4 | 3+5 2 |
| | 6 2 | 5 2 | 3+4 2 | 3+4 2 | 2+2 4 | 2+4 2 |
| | 6 2 | 5 2 | 5+6 2 | 5+6 2 | 2+2 4 | 2+4 2 |
| | 6 2 | 5 2 | 5+6 2 | 3+4 2 | 2+2 4 | 3+5 2 |
| | 5 2 | 6 2 | 3+4 2 | 5+6 2 | 2+2 4 | 3+5 2 |
| | 5 2 | 5 2 | 5+6 2 | 5+6 2 | 2+2 4 | 2+4 2 |
| | 6 2 | 6 2 | 3+4 2 | 3+4 2 | | 2+4 2 |
| | 6 2 | 5 2 | 5+6 2 | 5+6 2 | | 3+5 2 |
| | 6 2 | 5 2 | 5+6 2 | 5+6 2 | | 3+5 2 |

Continued on next column

**TABLE XLVI– continued**

| TQC | ICSD | NoE | SF | | | |
|---|---|---|---|---|---|---|
| | 5 2 | 6 2 | 5+6 2 | 3+4 2 | | 2+4 2 |
| | 6 2 | 5 2 | 3+4 2 | 5+6 2 | | 3+5 2 |
| | 5 2 | 5 2 | 5+6 2 | 3+4 2 | | 2+4 2 |
| | 5 2 | 6 2 | 3+4 2 | 5+6 2 | | 2+4 2 |
| | 6 2 | 5 2 | 3+4 2 | 5+6 2 | | 3+5 2 |
| | 5 2 | 6 2 | 5+6 2 | 3+4 2 | | 2+4 2 |
| | 6 2 | 6 2 | 3+4 2 | 3+4 2 | | 3+5 2 |
| | 5 2 | 5 2 | 3+4 2 | 3+4 2 | | 3+5 2 |
| | 5 2 | 6 2 | 5+6 2 | 5+6 2 | | 2+4 2 |
| 74.2.23 | 72289 | 128 | Al1F7Na2Ni1 | | | |
| | -GM | -X | -S | -R | -T | -W |
| | 5 2 | 5 2 | 5+6 2 | 5+6 2 | 2+2 4 | 2+4 2 |
| | 5 2 | 5 2 | 3+4 2 | 5+6 2 | 2+2 4 | 3+5 2 |
| | 6 2 | 6 2 | 5+6 2 | 3+4 2 | 2+2 4 | 2+4 2 |
| | 6 2 | 6 2 | 3+4 2 | 3+4 2 | 2+2 4 | 3+5 2 |
| | 6 2 | 6 2 | 5+6 2 | 3+4 2 | 2+2 4 | 2+4 2 |
| | 6 2 | 6 2 | 3+4 2 | 3+4 2 | 2+2 4 | 3+5 2 |
| | 5 2 | 5 2 | 5+6 2 | 5+6 2 | 2+2 4 | 2+4 2 |
| | 5 2 | 5 2 | 3+4 2 | 5+6 2 | 2+2 4 | 3+5 2 |
| | 6 2 | 6 2 | 5+6 2 | 3+4 2 | 2+2 4 | 2+4 2 |
| | 6 2 | 6 2 | 3+4 2 | 3+4 2 | 2+2 4 | 3+5 2 |
| | 5 2 | 5 2 | 5+6 2 | 5+6 2 | 2+2 4 | 2+4 2 |
| | 5 2 | 5 2 | 3+4 2 | 5+6 2 | 2+2 4 | 3+5 2 |
| | 5 2 | 6 2 | 5+6 2 | 3+4 2 | 2+2 4 | 2+4 2 |
| | 6 2 | 5 2 | 3+4 2 | 5+6 2 | 2+2 4 | 3+5 2 |
| | 5 2 | 5 2 | 5+6 2 | 5+6 2 | 2+2 4 | 2+4 2 |
| | 5 2 | 5 2 | 3+4 2 | 5+6 2 | 2+2 4 | 3+5 2 |
| | 6 2 | 6 2 | 5+6 2 | 3+4 2 | 2+2 4 | 2+4 2 |
| | 5 2 | 6 2 | 3+4 2 | 3+4 2 | 2+2 4 | 3+5 2 |
| | 6 2 | 6 2 | 5+6 2 | 3+4 2 | 2+2 4 | 3+5 2 |
| | 6 2 | 6 2 | 3+4 2 | 3+4 2 | 2+2 4 | 2+4 2 |
| | 6 2 | 6 2 | 5+6 2 | 3+4 2 | 2+2 4 | 2+4 2 |
| | 5 2 | 6 2 | 5+6 2 | 5+6 2 | 2+2 4 | 3+5 2 |
| | 5 2 | 5 2 | 5+6 2 | 5+6 2 | 2+2 4 | 2+4 2 |
| | 6 2 | 6 2 | 3+4 2 | 3+4 2 | 2+2 4 | 3+5 2 |
| | 5 2 | 5 2 | 3+4 2 | 5+6 2 | 2+2 4 | 3+5 2 |
| | 5 2 | 5 2 | 3+4 2 | 5+6 2 | 2+2 4 | 2+4 2 |
| | 6 2 | 6 2 | 5+6 2 | 5+6 2 | 2+2 4 | 2+4 2 |
| | 5 2 | 5 2 | 3+4 2 | 5+6 2 | 2+2 4 | 3+5 2 |
| | 5 2 | 5 2 | 5+6 2 | 3+4 2 | 2+2 4 | 2+4 2 |
| | 6 2 | 5 2 | 5+6 2 | 5+6 2 | 2+2 4 | 3+5 2 |
| | 6 2 | 5 2 | 3+4 2 | 3+4 2 | 2+2 4 | 3+5 2 |
| | 5 2 | 6 2 | 5+6 2 | 3+4 2 | 2+2 4 | 2+4 2 |
| | 5 2 | 5 2 | 5+6 2 | 5+6 2 | | 3+5 2 |
| | 5 2 | 5 2 | 3+4 2 | 5+6 2 | | 2+4 2 |
| | 5 2 | 6 2 | 3+4 2 | 3+4 2 | | 2+4 2 |
| | 6 2 | 6 2 | 5+6 2 | 5+6 2 | | 3+5 2 |
| | 6 2 | 6 2 | 3+4 2 | 5+6 2 | | 2+4 2 |
| | 6 2 | 5 2 | 5+6 2 | 3+4 2 | | 3+5 2 |
| | 5 2 | 5 2 | 3+4 2 | 3+4 2 | | 3+5 2 |
| | 6 2 | 6 2 | 5+6 2 | 3+4 2 | | 2+4 2 |
| | 6 2 | 6 2 | 3+4 2 | 3+4 2 | | 2+4 2 |
| | 5 2 | 6 2 | 5+6 2 | 3+4 2 | | 3+5 2 |
| | 6 2 | 6 2 | 3+4 2 | 5+6 2 | | 3+5 2 |
| | 6 2 | 6 2 | 3+4 2 | 5+6 2 | | 2+4 2 |
| | 6 2 | 5 2 | 5+6 2 | 3+4 2 | | 3+5 2 |
| | 5 2 | 6 2 | 5+6 2 | 5+6 2 | | 2+4 2 |
| | 6 2 | 5 2 | 5+6 2 | 3+4 2 | | 3+5 2 |
| | 6 2 | 5 2 | 3+4 2 | 3+4 2 | | 2+4 2 |
| | 5 2 | 6 2 | 3+4 2 | 3+4 2 | | 2+4 2 |
| | 5 2 | 6 2 | 5+6 2 | 5+6 2 | | 3+5 2 |

Continued on next column

**TABLE XLVI– continued**

| TQC | ICSD | NoE | SF | | | |
|---|---|---|---|---|---|---|
| | 5 2 | 5 2 | 3+4 2 | 3+4 2 | | 2+4 2 |
| | 6 2 | 5 2 | 3+4 2 | 3+4 2 | | 3+5 2 |
| | 6 2 | 6 2 | 5+6 2 | 5+6 2 | | 2+4 2 |
| | 6 2 | 5 2 | 3+4 2 | 5+6 2 | | 3+5 2 |
| | 5 2 | 5 2 | 3+4 2 | 5+6 2 | | 3+5 2 |
| | 5 2 | 5 2 | 5+6 2 | 5+6 2 | | 2+4 2 |
| | 5 2 | 6 2 | 5+6 2 | 5+6 2 | | 2+4 2 |
| | 5 2 | 6 2 | 5+6 2 | 3+4 2 | | 3+5 2 |
| | 5 2 | 6 2 | 5+6 2 | 3+4 2 | | 2+4 2 |
| | 5 2 | 6 2 | 5+6 2 | 5+6 2 | | 3+5 2 |
| | 5 2 | 6 2 | 5+6 2 | 3+4 2 | | 3+5 2 |
| | 5 2 | 6 2 | 5+6 2 | 5+6 2 | | 2+4 2 |
| | 5 2 | 6 2 | 5+6 2 | 5+6 2 | | 3+5 2 |
| | 5 2 | 6 2 | 5+6 2 | 5+6 2 | | 2+4 2 |
| 74.2.24 | 417225 | 48 | Ag2Ca1 | | | |
| | -GM | -X | -S | -R | -T | -W |
| | 5 2 | 5 2 | 3+4 2 | 5+6 2 | 2+2 4 | 2+4 2 |
| | 5 2 | 6 2 | 5+6 2 | 3+4 2 | 2+2 4 | 3+5 2 |
| | 5 2 | 5 2 | 3+4 2 | 5+6 2 | 2+2 4 | 2+4 2 |
| | 6 2 | 6 2 | 3+4 2 | 5+6 2 | 2+2 4 | 3+5 2 |
| | 5 2 | 6 2 | 5+6 2 | 3+4 2 | 2+2 4 | 3+5 2 |
| | 6 2 | 6 2 | 3+4 2 | 3+4 2 | 2+2 4 | 2+4 2 |
| | 5 2 | 5 2 | 5+6 2 | 3+4 2 | 2+2 4 | 2+4 2 |
| | 6 2 | 5 2 | 3+4 2 | 5+6 2 | 2+2 4 | 3+5 2 |
| | 6 2 | 6 2 | 5+6 2 | 3+4 2 | 2+2 4 | 2+4 2 |
| | 6 2 | 5 2 | 3+4 2 | 3+4 2 | 2+2 4 | 3+5 2 |
| | 5 2 | 6 2 | 5+6 2 | 5+6 2 | 2+2 4 | 2+4 2 |
| | 6 2 | 6 2 | 5+6 2 | 5+6 2 | 2+2 4 | 3+5 2 |
| | 5 2 | 5 2 | 3+4 2 | 3+4 2 | | 2+4 2 |
| | 6 2 | 5 2 | 5+6 2 | 5+6 2 | | 3+5 2 |
| | 6 2 | 5 2 | 5+6 2 | 3+4 2 | | 3+5 2 |
| | 5 2 | 6 2 | 5+6 2 | 5+6 2 | | 2+4 2 |
| | 6 2 | 5 2 | 3+4 2 | 3+4 2 | | 3+5 2 |
| | 5 2 | 6 2 | 5+6 2 | 5+6 2 | | 2+4 2 |
| | 5 2 | 5 2 | 3+4 2 | 5+6 2 | | 3+5 2 |
| | 6 2 | 6 2 | 5+6 2 | 3+4 2 | | 2+4 2 |
| | 5 2 | 5 2 | 3+4 2 | 5+6 2 | | 2+4 2 |
| | 6 2 | 6 2 | 3+4 2 | 3+4 2 | | 3+5 2 |
| | 5 2 | 5 2 | 3+4 2 | 3+4 2 | | 3+5 2 |
| | 5 2 | 6 2 | 5+6 2 | 5+6 2 | | 2+4 2 |
| 74.2.25 | 417227 | 64 | Ag1Ca1Zn1 | | | |
| | -GM | -X | -S | -R | -T | -W |
| | 5 2 | 5 2 | 5+6 2 | 5+6 2 | 2+2 4 | 2+4 2 |
| | 6 2 | 6 2 | 3+4 2 | 3+4 2 | 2+2 4 | 3+5 2 |
| | 5 2 | 5 2 | 5+6 2 | 5+6 2 | 2+2 4 | 2+4 2 |
| | 6 2 | 6 2 | 3+4 2 | 3+4 2 | 2+2 4 | 3+5 2 |
| | 5 2 | 5 2 | 5+6 2 | 5+6 2 | 2+2 4 | 2+4 2 |
| | 5 2 | 6 2 | 3+4 2 | 3+4 2 | 2+2 4 | 3+5 2 |
| | 6 2 | 5 2 | 5+6 2 | 5+6 2 | 2+2 4 | 3+5 2 |
| | 6 2 | 6 2 | 3+4 2 | 3+4 2 | 2+2 4 | 2+4 2 |
| | 5 2 | 5 2 | 3+4 2 | 5+6 2 | 2+2 4 | 3+5 2 |
| | 5 2 | 6 2 | 5+6 2 | 3+4 2 | 2+2 4 | 2+4 2 |
| | 5 2 | 6 2 | 3+4 2 | 5+6 2 | 2+2 4 | 3+5 2 |
| | 5 2 | 5 2 | 3+4 2 | 5+6 2 | 2+2 4 | 2+4 2 |
| | 6 2 | 6 2 | 5+6 2 | 3+4 2 | 2+2 4 | 2+4 2 |
| | 6 2 | 6 2 | 3+4 2 | 3+4 2 | 2+2 4 | 3+5 2 |
| | 6 2 | 5 2 | 5+6 2 | 3+4 2 | 2+2 4 | 2+4 2 |
| | 5 2 | 5 2 | 3+4 2 | 5+6 2 | 2+2 4 | 3+5 2 |
| | 6 2 | 6 2 | 5+6 2 | 3+4 2 | | 3+5 2 |
| | 5 2 | 5 2 | 5+6 2 | 3+4 2 | | 2+4 2 |
| | 6 2 | 6 2 | 3+4 2 | 5+6 2 | | 3+5 2 |

Continued on next column

**TABLE XLVI– continued**

| TQC | ICSD | NoE | SF |
|---|---|---|---|
| | 6 2 6 2 | 5+6 2 5+6 2 | 2+4 2 |
| | 6 2 5 2 | 3+4 2 3+4 2 | 3+5 2 |
| | 5 2 5 2 | 5+6 2 5+6 2 | 2+4 2 |
| | 6 2 5 2 | 5+6 2 3+4 2 | 2+4 2 |
| | 5 2 6 2 | 5+6 2 5+6 2 | 3+5 2 |
| | 6 2 5 2 | 3+4 2 3+4 2 | 2+4 2 |
| | 5 2 6 2 | 5+6 2 5+6 2 | 3+5 2 |
| | 6 2 5 2 | 3+4 2 5+6 2 | 2+4 2 |
| | 5 2 6 2 | 5+6 2 5+6 2 | 3+5 2 |
| | 5 2 5 2 | 3+4 2 3+4 2 | 3+5 2 |
| | 6 2 6 2 | 3+4 2 3+4 2 | 2+4 2 |
| | 5 2 5 2 | 3+4 2 3+4 2 | 2+4 2 |
| | 5 2 6 2 | 3+4 2 5+6 2 | 3+5 2 |

The End

TABLE XLVII: The computed irreducible representations for the valence bands for the compounds of SEBR and NLC in space group 82. See the list of tables.

| TQC | ICSD | | NoE | | SF | |
|---|---|---|---|---|---|---|
| 82.2.1 | 81539 | | 72 | | Te1Zr3 | |
| | -GM | -M | -P | -PA | -X | -N |
| | 6+8 2 | 5+7 2 | 5 1 | 5 1 | 3+4 2 | 2+2 2 |
| | 5+7 2 | 5+7 2 | 8 1 | 8 1 | 3+4 2 | 2+2 2 |
| | 6+8 2 | 6+8 2 | 5 1 | 5 1 | 3+4 2 | 2+2 2 |
| | 5+7 2 | 6+8 2 | 7 1 | 7 1 | 3+4 2 | 2+2 2 |
| | 6+8 2 | 6+8 2 | 6 1 | 6 1 | 3+4 2 | 2+2 2 |
| | 5+7 2 | 5+7 2 | 7 1 | 7 1 | 3+4 2 | 2+2 2 |
| | 6+8 2 | 6+8 2 | 8 1 | 8 1 | 3+4 2 | 2+2 2 |
| | 5+7 2 | 5+7 2 | 6 1 | 6 1 | 3+4 2 | 2+2 2 |
| | 5+7 2 | 5+7 2 | 8 1 | 8 1 | 3+4 2 | 2+2 2 |
| | 6+8 2 | 5+7 2 | 6 1 | 6 1 | 3+4 2 | 2+2 2 |
| | 5+7 2 | 6+8 2 | 8 1 | 8 1 | 3+4 2 | 2+2 2 |
| | 5+7 2 | 6+8 2 | 5 1 | 5 1 | 3+4 2 | 2+2 2 |
| | 6+8 2 | 5+7 2 | 6 1 | 6 1 | 3+4 2 | 2+2 2 |
| | 6+8 2 | 6+8 2 | 7 1 | 7 1 | 3+4 2 | 2+2 2 |
| | 5+7 2 | 5+7 2 | 7 1 | 7 1 | 3+4 2 | 2+2 2 |
| | 6+8 2 | 6+8 2 | 5 1 | 5 1 | 3+4 2 | 2+2 2 |
| | 6+8 2 | 6+8 2 | 7 1 | 7 1 | 3+4 2 | 2+2 2 |
| | 6+8 2 | 5+7 2 | 6 1 | 6 1 | 3+4 2 | 2+2 2 |
| | 5+7 2 | 5+7 2 | 7 1 | 7 1 | 3+4 2 | 2+2 2 |
| | 6+8 2 | 6+8 2 | 8 1 | 8 1 | 3+4 2 | 2+2 2 |
| | 6+8 2 | 5+7 2 | 5 1 | 5 1 | 3+4 2 | 2+2 2 |
| | 5+7 2 | 6+8 2 | 5 1 | 5 1 | 3+4 2 | 2+2 2 |
| | 5+7 2 | 6+8 2 | 6 1 | 6 1 | 3+4 2 | 2+2 2 |
| | 5+7 2 | 5+7 2 | 8 1 | 8 1 | 3+4 2 | 2+2 2 |
| | 6+8 2 | 6+8 2 | 8 1 | 8 1 | 3+4 2 | 2+2 2 |
| | 5+7 2 | 6+8 2 | 5 1 | 5 1 | 3+4 2 | 2+2 2 |
| | 6+8 2 | 5+7 2 | 7 1 | 7 1 | 3+4 2 | 2+2 2 |
| | 6+8 2 | 5+7 2 | 6 1 | 6 1 | 3+4 2 | 2+2 2 |
| | 5+7 2 | 6+8 2 | 5 1 | 5 1 | 3+4 2 | 2+2 2 |
| | 5+7 2 | 5+7 2 | 6 1 | 6 1 | 3+4 2 | 2+2 2 |
| | 6+8 2 | 6+8 2 | 8 1 | 8 1 | 3+4 2 | 2+2 2 |
| | 6+8 2 | 6+8 2 | 7 1 | 7 1 | 3+4 2 | 2+2 2 |
| | 5+7 2 | 5+7 2 | 6 1 | 6 1 | 3+4 2 | 2+2 2 |
| | 6+8 2 | 6+8 2 | 8 1 | 8 1 | 3+4 2 | 2+2 2 |
| | 5+7 2 | 5+7 2 | 5 1 | 5 1 | 3+4 2 | 2+2 2 |
| | 6+8 2 | 6+8 2 | 7 1 | 7 1 | 3+4 2 | 2+2 2 |
| | | | 5 1 | 5 1 | | |
| | | | 8 1 | 8 1 | | |

Continued on next column

**TABLE XLVII– continued**

| TQC | ICSD | | NoE | | SF | |
|---|---|---|---|---|---|---|
| | | | 6 1 | 6 1 | | |
| | | | 7 1 | 7 1 | | |
| | | | 6 1 | 6 1 | | |
| | | | 8 1 | 8 1 | | |
| | | | 6 1 | 6 1 | | |
| | | | 7 1 | 7 1 | | |
| | | | 5 1 | 5 1 | | |
| | | | 7 1 | 7 1 | | |
| | | | 5 1 | 5 1 | | |
| | | | 8 1 | 8 1 | | |
| | | | 7 1 | 7 1 | | |
| | | | 6 1 | 6 1 | | |
| | | | 5 1 | 5 1 | | |
| | | | 7 1 | 7 1 | | |
| | | | 5 1 | 5 1 | | |
| | | | 8 1 | 8 1 | | |
| | | | 6 1 | 6 1 | | |
| | | | 8 1 | 8 1 | | |
| | | | 7 1 | 7 1 | | |
| | | | 5 1 | 5 1 | | |
| | | | 6 1 | 6 1 | | |
| | | | 7 1 | 7 1 | | |
| | | | 8 1 | 8 1 | | |
| | | | 5 1 | 5 1 | | |
| | | | 6 1 | 6 1 | | |
| | | | 8 1 | 8 1 | | |
| | | | 7 1 | 7 1 | | |
| | | | 6 1 | 6 1 | | |
| | | | 7 1 | 7 1 | | |
| | | | 5 1 | 5 1 | | |
| | | | 5 1 | 5 1 | | |
| | | | 8 1 | 8 1 | | |
| | | | 8 1 | 8 1 | | |
| | | | 5 1 | 5 1 | | |
| 82.2.2 | 51048 | | 72 | | Te1Zr3 | |
| | -GM | -M | -P | -PA | -X | -N |
| | 6+8 2 | 5+7 2 | 6 1 | 6 1 | 3+4 2 | 2+2 2 |
| | 5+7 2 | 5+7 2 | 7 1 | 7 1 | 3+4 2 | 2+2 2 |
| | 6+8 2 | 6+8 2 | 5 1 | 5 1 | 3+4 2 | 2+2 2 |
| | 5+7 2 | 6+8 2 | 7 1 | 7 1 | 3+4 2 | 2+2 2 |
| | 6+8 2 | 6+8 2 | 5 1 | 5 1 | 3+4 2 | 2+2 2 |
| | 5+7 2 | 5+7 2 | 8 1 | 8 1 | 3+4 2 | 2+2 2 |
| | 6+8 2 | 6+8 2 | 6 1 | 6 1 | 3+4 2 | 2+2 2 |
| | 5+7 2 | 5+7 2 | 8 1 | 8 1 | 3+4 2 | 2+2 2 |
| | 5+7 2 | 5+7 2 | 6 1 | 6 1 | 3+4 2 | 2+2 2 |
| | 6+8 2 | 5+7 2 | 8 1 | 8 1 | 3+4 2 | 2+2 2 |
| | 5+7 2 | 6+8 2 | 6 1 | 6 1 | 3+4 2 | 2+2 2 |
| | 5+7 2 | 6+8 2 | 7 1 | 7 1 | 3+4 2 | 2+2 2 |
| | 6+8 2 | 5+7 2 | 5 1 | 5 1 | 3+4 2 | 2+2 2 |
| | 6+8 2 | 6+8 2 | 8 1 | 8 1 | 3+4 2 | 2+2 2 |
| | 5+7 2 | 5+7 2 | 5 1 | 5 1 | 3+4 2 | 2+2 2 |
| | 6+8 2 | 6+8 2 | 7 1 | 7 1 | 3+4 2 | 2+2 2 |
| | 6+8 2 | 6+8 2 | 5 1 | 5 1 | 3+4 2 | 2+2 2 |
| | 5+7 2 | 5+7 2 | 8 1 | 8 1 | 3+4 2 | 2+2 2 |
| | 6+8 2 | 5+7 2 | 5 1 | 5 1 | 3+4 2 | 2+2 2 |
| | 6+8 2 | 6+8 2 | 6 1 | 6 1 | 3+4 2 | 2+2 2 |
| | 6+8 2 | 5+7 2 | 7 1 | 7 1 | 3+4 2 | 2+2 2 |
| | 5+7 2 | 6+8 2 | 7 1 | 7 1 | 3+4 2 | 2+2 2 |
| | 5+7 2 | 5+7 2 | 8 1 | 8 1 | 3+4 2 | 2+2 2 |
| | 5+7 2 | 6+8 2 | 6 1 | 6 1 | 3+4 2 | 2+2 2 |
| | 6+8 2 | 6+8 2 | 6 1 | 6 1 | 3+4 2 | 2+2 2 |

Continued on next column

**TABLE XLVII– continued**

| TQC | ICSD | | NoE | | SF | |
|---|---|---|---|---|---|---|
| | 5+7 2 | 6+8 2 | 7 1 | 7 1 | 3+4 2 | 2+2 2 |
| | 6+8 2 | 5+7 2 | 5 1 | 5 1 | 3+4 2 | 2+2 2 |
| | 5+7 2 | 5+7 2 | 8 1 | 8 1 | 3+4 2 | 2+2 2 |
| | 6+8 2 | 6+8 2 | 7 1 | 7 1 | 3+4 2 | 2+2 2 |
| | 5+7 2 | 5+7 2 | 8 1 | 8 1 | 3+4 2 | 2+2 2 |
| | 6+8 2 | 6+8 2 | 6 1 | 6 1 | 3+4 2 | 2+2 2 |
| | 6+8 2 | 6+8 2 | 5 1 | 5 1 | 3+4 2 | 2+2 2 |
| | 5+7 2 | 5+7 2 | 8 1 | 8 1 | 3+4 2 | 2+2 2 |
| | 6+8 2 | 6+8 2 | 6 1 | 6 1 | 3+4 2 | 2+2 2 |
| | 5+7 2 | 5+7 2 | 7 1 | 7 1 | 3+4 2 | 2+2 2 |
| | 6+8 2 | 6+8 2 | 5 1 | 5 1 | 3+4 2 | 2+2 2 |
| | | | 6 1 | 6 1 | | |
| | | | 7 1 | 7 1 | | |
| | | | 8 1 | 8 1 | | |
| | | | 5 1 | 5 1 | | |
| | | | 8 1 | 8 1 | | |
| | | | 6 1 | 6 1 | | |
| | | | 5 1 | 5 1 | | |
| | | | 8 1 | 8 1 | | |
| | | | 7 1 | 7 1 | | |
| | | | 5 1 | 5 1 | | |
| | | | 7 1 | 7 1 | | |
| | | | 6 1 | 6 1 | | |
| | | | 5 1 | 5 1 | | |
| | | | 8 1 | 8 1 | | |
| | | | 7 1 | 7 1 | | |
| | | | 5 1 | 5 1 | | |
| | | | 7 1 | 7 1 | | |
| | | | 6 1 | 6 1 | | |
| | | | 8 1 | 8 1 | | |
| | | | 6 1 | 6 1 | | |
| | | | 5 1 | 5 1 | | |
| | | | 7 1 | 7 1 | | |
| | | | 8 1 | 8 1 | | |
| | | | 5 1 | 5 1 | | |
| | | | 6 1 | 6 1 | | |
| | | | 7 1 | 7 1 | | |
| | | | 8 1 | 8 1 | | |
| | | | 5 1 | 5 1 | | |
| | | | 6 1 | 6 1 | | |
| | | | 8 1 | 8 1 | | |
| | | | 5 1 | 5 1 | | |
| | | | 7 1 | 7 1 | | |
| | | | 7 1 | 7 1 | | |
| | | | 6 1 | 6 1 | | |
| | | | 6 1 | 6 1 | | |
| | | | 7 1 | 7 1 | | |
| 82.2.3 | 195057 | | 68 | | Sb1Zr3 | |
| | -GM | -M | -P | -PA | -X | -N |
| | 6+8 2 | 5+7 2 | 5 1 | 5 1 | 3+4 2 | 2+2 2 |
| | 5+7 2 | 5+7 2 | 8 1 | 8 1 | 3+4 2 | 2+2 2 |
| | 6+8 2 | 6+8 2 | 5 1 | 5 1 | 3+4 2 | 2+2 2 |
| | 5+7 2 | 6+8 2 | 7 1 | 7 1 | 3+4 2 | 2+2 2 |
| | 6+8 2 | 6+8 2 | 6 1 | 6 1 | 3+4 2 | 2+2 2 |
| | 5+7 2 | 6+8 2 | 7 1 | 7 1 | 3+4 2 | 2+2 2 |
| | 6+8 2 | 5+7 2 | 8 1 | 8 1 | 3+4 2 | 2+2 2 |
| | 5+7 2 | 5+7 2 | 6 1 | 6 1 | 3+4 2 | 2+2 2 |
| | 6+8 2 | 5+7 2 | 8 1 | 8 1 | 3+4 2 | 2+2 2 |
| | 5+7 2 | 5+7 2 | 6 1 | 6 1 | 3+4 2 | 2+2 2 |
| | 5+7 2 | 6+8 2 | 8 1 | 8 1 | 3+4 2 | 2+2 2 |
| | 5+7 2 | 6+8 2 | 5 1 | 5 1 | 3+4 2 | 2+2 2 |

Continued on next column

**TABLE XLVII– continued**

| TQC | ICSD | | NoE | | SF | |
|---|---|---|---|---|---|---|
| | 6+8 2 | 5+7 2 | 7 1 | 7 1 | 3+4 2 | 2+2 2 |
| | 5+7 2 | 6+8 2 | 6 1 | 6 1 | 3+4 2 | 2+2 2 |
| | 6+8 2 | 5+7 2 | 7 1 | 7 1 | 3+4 2 | 2+2 2 |
| | 6+8 2 | 6+8 2 | 5 1 | 5 1 | 3+4 2 | 2+2 2 |
| | 6+8 2 | 6+8 2 | 7 1 | 7 1 | 3+4 2 | 2+2 2 |
| | 6+8 2 | 5+7 2 | 6 1 | 6 1 | 3+4 2 | 2+2 2 |
| | 5+7 2 | 5+7 2 | 8 1 | 8 1 | 3+4 2 | 2+2 2 |
| | 6+8 2 | 6+8 2 | 5 1 | 5 1 | 3+4 2 | 2+2 2 |
| | 5+7 2 | 5+7 2 | 7 1 | 7 1 | 3+4 2 | 2+2 2 |
| | 6+8 2 | 6+8 2 | 5 1 | 5 1 | 3+4 2 | 2+2 2 |
| | 5+7 2 | 6+8 2 | 6 1 | 6 1 | 3+4 2 | 2+2 2 |
| | 5+7 2 | 5+7 2 | 8 1 | 8 1 | 3+4 2 | 2+2 2 |
| | 6+8 2 | 6+8 2 | 8 1 | 8 1 | 3+4 2 | 2+2 2 |
| | 6+8 2 | 6+8 2 | 5 1 | 5 1 | 3+4 2 | 2+2 2 |
| | 5+7 2 | 5+7 2 | 7 1 | 7 1 | 3+4 2 | 2+2 2 |
| | 6+8 2 | 5+7 2 | 5 1 | 5 1 | 3+4 2 | 2+2 2 |
| | 5+7 2 | 6+8 2 | 6 1 | 6 1 | 3+4 2 | 2+2 2 |
| | 5+7 2 | 6+8 2 | 8 1 | 8 1 | 3+4 2 | 2+2 2 |
| | 6+8 2 | 5+7 2 | 6 1 | 6 1 | 3+4 2 | 2+2 2 |
| | 5+7 2 | 5+7 2 | 7 1 | 7 1 | 3+4 2 | 2+2 2 |
| | 6+8 2 | 6+8 2 | 6 1 | 6 1 | 3+4 2 | 2+2 2 |
| | 6+8 2 | 6+8 2 | 8 1 | 8 1 | 3+4 2 | 2+2 2 |
| | | | 5 1 | 5 1 | | |
| | | | 7 1 | 7 1 | | |
| | | | 8 1 | 8 1 | | |
| | | | 5 1 | 5 1 | | |
| | | | 6 1 | 6 1 | | |
| | | | 7 1 | 7 1 | | |
| | | | 6 1 | 6 1 | | |
| | | | 8 1 | 8 1 | | |
| | | | 7 1 | 7 1 | | |
| | | | 6 1 | 6 1 | | |
| | | | 5 1 | 5 1 | | |
| | | | 7 1 | 7 1 | | |
| | | | 6 1 | 6 1 | | |
| | | | 7 1 | 7 1 | | |
| | | | 5 1 | 5 1 | | |
| | | | 8 1 | 8 1 | | |
| | | | 5 1 | 5 1 | | |
| | | | 7 1 | 7 1 | | |
| | | | 5 1 | 5 1 | | |
| | | | 8 1 | 8 1 | | |
| | | | 6 1 | 6 1 | | |
| | | | 8 1 | 8 1 | | |
| | | | 7 1 | 7 1 | | |
| | | | 5 1 | 5 1 | | |
| | | | 6 1 | 6 1 | | |
| | | | 7 1 | 7 1 | | |
| | | | 8 1 | 8 1 | | |
| | | | 5 1 | 5 1 | | |
| | | | 6 1 | 6 1 | | |
| | | | 8 1 | 8 1 | | |
| | | | 7 1 | 7 1 | | |
| | | | 6 1 | 6 1 | | |
| | | | 7 1 | 7 1 | | |
| | | | 5 1 | 5 1 | | |
| 82.2.4 | 98373 | | 140 | | Ni3P1 | |
| | -GM | -M | -P | -PA | -X | -N |
| | 6+8 2 | 5+7 2 | 5 1 | 5 1 | 3+4 2 | 2+2 2 |
| | 5+7 2 | 5+7 2 | 8 1 | 8 1 | 3+4 2 | 2+2 2 |
| | 6+8 2 | 6+8 2 | 5 1 | 5 1 | 3+4 2 | 2+2 2 |

Continued on next column

**TABLE XLVII– continued**

| TQC | ICSD | NoE | SF |
|---|---|---|---|
| | 5+7 2 6+8 2 | 7 1 | 7 1 3+4 2 2+2 2 |
| | 6+8 2 6+8 2 | 6 1 | 6 1 3+4 2 2+2 2 |
| | 5+7 2 5+7 2 | 7 1 | 7 1 3+4 2 2+2 2 |
| | 5+7 2 6+8 2 | 8 1 | 8 1 3+4 2 2+2 2 |
| | 6+8 2 5+7 2 | 6 1 | 6 1 3+4 2 2+2 2 |
| | 5+7 2 5+7 2 | 6 1 | 6 1 3+4 2 2+2 2 |
| | 6+8 2 6+8 2 | 8 1 | 8 1 3+4 2 2+2 2 |
| | 5+7 2 5+7 2 | 6 1 | 6 1 3+4 2 2+2 2 |
| | 6+8 2 6+8 2 | 7 1 | 7 1 3+4 2 2+2 2 |
| | 5+7 2 6+8 2 | 5 1 | 5 1 3+4 2 2+2 2 |
| | 6+8 2 5+7 2 | 8 1 | 8 1 3+4 2 2+2 2 |
| | 5+7 2 5+7 2 | 5 1 | 5 1 3+4 2 2+2 2 |
| | 6+8 2 6+8 2 | 7 1 | 7 1 3+4 2 2+2 2 |
| | 6+8 2 6+8 2 | 6 1 | 6 1 3+4 2 2+2 2 |
| | 6+8 2 5+7 2 | 7 1 | 7 1 3+4 2 2+2 2 |
| | 5+7 2 5+7 2 | 5 1 | 5 1 3+4 2 2+2 2 |
| | 6+8 2 6+8 2 | 7 1 | 7 1 3+4 2 2+2 2 |
| | 5+7 2 5+7 2 | 8 1 | 8 1 3+4 2 2+2 2 |
| | 6+8 2 6+8 2 | 5 1 | 5 1 3+4 2 2+2 2 |
| | 5+7 2 5+7 2 | 6 1 | 6 1 3+4 2 2+2 2 |
| | 5+7 2 6+8 2 | 8 1 | 8 1 3+4 2 2+2 2 |
| | 6+8 2 5+7 2 | 5 1 | 5 1 3+4 2 2+2 2 |
| | 6+8 2 6+8 2 | 8 1 | 8 1 3+4 2 2+2 2 |
| | 5+7 2 6+8 2 | 5 1 | 5 1 3+4 2 2+2 2 |
| | 5+7 2 5+7 2 | 7 1 | 7 1 3+4 2 2+2 2 |
| | 6+8 2 6+8 2 | 8 1 | 8 1 3+4 2 2+2 2 |
| | 5+7 2 5+7 2 | 6 1 | 6 1 3+4 2 2+2 2 |
| | 5+7 2 6+8 2 | 7 1 | 7 1 3+4 2 2+2 2 |
| | 6+8 2 5+7 2 | 6 1 | 6 1 3+4 2 2+2 2 |
| | 6+8 2 6+8 2 | 8 1 | 8 1 3+4 2 2+2 2 |
| | 6+8 2 6+8 2 | 6 1 | 6 1 3+4 2 2+2 2 |
| | 5+7 2 5+7 2 | 8 1 | 8 1 3+4 2 2+2 2 |
| | 6+8 2 5+7 2 | 5 1 | 5 1 3+4 2 2+2 2 |
| | 5+7 2 5+7 2 | 7 1 | 7 1 3+4 2 2+2 2 |
| | 6+8 2 6+8 2 | 5 1 | 5 1 3+4 2 2+2 2 |
| | 6+8 2 5+7 2 | 7 1 | 7 1 3+4 2 2+2 2 |
| | 5+7 2 5+7 2 | 6 1 | 6 1 3+4 2 2+2 2 |
| | 5+7 2 6+8 2 | 7 1 | 7 1 3+4 2 2+2 2 |
| | 6+8 2 6+8 2 | 5 1 | 5 1 3+4 2 2+2 2 |
| | 6+8 2 5+7 2 | 6 1 | 6 1 3+4 2 2+2 2 |
| | 6+8 2 5+7 2 | 8 1 | 8 1 3+4 2 2+2 2 |
| | 5+7 2 6+8 2 | 6 1 | 6 1 3+4 2 2+2 2 |
| | 5+7 2 6+8 2 | 7 1 | 7 1 3+4 2 2+2 2 |
| | 6+8 2 5+7 2 | 5 1 | 5 1 3+4 2 2+2 2 |
| | 5+7 2 5+7 2 | 8 1 | 8 1 3+4 2 2+2 2 |
| | 6+8 2 6+8 2 | 6 1 | 6 1 3+4 2 2+2 2 |
| | 5+7 2 6+8 2 | 7 1 | 7 1 3+4 2 2+2 2 |
| | 5+7 2 5+7 2 | 5 1 | 5 1 3+4 2 2+2 2 |
| | 6+8 2 6+8 2 | 7 1 | 7 1 3+4 2 2+2 2 |
| | 5+7 2 6+8 2 | 6 1 | 6 1 3+4 2 2+2 2 |
| | 6+8 2 5+7 2 | 8 1 | 8 1 3+4 2 2+2 2 |
| | 5+7 2 6+8 2 | 8 1 | 8 1 3+4 2 2+2 2 |
| | 6+8 2 6+8 2 | 5 1 | 5 1 3+4 2 2+2 2 |
| | 5+7 2 5+7 2 | 6 1 | 6 1 3+4 2 2+2 2 |
| | 6+8 2 5+7 2 | 7 1 | 7 1 3+4 2 2+2 2 |
| | 5+7 2 6+8 2 | 5 1 | 5 1 3+4 2 2+2 2 |
| | 6+8 2 5+7 2 | 7 1 | 7 1 3+4 2 2+2 2 |
| | 5+7 2 6+8 2 | 5 1 | 5 1 3+4 2 2+2 2 |
| | 6+8 2 6+8 2 | 8 1 | 8 1 3+4 2 2+2 2 |
| | 6+8 2 5+7 2 | 6 1 | 6 1 3+4 2 2+2 2 |
| | 5+7 2 5+7 2 | 7 1 | 7 1 3+4 2 2+2 2 |

Continued on next column

**TABLE XLVII– continued**

| TQC | ICSD | NoE | SF |
|---|---|---|---|
| | 6+8 2 5+7 2 | 8 1 | 8 1 3+4 2 2+2 2 |
| | 5+7 2 6+8 2 | 5 1 | 5 1 3+4 2 2+2 2 |
| | 5+7 2 6+8 2 | 8 1 | 8 1 3+4 2 2+2 2 |
| | 5+7 2 5+7 2 | 6 1 | 6 1 3+4 2 2+2 2 |
| | 6+8 2 6+8 2 | 8 1 | 8 1 3+4 2 2+2 2 |
| | 6+8 2 5+7 2 | 5 1 | 5 1 3+4 2 2+2 2 |
| | | 6 1 | 6 1 |
| | | 5 1 | 5 1 |
| | | 7 1 | 7 1 |
| | | 8 1 | 8 1 |
| | | 6 1 | 6 1 |
| | | 7 1 | 7 1 |
| | | 5 1 | 5 1 |
| | | 7 1 | 7 1 |
| | | 8 1 | 8 1 |
| | | 5 1 | 5 1 |
| | | 6 1 | 6 1 |
| | | 8 1 | 8 1 |
| | | 7 1 | 7 1 |
| | | 6 1 | 6 1 |
| | | 6 1 | 6 1 |
| | | 8 1 | 8 1 |
| | | 5 1 | 5 1 |
| | | 7 1 | 7 1 |
| | | 8 1 | 8 1 |
| | | 6 1 | 6 1 |
| | | 7 1 | 7 1 |
| | | 5 1 | 5 1 |
| | | 8 1 | 8 1 |
| | | 5 1 | 5 1 |
| | | 5 1 | 5 1 |
| | | 8 1 | 8 1 |
| | | 7 1 | 7 1 |
| | | 6 1 | 6 1 |
| | | 5 1 | 5 1 |
| | | 7 1 | 7 1 |
| | | 6 1 | 6 1 |
| | | 8 1 | 8 1 |
| | | 5 1 | 5 1 |
| | | 8 1 | 8 1 |
| | | 7 1 | 7 1 |
| | | 6 1 | 6 1 |
| | | 7 1 | 7 1 |
| | | 8 1 | 8 1 |
| | | 6 1 | 6 1 |
| | | 5 1 | 5 1 |
| | | 7 1 | 7 1 |
| | | 6 1 | 6 1 |
| | | 5 1 | 5 1 |
| | | 7 1 | 7 1 |
| | | 6 1 | 6 1 |
| | | 8 1 | 8 1 |
| | | 5 1 | 5 1 |
| | | 6 1 | 6 1 |
| | | 7 1 | 7 1 |
| | | 8 1 | 8 1 |
| | | 5 1 | 5 1 |
| | | 7 1 | 7 1 |
| | | 8 1 | 8 1 |
| | | 6 1 | 6 1 |
| | | 8 1 | 8 1 |

Continued on next column

**TABLE XLVII– continued**

| TQC | ICSD | | NoE | | SF | |
|---|---|---|---|---|---|---|
| | | | 6 1 | 6 1 | | |
| | | | 7 1 | 7 1 | | |
| | | | 5 1 | 5 1 | | |
| | | | 6 1 | 6 1 | | |
| | | | 8 1 | 8 1 | | |
| | | | 5 1 | 5 1 | | |
| | | | 7 1 | 7 1 | | |
| | | | 8 1 | 8 1 | | |
| | | | 5 1 | 5 1 | | |
| | | | 7 1 | 7 1 | | |
| | | | 6 1 | 6 1 | | |
| | | | 6 1 | 6 1 | | |
| | | | 7 1 | 7 1 | | |
| | | | 8 1 | 8 1 | | |
| | | | 5 1 | 5 1 | | |
| 82.2.5 | 23612 | | 32 | | In1P1S4 | |
| | -GM | -M | -P | -PA | -X | -N |
| | 6+8 2 | 5+7 2 | 7 1 | 7 1 | 3+4 2 | 2+2 2 |
| | 5+7 2 | 6+8 2 | 6 1 | 6 1 | 3+4 2 | 2+2 2 |
| | 6+8 2 | 5+7 2 | 6 1 | 6 1 | 3+4 2 | 2+2 2 |
| | 5+7 2 | 6+8 2 | 8 1 | 8 1 | 3+4 2 | 2+2 2 |
| | 6+8 2 | 5+7 2 | 5 1 | 5 1 | 3+4 2 | 2+2 2 |
| | 6+8 2 | 6+8 2 | 8 1 | 8 1 | 3+4 2 | 2+2 2 |
| | 5+7 2 | 6+8 2 | 5 1 | 5 1 | 3+4 2 | 2+2 2 |
| | 6+8 2 | 5+7 2 | 7 1 | 7 1 | 3+4 2 | 2+2 2 |
| | 5+7 2 | 6+8 2 | 7 1 | 7 1 | 3+4 2 | 2+2 2 |
| | 5+7 2 | 5+7 2 | 6 1 | 6 1 | 3+4 2 | 2+2 2 |
| | 5+7 2 | 6+8 2 | 6 1 | 6 1 | 3+4 2 | 2+2 2 |
| | 6+8 2 | 5+7 2 | 8 1 | 8 1 | 3+4 2 | 2+2 2 |
| | 6+8 2 | 6+8 2 | 5 1 | 5 1 | 3+4 2 | 2+2 2 |
| | 6+8 2 | 5+7 2 | 8 1 | 8 1 | 3+4 2 | 2+2 2 |
| | 5+7 2 | 6+8 2 | 7 1 | 7 1 | 3+4 2 | 2+2 2 |
| | 6+8 2 | 5+7 2 | 5 1 | 5 1 | 3+4 2 | 2+2 2 |
| | | | 8 1 | 8 1 | | |
| | | | 6 1 | 6 1 | | |
| | | | 5 1 | 5 1 | | |
| | | | 7 1 | 7 1 | | |
| | | | 7 1 | 7 1 | | |
| | | | 6 1 | 6 1 | | |
| | | | 5 1 | 5 1 | | |
| | | | 8 1 | 8 1 | | |
| | | | 7 1 | 7 1 | | |
| | | | 6 1 | 6 1 | | |
| | | | 5 1 | 5 1 | | |
| | | | 8 1 | 8 1 | | |
| | | | 7 1 | 7 1 | | |
| | | | 5 1 | 5 1 | | |
| | | | 8 1 | 8 1 | | |
| | | | 6 1 | 6 1 | | |
| 82.2.6 | 56027 | | 76 | | Ge1Ta3 | |
| | -GM | -M | -P | -PA | -X | -N |
| | 6+8 2 | 5+7 2 | 5 1 | 5 1 | 3+4 2 | 2+2 2 |
| | 5+7 2 | 6+8 2 | 8 1 | 8 1 | 3+4 2 | 2+2 2 |
| | 6+8 2 | 5+7 2 | 5 1 | 5 1 | 3+4 2 | 2+2 2 |
| | 5+7 2 | 6+8 2 | 7 1 | 7 1 | 3+4 2 | 2+2 2 |
| | 6+8 2 | 6+8 2 | 7 1 | 7 1 | 3+4 2 | 2+2 2 |
| | 5+7 2 | 5+7 2 | 6 1 | 6 1 | 3+4 2 | 2+2 2 |
| | 5+7 2 | 6+8 2 | 8 1 | 8 1 | 3+4 2 | 2+2 2 |
| | 6+8 2 | 5+7 2 | 6 1 | 6 1 | 3+4 2 | 2+2 2 |
| | 5+7 2 | 5+7 2 | 8 1 | 8 1 | 3+4 2 | 2+2 2 |
| | 6+8 2 | 5+7 2 | 6 1 | 6 1 | 3+4 2 | 2+2 2 |

Continued on next column

**TABLE XLVII– continued**

| TQC | ICSD | | NoE | | SF | |
|---|---|---|---|---|---|---|
| | 5+7 2 | 6+8 2 | 7 1 | 7 1 | 3+4 2 | 2+2 2 |
| | 5+7 2 | 6+8 2 | 6 1 | 6 1 | 3+4 2 | 2+2 2 |
| | 6+8 2 | 5+7 2 | 5 1 | 5 1 | 3+4 2 | 2+2 2 |
| | 6+8 2 | 6+8 2 | 8 1 | 8 1 | 3+4 2 | 2+2 2 |
| | 5+7 2 | 5+7 2 | 7 1 | 7 1 | 3+4 2 | 2+2 2 |
| | 6+8 2 | 6+8 2 | 5 1 | 5 1 | 3+4 2 | 2+2 2 |
| | 6+8 2 | 5+7 2 | 6 1 | 6 1 | 3+4 2 | 2+2 2 |
| | 6+8 2 | 6+8 2 | 7 1 | 7 1 | 3+4 2 | 2+2 2 |
| | 6+8 2 | 5+7 2 | 8 1 | 8 1 | 3+4 2 | 2+2 2 |
| | 5+7 2 | 6+8 2 | 6 1 | 6 1 | 3+4 2 | 2+2 2 |
| | 5+7 2 | 5+7 2 | 5 1 | 5 1 | 3+4 2 | 2+2 2 |
| | 6+8 2 | 6+8 2 | 7 1 | 7 1 | 3+4 2 | 2+2 2 |
| | 6+8 2 | 6+8 2 | 8 1 | 8 1 | 3+4 2 | 2+2 2 |
| | 5+7 2 | 5+7 2 | 5 1 | 5 1 | 3+4 2 | 2+2 2 |
| | 5+7 2 | 6+8 2 | 5 1 | 5 1 | 3+4 2 | 2+2 2 |
| | 6+8 2 | 6+8 2 | 7 1 | 7 1 | 3+4 2 | 2+2 2 |
| | 5+7 2 | 5+7 2 | 5 1 | 5 1 | 3+4 2 | 2+2 2 |
| | 6+8 2 | 6+8 2 | 8 1 | 8 1 | 3+4 2 | 2+2 2 |
| | 5+7 2 | 5+7 2 | 8 1 | 8 1 | 3+4 2 | 2+2 2 |
| | 5+7 2 | 6+8 2 | 6 1 | 6 1 | 3+4 2 | 2+2 2 |
| | 6+8 2 | 5+7 2 | 7 1 | 7 1 | 3+4 2 | 2+2 2 |
| | 6+8 2 | 5+7 2 | 6 1 | 6 1 | 3+4 2 | 2+2 2 |
| | 5+7 2 | 6+8 2 | 5 1 | 5 1 | 3+4 2 | 2+2 2 |
| | 6+8 2 | 5+7 2 | 7 1 | 7 1 | 3+4 2 | 2+2 2 |
| | 5+7 2 | 6+8 2 | 6 1 | 6 1 | 3+4 2 | 2+2 2 |
| | 6+8 2 | 5+7 2 | 8 1 | 8 1 | 3+4 2 | 2+2 2 |
| | 5+7 2 | 5+7 2 | 5 1 | 5 1 | 3+4 2 | 2+2 2 |
| | 6+8 2 | 6+8 2 | 8 1 | 8 1 | 3+4 2 | 2+2 2 |
| | | | 7 1 | 7 1 | | |
| | | | 6 1 | 6 1 | | |
| | | | 8 1 | 8 1 | | |
| | | | 6 1 | 6 1 | | |
| | | | 7 1 | 7 1 | | |
| | | | 6 1 | 6 1 | | |
| | | | 5 1 | 5 1 | | |
| | | | 7 1 | 7 1 | | |
| | | | 5 1 | 5 1 | | |
| | | | 8 1 | 8 1 | | |
| | | | 6 1 | 6 1 | | |
| | | | 7 1 | 7 1 | | |
| | | | 7 1 | 7 1 | | |
| | | | 5 1 | 5 1 | | |
| | | | 5 1 | 5 1 | | |
| | | | 6 1 | 6 1 | | |
| | | | 8 1 | 8 1 | | |
| | | | 8 1 | 8 1 | | |
| | | | 7 1 | 7 1 | | |
| | | | 6 1 | 6 1 | | |
| | | | 5 1 | 5 1 | | |
| | | | 7 1 | 7 1 | | |
| | | | 8 1 | 8 1 | | |
| | | | 5 1 | 5 1 | | |
| | | | 8 1 | 8 1 | | |
| | | | 6 1 | 6 1 | | |
| | | | 7 1 | 7 1 | | |
| | | | 6 1 | 6 1 | | |
| | | | 5 1 | 5 1 | | |
| | | | 7 1 | 7 1 | | |
| | | | 8 1 | 8 1 | | |
| | | | 6 1 | 6 1 | | |
| | | | 5 1 | 5 1 | | |

Continued on next column

**TABLE XLVII– continued**

| TQC | ICSD | | NoE | | SF | |
|---|---|---|---|---|---|---|
| | | | 8 1 | 8 1 | | |
| | | | 5 1 | 5 1 | | |
| | | | 8 1 | 8 1 | | |
| | | | 6 1 | 6 1 | | |
| | | | 7 1 | 7 1 | | |
| 82.2.7 | 43365 | | 116 | | Fe3P1 | |
| | -GM | -M | -P | -PA | -X | -N |
| | 6+8 2 | 5+7 2 | 6 1 | 6 1 | 3+4 2 | 2+2 2 |
| | 5+7 2 | 5+7 2 | 7 1 | 7 1 | 3+4 2 | 2+2 2 |
| | 6+8 2 | 6+8 2 | 5 1 | 5 1 | 3+4 2 | 2+2 2 |
| | 5+7 2 | 6+8 2 | 7 1 | 7 1 | 3+4 2 | 2+2 2 |
| | 6+8 2 | 6+8 2 | 5 1 | 5 1 | 3+4 2 | 2+2 2 |
| | 5+7 2 | 6+8 2 | 8 1 | 8 1 | 3+4 2 | 2+2 2 |
| | 6+8 2 | 5+7 2 | 6 1 | 6 1 | 3+4 2 | 2+2 2 |
| | 5+7 2 | 5+7 2 | 8 1 | 8 1 | 3+4 2 | 2+2 2 |
| | 5+7 2 | 5+7 2 | 6 1 | 6 1 | 3+4 2 | 2+2 2 |
| | 6+8 2 | 5+7 2 | 8 1 | 8 1 | 3+4 2 | 2+2 2 |
| | 5+7 2 | 6+8 2 | 6 1 | 6 1 | 3+4 2 | 2+2 2 |
| | 6+8 2 | 6+8 2 | 7 1 | 7 1 | 3+4 2 | 2+2 2 |
| | 5+7 2 | 5+7 2 | 5 1 | 5 1 | 3+4 2 | 2+2 2 |
| | 6+8 2 | 6+8 2 | 8 1 | 8 1 | 3+4 2 | 2+2 2 |
| | 5+7 2 | 5+7 2 | 5 1 | 5 1 | 3+4 2 | 2+2 2 |
| | 6+8 2 | 6+8 2 | 7 1 | 7 1 | 3+4 2 | 2+2 2 |
| | 6+8 2 | 6+8 2 | 5 1 | 5 1 | 3+4 2 | 2+2 2 |
| | 5+7 2 | 5+7 2 | 8 1 | 8 1 | 3+4 2 | 2+2 2 |
| | 6+8 2 | 5+7 2 | 5 1 | 5 1 | 3+4 2 | 2+2 2 |
| | 5+7 2 | 6+8 2 | 7 1 | 7 1 | 3+4 2 | 2+2 2 |
| | 6+8 2 | 5+7 2 | 6 1 | 6 1 | 3+4 2 | 2+2 2 |
| | 6+8 2 | 6+8 2 | 7 1 | 7 1 | 3+4 2 | 2+2 2 |
| | 5+7 2 | 5+7 2 | 6 1 | 6 1 | 3+4 2 | 2+2 2 |
| | 6+8 2 | 6+8 2 | 8 1 | 8 1 | 3+4 2 | 2+2 2 |
| | 5+7 2 | 6+8 2 | 6 1 | 6 1 | 3+4 2 | 2+2 2 |
| | 6+8 2 | 5+7 2 | 7 1 | 7 1 | 3+4 2 | 2+2 2 |
| | 6+8 2 | 6+8 2 | 5 1 | 5 1 | 3+4 2 | 2+2 2 |
| | 5+7 2 | 5+7 2 | 7 1 | 7 1 | 3+4 2 | 2+2 2 |
| | 5+7 2 | 6+8 2 | 6 1 | 6 1 | 3+4 2 | 2+2 2 |
| | 5+7 2 | 5+7 2 | 8 1 | 8 1 | 3+4 2 | 2+2 2 |
| | 6+8 2 | 6+8 2 | 5 1 | 5 1 | 3+4 2 | 2+2 2 |
| | 6+8 2 | 5+7 2 | 8 1 | 8 1 | 3+4 2 | 2+2 2 |
| | 5+7 2 | 6+8 2 | 6 1 | 6 1 | 3+4 2 | 2+2 2 |
| | 6+8 2 | 6+8 2 | 8 1 | 8 1 | 3+4 2 | 2+2 2 |
| | 5+7 2 | 5+7 2 | 5 1 | 5 1 | 3+4 2 | 2+2 2 |
| | 6+8 2 | 5+7 2 | 7 1 | 7 1 | 3+4 2 | 2+2 2 |
| | 5+7 2 | 5+7 2 | 6 1 | 6 1 | 3+4 2 | 2+2 2 |
| | 6+8 2 | 5+7 2 | 7 1 | 7 1 | 3+4 2 | 2+2 2 |
| | 5+7 2 | 6+8 2 | 5 1 | 5 1 | 3+4 2 | 2+2 2 |
| | 6+8 2 | 6+8 2 | 8 1 | 8 1 | 3+4 2 | 2+2 2 |
| | 5+7 2 | 6+8 2 | 8 1 | 8 1 | 3+4 2 | 2+2 2 |
| | 6+8 2 | 5+7 2 | 6 1 | 6 1 | 3+4 2 | 2+2 2 |
| | 6+8 2 | 5+7 2 | 5 1 | 5 1 | 3+4 2 | 2+2 2 |
| | 6+8 2 | 5+7 2 | 7 1 | 7 1 | 3+4 2 | 2+2 2 |
| | 5+7 2 | 6+8 2 | 8 1 | 8 1 | 3+4 2 | 2+2 2 |
| | 5+7 2 | 6+8 2 | 5 1 | 5 1 | 3+4 2 | 2+2 2 |
| | 6+8 2 | 5+7 2 | 5 1 | 5 1 | 3+4 2 | 2+2 2 |
| | 5+7 2 | 6+8 2 | 7 1 | 7 1 | 3+4 2 | 2+2 2 |
| | 5+7 2 | 5+7 2 | 8 1 | 8 1 | 3+4 2 | 2+2 2 |
| | 6+8 2 | 6+8 2 | 5 1 | 5 1 | 3+4 2 | 2+2 2 |
| | 5+7 2 | 5+7 2 | 6 1 | 7 1 | 3+4 2 | 2+2 2 |
| | 6+8 2 | 6+8 2 | 7 1 | 6 1 | 3+4 2 | 2+2 2 |
| | 5+7 2 | 6+8 2 | 6 1 | 6 1 | 3+4 2 | 2+2 2 |
| | 6+8 2 | 5+7 2 | 7 1 | 7 1 | 3+4 2 | 2+2 2 |

Continued on next column

**TABLE XLVII– continued**

| TQC | ICSD | | NoE | | SF | |
|---|---|---|---|---|---|---|
| | 5+7 2 | 5+7 2 | 6 1 | 6 1 | 3+4 2 | 2+2 2 |
| | 6+8 2 | 6+8 2 | 8 1 | 8 1 | 3+4 2 | 2+2 2 |
| | 5+7 2 | 6+8 2 | 5 1 | 5 1 | 3+4 2 | 2+2 2 |
| | 6+8 2 | 5+7 2 | 8 1 | 8 1 | 3+4 2 | 2+2 2 |
| | | | 5 1 | 5 1 | | |
| | | | 7 1 | 7 1 | | |
| | | | 7 1 | 7 1 | | |
| | | | 6 1 | 6 1 | | |
| | | | 5 1 | 5 1 | | |
| | | | 8 1 | 8 1 | | |
| | | | 8 1 | 8 1 | | |
| | | | 6 1 | 6 1 | | |
| | | | 7 1 | 7 1 | | |
| | | | 6 1 | 6 1 | | |
| | | | 6 1 | 6 1 | | |
| | | | 5 1 | 5 1 | | |
| | | | 7 1 | 7 1 | | |
| | | | 8 1 | 8 1 | | |
| | | | 7 1 | 7 1 | | |
| | | | 5 1 | 5 1 | | |
| | | | 6 1 | 6 1 | | |
| | | | 7 1 | 7 1 | | |
| | | | 8 1 | 8 1 | | |
| | | | 5 1 | 5 1 | | |
| | | | 6 1 | 6 1 | | |
| | | | 7 1 | 7 1 | | |
| | | | 8 1 | 8 1 | | |
| | | | 6 1 | 6 1 | | |
| | | | 5 1 | 5 1 | | |
| | | | 8 1 | 8 1 | | |
| | | | 8 1 | 8 1 | | |
| | | | 6 1 | 6 1 | | |
| | | | 7 1 | 7 1 | | |
| | | | 5 1 | 5 1 | | |
| | | | 6 1 | 6 1 | | |
| | | | 8 1 | 8 1 | | |
| | | | 7 1 | 7 1 | | |
| | | | 5 1 | 5 1 | | |
| | | | 6 1 | 6 1 | | |
| | | | 7 1 | 7 1 | | |
| | | | 6 1 | 6 1 | | |
| | | | 7 1 | 7 1 | | |
| | | | 5 1 | 5 1 | | |
| | | | 8 1 | 8 1 | | |
| | | | 8 1 | 8 1 | | |
| | | | 6 1 | 6 1 | | |
| | | | 5 1 | 5 1 | | |
| | | | 7 1 | 7 1 | | |
| | | | 5 1 | 5 1 | | |
| | | | 7 1 | 7 1 | | |
| | | | 8 1 | 8 1 | | |
| | | | 6 1 | 6 1 | | |
| | | | 7 1 | 7 1 | | |
| | | | 6 1 | 6 1 | | |
| | | | 5 1 | 5 1 | | |
| | | | 8 1 | 8 1 | | |
| | | | 5 1 | 5 1 | | |
| | | | 8 1 | 8 1 | | |
| | | | 5 1 | 5 1 | | |
| | | | 7 1 | 7 1 | | |
| | | | 8 1 | 8 1 | | |

Continued on next column

**TABLE XLVII– continued**

| TQC | -GM | -M | -P | -PA | -X | -N |
|---|---|---|---|---|---|---|
| | | | 6 1 | 6 1 | | |
| 82.2.8 | 23560 | | 92 | | Cr3P1 | |
| | 6+8 2 | 5+7 2 | 6 1 | 6 1 | 3+4 2 | 2+2 2 |
| | 5+7 2 | 5+7 2 | 7 1 | 7 1 | 3+4 2 | 2+2 2 |
| | 6+8 2 | 6+8 2 | 5 1 | 5 1 | 3+4 2 | 2+2 2 |
| | 5+7 2 | 6+8 2 | 7 1 | 7 1 | 3+4 2 | 2+2 2 |
| | 6+8 2 | 6+8 2 | 5 1 | 5 1 | 3+4 2 | 2+2 2 |
| | 5+7 2 | 6+8 2 | 8 1 | 8 1 | 3+4 2 | 2+2 2 |
| | 6+8 2 | 5+7 2 | 6 1 | 6 1 | 3+4 2 | 2+2 2 |
| | 5+7 2 | 5+7 2 | 8 1 | 8 1 | 3+4 2 | 2+2 2 |
| | 6+8 2 | 5+7 2 | 6 1 | 6 1 | 3+4 2 | 2+2 2 |
| | 5+7 2 | 5+7 2 | 8 1 | 8 1 | 3+4 2 | 2+2 2 |
| | 5+7 2 | 6+8 2 | 5 1 | 5 1 | 3+4 2 | 2+2 2 |
| | 5+7 2 | 6+8 2 | 8 1 | 8 1 | 3+4 2 | 2+2 2 |
| | 6+8 2 | 5+7 2 | 6 1 | 6 1 | 3+4 2 | 2+2 2 |
| | 6+8 2 | 6+8 2 | 7 1 | 7 1 | 3+4 2 | 2+2 2 |
| | 5+7 2 | 5+7 2 | 5 1 | 5 1 | 3+4 2 | 2+2 2 |
| | 6+8 2 | 6+8 2 | 7 1 | 7 1 | 3+4 2 | 2+2 2 |
| | 6+8 2 | 6+8 2 | 5 1 | 5 1 | 3+4 2 | 2+2 2 |
| | 5+7 2 | 5+7 2 | 8 1 | 8 1 | 3+4 2 | 2+2 2 |
| | 6+8 2 | 5+7 2 | 5 1 | 5 1 | 3+4 2 | 2+2 2 |
| | 6+8 2 | 5+7 2 | 7 1 | 7 1 | 3+4 2 | 2+2 2 |
| | 5+7 2 | 6+8 2 | 6 1 | 6 1 | 3+4 2 | 2+2 2 |
| | 6+8 2 | 6+8 2 | 7 1 | 7 1 | 3+4 2 | 2+2 2 |
| | 5+7 2 | 6+8 2 | 6 1 | 6 1 | 3+4 2 | 2+2 2 |
| | 6+8 2 | 5+7 2 | 8 1 | 8 1 | 3+4 2 | 2+2 2 |
| | 5+7 2 | 6+8 2 | 7 1 | 7 1 | 3+4 2 | 2+2 2 |
| | 6+8 2 | 6+8 2 | 5 1 | 5 1 | 3+4 2 | 2+2 2 |
| | 5+7 2 | 5+7 2 | 6 1 | 6 1 | 3+4 2 | 2+2 2 |
| | 6+8 2 | 5+7 2 | 7 1 | 7 1 | 3+4 2 | 2+2 2 |
| | 5+7 2 | 6+8 2 | 8 1 | 8 1 | 3+4 2 | 2+2 2 |
| | 5+7 2 | 5+7 2 | 6 1 | 6 1 | 3+4 2 | 2+2 2 |
| | 6+8 2 | 6+8 2 | 8 1 | 8 1 | 3+4 2 | 2+2 2 |
| | 6+8 2 | 5+7 2 | 5 1 | 5 1 | 3+4 2 | 2+2 2 |
| | 5+7 2 | 6+8 2 | 6 1 | 6 1 | 3+4 2 | 2+2 2 |
| | 6+8 2 | 6+8 2 | 8 1 | 8 1 | 3+4 2 | 2+2 2 |
| | 5+7 2 | 5+7 2 | 5 1 | 5 1 | 3+4 2 | 2+2 2 |
| | 6+8 2 | 5+7 2 | 7 1 | 7 1 | 3+4 2 | 2+2 2 |
| | 5+7 2 | 5+7 2 | 6 1 | 6 1 | 3+4 2 | 2+2 2 |
| | 6+8 2 | 6+8 2 | 7 1 | 7 1 | 3+4 2 | 2+2 2 |
| | 5+7 2 | 5+7 2 | 5 1 | 5 1 | 3+4 2 | 2+2 2 |
| | 6+8 2 | 6+8 2 | 8 1 | 8 1 | 3+4 2 | 2+2 2 |
| | 5+7 2 | 6+8 2 | 8 1 | 8 1 | 3+4 2 | 2+2 2 |
| | 6+8 2 | 5+7 2 | 6 1 | 6 1 | 3+4 2 | 2+2 2 |
| | 6+8 2 | 5+7 2 | 5 1 | 5 1 | 3+4 2 | 2+2 2 |
| | 5+7 2 | 6+8 2 | 7 1 | 7 1 | 3+4 2 | 2+2 2 |
| | 6+8 2 | 5+7 2 | 8 1 | 8 1 | 3+4 2 | 2+2 2 |
| | 5+7 2 | 6+8 2 | 5 1 | 5 1 | 3+4 2 | 2+2 2 |
| | | | 6 1 | 6 1 | | |
| | | | 7 1 | 7 1 | | |
| | | | 5 1 | 5 1 | | |
| | | | 7 1 | 7 1 | | |
| | | | 5 1 | 5 1 | | |
| | | | 8 1 | 8 1 | | |
| | | | 6 1 | 6 1 | | |
| | | | 7 1 | 7 1 | | |
| | | | 6 1 | 6 1 | | |
| | | | 8 1 | 8 1 | | |
| | | | 5 1 | 5 1 | | |
| | | | 7 1 | 7 1 | | |

Continued on next column

**TABLE XLVII– continued**

| TQC | ICSD | NoE | | SF |
|---|---|---|---|---|
| | | 5 1 | 5 1 | |
| | | 8 1 | 8 1 | |
| | | 7 1 | 7 1 | |
| | | 6 1 | 6 1 | |
| | | 5 1 | 5 1 | |
| | | 8 1 | 8 1 | |
| | | 7 1 | 7 1 | |
| | | 6 1 | 6 1 | |
| | | 8 1 | 8 1 | |
| | | 6 1 | 6 1 | |
| | | 6 1 | 6 1 | |
| | | 8 1 | 8 1 | |
| | | 5 1 | 5 1 | |
| | | 7 1 | 7 1 | |
| | | 7 1 | 7 1 | |
| | | 5 1 | 5 1 | |
| | | 6 1 | 6 1 | |
| | | 7 1 | 7 1 | |
| | | 8 1 | 8 1 | |
| | | 5 1 | 5 1 | |
| | | 6 1 | 6 1 | |
| | | 8 1 | 8 1 | |
| | | 7 1 | 7 1 | |
| | | 6 1 | 6 1 | |
| | | 5 1 | 5 1 | |
| | | 8 1 | 8 1 | |
| | | 8 1 | 8 1 | |
| | | 6 1 | 6 1 | |
| | | 7 1 | 7 1 | |
| | | 5 1 | 5 1 | |
| | | 6 1 | 6 1 | |
| | | 8 1 | 8 1 | |
| | | 7 1 | 7 1 | |
| | | 5 1 | 5 1 | |

The End

TABLE XLVIII: The computed irreducible representations for the valence bands for the compounds of SEBR and NLC in space group 83. See the list of tables.

| TQC | -A | -GM | -M | -Z | -R | -X |
|---|---|---|---|---|---|---|
| 83.2.1 | 106354 | | 106 | Ca7Ni4Sn13 | | |
| | 5+7 2 | 6+8 2 | 5+7 2 | 10+12 2 | 5+6 2 | 5+6 2 |
| | 10+12 2 | 11+9 2 | 11+9 2 | 6+8 2 | 3+4 2 | 3+4 2 |
| | 5+7 2 | 10+12 2 | 10+12 2 | 6+8 2 | 5+6 2 | 3+4 2 |
| | 6+8 2 | 6+8 2 | 6+8 2 | 5+7 2 | 5+6 2 | 5+6 2 |
| | 11+9 2 | 5+7 2 | 6+8 2 | 11+9 2 | 3+4 2 | 3+4 2 |
| | 11+9 2 | 11+9 2 | 10+12 2 | 10+12 2 | 3+4 2 | 5+6 2 |
| | 10+12 2 | 10+12 2 | 11+9 2 | 11+9 2 | 3+4 2 | 3+4 2 |
| | 10+12 2 | 5+7 2 | 5+7 2 | 10+12 2 | 5+6 2 | 5+6 2 |
| | 11+9 2 | 6+8 2 | 6+8 2 | 5+7 2 | 3+4 2 | 5+6 2 |
| | 6+8 2 | 5+7 2 | 10+12 2 | 6+8 2 | 5+6 2 | 3+4 2 |
| | 5+7 2 | 11+9 2 | 11+9 2 | 11+9 2 | 3+4 2 | 3+4 2 |
| | 6+8 2 | 10+12 2 | 5+7 2 | 5+7 2 | 5+6 2 | 5+6 2 |
| | 10+12 2 | 6+8 2 | 6+8 2 | 10+12 2 | 5+6 2 | 3+4 2 |
| | 6+8 2 | 6+8 2 | 10+12 2 | 11+9 2 | 3+4 2 | 3+4 2 |
| | 6+8 2 | 11+9 2 | 11+9 2 | 5+7 2 | 5+6 2 | 5+6 2 |
| | 5+7 2 | 10+12 2 | 5+7 2 | 6+8 2 | 5+6 2 | 5+6 2 |
| | 10+12 2 | 5+7 2 | 10+12 2 | 10+12 2 | 3+4 2 | 5+6 2 |
| | 11+9 2 | 10+12 2 | 11+9 2 | 5+7 2 | 3+4 2 | 3+4 2 |

Continued on next column

**TABLE XLVIII– continued**

| TQC | ICSD | | NoE | SF | | |
|---|---|---|---|---|---|---|
| | 11+9 2 | 5+7 2 | 11+9 2 | 6+8 2 | 5+6 2 | 3+4 2 |
| | 10+12 2 | 6+8 2 | 6+8 2 | 11+9 2 | 5+6 2 | 3+4 2 |
| | 6+8 2 | 5+7 2 | 5+7 2 | 10+12 2 | 3+4 2 | 5+6 2 |
| | 10+12 2 | 6+8 2 | 6+8 2 | 11+9 2 | 3+4 2 | 5+6 2 |
| | 5+7 2 | 10+12 2 | 5+7 2 | 6+8 2 | 3+4 2 | 3+4 2 |
| | 6+8 2 | 11+9 2 | 6+8 2 | 5+7 2 | 5+6 2 | 5+6 2 |
| | 11+9 2 | 6+8 2 | 10+12 2 | 5+7 2 | 5+6 2 | 5+6 2 |
| | 5+7 2 | 11+9 2 | 10+12 2 | 6+8 2 | 3+4 2 | 3+4 2 |
| | 10+12 2 | 10+12 2 | 11+9 2 | 10+12 2 | 5+6 2 | 5+6 2 |
| | 6+8 2 | 11+9 2 | 5+7 2 | 11+9 2 | 3+4 2 | 3+4 2 |
| | 5+7 2 | 11+9 2 | 10+12 2 | 11+9 2 | 3+4 2 | 3+4 2 |
| | 10+12 2 | 6+8 2 | 11+9 2 | 6+8 2 | 5+6 2 | 5+6 2 |
| | 5+7 2 | 5+7 2 | 5+7 2 | 5+7 2 | 3+4 2 | 5+6 2 |
| | 11+9 2 | 10+12 2 | 10+12 2 | 6+8 2 | 5+6 2 | 3+4 2 |
| | 5+7 2 | 6+8 2 | 6+8 2 | 10+12 2 | 5+6 2 | 3+4 2 |
| | 6+8 2 | 5+7 2 | 6+8 2 | 6+8 2 | 3+4 2 | 5+6 2 |
| | 11+9 2 | 11+9 2 | 11+9 2 | 5+7 2 | 5+6 2 | 3+4 2 |
| | 5+7 2 | 10+12 2 | 5+7 2 | 10+12 2 | 3+4 2 | 5+6 2 |
| | 6+8 2 | 5+7 2 | 6+8 2 | 10+12 2 | 3+4 2 | 5+6 2 |
| | 11+9 2 | 11+9 2 | 11+9 2 | 11+9 2 | 5+6 2 | 3+4 2 |
| | 10+12 2 | 10+12 2 | 5+7 2 | 6+8 2 | 3+4 2 | 3+4 2 |
| | 5+7 2 | 10+12 2 | 6+8 2 | 6+8 2 | 5+6 2 | 5+6 2 |
| | 6+8 2 | 6+8 2 | 5+7 2 | 5+7 2 | 5+6 2 | 5+6 2 |
| | 11+9 2 | 5+7 2 | 10+12 2 | 11+9 2 | 3+4 2 | 5+6 2 |
| | 10+12 2 | 6+8 2 | 11+9 2 | 10+12 2 | 5+6 2 | 5+6 2 |
| | 5+7 2 | 5+7 2 | 5+7 2 | 5+7 2 | 3+4 2 | 3+4 2 |
| | 6+8 2 | 11+9 2 | 10+12 2 | 6+8 2 | 5+6 2 | 3+4 2 |
| | 10+12 2 | 10+12 2 | 11+9 2 | 5+7 2 | 3+4 2 | 5+6 2 |
| | 11+9 2 | 5+7 2 | 10+12 2 | 10+12 2 | 3+4 2 | 3+4 2 |
| | 11+9 2 | 6+8 2 | 11+9 2 | 11+9 2 | 5+6 2 | 5+6 2 |
| | 10+12 2 | 6+8 2 | 6+8 2 | 6+8 2 | 3+4 2 | 3+4 2 |
| | 6+8 2 | 11+9 2 | 10+12 2 | 5+7 2 | 3+4 2 | 3+4 2 |
| | 11+9 2 | 10+12 2 | 6+8 2 | 10+12 2 | 3+4 2 | 3+4 2 |
| | 10+12 2 | 6+8 2 | 10+12 2 | 11+9 2 | 3+4 2 | 5+6 2 |
| | 11+9 2 | 11+9 2 | 5+7 2 | 10+12 2 | 5+6 2 | 5+6 2 |

The End

TABLE XLIX: The computed irreducible representations for the valence bands for the compounds of SEBR and NLC in space group 84. See the list of tables.

| TQC | ICSD | | NoE | SF | | |
|---|---|---|---|---|---|---|
| 84.2.1 | 74390 | | 120 | P2S6U1 | | |
| | -A | -GM | -M | -Z | -R | -X |
| | 3+4 4 | 6+8 2 | 6+8 2 | 3+4 4 | 3+4 2 | 3+4 2 |
| | 3+4 4 | 11+9 2 | 11+9 2 | 3+4 4 | 5+6 2 | 5+6 2 |
| | 3+4 4 | 5+7 2 | 5+7 2 | 3+4 4 | 3+4 2 | 3+4 2 |
| | 3+4 4 | 10+12 2 | 10+12 2 | 3+4 4 | 5+6 2 | 5+6 2 |
| | 3+4 4 | 5+7 2 | 11+9 2 | 3+4 4 | 3+4 2 | 3+4 2 |
| | 3+4 4 | 11+9 2 | 5+7 2 | 3+4 4 | 3+4 2 | 5+6 2 |
| | 3+4 4 | 6+8 2 | 10+12 2 | 3+4 4 | 5+6 2 | 3+4 2 |
| | 3+4 4 | 10+12 2 | 6+8 2 | 3+4 4 | 5+6 2 | 5+6 2 |
| | 3+4 4 | 6+8 2 | 10+12 2 | 3+4 4 | 3+4 2 | 3+4 2 |
| | 3+4 4 | 5+7 2 | 11+9 2 | 3+4 4 | 3+4 2 | 5+6 2 |
| | 3+4 4 | 10+12 2 | 6+8 2 | 3+4 4 | 5+6 2 | 3+4 2 |
| | 3+4 4 | 11+9 2 | 5+7 2 | 3+4 4 | 5+6 2 | 5+6 2 |
| | 3+4 4 | 11+9 2 | 5+7 2 | 3+4 4 | 5+6 2 | 5+6 2 |
| | 3+4 4 | 10+12 2 | 6+8 2 | 3+4 4 | 3+4 2 | 3+4 2 |
| | 3+4 4 | 5+7 2 | 10+12 2 | 3+4 4 | 5+6 2 | 5+6 2 |
| | 3+4 4 | 11+9 2 | 11+9 2 | 3+4 4 | 5+6 2 | 3+4 2 |

Continued on next column

**TABLE XLIX– continued**

| TQC | ICSD | | NoE | SF | | |
|---|---|---|---|---|---|---|
| | 3+4 4 | 6+8 2 | 5+7 2 | 3+4 4 | 3+4 2 | 5+6 2 |
| | 3+4 4 | 10+12 2 | 10+12 2 | 3+4 4 | 5+6 2 | 3+4 2 |
| | 3+4 4 | 5+7 2 | 11+9 2 | 3+4 4 | 3+4 2 | 5+6 2 |
| | 3+4 4 | 6+8 2 | 6+8 2 | 3+4 4 | 3+4 2 | 3+4 2 |
| | 3+4 4 | 6+8 2 | 10+12 2 | 3+4 4 | 3+4 2 | 3+4 2 |
| | 3+4 4 | 5+7 2 | 11+9 2 | 3+4 4 | 3+4 2 | 5+6 2 |
| | 3+4 4 | 10+12 2 | 6+8 2 | 3+4 4 | 5+6 2 | 3+4 2 |
| | 3+4 4 | 11+9 2 | 5+7 2 | 3+4 4 | 5+6 2 | 5+6 2 |
| | 3+4 4 | 6+8 2 | 10+12 2 | 3+4 4 | 3+4 2 | 5+6 2 |
| | 3+4 4 | 10+12 2 | 11+9 2 | 3+4 4 | 3+4 2 | 3+4 2 |
| | 3+4 4 | 10+12 2 | 5+7 2 | 3+4 4 | 5+6 2 | 5+6 2 |
| | 3+4 4 | 5+7 2 | 5+7 2 | 3+4 4 | 5+6 2 | 3+4 2 |
| | 3+4 4 | 11+9 2 | 6+8 2 | 3+4 4 | 5+6 2 | 3+4 2 |
| | 3+4 4 | 11+9 2 | 6+8 2 | 3+4 4 | 5+6 2 | 5+6 2 |
| | | 5+7 2 | 10+12 2 | | 3+4 2 | 3+4 2 |
| | | 5+7 2 | 10+12 2 | | 3+4 2 | 3+4 2 |
| | | 6+8 2 | 11+9 2 | | 3+4 2 | 5+6 2 |
| | | 6+8 2 | 11+9 2 | | 3+4 2 | 5+6 2 |
| | | 11+9 2 | 11+9 2 | | 5+6 2 | 5+6 2 |
| | | 6+8 2 | 6+8 2 | | 3+4 2 | 3+4 2 |
| | | 5+7 2 | 5+7 2 | | 5+6 2 | 3+4 2 |
| | | 6+8 2 | 5+7 2 | | 5+6 2 | 3+4 2 |
| | | 10+12 2 | 10+12 2 | | 3+4 2 | 5+6 2 |
| | | 5+7 2 | 6+8 2 | | 5+6 2 | 3+4 2 |
| | | 11+9 2 | 5+7 2 | | 3+4 2 | 5+6 2 |
| | | 10+12 2 | 10+12 2 | | 5+6 2 | 3+4 2 |
| | | 6+8 2 | 11+9 2 | | 3+4 2 | 5+6 2 |
| | | 11+9 2 | 6+8 2 | | 3+4 2 | 5+6 2 |
| | | 5+7 2 | 5+7 2 | | 5+6 2 | 3+4 2 |
| | | 10+12 2 | 6+8 2 | | 5+6 2 | 5+6 2 |
| | | 10+12 2 | 5+7 2 | | 5+6 2 | 3+4 2 |
| | | 5+7 2 | 6+8 2 | | 3+4 2 | 3+4 2 |
| | | 11+9 2 | 10+12 2 | | 5+6 2 | 5+6 2 |
| | | 5+7 2 | 11+9 2 | | 3+4 2 | 3+4 2 |
| | | 10+12 2 | 5+7 2 | | 3+4 2 | 5+6 2 |
| | | 6+8 2 | 6+8 2 | | 5+6 2 | 3+4 2 |
| | | 11+9 2 | 10+12 2 | | 3+4 2 | 3+4 2 |
| | | 11+9 2 | 11+9 2 | | 5+6 2 | 5+6 2 |
| | | 10+12 2 | 10+12 2 | | 5+6 2 | 5+6 2 |
| | | 6+8 2 | 11+9 2 | | 3+4 2 | 5+6 2 |
| | | 6+8 2 | 10+12 2 | | 3+4 2 | 5+6 2 |
| | | 5+7 2 | 11+9 2 | | 3+4 2 | 3+4 2 |
| | | 11+9 2 | 6+8 2 | | 5+6 2 | 3+4 2 |
| | | 6+8 2 | 11+9 2 | | 5+6 2 | 5+6 2 |

The End

TABLE L: The computed irreducible representations for the valence bands for the compounds of SEBR and NLC in space group 85. See the list of tables.

| TQC | ICSD | | NoE | SF | | |
|---|---|---|---|---|---|---|
| 85.2.1 | 33884 | | 136 | As2Cl4F6 | | |
| | -A | -GM | -M | -Z | -R | -X |
| | 3+4 4 | 6+8 2 | 3+4 4 | 10+12 2 | 2+2 4 | 2+2 4 |
| | 3+4 4 | 10+12 2 | 3+4 4 | 6+8 2 | 2+2 4 | 2+2 4 |
| | 3+4 4 | 6+8 2 | 3+4 4 | 6+8 2 | 2+2 4 | 2+2 4 |
| | 3+4 4 | 11+9 2 | 3+4 4 | 11+9 2 | 2+2 4 | 2+2 4 |
| | 3+4 4 | 6+8 2 | 3+4 4 | 10+12 2 | 2+2 4 | 2+2 4 |
| | 3+4 4 | 5+7 2 | 3+4 4 | 5+7 2 | 2+2 4 | 2+2 4 |
| | 3+4 4 | 10+12 2 | 3+4 4 | 6+8 2 | 2+2 4 | 2+2 4 |

Continued on next column

**TABLE L– continued**

| TQC | ICSD | | NoE | | SF | |
|---|---|---|---|---|---|---|
| | 3+4 4 | 10+12 2 | 3+4 4 | 10+12 2 | 2+2 4 | 2+2 4 |
| | 3+4 4 | 11+9 2 | 3+4 4 | 5+7 2 | 2+2 4 | 2+2 4 |
| | 3+4 4 | 5+7 2 | 3+4 4 | 11+9 2 | 2+2 4 | 2+2 4 |
| | 3+4 4 | 10+12 2 | 3+4 4 | 6+8 2 | 2+2 4 | 2+2 4 |
| | 3+4 4 | 6+8 2 | 3+4 4 | 10+12 2 | 2+2 4 | 2+2 4 |
| | 3+4 4 | 10+12 2 | 3+4 4 | 6+8 2 | 2+2 4 | 2+2 4 |
| | 3+4 4 | 6+8 2 | 3+4 4 | 11+9 2 | 2+2 4 | 2+2 4 |
| | 3+4 4 | 11+9 2 | 3+4 4 | 10+12 2 | 2+2 4 | 2+2 4 |
| | 3+4 4 | 5+7 2 | 3+4 4 | 5+7 2 | 2+2 4 | 2+2 4 |
| | 3+4 4 | 6+8 2 | 3+4 4 | 6+8 2 | 2+2 4 | 2+2 4 |
| | 3+4 4 | 11+9 2 | 3+4 4 | 11+9 2 | 2+2 4 | 2+2 4 |
| | 3+4 4 | 10+12 2 | 3+4 4 | 5+7 2 | 2+2 4 | 2+2 4 |
| | 3+4 4 | 6+8 2 | 3+4 4 | 11+9 2 | 2+2 4 | 2+2 4 |
| | 3+4 4 | 5+7 2 | 3+4 4 | 10+12 2 | 2+2 4 | 2+2 4 |
| | 3+4 4 | 11+9 2 | 3+4 4 | 6+8 2 | 2+2 4 | 2+2 4 |
| | 3+4 4 | 10+12 2 | 3+4 4 | 10+12 2 | 2+2 4 | 2+2 4 |
| | 3+4 4 | 10+12 2 | 3+4 4 | 10+12 2 | 2+2 4 | 2+2 4 |
| | 3+4 4 | 11+9 2 | 3+4 4 | 11+9 2 | 2+2 4 | 2+2 4 |
| | 3+4 4 | 5+7 2 | 3+4 4 | 5+7 2 | 2+2 4 | 2+2 4 |
| | 3+4 4 | 6+8 2 | 3+4 4 | 6+8 2 | 2+2 4 | 2+2 4 |
| | 3+4 4 | 6+8 2 | 3+4 4 | 10+12 2 | 2+2 4 | 2+2 4 |
| | 3+4 4 | 5+7 2 | 3+4 4 | 5+7 2 | 2+2 4 | 2+2 4 |
| | 3+4 4 | 5+7 2 | 3+4 4 | 6+8 2 | 2+2 4 | 2+2 4 |
| | 3+4 4 | 10+12 2 | 3+4 4 | 6+8 2 | 2+2 4 | 2+2 4 |
| | 3+4 4 | 6+8 2 | 3+4 4 | 6+8 2 | 2+2 4 | 2+2 4 |
| | 3+4 4 | 11+9 2 | 3+4 4 | 5+7 2 | 2+2 4 | 2+2 4 |
| | 3+4 4 | 5+7 2 | 3+4 4 | 10+12 2 | 2+2 4 | 2+2 4 |
| | | 10+12 2 | | 11+9 2 | | |
| | | 6+8 2 | | 5+7 2 | | |
| | | 5+7 2 | | 11+9 2 | | |
| | | 11+9 2 | | 10+12 2 | | |
| | | 10+12 2 | | 10+12 2 | | |
| | | 11+9 2 | | 5+7 2 | | |
| | | 6+8 2 | | 11+9 2 | | |
| | | 11+9 2 | | 6+8 2 | | |
| | | 5+7 2 | | 5+7 2 | | |
| | | 10+12 2 | | 11+9 2 | | |
| | | 10+12 2 | | 10+12 2 | | |
| | | 5+7 2 | | 6+8 2 | | |
| | | 6+8 2 | | 5+7 2 | | |
| | | 5+7 2 | | 11+9 2 | | |
| | | 10+12 2 | | 10+12 2 | | |
| | | 11+9 2 | | 6+8 2 | | |
| | | 11+9 2 | | 5+7 2 | | |
| | | 6+8 2 | | 11+9 2 | | |
| | | 5+7 2 | | 10+12 2 | | |
| | | 6+8 2 | | 6+8 2 | | |
| | | 10+12 2 | | 10+12 2 | | |
| | | 11+9 2 | | 11+9 2 | | |
| | | 6+8 2 | | 5+7 2 | | |
| | | 10+12 2 | | 6+8 2 | | |
| | | 11+9 2 | | 11+9 2 | | |
| | | 6+8 2 | | 10+12 2 | | |
| | | 10+12 2 | | 5+7 2 | | |
| | | 6+8 2 | | 6+8 2 | | |
| | | 11+9 2 | | 11+9 2 | | |
| | | 5+7 2 | | 10+12 2 | | |
| | | 10+12 2 | | 6+8 2 | | |
| | | 5+7 2 | | 5+7 2 | | |
| | | 6+8 2 | | 10+12 2 | | |
| | | 6+8 2 | | 6+8 2 | | |

Continued on next column

**TABLE LI– continued**

| TQC | ICSD | NoE | SF |
|---|---|---|---|

The End

TABLE LI: The computed irreducible representations for the valence bands for the compounds of SEBR and NLC in space group 87. See the list of tables.

| TQC | ICSD | | NoE | SF | |
|---|---|---|---|---|---|
| 87.1.1 | 51047 | | 44 | Te4Zr5 | |
| | -GM | -M | -P | -X | -N |
| | 6+8 2 | 5+7 2 | 5+7 2 | 5+6 2 | 3+3 2 |
| | 11+9 2 | 11+9 2 | 5+7 2 | 3+4 2 | 2+2 2 |
| | 10+12 2 | 10+12 2 | 6+8 2 | 3+4 2 | 3+3 2 |
| | 5+7 2 | 6+8 2 | 6+8 2 | 5+6 2 | 2+2 2 |
| | 5+7 2 | 6+8 2 | 5+7 2 | 5+6 2 | 2+2 2 |
| | 6+8 2 | 5+7 2 | 5+7 2 | 3+4 2 | 3+3 2 |
| | 11+9 2 | 11+9 2 | 6+8 2 | 3+4 2 | 3+3 2 |
| | 10+12 2 | 10+12 2 | 6+8 2 | 3+4 2 | 2+2 2 |
| | 11+9 2 | 5+7 2 | 6+8 2 | 5+6 2 | 3+3 2 |
| | 6+8 2 | 10+12 2 | 6+8 2 | 5+6 2 | 2+2 2 |
| | 5+7 2 | 11+9 2 | 5+7 2 | 5+6 2 | 3+3 2 |
| | 10+12 2 | 6+8 2 | 5+7 2 | 3+4 2 | 2+2 2 |
| | 6+8 2 | 11+9 2 | 6+8 2 | 3+4 2 | 3+3 2 |
| | 11+9 2 | 5+7 2 | 5+7 2 | 5+6 2 | 3+3 2 |
| | 5+7 2 | 6+8 2 | 5+7 2 | 3+4 2 | 2+2 2 |
| | 6+8 2 | 6+8 2 | 6+8 2 | 3+4 2 | 2+2 2 |
| | 10+12 2 | 10+12 2 | 6+8 2 | 5+6 2 | 3+3 2 |
| | 10+12 2 | 5+7 2 | 5+7 2 | 5+6 2 | 3+3 2 |
| | 11+9 2 | 6+8 2 | 6+8 2 | 3+4 2 | 2+2 2 |
| | 5+7 2 | 11+9 2 | 5+7 2 | 5+6 2 | 3+3 2 |
| | 6+8 2 | 10+12 2 | 5+7 2 | 3+4 2 | 2+2 2 |
| | 6+8 2 | 10+12 2 | 5+7 2 | 3+4 2 | 3+3 2 |
| 87.1.2 | 409337 | | 44 | Te4Zr5 | |
| | -GM | -M | -P | -X | -N |
| | 6+8 2 | 5+7 2 | 5+7 2 | 5+6 2 | 3+3 2 |
| | 11+9 2 | 11+9 2 | 5+7 2 | 3+4 2 | 2+2 2 |
| | 10+12 2 | 10+12 2 | 6+8 2 | 3+4 2 | 3+3 2 |
| | 5+7 2 | 6+8 2 | 6+8 2 | 5+6 2 | 2+2 2 |
| | 5+7 2 | 6+8 2 | 5+7 2 | 5+6 2 | 2+2 2 |
| | 6+8 2 | 5+7 2 | 5+7 2 | 3+4 2 | 3+3 2 |
| | 11+9 2 | 11+9 2 | 6+8 2 | 3+4 2 | 3+3 2 |
| | 10+12 2 | 10+12 2 | 6+8 2 | 3+4 2 | 2+2 2 |
| | 11+9 2 | 5+7 2 | 6+8 2 | 5+6 2 | 3+3 2 |
| | 6+8 2 | 10+12 2 | 6+8 2 | 5+6 2 | 2+2 2 |
| | 5+7 2 | 11+9 2 | 5+7 2 | 5+6 2 | 3+3 2 |
| | 10+12 2 | 6+8 2 | 5+7 2 | 3+4 2 | 2+2 2 |
| | 6+8 2 | 11+9 2 | 6+8 2 | 3+4 2 | 3+3 2 |
| | 5+7 2 | 5+7 2 | 5+7 2 | 5+6 2 | 3+3 2 |
| | 11+9 2 | 6+8 2 | 5+7 2 | 3+4 2 | 2+2 2 |
| | 6+8 2 | 6+8 2 | 6+8 2 | 3+4 2 | 2+2 2 |
| | 10+12 2 | 10+12 2 | 6+8 2 | 5+6 2 | 3+3 2 |
| | 10+12 2 | 5+7 2 | 5+7 2 | 5+6 2 | 3+3 2 |
| | 11+9 2 | 6+8 2 | 6+8 2 | 3+4 2 | 2+2 2 |
| | 5+7 2 | 11+9 2 | 5+7 2 | 5+6 2 | 3+3 2 |
| | 6+8 2 | 10+12 2 | 5+7 2 | 3+4 2 | 2+2 2 |
| | 6+8 2 | 10+12 2 | 5+7 2 | 3+4 2 | 3+3 2 |
| 87.1.3 | 55075 | | 44 | Se4Ti5 | |
| | -GM | -M | -P | -X | -N |
| | 6+8 2 | 11+9 2 | 6+8 2 | 3+4 2 | 3+3 2 |
| | 11+9 2 | 10+12 2 | 5+7 2 | 5+6 2 | 3+3 2 |
| | 10+12 2 | 5+7 2 | 5+7 2 | 3+4 2 | 2+2 2 |

Continued on next column

**TABLE LI– continued**

| TQC | ICSD | | NoE | SF | |
|---|---|---|---|---|---|
| | 5+7 2 | 6+8 2 | 6+8 2 | 5+6 2 | 2+2 2 |
| | 5+7 2 | 6+8 2 | 5+7 2 | 5+6 2 | 2+2 2 |
| | 6+8 2 | 5+7 2 | 5+7 2 | 3+4 2 | 3+3 2 |
| | 11+9 2 | 11+9 2 | 6+8 2 | 3+4 2 | 3+3 2 |
| | 10+12 2 | 10+12 2 | 6+8 2 | 5+6 2 | 2+2 2 |
| | 6+8 2 | 5+7 2 | 6+8 2 | 3+4 2 | 3+3 2 |
| | 11+9 2 | 10+12 2 | 6+8 2 | 5+6 2 | 2+2 2 |
| | 5+7 2 | 11+9 2 | 5+7 2 | 5+6 2 | 3+3 2 |
| | 10+12 2 | 11+9 2 | 5+7 2 | 3+4 2 | 2+2 2 |
| | 6+8 2 | 6+8 2 | 6+8 2 | 3+4 2 | 3+3 2 |
| | 5+7 2 | 5+7 2 | 5+7 2 | 5+6 2 | 3+3 2 |
| | 11+9 2 | 6+8 2 | 5+7 2 | 3+4 2 | 2+2 2 |
| | 6+8 2 | 6+8 2 | 6+8 2 | 3+4 2 | 2+2 2 |
| | 10+12 2 | 10+12 2 | 6+8 2 | 5+6 2 | 3+3 2 |
| | 10+12 2 | 5+7 2 | 5+7 2 | 5+6 2 | 3+3 2 |
| | 11+9 2 | 11+9 2 | 5+7 2 | 3+4 2 | 2+2 2 |
| | 5+7 2 | 10+12 2 | 6+8 2 | 3+4 2 | 2+2 2 |
| | 6+8 2 | 6+8 2 | 5+7 2 | 5+6 2 | 3+3 2 |
| | 6+8 2 | 10+12 2 | 5+7 2 | 3+4 2 | 2+2 2 |
| 87.1.4 | 157026 | | 72 | Mo1Ni1O6Sr2 | |
| | -GM | -M | -P | -X | -N |
| | 6+8 2 | 10+12 2 | 5+7 2 | 5+6 2 | 2+2 2 |
| | 11+9 2 | 5+7 2 | 6+8 2 | 3+4 2 | 3+3 2 |
| | 6+8 2 | 6+8 2 | 6+8 2 | 3+4 2 | 3+3 2 |
| | 5+7 2 | 6+8 2 | 5+7 2 | 5+6 2 | 2+2 2 |
| | 6+8 2 | 11+9 2 | 6+8 2 | 3+4 2 | 3+3 2 |
| | 10+12 2 | 10+12 2 | 5+7 2 | 5+6 2 | 3+3 2 |
| | 10+12 2 | 5+7 2 | 6+8 2 | 3+4 2 | 2+2 2 |
| | 11+9 2 | 10+12 2 | 5+7 2 | 5+6 2 | 2+2 2 |
| | 5+7 2 | 6+8 2 | 6+8 2 | 3+4 2 | 3+3 2 |
| | 10+12 2 | 11+9 2 | 5+7 2 | 5+6 2 | 2+2 2 |
| | 6+8 2 | 5+7 2 | 6+8 2 | 3+4 2 | 3+3 2 |
| | 5+7 2 | 11+9 2 | 5+7 2 | 5+6 2 | 2+2 2 |
| | 10+12 2 | 10+12 2 | 5+7 2 | 3+4 2 | 3+3 2 |
| | 11+9 2 | 6+8 2 | 6+8 2 | 5+6 2 | 2+2 2 |
| | 6+8 2 | 6+8 2 | 5+7 2 | 3+4 2 | 3+3 2 |
| | 6+8 2 | 5+7 2 | 6+8 2 | 3+4 2 | 3+3 2 |
| | 5+7 2 | 5+7 2 | 6+8 2 | 3+4 2 | 3+3 2 |
| | 5+7 2 | 6+8 2 | 5+7 2 | 3+4 2 | 2+2 2 |
| | 6+8 2 | 6+8 2 | 6+8 2 | 3+4 2 | 3+3 2 |
| | 5+7 2 | 5+7 2 | 5+7 2 | 3+4 2 | 3+3 2 |
| | 10+12 2 | 10+12 2 | 5+7 2 | 5+6 2 | 3+3 2 |
| | 11+9 2 | 11+9 2 | 5+7 2 | 5+6 2 | 2+2 2 |
| | 10+12 2 | 10+12 2 | 6+8 2 | 5+6 2 | 2+2 2 |
| | 10+12 2 | 6+8 2 | 6+8 2 | 5+6 2 | 2+2 2 |
| | 11+9 2 | 5+7 2 | 5+7 2 | 3+4 2 | 2+2 2 |
| | 11+9 2 | 10+12 2 | 6+8 2 | 3+4 2 | 2+2 2 |
| | 6+8 2 | 11+9 2 | 5+7 2 | 5+6 2 | 2+2 2 |
| | 10+12 2 | 11+9 2 | 6+8 2 | 5+6 2 | 3+3 2 |
| | 6+8 2 | 10+12 2 | 5+7 2 | 5+6 2 | 2+2 2 |
| | 10+12 2 | 10+12 2 | 6+8 2 | 5+6 2 | 2+2 2 |
| | 5+7 2 | 11+9 2 | 5+7 2 | 5+6 2 | 3+3 2 |
| | 11+9 2 | 6+8 2 | 6+8 2 | 3+4 2 | 3+3 2 |
| | 5+7 2 | 5+7 2 | 5+7 2 | 3+4 2 | 2+2 2 |
| | 6+8 2 | 6+8 2 | 6+8 2 | 3+4 2 | 2+2 2 |
| | 5+7 2 | 5+7 2 | 6+8 2 | 3+4 2 | 2+2 2 |
| | 5+7 2 | 5+7 2 | 5+7 2 | 3+4 2 | 2+2 2 |
| 87.1.5 | 157024 | | 72 | Mo1Ni1O6Sr2 | |
| | -GM | -M | -P | -X | -N |
| | 6+8 2 | 10+12 2 | 5+7 2 | 5+6 2 | 2+2 2 |
| | 11+9 2 | 5+7 2 | 6+8 2 | 3+4 2 | 3+3 2 |

Continued on next column

**TABLE LI– continued**

| TQC | ICSD | | NoE | SF | |
|---|---|---|---|---|---|
| | 6+8 2 | 6+8 2 | 6+8 2 | 3+4 2 | 3+3 2 |
| | 5+7 2 | 6+8 2 | 5+7 2 | 5+6 2 | 2+2 2 |
| | 6+8 2 | 11+9 2 | 6+8 2 | 3+4 2 | 3+3 2 |
| | 10+12 2 | 10+12 2 | 5+7 2 | 5+6 2 | 3+3 2 |
| | 10+12 2 | 10+12 2 | 6+8 2 | 3+4 2 | 2+2 2 |
| | 11+9 2 | 5+7 2 | 5+7 2 | 5+6 2 | 2+2 2 |
| | 5+7 2 | 6+8 2 | 6+8 2 | 3+4 2 | 3+3 2 |
| | 10+12 2 | 11+9 2 | 5+7 2 | 5+6 2 | 2+2 2 |
| | 6+8 2 | 5+7 2 | 6+8 2 | 3+4 2 | 3+3 2 |
| | 5+7 2 | 11+9 2 | 5+7 2 | 5+6 2 | 2+2 2 |
| | 10+12 2 | 10+12 2 | 6+8 2 | 3+4 2 | 3+3 2 |
| | 11+9 2 | 6+8 2 | 5+7 2 | 5+6 2 | 2+2 2 |
| | 6+8 2 | 6+8 2 | 5+7 2 | 3+4 2 | 3+3 2 |
| | 6+8 2 | 5+7 2 | 6+8 2 | 3+4 2 | 3+3 2 |
| | 5+7 2 | 5+7 2 | 6+8 2 | 3+4 2 | 3+3 2 |
| | 5+7 2 | 6+8 2 | 5+7 2 | 3+4 2 | 2+2 2 |
| | 6+8 2 | 6+8 2 | 6+8 2 | 3+4 2 | 3+3 2 |
| | 5+7 2 | 5+7 2 | 5+7 2 | 3+4 2 | 3+3 2 |
| | 10+12 2 | 10+12 2 | 5+7 2 | 5+6 2 | 3+3 2 |
| | 11+9 2 | 11+9 2 | 5+7 2 | 5+6 2 | 2+2 2 |
| | 10+12 2 | 10+12 2 | 6+8 2 | 5+6 2 | 2+2 2 |
| | 10+12 2 | 6+8 2 | 6+8 2 | 5+6 2 | 2+2 2 |
| | 11+9 2 | 5+7 2 | 5+7 2 | 3+4 2 | 2+2 2 |
| | 11+9 2 | 10+12 2 | 6+8 2 | 3+4 2 | 2+2 2 |
| | 6+8 2 | 11+9 2 | 5+7 2 | 5+6 2 | 2+2 2 |
| | 6+8 2 | 10+12 2 | 6+8 2 | 5+6 2 | 3+3 2 |
| | 10+12 2 | 11+9 2 | 5+7 2 | 5+6 2 | 2+2 2 |
| | 5+7 2 | 10+12 2 | 6+8 2 | 5+6 2 | 3+3 2 |
| | 10+12 2 | 11+9 2 | 5+7 2 | 5+6 2 | 2+2 2 |
| | 11+9 2 | 6+8 2 | 6+8 2 | 3+4 2 | 3+3 2 |
| | 5+7 2 | 5+7 2 | 5+7 2 | 3+4 2 | 2+2 2 |
| | 6+8 2 | 6+8 2 | 6+8 2 | 3+4 2 | 2+2 2 |
| | 5+7 2 | 5+7 2 | 6+8 2 | 3+4 2 | 2+2 2 |
| | 5+7 2 | 5+7 2 | 5+7 2 | 3+4 2 | 2+2 2 |
| 87.1.6 | 57076 | | 140 | Ge4Li4Yb5 | |
| | -GM | -M | -P | -X | -N |
| | 6+8 2 | 6+8 2 | 6+8 2 | 3+4 2 | 3+3 2 |
| | 5+7 2 | 11+9 2 | 5+7 2 | 5+6 2 | 2+2 2 |
| | 11+9 2 | 10+12 2 | 6+8 2 | 5+6 2 | 2+2 2 |
| | 10+12 2 | 5+7 2 | 5+7 2 | 3+4 2 | 2+2 2 |
| | 6+8 2 | 6+8 2 | 5+7 2 | 3+4 2 | 3+3 2 |
| | 10+12 2 | 10+12 2 | 6+8 2 | 5+6 2 | 3+3 2 |
| | 5+7 2 | 11+9 2 | 6+8 2 | 5+6 2 | 2+2 2 |
| | 6+8 2 | 5+7 2 | 6+8 2 | 3+4 2 | 3+3 2 |
| | 11+9 2 | 6+8 2 | 5+7 2 | 3+4 2 | 3+3 2 |
| | 10+12 2 | 10+12 2 | 5+7 2 | 5+6 2 | 2+2 2 |
| | 10+12 2 | 10+12 2 | 6+8 2 | 5+6 2 | 3+3 2 |
| | 11+9 2 | 11+9 2 | 5+7 2 | 5+6 2 | 3+3 2 |
| | 6+8 2 | 6+8 2 | 5+7 2 | 3+4 2 | 2+2 2 |
| | 5+7 2 | 10+12 2 | 5+7 2 | 5+6 2 | 3+3 2 |
| | 10+12 2 | 5+7 2 | 6+8 2 | 3+4 2 | 2+2 2 |
| | 5+7 2 | 11+9 2 | 6+8 2 | 3+4 2 | 3+3 2 |
| | 6+8 2 | 5+7 2 | 5+7 2 | 5+6 2 | 2+2 2 |
| | 11+9 2 | 6+8 2 | 6+8 2 | 3+4 2 | 2+2 2 |
| | 11+9 2 | 11+9 2 | 6+8 2 | 5+6 2 | 3+3 2 |
| | 10+12 2 | 10+12 2 | 5+7 2 | 5+6 2 | 3+3 2 |
| | 6+8 2 | 6+8 2 | 5+7 2 | 3+4 2 | 2+2 2 |
| | 11+9 2 | 10+12 2 | 6+8 2 | 5+6 2 | 2+2 2 |
| | 10+12 2 | 11+9 2 | 6+8 2 | 5+6 2 | 3+3 2 |
| | 5+7 2 | 5+7 2 | 5+7 2 | 3+4 2 | 3+3 2 |
| | 5+7 2 | 6+8 2 | 6+8 2 | 3+4 2 | 3+3 2 |

Continued on next column

**TABLE LI– continued**

| TQC | ICSD | | NoE | SF | |
|---|---|---|---|---|---|
| | 6+8 2 | 5+7 2 | 6+8 2 | 5+6 2 | 2+2 2 |
| | 10+12 2 | 11+9 2 | 5+7 2 | 3+4 2 | 2+2 2 |
| | 11+9 2 | 10+12 2 | 5+7 2 | 5+6 2 | 2+2 2 |
| | 6+8 2 | 10+12 2 | 5+7 2 | 3+4 2 | 3+3 2 |
| | 11+9 2 | 6+8 2 | 6+8 2 | 5+6 2 | 2+2 2 |
| | 10+12 2 | 5+7 2 | 5+7 2 | 3+4 2 | 3+3 2 |
| | 11+9 2 | 11+9 2 | 6+8 2 | 5+6 2 | 3+3 2 |
| | 6+8 2 | 6+8 2 | 5+7 2 | 3+4 2 | 2+2 2 |
| | 5+7 2 | 11+9 2 | 6+8 2 | 5+6 2 | 3+3 2 |
| | 11+9 2 | 10+12 2 | 5+7 2 | 5+6 2 | 3+3 2 |
| | 10+12 2 | 5+7 2 | 6+8 2 | 5+6 2 | 2+2 2 |
| | 5+7 2 | 11+9 2 | 5+7 2 | 3+4 2 | 3+3 2 |
| | 11+9 2 | 10+12 2 | 6+8 2 | 5+6 2 | 3+3 2 |
| | 6+8 2 | 11+9 2 | 6+8 2 | 3+4 2 | 2+2 2 |
| | 10+12 2 | 5+7 2 | 5+7 2 | 5+6 2 | 3+3 2 |
| | 5+7 2 | 6+8 2 | 5+7 2 | 3+4 2 | 2+2 2 |
| | 11+9 2 | 10+12 2 | 6+8 2 | 5+6 2 | 3+3 2 |
| | 10+12 2 | 11+9 2 | 5+7 2 | 5+6 2 | 3+3 2 |
| | 6+8 2 | 5+7 2 | 5+7 2 | 3+4 2 | 2+2 2 |
| | 11+9 2 | 10+12 2 | 5+7 2 | 5+6 2 | 2+2 2 |
| | 6+8 2 | 11+9 2 | 6+8 2 | 3+4 2 | 3+3 2 |
| | 5+7 2 | 6+8 2 | 6+8 2 | 3+4 2 | 3+3 2 |
| | 10+12 2 | 5+7 2 | 6+8 2 | 5+6 2 | 2+2 2 |
| | 5+7 2 | 6+8 2 | 6+8 2 | 3+4 2 | 3+3 2 |
| | 6+8 2 | 11+9 2 | 5+7 2 | 5+6 2 | 2+2 2 |
| | 11+9 2 | 10+12 2 | 5+7 2 | 5+6 2 | 3+3 2 |
| | 10+12 2 | 6+8 2 | 6+8 2 | 3+4 2 | 3+3 2 |
| | 11+9 2 | 10+12 2 | 5+7 2 | 5+6 2 | 3+3 2 |
| | 10+12 2 | 11+9 2 | 6+8 2 | 5+6 2 | 2+2 2 |
| | 10+12 2 | 5+7 2 | 5+7 2 | 5+6 2 | 3+3 2 |
| | 5+7 2 | 10+12 2 | 5+7 2 | 3+4 2 | 2+2 2 |
| | 11+9 2 | 5+7 2 | 5+7 2 | 5+6 2 | 3+3 2 |
| | 6+8 2 | 6+8 2 | 6+8 2 | 3+4 2 | 2+2 2 |
| | 5+7 2 | 11+9 2 | 5+7 2 | 5+6 2 | 3+3 2 |
| | 11+9 2 | 10+12 2 | 6+8 2 | 3+4 2 | 2+2 2 |
| | 10+12 2 | 5+7 2 | 6+8 2 | 5+6 2 | 3+3 2 |
| | 11+9 2 | 11+9 2 | 5+7 2 | 3+4 2 | 2+2 2 |
| | 10+12 2 | 6+8 2 | 6+8 2 | 5+6 2 | 3+3 2 |
| | 6+8 2 | 11+9 2 | 5+7 2 | 5+6 2 | 2+2 2 |
| | 11+9 2 | 10+12 2 | 5+7 2 | 3+4 2 | 3+3 2 |
| | 6+8 2 | 6+8 2 | 6+8 2 | 3+4 2 | 3+3 2 |
| | 5+7 2 | 11+9 2 | 6+8 2 | 5+6 2 | 2+2 2 |
| | 6+8 2 | 10+12 2 | 5+7 2 | 5+6 2 | 3+3 2 |
| | 5+7 2 | 5+7 2 | 5+7 2 | 3+4 2 | 2+2 2 |
| | 11+9 2 | 11+9 2 | 6+8 2 | 3+4 2 | 3+3 2 |
| 87.1.7 | 173488 | | 72 | Co1O6Re1Sr2 | |
| | -GM | -M | -P | -X | -N |
| | 6+8 2 | 10+12 2 | 5+7 2 | 5+6 2 | 2+2 2 |
| | 11+9 2 | 5+7 2 | 6+8 2 | 3+4 2 | 3+3 2 |
| | 6+8 2 | 6+8 2 | 5+7 2 | 3+4 2 | 2+2 2 |
| | 5+7 2 | 6+8 2 | 6+8 2 | 3+4 2 | 3+3 2 |
| | 6+8 2 | 5+7 2 | 5+7 2 | 5+6 2 | 2+2 2 |
| | 10+12 2 | 11+9 2 | 6+8 2 | 5+6 2 | 2+2 2 |
| | 11+9 2 | 10+12 2 | 5+7 2 | 3+4 2 | 3+3 2 |
| | 10+12 2 | 10+12 2 | 6+8 2 | 5+6 2 | 3+3 2 |
| | 5+7 2 | 6+8 2 | 5+7 2 | 3+4 2 | 2+2 2 |
| | 10+12 2 | 11+9 2 | 6+8 2 | 5+6 2 | 3+3 2 |
| | 6+8 2 | 5+7 2 | 5+7 2 | 3+4 2 | 2+2 2 |
| | 5+7 2 | 11+9 2 | 6+8 2 | 5+6 2 | 3+3 2 |
| | 10+12 2 | 10+12 2 | 5+7 2 | 3+4 2 | 2+2 2 |
| | 11+9 2 | 6+8 2 | 6+8 2 | 5+6 2 | 3+3 2 |

Continued on next column

**TABLE LI– continued**

| TQC | ICSD | | NoE | SF | |
|---|---|---|---|---|---|
| | 6+8 2 | 6+8 2 | 6+8 2 | 3+4 2 | 2+2 2 |
| | 5+7 2 | 5+7 2 | 5+7 2 | 3+4 2 | 2+2 2 |
| | 6+8 2 | 5+7 2 | 5+7 2 | 3+4 2 | 2+2 2 |
| | 5+7 2 | 6+8 2 | 6+8 2 | 3+4 2 | 2+2 2 |
| | 6+8 2 | 6+8 2 | 5+7 2 | 3+4 2 | 3+3 2 |
| | 5+7 2 | 5+7 2 | 6+8 2 | 3+4 2 | 2+2 2 |
| | 10+12 2 | 10+12 2 | 6+8 2 | 5+6 2 | 2+2 2 |
| | 10+12 2 | 11+9 2 | 6+8 2 | 5+6 2 | 3+3 2 |
| | 11+9 2 | 10+12 2 | 5+7 2 | 5+6 2 | 3+3 2 |
| | 11+9 2 | 10+12 2 | 6+8 2 | 5+6 2 | 3+3 2 |
| | 10+12 2 | 6+8 2 | 5+7 2 | 5+6 2 | 3+3 2 |
| | 11+9 2 | 5+7 2 | 5+7 2 | 3+4 2 | 3+3 2 |
| | 10+12 2 | 11+9 2 | 6+8 2 | 3+4 2 | 3+3 2 |
| | 10+12 2 | 11+9 2 | 5+7 2 | 5+6 2 | 3+3 2 |
| | 11+9 2 | 10+12 2 | 6+8 2 | 5+6 2 | 2+2 2 |
| | 6+8 2 | 10+12 2 | 5+7 2 | 5+6 2 | 3+3 2 |
| | 6+8 2 | 11+9 2 | 6+8 2 | 5+6 2 | 2+2 2 |
| | 5+7 2 | 6+8 2 | 5+7 2 | 3+4 2 | 2+2 2 |
| | 5+7 2 | 5+7 2 | 6+8 2 | 3+4 2 | 3+3 2 |
| | 6+8 2 | 6+8 2 | 5+7 2 | 3+4 2 | 3+3 2 |
| | 5+7 2 | 5+7 2 | 5+7 2 | 3+4 2 | 3+3 2 |
| | 5+7 2 | 5+7 2 | 5+7 2 | 3+4 2 | 2+2 2 |
| 87.1.8 | 60822 | | 54 | C4Cr4U1 | |
| | -GM | -M | -P | -X | -N |
| | 6+8 2 | 6+8 2 | 6+8 2 | 3+4 2 | 3+3 2 |
| | 10+12 2 | 10+12 2 | 5+7 2 | 5+6 2 | 2+2 2 |
| | 11+9 2 | 11+9 2 | 5+7 2 | 5+6 2 | 2+2 2 |
| | 10+12 2 | 10+12 2 | 6+8 2 | 5+6 2 | 2+2 2 |
| | 6+8 2 | 5+7 2 | 6+8 2 | 3+4 2 | 3+3 2 |
| | 5+7 2 | 6+8 2 | 5+7 2 | 5+6 2 | 3+3 2 |
| | 10+12 2 | 10+12 2 | 6+8 2 | 3+4 2 | 2+2 2 |
| | 11+9 2 | 11+9 2 | 5+7 2 | 5+6 2 | 2+2 2 |
| | 6+8 2 | 6+8 2 | 6+8 2 | 3+4 2 | 3+3 2 |
| | 5+7 2 | 5+7 2 | 5+7 2 | 5+6 2 | 2+2 2 |
| | 10+12 2 | 10+12 2 | 6+8 2 | 3+4 2 | 3+3 2 |
| | 11+9 2 | 11+9 2 | 5+7 2 | 5+6 2 | 3+3 2 |
| | 10+12 2 | 11+9 2 | 5+7 2 | 3+4 2 | 3+3 2 |
| | 6+8 2 | 5+7 2 | 6+8 2 | 3+4 2 | 2+2 2 |
| | 10+12 2 | 6+8 2 | 5+7 2 | 5+6 2 | 2+2 2 |
| | 5+7 2 | 6+8 2 | 6+8 2 | 3+4 2 | 3+3 2 |
| | 6+8 2 | 10+12 2 | 5+7 2 | 5+6 2 | 3+3 2 |
| | 5+7 2 | 5+7 2 | 5+7 2 | 5+6 2 | 2+2 2 |
| | 11+9 2 | 11+9 2 | 6+8 2 | 5+6 2 | 2+2 2 |
| | 11+9 2 | 10+12 2 | 6+8 2 | 3+4 2 | 2+2 2 |
| | 6+8 2 | 6+8 2 | 6+8 2 | 5+6 2 | 3+3 2 |
| | 5+7 2 | 5+7 2 | 5+7 2 | 3+4 2 | 2+2 2 |
| | 10+12 2 | 10+12 2 | 5+7 2 | 5+6 2 | 3+3 2 |
| | 6+8 2 | 11+9 2 | 6+8 2 | 3+4 2 | 2+2 2 |
| | 5+7 2 | 10+12 2 | 6+8 2 | 3+4 2 | 2+2 2 |
| | 11+9 2 | 11+9 2 | 6+8 2 | 5+6 2 | 3+3 2 |
| | 6+8 2 | 10+12 2 | 5+7 2 | 5+6 2 | 2+2 2 |
| 87.1.9 | 40319 | | 54 | C4Cr4U1 | |
| | -GM | -M | -P | -X | -N |
| | 6+8 2 | 6+8 2 | 6+8 2 | 3+4 2 | 3+3 2 |
| | 10+12 2 | 10+12 2 | 5+7 2 | 5+6 2 | 2+2 2 |
| | 11+9 2 | 11+9 2 | 5+7 2 | 5+6 2 | 2+2 2 |
| | 10+12 2 | 10+12 2 | 6+8 2 | 5+6 2 | 2+2 2 |
| | 6+8 2 | 5+7 2 | 6+8 2 | 3+4 2 | 3+3 2 |
| | 5+7 2 | 6+8 2 | 5+7 2 | 5+6 2 | 3+3 2 |
| | 10+12 2 | 10+12 2 | 6+8 2 | 3+4 2 | 2+2 2 |
| | 11+9 2 | 11+9 2 | 5+7 2 | 5+6 2 | 2+2 2 |

Continued on next column

**TABLE LI– continued**

| TQC | ICSD | | NoE | SF | |
|---|---|---|---|---|---|
| | 6+8 2 | 6+8 2 | 6+8 2 | 3+4 2 | 3+3 2 |
| | 5+7 2 | 5+7 2 | 5+7 2 | 5+6 2 | 3+3 2 |
| | 10+12 2 | 10+12 2 | 6+8 2 | 3+4 2 | 2+2 2 |
| | 11+9 2 | 11+9 2 | 5+7 2 | 5+6 2 | 3+3 2 |
| | 10+12 2 | 11+9 2 | 5+7 2 | 3+4 2 | 3+3 2 |
| | 6+8 2 | 5+7 2 | 6+8 2 | 3+4 2 | 2+2 2 |
| | 10+12 2 | 6+8 2 | 6+8 2 | 5+6 2 | 2+2 2 |
| | 5+7 2 | 6+8 2 | 5+7 2 | 3+4 2 | 3+3 2 |
| | 11+9 2 | 5+7 2 | 5+7 2 | 5+6 2 | 3+3 2 |
| | 6+8 2 | 10+12 2 | 5+7 2 | 5+6 2 | 2+2 2 |
| | 5+7 2 | 11+9 2 | 6+8 2 | 5+6 2 | 2+2 2 |
| | 11+9 2 | 10+12 2 | 6+8 2 | 3+4 2 | 2+2 2 |
| | 6+8 2 | 6+8 2 | 5+7 2 | 5+6 2 | 3+3 2 |
| | 5+7 2 | 5+7 2 | 6+8 2 | 3+4 2 | 2+2 2 |
| | 10+12 2 | 10+12 2 | 5+7 2 | 5+6 2 | 3+3 2 |
| | 6+8 2 | 11+9 2 | 6+8 2 | 3+4 2 | 2+2 2 |
| | 5+7 2 | 10+12 2 | 6+8 2 | 3+4 2 | 2+2 2 |
| | 11+9 2 | 11+9 2 | 5+7 2 | 5+6 2 | 3+3 2 |
| | 6+8 2 | 10+12 2 | 6+8 2 | 5+6 2 | 2+2 2 |
| 87.1.10 | 43186 | | 50 | As4Mo5 | |
| | -GM | -M | -P | -X | -N |
| | 6+8 2 | 5+7 2 | 5+7 2 | 5+6 2 | 3+3 2 |
| | 11+9 2 | 10+12 2 | 5+7 2 | 3+4 2 | 2+2 2 |
| | 10+12 2 | 11+9 2 | 6+8 2 | 3+4 2 | 2+2 2 |
| | 5+7 2 | 6+8 2 | 6+8 2 | 5+6 2 | 3+3 2 |
| | 5+7 2 | 5+7 2 | 5+7 2 | 5+6 2 | 2+2 2 |
| | 6+8 2 | 6+8 2 | 5+7 2 | 3+4 2 | 3+3 2 |
| | 11+9 2 | 5+7 2 | 6+8 2 | 3+4 2 | 3+3 2 |
| | 10+12 2 | 11+9 2 | 6+8 2 | 3+4 2 | 3+3 2 |
| | 6+8 2 | 10+12 2 | 6+8 2 | 5+6 2 | 2+2 2 |
| | 5+7 2 | 6+8 2 | 6+8 2 | 3+4 2 | 3+3 2 |
| | 10+12 2 | 10+12 2 | 5+7 2 | 5+6 2 | 3+3 2 |
| | 11+9 2 | 11+9 2 | 5+7 2 | 3+4 2 | 2+2 2 |
| | 11+9 2 | 11+9 2 | 6+8 2 | 5+6 2 | 2+2 2 |
| | 5+7 2 | 5+7 2 | 5+7 2 | 3+4 2 | 3+3 2 |
| | 6+8 2 | 6+8 2 | 6+8 2 | 5+6 2 | 2+2 2 |
| | 6+8 2 | 5+7 2 | 5+7 2 | 3+4 2 | 3+3 2 |
| | 10+12 2 | 6+8 2 | 5+7 2 | 5+6 2 | 2+2 2 |
| | 11+9 2 | 11+9 2 | 6+8 2 | 3+4 2 | 3+3 2 |
| | 5+7 2 | 10+12 2 | 6+8 2 | 3+4 2 | 3+3 2 |
| | 6+8 2 | 10+12 2 | 5+7 2 | 5+6 2 | 2+2 2 |
| | 6+8 2 | 6+8 2 | 5+7 2 | 3+4 2 | 2+2 2 |
| | 5+7 2 | 6+8 2 | 6+8 2 | 5+6 2 | 3+3 2 |
| | 10+12 2 | 5+7 2 | 5+7 2 | 3+4 2 | 2+2 2 |
| | 5+7 2 | 10+12 2 | 6+8 2 | 5+6 2 | 3+3 2 |
| | 10+12 2 | 11+9 2 | 5+7 2 | 5+6 2 | 2+2 2 |
| 87.1.11 | 66017 | | 118 | As4Ba1Cu8 | |
| | -GM | -M | -P | -X | -N |
| | 6+8 2 | 6+8 2 | 6+8 2 | 3+4 2 | 3+3 2 |
| | 10+12 2 | 10+12 2 | 5+7 2 | 5+6 2 | 2+2 2 |
| | 11+9 2 | 11+9 2 | 5+7 2 | 5+6 2 | 2+2 2 |
| | 10+12 2 | 10+12 2 | 6+8 2 | 5+6 2 | 2+2 2 |
| | 6+8 2 | 5+7 2 | 5+7 2 | 3+4 2 | 3+3 2 |
| | 5+7 2 | 10+12 2 | 6+8 2 | 5+6 2 | 3+3 2 |
| | 10+12 2 | 11+9 2 | 5+7 2 | 3+4 2 | 2+2 2 |
| | 11+9 2 | 6+8 2 | 6+8 2 | 5+6 2 | 2+2 2 |
| | 5+7 2 | 5+7 2 | 6+8 2 | 5+6 2 | 2+2 2 |
| | 6+8 2 | 10+12 2 | 5+7 2 | 3+4 2 | 3+3 2 |
| | 11+9 2 | 11+9 2 | 6+8 2 | 5+6 2 | 3+3 2 |
| | 10+12 2 | 6+8 2 | 5+7 2 | 3+4 2 | 2+2 2 |
| | 6+8 2 | 6+8 2 | 5+7 2 | 3+4 2 | 3+3 2 |

Continued on next column

**TABLE LI– continued**

| TQC | ICSD | | NoE | SF | |
|---|---|---|---|---|---|
| | 10+12 2 | 11+9 2 | 6+8 2 | 3+4 2 | 2+2 2 |
| | 11+9 2 | 5+7 2 | 6+8 2 | 5+6 2 | 3+3 2 |
| | 10+12 2 | 6+8 2 | 5+7 2 | 5+6 2 | 2+2 2 |
| | 5+7 2 | 5+7 2 | 5+7 2 | 5+6 2 | 3+3 2 |
| | 6+8 2 | 11+9 2 | 6+8 2 | 3+4 2 | 2+2 2 |
| | 5+7 2 | 10+12 2 | 6+8 2 | 5+6 2 | 3+3 2 |
| | 11+9 2 | 10+12 2 | 5+7 2 | 3+4 2 | 2+2 2 |
| | 6+8 2 | 6+8 2 | 5+7 2 | 5+6 2 | 2+2 2 |
| | 11+9 2 | 11+9 2 | 5+7 2 | 3+4 2 | 3+3 2 |
| | 6+8 2 | 5+7 2 | 6+8 2 | 3+4 2 | 3+3 2 |
| | 5+7 2 | 11+9 2 | 6+8 2 | 5+6 2 | 2+2 2 |
| | 10+12 2 | 10+12 2 | 5+7 2 | 3+4 2 | 3+3 2 |
| | 11+9 2 | 10+12 2 | 6+8 2 | 3+4 2 | 2+2 2 |
| | 10+12 2 | 5+7 2 | 5+7 2 | 5+6 2 | 3+3 2 |
| | 5+7 2 | 6+8 2 | 6+8 2 | 5+6 2 | 2+2 2 |
| | 10+12 2 | 10+12 2 | 5+7 2 | 3+4 2 | 2+2 2 |
| | 6+8 2 | 5+7 2 | 6+8 2 | 5+6 2 | 3+3 2 |
| | 11+9 2 | 6+8 2 | 6+8 2 | 5+6 2 | 3+3 2 |
| | 5+7 2 | 6+8 2 | 5+7 2 | 3+4 2 | 2+2 2 |
| | 6+8 2 | 10+12 2 | 6+8 2 | 5+6 2 | 3+3 2 |
| | 5+7 2 | 6+8 2 | 5+7 2 | 3+4 2 | 2+2 2 |
| | 11+9 2 | 5+7 2 | 5+7 2 | 5+6 2 | 3+3 2 |
| | 5+7 2 | 6+8 2 | 6+8 2 | 5+6 2 | 2+2 2 |
| | 6+8 2 | 11+9 2 | 6+8 2 | 5+6 2 | 3+3 2 |
| | 10+12 2 | 5+7 2 | 5+7 2 | 3+4 2 | 2+2 2 |
| | 11+9 2 | 11+9 2 | 5+7 2 | 3+4 2 | 3+3 2 |
| | 10+12 2 | 10+12 2 | 6+8 2 | 5+6 2 | 2+2 2 |
| | 6+8 2 | 5+7 2 | 6+8 2 | 3+4 2 | 2+2 2 |
| | 10+12 2 | 11+9 2 | 5+7 2 | 5+6 2 | 3+3 2 |
| | 5+7 2 | 6+8 2 | 5+7 2 | 3+4 2 | 3+3 2 |
| | 11+9 2 | 11+9 2 | 6+8 2 | 3+4 2 | 2+2 2 |
| | 6+8 2 | 10+12 2 | 6+8 2 | 5+6 2 | 3+3 2 |
| | 6+8 2 | 11+9 2 | 5+7 2 | 3+4 2 | 2+2 2 |
| | 11+9 2 | 10+12 2 | 5+7 2 | 3+4 2 | 3+3 2 |
| | 5+7 2 | 5+7 2 | 6+8 2 | 5+6 2 | 2+2 2 |
| | 5+7 2 | 11+9 2 | 5+7 2 | 3+4 2 | 3+3 2 |
| | 11+9 2 | 6+8 2 | 6+8 2 | 5+6 2 | 2+2 2 |
| | 10+12 2 | 10+12 2 | 5+7 2 | 5+6 2 | 3+3 2 |
| | 10+12 2 | 5+7 2 | 6+8 2 | 3+4 2 | 2+2 2 |
| | 6+8 2 | 10+12 2 | 5+7 2 | 3+4 2 | 2+2 2 |
| | 5+7 2 | 5+7 2 | 6+8 2 | 3+4 2 | 3+3 2 |
| | 11+9 2 | 11+9 2 | 5+7 2 | 5+6 2 | 3+3 2 |
| | 5+7 2 | 10+12 2 | 6+8 2 | 5+6 2 | 2+2 2 |
| | 6+8 2 | 6+8 2 | 6+8 2 | 5+6 2 | 3+3 2 |
| | 6+8 2 | 5+7 2 | 5+7 2 | 3+4 2 | 2+2 2 |
| | 10+12 2 | 6+8 2 | 6+8 2 | 5+6 2 | 2+2 2 |

The End

TABLE LII: The computed irreducible representations for the valence bands for the compounds of SEBR and NLC in space group 113. See the list of tables.

| TQC | ICSD | | NoE | | SF | |
|---|---|---|---|---|---|---|
| 113.2.1 | 246245 | | 44 | | Ca1Fe1O2 | |
| | -A | -GM | -M | -Z | -R | -X |
| | 6+7 4 | 7 2 | 6+7 4 | 7 2 | 2+5 2 | 2+5 2 |
| | 6+7 4 | 6 2 | 6+7 4 | 6 2 | 3+4 2 | 3+4 2 |
| | 6+7 4 | 7 2 | 6+7 4 | 7 2 | 2+5 2 | 2+5 2 |
| | 6+7 4 | 6 2 | 6+7 4 | 6 2 | 3+4 2 | 3+4 2 |
| | 6+7 4 | 7 2 | 6+7 4 | 7 2 | 2+5 2 | 2+5 2 |

Continued on next column

**TABLE LII– continued**

| TQC | ICSD | | NoE | | SF | |
|---|---|---|---|---|---|---|
| | 6+7 4 | 6 2 | 6+7 4 | 6 2 | 3+4 2 | 3+4 2 |
| | 6+7 4 | 6 2 | 6+7 4 | 6 2 | 2+5 2 | 2+5 2 |
| | 6+7 4 | 6 2 | 6+7 4 | 6 2 | 3+4 2 | 3+4 2 |
| | 6+7 4 | 7 2 | 6+7 4 | 6 2 | 3+4 2 | 3+4 2 |
| | 6+7 4 | 7 2 | 6+7 4 | 7 2 | 2+5 2 | 2+5 2 |
| | 6+7 4 | 6 2 | 6+7 4 | 6 2 | 3+4 2 | 3+4 2 |
| | | 7 2 | | 7 2 | 2+5 2 | 2+5 2 |
| | | 6 2 | | 7 2 | 2+5 2 | 3+4 2 |
| | | 6 2 | | 6 2 | 3+4 2 | 2+5 2 |
| | | 7 2 | | 7 2 | 2+5 2 | 2+5 2 |
| | | 7 2 | | 7 2 | 3+4 2 | 3+4 2 |
| | | 7 2 | | 6 2 | 3+4 2 | 2+5 2 |
| | | 6 2 | | 7 2 | 2+5 2 | 3+4 2 |
| | | 7 2 | | 6 2 | 2+5 2 | 3+4 2 |
| | | 7 2 | | 6 2 | 3+4 2 | 2+5 2 |
| | | 6 2 | | 7 2 | 3+4 2 | 2+5 2 |
| | | 6 2 | | 6 2 | 2+5 2 | 3+4 2 |
| 113.2.2 | 39120 | | 172 | Be2Gd2Ge1O7 | | |
| | -A | -GM | -M | -Z | -R | -X |
| | 6+7 4 | 7 2 | 6+7 4 | 6 2 | 2+5 2 | 2+5 2 |
| | 6+7 4 | 6 2 | 6+7 4 | 7 2 | 3+4 2 | 3+4 2 |
| | 6+7 4 | 6 2 | 6+7 4 | 6 2 | 2+5 2 | 2+5 2 |
| | 6+7 4 | 7 2 | 6+7 4 | 7 2 | 3+4 2 | 3+4 2 |
| | 6+7 4 | 6 2 | 6+7 4 | 7 2 | 2+5 2 | 3+4 2 |
| | 6+7 4 | 7 2 | 6+7 4 | 6 2 | 3+4 2 | 2+5 2 |
| | 6+7 4 | 7 2 | 6+7 4 | 6 2 | 3+4 2 | 2+5 2 |
| | 6+7 4 | 6 2 | 6+7 4 | 7 2 | 2+5 2 | 3+4 2 |
| | 6+7 4 | 7 2 | 6+7 4 | 7 2 | 3+4 2 | 2+5 2 |
| | 6+7 4 | 7 2 | 6+7 4 | 7 2 | 2+5 2 | 3+4 2 |
| | 6+7 4 | 6 2 | 6+7 4 | 7 2 | 3+4 2 | 2+5 2 |
| | 6+7 4 | 6 2 | 6+7 4 | 6 2 | 2+5 2 | 3+4 2 |
| | 6+7 4 | 7 2 | 6+7 4 | 6 2 | 2+5 2 | 3+4 2 |
| | 6+7 4 | 6 2 | 6+7 4 | 7 2 | 3+4 2 | 2+5 2 |
| | 6+7 4 | 7 2 | 6+7 4 | 6 2 | 2+5 2 | 3+4 2 |
| | 6+7 4 | 7 2 | 6+7 4 | 6 2 | 3+4 2 | 2+5 2 |
| | 6+7 4 | 6 2 | 6+7 4 | 7 2 | 2+5 2 | 2+5 2 |
| | 6+7 4 | 6 2 | 6+7 4 | 6 2 | 3+4 2 | 3+4 2 |
| | 6+7 4 | 6 2 | 6+7 4 | 7 2 | 2+5 2 | 3+4 2 |
| | 6+7 4 | 7 2 | 6+7 4 | 7 2 | 3+4 2 | 2+5 2 |
| | 6+7 4 | 7 2 | 6+7 4 | 6 2 | 2+5 2 | 3+4 2 |
| | 6+7 4 | 6 2 | 6+7 4 | 6 2 | 3+4 2 | 2+5 2 |
| | 6+7 4 | 7 2 | 6+7 4 | 7 2 | 3+4 2 | 2+5 2 |
| | 6+7 4 | 6 2 | 6+7 4 | 6 2 | 2+5 2 | 3+4 2 |
| | 6+7 4 | 7 2 | 6+7 4 | 6 2 | 3+4 2 | 2+5 2 |
| | 6+7 4 | 6 2 | 6+7 4 | 7 2 | 2+5 2 | 3+4 2 |
| | 6+7 4 | 7 2 | 6+7 4 | 6 2 | 2+5 2 | 2+5 2 |
| | 6+7 4 | 7 2 | 6+7 4 | 6 2 | 3+4 2 | 3+4 2 |
| | 6+7 4 | 6 2 | 6+7 4 | 7 2 | 2+5 2 | 2+5 2 |
| | 6+7 4 | 6 2 | 6+7 4 | 7 2 | 3+4 2 | 3+4 2 |
| | 6+7 4 | 7 2 | 6+7 4 | 7 2 | 2+5 2 | 3+4 2 |
| | 6+7 4 | 7 2 | 6+7 4 | 7 2 | 3+4 2 | 2+5 2 |
| | 6+7 4 | 6 2 | 6+7 4 | 6 2 | 3+4 2 | 3+4 2 |
| | 6+7 4 | 6 2 | 6+7 4 | 6 2 | 2+5 2 | 2+5 2 |
| | 6+7 4 | 7 2 | 6+7 4 | 7 2 | 3+4 2 | 2+5 2 |
| | 6+7 4 | 6 2 | 6+7 4 | 6 2 | 2+5 2 | 3+4 2 |
| | 6+7 4 | 7 2 | 6+7 4 | 6 2 | 2+5 2 | 2+5 2 |
| | 6+7 4 | 6 2 | 6+7 4 | 7 2 | 3+4 2 | 3+4 2 |
| | 6+7 4 | 7 2 | 6+7 4 | 6 2 | 2+5 2 | 2+5 2 |
| | 6+7 4 | 6 2 | 6+7 4 | 7 2 | 3+4 2 | 3+4 2 |
| | 6+7 4 | 7 2 | 6+7 4 | 7 2 | 2+5 2 | 3+4 2 |
| | 6+7 4 | 6 2 | 6+7 4 | 6 2 | 3+4 2 | 2+5 2 |
| | 6+7 4 | 7 2 | 6+7 4 | 7 2 | 2+5 2 | 2+5 2 |
| | | 6 2 | | 6 2 | 3+4 2 | 3+4 2 |
| | | 6 2 | | 7 2 | 3+4 2 | 3+4 2 |
| | | 7 2 | | 6 2 | 2+5 2 | 2+5 2 |
| | | 7 2 | | 7 2 | 3+4 2 | 3+4 2 |
| | | 6 2 | | 6 2 | 2+5 2 | 2+5 2 |
| | | 7 2 | | 7 2 | 2+5 2 | 2+5 2 |
| | | 6 2 | | 6 2 | 3+4 2 | 3+4 2 |
| | | 6 2 | | 6 2 | 2+5 2 | 3+4 2 |
| | | 6 2 | | 7 2 | 3+4 2 | 2+5 2 |
| | | 7 2 | | 6 2 | 2+5 2 | 2+5 2 |
| | | 6 2 | | 6 2 | 3+4 2 | 3+4 2 |
| | | 7 2 | | 7 2 | 3+4 2 | 2+5 2 |
| | | 6 2 | | 6 2 | 2+5 2 | 3+4 2 |
| | | 7 2 | | 6 2 | 2+5 2 | 2+5 2 |
| | | 7 2 | | 7 2 | 3+4 2 | 3+4 2 |
| | | 6 2 | | 7 2 | 3+4 2 | 2+5 2 |
| | | 7 2 | | 7 2 | 2+5 2 | 3+4 2 |
| | | 6 2 | | 7 2 | 3+4 2 | 2+5 2 |
| | | 7 2 | | 6 2 | 2+5 2 | 3+4 2 |
| | | 6 2 | | 7 2 | 3+4 2 | 2+5 2 |
| | | 7 2 | | 7 2 | 2+5 2 | 3+4 2 |
| | | 6 2 | | 6 2 | 2+5 2 | 3+4 2 |
| | | 7 2 | | 7 2 | 3+4 2 | 2+5 2 |
| | | 7 2 | | 7 2 | 3+4 2 | 2+5 2 |
| | | 6 2 | | 6 2 | 2+5 2 | 3+4 2 |
| | | 6 2 | | 6 2 | 2+5 2 | 2+5 2 |
| | | 7 2 | | 6 2 | 3+4 2 | 3+4 2 |
| | | 6 2 | | 7 2 | 2+5 2 | 3+4 2 |
| | | 7 2 | | 6 2 | 3+4 2 | 2+5 2 |
| | | 6 2 | | 7 2 | 3+4 2 | 2+5 2 |
| | | 7 2 | | 6 2 | 2+5 2 | 3+4 2 |
| | | 7 2 | | 6 2 | 2+5 2 | 3+4 2 |
| | | 6 2 | | 7 2 | 3+4 2 | 2+5 2 |
| | | 7 2 | | 6 2 | 2+5 2 | 3+4 2 |
| | | 6 2 | | 7 2 | 3+4 2 | 3+4 2 |
| | | 6 2 | | 6 2 | 2+5 2 | 2+5 2 |
| | | 7 2 | | 7 2 | 3+4 2 | 2+5 2 |
| | | 7 2 | | 6 2 | 2+5 2 | 3+4 2 |
| | | 6 2 | | 7 2 | 3+4 2 | 2+5 2 |
| | | 6 2 | | 7 2 | 3+4 2 | 3+4 2 |
| | | 7 2 | | 6 2 | 2+5 2 | 2+5 2 |
| | | 6 2 | | 6 2 | 2+5 2 | 2+5 2 |
| | | 6 2 | | 7 2 | 3+4 2 | 3+4 2 |
| 113.2.3 | 160121 | | 84 | Ag1Tb1Te2 | | |
| | -A | -GM | -M | -Z | -R | -X |
| | 6+7 4 | 6 2 | 6+7 4 | 6 2 | 2+5 2 | 2+5 2 |
| | 6+7 4 | 7 2 | 6+7 4 | 7 2 | 3+4 2 | 3+4 2 |
| | 6+7 4 | 7 2 | 6+7 4 | 7 2 | 3+4 2 | 2+5 2 |
| | 6+7 4 | 6 2 | 6+7 4 | 6 2 | 2+5 2 | 3+4 2 |
| | 6+7 4 | 6 2 | 6+7 4 | 6 2 | 2+5 2 | 2+5 2 |
| | 6+7 4 | 7 2 | 6+7 4 | 7 2 | 3+4 2 | 3+4 2 |
| | 6+7 4 | 7 2 | 6+7 4 | 6 2 | 3+4 2 | 2+5 2 |
| | 6+7 4 | 6 2 | 6+7 4 | 7 2 | 2+5 2 | 3+4 2 |
| | 6+7 4 | 7 2 | 6+7 4 | 6 2 | 3+4 2 | 2+5 2 |
| | 6+7 4 | 6 2 | 6+7 4 | 7 2 | 2+5 2 | 3+4 2 |
| | 6+7 4 | 7 2 | 6+7 4 | 7 2 | 3+4 2 | 3+4 2 |
| | 6+7 4 | 6 2 | 6+7 4 | 6 2 | 2+5 2 | 2+5 2 |
| | 6+7 4 | 6 2 | 6+7 4 | 6 2 | 3+4 2 | 2+5 2 |
| | 6+7 4 | 7 2 | 6+7 4 | 6 2 | 2+5 2 | 3+4 2 |
| | 6+7 4 | 7 2 | 6+7 4 | 7 2 | 2+5 2 | 2+5 2 |

Continued on next column

Continued on next column

**TABLE LII– continued**

| TQC | -A | -GM | -M | -Z | -R | -X |
|---|---|---|---|---|---|---|
| | 6+7 4 | 6 2 | 6+7 4 | 6 2 | 3+4 2 | 3+4 2 |
| | 6+7 4 | 6 2 | 6+7 4 | 6 2 | 3+4 2 | 3+4 2 |
| | 6+7 4 | 7 2 | 6+7 4 | 7 2 | 2+5 2 | 2+5 2 |
| | 6+7 4 | 7 2 | 6+7 4 | 7 2 | 2+5 2 | 3+4 2 |
| | 6+7 4 | 6 2 | 6+7 4 | 6 2 | 3+4 2 | 2+5 2 |
| | 6+7 4 | 6 2 | 6+7 4 | 7 2 | 2+5 2 | 3+4 2 |
| | | 6 2 | | 6 2 | 3+4 2 | 2+5 2 |
| | | 7 2 | | 7 2 | 3+4 2 | 2+5 2 |
| | | 6 2 | | 7 2 | 2+5 2 | 3+4 2 |
| | | 7 2 | | 7 2 | 2+5 2 | 3+4 2 |
| | | 7 2 | | 6 2 | 3+4 2 | 2+5 2 |
| | | 6 2 | | 7 2 | 2+5 2 | 3+4 2 |
| | | 6 2 | | 6 2 | 3+4 2 | 2+5 2 |
| | | 7 2 | | 6 2 | 3+4 2 | 2+5 2 |
| | | 7 2 | | 6 2 | 2+5 2 | 3+4 2 |
| | | 7 2 | | 7 2 | 2+5 2 | 2+5 2 |
| | | 6 2 | | 6 2 | 3+4 2 | 3+4 2 |
| | | 6 2 | | 7 2 | 3+4 2 | 3+4 2 |
| | | 6 2 | | 6 2 | 2+5 2 | 2+5 2 |
| | | 7 2 | | 7 2 | 3+4 2 | 3+4 2 |
| | | 6 2 | | 6 2 | 2+5 2 | 2+5 2 |
| | | 7 2 | | 6 2 | 2+5 2 | 2+5 2 |
| | | 6 2 | | 7 2 | 3+4 2 | 3+4 2 |
| | | 7 2 | | 7 2 | 2+5 2 | 3+4 2 |
| | | 7 2 | | 6 2 | 3+4 2 | 2+5 2 |
| | | 6 2 | | 7 2 | 2+5 2 | 2+5 2 |
| | | 7 2 | | 6 2 | 3+4 2 | 3+4 2 |
| 113.2.4 | 154793 | | 84 | | Ag1Tb1Te2 | |
| | -A | -GM | -M | -Z | -R | -X |
| | 6+7 4 | 6 2 | 6+7 4 | 6 2 | 2+5 2 | 2+5 2 |
| | 6+7 4 | 7 2 | 6+7 4 | 7 2 | 3+4 2 | 3+4 2 |
| | 6+7 4 | 7 2 | 6+7 4 | 6 2 | 3+4 2 | 2+5 2 |
| | 6+7 4 | 6 2 | 6+7 4 | 7 2 | 2+5 2 | 3+4 2 |
| | 6+7 4 | 6 2 | 6+7 4 | 6 2 | 2+5 2 | 2+5 2 |
| | 6+7 4 | 7 2 | 6+7 4 | 7 2 | 3+4 2 | 3+4 2 |
| | 6+7 4 | 7 2 | 6+7 4 | 6 2 | 3+4 2 | 2+5 2 |
| | 6+7 4 | 6 2 | 6+7 4 | 7 2 | 2+5 2 | 3+4 2 |
| | 6+7 4 | 7 2 | 6+7 4 | 6 2 | 3+4 2 | 2+5 2 |
| | 6+7 4 | 6 2 | 6+7 4 | 7 2 | 2+5 2 | 3+4 2 |
| | 6+7 4 | 7 2 | 6+7 4 | 6 2 | 3+4 2 | 3+4 2 |
| | 6+7 4 | 6 2 | 6+7 4 | 7 2 | 2+5 2 | 2+5 2 |
| | 6+7 4 | 6 2 | 6+7 4 | 6 2 | 3+4 2 | 2+5 2 |
| | 6+7 4 | 7 2 | 6+7 4 | 6 2 | 2+5 2 | 3+4 2 |
| | 6+7 4 | 7 2 | 6+7 4 | 7 2 | 2+5 2 | 2+5 2 |
| | 6+7 4 | 6 2 | 6+7 4 | 6 2 | 3+4 2 | 3+4 2 |
| | 6+7 4 | 6 2 | 6+7 4 | 6 2 | 3+4 2 | 3+4 2 |
| | 6+7 4 | 7 2 | 6+7 4 | 7 2 | 2+5 2 | 2+5 2 |
| | 6+7 4 | 7 2 | 6+7 4 | 7 2 | 2+5 2 | 3+4 2 |
| | 6+7 4 | 6 2 | 6+7 4 | 6 2 | 3+4 2 | 2+5 2 |
| | 6+7 4 | 6 2 | 6+7 4 | 7 2 | 2+5 2 | 3+4 2 |
| | | 6 2 | | 6 2 | 3+4 2 | 2+5 2 |
| | | 7 2 | | 7 2 | 3+4 2 | 2+5 2 |
| | | 6 2 | | 7 2 | 2+5 2 | 3+4 2 |
| | | 7 2 | | 7 2 | 2+5 2 | 3+4 2 |
| | | 7 2 | | 6 2 | 3+4 2 | 2+5 2 |
| | | 6 2 | | 7 2 | 2+5 2 | 3+4 2 |
| | | 6 2 | | 6 2 | 3+4 2 | 2+5 2 |
| | | 7 2 | | 6 2 | 3+4 2 | 2+5 2 |
| | | 7 2 | | 6 2 | 2+5 2 | 3+4 2 |
| | | 7 2 | | 7 2 | 2+5 2 | 2+5 2 |
| | | 6 2 | | 6 2 | 3+4 2 | 3+4 2 |

Continued on next column

**TABLE LII– continued**

| TQC | -A | -GM | -M | -Z | -R | -X |
|---|---|---|---|---|---|---|
| | | 6 2 | | 7 2 | 3+4 2 | 3+4 2 |
| | | 6 2 | | 6 2 | 2+5 2 | 2+5 2 |
| | | 7 2 | | 7 2 | 3+4 2 | 3+4 2 |
| | | 6 2 | | 6 2 | 2+5 2 | 2+5 2 |
| | | 7 2 | | 6 2 | 2+5 2 | 2+5 2 |
| | | 6 2 | | 7 2 | 3+4 2 | 3+4 2 |
| | | 7 2 | | 7 2 | 2+5 2 | 3+4 2 |
| | | 7 2 | | 6 2 | 3+4 2 | 2+5 2 |
| | | 6 2 | | 7 2 | 2+5 2 | 2+5 2 |
| | | 7 2 | | 6 2 | 3+4 2 | 3+4 2 |

The End

TABLE LIII: The computed irreducible representations for the valence bands for the compounds of SEBR and NLC in space group 115. See the list of tables.

| TQC | -A | -GM | -M | -Z | -R | -X |
|---|---|---|---|---|---|---|
| 115.1.1 | 93242 | | 34 | Cu1Sb3Zr2 | | |
| | -A | -GM | -M | -Z | -R | -X |
| | 6 2 | 7 2 | 6 2 | 6 2 | 5 2 | 5 2 |
| | 6 2 | 7 2 | 6 2 | 7 2 | 5 2 | 5 2 |
| | 7 2 | 6 2 | 7 2 | 7 2 | 5 2 | 5 2 |
| | 6 2 | 7 2 | 7 2 | 6 2 | 5 2 | 5 2 |
| | 7 2 | 6 2 | 6 2 | 6 2 | 5 2 | 5 2 |
| | 7 2 | 6 2 | 7 2 | 7 2 | 5 2 | 5 2 |
| | 6 2 | 7 2 | 6 2 | 6 2 | 5 2 | 5 2 |
| | 7 2 | 6 2 | 7 2 | 7 2 | 5 2 | 5 2 |
| | 6 2 | 6 2 | 6 2 | 6 2 | 5 2 | 5 2 |
| | 7 2 | 7 2 | 7 2 | 6 2 | 5 2 | 5 2 |
| | 6 2 | 7 2 | 6 2 | 6 2 | 5 2 | 5 2 |
| | 6 2 | 6 2 | 7 2 | 7 2 | 5 2 | 5 2 |
| | 7 2 | 7 2 | 6 2 | 6 2 | 5 2 | 5 2 |
| | 7 2 | 6 2 | 7 2 | 7 2 | 5 2 | 5 2 |
| | 7 2 | 6 2 | 7 2 | 7 2 | 5 2 | 5 2 |
| | 6 2 | 7 2 | 6 2 | 6 2 | 5 2 | 5 2 |
| | 7 2 | 6 2 | 6 2 | 7 2 | 5 2 | 5 2 |
| 115.1.2 | 426313 | | 34 | Cu1Sb3Zr2 | | |
| | -A | -GM | -M | -Z | -R | -X |
| | 7 2 | 7 2 | 7 2 | 6 2 | 5 2 | 5 2 |
| | 6 2 | 7 2 | 6 2 | 7 2 | 5 2 | 5 2 |
| | 7 2 | 6 2 | 7 2 | 7 2 | 5 2 | 5 2 |
| | 7 2 | 7 2 | 6 2 | 6 2 | 5 2 | 5 2 |
| | 6 2 | 6 2 | 7 2 | 6 2 | 5 2 | 5 2 |
| | 6 2 | 6 2 | 6 2 | 7 2 | 5 2 | 5 2 |
| | 7 2 | 7 2 | 7 2 | 6 2 | 5 2 | 5 2 |
| | 6 2 | 6 2 | 6 2 | 7 2 | 5 2 | 5 2 |
| | 7 2 | 6 2 | 7 2 | 6 2 | 5 2 | 5 2 |
| | 6 2 | 7 2 | 6 2 | 6 2 | 5 2 | 5 2 |
| | 7 2 | 7 2 | 7 2 | 6 2 | 5 2 | 5 2 |
| | 7 2 | 6 2 | 6 2 | 7 2 | 5 2 | 5 2 |
| | 6 2 | 7 2 | 7 2 | 6 2 | 5 2 | 5 2 |
| | 6 2 | 6 2 | 6 2 | 7 2 | 5 2 | 5 2 |
| | 6 2 | 6 2 | 6 2 | 7 2 | 5 2 | 5 2 |
| | 7 2 | 7 2 | 7 2 | 6 2 | 5 2 | 5 2 |
| | 6 2 | 6 2 | 7 2 | 7 2 | 5 2 | 5 2 |
| 115.1.3 | 195058 | | 34 | Cu1Sb3Zr2 | | |
| | -A | -GM | -M | -Z | -R | -X |
| | 7 2 | 7 2 | 7 2 | 6 2 | 5 2 | 5 2 |
| | 6 2 | 7 2 | 6 2 | 7 2 | 5 2 | 5 2 |
| | 7 2 | 6 2 | 7 2 | 7 2 | 5 2 | 5 2 |

Continued on next column

**TABLE LIII– continued**

| TQC | ICSD | | NoE | SF | | |
|---|---|---|---|---|---|---|
| | 7 2 | 7 2 | 6 2 | 6 2 | 5 2 | 5 2 |
| | 6 2 | 6 2 | 7 2 | 6 2 | 5 2 | 5 2 |
| | 6 2 | 6 2 | 6 2 | 7 2 | 5 2 | 5 2 |
| | 7 2 | 7 2 | 7 2 | 6 2 | 5 2 | 5 2 |
| | 6 2 | 6 2 | 6 2 | 7 2 | 5 2 | 5 2 |
| | 7 2 | 6 2 | 7 2 | 6 2 | 5 2 | 5 2 |
| | 6 2 | 7 2 | 6 2 | 6 2 | 5 2 | 5 2 |
| | 7 2 | 7 2 | 7 2 | 6 2 | 5 2 | 5 2 |
| | 7 2 | 6 2 | 6 2 | 7 2 | 5 2 | 5 2 |
| | 6 2 | 7 2 | 7 2 | 6 2 | 5 2 | 5 2 |
| | 6 2 | 6 2 | 6 2 | 7 2 | 5 2 | 5 2 |
| | 6 2 | 6 2 | 6 2 | 7 2 | 5 2 | 5 2 |
| | 7 2 | 7 2 | 7 2 | 6 2 | 5 2 | 5 2 |
| | 6 2 | 6 2 | 7 2 | 7 2 | 5 2 | 5 2 |
| 115.1.4 | 93241 | | 34 | Cu1Sb3Ti2 | | |
| | -A | -GM | -M | -Z | -R | -X |
| | 6 2 | 7 2 | 6 2 | 6 2 | 5 2 | 5 2 |
| | 6 2 | 7 2 | 6 2 | 7 2 | 5 2 | 5 2 |
| | 7 2 | 6 2 | 7 2 | 7 2 | 5 2 | 5 2 |
| | 7 2 | 7 2 | 7 2 | 6 2 | 5 2 | 5 2 |
| | 6 2 | 6 2 | 6 2 | 6 2 | 5 2 | 5 2 |
| | 7 2 | 6 2 | 7 2 | 7 2 | 5 2 | 5 2 |
| | 6 2 | 7 2 | 6 2 | 6 2 | 5 2 | 5 2 |
| | 7 2 | 6 2 | 7 2 | 7 2 | 5 2 | 5 2 |
| | 6 2 | 6 2 | 6 2 | 6 2 | 5 2 | 5 2 |
| | 7 2 | 7 2 | 7 2 | 6 2 | 5 2 | 5 2 |
| | 6 2 | 7 2 | 6 2 | 6 2 | 5 2 | 5 2 |
| | 6 2 | 6 2 | 7 2 | 7 2 | 5 2 | 5 2 |
| | 7 2 | 6 2 | 6 2 | 6 2 | 5 2 | 5 2 |
| | 7 2 | 7 2 | 7 2 | 7 2 | 5 2 | 5 2 |
| | 6 2 | 7 2 | 6 2 | 7 2 | 5 2 | 5 2 |
| | 7 2 | 6 2 | 7 2 | 6 2 | 5 2 | 5 2 |
| | 7 2 | 6 2 | 6 2 | 7 2 | 5 2 | 5 2 |
| 115.1.5 | 93243 | | 34 | Cu1Hf2Sb3 | | |
| | -A | -GM | -M | -Z | -R | -X |
| | 6 2 | 7 2 | 6 2 | 6 2 | 5 2 | 5 2 |
| | 6 2 | 7 2 | 7 2 | 7 2 | 5 2 | 5 2 |
| | 7 2 | 6 2 | 6 2 | 7 2 | 5 2 | 5 2 |
| | 7 2 | 7 2 | 7 2 | 6 2 | 5 2 | 5 2 |
| | 6 2 | 6 2 | 6 2 | 6 2 | 5 2 | 5 2 |
| | 7 2 | 6 2 | 7 2 | 7 2 | 5 2 | 5 2 |
| | 6 2 | 7 2 | 6 2 | 6 2 | 5 2 | 5 2 |
| | 7 2 | 6 2 | 7 2 | 7 2 | 5 2 | 5 2 |
| | 6 2 | 6 2 | 6 2 | 6 2 | 5 2 | 5 2 |
| | 7 2 | 7 2 | 7 2 | 6 2 | 5 2 | 5 2 |
| | 6 2 | 7 2 | 6 2 | 6 2 | 5 2 | 5 2 |
| | 7 2 | 6 2 | 7 2 | 7 2 | 5 2 | 5 2 |
| | 6 2 | 7 2 | 6 2 | 6 2 | 5 2 | 5 2 |
| | 7 2 | 6 2 | 7 2 | 7 2 | 5 2 | 5 2 |
| | 6 2 | 7 2 | 6 2 | 7 2 | 5 2 | 5 2 |
| | 7 2 | 6 2 | 7 2 | 6 2 | 5 2 | 5 2 |
| | 7 2 | 6 2 | 6 2 | 7 2 | 5 2 | 5 2 |

The End

TABLE LIV: The computed irreducible representations for the valence bands for the compounds of SEBR and NLC in space group 121. See the list of tables.

| TQC | ICSD | NoE | SF |
|---|---|---|---|
| 121.1.1 | 237706 | 92 | Mo3P1 |

Continued on next column

**TABLE LIV– continued**

| TQC | ICSD | | NoE | SF | | |
|---|---|---|---|---|---|---|
| | -GM | -M | -P | -PA | -X | -N |
| | 7 2 | 6 2 | 6 2 | 6 2 | 5 2 | 3+4 2 |
| | 6 2 | 7 2 | 7 2 | 7 2 | 5 2 | 3+4 2 |
| | 7 2 | 6 2 | 7 2 | 7 2 | 5 2 | 3+4 2 |
| | 6 2 | 7 2 | 6 2 | 6 2 | 5 2 | 3+4 2 |
| | 7 2 | 6 2 | 6 2 | 6 2 | 5 2 | 3+4 2 |
| | 6 2 | 7 2 | 7 2 | 7 2 | 5 2 | 3+4 2 |
| | 7 2 | 7 2 | 6 2 | 6 2 | 5 2 | 3+4 2 |
| | 6 2 | 6 2 | 7 2 | 7 2 | 5 2 | 3+4 2 |
| | 6 2 | 6 2 | 6 2 | 6 2 | 5 2 | 3+4 2 |
| | 6 2 | 6 2 | 6 2 | 6 2 | 5 2 | 3+4 2 |
| | 7 2 | 7 2 | 7 2 | 7 2 | 5 2 | 3+4 2 |
| | 6 2 | 7 2 | 7 2 | 7 2 | 5 2 | 3+4 2 |
| | 7 2 | 6 2 | 7 2 | 7 2 | 5 2 | 3+4 2 |
| | 6 2 | 6 2 | 7 2 | 7 2 | 5 2 | 3+4 2 |
| | 7 2 | 7 2 | 6 2 | 6 2 | 5 2 | 3+4 2 |
| | 7 2 | 7 2 | 6 2 | 6 2 | 5 2 | 3+4 2 |
| | 7 2 | 7 2 | 7 2 | 7 2 | 5 2 | 3+4 2 |
| | 7 2 | 6 2 | 6 2 | 6 2 | 5 2 | 3+4 2 |
| | 7 2 | 7 2 | 6 2 | 6 2 | 5 2 | 3+4 2 |
| | 6 2 | 6 2 | 7 2 | 7 2 | 5 2 | 3+4 2 |
| | 7 2 | 6 2 | 6 2 | 6 2 | 5 2 | 3+4 2 |
| | 6 2 | 7 2 | 7 2 | 7 2 | 5 2 | 3+4 2 |
| | 6 2 | 7 2 | 6 2 | 6 2 | 5 2 | 3+4 2 |
| | 7 2 | 6 2 | 7 2 | 7 2 | 5 2 | 3+4 2 |
| | 6 2 | 7 2 | 6 2 | 6 2 | 5 2 | 3+4 2 |
| | 7 2 | 7 2 | 6 2 | 6 2 | 5 2 | 3+4 2 |
| | 6 2 | 6 2 | 7 2 | 7 2 | 5 2 | 3+4 2 |
| | 6 2 | 6 2 | 7 2 | 7 2 | 5 2 | 3+4 2 |
| | 7 2 | 6 2 | 7 2 | 7 2 | 5 2 | 3+4 2 |
| | 7 2 | 7 2 | 6 2 | 6 2 | 5 2 | 3+4 2 |
| | 6 2 | 7 2 | 6 2 | 6 2 | 5 2 | 3+4 2 |
| | 6 2 | 7 2 | 7 2 | 7 2 | 5 2 | 3+4 2 |
| | 7 2 | 6 2 | 7 2 | 7 2 | 5 2 | 3+4 2 |
| | 6 2 | 6 2 | 6 2 | 6 2 | 5 2 | 3+4 2 |
| | 7 2 | 6 2 | 7 2 | 7 2 | 5 2 | 3+4 2 |
| | 6 2 | 6 2 | 6 2 | 6 2 | 5 2 | 3+4 2 |
| | 7 2 | 7 2 | 6 2 | 6 2 | 5 2 | 3+4 2 |
| | 6 2 | 7 2 | 7 2 | 7 2 | 5 2 | 3+4 2 |
| | 7 2 | 7 2 | 7 2 | 7 2 | 5 2 | 3+4 2 |
| | 7 2 | 6 2 | 6 2 | 6 2 | 5 2 | 3+4 2 |
| | 6 2 | 6 2 | 6 2 | 6 2 | 5 2 | 3+4 2 |
| | 7 2 | 6 2 | 7 2 | 7 2 | 5 2 | 3+4 2 |
| | 6 2 | 7 2 | 7 2 | 7 2 | 5 2 | 3+4 2 |
| | 7 2 | 6 2 | 7 2 | 7 2 | 5 2 | 3+4 2 |
| | 6 2 | 7 2 | 6 2 | 6 2 | 5 2 | 3+4 2 |
| | 7 2 | 7 2 | 6 2 | 6 2 | 5 2 | 3+4 2 |
| 121.1.2 | 104607 | | 84 | Ir1Zr3 | | |
| | -GM | -M | -P | -PA | -X | -N |
| | 7 2 | 6 2 | 7 2 | 7 2 | 5 2 | 3+4 2 |
| | 7 2 | 7 2 | 6 2 | 6 2 | 5 2 | 3+4 2 |
| | 6 2 | 6 2 | 7 2 | 7 2 | 5 2 | 3+4 2 |
| | 6 2 | 7 2 | 6 2 | 6 2 | 5 2 | 3+4 2 |
| | 7 2 | 7 2 | 7 2 | 7 2 | 5 2 | 3+4 2 |
| | 7 2 | 6 2 | 6 2 | 6 2 | 5 2 | 3+4 2 |
| | 6 2 | 6 2 | 7 2 | 7 2 | 5 2 | 3+4 2 |
| | 7 2 | 7 2 | 6 2 | 6 2 | 5 2 | 3+4 2 |
| | 6 2 | 6 2 | 6 2 | 6 2 | 5 2 | 3+4 2 |
| | 6 2 | 7 2 | 6 2 | 6 2 | 5 2 | 3+4 2 |
| | 7 2 | 6 2 | 7 2 | 7 2 | 5 2 | 3+4 2 |
| | 6 2 | 7 2 | 7 2 | 7 2 | 5 2 | 3+4 2 |

Continued on next column

**TABLE LIV– continued**

| TQC | ICSD | | NoE | SF | | |
|---|---|---|---|---|---|---|
| | 6 2 | 7 2 | 6 2 | 6 2 | 5 2 | 3+4 2 |
| | 7 2 | 6 2 | 7 2 | 7 2 | 5 2 | 3+4 2 |
| | 7 2 | 7 2 | 6 2 | 6 2 | 5 2 | 3+4 2 |
| | 6 2 | 6 2 | 7 2 | 7 2 | 5 2 | 3+4 2 |
| | 6 2 | 7 2 | 7 2 | 7 2 | 5 2 | 3+4 2 |
| | 7 2 | 6 2 | 6 2 | 6 2 | 5 2 | 3+4 2 |
| | 6 2 | 6 2 | 6 2 | 6 2 | 5 2 | 3+4 2 |
| | 7 2 | 7 2 | 7 2 | 7 2 | 5 2 | 3+4 2 |
| | 6 2 | 7 2 | 6 2 | 6 2 | 5 2 | 3+4 2 |
| | 7 2 | 6 2 | 7 2 | 7 2 | 5 2 | 3+4 2 |
| | 6 2 | 7 2 | 6 2 | 6 2 | 5 2 | 3+4 2 |
| | 7 2 | 6 2 | 6 2 | 6 2 | 5 2 | 3+4 2 |
| | 6 2 | 7 2 | 7 2 | 7 2 | 5 2 | 3+4 2 |
| | 7 2 | 7 2 | 7 2 | 7 2 | 5 2 | 3+4 2 |
| | 7 2 | 6 2 | 6 2 | 6 2 | 5 2 | 3+4 2 |
| | 6 2 | 6 2 | 7 2 | 7 2 | 5 2 | 3+4 2 |
| | 7 2 | 6 2 | 6 2 | 6 2 | 5 2 | 3+4 2 |
| | 6 2 | 6 2 | 7 2 | 7 2 | 5 2 | 3+4 2 |
| | 6 2 | 7 2 | 6 2 | 6 2 | 5 2 | 3+4 2 |
| | 7 2 | 7 2 | 7 2 | 7 2 | 5 2 | 3+4 2 |
| | 7 2 | 6 2 | 6 2 | 6 2 | 5 2 | 3+4 2 |
| | 7 2 | 7 2 | 7 2 | 7 2 | 5 2 | 3+4 2 |
| | 6 2 | 7 2 | 6 2 | 6 2 | 5 2 | 3+4 2 |
| | 6 2 | 6 2 | 7 2 | 7 2 | 5 2 | 3+4 2 |
| | 6 2 | 7 2 | 7 2 | 7 2 | 5 2 | 3+4 2 |
| | 7 2 | 6 2 | 6 2 | 6 2 | 5 2 | 3+4 2 |
| | 7 2 | 7 2 | 7 2 | 7 2 | 5 2 | 3+4 2 |
| | 6 2 | 6 2 | 6 2 | 6 2 | 5 2 | 3+4 2 |
| | 7 2 | 7 2 | 7 2 | 7 2 | 5 2 | 3+4 2 |
| | 7 2 | 6 2 | 6 2 | 6 2 | 5 2 | 3+4 2 |
| 121.1.3 | 400652 | | 62 | Cu3Sb1Se4 | | |
| | -GM | -M | -P | -PA | -X | -N |
| | 7 2 | 7 2 | 7 2 | 7 2 | 5 2 | 3+4 2 |
| | 6 2 | 6 2 | 6 2 | 6 2 | 5 2 | 3+4 2 |
| | 6 2 | 6 2 | 6 2 | 6 2 | 5 2 | 3+4 2 |
| | 7 2 | 7 2 | 7 2 | 7 2 | 5 2 | 3+4 2 |
| | 7 2 | 7 2 | 7 2 | 7 2 | 5 2 | 3+4 2 |
| | 6 2 | 6 2 | 6 2 | 6 2 | 5 2 | 3+4 2 |
| | 7 2 | 7 2 | 6 2 | 6 2 | 5 2 | 3+4 2 |
| | 7 2 | 6 2 | 7 2 | 7 2 | 5 2 | 3+4 2 |
| | 6 2 | 6 2 | 6 2 | 6 2 | 5 2 | 3+4 2 |
| | 7 2 | 7 2 | 7 2 | 7 2 | 5 2 | 3+4 2 |
| | 6 2 | 6 2 | 6 2 | 6 2 | 5 2 | 3+4 2 |
| | 6 2 | 6 2 | 7 2 | 7 2 | 5 2 | 3+4 2 |
| | 7 2 | 7 2 | 7 2 | 7 2 | 5 2 | 3+4 2 |
| | 6 2 | 7 2 | 6 2 | 6 2 | 5 2 | 3+4 2 |
| | 7 2 | 6 2 | 7 2 | 7 2 | 5 2 | 3+4 2 |
| | 6 2 | 7 2 | 6 2 | 6 2 | 5 2 | 3+4 2 |
| | 7 2 | 7 2 | 7 2 | 7 2 | 5 2 | 3+4 2 |
| | 6 2 | 7 2 | 6 2 | 6 2 | 5 2 | 3+4 2 |
| | 7 2 | 6 2 | 7 2 | 7 2 | 5 2 | 3+4 2 |
| | 6 2 | 7 2 | 6 2 | 6 2 | 5 2 | 3+4 2 |
| | 6 2 | 6 2 | 7 2 | 7 2 | 5 2 | 3+4 2 |
| | 7 2 | 6 2 | 6 2 | 6 2 | 5 2 | 3+4 2 |
| | 6 2 | 7 2 | 7 2 | 7 2 | 5 2 | 3+4 2 |
| | 7 2 | 6 2 | 6 2 | 6 2 | 5 2 | 3+4 2 |
| | 6 2 | 7 2 | 6 2 | 6 2 | 5 2 | 3+4 2 |
| | 7 2 | 7 2 | 7 2 | 7 2 | 5 2 | 3+4 2 |
| | 6 2 | 6 2 | 7 2 | 7 2 | 5 2 | 3+4 2 |
| | 6 2 | 6 2 | 6 2 | 6 2 | 5 2 | 3+4 2 |
| | 7 2 | 7 2 | 7 2 | 7 2 | 5 2 | 3+4 2 |

Continued on next column

**TABLE LIV– continued**

| TQC | ICSD | | NoE | SF | | |
|---|---|---|---|---|---|---|
| | 6 2 | 6 2 | 6 2 | 6 2 | 5 2 | 3+4 2 |
| | 7 2 | 7 2 | 7 2 | 7 2 | 5 2 | 3+4 2 |
| 121.1.4 | 412239 | | 62 | Cu3S4Sb1 | | |
| | -GM | -M | -P | -PA | -X | -N |
| | 7 2 | 7 2 | 7 2 | 7 2 | 5 2 | 3+4 2 |
| | 6 2 | 6 2 | 6 2 | 6 2 | 5 2 | 3+4 2 |
| | 6 2 | 6 2 | 7 2 | 7 2 | 5 2 | 3+4 2 |
| | 7 2 | 7 2 | 6 2 | 6 2 | 5 2 | 3+4 2 |
| | 7 2 | 7 2 | 7 2 | 7 2 | 5 2 | 3+4 2 |
| | 6 2 | 6 2 | 6 2 | 6 2 | 5 2 | 3+4 2 |
| | 7 2 | 6 2 | 6 2 | 6 2 | 5 2 | 3+4 2 |
| | 6 2 | 7 2 | 7 2 | 7 2 | 5 2 | 3+4 2 |
| | 7 2 | 6 2 | 6 2 | 6 2 | 5 2 | 3+4 2 |
| | 7 2 | 7 2 | 7 2 | 7 2 | 5 2 | 3+4 2 |
| | 6 2 | 6 2 | 6 2 | 6 2 | 5 2 | 3+4 2 |
| | 6 2 | 6 2 | 7 2 | 7 2 | 5 2 | 3+4 2 |
| | 7 2 | 7 2 | 7 2 | 7 2 | 5 2 | 3+4 2 |
| | 6 2 | 7 2 | 6 2 | 6 2 | 5 2 | 3+4 2 |
| | 7 2 | 7 2 | 7 2 | 7 2 | 5 2 | 3+4 2 |
| | 6 2 | 6 2 | 6 2 | 6 2 | 5 2 | 3+4 2 |
| | 7 2 | 7 2 | 7 2 | 7 2 | 5 2 | 3+4 2 |
| | 6 2 | 7 2 | 6 2 | 6 2 | 5 2 | 3+4 2 |
| | 7 2 | 6 2 | 7 2 | 7 2 | 5 2 | 3+4 2 |
| | 6 2 | 6 2 | 6 2 | 6 2 | 5 2 | 3+4 2 |
| | 6 2 | 7 2 | 7 2 | 7 2 | 5 2 | 3+4 2 |
| | 7 2 | 7 2 | 6 2 | 6 2 | 5 2 | 3+4 2 |
| | 6 2 | 6 2 | 7 2 | 7 2 | 5 2 | 3+4 2 |
| | 7 2 | 6 2 | 6 2 | 6 2 | 5 2 | 3+4 2 |
| | 6 2 | 7 2 | 6 2 | 6 2 | 5 2 | 3+4 2 |
| | 7 2 | 7 2 | 7 2 | 7 2 | 5 2 | 3+4 2 |
| | 6 2 | 7 2 | 7 2 | 7 2 | 5 2 | 3+4 2 |
| | 6 2 | 6 2 | 6 2 | 6 2 | 5 2 | 3+4 2 |
| | 7 2 | 6 2 | 7 2 | 7 2 | 5 2 | 3+4 2 |
| | 6 2 | 6 2 | 6 2 | 6 2 | 5 2 | 3+4 2 |
| | 7 2 | 7 2 | 7 2 | 7 2 | 5 2 | 3+4 2 |
| 121.1.5 | 2857 | | 62 | Cu3S4Sb1 | | |
| | -GM | -M | -P | -PA | -X | -N |
| | 7 2 | 7 2 | 7 2 | 7 2 | 5 2 | 3+4 2 |
| | 6 2 | 6 2 | 6 2 | 6 2 | 5 2 | 3+4 2 |
| | 6 2 | 6 2 | 7 2 | 7 2 | 5 2 | 3+4 2 |
| | 7 2 | 7 2 | 6 2 | 6 2 | 5 2 | 3+4 2 |
| | 7 2 | 7 2 | 7 2 | 7 2 | 5 2 | 3+4 2 |
| | 6 2 | 6 2 | 6 2 | 6 2 | 5 2 | 3+4 2 |
| | 7 2 | 6 2 | 6 2 | 6 2 | 5 2 | 3+4 2 |
| | 6 2 | 7 2 | 7 2 | 7 2 | 5 2 | 3+4 2 |
| | 7 2 | 6 2 | 6 2 | 6 2 | 5 2 | 3+4 2 |
| | 7 2 | 7 2 | 7 2 | 7 2 | 5 2 | 3+4 2 |
| | 6 2 | 6 2 | 6 2 | 6 2 | 5 2 | 3+4 2 |
| | 6 2 | 6 2 | 7 2 | 7 2 | 5 2 | 3+4 2 |
| | 7 2 | 7 2 | 7 2 | 7 2 | 5 2 | 3+4 2 |
| | 6 2 | 7 2 | 6 2 | 6 2 | 5 2 | 3+4 2 |
| | 7 2 | 7 2 | 6 2 | 6 2 | 5 2 | 3+4 2 |
| | 6 2 | 6 2 | 7 2 | 7 2 | 5 2 | 3+4 2 |
| | 6 2 | 7 2 | 7 2 | 7 2 | 5 2 | 3+4 2 |
| | 7 2 | 6 2 | 6 2 | 6 2 | 5 2 | 3+4 2 |
| | 7 2 | 7 2 | 7 2 | 7 2 | 5 2 | 3+4 2 |
| | 6 2 | 6 2 | 7 2 | 7 2 | 5 2 | 3+4 2 |
| | 6 2 | 7 2 | 6 2 | 6 2 | 5 2 | 3+4 2 |
| | 6 2 | 7 2 | 6 2 | 6 2 | 5 2 | 3+4 2 |
| | 7 2 | 6 2 | 7 2 | 7 2 | 5 2 | 3+4 2 |
| | 7 2 | 6 2 | 6 2 | 6 2 | 5 2 | 3+4 2 |

Continued on next column

**TABLE LIV– continued**

| TQC | ICSD | NoE | SF |
|---|---|---|---|
| | 6 2 7 2 | 6 2 | 6 2 5 2 3+4 2 |
| | 7 2 7 2 | 7 2 | 7 2 5 2 3+4 2 |
| | 6 2 6 2 | 7 2 | 7 2 5 2 3+4 2 |
| | 6 2 7 2 | 6 2 | 6 2 5 2 3+4 2 |
| | 7 2 6 2 | 7 2 | 7 2 5 2 3+4 2 |
| | 7 2 6 2 | 6 2 | 6 2 5 2 3+4 2 |
| | 6 2 7 2 | 7 2 | 7 2 5 2 3+4 2 |
| 121.1.6 | 95119 | 62 | Cu2Hg1Se4Sn1 |
| | -GM -M | -P | -PA -X -N |
| | 7 2 7 2 | 6 2 | 6 2 5 2 3+4 2 |
| | 6 2 6 2 | 7 2 | 7 2 5 2 3+4 2 |
| | 6 2 6 2 | 6 2 | 6 2 5 2 3+4 2 |
| | 7 2 7 2 | 7 2 | 7 2 5 2 3+4 2 |
| | 7 2 7 2 | 6 2 | 6 2 5 2 3+4 2 |
| | 6 2 6 2 | 7 2 | 7 2 5 2 3+4 2 |
| | 7 2 7 2 | 6 2 | 6 2 5 2 3+4 2 |
| | 6 2 6 2 | 6 2 | 6 2 5 2 3+4 2 |
| | 6 2 6 2 | 7 2 | 7 2 5 2 3+4 2 |
| | 7 2 7 2 | 6 2 | 6 2 5 2 3+4 2 |
| | 7 2 7 2 | 7 2 | 7 2 5 2 3+4 2 |
| | 6 2 6 2 | 7 2 | 7 2 5 2 3+4 2 |
| | 7 2 6 2 | 6 2 | 6 2 5 2 3+4 2 |
| | 6 2 6 2 | 7 2 | 7 2 5 2 3+4 2 |
| | 7 2 7 2 | 6 2 | 6 2 5 2 3+4 2 |
| | 6 2 6 2 | 7 2 | 7 2 5 2 3+4 2 |
| | 7 2 7 2 | 7 2 | 7 2 5 2 3+4 2 |
| | 6 2 7 2 | 6 2 | 6 2 5 2 3+4 2 |
| | 6 2 6 2 | 6 2 | 6 2 5 2 3+4 2 |
| | 6 2 7 2 | 7 2 | 7 2 5 2 3+4 2 |
| | 7 2 6 2 | 6 2 | 6 2 5 2 3+4 2 |
| | 7 2 7 2 | 7 2 | 7 2 5 2 3+4 2 |
| | 7 2 7 2 | 6 2 | 6 2 5 2 3+4 2 |
| | 6 2 6 2 | 7 2 | 7 2 5 2 3+4 2 |
| | 6 2 6 2 | 6 2 | 6 2 5 2 3+4 2 |
| | 7 2 7 2 | 7 2 | 7 2 5 2 3+4 2 |
| | 6 2 6 2 | 6 2 | 6 2 5 2 3+4 2 |
| | 6 2 7 2 | 6 2 | 6 2 5 2 3+4 2 |
| | 7 2 7 2 | 7 2 | 7 2 5 2 3+4 2 |
| | 6 2 6 2 | 7 2 | 7 2 5 2 3+4 2 |
| | 7 2 7 2 | 6 2 | 6 2 5 2 3+4 2 |
| 121.1.7 | 152761 | 62 | Cu2Ge1Hg1Se4 |
| | -GM -M | -P | -PA -X -N |
| | 7 2 7 2 | 6 2 | 6 2 5 2 3+4 2 |
| | 6 2 6 2 | 7 2 | 7 2 5 2 3+4 2 |
| | 6 2 6 2 | 6 2 | 6 2 5 2 3+4 2 |
| | 7 2 7 2 | 7 2 | 7 2 5 2 3+4 2 |
| | 7 2 7 2 | 6 2 | 6 2 5 2 3+4 2 |
| | 6 2 6 2 | 7 2 | 7 2 5 2 3+4 2 |
| | 7 2 7 2 | 6 2 | 6 2 5 2 3+4 2 |
| | 6 2 6 2 | 6 2 | 6 2 5 2 3+4 2 |
| | 6 2 6 2 | 7 2 | 7 2 5 2 3+4 2 |
| | 7 2 7 2 | 6 2 | 6 2 5 2 3+4 2 |
| | 7 2 7 2 | 7 2 | 7 2 5 2 3+4 2 |
| | 6 2 6 2 | 7 2 | 7 2 5 2 3+4 2 |
| | 7 2 6 2 | 6 2 | 6 2 5 2 3+4 2 |
| | 6 2 7 2 | 7 2 | 7 2 5 2 3+4 2 |
| | 7 2 6 2 | 6 2 | 6 2 5 2 3+4 2 |
| | 6 2 6 2 | 7 2 | 7 2 5 2 3+4 2 |
| | 7 2 7 2 | 7 2 | 7 2 5 2 3+4 2 |
| | 6 2 7 2 | 6 2 | 6 2 5 2 3+4 2 |
| | 6 2 6 2 | 6 2 | 6 2 5 2 3+4 2 |

Continued on next column

**TABLE LIV– continued**

| TQC | ICSD | NoE | SF |
|---|---|---|---|
| | 6 2 7 2 | 7 2 | 7 2 5 2 3+4 2 |
| | 7 2 6 2 | 6 2 | 6 2 5 2 3+4 2 |
| | 7 2 7 2 | 6 2 | 6 2 5 2 3+4 2 |
| | 7 2 7 2 | 7 2 | 7 2 5 2 3+4 2 |
| | 6 2 6 2 | 7 2 | 7 2 5 2 3+4 2 |
| | 6 2 6 2 | 6 2 | 6 2 5 2 3+4 2 |
| | 7 2 7 2 | 6 2 | 6 2 5 2 3+4 2 |
| | 6 2 6 2 | 7 2 | 7 2 5 2 3+4 2 |
| | 6 2 7 2 | 6 2 | 6 2 5 2 3+4 2 |
| | 7 2 7 2 | 7 2 | 7 2 5 2 3+4 2 |
| | 6 2 6 2 | 7 2 | 7 2 5 2 3+4 2 |
| | 7 2 7 2 | 6 2 | 6 2 5 2 3+4 2 |
| 121.1.8 | 85126 | 58 | Cu2Fe1Se4Sn1 |
| | -GM -M | -P | -PA -X -N |
| | 7 2 7 2 | 6 2 | 6 2 5 2 3+4 2 |
| | 6 2 6 2 | 7 2 | 7 2 5 2 3+4 2 |
| | 7 2 6 2 | 6 2 | 6 2 5 2 3+4 2 |
| | 6 2 7 2 | 7 2 | 7 2 5 2 3+4 2 |
| | 7 2 7 2 | 6 2 | 6 2 5 2 3+4 2 |
| | 6 2 6 2 | 7 2 | 7 2 5 2 3+4 2 |
| | 7 2 7 2 | 6 2 | 6 2 5 2 3+4 2 |
| | 7 2 6 2 | 7 2 | 7 2 5 2 3+4 2 |
| | 6 2 6 2 | 7 2 | 7 2 5 2 3+4 2 |
| | 6 2 7 2 | 6 2 | 6 2 5 2 3+4 2 |
| | 7 2 6 2 | 6 2 | 6 2 5 2 3+4 2 |
| | 6 2 7 2 | 7 2 | 7 2 5 2 3+4 2 |
| | 7 2 6 2 | 7 2 | 7 2 5 2 3+4 2 |
| | 6 2 7 2 | 6 2 | 6 2 5 2 3+4 2 |
| | 7 2 6 2 | 7 2 | 7 2 5 2 3+4 2 |
| | 6 2 7 2 | 6 2 | 6 2 5 2 3+4 2 |
| | 7 2 7 2 | 7 2 | 7 2 5 2 3+4 2 |
| | 6 2 7 2 | 6 2 | 6 2 5 2 3+4 2 |
| | 6 2 6 2 | 6 2 | 6 2 5 2 3+4 2 |
| | 7 2 6 2 | 7 2 | 7 2 5 2 3+4 2 |
| | 7 2 7 2 | 6 2 | 6 2 5 2 3+4 2 |
| | 6 2 6 2 | 7 2 | 7 2 5 2 3+4 2 |
| | 6 2 7 2 | 7 2 | 7 2 5 2 3+4 2 |
| | 6 2 7 2 | 6 2 | 6 2 5 2 3+4 2 |
| | 7 2 7 2 | 6 2 | 6 2 5 2 3+4 2 |
| | 6 2 6 2 | 7 2 | 7 2 5 2 3+4 2 |
| | 7 2 7 2 | 7 2 | 7 2 5 2 3+4 2 |
| | 6 2 6 2 | 6 2 | 6 2 5 2 3+4 2 |
| | 7 2 6 2 | 6 2 | 6 2 5 2 3+4 2 |
| 121.1.9 | 165094 | 62 | Cd1Cu2Ge1Te4 |
| | -GM -M | -P | -PA -X -N |
| | 7 2 7 2 | 6 2 | 6 2 5 2 3+4 2 |
| | 7 2 6 2 | 7 2 | 7 2 5 2 3+4 2 |
| | 6 2 7 2 | 6 2 | 6 2 5 2 3+4 2 |
| | 6 2 6 2 | 7 2 | 7 2 5 2 3+4 2 |
| | 7 2 6 2 | 6 2 | 6 2 5 2 3+4 2 |
| | 6 2 7 2 | 7 2 | 7 2 5 2 3+4 2 |
| | 7 2 7 2 | 6 2 | 6 2 5 2 3+4 2 |
| | 6 2 6 2 | 7 2 | 7 2 5 2 3+4 2 |
| | 6 2 6 2 | 6 2 | 6 2 5 2 3+4 2 |
| | 7 2 7 2 | 6 2 | 6 2 5 2 3+4 2 |
| | 7 2 6 2 | 7 2 | 7 2 5 2 3+4 2 |
| | 6 2 7 2 | 6 2 | 6 2 5 2 3+4 2 |
| | 7 2 6 2 | 7 2 | 7 2 5 2 3+4 2 |
| | 7 2 7 2 | 7 2 | 7 2 5 2 3+4 2 |
| | 6 2 6 2 | 6 2 | 6 2 5 2 3+4 2 |
| | 6 2 6 2 | 7 2 | 7 2 5 2 3+4 2 |

Continued on next column

**TABLE LIV– continued**

| TQC | ICSD | NoE | SF |
|---|---|---|---|
| | 6 2 7 2 | 6 2 | 6 2 5 2 3+4 2 |
| | 7 2 7 2 | 7 2 | 7 2 5 2 3+4 2 |
| | 6 2 6 2 | 6 2 | 6 2 5 2 3+4 2 |
| | 7 2 7 2 | 7 2 | 7 2 5 2 3+4 2 |
| | 6 2 6 2 | 6 2 | 6 2 5 2 3+4 2 |
| | 7 2 7 2 | 6 2 | 6 2 5 2 3+4 2 |
| | 6 2 6 2 | 7 2 | 7 2 5 2 3+4 2 |
| | 7 2 7 2 | 6 2 | 6 2 5 2 3+4 2 |
| | 6 2 6 2 | 7 2 | 7 2 5 2 3+4 2 |
| | 7 2 6 2 | 7 2 | 7 2 5 2 3+4 2 |
| | 6 2 7 2 | 6 2 | 6 2 5 2 3+4 2 |
| | 6 2 7 2 | 6 2 | 6 2 5 2 3+4 2 |
| | 6 2 7 2 | 7 2 | 7 2 5 2 3+4 2 |
| | 7 2 6 2 | 7 2 | 7 2 5 2 3+4 2 |
| | 7 2 7 2 | 6 2 | 6 2 5 2 3+4 2 |

The End

TABLE LV: The computed irreducible representations for the valence bands for the compounds of SEBR and NLC in space group 122. See the list of tables.

| TQC | ICSD | NoE | SF | | | |
|---|---|---|---|---|---|---|
| 122.1.1 | 248539 | 144 | O6Rb4 | | | |
| | -GM | -M | -PA | -P | -X | -N |
| | 7 2 | 6+7 4 | 3 1 | 3 1 | 2+3 2 | 3+4 2 |
| | 6 2 | 6+7 4 | 5 1 | 5 1 | 4+5 2 | 3+4 2 |
| | 6 2 | 6+7 4 | 7 2 | 7 2 | 2+3 2 | 3+4 2 |
| | 7 2 | 6+7 4 | 4 1 | 4 1 | 4+5 2 | 3+4 2 |
| | 6 2 | 6+7 4 | 6 1 | 6 1 | 2+3 2 | 3+4 2 |
| | 7 2 | 6+7 4 | 7 2 | 7 2 | 4+5 2 | 3+4 2 |
| | 6 2 | 6+7 4 | 3 1 | 3 1 | 2+3 2 | 3+4 2 |
| | 7 2 | 6+7 4 | 5 1 | 5 1 | 4+5 2 | 3+4 2 |
| | 6 2 | 6+7 4 | 7 2 | 7 2 | 2+3 2 | 3+4 2 |
| | 6 2 | 6+7 4 | 4 1 | 4 1 | 4+5 2 | 3+4 2 |
| | 7 2 | 6+7 4 | 6 1 | 6 1 | 2+3 2 | 3+4 2 |
| | 7 2 | 6+7 4 | 7 2 | 7 2 | 4+5 2 | 3+4 2 |
| | 7 2 | 6+7 4 | 4 1 | 4 1 | 2+3 2 | 3+4 2 |
| | 7 2 | 6+7 4 | 6 1 | 6 1 | 4+5 2 | 3+4 2 |
| | 6 2 | 6+7 4 | 7 2 | 7 2 | 4+5 2 | 3+4 2 |
| | 6 2 | 6+7 4 | 3 1 | 3 1 | 2+3 2 | 3+4 2 |
| | 7 2 | 6+7 4 | 5 1 | 5 1 | 2+3 2 | 3+4 2 |
| | 6 2 | 6+7 4 | 7 2 | 7 2 | 4+5 2 | 3+4 2 |
| | 7 2 | 6+7 4 | 3 1 | 3 1 | 2+3 2 | 3+4 2 |
| | 6 2 | 6+7 4 | 5 1 | 5 1 | 4+5 2 | 3+4 2 |
| | 6 2 | 6+7 4 | 7 2 | 7 2 | 2+3 2 | 3+4 2 |
| | 7 2 | 6+7 4 | 4 1 | 4 1 | 4+5 2 | 3+4 2 |
| | 6 2 | 6+7 4 | 6 1 | 6 1 | 2+3 2 | 3+4 2 |
| | 7 2 | 6+7 4 | 7 2 | 7 2 | 4+5 2 | 3+4 2 |
| | 7 2 | 6+7 4 | 6 1 | 6 1 | 4+5 2 | 3+4 2 |
| | 7 2 | 6+7 4 | 7 2 | 7 2 | 2+3 2 | 3+4 2 |
| | 6 2 | 6+7 4 | 4 1 | 4 1 | 4+5 2 | 3+4 2 |
| | 6 2 | 6+7 4 | 5 1 | 5 1 | 2+3 2 | 3+4 2 |
| | 6 2 | 6+7 4 | 7 2 | 7 2 | 4+5 2 | 3+4 2 |
| | 6 2 | 6+7 4 | 3 1 | 3 1 | 2+3 2 | 3+4 2 |
| | 7 2 | 6+7 4 | 7 2 | 7 2 | 2+3 2 | 3+4 2 |
| | 7 2 | 6+7 4 | 3 1 | 3 1 | 4+5 2 | 3+4 2 |
| | 6 2 | 6+7 4 | 5 1 | 5 1 | 2+3 2 | 3+4 2 |
| | 6 2 | 6+7 4 | 6 1 | 6 1 | 4+5 2 | 3+4 2 |
| | 7 2 | 6+7 4 | 4 1 | 4 1 | 2+3 2 | 3+4 2 |
| | 6 2 | 6+7 4 | 7 2 | 7 2 | 4+5 2 | 3+4 2 |

Continued on next column

**TABLE LV– continued**

| TQC | ICSD | | NoE | SF | | |
|---|---|---|---|---|---|---|
| | 7 2 | | 7 2 | 7 2 | 2+3 2 | 3+4 2 |
| | 7 2 | | 4 1 | 4 1 | 4+5 2 | 3+4 2 |
| | 6 2 | | 6 1 | 6 1 | 2+3 2 | 3+4 2 |
| | 7 2 | | 5 1 | 5 1 | 4+5 2 | 3+4 2 |
| | 7 2 | | 3 1 | 3 1 | 4+5 2 | 3+4 2 |
| | 7 2 | | 7 2 | 7 2 | 2+3 2 | 3+4 2 |
| | 6 2 | | 4 1 | 4 1 | 4+5 2 | 3+4 2 |
| | 6 2 | | 7 2 | 7 2 | 2+3 2 | 3+4 2 |
| | 7 2 | | 6 1 | 6 1 | 2+3 2 | 3+4 2 |
| | 6 2 | | 5 1 | 5 1 | 4+5 2 | 3+4 2 |
| | 7 2 | | 3 1 | 3 1 | 4+5 2 | 3+4 2 |
| | 6 2 | | 7 2 | 7 2 | 2+3 2 | 3+4 2 |
| | 7 2 | | 6 1 | 6 1 | 2+3 2 | 3+4 2 |
| | 6 2 | | 7 2 | 7 2 | 4+5 2 | 3+4 2 |
| | 7 2 | | 3 1 | 3 1 | 4+5 2 | 3+4 2 |
| | 7 2 | | 7 2 | 7 2 | 2+3 2 | 3+4 2 |
| | 6 2 | | 5 1 | 5 1 | 4+5 2 | 3+4 2 |
| | 6 2 | | 4 1 | 4 1 | 2+3 2 | 3+4 2 |
| | 6 2 | | 7 2 | 7 2 | 2+3 2 | 3+4 2 |
| | 7 2 | | 6 1 | 6 1 | 4+5 2 | 3+4 2 |
| | 7 2 | | 4 1 | 4 1 | 2+3 2 | 3+4 2 |
| | 7 2 | | 5 1 | 5 1 | 4+5 2 | 3+4 2 |
| | 6 2 | | 3 1 | 3 1 | 4+5 2 | 3+4 2 |
| | 7 2 | | 7 2 | 7 2 | 2+3 2 | 3+4 2 |
| | 7 2 | | 3 1 | 3 1 | 2+3 2 | 3+4 2 |
| | 6 2 | | 7 2 | 7 2 | 4+5 2 | 3+4 2 |
| | 7 2 | | 5 1 | 5 1 | 4+5 2 | 3+4 2 |
| | 6 2 | | 7 2 | 7 2 | 2+3 2 | 3+4 2 |
| | 7 2 | | 4 1 | 4 1 | 2+3 2 | 3+4 2 |
| | 6 2 | | 6 1 | 6 1 | 4+5 2 | 3+4 2 |
| | 7 2 | | 4 1 | 4 1 | 2+3 2 | 3+4 2 |
| | 6 2 | | 6 1 | 6 1 | 4+5 2 | 3+4 2 |
| | 6 2 | | 7 2 | 7 2 | 4+5 2 | 3+4 2 |
| | 7 2 | | 3 1 | 3 1 | 2+3 2 | 3+4 2 |
| | 6 2 | | 7 2 | 7 2 | 2+3 2 | 3+4 2 |
| | 7 2 | | 5 1 | 5 1 | 4+5 2 | 3+4 2 |
| | | | 3 1 | 3 1 | | |
| | | | 5 1 | 5 1 | | |
| | | | 7 2 | 7 2 | | |
| | | | 7 2 | 7 2 | | |
| | | | 4 1 | 4 1 | | |
| | | | 6 1 | 6 1 | | |
| | | | 7 2 | 7 2 | | |
| | | | 4 1 | 4 1 | | |
| | | | 5 1 | 5 1 | | |
| | | | 3 1 | 3 1 | | |
| | | | 6 1 | 6 1 | | |
| | | | 7 2 | 7 2 | | |
| | | | 3 1 | 3 1 | | |
| | | | 7 2 | 7 2 | | |
| | | | 5 1 | 5 1 | | |
| | | | 4 1 | 4 1 | | |
| | | | 6 1 | 6 1 | | |
| | | | 7 2 | 7 2 | | |
| | | | 5 1 | 5 1 | | |
| | | | 7 2 | 7 2 | | |
| | | | 3 1 | 3 1 | | |
| | | | 7 2 | 7 2 | | |
| | | | 4 1 | 4 1 | | |
| | | | 6 1 | 6 1 | | |
| | | | 5 1 | 5 1 | | |

Continued on next column

**TABLE LV– continued**

| TQC | ICSD | | NoE | | SF | |
|---|---|---|---|---|---|---|
| | | | 7 2 | 7 2 | | |
| | | | 3 1 | 3 1 | | |
| | | | 3 1 | 3 1 | | |
| | | | 5 1 | 5 1 | | |
| | | | 7 2 | 7 2 | | |
| | | | 4 1 | 4 1 | | |
| | | | 7 2 | 7 2 | | |
| | | | 6 1 | 6 1 | | |
| | | | 5 1 | 5 1 | | |
| | | | 3 1 | 3 1 | | |
| | | | 7 2 | 7 2 | | |
| 122.1.2 | 659 | | 196 | | F5U1 | |
| | -GM | -M | -P | -PA | -X | -N |
| | 7 2 | 6+7 4 | 4 1 | 4 1 | 2+3 2 | 3+4 2 |
| | 6 2 | 6+7 4 | 6 1 | 6 1 | 4+5 2 | 3+4 2 |
| | 6 2 | 6+7 4 | 7 2 | 7 2 | 2+3 2 | 3+4 2 |
| | 7 2 | 6+7 4 | 3 1 | 3 1 | 4+5 2 | 3+4 2 |
| | 7 2 | 6+7 4 | 5 1 | 5 1 | 2+3 2 | 3+4 2 |
| | 7 2 | 6+7 4 | 7 2 | 7 2 | 4+5 2 | 3+4 2 |
| | 6 2 | 6+7 4 | 6 1 | 6 1 | 4+5 2 | 3+4 2 |
| | 6 2 | 6+7 4 | 4 1 | 4 1 | 2+3 2 | 3+4 2 |
| | 7 2 | 6+7 4 | 7 2 | 7 2 | 4+5 2 | 3+4 2 |
| | 6 2 | 6+7 4 | 5 1 | 5 1 | 2+3 2 | 3+4 2 |
| | 7 2 | 6+7 4 | 3 1 | 3 1 | 4+5 2 | 3+4 2 |
| | 6 2 | 6+7 4 | 7 2 | 7 2 | 2+3 2 | 3+4 2 |
| | 7 2 | 6+7 4 | 7 2 | 7 2 | 2+3 2 | 3+4 2 |
| | 6 2 | 6+7 4 | 4 1 | 4 1 | 4+5 2 | 3+4 2 |
| | 7 2 | 6+7 4 | 6 1 | 6 1 | 2+3 2 | 3+4 2 |
| | 6 2 | 6+7 4 | 7 2 | 7 2 | 4+5 2 | 3+4 2 |
| | 7 2 | 6+7 4 | 3 1 | 3 1 | 2+3 2 | 3+4 2 |
| | 7 2 | 6+7 4 | 5 1 | 5 1 | 4+5 2 | 3+4 2 |
| | 6 2 | 6+7 4 | 7 2 | 7 2 | 2+3 2 | 3+4 2 |
| | 7 2 | 6+7 4 | 6 1 | 6 1 | 4+5 2 | 3+4 2 |
| | 6 2 | 6+7 4 | 5 1 | 5 1 | 2+3 2 | 3+4 2 |
| | 6 2 | 6+7 4 | 4 1 | 4 1 | 4+5 2 | 3+4 2 |
| | 6 2 | 6+7 4 | 3 1 | 3 1 | 2+3 2 | 3+4 2 |
| | 7 2 | 6+7 4 | 7 2 | 7 2 | 4+5 2 | 3+4 2 |
| | 6 2 | 6+7 4 | 6 1 | 6 1 | 2+3 2 | 3+4 2 |
| | 6 2 | 6+7 4 | 7 2 | 7 2 | 4+5 2 | 3+4 2 |
| | 7 2 | 6+7 4 | 4 1 | 4 1 | 4+5 2 | 3+4 2 |
| | 7 2 | 6+7 4 | 5 1 | 5 1 | 2+3 2 | 3+4 2 |
| | 7 2 | 6+7 4 | 3 1 | 3 1 | 2+3 2 | 3+4 2 |
| | 6 2 | 6+7 4 | 7 2 | 7 2 | 4+5 2 | 3+4 2 |
| | 6 2 | 6+7 4 | 7 2 | 7 2 | 4+5 2 | 3+4 2 |
| | 7 2 | 6+7 4 | 6 1 | 6 1 | 2+3 2 | 3+4 2 |
| | 7 2 | 6+7 4 | 4 1 | 4 1 | 2+3 2 | 3+4 2 |
| | 6 2 | 6+7 4 | 5 1 | 5 1 | 4+5 2 | 3+4 2 |
| | 6 2 | 6+7 4 | 3 1 | 3 1 | 4+5 2 | 3+4 2 |
| | 7 2 | 6+7 4 | 7 2 | 7 2 | 2+3 2 | 3+4 2 |
| | 6 2 | 6+7 4 | 5 1 | 5 1 | 2+3 2 | 3+4 2 |
| | 6 2 | 6+7 4 | 3 1 | 3 1 | 4+5 2 | 3+4 2 |
| | 7 2 | 6+7 4 | 7 2 | 7 2 | 4+5 2 | 3+4 2 |
| | 7 2 | 6+7 4 | 7 2 | 7 2 | 2+3 2 | 3+4 2 |
| | 6 2 | 6+7 4 | 6 1 | 6 1 | 2+3 2 | 3+4 2 |
| | 6 2 | 6+7 4 | 4 1 | 4 1 | 4+5 2 | 3+4 2 |
| | 6 2 | 6+7 4 | 7 2 | 7 2 | 2+3 2 | 3+4 2 |
| | 7 2 | 6+7 4 | 4 1 | 4 1 | 4+5 2 | 3+4 2 |
| | 7 2 | 6+7 4 | 6 1 | 6 1 | 4+5 2 | 3+4 2 |
| | 7 2 | 6+7 4 | 5 1 | 5 1 | 2+3 2 | 3+4 2 |
| | 7 2 | 6+7 4 | 3 1 | 3 1 | 2+3 2 | 3+4 2 |
| | 6 2 | 6+7 4 | 7 2 | 7 2 | 4+5 2 | 3+4 2 |

Continued on next column

**TABLE LV– continued**

| TQC | ICSD | | NoE | | SF | |
|---|---|---|---|---|---|---|
| | 7 2 | 6+7 4 | 3 1 | 3 1 | 2+3 2 | 3+4 2 |
| | 7 2 | | 5 1 | 5 1 | 4+5 2 | 3+4 2 |
| | 6 2 | | 7 2 | 7 2 | 4+5 2 | 3+4 2 |
| | 7 2 | | 6 1 | 6 1 | 2+3 2 | 3+4 2 |
| | 6 2 | | 4 1 | 4 1 | 2+3 2 | 3+4 2 |
| | 6 2 | | 7 2 | 7 2 | 4+5 2 | 3+4 2 |
| | 7 2 | | 7 2 | 7 2 | 4+5 2 | 3+4 2 |
| | 7 2 | | 5 1 | 5 1 | 2+3 2 | 3+4 2 |
| | 6 2 | | 3 1 | 3 1 | 2+3 2 | 3+4 2 |
| | 6 2 | | 7 2 | 7 2 | 4+5 2 | 3+4 2 |
| | 7 2 | | 4 1 | 4 1 | 2+3 2 | 3+4 2 |
| | 6 2 | | 6 1 | 6 1 | 2+3 2 | 3+4 2 |
| | 7 2 | | 7 2 | 7 2 | 4+5 2 | 3+4 2 |
| | 6 2 | | 6 1 | 6 1 | 4+5 2 | 3+4 2 |
| | 7 2 | | 4 1 | 4 1 | 2+3 2 | 3+4 2 |
| | 6 2 | | 7 2 | 7 2 | 4+5 2 | 3+4 2 |
| | 7 2 | | 5 1 | 5 1 | 4+5 2 | 3+4 2 |
| | 6 2 | | 3 1 | 3 1 | 2+3 2 | 3+4 2 |
| | 7 2 | | 6 1 | 6 1 | 2+3 2 | 3+4 2 |
| | 7 2 | | 4 1 | 4 1 | 4+5 2 | 3+4 2 |
| | 6 2 | | 7 2 | 7 2 | 2+3 2 | 3+4 2 |
| | 7 2 | | 7 2 | 7 2 | 4+5 2 | 3+4 2 |
| | 6 2 | | 3 1 | 3 1 | 2+3 2 | 3+4 2 |
| | 7 2 | | 5 1 | 5 1 | 4+5 2 | 3+4 2 |
| | 7 2 | | 7 2 | 7 2 | 4+5 2 | 3+4 2 |
| | 6 2 | | 4 1 | 4 1 | 2+3 2 | 3+4 2 |
| | 6 2 | | 6 1 | 6 1 | 4+5 2 | 3+4 2 |
| | 7 2 | | 4 1 | 4 1 | 2+3 2 | 3+4 2 |
| | 6 2 | | 3 1 | 3 1 | 4+5 2 | 3+4 2 |
| | 7 2 | | 7 2 | 7 2 | 2+3 2 | 3+4 2 |
| | 6 2 | | 6 1 | 6 1 | 4+5 2 | 3+4 2 |
| | 6 2 | | 7 2 | 7 2 | 2+3 2 | 3+4 2 |
| | 7 2 | | 5 1 | 5 1 | 2+3 2 | 3+4 2 |
| | 6 2 | | 5 1 | 5 1 | 4+5 2 | 3+4 2 |
| | 7 2 | | 3 1 | 3 1 | 4+5 2 | 3+4 2 |
| | 6 2 | | 7 2 | 7 2 | 2+3 2 | 3+4 2 |
| | 7 2 | | 4 1 | 4 1 | 4+5 2 | 3+4 2 |
| | 6 2 | | 6 1 | 6 1 | 2+3 2 | 3+4 2 |
| | 7 2 | | 3 1 | 3 1 | 2+3 2 | 3+4 2 |
| | 6 2 | | 5 1 | 5 1 | 4+5 2 | 3+4 2 |
| | 6 2 | | 7 2 | 7 2 | 2+3 2 | 3+4 2 |
| | 7 2 | | 7 2 | 7 2 | 4+5 2 | 3+4 2 |
| | 6 2 | | 6 1 | 6 1 | 2+3 2 | 3+4 2 |
| | 7 2 | | 7 2 | 7 2 | 4+5 2 | 3+4 2 |
| | 6 2 | | 4 1 | 4 1 | 4+5 2 | 3+4 2 |
| | 7 2 | | 7 2 | 7 2 | 2+3 2 | 3+4 2 |
| | 7 2 | | 3 1 | 3 1 | 4+5 2 | 3+4 2 |
| | 6 2 | | 5 1 | 5 1 | 2+3 2 | 3+4 2 |
| | 6 2 | | 5 1 | 5 1 | 4+5 2 | 3+4 2 |
| | 7 2 | | 6 1 | 6 1 | 2+3 2 | 3+4 2 |
| | | | 7 2 | 7 2 | | |
| | | | 4 1 | 4 1 | | |
| | | | 3 1 | 3 1 | | |
| | | | 7 2 | 7 2 | | |
| | | | 7 2 | 7 2 | | |
| | | | 5 1 | 5 1 | | |
| | | | 7 2 | 7 2 | | |
| | | | 6 1 | 6 1 | | |
| | | | 3 1 | 3 1 | | |
| | | | 4 1 | 4 1 | | |
| | | | 6 1 | 6 1 | | |

Continued on next column

**TABLE LV– continued**

| TQC | ICSD | | NoE | | SF | |
|---|---|---|---|---|---|---|
| | | | 7 2 | 7 2 | | |
| | | | 4 1 | 4 1 | | |
| | | | 5 1 | 5 1 | | |
| | | | 7 2 | 7 2 | | |
| | | | 3 1 | 3 1 | | |
| | | | 6 1 | 6 1 | | |
| | | | 4 1 | 4 1 | | |
| | | | 7 2 | 7 2 | | |
| | | | 7 2 | 7 2 | | |
| | | | 3 1 | 3 1 | | |
| | | | 7 2 | 7 2 | | |
| | | | 5 1 | 5 1 | | |
| | | | 4 1 | 4 1 | | |
| | | | 6 1 | 6 1 | | |
| | | | 5 1 | 5 1 | | |
| | | | 3 1 | 3 1 | | |
| | | | 7 2 | 7 2 | | |
| | | | 6 1 | 6 1 | | |
| | | | 7 2 | 7 2 | | |
| | | | 4 1 | 4 1 | | |
| | | | 3 1 | 3 1 | | |
| | | | 5 1 | 5 1 | | |
| | | | 7 2 | 7 2 | | |
| | | | 4 1 | 4 1 | | |
| | | | 6 1 | 6 1 | | |
| | | | 7 2 | 7 2 | | |
| | | | 3 1 | 3 1 | | |
| | | | 5 1 | 5 1 | | |
| | | | 7 2 | 7 2 | | |
| | | | 7 2 | 7 2 | | |
| | | | 7 2 | 7 2 | | |
| | | | 3 1 | 3 1 | | |
| | | | 4 1 | 4 1 | | |
| | | | 5 1 | 5 1 | | |
| | | | 6 1 | 6 1 | | |
| | | | 3 1 | 5 1 | | |
| | | | 5 1 | 3 1 | | |
| | | | 7 2 | 7 2 | | |
| 122.1.3 | 16427 | | 196 | | F5U1 | |
| | -GM | -M | -P | -PA | -X | -N |
| | 7 2 | 6+7 4 | 4 1 | 4 1 | 2+3 2 | 3+4 2 |
| | 6 2 | 6+7 4 | 6 1 | 6 1 | 4+5 2 | 3+4 2 |
| | 7 2 | 6+7 4 | 7 2 | 7 2 | 2+3 2 | 3+4 2 |
| | 6 2 | 6+7 4 | 3 1 | 3 1 | 4+5 2 | 3+4 2 |
| | 7 2 | 6+7 4 | 5 1 | 5 1 | 2+3 2 | 3+4 2 |
| | 7 2 | 6+7 4 | 7 2 | 7 2 | 4+5 2 | 3+4 2 |
| | 6 2 | 6+7 4 | 6 1 | 6 1 | 2+3 2 | 3+4 2 |
| | 6 2 | 6+7 4 | 4 1 | 4 1 | 4+5 2 | 3+4 2 |
| | 7 2 | 6+7 4 | 7 2 | 7 2 | 4+5 2 | 3+4 2 |
| | 6 2 | 6+7 4 | 5 1 | 5 1 | 2+3 2 | 3+4 2 |
| | 7 2 | 6+7 4 | 3 1 | 3 1 | 4+5 2 | 3+4 2 |
| | 6 2 | 6+7 4 | 7 2 | 7 2 | 2+3 2 | 3+4 2 |
| | 7 2 | 6+7 4 | 7 2 | 7 2 | 2+3 2 | 3+4 2 |
| | 6 2 | 6+7 4 | 4 1 | 4 1 | 4+5 2 | 3+4 2 |
| | 7 2 | 6+7 4 | 6 1 | 6 1 | 2+3 2 | 3+4 2 |
| | 7 2 | 6+7 4 | 7 2 | 7 2 | 4+5 2 | 3+4 2 |
| | 6 2 | 6+7 4 | 3 1 | 3 1 | 2+3 2 | 3+4 2 |
| | 6 2 | 6+7 4 | 5 1 | 5 1 | 4+5 2 | 3+4 2 |
| | 7 2 | 6+7 4 | 7 2 | 7 2 | 2+3 2 | 3+4 2 |
| | 6 2 | 6+7 4 | 5 1 | 5 1 | 4+5 2 | 3+4 2 |
| | 7 2 | 6+7 4 | 3 1 | 3 1 | 2+3 2 | 3+4 2 |

Continued on next column

**TABLE LV– continued**

| TQC | ICSD | | NoE | | SF | |
|---|---|---|---|---|---|---|
| | 6 2 | 6+7 4 | 6 1 | 6 1 | 4+5 2 | 3+4 2 |
| | 6 2 | 6+7 4 | 4 1 | 4 1 | 2+3 2 | 3+4 2 |
| | 7 2 | 6+7 4 | 7 2 | 7 2 | 4+5 2 | 3+4 2 |
| | 6 2 | 6+7 4 | 7 2 | 7 2 | 2+3 2 | 3+4 2 |
| | 6 2 | 6+7 4 | 6 1 | 6 1 | 4+5 2 | 3+4 2 |
| | 7 2 | 6+7 4 | 4 1 | 4 1 | 4+5 2 | 3+4 2 |
| | 7 2 | 6+7 4 | 3 1 | 3 1 | 2+3 2 | 3+4 2 |
| | 7 2 | 6+7 4 | 5 1 | 5 1 | 2+3 2 | 3+4 2 |
| | 6 2 | 6+7 4 | 7 2 | 7 2 | 4+5 2 | 3+4 2 |
| | 6 2 | 6+7 4 | 7 2 | 7 2 | 4+5 2 | 3+4 2 |
| | 7 2 | 6+7 4 | 6 1 | 6 1 | 2+3 2 | 3+4 2 |
| | 7 2 | 6+7 4 | 4 1 | 4 1 | 2+3 2 | 3+4 2 |
| | 6 2 | 6+7 4 | 5 1 | 5 1 | 4+5 2 | 3+4 2 |
| | 6 2 | 6+7 4 | 3 1 | 3 1 | 4+5 2 | 3+4 2 |
| | 7 2 | 6+7 4 | 7 2 | 7 2 | 2+3 2 | 3+4 2 |
| | 6 2 | 6+7 4 | 5 1 | 5 1 | 2+3 2 | 3+4 2 |
| | 6 2 | 6+7 4 | 3 1 | 3 1 | 4+5 2 | 3+4 2 |
| | 7 2 | 6+7 4 | 7 2 | 7 2 | 4+5 2 | 3+4 2 |
| | 7 2 | 6+7 4 | 7 2 | 7 2 | 2+3 2 | 3+4 2 |
| | 6 2 | 6+7 4 | 6 1 | 6 1 | 2+3 2 | 3+4 2 |
| | 7 2 | 6+7 4 | 4 1 | 4 1 | 4+5 2 | 3+4 2 |
| | 6 2 | 6+7 4 | 7 2 | 7 2 | 2+3 2 | 3+4 2 |
| | 6 2 | 6+7 4 | 4 1 | 4 1 | 4+5 2 | 3+4 2 |
| | 7 2 | 6+7 4 | 6 1 | 6 1 | 4+5 2 | 3+4 2 |
| | 7 2 | 6+7 4 | 5 1 | 5 1 | 2+3 2 | 3+4 2 |
| | 7 2 | 6+7 4 | 3 1 | 3 1 | 2+3 2 | 3+4 2 |
| | 6 2 | 6+7 4 | 7 2 | 7 2 | 4+5 2 | 3+4 2 |
| | 7 2 | 6+7 4 | 3 1 | 3 1 | 2+3 2 | 3+4 2 |
| | 7 2 | | 5 1 | 5 1 | 4+5 2 | 3+4 2 |
| | 6 2 | | 7 2 | 7 2 | 4+5 2 | 3+4 2 |
| | 7 2 | | 6 1 | 6 1 | 2+3 2 | 3+4 2 |
| | 6 2 | | 4 1 | 4 1 | 2+3 2 | 3+4 2 |
| | 7 2 | | 7 2 | 7 2 | 4+5 2 | 3+4 2 |
| | 6 2 | | 7 2 | 7 2 | 4+5 2 | 3+4 2 |
| | 7 2 | | 5 1 | 5 1 | 2+3 2 | 3+4 2 |
| | 6 2 | | 3 1 | 3 1 | 2+3 2 | 3+4 2 |
| | 6 2 | | 7 2 | 7 2 | 4+5 2 | 3+4 2 |
| | 7 2 | | 4 1 | 4 1 | 2+3 2 | 3+4 2 |
| | 7 2 | | 6 1 | 6 1 | 2+3 2 | 3+4 2 |
| | 6 2 | | 7 2 | 7 2 | 4+5 2 | 3+4 2 |
| | 6 2 | | 6 1 | 6 1 | 4+5 2 | 3+4 2 |
| | 7 2 | | 4 1 | 4 1 | 2+3 2 | 3+4 2 |
| | 6 2 | | 7 2 | 7 2 | 4+5 2 | 3+4 2 |
| | 7 2 | | 5 1 | 5 1 | 4+5 2 | 3+4 2 |
| | 6 2 | | 6 1 | 6 1 | 2+3 2 | 3+4 2 |
| | 7 2 | | 3 1 | 3 1 | 2+3 2 | 3+4 2 |
| | 7 2 | | 4 1 | 4 1 | 4+5 2 | 3+4 2 |
| | 7 2 | | 7 2 | 7 2 | 2+3 2 | 3+4 2 |
| | 6 2 | | 7 2 | 7 2 | 4+5 2 | 3+4 2 |
| | 6 2 | | 3 1 | 3 1 | 2+3 2 | 3+4 2 |
| | 7 2 | | 5 1 | 5 1 | 4+5 2 | 3+4 2 |
| | 7 2 | | 7 2 | 7 2 | 4+5 2 | 3+4 2 |
| | 6 2 | | 4 1 | 4 1 | 2+3 2 | 3+4 2 |
| | 6 2 | | 6 1 | 6 1 | 4+5 2 | 3+4 2 |
| | 7 2 | | 4 1 | 4 1 | 2+3 2 | 3+4 2 |
| | 6 2 | | 3 1 | 3 1 | 4+5 2 | 3+4 2 |
| | 7 2 | | 7 2 | 7 2 | 2+3 2 | 3+4 2 |
| | 6 2 | | 6 1 | 6 1 | 4+5 2 | 3+4 2 |
| | 6 2 | | 7 2 | 7 2 | 2+3 2 | 3+4 2 |
| | 7 2 | | 5 1 | 5 1 | 4+5 2 | 3+4 2 |
| | 6 2 | | 5 1 | 5 1 | 2+3 2 | 3+4 2 |

Continued on next column

**TABLE LV– continued**

| TQC | ICSD | | NoE | | SF | |
|---|---|---|---|---|---|---|
| | 7 2 | | 3 1 | 3 1 | 4+5 2 | 3+4 2 |
| | 6 2 | | 7 2 | 7 2 | 2+3 2 | 3+4 2 |
| | 6 2 | | 4 1 | 4 1 | 4+5 2 | 3+4 2 |
| | 7 2 | | 6 1 | 6 1 | 2+3 2 | 3+4 2 |
| | 6 2 | | 3 1 | 3 1 | 2+3 2 | 3+4 2 |
| | 7 2 | | 7 2 | 7 2 | 4+5 2 | 3+4 2 |
| | 6 2 | | 5 1 | 5 1 | 2+3 2 | 3+4 2 |
| | 7 2 | | 7 2 | 7 2 | 4+5 2 | 3+4 2 |
| | 6 2 | | 6 1 | 6 1 | 4+5 2 | 3+4 2 |
| | 7 2 | | 7 2 | 7 2 | 2+3 2 | 3+4 2 |
| | 6 2 | | 7 2 | 7 2 | 4+5 2 | 3+4 2 |
| | 7 2 | | 4 1 | 4 1 | 2+3 2 | 3+4 2 |
| | 7 2 | | 3 1 | 3 1 | 2+3 2 | 3+4 2 |
| | 6 2 | | 5 1 | 5 1 | 4+5 2 | 3+4 2 |
| | 6 2 | | 5 1 | 5 1 | 4+5 2 | 3+4 2 |
| | 7 2 | | 6 1 | 6 1 | 2+3 2 | 3+4 2 |
| | | | 7 2 | 7 2 | | |
| | | | 4 1 | 4 1 | | |
| | | | 3 1 | 3 1 | | |
| | | | 7 2 | 7 2 | | |
| | | | 7 2 | 7 2 | | |
| | | | 7 2 | 7 2 | | |
| | | | 5 1 | 5 1 | | |
| | | | 6 1 | 6 1 | | |
| | | | 4 1 | 4 1 | | |
| | | | 3 1 | 3 1 | | |
| | | | 6 1 | 6 1 | | |
| | | | 7 2 | 7 2 | | |
| | | | 4 1 | 4 1 | | |
| | | | 5 1 | 5 1 | | |
| | | | 3 1 | 3 1 | | |
| | | | 7 2 | 7 2 | | |
| | | | 6 1 | 6 1 | | |
| | | | 4 1 | 4 1 | | |
| | | | 7 2 | 7 2 | | |
| | | | 7 2 | 7 2 | | |
| | | | 3 1 | 3 1 | | |
| | | | 7 2 | 7 2 | | |
| | | | 5 1 | 5 1 | | |
| | | | 6 1 | 6 1 | | |
| | | | 4 1 | 4 1 | | |
| | | | 5 1 | 5 1 | | |
| | | | 3 1 | 3 1 | | |
| | | | 7 2 | 7 2 | | |
| | | | 7 2 | 7 2 | | |
| | | | 6 1 | 6 1 | | |
| | | | 3 1 | 3 1 | | |
| | | | 4 1 | 4 1 | | |
| | | | 7 2 | 7 2 | | |
| | | | 5 1 | 5 1 | | |
| | | | 3 1 | 3 1 | | |
| | | | 4 1 | 4 1 | | |
| | | | 6 1 | 6 1 | | |
| | | | 7 2 | 7 2 | | |
| | | | 7 2 | 7 2 | | |
| | | | 5 1 | 5 1 | | |
| | | | 7 2 | 7 2 | | |
| | | | 7 2 | 7 2 | | |
| | | | 3 1 | 3 1 | | |
| | | | 4 1 | 4 1 | | |
| | | | 5 1 | 5 1 | | |

Continued on next column

**TABLE LV– continued**

| TQC | ICSD | | NoE | | SF | |
|---|---|---|---|---|---|---|
| | | | 6 1 | 6 1 | | |
| | | | 3 1 | 3 1 | | |
| | | | 5 1 | 5 1 | | |
| | | | 7 2 | 7 2 | | |
| 122.1.4 | 16426 | | 196 | | F5U1 | |
| | -GM | -M | -P | -PA | -X | -N |
| | 7 2 | 6+7 4 | 4 1 | 6 1 | 2+3 2 | 3+4 2 |
| | 6 2 | 6+7 4 | 7 2 | 7 2 | 4+5 2 | 3+4 2 |
| | 7 2 | 6+7 4 | 5 1 | 3 1 | 2+3 2 | 3+4 2 |
| | 6 2 | 6+7 4 | 6 1 | 4 1 | 4+5 2 | 3+4 2 |
| | 7 2 | 6+7 4 | 7 2 | 7 2 | 2+3 2 | 3+4 2 |
| | 7 2 | 6+7 4 | 3 1 | 5 1 | 4+5 2 | 3+4 2 |
| | 6 2 | 6+7 4 | 6 1 | 4 1 | 2+3 2 | 3+4 2 |
| | 6 2 | 6+7 4 | 7 2 | 6 1 | 4+5 2 | 3+4 2 |
| | 7 2 | 6+7 4 | 4 1 | 7 2 | 4+5 2 | 3+4 2 |
| | 6 2 | 6+7 4 | 3 1 | 5 1 | 2+3 2 | 3+4 2 |
| | 7 2 | 6+7 4 | 5 1 | 3 1 | 4+5 2 | 3+4 2 |
| | 6 2 | 6+7 4 | 7 2 | 7 2 | 2+3 2 | 3+4 2 |
| | 7 2 | 6+7 4 | 7 2 | 7 2 | 2+3 2 | 3+4 2 |
| | 7 2 | 6+7 4 | 4 1 | 4 1 | 4+5 2 | 3+4 2 |
| | 6 2 | 6+7 4 | 6 1 | 6 1 | 4+5 2 | 3+4 2 |
| | 6 2 | 6+7 4 | 3 1 | 3 1 | 2+3 2 | 3+4 2 |
| | 7 2 | 6+7 4 | 5 1 | 5 1 | 2+3 2 | 3+4 2 |
| | 6 2 | 6+7 4 | 7 2 | 7 2 | 4+5 2 | 3+4 2 |
| | 7 2 | 6+7 4 | 7 2 | 7 2 | 2+3 2 | 3+4 2 |
| | 6 2 | 6+7 4 | 4 1 | 4 1 | 4+5 2 | 3+4 2 |
| | 7 2 | 6+7 4 | 6 1 | 6 1 | 2+3 2 | 3+4 2 |
| | 6 2 | 6+7 4 | 7 2 | 7 2 | 4+5 2 | 3+4 2 |
| | 6 2 | 6+7 4 | 3 1 | 3 1 | 2+3 2 | 3+4 2 |
| | 7 2 | 6+7 4 | 5 1 | 5 1 | 4+5 2 | 3+4 2 |
| | 6 2 | 6+7 4 | 5 1 | 5 1 | 2+3 2 | 3+4 2 |
| | 6 2 | 6+7 4 | 6 1 | 6 1 | 4+5 2 | 3+4 2 |
| | 7 2 | 6+7 4 | 7 2 | 7 2 | 4+5 2 | 3+4 2 |
| | 7 2 | 6+7 4 | 3 1 | 3 1 | 2+3 2 | 3+4 2 |
| | 7 2 | 6+7 4 | 4 1 | 4 1 | 2+3 2 | 3+4 2 |
| | 6 2 | 6+7 4 | 7 2 | 7 2 | 4+5 2 | 3+4 2 |
| | 6 2 | 6+7 4 | 7 2 | 7 2 | 4+5 2 | 3+4 2 |
| | 7 2 | 6+7 4 | 6 1 | 6 1 | 2+3 2 | 3+4 2 |
| | 7 2 | 6+7 4 | 4 1 | 4 1 | 2+3 2 | 3+4 2 |
| | 6 2 | 6+7 4 | 5 1 | 5 1 | 4+5 2 | 3+4 2 |
| | 6 2 | 6+7 4 | 3 1 | 3 1 | 4+5 2 | 3+4 2 |
| | 7 2 | 6+7 4 | 7 2 | 7 2 | 2+3 2 | 3+4 2 |
| | 6 2 | 6+7 4 | 3 1 | 5 1 | 2+3 2 | 3+4 2 |
| | 6 2 | 6+7 4 | 5 1 | 3 1 | 4+5 2 | 3+4 2 |
| | 7 2 | 6+7 4 | 7 2 | 7 2 | 2+3 2 | 3+4 2 |
| | 6 2 | 6+7 4 | 7 2 | 7 2 | 4+5 2 | 3+4 2 |
| | 7 2 | 6+7 4 | 6 1 | 4 1 | 4+5 2 | 3+4 2 |
| | 6 2 | 6+7 4 | 4 1 | 6 1 | 2+3 2 | 3+4 2 |
| | 7 2 | 6+7 4 | 7 2 | 7 2 | 4+5 2 | 3+4 2 |
| | 7 2 | 6+7 4 | 4 1 | 4 1 | 2+3 2 | 3+4 2 |
| | 6 2 | 6+7 4 | 6 1 | 6 1 | 4+5 2 | 3+4 2 |
| | 7 2 | 6+7 4 | 3 1 | 5 1 | 2+3 2 | 3+4 2 |
| | 7 2 | 6+7 4 | 5 1 | 3 1 | 2+3 2 | 3+4 2 |
| | 6 2 | 6+7 4 | 7 2 | 7 2 | 4+5 2 | 3+4 2 |
| | 7 2 | 6+7 4 | 5 1 | 3 1 | 2+3 2 | 3+4 2 |
| | 7 2 | | 3 1 | 5 1 | 4+5 2 | 3+4 2 |
| | 6 2 | | 7 2 | 7 2 | 4+5 2 | 3+4 2 |
| | 7 2 | | 6 1 | 6 1 | 2+3 2 | 3+4 2 |
| | 6 2 | | 4 1 | 4 1 | 2+3 2 | 3+4 2 |
| | 7 2 | | 7 2 | 7 2 | 4+5 2 | 3+4 2 |
| | 6 2 | | 7 2 | 7 2 | 4+5 2 | 3+4 2 |

Continued on next column

**TABLE LV– continued**

| TQC | ICSD | | NoE | | SF | |
|---|---|---|---|---|---|---|
| | 7 2 | | 5 1 | 5 1 | 2+3 2 | 3+4 2 |
| | 6 2 | | 3 1 | 3 1 | 4+5 2 | 3+4 2 |
| | 6 2 | | 7 2 | 7 2 | 2+3 2 | 3+4 2 |
| | 6 2 | | 6 1 | 6 1 | 2+3 2 | 3+4 2 |
| | 7 2 | | 4 1 | 4 1 | 4+5 2 | 3+4 2 |
| | 7 2 | | 7 2 | 7 2 | 4+5 2 | 3+4 2 |
| | 6 2 | | 6 1 | 6 1 | 2+3 2 | 3+4 2 |
| | 6 2 | | 4 1 | 4 1 | 2+3 2 | 3+4 2 |
| | 7 2 | | 5 1 | 5 1 | 4+5 2 | 3+4 2 |
| | 7 2 | | 7 2 | 7 2 | 2+3 2 | 3+4 2 |
| | 7 2 | | 3 1 | 3 1 | 4+5 2 | 3+4 2 |
| | 6 2 | | 6 1 | 4 1 | 2+3 2 | 3+4 2 |
| | 7 2 | | 7 2 | 7 2 | 4+5 2 | 3+4 2 |
| | 7 2 | | 4 1 | 6 1 | 2+3 2 | 3+4 2 |
| | 6 2 | | 7 2 | 7 2 | 4+5 2 | 3+4 2 |
| | 7 2 | | 3 1 | 3 1 | 2+3 2 | 3+4 2 |
| | 6 2 | | 5 1 | 5 1 | 4+5 2 | 3+4 2 |
| | 7 2 | | 4 1 | 4 1 | 4+5 2 | 3+4 2 |
| | 6 2 | | 6 1 | 6 1 | 2+3 2 | 3+4 2 |
| | 6 2 | | 7 2 | 7 2 | 4+5 2 | 3+4 2 |
| | 7 2 | | 3 1 | 3 1 | 2+3 2 | 3+4 2 |
| | 6 2 | | 7 2 | 7 2 | 4+5 2 | 3+4 2 |
| | 7 2 | | 5 1 | 5 1 | 2+3 2 | 3+4 2 |
| | 6 2 | | 3 1 | 3 1 | 4+5 2 | 3+4 2 |
| | 7 2 | | 5 1 | 5 1 | 2+3 2 | 3+4 2 |
| | 6 2 | | 7 2 | 4 1 | 4+5 2 | 3+4 2 |
| | 6 2 | | 4 1 | 7 2 | 2+3 2 | 3+4 2 |
| | 7 2 | | 6 1 | 6 1 | 4+5 2 | 3+4 2 |
| | 6 2 | | 7 2 | 7 2 | 2+3 2 | 3+4 2 |
| | 7 2 | | 4 1 | 4 1 | 4+5 2 | 3+4 2 |
| | 6 2 | | 6 1 | 6 1 | 2+3 2 | 3+4 2 |
| | 6 2 | | 7 2 | 7 2 | 2+3 2 | 3+4 2 |
| | 7 2 | | 3 1 | 3 1 | 4+5 2 | 3+4 2 |
| | 6 2 | | 5 1 | 5 1 | 2+3 2 | 3+4 2 |
| | 7 2 | | 7 2 | 7 2 | 4+5 2 | 3+4 2 |
| | 6 2 | | 6 1 | 6 1 | 4+5 2 | 3+4 2 |
| | 7 2 | | 4 1 | 4 1 | 2+3 2 | 3+4 2 |
| | 6 2 | | 7 2 | 7 2 | 4+5 2 | 3+4 2 |
| | 7 2 | | 7 2 | 7 2 | 2+3 2 | 3+4 2 |
| | 7 2 | | 3 1 | 3 1 | 2+3 2 | 3+4 2 |
| | 6 2 | | 5 1 | 5 1 | 4+5 2 | 3+4 2 |
| | 6 2 | | 5 1 | 5 1 | 4+5 2 | 3+4 2 |
| | 7 2 | | 7 2 | 7 2 | 2+3 2 | 3+4 2 |
| | | | 6 1 | 6 1 | | |
| | | | 4 1 | 4 1 | | |
| | | | 3 1 | 3 1 | | |
| | | | 7 2 | 7 2 | | |
| | | | 7 2 | 7 2 | | |
| | | | 6 1 | 6 1 | | |
| | | | 7 2 | 7 2 | | |
| | | | 4 1 | 4 1 | | |
| | | | 5 1 | 5 1 | | |
| | | | 3 1 | 3 1 | | |
| | | | 7 2 | 7 2 | | |
| | | | 4 1 | 4 1 | | |
| | | | 6 1 | 6 1 | | |
| | | | 5 1 | 5 1 | | |
| | | | 3 1 | 3 1 | | |
| | | | 7 2 | 7 2 | | |
| | | | 6 1 | 6 1 | | |
| | | | 4 1 | 4 1 | | |

Continued on next column

**TABLE LV– continued**

| TQC | ICSD | | NoE | | SF | |
|---|---|---|---|---|---|---|
| | | | 7 2 | 7 2 | | |
| | | | 7 2 | 7 2 | | |
| | | | 5 1 | 5 1 | | |
| | | | 3 1 | 3 1 | | |
| | | | 6 1 | 6 1 | | |
| | | | 7 2 | 7 2 | | |
| | | | 4 1 | 4 1 | | |
| | | | 5 1 | 5 1 | | |
| | | | 3 1 | 3 1 | | |
| | | | 7 2 | 7 2 | | |
| | | | 7 2 | 7 2 | | |
| | | | 6 1 | 6 1 | | |
| | | | 3 1 | 3 1 | | |
| | | | 4 1 | 4 1 | | |
| | | | 5 1 | 5 1 | | |
| | | | 7 2 | 7 2 | | |
| | | | 3 1 | 3 1 | | |
| | | | 7 2 | 4 1 | | |
| | | | 4 1 | 7 2 | | |
| | | | 6 1 | 6 1 | | |
| | | | 7 2 | 7 2 | | |
| | | | 5 1 | 5 1 | | |
| | | | 7 2 | 7 2 | | |
| | | | 7 2 | 7 2 | | |
| | | | 3 1 | 3 1 | | |
| | | | 4 1 | 4 1 | | |
| | | | 5 1 | 5 1 | | |
| | | | 6 1 | 6 1 | | |
| | | | 5 1 | 3 1 | | |
| | | | 3 1 | 7 2 | | |
| | | | 7 2 | 5 1 | | |
| 122.1.5 | 16425 | | 196 | | F5U1 | |
| | -GM | -M | -P | -PA | -X | -N |
| | 7 2 | 6+7 4 | 4 1 | 4 1 | 2+3 2 | 3+4 2 |
| | 6 2 | 6+7 4 | 6 1 | 6 1 | 4+5 2 | 3+4 2 |
| | 7 2 | 6+7 4 | 7 2 | 7 2 | 2+3 2 | 3+4 2 |
| | 6 2 | 6+7 4 | 3 1 | 3 1 | 4+5 2 | 3+4 2 |
| | 7 2 | 6+7 4 | 5 1 | 5 1 | 2+3 2 | 3+4 2 |
| | 7 2 | 6+7 4 | 7 2 | 7 2 | 4+5 2 | 3+4 2 |
| | 6 2 | 6+7 4 | 6 1 | 6 1 | 2+3 2 | 3+4 2 |
| | 6 2 | 6+7 4 | 4 1 | 4 1 | 4+5 2 | 3+4 2 |
| | 7 2 | 6+7 4 | 7 2 | 7 2 | 4+5 2 | 3+4 2 |
| | 6 2 | 6+7 4 | 5 1 | 5 1 | 2+3 2 | 3+4 2 |
| | 7 2 | 6+7 4 | 3 1 | 3 1 | 4+5 2 | 3+4 2 |
| | 6 2 | 6+7 4 | 7 2 | 7 2 | 2+3 2 | 3+4 2 |
| | 7 2 | 6+7 4 | 4 1 | 4 1 | 2+3 2 | 3+4 2 |
| | 6 2 | 6+7 4 | 7 2 | 7 2 | 4+5 2 | 3+4 2 |
| | 7 2 | 6+7 4 | 6 1 | 6 1 | 4+5 2 | 3+4 2 |
| | 6 2 | 6+7 4 | 7 2 | 7 2 | 2+3 2 | 3+4 2 |
| | 7 2 | 6+7 4 | 3 1 | 3 1 | 2+3 2 | 3+4 2 |
| | 6 2 | 6+7 4 | 5 1 | 5 1 | 4+5 2 | 3+4 2 |
| | 7 2 | 6+7 4 | 7 2 | 7 2 | 2+3 2 | 3+4 2 |
| | 6 2 | 6+7 4 | 6 1 | 6 1 | 4+5 2 | 3+4 2 |
| | 7 2 | 6+7 4 | 4 1 | 4 1 | 2+3 2 | 3+4 2 |
| | 6 2 | 6+7 4 | 3 1 | 3 1 | 4+5 2 | 3+4 2 |
| | 6 2 | 6+7 4 | 7 2 | 7 2 | 2+3 2 | 3+4 2 |
| | 7 2 | 6+7 4 | 5 1 | 5 1 | 4+5 2 | 3+4 2 |
| | 6 2 | 6+7 4 | 5 1 | 5 1 | 2+3 2 | 3+4 2 |
| | 6 2 | 6+7 4 | 7 2 | 7 2 | 4+5 2 | 3+4 2 |
| | 7 2 | 6+7 4 | 6 1 | 6 1 | 4+5 2 | 3+4 2 |
| | 7 2 | 6+7 4 | 4 1 | 4 1 | 2+3 2 | 3+4 2 |

Continued on next column

**TABLE LV– continued**

| TQC | ICSD | | NoE | | SF | |
|---|---|---|---|---|---|---|
| | 7 2 | 6+7 4 | 3 1 | 3 1 | 2+3 2 | 3+4 2 |
| | 6 2 | 6+7 4 | 7 2 | 7 2 | 4+5 2 | 3+4 2 |
| | 6 2 | 6+7 4 | 7 2 | 7 2 | 4+5 2 | 3+4 2 |
| | 7 2 | 6+7 4 | 6 1 | 6 1 | 2+3 2 | 3+4 2 |
| | 7 2 | 6+7 4 | 4 1 | 4 1 | 2+3 2 | 3+4 2 |
| | 6 2 | 6+7 4 | 5 1 | 5 1 | 4+5 2 | 3+4 2 |
| | 6 2 | 6+7 4 | 3 1 | 3 1 | 4+5 2 | 3+4 2 |
| | 7 2 | 6+7 4 | 7 2 | 7 2 | 2+3 2 | 3+4 2 |
| | 6 2 | 6+7 4 | 5 1 | 5 1 | 2+3 2 | 3+4 2 |
| | 6 2 | 6+7 4 | 3 1 | 3 1 | 4+5 2 | 3+4 2 |
| | 7 2 | 6+7 4 | 7 2 | 7 2 | 2+3 2 | 3+4 2 |
| | 6 2 | 6+7 4 | 7 2 | 7 2 | 4+5 2 | 3+4 2 |
| | 6 2 | 6+7 4 | 6 1 | 6 1 | 4+5 2 | 3+4 2 |
| | 7 2 | 6+7 4 | 4 1 | 4 1 | 2+3 2 | 3+4 2 |
| | 7 2 | 6+7 4 | 7 2 | 7 2 | 2+3 2 | 3+4 2 |
| | 7 2 | 6+7 4 | 4 1 | 4 1 | 4+5 2 | 3+4 2 |
| | 6 2 | 6+7 4 | 6 1 | 6 1 | 4+5 2 | 3+4 2 |
| | 7 2 | 6+7 4 | 3 1 | 3 1 | 2+3 2 | 3+4 2 |
| | 7 2 | 6+7 4 | 5 1 | 5 1 | 2+3 2 | 3+4 2 |
| | 6 2 | 6+7 4 | 7 2 | 7 2 | 4+5 2 | 3+4 2 |
| | 7 2 | 6+7 4 | 3 1 | 3 1 | 2+3 2 | 3+4 2 |
| | 7 2 | | 5 1 | 5 1 | 4+5 2 | 3+4 2 |
| | 6 2 | | 7 2 | 7 2 | 4+5 2 | 3+4 2 |
| | 7 2 | | 6 1 | 6 1 | 2+3 2 | 3+4 2 |
| | 6 2 | | 4 1 | 4 1 | 2+3 2 | 3+4 2 |
| | 7 2 | | 7 2 | 7 2 | 4+5 2 | 3+4 2 |
| | 7 2 | | 7 2 | 7 2 | 4+5 2 | 3+4 2 |
| | 6 2 | | 5 1 | 5 1 | 2+3 2 | 3+4 2 |
| | 6 2 | | 3 1 | 3 1 | 2+3 2 | 3+4 2 |
| | 6 2 | | 7 2 | 7 2 | 4+5 2 | 3+4 2 |
| | 6 2 | | 4 1 | 4 1 | 2+3 2 | 3+4 2 |
| | 7 2 | | 6 1 | 6 1 | 4+5 2 | 3+4 2 |
| | 7 2 | | 7 2 | 7 2 | 2+3 2 | 3+4 2 |
| | 6 2 | | 6 1 | 6 1 | 4+5 2 | 3+4 2 |
| | 7 2 | | 4 1 | 4 1 | 2+3 2 | 3+4 2 |
| | 6 2 | | 5 1 | 5 1 | 4+5 2 | 3+4 2 |
| | 7 2 | | 7 2 | 7 2 | 4+5 2 | 3+4 2 |
| | 7 2 | | 6 1 | 6 1 | 2+3 2 | 3+4 2 |
| | 6 2 | | 3 1 | 3 1 | 2+3 2 | 3+4 2 |
| | 7 2 | | 4 1 | 4 1 | 4+5 2 | 3+4 2 |
| | 7 2 | | 7 2 | 7 2 | 2+3 2 | 3+4 2 |
| | 6 2 | | 7 2 | 7 2 | 4+5 2 | 3+4 2 |
| | 7 2 | | 3 1 | 3 1 | 2+3 2 | 3+4 2 |
| | 6 2 | | 5 1 | 5 1 | 4+5 2 | 3+4 2 |
| | 7 2 | | 7 2 | 7 2 | 4+5 2 | 3+4 2 |
| | 6 2 | | 4 1 | 4 1 | 2+3 2 | 3+4 2 |
| | 6 2 | | 6 1 | 6 1 | 4+5 2 | 3+4 2 |
| | 7 2 | | 3 1 | 3 1 | 2+3 2 | 3+4 2 |
| | 6 2 | | 7 2 | 7 2 | 4+5 2 | 3+4 2 |
| | 7 2 | | 5 1 | 5 1 | 2+3 2 | 3+4 2 |
| | 6 2 | | 3 1 | 3 1 | 4+5 2 | 3+4 2 |
| | 7 2 | | 4 1 | 4 1 | 2+3 2 | 3+4 2 |
| | 6 2 | | 6 1 | 6 1 | 4+5 2 | 3+4 2 |
| | 6 2 | | 5 1 | 5 1 | 2+3 2 | 3+4 2 |
| | 7 2 | | 7 2 | 7 2 | 4+5 2 | 3+4 2 |
| | 6 2 | | 7 2 | 7 2 | 2+3 2 | 3+4 2 |
| | 6 2 | | 4 1 | 4 1 | 2+3 2 | 3+4 2 |
| | 7 2 | | 6 1 | 6 1 | 4+5 2 | 3+4 2 |
| | 6 2 | | 7 2 | 7 2 | 2+3 2 | 3+4 2 |
| | 7 2 | | 3 1 | 3 1 | 4+5 2 | 3+4 2 |
| | 6 2 | | 5 1 | 5 1 | 2+3 2 | 3+4 2 |

Continued on next column

**TABLE LV– continued**

| TQC | ICSD | | NoE | | SF | |
|---|---|---|---|---|---|---|
| | 7 2 | | 7 2 | 7 2 | 4+5 2 | 3+4 2 |
| | 6 2 | | 6 1 | 6 1 | 4+5 2 | 3+4 2 |
| | 7 2 | | 4 1 | 4 1 | 2+3 2 | 3+4 2 |
| | 7 2 | | 7 2 | 7 2 | 4+5 2 | 3+4 2 |
| | 6 2 | | 7 2 | 7 2 | 2+3 2 | 3+4 2 |
| | 7 2 | | 3 1 | 3 1 | 2+3 2 | 3+4 2 |
| | 6 2 | | 5 1 | 5 1 | 4+5 2 | 3+4 2 |
| | 6 2 | | 5 1 | 5 1 | 4+5 2 | 3+4 2 |
| | 7 2 | | 6 1 | 6 1 | 2+3 2 | 3+4 2 |
| | | | 7 2 | 7 2 | | |
| | | | 4 1 | 4 1 | | |
| | | | 3 1 | 3 1 | | |
| | | | 7 2 | 7 2 | | |
| | | | 7 2 | 7 2 | | |
| | | | 6 1 | 6 1 | | |
| | | | 7 2 | 7 2 | | |
| | | | 4 1 | 4 1 | | |
| | | | 5 1 | 5 1 | | |
| | | | 4 1 | 4 1 | | |
| | | | 3 1 | 3 1 | | |
| | | | 6 1 | 6 1 | | |
| | | | 7 2 | 7 2 | | |
| | | | 5 1 | 5 1 | | |
| | | | 7 2 | 7 2 | | |
| | | | 6 1 | 6 1 | | |
| | | | 3 1 | 3 1 | | |
| | | | 4 1 | 4 1 | | |
| | | | 7 2 | 7 2 | | |
| | | | 7 2 | 7 2 | | |
| | | | 3 1 | 3 1 | | |
| | | | 5 1 | 5 1 | | |
| | | | 6 1 | 6 1 | | |
| | | | 7 2 | 7 2 | | |
| | | | 4 1 | 4 1 | | |
| | | | 5 1 | 5 1 | | |
| | | | 3 1 | 3 1 | | |
| | | | 7 2 | 7 2 | | |
| | | | 7 2 | 7 2 | | |
| | | | 3 1 | 3 1 | | |
| | | | 6 1 | 6 1 | | |
| | | | 4 1 | 4 1 | | |
| | | | 7 2 | 7 2 | | |
| | | | 5 1 | 5 1 | | |
| | | | 3 1 | 3 1 | | |
| | | | 7 2 | 7 2 | | |
| | | | 4 1 | 4 1 | | |
| | | | 6 1 | 6 1 | | |
| | | | 7 2 | 7 2 | | |
| | | | 5 1 | 5 1 | | |
| | | | 7 2 | 7 2 | | |
| | | | 7 2 | 7 2 | | |
| | | | 3 1 | 3 1 | | |
| | | | 4 1 | 4 1 | | |
| | | | 5 1 | 5 1 | | |
| | | | 6 1 | 6 1 | | |
| | | | 3 1 | 5 1 | | |
| | | | 5 1 | 3 1 | | |
| | | | 7 2 | 7 2 | | |
| 122.1.6 | 16424 | | 196 | | F5U1 | |
| | -GM | -M | -P | -PA | -X | -N |
| | 7 2 | 6+7 4 | 4 1 | 6 1 | 2+3 2 | 3+4 2 |

Continued on next column

**TABLE LV– continued**

| TQC | ICSD | NoE | SF |
|---|---|---|---|
| | 6 2 6+7 4 | 7 2 | 7 2 4+5 2 3+4 2 |
| | 7 2 6+7 4 | 5 1 | 3 1 2+3 2 3+4 2 |
| | 6 2 6+7 4 | 6 1 | 4 1 4+5 2 3+4 2 |
| | 7 2 6+7 4 | 7 2 | 7 2 2+3 2 3+4 2 |
| | 7 2 6+7 4 | 3 1 | 5 1 4+5 2 3+4 2 |
| | 6 2 6+7 4 | 6 1 | 6 1 2+3 2 3+4 2 |
| | 6 2 6+7 4 | 4 1 | 4 1 4+5 2 3+4 2 |
| | 7 2 6+7 4 | 7 2 | 7 2 4+5 2 3+4 2 |
| | 6 2 6+7 4 | 5 1 | 5 1 2+3 2 3+4 2 |
| | 7 2 6+7 4 | 3 1 | 3 1 4+5 2 3+4 2 |
| | 6 2 6+7 4 | 7 2 | 7 2 2+3 2 3+4 2 |
| | 7 2 6+7 4 | 7 2 | 7 2 2+3 2 3+4 2 |
| | 6 2 6+7 4 | 4 1 | 4 1 4+5 2 3+4 2 |
| | 7 2 6+7 4 | 6 1 | 6 1 2+3 2 3+4 2 |
| | 7 2 6+7 4 | 7 2 | 7 2 4+5 2 3+4 2 |
| | 6 2 6+7 4 | 3 1 | 3 1 2+3 2 3+4 2 |
| | 6 2 6+7 4 | 5 1 | 5 1 4+5 2 3+4 2 |
| | 7 2 6+7 4 | 5 1 | 5 1 2+3 2 3+4 2 |
| | 7 2 6+7 4 | 7 2 | 7 2 4+5 2 3+4 2 |
| | 6 2 6+7 4 | 3 1 | 3 1 2+3 2 3+4 2 |
| | 6 2 6+7 4 | 6 1 | 6 1 4+5 2 3+4 2 |
| | 6 2 6+7 4 | 4 1 | 4 1 2+3 2 3+4 2 |
| | 7 2 6+7 4 | 7 2 | 7 2 4+5 2 3+4 2 |
| | 6 2 6+7 4 | 7 2 | 7 2 2+3 2 3+4 2 |
| | 6 2 6+7 4 | 6 1 | 6 1 4+5 2 3+4 2 |
| | 7 2 6+7 4 | 4 1 | 4 1 4+5 2 3+4 2 |
| | 7 2 6+7 4 | 3 1 | 3 1 2+3 2 3+4 2 |
| | 7 2 6+7 4 | 5 1 | 5 1 2+3 2 3+4 2 |
| | 6 2 6+7 4 | 7 2 | 7 2 4+5 2 3+4 2 |
| | 6 2 6+7 4 | 7 2 | 7 2 4+5 2 3+4 2 |
| | 7 2 6+7 4 | 6 1 | 6 1 2+3 2 3+4 2 |
| | 7 2 6+7 4 | 4 1 | 4 1 2+3 2 3+4 2 |
| | 6 2 6+7 4 | 5 1 | 5 1 4+5 2 3+4 2 |
| | 6 2 6+7 4 | 3 1 | 3 1 4+5 2 3+4 2 |
| | 7 2 6+7 4 | 7 2 | 7 2 2+3 2 3+4 2 |
| | 6 2 6+7 4 | 5 1 | 5 1 2+3 2 3+4 2 |
| | 6 2 6+7 4 | 3 1 | 3 1 4+5 2 3+4 2 |
| | 7 2 6+7 4 | 7 2 | 7 2 4+5 2 3+4 2 |
| | 7 2 6+7 4 | 7 2 | 7 2 2+3 2 3+4 2 |
| | 6 2 6+7 4 | 6 1 | 6 1 2+3 2 3+4 2 |
| | 6 2 6+7 4 | 4 1 | 4 1 4+5 2 3+4 2 |
| | 7 2 6+7 4 | 7 2 | 7 2 2+3 2 3+4 2 |
| | 6 2 6+7 4 | 4 1 | 4 1 4+5 2 3+4 2 |
| | 7 2 6+7 4 | 6 1 | 6 1 4+5 2 3+4 2 |
| | 7 2 6+7 4 | 5 1 | 5 1 2+3 2 3+4 2 |
| | 7 2 6+7 4 | 3 1 | 3 1 2+3 2 3+4 2 |
| | 6 2 6+7 4 | 7 2 | 7 2 4+5 2 3+4 2 |
| | 7 2 6+7 4 | 3 1 | 3 1 2+3 2 3+4 2 |
| | 7 2 | 5 1 | 5 1 4+5 2 3+4 2 |
| | 6 2 | 7 2 | 7 2 4+5 2 3+4 2 |
| | 7 2 | 6 1 | 6 1 2+3 2 3+4 2 |
| | 6 2 | 4 1 | 4 1 2+3 2 3+4 2 |
| | 7 2 | 7 2 | 7 2 4+5 2 3+4 2 |
| | 7 2 | 7 2 | 7 2 4+5 2 3+4 2 |
| | 6 2 | 5 1 | 5 1 2+3 2 3+4 2 |
| | 6 2 | 3 1 | 3 1 2+3 2 3+4 2 |
| | 7 2 | 7 2 | 7 2 4+5 2 3+4 2 |
| | 6 2 | 4 1 | 4 1 2+3 2 3+4 2 |
| | 7 2 | 6 1 | 6 1 2+3 2 3+4 2 |
| | 6 2 | 7 2 | 7 2 4+5 2 3+4 2 |
| | 6 2 | 6 1 | 6 1 4+5 2 3+4 2 |

Continued on next column

**TABLE LV– continued**

| TQC | ICSD | NoE | SF |
|---|---|---|---|
| | 7 2 | 4 1 | 4 1 2+3 2 3+4 2 |
| | 6 2 | 7 2 | 7 2 4+5 2 3+4 2 |
| | 7 2 | 5 1 | 5 1 2+3 2 3+4 2 |
| | 6 2 | 3 1 | 3 1 4+5 2 3+4 2 |
| | 7 2 | 6 1 | 6 1 2+3 2 3+4 2 |
| | 7 2 | 4 1 | 4 1 4+5 2 3+4 2 |
| | 6 2 | 7 2 | 7 2 2+3 2 3+4 2 |
| | 7 2 | 7 2 | 7 2 4+5 2 3+4 2 |
| | 6 2 | 3 1 | 3 1 4+5 2 3+4 2 |
| | 7 2 | 5 1 | 5 1 2+3 2 3+4 2 |
| | 7 2 | 7 2 | 7 2 4+5 2 3+4 2 |
| | 6 2 | 4 1 | 4 1 2+3 2 3+4 2 |
| | 6 2 | 6 1 | 6 1 4+5 2 3+4 2 |
| | 7 2 | 7 2 | 7 2 2+3 2 3+4 2 |
| | 6 2 | 3 1 | 3 1 4+5 2 3+4 2 |
| | 7 2 | 4 1 | 4 1 2+3 2 3+4 2 |
| | 6 2 | 6 1 | 6 1 4+5 2 3+4 2 |
| | 6 2 | 7 2 | 7 2 2+3 2 3+4 2 |
| | 7 2 | 5 1 | 5 1 4+5 2 3+4 2 |
| | 6 2 | 5 1 | 5 1 2+3 2 3+4 2 |
| | 7 2 | 3 1 | 3 1 4+5 2 3+4 2 |
| | 6 2 | 7 2 | 7 2 2+3 2 3+4 2 |
| | 6 2 | 4 1 | 4 1 4+5 2 3+4 2 |
| | 7 2 | 6 1 | 6 1 2+3 2 3+4 2 |
| | 7 2 | 3 1 | 3 1 2+3 2 3+4 2 |
| | 6 2 | 5 1 | 5 1 4+5 2 3+4 2 |
| | 6 2 | 7 2 | 7 2 2+3 2 3+4 2 |
| | 7 2 | 7 2 | 7 2 4+5 2 3+4 2 |
| | 6 2 | 6 1 | 6 1 2+3 2 3+4 2 |
| | 7 2 | 7 2 | 7 2 4+5 2 3+4 2 |
| | 6 2 | 7 2 | 7 2 4+5 2 3+4 2 |
| | 7 2 | 4 1 | 4 1 2+3 2 3+4 2 |
| | 7 2 | 3 1 | 3 1 4+5 2 3+4 2 |
| | 6 2 | 5 1 | 5 1 2+3 2 3+4 2 |
| | 6 2 | 6 1 | 6 1 4+5 2 3+4 2 |
| | 7 2 | 5 1 | 5 1 2+3 2 3+4 2 |
| | | 7 2 | 7 2 |
| | | 4 1 | 4 1 |
| | | 3 1 | 3 1 |
| | | 7 2 | 7 2 |
| | | 7 2 | 7 2 |
| | | 5 1 | 5 1 |
| | | 7 2 | 7 2 |
| | | 6 1 | 6 1 |
| | | 4 1 | 4 1 |
| | | 3 1 | 3 1 |
| | | 6 1 | 6 1 |
| | | 7 2 | 7 2 |
| | | 4 1 | 4 1 |
| | | 5 1 | 5 1 |
| | | 7 2 | 7 2 |
| | | 3 1 | 3 1 |
| | | 6 1 | 6 1 |
| | | 4 1 | 4 1 |
| | | 7 2 | 7 2 |
| | | 7 2 | 7 2 |
| | | 3 1 | 3 1 |
| | | 7 2 | 7 2 |
| | | 5 1 | 5 1 |
| | | 6 1 | 6 1 |
| | | 4 1 | 4 1 |

Continued on next column

**TABLE LV– continued**

| TQC | ICSD | | NoE | | SF | |
|---|---|---|---|---|---|---|
| | | | 5 1 | 5 1 | | |
| | | | 3 1 | 3 1 | | |
| | | | 7 2 | 7 2 | | |
| | | | 7 2 | 7 2 | | |
| | | | 6 1 | 6 1 | | |
| | | | 4 1 | 4 1 | | |
| | | | 3 1 | 3 1 | | |
| | | | 7 2 | 7 2 | | |
| | | | 5 1 | 5 1 | | |
| | | | 4 1 | 4 1 | | |
| | | | 6 1 | 6 1 | | |
| | | | 7 2 | 7 2 | | |
| | | | 3 1 | 3 1 | | |
| | | | 5 1 | 5 1 | | |
| | | | 7 2 | 7 2 | | |
| | | | 7 2 | 7 2 | | |
| | | | 7 2 | 7 2 | | |
| | | | 3 1 | 3 1 | | |
| | | | 4 1 | 4 1 | | |
| | | | 5 1 | 5 1 | | |
| | | | 6 1 | 6 1 | | |
| | | | 3 1 | 5 1 | | |
| | | | 5 1 | 3 1 | | |
| | | | 7 2 | 7 2 | | |
| 122.1.7 | 84027 | | 52 | | Cu1In1Se2 | |
| | -GM | -M | -PA | -P | -X | -N |
| | 7 2 | 6+7 4 | 7 2 | 7 2 | 2+3 2 | 3+4 2 |
| | 6 2 | 6+7 4 | 3 1 | 3 1 | 4+5 2 | 3+4 2 |
| | 6 2 | 6+7 4 | 5 1 | 5 1 | 2+3 2 | 3+4 2 |
| | 7 2 | 6+7 4 | 6 1 | 6 1 | 4+5 2 | 3+4 2 |
| | 7 2 | 6+7 4 | 4 1 | 4 1 | 4+5 2 | 3+4 2 |
| | 7 2 | 6+7 4 | 7 2 | 7 2 | 2+3 2 | 3+4 2 |
| | 7 2 | 6+7 4 | 3 1 | 3 1 | 4+5 2 | 3+4 2 |
| | 6 2 | 6+7 4 | 7 2 | 7 2 | 2+3 2 | 3+4 2 |
| | 6 2 | 6+7 4 | 5 1 | 5 1 | 2+3 2 | 3+4 2 |
| | 6 2 | 6+7 4 | 4 1 | 4 1 | 4+5 2 | 3+4 2 |
| | 7 2 | 6+7 4 | 7 2 | 7 2 | 4+5 2 | 3+4 2 |
| | 7 2 | 6+7 4 | 6 1 | 6 1 | 2+3 2 | 3+4 2 |
| | 6 2 | 6+7 4 | 5 1 | 5 1 | 2+3 2 | 3+4 2 |
| | 6 2 | | 7 2 | 7 2 | 4+5 2 | 3+4 2 |
| | 7 2 | | 3 1 | 3 1 | 2+3 2 | 3+4 2 |
| | 6 2 | | 4 1 | 4 1 | 4+5 2 | 3+4 2 |
| | 7 2 | | 7 2 | 7 2 | 4+5 2 | 3+4 2 |
| | 6 2 | | 6 1 | 6 1 | 2+3 2 | 3+4 2 |
| | 7 2 | | 7 2 | 7 2 | 2+3 2 | 3+4 2 |
| | 6 2 | | 5 1 | 5 1 | 4+5 2 | 3+4 2 |
| | 7 2 | | 3 1 | 3 1 | 2+3 2 | 3+4 2 |
| | 6 2 | | 6 1 | 6 1 | 4+5 2 | 3+4 2 |
| | 6 2 | | 4 1 | 4 1 | 4+5 2 | 3+4 2 |
| | 7 2 | | 7 2 | 7 2 | 2+3 2 | 3+4 2 |
| | 6 2 | | 7 2 | 7 2 | 2+3 2 | 3+4 2 |
| | 7 2 | | 3 1 | 3 1 | 4+5 2 | 3+4 2 |
| | | | 5 1 | 5 1 | | |
| | | | 6 1 | 6 1 | | |
| | | | 4 1 | 4 1 | | |
| | | | 7 2 | 7 2 | | |
| | | | 3 1 | 3 1 | | |
| | | | 5 1 | 5 1 | | |
| | | | 7 2 | 7 2 | | |
| | | | 6 1 | 6 1 | | |
| | | | 4 1 | 4 1 | | |

Continued on next column

**TABLE LV– continued**

| TQC | ICSD | | NoE | | SF | |
|---|---|---|---|---|---|---|
| | | | 7 2 | 7 2 | | |
| | | | 7 2 | 7 2 | | |
| | | | 5 1 | 5 1 | | |
| | | | 3 1 | 3 1 | | |
| 122.1.8 | 73351 | | 52 | | Cu1In1Se2 | |
| | -GM | -M | -PA | -P | -X | -N |
| | 7 2 | 6+7 4 | 7 2 | 7 2 | 2+3 2 | 3+4 2 |
| | 6 2 | 6+7 4 | 3 1 | 3 1 | 4+5 2 | 3+4 2 |
| | 6 2 | 6+7 4 | 5 1 | 5 1 | 2+3 2 | 3+4 2 |
| | 7 2 | 6+7 4 | 6 1 | 6 1 | 4+5 2 | 3+4 2 |
| | 7 2 | 6+7 4 | 4 1 | 4 1 | 4+5 2 | 3+4 2 |
| | 7 2 | 6+7 4 | 7 2 | 7 2 | 2+3 2 | 3+4 2 |
| | 7 2 | 6+7 4 | 3 1 | 3 1 | 4+5 2 | 3+4 2 |
| | 6 2 | 6+7 4 | 7 2 | 7 2 | 2+3 2 | 3+4 2 |
| | 6 2 | 6+7 4 | 5 1 | 5 1 | 2+3 2 | 3+4 2 |
| | 6 2 | 6+7 4 | 4 1 | 4 1 | 4+5 2 | 3+4 2 |
| | 7 2 | 6+7 4 | 7 2 | 7 2 | 4+5 2 | 3+4 2 |
| | 7 2 | 6+7 4 | 6 1 | 6 1 | 2+3 2 | 3+4 2 |
| | 6 2 | 6+7 4 | 5 1 | 5 1 | 2+3 2 | 3+4 2 |
| | 6 2 | | 7 2 | 7 2 | 4+5 2 | 3+4 2 |
| | 7 2 | | 3 1 | 3 1 | 2+3 2 | 3+4 2 |
| | 6 2 | | 4 1 | 4 1 | 4+5 2 | 3+4 2 |
| | 7 2 | | 7 2 | 7 2 | 4+5 2 | 3+4 2 |
| | 6 2 | | 6 1 | 6 1 | 2+3 2 | 3+4 2 |
| | 7 2 | | 7 2 | 7 2 | 2+3 2 | 3+4 2 |
| | 6 2 | | 5 1 | 5 1 | 4+5 2 | 3+4 2 |
| | 7 2 | | 3 1 | 3 1 | 2+3 2 | 3+4 2 |
| | 6 2 | | 6 1 | 6 1 | 4+5 2 | 3+4 2 |
| | 6 2 | | 4 1 | 4 1 | 4+5 2 | 3+4 2 |
| | 7 2 | | 7 2 | 7 2 | 2+3 2 | 3+4 2 |
| | 6 2 | | 7 2 | 7 2 | 2+3 2 | 3+4 2 |
| | 7 2 | | 3 1 | 3 1 | 4+5 2 | 3+4 2 |
| | | | 5 1 | 5 1 | | |
| | | | 6 1 | 6 1 | | |
| | | | 4 1 | 4 1 | | |
| | | | 7 2 | 7 2 | | |
| | | | 3 1 | 3 1 | | |
| | | | 5 1 | 5 1 | | |
| | | | 7 2 | 7 2 | | |
| | | | 6 1 | 6 1 | | |
| | | | 4 1 | 4 1 | | |
| | | | 7 2 | 7 2 | | |
| | | | 7 2 | 7 2 | | |
| | | | 5 1 | 5 1 | | |
| | | | 3 1 | 3 1 | | |
| 122.1.9 | 154409 | | 52 | | Cu1In1Se2 | |
| | -GM | -M | -PA | -P | -X | -N |
| | 7 2 | 6+7 4 | 7 2 | 7 2 | 2+3 2 | 3+4 2 |
| | 6 2 | 6+7 4 | 3 1 | 3 1 | 4+5 2 | 3+4 2 |
| | 6 2 | 6+7 4 | 5 1 | 5 1 | 2+3 2 | 3+4 2 |
| | 7 2 | 6+7 4 | 6 1 | 6 1 | 4+5 2 | 3+4 2 |
| | 7 2 | 6+7 4 | 4 1 | 4 1 | 4+5 2 | 3+4 2 |
| | 7 2 | 6+7 4 | 7 2 | 7 2 | 2+3 2 | 3+4 2 |
| | 7 2 | 6+7 4 | 3 1 | 3 1 | 4+5 2 | 3+4 2 |
| | 6 2 | 6+7 4 | 7 2 | 7 2 | 2+3 2 | 3+4 2 |
| | 6 2 | 6+7 4 | 5 1 | 5 1 | 2+3 2 | 3+4 2 |
| | 6 2 | 6+7 4 | 4 1 | 4 1 | 4+5 2 | 3+4 2 |
| | 7 2 | 6+7 4 | 7 2 | 7 2 | 4+5 2 | 3+4 2 |
| | 7 2 | 6+7 4 | 6 1 | 6 1 | 2+3 2 | 3+4 2 |
| | 6 2 | 6+7 4 | 5 1 | 5 1 | 2+3 2 | 3+4 2 |
| | 6 2 | | 7 2 | 7 2 | 4+5 2 | 3+4 2 |

Continued on next column

**TABLE LV– continued**

| TQC | ICSD | | NoE | SF | | |
|---|---|---|---|---|---|---|
| | 7 2 | | 3 1 | 3 1 | 4+5 2 | 3+4 2 |
| | 6 2 | | 4 1 | 4 1 | 2+3 2 | 3+4 2 |
| | 7 2 | | 7 2 | 7 2 | 4+5 2 | 3+4 2 |
| | 6 2 | | 6 1 | 6 1 | 2+3 2 | 3+4 2 |
| | 7 2 | | 7 2 | 7 2 | 2+3 2 | 3+4 2 |
| | 6 2 | | 5 1 | 5 1 | 4+5 2 | 3+4 2 |
| | 7 2 | | 3 1 | 3 1 | 2+3 2 | 3+4 2 |
| | 6 2 | | 6 1 | 6 1 | 4+5 2 | 3+4 2 |
| | 6 2 | | 4 1 | 4 1 | 4+5 2 | 3+4 2 |
| | 7 2 | | 7 2 | 7 2 | 2+3 2 | 3+4 2 |
| | 6 2 | | 7 2 | 7 2 | 2+3 2 | 3+4 2 |
| | 7 2 | | 3 1 | 3 1 | 4+5 2 | 3+4 2 |
| | | | 5 1 | 5 1 | | |
| | | | 6 1 | 6 1 | | |
| | | | 4 1 | 4 1 | | |
| | | | 7 2 | 7 2 | | |
| | | | 3 1 | 3 1 | | |
| | | | 5 1 | 5 1 | | |
| | | | 7 2 | 7 2 | | |
| | | | 4 1 | 4 1 | | |
| | | | 6 1 | 6 1 | | |
| | | | 7 2 | 7 2 | | |
| | | | 7 2 | 7 2 | | |
| | | | 5 1 | 5 1 | | |
| | | | 3 1 | 3 1 | | |
| 122.1.10 | 154408 | | 52 | Cu1In1Se2 | | |
| | -GM | -M | -PA | -P | -X | -N |
| | 7 2 | 6+7 4 | 7 2 | 7 2 | 2+3 2 | 3+4 2 |
| | 6 2 | 6+7 4 | 3 1 | 3 1 | 4+5 2 | 3+4 2 |
| | 6 2 | 6+7 4 | 5 1 | 5 1 | 2+3 2 | 3+4 2 |
| | 7 2 | 6+7 4 | 6 1 | 6 1 | 4+5 2 | 3+4 2 |
| | 7 2 | 6+7 4 | 4 1 | 4 1 | 4+5 2 | 3+4 2 |
| | 7 2 | 6+7 4 | 7 2 | 7 2 | 2+3 2 | 3+4 2 |
| | 7 2 | 6+7 4 | 3 1 | 3 1 | 4+5 2 | 3+4 2 |
| | 6 2 | 6+7 4 | 7 2 | 7 2 | 2+3 2 | 3+4 2 |
| | 6 2 | 6+7 4 | 5 1 | 5 1 | 2+3 2 | 3+4 2 |
| | 6 2 | 6+7 4 | 4 1 | 4 1 | 4+5 2 | 3+4 2 |
| | 7 2 | 6+7 4 | 7 2 | 7 2 | 4+5 2 | 3+4 2 |
| | 7 2 | 6+7 4 | 6 1 | 6 1 | 2+3 2 | 3+4 2 |
| | 6 2 | 6+7 4 | 5 1 | 5 1 | 2+3 2 | 3+4 2 |
| | 6 2 | | 7 2 | 7 2 | 4+5 2 | 3+4 2 |
| | 7 2 | | 3 1 | 3 1 | 4+5 2 | 3+4 2 |
| | 6 2 | | 4 1 | 4 1 | 2+3 2 | 3+4 2 |
| | 7 2 | | 7 2 | 7 2 | 4+5 2 | 3+4 2 |
| | 6 2 | | 6 1 | 6 1 | 2+3 2 | 3+4 2 |
| | 7 2 | | 7 2 | 7 2 | 2+3 2 | 3+4 2 |
| | 6 2 | | 5 1 | 5 1 | 4+5 2 | 3+4 2 |
| | 7 2 | | 3 1 | 3 1 | 2+3 2 | 3+4 2 |
| | 6 2 | | 6 1 | 6 1 | 4+5 2 | 3+4 2 |
| | 6 2 | | 4 1 | 4 1 | 4+5 2 | 3+4 2 |
| | 7 2 | | 7 2 | 7 2 | 2+3 2 | 3+4 2 |
| | 6 2 | | 7 2 | 7 2 | 2+3 2 | 3+4 2 |
| | 7 2 | | 3 1 | 3 1 | 4+5 2 | 3+4 2 |
| | | | 5 1 | 5 1 | | |
| | | | 6 1 | 6 1 | | |
| | | | 4 1 | 4 1 | | |
| | | | 7 2 | 7 2 | | |
| | | | 3 1 | 3 1 | | |
| | | | 5 1 | 5 1 | | |
| | | | 7 2 | 7 2 | | |
| | | | 4 1 | 4 1 | | |

Continued on next column

**TABLE LV– continued**

| TQC | ICSD | | NoE | SF | | |
|---|---|---|---|---|---|---|
| | | | 6 1 | 6 1 | | |
| | | | 7 2 | 7 2 | | |
| | | | 7 2 | 7 2 | | |
| | | | 5 1 | 5 1 | | |
| | | | 3 1 | 3 1 | | |
| 122.1.11 | 154407 | | 52 | Cu1In1Se2 | | |
| | -GM | -M | -PA | -P | -X | -N |
| | 7 2 | 6+7 4 | 7 2 | 7 2 | 2+3 2 | 3+4 2 |
| | 6 2 | 6+7 4 | 4 1 | 4 1 | 4+5 2 | 3+4 2 |
| | 6 2 | 6+7 4 | 6 1 | 6 1 | 2+3 2 | 3+4 2 |
| | 7 2 | 6+7 4 | 5 1 | 5 1 | 4+5 2 | 3+4 2 |
| | 7 2 | 6+7 4 | 3 1 | 3 1 | 4+5 2 | 3+4 2 |
| | 7 2 | 6+7 4 | 7 2 | 7 2 | 2+3 2 | 3+4 2 |
| | 7 2 | 6+7 4 | 4 1 | 4 1 | 4+5 2 | 3+4 2 |
| | 6 2 | 6+7 4 | 7 2 | 7 2 | 2+3 2 | 3+4 2 |
| | 6 2 | 6+7 4 | 6 1 | 6 1 | 2+3 2 | 3+4 2 |
| | 6 2 | 6+7 4 | 3 1 | 3 1 | 4+5 2 | 3+4 2 |
| | 7 2 | 6+7 4 | 7 2 | 7 2 | 4+5 2 | 3+4 2 |
| | 7 2 | 6+7 4 | 5 1 | 5 1 | 2+3 2 | 3+4 2 |
| | 6 2 | 6+7 4 | 6 1 | 6 1 | 2+3 2 | 3+4 2 |
| | 6 2 | | 7 2 | 7 2 | 4+5 2 | 3+4 2 |
| | 7 2 | | 4 1 | 4 1 | 4+5 2 | 3+4 2 |
| | 6 2 | | 3 1 | 3 1 | 2+3 2 | 3+4 2 |
| | 7 2 | | 7 2 | 7 2 | 4+5 2 | 3+4 2 |
| | 6 2 | | 5 1 | 5 1 | 2+3 2 | 3+4 2 |
| | 7 2 | | 7 2 | 7 2 | 2+3 2 | 3+4 2 |
| | 6 2 | | 6 1 | 6 1 | 4+5 2 | 3+4 2 |
| | 7 2 | | 4 1 | 4 1 | 2+3 2 | 3+4 2 |
| | 6 2 | | 5 1 | 5 1 | 4+5 2 | 3+4 2 |
| | 6 2 | | 3 1 | 3 1 | 4+5 2 | 3+4 2 |
| | 7 2 | | 7 2 | 7 2 | 2+3 2 | 3+4 2 |
| | 6 2 | | 7 2 | 7 2 | 2+3 2 | 3+4 2 |
| | 7 2 | | 4 1 | 4 1 | 4+5 2 | 3+4 2 |
| | | | 6 1 | 6 1 | | |
| | | | 5 1 | 5 1 | | |
| | | | 3 1 | 3 1 | | |
| | | | 7 2 | 7 2 | | |
| | | | 4 1 | 4 1 | | |
| | | | 6 1 | 6 1 | | |
| | | | 7 2 | 7 2 | | |
| | | | 3 1 | 3 1 | | |
| | | | 5 1 | 5 1 | | |
| | | | 7 2 | 7 2 | | |
| | | | 7 2 | 7 2 | | |
| | | | 6 1 | 6 1 | | |
| | | | 4 1 | 4 1 | | |
| 122.1.12 | 154406 | | 52 | Cu1In1Se2 | | |
| | -GM | -M | -PA | -P | -X | -N |
| | 7 2 | 6+7 4 | 7 2 | 7 2 | 2+3 2 | 3+4 2 |
| | 6 2 | 6+7 4 | 3 1 | 3 1 | 4+5 2 | 3+4 2 |
| | 6 2 | 6+7 4 | 5 1 | 5 1 | 2+3 2 | 3+4 2 |
| | 7 2 | 6+7 4 | 6 1 | 6 1 | 4+5 2 | 3+4 2 |
| | 7 2 | 6+7 4 | 4 1 | 4 1 | 4+5 2 | 3+4 2 |
| | 7 2 | 6+7 4 | 7 2 | 7 2 | 2+3 2 | 3+4 2 |
| | 7 2 | 6+7 4 | 3 1 | 3 1 | 4+5 2 | 3+4 2 |
| | 6 2 | 6+7 4 | 7 2 | 7 2 | 2+3 2 | 3+4 2 |
| | 6 2 | 6+7 4 | 5 1 | 5 1 | 2+3 2 | 3+4 2 |
| | 6 2 | 6+7 4 | 4 1 | 4 1 | 4+5 2 | 3+4 2 |
| | 7 2 | 6+7 4 | 7 2 | 7 2 | 4+5 2 | 3+4 2 |
| | 7 2 | 6+7 4 | 6 1 | 6 1 | 2+3 2 | 3+4 2 |
| | 6 2 | 6+7 4 | 5 1 | 5 1 | 2+3 2 | 3+4 2 |

Continued on next column

**TABLE LV– continued**

| TQC | ICSD | | NoE | | SF | |
|---|---|---|---|---|---|---|
| | 6 2 | | 7 2 | 7 2 | 4+5 2 | 3+4 2 |
| | 7 2 | | 3 1 | 3 1 | 4+5 2 | 3+4 2 |
| | 6 2 | | 4 1 | 4 1 | 2+3 2 | 3+4 2 |
| | 7 2 | | 7 2 | 7 2 | 4+5 2 | 3+4 2 |
| | 6 2 | | 6 1 | 6 1 | 2+3 2 | 3+4 2 |
| | 7 2 | | 7 2 | 7 2 | 2+3 2 | 3+4 2 |
| | 6 2 | | 5 1 | 5 1 | 4+5 2 | 3+4 2 |
| | 7 2 | | 3 1 | 3 1 | 2+3 2 | 3+4 2 |
| | 6 2 | | 6 1 | 6 1 | 4+5 2 | 3+4 2 |
| | 6 2 | | 4 1 | 4 1 | 4+5 2 | 3+4 2 |
| | 7 2 | | 7 2 | 7 2 | 2+3 2 | 3+4 2 |
| | 6 2 | | 7 2 | 7 2 | 2+3 2 | 3+4 2 |
| | 7 2 | | 3 1 | 3 1 | 4+5 2 | 3+4 2 |
| | | | 5 1 | 5 1 | | |
| | | | 6 1 | 6 1 | | |
| | | | 4 1 | 4 1 | | |
| | | | 7 2 | 7 2 | | |
| | | | 3 1 | 3 1 | | |
| | | | 5 1 | 5 1 | | |
| | | | 7 2 | 7 2 | | |
| | | | 4 1 | 4 1 | | |
| | | | 6 1 | 6 1 | | |
| | | | 7 2 | 7 2 | | |
| | | | 7 2 | 7 2 | | |
| | | | 5 1 | 5 1 | | |
| | | | 3 1 | 3 1 | | |
| 122.1.13 | 154405 | | 52 | | Cu1In1Se2 | |
| | -GM | -M | -PA | -P | -X | -N |
| | 7 2 | 6+7 4 | 7 2 | 7 2 | 2+3 2 | 3+4 2 |
| | 6 2 | 6+7 4 | 3 1 | 3 1 | 4+5 2 | 3+4 2 |
| | 6 2 | 6+7 4 | 5 1 | 5 1 | 2+3 2 | 3+4 2 |
| | 7 2 | 6+7 4 | 6 1 | 6 1 | 4+5 2 | 3+4 2 |
| | 7 2 | 6+7 4 | 4 1 | 4 1 | 4+5 2 | 3+4 2 |
| | 7 2 | 6+7 4 | 7 2 | 7 2 | 2+3 2 | 3+4 2 |
| | 7 2 | 6+7 4 | 3 1 | 3 1 | 4+5 2 | 3+4 2 |
| | 6 2 | 6+7 4 | 7 2 | 7 2 | 2+3 2 | 3+4 2 |
| | 6 2 | 6+7 4 | 5 1 | 5 1 | 2+3 2 | 3+4 2 |
| | 6 2 | 6+7 4 | 4 1 | 4 1 | 4+5 2 | 3+4 2 |
| | 7 2 | 6+7 4 | 7 2 | 7 2 | 4+5 2 | 3+4 2 |
| | 7 2 | 6+7 4 | 6 1 | 6 1 | 2+3 2 | 3+4 2 |
| | 6 2 | 6+7 4 | 5 1 | 5 1 | 2+3 2 | 3+4 2 |
| | 6 2 | | 7 2 | 7 2 | 4+5 2 | 3+4 2 |
| | 7 2 | | 3 1 | 3 1 | 2+3 2 | 3+4 2 |
| | 6 2 | | 4 1 | 4 1 | 4+5 2 | 3+4 2 |
| | 7 2 | | 7 2 | 7 2 | 4+5 2 | 3+4 2 |
| | 6 2 | | 6 1 | 6 1 | 2+3 2 | 3+4 2 |
| | 7 2 | | 7 2 | 7 2 | 2+3 2 | 3+4 2 |
| | 6 2 | | 5 1 | 5 1 | 4+5 2 | 3+4 2 |
| | 7 2 | | 3 1 | 3 1 | 2+3 2 | 3+4 2 |
| | 6 2 | | 6 1 | 6 1 | 4+5 2 | 3+4 2 |
| | 6 2 | | 4 1 | 4 1 | 4+5 2 | 3+4 2 |
| | 7 2 | | 7 2 | 7 2 | 2+3 2 | 3+4 2 |
| | 6 2 | | 7 2 | 7 2 | 2+3 2 | 3+4 2 |
| | 7 2 | | 3 1 | 3 1 | 4+5 2 | 3+4 2 |
| | | | 5 1 | 5 1 | | |
| | | | 6 1 | 6 1 | | |
| | | | 4 1 | 4 1 | | |
| | | | 7 2 | 7 2 | | |
| | | | 3 1 | 3 1 | | |
| | | | 5 1 | 5 1 | | |
| | | | 7 2 | 7 2 | | |

Continued on next column

**TABLE LV– continued**

| TQC | ICSD | | NoE | | SF | |
|---|---|---|---|---|---|---|
| | | | 6 1 | 6 1 | | |
| | | | 4 1 | 4 1 | | |
| | | | 7 2 | 7 2 | | |
| | | | 7 2 | 7 2 | | |
| | | | 5 1 | 5 1 | | |
| | | | 3 1 | 3 1 | | |
| 122.1.14 | 68325 | | 52 | | As2Sn1Zn1 | |
| | -GM | -M | -PA | -P | -X | -N |
| | 7 2 | 6+7 4 | 7 2 | 7 2 | 2+3 2 | 3+4 2 |
| | 6 2 | 6+7 4 | 3 1 | 3 1 | 4+5 2 | 3+4 2 |
| | 6 2 | 6+7 4 | 5 1 | 5 1 | 2+3 2 | 3+4 2 |
| | 7 2 | 6+7 4 | 6 1 | 6 1 | 4+5 2 | 3+4 2 |
| | 7 2 | 6+7 4 | 4 1 | 4 1 | 4+5 2 | 3+4 2 |
| | 6 2 | 6+7 4 | 7 2 | 7 2 | 2+3 2 | 3+4 2 |
| | 7 2 | 6+7 4 | 7 2 | 7 2 | 2+3 2 | 3+4 2 |
| | 6 2 | 6+7 4 | 3 1 | 3 1 | 4+5 2 | 3+4 2 |
| | 7 2 | 6+7 4 | 5 1 | 5 1 | 2+3 2 | 3+4 2 |
| | 6 2 | 6+7 4 | 7 2 | 7 2 | 4+5 2 | 3+4 2 |
| | 6 2 | 6+7 4 | 6 1 | 6 1 | 4+5 2 | 3+4 2 |
| | 7 2 | 6+7 4 | 4 1 | 4 1 | 2+3 2 | 3+4 2 |
| | 6 2 | 6+7 4 | 5 1 | 5 1 | 2+3 2 | 3+4 2 |
| | 7 2 | | 3 1 | 3 1 | 4+5 2 | 3+4 2 |
| | 6 2 | | 7 2 | 7 2 | 4+5 2 | 3+4 2 |
| | 7 2 | | 7 2 | 7 2 | 2+3 2 | 3+4 2 |
| | 7 2 | | 3 1 | 3 1 | 4+5 2 | 3+4 2 |
| | 6 2 | | 4 1 | 4 1 | 2+3 2 | 3+4 2 |
| | 7 2 | | 5 1 | 5 1 | 4+5 2 | 3+4 2 |
| | 6 2 | | 6 1 | 6 1 | 2+3 2 | 3+4 2 |
| | 6 2 | | 7 2 | 7 2 | 2+3 2 | 3+4 2 |
| | 7 2 | | 7 2 | 7 2 | 4+5 2 | 3+4 2 |
| | 6 2 | | 3 1 | 3 1 | 2+3 2 | 3+4 2 |
| | 6 2 | | 5 1 | 5 1 | 4+5 2 | 3+4 2 |
| | 7 2 | | 4 1 | 4 1 | 4+5 2 | 3+4 2 |
| | 7 2 | | 6 1 | 6 1 | 2+3 2 | 3+4 2 |
| | | | 7 2 | 7 2 | | |
| | | | 4 1 | 4 1 | | |
| | | | 6 1 | 6 1 | | |
| | | | 7 2 | 7 2 | | |
| | | | 7 2 | 7 2 | | |
| | | | 5 1 | 5 1 | | |
| | | | 3 1 | 3 1 | | |
| | | | 6 1 | 6 1 | | |
| | | | 7 2 | 7 2 | | |
| | | | 4 1 | 4 1 | | |
| | | | 3 1 | 3 1 | | |
| | | | 7 2 | 7 2 | | |
| | | | 5 1 | 5 1 | | |
| 122.1.15 | 42098 | | 52 | | As2Cd1Ge1 | |
| | -GM | -M | -PA | -P | -X | -N |
| | 7 2 | 6+7 4 | 3 1 | 3 1 | 4+5 2 | 3+4 2 |
| | 6 2 | 6+7 4 | 7 2 | 7 2 | 2+3 2 | 3+4 2 |
| | 7 2 | 6+7 4 | 5 1 | 5 1 | 2+3 2 | 3+4 2 |
| | 6 2 | 6+7 4 | 6 1 | 6 1 | 4+5 2 | 3+4 2 |
| | 7 2 | 6+7 4 | 4 1 | 4 1 | 2+3 2 | 3+4 2 |
| | 6 2 | 6+7 4 | 7 2 | 7 2 | 4+5 2 | 3+4 2 |
| | 7 2 | 6+7 4 | 7 2 | 7 2 | 4+5 2 | 3+4 2 |
| | 6 2 | 6+7 4 | 3 1 | 3 1 | 2+3 2 | 3+4 2 |
| | 7 2 | 6+7 4 | 5 1 | 5 1 | 2+3 2 | 3+4 2 |
| | 6 2 | 6+7 4 | 7 2 | 7 2 | 4+5 2 | 3+4 2 |
| | 6 2 | 6+7 4 | 4 1 | 4 1 | 4+5 2 | 3+4 2 |
| | 7 2 | 6+7 4 | 6 1 | 6 1 | 2+3 2 | 3+4 2 |

Continued on next column

**TABLE LV– continued**

| TQC | ICSD | NoE | SF |
|---|---|---|---|
| | 7 2 6+7 4 | 5 1 | 5 1 2+3 2 3+4 2 |
| | 6 2 | 3 1 | 3 1 4+5 2 3+4 2 |
| | 6 2 | 7 2 | 7 2 4+5 2 3+4 2 |
| | 7 2 | 3 1 | 3 1 2+3 2 3+4 2 |
| | 7 2 | 7 2 | 7 2 4+5 2 3+4 2 |
| | 6 2 | 5 1 | 5 1 2+3 2 3+4 2 |
| | 7 2 | 6 1 | 6 1 2+3 2 3+4 2 |
| | 6 2 | 4 1 | 4 1 4+5 2 3+4 2 |
| | 6 2 | 7 2 | 7 2 2+3 2 3+4 2 |
| | 6 2 | 7 2 | 7 2 4+5 2 3+4 2 |
| | 7 2 | 3 1 | 3 1 2+3 2 3+4 2 |
| | 6 2 | 5 1 | 5 1 4+5 2 3+4 2 |
| | 7 2 | 4 1 | 4 1 4+5 2 3+4 2 |
| | 7 2 | 7 2 | 7 2 2+3 2 3+4 2 |
| | | 6 1 | 6 1 |
| | | 4 1 | 4 1 |
| | | 6 1 | 6 1 |
| | | 7 2 | 7 2 |
| | | 5 1 | 5 1 |
| | | 7 2 | 7 2 |
| | | 3 1 | 3 1 |
| | | 7 2 | 7 2 |
| | | 6 1 | 6 1 |
| | | 3 1 | 3 1 |
| | | 7 2 | 7 2 |
| | | 5 1 | 5 1 |
| | | 4 1 | 4 1 |

The End

TABLE LVI: The computed irreducible representations for the valence bands for the compounds of SEBR and NLC in space group 123. See the list of tables.

| TQC | ICSD | | NoE | | SF | |
|---|---|---|---|---|---|---|
| 123.1.1 | 193701 | | 38 | | In5Rh1U1 | |
| | -A | -GM | -M | -Z | -R | -X |
| | 7 2 | 7 2 | 7 2 | 7 2 | 5 2 | 5 2 |
| | 9 2 | 9 2 | 9 2 | 9 2 | 6 2 | 6 2 |
| | 8 2 | 9 2 | 8 2 | 9 2 | 6 2 | 6 2 |
| | 9 2 | 8 2 | 9 2 | 8 2 | 6 2 | 6 2 |
| | 6 2 | 7 2 | 6 2 | 9 2 | 6 2 | 5 2 |
| | 7 2 | 9 2 | 9 2 | 7 2 | 5 2 | 6 2 |
| | 6 2 | 7 2 | 8 2 | 8 2 | 6 2 | 5 2 |
| | 9 2 | 6 2 | 7 2 | 7 2 | 5 2 | 6 2 |
| | 8 2 | 7 2 | 6 2 | 9 2 | 6 2 | 6 2 |
| | 8 2 | 6 2 | 7 2 | 9 2 | 6 2 | 5 2 |
| | 9 2 | 7 2 | 6 2 | 8 2 | 6 2 | 5 2 |
| | 7 2 | 6 2 | 8 2 | 6 2 | 6 2 | 5 2 |
| | 8 2 | 6 2 | 6 2 | 8 2 | 6 2 | 5 2 |
| | 9 2 | 9 2 | 7 2 | 6 2 | 5 2 | 5 2 |
| | 9 2 | 8 2 | 9 2 | 7 2 | 5 2 | 5 2 |
| | 8 2 | 8 2 | 7 2 | 9 2 | 5 2 | 5 2 |
| | 8 2 | 8 2 | 8 2 | 8 2 | 5 2 | 6 2 |
| | 7 2 | 9 2 | 6 2 | 8 2 | 6 2 | 6 2 |
| | 6 2 | 9 2 | 9 2 | 9 2 | 6 2 | 6 2 |
| 123.1.2 | 103639 | | 38 | | Ga5Ir1U1 | |
| | -A | -GM | -M | -Z | -R | -X |
| | 7 2 | 7 2 | 7 2 | 7 2 | 5 2 | 5 2 |
| | 9 2 | 9 2 | 9 2 | 9 2 | 6 2 | 6 2 |
| | 8 2 | 9 2 | 8 2 | 9 2 | 6 2 | 6 2 |

Continued on next column

**TABLE LVI– continued**

| TQC | ICSD | | NoE | | SF | |
|---|---|---|---|---|---|---|
| | 9 2 | 8 2 | 9 2 | 8 2 | 6 2 | 6 2 |
| | 7 2 | 7 2 | 9 2 | 9 2 | 6 2 | 5 2 |
| | 6 2 | 7 2 | 6 2 | 7 2 | 5 2 | 6 2 |
| | 6 2 | 6 2 | 8 2 | 8 2 | 6 2 | 5 2 |
| | 9 2 | 9 2 | 7 2 | 9 2 | 5 2 | 5 2 |
| | 8 2 | 7 2 | 7 2 | 8 2 | 6 2 | 6 2 |
| | 8 2 | 6 2 | 6 2 | 7 2 | 6 2 | 6 2 |
| | 9 2 | 7 2 | 6 2 | 9 2 | 6 2 | 5 2 |
| | 8 2 | 6 2 | 6 2 | 6 2 | 6 2 | 5 2 |
| | 7 2 | 6 2 | 7 2 | 8 2 | 6 2 | 5 2 |
| | 9 2 | 9 2 | 8 2 | 7 2 | 5 2 | 5 2 |
| | 9 2 | 8 2 | 9 2 | 6 2 | 5 2 | 5 2 |
| | 8 2 | 8 2 | 7 2 | 9 2 | 5 2 | 5 2 |
| | 7 2 | 8 2 | 6 2 | 8 2 | 5 2 | 6 2 |
| | 6 2 | 9 2 | 8 2 | 8 2 | 6 2 | 6 2 |
| | 8 2 | 9 2 | 9 2 | 9 2 | 6 2 | 6 2 |
| 123.1.3 | 5257 | | 16 | | Fe1Si2 | |
| | -A | -GM | -M | -Z | -R | -X |
| | 7 2 | 7 2 | 7 2 | 7 2 | 6 2 | 5 2 |
| | 6 2 | 9 2 | 6 2 | 9 2 | 5 2 | 6 2 |
| | 7 2 | 7 2 | 7 2 | 7 2 | 6 2 | 6 2 |
| | 9 2 | 6 2 | 8 2 | 6 2 | 5 2 | 5 2 |
| | 8 2 | 7 2 | 9 2 | 9 2 | 6 2 | 5 2 |
| | 9 2 | 7 2 | 7 2 | 7 2 | 6 2 | 6 2 |
| | 6 2 | 6 2 | 6 2 | 6 2 | 6 2 | 6 2 |
| | 7 2 | 6 2 | 7 2 | 6 2 | 6 2 | 6 2 |
| 123.1.4 | 106640 | | 78 | | Fe1Ga8Tm2 | |
| | -A | -GM | -M | -Z | -R | -X |
| | 7 2 | 7 2 | 7 2 | 7 2 | 5 2 | 5 2 |
| | 9 2 | 9 2 | 9 2 | 9 2 | 6 2 | 6 2 |
| | 9 2 | 7 2 | 7 2 | 9 2 | 6 2 | 5 2 |
| | 7 2 | 9 2 | 9 2 | 7 2 | 5 2 | 6 2 |
| | 6 2 | 7 2 | 8 2 | 9 2 | 6 2 | 5 2 |
| | 8 2 | 9 2 | 6 2 | 7 2 | 5 2 | 6 2 |
| | 9 2 | 8 2 | 7 2 | 6 2 | 6 2 | 5 2 |
| | 7 2 | 6 2 | 9 2 | 8 2 | 5 2 | 6 2 |
| | 9 2 | 7 2 | 9 2 | 7 2 | 6 2 | 6 2 |
| | 8 2 | 7 2 | 8 2 | 9 2 | 5 2 | 5 2 |
| | 8 2 | 9 2 | 6 2 | 9 2 | 5 2 | 5 2 |
| | 6 2 | 7 2 | 8 2 | 7 2 | 6 2 | 6 2 |
| | 7 2 | 6 2 | 9 2 | 6 2 | 5 2 | 5 2 |
| | 6 2 | 9 2 | 8 2 | 9 2 | 6 2 | 6 2 |
| | 7 2 | 6 2 | 7 2 | 8 2 | 6 2 | 6 2 |
| | 6 2 | 7 2 | 6 2 | 7 2 | 5 2 | 5 2 |
| | 7 2 | 6 2 | 7 2 | 7 2 | 5 2 | 5 2 |
| | 9 2 | 8 2 | 6 2 | 6 2 | 5 2 | 6 2 |
| | 6 2 | 9 2 | 7 2 | 8 2 | 5 2 | 5 2 |
| | 8 2 | 7 2 | 9 2 | 9 2 | 5 2 | 5 2 |
| | 9 2 | 8 2 | 9 2 | 8 2 | 6 2 | 5 2 |
| | 8 2 | 9 2 | 8 2 | 7 2 | 6 2 | 6 2 |
| | 9 2 | 8 2 | 6 2 | 6 2 | 5 2 | 6 2 |
| | 7 2 | 6 2 | 8 2 | 6 2 | 6 2 | 5 2 |
| | 6 2 | 8 2 | 7 2 | 8 2 | 5 2 | 5 2 |
| | 6 2 | 7 2 | 6 2 | 9 2 | 6 2 | 6 2 |
| | 8 2 | 6 2 | 9 2 | 7 2 | 5 2 | 5 2 |
| | 9 2 | 9 2 | 8 2 | 6 2 | 5 2 | 5 2 |
| | 8 2 | 8 2 | 6 2 | 6 2 | 6 2 | 5 2 |
| | 6 2 | 6 2 | 7 2 | 6 2 | 6 2 | 5 2 |
| | 7 2 | 9 2 | 6 2 | 8 2 | 5 2 | 6 2 |
| | 6 2 | 6 2 | 8 2 | 7 2 | 5 2 | 6 2 |
| | 7 2 | 7 2 | 6 2 | 9 2 | 6 2 | 6 2 |

Continued on next column

**TABLE LVI– continued**

| TQC | ICSD | NoE | SF | | | |
|---|---|---|---|---|---|---|
| | 6 2 | 7 2 | 7 2 | 6 2 | 6 2 | 5 2 |
| | 9 2 | 8 2 | 7 2 | 7 2 | 5 2 | 5 2 |
| | 8 2 | 6 2 | 8 2 | 8 2 | 6 2 | 6 2 |
| | 9 2 | 8 2 | 6 2 | 9 2 | 5 2 | 6 2 |
| | 6 2 | 9 2 | 9 2 | 6 2 | 6 2 | 5 2 |
| | 7 2 | 9 2 | 7 2 | 7 2 | 5 2 | 5 2 |
| 123.1.5 | 102455 | 38 | Co1Ga5U1 | | | |
| | -A | -GM | -M | -Z | -R | -X |
| | 7 2 | 7 2 | 7 2 | 7 2 | 5 2 | 5 2 |
| | 9 2 | 9 2 | 9 2 | 9 2 | 6 2 | 6 2 |
| | 8 2 | 9 2 | 8 2 | 9 2 | 6 2 | 6 2 |
| | 9 2 | 8 2 | 9 2 | 8 2 | 6 2 | 6 2 |
| | 7 2 | 7 2 | 6 2 | 9 2 | 5 2 | 5 2 |
| | 6 2 | 7 2 | 9 2 | 7 2 | 6 2 | 6 2 |
| | 6 2 | 9 2 | 8 2 | 8 2 | 6 2 | 5 2 |
| | 9 2 | 6 2 | 7 2 | 7 2 | 5 2 | 6 2 |
| | 8 2 | 7 2 | 6 2 | 6 2 | 6 2 | 6 2 |
| | 8 2 | 6 2 | 7 2 | 9 2 | 6 2 | 5 2 |
| | 9 2 | 6 2 | 8 2 | 9 2 | 5 2 | 5 2 |
| | 7 2 | 7 2 | 6 2 | 8 2 | 6 2 | 5 2 |
| | 8 2 | 9 2 | 9 2 | 6 2 | 5 2 | 5 2 |
| | 9 2 | 8 2 | 6 2 | 7 2 | 6 2 | 5 2 |
| | 7 2 | 8 2 | 7 2 | 8 2 | 5 2 | 5 2 |
| | 6 2 | 6 2 | 9 2 | 9 2 | 6 2 | 5 2 |
| | 9 2 | 8 2 | 7 2 | 8 2 | 5 2 | 6 2 |
| | 8 2 | 9 2 | 8 2 | 8 2 | 6 2 | 6 2 |
| | 8 2 | 9 2 | 6 2 | 9 2 | 6 2 | 6 2 |
| 123.1.6 | 192621 | 58 | Ce2In8Pt1 | | | |
| | -A | -GM | -M | -Z | -R | -X |
| | 9 2 | 7 2 | 7 2 | 7 2 | 6 2 | 5 2 |
| | 7 2 | 9 2 | 9 2 | 9 2 | 5 2 | 6 2 |
| | 9 2 | 7 2 | 7 2 | 9 2 | 6 2 | 5 2 |
| | 7 2 | 9 2 | 9 2 | 7 2 | 5 2 | 6 2 |
| | 6 2 | 7 2 | 8 2 | 9 2 | 6 2 | 5 2 |
| | 8 2 | 9 2 | 6 2 | 7 2 | 5 2 | 6 2 |
| | 9 2 | 8 2 | 7 2 | 6 2 | 6 2 | 5 2 |
| | 7 2 | 6 2 | 9 2 | 8 2 | 5 2 | 6 2 |
| | 9 2 | 7 2 | 6 2 | 7 2 | 5 2 | 5 2 |
| | 8 2 | 7 2 | 9 2 | 9 2 | 6 2 | 6 2 |
| | 8 2 | 9 2 | 8 2 | 9 2 | 5 2 | 5 2 |
| | 6 2 | 6 2 | 7 2 | 7 2 | 5 2 | 6 2 |
| | 7 2 | 7 2 | 8 2 | 6 2 | 6 2 | 5 2 |
| | 7 2 | 9 2 | 7 2 | 7 2 | 6 2 | 6 2 |
| | 6 2 | 7 2 | 6 2 | 6 2 | 5 2 | 5 2 |
| | 7 2 | 6 2 | 9 2 | 7 2 | 6 2 | 5 2 |
| | 6 2 | 7 2 | 8 2 | 9 2 | 5 2 | 5 2 |
| | 6 2 | 6 2 | 6 2 | 6 2 | 5 2 | 5 2 |
| | 6 2 | 6 2 | 6 2 | 8 2 | 5 2 | 6 2 |
| | 7 2 | 8 2 | 7 2 | 9 2 | 5 2 | 5 2 |
| | 9 2 | 9 2 | 6 2 | 8 2 | 5 2 | 5 2 |
| | 8 2 | 8 2 | 9 2 | 6 2 | 6 2 | 5 2 |
| | 7 2 | 8 2 | 7 2 | 7 2 | 6 2 | 6 2 |
| | 6 2 | 9 2 | 7 2 | 8 2 | 6 2 | 5 2 |
| | 9 2 | 6 2 | 6 2 | 6 2 | 5 2 | 6 2 |
| | 8 2 | 7 2 | 9 2 | 7 2 | 5 2 | 6 2 |
| | 9 2 | 9 2 | 8 2 | 6 2 | 6 2 | 6 2 |
| | 6 2 | 6 2 | 8 2 | 7 2 | 5 2 | 5 2 |
| | 9 2 | 8 2 | 7 2 | 6 2 | 5 2 | 6 2 |
| 123.1.7 | 150226 | 36 | Ce1In5Rh1 | | | |
| | -A | -GM | -M | -Z | -R | -X |
| | 7 2 | 7 2 | 7 2 | 7 2 | 5 2 | 5 2 |

Continued on next column

**TABLE LVI– continued**

| TQC | ICSD | NoE | SF | | | |
|---|---|---|---|---|---|---|
| | 9 2 | 9 2 | 9 2 | 9 2 | 6 2 | 6 2 |
| | 8 2 | 9 2 | 8 2 | 9 2 | 6 2 | 6 2 |
| | 9 2 | 8 2 | 9 2 | 8 2 | 6 2 | 6 2 |
| | 7 2 | 7 2 | 6 2 | 9 2 | 6 2 | 5 2 |
| | 6 2 | 9 2 | 9 2 | 7 2 | 5 2 | 6 2 |
| | 6 2 | 7 2 | 8 2 | 8 2 | 6 2 | 5 2 |
| | 9 2 | 6 2 | 7 2 | 7 2 | 5 2 | 6 2 |
| | 8 2 | 7 2 | 6 2 | 9 2 | 6 2 | 6 2 |
| | 8 2 | 6 2 | 7 2 | 8 2 | 6 2 | 5 2 |
| | 9 2 | 7 2 | 6 2 | 9 2 | 6 2 | 5 2 |
| | 7 2 | 6 2 | 8 2 | 6 2 | 6 2 | 5 2 |
| | 8 2 | 6 2 | 6 2 | 8 2 | 6 2 | 5 2 |
| | 9 2 | 8 2 | 7 2 | 6 2 | 5 2 | 5 2 |
| | 9 2 | 9 2 | 9 2 | 7 2 | 5 2 | 5 2 |
| | 8 2 | 8 2 | 8 2 | 9 2 | 5 2 | 5 2 |
| | 8 2 | 8 2 | 9 2 | 8 2 | 5 2 | 6 2 |
| | 7 2 | 9 2 | 7 2 | 8 2 | 6 2 | 6 2 |
| 123.1.8 | 110780 | 36 | Ce1In5Rh1 | | | |
| | -A | -GM | -M | -Z | -R | -X |
| | 7 2 | 7 2 | 7 2 | 7 2 | 5 2 | 5 2 |
| | 9 2 | 9 2 | 9 2 | 9 2 | 6 2 | 6 2 |
| | 8 2 | 9 2 | 8 2 | 9 2 | 6 2 | 6 2 |
| | 9 2 | 8 2 | 9 2 | 8 2 | 6 2 | 6 2 |
| | 6 2 | 7 2 | 6 2 | 9 2 | 6 2 | 5 2 |
| | 7 2 | 9 2 | 9 2 | 7 2 | 5 2 | 6 2 |
| | 6 2 | 7 2 | 8 2 | 8 2 | 6 2 | 5 2 |
| | 9 2 | 6 2 | 7 2 | 7 2 | 5 2 | 6 2 |
| | 8 2 | 7 2 | 6 2 | 9 2 | 6 2 | 6 2 |
| | 8 2 | 6 2 | 7 2 | 8 2 | 6 2 | 5 2 |
| | 9 2 | 7 2 | 6 2 | 9 2 | 6 2 | 5 2 |
| | 7 2 | 6 2 | 8 2 | 6 2 | 6 2 | 5 2 |
| | 8 2 | 6 2 | 6 2 | 8 2 | 6 2 | 5 2 |
| | 9 2 | 8 2 | 7 2 | 6 2 | 5 2 | 5 2 |
| | 9 2 | 9 2 | 9 2 | 7 2 | 5 2 | 5 2 |
| | 8 2 | 8 2 | 8 2 | 9 2 | 5 2 | 5 2 |
| | 8 2 | 8 2 | 7 2 | 8 2 | 5 2 | 6 2 |
| | 7 2 | 9 2 | 9 2 | 8 2 | 6 2 | 6 2 |
| 123.1.9 | 110779 | 36 | Ce1In5Rh1 | | | |
| | -A | -GM | -M | -Z | -R | -X |
| | 7 2 | 7 2 | 7 2 | 7 2 | 5 2 | 5 2 |
| | 9 2 | 9 2 | 9 2 | 9 2 | 6 2 | 6 2 |
| | 8 2 | 9 2 | 8 2 | 9 2 | 6 2 | 6 2 |
| | 9 2 | 8 2 | 9 2 | 8 2 | 6 2 | 6 2 |
| | 7 2 | 7 2 | 6 2 | 9 2 | 6 2 | 5 2 |
| | 6 2 | 9 2 | 9 2 | 7 2 | 5 2 | 6 2 |
| | 6 2 | 7 2 | 8 2 | 8 2 | 6 2 | 5 2 |
| | 9 2 | 6 2 | 7 2 | 7 2 | 5 2 | 6 2 |
| | 8 2 | 7 2 | 6 2 | 9 2 | 6 2 | 6 2 |
| | 8 2 | 6 2 | 7 2 | 8 2 | 6 2 | 5 2 |
| | 9 2 | 7 2 | 6 2 | 9 2 | 6 2 | 5 2 |
| | 7 2 | 6 2 | 8 2 | 6 2 | 6 2 | 5 2 |
| | 8 2 | 6 2 | 6 2 | 8 2 | 6 2 | 5 2 |
| | 9 2 | 8 2 | 7 2 | 6 2 | 5 2 | 5 2 |
| | 9 2 | 9 2 | 9 2 | 7 2 | 5 2 | 5 2 |
| | 8 2 | 8 2 | 8 2 | 9 2 | 5 2 | 5 2 |
| | 8 2 | 8 2 | 9 2 | 8 2 | 5 2 | 6 2 |
| | 7 2 | 9 2 | 7 2 | 8 2 | 6 2 | 6 2 |
| 123.1.10 | 110778 | 36 | Ce1In5Rh1 | | | |
| | -A | -GM | -M | -Z | -R | -X |
| | 7 2 | 7 2 | 7 2 | 7 2 | 5 2 | 5 2 |
| | 9 2 | 9 2 | 9 2 | 9 2 | 6 2 | 6 2 |

Continued on next column

**TABLE LVI– continued**

| TQC | ICSD | NoE | SF | | |
|---|---|---|---|---|---|
| | 8 2 | 9 2 | 8 2 9 2 6 2 | | 6 2 |
| | 9 2 | 8 2 | 9 2 8 2 6 2 | | 6 2 |
| | 6 2 | 7 2 | 6 2 9 2 6 2 | | 5 2 |
| | 7 2 | 9 2 | 9 2 7 2 5 2 | | 6 2 |
| | 6 2 | 7 2 | 8 2 8 2 6 2 | | 5 2 |
| | 9 2 | 6 2 | 7 2 7 2 5 2 | | 6 2 |
| | 8 2 | 7 2 | 6 2 9 2 6 2 | | 6 2 |
| | 8 2 | 6 2 | 7 2 8 2 6 2 | | 5 2 |
| | 9 2 | 7 2 | 6 2 9 2 6 2 | | 5 2 |
| | 7 2 | 6 2 | 8 2 6 2 6 2 | | 5 2 |
| | 8 2 | 6 2 | 6 2 8 2 6 2 | | 5 2 |
| | 9 2 | 8 2 | 7 2 6 2 5 2 | | 5 2 |
| | 9 2 | 9 2 | 9 2 7 2 5 2 | | 5 2 |
| | 8 2 | 8 2 | 8 2 9 2 5 2 | | 5 2 |
| | 8 2 | 8 2 | 9 2 8 2 5 2 | | 6 2 |
| | 7 2 | 9 2 | 7 2 8 2 6 2 | | 6 2 |
| 123.1.11 | 110777 | 36 | Ce1In5Rh1 | | |
| | -A | -GM | -M -Z -R | | -X |
| | 7 2 | 7 2 | 7 2 7 2 5 2 | | 5 2 |
| | 9 2 | 9 2 | 9 2 9 2 6 2 | | 6 2 |
| | 8 2 | 9 2 | 8 2 9 2 6 2 | | 6 2 |
| | 9 2 | 8 2 | 9 2 8 2 6 2 | | 6 2 |
| | 6 2 | 7 2 | 6 2 9 2 6 2 | | 5 2 |
| | 7 2 | 9 2 | 9 2 7 2 5 2 | | 6 2 |
| | 6 2 | 7 2 | 8 2 8 2 6 2 | | 5 2 |
| | 9 2 | 6 2 | 7 2 7 2 5 2 | | 6 2 |
| | 8 2 | 7 2 | 6 2 9 2 6 2 | | 6 2 |
| | 8 2 | 6 2 | 7 2 8 2 6 2 | | 5 2 |
| | 9 2 | 7 2 | 6 2 9 2 6 2 | | 5 2 |
| | 7 2 | 6 2 | 8 2 6 2 6 2 | | 5 2 |
| | 8 2 | 6 2 | 6 2 8 2 6 2 | | 5 2 |
| | 9 2 | 8 2 | 7 2 6 2 5 2 | | 5 2 |
| | 9 2 | 9 2 | 9 2 7 2 5 2 | | 5 2 |
| | 8 2 | 8 2 | 8 2 9 2 5 2 | | 5 2 |
| | 8 2 | 8 2 | 9 2 8 2 5 2 | | 6 2 |
| | 7 2 | 9 2 | 7 2 8 2 6 2 | | 6 2 |
| 123.1.12 | 150225 | 36 | Ce1In5Ir1 | | |
| | -A | -GM | -M -Z -R | | -X |
| | 7 2 | 7 2 | 7 2 7 2 5 2 | | 5 2 |
| | 9 2 | 9 2 | 9 2 9 2 6 2 | | 6 2 |
| | 8 2 | 9 2 | 8 2 9 2 6 2 | | 6 2 |
| | 9 2 | 8 2 | 9 2 8 2 6 2 | | 6 2 |
| | 7 2 | 7 2 | 6 2 9 2 6 2 | | 5 2 |
| | 6 2 | 9 2 | 9 2 7 2 5 2 | | 6 2 |
| | 6 2 | 7 2 | 8 2 8 2 6 2 | | 5 2 |
| | 9 2 | 6 2 | 7 2 9 2 5 2 | | 6 2 |
| | 8 2 | 7 2 | 7 2 7 2 6 2 | | 6 2 |
| | 9 2 | 6 2 | 6 2 8 2 6 2 | | 5 2 |
| | 8 2 | 7 2 | 6 2 9 2 6 2 | | 5 2 |
| | 7 2 | 6 2 | 8 2 6 2 6 2 | | 5 2 |
| | 8 2 | 6 2 | 6 2 8 2 6 2 | | 5 2 |
| | 9 2 | 9 2 | 7 2 6 2 5 2 | | 5 2 |
| | 9 2 | 8 2 | 9 2 7 2 5 2 | | 5 2 |
| | 8 2 | 8 2 | 8 2 9 2 5 2 | | 5 2 |
| | 8 2 | 8 2 | 9 2 8 2 5 2 | | 6 2 |
| | 7 2 | 9 2 | 7 2 8 2 6 2 | | 6 2 |
| 123.1.13 | 92435 | 64 | Bi2I1O4Tm1 | | |
| | -A | -GM | -M -Z -R | | -X |
| | 7 2 | 7 2 | 7 2 7 2 5 2 | | 5 2 |
| | 9 2 | 9 2 | 9 2 9 2 6 2 | | 6 2 |
| | 8 2 | 9 2 | 8 2 9 2 6 2 | | 6 2 |

Continued on next column

**TABLE LVI– continued**

| TQC | ICSD | NoE | SF | | |
|---|---|---|---|---|---|
| | 9 2 | 8 2 | 9 2 8 2 6 2 | | 6 2 |
| | 6 2 | 7 2 | 6 2 7 2 5 2 | | 5 2 |
| | 7 2 | 9 2 | 7 2 9 2 5 2 | | 5 2 |
| | 9 2 | 6 2 | 9 2 6 2 6 2 | | 6 2 |
| | 8 2 | 8 2 | 8 2 8 2 6 2 | | 6 2 |
| | 9 2 | 7 2 | 7 2 9 2 6 2 | | 5 2 |
| | 6 2 | 9 2 | 6 2 7 2 5 2 | | 6 2 |
| | 8 2 | 7 2 | 8 2 9 2 6 2 | | 5 2 |
| | 7 2 | 7 2 | 7 2 7 2 5 2 | | 5 2 |
| | 8 2 | 9 2 | 8 2 9 2 5 2 | | 6 2 |
| | 6 2 | 6 2 | 9 2 6 2 5 2 | | 6 2 |
| | 7 2 | 8 2 | 6 2 8 2 6 2 | | 5 2 |
| | 9 2 | 6 2 | 9 2 7 2 6 2 | | 6 2 |
| | 6 2 | 7 2 | 6 2 6 2 6 2 | | 5 2 |
| | 8 2 | 9 2 | 9 2 7 2 5 2 | | 5 2 |
| | 9 2 | 7 2 | 8 2 7 2 5 2 | | 5 2 |
| | 7 2 | 6 2 | 7 2 6 2 5 2 | | 6 2 |
| | 6 2 | 8 2 | 8 2 8 2 6 2 | | 6 2 |
| | 8 2 | 9 2 | 8 2 9 2 5 2 | | 6 2 |
| | 7 2 | 8 2 | 9 2 8 2 6 2 | | 6 2 |
| | 6 2 | 9 2 | 6 2 9 2 6 2 | | 6 2 |
| | 7 2 | 9 2 | 7 2 7 2 6 2 | | 6 2 |
| | 9 2 | 8 2 | 9 2 8 2 6 2 | | 5 2 |
| | 9 2 | 8 2 | 9 2 8 2 6 2 | | 6 2 |
| | 8 2 | 9 2 | 8 2 9 2 5 2 | | 6 2 |
| | 8 2 | 8 2 | 8 2 6 2 5 2 | | 6 2 |
| | 8 2 | 9 2 | 8 2 9 2 6 2 | | 6 2 |
| | 9 2 | 8 2 | 9 2 8 2 6 2 | | 6 2 |
| | 9 2 | 8 2 | 9 2 8 2 6 2 | | 6 2 |
| 123.1.14 | 92408 | 64 | Bi2Cl1O4Tm1 | | |
| | -A | -GM | -M -Z -R | | -X |
| | 7 2 | 7 2 | 7 2 7 2 5 2 | | 5 2 |
| | 9 2 | 9 2 | 9 2 9 2 6 2 | | 6 2 |
| | 8 2 | 9 2 | 8 2 9 2 6 2 | | 6 2 |
| | 9 2 | 8 2 | 9 2 8 2 6 2 | | 6 2 |
| | 6 2 | 7 2 | 6 2 7 2 5 2 | | 5 2 |
| | 7 2 | 9 2 | 7 2 9 2 5 2 | | 5 2 |
| | 9 2 | 6 2 | 9 2 6 2 6 2 | | 6 2 |
| | 8 2 | 8 2 | 8 2 8 2 6 2 | | 6 2 |
| | 9 2 | 7 2 | 7 2 9 2 6 2 | | 5 2 |
| | 6 2 | 7 2 | 6 2 7 2 5 2 | | 6 2 |
| | 8 2 | 9 2 | 8 2 9 2 6 2 | | 5 2 |
| | 7 2 | 7 2 | 7 2 7 2 5 2 | | 5 2 |
| | 6 2 | 9 2 | 8 2 9 2 5 2 | | 6 2 |
| | 7 2 | 6 2 | 9 2 6 2 5 2 | | 6 2 |
| | 8 2 | 8 2 | 6 2 8 2 6 2 | | 5 2 |
| | 9 2 | 6 2 | 9 2 6 2 6 2 | | 6 2 |
| | 8 2 | 7 2 | 6 2 7 2 6 2 | | 5 2 |
| | 6 2 | 9 2 | 7 2 7 2 5 2 | | 5 2 |
| | 9 2 | 9 2 | 9 2 6 2 5 2 | | 5 2 |
| | 7 2 | 8 2 | 8 2 7 2 5 2 | | 6 2 |
| | 6 2 | 7 2 | 8 2 7 2 5 2 | | 6 2 |
| | 7 2 | 8 2 | 9 2 8 2 6 2 | | 6 2 |
| | 8 2 | 6 2 | 8 2 6 2 5 2 | | 6 2 |
| | 6 2 | 9 2 | 6 2 9 2 6 2 | | 6 2 |
| | 7 2 | 8 2 | 7 2 8 2 6 2 | | 6 2 |
| | 9 2 | 9 2 | 9 2 9 2 6 2 | | 6 2 |
| | 9 2 | 8 2 | 9 2 8 2 6 2 | | 6 2 |
| | 8 2 | 8 2 | 8 2 8 2 6 2 | | 5 2 |
| | 8 2 | 9 2 | 8 2 9 2 5 2 | | 6 2 |
| | 9 2 | 9 2 | 9 2 9 2 6 2 | | 6 2 |

Continued on next column

**TABLE LVI– continued**

| TQC | ICSD | NoE | SF | | | |
|---|---|---|---|---|---|---|
| | 8 2 | 8 2 | 8 2 | 8 2 | 6 2 | 6 2 |
| | 9 2 | 8 2 | 9 2 | 8 2 | 6 2 | 6 2 |
| 123.1.15 | 92421 | 64 | Bi2Br1O4Tm1 | | | |
| | -A | -GM | -M | -Z | -R | -X |
| | 7 2 | 7 2 | 7 2 | 7 2 | 5 2 | 5 2 |
| | 9 2 | 9 2 | 9 2 | 9 2 | 6 2 | 6 2 |
| | 8 2 | 9 2 | 8 2 | 9 2 | 6 2 | 6 2 |
| | 9 2 | 8 2 | 9 2 | 8 2 | 6 2 | 6 2 |
| | 6 2 | 7 2 | 6 2 | 7 2 | 5 2 | 5 2 |
| | 7 2 | 9 2 | 7 2 | 9 2 | 5 2 | 5 2 |
| | 9 2 | 6 2 | 9 2 | 6 2 | 6 2 | 6 2 |
| | 8 2 | 8 2 | 8 2 | 8 2 | 6 2 | 6 2 |
| | 9 2 | 7 2 | 7 2 | 9 2 | 6 2 | 5 2 |
| | 6 2 | 7 2 | 6 2 | 7 2 | 5 2 | 6 2 |
| | 8 2 | 9 2 | 8 2 | 9 2 | 6 2 | 5 2 |
| | 7 2 | 7 2 | 7 2 | 7 2 | 5 2 | 5 2 |
| | 6 2 | 9 2 | 8 2 | 9 2 | 5 2 | 6 2 |
| | 8 2 | 6 2 | 9 2 | 6 2 | 5 2 | 6 2 |
| | 7 2 | 8 2 | 6 2 | 8 2 | 6 2 | 5 2 |
| | 9 2 | 6 2 | 9 2 | 6 2 | 6 2 | 6 2 |
| | 6 2 | 7 2 | 6 2 | 7 2 | 6 2 | 5 2 |
| | 8 2 | 9 2 | 7 2 | 7 2 | 5 2 | 5 2 |
| | 9 2 | 7 2 | 9 2 | 7 2 | 5 2 | 5 2 |
| | 7 2 | 9 2 | 8 2 | 8 2 | 5 2 | 6 2 |
| | 6 2 | 8 2 | 8 2 | 6 2 | 5 2 | 6 2 |
| | 7 2 | 6 2 | 9 2 | 7 2 | 6 2 | 6 2 |
| | 8 2 | 9 2 | 8 2 | 9 2 | 6 2 | 6 2 |
| | 6 2 | 8 2 | 6 2 | 8 2 | 6 2 | 6 2 |
| | 7 2 | 8 2 | 7 2 | 6 2 | 5 2 | 6 2 |
| | 9 2 | 9 2 | 9 2 | 9 2 | 6 2 | 6 2 |
| | 9 2 | 8 2 | 9 2 | 8 2 | 6 2 | 5 2 |
| | 8 2 | 8 2 | 8 2 | 8 2 | 6 2 | 6 2 |
| | 8 2 | 9 2 | 8 2 | 9 2 | 5 2 | 6 2 |
| | 9 2 | 9 2 | 9 2 | 9 2 | 6 2 | 6 2 |
| | 8 2 | 8 2 | 8 2 | 8 2 | 6 2 | 6 2 |
| | 9 2 | 8 2 | 9 2 | 8 2 | 6 2 | 6 2 |
| 123.1.16 | 68047 | 110 | Ba2Cu3Ho1O6 | | | |
| | -A | -GM | -M | -Z | -R | -X |
| | 8 2 | 7 2 | 6 2 | 9 2 | 5 2 | 6 2 |
| | 6 2 | 7 2 | 6 2 | 7 2 | 6 2 | 6 2 |
| | 8 2 | 9 2 | 8 2 | 9 2 | 5 2 | 5 2 |
| | 6 2 | 9 2 | 8 2 | 7 2 | 6 2 | 5 2 |
| | 7 2 | 9 2 | 9 2 | 7 2 | 6 2 | 5 2 |
| | 6 2 | 8 2 | 8 2 | 6 2 | 6 2 | 5 2 |
| | 8 2 | 7 2 | 6 2 | 9 2 | 5 2 | 6 2 |
| | 9 2 | 9 2 | 7 2 | 7 2 | 5 2 | 6 2 |
| | 6 2 | 6 2 | 8 2 | 8 2 | 6 2 | 5 2 |
| | 7 2 | 8 2 | 7 2 | 6 2 | 5 2 | 5 2 |
| | 7 2 | 7 2 | 9 2 | 7 2 | 6 2 | 5 2 |
| | 9 2 | 9 2 | 9 2 | 9 2 | 6 2 | 6 2 |
| | 6 2 | 7 2 | 6 2 | 7 2 | 6 2 | 6 2 |
| | 8 2 | 9 2 | 8 2 | 9 2 | 5 2 | 5 2 |
| | 9 2 | 7 2 | 9 2 | 7 2 | 6 2 | 6 2 |
| | 7 2 | 9 2 | 7 2 | 9 2 | 5 2 | 5 2 |
| | 6 2 | 8 2 | 6 2 | 8 2 | 5 2 | 5 2 |
| | 8 2 | 6 2 | 8 2 | 6 2 | 6 2 | 6 2 |
| | 9 2 | 7 2 | 7 2 | 9 2 | 6 2 | 5 2 |
| | 7 2 | 6 2 | 9 2 | 8 2 | 5 2 | 6 2 |
| | 8 2 | 7 2 | 6 2 | 9 2 | 5 2 | 6 2 |
| | 8 2 | 9 2 | 6 2 | 7 2 | 5 2 | 5 2 |
| | 6 2 | 8 2 | 8 2 | 7 2 | 5 2 | 6 2 |
| | 6 2 | 7 2 | 8 2 | 6 2 | 5 2 | 6 2 |
| | 7 2 | 6 2 | 7 2 | 8 2 | 6 2 | 5 2 |
| | 6 2 | 9 2 | 8 2 | 9 2 | 6 2 | 5 2 |
| | 7 2 | 7 2 | 9 2 | 7 2 | 6 2 | 5 2 |
| | 9 2 | 9 2 | 7 2 | 7 2 | 6 2 | 5 2 |
| | 8 2 | 6 2 | 6 2 | 6 2 | 5 2 | 6 2 |
| | 9 2 | 8 2 | 9 2 | 6 2 | 5 2 | 6 2 |
| | 6 2 | 6 2 | 6 2 | 8 2 | 6 2 | 5 2 |
| | 7 2 | 8 2 | 7 2 | 6 2 | 5 2 | 5 2 |
| | 9 2 | 7 2 | 7 2 | 9 2 | 5 2 | 6 2 |
| | 8 2 | 9 2 | 8 2 | 7 2 | 6 2 | 6 2 |
| | 7 2 | 6 2 | 9 2 | 8 2 | 6 2 | 5 2 |
| | 6 2 | 8 2 | 6 2 | 6 2 | 6 2 | 6 2 |
| | 9 2 | 7 2 | 9 2 | 7 2 | 5 2 | 5 2 |
| | 9 2 | 9 2 | 7 2 | 9 2 | 5 2 | 5 2 |
| | 7 2 | 7 2 | 6 2 | 8 2 | 6 2 | 6 2 |
| | 6 2 | 8 2 | 9 2 | 6 2 | 5 2 | 5 2 |
| | 7 2 | 6 2 | 6 2 | 7 2 | 5 2 | 5 2 |
| | 8 2 | 6 2 | 8 2 | 6 2 | 6 2 | 6 2 |
| | 6 2 | 7 2 | 9 2 | 8 2 | 5 2 | 6 2 |
| | 6 2 | 6 2 | 8 2 | 7 2 | 5 2 | 5 2 |
| | 6 2 | 6 2 | 7 2 | 6 2 | 6 2 | 5 2 |
| | 7 2 | 9 2 | 8 2 | 9 2 | 6 2 | 6 2 |
| | 9 2 | 8 2 | 6 2 | 6 2 | 5 2 | 5 2 |
| | 8 2 | 8 2 | 9 2 | 8 2 | 6 2 | 5 2 |
| | 7 2 | 9 2 | 7 2 | 9 2 | 6 2 | 5 2 |
| | 7 2 | 8 2 | 9 2 | 6 2 | 6 2 | 5 2 |
| | 6 2 | 8 2 | 6 2 | 6 2 | 5 2 | 5 2 |
| | 7 2 | 9 2 | 8 2 | 7 2 | 6 2 | 5 2 |
| | 6 2 | 7 2 | 7 2 | 7 2 | 5 2 | 5 2 |
| | 7 2 | 6 2 | 9 2 | 6 2 | 5 2 | 6 2 |
| | 6 2 | 9 2 | 8 2 | 7 2 | 6 2 | 5 2 |
| 123.1.17 | 74163 | 98 | Ba2Cu2O7Tl1Y1 | | | |
| | -A | -GM | -M | -Z | -R | -X |
| | 8 2 | 7 2 | 6 2 | 9 2 | 5 2 | 6 2 |
| | 6 2 | 7 2 | 6 2 | 7 2 | 6 2 | 6 2 |
| | 8 2 | 9 2 | 8 2 | 9 2 | 5 2 | 5 2 |
| | 6 2 | 9 2 | 8 2 | 7 2 | 6 2 | 5 2 |
| | 7 2 | 9 2 | 9 2 | 7 2 | 6 2 | 5 2 |
| | 6 2 | 8 2 | 8 2 | 6 2 | 6 2 | 5 2 |
| | 9 2 | 7 2 | 9 2 | 9 2 | 6 2 | 6 2 |
| | 9 2 | 9 2 | 7 2 | 9 2 | 5 2 | 6 2 |
| | 8 2 | 6 2 | 6 2 | 7 2 | 5 2 | 6 2 |
| | 7 2 | 7 2 | 7 2 | 8 2 | 5 2 | 5 2 |
| | 6 2 | 9 2 | 8 2 | 7 2 | 6 2 | 5 2 |
| | 7 2 | 8 2 | 9 2 | 6 2 | 6 2 | 5 2 |
| | 6 2 | 7 2 | 6 2 | 7 2 | 6 2 | 6 2 |
| | 8 2 | 9 2 | 8 2 | 9 2 | 5 2 | 5 2 |
| | 6 2 | 7 2 | 6 2 | 7 2 | 6 2 | 6 2 |
| | 9 2 | 9 2 | 9 2 | 9 2 | 5 2 | 5 2 |
| | 7 2 | 8 2 | 7 2 | 8 2 | 6 2 | 6 2 |
| | 8 2 | 6 2 | 8 2 | 6 2 | 5 2 | 5 2 |
| | 6 2 | 7 2 | 6 2 | 7 2 | 6 2 | 6 2 |
| | 9 2 | 7 2 | 7 2 | 7 2 | 5 2 | 5 2 |
| | 7 2 | 7 2 | 9 2 | 9 2 | 6 2 | 5 2 |
| | 8 2 | 7 2 | 6 2 | 9 2 | 5 2 | 6 2 |
| | 8 2 | 6 2 | 6 2 | 8 2 | 5 2 | 6 2 |
| | 6 2 | 9 2 | 8 2 | 7 2 | 5 2 | 6 2 |
| | 7 2 | 8 2 | 7 2 | 6 2 | 5 2 | 6 2 |
| | 6 2 | 9 2 | 8 2 | 9 2 | 6 2 | 6 2 |
| | 6 2 | 6 2 | 9 2 | 8 2 | 6 2 | 5 2 |

Continued on next column

Continued on next column

**TABLE LVI– continued**

| TQC | ICSD | NoE | SF | | | |
|---|---|---|---|---|---|---|
| | 7 2 | 6 2 | 8 2 | 8 2 | 6 2 | 5 2 |
| | 9 2 | 7 2 | 9 2 | 6 2 | 6 2 | 5 2 |
| | 9 2 | 8 2 | 9 2 | 7 2 | 6 2 | 5 2 |
| | 8 2 | 9 2 | 8 2 | 9 2 | 5 2 | 6 2 |
| | 9 2 | 9 2 | 7 2 | 9 2 | 6 2 | 6 2 |
| | 8 2 | 8 2 | 6 2 | 6 2 | 6 2 | 6 2 |
| | 9 2 | 9 2 | 7 2 | 8 2 | 5 2 | 5 2 |
| | 8 2 | 8 2 | 8 2 | 7 2 | 5 2 | 5 2 |
| | 6 2 | 7 2 | 6 2 | 9 2 | 5 2 | 5 2 |
| | 9 2 | 6 2 | 9 2 | 8 2 | 6 2 | 6 2 |
| | 7 2 | 6 2 | 7 2 | 6 2 | 5 2 | 5 2 |
| | 9 2 | 8 2 | 7 2 | 8 2 | 6 2 | 5 2 |
| | 7 2 | 7 2 | 8 2 | 9 2 | 6 2 | 6 2 |
| | 8 2 | 6 2 | 9 2 | 7 2 | 5 2 | 6 2 |
| | 8 2 | 7 2 | 6 2 | 8 2 | 5 2 | 5 2 |
| | 7 2 | 9 2 | 9 2 | 9 2 | 5 2 | 6 2 |
| | 6 2 | 9 2 | 8 2 | 6 2 | 5 2 | 5 2 |
| | 6 2 | 8 2 | 9 2 | 7 2 | 5 2 | 6 2 |
| | 7 2 | 6 2 | 8 2 | 8 2 | 5 2 | 5 2 |
| | 9 2 | 9 2 | 7 2 | 7 2 | 6 2 | 5 2 |
| | 8 2 | 8 2 | 6 2 | 6 2 | 6 2 | 5 2 |
| | 7 2 | 8 2 | 7 2 | 6 2 | 6 2 | 5 2 |
| 123.1.18 | 430061 | 34 | Ba1O1Sb2Ti2 | | | |
| | -A | -GM | -M | -Z | -R | -X |
| | 8 2 | 7 2 | 6 2 | 9 2 | 5 2 | 6 2 |
| | 6 2 | 7 2 | 6 2 | 7 2 | 6 2 | 6 2 |
| | 6 2 | 9 2 | 8 2 | 7 2 | 6 2 | 5 2 |
| | 7 2 | 9 2 | 9 2 | 7 2 | 6 2 | 5 2 |
| | 6 2 | 8 2 | 8 2 | 6 2 | 6 2 | 5 2 |
| | 9 2 | 7 2 | 7 2 | 7 2 | 5 2 | 5 2 |
| | 7 2 | 9 2 | 9 2 | 9 2 | 6 2 | 6 2 |
| | 8 2 | 9 2 | 8 2 | 9 2 | 5 2 | 5 2 |
| | 9 2 | 8 2 | 9 2 | 8 2 | 5 2 | 5 2 |
| | 8 2 | 7 2 | 8 2 | 9 2 | 5 2 | 5 2 |
| | 9 2 | 9 2 | 7 2 | 7 2 | 6 2 | 6 2 |
| | 8 2 | 7 2 | 9 2 | 9 2 | 5 2 | 5 2 |
| | 9 2 | 6 2 | 8 2 | 7 2 | 5 2 | 5 2 |
| | 7 2 | 6 2 | 7 2 | 6 2 | 6 2 | 6 2 |
| | 6 2 | 7 2 | 6 2 | 6 2 | 5 2 | 5 2 |
| | 7 2 | 9 2 | 9 2 | 9 2 | 6 2 | 6 2 |
| | 8 2 | 6 2 | 8 2 | 8 2 | 6 2 | 5 2 |
| 123.1.19 | 430060 | 34 | Ba1O1Sb2Ti2 | | | |
| | -A | -GM | -M | -Z | -R | -X |
| | 8 2 | 7 2 | 6 2 | 9 2 | 5 2 | 6 2 |
| | 6 2 | 7 2 | 6 2 | 7 2 | 6 2 | 6 2 |
| | 6 2 | 9 2 | 8 2 | 7 2 | 6 2 | 5 2 |
| | 7 2 | 9 2 | 9 2 | 7 2 | 6 2 | 5 2 |
| | 6 2 | 8 2 | 8 2 | 6 2 | 6 2 | 5 2 |
| | 9 2 | 7 2 | 7 2 | 7 2 | 5 2 | 5 2 |
| | 7 2 | 9 2 | 9 2 | 9 2 | 6 2 | 6 2 |
| | 8 2 | 9 2 | 8 2 | 9 2 | 5 2 | 5 2 |
| | 9 2 | 8 2 | 9 2 | 8 2 | 5 2 | 5 2 |
| | 8 2 | 7 2 | 8 2 | 9 2 | 5 2 | 5 2 |
| | 9 2 | 9 2 | 7 2 | 7 2 | 6 2 | 6 2 |
| | 8 2 | 7 2 | 9 2 | 9 2 | 5 2 | 5 2 |
| | 9 2 | 6 2 | 8 2 | 7 2 | 5 2 | 5 2 |
| | 7 2 | 6 2 | 7 2 | 6 2 | 6 2 | 6 2 |
| | 6 2 | 7 2 | 6 2 | 6 2 | 5 2 | 5 2 |
| | 7 2 | 9 2 | 9 2 | 9 2 | 6 2 | 6 2 |
| | 8 2 | 6 2 | 8 2 | 8 2 | 6 2 | 5 2 |
| 123.1.20 | 237789 | 34 | Ba1O1Sb2Ti2 | | | |

Continued on next column

**TABLE LVI– continued**

| TQC | ICSD | NoE | SF | | | |
|---|---|---|---|---|---|---|
| | -A | -GM | -M | -Z | -R | -X |
| | 7 2 | 7 2 | 7 2 | 7 2 | 5 2 | 5 2 |
| | 9 2 | 7 2 | 7 2 | 9 2 | 6 2 | 5 2 |
| | 9 2 | 9 2 | 9 2 | 9 2 | 6 2 | 6 2 |
| | 8 2 | 9 2 | 8 2 | 9 2 | 6 2 | 6 2 |
| | 9 2 | 8 2 | 9 2 | 8 2 | 6 2 | 6 2 |
| | 6 2 | 7 2 | 6 2 | 9 2 | 5 2 | 6 2 |
| | 8 2 | 9 2 | 8 2 | 7 2 | 6 2 | 5 2 |
| | 7 2 | 9 2 | 9 2 | 7 2 | 5 2 | 6 2 |
| | 6 2 | 8 2 | 8 2 | 6 2 | 5 2 | 6 2 |
| | 7 2 | 7 2 | 9 2 | 7 2 | 5 2 | 6 2 |
| | 6 2 | 9 2 | 6 2 | 9 2 | 6 2 | 5 2 |
| | 7 2 | 7 2 | 8 2 | 7 2 | 5 2 | 6 2 |
| | 6 2 | 6 2 | 9 2 | 9 2 | 5 2 | 6 2 |
| | 8 2 | 6 2 | 6 2 | 8 2 | 6 2 | 5 2 |
| | 9 2 | 7 2 | 7 2 | 8 2 | 5 2 | 6 2 |
| | 8 2 | 9 2 | 8 2 | 7 2 | 6 2 | 5 2 |
| | 7 2 | 6 2 | 9 2 | 6 2 | 6 2 | 6 2 |
| 123.1.21 | 165631 | 30 | Ba1Ge3Mg4 | | | |
| | -A | -GM | -M | -Z | -R | -X |
| | 7 2 | 7 2 | 7 2 | 7 2 | 5 2 | 5 2 |
| | 9 2 | 9 2 | 9 2 | 9 2 | 6 2 | 6 2 |
| | 8 2 | 9 2 | 8 2 | 9 2 | 6 2 | 6 2 |
| | 9 2 | 8 2 | 9 2 | 8 2 | 6 2 | 6 2 |
| | 6 2 | 7 2 | 6 2 | 7 2 | 6 2 | 6 2 |
| | 9 2 | 7 2 | 7 2 | 9 2 | 6 2 | 5 2 |
| | 8 2 | 9 2 | 8 2 | 9 2 | 5 2 | 5 2 |
| | 7 2 | 7 2 | 9 2 | 7 2 | 5 2 | 6 2 |
| | 6 2 | 9 2 | 8 2 | 7 2 | 5 2 | 6 2 |
| | 9 2 | 9 2 | 6 2 | 6 2 | 6 2 | 5 2 |
| | 8 2 | 8 2 | 7 2 | 9 2 | 6 2 | 5 2 |
| | 7 2 | 7 2 | 6 2 | 8 2 | 5 2 | 6 2 |
| | 6 2 | 6 2 | 9 2 | 7 2 | 5 2 | 6 2 |
| | 6 2 | 9 2 | 9 2 | 9 2 | 5 2 | 6 2 |
| | 7 2 | 6 2 | 8 2 | 6 2 | 6 2 | 6 2 |
| 123.1.22 | 262038 | 72 | As3Rb1Zn4 | | | |
| | -A | -GM | -M | -Z | -R | -X |
| | 9 2 | 7 2 | 7 2 | 9 2 | 6 2 | 5 2 |
| | 7 2 | 7 2 | 9 2 | 7 2 | 5 2 | 6 2 |
| | 6 2 | 9 2 | 8 2 | 7 2 | 5 2 | 6 2 |
| | 7 2 | 9 2 | 9 2 | 9 2 | 5 2 | 5 2 |
| | 7 2 | 8 2 | 7 2 | 7 2 | 5 2 | 6 2 |
| | 8 2 | 7 2 | 6 2 | 6 2 | 6 2 | 5 2 |
| | 6 2 | 9 2 | 8 2 | 7 2 | 5 2 | 6 2 |
| | 9 2 | 9 2 | 9 2 | 9 2 | 6 2 | 6 2 |
| | 8 2 | 8 2 | 8 2 | 8 2 | 6 2 | 6 2 |
| | 7 2 | 6 2 | 7 2 | 6 2 | 5 2 | 5 2 |
| | 6 2 | 8 2 | 6 2 | 8 2 | 5 2 | 5 2 |
| | 9 2 | 7 2 | 9 2 | 7 2 | 6 2 | 6 2 |
| | 8 2 | 6 2 | 8 2 | 9 2 | 6 2 | 6 2 |
| | 9 2 | 9 2 | 9 2 | 6 2 | 5 2 | 5 2 |
| | 7 2 | 7 2 | 7 2 | 9 2 | 5 2 | 5 2 |
| | 6 2 | 9 2 | 6 2 | 7 2 | 6 2 | 6 2 |
| | 8 2 | 7 2 | 8 2 | 9 2 | 6 2 | 6 2 |
| | 6 2 | 9 2 | 6 2 | 7 2 | 5 2 | 5 2 |
| | 9 2 | 8 2 | 9 2 | 8 2 | 5 2 | 5 2 |
| | 7 2 | 6 2 | 7 2 | 6 2 | 6 2 | 6 2 |
| | 8 2 | 7 2 | 8 2 | 7 2 | 5 2 | 5 2 |
| | 9 2 | 6 2 | 6 2 | 6 2 | 5 2 | 5 2 |
| | 6 2 | 8 2 | 9 2 | 8 2 | 6 2 | 6 2 |
| | 7 2 | 8 2 | 7 2 | 8 2 | 6 2 | 6 2 |

Continued on next column

**TABLE LVI– continued**

| TQC | ICSD | NoE | SF | | | |
|---|---|---|---|---|---|---|
| | 6 2 | 6 2 | 6 2 | 6 2 | 6 2 | 6 2 |
| | 7 2 | 9 2 | 7 2 | 9 2 | 5 2 | 5 2 |
| | 8 2 | 7 2 | 8 2 | 7 2 | 5 2 | 5 2 |
| | 9 2 | 9 2 | 9 2 | 9 2 | 6 2 | 6 2 |
| | 8 2 | 7 2 | 8 2 | 9 2 | 6 2 | 6 2 |
| | 6 2 | 9 2 | 6 2 | 8 2 | 5 2 | 5 2 |
| | 7 2 | 8 2 | 7 2 | 7 2 | 5 2 | 5 2 |
| | 9 2 | 7 2 | 9 2 | 7 2 | 6 2 | 6 2 |
| | 8 2 | 6 2 | 6 2 | 6 2 | 6 2 | 6 2 |
| | 9 2 | 9 2 | 8 2 | 9 2 | 6 2 | 6 2 |
| | 8 2 | 6 2 | 9 2 | 9 2 | 5 2 | 5 2 |
| | 6 2 | 9 2 | 8 2 | 6 2 | 6 2 | 6 2 |
| 123.1.23 | 262030 | 72 | As3Cs1Zn4 | | | |
| | -A | -GM | -M | -Z | -R | -X |
| | 9 2 | 7 2 | 7 2 | 9 2 | 6 2 | 5 2 |
| | 7 2 | 7 2 | 7 2 | 7 2 | 5 2 | 5 2 |
| | 8 2 | 9 2 | 6 2 | 9 2 | 5 2 | 6 2 |
| | 6 2 | 7 2 | 8 2 | 7 2 | 6 2 | 5 2 |
| | 7 2 | 9 2 | 9 2 | 7 2 | 5 2 | 6 2 |
| | 6 2 | 9 2 | 8 2 | 7 2 | 5 2 | 6 2 |
| | 7 2 | 8 2 | 9 2 | 6 2 | 5 2 | 6 2 |
| | 9 2 | 9 2 | 9 2 | 9 2 | 6 2 | 6 2 |
| | 8 2 | 8 2 | 8 2 | 8 2 | 6 2 | 6 2 |
| | 7 2 | 6 2 | 7 2 | 6 2 | 5 2 | 5 2 |
| | 6 2 | 8 2 | 6 2 | 8 2 | 5 2 | 5 2 |
| | 9 2 | 7 2 | 9 2 | 7 2 | 6 2 | 6 2 |
| | 8 2 | 6 2 | 8 2 | 9 2 | 6 2 | 6 2 |
| | 9 2 | 9 2 | 9 2 | 6 2 | 5 2 | 5 2 |
| | 7 2 | 7 2 | 7 2 | 9 2 | 5 2 | 5 2 |
| | 6 2 | 9 2 | 6 2 | 7 2 | 6 2 | 6 2 |
| | 8 2 | 7 2 | 8 2 | 9 2 | 6 2 | 6 2 |
| | 6 2 | 8 2 | 6 2 | 8 2 | 5 2 | 5 2 |
| | 9 2 | 9 2 | 9 2 | 6 2 | 6 2 | 5 2 |
| | 7 2 | 6 2 | 7 2 | 7 2 | 5 2 | 6 2 |
| | 8 2 | 7 2 | 8 2 | 7 2 | 5 2 | 5 2 |
| | 9 2 | 6 2 | 6 2 | 6 2 | 5 2 | 5 2 |
| | 6 2 | 8 2 | 9 2 | 8 2 | 6 2 | 6 2 |
| | 7 2 | 8 2 | 7 2 | 8 2 | 6 2 | 6 2 |
| | 6 2 | 6 2 | 6 2 | 6 2 | 6 2 | 6 2 |
| | 7 2 | 9 2 | 7 2 | 9 2 | 5 2 | 5 2 |
| | 8 2 | 7 2 | 8 2 | 7 2 | 5 2 | 5 2 |
| | 9 2 | 9 2 | 9 2 | 9 2 | 6 2 | 6 2 |
| | 8 2 | 7 2 | 8 2 | 9 2 | 6 2 | 6 2 |
| | 6 2 | 9 2 | 6 2 | 8 2 | 5 2 | 5 2 |
| | 7 2 | 8 2 | 7 2 | 7 2 | 5 2 | 5 2 |
| | 9 2 | 7 2 | 9 2 | 7 2 | 6 2 | 6 2 |
| | 8 2 | 6 2 | 6 2 | 6 2 | 6 2 | 6 2 |
| | 9 2 | 9 2 | 8 2 | 9 2 | 6 2 | 6 2 |
| | 8 2 | 6 2 | 9 2 | 9 2 | 5 2 | 5 2 |
| | 6 2 | 9 2 | 8 2 | 6 2 | 6 2 | 6 2 |
| 123.1.24 | 170662 | 28 | Ag1Sb1Te2 | | | |
| | -A | -GM | -M | -Z | -R | -X |
| | 8 2 | 7 2 | 7 2 | 9 2 | 5 2 | 6 2 |
| | 7 2 | 7 2 | 6 2 | 7 2 | 5 2 | 5 2 |
| | 9 2 | 7 2 | 7 2 | 9 2 | 6 2 | 5 2 |
| | 6 2 | 9 2 | 8 2 | 7 2 | 6 2 | 6 2 |
| | 7 2 | 6 2 | 9 2 | 7 2 | 6 2 | 6 2 |
| | 6 2 | 7 2 | 6 2 | 6 2 | 6 2 | 6 2 |
| | 7 2 | 9 2 | 9 2 | 6 2 | 6 2 | 6 2 |
| | 6 2 | 6 2 | 7 2 | 6 2 | 6 2 | 6 2 |
| | 7 2 | 6 2 | 6 2 | 7 2 | 6 2 | 6 2 |

Continued on next column

**TABLE LVI– continued**

| TQC | ICSD | NoE | SF | | | |
|---|---|---|---|---|---|---|
| | 9 2 | 7 2 | 7 2 | 6 2 | 6 2 | 5 2 |
| | 7 2 | 8 2 | 7 2 | 7 2 | 6 2 | 6 2 |
| | 8 2 | 9 2 | 9 2 | 9 2 | 5 2 | 5 2 |
| | 6 2 | 9 2 | 8 2 | 9 2 | 5 2 | 6 2 |
| | 9 2 | 8 2 | 8 2 | 8 2 | 5 2 | 5 2 |
| 123.2.1 | 98585 | 36 | Nd1Ni1O2 | | | |
| | -A | -GM | -M | -Z | -R | -X |
| | 8 2 | 7 2 | 6 2 | 9 2 | 5 2 | 6 2 |
| | 6 2 | 9 2 | 8 2 | 7 2 | 6 2 | 5 2 |
| | 9 2 | 7 2 | 9 2 | 7 2 | 5 2 | 5 2 |
| | 8 2 | 6 2 | 8 2 | 6 2 | 6 2 | 6 2 |
| | 6 2 | 8 2 | 8 2 | 6 2 | 6 2 | 5 2 |
| | 7 2 | 9 2 | 9 2 | 7 2 | 6 2 | 5 2 |
| | 7 2 | 9 2 | 7 2 | 9 2 | 5 2 | 5 2 |
| | 6 2 | 8 2 | 6 2 | 8 2 | 5 2 | 6 2 |
| | 6 2 | 9 2 | 6 2 | 9 2 | 5 2 | 6 2 |
| | 6 2 | 8 2 | 7 2 | 8 2 | 6 2 | 5 2 |
| | 7 2 | 9 2 | 6 2 | 9 2 | 6 2 | 6 2 |
| | 7 2 | 8 2 | 7 2 | 8 2 | 6 2 | 5 2 |
| | 9 2 | 7 2 | 7 2 | 6 2 | 5 2 | 5 2 |
| | 7 2 | 6 2 | 6 2 | 7 2 | 5 2 | 5 2 |
| | 6 2 | 7 2 | 7 2 | 6 2 | 5 2 | 5 2 |
| | 7 2 | 6 2 | 6 2 | 6 2 | 5 2 | 5 2 |
| | 6 2 | 6 2 | 9 2 | 7 2 | 6 2 | 5 2 |
| | 6 2 | 7 2 | 8 2 | 7 2 | 6 2 | 5 2 |
| 123.2.2 | 428960 | 16 | Mg1Ru1Sc2 | | | |
| | -A | -GM | -M | -Z | -R | -X |
| | 6 2 | 7 2 | 6 2 | 9 2 | 5 2 | 5 2 |
| | 7 2 | 7 2 | 7 2 | 7 2 | 5 2 | 5 2 |
| | 7 2 | 6 2 | 7 2 | 6 2 | 6 2 | 5 2 |
| | 6 2 | 9 2 | 6 2 | 7 2 | 5 2 | 6 2 |
| | 9 2 | 7 2 | 9 2 | 7 2 | 5 2 | 5 2 |
| | 9 2 | 7 2 | 9 2 | 9 2 | 6 2 | 5 2 |
| | 8 2 | 6 2 | 8 2 | 6 2 | 5 2 | 5 2 |
| | 6 2 | 6 2 | 6 2 | 6 2 | 5 2 | 6 2 |
| 123.2.3 | 157999 | 66 | Fe5Si3U1 | | | |
| | -A | -GM | -M | -Z | -R | -X |
| | 7 2 | 7 2 | 7 2 | 7 2 | 5 2 | 5 2 |
| | 9 2 | 9 2 | 9 2 | 9 2 | 6 2 | 6 2 |
| | 8 2 | 9 2 | 8 2 | 9 2 | 6 2 | 6 2 |
| | 9 2 | 8 2 | 9 2 | 8 2 | 6 2 | 6 2 |
| | 6 2 | 7 2 | 6 2 | 7 2 | 6 2 | 5 2 |
| | 9 2 | 7 2 | 7 2 | 9 2 | 6 2 | 6 2 |
| | 8 2 | 9 2 | 8 2 | 9 2 | 5 2 | 5 2 |
| | 6 2 | 7 2 | 9 2 | 7 2 | 5 2 | 6 2 |
| | 7 2 | 9 2 | 8 2 | 7 2 | 5 2 | 5 2 |
| | 9 2 | 9 2 | 7 2 | 6 2 | 6 2 | 6 2 |
| | 8 2 | 8 2 | 6 2 | 7 2 | 5 2 | 6 2 |
| | 8 2 | 6 2 | 6 2 | 8 2 | 5 2 | 6 2 |
| | 6 2 | 6 2 | 6 2 | 8 2 | 5 2 | 5 2 |
| | 7 2 | 7 2 | 9 2 | 9 2 | 5 2 | 6 2 |
| | 7 2 | 6 2 | 7 2 | 9 2 | 6 2 | 6 2 |
| | 6 2 | 7 2 | 9 2 | 8 2 | 6 2 | 6 2 |
| | 9 2 | 9 2 | 8 2 | 9 2 | 5 2 | 6 2 |
| | 7 2 | 8 2 | 7 2 | 6 2 | 5 2 | 5 2 |
| | 9 2 | 8 2 | 7 2 | 6 2 | 5 2 | 5 2 |
| | 7 2 | 9 2 | 6 2 | 7 2 | 6 2 | 5 2 |
| | 6 2 | 7 2 | 9 2 | 6 2 | 6 2 | 5 2 |
| | 8 2 | 6 2 | 8 2 | 9 2 | 5 2 | 6 2 |
| | 8 2 | 6 2 | 6 2 | 8 2 | 6 2 | 5 2 |
| | 9 2 | 8 2 | 8 2 | 8 2 | 6 2 | 6 2 |

Continued on next column

**TABLE LVI– continued**

| TQC | ICSD | | NoE | | SF | |
|---|---|---|---|---|---|---|
| | 6 2 | 7 2 | 7 2 | 9 2 | 5 2 | 6 2 |
| | 8 2 | 7 2 | 9 2 | 8 2 | 6 2 | 5 2 |
| | 9 2 | 9 2 | 6 2 | 7 2 | 6 2 | 5 2 |
| | 6 2 | 6 2 | 9 2 | 6 2 | 5 2 | 5 2 |
| | 7 2 | 6 2 | 9 2 | 7 2 | 6 2 | 6 2 |
| | 8 2 | 8 2 | 6 2 | 8 2 | 6 2 | 5 2 |
| | 7 2 | 6 2 | 8 2 | 6 2 | 6 2 | 6 2 |
| | 9 2 | 8 2 | 8 2 | 8 2 | 5 2 | 6 2 |
| | 6 2 | 8 2 | 7 2 | 8 2 | 5 2 | 5 2 |
| 123.2.4 | 427123 | | 58 | | Er2Mg3Ru1 | |
| | -A | -GM | -M | -Z | -R | -X |
| | 6 2 | 7 2 | 6 2 | 7 2 | 6 2 | 6 2 |
| | 8 2 | 9 2 | 8 2 | 9 2 | 5 2 | 5 2 |
| | 6 2 | 7 2 | 6 2 | 7 2 | 6 2 | 6 2 |
| | 8 2 | 9 2 | 8 2 | 9 2 | 5 2 | 5 2 |
| | 9 2 | 7 2 | 9 2 | 7 2 | 6 2 | 6 2 |
| | 7 2 | 9 2 | 7 2 | 9 2 | 5 2 | 5 2 |
| | 6 2 | 8 2 | 6 2 | 8 2 | 6 2 | 6 2 |
| | 8 2 | 6 2 | 8 2 | 6 2 | 5 2 | 5 2 |
| | 6 2 | 7 2 | 6 2 | 9 2 | 5 2 | 5 2 |
| | 7 2 | 9 2 | 7 2 | 7 2 | 5 2 | 6 2 |
| | 6 2 | 7 2 | 6 2 | 7 2 | 6 2 | 5 2 |
| | 7 2 | 6 2 | 7 2 | 9 2 | 5 2 | 5 2 |
| | 9 2 | 7 2 | 9 2 | 6 2 | 6 2 | 6 2 |
| | 7 2 | 9 2 | 7 2 | 7 2 | 5 2 | 5 2 |
| | 8 2 | 6 2 | 8 2 | 6 2 | 5 2 | 5 2 |
| | 6 2 | 7 2 | 6 2 | 6 2 | 5 2 | 5 2 |
| | 9 2 | 6 2 | 9 2 | 8 2 | 5 2 | 6 2 |
| | 7 2 | 8 2 | 7 2 | 7 2 | 6 2 | 5 2 |
| | 9 2 | 8 2 | 9 2 | 8 2 | 6 2 | 6 2 |
| | 6 2 | 9 2 | 9 2 | 9 2 | 5 2 | 6 2 |
| | 7 2 | 7 2 | 8 2 | 7 2 | 6 2 | 5 2 |
| | 9 2 | 7 2 | 9 2 | 6 2 | 5 2 | 6 2 |
| | 7 2 | 6 2 | 7 2 | 6 2 | 5 2 | 5 2 |
| | 8 2 | 6 2 | 8 2 | 9 2 | 6 2 | 5 2 |
| | 6 2 | 9 2 | 6 2 | 7 2 | 6 2 | 5 2 |
| | 8 2 | 8 2 | 8 2 | 9 2 | 6 2 | 6 2 |
| | 9 2 | 7 2 | 6 2 | 8 2 | 5 2 | 5 2 |
| | 7 2 | 8 2 | 7 2 | 8 2 | 6 2 | 6 2 |
| | 6 2 | 9 2 | 9 2 | 9 2 | 6 2 | 6 2 |
| 123.2.5 | 74164 | | 98 | | Cu2O7Sr2Tl1Y1 | |
| | -A | -GM | -M | -Z | -R | -X |
| | 8 2 | 7 2 | 6 2 | 9 2 | 5 2 | 6 2 |
| | 6 2 | 7 2 | 6 2 | 7 2 | 6 2 | 6 2 |
| | 8 2 | 9 2 | 8 2 | 9 2 | 5 2 | 5 2 |
| | 6 2 | 9 2 | 8 2 | 7 2 | 6 2 | 5 2 |
| | 7 2 | 9 2 | 9 2 | 7 2 | 6 2 | 5 2 |
| | 6 2 | 8 2 | 8 2 | 6 2 | 6 2 | 5 2 |
| | 9 2 | 7 2 | 9 2 | 7 2 | 6 2 | 6 2 |
| | 6 2 | 9 2 | 6 2 | 9 2 | 5 2 | 5 2 |
| | 7 2 | 7 2 | 7 2 | 9 2 | 6 2 | 6 2 |
| | 8 2 | 7 2 | 6 2 | 7 2 | 5 2 | 6 2 |
| | 9 2 | 9 2 | 8 2 | 9 2 | 5 2 | 6 2 |
| | 6 2 | 6 2 | 7 2 | 7 2 | 6 2 | 5 2 |
| | 8 2 | 9 2 | 6 2 | 8 2 | 5 2 | 5 2 |
| | 7 2 | 7 2 | 9 2 | 7 2 | 6 2 | 5 2 |
| | 6 2 | 8 2 | 8 2 | 6 2 | 6 2 | 6 2 |
| | 9 2 | 9 2 | 7 2 | 9 2 | 5 2 | 5 2 |
| | 7 2 | 8 2 | 9 2 | 8 2 | 6 2 | 6 2 |
| | 8 2 | 6 2 | 8 2 | 6 2 | 5 2 | 5 2 |
| | 6 2 | 7 2 | 6 2 | 7 2 | 6 2 | 6 2 |

Continued on next column

**TABLE LVI– continued**

| TQC | ICSD | | NoE | | SF | |
|---|---|---|---|---|---|---|
| | 9 2 | 7 2 | 7 2 | 7 2 | 5 2 | 5 2 |
| | 7 2 | 9 2 | 7 2 | 9 2 | 6 2 | 6 2 |
| | 7 2 | 7 2 | 9 2 | 9 2 | 5 2 | 5 2 |
| | 8 2 | 7 2 | 6 2 | 9 2 | 6 2 | 6 2 |
| | 9 2 | 6 2 | 9 2 | 8 2 | 6 2 | 6 2 |
| | 9 2 | 9 2 | 9 2 | 7 2 | 5 2 | 5 2 |
| | 8 2 | 8 2 | 8 2 | 6 2 | 5 2 | 6 2 |
| | 8 2 | 9 2 | 6 2 | 9 2 | 6 2 | 6 2 |
| | 6 2 | 6 2 | 8 2 | 8 2 | 5 2 | 6 2 |
| | 6 2 | 8 2 | 8 2 | 8 2 | 6 2 | 5 2 |
| | 6 2 | 9 2 | 8 2 | 7 2 | 5 2 | 6 2 |
| | 7 2 | 7 2 | 9 2 | 7 2 | 6 2 | 5 2 |
| | 9 2 | 6 2 | 6 2 | 6 2 | 5 2 | 5 2 |
| | 8 2 | 9 2 | 7 2 | 9 2 | 6 2 | 5 2 |
| | 6 2 | 8 2 | 8 2 | 8 2 | 5 2 | 6 2 |
| | 7 2 | 8 2 | 9 2 | 6 2 | 5 2 | 6 2 |
| | 9 2 | 9 2 | 7 2 | 9 2 | 6 2 | 5 2 |
| | 8 2 | 6 2 | 6 2 | 8 2 | 5 2 | 5 2 |
| | 8 2 | 8 2 | 8 2 | 6 2 | 5 2 | 5 2 |
| | 9 2 | 7 2 | 7 2 | 9 2 | 5 2 | 5 2 |
| | 7 2 | 6 2 | 9 2 | 8 2 | 6 2 | 5 2 |
| | 7 2 | 7 2 | 8 2 | 7 2 | 5 2 | 6 2 |
| | 6 2 | 9 2 | 7 2 | 6 2 | 5 2 | 6 2 |
| | 9 2 | 8 2 | 6 2 | 7 2 | 6 2 | 5 2 |
| | 8 2 | 6 2 | 9 2 | 9 2 | 5 2 | 6 2 |
| | 9 2 | 7 2 | 7 2 | 8 2 | 6 2 | 5 2 |
| | 8 2 | 6 2 | 6 2 | 8 2 | 5 2 | 5 2 |
| | 6 2 | 9 2 | 8 2 | 7 2 | 6 2 | 6 2 |
| | 7 2 | 8 2 | 9 2 | 9 2 | 6 2 | 5 2 |
| | 7 2 | 9 2 | 7 2 | 6 2 | 6 2 | 5 2 |
| 123.2.6 | 85244 | | 116 | | Cu2Gd1O8Ru1Sr2 | |
| | -A | -GM | -M | -Z | -R | -X |
| | 8 2 | 7 2 | 6 2 | 9 2 | 5 2 | 6 2 |
| | 6 2 | 7 2 | 6 2 | 7 2 | 6 2 | 5 2 |
| | 8 2 | 9 2 | 8 2 | 9 2 | 5 2 | 6 2 |
| | 6 2 | 9 2 | 8 2 | 7 2 | 6 2 | 5 2 |
| | 7 2 | 9 2 | 9 2 | 7 2 | 6 2 | 5 2 |
| | 6 2 | 7 2 | 8 2 | 7 2 | 5 2 | 5 2 |
| | 9 2 | 6 2 | 9 2 | 6 2 | 6 2 | 5 2 |
| | 8 2 | 7 2 | 8 2 | 9 2 | 6 2 | 6 2 |
| | 8 2 | 8 2 | 6 2 | 6 2 | 5 2 | 6 2 |
| | 9 2 | 6 2 | 7 2 | 8 2 | 5 2 | 6 2 |
| | 7 2 | 8 2 | 7 2 | 6 2 | 6 2 | 5 2 |
| | 6 2 | 7 2 | 8 2 | 7 2 | 5 2 | 6 2 |
| | 9 2 | 9 2 | 9 2 | 7 2 | 6 2 | 5 2 |
| | 8 2 | 9 2 | 6 2 | 9 2 | 6 2 | 5 2 |
| | 6 2 | 7 2 | 8 2 | 9 2 | 5 2 | 5 2 |
| | 7 2 | 9 2 | 9 2 | 7 2 | 6 2 | 6 2 |
| | 6 2 | 9 2 | 6 2 | 9 2 | 5 2 | 5 2 |
| | 8 2 | 8 2 | 7 2 | 8 2 | 6 2 | 6 2 |
| | 9 2 | 6 2 | 8 2 | 6 2 | 5 2 | 5 2 |
| | 7 2 | 7 2 | 9 2 | 7 2 | 6 2 | 6 2 |
| | 6 2 | 7 2 | 6 2 | 7 2 | 5 2 | 5 2 |
| | 7 2 | 7 2 | 7 2 | 9 2 | 6 2 | 5 2 |
| | 9 2 | 6 2 | 7 2 | 8 2 | 5 2 | 5 2 |
| | 6 2 | 7 2 | 6 2 | 9 2 | 5 2 | 6 2 |
| | 7 2 | 9 2 | 9 2 | 9 2 | 5 2 | 5 2 |
| | 8 2 | 9 2 | 7 2 | 7 2 | 5 2 | 5 2 |
| | 7 2 | 8 2 | 6 2 | 6 2 | 5 2 | 6 2 |
| | 6 2 | 8 2 | 6 2 | 8 2 | 5 2 | 6 2 |
| | 8 2 | 9 2 | 6 2 | 9 2 | 5 2 | 6 2 |

Continued on next column

**TABLE LVI– continued**

| TQC | ICSD | NoE | SF | | | |
|---|---|---|---|---|---|---|
| | 6 2 | 6 2 | 8 2 | 8 2 | 6 2 | 6 2 |
| | 7 2 | 7 2 | 7 2 | 7 2 | 6 2 | 5 2 |
| | 6 2 | 6 2 | 8 2 | 6 2 | 6 2 | 6 2 |
| | 6 2 | 9 2 | 8 2 | 9 2 | 6 2 | 5 2 |
| | 7 2 | 8 2 | 9 2 | 7 2 | 6 2 | 5 2 |
| | 9 2 | 9 2 | 7 2 | 8 2 | 5 2 | 5 2 |
| | 8 2 | 8 2 | 6 2 | 6 2 | 6 2 | 6 2 |
| | 9 2 | 9 2 | 9 2 | 8 2 | 5 2 | 5 2 |
| | 8 2 | 6 2 | 7 2 | 9 2 | 6 2 | 6 2 |
| | 9 2 | 8 2 | 8 2 | 6 2 | 5 2 | 6 2 |
| | 7 2 | 6 2 | 9 2 | 8 2 | 6 2 | 6 2 |
| | 9 2 | 8 2 | 9 2 | 6 2 | 6 2 | 6 2 |
| | 9 2 | 7 2 | 7 2 | 7 2 | 6 2 | 5 2 |
| | 7 2 | 9 2 | 7 2 | 9 2 | 5 2 | 6 2 |
| | 7 2 | 8 2 | 6 2 | 8 2 | 6 2 | 5 2 |
| | 8 2 | 6 2 | 9 2 | 6 2 | 5 2 | 5 2 |
| | 6 2 | 7 2 | 7 2 | 9 2 | 5 2 | 5 2 |
| | 7 2 | 9 2 | 6 2 | 8 2 | 6 2 | 6 2 |
| | 7 2 | 8 2 | 8 2 | 7 2 | 5 2 | 6 2 |
| | 6 2 | 6 2 | 7 2 | 6 2 | 5 2 | 6 2 |
| | 9 2 | 6 2 | 8 2 | 8 2 | 6 2 | 5 2 |
| | 6 2 | 7 2 | 6 2 | 6 2 | 6 2 | 5 2 |
| | 8 2 | 8 2 | 9 2 | 7 2 | 6 2 | 5 2 |
| | 6 2 | 9 2 | 8 2 | 9 2 | 6 2 | 6 2 |
| | 7 2 | 8 2 | 9 2 | 8 2 | 6 2 | 5 2 |
| | 7 2 | 9 2 | 9 2 | 9 2 | 5 2 | 5 2 |
| | 6 2 | 8 2 | 8 2 | 6 2 | 6 2 | 5 2 |
| | 7 2 | 8 2 | 9 2 | 6 2 | 5 2 | 5 2 |
| | 6 2 | 9 2 | 8 2 | 7 2 | 6 2 | 6 2 |
| 123.2.7 | 240161 | 40 | Ce1Ga6Pd1 | | | |
| | -A | -GM | -M | -Z | -R | -X |
| | 7 2 | 7 2 | 7 2 | 7 2 | 5 2 | 5 2 |
| | 9 2 | 9 2 | 9 2 | 9 2 | 6 2 | 6 2 |
| | 8 2 | 9 2 | 8 2 | 9 2 | 6 2 | 6 2 |
| | 9 2 | 8 2 | 9 2 | 8 2 | 6 2 | 6 2 |
| | 6 2 | 7 2 | 6 2 | 9 2 | 5 2 | 6 2 |
| | 7 2 | 9 2 | 9 2 | 7 2 | 6 2 | 5 2 |
| | 6 2 | 7 2 | 8 2 | 8 2 | 6 2 | 5 2 |
| | 8 2 | 6 2 | 7 2 | 7 2 | 5 2 | 6 2 |
| | 9 2 | 7 2 | 6 2 | 9 2 | 6 2 | 6 2 |
| | 8 2 | 6 2 | 8 2 | 9 2 | 6 2 | 5 2 |
| | 9 2 | 7 2 | 7 2 | 8 2 | 6 2 | 5 2 |
| | 8 2 | 6 2 | 6 2 | 8 2 | 6 2 | 5 2 |
| | 8 2 | 9 2 | 6 2 | 9 2 | 5 2 | 5 2 |
| | 9 2 | 8 2 | 7 2 | 6 2 | 6 2 | 5 2 |
| | 7 2 | 6 2 | 7 2 | 6 2 | 6 2 | 5 2 |
| | 9 2 | 8 2 | 6 2 | 7 2 | 5 2 | 5 2 |
| | 8 2 | 9 2 | 9 2 | 8 2 | 5 2 | 6 2 |
| | 6 2 | 7 2 | 6 2 | 9 2 | 6 2 | 6 2 |
| | 7 2 | 7 2 | 9 2 | 8 2 | 5 2 | 6 2 |
| | 6 2 | 6 2 | 8 2 | 9 2 | 6 2 | 6 2 |
| 123.2.8 | 102108 | 36 | Ce1Co1In5 | | | |
| | -A | -GM | -M | -Z | -R | -X |
| | 7 2 | 7 2 | 7 2 | 7 2 | 5 2 | 5 2 |
| | 9 2 | 9 2 | 9 2 | 9 2 | 6 2 | 6 2 |
| | 8 2 | 9 2 | 8 2 | 9 2 | 6 2 | 6 2 |
| | 9 2 | 8 2 | 9 2 | 8 2 | 6 2 | 6 2 |
| | 7 2 | 7 2 | 6 2 | 9 2 | 5 2 | 5 2 |
| | 6 2 | 9 2 | 9 2 | 7 2 | 6 2 | 6 2 |
| | 6 2 | 7 2 | 8 2 | 8 2 | 6 2 | 5 2 |
| | 9 2 | 6 2 | 7 2 | 7 2 | 5 2 | 6 2 |

Continued on next column

**TABLE LVI– continued**

| TQC | ICSD | NoE | SF | | | |
|---|---|---|---|---|---|---|
| | 8 2 | 6 2 | 7 2 | 9 2 | 6 2 | 6 2 |
| | 9 2 | 7 2 | 6 2 | 6 2 | 5 2 | 5 2 |
| | 8 2 | 8 2 | 8 2 | 6 2 | 6 2 | 5 2 |
| | 7 2 | 7 2 | 6 2 | 8 2 | 5 2 | 5 2 |
| | 8 2 | 9 2 | 9 2 | 9 2 | 6 2 | 5 2 |
| | 9 2 | 6 2 | 9 2 | 7 2 | 6 2 | 5 2 |
| | 7 2 | 6 2 | 8 2 | 8 2 | 5 2 | 5 2 |
| | 6 2 | 8 2 | 6 2 | 9 2 | 6 2 | 5 2 |
| | 8 2 | 8 2 | 7 2 | 8 2 | 5 2 | 6 2 |
| | 9 2 | 9 2 | 7 2 | 8 2 | 6 2 | 6 2 |
| 123.2.9 | 92425 | 54 | Bi2I1O4Pr1 | | | |
| | -A | -GM | -M | -Z | -R | -X |
| | 7 2 | 7 2 | 7 2 | 7 2 | 5 2 | 5 2 |
| | 9 2 | 9 2 | 9 2 | 9 2 | 6 2 | 6 2 |
| | 8 2 | 7 2 | 8 2 | 7 2 | 6 2 | 6 2 |
| | 6 2 | 9 2 | 6 2 | 9 2 | 5 2 | 5 2 |
| | 7 2 | 6 2 | 7 2 | 6 2 | 5 2 | 5 2 |
| | 9 2 | 8 2 | 9 2 | 8 2 | 6 2 | 6 2 |
| | 9 2 | 8 2 | 9 2 | 8 2 | 6 2 | 6 2 |
| | 8 2 | 9 2 | 8 2 | 9 2 | 6 2 | 6 2 |
| | 9 2 | 7 2 | 7 2 | 9 2 | 6 2 | 5 2 |
| | 6 2 | 9 2 | 6 2 | 7 2 | 5 2 | 6 2 |
| | 8 2 | 7 2 | 8 2 | 9 2 | 6 2 | 5 2 |
| | 7 2 | 7 2 | 6 2 | 7 2 | 5 2 | 5 2 |
| | 6 2 | 6 2 | 7 2 | 6 2 | 5 2 | 6 2 |
| | 7 2 | 9 2 | 8 2 | 9 2 | 6 2 | 5 2 |
| | 8 2 | 8 2 | 9 2 | 8 2 | 5 2 | 6 2 |
| | 6 2 | 6 2 | 6 2 | 6 2 | 6 2 | 6 2 |
| | 9 2 | 7 2 | 7 2 | 7 2 | 5 2 | 5 2 |
| | 8 2 | 9 2 | 8 2 | 7 2 | 6 2 | 5 2 |
| | 9 2 | 7 2 | 9 2 | 7 2 | 5 2 | 5 2 |
| | 7 2 | 6 2 | 9 2 | 6 2 | 5 2 | 6 2 |
| | 6 2 | 8 2 | 8 2 | 8 2 | 6 2 | 6 2 |
| | 7 2 | 9 2 | 7 2 | 9 2 | 5 2 | 6 2 |
| | 6 2 | 8 2 | 6 2 | 8 2 | 6 2 | 6 2 |
| | 9 2 | 9 2 | 9 2 | 9 2 | 6 2 | 5 2 |
| | 7 2 | 9 2 | 8 2 | 7 2 | 5 2 | 6 2 |
| | 8 2 | 8 2 | 9 2 | 6 2 | 5 2 | 6 2 |
| | 9 2 | 7 2 | 9 2 | 9 2 | 6 2 | 6 2 |
| 123.2.10 | 92403 | 60 | Bi2Cl1O4Tb1 | | | |
| | -A | -GM | -M | -Z | -R | -X |
| | 7 2 | 7 2 | 7 2 | 7 2 | 5 2 | 5 2 |
| | 9 2 | 9 2 | 9 2 | 9 2 | 6 2 | 6 2 |
| | 8 2 | 9 2 | 8 2 | 9 2 | 6 2 | 6 2 |
| | 9 2 | 7 2 | 9 2 | 7 2 | 6 2 | 6 2 |
| | 6 2 | 8 2 | 6 2 | 8 2 | 5 2 | 5 2 |
| | 7 2 | 9 2 | 7 2 | 9 2 | 5 2 | 5 2 |
| | 9 2 | 6 2 | 9 2 | 6 2 | 6 2 | 6 2 |
| | 8 2 | 8 2 | 8 2 | 8 2 | 6 2 | 6 2 |
| | 9 2 | 7 2 | 7 2 | 9 2 | 6 2 | 5 2 |
| | 6 2 | 7 2 | 6 2 | 7 2 | 5 2 | 6 2 |
| | 8 2 | 9 2 | 8 2 | 9 2 | 6 2 | 5 2 |
| | 7 2 | 7 2 | 8 2 | 7 2 | 5 2 | 5 2 |
| | 6 2 | 9 2 | 7 2 | 6 2 | 5 2 | 6 2 |
| | 7 2 | 6 2 | 9 2 | 9 2 | 6 2 | 5 2 |
| | 8 2 | 8 2 | 6 2 | 8 2 | 5 2 | 6 2 |
| | 9 2 | 6 2 | 9 2 | 6 2 | 6 2 | 6 2 |
| | 8 2 | 7 2 | 6 2 | 7 2 | 6 2 | 5 2 |
| | 6 2 | 9 2 | 7 2 | 7 2 | 5 2 | 5 2 |
| | 9 2 | 9 2 | 9 2 | 7 2 | 5 2 | 5 2 |
| | 7 2 | 8 2 | 8 2 | 6 2 | 5 2 | 6 2 |

Continued on next column

**TABLE LVI– continued**

| TQC | ICSD | NoE | SF | | |
|---|---|---|---|---|---|
| 6 2 | 7 2 | 8 2 | 7 2 | 5 2 | 6 2 |
| 7 2 | 8 2 | 9 2 | 8 2 | 6 2 | 6 2 |
| 7 2 | 6 2 | 7 2 | 6 2 | 5 2 | 6 2 |
| 9 2 | 9 2 | 9 2 | 9 2 | 6 2 | 6 2 |
| 6 2 | 8 2 | 6 2 | 8 2 | 6 2 | 6 2 |
| 8 2 | 9 2 | 8 2 | 9 2 | 5 2 | 5 2 |
| 9 2 | 8 2 | 9 2 | 8 2 | 6 2 | 6 2 |
| 8 2 | 8 2 | 8 2 | 8 2 | 6 2 | 6 2 |
| 8 2 | 9 2 | 8 2 | 9 2 | 6 2 | 6 2 |
| 9 2 | 7 2 | 9 2 | 9 2 | 6 2 | 6 2 |
| 123.2.11 | 92398 | 54 | Bi2Cl1O4Pr1 | | |
| -A | -GM | -M | -Z | -R | -X |
| 7 2 | 7 2 | 7 2 | 7 2 | 5 2 | 5 2 |
| 9 2 | 9 2 | 9 2 | 9 2 | 6 2 | 6 2 |
| 8 2 | 7 2 | 8 2 | 7 2 | 6 2 | 6 2 |
| 6 2 | 9 2 | 6 2 | 9 2 | 5 2 | 5 2 |
| 7 2 | 6 2 | 7 2 | 6 2 | 5 2 | 5 2 |
| 9 2 | 8 2 | 9 2 | 8 2 | 6 2 | 6 2 |
| 9 2 | 8 2 | 9 2 | 8 2 | 6 2 | 6 2 |
| 8 2 | 9 2 | 8 2 | 9 2 | 6 2 | 6 2 |
| 9 2 | 7 2 | 7 2 | 9 2 | 6 2 | 5 2 |
| 8 2 | 7 2 | 6 2 | 7 2 | 5 2 | 6 2 |
| 6 2 | 9 2 | 8 2 | 9 2 | 6 2 | 5 2 |
| 6 2 | 7 2 | 8 2 | 7 2 | 5 2 | 5 2 |
| 7 2 | 6 2 | 6 2 | 6 2 | 5 2 | 6 2 |
| 7 2 | 9 2 | 7 2 | 9 2 | 6 2 | 5 2 |
| 8 2 | 8 2 | 9 2 | 8 2 | 5 2 | 6 2 |
| 9 2 | 6 2 | 6 2 | 6 2 | 6 2 | 6 2 |
| 6 2 | 7 2 | 7 2 | 7 2 | 5 2 | 5 2 |
| 8 2 | 9 2 | 9 2 | 7 2 | 6 2 | 5 2 |
| 9 2 | 9 2 | 8 2 | 7 2 | 5 2 | 5 2 |
| 7 2 | 8 2 | 9 2 | 6 2 | 5 2 | 6 2 |
| 6 2 | 7 2 | 8 2 | 7 2 | 5 2 | 6 2 |
| 7 2 | 8 2 | 9 2 | 8 2 | 6 2 | 6 2 |
| 7 2 | 6 2 | 7 2 | 6 2 | 5 2 | 6 2 |
| 6 2 | 9 2 | 6 2 | 9 2 | 6 2 | 6 2 |
| 9 2 | 8 2 | 9 2 | 9 2 | 6 2 | 5 2 |
| 8 2 | 9 2 | 8 2 | 8 2 | 5 2 | 6 2 |
| 9 2 | 7 2 | 9 2 | 9 2 | 6 2 | 6 2 |
| 123.2.12 | 92401 | 58 | Bi2Cl1Eu1O4 | | |
| -A | -GM | -M | -Z | -R | -X |
| 7 2 | 7 2 | 7 2 | 7 2 | 5 2 | 5 2 |
| 9 2 | 9 2 | 9 2 | 9 2 | 6 2 | 6 2 |
| 8 2 | 9 2 | 8 2 | 9 2 | 6 2 | 6 2 |
| 9 2 | 7 2 | 9 2 | 7 2 | 6 2 | 6 2 |
| 6 2 | 8 2 | 6 2 | 8 2 | 5 2 | 5 2 |
| 7 2 | 6 2 | 7 2 | 6 2 | 5 2 | 5 2 |
| 9 2 | 8 2 | 9 2 | 8 2 | 6 2 | 6 2 |
| 8 2 | 9 2 | 8 2 | 9 2 | 6 2 | 6 2 |
| 9 2 | 7 2 | 7 2 | 9 2 | 6 2 | 5 2 |
| 8 2 | 7 2 | 6 2 | 7 2 | 5 2 | 6 2 |
| 6 2 | 9 2 | 8 2 | 9 2 | 6 2 | 5 2 |
| 6 2 | 7 2 | 8 2 | 7 2 | 5 2 | 5 2 |
| 7 2 | 9 2 | 6 2 | 6 2 | 5 2 | 6 2 |
| 7 2 | 6 2 | 7 2 | 9 2 | 6 2 | 6 2 |
| 8 2 | 8 2 | 9 2 | 8 2 | 5 2 | 5 2 |
| 9 2 | 6 2 | 6 2 | 6 2 | 6 2 | 6 2 |
| 8 2 | 7 2 | 9 2 | 7 2 | 6 2 | 5 2 |
| 6 2 | 9 2 | 7 2 | 7 2 | 5 2 | 5 2 |
| 9 2 | 9 2 | 8 2 | 7 2 | 5 2 | 5 2 |
| 7 2 | 8 2 | 9 2 | 6 2 | 5 2 | 6 2 |

Continued on next column

**TABLE LVI– continued**

| TQC | ICSD | NoE | SF | | |
|---|---|---|---|---|---|
| 6 2 | 7 2 | 8 2 | 7 2 | 5 2 | 6 2 |
| 7 2 | 8 2 | 9 2 | 8 2 | 5 2 | 6 2 |
| 7 2 | 6 2 | 7 2 | 6 2 | 6 2 | 6 2 |
| 9 2 | 9 2 | 9 2 | 9 2 | 6 2 | 6 2 |
| 6 2 | 8 2 | 6 2 | 9 2 | 6 2 | 6 2 |
| 8 2 | 9 2 | 8 2 | 8 2 | 5 2 | 5 2 |
| 9 2 | 8 2 | 9 2 | 8 2 | 6 2 | 6 2 |
| 8 2 | 8 2 | 8 2 | 8 2 | 6 2 | 6 2 |
| 8 2 | 7 2 | 8 2 | 9 2 | 6 2 | 6 2 |
| 123.2.13 | 92416 | 60 | Bi2Br1O4Tb1 | | |
| -A | -GM | -M | -Z | -R | -X |
| 7 2 | 7 2 | 7 2 | 7 2 | 5 2 | 5 2 |
| 9 2 | 9 2 | 9 2 | 9 2 | 6 2 | 6 2 |
| 8 2 | 9 2 | 8 2 | 9 2 | 6 2 | 6 2 |
| 9 2 | 7 2 | 9 2 | 7 2 | 6 2 | 6 2 |
| 6 2 | 8 2 | 6 2 | 8 2 | 5 2 | 5 2 |
| 7 2 | 9 2 | 7 2 | 9 2 | 5 2 | 5 2 |
| 9 2 | 6 2 | 9 2 | 6 2 | 6 2 | 6 2 |
| 8 2 | 8 2 | 8 2 | 8 2 | 6 2 | 6 2 |
| 9 2 | 7 2 | 7 2 | 9 2 | 6 2 | 5 2 |
| 6 2 | 7 2 | 6 2 | 7 2 | 5 2 | 6 2 |
| 8 2 | 9 2 | 8 2 | 9 2 | 6 2 | 5 2 |
| 7 2 | 7 2 | 7 2 | 7 2 | 5 2 | 5 2 |
| 6 2 | 9 2 | 8 2 | 9 2 | 5 2 | 6 2 |
| 7 2 | 6 2 | 9 2 | 6 2 | 6 2 | 5 2 |
| 8 2 | 8 2 | 6 2 | 8 2 | 5 2 | 6 2 |
| 9 2 | 6 2 | 9 2 | 6 2 | 6 2 | 6 2 |
| 6 2 | 7 2 | 6 2 | 7 2 | 6 2 | 5 2 |
| 8 2 | 9 2 | 7 2 | 7 2 | 5 2 | 5 2 |
| 9 2 | 7 2 | 9 2 | 7 2 | 5 2 | 5 2 |
| 7 2 | 6 2 | 8 2 | 6 2 | 5 2 | 6 2 |
| 6 2 | 8 2 | 8 2 | 8 2 | 5 2 | 6 2 |
| 7 2 | 9 2 | 9 2 | 7 2 | 6 2 | 6 2 |
| 7 2 | 9 2 | 7 2 | 9 2 | 6 2 | 6 2 |
| 9 2 | 8 2 | 9 2 | 6 2 | 6 2 | 6 2 |
| 6 2 | 8 2 | 6 2 | 8 2 | 5 2 | 5 2 |
| 8 2 | 9 2 | 8 2 | 9 2 | 5 2 | 6 2 |
| 9 2 | 8 2 | 9 2 | 8 2 | 6 2 | 6 2 |
| 8 2 | 8 2 | 8 2 | 8 2 | 6 2 | 6 2 |
| 8 2 | 9 2 | 8 2 | 9 2 | 6 2 | 6 2 |
| 9 2 | 7 2 | 9 2 | 9 2 | 6 2 | 6 2 |
| 123.2.14 | 92411 | 54 | Bi2Br1O4Pr1 | | |
| -A | -GM | -M | -Z | -R | -X |
| 7 2 | 7 2 | 7 2 | 7 2 | 5 2 | 5 2 |
| 9 2 | 9 2 | 9 2 | 9 2 | 6 2 | 6 2 |
| 8 2 | 7 2 | 8 2 | 7 2 | 6 2 | 6 2 |
| 6 2 | 9 2 | 6 2 | 9 2 | 5 2 | 5 2 |
| 7 2 | 6 2 | 7 2 | 6 2 | 5 2 | 5 2 |
| 9 2 | 8 2 | 9 2 | 8 2 | 6 2 | 6 2 |
| 9 2 | 8 2 | 9 2 | 8 2 | 6 2 | 6 2 |
| 8 2 | 9 2 | 8 2 | 9 2 | 6 2 | 6 2 |
| 9 2 | 7 2 | 7 2 | 9 2 | 6 2 | 5 2 |
| 6 2 | 7 2 | 6 2 | 7 2 | 5 2 | 6 2 |
| 8 2 | 9 2 | 8 2 | 9 2 | 6 2 | 5 2 |
| 7 2 | 7 2 | 7 2 | 7 2 | 5 2 | 5 2 |
| 6 2 | 6 2 | 8 2 | 6 2 | 5 2 | 6 2 |
| 7 2 | 9 2 | 6 2 | 9 2 | 6 2 | 5 2 |
| 8 2 | 8 2 | 9 2 | 8 2 | 5 2 | 6 2 |
| 9 2 | 6 2 | 6 2 | 6 2 | 6 2 | 6 2 |
| 6 2 | 7 2 | 7 2 | 7 2 | 5 2 | 5 2 |
| 8 2 | 9 2 | 9 2 | 7 2 | 6 2 | 5 2 |

Continued on next column

**TABLE LVI– continued**

| TQC | ICSD | NoE | SF | | | |
|---|---|---|---|---|---|---|
| | 9 2 | 7 2 | 8 2 | 7 2 | 5 2 | 5 2 |
| | 7 2 | 6 2 | 9 2 | 6 2 | 5 2 | 6 2 |
| | 6 2 | 8 2 | 8 2 | 8 2 | 5 2 | 6 2 |
| | 7 2 | 9 2 | 9 2 | 7 2 | 6 2 | 6 2 |
| | 7 2 | 8 2 | 7 2 | 9 2 | 6 2 | 6 2 |
| | 9 2 | 9 2 | 9 2 | 8 2 | 6 2 | 6 2 |
| | 6 2 | 8 2 | 6 2 | 9 2 | 5 2 | 5 2 |
| | 8 2 | 9 2 | 8 2 | 6 2 | 5 2 | 6 2 |
| | 9 2 | 7 2 | 9 2 | 9 2 | 6 2 | 6 2 |
| 123.2.15 | 424316 | 134 | Ba3O5Te1Yb2 | | | |
| | -A | -GM | -M | -Z | -R | -X |
| | 6 2 | 7 2 | 6 2 | 7 2 | 6 2 | 6 2 |
| | 8 2 | 9 2 | 8 2 | 9 2 | 5 2 | 5 2 |
| | 6 2 | 7 2 | 6 2 | 7 2 | 6 2 | 6 2 |
| | 8 2 | 9 2 | 8 2 | 9 2 | 5 2 | 5 2 |
| | 9 2 | 7 2 | 9 2 | 7 2 | 6 2 | 6 2 |
| | 7 2 | 9 2 | 7 2 | 9 2 | 5 2 | 5 2 |
| | 6 2 | 7 2 | 6 2 | 7 2 | 6 2 | 6 2 |
| | 7 2 | 6 2 | 7 2 | 6 2 | 5 2 | 5 2 |
| | 9 2 | 8 2 | 9 2 | 8 2 | 6 2 | 6 2 |
| | 7 2 | 7 2 | 7 2 | 7 2 | 5 2 | 5 2 |
| | 8 2 | 9 2 | 8 2 | 9 2 | 5 2 | 5 2 |
| | 6 2 | 7 2 | 6 2 | 7 2 | 6 2 | 6 2 |
| | 8 2 | 8 2 | 8 2 | 8 2 | 5 2 | 5 2 |
| | 9 2 | 6 2 | 9 2 | 6 2 | 6 2 | 6 2 |
| | 7 2 | 9 2 | 7 2 | 9 2 | 5 2 | 5 2 |
| | 6 2 | 7 2 | 6 2 | 7 2 | 6 2 | 6 2 |
| | 9 2 | 9 2 | 9 2 | 9 2 | 6 2 | 6 2 |
| | 7 2 | 7 2 | 7 2 | 7 2 | 5 2 | 5 2 |
| | 9 2 | 8 2 | 9 2 | 8 2 | 6 2 | 6 2 |
| | 8 2 | 9 2 | 8 2 | 9 2 | 6 2 | 6 2 |
| | 7 2 | 6 2 | 7 2 | 6 2 | 5 2 | 5 2 |
| | 6 2 | 9 2 | 6 2 | 9 2 | 5 2 | 5 2 |
| | 8 2 | 8 2 | 8 2 | 8 2 | 5 2 | 6 2 |
| | 9 2 | 7 2 | 9 2 | 7 2 | 6 2 | 5 2 |
| | 6 2 | 9 2 | 6 2 | 9 2 | 6 2 | 6 2 |
| | 9 2 | 6 2 | 9 2 | 6 2 | 5 2 | 5 2 |
| | 8 2 | 8 2 | 8 2 | 8 2 | 6 2 | 6 2 |
| | 7 2 | 7 2 | 7 2 | 7 2 | 5 2 | 5 2 |
| | 9 2 | 9 2 | 9 2 | 9 2 | 6 2 | 6 2 |
| | 7 2 | 9 2 | 7 2 | 9 2 | 5 2 | 5 2 |
| | 8 2 | 7 2 | 8 2 | 8 2 | 6 2 | 6 2 |
| | 6 2 | 8 2 | 6 2 | 9 2 | 5 2 | 5 2 |
| | 9 2 | 9 2 | 9 2 | 7 2 | 5 2 | 5 2 |
| | 8 2 | 6 2 | 8 2 | 6 2 | 5 2 | 6 2 |
| | 7 2 | 9 2 | 7 2 | 9 2 | 6 2 | 6 2 |
| | 6 2 | 7 2 | 6 2 | 7 2 | 6 2 | 5 2 |
| | 9 2 | 8 2 | 9 2 | 8 2 | 5 2 | 5 2 |
| | 9 2 | 9 2 | 8 2 | 9 2 | 6 2 | 6 2 |
| | 8 2 | 7 2 | 9 2 | 9 2 | 6 2 | 6 2 |
| | 8 2 | 6 2 | 8 2 | 6 2 | 5 2 | 5 2 |
| | 8 2 | 8 2 | 6 2 | 8 2 | 5 2 | 6 2 |
| | 6 2 | 6 2 | 7 2 | 6 2 | 6 2 | 6 2 |
| | 7 2 | 8 2 | 9 2 | 8 2 | 5 2 | 5 2 |
| | 9 2 | 7 2 | 6 2 | 7 2 | 6 2 | 6 2 |
| | 6 2 | 6 2 | 8 2 | 6 2 | 6 2 | 6 2 |
| | 7 2 | 9 2 | 7 2 | 7 2 | 5 2 | 5 2 |
| | 9 2 | 7 2 | 9 2 | 6 2 | 5 2 | 5 2 |
| | 8 2 | 6 2 | 8 2 | 8 2 | 6 2 | 6 2 |
| | 6 2 | 8 2 | 8 2 | 9 2 | 6 2 | 5 2 |
| | 8 2 | 8 2 | 6 2 | 8 2 | 6 2 | 6 2 |

Continued on next column

**TABLE LVI– continued**

| TQC | ICSD | NoE | SF | | | |
|---|---|---|---|---|---|---|
| | 7 2 | 6 2 | 6 2 | 6 2 | 5 2 | 5 2 |
| | 6 2 | 9 2 | 9 2 | 7 2 | 6 2 | 5 2 |
| | 7 2 | 7 2 | 7 2 | 6 2 | 6 2 | 6 2 |
| | 6 2 | 6 2 | 8 2 | 7 2 | 5 2 | 5 2 |
| | 7 2 | 8 2 | 7 2 | 9 2 | 6 2 | 6 2 |
| | 9 2 | 6 2 | 9 2 | 8 2 | 5 2 | 5 2 |
| | 8 2 | 9 2 | 8 2 | 6 2 | 6 2 | 6 2 |
| | 9 2 | 7 2 | 9 2 | 7 2 | 5 2 | 5 2 |
| | 8 2 | 8 2 | 6 2 | 9 2 | 5 2 | 5 2 |
| | 6 2 | 6 2 | 6 2 | 8 2 | 6 2 | 5 2 |
| | 7 2 | 7 2 | 7 2 | 6 2 | 5 2 | 5 2 |
| | 6 2 | 9 2 | 7 2 | 7 2 | 6 2 | 6 2 |
| | 7 2 | 9 2 | 7 2 | 7 2 | 5 2 | 6 2 |
| | 9 2 | 8 2 | 7 2 | 9 2 | 6 2 | 6 2 |
| | 8 2 | 9 2 | 6 2 | 6 2 | 6 2 | 5 2 |
| | 7 2 | 9 2 | 8 2 | 9 2 | 5 2 | 6 2 |
| | 7 2 | 8 2 | 9 2 | 8 2 | 5 2 | 6 2 |
| 123.2.16 | 84739 | 100 | Ba2Cu3O6Y1 | | | |
| | -A | -GM | -M | -Z | -R | -X |
| | 8 2 | 7 2 | 6 2 | 9 2 | 5 2 | 6 2 |
| | 6 2 | 7 2 | 6 2 | 7 2 | 6 2 | 6 2 |
| | 8 2 | 9 2 | 8 2 | 9 2 | 5 2 | 5 2 |
| | 6 2 | 9 2 | 8 2 | 7 2 | 6 2 | 5 2 |
| | 7 2 | 9 2 | 9 2 | 7 2 | 6 2 | 5 2 |
| | 6 2 | 8 2 | 8 2 | 6 2 | 6 2 | 5 2 |
| | 8 2 | 7 2 | 6 2 | 9 2 | 5 2 | 6 2 |
| | 9 2 | 9 2 | 7 2 | 7 2 | 5 2 | 6 2 |
| | 6 2 | 6 2 | 8 2 | 8 2 | 6 2 | 5 2 |
| | 7 2 | 8 2 | 9 2 | 6 2 | 6 2 | 5 2 |
| | 7 2 | 7 2 | 7 2 | 7 2 | 5 2 | 5 2 |
| | 9 2 | 9 2 | 9 2 | 9 2 | 6 2 | 6 2 |
| | 6 2 | 7 2 | 6 2 | 7 2 | 6 2 | 6 2 |
| | 8 2 | 9 2 | 8 2 | 9 2 | 5 2 | 5 2 |
| | 9 2 | 7 2 | 9 2 | 7 2 | 6 2 | 6 2 |
| | 7 2 | 9 2 | 7 2 | 9 2 | 5 2 | 5 2 |
| | 6 2 | 8 2 | 6 2 | 8 2 | 5 2 | 5 2 |
| | 8 2 | 6 2 | 8 2 | 6 2 | 6 2 | 6 2 |
| | 9 2 | 7 2 | 7 2 | 9 2 | 6 2 | 5 2 |
| | 7 2 | 6 2 | 9 2 | 8 2 | 5 2 | 6 2 |
| | 8 2 | 7 2 | 6 2 | 9 2 | 5 2 | 6 2 |
| | 8 2 | 9 2 | 6 2 | 7 2 | 5 2 | 5 2 |
| | 6 2 | 8 2 | 8 2 | 7 2 | 5 2 | 6 2 |
| | 6 2 | 7 2 | 8 2 | 6 2 | 5 2 | 6 2 |
| | 7 2 | 6 2 | 7 2 | 8 2 | 6 2 | 5 2 |
| | 6 2 | 9 2 | 8 2 | 9 2 | 6 2 | 5 2 |
| | 7 2 | 7 2 | 9 2 | 7 2 | 6 2 | 5 2 |
| | 9 2 | 6 2 | 7 2 | 7 2 | 6 2 | 5 2 |
| | 8 2 | 9 2 | 6 2 | 6 2 | 5 2 | 6 2 |
| | 9 2 | 6 2 | 9 2 | 8 2 | 5 2 | 5 2 |
| | 6 2 | 8 2 | 6 2 | 6 2 | 5 2 | 6 2 |
| | 7 2 | 7 2 | 7 2 | 9 2 | 6 2 | 5 2 |
| | 9 2 | 8 2 | 7 2 | 6 2 | 5 2 | 6 2 |
| | 8 2 | 9 2 | 8 2 | 7 2 | 6 2 | 6 2 |
| | 6 2 | 6 2 | 6 2 | 8 2 | 6 2 | 6 2 |
| | 9 2 | 8 2 | 9 2 | 6 2 | 6 2 | 5 2 |
| | 9 2 | 7 2 | 7 2 | 7 2 | 5 2 | 5 2 |
| | 7 2 | 7 2 | 6 2 | 9 2 | 5 2 | 5 2 |
| | 6 2 | 9 2 | 6 2 | 8 2 | 6 2 | 6 2 |
| | 8 2 | 6 2 | 9 2 | 6 2 | 5 2 | 5 2 |
| | 7 2 | 6 2 | 9 2 | 7 2 | 5 2 | 5 2 |
| | 7 2 | 8 2 | 8 2 | 8 2 | 5 2 | 6 2 |

Continued on next column

**TABLE LVI– continued**

| TQC | ICSD | NoE | SF | | | |
|---|---|---|---|---|---|---|
| | 6 2 | 7 2 | 9 2 | 7 2 | 6 2 | 6 2 |
| | 6 2 | 6 2 | 7 2 | 6 2 | 5 2 | 6 2 |
| | 9 2 | 6 2 | 8 2 | 9 2 | 5 2 | 5 2 |
| | 8 2 | 9 2 | 6 2 | 6 2 | 6 2 | 5 2 |
| | 7 2 | 8 2 | 7 2 | 8 2 | 6 2 | 5 2 |
| | 6 2 | 8 2 | 6 2 | 9 2 | 6 2 | 5 2 |
| | 7 2 | 9 2 | 7 2 | 6 2 | 5 2 | 5 2 |
| | 8 2 | 7 2 | 6 2 | 7 2 | 5 2 | 5 2 |
| 123.2.17 | 72254 | 100 | Ba2Cu3O6Y1 | | | |
| | -A | -GM | -M | -Z | -R | -X |
| | 8 2 | 7 2 | 6 2 | 9 2 | 5 2 | 6 2 |
| | 6 2 | 7 2 | 6 2 | 7 2 | 6 2 | 6 2 |
| | 8 2 | 9 2 | 8 2 | 9 2 | 5 2 | 5 2 |
| | 6 2 | 9 2 | 8 2 | 7 2 | 6 2 | 5 2 |
| | 7 2 | 9 2 | 9 2 | 7 2 | 6 2 | 5 2 |
| | 6 2 | 8 2 | 8 2 | 6 2 | 6 2 | 5 2 |
| | 8 2 | 7 2 | 6 2 | 9 2 | 5 2 | 6 2 |
| | 9 2 | 9 2 | 7 2 | 7 2 | 5 2 | 6 2 |
| | 6 2 | 6 2 | 8 2 | 8 2 | 6 2 | 5 2 |
| | 7 2 | 8 2 | 9 2 | 6 2 | 6 2 | 5 2 |
| | 7 2 | 7 2 | 7 2 | 7 2 | 5 2 | 5 2 |
| | 9 2 | 9 2 | 9 2 | 9 2 | 6 2 | 6 2 |
| | 6 2 | 7 2 | 6 2 | 7 2 | 6 2 | 6 2 |
| | 8 2 | 9 2 | 8 2 | 9 2 | 5 2 | 5 2 |
| | 9 2 | 7 2 | 9 2 | 7 2 | 6 2 | 6 2 |
| | 7 2 | 9 2 | 7 2 | 9 2 | 5 2 | 5 2 |
| | 6 2 | 8 2 | 6 2 | 8 2 | 5 2 | 5 2 |
| | 8 2 | 6 2 | 8 2 | 6 2 | 6 2 | 6 2 |
| | 9 2 | 7 2 | 7 2 | 9 2 | 6 2 | 5 2 |
| | 7 2 | 6 2 | 9 2 | 8 2 | 5 2 | 6 2 |
| | 8 2 | 7 2 | 6 2 | 9 2 | 5 2 | 6 2 |
| | 8 2 | 9 2 | 6 2 | 7 2 | 5 2 | 5 2 |
| | 6 2 | 8 2 | 8 2 | 7 2 | 5 2 | 6 2 |
| | 6 2 | 7 2 | 8 2 | 6 2 | 5 2 | 6 2 |
| | 7 2 | 6 2 | 7 2 | 8 2 | 6 2 | 5 2 |
| | 6 2 | 9 2 | 8 2 | 9 2 | 6 2 | 5 2 |
| | 7 2 | 7 2 | 9 2 | 7 2 | 6 2 | 5 2 |
| | 9 2 | 6 2 | 7 2 | 7 2 | 6 2 | 5 2 |
| | 8 2 | 9 2 | 6 2 | 6 2 | 5 2 | 6 2 |
| | 9 2 | 6 2 | 9 2 | 8 2 | 5 2 | 5 2 |
| | 6 2 | 8 2 | 6 2 | 6 2 | 5 2 | 6 2 |
| | 7 2 | 7 2 | 7 2 | 9 2 | 6 2 | 5 2 |
| | 9 2 | 8 2 | 7 2 | 6 2 | 5 2 | 6 2 |
| | 8 2 | 9 2 | 8 2 | 7 2 | 6 2 | 6 2 |
| | 6 2 | 6 2 | 6 2 | 8 2 | 6 2 | 6 2 |
| | 9 2 | 8 2 | 9 2 | 6 2 | 6 2 | 5 2 |
| | 9 2 | 7 2 | 7 2 | 7 2 | 5 2 | 5 2 |
| | 7 2 | 7 2 | 6 2 | 9 2 | 5 2 | 5 2 |
| | 6 2 | 9 2 | 6 2 | 8 2 | 6 2 | 6 2 |
| | 8 2 | 6 2 | 9 2 | 6 2 | 5 2 | 5 2 |
| | 7 2 | 6 2 | 9 2 | 7 2 | 5 2 | 5 2 |
| | 7 2 | 8 2 | 8 2 | 8 2 | 5 2 | 6 2 |
| | 6 2 | 7 2 | 9 2 | 7 2 | 6 2 | 6 2 |
| | 6 2 | 6 2 | 7 2 | 6 2 | 5 2 | 6 2 |
| | 9 2 | 6 2 | 8 2 | 9 2 | 5 2 | 5 2 |
| | 8 2 | 9 2 | 6 2 | 6 2 | 6 2 | 5 2 |
| | 7 2 | 8 2 | 7 2 | 8 2 | 6 2 | 5 2 |
| | 6 2 | 8 2 | 6 2 | 9 2 | 6 2 | 5 2 |
| | 7 2 | 9 2 | 7 2 | 6 2 | 5 2 | 5 2 |
| | 8 2 | 7 2 | 6 2 | 7 2 | 5 2 | 5 2 |
| 123.2.18 | 68290 | 100 | Ba2Cu3O6Y1 | | | |
| | -A | -GM | -M | -Z | -R | -X |
| | 8 2 | 7 2 | 6 2 | 9 2 | 5 2 | 6 2 |
| | 6 2 | 7 2 | 6 2 | 7 2 | 6 2 | 6 2 |
| | 8 2 | 9 2 | 8 2 | 9 2 | 5 2 | 5 2 |
| | 6 2 | 9 2 | 8 2 | 7 2 | 6 2 | 5 2 |
| | 7 2 | 9 2 | 9 2 | 7 2 | 6 2 | 5 2 |
| | 6 2 | 8 2 | 8 2 | 6 2 | 6 2 | 5 2 |
| | 8 2 | 7 2 | 6 2 | 9 2 | 5 2 | 6 2 |
| | 9 2 | 9 2 | 7 2 | 7 2 | 5 2 | 6 2 |
| | 6 2 | 6 2 | 8 2 | 8 2 | 6 2 | 5 2 |
| | 7 2 | 8 2 | 9 2 | 6 2 | 6 2 | 5 2 |
| | 7 2 | 7 2 | 7 2 | 7 2 | 5 2 | 5 2 |
| | 9 2 | 9 2 | 9 2 | 9 2 | 6 2 | 6 2 |
| | 6 2 | 7 2 | 6 2 | 7 2 | 6 2 | 6 2 |
| | 8 2 | 9 2 | 8 2 | 9 2 | 5 2 | 5 2 |
| | 9 2 | 7 2 | 9 2 | 7 2 | 6 2 | 6 2 |
| | 7 2 | 9 2 | 7 2 | 9 2 | 5 2 | 5 2 |
| | 6 2 | 8 2 | 6 2 | 8 2 | 5 2 | 5 2 |
| | 8 2 | 6 2 | 8 2 | 6 2 | 6 2 | 6 2 |
| | 9 2 | 7 2 | 7 2 | 9 2 | 6 2 | 5 2 |
| | 7 2 | 6 2 | 9 2 | 8 2 | 5 2 | 6 2 |
| | 8 2 | 7 2 | 6 2 | 9 2 | 5 2 | 6 2 |
| | 8 2 | 9 2 | 6 2 | 7 2 | 5 2 | 5 2 |
| | 6 2 | 8 2 | 8 2 | 7 2 | 5 2 | 6 2 |
| | 6 2 | 7 2 | 8 2 | 6 2 | 5 2 | 6 2 |
| | 7 2 | 6 2 | 7 2 | 8 2 | 6 2 | 5 2 |
| | 6 2 | 9 2 | 8 2 | 9 2 | 6 2 | 5 2 |
| | 7 2 | 7 2 | 9 2 | 7 2 | 6 2 | 5 2 |
| | 9 2 | 6 2 | 7 2 | 7 2 | 6 2 | 5 2 |
| | 8 2 | 9 2 | 6 2 | 6 2 | 5 2 | 6 2 |
| | 9 2 | 6 2 | 9 2 | 8 2 | 5 2 | 5 2 |
| | 6 2 | 8 2 | 6 2 | 6 2 | 5 2 | 6 2 |
| | 7 2 | 7 2 | 7 2 | 9 2 | 6 2 | 5 2 |
| | 9 2 | 8 2 | 7 2 | 6 2 | 5 2 | 6 2 |
| | 8 2 | 9 2 | 8 2 | 7 2 | 6 2 | 6 2 |
| | 6 2 | 6 2 | 6 2 | 8 2 | 6 2 | 6 2 |
| | 9 2 | 8 2 | 9 2 | 6 2 | 6 2 | 5 2 |
| | 9 2 | 7 2 | 7 2 | 7 2 | 5 2 | 5 2 |
| | 7 2 | 7 2 | 6 2 | 9 2 | 5 2 | 5 2 |
| | 6 2 | 9 2 | 6 2 | 8 2 | 6 2 | 6 2 |
| | 8 2 | 6 2 | 9 2 | 6 2 | 5 2 | 5 2 |
| | 7 2 | 6 2 | 9 2 | 7 2 | 5 2 | 5 2 |
| | 7 2 | 8 2 | 8 2 | 8 2 | 5 2 | 6 2 |
| | 6 2 | 7 2 | 9 2 | 7 2 | 6 2 | 6 2 |
| | 6 2 | 6 2 | 7 2 | 6 2 | 5 2 | 6 2 |
| | 9 2 | 6 2 | 8 2 | 9 2 | 5 2 | 5 2 |
| | 8 2 | 9 2 | 6 2 | 6 2 | 6 2 | 5 2 |
| | 7 2 | 8 2 | 7 2 | 8 2 | 6 2 | 5 2 |
| | 6 2 | 8 2 | 6 2 | 9 2 | 6 2 | 5 2 |
| | 7 2 | 9 2 | 7 2 | 6 2 | 5 2 | 5 2 |
| | 8 2 | 7 2 | 6 2 | 7 2 | 5 2 | 5 2 |
| 123.2.19 | 68046 | 100 | Ba2Cu3O6Y1 | | | |
| | -A | -GM | -M | -Z | -R | -X |
| | 8 2 | 7 2 | 6 2 | 9 2 | 5 2 | 6 2 |
| | 6 2 | 7 2 | 6 2 | 7 2 | 6 2 | 6 2 |
| | 8 2 | 9 2 | 8 2 | 9 2 | 5 2 | 5 2 |
| | 6 2 | 9 2 | 8 2 | 7 2 | 6 2 | 5 2 |
| | 7 2 | 9 2 | 9 2 | 7 2 | 6 2 | 5 2 |
| | 6 2 | 8 2 | 8 2 | 6 2 | 6 2 | 5 2 |
| | 8 2 | 7 2 | 6 2 | 9 2 | 5 2 | 6 2 |
| | 9 2 | 9 2 | 7 2 | 7 2 | 5 2 | 6 2 |

Continued on next column

Continued on next column

**TABLE LVI– continued**

| TQC | ICSD | NoE | SF | | | |
|---|---|---|---|---|---|---|
| | 6 2 | 6 2 | 8 2 | 8 2 | 6 2 | 5 2 |
| | 7 2 | 8 2 | 9 2 | 6 2 | 6 2 | 5 2 |
| | 7 2 | 7 2 | 7 2 | 7 2 | 5 2 | 5 2 |
| | 9 2 | 9 2 | 9 2 | 9 2 | 6 2 | 6 2 |
| | 6 2 | 7 2 | 6 2 | 7 2 | 6 2 | 6 2 |
| | 8 2 | 9 2 | 8 2 | 9 2 | 5 2 | 5 2 |
| | 9 2 | 7 2 | 9 2 | 7 2 | 6 2 | 6 2 |
| | 7 2 | 9 2 | 7 2 | 9 2 | 5 2 | 5 2 |
| | 6 2 | 8 2 | 6 2 | 8 2 | 5 2 | 5 2 |
| | 8 2 | 6 2 | 8 2 | 6 2 | 6 2 | 6 2 |
| | 9 2 | 7 2 | 7 2 | 9 2 | 6 2 | 5 2 |
| | 7 2 | 6 2 | 9 2 | 8 2 | 5 2 | 6 2 |
| | 8 2 | 7 2 | 6 2 | 9 2 | 5 2 | 6 2 |
| | 8 2 | 9 2 | 6 2 | 7 2 | 5 2 | 5 2 |
| | 6 2 | 8 2 | 8 2 | 7 2 | 5 2 | 6 2 |
| | 6 2 | 7 2 | 8 2 | 6 2 | 5 2 | 6 2 |
| | 7 2 | 6 2 | 7 2 | 8 2 | 6 2 | 5 2 |
| | 6 2 | 9 2 | 8 2 | 9 2 | 6 2 | 5 2 |
| | 7 2 | 7 2 | 9 2 | 7 2 | 6 2 | 5 2 |
| | 9 2 | 6 2 | 7 2 | 7 2 | 6 2 | 5 2 |
| | 8 2 | 9 2 | 6 2 | 6 2 | 5 2 | 6 2 |
| | 9 2 | 6 2 | 9 2 | 8 2 | 5 2 | 5 2 |
| | 6 2 | 8 2 | 6 2 | 6 2 | 5 2 | 6 2 |
| | 7 2 | 7 2 | 7 2 | 9 2 | 6 2 | 5 2 |
| | 9 2 | 8 2 | 7 2 | 6 2 | 5 2 | 6 2 |
| | 8 2 | 9 2 | 8 2 | 7 2 | 6 2 | 6 2 |
| | 6 2 | 6 2 | 6 2 | 8 2 | 6 2 | 6 2 |
| | 9 2 | 8 2 | 9 2 | 6 2 | 6 2 | 5 2 |
| | 9 2 | 7 2 | 7 2 | 7 2 | 5 2 | 5 2 |
| | 7 2 | 7 2 | 6 2 | 9 2 | 5 2 | 5 2 |
| | 6 2 | 9 2 | 6 2 | 8 2 | 6 2 | 6 2 |
| | 8 2 | 6 2 | 9 2 | 6 2 | 5 2 | 5 2 |
| | 7 2 | 6 2 | 9 2 | 7 2 | 5 2 | 5 2 |
| | 7 2 | 8 2 | 8 2 | 8 2 | 5 2 | 6 2 |
| | 6 2 | 7 2 | 9 2 | 7 2 | 6 2 | 6 2 |
| | 6 2 | 6 2 | 7 2 | 6 2 | 5 2 | 6 2 |
| | 9 2 | 6 2 | 8 2 | 9 2 | 5 2 | 5 2 |
| | 8 2 | 9 2 | 6 2 | 6 2 | 6 2 | 5 2 |
| | 7 2 | 8 2 | 7 2 | 8 2 | 6 2 | 5 2 |
| | 6 2 | 8 2 | 6 2 | 9 2 | 6 2 | 5 2 |
| | 7 2 | 9 2 | 7 2 | 6 2 | 5 2 | 5 2 |
| | 8 2 | 7 2 | 6 2 | 7 2 | 5 2 | 5 2 |
| 123.2.20 | 68008 | 100 | Ba2Cu3O6Y1 | | | |
| | -A | -GM | -M | -Z | -R | -X |
| | 8 2 | 7 2 | 6 2 | 9 2 | 5 2 | 6 2 |
| | 6 2 | 7 2 | 6 2 | 7 2 | 6 2 | 6 2 |
| | 8 2 | 9 2 | 8 2 | 9 2 | 5 2 | 5 2 |
| | 6 2 | 9 2 | 8 2 | 7 2 | 6 2 | 5 2 |
| | 7 2 | 9 2 | 9 2 | 7 2 | 6 2 | 5 2 |
| | 6 2 | 8 2 | 8 2 | 6 2 | 6 2 | 5 2 |
| | 8 2 | 7 2 | 6 2 | 9 2 | 5 2 | 6 2 |
| | 9 2 | 9 2 | 7 2 | 7 2 | 5 2 | 6 2 |
| | 6 2 | 6 2 | 8 2 | 8 2 | 6 2 | 5 2 |
| | 7 2 | 8 2 | 9 2 | 6 2 | 5 2 | 5 2 |
| | 7 2 | 7 2 | 7 2 | 7 2 | 6 2 | 5 2 |
| | 9 2 | 9 2 | 9 2 | 9 2 | 6 2 | 6 2 |
| | 6 2 | 7 2 | 6 2 | 7 2 | 6 2 | 6 2 |
| | 8 2 | 9 2 | 8 2 | 9 2 | 5 2 | 5 2 |
| | 9 2 | 7 2 | 9 2 | 7 2 | 6 2 | 6 2 |
| | 7 2 | 9 2 | 7 2 | 9 2 | 5 2 | 5 2 |
| | 6 2 | 8 2 | 6 2 | 8 2 | 5 2 | 5 2 |
| | 8 2 | 6 2 | 8 2 | 6 2 | 6 2 | 6 2 |
| | 9 2 | 7 2 | 7 2 | 9 2 | 6 2 | 5 2 |
| | 7 2 | 6 2 | 9 2 | 8 2 | 5 2 | 6 2 |
| | 8 2 | 7 2 | 6 2 | 9 2 | 5 2 | 6 2 |
| | 8 2 | 9 2 | 6 2 | 7 2 | 5 2 | 5 2 |
| | 6 2 | 7 2 | 8 2 | 7 2 | 5 2 | 6 2 |
| | 6 2 | 8 2 | 8 2 | 6 2 | 5 2 | 6 2 |
| | 7 2 | 6 2 | 7 2 | 8 2 | 6 2 | 5 2 |
| | 6 2 | 9 2 | 8 2 | 9 2 | 6 2 | 5 2 |
| | 7 2 | 7 2 | 9 2 | 7 2 | 6 2 | 5 2 |
| | 9 2 | 6 2 | 7 2 | 7 2 | 6 2 | 5 2 |
| | 8 2 | 9 2 | 6 2 | 6 2 | 5 2 | 6 2 |
| | 9 2 | 6 2 | 9 2 | 8 2 | 5 2 | 5 2 |
| | 6 2 | 8 2 | 6 2 | 6 2 | 5 2 | 6 2 |
| | 7 2 | 7 2 | 7 2 | 9 2 | 6 2 | 5 2 |
| | 9 2 | 8 2 | 7 2 | 6 2 | 5 2 | 6 2 |
| | 8 2 | 6 2 | 8 2 | 7 2 | 6 2 | 6 2 |
| | 6 2 | 9 2 | 6 2 | 8 2 | 6 2 | 6 2 |
| | 9 2 | 8 2 | 9 2 | 6 2 | 6 2 | 5 2 |
| | 9 2 | 7 2 | 7 2 | 7 2 | 5 2 | 5 2 |
| | 7 2 | 7 2 | 6 2 | 9 2 | 5 2 | 5 2 |
| | 6 2 | 9 2 | 6 2 | 8 2 | 6 2 | 6 2 |
| | 8 2 | 6 2 | 9 2 | 6 2 | 5 2 | 5 2 |
| | 7 2 | 6 2 | 9 2 | 7 2 | 5 2 | 5 2 |
| | 7 2 | 8 2 | 8 2 | 8 2 | 5 2 | 6 2 |
| | 6 2 | 7 2 | 9 2 | 7 2 | 6 2 | 6 2 |
| | 6 2 | 6 2 | 7 2 | 9 2 | 5 2 | 6 2 |
| | 9 2 | 6 2 | 8 2 | 6 2 | 5 2 | 5 2 |
| | 8 2 | 9 2 | 6 2 | 6 2 | 6 2 | 5 2 |
| | 7 2 | 8 2 | 7 2 | 8 2 | 6 2 | 5 2 |
| | 6 2 | 8 2 | 6 2 | 9 2 | 6 2 | 5 2 |
| | 7 2 | 9 2 | 7 2 | 6 2 | 5 2 | 5 2 |
| | 8 2 | 7 2 | 6 2 | 7 2 | 5 2 | 5 2 |
| 123.2.21 | 67016 | 100 | Ba2Cu3O6Y1 | | | |
| | -A | -GM | -M | -Z | -R | -X |
| | 8 2 | 7 2 | 6 2 | 9 2 | 5 2 | 6 2 |
| | 6 2 | 7 2 | 6 2 | 7 2 | 6 2 | 6 2 |
| | 8 2 | 9 2 | 8 2 | 9 2 | 5 2 | 5 2 |
| | 6 2 | 9 2 | 8 2 | 7 2 | 6 2 | 5 2 |
| | 7 2 | 9 2 | 9 2 | 7 2 | 6 2 | 5 2 |
| | 6 2 | 8 2 | 8 2 | 6 2 | 6 2 | 5 2 |
| | 8 2 | 7 2 | 6 2 | 9 2 | 5 2 | 6 2 |
| | 9 2 | 9 2 | 7 2 | 7 2 | 5 2 | 6 2 |
| | 6 2 | 6 2 | 8 2 | 8 2 | 6 2 | 5 2 |
| | 7 2 | 8 2 | 9 2 | 6 2 | 5 2 | 5 2 |
| | 7 2 | 7 2 | 7 2 | 7 2 | 6 2 | 5 2 |
| | 9 2 | 9 2 | 9 2 | 9 2 | 6 2 | 6 2 |
| | 6 2 | 7 2 | 6 2 | 7 2 | 6 2 | 6 2 |
| | 8 2 | 9 2 | 8 2 | 9 2 | 5 2 | 5 2 |
| | 9 2 | 7 2 | 9 2 | 7 2 | 6 2 | 6 2 |
| | 7 2 | 9 2 | 7 2 | 9 2 | 5 2 | 5 2 |
| | 6 2 | 8 2 | 6 2 | 8 2 | 5 2 | 5 2 |
| | 8 2 | 6 2 | 8 2 | 6 2 | 6 2 | 6 2 |
| | 9 2 | 7 2 | 7 2 | 9 2 | 6 2 | 5 2 |
| | 7 2 | 6 2 | 9 2 | 8 2 | 5 2 | 6 2 |
| | 8 2 | 7 2 | 6 2 | 9 2 | 5 2 | 6 2 |
| | 8 2 | 9 2 | 6 2 | 7 2 | 5 2 | 5 2 |
| | 6 2 | 7 2 | 8 2 | 7 2 | 5 2 | 6 2 |
| | 6 2 | 8 2 | 8 2 | 6 2 | 5 2 | 6 2 |
| | 7 2 | 6 2 | 7 2 | 8 2 | 6 2 | 5 2 |
| | 6 2 | 9 2 | 8 2 | 9 2 | 6 2 | 5 2 |

Continued on next column

Continued on next column

**TABLE LVI– continued**

| TQC | ICSD | | NoE | | SF | |
|---|---|---|---|---|---|---|
| | 7 2 | 7 2 | 9 2 | 7 2 | 6 2 | 5 2 |
| | 9 2 | 6 2 | 7 2 | 7 2 | 6 2 | 5 2 |
| | 8 2 | 9 2 | 6 2 | 6 2 | 5 2 | 6 2 |
| | 9 2 | 6 2 | 9 2 | 8 2 | 5 2 | 5 2 |
| | 6 2 | 8 2 | 6 2 | 6 2 | 5 2 | 6 2 |
| | 7 2 | 7 2 | 7 2 | 9 2 | 6 2 | 5 2 |
| | 9 2 | 8 2 | 7 2 | 6 2 | 5 2 | 6 2 |
| | 8 2 | 6 2 | 8 2 | 7 2 | 6 2 | 6 2 |
| | 6 2 | 9 2 | 6 2 | 8 2 | 6 2 | 6 2 |
| | 9 2 | 8 2 | 9 2 | 6 2 | 6 2 | 5 2 |
| | 9 2 | 7 2 | 7 2 | 7 2 | 5 2 | 5 2 |
| | 7 2 | 7 2 | 6 2 | 9 2 | 5 2 | 5 2 |
| | 6 2 | 6 2 | 6 2 | 8 2 | 6 2 | 6 2 |
| | 8 2 | 9 2 | 9 2 | 6 2 | 5 2 | 5 2 |
| | 7 2 | 6 2 | 9 2 | 7 2 | 5 2 | 5 2 |
| | 6 2 | 8 2 | 8 2 | 8 2 | 5 2 | 6 2 |
| | 7 2 | 7 2 | 9 2 | 6 2 | 6 2 | 6 2 |
| | 6 2 | 6 2 | 7 2 | 7 2 | 5 2 | 6 2 |
| | 9 2 | 6 2 | 8 2 | 9 2 | 5 2 | 5 2 |
| | 8 2 | 9 2 | 6 2 | 6 2 | 6 2 | 5 2 |
| | 7 2 | 8 2 | 7 2 | 8 2 | 6 2 | 5 2 |
| | 6 2 | 8 2 | 6 2 | 9 2 | 6 2 | 5 2 |
| | 7 2 | 9 2 | 7 2 | 6 2 | 5 2 | 5 2 |
| | 8 2 | 7 2 | 6 2 | 7 2 | 5 2 | 5 2 |
| 123.2.22 | 66903 | | 100 | Ba2Cu3O6Y1 | | |
| | -A | -GM | -M | -Z | -R | -X |
| | 8 2 | 7 2 | 6 2 | 9 2 | 5 2 | 6 2 |
| | 6 2 | 7 2 | 6 2 | 7 2 | 6 2 | 6 2 |
| | 8 2 | 9 2 | 8 2 | 9 2 | 5 2 | 5 2 |
| | 6 2 | 9 2 | 8 2 | 7 2 | 6 2 | 5 2 |
| | 7 2 | 9 2 | 9 2 | 7 2 | 6 2 | 5 2 |
| | 6 2 | 8 2 | 8 2 | 6 2 | 6 2 | 5 2 |
| | 8 2 | 7 2 | 6 2 | 9 2 | 5 2 | 6 2 |
| | 9 2 | 9 2 | 7 2 | 7 2 | 5 2 | 6 2 |
| | 6 2 | 6 2 | 8 2 | 8 2 | 6 2 | 5 2 |
| | 7 2 | 8 2 | 9 2 | 6 2 | 5 2 | 5 2 |
| | 7 2 | 7 2 | 7 2 | 7 2 | 6 2 | 5 2 |
| | 9 2 | 9 2 | 9 2 | 9 2 | 6 2 | 6 2 |
| | 6 2 | 7 2 | 6 2 | 7 2 | 6 2 | 6 2 |
| | 8 2 | 9 2 | 8 2 | 9 2 | 5 2 | 5 2 |
| | 9 2 | 7 2 | 9 2 | 7 2 | 6 2 | 6 2 |
| | 7 2 | 9 2 | 7 2 | 9 2 | 5 2 | 5 2 |
| | 6 2 | 8 2 | 6 2 | 8 2 | 5 2 | 5 2 |
| | 8 2 | 6 2 | 8 2 | 6 2 | 6 2 | 6 2 |
| | 9 2 | 7 2 | 7 2 | 9 2 | 6 2 | 5 2 |
| | 7 2 | 6 2 | 9 2 | 8 2 | 5 2 | 6 2 |
| | 8 2 | 7 2 | 6 2 | 9 2 | 5 2 | 6 2 |
| | 8 2 | 9 2 | 6 2 | 7 2 | 5 2 | 5 2 |
| | 6 2 | 7 2 | 8 2 | 7 2 | 5 2 | 6 2 |
| | 6 2 | 8 2 | 8 2 | 6 2 | 5 2 | 6 2 |
| | 7 2 | 6 2 | 7 2 | 8 2 | 6 2 | 5 2 |
| | 6 2 | 9 2 | 8 2 | 9 2 | 6 2 | 5 2 |
| | 7 2 | 7 2 | 9 2 | 7 2 | 6 2 | 5 2 |
| | 9 2 | 6 2 | 7 2 | 6 2 | 6 2 | 5 2 |
| | 8 2 | 9 2 | 6 2 | 7 2 | 5 2 | 6 2 |
| | 9 2 | 6 2 | 9 2 | 8 2 | 5 2 | 5 2 |
| | 6 2 | 8 2 | 6 2 | 6 2 | 5 2 | 6 2 |
| | 7 2 | 7 2 | 7 2 | 9 2 | 6 2 | 5 2 |
| | 9 2 | 8 2 | 7 2 | 6 2 | 5 2 | 6 2 |
| | 8 2 | 6 2 | 8 2 | 7 2 | 6 2 | 6 2 |
| | 6 2 | 9 2 | 9 2 | 8 2 | 6 2 | 6 2 |

Continued on next column

**TABLE LVI– continued**

| TQC | ICSD | | NoE | | SF | |
|---|---|---|---|---|---|---|
| | 9 2 | 8 2 | 6 2 | 6 2 | 6 2 | 5 2 |
| | 9 2 | 7 2 | 7 2 | 7 2 | 5 2 | 5 2 |
| | 7 2 | 7 2 | 6 2 | 9 2 | 5 2 | 5 2 |
| | 6 2 | 9 2 | 6 2 | 8 2 | 6 2 | 6 2 |
| | 8 2 | 6 2 | 9 2 | 6 2 | 5 2 | 5 2 |
| | 7 2 | 8 2 | 9 2 | 7 2 | 5 2 | 5 2 |
| | 7 2 | 6 2 | 8 2 | 8 2 | 5 2 | 6 2 |
| | 6 2 | 7 2 | 9 2 | 9 2 | 6 2 | 6 2 |
| | 6 2 | 6 2 | 7 2 | 7 2 | 5 2 | 6 2 |
| | 9 2 | 6 2 | 8 2 | 6 2 | 5 2 | 5 2 |
| | 8 2 | 9 2 | 6 2 | 6 2 | 6 2 | 5 2 |
| | 7 2 | 8 2 | 7 2 | 8 2 | 6 2 | 5 2 |
| | 6 2 | 8 2 | 6 2 | 9 2 | 6 2 | 5 2 |
| | 7 2 | 9 2 | 7 2 | 6 2 | 5 2 | 5 2 |
| | 8 2 | 7 2 | 6 2 | 7 2 | 5 2 | 5 2 |
| 123.2.23 | 63269 | | 100 | Ba2Cu3O6Y1 | | |
| | -A | -GM | -M | -Z | -R | -X |
| | 8 2 | 7 2 | 6 2 | 9 2 | 5 2 | 6 2 |
| | 6 2 | 7 2 | 6 2 | 7 2 | 6 2 | 6 2 |
| | 8 2 | 9 2 | 8 2 | 9 2 | 5 2 | 5 2 |
| | 6 2 | 9 2 | 8 2 | 7 2 | 6 2 | 5 2 |
| | 7 2 | 9 2 | 9 2 | 7 2 | 6 2 | 5 2 |
| | 6 2 | 8 2 | 8 2 | 6 2 | 6 2 | 5 2 |
| | 8 2 | 7 2 | 6 2 | 9 2 | 5 2 | 6 2 |
| | 9 2 | 9 2 | 7 2 | 7 2 | 5 2 | 6 2 |
| | 6 2 | 6 2 | 8 2 | 8 2 | 6 2 | 5 2 |
| | 7 2 | 8 2 | 9 2 | 6 2 | 5 2 | 5 2 |
| | 7 2 | 7 2 | 7 2 | 7 2 | 6 2 | 5 2 |
| | 9 2 | 9 2 | 9 2 | 9 2 | 6 2 | 6 2 |
| | 6 2 | 7 2 | 6 2 | 7 2 | 6 2 | 6 2 |
| | 8 2 | 9 2 | 8 2 | 9 2 | 5 2 | 5 2 |
| | 9 2 | 7 2 | 9 2 | 7 2 | 6 2 | 6 2 |
| | 7 2 | 9 2 | 7 2 | 9 2 | 5 2 | 5 2 |
| | 6 2 | 8 2 | 6 2 | 8 2 | 5 2 | 5 2 |
| | 8 2 | 6 2 | 8 2 | 6 2 | 6 2 | 6 2 |
| | 9 2 | 7 2 | 7 2 | 9 2 | 6 2 | 5 2 |
| | 7 2 | 6 2 | 9 2 | 8 2 | 5 2 | 6 2 |
| | 8 2 | 7 2 | 6 2 | 9 2 | 5 2 | 6 2 |
| | 8 2 | 9 2 | 6 2 | 7 2 | 5 2 | 5 2 |
| | 6 2 | 7 2 | 8 2 | 7 2 | 5 2 | 6 2 |
| | 6 2 | 8 2 | 8 2 | 6 2 | 5 2 | 6 2 |
| | 7 2 | 6 2 | 7 2 | 8 2 | 6 2 | 5 2 |
| | 6 2 | 9 2 | 8 2 | 9 2 | 6 2 | 5 2 |
| | 7 2 | 7 2 | 9 2 | 7 2 | 6 2 | 5 2 |
| | 9 2 | 6 2 | 7 2 | 7 2 | 6 2 | 5 2 |
| | 8 2 | 9 2 | 6 2 | 6 2 | 5 2 | 6 2 |
| | 9 2 | 6 2 | 9 2 | 8 2 | 5 2 | 5 2 |
| | 6 2 | 8 2 | 6 2 | 6 2 | 5 2 | 6 2 |
| | 7 2 | 7 2 | 7 2 | 9 2 | 6 2 | 5 2 |
| | 9 2 | 8 2 | 7 2 | 6 2 | 5 2 | 6 2 |
| | 8 2 | 6 2 | 8 2 | 7 2 | 6 2 | 6 2 |
| | 6 2 | 9 2 | 6 2 | 8 2 | 6 2 | 6 2 |
| | 9 2 | 8 2 | 9 2 | 6 2 | 6 2 | 5 2 |
| | 9 2 | 7 2 | 7 2 | 7 2 | 5 2 | 5 2 |
| | 7 2 | 7 2 | 6 2 | 9 2 | 5 2 | 5 2 |
| | 6 2 | 9 2 | 6 2 | 8 2 | 6 2 | 6 2 |
| | 8 2 | 6 2 | 9 2 | 6 2 | 5 2 | 5 2 |
| | 7 2 | 6 2 | 9 2 | 7 2 | 5 2 | 5 2 |
| | 7 2 | 8 2 | 8 2 | 8 2 | 5 2 | 6 2 |
| | 6 2 | 7 2 | 9 2 | 7 2 | 6 2 | 6 2 |
| | 6 2 | 6 2 | 7 2 | 9 2 | 5 2 | 6 2 |

Continued on next column

**TABLE LVI– continued**

| TQC | ICSD | NoE | SF | | | |
|---|---|---|---|---|---|---|
| | 9 2 | 6 2 | 8 2 | 6 2 | 5 2 | 5 2 |
| | 8 2 | 9 2 | 6 2 | 6 2 | 6 2 | 5 2 |
| | 7 2 | 8 2 | 7 2 | 8 2 | 6 2 | 5 2 |
| | 6 2 | 8 2 | 6 2 | 9 2 | 6 2 | 5 2 |
| | 7 2 | 9 2 | 7 2 | 6 2 | 5 2 | 5 2 |
| | 8 2 | 7 2 | 6 2 | 7 2 | 5 2 | 5 2 |
| 123.2.24 | 203116 | 100 | Ba2Cu3O6Y1 | | | |
| | -A | -GM | -M | -Z | -R | -X |
| | 8 2 | 7 2 | 6 2 | 9 2 | 5 2 | 6 2 |
| | 6 2 | 7 2 | 6 2 | 7 2 | 6 2 | 6 2 |
| | 8 2 | 9 2 | 8 2 | 9 2 | 5 2 | 5 2 |
| | 6 2 | 9 2 | 8 2 | 7 2 | 6 2 | 5 2 |
| | 7 2 | 9 2 | 9 2 | 7 2 | 6 2 | 5 2 |
| | 6 2 | 8 2 | 8 2 | 6 2 | 6 2 | 5 2 |
| | 8 2 | 7 2 | 6 2 | 9 2 | 5 2 | 6 2 |
| | 9 2 | 9 2 | 7 2 | 7 2 | 5 2 | 6 2 |
| | 6 2 | 6 2 | 8 2 | 8 2 | 6 2 | 5 2 |
| | 7 2 | 8 2 | 7 2 | 6 2 | 5 2 | 5 2 |
| | 7 2 | 7 2 | 9 2 | 7 2 | 6 2 | 5 2 |
| | 9 2 | 9 2 | 9 2 | 9 2 | 6 2 | 6 2 |
| | 6 2 | 7 2 | 6 2 | 7 2 | 6 2 | 6 2 |
| | 8 2 | 9 2 | 8 2 | 9 2 | 5 2 | 5 2 |
| | 9 2 | 7 2 | 9 2 | 7 2 | 6 2 | 6 2 |
| | 7 2 | 9 2 | 7 2 | 9 2 | 5 2 | 5 2 |
| | 6 2 | 8 2 | 6 2 | 8 2 | 5 2 | 5 2 |
| | 8 2 | 6 2 | 8 2 | 6 2 | 6 2 | 6 2 |
| | 9 2 | 7 2 | 7 2 | 9 2 | 6 2 | 5 2 |
| | 7 2 | 6 2 | 9 2 | 8 2 | 5 2 | 6 2 |
| | 8 2 | 7 2 | 6 2 | 7 2 | 5 2 | 5 2 |
| | 8 2 | 7 2 | 6 2 | 9 2 | 5 2 | 6 2 |
| | 6 2 | 9 2 | 8 2 | 7 2 | 5 2 | 6 2 |
| | 6 2 | 8 2 | 8 2 | 6 2 | 5 2 | 6 2 |
| | 7 2 | 6 2 | 7 2 | 8 2 | 6 2 | 5 2 |
| | 6 2 | 9 2 | 8 2 | 9 2 | 6 2 | 5 2 |
| | 7 2 | 7 2 | 9 2 | 7 2 | 6 2 | 5 2 |
| | 9 2 | 6 2 | 7 2 | 6 2 | 6 2 | 5 2 |
| | 8 2 | 9 2 | 6 2 | 7 2 | 5 2 | 5 2 |
| | 9 2 | 6 2 | 9 2 | 8 2 | 5 2 | 6 2 |
| | 6 2 | 8 2 | 6 2 | 6 2 | 5 2 | 6 2 |
| | 7 2 | 7 2 | 7 2 | 9 2 | 6 2 | 5 2 |
| | 9 2 | 8 2 | 7 2 | 6 2 | 5 2 | 6 2 |
| | 8 2 | 9 2 | 8 2 | 7 2 | 6 2 | 6 2 |
| | 6 2 | 6 2 | 6 2 | 8 2 | 6 2 | 6 2 |
| | 9 2 | 7 2 | 9 2 | 6 2 | 6 2 | 5 2 |
| | 9 2 | 8 2 | 7 2 | 7 2 | 5 2 | 5 2 |
| | 7 2 | 6 2 | 6 2 | 9 2 | 5 2 | 5 2 |
| | 6 2 | 7 2 | 6 2 | 8 2 | 6 2 | 6 2 |
| | 8 2 | 9 2 | 9 2 | 6 2 | 5 2 | 5 2 |
| | 7 2 | 8 2 | 9 2 | 7 2 | 5 2 | 5 2 |
| | 7 2 | 6 2 | 8 2 | 6 2 | 5 2 | 6 2 |
| | 6 2 | 7 2 | 9 2 | 8 2 | 6 2 | 6 2 |
| | 6 2 | 6 2 | 7 2 | 9 2 | 5 2 | 5 2 |
| | 9 2 | 6 2 | 8 2 | 7 2 | 5 2 | 6 2 |
| | 8 2 | 9 2 | 6 2 | 6 2 | 6 2 | 5 2 |
| | 7 2 | 8 2 | 7 2 | 8 2 | 6 2 | 5 2 |
| | 6 2 | 8 2 | 6 2 | 9 2 | 6 2 | 5 2 |
| | 7 2 | 9 2 | 7 2 | 6 2 | 5 2 | 5 2 |
| | 8 2 | 7 2 | 6 2 | 7 2 | 5 2 | 5 2 |
| 123.2.25 | 98114 | 114 | Ba2Cu3Lu1O6 | | | |
| | -A | -GM | -M | -Z | -R | -X |
| | 8 2 | 7 2 | 6 2 | 9 2 | 5 2 | 6 2 |

Continued on next column

**TABLE LVI– continued**

| TQC | ICSD | NoE | SF | | | |
|---|---|---|---|---|---|---|
| | 6 2 | 9 2 | 8 2 | 7 2 | 6 2 | 5 2 |
| | 6 2 | 7 2 | 6 2 | 7 2 | 6 2 | 6 2 |
| | 8 2 | 9 2 | 8 2 | 9 2 | 5 2 | 5 2 |
| | 7 2 | 9 2 | 9 2 | 7 2 | 6 2 | 5 2 |
| | 6 2 | 8 2 | 8 2 | 6 2 | 6 2 | 5 2 |
| | 8 2 | 7 2 | 6 2 | 9 2 | 5 2 | 6 2 |
| | 9 2 | 9 2 | 7 2 | 7 2 | 5 2 | 6 2 |
| | 6 2 | 6 2 | 8 2 | 8 2 | 6 2 | 5 2 |
| | 7 2 | 8 2 | 9 2 | 6 2 | 6 2 | 5 2 |
| | 7 2 | 7 2 | 7 2 | 7 2 | 5 2 | 5 2 |
| | 9 2 | 9 2 | 9 2 | 9 2 | 6 2 | 6 2 |
| | 6 2 | 7 2 | 6 2 | 7 2 | 6 2 | 6 2 |
| | 8 2 | 9 2 | 8 2 | 9 2 | 5 2 | 5 2 |
| | 9 2 | 7 2 | 9 2 | 7 2 | 6 2 | 6 2 |
| | 7 2 | 9 2 | 7 2 | 9 2 | 5 2 | 5 2 |
| | 6 2 | 8 2 | 6 2 | 8 2 | 5 2 | 5 2 |
| | 8 2 | 6 2 | 8 2 | 6 2 | 6 2 | 6 2 |
| | 9 2 | 7 2 | 7 2 | 9 2 | 6 2 | 5 2 |
| | 7 2 | 7 2 | 9 2 | 9 2 | 5 2 | 6 2 |
| | 8 2 | 6 2 | 6 2 | 8 2 | 5 2 | 6 2 |
| | 8 2 | 9 2 | 6 2 | 7 2 | 5 2 | 6 2 |
| | 6 2 | 8 2 | 8 2 | 6 2 | 5 2 | 5 2 |
| | 6 2 | 7 2 | 8 2 | 7 2 | 5 2 | 6 2 |
| | 7 2 | 8 2 | 7 2 | 6 2 | 6 2 | 5 2 |
| | 7 2 | 9 2 | 9 2 | 7 2 | 6 2 | 5 2 |
| | 6 2 | 8 2 | 8 2 | 6 2 | 6 2 | 5 2 |
| | 7 2 | 6 2 | 9 2 | 8 2 | 6 2 | 5 2 |
| | 6 2 | 9 2 | 8 2 | 9 2 | 6 2 | 5 2 |
| | 7 2 | 9 2 | 9 2 | 7 2 | 6 2 | 5 2 |
| | 9 2 | 8 2 | 7 2 | 6 2 | 5 2 | 6 2 |
| | 8 2 | 7 2 | 6 2 | 7 2 | 6 2 | 5 2 |
| | 9 2 | 8 2 | 9 2 | 6 2 | 5 2 | 5 2 |
| | 6 2 | 6 2 | 6 2 | 7 2 | 5 2 | 6 2 |
| | 7 2 | 9 2 | 7 2 | 6 2 | 6 2 | 5 2 |
| | 7 2 | 6 2 | 9 2 | 8 2 | 6 2 | 5 2 |
| | 6 2 | 8 2 | 8 2 | 6 2 | 5 2 | 6 2 |
| | 6 2 | 7 2 | 8 2 | 7 2 | 6 2 | 5 2 |
| | 7 2 | 9 2 | 9 2 | 9 2 | 6 2 | 6 2 |
| | 9 2 | 6 2 | 7 2 | 8 2 | 6 2 | 6 2 |
| | 8 2 | 9 2 | 8 2 | 7 2 | 6 2 | 5 2 |
| | 6 2 | 8 2 | 6 2 | 6 2 | 6 2 | 5 2 |
| | 9 2 | 7 2 | 9 2 | 6 2 | 6 2 | 5 2 |
| | 9 2 | 8 2 | 7 2 | 7 2 | 5 2 | 5 2 |
| | 7 2 | 7 2 | 6 2 | 9 2 | 5 2 | 5 2 |
| | 6 2 | 6 2 | 6 2 | 8 2 | 6 2 | 6 2 |
| | 8 2 | 9 2 | 9 2 | 6 2 | 5 2 | 5 2 |
| | 7 2 | 6 2 | 8 2 | 7 2 | 5 2 | 5 2 |
| | 6 2 | 7 2 | 8 2 | 8 2 | 5 2 | 6 2 |
| | 6 2 | 8 2 | 9 2 | 9 2 | 6 2 | 6 2 |
| | 7 2 | 6 2 | 7 2 | 7 2 | 5 2 | 6 2 |
| | 9 2 | 6 2 | 9 2 | 6 2 | 5 2 | 5 2 |
| | 8 2 | 9 2 | 6 2 | 6 2 | 6 2 | 5 2 |
| | 7 2 | 8 2 | 7 2 | 8 2 | 6 2 | 5 2 |
| | 6 2 | 8 2 | 6 2 | 9 2 | 6 2 | 5 2 |
| | 7 2 | 9 2 | 7 2 | 6 2 | 5 2 | 5 2 |
| | 8 2 | 7 2 | 6 2 | 7 2 | 5 2 | 5 2 |
| 123.2.26 | 98113 | 114 | Ba2Cu3Lu1O6 | | | |
| | -A | -GM | -M | -Z | -R | -X |
| | 8 2 | 7 2 | 6 2 | 9 2 | 5 2 | 6 2 |
| | 6 2 | 9 2 | 8 2 | 7 2 | 6 2 | 5 2 |
| | 6 2 | 7 2 | 6 2 | 7 2 | 6 2 | 6 2 |

Continued on next column

**TABLE LVI– continued**

| TQC | ICSD | NoE | SF | | | |
|---|---|---|---|---|---|---|
| | 8 2 | 9 2 | 8 2 | 9 2 | 5 2 | 5 2 |
| | 7 2 | 9 2 | 9 2 | 7 2 | 6 2 | 5 2 |
| | 6 2 | 8 2 | 8 2 | 6 2 | 6 2 | 5 2 |
| | 8 2 | 7 2 | 6 2 | 9 2 | 5 2 | 6 2 |
| | 9 2 | 9 2 | 7 2 | 7 2 | 5 2 | 6 2 |
| | 6 2 | 6 2 | 8 2 | 8 2 | 6 2 | 5 2 |
| | 7 2 | 8 2 | 9 2 | 6 2 | 6 2 | 5 2 |
| | 7 2 | 7 2 | 7 2 | 7 2 | 5 2 | 5 2 |
| | 9 2 | 9 2 | 9 2 | 9 2 | 6 2 | 6 2 |
| | 6 2 | 7 2 | 6 2 | 7 2 | 6 2 | 6 2 |
| | 8 2 | 9 2 | 8 2 | 9 2 | 5 2 | 5 2 |
| | 9 2 | 7 2 | 9 2 | 7 2 | 6 2 | 6 2 |
| | 7 2 | 9 2 | 7 2 | 9 2 | 5 2 | 5 2 |
| | 6 2 | 8 2 | 6 2 | 8 2 | 5 2 | 5 2 |
| | 8 2 | 6 2 | 8 2 | 6 2 | 6 2 | 6 2 |
| | 9 2 | 7 2 | 7 2 | 9 2 | 6 2 | 5 2 |
| | 7 2 | 7 2 | 9 2 | 8 2 | 5 2 | 6 2 |
| | 8 2 | 6 2 | 6 2 | 9 2 | 5 2 | 6 2 |
| | 8 2 | 9 2 | 6 2 | 7 2 | 5 2 | 6 2 |
| | 6 2 | 8 2 | 8 2 | 6 2 | 5 2 | 5 2 |
| | 6 2 | 7 2 | 8 2 | 7 2 | 5 2 | 6 2 |
| | 7 2 | 8 2 | 7 2 | 6 2 | 6 2 | 5 2 |
| | 7 2 | 9 2 | 9 2 | 7 2 | 6 2 | 5 2 |
| | 6 2 | 8 2 | 8 2 | 6 2 | 6 2 | 5 2 |
| | 7 2 | 6 2 | 9 2 | 8 2 | 6 2 | 5 2 |
| | 6 2 | 9 2 | 8 2 | 9 2 | 6 2 | 5 2 |
| | 7 2 | 9 2 | 9 2 | 7 2 | 6 2 | 5 2 |
| | 9 2 | 8 2 | 7 2 | 6 2 | 5 2 | 6 2 |
| | 8 2 | 8 2 | 6 2 | 6 2 | 6 2 | 5 2 |
| | 9 2 | 7 2 | 9 2 | 7 2 | 5 2 | 5 2 |
| | 6 2 | 9 2 | 6 2 | 7 2 | 5 2 | 6 2 |
| | 7 2 | 6 2 | 9 2 | 6 2 | 6 2 | 5 2 |
| | 7 2 | 6 2 | 7 2 | 8 2 | 6 2 | 5 2 |
| | 6 2 | 8 2 | 8 2 | 6 2 | 5 2 | 6 2 |
| | 6 2 | 7 2 | 8 2 | 7 2 | 6 2 | 5 2 |
| | 7 2 | 9 2 | 9 2 | 9 2 | 6 2 | 6 2 |
| | 9 2 | 6 2 | 7 2 | 8 2 | 6 2 | 5 2 |
| | 8 2 | 9 2 | 8 2 | 7 2 | 6 2 | 5 2 |
| | 6 2 | 8 2 | 6 2 | 6 2 | 6 2 | 6 2 |
| | 9 2 | 7 2 | 7 2 | 6 2 | 6 2 | 5 2 |
| | 9 2 | 8 2 | 9 2 | 7 2 | 5 2 | 5 2 |
| | 7 2 | 7 2 | 6 2 | 9 2 | 5 2 | 5 2 |
| | 6 2 | 6 2 | 6 2 | 8 2 | 6 2 | 6 2 |
| | 8 2 | 9 2 | 9 2 | 6 2 | 5 2 | 5 2 |
| | 7 2 | 6 2 | 8 2 | 7 2 | 5 2 | 5 2 |
| | 6 2 | 7 2 | 8 2 | 8 2 | 5 2 | 6 2 |
| | 6 2 | 8 2 | 9 2 | 7 2 | 6 2 | 6 2 |
| | 7 2 | 6 2 | 7 2 | 9 2 | 5 2 | 6 2 |
| | 9 2 | 6 2 | 9 2 | 6 2 | 5 2 | 5 2 |
| | 8 2 | 9 2 | 6 2 | 6 2 | 6 2 | 5 2 |
| | 7 2 | 8 2 | 7 2 | 8 2 | 6 2 | 5 2 |
| | 6 2 | 8 2 | 6 2 | 9 2 | 6 2 | 5 2 |
| | 7 2 | 9 2 | 7 2 | 6 2 | 5 2 | 5 2 |
| | 8 2 | 7 2 | 6 2 | 7 2 | 5 2 | 5 2 |
| 123.2.27 | 64639 | 106 | Ba2Cu2La1O8Ta1 | | | |
| | -A | -GM | -M | -Z | -R | -X |
| | 8 2 | 7 2 | 6 2 | 9 2 | 5 2 | 6 2 |
| | 6 2 | 7 2 | 6 2 | 7 2 | 6 2 | 6 2 |
| | 8 2 | 9 2 | 8 2 | 9 2 | 5 2 | 5 2 |
| | 9 2 | 7 2 | 9 2 | 7 2 | 5 2 | 5 2 |
| | 8 2 | 6 2 | 8 2 | 6 2 | 6 2 | 6 2 |
| | 6 2 | 9 2 | 8 2 | 7 2 | 6 2 | 5 2 |
| | 7 2 | 7 2 | 9 2 | 9 2 | 5 2 | 6 2 |
| | 7 2 | 9 2 | 7 2 | 7 2 | 6 2 | 5 2 |
| | 9 2 | 7 2 | 9 2 | 9 2 | 6 2 | 5 2 |
| | 8 2 | 6 2 | 6 2 | 8 2 | 5 2 | 6 2 |
| | 9 2 | 8 2 | 7 2 | 6 2 | 5 2 | 6 2 |
| | 6 2 | 9 2 | 8 2 | 7 2 | 6 2 | 5 2 |
| | 6 2 | 8 2 | 8 2 | 6 2 | 6 2 | 5 2 |
| | 7 2 | 9 2 | 9 2 | 7 2 | 6 2 | 5 2 |
| | 6 2 | 9 2 | 6 2 | 9 2 | 5 2 | 6 2 |
| | 8 2 | 7 2 | 8 2 | 7 2 | 6 2 | 5 2 |
| | 7 2 | 9 2 | 7 2 | 9 2 | 5 2 | 6 2 |
| | 9 2 | 7 2 | 6 2 | 7 2 | 6 2 | 5 2 |
| | 6 2 | 8 2 | 9 2 | 8 2 | 5 2 | 6 2 |
| | 8 2 | 6 2 | 8 2 | 6 2 | 6 2 | 5 2 |
| | 7 2 | 9 2 | 7 2 | 9 2 | 5 2 | 5 2 |
| | 6 2 | 7 2 | 6 2 | 7 2 | 5 2 | 5 2 |
| | 6 2 | 9 2 | 6 2 | 9 2 | 5 2 | 5 2 |
| | 9 2 | 8 2 | 7 2 | 8 2 | 6 2 | 6 2 |
| | 7 2 | 7 2 | 9 2 | 9 2 | 5 2 | 5 2 |
| | 6 2 | 6 2 | 6 2 | 8 2 | 5 2 | 5 2 |
| | 7 2 | 8 2 | 7 2 | 8 2 | 6 2 | 6 2 |
| | 7 2 | 9 2 | 7 2 | 9 2 | 6 2 | 6 2 |
| | 8 2 | 9 2 | 6 2 | 7 2 | 6 2 | 6 2 |
| | 6 2 | 8 2 | 8 2 | 6 2 | 5 2 | 6 2 |
| | 8 2 | 7 2 | 9 2 | 9 2 | 6 2 | 5 2 |
| | 9 2 | 9 2 | 6 2 | 9 2 | 5 2 | 5 2 |
| | 6 2 | 8 2 | 8 2 | 8 2 | 6 2 | 6 2 |
| | 8 2 | 7 2 | 6 2 | 7 2 | 5 2 | 6 2 |
| | 9 2 | 6 2 | 7 2 | 9 2 | 6 2 | 6 2 |
| | 6 2 | 9 2 | 8 2 | 6 2 | 5 2 | 6 2 |
| | 7 2 | 6 2 | 9 2 | 8 2 | 6 2 | 5 2 |
| | 7 2 | 8 2 | 7 2 | 8 2 | 6 2 | 5 2 |
| | 8 2 | 6 2 | 7 2 | 6 2 | 5 2 | 6 2 |
| | 9 2 | 8 2 | 8 2 | 8 2 | 5 2 | 6 2 |
| | 6 2 | 9 2 | 6 2 | 7 2 | 6 2 | 5 2 |
| | 7 2 | 8 2 | 7 2 | 6 2 | 5 2 | 6 2 |
| | 9 2 | 7 2 | 9 2 | 9 2 | 6 2 | 5 2 |
| | 8 2 | 6 2 | 6 2 | 7 2 | 6 2 | 5 2 |
| | 9 2 | 6 2 | 7 2 | 8 2 | 5 2 | 5 2 |
| | 7 2 | 9 2 | 9 2 | 6 2 | 5 2 | 6 2 |
| | 7 2 | 8 2 | 7 2 | 8 2 | 6 2 | 6 2 |
| | 8 2 | 8 2 | 6 2 | 6 2 | 5 2 | 6 2 |
| | 9 2 | 7 2 | 8 2 | 7 2 | 5 2 | 5 2 |
| | 6 2 | 6 2 | 9 2 | 6 2 | 6 2 | 6 2 |
| | 6 2 | 7 2 | 8 2 | 9 2 | 6 2 | 5 2 |
| | 7 2 | 9 2 | 9 2 | 7 2 | 6 2 | 5 2 |
| | 8 2 | 8 2 | 6 2 | 8 2 | 6 2 | 5 2 |
| 123.2.28 | 165630 | 30 | Ba1Mg4Si3 | | | |
| | -A | -GM | -M | -Z | -R | -X |
| | 7 2 | 7 2 | 7 2 | 7 2 | 5 2 | 5 2 |
| | 9 2 | 9 2 | 9 2 | 9 2 | 6 2 | 6 2 |
| | 8 2 | 9 2 | 8 2 | 9 2 | 6 2 | 6 2 |
| | 9 2 | 8 2 | 9 2 | 8 2 | 6 2 | 6 2 |
| | 6 2 | 7 2 | 6 2 | 7 2 | 6 2 | 6 2 |
| | 9 2 | 7 2 | 7 2 | 9 2 | 6 2 | 5 2 |
| | 8 2 | 9 2 | 8 2 | 9 2 | 5 2 | 5 2 |
| | 7 2 | 7 2 | 9 2 | 7 2 | 5 2 | 6 2 |
| | 6 2 | 9 2 | 8 2 | 7 2 | 5 2 | 6 2 |
| | 9 2 | 9 2 | 6 2 | 6 2 | 6 2 | 5 2 |
| | 7 2 | 8 2 | 6 2 | 9 2 | 5 2 | 5 2 |

Continued on next column

**TABLE LVI– continued**

| TQC | ICSD | NoE | SF | | |
|---|---|---|---|---|---|
| | 8 2 | 7 2 | 7 2 8 2 | 6 2 | 6 2 |
| | 6 2 | 6 2 | 9 2 7 2 | 5 2 | 6 2 |
| | 7 2 | 9 2 | 9 2 6 2 | 5 2 | 6 2 |
| | 6 2 | 6 2 | 8 2 7 2 | 6 2 | 6 2 |
| 123.2.29 | 169074 | 34 | As2Ba1O1Ti2 | | |
| | -A | -GM | -M -Z | -R | -X |
| | 8 2 | 7 2 | 6 2 9 2 | 5 2 | 6 2 |
| | 6 2 | 7 2 | 6 2 7 2 | 6 2 | 6 2 |
| | 6 2 | 9 2 | 8 2 7 2 | 6 2 | 5 2 |
| | 7 2 | 9 2 | 9 2 7 2 | 6 2 | 5 2 |
| | 6 2 | 8 2 | 8 2 6 2 | 6 2 | 5 2 |
| | 9 2 | 7 2 | 7 2 7 2 | 5 2 | 5 2 |
| | 7 2 | 9 2 | 9 2 9 2 | 6 2 | 6 2 |
| | 8 2 | 8 2 | 8 2 8 2 | 5 2 | 5 2 |
| | 9 2 | 9 2 | 9 2 9 2 | 5 2 | 5 2 |
| | 8 2 | 7 2 | 8 2 9 2 | 5 2 | 5 2 |
| | 9 2 | 9 2 | 7 2 7 2 | 6 2 | 6 2 |
| | 8 2 | 7 2 | 9 2 7 2 | 5 2 | 5 2 |
| | 9 2 | 6 2 | 8 2 6 2 | 5 2 | 5 2 |
| | 7 2 | 6 2 | 7 2 9 2 | 5 2 | 6 2 |
| | 6 2 | 7 2 | 6 2 6 2 | 6 2 | 5 2 |
| | 7 2 | 9 2 | 9 2 9 2 | 6 2 | 6 2 |
| | 8 2 | 8 2 | 8 2 8 2 | 6 2 | 6 2 |
| 123.2.30 | 174342 | 36 | Al2B1La1Ru2 | | |
| | -A | -GM | -M -Z | -R | -X |
| | 9 2 | 7 2 | 7 2 9 2 | 6 2 | 5 2 |
| | 7 2 | 9 2 | 9 2 7 2 | 5 2 | 6 2 |
| | 6 2 | 9 2 | 8 2 7 2 | 5 2 | 6 2 |
| | 7 2 | 8 2 | 9 2 6 2 | 5 2 | 6 2 |
| | 8 2 | 7 2 | 7 2 7 2 | 5 2 | 5 2 |
| | 8 2 | 7 2 | 6 2 7 2 | 6 2 | 6 2 |
| | 9 2 | 7 2 | 8 2 9 2 | 5 2 | 5 2 |
| | 7 2 | 6 2 | 9 2 6 2 | 6 2 | 6 2 |
| | 9 2 | 6 2 | 8 2 6 2 | 6 2 | 5 2 |
| | 6 2 | 7 2 | 9 2 7 2 | 6 2 | 6 2 |
| | 9 2 | 9 2 | 9 2 7 2 | 6 2 | 6 2 |
| | 8 2 | 6 2 | 8 2 9 2 | 5 2 | 6 2 |
| | 9 2 | 7 2 | 6 2 8 2 | 5 2 | 5 2 |
| | 8 2 | 8 2 | 9 2 9 2 | 6 2 | 6 2 |
| | 8 2 | 9 2 | 8 2 6 2 | 5 2 | 5 2 |
| | 9 2 | 6 2 | 7 2 7 2 | 6 2 | 5 2 |
| | 8 2 | 7 2 | 6 2 6 2 | 5 2 | 6 2 |
| | 7 2 | 6 2 | 9 2 6 2 | 5 2 | 5 2 |
| 123.2.31 | 251834 | 36 | Al2B1La1Os2 | | |
| | -A | -GM | -M -Z | -R | -X |
| | 9 2 | 7 2 | 7 2 9 2 | 6 2 | 5 2 |
| | 7 2 | 9 2 | 9 2 7 2 | 5 2 | 6 2 |
| | 6 2 | 9 2 | 8 2 7 2 | 5 2 | 6 2 |
| | 7 2 | 8 2 | 9 2 6 2 | 5 2 | 6 2 |
| | 8 2 | 7 2 | 8 2 7 2 | 5 2 | 5 2 |
| | 9 2 | 7 2 | 9 2 7 2 | 6 2 | 6 2 |
| | 8 2 | 7 2 | 7 2 9 2 | 5 2 | 6 2 |
| | 7 2 | 6 2 | 6 2 6 2 | 6 2 | 5 2 |
| | 9 2 | 6 2 | 8 2 6 2 | 6 2 | 5 2 |
| | 6 2 | 7 2 | 9 2 7 2 | 6 2 | 6 2 |
| | 9 2 | 9 2 | 9 2 7 2 | 5 2 | 6 2 |
| | 8 2 | 6 2 | 8 2 9 2 | 6 2 | 6 2 |
| | 9 2 | 8 2 | 6 2 8 2 | 5 2 | 5 2 |
| | 8 2 | 7 2 | 9 2 9 2 | 6 2 | 6 2 |
| | 8 2 | 9 2 | 8 2 6 2 | 5 2 | 5 2 |
| | 7 2 | 6 2 | 7 2 7 2 | 6 2 | 5 2 |

Continued on next column

**TABLE LVI– continued**

| TQC | ICSD | NoE | SF | | |
|---|---|---|---|---|---|
| | 9 2 | 7 2 | 6 2 6 2 | 5 2 | 6 2 |
| | 6 2 | 6 2 | 9 2 8 2 | 5 2 | 5 2 |

The End

TABLE LVII: The computed irreducible representations for the valence bands for the compounds of SEBR and NLC in space group 125. See the list of tables.

| TQC | ICSD | NoE | SF | | |
|---|---|---|---|---|---|
| 125.1.1 | 412794 | 92 | Pr1Rb1Te4 | | |
| | -A | -GM | -M -Z | -R | -X |
| | 5 4 | 7 2 | 5 4 9 2 | 3+4 4 | 3+4 4 |
| | 5 4 | 9 2 | 5 4 7 2 | 3+4 4 | 3+4 4 |
| | 5 4 | 7 2 | 5 4 7 2 | 3+4 4 | 3+4 4 |
| | 5 4 | 9 2 | 5 4 9 2 | 3+4 4 | 3+4 4 |
| | 5 4 | 7 2 | 5 4 9 2 | 3+4 4 | 3+4 4 |
| | 5 4 | 9 2 | 5 4 7 2 | 3+4 4 | 3+4 4 |
| | 5 4 | 9 2 | 5 4 7 2 | 3+4 4 | 3+4 4 |
| | 5 4 | 6 2 | 5 4 8 2 | 3+4 4 | 3+4 4 |
| | 5 4 | 7 2 | 5 4 9 2 | 3+4 4 | 3+4 4 |
| | 5 4 | 8 2 | 5 4 6 2 | 3+4 4 | 3+4 4 |
| | 5 4 | 7 2 | 5 4 9 2 | 3+4 4 | 3+4 4 |
| | 5 4 | 9 2 | 5 4 7 2 | 3+4 4 | 3+4 4 |
| | 5 4 | 7 2 | 5 4 7 2 | 3+4 4 | 3+4 4 |
| | 5 4 | 9 2 | 5 4 9 2 | 3+4 4 | 3+4 4 |
| | 5 4 | 8 2 | 5 4 6 2 | 3+4 4 | 3+4 4 |
| | 5 4 | 6 2 | 5 4 8 2 | 3+4 4 | 3+4 4 |
| | 5 4 | 6 2 | 5 4 6 2 | 3+4 4 | 3+4 4 |
| | 5 4 | 7 2 | 5 4 7 2 | 3+4 4 | 3+4 4 |
| | 5 4 | 8 2 | 5 4 8 2 | 3+4 4 | 3+4 4 |
| | 5 4 | 9 2 | 5 4 9 2 | 3+4 4 | 3+4 4 |
| | 5 4 | 9 2 | 5 4 9 2 | 3+4 4 | 3+4 4 |
| | 5 4 | 8 2 | 5 4 7 2 | 3+4 4 | 3+4 4 |
| | 5 4 | 7 2 | 5 4 8 2 | 3+4 4 | 3+4 4 |
| | | 6 2 | 6 2 | | |
| | | 6 2 | 8 2 | | |
| | | 7 2 | 6 2 | | |
| | | 8 2 | 9 2 | | |
| | | 9 2 | 7 2 | | |
| | | 7 2 | 8 2 | | |
| | | 6 2 | 9 2 | | |
| | | 8 2 | 6 2 | | |
| | | 7 2 | 9 2 | | |
| | | 9 2 | 7 2 | | |
| | | 9 2 | 7 2 | | |
| | | 8 2 | 6 2 | | |
| | | 7 2 | 9 2 | | |
| | | 9 2 | 7 2 | | |
| | | 6 2 | 8 2 | | |
| | | 7 2 | 9 2 | | |
| | | 6 2 | 8 2 | | |
| | | 8 2 | 6 2 | | |
| | | 9 2 | 7 2 | | |
| | | 6 2 | 8 2 | | |
| | | 8 2 | 6 2 | | |
| | | 6 2 | 9 2 | | |
| | | 7 2 | 8 2 | | |
| 125.1.2 | 391204 | 92 | K1Pr1Te4 | | |
| | -A | -GM | -M -Z | -R | -X |
| | 5 4 | 7 2 | 5 4 9 2 | 3+4 4 | 3+4 4 |

Continued on next column

**TABLE LVII– continued**

| TQC | ICSD | | NoE | | SF | |
|---|---|---|---|---|---|---|
| | 5 4 | 9 2 | 5 4 | 7 2 | 3+4 4 | 3+4 4 |
| | 5 4 | 7 2 | 5 4 | 7 2 | 3+4 4 | 3+4 4 |
| | 5 4 | 9 2 | 5 4 | 9 2 | 3+4 4 | 3+4 4 |
| | 5 4 | 7 2 | 5 4 | 9 2 | 3+4 4 | 3+4 4 |
| | 5 4 | 9 2 | 5 4 | 7 2 | 3+4 4 | 3+4 4 |
| | 5 4 | 9 2 | 5 4 | 7 2 | 3+4 4 | 3+4 4 |
| | 5 4 | 6 2 | 5 4 | 8 2 | 3+4 4 | 3+4 4 |
| | 5 4 | 7 2 | 5 4 | 9 2 | 3+4 4 | 3+4 4 |
| | 5 4 | 8 2 | 5 4 | 6 2 | 3+4 4 | 3+4 4 |
| | 5 4 | 7 2 | 5 4 | 7 2 | 3+4 4 | 3+4 4 |
| | 5 4 | 9 2 | 5 4 | 9 2 | 3+4 4 | 3+4 4 |
| | 5 4 | 9 2 | 5 4 | 9 2 | 3+4 4 | 3+4 4 |
| | 5 4 | 6 2 | 5 4 | 6 2 | 3+4 4 | 3+4 4 |
| | 5 4 | 7 2 | 5 4 | 7 2 | 3+4 4 | 3+4 4 |
| | 5 4 | 8 2 | 5 4 | 8 2 | 3+4 4 | 3+4 4 |
| | 5 4 | 7 2 | 5 4 | 9 2 | 3+4 4 | 3+4 4 |
| | 5 4 | 9 2 | 5 4 | 7 2 | 3+4 4 | 3+4 4 |
| | 5 4 | 8 2 | 5 4 | 6 2 | 3+4 4 | 3+4 4 |
| | 5 4 | 6 2 | 5 4 | 8 2 | 3+4 4 | 3+4 4 |
| | 5 4 | 8 2 | 5 4 | 6 2 | 3+4 4 | 3+4 4 |
| | 5 4 | 9 2 | 5 4 | 7 2 | 3+4 4 | 3+4 4 |
| | 5 4 | 7 2 | 5 4 | 9 2 | 3+4 4 | 3+4 4 |
| | | 6 2 | | 8 2 | | |
| | | 6 2 | | 8 2 | | |
| | | 7 2 | | 6 2 | | |
| | | 8 2 | | 9 2 | | |
| | | 9 2 | | 7 2 | | |
| | | 7 2 | | 8 2 | | |
| | | 6 2 | | 9 2 | | |
| | | 8 2 | | 6 2 | | |
| | | 7 2 | | 9 2 | | |
| | | 9 2 | | 7 2 | | |
| | | 9 2 | | 7 2 | | |
| | | 8 2 | | 6 2 | | |
| | | 7 2 | | 9 2 | | |
| | | 9 2 | | 7 2 | | |
| | | 6 2 | | 8 2 | | |
| | | 7 2 | | 9 2 | | |
| | | 6 2 | | 8 2 | | |
| | | 8 2 | | 6 2 | | |
| | | 9 2 | | 7 2 | | |
| | | 6 2 | | 8 2 | | |
| | | 8 2 | | 6 2 | | |
| | | 7 2 | | 9 2 | | |
| | | 6 2 | | 8 2 | | |
| 125.1.3 | 152562 | | 60 | | Ce1Ga6 | |
| | -A | -GM | -M | -Z | -R | -X |
| | 5 4 | 7 2 | 5 4 | 7 2 | 3+4 4 | 3+4 4 |
| | 5 4 | 8 2 | 5 4 | 8 2 | 3+4 4 | 3+4 4 |
| | 5 4 | 6 2 | 5 4 | 6 2 | 3+4 4 | 3+4 4 |
| | 5 4 | 9 2 | 5 4 | 9 2 | 3+4 4 | 3+4 4 |
| | 5 4 | 7 2 | 5 4 | 7 2 | 3+4 4 | 3+4 4 |
| | 5 4 | 6 2 | 5 4 | 6 2 | 3+4 4 | 3+4 4 |
| | 5 4 | 9 2 | 5 4 | 9 2 | 3+4 4 | 3+4 4 |
| | 5 4 | 8 2 | 5 4 | 8 2 | 3+4 4 | 3+4 4 |
| | 5 4 | 7 2 | 5 4 | 9 2 | 3+4 4 | 3+4 4 |
| | 5 4 | 9 2 | 5 4 | 7 2 | 3+4 4 | 3+4 4 |
| | 5 4 | 9 2 | 5 4 | 9 2 | 3+4 4 | 3+4 4 |
| | 5 4 | 7 2 | 5 4 | 9 2 | 3+4 4 | 3+4 4 |
| | 5 4 | 6 2 | 5 4 | 8 2 | 3+4 4 | 3+4 4 |
| | 5 4 | 7 2 | 5 4 | 7 2 | 3+4 4 | 3+4 4 |

Continued on next column

**TABLE LVII– continued**

| TQC | ICSD | | NoE | | SF | |
|---|---|---|---|---|---|---|
| | 5 4 | 7 2 | 5 4 | 7 2 | 3+4 4 | 3+4 4 |
| | | 9 2 | | 6 2 | | |
| | | 8 2 | | 7 2 | | |
| | | 8 2 | | 6 2 | | |
| | | 6 2 | | 9 2 | | |
| | | 9 2 | | 8 2 | | |
| | | 9 2 | | 8 2 | | |
| | | 8 2 | | 6 2 | | |
| | | 6 2 | | 9 2 | | |
| | | 7 2 | | 7 2 | | |
| | | 6 2 | | 9 2 | | |
| | | 7 2 | | 8 2 | | |
| | | 6 2 | | 6 2 | | |
| | | 7 2 | | 9 2 | | |
| | | 8 2 | | 8 2 | | |
| | | 9 2 | | 7 2 | | |
| 125.1.4 | 102167 | | 60 | | Ce1Ga6 | |
| | -A | -GM | -M | -Z | -R | -X |
| | 5 4 | 7 2 | 5 4 | 7 2 | 3+4 4 | 3+4 4 |
| | 5 4 | 8 2 | 5 4 | 8 2 | 3+4 4 | 3+4 4 |
| | 5 4 | 6 2 | 5 4 | 6 2 | 3+4 4 | 3+4 4 |
| | 5 4 | 9 2 | 5 4 | 9 2 | 3+4 4 | 3+4 4 |
| | 5 4 | 7 2 | 5 4 | 7 2 | 3+4 4 | 3+4 4 |
| | 5 4 | 6 2 | 5 4 | 6 2 | 3+4 4 | 3+4 4 |
| | 5 4 | 9 2 | 5 4 | 9 2 | 3+4 4 | 3+4 4 |
| | 5 4 | 8 2 | 5 4 | 8 2 | 3+4 4 | 3+4 4 |
| | 5 4 | 7 2 | 5 4 | 9 2 | 3+4 4 | 3+4 4 |
| | 5 4 | 9 2 | 5 4 | 7 2 | 3+4 4 | 3+4 4 |
| | 5 4 | 9 2 | 5 4 | 9 2 | 3+4 4 | 3+4 4 |
| | 5 4 | 7 2 | 5 4 | 9 2 | 3+4 4 | 3+4 4 |
| | 5 4 | 6 2 | 5 4 | 8 2 | 3+4 4 | 3+4 4 |
| | 5 4 | 7 2 | 5 4 | 7 2 | 3+4 4 | 3+4 4 |
| | 5 4 | 7 2 | 5 4 | 7 2 | 3+4 4 | 3+4 4 |
| | | 9 2 | | 6 2 | | |
| | | 8 2 | | 7 2 | | |
| | | 8 2 | | 9 2 | | |
| | | 6 2 | | 6 2 | | |
| | | 9 2 | | 8 2 | | |
| | | 8 2 | | 8 2 | | |
| | | 9 2 | | 6 2 | | |
| | | 6 2 | | 9 2 | | |
| | | 7 2 | | 7 2 | | |
| | | 7 2 | | 9 2 | | |
| | | 6 2 | | 8 2 | | |
| | | 6 2 | | 6 2 | | |
| | | 7 2 | | 9 2 | | |
| | | 8 2 | | 8 2 | | |
| | | 9 2 | | 7 2 | | |
| 125.2.1 | 400461 | | 80 | | H8I1N4P1 | |
| | -A | -GM | -M | -Z | -R | -X |
| | 5 4 | 7 2 | 5 4 | 9 2 | 3+4 4 | 3+4 4 |
| | 5 4 | 8 2 | 5 4 | 6 2 | 3+4 4 | 3+4 4 |
| | 5 4 | 6 2 | 5 4 | 8 2 | 3+4 4 | 3+4 4 |
| | 5 4 | 7 2 | 5 4 | 9 2 | 3+4 4 | 3+4 4 |
| | 5 4 | 9 2 | 5 4 | 7 2 | 3+4 4 | 3+4 4 |
| | 5 4 | 8 2 | 5 4 | 6 2 | 3+4 4 | 3+4 4 |
| | 5 4 | 6 2 | 5 4 | 8 2 | 3+4 4 | 3+4 4 |
| | 5 4 | 9 2 | 5 4 | 7 2 | 3+4 4 | 3+4 4 |
| | 5 4 | 7 2 | 5 4 | 7 2 | 3+4 4 | 3+4 4 |
| | 5 4 | 9 2 | 5 4 | 9 2 | 3+4 4 | 3+4 4 |
| | 5 4 | 6 2 | 5 4 | 8 2 | 3+4 4 | 3+4 4 |

Continued on next column

**TABLE LVII– continued**

| TQC | ICSD | | NoE | | SF | |
|---|---|---|---|---|---|---|
| | 5 4 | 9 2 | 5 4 | 7 2 | 3+4 4 | 3+4 4 |
| | 5 4 | 9 2 | 5 4 | 7 2 | 3+4 4 | 3+4 4 |
| | 5 4 | 8 2 | 5 4 | 6 2 | 3+4 4 | 3+4 4 |
| | 5 4 | 7 2 | 5 4 | 9 2 | 3+4 4 | 3+4 4 |
| | 5 4 | 6 2 | 5 4 | 8 2 | 3+4 4 | 3+4 4 |
| | 5 4 | 7 2 | 5 4 | 9 2 | 3+4 4 | 3+4 4 |
| | 5 4 | 8 2 | 5 4 | 6 2 | 3+4 4 | 3+4 4 |
| | 5 4 | 9 2 | 5 4 | 7 2 | 3+4 4 | 3+4 4 |
| | 5 4 | 6 2 | 5 4 | 8 2 | 3+4 4 | 3+4 4 |
| | | 7 2 | | 9 2 | | |
| | | 6 2 | | 6 2 | | |
| | | 8 2 | | 8 2 | | |
| | | 7 2 | | 9 2 | | |
| | | 9 2 | | 7 2 | | |
| | | 8 2 | | 7 2 | | |
| | | 9 2 | | 6 2 | | |
| | | 7 2 | | 8 2 | | |
| | | 8 2 | | 9 2 | | |
| | | 6 2 | | 7 2 | | |
| | | 9 2 | | 9 2 | | |
| | | 7 2 | | 6 2 | | |
| | | 8 2 | | 6 2 | | |
| | | 6 2 | | 8 2 | | |
| | | 7 2 | | 7 2 | | |
| | | 9 2 | | 9 2 | | |
| | | 8 2 | | 6 2 | | |
| | | 7 2 | | 9 2 | | |
| | | 9 2 | | 7 2 | | |
| | | 6 2 | | 6 2 | | |

The End

TABLE LVIII: The computed irreducible representations for the valence bands for the compounds of SEBR and NLC in space group 127. See the list of tables.

| TQC | ICSD | | NoE | | SF | |
|---|---|---|---|---|---|---|
| 127.2.1 | 105627 | | 64 | | Pb3Sr2 | |
| | -A | -GM | -M | -Z | -R | -X |
| | 8+9 4 | 7 2 | 8+9 4 | 7 2 | 3+4 4 | 3+4 4 |
| | 6+7 4 | 6 2 | 6+7 4 | 6 2 | 3+4 4 | 3+4 4 |
| | 8+9 4 | 8 2 | 8+9 4 | 8 2 | 3+4 4 | 3+4 4 |
| | 6+7 4 | 9 2 | 6+7 4 | 9 2 | 3+4 4 | 3+4 4 |
| | 6+7 4 | 6 2 | 8+9 4 | 6 2 | 3+4 4 | 3+4 4 |
| | 8+9 4 | 8 2 | 6+7 4 | 8 2 | 3+4 4 | 3+4 4 |
| | 8+9 4 | 7 2 | 8+9 4 | 7 2 | 3+4 4 | 3+4 4 |
| | 6+7 4 | 9 2 | 6+7 4 | 9 2 | 3+4 4 | 3+4 4 |
| | 6+7 4 | 6 2 | 8+9 4 | 6 2 | 3+4 4 | 3+4 4 |
| | 8+9 4 | 9 2 | 6+7 4 | 9 2 | 3+4 4 | 3+4 4 |
| | 8+9 4 | 7 2 | 6+7 4 | 7 2 | 3+4 4 | 3+4 4 |
| | 8+9 4 | 8 2 | 8+9 4 | 8 2 | 3+4 4 | 3+4 4 |
| | 6+7 4 | 7 2 | 8+9 4 | 9 2 | 3+4 4 | 3+4 4 |
| | 6+7 4 | 9 2 | 6+7 4 | 7 2 | 3+4 4 | 3+4 4 |
| | 6+7 4 | 6 2 | 8+9 4 | 6 2 | 3+4 4 | 3+4 4 |
| | 6+7 4 | 8 2 | 6+7 4 | 8 2 | 3+4 4 | 3+4 4 |
| | | 7 2 | | 9 2 | | |
| | | 6 2 | | 8 2 | | |
| | | 7 2 | | 9 2 | | |
| | | 9 2 | | 7 2 | | |
| | | 8 2 | | 6 2 | | |
| | | 7 2 | | 9 2 | | |

Continued on next column

**TABLE LVIII– continued**

| TQC | ICSD | | NoE | | SF | |
|---|---|---|---|---|---|---|
| | | 9 2 | | 7 2 | | |
| | | 8 2 | | 7 2 | | |
| | | 6 2 | | 6 2 | | |
| | | 8 2 | | 8 2 | | |
| | | 9 2 | | 9 2 | | |
| | | 7 2 | | 6 2 | | |
| | | 9 2 | | 7 2 | | |
| | | 6 2 | | 8 2 | | |
| | | 7 2 | | 7 2 | | |
| | | 9 2 | | 9 2 | | |
| 127.2.2 | 427282 | | 124 | | Ni2Sn1Tb2 | |
| | -A | -GM | -M | -Z | -R | -X |
| | 6+7 4 | 7 2 | 8+9 4 | 9 2 | 3+4 4 | 3+4 4 |
| | 8+9 4 | 6 2 | 6+7 4 | 6 2 | 3+4 4 | 3+4 4 |
| | 6+7 4 | 8 2 | 8+9 4 | 7 2 | 3+4 4 | 3+4 4 |
| | 8+9 4 | 9 2 | 6+7 4 | 8 2 | 3+4 4 | 3+4 4 |
| | 8+9 4 | 8 2 | 8+9 4 | 6 2 | 3+4 4 | 3+4 4 |
| | 6+7 4 | 6 2 | 6+7 4 | 9 2 | 3+4 4 | 3+4 4 |
| | 6+7 4 | 7 2 | 8+9 4 | 8 2 | 3+4 4 | 3+4 4 |
| | 8+9 4 | 9 2 | 6+7 4 | 7 2 | 3+4 4 | 3+4 4 |
| | 6+7 4 | 6 2 | 6+7 4 | 8 2 | 3+4 4 | 3+4 4 |
| | 8+9 4 | 9 2 | 8+9 4 | 9 2 | 3+4 4 | 3+4 4 |
| | 8+9 4 | 7 2 | 6+7 4 | 7 2 | 3+4 4 | 3+4 4 |
| | 6+7 4 | 8 2 | 8+9 4 | 6 2 | 3+4 4 | 3+4 4 |
| | 8+9 4 | 7 2 | 6+7 4 | 7 2 | 3+4 4 | 3+4 4 |
| | 8+9 4 | 6 2 | 8+9 4 | 9 2 | 3+4 4 | 3+4 4 |
| | 6+7 4 | 9 2 | 8+9 4 | 8 2 | 3+4 4 | 3+4 4 |
| | 6+7 4 | 8 2 | 6+7 4 | 6 2 | 3+4 4 | 3+4 4 |
| | 8+9 4 | 7 2 | 8+9 4 | 7 2 | 3+4 4 | 3+4 4 |
| | 8+9 4 | 7 2 | 6+7 4 | 7 2 | 3+4 4 | 3+4 4 |
| | 6+7 4 | 9 2 | 6+7 4 | 9 2 | 3+4 4 | 3+4 4 |
| | 6+7 4 | 8 2 | 8+9 4 | 6 2 | 3+4 4 | 3+4 4 |
| | 8+9 4 | 7 2 | 6+7 4 | 9 2 | 3+4 4 | 3+4 4 |
| | 6+7 4 | 6 2 | 8+9 4 | 8 2 | 3+4 4 | 3+4 4 |
| | 8+9 4 | 7 2 | 8+9 4 | 8 2 | 3+4 4 | 3+4 4 |
| | 6+7 4 | 9 2 | 6+7 4 | 9 2 | 3+4 4 | 3+4 4 |
| | 8+9 4 | 8 2 | 8+9 4 | 7 2 | 3+4 4 | 3+4 4 |
| | 6+7 4 | 6 2 | 6+7 4 | 9 2 | 3+4 4 | 3+4 4 |
| | 8+9 4 | 9 2 | 6+7 4 | 8 2 | 3+4 4 | 3+4 4 |
| | 6+7 4 | 8 2 | 8+9 4 | 6 2 | 3+4 4 | 3+4 4 |
| | 8+9 4 | 9 2 | 8+9 4 | 7 2 | 3+4 4 | 3+4 4 |
| | 6+7 4 | 6 2 | 6+7 4 | 9 2 | 3+4 4 | 3+4 4 |
| | 8+9 4 | 9 2 | 8+9 4 | 8 2 | 3+4 4 | 3+4 4 |
| | | 8 2 | | 6 2 | | |
| | | 9 2 | | 7 2 | | |
| | | 8 2 | | 6 2 | | |
| | | 7 2 | | 9 2 | | |
| | | 6 2 | | 8 2 | | |
| | | 8 2 | | 7 2 | | |
| | | 6 2 | | 6 2 | | |
| | | 6 2 | | 9 2 | | |
| | | 7 2 | | 6 2 | | |
| | | 7 2 | | 9 2 | | |
| | | 8 2 | | 8 2 | | |
| | | 9 2 | | 7 2 | | |
| | | 7 2 | | 8 2 | | |
| | | 6 2 | | 9 2 | | |
| | | 7 2 | | 7 2 | | |
| | | 7 2 | | 6 2 | | |
| | | 9 2 | | 6 2 | | |
| | | 9 2 | | 8 2 | | |

Continued on next column

**TABLE LVIII– continued**

| TQC | ICSD | | NoE | | SF | |
|---|---|---|---|---|---|---|
| | | 8 2 | | 6 2 | | |
| | | 6 2 | | 9 2 | | |
| | | 6 2 | | 9 2 | | |
| | | 7 2 | | 8 2 | | |
| | | 8 2 | | 7 2 | | |
| | | 9 2 | | 9 2 | | |
| | | 8 2 | | 7 2 | | |
| | | 6 2 | | 7 2 | | |
| | | 7 2 | | 8 2 | | |
| | | 9 2 | | 9 2 | | |
| | | 6 2 | | 7 2 | | |
| | | 9 2 | | 6 2 | | |
| | | 7 2 | | 6 2 | | |
| 127.2.3 | 391441 | | 116 | | Mg1Si2Yb2 | |
| | -A | -GM | -M | -Z | -R | -X |
| | 6+7 4 | 7 2 | 8+9 4 | 9 2 | 3+4 4 | 3+4 4 |
| | 8+9 4 | 6 2 | 6+7 4 | 6 2 | 3+4 4 | 3+4 4 |
| | 6+7 4 | 8 2 | 8+9 4 | 7 2 | 3+4 4 | 3+4 4 |
| | 8+9 4 | 9 2 | 6+7 4 | 8 2 | 3+4 4 | 3+4 4 |
| | 8+9 4 | 8 2 | 8+9 4 | 6 2 | 3+4 4 | 3+4 4 |
| | 6+7 4 | 6 2 | 6+7 4 | 8 2 | 3+4 4 | 3+4 4 |
| | 6+7 4 | 7 2 | 8+9 4 | 9 2 | 3+4 4 | 3+4 4 |
| | 8+9 4 | 9 2 | 6+7 4 | 7 2 | 3+4 4 | 3+4 4 |
| | 8+9 4 | 6 2 | 8+9 4 | 8 2 | 3+4 4 | 3+4 4 |
| | 6+7 4 | 9 2 | 6+7 4 | 7 2 | 3+4 4 | 3+4 4 |
| | 6+7 4 | 7 2 | 6+7 4 | 9 2 | 3+4 4 | 3+4 4 |
| | 8+9 4 | 8 2 | 8+9 4 | 6 2 | 3+4 4 | 3+4 4 |
| | 6+7 4 | 7 2 | 6+7 4 | 7 2 | 3+4 4 | 3+4 4 |
| | 8+9 4 | 9 2 | 8+9 4 | 9 2 | 3+4 4 | 3+4 4 |
| | 8+9 4 | 6 2 | 6+7 4 | 8 2 | 3+4 4 | 3+4 4 |
| | 6+7 4 | 8 2 | 8+9 4 | 6 2 | 3+4 4 | 3+4 4 |
| | 8+9 4 | 7 2 | 6+7 4 | 7 2 | 3+4 4 | 3+4 4 |
| | 6+7 4 | 6 2 | 8+9 4 | 6 2 | 3+4 4 | 3+4 4 |
| | 8+9 4 | 8 2 | 6+7 4 | 8 2 | 3+4 4 | 3+4 4 |
| | 6+7 4 | 9 2 | 8+9 4 | 9 2 | 3+4 4 | 3+4 4 |
| | 8+9 4 | 7 2 | 6+7 4 | 9 2 | 3+4 4 | 3+4 4 |
| | 8+9 4 | 7 2 | 8+9 4 | 7 2 | 3+4 4 | 3+4 4 |
| | 6+7 4 | 8 2 | 6+7 4 | 7 2 | 3+4 4 | 3+4 4 |
| | 8+9 4 | 6 2 | 8+9 4 | 8 2 | 3+4 4 | 3+4 4 |
| | 6+7 4 | 9 2 | 6+7 4 | 9 2 | 3+4 4 | 3+4 4 |
| | 6+7 4 | 8 2 | 8+9 4 | 6 2 | 3+4 4 | 3+4 4 |
| | 8+9 4 | 9 2 | 6+7 4 | 8 2 | 3+4 4 | 3+4 4 |
| | 6+7 4 | 7 2 | 6+7 4 | 6 2 | 3+4 4 | 3+4 4 |
| | 8+9 4 | 6 2 | 8+9 4 | 7 2 | 3+4 4 | 3+4 4 |
| | | 8 2 | | 6 2 | | |
| | | 8 2 | | 9 2 | | |
| | | 7 2 | | 7 2 | | |
| | | 9 2 | | 8 2 | | |
| | | 6 2 | | 7 2 | | |
| | | 6 2 | | 9 2 | | |
| | | 7 2 | | 8 2 | | |
| | | 9 2 | | 7 2 | | |
| | | 8 2 | | 9 2 | | |
| | | 9 2 | | 8 2 | | |
| | | 6 2 | | 6 2 | | |
| | | 7 2 | | 6 2 | | |
| | | 7 2 | | 9 2 | | |
| | | 6 2 | | 6 2 | | |
| | | 9 2 | | 9 2 | | |
| | | 8 2 | | 7 2 | | |
| | | 8 2 | | 6 2 | | |

Continued on next column

**TABLE LVIII– continued**

| TQC | ICSD | | NoE | | SF | |
|---|---|---|---|---|---|---|
| | | 6 2 | | 8 2 | | |
| | | 7 2 | | 9 2 | | |
| | | 9 2 | | 6 2 | | |
| | | 7 2 | | 7 2 | | |
| | | 8 2 | | 8 2 | | |
| | | 6 2 | | 7 2 | | |
| | | 7 2 | | 8 2 | | |
| | | 9 2 | | 7 2 | | |
| | | 8 2 | | 8 2 | | |
| | | 9 2 | | 9 2 | | |
| | | 6 2 | | 9 2 | | |
| | | 7 2 | | 6 2 | | |
| 127.2.4 | 26200 | | 44 | | Li2Sn5 | |
| | -A | -GM | -M | -Z | -R | -X |
| | 8+9 4 | 7 2 | 8+9 4 | 7 2 | 3+4 4 | 3+4 4 |
| | 6+7 4 | 6 2 | 6+7 4 | 6 2 | 3+4 4 | 3+4 4 |
| | 8+9 4 | 7 2 | 8+9 4 | 7 2 | 3+4 4 | 3+4 4 |
| | 6+7 4 | 9 2 | 6+7 4 | 9 2 | 3+4 4 | 3+4 4 |
| | 8+9 4 | 8 2 | 8+9 4 | 8 2 | 3+4 4 | 3+4 4 |
| | 6+7 4 | 8 2 | 6+7 4 | 8 2 | 3+4 4 | 3+4 4 |
| | 8+9 4 | 9 2 | 6+7 4 | 9 2 | 3+4 4 | 3+4 4 |
| | 6+7 4 | 7 2 | 8+9 4 | 7 2 | 3+4 4 | 3+4 4 |
| | 6+7 4 | 6 2 | 6+7 4 | 9 2 | 3+4 4 | 3+4 4 |
| | 6+7 4 | 6 2 | 8+9 4 | 8 2 | 3+4 4 | 3+4 4 |
| | 8+9 4 | 8 2 | 8+9 4 | 6 2 | 3+4 4 | 3+4 4 |
| | | 9 2 | | 9 2 | | |
| | | 6 2 | | 7 2 | | |
| | | 7 2 | | 6 2 | | |
| | | 8 2 | | 6 2 | | |
| | | 6 2 | | 9 2 | | |
| | | 9 2 | | 7 2 | | |
| | | 9 2 | | 6 2 | | |
| | | 8 2 | | 8 2 | | |
| | | 7 2 | | 8 2 | | |
| | | 9 2 | | 9 2 | | |
| | | 8 2 | | 8 2 | | |
| 127.2.5 | 415117 | | 100 | | Dy2Mg1Si2 | |
| | -A | -GM | -M | -Z | -R | -X |
| | 6+7 4 | 7 2 | 8+9 4 | 9 2 | 3+4 4 | 3+4 4 |
| | 8+9 4 | 8 2 | 6+7 4 | 6 2 | 3+4 4 | 3+4 4 |
| | 6+7 4 | 9 2 | 8+9 4 | 7 2 | 3+4 4 | 3+4 4 |
| | 8+9 4 | 6 2 | 6+7 4 | 8 2 | 3+4 4 | 3+4 4 |
| | 8+9 4 | 8 2 | 8+9 4 | 6 2 | 3+4 4 | 3+4 4 |
| | 6+7 4 | 6 2 | 6+7 4 | 8 2 | 3+4 4 | 3+4 4 |
| | 6+7 4 | 7 2 | 8+9 4 | 9 2 | 3+4 4 | 3+4 4 |
| | 8+9 4 | 9 2 | 6+7 4 | 7 2 | 3+4 4 | 3+4 4 |
| | 8+9 4 | 6 2 | 8+9 4 | 8 2 | 3+4 4 | 3+4 4 |
| | 6+7 4 | 9 2 | 6+7 4 | 7 2 | 3+4 4 | 3+4 4 |
| | 6+7 4 | 7 2 | 6+7 4 | 9 2 | 3+4 4 | 3+4 4 |
| | 6+7 4 | 8 2 | 8+9 4 | 6 2 | 3+4 4 | 3+4 4 |
| | 8+9 4 | 7 2 | 6+7 4 | 7 2 | 3+4 4 | 3+4 4 |
| | 8+9 4 | 9 2 | 8+9 4 | 9 2 | 3+4 4 | 3+4 4 |
| | 8+9 4 | 6 2 | 8+9 4 | 8 2 | 3+4 4 | 3+4 4 |
| | 8+9 4 | 8 2 | 6+7 4 | 6 2 | 3+4 4 | 3+4 4 |
| | 6+7 4 | 7 2 | 6+7 4 | 6 2 | 3+4 4 | 3+4 4 |
| | 6+7 4 | 6 2 | 8+9 4 | 7 2 | 3+4 4 | 3+4 4 |
| | 8+9 4 | 8 2 | 6+7 4 | 8 2 | 3+4 4 | 3+4 4 |
| | 6+7 4 | 9 2 | 8+9 4 | 9 2 | 3+4 4 | 3+4 4 |
| | 8+9 4 | 7 2 | 6+7 4 | 9 2 | 3+4 4 | 3+4 4 |
| | 8+9 4 | 7 2 | 8+9 4 | 7 2 | 3+4 4 | 3+4 4 |
| | 6+7 4 | 8 2 | 6+7 4 | 7 2 | 3+4 4 | 3+4 4 |

Continued on next column

**TABLE LVIII– continued**

| TQC | ICSD | | NoE | | SF | |
|---|---|---|---|---|---|---|
| | 8+9 4 | 6 2 | 6+7 4 | 8 2 | 3+4 4 | 3+4 4 |
| | 6+7 4 | 9 2 | 8+9 4 | 9 2 | 3+4 4 | 3+4 4 |
| | | 8 2 | | 6 2 | | |
| | | 9 2 | | 6 2 | | |
| | | 7 2 | | 7 2 | | |
| | | 6 2 | | 8 2 | | |
| | | 8 2 | | 6 2 | | |
| | | 7 2 | | 9 2 | | |
| | | 8 2 | | 8 2 | | |
| | | 6 2 | | 7 2 | | |
| | | 9 2 | | 7 2 | | |
| | | 9 2 | | 8 2 | | |
| | | 6 2 | | 6 2 | | |
| | | 7 2 | | 9 2 | | |
| | | 8 2 | | 7 2 | | |
| | | 7 2 | | 9 2 | | |
| | | 9 2 | | 8 2 | | |
| | | 6 2 | | 6 2 | | |
| | | 7 2 | | 9 2 | | |
| | | 6 2 | | 6 2 | | |
| | | 9 2 | | 9 2 | | |
| | | 8 2 | | 6 2 | | |
| | | 8 2 | | 7 2 | | |
| | | 7 2 | | 8 2 | | |
| | | 7 2 | | 9 2 | | |
| | | 6 2 | | 7 2 | | |
| | | 9 2 | | 8 2 | | |
| 127.2.6 | 423452 | | 100 | | Dy2Ge2Mg1 | |
| | -A | -GM | -M | -Z | -R | -X |
| | 6+7 4 | 7 2 | 8+9 4 | 9 2 | 3+4 4 | 3+4 4 |
| | 8+9 4 | 6 2 | 6+7 4 | 6 2 | 3+4 4 | 3+4 4 |
| | 6+7 4 | 8 2 | 8+9 4 | 7 2 | 3+4 4 | 3+4 4 |
| | 8+9 4 | 9 2 | 6+7 4 | 8 2 | 3+4 4 | 3+4 4 |
| | 8+9 4 | 8 2 | 8+9 4 | 6 2 | 3+4 4 | 3+4 4 |
| | 6+7 4 | 6 2 | 6+7 4 | 8 2 | 3+4 4 | 3+4 4 |
| | 6+7 4 | 7 2 | 8+9 4 | 9 2 | 3+4 4 | 3+4 4 |
| | 8+9 4 | 9 2 | 6+7 4 | 7 2 | 3+4 4 | 3+4 4 |
| | 8+9 4 | 6 2 | 8+9 4 | 8 2 | 3+4 4 | 3+4 4 |
| | 6+7 4 | 9 2 | 6+7 4 | 7 2 | 3+4 4 | 3+4 4 |
| | 6+7 4 | 7 2 | 6+7 4 | 9 2 | 3+4 4 | 3+4 4 |
| | 6+7 4 | 8 2 | 8+9 4 | 6 2 | 3+4 4 | 3+4 4 |
| | 8+9 4 | 7 2 | 6+7 4 | 7 2 | 3+4 4 | 3+4 4 |
| | 8+9 4 | 9 2 | 8+9 4 | 9 2 | 3+4 4 | 3+4 4 |
| | 8+9 4 | 6 2 | 8+9 4 | 8 2 | 3+4 4 | 3+4 4 |
| | 8+9 4 | 8 2 | 6+7 4 | 6 2 | 3+4 4 | 3+4 4 |
| | 6+7 4 | 7 2 | 6+7 4 | 6 2 | 3+4 4 | 3+4 4 |
| | 6+7 4 | 6 2 | 8+9 4 | 7 2 | 3+4 4 | 3+4 4 |
| | 8+9 4 | 8 2 | 6+7 4 | 8 2 | 3+4 4 | 3+4 4 |
| | 6+7 4 | 9 2 | 8+9 4 | 9 2 | 3+4 4 | 3+4 4 |
| | 8+9 4 | 7 2 | 6+7 4 | 9 2 | 3+4 4 | 3+4 4 |
| | 8+9 4 | 7 2 | 8+9 4 | 7 2 | 3+4 4 | 3+4 4 |
| | 6+7 4 | 6 2 | 6+7 4 | 7 2 | 3+4 4 | 3+4 4 |
| | 8+9 4 | 8 2 | 6+7 4 | 8 2 | 3+4 4 | 3+4 4 |
| | 6+7 4 | 9 2 | 8+9 4 | 9 2 | 3+4 4 | 3+4 4 |
| | | 8 2 | | 6 2 | | |
| | | 9 2 | | 6 2 | | |
| | | 7 2 | | 7 2 | | |
| | | 6 2 | | 8 2 | | |
| | | 7 2 | | 6 2 | | |
| | | 8 2 | | 9 2 | | |
| | | 8 2 | | 8 2 | | |

Continued on next column

| TQC | ICSD | | NoE | | SF | |
|---|---|---|---|---|---|---|
| | | 9 2 | | 7 2 | | |
| | | 6 2 | | 7 2 | | |
| | | 6 2 | | 8 2 | | |
| | | 9 2 | | 6 2 | | |
| | | 7 2 | | 9 2 | | |
| | | 8 2 | | 7 2 | | |
| | | 7 2 | | 9 2 | | |
| | | 9 2 | | 8 2 | | |
| | | 6 2 | | 6 2 | | |
| | | 7 2 | | 9 2 | | |
| | | 6 2 | | 6 2 | | |
| | | 9 2 | | 9 2 | | |
| | | 8 2 | | 8 2 | | |
| | | 7 2 | | 6 2 | | |
| | | 8 2 | | 7 2 | | |
| | | 6 2 | | 7 2 | | |
| | | 7 2 | | 8 2 | | |
| | | 9 2 | | 9 2 | | |
| 127.2.7 | 262925 | | 68 | | Cs1I3Sn1 | |
| | -A | -GM | -M | -Z | -R | -X |
| | 8+9 4 | 6 2 | 8+9 4 | 6 2 | 3+4 4 | 3+4 4 |
| | 8+9 4 | 7 2 | 6+7 4 | 7 2 | 3+4 4 | 3+4 4 |
| | 6+7 4 | 7 2 | 8+9 4 | 9 2 | 3+4 4 | 3+4 4 |
| | 6+7 4 | 7 2 | 6+7 4 | 7 2 | 3+4 4 | 3+4 4 |
| | 6+7 4 | 9 2 | 6+7 4 | 7 2 | 3+4 4 | 3+4 4 |
| | 6+7 4 | 8 2 | 6+7 4 | 7 2 | 3+4 4 | 3+4 4 |
| | 6+7 4 | 6 2 | 6+7 4 | 6 2 | 3+4 4 | 3+4 4 |
| | 8+9 4 | 7 2 | 6+7 4 | 8 2 | 3+4 4 | 3+4 4 |
| | 6+7 4 | 9 2 | 8+9 4 | 8 2 | 3+4 4 | 3+4 4 |
| | 8+9 4 | 8 2 | 8+9 4 | 9 2 | 3+4 4 | 3+4 4 |
| | 6+7 4 | 9 2 | 8+9 4 | 9 2 | 3+4 4 | 3+4 4 |
| | 8+9 4 | 8 2 | 6+7 4 | 8 2 | 3+4 4 | 3+4 4 |
| | 8+9 4 | 9 2 | 8+9 4 | 8 2 | 3+4 4 | 3+4 4 |
| | 6+7 4 | 8 2 | 8+9 4 | 9 2 | 3+4 4 | 3+4 4 |
| | 8+9 4 | 7 2 | 6+7 4 | 9 2 | 3+4 4 | 3+4 4 |
| | 8+9 4 | 7 2 | 8+9 4 | 9 2 | 3+4 4 | 3+4 4 |
| | 8+9 4 | 9 2 | 6+7 4 | 6 2 | 3+4 4 | 3+4 4 |
| | | 8 2 | | 7 2 | | |
| | | 9 2 | | 8 2 | | |
| | | 9 2 | | 9 2 | | |
| | | 6 2 | | 7 2 | | |
| | | 9 2 | | 8 2 | | |
| | | 7 2 | | 9 2 | | |
| | | 9 2 | | 6 2 | | |
| | | 6 2 | | 8 2 | | |
| | | 8 2 | | 9 2 | | |
| | | 9 2 | | 7 2 | | |
| | | 8 2 | | 8 2 | | |
| | | 6 2 | | 9 2 | | |
| | | 8 2 | | 6 2 | | |
| | | 9 2 | | 9 2 | | |
| | | 8 2 | | 8 2 | | |
| | | 7 2 | | 9 2 | | |
| | | 7 2 | | 7 2 | | |
| 127.2.8 | 413849 | | 68 | | Ce2Ge2Mg1 | |
| | -A | -GM | -M | -Z | -R | -X |
| | 6+7 4 | 7 2 | 8+9 4 | 9 2 | 3+4 4 | 3+4 4 |
| | 8+9 4 | 8 2 | 6+7 4 | 6 2 | 3+4 4 | 3+4 4 |
| | 6+7 4 | 9 2 | 8+9 4 | 7 2 | 3+4 4 | 3+4 4 |
| | 8+9 4 | 6 2 | 6+7 4 | 8 2 | 3+4 4 | 3+4 4 |
| | 8+9 4 | 6 2 | 8+9 4 | 6 2 | 3+4 4 | 3+4 4 |

Continued on next column

**TABLE LVIII– continued**

| TQC | ICSD | | NoE | | SF | |
|---|---|---|---|---|---|---|
| | 6+7 4 | 8 2 | 6+7 4 | 8 2 | 3+4 4 | 3+4 4 |
| | 6+7 4 | 7 2 | 8+9 4 | 9 2 | 3+4 4 | 3+4 4 |
| | 8+9 4 | 9 2 | 6+7 4 | 7 2 | 3+4 4 | 3+4 4 |
| | 8+9 4 | 6 2 | 8+9 4 | 8 2 | 3+4 4 | 3+4 4 |
| | 6+7 4 | 9 2 | 6+7 4 | 7 2 | 3+4 4 | 3+4 4 |
| | 6+7 4 | 7 2 | 6+7 4 | 9 2 | 3+4 4 | 3+4 4 |
| | 6+7 4 | 8 2 | 8+9 4 | 6 2 | 3+4 4 | 3+4 4 |
| | 8+9 4 | 7 2 | 6+7 4 | 7 2 | 3+4 4 | 3+4 4 |
| | 8+9 4 | 9 2 | 8+9 4 | 9 2 | 3+4 4 | 3+4 4 |
| | 8+9 4 | 6 2 | 8+9 4 | 8 2 | 3+4 4 | 3+4 4 |
| | 8+9 4 | 8 2 | 6+7 4 | 6 2 | 3+4 4 | 3+4 4 |
| | 6+7 4 | 7 2 | 8+9 4 | 6 2 | 3+4 4 | 3+4 4 |
| | | 6 2 | | 7 2 | | |
| | | 8 2 | | 8 2 | | |
| | | 9 2 | | 9 2 | | |
| | | 7 2 | | 7 2 | | |
| | | 7 2 | | 9 2 | | |
| | | 6 2 | | 7 2 | | |
| | | 8 2 | | 8 2 | | |
| | | 9 2 | | 9 2 | | |
| | | 8 2 | | 6 2 | | |
| | | 7 2 | | 6 2 | | |
| | | 6 2 | | 7 2 | | |
| | | 9 2 | | 8 2 | | |
| | | 7 2 | | 9 2 | | |
| | | 6 2 | | 6 2 | | |
| | | 7 2 | | 9 2 | | |
| | | 8 2 | | 8 2 | | |
| | | 9 2 | | 7 2 | | |
| 127.2.9 | 157389 | | 160 | | Cd1Pd2Yb2 | |
| | -A | -GM | -M | -Z | -R | -X |
| | 6+7 4 | 7 2 | 8+9 4 | 9 2 | 3+4 4 | 3+4 4 |
| | 8+9 4 | 6 2 | 6+7 4 | 8 2 | 3+4 4 | 3+4 4 |
| | 6+7 4 | 8 2 | 8+9 4 | 6 2 | 3+4 4 | 3+4 4 |
| | 8+9 4 | 9 2 | 6+7 4 | 7 2 | 3+4 4 | 3+4 4 |
| | 6+7 4 | 6 2 | 8+9 4 | 9 2 | 3+4 4 | 3+4 4 |
| | 8+9 4 | 7 2 | 6+7 4 | 8 2 | 3+4 4 | 3+4 4 |
| | 6+7 4 | 8 2 | 8+9 4 | 6 2 | 3+4 4 | 3+4 4 |
| | 8+9 4 | 9 2 | 6+7 4 | 7 2 | 3+4 4 | 3+4 4 |
| | 6+7 4 | 6 2 | 6+7 4 | 8 2 | 3+4 4 | 3+4 4 |
| | 6+7 4 | 7 2 | 6+7 4 | 7 2 | 3+4 4 | 3+4 4 |
| | 6+7 4 | 9 2 | 6+7 4 | 9 2 | 3+4 4 | 3+4 4 |
| | 6+7 4 | 8 2 | 6+7 4 | 6 2 | 3+4 4 | 3+4 4 |
| | 6+7 4 | 7 2 | 6+7 4 | 9 2 | 3+4 4 | 3+4 4 |
| | 6+7 4 | 6 2 | 6+7 4 | 7 2 | 3+4 4 | 3+4 4 |
| | 8+9 4 | 8 2 | 8+9 4 | 8 2 | 3+4 4 | 3+4 4 |
| | 6+7 4 | 9 2 | 8+9 4 | 6 2 | 3+4 4 | 3+4 4 |
| | 8+9 4 | 7 2 | 6+7 4 | 6 2 | 3+4 4 | 3+4 4 |
| | 6+7 4 | 7 2 | 6+7 4 | 7 2 | 3+4 4 | 3+4 4 |
| | 8+9 4 | 6 2 | 8+9 4 | 7 2 | 3+4 4 | 3+4 4 |
| | 6+7 4 | 6 2 | 6+7 4 | 6 2 | 3+4 4 | 3+4 4 |
| | 8+9 4 | 7 2 | 8+9 4 | 6 2 | 3+4 4 | 3+4 4 |
| | 8+9 4 | 6 2 | 6+7 4 | 6 2 | 3+4 4 | 3+4 4 |
| | 6+7 4 | 7 2 | 8+9 4 | 6 2 | 3+4 4 | 3+4 4 |
| | 6+7 4 | 6 2 | 8+9 4 | 6 2 | 3+4 4 | 3+4 4 |
| | 8+9 4 | 6 2 | 6+7 4 | 7 2 | 3+4 4 | 3+4 4 |
| | 6+7 4 | 6 2 | 8+9 4 | 7 2 | 3+4 4 | 3+4 4 |
| | 8+9 4 | 7 2 | 6+7 4 | 7 2 | 3+4 4 | 3+4 4 |
| | 8+9 4 | 7 2 | 6+7 4 | 8 2 | 3+4 4 | 3+4 4 |
| | 6+7 4 | 7 2 | 8+9 4 | 6 2 | 3+4 4 | 3+4 4 |
| | 6+7 4 | 8 2 | 8+9 4 | 7 2 | 3+4 4 | 3+4 4 |

Continued on next column

**TABLE LVIII– continued**

| TQC | ICSD | | NoE | | SF | |
|---|---|---|---|---|---|---|
| | 8+9 4 | 9 2 | 6+7 4 | 8 2 | 3+4 4 | 3+4 4 |
| | 6+7 4 | 7 2 | 8+9 4 | 9 2 | 3+4 4 | 3+4 4 |
| | 8+9 4 | 9 2 | 6+7 4 | 7 2 | 3+4 4 | 3+4 4 |
| | 8+9 4 | 6 2 | 8+9 4 | 6 2 | 3+4 4 | 3+4 4 |
| | 6+7 4 | 8 2 | 6+7 4 | 9 2 | 3+4 4 | 3+4 4 |
| | 8+9 4 | 9 2 | 6+7 4 | 7 2 | 3+4 4 | 3+4 4 |
| | 6+7 4 | 6 2 | 8+9 4 | 9 2 | 3+4 4 | 3+4 4 |
| | 6+7 4 | 8 2 | 6+7 4 | 8 2 | 3+4 4 | 3+4 4 |
| | 8+9 4 | 9 2 | 8+9 4 | 9 2 | 3+4 4 | 3+4 4 |
| | 6+7 4 | 8 2 | 6+7 4 | 9 2 | 3+4 4 | 3+4 4 |
| | | 7 2 | | 8 2 | | |
| | | 6 2 | | 7 2 | | |
| | | 9 2 | | 6 2 | | |
| | | 8 2 | | 6 2 | | |
| | | 9 2 | | 6 2 | | |
| | | 6 2 | | 8 2 | | |
| | | 8 2 | | 9 2 | | |
| | | 6 2 | | 7 2 | | |
| | | 7 2 | | 7 2 | | |
| | | 6 2 | | 9 2 | | |
| | | 7 2 | | 6 2 | | |
| | | 9 2 | | 8 2 | | |
| | | 7 2 | | 6 2 | | |
| | | 8 2 | | 8 2 | | |
| | | 6 2 | | 6 2 | | |
| | | 9 2 | | 9 2 | | |
| | | 7 2 | | 9 2 | | |
| | | 9 2 | | 8 2 | | |
| | | 6 2 | | 7 2 | | |
| | | 7 2 | | 7 2 | | |
| | | 6 2 | | 6 2 | | |
| | | 8 2 | | 7 2 | | |
| | | 8 2 | | 8 2 | | |
| | | 7 2 | | 9 2 | | |
| | | 9 2 | | 9 2 | | |
| | | 8 2 | | 9 2 | | |
| | | 6 2 | | 6 2 | | |
| | | 7 2 | | 8 2 | | |
| | | 6 2 | | 7 2 | | |
| | | 9 2 | | 7 2 | | |
| | | 9 2 | | 6 2 | | |
| | | 8 2 | | 9 2 | | |
| | | 7 2 | | 8 2 | | |
| | | 6 2 | | 8 2 | | |
| | | 9 2 | | 9 2 | | |
| | | 7 2 | | 6 2 | | |
| | | 8 2 | | 7 2 | | |
| | | 6 2 | | 7 2 | | |
| | | 8 2 | | 8 2 | | |
| | | 7 2 | | 6 2 | | |
| 127.2.10 | 156063 | | 160 | | Cd1Pd2Yb2 | |
| | -A | -GM | -M | -Z | -R | -X |
| | 6+7 4 | 7 2 | 8+9 4 | 9 2 | 3+4 4 | 3+4 4 |
| | 8+9 4 | 6 2 | 6+7 4 | 8 2 | 3+4 4 | 3+4 4 |
| | 6+7 4 | 8 2 | 8+9 4 | 6 2 | 3+4 4 | 3+4 4 |
| | 8+9 4 | 9 2 | 6+7 4 | 7 2 | 3+4 4 | 3+4 4 |
| | 6+7 4 | 6 2 | 8+9 4 | 9 2 | 3+4 4 | 3+4 4 |
| | 8+9 4 | 7 2 | 6+7 4 | 8 2 | 3+4 4 | 3+4 4 |
| | 6+7 4 | 8 2 | 8+9 4 | 6 2 | 3+4 4 | 3+4 4 |
| | 8+9 4 | 9 2 | 6+7 4 | 7 2 | 3+4 4 | 3+4 4 |
| | 6+7 4 | 6 2 | 6+7 4 | 8 2 | 3+4 4 | 3+4 4 |

Continued on next column

**TABLE LVIII– continued**

| TQC | ICSD | | NoE | | SF | |
|---|---|---|---|---|---|---|
| | 6+7 4 | 7 2 | 6+7 4 | 7 2 | 3+4 4 | 3+4 4 |
| | 6+7 4 | 9 2 | 6+7 4 | 9 2 | 3+4 4 | 3+4 4 |
| | 6+7 4 | 8 2 | 6+7 4 | 6 2 | 3+4 4 | 3+4 4 |
| | 6+7 4 | 7 2 | 6+7 4 | 9 2 | 3+4 4 | 3+4 4 |
| | 6+7 4 | 6 2 | 6+7 4 | 7 2 | 3+4 4 | 3+4 4 |
| | 8+9 4 | 8 2 | 8+9 4 | 8 2 | 3+4 4 | 3+4 4 |
| | 6+7 4 | 9 2 | 8+9 4 | 6 2 | 3+4 4 | 3+4 4 |
| | 8+9 4 | 7 2 | 6+7 4 | 6 2 | 3+4 4 | 3+4 4 |
| | 6+7 4 | 7 2 | 6+7 4 | 7 2 | 3+4 4 | 3+4 4 |
| | 8+9 4 | 6 2 | 8+9 4 | 7 2 | 3+4 4 | 3+4 4 |
| | 6+7 4 | 6 2 | 6+7 4 | 6 2 | 3+4 4 | 3+4 4 |
| | 8+9 4 | 7 2 | 8+9 4 | 6 2 | 3+4 4 | 3+4 4 |
| | 8+9 4 | 6 2 | 6+7 4 | 6 2 | 3+4 4 | 3+4 4 |
| | 6+7 4 | 7 2 | 8+9 4 | 6 2 | 3+4 4 | 3+4 4 |
| | 6+7 4 | 6 2 | 8+9 4 | 6 2 | 3+4 4 | 3+4 4 |
| | 8+9 4 | 6 2 | 6+7 4 | 7 2 | 3+4 4 | 3+4 4 |
| | 6+7 4 | 6 2 | 8+9 4 | 7 2 | 3+4 4 | 3+4 4 |
| | 8+9 4 | 7 2 | 6+7 4 | 7 2 | 3+4 4 | 3+4 4 |
| | 8+9 4 | 7 2 | 6+7 4 | 8 2 | 3+4 4 | 3+4 4 |
| | 6+7 4 | 7 2 | 8+9 4 | 6 2 | 3+4 4 | 3+4 4 |
| | 6+7 4 | 8 2 | 8+9 4 | 7 2 | 3+4 4 | 3+4 4 |
| | 8+9 4 | 9 2 | 6+7 4 | 8 2 | 3+4 4 | 3+4 4 |
| | 6+7 4 | 7 2 | 8+9 4 | 9 2 | 3+4 4 | 3+4 4 |
| | 8+9 4 | 9 2 | 6+7 4 | 7 2 | 3+4 4 | 3+4 4 |
| | 8+9 4 | 6 2 | 8+9 4 | 6 2 | 3+4 4 | 3+4 4 |
| | 6+7 4 | 8 2 | 6+7 4 | 9 2 | 3+4 4 | 3+4 4 |
| | 8+9 4 | 9 2 | 6+7 4 | 7 2 | 3+4 4 | 3+4 4 |
| | 6+7 4 | 6 2 | 8+9 4 | 9 2 | 3+4 4 | 3+4 4 |
| | 6+7 4 | 8 2 | 6+7 4 | 8 2 | 3+4 4 | 3+4 4 |
| | 8+9 4 | 9 2 | 8+9 4 | 9 2 | 3+4 4 | 3+4 4 |
| | 6+7 4 | 7 2 | 6+7 4 | 9 2 | 3+4 4 | 3+4 4 |
| | | 8 2 | | 8 2 | | |
| | | 6 2 | | 7 2 | | |
| | | 9 2 | | 6 2 | | |
| | | 8 2 | | 6 2 | | |
| | | 9 2 | | 6 2 | | |
| | | 6 2 | | 8 2 | | |
| | | 8 2 | | 9 2 | | |
| | | 6 2 | | 7 2 | | |
| | | 7 2 | | 7 2 | | |
| | | 6 2 | | 9 2 | | |
| | | 7 2 | | 6 2 | | |
| | | 9 2 | | 8 2 | | |
| | | 7 2 | | 6 2 | | |
| | | 8 2 | | 8 2 | | |
| | | 6 2 | | 6 2 | | |
| | | 9 2 | | 9 2 | | |
| | | 7 2 | | 9 2 | | |
| | | 9 2 | | 8 2 | | |
| | | 6 2 | | 7 2 | | |
| | | 7 2 | | 7 2 | | |
| | | 6 2 | | 6 2 | | |
| | | 8 2 | | 7 2 | | |
| | | 8 2 | | 8 2 | | |
| | | 7 2 | | 9 2 | | |
| | | 9 2 | | 9 2 | | |
| | | 8 2 | | 9 2 | | |
| | | 6 2 | | 6 2 | | |
| | | 7 2 | | 8 2 | | |
| | | 6 2 | | 7 2 | | |
| | | 9 2 | | 7 2 | | |
| | | 9 2 | | 6 2 | | |
| | | 8 2 | | 9 2 | | |
| | | 7 2 | | 8 2 | | |
| | | 6 2 | | 8 2 | | |
| | | 9 2 | | 9 2 | | |
| | | 7 2 | | 6 2 | | |
| | | 8 2 | | 7 2 | | |
| | | 6 2 | | 7 2 | | |
| | | 8 2 | | 8 2 | | |
| | | 7 2 | | 6 2 | | |
| 127.2.11 | 157385 | | 116 | | Cd1Pd2Pr2 | |
| | -A | -GM | -M | -Z | -R | -X |
| | 6+7 4 | 7 2 | 8+9 4 | 9 2 | 3+4 4 | 3+4 4 |
| | 8+9 4 | 6 2 | 6+7 4 | 6 2 | 3+4 4 | 3+4 4 |
| | 6+7 4 | 8 2 | 8+9 4 | 7 2 | 3+4 4 | 3+4 4 |
| | 8+9 4 | 9 2 | 6+7 4 | 8 2 | 3+4 4 | 3+4 4 |
| | 6+7 4 | 6 2 | 8+9 4 | 9 2 | 3+4 4 | 3+4 4 |
| | 8+9 4 | 8 2 | 6+7 4 | 6 2 | 3+4 4 | 3+4 4 |
| | 6+7 4 | 7 2 | 8+9 4 | 8 2 | 3+4 4 | 3+4 4 |
| | 8+9 4 | 9 2 | 6+7 4 | 7 2 | 3+4 4 | 3+4 4 |
| | 6+7 4 | 6 2 | 6+7 4 | 8 2 | 3+4 4 | 3+4 4 |
| | 6+7 4 | 9 2 | 6+7 4 | 9 2 | 3+4 4 | 3+4 4 |
| | 6+7 4 | 7 2 | 6+7 4 | 6 2 | 3+4 4 | 3+4 4 |
| | 6+7 4 | 8 2 | 6+7 4 | 7 2 | 3+4 4 | 3+4 4 |
| | 6+7 4 | 7 2 | 6+7 4 | 9 2 | 3+4 4 | 3+4 4 |
| | 6+7 4 | 6 2 | 6+7 4 | 7 2 | 3+4 4 | 3+4 4 |
| | 8+9 4 | 8 2 | 8+9 4 | 8 2 | 3+4 4 | 3+4 4 |
| | 6+7 4 | 9 2 | 8+9 4 | 6 2 | 3+4 4 | 3+4 4 |
| | 8+9 4 | 7 2 | 6+7 4 | 6 2 | 3+4 4 | 3+4 4 |
| | 6+7 4 | 6 2 | 6+7 4 | 7 2 | 3+4 4 | 3+4 4 |
| | 6+7 4 | 7 2 | 6+7 4 | 6 2 | 3+4 4 | 3+4 4 |
| | 8+9 4 | 6 2 | 8+9 4 | 7 2 | 3+4 4 | 3+4 4 |
| | 8+9 4 | 7 2 | 8+9 4 | 6 2 | 3+4 4 | 3+4 4 |
| | 8+9 4 | 6 2 | 6+7 4 | 6 2 | 3+4 4 | 3+4 4 |
| | 6+7 4 | 7 2 | 8+9 4 | 6 2 | 3+4 4 | 3+4 4 |
| | 8+9 4 | 6 2 | 8+9 4 | 6 2 | 3+4 4 | 3+4 4 |
| | 6+7 4 | 6 2 | 6+7 4 | 7 2 | 3+4 4 | 3+4 4 |
| | 8+9 4 | 6 2 | 6+7 4 | 7 2 | 3+4 4 | 3+4 4 |
| | 8+9 4 | 7 2 | 8+9 4 | 6 2 | 3+4 4 | 3+4 4 |
| | 6+7 4 | 7 2 | 8+9 4 | 8 2 | 3+4 4 | 3+4 4 |
| | 6+7 4 | 7 2 | 6+7 4 | 7 2 | 3+4 4 | 3+4 4 |
| | | 9 2 | | 7 2 | | |
| | | 7 2 | | 8 2 | | |
| | | 6 2 | | 7 2 | | |
| | | 8 2 | | 9 2 | | |
| | | 9 2 | | 8 2 | | |
| | | 8 2 | | 6 2 | | |
| | | 9 2 | | 9 2 | | |
| | | 6 2 | | 7 2 | | |
| | | 8 2 | | 9 2 | | |
| | | 7 2 | | 9 2 | | |
| | | 9 2 | | 8 2 | | |
| | | 6 2 | | 6 2 | | |
| | | 8 2 | | 7 2 | | |
| | | 9 2 | | 6 2 | | |
| | | 8 2 | | 6 2 | | |
| | | 6 2 | | 8 2 | | |
| | | 7 2 | | 9 2 | | |
| | | 8 2 | | 9 2 | | |
| | | 9 2 | | 7 2 | | |
| | | 6 2 | | 6 2 | | |

Continued on next column

**TABLE LVIII– continued**

| TQC | ICSD | | NoE | | SF | |
|---|---|---|---|---|---|---|
| | | 6 2 | | 7 2 | | |
| | | 7 2 | | 9 2 | | |
| | | 7 2 | | 8 2 | | |
| | | 8 2 | | 7 2 | | |
| | | 9 2 | | 9 2 | | |
| | | 6 2 | | 6 2 | | |
| | | 7 2 | | 9 2 | | |
| | | 7 2 | | 7 2 | | |
| | | 9 2 | | 8 2 | | |
| 127.2.12 | 414596 | | 116 | | Cd1Ni2Pr2 | |
| | -A | -GM | -M | -Z | -R | -X |
| | 6+7 4 | 7 2 | 8+9 4 | 9 2 | 3+4 4 | 3+4 4 |
| | 8+9 4 | 8 2 | 6+7 4 | 6 2 | 3+4 4 | 3+4 4 |
| | 6+7 4 | 9 2 | 8+9 4 | 7 2 | 3+4 4 | 3+4 4 |
| | 8+9 4 | 6 2 | 6+7 4 | 8 2 | 3+4 4 | 3+4 4 |
| | 6+7 4 | 6 2 | 8+9 4 | 6 2 | 3+4 4 | 3+4 4 |
| | 8+9 4 | 8 2 | 6+7 4 | 8 2 | 3+4 4 | 3+4 4 |
| | 6+7 4 | 7 2 | 8+9 4 | 9 2 | 3+4 4 | 3+4 4 |
| | 8+9 4 | 9 2 | 6+7 4 | 7 2 | 3+4 4 | 3+4 4 |
| | 6+7 4 | 6 2 | 6+7 4 | 8 2 | 3+4 4 | 3+4 4 |
| | 6+7 4 | 9 2 | 6+7 4 | 9 2 | 3+4 4 | 3+4 4 |
| | 6+7 4 | 7 2 | 6+7 4 | 6 2 | 3+4 4 | 3+4 4 |
| | 6+7 4 | 8 2 | 6+7 4 | 7 2 | 3+4 4 | 3+4 4 |
| | 6+7 4 | 7 2 | 6+7 4 | 7 2 | 3+4 4 | 3+4 4 |
| | 6+7 4 | 9 2 | 6+7 4 | 9 2 | 3+4 4 | 3+4 4 |
| | 8+9 4 | 6 2 | 8+9 4 | 8 2 | 3+4 4 | 3+4 4 |
| | 6+7 4 | 8 2 | 8+9 4 | 6 2 | 3+4 4 | 3+4 4 |
| | 8+9 4 | 7 2 | 6+7 4 | 6 2 | 3+4 4 | 3+4 4 |
| | 6+7 4 | 7 2 | 6+7 4 | 7 2 | 3+4 4 | 3+4 4 |
| | 8+9 4 | 6 2 | 6+7 4 | 6 2 | 3+4 4 | 3+4 4 |
| | 6+7 4 | 6 2 | 8+9 4 | 7 2 | 3+4 4 | 3+4 4 |
| | 8+9 4 | 7 2 | 8+9 4 | 6 2 | 3+4 4 | 3+4 4 |
| | 8+9 4 | 6 2 | 6+7 4 | 6 2 | 3+4 4 | 3+4 4 |
| | 8+9 4 | 7 2 | 8+9 4 | 6 2 | 3+4 4 | 3+4 4 |
| | 6+7 4 | 6 2 | 8+9 4 | 6 2 | 3+4 4 | 3+4 4 |
| | 6+7 4 | 6 2 | 6+7 4 | 7 2 | 3+4 4 | 3+4 4 |
| | 8+9 4 | 6 2 | 6+7 4 | 7 2 | 3+4 4 | 3+4 4 |
| | 8+9 4 | 7 2 | 8+9 4 | 7 2 | 3+4 4 | 3+4 4 |
| | 6+7 4 | 7 2 | 8+9 4 | 6 2 | 3+4 4 | 3+4 4 |
| | 6+7 4 | 7 2 | 6+7 4 | 7 2 | 3+4 4 | 3+4 4 |
| | | 8 2 | | 9 2 | | |
| | | 9 2 | | 8 2 | | |
| | | 7 2 | | 9 2 | | |
| | | 6 2 | | 8 2 | | |
| | | 9 2 | | 7 2 | | |
| | | 8 2 | | 8 2 | | |
| | | 9 2 | | 9 2 | | |
| | | 6 2 | | 8 2 | | |
| | | 8 2 | | 9 2 | | |
| | | 6 2 | | 6 2 | | |
| | | 9 2 | | 7 2 | | |
| | | 8 2 | | 9 2 | | |
| | | 9 2 | | 6 2 | | |
| | | 7 2 | | 7 2 | | |
| | | 6 2 | | 6 2 | | |
| | | 8 2 | | 6 2 | | |
| | | 9 2 | | 8 2 | | |
| | | 8 2 | | 9 2 | | |
| | | 6 2 | | 7 2 | | |
| | | 7 2 | | 6 2 | | |
| | | 6 2 | | 9 2 | | |

Continued on next column

**TABLE LVIII– continued**

| TQC | ICSD | | NoE | | SF | |
|---|---|---|---|---|---|---|
| | | 7 2 | | 8 2 | | |
| | | 7 2 | | 7 2 | | |
| | | 8 2 | | 7 2 | | |
| | | 9 2 | | 9 2 | | |
| | | 6 2 | | 6 2 | | |
| | | 7 2 | | 9 2 | | |
| | | 7 2 | | 7 2 | | |
| | | 9 2 | | 8 2 | | |
| 127.2.13 | 107389 | | 104 | | Cd1La2Rh2 | |
| | -A | -GM | -M | -Z | -R | -X |
| | 6+7 4 | 7 2 | 8+9 4 | 9 2 | 3+4 4 | 3+4 4 |
| | 8+9 4 | 8 2 | 6+7 4 | 6 2 | 3+4 4 | 3+4 4 |
| | 6+7 4 | 9 2 | 8+9 4 | 7 2 | 3+4 4 | 3+4 4 |
| | 8+9 4 | 6 2 | 6+7 4 | 8 2 | 3+4 4 | 3+4 4 |
| | 6+7 4 | 8 2 | 8+9 4 | 6 2 | 3+4 4 | 3+4 4 |
| | 8+9 4 | 6 2 | 6+7 4 | 9 2 | 3+4 4 | 3+4 4 |
| | 6+7 4 | 7 2 | 8+9 4 | 8 2 | 3+4 4 | 3+4 4 |
| | 8+9 4 | 9 2 | 6+7 4 | 7 2 | 3+4 4 | 3+4 4 |
| | 6+7 4 | 6 2 | 6+7 4 | 8 2 | 3+4 4 | 3+4 4 |
| | 6+7 4 | 9 2 | 6+7 4 | 9 2 | 3+4 4 | 3+4 4 |
| | 6+7 4 | 7 2 | 6+7 4 | 6 2 | 3+4 4 | 3+4 4 |
| | 6+7 4 | 8 2 | 6+7 4 | 7 2 | 3+4 4 | 3+4 4 |
| | 6+7 4 | 7 2 | 6+7 4 | 7 2 | 3+4 4 | 3+4 4 |
| | 6+7 4 | 6 2 | 6+7 4 | 9 2 | 3+4 4 | 3+4 4 |
| | 8+9 4 | 8 2 | 8+9 4 | 8 2 | 3+4 4 | 3+4 4 |
| | 6+7 4 | 9 2 | 6+7 4 | 6 2 | 3+4 4 | 3+4 4 |
| | 6+7 4 | 7 2 | 6+7 4 | 6 2 | 3+4 4 | 3+4 4 |
| | 8+9 4 | 7 2 | 8+9 4 | 7 2 | 3+4 4 | 3+4 4 |
| | 8+9 4 | 6 2 | 8+9 4 | 6 2 | 3+4 4 | 3+4 4 |
| | 6+7 4 | 6 2 | 6+7 4 | 7 2 | 3+4 4 | 3+4 4 |
| | 8+9 4 | 7 2 | 8+9 4 | 6 2 | 3+4 4 | 3+4 4 |
| | 8+9 4 | 7 2 | 6+7 4 | 6 2 | 3+4 4 | 3+4 4 |
| | 8+9 4 | 6 2 | 8+9 4 | 6 2 | 3+4 4 | 3+4 4 |
| | 6+7 4 | 6 2 | 6+7 4 | 7 2 | 3+4 4 | 3+4 4 |
| | 6+7 4 | 6 2 | 8+9 4 | 6 2 | 3+4 4 | 3+4 4 |
| | 8+9 4 | 6 2 | 6+7 4 | 7 2 | 3+4 4 | 3+4 4 |
| | | 7 2 | | 7 2 | | |
| | | 7 2 | | 6 2 | | |
| | | 7 2 | | 8 2 | | |
| | | 7 2 | | 7 2 | | |
| | | 9 2 | | 8 2 | | |
| | | 6 2 | | 9 2 | | |
| | | 8 2 | | 7 2 | | |
| | | 9 2 | | 9 2 | | |
| | | 8 2 | | 9 2 | | |
| | | 8 2 | | 8 2 | | |
| | | 9 2 | | 8 2 | | |
| | | 6 2 | | 6 2 | | |
| | | 7 2 | | 7 2 | | |
| | | 9 2 | | 9 2 | | |
| | | 6 2 | | 6 2 | | |
| | | 8 2 | | 9 2 | | |
| | | 9 2 | | 6 2 | | |
| | | 8 2 | | 7 2 | | |
| | | 8 2 | | 6 2 | | |
| | | 9 2 | | 7 2 | | |
| | | 6 2 | | 8 2 | | |
| | | 6 2 | | 9 2 | | |
| | | 7 2 | | 6 2 | | |
| | | 6 2 | | 9 2 | | |
| | | 7 2 | | 8 2 | | |

Continued on next column

**TABLE LVIII– continued**

| TQC | ICSD | NoE | SF | | | |
|---|---|---|---|---|---|---|
| | | 7 2 | | 7 2 | | |
| 127.2.14 | 424179 | 136 | Cd1Ge2Yb2 | | | |
| | -A | -GM | -M | -Z | -R | -X |
| | 6+7 4 | 7 2 | 8+9 4 | 9 2 | 3+4 4 | 3+4 4 |
| | 8+9 4 | 6 2 | 6+7 4 | 6 2 | 3+4 4 | 3+4 4 |
| | 6+7 4 | 8 2 | 8+9 4 | 7 2 | 3+4 4 | 3+4 4 |
| | 8+9 4 | 9 2 | 6+7 4 | 8 2 | 3+4 4 | 3+4 4 |
| | 8+9 4 | 8 2 | 8+9 4 | 6 2 | 3+4 4 | 3+4 4 |
| | 6+7 4 | 6 2 | 6+7 4 | 8 2 | 3+4 4 | 3+4 4 |
| | 6+7 4 | 7 2 | 8+9 4 | 9 2 | 3+4 4 | 3+4 4 |
| | 8+9 4 | 9 2 | 6+7 4 | 7 2 | 3+4 4 | 3+4 4 |
| | 6+7 4 | 6 2 | 8+9 4 | 8 2 | 3+4 4 | 3+4 4 |
| | 8+9 4 | 9 2 | 6+7 4 | 7 2 | 3+4 4 | 3+4 4 |
| | 6+7 4 | 7 2 | 6+7 4 | 9 2 | 3+4 4 | 3+4 4 |
| | 6+7 4 | 8 2 | 6+7 4 | 6 2 | 3+4 4 | 3+4 4 |
| | 6+7 4 | 7 2 | 6+7 4 | 7 2 | 3+4 4 | 3+4 4 |
| | 6+7 4 | 9 2 | 6+7 4 | 9 2 | 3+4 4 | 3+4 4 |
| | 6+7 4 | 6 2 | 6+7 4 | 8 2 | 3+4 4 | 3+4 4 |
| | 6+7 4 | 8 2 | 6+7 4 | 6 2 | 3+4 4 | 3+4 4 |
| | 6+7 4 | 7 2 | 8+9 4 | 6 2 | 3+4 4 | 3+4 4 |
| | 8+9 4 | 6 2 | 8+9 4 | 7 2 | 3+4 4 | 3+4 4 |
| | 8+9 4 | 6 2 | 6+7 4 | 6 2 | 3+4 4 | 3+4 4 |
| | 8+9 4 | 7 2 | 8+9 4 | 7 2 | 3+4 4 | 3+4 4 |
| | 6+7 4 | 7 2 | 6+7 4 | 7 2 | 3+4 4 | 3+4 4 |
| | 8+9 4 | 6 2 | 6+7 4 | 6 2 | 3+4 4 | 3+4 4 |
| | 6+7 4 | 6 2 | 8+9 4 | 6 2 | 3+4 4 | 3+4 4 |
| | 8+9 4 | 6 2 | 8+9 4 | 6 2 | 3+4 4 | 3+4 4 |
| | 6+7 4 | 7 2 | 6+7 4 | 6 2 | 3+4 4 | 3+4 4 |
| | 8+9 4 | 6 2 | 8+9 4 | 7 2 | 3+4 4 | 3+4 4 |
| | 8+9 4 | 7 2 | 6+7 4 | 7 2 | 3+4 4 | 3+4 4 |
| | 6+7 4 | 6 2 | 6+7 4 | 6 2 | 3+4 4 | 3+4 4 |
| | 8+9 4 | 8 2 | 8+9 4 | 9 2 | 3+4 4 | 3+4 4 |
| | 6+7 4 | 9 2 | 6+7 4 | 8 2 | 3+4 4 | 3+4 4 |
| | 6+7 4 | 7 2 | 8+9 4 | 9 2 | 3+4 4 | 3+4 4 |
| | 8+9 4 | 7 2 | 6+7 4 | 7 2 | 3+4 4 | 3+4 4 |
| | 8+9 4 | 8 2 | 8+9 4 | 7 2 | 3+4 4 | 3+4 4 |
| | 6+7 4 | 9 2 | 6+7 4 | 9 2 | 3+4 4 | 3+4 4 |
| | | 6 2 | | 8 2 | | |
| | | 8 2 | | 8 2 | | |
| | | 9 2 | | 6 2 | | |
| | | 7 2 | | 6 2 | | |
| | | 8 2 | | 7 2 | | |
| | | 6 2 | | 6 2 | | |
| | | 8 2 | | 9 2 | | |
| | | 7 2 | | 7 2 | | |
| | | 9 2 | | 8 2 | | |
| | | 6 2 | | 7 2 | | |
| | | 6 2 | | 9 2 | | |
| | | 7 2 | | 8 2 | | |
| | | 9 2 | | 8 2 | | |
| | | 8 2 | | 9 2 | | |
| | | 9 2 | | 7 2 | | |
| | | 6 2 | | 6 2 | | |
| | | 7 2 | | 6 2 | | |
| | | 9 2 | | 9 2 | | |
| | | 8 2 | | 9 2 | | |
| | | 6 2 | | 6 2 | | |
| | | 8 2 | | 7 2 | | |
| | | 9 2 | | 8 2 | | |
| | | 6 2 | | 9 2 | | |
| | | 7 2 | | 6 2 | | |

Continued on next column

| TQC | ICSD | NoE | SF | | | |
|---|---|---|---|---|---|---|
| | | 7 2 | | 7 2 | | |
| | | 8 2 | | 8 2 | | |
| | | 7 2 | | 6 2 | | |
| | | 6 2 | | 8 2 | | |
| | | 9 2 | | 7 2 | | |
| | | 8 2 | | 8 2 | | |
| | | 7 2 | | 9 2 | | |
| | | 9 2 | | 9 2 | | |
| | | 7 2 | | 7 2 | | |
| | | 6 2 | | 6 2 | | |
| 127.2.15 | 424169 | 84 | Cd1Ge2Y2 | | | |
| | -A | -GM | -M | -Z | -R | -X |
| | 6+7 4 | 7 2 | 8+9 4 | 9 2 | 3+4 4 | 3+4 4 |
| | 8+9 4 | 8 2 | 6+7 4 | 6 2 | 3+4 4 | 3+4 4 |
| | 6+7 4 | 9 2 | 8+9 4 | 7 2 | 3+4 4 | 3+4 4 |
| | 8+9 4 | 6 2 | 6+7 4 | 8 2 | 3+4 4 | 3+4 4 |
| | 8+9 4 | 8 2 | 8+9 4 | 6 2 | 3+4 4 | 3+4 4 |
| | 6+7 4 | 6 2 | 6+7 4 | 8 2 | 3+4 4 | 3+4 4 |
| | 6+7 4 | 7 2 | 8+9 4 | 9 2 | 3+4 4 | 3+4 4 |
| | 8+9 4 | 9 2 | 6+7 4 | 7 2 | 3+4 4 | 3+4 4 |
| | 8+9 4 | 6 2 | 8+9 4 | 8 2 | 3+4 4 | 3+4 4 |
| | 6+7 4 | 9 2 | 6+7 4 | 7 2 | 3+4 4 | 3+4 4 |
| | 6+7 4 | 7 2 | 6+7 4 | 9 2 | 3+4 4 | 3+4 4 |
| | 6+7 4 | 8 2 | 6+7 4 | 6 2 | 3+4 4 | 3+4 4 |
| | 6+7 4 | 7 2 | 6+7 4 | 7 2 | 3+4 4 | 3+4 4 |
| | 6+7 4 | 9 2 | 6+7 4 | 9 2 | 3+4 4 | 3+4 4 |
| | 6+7 4 | 6 2 | 6+7 4 | 8 2 | 3+4 4 | 3+4 4 |
| | 6+7 4 | 8 2 | 6+7 4 | 6 2 | 3+4 4 | 3+4 4 |
| | 6+7 4 | 7 2 | 8+9 4 | 6 2 | 3+4 4 | 3+4 4 |
| | 8+9 4 | 6 2 | 8+9 4 | 7 2 | 3+4 4 | 3+4 4 |
| | 8+9 4 | 6 2 | 6+7 4 | 6 2 | 3+4 4 | 3+4 4 |
| | 8+9 4 | 7 2 | 6+7 4 | 7 2 | 3+4 4 | 3+4 4 |
| | 8+9 4 | 7 2 | 8+9 4 | 7 2 | 3+4 4 | 3+4 4 |
| | | 6 2 | | 6 2 | | |
| | | 6 2 | | 6 2 | | |
| | | 6 2 | | 6 2 | | |
| | | 7 2 | | 6 2 | | |
| | | 6 2 | | 7 2 | | |
| | | 7 2 | | 7 2 | | |
| | | 6 2 | | 6 2 | | |
| | | 8 2 | | 9 2 | | |
| | | 9 2 | | 8 2 | | |
| | | 7 2 | | 7 2 | | |
| | | 7 2 | | 9 2 | | |
| | | 9 2 | | 7 2 | | |
| | | 8 2 | | 9 2 | | |
| | | 6 2 | | 8 2 | | |
| | | 8 2 | | 8 2 | | |
| | | 9 2 | | 6 2 | | |
| | | 7 2 | | 7 2 | | |
| | | 6 2 | | 6 2 | | |
| | | 7 2 | | 9 2 | | |
| | | 6 2 | | 9 2 | | |
| | | 9 2 | | 6 2 | | |
| 127.2.16 | 424178 | 132 | Cd1Ge2Tm2 | | | |
| | -A | -GM | -M | -Z | -R | -X |
| | 6+7 4 | 7 2 | 8+9 4 | 9 2 | 3+4 4 | 3+4 4 |
| | 8+9 4 | 6 2 | 6+7 4 | 6 2 | 3+4 4 | 3+4 4 |
| | 6+7 4 | 8 2 | 8+9 4 | 7 2 | 3+4 4 | 3+4 4 |
| | 8+9 4 | 9 2 | 6+7 4 | 8 2 | 3+4 4 | 3+4 4 |
| | 8+9 4 | 8 2 | 8+9 4 | 6 2 | 3+4 4 | 3+4 4 |

Continued on next column

**TABLE LVIII– continued**

| TQC | ICSD | | NoE | | SF | |
|---|---|---|---|---|---|---|
| | 6+7 4 | 6 2 | 6+7 4 | 8 2 | 3+4 4 | 3+4 4 |
| | 6+7 4 | 7 2 | 8+9 4 | 9 2 | 3+4 4 | 3+4 4 |
| | 8+9 4 | 9 2 | 6+7 4 | 7 2 | 3+4 4 | 3+4 4 |
| | 8+9 4 | 6 2 | 8+9 4 | 8 2 | 3+4 4 | 3+4 4 |
| | 6+7 4 | 9 2 | 6+7 4 | 7 2 | 3+4 4 | 3+4 4 |
| | 6+7 4 | 7 2 | 6+7 4 | 9 2 | 3+4 4 | 3+4 4 |
| | 6+7 4 | 8 2 | 6+7 4 | 6 2 | 3+4 4 | 3+4 4 |
| | 6+7 4 | 7 2 | 6+7 4 | 7 2 | 3+4 4 | 3+4 4 |
| | 6+7 4 | 9 2 | 6+7 4 | 9 2 | 3+4 4 | 3+4 4 |
| | 6+7 4 | 6 2 | 6+7 4 | 8 2 | 3+4 4 | 3+4 4 |
| | 6+7 4 | 8 2 | 6+7 4 | 6 2 | 3+4 4 | 3+4 4 |
| | 6+7 4 | 7 2 | 8+9 4 | 6 2 | 3+4 4 | 3+4 4 |
| | 8+9 4 | 6 2 | 8+9 4 | 7 2 | 3+4 4 | 3+4 4 |
| | 8+9 4 | 6 2 | 6+7 4 | 6 2 | 3+4 4 | 3+4 4 |
| | 8+9 4 | 7 2 | 8+9 4 | 7 2 | 3+4 4 | 3+4 4 |
| | 8+9 4 | 7 2 | 6+7 4 | 7 2 | 3+4 4 | 3+4 4 |
| | 6+7 4 | 6 2 | 8+9 4 | 6 2 | 3+4 4 | 3+4 4 |
| | 6+7 4 | 6 2 | 6+7 4 | 6 2 | 3+4 4 | 3+4 4 |
| | 8+9 4 | 6 2 | 8+9 4 | 6 2 | 3+4 4 | 3+4 4 |
| | 6+7 4 | 7 2 | 6+7 4 | 6 2 | 3+4 4 | 3+4 4 |
| | 8+9 4 | 6 2 | 6+7 4 | 7 2 | 3+4 4 | 3+4 4 |
| | 8+9 4 | 7 2 | 8+9 4 | 7 2 | 3+4 4 | 3+4 4 |
| | 6+7 4 | 6 2 | 6+7 4 | 6 2 | 3+4 4 | 3+4 4 |
| | 8+9 4 | 8 2 | 8+9 4 | 9 2 | 3+4 4 | 3+4 4 |
| | 6+7 4 | 9 2 | 6+7 4 | 8 2 | 3+4 4 | 3+4 4 |
| | 6+7 4 | 7 2 | 8+9 4 | 9 2 | 3+4 4 | 3+4 4 |
| | 8+9 4 | 7 2 | 6+7 4 | 7 2 | 3+4 4 | 3+4 4 |
| | 8+9 4 | 8 2 | 8+9 4 | 7 2 | 3+4 4 | 3+4 4 |
| | | 9 2 | | 9 2 | | |
| | | 6 2 | | 8 2 | | |
| | | 8 2 | | 8 2 | | |
| | | 9 2 | | 6 2 | | |
| | | 7 2 | | 7 2 | | |
| | | 6 2 | | 6 2 | | |
| | | 8 2 | | 9 2 | | |
| | | 7 2 | | 6 2 | | |
| | | 8 2 | | 8 2 | | |
| | | 9 2 | | 7 2 | | |
| | | 6 2 | | 7 2 | | |
| | | 6 2 | | 8 2 | | |
| | | 9 2 | | 9 2 | | |
| | | 7 2 | | 7 2 | | |
| | | 8 2 | | 9 2 | | |
| | | 9 2 | | 8 2 | | |
| | | 6 2 | | 6 2 | | |
| | | 7 2 | | 6 2 | | |
| | | 7 2 | | 9 2 | | |
| | | 9 2 | | 9 2 | | |
| | | 6 2 | | 6 2 | | |
| | | 8 2 | | 8 2 | | |
| | | 8 2 | | 7 2 | | |
| | | 9 2 | | 9 2 | | |
| | | 6 2 | | 6 2 | | |
| | | 7 2 | | 7 2 | | |
| | | 7 2 | | 8 2 | | |
| | | 8 2 | | 6 2 | | |
| | | 6 2 | | 7 2 | | |
| | | 7 2 | | 8 2 | | |
| | | 9 2 | | 9 2 | | |
| | | 8 2 | | 8 2 | | |
| | | 9 2 | | 9 2 | | |

Continued on next column

**TABLE LVIII– continued**

| TQC | ICSD | | NoE | | SF | |
|---|---|---|---|---|---|---|
| 127.2.17 | 424177 | | 128 | | Cd1Er2Ge2 | |
| | -A | -GM | -M | -Z | -R | -X |
| | 6+7 4 | 7 2 | 8+9 4 | 9 2 | 3+4 4 | 3+4 4 |
| | 8+9 4 | 6 2 | 6+7 4 | 6 2 | 3+4 4 | 3+4 4 |
| | 6+7 4 | 8 2 | 8+9 4 | 7 2 | 3+4 4 | 3+4 4 |
| | 8+9 4 | 9 2 | 6+7 4 | 8 2 | 3+4 4 | 3+4 4 |
| | 8+9 4 | 8 2 | 8+9 4 | 6 2 | 3+4 4 | 3+4 4 |
| | 6+7 4 | 6 2 | 6+7 4 | 8 2 | 3+4 4 | 3+4 4 |
| | 6+7 4 | 7 2 | 8+9 4 | 9 2 | 3+4 4 | 3+4 4 |
| | 8+9 4 | 9 2 | 6+7 4 | 7 2 | 3+4 4 | 3+4 4 |
| | 8+9 4 | 6 2 | 8+9 4 | 8 2 | 3+4 4 | 3+4 4 |
| | 6+7 4 | 9 2 | 6+7 4 | 7 2 | 3+4 4 | 3+4 4 |
| | 6+7 4 | 7 2 | 6+7 4 | 9 2 | 3+4 4 | 3+4 4 |
| | 6+7 4 | 8 2 | 6+7 4 | 6 2 | 3+4 4 | 3+4 4 |
| | 6+7 4 | 7 2 | 6+7 4 | 7 2 | 3+4 4 | 3+4 4 |
| | 6+7 4 | 9 2 | 6+7 4 | 9 2 | 3+4 4 | 3+4 4 |
| | 6+7 4 | 6 2 | 6+7 4 | 8 2 | 3+4 4 | 3+4 4 |
| | 6+7 4 | 8 2 | 6+7 4 | 6 2 | 3+4 4 | 3+4 4 |
| | 6+7 4 | 7 2 | 8+9 4 | 6 2 | 3+4 4 | 3+4 4 |
| | 8+9 4 | 6 2 | 8+9 4 | 7 2 | 3+4 4 | 3+4 4 |
| | 8+9 4 | 6 2 | 6+7 4 | 6 2 | 3+4 4 | 3+4 4 |
| | 8+9 4 | 7 2 | 8+9 4 | 7 2 | 3+4 4 | 3+4 4 |
| | 8+9 4 | 7 2 | 6+7 4 | 7 2 | 3+4 4 | 3+4 4 |
| | 6+7 4 | 6 2 | 8+9 4 | 6 2 | 3+4 4 | 3+4 4 |
| | 6+7 4 | 6 2 | 6+7 4 | 6 2 | 3+4 4 | 3+4 4 |
| | 8+9 4 | 6 2 | 8+9 4 | 6 2 | 3+4 4 | 3+4 4 |
| | 6+7 4 | 7 2 | 6+7 4 | 6 2 | 3+4 4 | 3+4 4 |
| | 8+9 4 | 6 2 | 6+7 4 | 7 2 | 3+4 4 | 3+4 4 |
| | 8+9 4 | 7 2 | 8+9 4 | 7 2 | 3+4 4 | 3+4 4 |
| | 6+7 4 | 6 2 | 6+7 4 | 6 2 | 3+4 4 | 3+4 4 |
| | 8+9 4 | 8 2 | 8+9 4 | 9 2 | 3+4 4 | 3+4 4 |
| | 6+7 4 | 9 2 | 6+7 4 | 8 2 | 3+4 4 | 3+4 4 |
| | 6+7 4 | 7 2 | 8+9 4 | 9 2 | 3+4 4 | 3+4 4 |
| | 8+9 4 | 7 2 | 6+7 4 | 7 2 | 3+4 4 | 3+4 4 |
| | | 8 2 | | 7 2 | | |
| | | 9 2 | | 9 2 | | |
| | | 6 2 | | 8 2 | | |
| | | 8 2 | | 6 2 | | |
| | | 9 2 | | 8 2 | | |
| | | 7 2 | | 7 2 | | |
| | | 6 2 | | 6 2 | | |
| | | 8 2 | | 9 2 | | |
| | | 7 2 | | 6 2 | | |
| | | 8 2 | | 8 2 | | |
| | | 9 2 | | 7 2 | | |
| | | 6 2 | | 7 2 | | |
| | | 6 2 | | 8 2 | | |
| | | 9 2 | | 9 2 | | |
| | | 7 2 | | 7 2 | | |
| | | 8 2 | | 9 2 | | |
| | | 9 2 | | 8 2 | | |
| | | 6 2 | | 6 2 | | |
| | | 7 2 | | 6 2 | | |
| | | 7 2 | | 9 2 | | |
| | | 9 2 | | 9 2 | | |
| | | 6 2 | | 6 2 | | |
| | | 8 2 | | 8 2 | | |
| | | 8 2 | | 7 2 | | |
| | | 9 2 | | 9 2 | | |
| | | 6 2 | | 6 2 | | |
| | | 7 2 | | 7 2 | | |

Continued on next column

**TABLE LVIII– continued**

| TQC | ICSD | | NoE | | SF | |
|---|---|---|---|---|---|---|
| | | 7 2 | | 8 2 | | |
| | | 8 2 | | 6 2 | | |
| | | 6 2 | | 7 2 | | |
| | | 9 2 | | 8 2 | | |
| | | 7 2 | | 9 2 | | |
| 127.2.18 | 424175 | | 120 | | Cd1Dy2Ge2 | |
| | -A | -GM | -M | -Z | -R | -X |
| | 6+7 4 | 7 2 | 8+9 4 | 9 2 | 3+4 4 | 3+4 4 |
| | 8+9 4 | 6 2 | 6+7 4 | 6 2 | 3+4 4 | 3+4 4 |
| | 6+7 4 | 8 2 | 8+9 4 | 7 2 | 3+4 4 | 3+4 4 |
| | 8+9 4 | 9 2 | 6+7 4 | 8 2 | 3+4 4 | 3+4 4 |
| | 8+9 4 | 8 2 | 8+9 4 | 6 2 | 3+4 4 | 3+4 4 |
| | 6+7 4 | 6 2 | 6+7 4 | 8 2 | 3+4 4 | 3+4 4 |
| | 6+7 4 | 7 2 | 8+9 4 | 9 2 | 3+4 4 | 3+4 4 |
| | 8+9 4 | 9 2 | 6+7 4 | 7 2 | 3+4 4 | 3+4 4 |
| | 6+7 4 | 6 2 | 8+9 4 | 8 2 | 3+4 4 | 3+4 4 |
| | 8+9 4 | 9 2 | 6+7 4 | 7 2 | 3+4 4 | 3+4 4 |
| | 6+7 4 | 7 2 | 6+7 4 | 9 2 | 3+4 4 | 3+4 4 |
| | 6+7 4 | 8 2 | 6+7 4 | 6 2 | 3+4 4 | 3+4 4 |
| | 6+7 4 | 7 2 | 6+7 4 | 7 2 | 3+4 4 | 3+4 4 |
| | 6+7 4 | 9 2 | 6+7 4 | 9 2 | 3+4 4 | 3+4 4 |
| | 6+7 4 | 6 2 | 6+7 4 | 8 2 | 3+4 4 | 3+4 4 |
| | 6+7 4 | 8 2 | 6+7 4 | 6 2 | 3+4 4 | 3+4 4 |
| | 6+7 4 | 7 2 | 8+9 4 | 6 2 | 3+4 4 | 3+4 4 |
| | 8+9 4 | 6 2 | 8+9 4 | 7 2 | 3+4 4 | 3+4 4 |
| | 8+9 4 | 6 2 | 6+7 4 | 6 2 | 3+4 4 | 3+4 4 |
| | 8+9 4 | 7 2 | 8+9 4 | 7 2 | 3+4 4 | 3+4 4 |
| | 8+9 4 | 7 2 | 6+7 4 | 7 2 | 3+4 4 | 3+4 4 |
| | 6+7 4 | 6 2 | 8+9 4 | 6 2 | 3+4 4 | 3+4 4 |
| | 8+9 4 | 6 2 | 6+7 4 | 6 2 | 3+4 4 | 3+4 4 |
| | 6+7 4 | 6 2 | 6+7 4 | 6 2 | 3+4 4 | 3+4 4 |
| | 8+9 4 | 7 2 | 8+9 4 | 6 2 | 3+4 4 | 3+4 4 |
| | 6+7 4 | 6 2 | 6+7 4 | 7 2 | 3+4 4 | 3+4 4 |
| | 8+9 4 | 7 2 | 8+9 4 | 7 2 | 3+4 4 | 3+4 4 |
| | 6+7 4 | 6 2 | 6+7 4 | 6 2 | 3+4 4 | 3+4 4 |
| | 8+9 4 | 8 2 | 8+9 4 | 9 2 | 3+4 4 | 3+4 4 |
| | 6+7 4 | 9 2 | 6+7 4 | 8 2 | 3+4 4 | 3+4 4 |
| | | 7 2 | | 7 2 | | |
| | | 7 2 | | 9 2 | | |
| | | 8 2 | | 7 2 | | |
| | | 9 2 | | 9 2 | | |
| | | 6 2 | | 8 2 | | |
| | | 8 2 | | 6 2 | | |
| | | 9 2 | | 8 2 | | |
| | | 7 2 | | 7 2 | | |
| | | 6 2 | | 6 2 | | |
| | | 8 2 | | 9 2 | | |
| | | 7 2 | | 6 2 | | |
| | | 8 2 | | 8 2 | | |
| | | 9 2 | | 7 2 | | |
| | | 6 2 | | 7 2 | | |
| | | 9 2 | | 8 2 | | |
| | | 6 2 | | 9 2 | | |
| | | 7 2 | | 7 2 | | |
| | | 8 2 | | 9 2 | | |
| | | 9 2 | | 6 2 | | |
| | | 6 2 | | 8 2 | | |
| | | 7 2 | | 6 2 | | |
| | | 7 2 | | 9 2 | | |
| | | 9 2 | | 9 2 | | |
| | | 6 2 | | 6 2 | | |

Continued on next column

**TABLE LVIII– continued**

| TQC | ICSD | | NoE | | SF | |
|---|---|---|---|---|---|---|
| | | 8 2 | | 8 2 | | |
| | | 8 2 | | 7 2 | | |
| | | 9 2 | | 6 2 | | |
| | | 7 2 | | 9 2 | | |
| | | 6 2 | | 7 2 | | |
| | | 7 2 | | 8 2 | | |
| 127.2.19 | 193226 | | 32 | | Ca2Ge2Sn1 | |
| | -A | -GM | -M | -Z | -R | -X |
| | 8+9 4 | 7 2 | 8+9 4 | 7 2 | 3+4 4 | 3+4 4 |
| | 6+7 4 | 6 2 | 6+7 4 | 6 2 | 3+4 4 | 3+4 4 |
| | 6+7 4 | 7 2 | 6+7 4 | 7 2 | 3+4 4 | 3+4 4 |
| | 6+7 4 | 9 2 | 8+9 4 | 9 2 | 3+4 4 | 3+4 4 |
| | 8+9 4 | 8 2 | 8+9 4 | 8 2 | 3+4 4 | 3+4 4 |
| | 8+9 4 | 7 2 | 6+7 4 | 7 2 | 3+4 4 | 3+4 4 |
| | 8+9 4 | 9 2 | 6+7 4 | 9 2 | 3+4 4 | 3+4 4 |
| | 8+9 4 | 8 2 | 8+9 4 | 9 2 | 3+4 4 | 3+4 4 |
| | | 6 2 | | 8 2 | | |
| | | 8 2 | | 8 2 | | |
| | | 9 2 | | 6 2 | | |
| | | 7 2 | | 9 2 | | |
| | | 9 2 | | 6 2 | | |
| | | 6 2 | | 7 2 | | |
| | | 7 2 | | 9 2 | | |
| | | 7 2 | | 7 2 | | |
| 127.2.20 | 79239 | | 184 | | C5Ru6Th2 | |
| | -A | -GM | -M | -Z | -R | -X |
| | 6+7 4 | 7 2 | 8+9 4 | 9 2 | 3+4 4 | 3+4 4 |
| | 8+9 4 | 6 2 | 6+7 4 | 6 2 | 3+4 4 | 3+4 4 |
| | 6+7 4 | 8 2 | 8+9 4 | 7 2 | 3+4 4 | 3+4 4 |
| | 8+9 4 | 9 2 | 6+7 4 | 8 2 | 3+4 4 | 3+4 4 |
| | 6+7 4 | 6 2 | 8+9 4 | 8 2 | 3+4 4 | 3+4 4 |
| | 8+9 4 | 8 2 | 6+7 4 | 6 2 | 3+4 4 | 3+4 4 |
| | 8+9 4 | 7 2 | 6+7 4 | 9 2 | 3+4 4 | 3+4 4 |
| | 6+7 4 | 9 2 | 8+9 4 | 7 2 | 3+4 4 | 3+4 4 |
| | 8+9 4 | 6 2 | 8+9 4 | 8 2 | 3+4 4 | 3+4 4 |
| | 6+7 4 | 9 2 | 6+7 4 | 7 2 | 3+4 4 | 3+4 4 |
| | 6+7 4 | 8 2 | 6+7 4 | 6 2 | 3+4 4 | 3+4 4 |
| | 8+9 4 | 7 2 | 8+9 4 | 9 2 | 3+4 4 | 3+4 4 |
| | 6+7 4 | 7 2 | 6+7 4 | 7 2 | 3+4 4 | 3+4 4 |
| | 8+9 4 | 8 2 | 8+9 4 | 9 2 | 3+4 4 | 3+4 4 |
| | 8+9 4 | 7 2 | 6+7 4 | 8 2 | 3+4 4 | 3+4 4 |
| | 8+9 4 | 6 2 | 8+9 4 | 6 2 | 3+4 4 | 3+4 4 |
| | 8+9 4 | 9 2 | 6+7 4 | 9 2 | 3+4 4 | 3+4 4 |
| | 6+7 4 | 8 2 | 8+9 4 | 6 2 | 3+4 4 | 3+4 4 |
| | 6+7 4 | 9 2 | 8+9 4 | 7 2 | 3+4 4 | 3+4 4 |
| | 8+9 4 | 6 2 | 6+7 4 | 8 2 | 3+4 4 | 3+4 4 |
| | 8+9 4 | 7 2 | 8+9 4 | 7 2 | 3+4 4 | 3+4 4 |
| | 6+7 4 | 7 2 | 8+9 4 | 7 2 | 3+4 4 | 3+4 4 |
| | 8+9 4 | 6 2 | 8+9 4 | 6 2 | 3+4 4 | 3+4 4 |
| | 8+9 4 | 7 2 | 8+9 4 | 9 2 | 3+4 4 | 3+4 4 |
| | 6+7 4 | 8 2 | 6+7 4 | 8 2 | 3+4 4 | 3+4 4 |
| | 6+7 4 | 9 2 | 6+7 4 | 7 2 | 3+4 4 | 3+4 4 |
| | 8+9 4 | 6 2 | 6+7 4 | 9 2 | 3+4 4 | 3+4 4 |
| | 6+7 4 | 7 2 | 6+7 4 | 9 2 | 3+4 4 | 3+4 4 |
| | 8+9 4 | 8 2 | 8+9 4 | 7 2 | 3+4 4 | 3+4 4 |
| | 8+9 4 | 9 2 | 8+9 4 | 6 2 | 3+4 4 | 3+4 4 |
| | 8+9 4 | 6 2 | 8+9 4 | 8 2 | 3+4 4 | 3+4 4 |
| | 8+9 4 | 6 2 | 6+7 4 | 8 2 | 3+4 4 | 3+4 4 |
| | 6+7 4 | 8 2 | 6+7 4 | 9 2 | 3+4 4 | 3+4 4 |
| | 6+7 4 | 9 2 | 6+7 4 | 6 2 | 3+4 4 | 3+4 4 |
| | 8+9 4 | 7 2 | 8+9 4 | 7 2 | 3+4 4 | 3+4 4 |

Continued on next column

**TABLE LVIII– continued**

| TQC | ICSD | | NoE | | SF | |
|---|---|---|---|---|---|---|
| | 8+9 4 | 8 2 | 8+9 4 | 8 2 | 3+4 4 | 3+4 4 |
| | 8+9 4 | 9 2 | 6+7 4 | 8 2 | 3+4 4 | 3+4 4 |
| | 6+7 4 | 7 2 | 8+9 4 | 9 2 | 3+4 4 | 3+4 4 |
| | 6+7 4 | 7 2 | 8+9 4 | 7 2 | 3+4 4 | 3+4 4 |
| | 8+9 4 | 9 2 | 6+7 4 | 6 2 | 3+4 4 | 3+4 4 |
| | 8+9 4 | 6 2 | 8+9 4 | 8 2 | 3+4 4 | 3+4 4 |
| | 6+7 4 | 7 2 | 6+7 4 | 9 2 | 3+4 4 | 3+4 4 |
| | 8+9 4 | 8 2 | 6+7 4 | 7 2 | 3+4 4 | 3+4 4 |
| | 6+7 4 | 9 2 | 8+9 4 | 6 2 | 3+4 4 | 3+4 4 |
| | 8+9 4 | 8 2 | 6+7 4 | 9 2 | 3+4 4 | 3+4 4 |
| | 6+7 4 | 9 2 | 6+7 4 | 7 2 | 3+4 4 | 3+4 4 |
| | | 6 2 | | 6 2 | | |
| | | 7 2 | | 9 2 | | |
| | | 6 2 | | 6 2 | | |
| | | 9 2 | | 9 2 | | |
| | | 6 2 | | 8 2 | | |
| | | 7 2 | | 7 2 | | |
| | | 7 2 | | 8 2 | | |
| | | 8 2 | | 9 2 | | |
| | | 8 2 | | 8 2 | | |
| | | 9 2 | | 6 2 | | |
| | | 6 2 | | 9 2 | | |
| | | 9 2 | | 7 2 | | |
| | | 7 2 | | 6 2 | | |
| | | 6 2 | | 7 2 | | |
| | | 7 2 | | 8 2 | | |
| | | 7 2 | | 7 2 | | |
| | | 6 2 | | 9 2 | | |
| | | 8 2 | | 8 2 | | |
| | | 9 2 | | 6 2 | | |
| | | 6 2 | | 9 2 | | |
| | | 8 2 | | 8 2 | | |
| | | 7 2 | | 7 2 | | |
| | | 8 2 | | 9 2 | | |
| | | 6 2 | | 8 2 | | |
| | | 9 2 | | 7 2 | | |
| | | 8 2 | | 9 2 | | |
| | | 9 2 | | 8 2 | | |
| | | 7 2 | | 7 2 | | |
| | | 6 2 | | 6 2 | | |
| | | 6 2 | | 6 2 | | |
| | | 7 2 | | 6 2 | | |
| | | 7 2 | | 6 2 | | |
| | | 6 2 | | 8 2 | | |
| | | 6 2 | | 7 2 | | |
| | | 7 2 | | 7 2 | | |
| | | 6 2 | | 8 2 | | |
| | | 6 2 | | 9 2 | | |
| | | 9 2 | | 9 2 | | |
| | | 7 2 | | 9 2 | | |
| | | 8 2 | | 7 2 | | |
| | | 6 2 | | 6 2 | | |
| | | 8 2 | | 6 2 | | |
| | | 9 2 | | 6 2 | | |
| | | 7 2 | | 8 2 | | |
| | | 6 2 | | 8 2 | | |
| | | 6 2 | | 6 2 | | |
| 127.2.21 | 39204 | | 184 | | C5Ru6Th2 | |
| | -A | -GM | -M | -Z | -R | -X |
| | 6+7 4 | 7 2 | 8+9 4 | 7 2 | 3+4 4 | 3+4 4 |
| | 8+9 4 | 6 2 | 6+7 4 | 6 2 | 3+4 4 | 3+4 4 |

Continued on next column

| TQC | ICSD | | NoE | | SF | |
|---|---|---|---|---|---|---|
| | 8+9 4 | 8 2 | 8+9 4 | 8 2 | 3+4 4 | 3+4 4 |
| | 6+7 4 | 9 2 | 6+7 4 | 9 2 | 3+4 4 | 3+4 4 |
| | 8+9 4 | 6 2 | 8+9 4 | 6 2 | 3+4 4 | 3+4 4 |
| | 6+7 4 | 8 2 | 6+7 4 | 8 2 | 3+4 4 | 3+4 4 |
| | 6+7 4 | 7 2 | 6+7 4 | 7 2 | 3+4 4 | 3+4 4 |
| | 8+9 4 | 9 2 | 8+9 4 | 9 2 | 3+4 4 | 3+4 4 |
| | 6+7 4 | 6 2 | 8+9 4 | 6 2 | 3+4 4 | 3+4 4 |
| | 8+9 4 | 9 2 | 6+7 4 | 9 2 | 3+4 4 | 3+4 4 |
| | 8+9 4 | 8 2 | 6+7 4 | 8 2 | 3+4 4 | 3+4 4 |
| | 6+7 4 | 7 2 | 8+9 4 | 7 2 | 3+4 4 | 3+4 4 |
| | 8+9 4 | 7 2 | 6+7 4 | 9 2 | 3+4 4 | 3+4 4 |
| | 6+7 4 | 8 2 | 8+9 4 | 7 2 | 3+4 4 | 3+4 4 |
| | 6+7 4 | 9 2 | 6+7 4 | 6 2 | 3+4 4 | 3+4 4 |
| | 6+7 4 | 6 2 | 8+9 4 | 8 2 | 3+4 4 | 3+4 4 |
| | 6+7 4 | 7 2 | 6+7 4 | 7 2 | 3+4 4 | 3+4 4 |
| | 8+9 4 | 8 2 | 8+9 4 | 8 2 | 3+4 4 | 3+4 4 |
| | 8+9 4 | 9 2 | 8+9 4 | 9 2 | 3+4 4 | 3+4 4 |
| | 6+7 4 | 6 2 | 6+7 4 | 6 2 | 3+4 4 | 3+4 4 |
| | 6+7 4 | 7 2 | 8+9 4 | 9 2 | 3+4 4 | 3+4 4 |
| | 8+9 4 | 7 2 | 8+9 4 | 9 2 | 3+4 4 | 3+4 4 |
| | 6+7 4 | 6 2 | 8+9 4 | 8 2 | 3+4 4 | 3+4 4 |
| | 6+7 4 | 7 2 | 6+7 4 | 7 2 | 3+4 4 | 3+4 4 |
| | 8+9 4 | 8 2 | 8+9 4 | 6 2 | 3+4 4 | 3+4 4 |
| | 8+9 4 | 9 2 | 6+7 4 | 9 2 | 3+4 4 | 3+4 4 |
| | 8+9 4 | 6 2 | 6+7 4 | 7 2 | 3+4 4 | 3+4 4 |
| | 6+7 4 | 8 2 | 6+7 4 | 7 2 | 3+4 4 | 3+4 4 |
| | 6+7 4 | 9 2 | 8+9 4 | 6 2 | 3+4 4 | 3+4 4 |
| | 6+7 4 | 7 2 | 8+9 4 | 9 2 | 3+4 4 | 3+4 4 |
| | 8+9 4 | 6 2 | 8+9 4 | 7 2 | 3+4 4 | 3+4 4 |
| | 6+7 4 | 6 2 | 6+7 4 | 8 2 | 3+4 4 | 3+4 4 |
| | 6+7 4 | 8 2 | 6+7 4 | 6 2 | 3+4 4 | 3+4 4 |
| | 6+7 4 | 9 2 | 6+7 4 | 8 2 | 3+4 4 | 3+4 4 |
| | 8+9 4 | 7 2 | 8+9 4 | 6 2 | 3+4 4 | 3+4 4 |
| | 6+7 4 | 9 2 | 8+9 4 | 6 2 | 3+4 4 | 3+4 4 |
| | 8+9 4 | 8 2 | 6+7 4 | 9 2 | 3+4 4 | 3+4 4 |
| | 6+7 4 | 7 2 | 8+9 4 | 7 2 | 3+4 4 | 3+4 4 |
| | 8+9 4 | 9 2 | 8+9 4 | 9 2 | 3+4 4 | 3+4 4 |
| | 6+7 4 | 6 2 | 8+9 4 | 8 2 | 3+4 4 | 3+4 4 |
| | 6+7 4 | 7 2 | 6+7 4 | 6 2 | 3+4 4 | 3+4 4 |
| | 8+9 4 | 7 2 | 6+7 4 | 9 2 | 3+4 4 | 3+4 4 |
| | 6+7 4 | 8 2 | 6+7 4 | 8 2 | 3+4 4 | 3+4 4 |
| | 8+9 4 | 9 2 | 8+9 4 | 7 2 | 3+4 4 | 3+4 4 |
| | 6+7 4 | 8 2 | 6+7 4 | 9 2 | 3+4 4 | 3+4 4 |
| | 8+9 4 | 9 2 | 6+7 4 | 8 2 | 3+4 4 | 3+4 4 |
| | | 6 2 | | 7 2 | | |
| | | 7 2 | | 7 2 | | |
| | | 6 2 | | 6 2 | | |
| | | 6 2 | | 7 2 | | |
| | | 9 2 | | 8 2 | | |
| | | 7 2 | | 7 2 | | |
| | | 7 2 | | 9 2 | | |
| | | 8 2 | | 6 2 | | |
| | | 9 2 | | 6 2 | | |
| | | 8 2 | | 8 2 | | |
| | | 9 2 | | 9 2 | | |
| | | 6 2 | | 9 2 | | |
| | | 7 2 | | 7 2 | | |
| | | 6 2 | | 6 2 | | |
| | | 8 2 | | 7 2 | | |
| | | 7 2 | | 9 2 | | |
| | | 7 2 | | 8 2 | | |

Continued on next column

**TABLE LVIII– continued**

| TQC | ICSD | | NoE | | SF | |
|---|---|---|---|---|---|---|
| | | 6 2 | | 8 2 | | |
| | | 8 2 | | 6 2 | | |
| | | 9 2 | | 7 2 | | |
| | | 6 2 | | 9 2 | | |
| | | 8 2 | | 6 2 | | |
| | | 7 2 | | 7 2 | | |
| | | 6 2 | | 6 2 | | |
| | | 9 2 | | 9 2 | | |
| | | 8 2 | | 7 2 | | |
| | | 9 2 | | 6 2 | | |
| | | 7 2 | | 8 2 | | |
| | | 6 2 | | 9 2 | | |
| | | 6 2 | | 8 2 | | |
| | | 7 2 | | 8 2 | | |
| | | 6 2 | | 8 2 | | |
| | | 7 2 | | 6 2 | | |
| | | 6 2 | | 9 2 | | |
| | | 7 2 | | 7 2 | | |
| | | 6 2 | | 9 2 | | |
| | | 6 2 | | 6 2 | | |
| | | 9 2 | | 7 2 | | |
| | | 9 2 | | 7 2 | | |
| | | 7 2 | | 9 2 | | |
| | | 6 2 | | 8 2 | | |
| | | 7 2 | | 8 2 | | |
| | | 6 2 | | 8 2 | | |
| | | 8 2 | | 6 2 | | |
| | | 8 2 | | 6 2 | | |
| | | 6 2 | | 8 2 | | |
| 127.2.22 | 427156 | | 64 | | C2B2Gd1 | |
| | -A | -GM | -M | -Z | -R | -X |
| | 6+7 4 | 7 2 | 6+7 4 | 7 2 | 3+4 4 | 3+4 4 |
| | 8+9 4 | 7 2 | 8+9 4 | 7 2 | 3+4 4 | 3+4 4 |
| | 8+9 4 | 9 2 | 8+9 4 | 9 2 | 3+4 4 | 3+4 4 |
| | 8+9 4 | 9 2 | 8+9 4 | 9 2 | 3+4 4 | 3+4 4 |
| | 6+7 4 | 8 2 | 8+9 4 | 9 2 | 3+4 4 | 3+4 4 |
| | 8+9 4 | 9 2 | 6+7 4 | 9 2 | 3+4 4 | 3+4 4 |
| | 6+7 4 | 9 2 | 8+9 4 | 8 2 | 3+4 4 | 3+4 4 |
| | 8+9 4 | 8 2 | 6+7 4 | 8 2 | 3+4 4 | 3+4 4 |
| | 6+7 4 | 7 2 | 6+7 4 | 9 2 | 3+4 4 | 3+4 4 |
| | 8+9 4 | 6 2 | 8+9 4 | 8 2 | 3+4 4 | 3+4 4 |
| | 6+7 4 | 9 2 | 6+7 4 | 7 2 | 3+4 4 | 3+4 4 |
| | 8+9 4 | 8 2 | 8+9 4 | 6 2 | 3+4 4 | 3+4 4 |
| | 8+9 4 | 6 2 | 8+9 4 | 8 2 | 3+4 4 | 3+4 4 |
| | 8+9 4 | 7 2 | 8+9 4 | 9 2 | 3+4 4 | 3+4 4 |
| | 8+9 4 | 7 2 | 8+9 4 | 7 2 | 3+4 4 | 3+4 4 |
| | 8+9 4 | 9 2 | 8+9 4 | 6 2 | 3+4 4 | 3+4 4 |
| | | 8 2 | | 7 2 | | |
| | | 9 2 | | 9 2 | | |
| | | 7 2 | | 6 2 | | |
| | | 6 2 | | 9 2 | | |
| | | 6 2 | | 6 2 | | |
| | | 8 2 | | 8 2 | | |
| | | 8 2 | | 8 2 | | |
| | | 9 2 | | 7 2 | | |
| | | 8 2 | | 6 2 | | |
| | | 8 2 | | 8 2 | | |
| | | 9 2 | | 8 2 | | |
| | | 8 2 | | 9 2 | | |
| | | 9 2 | | 9 2 | | |
| | | 8 2 | | 8 2 | | |

Continued on next column

**TABLE LVIII– continued**

| TQC | ICSD | | NoE | | SF | |
|---|---|---|---|---|---|---|
| | | 7 2 | | 8 2 | | |
| | | 8 2 | | 9 2 | | |
| 127.2.23 | 94042 | | 72 | | C2B2Er1 | |
| | -A | -GM | -M | -Z | -R | -X |
| | 6+7 4 | 7 2 | 6+7 4 | 7 2 | 3+4 4 | 3+4 4 |
| | 8+9 4 | 7 2 | 8+9 4 | 7 2 | 3+4 4 | 3+4 4 |
| | 8+9 4 | 9 2 | 8+9 4 | 9 2 | 3+4 4 | 3+4 4 |
| | 8+9 4 | 9 2 | 8+9 4 | 9 2 | 3+4 4 | 3+4 4 |
| | 6+7 4 | 8 2 | 8+9 4 | 9 2 | 3+4 4 | 3+4 4 |
| | 8+9 4 | 9 2 | 6+7 4 | 9 2 | 3+4 4 | 3+4 4 |
| | 6+7 4 | 9 2 | 8+9 4 | 8 2 | 3+4 4 | 3+4 4 |
| | 8+9 4 | 8 2 | 6+7 4 | 8 2 | 3+4 4 | 3+4 4 |
| | 6+7 4 | 7 2 | 8+9 4 | 9 2 | 3+4 4 | 3+4 4 |
| | 8+9 4 | 6 2 | 6+7 4 | 8 2 | 3+4 4 | 3+4 4 |
| | 6+7 4 | 9 2 | 6+7 4 | 7 2 | 3+4 4 | 3+4 4 |
| | 8+9 4 | 8 2 | 8+9 4 | 6 2 | 3+4 4 | 3+4 4 |
| | 8+9 4 | 6 2 | 8+9 4 | 8 2 | 3+4 4 | 3+4 4 |
| | 8+9 4 | 7 2 | 8+9 4 | 7 2 | 3+4 4 | 3+4 4 |
| | 8+9 4 | 7 2 | 8+9 4 | 9 2 | 3+4 4 | 3+4 4 |
| | 8+9 4 | 9 2 | 8+9 4 | 7 2 | 3+4 4 | 3+4 4 |
| | 8+9 4 | 8 2 | 8+9 4 | 6 2 | 3+4 4 | 3+4 4 |
| | 8+9 4 | 9 2 | 8+9 4 | 9 2 | 3+4 4 | 3+4 4 |
| | | 7 2 | | 6 2 | | |
| | | 6 2 | | 9 2 | | |
| | | 6 2 | | 8 2 | | |
| | | 8 2 | | 8 2 | | |
| | | 8 2 | | 6 2 | | |
| | | 9 2 | | 7 2 | | |
| | | 8 2 | | 6 2 | | |
| | | 8 2 | | 8 2 | | |
| | | 9 2 | | 8 2 | | |
| | | 8 2 | | 9 2 | | |
| | | 9 2 | | 9 2 | | |
| | | 8 2 | | 8 2 | | |
| | | 8 2 | | 8 2 | | |
| | | 8 2 | | 9 2 | | |
| | | 9 2 | | 8 2 | | |
| | | 8 2 | | 9 2 | | |
| | | 9 2 | | 9 2 | | |
| | | 7 2 | | 8 2 | | |
| 127.2.24 | 94040 | | 68 | | C2B2Dy1 | |
| | -A | -GM | -M | -Z | -R | -X |
| | 6+7 4 | 7 2 | 6+7 4 | 7 2 | 3+4 4 | 3+4 4 |
| | 8+9 4 | 7 2 | 8+9 4 | 7 2 | 3+4 4 | 3+4 4 |
| | 8+9 4 | 9 2 | 8+9 4 | 9 2 | 3+4 4 | 3+4 4 |
| | 8+9 4 | 9 2 | 8+9 4 | 9 2 | 3+4 4 | 3+4 4 |
| | 6+7 4 | 8 2 | 8+9 4 | 9 2 | 3+4 4 | 3+4 4 |
| | 8+9 4 | 9 2 | 6+7 4 | 9 2 | 3+4 4 | 3+4 4 |
| | 6+7 4 | 9 2 | 8+9 4 | 8 2 | 3+4 4 | 3+4 4 |
| | 8+9 4 | 8 2 | 6+7 4 | 8 2 | 3+4 4 | 3+4 4 |
| | 6+7 4 | 7 2 | 6+7 4 | 9 2 | 3+4 4 | 3+4 4 |
| | 8+9 4 | 6 2 | 8+9 4 | 8 2 | 3+4 4 | 3+4 4 |
| | 6+7 4 | 9 2 | 6+7 4 | 7 2 | 3+4 4 | 3+4 4 |
| | 8+9 4 | 8 2 | 8+9 4 | 6 2 | 3+4 4 | 3+4 4 |
| | 8+9 4 | 6 2 | 8+9 4 | 8 2 | 3+4 4 | 3+4 4 |
| | 8+9 4 | 7 2 | 8+9 4 | 9 2 | 3+4 4 | 3+4 4 |
| | 8+9 4 | 7 2 | 8+9 4 | 7 2 | 3+4 4 | 3+4 4 |
| | 8+9 4 | 9 2 | 8+9 4 | 6 2 | 3+4 4 | 3+4 4 |
| | 8+9 4 | 8 2 | 8+9 4 | 7 2 | 3+4 4 | 3+4 4 |
| | | 9 2 | | 9 2 | | |
| | | 7 2 | | 6 2 | | |

Continued on next column

**TABLE LVIII– continued**

| TQC | ICSD | NoE | SF | | | |
|---|---|---|---|---|---|---|
| | | 6 2 | | 9 2 | | |
| | | 6 2 | | 8 2 | | |
| | | 8 2 | | 6 2 | | |
| | | 8 2 | | 8 2 | | |
| | | 9 2 | | 7 2 | | |
| | | 8 2 | | 6 2 | | |
| | | 8 2 | | 8 2 | | |
| | | 9 2 | | 8 2 | | |
| | | 8 2 | | 9 2 | | |
| | | 9 2 | | 9 2 | | |
| | | 8 2 | | 8 2 | | |
| | | 8 2 | | 8 2 | | |
| | | 7 2 | | 9 2 | | |
| | | 8 2 | | 8 2 | | |
| | | 9 2 | | 9 2 | | |
| 127.2.25 | 94036 | 52 | C2B2Ce1 | | | |
| | -A | -GM | -M | -Z | -R | -X |
| | 6+7 4 | 7 2 | 6+7 4 | 7 2 | 3+4 4 | 3+4 4 |
| | 8+9 4 | 7 2 | 8+9 4 | 7 2 | 3+4 4 | 3+4 4 |
| | 8+9 4 | 9 2 | 8+9 4 | 9 2 | 3+4 4 | 3+4 4 |
| | 8+9 4 | 9 2 | 8+9 4 | 9 2 | 3+4 4 | 3+4 4 |
| | 6+7 4 | 8 2 | 8+9 4 | 9 2 | 3+4 4 | 3+4 4 |
| | 8+9 4 | 9 2 | 6+7 4 | 9 2 | 3+4 4 | 3+4 4 |
| | 6+7 4 | 9 2 | 8+9 4 | 8 2 | 3+4 4 | 3+4 4 |
| | 8+9 4 | 8 2 | 6+7 4 | 8 2 | 3+4 4 | 3+4 4 |
| | 6+7 4 | 7 2 | 6+7 4 | 9 2 | 3+4 4 | 3+4 4 |
| | 8+9 4 | 6 2 | 8+9 4 | 8 2 | 3+4 4 | 3+4 4 |
| | 6+7 4 | 9 2 | 6+7 4 | 7 2 | 3+4 4 | 3+4 4 |
| | 8+9 4 | 8 2 | 8+9 4 | 6 2 | 3+4 4 | 3+4 4 |
| | 8+9 4 | 6 2 | 6+7 4 | 8 2 | 3+4 4 | 3+4 4 |
| | | 7 2 | | 7 2 | | |
| | | 7 2 | | 6 2 | | |
| | | 9 2 | | 9 2 | | |
| | | 8 2 | | 7 2 | | |
| | | 9 2 | | 9 2 | | |
| | | 7 2 | | 6 2 | | |
| | | 6 2 | | 6 2 | | |
| | | 8 2 | | 8 2 | | |
| | | 6 2 | | 9 2 | | |
| | | 9 2 | | 7 2 | | |
| | | 8 2 | | 6 2 | | |
| | | 7 2 | | 8 2 | | |
| | | 8 2 | | 9 2 | | |
| 127.2.26 | 88561 | 52 | C2B2Ce1 | | | |
| | -A | -GM | -M | -Z | -R | -X |
| | 6+7 4 | 7 2 | 6+7 4 | 7 2 | 3+4 4 | 3+4 4 |
| | 8+9 4 | 7 2 | 8+9 4 | 7 2 | 3+4 4 | 3+4 4 |
| | 8+9 4 | 9 2 | 8+9 4 | 9 2 | 3+4 4 | 3+4 4 |
| | 8+9 4 | 9 2 | 8+9 4 | 9 2 | 3+4 4 | 3+4 4 |
| | 6+7 4 | 8 2 | 8+9 4 | 9 2 | 3+4 4 | 3+4 4 |
| | 8+9 4 | 9 2 | 6+7 4 | 9 2 | 3+4 4 | 3+4 4 |
| | 6+7 4 | 9 2 | 8+9 4 | 8 2 | 3+4 4 | 3+4 4 |
| | 8+9 4 | 8 2 | 6+7 4 | 8 2 | 3+4 4 | 3+4 4 |
| | 6+7 4 | 7 2 | 6+7 4 | 9 2 | 3+4 4 | 3+4 4 |
| | 8+9 4 | 6 2 | 8+9 4 | 8 2 | 3+4 4 | 3+4 4 |
| | 6+7 4 | 9 2 | 6+7 4 | 7 2 | 3+4 4 | 3+4 4 |
| | 8+9 4 | 8 2 | 8+9 4 | 6 2 | 3+4 4 | 3+4 4 |
| | 8+9 4 | 6 2 | 6+7 4 | 8 2 | 3+4 4 | 3+4 4 |
| | | 7 2 | | 7 2 | | |
| | | 7 2 | | 6 2 | | |
| | | 9 2 | | 9 2 | | |

Continued on next column

**TABLE LVIII– continued**

| TQC | ICSD | NoE | SF | | | |
|---|---|---|---|---|---|---|
| | | 8 2 | | 7 2 | | |
| | | 9 2 | | 9 2 | | |
| | | 7 2 | | 6 2 | | |
| | | 6 2 | | 6 2 | | |
| | | 8 2 | | 8 2 | | |
| | | 6 2 | | 9 2 | | |
| | | 9 2 | | 7 2 | | |
| | | 8 2 | | 6 2 | | |
| | | 7 2 | | 8 2 | | |
| | | 8 2 | | 9 2 | | |
| 127.2.27 | 88560 | 52 | C2B2Ce1 | | | |
| | -A | -GM | -M | -Z | -R | -X |
| | 6+7 4 | 7 2 | 6+7 4 | 7 2 | 3+4 4 | 3+4 4 |
| | 8+9 4 | 7 2 | 8+9 4 | 7 2 | 3+4 4 | 3+4 4 |
| | 8+9 4 | 9 2 | 8+9 4 | 9 2 | 3+4 4 | 3+4 4 |
| | 8+9 4 | 9 2 | 8+9 4 | 9 2 | 3+4 4 | 3+4 4 |
| | 6+7 4 | 8 2 | 8+9 4 | 9 2 | 3+4 4 | 3+4 4 |
| | 8+9 4 | 9 2 | 6+7 4 | 9 2 | 3+4 4 | 3+4 4 |
| | 6+7 4 | 9 2 | 8+9 4 | 8 2 | 3+4 4 | 3+4 4 |
| | 8+9 4 | 8 2 | 6+7 4 | 8 2 | 3+4 4 | 3+4 4 |
| | 6+7 4 | 7 2 | 6+7 4 | 9 2 | 3+4 4 | 3+4 4 |
| | 8+9 4 | 6 2 | 8+9 4 | 8 2 | 3+4 4 | 3+4 4 |
| | 6+7 4 | 9 2 | 6+7 4 | 7 2 | 3+4 4 | 3+4 4 |
| | 8+9 4 | 8 2 | 8+9 4 | 6 2 | 3+4 4 | 3+4 4 |
| | 8+9 4 | 6 2 | 6+7 4 | 8 2 | 3+4 4 | 3+4 4 |
| | | 7 2 | | 7 2 | | |
| | | 7 2 | | 6 2 | | |
| | | 9 2 | | 9 2 | | |
| | | 8 2 | | 7 2 | | |
| | | 9 2 | | 9 2 | | |
| | | 7 2 | | 6 2 | | |
| | | 6 2 | | 6 2 | | |
| | | 8 2 | | 8 2 | | |
| | | 6 2 | | 9 2 | | |
| | | 9 2 | | 7 2 | | |
| | | 8 2 | | 6 2 | | |
| | | 7 2 | | 8 2 | | |
| | | 8 2 | | 9 2 | | |
| 127.2.28 | 280180 | 52 | C2B2Ce1 | | | |
| | -A | -GM | -M | -Z | -R | -X |
| | 6+7 4 | 7 2 | 6+7 4 | 7 2 | 3+4 4 | 3+4 4 |
| | 8+9 4 | 7 2 | 8+9 4 | 7 2 | 3+4 4 | 3+4 4 |
| | 8+9 4 | 9 2 | 8+9 4 | 9 2 | 3+4 4 | 3+4 4 |
| | 8+9 4 | 9 2 | 8+9 4 | 9 2 | 3+4 4 | 3+4 4 |
| | 6+7 4 | 8 2 | 8+9 4 | 9 2 | 3+4 4 | 3+4 4 |
| | 8+9 4 | 9 2 | 6+7 4 | 9 2 | 3+4 4 | 3+4 4 |
| | 6+7 4 | 9 2 | 8+9 4 | 8 2 | 3+4 4 | 3+4 4 |
| | 8+9 4 | 8 2 | 6+7 4 | 8 2 | 3+4 4 | 3+4 4 |
| | 6+7 4 | 7 2 | 6+7 4 | 9 2 | 3+4 4 | 3+4 4 |
| | 8+9 4 | 6 2 | 8+9 4 | 8 2 | 3+4 4 | 3+4 4 |
| | 6+7 4 | 9 2 | 6+7 4 | 7 2 | 3+4 4 | 3+4 4 |
| | 8+9 4 | 8 2 | 8+9 4 | 6 2 | 3+4 4 | 3+4 4 |
| | 8+9 4 | 6 2 | 6+7 4 | 8 2 | 3+4 4 | 3+4 4 |
| | | 7 2 | | 7 2 | | |
| | | 7 2 | | 6 2 | | |
| | | 9 2 | | 9 2 | | |
| | | 8 2 | | 7 2 | | |
| | | 9 2 | | 9 2 | | |
| | | 7 2 | | 6 2 | | |
| | | 6 2 | | 6 2 | | |
| | | 8 2 | | 8 2 | | |

Continued on next column

**TABLE LVIII– continued**

| TQC | ICSD | | NoE | | SF | |
|---|---|---|---|---|---|---|
| | | 6 2 | | 9 2 | | |
| | | 9 2 | | 7 2 | | |
| | | 8 2 | | 6 2 | | |
| | | 7 2 | | 8 2 | | |
| | | 8 2 | | 9 2 | | |
| 127.2.29 | 202173 | | 176 | | Ba1O5Pd1Tb2 | |
| | -A | -GM | -M | -Z | -R | -X |
| | 6+7 4 | 7 2 | 8+9 4 | 9 2 | 3+4 4 | 3+4 4 |
| | 8+9 4 | 6 2 | 6+7 4 | 6 2 | 3+4 4 | 3+4 4 |
| | 6+7 4 | 8 2 | 6+7 4 | 7 2 | 3+4 4 | 3+4 4 |
| | 6+7 4 | 9 2 | 8+9 4 | 8 2 | 3+4 4 | 3+4 4 |
| | 8+9 4 | 7 2 | 6+7 4 | 7 2 | 3+4 4 | 3+4 4 |
| | 8+9 4 | 7 2 | 8+9 4 | 7 2 | 3+4 4 | 3+4 4 |
| | 6+7 4 | 6 2 | 6+7 4 | 8 2 | 3+4 4 | 3+4 4 |
| | 8+9 4 | 8 2 | 6+7 4 | 6 2 | 3+4 4 | 3+4 4 |
| | 6+7 4 | 7 2 | 8+9 4 | 9 2 | 3+4 4 | 3+4 4 |
| | 8+9 4 | 9 2 | 8+9 4 | 7 2 | 3+4 4 | 3+4 4 |
| | 6+7 4 | 6 2 | 8+9 4 | 8 2 | 3+4 4 | 3+4 4 |
| | 8+9 4 | 9 2 | 6+7 4 | 7 2 | 3+4 4 | 3+4 4 |
| | 6+7 4 | 7 2 | 6+7 4 | 7 2 | 3+4 4 | 3+4 4 |
| | 8+9 4 | 9 2 | 6+7 4 | 9 2 | 3+4 4 | 3+4 4 |
| | 8+9 4 | 8 2 | 8+9 4 | 6 2 | 3+4 4 | 3+4 4 |
| | 8+9 4 | 7 2 | 8+9 4 | 9 2 | 3+4 4 | 3+4 4 |
| | 8+9 4 | 6 2 | 8+9 4 | 8 2 | 3+4 4 | 3+4 4 |
| | 8+9 4 | 8 2 | 8+9 4 | 6 2 | 3+4 4 | 3+4 4 |
| | 8+9 4 | 7 2 | 8+9 4 | 6 2 | 3+4 4 | 3+4 4 |
| | 8+9 4 | 8 2 | 8+9 4 | 9 2 | 3+4 4 | 3+4 4 |
| | 8+9 4 | 6 2 | 8+9 4 | 6 2 | 3+4 4 | 3+4 4 |
| | 6+7 4 | 7 2 | 6+7 4 | 9 2 | 3+4 4 | 3+4 4 |
| | 8+9 4 | 8 2 | 6+7 4 | 8 2 | 3+4 4 | 3+4 4 |
| | 6+7 4 | 9 2 | 6+7 4 | 7 2 | 3+4 4 | 3+4 4 |
| | 6+7 4 | 6 2 | 8+9 4 | 7 2 | 3+4 4 | 3+4 4 |
| | 6+7 4 | 7 2 | 6+7 4 | 9 2 | 3+4 4 | 3+4 4 |
| | 6+7 4 | 9 2 | 8+9 4 | 8 2 | 3+4 4 | 3+4 4 |
| | 6+7 4 | 7 2 | 6+7 4 | 9 2 | 3+4 4 | 3+4 4 |
| | 6+7 4 | 9 2 | 6+7 4 | 9 2 | 3+4 4 | 3+4 4 |
| | 8+9 4 | 9 2 | 8+9 4 | 9 2 | 3+4 4 | 3+4 4 |
| | 6+7 4 | 8 2 | 8+9 4 | 8 2 | 3+4 4 | 3+4 4 |
| | 8+9 4 | 9 2 | 8+9 4 | 8 2 | 3+4 4 | 3+4 4 |
| | 6+7 4 | 9 2 | 8+9 4 | 9 2 | 3+4 4 | 3+4 4 |
| | 8+9 4 | 8 2 | 8+9 4 | 9 2 | 3+4 4 | 3+4 4 |
| | 8+9 4 | 6 2 | 8+9 4 | 6 2 | 3+4 4 | 3+4 4 |
| | 8+9 4 | 7 2 | 8+9 4 | 7 2 | 3+4 4 | 3+4 4 |
| | 6+7 4 | 6 2 | 6+7 4 | 7 2 | 3+4 4 | 3+4 4 |
| | 8+9 4 | 7 2 | 6+7 4 | 6 2 | 3+4 4 | 3+4 4 |
| | 8+9 4 | 7 2 | 8+9 4 | 6 2 | 3+4 4 | 3+4 4 |
| | 6+7 4 | 6 2 | 6+7 4 | 6 2 | 3+4 4 | 3+4 4 |
| | 6+7 4 | 7 2 | 8+9 4 | 7 2 | 3+4 4 | 3+4 4 |
| | 8+9 4 | 6 2 | 8+9 4 | 7 2 | 3+4 4 | 3+4 4 |
| | 8+9 4 | 6 2 | 6+7 4 | 7 2 | 3+4 4 | 3+4 4 |
| | 6+7 4 | 9 2 | 8+9 4 | 8 2 | 3+4 4 | 3+4 4 |
| | | 9 2 | | 6 2 | | |
| | | 8 2 | | 9 2 | | |
| | | 7 2 | | 9 2 | | |
| | | 8 2 | | 8 2 | | |
| | | 9 2 | | 9 2 | | |
| | | 8 2 | | 7 2 | | |
| | | 9 2 | | 6 2 | | |
| | | 8 2 | | 8 2 | | |
| | | 8 2 | | 7 2 | | |
| | | 9 2 | | 9 2 | | |

Continued on next column

**TABLE LVIII– continued**

| TQC | ICSD | | NoE | | SF | |
|---|---|---|---|---|---|---|
| | | 8 2 | | 8 2 | | |
| | | 6 2 | | 6 2 | | |
| | | 7 2 | | 7 2 | | |
| | | 9 2 | | 8 2 | | |
| | | 8 2 | | 6 2 | | |
| | | 9 2 | | 7 2 | | |
| | | 7 2 | | 9 2 | | |
| | | 6 2 | | 9 2 | | |
| | | 9 2 | | 8 2 | | |
| | | 9 2 | | 7 2 | | |
| | | 8 2 | | 6 2 | | |
| | | 6 2 | | 7 2 | | |
| | | 9 2 | | 6 2 | | |
| | | 7 2 | | 7 2 | | |
| | | 6 2 | | 7 2 | | |
| | | 7 2 | | 6 2 | | |
| | | 6 2 | | 6 2 | | |
| | | 7 2 | | 7 2 | | |
| | | 6 2 | | 8 2 | | |
| | | 7 2 | | 9 2 | | |
| | | 9 2 | | 8 2 | | |
| | | 6 2 | | 8 2 | | |
| | | 8 2 | | 9 2 | | |
| | | 7 2 | | 7 2 | | |
| | | 9 2 | | 6 2 | | |
| | | 6 2 | | 9 2 | | |
| | | 8 2 | | 7 2 | | |
| | | 7 2 | | 6 2 | | |
| | | 9 2 | | 7 2 | | |
| | | 8 2 | | 6 2 | | |
| | | 6 2 | | 8 2 | | |
| | | 7 2 | | 8 2 | | |
| | | 8 2 | | 9 2 | | |
| | | 6 2 | | 6 2 | | |
| 127.2.30 | 404497 | | 184 | | Ba1Ho2O5Pd1 | |
| | -A | -GM | -M | -Z | -R | -X |
| | 6+7 4 | 7 2 | 8+9 4 | 9 2 | 3+4 4 | 3+4 4 |
| | 8+9 4 | 6 2 | 6+7 4 | 6 2 | 3+4 4 | 3+4 4 |
| | 6+7 4 | 8 2 | 6+7 4 | 7 2 | 3+4 4 | 3+4 4 |
| | 6+7 4 | 9 2 | 8+9 4 | 8 2 | 3+4 4 | 3+4 4 |
| | 8+9 4 | 7 2 | 6+7 4 | 7 2 | 3+4 4 | 3+4 4 |
| | 8+9 4 | 7 2 | 8+9 4 | 7 2 | 3+4 4 | 3+4 4 |
| | 6+7 4 | 6 2 | 6+7 4 | 8 2 | 3+4 4 | 3+4 4 |
| | 8+9 4 | 8 2 | 6+7 4 | 6 2 | 3+4 4 | 3+4 4 |
| | 6+7 4 | 7 2 | 8+9 4 | 9 2 | 3+4 4 | 3+4 4 |
| | 8+9 4 | 9 2 | 8+9 4 | 7 2 | 3+4 4 | 3+4 4 |
| | 6+7 4 | 6 2 | 8+9 4 | 8 2 | 3+4 4 | 3+4 4 |
| | 8+9 4 | 9 2 | 6+7 4 | 7 2 | 3+4 4 | 3+4 4 |
| | 6+7 4 | 7 2 | 6+7 4 | 9 2 | 3+4 4 | 3+4 4 |
| | 8+9 4 | 9 2 | 6+7 4 | 7 2 | 3+4 4 | 3+4 4 |
| | 8+9 4 | 8 2 | 8+9 4 | 6 2 | 3+4 4 | 3+4 4 |
| | 8+9 4 | 7 2 | 8+9 4 | 9 2 | 3+4 4 | 3+4 4 |
| | 8+9 4 | 6 2 | 8+9 4 | 8 2 | 3+4 4 | 3+4 4 |
| | 8+9 4 | 8 2 | 8+9 4 | 6 2 | 3+4 4 | 3+4 4 |
| | 8+9 4 | 7 2 | 8+9 4 | 6 2 | 3+4 4 | 3+4 4 |
| | 8+9 4 | 8 2 | 8+9 4 | 9 2 | 3+4 4 | 3+4 4 |
| | 6+7 4 | 6 2 | 6+7 4 | 6 2 | 3+4 4 | 3+4 4 |
| | 8+9 4 | 7 2 | 8+9 4 | 9 2 | 3+4 4 | 3+4 4 |
| | 6+7 4 | 8 2 | 6+7 4 | 8 2 | 3+4 4 | 3+4 4 |
| | 8+9 4 | 6 2 | 6+7 4 | 7 2 | 3+4 4 | 3+4 4 |
| | 6+7 4 | 9 2 | 8+9 4 | 7 2 | 3+4 4 | 3+4 4 |

Continued on next column

**TABLE LVIII– continued**

| TQC | ICSD | | NoE | | SF | |
|---|---|---|---|---|---|---|
| | 6+7 4 | 9 2 | 6+7 4 | 8 2 | 3+4 4 | 3+4 4 |
| | 6+7 4 | 7 2 | 8+9 4 | 9 2 | 3+4 4 | 3+4 4 |
| | 6+7 4 | 7 2 | 6+7 4 | 9 2 | 3+4 4 | 3+4 4 |
| | 8+9 4 | 9 2 | 6+7 4 | 9 2 | 3+4 4 | 3+4 4 |
| | 6+7 4 | 9 2 | 8+9 4 | 9 2 | 3+4 4 | 3+4 4 |
| | 8+9 4 | 8 2 | 8+9 4 | 8 2 | 3+4 4 | 3+4 4 |
| | 6+7 4 | 9 2 | 8+9 4 | 8 2 | 3+4 4 | 3+4 4 |
| | 6+7 4 | 8 2 | 8+9 4 | 9 2 | 3+4 4 | 3+4 4 |
| | 8+9 4 | 9 2 | 8+9 4 | 9 2 | 3+4 4 | 3+4 4 |
| | 8+9 4 | 6 2 | 8+9 4 | 6 2 | 3+4 4 | 3+4 4 |
| | 8+9 4 | 7 2 | 6+7 4 | 7 2 | 3+4 4 | 3+4 4 |
| | 6+7 4 | 6 2 | 6+7 4 | 7 2 | 3+4 4 | 3+4 4 |
| | 8+9 4 | 7 2 | 8+9 4 | 6 2 | 3+4 4 | 3+4 4 |
| | 6+7 4 | 7 2 | 8+9 4 | 6 2 | 3+4 4 | 3+4 4 |
| | 8+9 4 | 6 2 | 6+7 4 | 6 2 | 3+4 4 | 3+4 4 |
| | 6+7 4 | 7 2 | 8+9 4 | 7 2 | 3+4 4 | 3+4 4 |
| | 8+9 4 | 6 2 | 8+9 4 | 7 2 | 3+4 4 | 3+4 4 |
| | 8+9 4 | 9 2 | 6+7 4 | 7 2 | 3+4 4 | 3+4 4 |
| | 6+7 4 | 6 2 | 8+9 4 | 8 2 | 3+4 4 | 3+4 4 |
| | 8+9 4 | 8 2 | 6+7 4 | 6 2 | 3+4 4 | 3+4 4 |
| | 6+7 4 | 9 2 | 6+7 4 | 9 2 | 3+4 4 | 3+4 4 |
| | | 7 2 | | 9 2 | | |
| | | 8 2 | | 8 2 | | |
| | | 9 2 | | 9 2 | | |
| | | 8 2 | | 7 2 | | |
| | | 9 2 | | 6 2 | | |
| | | 9 2 | | 8 2 | | |
| | | 8 2 | | 8 2 | | |
| | | 8 2 | | 7 2 | | |
| | | 6 2 | | 9 2 | | |
| | | 8 2 | | 6 2 | | |
| | | 7 2 | | 7 2 | | |
| | | 9 2 | | 8 2 | | |
| | | 8 2 | | 6 2 | | |
| | | 9 2 | | 9 2 | | |
| | | 6 2 | | 7 2 | | |
| | | 7 2 | | 9 2 | | |
| | | 9 2 | | 8 2 | | |
| | | 6 2 | | 7 2 | | |
| | | 8 2 | | 6 2 | | |
| | | 9 2 | | 7 2 | | |
| | | 7 2 | | 6 2 | | |
| | | 6 2 | | 7 2 | | |
| | | 7 2 | | 7 2 | | |
| | | 6 2 | | 8 2 | | |
| | | 9 2 | | 6 2 | | |
| | | 7 2 | | 9 2 | | |
| | | 6 2 | | 7 2 | | |
| | | 7 2 | | 8 2 | | |
| | | 9 2 | | 6 2 | | |
| | | 6 2 | | 9 2 | | |
| | | 7 2 | | 8 2 | | |
| | | 8 2 | | 9 2 | | |
| | | 8 2 | | 7 2 | | |
| | | 9 2 | | 6 2 | | |
| | | 6 2 | | 6 2 | | |
| | | 7 2 | | 6 2 | | |
| | | 9 2 | | 7 2 | | |
| | | 8 2 | | 7 2 | | |
| | | 6 2 | | 8 2 | | |
| | | 7 2 | | 8 2 | | |

Continued on next column

**TABLE LVIII– continued**

| TQC | ICSD | | NoE | | SF | |
|---|---|---|---|---|---|---|
| | | 8 2 | | 9 2 | | |
| | | 6 2 | | 6 2 | | |
| | | 6 2 | | 8 2 | | |
| | | 7 2 | | 9 2 | | |
| | | 9 2 | | 6 2 | | |
| | | 8 2 | | 9 2 | | |
| 127.2.31 | 404496 | | 180 | Ba1Dy2O5Pd1 | | |
| | -A | -GM | -M | -Z | -R | -X |
| | 6+7 4 | 7 2 | 8+9 4 | 9 2 | 3+4 4 | 3+4 4 |
| | 8+9 4 | 6 2 | 6+7 4 | 6 2 | 3+4 4 | 3+4 4 |
| | 6+7 4 | 8 2 | 6+7 4 | 7 2 | 3+4 4 | 3+4 4 |
| | 6+7 4 | 9 2 | 8+9 4 | 8 2 | 3+4 4 | 3+4 4 |
| | 8+9 4 | 7 2 | 6+7 4 | 7 2 | 3+4 4 | 3+4 4 |
| | 8+9 4 | 7 2 | 8+9 4 | 7 2 | 3+4 4 | 3+4 4 |
| | 6+7 4 | 6 2 | 6+7 4 | 8 2 | 3+4 4 | 3+4 4 |
| | 8+9 4 | 8 2 | 6+7 4 | 6 2 | 3+4 4 | 3+4 4 |
| | 6+7 4 | 7 2 | 8+9 4 | 9 2 | 3+4 4 | 3+4 4 |
| | 8+9 4 | 9 2 | 8+9 4 | 7 2 | 3+4 4 | 3+4 4 |
| | 6+7 4 | 6 2 | 8+9 4 | 8 2 | 3+4 4 | 3+4 4 |
| | 8+9 4 | 9 2 | 6+7 4 | 7 2 | 3+4 4 | 3+4 4 |
| | 6+7 4 | 7 2 | 6+7 4 | 9 2 | 3+4 4 | 3+4 4 |
| | 8+9 4 | 9 2 | 6+7 4 | 7 2 | 3+4 4 | 3+4 4 |
| | 8+9 4 | 8 2 | 8+9 4 | 6 2 | 3+4 4 | 3+4 4 |
| | 8+9 4 | 7 2 | 8+9 4 | 9 2 | 3+4 4 | 3+4 4 |
| | 8+9 4 | 6 2 | 8+9 4 | 8 2 | 3+4 4 | 3+4 4 |
| | 8+9 4 | 8 2 | 8+9 4 | 6 2 | 3+4 4 | 3+4 4 |
| | 8+9 4 | 7 2 | 8+9 4 | 6 2 | 3+4 4 | 3+4 4 |
| | 8+9 4 | 8 2 | 8+9 4 | 9 2 | 3+4 4 | 3+4 4 |
| | 6+7 4 | 6 2 | 6+7 4 | 6 2 | 3+4 4 | 3+4 4 |
| | 8+9 4 | 7 2 | 8+9 4 | 9 2 | 3+4 4 | 3+4 4 |
| | 6+7 4 | 8 2 | 6+7 4 | 8 2 | 3+4 4 | 3+4 4 |
| | 8+9 4 | 9 2 | 6+7 4 | 7 2 | 3+4 4 | 3+4 4 |
| | 6+7 4 | 6 2 | 8+9 4 | 7 2 | 3+4 4 | 3+4 4 |
| | 6+7 4 | 9 2 | 6+7 4 | 9 2 | 3+4 4 | 3+4 4 |
| | 6+7 4 | 7 2 | 8+9 4 | 8 2 | 3+4 4 | 3+4 4 |
| | 6+7 4 | 7 2 | 6+7 4 | 9 2 | 3+4 4 | 3+4 4 |
| | 8+9 4 | 9 2 | 6+7 4 | 9 2 | 3+4 4 | 3+4 4 |
| | 6+7 4 | 9 2 | 8+9 4 | 9 2 | 3+4 4 | 3+4 4 |
| | 8+9 4 | 8 2 | 8+9 4 | 8 2 | 3+4 4 | 3+4 4 |
| | 6+7 4 | 9 2 | 8+9 4 | 8 2 | 3+4 4 | 3+4 4 |
| | 6+7 4 | 8 2 | 8+9 4 | 9 2 | 3+4 4 | 3+4 4 |
| | 8+9 4 | 9 2 | 8+9 4 | 9 2 | 3+4 4 | 3+4 4 |
| | 8+9 4 | 6 2 | 8+9 4 | 6 2 | 3+4 4 | 3+4 4 |
| | 8+9 4 | 7 2 | 8+9 4 | 7 2 | 3+4 4 | 3+4 4 |
| | 6+7 4 | 6 2 | 6+7 4 | 7 2 | 3+4 4 | 3+4 4 |
| | 8+9 4 | 7 2 | 6+7 4 | 6 2 | 3+4 4 | 3+4 4 |
| | 8+9 4 | 7 2 | 8+9 4 | 6 2 | 3+4 4 | 3+4 4 |
| | 6+7 4 | 6 2 | 6+7 4 | 6 2 | 3+4 4 | 3+4 4 |
| | 6+7 4 | 7 2 | 8+9 4 | 7 2 | 3+4 4 | 3+4 4 |
| | 8+9 4 | 9 2 | 8+9 4 | 7 2 | 3+4 4 | 3+4 4 |
| | 8+9 4 | 6 2 | 6+7 4 | 7 2 | 3+4 4 | 3+4 4 |
| | 6+7 4 | 6 2 | 8+9 4 | 8 2 | 3+4 4 | 3+4 4 |
| | 8+9 4 | 9 2 | 6+7 4 | 6 2 | 3+4 4 | 3+4 4 |
| | | 8 2 | | 9 2 | | |
| | | 7 2 | | 9 2 | | |
| | | 8 2 | | 8 2 | | |
| | | 9 2 | | 7 2 | | |
| | | 8 2 | | 6 2 | | |
| | | 9 2 | | 9 2 | | |
| | | 9 2 | | 8 2 | | |
| | | 8 2 | | 8 2 | | |

Continued on next column

**TABLE LVIII– continued**

| TQC | ICSD | | NoE | | SF | |
|---|---|---|---|---|---|---|
| | | 8 2 | | 7 2 | | |
| | | 8 2 | | 9 2 | | |
| | | 6 2 | | 6 2 | | |
| | | 7 2 | | 7 2 | | |
| | | 9 2 | | 8 2 | | |
| | | 8 2 | | 6 2 | | |
| | | 9 2 | | 9 2 | | |
| | | 7 2 | | 7 2 | | |
| | | 6 2 | | 9 2 | | |
| | | 9 2 | | 8 2 | | |
| | | 6 2 | | 7 2 | | |
| | | 8 2 | | 6 2 | | |
| | | 9 2 | | 7 2 | | |
| | | 7 2 | | 6 2 | | |
| | | 6 2 | | 7 2 | | |
| | | 7 2 | | 7 2 | | |
| | | 6 2 | | 6 2 | | |
| | | 9 2 | | 6 2 | | |
| | | 7 2 | | 8 2 | | |
| | | 6 2 | | 7 2 | | |
| | | 7 2 | | 9 2 | | |
| | | 9 2 | | 8 2 | | |
| | | 6 2 | | 9 2 | | |
| | | 7 2 | | 8 2 | | |
| | | 8 2 | | 7 2 | | |
| | | 8 2 | | 9 2 | | |
| | | 9 2 | | 6 2 | | |
| | | 6 2 | | 7 2 | | |
| | | 7 2 | | 6 2 | | |
| | | 9 2 | | 6 2 | | |
| | | 8 2 | | 7 2 | | |
| | | 6 2 | | 8 2 | | |
| | | 7 2 | | 8 2 | | |
| | | 8 2 | | 6 2 | | |
| | | 6 2 | | 9 2 | | |
| | | 6 2 | | 8 2 | | |
| | | 7 2 | | 9 2 | | |
| 127.2.32 | 67222 | | 92 | | B4Y1 | |
| | -A | -GM | -M | -Z | -R | -X |
| | 8+9 4 | 7 2 | 8+9 4 | 7 2 | 3+4 4 | 3+4 4 |
| | 6+7 4 | 8 2 | 6+7 4 | 8 2 | 3+4 4 | 3+4 4 |
| | 8+9 4 | 9 2 | 8+9 4 | 9 2 | 3+4 4 | 3+4 4 |
| | 6+7 4 | 6 2 | 6+7 4 | 6 2 | 3+4 4 | 3+4 4 |
| | 6+7 4 | 7 2 | 6+7 4 | 7 2 | 3+4 4 | 3+4 4 |
| | 8+9 4 | 8 2 | 8+9 4 | 8 2 | 3+4 4 | 3+4 4 |
| | 8+9 4 | 6 2 | 8+9 4 | 6 2 | 3+4 4 | 3+4 4 |
| | 6+7 4 | 9 2 | 6+7 4 | 9 2 | 3+4 4 | 3+4 4 |
| | 8+9 4 | 7 2 | 6+7 4 | 7 2 | 3+4 4 | 3+4 4 |
| | 6+7 4 | 9 2 | 8+9 4 | 9 2 | 3+4 4 | 3+4 4 |
| | 6+7 4 | 6 2 | 8+9 4 | 6 2 | 3+4 4 | 3+4 4 |
| | 6+7 4 | 8 2 | 6+7 4 | 8 2 | 3+4 4 | 3+4 4 |
| | 8+9 4 | 6 2 | 6+7 4 | 9 2 | 3+4 4 | 3+4 4 |
| | 6+7 4 | 9 2 | 8+9 4 | 6 2 | 3+4 4 | 3+4 4 |
| | 8+9 4 | 7 2 | 8+9 4 | 7 2 | 3+4 4 | 3+4 4 |
| | 8+9 4 | 8 2 | 6+7 4 | 8 2 | 3+4 4 | 3+4 4 |
| | 8+9 4 | 7 2 | 6+7 4 | 9 2 | 3+4 4 | 3+4 4 |
| | 8+9 4 | 7 2 | 8+9 4 | 9 2 | 3+4 4 | 3+4 4 |
| | 6+7 4 | 8 2 | 6+7 4 | 7 2 | 3+4 4 | 3+4 4 |
| | 6+7 4 | 9 2 | 6+7 4 | 7 2 | 3+4 4 | 3+4 4 |
| | 8+9 4 | 7 2 | 8+9 4 | 8 2 | 3+4 4 | 3+4 4 |
| | 6+7 4 | 6 2 | 8+9 4 | 6 2 | 3+4 4 | 3+4 4 |
| | 6+7 4 | 8 2 | 6+7 4 | 7 2 | 3+4 4 | 3+4 4 |
| | | 9 2 | | 9 2 | | |
| | | 7 2 | | 6 2 | | |
| | | 9 2 | | 7 2 | | |
| | | 7 2 | | 8 2 | | |
| | | 9 2 | | 9 2 | | |
| | | 6 2 | | 8 2 | | |
| | | 9 2 | | 6 2 | | |
| | | 8 2 | | 7 2 | | |
| | | 6 2 | | 8 2 | | |
| | | 6 2 | | 9 2 | | |
| | | 7 2 | | 9 2 | | |
| | | 7 2 | | 8 2 | | |
| | | 8 2 | | 9 2 | | |
| | | 9 2 | | 7 2 | | |
| | | 7 2 | | 6 2 | | |
| | | 8 2 | | 7 2 | | |
| | | 6 2 | | 6 2 | | |
| | | 6 2 | | 8 2 | | |
| | | 7 2 | | 7 2 | | |
| | | 6 2 | | 7 2 | | |
| | | 8 2 | | 8 2 | | |
| | | 9 2 | | 6 2 | | |
| | | 9 2 | | 9 2 | | |
| 127.2.33 | 26777 | | 92 | | B4Y1 | |
| | -A | -GM | -M | -Z | -R | -X |
| | 8+9 4 | 7 2 | 8+9 4 | 7 2 | 3+4 4 | 3+4 4 |
| | 6+7 4 | 8 2 | 6+7 4 | 8 2 | 3+4 4 | 3+4 4 |
| | 8+9 4 | 9 2 | 8+9 4 | 9 2 | 3+4 4 | 3+4 4 |
| | 6+7 4 | 6 2 | 6+7 4 | 6 2 | 3+4 4 | 3+4 4 |
| | 6+7 4 | 8 2 | 6+7 4 | 7 2 | 3+4 4 | 3+4 4 |
| | 8+9 4 | 7 2 | 8+9 4 | 8 2 | 3+4 4 | 3+4 4 |
| | 8+9 4 | 6 2 | 8+9 4 | 6 2 | 3+4 4 | 3+4 4 |
| | 6+7 4 | 9 2 | 6+7 4 | 9 2 | 3+4 4 | 3+4 4 |
| | 8+9 4 | 7 2 | 6+7 4 | 7 2 | 3+4 4 | 3+4 4 |
| | 6+7 4 | 9 2 | 8+9 4 | 9 2 | 3+4 4 | 3+4 4 |
| | 6+7 4 | 6 2 | 6+7 4 | 6 2 | 3+4 4 | 3+4 4 |
| | 8+9 4 | 8 2 | 8+9 4 | 8 2 | 3+4 4 | 3+4 4 |
| | 6+7 4 | 9 2 | 6+7 4 | 9 2 | 3+4 4 | 3+4 4 |
| | 6+7 4 | 7 2 | 8+9 4 | 7 2 | 3+4 4 | 3+4 4 |
| | 8+9 4 | 6 2 | 8+9 4 | 6 2 | 3+4 4 | 3+4 4 |
| | 8+9 4 | 8 2 | 6+7 4 | 8 2 | 3+4 4 | 3+4 4 |
| | 8+9 4 | 7 2 | 8+9 4 | 9 2 | 3+4 4 | 3+4 4 |
| | 6+7 4 | 7 2 | 6+7 4 | 9 2 | 3+4 4 | 3+4 4 |
| | 8+9 4 | 8 2 | 6+7 4 | 7 2 | 3+4 4 | 3+4 4 |
| | 6+7 4 | 9 2 | 6+7 4 | 8 2 | 3+4 4 | 3+4 4 |
| | 8+9 4 | 7 2 | 8+9 4 | 6 2 | 3+4 4 | 3+4 4 |
| | 6+7 4 | 6 2 | 8+9 4 | 7 2 | 3+4 4 | 3+4 4 |
| | 6+7 4 | 8 2 | 8+9 4 | 7 2 | 3+4 4 | 3+4 4 |
| | | 9 2 | | 9 2 | | |
| | | 7 2 | | 6 2 | | |
| | | 6 2 | | 7 2 | | |
| | | 7 2 | | 8 2 | | |
| | | 9 2 | | 9 2 | | |
| | | 9 2 | | 8 2 | | |
| | | 9 2 | | 9 2 | | |
| | | 8 2 | | 6 2 | | |
| | | 7 2 | | 7 2 | | |
| | | 6 2 | | 8 2 | | |
| | | 6 2 | | 8 2 | | |
| | | 7 2 | | 9 2 | | |

Continued on next column

**TABLE LVIII– continued**

| TQC | ICSD | | NoE | | SF | |
|---|---|---|---|---|---|---|
| | | 8 2 | | 7 2 | | |
| | | 8 2 | | 9 2 | | |
| | | 9 2 | | 6 2 | | |
| | | 6 2 | | 6 2 | | |
| | | 7 2 | | 7 2 | | |
| | | 6 2 | | 8 2 | | |
| | | 7 2 | | 7 2 | | |
| | | 6 2 | | 7 2 | | |
| | | 9 2 | | 8 2 | | |
| | | 8 2 | | 6 2 | | |
| | | 9 2 | | 9 2 | | |
| 127.2.34 | 2360 | | 92 | | B4La1 | |
| | -A | -GM | -M | -Z | -R | -X |
| | 8+9 4 | 7 2 | 8+9 4 | 7 2 | 3+4 4 | 3+4 4 |
| | 6+7 4 | 8 2 | 6+7 4 | 8 2 | 3+4 4 | 3+4 4 |
| | 8+9 4 | 9 2 | 6+7 4 | 9 2 | 3+4 4 | 3+4 4 |
| | 6+7 4 | 6 2 | 8+9 4 | 6 2 | 3+4 4 | 3+4 4 |
| | 6+7 4 | 7 2 | 6+7 4 | 7 2 | 3+4 4 | 3+4 4 |
| | 8+9 4 | 8 2 | 8+9 4 | 8 2 | 3+4 4 | 3+4 4 |
| | 8+9 4 | 6 2 | 8+9 4 | 9 2 | 3+4 4 | 3+4 4 |
| | 6+7 4 | 9 2 | 6+7 4 | 6 2 | 3+4 4 | 3+4 4 |
| | 8+9 4 | 7 2 | 6+7 4 | 7 2 | 3+4 4 | 3+4 4 |
| | 6+7 4 | 9 2 | 8+9 4 | 9 2 | 3+4 4 | 3+4 4 |
| | 6+7 4 | 6 2 | 8+9 4 | 6 2 | 3+4 4 | 3+4 4 |
| | 6+7 4 | 8 2 | 6+7 4 | 8 2 | 3+4 4 | 3+4 4 |
| | 8+9 4 | 7 2 | 8+9 4 | 9 2 | 3+4 4 | 3+4 4 |
| | 6+7 4 | 9 2 | 8+9 4 | 7 2 | 3+4 4 | 3+4 4 |
| | 8+9 4 | 6 2 | 6+7 4 | 6 2 | 3+4 4 | 3+4 4 |
| | 8+9 4 | 8 2 | 6+7 4 | 8 2 | 3+4 4 | 3+4 4 |
| | 8+9 4 | 7 2 | 6+7 4 | 9 2 | 3+4 4 | 3+4 4 |
| | 8+9 4 | 7 2 | 6+7 4 | 9 2 | 3+4 4 | 3+4 4 |
| | 6+7 4 | 8 2 | 8+9 4 | 7 2 | 3+4 4 | 3+4 4 |
| | 6+7 4 | 9 2 | 6+7 4 | 8 2 | 3+4 4 | 3+4 4 |
| | 6+7 4 | 7 2 | 8+9 4 | 6 2 | 3+4 4 | 3+4 4 |
| | 8+9 4 | 6 2 | 8+9 4 | 7 2 | 3+4 4 | 3+4 4 |
| | 6+7 4 | 8 2 | 8+9 4 | 7 2 | 3+4 4 | 3+4 4 |
| | | 9 2 | | 9 2 | | |
| | | 9 2 | | 6 2 | | |
| | | 7 2 | | 7 2 | | |
| | | 9 2 | | 8 2 | | |
| | | 7 2 | | 9 2 | | |
| | | 6 2 | | 6 2 | | |
| | | 9 2 | | 7 2 | | |
| | | 8 2 | | 9 2 | | |
| | | 6 2 | | 8 2 | | |
| | | 6 2 | | 8 2 | | |
| | | 7 2 | | 9 2 | | |
| | | 7 2 | | 7 2 | | |
| | | 9 2 | | 8 2 | | |
| | | 8 2 | | 9 2 | | |
| | | 7 2 | | 6 2 | | |
| | | 8 2 | | 7 2 | | |
| | | 6 2 | | 8 2 | | |
| | | 6 2 | | 7 2 | | |
| | | 7 2 | | 7 2 | | |
| | | 6 2 | | 8 2 | | |
| | | 8 2 | | 6 2 | | |
| | | 9 2 | | 6 2 | | |
| | | 9 2 | | 9 2 | | |
| 127.2.35 | 182659 | | 132 | | B4Ho1 | |
| | -A | -GM | -M | -Z | -R | -X |

Continued on next column

**TABLE LVIII– continued**

| TQC | ICSD | | NoE | | SF | |
|---|---|---|---|---|---|---|
| | 8+9 4 | 7 2 | 8+9 4 | 7 2 | 3+4 4 | 3+4 4 |
| | 6+7 4 | 8 2 | 6+7 4 | 8 2 | 3+4 4 | 3+4 4 |
| | 8+9 4 | 9 2 | 8+9 4 | 9 2 | 3+4 4 | 3+4 4 |
| | 6+7 4 | 6 2 | 6+7 4 | 6 2 | 3+4 4 | 3+4 4 |
| | 6+7 4 | 8 2 | 6+7 4 | 8 2 | 3+4 4 | 3+4 4 |
| | 8+9 4 | 7 2 | 8+9 4 | 7 2 | 3+4 4 | 3+4 4 |
| | 8+9 4 | 6 2 | 8+9 4 | 6 2 | 3+4 4 | 3+4 4 |
| | 6+7 4 | 9 2 | 6+7 4 | 9 2 | 3+4 4 | 3+4 4 |
| | 8+9 4 | 7 2 | 6+7 4 | 7 2 | 3+4 4 | 3+4 4 |
| | 6+7 4 | 9 2 | 8+9 4 | 9 2 | 3+4 4 | 3+4 4 |
| | 6+7 4 | 6 2 | 6+7 4 | 6 2 | 3+4 4 | 3+4 4 |
| | 6+7 4 | 8 2 | 8+9 4 | 8 2 | 3+4 4 | 3+4 4 |
| | 6+7 4 | 9 2 | 8+9 4 | 9 2 | 3+4 4 | 3+4 4 |
| | 8+9 4 | 7 2 | 6+7 4 | 7 2 | 3+4 4 | 3+4 4 |
| | 8+9 4 | 6 2 | 8+9 4 | 6 2 | 3+4 4 | 3+4 4 |
| | 8+9 4 | 8 2 | 6+7 4 | 8 2 | 3+4 4 | 3+4 4 |
| | 8+9 4 | 7 2 | 8+9 4 | 9 2 | 3+4 4 | 3+4 4 |
| | 8+9 4 | 7 2 | 6+7 4 | 9 2 | 3+4 4 | 3+4 4 |
| | 6+7 4 | 8 2 | 6+7 4 | 7 2 | 3+4 4 | 3+4 4 |
| | 6+7 4 | 9 2 | 6+7 4 | 7 2 | 3+4 4 | 3+4 4 |
| | 6+7 4 | 7 2 | 8+9 4 | 6 2 | 3+4 4 | 3+4 4 |
| | 8+9 4 | 6 2 | 6+7 4 | 7 2 | 3+4 4 | 3+4 4 |
| | 6+7 4 | 8 2 | 8+9 4 | 8 2 | 3+4 4 | 3+4 4 |
| | 8+9 4 | 9 2 | 6+7 4 | 9 2 | 3+4 4 | 3+4 4 |
| | 6+7 4 | 7 2 | 8+9 4 | 6 2 | 3+4 4 | 3+4 4 |
| | 8+9 4 | 9 2 | 6+7 4 | 7 2 | 3+4 4 | 3+4 4 |
| | 8+9 4 | 7 2 | 8+9 4 | 8 2 | 3+4 4 | 3+4 4 |
| | 6+7 4 | 9 2 | 6+7 4 | 9 2 | 3+4 4 | 3+4 4 |
| | 6+7 4 | 8 2 | 8+9 4 | 8 2 | 3+4 4 | 3+4 4 |
| | 8+9 4 | 9 2 | 6+7 4 | 6 2 | 3+4 4 | 3+4 4 |
| | 6+7 4 | 6 2 | 8+9 4 | 7 2 | 3+4 4 | 3+4 4 |
| | 8+9 4 | 7 2 | 8+9 4 | 9 2 | 3+4 4 | 3+4 4 |
| | 8+9 4 | 6 2 | 6+7 4 | 8 2 | 3+4 4 | 3+4 4 |
| | | 7 2 | | 8 2 | | |
| | | 9 2 | | 9 2 | | |
| | | 8 2 | | 9 2 | | |
| | | 6 2 | | 7 2 | | |
| | | 7 2 | | 7 2 | | |
| | | 6 2 | | 6 2 | | |
| | | 6 2 | | 6 2 | | |
| | | 7 2 | | 8 2 | | |
| | | 8 2 | | 7 2 | | |
| | | 6 2 | | 8 2 | | |
| | | 9 2 | | 7 2 | | |
| | | 8 2 | | 9 2 | | |
| | | 9 2 | | 8 2 | | |
| | | 8 2 | | 6 2 | | |
| | | 7 2 | | 9 2 | | |
| | | 9 2 | | 6 2 | | |
| | | 6 2 | | 9 2 | | |
| | | 7 2 | | 8 2 | | |
| | | 8 2 | | 7 2 | | |
| | | 6 2 | | 6 2 | | |
| | | 9 2 | | 7 2 | | |
| | | 7 2 | | 7 2 | | |
| | | 6 2 | | 8 2 | | |
| | | 8 2 | | 9 2 | | |
| | | 9 2 | | 6 2 | | |
| | | 7 2 | | 9 2 | | |
| | | 8 2 | | 8 2 | | |
| | | 9 2 | | 6 2 | | |

Continued on next column

**TABLE LVIII– continued**

| TQC | ICSD | | NoE | | SF | |
|---|---|---|---|---|---|---|
| | | 8 2 | | 8 2 | | |
| | | 6 2 | | 7 2 | | |
| | | 6 2 | | 6 2 | | |
| | | 7 2 | | 9 2 | | |
| | | 8 2 | | 8 2 | | |
| 127.2.36 | 51434 | | 120 | | B2Rh5Sc3 | |
| | -A | -GM | -M | -Z | -R | -X |
| | 6+7 4 | 7 2 | 8+9 4 | 8 2 | 3+4 4 | 3+4 4 |
| | 8+9 4 | 6 2 | 6+7 4 | 9 2 | 3+4 4 | 3+4 4 |
| | 6+7 4 | 8 2 | 8+9 4 | 7 2 | 3+4 4 | 3+4 4 |
| | 6+7 4 | 9 2 | 6+7 4 | 6 2 | 3+4 4 | 3+4 4 |
| | 8+9 4 | 6 2 | 8+9 4 | 9 2 | 3+4 4 | 3+4 4 |
| | 6+7 4 | 7 2 | 8+9 4 | 8 2 | 3+4 4 | 3+4 4 |
| | 8+9 4 | 7 2 | 8+9 4 | 6 2 | 3+4 4 | 3+4 4 |
| | 6+7 4 | 6 2 | 6+7 4 | 7 2 | 3+4 4 | 3+4 4 |
| | 8+9 4 | 8 2 | 8+9 4 | 9 2 | 3+4 4 | 3+4 4 |
| | 6+7 4 | 9 2 | 6+7 4 | 8 2 | 3+4 4 | 3+4 4 |
| | 6+7 4 | 7 2 | 6+7 4 | 6 2 | 3+4 4 | 3+4 4 |
| | 8+9 4 | 7 2 | 8+9 4 | 8 2 | 3+4 4 | 3+4 4 |
| | 6+7 4 | 6 2 | 6+7 4 | 9 2 | 3+4 4 | 3+4 4 |
| | 8+9 4 | 7 2 | 8+9 4 | 7 2 | 3+4 4 | 3+4 4 |
| | 6+7 4 | 6 2 | 6+7 4 | 9 2 | 3+4 4 | 3+4 4 |
| | 8+9 4 | 9 2 | 8+9 4 | 6 2 | 3+4 4 | 3+4 4 |
| | 6+7 4 | 8 2 | 6+7 4 | 7 2 | 3+4 4 | 3+4 4 |
| | 6+7 4 | 7 2 | 8+9 4 | 8 2 | 3+4 4 | 3+4 4 |
| | 8+9 4 | 6 2 | 6+7 4 | 8 2 | 3+4 4 | 3+4 4 |
| | 6+7 4 | 6 2 | 8+9 4 | 9 2 | 3+4 4 | 3+4 4 |
| | 6+7 4 | 9 2 | 6+7 4 | 7 2 | 3+4 4 | 3+4 4 |
| | 8+9 4 | 8 2 | 8+9 4 | 6 2 | 3+4 4 | 3+4 4 |
| | 8+9 4 | 9 2 | 6+7 4 | 9 2 | 3+4 4 | 3+4 4 |
| | 6+7 4 | 8 2 | 8+9 4 | 8 2 | 3+4 4 | 3+4 4 |
| | 8+9 4 | 9 2 | 8+9 4 | 8 2 | 3+4 4 | 3+4 4 |
| | 6+7 4 | 6 2 | 6+7 4 | 6 2 | 3+4 4 | 3+4 4 |
| | 6+7 4 | 8 2 | 6+7 4 | 7 2 | 3+4 4 | 3+4 4 |
| | 8+9 4 | 7 2 | 8+9 4 | 9 2 | 3+4 4 | 3+4 4 |
| | 6+7 4 | 7 2 | 8+9 4 | 6 2 | 3+4 4 | 3+4 4 |
| | 8+9 4 | 9 2 | 8+9 4 | 7 2 | 3+4 4 | 3+4 4 |
| | | 7 2 | | 7 2 | | |
| | | 8 2 | | 9 2 | | |
| | | 6 2 | | 8 2 | | |
| | | 7 2 | | 9 2 | | |
| | | 6 2 | | 6 2 | | |
| | | 7 2 | | 7 2 | | |
| | | 7 2 | | 9 2 | | |
| | | 6 2 | | 8 2 | | |
| | | 8 2 | | 6 2 | | |
| | | 6 2 | | 7 2 | | |
| | | 7 2 | | 9 2 | | |
| | | 8 2 | | 8 2 | | |
| | | 9 2 | | 6 2 | | |
| | | 8 2 | | 9 2 | | |
| | | 9 2 | | 8 2 | | |
| | | 6 2 | | 6 2 | | |
| | | 7 2 | | 7 2 | | |
| | | 9 2 | | 7 2 | | |
| | | 8 2 | | 8 2 | | |
| | | 6 2 | | 6 2 | | |
| | | 9 2 | | 9 2 | | |
| | | 6 2 | | 6 2 | | |
| | | 7 2 | | 8 2 | | |
| | | 8 2 | | 7 2 | | |

Continued on next column

**TABLE LVIII– continued**

| TQC | ICSD | | NoE | | SF | |
|---|---|---|---|---|---|---|
| | | 6 2 | | 9 2 | | |
| | | 7 2 | | 8 2 | | |
| | | 6 2 | | 7 2 | | |
| | | 9 2 | | 6 2 | | |
| | | 8 2 | | 8 2 | | |
| | | 7 2 | | 9 2 | | |
| 127.2.37 | 107439 | | 132 | | Al1Pt3 | |
| | -A | -GM | -M | -Z | -R | -X |
| | 6+7 4 | 7 2 | 8+9 4 | 9 2 | 3+4 4 | 3+4 4 |
| | 8+9 4 | 7 2 | 6+7 4 | 7 2 | 3+4 4 | 3+4 4 |
| | 8+9 4 | 8 2 | 6+7 4 | 6 2 | 3+4 4 | 3+4 4 |
| | 6+7 4 | 9 2 | 8+9 4 | 8 2 | 3+4 4 | 3+4 4 |
| | 8+9 4 | 7 2 | 8+9 4 | 7 2 | 3+4 4 | 3+4 4 |
| | 8+9 4 | 6 2 | 6+7 4 | 9 2 | 3+4 4 | 3+4 4 |
| | 6+7 4 | 9 2 | 6+7 4 | 7 2 | 3+4 4 | 3+4 4 |
| | 6+7 4 | 8 2 | 6+7 4 | 6 2 | 3+4 4 | 3+4 4 |
| | 8+9 4 | 9 2 | 8+9 4 | 9 2 | 3+4 4 | 3+4 4 |
| | 8+9 4 | 9 2 | 6+7 4 | 9 2 | 3+4 4 | 3+4 4 |
| | 6+7 4 | 8 2 | 8+9 4 | 8 2 | 3+4 4 | 3+4 4 |
| | 6+7 4 | 7 2 | 8+9 4 | 7 2 | 3+4 4 | 3+4 4 |
| | 6+7 4 | 6 2 | 6+7 4 | 9 2 | 3+4 4 | 3+4 4 |
| | 8+9 4 | 7 2 | 8+9 4 | 7 2 | 3+4 4 | 3+4 4 |
| | 6+7 4 | 6 2 | 8+9 4 | 8 2 | 3+4 4 | 3+4 4 |
| | 8+9 4 | 7 2 | 6+7 4 | 9 2 | 3+4 4 | 3+4 4 |
| | 8+9 4 | 6 2 | 8+9 4 | 6 2 | 3+4 4 | 3+4 4 |
| | 6+7 4 | 8 2 | 6+7 4 | 8 2 | 3+4 4 | 3+4 4 |
| | 6+7 4 | 6 2 | 8+9 4 | 6 2 | 3+4 4 | 3+4 4 |
| | 8+9 4 | 6 2 | 6+7 4 | 7 2 | 3+4 4 | 3+4 4 |
| | 6+7 4 | 7 2 | 8+9 4 | 8 2 | 3+4 4 | 3+4 4 |
| | 8+9 4 | 9 2 | 8+9 4 | 6 2 | 3+4 4 | 3+4 4 |
| | 6+7 4 | 7 2 | 6+7 4 | 7 2 | 3+4 4 | 3+4 4 |
| | 8+9 4 | 8 2 | 8+9 4 | 6 2 | 3+4 4 | 3+4 4 |
| | 6+7 4 | 9 2 | 6+7 4 | 8 2 | 3+4 4 | 3+4 4 |
| | 8+9 4 | 8 2 | 8+9 4 | 9 2 | 3+4 4 | 3+4 4 |
| | 6+7 4 | 9 2 | 6+7 4 | 6 2 | 3+4 4 | 3+4 4 |
| | 8+9 4 | 8 2 | 6+7 4 | 8 2 | 3+4 4 | 3+4 4 |
| | 6+7 4 | 9 2 | 8+9 4 | 7 2 | 3+4 4 | 3+4 4 |
| | 8+9 4 | 7 2 | 8+9 4 | 9 2 | 3+4 4 | 3+4 4 |
| | 8+9 4 | 8 2 | 6+7 4 | 8 2 | 3+4 4 | 3+4 4 |
| | 6+7 4 | 7 2 | 6+7 4 | 6 2 | 3+4 4 | 3+4 4 |
| | 8+9 4 | 6 2 | 6+7 4 | 6 2 | 3+4 4 | 3+4 4 |
| | | 8 2 | | 9 2 | | |
| | | 8 2 | | 7 2 | | |
| | | 6 2 | | 9 2 | | |
| | | 7 2 | | 8 2 | | |
| | | 9 2 | | 6 2 | | |
| | | 9 2 | | 7 2 | | |
| | | 6 2 | | 8 2 | | |
| | | 8 2 | | 7 2 | | |
| | | 9 2 | | 8 2 | | |
| | | 7 2 | | 6 2 | | |
| | | 6 2 | | 9 2 | | |
| | | 8 2 | | 7 2 | | |
| | | 7 2 | | 8 2 | | |
| | | 9 2 | | 6 2 | | |
| | | 6 2 | | 9 2 | | |
| | | 9 2 | | 7 2 | | |
| | | 7 2 | | 9 2 | | |
| | | 6 2 | | 8 2 | | |
| | | 6 2 | | 9 2 | | |
| | | 8 2 | | 7 2 | | |

Continued on next column

**TABLE LVIII– continued**

| TQC | ICSD | NoE | SF |
|---|---|---|---|
| | 6 2 | 7 2 | |
| | 6 2 | 9 2 | |
| | 8 2 | 8 2 | |
| | 9 2 | 6 2 | |
| | 7 2 | 9 2 | |
| | 7 2 | 7 2 | |
| | 6 2 | 6 2 | |
| | 8 2 | 8 2 | |
| | 6 2 | 8 2 | |
| | 9 2 | 6 2 | |
| | 6 2 | 6 2 | |
| | 8 2 | 8 2 | |
| | 9 2 | 6 2 | |

The End

TABLE LIX: The computed irreducible representations for the valence bands for the compounds of SEBR and NLC in space group 128. See the list of tables.

| TQC | ICSD | | NoE | SF | | |
|---|---|---|---|---|---|---|
| 128.1.1 | 425655 | | 160 | B2Nb2Os1 | | |
| | -A | -GM | -M | -Z | -R | -X |
| | 5+7 4 | 7 2 | 8+9 4 | 6+8 4 | 2+5 2 | 3+4 4 |
| | 6+8 4 | 7 2 | 8+9 4 | 5+7 4 | 3+4 2 | 3+4 4 |
| | 6+8 4 | 8 2 | 6+7 4 | 6+8 4 | 6+9 2 | 3+4 4 |
| | 5+7 4 | 9 2 | 6+7 4 | 5+7 4 | 7+8 2 | 3+4 4 |
| | 6+8 4 | 6 2 | 8+9 4 | 5+7 4 | 6+9 2 | 3+4 4 |
| | 5+7 4 | 8 2 | 8+9 4 | 5+7 4 | 7+8 2 | 3+4 4 |
| | 6+8 4 | 9 2 | 6+7 4 | 6+8 4 | 2+5 2 | 3+4 4 |
| | 5+7 4 | 6 2 | 6+7 4 | 6+8 4 | 3+4 2 | 3+4 4 |
| | 5+7 4 | 8 2 | 8+9 4 | 5+7 4 | 6+9 2 | 3+4 4 |
| | 6+8 4 | 6 2 | 6+7 4 | 6+8 4 | 7+8 2 | 3+4 4 |
| | 6+8 4 | 6 2 | 8+9 4 | 6+8 4 | 3+4 2 | 3+4 4 |
| | 5+7 4 | 7 2 | 6+7 4 | 5+7 4 | 2+5 2 | 3+4 4 |
| | 5+7 4 | 8 2 | 8+9 4 | 6+8 4 | 2+5 2 | 3+4 4 |
| | 6+8 4 | 7 2 | 8+9 4 | 6+8 4 | 3+4 2 | 3+4 4 |
| | 5+7 4 | 9 2 | 6+7 4 | 5+7 4 | 7+8 2 | 3+4 4 |
| | 6+8 4 | 9 2 | 6+7 4 | 5+7 4 | 6+9 2 | 3+4 4 |
| | 5+7 4 | 6 2 | 8+9 4 | 6+8 4 | 2+5 2 | 3+4 4 |
| | 6+8 4 | 6 2 | 6+7 4 | 5+7 4 | 3+4 2 | 3+4 4 |
| | 6+8 4 | 7 2 | 6+7 4 | 6+8 4 | 7+8 2 | 3+4 4 |
| | 5+7 4 | 9 2 | 8+9 4 | 5+7 4 | 6+9 2 | 3+4 4 |
| | 6+8 4 | 9 2 | 6+7 4 | 6+8 4 | 2+5 2 | 3+4 4 |
| | 5+7 4 | 8 2 | 8+9 4 | 5+7 4 | 3+4 2 | 3+4 4 |
| | 6+8 4 | 7 2 | 8+9 4 | 6+8 4 | 6+9 2 | 3+4 4 |
| | 5+7 4 | 8 2 | 6+7 4 | 6+8 4 | 7+8 2 | 3+4 4 |
| | 5+7 4 | 9 2 | 6+7 4 | 5+7 4 | 6+9 2 | 3+4 4 |
| | 6+8 4 | 7 2 | 8+9 4 | 6+8 4 | 7+8 2 | 3+4 4 |
| | 5+7 4 | 7 2 | 8+9 4 | 5+7 4 | 3+4 2 | 3+4 4 |
| | 6+8 4 | 9 2 | 6+7 4 | 6+8 4 | 2+5 2 | 3+4 4 |
| | 6+8 4 | 6 2 | 6+7 4 | 6+8 4 | 2+5 2 | 3+4 4 |
| | 5+7 4 | 6 2 | 8+9 4 | 5+7 4 | 3+4 2 | 3+4 4 |
| | 5+7 4 | 8 2 | 6+7 4 | 5+7 4 | 7+8 2 | 3+4 4 |
| | 6+8 4 | 8 2 | 8+9 4 | 6+8 4 | 6+9 2 | 3+4 4 |
| | 6+8 4 | 7 2 | 6+7 4 | 5+7 4 | 6+9 2 | 3+4 4 |
| | 5+7 4 | 8 2 | 6+7 4 | 6+8 4 | 7+8 2 | 3+4 4 |
| | 5+7 4 | 9 2 | 8+9 4 | 5+7 4 | 2+5 2 | 3+4 4 |
| | 6+8 4 | 6 2 | 8+9 4 | 5+7 4 | 3+4 2 | 3+4 4 |
| | 6+8 4 | 9 2 | 6+7 4 | 6+8 4 | 6+9 2 | 3+4 4 |
| | 5+7 4 | 8 2 | 8+9 4 | 5+7 4 | 7+8 2 | 3+4 4 |
| | 5+7 4 | 7 2 | 6+7 4 | 6+8 4 | 3+4 2 | 3+4 4 |
| | 6+8 4 | 9 2 | 6+7 4 | 5+7 4 | 2+5 2 | 3+4 4 |
| | | 7 2 | | | 7+8 2 | |
| | | 6 2 | | | 6+9 2 | |
| | | 6 2 | | | 2+5 2 | |
| | | 7 2 | | | 3+4 2 | |
| | | 9 2 | | | 2+5 2 | |
| | | 9 2 | | | 3+4 2 | |
| | | 7 2 | | | 6+9 2 | |
| | | 8 2 | | | 7+8 2 | |
| | | 6 2 | | | 3+4 2 | |
| | | 8 2 | | | 2+5 2 | |
| | | 8 2 | | | 7+8 2 | |
| | | 7 2 | | | 3+4 2 | |
| | | 8 2 | | | 6+9 2 | |
| | | 7 2 | | | 2+5 2 | |
| | | 6 2 | | | 6+9 2 | |
| | | 9 2 | | | 7+8 2 | |
| | | 8 2 | | | 2+5 2 | |
| | | 7 2 | | | 3+4 2 | |
| | | 9 2 | | | 7+8 2 | |
| | | 8 2 | | | 6+9 2 | |
| | | 6 2 | | | 2+5 2 | |
| | | 7 2 | | | 7+8 2 | |
| | | 8 2 | | | 6+9 2 | |
| | | 6 2 | | | 3+4 2 | |
| | | 7 2 | | | 3+4 2 | |
| | | 6 2 | | | 2+5 2 | |
| | | 6 2 | | | 7+8 2 | |
| | | 7 2 | | | 6+9 2 | |
| | | 9 2 | | | 6+9 2 | |
| | | 8 2 | | | 7+8 2 | |
| | | 7 2 | | | 2+5 2 | |
| | | 8 2 | | | 3+4 2 | |
| | | 9 2 | | | 2+5 2 | |
| | | 6 2 | | | 3+4 2 | |
| | | 9 2 | | | 7+8 2 | |
| | | 9 2 | | | 6+9 2 | |
| | | 8 2 | | | 6+9 2 | |
| | | 6 2 | | | 7+8 2 | |
| | | 9 2 | | | 3+4 2 | |
| | | 6 2 | | | 2+5 2 | |

Continued on next column

**TABLE LIX– continued**

The End

TABLE LX: The computed irreducible representations for the valence bands for the compounds of SEBR and NLC in space group 129. See the list of tables.

| TQC | ICSD | NoE | | SF | | |
|---|---|---|---|---|---|---|
| 129.1.1 | 80190 | 28 | | Sn1Te1Zr1 | | |
| | -A | -GM | -M | -Z | -R | -X |
| | 5 4 | 7 2 | 5 4 | 9 2 | 3+4 4 | 3+4 4 |
| | 5 4 | 9 2 | 5 4 | 7 2 | 3+4 4 | 3+4 4 |
| | 5 4 | 7 2 | 5 4 | 7 2 | 3+4 4 | 3+4 4 |
| | 5 4 | 7 2 | 5 4 | 9 2 | 3+4 4 | 3+4 4 |
| | 5 4 | 9 2 | 5 4 | 9 2 | 3+4 4 | 3+4 4 |
| | 5 4 | 7 2 | 5 4 | 8 2 | 3+4 4 | 3+4 4 |
| | 5 4 | 8 2 | 5 4 | 8 2 | 3+4 4 | 3+4 4 |
| | | 6 2 | | 6 2 | | |
| | | 6 2 | | 7 2 | | |

Continued on next column

**TABLE LX– continued**

| TQC | ICSD | NoE | SF | | | |
|---|---|---|---|---|---|---|
| | | 7 2 | | 9 2 | | |
| | | 6 2 | | 7 2 | | |
| | | 9 2 | | 6 2 | | |
| | | 8 2 | | 7 2 | | |
| | | 9 2 | | 6 2 | | |
| 129.1.2 | 26273 | 48 | Sn1Te1U1 | | | |
| | -A | -GM | -M | -Z | -R | -X |
| | 5 4 | 7 2 | 5 4 | 7 2 | 3+4 4 | 3+4 4 |
| | 5 4 | 9 2 | 5 4 | 9 2 | 3+4 4 | 3+4 4 |
| | 5 4 | 7 2 | 5 4 | 9 2 | 3+4 4 | 3+4 4 |
| | 5 4 | 9 2 | 5 4 | 7 2 | 3+4 4 | 3+4 4 |
| | 5 4 | 7 2 | 5 4 | 9 2 | 3+4 4 | 3+4 4 |
| | 5 4 | 9 2 | 5 4 | 7 2 | 3+4 4 | 3+4 4 |
| | 5 4 | 6 2 | 5 4 | 8 2 | 3+4 4 | 3+4 4 |
| | 5 4 | 8 2 | 5 4 | 6 2 | 3+4 4 | 3+4 4 |
| | 5 4 | 7 2 | 5 4 | 9 2 | 3+4 4 | 3+4 4 |
| | 5 4 | 9 2 | 5 4 | 7 2 | 3+4 4 | 3+4 4 |
| | 5 4 | 7 2 | 5 4 | 7 2 | 3+4 4 | 3+4 4 |
| | 5 4 | 7 2 | 5 4 | 9 2 | 3+4 4 | 3+4 4 |
| | | 9 2 | | 8 2 | | |
| | | 8 2 | | 9 2 | | |
| | | 7 2 | | 8 2 | | |
| | | 6 2 | | 7 2 | | |
| | | 9 2 | | 6 2 | | |
| | | 6 2 | | 9 2 | | |
| | | 7 2 | | 7 2 | | |
| | | 8 2 | | 7 2 | | |
| | | 6 2 | | 6 2 | | |
| | | 8 2 | | 6 2 | | |
| | | 6 2 | | 6 2 | | |
| | | 9 2 | | 8 2 | | |
| 129.1.3 | 74522 | 28 | Si1Te1Zr1 | | | |
| | -A | -GM | -M | -Z | -R | -X |
| | 5 4 | 7 2 | 5 4 | 7 2 | 3+4 4 | 3+4 4 |
| | 5 4 | 9 2 | 5 4 | 9 2 | 3+4 4 | 3+4 4 |
| | 5 4 | 7 2 | 5 4 | 9 2 | 3+4 4 | 3+4 4 |
| | 5 4 | 7 2 | 5 4 | 7 2 | 3+4 4 | 3+4 4 |
| | 5 4 | 9 2 | 5 4 | 6 2 | 3+4 4 | 3+4 4 |
| | 5 4 | 8 2 | 5 4 | 9 2 | 3+4 4 | 3+4 4 |
| | 5 4 | 7 2 | 5 4 | 8 2 | 3+4 4 | 3+4 4 |
| | | 6 2 | | 9 2 | | |
| | | 6 2 | | 7 2 | | |
| | | 7 2 | | 7 2 | | |
| | | 9 2 | | 8 2 | | |
| | | 6 2 | | 6 2 | | |
| | | 9 2 | | 9 2 | | |
| | | 8 2 | | 8 2 | | |
| 129.1.4 | 25734 | 28 | Si1Te1Zr1 | | | |
| | -A | -GM | -M | -Z | -R | -X |
| | 5 4 | 7 2 | 5 4 | 9 2 | 3+4 4 | 3+4 4 |
| | 5 4 | 9 2 | 5 4 | 7 2 | 3+4 4 | 3+4 4 |
| | 5 4 | 7 2 | 5 4 | 7 2 | 3+4 4 | 3+4 4 |
| | 5 4 | 7 2 | 5 4 | 9 2 | 3+4 4 | 3+4 4 |
| | 5 4 | 9 2 | 5 4 | 8 2 | 3+4 4 | 3+4 4 |
| | 5 4 | 8 2 | 5 4 | 9 2 | 3+4 4 | 3+4 4 |
| | 5 4 | 7 2 | 5 4 | 7 2 | 3+4 4 | 3+4 4 |
| | | 6 2 | | 6 2 | | |
| | | 6 2 | | 7 2 | | |
| | | 7 2 | | 9 2 | | |
| | | 9 2 | | 6 2 | | |
| | | 6 2 | | 8 2 | | |

Continued on next column

**TABLE LX– continued**

| TQC | ICSD | NoE | SF | | | |
|---|---|---|---|---|---|---|
| | | 9 2 | | 7 2 | | |
| | | 8 2 | | 6 2 | | |
| 129.1.5 | 15572 | 28 | Si1Te1Zr1 | | | |
| | -A | -GM | -M | -Z | -R | -X |
| | 5 4 | 7 2 | 5 4 | 9 2 | 3+4 4 | 3+4 4 |
| | 5 4 | 9 2 | 5 4 | 7 2 | 3+4 4 | 3+4 4 |
| | 5 4 | 7 2 | 5 4 | 7 2 | 3+4 4 | 3+4 4 |
| | 5 4 | 7 2 | 5 4 | 9 2 | 3+4 4 | 3+4 4 |
| | 5 4 | 9 2 | 5 4 | 8 2 | 3+4 4 | 3+4 4 |
| | 5 4 | 8 2 | 5 4 | 7 2 | 3+4 4 | 3+4 4 |
| | 5 4 | 7 2 | 5 4 | 6 2 | 3+4 4 | 3+4 4 |
| | | 6 2 | | 9 2 | | |
| | | 6 2 | | 7 2 | | |
| | | 7 2 | | 9 2 | | |
| | | 9 2 | | 6 2 | | |
| | | 6 2 | | 8 2 | | |
| | | 9 2 | | 7 2 | | |
| | | 8 2 | | 6 2 | | |
| 129.1.6 | 25733 | 28 | Se1Si1Zr1 | | | |
| | -A | -GM | -M | -Z | -R | -X |
| | 5 4 | 7 2 | 5 4 | 9 2 | 3+4 4 | 3+4 4 |
| | 5 4 | 9 2 | 5 4 | 7 2 | 3+4 4 | 3+4 4 |
| | 5 4 | 7 2 | 5 4 | 7 2 | 3+4 4 | 3+4 4 |
| | 5 4 | 7 2 | 5 4 | 9 2 | 3+4 4 | 3+4 4 |
| | 5 4 | 9 2 | 5 4 | 9 2 | 3+4 4 | 3+4 4 |
| | 5 4 | 7 2 | 5 4 | 8 2 | 3+4 4 | 3+4 4 |
| | 5 4 | 6 2 | 5 4 | 8 2 | 3+4 4 | 3+4 4 |
| | | 8 2 | | 9 2 | | |
| | | 7 2 | | 7 2 | | |
| | | 6 2 | | 6 2 | | |
| | | 6 2 | | 7 2 | | |
| | | 9 2 | | 6 2 | | |
| | | 8 2 | | 7 2 | | |
| | | 9 2 | | 6 2 | | |
| 129.1.7 | 15571 | 28 | Se1Si1Zr1 | | | |
| | -A | -GM | -M | -Z | -R | -X |
| | 5 4 | 7 2 | 5 4 | 9 2 | 3+4 4 | 3+4 4 |
| | 5 4 | 9 2 | 5 4 | 7 2 | 3+4 4 | 3+4 4 |
| | 5 4 | 7 2 | 5 4 | 7 2 | 3+4 4 | 3+4 4 |
| | 5 4 | 7 2 | 5 4 | 9 2 | 3+4 4 | 3+4 4 |
| | 5 4 | 9 2 | 5 4 | 9 2 | 3+4 4 | 3+4 4 |
| | 5 4 | 7 2 | 5 4 | 8 2 | 3+4 4 | 3+4 4 |
| | 5 4 | 6 2 | 5 4 | 8 2 | 3+4 4 | 3+4 4 |
| | | 8 2 | | 9 2 | | |
| | | 7 2 | | 7 2 | | |
| | | 6 2 | | 6 2 | | |
| | | 6 2 | | 7 2 | | |
| | | 9 2 | | 6 2 | | |
| | | 8 2 | | 7 2 | | |
| | | 9 2 | | 6 2 | | |
| 129.1.8 | 25732 | 28 | S1Si1Zr1 | | | |
| | -A | -GM | -M | -Z | -R | -X |
| | 5 4 | 7 2 | 5 4 | 9 2 | 3+4 4 | 3+4 4 |
| | 5 4 | 9 2 | 5 4 | 7 2 | 3+4 4 | 3+4 4 |
| | 5 4 | 7 2 | 5 4 | 7 2 | 3+4 4 | 3+4 4 |
| | 5 4 | 7 2 | 5 4 | 9 2 | 3+4 4 | 3+4 4 |
| | 5 4 | 9 2 | 5 4 | 9 2 | 3+4 4 | 3+4 4 |
| | 5 4 | 7 2 | 5 4 | 8 2 | 3+4 4 | 3+4 4 |
| | 5 4 | 6 2 | 5 4 | 9 2 | 3+4 4 | 3+4 4 |
| | | 8 2 | | 8 2 | | |
| | | 7 2 | | 7 2 | | |

Continued on next column

**TABLE LX– continued**

| TQC | ICSD | NoE | SF | | | |
|---|---|---|---|---|---|---|
| | | 6 2 | | 6 2 | | |
| | | 6 2 | | 7 2 | | |
| | | 9 2 | | 6 2 | | |
| | | 8 2 | | 7 2 | | |
| | | 9 2 | | 6 2 | | |
| 129.1.9 | 15570 | | 28 | S1Si1Zr1 | | |
| | -A | -GM | -M | -Z | -R | -X |
| | 5 4 | 7 2 | 5 4 | 9 2 | 3+4 4 | 3+4 4 |
| | 5 4 | 9 2 | 5 4 | 7 2 | 3+4 4 | 3+4 4 |
| | 5 4 | 7 2 | 5 4 | 7 2 | 3+4 4 | 3+4 4 |
| | 5 4 | 7 2 | 5 4 | 9 2 | 3+4 4 | 3+4 4 |
| | 5 4 | 9 2 | 5 4 | 9 2 | 3+4 4 | 3+4 4 |
| | 5 4 | 7 2 | 5 4 | 8 2 | 3+4 4 | 3+4 4 |
| | 5 4 | 6 2 | 5 4 | 9 2 | 3+4 4 | 3+4 4 |
| | | 8 2 | | 8 2 | | |
| | | 7 2 | | 7 2 | | |
| | | 6 2 | | 6 2 | | |
| | | 6 2 | | 7 2 | | |
| | | 9 2 | | 6 2 | | |
| | | 8 2 | | 7 2 | | |
| | | 9 2 | | 6 2 | | |
| 129.1.10 | 15569 | | 28 | S1Si1Zr1 | | |
| | -A | -GM | -M | -Z | -R | -X |
| | 5 4 | 7 2 | 5 4 | 9 2 | 3+4 4 | 3+4 4 |
| | 5 4 | 9 2 | 5 4 | 7 2 | 3+4 4 | 3+4 4 |
| | 5 4 | 7 2 | 5 4 | 7 2 | 3+4 4 | 3+4 4 |
| | 5 4 | 7 2 | 5 4 | 9 2 | 3+4 4 | 3+4 4 |
| | 5 4 | 9 2 | 5 4 | 9 2 | 3+4 4 | 3+4 4 |
| | 5 4 | 7 2 | 5 4 | 8 2 | 3+4 4 | 3+4 4 |
| | 5 4 | 6 2 | 5 4 | 9 2 | 3+4 4 | 3+4 4 |
| | | 8 2 | | 8 2 | | |
| | | 7 2 | | 7 2 | | |
| | | 6 2 | | 6 2 | | |
| | | 6 2 | | 7 2 | | |
| | | 9 2 | | 6 2 | | |
| | | 8 2 | | 7 2 | | |
| | | 9 2 | | 6 2 | | |
| 129.1.11 | 26271 | | 48 | S1Si1U1 | | |
| | -A | -GM | -M | -Z | -R | -X |
| | 5 4 | 7 2 | 5 4 | 9 2 | 3+4 4 | 3+4 4 |
| | 5 4 | 9 2 | 5 4 | 7 2 | 3+4 4 | 3+4 4 |
| | 5 4 | 7 2 | 5 4 | 9 2 | 3+4 4 | 3+4 4 |
| | 5 4 | 9 2 | 5 4 | 7 2 | 3+4 4 | 3+4 4 |
| | 5 4 | 7 2 | 5 4 | 7 2 | 3+4 4 | 3+4 4 |
| | 5 4 | 9 2 | 5 4 | 9 2 | 3+4 4 | 3+4 4 |
| | 5 4 | 6 2 | 5 4 | 8 2 | 3+4 4 | 3+4 4 |
| | 5 4 | 8 2 | 5 4 | 6 2 | 3+4 4 | 3+4 4 |
| | 5 4 | 7 2 | 5 4 | 7 2 | 3+4 4 | 3+4 4 |
| | 5 4 | 8 2 | 5 4 | 8 2 | 3+4 4 | 3+4 4 |
| | 5 4 | 7 2 | 5 4 | 9 2 | 3+4 4 | 3+4 4 |
| | 5 4 | 9 2 | 5 4 | 7 2 | 3+4 4 | 3+4 4 |
| | | 7 2 | | 7 2 | | |
| | | 6 2 | | 6 2 | | |
| | | 9 2 | | 9 2 | | |
| | | 6 2 | | 6 2 | | |
| | | 7 2 | | 9 2 | | |
| | | 8 2 | | 8 2 | | |
| | | 6 2 | | 9 2 | | |
| | | 7 2 | | 7 2 | | |
| | | 8 2 | | 6 2 | | |
| | | 9 2 | | 6 2 | | |

Continued on next column

**TABLE LX– continued**

| TQC | ICSD | NoE | SF | | | |
|---|---|---|---|---|---|---|
| | | 9 2 | | 8 2 | | |
| | | 6 2 | | 7 2 | | |
| 129.1.12 | 50351 | | 64 | Ru1Sb2U1 | | |
| | -A | -GM | -M | -Z | -R | -X |
| | 5 4 | 7 2 | 5 4 | 7 2 | 3+4 4 | 3+4 4 |
| | 5 4 | 9 2 | 5 4 | 9 2 | 3+4 4 | 3+4 4 |
| | 5 4 | 7 2 | 5 4 | 7 2 | 3+4 4 | 3+4 4 |
| | 5 4 | 9 2 | 5 4 | 9 2 | 3+4 4 | 3+4 4 |
| | 5 4 | 7 2 | 5 4 | 7 2 | 3+4 4 | 3+4 4 |
| | 5 4 | 9 2 | 5 4 | 9 2 | 3+4 4 | 3+4 4 |
| | 5 4 | 8 2 | 5 4 | 6 2 | 3+4 4 | 3+4 4 |
| | 5 4 | 6 2 | 5 4 | 8 2 | 3+4 4 | 3+4 4 |
| | 5 4 | 7 2 | 5 4 | 7 2 | 3+4 4 | 3+4 4 |
| | 5 4 | 9 2 | 5 4 | 9 2 | 3+4 4 | 3+4 4 |
| | 5 4 | 7 2 | 5 4 | 7 2 | 3+4 4 | 3+4 4 |
| | 5 4 | 8 2 | 5 4 | 8 2 | 3+4 4 | 3+4 4 |
| | 5 4 | 9 2 | 5 4 | 9 2 | 3+4 4 | 3+4 4 |
| | 5 4 | 7 2 | 5 4 | 7 2 | 3+4 4 | 3+4 4 |
| | 5 4 | 9 2 | 5 4 | 7 2 | 3+4 4 | 3+4 4 |
| | 5 4 | 8 2 | 5 4 | 6 2 | 3+4 4 | 3+4 4 |
| | | 6 2 | | 8 2 | | |
| | | 6 2 | | 6 2 | | |
| | | 7 2 | | 9 2 | | |
| | | 7 2 | | 6 2 | | |
| | | 6 2 | | 7 2 | | |
| | | 8 2 | | 8 2 | | |
| | | 6 2 | | 9 2 | | |
| | | 9 2 | | 9 2 | | |
| | | 7 2 | | 8 2 | | |
| | | 6 2 | | 8 2 | | |
| | | 6 2 | | 6 2 | | |
| | | 7 2 | | 9 2 | | |
| | | 8 2 | | 8 2 | | |
| | | 9 2 | | 6 2 | | |
| | | 9 2 | | 7 2 | | |
| | | 6 2 | | 7 2 | | |
| 129.1.13 | 50350 | | 64 | Ru1Sb2U1 | | |
| | -A | -GM | -M | -Z | -R | -X |
| | 5 4 | 7 2 | 5 4 | 7 2 | 3+4 4 | 3+4 4 |
| | 5 4 | 9 2 | 5 4 | 9 2 | 3+4 4 | 3+4 4 |
| | 5 4 | 7 2 | 5 4 | 7 2 | 3+4 4 | 3+4 4 |
| | 5 4 | 9 2 | 5 4 | 9 2 | 3+4 4 | 3+4 4 |
| | 5 4 | 7 2 | 5 4 | 7 2 | 3+4 4 | 3+4 4 |
| | 5 4 | 9 2 | 5 4 | 9 2 | 3+4 4 | 3+4 4 |
| | 5 4 | 8 2 | 5 4 | 6 2 | 3+4 4 | 3+4 4 |
| | 5 4 | 6 2 | 5 4 | 8 2 | 3+4 4 | 3+4 4 |
| | 5 4 | 7 2 | 5 4 | 7 2 | 3+4 4 | 3+4 4 |
| | 5 4 | 9 2 | 5 4 | 9 2 | 3+4 4 | 3+4 4 |
| | 5 4 | 7 2 | 5 4 | 7 2 | 3+4 4 | 3+4 4 |
| | 5 4 | 8 2 | 5 4 | 8 2 | 3+4 4 | 3+4 4 |
| | 5 4 | 9 2 | 5 4 | 9 2 | 3+4 4 | 3+4 4 |
| | 5 4 | 7 2 | 5 4 | 7 2 | 3+4 4 | 3+4 4 |
| | 5 4 | 9 2 | 5 4 | 7 2 | 3+4 4 | 3+4 4 |
| | 5 4 | 8 2 | 5 4 | 6 2 | 3+4 4 | 3+4 4 |
| | | 6 2 | | 8 2 | | |
| | | 6 2 | | 6 2 | | |
| | | 7 2 | | 9 2 | | |
| | | 7 2 | | 7 2 | | |
| | | 6 2 | | 6 2 | | |
| | | 8 2 | | 9 2 | | |
| | | 6 2 | | 8 2 | | |

Continued on next column

**TABLE LX– continued**

| TQC | -A | -GM | -M | -Z | -R | -X |
|---|---|---|---|---|---|---|
| | | 9 2 | | 9 2 | | |
| | | 7 2 | | 8 2 | | |
| | | 6 2 | | 6 2 | | |
| | | 7 2 | | 8 2 | | |
| | | 6 2 | | 9 2 | | |
| | | 8 2 | | 8 2 | | |
| | | 9 2 | | 6 2 | | |
| | | 9 2 | | 7 2 | | |
| | | 6 2 | | 7 2 | | |
| 129.1.14 | ICSD: 61194 | NoE: 80 | SF: Pd2Sb2Sr1 | | | |
| | -A | -GM | -M | -Z | -R | -X |
| | 5 4 | 7 2 | 5 4 | 7 2 | 3+4 4 | 3+4 4 |
| | 5 4 | 9 2 | 5 4 | 9 2 | 3+4 4 | 3+4 4 |
| | 5 4 | 7 2 | 5 4 | 7 2 | 3+4 4 | 3+4 4 |
| | 5 4 | 9 2 | 5 4 | 9 2 | 3+4 4 | 3+4 4 |
| | 5 4 | 7 2 | 5 4 | 9 2 | 3+4 4 | 3+4 4 |
| | 5 4 | 9 2 | 5 4 | 7 2 | 3+4 4 | 3+4 4 |
| | 5 4 | 6 2 | 5 4 | 6 2 | 3+4 4 | 3+4 4 |
| | 5 4 | 8 2 | 5 4 | 8 2 | 3+4 4 | 3+4 4 |
| | 5 4 | 7 2 | 5 4 | 9 2 | 3+4 4 | 3+4 4 |
| | 5 4 | 7 2 | 5 4 | 7 2 | 3+4 4 | 3+4 4 |
| | 5 4 | 9 2 | 5 4 | 9 2 | 3+4 4 | 3+4 4 |
| | 5 4 | 8 2 | 5 4 | 6 2 | 3+4 4 | 3+4 4 |
| | 5 4 | 9 2 | 5 4 | 7 2 | 3+4 4 | 3+4 4 |
| | 5 4 | 6 2 | 5 4 | 8 2 | 3+4 4 | 3+4 4 |
| | 5 4 | 7 2 | 5 4 | 9 2 | 3+4 4 | 3+4 4 |
| | 5 4 | 7 2 | 5 4 | 7 2 | 3+4 4 | 3+4 4 |
| | 5 4 | 9 2 | 5 4 | 9 2 | 3+4 4 | 3+4 4 |
| | 5 4 | 9 2 | 5 4 | 8 2 | 3+4 4 | 3+4 4 |
| | 5 4 | 8 2 | 5 4 | 8 2 | 3+4 4 | 3+4 4 |
| | 5 4 | 6 2 | 5 4 | 9 2 | 3+4 4 | 3+4 4 |
| | | 6 2 | | 6 2 | | |
| | | 9 2 | | 7 2 | | |
| | | 7 2 | | 6 2 | | |
| | | 8 2 | | 7 2 | | |
| | | 6 2 | | 8 2 | | |
| | | 6 2 | | 9 2 | | |
| | | 7 2 | | 6 2 | | |
| | | 8 2 | | 6 2 | | |
| | | 8 2 | | 8 2 | | |
| | | 6 2 | | 7 2 | | |
| | | 9 2 | | 6 2 | | |
| | | 7 2 | | 9 2 | | |
| | | 9 2 | | 7 2 | | |
| | | 6 2 | | 9 2 | | |
| | | 7 2 | | 6 2 | | |
| | | 8 2 | | 8 2 | | |
| | | 7 2 | | 6 2 | | |
| | | 6 2 | | 7 2 | | |
| | | 6 2 | | 9 2 | | |
| | | 9 2 | | 8 2 | | |
| 129.1.15 | ICSD: 50349 | NoE: 68 | SF: Pd1Sb2U1 | | | |
| | -A | -GM | -M | -Z | -R | -X |
| | 5 4 | 7 2 | 5 4 | 7 2 | 3+4 4 | 3+4 4 |
| | 5 4 | 9 2 | 5 4 | 9 2 | 3+4 4 | 3+4 4 |
| | 5 4 | 7 2 | 5 4 | 7 2 | 3+4 4 | 3+4 4 |
| | 5 4 | 9 2 | 5 4 | 9 2 | 3+4 4 | 3+4 4 |
| | 5 4 | 7 2 | 5 4 | 7 2 | 3+4 4 | 3+4 4 |
| | 5 4 | 9 2 | 5 4 | 9 2 | 3+4 4 | 3+4 4 |
| | 5 4 | 6 2 | 5 4 | 6 2 | 3+4 4 | 3+4 4 |
| | 5 4 | 8 2 | 5 4 | 8 2 | 3+4 4 | 3+4 4 |

Continued on next column

**TABLE LX– continued**

| TQC | -A | -GM | -M | -Z | -R | -X |
|---|---|---|---|---|---|---|
| | 5 4 | 7 2 | 5 4 | 7 2 | 3+4 4 | 3+4 4 |
| | 5 4 | 9 2 | 5 4 | 9 2 | 3+4 4 | 3+4 4 |
| | 5 4 | 7 2 | 5 4 | 7 2 | 3+4 4 | 3+4 4 |
| | 5 4 | 8 2 | 5 4 | 8 2 | 3+4 4 | 3+4 4 |
| | 5 4 | 9 2 | 5 4 | 9 2 | 3+4 4 | 3+4 4 |
| | 5 4 | 7 2 | 5 4 | 7 2 | 3+4 4 | 3+4 4 |
| | 5 4 | 9 2 | 5 4 | 7 2 | 3+4 4 | 3+4 4 |
| | 5 4 | 8 2 | 5 4 | 6 2 | 3+4 4 | 3+4 4 |
| | 5 4 | 6 2 | 5 4 | 8 2 | 3+4 4 | 3+4 4 |
| | | 6 2 | | 6 2 | | |
| | | 7 2 | | 6 2 | | |
| | | 8 2 | | 9 2 | | |
| | | 7 2 | | 7 2 | | |
| | | 6 2 | | 9 2 | | |
| | | 9 2 | | 8 2 | | |
| | | 7 2 | | 8 2 | | |
| | | 6 2 | | 9 2 | | |
| | | 6 2 | | 9 2 | | |
| | | 7 2 | | 6 2 | | |
| | | 6 2 | | 8 2 | | |
| | | 9 2 | | 7 2 | | |
| | | 8 2 | | 7 2 | | |
| | | 9 2 | | 8 2 | | |
| | | 9 2 | | 6 2 | | |
| | | 6 2 | | 7 2 | | |
| | | 7 2 | | 9 2 | | |
| 129.1.16 | ICSD: 50348 | NoE: 68 | SF: Pd1Sb2U1 | | | |
| | -A | -GM | -M | -Z | -R | -X |
| | 5 4 | 7 2 | 5 4 | 7 2 | 3+4 4 | 3+4 4 |
| | 5 4 | 9 2 | 5 4 | 9 2 | 3+4 4 | 3+4 4 |
| | 5 4 | 7 2 | 5 4 | 7 2 | 3+4 4 | 3+4 4 |
| | 5 4 | 9 2 | 5 4 | 9 2 | 3+4 4 | 3+4 4 |
| | 5 4 | 7 2 | 5 4 | 7 2 | 3+4 4 | 3+4 4 |
| | 5 4 | 9 2 | 5 4 | 9 2 | 3+4 4 | 3+4 4 |
| | 5 4 | 6 2 | 5 4 | 6 2 | 3+4 4 | 3+4 4 |
| | 5 4 | 8 2 | 5 4 | 8 2 | 3+4 4 | 3+4 4 |
| | 5 4 | 7 2 | 5 4 | 7 2 | 3+4 4 | 3+4 4 |
| | 5 4 | 9 2 | 5 4 | 9 2 | 3+4 4 | 3+4 4 |
| | 5 4 | 7 2 | 5 4 | 7 2 | 3+4 4 | 3+4 4 |
| | 5 4 | 8 2 | 5 4 | 8 2 | 3+4 4 | 3+4 4 |
| | 5 4 | 9 2 | 5 4 | 9 2 | 3+4 4 | 3+4 4 |
| | 5 4 | 7 2 | 5 4 | 7 2 | 3+4 4 | 3+4 4 |
| | 5 4 | 9 2 | 5 4 | 7 2 | 3+4 4 | 3+4 4 |
| | 5 4 | 8 2 | 5 4 | 6 2 | 3+4 4 | 3+4 4 |
| | 5 4 | 6 2 | 5 4 | 8 2 | 3+4 4 | 3+4 4 |
| | | 6 2 | | 6 2 | | |
| | | 7 2 | | 6 2 | | |
| | | 8 2 | | 9 2 | | |
| | | 7 2 | | 7 2 | | |
| | | 6 2 | | 9 2 | | |
| | | 9 2 | | 8 2 | | |
| | | 7 2 | | 8 2 | | |
| | | 6 2 | | 9 2 | | |
| | | 6 2 | | 9 2 | | |
| | | 7 2 | | 6 2 | | |
| | | 6 2 | | 8 2 | | |
| | | 9 2 | | 7 2 | | |
| | | 8 2 | | 7 2 | | |
| | | 9 2 | | 8 2 | | |
| | | 9 2 | | 6 2 | | |
| | | 6 2 | | 7 2 | | |

Continued on next column

**TABLE LX– continued**

| TQC | ICSD | NoE | SF | | | |
|---|---|---|---|---|---|---|
| | | 7 2 | | 9 2 | | |
| 129.1.17 | 77853 | 48 | P2U1 | | | |
| | -A | -GM | -M | -Z | -R | -X |
| | 5 4 | 7 2 | 5 4 | 9 2 | 3+4 4 | 3+4 4 |
| | 5 4 | 9 2 | 5 4 | 7 2 | 3+4 4 | 3+4 4 |
| | 5 4 | 7 2 | 5 4 | 9 2 | 3+4 4 | 3+4 4 |
| | 5 4 | 9 2 | 5 4 | 7 2 | 3+4 4 | 3+4 4 |
| | 5 4 | 7 2 | 5 4 | 9 2 | 3+4 4 | 3+4 4 |
| | 5 4 | 9 2 | 5 4 | 7 2 | 3+4 4 | 3+4 4 |
| | 5 4 | 6 2 | 5 4 | 8 2 | 3+4 4 | 3+4 4 |
| | 5 4 | 8 2 | 5 4 | 6 2 | 3+4 4 | 3+4 4 |
| | 5 4 | 7 2 | 5 4 | 7 2 | 3+4 4 | 3+4 4 |
| | 5 4 | 9 2 | 5 4 | 9 2 | 3+4 4 | 3+4 4 |
| | 5 4 | 7 2 | 5 4 | 7 2 | 3+4 4 | 3+4 4 |
| | 5 4 | 8 2 | 5 4 | 8 2 | 3+4 4 | 3+4 4 |
| | | 9 2 | | 9 2 | | |
| | | 7 2 | | 7 2 | | |
| | | 6 2 | | 6 2 | | |
| | | 7 2 | | 6 2 | | |
| | | 6 2 | | 9 2 | | |
| | | 7 2 | | 9 2 | | |
| | | 6 2 | | 8 2 | | |
| | | 8 2 | | 7 2 | | |
| | | 9 2 | | 6 2 | | |
| | | 8 2 | | 7 2 | | |
| | | 9 2 | | 6 2 | | |
| | | 6 2 | | 8 2 | | |
| 129.1.18 | 25731 | 28 | O1Si1Zr1 | | | |
| | -A | -GM | -M | -Z | -R | -X |
| | 5 4 | 7 2 | 5 4 | 9 2 | 3+4 4 | 3+4 4 |
| | 5 4 | 9 2 | 5 4 | 7 2 | 3+4 4 | 3+4 4 |
| | 5 4 | 7 2 | 5 4 | 7 2 | 3+4 4 | 3+4 4 |
| | 5 4 | 9 2 | 5 4 | 9 2 | 3+4 4 | 3+4 4 |
| | 5 4 | 7 2 | 5 4 | 9 2 | 3+4 4 | 3+4 4 |
| | 5 4 | 7 2 | 5 4 | 9 2 | 3+4 4 | 3+4 4 |
| | 5 4 | 6 2 | 5 4 | 8 2 | 3+4 4 | 3+4 4 |
| | | 8 2 | | 7 2 | | |
| | | 7 2 | | 8 2 | | |
| | | 6 2 | | 6 2 | | |
| | | 9 2 | | 7 2 | | |
| | | 8 2 | | 7 2 | | |
| | | 9 2 | | 6 2 | | |
| | | 6 2 | | 6 2 | | |
| 129.1.19 | 85856 | 52 | Nd1Ru1Si1 | | | |
| | -A | -GM | -M | -Z | -R | -X |
| | 5 4 | 7 2 | 5 4 | 9 2 | 3+4 4 | 3+4 4 |
| | 5 4 | 9 2 | 5 4 | 7 2 | 3+4 4 | 3+4 4 |
| | 5 4 | 7 2 | 5 4 | 9 2 | 3+4 4 | 3+4 4 |
| | 5 4 | 9 2 | 5 4 | 7 2 | 3+4 4 | 3+4 4 |
| | 5 4 | 7 2 | 5 4 | 9 2 | 3+4 4 | 3+4 4 |
| | 5 4 | 6 2 | 5 4 | 8 2 | 3+4 4 | 3+4 4 |
| | 5 4 | 9 2 | 5 4 | 7 2 | 3+4 4 | 3+4 4 |
| | 5 4 | 8 2 | 5 4 | 6 2 | 3+4 4 | 3+4 4 |
| | 5 4 | 7 2 | 5 4 | 7 2 | 3+4 4 | 3+4 4 |
| | 5 4 | 9 2 | 5 4 | 9 2 | 3+4 4 | 3+4 4 |
| | 5 4 | 7 2 | 5 4 | 7 2 | 3+4 4 | 3+4 4 |
| | 5 4 | 8 2 | 5 4 | 9 2 | 3+4 4 | 3+4 4 |
| | 5 4 | 9 2 | 5 4 | 8 2 | 3+4 4 | 3+4 4 |
| | | 6 2 | | 9 2 | | |
| | | 9 2 | | 6 2 | | |
| | | 8 2 | | 8 2 | | |

Continued on next column

**TABLE LX– continued**

| TQC | ICSD | NoE | SF | | | |
|---|---|---|---|---|---|---|
| | | 7 2 | | 7 2 | | |
| | | 6 2 | | 6 2 | | |
| | | 6 2 | | 6 2 | | |
| | | 7 2 | | 7 2 | | |
| | | 7 2 | | 6 2 | | |
| | | 6 2 | | 7 2 | | |
| | | 8 2 | | 6 2 | | |
| | | 9 2 | | 8 2 | | |
| | | 8 2 | | 9 2 | | |
| | | 6 2 | | 8 2 | | |
| 129.1.20 | 41250 | 52 | Nd1Ru1Si1 | | | |
| | -A | -GM | -M | -Z | -R | -X |
| | 5 4 | 7 2 | 5 4 | 9 2 | 3+4 4 | 3+4 4 |
| | 5 4 | 9 2 | 5 4 | 7 2 | 3+4 4 | 3+4 4 |
| | 5 4 | 7 2 | 5 4 | 9 2 | 3+4 4 | 3+4 4 |
| | 5 4 | 9 2 | 5 4 | 7 2 | 3+4 4 | 3+4 4 |
| | 5 4 | 7 2 | 5 4 | 9 2 | 3+4 4 | 3+4 4 |
| | 5 4 | 6 2 | 5 4 | 8 2 | 3+4 4 | 3+4 4 |
| | 5 4 | 9 2 | 5 4 | 7 2 | 3+4 4 | 3+4 4 |
| | 5 4 | 8 2 | 5 4 | 6 2 | 3+4 4 | 3+4 4 |
| | 5 4 | 7 2 | 5 4 | 7 2 | 3+4 4 | 3+4 4 |
| | 5 4 | 9 2 | 5 4 | 9 2 | 3+4 4 | 3+4 4 |
| | 5 4 | 7 2 | 5 4 | 7 2 | 3+4 4 | 3+4 4 |
| | 5 4 | 8 2 | 5 4 | 9 2 | 3+4 4 | 3+4 4 |
| | 5 4 | 9 2 | 5 4 | 6 2 | 3+4 4 | 3+4 4 |
| | | 6 2 | | 8 2 | | |
| | | 9 2 | | 9 2 | | |
| | | 8 2 | | 8 2 | | |
| | | 7 2 | | 7 2 | | |
| | | 6 2 | | 6 2 | | |
| | | 6 2 | | 6 2 | | |
| | | 7 2 | | 7 2 | | |
| | | 7 2 | | 6 2 | | |
| | | 6 2 | | 7 2 | | |
| | | 8 2 | | 6 2 | | |
| | | 9 2 | | 8 2 | | |
| | | 8 2 | | 9 2 | | |
| | | 6 2 | | 8 2 | | |
| 129.1.21 | 76235 | 44 | Mn1Si1Y1 | | | |
| | -A | -GM | -M | -Z | -R | -X |
| | 5 4 | 7 2 | 5 4 | 9 2 | 3+4 4 | 3+4 4 |
| | 5 4 | 9 2 | 5 4 | 7 2 | 3+4 4 | 3+4 4 |
| | 5 4 | 7 2 | 5 4 | 9 2 | 3+4 4 | 3+4 4 |
| | 5 4 | 9 2 | 5 4 | 7 2 | 3+4 4 | 3+4 4 |
| | 5 4 | 7 2 | 5 4 | 9 2 | 3+4 4 | 3+4 4 |
| | 5 4 | 6 2 | 5 4 | 8 2 | 3+4 4 | 3+4 4 |
| | 5 4 | 8 2 | 5 4 | 6 2 | 3+4 4 | 3+4 4 |
| | 5 4 | 9 2 | 5 4 | 7 2 | 3+4 4 | 3+4 4 |
| | 5 4 | 7 2 | 5 4 | 7 2 | 3+4 4 | 3+4 4 |
| | 5 4 | 9 2 | 5 4 | 9 2 | 3+4 4 | 3+4 4 |
| | 5 4 | 7 2 | 5 4 | 9 2 | 3+4 4 | 3+4 4 |
| | | 9 2 | | 7 2 | | |
| | | 8 2 | | 8 2 | | |
| | | 9 2 | | 9 2 | | |
| | | 6 2 | | 6 2 | | |
| | | 8 2 | | 7 2 | | |
| | | 7 2 | | 6 2 | | |
| | | 6 2 | | 8 2 | | |
| | | 7 2 | | 6 2 | | |
| | | 9 2 | | 7 2 | | |
| | | 7 2 | | 9 2 | | |

Continued on next column

**TABLE LX– continued**

| TQC | ICSD | NoE | SF | | | |
|---|---|---|---|---|---|---|
| | | 6 2 | | 7 2 | | |
| 129.1.22 | 73039 | 84 | Ir2Pu1Si2 | | | |
| | -A | -GM | -M | -Z | -R | -X |
| | 5 4 | 7 2 | 5 4 | 7 2 | 3+4 4 | 3+4 4 |
| | 5 4 | 9 2 | 5 4 | 9 2 | 3+4 4 | 3+4 4 |
| | 5 4 | 7 2 | 5 4 | 7 2 | 3+4 4 | 3+4 4 |
| | 5 4 | 9 2 | 5 4 | 9 2 | 3+4 4 | 3+4 4 |
| | 5 4 | 7 2 | 5 4 | 9 2 | 3+4 4 | 3+4 4 |
| | 5 4 | 9 2 | 5 4 | 7 2 | 3+4 4 | 3+4 4 |
| | 5 4 | 8 2 | 5 4 | 8 2 | 3+4 4 | 3+4 4 |
| | 5 4 | 6 2 | 5 4 | 6 2 | 3+4 4 | 3+4 4 |
| | 5 4 | 7 2 | 5 4 | 9 2 | 3+4 4 | 3+4 4 |
| | 5 4 | 7 2 | 5 4 | 7 2 | 3+4 4 | 3+4 4 |
| | 5 4 | 9 2 | 5 4 | 9 2 | 3+4 4 | 3+4 4 |
| | 5 4 | 8 2 | 5 4 | 6 2 | 3+4 4 | 3+4 4 |
| | 5 4 | 9 2 | 5 4 | 7 2 | 3+4 4 | 3+4 4 |
| | 5 4 | 7 2 | 5 4 | 9 2 | 3+4 4 | 3+4 4 |
| | 5 4 | 7 2 | 5 4 | 7 2 | 3+4 4 | 3+4 4 |
| | 5 4 | 6 2 | 5 4 | 8 2 | 3+4 4 | 3+4 4 |
| | 5 4 | 9 2 | 5 4 | 9 2 | 3+4 4 | 3+4 4 |
| | 5 4 | 6 2 | 5 4 | 8 2 | 3+4 4 | 3+4 4 |
| | 5 4 | 8 2 | 5 4 | 6 2 | 3+4 4 | 3+4 4 |
| | 5 4 | 6 2 | 5 4 | 8 2 | 3+4 4 | 3+4 4 |
| | 5 4 | 9 2 | 5 4 | 9 2 | 3+4 4 | 3+4 4 |
| | | 6 2 | | 7 2 | | |
| | | 7 2 | | 8 2 | | |
| | | 9 2 | | 6 2 | | |
| | | 8 2 | | 7 2 | | |
| | | 8 2 | | 9 2 | | |
| | | 7 2 | | 8 2 | | |
| | | 6 2 | | 6 2 | | |
| | | 8 2 | | 7 2 | | |
| | | 9 2 | | 6 2 | | |
| | | 7 2 | | 6 2 | | |
| | | 6 2 | | 9 2 | | |
| | | 7 2 | | 7 2 | | |
| | | 6 2 | | 8 2 | | |
| | | 6 2 | | 6 2 | | |
| | | 7 2 | | 9 2 | | |
| | | 6 2 | | 6 2 | | |
| | | 8 2 | | 6 2 | | |
| | | 9 2 | | 7 2 | | |
| | | 8 2 | | 8 2 | | |
| | | 7 2 | | 9 2 | | |
| | | 6 2 | | 6 2 | | |
| 129.1.23 | 15420 | 72 | I1O1Tm1 | | | |
| | -A | -GM | -M | -Z | -R | -X |
| | 5 4 | 7 2 | 5 4 | 7 2 | 3+4 4 | 3+4 4 |
| | 5 4 | 9 2 | 5 4 | 9 2 | 3+4 4 | 3+4 4 |
| | 5 4 | 7 2 | 5 4 | 7 2 | 3+4 4 | 3+4 4 |
| | 5 4 | 9 2 | 5 4 | 9 2 | 3+4 4 | 3+4 4 |
| | 5 4 | 7 2 | 5 4 | 7 2 | 3+4 4 | 3+4 4 |
| | 5 4 | 6 2 | 5 4 | 6 2 | 3+4 4 | 3+4 4 |
| | 5 4 | 9 2 | 5 4 | 9 2 | 3+4 4 | 3+4 4 |
| | 5 4 | 8 2 | 5 4 | 8 2 | 3+4 4 | 3+4 4 |
| | 5 4 | 7 2 | 5 4 | 7 2 | 3+4 4 | 3+4 4 |
| | 5 4 | 8 2 | 5 4 | 8 2 | 3+4 4 | 3+4 4 |
| | 5 4 | 7 2 | 5 4 | 9 2 | 3+4 4 | 3+4 4 |
| | 5 4 | 9 2 | 5 4 | 7 2 | 3+4 4 | 3+4 4 |
| | 5 4 | 9 2 | 5 4 | 9 2 | 3+4 4 | 3+4 4 |
| | 5 4 | 7 2 | 5 4 | 7 2 | 3+4 4 | 3+4 4 |

Continued on next column

**TABLE LX– continued**

| TQC | ICSD | NoE | SF | | | |
|---|---|---|---|---|---|---|
| | 5 4 | 6 2 | 5 4 | 6 2 | 3+4 4 | 3+4 4 |
| | 5 4 | 6 2 | 5 4 | 6 2 | 3+4 4 | 3+4 4 |
| | 5 4 | 7 2 | 5 4 | 9 2 | 3+4 4 | 3+4 4 |
| | 5 4 | 8 2 | 5 4 | 8 2 | 3+4 4 | 3+4 4 |
| | | 9 2 | | 7 2 | | |
| | | 6 2 | | 9 2 | | |
| | | 7 2 | | 6 2 | | |
| | | 6 2 | | 6 2 | | |
| | | 8 2 | | 8 2 | | |
| | | 9 2 | | 9 2 | | |
| | | 9 2 | | 7 2 | | |
| | | 7 2 | | 8 2 | | |
| | | 8 2 | | 7 2 | | |
| | | 6 2 | | 9 2 | | |
| | | 8 2 | | 8 2 | | |
| | | 9 2 | | 6 2 | | |
| | | 6 2 | | 6 2 | | |
| | | 7 2 | | 8 2 | | |
| | | 8 2 | | 7 2 | | |
| | | 6 2 | | 9 2 | | |
| | | 7 2 | | 7 2 | | |
| | | 9 2 | | 6 2 | | |
| 129.1.24 | 25737 | 28 | Hf1Si1Te1 | | | |
| | -A | -GM | -M | -Z | -R | -X |
| | 5 4 | 7 2 | 5 4 | 9 2 | 3+4 4 | 3+4 4 |
| | 5 4 | 9 2 | 5 4 | 7 2 | 3+4 4 | 3+4 4 |
| | 5 4 | 7 2 | 5 4 | 7 2 | 3+4 4 | 3+4 4 |
| | 5 4 | 7 2 | 5 4 | 9 2 | 3+4 4 | 3+4 4 |
| | 5 4 | 9 2 | 5 4 | 8 2 | 3+4 4 | 3+4 4 |
| | 5 4 | 8 2 | 5 4 | 7 2 | 3+4 4 | 3+4 4 |
| | 5 4 | 7 2 | 5 4 | 9 2 | 3+4 4 | 3+4 4 |
| | | 6 2 | | 7 2 | | |
| | | 6 2 | | 6 2 | | |
| | | 7 2 | | 9 2 | | |
| | | 9 2 | | 6 2 | | |
| | | 9 2 | | 8 2 | | |
| | | 6 2 | | 7 2 | | |
| | | 8 2 | | 6 2 | | |
| 129.1.25 | 25736 | 28 | Hf1Se1Si1 | | | |
| | -A | -GM | -M | -Z | -R | -X |
| | 5 4 | 7 2 | 5 4 | 9 2 | 3+4 4 | 3+4 4 |
| | 5 4 | 9 2 | 5 4 | 7 2 | 3+4 4 | 3+4 4 |
| | 5 4 | 7 2 | 5 4 | 7 2 | 3+4 4 | 3+4 4 |
| | 5 4 | 7 2 | 5 4 | 9 2 | 3+4 4 | 3+4 4 |
| | 5 4 | 9 2 | 5 4 | 9 2 | 3+4 4 | 3+4 4 |
| | 5 4 | 7 2 | 5 4 | 8 2 | 3+4 4 | 3+4 4 |
| | 5 4 | 6 2 | 5 4 | 9 2 | 3+4 4 | 3+4 4 |
| | | 8 2 | | 7 2 | | |
| | | 7 2 | | 8 2 | | |
| | | 6 2 | | 7 2 | | |
| | | 6 2 | | 6 2 | | |
| | | 9 2 | | 6 2 | | |
| | | 8 2 | | 7 2 | | |
| | | 9 2 | | 6 2 | | |
| 129.1.26 | 25735 | 28 | Hf1S1Si1 | | | |
| | -A | -GM | -M | -Z | -R | -X |
| | 5 4 | 7 2 | 5 4 | 9 2 | 3+4 4 | 3+4 4 |
| | 5 4 | 9 2 | 5 4 | 7 2 | 3+4 4 | 3+4 4 |
| | 5 4 | 7 2 | 5 4 | 7 2 | 3+4 4 | 3+4 4 |
| | 5 4 | 7 2 | 5 4 | 9 2 | 3+4 4 | 3+4 4 |
| | 5 4 | 9 2 | 5 4 | 9 2 | 3+4 4 | 3+4 4 |

Continued on next column

**TABLE LX– continued**

| TQC | ICSD | | NoE | | SF | |
|---|---|---|---|---|---|---|
| | 5 4 | 7 2 | 5 4 | 8 2 | 3+4 4 | 3+4 4 |
| | 5 4 | 6 2 | 5 4 | 9 2 | 3+4 4 | 3+4 4 |
| | | 8 2 | | 7 2 | | |
| | | 7 2 | | 8 2 | | |
| | | 6 2 | | 7 2 | | |
| | | 9 2 | | 6 2 | | |
| | | 6 2 | | 6 2 | | |
| | | 8 2 | | 7 2 | | |
| | | 9 2 | | 6 2 | | |
| 129.1.27 | 25740 | | 28 | | Ge1Te1Zr1 | |
| | -A | -GM | -M | -Z | -R | -X |
| | 5 4 | 7 2 | 5 4 | 9 2 | 3+4 4 | 3+4 4 |
| | 5 4 | 9 2 | 5 4 | 7 2 | 3+4 4 | 3+4 4 |
| | 5 4 | 7 2 | 5 4 | 7 2 | 3+4 4 | 3+4 4 |
| | 5 4 | 7 2 | 5 4 | 9 2 | 3+4 4 | 3+4 4 |
| | 5 4 | 8 2 | 5 4 | 8 2 | 3+4 4 | 3+4 4 |
| | 5 4 | 9 2 | 5 4 | 9 2 | 3+4 4 | 3+4 4 |
| | 5 4 | 7 2 | 5 4 | 8 2 | 3+4 4 | 3+4 4 |
| | | 6 2 | | 6 2 | | |
| | | 7 2 | | 7 2 | | |
| | | 6 2 | | 6 2 | | |
| | | 6 2 | | 9 2 | | |
| | | 9 2 | | 7 2 | | |
| | | 8 2 | | 7 2 | | |
| | | 9 2 | | 6 2 | | |
| 129.1.28 | 25739 | | 28 | | Ge1Se1Zr1 | |
| | -A | -GM | -M | -Z | -R | -X |
| | 5 4 | 7 2 | 5 4 | 9 2 | 3+4 4 | 3+4 4 |
| | 5 4 | 9 2 | 5 4 | 7 2 | 3+4 4 | 3+4 4 |
| | 5 4 | 7 2 | 5 4 | 7 2 | 3+4 4 | 3+4 4 |
| | 5 4 | 7 2 | 5 4 | 9 2 | 3+4 4 | 3+4 4 |
| | 5 4 | 9 2 | 5 4 | 9 2 | 3+4 4 | 3+4 4 |
| | 5 4 | 7 2 | 5 4 | 8 2 | 3+4 4 | 3+4 4 |
| | 5 4 | 8 2 | 5 4 | 8 2 | 3+4 4 | 3+4 4 |
| | | 6 2 | | 9 2 | | |
| | | 6 2 | | 7 2 | | |
| | | 7 2 | | 6 2 | | |
| | | 6 2 | | 7 2 | | |
| | | 9 2 | | 6 2 | | |
| | | 8 2 | | 7 2 | | |
| | | 9 2 | | 6 2 | | |
| 129.1.29 | 15574 | | 28 | | Ge1Se1Zr1 | |
| | -A | -GM | -M | -Z | -R | -X |
| | 5 4 | 7 2 | 5 4 | 9 2 | 3+4 4 | 3+4 4 |
| | 5 4 | 9 2 | 5 4 | 7 2 | 3+4 4 | 3+4 4 |
| | 5 4 | 7 2 | 5 4 | 7 2 | 3+4 4 | 3+4 4 |
| | 5 4 | 7 2 | 5 4 | 9 2 | 3+4 4 | 3+4 4 |
| | 5 4 | 9 2 | 5 4 | 8 2 | 3+4 4 | 3+4 4 |
| | 5 4 | 8 2 | 5 4 | 9 2 | 3+4 4 | 3+4 4 |
| | 5 4 | 7 2 | 5 4 | 8 2 | 3+4 4 | 3+4 4 |
| | | 6 2 | | 9 2 | | |
| | | 6 2 | | 6 2 | | |
| | | 7 2 | | 7 2 | | |
| | | 6 2 | | 7 2 | | |
| | | 9 2 | | 6 2 | | |
| | | 8 2 | | 7 2 | | |
| | | 9 2 | | 6 2 | | |
| 129.1.30 | 25738 | | 28 | | Ge1S1Zr1 | |
| | -A | -GM | -M | -Z | -R | -X |
| | 5 4 | 7 2 | 5 4 | 9 2 | 3+4 4 | 3+4 4 |
| | 5 4 | 9 2 | 5 4 | 7 2 | 3+4 4 | 3+4 4 |

Continued on next column

**TABLE LX– continued**

| TQC | ICSD | | NoE | | SF | |
|---|---|---|---|---|---|---|
| | 5 4 | 7 2 | 5 4 | 7 2 | 3+4 4 | 3+4 4 |
| | 5 4 | 7 2 | 5 4 | 9 2 | 3+4 4 | 3+4 4 |
| | 5 4 | 9 2 | 5 4 | 9 2 | 3+4 4 | 3+4 4 |
| | 5 4 | 7 2 | 5 4 | 8 2 | 3+4 4 | 3+4 4 |
| | 5 4 | 6 2 | 5 4 | 8 2 | 3+4 4 | 3+4 4 |
| | | 8 2 | | 9 2 | | |
| | | 6 2 | | 7 2 | | |
| | | 7 2 | | 6 2 | | |
| | | 6 2 | | 7 2 | | |
| | | 9 2 | | 6 2 | | |
| | | 8 2 | | 7 2 | | |
| | | 9 2 | | 6 2 | | |
| 129.1.31 | 15573 | | 28 | | Ge1S1Zr1 | |
| | -A | -GM | -M | -Z | -R | -X |
| | 5 4 | 7 2 | 5 4 | 9 2 | 3+4 4 | 3+4 4 |
| | 5 4 | 9 2 | 5 4 | 7 2 | 3+4 4 | 3+4 4 |
| | 5 4 | 7 2 | 5 4 | 7 2 | 3+4 4 | 3+4 4 |
| | 5 4 | 7 2 | 5 4 | 9 2 | 3+4 4 | 3+4 4 |
| | 5 4 | 9 2 | 5 4 | 9 2 | 3+4 4 | 3+4 4 |
| | 5 4 | 7 2 | 5 4 | 8 2 | 3+4 4 | 3+4 4 |
| | 5 4 | 6 2 | 5 4 | 8 2 | 3+4 4 | 3+4 4 |
| | | 8 2 | | 9 2 | | |
| | | 6 2 | | 7 2 | | |
| | | 7 2 | | 6 2 | | |
| | | 9 2 | | 7 2 | | |
| | | 6 2 | | 6 2 | | |
| | | 8 2 | | 7 2 | | |
| | | 9 2 | | 6 2 | | |
| 129.1.32 | 87315 | | 48 | | Ge1S1U1 | |
| | -A | -GM | -M | -Z | -R | -X |
| | 5 4 | 7 2 | 5 4 | 9 2 | 3+4 4 | 3+4 4 |
| | 5 4 | 9 2 | 5 4 | 7 2 | 3+4 4 | 3+4 4 |
| | 5 4 | 7 2 | 5 4 | 9 2 | 3+4 4 | 3+4 4 |
| | 5 4 | 9 2 | 5 4 | 7 2 | 3+4 4 | 3+4 4 |
| | 5 4 | 7 2 | 5 4 | 9 2 | 3+4 4 | 3+4 4 |
| | 5 4 | 9 2 | 5 4 | 7 2 | 3+4 4 | 3+4 4 |
| | 5 4 | 6 2 | 5 4 | 8 2 | 3+4 4 | 3+4 4 |
| | 5 4 | 8 2 | 5 4 | 6 2 | 3+4 4 | 3+4 4 |
| | 5 4 | 7 2 | 5 4 | 9 2 | 3+4 4 | 3+4 4 |
| | 5 4 | 9 2 | 5 4 | 7 2 | 3+4 4 | 3+4 4 |
| | 5 4 | 7 2 | 5 4 | 7 2 | 3+4 4 | 3+4 4 |
| | 5 4 | 7 2 | 5 4 | 9 2 | 3+4 4 | 3+4 4 |
| | | 9 2 | | 9 2 | | |
| | | 8 2 | | 8 2 | | |
| | | 7 2 | | 8 2 | | |
| | | 6 2 | | 7 2 | | |
| | | 9 2 | | 7 2 | | |
| | | 8 2 | | 6 2 | | |
| | | 7 2 | | 9 2 | | |
| | | 6 2 | | 6 2 | | |
| | | 6 2 | | 6 2 | | |
| | | 8 2 | | 7 2 | | |
| | | 9 2 | | 8 2 | | |
| | | 6 2 | | 6 2 | | |
| 129.1.33 | 26272 | | 48 | | Ge1S1U1 | |
| | -A | -GM | -M | -Z | -R | -X |
| | 5 4 | 7 2 | 5 4 | 9 2 | 3+4 4 | 3+4 4 |
| | 5 4 | 9 2 | 5 4 | 7 2 | 3+4 4 | 3+4 4 |
| | 5 4 | 7 2 | 5 4 | 9 2 | 3+4 4 | 3+4 4 |
| | 5 4 | 9 2 | 5 4 | 7 2 | 3+4 4 | 3+4 4 |
| | 5 4 | 7 2 | 5 4 | 7 2 | 3+4 4 | 3+4 4 |

Continued on next column

**TABLE LX– continued**

| TQC | ICSD | | NoE | | SF | |
|---|---|---|---|---|---|---|
| | 5 4 | 9 2 | 5 4 | 9 2 | 3+4 4 | 3+4 4 |
| | 5 4 | 6 2 | 5 4 | 8 2 | 3+4 4 | 3+4 4 |
| | 5 4 | 8 2 | 5 4 | 6 2 | 3+4 4 | 3+4 4 |
| | 5 4 | 7 2 | 5 4 | 7 2 | 3+4 4 | 3+4 4 |
| | 5 4 | 8 2 | 5 4 | 8 2 | 3+4 4 | 3+4 4 |
| | 5 4 | 7 2 | 5 4 | 9 2 | 3+4 4 | 3+4 4 |
| | 5 4 | 9 2 | 5 4 | 7 2 | 3+4 4 | 3+4 4 |
| | | 6 2 | | 7 2 | | |
| | | 7 2 | | 6 2 | | |
| | | 9 2 | | 9 2 | | |
| | | 6 2 | | 6 2 | | |
| | | 7 2 | | 9 2 | | |
| | | 8 2 | | 9 2 | | |
| | | 7 2 | | 8 2 | | |
| | | 6 2 | | 7 2 | | |
| | | 8 2 | | 6 2 | | |
| | | 9 2 | | 6 2 | | |
| | | 9 2 | | 8 2 | | |
| | | 6 2 | | 7 2 | | |
| 129.1.34 | 85863 | | 52 | | Ge1Nd1Ru1 | |
| | -A | -GM | -M | -Z | -R | -X |
| | 5 4 | 7 2 | 5 4 | 9 2 | 3+4 4 | 3+4 4 |
| | 5 4 | 9 2 | 5 4 | 7 2 | 3+4 4 | 3+4 4 |
| | 5 4 | 7 2 | 5 4 | 9 2 | 3+4 4 | 3+4 4 |
| | 5 4 | 9 2 | 5 4 | 7 2 | 3+4 4 | 3+4 4 |
| | 5 4 | 7 2 | 5 4 | 9 2 | 3+4 4 | 3+4 4 |
| | 5 4 | 6 2 | 5 4 | 8 2 | 3+4 4 | 3+4 4 |
| | 5 4 | 9 2 | 5 4 | 7 2 | 3+4 4 | 3+4 4 |
| | 5 4 | 8 2 | 5 4 | 6 2 | 3+4 4 | 3+4 4 |
| | 5 4 | 7 2 | 5 4 | 7 2 | 3+4 4 | 3+4 4 |
| | 5 4 | 9 2 | 5 4 | 9 2 | 3+4 4 | 3+4 4 |
| | 5 4 | 7 2 | 5 4 | 7 2 | 3+4 4 | 3+4 4 |
| | 5 4 | 9 2 | 5 4 | 9 2 | 3+4 4 | 3+4 4 |
| | 5 4 | 8 2 | 5 4 | 9 2 | 3+4 4 | 3+4 4 |
| | | 9 2 | | 8 2 | | |
| | | 6 2 | | 6 2 | | |
| | | 8 2 | | 8 2 | | |
| | | 7 2 | | 7 2 | | |
| | | 6 2 | | 6 2 | | |
| | | 7 2 | | 7 2 | | |
| | | 6 2 | | 6 2 | | |
| | | 7 2 | | 7 2 | | |
| | | 6 2 | | 6 2 | | |
| | | 8 2 | | 6 2 | | |
| | | 9 2 | | 8 2 | | |
| | | 8 2 | | 9 2 | | |
| | | 6 2 | | 8 2 | | |
| 129.1.35 | 25743 | | 28 | | Ge1Hf1Te1 | |
| | -A | -GM | -M | -Z | -R | -X |
| | 5 4 | 7 2 | 5 4 | 9 2 | 3+4 4 | 3+4 4 |
| | 5 4 | 9 2 | 5 4 | 7 2 | 3+4 4 | 3+4 4 |
| | 5 4 | 7 2 | 5 4 | 7 2 | 3+4 4 | 3+4 4 |
| | 5 4 | 7 2 | 5 4 | 9 2 | 3+4 4 | 3+4 4 |
| | 5 4 | 9 2 | 5 4 | 8 2 | 3+4 4 | 3+4 4 |
| | 5 4 | 8 2 | 5 4 | 9 2 | 3+4 4 | 3+4 4 |
| | 5 4 | 7 2 | 5 4 | 8 2 | 3+4 4 | 3+4 4 |
| | | 6 2 | | 6 2 | | |
| | | 7 2 | | 7 2 | | |
| | | 6 2 | | 7 2 | | |
| | | 6 2 | | 6 2 | | |
| | | 9 2 | | 9 2 | | |
| | | 8 2 | | 7 2 | | |
| | | 9 2 | | 6 2 | | |
| 129.1.36 | 25741 | | 28 | | Ge1Hf1Se1 | |
| | -A | -GM | -M | -Z | -R | -X |
| | 5 4 | 7 2 | 5 4 | 9 2 | 3+4 4 | 3+4 4 |
| | 5 4 | 9 2 | 5 4 | 7 2 | 3+4 4 | 3+4 4 |
| | 5 4 | 7 2 | 5 4 | 7 2 | 3+4 4 | 3+4 4 |
| | 5 4 | 7 2 | 5 4 | 9 2 | 3+4 4 | 3+4 4 |
| | 5 4 | 9 2 | 5 4 | 9 2 | 3+4 4 | 3+4 4 |
| | 5 4 | 7 2 | 5 4 | 8 2 | 3+4 4 | 3+4 4 |
| | 5 4 | 8 2 | 5 4 | 8 2 | 3+4 4 | 3+4 4 |
| | | 6 2 | | 9 2 | | |
| | | 6 2 | | 7 2 | | |
| | | 7 2 | | 6 2 | | |
| | | 6 2 | | 7 2 | | |
| | | 9 2 | | 6 2 | | |
| | | 8 2 | | 7 2 | | |
| | | 9 2 | | 6 2 | | |
| 129.1.37 | 25742 | | 28 | | Ge1Hf1S1 | |
| | -A | -GM | -M | -Z | -R | -X |
| | 5 4 | 7 2 | 5 4 | 9 2 | 3+4 4 | 3+4 4 |
| | 5 4 | 9 2 | 5 4 | 7 2 | 3+4 4 | 3+4 4 |
| | 5 4 | 7 2 | 5 4 | 7 2 | 3+4 4 | 3+4 4 |
| | 5 4 | 7 2 | 5 4 | 9 2 | 3+4 4 | 3+4 4 |
| | 5 4 | 9 2 | 5 4 | 9 2 | 3+4 4 | 3+4 4 |
| | 5 4 | 7 2 | 5 4 | 8 2 | 3+4 4 | 3+4 4 |
| | 5 4 | 6 2 | 5 4 | 8 2 | 3+4 4 | 3+4 4 |
| | | 8 2 | | 9 2 | | |
| | | 6 2 | | 7 2 | | |
| | | 7 2 | | 6 2 | | |
| | | 9 2 | | 7 2 | | |
| | | 8 2 | | 6 2 | | |
| | | 6 2 | | 7 2 | | |
| | | 9 2 | | 6 2 | | |
| 129.1.38 | 106758 | | 92 | | Gd1Pt2Si2 | |
| | -A | -GM | -M | -Z | -R | -X |
| | 5 4 | 7 2 | 5 4 | 7 2 | 3+4 4 | 3+4 4 |
| | 5 4 | 9 2 | 5 4 | 9 2 | 3+4 4 | 3+4 4 |
| | 5 4 | 7 2 | 5 4 | 7 2 | 3+4 4 | 3+4 4 |
| | 5 4 | 9 2 | 5 4 | 9 2 | 3+4 4 | 3+4 4 |
| | 5 4 | 7 2 | 5 4 | 9 2 | 3+4 4 | 3+4 4 |
| | 5 4 | 9 2 | 5 4 | 7 2 | 3+4 4 | 3+4 4 |
| | 5 4 | 6 2 | 5 4 | 8 2 | 3+4 4 | 3+4 4 |
| | 5 4 | 8 2 | 5 4 | 6 2 | 3+4 4 | 3+4 4 |
| | 5 4 | 7 2 | 5 4 | 9 2 | 3+4 4 | 3+4 4 |
| | 5 4 | 7 2 | 5 4 | 7 2 | 3+4 4 | 3+4 4 |
| | 5 4 | 9 2 | 5 4 | 9 2 | 3+4 4 | 3+4 4 |
| | 5 4 | 8 2 | 5 4 | 6 2 | 3+4 4 | 3+4 4 |
| | 5 4 | 9 2 | 5 4 | 7 2 | 3+4 4 | 3+4 4 |
| | 5 4 | 7 2 | 5 4 | 9 2 | 3+4 4 | 3+4 4 |
| | 5 4 | 7 2 | 5 4 | 7 2 | 3+4 4 | 3+4 4 |
| | 5 4 | 6 2 | 5 4 | 8 2 | 3+4 4 | 3+4 4 |
| | 5 4 | 9 2 | 5 4 | 9 2 | 3+4 4 | 3+4 4 |
| | 5 4 | 6 2 | 5 4 | 8 2 | 3+4 4 | 3+4 4 |
| | 5 4 | 8 2 | 5 4 | 6 2 | 3+4 4 | 3+4 4 |
| | 5 4 | 6 2 | 5 4 | 8 2 | 3+4 4 | 3+4 4 |
| | 5 4 | 9 2 | 5 4 | 9 2 | 3+4 4 | 3+4 4 |
| | 5 4 | 9 2 | 5 4 | 7 2 | 3+4 4 | 3+4 4 |
| | 5 4 | 6 2 | 5 4 | 8 2 | 3+4 4 | 3+4 4 |
| | | 7 2 | | 6 2 | | |
| | | 8 2 | | 7 2 | | |

Continued on next column

**TABLE LX– continued**

| TQC | ICSD | | NoE | | SF | |
|---|---|---|---|---|---|---|
| | | 7 2 | | 9 2 | | |
| | | 8 2 | | 6 2 | | |
| | | 6 2 | | 8 2 | | |
| | | 8 2 | | 6 2 | | |
| | | 9 2 | | 7 2 | | |
| | | 6 2 | | 6 2 | | |
| | | 7 2 | | 9 2 | | |
| | | 6 2 | | 6 2 | | |
| | | 7 2 | | 7 2 | | |
| | | 9 2 | | 8 2 | | |
| | | 7 2 | | 9 2 | | |
| | | 6 2 | | 9 2 | | |
| | | 8 2 | | 6 2 | | |
| | | 9 2 | | 6 2 | | |
| | | 6 2 | | 7 2 | | |
| | | 8 2 | | 8 2 | | |
| | | 7 2 | | 9 2 | | |
| | | 6 2 | | 8 2 | | |
| | | 9 2 | | 7 2 | | |
| | | 8 2 | | 6 2 | | |
| | | 8 2 | | 6 2 | | |
| 129.1.39 | 180602 | | 28 | | Fe1Te1 | |
| | -A | -GM | -M | -Z | -R | -X |
| | 5 4 | 7 2 | 5 4 | 9 2 | 3+4 4 | 3+4 4 |
| | 5 4 | 9 2 | 5 4 | 7 2 | 3+4 4 | 3+4 4 |
| | 5 4 | 7 2 | 5 4 | 9 2 | 3+4 4 | 3+4 4 |
| | 5 4 | 9 2 | 5 4 | 7 2 | 3+4 4 | 3+4 4 |
| | 5 4 | 9 2 | 5 4 | 9 2 | 3+4 4 | 3+4 4 |
| | 5 4 | 8 2 | 5 4 | 8 2 | 3+4 4 | 3+4 4 |
| | 5 4 | 7 2 | 5 4 | 6 2 | 3+4 4 | 3+4 4 |
| | | 6 2 | | 7 2 | | |
| | | 6 2 | | 6 2 | | |
| | | 7 2 | | 9 2 | | |
| | | 8 2 | | 7 2 | | |
| | | 7 2 | | 8 2 | | |
| | | 9 2 | | 7 2 | | |
| | | 6 2 | | 6 2 | | |
| 129.1.40 | 169271 | | 28 | | Fe1Se1 | |
| | -A | -GM | -M | -Z | -R | -X |
| | 5 4 | 7 2 | 5 4 | 9 2 | 3+4 4 | 3+4 4 |
| | 5 4 | 9 2 | 5 4 | 7 2 | 3+4 4 | 3+4 4 |
| | 5 4 | 7 2 | 5 4 | 9 2 | 3+4 4 | 3+4 4 |
| | 5 4 | 9 2 | 5 4 | 7 2 | 3+4 4 | 3+4 4 |
| | 5 4 | 9 2 | 5 4 | 9 2 | 3+4 4 | 3+4 4 |
| | 5 4 | 8 2 | 5 4 | 8 2 | 3+4 4 | 3+4 4 |
| | 5 4 | 7 2 | 5 4 | 6 2 | 3+4 4 | 3+4 4 |
| | | 6 2 | | 7 2 | | |
| | | 6 2 | | 6 2 | | |
| | | 7 2 | | 7 2 | | |
| | | 8 2 | | 8 2 | | |
| | | 9 2 | | 9 2 | | |
| | | 7 2 | | 6 2 | | |
| | | 6 2 | | 7 2 | | |
| 129.1.41 | 169269 | | 28 | | Fe1Se1 | |
| | -A | -GM | -M | -Z | -R | -X |
| | 5 4 | 7 2 | 5 4 | 9 2 | 3+4 4 | 3+4 4 |
| | 5 4 | 9 2 | 5 4 | 7 2 | 3+4 4 | 3+4 4 |
| | 5 4 | 7 2 | 5 4 | 9 2 | 3+4 4 | 3+4 4 |
| | 5 4 | 9 2 | 5 4 | 7 2 | 3+4 4 | 3+4 4 |
| | 5 4 | 9 2 | 5 4 | 9 2 | 3+4 4 | 3+4 4 |
| | 5 4 | 7 2 | 5 4 | 8 2 | 3+4 4 | 3+4 4 |

Continued on next column

**TABLE LX– continued**

| TQC | ICSD | | NoE | | SF | |
|---|---|---|---|---|---|---|
| | 5 4 | 8 2 | 5 4 | 6 2 | 3+4 4 | 3+4 4 |
| | | 6 2 | | 7 2 | | |
| | | 6 2 | | 6 2 | | |
| | | 7 2 | | 7 2 | | |
| | | 9 2 | | 8 2 | | |
| | | 8 2 | | 9 2 | | |
| | | 7 2 | | 6 2 | | |
| | | 6 2 | | 7 2 | | |
| 129.1.42 | 169267 | | 28 | | Fe1Se1 | |
| | -A | -GM | -M | -Z | -R | -X |
| | 5 4 | 7 2 | 5 4 | 9 2 | 3+4 4 | 3+4 4 |
| | 5 4 | 9 2 | 5 4 | 7 2 | 3+4 4 | 3+4 4 |
| | 5 4 | 7 2 | 5 4 | 9 2 | 3+4 4 | 3+4 4 |
| | 5 4 | 9 2 | 5 4 | 7 2 | 3+4 4 | 3+4 4 |
| | 5 4 | 9 2 | 5 4 | 9 2 | 3+4 4 | 3+4 4 |
| | 5 4 | 7 2 | 5 4 | 8 2 | 3+4 4 | 3+4 4 |
| | 5 4 | 8 2 | 5 4 | 6 2 | 3+4 4 | 3+4 4 |
| | | 6 2 | | 7 2 | | |
| | | 6 2 | | 6 2 | | |
| | | 7 2 | | 7 2 | | |
| | | 9 2 | | 8 2 | | |
| | | 8 2 | | 9 2 | | |
| | | 7 2 | | 6 2 | | |
| | | 6 2 | | 7 2 | | |
| 129.1.43 | 169265 | | 28 | | Fe1Se1 | |
| | -A | -GM | -M | -Z | -R | -X |
| | 5 4 | 7 2 | 5 4 | 9 2 | 3+4 4 | 3+4 4 |
| | 5 4 | 9 2 | 5 4 | 7 2 | 3+4 4 | 3+4 4 |
| | 5 4 | 7 2 | 5 4 | 9 2 | 3+4 4 | 3+4 4 |
| | 5 4 | 9 2 | 5 4 | 7 2 | 3+4 4 | 3+4 4 |
| | 5 4 | 9 2 | 5 4 | 9 2 | 3+4 4 | 3+4 4 |
| | 5 4 | 8 2 | 5 4 | 8 2 | 3+4 4 | 3+4 4 |
| | 5 4 | 7 2 | 5 4 | 6 2 | 3+4 4 | 3+4 4 |
| | | 6 2 | | 7 2 | | |
| | | 6 2 | | 6 2 | | |
| | | 7 2 | | 7 2 | | |
| | | 8 2 | | 8 2 | | |
| | | 9 2 | | 9 2 | | |
| | | 7 2 | | 6 2 | | |
| | | 6 2 | | 7 2 | | |
| 129.1.44 | 169263 | | 28 | | Fe1Se1 | |
| | -A | -GM | -M | -Z | -R | -X |
| | 5 4 | 7 2 | 5 4 | 9 2 | 3+4 4 | 3+4 4 |
| | 5 4 | 9 2 | 5 4 | 7 2 | 3+4 4 | 3+4 4 |
| | 5 4 | 7 2 | 5 4 | 9 2 | 3+4 4 | 3+4 4 |
| | 5 4 | 9 2 | 5 4 | 7 2 | 3+4 4 | 3+4 4 |
| | 5 4 | 9 2 | 5 4 | 9 2 | 3+4 4 | 3+4 4 |
| | 5 4 | 8 2 | 5 4 | 8 2 | 3+4 4 | 3+4 4 |
| | 5 4 | 7 2 | 5 4 | 6 2 | 3+4 4 | 3+4 4 |
| | | 6 2 | | 7 2 | | |
| | | 6 2 | | 6 2 | | |
| | | 7 2 | | 7 2 | | |
| | | 8 2 | | 8 2 | | |
| | | 9 2 | | 9 2 | | |
| | | 7 2 | | 6 2 | | |
| | | 6 2 | | 7 2 | | |
| 129.1.45 | 169261 | | 28 | | Fe1Se1 | |
| | -A | -GM | -M | -Z | -R | -X |
| | 5 4 | 7 2 | 5 4 | 9 2 | 3+4 4 | 3+4 4 |
| | 5 4 | 9 2 | 5 4 | 7 2 | 3+4 4 | 3+4 4 |
| | 5 4 | 7 2 | 5 4 | 9 2 | 3+4 4 | 3+4 4 |

Continued on next column

**TABLE LX– continued**

| TQC | ICSD | | NoE | | SF | |
|---|---|---|---|---|---|---|
| | 5 4 | 9 2 | 5 4 | 7 2 | 3+4 4 | 3+4 4 |
| | 5 4 | 9 2 | 5 4 | 9 2 | 3+4 4 | 3+4 4 |
| | 5 4 | 8 2 | 5 4 | 8 2 | 3+4 4 | 3+4 4 |
| | 5 4 | 7 2 | 5 4 | 7 2 | 3+4 4 | 3+4 4 |
| | | 6 2 | | 6 2 | | |
| | | 6 2 | | 6 2 | | |
| | | 7 2 | | 7 2 | | |
| | | 8 2 | | 8 2 | | |
| | | 9 2 | | 9 2 | | |
| | | 7 2 | | 6 2 | | |
| | | 6 2 | | 7 2 | | |
| 129.1.46 | 169259 | | 28 | | Fe1Se1 | |
| | -A | -GM | -M | -Z | -R | -X |
| | 5 4 | 7 2 | 5 4 | 9 2 | 3+4 4 | 3+4 4 |
| | 5 4 | 9 2 | 5 4 | 7 2 | 3+4 4 | 3+4 4 |
| | 5 4 | 7 2 | 5 4 | 9 2 | 3+4 4 | 3+4 4 |
| | 5 4 | 9 2 | 5 4 | 7 2 | 3+4 4 | 3+4 4 |
| | 5 4 | 9 2 | 5 4 | 9 2 | 3+4 4 | 3+4 4 |
| | 5 4 | 8 2 | 5 4 | 8 2 | 3+4 4 | 3+4 4 |
| | 5 4 | 7 2 | 5 4 | 6 2 | 3+4 4 | 3+4 4 |
| | | 6 2 | | 7 2 | | |
| | | 6 2 | | 6 2 | | |
| | | 7 2 | | 7 2 | | |
| | | 8 2 | | 8 2 | | |
| | | 9 2 | | 9 2 | | |
| | | 7 2 | | 6 2 | | |
| | | 6 2 | | 7 2 | | |
| 129.1.47 | 169257 | | 28 | | Fe1Se1 | |
| | -A | -GM | -M | -Z | -R | -X |
| | 5 4 | 7 2 | 5 4 | 9 2 | 3+4 4 | 3+4 4 |
| | 5 4 | 9 2 | 5 4 | 7 2 | 3+4 4 | 3+4 4 |
| | 5 4 | 7 2 | 5 4 | 9 2 | 3+4 4 | 3+4 4 |
| | 5 4 | 9 2 | 5 4 | 7 2 | 3+4 4 | 3+4 4 |
| | 5 4 | 9 2 | 5 4 | 9 2 | 3+4 4 | 3+4 4 |
| | 5 4 | 8 2 | 5 4 | 8 2 | 3+4 4 | 3+4 4 |
| | 5 4 | 7 2 | 5 4 | 7 2 | 3+4 4 | 3+4 4 |
| | | 6 2 | | 6 2 | | |
| | | 6 2 | | 6 2 | | |
| | | 7 2 | | 7 2 | | |
| | | 8 2 | | 8 2 | | |
| | | 9 2 | | 9 2 | | |
| | | 7 2 | | 6 2 | | |
| | | 6 2 | | 7 2 | | |
| 129.1.48 | 169255 | | 28 | | Fe1Se1 | |
| | -A | -GM | -M | -Z | -R | -X |
| | 5 4 | 7 2 | 5 4 | 9 2 | 3+4 4 | 3+4 4 |
| | 5 4 | 9 2 | 5 4 | 7 2 | 3+4 4 | 3+4 4 |
| | 5 4 | 7 2 | 5 4 | 9 2 | 3+4 4 | 3+4 4 |
| | 5 4 | 9 2 | 5 4 | 7 2 | 3+4 4 | 3+4 4 |
| | 5 4 | 9 2 | 5 4 | 9 2 | 3+4 4 | 3+4 4 |
| | 5 4 | 8 2 | 5 4 | 8 2 | 3+4 4 | 3+4 4 |
| | 5 4 | 7 2 | 5 4 | 7 2 | 3+4 4 | 3+4 4 |
| | | 6 2 | | 6 2 | | |
| | | 6 2 | | 6 2 | | |
| | | 7 2 | | 7 2 | | |
| | | 8 2 | | 8 2 | | |
| | | 9 2 | | 9 2 | | |
| | | 7 2 | | 6 2 | | |
| | | 6 2 | | 7 2 | | |
| 129.1.49 | 169253 | | 28 | | Fe1Se1 | |
| | -A | -GM | -M | -Z | -R | -X |

Continued on next column

**TABLE LX– continued**

| TQC | ICSD | | NoE | | SF | |
|---|---|---|---|---|---|---|
| | 5 4 | 7 2 | 5 4 | 9 2 | 3+4 4 | 3+4 4 |
| | 5 4 | 9 2 | 5 4 | 7 2 | 3+4 4 | 3+4 4 |
| | 5 4 | 7 2 | 5 4 | 9 2 | 3+4 4 | 3+4 4 |
| | 5 4 | 9 2 | 5 4 | 7 2 | 3+4 4 | 3+4 4 |
| | 5 4 | 9 2 | 5 4 | 9 2 | 3+4 4 | 3+4 4 |
| | 5 4 | 8 2 | 5 4 | 8 2 | 3+4 4 | 3+4 4 |
| | 5 4 | 7 2 | 5 4 | 7 2 | 3+4 4 | 3+4 4 |
| | | 6 2 | | 6 2 | | |
| | | 6 2 | | 6 2 | | |
| | | 7 2 | | 7 2 | | |
| | | 8 2 | | 8 2 | | |
| | | 9 2 | | 9 2 | | |
| | | 7 2 | | 6 2 | | |
| | | 6 2 | | 7 2 | | |
| 129.1.50 | 94475 | | 68 | | F1Ho1S1 | |
| | -A | -GM | -M | -Z | -R | -X |
| | 5 4 | 7 2 | 5 4 | 7 2 | 3+4 4 | 3+4 4 |
| | 5 4 | 9 2 | 5 4 | 9 2 | 3+4 4 | 3+4 4 |
| | 5 4 | 7 2 | 5 4 | 7 2 | 3+4 4 | 3+4 4 |
| | 5 4 | 9 2 | 5 4 | 9 2 | 3+4 4 | 3+4 4 |
| | 5 4 | 7 2 | 5 4 | 7 2 | 3+4 4 | 3+4 4 |
| | 5 4 | 8 2 | 5 4 | 8 2 | 3+4 4 | 3+4 4 |
| | 5 4 | 6 2 | 5 4 | 9 2 | 3+4 4 | 3+4 4 |
| | 5 4 | 9 2 | 5 4 | 6 2 | 3+4 4 | 3+4 4 |
| | 5 4 | 8 2 | 5 4 | 8 2 | 3+4 4 | 3+4 4 |
| | 5 4 | 7 2 | 5 4 | 7 2 | 3+4 4 | 3+4 4 |
| | 5 4 | 7 2 | 5 4 | 9 2 | 3+4 4 | 3+4 4 |
| | 5 4 | 9 2 | 5 4 | 7 2 | 3+4 4 | 3+4 4 |
| | 5 4 | 7 2 | 5 4 | 7 2 | 3+4 4 | 3+4 4 |
| | 5 4 | 6 2 | 5 4 | 6 2 | 3+4 4 | 3+4 4 |
| | 5 4 | 9 2 | 5 4 | 9 2 | 3+4 4 | 3+4 4 |
| | 5 4 | 6 2 | 5 4 | 6 2 | 3+4 4 | 3+4 4 |
| | 5 4 | 9 2 | 5 4 | 9 2 | 3+4 4 | 3+4 4 |
| | | 8 2 | | 8 2 | | |
| | | 7 2 | | 9 2 | | |
| | | 7 2 | | 9 2 | | |
| | | 6 2 | | 8 2 | | |
| | | 8 2 | | 7 2 | | |
| | | 6 2 | | 6 2 | | |
| | | 8 2 | | 8 2 | | |
| | | 7 2 | | 6 2 | | |
| | | 9 2 | | 7 2 | | |
| | | 6 2 | | 9 2 | | |
| | | 9 2 | | 8 2 | | |
| | | 9 2 | | 7 2 | | |
| | | 8 2 | | 6 2 | | |
| | | 6 2 | | 8 2 | | |
| | | 7 2 | | 6 2 | | |
| | | 8 2 | | 9 2 | | |
| | | 6 2 | | 8 2 | | |
| 129.1.51 | 42969 | | 68 | | Er1Se2 | |
| | -A | -GM | -M | -Z | -R | -X |
| | 5 4 | 7 2 | 5 4 | 7 2 | 3+4 4 | 3+4 4 |
| | 5 4 | 9 2 | 5 4 | 9 2 | 3+4 4 | 3+4 4 |
| | 5 4 | 7 2 | 5 4 | 9 2 | 3+4 4 | 3+4 4 |
| | 5 4 | 9 2 | 5 4 | 7 2 | 3+4 4 | 3+4 4 |
| | 5 4 | 7 2 | 5 4 | 7 2 | 3+4 4 | 3+4 4 |
| | 5 4 | 9 2 | 5 4 | 9 2 | 3+4 4 | 3+4 4 |
| | 5 4 | 6 2 | 5 4 | 8 2 | 3+4 4 | 3+4 4 |
| | 5 4 | 8 2 | 5 4 | 6 2 | 3+4 4 | 3+4 4 |
| | 5 4 | 7 2 | 5 4 | 7 2 | 3+4 4 | 3+4 4 |

Continued on next column

**TABLE LX– continued**

| TQC | ICSD | | NoE | | SF | |
|---|---|---|---|---|---|---|
| | 5 4 | 7 2 | 5 4 | 9 2 | 3+4 4 | 3+4 4 |
| | 5 4 | 9 2 | 5 4 | 7 2 | 3+4 4 | 3+4 4 |
| | 5 4 | 8 2 | 5 4 | 8 2 | 3+4 4 | 3+4 4 |
| | 5 4 | 9 2 | 5 4 | 9 2 | 3+4 4 | 3+4 4 |
| | 5 4 | 6 2 | 5 4 | 6 2 | 3+4 4 | 3+4 4 |
| | 5 4 | 7 2 | 5 4 | 7 2 | 3+4 4 | 3+4 4 |
| | 5 4 | 7 2 | 5 4 | 9 2 | 3+4 4 | 3+4 4 |
| | 5 4 | 7 2 | 5 4 | 9 2 | 3+4 4 | 3+4 4 |
| | | 6 2 | | 6 2 | | |
| | | 6 2 | | 8 2 | | |
| | | 8 2 | | 7 2 | | |
| | | 9 2 | | 6 2 | | |
| | | 7 2 | | 7 2 | | |
| | | 8 2 | | 9 2 | | |
| | | 6 2 | | 8 2 | | |
| | | 9 2 | | 6 2 | | |
| | | 6 2 | | 8 2 | | |
| | | 8 2 | | 7 2 | | |
| | | 9 2 | | 6 2 | | |
| | | 7 2 | | 9 2 | | |
| | | 8 2 | | 8 2 | | |
| | | 6 2 | | 6 2 | | |
| | | 6 2 | | 8 2 | | |
| | | 9 2 | | 7 2 | | |
| | | 7 2 | | 9 2 | | |
| 129.1.52 | 59594 | | 48 | | Cu1P1Si1Zr1 | |
| | -A | -GM | -M | -Z | -R | -X |
| | 5 4 | 7 2 | 5 4 | 9 2 | 3+4 4 | 3+4 4 |
| | 5 4 | 9 2 | 5 4 | 7 2 | 3+4 4 | 3+4 4 |
| | 5 4 | 7 2 | 5 4 | 7 2 | 3+4 4 | 3+4 4 |
| | 5 4 | 7 2 | 5 4 | 9 2 | 3+4 4 | 3+4 4 |
| | 5 4 | 9 2 | 5 4 | 7 2 | 3+4 4 | 3+4 4 |
| | 5 4 | 8 2 | 5 4 | 6 2 | 3+4 4 | 3+4 4 |
| | 5 4 | 9 2 | 5 4 | 7 2 | 3+4 4 | 3+4 4 |
| | 5 4 | 6 2 | 5 4 | 8 2 | 3+4 4 | 3+4 4 |
| | 5 4 | 9 2 | 5 4 | 9 2 | 3+4 4 | 3+4 4 |
| | 5 4 | 8 2 | 5 4 | 6 2 | 3+4 4 | 3+4 4 |
| | 5 4 | 7 2 | 5 4 | 9 2 | 3+4 4 | 3+4 4 |
| | 5 4 | 7 2 | 5 4 | 8 2 | 3+4 4 | 3+4 4 |
| | | 6 2 | | 7 2 | | |
| | | 6 2 | | 8 2 | | |
| | | 7 2 | | 8 2 | | |
| | | 6 2 | | 9 2 | | |
| | | 8 2 | | 6 2 | | |
| | | 9 2 | | 7 2 | | |
| | | 6 2 | | 9 2 | | |
| | | 9 2 | | 7 2 | | |
| | | 7 2 | | 8 2 | | |
| | | 6 2 | | 6 2 | | |
| | | 9 2 | | 7 2 | | |
| | | 8 2 | | 6 2 | | |
| 129.1.53 | 91199 | | 124 | | Cu1Mn1O3S1Sr2 | |
| | -A | -GM | -M | -Z | -R | -X |
| | 5 4 | 7 2 | 5 4 | 9 2 | 3+4 4 | 3+4 4 |
| | 5 4 | 9 2 | 5 4 | 7 2 | 3+4 4 | 3+4 4 |
| | 5 4 | 7 2 | 5 4 | 9 2 | 3+4 4 | 3+4 4 |
| | 5 4 | 9 2 | 5 4 | 7 2 | 3+4 4 | 3+4 4 |
| | 5 4 | 9 2 | 5 4 | 7 2 | 3+4 4 | 3+4 4 |
| | 5 4 | 7 2 | 5 4 | 9 2 | 3+4 4 | 3+4 4 |
| | 5 4 | 6 2 | 5 4 | 6 2 | 3+4 4 | 3+4 4 |
| | 5 4 | 8 2 | 5 4 | 8 2 | 3+4 4 | 3+4 4 |
| | 5 4 | 7 2 | 5 4 | 9 2 | 3+4 4 | 3+4 4 |
| | 5 4 | 9 2 | 5 4 | 7 2 | 3+4 4 | 3+4 4 |
| | 5 4 | 7 2 | 5 4 | 9 2 | 3+4 4 | 3+4 4 |
| | 5 4 | 9 2 | 5 4 | 7 2 | 3+4 4 | 3+4 4 |
| | 5 4 | 7 2 | 5 4 | 9 2 | 3+4 4 | 3+4 4 |
| | 5 4 | 9 2 | 5 4 | 7 2 | 3+4 4 | 3+4 4 |
| | 5 4 | 6 2 | 5 4 | 8 2 | 3+4 4 | 3+4 4 |
| | 5 4 | 7 2 | 5 4 | 6 2 | 3+4 4 | 3+4 4 |
| | 5 4 | 8 2 | 5 4 | 9 2 | 3+4 4 | 3+4 4 |
| | 5 4 | 6 2 | 5 4 | 8 2 | 3+4 4 | 3+4 4 |
| | 5 4 | 9 2 | 5 4 | 7 2 | 3+4 4 | 3+4 4 |
| | 5 4 | 8 2 | 5 4 | 6 2 | 3+4 4 | 3+4 4 |
| | 5 4 | 7 2 | 5 4 | 9 2 | 3+4 4 | 3+4 4 |
| | 5 4 | 9 2 | 5 4 | 7 2 | 3+4 4 | 3+4 4 |
| | 5 4 | 7 2 | 5 4 | 7 2 | 3+4 4 | 3+4 4 |
| | 5 4 | 9 2 | 5 4 | 9 2 | 3+4 4 | 3+4 4 |
| | 5 4 | 7 2 | 5 4 | 7 2 | 3+4 4 | 3+4 4 |
| | 5 4 | 9 2 | 5 4 | 9 2 | 3+4 4 | 3+4 4 |
| | 5 4 | 9 2 | 5 4 | 9 2 | 3+4 4 | 3+4 4 |
| | 5 4 | 7 2 | 5 4 | 7 2 | 3+4 4 | 3+4 4 |
| | 5 4 | 8 2 | 5 4 | 8 2 | 3+4 4 | 3+4 4 |
| | 5 4 | 6 2 | 5 4 | 6 2 | 3+4 4 | 3+4 4 |
| | 5 4 | 7 2 | 5 4 | 7 2 | 3+4 4 | 3+4 4 |
| | | 8 2 | | 8 2 | | |
| | | 9 2 | | 9 2 | | |
| | | 9 2 | | 9 2 | | |
| | | 7 2 | | 9 2 | | |
| | | 8 2 | | 6 2 | | |
| | | 7 2 | | 7 2 | | |
| | | 6 2 | | 8 2 | | |
| | | 6 2 | | 6 2 | | |
| | | 9 2 | | 7 2 | | |
| | | 6 2 | | 6 2 | | |
| | | 7 2 | | 9 2 | | |
| | | 6 2 | | 8 2 | | |
| | | 9 2 | | 7 2 | | |
| | | 8 2 | | 6 2 | | |
| | | 7 2 | | 9 2 | | |
| | | 6 2 | | 8 2 | | |
| | | 9 2 | | 7 2 | | |
| | | 8 2 | | 6 2 | | |
| | | 7 2 | | 7 2 | | |
| | | 8 2 | | 8 2 | | |
| | | 6 2 | | 6 2 | | |
| | | 7 2 | | 7 2 | | |
| | | 6 2 | | 6 2 | | |
| | | 9 2 | | 9 2 | | |
| | | 9 2 | | 9 2 | | |
| | | 6 2 | | 6 2 | | |
| | | 8 2 | | 8 2 | | |
| | | 8 2 | | 8 2 | | |
| | | 9 2 | | 9 2 | | |
| | | 7 2 | | 9 2 | | |
| | | 6 2 | | 8 2 | | |
| 129.1.54 | 27230 | | 164 | | Cs1F4Ti1 | |
| | -A | -GM | -M | -Z | -R | -X |
| | 5 4 | 7 2 | 5 4 | 7 2 | 3+4 4 | 3+4 4 |
| | 5 4 | 6 2 | 5 4 | 6 2 | 3+4 4 | 3+4 4 |
| | 5 4 | 7 2 | 5 4 | 7 2 | 3+4 4 | 3+4 4 |
| | 5 4 | 6 2 | 5 4 | 6 2 | 3+4 4 | 3+4 4 |
| | 5 4 | 9 2 | 5 4 | 9 2 | 3+4 4 | 3+4 4 |

Continued on next column

**TABLE LX– continued**

| TQC | ICSD | | NoE | | SF | |
|---|---|---|---|---|---|---|
| | 5 4 | 8 2 | 5 4 | 8 2 | 3+4 4 | 3+4 4 |
| | 5 4 | 8 2 | 5 4 | 8 2 | 3+4 4 | 3+4 4 |
| | 5 4 | 9 2 | 5 4 | 9 2 | 3+4 4 | 3+4 4 |
| | 5 4 | 7 2 | 5 4 | 7 2 | 3+4 4 | 3+4 4 |
| | 5 4 | 6 2 | 5 4 | 6 2 | 3+4 4 | 3+4 4 |
| | 5 4 | 6 2 | 5 4 | 6 2 | 3+4 4 | 3+4 4 |
| | 5 4 | 7 2 | 5 4 | 7 2 | 3+4 4 | 3+4 4 |
| | 5 4 | 9 2 | 5 4 | 9 2 | 3+4 4 | 3+4 4 |
| | 5 4 | 8 2 | 5 4 | 8 2 | 3+4 4 | 3+4 4 |
| | 5 4 | 9 2 | 5 4 | 9 2 | 3+4 4 | 3+4 4 |
| | 5 4 | 8 2 | 5 4 | 8 2 | 3+4 4 | 3+4 4 |
| | 5 4 | 8 2 | 5 4 | 6 2 | 3+4 4 | 3+4 4 |
| | 5 4 | 7 2 | 5 4 | 9 2 | 3+4 4 | 3+4 4 |
| | 5 4 | 9 2 | 5 4 | 7 2 | 3+4 4 | 3+4 4 |
| | 5 4 | 7 2 | 5 4 | 9 2 | 3+4 4 | 3+4 4 |
| | 5 4 | 9 2 | 5 4 | 7 2 | 3+4 4 | 3+4 4 |
| | 5 4 | 6 2 | 5 4 | 8 2 | 3+4 4 | 3+4 4 |
| | 5 4 | 7 2 | 5 4 | 9 2 | 3+4 4 | 3+4 4 |
| | 5 4 | 9 2 | 5 4 | 7 2 | 3+4 4 | 3+4 4 |
| | 5 4 | 8 2 | 5 4 | 6 2 | 3+4 4 | 3+4 4 |
| | 5 4 | 6 2 | 5 4 | 8 2 | 3+4 4 | 3+4 4 |
| | 5 4 | 7 2 | 5 4 | 9 2 | 3+4 4 | 3+4 4 |
| | 5 4 | 9 2 | 5 4 | 7 2 | 3+4 4 | 3+4 4 |
| | 5 4 | 9 2 | 5 4 | 7 2 | 3+4 4 | 3+4 4 |
| | 5 4 | 6 2 | 5 4 | 6 2 | 3+4 4 | 3+4 4 |
| | 5 4 | 6 2 | 5 4 | 6 2 | 3+4 4 | 3+4 4 |
| | 5 4 | 7 2 | 5 4 | 7 2 | 3+4 4 | 3+4 4 |
| | 5 4 | 8 2 | 5 4 | 9 2 | 3+4 4 | 3+4 4 |
| | 5 4 | 7 2 | 5 4 | 8 2 | 3+4 4 | 3+4 4 |
| | 5 4 | 6 2 | 5 4 | 6 2 | 3+4 4 | 3+4 4 |
| | 5 4 | 7 2 | 5 4 | 7 2 | 3+4 4 | 3+4 4 |
| | 5 4 | 7 2 | 5 4 | 7 2 | 3+4 4 | 3+4 4 |
| | 5 4 | 6 2 | 5 4 | 6 2 | 3+4 4 | 3+4 4 |
| | 5 4 | 7 2 | 5 4 | 7 2 | 3+4 4 | 3+4 4 |
| | 5 4 | 7 2 | 5 4 | 9 2 | 3+4 4 | 3+4 4 |
| | 5 4 | 6 2 | 5 4 | 8 2 | 3+4 4 | 3+4 4 |
| | | 9 2 | | 7 2 | | |
| | | 8 2 | | 8 2 | | |
| | | 7 2 | | 6 2 | | |
| | | 8 2 | | 9 2 | | |
| | | 6 2 | | 6 2 | | |
| | | 9 2 | | 9 2 | | |
| | | 6 2 | | 7 2 | | |
| | | 7 2 | | 7 2 | | |
| | | 9 2 | | 6 2 | | |
| | | 8 2 | | 8 2 | | |
| | | 7 2 | | 9 2 | | |
| | | 6 2 | | 7 2 | | |
| | | 9 2 | | 6 2 | | |
| | | 9 2 | | 8 2 | | |
| | | 8 2 | | 8 2 | | |
| | | 8 2 | | 9 2 | | |
| | | 7 2 | | 7 2 | | |
| | | 6 2 | | 8 2 | | |
| | | 6 2 | | 6 2 | | |
| | | 7 2 | | 8 2 | | |
| | | 9 2 | | 9 2 | | |
| | | 8 2 | | 9 2 | | |
| | | 8 2 | | 9 2 | | |
| | | 9 2 | | 8 2 | | |
| | | 8 2 | | 7 2 | | |

Continued on next column

**TABLE LX– continued**

| TQC | ICSD | | NoE | | SF | |
|---|---|---|---|---|---|---|
| | | 9 2 | | 6 2 | | |
| | | 9 2 | | 9 2 | | |
| | | 8 2 | | 8 2 | | |
| | | 9 2 | | 8 2 | | |
| | | 8 2 | | 9 2 | | |
| | | 8 2 | | 9 2 | | |
| | | 6 2 | | 8 2 | | |
| | | 9 2 | | 8 2 | | |
| | | 6 2 | | 9 2 | | |
| | | 7 2 | | 6 2 | | |
| | | 8 2 | | 6 2 | | |
| | | 6 2 | | 7 2 | | |
| | | 7 2 | | 7 2 | | |
| | | 9 2 | | 6 2 | | |
| | | 6 2 | | 6 2 | | |
| | | 7 2 | | 7 2 | | |
| 129.1.55 | 62696 | | 180 | | Cs1F4Fe1 | |
| | -A | -GM | -M | -Z | -R | -X |
| | 5 4 | 7 2 | 5 4 | 7 2 | 3+4 4 | 3+4 4 |
| | 5 4 | 6 2 | 5 4 | 6 2 | 3+4 4 | 3+4 4 |
| | 5 4 | 7 2 | 5 4 | 7 2 | 3+4 4 | 3+4 4 |
| | 5 4 | 9 2 | 5 4 | 9 2 | 3+4 4 | 3+4 4 |
| | 5 4 | 8 2 | 5 4 | 8 2 | 3+4 4 | 3+4 4 |
| | 5 4 | 8 2 | 5 4 | 8 2 | 3+4 4 | 3+4 4 |
| | 5 4 | 9 2 | 5 4 | 9 2 | 3+4 4 | 3+4 4 |
| | 5 4 | 6 2 | 5 4 | 6 2 | 3+4 4 | 3+4 4 |
| | 5 4 | 7 2 | 5 4 | 9 2 | 3+4 4 | 3+4 4 |
| | 5 4 | 6 2 | 5 4 | 6 2 | 3+4 4 | 3+4 4 |
| | 5 4 | 6 2 | 5 4 | 8 2 | 3+4 4 | 3+4 4 |
| | 5 4 | 7 2 | 5 4 | 9 2 | 3+4 4 | 3+4 4 |
| | 5 4 | 9 2 | 5 4 | 6 2 | 3+4 4 | 3+4 4 |
| | 5 4 | 8 2 | 5 4 | 7 2 | 3+4 4 | 3+4 4 |
| | 5 4 | 9 2 | 5 4 | 7 2 | 3+4 4 | 3+4 4 |
| | 5 4 | 8 2 | 5 4 | 8 2 | 3+4 4 | 3+4 4 |
| | 5 4 | 7 2 | 5 4 | 9 2 | 3+4 4 | 3+4 4 |
| | 5 4 | 8 2 | 5 4 | 6 2 | 3+4 4 | 3+4 4 |
| | 5 4 | 9 2 | 5 4 | 7 2 | 3+4 4 | 3+4 4 |
| | 5 4 | 7 2 | 5 4 | 9 2 | 3+4 4 | 3+4 4 |
| | 5 4 | 9 2 | 5 4 | 7 2 | 3+4 4 | 3+4 4 |
| | 5 4 | 6 2 | 5 4 | 8 2 | 3+4 4 | 3+4 4 |
| | 5 4 | 7 2 | 5 4 | 9 2 | 3+4 4 | 3+4 4 |
| | 5 4 | 9 2 | 5 4 | 7 2 | 3+4 4 | 3+4 4 |
| | 5 4 | 6 2 | 5 4 | 6 2 | 3+4 4 | 3+4 4 |
| | 5 4 | 8 2 | 5 4 | 6 2 | 3+4 4 | 3+4 4 |
| | 5 4 | 7 2 | 5 4 | 6 2 | 3+4 4 | 3+4 4 |
| | 5 4 | 6 2 | 5 4 | 7 2 | 3+4 4 | 3+4 4 |
| | 5 4 | 7 2 | 5 4 | 9 2 | 3+4 4 | 3+4 4 |
| | 5 4 | 6 2 | 5 4 | 8 2 | 3+4 4 | 3+4 4 |
| | 5 4 | 7 2 | 5 4 | 7 2 | 3+4 4 | 3+4 4 |
| | 5 4 | 9 2 | 5 4 | 7 2 | 3+4 4 | 3+4 4 |
| | 5 4 | 9 2 | 5 4 | 7 2 | 3+4 4 | 3+4 4 |
| | 5 4 | 8 2 | 5 4 | 6 2 | 3+4 4 | 3+4 4 |
| | 5 4 | 7 2 | 5 4 | 8 2 | 3+4 4 | 3+4 4 |
| | 5 4 | 6 2 | 5 4 | 9 2 | 3+4 4 | 3+4 4 |
| | 5 4 | 9 2 | 5 4 | 9 2 | 3+4 4 | 3+4 4 |
| | 5 4 | 7 2 | 5 4 | 8 2 | 3+4 4 | 3+4 4 |
| | 5 4 | 8 2 | 5 4 | 8 2 | 3+4 4 | 3+4 4 |
| | 5 4 | 6 2 | 5 4 | 9 2 | 3+4 4 | 3+4 4 |
| | 5 4 | 8 2 | 5 4 | 7 2 | 3+4 4 | 3+4 4 |
| | 5 4 | 7 2 | 5 4 | 7 2 | 3+4 4 | 3+4 4 |
| | 5 4 | 9 2 | 5 4 | 6 2 | 3+4 4 | 3+4 4 |

Continued on next column

**TABLE LX– continued**

| TQC | ICSD | | NoE | | SF | |
|---|---|---|---|---|---|---|
| | 5 4 | 6 2 | 5 4 | 6 2 | 3+4 4 | 3+4 4 |
| | 5 4 | 7 2 | 5 4 | 7 2 | 3+4 4 | 3+4 4 |
| | | 7 2 | | 9 2 | | |
| | | 6 2 | | 6 2 | | |
| | | 6 2 | | 7 2 | | |
| | | 9 2 | | 7 2 | | |
| | | 7 2 | | 6 2 | | |
| | | 7 2 | | 8 2 | | |
| | | 6 2 | | 9 2 | | |
| | | 9 2 | | 8 2 | | |
| | | 8 2 | | 7 2 | | |
| | | 9 2 | | 6 2 | | |
| | | 8 2 | | 8 2 | | |
| | | 8 2 | | 9 2 | | |
| | | 7 2 | | 7 2 | | |
| | | 8 2 | | 6 2 | | |
| | | 6 2 | | 8 2 | | |
| | | 6 2 | | 9 2 | | |
| | | 7 2 | | 9 2 | | |
| | | 9 2 | | 8 2 | | |
| | | 8 2 | | 7 2 | | |
| | | 8 2 | | 8 2 | | |
| | | 9 2 | | 9 2 | | |
| | | 9 2 | | 8 2 | | |
| | | 8 2 | | 9 2 | | |
| | | 9 2 | | 6 2 | | |
| | | 8 2 | | 9 2 | | |
| | | 9 2 | | 8 2 | | |
| | | 9 2 | | 9 2 | | |
| | | 6 2 | | 6 2 | | |
| | | 8 2 | | 8 2 | | |
| | | 9 2 | | 8 2 | | |
| | | 6 2 | | 9 2 | | |
| | | 7 2 | | 6 2 | | |
| | | 8 2 | | 7 2 | | |
| | | 7 2 | | 7 2 | | |
| | | 6 2 | | 6 2 | | |
| | | 6 2 | | 6 2 | | |
| | | 6 2 | | 6 2 | | |
| | | 7 2 | | 7 2 | | |
| | | 6 2 | | 6 2 | | |
| | | 7 2 | | 7 2 | | |
| | | 6 2 | | 6 2 | | |
| | | 6 2 | | 6 2 | | |
| | | 7 2 | | 7 2 | | |
| | | 6 2 | | 6 2 | | |
| | | 7 2 | | 7 2 | | |
| 129.1.56 | 6043 | | 180 | | Cs1F4Fe1 | |
| | -A | -GM | -M | -Z | -R | -X |
| | 5 4 | 7 2 | 5 4 | 7 2 | 3+4 4 | 3+4 4 |
| | 5 4 | 6 2 | 5 4 | 6 2 | 3+4 4 | 3+4 4 |
| | 5 4 | 7 2 | 5 4 | 7 2 | 3+4 4 | 3+4 4 |
| | 5 4 | 9 2 | 5 4 | 9 2 | 3+4 4 | 3+4 4 |
| | 5 4 | 8 2 | 5 4 | 8 2 | 3+4 4 | 3+4 4 |
| | 5 4 | 8 2 | 5 4 | 8 2 | 3+4 4 | 3+4 4 |
| | 5 4 | 9 2 | 5 4 | 9 2 | 3+4 4 | 3+4 4 |
| | 5 4 | 6 2 | 5 4 | 6 2 | 3+4 4 | 3+4 4 |
| | 5 4 | 6 2 | 5 4 | 9 2 | 3+4 4 | 3+4 4 |
| | 5 4 | 7 2 | 5 4 | 6 2 | 3+4 4 | 3+4 4 |
| | 5 4 | 6 2 | 5 4 | 6 2 | 3+4 4 | 3+4 4 |
| | 5 4 | 7 2 | 5 4 | 7 2 | 3+4 4 | 3+4 4 |

Continued on next column

**TABLE LX– continued**

| TQC | ICSD | | NoE | | SF | |
|---|---|---|---|---|---|---|
| | 5 4 | 9 2 | 5 4 | 7 2 | 3+4 4 | 3+4 4 |
| | 5 4 | 8 2 | 5 4 | 8 2 | 3+4 4 | 3+4 4 |
| | 5 4 | 9 2 | 5 4 | 9 2 | 3+4 4 | 3+4 4 |
| | 5 4 | 8 2 | 5 4 | 8 2 | 3+4 4 | 3+4 4 |
| | 5 4 | 7 2 | 5 4 | 9 2 | 3+4 4 | 3+4 4 |
| | 5 4 | 8 2 | 5 4 | 6 2 | 3+4 4 | 3+4 4 |
| | 5 4 | 9 2 | 5 4 | 7 2 | 3+4 4 | 3+4 4 |
| | 5 4 | 7 2 | 5 4 | 9 2 | 3+4 4 | 3+4 4 |
| | 5 4 | 9 2 | 5 4 | 7 2 | 3+4 4 | 3+4 4 |
| | 5 4 | 6 2 | 5 4 | 8 2 | 3+4 4 | 3+4 4 |
| | 5 4 | 7 2 | 5 4 | 9 2 | 3+4 4 | 3+4 4 |
| | 5 4 | 9 2 | 5 4 | 7 2 | 3+4 4 | 3+4 4 |
| | 5 4 | 6 2 | 5 4 | 6 2 | 3+4 4 | 3+4 4 |
| | 5 4 | 8 2 | 5 4 | 6 2 | 3+4 4 | 3+4 4 |
| | 5 4 | 7 2 | 5 4 | 6 2 | 3+4 4 | 3+4 4 |
| | 5 4 | 6 2 | 5 4 | 7 2 | 3+4 4 | 3+4 4 |
| | 5 4 | 6 2 | 5 4 | 9 2 | 3+4 4 | 3+4 4 |
| | 5 4 | 7 2 | 5 4 | 8 2 | 3+4 4 | 3+4 4 |
| | 5 4 | 7 2 | 5 4 | 7 2 | 3+4 4 | 3+4 4 |
| | 5 4 | 9 2 | 5 4 | 7 2 | 3+4 4 | 3+4 4 |
| | 5 4 | 9 2 | 5 4 | 7 2 | 3+4 4 | 3+4 4 |
| | 5 4 | 8 2 | 5 4 | 6 2 | 3+4 4 | 3+4 4 |
| | 5 4 | 7 2 | 5 4 | 9 2 | 3+4 4 | 3+4 4 |
| | 5 4 | 6 2 | 5 4 | 8 2 | 3+4 4 | 3+4 4 |
| | 5 4 | 9 2 | 5 4 | 9 2 | 3+4 4 | 3+4 4 |
| | 5 4 | 6 2 | 5 4 | 8 2 | 3+4 4 | 3+4 4 |
| | 5 4 | 7 2 | 5 4 | 8 2 | 3+4 4 | 3+4 4 |
| | 5 4 | 8 2 | 5 4 | 9 2 | 3+4 4 | 3+4 4 |
| | 5 4 | 8 2 | 5 4 | 7 2 | 3+4 4 | 3+4 4 |
| | 5 4 | 7 2 | 5 4 | 7 2 | 3+4 4 | 3+4 4 |
| | 5 4 | 9 2 | 5 4 | 6 2 | 3+4 4 | 3+4 4 |
| | 5 4 | 6 2 | 5 4 | 6 2 | 3+4 4 | 3+4 4 |
| | 5 4 | 7 2 | 5 4 | 7 2 | 3+4 4 | 3+4 4 |
| | | 7 2 | | 6 2 | | |
| | | 6 2 | | 7 2 | | |
| | | 6 2 | | 9 2 | | |
| | | 7 2 | | 7 2 | | |
| | | 9 2 | | 6 2 | | |
| | | 9 2 | | 8 2 | | |
| | | 8 2 | | 9 2 | | |
| | | 7 2 | | 8 2 | | |
| | | 6 2 | | 7 2 | | |
| | | 8 2 | | 6 2 | | |
| | | 9 2 | | 8 2 | | |
| | | 7 2 | | 9 2 | | |
| | | 8 2 | | 7 2 | | |
| | | 6 2 | | 6 2 | | |
| | | 6 2 | | 9 2 | | |
| | | 7 2 | | 8 2 | | |
| | | 8 2 | | 7 2 | | |
| | | 9 2 | | 8 2 | | |
| | | 8 2 | | 9 2 | | |
| | | 8 2 | | 9 2 | | |
| | | 9 2 | | 8 2 | | |
| | | 8 2 | | 6 2 | | |
| | | 9 2 | | 9 2 | | |
| | | 9 2 | | 8 2 | | |
| | | 9 2 | | 9 2 | | |
| | | 8 2 | | 8 2 | | |
| | | 9 2 | | 9 2 | | |
| | | 8 2 | | 8 2 | | |

Continued on next column

**TABLE LX– continued**

| TQC | ICSD | NoE | SF | | | |
|---|---|---|---|---|---|---|
| | | 6 2 | | 8 2 | | |
| | | 9 2 | | 9 2 | | |
| | | 6 2 | | 6 2 | | |
| | | 7 2 | | 6 2 | | |
| | | 8 2 | | 7 2 | | |
| | | 7 2 | | 7 2 | | |
| | | 6 2 | | 6 2 | | |
| | | 6 2 | | 6 2 | | |
| | | 6 2 | | 6 2 | | |
| | | 7 2 | | 7 2 | | |
| | | 6 2 | | 6 2 | | |
| | | 7 2 | | 7 2 | | |
| | | 6 2 | | 6 2 | | |
| | | 6 2 | | 6 2 | | |
| | | 7 2 | | 7 2 | | |
| | | 6 2 | | 6 2 | | |
| | | 7 2 | | 7 2 | | |
| 129.1.57 | 16496 | 48 | Cr1N1Nb1 | | | |
| | -A | -GM | -M | -Z | -R | -X |
| | 5 4 | 7 2 | 5 4 | 9 2 | 3+4 4 | 3+4 4 |
| | 5 4 | 9 2 | 5 4 | 7 2 | 3+4 4 | 3+4 4 |
| | 5 4 | 7 2 | 5 4 | 9 2 | 3+4 4 | 3+4 4 |
| | 5 4 | 9 2 | 5 4 | 7 2 | 3+4 4 | 3+4 4 |
| | 5 4 | 7 2 | 5 4 | 9 2 | 3+4 4 | 3+4 4 |
| | 5 4 | 6 2 | 5 4 | 8 2 | 3+4 4 | 3+4 4 |
| | 5 4 | 8 2 | 5 4 | 6 2 | 3+4 4 | 3+4 4 |
| | 5 4 | 9 2 | 5 4 | 7 2 | 3+4 4 | 3+4 4 |
| | 5 4 | 7 2 | 5 4 | 9 2 | 3+4 4 | 3+4 4 |
| | 5 4 | 9 2 | 5 4 | 7 2 | 3+4 4 | 3+4 4 |
| | 5 4 | 7 2 | 5 4 | 9 2 | 3+4 4 | 3+4 4 |
| | 5 4 | 7 2 | 5 4 | 7 2 | 3+4 4 | 3+4 4 |
| | | 9 2 | | 6 2 | | |
| | | 8 2 | | 9 2 | | |
| | | 7 2 | | 8 2 | | |
| | | 6 2 | | 8 2 | | |
| | | 9 2 | | 7 2 | | |
| | | 9 2 | | 7 2 | | |
| | | 6 2 | | 6 2 | | |
| | | 8 2 | | 9 2 | | |
| | | 8 2 | | 8 2 | | |
| | | 7 2 | | 9 2 | | |
| | | 6 2 | | 8 2 | | |
| | | 6 2 | | 6 2 | | |
| 129.1.58 | 78578 | 84 | Co2P2U1 | | | |
| | -A | -GM | -M | -Z | -R | -X |
| | 5 4 | 7 2 | 5 4 | 7 2 | 3+4 4 | 3+4 4 |
| | 5 4 | 9 2 | 5 4 | 9 2 | 3+4 4 | 3+4 4 |
| | 5 4 | 7 2 | 5 4 | 9 2 | 3+4 4 | 3+4 4 |
| | 5 4 | 9 2 | 5 4 | 7 2 | 3+4 4 | 3+4 4 |
| | 5 4 | 7 2 | 5 4 | 9 2 | 3+4 4 | 3+4 4 |
| | 5 4 | 9 2 | 5 4 | 7 2 | 3+4 4 | 3+4 4 |
| | 5 4 | 6 2 | 5 4 | 8 2 | 3+4 4 | 3+4 4 |
| | 5 4 | 8 2 | 5 4 | 6 2 | 3+4 4 | 3+4 4 |
| | 5 4 | 7 2 | 5 4 | 9 2 | 3+4 4 | 3+4 4 |
| | 5 4 | 9 2 | 5 4 | 7 2 | 3+4 4 | 3+4 4 |
| | 5 4 | 8 2 | 5 4 | 6 2 | 3+4 4 | 3+4 4 |
| | 5 4 | 7 2 | 5 4 | 9 2 | 3+4 4 | 3+4 4 |
| | 5 4 | 7 2 | 5 4 | 9 2 | 3+4 4 | 3+4 4 |
| | 5 4 | 6 2 | 5 4 | 8 2 | 3+4 4 | 3+4 4 |
| | 5 4 | 7 2 | 5 4 | 7 2 | 3+4 4 | 3+4 4 |
| | 5 4 | 9 2 | 5 4 | 8 2 | 3+4 4 | 3+4 4 |

Continued on next column

**TABLE LX– continued**

| TQC | ICSD | NoE | SF | | | |
|---|---|---|---|---|---|---|
| | 5 4 | 6 2 | 5 4 | 7 2 | 3+4 4 | 3+4 4 |
| | 5 4 | 9 2 | 5 4 | 7 2 | 3+4 4 | 3+4 4 |
| | 5 4 | 6 2 | 5 4 | 6 2 | 3+4 4 | 3+4 4 |
| | 5 4 | 9 2 | 5 4 | 6 2 | 3+4 4 | 3+4 4 |
| | 5 4 | 8 2 | 5 4 | 9 2 | 3+4 4 | 3+4 4 |
| | | 8 2 | | 7 2 | | |
| | | 7 2 | | 6 2 | | |
| | | 9 2 | | 9 2 | | |
| | | 6 2 | | 8 2 | | |
| | | 9 2 | | 9 2 | | |
| | | 7 2 | | 8 2 | | |
| | | 6 2 | | 7 2 | | |
| | | 8 2 | | 7 2 | | |
| | | 9 2 | | 6 2 | | |
| | | 8 2 | | 9 2 | | |
| | | 9 2 | | 8 2 | | |
| | | 8 2 | | 8 2 | | |
| | | 7 2 | | 7 2 | | |
| | | 6 2 | | 6 2 | | |
| | | 6 2 | | 6 2 | | |
| | | 8 2 | | 8 2 | | |
| | | 7 2 | | 9 2 | | |
| | | 6 2 | | 9 2 | | |
| | | 9 2 | | 6 2 | | |
| | | 7 2 | | 7 2 | | |
| | | 6 2 | | 6 2 | | |
| 129.1.59 | 15770 | 60 | Co1K1O2 | | | |
| | -A | -GM | -M | -Z | -R | -X |
| | 5 4 | 7 2 | 5 4 | 9 2 | 3+4 4 | 3+4 4 |
| | 5 4 | 9 2 | 5 4 | 7 2 | 3+4 4 | 3+4 4 |
| | 5 4 | 7 2 | 5 4 | 7 2 | 3+4 4 | 3+4 4 |
| | 5 4 | 8 2 | 5 4 | 8 2 | 3+4 4 | 3+4 4 |
| | 5 4 | 9 2 | 5 4 | 7 2 | 3+4 4 | 3+4 4 |
| | 5 4 | 7 2 | 5 4 | 9 2 | 3+4 4 | 3+4 4 |
| | 5 4 | 7 2 | 5 4 | 9 2 | 3+4 4 | 3+4 4 |
| | 5 4 | 9 2 | 5 4 | 7 2 | 3+4 4 | 3+4 4 |
| | 5 4 | 7 2 | 5 4 | 9 2 | 3+4 4 | 3+4 4 |
| | 5 4 | 6 2 | 5 4 | 8 2 | 3+4 4 | 3+4 4 |
| | 5 4 | 9 2 | 5 4 | 7 2 | 3+4 4 | 3+4 4 |
| | 5 4 | 8 2 | 5 4 | 6 2 | 3+4 4 | 3+4 4 |
| | 5 4 | 7 2 | 5 4 | 7 2 | 3+4 4 | 3+4 4 |
| | 5 4 | 6 2 | 5 4 | 6 2 | 3+4 4 | 3+4 4 |
| | 5 4 | 9 2 | 5 4 | 9 2 | 3+4 4 | 3+4 4 |
| | | 6 2 | | 6 2 | | |
| | | 9 2 | | 9 2 | | |
| | | 8 2 | | 8 2 | | |
| | | 7 2 | | 7 2 | | |
| | | 9 2 | | 9 2 | | |
| | | 6 2 | | 6 2 | | |
| | | 7 2 | | 7 2 | | |
| | | 9 2 | | 6 2 | | |
| | | 6 2 | | 9 2 | | |
| | | 8 2 | | 8 2 | | |
| | | 8 2 | | 8 2 | | |
| | | 8 2 | | 8 2 | | |
| | | 7 2 | | 7 2 | | |
| | | 9 2 | | 9 2 | | |
| | | 8 2 | | 8 2 | | |
| 129.1.60 | 97019 | 104 | Cl1Mn1O3Sr2 | | | |
| | -A | -GM | -M | -Z | -R | -X |
| | 5 4 | 7 2 | 5 4 | 7 2 | 3+4 4 | 3+4 4 |

Continued on next column

**TABLE LX– continued**

| TQC | ICSD | | NoE | | SF | |
|---|---|---|---|---|---|---|
| | 5 4 | 9 2 | 5 4 | 9 2 | 3+4 4 | 3+4 4 |
| | 5 4 | 7 2 | 5 4 | 7 2 | 3+4 4 | 3+4 4 |
| | 5 4 | 9 2 | 5 4 | 9 2 | 3+4 4 | 3+4 4 |
| | 5 4 | 9 2 | 5 4 | 9 2 | 3+4 4 | 3+4 4 |
| | 5 4 | 7 2 | 5 4 | 7 2 | 3+4 4 | 3+4 4 |
| | 5 4 | 6 2 | 5 4 | 6 2 | 3+4 4 | 3+4 4 |
| | 5 4 | 8 2 | 5 4 | 8 2 | 3+4 4 | 3+4 4 |
| | 5 4 | 7 2 | 5 4 | 7 2 | 3+4 4 | 3+4 4 |
| | 5 4 | 9 2 | 5 4 | 9 2 | 3+4 4 | 3+4 4 |
| | 5 4 | 7 2 | 5 4 | 9 2 | 3+4 4 | 3+4 4 |
| | 5 4 | 9 2 | 5 4 | 7 2 | 3+4 4 | 3+4 4 |
| | 5 4 | 7 2 | 5 4 | 7 2 | 3+4 4 | 3+4 4 |
| | 5 4 | 7 2 | 5 4 | 9 2 | 3+4 4 | 3+4 4 |
| | 5 4 | 9 2 | 5 4 | 9 2 | 3+4 4 | 3+4 4 |
| | 5 4 | 9 2 | 5 4 | 7 2 | 3+4 4 | 3+4 4 |
| | 5 4 | 6 2 | 5 4 | 8 2 | 3+4 4 | 3+4 4 |
| | 5 4 | 8 2 | 5 4 | 6 2 | 3+4 4 | 3+4 4 |
| | 5 4 | 6 2 | 5 4 | 6 2 | 3+4 4 | 3+4 4 |
| | 5 4 | 8 2 | 5 4 | 8 2 | 3+4 4 | 3+4 4 |
| | 5 4 | 7 2 | 5 4 | 7 2 | 3+4 4 | 3+4 4 |
| | 5 4 | 9 2 | 5 4 | 9 2 | 3+4 4 | 3+4 4 |
| | 5 4 | 9 2 | 5 4 | 7 2 | 3+4 4 | 3+4 4 |
| | 5 4 | 7 2 | 5 4 | 9 2 | 3+4 4 | 3+4 4 |
| | 5 4 | 7 2 | 5 4 | 9 2 | 3+4 4 | 3+4 4 |
| | 5 4 | 9 2 | 5 4 | 7 2 | 3+4 4 | 3+4 4 |
| | | 7 2 | | 9 2 | | |
| | | 9 2 | | 7 2 | | |
| | | 6 2 | | 8 2 | | |
| | | 8 2 | | 6 2 | | |
| | | 7 2 | | 9 2 | | |
| | | 6 2 | | 9 2 | | |
| | | 7 2 | | 8 2 | | |
| | | 9 2 | | 7 2 | | |
| | | 7 2 | | 8 2 | | |
| | | 8 2 | | 6 2 | | |
| | | 6 2 | | 7 2 | | |
| | | 9 2 | | 6 2 | | |
| | | 8 2 | | 9 2 | | |
| | | 9 2 | | 7 2 | | |
| | | 7 2 | | 7 2 | | |
| | | 6 2 | | 9 2 | | |
| | | 9 2 | | 6 2 | | |
| | | 8 2 | | 8 2 | | |
| | | 7 2 | | 7 2 | | |
| | | 6 2 | | 6 2 | | |
| | | 9 2 | | 9 2 | | |
| | | 8 2 | | 8 2 | | |
| | | 6 2 | | 6 2 | | |
| | | 8 2 | | 8 2 | | |
| | | 7 2 | | 7 2 | | |
| | | 6 2 | | 6 2 | | |
| 129.1.61 | 97018 | | 104 | | Cl1Mn1O3Sr2 | |
| | -A | -GM | -M | -Z | -R | -X |
| | 5 4 | 7 2 | 5 4 | 7 2 | 3+4 4 | 3+4 4 |
| | 5 4 | 9 2 | 5 4 | 9 2 | 3+4 4 | 3+4 4 |
| | 5 4 | 7 2 | 5 4 | 7 2 | 3+4 4 | 3+4 4 |
| | 5 4 | 9 2 | 5 4 | 9 2 | 3+4 4 | 3+4 4 |
| | 5 4 | 9 2 | 5 4 | 9 2 | 3+4 4 | 3+4 4 |
| | 5 4 | 7 2 | 5 4 | 7 2 | 3+4 4 | 3+4 4 |
| | 5 4 | 6 2 | 5 4 | 6 2 | 3+4 4 | 3+4 4 |
| | 5 4 | 8 2 | 5 4 | 8 2 | 3+4 4 | 3+4 4 |

Continued on next column

**TABLE LX– continued**

| TQC | ICSD | | NoE | | SF | |
|---|---|---|---|---|---|---|
| | 5 4 | 7 2 | 5 4 | 7 2 | 3+4 4 | 3+4 4 |
| | 5 4 | 9 2 | 5 4 | 9 2 | 3+4 4 | 3+4 4 |
| | 5 4 | 7 2 | 5 4 | 7 2 | 3+4 4 | 3+4 4 |
| | 5 4 | 9 2 | 5 4 | 9 2 | 3+4 4 | 3+4 4 |
| | 5 4 | 7 2 | 5 4 | 9 2 | 3+4 4 | 3+4 4 |
| | 5 4 | 7 2 | 5 4 | 7 2 | 3+4 4 | 3+4 4 |
| | 5 4 | 9 2 | 5 4 | 9 2 | 3+4 4 | 3+4 4 |
| | 5 4 | 9 2 | 5 4 | 7 2 | 3+4 4 | 3+4 4 |
| | 5 4 | 6 2 | 5 4 | 8 2 | 3+4 4 | 3+4 4 |
| | 5 4 | 8 2 | 5 4 | 6 2 | 3+4 4 | 3+4 4 |
| | 5 4 | 6 2 | 5 4 | 6 2 | 3+4 4 | 3+4 4 |
| | 5 4 | 7 2 | 5 4 | 7 2 | 3+4 4 | 3+4 4 |
| | 5 4 | 8 2 | 5 4 | 8 2 | 3+4 4 | 3+4 4 |
| | 5 4 | 9 2 | 5 4 | 9 2 | 3+4 4 | 3+4 4 |
| | 5 4 | 9 2 | 5 4 | 7 2 | 3+4 4 | 3+4 4 |
| | 5 4 | 7 2 | 5 4 | 9 2 | 3+4 4 | 3+4 4 |
| | 5 4 | 7 2 | 5 4 | 9 2 | 3+4 4 | 3+4 4 |
| | 5 4 | 9 2 | 5 4 | 7 2 | 3+4 4 | 3+4 4 |
| | | 7 2 | | 9 2 | | |
| | | 9 2 | | 7 2 | | |
| | | 6 2 | | 8 2 | | |
| | | 8 2 | | 6 2 | | |
| | | 7 2 | | 9 2 | | |
| | | 6 2 | | 8 2 | | |
| | | 7 2 | | 9 2 | | |
| | | 9 2 | | 7 2 | | |
| | | 7 2 | | 7 2 | | |
| | | 9 2 | | 8 2 | | |
| | | 8 2 | | 6 2 | | |
| | | 6 2 | | 6 2 | | |
| | | 8 2 | | 9 2 | | |
| | | 9 2 | | 7 2 | | |
| | | 7 2 | | 7 2 | | |
| | | 6 2 | | 6 2 | | |
| | | 9 2 | | 9 2 | | |
| | | 8 2 | | 8 2 | | |
| | | 7 2 | | 7 2 | | |
| | | 6 2 | | 6 2 | | |
| | | 9 2 | | 9 2 | | |
| | | 8 2 | | 8 2 | | |
| | | 6 2 | | 6 2 | | |
| | | 8 2 | | 8 2 | | |
| | | 7 2 | | 7 2 | | |
| | | 6 2 | | 6 2 | | |
| 129.1.62 | 96044 | | 104 | | Cl1Mn1O3Sr2 | |
| | -A | -GM | -M | -Z | -R | -X |
| | 5 4 | 7 2 | 5 4 | 7 2 | 3+4 4 | 3+4 4 |
| | 5 4 | 9 2 | 5 4 | 9 2 | 3+4 4 | 3+4 4 |
| | 5 4 | 7 2 | 5 4 | 7 2 | 3+4 4 | 3+4 4 |
| | 5 4 | 9 2 | 5 4 | 9 2 | 3+4 4 | 3+4 4 |
| | 5 4 | 9 2 | 5 4 | 9 2 | 3+4 4 | 3+4 4 |
| | 5 4 | 7 2 | 5 4 | 7 2 | 3+4 4 | 3+4 4 |
| | 5 4 | 6 2 | 5 4 | 6 2 | 3+4 4 | 3+4 4 |
| | 5 4 | 8 2 | 5 4 | 8 2 | 3+4 4 | 3+4 4 |
| | 5 4 | 7 2 | 5 4 | 7 2 | 3+4 4 | 3+4 4 |
| | 5 4 | 9 2 | 5 4 | 9 2 | 3+4 4 | 3+4 4 |
| | 5 4 | 7 2 | 5 4 | 7 2 | 3+4 4 | 3+4 4 |
| | 5 4 | 9 2 | 5 4 | 9 2 | 3+4 4 | 3+4 4 |
| | 5 4 | 7 2 | 5 4 | 9 2 | 3+4 4 | 3+4 4 |
| | 5 4 | 7 2 | 5 4 | 7 2 | 3+4 4 | 3+4 4 |
| | 5 4 | 9 2 | 5 4 | 9 2 | 3+4 4 | 3+4 4 |

Continued on next column

**TABLE LX– continued**

| TQC | ICSD | | NoE | | SF | |
|---|---|---|---|---|---|---|
| | 5 4 | 9 2 | 5 4 | 7 2 | 3+4 4 | 3+4 4 |
| | 5 4 | 6 2 | 5 4 | 8 2 | 3+4 4 | 3+4 4 |
| | 5 4 | 8 2 | 5 4 | 6 2 | 3+4 4 | 3+4 4 |
| | 5 4 | 6 2 | 5 4 | 6 2 | 3+4 4 | 3+4 4 |
| | 5 4 | 8 2 | 5 4 | 8 2 | 3+4 4 | 3+4 4 |
| | 5 4 | 7 2 | 5 4 | 7 2 | 3+4 4 | 3+4 4 |
| | 5 4 | 9 2 | 5 4 | 9 2 | 3+4 4 | 3+4 4 |
| | 5 4 | 9 2 | 5 4 | 7 2 | 3+4 4 | 3+4 4 |
| | 5 4 | 7 2 | 5 4 | 9 2 | 3+4 4 | 3+4 4 |
| | 5 4 | 7 2 | 5 4 | 9 2 | 3+4 4 | 3+4 4 |
| | 5 4 | 9 2 | 5 4 | 7 2 | 3+4 4 | 3+4 4 |
| | | 7 2 | | 9 2 | | |
| | | 9 2 | | 7 2 | | |
| | | 6 2 | | 8 2 | | |
| | | 8 2 | | 6 2 | | |
| | | 7 2 | | 9 2 | | |
| | | 6 2 | | 9 2 | | |
| | | 7 2 | | 8 2 | | |
| | | 9 2 | | 7 2 | | |
| | | 7 2 | | 8 2 | | |
| | | 8 2 | | 6 2 | | |
| | | 6 2 | | 7 2 | | |
| | | 9 2 | | 6 2 | | |
| | | 8 2 | | 9 2 | | |
| | | 9 2 | | 7 2 | | |
| | | 7 2 | | 7 2 | | |
| | | 6 2 | | 6 2 | | |
| | | 9 2 | | 9 2 | | |
| | | 8 2 | | 8 2 | | |
| | | 7 2 | | 7 2 | | |
| | | 6 2 | | 6 2 | | |
| | | 9 2 | | 9 2 | | |
| | | 8 2 | | 8 2 | | |
| | | 6 2 | | 6 2 | | |
| | | 8 2 | | 8 2 | | |
| | | 7 2 | | 7 2 | | |
| | | 6 2 | | 6 2 | | |
| 129.1.63 | 94745 | | 104 | | Cl1Mn1O3Sr2 | |
| | -A | -GM | -M | -Z | -R | -X |
| | 5 4 | 7 2 | 5 4 | 7 2 | 3+4 4 | 3+4 4 |
| | 5 4 | 9 2 | 5 4 | 9 2 | 3+4 4 | 3+4 4 |
| | 5 4 | 7 2 | 5 4 | 7 2 | 3+4 4 | 3+4 4 |
| | 5 4 | 9 2 | 5 4 | 9 2 | 3+4 4 | 3+4 4 |
| | 5 4 | 9 2 | 5 4 | 9 2 | 3+4 4 | 3+4 4 |
| | 5 4 | 7 2 | 5 4 | 7 2 | 3+4 4 | 3+4 4 |
| | 5 4 | 6 2 | 5 4 | 6 2 | 3+4 4 | 3+4 4 |
| | 5 4 | 8 2 | 5 4 | 8 2 | 3+4 4 | 3+4 4 |
| | 5 4 | 7 2 | 5 4 | 7 2 | 3+4 4 | 3+4 4 |
| | 5 4 | 9 2 | 5 4 | 9 2 | 3+4 4 | 3+4 4 |
| | 5 4 | 7 2 | 5 4 | 7 2 | 3+4 4 | 3+4 4 |
| | 5 4 | 9 2 | 5 4 | 9 2 | 3+4 4 | 3+4 4 |
| | 5 4 | 7 2 | 5 4 | 9 2 | 3+4 4 | 3+4 4 |
| | 5 4 | 7 2 | 5 4 | 7 2 | 3+4 4 | 3+4 4 |
| | 5 4 | 9 2 | 5 4 | 9 2 | 3+4 4 | 3+4 4 |
| | 5 4 | 9 2 | 5 4 | 7 2 | 3+4 4 | 3+4 4 |
| | 5 4 | 6 2 | 5 4 | 8 2 | 3+4 4 | 3+4 4 |
| | 5 4 | 8 2 | 5 4 | 6 2 | 3+4 4 | 3+4 4 |
| | 5 4 | 6 2 | 5 4 | 6 2 | 3+4 4 | 3+4 4 |
| | 5 4 | 8 2 | 5 4 | 8 2 | 3+4 4 | 3+4 4 |
| | 5 4 | 7 2 | 5 4 | 7 2 | 3+4 4 | 3+4 4 |
| | 5 4 | 9 2 | 5 4 | 9 2 | 3+4 4 | 3+4 4 |

Continued on next column

**TABLE LX– continued**

| TQC | ICSD | | NoE | | SF | |
|---|---|---|---|---|---|---|
| | 5 4 | 9 2 | 5 4 | 7 2 | 3+4 4 | 3+4 4 |
| | 5 4 | 7 2 | 5 4 | 9 2 | 3+4 4 | 3+4 4 |
| | 5 4 | 7 2 | 5 4 | 9 2 | 3+4 4 | 3+4 4 |
| | 5 4 | 9 2 | 5 4 | 7 2 | 3+4 4 | 3+4 4 |
| | | 7 2 | | 9 2 | | |
| | | 9 2 | | 7 2 | | |
| | | 6 2 | | 8 2 | | |
| | | 8 2 | | 6 2 | | |
| | | 7 2 | | 9 2 | | |
| | | 6 2 | | 9 2 | | |
| | | 7 2 | | 8 2 | | |
| | | 9 2 | | 7 2 | | |
| | | 7 2 | | 8 2 | | |
| | | 8 2 | | 6 2 | | |
| | | 6 2 | | 7 2 | | |
| | | 9 2 | | 6 2 | | |
| | | 8 2 | | 9 2 | | |
| | | 9 2 | | 7 2 | | |
| | | 7 2 | | 7 2 | | |
| | | 6 2 | | 6 2 | | |
| | | 9 2 | | 9 2 | | |
| | | 8 2 | | 8 2 | | |
| | | 7 2 | | 7 2 | | |
| | | 6 2 | | 6 2 | | |
| | | 9 2 | | 9 2 | | |
| | | 8 2 | | 8 2 | | |
| | | 6 2 | | 6 2 | | |
| | | 8 2 | | 8 2 | | |
| | | 7 2 | | 7 2 | | |
| | | 6 2 | | 6 2 | | |
| 129.1.64 | 76171 | | 68 | | Cl1Ho1O1 | |
| | -A | -GM | -M | -Z | -R | -X |
| | 5 4 | 7 2 | 5 4 | 7 2 | 3+4 4 | 3+4 4 |
| | 5 4 | 9 2 | 5 4 | 9 2 | 3+4 4 | 3+4 4 |
| | 5 4 | 7 2 | 5 4 | 7 2 | 3+4 4 | 3+4 4 |
| | 5 4 | 9 2 | 5 4 | 9 2 | 3+4 4 | 3+4 4 |
| | 5 4 | 7 2 | 5 4 | 7 2 | 3+4 4 | 3+4 4 |
| | 5 4 | 6 2 | 5 4 | 6 2 | 3+4 4 | 3+4 4 |
| | 5 4 | 9 2 | 5 4 | 9 2 | 3+4 4 | 3+4 4 |
| | 5 4 | 8 2 | 5 4 | 8 2 | 3+4 4 | 3+4 4 |
| | 5 4 | 8 2 | 5 4 | 8 2 | 3+4 4 | 3+4 4 |
| | 5 4 | 7 2 | 5 4 | 7 2 | 3+4 4 | 3+4 4 |
| | 5 4 | 7 2 | 5 4 | 9 2 | 3+4 4 | 3+4 4 |
| | 5 4 | 9 2 | 5 4 | 7 2 | 3+4 4 | 3+4 4 |
| | 5 4 | 9 2 | 5 4 | 9 2 | 3+4 4 | 3+4 4 |
| | 5 4 | 7 2 | 5 4 | 7 2 | 3+4 4 | 3+4 4 |
| | 5 4 | 6 2 | 5 4 | 6 2 | 3+4 4 | 3+4 4 |
| | 5 4 | 7 2 | 5 4 | 6 2 | 3+4 4 | 3+4 4 |
| | 5 4 | 6 2 | 5 4 | 9 2 | 3+4 4 | 3+4 4 |
| | | 7 2 | | 8 2 | | |
| | | 6 2 | | 7 2 | | |
| | | 8 2 | | 9 2 | | |
| | | 9 2 | | 7 2 | | |
| | | 9 2 | | 8 2 | | |
| | | 9 2 | | 9 2 | | |
| | | 8 2 | | 6 2 | | |
| | | 6 2 | | 6 2 | | |
| | | 8 2 | | 8 2 | | |
| | | 6 2 | | 7 2 | | |
| | | 7 2 | | 6 2 | | |
| | | 9 2 | | 9 2 | | |

Continued on next column

**TABLE LX– continued**

| TQC | ICSD | NoE | SF | | | |
|---|---|---|---|---|---|---|
| | | 8 2 | | 8 2 | | |
| | | 6 2 | | 6 2 | | |
| | | 8 2 | | 8 2 | | |
| | | 7 2 | | 6 2 | | |
| | | 6 2 | | 7 2 | | |
| 129.1.65 | 91750 | 108 | Cl1Co1O3Sr2 | | | |
| | -A | -GM | -M | -Z | -R | -X |
| | 5 4 | 7 2 | 5 4 | 7 2 | 3+4 4 | 3+4 4 |
| | 5 4 | 9 2 | 5 4 | 9 2 | 3+4 4 | 3+4 4 |
| | 5 4 | 7 2 | 5 4 | 7 2 | 3+4 4 | 3+4 4 |
| | 5 4 | 9 2 | 5 4 | 9 2 | 3+4 4 | 3+4 4 |
| | 5 4 | 9 2 | 5 4 | 9 2 | 3+4 4 | 3+4 4 |
| | 5 4 | 7 2 | 5 4 | 7 2 | 3+4 4 | 3+4 4 |
| | 5 4 | 7 2 | 5 4 | 7 2 | 3+4 4 | 3+4 4 |
| | 5 4 | 9 2 | 5 4 | 9 2 | 3+4 4 | 3+4 4 |
| | 5 4 | 6 2 | 5 4 | 6 2 | 3+4 4 | 3+4 4 |
| | 5 4 | 8 2 | 5 4 | 8 2 | 3+4 4 | 3+4 4 |
| | 5 4 | 7 2 | 5 4 | 9 2 | 3+4 4 | 3+4 4 |
| | 5 4 | 9 2 | 5 4 | 7 2 | 3+4 4 | 3+4 4 |
| | 5 4 | 7 2 | 5 4 | 7 2 | 3+4 4 | 3+4 4 |
| | 5 4 | 9 2 | 5 4 | 9 2 | 3+4 4 | 3+4 4 |
| | 5 4 | 7 2 | 5 4 | 9 2 | 3+4 4 | 3+4 4 |
| | 5 4 | 9 2 | 5 4 | 7 2 | 3+4 4 | 3+4 4 |
| | 5 4 | 6 2 | 5 4 | 8 2 | 3+4 4 | 3+4 4 |
| | 5 4 | 8 2 | 5 4 | 6 2 | 3+4 4 | 3+4 4 |
| | 5 4 | 6 2 | 5 4 | 6 2 | 3+4 4 | 3+4 4 |
| | 5 4 | 8 2 | 5 4 | 8 2 | 3+4 4 | 3+4 4 |
| | 5 4 | 7 2 | 5 4 | 7 2 | 3+4 4 | 3+4 4 |
| | 5 4 | 9 2 | 5 4 | 9 2 | 3+4 4 | 3+4 4 |
| | 5 4 | 9 2 | 5 4 | 7 2 | 3+4 4 | 3+4 4 |
| | 5 4 | 7 2 | 5 4 | 9 2 | 3+4 4 | 3+4 4 |
| | 5 4 | 7 2 | 5 4 | 9 2 | 3+4 4 | 3+4 4 |
| | 5 4 | 9 2 | 5 4 | 7 2 | 3+4 4 | 3+4 4 |
| | 5 4 | 7 2 | 5 4 | 9 2 | 3+4 4 | 3+4 4 |
| | | 6 2 | | 7 2 | | |
| | | 8 2 | | 6 2 | | |
| | | 9 2 | | 8 2 | | |
| | | 7 2 | | 7 2 | | |
| | | 9 2 | | 9 2 | | |
| | | 7 2 | | 9 2 | | |
| | | 6 2 | | 8 2 | | |
| | | 7 2 | | 9 2 | | |
| | | 8 2 | | 8 2 | | |
| | | 6 2 | | 6 2 | | |
| | | 9 2 | | 7 2 | | |
| | | 8 2 | | 6 2 | | |
| | | 7 2 | | 7 2 | | |
| | | 6 2 | | 6 2 | | |
| | | 9 2 | | 9 2 | | |
| | | 8 2 | | 8 2 | | |
| | | 9 2 | | 7 2 | | |
| | | 7 2 | | 7 2 | | |
| | | 6 2 | | 6 2 | | |
| | | 9 2 | | 9 2 | | |
| | | 8 2 | | 8 2 | | |
| | | 6 2 | | 6 2 | | |
| | | 8 2 | | 8 2 | | |
| | | 7 2 | | 7 2 | | |
| | | 6 2 | | 6 2 | | |
| | | 8 2 | | 8 2 | | |
| | | 6 2 | | 6 2 | | |

Continued on next column

**TABLE LX– continued**

| TQC | ICSD | NoE | SF | | | |
|---|---|---|---|---|---|---|
| 129.1.66 | 90122 | 108 | Cl1Co1O3Sr2 | | | |
| | -A | -GM | -M | -Z | -R | -X |
| | 5 4 | 7 2 | 5 4 | 7 2 | 3+4 4 | 3+4 4 |
| | 5 4 | 9 2 | 5 4 | 9 2 | 3+4 4 | 3+4 4 |
| | 5 4 | 7 2 | 5 4 | 7 2 | 3+4 4 | 3+4 4 |
| | 5 4 | 9 2 | 5 4 | 9 2 | 3+4 4 | 3+4 4 |
| | 5 4 | 9 2 | 5 4 | 9 2 | 3+4 4 | 3+4 4 |
| | 5 4 | 7 2 | 5 4 | 7 2 | 3+4 4 | 3+4 4 |
| | 5 4 | 7 2 | 5 4 | 7 2 | 3+4 4 | 3+4 4 |
| | 5 4 | 6 2 | 5 4 | 9 2 | 3+4 4 | 3+4 4 |
| | 5 4 | 8 2 | 5 4 | 6 2 | 3+4 4 | 3+4 4 |
| | 5 4 | 9 2 | 5 4 | 8 2 | 3+4 4 | 3+4 4 |
| | 5 4 | 7 2 | 5 4 | 9 2 | 3+4 4 | 3+4 4 |
| | 5 4 | 9 2 | 5 4 | 7 2 | 3+4 4 | 3+4 4 |
| | 5 4 | 7 2 | 5 4 | 7 2 | 3+4 4 | 3+4 4 |
| | 5 4 | 9 2 | 5 4 | 9 2 | 3+4 4 | 3+4 4 |
| | 5 4 | 7 2 | 5 4 | 9 2 | 3+4 4 | 3+4 4 |
| | 5 4 | 9 2 | 5 4 | 7 2 | 3+4 4 | 3+4 4 |
| | 5 4 | 6 2 | 5 4 | 8 2 | 3+4 4 | 3+4 4 |
| | 5 4 | 8 2 | 5 4 | 6 2 | 3+4 4 | 3+4 4 |
| | 5 4 | 6 2 | 5 4 | 6 2 | 3+4 4 | 3+4 4 |
| | 5 4 | 8 2 | 5 4 | 8 2 | 3+4 4 | 3+4 4 |
| | 5 4 | 7 2 | 5 4 | 7 2 | 3+4 4 | 3+4 4 |
| | 5 4 | 9 2 | 5 4 | 9 2 | 3+4 4 | 3+4 4 |
| | 5 4 | 9 2 | 5 4 | 7 2 | 3+4 4 | 3+4 4 |
| | 5 4 | 7 2 | 5 4 | 9 2 | 3+4 4 | 3+4 4 |
| | 5 4 | 7 2 | 5 4 | 9 2 | 3+4 4 | 3+4 4 |
| | 5 4 | 9 2 | 5 4 | 7 2 | 3+4 4 | 3+4 4 |
| | 5 4 | 7 2 | 5 4 | 9 2 | 3+4 4 | 3+4 4 |
| | | 6 2 | | 6 2 | | |
| | | 8 2 | | 8 2 | | |
| | | 9 2 | | 7 2 | | |
| | | 7 2 | | 7 2 | | |
| | | 9 2 | | 9 2 | | |
| | | 7 2 | | 9 2 | | |
| | | 6 2 | | 8 2 | | |
| | | 7 2 | | 9 2 | | |
| | | 8 2 | | 8 2 | | |
| | | 6 2 | | 6 2 | | |
| | | 9 2 | | 7 2 | | |
| | | 8 2 | | 6 2 | | |
| | | 7 2 | | 7 2 | | |
| | | 6 2 | | 6 2 | | |
| | | 9 2 | | 7 2 | | |
| | | 9 2 | | 9 2 | | |
| | | 8 2 | | 8 2 | | |
| | | 7 2 | | 7 2 | | |
| | | 6 2 | | 6 2 | | |
| | | 9 2 | | 9 2 | | |
| | | 8 2 | | 8 2 | | |
| | | 6 2 | | 6 2 | | |
| | | 8 2 | | 8 2 | | |
| | | 7 2 | | 7 2 | | |
| | | 6 2 | | 6 2 | | |
| | | 8 2 | | 8 2 | | |
| | | 6 2 | | 6 2 | | |
| 129.1.67 | 84529 | 64 | Ce1Ni1Sb2 | | | |
| | -A | -GM | -M | -Z | -R | -X |
| | 5 4 | 7 2 | 5 4 | 7 2 | 3+4 4 | 3+4 4 |
| | 5 4 | 9 2 | 5 4 | 9 2 | 3+4 4 | 3+4 4 |
| | 5 4 | 7 2 | 5 4 | 9 2 | 3+4 4 | 3+4 4 |

Continued on next column

**TABLE LX– continued**

| TQC | ICSD | | NoE | | SF | |
|---|---|---|---|---|---|---|
| | 5 4 | 9 2 | 5 4 | 7 2 | 3+4 4 | 3+4 4 |
| | 5 4 | 7 2 | 5 4 | 9 2 | 3+4 4 | 3+4 4 |
| | 5 4 | 9 2 | 5 4 | 7 2 | 3+4 4 | 3+4 4 |
| | 5 4 | 6 2 | 5 4 | 8 2 | 3+4 4 | 3+4 4 |
| | 5 4 | 8 2 | 5 4 | 6 2 | 3+4 4 | 3+4 4 |
| | 5 4 | 7 2 | 5 4 | 9 2 | 3+4 4 | 3+4 4 |
| | 5 4 | 7 2 | 5 4 | 7 2 | 3+4 4 | 3+4 4 |
| | 5 4 | 9 2 | 5 4 | 9 2 | 3+4 4 | 3+4 4 |
| | 5 4 | 8 2 | 5 4 | 6 2 | 3+4 4 | 3+4 4 |
| | 5 4 | 9 2 | 5 4 | 7 2 | 3+4 4 | 3+4 4 |
| | 5 4 | 7 2 | 5 4 | 8 2 | 3+4 4 | 3+4 4 |
| | 5 4 | 6 2 | 5 4 | 9 2 | 3+4 4 | 3+4 4 |
| | 5 4 | 7 2 | 5 4 | 9 2 | 3+4 4 | 3+4 4 |
| | | 9 2 | | 7 2 | | |
| | | 8 2 | | 8 2 | | |
| | | 7 2 | | 9 2 | | |
| | | 6 2 | | 6 2 | | |
| | | 6 2 | | 7 2 | | |
| | | 9 2 | | 6 2 | | |
| | | 8 2 | | 8 2 | | |
| | | 6 2 | | 7 2 | | |
| | | 7 2 | | 6 2 | | |
| | | 6 2 | | 6 2 | | |
| | | 6 2 | | 7 2 | | |
| | | 7 2 | | 8 2 | | |
| | | 9 2 | | 9 2 | | |
| | | 8 2 | | 6 2 | | |
| | | 9 2 | | 8 2 | | |
| | | 6 2 | | 9 2 | | |
| 129.1.68 | 411678 | | 76 | | Ce1Ir2Si2 | |
| | -A | -GM | -M | -Z | -R | -X |
| | 5 4 | 7 2 | 5 4 | 7 2 | 3+4 4 | 3+4 4 |
| | 5 4 | 9 2 | 5 4 | 9 2 | 3+4 4 | 3+4 4 |
| | 5 4 | 7 2 | 5 4 | 7 2 | 3+4 4 | 3+4 4 |
| | 5 4 | 9 2 | 5 4 | 9 2 | 3+4 4 | 3+4 4 |
| | 5 4 | 7 2 | 5 4 | 7 2 | 3+4 4 | 3+4 4 |
| | 5 4 | 9 2 | 5 4 | 9 2 | 3+4 4 | 3+4 4 |
| | 5 4 | 8 2 | 5 4 | 6 2 | 3+4 4 | 3+4 4 |
| | 5 4 | 6 2 | 5 4 | 8 2 | 3+4 4 | 3+4 4 |
| | 5 4 | 7 2 | 5 4 | 7 2 | 3+4 4 | 3+4 4 |
| | 5 4 | 7 2 | 5 4 | 9 2 | 3+4 4 | 3+4 4 |
| | 5 4 | 9 2 | 5 4 | 7 2 | 3+4 4 | 3+4 4 |
| | 5 4 | 8 2 | 5 4 | 8 2 | 3+4 4 | 3+4 4 |
| | 5 4 | 9 2 | 5 4 | 9 2 | 3+4 4 | 3+4 4 |
| | 5 4 | 7 2 | 5 4 | 7 2 | 3+4 4 | 3+4 4 |
| | 5 4 | 7 2 | 5 4 | 6 2 | 3+4 4 | 3+4 4 |
| | 5 4 | 6 2 | 5 4 | 7 2 | 3+4 4 | 3+4 4 |
| | 5 4 | 6 2 | 5 4 | 9 2 | 3+4 4 | 3+4 4 |
| | 5 4 | 8 2 | 5 4 | 8 2 | 3+4 4 | 3+4 4 |
| | 5 4 | 9 2 | 5 4 | 6 2 | 3+4 4 | 3+4 4 |
| | | 6 2 | | 6 2 | | |
| | | 9 2 | | 7 2 | | |
| | | 6 2 | | 9 2 | | |
| | | 7 2 | | 6 2 | | |
| | | 9 2 | | 8 2 | | |
| | | 8 2 | | 9 2 | | |
| | | 8 2 | | 7 2 | | |
| | | 7 2 | | 6 2 | | |
| | | 6 2 | | 8 2 | | |
| | | 8 2 | | 9 2 | | |
| | | 9 2 | | 8 2 | | |

Continued on next column

**TABLE LX– continued**

| TQC | ICSD | | NoE | | SF | |
|---|---|---|---|---|---|---|
| | | 7 2 | | 8 2 | | |
| | | 6 2 | | 9 2 | | |
| | | 6 2 | | 7 2 | | |
| | | 7 2 | | 8 2 | | |
| | | 7 2 | | 7 2 | | |
| | | 6 2 | | 6 2 | | |
| | | 6 2 | | 8 2 | | |
| | | 9 2 | | 9 2 | | |
| 129.1.69 | 61371 | | 80 | | Ce1Ge2Pt2 | |
| | -A | -GM | -M | -Z | -R | -X |
| | 5 4 | 7 2 | 5 4 | 7 2 | 3+4 4 | 3+4 4 |
| | 5 4 | 9 2 | 5 4 | 9 2 | 3+4 4 | 3+4 4 |
| | 5 4 | 7 2 | 5 4 | 9 2 | 3+4 4 | 3+4 4 |
| | 5 4 | 9 2 | 5 4 | 7 2 | 3+4 4 | 3+4 4 |
| | 5 4 | 7 2 | 5 4 | 9 2 | 3+4 4 | 3+4 4 |
| | 5 4 | 9 2 | 5 4 | 7 2 | 3+4 4 | 3+4 4 |
| | 5 4 | 6 2 | 5 4 | 8 2 | 3+4 4 | 3+4 4 |
| | 5 4 | 8 2 | 5 4 | 6 2 | 3+4 4 | 3+4 4 |
| | 5 4 | 7 2 | 5 4 | 9 2 | 3+4 4 | 3+4 4 |
| | 5 4 | 7 2 | 5 4 | 7 2 | 3+4 4 | 3+4 4 |
| | 5 4 | 9 2 | 5 4 | 9 2 | 3+4 4 | 3+4 4 |
| | 5 4 | 8 2 | 5 4 | 6 2 | 3+4 4 | 3+4 4 |
| | 5 4 | 9 2 | 5 4 | 7 2 | 3+4 4 | 3+4 4 |
| | 5 4 | 7 2 | 5 4 | 9 2 | 3+4 4 | 3+4 4 |
| | 5 4 | 6 2 | 5 4 | 8 2 | 3+4 4 | 3+4 4 |
| | 5 4 | 7 2 | 5 4 | 7 2 | 3+4 4 | 3+4 4 |
| | 5 4 | 9 2 | 5 4 | 8 2 | 3+4 4 | 3+4 4 |
| | 5 4 | 6 2 | 5 4 | 9 2 | 3+4 4 | 3+4 4 |
| | 5 4 | 8 2 | 5 4 | 8 2 | 3+4 4 | 3+4 4 |
| | 5 4 | 6 2 | 5 4 | 6 2 | 3+4 4 | 3+4 4 |
| | | 9 2 | | 9 2 | | |
| | | 6 2 | | 7 2 | | |
| | | 9 2 | | 6 2 | | |
| | | 8 2 | | 8 2 | | |
| | | 7 2 | | 7 2 | | |
| | | 8 2 | | 6 2 | | |
| | | 7 2 | | 9 2 | | |
| | | 6 2 | | 8 2 | | |
| | | 8 2 | | 6 2 | | |
| | | 9 2 | | 7 2 | | |
| | | 6 2 | | 6 2 | | |
| | | 7 2 | | 9 2 | | |
| | | 7 2 | | 6 2 | | |
| | | 6 2 | | 7 2 | | |
| | | 7 2 | | 9 2 | | |
| | | 9 2 | | 8 2 | | |
| | | 6 2 | | 9 2 | | |
| | | 8 2 | | 6 2 | | |
| | | 8 2 | | 7 2 | | |
| | | 9 2 | | 6 2 | | |
| 129.1.70 | 106393 | | 76 | | Ce1Ge2Ir2 | |
| | -A | -GM | -M | -Z | -R | -X |
| | 5 4 | 7 2 | 5 4 | 7 2 | 3+4 4 | 3+4 4 |
| | 5 4 | 9 2 | 5 4 | 9 2 | 3+4 4 | 3+4 4 |
| | 5 4 | 7 2 | 5 4 | 9 2 | 3+4 4 | 3+4 4 |
| | 5 4 | 9 2 | 5 4 | 7 2 | 3+4 4 | 3+4 4 |
| | 5 4 | 7 2 | 5 4 | 9 2 | 3+4 4 | 3+4 4 |
| | 5 4 | 9 2 | 5 4 | 7 2 | 3+4 4 | 3+4 4 |
| | 5 4 | 6 2 | 5 4 | 8 2 | 3+4 4 | 3+4 4 |
| | 5 4 | 8 2 | 5 4 | 6 2 | 3+4 4 | 3+4 4 |
| | 5 4 | 7 2 | 5 4 | 9 2 | 3+4 4 | 3+4 4 |

Continued on next column

**TABLE LX– continued**

| TQC | ICSD | NoE | SF | | | |
|---|---|---|---|---|---|---|
| | 5 4 | 7 2 | 5 4 | 7 2 | 3+4 4 | 3+4 4 |
| | 5 4 | 9 2 | 5 4 | 9 2 | 3+4 4 | 3+4 4 |
| | 5 4 | 8 2 | 5 4 | 6 2 | 3+4 4 | 3+4 4 |
| | 5 4 | 9 2 | 5 4 | 7 2 | 3+4 4 | 3+4 4 |
| | 5 4 | 7 2 | 5 4 | 9 2 | 3+4 4 | 3+4 4 |
| | 5 4 | 7 2 | 5 4 | 8 2 | 3+4 4 | 3+4 4 |
| | 5 4 | 6 2 | 5 4 | 7 2 | 3+4 4 | 3+4 4 |
| | 5 4 | 9 2 | 5 4 | 9 2 | 3+4 4 | 3+4 4 |
| | 5 4 | 6 2 | 5 4 | 8 2 | 3+4 4 | 3+4 4 |
| | 5 4 | 8 2 | 5 4 | 6 2 | 3+4 4 | 3+4 4 |
| | | 6 2 | | 8 2 | | |
| | | 9 2 | | 9 2 | | |
| | | 6 2 | | 7 2 | | |
| | | 9 2 | | 6 2 | | |
| | | 7 2 | | 8 2 | | |
| | | 8 2 | | 7 2 | | |
| | | 7 2 | | 9 2 | | |
| | | 8 2 | | 6 2 | | |
| | | 8 2 | | 8 2 | | |
| | | 6 2 | | 7 2 | | |
| | | 9 2 | | 6 2 | | |
| | | 7 2 | | 9 2 | | |
| | | 6 2 | | 6 2 | | |
| | | 7 2 | | 7 2 | | |
| | | 6 2 | | 9 2 | | |
| | | 7 2 | | 8 2 | | |
| | | 6 2 | | 6 2 | | |
| | | 6 2 | | 6 2 | | |
| | | 9 2 | | 7 2 | | |
| 129.1.71 | 191053 | 76 | Ce1Ga2Pd2 | | | |
| | -A | -GM | -M | -Z | -R | -X |
| | 5 4 | 7 2 | 5 4 | 7 2 | 3+4 4 | 3+4 4 |
| | 5 4 | 9 2 | 5 4 | 9 2 | 3+4 4 | 3+4 4 |
| | 5 4 | 7 2 | 5 4 | 9 2 | 3+4 4 | 3+4 4 |
| | 5 4 | 9 2 | 5 4 | 7 2 | 3+4 4 | 3+4 4 |
| | 5 4 | 7 2 | 5 4 | 9 2 | 3+4 4 | 3+4 4 |
| | 5 4 | 9 2 | 5 4 | 7 2 | 3+4 4 | 3+4 4 |
| | 5 4 | 6 2 | 5 4 | 8 2 | 3+4 4 | 3+4 4 |
| | 5 4 | 8 2 | 5 4 | 6 2 | 3+4 4 | 3+4 4 |
| | 5 4 | 7 2 | 5 4 | 9 2 | 3+4 4 | 3+4 4 |
| | 5 4 | 7 2 | 5 4 | 7 2 | 3+4 4 | 3+4 4 |
| | 5 4 | 9 2 | 5 4 | 9 2 | 3+4 4 | 3+4 4 |
| | 5 4 | 8 2 | 5 4 | 6 2 | 3+4 4 | 3+4 4 |
| | 5 4 | 9 2 | 5 4 | 7 2 | 3+4 4 | 3+4 4 |
| | 5 4 | 7 2 | 5 4 | 9 2 | 3+4 4 | 3+4 4 |
| | 5 4 | 7 2 | 5 4 | 8 2 | 3+4 4 | 3+4 4 |
| | 5 4 | 6 2 | 5 4 | 7 2 | 3+4 4 | 3+4 4 |
| | 5 4 | 8 2 | 5 4 | 9 2 | 3+4 4 | 3+4 4 |
| | 5 4 | 9 2 | 5 4 | 6 2 | 3+4 4 | 3+4 4 |
| | 5 4 | 6 2 | 5 4 | 8 2 | 3+4 4 | 3+4 4 |
| | | 9 2 | | 9 2 | | |
| | | 6 2 | | 8 2 | | |
| | | 6 2 | | 7 2 | | |
| | | 8 2 | | 8 2 | | |
| | | 9 2 | | 6 2 | | |
| | | 7 2 | | 6 2 | | |
| | | 8 2 | | 7 2 | | |
| | | 7 2 | | 9 2 | | |
| | | 8 2 | | 8 2 | | |
| | | 6 2 | | 6 2 | | |
| | | 6 2 | | 6 2 | | |

Continued on next column

**TABLE LX– continued**

| TQC | ICSD | NoE | SF | | | |
|---|---|---|---|---|---|---|
| | | 9 2 | | 7 2 | | |
| | | 7 2 | | 9 2 | | |
| | | 9 2 | | 6 2 | | |
| | | 6 2 | | 7 2 | | |
| | | 7 2 | | 9 2 | | |
| | | 7 2 | | 8 2 | | |
| | | 6 2 | | 6 2 | | |
| | | 8 2 | | 9 2 | | |
| 129.1.72 | 92319 | 180 | Br1Nd1O3Te1 | | | |
| | -A | -GM | -M | -Z | -R | -X |
| | 5 4 | 7 2 | 5 4 | 7 2 | 3+4 4 | 3+4 4 |
| | 5 4 | 9 2 | 5 4 | 9 2 | 3+4 4 | 3+4 4 |
| | 5 4 | 7 2 | 5 4 | 7 2 | 3+4 4 | 3+4 4 |
| | 5 4 | 9 2 | 5 4 | 9 2 | 3+4 4 | 3+4 4 |
| | 5 4 | 7 2 | 5 4 | 7 2 | 3+4 4 | 3+4 4 |
| | 5 4 | 9 2 | 5 4 | 9 2 | 3+4 4 | 3+4 4 |
| | 5 4 | 7 2 | 5 4 | 7 2 | 3+4 4 | 3+4 4 |
| | 5 4 | 9 2 | 5 4 | 9 2 | 3+4 4 | 3+4 4 |
| | 5 4 | 7 2 | 5 4 | 7 2 | 3+4 4 | 3+4 4 |
| | 5 4 | 9 2 | 5 4 | 9 2 | 3+4 4 | 3+4 4 |
| | 5 4 | 7 2 | 5 4 | 7 2 | 3+4 4 | 3+4 4 |
| | 5 4 | 7 2 | 5 4 | 7 2 | 3+4 4 | 3+4 4 |
| | 5 4 | 9 2 | 5 4 | 9 2 | 3+4 4 | 3+4 4 |
| | 5 4 | 6 2 | 5 4 | 6 2 | 3+4 4 | 3+4 4 |
| | 5 4 | 6 2 | 5 4 | 6 2 | 3+4 4 | 3+4 4 |
| | 5 4 | 8 2 | 5 4 | 8 2 | 3+4 4 | 3+4 4 |
| | 5 4 | 8 2 | 5 4 | 8 2 | 3+4 4 | 3+4 4 |
| | 5 4 | 9 2 | 5 4 | 9 2 | 3+4 4 | 3+4 4 |
| | 5 4 | 8 2 | 5 4 | 8 2 | 3+4 4 | 3+4 4 |
| | 5 4 | 7 2 | 5 4 | 7 2 | 3+4 4 | 3+4 4 |
| | 5 4 | 9 2 | 5 4 | 9 2 | 3+4 4 | 3+4 4 |
| | 5 4 | 6 2 | 5 4 | 6 2 | 3+4 4 | 3+4 4 |
| | 5 4 | 8 2 | 5 4 | 8 2 | 3+4 4 | 3+4 4 |
| | 5 4 | 6 2 | 5 4 | 6 2 | 3+4 4 | 3+4 4 |
| | 5 4 | 8 2 | 5 4 | 8 2 | 3+4 4 | 3+4 4 |
| | 5 4 | 7 2 | 5 4 | 7 2 | 3+4 4 | 3+4 4 |
| | 5 4 | 7 2 | 5 4 | 9 2 | 3+4 4 | 3+4 4 |
| | 5 4 | 9 2 | 5 4 | 7 2 | 3+4 4 | 3+4 4 |
| | 5 4 | 9 2 | 5 4 | 9 2 | 3+4 4 | 3+4 4 |
| | 5 4 | 7 2 | 5 4 | 7 2 | 3+4 4 | 3+4 4 |
| | 5 4 | 7 2 | 5 4 | 9 2 | 3+4 4 | 3+4 4 |
| | 5 4 | 9 2 | 5 4 | 7 2 | 3+4 4 | 3+4 4 |
| | 5 4 | 9 2 | 5 4 | 7 2 | 3+4 4 | 3+4 4 |
| | 5 4 | 7 2 | 5 4 | 9 2 | 3+4 4 | 3+4 4 |
| | 5 4 | 7 2 | 5 4 | 7 2 | 3+4 4 | 3+4 4 |
| | 5 4 | 9 2 | 5 4 | 9 2 | 3+4 4 | 3+4 4 |
| | 5 4 | 7 2 | 5 4 | 7 2 | 3+4 4 | 3+4 4 |
| | 5 4 | 9 2 | 5 4 | 9 2 | 3+4 4 | 3+4 4 |
| | 5 4 | 6 2 | 5 4 | 6 2 | 3+4 4 | 3+4 4 |
| | 5 4 | 8 2 | 5 4 | 8 2 | 3+4 4 | 3+4 4 |
| | 5 4 | 9 2 | 5 4 | 9 2 | 3+4 4 | 3+4 4 |
| | 5 4 | 7 2 | 5 4 | 7 2 | 3+4 4 | 3+4 4 |
| | 5 4 | 7 2 | 5 4 | 9 2 | 3+4 4 | 3+4 4 |
| | 5 4 | 9 2 | 5 4 | 7 2 | 3+4 4 | 3+4 4 |
| | 5 4 | 6 2 | 5 4 | 8 2 | 3+4 4 | 3+4 4 |
| | | 8 2 | | 6 2 | | |
| | | 9 2 | | 9 2 | | |
| | | 7 2 | | 7 2 | | |
| | | 6 2 | | 6 2 | | |
| | | 6 2 | | 6 2 | | |
| | | 6 2 | | 6 2 | | |

Continued on next column

**TABLE LX– continued**

| TQC | ICSD | | NoE | | SF | |
|---|---|---|---|---|---|---|
| | | 8 2 | | 8 2 | | |
| | | 9 2 | | 7 2 | | |
| | | 7 2 | | 9 2 | | |
| | | 7 2 | | 7 2 | | |
| | | 9 2 | | 9 2 | | |
| | | 8 2 | | 8 2 | | |
| | | 7 2 | | 7 2 | | |
| | | 7 2 | | 9 2 | | |
| | | 9 2 | | 6 2 | | |
| | | 6 2 | | 7 2 | | |
| | | 8 2 | | 8 2 | | |
| | | 7 2 | | 9 2 | | |
| | | 6 2 | | 6 2 | | |
| | | 9 2 | | 9 2 | | |
| | | 9 2 | | 9 2 | | |
| | | 8 2 | | 8 2 | | |
| | | 7 2 | | 7 2 | | |
| | | 6 2 | | 6 2 | | |
| | | 8 2 | | 8 2 | | |
| | | 6 2 | | 9 2 | | |
| | | 7 2 | | 6 2 | | |
| | | 9 2 | | 8 2 | | |
| | | 6 2 | | 7 2 | | |
| | | 9 2 | | 9 2 | | |
| | | 8 2 | | 8 2 | | |
| | | 7 2 | | 7 2 | | |
| | | 9 2 | | 9 2 | | |
| | | 8 2 | | 6 2 | | |
| | | 6 2 | | 8 2 | | |
| | | 7 2 | | 9 2 | | |
| | | 9 2 | | 7 2 | | |
| | | 9 2 | | 7 2 | | |
| | | 8 2 | | 6 2 | | |
| | | 7 2 | | 9 2 | | |
| | | 6 2 | | 6 2 | | |
| | | 8 2 | | 8 2 | | |
| | | 6 2 | | 6 2 | | |
| | | 7 2 | | 7 2 | | |
| | | 9 2 | | 7 2 | | |
| 129.1.73 | 92318 | | 144 | | Bi1I1O3Te1 | |
| | -A | -GM | -M | -Z | -R | -X |
| | 5 4 | 7 2 | 5 4 | 7 2 | 3+4 4 | 3+4 4 |
| | 5 4 | 9 2 | 5 4 | 9 2 | 3+4 4 | 3+4 4 |
| | 5 4 | 7 2 | 5 4 | 7 2 | 3+4 4 | 3+4 4 |
| | 5 4 | 7 2 | 5 4 | 7 2 | 3+4 4 | 3+4 4 |
| | 5 4 | 9 2 | 5 4 | 9 2 | 3+4 4 | 3+4 4 |
| | 5 4 | 8 2 | 5 4 | 8 2 | 3+4 4 | 3+4 4 |
| | 5 4 | 6 2 | 5 4 | 6 2 | 3+4 4 | 3+4 4 |
| | 5 4 | 8 2 | 5 4 | 8 2 | 3+4 4 | 3+4 4 |
| | 5 4 | 6 2 | 5 4 | 6 2 | 3+4 4 | 3+4 4 |
| | 5 4 | 8 2 | 5 4 | 8 2 | 3+4 4 | 3+4 4 |
| | 5 4 | 7 2 | 5 4 | 7 2 | 3+4 4 | 3+4 4 |
| | 5 4 | 9 2 | 5 4 | 9 2 | 3+4 4 | 3+4 4 |
| | 5 4 | 7 2 | 5 4 | 9 2 | 3+4 4 | 3+4 4 |
| | 5 4 | 9 2 | 5 4 | 9 2 | 3+4 4 | 3+4 4 |
| | 5 4 | 7 2 | 5 4 | 7 2 | 3+4 4 | 3+4 4 |
| | 5 4 | 9 2 | 5 4 | 7 2 | 3+4 4 | 3+4 4 |
| | 5 4 | 9 2 | 5 4 | 9 2 | 3+4 4 | 3+4 4 |
| | 5 4 | 7 2 | 5 4 | 9 2 | 3+4 4 | 3+4 4 |
| | 5 4 | 9 2 | 5 4 | 7 2 | 3+4 4 | 3+4 4 |
| | 5 4 | 7 2 | 5 4 | 7 2 | 3+4 4 | 3+4 4 |

Continued on next column

**TABLE LX– continued**

| TQC | ICSD | | NoE | | SF | |
|---|---|---|---|---|---|---|
| | 5 4 | 9 2 | 5 4 | 9 2 | 3+4 4 | 3+4 4 |
| | 5 4 | 7 2 | 5 4 | 7 2 | 3+4 4 | 3+4 4 |
| | 5 4 | 7 2 | 5 4 | 9 2 | 3+4 4 | 3+4 4 |
| | 5 4 | 9 2 | 5 4 | 7 2 | 3+4 4 | 3+4 4 |
| | 5 4 | 7 2 | 5 4 | 7 2 | 3+4 4 | 3+4 4 |
| | 5 4 | 9 2 | 5 4 | 9 2 | 3+4 4 | 3+4 4 |
| | 5 4 | 6 2 | 5 4 | 6 2 | 3+4 4 | 3+4 4 |
| | 5 4 | 8 2 | 5 4 | 8 2 | 3+4 4 | 3+4 4 |
| | 5 4 | 7 2 | 5 4 | 7 2 | 3+4 4 | 3+4 4 |
| | 5 4 | 6 2 | 5 4 | 6 2 | 3+4 4 | 3+4 4 |
| | 5 4 | 7 2 | 5 4 | 7 2 | 3+4 4 | 3+4 4 |
| | 5 4 | 9 2 | 5 4 | 9 2 | 3+4 4 | 3+4 4 |
| | 5 4 | 6 2 | 5 4 | 6 2 | 3+4 4 | 3+4 4 |
| | 5 4 | 8 2 | 5 4 | 8 2 | 3+4 4 | 3+4 4 |
| | 5 4 | 7 2 | 5 4 | 9 2 | 3+4 4 | 3+4 4 |
| | 5 4 | 9 2 | 5 4 | 7 2 | 3+4 4 | 3+4 4 |
| | | 9 2 | | 9 2 | | |
| | | 7 2 | | 7 2 | | |
| | | 9 2 | | 9 2 | | |
| | | 6 2 | | 6 2 | | |
| | | 8 2 | | 8 2 | | |
| | | 8 2 | | 8 2 | | |
| | | 6 2 | | 6 2 | | |
| | | 6 2 | | 6 2 | | |
| | | 8 2 | | 8 2 | | |
| | | 9 2 | | 9 2 | | |
| | | 7 2 | | 7 2 | | |
| | | 9 2 | | 9 2 | | |
| | | 7 2 | | 9 2 | | |
| | | 9 2 | | 9 2 | | |
| | | 7 2 | | 7 2 | | |
| | | 9 2 | | 7 2 | | |
| | | 7 2 | | 9 2 | | |
| | | 9 2 | | 9 2 | | |
| | | 7 2 | | 7 2 | | |
| | | 6 2 | | 8 2 | | |
| | | 8 2 | | 6 2 | | |
| | | 7 2 | | 9 2 | | |
| | | 9 2 | | 8 2 | | |
| | | 6 2 | | 9 2 | | |
| | | 6 2 | | 6 2 | | |
| | | 8 2 | | 8 2 | | |
| | | 7 2 | | 7 2 | | |
| | | 8 2 | | 8 2 | | |
| | | 9 2 | | 7 2 | | |
| | | 6 2 | | 6 2 | | |
| | | 9 2 | | 7 2 | | |
| | | 7 2 | | 9 2 | | |
| | | 9 2 | | 7 2 | | |
| | | 8 2 | | 6 2 | | |
| | | 7 2 | | 9 2 | | |
| | | 6 2 | | 8 2 | | |
| 129.1.74 | 246945 | | 64 | | Ba1Ni1S2 | |
| | -A | -GM | -M | -Z | -R | -X |
| | 5 4 | 7 2 | 5 4 | 7 2 | 3+4 4 | 3+4 4 |
| | 5 4 | 9 2 | 5 4 | 9 2 | 3+4 4 | 3+4 4 |
| | 5 4 | 7 2 | 5 4 | 9 2 | 3+4 4 | 3+4 4 |
| | 5 4 | 7 2 | 5 4 | 7 2 | 3+4 4 | 3+4 4 |
| | 5 4 | 9 2 | 5 4 | 9 2 | 3+4 4 | 3+4 4 |
| | 5 4 | 8 2 | 5 4 | 6 2 | 3+4 4 | 3+4 4 |
| | 5 4 | 7 2 | 5 4 | 7 2 | 3+4 4 | 3+4 4 |

Continued on next column

**TABLE LX– continued**

| TQC | ICSD | NoE | SF | | | |
|---|---|---|---|---|---|---|
| | 5 4 | 9 2 | 5 4 | 9 2 | 3+4 4 | 3+4 4 |
| | 5 4 | 6 2 | 5 4 | 6 2 | 3+4 4 | 3+4 4 |
| | 5 4 | 8 2 | 5 4 | 8 2 | 3+4 4 | 3+4 4 |
| | 5 4 | 9 2 | 5 4 | 7 2 | 3+4 4 | 3+4 4 |
| | 5 4 | 7 2 | 5 4 | 9 2 | 3+4 4 | 3+4 4 |
| | 5 4 | 6 2 | 5 4 | 9 2 | 3+4 4 | 3+4 4 |
| | 5 4 | 7 2 | 5 4 | 8 2 | 3+4 4 | 3+4 4 |
| | 5 4 | 9 2 | 5 4 | 7 2 | 3+4 4 | 3+4 4 |
| | 5 4 | 6 2 | 5 4 | 8 2 | 3+4 4 | 3+4 4 |
| | | 7 2 | | 7 2 | | |
| | | 9 2 | | 9 2 | | |
| | | 9 2 | | 7 2 | | |
| | | 8 2 | | 6 2 | | |
| | | 6 2 | | 8 2 | | |
| | | 7 2 | | 9 2 | | |
| | | 6 2 | | 8 2 | | |
| | | 9 2 | | 7 2 | | |
| | | 8 2 | | 6 2 | | |
| | | 8 2 | | 6 2 | | |
| | | 7 2 | | 6 2 | | |
| | | 8 2 | | 9 2 | | |
| | | 7 2 | | 9 2 | | |
| | | 6 2 | | 8 2 | | |
| | | 6 2 | | 8 2 | | |
| | | 9 2 | | 7 2 | | |
| 129.1.75 | 15289 | 64 | Ba1Ni1S2 | | | |
| | -A | -GM | -M | -Z | -R | -X |
| | 5 4 | 7 2 | 5 4 | 7 2 | 3+4 4 | 3+4 4 |
| | 5 4 | 9 2 | 5 4 | 9 2 | 3+4 4 | 3+4 4 |
| | 5 4 | 7 2 | 5 4 | 9 2 | 3+4 4 | 3+4 4 |
| | 5 4 | 7 2 | 5 4 | 7 2 | 3+4 4 | 3+4 4 |
| | 5 4 | 9 2 | 5 4 | 9 2 | 3+4 4 | 3+4 4 |
| | 5 4 | 8 2 | 5 4 | 6 2 | 3+4 4 | 3+4 4 |
| | 5 4 | 7 2 | 5 4 | 7 2 | 3+4 4 | 3+4 4 |
| | 5 4 | 9 2 | 5 4 | 9 2 | 3+4 4 | 3+4 4 |
| | 5 4 | 6 2 | 5 4 | 6 2 | 3+4 4 | 3+4 4 |
| | 5 4 | 8 2 | 5 4 | 8 2 | 3+4 4 | 3+4 4 |
| | 5 4 | 9 2 | 5 4 | 9 2 | 3+4 4 | 3+4 4 |
| | 5 4 | 7 2 | 5 4 | 7 2 | 3+4 4 | 3+4 4 |
| | 5 4 | 6 2 | 5 4 | 9 2 | 3+4 4 | 3+4 4 |
| | 5 4 | 7 2 | 5 4 | 8 2 | 3+4 4 | 3+4 4 |
| | 5 4 | 9 2 | 5 4 | 7 2 | 3+4 4 | 3+4 4 |
| | 5 4 | 6 2 | 5 4 | 8 2 | 3+4 4 | 3+4 4 |
| | | 7 2 | | 7 2 | | |
| | | 9 2 | | 9 2 | | |
| | | 9 2 | | 7 2 | | |
| | | 8 2 | | 6 2 | | |
| | | 6 2 | | 8 2 | | |
| | | 7 2 | | 9 2 | | |
| | | 6 2 | | 8 2 | | |
| | | 8 2 | | 7 2 | | |
| | | 9 2 | | 6 2 | | |
| | | 8 2 | | 6 2 | | |
| | | 7 2 | | 6 2 | | |
| | | 8 2 | | 9 2 | | |
| | | 7 2 | | 9 2 | | |
| | | 6 2 | | 8 2 | | |
| | | 6 2 | | 8 2 | | |
| | | 9 2 | | 7 2 | | |
| 129.1.76 | 41915 | 88 | Ba1Mg2Sn2 | | | |
| | -A | -GM | -M | -Z | -R | -X |
| | 5 4 | 7 2 | 5 4 | 7 2 | 3+4 4 | 3+4 4 |
| | 5 4 | 9 2 | 5 4 | 9 2 | 3+4 4 | 3+4 4 |
| | 5 4 | 7 2 | 5 4 | 7 2 | 3+4 4 | 3+4 4 |
| | 5 4 | 9 2 | 5 4 | 9 2 | 3+4 4 | 3+4 4 |
| | 5 4 | 7 2 | 5 4 | 7 2 | 3+4 4 | 3+4 4 |
| | 5 4 | 9 2 | 5 4 | 9 2 | 3+4 4 | 3+4 4 |
| | 5 4 | 7 2 | 5 4 | 9 2 | 3+4 4 | 3+4 4 |
| | 5 4 | 9 2 | 5 4 | 7 2 | 3+4 4 | 3+4 4 |
| | 5 4 | 7 2 | 5 4 | 7 2 | 3+4 4 | 3+4 4 |
| | 5 4 | 9 2 | 5 4 | 9 2 | 3+4 4 | 3+4 4 |
| | 5 4 | 7 2 | 5 4 | 9 2 | 3+4 4 | 3+4 4 |
| | 5 4 | 9 2 | 5 4 | 7 2 | 3+4 4 | 3+4 4 |
| | 5 4 | 6 2 | 5 4 | 6 2 | 3+4 4 | 3+4 4 |
| | 5 4 | 8 2 | 5 4 | 8 2 | 3+4 4 | 3+4 4 |
| | 5 4 | 6 2 | 5 4 | 8 2 | 3+4 4 | 3+4 4 |
| | 5 4 | 8 2 | 5 4 | 6 2 | 3+4 4 | 3+4 4 |
| | 5 4 | 7 2 | 5 4 | 9 2 | 3+4 4 | 3+4 4 |
| | 5 4 | 7 2 | 5 4 | 7 2 | 3+4 4 | 3+4 4 |
| | 5 4 | 9 2 | 5 4 | 9 2 | 3+4 4 | 3+4 4 |
| | 5 4 | 7 2 | 5 4 | 7 2 | 3+4 4 | 3+4 4 |
| | 5 4 | 9 2 | 5 4 | 9 2 | 3+4 4 | 3+4 4 |
| | 5 4 | 7 2 | 5 4 | 7 2 | 3+4 4 | 3+4 4 |
| | | 9 2 | | 9 2 | | |
| | | 8 2 | | 6 2 | | |
| | | 9 2 | | 7 2 | | |
| | | 7 2 | | 7 2 | | |
| | | 9 2 | | 9 2 | | |
| | | 6 2 | | 8 2 | | |
| | | 7 2 | | 9 2 | | |
| | | 7 2 | | 7 2 | | |
| | | 7 2 | | 9 2 | | |
| | | 9 2 | | 7 2 | | |
| | | 7 2 | | 9 2 | | |
| | | 6 2 | | 6 2 | | |
| | | 8 2 | | 7 2 | | |
| | | 6 2 | | 8 2 | | |
| | | 9 2 | | 6 2 | | |
| | | 6 2 | | 8 2 | | |
| | | 9 2 | | 9 2 | | |
| | | 8 2 | | 8 2 | | |
| | | 7 2 | | 7 2 | | |
| | | 6 2 | | 6 2 | | |
| | | 8 2 | | 8 2 | | |
| | | 9 2 | | 9 2 | | |
| 129.1.77 | 25338 | 44 | Ba1Mg2Pb2 | | | |
| | -A | -GM | -M | -Z | -R | -X |
| | 5 4 | 7 2 | 5 4 | 7 2 | 3+4 4 | 3+4 4 |
| | 5 4 | 9 2 | 5 4 | 9 2 | 3+4 4 | 3+4 4 |
| | 5 4 | 7 2 | 5 4 | 9 2 | 3+4 4 | 3+4 4 |
| | 5 4 | 9 2 | 5 4 | 7 2 | 3+4 4 | 3+4 4 |
| | 5 4 | 7 2 | 5 4 | 9 2 | 3+4 4 | 3+4 4 |
| | 5 4 | 9 2 | 5 4 | 7 2 | 3+4 4 | 3+4 4 |
| | 5 4 | 6 2 | 5 4 | 8 2 | 3+4 4 | 3+4 4 |
| | 5 4 | 8 2 | 5 4 | 6 2 | 3+4 4 | 3+4 4 |
| | 5 4 | 7 2 | 5 4 | 9 2 | 3+4 4 | 3+4 4 |
| | 5 4 | 7 2 | 5 4 | 7 2 | 3+4 4 | 3+4 4 |
| | 5 4 | 9 2 | 5 4 | 9 2 | 3+4 4 | 3+4 4 |
| | | 8 2 | | 6 2 | | |
| | | 9 2 | | 7 2 | | |
| | | 7 2 | | 8 2 | | |
| | | 6 2 | | 7 2 | | |

Continued on next column

**TABLE LX– continued**

| TQC | ICSD | NoE | SF | | | |
|---|---|---|---|---|---|---|
| | | 7 2 | | 9 2 | | |
| | | 7 2 | | 9 2 | | |
| | | 6 2 | | 7 2 | | |
| | | 9 2 | | 6 2 | | |
| | | 6 2 | | 8 2 | | |
| | | 9 2 | | 9 2 | | |
| | | 8 2 | | 8 2 | | |
| 129.1.78 | 370004 | 76 | Au2Ga2Sr1 | | | |
| | -A | -GM | -M | -Z | -R | -X |
| | 5 4 | 7 2 | 5 4 | 7 2 | 3+4 4 | 3+4 4 |
| | 5 4 | 9 2 | 5 4 | 9 2 | 3+4 4 | 3+4 4 |
| | 5 4 | 7 2 | 5 4 | 7 2 | 3+4 4 | 3+4 4 |
| | 5 4 | 9 2 | 5 4 | 9 2 | 3+4 4 | 3+4 4 |
| | 5 4 | 7 2 | 5 4 | 7 2 | 3+4 4 | 3+4 4 |
| | 5 4 | 9 2 | 5 4 | 9 2 | 3+4 4 | 3+4 4 |
| | 5 4 | 6 2 | 5 4 | 6 2 | 3+4 4 | 3+4 4 |
| | 5 4 | 8 2 | 5 4 | 8 2 | 3+4 4 | 3+4 4 |
| | 5 4 | 7 2 | 5 4 | 9 2 | 3+4 4 | 3+4 4 |
| | 5 4 | 7 2 | 5 4 | 7 2 | 3+4 4 | 3+4 4 |
| | 5 4 | 9 2 | 5 4 | 9 2 | 3+4 4 | 3+4 4 |
| | 5 4 | 8 2 | 5 4 | 6 2 | 3+4 4 | 3+4 4 |
| | 5 4 | 7 2 | 5 4 | 9 2 | 3+4 4 | 3+4 4 |
| | 5 4 | 6 2 | 5 4 | 7 2 | 3+4 4 | 3+4 4 |
| | 5 4 | 9 2 | 5 4 | 6 2 | 3+4 4 | 3+4 4 |
| | 5 4 | 8 2 | 5 4 | 8 2 | 3+4 4 | 3+4 4 |
| | 5 4 | 6 2 | 5 4 | 7 2 | 3+4 4 | 3+4 4 |
| | 5 4 | 9 2 | 5 4 | 8 2 | 3+4 4 | 3+4 4 |
| | 5 4 | 7 2 | 5 4 | 9 2 | 3+4 4 | 3+4 4 |
| | | 9 2 | | 9 2 | | |
| | | 6 2 | | 7 2 | | |
| | | 8 2 | | 8 2 | | |
| | | 7 2 | | 6 2 | | |
| | | 9 2 | | 7 2 | | |
| | | 6 2 | | 8 2 | | |
| | | 6 2 | | 6 2 | | |
| | | 7 2 | | 6 2 | | |
| | | 8 2 | | 8 2 | | |
| | | 8 2 | | 9 2 | | |
| | | 6 2 | | 7 2 | | |
| | | 9 2 | | 6 2 | | |
| | | 9 2 | | 9 2 | | |
| | | 7 2 | | 6 2 | | |
| | | 6 2 | | 9 2 | | |
| | | 7 2 | | 7 2 | | |
| | | 7 2 | | 8 2 | | |
| | | 6 2 | | 6 2 | | |
| | | 8 2 | | 9 2 | | |
| 129.1.79 | 238706 | 68 | Au1Bi2Pr1 | | | |
| | -A | -GM | -M | -Z | -R | -X |
| | 5 4 | 7 2 | 5 4 | 7 2 | 3+4 4 | 3+4 4 |
| | 5 4 | 9 2 | 5 4 | 9 2 | 3+4 4 | 3+4 4 |
| | 5 4 | 7 2 | 5 4 | 9 2 | 3+4 4 | 3+4 4 |
| | 5 4 | 9 2 | 5 4 | 7 2 | 3+4 4 | 3+4 4 |
| | 5 4 | 7 2 | 5 4 | 9 2 | 3+4 4 | 3+4 4 |
| | 5 4 | 9 2 | 5 4 | 7 2 | 3+4 4 | 3+4 4 |
| | 5 4 | 6 2 | 5 4 | 8 2 | 3+4 4 | 3+4 4 |
| | 5 4 | 8 2 | 5 4 | 6 2 | 3+4 4 | 3+4 4 |
| | 5 4 | 7 2 | 5 4 | 9 2 | 3+4 4 | 3+4 4 |
| | 5 4 | 7 2 | 5 4 | 7 2 | 3+4 4 | 3+4 4 |
| | 5 4 | 9 2 | 5 4 | 9 2 | 3+4 4 | 3+4 4 |
| | 5 4 | 8 2 | 5 4 | 6 2 | 3+4 4 | 3+4 4 |

Continued on next column

**TABLE LX– continued**

| TQC | ICSD | NoE | SF | | | |
|---|---|---|---|---|---|---|
| | 5 4 | 7 2 | 5 4 | 7 2 | 3+4 4 | 3+4 4 |
| | 5 4 | 9 2 | 5 4 | 8 2 | 3+4 4 | 3+4 4 |
| | 5 4 | 8 2 | 5 4 | 6 2 | 3+4 4 | 3+4 4 |
| | 5 4 | 6 2 | 5 4 | 9 2 | 3+4 4 | 3+4 4 |
| | 5 4 | 9 2 | 5 4 | 9 2 | 3+4 4 | 3+4 4 |
| | | 7 2 | | 7 2 | | |
| | | 8 2 | | 8 2 | | |
| | | 6 2 | | 6 2 | | |
| | | 6 2 | | 7 2 | | |
| | | 6 2 | | 8 2 | | |
| | | 7 2 | | 6 2 | | |
| | | 9 2 | | 9 2 | | |
| | | 9 2 | | 7 2 | | |
| | | 7 2 | | 9 2 | | |
| | | 7 2 | | 7 2 | | |
| | | 6 2 | | 6 2 | | |
| | | 6 2 | | 8 2 | | |
| | | 9 2 | | 9 2 | | |
| | | 8 2 | | 6 2 | | |
| | | 8 2 | | 8 2 | | |
| | | 9 2 | | 9 2 | | |
| | | 6 2 | | 8 2 | | |
| 129.1.80 | 238707 | 68 | Au1Bi2Nd1 | | | |
| | -A | -GM | -M | -Z | -R | -X |
| | 5 4 | 7 2 | 5 4 | 7 2 | 3+4 4 | 3+4 4 |
| | 5 4 | 9 2 | 5 4 | 9 2 | 3+4 4 | 3+4 4 |
| | 5 4 | 7 2 | 5 4 | 9 2 | 3+4 4 | 3+4 4 |
| | 5 4 | 9 2 | 5 4 | 7 2 | 3+4 4 | 3+4 4 |
| | 5 4 | 7 2 | 5 4 | 9 2 | 3+4 4 | 3+4 4 |
| | 5 4 | 9 2 | 5 4 | 7 2 | 3+4 4 | 3+4 4 |
| | 5 4 | 6 2 | 5 4 | 8 2 | 3+4 4 | 3+4 4 |
| | 5 4 | 8 2 | 5 4 | 6 2 | 3+4 4 | 3+4 4 |
| | 5 4 | 7 2 | 5 4 | 9 2 | 3+4 4 | 3+4 4 |
| | 5 4 | 7 2 | 5 4 | 7 2 | 3+4 4 | 3+4 4 |
| | 5 4 | 9 2 | 5 4 | 9 2 | 3+4 4 | 3+4 4 |
| | 5 4 | 8 2 | 5 4 | 6 2 | 3+4 4 | 3+4 4 |
| | 5 4 | 7 2 | 5 4 | 7 2 | 3+4 4 | 3+4 4 |
| | 5 4 | 9 2 | 5 4 | 8 2 | 3+4 4 | 3+4 4 |
| | 5 4 | 8 2 | 5 4 | 6 2 | 3+4 4 | 3+4 4 |
| | 5 4 | 6 2 | 5 4 | 9 2 | 3+4 4 | 3+4 4 |
| | 5 4 | 9 2 | 5 4 | 9 2 | 3+4 4 | 3+4 4 |
| | | 7 2 | | 7 2 | | |
| | | 8 2 | | 8 2 | | |
| | | 6 2 | | 6 2 | | |
| | | 6 2 | | 7 2 | | |
| | | 6 2 | | 8 2 | | |
| | | 7 2 | | 6 2 | | |
| | | 9 2 | | 9 2 | | |
| | | 9 2 | | 7 2 | | |
| | | 7 2 | | 9 2 | | |
| | | 7 2 | | 7 2 | | |
| | | 6 2 | | 6 2 | | |
| | | 6 2 | | 8 2 | | |
| | | 9 2 | | 9 2 | | |
| | | 8 2 | | 6 2 | | |
| | | 8 2 | | 8 2 | | |
| | | 9 2 | | 9 2 | | |
| | | 6 2 | | 8 2 | | |
| 129.1.81 | 66765 | 48 | As2U1 | | | |
| | -A | -GM | -M | -Z | -R | -X |
| | 5 4 | 7 2 | 5 4 | 9 2 | 3+4 4 | 3+4 4 |

Continued on next column

**TABLE LX– continued**

| TQC | ICSD | | NoE | | SF | |
|---|---|---|---|---|---|---|
| | 5 4 | 9 2 | 5 4 | 7 2 | 3+4 4 | 3+4 4 |
| | 5 4 | 7 2 | 5 4 | 9 2 | 3+4 4 | 3+4 4 |
| | 5 4 | 9 2 | 5 4 | 7 2 | 3+4 4 | 3+4 4 |
| | 5 4 | 7 2 | 5 4 | 9 2 | 3+4 4 | 3+4 4 |
| | 5 4 | 9 2 | 5 4 | 7 2 | 3+4 4 | 3+4 4 |
| | 5 4 | 6 2 | 5 4 | 8 2 | 3+4 4 | 3+4 4 |
| | 5 4 | 8 2 | 5 4 | 6 2 | 3+4 4 | 3+4 4 |
| | 5 4 | 7 2 | 5 4 | 7 2 | 3+4 4 | 3+4 4 |
| | 5 4 | 9 2 | 5 4 | 9 2 | 3+4 4 | 3+4 4 |
| | 5 4 | 7 2 | 5 4 | 7 2 | 3+4 4 | 3+4 4 |
| | 5 4 | 8 2 | 5 4 | 8 2 | 3+4 4 | 3+4 4 |
| | | 9 2 | | 9 2 | | |
| | | 6 2 | | 6 2 | | |
| | | 7 2 | | 7 2 | | |
| | | 7 2 | | 6 2 | | |
| | | 6 2 | | 9 2 | | |
| | | 7 2 | | 9 2 | | |
| | | 6 2 | | 8 2 | | |
| | | 9 2 | | 7 2 | | |
| | | 8 2 | | 6 2 | | |
| | | 8 2 | | 7 2 | | |
| | | 9 2 | | 6 2 | | |
| | | 6 2 | | 8 2 | | |
| 129.1.82 | 44081 | | 48 | | As2U1 | |
| | -A | -GM | -M | -Z | -R | -X |
| | 5 4 | 7 2 | 5 4 | 9 2 | 3+4 4 | 3+4 4 |
| | 5 4 | 9 2 | 5 4 | 7 2 | 3+4 4 | 3+4 4 |
| | 5 4 | 7 2 | 5 4 | 9 2 | 3+4 4 | 3+4 4 |
| | 5 4 | 9 2 | 5 4 | 7 2 | 3+4 4 | 3+4 4 |
| | 5 4 | 7 2 | 5 4 | 7 2 | 3+4 4 | 3+4 4 |
| | 5 4 | 9 2 | 5 4 | 9 2 | 3+4 4 | 3+4 4 |
| | 5 4 | 6 2 | 5 4 | 8 2 | 3+4 4 | 3+4 4 |
| | 5 4 | 8 2 | 5 4 | 6 2 | 3+4 4 | 3+4 4 |
| | 5 4 | 7 2 | 5 4 | 7 2 | 3+4 4 | 3+4 4 |
| | 5 4 | 9 2 | 5 4 | 9 2 | 3+4 4 | 3+4 4 |
| | 5 4 | 7 2 | 5 4 | 7 2 | 3+4 4 | 3+4 4 |
| | 5 4 | 8 2 | 5 4 | 8 2 | 3+4 4 | 3+4 4 |
| | | 9 2 | | 9 2 | | |
| | | 6 2 | | 6 2 | | |
| | | 7 2 | | 7 2 | | |
| | | 7 2 | | 6 2 | | |
| | | 6 2 | | 9 2 | | |
| | | 7 2 | | 9 2 | | |
| | | 6 2 | | 8 2 | | |
| | | 9 2 | | 7 2 | | |
| | | 8 2 | | 6 2 | | |
| | | 8 2 | | 7 2 | | |
| | | 9 2 | | 6 2 | | |
| | | 6 2 | | 8 2 | | |
| 129.1.83 | 107523 | | 48 | | As2U1 | |
| | -A | -GM | -M | -Z | -R | -X |
| | 5 4 | 7 2 | 5 4 | 9 2 | 3+4 4 | 3+4 4 |
| | 5 4 | 9 2 | 5 4 | 7 2 | 3+4 4 | 3+4 4 |
| | 5 4 | 7 2 | 5 4 | 9 2 | 3+4 4 | 3+4 4 |
| | 5 4 | 9 2 | 5 4 | 7 2 | 3+4 4 | 3+4 4 |
| | 5 4 | 7 2 | 5 4 | 9 2 | 3+4 4 | 3+4 4 |
| | 5 4 | 9 2 | 5 4 | 7 2 | 3+4 4 | 3+4 4 |
| | 5 4 | 6 2 | 5 4 | 8 2 | 3+4 4 | 3+4 4 |
| | 5 4 | 8 2 | 5 4 | 6 2 | 3+4 4 | 3+4 4 |
| | 5 4 | 7 2 | 5 4 | 7 2 | 3+4 4 | 3+4 4 |
| | 5 4 | 9 2 | 5 4 | 9 2 | 3+4 4 | 3+4 4 |

Continued on next column

**TABLE LX– continued**

| TQC | ICSD | | NoE | | SF | |
|---|---|---|---|---|---|---|
| | 5 4 | 7 2 | 5 4 | 7 2 | 3+4 4 | 3+4 4 |
| | 5 4 | 8 2 | 5 4 | 8 2 | 3+4 4 | 3+4 4 |
| | | 9 2 | | 9 2 | | |
| | | 6 2 | | 6 2 | | |
| | | 7 2 | | 7 2 | | |
| | | 7 2 | | 6 2 | | |
| | | 6 2 | | 9 2 | | |
| | | 7 2 | | 9 2 | | |
| | | 6 2 | | 8 2 | | |
| | | 8 2 | | 7 2 | | |
| | | 9 2 | | 6 2 | | |
| | | 8 2 | | 7 2 | | |
| | | 9 2 | | 6 2 | | |
| | | 6 2 | | 8 2 | | |
| 129.1.84 | 56976 | | 68 | | As2Pd1U1 | |
| | -A | -GM | -M | -Z | -R | -X |
| | 5 4 | 7 2 | 5 4 | 7 2 | 3+4 4 | 3+4 4 |
| | 5 4 | 9 2 | 5 4 | 9 2 | 3+4 4 | 3+4 4 |
| | 5 4 | 7 2 | 5 4 | 7 2 | 3+4 4 | 3+4 4 |
| | 5 4 | 9 2 | 5 4 | 9 2 | 3+4 4 | 3+4 4 |
| | 5 4 | 7 2 | 5 4 | 7 2 | 3+4 4 | 3+4 4 |
| | 5 4 | 9 2 | 5 4 | 9 2 | 3+4 4 | 3+4 4 |
| | 5 4 | 6 2 | 5 4 | 6 2 | 3+4 4 | 3+4 4 |
| | 5 4 | 8 2 | 5 4 | 8 2 | 3+4 4 | 3+4 4 |
| | 5 4 | 7 2 | 5 4 | 7 2 | 3+4 4 | 3+4 4 |
| | 5 4 | 9 2 | 5 4 | 9 2 | 3+4 4 | 3+4 4 |
| | 5 4 | 7 2 | 5 4 | 7 2 | 3+4 4 | 3+4 4 |
| | 5 4 | 8 2 | 5 4 | 8 2 | 3+4 4 | 3+4 4 |
| | 5 4 | 7 2 | 5 4 | 9 2 | 3+4 4 | 3+4 4 |
| | 5 4 | 9 2 | 5 4 | 7 2 | 3+4 4 | 3+4 4 |
| | 5 4 | 9 2 | 5 4 | 7 2 | 3+4 4 | 3+4 4 |
| | 5 4 | 8 2 | 5 4 | 6 2 | 3+4 4 | 3+4 4 |
| | 5 4 | 6 2 | 5 4 | 8 2 | 3+4 4 | 3+4 4 |
| | | 6 2 | | 9 2 | | |
| | | 7 2 | | 6 2 | | |
| | | 8 2 | | 7 2 | | |
| | | 7 2 | | 6 2 | | |
| | | 9 2 | | 9 2 | | |
| | | 6 2 | | 8 2 | | |
| | | 7 2 | | 9 2 | | |
| | | 6 2 | | 6 2 | | |
| | | 6 2 | | 9 2 | | |
| | | 7 2 | | 8 2 | | |
| | | 6 2 | | 7 2 | | |
| | | 9 2 | | 8 2 | | |
| | | 8 2 | | 8 2 | | |
| | | 9 2 | | 6 2 | | |
| | | 9 2 | | 7 2 | | |
| | | 6 2 | | 7 2 | | |
| | | 7 2 | | 9 2 | | |
| 129.1.85 | 412554 | | 88 | | As2Cu1Tm1 | |
| | -A | -GM | -M | -Z | -R | -X |
| | 5 4 | 7 2 | 5 4 | 7 2 | 3+4 4 | 3+4 4 |
| | 5 4 | 9 2 | 5 4 | 9 2 | 3+4 4 | 3+4 4 |
| | 5 4 | 7 2 | 5 4 | 7 2 | 3+4 4 | 3+4 4 |
| | 5 4 | 9 2 | 5 4 | 9 2 | 3+4 4 | 3+4 4 |
| | 5 4 | 7 2 | 5 4 | 7 2 | 3+4 4 | 3+4 4 |
| | 5 4 | 9 2 | 5 4 | 9 2 | 3+4 4 | 3+4 4 |
| | 5 4 | 6 2 | 5 4 | 6 2 | 3+4 4 | 3+4 4 |
| | 5 4 | 8 2 | 5 4 | 8 2 | 3+4 4 | 3+4 4 |
| | 5 4 | 7 2 | 5 4 | 7 2 | 3+4 4 | 3+4 4 |

Continued on next column

**TABLE LX– continued**

| TQC | ICSD | | NoE | | SF | |
|---|---|---|---|---|---|---|
| | 5 4 | 7 2 | 5 4 | 9 2 | 3+4 4 | 3+4 4 |
| | 5 4 | 9 2 | 5 4 | 7 2 | 3+4 4 | 3+4 4 |
| | 5 4 | 8 2 | 5 4 | 8 2 | 3+4 4 | 3+4 4 |
| | 5 4 | 7 2 | 5 4 | 9 2 | 3+4 4 | 3+4 4 |
| | 5 4 | 9 2 | 5 4 | 7 2 | 3+4 4 | 3+4 4 |
| | 5 4 | 8 2 | 5 4 | 6 2 | 3+4 4 | 3+4 4 |
| | 5 4 | 9 2 | 5 4 | 7 2 | 3+4 4 | 3+4 4 |
| | 5 4 | 6 2 | 5 4 | 8 2 | 3+4 4 | 3+4 4 |
| | 5 4 | 6 2 | 5 4 | 9 2 | 3+4 4 | 3+4 4 |
| | 5 4 | 9 2 | 5 4 | 6 2 | 3+4 4 | 3+4 4 |
| | 5 4 | 7 2 | 5 4 | 7 2 | 3+4 4 | 3+4 4 |
| | 5 4 | 7 2 | 5 4 | 9 2 | 3+4 4 | 3+4 4 |
| | 5 4 | 8 2 | 5 4 | 6 2 | 3+4 4 | 3+4 4 |
| | | 6 2 | | 8 2 | | |
| | | 7 2 | | 9 2 | | |
| | | 6 2 | | 8 2 | | |
| | | 9 2 | | 7 2 | | |
| | | 7 2 | | 9 2 | | |
| | | 6 2 | | 8 2 | | |
| | | 8 2 | | 7 2 | | |
| | | 6 2 | | 6 2 | | |
| | | 9 2 | | 6 2 | | |
| | | 8 2 | | 8 2 | | |
| | | 7 2 | | 9 2 | | |
| | | 6 2 | | 7 2 | | |
| | | 9 2 | | 8 2 | | |
| | | 6 2 | | 6 2 | | |
| | | 9 2 | | 7 2 | | |
| | | 8 2 | | 6 2 | | |
| | | 6 2 | | 8 2 | | |
| | | 7 2 | | 9 2 | | |
| | | 8 2 | | 8 2 | | |
| | | 9 2 | | 7 2 | | |
| | | 6 2 | | 6 2 | | |
| | | 7 2 | | 9 2 | | |
| 129.1.86 | 412551 | | 80 | | As2Cu1Tb1 | |
| | -A | -GM | -M | -Z | -R | -X |
| | 5 4 | 7 2 | 5 4 | 7 2 | 3+4 4 | 3+4 4 |
| | 5 4 | 9 2 | 5 4 | 9 2 | 3+4 4 | 3+4 4 |
| | 5 4 | 7 2 | 5 4 | 7 2 | 3+4 4 | 3+4 4 |
| | 5 4 | 9 2 | 5 4 | 9 2 | 3+4 4 | 3+4 4 |
| | 5 4 | 7 2 | 5 4 | 7 2 | 3+4 4 | 3+4 4 |
| | 5 4 | 9 2 | 5 4 | 9 2 | 3+4 4 | 3+4 4 |
| | 5 4 | 6 2 | 5 4 | 6 2 | 3+4 4 | 3+4 4 |
| | 5 4 | 8 2 | 5 4 | 8 2 | 3+4 4 | 3+4 4 |
| | 5 4 | 7 2 | 5 4 | 7 2 | 3+4 4 | 3+4 4 |
| | 5 4 | 7 2 | 5 4 | 9 2 | 3+4 4 | 3+4 4 |
| | 5 4 | 9 2 | 5 4 | 7 2 | 3+4 4 | 3+4 4 |
| | 5 4 | 8 2 | 5 4 | 8 2 | 3+4 4 | 3+4 4 |
| | 5 4 | 7 2 | 5 4 | 9 2 | 3+4 4 | 3+4 4 |
| | 5 4 | 9 2 | 5 4 | 7 2 | 3+4 4 | 3+4 4 |
| | 5 4 | 8 2 | 5 4 | 6 2 | 3+4 4 | 3+4 4 |
| | 5 4 | 9 2 | 5 4 | 7 2 | 3+4 4 | 3+4 4 |
| | 5 4 | 6 2 | 5 4 | 8 2 | 3+4 4 | 3+4 4 |
| | 5 4 | 6 2 | 5 4 | 9 2 | 3+4 4 | 3+4 4 |
| | 5 4 | 9 2 | 5 4 | 6 2 | 3+4 4 | 3+4 4 |
| | 5 4 | 7 2 | 5 4 | 7 2 | 3+4 4 | 3+4 4 |
| | | 7 2 | | 9 2 | | |
| | | 8 2 | | 6 2 | | |
| | | 6 2 | | 8 2 | | |
| | | 7 2 | | 9 2 | | |

Continued on next column

**TABLE LX– continued**

| TQC | ICSD | | NoE | | SF | |
|---|---|---|---|---|---|---|
| | | 6 2 | | 8 2 | | |
| | | 9 2 | | 7 2 | | |
| | | 7 2 | | 9 2 | | |
| | | 6 2 | | 8 2 | | |
| | | 6 2 | | 7 2 | | |
| | | 8 2 | | 6 2 | | |
| | | 9 2 | | 6 2 | | |
| | | 9 2 | | 8 2 | | |
| | | 8 2 | | 9 2 | | |
| | | 6 2 | | 7 2 | | |
| | | 7 2 | | 8 2 | | |
| | | 6 2 | | 6 2 | | |
| | | 9 2 | | 7 2 | | |
| | | 8 2 | | 6 2 | | |
| | | 6 2 | | 8 2 | | |
| | | 7 2 | | 9 2 | | |
| 129.1.87 | 246061 | | 64 | | As1La1Ni1O1 | |
| | -A | -GM | -M | -Z | -R | -X |
| | 5 4 | 7 2 | 5 4 | 7 2 | 3+4 4 | 3+4 4 |
| | 5 4 | 9 2 | 5 4 | 9 2 | 3+4 4 | 3+4 4 |
| | 5 4 | 7 2 | 5 4 | 7 2 | 3+4 4 | 3+4 4 |
| | 5 4 | 8 2 | 5 4 | 8 2 | 3+4 4 | 3+4 4 |
| | 5 4 | 9 2 | 5 4 | 9 2 | 3+4 4 | 3+4 4 |
| | 5 4 | 7 2 | 5 4 | 7 2 | 3+4 4 | 3+4 4 |
| | 5 4 | 6 2 | 5 4 | 6 2 | 3+4 4 | 3+4 4 |
| | 5 4 | 9 2 | 5 4 | 9 2 | 3+4 4 | 3+4 4 |
| | 5 4 | 8 2 | 5 4 | 8 2 | 3+4 4 | 3+4 4 |
| | 5 4 | 7 2 | 5 4 | 7 2 | 3+4 4 | 3+4 4 |
| | 5 4 | 7 2 | 5 4 | 9 2 | 3+4 4 | 3+4 4 |
| | 5 4 | 9 2 | 5 4 | 7 2 | 3+4 4 | 3+4 4 |
| | 5 4 | 9 2 | 5 4 | 9 2 | 3+4 4 | 3+4 4 |
| | 5 4 | 6 2 | 5 4 | 6 2 | 3+4 4 | 3+4 4 |
| | 5 4 | 7 2 | 5 4 | 7 2 | 3+4 4 | 3+4 4 |
| | 5 4 | 7 2 | 5 4 | 6 2 | 3+4 4 | 3+4 4 |
| | | 6 2 | | 7 2 | | |
| | | 9 2 | | 9 2 | | |
| | | 8 2 | | 8 2 | | |
| | | 9 2 | | 7 2 | | |
| | | 7 2 | | 6 2 | | |
| | | 6 2 | | 9 2 | | |
| | | 9 2 | | 8 2 | | |
| | | 6 2 | | 9 2 | | |
| | | 8 2 | | 8 2 | | |
| | | 8 2 | | 6 2 | | |
| | | 6 2 | | 8 2 | | |
| | | 7 2 | | 9 2 | | |
| | | 6 2 | | 8 2 | | |
| | | 7 2 | | 7 2 | | |
| | | 9 2 | | 9 2 | | |
| | | 9 2 | | 7 2 | | |
| 129.1.88 | 248695 | | 112 | | As1Fe1O3Sr2V1 | |
| | -A | -GM | -M | -Z | -R | -X |
| | 5 4 | 7 2 | 5 4 | 9 2 | 3+4 4 | 3+4 4 |
| | 5 4 | 9 2 | 5 4 | 7 2 | 3+4 4 | 3+4 4 |
| | 5 4 | 7 2 | 5 4 | 7 2 | 3+4 4 | 3+4 4 |
| | 5 4 | 9 2 | 5 4 | 9 2 | 3+4 4 | 3+4 4 |
| | 5 4 | 9 2 | 5 4 | 7 2 | 3+4 4 | 3+4 4 |
| | 5 4 | 7 2 | 5 4 | 9 2 | 3+4 4 | 3+4 4 |
| | 5 4 | 7 2 | 5 4 | 9 2 | 3+4 4 | 3+4 4 |
| | 5 4 | 9 2 | 5 4 | 7 2 | 3+4 4 | 3+4 4 |
| | 5 4 | 6 2 | 5 4 | 6 2 | 3+4 4 | 3+4 4 |

Continued on next column

**TABLE LX– continued**

| TQC | ICSD | | NoE | | SF | |
|---|---|---|---|---|---|---|
| | 5 4 | 8 2 | 5 4 | 8 2 | 3+4 4 | 3+4 4 |
| | 5 4 | 7 2 | 5 4 | 9 2 | 3+4 4 | 3+4 4 |
| | 5 4 | 9 2 | 5 4 | 7 2 | 3+4 4 | 3+4 4 |
| | 5 4 | 7 2 | 5 4 | 9 2 | 3+4 4 | 3+4 4 |
| | 5 4 | 9 2 | 5 4 | 7 2 | 3+4 4 | 3+4 4 |
| | 5 4 | 6 2 | 5 4 | 8 2 | 3+4 4 | 3+4 4 |
| | 5 4 | 8 2 | 5 4 | 6 2 | 3+4 4 | 3+4 4 |
| | 5 4 | 7 2 | 5 4 | 9 2 | 3+4 4 | 3+4 4 |
| | 5 4 | 9 2 | 5 4 | 7 2 | 3+4 4 | 3+4 4 |
| | 5 4 | 6 2 | 5 4 | 6 2 | 3+4 4 | 3+4 4 |
| | 5 4 | 8 2 | 5 4 | 8 2 | 3+4 4 | 3+4 4 |
| | 5 4 | 7 2 | 5 4 | 7 2 | 3+4 4 | 3+4 4 |
| | 5 4 | 9 2 | 5 4 | 9 2 | 3+4 4 | 3+4 4 |
| | 5 4 | 7 2 | 5 4 | 7 2 | 3+4 4 | 3+4 4 |
| | 5 4 | 9 2 | 5 4 | 9 2 | 3+4 4 | 3+4 4 |
| | 5 4 | 7 2 | 5 4 | 9 2 | 3+4 4 | 3+4 4 |
| | 5 4 | 7 2 | 5 4 | 7 2 | 3+4 4 | 3+4 4 |
| | 5 4 | 9 2 | 5 4 | 9 2 | 3+4 4 | 3+4 4 |
| | 5 4 | 6 2 | 5 4 | 7 2 | 3+4 4 | 3+4 4 |
| | | 8 2 | | 8 2 | | |
| | | 9 2 | | 6 2 | | |
| | | 7 2 | | 7 2 | | |
| | | 9 2 | | 9 2 | | |
| | | 7 2 | | 9 2 | | |
| | | 6 2 | | 8 2 | | |
| | | 9 2 | | 7 2 | | |
| | | 8 2 | | 6 2 | | |
| | | 8 2 | | 6 2 | | |
| | | 6 2 | | 8 2 | | |
| | | 7 2 | | 7 2 | | |
| | | 7 2 | | 9 2 | | |
| | | 9 2 | | 7 2 | | |
| | | 6 2 | | 8 2 | | |
| | | 8 2 | | 6 2 | | |
| | | 9 2 | | 9 2 | | |
| | | 9 2 | | 9 2 | | |
| | | 8 2 | | 8 2 | | |
| | | 7 2 | | 7 2 | | |
| | | 6 2 | | 6 2 | | |
| | | 6 2 | | 6 2 | | |
| | | 6 2 | | 8 2 | | |
| | | 8 2 | | 6 2 | | |
| | | 8 2 | | 8 2 | | |
| | | 7 2 | | 7 2 | | |
| | | 9 2 | | 9 2 | | |
| | | 7 2 | | 9 2 | | |
| | | 6 2 | | 8 2 | | |
| 129.1.89 | 422004 | | 76 | | As1Fe1O1Tb1 | |
| | -A | -GM | -M | -Z | -R | -X |
| | 5 4 | 7 2 | 5 4 | 7 2 | 3+4 4 | 3+4 4 |
| | 5 4 | 9 2 | 5 4 | 9 2 | 3+4 4 | 3+4 4 |
| | 5 4 | 7 2 | 5 4 | 7 2 | 3+4 4 | 3+4 4 |
| | 5 4 | 9 2 | 5 4 | 9 2 | 3+4 4 | 3+4 4 |
| | 5 4 | 7 2 | 5 4 | 7 2 | 3+4 4 | 3+4 4 |
| | 5 4 | 6 2 | 5 4 | 6 2 | 3+4 4 | 3+4 4 |
| | 5 4 | 9 2 | 5 4 | 9 2 | 3+4 4 | 3+4 4 |
| | 5 4 | 8 2 | 5 4 | 8 2 | 3+4 4 | 3+4 4 |
| | 5 4 | 8 2 | 5 4 | 8 2 | 3+4 4 | 3+4 4 |
| | 5 4 | 7 2 | 5 4 | 7 2 | 3+4 4 | 3+4 4 |
| | 5 4 | 7 2 | 5 4 | 9 2 | 3+4 4 | 3+4 4 |
| | 5 4 | 9 2 | 5 4 | 7 2 | 3+4 4 | 3+4 4 |

Continued on next column

**TABLE LX– continued**

| TQC | ICSD | | NoE | | SF | |
|---|---|---|---|---|---|---|
| | 5 4 | 9 2 | 5 4 | 9 2 | 3+4 4 | 3+4 4 |
| | 5 4 | 7 2 | 5 4 | 7 2 | 3+4 4 | 3+4 4 |
| | 5 4 | 6 2 | 5 4 | 6 2 | 3+4 4 | 3+4 4 |
| | 5 4 | 6 2 | 5 4 | 6 2 | 3+4 4 | 3+4 4 |
| | 5 4 | 7 2 | 5 4 | 7 2 | 3+4 4 | 3+4 4 |
| | 5 4 | 9 2 | 5 4 | 9 2 | 3+4 4 | 3+4 4 |
| | 5 4 | 8 2 | 5 4 | 7 2 | 3+4 4 | 3+4 4 |
| | | 9 2 | | 9 2 | | |
| | | 9 2 | | 8 2 | | |
| | | 8 2 | | 6 2 | | |
| | | 7 2 | | 8 2 | | |
| | | 6 2 | | 9 2 | | |
| | | 6 2 | | 8 2 | | |
| | | 8 2 | | 6 2 | | |
| | | 6 2 | | 6 2 | | |
| | | 6 2 | | 8 2 | | |
| | | 7 2 | | 7 2 | | |
| | | 9 2 | | 9 2 | | |
| | | 8 2 | | 8 2 | | |
| | | 8 2 | | 6 2 | | |
| | | 7 2 | | 9 2 | | |
| | | 6 2 | | 7 2 | | |
| | | 7 2 | | 8 2 | | |
| | | 6 2 | | 7 2 | | |
| | | 9 2 | | 6 2 | | |
| | | 8 2 | | 6 2 | | |
| 129.1.90 | 42165 | | 48 | | As1Cu1Si1Zr1 | |
| | -A | -GM | -M | -Z | -R | -X |
| | 5 4 | 7 2 | 5 4 | 9 2 | 3+4 4 | 3+4 4 |
| | 5 4 | 9 2 | 5 4 | 7 2 | 3+4 4 | 3+4 4 |
| | 5 4 | 7 2 | 5 4 | 7 2 | 3+4 4 | 3+4 4 |
| | 5 4 | 7 2 | 5 4 | 9 2 | 3+4 4 | 3+4 4 |
| | 5 4 | 9 2 | 5 4 | 7 2 | 3+4 4 | 3+4 4 |
| | 5 4 | 8 2 | 5 4 | 6 2 | 3+4 4 | 3+4 4 |
| | 5 4 | 9 2 | 5 4 | 7 2 | 3+4 4 | 3+4 4 |
| | 5 4 | 6 2 | 5 4 | 8 2 | 3+4 4 | 3+4 4 |
| | 5 4 | 9 2 | 5 4 | 9 2 | 3+4 4 | 3+4 4 |
| | 5 4 | 8 2 | 5 4 | 6 2 | 3+4 4 | 3+4 4 |
| | 5 4 | 7 2 | 5 4 | 9 2 | 3+4 4 | 3+4 4 |
| | 5 4 | 7 2 | 5 4 | 8 2 | 3+4 4 | 3+4 4 |
| | | 6 2 | | 8 2 | | |
| | | 8 2 | | 7 2 | | |
| | | 6 2 | | 8 2 | | |
| | | 7 2 | | 9 2 | | |
| | | 6 2 | | 6 2 | | |
| | | 9 2 | | 7 2 | | |
| | | 6 2 | | 7 2 | | |
| | | 7 2 | | 9 2 | | |
| | | 9 2 | | 8 2 | | |
| | | 6 2 | | 6 2 | | |
| | | 9 2 | | 7 2 | | |
| | | 8 2 | | 6 2 | | |
| 129.1.91 | 423103 | | 48 | | As1Cu1Ge1Zr1 | |
| | -A | -GM | -M | -Z | -R | -X |
| | 5 4 | 7 2 | 5 4 | 9 2 | 3+4 4 | 3+4 4 |
| | 5 4 | 9 2 | 5 4 | 7 2 | 3+4 4 | 3+4 4 |
| | 5 4 | 7 2 | 5 4 | 7 2 | 3+4 4 | 3+4 4 |
| | 5 4 | 7 2 | 5 4 | 9 2 | 3+4 4 | 3+4 4 |
| | 5 4 | 9 2 | 5 4 | 7 2 | 3+4 4 | 3+4 4 |
| | 5 4 | 8 2 | 5 4 | 6 2 | 3+4 4 | 3+4 4 |
| | 5 4 | 9 2 | 5 4 | 7 2 | 3+4 4 | 3+4 4 |

Continued on next column

**TABLE LX– continued**

| TQC | ICSD | | NoE | SF | | |
|---|---|---|---|---|---|---|
| | 5 4 | 6 2 | 5 4 | 8 2 | 3+4 4 | 3+4 4 |
| | 5 4 | 8 2 | 5 4 | 9 2 | 3+4 4 | 3+4 4 |
| | 5 4 | 9 2 | 5 4 | 8 2 | 3+4 4 | 3+4 4 |
| | 5 4 | 7 2 | 5 4 | 6 2 | 3+4 4 | 3+4 4 |
| | 5 4 | 7 2 | 5 4 | 9 2 | 3+4 4 | 3+4 4 |
| | | 6 2 | | 8 2 | | |
| | | 8 2 | | 7 2 | | |
| | | 6 2 | | 8 2 | | |
| | | 7 2 | | 6 2 | | |
| | | 6 2 | | 9 2 | | |
| | | 9 2 | | 7 2 | | |
| | | 6 2 | | 7 2 | | |
| | | 7 2 | | 9 2 | | |
| | | 6 2 | | 8 2 | | |
| | | 9 2 | | 6 2 | | |
| | | 9 2 | | 7 2 | | |
| | | 8 2 | | 6 2 | | |
| 129.1.92 | 42750 | | 28 | Al1Ge1Mn1 | | |
| | -A | -GM | -M | -Z | -R | -X |
| | 5 4 | 7 2 | 5 4 | 9 2 | 3+4 4 | 3+4 4 |
| | 5 4 | 9 2 | 5 4 | 7 2 | 3+4 4 | 3+4 4 |
| | 5 4 | 7 2 | 5 4 | 9 2 | 3+4 4 | 3+4 4 |
| | 5 4 | 9 2 | 5 4 | 7 2 | 3+4 4 | 3+4 4 |
| | 5 4 | 9 2 | 5 4 | 9 2 | 3+4 4 | 3+4 4 |
| | 5 4 | 8 2 | 5 4 | 8 2 | 3+4 4 | 3+4 4 |
| | 5 4 | 7 2 | 5 4 | 6 2 | 3+4 4 | 3+4 4 |
| | | 6 2 | | 9 2 | | |
| | | 6 2 | | 7 2 | | |
| | | 9 2 | | 7 2 | | |
| | | 7 2 | | 6 2 | | |
| | | 8 2 | | 8 2 | | |
| | | 7 2 | | 7 2 | | |
| | | 6 2 | | 6 2 | | |
| 129.1.93 | 190195 | | 68 | Ag1Pr1Sb2 | | |
| | -A | -GM | -M | -Z | -R | -X |
| | 5 4 | 7 2 | 5 4 | 7 2 | 3+4 4 | 3+4 4 |
| | 5 4 | 9 2 | 5 4 | 9 2 | 3+4 4 | 3+4 4 |
| | 5 4 | 7 2 | 5 4 | 7 2 | 3+4 4 | 3+4 4 |
| | 5 4 | 9 2 | 5 4 | 9 2 | 3+4 4 | 3+4 4 |
| | 5 4 | 7 2 | 5 4 | 7 2 | 3+4 4 | 3+4 4 |
| | 5 4 | 9 2 | 5 4 | 9 2 | 3+4 4 | 3+4 4 |
| | 5 4 | 6 2 | 5 4 | 6 2 | 3+4 4 | 3+4 4 |
| | 5 4 | 8 2 | 5 4 | 8 2 | 3+4 4 | 3+4 4 |
| | 5 4 | 7 2 | 5 4 | 7 2 | 3+4 4 | 3+4 4 |
| | 5 4 | 7 2 | 5 4 | 9 2 | 3+4 4 | 3+4 4 |
| | 5 4 | 9 2 | 5 4 | 7 2 | 3+4 4 | 3+4 4 |
| | 5 4 | 8 2 | 5 4 | 9 2 | 3+4 4 | 3+4 4 |
| | 5 4 | 9 2 | 5 4 | 8 2 | 3+4 4 | 3+4 4 |
| | 5 4 | 7 2 | 5 4 | 7 2 | 3+4 4 | 3+4 4 |
| | 5 4 | 6 2 | 5 4 | 8 2 | 3+4 4 | 3+4 4 |
| | 5 4 | 8 2 | 5 4 | 6 2 | 3+4 4 | 3+4 4 |
| | 5 4 | 9 2 | 5 4 | 7 2 | 3+4 4 | 3+4 4 |
| | | 7 2 | | 9 2 | | |
| | | 8 2 | | 6 2 | | |
| | | 6 2 | | 8 2 | | |
| | | 6 2 | | 9 2 | | |
| | | 7 2 | | 8 2 | | |
| | | 9 2 | | 7 2 | | |
| | | 6 2 | | 6 2 | | |
| | | 9 2 | | 7 2 | | |
| | | 7 2 | | 9 2 | | |

Continued on next column

**TABLE LX– continued**

| TQC | ICSD | | NoE | SF | | |
|---|---|---|---|---|---|---|
| | | 7 2 | | 9 2 | | |
| | | 6 2 | | 7 2 | | |
| | | 6 2 | | 8 2 | | |
| | | 9 2 | | 6 2 | | |
| | | 9 2 | | 7 2 | | |
| | | 8 2 | | 6 2 | | |
| | | 8 2 | | 8 2 | | |
| | | 6 2 | | 6 2 | | |
| 129.2.1 | 52412 | | 12 | Sc1 | | |
| | -A | -GM | -M | -Z | -R | -X |
| | 5 4 | 7 2 | 5 4 | 7 2 | 3+4 4 | 3+4 4 |
| | 5 4 | 6 2 | 5 4 | 9 2 | 3+4 4 | 3+4 4 |
| | 5 4 | 9 2 | 5 4 | 9 2 | 3+4 4 | 3+4 4 |
| | | 8 2 | | 8 2 | | |
| | | 7 2 | | 6 2 | | |
| | | 6 2 | | 6 2 | | |
| 129.2.2 | 54346 | | 76 | Pt2Si2U1 | | |
| | -A | -GM | -M | -Z | -R | -X |
| | 5 4 | 7 2 | 5 4 | 9 2 | 3+4 4 | 3+4 4 |
| | 5 4 | 7 2 | 5 4 | 7 2 | 3+4 4 | 3+4 4 |
| | 5 4 | 9 2 | 5 4 | 9 2 | 3+4 4 | 3+4 4 |
| | 5 4 | 8 2 | 5 4 | 6 2 | 3+4 4 | 3+4 4 |
| | 5 4 | 9 2 | 5 4 | 7 2 | 3+4 4 | 3+4 4 |
| | 5 4 | 7 2 | 5 4 | 9 2 | 3+4 4 | 3+4 4 |
| | 5 4 | 9 2 | 5 4 | 7 2 | 3+4 4 | 3+4 4 |
| | 5 4 | 7 2 | 5 4 | 9 2 | 3+4 4 | 3+4 4 |
| | 5 4 | 6 2 | 5 4 | 8 2 | 3+4 4 | 3+4 4 |
| | 5 4 | 6 2 | 5 4 | 6 2 | 3+4 4 | 3+4 4 |
| | 5 4 | 8 2 | 5 4 | 8 2 | 3+4 4 | 3+4 4 |
| | 5 4 | 6 2 | 5 4 | 8 2 | 3+4 4 | 3+4 4 |
| | 5 4 | 9 2 | 5 4 | 9 2 | 3+4 4 | 3+4 4 |
| | 5 4 | 6 2 | 5 4 | 7 2 | 3+4 4 | 3+4 4 |
| | 5 4 | 9 2 | 5 4 | 8 2 | 3+4 4 | 3+4 4 |
| | 5 4 | 7 2 | 5 4 | 6 2 | 3+4 4 | 3+4 4 |
| | 5 4 | 8 2 | 5 4 | 7 2 | 3+4 4 | 3+4 4 |
| | 5 4 | 7 2 | 5 4 | 9 2 | 3+4 4 | 3+4 4 |
| | 5 4 | 8 2 | 5 4 | 6 2 | 3+4 4 | 3+4 4 |
| | | 6 2 | | 8 2 | | |
| | | 8 2 | | 6 2 | | |
| | | 9 2 | | 7 2 | | |
| | | 7 2 | | 7 2 | | |
| | | 6 2 | | 9 2 | | |
| | | 6 2 | | 6 2 | | |
| | | 7 2 | | 6 2 | | |
| | | 9 2 | | 8 2 | | |
| | | 6 2 | | 6 2 | | |
| | | 8 2 | | 9 2 | | |
| | | 7 2 | | 8 2 | | |
| | | 8 2 | | 7 2 | | |
| | | 6 2 | | 6 2 | | |
| | | 9 2 | | 9 2 | | |
| | | 6 2 | | 8 2 | | |
| | | 8 2 | | 7 2 | | |
| | | 9 2 | | 9 2 | | |
| | | 7 2 | | 6 2 | | |
| | | 8 2 | | 8 2 | | |
| 129.2.3 | 38176 | | 52 | O1Te1U1 | | |
| | -A | -GM | -M | -Z | -R | -X |
| | 5 4 | 7 2 | 5 4 | 7 2 | 3+4 4 | 3+4 4 |
| | 5 4 | 9 2 | 5 4 | 9 2 | 3+4 4 | 3+4 4 |
| | 5 4 | 7 2 | 5 4 | 7 2 | 3+4 4 | 3+4 4 |

Continued on next column

**TABLE LX– continued**

| TQC | ICSD | | NoE | | SF | |
|---|---|---|---|---|---|---|
| | 5 4 | 9 2 | 5 4 | 9 2 | 3+4 4 | 3+4 4 |
| | 5 4 | 7 2 | 5 4 | 7 2 | 3+4 4 | 3+4 4 |
| | 5 4 | 8 2 | 5 4 | 8 2 | 3+4 4 | 3+4 4 |
| | 5 4 | 6 2 | 5 4 | 6 2 | 3+4 4 | 3+4 4 |
| | 5 4 | 9 2 | 5 4 | 9 2 | 3+4 4 | 3+4 4 |
| | 5 4 | 8 2 | 5 4 | 8 2 | 3+4 4 | 3+4 4 |
| | 5 4 | 7 2 | 5 4 | 7 2 | 3+4 4 | 3+4 4 |
| | 5 4 | 7 2 | 5 4 | 9 2 | 3+4 4 | 3+4 4 |
| | 5 4 | 9 2 | 5 4 | 7 2 | 3+4 4 | 3+4 4 |
| | 5 4 | 7 2 | 5 4 | 9 2 | 3+4 4 | 3+4 4 |
| | | 6 2 | | 7 2 | | |
| | | 9 2 | | 6 2 | | |
| | | 6 2 | | 6 2 | | |
| | | 9 2 | | 9 2 | | |
| | | 8 2 | | 8 2 | | |
| | | 7 2 | | 9 2 | | |
| | | 7 2 | | 9 2 | | |
| | | 6 2 | | 7 2 | | |
| | | 9 2 | | 8 2 | | |
| | | 8 2 | | 7 2 | | |
| | | 9 2 | | 6 2 | | |
| | | 6 2 | | 6 2 | | |
| | | 6 2 | | 8 2 | | |
| 129.2.4 | 22204 | | 52 | | O1Te1U1 | |
| | -A | -GM | -M | -Z | -R | -X |
| | 5 4 | 7 2 | 5 4 | 7 2 | 3+4 4 | 3+4 4 |
| | 5 4 | 9 2 | 5 4 | 9 2 | 3+4 4 | 3+4 4 |
| | 5 4 | 7 2 | 5 4 | 7 2 | 3+4 4 | 3+4 4 |
| | 5 4 | 9 2 | 5 4 | 9 2 | 3+4 4 | 3+4 4 |
| | 5 4 | 7 2 | 5 4 | 7 2 | 3+4 4 | 3+4 4 |
| | 5 4 | 8 2 | 5 4 | 8 2 | 3+4 4 | 3+4 4 |
| | 5 4 | 6 2 | 5 4 | 6 2 | 3+4 4 | 3+4 4 |
| | 5 4 | 9 2 | 5 4 | 9 2 | 3+4 4 | 3+4 4 |
| | 5 4 | 8 2 | 5 4 | 8 2 | 3+4 4 | 3+4 4 |
| | 5 4 | 7 2 | 5 4 | 7 2 | 3+4 4 | 3+4 4 |
| | 5 4 | 7 2 | 5 4 | 9 2 | 3+4 4 | 3+4 4 |
| | 5 4 | 9 2 | 5 4 | 7 2 | 3+4 4 | 3+4 4 |
| | 5 4 | 7 2 | 5 4 | 9 2 | 3+4 4 | 3+4 4 |
| | | 6 2 | | 7 2 | | |
| | | 9 2 | | 6 2 | | |
| | | 6 2 | | 6 2 | | |
| | | 9 2 | | 9 2 | | |
| | | 8 2 | | 8 2 | | |
| | | 7 2 | | 9 2 | | |
| | | 7 2 | | 9 2 | | |
| | | 6 2 | | 8 2 | | |
| | | 9 2 | | 7 2 | | |
| | | 8 2 | | 7 2 | | |
| | | 9 2 | | 6 2 | | |
| | | 6 2 | | 6 2 | | |
| | | 6 2 | | 8 2 | | |
| 129.2.5 | 419464 | | 64 | | O1Os1P1Pr1 | |
| | -A | -GM | -M | -Z | -R | -X |
| | 5 4 | 7 2 | 5 4 | 7 2 | 3+4 4 | 3+4 4 |
| | 5 4 | 9 2 | 5 4 | 9 2 | 3+4 4 | 3+4 4 |
| | 5 4 | 7 2 | 5 4 | 7 2 | 3+4 4 | 3+4 4 |
| | 5 4 | 9 2 | 5 4 | 9 2 | 3+4 4 | 3+4 4 |
| | 5 4 | 7 2 | 5 4 | 7 2 | 3+4 4 | 3+4 4 |
| | 5 4 | 8 2 | 5 4 | 8 2 | 3+4 4 | 3+4 4 |
| | 5 4 | 6 2 | 5 4 | 6 2 | 3+4 4 | 3+4 4 |
| | 5 4 | 9 2 | 5 4 | 9 2 | 3+4 4 | 3+4 4 |

Continued on next column

**TABLE LX– continued**

| TQC | ICSD | | NoE | | SF | |
|---|---|---|---|---|---|---|
| | 5 4 | 8 2 | 5 4 | 8 2 | 3+4 4 | 3+4 4 |
| | 5 4 | 7 2 | 5 4 | 7 2 | 3+4 4 | 3+4 4 |
| | 5 4 | 7 2 | 5 4 | 9 2 | 3+4 4 | 3+4 4 |
| | 5 4 | 9 2 | 5 4 | 7 2 | 3+4 4 | 3+4 4 |
| | 5 4 | 9 2 | 5 4 | 9 2 | 3+4 4 | 3+4 4 |
| | 5 4 | 7 2 | 5 4 | 7 2 | 3+4 4 | 3+4 4 |
| | 5 4 | 8 2 | 5 4 | 6 2 | 3+4 4 | 3+4 4 |
| | 5 4 | 9 2 | 5 4 | 7 2 | 3+4 4 | 3+4 4 |
| | | 7 2 | | 6 2 | | |
| | | 6 2 | | 8 2 | | |
| | | 6 2 | | 7 2 | | |
| | | 6 2 | | 6 2 | | |
| | | 9 2 | | 9 2 | | |
| | | 6 2 | | 8 2 | | |
| | | 9 2 | | 9 2 | | |
| | | 8 2 | | 8 2 | | |
| | | 7 2 | | 9 2 | | |
| | | 8 2 | | 6 2 | | |
| | | 7 2 | | 9 2 | | |
| | | 6 2 | | 8 2 | | |
| | | 6 2 | | 7 2 | | |
| | | 6 2 | | 8 2 | | |
| | | 9 2 | | 6 2 | | |
| | | 7 2 | | 9 2 | | |
| 129.2.6 | 74786 | | 84 | | Ni2P2Th1 | |
| | -A | -GM | -M | -Z | -R | -X |
| | 5 4 | 7 2 | 5 4 | 7 2 | 3+4 4 | 3+4 4 |
| | 5 4 | 9 2 | 5 4 | 9 2 | 3+4 4 | 3+4 4 |
| | 5 4 | 7 2 | 5 4 | 9 2 | 3+4 4 | 3+4 4 |
| | 5 4 | 9 2 | 5 4 | 7 2 | 3+4 4 | 3+4 4 |
| | 5 4 | 7 2 | 5 4 | 9 2 | 3+4 4 | 3+4 4 |
| | 5 4 | 9 2 | 5 4 | 7 2 | 3+4 4 | 3+4 4 |
| | 5 4 | 6 2 | 5 4 | 8 2 | 3+4 4 | 3+4 4 |
| | 5 4 | 8 2 | 5 4 | 6 2 | 3+4 4 | 3+4 4 |
| | 5 4 | 7 2 | 5 4 | 9 2 | 3+4 4 | 3+4 4 |
| | 5 4 | 7 2 | 5 4 | 7 2 | 3+4 4 | 3+4 4 |
| | 5 4 | 9 2 | 5 4 | 9 2 | 3+4 4 | 3+4 4 |
| | 5 4 | 8 2 | 5 4 | 6 2 | 3+4 4 | 3+4 4 |
| | 5 4 | 9 2 | 5 4 | 7 2 | 3+4 4 | 3+4 4 |
| | 5 4 | 7 2 | 5 4 | 9 2 | 3+4 4 | 3+4 4 |
| | 5 4 | 6 2 | 5 4 | 8 2 | 3+4 4 | 3+4 4 |
| | 5 4 | 7 2 | 5 4 | 7 2 | 3+4 4 | 3+4 4 |
| | 5 4 | 6 2 | 5 4 | 8 2 | 3+4 4 | 3+4 4 |
| | 5 4 | 9 2 | 5 4 | 9 2 | 3+4 4 | 3+4 4 |
| | 5 4 | 9 2 | 5 4 | 9 2 | 3+4 4 | 3+4 4 |
| | 5 4 | 8 2 | 5 4 | 7 2 | 3+4 4 | 3+4 4 |
| | 5 4 | 7 2 | 5 4 | 8 2 | 3+4 4 | 3+4 4 |
| | | 6 2 | | 6 2 | | |
| | | 6 2 | | 8 2 | | |
| | | 9 2 | | 6 2 | | |
| | | 6 2 | | 7 2 | | |
| | | 8 2 | | 6 2 | | |
| | | 8 2 | | 9 2 | | |
| | | 7 2 | | 8 2 | | |
| | | 8 2 | | 6 2 | | |
| | | 9 2 | | 7 2 | | |
| | | 7 2 | | 9 2 | | |
| | | 6 2 | | 8 2 | | |
| | | 7 2 | | 7 2 | | |
| | | 7 2 | | 6 2 | | |
| | | 6 2 | | 6 2 | | |

Continued on next column

**TABLE LX– continued**

| TQC | ICSD | NoE | SF | | | |
|---|---|---|---|---|---|---|
| | | 6 2 | | 6 2 | | |
| | | 8 2 | | 9 2 | | |
| | | 6 2 | | 6 2 | | |
| | | 9 2 | | 9 2 | | |
| | | 8 2 | | 7 2 | | |
| | | 9 2 | | 8 2 | | |
| | | 8 2 | | 8 2 | | |
| 129.2.7 | 54276 | 60 | Mn1Si1Tb1 | | | |
| | -A | -GM | -M | -Z | -R | -X |
| | 5 4 | 7 2 | 5 4 | 9 2 | 3+4 4 | 3+4 4 |
| | 5 4 | 9 2 | 5 4 | 7 2 | 3+4 4 | 3+4 4 |
| | 5 4 | 7 2 | 5 4 | 9 2 | 3+4 4 | 3+4 4 |
| | 5 4 | 9 2 | 5 4 | 7 2 | 3+4 4 | 3+4 4 |
| | 5 4 | 7 2 | 5 4 | 9 2 | 3+4 4 | 3+4 4 |
| | 5 4 | 6 2 | 5 4 | 8 2 | 3+4 4 | 3+4 4 |
| | 5 4 | 9 2 | 5 4 | 7 2 | 3+4 4 | 3+4 4 |
| | 5 4 | 8 2 | 5 4 | 6 2 | 3+4 4 | 3+4 4 |
| | 5 4 | 7 2 | 5 4 | 7 2 | 3+4 4 | 3+4 4 |
| | 5 4 | 9 2 | 5 4 | 9 2 | 3+4 4 | 3+4 4 |
| | 5 4 | 7 2 | 5 4 | 7 2 | 3+4 4 | 3+4 4 |
| | 5 4 | 9 2 | 5 4 | 9 2 | 3+4 4 | 3+4 4 |
| | 5 4 | 9 2 | 5 4 | 9 2 | 3+4 4 | 3+4 4 |
| | 5 4 | 8 2 | 5 4 | 8 2 | 3+4 4 | 3+4 4 |
| | 5 4 | 6 2 | 5 4 | 6 2 | 3+4 4 | 3+4 4 |
| | | 8 2 | | 8 2 | | |
| | | 6 2 | | 6 2 | | |
| | | 7 2 | | 7 2 | | |
| | | 7 2 | | 7 2 | | |
| | | 9 2 | | 8 2 | | |
| | | 8 2 | | 6 2 | | |
| | | 6 2 | | 9 2 | | |
| | | 8 2 | | 8 2 | | |
| | | 7 2 | | 6 2 | | |
| | | 6 2 | | 7 2 | | |
| | | 9 2 | | 6 2 | | |
| | | 6 2 | | 9 2 | | |
| | | 8 2 | | 9 2 | | |
| | | 9 2 | | 6 2 | | |
| | | 8 2 | | 8 2 | | |
| 129.2.8 | 75054 | 48 | Mn1Pr1Si1 | | | |
| | -A | -GM | -M | -Z | -R | -X |
| | 5 4 | 7 2 | 5 4 | 9 2 | 3+4 4 | 3+4 4 |
| | 5 4 | 9 2 | 5 4 | 7 2 | 3+4 4 | 3+4 4 |
| | 5 4 | 7 2 | 5 4 | 9 2 | 3+4 4 | 3+4 4 |
| | 5 4 | 9 2 | 5 4 | 7 2 | 3+4 4 | 3+4 4 |
| | 5 4 | 7 2 | 5 4 | 9 2 | 3+4 4 | 3+4 4 |
| | 5 4 | 6 2 | 5 4 | 8 2 | 3+4 4 | 3+4 4 |
| | 5 4 | 9 2 | 5 4 | 7 2 | 3+4 4 | 3+4 4 |
| | 5 4 | 8 2 | 5 4 | 6 2 | 3+4 4 | 3+4 4 |
| | 5 4 | 7 2 | 5 4 | 7 2 | 3+4 4 | 3+4 4 |
| | 5 4 | 9 2 | 5 4 | 9 2 | 3+4 4 | 3+4 4 |
| | 5 4 | 7 2 | 5 4 | 7 2 | 3+4 4 | 3+4 4 |
| | 5 4 | 9 2 | 5 4 | 9 2 | 3+4 4 | 3+4 4 |
| | | 9 2 | | 9 2 | | |
| | | 8 2 | | 8 2 | | |
| | | 8 2 | | 6 2 | | |
| | | 6 2 | | 8 2 | | |
| | | 7 2 | | 7 2 | | |
| | | 6 2 | | 6 2 | | |
| | | 9 2 | | 6 2 | | |
| | | 7 2 | | 7 2 | | |

Continued on next column

**TABLE LX– continued**

| TQC | ICSD | NoE | SF | | | |
|---|---|---|---|---|---|---|
| | | 8 2 | | 8 2 | | |
| | | 8 2 | | 6 2 | | |
| | | 6 2 | | 7 2 | | |
| | | 7 2 | | 9 2 | | |
| 129.2.9 | 75053 | 48 | Mn1Pr1Si1 | | | |
| | -A | -GM | -M | -Z | -R | -X |
| | 5 4 | 7 2 | 5 4 | 9 2 | 3+4 4 | 3+4 4 |
| | 5 4 | 9 2 | 5 4 | 7 2 | 3+4 4 | 3+4 4 |
| | 5 4 | 7 2 | 5 4 | 9 2 | 3+4 4 | 3+4 4 |
| | 5 4 | 9 2 | 5 4 | 7 2 | 3+4 4 | 3+4 4 |
| | 5 4 | 7 2 | 5 4 | 9 2 | 3+4 4 | 3+4 4 |
| | 5 4 | 6 2 | 5 4 | 8 2 | 3+4 4 | 3+4 4 |
| | 5 4 | 9 2 | 5 4 | 7 2 | 3+4 4 | 3+4 4 |
| | 5 4 | 8 2 | 5 4 | 6 2 | 3+4 4 | 3+4 4 |
| | 5 4 | 7 2 | 5 4 | 7 2 | 3+4 4 | 3+4 4 |
| | 5 4 | 9 2 | 5 4 | 9 2 | 3+4 4 | 3+4 4 |
| | 5 4 | 7 2 | 5 4 | 7 2 | 3+4 4 | 3+4 4 |
| | 5 4 | 9 2 | 5 4 | 9 2 | 3+4 4 | 3+4 4 |
| | | 9 2 | | 9 2 | | |
| | | 8 2 | | 8 2 | | |
| | | 8 2 | | 6 2 | | |
| | | 6 2 | | 8 2 | | |
| | | 7 2 | | 7 2 | | |
| | | 6 2 | | 6 2 | | |
| | | 9 2 | | 6 2 | | |
| | | 7 2 | | 7 2 | | |
| | | 8 2 | | 8 2 | | |
| | | 8 2 | | 6 2 | | |
| | | 6 2 | | 7 2 | | |
| | | 7 2 | | 9 2 | | |
| 129.2.10 | 75052 | 48 | Mn1Pr1Si1 | | | |
| | -A | -GM | -M | -Z | -R | -X |
| | 5 4 | 7 2 | 5 4 | 9 2 | 3+4 4 | 3+4 4 |
| | 5 4 | 9 2 | 5 4 | 7 2 | 3+4 4 | 3+4 4 |
| | 5 4 | 7 2 | 5 4 | 9 2 | 3+4 4 | 3+4 4 |
| | 5 4 | 9 2 | 5 4 | 7 2 | 3+4 4 | 3+4 4 |
| | 5 4 | 7 2 | 5 4 | 9 2 | 3+4 4 | 3+4 4 |
| | 5 4 | 6 2 | 5 4 | 8 2 | 3+4 4 | 3+4 4 |
| | 5 4 | 9 2 | 5 4 | 7 2 | 3+4 4 | 3+4 4 |
| | 5 4 | 8 2 | 5 4 | 6 2 | 3+4 4 | 3+4 4 |
| | 5 4 | 7 2 | 5 4 | 7 2 | 3+4 4 | 3+4 4 |
| | 5 4 | 9 2 | 5 4 | 9 2 | 3+4 4 | 3+4 4 |
| | 5 4 | 7 2 | 5 4 | 7 2 | 3+4 4 | 3+4 4 |
| | 5 4 | 9 2 | 5 4 | 9 2 | 3+4 4 | 3+4 4 |
| | | 9 2 | | 9 2 | | |
| | | 8 2 | | 8 2 | | |
| | | 8 2 | | 6 2 | | |
| | | 6 2 | | 8 2 | | |
| | | 7 2 | | 7 2 | | |
| | | 6 2 | | 6 2 | | |
| | | 9 2 | | 6 2 | | |
| | | 7 2 | | 7 2 | | |
| | | 8 2 | | 8 2 | | |
| | | 8 2 | | 6 2 | | |
| | | 6 2 | | 7 2 | | |
| | | 7 2 | | 9 2 | | |
| 129.2.11 | 50243 | 64 | Mn1Nd1O1Sb1 | | | |
| | -A | -GM | -M | -Z | -R | -X |
| | 5 4 | 7 2 | 5 4 | 7 2 | 3+4 4 | 3+4 4 |
| | 5 4 | 9 2 | 5 4 | 9 2 | 3+4 4 | 3+4 4 |
| | 5 4 | 7 2 | 5 4 | 7 2 | 3+4 4 | 3+4 4 |

Continued on next column

**TABLE LX– continued**

| TQC | ICSD | | NoE | | SF | |
|---|---|---|---|---|---|---|
| | 5 4 | 9 2 | 5 4 | 9 2 | 3+4 4 | 3+4 4 |
| | 5 4 | 7 2 | 5 4 | 7 2 | 3+4 4 | 3+4 4 |
| | 5 4 | 8 2 | 5 4 | 8 2 | 3+4 4 | 3+4 4 |
| | 5 4 | 6 2 | 5 4 | 6 2 | 3+4 4 | 3+4 4 |
| | 5 4 | 9 2 | 5 4 | 9 2 | 3+4 4 | 3+4 4 |
| | 5 4 | 8 2 | 5 4 | 8 2 | 3+4 4 | 3+4 4 |
| | 5 4 | 7 2 | 5 4 | 7 2 | 3+4 4 | 3+4 4 |
| | 5 4 | 7 2 | 5 4 | 9 2 | 3+4 4 | 3+4 4 |
| | 5 4 | 9 2 | 5 4 | 7 2 | 3+4 4 | 3+4 4 |
| | 5 4 | 9 2 | 5 4 | 9 2 | 3+4 4 | 3+4 4 |
| | 5 4 | 7 2 | 5 4 | 7 2 | 3+4 4 | 3+4 4 |
| | 5 4 | 6 2 | 5 4 | 6 2 | 3+4 4 | 3+4 4 |
| | 5 4 | 6 2 | 5 4 | 6 2 | 3+4 4 | 3+4 4 |
| | | 7 2 | | 9 2 | | |
| | | 9 2 | | 8 2 | | |
| | | 8 2 | | 9 2 | | |
| | | 9 2 | | 7 2 | | |
| | | 9 2 | | 7 2 | | |
| | | 8 2 | | 6 2 | | |
| | | 6 2 | | 8 2 | | |
| | | 7 2 | | 9 2 | | |
| | | 6 2 | | 8 2 | | |
| | | 7 2 | | 7 2 | | |
| | | 8 2 | | 6 2 | | |
| | | 9 2 | | 9 2 | | |
| | | 6 2 | | 6 2 | | |
| | | 6 2 | | 8 2 | | |
| | | 8 2 | | 6 2 | | |
| | | 7 2 | | 9 2 | | |
| 129.2.12 | 50241 | | 64 | | Mn1Nd1O1P1 | |
| | -A | -GM | -M | -Z | -R | -X |
| | 5 4 | 7 2 | 5 4 | 7 2 | 3+4 4 | 3+4 4 |
| | 5 4 | 9 2 | 5 4 | 9 2 | 3+4 4 | 3+4 4 |
| | 5 4 | 7 2 | 5 4 | 7 2 | 3+4 4 | 3+4 4 |
| | 5 4 | 9 2 | 5 4 | 9 2 | 3+4 4 | 3+4 4 |
| | 5 4 | 7 2 | 5 4 | 7 2 | 3+4 4 | 3+4 4 |
| | 5 4 | 8 2 | 5 4 | 8 2 | 3+4 4 | 3+4 4 |
| | 5 4 | 6 2 | 5 4 | 6 2 | 3+4 4 | 3+4 4 |
| | 5 4 | 9 2 | 5 4 | 9 2 | 3+4 4 | 3+4 4 |
| | 5 4 | 8 2 | 5 4 | 8 2 | 3+4 4 | 3+4 4 |
| | 5 4 | 7 2 | 5 4 | 7 2 | 3+4 4 | 3+4 4 |
| | 5 4 | 7 2 | 5 4 | 9 2 | 3+4 4 | 3+4 4 |
| | 5 4 | 9 2 | 5 4 | 7 2 | 3+4 4 | 3+4 4 |
| | 5 4 | 9 2 | 5 4 | 9 2 | 3+4 4 | 3+4 4 |
| | 5 4 | 7 2 | 5 4 | 7 2 | 3+4 4 | 3+4 4 |
| | 5 4 | 6 2 | 5 4 | 6 2 | 3+4 4 | 3+4 4 |
| | 5 4 | 6 2 | 5 4 | 6 2 | 3+4 4 | 3+4 4 |
| | | 7 2 | | 7 2 | | |
| | | 9 2 | | 9 2 | | |
| | | 8 2 | | 8 2 | | |
| | | 9 2 | | 9 2 | | |
| | | 9 2 | | 7 2 | | |
| | | 8 2 | | 6 2 | | |
| | | 7 2 | | 9 2 | | |
| | | 6 2 | | 8 2 | | |
| | | 6 2 | | 8 2 | | |
| | | 8 2 | | 6 2 | | |
| | | 7 2 | | 9 2 | | |
| | | 6 2 | | 7 2 | | |
| | | 6 2 | | 6 2 | | |
| | | 9 2 | | 8 2 | | |

Continued on next column

**TABLE LX– continued**

| TQC | ICSD | | NoE | | SF | |
|---|---|---|---|---|---|---|
| | | 8 2 | | 6 2 | | |
| | | 7 2 | | 9 2 | | |
| 129.2.13 | 423892 | | 32 | | La1N1 | |
| | -A | -GM | -M | -Z | -R | -X |
| | 5 4 | 7 2 | 5 4 | 9 2 | 3+4 4 | 3+4 4 |
| | 5 4 | 9 2 | 5 4 | 7 2 | 3+4 4 | 3+4 4 |
| | 5 4 | 7 2 | 5 4 | 7 2 | 3+4 4 | 3+4 4 |
| | 5 4 | 6 2 | 5 4 | 9 2 | 3+4 4 | 3+4 4 |
| | 5 4 | 9 2 | 5 4 | 9 2 | 3+4 4 | 3+4 4 |
| | 5 4 | 7 2 | 5 4 | 8 2 | 3+4 4 | 3+4 4 |
| | 5 4 | 8 2 | 5 4 | 7 2 | 3+4 4 | 3+4 4 |
| | 5 4 | 8 2 | 5 4 | 6 2 | 3+4 4 | 3+4 4 |
| | | 9 2 | | 8 2 | | |
| | | 7 2 | | 7 2 | | |
| | | 9 2 | | 6 2 | | |
| | | 7 2 | | 7 2 | | |
| | | 6 2 | | 6 2 | | |
| | | 6 2 | | 9 2 | | |
| | | 6 2 | | 9 2 | | |
| | | 9 2 | | 8 2 | | |
| 129.2.14 | 62716 | | 80 | | Ir2Si2U1 | |
| | -A | -GM | -M | -Z | -R | -X |
| | 5 4 | 7 2 | 5 4 | 7 2 | 3+4 4 | 3+4 4 |
| | 5 4 | 9 2 | 5 4 | 9 2 | 3+4 4 | 3+4 4 |
| | 5 4 | 7 2 | 5 4 | 7 2 | 3+4 4 | 3+4 4 |
| | 5 4 | 9 2 | 5 4 | 9 2 | 3+4 4 | 3+4 4 |
| | 5 4 | 7 2 | 5 4 | 9 2 | 3+4 4 | 3+4 4 |
| | 5 4 | 9 2 | 5 4 | 7 2 | 3+4 4 | 3+4 4 |
| | 5 4 | 8 2 | 5 4 | 8 2 | 3+4 4 | 3+4 4 |
| | 5 4 | 6 2 | 5 4 | 6 2 | 3+4 4 | 3+4 4 |
| | 5 4 | 7 2 | 5 4 | 7 2 | 3+4 4 | 3+4 4 |
| | 5 4 | 7 2 | 5 4 | 9 2 | 3+4 4 | 3+4 4 |
| | 5 4 | 9 2 | 5 4 | 9 2 | 3+4 4 | 3+4 4 |
| | 5 4 | 8 2 | 5 4 | 6 2 | 3+4 4 | 3+4 4 |
| | 5 4 | 9 2 | 5 4 | 7 2 | 3+4 4 | 3+4 4 |
| | 5 4 | 6 2 | 5 4 | 6 2 | 3+4 4 | 3+4 4 |
| | 5 4 | 7 2 | 5 4 | 8 2 | 3+4 4 | 3+4 4 |
| | 5 4 | 8 2 | 5 4 | 9 2 | 3+4 4 | 3+4 4 |
| | 5 4 | 9 2 | 5 4 | 9 2 | 3+4 4 | 3+4 4 |
| | 5 4 | 7 2 | 5 4 | 7 2 | 3+4 4 | 3+4 4 |
| | 5 4 | 6 2 | 5 4 | 8 2 | 3+4 4 | 3+4 4 |
| | 5 4 | 6 2 | 5 4 | 9 2 | 3+4 4 | 3+4 4 |
| | | 9 2 | | 8 2 | | |
| | | 7 2 | | 7 2 | | |
| | | 6 2 | | 6 2 | | |
| | | 6 2 | | 8 2 | | |
| | | 8 2 | | 7 2 | | |
| | | 9 2 | | 8 2 | | |
| | | 7 2 | | 9 2 | | |
| | | 8 2 | | 6 2 | | |
| | | 8 2 | | 7 2 | | |
| | | 9 2 | | 6 2 | | |
| | | 6 2 | | 6 2 | | |
| | | 7 2 | | 6 2 | | |
| | | 6 2 | | 7 2 | | |
| | | 7 2 | | 9 2 | | |
| | | 7 2 | | 8 2 | | |
| | | 6 2 | | 9 2 | | |
| | | 6 2 | | 6 2 | | |
| | | 9 2 | | 9 2 | | |
| | | 9 2 | | 7 2 | | |

Continued on next column

**TABLE LX– continued**

| TQC | ICSD | | NoE | | SF | |
|---|---|---|---|---|---|---|
| | | 8 2 | | 6 2 | | |
| 129.2.15 | 200099 | | 56 | | I1Np1O1 | |
| | -A | -GM | -M | -Z | -R | -X |
| | 5 4 | 7 2 | 5 4 | 7 2 | 3+4 4 | 3+4 4 |
| | 5 4 | 9 2 | 5 4 | 9 2 | 3+4 4 | 3+4 4 |
| | 5 4 | 7 2 | 5 4 | 7 2 | 3+4 4 | 3+4 4 |
| | 5 4 | 9 2 | 5 4 | 9 2 | 3+4 4 | 3+4 4 |
| | 5 4 | 7 2 | 5 4 | 7 2 | 3+4 4 | 3+4 4 |
| | 5 4 | 8 2 | 5 4 | 8 2 | 3+4 4 | 3+4 4 |
| | 5 4 | 6 2 | 5 4 | 6 2 | 3+4 4 | 3+4 4 |
| | 5 4 | 9 2 | 5 4 | 9 2 | 3+4 4 | 3+4 4 |
| | 5 4 | 8 2 | 5 4 | 8 2 | 3+4 4 | 3+4 4 |
| | 5 4 | 7 2 | 5 4 | 7 2 | 3+4 4 | 3+4 4 |
| | 5 4 | 7 2 | 5 4 | 9 2 | 3+4 4 | 3+4 4 |
| | 5 4 | 9 2 | 5 4 | 7 2 | 3+4 4 | 3+4 4 |
| | 5 4 | 9 2 | 5 4 | 9 2 | 3+4 4 | 3+4 4 |
| | 5 4 | 7 2 | 5 4 | 7 2 | 3+4 4 | 3+4 4 |
| | | 6 2 | | 6 2 | | |
| | | 6 2 | | 6 2 | | |
| | | 7 2 | | 9 2 | | |
| | | 9 2 | | 8 2 | | |
| | | 8 2 | | 9 2 | | |
| | | 9 2 | | 7 2 | | |
| | | 7 2 | | 9 2 | | |
| | | 6 2 | | 7 2 | | |
| | | 9 2 | | 8 2 | | |
| | | 8 2 | | 6 2 | | |
| | | 8 2 | | 6 2 | | |
| | | 6 2 | | 8 2 | | |
| | | 6 2 | | 6 2 | | |
| | | 7 2 | | 9 2 | | |
| 129.2.16 | 87913 | | 48 | | Ge1Ni1Pd1 | |
| | -A | -GM | -M | -Z | -R | -X |
| | 5 4 | 7 2 | 5 4 | 7 2 | 3+4 4 | 3+4 4 |
| | 5 4 | 9 2 | 5 4 | 9 2 | 3+4 4 | 3+4 4 |
| | 5 4 | 7 2 | 5 4 | 7 2 | 3+4 4 | 3+4 4 |
| | 5 4 | 8 2 | 5 4 | 9 2 | 3+4 4 | 3+4 4 |
| | 5 4 | 9 2 | 5 4 | 6 2 | 3+4 4 | 3+4 4 |
| | 5 4 | 9 2 | 5 4 | 9 2 | 3+4 4 | 3+4 4 |
| | 5 4 | 8 2 | 5 4 | 8 2 | 3+4 4 | 3+4 4 |
| | 5 4 | 9 2 | 5 4 | 8 2 | 3+4 4 | 3+4 4 |
| | 5 4 | 6 2 | 5 4 | 7 2 | 3+4 4 | 3+4 4 |
| | 5 4 | 7 2 | 5 4 | 6 2 | 3+4 4 | 3+4 4 |
| | 5 4 | 6 2 | 5 4 | 8 2 | 3+4 4 | 3+4 4 |
| | 5 4 | 6 2 | 5 4 | 6 2 | 3+4 4 | 3+4 4 |
| | | 6 2 | | 9 2 | | |
| | | 7 2 | | 9 2 | | |
| | | 8 2 | | 7 2 | | |
| | | 7 2 | | 8 2 | | |
| | | 7 2 | | 6 2 | | |
| | | 9 2 | | 7 2 | | |
| | | 8 2 | | 6 2 | | |
| | | 6 2 | | 9 2 | | |
| | | 8 2 | | 8 2 | | |
| | | 6 2 | | 6 2 | | |
| | | 7 2 | | 9 2 | | |
| | | 9 2 | | 7 2 | | |
| 129.2.17 | 42369 | | 48 | | Ge1Ni1Pd1 | |
| | -A | -GM | -M | -Z | -R | -X |
| | 5 4 | 7 2 | 5 4 | 7 2 | 3+4 4 | 3+4 4 |
| | 5 4 | 9 2 | 5 4 | 9 2 | 3+4 4 | 3+4 4 |

Continued on next column

**TABLE LX– continued**

| TQC | ICSD | | NoE | | SF | |
|---|---|---|---|---|---|---|
| | 5 4 | 7 2 | 5 4 | 7 2 | 3+4 4 | 3+4 4 |
| | 5 4 | 8 2 | 5 4 | 9 2 | 3+4 4 | 3+4 4 |
| | 5 4 | 9 2 | 5 4 | 6 2 | 3+4 4 | 3+4 4 |
| | 5 4 | 9 2 | 5 4 | 9 2 | 3+4 4 | 3+4 4 |
| | 5 4 | 8 2 | 5 4 | 8 2 | 3+4 4 | 3+4 4 |
| | 5 4 | 9 2 | 5 4 | 8 2 | 3+4 4 | 3+4 4 |
| | 5 4 | 6 2 | 5 4 | 7 2 | 3+4 4 | 3+4 4 |
| | 5 4 | 7 2 | 5 4 | 6 2 | 3+4 4 | 3+4 4 |
| | 5 4 | 6 2 | 5 4 | 8 2 | 3+4 4 | 3+4 4 |
| | 5 4 | 6 2 | 5 4 | 6 2 | 3+4 4 | 3+4 4 |
| | | 6 2 | | 9 2 | | |
| | | 7 2 | | 9 2 | | |
| | | 8 2 | | 7 2 | | |
| | | 7 2 | | 8 2 | | |
| | | 7 2 | | 6 2 | | |
| | | 9 2 | | 7 2 | | |
| | | 8 2 | | 6 2 | | |
| | | 6 2 | | 9 2 | | |
| | | 8 2 | | 8 2 | | |
| | | 6 2 | | 6 2 | | |
| | | 7 2 | | 9 2 | | |
| | | 9 2 | | 7 2 | | |
| 129.2.18 | 85859 | | 48 | | Ge1Mn1Pr1 | |
| | -A | -GM | -M | -Z | -R | -X |
| | 5 4 | 7 2 | 5 4 | 9 2 | 3+4 4 | 3+4 4 |
| | 5 4 | 9 2 | 5 4 | 7 2 | 3+4 4 | 3+4 4 |
| | 5 4 | 7 2 | 5 4 | 9 2 | 3+4 4 | 3+4 4 |
| | 5 4 | 9 2 | 5 4 | 7 2 | 3+4 4 | 3+4 4 |
| | 5 4 | 7 2 | 5 4 | 9 2 | 3+4 4 | 3+4 4 |
| | 5 4 | 6 2 | 5 4 | 8 2 | 3+4 4 | 3+4 4 |
| | 5 4 | 9 2 | 5 4 | 7 2 | 3+4 4 | 3+4 4 |
| | 5 4 | 8 2 | 5 4 | 6 2 | 3+4 4 | 3+4 4 |
| | 5 4 | 7 2 | 5 4 | 7 2 | 3+4 4 | 3+4 4 |
| | 5 4 | 9 2 | 5 4 | 9 2 | 3+4 4 | 3+4 4 |
| | 5 4 | 7 2 | 5 4 | 7 2 | 3+4 4 | 3+4 4 |
| | 5 4 | 9 2 | 5 4 | 9 2 | 3+4 4 | 3+4 4 |
| | | 9 2 | | 9 2 | | |
| | | 8 2 | | 8 2 | | |
| | | 8 2 | | 6 2 | | |
| | | 6 2 | | 7 2 | | |
| | | 7 2 | | 6 2 | | |
| | | 6 2 | | 8 2 | | |
| | | 9 2 | | 6 2 | | |
| | | 7 2 | | 7 2 | | |
| | | 8 2 | | 8 2 | | |
| | | 8 2 | | 6 2 | | |
| | | 6 2 | | 7 2 | | |
| | | 7 2 | | 9 2 | | |
| 129.2.19 | 80012 | | 48 | | Ge1Mn1Pr1 | |
| | -A | -GM | -M | -Z | -R | -X |
| | 5 4 | 7 2 | 5 4 | 9 2 | 3+4 4 | 3+4 4 |
| | 5 4 | 9 2 | 5 4 | 7 2 | 3+4 4 | 3+4 4 |
| | 5 4 | 7 2 | 5 4 | 9 2 | 3+4 4 | 3+4 4 |
| | 5 4 | 9 2 | 5 4 | 7 2 | 3+4 4 | 3+4 4 |
| | 5 4 | 7 2 | 5 4 | 9 2 | 3+4 4 | 3+4 4 |
| | 5 4 | 6 2 | 5 4 | 8 2 | 3+4 4 | 3+4 4 |
| | 5 4 | 9 2 | 5 4 | 7 2 | 3+4 4 | 3+4 4 |
| | 5 4 | 8 2 | 5 4 | 6 2 | 3+4 4 | 3+4 4 |
| | 5 4 | 7 2 | 5 4 | 7 2 | 3+4 4 | 3+4 4 |
| | 5 4 | 9 2 | 5 4 | 9 2 | 3+4 4 | 3+4 4 |
| | 5 4 | 7 2 | 5 4 | 7 2 | 3+4 4 | 3+4 4 |

Continued on next column

**TABLE LX– continued**

| TQC | ICSD | | NoE | | SF | |
|---|---|---|---|---|---|---|
| | 5 4 | 9 2 | 5 4 | 9 2 | 3+4 4 | 3+4 4 |
| | | 9 2 | | 9 2 | | |
| | | 8 2 | | 8 2 | | |
| | | 8 2 | | 6 2 | | |
| | | 6 2 | | 7 2 | | |
| | | 7 2 | | 6 2 | | |
| | | 6 2 | | 8 2 | | |
| | | 9 2 | | 6 2 | | |
| | | 7 2 | | 7 2 | | |
| | | 8 2 | | 8 2 | | |
| | | 8 2 | | 6 2 | | |
| | | 6 2 | | 7 2 | | |
| | | 7 2 | | 9 2 | | |
| 129.2.20 | 80011 | | 48 | | Ge1Mn1Pr1 | |
| | -A | -GM | -M | -Z | -R | -X |
| | 5 4 | 7 2 | 5 4 | 9 2 | 3+4 4 | 3+4 4 |
| | 5 4 | 9 2 | 5 4 | 7 2 | 3+4 4 | 3+4 4 |
| | 5 4 | 7 2 | 5 4 | 9 2 | 3+4 4 | 3+4 4 |
| | 5 4 | 9 2 | 5 4 | 7 2 | 3+4 4 | 3+4 4 |
| | 5 4 | 7 2 | 5 4 | 9 2 | 3+4 4 | 3+4 4 |
| | 5 4 | 6 2 | 5 4 | 8 2 | 3+4 4 | 3+4 4 |
| | 5 4 | 9 2 | 5 4 | 7 2 | 3+4 4 | 3+4 4 |
| | 5 4 | 8 2 | 5 4 | 6 2 | 3+4 4 | 3+4 4 |
| | 5 4 | 7 2 | 5 4 | 7 2 | 3+4 4 | 3+4 4 |
| | 5 4 | 9 2 | 5 4 | 9 2 | 3+4 4 | 3+4 4 |
| | 5 4 | 7 2 | 5 4 | 7 2 | 3+4 4 | 3+4 4 |
| | 5 4 | 9 2 | 5 4 | 9 2 | 3+4 4 | 3+4 4 |
| | | 9 2 | | 9 2 | | |
| | | 8 2 | | 8 2 | | |
| | | 8 2 | | 6 2 | | |
| | | 6 2 | | 7 2 | | |
| | | 7 2 | | 6 2 | | |
| | | 6 2 | | 8 2 | | |
| | | 9 2 | | 6 2 | | |
| | | 7 2 | | 7 2 | | |
| | | 8 2 | | 8 2 | | |
| | | 8 2 | | 6 2 | | |
| | | 6 2 | | 7 2 | | |
| | | 7 2 | | 9 2 | | |
| 129.2.21 | 183310 | | 76 | | Gd1Sn2Zn1 | |
| | -A | -GM | -M | -Z | -R | -X |
| | 5 4 | 7 2 | 5 4 | 7 2 | 3+4 4 | 3+4 4 |
| | 5 4 | 9 2 | 5 4 | 9 2 | 3+4 4 | 3+4 4 |
| | 5 4 | 7 2 | 5 4 | 7 2 | 3+4 4 | 3+4 4 |
| | 5 4 | 9 2 | 5 4 | 9 2 | 3+4 4 | 3+4 4 |
| | 5 4 | 7 2 | 5 4 | 7 2 | 3+4 4 | 3+4 4 |
| | 5 4 | 9 2 | 5 4 | 9 2 | 3+4 4 | 3+4 4 |
| | 5 4 | 6 2 | 5 4 | 6 2 | 3+4 4 | 3+4 4 |
| | 5 4 | 8 2 | 5 4 | 8 2 | 3+4 4 | 3+4 4 |
| | 5 4 | 7 2 | 5 4 | 7 2 | 3+4 4 | 3+4 4 |
| | 5 4 | 7 2 | 5 4 | 9 2 | 3+4 4 | 3+4 4 |
| | 5 4 | 9 2 | 5 4 | 7 2 | 3+4 4 | 3+4 4 |
| | 5 4 | 6 2 | 5 4 | 8 2 | 3+4 4 | 3+4 4 |
| | 5 4 | 8 2 | 5 4 | 6 2 | 3+4 4 | 3+4 4 |
| | 5 4 | 7 2 | 5 4 | 9 2 | 3+4 4 | 3+4 4 |
| | 5 4 | 9 2 | 5 4 | 7 2 | 3+4 4 | 3+4 4 |
| | 5 4 | 9 2 | 5 4 | 9 2 | 3+4 4 | 3+4 4 |
| | 5 4 | 7 2 | 5 4 | 8 2 | 3+4 4 | 3+4 4 |
| | 5 4 | 6 2 | 5 4 | 6 2 | 3+4 4 | 3+4 4 |
| | 5 4 | 8 2 | 5 4 | 8 2 | 3+4 4 | 3+4 4 |
| | | 6 2 | | 7 2 | | |

Continued on next column

**TABLE LX– continued**

| TQC | ICSD | | NoE | | SF | |
|---|---|---|---|---|---|---|
| | | 9 2 | | 7 2 | | |
| | | 7 2 | | 9 2 | | |
| | | 8 2 | | 8 2 | | |
| | | 9 2 | | 6 2 | | |
| | | 6 2 | | 7 2 | | |
| | | 7 2 | | 9 2 | | |
| | | 7 2 | | 8 2 | | |
| | | 6 2 | | 6 2 | | |
| | | 8 2 | | 9 2 | | |
| | | 8 2 | | 6 2 | | |
| | | 6 2 | | 7 2 | | |
| | | 9 2 | | 6 2 | | |
| | | 7 2 | | 8 2 | | |
| | | 6 2 | | 9 2 | | |
| | | 6 2 | | 8 2 | | |
| | | 8 2 | | 7 2 | | |
| | | 9 2 | | 7 2 | | |
| | | 7 2 | | 6 2 | | |
| 129.2.22 | 41255 | | 60 | | Gd1Ru1Si1 | |
| | -A | -GM | -M | -Z | -R | -X |
| | 5 4 | 7 2 | 5 4 | 9 2 | 3+4 4 | 3+4 4 |
| | 5 4 | 9 2 | 5 4 | 7 2 | 3+4 4 | 3+4 4 |
| | 5 4 | 7 2 | 5 4 | 9 2 | 3+4 4 | 3+4 4 |
| | 5 4 | 9 2 | 5 4 | 7 2 | 3+4 4 | 3+4 4 |
| | 5 4 | 7 2 | 5 4 | 9 2 | 3+4 4 | 3+4 4 |
| | 5 4 | 6 2 | 5 4 | 8 2 | 3+4 4 | 3+4 4 |
| | 5 4 | 9 2 | 5 4 | 7 2 | 3+4 4 | 3+4 4 |
| | 5 4 | 8 2 | 5 4 | 6 2 | 3+4 4 | 3+4 4 |
| | 5 4 | 7 2 | 5 4 | 7 2 | 3+4 4 | 3+4 4 |
| | 5 4 | 9 2 | 5 4 | 9 2 | 3+4 4 | 3+4 4 |
| | 5 4 | 7 2 | 5 4 | 9 2 | 3+4 4 | 3+4 4 |
| | 5 4 | 6 2 | 5 4 | 7 2 | 3+4 4 | 3+4 4 |
| | 5 4 | 9 2 | 5 4 | 6 2 | 3+4 4 | 3+4 4 |
| | 5 4 | 8 2 | 5 4 | 8 2 | 3+4 4 | 3+4 4 |
| | 5 4 | 9 2 | 5 4 | 9 2 | 3+4 4 | 3+4 4 |
| | | 8 2 | | 6 2 | | |
| | | 7 2 | | 7 2 | | |
| | | 6 2 | | 8 2 | | |
| | | 7 2 | | 7 2 | | |
| | | 6 2 | | 6 2 | | |
| | | 8 2 | | 8 2 | | |
| | | 9 2 | | 9 2 | | |
| | | 8 2 | | 8 2 | | |
| | | 7 2 | | 6 2 | | |
| | | 6 2 | | 7 2 | | |
| | | 6 2 | | 6 2 | | |
| | | 6 2 | | 6 2 | | |
| | | 9 2 | | 6 2 | | |
| | | 7 2 | | 7 2 | | |
| | | 8 2 | | 9 2 | | |
| 129.2.23 | 80200 | | 64 | | Fe1O1P1Pr1 | |
| | -A | -GM | -M | -Z | -R | -X |
| | 5 4 | 7 2 | 5 4 | 7 2 | 3+4 4 | 3+4 4 |
| | 5 4 | 9 2 | 5 4 | 9 2 | 3+4 4 | 3+4 4 |
| | 5 4 | 7 2 | 5 4 | 7 2 | 3+4 4 | 3+4 4 |
| | 5 4 | 9 2 | 5 4 | 9 2 | 3+4 4 | 3+4 4 |
| | 5 4 | 7 2 | 5 4 | 7 2 | 3+4 4 | 3+4 4 |
| | 5 4 | 8 2 | 5 4 | 8 2 | 3+4 4 | 3+4 4 |
| | 5 4 | 6 2 | 5 4 | 6 2 | 3+4 4 | 3+4 4 |
| | 5 4 | 9 2 | 5 4 | 9 2 | 3+4 4 | 3+4 4 |
| | 5 4 | 8 2 | 5 4 | 8 2 | 3+4 4 | 3+4 4 |

Continued on next column

**TABLE LX– continued**

| TQC | ICSD | | NoE | | SF | |
|---|---|---|---|---|---|---|
| | 5 4 | 7 2 | 5 4 | 7 2 | 3+4 4 | 3+4 4 |
| | 5 4 | 7 2 | 5 4 | 9 2 | 3+4 4 | 3+4 4 |
| | 5 4 | 9 2 | 5 4 | 7 2 | 3+4 4 | 3+4 4 |
| | 5 4 | 9 2 | 5 4 | 9 2 | 3+4 4 | 3+4 4 |
| | 5 4 | 7 2 | 5 4 | 7 2 | 3+4 4 | 3+4 4 |
| | 5 4 | 6 2 | 5 4 | 6 2 | 3+4 4 | 3+4 4 |
| | 5 4 | 6 2 | 5 4 | 6 2 | 3+4 4 | 3+4 4 |
| | | 7 2 | | 7 2 | | |
| | | 9 2 | | 9 2 | | |
| | | 8 2 | | 8 2 | | |
| | | 9 2 | | 9 2 | | |
| | | 9 2 | | 7 2 | | |
| | | 8 2 | | 6 2 | | |
| | | 7 2 | | 9 2 | | |
| | | 6 2 | | 8 2 | | |
| | | 6 2 | | 8 2 | | |
| | | 8 2 | | 6 2 | | |
| | | 7 2 | | 9 2 | | |
| | | 6 2 | | 8 2 | | |
| | | 6 2 | | 8 2 | | |
| | | 7 2 | | 7 2 | | |
| | | 6 2 | | 6 2 | | |
| | | 9 2 | | 9 2 | | |
| 129.2.24 | 85854 | | 52 | | Fe1Nd1Si1 | |
| | -A | -GM | -M | -Z | -R | -X |
| | 5 4 | 7 2 | 5 4 | 9 2 | 3+4 4 | 3+4 4 |
| | 5 4 | 9 2 | 5 4 | 7 2 | 3+4 4 | 3+4 4 |
| | 5 4 | 7 2 | 5 4 | 9 2 | 3+4 4 | 3+4 4 |
| | 5 4 | 9 2 | 5 4 | 7 2 | 3+4 4 | 3+4 4 |
| | 5 4 | 7 2 | 5 4 | 9 2 | 3+4 4 | 3+4 4 |
| | 5 4 | 6 2 | 5 4 | 8 2 | 3+4 4 | 3+4 4 |
| | 5 4 | 9 2 | 5 4 | 7 2 | 3+4 4 | 3+4 4 |
| | 5 4 | 8 2 | 5 4 | 6 2 | 3+4 4 | 3+4 4 |
| | 5 4 | 7 2 | 5 4 | 7 2 | 3+4 4 | 3+4 4 |
| | 5 4 | 9 2 | 5 4 | 9 2 | 3+4 4 | 3+4 4 |
| | 5 4 | 7 2 | 5 4 | 7 2 | 3+4 4 | 3+4 4 |
| | 5 4 | 9 2 | 5 4 | 9 2 | 3+4 4 | 3+4 4 |
| | 5 4 | 9 2 | 5 4 | 9 2 | 3+4 4 | 3+4 4 |
| | | 8 2 | | 8 2 | | |
| | | 6 2 | | 6 2 | | |
| | | 8 2 | | 7 2 | | |
| | | 7 2 | | 6 2 | | |
| | | 6 2 | | 8 2 | | |
| | | 7 2 | | 6 2 | | |
| | | 7 2 | | 7 2 | | |
| | | 6 2 | | 7 2 | | |
| | | 9 2 | | 6 2 | | |
| | | 8 2 | | 8 2 | | |
| | | 8 2 | | 6 2 | | |
| | | 6 2 | | 9 2 | | |
| | | 6 2 | | 6 2 | | |
| 129.2.25 | 57235 | | 52 | | Fe1Nd1Si1 | |
| | -A | -GM | -M | -Z | -R | -X |
| | 5 4 | 7 2 | 5 4 | 9 2 | 3+4 4 | 3+4 4 |
| | 5 4 | 9 2 | 5 4 | 7 2 | 3+4 4 | 3+4 4 |
| | 5 4 | 7 2 | 5 4 | 9 2 | 3+4 4 | 3+4 4 |
| | 5 4 | 9 2 | 5 4 | 7 2 | 3+4 4 | 3+4 4 |
| | 5 4 | 7 2 | 5 4 | 9 2 | 3+4 4 | 3+4 4 |
| | 5 4 | 6 2 | 5 4 | 8 2 | 3+4 4 | 3+4 4 |
| | 5 4 | 9 2 | 5 4 | 7 2 | 3+4 4 | 3+4 4 |
| | 5 4 | 8 2 | 5 4 | 6 2 | 3+4 4 | 3+4 4 |

Continued on next column

**TABLE LX– continued**

| TQC | ICSD | | NoE | | SF | |
|---|---|---|---|---|---|---|
| | 5 4 | 7 2 | 5 4 | 7 2 | 3+4 4 | 3+4 4 |
| | 5 4 | 9 2 | 5 4 | 9 2 | 3+4 4 | 3+4 4 |
| | 5 4 | 7 2 | 5 4 | 7 2 | 3+4 4 | 3+4 4 |
| | 5 4 | 9 2 | 5 4 | 9 2 | 3+4 4 | 3+4 4 |
| | 5 4 | 9 2 | 5 4 | 9 2 | 3+4 4 | 3+4 4 |
| | | 8 2 | | 8 2 | | |
| | | 6 2 | | 6 2 | | |
| | | 8 2 | | 7 2 | | |
| | | 7 2 | | 6 2 | | |
| | | 6 2 | | 8 2 | | |
| | | 7 2 | | 6 2 | | |
| | | 7 2 | | 7 2 | | |
| | | 6 2 | | 7 2 | | |
| | | 9 2 | | 6 2 | | |
| | | 8 2 | | 8 2 | | |
| | | 8 2 | | 6 2 | | |
| | | 6 2 | | 9 2 | | |
| | | 6 2 | | 6 2 | | |
| 129.2.26 | 57234 | | 52 | | Fe1Nd1Si1 | |
| | -A | -GM | -M | -Z | -R | -X |
| | 5 4 | 7 2 | 5 4 | 9 2 | 3+4 4 | 3+4 4 |
| | 5 4 | 9 2 | 5 4 | 7 2 | 3+4 4 | 3+4 4 |
| | 5 4 | 7 2 | 5 4 | 9 2 | 3+4 4 | 3+4 4 |
| | 5 4 | 9 2 | 5 4 | 7 2 | 3+4 4 | 3+4 4 |
| | 5 4 | 7 2 | 5 4 | 9 2 | 3+4 4 | 3+4 4 |
| | 5 4 | 6 2 | 5 4 | 8 2 | 3+4 4 | 3+4 4 |
| | 5 4 | 9 2 | 5 4 | 7 2 | 3+4 4 | 3+4 4 |
| | 5 4 | 8 2 | 5 4 | 6 2 | 3+4 4 | 3+4 4 |
| | 5 4 | 7 2 | 5 4 | 7 2 | 3+4 4 | 3+4 4 |
| | 5 4 | 9 2 | 5 4 | 9 2 | 3+4 4 | 3+4 4 |
| | 5 4 | 7 2 | 5 4 | 7 2 | 3+4 4 | 3+4 4 |
| | 5 4 | 9 2 | 5 4 | 9 2 | 3+4 4 | 3+4 4 |
| | 5 4 | 9 2 | 5 4 | 9 2 | 3+4 4 | 3+4 4 |
| | | 8 2 | | 8 2 | | |
| | | 6 2 | | 6 2 | | |
| | | 8 2 | | 7 2 | | |
| | | 7 2 | | 6 2 | | |
| | | 6 2 | | 8 2 | | |
| | | 7 2 | | 6 2 | | |
| | | 7 2 | | 7 2 | | |
| | | 6 2 | | 7 2 | | |
| | | 9 2 | | 6 2 | | |
| | | 8 2 | | 8 2 | | |
| | | 8 2 | | 6 2 | | |
| | | 6 2 | | 9 2 | | |
| | | 6 2 | | 6 2 | | |
| 129.2.27 | 95081 | | 52 | | Eu1P1Sn1 | |
| | -A | -GM | -M | -Z | -R | -X |
| | 5 4 | 7 2 | 5 4 | 7 2 | 3+4 4 | 3+4 4 |
| | 5 4 | 9 2 | 5 4 | 9 2 | 3+4 4 | 3+4 4 |
| | 5 4 | 7 2 | 5 4 | 7 2 | 3+4 4 | 3+4 4 |
| | 5 4 | 9 2 | 5 4 | 9 2 | 3+4 4 | 3+4 4 |
| | 5 4 | 7 2 | 5 4 | 7 2 | 3+4 4 | 3+4 4 |
| | 5 4 | 9 2 | 5 4 | 9 2 | 3+4 4 | 3+4 4 |
| | 5 4 | 6 2 | 5 4 | 6 2 | 3+4 4 | 3+4 4 |
| | 5 4 | 8 2 | 5 4 | 8 2 | 3+4 4 | 3+4 4 |
| | 5 4 | 7 2 | 5 4 | 9 2 | 3+4 4 | 3+4 4 |
| | 5 4 | 9 2 | 5 4 | 7 2 | 3+4 4 | 3+4 4 |
| | 5 4 | 7 2 | 5 4 | 9 2 | 3+4 4 | 3+4 4 |
| | 5 4 | 9 2 | 5 4 | 7 2 | 3+4 4 | 3+4 4 |
| | 5 4 | 7 2 | 5 4 | 9 2 | 3+4 4 | 3+4 4 |

Continued on next column

**TABLE LX– continued**

| TQC | ICSD | NoE | SF | | | |
|---|---|---|---|---|---|---|
| | | 7 2 | | 9 2 | | |
| | | 6 2 | | 8 2 | | |
| | | 9 2 | | 7 2 | | |
| | | 9 2 | | 6 2 | | |
| | | 8 2 | | 7 2 | | |
| | | 7 2 | | 9 2 | | |
| | | 6 2 | | 8 2 | | |
| | | 8 2 | | 6 2 | | |
| | | 6 2 | | 8 2 | | |
| | | 7 2 | | 7 2 | | |
| | | 9 2 | | 6 2 | | |
| | | 8 2 | | 9 2 | | |
| | | 8 2 | | 8 2 | | |
| 129.2.28 | 88118 | 60 | Er1Si1Ti1 | | | |
| | -A | -GM | -M | -Z | -R | -X |
| | 5 4 | 7 2 | 5 4 | 9 2 | 3+4 4 | 3+4 4 |
| | 5 4 | 9 2 | 5 4 | 7 2 | 3+4 4 | 3+4 4 |
| | 5 4 | 7 2 | 5 4 | 9 2 | 3+4 4 | 3+4 4 |
| | 5 4 | 9 2 | 5 4 | 7 2 | 3+4 4 | 3+4 4 |
| | 5 4 | 7 2 | 5 4 | 9 2 | 3+4 4 | 3+4 4 |
| | 5 4 | 6 2 | 5 4 | 8 2 | 3+4 4 | 3+4 4 |
| | 5 4 | 9 2 | 5 4 | 7 2 | 3+4 4 | 3+4 4 |
| | 5 4 | 8 2 | 5 4 | 6 2 | 3+4 4 | 3+4 4 |
| | 5 4 | 7 2 | 5 4 | 9 2 | 3+4 4 | 3+4 4 |
| | 5 4 | 9 2 | 5 4 | 7 2 | 3+4 4 | 3+4 4 |
| | 5 4 | 7 2 | 5 4 | 7 2 | 3+4 4 | 3+4 4 |
| | 5 4 | 9 2 | 5 4 | 9 2 | 3+4 4 | 3+4 4 |
| | 5 4 | 8 2 | 5 4 | 8 2 | 3+4 4 | 3+4 4 |
| | 5 4 | 9 2 | 5 4 | 9 2 | 3+4 4 | 3+4 4 |
| | 5 4 | 6 2 | 5 4 | 6 2 | 3+4 4 | 3+4 4 |
| | | 9 2 | | 7 2 | | |
| | | 7 2 | | 8 2 | | |
| | | 8 2 | | 6 2 | | |
| | | 6 2 | | 9 2 | | |
| | | 8 2 | | 8 2 | | |
| | | 8 2 | | 6 2 | | |
| | | 9 2 | | 7 2 | | |
| | | 6 2 | | 8 2 | | |
| | | 7 2 | | 9 2 | | |
| | | 7 2 | | 6 2 | | |
| | | 6 2 | | 7 2 | | |
| | | 8 2 | | 9 2 | | |
| | | 8 2 | | 6 2 | | |
| | | 9 2 | | 8 2 | | |
| | | 6 2 | | 6 2 | | |
| 129.2.29 | 88116 | 56 | Dy1Si1Ti1 | | | |
| | -A | -GM | -M | -Z | -R | -X |
| | 5 4 | 7 2 | 5 4 | 9 2 | 3+4 4 | 3+4 4 |
| | 5 4 | 9 2 | 5 4 | 7 2 | 3+4 4 | 3+4 4 |
| | 5 4 | 7 2 | 5 4 | 9 2 | 3+4 4 | 3+4 4 |
| | 5 4 | 9 2 | 5 4 | 7 2 | 3+4 4 | 3+4 4 |
| | 5 4 | 7 2 | 5 4 | 9 2 | 3+4 4 | 3+4 4 |
| | 5 4 | 6 2 | 5 4 | 8 2 | 3+4 4 | 3+4 4 |
| | 5 4 | 9 2 | 5 4 | 7 2 | 3+4 4 | 3+4 4 |
| | 5 4 | 8 2 | 5 4 | 6 2 | 3+4 4 | 3+4 4 |
| | 5 4 | 7 2 | 5 4 | 7 2 | 3+4 4 | 3+4 4 |
| | 5 4 | 9 2 | 5 4 | 9 2 | 3+4 4 | 3+4 4 |
| | 5 4 | 7 2 | 5 4 | 7 2 | 3+4 4 | 3+4 4 |
| | 5 4 | 9 2 | 5 4 | 9 2 | 3+4 4 | 3+4 4 |
| | 5 4 | 8 2 | 5 4 | 9 2 | 3+4 4 | 3+4 4 |
| | 5 4 | 9 2 | 5 4 | 8 2 | 3+4 4 | 3+4 4 |

Continued on next column

**TABLE LX– continued**

| TQC | ICSD | NoE | SF | | | |
|---|---|---|---|---|---|---|
| | | 9 2 | | 6 2 | | |
| | | 6 2 | | 7 2 | | |
| | | 7 2 | | 8 2 | | |
| | | 6 2 | | 6 2 | | |
| | | 8 2 | | 9 2 | | |
| | | 8 2 | | 8 2 | | |
| | | 8 2 | | 6 2 | | |
| | | 9 2 | | 7 2 | | |
| | | 6 2 | | 6 2 | | |
| | | 7 2 | | 8 2 | | |
| | | 7 2 | | 9 2 | | |
| | | 6 2 | | 7 2 | | |
| | | 8 2 | | 6 2 | | |
| | | 9 2 | | 9 2 | | |
| 129.2.30 | 57233 | 64 | Dy1Fe1Si1 | | | |
| | -A | -GM | -M | -Z | -R | -X |
| | 5 4 | 7 2 | 5 4 | 9 2 | 3+4 4 | 3+4 4 |
| | 5 4 | 9 2 | 5 4 | 7 2 | 3+4 4 | 3+4 4 |
| | 5 4 | 7 2 | 5 4 | 9 2 | 3+4 4 | 3+4 4 |
| | 5 4 | 9 2 | 5 4 | 7 2 | 3+4 4 | 3+4 4 |
| | 5 4 | 7 2 | 5 4 | 9 2 | 3+4 4 | 3+4 4 |
| | 5 4 | 6 2 | 5 4 | 8 2 | 3+4 4 | 3+4 4 |
| | 5 4 | 9 2 | 5 4 | 7 2 | 3+4 4 | 3+4 4 |
| | 5 4 | 8 2 | 5 4 | 6 2 | 3+4 4 | 3+4 4 |
| | 5 4 | 7 2 | 5 4 | 7 2 | 3+4 4 | 3+4 4 |
| | 5 4 | 9 2 | 5 4 | 9 2 | 3+4 4 | 3+4 4 |
| | 5 4 | 7 2 | 5 4 | 7 2 | 3+4 4 | 3+4 4 |
| | 5 4 | 9 2 | 5 4 | 9 2 | 3+4 4 | 3+4 4 |
| | 5 4 | 9 2 | 5 4 | 9 2 | 3+4 4 | 3+4 4 |
| | 5 4 | 8 2 | 5 4 | 8 2 | 3+4 4 | 3+4 4 |
| | 5 4 | 6 2 | 5 4 | 6 2 | 3+4 4 | 3+4 4 |
| | 5 4 | 8 2 | 5 4 | 8 2 | 3+4 4 | 3+4 4 |
| | | 6 2 | | 6 2 | | |
| | | 7 2 | | 7 2 | | |
| | | 7 2 | | 7 2 | | |
| | | 9 2 | | 8 2 | | |
| | | 8 2 | | 6 2 | | |
| | | 6 2 | | 9 2 | | |
| | | 8 2 | | 8 2 | | |
| | | 7 2 | | 6 2 | | |
| | | 6 2 | | 7 2 | | |
| | | 9 2 | | 6 2 | | |
| | | 6 2 | | 6 2 | | |
| | | 8 2 | | 9 2 | | |
| | | 8 2 | | 9 2 | | |
| | | 9 2 | | 7 2 | | |
| | | 6 2 | | 8 2 | | |
| | | 7 2 | | 6 2 | | |
| 129.2.31 | 57218 | 64 | Dy1Fe1Si1 | | | |
| | -A | -GM | -M | -Z | -R | -X |
| | 5 4 | 7 2 | 5 4 | 9 2 | 3+4 4 | 3+4 4 |
| | 5 4 | 9 2 | 5 4 | 7 2 | 3+4 4 | 3+4 4 |
| | 5 4 | 7 2 | 5 4 | 9 2 | 3+4 4 | 3+4 4 |
| | 5 4 | 9 2 | 5 4 | 7 2 | 3+4 4 | 3+4 4 |
| | 5 4 | 7 2 | 5 4 | 9 2 | 3+4 4 | 3+4 4 |
| | 5 4 | 6 2 | 5 4 | 8 2 | 3+4 4 | 3+4 4 |
| | 5 4 | 9 2 | 5 4 | 7 2 | 3+4 4 | 3+4 4 |
| | 5 4 | 8 2 | 5 4 | 6 2 | 3+4 4 | 3+4 4 |
| | 5 4 | 7 2 | 5 4 | 7 2 | 3+4 4 | 3+4 4 |
| | 5 4 | 9 2 | 5 4 | 9 2 | 3+4 4 | 3+4 4 |
| | 5 4 | 7 2 | 5 4 | 7 2 | 3+4 4 | 3+4 4 |

Continued on next column

**TABLE LX– continued**

| TQC | ICSD | | NoE | | SF | |
|---|---|---|---|---|---|---|
| | 5 4 | 9 2 | 5 4 | 9 2 | 3+4 4 | 3+4 4 |
| | 5 4 | 9 2 | 5 4 | 9 2 | 3+4 4 | 3+4 4 |
| | 5 4 | 8 2 | 5 4 | 8 2 | 3+4 4 | 3+4 4 |
| | 5 4 | 6 2 | 5 4 | 6 2 | 3+4 4 | 3+4 4 |
| | 5 4 | 8 2 | 5 4 | 8 2 | 3+4 4 | 3+4 4 |
| | | 6 2 | | 6 2 | | |
| | | 7 2 | | 7 2 | | |
| | | 7 2 | | 7 2 | | |
| | | 9 2 | | 8 2 | | |
| | | 8 2 | | 6 2 | | |
| | | 6 2 | | 9 2 | | |
| | | 8 2 | | 8 2 | | |
| | | 6 2 | | 6 2 | | |
| | | 7 2 | | 7 2 | | |
| | | 9 2 | | 6 2 | | |
| | | 6 2 | | 6 2 | | |
| | | 8 2 | | 9 2 | | |
| | | 6 2 | | 9 2 | | |
| | | 8 2 | | 7 2 | | |
| | | 7 2 | | 8 2 | | |
| | | 9 2 | | 6 2 | | |
| 129.2.32 | 428492 | | 140 | | Cu4La1P3 | |
| | -A | -GM | -M | -Z | -R | -X |
| | 5 4 | 7 2 | 5 4 | 7 2 | 3+4 4 | 3+4 4 |
| | 5 4 | 9 2 | 5 4 | 9 2 | 3+4 4 | 3+4 4 |
| | 5 4 | 7 2 | 5 4 | 7 2 | 3+4 4 | 3+4 4 |
| | 5 4 | 9 2 | 5 4 | 9 2 | 3+4 4 | 3+4 4 |
| | 5 4 | 6 2 | 5 4 | 6 2 | 3+4 4 | 3+4 4 |
| | 5 4 | 9 2 | 5 4 | 9 2 | 3+4 4 | 3+4 4 |
| | 5 4 | 7 2 | 5 4 | 7 2 | 3+4 4 | 3+4 4 |
| | 5 4 | 8 2 | 5 4 | 8 2 | 3+4 4 | 3+4 4 |
| | 5 4 | 7 2 | 5 4 | 7 2 | 3+4 4 | 3+4 4 |
| | 5 4 | 8 2 | 5 4 | 8 2 | 3+4 4 | 3+4 4 |
| | 5 4 | 7 2 | 5 4 | 9 2 | 3+4 4 | 3+4 4 |
| | 5 4 | 9 2 | 5 4 | 7 2 | 3+4 4 | 3+4 4 |
| | 5 4 | 9 2 | 5 4 | 9 2 | 3+4 4 | 3+4 4 |
| | 5 4 | 6 2 | 5 4 | 6 2 | 3+4 4 | 3+4 4 |
| | 5 4 | 7 2 | 5 4 | 7 2 | 3+4 4 | 3+4 4 |
| | 5 4 | 6 2 | 5 4 | 8 2 | 3+4 4 | 3+4 4 |
| | 5 4 | 7 2 | 5 4 | 9 2 | 3+4 4 | 3+4 4 |
| | 5 4 | 9 2 | 5 4 | 7 2 | 3+4 4 | 3+4 4 |
| | 5 4 | 9 2 | 5 4 | 7 2 | 3+4 4 | 3+4 4 |
| | 5 4 | 8 2 | 5 4 | 8 2 | 3+4 4 | 3+4 4 |
| | 5 4 | 8 2 | 5 4 | 6 2 | 3+4 4 | 3+4 4 |
| | 5 4 | 8 2 | 5 4 | 6 2 | 3+4 4 | 3+4 4 |
| | 5 4 | 9 2 | 5 4 | 7 2 | 3+4 4 | 3+4 4 |
| | 5 4 | 6 2 | 5 4 | 9 2 | 3+4 4 | 3+4 4 |
| | 5 4 | 7 2 | 5 4 | 8 2 | 3+4 4 | 3+4 4 |
| | 5 4 | 9 2 | 5 4 | 9 2 | 3+4 4 | 3+4 4 |
| | 5 4 | 8 2 | 5 4 | 6 2 | 3+4 4 | 3+4 4 |
| | 5 4 | 6 2 | 5 4 | 7 2 | 3+4 4 | 3+4 4 |
| | 5 4 | 7 2 | 5 4 | 9 2 | 3+4 4 | 3+4 4 |
| | 5 4 | 6 2 | 5 4 | 6 2 | 3+4 4 | 3+4 4 |
| | 5 4 | 7 2 | 5 4 | 8 2 | 3+4 4 | 3+4 4 |
| | 5 4 | 8 2 | 5 4 | 9 2 | 3+4 4 | 3+4 4 |
| | 5 4 | 6 2 | 5 4 | 8 2 | 3+4 4 | 3+4 4 |
| | 5 4 | 7 2 | 5 4 | 6 2 | 3+4 4 | 3+4 4 |
| | 5 4 | 9 2 | 5 4 | 8 2 | 3+4 4 | 3+4 4 |
| | | 8 2 | | 9 2 | | |
| | | 6 2 | | 8 2 | | |
| | | 7 2 | | 9 2 | | |

Continued on next column

**TABLE LX– continued**

| TQC | ICSD | | NoE | | SF | |
|---|---|---|---|---|---|---|
| | | 9 2 | | 9 2 | | |
| | | 8 2 | | 6 2 | | |
| | | 9 2 | | 7 2 | | |
| | | 9 2 | | 6 2 | | |
| | | 8 2 | | 7 2 | | |
| | | 7 2 | | 7 2 | | |
| | | 8 2 | | 9 2 | | |
| | | 9 2 | | 6 2 | | |
| | | 7 2 | | 7 2 | | |
| | | 6 2 | | 8 2 | | |
| | | 6 2 | | 6 2 | | |
| | | 7 2 | | 8 2 | | |
| | | 6 2 | | 9 2 | | |
| | | 8 2 | | 7 2 | | |
| | | 6 2 | | 8 2 | | |
| | | 8 2 | | 6 2 | | |
| | | 7 2 | | 6 2 | | |
| | | 9 2 | | 7 2 | | |
| | | 8 2 | | 8 2 | | |
| | | 7 2 | | 9 2 | | |
| | | 9 2 | | 7 2 | | |
| | | 7 2 | | 9 2 | | |
| | | 6 2 | | 8 2 | | |
| | | 9 2 | | 7 2 | | |
| | | 8 2 | | 8 2 | | |
| | | 6 2 | | 6 2 | | |
| | | 7 2 | | 9 2 | | |
| | | 6 2 | | 9 2 | | |
| | | 6 2 | | 8 2 | | |
| | | 7 2 | | 6 2 | | |
| | | 9 2 | | 6 2 | | |
| | | 8 2 | | 7 2 | | |
| 129.2.33 | 153572 | | 64 | | Cu1Sb2Y1 | |
| | -A | -GM | -M | -Z | -R | -X |
| | 5 4 | 7 2 | 5 4 | 7 2 | 3+4 4 | 3+4 4 |
| | 5 4 | 9 2 | 5 4 | 9 2 | 3+4 4 | 3+4 4 |
| | 5 4 | 7 2 | 5 4 | 7 2 | 3+4 4 | 3+4 4 |
| | 5 4 | 9 2 | 5 4 | 9 2 | 3+4 4 | 3+4 4 |
| | 5 4 | 7 2 | 5 4 | 7 2 | 3+4 4 | 3+4 4 |
| | 5 4 | 9 2 | 5 4 | 9 2 | 3+4 4 | 3+4 4 |
| | 5 4 | 6 2 | 5 4 | 6 2 | 3+4 4 | 3+4 4 |
| | 5 4 | 8 2 | 5 4 | 8 2 | 3+4 4 | 3+4 4 |
| | 5 4 | 7 2 | 5 4 | 7 2 | 3+4 4 | 3+4 4 |
| | 5 4 | 7 2 | 5 4 | 9 2 | 3+4 4 | 3+4 4 |
| | 5 4 | 9 2 | 5 4 | 7 2 | 3+4 4 | 3+4 4 |
| | 5 4 | 8 2 | 5 4 | 8 2 | 3+4 4 | 3+4 4 |
| | 5 4 | 7 2 | 5 4 | 9 2 | 3+4 4 | 3+4 4 |
| | 5 4 | 9 2 | 5 4 | 7 2 | 3+4 4 | 3+4 4 |
| | 5 4 | 9 2 | 5 4 | 6 2 | 3+4 4 | 3+4 4 |
| | 5 4 | 8 2 | 5 4 | 7 2 | 3+4 4 | 3+4 4 |
| | | 6 2 | | 6 2 | | |
| | | 6 2 | | 8 2 | | |
| | | 7 2 | | 9 2 | | |
| | | 9 2 | | 9 2 | | |
| | | 7 2 | | 7 2 | | |
| | | 8 2 | | 6 2 | | |
| | | 6 2 | | 8 2 | | |
| | | 7 2 | | 8 2 | | |
| | | 6 2 | | 7 2 | | |
| | | 9 2 | | 9 2 | | |
| | | 7 2 | | 9 2 | | |

Continued on next column

**TABLE LX– continued**

| TQC | ICSD NoE SF | | | | | |
|---|---|---|---|---|---|---|
| | | 6 2 | | 8 2 | | |
| | | 6 2 | | 6 2 | | |
| | | 9 2 | | 7 2 | | |
| | | 8 2 | | 8 2 | | |
| | | 9 2 | | 7 2 | | |
| 129.2.34 | 79579 72 Cu1O1P1U1 | | | | | |
| | -A | -GM | -M | -Z | -R | -X |
| | 5 4 | 7 2 | 5 4 | 9 2 | 3+4 4 | 3+4 4 |
| | 5 4 | 9 2 | 5 4 | 7 2 | 3+4 4 | 3+4 4 |
| | 5 4 | 7 2 | 5 4 | 9 2 | 3+4 4 | 3+4 4 |
| | 5 4 | 9 2 | 5 4 | 7 2 | 3+4 4 | 3+4 4 |
| | 5 4 | 7 2 | 5 4 | 9 2 | 3+4 4 | 3+4 4 |
| | 5 4 | 8 2 | 5 4 | 6 2 | 3+4 4 | 3+4 4 |
| | 5 4 | 6 2 | 5 4 | 8 2 | 3+4 4 | 3+4 4 |
| | 5 4 | 9 2 | 5 4 | 7 2 | 3+4 4 | 3+4 4 |
| | 5 4 | 8 2 | 5 4 | 6 2 | 3+4 4 | 3+4 4 |
| | 5 4 | 7 2 | 5 4 | 9 2 | 3+4 4 | 3+4 4 |
| | 5 4 | 7 2 | 5 4 | 7 2 | 3+4 4 | 3+4 4 |
| | 5 4 | 9 2 | 5 4 | 9 2 | 3+4 4 | 3+4 4 |
| | 5 4 | 9 2 | 5 4 | 7 2 | 3+4 4 | 3+4 4 |
| | 5 4 | 7 2 | 5 4 | 9 2 | 3+4 4 | 3+4 4 |
| | 5 4 | 6 2 | 5 4 | 8 2 | 3+4 4 | 3+4 4 |
| | 5 4 | 6 2 | 5 4 | 8 2 | 3+4 4 | 3+4 4 |
| | 5 4 | 7 2 | 5 4 | 9 2 | 3+4 4 | 3+4 4 |
| | 5 4 | 8 2 | 5 4 | 7 2 | 3+4 4 | 3+4 4 |
| | | 9 2 | | 7 2 | | |
| | | 9 2 | | 6 2 | | |
| | | 6 2 | | 9 2 | | |
| | | 9 2 | | 8 2 | | |
| | | 8 2 | | 6 2 | | |
| | | 7 2 | | 7 2 | | |
| | | 6 2 | | 6 2 | | |
| | | 8 2 | | 8 2 | | |
| | | 7 2 | | 7 2 | | |
| | | 6 2 | | 6 2 | | |
| | | 9 2 | | 9 2 | | |
| | | 7 2 | | 7 2 | | |
| | | 6 2 | | 6 2 | | |
| | | 9 2 | | 9 2 | | |
| | | 6 2 | | 6 2 | | |
| | | 8 2 | | 8 2 | | |
| | | 6 2 | | 9 2 | | |
| | | 9 2 | | 8 2 | | |
| 129.2.35 | 261277 72 Cu1O1P1U1 | | | | | |
| | -A | -GM | -M | -Z | -R | -X |
| | 5 4 | 7 2 | 5 4 | 9 2 | 3+4 4 | 3+4 4 |
| | 5 4 | 9 2 | 5 4 | 7 2 | 3+4 4 | 3+4 4 |
| | 5 4 | 7 2 | 5 4 | 9 2 | 3+4 4 | 3+4 4 |
| | 5 4 | 9 2 | 5 4 | 7 2 | 3+4 4 | 3+4 4 |
| | 5 4 | 7 2 | 5 4 | 9 2 | 3+4 4 | 3+4 4 |
| | 5 4 | 8 2 | 5 4 | 6 2 | 3+4 4 | 3+4 4 |
| | 5 4 | 6 2 | 5 4 | 8 2 | 3+4 4 | 3+4 4 |
| | 5 4 | 9 2 | 5 4 | 7 2 | 3+4 4 | 3+4 4 |
| | 5 4 | 8 2 | 5 4 | 6 2 | 3+4 4 | 3+4 4 |
| | 5 4 | 7 2 | 5 4 | 9 2 | 3+4 4 | 3+4 4 |
| | 5 4 | 7 2 | 5 4 | 7 2 | 3+4 4 | 3+4 4 |
| | 5 4 | 9 2 | 5 4 | 9 2 | 3+4 4 | 3+4 4 |
| | 5 4 | 9 2 | 5 4 | 7 2 | 3+4 4 | 3+4 4 |
| | 5 4 | 7 2 | 5 4 | 9 2 | 3+4 4 | 3+4 4 |
| | 5 4 | 6 2 | 5 4 | 8 2 | 3+4 4 | 3+4 4 |
| | 5 4 | 6 2 | 5 4 | 8 2 | 3+4 4 | 3+4 4 |
| | 5 4 | 7 2 | 5 4 | 9 2 | 3+4 4 | 3+4 4 |
| | 5 4 | 8 2 | 5 4 | 7 2 | 3+4 4 | 3+4 4 |
| | | 9 2 | | 7 2 | | |
| | | 9 2 | | 6 2 | | |
| | | 6 2 | | 9 2 | | |
| | | 9 2 | | 8 2 | | |
| | | 8 2 | | 6 2 | | |
| | | 7 2 | | 7 2 | | |
| | | 6 2 | | 6 2 | | |
| | | 8 2 | | 8 2 | | |
| | | 7 2 | | 7 2 | | |
| | | 6 2 | | 6 2 | | |
| | | 9 2 | | 9 2 | | |
| | | 7 2 | | 7 2 | | |
| | | 6 2 | | 6 2 | | |
| | | 9 2 | | 9 2 | | |
| | | 6 2 | | 6 2 | | |
| | | 8 2 | | 8 2 | | |
| | | 6 2 | | 9 2 | | |
| | | 9 2 | | 8 2 | | |
| 129.2.36 | 182310 72 Cu1O1P1U1 | | | | | |
| | -A | -GM | -M | -Z | -R | -X |
| | 5 4 | 7 2 | 5 4 | 9 2 | 3+4 4 | 3+4 4 |
| | 5 4 | 9 2 | 5 4 | 7 2 | 3+4 4 | 3+4 4 |
| | 5 4 | 7 2 | 5 4 | 9 2 | 3+4 4 | 3+4 4 |
| | 5 4 | 9 2 | 5 4 | 7 2 | 3+4 4 | 3+4 4 |
| | 5 4 | 7 2 | 5 4 | 9 2 | 3+4 4 | 3+4 4 |
| | 5 4 | 8 2 | 5 4 | 6 2 | 3+4 4 | 3+4 4 |
| | 5 4 | 6 2 | 5 4 | 8 2 | 3+4 4 | 3+4 4 |
| | 5 4 | 9 2 | 5 4 | 7 2 | 3+4 4 | 3+4 4 |
| | 5 4 | 8 2 | 5 4 | 6 2 | 3+4 4 | 3+4 4 |
| | 5 4 | 7 2 | 5 4 | 9 2 | 3+4 4 | 3+4 4 |
| | 5 4 | 7 2 | 5 4 | 7 2 | 3+4 4 | 3+4 4 |
| | 5 4 | 9 2 | 5 4 | 9 2 | 3+4 4 | 3+4 4 |
| | 5 4 | 9 2 | 5 4 | 7 2 | 3+4 4 | 3+4 4 |
| | 5 4 | 7 2 | 5 4 | 9 2 | 3+4 4 | 3+4 4 |
| | 5 4 | 6 2 | 5 4 | 8 2 | 3+4 4 | 3+4 4 |
| | 5 4 | 6 2 | 5 4 | 8 2 | 3+4 4 | 3+4 4 |
| | 5 4 | 7 2 | 5 4 | 9 2 | 3+4 4 | 3+4 4 |
| | 5 4 | 8 2 | 5 4 | 7 2 | 3+4 4 | 3+4 4 |
| | | 9 2 | | 7 2 | | |
| | | 9 2 | | 9 2 | | |
| | | 6 2 | | 6 2 | | |
| | | 9 2 | | 8 2 | | |
| | | 8 2 | | 6 2 | | |
| | | 7 2 | | 7 2 | | |
| | | 6 2 | | 6 2 | | |
| | | 8 2 | | 8 2 | | |
| | | 7 2 | | 6 2 | | |
| | | 6 2 | | 7 2 | | |
| | | 9 2 | | 9 2 | | |
| | | 7 2 | | 7 2 | | |
| | | 6 2 | | 6 2 | | |
| | | 9 2 | | 9 2 | | |
| | | 6 2 | | 6 2 | | |
| | | 8 2 | | 8 2 | | |
| | | 6 2 | | 9 2 | | |
| | | 9 2 | | 8 2 | | |
| 129.2.37 | 182311 68 Cu1O1P1Th1 | | | | | |
| | -A | -GM | -M | -Z | -R | -X |
| | 5 4 | 7 2 | 5 4 | 9 2 | 3+4 4 | 3+4 4 |

Continued on next column

**TABLE LX– continued**

| TQC | ICSD | NoE | SF | | | |
|---|---|---|---|---|---|---|
| | 5 4 | 9 2 | 5 4 | 7 2 | 3+4 4 | 3+4 4 |
| | 5 4 | 7 2 | 5 4 | 9 2 | 3+4 4 | 3+4 4 |
| | 5 4 | 9 2 | 5 4 | 7 2 | 3+4 4 | 3+4 4 |
| | 5 4 | 7 2 | 5 4 | 9 2 | 3+4 4 | 3+4 4 |
| | 5 4 | 8 2 | 5 4 | 6 2 | 3+4 4 | 3+4 4 |
| | 5 4 | 6 2 | 5 4 | 8 2 | 3+4 4 | 3+4 4 |
| | 5 4 | 9 2 | 5 4 | 7 2 | 3+4 4 | 3+4 4 |
| | 5 4 | 8 2 | 5 4 | 6 2 | 3+4 4 | 3+4 4 |
| | 5 4 | 7 2 | 5 4 | 9 2 | 3+4 4 | 3+4 4 |
| | 5 4 | 7 2 | 5 4 | 7 2 | 3+4 4 | 3+4 4 |
| | 5 4 | 9 2 | 5 4 | 9 2 | 3+4 4 | 3+4 4 |
| | 5 4 | 9 2 | 5 4 | 7 2 | 3+4 4 | 3+4 4 |
| | 5 4 | 6 2 | 5 4 | 8 2 | 3+4 4 | 3+4 4 |
| | 5 4 | 7 2 | 5 4 | 9 2 | 3+4 4 | 3+4 4 |
| | 5 4 | 6 2 | 5 4 | 8 2 | 3+4 4 | 3+4 4 |
| | 5 4 | 7 2 | 5 4 | 9 2 | 3+4 4 | 3+4 4 |
| | | 8 2 | | 7 2 | | |
| | | 9 2 | | 7 2 | | |
| | | 9 2 | | 6 2 | | |
| | | 6 2 | | 9 2 | | |
| | | 9 2 | | 8 2 | | |
| | | 8 2 | | 6 2 | | |
| | | 7 2 | | 7 2 | | |
| | | 6 2 | | 6 2 | | |
| | | 8 2 | | 8 2 | | |
| | | 7 2 | | 6 2 | | |
| | | 6 2 | | 7 2 | | |
| | | 9 2 | | 9 2 | | |
| | | 7 2 | | 7 2 | | |
| | | 6 2 | | 6 2 | | |
| | | 9 2 | | 9 2 | | |
| | | 8 2 | | 6 2 | | |
| | | 9 2 | | 9 2 | | |
| 129.2.38 | 172798 | 68 | Cu1F1Se1Sr1 | | | |
| | -A | -GM | -M | -Z | -R | -X |
| | 5 4 | 7 2 | 5 4 | 7 2 | 3+4 4 | 3+4 4 |
| | 5 4 | 9 2 | 5 4 | 9 2 | 3+4 4 | 3+4 4 |
| | 5 4 | 7 2 | 5 4 | 9 2 | 3+4 4 | 3+4 4 |
| | 5 4 | 8 2 | 5 4 | 6 2 | 3+4 4 | 3+4 4 |
| | 5 4 | 7 2 | 5 4 | 7 2 | 3+4 4 | 3+4 4 |
| | 5 4 | 9 2 | 5 4 | 9 2 | 3+4 4 | 3+4 4 |
| | 5 4 | 7 2 | 5 4 | 7 2 | 3+4 4 | 3+4 4 |
| | 5 4 | 6 2 | 5 4 | 6 2 | 3+4 4 | 3+4 4 |
| | 5 4 | 9 2 | 5 4 | 9 2 | 3+4 4 | 3+4 4 |
| | 5 4 | 8 2 | 5 4 | 8 2 | 3+4 4 | 3+4 4 |
| | 5 4 | 9 2 | 5 4 | 7 2 | 3+4 4 | 3+4 4 |
| | 5 4 | 7 2 | 5 4 | 9 2 | 3+4 4 | 3+4 4 |
| | 5 4 | 7 2 | 5 4 | 7 2 | 3+4 4 | 3+4 4 |
| | 5 4 | 7 2 | 5 4 | 9 2 | 3+4 4 | 3+4 4 |
| | 5 4 | 6 2 | 5 4 | 8 2 | 3+4 4 | 3+4 4 |
| | 5 4 | 6 2 | 5 4 | 6 2 | 3+4 4 | 3+4 4 |
| | 5 4 | 9 2 | 5 4 | 9 2 | 3+4 4 | 3+4 4 |
| | | 7 2 | | 7 2 | | |
| | | 8 2 | | 8 2 | | |
| | | 9 2 | | 9 2 | | |
| | | 8 2 | | 8 2 | | |
| | | 9 2 | | 7 2 | | |
| | | 7 2 | | 6 2 | | |
| | | 6 2 | | 8 2 | | |
| | | 6 2 | | 6 2 | | |
| | | 6 2 | | 7 2 | | |

Continued on next column

**TABLE LX– continued**

| TQC | ICSD | NoE | SF | | | |
|---|---|---|---|---|---|---|
| | | 9 2 | | 7 2 | | |
| | | 8 2 | | 6 2 | | |
| | | 9 2 | | 9 2 | | |
| | | 9 2 | | 9 2 | | |
| | | 9 2 | | 7 2 | | |
| | | 7 2 | | 6 2 | | |
| | | 8 2 | | 9 2 | | |
| | | 7 2 | | 7 2 | | |
| 129.2.39 | 172796 | 68 | Cu1F1S1Sr1 | | | |
| | -A | -GM | -M | -Z | -R | -X |
| | 5 4 | 7 2 | 5 4 | 7 2 | 3+4 4 | 3+4 4 |
| | 5 4 | 9 2 | 5 4 | 9 2 | 3+4 4 | 3+4 4 |
| | 5 4 | 7 2 | 5 4 | 9 2 | 3+4 4 | 3+4 4 |
| | 5 4 | 8 2 | 5 4 | 6 2 | 3+4 4 | 3+4 4 |
| | 5 4 | 7 2 | 5 4 | 7 2 | 3+4 4 | 3+4 4 |
| | 5 4 | 9 2 | 5 4 | 9 2 | 3+4 4 | 3+4 4 |
| | 5 4 | 7 2 | 5 4 | 7 2 | 3+4 4 | 3+4 4 |
| | 5 4 | 6 2 | 5 4 | 6 2 | 3+4 4 | 3+4 4 |
| | 5 4 | 9 2 | 5 4 | 9 2 | 3+4 4 | 3+4 4 |
| | 5 4 | 8 2 | 5 4 | 8 2 | 3+4 4 | 3+4 4 |
| | 5 4 | 9 2 | 5 4 | 7 2 | 3+4 4 | 3+4 4 |
| | 5 4 | 7 2 | 5 4 | 9 2 | 3+4 4 | 3+4 4 |
| | 5 4 | 7 2 | 5 4 | 7 2 | 3+4 4 | 3+4 4 |
| | 5 4 | 6 2 | 5 4 | 6 2 | 3+4 4 | 3+4 4 |
| | 5 4 | 6 2 | 5 4 | 8 2 | 3+4 4 | 3+4 4 |
| | 5 4 | 7 2 | 5 4 | 9 2 | 3+4 4 | 3+4 4 |
| | 5 4 | 9 2 | 5 4 | 9 2 | 3+4 4 | 3+4 4 |
| | | 7 2 | | 7 2 | | |
| | | 8 2 | | 8 2 | | |
| | | 9 2 | | 9 2 | | |
| | | 8 2 | | 8 2 | | |
| | | 7 2 | | 7 2 | | |
| | | 9 2 | | 6 2 | | |
| | | 6 2 | | 6 2 | | |
| | | 6 2 | | 8 2 | | |
| | | 6 2 | | 7 2 | | |
| | | 9 2 | | 7 2 | | |
| | | 9 2 | | 9 2 | | |
| | | 8 2 | | 6 2 | | |
| | | 9 2 | | 9 2 | | |
| | | 7 2 | | 7 2 | | |
| | | 9 2 | | 6 2 | | |
| | | 8 2 | | 7 2 | | |
| | | 7 2 | | 9 2 | | |
| 129.2.40 | 67932 | 84 | Co2P2U1 | | | |
| | -A | -GM | -M | -Z | -R | -X |
| | 5 4 | 7 2 | 5 4 | 7 2 | 3+4 4 | 3+4 4 |
| | 5 4 | 9 2 | 5 4 | 9 2 | 3+4 4 | 3+4 4 |
| | 5 4 | 7 2 | 5 4 | 7 2 | 3+4 4 | 3+4 4 |
| | 5 4 | 9 2 | 5 4 | 9 2 | 3+4 4 | 3+4 4 |
| | 5 4 | 7 2 | 5 4 | 7 2 | 3+4 4 | 3+4 4 |
| | 5 4 | 9 2 | 5 4 | 9 2 | 3+4 4 | 3+4 4 |
| | 5 4 | 6 2 | 5 4 | 6 2 | 3+4 4 | 3+4 4 |
| | 5 4 | 8 2 | 5 4 | 8 2 | 3+4 4 | 3+4 4 |
| | 5 4 | 7 2 | 5 4 | 7 2 | 3+4 4 | 3+4 4 |
| | 5 4 | 7 2 | 5 4 | 9 2 | 3+4 4 | 3+4 4 |
| | 5 4 | 9 2 | 5 4 | 7 2 | 3+4 4 | 3+4 4 |
| | 5 4 | 8 2 | 5 4 | 8 2 | 3+4 4 | 3+4 4 |
| | 5 4 | 9 2 | 5 4 | 9 2 | 3+4 4 | 3+4 4 |
| | 5 4 | 7 2 | 5 4 | 7 2 | 3+4 4 | 3+4 4 |
| | 5 4 | 7 2 | 5 4 | 6 2 | 3+4 4 | 3+4 4 |

Continued on next column

**TABLE LX– continued**

| TQC | ICSD | | NoE | | SF | |
|---|---|---|---|---|---|---|
| | 5 4 | 6 2 | 5 4 | 7 2 | 3+4 4 | 3+4 4 |
| | 5 4 | 6 2 | 5 4 | 9 2 | 3+4 4 | 3+4 4 |
| | 5 4 | 9 2 | 5 4 | 6 2 | 3+4 4 | 3+4 4 |
| | 5 4 | 7 2 | 5 4 | 9 2 | 3+4 4 | 3+4 4 |
| | 5 4 | 8 2 | 5 4 | 8 2 | 3+4 4 | 3+4 4 |
| | 5 4 | 6 2 | 5 4 | 7 2 | 3+4 4 | 3+4 4 |
| | | 9 2 | | 6 2 | | |
| | | 6 2 | | 8 2 | | |
| | | 8 2 | | 7 2 | | |
| | | 7 2 | | 6 2 | | |
| | | 6 2 | | 6 2 | | |
| | | 8 2 | | 8 2 | | |
| | | 9 2 | | 9 2 | | |
| | | 8 2 | | 8 2 | | |
| | | 9 2 | | 9 2 | | |
| | | 7 2 | | 8 2 | | |
| | | 6 2 | | 9 2 | | |
| | | 6 2 | | 8 2 | | |
| | | 7 2 | | 8 2 | | |
| | | 7 2 | | 7 2 | | |
| | | 6 2 | | 6 2 | | |
| | | 8 2 | | 7 2 | | |
| | | 6 2 | | 8 2 | | |
| | | 9 2 | | 7 2 | | |
| | | 8 2 | | 8 2 | | |
| | | 9 2 | | 9 2 | | |
| | | 9 2 | | 9 2 | | |
| 129.2.41 | 106477 | | 52 | | Co1Ge1Pr1 | |
| | -A | -GM | -M | -Z | -R | -X |
| | 5 4 | 7 2 | 5 4 | 9 2 | 3+4 4 | 3+4 4 |
| | 5 4 | 9 2 | 5 4 | 7 2 | 3+4 4 | 3+4 4 |
| | 5 4 | 7 2 | 5 4 | 9 2 | 3+4 4 | 3+4 4 |
| | 5 4 | 9 2 | 5 4 | 7 2 | 3+4 4 | 3+4 4 |
| | 5 4 | 7 2 | 5 4 | 9 2 | 3+4 4 | 3+4 4 |
| | 5 4 | 6 2 | 5 4 | 8 2 | 3+4 4 | 3+4 4 |
| | 5 4 | 9 2 | 5 4 | 7 2 | 3+4 4 | 3+4 4 |
| | 5 4 | 8 2 | 5 4 | 6 2 | 3+4 4 | 3+4 4 |
| | 5 4 | 7 2 | 5 4 | 7 2 | 3+4 4 | 3+4 4 |
| | 5 4 | 9 2 | 5 4 | 9 2 | 3+4 4 | 3+4 4 |
| | 5 4 | 7 2 | 5 4 | 7 2 | 3+4 4 | 3+4 4 |
| | 5 4 | 9 2 | 5 4 | 9 2 | 3+4 4 | 3+4 4 |
| | 5 4 | 9 2 | 5 4 | 9 2 | 3+4 4 | 3+4 4 |
| | | 8 2 | | 8 2 | | |
| | | 6 2 | | 6 2 | | |
| | | 7 2 | | 7 2 | | |
| | | 8 2 | | 6 2 | | |
| | | 6 2 | | 8 2 | | |
| | | 7 2 | | 6 2 | | |
| | | 6 2 | | 7 2 | | |
| | | 6 2 | | 6 2 | | |
| | | 7 2 | | 7 2 | | |
| | | 9 2 | | 6 2 | | |
| | | 8 2 | | 9 2 | | |
| | | 9 2 | | 8 2 | | |
| | | 8 2 | | 9 2 | | |
| 129.2.42 | 106471 | | 48 | | Co1Ge1La1 | |
| | -A | -GM | -M | -Z | -R | -X |
| | 5 4 | 7 2 | 5 4 | 9 2 | 3+4 4 | 3+4 4 |
| | 5 4 | 9 2 | 5 4 | 7 2 | 3+4 4 | 3+4 4 |
| | 5 4 | 7 2 | 5 4 | 9 2 | 3+4 4 | 3+4 4 |
| | 5 4 | 9 2 | 5 4 | 7 2 | 3+4 4 | 3+4 4 |

Continued on next column

**TABLE LX– continued**

| TQC | ICSD | | NoE | | SF | |
|---|---|---|---|---|---|---|
| | 5 4 | 7 2 | 5 4 | 9 2 | 3+4 4 | 3+4 4 |
| | 5 4 | 6 2 | 5 4 | 8 2 | 3+4 4 | 3+4 4 |
| | 5 4 | 9 2 | 5 4 | 7 2 | 3+4 4 | 3+4 4 |
| | 5 4 | 8 2 | 5 4 | 6 2 | 3+4 4 | 3+4 4 |
| | 5 4 | 7 2 | 5 4 | 7 2 | 3+4 4 | 3+4 4 |
| | 5 4 | 9 2 | 5 4 | 9 2 | 3+4 4 | 3+4 4 |
| | 5 4 | 7 2 | 5 4 | 7 2 | 3+4 4 | 3+4 4 |
| | 5 4 | 9 2 | 5 4 | 9 2 | 3+4 4 | 3+4 4 |
| | | 9 2 | | 9 2 | | |
| | | 8 2 | | 8 2 | | |
| | | 6 2 | | 6 2 | | |
| | | 8 2 | | 7 2 | | |
| | | 7 2 | | 6 2 | | |
| | | 6 2 | | 8 2 | | |
| | | 7 2 | | 6 2 | | |
| | | 6 2 | | 7 2 | | |
| | | 6 2 | | 6 2 | | |
| | | 7 2 | | 7 2 | | |
| | | 9 2 | | 6 2 | | |
| | | 8 2 | | 9 2 | | |
| 129.2.43 | 61370 | | 80 | | Ce1Pt2Si2 | |
| | -A | -GM | -M | -Z | -R | -X |
| | 5 4 | 7 2 | 5 4 | 7 2 | 3+4 4 | 3+4 4 |
| | 5 4 | 9 2 | 5 4 | 9 2 | 3+4 4 | 3+4 4 |
| | 5 4 | 7 2 | 5 4 | 9 2 | 3+4 4 | 3+4 4 |
| | 5 4 | 9 2 | 5 4 | 7 2 | 3+4 4 | 3+4 4 |
| | 5 4 | 7 2 | 5 4 | 9 2 | 3+4 4 | 3+4 4 |
| | 5 4 | 9 2 | 5 4 | 7 2 | 3+4 4 | 3+4 4 |
| | 5 4 | 8 2 | 5 4 | 8 2 | 3+4 4 | 3+4 4 |
| | 5 4 | 6 2 | 5 4 | 6 2 | 3+4 4 | 3+4 4 |
| | 5 4 | 7 2 | 5 4 | 9 2 | 3+4 4 | 3+4 4 |
| | 5 4 | 7 2 | 5 4 | 7 2 | 3+4 4 | 3+4 4 |
| | 5 4 | 9 2 | 5 4 | 9 2 | 3+4 4 | 3+4 4 |
| | 5 4 | 8 2 | 5 4 | 6 2 | 3+4 4 | 3+4 4 |
| | 5 4 | 9 2 | 5 4 | 7 2 | 3+4 4 | 3+4 4 |
| | 5 4 | 7 2 | 5 4 | 9 2 | 3+4 4 | 3+4 4 |
| | 5 4 | 6 2 | 5 4 | 8 2 | 3+4 4 | 3+4 4 |
| | 5 4 | 7 2 | 5 4 | 7 2 | 3+4 4 | 3+4 4 |
| | 5 4 | 9 2 | 5 4 | 8 2 | 3+4 4 | 3+4 4 |
| | 5 4 | 6 2 | 5 4 | 9 2 | 3+4 4 | 3+4 4 |
| | 5 4 | 8 2 | 5 4 | 6 2 | 3+4 4 | 3+4 4 |
| | 5 4 | 6 2 | 5 4 | 8 2 | 3+4 4 | 3+4 4 |
| | | 9 2 | | 9 2 | | |
| | | 6 2 | | 8 2 | | |
| | | 9 2 | | 7 2 | | |
| | | 8 2 | | 6 2 | | |
| | | 7 2 | | 7 2 | | |
| | | 7 2 | | 9 2 | | |
| | | 8 2 | | 6 2 | | |
| | | 8 2 | | 8 2 | | |
| | | 6 2 | | 6 2 | | |
| | | 9 2 | | 7 2 | | |
| | | 6 2 | | 6 2 | | |
| | | 7 2 | | 9 2 | | |
| | | 6 2 | | 6 2 | | |
| | | 7 2 | | 7 2 | | |
| | | 7 2 | | 9 2 | | |
| | | 9 2 | | 8 2 | | |
| | | 6 2 | | 9 2 | | |
| | | 6 2 | | 6 2 | | |
| | | 8 2 | | 7 2 | | |

Continued on next column

**TABLE LX– continued**

| TQC | ICSD | NoE | SF | | | |
|---|---|---|---|---|---|---|
| | | 9 2 | | 6 2 | | |
| 129.2.44 | 40764 | 80 | Ce1P2Rh2 | | | |
| | -A | -GM | -M | -Z | -R | -X |
| | 5 4 | 7 2 | 5 4 | 7 2 | 3+4 4 | 3+4 4 |
| | 5 4 | 9 2 | 5 4 | 9 2 | 3+4 4 | 3+4 4 |
| | 5 4 | 7 2 | 5 4 | 9 2 | 3+4 4 | 3+4 4 |
| | 5 4 | 9 2 | 5 4 | 7 2 | 3+4 4 | 3+4 4 |
| | 5 4 | 7 2 | 5 4 | 9 2 | 3+4 4 | 3+4 4 |
| | 5 4 | 9 2 | 5 4 | 7 2 | 3+4 4 | 3+4 4 |
| | 5 4 | 8 2 | 5 4 | 8 2 | 3+4 4 | 3+4 4 |
| | 5 4 | 6 2 | 5 4 | 6 2 | 3+4 4 | 3+4 4 |
| | 5 4 | 7 2 | 5 4 | 9 2 | 3+4 4 | 3+4 4 |
| | 5 4 | 7 2 | 5 4 | 7 2 | 3+4 4 | 3+4 4 |
| | 5 4 | 9 2 | 5 4 | 9 2 | 3+4 4 | 3+4 4 |
| | 5 4 | 8 2 | 5 4 | 6 2 | 3+4 4 | 3+4 4 |
| | 5 4 | 7 2 | 5 4 | 9 2 | 3+4 4 | 3+4 4 |
| | 5 4 | 6 2 | 5 4 | 8 2 | 3+4 4 | 3+4 4 |
| | 5 4 | 9 2 | 5 4 | 7 2 | 3+4 4 | 3+4 4 |
| | 5 4 | 7 2 | 5 4 | 7 2 | 3+4 4 | 3+4 4 |
| | 5 4 | 6 2 | 5 4 | 8 2 | 3+4 4 | 3+4 4 |
| | 5 4 | 9 2 | 5 4 | 9 2 | 3+4 4 | 3+4 4 |
| | 5 4 | 8 2 | 5 4 | 8 2 | 3+4 4 | 3+4 4 |
| | 5 4 | 9 2 | 5 4 | 9 2 | 3+4 4 | 3+4 4 |
| | | 6 2 | | 6 2 | | |
| | | 6 2 | | 7 2 | | |
| | | 9 2 | | 8 2 | | |
| | | 7 2 | | 6 2 | | |
| | | 6 2 | | 7 2 | | |
| | | 8 2 | | 6 2 | | |
| | | 7 2 | | 6 2 | | |
| | | 8 2 | | 9 2 | | |
| | | 8 2 | | 8 2 | | |
| | | 9 2 | | 7 2 | | |
| | | 7 2 | | 6 2 | | |
| | | 6 2 | | 7 2 | | |
| | | 7 2 | | 9 2 | | |
| | | 6 2 | | 6 2 | | |
| | | 7 2 | | 8 2 | | |
| | | 6 2 | | 9 2 | | |
| | | 9 2 | | 6 2 | | |
| | | 8 2 | | 9 2 | | |
| | | 9 2 | | 7 2 | | |
| | | 6 2 | | 6 2 | | |
| 129.2.45 | 419356 | 60 | Ce1Mn1O1Sb1 | | | |
| | -A | -GM | -M | -Z | -R | -X |
| | 5 4 | 7 2 | 5 4 | 7 2 | 3+4 4 | 3+4 4 |
| | 5 4 | 9 2 | 5 4 | 9 2 | 3+4 4 | 3+4 4 |
| | 5 4 | 7 2 | 5 4 | 7 2 | 3+4 4 | 3+4 4 |
| | 5 4 | 9 2 | 5 4 | 9 2 | 3+4 4 | 3+4 4 |
| | 5 4 | 8 2 | 5 4 | 8 2 | 3+4 4 | 3+4 4 |
| | 5 4 | 7 2 | 5 4 | 7 2 | 3+4 4 | 3+4 4 |
| | 5 4 | 6 2 | 5 4 | 6 2 | 3+4 4 | 3+4 4 |
| | 5 4 | 9 2 | 5 4 | 9 2 | 3+4 4 | 3+4 4 |
| | 5 4 | 8 2 | 5 4 | 8 2 | 3+4 4 | 3+4 4 |
| | 5 4 | 7 2 | 5 4 | 7 2 | 3+4 4 | 3+4 4 |
| | 5 4 | 7 2 | 5 4 | 9 2 | 3+4 4 | 3+4 4 |
| | 5 4 | 9 2 | 5 4 | 7 2 | 3+4 4 | 3+4 4 |
| | 5 4 | 9 2 | 5 4 | 9 2 | 3+4 4 | 3+4 4 |
| | 5 4 | 6 2 | 5 4 | 6 2 | 3+4 4 | 3+4 4 |
| | 5 4 | 7 2 | 5 4 | 7 2 | 3+4 4 | 3+4 4 |
| | | 6 2 | | 6 2 | | |

Continued on next column

**TABLE LX– continued**

| TQC | ICSD | NoE | SF | | | |
|---|---|---|---|---|---|---|
| | | 9 2 | | 9 2 | | |
| | | 8 2 | | 8 2 | | |
| | | 7 2 | | 9 2 | | |
| | | 9 2 | | 7 2 | | |
| | | 9 2 | | 7 2 | | |
| | | 8 2 | | 6 2 | | |
| | | 7 2 | | 8 2 | | |
| | | 6 2 | | 9 2 | | |
| | | 6 2 | | 8 2 | | |
| | | 7 2 | | 7 2 | | |
| | | 8 2 | | 9 2 | | |
| | | 9 2 | | 6 2 | | |
| | | 6 2 | | 6 2 | | |
| | | 6 2 | | 8 2 | | |
| 129.2.46 | 106407 | 76 | Ce1Ir2Sn2 | | | |
| | -A | -GM | -M | -Z | -R | -X |
| | 5 4 | 7 2 | 5 4 | 7 2 | 3+4 4 | 3+4 4 |
| | 5 4 | 9 2 | 5 4 | 9 2 | 3+4 4 | 3+4 4 |
| | 5 4 | 7 2 | 5 4 | 7 2 | 3+4 4 | 3+4 4 |
| | 5 4 | 9 2 | 5 4 | 9 2 | 3+4 4 | 3+4 4 |
| | 5 4 | 7 2 | 5 4 | 7 2 | 3+4 4 | 3+4 4 |
| | 5 4 | 9 2 | 5 4 | 9 2 | 3+4 4 | 3+4 4 |
| | 5 4 | 6 2 | 5 4 | 8 2 | 3+4 4 | 3+4 4 |
| | 5 4 | 8 2 | 5 4 | 6 2 | 3+4 4 | 3+4 4 |
| | 5 4 | 7 2 | 5 4 | 7 2 | 3+4 4 | 3+4 4 |
| | 5 4 | 7 2 | 5 4 | 9 2 | 3+4 4 | 3+4 4 |
| | 5 4 | 9 2 | 5 4 | 9 2 | 3+4 4 | 3+4 4 |
| | 5 4 | 8 2 | 5 4 | 6 2 | 3+4 4 | 3+4 4 |
| | 5 4 | 9 2 | 5 4 | 7 2 | 3+4 4 | 3+4 4 |
| | 5 4 | 7 2 | 5 4 | 9 2 | 3+4 4 | 3+4 4 |
| | 5 4 | 6 2 | 5 4 | 8 2 | 3+4 4 | 3+4 4 |
| | 5 4 | 8 2 | 5 4 | 9 2 | 3+4 4 | 3+4 4 |
| | 5 4 | 7 2 | 5 4 | 6 2 | 3+4 4 | 3+4 4 |
| | 5 4 | 6 2 | 5 4 | 7 2 | 3+4 4 | 3+4 4 |
| | 5 4 | 9 2 | 5 4 | 8 2 | 3+4 4 | 3+4 4 |
| | | 9 2 | | 9 2 | | |
| | | 6 2 | | 8 2 | | |
| | | 7 2 | | 7 2 | | |
| | | 6 2 | | 6 2 | | |
| | | 8 2 | | 7 2 | | |
| | | 9 2 | | 8 2 | | |
| | | 8 2 | | 6 2 | | |
| | | 6 2 | | 9 2 | | |
| | | 7 2 | | 8 2 | | |
| | | 8 2 | | 6 2 | | |
| | | 6 2 | | 7 2 | | |
| | | 7 2 | | 6 2 | | |
| | | 6 2 | | 6 2 | | |
| | | 9 2 | | 7 2 | | |
| | | 7 2 | | 9 2 | | |
| | | 9 2 | | 9 2 | | |
| | | 7 2 | | 8 2 | | |
| | | 6 2 | | 6 2 | | |
| | | 6 2 | | 9 2 | | |
| 129.2.47 | 102076 | 32 | Cd1Ti1 | | | |
| | -A | -GM | -M | -Z | -R | -X |
| | 5 4 | 8 2 | 5 4 | 8 2 | 3+4 4 | 3+4 4 |
| | 5 4 | 6 2 | 5 4 | 6 2 | 3+4 4 | 3+4 4 |
| | 5 4 | 9 2 | 5 4 | 9 2 | 3+4 4 | 3+4 4 |
| | 5 4 | 7 2 | 5 4 | 7 2 | 3+4 4 | 3+4 4 |
| | 5 4 | 7 2 | 5 4 | 7 2 | 3+4 4 | 3+4 4 |

Continued on next column

**TABLE LX– continued**

| TQC | ICSD | | NoE | | SF | |
|---|---|---|---|---|---|---|
| | 5 4 | 6 2 | 5 4 | 6 2 | 3+4 4 | 3+4 4 |
| | 5 4 | 9 2 | 5 4 | 9 2 | 3+4 4 | 3+4 4 |
| | 5 4 | 7 2 | 5 4 | 8 2 | 3+4 4 | 3+4 4 |
| | | 8 2 | | 7 2 | | |
| | | 6 2 | | 6 2 | | |
| | | 8 2 | | 8 2 | | |
| | | 9 2 | | 9 2 | | |
| | | 7 2 | | 7 2 | | |
| | | 9 2 | | 9 2 | | |
| | | 8 2 | | 6 2 | | |
| | | 7 2 | | 7 2 | | |
| 129.2.48 | 2027 | | 20 | | Ca1Ga1N1 | |
| | -A | -GM | -M | -Z | -R | -X |
| | 5 4 | 7 2 | 5 4 | 9 2 | 3+4 4 | 3+4 4 |
| | 5 4 | 9 2 | 5 4 | 7 2 | 3+4 4 | 3+4 4 |
| | 5 4 | 7 2 | 5 4 | 7 2 | 3+4 4 | 3+4 4 |
| | 5 4 | 9 2 | 5 4 | 9 2 | 3+4 4 | 3+4 4 |
| | 5 4 | 7 2 | 5 4 | 7 2 | 3+4 4 | 3+4 4 |
| | | 7 2 | | 7 2 | | |
| | | 6 2 | | 6 2 | | |
| | | 9 2 | | 9 2 | | |
| | | 8 2 | | 8 2 | | |
| | | 7 2 | | 9 2 | | |
| 129.2.49 | 62598 | | 40 | | C2Co1Sc1 | |
| | -A | -GM | -M | -Z | -R | -X |
| | 5 4 | 7 2 | 5 4 | 7 2 | 3+4 4 | 3+4 4 |
| | 5 4 | 9 2 | 5 4 | 9 2 | 3+4 4 | 3+4 4 |
| | 5 4 | 7 2 | 5 4 | 7 2 | 3+4 4 | 3+4 4 |
| | 5 4 | 9 2 | 5 4 | 9 2 | 3+4 4 | 3+4 4 |
| | 5 4 | 7 2 | 5 4 | 9 2 | 3+4 4 | 3+4 4 |
| | 5 4 | 8 2 | 5 4 | 8 2 | 3+4 4 | 3+4 4 |
| | 5 4 | 9 2 | 5 4 | 9 2 | 3+4 4 | 3+4 4 |
| | 5 4 | 9 2 | 5 4 | 7 2 | 3+4 4 | 3+4 4 |
| | 5 4 | 7 2 | 5 4 | 6 2 | 3+4 4 | 3+4 4 |
| | 5 4 | 6 2 | 5 4 | 7 2 | 3+4 4 | 3+4 4 |
| | | 6 2 | | 6 2 | | |
| | | 7 2 | | 9 2 | | |
| | | 9 2 | | 7 2 | | |
| | | 8 2 | | 8 2 | | |
| | | 8 2 | | 9 2 | | |
| | | 9 2 | | 8 2 | | |
| | | 7 2 | | 6 2 | | |
| | | 6 2 | | 7 2 | | |
| | | 8 2 | | 6 2 | | |
| | | 6 2 | | 6 2 | | |
| 129.2.50 | 85283 | | 56 | | C1B1Ni1Y1 | |
| | -A | -GM | -M | -Z | -R | -X |
| | 5 4 | 7 2 | 5 4 | 7 2 | 3+4 4 | 3+4 4 |
| | 5 4 | 9 2 | 5 4 | 9 2 | 3+4 4 | 3+4 4 |
| | 5 4 | 7 2 | 5 4 | 7 2 | 3+4 4 | 3+4 4 |
| | 5 4 | 9 2 | 5 4 | 9 2 | 3+4 4 | 3+4 4 |
| | 5 4 | 7 2 | 5 4 | 7 2 | 3+4 4 | 3+4 4 |
| | 5 4 | 6 2 | 5 4 | 6 2 | 3+4 4 | 3+4 4 |
| | 5 4 | 9 2 | 5 4 | 9 2 | 3+4 4 | 3+4 4 |
| | 5 4 | 8 2 | 5 4 | 8 2 | 3+4 4 | 3+4 4 |
| | 5 4 | 7 2 | 5 4 | 9 2 | 3+4 4 | 3+4 4 |
| | 5 4 | 9 2 | 5 4 | 7 2 | 3+4 4 | 3+4 4 |
| | 5 4 | 7 2 | 5 4 | 9 2 | 3+4 4 | 3+4 4 |
| | 5 4 | 9 2 | 5 4 | 7 2 | 3+4 4 | 3+4 4 |
| | 5 4 | 8 2 | 5 4 | 7 2 | 3+4 4 | 3+4 4 |
| | 5 4 | 7 2 | 5 4 | 6 2 | 3+4 4 | 3+4 4 |

Continued on next column

**TABLE LX– continued**

| TQC | ICSD | | NoE | | SF | |
|---|---|---|---|---|---|---|
| | | 6 2 | | 9 2 | | |
| | | 9 2 | | 8 2 | | |
| | | 7 2 | | 7 2 | | |
| | | 9 2 | | 9 2 | | |
| | | 6 2 | | 8 2 | | |
| | | 8 2 | | 6 2 | | |
| | | 7 2 | | 9 2 | | |
| | | 8 2 | | 8 2 | | |
| | | 7 2 | | 6 2 | | |
| | | 9 2 | | 9 2 | | |
| | | 6 2 | | 7 2 | | |
| | | 6 2 | | 8 2 | | |
| | | 9 2 | | 7 2 | | |
| | | 8 2 | | 8 2 | | |
| 129.2.51 | 415728 | | 44 | | Bi2La1Li1 | |
| | -A | -GM | -M | -Z | -R | -X |
| | 5 4 | 7 2 | 5 4 | 7 2 | 3+4 4 | 3+4 4 |
| | 5 4 | 9 2 | 5 4 | 9 2 | 3+4 4 | 3+4 4 |
| | 5 4 | 7 2 | 5 4 | 9 2 | 3+4 4 | 3+4 4 |
| | 5 4 | 9 2 | 5 4 | 7 2 | 3+4 4 | 3+4 4 |
| | 5 4 | 7 2 | 5 4 | 9 2 | 3+4 4 | 3+4 4 |
| | 5 4 | 9 2 | 5 4 | 7 2 | 3+4 4 | 3+4 4 |
| | 5 4 | 6 2 | 5 4 | 8 2 | 3+4 4 | 3+4 4 |
| | 5 4 | 8 2 | 5 4 | 6 2 | 3+4 4 | 3+4 4 |
| | 5 4 | 7 2 | 5 4 | 9 2 | 3+4 4 | 3+4 4 |
| | 5 4 | 7 2 | 5 4 | 7 2 | 3+4 4 | 3+4 4 |
| | 5 4 | 9 2 | 5 4 | 9 2 | 3+4 4 | 3+4 4 |
| | | 8 2 | | 6 2 | | |
| | | 6 2 | | 7 2 | | |
| | | 9 2 | | 8 2 | | |
| | | 7 2 | | 9 2 | | |
| | | 7 2 | | 7 2 | | |
| | | 7 2 | | 9 2 | | |
| | | 9 2 | | 7 2 | | |
| | | 6 2 | | 8 2 | | |
| | | 6 2 | | 6 2 | | |
| | | 9 2 | | 9 2 | | |
| | | 8 2 | | 6 2 | | |
| 129.2.52 | 426715 | | 76 | | Bi1N1Nd2 | |
| | -A | -GM | -M | -Z | -R | -X |
| | 5 4 | 7 2 | 5 4 | 7 2 | 3+4 4 | 3+4 4 |
| | 5 4 | 9 2 | 5 4 | 9 2 | 3+4 4 | 3+4 4 |
| | 5 4 | 7 2 | 5 4 | 7 2 | 3+4 4 | 3+4 4 |
| | 5 4 | 8 2 | 5 4 | 8 2 | 3+4 4 | 3+4 4 |
| | 5 4 | 6 2 | 5 4 | 6 2 | 3+4 4 | 3+4 4 |
| | 5 4 | 9 2 | 5 4 | 9 2 | 3+4 4 | 3+4 4 |
| | 5 4 | 7 2 | 5 4 | 7 2 | 3+4 4 | 3+4 4 |
| | 5 4 | 9 2 | 5 4 | 9 2 | 3+4 4 | 3+4 4 |
| | 5 4 | 7 2 | 5 4 | 7 2 | 3+4 4 | 3+4 4 |
| | 5 4 | 9 2 | 5 4 | 9 2 | 3+4 4 | 3+4 4 |
| | 5 4 | 7 2 | 5 4 | 7 2 | 3+4 4 | 3+4 4 |
| | 5 4 | 6 2 | 5 4 | 6 2 | 3+4 4 | 3+4 4 |
| | 5 4 | 8 2 | 5 4 | 8 2 | 3+4 4 | 3+4 4 |
| | 5 4 | 6 2 | 5 4 | 6 2 | 3+4 4 | 3+4 4 |
| | 5 4 | 9 2 | 5 4 | 9 2 | 3+4 4 | 3+4 4 |
| | 5 4 | 8 2 | 5 4 | 8 2 | 3+4 4 | 3+4 4 |
| | 5 4 | 7 2 | 5 4 | 7 2 | 3+4 4 | 3+4 4 |
| | 5 4 | 9 2 | 5 4 | 9 2 | 3+4 4 | 3+4 4 |
| | 5 4 | 7 2 | 5 4 | 9 2 | 3+4 4 | 3+4 4 |
| | | 9 2 | | 7 2 | | |
| | | 7 2 | | 9 2 | | |

Continued on next column

**TABLE LX– continued**

| TQC | ICSD | NoE | SF | | | |
|---|---|---|---|---|---|---|
| | | 9 2 | | 8 2 | | |
| | | 9 2 | | 9 2 | | |
| | | 8 2 | | 7 2 | | |
| | | 6 2 | | 9 2 | | |
| | | 7 2 | | 6 2 | | |
| | | 7 2 | | 7 2 | | |
| | | 9 2 | | 7 2 | | |
| | | 7 2 | | 9 2 | | |
| | | 9 2 | | 7 2 | | |
| | | 6 2 | | 8 2 | | |
| | | 8 2 | | 6 2 | | |
| | | 6 2 | | 6 2 | | |
| | | 8 2 | | 6 2 | | |
| | | 8 2 | | 8 2 | | |
| | | 6 2 | | 8 2 | | |
| | | 8 2 | | 7 2 | | |
| | | 7 2 | | 6 2 | | |
| 129.2.53 | 252466 | 68 | Bi1La1O1S2 | | | |
| | -A | -GM | -M | -Z | -R | -X |
| | 5 4 | 7 2 | 5 4 | 7 2 | 3+4 4 | 3+4 4 |
| | 5 4 | 9 2 | 5 4 | 9 2 | 3+4 4 | 3+4 4 |
| | 5 4 | 7 2 | 5 4 | 7 2 | 3+4 4 | 3+4 4 |
| | 5 4 | 9 2 | 5 4 | 9 2 | 3+4 4 | 3+4 4 |
| | 5 4 | 7 2 | 5 4 | 9 2 | 3+4 4 | 3+4 4 |
| | 5 4 | 7 2 | 5 4 | 7 2 | 3+4 4 | 3+4 4 |
| | 5 4 | 6 2 | 5 4 | 6 2 | 3+4 4 | 3+4 4 |
| | 5 4 | 8 2 | 5 4 | 6 2 | 3+4 4 | 3+4 4 |
| | 5 4 | 9 2 | 5 4 | 9 2 | 3+4 4 | 3+4 4 |
| | 5 4 | 8 2 | 5 4 | 8 2 | 3+4 4 | 3+4 4 |
| | 5 4 | 9 2 | 5 4 | 7 2 | 3+4 4 | 3+4 4 |
| | 5 4 | 7 2 | 5 4 | 9 2 | 3+4 4 | 3+4 4 |
| | 5 4 | 9 2 | 5 4 | 7 2 | 3+4 4 | 3+4 4 |
| | 5 4 | 7 2 | 5 4 | 9 2 | 3+4 4 | 3+4 4 |
| | 5 4 | 9 2 | 5 4 | 7 2 | 3+4 4 | 3+4 4 |
| | 5 4 | 7 2 | 5 4 | 9 2 | 3+4 4 | 3+4 4 |
| | 5 4 | 7 2 | 5 4 | 9 2 | 3+4 4 | 3+4 4 |
| | | 9 2 | | 7 2 | | |
| | | 6 2 | | 8 2 | | |
| | | 7 2 | | 9 2 | | |
| | | 9 2 | | 7 2 | | |
| | | 8 2 | | 6 2 | | |
| | | 9 2 | | 7 2 | | |
| | | 7 2 | | 9 2 | | |
| | | 6 2 | | 8 2 | | |
| | | 9 2 | | 7 2 | | |
| | | 8 2 | | 6 2 | | |
| | | 6 2 | | 8 2 | | |
| | | 7 2 | | 9 2 | | |
| | | 8 2 | | 8 2 | | |
| | | 9 2 | | 7 2 | | |
| | | 8 2 | | 6 2 | | |
| | | 7 2 | | 9 2 | | |
| | | 6 2 | | 8 2 | | |
| 129.2.54 | 426713 | 68 | Bi1Ce2N1 | | | |
| | -A | -GM | -M | -Z | -R | -X |
| | 5 4 | 7 2 | 5 4 | 7 2 | 3+4 4 | 3+4 4 |
| | 5 4 | 9 2 | 5 4 | 9 2 | 3+4 4 | 3+4 4 |
| | 5 4 | 7 2 | 5 4 | 7 2 | 3+4 4 | 3+4 4 |
| | 5 4 | 8 2 | 5 4 | 8 2 | 3+4 4 | 3+4 4 |
| | 5 4 | 6 2 | 5 4 | 6 2 | 3+4 4 | 3+4 4 |
| | 5 4 | 9 2 | 5 4 | 9 2 | 3+4 4 | 3+4 4 |

Continued on next column

**TABLE LX– continued**

| TQC | ICSD | NoE | SF | | | |
|---|---|---|---|---|---|---|
| | 5 4 | 7 2 | 5 4 | 7 2 | 3+4 4 | 3+4 4 |
| | 5 4 | 9 2 | 5 4 | 9 2 | 3+4 4 | 3+4 4 |
| | 5 4 | 7 2 | 5 4 | 7 2 | 3+4 4 | 3+4 4 |
| | 5 4 | 9 2 | 5 4 | 9 2 | 3+4 4 | 3+4 4 |
| | 5 4 | 7 2 | 5 4 | 7 2 | 3+4 4 | 3+4 4 |
| | 5 4 | 6 2 | 5 4 | 8 2 | 3+4 4 | 3+4 4 |
| | 5 4 | 8 2 | 5 4 | 6 2 | 3+4 4 | 3+4 4 |
| | 5 4 | 6 2 | 5 4 | 6 2 | 3+4 4 | 3+4 4 |
| | 5 4 | 9 2 | 5 4 | 9 2 | 3+4 4 | 3+4 4 |
| | 5 4 | 8 2 | 5 4 | 8 2 | 3+4 4 | 3+4 4 |
| | 5 4 | 7 2 | 5 4 | 7 2 | 3+4 4 | 3+4 4 |
| | | 9 2 | | 9 2 | | |
| | | 7 2 | | 9 2 | | |
| | | 9 2 | | 7 2 | | |
| | | 9 2 | | 9 2 | | |
| | | 7 2 | | 8 2 | | |
| | | 9 2 | | 9 2 | | |
| | | 8 2 | | 7 2 | | |
| | | 7 2 | | 9 2 | | |
| | | 6 2 | | 7 2 | | |
| | | 7 2 | | 6 2 | | |
| | | 9 2 | | 7 2 | | |
| | | 7 2 | | 9 2 | | |
| | | 6 2 | | 7 2 | | |
| | | 8 2 | | 6 2 | | |
| | | 6 2 | | 8 2 | | |
| | | 9 2 | | 6 2 | | |
| | | 8 2 | | 6 2 | | |
| 129.2.55 | 25337 | 28 | Be2Ca1Ge2 | | | |
| | -A | -GM | -M | -Z | -R | -X |
| | 5 4 | 7 2 | 5 4 | 7 2 | 3+4 4 | 3+4 4 |
| | 5 4 | 7 2 | 5 4 | 9 2 | 3+4 4 | 3+4 4 |
| | 5 4 | 9 2 | 5 4 | 7 2 | 3+4 4 | 3+4 4 |
| | 5 4 | 9 2 | 5 4 | 9 2 | 3+4 4 | 3+4 4 |
| | 5 4 | 8 2 | 5 4 | 8 2 | 3+4 4 | 3+4 4 |
| | 5 4 | 6 2 | 5 4 | 6 2 | 3+4 4 | 3+4 4 |
| | 5 4 | 7 2 | 5 4 | 7 2 | 3+4 4 | 3+4 4 |
| | | 7 2 | | 9 2 | | |
| | | 7 2 | | 8 2 | | |
| | | 6 2 | | 9 2 | | |
| | | 9 2 | | 7 2 | | |
| | | 6 2 | | 6 2 | | |
| | | 9 2 | | 9 2 | | |
| | | 9 2 | | 7 2 | | |
| 129.2.56 | 25339 | 84 | Ba1Sn2Zn2 | | | |
| | -A | -GM | -M | -Z | -R | -X |
| | 5 4 | 7 2 | 5 4 | 7 2 | 3+4 4 | 3+4 4 |
| | 5 4 | 9 2 | 5 4 | 9 2 | 3+4 4 | 3+4 4 |
| | 5 4 | 7 2 | 5 4 | 9 2 | 3+4 4 | 3+4 4 |
| | 5 4 | 9 2 | 5 4 | 7 2 | 3+4 4 | 3+4 4 |
| | 5 4 | 7 2 | 5 4 | 9 2 | 3+4 4 | 3+4 4 |
| | 5 4 | 9 2 | 5 4 | 7 2 | 3+4 4 | 3+4 4 |
| | 5 4 | 6 2 | 5 4 | 8 2 | 3+4 4 | 3+4 4 |
| | 5 4 | 8 2 | 5 4 | 6 2 | 3+4 4 | 3+4 4 |
| | 5 4 | 7 2 | 5 4 | 9 2 | 3+4 4 | 3+4 4 |
| | 5 4 | 7 2 | 5 4 | 7 2 | 3+4 4 | 3+4 4 |
| | 5 4 | 9 2 | 5 4 | 6 2 | 3+4 4 | 3+4 4 |
| | 5 4 | 8 2 | 5 4 | 9 2 | 3+4 4 | 3+4 4 |
| | 5 4 | 7 2 | 5 4 | 8 2 | 3+4 4 | 3+4 4 |
| | 5 4 | 6 2 | 5 4 | 9 2 | 3+4 4 | 3+4 4 |
| | 5 4 | 9 2 | 5 4 | 7 2 | 3+4 4 | 3+4 4 |

Continued on next column

**TABLE LX– continued**

| TQC | ICSD | | NoE | | SF | |
|---|---|---|---|---|---|---|
| | 5 4 | 6 2 | 5 4 | 8 2 | 3+4 4 | 3+4 4 |
| | 5 4 | 8 2 | 5 4 | 6 2 | 3+4 4 | 3+4 4 |
| | 5 4 | 9 2 | 5 4 | 7 2 | 3+4 4 | 3+4 4 |
| | 5 4 | 7 2 | 5 4 | 9 2 | 3+4 4 | 3+4 4 |
| | 5 4 | 6 2 | 5 4 | 8 2 | 3+4 4 | 3+4 4 |
| | 5 4 | 6 2 | 5 4 | 6 2 | 3+4 4 | 3+4 4 |
| | | 8 2 | | 8 2 | | |
| | | 8 2 | | 6 2 | | |
| | | 7 2 | | 7 2 | | |
| | | 9 2 | | 9 2 | | |
| | | 6 2 | | 6 2 | | |
| | | 7 2 | | 7 2 | | |
| | | 8 2 | | 8 2 | | |
| | | 6 2 | | 6 2 | | |
| | | 9 2 | | 9 2 | | |
| | | 9 2 | | 9 2 | | |
| | | 9 2 | | 7 2 | | |
| | | 8 2 | | 6 2 | | |
| | | 7 2 | | 8 2 | | |
| | | 6 2 | | 9 2 | | |
| | | 7 2 | | 7 2 | | |
| | | 7 2 | | 7 2 | | |
| | | 6 2 | | 9 2 | | |
| | | 8 2 | | 6 2 | | |
| | | 6 2 | | 8 2 | | |
| | | 9 2 | | 7 2 | | |
| | | 9 2 | | 8 2 | | |
| 129.2.57 | 42460 | | 32 | | Ba1Ge1Mg1 | |
| | -A | -GM | -M | -Z | -R | -X |
| | 5 4 | 7 2 | 5 4 | 7 2 | 3+4 4 | 3+4 4 |
| | 5 4 | 9 2 | 5 4 | 9 2 | 3+4 4 | 3+4 4 |
| | 5 4 | 7 2 | 5 4 | 7 2 | 3+4 4 | 3+4 4 |
| | 5 4 | 9 2 | 5 4 | 9 2 | 3+4 4 | 3+4 4 |
| | 5 4 | 7 2 | 5 4 | 7 2 | 3+4 4 | 3+4 4 |
| | 5 4 | 6 2 | 5 4 | 6 2 | 3+4 4 | 3+4 4 |
| | 5 4 | 9 2 | 5 4 | 9 2 | 3+4 4 | 3+4 4 |
| | 5 4 | 8 2 | 5 4 | 8 2 | 3+4 4 | 3+4 4 |
| | | 7 2 | | 9 2 | | |
| | | 9 2 | | 7 2 | | |
| | | 7 2 | | 9 2 | | |
| | | 7 2 | | 7 2 | | |
| | | 6 2 | | 9 2 | | |
| | | 9 2 | | 8 2 | | |
| | | 8 2 | | 8 2 | | |
| | | 9 2 | | 7 2 | | |
| 129.2.58 | 236308 | | 84 | | Ba1Cu2Sb2 | |
| | -A | -GM | -M | -Z | -R | -X |
| | 5 4 | 7 2 | 5 4 | 7 2 | 3+4 4 | 3+4 4 |
| | 5 4 | 9 2 | 5 4 | 9 2 | 3+4 4 | 3+4 4 |
| | 5 4 | 7 2 | 5 4 | 9 2 | 3+4 4 | 3+4 4 |
| | 5 4 | 9 2 | 5 4 | 7 2 | 3+4 4 | 3+4 4 |
| | 5 4 | 7 2 | 5 4 | 9 2 | 3+4 4 | 3+4 4 |
| | 5 4 | 9 2 | 5 4 | 7 2 | 3+4 4 | 3+4 4 |
| | 5 4 | 6 2 | 5 4 | 6 2 | 3+4 4 | 3+4 4 |
| | 5 4 | 8 2 | 5 4 | 8 2 | 3+4 4 | 3+4 4 |
| | 5 4 | 7 2 | 5 4 | 9 2 | 3+4 4 | 3+4 4 |
| | 5 4 | 7 2 | 5 4 | 7 2 | 3+4 4 | 3+4 4 |
| | 5 4 | 9 2 | 5 4 | 9 2 | 3+4 4 | 3+4 4 |
| | 5 4 | 8 2 | 5 4 | 6 2 | 3+4 4 | 3+4 4 |
| | 5 4 | 9 2 | 5 4 | 7 2 | 3+4 4 | 3+4 4 |
| | 5 4 | 7 2 | 5 4 | 8 2 | 3+4 4 | 3+4 4 |

Continued on next column

**TABLE LX– continued**

| TQC | ICSD | | NoE | | SF | |
|---|---|---|---|---|---|---|
| | 5 4 | 6 2 | 5 4 | 9 2 | 3+4 4 | 3+4 4 |
| | 5 4 | 7 2 | 5 4 | 7 2 | 3+4 4 | 3+4 4 |
| | 5 4 | 9 2 | 5 4 | 9 2 | 3+4 4 | 3+4 4 |
| | 5 4 | 6 2 | 5 4 | 8 2 | 3+4 4 | 3+4 4 |
| | 5 4 | 9 2 | 5 4 | 8 2 | 3+4 4 | 3+4 4 |
| | 5 4 | 8 2 | 5 4 | 9 2 | 3+4 4 | 3+4 4 |
| | 5 4 | 9 2 | 5 4 | 7 2 | 3+4 4 | 3+4 4 |
| | | 6 2 | | 6 2 | | |
| | | 8 2 | | 6 2 | | |
| | | 7 2 | | 8 2 | | |
| | | 6 2 | | 7 2 | | |
| | | 7 2 | | 9 2 | | |
| | | 8 2 | | 6 2 | | |
| | | 6 2 | | 6 2 | | |
| | | 8 2 | | 7 2 | | |
| | | 9 2 | | 8 2 | | |
| | | 6 2 | | 6 2 | | |
| | | 8 2 | | 6 2 | | |
| | | 7 2 | | 9 2 | | |
| | | 9 2 | | 8 2 | | |
| | | 7 2 | | 9 2 | | |
| | | 6 2 | | 7 2 | | |
| | | 7 2 | | 6 2 | | |
| | | 6 2 | | 7 2 | | |
| | | 8 2 | | 9 2 | | |
| | | 6 2 | | 8 2 | | |
| | | 9 2 | | 9 2 | | |
| | | 9 2 | | 6 2 | | |
| 129.2.59 | 428691 | | 80 | | Au2Sn2Sr1 | |
| | -A | -GM | -M | -Z | -R | -X |
| | 5 4 | 7 2 | 5 4 | 7 2 | 3+4 4 | 3+4 4 |
| | 5 4 | 9 2 | 5 4 | 9 2 | 3+4 4 | 3+4 4 |
| | 5 4 | 7 2 | 5 4 | 7 2 | 3+4 4 | 3+4 4 |
| | 5 4 | 9 2 | 5 4 | 9 2 | 3+4 4 | 3+4 4 |
| | 5 4 | 7 2 | 5 4 | 7 2 | 3+4 4 | 3+4 4 |
| | 5 4 | 9 2 | 5 4 | 9 2 | 3+4 4 | 3+4 4 |
| | 5 4 | 6 2 | 5 4 | 6 2 | 3+4 4 | 3+4 4 |
| | 5 4 | 8 2 | 5 4 | 8 2 | 3+4 4 | 3+4 4 |
| | 5 4 | 7 2 | 5 4 | 7 2 | 3+4 4 | 3+4 4 |
| | 5 4 | 7 2 | 5 4 | 9 2 | 3+4 4 | 3+4 4 |
| | 5 4 | 9 2 | 5 4 | 7 2 | 3+4 4 | 3+4 4 |
| | 5 4 | 8 2 | 5 4 | 8 2 | 3+4 4 | 3+4 4 |
| | 5 4 | 7 2 | 5 4 | 9 2 | 3+4 4 | 3+4 4 |
| | 5 4 | 9 2 | 5 4 | 7 2 | 3+4 4 | 3+4 4 |
| | 5 4 | 9 2 | 5 4 | 9 2 | 3+4 4 | 3+4 4 |
| | 5 4 | 6 2 | 5 4 | 6 2 | 3+4 4 | 3+4 4 |
| | 5 4 | 8 2 | 5 4 | 6 2 | 3+4 4 | 3+4 4 |
| | 5 4 | 6 2 | 5 4 | 8 2 | 3+4 4 | 3+4 4 |
| | 5 4 | 7 2 | 5 4 | 7 2 | 3+4 4 | 3+4 4 |
| | 5 4 | 9 2 | 5 4 | 7 2 | 3+4 4 | 3+4 4 |
| | | 6 2 | | 9 2 | | |
| | | 7 2 | | 6 2 | | |
| | | 8 2 | | 8 2 | | |
| | | 9 2 | | 9 2 | | |
| | | 6 2 | | 6 2 | | |
| | | 7 2 | | 7 2 | | |
| | | 6 2 | | 8 2 | | |
| | | 8 2 | | 8 2 | | |
| | | 8 2 | | 6 2 | | |
| | | 6 2 | | 9 2 | | |
| | | 9 2 | | 8 2 | | |

Continued on next column

**TABLE LX– continued**

| TQC | ICSD | | NoE | | SF | |
|---|---|---|---|---|---|---|
| | | 9 2 | | 7 2 | | |
| | | 7 2 | | 8 2 | | |
| | | 8 2 | | 7 2 | | |
| | | 7 2 | | 8 2 | | |
| | | 6 2 | | 6 2 | | |
| | | 6 2 | | 9 2 | | |
| | | 7 2 | | 9 2 | | |
| | | 8 2 | | 7 2 | | |
| | | 9 2 | | 6 2 | | |
| 129.2.60 | 427711 | | 80 | | Au2Sn2Sr1 | |
| | -A | -GM | -M | -Z | -R | -X |
| | 5 4 | 7 2 | 5 4 | 7 2 | 3+4 4 | 3+4 4 |
| | 5 4 | 9 2 | 5 4 | 9 2 | 3+4 4 | 3+4 4 |
| | 5 4 | 7 2 | 5 4 | 7 2 | 3+4 4 | 3+4 4 |
| | 5 4 | 9 2 | 5 4 | 9 2 | 3+4 4 | 3+4 4 |
| | 5 4 | 7 2 | 5 4 | 7 2 | 3+4 4 | 3+4 4 |
| | 5 4 | 9 2 | 5 4 | 9 2 | 3+4 4 | 3+4 4 |
| | 5 4 | 6 2 | 5 4 | 6 2 | 3+4 4 | 3+4 4 |
| | 5 4 | 8 2 | 5 4 | 8 2 | 3+4 4 | 3+4 4 |
| | 5 4 | 7 2 | 5 4 | 7 2 | 3+4 4 | 3+4 4 |
| | 5 4 | 7 2 | 5 4 | 9 2 | 3+4 4 | 3+4 4 |
| | 5 4 | 9 2 | 5 4 | 7 2 | 3+4 4 | 3+4 4 |
| | 5 4 | 8 2 | 5 4 | 8 2 | 3+4 4 | 3+4 4 |
| | 5 4 | 7 2 | 5 4 | 9 2 | 3+4 4 | 3+4 4 |
| | 5 4 | 9 2 | 5 4 | 7 2 | 3+4 4 | 3+4 4 |
| | 5 4 | 9 2 | 5 4 | 9 2 | 3+4 4 | 3+4 4 |
| | 5 4 | 6 2 | 5 4 | 6 2 | 3+4 4 | 3+4 4 |
| | 5 4 | 8 2 | 5 4 | 6 2 | 3+4 4 | 3+4 4 |
| | 5 4 | 6 2 | 5 4 | 8 2 | 3+4 4 | 3+4 4 |
| | 5 4 | 7 2 | 5 4 | 7 2 | 3+4 4 | 3+4 4 |
| | 5 4 | 9 2 | 5 4 | 9 2 | 3+4 4 | 3+4 4 |
| | | 6 2 | | 7 2 | | |
| | | 8 2 | | 6 2 | | |
| | | 7 2 | | 8 2 | | |
| | | 9 2 | | 9 2 | | |
| | | 6 2 | | 6 2 | | |
| | | 7 2 | | 7 2 | | |
| | | 8 2 | | 8 2 | | |
| | | 6 2 | | 8 2 | | |
| | | 8 2 | | 6 2 | | |
| | | 6 2 | | 9 2 | | |
| | | 9 2 | | 8 2 | | |
| | | 9 2 | | 7 2 | | |
| | | 7 2 | | 8 2 | | |
| | | 8 2 | | 7 2 | | |
| | | 7 2 | | 8 2 | | |
| | | 6 2 | | 6 2 | | |
| | | 6 2 | | 9 2 | | |
| | | 7 2 | | 9 2 | | |
| | | 8 2 | | 7 2 | | |
| | | 9 2 | | 6 2 | | |
| 129.2.61 | 107956 | | 88 | | As2Nd1Ni2 | |
| | -A | -GM | -M | -Z | -R | -X |
| | 5 4 | 7 2 | 5 4 | 7 2 | 3+4 4 | 3+4 4 |
| | 5 4 | 9 2 | 5 4 | 9 2 | 3+4 4 | 3+4 4 |
| | 5 4 | 7 2 | 5 4 | 9 2 | 3+4 4 | 3+4 4 |
| | 5 4 | 9 2 | 5 4 | 7 2 | 3+4 4 | 3+4 4 |
| | 5 4 | 7 2 | 5 4 | 9 2 | 3+4 4 | 3+4 4 |
| | 5 4 | 9 2 | 5 4 | 7 2 | 3+4 4 | 3+4 4 |
| | 5 4 | 6 2 | 5 4 | 8 2 | 3+4 4 | 3+4 4 |
| | 5 4 | 8 2 | 5 4 | 6 2 | 3+4 4 | 3+4 4 |

Continued on next column

| TQC | ICSD | | NoE | | SF | |
|---|---|---|---|---|---|---|
| | 5 4 | 7 2 | 5 4 | 9 2 | 3+4 4 | 3+4 4 |
| | 5 4 | 7 2 | 5 4 | 7 2 | 3+4 4 | 3+4 4 |
| | 5 4 | 9 2 | 5 4 | 9 2 | 3+4 4 | 3+4 4 |
| | 5 4 | 8 2 | 5 4 | 6 2 | 3+4 4 | 3+4 4 |
| | 5 4 | 9 2 | 5 4 | 7 2 | 3+4 4 | 3+4 4 |
| | 5 4 | 7 2 | 5 4 | 8 2 | 3+4 4 | 3+4 4 |
| | 5 4 | 6 2 | 5 4 | 9 2 | 3+4 4 | 3+4 4 |
| | 5 4 | 7 2 | 5 4 | 7 2 | 3+4 4 | 3+4 4 |
| | 5 4 | 9 2 | 5 4 | 9 2 | 3+4 4 | 3+4 4 |
| | 5 4 | 6 2 | 5 4 | 8 2 | 3+4 4 | 3+4 4 |
| | 5 4 | 9 2 | 5 4 | 9 2 | 3+4 4 | 3+4 4 |
| | 5 4 | 8 2 | 5 4 | 7 2 | 3+4 4 | 3+4 4 |
| | 5 4 | 7 2 | 5 4 | 8 2 | 3+4 4 | 3+4 4 |
| | 5 4 | 6 2 | 5 4 | 6 2 | 3+4 4 | 3+4 4 |
| | | 6 2 | | 6 2 | | |
| | | 9 2 | | 8 2 | | |
| | | 6 2 | | 7 2 | | |
| | | 8 2 | | 6 2 | | |
| | | 7 2 | | 9 2 | | |
| | | 8 2 | | 8 2 | | |
| | | 8 2 | | 6 2 | | |
| | | 9 2 | | 7 2 | | |
| | | 7 2 | | 7 2 | | |
| | | 6 2 | | 6 2 | | |
| | | 6 2 | | 9 2 | | |
| | | 7 2 | | 8 2 | | |
| | | 7 2 | | 6 2 | | |
| | | 6 2 | | 6 2 | | |
| | | 8 2 | | 9 2 | | |
| | | 9 2 | | 9 2 | | |
| | | 8 2 | | 7 2 | | |
| | | 9 2 | | 6 2 | | |
| | | 6 2 | | 8 2 | | |
| | | 8 2 | | 6 2 | | |
| | | 6 2 | | 7 2 | | |
| | | 7 2 | | 8 2 | | |
| 129.2.62 | 412550 | | 64 | | As2Cu1Y1 | |
| | -A | -GM | -M | -Z | -R | -X |
| | 5 4 | 7 2 | 5 4 | 7 2 | 3+4 4 | 3+4 4 |
| | 5 4 | 9 2 | 5 4 | 9 2 | 3+4 4 | 3+4 4 |
| | 5 4 | 7 2 | 5 4 | 7 2 | 3+4 4 | 3+4 4 |
| | 5 4 | 9 2 | 5 4 | 9 2 | 3+4 4 | 3+4 4 |
| | 5 4 | 7 2 | 5 4 | 7 2 | 3+4 4 | 3+4 4 |
| | 5 4 | 9 2 | 5 4 | 9 2 | 3+4 4 | 3+4 4 |
| | 5 4 | 6 2 | 5 4 | 6 2 | 3+4 4 | 3+4 4 |
| | 5 4 | 8 2 | 5 4 | 8 2 | 3+4 4 | 3+4 4 |
| | 5 4 | 7 2 | 5 4 | 7 2 | 3+4 4 | 3+4 4 |
| | 5 4 | 7 2 | 5 4 | 9 2 | 3+4 4 | 3+4 4 |
| | 5 4 | 9 2 | 5 4 | 7 2 | 3+4 4 | 3+4 4 |
| | 5 4 | 8 2 | 5 4 | 8 2 | 3+4 4 | 3+4 4 |
| | 5 4 | 7 2 | 5 4 | 9 2 | 3+4 4 | 3+4 4 |
| | 5 4 | 9 2 | 5 4 | 7 2 | 3+4 4 | 3+4 4 |
| | 5 4 | 8 2 | 5 4 | 6 2 | 3+4 4 | 3+4 4 |
| | 5 4 | 9 2 | 5 4 | 7 2 | 3+4 4 | 3+4 4 |
| | | 6 2 | | 9 2 | | |
| | | 6 2 | | 8 2 | | |
| | | 9 2 | | 6 2 | | |
| | | 7 2 | | 7 2 | | |
| | | 7 2 | | 9 2 | | |
| | | 8 2 | | 6 2 | | |
| | | 6 2 | | 8 2 | | |

Continued on next column

**TABLE LX– continued**

| TQC | ICSD | NoE | SF | | | |
|---|---|---|---|---|---|---|
| | | 7 2 | | 9 2 | | |
| | | 6 2 | | 8 2 | | |
| | | 9 2 | | 7 2 | | |
| | | 7 2 | | 9 2 | | |
| | | 6 2 | | 8 2 | | |
| | | 6 2 | | 7 2 | | |
| | | 9 2 | | 6 2 | | |
| | | 8 2 | | 8 2 | | |
| | | 9 2 | | 7 2 | | |
| 129.2.63 | 174358 | 92 | As2Cu1Lu1 | | | |
| | -A | -GM | -M | -Z | -R | -X |
| | 5 4 | 7 2 | 5 4 | 7 2 | 3+4 4 | 3+4 4 |
| | 5 4 | 9 2 | 5 4 | 9 2 | 3+4 4 | 3+4 4 |
| | 5 4 | 7 2 | 5 4 | 7 2 | 3+4 4 | 3+4 4 |
| | 5 4 | 9 2 | 5 4 | 9 2 | 3+4 4 | 3+4 4 |
| | 5 4 | 7 2 | 5 4 | 7 2 | 3+4 4 | 3+4 4 |
| | 5 4 | 9 2 | 5 4 | 9 2 | 3+4 4 | 3+4 4 |
| | 5 4 | 6 2 | 5 4 | 6 2 | 3+4 4 | 3+4 4 |
| | 5 4 | 8 2 | 5 4 | 8 2 | 3+4 4 | 3+4 4 |
| | 5 4 | 7 2 | 5 4 | 7 2 | 3+4 4 | 3+4 4 |
| | 5 4 | 7 2 | 5 4 | 9 2 | 3+4 4 | 3+4 4 |
| | 5 4 | 9 2 | 5 4 | 7 2 | 3+4 4 | 3+4 4 |
| | 5 4 | 8 2 | 5 4 | 8 2 | 3+4 4 | 3+4 4 |
| | 5 4 | 7 2 | 5 4 | 9 2 | 3+4 4 | 3+4 4 |
| | 5 4 | 9 2 | 5 4 | 6 2 | 3+4 4 | 3+4 4 |
| | 5 4 | 6 2 | 5 4 | 7 2 | 3+4 4 | 3+4 4 |
| | 5 4 | 6 2 | 5 4 | 9 2 | 3+4 4 | 3+4 4 |
| | 5 4 | 8 2 | 5 4 | 6 2 | 3+4 4 | 3+4 4 |
| | 5 4 | 8 2 | 5 4 | 8 2 | 3+4 4 | 3+4 4 |
| | 5 4 | 7 2 | 5 4 | 8 2 | 3+4 4 | 3+4 4 |
| | 5 4 | 9 2 | 5 4 | 7 2 | 3+4 4 | 3+4 4 |
| | 5 4 | 9 2 | 5 4 | 7 2 | 3+4 4 | 3+4 4 |
| | 5 4 | 6 2 | 5 4 | 6 2 | 3+4 4 | 3+4 4 |
| | 5 4 | 7 2 | 5 4 | 9 2 | 3+4 4 | 3+4 4 |
| | | 8 2 | | 9 2 | | |
| | | 7 2 | | 7 2 | | |
| | | 6 2 | | 6 2 | | |
| | | 9 2 | | 8 2 | | |
| | | 8 2 | | 8 2 | | |
| | | 8 2 | | 6 2 | | |
| | | 9 2 | | 7 2 | | |
| | | 6 2 | | 9 2 | | |
| | | 9 2 | | 8 2 | | |
| | | 6 2 | | 6 2 | | |
| | | 7 2 | | 7 2 | | |
| | | 7 2 | | 9 2 | | |
| | | 8 2 | | 6 2 | | |
| | | 6 2 | | 8 2 | | |
| | | 7 2 | | 9 2 | | |
| | | 6 2 | | 8 2 | | |
| | | 9 2 | | 7 2 | | |
| | | 7 2 | | 7 2 | | |
| | | 6 2 | | 9 2 | | |
| | | 6 2 | | 8 2 | | |
| | | 9 2 | | 6 2 | | |
| | | 9 2 | | 8 2 | | |
| | | 8 2 | | 7 2 | | |
| 129.2.64 | 94440 | 84 | As2Cu1Ho1 | | | |
| | -A | -GM | -M | -Z | -R | -X |
| | 5 4 | 7 2 | 5 4 | 7 2 | 3+4 4 | 3+4 4 |
| | 5 4 | 9 2 | 5 4 | 9 2 | 3+4 4 | 3+4 4 |

Continued on next column

**TABLE LX– continued**

| TQC | ICSD | NoE | SF | | | |
|---|---|---|---|---|---|---|
| | 5 4 | 7 2 | 5 4 | 7 2 | 3+4 4 | 3+4 4 |
| | 5 4 | 9 2 | 5 4 | 9 2 | 3+4 4 | 3+4 4 |
| | 5 4 | 7 2 | 5 4 | 7 2 | 3+4 4 | 3+4 4 |
| | 5 4 | 9 2 | 5 4 | 9 2 | 3+4 4 | 3+4 4 |
| | 5 4 | 6 2 | 5 4 | 6 2 | 3+4 4 | 3+4 4 |
| | 5 4 | 8 2 | 5 4 | 8 2 | 3+4 4 | 3+4 4 |
| | 5 4 | 7 2 | 5 4 | 7 2 | 3+4 4 | 3+4 4 |
| | 5 4 | 7 2 | 5 4 | 9 2 | 3+4 4 | 3+4 4 |
| | 5 4 | 9 2 | 5 4 | 7 2 | 3+4 4 | 3+4 4 |
| | 5 4 | 8 2 | 5 4 | 8 2 | 3+4 4 | 3+4 4 |
| | 5 4 | 7 2 | 5 4 | 9 2 | 3+4 4 | 3+4 4 |
| | 5 4 | 9 2 | 5 4 | 7 2 | 3+4 4 | 3+4 4 |
| | 5 4 | 8 2 | 5 4 | 6 2 | 3+4 4 | 3+4 4 |
| | 5 4 | 9 2 | 5 4 | 7 2 | 3+4 4 | 3+4 4 |
| | 5 4 | 6 2 | 5 4 | 8 2 | 3+4 4 | 3+4 4 |
| | 5 4 | 6 2 | 5 4 | 9 2 | 3+4 4 | 3+4 4 |
| | 5 4 | 9 2 | 5 4 | 6 2 | 3+4 4 | 3+4 4 |
| | 5 4 | 7 2 | 5 4 | 7 2 | 3+4 4 | 3+4 4 |
| | 5 4 | 7 2 | 5 4 | 9 2 | 3+4 4 | 3+4 4 |
| | | 8 2 | | 6 2 | | |
| | | 6 2 | | 8 2 | | |
| | | 7 2 | | 9 2 | | |
| | | 6 2 | | 8 2 | | |
| | | 9 2 | | 7 2 | | |
| | | 7 2 | | 9 2 | | |
| | | 6 2 | | 8 2 | | |
| | | 8 2 | | 7 2 | | |
| | | 6 2 | | 6 2 | | |
| | | 9 2 | | 6 2 | | |
| | | 8 2 | | 8 2 | | |
| | | 7 2 | | 9 2 | | |
| | | 6 2 | | 7 2 | | |
| | | 9 2 | | 8 2 | | |
| | | 6 2 | | 6 2 | | |
| | | 9 2 | | 7 2 | | |
| | | 8 2 | | 6 2 | | |
| | | 6 2 | | 8 2 | | |
| | | 7 2 | | 9 2 | | |
| | | 8 2 | | 8 2 | | |
| | | 9 2 | | 7 2 | | |
| 129.2.65 | 412178 | 84 | As2Cu1Ho1 | | | |
| | -A | -GM | -M | -Z | -R | -X |
| | 5 4 | 7 2 | 5 4 | 7 2 | 3+4 4 | 3+4 4 |
| | 5 4 | 9 2 | 5 4 | 9 2 | 3+4 4 | 3+4 4 |
| | 5 4 | 7 2 | 5 4 | 7 2 | 3+4 4 | 3+4 4 |
| | 5 4 | 9 2 | 5 4 | 9 2 | 3+4 4 | 3+4 4 |
| | 5 4 | 7 2 | 5 4 | 7 2 | 3+4 4 | 3+4 4 |
| | 5 4 | 9 2 | 5 4 | 9 2 | 3+4 4 | 3+4 4 |
| | 5 4 | 6 2 | 5 4 | 6 2 | 3+4 4 | 3+4 4 |
| | 5 4 | 8 2 | 5 4 | 8 2 | 3+4 4 | 3+4 4 |
| | 5 4 | 7 2 | 5 4 | 7 2 | 3+4 4 | 3+4 4 |
| | 5 4 | 7 2 | 5 4 | 9 2 | 3+4 4 | 3+4 4 |
| | 5 4 | 9 2 | 5 4 | 7 2 | 3+4 4 | 3+4 4 |
| | 5 4 | 8 2 | 5 4 | 8 2 | 3+4 4 | 3+4 4 |
| | 5 4 | 7 2 | 5 4 | 9 2 | 3+4 4 | 3+4 4 |
| | 5 4 | 9 2 | 5 4 | 7 2 | 3+4 4 | 3+4 4 |
| | 5 4 | 8 2 | 5 4 | 6 2 | 3+4 4 | 3+4 4 |
| | 5 4 | 9 2 | 5 4 | 7 2 | 3+4 4 | 3+4 4 |
| | 5 4 | 6 2 | 5 4 | 8 2 | 3+4 4 | 3+4 4 |
| | 5 4 | 6 2 | 5 4 | 9 2 | 3+4 4 | 3+4 4 |
| | 5 4 | 9 2 | 5 4 | 6 2 | 3+4 4 | 3+4 4 |

Continued on next column

**TABLE LX– continued**

| TQC | ICSD | | NoE | | SF | |
|---|---|---|---|---|---|---|
| | 5 4 | 7 2 | 5 4 | 7 2 | 3+4 4 | 3+4 4 |
| | 5 4 | 7 2 | 5 4 | 9 2 | 3+4 4 | 3+4 4 |
| | | 8 2 | | 6 2 | | |
| | | 6 2 | | 8 2 | | |
| | | 7 2 | | 9 2 | | |
| | | 6 2 | | 8 2 | | |
| | | 9 2 | | 7 2 | | |
| | | 7 2 | | 9 2 | | |
| | | 6 2 | | 8 2 | | |
| | | 8 2 | | 7 2 | | |
| | | 6 2 | | 6 2 | | |
| | | 9 2 | | 6 2 | | |
| | | 8 2 | | 8 2 | | |
| | | 7 2 | | 9 2 | | |
| | | 6 2 | | 7 2 | | |
| | | 9 2 | | 8 2 | | |
| | | 6 2 | | 6 2 | | |
| | | 9 2 | | 7 2 | | |
| | | 8 2 | | 6 2 | | |
| | | 6 2 | | 8 2 | | |
| | | 7 2 | | 9 2 | | |
| | | 8 2 | | 8 2 | | |
| | | 9 2 | | 7 2 | | |
| 129.2.66 | 44038 | | 52 | | As1Np1Te1 | |
| | -A | -GM | -M | -Z | -R | -X |
| | 5 4 | 7 2 | 5 4 | 7 2 | 3+4 4 | 3+4 4 |
| | 5 4 | 9 2 | 5 4 | 9 2 | 3+4 4 | 3+4 4 |
| | 5 4 | 7 2 | 5 4 | 9 2 | 3+4 4 | 3+4 4 |
| | 5 4 | 9 2 | 5 4 | 7 2 | 3+4 4 | 3+4 4 |
| | 5 4 | 7 2 | 5 4 | 9 2 | 3+4 4 | 3+4 4 |
| | 5 4 | 9 2 | 5 4 | 7 2 | 3+4 4 | 3+4 4 |
| | 5 4 | 6 2 | 5 4 | 8 2 | 3+4 4 | 3+4 4 |
| | 5 4 | 8 2 | 5 4 | 6 2 | 3+4 4 | 3+4 4 |
| | 5 4 | 7 2 | 5 4 | 7 2 | 3+4 4 | 3+4 4 |
| | 5 4 | 7 2 | 5 4 | 9 2 | 3+4 4 | 3+4 4 |
| | 5 4 | 9 2 | 5 4 | 7 2 | 3+4 4 | 3+4 4 |
| | 5 4 | 8 2 | 5 4 | 8 2 | 3+4 4 | 3+4 4 |
| | 5 4 | 7 2 | 5 4 | 9 2 | 3+4 4 | 3+4 4 |
| | | 9 2 | | 6 2 | | |
| | | 6 2 | | 9 2 | | |
| | | 7 2 | | 7 2 | | |
| | | 7 2 | | 9 2 | | |
| | | 6 2 | | 8 2 | | |
| | | 9 2 | | 7 2 | | |
| | | 6 2 | | 6 2 | | |
| | | 8 2 | | 7 2 | | |
| | | 8 2 | | 6 2 | | |
| | | 9 2 | | 6 2 | | |
| | | 6 2 | | 8 2 | | |
| | | 7 2 | | 9 2 | | |
| | | 8 2 | | 8 2 | | |
| 129.2.67 | 50242 | | 64 | | As1Mn1Nd1O1 | |
| | -A | -GM | -M | -Z | -R | -X |
| | 5 4 | 7 2 | 5 4 | 7 2 | 3+4 4 | 3+4 4 |
| | 5 4 | 9 2 | 5 4 | 9 2 | 3+4 4 | 3+4 4 |
| | 5 4 | 7 2 | 5 4 | 7 2 | 3+4 4 | 3+4 4 |
| | 5 4 | 9 2 | 5 4 | 9 2 | 3+4 4 | 3+4 4 |
| | 5 4 | 7 2 | 5 4 | 7 2 | 3+4 4 | 3+4 4 |
| | 5 4 | 8 2 | 5 4 | 8 2 | 3+4 4 | 3+4 4 |
| | 5 4 | 6 2 | 5 4 | 6 2 | 3+4 4 | 3+4 4 |
| | 5 4 | 9 2 | 5 4 | 9 2 | 3+4 4 | 3+4 4 |

Continued on next column

**TABLE LX– continued**

| TQC | ICSD | | NoE | | SF | |
|---|---|---|---|---|---|---|
| | 5 4 | 8 2 | 5 4 | 8 2 | 3+4 4 | 3+4 4 |
| | 5 4 | 7 2 | 5 4 | 7 2 | 3+4 4 | 3+4 4 |
| | 5 4 | 7 2 | 5 4 | 9 2 | 3+4 4 | 3+4 4 |
| | 5 4 | 9 2 | 5 4 | 7 2 | 3+4 4 | 3+4 4 |
| | 5 4 | 9 2 | 5 4 | 9 2 | 3+4 4 | 3+4 4 |
| | 5 4 | 7 2 | 5 4 | 7 2 | 3+4 4 | 3+4 4 |
| | 5 4 | 6 2 | 5 4 | 6 2 | 3+4 4 | 3+4 4 |
| | 5 4 | 6 2 | 5 4 | 6 2 | 3+4 4 | 3+4 4 |
| | | 7 2 | | 9 2 | | |
| | | 9 2 | | 8 2 | | |
| | | 8 2 | | 9 2 | | |
| | | 9 2 | | 7 2 | | |
| | | 9 2 | | 7 2 | | |
| | | 8 2 | | 6 2 | | |
| | | 7 2 | | 9 2 | | |
| | | 6 2 | | 8 2 | | |
| | | 6 2 | | 8 2 | | |
| | | 8 2 | | 6 2 | | |
| | | 7 2 | | 9 2 | | |
| | | 6 2 | | 7 2 | | |
| | | 9 2 | | 6 2 | | |
| | | 6 2 | | 8 2 | | |
| | | 7 2 | | 6 2 | | |
| | | 8 2 | | 9 2 | | |
| 129.2.68 | 262831 | | 60 | | As1La1O1Ru1 | |
| | -A | -GM | -M | -Z | -R | -X |
| | 5 4 | 7 2 | 5 4 | 7 2 | 3+4 4 | 3+4 4 |
| | 5 4 | 9 2 | 5 4 | 9 2 | 3+4 4 | 3+4 4 |
| | 5 4 | 7 2 | 5 4 | 9 2 | 3+4 4 | 3+4 4 |
| | 5 4 | 9 2 | 5 4 | 9 2 | 3+4 4 | 3+4 4 |
| | 5 4 | 7 2 | 5 4 | 7 2 | 3+4 4 | 3+4 4 |
| | 5 4 | 8 2 | 5 4 | 6 2 | 3+4 4 | 3+4 4 |
| | 5 4 | 7 2 | 5 4 | 7 2 | 3+4 4 | 3+4 4 |
| | 5 4 | 9 2 | 5 4 | 9 2 | 3+4 4 | 3+4 4 |
| | 5 4 | 8 2 | 5 4 | 8 2 | 3+4 4 | 3+4 4 |
| | 5 4 | 6 2 | 5 4 | 6 2 | 3+4 4 | 3+4 4 |
| | 5 4 | 9 2 | 5 4 | 7 2 | 3+4 4 | 3+4 4 |
| | 5 4 | 7 2 | 5 4 | 9 2 | 3+4 4 | 3+4 4 |
| | 5 4 | 6 2 | 5 4 | 6 2 | 3+4 4 | 3+4 4 |
| | 5 4 | 7 2 | 5 4 | 9 2 | 3+4 4 | 3+4 4 |
| | 5 4 | 6 2 | 5 4 | 8 2 | 3+4 4 | 3+4 4 |
| | | 9 2 | | 7 2 | | |
| | | 7 2 | | 9 2 | | |
| | | 8 2 | | 8 2 | | |
| | | 6 2 | | 7 2 | | |
| | | 9 2 | | 7 2 | | |
| | | 6 2 | | 6 2 | | |
| | | 8 2 | | 6 2 | | |
| | | 7 2 | | 8 2 | | |
| | | 7 2 | | 7 2 | | |
| | | 9 2 | | 7 2 | | |
| | | 6 2 | | 6 2 | | |
| | | 9 2 | | 9 2 | | |
| | | 8 2 | | 8 2 | | |
| | | 9 2 | | 9 2 | | |
| | | 9 2 | | 7 2 | | |
| 129.2.69 | 90316 | | 64 | | As1Fe1O1Pr1 | |
| | -A | -GM | -M | -Z | -R | -X |
| | 5 4 | 7 2 | 5 4 | 7 2 | 3+4 4 | 3+4 4 |
| | 5 4 | 9 2 | 5 4 | 9 2 | 3+4 4 | 3+4 4 |
| | 5 4 | 7 2 | 5 4 | 7 2 | 3+4 4 | 3+4 4 |

Continued on next column

**TABLE LX– continued**

| TQC | ICSD | | NoE | | SF | |
|---|---|---|---|---|---|---|
| | 5 4 | 9 2 | 5 4 | 9 2 | 3+4 4 | 3+4 4 |
| | 5 4 | 7 2 | 5 4 | 7 2 | 3+4 4 | 3+4 4 |
| | 5 4 | 8 2 | 5 4 | 8 2 | 3+4 4 | 3+4 4 |
| | 5 4 | 6 2 | 5 4 | 6 2 | 3+4 4 | 3+4 4 |
| | 5 4 | 9 2 | 5 4 | 9 2 | 3+4 4 | 3+4 4 |
| | 5 4 | 8 2 | 5 4 | 8 2 | 3+4 4 | 3+4 4 |
| | 5 4 | 7 2 | 5 4 | 7 2 | 3+4 4 | 3+4 4 |
| | 5 4 | 7 2 | 5 4 | 9 2 | 3+4 4 | 3+4 4 |
| | 5 4 | 9 2 | 5 4 | 7 2 | 3+4 4 | 3+4 4 |
| | 5 4 | 9 2 | 5 4 | 9 2 | 3+4 4 | 3+4 4 |
| | 5 4 | 6 2 | 5 4 | 6 2 | 3+4 4 | 3+4 4 |
| | 5 4 | 7 2 | 5 4 | 7 2 | 3+4 4 | 3+4 4 |
| | 5 4 | 6 2 | 5 4 | 6 2 | 3+4 4 | 3+4 4 |
| | | 7 2 | | 9 2 | | |
| | | 9 2 | | 8 2 | | |
| | | 8 2 | | 7 2 | | |
| | | 9 2 | | 9 2 | | |
| | | 9 2 | | 7 2 | | |
| | | 8 2 | | 6 2 | | |
| | | 7 2 | | 9 2 | | |
| | | 6 2 | | 8 2 | | |
| | | 6 2 | | 8 2 | | |
| | | 8 2 | | 6 2 | | |
| | | 7 2 | | 9 2 | | |
| | | 6 2 | | 7 2 | | |
| | | 6 2 | | 8 2 | | |
| | | 9 2 | | 6 2 | | |
| | | 7 2 | | 9 2 | | |
| | | 6 2 | | 8 2 | | |
| 129.2.70 | 422000 | | 64 | | As1Fe1O1Pr1 | |
| | -A | -GM | -M | -Z | -R | -X |
| | 5 4 | 7 2 | 5 4 | 7 2 | 3+4 4 | 3+4 4 |
| | 5 4 | 9 2 | 5 4 | 9 2 | 3+4 4 | 3+4 4 |
| | 5 4 | 7 2 | 5 4 | 7 2 | 3+4 4 | 3+4 4 |
| | 5 4 | 9 2 | 5 4 | 9 2 | 3+4 4 | 3+4 4 |
| | 5 4 | 7 2 | 5 4 | 7 2 | 3+4 4 | 3+4 4 |
| | 5 4 | 8 2 | 5 4 | 8 2 | 3+4 4 | 3+4 4 |
| | 5 4 | 6 2 | 5 4 | 6 2 | 3+4 4 | 3+4 4 |
| | 5 4 | 9 2 | 5 4 | 9 2 | 3+4 4 | 3+4 4 |
| | 5 4 | 8 2 | 5 4 | 8 2 | 3+4 4 | 3+4 4 |
| | 5 4 | 7 2 | 5 4 | 7 2 | 3+4 4 | 3+4 4 |
| | 5 4 | 7 2 | 5 4 | 9 2 | 3+4 4 | 3+4 4 |
| | 5 4 | 9 2 | 5 4 | 7 2 | 3+4 4 | 3+4 4 |
| | 5 4 | 9 2 | 5 4 | 9 2 | 3+4 4 | 3+4 4 |
| | 5 4 | 6 2 | 5 4 | 6 2 | 3+4 4 | 3+4 4 |
| | 5 4 | 7 2 | 5 4 | 7 2 | 3+4 4 | 3+4 4 |
| | 5 4 | 6 2 | 5 4 | 6 2 | 3+4 4 | 3+4 4 |
| | | 7 2 | | 9 2 | | |
| | | 9 2 | | 8 2 | | |
| | | 8 2 | | 7 2 | | |
| | | 9 2 | | 9 2 | | |
| | | 9 2 | | 7 2 | | |
| | | 8 2 | | 6 2 | | |
| | | 7 2 | | 9 2 | | |
| | | 6 2 | | 8 2 | | |
| | | 6 2 | | 8 2 | | |
| | | 8 2 | | 6 2 | | |
| | | 7 2 | | 9 2 | | |
| | | 6 2 | | 7 2 | | |
| | | 6 2 | | 8 2 | | |
| | | 9 2 | | 6 2 | | |

Continued on next column

**TABLE LX– continued**

| TQC | ICSD | | NoE | | SF | |
|---|---|---|---|---|---|---|
| | | 7 2 | | 9 2 | | |
| | | 6 2 | | 8 2 | | |
| 129.2.71 | 246832 | | 64 | | As1Fe1O1Pr1 | |
| | -A | -GM | -M | -Z | -R | -X |
| | 5 4 | 7 2 | 5 4 | 7 2 | 3+4 4 | 3+4 4 |
| | 5 4 | 9 2 | 5 4 | 9 2 | 3+4 4 | 3+4 4 |
| | 5 4 | 7 2 | 5 4 | 7 2 | 3+4 4 | 3+4 4 |
| | 5 4 | 9 2 | 5 4 | 9 2 | 3+4 4 | 3+4 4 |
| | 5 4 | 7 2 | 5 4 | 7 2 | 3+4 4 | 3+4 4 |
| | 5 4 | 8 2 | 5 4 | 8 2 | 3+4 4 | 3+4 4 |
| | 5 4 | 6 2 | 5 4 | 6 2 | 3+4 4 | 3+4 4 |
| | 5 4 | 9 2 | 5 4 | 9 2 | 3+4 4 | 3+4 4 |
| | 5 4 | 8 2 | 5 4 | 8 2 | 3+4 4 | 3+4 4 |
| | 5 4 | 7 2 | 5 4 | 7 2 | 3+4 4 | 3+4 4 |
| | 5 4 | 7 2 | 5 4 | 9 2 | 3+4 4 | 3+4 4 |
| | 5 4 | 9 2 | 5 4 | 7 2 | 3+4 4 | 3+4 4 |
| | 5 4 | 9 2 | 5 4 | 9 2 | 3+4 4 | 3+4 4 |
| | 5 4 | 6 2 | 5 4 | 6 2 | 3+4 4 | 3+4 4 |
| | 5 4 | 7 2 | 5 4 | 7 2 | 3+4 4 | 3+4 4 |
| | 5 4 | 6 2 | 5 4 | 6 2 | 3+4 4 | 3+4 4 |
| | | 7 2 | | 9 2 | | |
| | | 9 2 | | 8 2 | | |
| | | 8 2 | | 7 2 | | |
| | | 9 2 | | 9 2 | | |
| | | 9 2 | | 7 2 | | |
| | | 8 2 | | 6 2 | | |
| | | 7 2 | | 9 2 | | |
| | | 6 2 | | 8 2 | | |
| | | 6 2 | | 8 2 | | |
| | | 8 2 | | 6 2 | | |
| | | 7 2 | | 9 2 | | |
| | | 6 2 | | 7 2 | | |
| | | 6 2 | | 8 2 | | |
| | | 9 2 | | 6 2 | | |
| | | 7 2 | | 9 2 | | |
| | | 6 2 | | 8 2 | | |
| 129.2.72 | 162830 | | 64 | | As1Fe1O1Pr1 | |
| | -A | -GM | -M | -Z | -R | -X |
| | 5 4 | 7 2 | 5 4 | 7 2 | 3+4 4 | 3+4 4 |
| | 5 4 | 9 2 | 5 4 | 9 2 | 3+4 4 | 3+4 4 |
| | 5 4 | 7 2 | 5 4 | 7 2 | 3+4 4 | 3+4 4 |
| | 5 4 | 9 2 | 5 4 | 9 2 | 3+4 4 | 3+4 4 |
| | 5 4 | 7 2 | 5 4 | 7 2 | 3+4 4 | 3+4 4 |
| | 5 4 | 8 2 | 5 4 | 8 2 | 3+4 4 | 3+4 4 |
| | 5 4 | 6 2 | 5 4 | 6 2 | 3+4 4 | 3+4 4 |
| | 5 4 | 9 2 | 5 4 | 9 2 | 3+4 4 | 3+4 4 |
| | 5 4 | 8 2 | 5 4 | 8 2 | 3+4 4 | 3+4 4 |
| | 5 4 | 7 2 | 5 4 | 7 2 | 3+4 4 | 3+4 4 |
| | 5 4 | 7 2 | 5 4 | 9 2 | 3+4 4 | 3+4 4 |
| | 5 4 | 9 2 | 5 4 | 7 2 | 3+4 4 | 3+4 4 |
| | 5 4 | 9 2 | 5 4 | 9 2 | 3+4 4 | 3+4 4 |
| | 5 4 | 6 2 | 5 4 | 6 2 | 3+4 4 | 3+4 4 |
| | 5 4 | 7 2 | 5 4 | 7 2 | 3+4 4 | 3+4 4 |
| | 5 4 | 6 2 | 5 4 | 6 2 | 3+4 4 | 3+4 4 |
| | | 7 2 | | 9 2 | | |
| | | 9 2 | | 8 2 | | |
| | | 8 2 | | 7 2 | | |
| | | 9 2 | | 9 2 | | |
| | | 9 2 | | 7 2 | | |
| | | 8 2 | | 6 2 | | |
| | | 7 2 | | 9 2 | | |

Continued on next column

**TABLE LX– continued**

| TQC | ICSD | NoE | SF | | | |
|---|---|---|---|---|---|---|
| | | 6 2 | | 8 2 | | |
| | | 6 2 | | 8 2 | | |
| | | 8 2 | | 6 2 | | |
| | | 7 2 | | 9 2 | | |
| | | 6 2 | | 7 2 | | |
| | | 9 2 | | 8 2 | | |
| | | 6 2 | | 6 2 | | |
| | | 7 2 | | 9 2 | | |
| | | 6 2 | | 8 2 | | |
| 129.2.73 | 93238 | 40 | As1Fe1Mn1 | | | |
| | -A | -GM | -M | -Z | -R | -X |
| | 5 4 | 7 2 | 5 4 | 7 2 | 3+4 4 | 3+4 4 |
| | 5 4 | 9 2 | 5 4 | 9 2 | 3+4 4 | 3+4 4 |
| | 5 4 | 7 2 | 5 4 | 7 2 | 3+4 4 | 3+4 4 |
| | 5 4 | 9 2 | 5 4 | 9 2 | 3+4 4 | 3+4 4 |
| | 5 4 | 9 2 | 5 4 | 9 2 | 3+4 4 | 3+4 4 |
| | 5 4 | 8 2 | 5 4 | 8 2 | 3+4 4 | 3+4 4 |
| | 5 4 | 6 2 | 5 4 | 7 2 | 3+4 4 | 3+4 4 |
| | 5 4 | 8 2 | 5 4 | 6 2 | 3+4 4 | 3+4 4 |
| | 5 4 | 7 2 | 5 4 | 6 2 | 3+4 4 | 3+4 4 |
| | 5 4 | 6 2 | 5 4 | 8 2 | 3+4 4 | 3+4 4 |
| | | 9 2 | | 6 2 | | |
| | | 7 2 | | 7 2 | | |
| | | 7 2 | | 9 2 | | |
| | | 6 2 | | 9 2 | | |
| | | 6 2 | | 8 2 | | |
| | | 6 2 | | 7 2 | | |
| | | 8 2 | | 6 2 | | |
| | | 7 2 | | 8 2 | | |
| | | 7 2 | | 9 2 | | |
| | | 6 2 | | 6 2 | | |
| 129.2.74 | 93237 | 40 | As1Fe1Mn1 | | | |
| | -A | -GM | -M | -Z | -R | -X |
| | 5 4 | 7 2 | 5 4 | 7 2 | 3+4 4 | 3+4 4 |
| | 5 4 | 9 2 | 5 4 | 9 2 | 3+4 4 | 3+4 4 |
| | 5 4 | 7 2 | 5 4 | 7 2 | 3+4 4 | 3+4 4 |
| | 5 4 | 9 2 | 5 4 | 9 2 | 3+4 4 | 3+4 4 |
| | 5 4 | 9 2 | 5 4 | 9 2 | 3+4 4 | 3+4 4 |
| | 5 4 | 8 2 | 5 4 | 8 2 | 3+4 4 | 3+4 4 |
| | 5 4 | 6 2 | 5 4 | 7 2 | 3+4 4 | 3+4 4 |
| | 5 4 | 7 2 | 5 4 | 6 2 | 3+4 4 | 3+4 4 |
| | 5 4 | 8 2 | 5 4 | 6 2 | 3+4 4 | 3+4 4 |
| | 5 4 | 6 2 | 5 4 | 8 2 | 3+4 4 | 3+4 4 |
| | | 9 2 | | 6 2 | | |
| | | 7 2 | | 7 2 | | |
| | | 7 2 | | 9 2 | | |
| | | 6 2 | | 9 2 | | |
| | | 6 2 | | 8 2 | | |
| | | 6 2 | | 7 2 | | |
| | | 8 2 | | 6 2 | | |
| | | 7 2 | | 8 2 | | |
| | | 7 2 | | 9 2 | | |
| | | 6 2 | | 6 2 | | |
| 129.2.75 | 180774 | 68 | As1Co1Nd1O1 | | | |
| | -A | -GM | -M | -Z | -R | -X |
| | 5 4 | 7 2 | 5 4 | 7 2 | 3+4 4 | 3+4 4 |
| | 5 4 | 9 2 | 5 4 | 9 2 | 3+4 4 | 3+4 4 |
| | 5 4 | 7 2 | 5 4 | 7 2 | 3+4 4 | 3+4 4 |
| | 5 4 | 9 2 | 5 4 | 9 2 | 3+4 4 | 3+4 4 |
| | 5 4 | 7 2 | 5 4 | 7 2 | 3+4 4 | 3+4 4 |
| | 5 4 | 8 2 | 5 4 | 8 2 | 3+4 4 | 3+4 4 |

Continued on next column

**TABLE LX– continued**

| TQC | ICSD | NoE | SF | | | |
|---|---|---|---|---|---|---|
| | 5 4 | 6 2 | 5 4 | 6 2 | 3+4 4 | 3+4 4 |
| | 5 4 | 9 2 | 5 4 | 9 2 | 3+4 4 | 3+4 4 |
| | 5 4 | 8 2 | 5 4 | 8 2 | 3+4 4 | 3+4 4 |
| | 5 4 | 7 2 | 5 4 | 7 2 | 3+4 4 | 3+4 4 |
| | 5 4 | 7 2 | 5 4 | 9 2 | 3+4 4 | 3+4 4 |
| | 5 4 | 9 2 | 5 4 | 7 2 | 3+4 4 | 3+4 4 |
| | 5 4 | 9 2 | 5 4 | 9 2 | 3+4 4 | 3+4 4 |
| | 5 4 | 7 2 | 5 4 | 7 2 | 3+4 4 | 3+4 4 |
| | 5 4 | 6 2 | 5 4 | 6 2 | 3+4 4 | 3+4 4 |
| | 5 4 | 6 2 | 5 4 | 6 2 | 3+4 4 | 3+4 4 |
| | 5 4 | 7 2 | 5 4 | 7 2 | 3+4 4 | 3+4 4 |
| | | 9 2 | | 9 2 | | |
| | | 8 2 | | 8 2 | | |
| | | 9 2 | | 9 2 | | |
| | | 9 2 | | 7 2 | | |
| | | 8 2 | | 6 2 | | |
| | | 7 2 | | 9 2 | | |
| | | 6 2 | | 8 2 | | |
| | | 6 2 | | 8 2 | | |
| | | 8 2 | | 6 2 | | |
| | | 7 2 | | 8 2 | | |
| | | 6 2 | | 9 2 | | |
| | | 7 2 | | 9 2 | | |
| | | 6 2 | | 8 2 | | |
| | | 9 2 | | 7 2 | | |
| | | 6 2 | | 7 2 | | |
| | | 6 2 | | 6 2 | | |
| | | 9 2 | | 8 2 | | |
| 129.2.76 | 180773 | 68 | As1Co1Nd1O1 | | | |
| | -A | -GM | -M | -Z | -R | -X |
| | 5 4 | 7 2 | 5 4 | 7 2 | 3+4 4 | 3+4 4 |
| | 5 4 | 9 2 | 5 4 | 9 2 | 3+4 4 | 3+4 4 |
| | 5 4 | 7 2 | 5 4 | 7 2 | 3+4 4 | 3+4 4 |
| | 5 4 | 9 2 | 5 4 | 9 2 | 3+4 4 | 3+4 4 |
| | 5 4 | 7 2 | 5 4 | 7 2 | 3+4 4 | 3+4 4 |
| | 5 4 | 8 2 | 5 4 | 8 2 | 3+4 4 | 3+4 4 |
| | 5 4 | 6 2 | 5 4 | 6 2 | 3+4 4 | 3+4 4 |
| | 5 4 | 9 2 | 5 4 | 9 2 | 3+4 4 | 3+4 4 |
| | 5 4 | 8 2 | 5 4 | 8 2 | 3+4 4 | 3+4 4 |
| | 5 4 | 7 2 | 5 4 | 7 2 | 3+4 4 | 3+4 4 |
| | 5 4 | 7 2 | 5 4 | 9 2 | 3+4 4 | 3+4 4 |
| | 5 4 | 9 2 | 5 4 | 7 2 | 3+4 4 | 3+4 4 |
| | 5 4 | 9 2 | 5 4 | 9 2 | 3+4 4 | 3+4 4 |
| | 5 4 | 7 2 | 5 4 | 7 2 | 3+4 4 | 3+4 4 |
| | 5 4 | 6 2 | 5 4 | 6 2 | 3+4 4 | 3+4 4 |
| | 5 4 | 6 2 | 5 4 | 6 2 | 3+4 4 | 3+4 4 |
| | 5 4 | 7 2 | 5 4 | 7 2 | 3+4 4 | 3+4 4 |
| | | 9 2 | | 9 2 | | |
| | | 8 2 | | 8 2 | | |
| | | 9 2 | | 9 2 | | |
| | | 9 2 | | 7 2 | | |
| | | 8 2 | | 6 2 | | |
| | | 7 2 | | 9 2 | | |
| | | 6 2 | | 8 2 | | |
| | | 6 2 | | 8 2 | | |
| | | 8 2 | | 6 2 | | |
| | | 7 2 | | 8 2 | | |
| | | 6 2 | | 9 2 | | |
| | | 7 2 | | 9 2 | | |
| | | 6 2 | | 8 2 | | |
| | | 9 2 | | 7 2 | | |

Continued on next column

**TABLE LX– continued**

| TQC | ICSD | | NoE | SF | | |
|---|---|---|---|---|---|---|
| | | 6 2 | | 7 2 | | |
| | | 6 2 | | 6 2 | | |
| | | 9 2 | | 8 2 | | |
| 129.2.77 | 180772 | | 68 | As1Co1Nd1O1 | | |
| | -A | -GM | -M | -Z | -R | -X |
| | 5 4 | 7 2 | 5 4 | 7 2 | 3+4 4 | 3+4 4 |
| | 5 4 | 9 2 | 5 4 | 9 2 | 3+4 4 | 3+4 4 |
| | 5 4 | 7 2 | 5 4 | 7 2 | 3+4 4 | 3+4 4 |
| | 5 4 | 9 2 | 5 4 | 9 2 | 3+4 4 | 3+4 4 |
| | 5 4 | 7 2 | 5 4 | 7 2 | 3+4 4 | 3+4 4 |
| | 5 4 | 8 2 | 5 4 | 8 2 | 3+4 4 | 3+4 4 |
| | 5 4 | 6 2 | 5 4 | 6 2 | 3+4 4 | 3+4 4 |
| | 5 4 | 9 2 | 5 4 | 9 2 | 3+4 4 | 3+4 4 |
| | 5 4 | 8 2 | 5 4 | 8 2 | 3+4 4 | 3+4 4 |
| | 5 4 | 7 2 | 5 4 | 7 2 | 3+4 4 | 3+4 4 |
| | 5 4 | 7 2 | 5 4 | 9 2 | 3+4 4 | 3+4 4 |
| | 5 4 | 9 2 | 5 4 | 7 2 | 3+4 4 | 3+4 4 |
| | 5 4 | 9 2 | 5 4 | 9 2 | 3+4 4 | 3+4 4 |
| | 5 4 | 7 2 | 5 4 | 7 2 | 3+4 4 | 3+4 4 |
| | 5 4 | 6 2 | 5 4 | 6 2 | 3+4 4 | 3+4 4 |
| | 5 4 | 6 2 | 5 4 | 6 2 | 3+4 4 | 3+4 4 |
| | 5 4 | 7 2 | 5 4 | 7 2 | 3+4 4 | 3+4 4 |
| | | 9 2 | | 9 2 | | |
| | | 8 2 | | 8 2 | | |
| | | 9 2 | | 9 2 | | |
| | | 9 2 | | 7 2 | | |
| | | 8 2 | | 6 2 | | |
| | | 7 2 | | 9 2 | | |
| | | 6 2 | | 8 2 | | |
| | | 6 2 | | 8 2 | | |
| | | 8 2 | | 6 2 | | |
| | | 7 2 | | 8 2 | | |
| | | 6 2 | | 9 2 | | |
| | | 7 2 | | 9 2 | | |
| | | 6 2 | | 8 2 | | |
| | | 9 2 | | 7 2 | | |
| | | 6 2 | | 7 2 | | |
| | | 6 2 | | 6 2 | | |
| | | 9 2 | | 8 2 | | |
| 129.2.78 | 169027 | | 68 | As1Co1Nd1O1 | | |
| | -A | -GM | -M | -Z | -R | -X |
| | 5 4 | 7 2 | 5 4 | 7 2 | 3+4 4 | 3+4 4 |
| | 5 4 | 9 2 | 5 4 | 9 2 | 3+4 4 | 3+4 4 |
| | 5 4 | 7 2 | 5 4 | 7 2 | 3+4 4 | 3+4 4 |
| | 5 4 | 9 2 | 5 4 | 9 2 | 3+4 4 | 3+4 4 |
| | 5 4 | 7 2 | 5 4 | 7 2 | 3+4 4 | 3+4 4 |
| | 5 4 | 8 2 | 5 4 | 8 2 | 3+4 4 | 3+4 4 |
| | 5 4 | 6 2 | 5 4 | 6 2 | 3+4 4 | 3+4 4 |
| | 5 4 | 9 2 | 5 4 | 9 2 | 3+4 4 | 3+4 4 |
| | 5 4 | 8 2 | 5 4 | 8 2 | 3+4 4 | 3+4 4 |
| | 5 4 | 7 2 | 5 4 | 7 2 | 3+4 4 | 3+4 4 |
| | 5 4 | 7 2 | 5 4 | 9 2 | 3+4 4 | 3+4 4 |
| | 5 4 | 9 2 | 5 4 | 7 2 | 3+4 4 | 3+4 4 |
| | 5 4 | 9 2 | 5 4 | 9 2 | 3+4 4 | 3+4 4 |
| | 5 4 | 7 2 | 5 4 | 7 2 | 3+4 4 | 3+4 4 |
| | 5 4 | 6 2 | 5 4 | 6 2 | 3+4 4 | 3+4 4 |
| | 5 4 | 6 2 | 5 4 | 6 2 | 3+4 4 | 3+4 4 |
| | 5 4 | 7 2 | 5 4 | 7 2 | 3+4 4 | 3+4 4 |
| | | 9 2 | | 9 2 | | |
| | | 8 2 | | 8 2 | | |
| | | 9 2 | | 9 2 | | |

Continued on next column

**TABLE LX– continued**

| TQC | ICSD | | NoE | SF | | |
|---|---|---|---|---|---|---|
| | | 9 2 | | 7 2 | | |
| | | 8 2 | | 6 2 | | |
| | | 7 2 | | 9 2 | | |
| | | 6 2 | | 8 2 | | |
| | | 6 2 | | 8 2 | | |
| | | 8 2 | | 6 2 | | |
| | | 7 2 | | 8 2 | | |
| | | 6 2 | | 9 2 | | |
| | | 7 2 | | 9 2 | | |
| | | 6 2 | | 8 2 | | |
| | | 9 2 | | 7 2 | | |
| | | 6 2 | | 7 2 | | |
| | | 6 2 | | 6 2 | | |
| | | 9 2 | | 8 2 | | |
| 129.2.79 | 195499 | | 60 | As1Ce1Mn1O1 | | |
| | -A | -GM | -M | -Z | -R | -X |
| | 5 4 | 7 2 | 5 4 | 7 2 | 3+4 4 | 3+4 4 |
| | 5 4 | 9 2 | 5 4 | 9 2 | 3+4 4 | 3+4 4 |
| | 5 4 | 7 2 | 5 4 | 7 2 | 3+4 4 | 3+4 4 |
| | 5 4 | 9 2 | 5 4 | 9 2 | 3+4 4 | 3+4 4 |
| | 5 4 | 8 2 | 5 4 | 8 2 | 3+4 4 | 3+4 4 |
| | 5 4 | 7 2 | 5 4 | 7 2 | 3+4 4 | 3+4 4 |
| | 5 4 | 6 2 | 5 4 | 6 2 | 3+4 4 | 3+4 4 |
| | 5 4 | 9 2 | 5 4 | 9 2 | 3+4 4 | 3+4 4 |
| | 5 4 | 8 2 | 5 4 | 8 2 | 3+4 4 | 3+4 4 |
| | 5 4 | 7 2 | 5 4 | 7 2 | 3+4 4 | 3+4 4 |
| | 5 4 | 7 2 | 5 4 | 9 2 | 3+4 4 | 3+4 4 |
| | 5 4 | 9 2 | 5 4 | 7 2 | 3+4 4 | 3+4 4 |
| | 5 4 | 9 2 | 5 4 | 9 2 | 3+4 4 | 3+4 4 |
| | 5 4 | 6 2 | 5 4 | 6 2 | 3+4 4 | 3+4 4 |
| | 5 4 | 7 2 | 5 4 | 7 2 | 3+4 4 | 3+4 4 |
| | | 6 2 | | 6 2 | | |
| | | 9 2 | | 9 2 | | |
| | | 8 2 | | 8 2 | | |
| | | 7 2 | | 9 2 | | |
| | | 9 2 | | 7 2 | | |
| | | 9 2 | | 7 2 | | |
| | | 8 2 | | 6 2 | | |
| | | 7 2 | | 9 2 | | |
| | | 6 2 | | 8 2 | | |
| | | 6 2 | | 8 2 | | |
| | | 8 2 | | 6 2 | | |
| | | 7 2 | | 9 2 | | |
| | | 9 2 | | 7 2 | | |
| | | 6 2 | | 6 2 | | |
| | | 6 2 | | 8 2 | | |
| 129.2.80 | 195498 | | 60 | As1Ce1Mn1O1 | | |
| | -A | -GM | -M | -Z | -R | -X |
| | 5 4 | 7 2 | 5 4 | 7 2 | 3+4 4 | 3+4 4 |
| | 5 4 | 9 2 | 5 4 | 9 2 | 3+4 4 | 3+4 4 |
| | 5 4 | 7 2 | 5 4 | 7 2 | 3+4 4 | 3+4 4 |
| | 5 4 | 9 2 | 5 4 | 9 2 | 3+4 4 | 3+4 4 |
| | 5 4 | 8 2 | 5 4 | 8 2 | 3+4 4 | 3+4 4 |
| | 5 4 | 7 2 | 5 4 | 7 2 | 3+4 4 | 3+4 4 |
| | 5 4 | 6 2 | 5 4 | 6 2 | 3+4 4 | 3+4 4 |
| | 5 4 | 9 2 | 5 4 | 9 2 | 3+4 4 | 3+4 4 |
| | 5 4 | 8 2 | 5 4 | 8 2 | 3+4 4 | 3+4 4 |
| | 5 4 | 7 2 | 5 4 | 7 2 | 3+4 4 | 3+4 4 |
| | 5 4 | 7 2 | 5 4 | 9 2 | 3+4 4 | 3+4 4 |
| | 5 4 | 9 2 | 5 4 | 7 2 | 3+4 4 | 3+4 4 |
| | 5 4 | 9 2 | 5 4 | 9 2 | 3+4 4 | 3+4 4 |

Continued on next column

**TABLE LX– continued**

| TQC | ICSD | | NoE | | SF | |
|---|---|---|---|---|---|---|
| | 5 4 | 6 2 | 5 4 | 6 2 | 3+4 4 | 3+4 4 |
| | 5 4 | 7 2 | 5 4 | 7 2 | 3+4 4 | 3+4 4 |
| | | 6 2 | | 6 2 | | |
| | | 9 2 | | 9 2 | | |
| | | 8 2 | | 8 2 | | |
| | | 7 2 | | 9 2 | | |
| | | 9 2 | | 7 2 | | |
| | | 9 2 | | 7 2 | | |
| | | 8 2 | | 6 2 | | |
| | | 7 2 | | 9 2 | | |
| | | 6 2 | | 8 2 | | |
| | | 6 2 | | 8 2 | | |
| | | 8 2 | | 6 2 | | |
| | | 7 2 | | 9 2 | | |
| | | 9 2 | | 7 2 | | |
| | | 6 2 | | 6 2 | | |
| | | 6 2 | | 8 2 | | |
| 129.2.81 | 194698 | | 76 | | Al2Ce1Pd2 | |
| | -A | -GM | -M | -Z | -R | -X |
| | 5 4 | 7 2 | 5 4 | 7 2 | 3+4 4 | 3+4 4 |
| | 5 4 | 9 2 | 5 4 | 9 2 | 3+4 4 | 3+4 4 |
| | 5 4 | 7 2 | 5 4 | 7 2 | 3+4 4 | 3+4 4 |
| | 5 4 | 9 2 | 5 4 | 9 2 | 3+4 4 | 3+4 4 |
| | 5 4 | 7 2 | 5 4 | 9 2 | 3+4 4 | 3+4 4 |
| | 5 4 | 9 2 | 5 4 | 7 2 | 3+4 4 | 3+4 4 |
| | 5 4 | 6 2 | 5 4 | 6 2 | 3+4 4 | 3+4 4 |
| | 5 4 | 8 2 | 5 4 | 8 2 | 3+4 4 | 3+4 4 |
| | 5 4 | 7 2 | 5 4 | 7 2 | 3+4 4 | 3+4 4 |
| | 5 4 | 7 2 | 5 4 | 9 2 | 3+4 4 | 3+4 4 |
| | 5 4 | 9 2 | 5 4 | 7 2 | 3+4 4 | 3+4 4 |
| | 5 4 | 8 2 | 5 4 | 8 2 | 3+4 4 | 3+4 4 |
| | 5 4 | 9 2 | 5 4 | 9 2 | 3+4 4 | 3+4 4 |
| | 5 4 | 7 2 | 5 4 | 8 2 | 3+4 4 | 3+4 4 |
| | 5 4 | 6 2 | 5 4 | 7 2 | 3+4 4 | 3+4 4 |
| | 5 4 | 8 2 | 5 4 | 6 2 | 3+4 4 | 3+4 4 |
| | 5 4 | 7 2 | 5 4 | 9 2 | 3+4 4 | 3+4 4 |
| | 5 4 | 6 2 | 5 4 | 6 2 | 3+4 4 | 3+4 4 |
| | 5 4 | 9 2 | 5 4 | 7 2 | 3+4 4 | 3+4 4 |
| | | 9 2 | | 7 2 | | |
| | | 6 2 | | 6 2 | | |
| | | 6 2 | | 6 2 | | |
| | | 7 2 | | 9 2 | | |
| | | 9 2 | | 8 2 | | |
| | | 8 2 | | 8 2 | | |
| | | 8 2 | | 9 2 | | |
| | | 7 2 | | 6 2 | | |
| | | 8 2 | | 7 2 | | |
| | | 6 2 | | 8 2 | | |
| | | 6 2 | | 8 2 | | |
| | | 9 2 | | 9 2 | | |
| | | 7 2 | | 7 2 | | |
| | | 9 2 | | 8 2 | | |
| | | 6 2 | | 9 2 | | |
| | | 7 2 | | 7 2 | | |
| | | 7 2 | | 6 2 | | |
| | | 6 2 | | 8 2 | | |
| | | 6 2 | | 7 2 | | |
| 129.2.82 | 191050 | | 76 | | Al2Ce1Pd2 | |
| | -A | -GM | -M | -Z | -R | -X |
| | 5 4 | 7 2 | 5 4 | 7 2 | 3+4 4 | 3+4 4 |
| | 5 4 | 9 2 | 5 4 | 9 2 | 3+4 4 | 3+4 4 |
| | 5 4 | 7 2 | 5 4 | 7 2 | 3+4 4 | 3+4 4 |
| | 5 4 | 9 2 | 5 4 | 9 2 | 3+4 4 | 3+4 4 |
| | 5 4 | 7 2 | 5 4 | 7 2 | 3+4 4 | 3+4 4 |
| | 5 4 | 9 2 | 5 4 | 9 2 | 3+4 4 | 3+4 4 |
| | 5 4 | 6 2 | 5 4 | 8 2 | 3+4 4 | 3+4 4 |
| | 5 4 | 8 2 | 5 4 | 6 2 | 3+4 4 | 3+4 4 |
| | 5 4 | 7 2 | 5 4 | 9 2 | 3+4 4 | 3+4 4 |
| | 5 4 | 7 2 | 5 4 | 7 2 | 3+4 4 | 3+4 4 |
| | 5 4 | 9 2 | 5 4 | 9 2 | 3+4 4 | 3+4 4 |
| | 5 4 | 8 2 | 5 4 | 6 2 | 3+4 4 | 3+4 4 |
| | 5 4 | 9 2 | 5 4 | 7 2 | 3+4 4 | 3+4 4 |
| | 5 4 | 7 2 | 5 4 | 9 2 | 3+4 4 | 3+4 4 |
| | 5 4 | 6 2 | 5 4 | 8 2 | 3+4 4 | 3+4 4 |
| | 5 4 | 7 2 | 5 4 | 6 2 | 3+4 4 | 3+4 4 |
| | 5 4 | 6 2 | 5 4 | 7 2 | 3+4 4 | 3+4 4 |
| | 5 4 | 8 2 | 5 4 | 8 2 | 3+4 4 | 3+4 4 |
| | 5 4 | 9 2 | 5 4 | 9 2 | 3+4 4 | 3+4 4 |
| | | 9 2 | | 9 2 | | |
| | | 6 2 | | 8 2 | | |
| | | 6 2 | | 8 2 | | |
| | | 7 2 | | 7 2 | | |
| | | 9 2 | | 6 2 | | |
| | | 8 2 | | 6 2 | | |
| | | 8 2 | | 7 2 | | |
| | | 7 2 | | 9 2 | | |
| | | 8 2 | | 8 2 | | |
| | | 6 2 | | 6 2 | | |
| | | 6 2 | | 6 2 | | |
| | | 9 2 | | 7 2 | | |
| | | 7 2 | | 9 2 | | |
| | | 9 2 | | 6 2 | | |
| | | 6 2 | | 7 2 | | |
| | | 7 2 | | 9 2 | | |
| | | 7 2 | | 8 2 | | |
| | | 6 2 | | 6 2 | | |
| | | 6 2 | | 9 2 | | |
| 129.2.83 | 190094 | | 64 | | Ag1Sb2Y1 | |
| | -A | -GM | -M | -Z | -R | -X |
| | 5 4 | 7 2 | 5 4 | 7 2 | 3+4 4 | 3+4 4 |
| | 5 4 | 9 2 | 5 4 | 9 2 | 3+4 4 | 3+4 4 |
| | 5 4 | 7 2 | 5 4 | 7 2 | 3+4 4 | 3+4 4 |
| | 5 4 | 9 2 | 5 4 | 9 2 | 3+4 4 | 3+4 4 |
| | 5 4 | 7 2 | 5 4 | 7 2 | 3+4 4 | 3+4 4 |
| | 5 4 | 9 2 | 5 4 | 9 2 | 3+4 4 | 3+4 4 |
| | 5 4 | 6 2 | 5 4 | 6 2 | 3+4 4 | 3+4 4 |
| | 5 4 | 8 2 | 5 4 | 8 2 | 3+4 4 | 3+4 4 |
| | 5 4 | 7 2 | 5 4 | 7 2 | 3+4 4 | 3+4 4 |
| | 5 4 | 7 2 | 5 4 | 9 2 | 3+4 4 | 3+4 4 |
| | 5 4 | 9 2 | 5 4 | 7 2 | 3+4 4 | 3+4 4 |
| | 5 4 | 9 2 | 5 4 | 9 2 | 3+4 4 | 3+4 4 |
| | 5 4 | 8 2 | 5 4 | 8 2 | 3+4 4 | 3+4 4 |
| | 5 4 | 7 2 | 5 4 | 7 2 | 3+4 4 | 3+4 4 |
| | 5 4 | 6 2 | 5 4 | 8 2 | 3+4 4 | 3+4 4 |
| | 5 4 | 8 2 | 5 4 | 6 2 | 3+4 4 | 3+4 4 |
| | | 9 2 | | 7 2 | | |
| | | 7 2 | | 9 2 | | |
| | | 8 2 | | 6 2 | | |
| | | 6 2 | | 9 2 | | |
| | | 9 2 | | 8 2 | | |
| | | 6 2 | | 8 2 | | |
| | | 7 2 | | 6 2 | | |

Continued on next column

**TABLE LX– continued**

| TQC | -A | -GM | -M | -Z | -R | -X |
|---|---|---|---|---|---|---|
| | | 6 2 | | 7 2 | | |
| | | 9 2 | | 7 2 | | |
| | | 7 2 | | 9 2 | | |
| | | 7 2 | | 7 2 | | |
| | | 6 2 | | 9 2 | | |
| | | 6 2 | | 8 2 | | |
| | | 9 2 | | 6 2 | | |
| | | 9 2 | | 7 2 | | |
| | | 8 2 | | 8 2 | | |
| 129.2.84 | ICSD: 251149 | NoE: 168 | SF: Ag1Cu1S1 | | | |
| | -A | -GM | -M | -Z | -R | -X |
| | 5 4 | 7 2 | 5 4 | 9 2 | 3+4 4 | 3+4 4 |
| | 5 4 | 9 2 | 5 4 | 7 2 | 3+4 4 | 3+4 4 |
| | 5 4 | 7 2 | 5 4 | 7 2 | 3+4 4 | 3+4 4 |
| | 5 4 | 9 2 | 5 4 | 9 2 | 3+4 4 | 3+4 4 |
| | 5 4 | 7 2 | 5 4 | 7 2 | 3+4 4 | 3+4 4 |
| | 5 4 | 9 2 | 5 4 | 9 2 | 3+4 4 | 3+4 4 |
| | 5 4 | 7 2 | 5 4 | 7 2 | 3+4 4 | 3+4 4 |
| | 5 4 | 9 2 | 5 4 | 9 2 | 3+4 4 | 3+4 4 |
| | 5 4 | 7 2 | 5 4 | 7 2 | 3+4 4 | 3+4 4 |
| | 5 4 | 7 2 | 5 4 | 9 2 | 3+4 4 | 3+4 4 |
| | 5 4 | 9 2 | 5 4 | 7 2 | 3+4 4 | 3+4 4 |
| | 5 4 | 9 2 | 5 4 | 9 2 | 3+4 4 | 3+4 4 |
| | 5 4 | 9 2 | 5 4 | 7 2 | 3+4 4 | 3+4 4 |
| | 5 4 | 8 2 | 5 4 | 6 2 | 3+4 4 | 3+4 4 |
| | 5 4 | 9 2 | 5 4 | 9 2 | 3+4 4 | 3+4 4 |
| | 5 4 | 8 2 | 5 4 | 8 2 | 3+4 4 | 3+4 4 |
| | 5 4 | 7 2 | 5 4 | 7 2 | 3+4 4 | 3+4 4 |
| | 5 4 | 6 2 | 5 4 | 9 2 | 3+4 4 | 3+4 4 |
| | 5 4 | 7 2 | 5 4 | 6 2 | 3+4 4 | 3+4 4 |
| | 5 4 | 8 2 | 5 4 | 8 2 | 3+4 4 | 3+4 4 |
| | 5 4 | 6 2 | 5 4 | 6 2 | 3+4 4 | 3+4 4 |
| | 5 4 | 8 2 | 5 4 | 8 2 | 3+4 4 | 3+4 4 |
| | 5 4 | 6 2 | 5 4 | 6 2 | 3+4 4 | 3+4 4 |
| | 5 4 | 7 2 | 5 4 | 7 2 | 3+4 4 | 3+4 4 |
| | 5 4 | 6 2 | 5 4 | 9 2 | 3+4 4 | 3+4 4 |
| | 5 4 | 9 2 | 5 4 | 8 2 | 3+4 4 | 3+4 4 |
| | 5 4 | 8 2 | 5 4 | 6 2 | 3+4 4 | 3+4 4 |
| | 5 4 | 6 2 | 5 4 | 8 2 | 3+4 4 | 3+4 4 |
| | 5 4 | 7 2 | 5 4 | 9 2 | 3+4 4 | 3+4 4 |
| | 5 4 | 6 2 | 5 4 | 6 2 | 3+4 4 | 3+4 4 |
| | 5 4 | 8 2 | 5 4 | 7 2 | 3+4 4 | 3+4 4 |
| | 5 4 | 9 2 | 5 4 | 8 2 | 3+4 4 | 3+4 4 |
| | 5 4 | 7 2 | 5 4 | 9 2 | 3+4 4 | 3+4 4 |
| | 5 4 | 9 2 | 5 4 | 8 2 | 3+4 4 | 3+4 4 |
| | 5 4 | 8 2 | 5 4 | 9 2 | 3+4 4 | 3+4 4 |
| | 5 4 | 6 2 | 5 4 | 8 2 | 3+4 4 | 3+4 4 |
| | 5 4 | 8 2 | 5 4 | 8 2 | 3+4 4 | 3+4 4 |
| | 5 4 | 9 2 | 5 4 | 7 2 | 3+4 4 | 3+4 4 |
| | 5 4 | 6 2 | 5 4 | 6 2 | 3+4 4 | 3+4 4 |
| | 5 4 | 7 2 | 5 4 | 8 2 | 3+4 4 | 3+4 4 |
| | 5 4 | 6 2 | 5 4 | 9 2 | 3+4 4 | 3+4 4 |
| | 5 4 | 7 2 | 5 4 | 7 2 | 3+4 4 | 3+4 4 |
| | | 6 2 | | 6 2 | | |
| | | 8 2 | | 6 2 | | |
| | | 6 2 | | 8 2 | | |
| | | 8 2 | | 6 2 | | |
| | | 9 2 | | 9 2 | | |
| | | 6 2 | | 7 2 | | |
| | | 8 2 | | 6 2 | | |
| | | 9 2 | | 8 2 | | |

Continued on next column

**TABLE LX– continued**

| TQC | -A | -GM | -M | -Z | -R | -X |
|---|---|---|---|---|---|---|
| | | 8 2 | | 6 2 | | |
| | | 7 2 | | 6 2 | | |
| | | 6 2 | | 7 2 | | |
| | | 7 2 | | 7 2 | | |
| | | 9 2 | | 9 2 | | |
| | | 6 2 | | 6 2 | | |
| | | 8 2 | | 8 2 | | |
| | | 9 2 | | 9 2 | | |
| | | 7 2 | | 7 2 | | |
| | | 7 2 | | 9 2 | | |
| | | 8 2 | | 6 2 | | |
| | | 6 2 | | 8 2 | | |
| | | 6 2 | | 6 2 | | |
| | | 8 2 | | 8 2 | | |
| | | 8 2 | | 8 2 | | |
| | | 7 2 | | 7 2 | | |
| | | 6 2 | | 6 2 | | |
| | | 6 2 | | 7 2 | | |
| | | 9 2 | | 9 2 | | |
| | | 7 2 | | 6 2 | | |
| | | 9 2 | | 8 2 | | |
| | | 7 2 | | 7 2 | | |
| | | 8 2 | | 9 2 | | |
| | | 9 2 | | 9 2 | | |
| | | 9 2 | | 9 2 | | |
| | | 7 2 | | 7 2 | | |
| | | 7 2 | | 7 2 | | |
| | | 6 2 | | 6 2 | | |
| | | 7 2 | | 8 2 | | |
| | | 6 2 | | 9 2 | | |
| | | 9 2 | | 7 2 | | |
| | | 8 2 | | 6 2 | | |
| | | 9 2 | | 9 2 | | |
| | | 8 2 | | 8 2 | | |

The End

TABLE LXI: The computed irreducible representations for the valence bands for the compounds of SEBR and NLC in space group 131. See the list of tables.

| TQC | -A | -GM | -M | -Z | -R | -X |
|---|---|---|---|---|---|---|
| 131.2.1 | ICSD: 50987 | NoE: 92 | SF: I1Pd1Te1 | | | |
| | -A | -GM | -M | -Z | -R | -X |
| | 5 4 | 7 2 | 6 2 | 5 4 | 6 2 | 5 2 |
| | 5 4 | 6 2 | 8 2 | 5 4 | 6 2 | 6 2 |
| | 5 4 | 8 2 | 9 2 | 5 4 | 5 2 | 5 2 |
| | 5 4 | 9 2 | 7 2 | 5 4 | 5 2 | 6 2 |
| | 5 4 | 7 2 | 7 2 | 5 4 | 5 2 | 5 2 |
| | 5 4 | 8 2 | 8 2 | 5 4 | 6 2 | 6 2 |
| | 5 4 | 6 2 | 9 2 | 5 4 | 5 2 | 6 2 |
| | 5 4 | 9 2 | 6 2 | 5 4 | 6 2 | 5 2 |
| | 5 4 | 8 2 | 8 2 | 5 4 | 5 2 | 6 2 |
| | 5 4 | 7 2 | 7 2 | 5 4 | 5 2 | 6 2 |
| | 5 4 | 9 2 | 6 2 | 5 4 | 6 2 | 5 2 |
| | 5 4 | 8 2 | 9 2 | 5 4 | 6 2 | 5 2 |
| | 5 4 | 6 2 | 8 2 | 5 4 | 6 2 | 6 2 |
| | 5 4 | 6 2 | 7 2 | 5 4 | 5 2 | 5 2 |
| | 5 4 | 7 2 | 9 2 | 5 4 | 6 2 | 5 2 |
| | 5 4 | 7 2 | 8 2 | 5 4 | 5 2 | 6 2 |
| | 5 4 | 6 2 | 6 2 | 5 4 | 6 2 | 6 2 |
| | 5 4 | 7 2 | 7 2 | 5 4 | 6 2 | 5 2 |

Continued on next column

**TABLE LXI– continued**

| TQC | ICSD NoE SF |
|---|---|
| | 5 4 9 2 9 2 5 4 5 2 5 2 |
| | 5 4 8 2 8 2 5 4 6 2 6 2 |
| | 5 4 6 2 6 2 5 4 5 2 5 2 |
| | 5 4 8 2 7 2 5 4 5 2 5 2 |
| | 5 4 9 2 6 2 5 4 6 2 6 2 |
| | 9 2 9 2 5 2 5 2 |
| | 7 2 7 2 6 2 6 2 |
| | 6 2 6 2 5 2 5 2 |
| | 6 2 6 2 5 2 6 2 |
| | 9 2 8 2 6 2 5 2 |
| | 8 2 9 2 6 2 6 2 |
| | 7 2 9 2 5 2 5 2 |
| | 8 2 8 2 6 2 6 2 |
| | 7 2 6 2 5 2 6 2 |
| | 6 2 7 2 6 2 5 2 |
| | 8 2 8 2 5 2 6 2 |
| | 9 2 9 2 6 2 5 2 |
| | 6 2 6 2 6 2 6 2 |
| | 7 2 7 2 5 2 5 2 |
| | 8 2 8 2 6 2 6 2 |
| | 7 2 9 2 5 2 5 2 |
| | 6 2 7 2 6 2 5 2 |
| | 9 2 8 2 5 2 6 2 |
| | 9 2 7 2 6 2 6 2 |
| | 8 2 9 2 5 2 5 2 |
| | 6 2 6 2 5 2 5 2 |
| | 7 2 8 2 5 2 6 2 |
| | 9 2 7 2 5 2 6 2 |
| 131.2.2 | 6159 92 Ca1O4Pt2 |
| | -A -GM -M -Z -R -X |
| | 5 4 7 2 7 2 5 4 5 2 5 2 |
| | 5 4 8 2 9 2 5 4 6 2 6 2 |
| | 5 4 9 2 8 2 5 4 5 2 5 2 |
| | 5 4 6 2 9 2 5 4 6 2 6 2 |
| | 5 4 7 2 6 2 5 4 6 2 6 2 |
| | 5 4 8 2 6 2 5 4 6 2 5 2 |
| | 5 4 9 2 7 2 5 4 5 2 6 2 |
| | 5 4 6 2 8 2 5 4 5 2 5 2 |
| | 5 4 6 2 6 2 5 4 5 2 5 2 |
| | 5 4 9 2 8 2 5 4 6 2 6 2 |
| | 5 4 8 2 7 2 5 4 6 2 6 2 |
| | 5 4 7 2 9 2 5 4 6 2 5 2 |
| | 5 4 8 2 6 2 5 4 5 2 5 2 |
| | 5 4 6 2 8 2 5 4 6 2 6 2 |
| | 5 4 7 2 9 2 5 4 5 2 6 2 |
| | 5 4 8 2 8 2 5 4 5 2 5 2 |
| | 5 4 6 2 7 2 5 4 6 2 6 2 |
| | 5 4 7 2 7 2 5 4 5 2 5 2 |
| | 5 4 9 2 6 2 5 4 5 2 6 2 |
| | 5 4 8 2 9 2 5 4 6 2 5 2 |
| | 5 4 6 2 8 2 5 4 5 2 5 2 |
| | 5 4 9 2 9 2 5 4 6 2 6 2 |
| | 5 4 7 2 8 2 5 4 6 2 6 2 |
| | 9 2 6 2 5 2 6 2 |
| | 6 2 7 2 5 2 5 2 |
| | 7 2 9 2 6 2 5 2 |
| | 7 2 7 2 6 2 5 2 |
| | 8 2 6 2 6 2 5 2 |
| | 7 2 7 2 5 2 6 2 |
| | 6 2 9 2 6 2 5 2 |
| | 9 2 8 2 5 2 6 2 |

Continued on next column

**TABLE LXI– continued**

| TQC | ICSD NoE SF |
|---|---|
| | 6 2 6 2 5 2 5 2 |
| | 8 2 8 2 5 2 5 2 |
| | 7 2 7 2 5 2 6 2 |
| | 9 2 6 2 6 2 5 2 |
| | 8 2 7 2 5 2 6 2 |
| | 9 2 9 2 6 2 6 2 |
| | 9 2 8 2 6 2 6 2 |
| | 6 2 9 2 6 2 6 2 |
| | 8 2 6 2 5 2 5 2 |
| | 8 2 6 2 6 2 6 2 |
| | 7 2 8 2 5 2 5 2 |
| | 6 2 7 2 5 2 5 2 |
| | 9 2 9 2 6 2 6 2 |
| | 6 2 8 2 6 2 6 2 |
| | 7 2 7 2 5 2 5 2 |

The End

TABLE LXII: The computed irreducible representations for the valence bands for the compounds of SEBR and NLC in space group 136. See the list of tables.

| TQC | ICSD NoE SF |
|---|---|
| 136.1.1 | 76166 420 U1 |
| | -A -GM -M -Z -R -X |
| | 5 4 7 2 8+9 4 5 4 6+9 2 3+4 4 |
| | 5 4 8 2 6+7 4 5 4 7+8 2 3+4 4 |
| | 5 4 9 2 8+9 4 5 4 3+4 2 3+4 4 |
| | 5 4 6 2 6+7 4 5 4 2+5 2 3+4 4 |
| | 5 4 7 2 8+9 4 5 4 7+8 2 3+4 4 |
| | 5 4 6 2 6+7 4 5 4 6+9 2 3+4 4 |
| | 5 4 8 2 6+7 4 5 4 2+5 2 3+4 4 |
| | 5 4 9 2 8+9 4 5 4 3+4 2 3+4 4 |
| | 5 4 7 2 8+9 4 5 4 6+9 2 3+4 4 |
| | 5 4 6 2 6+7 4 5 4 7+8 2 3+4 4 |
| | 5 4 7 2 6+7 4 5 4 3+4 2 3+4 4 |
| | 5 4 6 2 8+9 4 5 4 2+5 2 3+4 4 |
| | 5 4 8 2 6+7 4 5 4 6+9 2 3+4 4 |
| | 5 4 9 2 6+7 4 5 4 7+8 2 3+4 4 |
| | 5 4 7 2 8+9 4 5 4 3+4 2 3+4 4 |
| | 5 4 8 2 8+9 4 5 4 2+5 2 3+4 4 |
| | 5 4 9 2 6+7 4 5 4 2+5 2 3+4 4 |
| | 5 4 6 2 8+9 4 5 4 3+4 2 3+4 4 |
| | 5 4 6 2 6+7 4 5 4 6+9 2 3+4 4 |
| | 5 4 8 2 8+9 4 5 4 7+8 2 3+4 4 |
| | 5 4 9 2 6+7 4 5 4 6+9 2 3+4 4 |
| | 5 4 8 2 8+9 4 5 4 7+8 2 3+4 4 |
| | 5 4 9 2 8+9 4 5 4 3+4 2 3+4 4 |
| | 5 4 6 2 6+7 4 5 4 2+5 2 3+4 4 |
| | 5 4 7 2 6+7 4 5 4 6+9 2 3+4 4 |
| | 5 4 7 2 8+9 4 5 4 7+8 2 3+4 4 |
| | 5 4 6 2 8+9 4 5 4 2+5 2 3+4 4 |
| | 5 4 7 2 6+7 4 5 4 7+8 2 3+4 4 |
| | 5 4 8 2 8+9 4 5 4 6+9 2 3+4 4 |
| | 5 4 9 2 6+7 4 5 4 3+4 2 3+4 4 |
| | 5 4 7 2 8+9 4 5 4 3+4 2 3+4 4 |
| | 5 4 8 2 6+7 4 5 4 2+5 2 3+4 4 |
| | 5 4 9 2 8+9 4 5 4 2+5 2 3+4 4 |
| | 5 4 6 2 6+7 4 5 4 6+9 2 3+4 4 |
| | 5 4 9 2 8+9 4 5 4 3+4 2 3+4 4 |
| | 5 4 7 2 6+7 4 5 4 7+8 2 3+4 4 |

Continued on next column

**TABLE LXII– continued**

| TQC | ICSD | NoE | SF |
|---|---|---|---|
| | 5 4 9 2 | 8+9 4 | 5 4 3+4 2 3+4 4 |
| | 5 4 6 2 | 6+7 4 | 5 4 7+8 2 3+4 4 |
| | 5 4 8 2 | 8+9 4 | 5 4 6+9 2 3+4 4 |
| | 5 4 6 2 | 6+7 4 | 5 4 2+5 2 3+4 4 |
| | 5 4 7 2 | 8+9 4 | 5 4 3+4 2 3+4 4 |
| | 5 4 8 2 | 8+9 4 | 5 4 2+5 2 3+4 4 |
| | 5 4 7 2 | 6+7 4 | 5 4 7+8 2 3+4 4 |
| | 5 4 8 2 | 8+9 4 | 5 4 6+9 2 3+4 4 |
| | 5 4 6 2 | 6+7 4 | 5 4 7+8 2 3+4 4 |
| | 5 4 9 2 | 6+7 4 | 5 4 6+9 2 3+4 4 |
| | 5 4 6 2 | 8+9 4 | 5 4 2+5 2 3+4 4 |
| | 5 4 7 2 | 6+7 4 | 5 4 3+4 2 3+4 4 |
| | 5 4 8 2 | 8+9 4 | 5 4 6+9 2 3+4 4 |
| | 5 4 9 2 | 8+9 4 | 5 4 7+8 2 3+4 4 |
| | 5 4 8 2 | 6+7 4 | 5 4 2+5 2 3+4 4 |
| | 5 4 9 2 | 8+9 4 | 5 4 7+8 2 3+4 4 |
| | 5 4 6 2 | 8+9 4 | 5 4 6+9 2 3+4 4 |
| | 5 4 7 2 | 6+7 4 | 5 4 3+4 2 3+4 4 |
| | 5 4 8 2 | 6+7 4 | 5 4 3+4 2 3+4 4 |
| | 5 4 9 2 | 8+9 4 | 5 4 2+5 2 3+4 4 |
| | 5 4 8 2 | 8+9 4 | 5 4 7+8 2 3+4 4 |
| | 5 4 7 2 | 6+7 4 | 5 4 6+9 2 3+4 4 |
| | 5 4 6 2 | 6+7 4 | 5 4 2+5 2 3+4 4 |
| | 5 4 9 2 | 8+9 4 | 5 4 3+4 2 3+4 4 |
| | 5 4 9 2 | 6+7 4 | 5 4 2+5 2 3+4 4 |
| | 5 4 6 2 | 6+7 4 | 5 4 3+4 2 3+4 4 |
| | 5 4 6 2 | 6+7 4 | 5 4 7+8 2 3+4 4 |
| | 5 4 8 2 | 8+9 4 | 5 4 2+5 2 3+4 4 |
| | 5 4 7 2 | 8+9 4 | 5 4 6+9 2 3+4 4 |
| | 5 4 9 2 | 6+7 4 | 5 4 3+4 2 3+4 4 |
| | 5 4 6 2 | 6+7 4 | 5 4 2+5 2 3+4 4 |
| | 5 4 8 2 | 8+9 4 | 5 4 3+4 2 3+4 4 |
| | 5 4 9 2 | 8+9 4 | 5 4 6+9 2 3+4 4 |
| | 5 4 7 2 | 8+9 4 | 5 4 7+8 2 3+4 4 |
| | 5 4 8 2 | 6+7 4 | 5 4 6+9 2 3+4 4 |
| | 5 4 8 2 | 8+9 4 | 5 4 7+8 2 3+4 4 |
| | 5 4 7 2 | 6+7 4 | 5 4 3+4 2 3+4 4 |
| | 5 4 6 2 | 6+7 4 | 5 4 2+5 2 3+4 4 |
| | 5 4 7 2 | 8+9 4 | 5 4 6+9 2 3+4 4 |
| | 5 4 9 2 | 6+7 4 | 5 4 2+5 2 3+4 4 |
| | 5 4 7 2 | 8+9 4 | 5 4 3+4 2 3+4 4 |
| | 5 4 8 2 | 6+7 4 | 5 4 7+8 2 3+4 4 |
| | 5 4 9 2 | 8+9 4 | 5 4 7+8 2 3+4 4 |
| | 5 4 6 2 | 6+7 4 | 5 4 3+4 2 3+4 4 |
| | 5 4 8 2 | 8+9 4 | 5 4 6+9 2 3+4 4 |
| | 5 4 9 2 | 6+7 4 | 5 4 2+5 2 3+4 4 |
| | 5 4 6 2 | 8+9 4 | 5 4 6+9 2 3+4 4 |
| | 5 4 9 2 | 6+7 4 | 5 4 7+8 2 3+4 4 |
| | 5 4 6 2 | 6+7 4 | 5 4 3+4 2 3+4 4 |
| | 5 4 8 2 | 6+7 4 | 5 4 2+5 2 3+4 4 |
| | 5 4 7 2 | 8+9 4 | 5 4 7+8 2 3+4 4 |
| | 5 4 7 2 | 8+9 4 | 5 4 6+9 2 3+4 4 |
| | 5 4 8 2 | 6+7 4 | 5 4 3+4 2 3+4 4 |
| | 5 4 9 2 | 8+9 4 | 5 4 2+5 2 3+4 4 |
| | 5 4 7 2 | 8+9 4 | 5 4 6+9 2 3+4 4 |
| | 5 4 6 2 | 6+7 4 | 5 4 7+8 2 3+4 4 |
| | 5 4 9 2 | 6+7 4 | 5 4 3+4 2 3+4 4 |
| | 5 4 8 2 | 8+9 4 | 5 4 2+5 2 3+4 4 |
| | 5 4 6 2 | 8+9 4 | 5 4 3+4 2 3+4 4 |
| | 5 4 7 2 | 6+7 4 | 5 4 2+5 2 3+4 4 |
| | 5 4 8 2 | 8+9 4 | 5 4 7+8 2 3+4 4 |

Continued on next column

**TABLE LXII– continued**

| TQC | ICSD | NoE | SF |
|---|---|---|---|
| | 5 4 7 2 | 8+9 4 | 5 4 2+5 2 3+4 4 |
| | 5 4 6 2 | 8+9 4 | 5 4 6+9 2 3+4 4 |
| | 5 4 7 2 | 6+7 4 | 5 4 7+8 2 3+4 4 |
| | 5 4 9 2 | 8+9 4 | 5 4 6+9 2 3+4 4 |
| | 5 4 8 2 | 6+7 4 | 5 4 3+4 2 3+4 4 |
| | 5 4 6 2 | 8+9 4 | 5 4 2+5 2 3+4 4 |
| | 5 4 9 2 | 6+7 4 | 5 4 3+4 2 3+4 4 |
| | 5 4 8 2 | 6+7 4 | 5 4 7+8 2 3+4 4 |
| | 7 2 | | 6+9 2 |
| | 9 2 | | 2+5 2 |
| | 8 2 | | 3+4 2 |
| | 6 2 | | 7+8 2 |
| | 7 2 | | 6+9 2 |
| | 9 2 | | 2+5 2 |
| | 6 2 | | 3+4 2 |
| | 9 2 | | 7+8 2 |
| | 8 2 | | 6+9 2 |
| | 9 2 | | 2+5 2 |
| | 8 2 | | 3+4 2 |
| | 6 2 | | 7+8 2 |
| | 7 2 | | 6+9 2 |
| | 8 2 | | 3+4 2 |
| | 9 2 | | 2+5 2 |
| | 7 2 | | 7+8 2 |
| | 7 2 | | 6+9 2 |
| | 9 2 | | 6+9 2 |
| | 6 2 | | 7+8 2 |
| | 8 2 | | 3+4 2 |
| | 9 2 | | 2+5 2 |
| | 8 2 | | 3+4 2 |
| | 6 2 | | 2+5 2 |
| | 9 2 | | 7+8 2 |
| | 7 2 | | 6+9 2 |
| | 8 2 | | 3+4 2 |
| | 6 2 | | 6+9 2 |
| | 7 2 | | 7+8 2 |
| | 9 2 | | 2+5 2 |
| | 6 2 | | 7+8 2 |
| | 7 2 | | 6+9 2 |
| | 6 2 | | 3+4 2 |
| | 8 2 | | 2+5 2 |
| | 9 2 | | 6+9 2 |
| | 8 2 | | 7+8 2 |
| | 6 2 | | 2+5 2 |
| | 7 2 | | 6+9 2 |
| | 6 2 | | 3+4 2 |
| | 8 2 | | 7+8 2 |
| | 7 2 | | 2+5 2 |
| | 8 2 | | 3+4 2 |
| | 9 2 | | 7+8 2 |
| | 6 2 | | 3+4 2 |
| | 9 2 | | 2+5 2 |
| | 7 2 | | 6+9 2 |
| | 7 2 | | 7+8 2 |
| | 9 2 | | 6+9 2 |
| | 6 2 | | 6+9 2 |
| | 8 2 | | 7+8 2 |
| | 6 2 | | 3+4 2 |
| | 7 2 | | 7+8 2 |
| | 6 2 | | 6+9 2 |
| | 8 2 | | 3+4 2 |

Continued on next column

**TABLE LXII– continued**

| TQC | ICSD | | NoE | | SF | |
|---|---|---|---|---|---|---|
| | | 9 2 | | | 2+5 2 | |
| | | 8 2 | | | 2+5 2 | |
| | | 9 2 | | | 6+9 2 | |
| | | 9 2 | | | 3+4 2 | |
| | | 7 2 | | | 2+5 2 | |
| | | 8 2 | | | 7+8 2 | |
| | | 7 2 | | | 2+5 2 | |
| | | 6 2 | | | 3+4 2 | |
| | | 7 2 | | | 2+5 2 | |
| | | 8 2 | | | 6+9 2 | |
| | | 9 2 | | | 7+8 2 | |
| | | 8 2 | | | 3+4 2 | |
| | | 9 2 | | | 3+4 2 | |
| | | 6 2 | | | 6+9 2 | |
| | | 8 2 | | | 7+8 2 | |
| | | 9 2 | | | 2+5 2 | |
| | | 8 2 | | | 7+8 2 | |
| | | 7 2 | | | 6+9 2 | |
| | | 6 2 | | | 2+5 2 | |
| | | 9 2 | | | 3+4 2 | |
| | | 7 2 | | | 7+8 2 | |
| | | 6 2 | | | 6+9 2 | |
| | | 6 2 | | | 2+5 2 | |
| | | 7 2 | | | 3+4 2 | |
| | | 8 2 | | | 6+9 2 | |
| | | 9 2 | | | 7+8 2 | |
| | | 6 2 | | | 3+4 2 | |
| | | 7 2 | | | 6+9 2 | |
| | | 6 2 | | | 2+5 2 | |
| | | 7 2 | | | 7+8 2 | |
| | | 6 2 | | | 2+5 2 | |
| | | 8 2 | | | 3+4 2 | |
| | | 9 2 | | | 6+9 2 | |
| | | 7 2 | | | 7+8 2 | |
| | | 8 2 | | | 2+5 2 | |
| | | 9 2 | | | 7+8 2 | |
| | | 7 2 | | | 3+4 2 | |
| | | 9 2 | | | 6+9 2 | |
| | | 8 2 | | | 2+5 2 | |
| | | 6 2 | | | 6+9 2 | |
| | | 8 2 | | | 7+8 2 | |
| | | 9 2 | | | 3+4 2 | |
| | | 9 2 | | | 2+5 2 | |
| | | 6 2 | | | 3+4 2 | |
| | | 8 2 | | | 3+4 2 | |
| | | 7 2 | | | 6+9 2 | |
| | | 8 2 | | | 7+8 2 | |
| | | 7 2 | | | 2+5 2 | |
| | | 9 2 | | | 7+8 2 | |
| | | 8 2 | | | 6+9 2 | |
| | | 8 2 | | | 2+5 2 | |
| | | 7 2 | | | 3+4 2 | |
| 136.1.2 | 90270 | | 148 | | Ni4P2Yb1 | |
| | -A | -GM | -M | -Z | -R | -X |
| | 5 4 | 6 2 | 6+7 4 | 5 4 | 6+9 2 | 3+4 4 |
| | 5 4 | 7 2 | 8+9 4 | 5 4 | 7+8 2 | 3+4 4 |
| | 5 4 | 8 2 | 8+9 4 | 5 4 | 2+5 2 | 3+4 4 |
| | 5 4 | 9 2 | 8+9 4 | 5 4 | 3+4 2 | 3+4 4 |
| | 5 4 | 9 2 | 6+7 4 | 5 4 | 2+5 2 | 3+4 4 |
| | 5 4 | 8 2 | 8+9 4 | 5 4 | 3+4 2 | 3+4 4 |
| | 5 4 | 8 2 | 8+9 4 | 5 4 | 3+4 2 | 3+4 4 |

Continued on next column

**TABLE LXII– continued**

| TQC | ICSD | | NoE | | SF | |
|---|---|---|---|---|---|---|
| | 5 4 | 9 2 | 6+7 4 | 5 4 | 2+5 2 | 3+4 4 |
| | 5 4 | 7 2 | 6+7 4 | 5 4 | 6+9 2 | 3+4 4 |
| | 5 4 | 8 2 | 6+7 4 | 5 4 | 7+8 2 | 3+4 4 |
| | 5 4 | 9 2 | 8+9 4 | 5 4 | 2+5 2 | 3+4 4 |
| | 5 4 | 6 2 | 8+9 4 | 5 4 | 3+4 2 | 3+4 4 |
| | 5 4 | 7 2 | 8+9 4 | 5 4 | 6+9 2 | 3+4 4 |
| | 5 4 | 6 2 | 6+7 4 | 5 4 | 7+8 2 | 3+4 4 |
| | 5 4 | 9 2 | 6+7 4 | 5 4 | 3+4 2 | 3+4 4 |
| | 5 4 | 8 2 | 6+7 4 | 5 4 | 2+5 2 | 3+4 4 |
| | 5 4 | 7 2 | 6+7 4 | 5 4 | 3+4 2 | 3+4 4 |
| | 5 4 | 9 2 | 8+9 4 | 5 4 | 2+5 2 | 3+4 4 |
| | 5 4 | 6 2 | 8+9 4 | 5 4 | 7+8 2 | 3+4 4 |
| | 5 4 | 6 2 | 8+9 4 | 5 4 | 6+9 2 | 3+4 4 |
| | 5 4 | 7 2 | 8+9 4 | 5 4 | 7+8 2 | 3+4 4 |
| | 5 4 | 9 2 | 6+7 4 | 5 4 | 6+9 2 | 3+4 4 |
| | 5 4 | 8 2 | 8+9 4 | 5 4 | 3+4 2 | 3+4 4 |
| | 5 4 | 8 2 | 8+9 4 | 5 4 | 2+5 2 | 3+4 4 |
| | 5 4 | 6 2 | 8+9 4 | 5 4 | 7+8 2 | 3+4 4 |
| | 5 4 | 7 2 | 8+9 4 | 5 4 | 6+9 2 | 3+4 4 |
| | 5 4 | 7 2 | 8+9 4 | 5 4 | 3+4 2 | 3+4 4 |
| | 5 4 | 9 2 | 8+9 4 | 5 4 | 2+5 2 | 3+4 4 |
| | 5 4 | 9 2 | 6+7 4 | 5 4 | 7+8 2 | 3+4 4 |
| | 5 4 | 8 2 | 8+9 4 | 5 4 | 6+9 2 | 3+4 4 |
| | 5 4 | 8 2 | 6+7 4 | 5 4 | 6+9 2 | 3+4 4 |
| | 5 4 | 8 2 | 6+7 4 | 5 4 | 7+8 2 | 3+4 4 |
| | 5 4 | 9 2 | 8+9 4 | 5 4 | 3+4 2 | 3+4 4 |
| | 5 4 | 7 2 | 8+9 4 | 5 4 | 2+5 2 | 3+4 4 |
| | 5 4 | 6 2 | 8+9 4 | 5 4 | 3+4 2 | 3+4 4 |
| | 5 4 | 7 2 | 8+9 4 | 5 4 | 2+5 2 | 3+4 4 |
| | 5 4 | 8 2 | 6+7 4 | 5 4 | 2+5 2 | 3+4 4 |
| | | 6 2 | | | 3+4 2 | |
| | | 7 2 | | | 6+9 2 | |
| | | 9 2 | | | 7+8 2 | |
| | | 6 2 | | | 7+8 2 | |
| | | 9 2 | | | 6+9 2 | |
| | | 8 2 | | | 2+5 2 | |
| | | 6 2 | | | 3+4 2 | |
| | | 6 2 | | | 2+5 2 | |
| | | 9 2 | | | 6+9 2 | |
| | | 7 2 | | | 3+4 2 | |
| | | 8 2 | | | 7+8 2 | |
| | | 8 2 | | | 7+8 2 | |
| | | 9 2 | | | 6+9 2 | |
| | | 9 2 | | | 2+5 2 | |
| | | 8 2 | | | 3+4 2 | |
| | | 6 2 | | | 3+4 2 | |
| | | 7 2 | | | 2+5 2 | |
| | | 8 2 | | | 3+4 2 | |
| | | 6 2 | | | 2+5 2 | |
| | | 9 2 | | | 3+4 2 | |
| | | 7 2 | | | 7+8 2 | |
| | | 7 2 | | | 2+5 2 | |
| | | 9 2 | | | 6+9 2 | |
| | | 8 2 | | | 2+5 2 | |
| | | 6 2 | | | 7+8 2 | |
| | | 8 2 | | | 6+9 2 | |
| | | 9 2 | | | 3+4 2 | |
| | | 8 2 | | | 2+5 2 | |
| | | 8 2 | | | 3+4 2 | |
| | | 8 2 | | | 3+4 2 | |
| | | 9 2 | | | 2+5 2 | |

Continued on next column

**TABLE LXII– continued**

| TQC | ICSD | | NoE | | SF | |
|---|---|---|---|---|---|---|
| | | 9 2 | | | 2+5 2 | |
| | | 9 2 | | | 3+4 2 | |
| | | 7 2 | | | 2+5 2 | |
| | | 8 2 | | | 3+4 2 | |
| | | 9 2 | | | 3+4 2 | |
| | | 8 2 | | | 2+5 2 | |
| 136.1.3 | 421229 | | 128 | | Li2Sr3 | |
| | -A | -GM | -M | -Z | -R | -X |
| | 5 4 | 7 2 | 8+9 4 | 5 4 | 2+5 2 | 3+4 4 |
| | 5 4 | 7 2 | 6+7 4 | 5 4 | 3+4 2 | 3+4 4 |
| | 5 4 | 8 2 | 6+7 4 | 5 4 | 6+9 2 | 3+4 4 |
| | 5 4 | 8 2 | 8+9 4 | 5 4 | 7+8 2 | 3+4 4 |
| | 5 4 | 6 2 | 6+7 4 | 5 4 | 2+5 2 | 3+4 4 |
| | 5 4 | 7 2 | 8+9 4 | 5 4 | 3+4 2 | 3+4 4 |
| | 5 4 | 8 2 | 8+9 4 | 5 4 | 6+9 2 | 3+4 4 |
| | 5 4 | 9 2 | 6+7 4 | 5 4 | 7+8 2 | 3+4 4 |
| | 5 4 | 6 2 | 8+9 4 | 5 4 | 2+5 2 | 3+4 4 |
| | 5 4 | 7 2 | 6+7 4 | 5 4 | 3+4 2 | 3+4 4 |
| | 5 4 | 8 2 | 6+7 4 | 5 4 | 6+9 2 | 3+4 4 |
| | 5 4 | 9 2 | 8+9 4 | 5 4 | 7+8 2 | 3+4 4 |
| | 5 4 | 6 2 | 6+7 4 | 5 4 | 2+5 2 | 3+4 4 |
| | 5 4 | 9 2 | 8+9 4 | 5 4 | 3+4 2 | 3+4 4 |
| | 5 4 | 6 2 | 8+9 4 | 5 4 | 6+9 2 | 3+4 4 |
| | 5 4 | 9 2 | 6+7 4 | 5 4 | 7+8 2 | 3+4 4 |
| | 5 4 | 6 2 | 6+7 4 | 5 4 | 6+9 2 | 3+4 4 |
| | 5 4 | 7 2 | 8+9 4 | 5 4 | 7+8 2 | 3+4 4 |
| | 5 4 | 8 2 | 6+7 4 | 5 4 | 2+5 2 | 3+4 4 |
| | 5 4 | 9 2 | 8+9 4 | 5 4 | 3+4 2 | 3+4 4 |
| | 5 4 | 6 2 | 6+7 4 | 5 4 | 2+5 2 | 3+4 4 |
| | 5 4 | 7 2 | 8+9 4 | 5 4 | 3+4 2 | 3+4 4 |
| | 5 4 | 8 2 | 8+9 4 | 5 4 | 6+9 2 | 3+4 4 |
| | 5 4 | 9 2 | 6+7 4 | 5 4 | 7+8 2 | 3+4 4 |
| | 5 4 | 6 2 | 8+9 4 | 5 4 | 3+4 2 | 3+4 4 |
| | 5 4 | 8 2 | 6+7 4 | 5 4 | 7+8 2 | 3+4 4 |
| | 5 4 | 7 2 | 6+7 4 | 5 4 | 6+9 2 | 3+4 4 |
| | 5 4 | 6 2 | 6+7 4 | 5 4 | 2+5 2 | 3+4 4 |
| | 5 4 | 9 2 | 8+9 4 | 5 4 | 2+5 2 | 3+4 4 |
| | 5 4 | 8 2 | 8+9 4 | 5 4 | 3+4 2 | 3+4 4 |
| | 5 4 | 7 2 | 8+9 4 | 5 4 | 6+9 2 | 3+4 4 |
| | 5 4 | 9 2 | 8+9 4 | 5 4 | 7+8 2 | 3+4 4 |
| | | 6 2 | | | 2+5 2 | |
| | | 7 2 | | | 3+4 2 | |
| | | 9 2 | | | 6+9 2 | |
| | | 8 2 | | | 2+5 2 | |
| | | 6 2 | | | 6+9 2 | |
| | | 7 2 | | | 7+8 2 | |
| | | 8 2 | | | 3+4 2 | |
| | | 9 2 | | | 7+8 2 | |
| | | 6 2 | | | 2+5 2 | |
| | | 7 2 | | | 3+4 2 | |
| | | 6 2 | | | 7+8 2 | |
| | | 7 2 | | | 3+4 2 | |
| | | 9 2 | | | 6+9 2 | |
| | | 8 2 | | | 2+5 2 | |
| | | 8 2 | | | 6+9 2 | |
| | | 9 2 | | | 7+8 2 | |
| | | 7 2 | | | 2+5 2 | |
| | | 6 2 | | | 3+4 2 | |
| | | 9 2 | | | 6+9 2 | |
| | | 8 2 | | | 7+8 2 | |
| | | 7 2 | | | 7+8 2 | |
| | | 6 2 | | | 6+9 2 | |
| | | 8 2 | | | 2+5 2 | |
| | | 7 2 | | | 3+4 2 | |
| | | 9 2 | | | 2+5 2 | |
| | | 6 2 | | | 3+4 2 | |
| | | 8 2 | | | 6+9 2 | |
| | | 9 2 | | | 7+8 2 | |
| | | 8 2 | | | 6+9 2 | |
| | | 8 2 | | | 7+8 2 | |
| | | 9 2 | | | 3+4 2 | |
| | | 7 2 | | | 2+5 2 | |
| 136.1.4 | 93213 | | 116 | | Fe2O6Te1 | |
| | -A | -GM | -M | -Z | -R | -X |
| | 5 4 | 7 2 | 6+7 4 | 5 4 | 2+5 2 | 3+4 4 |
| | 5 4 | 6 2 | 8+9 4 | 5 4 | 3+4 2 | 3+4 4 |
| | 5 4 | 8 2 | 8+9 4 | 5 4 | 7+8 2 | 3+4 4 |
| | 5 4 | 9 2 | 8+9 4 | 5 4 | 6+9 2 | 3+4 4 |
| | 5 4 | 8 2 | 6+7 4 | 5 4 | 6+9 2 | 3+4 4 |
| | 5 4 | 9 2 | 6+7 4 | 5 4 | 7+8 2 | 3+4 4 |
| | 5 4 | 8 2 | 6+7 4 | 5 4 | 7+8 2 | 3+4 4 |
| | 5 4 | 9 2 | 8+9 4 | 5 4 | 6+9 2 | 3+4 4 |
| | 5 4 | 6 2 | 8+9 4 | 5 4 | 2+5 2 | 3+4 4 |
| | 5 4 | 7 2 | 6+7 4 | 5 4 | 3+4 2 | 3+4 4 |
| | 5 4 | 7 2 | 8+9 4 | 5 4 | 2+5 2 | 3+4 4 |
| | 5 4 | 6 2 | 6+7 4 | 5 4 | 3+4 2 | 3+4 4 |
| | 5 4 | 7 2 | 8+9 4 | 5 4 | 2+5 2 | 3+4 4 |
| | 5 4 | 6 2 | 6+7 4 | 5 4 | 3+4 2 | 3+4 4 |
| | 5 4 | 9 2 | 8+9 4 | 5 4 | 6+9 2 | 3+4 4 |
| | 5 4 | 8 2 | 6+7 4 | 5 4 | 7+8 2 | 3+4 4 |
| | 5 4 | 9 2 | 8+9 4 | 5 4 | 7+8 2 | 3+4 4 |
| | 5 4 | 8 2 | 8+9 4 | 5 4 | 6+9 2 | 3+4 4 |
| | 5 4 | 8 2 | 6+7 4 | 5 4 | 6+9 2 | 3+4 4 |
| | 5 4 | 9 2 | 6+7 4 | 5 4 | 7+8 2 | 3+4 4 |
| | 5 4 | 6 2 | 8+9 4 | 5 4 | 3+4 2 | 3+4 4 |
| | 5 4 | 7 2 | 8+9 4 | 5 4 | 2+5 2 | 3+4 4 |
| | 5 4 | 9 2 | 6+7 4 | 5 4 | 2+5 2 | 3+4 4 |
| | 5 4 | 7 2 | 6+7 4 | 5 4 | 3+4 2 | 3+4 4 |
| | 5 4 | 7 2 | 6+7 4 | 5 4 | 6+9 2 | 3+4 4 |
| | 5 4 | 8 2 | 6+7 4 | 5 4 | 7+8 2 | 3+4 4 |
| | 5 4 | 6 2 | 8+9 4 | 5 4 | 7+8 2 | 3+4 4 |
| | 5 4 | 7 2 | 8+9 4 | 5 4 | 6+9 2 | 3+4 4 |
| | 5 4 | 6 2 | 8+9 4 | 5 4 | 3+4 2 | 3+4 4 |
| | | 6 2 | | | 2+5 2 | |
| | | 8 2 | | | 2+5 2 | |
| | | 9 2 | | | 3+4 2 | |
| | | 8 2 | | | 6+9 2 | |
| | | 7 2 | | | 7+8 2 | |
| | | 6 2 | | | 2+5 2 | |
| | | 8 2 | | | 3+4 2 | |
| | | 8 2 | | | 7+8 2 | |
| | | 9 2 | | | 6+9 2 | |
| | | 6 2 | | | 3+4 2 | |
| | | 7 2 | | | 2+5 2 | |
| | | 9 2 | | | 2+5 2 | |
| | | 6 2 | | | 3+4 2 | |
| | | 7 2 | | | 6+9 2 | |
| | | 9 2 | | | 7+8 2 | |
| | | 7 2 | | | 6+9 2 | |
| | | 9 2 | | | 7+8 2 | |
| | | 8 2 | | | 3+4 2 | |
| | | 6 2 | | | 2+5 2 | |

Continued on next column

**TABLE LXII– continued**

| TQC | ICSD | | NoE | | SF | |
|---|---|---|---|---|---|---|
| | | 6 2 | | | 2+5 2 | |
| | | 9 2 | | | 3+4 2 | |
| | | 7 2 | | | 6+9 2 | |
| | | 6 2 | | | 7+8 2 | |
| | | 9 2 | | | 6+9 2 | |
| | | 6 2 | | | 7+8 2 | |
| | | 7 2 | | | 3+4 2 | |
| | | 8 2 | | | 2+5 2 | |
| | | 8 2 | | | 7+8 2 | |
| | | 9 2 | | | 6+9 2 | |
| 136.1.5 | 33797 | | 204 | | Co2K4O5Rb2 | |
| | -A | -GM | -M | -Z | -R | -X |
| | 5 4 | 7 2 | 6+7 4 | 5 4 | 2+5 2 | 3+4 4 |
| | 5 4 | 8 2 | 8+9 4 | 5 4 | 3+4 2 | 3+4 4 |
| | 5 4 | 7 2 | 6+7 4 | 5 4 | 6+9 2 | 3+4 4 |
| | 5 4 | 8 2 | 8+9 4 | 5 4 | 7+8 2 | 3+4 4 |
| | 5 4 | 6 2 | 6+7 4 | 5 4 | 2+5 2 | 3+4 4 |
| | 5 4 | 9 2 | 8+9 4 | 5 4 | 3+4 2 | 3+4 4 |
| | 5 4 | 7 2 | 6+7 4 | 5 4 | 6+9 2 | 3+4 4 |
| | 5 4 | 8 2 | 8+9 4 | 5 4 | 7+8 2 | 3+4 4 |
| | 5 4 | 6 2 | 6+7 4 | 5 4 | 2+5 2 | 3+4 4 |
| | 5 4 | 7 2 | 8+9 4 | 5 4 | 3+4 2 | 3+4 4 |
| | 5 4 | 8 2 | 6+7 4 | 5 4 | 6+9 2 | 3+4 4 |
| | 5 4 | 9 2 | 6+7 4 | 5 4 | 7+8 2 | 3+4 4 |
| | 5 4 | 6 2 | 8+9 4 | 5 4 | 6+9 2 | 3+4 4 |
| | 5 4 | 7 2 | 8+9 4 | 5 4 | 7+8 2 | 3+4 4 |
| | 5 4 | 8 2 | 6+7 4 | 5 4 | 2+5 2 | 3+4 4 |
| | 5 4 | 9 2 | 8+9 4 | 5 4 | 3+4 2 | 3+4 4 |
| | 5 4 | 6 2 | 6+7 4 | 5 4 | 6+9 2 | 3+4 4 |
| | 5 4 | 7 2 | 8+9 4 | 5 4 | 7+8 2 | 3+4 4 |
| | 5 4 | 6 2 | 6+7 4 | 5 4 | 6+9 2 | 3+4 4 |
| | 5 4 | 7 2 | 8+9 4 | 5 4 | 7+8 2 | 3+4 4 |
| | 5 4 | 9 2 | 6+7 4 | 5 4 | 2+5 2 | 3+4 4 |
| | 5 4 | 8 2 | 8+9 4 | 5 4 | 3+4 2 | 3+4 4 |
| | 5 4 | 9 2 | 6+7 4 | 5 4 | 2+5 2 | 3+4 4 |
| | 5 4 | 6 2 | 6+7 4 | 5 4 | 6+9 2 | 3+4 4 |
| | 5 4 | 9 2 | 8+9 4 | 5 4 | 3+4 2 | 3+4 4 |
| | 5 4 | 6 2 | 6+7 4 | 5 4 | 7+8 2 | 3+4 4 |
| | 5 4 | 7 2 | 8+9 4 | 5 4 | 3+4 2 | 3+4 4 |
| | 5 4 | 8 2 | 6+7 4 | 5 4 | 7+8 2 | 3+4 4 |
| | 5 4 | 8 2 | 8+9 4 | 5 4 | 2+5 2 | 3+4 4 |
| | 5 4 | 7 2 | 8+9 4 | 5 4 | 6+9 2 | 3+4 4 |
| | 5 4 | 6 2 | 6+7 4 | 5 4 | 2+5 2 | 3+4 4 |
| | 5 4 | 9 2 | 8+9 4 | 5 4 | 3+4 2 | 3+4 4 |
| | 5 4 | 9 2 | 8+9 4 | 5 4 | 6+9 2 | 3+4 4 |
| | 5 4 | 6 2 | 6+7 4 | 5 4 | 7+8 2 | 3+4 4 |
| | 5 4 | 8 2 | 8+9 4 | 5 4 | 2+5 2 | 3+4 4 |
| | 5 4 | 9 2 | 8+9 4 | 5 4 | 3+4 2 | 3+4 4 |
| | 5 4 | 6 2 | 6+7 4 | 5 4 | 6+9 2 | 3+4 4 |
| | 5 4 | 7 2 | 6+7 4 | 5 4 | 7+8 2 | 3+4 4 |
| | 5 4 | 9 2 | 8+9 4 | 5 4 | 3+4 2 | 3+4 4 |
| | 5 4 | 8 2 | 8+9 4 | 5 4 | 2+5 2 | 3+4 4 |
| | 5 4 | 7 2 | 8+9 4 | 5 4 | 7+8 2 | 3+4 4 |
| | 5 4 | 8 2 | 6+7 4 | 5 4 | 6+9 2 | 3+4 4 |
| | 5 4 | 6 2 | 6+7 4 | 5 4 | 2+5 2 | 3+4 4 |
| | 5 4 | 9 2 | 8+9 4 | 5 4 | 3+4 2 | 3+4 4 |
| | 5 4 | 7 2 | 8+9 4 | 5 4 | 6+9 2 | 3+4 4 |
| | 5 4 | 6 2 | 6+7 4 | 5 4 | 7+8 2 | 3+4 4 |
| | 5 4 | 7 2 | 6+7 4 | 5 4 | 2+5 2 | 3+4 4 |
| | 5 4 | 6 2 | 6+7 4 | 5 4 | 3+4 2 | 3+4 4 |
| | 5 4 | 8 2 | 6+7 4 | 5 4 | 6+9 2 | 3+4 4 |
| | 5 4 | 9 2 | 8+9 4 | 5 4 | 7+8 2 | 3+4 4 |
| | 5 4 | 6 2 | 6+7 4 | 5 4 | 2+5 2 | 3+4 4 |
| | | 7 2 | | | 3+4 2 | |
| | | 8 2 | | | 6+9 2 | |
| | | 9 2 | | | 7+8 2 | |
| | | 7 2 | | | 3+4 2 | |
| | | 6 2 | | | 2+5 2 | |
| | | 8 2 | | | 6+9 2 | |
| | | 9 2 | | | 7+8 2 | |
| | | 8 2 | | | 2+5 2 | |
| | | 9 2 | | | 3+4 2 | |
| | | 6 2 | | | 3+4 2 | |
| | | 7 2 | | | 2+5 2 | |
| | | 8 2 | | | 7+8 2 | |
| | | 9 2 | | | 6+9 2 | |
| | | 8 2 | | | 2+5 2 | |
| | | 9 2 | | | 3+4 2 | |
| | | 6 2 | | | 6+9 2 | |
| | | 7 2 | | | 7+8 2 | |
| | | 9 2 | | | 2+5 2 | |
| | | 8 2 | | | 3+4 2 | |
| | | 9 2 | | | 2+5 2 | |
| | | 8 2 | | | 3+4 2 | |
| | | 6 2 | | | 7+8 2 | |
| | | 7 2 | | | 6+9 2 | |
| | | 8 2 | | | 6+9 2 | |
| | | 9 2 | | | 2+5 2 | |
| | | 6 2 | | | 3+4 2 | |
| | | 7 2 | | | 7+8 2 | |
| | | 7 2 | | | 2+5 2 | |
| | | 9 2 | | | 3+4 2 | |
| | | 8 2 | | | 6+9 2 | |
| | | 8 2 | | | 7+8 2 | |
| | | 9 2 | | | 2+5 2 | |
| | | 6 2 | | | 3+4 2 | |
| | | 6 2 | | | 7+8 2 | |
| | | 7 2 | | | 6+9 2 | |
| | | 8 2 | | | 2+5 2 | |
| | | 9 2 | | | 3+4 2 | |
| | | 7 2 | | | 3+4 2 | |
| | | 6 2 | | | 2+5 2 | |
| | | 7 2 | | | 6+9 2 | |
| | | 6 2 | | | 7+8 2 | |
| | | 8 2 | | | 6+9 2 | |
| | | 9 2 | | | 7+8 2 | |
| | | 6 2 | | | 7+8 2 | |
| | | 7 2 | | | 6+9 2 | |
| | | 8 2 | | | 6+9 2 | |
| | | 7 2 | | | 7+8 2 | |
| | | 9 2 | | | 3+4 2 | |
| | | 6 2 | | | 2+5 2 | |
| | | 8 2 | | | 7+8 2 | |
| | | 9 2 | | | 6+9 2 | |
| 136.1.6 | 237545 | | 156 | | Cd4Pt2Sr1 | |
| | -A | -GM | -M | -Z | -R | -X |
| | 5 4 | 6 2 | 6+7 4 | 5 4 | 2+5 2 | 3+4 4 |
| | 5 4 | 7 2 | 8+9 4 | 5 4 | 3+4 2 | 3+4 4 |
| | 5 4 | 8 2 | 8+9 4 | 5 4 | 7+8 2 | 3+4 4 |
| | 5 4 | 9 2 | 8+9 4 | 5 4 | 6+9 2 | 3+4 4 |
| | 5 4 | 9 2 | 8+9 4 | 5 4 | 6+9 2 | 3+4 4 |
| | 5 4 | 8 2 | 6+7 4 | 5 4 | 7+8 2 | 3+4 4 |

Continued on next column

**TABLE LXII– continued**

| TQC | ICSD | NoE | SF |
|---|---|---|---|
| | 5 4 8 2 6+7 4 5 4 6+9 2 3+4 4 | | |
| | 5 4 9 2 6+7 4 5 4 7+8 2 3+4 4 | | |
| | 5 4 8 2 8+9 4 5 4 6+9 2 3+4 4 | | |
| | 5 4 7 2 6+7 4 5 4 2+5 2 3+4 4 | | |
| | 5 4 6 2 8+9 4 5 4 7+8 2 3+4 4 | | |
| | 5 4 8 2 8+9 4 5 4 3+4 2 3+4 4 | | |
| | 5 4 9 2 6+7 4 5 4 2+5 2 3+4 4 | | |
| | 5 4 8 2 8+9 4 5 4 3+4 2 3+4 4 | | |
| | 5 4 9 2 6+7 4 5 4 7+8 2 3+4 4 | | |
| | 5 4 9 2 8+9 4 5 4 6+9 2 3+4 4 | | |
| | 5 4 7 2 8+9 4 5 4 6+9 2 3+4 4 | | |
| | 5 4 8 2 8+9 4 5 4 3+4 2 3+4 4 | | |
| | 5 4 9 2 6+7 4 5 4 2+5 2 3+4 4 | | |
| | 5 4 7 2 8+9 4 5 4 7+8 2 3+4 4 | | |
| | 5 4 6 2 6+7 4 5 4 3+4 2 3+4 4 | | |
| | 5 4 6 2 8+9 4 5 4 7+8 2 3+4 4 | | |
| | 5 4 6 2 6+7 4 5 4 2+5 2 3+4 4 | | |
| | 5 4 7 2 6+7 4 5 4 6+9 2 3+4 4 | | |
| | 5 4 8 2 8+9 4 5 4 3+4 2 3+4 4 | | |
| | 5 4 7 2 6+7 4 5 4 7+8 2 3+4 4 | | |
| | 5 4 6 2 8+9 4 5 4 6+9 2 3+4 4 | | |
| | 5 4 9 2 6+7 4 5 4 2+5 2 3+4 4 | | |
| | 5 4 6 2 6+7 4 5 4 3+4 2 3+4 4 | | |
| | 5 4 9 2 8+9 4 5 4 7+8 2 3+4 4 | | |
| | 5 4 9 2 6+7 4 5 4 2+5 2 3+4 4 | | |
| | 5 4 6 2 8+9 4 5 4 6+9 2 3+4 4 | | |
| | 5 4 7 2 6+7 4 5 4 2+5 2 3+4 4 | | |
| | 5 4 7 2 8+9 4 5 4 3+4 2 3+4 4 | | |
| | 5 4 8 2 6+7 4 5 4 6+9 2 3+4 4 | | |
| | 5 4 6 2 8+9 4 5 4 7+8 2 3+4 4 | | |
| | 5 4 6 2 8+9 4 5 4 3+4 2 3+4 4 | | |
| | 5 4 7 2 6+7 4 5 4 6+9 2 3+4 4 | | |
| | 5 4 8 2 6+7 4 5 4 7+8 2 3+4 4 | | |
| | 7 2 2+5 2 | | |
| | 9 2 2+5 2 | | |
| | 6 2 6+9 2 | | |
| | 8 2 3+4 2 | | |
| | 9 2 7+8 2 | | |
| | 7 2 7+8 2 | | |
| | 8 2 2+5 2 | | |
| | 7 2 6+9 2 | | |
| | 9 2 3+4 2 | | |
| | 8 2 6+9 2 | | |
| | 7 2 2+5 2 | | |
| | 8 2 7+8 2 | | |
| | 9 2 3+4 2 | | |
| | 6 2 3+4 2 | | |
| | 7 2 2+5 2 | | |
| | 8 2 6+9 2 | | |
| | 6 2 7+8 2 | | |
| | 7 2 2+5 2 | | |
| | 9 2 3+4 2 | | |
| | 6 2 6+9 2 | | |
| | 9 2 7+8 2 | | |
| | 8 2 2+5 2 | | |
| | 6 2 3+4 2 | | |
| | 8 2 7+8 2 | | |
| | 7 2 6+9 2 | | |
| | 6 2 3+4 2 | | |
| | 8 2 7+8 2 | | |
| | 7 2 2+5 2 | | |
| | 6 2 6+9 2 | | |
| | 9 2 2+5 2 | | |
| | 9 2 7+8 2 | | |
| | 8 2 6+9 2 | | |
| | 7 2 3+4 2 | | |
| | 9 2 2+5 2 | | |
| | 9 2 7+8 2 | | |
| | 6 2 3+4 2 | | |
| | 8 2 6+9 2 | | |
| | 7 2 6+9 2 | | |
| | 8 2 7+8 2 | | |
| 136.1.7 | 62553 | 112 | Ca1Cu4P2 |
| | -A -GM -M -Z -R -X | | |
| | 5 4 7 2 8+9 4 5 4 2+5 2 3+4 4 | | |
| | 5 4 9 2 6+7 4 5 4 3+4 2 3+4 4 | | |
| | 5 4 8 2 8+9 4 5 4 6+9 2 3+4 4 | | |
| | 5 4 6 2 6+7 4 5 4 7+8 2 3+4 4 | | |
| | 5 4 7 2 6+7 4 5 4 3+4 2 3+4 4 | | |
| | 5 4 9 2 6+7 4 5 4 2+5 2 3+4 4 | | |
| | 5 4 8 2 8+9 4 5 4 6+9 2 3+4 4 | | |
| | 5 4 6 2 8+9 4 5 4 7+8 2 3+4 4 | | |
| | 5 4 9 2 6+7 4 5 4 6+9 2 3+4 4 | | |
| | 5 4 7 2 8+9 4 5 4 7+8 2 3+4 4 | | |
| | 5 4 6 2 6+7 4 5 4 2+5 2 3+4 4 | | |
| | 5 4 6 2 6+7 4 5 4 3+4 2 3+4 4 | | |
| | 5 4 7 2 8+9 4 5 4 2+5 2 3+4 4 | | |
| | 5 4 8 2 8+9 4 5 4 3+4 2 3+4 4 | | |
| | 5 4 8 2 6+7 4 5 4 7+8 2 3+4 4 | | |
| | 5 4 9 2 8+9 4 5 4 6+9 2 3+4 4 | | |
| | 5 4 6 2 6+7 4 5 4 6+9 2 3+4 4 | | |
| | 5 4 7 2 8+9 4 5 4 7+8 2 3+4 4 | | |
| | 5 4 8 2 8+9 4 5 4 2+5 2 3+4 4 | | |
| | 5 4 9 2 8+9 4 5 4 3+4 2 3+4 4 | | |
| | 5 4 9 2 8+9 4 5 4 3+4 2 3+4 4 | | |
| | 5 4 8 2 6+7 4 5 4 2+5 2 3+4 4 | | |
| | 5 4 7 2 6+7 4 5 4 6+9 2 3+4 4 | | |
| | 5 4 7 2 8+9 4 5 4 7+8 2 3+4 4 | | |
| | 5 4 8 2 6+7 4 5 4 6+9 2 3+4 4 | | |
| | 5 4 6 2 6+7 4 5 4 7+8 2 3+4 4 | | |
| | 5 4 9 2 6+7 4 5 4 2+5 2 3+4 4 | | |
| | 5 4 7 2 8+9 4 5 4 3+4 2 3+4 4 | | |
| | 6 2 3+4 2 | | |
| | 7 2 7+8 2 | | |
| | 9 2 6+9 2 | | |
| | 6 2 2+5 2 | | |
| | 9 2 2+5 2 | | |
| | 8 2 7+8 2 | | |
| | 6 2 3+4 2 | | |
| | 8 2 6+9 2 | | |
| | 6 2 6+9 2 | | |
| | 6 2 7+8 2 | | |
| | 7 2 3+4 2 | | |
| | 8 2 2+5 2 | | |
| | 9 2 2+5 2 | | |
| | 6 2 3+4 2 | | |
| | 7 2 2+5 2 | | |
| | 7 2 6+9 2 | | |
| | 9 2 7+8 2 | | |
| | 8 2 3+4 2 | | |
| | 7 2 6+9 2 | | |
| | 8 2 7+8 2 | | |

Continued on next column

**TABLE LXII– continued**

| TQC | ICSD | | NoE | | SF | |
|---|---|---|---|---|---|---|
| | | 8 2 | | | 7+8 2 | |
| | | 9 2 | | | 2+5 2 | |
| | | 6 2 | | | 3+4 2 | |
| | | 7 2 | | | 6+9 2 | |
| | | 8 2 | | | 2+5 2 | |
| | | 9 2 | | | 3+4 2 | |
| | | 6 2 | | | 6+9 2 | |
| | | 7 2 | | | 7+8 2 | |
| 136.1.8 | 417667 | | 168 | | C2Nd2W1 | |
| | -A | -GM | -M | -Z | -R | -X |
| | 5 4 | 7 2 | 6+7 4 | 5 4 | 6+9 2 | 3+4 4 |
| | 5 4 | 8 2 | 8+9 4 | 5 4 | 7+8 2 | 3+4 4 |
| | 5 4 | 6 2 | 8+9 4 | 5 4 | 2+5 2 | 3+4 4 |
| | 5 4 | 8 2 | 6+7 4 | 5 4 | 3+4 2 | 3+4 4 |
| | 5 4 | 9 2 | 8+9 4 | 5 4 | 2+5 2 | 3+4 4 |
| | 5 4 | 9 2 | 6+7 4 | 5 4 | 3+4 2 | 3+4 4 |
| | 5 4 | 6 2 | 6+7 4 | 5 4 | 6+9 2 | 3+4 4 |
| | 5 4 | 7 2 | 8+9 4 | 5 4 | 7+8 2 | 3+4 4 |
| | 5 4 | 9 2 | 8+9 4 | 5 4 | 2+5 2 | 3+4 4 |
| | 5 4 | 7 2 | 6+7 4 | 5 4 | 3+4 2 | 3+4 4 |
| | 5 4 | 6 2 | 6+7 4 | 5 4 | 6+9 2 | 3+4 4 |
| | 5 4 | 8 2 | 8+9 4 | 5 4 | 7+8 2 | 3+4 4 |
| | 5 4 | 7 2 | 8+9 4 | 5 4 | 7+8 2 | 3+4 4 |
| | 5 4 | 8 2 | 6+7 4 | 5 4 | 6+9 2 | 3+4 4 |
| | 5 4 | 6 2 | 8+9 4 | 5 4 | 2+5 2 | 3+4 4 |
| | 5 4 | 9 2 | 6+7 4 | 5 4 | 3+4 2 | 3+4 4 |
| | 5 4 | 9 2 | 6+7 4 | 5 4 | 3+4 2 | 3+4 4 |
| | 5 4 | 6 2 | 8+9 4 | 5 4 | 7+8 2 | 3+4 4 |
| | 5 4 | 7 2 | 8+9 4 | 5 4 | 6+9 2 | 3+4 4 |
| | 5 4 | 7 2 | 8+9 4 | 5 4 | 2+5 2 | 3+4 4 |
| | 5 4 | 8 2 | 6+7 4 | 5 4 | 6+9 2 | 3+4 4 |
| | 5 4 | 8 2 | 6+7 4 | 5 4 | 2+5 2 | 3+4 4 |
| | 5 4 | 6 2 | 6+7 4 | 5 4 | 7+8 2 | 3+4 4 |
| | 5 4 | 9 2 | 8+9 4 | 5 4 | 2+5 2 | 3+4 4 |
| | 5 4 | 8 2 | 8+9 4 | 5 4 | 3+4 2 | 3+4 4 |
| | 5 4 | 9 2 | 6+7 4 | 5 4 | 3+4 2 | 3+4 4 |
| | 5 4 | 6 2 | 6+7 4 | 5 4 | 6+9 2 | 3+4 4 |
| | 5 4 | 7 2 | 8+9 4 | 5 4 | 7+8 2 | 3+4 4 |
| | 5 4 | 6 2 | 6+7 4 | 5 4 | 3+4 2 | 3+4 4 |
| | 5 4 | 8 2 | 6+7 4 | 5 4 | 7+8 2 | 3+4 4 |
| | 5 4 | 9 2 | 6+7 4 | 5 4 | 6+9 2 | 3+4 4 |
| | 5 4 | 7 2 | 6+7 4 | 5 4 | 2+5 2 | 3+4 4 |
| | 5 4 | 7 2 | 8+9 4 | 5 4 | 2+5 2 | 3+4 4 |
| | 5 4 | 6 2 | 8+9 4 | 5 4 | 3+4 2 | 3+4 4 |
| | 5 4 | 8 2 | 6+7 4 | 5 4 | 6+9 2 | 3+4 4 |
| | 5 4 | 6 2 | 8+9 4 | 5 4 | 7+8 2 | 3+4 4 |
| | 5 4 | 7 2 | 6+7 4 | 5 4 | 6+9 2 | 3+4 4 |
| | 5 4 | 9 2 | 8+9 4 | 5 4 | 2+5 2 | 3+4 4 |
| | 5 4 | 6 2 | 6+7 4 | 5 4 | 7+8 2 | 3+4 4 |
| | 5 4 | 7 2 | 8+9 4 | 5 4 | 3+4 2 | 3+4 4 |
| | 5 4 | 8 2 | 8+9 4 | 5 4 | 2+5 2 | 3+4 4 |
| | 5 4 | 9 2 | 6+7 4 | 5 4 | 3+4 2 | 3+4 4 |
| | | 6 2 | | | 6+9 2 | |
| | | 9 2 | | | 7+8 2 | |
| | | 8 2 | | | 7+8 2 | |
| | | 7 2 | | | 6+9 2 | |
| | | 8 2 | | | 3+4 2 | |
| | | 8 2 | | | 2+5 2 | |
| | | 9 2 | | | 3+4 2 | |
| | | 9 2 | | | 2+5 2 | |
| | | 6 2 | | | 6+9 2 | |

Continued on next column

**TABLE LXII– continued**

| TQC | ICSD | | NoE | | SF | |
|---|---|---|---|---|---|---|
| | | 7 2 | | | 7+8 2 | |
| | | 9 2 | | | 2+5 2 | |
| | | 9 2 | | | 7+8 2 | |
| | | 7 2 | | | 3+4 2 | |
| | | 8 2 | | | 6+9 2 | |
| | | 8 2 | | | 2+5 2 | |
| | | 6 2 | | | 6+9 2 | |
| | | 9 2 | | | 3+4 2 | |
| | | 8 2 | | | 7+8 2 | |
| | | 8 2 | | | 7+8 2 | |
| | | 9 2 | | | 6+9 2 | |
| | | 7 2 | | | 3+4 2 | |
| | | 6 2 | | | 2+5 2 | |
| | | 9 2 | | | 7+8 2 | |
| | | 8 2 | | | 6+9 2 | |
| | | 7 2 | | | 3+4 2 | |
| | | 6 2 | | | 2+5 2 | |
| | | 7 2 | | | 2+5 2 | |
| | | 6 2 | | | 3+4 2 | |
| | | 8 2 | | | 6+9 2 | |
| | | 9 2 | | | 7+8 2 | |
| | | 7 2 | | | 3+4 2 | |
| | | 7 2 | | | 2+5 2 | |
| | | 6 2 | | | 7+8 2 | |
| | | 9 2 | | | 6+9 2 | |
| | | 7 2 | | | 3+4 2 | |
| | | 6 2 | | | 2+5 2 | |
| | | 8 2 | | | 7+8 2 | |
| | | 8 2 | | | 6+9 2 | |
| | | 7 2 | | | 3+4 2 | |
| | | 6 2 | | | 6+9 2 | |
| | | 6 2 | | | 2+5 2 | |
| | | 9 2 | | | 7+8 2 | |
| 136.1.9 | 413345 | | 160 | | C2Mo1Pr2 | |
| | -A | -GM | -M | -Z | -R | -X |
| | 5 4 | 7 2 | 6+7 4 | 5 4 | 6+9 2 | 3+4 4 |
| | 5 4 | 8 2 | 8+9 4 | 5 4 | 7+8 2 | 3+4 4 |
| | 5 4 | 6 2 | 8+9 4 | 5 4 | 3+4 2 | 3+4 4 |
| | 5 4 | 8 2 | 6+7 4 | 5 4 | 2+5 2 | 3+4 4 |
| | 5 4 | 9 2 | 8+9 4 | 5 4 | 2+5 2 | 3+4 4 |
| | 5 4 | 9 2 | 6+7 4 | 5 4 | 3+4 2 | 3+4 4 |
| | 5 4 | 6 2 | 6+7 4 | 5 4 | 6+9 2 | 3+4 4 |
| | 5 4 | 7 2 | 8+9 4 | 5 4 | 7+8 2 | 3+4 4 |
| | 5 4 | 9 2 | 8+9 4 | 5 4 | 2+5 2 | 3+4 4 |
| | 5 4 | 7 2 | 6+7 4 | 5 4 | 6+9 2 | 3+4 4 |
| | 5 4 | 6 2 | 6+7 4 | 5 4 | 3+4 2 | 3+4 4 |
| | 5 4 | 8 2 | 8+9 4 | 5 4 | 7+8 2 | 3+4 4 |
| | 5 4 | 7 2 | 8+9 4 | 5 4 | 7+8 2 | 3+4 4 |
| | 5 4 | 8 2 | 6+7 4 | 5 4 | 6+9 2 | 3+4 4 |
| | 5 4 | 6 2 | 8+9 4 | 5 4 | 2+5 2 | 3+4 4 |
| | 5 4 | 9 2 | 6+7 4 | 5 4 | 3+4 2 | 3+4 4 |
| | 5 4 | 9 2 | 6+7 4 | 5 4 | 3+4 2 | 3+4 4 |
| | 5 4 | 6 2 | 8+9 4 | 5 4 | 7+8 2 | 3+4 4 |
| | 5 4 | 7 2 | 8+9 4 | 5 4 | 6+9 2 | 3+4 4 |
| | 5 4 | 7 2 | 8+9 4 | 5 4 | 2+5 2 | 3+4 4 |
| | 5 4 | 8 2 | 6+7 4 | 5 4 | 6+9 2 | 3+4 4 |
| | 5 4 | 8 2 | 6+7 4 | 5 4 | 7+8 2 | 3+4 4 |
| | 5 4 | 6 2 | 6+7 4 | 5 4 | 2+5 2 | 3+4 4 |
| | 5 4 | 9 2 | 8+9 4 | 5 4 | 2+5 2 | 3+4 4 |
| | 5 4 | 8 2 | 6+7 4 | 5 4 | 3+4 2 | 3+4 4 |
| | 5 4 | 9 2 | 8+9 4 | 5 4 | 3+4 2 | 3+4 4 |

Continued on next column

**TABLE LXII– continued**

| TQC | ICSD | | NoE | | SF | |
|---|---|---|---|---|---|---|
| | 5 4 | 6 2 | 6+7 4 | 5 4 | 6+9 2 | 3+4 4 |
| | 5 4 | 7 2 | 8+9 4 | 5 4 | 7+8 2 | 3+4 4 |
| | 5 4 | 6 2 | 6+7 4 | 5 4 | 3+4 2 | 3+4 4 |
| | 5 4 | 8 2 | 6+7 4 | 5 4 | 7+8 2 | 3+4 4 |
| | 5 4 | 9 2 | 6+7 4 | 5 4 | 6+9 2 | 3+4 4 |
| | 5 4 | 7 2 | 6+7 4 | 5 4 | 2+5 2 | 3+4 4 |
| | 5 4 | 7 2 | 8+9 4 | 5 4 | 2+5 2 | 3+4 4 |
| | 5 4 | 8 2 | 8+9 4 | 5 4 | 3+4 2 | 3+4 4 |
| | 5 4 | 6 2 | 6+7 4 | 5 4 | 6+9 2 | 3+4 4 |
| | 5 4 | 7 2 | 8+9 4 | 5 4 | 7+8 2 | 3+4 4 |
| | 5 4 | 6 2 | 6+7 4 | 5 4 | 2+5 2 | 3+4 4 |
| | 5 4 | 9 2 | 8+9 4 | 5 4 | 3+4 2 | 3+4 4 |
| | 5 4 | 7 2 | 6+7 4 | 5 4 | 6+9 2 | 3+4 4 |
| | 5 4 | 6 2 | 8+9 4 | 5 4 | 7+8 2 | 3+4 4 |
| | | 8 2 | | | 3+4 2 | |
| | | 9 2 | | | 2+5 2 | |
| | | 6 2 | | | 7+8 2 | |
| | | 9 2 | | | 6+9 2 | |
| | | 8 2 | | | 7+8 2 | |
| | | 8 2 | | | 6+9 2 | |
| | | 7 2 | | | 3+4 2 | |
| | | 8 2 | | | 2+5 2 | |
| | | 9 2 | | | 6+9 2 | |
| | | 9 2 | | | 2+5 2 | |
| | | 6 2 | | | 3+4 2 | |
| | | 7 2 | | | 7+8 2 | |
| | | 9 2 | | | 2+5 2 | |
| | | 9 2 | | | 3+4 2 | |
| | | 8 2 | | | 7+8 2 | |
| | | 8 2 | | | 6+9 2 | |
| | | 7 2 | | | 2+5 2 | |
| | | 6 2 | | | 6+9 2 | |
| | | 9 2 | | | 3+4 2 | |
| | | 8 2 | | | 7+8 2 | |
| | | 8 2 | | | 6+9 2 | |
| | | 9 2 | | | 7+8 2 | |
| | | 6 2 | | | 3+4 2 | |
| | | 7 2 | | | 2+5 2 | |
| | | 9 2 | | | 7+8 2 | |
| | | 8 2 | | | 6+9 2 | |
| | | 7 2 | | | 3+4 2 | |
| | | 6 2 | | | 2+5 2 | |
| | | 7 2 | | | 2+5 2 | |
| | | 6 2 | | | 3+4 2 | |
| | | 8 2 | | | 6+9 2 | |
| | | 9 2 | | | 7+8 2 | |
| | | 7 2 | | | 2+5 2 | |
| | | 6 2 | | | 3+4 2 | |
| | | 7 2 | | | 7+8 2 | |
| | | 7 2 | | | 6+9 2 | |
| | | 9 2 | | | 3+4 2 | |
| | | 6 2 | | | 2+5 2 | |
| | | 8 2 | | | 7+8 2 | |
| | | 6 2 | | | 6+9 2 | |
| 136.1.10 | 260392 | | 160 | | Ba1Hg2Tl2 | |
| | -A | -GM | -M | -Z | -R | -X |
| | 5 4 | 6 2 | 6+7 4 | 5 4 | 2+5 2 | 3+4 4 |
| | 5 4 | 7 2 | 8+9 4 | 5 4 | 3+4 2 | 3+4 4 |
| | 5 4 | 8 2 | 6+7 4 | 5 4 | 6+9 2 | 3+4 4 |
| | 5 4 | 9 2 | 8+9 4 | 5 4 | 7+8 2 | 3+4 4 |
| | 5 4 | 6 2 | 6+7 4 | 5 4 | 2+5 2 | 3+4 4 |

Continued on next column

**TABLE LXII– continued**

| TQC | ICSD | | NoE | | SF | |
|---|---|---|---|---|---|---|
| | 5 4 | 7 2 | 8+9 4 | 5 4 | 3+4 2 | 3+4 4 |
| | 5 4 | 8 2 | 6+7 4 | 5 4 | 7+8 2 | 3+4 4 |
| | 5 4 | 9 2 | 8+9 4 | 5 4 | 6+9 2 | 3+4 4 |
| | 5 4 | 6 2 | 8+9 4 | 5 4 | 6+9 2 | 3+4 4 |
| | 5 4 | 7 2 | 6+7 4 | 5 4 | 3+4 2 | 3+4 4 |
| | 5 4 | 9 2 | 8+9 4 | 5 4 | 7+8 2 | 3+4 4 |
| | 5 4 | 8 2 | 6+7 4 | 5 4 | 2+5 2 | 3+4 4 |
| | 5 4 | 7 2 | 8+9 4 | 5 4 | 2+5 2 | 3+4 4 |
| | 5 4 | 6 2 | 6+7 4 | 5 4 | 3+4 2 | 3+4 4 |
| | 5 4 | 9 2 | 6+7 4 | 5 4 | 7+8 2 | 3+4 4 |
| | 5 4 | 8 2 | 8+9 4 | 5 4 | 6+9 2 | 3+4 4 |
| | 5 4 | 7 2 | 8+9 4 | 5 4 | 7+8 2 | 3+4 4 |
| | 5 4 | 7 2 | 6+7 4 | 5 4 | 6+9 2 | 3+4 4 |
| | 5 4 | 6 2 | 8+9 4 | 5 4 | 3+4 2 | 3+4 4 |
| | 5 4 | 7 2 | 6+7 4 | 5 4 | 2+5 2 | 3+4 4 |
| | 5 4 | 9 2 | 6+7 4 | 5 4 | 6+9 2 | 3+4 4 |
| | 5 4 | 8 2 | 8+9 4 | 5 4 | 7+8 2 | 3+4 4 |
| | 5 4 | 7 2 | 6+7 4 | 5 4 | 2+5 2 | 3+4 4 |
| | 5 4 | 6 2 | 8+9 4 | 5 4 | 3+4 2 | 3+4 4 |
| | 5 4 | 8 2 | 8+9 4 | 5 4 | 3+4 2 | 3+4 4 |
| | 5 4 | 9 2 | 6+7 4 | 5 4 | 6+9 2 | 3+4 4 |
| | 5 4 | 6 2 | 6+7 4 | 5 4 | 2+5 2 | 3+4 4 |
| | 5 4 | 8 2 | 8+9 4 | 5 4 | 7+8 2 | 3+4 4 |
| | 5 4 | 9 2 | 8+9 4 | 5 4 | 3+4 2 | 3+4 4 |
| | 5 4 | 6 2 | 6+7 4 | 5 4 | 7+8 2 | 3+4 4 |
| | 5 4 | 8 2 | 8+9 4 | 5 4 | 6+9 2 | 3+4 4 |
| | 5 4 | 7 2 | 6+7 4 | 5 4 | 2+5 2 | 3+4 4 |
| | 5 4 | 9 2 | 8+9 4 | 5 4 | 2+5 2 | 3+4 4 |
| | 5 4 | 7 2 | 6+7 4 | 5 4 | 6+9 2 | 3+4 4 |
| | 5 4 | 8 2 | 6+7 4 | 5 4 | 7+8 2 | 3+4 4 |
| | 5 4 | 9 2 | 8+9 4 | 5 4 | 3+4 2 | 3+4 4 |
| | 5 4 | 9 2 | 6+7 4 | 5 4 | 7+8 2 | 3+4 4 |
| | 5 4 | 6 2 | 8+9 4 | 5 4 | 6+9 2 | 3+4 4 |
| | 5 4 | 8 2 | 8+9 4 | 5 4 | 2+5 2 | 3+4 4 |
| | 5 4 | 6 2 | 6+7 4 | 5 4 | 7+8 2 | 3+4 4 |
| | | 7 2 | | | 3+4 2 | |
| | | 9 2 | | | 6+9 2 | |
| | | 7 2 | | | 3+4 2 | |
| | | 6 2 | | | 2+5 2 | |
| | | 8 2 | | | 2+5 2 | |
| | | 8 2 | | | 7+8 2 | |
| | | 7 2 | | | 6+9 2 | |
| | | 9 2 | | | 3+4 2 | |
| | | 8 2 | | | 3+4 2 | |
| | | 7 2 | | | 2+5 2 | |
| | | 6 2 | | | 6+9 2 | |
| | | 9 2 | | | 7+8 2 | |
| | | 6 2 | | | 6+9 2 | |
| | | 8 2 | | | 7+8 2 | |
| | | 6 2 | | | 3+4 2 | |
| | | 9 2 | | | 2+5 2 | |
| | | 7 2 | | | 7+8 2 | |
| | | 8 2 | | | 6+9 2 | |
| | | 9 2 | | | 2+5 2 | |
| | | 7 2 | | | 3+4 2 | |
| | | 6 2 | | | 6+9 2 | |
| | | 7 2 | | | 2+5 2 | |
| | | 8 2 | | | 3+4 2 | |
| | | 9 2 | | | 7+8 2 | |
| | | 6 2 | | | 6+9 2 | |
| | | 6 2 | | | 7+8 2 | |

Continued on next column

**TABLE LXII– continued**

| TQC | ICSD | | NoE | | SF | |
|---|---|---|---|---|---|---|
| | | 9 2 | | | 2+5 2 | |
| | | 8 2 | | | 3+4 2 | |
| | | 6 2 | | | 7+8 2 | |
| | | 7 2 | | | 6+9 2 | |
| | | 8 2 | | | 3+4 2 | |
| | | 9 2 | | | 2+5 2 | |
| | | 6 2 | | | 6+9 2 | |
| | | 8 2 | | | 7+8 2 | |
| | | 8 2 | | | 2+5 2 | |
| | | 7 2 | | | 3+4 2 | |
| | | 7 2 | | | 7+8 2 | |
| | | 9 2 | | | 6+9 2 | |
| | | 6 2 | | | 3+4 2 | |
| | | 7 2 | | | 2+5 2 | |
| 136.1.11 | 260391 | | 160 | | Ba1Hg2Tl2 | |
| | -A | -GM | -M | -Z | -R | -X |
| | 5 4 | 6 2 | 6+7 4 | 5 4 | 6+9 2 | 3+4 4 |
| | 5 4 | 7 2 | 8+9 4 | 5 4 | 7+8 2 | 3+4 4 |
| | 5 4 | 8 2 | 6+7 4 | 5 4 | 2+5 2 | 3+4 4 |
| | 5 4 | 9 2 | 8+9 4 | 5 4 | 3+4 2 | 3+4 4 |
| | 5 4 | 6 2 | 6+7 4 | 5 4 | 6+9 2 | 3+4 4 |
| | 5 4 | 7 2 | 8+9 4 | 5 4 | 7+8 2 | 3+4 4 |
| | 5 4 | 8 2 | 6+7 4 | 5 4 | 3+4 2 | 3+4 4 |
| | 5 4 | 9 2 | 8+9 4 | 5 4 | 2+5 2 | 3+4 4 |
| | 5 4 | 6 2 | 8+9 4 | 5 4 | 2+5 2 | 3+4 4 |
| | 5 4 | 7 2 | 6+7 4 | 5 4 | 7+8 2 | 3+4 4 |
| | 5 4 | 9 2 | 8+9 4 | 5 4 | 3+4 2 | 3+4 4 |
| | 5 4 | 8 2 | 6+7 4 | 5 4 | 6+9 2 | 3+4 4 |
| | 5 4 | 7 2 | 8+9 4 | 5 4 | 6+9 2 | 3+4 4 |
| | 5 4 | 6 2 | 6+7 4 | 5 4 | 7+8 2 | 3+4 4 |
| | 5 4 | 9 2 | 6+7 4 | 5 4 | 3+4 2 | 3+4 4 |
| | 5 4 | 8 2 | 8+9 4 | 5 4 | 2+5 2 | 3+4 4 |
| | 5 4 | 7 2 | 8+9 4 | 5 4 | 3+4 2 | 3+4 4 |
| | 5 4 | 7 2 | 6+7 4 | 5 4 | 2+5 2 | 3+4 4 |
| | 5 4 | 9 2 | 8+9 4 | 5 4 | 7+8 2 | 3+4 4 |
| | 5 4 | 6 2 | 6+7 4 | 5 4 | 6+9 2 | 3+4 4 |
| | 5 4 | 7 2 | 8+9 4 | 5 4 | 2+5 2 | 3+4 4 |
| | 5 4 | 8 2 | 6+7 4 | 5 4 | 3+4 2 | 3+4 4 |
| | 5 4 | 7 2 | 6+7 4 | 5 4 | 6+9 2 | 3+4 4 |
| | 5 4 | 6 2 | 8+9 4 | 5 4 | 7+8 2 | 3+4 4 |
| | 5 4 | 8 2 | 8+9 4 | 5 4 | 7+8 2 | 3+4 4 |
| | 5 4 | 9 2 | 6+7 4 | 5 4 | 2+5 2 | 3+4 4 |
| | 5 4 | 6 2 | 6+7 4 | 5 4 | 6+9 2 | 3+4 4 |
| | 5 4 | 8 2 | 8+9 4 | 5 4 | 3+4 2 | 3+4 4 |
| | 5 4 | 9 2 | 8+9 4 | 5 4 | 7+8 2 | 3+4 4 |
| | 5 4 | 6 2 | 6+7 4 | 5 4 | 3+4 2 | 3+4 4 |
| | 5 4 | 7 2 | 8+9 4 | 5 4 | 2+5 2 | 3+4 4 |
| | 5 4 | 8 2 | 6+7 4 | 5 4 | 6+9 2 | 3+4 4 |
| | 5 4 | 9 2 | 8+9 4 | 5 4 | 6+9 2 | 3+4 4 |
| | 5 4 | 7 2 | 6+7 4 | 5 4 | 2+5 2 | 3+4 4 |
| | 5 4 | 8 2 | 6+7 4 | 5 4 | 3+4 2 | 3+4 4 |
| | 5 4 | 9 2 | 8+9 4 | 5 4 | 7+8 2 | 3+4 4 |
| | 5 4 | 9 2 | 6+7 4 | 5 4 | 3+4 2 | 3+4 4 |
| | 5 4 | 6 2 | 8+9 4 | 5 4 | 2+5 2 | 3+4 4 |
| | 5 4 | 8 2 | 8+9 4 | 5 4 | 6+9 2 | 3+4 4 |
| | 5 4 | 6 2 | 6+7 4 | 5 4 | 3+4 2 | 3+4 4 |
| | | 7 2 | | | 7+8 2 | |
| | | 9 2 | | | 2+5 2 | |
| | | 7 2 | | | 7+8 2 | |
| | | 6 2 | | | 6+9 2 | |
| | | 8 2 | | | 6+9 2 | |

Continued on next column

**TABLE LXII– continued**

| TQC | ICSD | | NoE | | SF | |
|---|---|---|---|---|---|---|
| | | 7 2 | | | 3+4 2 | |
| | | 8 2 | | | 2+5 2 | |
| | | 9 2 | | | 7+8 2 | |
| | | 8 2 | | | 7+8 2 | |
| | | 7 2 | | | 6+9 2 | |
| | | 6 2 | | | 2+5 2 | |
| | | 9 2 | | | 3+4 2 | |
| | | 6 2 | | | 2+5 2 | |
| | | 8 2 | | | 3+4 2 | |
| | | 6 2 | | | 7+8 2 | |
| | | 9 2 | | | 6+9 2 | |
| | | 7 2 | | | 3+4 2 | |
| | | 8 2 | | | 2+5 2 | |
| | | 9 2 | | | 6+9 2 | |
| | | 7 2 | | | 7+8 2 | |
| | | 6 2 | | | 2+5 2 | |
| | | 7 2 | | | 6+9 2 | |
| | | 8 2 | | | 7+8 2 | |
| | | 9 2 | | | 3+4 2 | |
| | | 6 2 | | | 2+5 2 | |
| | | 6 2 | | | 3+4 2 | |
| | | 9 2 | | | 6+9 2 | |
| | | 8 2 | | | 7+8 2 | |
| | | 6 2 | | | 3+4 2 | |
| | | 7 2 | | | 2+5 2 | |
| | | 8 2 | | | 7+8 2 | |
| | | 9 2 | | | 6+9 2 | |
| | | 6 2 | | | 2+5 2 | |
| | | 8 2 | | | 3+4 2 | |
| | | 8 2 | | | 6+9 2 | |
| | | 7 2 | | | 7+8 2 | |
| | | 7 2 | | | 3+4 2 | |
| | | 9 2 | | | 2+5 2 | |
| | | 6 2 | | | 7+8 2 | |
| | | 7 2 | | | 6+9 2 | |
| 136.1.12 | 237544 | | 156 | | Ba1Cd4Pt2 | |
| | -A | -GM | -M | -Z | -R | -X |
| | 5 4 | 7 2 | 6+7 4 | 5 4 | 2+5 2 | 3+4 4 |
| | 5 4 | 6 2 | 8+9 4 | 5 4 | 3+4 2 | 3+4 4 |
| | 5 4 | 8 2 | 8+9 4 | 5 4 | 7+8 2 | 3+4 4 |
| | 5 4 | 9 2 | 8+9 4 | 5 4 | 6+9 2 | 3+4 4 |
| | 5 4 | 9 2 | 8+9 4 | 5 4 | 6+9 2 | 3+4 4 |
| | 5 4 | 8 2 | 6+7 4 | 5 4 | 7+8 2 | 3+4 4 |
| | 5 4 | 8 2 | 6+7 4 | 5 4 | 7+8 2 | 3+4 4 |
| | 5 4 | 9 2 | 6+7 4 | 5 4 | 6+9 2 | 3+4 4 |
| | 5 4 | 8 2 | 8+9 4 | 5 4 | 6+9 2 | 3+4 4 |
| | 5 4 | 7 2 | 6+7 4 | 5 4 | 2+5 2 | 3+4 4 |
| | 5 4 | 6 2 | 8+9 4 | 5 4 | 7+8 2 | 3+4 4 |
| | 5 4 | 8 2 | 8+9 4 | 5 4 | 3+4 2 | 3+4 4 |
| | 5 4 | 9 2 | 6+7 4 | 5 4 | 2+5 2 | 3+4 4 |
| | 5 4 | 8 2 | 8+9 4 | 5 4 | 3+4 2 | 3+4 4 |
| | 5 4 | 9 2 | 6+7 4 | 5 4 | 7+8 2 | 3+4 4 |
| | 5 4 | 9 2 | 8+9 4 | 5 4 | 6+9 2 | 3+4 4 |
| | 5 4 | 7 2 | 8+9 4 | 5 4 | 2+5 2 | 3+4 4 |
| | 5 4 | 7 2 | 6+7 4 | 5 4 | 3+4 2 | 3+4 4 |
| | 5 4 | 9 2 | 8+9 4 | 5 4 | 6+9 2 | 3+4 4 |
| | 5 4 | 6 2 | 8+9 4 | 5 4 | 7+8 2 | 3+4 4 |
| | 5 4 | 8 2 | 6+7 4 | 5 4 | 3+4 2 | 3+4 4 |
| | 5 4 | 6 2 | 8+9 4 | 5 4 | 2+5 2 | 3+4 4 |
| | 5 4 | 6 2 | 6+7 4 | 5 4 | 7+8 2 | 3+4 4 |
| | 5 4 | 7 2 | 6+7 4 | 5 4 | 6+9 2 | 3+4 4 |

Continued on next column

**TABLE LXII– continued**

| TQC | ICSD | | NoE | | SF | |
|---|---|---|---|---|---|---|
| | 5 4 | 8 2 | 8+9 4 | 5 4 | 3+4 2 | 3+4 4 |
| | 5 4 | 7 2 | 6+7 4 | 5 4 | 7+8 2 | 3+4 4 |
| | 5 4 | 6 2 | 8+9 4 | 5 4 | 6+9 2 | 3+4 4 |
| | 5 4 | 9 2 | 6+7 4 | 5 4 | 2+5 2 | 3+4 4 |
| | 5 4 | 6 2 | 6+7 4 | 5 4 | 3+4 2 | 3+4 4 |
| | 5 4 | 9 2 | 8+9 4 | 5 4 | 2+5 2 | 3+4 4 |
| | 5 4 | 9 2 | 6+7 4 | 5 4 | 7+8 2 | 3+4 4 |
| | 5 4 | 6 2 | 8+9 4 | 5 4 | 6+9 2 | 3+4 4 |
| | 5 4 | 7 2 | 6+7 4 | 5 4 | 3+4 2 | 3+4 4 |
| | 5 4 | 7 2 | 6+7 4 | 5 4 | 2+5 2 | 3+4 4 |
| | 5 4 | 8 2 | 8+9 4 | 5 4 | 6+9 2 | 3+4 4 |
| | 5 4 | 6 2 | 8+9 4 | 5 4 | 7+8 2 | 3+4 4 |
| | 5 4 | 6 2 | 8+9 4 | 5 4 | 3+4 2 | 3+4 4 |
| | 5 4 | 7 2 | 6+7 4 | 5 4 | 6+9 2 | 3+4 4 |
| | 5 4 | 8 2 | 6+7 4 | 5 4 | 2+5 2 | 3+4 4 |
| | | 7 2 | | | 7+8 2 | |
| | | 6 2 | | | 2+5 2 | |
| | | 9 2 | | | 3+4 2 | |
| | | 8 2 | | | 6+9 2 | |
| | | 9 2 | | | 7+8 2 | |
| | | 8 2 | | | 2+5 2 | |
| | | 7 2 | | | 7+8 2 | |
| | | 8 2 | | | 6+9 2 | |
| | | 9 2 | | | 3+4 2 | |
| | | 7 2 | | | 2+5 2 | |
| | | 7 2 | | | 3+4 2 | |
| | | 8 2 | | | 7+8 2 | |
| | | 9 2 | | | 6+9 2 | |
| | | 6 2 | | | 3+4 2 | |
| | | 7 2 | | | 2+5 2 | |
| | | 6 2 | | | 6+9 2 | |
| | | 8 2 | | | 7+8 2 | |
| | | 7 2 | | | 2+5 2 | |
| | | 9 2 | | | 3+4 2 | |
| | | 9 2 | | | 6+9 2 | |
| | | 6 2 | | | 7+8 2 | |
| | | 8 2 | | | 3+4 2 | |
| | | 6 2 | | | 2+5 2 | |
| | | 8 2 | | | 7+8 2 | |
| | | 7 2 | | | 3+4 2 | |
| | | 6 2 | | | 2+5 2 | |
| | | 6 2 | | | 6+9 2 | |
| | | 8 2 | | | 6+9 2 | |
| | | 7 2 | | | 7+8 2 | |
| | | 9 2 | | | 2+5 2 | |
| | | 9 2 | | | 3+4 2 | |
| | | 8 2 | | | 7+8 2 | |
| | | 7 2 | | | 6+9 2 | |
| | | 9 2 | | | 2+5 2 | |
| | | 9 2 | | | 7+8 2 | |
| | | 8 2 | | | 3+4 2 | |
| | | 6 2 | | | 6+9 2 | |
| | | 7 2 | | | 6+9 2 | |
| | | 8 2 | | | 7+8 2 | |
| 136.1.13 | 173224 | | 144 | | As2Er1Ni4 | |
| | -A | -GM | -M | -Z | -R | -X |
| | 5 4 | 6 2 | 6+7 4 | 5 4 | 6+9 2 | 3+4 4 |
| | 5 4 | 7 2 | 8+9 4 | 5 4 | 7+8 2 | 3+4 4 |
| | 5 4 | 8 2 | 8+9 4 | 5 4 | 2+5 2 | 3+4 4 |
| | 5 4 | 9 2 | 8+9 4 | 5 4 | 3+4 2 | 3+4 4 |
| | 5 4 | 9 2 | 6+7 4 | 5 4 | 2+5 2 | 3+4 4 |
| | 5 4 | 8 2 | 8+9 4 | 5 4 | 3+4 2 | 3+4 4 |
| | 5 4 | 8 2 | 8+9 4 | 5 4 | 3+4 2 | 3+4 4 |
| | 5 4 | 9 2 | 6+7 4 | 5 4 | 2+5 2 | 3+4 4 |
| | 5 4 | 7 2 | 6+7 4 | 5 4 | 6+9 2 | 3+4 4 |
| | 5 4 | 8 2 | 6+7 4 | 5 4 | 7+8 2 | 3+4 4 |
| | 5 4 | 9 2 | 8+9 4 | 5 4 | 2+5 2 | 3+4 4 |
| | 5 4 | 6 2 | 8+9 4 | 5 4 | 3+4 2 | 3+4 4 |
| | 5 4 | 7 2 | 8+9 4 | 5 4 | 7+8 2 | 3+4 4 |
| | 5 4 | 6 2 | 6+7 4 | 5 4 | 6+9 2 | 3+4 4 |
| | 5 4 | 9 2 | 6+7 4 | 5 4 | 3+4 2 | 3+4 4 |
| | 5 4 | 8 2 | 6+7 4 | 5 4 | 2+5 2 | 3+4 4 |
| | 5 4 | 7 2 | 8+9 4 | 5 4 | 2+5 2 | 3+4 4 |
| | 5 4 | 6 2 | 8+9 4 | 5 4 | 3+4 2 | 3+4 4 |
| | 5 4 | 9 2 | 6+7 4 | 5 4 | 6+9 2 | 3+4 4 |
| | 5 4 | 6 2 | 8+9 4 | 5 4 | 7+8 2 | 3+4 4 |
| | 5 4 | 7 2 | 6+7 4 | 5 4 | 6+9 2 | 3+4 4 |
| | 5 4 | 8 2 | 8+9 4 | 5 4 | 7+8 2 | 3+4 4 |
| | 5 4 | 9 2 | 8+9 4 | 5 4 | 3+4 2 | 3+4 4 |
| | 5 4 | 8 2 | 8+9 4 | 5 4 | 2+5 2 | 3+4 4 |
| | 5 4 | 6 2 | 8+9 4 | 5 4 | 7+8 2 | 3+4 4 |
| | 5 4 | 8 2 | 8+9 4 | 5 4 | 6+9 2 | 3+4 4 |
| | 5 4 | 7 2 | 8+9 4 | 5 4 | 3+4 2 | 3+4 4 |
| | 5 4 | 7 2 | 6+7 4 | 5 4 | 2+5 2 | 3+4 4 |
| | 5 4 | 9 2 | 8+9 4 | 5 4 | 7+8 2 | 3+4 4 |
| | 5 4 | 9 2 | 8+9 4 | 5 4 | 6+9 2 | 3+4 4 |
| | 5 4 | 8 2 | 6+7 4 | 5 4 | 7+8 2 | 3+4 4 |
| | 5 4 | 8 2 | 6+7 4 | 5 4 | 6+9 2 | 3+4 4 |
| | 5 4 | 9 2 | 8+9 4 | 5 4 | 2+5 2 | 3+4 4 |
| | 5 4 | 7 2 | 6+7 4 | 5 4 | 3+4 2 | 3+4 4 |
| | 5 4 | 7 2 | 8+9 4 | 5 4 | 2+5 2 | 3+4 4 |
| | 5 4 | 6 2 | 8+9 4 | 5 4 | 3+4 2 | 3+4 4 |
| | | 9 2 | | | 6+9 2 | |
| | | 8 2 | | | 3+4 2 | |
| | | 7 2 | | | 7+8 2 | |
| | | 6 2 | | | 2+5 2 | |
| | | 7 2 | | | 2+5 2 | |
| | | 6 2 | | | 7+8 2 | |
| | | 9 2 | | | 6+9 2 | |
| | | 8 2 | | | 3+4 2 | |
| | | 6 2 | | | 6+9 2 | |
| | | 6 2 | | | 7+8 2 | |
| | | 9 2 | | | 2+5 2 | |
| | | 8 2 | | | 3+4 2 | |
| | | 8 2 | | | 6+9 2 | |
| | | 6 2 | | | 7+8 2 | |
| | | 9 2 | | | 3+4 2 | |
| | | 7 2 | | | 2+5 2 | |
| | | 9 2 | | | 2+5 2 | |
| | | 6 2 | | | 3+4 2 | |
| | | 8 2 | | | 3+4 2 | |
| | | 8 2 | | | 2+5 2 | |
| | | 7 2 | | | 7+8 2 | |
| | | 9 2 | | | 3+4 2 | |
| | | 7 2 | | | 6+9 2 | |
| | | 9 2 | | | 2+5 2 | |
| | | 8 2 | | | 2+5 2 | |
| | | 6 2 | | | 7+8 2 | |
| | | 8 2 | | | 6+9 2 | |
| | | 7 2 | | | 3+4 2 | |
| | | 9 2 | | | 3+4 2 | |
| | | 8 2 | | | 2+5 2 | |

Continued on next column

**TABLE LXII– continued**

| TQC | ICSD | | NoE | | SF | |
|---|---|---|---|---|---|---|
| | | 8 2 | | | 3+4 2 | |
| | | 8 2 | | | 6+9 2 | |
| | | 9 2 | | | 7+8 2 | |
| | | 9 2 | | | 2+5 2 | |
| | | 9 2 | | | 3+4 2 | |
| | | 8 2 | | | 2+5 2 | |
| 136.1.14 | 68526 | | 140 | | As2Dy1Ni4 | |
| | -A | -GM | -M | -Z | -R | -X |
| | 5 4 | 6 2 | 6+7 4 | 5 4 | 6+9 2 | 3+4 4 |
| | 5 4 | 7 2 | 8+9 4 | 5 4 | 7+8 2 | 3+4 4 |
| | 5 4 | 8 2 | 8+9 4 | 5 4 | 2+5 2 | 3+4 4 |
| | 5 4 | 9 2 | 8+9 4 | 5 4 | 3+4 2 | 3+4 4 |
| | 5 4 | 9 2 | 6+7 4 | 5 4 | 3+4 2 | 3+4 4 |
| | 5 4 | 8 2 | 8+9 4 | 5 4 | 2+5 2 | 3+4 4 |
| | 5 4 | 8 2 | 8+9 4 | 5 4 | 3+4 2 | 3+4 4 |
| | 5 4 | 9 2 | 6+7 4 | 5 4 | 2+5 2 | 3+4 4 |
| | 5 4 | 7 2 | 6+7 4 | 5 4 | 6+9 2 | 3+4 4 |
| | 5 4 | 8 2 | 6+7 4 | 5 4 | 7+8 2 | 3+4 4 |
| | 5 4 | 9 2 | 8+9 4 | 5 4 | 2+5 2 | 3+4 4 |
| | 5 4 | 6 2 | 8+9 4 | 5 4 | 3+4 2 | 3+4 4 |
| | 5 4 | 7 2 | 8+9 4 | 5 4 | 7+8 2 | 3+4 4 |
| | 5 4 | 6 2 | 6+7 4 | 5 4 | 6+9 2 | 3+4 4 |
| | 5 4 | 9 2 | 6+7 4 | 5 4 | 3+4 2 | 3+4 4 |
| | 5 4 | 8 2 | 6+7 4 | 5 4 | 2+5 2 | 3+4 4 |
| | 5 4 | 7 2 | 8+9 4 | 5 4 | 2+5 2 | 3+4 4 |
| | 5 4 | 6 2 | 8+9 4 | 5 4 | 3+4 2 | 3+4 4 |
| | 5 4 | 9 2 | 6+7 4 | 5 4 | 6+9 2 | 3+4 4 |
| | 5 4 | 6 2 | 8+9 4 | 5 4 | 7+8 2 | 3+4 4 |
| | 5 4 | 7 2 | 6+7 4 | 5 4 | 6+9 2 | 3+4 4 |
| | 5 4 | 9 2 | 8+9 4 | 5 4 | 7+8 2 | 3+4 4 |
| | 5 4 | 8 2 | 8+9 4 | 5 4 | 3+4 2 | 3+4 4 |
| | 5 4 | 8 2 | 8+9 4 | 5 4 | 2+5 2 | 3+4 4 |
| | 5 4 | 6 2 | 8+9 4 | 5 4 | 7+8 2 | 3+4 4 |
| | 5 4 | 7 2 | 6+7 4 | 5 4 | 6+9 2 | 3+4 4 |
| | 5 4 | 8 2 | 8+9 4 | 5 4 | 3+4 2 | 3+4 4 |
| | 5 4 | 7 2 | 8+9 4 | 5 4 | 2+5 2 | 3+4 4 |
| | 5 4 | 9 2 | 6+7 4 | 5 4 | 7+8 2 | 3+4 4 |
| | 5 4 | 9 2 | 8+9 4 | 5 4 | 6+9 2 | 3+4 4 |
| | 5 4 | 8 2 | 8+9 4 | 5 4 | 7+8 2 | 3+4 4 |
| | 5 4 | 8 2 | 6+7 4 | 5 4 | 6+9 2 | 3+4 4 |
| | 5 4 | 9 2 | 6+7 4 | 5 4 | 3+4 2 | 3+4 4 |
| | 5 4 | 7 2 | 8+9 4 | 5 4 | 2+5 2 | 3+4 4 |
| | 5 4 | 7 2 | 8+9 4 | 5 4 | 2+5 2 | 3+4 4 |
| | | 6 2 | | | 3+4 2 | |
| | | 8 2 | | | 6+9 2 | |
| | | 7 2 | | | 3+4 2 | |
| | | 6 2 | | | 7+8 2 | |
| | | 9 2 | | | 2+5 2 | |
| | | 7 2 | | | 7+8 2 | |
| | | 6 2 | | | 2+5 2 | |
| | | 9 2 | | | 6+9 2 | |
| | | 8 2 | | | 3+4 2 | |
| | | 6 2 | | | 6+9 2 | |
| | | 6 2 | | | 7+8 2 | |
| | | 9 2 | | | 2+5 2 | |
| | | 8 2 | | | 3+4 2 | |
| | | 6 2 | | | 6+9 2 | |
| | | 7 2 | | | 7+8 2 | |
| | | 6 2 | | | 3+4 2 | |
| | | 7 2 | | | 2+5 2 | |
| | | 9 2 | | | 7+8 2 | |

Continued on next column

**TABLE LXII– continued**

| TQC | ICSD | | NoE | | SF | |
|---|---|---|---|---|---|---|
| | | 8 2 | | | 3+4 2 | |
| | | 7 2 | | | 6+9 2 | |
| | | 8 2 | | | 2+5 2 | |
| | | 9 2 | | | 3+4 2 | |
| | | 8 2 | | | 2+5 2 | |
| | | 9 2 | | | 3+4 2 | |
| | | 9 2 | | | 2+5 2 | |
| | | 8 2 | | | 2+5 2 | |
| | | 6 2 | | | 7+8 2 | |
| | | 8 2 | | | 6+9 2 | |
| | | 7 2 | | | 3+4 2 | |
| | | 9 2 | | | 2+5 2 | |
| | | 8 2 | | | 3+4 2 | |
| | | 8 2 | | | 6+9 2 | |
| | | 9 2 | | | 7+8 2 | |
| | | 8 2 | | | 3+4 2 | |
| | | 9 2 | | | 2+5 2 | |

The End

TABLE LXIII: The computed irreducible representations for the valence bands for the compounds of SEBR and NLC in space group 137. See the list of tables.

| TQC | ICSD | | NoE | | SF | |
|---|---|---|---|---|---|---|
| 137.1.1 | 250760 | | 120 | | Np2Se5 | |
| | -A | -GM | -M | -Z | -R | -X |
| | 5+7 4 | 7 2 | 5 4 | 5 4 | 3+4 4 | 3+4 4 |
| | 6+8 4 | 8 2 | 5 4 | 5 4 | 3+4 4 | 3+4 4 |
| | 5+7 4 | 6 2 | 5 4 | 5 4 | 3+4 4 | 3+4 4 |
| | 6+8 4 | 9 2 | 5 4 | 5 4 | 3+4 4 | 3+4 4 |
| | 6+8 4 | 7 2 | 5 4 | 5 4 | 3+4 4 | 3+4 4 |
| | 5+7 4 | 8 2 | 5 4 | 5 4 | 3+4 4 | 3+4 4 |
| | 5+7 4 | 6 2 | 5 4 | 5 4 | 3+4 4 | 3+4 4 |
| | 6+8 4 | 9 2 | 5 4 | 5 4 | 3+4 4 | 3+4 4 |
| | 5+7 4 | 6 2 | 5 4 | 5 4 | 3+4 4 | 3+4 4 |
| | 6+8 4 | 8 2 | 5 4 | 5 4 | 3+4 4 | 3+4 4 |
| | 5+7 4 | 6 2 | 5 4 | 5 4 | 3+4 4 | 3+4 4 |
| | 6+8 4 | 7 2 | 5 4 | 5 4 | 3+4 4 | 3+4 4 |
| | 5+7 4 | 9 2 | 5 4 | 5 4 | 3+4 4 | 3+4 4 |
| | 6+8 4 | 9 2 | 5 4 | 5 4 | 3+4 4 | 3+4 4 |
| | 6+8 4 | 7 2 | 5 4 | 5 4 | 3+4 4 | 3+4 4 |
| | 5+7 4 | 8 2 | 5 4 | 5 4 | 3+4 4 | 3+4 4 |
| | 5+7 4 | 7 2 | 5 4 | 5 4 | 3+4 4 | 3+4 4 |
| | 6+8 4 | 8 2 | 5 4 | 5 4 | 3+4 4 | 3+4 4 |
| | 6+8 4 | 9 2 | 5 4 | 5 4 | 3+4 4 | 3+4 4 |
| | 5+7 4 | 6 2 | 5 4 | 5 4 | 3+4 4 | 3+4 4 |
| | 6+8 4 | 8 2 | 5 4 | 5 4 | 3+4 4 | 3+4 4 |
| | 6+8 4 | 7 2 | 5 4 | 5 4 | 3+4 4 | 3+4 4 |
| | 5+7 4 | 7 2 | 5 4 | 5 4 | 3+4 4 | 3+4 4 |
| | 5+7 4 | 6 2 | 5 4 | 5 4 | 3+4 4 | 3+4 4 |
| | 5+7 4 | 8 2 | 5 4 | 5 4 | 3+4 4 | 3+4 4 |
| | 5+7 4 | 9 2 | 5 4 | 5 4 | 3+4 4 | 3+4 4 |
| | 6+8 4 | 9 2 | 5 4 | 5 4 | 3+4 4 | 3+4 4 |
| | 6+8 4 | 7 2 | 5 4 | 5 4 | 3+4 4 | 3+4 4 |
| | 6+8 4 | 8 2 | 5 4 | 5 4 | 3+4 4 | 3+4 4 |
| | 5+7 4 | 6 2 | 5 4 | 5 4 | 3+4 4 | 3+4 4 |
| | | 7 2 | | | | |
| | | 6 2 | | | | |
| | | 8 2 | | | | |
| | | 6 2 | | | | |

Continued on next column

**TABLE LXIII– continued**

| TQC | ICSD | | NoE | | SF | |
|---|---|---|---|---|---|---|
| | | 7 2 | | | | |
| | | 8 2 | | | | |
| | | 6 2 | | | | |
| | | 9 2 | | | | |
| | | 7 2 | | | | |
| | | 6 2 | | | | |
| | | 9 2 | | | | |
| | | 7 2 | | | | |
| | | 6 2 | | | | |
| | | 9 2 | | | | |
| | | 8 2 | | | | |
| | | 9 2 | | | | |
| | | 6 2 | | | | |
| | | 9 2 | | | | |
| | | 7 2 | | | | |
| | | 8 2 | | | | |
| | | 8 2 | | | | |
| | | 9 2 | | | | |
| | | 6 2 | | | | |
| | | 7 2 | | | | |
| | | 9 2 | | | | |
| | | 6 2 | | | | |
| | | 9 2 | | | | |
| | | 7 2 | | | | |
| | | 6 2 | | | | |
| | | 8 2 | | | | |
| 137.1.2 | 248874 | | 140 | | Ba2S6U1 | |
| | -A | -GM | -M | -Z | -R | -X |
| | 6+8 4 | 7 2 | 5 4 | 5 4 | 3+4 4 | 3+4 4 |
| | 5+7 4 | 8 2 | 5 4 | 5 4 | 3+4 4 | 3+4 4 |
| | 6+8 4 | 7 2 | 5 4 | 5 4 | 3+4 4 | 3+4 4 |
| | 5+7 4 | 8 2 | 5 4 | 5 4 | 3+4 4 | 3+4 4 |
| | 5+7 4 | 6 2 | 5 4 | 5 4 | 3+4 4 | 3+4 4 |
| | 6+8 4 | 9 2 | 5 4 | 5 4 | 3+4 4 | 3+4 4 |
| | 6+8 4 | 6 2 | 5 4 | 5 4 | 3+4 4 | 3+4 4 |
| | 5+7 4 | 9 2 | 5 4 | 5 4 | 3+4 4 | 3+4 4 |
| | 6+8 4 | 6 2 | 5 4 | 5 4 | 3+4 4 | 3+4 4 |
| | 5+7 4 | 9 2 | 5 4 | 5 4 | 3+4 4 | 3+4 4 |
| | 5+7 4 | 7 2 | 5 4 | 5 4 | 3+4 4 | 3+4 4 |
| | 6+8 4 | 8 2 | 5 4 | 5 4 | 3+4 4 | 3+4 4 |
| | 5+7 4 | 7 2 | 5 4 | 5 4 | 3+4 4 | 3+4 4 |
| | 6+8 4 | 8 2 | 5 4 | 5 4 | 3+4 4 | 3+4 4 |
| | 5+7 4 | 6 2 | 5 4 | 5 4 | 3+4 4 | 3+4 4 |
| | 6+8 4 | 9 2 | 5 4 | 5 4 | 3+4 4 | 3+4 4 |
| | 5+7 4 | 7 2 | 5 4 | 5 4 | 3+4 4 | 3+4 4 |
| | 6+8 4 | 8 2 | 5 4 | 5 4 | 3+4 4 | 3+4 4 |
| | 6+8 4 | 6 2 | 5 4 | 5 4 | 3+4 4 | 3+4 4 |
| | 6+8 4 | 9 2 | 5 4 | 5 4 | 3+4 4 | 3+4 4 |
| | 5+7 4 | 7 2 | 5 4 | 5 4 | 3+4 4 | 3+4 4 |
| | 5+7 4 | 8 2 | 5 4 | 5 4 | 3+4 4 | 3+4 4 |
| | 6+8 4 | 6 2 | 5 4 | 5 4 | 3+4 4 | 3+4 4 |
| | 5+7 4 | 7 2 | 5 4 | 5 4 | 3+4 4 | 3+4 4 |
| | 5+7 4 | 9 2 | 5 4 | 5 4 | 3+4 4 | 3+4 4 |
| | 6+8 4 | 6 2 | 5 4 | 5 4 | 3+4 4 | 3+4 4 |
| | 5+7 4 | 8 2 | 5 4 | 5 4 | 3+4 4 | 3+4 4 |
| | 5+7 4 | 9 2 | 5 4 | 5 4 | 3+4 4 | 3+4 4 |
| | 6+8 4 | 6 2 | 5 4 | 5 4 | 3+4 4 | 3+4 4 |
| | 6+8 4 | 9 2 | 5 4 | 5 4 | 3+4 4 | 3+4 4 |
| | 6+8 4 | 8 2 | 5 4 | 5 4 | 3+4 4 | 3+4 4 |
| | 5+7 4 | 7 2 | 5 4 | 5 4 | 3+4 4 | 3+4 4 |
| | 5+7 4 | 9 2 | 5 4 | 5 4 | 3+4 4 | 3+4 4 |

Continued on next column

**TABLE LXIII– continued**

| TQC | ICSD | | NoE | | SF | |
|---|---|---|---|---|---|---|
| | 6+8 4 | 6 2 | 5 4 | 5 4 | 3+4 4 | 3+4 4 |
| | 6+8 4 | 8 2 | 5 4 | 5 4 | 3+4 4 | 3+4 4 |
| | | 7 2 | | | | |
| | | 7 2 | | | | |
| | | 8 2 | | | | |
| | | 9 2 | | | | |
| | | 6 2 | | | | |
| | | 7 2 | | | | |
| | | 8 2 | | | | |
| | | 6 2 | | | | |
| | | 7 2 | | | | |
| | | 6 2 | | | | |
| | | 9 2 | | | | |
| | | 8 2 | | | | |
| | | 9 2 | | | | |
| | | 6 2 | | | | |
| | | 7 2 | | | | |
| | | 9 2 | | | | |
| | | 8 2 | | | | |
| | | 7 2 | | | | |
| | | 8 2 | | | | |
| | | 6 2 | | | | |
| | | 7 2 | | | | |
| | | 6 2 | | | | |
| | | 6 2 | | | | |
| | | 9 2 | | | | |
| | | 7 2 | | | | |
| | | 9 2 | | | | |
| | | 8 2 | | | | |
| | | 6 2 | | | | |
| | | 7 2 | | | | |
| | | 9 2 | | | | |
| | | 9 2 | | | | |
| | | 8 2 | | | | |
| | | 8 2 | | | | |
| | | 6 2 | | | | |
| | | 7 2 | | | | |
| 137.1.3 | 49631 | | 140 | | B4Er1Rh4 | |
| | -A | -GM | -M | -Z | -R | -X |
| | 5+7 4 | 7 2 | 5 4 | 5 4 | 3+4 4 | 3+4 4 |
| | 6+8 4 | 8 2 | 5 4 | 5 4 | 3+4 4 | 3+4 4 |
| | 6+8 4 | 6 2 | 5 4 | 5 4 | 3+4 4 | 3+4 4 |
| | 5+7 4 | 9 2 | 5 4 | 5 4 | 3+4 4 | 3+4 4 |
| | 6+8 4 | 6 2 | 5 4 | 5 4 | 3+4 4 | 3+4 4 |
| | 5+7 4 | 9 2 | 5 4 | 5 4 | 3+4 4 | 3+4 4 |
| | 6+8 4 | 7 2 | 5 4 | 5 4 | 3+4 4 | 3+4 4 |
| | 5+7 4 | 8 2 | 5 4 | 5 4 | 3+4 4 | 3+4 4 |
| | 6+8 4 | 7 2 | 5 4 | 5 4 | 3+4 4 | 3+4 4 |
| | 5+7 4 | 6 2 | 5 4 | 5 4 | 3+4 4 | 3+4 4 |
| | 5+7 4 | 9 2 | 5 4 | 5 4 | 3+4 4 | 3+4 4 |
| | 6+8 4 | 8 2 | 5 4 | 5 4 | 3+4 4 | 3+4 4 |
| | 6+8 4 | 9 2 | 5 4 | 5 4 | 3+4 4 | 3+4 4 |
| | 5+7 4 | 8 2 | 5 4 | 5 4 | 3+4 4 | 3+4 4 |
| | 5+7 4 | 8 2 | 5 4 | 5 4 | 3+4 4 | 3+4 4 |
| | 6+8 4 | 6 2 | 5 4 | 5 4 | 3+4 4 | 3+4 4 |
| | 6+8 4 | 7 2 | 5 4 | 5 4 | 3+4 4 | 3+4 4 |
| | 5+7 4 | 6 2 | 5 4 | 5 4 | 3+4 4 | 3+4 4 |
| | 5+7 4 | 9 2 | 5 4 | 5 4 | 3+4 4 | 3+4 4 |
| | 6+8 4 | 6 2 | 5 4 | 5 4 | 3+4 4 | 3+4 4 |
| | 5+7 4 | 7 2 | 5 4 | 5 4 | 3+4 4 | 3+4 4 |
| | 6+8 4 | 7 2 | 5 4 | 5 4 | 3+4 4 | 3+4 4 |

Continued on next column

**TABLE LXIII– continued**

| TQC | ICSD | NoE | SF |
|---|---|---|---|
| | 5+7 4 | 9 2 | 5 4 5 4 3+4 4 3+4 4 |
| | 6+8 4 | 7 2 | 5 4 5 4 3+4 4 3+4 4 |
| | 6+8 4 | 7 2 | 5 4 5 4 3+4 4 3+4 4 |
| | 5+7 4 | 6 2 | 5 4 5 4 3+4 4 3+4 4 |
| | 6+8 4 | 8 2 | 5 4 5 4 3+4 4 3+4 4 |
| | 5+7 4 | 9 2 | 5 4 5 4 3+4 4 3+4 4 |
| | 5+7 4 | 6 2 | 5 4 5 4 3+4 4 3+4 4 |
| | 6+8 4 | 8 2 | 5 4 5 4 3+4 4 3+4 4 |
| | 5+7 4 | 9 2 | 5 4 5 4 3+4 4 3+4 4 |
| | 6+8 4 | 7 2 | 5 4 5 4 3+4 4 3+4 4 |
| | 5+7 4 | 6 2 | 5 4 5 4 3+4 4 3+4 4 |
| | 6+8 4 | 9 2 | 5 4 5 4 3+4 4 3+4 4 |
| | 5+7 4 | 8 2 | 5 4 5 4 3+4 4 3+4 4 |
| | | 8 2 | |
| | | 6 2 | |
| | | 7 2 | |
| | | 7 2 | |
| | | 6 2 | |
| | | 8 2 | |
| | | 7 2 | |
| | | 8 2 | |
| | | 9 2 | |
| | | 8 2 | |
| | | 6 2 | |
| | | 9 2 | |
| | | 7 2 | |
| | | 6 2 | |
| | | 9 2 | |
| | | 8 2 | |
| | | 8 2 | |
| | | 6 2 | |
| | | 7 2 | |
| | | 7 2 | |
| | | 6 2 | |
| | | 9 2 | |
| | | 8 2 | |
| | | 7 2 | |
| | | 6 2 | |
| | | 8 2 | |
| | | 7 2 | |
| | | 6 2 | |
| | | 7 2 | |
| | | 9 2 | |
| | | 9 2 | |
| | | 8 2 | |
| | | 9 2 | |
| | | 7 2 | |
| | | 6 2 | |

The End

TABLE LXIV: The computed irreducible representations for the valence bands for the compounds of SEBR and NLC in space group 139. See the list of tables.

| TQC | ICSD | NoE | SF |
|---|---|---|---|
| 139.1.1 | 263005 | 92 | Si3Y1 |
| | -GM -M | -P -X | -N |
| | 7 2 7 2 | 6 2 5 2 | 3+4 2 |
| | 9 2 9 2 | 7 2 6 2 | 5+6 2 |
| | 7 2 6 2 | 6 2 5 2 | 3+4 2 |
| | 8 2 9 2 | 7 2 6 2 | 5+6 2 |
| | 7 2 7 2 | 6 2 5 2 | 3+4 2 |
| | 9 2 9 2 | 7 2 6 2 | 5+6 2 |
| | 6 2 8 2 | 6 2 6 2 | 3+4 2 |
| | 9 2 7 2 | 7 2 5 2 | 5+6 2 |
| | 7 2 8 2 | 6 2 5 2 | 3+4 2 |
| | 8 2 6 2 | 6 2 6 2 | 5+6 2 |
| | 6 2 7 2 | 7 2 5 2 | 3+4 2 |
| | 9 2 9 2 | 7 2 6 2 | 5+6 2 |
| | 9 2 7 2 | 6 2 6 2 | 3+4 2 |
| | 6 2 8 2 | 7 2 5 2 | 5+6 2 |
| | 7 2 9 2 | 6 2 5 2 | 3+4 2 |
| | 8 2 6 2 | 7 2 6 2 | 5+6 2 |
| | 7 2 7 2 | 7 2 6 2 | 5+6 2 |
| | 9 2 8 2 | 7 2 5 2 | 3+4 2 |
| | 8 2 9 2 | 6 2 6 2 | 3+4 2 |
| | 6 2 6 2 | 6 2 5 2 | 5+6 2 |
| | 6 2 9 2 | 6 2 6 2 | 5+6 2 |
| | 9 2 7 2 | 6 2 5 2 | 3+4 2 |
| | 7 2 6 2 | 7 2 6 2 | 3+4 2 |
| | 6 2 8 2 | 6 2 6 2 | 5+6 2 |
| | 7 2 9 2 | 7 2 5 2 | 5+6 2 |
| | 9 2 7 2 | 7 2 5 2 | 3+4 2 |
| | 8 2 6 2 | 7 2 6 2 | 5+6 2 |
| | 7 2 8 2 | 6 2 5 2 | 3+4 2 |
| | 8 2 7 2 | 7 2 5 2 | 5+6 2 |
| | 9 2 9 2 | 7 2 6 2 | 3+4 2 |
| | 7 2 8 2 | 6 2 6 2 | 3+4 2 |
| | 8 2 9 2 | 6 2 5 2 | 5+6 2 |
| | 6 2 8 2 | 7 2 6 2 | 3+4 2 |
| | 7 2 7 2 | 7 2 5 2 | 5+6 2 |
| | 6 2 6 2 | 6 2 5 2 | 3+4 2 |
| | 6 2 7 2 | 7 2 6 2 | 5+6 2 |
| | 9 2 6 2 | 6 2 6 2 | 3+4 2 |
| | 8 2 9 2 | 6 2 5 2 | 5+6 2 |
| | 7 2 9 2 | 7 2 6 2 | 5+6 2 |
| | 9 2 8 2 | 6 2 6 2 | 3+4 2 |
| | 8 2 7 2 | 6 2 5 2 | 3+4 2 |
| | 7 2 6 2 | 7 2 6 2 | 5+6 2 |
| | 6 2 8 2 | 6 2 5 2 | 5+6 2 |
| | 8 2 7 2 | 7 2 6 2 | 3+4 2 |
| | 7 2 6 2 | 6 2 5 2 | 3+4 2 |
| | 9 2 7 2 | 7 2 6 2 | 5+6 2 |
| 139.1.2 | 42119 | 50 | Sb1Sr2 |
| | -GM -M | -P -X | -N |
| | 7 2 9 2 | 6 2 5 2 | 3+4 2 |
| | 9 2 7 2 | 7 2 6 2 | 5+6 2 |
| | 7 2 7 2 | 6 2 6 2 | 3+4 2 |
| | 6 2 6 2 | 7 2 6 2 | 5+6 2 |
| | 7 2 9 2 | 7 2 5 2 | 3+4 2 |
| | 9 2 7 2 | 6 2 6 2 | 5+6 2 |
| | 8 2 8 2 | 6 2 5 2 | 5+6 2 |
| | 9 2 9 2 | 7 2 5 2 | 3+4 2 |
| | 9 2 9 2 | 6 2 5 2 | 5+6 2 |
| | 7 2 8 2 | 7 2 6 2 | 5+6 2 |
| | 6 2 9 2 | 6 2 5 2 | 3+4 2 |
| | 9 2 7 2 | 7 2 6 2 | 5+6 2 |
| | 8 2 6 2 | 6 2 5 2 | 3+4 2 |
| | 8 2 8 2 | 7 2 5 2 | 5+6 2 |
| | 9 2 9 2 | 7 2 5 2 | 5+6 2 |
| | 8 2 8 2 | 6 2 5 2 | 3+4 2 |

Continued on next column

**TABLE LXIV– continued**

| TQC | ICSD | NoE | SF |
|---|---|---|---|
| | 7 2 7 2 | 7 2 6 2 | 3+4 2 |
| | 9 2 9 2 | 6 2 5 2 | 5+6 2 |
| | 7 2 7 2 | 7 2 6 2 | 5+6 2 |
| | 7 2 9 2 | 6 2 6 2 | 3+4 2 |
| | 7 2 7 2 | 6 2 5 2 | 3+4 2 |
| | 9 2 6 2 | 7 2 5 2 | 5+6 2 |
| | 6 2 7 2 | 7 2 6 2 | 3+4 2 |
| | 9 2 9 2 | 6 2 5 2 | 5+6 2 |
| | 8 2 8 2 | 6 2 6 2 | 5+6 2 |
| 139.1.3 | 416747 | 88 | Ru1Zn3 |
| | -GM -M | -P -X | -N |
| | 7 2 7 2 | 6 2 6 2 | 5+6 2 |
| | 7 2 9 2 | 6 2 6 2 | 3+4 2 |
| | 7 2 7 2 | 7 2 6 2 | 5+6 2 |
| | 6 2 6 2 | 7 2 5 2 | 5+6 2 |
| | 7 2 6 2 | 7 2 6 2 | 3+4 2 |
| | 9 2 7 2 | 6 2 5 2 | 3+4 2 |
| | 6 2 6 2 | 6 2 6 2 | 3+4 2 |
| | 7 2 7 2 | 7 2 6 2 | 5+6 2 |
| | 6 2 9 2 | 6 2 5 2 | 3+4 2 |
| | 9 2 8 2 | 7 2 6 2 | 5+6 2 |
| | 8 2 7 2 | 7 2 6 2 | 3+4 2 |
| | 6 2 8 2 | 6 2 5 2 | 5+6 2 |
| | 8 2 6 2 | 7 2 5 2 | 5+6 2 |
| | 9 2 9 2 | 6 2 6 2 | 3+4 2 |
| | 8 2 7 2 | 6 2 6 2 | 3+4 2 |
| | 7 2 6 2 | 7 2 6 2 | 3+4 2 |
| | 6 2 8 2 | 7 2 6 2 | 5+6 2 |
| | 6 2 7 2 | 6 2 5 2 | 3+4 2 |
| | 7 2 8 2 | 7 2 6 2 | 5+6 2 |
| | 6 2 6 2 | 6 2 5 2 | 5+6 2 |
| | 7 2 9 2 | 6 2 6 2 | 3+4 2 |
| | 9 2 7 2 | 7 2 6 2 | 5+6 2 |
| | 6 2 6 2 | 6 2 6 2 | 3+4 2 |
| | 8 2 9 2 | 7 2 5 2 | 5+6 2 |
| | 7 2 8 2 | 6 2 6 2 | 3+4 2 |
| | 6 2 6 2 | 7 2 5 2 | 5+6 2 |
| | 6 2 7 2 | 7 2 6 2 | 3+4 2 |
| | 7 2 8 2 | 6 2 6 2 | 5+6 2 |
| | 8 2 6 2 | 6 2 5 2 | 3+4 2 |
| | 9 2 6 2 | 7 2 6 2 | 5+6 2 |
| | 6 2 7 2 | 7 2 5 2 | 3+4 2 |
| | 7 2 7 2 | 6 2 6 2 | 5+6 2 |
| | 9 2 9 2 | 7 2 5 2 | 5+6 2 |
| | 7 2 7 2 | 6 2 6 2 | 3+4 2 |
| | 6 2 8 2 | 6 2 6 2 | 5+6 2 |
| | 8 2 6 2 | 6 2 5 2 | 3+4 2 |
| | 9 2 9 2 | 7 2 6 2 | 3+4 2 |
| | 8 2 8 2 | 7 2 5 2 | 3+4 2 |
| | 7 2 7 2 | 6 2 5 2 | 5+6 2 |
| | 6 2 7 2 | 7 2 6 2 | 5+6 2 |
| | 9 2 6 2 | 6 2 6 2 | 5+6 2 |
| | 8 2 8 2 | 7 2 5 2 | 3+4 2 |
| | 6 2 6 2 | 6 2 5 2 | 3+4 2 |
| | 9 2 9 2 | 7 2 5 2 | 3+4 2 |
| 139.1.4 | 249448 | 88 | Ru1Zn3 |
| | -GM -M | -P -X | -N |
| | 7 2 7 2 | 6 2 6 2 | 5+6 2 |
| | 7 2 9 2 | 6 2 6 2 | 3+4 2 |
| | 7 2 7 2 | 7 2 6 2 | 5+6 2 |
| | 6 2 6 2 | 7 2 5 2 | 5+6 2 |

Continued on next column

**TABLE LXIV– continued**

| TQC | ICSD | NoE | SF |
|---|---|---|---|
| | 7 2 6 2 | 7 2 6 2 | 3+4 2 |
| | 9 2 7 2 | 6 2 5 2 | 3+4 2 |
| | 6 2 6 2 | 6 2 6 2 | 3+4 2 |
| | 7 2 7 2 | 7 2 6 2 | 5+6 2 |
| | 6 2 9 2 | 6 2 5 2 | 3+4 2 |
| | 9 2 8 2 | 7 2 6 2 | 5+6 2 |
| | 8 2 7 2 | 7 2 6 2 | 3+4 2 |
| | 6 2 8 2 | 6 2 5 2 | 5+6 2 |
| | 8 2 6 2 | 7 2 5 2 | 5+6 2 |
| | 9 2 9 2 | 6 2 6 2 | 3+4 2 |
| | 8 2 7 2 | 6 2 6 2 | 3+4 2 |
| | 7 2 6 2 | 7 2 6 2 | 3+4 2 |
| | 6 2 8 2 | 7 2 6 2 | 5+6 2 |
| | 6 2 7 2 | 6 2 5 2 | 3+4 2 |
| | 7 2 8 2 | 7 2 6 2 | 5+6 2 |
| | 6 2 6 2 | 6 2 5 2 | 5+6 2 |
| | 7 2 9 2 | 6 2 6 2 | 3+4 2 |
| | 9 2 7 2 | 7 2 6 2 | 5+6 2 |
| | 6 2 6 2 | 6 2 6 2 | 3+4 2 |
| | 8 2 9 2 | 7 2 5 2 | 5+6 2 |
| | 7 2 8 2 | 6 2 6 2 | 3+4 2 |
| | 6 2 6 2 | 7 2 5 2 | 5+6 2 |
| | 6 2 7 2 | 7 2 6 2 | 3+4 2 |
| | 7 2 8 2 | 6 2 6 2 | 5+6 2 |
| | 8 2 6 2 | 6 2 5 2 | 3+4 2 |
| | 9 2 6 2 | 7 2 6 2 | 5+6 2 |
| | 6 2 7 2 | 7 2 5 2 | 3+4 2 |
| | 7 2 7 2 | 6 2 6 2 | 5+6 2 |
| | 9 2 9 2 | 7 2 5 2 | 5+6 2 |
| | 7 2 7 2 | 6 2 6 2 | 3+4 2 |
| | 6 2 8 2 | 6 2 6 2 | 5+6 2 |
| | 8 2 6 2 | 6 2 5 2 | 3+4 2 |
| | 9 2 9 2 | 7 2 6 2 | 3+4 2 |
| | 8 2 8 2 | 7 2 5 2 | 3+4 2 |
| | 7 2 7 2 | 6 2 5 2 | 5+6 2 |
| | 6 2 7 2 | 7 2 6 2 | 5+6 2 |
| | 9 2 6 2 | 6 2 6 2 | 5+6 2 |
| | 8 2 8 2 | 7 2 5 2 | 3+4 2 |
| | 6 2 6 2 | 6 2 5 2 | 3+4 2 |
| | 9 2 9 2 | 7 2 5 2 | 3+4 2 |
| 139.1.5 | 155990 | 88 | Ru1Zn3 |
| | -GM -M | -P -X | -N |
| | 7 2 7 2 | 6 2 6 2 | 5+6 2 |
| | 7 2 9 2 | 6 2 6 2 | 3+4 2 |
| | 7 2 7 2 | 7 2 6 2 | 5+6 2 |
| | 6 2 6 2 | 7 2 5 2 | 5+6 2 |
| | 7 2 6 2 | 7 2 6 2 | 3+4 2 |
| | 9 2 7 2 | 6 2 5 2 | 3+4 2 |
| | 6 2 6 2 | 6 2 6 2 | 3+4 2 |
| | 7 2 7 2 | 7 2 6 2 | 5+6 2 |
| | 6 2 9 2 | 6 2 5 2 | 3+4 2 |
| | 9 2 8 2 | 7 2 6 2 | 5+6 2 |
| | 8 2 7 2 | 7 2 6 2 | 3+4 2 |
| | 6 2 8 2 | 6 2 5 2 | 5+6 2 |
| | 8 2 6 2 | 7 2 5 2 | 5+6 2 |
| | 9 2 9 2 | 6 2 6 2 | 3+4 2 |
| | 8 2 7 2 | 6 2 6 2 | 3+4 2 |
| | 7 2 6 2 | 7 2 6 2 | 3+4 2 |
| | 6 2 8 2 | 7 2 6 2 | 5+6 2 |
| | 6 2 7 2 | 6 2 5 2 | 3+4 2 |
| | 7 2 8 2 | 7 2 6 2 | 5+6 2 |

Continued on next column

**TABLE LXIV– continued**

| TQC | ICSD | NoE | SF |
|---|---|---|---|
| | 6 2 6 2 | 6 2 5 2 | 5+6 2 |
| | 7 2 9 2 | 6 2 6 2 | 3+4 2 |
| | 9 2 7 2 | 7 2 6 2 | 5+6 2 |
| | 6 2 6 2 | 6 2 6 2 | 3+4 2 |
| | 8 2 9 2 | 7 2 5 2 | 5+6 2 |
| | 7 2 8 2 | 6 2 6 2 | 3+4 2 |
| | 6 2 6 2 | 7 2 5 2 | 5+6 2 |
| | 6 2 7 2 | 7 2 6 2 | 3+4 2 |
| | 7 2 8 2 | 6 2 6 2 | 5+6 2 |
| | 8 2 6 2 | 6 2 5 2 | 3+4 2 |
| | 9 2 6 2 | 7 2 6 2 | 5+6 2 |
| | 6 2 7 2 | 7 2 5 2 | 3+4 2 |
| | 7 2 7 2 | 6 2 6 2 | 5+6 2 |
| | 9 2 9 2 | 7 2 5 2 | 5+6 2 |
| | 7 2 7 2 | 6 2 6 2 | 3+4 2 |
| | 6 2 8 2 | 6 2 6 2 | 5+6 2 |
| | 8 2 6 2 | 6 2 5 2 | 3+4 2 |
| | 9 2 9 2 | 7 2 6 2 | 3+4 2 |
| | 8 2 8 2 | 7 2 5 2 | 3+4 2 |
| | 7 2 7 2 | 6 2 5 2 | 5+6 2 |
| | 6 2 7 2 | 7 2 6 2 | 5+6 2 |
| | 9 2 6 2 | 6 2 6 2 | 5+6 2 |
| | 8 2 8 2 | 7 2 5 2 | 3+4 2 |
| | 6 2 6 2 | 6 2 5 2 | 3+4 2 |
| | 9 2 9 2 | 7 2 5 2 | 3+4 2 |
| 139.1.6 | 251818 | 64 | Pt2Th1 |
| | -GM -M | -P -X | -N |
| | 7 2 9 2 | 6 2 5 2 | 5+6 2 |
| | 9 2 7 2 | 7 2 6 2 | 3+4 2 |
| | 7 2 9 2 | 6 2 6 2 | 5+6 2 |
| | 9 2 7 2 | 7 2 5 2 | 3+4 2 |
| | 7 2 9 2 | 7 2 6 2 | 3+4 2 |
| | 6 2 7 2 | 6 2 5 2 | 5+6 2 |
| | 9 2 8 2 | 7 2 6 2 | 5+6 2 |
| | 8 2 6 2 | 6 2 5 2 | 3+4 2 |
| | 7 2 7 2 | 7 2 6 2 | 3+4 2 |
| | 6 2 6 2 | 6 2 6 2 | 5+6 2 |
| | 7 2 9 2 | 7 2 6 2 | 3+4 2 |
| | 6 2 7 2 | 6 2 5 2 | 5+6 2 |
| | 7 2 6 2 | 6 2 6 2 | 3+4 2 |
| | 6 2 7 2 | 6 2 5 2 | 5+6 2 |
| | 7 2 6 2 | 7 2 6 2 | 5+6 2 |
| | 8 2 7 2 | 7 2 5 2 | 3+4 2 |
| | 9 2 8 2 | 6 2 6 2 | 5+6 2 |
| | 8 2 9 2 | 6 2 5 2 | 3+4 2 |
| | 9 2 8 2 | 7 2 6 2 | 3+4 2 |
| | 6 2 8 2 | 7 2 6 2 | 5+6 2 |
| | 8 2 6 2 | 6 2 5 2 | 5+6 2 |
| | 7 2 7 2 | 6 2 6 2 | 3+4 2 |
| | 9 2 9 2 | 7 2 6 2 | 5+6 2 |
| | 6 2 6 2 | 7 2 5 2 | 5+6 2 |
| | 7 2 7 2 | 6 2 6 2 | 3+4 2 |
| | 7 2 6 2 | 7 2 6 2 | 3+4 2 |
| | 6 2 7 2 | 7 2 6 2 | 5+6 2 |
| | 6 2 6 2 | 6 2 6 2 | 3+4 2 |
| | 6 2 7 2 | 7 2 6 2 | 5+6 2 |
| | 7 2 6 2 | 6 2 5 2 | 3+4 2 |
| | 7 2 7 2 | 7 2 6 2 | 3+4 2 |
| | 9 2 9 2 | 6 2 5 2 | 5+6 2 |
| 139.1.7 | 423291 | 62 | Pr2Sb1 |
| | -GM -M | -P -X | -N |

Continued on next column

| TQC | ICSD | NoE | SF |
|---|---|---|---|
| | 7 2 9 2 | 6 2 5 2 | 3+4 2 |
| | 9 2 7 2 | 7 2 6 2 | 5+6 2 |
| | 7 2 7 2 | 6 2 6 2 | 5+6 2 |
| | 6 2 6 2 | 7 2 6 2 | 3+4 2 |
| | 8 2 8 2 | 7 2 5 2 | 5+6 2 |
| | 7 2 9 2 | 6 2 5 2 | 3+4 2 |
| | 9 2 7 2 | 7 2 6 2 | 3+4 2 |
| | 9 2 9 2 | 6 2 5 2 | 5+6 2 |
| | 9 2 9 2 | 7 2 5 2 | 3+4 2 |
| | 7 2 9 2 | 6 2 5 2 | 5+6 2 |
| | 6 2 8 2 | 6 2 6 2 | 3+4 2 |
| | 9 2 7 2 | 7 2 5 2 | 5+6 2 |
| | 8 2 6 2 | 6 2 6 2 | 5+6 2 |
| | 8 2 8 2 | 7 2 5 2 | 3+4 2 |
| | 9 2 9 2 | 6 2 5 2 | 3+4 2 |
| | 8 2 8 2 | 7 2 5 2 | 5+6 2 |
| | 7 2 7 2 | 6 2 6 2 | 5+6 2 |
| | 9 2 9 2 | 7 2 5 2 | 3+4 2 |
| | 7 2 7 2 | 6 2 6 2 | 3+4 2 |
| | 9 2 9 2 | 7 2 6 2 | 5+6 2 |
| | 7 2 7 2 | 6 2 6 2 | 5+6 2 |
| | 7 2 7 2 | 7 2 5 2 | 3+4 2 |
| | 6 2 6 2 | 6 2 5 2 | 5+6 2 |
| | 9 2 9 2 | 7 2 5 2 | 3+4 2 |
| | 8 2 8 2 | 7 2 6 2 | 3+4 2 |
| | 6 2 6 2 | 7 2 6 2 | 5+6 2 |
| | 8 2 6 2 | 6 2 5 2 | 5+6 2 |
| | 7 2 7 2 | 6 2 5 2 | 3+4 2 |
| | 6 2 9 2 | 7 2 5 2 | 3+4 2 |
| | 8 2 8 2 | 7 2 6 2 | 5+6 2 |
| | 9 2 8 2 | 6 2 5 2 | 3+4 2 |
| 139.1.8 | 247272 | 66 | Pd3Tl1 |
| | -GM -M | -P -X | -N |
| | 7 2 7 2 | 7 2 6 2 | 3+4 2 |
| | 9 2 9 2 | 6 2 5 2 | 5+6 2 |
| | 7 2 7 2 | 7 2 6 2 | 5+6 2 |
| | 7 2 9 2 | 6 2 6 2 | 3+4 2 |
| | 9 2 7 2 | 7 2 5 2 | 3+4 2 |
| | 7 2 6 2 | 6 2 6 2 | 5+6 2 |
| | 6 2 7 2 | 6 2 6 2 | 3+4 2 |
| | 6 2 6 2 | 7 2 6 2 | 5+6 2 |
| | 9 2 7 2 | 7 2 5 2 | 5+6 2 |
| | 8 2 6 2 | 6 2 6 2 | 5+6 2 |
| | 6 2 6 2 | 6 2 6 2 | 3+4 2 |
| | 7 2 8 2 | 7 2 5 2 | 3+4 2 |
| | 6 2 7 2 | 7 2 6 2 | 3+4 2 |
| | 8 2 9 2 | 6 2 6 2 | 5+6 2 |
| | 7 2 8 2 | 7 2 5 2 | 3+4 2 |
| | 6 2 8 2 | 7 2 6 2 | 5+6 2 |
| | 7 2 7 2 | 6 2 5 2 | 3+4 2 |
| | 9 2 6 2 | 6 2 5 2 | 5+6 2 |
| | 8 2 8 2 | 6 2 6 2 | 5+6 2 |
| | 6 2 6 2 | 7 2 6 2 | 3+4 2 |
| | 7 2 9 2 | 6 2 5 2 | 3+4 2 |
| | 8 2 6 2 | 7 2 6 2 | 5+6 2 |
| | 6 2 9 2 | 6 2 6 2 | 3+4 2 |
| | 9 2 8 2 | 6 2 6 2 | 5+6 2 |
| | 6 2 6 2 | 7 2 5 2 | 3+4 2 |
| | 6 2 8 2 | 7 2 6 2 | 5+6 2 |
| | 8 2 7 2 | 6 2 5 2 | 3+4 2 |
| | 7 2 7 2 | 7 2 5 2 | 5+6 2 |

Continued on next column

**TABLE LXIV– continued**

| TQC | ICSD | NoE | SF |
|---|---|---|---|
| | 7 2 9 2 | 6 2 6 2 | 3+4 2 |
| | 6 2 6 2 | 6 2 5 2 | 5+6 2 |
| | 6 2 7 2 | 7 2 6 2 | 3+4 2 |
| | 7 2 7 2 | 6 2 5 2 | 5+6 2 |
| | 9 2 6 2 | 7 2 6 2 | 5+6 2 |
| 139.1.9 | 61195 | 40 | Pd2Sb2Sr1 |
| | -GM -M | -P -X | -N |
| | 7 2 7 2 | 6 2 5 2 | 3+4 2 |
| | 9 2 9 2 | 7 2 6 2 | 5+6 2 |
| | 8 2 9 2 | 6 2 6 2 | 5+6 2 |
| | 9 2 8 2 | 7 2 6 2 | 5+6 2 |
| | 7 2 7 2 | 7 2 5 2 | 5+6 2 |
| | 9 2 9 2 | 6 2 6 2 | 3+4 2 |
| | 7 2 7 2 | 6 2 6 2 | 3+4 2 |
| | 9 2 6 2 | 7 2 5 2 | 5+6 2 |
| | 8 2 7 2 | 6 2 6 2 | 3+4 2 |
| | 6 2 8 2 | 7 2 5 2 | 5+6 2 |
| | 7 2 9 2 | 7 2 5 2 | 5+6 2 |
| | 9 2 9 2 | 6 2 5 2 | 3+4 2 |
| | 6 2 8 2 | 7 2 6 2 | 5+6 2 |
| | 8 2 6 2 | 6 2 5 2 | 3+4 2 |
| | 6 2 8 2 | 7 2 6 2 | 3+4 2 |
| | 7 2 7 2 | 6 2 5 2 | 5+6 2 |
| | 9 2 9 2 | 7 2 6 2 | 5+6 2 |
| | 7 2 8 2 | 7 2 6 2 | 3+4 2 |
| | 6 2 7 2 | 6 2 6 2 | 3+4 2 |
| | 6 2 6 2 | 6 2 5 2 | 5+6 2 |
| 139.1.10 | 105758 | 18 | Pd1Zr2 |
| | -GM -M | -P -X | -N |
| | 7 2 7 2 | 7 2 5 2 | 3+4 2 |
| | 6 2 6 2 | 6 2 5 2 | 3+4 2 |
| | 7 2 7 2 | 6 2 5 2 | 3+4 2 |
| | 7 2 9 2 | 7 2 5 2 | 3+4 2 |
| | 6 2 7 2 | 7 2 6 2 | 3+4 2 |
| | 6 2 6 2 | 7 2 5 2 | 5+6 2 |
| | 9 2 6 2 | 7 2 6 2 | 3+4 2 |
| | 8 2 6 2 | 6 2 5 2 | 5+6 2 |
| | 7 2 7 2 | 6 2 6 2 | 3+4 2 |
| 139.1.11 | 42966 | 36 | P2Ru2Sr1 |
| | -GM -M | -P -X | -N |
| | 7 2 7 2 | 6 2 5 2 | 3+4 2 |
| | 9 2 9 2 | 7 2 6 2 | 5+6 2 |
| | 9 2 9 2 | 6 2 6 2 | 5+6 2 |
| | 8 2 8 2 | 7 2 6 2 | 5+6 2 |
| | 7 2 9 2 | 7 2 5 2 | 5+6 2 |
| | 9 2 7 2 | 6 2 6 2 | 3+4 2 |
| | 7 2 7 2 | 7 2 5 2 | 3+4 2 |
| | 8 2 6 2 | 6 2 6 2 | 5+6 2 |
| | 9 2 7 2 | 7 2 6 2 | 3+4 2 |
| | 9 2 9 2 | 6 2 5 2 | 5+6 2 |
| | 6 2 8 2 | 7 2 5 2 | 5+6 2 |
| | 6 2 8 2 | 6 2 5 2 | 3+4 2 |
| | 7 2 9 2 | 7 2 6 2 | 3+4 2 |
| | 8 2 6 2 | 6 2 5 2 | 5+6 2 |
| | 7 2 8 2 | 7 2 6 2 | 3+4 2 |
| | 6 2 9 2 | 7 2 5 2 | 5+6 2 |
| | 7 2 7 2 | 6 2 6 2 | 5+6 2 |
| | 6 2 8 2 | 6 2 5 2 | 3+4 2 |
| 139.1.12 | 50186 | 38 | P2Rh2Sr1 |
| | -GM -M | -P -X | -N |
| | 7 2 7 2 | 6 2 5 2 | 3+4 2 |

Continued on next column

**TABLE LXIV– continued**

| TQC | ICSD | NoE | SF |
|---|---|---|---|
| | 9 2 9 2 | 7 2 6 2 | 5+6 2 |
| | 9 2 9 2 | 6 2 6 2 | 5+6 2 |
| | 8 2 8 2 | 7 2 6 2 | 5+6 2 |
| | 7 2 9 2 | 7 2 5 2 | 5+6 2 |
| | 9 2 7 2 | 6 2 6 2 | 3+4 2 |
| | 7 2 7 2 | 7 2 5 2 | 3+4 2 |
| | 8 2 6 2 | 6 2 6 2 | 5+6 2 |
| | 9 2 7 2 | 7 2 5 2 | 3+4 2 |
| | 9 2 9 2 | 6 2 6 2 | 5+6 2 |
| | 6 2 8 2 | 7 2 5 2 | 5+6 2 |
| | 7 2 9 2 | 6 2 5 2 | 3+4 2 |
| | 6 2 8 2 | 7 2 6 2 | 3+4 2 |
| | 8 2 6 2 | 6 2 5 2 | 5+6 2 |
| | 7 2 8 2 | 7 2 6 2 | 5+6 2 |
| | 6 2 7 2 | 7 2 5 2 | 3+4 2 |
| | 7 2 9 2 | 6 2 6 2 | 5+6 2 |
| | 6 2 8 2 | 7 2 5 2 | 3+4 2 |
| | 9 2 9 2 | 6 2 6 2 | 3+4 2 |
| 139.1.13 | 30849 | 44 | P2Pd2Yb1 |
| | -GM -M | -P -X | -N |
| | 7 2 7 2 | 6 2 5 2 | 3+4 2 |
| | 9 2 9 2 | 7 2 6 2 | 5+6 2 |
| | 8 2 8 2 | 6 2 6 2 | 5+6 2 |
| | 9 2 9 2 | 7 2 6 2 | 5+6 2 |
| | 7 2 9 2 | 6 2 5 2 | 3+4 2 |
| | 6 2 7 2 | 7 2 5 2 | 3+4 2 |
| | 9 2 8 2 | 6 2 6 2 | 5+6 2 |
| | 7 2 6 2 | 7 2 6 2 | 5+6 2 |
| | 8 2 7 2 | 7 2 6 2 | 3+4 2 |
| | 9 2 7 2 | 6 2 5 2 | 3+4 2 |
| | 8 2 9 2 | 7 2 6 2 | 5+6 2 |
| | 8 2 6 2 | 6 2 6 2 | 5+6 2 |
| | 9 2 9 2 | 6 2 6 2 | 3+4 2 |
| | 7 2 8 2 | 6 2 5 2 | 5+6 2 |
| | 8 2 8 2 | 7 2 6 2 | 5+6 2 |
| | 6 2 9 2 | 7 2 5 2 | 5+6 2 |
| | 6 2 8 2 | 6 2 6 2 | 5+6 2 |
| | 9 2 8 2 | 6 2 6 2 | 3+4 2 |
| | 8 2 8 2 | 7 2 6 2 | 5+6 2 |
| | 9 2 9 2 | 7 2 6 2 | 5+6 2 |
| | 8 2 9 2 | 6 2 6 2 | 5+6 2 |
| | 9 2 8 2 | 7 2 5 2 | 5+6 2 |
| 139.1.14 | 36373 | 40 | P2Pd2Sr1 |
| | -GM -M | -P -X | -N |
| | 7 2 7 2 | 6 2 5 2 | 3+4 2 |
| | 9 2 9 2 | 7 2 6 2 | 5+6 2 |
| | 9 2 9 2 | 6 2 6 2 | 5+6 2 |
| | 8 2 8 2 | 7 2 6 2 | 5+6 2 |
| | 7 2 7 2 | 7 2 5 2 | 5+6 2 |
| | 9 2 9 2 | 6 2 6 2 | 3+4 2 |
| | 7 2 7 2 | 7 2 6 2 | 3+4 2 |
| | 8 2 6 2 | 6 2 5 2 | 5+6 2 |
| | 9 2 7 2 | 7 2 5 2 | 3+4 2 |
| | 6 2 8 2 | 6 2 6 2 | 5+6 2 |
| | 7 2 9 2 | 7 2 5 2 | 5+6 2 |
| | 9 2 8 2 | 6 2 6 2 | 3+4 2 |
| | 6 2 9 2 | 7 2 5 2 | 5+6 2 |
| | 7 2 6 2 | 6 2 5 2 | 3+4 2 |
| | 8 2 8 2 | 7 2 6 2 | 3+4 2 |
| | 6 2 9 2 | 6 2 5 2 | 5+6 2 |
| | 6 2 8 2 | 7 2 6 2 | 5+6 2 |

Continued on next column

**TABLE LXIV– continued**

| TQC | ICSD | NoE | SF |
|---|---|---|---|
| | 7 2 7 2 | 7 2 6 2 | 3+4 2 |
| | 9 2 7 2 | 6 2 5 2 | 3+4 2 |
| | 8 2 6 2 | 6 2 6 2 | 5+6 2 |
| 139.1.15 | 71451 | 82 | O7Sr3V2 |
| | -GM -M | -P -X | -N |
| | 7 2 7 2 | 7 2 5 2 | 5+6 2 |
| | 7 2 9 2 | 6 2 5 2 | 3+4 2 |
| | 9 2 7 2 | 7 2 6 2 | 5+6 2 |
| | 9 2 9 2 | 6 2 6 2 | 3+4 2 |
| | 7 2 7 2 | 7 2 6 2 | 5+6 2 |
| | 6 2 6 2 | 7 2 5 2 | 3+4 2 |
| | 8 2 8 2 | 6 2 5 2 | 5+6 2 |
| | 7 2 7 2 | 6 2 5 2 | 3+4 2 |
| | 7 2 9 2 | 7 2 6 2 | 3+4 2 |
| | 9 2 7 2 | 6 2 5 2 | 5+6 2 |
| | 9 2 9 2 | 6 2 6 2 | 3+4 2 |
| | 7 2 9 2 | 7 2 5 2 | 5+6 2 |
| | 9 2 7 2 | 6 2 6 2 | 3+4 2 |
| | 6 2 8 2 | 7 2 6 2 | 3+4 2 |
| | 8 2 9 2 | 6 2 6 2 | 5+6 2 |
| | 9 2 8 2 | 6 2 5 2 | 3+4 2 |
| | 8 2 6 2 | 7 2 5 2 | 3+4 2 |
| | 7 2 7 2 | 7 2 6 2 | 5+6 2 |
| | 9 2 9 2 | 6 2 6 2 | 3+4 2 |
| | 9 2 9 2 | 7 2 6 2 | 5+6 2 |
| | 9 2 7 2 | 6 2 5 2 | 3+4 2 |
| | 7 2 9 2 | 7 2 5 2 | 5+6 2 |
| | 6 2 6 2 | 7 2 5 2 | 5+6 2 |
| | 8 2 8 2 | 6 2 6 2 | 3+4 2 |
| | 7 2 7 2 | 6 2 6 2 | 5+6 2 |
| | 9 2 9 2 | 7 2 6 2 | 5+6 2 |
| | 6 2 6 2 | 6 2 6 2 | 3+4 2 |
| | 9 2 9 2 | 7 2 6 2 | 5+6 2 |
| | 8 2 8 2 | 7 2 5 2 | 3+4 2 |
| | 7 2 8 2 | 6 2 5 2 | 3+4 2 |
| | 8 2 7 2 | 6 2 5 2 | 5+6 2 |
| | 7 2 9 2 | 7 2 6 2 | 5+6 2 |
| | 6 2 7 2 | 6 2 6 2 | 3+4 2 |
| | 9 2 8 2 | 7 2 5 2 | 5+6 2 |
| | 8 2 6 2 | 7 2 6 2 | 3+4 2 |
| | 7 2 9 2 | 6 2 5 2 | 3+4 2 |
| | 6 2 7 2 | 6 2 6 2 | 3+4 2 |
| | 9 2 6 2 | 7 2 5 2 | 5+6 2 |
| | 8 2 9 2 | 7 2 6 2 | 5+6 2 |
| | 9 2 8 2 | 6 2 6 2 | 5+6 2 |
| | 6 2 6 2 | 6 2 5 2 | 3+4 2 |
| 139.1.16 | 71320 | 82 | O7Sr3V2 |
| | -GM -M | -P -X | -N |
| | 7 2 9 2 | 6 2 5 2 | 3+4 2 |
| | 9 2 7 2 | 7 2 6 2 | 5+6 2 |
| | 7 2 7 2 | 7 2 5 2 | 5+6 2 |
| | 9 2 9 2 | 6 2 6 2 | 3+4 2 |
| | 7 2 7 2 | 7 2 6 2 | 5+6 2 |
| | 6 2 6 2 | 7 2 5 2 | 3+4 2 |
| | 8 2 8 2 | 6 2 5 2 | 5+6 2 |
| | 7 2 9 2 | 6 2 5 2 | 3+4 2 |
| | 9 2 7 2 | 7 2 6 2 | 5+6 2 |
| | 7 2 7 2 | 6 2 5 2 | 3+4 2 |
| | 7 2 9 2 | 6 2 6 2 | 3+4 2 |
| | 9 2 7 2 | 7 2 5 2 | 5+6 2 |
| | 9 2 9 2 | 6 2 6 2 | 3+4 2 |

Continued on next column

**TABLE LXIV– continued**

| TQC | ICSD | NoE | SF |
|---|---|---|---|
| | 6 2 8 2 | 7 2 6 2 | 3+4 2 |
| | 8 2 9 2 | 6 2 6 2 | 5+6 2 |
| | 7 2 6 2 | 6 2 5 2 | 3+4 2 |
| | 9 2 8 2 | 7 2 5 2 | 5+6 2 |
| | 8 2 7 2 | 6 2 6 2 | 3+4 2 |
| | 9 2 9 2 | 7 2 6 2 | 3+4 2 |
| | 9 2 9 2 | 7 2 6 2 | 5+6 2 |
| | 7 2 7 2 | 6 2 5 2 | 3+4 2 |
| | 9 2 6 2 | 7 2 5 2 | 5+6 2 |
| | 6 2 9 2 | 7 2 5 2 | 5+6 2 |
| | 8 2 8 2 | 6 2 6 2 | 3+4 2 |
| | 7 2 7 2 | 6 2 6 2 | 5+6 2 |
| | 9 2 9 2 | 7 2 6 2 | 5+6 2 |
| | 7 2 9 2 | 6 2 6 2 | 3+4 2 |
| | 9 2 6 2 | 7 2 6 2 | 5+6 2 |
| | 8 2 8 2 | 7 2 5 2 | 3+4 2 |
| | 6 2 7 2 | 6 2 5 2 | 3+4 2 |
| | 7 2 8 2 | 6 2 5 2 | 5+6 2 |
| | 6 2 7 2 | 7 2 6 2 | 5+6 2 |
| | 8 2 6 2 | 6 2 6 2 | 5+6 2 |
| | 9 2 9 2 | 7 2 5 2 | 3+4 2 |
| | 8 2 8 2 | 7 2 6 2 | 3+4 2 |
| | 7 2 9 2 | 6 2 5 2 | 3+4 2 |
| | 6 2 7 2 | 7 2 6 2 | 3+4 2 |
| | 9 2 6 2 | 6 2 5 2 | 5+6 2 |
| | 9 2 9 2 | 6 2 6 2 | 5+6 2 |
| | 8 2 8 2 | 7 2 6 2 | 5+6 2 |
| | 6 2 6 2 | 6 2 5 2 | 3+4 2 |
| 139.1.17 | 56577 | 88 | O7Ru2Sr3 |
| | -GM -M | -P -X | -N |
| | 7 2 7 2 | 7 2 5 2 | 5+6 2 |
| | 7 2 9 2 | 6 2 5 2 | 3+4 2 |
| | 9 2 7 2 | 7 2 6 2 | 5+6 2 |
| | 9 2 9 2 | 6 2 6 2 | 3+4 2 |
| | 7 2 7 2 | 7 2 6 2 | 5+6 2 |
| | 6 2 6 2 | 7 2 5 2 | 3+4 2 |
| | 8 2 8 2 | 6 2 5 2 | 5+6 2 |
| | 7 2 7 2 | 6 2 5 2 | 3+4 2 |
| | 7 2 9 2 | 7 2 6 2 | 3+4 2 |
| | 9 2 7 2 | 6 2 5 2 | 5+6 2 |
| | 9 2 9 2 | 6 2 6 2 | 3+4 2 |
| | 7 2 9 2 | 7 2 5 2 | 5+6 2 |
| | 9 2 7 2 | 6 2 6 2 | 3+4 2 |
| | 8 2 8 2 | 7 2 6 2 | 3+4 2 |
| | 9 2 9 2 | 6 2 6 2 | 3+4 2 |
| | 6 2 8 2 | 6 2 5 2 | 5+6 2 |
| | 8 2 6 2 | 7 2 6 2 | 3+4 2 |
| | 7 2 9 2 | 7 2 5 2 | 5+6 2 |
| | 9 2 7 2 | 6 2 6 2 | 3+4 2 |
| | 9 2 9 2 | 7 2 5 2 | 5+6 2 |
| | 7 2 7 2 | 6 2 6 2 | 3+4 2 |
| | 9 2 9 2 | 6 2 5 2 | 5+6 2 |
| | 6 2 6 2 | 7 2 6 2 | 5+6 2 |
| | 8 2 8 2 | 7 2 5 2 | 3+4 2 |
| | 7 2 7 2 | 6 2 6 2 | 5+6 2 |
| | 9 2 9 2 | 7 2 6 2 | 3+4 2 |
| | 9 2 9 2 | 6 2 6 2 | 5+6 2 |
| | 8 2 8 2 | 7 2 6 2 | 3+4 2 |
| | 7 2 6 2 | 6 2 5 2 | 5+6 2 |
| | 6 2 7 2 | 7 2 5 2 | 3+4 2 |
| | 7 2 7 2 | 6 2 5 2 | 5+6 2 |

Continued on next column

**TABLE LXIV– continued**

| TQC | ICSD | NoE | SF |
|---|---|---|---|
| | 8 2 8 2 | 7 2 6 2 | 5+6 2 |
| | 6 2 6 2 | 6 2 6 2 | 3+4 2 |
| | 9 2 9 2 | 7 2 5 2 | 3+4 2 |
| | 7 2 8 2 | 7 2 6 2 | 3+4 2 |
| | 8 2 7 2 | 6 2 5 2 | 5+6 2 |
| | 6 2 6 2 | 6 2 5 2 | 3+4 2 |
| | 6 2 9 2 | 7 2 6 2 | 5+6 2 |
| | 8 2 6 2 | 7 2 6 2 | 5+6 2 |
| | 9 2 8 2 | 6 2 6 2 | 5+6 2 |
| | 9 2 9 2 | 6 2 5 2 | 3+4 2 |
| | 8 2 8 2 | 7 2 5 2 | 5+6 2 |
| | 7 2 7 2 | 7 2 6 2 | 5+6 2 |
| | 6 2 6 2 | 6 2 6 2 | 3+4 2 |
| 139.1.18 | 79214 | 98 | O4Te3U4 |
| | -GM -M | -P -X | -N |
| | 7 2 7 2 | 7 2 5 2 | 5+6 2 |
| | 9 2 9 2 | 6 2 6 2 | 3+4 2 |
| | 7 2 7 2 | 6 2 5 2 | 3+4 2 |
| | 9 2 9 2 | 7 2 6 2 | 5+6 2 |
| | 7 2 7 2 | 7 2 5 2 | 5+6 2 |
| | 9 2 9 2 | 6 2 6 2 | 3+4 2 |
| | 7 2 7 2 | 6 2 5 2 | 3+4 2 |
| | 9 2 9 2 | 7 2 6 2 | 5+6 2 |
| | 7 2 7 2 | 7 2 6 2 | 5+6 2 |
| | 9 2 9 2 | 6 2 5 2 | 3+4 2 |
| | 6 2 6 2 | 7 2 5 2 | 3+4 2 |
| | 8 2 8 2 | 6 2 6 2 | 5+6 2 |
| | 8 2 8 2 | 7 2 5 2 | 5+6 2 |
| | 6 2 6 2 | 7 2 6 2 | 3+4 2 |
| | 7 2 7 2 | 6 2 5 2 | 5+6 2 |
| | 9 2 9 2 | 6 2 6 2 | 3+4 2 |
| | 8 2 8 2 | 7 2 5 2 | 5+6 2 |
| | 6 2 6 2 | 6 2 6 2 | 3+4 2 |
| | 7 2 7 2 | 7 2 6 2 | 3+4 2 |
| | 9 2 9 2 | 6 2 5 2 | 5+6 2 |
| | 7 2 9 2 | 6 2 5 2 | 3+4 2 |
| | 7 2 7 2 | 7 2 5 2 | 3+4 2 |
| | 9 2 7 2 | 6 2 6 2 | 5+6 2 |
| | 7 2 9 2 | 7 2 6 2 | 5+6 2 |
| | 9 2 7 2 | 6 2 5 2 | 3+4 2 |
| | 9 2 9 2 | 7 2 5 2 | 5+6 2 |
| | 7 2 7 2 | 6 2 6 2 | 3+4 2 |
| | 6 2 6 2 | 6 2 5 2 | 5+6 2 |
| | 8 2 8 2 | 7 2 6 2 | 3+4 2 |
| | 8 2 8 2 | 7 2 5 2 | 5+6 2 |
| | 6 2 6 2 | 6 2 6 2 | 5+6 2 |
| | 9 2 9 2 | 7 2 6 2 | 3+4 2 |
| | 7 2 8 2 | 6 2 5 2 | 3+4 2 |
| | 8 2 7 2 | 7 2 6 2 | 5+6 2 |
| | 6 2 6 2 | 6 2 5 2 | 3+4 2 |
| | 7 2 9 2 | 7 2 5 2 | 5+6 2 |
| | 9 2 9 2 | 6 2 6 2 | 3+4 2 |
| | 7 2 9 2 | 6 2 6 2 | 5+6 2 |
| | 9 2 8 2 | 7 2 6 2 | 5+6 2 |
| | 6 2 9 2 | 7 2 6 2 | 3+4 2 |
| | 8 2 7 2 | 6 2 5 2 | 5+6 2 |
| | 9 2 8 2 | 7 2 6 2 | 5+6 2 |
| | 9 2 7 2 | 6 2 5 2 | 3+4 2 |
| | 8 2 6 2 | 7 2 6 2 | 5+6 2 |
| | 6 2 6 2 | 6 2 5 2 | 3+4 2 |
| | 8 2 6 2 | 7 2 6 2 | 3+4 2 |

Continued on next column

**TABLE LXIV– continued**

| TQC | ICSD | NoE | SF |
|---|---|---|---|
| | 6 2 8 2 | 7 2 5 2 | 5+6 2 |
| | 7 2 7 2 | 6 2 6 2 | 3+4 2 |
| | 8 2 8 2 | 7 2 5 2 | 5+6 2 |
| 139.1.19 | 68211 | 40 | Ni2Si2Th1 |
| | -GM -M | -P -X | -N |
| | 7 2 7 2 | 6 2 5 2 | 3+4 2 |
| | 9 2 9 2 | 7 2 6 2 | 5+6 2 |
| | 9 2 9 2 | 6 2 6 2 | 5+6 2 |
| | 8 2 8 2 | 7 2 6 2 | 5+6 2 |
| | 7 2 7 2 | 7 2 5 2 | 5+6 2 |
| | 9 2 9 2 | 6 2 6 2 | 3+4 2 |
| | 7 2 7 2 | 7 2 5 2 | 3+4 2 |
| | 9 2 6 2 | 6 2 6 2 | 5+6 2 |
| | 8 2 7 2 | 6 2 5 2 | 3+4 2 |
| | 6 2 6 2 | 7 2 5 2 | 5+6 2 |
| | 7 2 8 2 | 7 2 6 2 | 3+4 2 |
| | 6 2 9 2 | 6 2 5 2 | 5+6 2 |
| | 8 2 9 2 | 6 2 5 2 | 3+4 2 |
| | 9 2 8 2 | 7 2 6 2 | 3+4 2 |
| | 7 2 8 2 | 7 2 6 2 | 5+6 2 |
| | 6 2 9 2 | 6 2 5 2 | 5+6 2 |
| | 6 2 7 2 | 7 2 5 2 | 5+6 2 |
| | 7 2 8 2 | 7 2 6 2 | 3+4 2 |
| | 6 2 6 2 | 6 2 5 2 | 3+4 2 |
| | 9 2 7 2 | 7 2 6 2 | 5+6 2 |
| 139.1.20 | 41771 | 70 | Ni2Sc1Si3 |
| | -GM -M | -P -X | -N |
| | 7 2 7 2 | 6 2 6 2 | 5+6 2 |
| | 9 2 9 2 | 7 2 6 2 | 3+4 2 |
| | 7 2 9 2 | 7 2 5 2 | 3+4 2 |
| | 9 2 7 2 | 6 2 6 2 | 5+6 2 |
| | 7 2 7 2 | 7 2 5 2 | 3+4 2 |
| | 6 2 6 2 | 7 2 5 2 | 5+6 2 |
| | 8 2 9 2 | 6 2 5 2 | 3+4 2 |
| | 7 2 8 2 | 6 2 5 2 | 5+6 2 |
| | 9 2 9 2 | 6 2 6 2 | 5+6 2 |
| | 8 2 8 2 | 6 2 5 2 | 3+4 2 |
| | 9 2 7 2 | 7 2 6 2 | 3+4 2 |
| | 7 2 7 2 | 7 2 6 2 | 5+6 2 |
| | 6 2 6 2 | 6 2 6 2 | 5+6 2 |
| | 8 2 6 2 | 7 2 5 2 | 3+4 2 |
| | 9 2 7 2 | 7 2 5 2 | 5+6 2 |
| | 9 2 9 2 | 6 2 6 2 | 3+4 2 |
| | 6 2 6 2 | 7 2 6 2 | 3+4 2 |
| | 6 2 8 2 | 6 2 6 2 | 3+4 2 |
| | 8 2 6 2 | 7 2 5 2 | 5+6 2 |
| | 6 2 8 2 | 6 2 6 2 | 5+6 2 |
| | 7 2 7 2 | 6 2 5 2 | 3+4 2 |
| | 7 2 9 2 | 6 2 5 2 | 5+6 2 |
| | 6 2 8 2 | 7 2 6 2 | 3+4 2 |
| | 8 2 6 2 | 7 2 6 2 | 5+6 2 |
| | 9 2 9 2 | 7 2 5 2 | 3+4 2 |
| | 8 2 9 2 | 6 2 5 2 | 5+6 2 |
| | 6 2 8 2 | 6 2 6 2 | 3+4 2 |
| | 9 2 8 2 | 7 2 5 2 | 5+6 2 |
| | 7 2 7 2 | 6 2 5 2 | 3+4 2 |
| | 7 2 6 2 | 7 2 6 2 | 5+6 2 |
| | 6 2 7 2 | 6 2 6 2 | 3+4 2 |
| | 9 2 9 2 | 6 2 5 2 | 3+4 2 |
| | 7 2 8 2 | 7 2 6 2 | 5+6 2 |
| | 6 2 7 2 | 7 2 6 2 | 5+6 2 |

Continued on next column

**TABLE LXIV– continued**

| TQC | ICSD | NoE | SF |
|---|---|---|---|
| | 6 2 8 2 | 6 2 5 2 | 5+6 2 |
| 139.1.21 | 89560 | 46 | Nd2O2Te1 |
| | -GM -M | -P -X | -N |
| | 7 2 9 2 | 7 2 5 2 | 5+6 2 |
| | 9 2 7 2 | 6 2 6 2 | 3+4 2 |
| | 7 2 9 2 | 7 2 5 2 | 5+6 2 |
| | 9 2 7 2 | 6 2 6 2 | 3+4 2 |
| | 7 2 9 2 | 7 2 5 2 | 5+6 2 |
| | 8 2 6 2 | 6 2 6 2 | 3+4 2 |
| | 6 2 8 2 | 7 2 5 2 | 5+6 2 |
| | 8 2 7 2 | 6 2 6 2 | 3+4 2 |
| | 9 2 6 2 | 7 2 6 2 | 3+4 2 |
| | 7 2 9 2 | 6 2 5 2 | 5+6 2 |
| | 7 2 7 2 | 6 2 5 2 | 3+4 2 |
| | 9 2 7 2 | 7 2 6 2 | 5+6 2 |
| | 7 2 9 2 | 6 2 5 2 | 3+4 2 |
| | 6 2 8 2 | 6 2 5 2 | 5+6 2 |
| | 6 2 8 2 | 7 2 6 2 | 5+6 2 |
| | 9 2 7 2 | 7 2 6 2 | 3+4 2 |
| | 8 2 6 2 | 6 2 6 2 | 3+4 2 |
| | 9 2 9 2 | 7 2 5 2 | 5+6 2 |
| | 9 2 9 2 | 6 2 6 2 | 5+6 2 |
| | 8 2 8 2 | 7 2 6 2 | 5+6 2 |
| | 6 2 6 2 | 6 2 5 2 | 3+4 2 |
| | 6 2 8 2 | 7 2 6 2 | 3+4 2 |
| | 7 2 7 2 | 6 2 5 2 | 5+6 2 |
| 139.1.22 | 93412 | 42 | Nd1Ni2Si2 |
| | -GM -M | -P -X | -N |
| | 7 2 7 2 | 6 2 5 2 | 3+4 2 |
| | 9 2 9 2 | 7 2 6 2 | 5+6 2 |
| | 9 2 9 2 | 6 2 6 2 | 5+6 2 |
| | 8 2 8 2 | 7 2 6 2 | 5+6 2 |
| | 7 2 7 2 | 7 2 5 2 | 5+6 2 |
| | 9 2 9 2 | 6 2 6 2 | 3+4 2 |
| | 7 2 7 2 | 7 2 5 2 | 3+4 2 |
| | 8 2 6 2 | 6 2 6 2 | 5+6 2 |
| | 9 2 7 2 | 7 2 5 2 | 3+4 2 |
| | 6 2 8 2 | 6 2 6 2 | 5+6 2 |
| | 7 2 9 2 | 7 2 5 2 | 5+6 2 |
| | 6 2 6 2 | 6 2 5 2 | 3+4 2 |
| | 9 2 9 2 | 7 2 6 2 | 3+4 2 |
| | 8 2 8 2 | 6 2 5 2 | 5+6 2 |
| | 7 2 8 2 | 7 2 6 2 | 3+4 2 |
| | 6 2 9 2 | 6 2 5 2 | 5+6 2 |
| | 7 2 7 2 | 7 2 6 2 | 5+6 2 |
| | 6 2 8 2 | 7 2 5 2 | 3+4 2 |
| | 9 2 8 2 | 6 2 6 2 | 3+4 2 |
| | 6 2 9 2 | 7 2 5 2 | 5+6 2 |
| | 8 2 8 2 | 6 2 6 2 | 5+6 2 |
| 139.1.23 | 91220 | 26 | Na2O1Sb2Ti2 |
| | -GM -M | -P -X | -N |
| | 7 2 7 2 | 7 2 5 2 | 5+6 2 |
| | 7 2 7 2 | 6 2 5 2 | 3+4 2 |
| | 9 2 9 2 | 7 2 6 2 | 5+6 2 |
| | 9 2 9 2 | 6 2 6 2 | 3+4 2 |
| | 8 2 8 2 | 7 2 6 2 | 3+4 2 |
| | 7 2 9 2 | 6 2 6 2 | 3+4 2 |
| | 9 2 7 2 | 7 2 6 2 | 5+6 2 |
| | 7 2 9 2 | 6 2 6 2 | 3+4 2 |
| | 6 2 7 2 | 6 2 5 2 | 3+4 2 |
| | 6 2 6 2 | 7 2 5 2 | 5+6 2 |

Continued on next column

**TABLE LXIV– continued**

| TQC | ICSD | NoE | SF |
|---|---|---|---|
| | 7 2 6 2 | 6 2 5 2 | 3+4 2 |
| | 9 2 9 2 | 7 2 6 2 | 5+6 2 |
| | 9 2 7 2 | 6 2 6 2 | 3+4 2 |
| 139.1.24 | 91219 | 26 | Na2O1Sb2Ti2 |
| | -GM -M | -P -X | -N |
| | 7 2 7 2 | 7 2 5 2 | 5+6 2 |
| | 7 2 7 2 | 6 2 5 2 | 3+4 2 |
| | 9 2 9 2 | 7 2 6 2 | 5+6 2 |
| | 9 2 9 2 | 6 2 6 2 | 3+4 2 |
| | 8 2 8 2 | 7 2 6 2 | 3+4 2 |
| | 7 2 9 2 | 6 2 6 2 | 3+4 2 |
| | 9 2 7 2 | 7 2 6 2 | 5+6 2 |
| | 7 2 9 2 | 6 2 6 2 | 3+4 2 |
| | 6 2 7 2 | 6 2 5 2 | 3+4 2 |
| | 6 2 6 2 | 7 2 5 2 | 5+6 2 |
| | 7 2 6 2 | 6 2 5 2 | 3+4 2 |
| | 9 2 9 2 | 7 2 6 2 | 5+6 2 |
| | 9 2 7 2 | 6 2 6 2 | 3+4 2 |
| 139.1.25 | 91218 | 26 | Na2O1Sb2Ti2 |
| | -GM -M | -P -X | -N |
| | 7 2 7 2 | 7 2 5 2 | 5+6 2 |
| | 7 2 7 2 | 6 2 5 2 | 3+4 2 |
| | 9 2 9 2 | 7 2 6 2 | 5+6 2 |
| | 9 2 9 2 | 6 2 6 2 | 3+4 2 |
| | 8 2 8 2 | 7 2 6 2 | 3+4 2 |
| | 7 2 9 2 | 6 2 6 2 | 3+4 2 |
| | 9 2 7 2 | 7 2 6 2 | 5+6 2 |
| | 7 2 9 2 | 6 2 6 2 | 3+4 2 |
| | 6 2 7 2 | 6 2 5 2 | 3+4 2 |
| | 6 2 6 2 | 7 2 5 2 | 5+6 2 |
| | 7 2 6 2 | 6 2 5 2 | 3+4 2 |
| | 9 2 9 2 | 7 2 6 2 | 5+6 2 |
| | 9 2 7 2 | 6 2 6 2 | 3+4 2 |
| 139.1.26 | 91217 | 26 | Na2O1Sb2Ti2 |
| | -GM -M | -P -X | -N |
| | 7 2 7 2 | 7 2 5 2 | 5+6 2 |
| | 7 2 7 2 | 6 2 5 2 | 3+4 2 |
| | 9 2 9 2 | 7 2 6 2 | 5+6 2 |
| | 9 2 9 2 | 6 2 6 2 | 3+4 2 |
| | 8 2 8 2 | 7 2 6 2 | 3+4 2 |
| | 7 2 9 2 | 6 2 6 2 | 3+4 2 |
| | 9 2 7 2 | 7 2 6 2 | 5+6 2 |
| | 7 2 9 2 | 6 2 6 2 | 3+4 2 |
| | 6 2 7 2 | 6 2 5 2 | 3+4 2 |
| | 6 2 6 2 | 7 2 5 2 | 5+6 2 |
| | 7 2 6 2 | 6 2 5 2 | 3+4 2 |
| | 9 2 9 2 | 7 2 6 2 | 5+6 2 |
| | 9 2 7 2 | 6 2 6 2 | 3+4 2 |
| 139.1.27 | 91216 | 26 | Na2O1Sb2Ti2 |
| | -GM -M | -P -X | -N |
| | 7 2 7 2 | 7 2 5 2 | 5+6 2 |
| | 7 2 7 2 | 6 2 5 2 | 3+4 2 |
| | 9 2 9 2 | 7 2 6 2 | 5+6 2 |
| | 9 2 9 2 | 6 2 6 2 | 3+4 2 |
| | 8 2 8 2 | 7 2 6 2 | 3+4 2 |
| | 7 2 9 2 | 6 2 6 2 | 3+4 2 |
| | 9 2 7 2 | 7 2 6 2 | 5+6 2 |
| | 7 2 9 2 | 6 2 6 2 | 3+4 2 |
| | 6 2 7 2 | 6 2 5 2 | 3+4 2 |
| | 6 2 6 2 | 7 2 5 2 | 5+6 2 |
| | 7 2 6 2 | 6 2 5 2 | 3+4 2 |

Continued on next column

**TABLE LXIV– continued**

| TQC | ICSD | NoE | SF |
|---|---|---|---|
| | 9 2 9 2 | 7 2 6 2 | 5+6 2 |
| | 9 2 7 2 | 6 2 6 2 | 3+4 2 |
| 139.1.28 | 91215 | 26 | Na2O1Sb2Ti2 |
| | -GM -M | -P -X | -N |
| | 7 2 7 2 | 7 2 5 2 | 5+6 2 |
| | 7 2 7 2 | 6 2 5 2 | 3+4 2 |
| | 9 2 9 2 | 7 2 6 2 | 5+6 2 |
| | 9 2 9 2 | 6 2 6 2 | 3+4 2 |
| | 8 2 8 2 | 7 2 6 2 | 3+4 2 |
| | 7 2 9 2 | 6 2 6 2 | 3+4 2 |
| | 9 2 7 2 | 7 2 6 2 | 5+6 2 |
| | 7 2 9 2 | 6 2 6 2 | 3+4 2 |
| | 6 2 7 2 | 6 2 5 2 | 3+4 2 |
| | 6 2 6 2 | 7 2 5 2 | 5+6 2 |
| | 7 2 6 2 | 6 2 5 2 | 3+4 2 |
| | 9 2 9 2 | 7 2 6 2 | 5+6 2 |
| | 9 2 7 2 | 6 2 6 2 | 3+4 2 |
| 139.1.29 | 91214 | 26 | Na2O1Sb2Ti2 |
| | -GM -M | -P -X | -N |
| | 7 2 7 2 | 7 2 5 2 | 5+6 2 |
| | 7 2 7 2 | 6 2 5 2 | 3+4 2 |
| | 9 2 9 2 | 7 2 6 2 | 5+6 2 |
| | 9 2 9 2 | 6 2 6 2 | 3+4 2 |
| | 8 2 8 2 | 7 2 6 2 | 3+4 2 |
| | 7 2 9 2 | 6 2 6 2 | 3+4 2 |
| | 9 2 7 2 | 7 2 6 2 | 5+6 2 |
| | 7 2 9 2 | 6 2 6 2 | 3+4 2 |
| | 6 2 7 2 | 6 2 5 2 | 3+4 2 |
| | 6 2 6 2 | 7 2 5 2 | 5+6 2 |
| | 7 2 6 2 | 6 2 5 2 | 3+4 2 |
| | 9 2 9 2 | 7 2 6 2 | 5+6 2 |
| | 9 2 7 2 | 6 2 6 2 | 3+4 2 |
| 139.1.30 | 91213 | 26 | Na2O1Sb2Ti2 |
| | -GM -M | -P -X | -N |
| | 7 2 7 2 | 7 2 5 2 | 5+6 2 |
| | 7 2 7 2 | 6 2 5 2 | 3+4 2 |
| | 9 2 9 2 | 7 2 6 2 | 5+6 2 |
| | 9 2 9 2 | 6 2 6 2 | 3+4 2 |
| | 8 2 8 2 | 7 2 6 2 | 3+4 2 |
| | 7 2 9 2 | 6 2 6 2 | 3+4 2 |
| | 9 2 7 2 | 7 2 6 2 | 5+6 2 |
| | 7 2 9 2 | 6 2 6 2 | 3+4 2 |
| | 6 2 7 2 | 6 2 5 2 | 3+4 2 |
| | 6 2 6 2 | 7 2 5 2 | 5+6 2 |
| | 7 2 6 2 | 6 2 5 2 | 3+4 2 |
| | 9 2 9 2 | 7 2 6 2 | 5+6 2 |
| | 9 2 7 2 | 6 2 6 2 | 3+4 2 |
| 139.1.31 | 91212 | 26 | Na2O1Sb2Ti2 |
| | -GM -M | -P -X | -N |
| | 7 2 7 2 | 7 2 5 2 | 5+6 2 |
| | 7 2 7 2 | 6 2 5 2 | 3+4 2 |
| | 9 2 9 2 | 7 2 6 2 | 5+6 2 |
| | 9 2 9 2 | 6 2 6 2 | 3+4 2 |
| | 8 2 8 2 | 7 2 6 2 | 3+4 2 |
| | 7 2 9 2 | 6 2 6 2 | 3+4 2 |
| | 9 2 7 2 | 7 2 6 2 | 5+6 2 |
| | 7 2 9 2 | 6 2 6 2 | 3+4 2 |
| | 6 2 7 2 | 6 2 5 2 | 3+4 2 |
| | 6 2 6 2 | 7 2 5 2 | 5+6 2 |
| | 7 2 6 2 | 6 2 5 2 | 3+4 2 |
| | 9 2 9 2 | 7 2 6 2 | 5+6 2 |

Continued on next column

**TABLE LXIV– continued**

| TQC | ICSD | NoE | SF |
|---|---|---|---|
| | 9 2 7 2 | 6 2 6 2 | 3+4 2 |
| 139.1.32 | 91211 | 26 | Na2O1Sb2Ti2 |
| | -GM -M | -P -X | -N |
| | 7 2 7 2 | 7 2 5 2 | 5+6 2 |
| | 7 2 7 2 | 6 2 5 2 | 3+4 2 |
| | 9 2 9 2 | 7 2 6 2 | 5+6 2 |
| | 9 2 9 2 | 6 2 6 2 | 3+4 2 |
| | 8 2 8 2 | 7 2 6 2 | 3+4 2 |
| | 7 2 9 2 | 6 2 6 2 | 3+4 2 |
| | 9 2 7 2 | 7 2 6 2 | 5+6 2 |
| | 7 2 9 2 | 6 2 6 2 | 3+4 2 |
| | 6 2 7 2 | 6 2 5 2 | 3+4 2 |
| | 6 2 6 2 | 7 2 5 2 | 5+6 2 |
| | 7 2 6 2 | 6 2 5 2 | 3+4 2 |
| | 9 2 9 2 | 7 2 6 2 | 5+6 2 |
| | 9 2 7 2 | 6 2 6 2 | 3+4 2 |
| 139.1.33 | 91210 | 26 | Na2O1Sb2Ti2 |
| | -GM -M | -P -X | -N |
| | 7 2 7 2 | 7 2 5 2 | 5+6 2 |
| | 7 2 7 2 | 6 2 5 2 | 3+4 2 |
| | 9 2 9 2 | 7 2 6 2 | 5+6 2 |
| | 9 2 9 2 | 6 2 6 2 | 3+4 2 |
| | 8 2 8 2 | 7 2 6 2 | 3+4 2 |
| | 7 2 9 2 | 6 2 6 2 | 3+4 2 |
| | 9 2 7 2 | 7 2 6 2 | 5+6 2 |
| | 7 2 9 2 | 6 2 6 2 | 3+4 2 |
| | 6 2 7 2 | 6 2 5 2 | 3+4 2 |
| | 6 2 6 2 | 7 2 5 2 | 5+6 2 |
| | 7 2 6 2 | 6 2 5 2 | 3+4 2 |
| | 9 2 9 2 | 7 2 6 2 | 5+6 2 |
| | 9 2 7 2 | 6 2 6 2 | 3+4 2 |
| 139.1.34 | 91209 | 26 | Na2O1Sb2Ti2 |
| | -GM -M | -P -X | -N |
| | 7 2 7 2 | 7 2 5 2 | 5+6 2 |
| | 7 2 7 2 | 6 2 5 2 | 3+4 2 |
| | 9 2 9 2 | 7 2 6 2 | 5+6 2 |
| | 9 2 9 2 | 6 2 6 2 | 3+4 2 |
| | 8 2 8 2 | 7 2 6 2 | 3+4 2 |
| | 7 2 9 2 | 6 2 6 2 | 3+4 2 |
| | 9 2 7 2 | 7 2 6 2 | 5+6 2 |
| | 7 2 9 2 | 6 2 6 2 | 3+4 2 |
| | 6 2 7 2 | 6 2 5 2 | 3+4 2 |
| | 6 2 6 2 | 7 2 5 2 | 5+6 2 |
| | 7 2 6 2 | 6 2 5 2 | 3+4 2 |
| | 9 2 9 2 | 7 2 6 2 | 5+6 2 |
| | 9 2 7 2 | 6 2 6 2 | 3+4 2 |
| 139.1.35 | 91208 | 26 | Na2O1Sb2Ti2 |
| | -GM -M | -P -X | -N |
| | 7 2 7 2 | 7 2 5 2 | 5+6 2 |
| | 7 2 7 2 | 6 2 5 2 | 3+4 2 |
| | 9 2 9 2 | 7 2 6 2 | 5+6 2 |
| | 9 2 9 2 | 6 2 6 2 | 3+4 2 |
| | 8 2 8 2 | 7 2 6 2 | 3+4 2 |
| | 7 2 9 2 | 6 2 6 2 | 3+4 2 |
| | 9 2 7 2 | 7 2 6 2 | 5+6 2 |
| | 7 2 9 2 | 6 2 6 2 | 3+4 2 |
| | 6 2 7 2 | 6 2 5 2 | 3+4 2 |
| | 6 2 6 2 | 7 2 5 2 | 5+6 2 |
| | 7 2 6 2 | 6 2 5 2 | 3+4 2 |
| | 9 2 9 2 | 7 2 6 2 | 5+6 2 |

Continued on next column

**TABLE LXIV– continued**

| TQC | ICSD NoE SF |
|---|---|
| | 9 2 7 2 6 2 6 2 3+4 2 |
| 139.1.36 | 91207 26 Na2O1Sb2Ti2 |
| | -GM -M -P -X -N |
| | 7 2 7 2 7 2 5 2 5+6 2 |
| | 7 2 7 2 6 2 5 2 3+4 2 |
| | 9 2 9 2 7 2 6 2 5+6 2 |
| | 9 2 9 2 6 2 6 2 3+4 2 |
| | 8 2 8 2 7 2 6 2 3+4 2 |
| | 7 2 9 2 6 2 6 2 3+4 2 |
| | 9 2 7 2 7 2 6 2 5+6 2 |
| | 7 2 9 2 6 2 6 2 3+4 2 |
| | 6 2 7 2 6 2 5 2 3+4 2 |
| | 6 2 6 2 7 2 5 2 5+6 2 |
| | 7 2 6 2 6 2 5 2 3+4 2 |
| | 9 2 9 2 7 2 6 2 5+6 2 |
| | 9 2 7 2 6 2 6 2 3+4 2 |
| 139.1.37 | 91206 26 Na2O1Sb2Ti2 |
| | -GM -M -P -X -N |
| | 7 2 7 2 7 2 5 2 5+6 2 |
| | 7 2 7 2 6 2 5 2 3+4 2 |
| | 9 2 9 2 7 2 6 2 5+6 2 |
| | 9 2 9 2 6 2 6 2 3+4 2 |
| | 8 2 8 2 7 2 6 2 3+4 2 |
| | 7 2 9 2 6 2 6 2 3+4 2 |
| | 9 2 7 2 7 2 6 2 5+6 2 |
| | 7 2 9 2 6 2 6 2 3+4 2 |
| | 6 2 7 2 6 2 5 2 3+4 2 |
| | 6 2 6 2 7 2 5 2 5+6 2 |
| | 7 2 6 2 6 2 5 2 3+4 2 |
| | 9 2 9 2 7 2 6 2 5+6 2 |
| | 9 2 7 2 6 2 6 2 3+4 2 |
| 139.1.38 | 91205 26 Na2O1Sb2Ti2 |
| | -GM -M -P -X -N |
| | 7 2 7 2 7 2 5 2 5+6 2 |
| | 7 2 7 2 6 2 5 2 3+4 2 |
| | 9 2 9 2 7 2 6 2 5+6 2 |
| | 9 2 9 2 6 2 6 2 3+4 2 |
| | 8 2 8 2 7 2 6 2 3+4 2 |
| | 7 2 9 2 6 2 6 2 3+4 2 |
| | 9 2 7 2 7 2 6 2 5+6 2 |
| | 7 2 9 2 6 2 6 2 3+4 2 |
| | 6 2 7 2 6 2 5 2 3+4 2 |
| | 6 2 6 2 7 2 5 2 5+6 2 |
| | 7 2 6 2 6 2 5 2 3+4 2 |
| | 9 2 9 2 7 2 6 2 5+6 2 |
| | 9 2 7 2 6 2 6 2 3+4 2 |
| 139.1.39 | 91204 26 Na2O1Sb2Ti2 |
| | -GM -M -P -X -N |
| | 7 2 7 2 7 2 5 2 5+6 2 |
| | 7 2 7 2 6 2 5 2 3+4 2 |
| | 9 2 9 2 7 2 6 2 5+6 2 |
| | 9 2 9 2 6 2 6 2 3+4 2 |
| | 8 2 8 2 7 2 6 2 3+4 2 |
| | 7 2 9 2 6 2 6 2 3+4 2 |
| | 9 2 7 2 7 2 6 2 5+6 2 |
| | 7 2 9 2 6 2 6 2 3+4 2 |
| | 6 2 7 2 6 2 5 2 3+4 2 |
| | 6 2 6 2 7 2 5 2 5+6 2 |
| | 7 2 6 2 6 2 5 2 3+4 2 |
| | 9 2 9 2 7 2 6 2 5+6 2 |

Continued on next column

**TABLE LXIV– continued**

| TQC | ICSD NoE SF |
|---|---|
| | 9 2 7 2 6 2 6 2 3+4 2 |
| 139.1.40 | 91203 26 Na2O1Sb2Ti2 |
| | -GM -M -P -X -N |
| | 7 2 7 2 7 2 5 2 5+6 2 |
| | 7 2 7 2 6 2 5 2 3+4 2 |
| | 9 2 9 2 7 2 6 2 5+6 2 |
| | 9 2 9 2 6 2 6 2 3+4 2 |
| | 8 2 8 2 7 2 6 2 3+4 2 |
| | 7 2 9 2 6 2 6 2 3+4 2 |
| | 9 2 7 2 7 2 6 2 5+6 2 |
| | 7 2 9 2 6 2 6 2 3+4 2 |
| | 6 2 7 2 6 2 5 2 3+4 2 |
| | 6 2 6 2 7 2 5 2 5+6 2 |
| | 7 2 6 2 6 2 5 2 3+4 2 |
| | 9 2 9 2 7 2 6 2 5+6 2 |
| | 9 2 7 2 6 2 6 2 3+4 2 |
| 139.1.41 | 91202 26 Na2O1Sb2Ti2 |
| | -GM -M -P -X -N |
| | 7 2 7 2 7 2 5 2 5+6 2 |
| | 7 2 7 2 6 2 5 2 3+4 2 |
| | 9 2 9 2 7 2 6 2 5+6 2 |
| | 9 2 9 2 6 2 6 2 3+4 2 |
| | 8 2 8 2 7 2 6 2 3+4 2 |
| | 7 2 9 2 6 2 6 2 3+4 2 |
| | 9 2 7 2 7 2 6 2 5+6 2 |
| | 7 2 9 2 6 2 6 2 3+4 2 |
| | 6 2 7 2 6 2 5 2 3+4 2 |
| | 6 2 6 2 7 2 5 2 5+6 2 |
| | 7 2 6 2 6 2 5 2 3+4 2 |
| | 9 2 9 2 7 2 6 2 5+6 2 |
| | 9 2 7 2 6 2 6 2 3+4 2 |
| 139.1.42 | 69648 26 Na2O1Sb2Ti2 |
| | -GM -M -P -X -N |
| | 7 2 7 2 7 2 5 2 5+6 2 |
| | 7 2 7 2 6 2 5 2 3+4 2 |
| | 9 2 9 2 7 2 6 2 5+6 2 |
| | 9 2 9 2 6 2 6 2 3+4 2 |
| | 8 2 8 2 7 2 6 2 3+4 2 |
| | 7 2 9 2 6 2 6 2 3+4 2 |
| | 9 2 7 2 7 2 6 2 5+6 2 |
| | 7 2 9 2 6 2 6 2 3+4 2 |
| | 6 2 7 2 6 2 5 2 3+4 2 |
| | 6 2 6 2 7 2 5 2 5+6 2 |
| | 7 2 6 2 6 2 5 2 3+4 2 |
| | 9 2 9 2 7 2 6 2 5+6 2 |
| | 9 2 7 2 6 2 6 2 3+4 2 |
| 139.1.43 | 82464 102 Na2Nd2O10Ti3 |
| | -GM -M -P -X -N |
| | 7 2 7 2 7 2 5 2 5+6 2 |
| | 9 2 9 2 6 2 6 2 3+4 2 |
| | 7 2 7 2 6 2 6 2 3+4 2 |
| | 9 2 9 2 7 2 5 2 5+6 2 |
| | 7 2 7 2 7 2 6 2 5+6 2 |
| | 9 2 9 2 6 2 5 2 3+4 2 |
| | 7 2 7 2 7 2 5 2 3+4 2 |
| | 6 2 6 2 6 2 6 2 5+6 2 |
| | 8 2 8 2 6 2 5 2 3+4 2 |
| | 7 2 9 2 7 2 6 2 3+4 2 |
| | 6 2 6 2 7 2 6 2 5+6 2 |
| | 9 2 7 2 6 2 6 2 5+6 2 |
| | 7 2 7 2 6 2 5 2 3+4 2 |

Continued on next column

**TABLE LXIV– continued**

| TQC | ICSD | NoE | SF |
|---|---|---|---|
| | 9 2 9 2 | 7 2 6 2 | 5+6 2 |
| | 8 2 8 2 | 7 2 5 2 | 3+4 2 |
| | 6 2 6 2 | 6 2 6 2 | 5+6 2 |
| | 9 2 9 2 | 6 2 5 2 | 3+4 2 |
| | 7 2 7 2 | 7 2 6 2 | 5+6 2 |
| | 7 2 7 2 | 7 2 5 2 | 3+4 2 |
| | 9 2 9 2 | 6 2 5 2 | 3+4 2 |
| | 7 2 9 2 | 7 2 6 2 | 3+4 2 |
| | 8 2 7 2 | 6 2 5 2 | 5+6 2 |
| | 9 2 8 2 | 6 2 6 2 | 5+6 2 |
| | 6 2 6 2 | 7 2 5 2 | 3+4 2 |
| | 9 2 9 2 | 7 2 5 2 | 3+4 2 |
| | 8 2 8 2 | 7 2 5 2 | 5+6 2 |
| | 7 2 9 2 | 6 2 5 2 | 3+4 2 |
| | 8 2 8 2 | 7 2 6 2 | 5+6 2 |
| | 9 2 6 2 | 6 2 6 2 | 3+4 2 |
| | 6 2 9 2 | 7 2 5 2 | 5+6 2 |
| | 7 2 7 2 | 6 2 5 2 | 5+6 2 |
| | 6 2 9 2 | 6 2 6 2 | 3+4 2 |
| | 9 2 8 2 | 7 2 5 2 | 5+6 2 |
| | 9 2 7 2 | 6 2 5 2 | 3+4 2 |
| | 8 2 6 2 | 7 2 6 2 | 5+6 2 |
| | 7 2 7 2 | 6 2 5 2 | 5+6 2 |
| | 6 2 6 2 | 6 2 6 2 | 3+4 2 |
| | 7 2 7 2 | 7 2 5 2 | 3+4 2 |
| | 8 2 8 2 | 7 2 6 2 | 5+6 2 |
| | 9 2 9 2 | 6 2 5 2 | 5+6 2 |
| | 8 2 8 2 | 7 2 6 2 | 3+4 2 |
| | 7 2 7 2 | 6 2 6 2 | 5+6 2 |
| | 9 2 9 2 | 7 2 6 2 | 3+4 2 |
| | 8 2 8 2 | 6 2 5 2 | 3+4 2 |
| | 6 2 6 2 | 7 2 6 2 | 5+6 2 |
| | 9 2 9 2 | 6 2 5 2 | 5+6 2 |
| | 9 2 9 2 | 7 2 5 2 | 5+6 2 |
| | 8 2 8 2 | 6 2 5 2 | 3+4 2 |
| | 6 2 6 2 | 7 2 6 2 | 5+6 2 |
| | 8 2 8 2 | 6 2 6 2 | 3+4 2 |
| | 7 2 7 2 | 6 2 5 2 | 3+4 2 |
| 139.1.44 | 75607 | 102 | Na2Nd2O10Ti3 |
| | -GM -M | -P -X | -N |
| | 7 2 7 2 | 7 2 5 2 | 5+6 2 |
| | 9 2 9 2 | 6 2 6 2 | 3+4 2 |
| | 7 2 7 2 | 6 2 6 2 | 3+4 2 |
| | 9 2 9 2 | 7 2 5 2 | 5+6 2 |
| | 7 2 7 2 | 7 2 6 2 | 3+4 2 |
| | 7 2 7 2 | 6 2 5 2 | 5+6 2 |
| | 9 2 9 2 | 6 2 5 2 | 3+4 2 |
| | 6 2 6 2 | 7 2 5 2 | 5+6 2 |
| | 8 2 6 2 | 6 2 6 2 | 3+4 2 |
| | 6 2 8 2 | 7 2 6 2 | 5+6 2 |
| | 7 2 7 2 | 6 2 6 2 | 3+4 2 |
| | 9 2 9 2 | 7 2 6 2 | 5+6 2 |
| | 8 2 8 2 | 7 2 5 2 | 3+4 2 |
| | 6 2 6 2 | 6 2 6 2 | 5+6 2 |
| | 7 2 9 2 | 6 2 5 2 | 3+4 2 |
| | 9 2 9 2 | 6 2 5 2 | 5+6 2 |
| | 9 2 7 2 | 7 2 6 2 | 3+4 2 |
| | 7 2 7 2 | 7 2 6 2 | 5+6 2 |
| | 7 2 7 2 | 7 2 5 2 | 3+4 2 |
| | 9 2 9 2 | 6 2 5 2 | 3+4 2 |
| | 9 2 9 2 | 6 2 5 2 | 3+4 2 |

Continued on next column

**TABLE LXIV– continued**

| TQC | ICSD | NoE | SF |
|---|---|---|---|
| | 7 2 7 2 | 7 2 6 2 | 5+6 2 |
| | 8 2 8 2 | 7 2 5 2 | 5+6 2 |
| | 6 2 6 2 | 6 2 6 2 | 3+4 2 |
| | 9 2 9 2 | 7 2 5 2 | 3+4 2 |
| | 8 2 8 2 | 7 2 5 2 | 5+6 2 |
| | 9 2 9 2 | 6 2 5 2 | 3+4 2 |
| | 8 2 8 2 | 6 2 5 2 | 5+6 2 |
| | 7 2 7 2 | 7 2 6 2 | 3+4 2 |
| | 6 2 6 2 | 6 2 5 2 | 5+6 2 |
| | 6 2 6 2 | 6 2 6 2 | 3+4 2 |
| | 7 2 7 2 | 7 2 5 2 | 5+6 2 |
| | 8 2 9 2 | 6 2 6 2 | 5+6 2 |
| | 7 2 8 2 | 7 2 5 2 | 5+6 2 |
| | 9 2 9 2 | 7 2 6 2 | 3+4 2 |
| | 8 2 8 2 | 6 2 6 2 | 3+4 2 |
| | 9 2 7 2 | 7 2 6 2 | 5+6 2 |
| | 7 2 9 2 | 6 2 5 2 | 3+4 2 |
| | 6 2 8 2 | 7 2 5 2 | 5+6 2 |
| | 9 2 7 2 | 7 2 6 2 | 5+6 2 |
| | 8 2 6 2 | 6 2 6 2 | 3+4 2 |
| | 7 2 9 2 | 7 2 5 2 | 5+6 2 |
| | 6 2 8 2 | 6 2 5 2 | 3+4 2 |
| | 9 2 9 2 | 6 2 6 2 | 5+6 2 |
| | 8 2 7 2 | 7 2 5 2 | 3+4 2 |
| | 9 2 6 2 | 6 2 6 2 | 5+6 2 |
| | 8 2 9 2 | 6 2 5 2 | 5+6 2 |
| | 9 2 8 2 | 7 2 5 2 | 3+4 2 |
| | 6 2 6 2 | 7 2 6 2 | 5+6 2 |
| | 7 2 7 2 | 6 2 5 2 | 3+4 2 |
| | 8 2 8 2 | 6 2 6 2 | 3+4 2 |
| 139.1.45 | 75149 | 102 | Na2Nd2O10Ti3 |
| | -GM -M | -P -X | -N |
| | 7 2 7 2 | 7 2 5 2 | 5+6 2 |
| | 9 2 9 2 | 6 2 6 2 | 3+4 2 |
| | 7 2 7 2 | 6 2 6 2 | 3+4 2 |
| | 9 2 9 2 | 7 2 5 2 | 5+6 2 |
| | 7 2 7 2 | 7 2 6 2 | 3+4 2 |
| | 7 2 7 2 | 6 2 5 2 | 5+6 2 |
| | 9 2 9 2 | 6 2 5 2 | 3+4 2 |
| | 6 2 6 2 | 7 2 5 2 | 5+6 2 |
| | 8 2 6 2 | 6 2 6 2 | 3+4 2 |
| | 6 2 8 2 | 7 2 6 2 | 5+6 2 |
| | 7 2 7 2 | 6 2 6 2 | 3+4 2 |
| | 9 2 9 2 | 7 2 6 2 | 5+6 2 |
| | 8 2 8 2 | 7 2 5 2 | 3+4 2 |
| | 6 2 6 2 | 6 2 6 2 | 5+6 2 |
| | 7 2 9 2 | 6 2 5 2 | 3+4 2 |
| | 9 2 9 2 | 6 2 5 2 | 5+6 2 |
| | 9 2 7 2 | 7 2 6 2 | 3+4 2 |
| | 7 2 7 2 | 7 2 6 2 | 5+6 2 |
| | 7 2 7 2 | 7 2 5 2 | 3+4 2 |
| | 9 2 9 2 | 6 2 5 2 | 3+4 2 |
| | 9 2 9 2 | 6 2 5 2 | 3+4 2 |
| | 7 2 7 2 | 7 2 6 2 | 5+6 2 |
| | 8 2 8 2 | 7 2 5 2 | 5+6 2 |
| | 6 2 6 2 | 6 2 6 2 | 3+4 2 |
| | 9 2 9 2 | 7 2 5 2 | 3+4 2 |
| | 8 2 8 2 | 7 2 5 2 | 5+6 2 |
| | 9 2 9 2 | 6 2 5 2 | 3+4 2 |
| | 8 2 8 2 | 6 2 5 2 | 5+6 2 |
| | 7 2 7 2 | 7 2 6 2 | 3+4 2 |

Continued on next column

**TABLE LXIV– continued**

| TQC | ICSD | NoE | SF |
|---|---|---|---|
| | 6 2 6 2 | 6 2 5 2 | 5+6 2 |
| | 6 2 6 2 | 6 2 6 2 | 3+4 2 |
| | 7 2 7 2 | 7 2 5 2 | 5+6 2 |
| | 8 2 9 2 | 6 2 6 2 | 5+6 2 |
| | 7 2 8 2 | 7 2 5 2 | 5+6 2 |
| | 9 2 9 2 | 7 2 6 2 | 3+4 2 |
| | 8 2 8 2 | 6 2 6 2 | 3+4 2 |
| | 9 2 7 2 | 7 2 6 2 | 5+6 2 |
| | 7 2 9 2 | 6 2 5 2 | 3+4 2 |
| | 6 2 8 2 | 7 2 5 2 | 5+6 2 |
| | 9 2 7 2 | 7 2 6 2 | 5+6 2 |
| | 8 2 6 2 | 6 2 6 2 | 3+4 2 |
| | 7 2 9 2 | 7 2 5 2 | 5+6 2 |
| | 6 2 8 2 | 6 2 5 2 | 3+4 2 |
| | 9 2 9 2 | 6 2 6 2 | 5+6 2 |
| | 8 2 7 2 | 7 2 5 2 | 3+4 2 |
| | 9 2 6 2 | 6 2 6 2 | 5+6 2 |
| | 8 2 9 2 | 6 2 5 2 | 5+6 2 |
| | 9 2 8 2 | 7 2 5 2 | 3+4 2 |
| | 6 2 6 2 | 7 2 6 2 | 5+6 2 |
| | 7 2 7 2 | 6 2 5 2 | 3+4 2 |
| | 8 2 8 2 | 6 2 6 2 | 3+4 2 |
| 139.1.46 | 152123 | 50 | Mo1O4Sr2 |
| | -GM -M | -P -X | -N |
| | 7 2 9 2 | 7 2 5 2 | 5+6 2 |
| | 9 2 7 2 | 6 2 6 2 | 3+4 2 |
| | 7 2 7 2 | 6 2 6 2 | 3+4 2 |
| | 6 2 6 2 | 7 2 6 2 | 5+6 2 |
| | 7 2 9 2 | 7 2 5 2 | 3+4 2 |
| | 9 2 7 2 | 6 2 6 2 | 5+6 2 |
| | 7 2 9 2 | 7 2 5 2 | 3+4 2 |
| | 9 2 7 2 | 6 2 6 2 | 5+6 2 |
| | 6 2 8 2 | 6 2 5 2 | 5+6 2 |
| | 8 2 6 2 | 7 2 6 2 | 3+4 2 |
| | 9 2 9 2 | 7 2 5 2 | 3+4 2 |
| | 7 2 7 2 | 6 2 6 2 | 5+6 2 |
| | 9 2 9 2 | 7 2 5 2 | 3+4 2 |
| | 9 2 8 2 | 6 2 5 2 | 3+4 2 |
| | 8 2 9 2 | 7 2 5 2 | 3+4 2 |
| | 7 2 8 2 | 7 2 5 2 | 5+6 2 |
| | 8 2 7 2 | 6 2 5 2 | 5+6 2 |
| | 9 2 9 2 | 6 2 5 2 | 5+6 2 |
| | 8 2 8 2 | 6 2 6 2 | 3+4 2 |
| | 7 2 7 2 | 7 2 5 2 | 3+4 2 |
| | 6 2 6 2 | 6 2 5 2 | 5+6 2 |
| | 9 2 9 2 | 7 2 6 2 | 5+6 2 |
| | 9 2 9 2 | 6 2 6 2 | 3+4 2 |
| | 8 2 8 2 | 7 2 5 2 | 5+6 2 |
| | 6 2 6 2 | 6 2 5 2 | 3+4 2 |
| 139.1.47 | 290825 | 46 | Mn2Si2Yb1 |
| | -GM -M | -P -X | -N |
| | 7 2 7 2 | 6 2 5 2 | 3+4 2 |
| | 9 2 9 2 | 7 2 6 2 | 5+6 2 |
| | 8 2 8 2 | 6 2 6 2 | 5+6 2 |
| | 9 2 9 2 | 7 2 6 2 | 5+6 2 |
| | 7 2 7 2 | 7 2 5 2 | 5+6 2 |
| | 9 2 9 2 | 6 2 6 2 | 3+4 2 |
| | 7 2 7 2 | 7 2 6 2 | 3+4 2 |
| | 9 2 6 2 | 6 2 5 2 | 5+6 2 |
| | 8 2 7 2 | 7 2 5 2 | 3+4 2 |
| | 9 2 9 2 | 7 2 6 2 | 5+6 2 |

Continued on next column

**TABLE LXIV– continued**

| TQC | ICSD | NoE | SF |
|---|---|---|---|
| | 6 2 8 2 | 6 2 5 2 | 5+6 2 |
| | 8 2 8 2 | 6 2 5 2 | 3+4 2 |
| | 8 2 8 2 | 7 2 6 2 | 5+6 2 |
| | 9 2 9 2 | 6 2 6 2 | 5+6 2 |
| | 7 2 9 2 | 6 2 6 2 | 5+6 2 |
| | 6 2 6 2 | 7 2 6 2 | 3+4 2 |
| | 7 2 8 2 | 6 2 6 2 | 5+6 2 |
| | 8 2 7 2 | 7 2 5 2 | 5+6 2 |
| | 9 2 8 2 | 6 2 6 2 | 5+6 2 |
| | 8 2 8 2 | 7 2 6 2 | 3+4 2 |
| | 9 2 9 2 | 6 2 6 2 | 3+4 2 |
| | 8 2 9 2 | 7 2 6 2 | 5+6 2 |
| | 6 2 8 2 | 7 2 5 2 | 5+6 2 |
| 139.1.48 | 290824 | 46 | Mn2Si2Yb1 |
| | -GM -M | -P -X | -N |
| | 7 2 7 2 | 6 2 5 2 | 3+4 2 |
| | 9 2 9 2 | 7 2 6 2 | 5+6 2 |
| | 8 2 8 2 | 6 2 6 2 | 5+6 2 |
| | 9 2 9 2 | 7 2 6 2 | 5+6 2 |
| | 7 2 7 2 | 7 2 5 2 | 5+6 2 |
| | 9 2 9 2 | 6 2 6 2 | 3+4 2 |
| | 7 2 7 2 | 7 2 6 2 | 3+4 2 |
| | 9 2 6 2 | 6 2 5 2 | 5+6 2 |
| | 8 2 7 2 | 7 2 5 2 | 3+4 2 |
| | 9 2 9 2 | 7 2 6 2 | 5+6 2 |
| | 6 2 8 2 | 6 2 5 2 | 5+6 2 |
| | 8 2 8 2 | 6 2 5 2 | 3+4 2 |
| | 8 2 8 2 | 7 2 6 2 | 5+6 2 |
| | 9 2 9 2 | 6 2 6 2 | 5+6 2 |
| | 7 2 9 2 | 6 2 6 2 | 5+6 2 |
| | 6 2 6 2 | 7 2 6 2 | 3+4 2 |
| | 7 2 8 2 | 6 2 6 2 | 5+6 2 |
| | 8 2 8 2 | 7 2 5 2 | 5+6 2 |
| | 9 2 7 2 | 6 2 6 2 | 3+4 2 |
| | 8 2 8 2 | 7 2 6 2 | 5+6 2 |
| | 9 2 9 2 | 6 2 6 2 | 5+6 2 |
| | 8 2 9 2 | 7 2 6 2 | 3+4 2 |
| | 6 2 8 2 | 7 2 5 2 | 5+6 2 |
| 139.1.49 | 290822 | 46 | Mn2Si2Yb1 |
| | -GM -M | -P -X | -N |
| | 7 2 7 2 | 6 2 5 2 | 3+4 2 |
| | 9 2 9 2 | 7 2 6 2 | 5+6 2 |
| | 8 2 8 2 | 6 2 6 2 | 5+6 2 |
| | 9 2 9 2 | 7 2 6 2 | 5+6 2 |
| | 7 2 7 2 | 7 2 5 2 | 5+6 2 |
| | 9 2 9 2 | 6 2 6 2 | 3+4 2 |
| | 7 2 7 2 | 7 2 6 2 | 3+4 2 |
| | 9 2 6 2 | 6 2 5 2 | 5+6 2 |
| | 8 2 7 2 | 7 2 5 2 | 3+4 2 |
| | 9 2 9 2 | 7 2 6 2 | 5+6 2 |
| | 6 2 8 2 | 6 2 5 2 | 5+6 2 |
| | 8 2 8 2 | 6 2 5 2 | 3+4 2 |
| | 8 2 8 2 | 7 2 6 2 | 5+6 2 |
| | 9 2 9 2 | 6 2 6 2 | 5+6 2 |
| | 7 2 9 2 | 6 2 6 2 | 5+6 2 |
| | 6 2 6 2 | 7 2 6 2 | 3+4 2 |
| | 7 2 8 2 | 6 2 6 2 | 5+6 2 |
| | 8 2 7 2 | 7 2 5 2 | 5+6 2 |
| | 9 2 8 2 | 6 2 6 2 | 5+6 2 |
| | 8 2 8 2 | 7 2 6 2 | 3+4 2 |
| | 9 2 9 2 | 6 2 6 2 | 3+4 2 |

Continued on next column

**TABLE LXIV– continued**

| TQC | ICSD | NoE | SF |
|---|---|---|---|
| | 8 2 9 2 | 7 2 6 2 | 5+6 2 |
| | 6 2 8 2 | 7 2 5 2 | 5+6 2 |
| 139.1.50 | 104941 | 74 | Mn1Pd3 |
| | -GM -M | -P -X | -N |
| | 7 2 7 2 | 7 2 6 2 | 5+6 2 |
| | 7 2 9 2 | 6 2 6 2 | 3+4 2 |
| | 7 2 7 2 | 6 2 5 2 | 5+6 2 |
| | 9 2 6 2 | 7 2 6 2 | 3+4 2 |
| | 6 2 9 2 | 6 2 6 2 | 5+6 2 |
| | 7 2 6 2 | 6 2 5 2 | 5+6 2 |
| | 9 2 7 2 | 7 2 6 2 | 3+4 2 |
| | 6 2 7 2 | 7 2 6 2 | 3+4 2 |
| | 8 2 6 2 | 6 2 5 2 | 3+4 2 |
| | 7 2 8 2 | 7 2 6 2 | 5+6 2 |
| | 9 2 7 2 | 6 2 6 2 | 3+4 2 |
| | 8 2 6 2 | 7 2 5 2 | 5+6 2 |
| | 6 2 8 2 | 6 2 6 2 | 3+4 2 |
| | 6 2 7 2 | 7 2 5 2 | 5+6 2 |
| | 8 2 9 2 | 7 2 6 2 | 3+4 2 |
| | 7 2 8 2 | 7 2 5 2 | 3+4 2 |
| | 6 2 6 2 | 6 2 6 2 | 5+6 2 |
| | 6 2 8 2 | 6 2 6 2 | 3+4 2 |
| | 7 2 7 2 | 6 2 5 2 | 5+6 2 |
| | 9 2 6 2 | 7 2 6 2 | 5+6 2 |
| | 7 2 9 2 | 7 2 6 2 | 3+4 2 |
| | 6 2 8 2 | 6 2 5 2 | 5+6 2 |
| | 8 2 9 2 | 6 2 6 2 | 5+6 2 |
| | 9 2 6 2 | 7 2 6 2 | 3+4 2 |
| | 6 2 7 2 | 6 2 5 2 | 3+4 2 |
| | 7 2 6 2 | 7 2 5 2 | 3+4 2 |
| | 7 2 8 2 | 7 2 6 2 | 5+6 2 |
| | 6 2 7 2 | 6 2 6 2 | 3+4 2 |
| | 6 2 6 2 | 6 2 5 2 | 5+6 2 |
| | 8 2 6 2 | 7 2 5 2 | 5+6 2 |
| | 6 2 9 2 | 6 2 6 2 | 5+6 2 |
| | 9 2 7 2 | 7 2 6 2 | 3+4 2 |
| | 8 2 9 2 | 6 2 6 2 | 3+4 2 |
| | 7 2 7 2 | 7 2 5 2 | 5+6 2 |
| | 9 2 8 2 | 7 2 5 2 | 5+6 2 |
| | 8 2 8 2 | 6 2 5 2 | 3+4 2 |
| | 9 2 7 2 | 7 2 6 2 | 5+6 2 |
| 139.1.51 | 150228 | 64 | Mg1Pd3 |
| | -GM -M | -P -X | -N |
| | 7 2 9 2 | 7 2 6 2 | 5+6 2 |
| | 7 2 7 2 | 6 2 5 2 | 3+4 2 |
| | 9 2 7 2 | 7 2 6 2 | 3+4 2 |
| | 7 2 9 2 | 6 2 5 2 | 5+6 2 |
| | 9 2 7 2 | 6 2 6 2 | 5+6 2 |
| | 6 2 6 2 | 7 2 6 2 | 5+6 2 |
| | 7 2 7 2 | 6 2 6 2 | 3+4 2 |
| | 8 2 6 2 | 7 2 6 2 | 3+4 2 |
| | 6 2 8 2 | 6 2 5 2 | 3+4 2 |
| | 7 2 6 2 | 7 2 6 2 | 5+6 2 |
| | 9 2 7 2 | 6 2 6 2 | 3+4 2 |
| | 6 2 6 2 | 7 2 5 2 | 5+6 2 |
| | 8 2 7 2 | 6 2 6 2 | 3+4 2 |
| | 6 2 8 2 | 7 2 5 2 | 5+6 2 |
| | 8 2 9 2 | 7 2 6 2 | 3+4 2 |
| | 7 2 8 2 | 7 2 5 2 | 5+6 2 |
| | 6 2 6 2 | 6 2 6 2 | 3+4 2 |
| | 7 2 8 2 | 6 2 5 2 | 5+6 2 |

Continued on next column

**TABLE LXIV– continued**

| TQC | ICSD | NoE | SF |
|---|---|---|---|
| | 6 2 7 2 | 6 2 6 2 | 3+4 2 |
| | 9 2 6 2 | 7 2 6 2 | 5+6 2 |
| | 7 2 9 2 | 6 2 6 2 | 3+4 2 |
| | 6 2 6 2 | 7 2 5 2 | 5+6 2 |
| | 8 2 9 2 | 6 2 6 2 | 3+4 2 |
| | 9 2 8 2 | 7 2 5 2 | 3+4 2 |
| | 7 2 7 2 | 6 2 6 2 | 5+6 2 |
| | 6 2 6 2 | 7 2 6 2 | 3+4 2 |
| | 6 2 8 2 | 7 2 6 2 | 5+6 2 |
| | 7 2 6 2 | 6 2 5 2 | 5+6 2 |
| | 8 2 7 2 | 6 2 5 2 | 3+4 2 |
| | 6 2 6 2 | 7 2 6 2 | 5+6 2 |
| | 6 2 7 2 | 7 2 6 2 | 3+4 2 |
| | 9 2 9 2 | 6 2 5 2 | 5+6 2 |
| 139.1.52 | 42733 | 54 | La2Sb1 |
| | -GM -M | -P -X | -N |
| | 7 2 9 2 | 6 2 5 2 | 3+4 2 |
| | 9 2 7 2 | 7 2 6 2 | 5+6 2 |
| | 7 2 7 2 | 6 2 6 2 | 3+4 2 |
| | 6 2 6 2 | 7 2 6 2 | 5+6 2 |
| | 8 2 8 2 | 6 2 5 2 | 3+4 2 |
| | 7 2 9 2 | 7 2 5 2 | 5+6 2 |
| | 9 2 7 2 | 6 2 6 2 | 5+6 2 |
| | 9 2 9 2 | 7 2 5 2 | 3+4 2 |
| | 9 2 9 2 | 6 2 5 2 | 5+6 2 |
| | 7 2 9 2 | 7 2 5 2 | 3+4 2 |
| | 6 2 8 2 | 7 2 6 2 | 3+4 2 |
| | 9 2 7 2 | 6 2 5 2 | 5+6 2 |
| | 8 2 6 2 | 7 2 6 2 | 3+4 2 |
| | 8 2 8 2 | 6 2 5 2 | 5+6 2 |
| | 9 2 9 2 | 7 2 5 2 | 5+6 2 |
| | 8 2 8 2 | 6 2 5 2 | 3+4 2 |
| | 7 2 7 2 | 7 2 6 2 | 3+4 2 |
| | 9 2 9 2 | 6 2 5 2 | 5+6 2 |
| | 7 2 7 2 | 7 2 6 2 | 5+6 2 |
| | 7 2 7 2 | 6 2 6 2 | 3+4 2 |
| | 9 2 7 2 | 7 2 6 2 | 3+4 2 |
| | 7 2 9 2 | 6 2 5 2 | 5+6 2 |
| | 6 2 6 2 | 6 2 5 2 | 3+4 2 |
| | 9 2 9 2 | 7 2 5 2 | 5+6 2 |
| | 8 2 6 2 | 6 2 6 2 | 5+6 2 |
| | 6 2 6 2 | 6 2 6 2 | 3+4 2 |
| | 8 2 8 2 | 7 2 5 2 | 3+4 2 |
| 139.1.53 | 423290 | 54 | La2Sb1 |
| | -GM -M | -P -X | -N |
| | 7 2 9 2 | 6 2 5 2 | 3+4 2 |
| | 9 2 7 2 | 7 2 6 2 | 5+6 2 |
| | 7 2 7 2 | 6 2 6 2 | 5+6 2 |
| | 6 2 6 2 | 7 2 6 2 | 3+4 2 |
| | 8 2 8 2 | 7 2 5 2 | 5+6 2 |
| | 7 2 9 2 | 6 2 5 2 | 3+4 2 |
| | 9 2 7 2 | 7 2 6 2 | 3+4 2 |
| | 9 2 9 2 | 6 2 5 2 | 5+6 2 |
| | 9 2 9 2 | 7 2 5 2 | 3+4 2 |
| | 7 2 9 2 | 6 2 5 2 | 5+6 2 |
| | 6 2 8 2 | 6 2 6 2 | 3+4 2 |
| | 9 2 7 2 | 7 2 5 2 | 5+6 2 |
| | 8 2 6 2 | 6 2 6 2 | 5+6 2 |
| | 8 2 8 2 | 7 2 5 2 | 3+4 2 |
| | 9 2 9 2 | 6 2 5 2 | 3+4 2 |
| | 8 2 8 2 | 7 2 5 2 | 5+6 2 |

Continued on next column

**TABLE LXIV– continued**

| TQC | ICSD | NoE | SF |
|---|---|---|---|
| | 7 2 7 2 | 6 2 6 2 | 5+6 2 |
| | 9 2 9 2 | 7 2 5 2 | 3+4 2 |
| | 7 2 7 2 | 6 2 6 2 | 3+4 2 |
| | 7 2 7 2 | 7 2 6 2 | 5+6 2 |
| | 9 2 7 2 | 6 2 6 2 | 5+6 2 |
| | 7 2 9 2 | 7 2 5 2 | 3+4 2 |
| | 6 2 6 2 | 7 2 5 2 | 5+6 2 |
| | 9 2 9 2 | 6 2 5 2 | 3+4 2 |
| | 8 2 6 2 | 7 2 6 2 | 3+4 2 |
| | 6 2 6 2 | 7 2 6 2 | 5+6 2 |
| | 8 2 8 2 | 6 2 5 2 | 5+6 2 |
| 139.1.54 | 10442 | 54 | La2Sb1 |
| | -GM -M | -P -X | -N |
| | 7 2 9 2 | 6 2 5 2 | 3+4 2 |
| | 9 2 7 2 | 7 2 6 2 | 5+6 2 |
| | 7 2 7 2 | 6 2 6 2 | 3+4 2 |
| | 6 2 6 2 | 7 2 6 2 | 5+6 2 |
| | 8 2 8 2 | 6 2 5 2 | 3+4 2 |
| | 7 2 9 2 | 7 2 5 2 | 5+6 2 |
| | 9 2 7 2 | 6 2 6 2 | 5+6 2 |
| | 9 2 9 2 | 7 2 5 2 | 3+4 2 |
| | 9 2 9 2 | 6 2 5 2 | 5+6 2 |
| | 7 2 9 2 | 7 2 5 2 | 3+4 2 |
| | 6 2 8 2 | 7 2 6 2 | 3+4 2 |
| | 9 2 7 2 | 6 2 5 2 | 5+6 2 |
| | 8 2 6 2 | 7 2 6 2 | 3+4 2 |
| | 8 2 8 2 | 6 2 5 2 | 5+6 2 |
| | 9 2 9 2 | 7 2 5 2 | 5+6 2 |
| | 8 2 8 2 | 6 2 5 2 | 3+4 2 |
| | 7 2 7 2 | 7 2 6 2 | 3+4 2 |
| | 9 2 9 2 | 6 2 5 2 | 5+6 2 |
| | 7 2 7 2 | 7 2 6 2 | 5+6 2 |
| | 7 2 7 2 | 6 2 6 2 | 3+4 2 |
| | 9 2 7 2 | 7 2 6 2 | 3+4 2 |
| | 7 2 9 2 | 6 2 5 2 | 5+6 2 |
| | 6 2 6 2 | 6 2 5 2 | 3+4 2 |
| | 9 2 9 2 | 7 2 5 2 | 5+6 2 |
| | 8 2 6 2 | 6 2 6 2 | 5+6 2 |
| | 6 2 6 2 | 6 2 6 2 | 3+4 2 |
| | 8 2 8 2 | 7 2 5 2 | 3+4 2 |
| 139.1.55 | 40262 | 56 | La2O4Pd1 |
| | -GM -M | -P -X | -N |
| | 7 2 9 2 | 7 2 5 2 | 5+6 2 |
| | 9 2 7 2 | 6 2 6 2 | 3+4 2 |
| | 7 2 7 2 | 6 2 6 2 | 3+4 2 |
| | 6 2 6 2 | 7 2 5 2 | 5+6 2 |
| | 7 2 9 2 | 7 2 6 2 | 5+6 2 |
| | 8 2 6 2 | 6 2 6 2 | 3+4 2 |
| | 9 2 7 2 | 7 2 5 2 | 5+6 2 |
| | 7 2 9 2 | 6 2 6 2 | 3+4 2 |
| | 6 2 8 2 | 7 2 5 2 | 5+6 2 |
| | 8 2 6 2 | 6 2 6 2 | 3+4 2 |
| | 9 2 7 2 | 6 2 5 2 | 5+6 2 |
| | 7 2 9 2 | 7 2 6 2 | 3+4 2 |
| | 9 2 9 2 | 7 2 5 2 | 3+4 2 |
| | 8 2 8 2 | 6 2 5 2 | 3+4 2 |
| | 9 2 8 2 | 7 2 5 2 | 3+4 2 |
| | 6 2 7 2 | 7 2 5 2 | 5+6 2 |
| | 7 2 9 2 | 6 2 5 2 | 5+6 2 |
| | 8 2 6 2 | 6 2 5 2 | 5+6 2 |
| | 6 2 9 2 | 7 2 6 2 | 3+4 2 |

Continued on next column

**TABLE LXIV– continued**

| TQC | ICSD | NoE | SF |
|---|---|---|---|
| | 6 2 8 2 | 6 2 5 2 | 3+4 2 |
| | 7 2 7 2 | 6 2 5 2 | 5+6 2 |
| | 9 2 6 2 | 7 2 6 2 | 5+6 2 |
| | 6 2 9 2 | 6 2 5 2 | 3+4 2 |
| | 8 2 8 2 | 7 2 6 2 | 3+4 2 |
| | 7 2 7 2 | 6 2 5 2 | 5+6 2 |
| | 9 2 8 2 | 6 2 5 2 | 3+4 2 |
| | 6 2 6 2 | 7 2 5 2 | 3+4 2 |
| | 9 2 7 2 | 7 2 5 2 | 3+4 2 |
| 139.1.56 | 69172 | 56 | La2Ni1O4 |
| | -GM -M | -P -X | -N |
| | 7 2 9 2 | 7 2 5 2 | 5+6 2 |
| | 9 2 7 2 | 6 2 6 2 | 3+4 2 |
| | 7 2 7 2 | 6 2 6 2 | 3+4 2 |
| | 7 2 9 2 | 7 2 6 2 | 5+6 2 |
| | 9 2 6 2 | 7 2 5 2 | 5+6 2 |
| | 6 2 7 2 | 6 2 6 2 | 3+4 2 |
| | 9 2 9 2 | 7 2 5 2 | 5+6 2 |
| | 7 2 7 2 | 6 2 6 2 | 3+4 2 |
| | 6 2 8 2 | 6 2 5 2 | 5+6 2 |
| | 8 2 6 2 | 7 2 5 2 | 3+4 2 |
| | 9 2 9 2 | 7 2 6 2 | 5+6 2 |
| | 7 2 7 2 | 6 2 6 2 | 3+4 2 |
| | 9 2 9 2 | 6 2 5 2 | 3+4 2 |
| | 8 2 8 2 | 7 2 5 2 | 3+4 2 |
| | 9 2 9 2 | 7 2 5 2 | 5+6 2 |
| | 7 2 8 2 | 7 2 5 2 | 3+4 2 |
| | 8 2 7 2 | 6 2 5 2 | 5+6 2 |
| | 7 2 9 2 | 6 2 6 2 | 5+6 2 |
| | 6 2 8 2 | 7 2 5 2 | 3+4 2 |
| | 9 2 9 2 | 6 2 6 2 | 3+4 2 |
| | 8 2 7 2 | 7 2 5 2 | 5+6 2 |
| | 9 2 6 2 | 6 2 6 2 | 5+6 2 |
| | 8 2 9 2 | 7 2 5 2 | 3+4 2 |
| | 9 2 8 2 | 6 2 5 2 | 5+6 2 |
| | 6 2 6 2 | 7 2 5 2 | 3+4 2 |
| | 7 2 7 2 | 6 2 5 2 | 3+4 2 |
| | 6 2 6 2 | 7 2 5 2 | 3+4 2 |
| | 6 2 6 2 | 6 2 5 2 | 3+4 2 |
| 139.1.57 | 33536 | 56 | La2Ni1O4 |
| | -GM -M | -P -X | -N |
| | 7 2 7 2 | 7 2 5 2 | 5+6 2 |
| | 9 2 9 2 | 6 2 6 2 | 3+4 2 |
| | 7 2 7 2 | 6 2 6 2 | 3+4 2 |
| | 7 2 9 2 | 7 2 6 2 | 5+6 2 |
| | 9 2 6 2 | 7 2 5 2 | 5+6 2 |
| | 6 2 7 2 | 6 2 6 2 | 3+4 2 |
| | 9 2 9 2 | 7 2 5 2 | 3+4 2 |
| | 7 2 7 2 | 6 2 6 2 | 5+6 2 |
| | 6 2 8 2 | 6 2 5 2 | 5+6 2 |
| | 8 2 9 2 | 7 2 5 2 | 3+4 2 |
| | 9 2 6 2 | 7 2 6 2 | 5+6 2 |
| | 7 2 7 2 | 6 2 6 2 | 3+4 2 |
| | 9 2 9 2 | 6 2 5 2 | 3+4 2 |
| | 8 2 8 2 | 7 2 5 2 | 3+4 2 |
| | 7 2 7 2 | 7 2 5 2 | 5+6 2 |
| | 9 2 9 2 | 7 2 5 2 | 5+6 2 |
| | 8 2 8 2 | 6 2 5 2 | 3+4 2 |
| | 7 2 9 2 | 6 2 5 2 | 3+4 2 |
| | 6 2 8 2 | 7 2 6 2 | 3+4 2 |
| | 9 2 7 2 | 6 2 6 2 | 5+6 2 |

Continued on next column

**TABLE LXIV– continued**

| TQC | ICSD | NoE | SF |
|---|---|---|---|
| | 8 2 6 2 | 7 2 5 2 | 5+6 2 |
| | 9 2 9 2 | 6 2 6 2 | 5+6 2 |
| | 9 2 9 2 | 7 2 5 2 | 3+4 2 |
| | 8 2 8 2 | 6 2 5 2 | 5+6 2 |
| | 6 2 6 2 | 6 2 5 2 | 3+4 2 |
| | 7 2 7 2 | 7 2 5 2 | 3+4 2 |
| | 6 2 6 2 | 7 2 5 2 | 3+4 2 |
| | 6 2 6 2 | 6 2 5 2 | 3+4 2 |
| 139.1.58 | 181385 | 66 | La2Mn2O3Se2 |
| | -GM -M | -P -X | -N |
| | 7 2 9 2 | 6 2 5 2 | 3+4 2 |
| | 9 2 7 2 | 7 2 6 2 | 5+6 2 |
| | 7 2 9 2 | 6 2 5 2 | 3+4 2 |
| | 7 2 7 2 | 7 2 6 2 | 5+6 2 |
| | 8 2 6 2 | 7 2 5 2 | 5+6 2 |
| | 9 2 7 2 | 6 2 6 2 | 3+4 2 |
| | 7 2 9 2 | 7 2 5 2 | 5+6 2 |
| | 6 2 8 2 | 6 2 6 2 | 3+4 2 |
| | 9 2 7 2 | 7 2 5 2 | 5+6 2 |
| | 8 2 6 2 | 6 2 5 2 | 3+4 2 |
| | 7 2 9 2 | 7 2 6 2 | 5+6 2 |
| | 7 2 7 2 | 6 2 5 2 | 3+4 2 |
| | 9 2 9 2 | 7 2 6 2 | 5+6 2 |
| | 9 2 7 2 | 7 2 6 2 | 3+4 2 |
| | 9 2 9 2 | 6 2 6 2 | 3+4 2 |
| | 7 2 8 2 | 6 2 6 2 | 3+4 2 |
| | 8 2 8 2 | 7 2 5 2 | 5+6 2 |
| | 6 2 9 2 | 6 2 6 2 | 3+4 2 |
| | 7 2 8 2 | 7 2 6 2 | 5+6 2 |
| | 6 2 9 2 | 6 2 6 2 | 3+4 2 |
| | 9 2 7 2 | 6 2 5 2 | 5+6 2 |
| | 9 2 7 2 | 7 2 5 2 | 3+4 2 |
| | 8 2 6 2 | 6 2 5 2 | 3+4 2 |
| | 7 2 7 2 | 7 2 6 2 | 5+6 2 |
| | 6 2 6 2 | 6 2 5 2 | 3+4 2 |
| | 9 2 9 2 | 6 2 6 2 | 5+6 2 |
| | 6 2 9 2 | 7 2 5 2 | 5+6 2 |
| | 8 2 8 2 | 7 2 6 2 | 3+4 2 |
| | 9 2 6 2 | 6 2 6 2 | 5+6 2 |
| | 7 2 7 2 | 7 2 6 2 | 3+4 2 |
| | 6 2 6 2 | 6 2 6 2 | 5+6 2 |
| | 7 2 7 2 | 7 2 6 2 | 5+6 2 |
| | 6 2 6 2 | 7 2 6 2 | 3+4 2 |
| 139.1.59 | 427126 | 38 | La1Sb1Sc1 |
| | -GM -M | -P -X | -N |
| | 7 2 9 2 | 6 2 5 2 | 3+4 2 |
| | 9 2 7 2 | 7 2 6 2 | 5+6 2 |
| | 7 2 9 2 | 6 2 5 2 | 3+4 2 |
| | 9 2 7 2 | 7 2 6 2 | 5+6 2 |
| | 7 2 9 2 | 6 2 6 2 | 3+4 2 |
| | 6 2 8 2 | 7 2 5 2 | 5+6 2 |
| | 9 2 7 2 | 6 2 6 2 | 3+4 2 |
| | 8 2 6 2 | 7 2 5 2 | 5+6 2 |
| | 7 2 7 2 | 7 2 5 2 | 5+6 2 |
| | 9 2 9 2 | 6 2 6 2 | 3+4 2 |
| | 7 2 7 2 | 6 2 6 2 | 3+4 2 |
| | 9 2 7 2 | 7 2 6 2 | 5+6 2 |
| | 7 2 9 2 | 7 2 5 2 | 5+6 2 |
| | 6 2 6 2 | 6 2 5 2 | 5+6 2 |
| | 7 2 7 2 | 7 2 6 2 | 3+4 2 |
| | 9 2 9 2 | 6 2 5 2 | 3+4 2 |

Continued on next column

**TABLE LXIV– continued**

| TQC | ICSD | NoE | SF |
|---|---|---|---|
| | 8 2 6 2 | 7 2 6 2 | 3+4 2 |
| | 8 2 8 2 | 6 2 6 2 | 5+6 2 |
| | 6 2 6 2 | 7 2 5 2 | 5+6 2 |
| 139.1.60 | 421342 | 40 | K2Mg5Sn3 |
| | -GM -M | -P -X | -N |
| | 7 2 9 2 | 6 2 5 2 | 3+4 2 |
| | 9 2 7 2 | 7 2 6 2 | 5+6 2 |
| | 7 2 9 2 | 6 2 5 2 | 3+4 2 |
| | 9 2 7 2 | 7 2 6 2 | 5+6 2 |
| | 7 2 9 2 | 6 2 5 2 | 3+4 2 |
| | 6 2 8 2 | 7 2 6 2 | 5+6 2 |
| | 9 2 7 2 | 6 2 5 2 | 3+4 2 |
| | 8 2 6 2 | 7 2 6 2 | 5+6 2 |
| | 7 2 7 2 | 6 2 5 2 | 3+4 2 |
| | 9 2 9 2 | 7 2 5 2 | 5+6 2 |
| | 7 2 7 2 | 6 2 6 2 | 3+4 2 |
| | 9 2 9 2 | 7 2 6 2 | 5+6 2 |
| | 7 2 7 2 | 6 2 6 2 | 5+6 2 |
| | 9 2 9 2 | 7 2 5 2 | 3+4 2 |
| | 8 2 8 2 | 7 2 5 2 | 3+4 2 |
| | 7 2 9 2 | 6 2 5 2 | 5+6 2 |
| | 6 2 7 2 | 7 2 6 2 | 5+6 2 |
| | 7 2 6 2 | 6 2 6 2 | 3+4 2 |
| | 9 2 6 2 | 7 2 6 2 | 5+6 2 |
| | 6 2 7 2 | 6 2 6 2 | 5+6 2 |
| 139.1.61 | 420251 | 66 | In1Pd3 |
| | -GM -M | -P -X | -N |
| | 7 2 7 2 | 7 2 6 2 | 3+4 2 |
| | 9 2 9 2 | 6 2 5 2 | 5+6 2 |
| | 7 2 7 2 | 6 2 6 2 | 5+6 2 |
| | 7 2 9 2 | 7 2 5 2 | 3+4 2 |
| | 9 2 7 2 | 7 2 6 2 | 3+4 2 |
| | 6 2 6 2 | 6 2 6 2 | 5+6 2 |
| | 8 2 7 2 | 6 2 6 2 | 3+4 2 |
| | 9 2 6 2 | 7 2 6 2 | 5+6 2 |
| | 7 2 8 2 | 6 2 5 2 | 5+6 2 |
| | 6 2 6 2 | 7 2 6 2 | 3+4 2 |
| | 7 2 7 2 | 6 2 5 2 | 5+6 2 |
| | 6 2 6 2 | 7 2 6 2 | 3+4 2 |
| | 8 2 7 2 | 6 2 6 2 | 3+4 2 |
| | 6 2 9 2 | 7 2 5 2 | 5+6 2 |
| | 7 2 8 2 | 7 2 6 2 | 3+4 2 |
| | 6 2 8 2 | 6 2 5 2 | 5+6 2 |
| | 7 2 6 2 | 7 2 6 2 | 3+4 2 |
| | 8 2 7 2 | 6 2 5 2 | 5+6 2 |
| | 9 2 8 2 | 6 2 6 2 | 3+4 2 |
| | 6 2 6 2 | 7 2 6 2 | 5+6 2 |
| | 7 2 9 2 | 6 2 5 2 | 3+4 2 |
| | 8 2 6 2 | 7 2 6 2 | 5+6 2 |
| | 6 2 8 2 | 6 2 6 2 | 3+4 2 |
| | 9 2 9 2 | 7 2 6 2 | 5+6 2 |
| | 6 2 7 2 | 6 2 5 2 | 3+4 2 |
| | 6 2 6 2 | 7 2 6 2 | 5+6 2 |
| | 7 2 8 2 | 7 2 5 2 | 3+4 2 |
| | 7 2 9 2 | 6 2 6 2 | 5+6 2 |
| | 8 2 6 2 | 6 2 5 2 | 3+4 2 |
| | 6 2 7 2 | 6 2 5 2 | 5+6 2 |
| | 6 2 7 2 | 7 2 5 2 | 3+4 2 |
| | 9 2 7 2 | 6 2 6 2 | 5+6 2 |
| | 7 2 6 2 | 7 2 6 2 | 5+6 2 |
| 139.1.62 | 72190 | 28 | I2Nd1 |

Continued on next column

**TABLE LXIV– continued**

| TQC | ICSD | NoE | SF |
|---|---|---|---|
| | -GM -M | -P -X | -N |
| | 7 2 7 2 | 6 2 5 2 | 3+4 2 |
| | 9 2 9 2 | 7 2 6 2 | 5+6 2 |
| | 9 2 9 2 | 6 2 6 2 | 5+6 2 |
| | 8 2 8 2 | 7 2 6 2 | 5+6 2 |
| | 7 2 7 2 | 7 2 5 2 | 3+4 2 |
| | 9 2 9 2 | 6 2 6 2 | 5+6 2 |
| | 7 2 7 2 | 7 2 5 2 | 3+4 2 |
| | 7 2 9 2 | 6 2 6 2 | 5+6 2 |
| | 9 2 7 2 | 6 2 5 2 | 5+6 2 |
| | 6 2 9 2 | 7 2 5 2 | 3+4 2 |
| | 8 2 6 2 | 7 2 6 2 | 3+4 2 |
| | 9 2 8 2 | 6 2 6 2 | 5+6 2 |
| | 6 2 6 2 | 7 2 6 2 | 3+4 2 |
| | 8 2 8 2 | 6 2 6 2 | 5+6 2 |
| 139.1.63 | 710068 | 132 | Ho1Si3 |
| | -GM -M | -P -X | -N |
| | 7 2 7 2 | 6 2 5 2 | 3+4 2 |
| | 9 2 9 2 | 7 2 6 2 | 5+6 2 |
| | 7 2 6 2 | 6 2 5 2 | 3+4 2 |
| | 8 2 9 2 | 7 2 6 2 | 5+6 2 |
| | 7 2 7 2 | 6 2 5 2 | 3+4 2 |
| | 9 2 9 2 | 7 2 6 2 | 5+6 2 |
| | 6 2 8 2 | 6 2 6 2 | 3+4 2 |
| | 9 2 7 2 | 7 2 5 2 | 5+6 2 |
| | 7 2 8 2 | 6 2 5 2 | 3+4 2 |
| | 8 2 6 2 | 6 2 6 2 | 5+6 2 |
| | 6 2 7 2 | 7 2 5 2 | 3+4 2 |
| | 9 2 9 2 | 7 2 6 2 | 5+6 2 |
| | 9 2 7 2 | 6 2 6 2 | 3+4 2 |
| | 6 2 8 2 | 7 2 5 2 | 5+6 2 |
| | 7 2 9 2 | 6 2 5 2 | 3+4 2 |
| | 8 2 6 2 | 7 2 6 2 | 5+6 2 |
| | 7 2 7 2 | 7 2 6 2 | 5+6 2 |
| | 9 2 8 2 | 7 2 5 2 | 3+4 2 |
| | 8 2 9 2 | 6 2 6 2 | 3+4 2 |
| | 6 2 6 2 | 6 2 5 2 | 5+6 2 |
| | 6 2 9 2 | 6 2 6 2 | 5+6 2 |
| | 9 2 7 2 | 6 2 5 2 | 3+4 2 |
| | 7 2 6 2 | 7 2 6 2 | 3+4 2 |
| | 6 2 9 2 | 6 2 6 2 | 5+6 2 |
| | 7 2 8 2 | 7 2 5 2 | 5+6 2 |
| | 9 2 7 2 | 7 2 5 2 | 3+4 2 |
| | 8 2 6 2 | 7 2 6 2 | 5+6 2 |
| | 7 2 8 2 | 6 2 5 2 | 3+4 2 |
| | 8 2 7 2 | 7 2 5 2 | 5+6 2 |
| | 9 2 9 2 | 7 2 6 2 | 3+4 2 |
| | 7 2 8 2 | 6 2 6 2 | 3+4 2 |
| | 8 2 9 2 | 6 2 5 2 | 5+6 2 |
| | 6 2 8 2 | 7 2 6 2 | 3+4 2 |
| | 7 2 7 2 | 7 2 5 2 | 5+6 2 |
| | 6 2 7 2 | 6 2 5 2 | 3+4 2 |
| | 9 2 6 2 | 6 2 6 2 | 5+6 2 |
| | 6 2 6 2 | 7 2 6 2 | 3+4 2 |
| | 8 2 9 2 | 6 2 5 2 | 5+6 2 |
| | 7 2 8 2 | 7 2 6 2 | 5+6 2 |
| | 7 2 9 2 | 6 2 6 2 | 3+4 2 |
| | 9 2 6 2 | 7 2 5 2 | 3+4 2 |
| | 6 2 7 2 | 6 2 6 2 | 5+6 2 |
| | 8 2 9 2 | 6 2 5 2 | 5+6 2 |
| | 8 2 8 2 | 7 2 6 2 | 3+4 2 |

Continued on next column

**TABLE LXIV– continued**

| TQC | ICSD | NoE | SF |
|---|---|---|---|
| | 6 2 8 2 | 6 2 5 2 | 5+6 2 |
| | 9 2 9 2 | 7 2 6 2 | 3+4 2 |
| | 7 2 6 2 | 6 2 5 2 | 3+4 2 |
| | 6 2 7 2 | 7 2 6 2 | 3+4 2 |
| | 8 2 6 2 | 6 2 5 2 | 5+6 2 |
| | 7 2 6 2 | 7 2 5 2 | 5+6 2 |
| | 9 2 7 2 | 6 2 6 2 | 3+4 2 |
| | 8 2 8 2 | 7 2 6 2 | 5+6 2 |
| | 8 2 9 2 | 6 2 5 2 | 3+4 2 |
| | 7 2 8 2 | 7 2 6 2 | 5+6 2 |
| | 6 2 6 2 | 6 2 5 2 | 5+6 2 |
| | 8 2 7 2 | 6 2 6 2 | 3+4 2 |
| | 7 2 8 2 | 7 2 5 2 | 5+6 2 |
| | 8 2 9 2 | 7 2 6 2 | 3+4 2 |
| | 6 2 7 2 | 6 2 5 2 | 3+4 2 |
| | 9 2 6 2 | 7 2 6 2 | 5+6 2 |
| | 7 2 9 2 | 6 2 5 2 | 3+4 2 |
| | 9 2 8 2 | 6 2 6 2 | 5+6 2 |
| | 6 2 6 2 | 7 2 5 2 | 3+4 2 |
| | 9 2 9 2 | 6 2 5 2 | 5+6 2 |
| | 6 2 7 2 | 7 2 5 2 | 5+6 2 |
| | 7 2 8 2 | 6 2 6 2 | 3+4 2 |
| 139.1.64 | 427211 | 40 | Hg2In2Sr1 |
| | -GM -M | -P -X | -N |
| | 7 2 7 2 | 6 2 5 2 | 3+4 2 |
| | 9 2 9 2 | 7 2 6 2 | 5+6 2 |
| | 9 2 9 2 | 6 2 6 2 | 5+6 2 |
| | 8 2 8 2 | 7 2 6 2 | 5+6 2 |
| | 7 2 7 2 | 7 2 5 2 | 5+6 2 |
| | 7 2 9 2 | 6 2 6 2 | 3+4 2 |
| | 9 2 9 2 | 7 2 6 2 | 3+4 2 |
| | 8 2 8 2 | 6 2 5 2 | 5+6 2 |
| | 6 2 6 2 | 7 2 5 2 | 5+6 2 |
| | 8 2 8 2 | 7 2 6 2 | 3+4 2 |
| | 7 2 7 2 | 6 2 5 2 | 3+4 2 |
| | 6 2 6 2 | 6 2 5 2 | 5+6 2 |
| | 8 2 7 2 | 7 2 6 2 | 5+6 2 |
| | 9 2 8 2 | 6 2 6 2 | 3+4 2 |
| | 6 2 9 2 | 7 2 5 2 | 3+4 2 |
| | 9 2 6 2 | 6 2 6 2 | 5+6 2 |
| | 7 2 9 2 | 7 2 6 2 | 5+6 2 |
| | 7 2 8 2 | 6 2 5 2 | 3+4 2 |
| | 8 2 6 2 | 7 2 5 2 | 5+6 2 |
| | 6 2 7 2 | 6 2 6 2 | 3+4 2 |
| 139.1.65 | 75936 | 24 | Hf3Te2 |
| | -GM -M | -P -X | -N |
| | 7 2 9 2 | 6 2 5 2 | 3+4 2 |
| | 9 2 7 2 | 7 2 6 2 | 5+6 2 |
| | 7 2 9 2 | 6 2 5 2 | 3+4 2 |
| | 7 2 7 2 | 7 2 6 2 | 5+6 2 |
| | 9 2 7 2 | 6 2 5 2 | 3+4 2 |
| | 6 2 6 2 | 7 2 6 2 | 5+6 2 |
| | 7 2 9 2 | 6 2 5 2 | 3+4 2 |
| | 9 2 7 2 | 7 2 6 2 | 3+4 2 |
| | 6 2 8 2 | 7 2 5 2 | 5+6 2 |
| | 8 2 8 2 | 7 2 5 2 | 3+4 2 |
| | 8 2 6 2 | 6 2 5 2 | 5+6 2 |
| | 7 2 7 2 | 7 2 6 2 | 5+6 2 |
| 139.1.66 | 415033 | 108 | H2Ba8Bi4O1 |
| | -GM -M | -P -X | -N |
| | 7 2 9 2 | 6 2 5 2 | 3+4 2 |

Continued on next column

**TABLE LXIV– continued**

| TQC | ICSD | NoE | SF |
|---|---|---|---|
| | 7 2 7 2 | 7 2 6 2 | 5+6 2 |
| | 9 2 9 2 | 6 2 5 2 | 3+4 2 |
| | 7 2 7 2 | 7 2 6 2 | 5+6 2 |
| | 9 2 9 2 | 6 2 5 2 | 3+4 2 |
| | 8 2 6 2 | 7 2 6 2 | 5+6 2 |
| | 7 2 7 2 | 6 2 6 2 | 3+4 2 |
| | 6 2 6 2 | 7 2 6 2 | 5+6 2 |
| | 7 2 7 2 | 6 2 5 2 | 3+4 2 |
| | 6 2 8 2 | 6 2 5 2 | 5+6 2 |
| | 9 2 9 2 | 7 2 6 2 | 3+4 2 |
| | 7 2 7 2 | 6 2 5 2 | 3+4 2 |
| | 9 2 7 2 | 7 2 6 2 | 5+6 2 |
| | 7 2 9 2 | 6 2 5 2 | 3+4 2 |
| | 8 2 8 2 | 7 2 6 2 | 5+6 2 |
| | 9 2 7 2 | 7 2 5 2 | 5+6 2 |
| | 9 2 9 2 | 6 2 5 2 | 3+4 2 |
| | 7 2 9 2 | 6 2 5 2 | 3+4 2 |
| | 9 2 9 2 | 7 2 6 2 | 5+6 2 |
| | 9 2 9 2 | 7 2 5 2 | 5+6 2 |
| | 7 2 7 2 | 7 2 6 2 | 3+4 2 |
| | 9 2 7 2 | 6 2 6 2 | 5+6 2 |
| | 6 2 6 2 | 6 2 5 2 | 3+4 2 |
| | 8 2 8 2 | 7 2 5 2 | 5+6 2 |
| | 6 2 8 2 | 6 2 6 2 | 3+4 2 |
| | 7 2 9 2 | 7 2 5 2 | 5+6 2 |
| | 8 2 6 2 | 6 2 6 2 | 3+4 2 |
| | 6 2 8 2 | 7 2 5 2 | 3+4 2 |
| | 9 2 7 2 | 6 2 5 2 | 5+6 2 |
| | 8 2 8 2 | 7 2 6 2 | 5+6 2 |
| | 8 2 6 2 | 6 2 5 2 | 3+4 2 |
| | 9 2 9 2 | 7 2 5 2 | 5+6 2 |
| | 8 2 8 2 | 6 2 5 2 | 3+4 2 |
| | 7 2 7 2 | 6 2 6 2 | 5+6 2 |
| | 7 2 9 2 | 6 2 5 2 | 3+4 2 |
| | 9 2 7 2 | 7 2 6 2 | 3+4 2 |
| | 9 2 9 2 | 7 2 5 2 | 5+6 2 |
| | 7 2 9 2 | 6 2 6 2 | 5+6 2 |
| | 9 2 7 2 | 7 2 6 2 | 3+4 2 |
| | 9 2 9 2 | 7 2 6 2 | 5+6 2 |
| | 7 2 7 2 | 6 2 6 2 | 5+6 2 |
| | 9 2 7 2 | 7 2 5 2 | 3+4 2 |
| | 8 2 9 2 | 6 2 6 2 | 5+6 2 |
| | 9 2 8 2 | 6 2 5 2 | 3+4 2 |
| | 7 2 9 2 | 7 2 6 2 | 5+6 2 |
| | 9 2 9 2 | 7 2 5 2 | 5+6 2 |
| | 7 2 7 2 | 7 2 6 2 | 5+6 2 |
| | 9 2 7 2 | 6 2 5 2 | 3+4 2 |
| | 6 2 6 2 | 7 2 6 2 | 5+6 2 |
| | 8 2 6 2 | 6 2 6 2 | 3+4 2 |
| | 7 2 9 2 | 6 2 5 2 | 5+6 2 |
| | 9 2 9 2 | 7 2 6 2 | 3+4 2 |
| | 6 2 8 2 | 6 2 5 2 | 5+6 2 |
| | 8 2 6 2 | 7 2 5 2 | 3+4 2 |
| 139.1.67 | 411 | 42 | Ge2Sr1Zn2 |
| | -GM -M | -P -X | -N |
| | 7 2 7 2 | 6 2 5 2 | 3+4 2 |
| | 9 2 9 2 | 7 2 6 2 | 5+6 2 |
| | 9 2 9 2 | 6 2 6 2 | 5+6 2 |
| | 8 2 8 2 | 7 2 6 2 | 5+6 2 |
| | 7 2 7 2 | 7 2 5 2 | 5+6 2 |
| | 9 2 9 2 | 6 2 6 2 | 3+4 2 |

Continued on next column

**TABLE LXIV– continued**

| TQC | ICSD | NoE | SF |
|---|---|---|---|
| | 6 2 8 2 | 7 2 5 2 | 5+6 2 |
| | 7 2 7 2 | 6 2 6 2 | 3+4 2 |
| | 8 2 6 2 | 7 2 5 2 | 5+6 2 |
| | 9 2 9 2 | 6 2 6 2 | 3+4 2 |
| | 7 2 7 2 | 6 2 5 2 | 5+6 2 |
| | 6 2 8 2 | 7 2 6 2 | 3+4 2 |
| | 8 2 9 2 | 6 2 5 2 | 3+4 2 |
| | 6 2 6 2 | 7 2 6 2 | 5+6 2 |
| | 9 2 8 2 | 7 2 5 2 | 5+6 2 |
| | 9 2 7 2 | 6 2 6 2 | 3+4 2 |
| | 7 2 9 2 | 7 2 6 2 | 5+6 2 |
| | 7 2 8 2 | 6 2 5 2 | 3+4 2 |
| | 6 2 6 2 | 7 2 5 2 | 3+4 2 |
| | 9 2 7 2 | 7 2 6 2 | 5+6 2 |
| | 8 2 9 2 | 6 2 5 2 | 5+6 2 |
| 139.1.68 | 53731 | 42 | Ge2Nd1Pd2 |
| | -GM -M | -P -X | -N |
| | 7 2 7 2 | 6 2 5 2 | 3+4 2 |
| | 9 2 9 2 | 7 2 6 2 | 5+6 2 |
| | 9 2 9 2 | 6 2 6 2 | 5+6 2 |
| | 8 2 8 2 | 7 2 6 2 | 5+6 2 |
| | 7 2 7 2 | 7 2 5 2 | 5+6 2 |
| | 9 2 9 2 | 6 2 6 2 | 3+4 2 |
| | 7 2 7 2 | 7 2 5 2 | 3+4 2 |
| | 8 2 6 2 | 6 2 6 2 | 5+6 2 |
| | 9 2 7 2 | 6 2 5 2 | 3+4 2 |
| | 6 2 8 2 | 7 2 6 2 | 5+6 2 |
| | 7 2 9 2 | 6 2 5 2 | 3+4 2 |
| | 6 2 9 2 | 7 2 6 2 | 5+6 2 |
| | 9 2 6 2 | 7 2 5 2 | 3+4 2 |
| | 8 2 8 2 | 6 2 5 2 | 5+6 2 |
| | 6 2 8 2 | 6 2 6 2 | 3+4 2 |
| | 7 2 7 2 | 7 2 5 2 | 5+6 2 |
| | 9 2 9 2 | 7 2 6 2 | 5+6 2 |
| | 6 2 8 2 | 7 2 6 2 | 3+4 2 |
| | 7 2 8 2 | 6 2 5 2 | 3+4 2 |
| | 8 2 9 2 | 7 2 6 2 | 5+6 2 |
| | 6 2 8 2 | 6 2 5 2 | 5+6 2 |
| 139.1.69 | 81753 | 42 | Ge2Nd1Ni2 |
| | -GM -M | -P -X | -N |
| | 7 2 7 2 | 6 2 5 2 | 3+4 2 |
| | 9 2 9 2 | 7 2 6 2 | 5+6 2 |
| | 9 2 9 2 | 6 2 6 2 | 5+6 2 |
| | 8 2 8 2 | 7 2 6 2 | 5+6 2 |
| | 7 2 7 2 | 7 2 5 2 | 5+6 2 |
| | 9 2 9 2 | 6 2 6 2 | 3+4 2 |
| | 7 2 7 2 | 7 2 5 2 | 3+4 2 |
| | 9 2 6 2 | 6 2 6 2 | 5+6 2 |
| | 8 2 7 2 | 7 2 5 2 | 3+4 2 |
| | 6 2 8 2 | 6 2 6 2 | 5+6 2 |
| | 7 2 9 2 | 7 2 5 2 | 5+6 2 |
| | 6 2 9 2 | 6 2 5 2 | 3+4 2 |
| | 9 2 8 2 | 7 2 6 2 | 3+4 2 |
| | 8 2 6 2 | 6 2 5 2 | 5+6 2 |
| | 7 2 8 2 | 7 2 6 2 | 3+4 2 |
| | 6 2 9 2 | 6 2 5 2 | 5+6 2 |
| | 7 2 8 2 | 7 2 6 2 | 5+6 2 |
| | 6 2 7 2 | 7 2 5 2 | 3+4 2 |
| | 9 2 8 2 | 6 2 6 2 | 5+6 2 |
| | 6 2 9 2 | 7 2 6 2 | 3+4 2 |
| | 8 2 8 2 | 6 2 5 2 | 5+6 2 |

Continued on next column

**TABLE LXIV– continued**

| TQC | ICSD | NoE | SF |
|---|---|---|---|
| 139.1.70 | 81744 | 36 | Ge2Mn2U1 |
| | -GM -M | -P -X | -N |
| | 7 2 7 2 | 6 2 5 2 | 3+4 2 |
| | 9 2 9 2 | 7 2 6 2 | 5+6 2 |
| | 9 2 9 2 | 6 2 6 2 | 5+6 2 |
| | 8 2 8 2 | 7 2 6 2 | 5+6 2 |
| | 7 2 7 2 | 7 2 5 2 | 5+6 2 |
| | 9 2 9 2 | 6 2 6 2 | 3+4 2 |
| | 7 2 7 2 | 7 2 5 2 | 3+4 2 |
| | 9 2 7 2 | 6 2 6 2 | 5+6 2 |
| | 8 2 6 2 | 7 2 5 2 | 3+4 2 |
| | 6 2 9 2 | 7 2 5 2 | 5+6 2 |
| | 9 2 6 2 | 6 2 5 2 | 3+4 2 |
| | 7 2 8 2 | 6 2 6 2 | 5+6 2 |
| | 6 2 9 2 | 7 2 5 2 | 3+4 2 |
| | 6 2 8 2 | 6 2 6 2 | 5+6 2 |
| | 8 2 7 2 | 7 2 6 2 | 5+6 2 |
| | 7 2 8 2 | 7 2 5 2 | 3+4 2 |
| | 8 2 9 2 | 6 2 6 2 | 5+6 2 |
| | 7 2 8 2 | 6 2 6 2 | 3+4 2 |
| 139.1.71 | 41976 | 34 | Ge2Mn2Th1 |
| | -GM -M | -P -X | -N |
| | 7 2 7 2 | 6 2 5 2 | 3+4 2 |
| | 9 2 9 2 | 7 2 6 2 | 5+6 2 |
| | 9 2 9 2 | 6 2 6 2 | 5+6 2 |
| | 8 2 8 2 | 7 2 6 2 | 5+6 2 |
| | 7 2 7 2 | 7 2 5 2 | 5+6 2 |
| | 9 2 9 2 | 6 2 6 2 | 3+4 2 |
| | 7 2 7 2 | 7 2 5 2 | 3+4 2 |
| | 9 2 7 2 | 6 2 6 2 | 5+6 2 |
| | 8 2 6 2 | 7 2 5 2 | 3+4 2 |
| | 9 2 6 2 | 7 2 5 2 | 5+6 2 |
| | 6 2 9 2 | 6 2 5 2 | 3+4 2 |
| | 7 2 8 2 | 6 2 6 2 | 5+6 2 |
| | 6 2 9 2 | 7 2 5 2 | 3+4 2 |
| | 6 2 8 2 | 6 2 6 2 | 5+6 2 |
| | 8 2 7 2 | 7 2 6 2 | 5+6 2 |
| | 7 2 9 2 | 7 2 5 2 | 3+4 2 |
| | 7 2 8 2 | 6 2 6 2 | 3+4 2 |
| 139.1.72 | 403 | 32 | Ge2Mn2Sr1 |
| | -GM -M | -P -X | -N |
| | 7 2 7 2 | 6 2 5 2 | 3+4 2 |
| | 9 2 9 2 | 7 2 6 2 | 5+6 2 |
| | 8 2 8 2 | 6 2 6 2 | 5+6 2 |
| | 9 2 9 2 | 7 2 6 2 | 5+6 2 |
| | 7 2 7 2 | 7 2 5 2 | 5+6 2 |
| | 9 2 9 2 | 6 2 6 2 | 3+4 2 |
| | 7 2 7 2 | 7 2 6 2 | 3+4 2 |
| | 9 2 7 2 | 6 2 5 2 | 5+6 2 |
| | 8 2 6 2 | 7 2 5 2 | 3+4 2 |
| | 9 2 9 2 | 7 2 6 2 | 5+6 2 |
| | 6 2 8 2 | 6 2 5 2 | 5+6 2 |
| | 7 2 9 2 | 6 2 5 2 | 3+4 2 |
| | 6 2 8 2 | 7 2 6 2 | 3+4 2 |
| | 7 2 6 2 | 7 2 6 2 | 5+6 2 |
| | 8 2 8 2 | 6 2 5 2 | 5+6 2 |
| | 6 2 9 2 | 7 2 6 2 | 3+4 2 |
| 139.1.73 | 165737 | 36 | Ge2Ir2Sr1 |
| | -GM -M | -P -X | -N |
| | 7 2 7 2 | 6 2 5 2 | 3+4 2 |
| | 9 2 9 2 | 7 2 6 2 | 5+6 2 |
| | 8 2 9 2 | 6 2 6 2 | 5+6 2 |
| | 9 2 8 2 | 7 2 6 2 | 5+6 2 |
| | 7 2 9 2 | 7 2 5 2 | 5+6 2 |
| | 9 2 7 2 | 6 2 6 2 | 3+4 2 |
| | 7 2 7 2 | 7 2 5 2 | 3+4 2 |
| | 6 2 6 2 | 6 2 6 2 | 5+6 2 |
| | 8 2 8 2 | 6 2 5 2 | 3+4 2 |
| | 9 2 7 2 | 7 2 6 2 | 5+6 2 |
| | 9 2 9 2 | 6 2 5 2 | 5+6 2 |
| | 7 2 9 2 | 7 2 6 2 | 3+4 2 |
| | 6 2 8 2 | 7 2 5 2 | 5+6 2 |
| | 8 2 6 2 | 6 2 5 2 | 3+4 2 |
| | 7 2 8 2 | 7 2 6 2 | 3+4 2 |
| | 6 2 9 2 | 6 2 6 2 | 5+6 2 |
| | 6 2 8 2 | 7 2 5 2 | 5+6 2 |
| | 7 2 7 2 | 7 2 6 2 | 3+4 2 |
| 139.1.74 | 26277 | 48 | Ge1Te1U1 |
| | -GM -M | -P -X | -N |
| | 7 2 7 2 | 6 2 5 2 | 3+4 2 |
| | 9 2 9 2 | 7 2 6 2 | 5+6 2 |
| | 7 2 7 2 | 6 2 5 2 | 3+4 2 |
| | 9 2 9 2 | 7 2 6 2 | 5+6 2 |
| | 7 2 7 2 | 6 2 6 2 | 3+4 2 |
| | 9 2 9 2 | 7 2 5 2 | 5+6 2 |
| | 8 2 8 2 | 6 2 5 2 | 3+4 2 |
| | 6 2 6 2 | 7 2 6 2 | 5+6 2 |
| | 7 2 9 2 | 7 2 5 2 | 3+4 2 |
| | 9 2 7 2 | 6 2 6 2 | 5+6 2 |
| | 7 2 7 2 | 7 2 6 2 | 5+6 2 |
| | 6 2 6 2 | 6 2 6 2 | 3+4 2 |
| | 7 2 9 2 | 6 2 5 2 | 5+6 2 |
| | 9 2 9 2 | 7 2 5 2 | 3+4 2 |
| | 7 2 8 2 | 7 2 6 2 | 3+4 2 |
| | 6 2 9 2 | 6 2 5 2 | 5+6 2 |
| | 9 2 7 2 | 6 2 6 2 | 3+4 2 |
| | 8 2 8 2 | 7 2 5 2 | 5+6 2 |
| | 8 2 8 2 | 6 2 6 2 | 3+4 2 |
| | 9 2 9 2 | 7 2 5 2 | 5+6 2 |
| | 8 2 7 2 | 6 2 5 2 | 5+6 2 |
| | 8 2 6 2 | 7 2 5 2 | 3+4 2 |
| | 6 2 6 2 | 6 2 6 2 | 3+4 2 |
| | 7 2 8 2 | 7 2 5 2 | 5+6 2 |
| 139.1.75 | 87316 | 48 | Ge1Se1U1 |
| | -GM -M | -P -X | -N |
| | 7 2 7 2 | 6 2 5 2 | 3+4 2 |
| | 9 2 9 2 | 7 2 6 2 | 5+6 2 |
| | 7 2 9 2 | 6 2 5 2 | 3+4 2 |
| | 9 2 7 2 | 7 2 6 2 | 5+6 2 |
| | 7 2 9 2 | 6 2 6 2 | 3+4 2 |
| | 9 2 7 2 | 7 2 5 2 | 5+6 2 |
| | 6 2 8 2 | 6 2 5 2 | 3+4 2 |
| | 8 2 6 2 | 7 2 6 2 | 5+6 2 |
| | 7 2 9 2 | 6 2 5 2 | 3+4 2 |
| | 9 2 7 2 | 7 2 6 2 | 5+6 2 |
| | 7 2 7 2 | 7 2 6 2 | 5+6 2 |
| | 7 2 9 2 | 6 2 6 2 | 3+4 2 |
| | 6 2 6 2 | 6 2 5 2 | 5+6 2 |
| | 9 2 9 2 | 7 2 5 2 | 3+4 2 |
| | 7 2 8 2 | 7 2 6 2 | 3+4 2 |
| | 6 2 7 2 | 6 2 5 2 | 5+6 2 |
| | 9 2 9 2 | 6 2 6 2 | 3+4 2 |

Continued on next column

**TABLE LXIV– continued**

| TQC | ICSD | NoE | SF |
|---|---|---|---|
| | 8 2 8 2 | 7 2 5 2 | 5+6 2 |
| | 9 2 9 2 | 7 2 5 2 | 3+4 2 |
| | 8 2 7 2 | 6 2 6 2 | 5+6 2 |
| | 8 2 6 2 | 6 2 5 2 | 5+6 2 |
| | 8 2 8 2 | 7 2 5 2 | 3+4 2 |
| | 6 2 6 2 | 6 2 6 2 | 3+4 2 |
| | 7 2 8 2 | 7 2 5 2 | 5+6 2 |
| 139.1.76 | 26276 | 48 | Ge1Se1U1 |
| | -GM -M | -P -X | -N |
| | 7 2 7 2 | 6 2 5 2 | 3+4 2 |
| | 9 2 9 2 | 7 2 6 2 | 5+6 2 |
| | 7 2 9 2 | 6 2 5 2 | 3+4 2 |
| | 9 2 7 2 | 7 2 6 2 | 5+6 2 |
| | 7 2 9 2 | 6 2 6 2 | 3+4 2 |
| | 9 2 7 2 | 7 2 5 2 | 5+6 2 |
| | 6 2 8 2 | 6 2 5 2 | 3+4 2 |
| | 8 2 6 2 | 7 2 6 2 | 5+6 2 |
| | 7 2 9 2 | 6 2 5 2 | 3+4 2 |
| | 9 2 7 2 | 7 2 6 2 | 5+6 2 |
| | 7 2 7 2 | 7 2 6 2 | 5+6 2 |
| | 7 2 9 2 | 6 2 6 2 | 3+4 2 |
| | 6 2 6 2 | 6 2 5 2 | 5+6 2 |
| | 9 2 9 2 | 7 2 5 2 | 3+4 2 |
| | 7 2 8 2 | 7 2 6 2 | 3+4 2 |
| | 6 2 7 2 | 6 2 5 2 | 5+6 2 |
| | 9 2 9 2 | 6 2 6 2 | 3+4 2 |
| | 8 2 7 2 | 7 2 5 2 | 5+6 2 |
| | 9 2 8 2 | 7 2 6 2 | 3+4 2 |
| | 8 2 9 2 | 6 2 5 2 | 5+6 2 |
| | 8 2 6 2 | 6 2 5 2 | 5+6 2 |
| | 8 2 8 2 | 7 2 5 2 | 3+4 2 |
| | 6 2 6 2 | 6 2 6 2 | 3+4 2 |
| | 7 2 8 2 | 7 2 5 2 | 5+6 2 |
| 139.1.77 | 89563 | 54 | Gd2O2Te1 |
| | -GM -M | -P -X | -N |
| | 7 2 9 2 | 7 2 5 2 | 5+6 2 |
| | 9 2 7 2 | 6 2 6 2 | 3+4 2 |
| | 7 2 9 2 | 7 2 5 2 | 5+6 2 |
| | 9 2 7 2 | 6 2 6 2 | 3+4 2 |
| | 7 2 9 2 | 7 2 5 2 | 5+6 2 |
| | 6 2 8 2 | 6 2 6 2 | 3+4 2 |
| | 9 2 7 2 | 7 2 5 2 | 5+6 2 |
| | 8 2 6 2 | 6 2 6 2 | 3+4 2 |
| | 8 2 6 2 | 7 2 6 2 | 5+6 2 |
| | 7 2 9 2 | 6 2 5 2 | 3+4 2 |
| | 7 2 7 2 | 6 2 5 2 | 3+4 2 |
| | 9 2 7 2 | 7 2 6 2 | 5+6 2 |
| | 7 2 9 2 | 6 2 5 2 | 5+6 2 |
| | 6 2 8 2 | 7 2 6 2 | 3+4 2 |
| | 6 2 8 2 | 6 2 6 2 | 5+6 2 |
| | 9 2 7 2 | 6 2 5 2 | 3+4 2 |
| | 8 2 6 2 | 7 2 6 2 | 3+4 2 |
| | 9 2 9 2 | 7 2 5 2 | 5+6 2 |
| | 9 2 9 2 | 6 2 6 2 | 5+6 2 |
| | 6 2 8 2 | 7 2 6 2 | 5+6 2 |
| | 8 2 8 2 | 7 2 5 2 | 3+4 2 |
| | 6 2 6 2 | 6 2 6 2 | 3+4 2 |
| | 7 2 7 2 | 6 2 5 2 | 5+6 2 |
| | 8 2 9 2 | 7 2 5 2 | 5+6 2 |
| | 8 2 6 2 | 6 2 6 2 | 3+4 2 |
| | 9 2 8 2 | 7 2 6 2 | 5+6 2 |

Continued on next column

**TABLE LXIV– continued**

| TQC | ICSD | NoE | SF |
|---|---|---|---|
| | 6 2 8 2 | 7 2 5 2 | 3+4 2 |
| 139.1.78 | 82466 | 110 | Gd2Na2O10Ti3 |
| | -GM -M | -P -X | -N |
| | 7 2 7 2 | 6 2 5 2 | 3+4 2 |
| | 9 2 9 2 | 7 2 6 2 | 5+6 2 |
| | 7 2 7 2 | 6 2 5 2 | 5+6 2 |
| | 9 2 9 2 | 7 2 6 2 | 3+4 2 |
| | 7 2 7 2 | 7 2 6 2 | 3+4 2 |
| | 9 2 9 2 | 6 2 5 2 | 5+6 2 |
| | 8 2 8 2 | 7 2 5 2 | 5+6 2 |
| | 6 2 6 2 | 6 2 6 2 | 3+4 2 |
| | 7 2 7 2 | 6 2 5 2 | 3+4 2 |
| | 9 2 9 2 | 7 2 6 2 | 5+6 2 |
| | 6 2 6 2 | 6 2 6 2 | 3+4 2 |
| | 7 2 8 2 | 7 2 6 2 | 5+6 2 |
| | 8 2 7 2 | 7 2 5 2 | 5+6 2 |
| | 6 2 6 2 | 6 2 5 2 | 3+4 2 |
| | 9 2 9 2 | 6 2 5 2 | 3+4 2 |
| | 7 2 7 2 | 6 2 6 2 | 5+6 2 |
| | 7 2 7 2 | 7 2 6 2 | 3+4 2 |
| | 9 2 9 2 | 7 2 6 2 | 5+6 2 |
| | 7 2 7 2 | 7 2 5 2 | 3+4 2 |
| | 9 2 9 2 | 6 2 5 2 | 3+4 2 |
| | 7 2 7 2 | 6 2 6 2 | 3+4 2 |
| | 9 2 9 2 | 7 2 6 2 | 5+6 2 |
| | 6 2 8 2 | 6 2 5 2 | 5+6 2 |
| | 8 2 6 2 | 7 2 5 2 | 3+4 2 |
| | 9 2 9 2 | 7 2 5 2 | 3+4 2 |
| | 8 2 8 2 | 7 2 5 2 | 3+4 2 |
| | 7 2 9 2 | 6 2 5 2 | 5+6 2 |
| | 9 2 9 2 | 6 2 6 2 | 5+6 2 |
| | 8 2 8 2 | 6 2 5 2 | 3+4 2 |
| | 6 2 6 2 | 7 2 6 2 | 5+6 2 |
| | 7 2 9 2 | 7 2 5 2 | 5+6 2 |
| | 6 2 7 2 | 6 2 6 2 | 3+4 2 |
| | 9 2 8 2 | 7 2 5 2 | 5+6 2 |
| | 9 2 7 2 | 6 2 5 2 | 3+4 2 |
| | 8 2 6 2 | 7 2 6 2 | 3+4 2 |
| | 7 2 7 2 | 6 2 5 2 | 5+6 2 |
| | 6 2 6 2 | 6 2 6 2 | 5+6 2 |
| | 7 2 7 2 | 7 2 6 2 | 3+4 2 |
| | 8 2 8 2 | 7 2 5 2 | 5+6 2 |
| | 8 2 8 2 | 6 2 5 2 | 5+6 2 |
| | 9 2 9 2 | 7 2 6 2 | 3+4 2 |
| | 7 2 7 2 | 7 2 6 2 | 5+6 2 |
| | 6 2 9 2 | 6 2 6 2 | 3+4 2 |
| | 9 2 6 2 | 6 2 5 2 | 3+4 2 |
| | 8 2 8 2 | 7 2 6 2 | 5+6 2 |
| | 9 2 9 2 | 6 2 5 2 | 5+6 2 |
| | 8 2 9 2 | 7 2 5 2 | 5+6 2 |
| | 9 2 8 2 | 6 2 5 2 | 3+4 2 |
| | 6 2 6 2 | 6 2 6 2 | 5+6 2 |
| | 7 2 7 2 | 7 2 6 2 | 3+4 2 |
| | 8 2 8 2 | 6 2 5 2 | 3+4 2 |
| | 6 2 6 2 | 7 2 6 2 | 5+6 2 |
| | 9 2 9 2 | 7 2 5 2 | 5+6 2 |
| | 8 2 8 2 | 6 2 5 2 | 3+4 2 |
| | 6 2 6 2 | 7 2 5 2 | 5+6 2 |
| 139.1.79 | 30844 | 38 | Gd1P2Pd2 |
| | -GM -M | -P -X | -N |
| | 7 2 7 2 | 6 2 5 2 | 3+4 2 |

Continued on next column

**TABLE LXIV– continued**

| TQC | ICSD | NoE | SF |
|---|---|---|---|
| | 9 2 9 2 | 7 2 6 2 | 5+6 2 |
| | 8 2 8 2 | 6 2 6 2 | 5+6 2 |
| | 9 2 9 2 | 7 2 6 2 | 5+6 2 |
| | 7 2 9 2 | 6 2 5 2 | 3+4 2 |
| | 6 2 7 2 | 7 2 5 2 | 3+4 2 |
| | 9 2 8 2 | 6 2 6 2 | 5+6 2 |
| | 7 2 6 2 | 7 2 6 2 | 5+6 2 |
| | 8 2 7 2 | 7 2 5 2 | 3+4 2 |
| | 9 2 7 2 | 6 2 6 2 | 3+4 2 |
| | 8 2 6 2 | 6 2 5 2 | 5+6 2 |
| | 7 2 9 2 | 7 2 6 2 | 5+6 2 |
| | 6 2 9 2 | 6 2 6 2 | 3+4 2 |
| | 8 2 8 2 | 7 2 5 2 | 5+6 2 |
| | 9 2 8 2 | 6 2 6 2 | 3+4 2 |
| | 8 2 8 2 | 7 2 6 2 | 5+6 2 |
| | 6 2 9 2 | 6 2 6 2 | 5+6 2 |
| | 9 2 8 2 | 6 2 5 2 | 5+6 2 |
| | 8 2 8 2 | 7 2 6 2 | 5+6 2 |
| 139.1.80 | 26823 | 102 | Gd1Mn12 |
| | -GM -M | -P -X | -N |
| | 7 2 7 2 | 6 2 5 2 | 3+4 2 |
| | 9 2 9 2 | 7 2 6 2 | 5+6 2 |
| | 8 2 8 2 | 7 2 6 2 | 5+6 2 |
| | 9 2 9 2 | 6 2 6 2 | 5+6 2 |
| | 7 2 6 2 | 7 2 6 2 | 5+6 2 |
| | 7 2 9 2 | 6 2 5 2 | 3+4 2 |
| | 6 2 8 2 | 7 2 6 2 | 3+4 2 |
| | 6 2 7 2 | 6 2 6 2 | 5+6 2 |
| | 9 2 8 2 | 7 2 5 2 | 5+6 2 |
| | 8 2 9 2 | 6 2 5 2 | 3+4 2 |
| | 7 2 8 2 | 7 2 5 2 | 3+4 2 |
| | 6 2 9 2 | 6 2 5 2 | 3+4 2 |
| | 6 2 8 2 | 7 2 5 2 | 3+4 2 |
| | 7 2 7 2 | 6 2 5 2 | 5+6 2 |
| | 7 2 9 2 | 7 2 6 2 | 5+6 2 |
| | 6 2 8 2 | 6 2 6 2 | 5+6 2 |
| | 7 2 9 2 | 7 2 6 2 | 3+4 2 |
| | 6 2 9 2 | 7 2 5 2 | 5+6 2 |
| | 8 2 8 2 | 6 2 6 2 | 3+4 2 |
| | 7 2 7 2 | 7 2 5 2 | 5+6 2 |
| | 6 2 6 2 | 6 2 5 2 | 3+4 2 |
| | 7 2 7 2 | 7 2 6 2 | 5+6 2 |
| | 9 2 9 2 | 6 2 6 2 | 5+6 2 |
| | 8 2 8 2 | 6 2 5 2 | 3+4 2 |
| | 6 2 9 2 | 6 2 6 2 | 5+6 2 |
| | 9 2 6 2 | 7 2 6 2 | 3+4 2 |
| | 7 2 8 2 | 7 2 6 2 | 3+4 2 |
| | 7 2 9 2 | 6 2 5 2 | 5+6 2 |
| | 6 2 6 2 | 7 2 5 2 | 5+6 2 |
| | 6 2 8 2 | 6 2 5 2 | 3+4 2 |
| | 7 2 7 2 | 7 2 6 2 | 3+4 2 |
| | 9 2 8 2 | 6 2 5 2 | 5+6 2 |
| | 8 2 9 2 | 7 2 5 2 | 3+4 2 |
| | 8 2 9 2 | 7 2 6 2 | 5+6 2 |
| | 9 2 6 2 | 6 2 5 2 | 5+6 2 |
| | 8 2 7 2 | 7 2 6 2 | 3+4 2 |
| | 9 2 6 2 | 6 2 6 2 | 3+4 2 |
| | 8 2 6 2 | 6 2 5 2 | 5+6 2 |
| | 6 2 7 2 | 7 2 6 2 | 3+4 2 |
| | 9 2 6 2 | 6 2 5 2 | 5+6 2 |
| | 8 2 7 2 | 7 2 6 2 | 5+6 2 |

Continued on next column

**TABLE LXIV– continued**

| TQC | ICSD | NoE | SF |
|---|---|---|---|
| | 8 2 8 2 | 6 2 6 2 | 3+4 2 |
| | 7 2 8 2 | 6 2 6 2 | 5+6 2 |
| | 9 2 9 2 | 7 2 5 2 | 3+4 2 |
| | 9 2 6 2 | 6 2 6 2 | 5+6 2 |
| | 8 2 9 2 | 7 2 6 2 | 5+6 2 |
| | 7 2 7 2 | 6 2 6 2 | 5+6 2 |
| | 6 2 8 2 | 7 2 6 2 | 5+6 2 |
| | 8 2 9 2 | 7 2 6 2 | 3+4 2 |
| | 9 2 7 2 | 6 2 5 2 | 5+6 2 |
| | 9 2 8 2 | 6 2 5 2 | 3+4 2 |
| 139.1.81 | 90392 | 52 | Gd1Ge1Ti1 |
| | -GM -M | -P -X | -N |
| | 7 2 9 2 | 6 2 5 2 | 3+4 2 |
| | 9 2 7 2 | 7 2 6 2 | 5+6 2 |
| | 7 2 9 2 | 6 2 5 2 | 3+4 2 |
| | 9 2 7 2 | 7 2 6 2 | 5+6 2 |
| | 7 2 9 2 | 6 2 6 2 | 3+4 2 |
| | 6 2 8 2 | 7 2 5 2 | 5+6 2 |
| | 9 2 7 2 | 6 2 6 2 | 3+4 2 |
| | 8 2 6 2 | 7 2 5 2 | 5+6 2 |
| | 7 2 7 2 | 6 2 5 2 | 3+4 2 |
| | 9 2 9 2 | 7 2 6 2 | 5+6 2 |
| | 7 2 7 2 | 6 2 6 2 | 5+6 2 |
| | 7 2 7 2 | 7 2 6 2 | 3+4 2 |
| | 7 2 7 2 | 6 2 5 2 | 3+4 2 |
| | 6 2 6 2 | 7 2 6 2 | 3+4 2 |
| | 9 2 9 2 | 7 2 5 2 | 5+6 2 |
| | 8 2 6 2 | 6 2 5 2 | 5+6 2 |
| | 6 2 6 2 | 6 2 6 2 | 5+6 2 |
| | 8 2 8 2 | 7 2 6 2 | 3+4 2 |
| | 6 2 7 2 | 6 2 6 2 | 5+6 2 |
| | 7 2 8 2 | 7 2 5 2 | 3+4 2 |
| | 9 2 6 2 | 6 2 5 2 | 5+6 2 |
| | 6 2 9 2 | 6 2 6 2 | 3+4 2 |
| | 9 2 6 2 | 7 2 5 2 | 5+6 2 |
| | 8 2 9 2 | 6 2 6 2 | 3+4 2 |
| | 6 2 7 2 | 7 2 6 2 | 5+6 2 |
| | 7 2 8 2 | 7 2 5 2 | 3+4 2 |
| 139.1.82 | 68209 | 36 | Fe2Si2Th1 |
| | -GM -M | -P -X | -N |
| | 7 2 7 2 | 6 2 5 2 | 3+4 2 |
| | 9 2 9 2 | 7 2 6 2 | 5+6 2 |
| | 9 2 9 2 | 6 2 6 2 | 5+6 2 |
| | 8 2 8 2 | 7 2 6 2 | 5+6 2 |
| | 7 2 7 2 | 7 2 5 2 | 5+6 2 |
| | 9 2 9 2 | 6 2 6 2 | 3+4 2 |
| | 7 2 7 2 | 7 2 5 2 | 3+4 2 |
| | 9 2 7 2 | 6 2 6 2 | 5+6 2 |
| | 8 2 6 2 | 7 2 5 2 | 3+4 2 |
| | 6 2 6 2 | 6 2 5 2 | 3+4 2 |
| | 7 2 8 2 | 7 2 5 2 | 5+6 2 |
| | 6 2 9 2 | 6 2 6 2 | 5+6 2 |
| | 9 2 9 2 | 7 2 5 2 | 3+4 2 |
| | 6 2 8 2 | 6 2 6 2 | 5+6 2 |
| | 8 2 8 2 | 7 2 6 2 | 3+4 2 |
| | 7 2 9 2 | 7 2 5 2 | 5+6 2 |
| | 7 2 8 2 | 6 2 6 2 | 5+6 2 |
| | 6 2 7 2 | 7 2 5 2 | 3+4 2 |
| 139.1.83 | 10467 | 36 | Fe2P2Sr1 |
| | -GM -M | -P -X | -N |
| | 7 2 7 2 | 6 2 5 2 | 3+4 2 |

Continued on next column

**TABLE LXIV– continued**

| TQC | ICSD | NoE | SF |
|---|---|---|---|
| | 9 2 9 2 | 7 2 6 2 | 5+6 2 |
| | 9 2 9 2 | 6 2 6 2 | 5+6 2 |
| | 8 2 8 2 | 7 2 6 2 | 5+6 2 |
| | 7 2 9 2 | 7 2 5 2 | 5+6 2 |
| | 9 2 7 2 | 6 2 6 2 | 3+4 2 |
| | 7 2 7 2 | 7 2 5 2 | 3+4 2 |
| | 9 2 9 2 | 6 2 6 2 | 5+6 2 |
| | 9 2 6 2 | 7 2 5 2 | 5+6 2 |
| | 8 2 7 2 | 7 2 5 2 | 3+4 2 |
| | 6 2 9 2 | 6 2 6 2 | 5+6 2 |
| | 7 2 8 2 | 6 2 5 2 | 3+4 2 |
| | 6 2 8 2 | 7 2 6 2 | 3+4 2 |
| | 7 2 6 2 | 6 2 5 2 | 5+6 2 |
| | 8 2 7 2 | 7 2 6 2 | 3+4 2 |
| | 6 2 8 2 | 6 2 6 2 | 5+6 2 |
| | 7 2 9 2 | 7 2 5 2 | 5+6 2 |
| | 6 2 8 2 | 6 2 6 2 | 3+4 2 |
| 139.1.84 | 74434 | 82 | Fe2O6Sr3 |
| | -GM -M | -P -X | -N |
| | 7 2 9 2 | 6 2 5 2 | 3+4 2 |
| | 9 2 7 2 | 7 2 6 2 | 5+6 2 |
| | 7 2 7 2 | 7 2 5 2 | 5+6 2 |
| | 9 2 9 2 | 6 2 6 2 | 3+4 2 |
| | 7 2 7 2 | 7 2 6 2 | 5+6 2 |
| | 7 2 9 2 | 7 2 5 2 | 3+4 2 |
| | 6 2 6 2 | 6 2 5 2 | 5+6 2 |
| | 9 2 8 2 | 6 2 6 2 | 3+4 2 |
| | 8 2 7 2 | 7 2 5 2 | 5+6 2 |
| | 7 2 9 2 | 6 2 6 2 | 3+4 2 |
| | 9 2 7 2 | 7 2 5 2 | 5+6 2 |
| | 9 2 9 2 | 6 2 6 2 | 3+4 2 |
| | 6 2 8 2 | 7 2 6 2 | 3+4 2 |
| | 8 2 6 2 | 6 2 5 2 | 5+6 2 |
| | 9 2 9 2 | 6 2 6 2 | 3+4 2 |
| | 7 2 7 2 | 7 2 5 2 | 5+6 2 |
| | 8 2 8 2 | 6 2 6 2 | 3+4 2 |
| | 9 2 9 2 | 7 2 6 2 | 3+4 2 |
| | 7 2 7 2 | 6 2 5 2 | 3+4 2 |
| | 6 2 6 2 | 7 2 6 2 | 5+6 2 |
| | 7 2 7 2 | 7 2 5 2 | 5+6 2 |
| | 9 2 9 2 | 7 2 5 2 | 3+4 2 |
| | 8 2 8 2 | 6 2 6 2 | 5+6 2 |
| | 7 2 9 2 | 6 2 6 2 | 5+6 2 |
| | 9 2 7 2 | 7 2 5 2 | 3+4 2 |
| | 6 2 6 2 | 6 2 5 2 | 3+4 2 |
| | 9 2 9 2 | 6 2 6 2 | 5+6 2 |
| | 8 2 8 2 | 7 2 5 2 | 3+4 2 |
| | 7 2 9 2 | 7 2 6 2 | 3+4 2 |
| | 6 2 8 2 | 6 2 6 2 | 5+6 2 |
| | 9 2 7 2 | 6 2 5 2 | 5+6 2 |
| | 8 2 6 2 | 7 2 6 2 | 3+4 2 |
| | 7 2 7 2 | 6 2 5 2 | 5+6 2 |
| | 6 2 6 2 | 7 2 6 2 | 5+6 2 |
| | 9 2 9 2 | 6 2 5 2 | 3+4 2 |
| | 8 2 8 2 | 7 2 6 2 | 3+4 2 |
| | 6 2 6 2 | 6 2 5 2 | 5+6 2 |
| | 8 2 8 2 | 7 2 6 2 | 5+6 2 |
| | 7 2 9 2 | 6 2 5 2 | 3+4 2 |
| | 6 2 8 2 | 7 2 5 2 | 5+6 2 |
| | 9 2 7 2 | 6 2 6 2 | 3+4 2 |
| 139.1.85 | 249015 | 82 | Fe2O6Sr3 |

Continued on next column

**TABLE LXIV– continued**

| TQC | ICSD | NoE | SF |
|---|---|---|---|
| | -GM -M | -P -X | -N |
| | 7 2 9 2 | 6 2 5 2 | 3+4 2 |
| | 9 2 7 2 | 7 2 6 2 | 5+6 2 |
| | 7 2 7 2 | 7 2 5 2 | 5+6 2 |
| | 9 2 9 2 | 6 2 6 2 | 3+4 2 |
| | 7 2 7 2 | 7 2 6 2 | 5+6 2 |
| | 7 2 9 2 | 7 2 5 2 | 3+4 2 |
| | 9 2 7 2 | 6 2 6 2 | 3+4 2 |
| | 6 2 6 2 | 6 2 5 2 | 5+6 2 |
| | 8 2 8 2 | 7 2 5 2 | 5+6 2 |
| | 7 2 9 2 | 6 2 6 2 | 3+4 2 |
| | 9 2 7 2 | 7 2 5 2 | 5+6 2 |
| | 9 2 9 2 | 7 2 6 2 | 3+4 2 |
| | 6 2 8 2 | 6 2 6 2 | 3+4 2 |
| | 8 2 6 2 | 6 2 5 2 | 5+6 2 |
| | 7 2 9 2 | 6 2 6 2 | 3+4 2 |
| | 9 2 7 2 | 7 2 5 2 | 5+6 2 |
| | 8 2 8 2 | 6 2 6 2 | 3+4 2 |
| | 9 2 9 2 | 7 2 6 2 | 3+4 2 |
| | 7 2 7 2 | 6 2 5 2 | 3+4 2 |
| | 6 2 6 2 | 7 2 6 2 | 5+6 2 |
| | 7 2 7 2 | 7 2 5 2 | 5+6 2 |
| | 9 2 9 2 | 7 2 5 2 | 3+4 2 |
| | 8 2 8 2 | 6 2 6 2 | 5+6 2 |
| | 7 2 9 2 | 6 2 6 2 | 3+4 2 |
| | 9 2 7 2 | 7 2 5 2 | 5+6 2 |
| | 6 2 9 2 | 6 2 6 2 | 3+4 2 |
| | 9 2 6 2 | 7 2 5 2 | 5+6 2 |
| | 7 2 9 2 | 6 2 5 2 | 3+4 2 |
| | 6 2 8 2 | 6 2 6 2 | 5+6 2 |
| | 9 2 7 2 | 7 2 6 2 | 3+4 2 |
| | 8 2 6 2 | 6 2 5 2 | 3+4 2 |
| | 8 2 8 2 | 7 2 6 2 | 5+6 2 |
| | 7 2 7 2 | 6 2 5 2 | 5+6 2 |
| | 6 2 6 2 | 7 2 6 2 | 5+6 2 |
| | 9 2 9 2 | 6 2 5 2 | 3+4 2 |
| | 8 2 8 2 | 7 2 6 2 | 3+4 2 |
| | 6 2 6 2 | 6 2 5 2 | 5+6 2 |
| | 8 2 8 2 | 7 2 6 2 | 5+6 2 |
| | 7 2 9 2 | 6 2 5 2 | 3+4 2 |
| | 6 2 8 2 | 7 2 5 2 | 3+4 2 |
| | 9 2 7 2 | 6 2 6 2 | 5+6 2 |
| 139.1.86 | 249014 | 82 | Fe2O6Sr3 |
| | -GM -M | -P -X | -N |
| | 7 2 9 2 | 6 2 5 2 | 3+4 2 |
| | 9 2 7 2 | 7 2 6 2 | 5+6 2 |
| | 7 2 7 2 | 7 2 5 2 | 5+6 2 |
| | 9 2 9 2 | 6 2 6 2 | 3+4 2 |
| | 7 2 7 2 | 7 2 6 2 | 5+6 2 |
| | 7 2 9 2 | 7 2 5 2 | 3+4 2 |
| | 6 2 6 2 | 6 2 5 2 | 5+6 2 |
| | 8 2 8 2 | 6 2 6 2 | 3+4 2 |
| | 9 2 7 2 | 7 2 5 2 | 5+6 2 |
| | 7 2 9 2 | 6 2 6 2 | 3+4 2 |
| | 9 2 7 2 | 7 2 5 2 | 5+6 2 |
| | 9 2 9 2 | 6 2 6 2 | 3+4 2 |
| | 6 2 8 2 | 7 2 6 2 | 3+4 2 |
| | 8 2 6 2 | 6 2 5 2 | 5+6 2 |
| | 9 2 9 2 | 6 2 6 2 | 3+4 2 |
| | 7 2 7 2 | 7 2 5 2 | 5+6 2 |
| | 8 2 8 2 | 6 2 6 2 | 3+4 2 |

Continued on next column

**TABLE LXIV– continued**

| TQC | ICSD | NoE | SF |
|---|---|---|---|
| | 9 2 9 2 | 7 2 6 2 | 3+4 2 |
| | 7 2 7 2 | 6 2 5 2 | 3+4 2 |
| | 6 2 6 2 | 7 2 6 2 | 5+6 2 |
| | 7 2 7 2 | 7 2 5 2 | 5+6 2 |
| | 9 2 9 2 | 7 2 5 2 | 3+4 2 |
| | 8 2 8 2 | 6 2 6 2 | 5+6 2 |
| | 7 2 9 2 | 6 2 6 2 | 3+4 2 |
| | 9 2 7 2 | 7 2 5 2 | 5+6 2 |
| | 6 2 6 2 | 6 2 5 2 | 5+6 2 |
| | 9 2 9 2 | 6 2 6 2 | 3+4 2 |
| | 8 2 8 2 | 7 2 6 2 | 3+4 2 |
| | 7 2 9 2 | 6 2 5 2 | 3+4 2 |
| | 6 2 8 2 | 7 2 6 2 | 5+6 2 |
| | 9 2 7 2 | 6 2 5 2 | 5+6 2 |
| | 8 2 6 2 | 7 2 6 2 | 5+6 2 |
| | 7 2 7 2 | 6 2 5 2 | 3+4 2 |
| | 6 2 6 2 | 7 2 6 2 | 3+4 2 |
| | 9 2 9 2 | 6 2 5 2 | 5+6 2 |
| | 8 2 8 2 | 7 2 6 2 | 3+4 2 |
| | 6 2 6 2 | 6 2 5 2 | 5+6 2 |
| | 8 2 8 2 | 7 2 6 2 | 3+4 2 |
| | 7 2 9 2 | 6 2 5 2 | 5+6 2 |
| | 6 2 8 2 | 7 2 5 2 | 5+6 2 |
| | 9 2 7 2 | 6 2 6 2 | 3+4 2 |
| 139.1.87 | 81746 | 38 | Fe2Ge2Nd1 |
| | -GM -M | -P -X | -N |
| | 7 2 7 2 | 6 2 5 2 | 3+4 2 |
| | 9 2 9 2 | 7 2 6 2 | 5+6 2 |
| | 9 2 9 2 | 6 2 6 2 | 5+6 2 |
| | 8 2 8 2 | 7 2 6 2 | 5+6 2 |
| | 7 2 7 2 | 7 2 5 2 | 5+6 2 |
| | 9 2 9 2 | 6 2 6 2 | 3+4 2 |
| | 7 2 7 2 | 7 2 5 2 | 3+4 2 |
| | 9 2 7 2 | 6 2 6 2 | 5+6 2 |
| | 8 2 6 2 | 7 2 5 2 | 3+4 2 |
| | 6 2 9 2 | 7 2 5 2 | 5+6 2 |
| | 9 2 8 2 | 6 2 6 2 | 5+6 2 |
| | 7 2 9 2 | 6 2 5 2 | 3+4 2 |
| | 6 2 6 2 | 7 2 6 2 | 3+4 2 |
| | 8 2 8 2 | 6 2 5 2 | 5+6 2 |
| | 7 2 9 2 | 7 2 6 2 | 5+6 2 |
| | 6 2 8 2 | 7 2 5 2 | 3+4 2 |
| | 7 2 8 2 | 6 2 6 2 | 5+6 2 |
| | 8 2 8 2 | 6 2 6 2 | 3+4 2 |
| | 6 2 7 2 | 7 2 5 2 | 5+6 2 |
| 139.1.88 | 74421 | 52 | Fe1O4Sr2 |
| | -GM -M | -P -X | -N |
| | 7 2 9 2 | 7 2 5 2 | 5+6 2 |
| | 9 2 7 2 | 6 2 6 2 | 3+4 2 |
| | 7 2 7 2 | 6 2 6 2 | 3+4 2 |
| | 6 2 6 2 | 7 2 6 2 | 5+6 2 |
| | 7 2 9 2 | 6 2 5 2 | 5+6 2 |
| | 9 2 7 2 | 7 2 6 2 | 3+4 2 |
| | 9 2 9 2 | 7 2 5 2 | 3+4 2 |
| | 7 2 7 2 | 6 2 6 2 | 5+6 2 |
| | 6 2 8 2 | 6 2 5 2 | 5+6 2 |
| | 8 2 9 2 | 7 2 5 2 | 3+4 2 |
| | 9 2 6 2 | 7 2 6 2 | 5+6 2 |
| | 7 2 7 2 | 6 2 6 2 | 3+4 2 |
| | 9 2 9 2 | 6 2 5 2 | 3+4 2 |
| | 8 2 8 2 | 7 2 5 2 | 3+4 2 |
| | 7 2 7 2 | 7 2 5 2 | 5+6 2 |
| | 9 2 9 2 | 7 2 5 2 | 3+4 2 |
| | 7 2 7 2 | 6 2 5 2 | 3+4 2 |
| | 8 2 8 2 | 6 2 5 2 | 5+6 2 |
| | 6 2 6 2 | 7 2 6 2 | 3+4 2 |
| | 9 2 9 2 | 6 2 6 2 | 5+6 2 |
| | 8 2 8 2 | 7 2 5 2 | 5+6 2 |
| | 6 2 9 2 | 6 2 6 2 | 5+6 2 |
| | 9 2 6 2 | 6 2 5 2 | 5+6 2 |
| | 9 2 9 2 | 7 2 5 2 | 3+4 2 |
| | 8 2 8 2 | 7 2 5 2 | 3+4 2 |
| | 6 2 6 2 | 6 2 5 2 | 3+4 2 |
| 139.1.89 | 74420 | 52 | Fe1O4Sr2 |
| | -GM -M | -P -X | -N |
| | 7 2 9 2 | 7 2 5 2 | 5+6 2 |
| | 9 2 7 2 | 6 2 6 2 | 3+4 2 |
| | 7 2 7 2 | 6 2 6 2 | 3+4 2 |
| | 6 2 6 2 | 7 2 6 2 | 5+6 2 |
| | 7 2 9 2 | 6 2 5 2 | 5+6 2 |
| | 9 2 7 2 | 7 2 6 2 | 3+4 2 |
| | 9 2 9 2 | 7 2 5 2 | 3+4 2 |
| | 7 2 7 2 | 6 2 6 2 | 5+6 2 |
| | 6 2 8 2 | 6 2 5 2 | 3+4 2 |
| | 8 2 9 2 | 7 2 5 2 | 5+6 2 |
| | 9 2 6 2 | 7 2 6 2 | 5+6 2 |
| | 7 2 7 2 | 6 2 6 2 | 3+4 2 |
| | 9 2 9 2 | 6 2 5 2 | 3+4 2 |
| | 8 2 8 2 | 7 2 5 2 | 3+4 2 |
| | 7 2 7 2 | 7 2 5 2 | 5+6 2 |
| | 9 2 9 2 | 7 2 5 2 | 3+4 2 |
| | 7 2 7 2 | 6 2 5 2 | 3+4 2 |
| | 8 2 6 2 | 6 2 5 2 | 5+6 2 |
| | 6 2 8 2 | 7 2 6 2 | 3+4 2 |
| | 9 2 9 2 | 6 2 6 2 | 5+6 2 |
| | 8 2 8 2 | 7 2 5 2 | 5+6 2 |
| | 6 2 9 2 | 6 2 6 2 | 5+6 2 |
| | 9 2 6 2 | 6 2 5 2 | 5+6 2 |
| | 9 2 9 2 | 7 2 5 2 | 3+4 2 |
| | 8 2 8 2 | 7 2 5 2 | 3+4 2 |
| | 6 2 6 2 | 6 2 5 2 | 3+4 2 |
| 139.1.90 | 74419 | 52 | Fe1O4Sr2 |
| | -GM -M | -P -X | -N |
| | 7 2 9 2 | 7 2 5 2 | 5+6 2 |
| | 9 2 7 2 | 6 2 6 2 | 3+4 2 |
| | 7 2 7 2 | 6 2 6 2 | 3+4 2 |
| | 6 2 6 2 | 7 2 6 2 | 5+6 2 |
| | 7 2 9 2 | 6 2 5 2 | 5+6 2 |
| | 9 2 7 2 | 7 2 6 2 | 3+4 2 |
| | 9 2 9 2 | 7 2 5 2 | 3+4 2 |
| | 7 2 7 2 | 6 2 6 2 | 5+6 2 |
| | 6 2 8 2 | 6 2 5 2 | 5+6 2 |
| | 8 2 9 2 | 7 2 5 2 | 3+4 2 |
| | 9 2 6 2 | 7 2 6 2 | 5+6 2 |
| | 7 2 7 2 | 6 2 6 2 | 3+4 2 |
| | 9 2 9 2 | 6 2 5 2 | 3+4 2 |
| | 8 2 8 2 | 7 2 5 2 | 3+4 2 |
| | 7 2 7 2 | 7 2 5 2 | 5+6 2 |
| | 9 2 9 2 | 7 2 5 2 | 3+4 2 |
| | 7 2 7 2 | 6 2 5 2 | 3+4 2 |
| | 8 2 8 2 | 6 2 5 2 | 5+6 2 |
| | 6 2 6 2 | 7 2 6 2 | 3+4 2 |

Continued on next column

Continued on next column

**TABLE LXIV– continued**

| TQC | ICSD | NoE | SF |
|---|---|---|---|
| | 9 2 9 2 | 6 2 6 2 | 5+6 2 |
| | 8 2 8 2 | 7 2 5 2 | 5+6 2 |
| | 6 2 9 2 | 6 2 6 2 | 5+6 2 |
| | 9 2 6 2 | 6 2 5 2 | 5+6 2 |
| | 9 2 9 2 | 7 2 5 2 | 3+4 2 |
| | 8 2 8 2 | 7 2 5 2 | 3+4 2 |
| | 6 2 6 2 | 6 2 5 2 | 3+4 2 |
| 139.1.91 | 69849 | 52 | Fe1O4Sr2 |
| | -GM -M | -P -X | -N |
| | 7 2 9 2 | 7 2 5 2 | 5+6 2 |
| | 9 2 7 2 | 6 2 6 2 | 3+4 2 |
| | 7 2 7 2 | 6 2 6 2 | 3+4 2 |
| | 6 2 6 2 | 7 2 6 2 | 5+6 2 |
| | 7 2 9 2 | 6 2 5 2 | 5+6 2 |
| | 9 2 7 2 | 7 2 6 2 | 3+4 2 |
| | 9 2 9 2 | 7 2 5 2 | 3+4 2 |
| | 7 2 7 2 | 6 2 6 2 | 5+6 2 |
| | 6 2 8 2 | 6 2 5 2 | 5+6 2 |
| | 8 2 9 2 | 7 2 5 2 | 3+4 2 |
| | 9 2 6 2 | 7 2 6 2 | 5+6 2 |
| | 7 2 7 2 | 6 2 6 2 | 3+4 2 |
| | 9 2 9 2 | 6 2 5 2 | 3+4 2 |
| | 8 2 8 2 | 7 2 5 2 | 3+4 2 |
| | 7 2 7 2 | 7 2 5 2 | 5+6 2 |
| | 9 2 9 2 | 7 2 5 2 | 3+4 2 |
| | 7 2 7 2 | 6 2 5 2 | 3+4 2 |
| | 8 2 8 2 | 6 2 5 2 | 5+6 2 |
| | 6 2 6 2 | 7 2 6 2 | 3+4 2 |
| | 9 2 9 2 | 6 2 6 2 | 5+6 2 |
| | 8 2 8 2 | 7 2 5 2 | 5+6 2 |
| | 6 2 9 2 | 6 2 6 2 | 5+6 2 |
| | 9 2 6 2 | 6 2 5 2 | 5+6 2 |
| | 9 2 9 2 | 7 2 5 2 | 3+4 2 |
| | 8 2 8 2 | 7 2 5 2 | 3+4 2 |
| | 6 2 6 2 | 6 2 5 2 | 3+4 2 |
| 139.1.92 | 69682 | 56 | F4Ni1Rb2 |
| | -GM -M | -P -X | -N |
| | 7 2 9 2 | 7 2 5 2 | 5+6 2 |
| | 9 2 7 2 | 6 2 6 2 | 3+4 2 |
| | 7 2 7 2 | 7 2 6 2 | 3+4 2 |
| | 6 2 6 2 | 6 2 6 2 | 5+6 2 |
| | 9 2 7 2 | 6 2 5 2 | 5+6 2 |
| | 7 2 9 2 | 7 2 6 2 | 3+4 2 |
| | 7 2 9 2 | 7 2 5 2 | 5+6 2 |
| | 9 2 7 2 | 6 2 6 2 | 3+4 2 |
| | 7 2 9 2 | 6 2 5 2 | 5+6 2 |
| | 6 2 8 2 | 7 2 5 2 | 3+4 2 |
| | 8 2 7 2 | 7 2 6 2 | 3+4 2 |
| | 9 2 6 2 | 6 2 6 2 | 5+6 2 |
| | 9 2 9 2 | 6 2 5 2 | 3+4 2 |
| | 8 2 8 2 | 7 2 5 2 | 5+6 2 |
| | 7 2 7 2 | 7 2 5 2 | 3+4 2 |
| | 9 2 9 2 | 6 2 5 2 | 3+4 2 |
| | 8 2 8 2 | 7 2 5 2 | 5+6 2 |
| | 9 2 9 2 | 6 2 5 2 | 3+4 2 |
| | 8 2 9 2 | 7 2 6 2 | 5+6 2 |
| | 7 2 7 2 | 6 2 6 2 | 3+4 2 |
| | 6 2 8 2 | 7 2 6 2 | 5+6 2 |
| | 9 2 6 2 | 6 2 5 2 | 5+6 2 |
| | 9 2 9 2 | 6 2 5 2 | 5+6 2 |
| | 8 2 8 2 | 7 2 5 2 | 3+4 2 |

Continued on next column

**TABLE LXIV– continued**

| TQC | ICSD | NoE | SF |
|---|---|---|---|
| | 6 2 6 2 | 6 2 5 2 | 3+4 2 |
| | 7 2 7 2 | 7 2 5 2 | 3+4 2 |
| | 6 2 6 2 | 7 2 5 2 | 3+4 2 |
| | 6 2 6 2 | 6 2 5 2 | 3+4 2 |
| 139.1.93 | 73450 | 56 | F4K2Ni1 |
| | -GM -M | -P -X | -N |
| | 7 2 9 2 | 7 2 5 2 | 5+6 2 |
| | 9 2 7 2 | 6 2 6 2 | 3+4 2 |
| | 7 2 7 2 | 7 2 6 2 | 3+4 2 |
| | 6 2 6 2 | 6 2 6 2 | 5+6 2 |
| | 7 2 9 2 | 6 2 5 2 | 3+4 2 |
| | 9 2 7 2 | 7 2 6 2 | 5+6 2 |
| | 7 2 9 2 | 7 2 5 2 | 5+6 2 |
| | 9 2 7 2 | 6 2 6 2 | 3+4 2 |
| | 7 2 9 2 | 7 2 5 2 | 5+6 2 |
| | 6 2 8 2 | 6 2 5 2 | 3+4 2 |
| | 8 2 7 2 | 7 2 6 2 | 3+4 2 |
| | 9 2 6 2 | 6 2 6 2 | 5+6 2 |
| | 8 2 8 2 | 6 2 5 2 | 3+4 2 |
| | 9 2 9 2 | 7 2 5 2 | 5+6 2 |
| | 9 2 7 2 | 7 2 5 2 | 3+4 2 |
| | 7 2 9 2 | 6 2 5 2 | 3+4 2 |
| | 8 2 8 2 | 6 2 5 2 | 5+6 2 |
| | 9 2 9 2 | 7 2 5 2 | 3+4 2 |
| | 8 2 8 2 | 7 2 6 2 | 5+6 2 |
| | 7 2 9 2 | 7 2 6 2 | 3+4 2 |
| | 6 2 7 2 | 6 2 6 2 | 5+6 2 |
| | 9 2 6 2 | 6 2 5 2 | 5+6 2 |
| | 9 2 9 2 | 6 2 5 2 | 5+6 2 |
| | 8 2 8 2 | 7 2 5 2 | 3+4 2 |
| | 6 2 6 2 | 6 2 5 2 | 3+4 2 |
| | 7 2 7 2 | 7 2 5 2 | 3+4 2 |
| | 6 2 6 2 | 7 2 5 2 | 3+4 2 |
| | 6 2 6 2 | 6 2 5 2 | 3+4 2 |
| 139.1.94 | 33520 | 56 | F4K2Ni1 |
| | -GM -M | -P -X | -N |
| | 7 2 9 2 | 7 2 5 2 | 5+6 2 |
| | 9 2 7 2 | 6 2 6 2 | 3+4 2 |
| | 7 2 7 2 | 7 2 6 2 | 3+4 2 |
| | 6 2 6 2 | 6 2 6 2 | 5+6 2 |
| | 7 2 9 2 | 6 2 5 2 | 3+4 2 |
| | 9 2 7 2 | 7 2 6 2 | 5+6 2 |
| | 7 2 9 2 | 7 2 5 2 | 5+6 2 |
| | 9 2 7 2 | 6 2 6 2 | 3+4 2 |
| | 7 2 9 2 | 7 2 5 2 | 5+6 2 |
| | 6 2 8 2 | 6 2 5 2 | 3+4 2 |
| | 8 2 7 2 | 7 2 6 2 | 3+4 2 |
| | 9 2 6 2 | 6 2 6 2 | 5+6 2 |
| | 8 2 8 2 | 6 2 5 2 | 3+4 2 |
| | 9 2 9 2 | 7 2 5 2 | 5+6 2 |
| | 9 2 7 2 | 7 2 5 2 | 3+4 2 |
| | 7 2 9 2 | 6 2 5 2 | 3+4 2 |
| | 8 2 8 2 | 6 2 5 2 | 5+6 2 |
| | 9 2 9 2 | 7 2 5 2 | 3+4 2 |
| | 8 2 8 2 | 7 2 6 2 | 5+6 2 |
| | 7 2 9 2 | 7 2 6 2 | 3+4 2 |
| | 6 2 7 2 | 6 2 6 2 | 5+6 2 |
| | 9 2 6 2 | 6 2 5 2 | 5+6 2 |
| | 9 2 9 2 | 6 2 5 2 | 5+6 2 |
| | 8 2 8 2 | 7 2 5 2 | 3+4 2 |
| | 6 2 6 2 | 6 2 5 2 | 3+4 2 |

Continued on next column

**TABLE LXIV– continued**

| TQC | ICSD | NoE | SF |
|---|---|---|---|
| | 7 2 7 2 | 7 2 5 2 | 3+4 2 |
| | 6 2 6 2 | 7 2 5 2 | 3+4 2 |
| | 6 2 6 2 | 6 2 5 2 | 3+4 2 |
| 139.1.95 | 167014 | 58 | F2O1Sb2Sr2Ti2 |
| | -GM -M | -P -X | -N |
| | 7 2 9 2 | 6 2 5 2 | 3+4 2 |
| | 9 2 7 2 | 7 2 6 2 | 5+6 2 |
| | 7 2 9 2 | 6 2 5 2 | 3+4 2 |
| | 8 2 6 2 | 7 2 6 2 | 5+6 2 |
| | 7 2 7 2 | 7 2 5 2 | 5+6 2 |
| | 7 2 9 2 | 6 2 5 2 | 3+4 2 |
| | 9 2 7 2 | 7 2 6 2 | 5+6 2 |
| | 6 2 8 2 | 6 2 5 2 | 3+4 2 |
| | 9 2 7 2 | 7 2 6 2 | 5+6 2 |
| | 8 2 6 2 | 6 2 5 2 | 3+4 2 |
| | 7 2 9 2 | 7 2 6 2 | 5+6 2 |
| | 7 2 7 2 | 6 2 5 2 | 3+4 2 |
| | 9 2 9 2 | 7 2 6 2 | 5+6 2 |
| | 7 2 7 2 | 6 2 6 2 | 3+4 2 |
| | 6 2 9 2 | 7 2 6 2 | 5+6 2 |
| | 9 2 8 2 | 6 2 5 2 | 3+4 2 |
| | 6 2 8 2 | 7 2 6 2 | 3+4 2 |
| | 9 2 7 2 | 6 2 5 2 | 5+6 2 |
| | 8 2 6 2 | 7 2 6 2 | 3+4 2 |
| | 9 2 9 2 | 6 2 5 2 | 5+6 2 |
| | 8 2 8 2 | 7 2 6 2 | 3+4 2 |
| | 9 2 9 2 | 6 2 6 2 | 3+4 2 |
| | 7 2 7 2 | 7 2 6 2 | 5+6 2 |
| | 7 2 7 2 | 6 2 6 2 | 3+4 2 |
| | 6 2 6 2 | 6 2 5 2 | 3+4 2 |
| | 6 2 9 2 | 7 2 5 2 | 5+6 2 |
| | 9 2 6 2 | 6 2 5 2 | 3+4 2 |
| | 7 2 7 2 | 7 2 6 2 | 5+6 2 |
| | 9 2 9 2 | 6 2 6 2 | 3+4 2 |
| 139.1.96 | 426633 | 56 | Er2Mg2Ru1 |
| | -GM -M | -P -X | -N |
| | 7 2 7 2 | 7 2 5 2 | 5+6 2 |
| | 9 2 9 2 | 6 2 6 2 | 3+4 2 |
| | 7 2 7 2 | 7 2 5 2 | 5+6 2 |
| | 9 2 9 2 | 6 2 6 2 | 3+4 2 |
| | 7 2 7 2 | 7 2 6 2 | 5+6 2 |
| | 9 2 9 2 | 6 2 5 2 | 3+4 2 |
| | 8 2 8 2 | 7 2 5 2 | 5+6 2 |
| | 6 2 6 2 | 6 2 6 2 | 3+4 2 |
| | 7 2 9 2 | 6 2 5 2 | 3+4 2 |
| | 7 2 7 2 | 7 2 5 2 | 3+4 2 |
| | 9 2 6 2 | 7 2 5 2 | 5+6 2 |
| | 6 2 7 2 | 6 2 5 2 | 3+4 2 |
| | 7 2 9 2 | 7 2 6 2 | 3+4 2 |
| | 6 2 7 2 | 6 2 5 2 | 5+6 2 |
| | 6 2 6 2 | 6 2 6 2 | 3+4 2 |
| | 7 2 6 2 | 7 2 5 2 | 3+4 2 |
| | 8 2 8 2 | 7 2 6 2 | 5+6 2 |
| | 8 2 8 2 | 6 2 5 2 | 5+6 2 |
| | 9 2 9 2 | 7 2 6 2 | 3+4 2 |
| | 7 2 7 2 | 7 2 6 2 | 5+6 2 |
| | 6 2 7 2 | 6 2 5 2 | 3+4 2 |
| | 9 2 6 2 | 7 2 6 2 | 3+4 2 |
| | 6 2 6 2 | 6 2 5 2 | 3+4 2 |
| | 7 2 9 2 | 6 2 6 2 | 5+6 2 |
| | 9 2 8 2 | 7 2 5 2 | 5+6 2 |

Continued on next column

**TABLE LXIV– continued**

| TQC | ICSD | NoE | SF |
|---|---|---|---|
| | 8 2 7 2 | 6 2 6 2 | 3+4 2 |
| | 8 2 8 2 | 7 2 5 2 | 5+6 2 |
| | 9 2 9 2 | 7 2 6 2 | 5+6 2 |
| 139.1.97 | 60667 | 80 | Er1O2S3U2 |
| | -GM -M | -P -X | -N |
| | 7 2 7 2 | 7 2 5 2 | 5+6 2 |
| | 7 2 9 2 | 6 2 5 2 | 3+4 2 |
| | 9 2 7 2 | 7 2 6 2 | 5+6 2 |
| | 9 2 9 2 | 6 2 6 2 | 3+4 2 |
| | 7 2 9 2 | 6 2 5 2 | 3+4 2 |
| | 9 2 7 2 | 7 2 6 2 | 5+6 2 |
| | 7 2 9 2 | 7 2 6 2 | 3+4 2 |
| | 9 2 9 2 | 6 2 6 2 | 3+4 2 |
| | 8 2 8 2 | 6 2 5 2 | 3+4 2 |
| | 8 2 6 2 | 7 2 6 2 | 5+6 2 |
| | 6 2 8 2 | 6 2 5 2 | 5+6 2 |
| | 9 2 7 2 | 7 2 6 2 | 3+4 2 |
| | 8 2 6 2 | 6 2 5 2 | 5+6 2 |
| | 7 2 9 2 | 7 2 6 2 | 3+4 2 |
| | 7 2 7 2 | 7 2 5 2 | 5+6 2 |
| | 9 2 7 2 | 6 2 6 2 | 3+4 2 |
| | 7 2 9 2 | 6 2 5 2 | 3+4 2 |
| | 9 2 7 2 | 7 2 5 2 | 5+6 2 |
| | 7 2 9 2 | 6 2 6 2 | 3+4 2 |
| | 6 2 8 2 | 6 2 5 2 | 5+6 2 |
| | 6 2 8 2 | 7 2 6 2 | 3+4 2 |
| | 9 2 7 2 | 6 2 5 2 | 5+6 2 |
| | 8 2 6 2 | 7 2 6 2 | 3+4 2 |
| | 7 2 9 2 | 7 2 5 2 | 5+6 2 |
| | 9 2 7 2 | 6 2 6 2 | 5+6 2 |
| | 7 2 9 2 | 6 2 6 2 | 3+4 2 |
| | 6 2 8 2 | 7 2 5 2 | 5+6 2 |
| | 9 2 7 2 | 7 2 5 2 | 5+6 2 |
| | 8 2 9 2 | 6 2 6 2 | 5+6 2 |
| | 9 2 9 2 | 6 2 6 2 | 3+4 2 |
| | 9 2 8 2 | 7 2 6 2 | 3+4 2 |
| | 8 2 6 2 | 7 2 6 2 | 5+6 2 |
| | 8 2 8 2 | 7 2 6 2 | 3+4 2 |
| | 9 2 9 2 | 6 2 6 2 | 3+4 2 |
| | 8 2 8 2 | 7 2 6 2 | 3+4 2 |
| | 8 2 6 2 | 6 2 5 2 | 3+4 2 |
| | 6 2 8 2 | 7 2 6 2 | 5+6 2 |
| | 6 2 8 2 | 6 2 6 2 | 3+4 2 |
| | 8 2 8 2 | 6 2 6 2 | 3+4 2 |
| | 9 2 9 2 | 7 2 6 2 | 3+4 2 |
| 139.1.98 | 245051 | 50 | Er1Ni2Si2 |
| | -GM -M | -P -X | -N |
| | 7 2 7 2 | 6 2 5 2 | 3+4 2 |
| | 9 2 9 2 | 7 2 6 2 | 5+6 2 |
| | 9 2 9 2 | 6 2 6 2 | 5+6 2 |
| | 8 2 8 2 | 7 2 6 2 | 5+6 2 |
| | 7 2 7 2 | 7 2 5 2 | 5+6 2 |
| | 9 2 9 2 | 6 2 6 2 | 3+4 2 |
| | 7 2 7 2 | 7 2 5 2 | 3+4 2 |
| | 9 2 6 2 | 6 2 6 2 | 5+6 2 |
| | 8 2 7 2 | 6 2 5 2 | 3+4 2 |
| | 6 2 8 2 | 7 2 6 2 | 5+6 2 |
| | 9 2 9 2 | 7 2 5 2 | 5+6 2 |
| | 7 2 9 2 | 6 2 6 2 | 3+4 2 |
| | 6 2 8 2 | 7 2 5 2 | 3+4 2 |
| | 8 2 6 2 | 7 2 6 2 | 5+6 2 |

Continued on next column

**TABLE LXIV– continued**

| TQC | ICSD | NoE | SF |
|---|---|---|---|
| | 7 2 8 2 | 6 2 5 2 | 3+4 2 |
| | 6 2 9 2 | 6 2 6 2 | 5+6 2 |
| | 8 2 8 2 | 7 2 5 2 | 5+6 2 |
| | 9 2 8 2 | 7 2 6 2 | 3+4 2 |
| | 8 2 7 2 | 6 2 6 2 | 5+6 2 |
| | 7 2 9 2 | 7 2 6 2 | 5+6 2 |
| | 6 2 8 2 | 6 2 6 2 | 5+6 2 |
| | 9 2 9 2 | 6 2 5 2 | 5+6 2 |
| | 9 2 8 2 | 6 2 6 2 | 5+6 2 |
| | 8 2 8 2 | 7 2 5 2 | 5+6 2 |
| | 8 2 9 2 | 6 2 6 2 | 3+4 2 |
| 139.1.99 | 109428 | 50 | Er1Ge2Ni2 |
| | -GM -M | -P -X | -N |
| | 7 2 7 2 | 6 2 5 2 | 3+4 2 |
| | 9 2 9 2 | 7 2 6 2 | 5+6 2 |
| | 9 2 9 2 | 6 2 6 2 | 5+6 2 |
| | 8 2 8 2 | 7 2 6 2 | 5+6 2 |
| | 7 2 7 2 | 7 2 5 2 | 5+6 2 |
| | 9 2 9 2 | 6 2 6 2 | 3+4 2 |
| | 7 2 7 2 | 7 2 6 2 | 3+4 2 |
| | 9 2 6 2 | 6 2 5 2 | 5+6 2 |
| | 8 2 7 2 | 7 2 5 2 | 3+4 2 |
| | 6 2 8 2 | 6 2 6 2 | 5+6 2 |
| | 9 2 9 2 | 7 2 5 2 | 5+6 2 |
| | 7 2 9 2 | 6 2 5 2 | 3+4 2 |
| | 6 2 8 2 | 7 2 6 2 | 3+4 2 |
| | 8 2 6 2 | 7 2 6 2 | 5+6 2 |
| | 7 2 8 2 | 6 2 5 2 | 3+4 2 |
| | 6 2 9 2 | 6 2 6 2 | 5+6 2 |
| | 8 2 8 2 | 7 2 5 2 | 5+6 2 |
| | 9 2 8 2 | 7 2 6 2 | 3+4 2 |
| | 7 2 9 2 | 6 2 6 2 | 5+6 2 |
| | 8 2 7 2 | 7 2 6 2 | 5+6 2 |
| | 6 2 8 2 | 6 2 6 2 | 5+6 2 |
| | 9 2 8 2 | 6 2 5 2 | 5+6 2 |
| | 9 2 9 2 | 6 2 6 2 | 5+6 2 |
| | 8 2 8 2 | 7 2 5 2 | 5+6 2 |
| | 8 2 9 2 | 6 2 6 2 | 5+6 2 |
| 139.1.100 | 30846 | 40 | Dy1P2Pd2 |
| | -GM -M | -P -X | -N |
| | 7 2 7 2 | 6 2 5 2 | 3+4 2 |
| | 9 2 9 2 | 7 2 6 2 | 5+6 2 |
| | 8 2 8 2 | 6 2 6 2 | 5+6 2 |
| | 9 2 9 2 | 7 2 6 2 | 5+6 2 |
| | 7 2 9 2 | 6 2 5 2 | 3+4 2 |
| | 6 2 7 2 | 7 2 5 2 | 3+4 2 |
| | 9 2 8 2 | 6 2 6 2 | 5+6 2 |
| | 7 2 6 2 | 7 2 6 2 | 5+6 2 |
| | 8 2 7 2 | 7 2 5 2 | 3+4 2 |
| | 9 2 7 2 | 6 2 6 2 | 3+4 2 |
| | 8 2 9 2 | 6 2 5 2 | 5+6 2 |
| | 7 2 6 2 | 7 2 6 2 | 5+6 2 |
| | 6 2 9 2 | 6 2 6 2 | 3+4 2 |
| | 8 2 8 2 | 7 2 5 2 | 5+6 2 |
| | 9 2 8 2 | 6 2 6 2 | 3+4 2 |
| | 8 2 9 2 | 7 2 6 2 | 5+6 2 |
| | 6 2 8 2 | 6 2 6 2 | 5+6 2 |
| | 9 2 8 2 | 6 2 5 2 | 5+6 2 |
| | 8 2 8 2 | 7 2 6 2 | 5+6 2 |
| | 9 2 9 2 | 7 2 6 2 | 5+6 2 |
| 139.1.101 | 88184 | 50 | Dy1Ni2P2 |

Continued on next column

**TABLE LXIV– continued**

| TQC | ICSD | NoE | SF |
|---|---|---|---|
| | -GM -M | -P -X | -N |
| | 7 2 7 2 | 6 2 5 2 | 3+4 2 |
| | 9 2 9 2 | 7 2 6 2 | 5+6 2 |
| | 9 2 9 2 | 6 2 6 2 | 5+6 2 |
| | 8 2 8 2 | 7 2 6 2 | 5+6 2 |
| | 7 2 7 2 | 7 2 5 2 | 5+6 2 |
| | 9 2 9 2 | 6 2 6 2 | 3+4 2 |
| | 7 2 7 2 | 7 2 5 2 | 3+4 2 |
| | 9 2 6 2 | 6 2 6 2 | 5+6 2 |
| | 8 2 7 2 | 7 2 5 2 | 3+4 2 |
| | 6 2 9 2 | 7 2 6 2 | 5+6 2 |
| | 7 2 8 2 | 6 2 5 2 | 5+6 2 |
| | 6 2 9 2 | 6 2 5 2 | 3+4 2 |
| | 9 2 8 2 | 7 2 6 2 | 3+4 2 |
| | 8 2 6 2 | 6 2 6 2 | 5+6 2 |
| | 7 2 8 2 | 7 2 5 2 | 3+4 2 |
| | 6 2 9 2 | 6 2 5 2 | 5+6 2 |
| | 7 2 8 2 | 7 2 6 2 | 5+6 2 |
| | 6 2 7 2 | 7 2 5 2 | 3+4 2 |
| | 8 2 7 2 | 6 2 6 2 | 5+6 2 |
| | 9 2 8 2 | 7 2 6 2 | 5+6 2 |
| | 9 2 9 2 | 6 2 6 2 | 3+4 2 |
| | 8 2 8 2 | 6 2 6 2 | 5+6 2 |
| | 6 2 9 2 | 6 2 6 2 | 5+6 2 |
| | 9 2 8 2 | 7 2 5 2 | 5+6 2 |
| | 8 2 8 2 | 6 2 6 2 | 5+6 2 |
| 139.1.102 | 88183 | 50 | Dy1Ni2P2 |
| | -GM -M | -P -X | -N |
| | 7 2 7 2 | 6 2 5 2 | 3+4 2 |
| | 9 2 9 2 | 7 2 6 2 | 5+6 2 |
| | 9 2 9 2 | 6 2 6 2 | 5+6 2 |
| | 8 2 8 2 | 7 2 6 2 | 5+6 2 |
| | 7 2 7 2 | 7 2 5 2 | 5+6 2 |
| | 9 2 9 2 | 6 2 6 2 | 3+4 2 |
| | 7 2 7 2 | 7 2 5 2 | 3+4 2 |
| | 9 2 6 2 | 6 2 6 2 | 5+6 2 |
| | 8 2 7 2 | 7 2 5 2 | 3+4 2 |
| | 6 2 9 2 | 7 2 6 2 | 5+6 2 |
| | 7 2 8 2 | 6 2 5 2 | 5+6 2 |
| | 6 2 9 2 | 6 2 5 2 | 3+4 2 |
| | 9 2 8 2 | 7 2 6 2 | 3+4 2 |
| | 8 2 6 2 | 6 2 6 2 | 5+6 2 |
| | 7 2 8 2 | 7 2 5 2 | 3+4 2 |
| | 6 2 9 2 | 6 2 5 2 | 5+6 2 |
| | 7 2 8 2 | 7 2 6 2 | 5+6 2 |
| | 6 2 7 2 | 7 2 5 2 | 3+4 2 |
| | 8 2 8 2 | 6 2 6 2 | 5+6 2 |
| | 9 2 7 2 | 7 2 6 2 | 3+4 2 |
| | 9 2 9 2 | 6 2 6 2 | 5+6 2 |
| | 8 2 8 2 | 6 2 6 2 | 5+6 2 |
| | 6 2 9 2 | 6 2 6 2 | 5+6 2 |
| | 9 2 8 2 | 7 2 5 2 | 5+6 2 |
| | 8 2 8 2 | 6 2 6 2 | 5+6 2 |
| 139.1.103 | 246933 | 44 | Cu2Si2U1 |
| | -GM -M | -P -X | -N |
| | 7 2 7 2 | 6 2 5 2 | 3+4 2 |
| | 9 2 9 2 | 7 2 6 2 | 5+6 2 |
| | 9 2 9 2 | 6 2 6 2 | 5+6 2 |
| | 8 2 8 2 | 7 2 6 2 | 5+6 2 |
| | 7 2 7 2 | 7 2 5 2 | 5+6 2 |
| | 9 2 9 2 | 6 2 6 2 | 3+4 2 |

Continued on next column

**TABLE LXIV– continued**

| TQC | ICSD | NoE | SF |
|---|---|---|---|
| | 7 2 7 2 | 7 2 5 2 | 3+4 2 |
| | 8 2 6 2 | 6 2 6 2 | 5+6 2 |
| | 9 2 7 2 | 6 2 5 2 | 3+4 2 |
| | 6 2 8 2 | 7 2 5 2 | 5+6 2 |
| | 7 2 9 2 | 7 2 5 2 | 3+4 2 |
| | 8 2 6 2 | 6 2 6 2 | 5+6 2 |
| | 6 2 9 2 | 7 2 6 2 | 3+4 2 |
| | 9 2 8 2 | 6 2 5 2 | 5+6 2 |
| | 7 2 8 2 | 6 2 6 2 | 3+4 2 |
| | 6 2 7 2 | 7 2 5 2 | 5+6 2 |
| | 9 2 9 2 | 7 2 6 2 | 5+6 2 |
| | 7 2 8 2 | 6 2 5 2 | 3+4 2 |
| | 6 2 7 2 | 7 2 5 2 | 3+4 2 |
| | 6 2 6 2 | 7 2 6 2 | 5+6 2 |
| | 8 2 8 2 | 6 2 6 2 | 5+6 2 |
| | 9 2 9 2 | 6 2 6 2 | 3+4 2 |
| 139.1.104 | 68212 | 42 | Cu2Si2Th1 |
| | -GM -M | -P -X | -N |
| | 7 2 7 2 | 6 2 5 2 | 3+4 2 |
| | 9 2 9 2 | 7 2 6 2 | 5+6 2 |
| | 9 2 9 2 | 6 2 6 2 | 5+6 2 |
| | 8 2 8 2 | 7 2 6 2 | 5+6 2 |
| | 7 2 7 2 | 7 2 5 2 | 5+6 2 |
| | 9 2 9 2 | 6 2 6 2 | 3+4 2 |
| | 7 2 7 2 | 7 2 5 2 | 3+4 2 |
| | 8 2 6 2 | 6 2 6 2 | 5+6 2 |
| | 9 2 7 2 | 6 2 5 2 | 3+4 2 |
| | 6 2 8 2 | 7 2 5 2 | 5+6 2 |
| | 7 2 9 2 | 7 2 6 2 | 3+4 2 |
| | 6 2 6 2 | 6 2 5 2 | 5+6 2 |
| | 8 2 9 2 | 6 2 5 2 | 3+4 2 |
| | 6 2 8 2 | 7 2 6 2 | 5+6 2 |
| | 9 2 8 2 | 6 2 6 2 | 3+4 2 |
| | 7 2 7 2 | 7 2 5 2 | 5+6 2 |
| | 9 2 9 2 | 7 2 6 2 | 5+6 2 |
| | 7 2 8 2 | 6 2 5 2 | 3+4 2 |
| | 6 2 6 2 | 7 2 5 2 | 3+4 2 |
| | 6 2 7 2 | 7 2 6 2 | 5+6 2 |
| | 9 2 9 2 | 6 2 6 2 | 3+4 2 |
| 139.1.105 | 88424 | 76 | Cu2Ni1O2S2Sr2 |
| | -GM -M | -P -X | -N |
| | 7 2 7 2 | 7 2 5 2 | 5+6 2 |
| | 9 2 9 2 | 6 2 6 2 | 3+4 2 |
| | 7 2 7 2 | 6 2 6 2 | 3+4 2 |
| | 6 2 6 2 | 7 2 6 2 | 5+6 2 |
| | 9 2 9 2 | 7 2 5 2 | 5+6 2 |
| | 7 2 7 2 | 6 2 6 2 | 3+4 2 |
| | 8 2 8 2 | 7 2 5 2 | 5+6 2 |
| | 9 2 9 2 | 6 2 5 2 | 3+4 2 |
| | 6 2 6 2 | 6 2 6 2 | 5+6 2 |
| | 7 2 7 2 | 7 2 6 2 | 3+4 2 |
| | 7 2 9 2 | 6 2 5 2 | 5+6 2 |
| | 9 2 7 2 | 7 2 6 2 | 3+4 2 |
| | 7 2 9 2 | 6 2 5 2 | 3+4 2 |
| | 9 2 9 2 | 6 2 5 2 | 3+4 2 |
| | 9 2 8 2 | 7 2 6 2 | 5+6 2 |
| | 8 2 7 2 | 7 2 5 2 | 3+4 2 |
| | 8 2 6 2 | 7 2 5 2 | 5+6 2 |
| | 9 2 7 2 | 6 2 6 2 | 5+6 2 |
| | 7 2 9 2 | 7 2 5 2 | 3+4 2 |
| | 6 2 8 2 | 7 2 6 2 | 3+4 2 |

Continued on next column

**TABLE LXIV– continued**

| TQC | ICSD | NoE | SF |
|---|---|---|---|
| | 9 2 9 2 | 6 2 5 2 | 5+6 2 |
| | 8 2 8 2 | 7 2 5 2 | 3+4 2 |
| | 6 2 8 2 | 6 2 5 2 | 5+6 2 |
| | 7 2 6 2 | 6 2 5 2 | 5+6 2 |
| | 9 2 9 2 | 7 2 6 2 | 3+4 2 |
| | 8 2 8 2 | 6 2 6 2 | 5+6 2 |
| | 8 2 9 2 | 7 2 5 2 | 3+4 2 |
| | 6 2 7 2 | 6 2 5 2 | 5+6 2 |
| | 6 2 6 2 | 7 2 6 2 | 3+4 2 |
| | 7 2 8 2 | 6 2 5 2 | 5+6 2 |
| | 6 2 9 2 | 7 2 6 2 | 3+4 2 |
| | 7 2 8 2 | 6 2 5 2 | 3+4 2 |
| | 9 2 7 2 | 6 2 5 2 | 5+6 2 |
| | 6 2 6 2 | 7 2 6 2 | 5+6 2 |
| | 6 2 7 2 | 6 2 5 2 | 3+4 2 |
| | 7 2 6 2 | 7 2 5 2 | 3+4 2 |
| | 9 2 7 2 | 6 2 5 2 | 3+4 2 |
| | 8 2 6 2 | 7 2 5 2 | 3+4 2 |
| 139.1.106 | 86002 | 44 | Cu2Ge2U1 |
| | -GM -M | -P -X | -N |
| | 7 2 7 2 | 6 2 5 2 | 3+4 2 |
| | 9 2 9 2 | 7 2 6 2 | 5+6 2 |
| | 9 2 9 2 | 6 2 6 2 | 5+6 2 |
| | 8 2 8 2 | 7 2 6 2 | 5+6 2 |
| | 7 2 7 2 | 7 2 5 2 | 5+6 2 |
| | 9 2 9 2 | 6 2 6 2 | 3+4 2 |
| | 7 2 7 2 | 7 2 5 2 | 3+4 2 |
| | 9 2 6 2 | 6 2 6 2 | 5+6 2 |
| | 8 2 7 2 | 7 2 5 2 | 3+4 2 |
| | 6 2 8 2 | 6 2 5 2 | 5+6 2 |
| | 7 2 9 2 | 7 2 5 2 | 3+4 2 |
| | 9 2 6 2 | 6 2 6 2 | 5+6 2 |
| | 6 2 9 2 | 7 2 6 2 | 5+6 2 |
| | 8 2 8 2 | 6 2 5 2 | 3+4 2 |
| | 7 2 8 2 | 7 2 6 2 | 3+4 2 |
| | 6 2 7 2 | 6 2 5 2 | 5+6 2 |
| | 9 2 9 2 | 7 2 6 2 | 5+6 2 |
| | 7 2 8 2 | 7 2 5 2 | 3+4 2 |
| | 6 2 7 2 | 6 2 5 2 | 3+4 2 |
| | 6 2 6 2 | 7 2 6 2 | 5+6 2 |
| | 8 2 9 2 | 6 2 6 2 | 5+6 2 |
| | 9 2 8 2 | 6 2 6 2 | 3+4 2 |
| 139.1.107 | 86001 | 44 | Cu2Ge2U1 |
| | -GM -M | -P -X | -N |
| | 7 2 7 2 | 6 2 5 2 | 3+4 2 |
| | 9 2 9 2 | 7 2 6 2 | 5+6 2 |
| | 9 2 9 2 | 6 2 6 2 | 5+6 2 |
| | 8 2 8 2 | 7 2 6 2 | 5+6 2 |
| | 7 2 7 2 | 7 2 5 2 | 5+6 2 |
| | 9 2 9 2 | 6 2 6 2 | 3+4 2 |
| | 7 2 7 2 | 7 2 5 2 | 3+4 2 |
| | 9 2 6 2 | 6 2 6 2 | 5+6 2 |
| | 8 2 7 2 | 7 2 5 2 | 3+4 2 |
| | 6 2 8 2 | 6 2 5 2 | 5+6 2 |
| | 7 2 9 2 | 7 2 5 2 | 3+4 2 |
| | 9 2 6 2 | 6 2 6 2 | 5+6 2 |
| | 6 2 9 2 | 7 2 6 2 | 5+6 2 |
| | 8 2 8 2 | 6 2 5 2 | 3+4 2 |
| | 7 2 8 2 | 7 2 6 2 | 3+4 2 |
| | 6 2 7 2 | 6 2 5 2 | 5+6 2 |
| | 9 2 9 2 | 7 2 6 2 | 5+6 2 |

Continued on next column

**TABLE LXIV– continued**

| TQC | ICSD | NoE | SF |
|---|---|---|---|
| | 7 2 8 2 7 2 5 2 3+4 2 | | |
| | 6 2 7 2 6 2 5 2 3+4 2 | | |
| | 6 2 6 2 7 2 6 2 5+6 2 | | |
| | 8 2 9 2 6 2 6 2 5+6 2 | | |
| | 9 2 8 2 6 2 6 2 3+4 2 | | |
| 139.1.108 | 86000 | 44 | Cu2Ge2U1 |
| | -GM -M -P -X -N | | |
| | 7 2 7 2 6 2 5 2 3+4 2 | | |
| | 9 2 9 2 7 2 6 2 5+6 2 | | |
| | 9 2 9 2 6 2 6 2 5+6 2 | | |
| | 8 2 8 2 7 2 6 2 5+6 2 | | |
| | 7 2 7 2 7 2 5 2 5+6 2 | | |
| | 9 2 9 2 6 2 6 2 3+4 2 | | |
| | 7 2 7 2 7 2 5 2 3+4 2 | | |
| | 9 2 6 2 6 2 6 2 5+6 2 | | |
| | 8 2 7 2 7 2 5 2 3+4 2 | | |
| | 6 2 8 2 6 2 5 2 5+6 2 | | |
| | 7 2 9 2 7 2 5 2 3+4 2 | | |
| | 9 2 6 2 6 2 6 2 5+6 2 | | |
| | 6 2 9 2 7 2 6 2 5+6 2 | | |
| | 8 2 8 2 6 2 5 2 3+4 2 | | |
| | 7 2 8 2 7 2 6 2 3+4 2 | | |
| | 6 2 7 2 6 2 5 2 5+6 2 | | |
| | 9 2 9 2 7 2 6 2 5+6 2 | | |
| | 7 2 8 2 6 2 5 2 3+4 2 | | |
| | 6 2 7 2 7 2 5 2 3+4 2 | | |
| | 6 2 6 2 7 2 6 2 5+6 2 | | |
| | 8 2 9 2 6 2 6 2 5+6 2 | | |
| | 9 2 8 2 6 2 6 2 3+4 2 | | |
| 139.1.109 | 85999 | 44 | Cu2Ge2U1 |
| | -GM -M -P -X -N | | |
| | 7 2 7 2 6 2 5 2 3+4 2 | | |
| | 9 2 9 2 7 2 6 2 5+6 2 | | |
| | 9 2 9 2 6 2 6 2 5+6 2 | | |
| | 8 2 8 2 7 2 6 2 5+6 2 | | |
| | 7 2 7 2 7 2 5 2 5+6 2 | | |
| | 9 2 9 2 6 2 6 2 3+4 2 | | |
| | 7 2 7 2 7 2 5 2 3+4 2 | | |
| | 9 2 6 2 6 2 6 2 5+6 2 | | |
| | 8 2 7 2 7 2 5 2 3+4 2 | | |
| | 6 2 8 2 6 2 5 2 5+6 2 | | |
| | 7 2 9 2 7 2 5 2 3+4 2 | | |
| | 9 2 6 2 6 2 6 2 5+6 2 | | |
| | 6 2 9 2 7 2 6 2 5+6 2 | | |
| | 8 2 8 2 6 2 5 2 3+4 2 | | |
| | 7 2 8 2 7 2 6 2 3+4 2 | | |
| | 6 2 7 2 6 2 5 2 5+6 2 | | |
| | 9 2 9 2 7 2 6 2 5+6 2 | | |
| | 7 2 8 2 6 2 5 2 3+4 2 | | |
| | 6 2 7 2 7 2 5 2 3+4 2 | | |
| | 6 2 6 2 7 2 6 2 5+6 2 | | |
| | 8 2 9 2 6 2 6 2 5+6 2 | | |
| | 9 2 8 2 6 2 6 2 3+4 2 | | |
| 139.1.110 | 85998 | 44 | Cu2Ge2U1 |
| | -GM -M -P -X -N | | |
| | 7 2 7 2 6 2 5 2 3+4 2 | | |
| | 9 2 9 2 7 2 6 2 5+6 2 | | |
| | 9 2 9 2 6 2 6 2 5+6 2 | | |
| | 8 2 8 2 7 2 6 2 5+6 2 | | |
| | 7 2 7 2 7 2 5 2 5+6 2 | | |
| | 9 2 9 2 6 2 6 2 3+4 2 | | |

Continued on next column

**TABLE LXIV– continued**

| TQC | ICSD | NoE | SF |
|---|---|---|---|
| | 7 2 7 2 7 2 5 2 3+4 2 | | |
| | 9 2 6 2 6 2 6 2 5+6 2 | | |
| | 8 2 7 2 7 2 5 2 3+4 2 | | |
| | 6 2 8 2 6 2 5 2 5+6 2 | | |
| | 7 2 9 2 7 2 5 2 3+4 2 | | |
| | 9 2 6 2 6 2 6 2 5+6 2 | | |
| | 6 2 9 2 7 2 6 2 5+6 2 | | |
| | 8 2 8 2 6 2 5 2 3+4 2 | | |
| | 7 2 8 2 7 2 6 2 3+4 2 | | |
| | 6 2 7 2 6 2 5 2 5+6 2 | | |
| | 9 2 9 2 7 2 6 2 5+6 2 | | |
| | 7 2 8 2 6 2 5 2 3+4 2 | | |
| | 6 2 7 2 7 2 5 2 3+4 2 | | |
| | 6 2 6 2 7 2 6 2 5+6 2 | | |
| | 8 2 9 2 6 2 6 2 5+6 2 | | |
| | 9 2 8 2 6 2 6 2 3+4 2 | | |
| 139.1.111 | 424224 | 40 | Cu2Ge2Sr1 |
| | -GM -M -P -X -N | | |
| | 7 2 7 2 6 2 5 2 3+4 2 | | |
| | 9 2 9 2 7 2 6 2 5+6 2 | | |
| | 9 2 9 2 6 2 6 2 5+6 2 | | |
| | 8 2 8 2 7 2 6 2 5+6 2 | | |
| | 7 2 7 2 7 2 5 2 5+6 2 | | |
| | 9 2 9 2 6 2 6 2 3+4 2 | | |
| | 7 2 7 2 7 2 6 2 3+4 2 | | |
| | 8 2 6 2 6 2 5 2 5+6 2 | | |
| | 9 2 7 2 6 2 6 2 3+4 2 | | |
| | 6 2 8 2 7 2 5 2 5+6 2 | | |
| | 7 2 9 2 6 2 5 2 5+6 2 | | |
| | 9 2 9 2 7 2 6 2 3+4 2 | | |
| | 6 2 8 2 7 2 5 2 5+6 2 | | |
| | 8 2 6 2 6 2 6 2 3+4 2 | | |
| | 6 2 8 2 6 2 5 2 3+4 2 | | |
| | 7 2 7 2 7 2 5 2 5+6 2 | | |
| | 9 2 9 2 7 2 6 2 5+6 2 | | |
| | 7 2 8 2 7 2 6 2 3+4 2 | | |
| | 6 2 6 2 6 2 5 2 3+4 2 | | |
| | 6 2 7 2 7 2 6 2 5+6 2 | | |
| 139.1.112 | 410 | 40 | Cu2Ge2Sr1 |
| | -GM -M -P -X -N | | |
| | 7 2 7 2 6 2 5 2 3+4 2 | | |
| | 9 2 9 2 7 2 6 2 5+6 2 | | |
| | 9 2 9 2 6 2 6 2 5+6 2 | | |
| | 8 2 8 2 7 2 6 2 5+6 2 | | |
| | 7 2 9 2 7 2 5 2 5+6 2 | | |
| | 9 2 7 2 6 2 6 2 3+4 2 | | |
| | 7 2 7 2 7 2 5 2 3+4 2 | | |
| | 8 2 6 2 6 2 6 2 5+6 2 | | |
| | 9 2 7 2 6 2 6 2 3+4 2 | | |
| | 6 2 8 2 7 2 5 2 5+6 2 | | |
| | 7 2 9 2 6 2 5 2 5+6 2 | | |
| | 9 2 9 2 7 2 6 2 3+4 2 | | |
| | 6 2 8 2 6 2 5 2 5+6 2 | | |
| | 8 2 6 2 7 2 6 2 3+4 2 | | |
| | 6 2 8 2 6 2 5 2 3+4 2 | | |
| | 9 2 7 2 7 2 6 2 5+6 2 | | |
| | 7 2 9 2 7 2 5 2 5+6 2 | | |
| | 7 2 8 2 7 2 6 2 3+4 2 | | |
| | 6 2 6 2 6 2 5 2 3+4 2 | | |
| | 6 2 7 2 7 2 6 2 5+6 2 | | |
| 139.1.113 | 154203 | 110 | Cu2Fe2O5Se2Sr3 |

Continued on next column

**TABLE LXIV– continued**

| TQC | ICSD NoE SF |
|---|---|
| | -GM -M -P -X -N |
| | 7 2 7 2 7 2 5 2 5+6 2 |
| | 7 2 7 2 6 2 5 2 3+4 2 |
| | 9 2 9 2 7 2 6 2 5+6 2 |
| | 7 2 7 2 6 2 5 2 3+4 2 |
| | 9 2 9 2 6 2 6 2 5+6 2 |
| | 7 2 7 2 7 2 5 2 3+4 2 |
| | 6 2 6 2 7 2 6 2 3+4 2 |
| | 8 2 8 2 6 2 5 2 5+6 2 |
| | 9 2 9 2 6 2 6 2 3+4 2 |
| | 9 2 9 2 7 2 5 2 3+4 2 |
| | 7 2 7 2 6 2 6 2 5+6 2 |
| | 8 2 8 2 7 2 6 2 3+4 2 |
| | 9 2 9 2 6 2 6 2 3+4 2 |
| | 6 2 6 2 7 2 5 2 5+6 2 |
| | 8 2 8 2 6 2 6 2 3+4 2 |
| | 7 2 9 2 6 2 5 2 5+6 2 |
| | 9 2 7 2 7 2 6 2 3+4 2 |
| | 7 2 9 2 6 2 5 2 3+4 2 |
| | 9 2 7 2 7 2 6 2 5+6 2 |
| | 9 2 9 2 7 2 6 2 5+6 2 |
| | 7 2 9 2 6 2 5 2 3+4 2 |
| | 9 2 7 2 7 2 6 2 5+6 2 |
| | 9 2 9 2 6 2 5 2 5+6 2 |
| | 7 2 7 2 7 2 6 2 3+4 2 |
| | 6 2 6 2 7 2 5 2 5+6 2 |
| | 8 2 8 2 7 2 5 2 5+6 2 |
| | 9 2 9 2 6 2 6 2 3+4 2 |
| | 8 2 8 2 6 2 6 2 3+4 2 |
| | 7 2 7 2 6 2 6 2 5+6 2 |
| | 9 2 6 2 7 2 6 2 5+6 2 |
| | 8 2 7 2 6 2 5 2 5+6 2 |
| | 9 2 9 2 7 2 6 2 3+4 2 |
| | 6 2 6 2 7 2 5 2 3+4 2 |
| | 7 2 9 2 6 2 5 2 3+4 2 |
| | 6 2 8 2 6 2 5 2 5+6 2 |
| | 8 2 8 2 7 2 6 2 5+6 2 |
| | 6 2 8 2 6 2 5 2 3+4 2 |
| | 7 2 7 2 7 2 6 2 5+6 2 |
| | 6 2 6 2 6 2 5 2 3+4 2 |
| | 9 2 9 2 7 2 6 2 5+6 2 |
| | 8 2 6 2 6 2 6 2 3+4 2 |
| | 7 2 8 2 7 2 6 2 5+6 2 |
| | 8 2 9 2 6 2 5 2 3+4 2 |
| | 6 2 8 2 7 2 6 2 5+6 2 |
| | 7 2 9 2 6 2 5 2 3+4 2 |
| | 6 2 8 2 7 2 6 2 5+6 2 |
| | 6 2 6 2 6 2 5 2 3+4 2 |
| | 9 2 7 2 7 2 6 2 3+4 2 |
| | 8 2 8 2 6 2 5 2 5+6 2 |
| | 7 2 7 2 7 2 6 2 5+6 2 |
| | 9 2 7 2 6 2 5 2 3+4 2 |
| | 6 2 6 2 6 2 5 2 3+4 2 |
| | 7 2 6 2 7 2 5 2 3+4 2 |
| | 6 2 8 2 7 2 5 2 5+6 2 |
| | 8 2 7 2 7 2 6 2 3+4 2 |
| 139.1.114 | 40763 70 Cu1P2U1 |
| | -GM -M -P -X -N |
| | 7 2 7 2 6 2 5 2 3+4 2 |
| | 9 2 9 2 7 2 6 2 5+6 2 |
| | 7 2 7 2 6 2 5 2 3+4 2 |

Continued on next column

**TABLE LXIV– continued**

| TQC | ICSD NoE SF |
|---|---|
| | 9 2 9 2 7 2 6 2 5+6 2 |
| | 7 2 7 2 6 2 6 2 3+4 2 |
| | 9 2 9 2 7 2 5 2 5+6 2 |
| | 8 2 8 2 6 2 5 2 3+4 2 |
| | 6 2 6 2 7 2 6 2 5+6 2 |
| | 7 2 7 2 7 2 5 2 5+6 2 |
| | 9 2 9 2 6 2 6 2 3+4 2 |
| | 7 2 7 2 7 2 6 2 3+4 2 |
| | 6 2 6 2 6 2 6 2 5+6 2 |
| | 7 2 9 2 7 2 5 2 5+6 2 |
| | 9 2 7 2 6 2 5 2 3+4 2 |
| | 8 2 6 2 7 2 5 2 3+4 2 |
| | 9 2 7 2 6 2 6 2 5+6 2 |
| | 6 2 9 2 6 2 6 2 5+6 2 |
| | 9 2 8 2 7 2 5 2 3+4 2 |
| | 8 2 8 2 6 2 5 2 5+6 2 |
| | 8 2 9 2 7 2 6 2 3+4 2 |
| | 7 2 9 2 7 2 6 2 3+4 2 |
| | 6 2 6 2 6 2 5 2 5+6 2 |
| | 7 2 8 2 6 2 6 2 3+4 2 |
| | 9 2 8 2 7 2 5 2 5+6 2 |
| | 6 2 8 2 7 2 6 2 3+4 2 |
| | 8 2 9 2 6 2 5 2 5+6 2 |
| | 7 2 7 2 6 2 5 2 5+6 2 |
| | 9 2 7 2 7 2 6 2 3+4 2 |
| | 6 2 9 2 6 2 5 2 3+4 2 |
| | 9 2 8 2 7 2 6 2 5+6 2 |
| | 6 2 6 2 6 2 5 2 3+4 2 |
| | 8 2 7 2 7 2 5 2 5+6 2 |
| | 9 2 8 2 6 2 6 2 3+4 2 |
| | 7 2 6 2 7 2 6 2 5+6 2 |
| | 8 2 7 2 6 2 5 2 3+4 2 |
| 139.1.115 | 290194 38 Cu1Ge1Hf1 |
| | -GM -M -P -X -N |
| | 7 2 7 2 6 2 5 2 3+4 2 |
| | 9 2 9 2 7 2 6 2 5+6 2 |
| | 7 2 7 2 6 2 6 2 5+6 2 |
| | 7 2 7 2 7 2 6 2 3+4 2 |
| | 6 2 6 2 6 2 6 2 3+4 2 |
| | 7 2 7 2 7 2 6 2 3+4 2 |
| | 6 2 6 2 6 2 6 2 5+6 2 |
| | 7 2 6 2 7 2 6 2 5+6 2 |
| | 6 2 7 2 6 2 5 2 5+6 2 |
| | 7 2 7 2 7 2 5 2 3+4 2 |
| | 6 2 6 2 7 2 6 2 5+6 2 |
| | 6 2 9 2 6 2 6 2 3+4 2 |
| | 7 2 6 2 6 2 5 2 5+6 2 |
| | 9 2 7 2 6 2 6 2 3+4 2 |
| | 8 2 6 2 7 2 6 2 3+4 2 |
| | 9 2 9 2 7 2 5 2 5+6 2 |
| | 8 2 8 2 6 2 6 2 5+6 2 |
| | 6 2 8 2 7 2 6 2 3+4 2 |
| | 7 2 9 2 7 2 6 2 5+6 2 |
| 139.1.116 | 67472 38 Cr2Gd1Si2 |
| | -GM -M -P -X -N |
| | 7 2 7 2 6 2 5 2 3+4 2 |
| | 9 2 9 2 7 2 6 2 5+6 2 |
| | 9 2 9 2 6 2 6 2 5+6 2 |
| | 8 2 8 2 7 2 6 2 5+6 2 |
| | 7 2 7 2 7 2 5 2 5+6 2 |
| | 9 2 9 2 6 2 6 2 3+4 2 |

Continued on next column

**TABLE LXIV– continued**

| TQC | ICSD | NoE | SF |
|---|---|---|---|
| | 7 2 7 2 | 7 2 5 2 | 3+4 2 |
| | 9 2 6 2 | 6 2 6 2 | 5+6 2 |
| | 8 2 7 2 | 7 2 5 2 | 3+4 2 |
| | 9 2 9 2 | 7 2 6 2 | 5+6 2 |
| | 6 2 8 2 | 6 2 5 2 | 3+4 2 |
| | 6 2 6 2 | 6 2 5 2 | 5+6 2 |
| | 7 2 8 2 | 7 2 6 2 | 5+6 2 |
| | 8 2 9 2 | 6 2 6 2 | 3+4 2 |
| | 8 2 8 2 | 7 2 5 2 | 5+6 2 |
| | 9 2 9 2 | 6 2 6 2 | 5+6 2 |
| | 7 2 8 2 | 6 2 6 2 | 5+6 2 |
| | 8 2 7 2 | 7 2 6 2 | 5+6 2 |
| | 6 2 8 2 | 7 2 5 2 | 3+4 2 |
| 139.1.117 | 245595 | 50 | Cr1O4Sr2 |
| | -GM -M | -P -X | -N |
| | 7 2 9 2 | 7 2 5 2 | 5+6 2 |
| | 9 2 7 2 | 6 2 6 2 | 3+4 2 |
| | 7 2 7 2 | 6 2 6 2 | 3+4 2 |
| | 6 2 6 2 | 7 2 6 2 | 5+6 2 |
| | 7 2 9 2 | 7 2 5 2 | 5+6 2 |
| | 9 2 7 2 | 6 2 6 2 | 3+4 2 |
| | 9 2 9 2 | 7 2 5 2 | 3+4 2 |
| | 7 2 7 2 | 6 2 6 2 | 5+6 2 |
| | 6 2 8 2 | 6 2 5 2 | 3+4 2 |
| | 8 2 9 2 | 7 2 5 2 | 5+6 2 |
| | 9 2 6 2 | 7 2 6 2 | 5+6 2 |
| | 7 2 7 2 | 6 2 6 2 | 3+4 2 |
| | 9 2 9 2 | 6 2 5 2 | 3+4 2 |
| | 8 2 8 2 | 7 2 5 2 | 3+4 2 |
| | 9 2 9 2 | 7 2 5 2 | 3+4 2 |
| | 7 2 7 2 | 7 2 5 2 | 5+6 2 |
| | 8 2 8 2 | 6 2 5 2 | 5+6 2 |
| | 9 2 9 2 | 6 2 5 2 | 3+4 2 |
| | 8 2 8 2 | 7 2 6 2 | 5+6 2 |
| | 7 2 7 2 | 6 2 6 2 | 3+4 2 |
| | 6 2 6 2 | 7 2 5 2 | 5+6 2 |
| | 9 2 9 2 | 6 2 5 2 | 5+6 2 |
| | 9 2 9 2 | 6 2 6 2 | 5+6 2 |
| | 8 2 8 2 | 7 2 5 2 | 3+4 2 |
| | 6 2 6 2 | 6 2 5 2 | 3+4 2 |
| 139.1.118 | 43091 | 106 | Co4Ge7U3 |
| | -GM -M | -P -X | -N |
| | 7 2 7 2 | 6 2 5 2 | 3+4 2 |
| | 9 2 9 2 | 7 2 6 2 | 5+6 2 |
| | 7 2 7 2 | 6 2 5 2 | 3+4 2 |
| | 7 2 9 2 | 6 2 5 2 | 3+4 2 |
| | 9 2 7 2 | 7 2 6 2 | 5+6 2 |
| | 9 2 9 2 | 7 2 6 2 | 5+6 2 |
| | 9 2 9 2 | 6 2 6 2 | 5+6 2 |
| | 7 2 9 2 | 7 2 6 2 | 5+6 2 |
| | 9 2 7 2 | 6 2 5 2 | 3+4 2 |
| | 8 2 8 2 | 7 2 6 2 | 5+6 2 |
| | 6 2 8 2 | 6 2 5 2 | 3+4 2 |
| | 8 2 6 2 | 7 2 6 2 | 5+6 2 |
| | 7 2 7 2 | 7 2 5 2 | 3+4 2 |
| | 9 2 9 2 | 6 2 5 2 | 3+4 2 |
| | 7 2 9 2 | 7 2 6 2 | 5+6 2 |
| | 9 2 7 2 | 7 2 5 2 | 5+6 2 |
| | 7 2 7 2 | 6 2 5 2 | 5+6 2 |
| | 6 2 6 2 | 7 2 6 2 | 3+4 2 |
| | 8 2 8 2 | 6 2 6 2 | 5+6 2 |

Continued on next column

**TABLE LXIV– continued**

| TQC | ICSD | NoE | SF |
|---|---|---|---|
| | 7 2 9 2 | 6 2 5 2 | 3+4 2 |
| | 9 2 6 2 | 6 2 6 2 | 3+4 2 |
| | 6 2 7 2 | 7 2 6 2 | 5+6 2 |
| | 7 2 7 2 | 7 2 5 2 | 3+4 2 |
| | 9 2 9 2 | 6 2 6 2 | 3+4 2 |
| | 8 2 8 2 | 7 2 5 2 | 5+6 2 |
| | 7 2 9 2 | 6 2 5 2 | 5+6 2 |
| | 9 2 6 2 | 7 2 6 2 | 3+4 2 |
| | 6 2 7 2 | 7 2 5 2 | 3+4 2 |
| | 9 2 7 2 | 6 2 6 2 | 3+4 2 |
| | 8 2 6 2 | 7 2 5 2 | 5+6 2 |
| | 6 2 8 2 | 6 2 6 2 | 3+4 2 |
| | 8 2 9 2 | 7 2 5 2 | 5+6 2 |
| | 7 2 6 2 | 6 2 6 2 | 5+6 2 |
| | 6 2 8 2 | 7 2 5 2 | 5+6 2 |
| | 8 2 8 2 | 6 2 6 2 | 3+4 2 |
| | 6 2 6 2 | 6 2 6 2 | 3+4 2 |
| | 8 2 8 2 | 7 2 5 2 | 5+6 2 |
| | 9 2 9 2 | 6 2 6 2 | 5+6 2 |
| | 9 2 9 2 | 6 2 5 2 | 3+4 2 |
| | 7 2 8 2 | 7 2 5 2 | 3+4 2 |
| | 8 2 7 2 | 7 2 6 2 | 5+6 2 |
| | 7 2 6 2 | 6 2 6 2 | 3+4 2 |
| | 6 2 9 2 | 6 2 6 2 | 5+6 2 |
| | 7 2 8 2 | 7 2 5 2 | 3+4 2 |
| | 6 2 8 2 | 6 2 6 2 | 3+4 2 |
| | 6 2 9 2 | 7 2 5 2 | 5+6 2 |
| | 8 2 7 2 | 7 2 6 2 | 5+6 2 |
| | 9 2 8 2 | 6 2 5 2 | 5+6 2 |
| | 6 2 9 2 | 7 2 6 2 | 3+4 2 |
| | 9 2 8 2 | 6 2 5 2 | 5+6 2 |
| | 9 2 7 2 | 7 2 5 2 | 3+4 2 |
| | 8 2 6 2 | 7 2 6 2 | 5+6 2 |
| | 7 2 7 2 | 6 2 6 2 | 5+6 2 |
| 139.1.119 | 290506 | 106 | Co4Ge7U3 |
| | -GM -M | -P -X | -N |
| | 7 2 7 2 | 6 2 5 2 | 3+4 2 |
| | 9 2 9 2 | 7 2 6 2 | 5+6 2 |
| | 7 2 7 2 | 6 2 5 2 | 3+4 2 |
| | 7 2 9 2 | 6 2 5 2 | 3+4 2 |
| | 9 2 7 2 | 7 2 6 2 | 5+6 2 |
| | 9 2 9 2 | 7 2 6 2 | 5+6 2 |
| | 9 2 9 2 | 6 2 6 2 | 5+6 2 |
| | 7 2 7 2 | 7 2 6 2 | 5+6 2 |
| | 9 2 9 2 | 6 2 5 2 | 3+4 2 |
| | 8 2 8 2 | 7 2 6 2 | 5+6 2 |
| | 6 2 6 2 | 6 2 5 2 | 3+4 2 |
| | 8 2 8 2 | 7 2 6 2 | 5+6 2 |
| | 7 2 7 2 | 7 2 5 2 | 3+4 2 |
| | 9 2 9 2 | 6 2 5 2 | 3+4 2 |
| | 7 2 9 2 | 7 2 6 2 | 5+6 2 |
| | 9 2 7 2 | 7 2 5 2 | 5+6 2 |
| | 7 2 7 2 | 6 2 5 2 | 5+6 2 |
| | 6 2 6 2 | 7 2 6 2 | 3+4 2 |
| | 8 2 8 2 | 6 2 6 2 | 5+6 2 |
| | 7 2 9 2 | 6 2 5 2 | 3+4 2 |
| | 9 2 6 2 | 6 2 6 2 | 3+4 2 |
| | 6 2 7 2 | 7 2 6 2 | 5+6 2 |
| | 7 2 9 2 | 7 2 5 2 | 3+4 2 |
| | 9 2 7 2 | 6 2 6 2 | 5+6 2 |
| | 8 2 8 2 | 7 2 5 2 | 3+4 2 |

Continued on next column

**TABLE LXIV– continued**

| TQC | ICSD | NoE | SF |
|---|---|---|---|
| | 7 2 9 2 | 6 2 5 2 | 5+6 2 |
| | 9 2 7 2 | 7 2 6 2 | 3+4 2 |
| | 6 2 6 2 | 7 2 5 2 | 5+6 2 |
| | 9 2 7 2 | 6 2 6 2 | 3+4 2 |
| | 8 2 6 2 | 7 2 5 2 | 3+4 2 |
| | 6 2 8 2 | 6 2 6 2 | 5+6 2 |
| | 8 2 9 2 | 7 2 5 2 | 3+4 2 |
| | 7 2 6 2 | 6 2 6 2 | 5+6 2 |
| | 8 2 8 2 | 7 2 5 2 | 5+6 2 |
| | 6 2 8 2 | 6 2 5 2 | 3+4 2 |
| | 6 2 8 2 | 6 2 6 2 | 3+4 2 |
| | 9 2 6 2 | 6 2 6 2 | 5+6 2 |
| | 8 2 9 2 | 7 2 5 2 | 5+6 2 |
| | 9 2 7 2 | 6 2 6 2 | 3+4 2 |
| | 8 2 6 2 | 7 2 5 2 | 5+6 2 |
| | 7 2 9 2 | 7 2 6 2 | 3+4 2 |
| | 7 2 8 2 | 6 2 6 2 | 3+4 2 |
| | 6 2 9 2 | 7 2 6 2 | 5+6 2 |
| | 7 2 8 2 | 6 2 5 2 | 3+4 2 |
| | 6 2 9 2 | 6 2 6 2 | 3+4 2 |
| | 8 2 7 2 | 7 2 5 2 | 5+6 2 |
| | 9 2 8 2 | 6 2 5 2 | 5+6 2 |
| | 6 2 8 2 | 7 2 6 2 | 5+6 2 |
| | 9 2 9 2 | 7 2 5 2 | 3+4 2 |
| | 6 2 8 2 | 7 2 6 2 | 5+6 2 |
| | 9 2 7 2 | 6 2 5 2 | 3+4 2 |
| | 8 2 6 2 | 7 2 6 2 | 5+6 2 |
| | 7 2 7 2 | 6 2 6 2 | 5+6 2 |
| 139.1.120 | 290505 | 106 | Co4Ge7U3 |
| | -GM -M | -P -X | -N |
| | 7 2 7 2 | 6 2 5 2 | 3+4 2 |
| | 9 2 9 2 | 7 2 6 2 | 5+6 2 |
| | 7 2 7 2 | 6 2 5 2 | 3+4 2 |
| | 7 2 9 2 | 6 2 5 2 | 3+4 2 |
| | 9 2 7 2 | 7 2 6 2 | 5+6 2 |
| | 9 2 9 2 | 7 2 6 2 | 5+6 2 |
| | 9 2 9 2 | 6 2 6 2 | 5+6 2 |
| | 7 2 7 2 | 7 2 6 2 | 5+6 2 |
| | 9 2 9 2 | 6 2 5 2 | 3+4 2 |
| | 8 2 8 2 | 7 2 6 2 | 5+6 2 |
| | 6 2 6 2 | 6 2 5 2 | 3+4 2 |
| | 8 2 8 2 | 7 2 6 2 | 5+6 2 |
| | 7 2 7 2 | 7 2 5 2 | 3+4 2 |
| | 9 2 9 2 | 6 2 5 2 | 3+4 2 |
| | 7 2 9 2 | 7 2 6 2 | 5+6 2 |
| | 9 2 7 2 | 7 2 5 2 | 5+6 2 |
| | 7 2 7 2 | 6 2 5 2 | 5+6 2 |
| | 6 2 6 2 | 7 2 6 2 | 3+4 2 |
| | 8 2 8 2 | 6 2 6 2 | 5+6 2 |
| | 7 2 9 2 | 6 2 5 2 | 3+4 2 |
| | 9 2 6 2 | 6 2 6 2 | 3+4 2 |
| | 6 2 7 2 | 7 2 6 2 | 5+6 2 |
| | 7 2 9 2 | 7 2 5 2 | 3+4 2 |
| | 9 2 7 2 | 6 2 6 2 | 5+6 2 |
| | 8 2 8 2 | 7 2 5 2 | 3+4 2 |
| | 7 2 9 2 | 6 2 5 2 | 5+6 2 |
| | 9 2 7 2 | 7 2 6 2 | 3+4 2 |
| | 6 2 6 2 | 7 2 5 2 | 5+6 2 |
| | 9 2 7 2 | 6 2 6 2 | 3+4 2 |
| | 8 2 6 2 | 7 2 5 2 | 3+4 2 |
| | 6 2 8 2 | 6 2 6 2 | 5+6 2 |

Continued on next column

**TABLE LXIV– continued**

| TQC | ICSD | NoE | SF |
|---|---|---|---|
| | 8 2 9 2 | 7 2 5 2 | 3+4 2 |
| | 7 2 6 2 | 6 2 6 2 | 5+6 2 |
| | 8 2 8 2 | 7 2 5 2 | 5+6 2 |
| | 6 2 8 2 | 6 2 5 2 | 3+4 2 |
| | 6 2 8 2 | 6 2 6 2 | 3+4 2 |
| | 9 2 6 2 | 6 2 6 2 | 5+6 2 |
| | 8 2 9 2 | 7 2 5 2 | 5+6 2 |
| | 9 2 7 2 | 6 2 6 2 | 3+4 2 |
| | 8 2 6 2 | 7 2 6 2 | 5+6 2 |
| | 7 2 9 2 | 7 2 5 2 | 3+4 2 |
| | 7 2 8 2 | 6 2 6 2 | 3+4 2 |
| | 6 2 9 2 | 7 2 6 2 | 5+6 2 |
| | 7 2 8 2 | 6 2 5 2 | 3+4 2 |
| | 6 2 9 2 | 6 2 6 2 | 3+4 2 |
| | 8 2 7 2 | 7 2 5 2 | 5+6 2 |
| | 9 2 8 2 | 6 2 5 2 | 5+6 2 |
| | 6 2 8 2 | 7 2 6 2 | 5+6 2 |
| | 9 2 9 2 | 7 2 5 2 | 3+4 2 |
| | 6 2 8 2 | 7 2 6 2 | 5+6 2 |
| | 9 2 7 2 | 6 2 5 2 | 3+4 2 |
| | 8 2 6 2 | 7 2 6 2 | 5+6 2 |
| | 7 2 7 2 | 6 2 6 2 | 5+6 2 |
| 139.1.121 | 290504 | 106 | Co4Ge7U3 |
| | -GM -M | -P -X | -N |
| | 7 2 7 2 | 6 2 5 2 | 3+4 2 |
| | 9 2 9 2 | 7 2 6 2 | 5+6 2 |
| | 7 2 7 2 | 6 2 5 2 | 3+4 2 |
| | 7 2 9 2 | 6 2 5 2 | 3+4 2 |
| | 9 2 7 2 | 7 2 6 2 | 5+6 2 |
| | 9 2 9 2 | 7 2 6 2 | 5+6 2 |
| | 9 2 9 2 | 6 2 6 2 | 5+6 2 |
| | 7 2 9 2 | 7 2 6 2 | 5+6 2 |
| | 9 2 7 2 | 6 2 5 2 | 3+4 2 |
| | 8 2 8 2 | 7 2 6 2 | 5+6 2 |
| | 6 2 8 2 | 6 2 5 2 | 3+4 2 |
| | 8 2 6 2 | 7 2 6 2 | 5+6 2 |
| | 7 2 7 2 | 7 2 5 2 | 3+4 2 |
| | 9 2 9 2 | 6 2 5 2 | 3+4 2 |
| | 7 2 9 2 | 7 2 6 2 | 5+6 2 |
| | 9 2 7 2 | 7 2 5 2 | 5+6 2 |
| | 7 2 7 2 | 6 2 5 2 | 3+4 2 |
| | 6 2 6 2 | 7 2 6 2 | 5+6 2 |
| | 8 2 8 2 | 6 2 6 2 | 5+6 2 |
| | 7 2 9 2 | 6 2 5 2 | 3+4 2 |
| | 9 2 6 2 | 6 2 6 2 | 3+4 2 |
| | 6 2 7 2 | 7 2 6 2 | 5+6 2 |
| | 7 2 7 2 | 7 2 5 2 | 3+4 2 |
| | 9 2 9 2 | 6 2 6 2 | 5+6 2 |
| | 8 2 8 2 | 7 2 5 2 | 3+4 2 |
| | 7 2 9 2 | 6 2 5 2 | 5+6 2 |
| | 9 2 7 2 | 7 2 6 2 | 3+4 2 |
| | 6 2 6 2 | 7 2 5 2 | 3+4 2 |
| | 9 2 7 2 | 6 2 6 2 | 5+6 2 |
| | 8 2 6 2 | 7 2 6 2 | 3+4 2 |
| | 8 2 8 2 | 6 2 5 2 | 3+4 2 |
| | 6 2 9 2 | 7 2 5 2 | 5+6 2 |
| | 7 2 6 2 | 6 2 6 2 | 5+6 2 |
| | 6 2 8 2 | 6 2 6 2 | 5+6 2 |
| | 8 2 8 2 | 7 2 5 2 | 3+4 2 |
| | 6 2 6 2 | 6 2 6 2 | 3+4 2 |
| | 8 2 8 2 | 7 2 5 2 | 5+6 2 |

Continued on next column

**TABLE LXIV– continued**

| TQC | ICSD | NoE | SF |
|---|---|---|---|
| | 9 2 9 2 6 2 5 2 5+6 2 | | |
| | 9 2 7 2 6 2 6 2 3+4 2 | | |
| | 7 2 6 2 7 2 5 2 3+4 2 | | |
| | 8 2 9 2 7 2 6 2 5+6 2 | | |
| | 7 2 8 2 6 2 6 2 3+4 2 | | |
| | 6 2 9 2 6 2 5 2 5+6 2 | | |
| | 7 2 8 2 7 2 6 2 3+4 2 | | |
| | 6 2 9 2 6 2 6 2 3+4 2 | | |
| | 8 2 8 2 7 2 5 2 5+6 2 | | |
| | 6 2 7 2 6 2 5 2 5+6 2 | | |
| | 9 2 8 2 7 2 6 2 5+6 2 | | |
| | 6 2 9 2 7 2 6 2 3+4 2 | | |
| | 9 2 8 2 7 2 5 2 5+6 2 | | |
| | 9 2 7 2 6 2 5 2 3+4 2 | | |
| | 8 2 6 2 7 2 6 2 5+6 2 | | |
| | 7 2 7 2 6 2 6 2 5+6 2 | | |
| 139.1.122 | 425470 | 36 | Co2Si2Sr1 |
| | -GM -M -P -X -N | | |
| | 7 2 7 2 6 2 5 2 3+4 2 | | |
| | 9 2 9 2 7 2 6 2 5+6 2 | | |
| | 9 2 9 2 6 2 6 2 5+6 2 | | |
| | 8 2 8 2 7 2 6 2 5+6 2 | | |
| | 7 2 9 2 7 2 5 2 5+6 2 | | |
| | 9 2 7 2 6 2 6 2 3+4 2 | | |
| | 7 2 7 2 7 2 5 2 3+4 2 | | |
| | 8 2 6 2 6 2 6 2 5+6 2 | | |
| | 9 2 7 2 7 2 5 2 3+4 2 | | |
| | 6 2 8 2 6 2 6 2 5+6 2 | | |
| | 9 2 9 2 7 2 5 2 5+6 2 | | |
| | 7 2 9 2 6 2 5 2 3+4 2 | | |
| | 6 2 8 2 7 2 6 2 3+4 2 | | |
| | 8 2 6 2 6 2 6 2 5+6 2 | | |
| | 7 2 8 2 7 2 5 2 3+4 2 | | |
| | 6 2 7 2 6 2 5 2 5+6 2 | | |
| | 7 2 9 2 7 2 6 2 5+6 2 | | |
| | 6 2 8 2 7 2 5 2 3+4 2 | | |
| 139.1.123 | 73653 | 42 | Co2Nd1P2 |
| | -GM -M -P -X -N | | |
| | 7 2 7 2 6 2 5 2 3+4 2 | | |
| | 9 2 9 2 7 2 6 2 5+6 2 | | |
| | 9 2 9 2 6 2 6 2 5+6 2 | | |
| | 8 2 8 2 7 2 6 2 5+6 2 | | |
| | 7 2 7 2 7 2 5 2 5+6 2 | | |
| | 9 2 9 2 6 2 6 2 3+4 2 | | |
| | 7 2 7 2 7 2 5 2 3+4 2 | | |
| | 9 2 6 2 6 2 6 2 5+6 2 | | |
| | 8 2 7 2 7 2 5 2 3+4 2 | | |
| | 6 2 9 2 7 2 5 2 5+6 2 | | |
| | 7 2 8 2 6 2 6 2 5+6 2 | | |
| | 6 2 9 2 6 2 5 2 3+4 2 | | |
| | 9 2 8 2 7 2 6 2 3+4 2 | | |
| | 8 2 6 2 6 2 5 2 5+6 2 | | |
| | 7 2 8 2 7 2 6 2 3+4 2 | | |
| | 6 2 9 2 7 2 5 2 5+6 2 | | |
| | 7 2 7 2 6 2 6 2 5+6 2 | | |
| | 6 2 8 2 6 2 5 2 3+4 2 | | |
| | 6 2 7 2 7 2 6 2 3+4 2 | | |
| | 8 2 8 2 6 2 6 2 5+6 2 | | |
| | 9 2 8 2 7 2 6 2 5+6 2 | | |
| 139.1.124 | 73652 | 42 | Co2Nd1P2 |
| | -GM -M -P -X -N | | |

Continued on next column

**TABLE LXIV– continued**

| TQC | ICSD | NoE | SF |
|---|---|---|---|
| | 7 2 7 2 6 2 5 2 3+4 2 | | |
| | 9 2 9 2 7 2 6 2 5+6 2 | | |
| | 9 2 9 2 6 2 6 2 5+6 2 | | |
| | 8 2 8 2 7 2 6 2 5+6 2 | | |
| | 7 2 7 2 7 2 5 2 5+6 2 | | |
| | 9 2 9 2 6 2 6 2 3+4 2 | | |
| | 7 2 7 2 7 2 5 2 3+4 2 | | |
| | 9 2 6 2 6 2 6 2 5+6 2 | | |
| | 8 2 7 2 7 2 5 2 3+4 2 | | |
| | 6 2 9 2 7 2 5 2 5+6 2 | | |
| | 7 2 8 2 6 2 6 2 5+6 2 | | |
| | 6 2 9 2 6 2 5 2 3+4 2 | | |
| | 9 2 8 2 7 2 6 2 3+4 2 | | |
| | 8 2 6 2 6 2 5 2 5+6 2 | | |
| | 7 2 8 2 7 2 6 2 3+4 2 | | |
| | 6 2 9 2 7 2 5 2 5+6 2 | | |
| | 7 2 8 2 6 2 6 2 5+6 2 | | |
| | 6 2 7 2 6 2 5 2 3+4 2 | | |
| | 6 2 7 2 7 2 6 2 3+4 2 | | |
| | 8 2 8 2 6 2 6 2 5+6 2 | | |
| | 9 2 8 2 7 2 6 2 5+6 2 | | |
| 139.1.125 | 73651 | 42 | Co2Nd1P2 |
| | -GM -M -P -X -N | | |
| | 7 2 7 2 6 2 5 2 3+4 2 | | |
| | 9 2 9 2 7 2 6 2 5+6 2 | | |
| | 9 2 9 2 6 2 6 2 5+6 2 | | |
| | 8 2 8 2 7 2 6 2 5+6 2 | | |
| | 7 2 7 2 7 2 5 2 5+6 2 | | |
| | 9 2 9 2 6 2 6 2 3+4 2 | | |
| | 7 2 7 2 7 2 5 2 3+4 2 | | |
| | 9 2 6 2 6 2 6 2 5+6 2 | | |
| | 8 2 7 2 7 2 5 2 3+4 2 | | |
| | 6 2 9 2 7 2 5 2 5+6 2 | | |
| | 7 2 8 2 6 2 6 2 5+6 2 | | |
| | 6 2 9 2 6 2 5 2 3+4 2 | | |
| | 9 2 8 2 7 2 6 2 3+4 2 | | |
| | 8 2 6 2 6 2 5 2 5+6 2 | | |
| | 7 2 8 2 7 2 6 2 3+4 2 | | |
| | 6 2 9 2 7 2 5 2 5+6 2 | | |
| | 7 2 8 2 6 2 6 2 5+6 2 | | |
| | 6 2 7 2 6 2 5 2 3+4 2 | | |
| | 6 2 7 2 7 2 6 2 3+4 2 | | |
| | 8 2 8 2 6 2 6 2 5+6 2 | | |
| | 9 2 8 2 7 2 6 2 5+6 2 | | |
| 139.1.126 | 261192 | 70 | Co2La2O3Se2 |
| | -GM -M -P -X -N | | |
| | 7 2 9 2 6 2 5 2 3+4 2 | | |
| | 9 2 7 2 7 2 6 2 5+6 2 | | |
| | 7 2 9 2 6 2 5 2 5+6 2 | | |
| | 7 2 7 2 7 2 6 2 3+4 2 | | |
| | 8 2 6 2 7 2 5 2 5+6 2 | | |
| | 9 2 7 2 6 2 6 2 3+4 2 | | |
| | 7 2 9 2 7 2 5 2 5+6 2 | | |
| | 6 2 8 2 6 2 6 2 3+4 2 | | |
| | 9 2 7 2 7 2 5 2 5+6 2 | | |
| | 8 2 6 2 6 2 6 2 3+4 2 | | |
| | 7 2 9 2 7 2 5 2 5+6 2 | | |
| | 7 2 7 2 6 2 5 2 3+4 2 | | |
| | 9 2 9 2 7 2 6 2 5+6 2 | | |
| | 9 2 7 2 7 2 6 2 3+4 2 | | |
| | 7 2 9 2 6 2 6 2 3+4 2 | | |

Continued on next column

**TABLE LXIV– continued**

| TQC | ICSD | NoE | SF |
|---|---|---|---|
| | 9 2 8 2 | 6 2 5 2 | 3+4 2 |
| | 8 2 8 2 | 7 2 6 2 | 5+6 2 |
| | 6 2 9 2 | 6 2 6 2 | 3+4 2 |
| | 7 2 8 2 | 7 2 6 2 | 5+6 2 |
| | 6 2 9 2 | 6 2 6 2 | 3+4 2 |
| | 9 2 7 2 | 6 2 5 2 | 5+6 2 |
| | 9 2 7 2 | 7 2 5 2 | 3+4 2 |
| | 7 2 6 2 | 6 2 5 2 | 3+4 2 |
| | 8 2 7 2 | 6 2 5 2 | 5+6 2 |
| | 6 2 6 2 | 7 2 6 2 | 3+4 2 |
| | 6 2 9 2 | 7 2 6 2 | 5+6 2 |
| | 9 2 6 2 | 6 2 5 2 | 5+6 2 |
| | 8 2 9 2 | 7 2 6 2 | 3+4 2 |
| | 9 2 8 2 | 6 2 6 2 | 5+6 2 |
| | 7 2 7 2 | 7 2 6 2 | 5+6 2 |
| | 6 2 6 2 | 6 2 6 2 | 3+4 2 |
| | 7 2 7 2 | 7 2 6 2 | 5+6 2 |
| | 6 2 6 2 | 7 2 6 2 | 3+4 2 |
| | 6 2 6 2 | 6 2 6 2 | 5+6 2 |
| | 7 2 7 2 | 6 2 6 2 | 3+4 2 |
| 139.1.127 | 261191 | 70 | Co2La2O3Se2 |
| | -GM -M | -P -X | -N |
| | 7 2 9 2 | 6 2 5 2 | 3+4 2 |
| | 9 2 7 2 | 7 2 6 2 | 5+6 2 |
| | 7 2 9 2 | 6 2 5 2 | 5+6 2 |
| | 7 2 7 2 | 7 2 6 2 | 3+4 2 |
| | 8 2 6 2 | 7 2 5 2 | 5+6 2 |
| | 9 2 7 2 | 6 2 6 2 | 3+4 2 |
| | 7 2 9 2 | 7 2 5 2 | 5+6 2 |
| | 6 2 8 2 | 6 2 6 2 | 3+4 2 |
| | 9 2 7 2 | 7 2 5 2 | 5+6 2 |
| | 8 2 6 2 | 6 2 6 2 | 3+4 2 |
| | 7 2 9 2 | 7 2 5 2 | 5+6 2 |
| | 7 2 7 2 | 6 2 5 2 | 3+4 2 |
| | 9 2 9 2 | 7 2 6 2 | 5+6 2 |
| | 9 2 7 2 | 7 2 6 2 | 3+4 2 |
| | 7 2 9 2 | 6 2 6 2 | 3+4 2 |
| | 9 2 8 2 | 6 2 5 2 | 3+4 2 |
| | 8 2 8 2 | 7 2 6 2 | 5+6 2 |
| | 6 2 9 2 | 6 2 6 2 | 3+4 2 |
| | 7 2 8 2 | 7 2 6 2 | 5+6 2 |
| | 6 2 9 2 | 6 2 6 2 | 3+4 2 |
| | 9 2 7 2 | 6 2 5 2 | 5+6 2 |
| | 9 2 7 2 | 7 2 5 2 | 3+4 2 |
| | 7 2 6 2 | 6 2 5 2 | 3+4 2 |
| | 8 2 7 2 | 6 2 5 2 | 5+6 2 |
| | 6 2 6 2 | 7 2 6 2 | 3+4 2 |
| | 6 2 9 2 | 7 2 6 2 | 5+6 2 |
| | 9 2 6 2 | 6 2 5 2 | 5+6 2 |
| | 8 2 9 2 | 7 2 6 2 | 3+4 2 |
| | 9 2 8 2 | 6 2 6 2 | 5+6 2 |
| | 7 2 7 2 | 7 2 6 2 | 5+6 2 |
| | 6 2 6 2 | 6 2 6 2 | 3+4 2 |
| | 7 2 7 2 | 7 2 6 2 | 5+6 2 |
| | 6 2 6 2 | 7 2 6 2 | 3+4 2 |
| | 6 2 6 2 | 6 2 6 2 | 5+6 2 |
| | 7 2 7 2 | 6 2 6 2 | 3+4 2 |
| 139.1.128 | 248239 | 70 | Co2La2O3Se2 |
| | -GM -M | -P -X | -N |
| | 7 2 9 2 | 6 2 5 2 | 3+4 2 |
| | 9 2 7 2 | 7 2 6 2 | 5+6 2 |

Continued on next column

**TABLE LXIV– continued**

| TQC | ICSD | NoE | SF |
|---|---|---|---|
| | 7 2 9 2 | 6 2 5 2 | 3+4 2 |
| | 7 2 7 2 | 7 2 6 2 | 5+6 2 |
| | 8 2 6 2 | 7 2 5 2 | 5+6 2 |
| | 9 2 7 2 | 6 2 6 2 | 3+4 2 |
| | 7 2 9 2 | 7 2 5 2 | 5+6 2 |
| | 6 2 8 2 | 6 2 6 2 | 3+4 2 |
| | 9 2 7 2 | 7 2 5 2 | 5+6 2 |
| | 8 2 6 2 | 6 2 6 2 | 3+4 2 |
| | 7 2 9 2 | 7 2 5 2 | 5+6 2 |
| | 7 2 7 2 | 6 2 5 2 | 3+4 2 |
| | 9 2 9 2 | 7 2 6 2 | 5+6 2 |
| | 9 2 7 2 | 7 2 6 2 | 3+4 2 |
| | 7 2 9 2 | 6 2 6 2 | 3+4 2 |
| | 9 2 8 2 | 6 2 5 2 | 3+4 2 |
| | 8 2 8 2 | 7 2 6 2 | 5+6 2 |
| | 6 2 9 2 | 6 2 6 2 | 3+4 2 |
| | 7 2 8 2 | 7 2 6 2 | 5+6 2 |
| | 6 2 9 2 | 6 2 6 2 | 3+4 2 |
| | 9 2 7 2 | 6 2 5 2 | 5+6 2 |
| | 9 2 7 2 | 7 2 5 2 | 3+4 2 |
| | 8 2 6 2 | 6 2 5 2 | 3+4 2 |
| | 7 2 7 2 | 6 2 5 2 | 5+6 2 |
| | 6 2 6 2 | 7 2 6 2 | 3+4 2 |
| | 6 2 9 2 | 7 2 6 2 | 5+6 2 |
| | 9 2 6 2 | 6 2 5 2 | 5+6 2 |
| | 8 2 9 2 | 7 2 6 2 | 3+4 2 |
| | 9 2 8 2 | 6 2 6 2 | 5+6 2 |
| | 7 2 7 2 | 7 2 6 2 | 5+6 2 |
| | 6 2 6 2 | 6 2 6 2 | 3+4 2 |
| | 7 2 7 2 | 7 2 6 2 | 5+6 2 |
| | 6 2 6 2 | 7 2 6 2 | 3+4 2 |
| | 6 2 6 2 | 6 2 6 2 | 5+6 2 |
| | 7 2 7 2 | 6 2 6 2 | 3+4 2 |
| 139.1.129 | 168330 | 70 | Co2La2O3Se2 |
| | -GM -M | -P -X | -N |
| | 7 2 9 2 | 6 2 5 2 | 3+4 2 |
| | 9 2 7 2 | 7 2 6 2 | 5+6 2 |
| | 7 2 9 2 | 6 2 5 2 | 5+6 2 |
| | 7 2 7 2 | 7 2 6 2 | 3+4 2 |
| | 8 2 6 2 | 7 2 5 2 | 5+6 2 |
| | 9 2 7 2 | 6 2 6 2 | 3+4 2 |
| | 7 2 9 2 | 7 2 5 2 | 5+6 2 |
| | 6 2 8 2 | 6 2 6 2 | 3+4 2 |
| | 9 2 7 2 | 7 2 5 2 | 5+6 2 |
| | 8 2 6 2 | 6 2 6 2 | 3+4 2 |
| | 7 2 9 2 | 7 2 5 2 | 5+6 2 |
| | 7 2 7 2 | 6 2 5 2 | 3+4 2 |
| | 9 2 9 2 | 7 2 6 2 | 5+6 2 |
| | 9 2 7 2 | 7 2 6 2 | 3+4 2 |
| | 7 2 9 2 | 6 2 6 2 | 3+4 2 |
| | 9 2 8 2 | 6 2 5 2 | 3+4 2 |
| | 8 2 8 2 | 7 2 6 2 | 5+6 2 |
| | 6 2 9 2 | 6 2 6 2 | 3+4 2 |
| | 7 2 8 2 | 7 2 6 2 | 5+6 2 |
| | 6 2 9 2 | 6 2 6 2 | 3+4 2 |
| | 9 2 7 2 | 6 2 5 2 | 5+6 2 |
| | 9 2 7 2 | 7 2 5 2 | 3+4 2 |
| | 7 2 6 2 | 6 2 5 2 | 3+4 2 |
| | 8 2 7 2 | 6 2 5 2 | 5+6 2 |
| | 6 2 6 2 | 7 2 6 2 | 3+4 2 |
| | 6 2 9 2 | 7 2 6 2 | 5+6 2 |

Continued on next column

**TABLE LXIV– continued**

| TQC | ICSD | NoE | SF |
|---|---|---|---|
| | 9 2 6 2 | 6 2 5 2 | 5+6 2 |
| | 8 2 9 2 | 7 2 6 2 | 3+4 2 |
| | 9 2 8 2 | 6 2 6 2 | 5+6 2 |
| | 7 2 7 2 | 7 2 6 2 | 5+6 2 |
| | 6 2 6 2 | 6 2 6 2 | 3+4 2 |
| | 7 2 7 2 | 7 2 6 2 | 5+6 2 |
| | 6 2 6 2 | 7 2 6 2 | 3+4 2 |
| | 6 2 6 2 | 6 2 6 2 | 5+6 2 |
| | 7 2 7 2 | 6 2 6 2 | 3+4 2 |
| 139.1.130 | 407 | 36 | Co2Ge2Sr1 |
| | -GM -M | -P -X | -N |
| | 7 2 7 2 | 6 2 5 2 | 3+4 2 |
| | 9 2 9 2 | 7 2 6 2 | 5+6 2 |
| | 8 2 9 2 | 6 2 6 2 | 5+6 2 |
| | 9 2 8 2 | 7 2 6 2 | 5+6 2 |
| | 7 2 9 2 | 7 2 5 2 | 5+6 2 |
| | 9 2 7 2 | 6 2 6 2 | 3+4 2 |
| | 7 2 7 2 | 7 2 5 2 | 3+4 2 |
| | 9 2 6 2 | 6 2 6 2 | 5+6 2 |
| | 8 2 7 2 | 7 2 5 2 | 3+4 2 |
| | 9 2 8 2 | 6 2 6 2 | 5+6 2 |
| | 6 2 9 2 | 7 2 5 2 | 5+6 2 |
| | 7 2 9 2 | 6 2 5 2 | 3+4 2 |
| | 6 2 8 2 | 7 2 6 2 | 3+4 2 |
| | 8 2 6 2 | 6 2 6 2 | 5+6 2 |
| | 7 2 8 2 | 7 2 5 2 | 3+4 2 |
| | 6 2 7 2 | 7 2 5 2 | 5+6 2 |
| | 7 2 9 2 | 6 2 6 2 | 5+6 2 |
| | 6 2 8 2 | 7 2 5 2 | 3+4 2 |
| 139.1.131 | 33524 | 94 | Co2F7K3 |
| | -GM -M | -P -X | -N |
| | 7 2 9 2 | 6 2 5 2 | 3+4 2 |
| | 9 2 7 2 | 7 2 6 2 | 5+6 2 |
| | 7 2 7 2 | 7 2 5 2 | 5+6 2 |
| | 7 2 7 2 | 7 2 6 2 | 3+4 2 |
| | 9 2 9 2 | 6 2 6 2 | 5+6 2 |
| | 6 2 6 2 | 6 2 5 2 | 3+4 2 |
| | 8 2 8 2 | 7 2 5 2 | 5+6 2 |
| | 7 2 7 2 | 6 2 5 2 | 3+4 2 |
| | 7 2 9 2 | 7 2 6 2 | 3+4 2 |
| | 9 2 7 2 | 6 2 5 2 | 5+6 2 |
| | 7 2 9 2 | 6 2 6 2 | 3+4 2 |
| | 9 2 7 2 | 7 2 5 2 | 5+6 2 |
| | 7 2 9 2 | 6 2 6 2 | 3+4 2 |
| | 6 2 8 2 | 7 2 5 2 | 5+6 2 |
| | 8 2 6 2 | 6 2 6 2 | 3+4 2 |
| | 9 2 7 2 | 7 2 5 2 | 5+6 2 |
| | 9 2 9 2 | 6 2 6 2 | 3+4 2 |
| | 9 2 9 2 | 7 2 6 2 | 3+4 2 |
| | 8 2 8 2 | 6 2 6 2 | 3+4 2 |
| | 9 2 9 2 | 7 2 6 2 | 5+6 2 |
| | 9 2 9 2 | 6 2 5 2 | 3+4 2 |
| | 8 2 8 2 | 7 2 5 2 | 5+6 2 |
| | 6 2 6 2 | 6 2 6 2 | 5+6 2 |
| | 7 2 7 2 | 7 2 6 2 | 3+4 2 |
| | 9 2 7 2 | 7 2 5 2 | 5+6 2 |
| | 7 2 9 2 | 6 2 6 2 | 3+4 2 |
| | 7 2 7 2 | 7 2 6 2 | 5+6 2 |
| | 6 2 6 2 | 6 2 6 2 | 3+4 2 |
| | 9 2 9 2 | 6 2 5 2 | 5+6 2 |
| | 8 2 8 2 | 7 2 5 2 | 5+6 2 |

Continued on next column

**TABLE LXIV– continued**

| TQC | ICSD | NoE | SF |
|---|---|---|---|
| | 8 2 8 2 | 6 2 5 2 | 3+4 2 |
| | 7 2 7 2 | 7 2 6 2 | 3+4 2 |
| | 6 2 9 2 | 7 2 6 2 | 5+6 2 |
| | 9 2 6 2 | 6 2 6 2 | 3+4 2 |
| | 8 2 8 2 | 6 2 5 2 | 5+6 2 |
| | 7 2 9 2 | 7 2 5 2 | 3+4 2 |
| | 6 2 7 2 | 6 2 5 2 | 5+6 2 |
| | 9 2 6 2 | 7 2 6 2 | 5+6 2 |
| | 9 2 9 2 | 7 2 6 2 | 3+4 2 |
| | 8 2 8 2 | 6 2 6 2 | 5+6 2 |
| | 6 2 6 2 | 6 2 5 2 | 3+4 2 |
| | 8 2 8 2 | 7 2 5 2 | 5+6 2 |
| | 7 2 7 2 | 7 2 6 2 | 3+4 2 |
| | 6 2 6 2 | 6 2 6 2 | 5+6 2 |
| | 9 2 9 2 | 7 2 5 2 | 3+4 2 |
| | 8 2 8 2 | 6 2 6 2 | 3+4 2 |
| | 6 2 6 2 | 6 2 5 2 | 5+6 2 |
| 139.1.132 | 30081 | 90 | Cl7Mn2Rb3 |
| | -GM -M | -P -X | -N |
| | 7 2 7 2 | 7 2 5 2 | 5+6 2 |
| | 7 2 9 2 | 6 2 5 2 | 3+4 2 |
| | 9 2 7 2 | 7 2 6 2 | 5+6 2 |
| | 9 2 9 2 | 6 2 6 2 | 3+4 2 |
| | 7 2 7 2 | 7 2 6 2 | 5+6 2 |
| | 6 2 6 2 | 6 2 5 2 | 3+4 2 |
| | 8 2 8 2 | 7 2 5 2 | 5+6 2 |
| | 7 2 9 2 | 6 2 5 2 | 3+4 2 |
| | 9 2 7 2 | 7 2 6 2 | 5+6 2 |
| | 7 2 7 2 | 6 2 5 2 | 3+4 2 |
| | 9 2 9 2 | 6 2 6 2 | 3+4 2 |
| | 7 2 9 2 | 6 2 5 2 | 3+4 2 |
| | 9 2 7 2 | 7 2 6 2 | 5+6 2 |
| | 9 2 9 2 | 6 2 6 2 | 3+4 2 |
| | 8 2 8 2 | 7 2 6 2 | 3+4 2 |
| | 6 2 8 2 | 6 2 5 2 | 3+4 2 |
| | 7 2 9 2 | 7 2 6 2 | 5+6 2 |
| | 8 2 6 2 | 6 2 5 2 | 3+4 2 |
| | 9 2 7 2 | 7 2 6 2 | 5+6 2 |
| | 6 2 6 2 | 7 2 5 2 | 5+6 2 |
| | 7 2 7 2 | 6 2 6 2 | 3+4 2 |
| | 8 2 8 2 | 7 2 6 2 | 5+6 2 |
| | 9 2 9 2 | 6 2 5 2 | 3+4 2 |
| | 9 2 9 2 | 7 2 6 2 | 5+6 2 |
| | 7 2 7 2 | 6 2 5 2 | 3+4 2 |
| | 7 2 9 2 | 7 2 6 2 | 5+6 2 |
| | 9 2 7 2 | 6 2 5 2 | 3+4 2 |
| | 7 2 9 2 | 6 2 6 2 | 5+6 2 |
| | 6 2 6 2 | 7 2 5 2 | 5+6 2 |
| | 6 2 8 2 | 6 2 6 2 | 3+4 2 |
| | 9 2 7 2 | 7 2 5 2 | 5+6 2 |
| | 8 2 8 2 | 6 2 6 2 | 3+4 2 |
| | 8 2 6 2 | 7 2 5 2 | 5+6 2 |
| | 9 2 9 2 | 7 2 6 2 | 3+4 2 |
| | 7 2 7 2 | 6 2 5 2 | 3+4 2 |
| | 8 2 8 2 | 7 2 5 2 | 5+6 2 |
| | 6 2 6 2 | 6 2 6 2 | 3+4 2 |
| | 9 2 9 2 | 7 2 6 2 | 5+6 2 |
| | 6 2 8 2 | 6 2 6 2 | 5+6 2 |
| | 8 2 6 2 | 7 2 6 2 | 5+6 2 |
| | 8 2 8 2 | 7 2 5 2 | 5+6 2 |
| | 9 2 9 2 | 6 2 5 2 | 3+4 2 |

Continued on next column

**TABLE LXIV– continued**

| TQC | ICSD | NoE | SF |
|---|---|---|---|
| | 7 2 7 2 | 7 2 5 2 | 5+6 2 |
| | 6 2 6 2 | 6 2 6 2 | 3+4 2 |
| | 9 2 9 2 | 7 2 6 2 | 5+6 2 |
| 139.1.133 | 2241 | 90 | Cl7Mn2Rb3 |
| | -GM -M | -P -X | -N |
| | 7 2 9 2 | 6 2 5 2 | 3+4 2 |
| | 9 2 7 2 | 7 2 6 2 | 5+6 2 |
| | 7 2 7 2 | 6 2 5 2 | 3+4 2 |
| | 7 2 7 2 | 6 2 6 2 | 5+6 2 |
| | 9 2 9 2 | 7 2 6 2 | 3+4 2 |
| | 6 2 6 2 | 6 2 5 2 | 5+6 2 |
| | 8 2 8 2 | 7 2 5 2 | 3+4 2 |
| | 7 2 9 2 | 7 2 5 2 | 5+6 2 |
| | 9 2 7 2 | 6 2 6 2 | 3+4 2 |
| | 7 2 7 2 | 7 2 5 2 | 5+6 2 |
| | 7 2 9 2 | 7 2 6 2 | 3+4 2 |
| | 9 2 7 2 | 6 2 5 2 | 5+6 2 |
| | 9 2 9 2 | 7 2 6 2 | 5+6 2 |
| | 7 2 9 2 | 7 2 6 2 | 3+4 2 |
| | 6 2 8 2 | 6 2 5 2 | 5+6 2 |
| | 8 2 6 2 | 6 2 6 2 | 3+4 2 |
| | 9 2 7 2 | 7 2 5 2 | 5+6 2 |
| | 9 2 9 2 | 6 2 6 2 | 5+6 2 |
| | 8 2 8 2 | 7 2 6 2 | 5+6 2 |
| | 8 2 8 2 | 6 2 6 2 | 3+4 2 |
| | 9 2 9 2 | 7 2 5 2 | 5+6 2 |
| | 9 2 9 2 | 6 2 5 2 | 3+4 2 |
| | 6 2 6 2 | 7 2 6 2 | 5+6 2 |
| | 7 2 7 2 | 6 2 5 2 | 3+4 2 |
| | 7 2 7 2 | 7 2 6 2 | 3+4 2 |
| | 9 2 9 2 | 6 2 6 2 | 5+6 2 |
| | 7 2 7 2 | 7 2 6 2 | 3+4 2 |
| | 9 2 9 2 | 6 2 6 2 | 5+6 2 |
| | 7 2 8 2 | 7 2 5 2 | 3+4 2 |
| | 6 2 6 2 | 6 2 5 2 | 3+4 2 |
| | 8 2 7 2 | 7 2 5 2 | 5+6 2 |
| | 6 2 8 2 | 6 2 6 2 | 5+6 2 |
| | 8 2 6 2 | 6 2 6 2 | 3+4 2 |
| | 9 2 9 2 | 7 2 6 2 | 5+6 2 |
| | 8 2 8 2 | 7 2 5 2 | 3+4 2 |
| | 7 2 7 2 | 6 2 5 2 | 5+6 2 |
| | 6 2 9 2 | 6 2 6 2 | 3+4 2 |
| | 9 2 6 2 | 7 2 5 2 | 5+6 2 |
| | 9 2 9 2 | 6 2 6 2 | 3+4 2 |
| | 8 2 8 2 | 7 2 6 2 | 3+4 2 |
| | 6 2 6 2 | 7 2 5 2 | 5+6 2 |
| | 8 2 8 2 | 6 2 5 2 | 3+4 2 |
| | 7 2 7 2 | 6 2 6 2 | 5+6 2 |
| | 6 2 6 2 | 7 2 6 2 | 3+4 2 |
| | 8 2 8 2 | 6 2 5 2 | 5+6 2 |
| 139.1.134 | 36127 | 90 | Cl7K3Mn2 |
| | -GM -M | -P -X | -N |
| | 7 2 9 2 | 6 2 5 2 | 3+4 2 |
| | 9 2 7 2 | 7 2 6 2 | 5+6 2 |
| | 7 2 7 2 | 6 2 5 2 | 3+4 2 |
| | 7 2 7 2 | 7 2 5 2 | 5+6 2 |
| | 9 2 9 2 | 6 2 6 2 | 3+4 2 |
| | 7 2 9 2 | 6 2 6 2 | 5+6 2 |
| | 9 2 7 2 | 7 2 6 2 | 3+4 2 |
| | 6 2 6 2 | 6 2 5 2 | 5+6 2 |
| | 8 2 8 2 | 7 2 5 2 | 3+4 2 |
| | 7 2 7 2 | 7 2 5 2 | 5+6 2 |
| | 7 2 9 2 | 6 2 5 2 | 3+4 2 |
| | 9 2 7 2 | 7 2 6 2 | 5+6 2 |
| | 6 2 8 2 | 6 2 5 2 | 3+4 2 |
| | 8 2 6 2 | 7 2 6 2 | 5+6 2 |
| | 7 2 9 2 | 6 2 5 2 | 3+4 2 |
| | 9 2 7 2 | 7 2 6 2 | 5+6 2 |
| | 9 2 9 2 | 7 2 6 2 | 5+6 2 |
| | 8 2 8 2 | 7 2 6 2 | 5+6 2 |
| | 9 2 9 2 | 6 2 6 2 | 5+6 2 |
| | 9 2 9 2 | 6 2 6 2 | 3+4 2 |
| | 8 2 8 2 | 7 2 5 2 | 5+6 2 |
| | 9 2 9 2 | 6 2 5 2 | 3+4 2 |
| | 7 2 6 2 | 7 2 6 2 | 3+4 2 |
| | 6 2 7 2 | 6 2 6 2 | 5+6 2 |
| | 7 2 7 2 | 7 2 6 2 | 5+6 2 |
| | 9 2 9 2 | 6 2 5 2 | 3+4 2 |
| | 9 2 9 2 | 7 2 6 2 | 3+4 2 |
| | 7 2 8 2 | 6 2 5 2 | 5+6 2 |
| | 8 2 7 2 | 7 2 5 2 | 3+4 2 |
| | 7 2 7 2 | 7 2 6 2 | 3+4 2 |
| | 6 2 6 2 | 6 2 5 2 | 3+4 2 |
| | 6 2 6 2 | 6 2 6 2 | 5+6 2 |
| | 8 2 8 2 | 6 2 6 2 | 5+6 2 |
| | 9 2 9 2 | 7 2 6 2 | 5+6 2 |
| | 8 2 8 2 | 7 2 5 2 | 5+6 2 |
| | 7 2 9 2 | 6 2 5 2 | 3+4 2 |
| | 9 2 7 2 | 6 2 6 2 | 3+4 2 |
| | 6 2 6 2 | 7 2 5 2 | 3+4 2 |
| | 9 2 9 2 | 7 2 6 2 | 5+6 2 |
| | 8 2 8 2 | 6 2 6 2 | 3+4 2 |
| | 6 2 6 2 | 7 2 5 2 | 5+6 2 |
| | 8 2 8 2 | 6 2 5 2 | 3+4 2 |
| | 7 2 7 2 | 6 2 6 2 | 5+6 2 |
| | 6 2 6 2 | 7 2 6 2 | 3+4 2 |
| | 8 2 8 2 | 6 2 5 2 | 5+6 2 |
| 139.1.135 | 99942 | 90 | Cl2Fe2O5Sr3 |
| | -GM -M | -P -X | -N |
| | 7 2 7 2 | 6 2 5 2 | 3+4 2 |
| | 7 2 7 2 | 6 2 5 2 | 3+4 2 |
| | 9 2 9 2 | 7 2 6 2 | 5+6 2 |
| | 7 2 7 2 | 7 2 5 2 | 5+6 2 |
| | 9 2 9 2 | 7 2 6 2 | 3+4 2 |
| | 7 2 7 2 | 6 2 5 2 | 5+6 2 |
| | 6 2 6 2 | 6 2 6 2 | 5+6 2 |
| | 8 2 8 2 | 7 2 5 2 | 3+4 2 |
| | 9 2 9 2 | 7 2 6 2 | 5+6 2 |
| | 7 2 9 2 | 6 2 5 2 | 5+6 2 |
| | 9 2 7 2 | 7 2 6 2 | 3+4 2 |
| | 8 2 9 2 | 6 2 6 2 | 5+6 2 |
| | 7 2 8 2 | 7 2 6 2 | 3+4 2 |
| | 9 2 9 2 | 6 2 5 2 | 5+6 2 |
| | 6 2 8 2 | 7 2 6 2 | 5+6 2 |
| | 8 2 6 2 | 7 2 5 2 | 3+4 2 |
| | 9 2 7 2 | 6 2 6 2 | 5+6 2 |
| | 9 2 7 2 | 6 2 5 2 | 3+4 2 |
| | 7 2 9 2 | 7 2 6 2 | 5+6 2 |
| | 9 2 9 2 | 6 2 6 2 | 3+4 2 |
| | 7 2 9 2 | 7 2 5 2 | 5+6 2 |
| | 9 2 7 2 | 6 2 6 2 | 3+4 2 |
| | 7 2 7 2 | 6 2 5 2 | 3+4 2 |

Continued on next column

**TABLE LXIV– continued**

| TQC | ICSD | NoE | SF |
|---|---|---|---|
| | 6 2 9 2 | 6 2 5 2 | 5+6 2 |
| | 8 2 6 2 | 7 2 6 2 | 3+4 2 |
| | 9 2 8 2 | 7 2 6 2 | 3+4 2 |
| | 9 2 9 2 | 6 2 5 2 | 5+6 2 |
| | 7 2 8 2 | 7 2 6 2 | 3+4 2 |
| | 8 2 9 2 | 7 2 6 2 | 5+6 2 |
| | 6 2 8 2 | 6 2 6 2 | 3+4 2 |
| | 7 2 6 2 | 7 2 5 2 | 3+4 2 |
| | 6 2 7 2 | 6 2 6 2 | 5+6 2 |
| | 9 2 9 2 | 6 2 5 2 | 5+6 2 |
| | 8 2 6 2 | 7 2 5 2 | 3+4 2 |
| | 8 2 8 2 | 7 2 6 2 | 5+6 2 |
| | 9 2 7 2 | 6 2 5 2 | 3+4 2 |
| | 7 2 7 2 | 7 2 6 2 | 3+4 2 |
| | 6 2 6 2 | 6 2 6 2 | 5+6 2 |
| | 9 2 9 2 | 7 2 5 2 | 5+6 2 |
| | 8 2 8 2 | 6 2 6 2 | 3+4 2 |
| | 6 2 6 2 | 7 2 5 2 | 5+6 2 |
| | 8 2 8 2 | 6 2 5 2 | 5+6 2 |
| | 7 2 7 2 | 7 2 5 2 | 5+6 2 |
| | 6 2 6 2 | 6 2 5 2 | 3+4 2 |
| | 7 2 7 2 | 6 2 6 2 | 5+6 2 |
| 139.1.136 | 69871 | 90 | Cl2Fe2O5Sr3 |
| | -GM -M | -P -X | -N |
| | 7 2 7 2 | 6 2 5 2 | 3+4 2 |
| | 7 2 7 2 | 6 2 5 2 | 3+4 2 |
| | 9 2 9 2 | 7 2 6 2 | 5+6 2 |
| | 7 2 7 2 | 7 2 5 2 | 5+6 2 |
| | 9 2 9 2 | 7 2 6 2 | 3+4 2 |
| | 7 2 7 2 | 6 2 5 2 | 5+6 2 |
| | 9 2 9 2 | 6 2 6 2 | 5+6 2 |
| | 6 2 6 2 | 7 2 5 2 | 3+4 2 |
| | 8 2 8 2 | 7 2 6 2 | 5+6 2 |
| | 7 2 9 2 | 6 2 5 2 | 5+6 2 |
| | 9 2 7 2 | 7 2 6 2 | 3+4 2 |
| | 8 2 8 2 | 6 2 6 2 | 5+6 2 |
| | 9 2 9 2 | 7 2 6 2 | 5+6 2 |
| | 7 2 9 2 | 6 2 5 2 | 3+4 2 |
| | 6 2 8 2 | 7 2 6 2 | 5+6 2 |
| | 8 2 6 2 | 7 2 5 2 | 3+4 2 |
| | 9 2 7 2 | 6 2 6 2 | 5+6 2 |
| | 9 2 7 2 | 6 2 5 2 | 3+4 2 |
| | 7 2 9 2 | 7 2 6 2 | 5+6 2 |
| | 9 2 9 2 | 6 2 6 2 | 3+4 2 |
| | 7 2 9 2 | 7 2 5 2 | 5+6 2 |
| | 7 2 7 2 | 6 2 6 2 | 3+4 2 |
| | 9 2 7 2 | 6 2 5 2 | 3+4 2 |
| | 6 2 9 2 | 6 2 5 2 | 5+6 2 |
| | 8 2 6 2 | 7 2 6 2 | 3+4 2 |
| | 9 2 8 2 | 7 2 6 2 | 3+4 2 |
| | 9 2 9 2 | 7 2 6 2 | 5+6 2 |
| | 8 2 8 2 | 6 2 6 2 | 3+4 2 |
| | 7 2 9 2 | 6 2 5 2 | 5+6 2 |
| | 6 2 8 2 | 7 2 6 2 | 3+4 2 |
| | 7 2 6 2 | 7 2 5 2 | 3+4 2 |
| | 6 2 9 2 | 6 2 6 2 | 5+6 2 |
| | 9 2 7 2 | 6 2 5 2 | 5+6 2 |
| | 8 2 6 2 | 7 2 5 2 | 3+4 2 |
| | 8 2 8 2 | 7 2 6 2 | 5+6 2 |
| | 9 2 7 2 | 6 2 5 2 | 3+4 2 |
| | 7 2 7 2 | 7 2 6 2 | 3+4 2 |

Continued on next column

**TABLE LXIV– continued**

| TQC | ICSD | NoE | SF |
|---|---|---|---|
| | 6 2 6 2 | 6 2 6 2 | 5+6 2 |
| | 9 2 9 2 | 7 2 5 2 | 5+6 2 |
| | 8 2 8 2 | 6 2 6 2 | 3+4 2 |
| | 6 2 6 2 | 7 2 5 2 | 5+6 2 |
| | 8 2 8 2 | 6 2 5 2 | 5+6 2 |
| | 7 2 7 2 | 7 2 5 2 | 5+6 2 |
| | 6 2 6 2 | 6 2 6 2 | 3+4 2 |
| | 7 2 7 2 | 6 2 5 2 | 5+6 2 |
| 139.1.137 | 33258 | 90 | Cl2Fe2O5Sr3 |
| | -GM -M | -P -X | -N |
| | 7 2 7 2 | 7 2 5 2 | 5+6 2 |
| | 7 2 7 2 | 6 2 5 2 | 3+4 2 |
| | 9 2 9 2 | 7 2 6 2 | 5+6 2 |
| | 7 2 7 2 | 6 2 5 2 | 3+4 2 |
| | 9 2 9 2 | 6 2 6 2 | 5+6 2 |
| | 7 2 7 2 | 7 2 5 2 | 3+4 2 |
| | 6 2 6 2 | 7 2 6 2 | 3+4 2 |
| | 8 2 8 2 | 6 2 5 2 | 5+6 2 |
| | 9 2 9 2 | 6 2 6 2 | 3+4 2 |
| | 7 2 9 2 | 7 2 5 2 | 3+4 2 |
| | 9 2 7 2 | 6 2 6 2 | 5+6 2 |
| | 8 2 8 2 | 7 2 6 2 | 3+4 2 |
| | 7 2 9 2 | 6 2 6 2 | 3+4 2 |
| | 9 2 9 2 | 7 2 5 2 | 5+6 2 |
| | 6 2 8 2 | 6 2 6 2 | 3+4 2 |
| | 8 2 6 2 | 6 2 5 2 | 5+6 2 |
| | 9 2 7 2 | 7 2 6 2 | 3+4 2 |
| | 9 2 7 2 | 7 2 5 2 | 5+6 2 |
| | 7 2 9 2 | 6 2 6 2 | 3+4 2 |
| | 9 2 9 2 | 7 2 6 2 | 5+6 2 |
| | 7 2 9 2 | 6 2 5 2 | 3+4 2 |
| | 7 2 7 2 | 7 2 6 2 | 5+6 2 |
| | 9 2 7 2 | 7 2 5 2 | 5+6 2 |
| | 6 2 9 2 | 7 2 5 2 | 3+4 2 |
| | 8 2 6 2 | 6 2 6 2 | 5+6 2 |
| | 9 2 8 2 | 6 2 6 2 | 5+6 2 |
| | 9 2 9 2 | 6 2 6 2 | 3+4 2 |
| | 8 2 8 2 | 7 2 5 2 | 5+6 2 |
| | 7 2 9 2 | 6 2 6 2 | 3+4 2 |
| | 6 2 8 2 | 7 2 6 2 | 5+6 2 |
| | 7 2 6 2 | 6 2 5 2 | 5+6 2 |
| | 6 2 9 2 | 7 2 6 2 | 3+4 2 |
| | 9 2 7 2 | 7 2 5 2 | 3+4 2 |
| | 8 2 6 2 | 6 2 5 2 | 5+6 2 |
| | 8 2 8 2 | 6 2 6 2 | 3+4 2 |
| | 9 2 7 2 | 7 2 5 2 | 5+6 2 |
| | 7 2 7 2 | 6 2 6 2 | 5+6 2 |
| | 6 2 6 2 | 7 2 6 2 | 3+4 2 |
| | 9 2 9 2 | 6 2 5 2 | 3+4 2 |
| | 8 2 8 2 | 7 2 6 2 | 5+6 2 |
| | 6 2 6 2 | 6 2 5 2 | 3+4 2 |
| | 8 2 8 2 | 7 2 5 2 | 3+4 2 |
| | 7 2 7 2 | 6 2 5 2 | 3+4 2 |
| | 6 2 6 2 | 7 2 6 2 | 5+6 2 |
| | 7 2 7 2 | 7 2 5 2 | 3+4 2 |
| 139.1.138 | 174532 | 84 | Cl2Fe2O4Sr3 |
| | -GM -M | -P -X | -N |
| | 7 2 7 2 | 6 2 5 2 | 3+4 2 |
| | 9 2 9 2 | 7 2 6 2 | 5+6 2 |
| | 7 2 7 2 | 6 2 5 2 | 3+4 2 |
| | 9 2 9 2 | 7 2 6 2 | 5+6 2 |

Continued on next column

**TABLE LXIV– continued**

| TQC | ICSD | NoE | SF |
|---|---|---|---|
| | 7 2 7 2 | 6 2 6 2 | 3+4 2 |
| | 6 2 6 2 | 6 2 5 2 | 5+6 2 |
| | 8 2 8 2 | 7 2 5 2 | 3+4 2 |
| | 7 2 9 2 | 7 2 6 2 | 5+6 2 |
| | 9 2 7 2 | 6 2 5 2 | 3+4 2 |
| | 7 2 9 2 | 7 2 6 2 | 5+6 2 |
| | 9 2 9 2 | 7 2 6 2 | 3+4 2 |
| | 6 2 8 2 | 6 2 6 2 | 5+6 2 |
| | 8 2 6 2 | 6 2 5 2 | 3+4 2 |
| | 9 2 7 2 | 7 2 5 2 | 5+6 2 |
| | 8 2 9 2 | 7 2 6 2 | 5+6 2 |
| | 7 2 8 2 | 6 2 5 2 | 5+6 2 |
| | 9 2 7 2 | 7 2 6 2 | 3+4 2 |
| | 9 2 9 2 | 6 2 6 2 | 5+6 2 |
| | 7 2 9 2 | 7 2 5 2 | 5+6 2 |
| | 7 2 7 2 | 6 2 6 2 | 3+4 2 |
| | 6 2 7 2 | 6 2 5 2 | 3+4 2 |
| | 9 2 6 2 | 7 2 5 2 | 5+6 2 |
| | 9 2 9 2 | 6 2 6 2 | 3+4 2 |
| | 8 2 8 2 | 6 2 5 2 | 3+4 2 |
| | 7 2 9 2 | 7 2 6 2 | 5+6 2 |
| | 7 2 8 2 | 7 2 6 2 | 3+4 2 |
| | 6 2 7 2 | 6 2 5 2 | 5+6 2 |
| | 6 2 6 2 | 7 2 6 2 | 5+6 2 |
| | 9 2 7 2 | 7 2 6 2 | 3+4 2 |
| | 8 2 6 2 | 6 2 5 2 | 3+4 2 |
| | 8 2 9 2 | 7 2 6 2 | 5+6 2 |
| | 9 2 8 2 | 6 2 5 2 | 3+4 2 |
| | 7 2 7 2 | 7 2 6 2 | 5+6 2 |
| | 6 2 6 2 | 6 2 5 2 | 5+6 2 |
| | 9 2 9 2 | 7 2 5 2 | 3+4 2 |
| | 8 2 8 2 | 6 2 6 2 | 5+6 2 |
| | 6 2 6 2 | 7 2 5 2 | 3+4 2 |
| | 8 2 8 2 | 6 2 6 2 | 5+6 2 |
| | 7 2 7 2 | 6 2 5 2 | 5+6 2 |
| | 9 2 9 2 | 7 2 6 2 | 3+4 2 |
| | 8 2 8 2 | 6 2 6 2 | 3+4 2 |
| | 7 2 7 2 | 7 2 5 2 | 5+6 2 |
| 139.1.139 | 91749 | 92 | Cl2Co2O5Sr3 |
| | -GM -M | -P -X | -N |
| | 7 2 7 2 | 6 2 5 2 | 3+4 2 |
| | 7 2 7 2 | 6 2 5 2 | 3+4 2 |
| | 9 2 9 2 | 7 2 6 2 | 5+6 2 |
| | 9 2 9 2 | 7 2 6 2 | 5+6 2 |
| | 7 2 7 2 | 6 2 5 2 | 3+4 2 |
| | 7 2 7 2 | 7 2 5 2 | 5+6 2 |
| | 6 2 6 2 | 6 2 6 2 | 5+6 2 |
| | 8 2 8 2 | 7 2 5 2 | 3+4 2 |
| | 9 2 9 2 | 7 2 6 2 | 5+6 2 |
| | 7 2 9 2 | 6 2 5 2 | 5+6 2 |
| | 9 2 7 2 | 7 2 6 2 | 3+4 2 |
| | 7 2 9 2 | 6 2 6 2 | 5+6 2 |
| | 8 2 8 2 | 7 2 6 2 | 3+4 2 |
| | 9 2 9 2 | 6 2 5 2 | 5+6 2 |
| | 6 2 8 2 | 7 2 6 2 | 5+6 2 |
| | 8 2 6 2 | 7 2 5 2 | 3+4 2 |
| | 9 2 7 2 | 6 2 6 2 | 5+6 2 |
| | 9 2 7 2 | 6 2 5 2 | 3+4 2 |
| | 7 2 9 2 | 7 2 6 2 | 5+6 2 |
| | 9 2 9 2 | 6 2 6 2 | 3+4 2 |
| | 7 2 9 2 | 7 2 5 2 | 5+6 2 |

Continued on next column

**TABLE LXIV– continued**

| TQC | ICSD | NoE | SF |
|---|---|---|---|
| | 7 2 7 2 | 6 2 5 2 | 3+4 2 |
| | 9 2 9 2 | 7 2 6 2 | 3+4 2 |
| | 6 2 6 2 | 6 2 6 2 | 5+6 2 |
| | 8 2 8 2 | 6 2 5 2 | 3+4 2 |
| | 9 2 7 2 | 7 2 6 2 | 3+4 2 |
| | 7 2 9 2 | 6 2 5 2 | 5+6 2 |
| | 9 2 8 2 | 7 2 6 2 | 3+4 2 |
| | 8 2 9 2 | 7 2 6 2 | 3+4 2 |
| | 6 2 8 2 | 6 2 6 2 | 5+6 2 |
| | 7 2 6 2 | 6 2 5 2 | 5+6 2 |
| | 6 2 9 2 | 7 2 5 2 | 3+4 2 |
| | 8 2 8 2 | 7 2 5 2 | 5+6 2 |
| | 9 2 7 2 | 6 2 6 2 | 5+6 2 |
| | 8 2 6 2 | 7 2 6 2 | 3+4 2 |
| | 9 2 7 2 | 6 2 5 2 | 3+4 2 |
| | 7 2 7 2 | 7 2 6 2 | 3+4 2 |
| | 6 2 6 2 | 6 2 6 2 | 5+6 2 |
| | 9 2 9 2 | 7 2 5 2 | 5+6 2 |
| | 6 2 8 2 | 6 2 6 2 | 3+4 2 |
| | 8 2 6 2 | 7 2 5 2 | 5+6 2 |
| | 8 2 8 2 | 7 2 5 2 | 5+6 2 |
| | 7 2 7 2 | 6 2 5 2 | 5+6 2 |
| | 6 2 6 2 | 6 2 6 2 | 3+4 2 |
| | 7 2 6 2 | 6 2 5 2 | 3+4 2 |
| | 6 2 7 2 | 7 2 6 2 | 5+6 2 |
| 139.1.140 | 90123 | 92 | Cl2Co2O5Sr3 |
| | -GM -M | -P -X | -N |
| | 7 2 7 2 | 6 2 5 2 | 3+4 2 |
| | 7 2 7 2 | 6 2 5 2 | 3+4 2 |
| | 9 2 9 2 | 7 2 6 2 | 5+6 2 |
| | 7 2 7 2 | 7 2 5 2 | 5+6 2 |
| | 9 2 9 2 | 7 2 6 2 | 3+4 2 |
| | 7 2 7 2 | 6 2 5 2 | 5+6 2 |
| | 6 2 6 2 | 6 2 6 2 | 5+6 2 |
| | 8 2 8 2 | 7 2 5 2 | 3+4 2 |
| | 9 2 9 2 | 7 2 6 2 | 5+6 2 |
| | 7 2 9 2 | 6 2 5 2 | 5+6 2 |
| | 9 2 7 2 | 7 2 6 2 | 3+4 2 |
| | 8 2 8 2 | 6 2 6 2 | 5+6 2 |
| | 7 2 9 2 | 7 2 6 2 | 3+4 2 |
| | 9 2 9 2 | 6 2 5 2 | 5+6 2 |
| | 6 2 8 2 | 7 2 6 2 | 5+6 2 |
| | 8 2 6 2 | 7 2 5 2 | 3+4 2 |
| | 9 2 7 2 | 6 2 6 2 | 5+6 2 |
| | 9 2 7 2 | 6 2 5 2 | 3+4 2 |
| | 7 2 9 2 | 7 2 6 2 | 5+6 2 |
| | 9 2 9 2 | 6 2 6 2 | 3+4 2 |
| | 7 2 9 2 | 7 2 5 2 | 5+6 2 |
| | 7 2 7 2 | 6 2 6 2 | 3+4 2 |
| | 9 2 9 2 | 6 2 5 2 | 3+4 2 |
| | 6 2 6 2 | 7 2 6 2 | 5+6 2 |
| | 8 2 7 2 | 6 2 5 2 | 3+4 2 |
| | 9 2 8 2 | 7 2 6 2 | 3+4 2 |
| | 9 2 9 2 | 7 2 6 2 | 5+6 2 |
| | 8 2 8 2 | 6 2 6 2 | 3+4 2 |
| | 7 2 9 2 | 6 2 5 2 | 5+6 2 |
| | 6 2 8 2 | 7 2 6 2 | 3+4 2 |
| | 7 2 6 2 | 7 2 5 2 | 3+4 2 |
| | 6 2 9 2 | 6 2 5 2 | 5+6 2 |
| | 8 2 7 2 | 7 2 6 2 | 5+6 2 |
| | 9 2 8 2 | 6 2 5 2 | 3+4 2 |

Continued on next column

**TABLE LXIV– continued**

| TQC | ICSD | NoE | SF |
|---|---|---|---|
| | 8 2 6 2 | 7 2 6 2 | 5+6 2 |
| | 9 2 7 2 | 6 2 5 2 | 3+4 2 |
| | 7 2 7 2 | 7 2 6 2 | 3+4 2 |
| | 6 2 6 2 | 6 2 6 2 | 5+6 2 |
| | 9 2 9 2 | 7 2 5 2 | 5+6 2 |
| | 8 2 8 2 | 6 2 6 2 | 3+4 2 |
| | 6 2 6 2 | 7 2 5 2 | 5+6 2 |
| | 8 2 8 2 | 7 2 5 2 | 5+6 2 |
| | 7 2 7 2 | 6 2 5 2 | 5+6 2 |
| | 6 2 6 2 | 6 2 6 2 | 3+4 2 |
| | 7 2 7 2 | 6 2 6 2 | 3+4 2 |
| | 6 2 6 2 | 7 2 5 2 | 5+6 2 |
| 139.1.141 | 251170 | 86 | Cl2Co2O4Sr3 |
| | -GM -M | -P -X | -N |
| | 7 2 7 2 | 6 2 5 2 | 3+4 2 |
| | 9 2 9 2 | 7 2 6 2 | 5+6 2 |
| | 7 2 7 2 | 6 2 5 2 | 3+4 2 |
| | 9 2 9 2 | 7 2 6 2 | 5+6 2 |
| | 7 2 7 2 | 6 2 6 2 | 3+4 2 |
| | 6 2 6 2 | 6 2 5 2 | 5+6 2 |
| | 8 2 8 2 | 7 2 5 2 | 3+4 2 |
| | 7 2 9 2 | 7 2 6 2 | 5+6 2 |
| | 9 2 7 2 | 6 2 5 2 | 3+4 2 |
| | 7 2 9 2 | 7 2 6 2 | 5+6 2 |
| | 9 2 9 2 | 6 2 6 2 | 3+4 2 |
| | 6 2 8 2 | 6 2 5 2 | 5+6 2 |
| | 8 2 6 2 | 7 2 6 2 | 3+4 2 |
| | 9 2 7 2 | 7 2 5 2 | 5+6 2 |
| | 9 2 9 2 | 7 2 6 2 | 5+6 2 |
| | 7 2 7 2 | 6 2 5 2 | 3+4 2 |
| | 8 2 8 2 | 7 2 6 2 | 5+6 2 |
| | 9 2 9 2 | 6 2 6 2 | 5+6 2 |
| | 7 2 9 2 | 7 2 5 2 | 5+6 2 |
| | 7 2 7 2 | 6 2 6 2 | 3+4 2 |
| | 6 2 7 2 | 6 2 5 2 | 3+4 2 |
| | 9 2 6 2 | 7 2 6 2 | 5+6 2 |
| | 9 2 9 2 | 6 2 5 2 | 3+4 2 |
| | 8 2 8 2 | 6 2 5 2 | 3+4 2 |
| | 7 2 9 2 | 7 2 6 2 | 5+6 2 |
| | 6 2 8 2 | 7 2 6 2 | 3+4 2 |
| | 7 2 7 2 | 6 2 5 2 | 5+6 2 |
| | 6 2 6 2 | 7 2 6 2 | 3+4 2 |
| | 9 2 7 2 | 7 2 6 2 | 3+4 2 |
| | 8 2 6 2 | 6 2 5 2 | 5+6 2 |
| | 8 2 9 2 | 7 2 6 2 | 5+6 2 |
| | 9 2 8 2 | 6 2 5 2 | 3+4 2 |
| | 7 2 7 2 | 7 2 6 2 | 5+6 2 |
| | 6 2 6 2 | 6 2 5 2 | 5+6 2 |
| | 9 2 9 2 | 7 2 5 2 | 3+4 2 |
| | 8 2 8 2 | 6 2 6 2 | 5+6 2 |
| | 6 2 6 2 | 7 2 5 2 | 3+4 2 |
| | 8 2 8 2 | 6 2 6 2 | 5+6 2 |
| | 7 2 7 2 | 6 2 5 2 | 5+6 2 |
| | 9 2 9 2 | 7 2 6 2 | 3+4 2 |
| | 8 2 8 2 | 6 2 6 2 | 3+4 2 |
| | 7 2 7 2 | 7 2 5 2 | 5+6 2 |
| | 6 2 6 2 | 7 2 6 2 | 3+4 2 |
| 139.1.142 | 409587 | 58 | Ce2Sb1 |
| | -GM -M | -P -X | -N |
| | 7 2 9 2 | 6 2 5 2 | 3+4 2 |
| | 9 2 7 2 | 7 2 6 2 | 5+6 2 |

Continued on next column

**TABLE LXIV– continued**

| TQC | ICSD | NoE | SF |
|---|---|---|---|
| | 7 2 7 2 | 6 2 6 2 | 5+6 2 |
| | 6 2 6 2 | 7 2 6 2 | 3+4 2 |
| | 8 2 8 2 | 7 2 5 2 | 5+6 2 |
| | 7 2 9 2 | 6 2 5 2 | 3+4 2 |
| | 9 2 7 2 | 7 2 6 2 | 3+4 2 |
| | 9 2 9 2 | 6 2 5 2 | 5+6 2 |
| | 9 2 9 2 | 7 2 5 2 | 3+4 2 |
| | 7 2 9 2 | 6 2 5 2 | 5+6 2 |
| | 6 2 8 2 | 6 2 6 2 | 5+6 2 |
| | 9 2 7 2 | 7 2 5 2 | 3+4 2 |
| | 8 2 6 2 | 6 2 6 2 | 5+6 2 |
| | 8 2 8 2 | 7 2 5 2 | 3+4 2 |
| | 9 2 9 2 | 6 2 5 2 | 3+4 2 |
| | 8 2 8 2 | 7 2 5 2 | 5+6 2 |
| | 7 2 7 2 | 6 2 6 2 | 5+6 2 |
| | 9 2 9 2 | 7 2 5 2 | 3+4 2 |
| | 7 2 7 2 | 6 2 6 2 | 3+4 2 |
| | 9 2 7 2 | 7 2 6 2 | 5+6 2 |
| | 7 2 7 2 | 6 2 6 2 | 5+6 2 |
| | 7 2 9 2 | 7 2 5 2 | 3+4 2 |
| | 6 2 6 2 | 7 2 5 2 | 5+6 2 |
| | 9 2 9 2 | 6 2 5 2 | 3+4 2 |
| | 8 2 6 2 | 7 2 6 2 | 5+6 2 |
| | 6 2 8 2 | 7 2 6 2 | 3+4 2 |
| | 8 2 6 2 | 6 2 5 2 | 5+6 2 |
| | 7 2 7 2 | 6 2 6 2 | 3+4 2 |
| | 6 2 8 2 | 7 2 5 2 | 3+4 2 |
| 139.1.143 | 150461 | 64 | Ce2Ga10Pd1 |
| | -GM -M | -P -X | -N |
| | 7 2 7 2 | 6 2 5 2 | 3+4 2 |
| | 9 2 9 2 | 7 2 6 2 | 5+6 2 |
| | 7 2 9 2 | 6 2 5 2 | 3+4 2 |
| | 9 2 7 2 | 7 2 6 2 | 5+6 2 |
| | 7 2 9 2 | 6 2 5 2 | 3+4 2 |
| | 9 2 7 2 | 7 2 6 2 | 5+6 2 |
| | 6 2 8 2 | 6 2 5 2 | 3+4 2 |
| | 8 2 6 2 | 7 2 6 2 | 5+6 2 |
| | 7 2 9 2 | 6 2 5 2 | 3+4 2 |
| | 7 2 7 2 | 7 2 6 2 | 5+6 2 |
| | 9 2 9 2 | 6 2 6 2 | 3+4 2 |
| | 9 2 7 2 | 7 2 6 2 | 5+6 2 |
| | 7 2 6 2 | 6 2 5 2 | 5+6 2 |
| | 6 2 9 2 | 7 2 6 2 | 3+4 2 |
| | 7 2 7 2 | 6 2 5 2 | 3+4 2 |
| | 6 2 7 2 | 7 2 5 2 | 5+6 2 |
| | 7 2 6 2 | 6 2 5 2 | 5+6 2 |
| | 6 2 7 2 | 7 2 6 2 | 5+6 2 |
| | 9 2 6 2 | 7 2 5 2 | 5+6 2 |
| | 8 2 6 2 | 6 2 5 2 | 5+6 2 |
| | 7 2 8 2 | 6 2 5 2 | 3+4 2 |
| | 6 2 9 2 | 7 2 5 2 | 5+6 2 |
| | 7 2 9 2 | 7 2 6 2 | 5+6 2 |
| | 9 2 8 2 | 7 2 5 2 | 3+4 2 |
| | 8 2 8 2 | 6 2 5 2 | 5+6 2 |
| | 8 2 6 2 | 6 2 5 2 | 3+4 2 |
| | 6 2 7 2 | 7 2 6 2 | 5+6 2 |
| | 7 2 7 2 | 6 2 6 2 | 3+4 2 |
| | 6 2 6 2 | 6 2 5 2 | 3+4 2 |
| | 6 2 9 2 | 6 2 6 2 | 5+6 2 |
| | 8 2 8 2 | 7 2 6 2 | 3+4 2 |
| | 9 2 6 2 | 6 2 5 2 | 5+6 2 |

Continued on next column

**TABLE LXIV– continued**

| TQC | ICSD | NoE | SF |
|---|---|---|---|
| 139.1.144 | 20525 | 64 | Ce2Ga10Ni1 |
| | -GM -M | -P -X | -N |
| | 7 2 7 2 | 6 2 5 2 | 3+4 2 |
| | 9 2 9 2 | 7 2 6 2 | 5+6 2 |
| | 7 2 9 2 | 6 2 5 2 | 3+4 2 |
| | 9 2 7 2 | 7 2 6 2 | 5+6 2 |
| | 7 2 9 2 | 6 2 5 2 | 3+4 2 |
| | 9 2 7 2 | 7 2 6 2 | 5+6 2 |
| | 6 2 8 2 | 6 2 5 2 | 3+4 2 |
| | 8 2 6 2 | 7 2 6 2 | 5+6 2 |
| | 7 2 9 2 | 6 2 5 2 | 3+4 2 |
| | 7 2 7 2 | 7 2 6 2 | 5+6 2 |
| | 9 2 9 2 | 6 2 6 2 | 3+4 2 |
| | 9 2 7 2 | 7 2 6 2 | 5+6 2 |
| | 7 2 9 2 | 6 2 5 2 | 5+6 2 |
| | 6 2 7 2 | 7 2 6 2 | 3+4 2 |
| | 7 2 6 2 | 6 2 5 2 | 3+4 2 |
| | 9 2 6 2 | 7 2 6 2 | 5+6 2 |
| | 8 2 7 2 | 7 2 5 2 | 3+4 2 |
| | 7 2 6 2 | 6 2 5 2 | 5+6 2 |
| | 6 2 8 2 | 7 2 5 2 | 5+6 2 |
| | 7 2 9 2 | 6 2 5 2 | 5+6 2 |
| | 6 2 7 2 | 7 2 6 2 | 5+6 2 |
| | 8 2 8 2 | 6 2 6 2 | 5+6 2 |
| | 9 2 9 2 | 7 2 6 2 | 5+6 2 |
| | 6 2 8 2 | 6 2 5 2 | 3+4 2 |
| | 8 2 6 2 | 6 2 6 2 | 3+4 2 |
| | 7 2 6 2 | 6 2 5 2 | 5+6 2 |
| | 6 2 7 2 | 7 2 6 2 | 5+6 2 |
| | 7 2 7 2 | 7 2 5 2 | 3+4 2 |
| | 6 2 6 2 | 6 2 5 2 | 3+4 2 |
| | 6 2 9 2 | 7 2 5 2 | 5+6 2 |
| | 8 2 8 2 | 6 2 5 2 | 3+4 2 |
| | 9 2 6 2 | 6 2 5 2 | 5+6 2 |
| 139.1.145 | 87159 | 38 | Ce1Sc1Si1 |
| | -GM -M | -P -X | -N |
| | 7 2 9 2 | 6 2 5 2 | 3+4 2 |
| | 9 2 7 2 | 7 2 6 2 | 5+6 2 |
| | 7 2 9 2 | 6 2 6 2 | 3+4 2 |
| | 9 2 7 2 | 7 2 5 2 | 5+6 2 |
| | 7 2 9 2 | 7 2 6 2 | 3+4 2 |
| | 6 2 8 2 | 6 2 5 2 | 5+6 2 |
| | 9 2 7 2 | 6 2 6 2 | 5+6 2 |
| | 8 2 6 2 | 7 2 5 2 | 3+4 2 |
| | 7 2 7 2 | 6 2 5 2 | 3+4 2 |
| | 9 2 9 2 | 7 2 6 2 | 5+6 2 |
| | 7 2 7 2 | 6 2 6 2 | 5+6 2 |
| | 7 2 7 2 | 7 2 6 2 | 3+4 2 |
| | 9 2 7 2 | 6 2 5 2 | 3+4 2 |
| | 7 2 6 2 | 7 2 6 2 | 3+4 2 |
| | 6 2 6 2 | 7 2 5 2 | 5+6 2 |
| | 8 2 9 2 | 6 2 5 2 | 5+6 2 |
| | 8 2 8 2 | 6 2 6 2 | 5+6 2 |
| | 9 2 9 2 | 7 2 5 2 | 3+4 2 |
| | 6 2 6 2 | 6 2 6 2 | 5+6 2 |
| 139.1.146 | 290821 | 38 | Ce1Sc1Si1 |
| | -GM -M | -P -X | -N |
| | 7 2 9 2 | 6 2 5 2 | 3+4 2 |
| | 9 2 7 2 | 7 2 6 2 | 5+6 2 |
| | 7 2 9 2 | 6 2 5 2 | 3+4 2 |
| | 9 2 7 2 | 7 2 6 2 | 5+6 2 |
| | 7 2 9 2 | 6 2 6 2 | 3+4 2 |
| | 6 2 8 2 | 7 2 5 2 | 5+6 2 |
| | 9 2 7 2 | 6 2 6 2 | 5+6 2 |
| | 8 2 6 2 | 7 2 5 2 | 3+4 2 |
| | 7 2 7 2 | 6 2 5 2 | 3+4 2 |
| | 9 2 9 2 | 7 2 6 2 | 5+6 2 |
| | 7 2 7 2 | 6 2 6 2 | 5+6 2 |
| | 7 2 7 2 | 7 2 6 2 | 3+4 2 |
| | 7 2 7 2 | 6 2 5 2 | 3+4 2 |
| | 9 2 6 2 | 7 2 6 2 | 3+4 2 |
| | 6 2 6 2 | 7 2 5 2 | 5+6 2 |
| | 8 2 9 2 | 6 2 5 2 | 5+6 2 |
| | 8 2 8 2 | 6 2 6 2 | 5+6 2 |
| | 9 2 9 2 | 7 2 5 2 | 3+4 2 |
| | 6 2 6 2 | 6 2 6 2 | 5+6 2 |
| 139.1.147 | 411677 | 38 | Ce1Ir2Si2 |
| | -GM -M | -P -X | -N |
| | 7 2 7 2 | 6 2 5 2 | 3+4 2 |
| | 9 2 9 2 | 7 2 6 2 | 5+6 2 |
| | 9 2 9 2 | 6 2 6 2 | 5+6 2 |
| | 8 2 8 2 | 7 2 6 2 | 5+6 2 |
| | 7 2 9 2 | 7 2 5 2 | 5+6 2 |
| | 9 2 7 2 | 6 2 6 2 | 3+4 2 |
| | 7 2 7 2 | 7 2 5 2 | 3+4 2 |
| | 6 2 6 2 | 6 2 6 2 | 5+6 2 |
| | 8 2 8 2 | 6 2 5 2 | 3+4 2 |
| | 9 2 7 2 | 7 2 5 2 | 5+6 2 |
| | 7 2 9 2 | 6 2 6 2 | 3+4 2 |
| | 9 2 6 2 | 7 2 6 2 | 5+6 2 |
| | 6 2 8 2 | 7 2 5 2 | 3+4 2 |
| | 8 2 9 2 | 6 2 5 2 | 5+6 2 |
| | 7 2 8 2 | 7 2 6 2 | 3+4 2 |
| | 6 2 9 2 | 6 2 5 2 | 5+6 2 |
| | 7 2 8 2 | 7 2 6 2 | 5+6 2 |
| | 6 2 7 2 | 7 2 5 2 | 3+4 2 |
| | 6 2 8 2 | 6 2 6 2 | 3+4 2 |
| 139.1.148 | 85893 | 40 | Ce1Co2P2 |
| | -GM -M | -P -X | -N |
| | 7 2 7 2 | 6 2 5 2 | 3+4 2 |
| | 9 2 9 2 | 7 2 6 2 | 5+6 2 |
| | 9 2 9 2 | 6 2 6 2 | 5+6 2 |
| | 8 2 8 2 | 7 2 6 2 | 5+6 2 |
| | 7 2 7 2 | 7 2 5 2 | 5+6 2 |
| | 9 2 9 2 | 6 2 6 2 | 3+4 2 |
| | 7 2 7 2 | 7 2 5 2 | 3+4 2 |
| | 9 2 6 2 | 6 2 6 2 | 5+6 2 |
| | 8 2 7 2 | 7 2 5 2 | 3+4 2 |
| | 6 2 9 2 | 7 2 5 2 | 5+6 2 |
| | 7 2 8 2 | 6 2 6 2 | 5+6 2 |
| | 6 2 9 2 | 6 2 5 2 | 3+4 2 |
| | 9 2 8 2 | 7 2 5 2 | 3+4 2 |
| | 8 2 6 2 | 6 2 6 2 | 5+6 2 |
| | 7 2 7 2 | 7 2 6 2 | 3+4 2 |
| | 6 2 8 2 | 7 2 5 2 | 5+6 2 |
| | 7 2 9 2 | 6 2 6 2 | 5+6 2 |
| | 6 2 8 2 | 6 2 5 2 | 3+4 2 |
| | 6 2 7 2 | 7 2 6 2 | 3+4 2 |
| | 9 2 8 2 | 6 2 6 2 | 5+6 2 |
| 139.1.149 | 370021 | 40 | Cd2Ga2Sr1 |
| | -GM -M | -P -X | -N |
| | 7 2 7 2 | 6 2 5 2 | 3+4 2 |

Continued on next column

**TABLE LXIV– continued**

| TQC | ICSD | NoE | SF |
|---|---|---|---|
| | 9 2 9 2 | 7 2 6 2 | 5+6 2 |
| | 9 2 9 2 | 6 2 6 2 | 5+6 2 |
| | 8 2 8 2 | 7 2 6 2 | 5+6 2 |
| | 9 2 7 2 | 7 2 5 2 | 5+6 2 |
| | 6 2 8 2 | 6 2 6 2 | 3+4 2 |
| | 7 2 9 2 | 6 2 5 2 | 3+4 2 |
| | 8 2 6 2 | 7 2 6 2 | 5+6 2 |
| | 7 2 7 2 | 7 2 5 2 | 5+6 2 |
| | 9 2 9 2 | 6 2 6 2 | 3+4 2 |
| | 6 2 8 2 | 7 2 5 2 | 3+4 2 |
| | 7 2 6 2 | 6 2 6 2 | 5+6 2 |
| | 8 2 8 2 | 6 2 5 2 | 3+4 2 |
| | 6 2 7 2 | 7 2 6 2 | 5+6 2 |
| | 9 2 9 2 | 7 2 5 2 | 5+6 2 |
| | 9 2 7 2 | 6 2 6 2 | 3+4 2 |
| | 7 2 8 2 | 7 2 6 2 | 5+6 2 |
| | 6 2 9 2 | 6 2 5 2 | 3+4 2 |
| | 7 2 6 2 | 7 2 5 2 | 3+4 2 |
| | 6 2 7 2 | 7 2 6 2 | 5+6 2 |
| 139.1.150 | 416139 | 140 | Ca2Pd3Sb4 |
| | -GM -M | -P -X | -N |
| | 7 2 7 2 | 6 2 5 2 | 3+4 2 |
| | 9 2 9 2 | 7 2 6 2 | 5+6 2 |
| | 7 2 7 2 | 6 2 5 2 | 3+4 2 |
| | 9 2 9 2 | 7 2 6 2 | 5+6 2 |
| | 7 2 7 2 | 6 2 5 2 | 3+4 2 |
| | 9 2 9 2 | 7 2 6 2 | 5+6 2 |
| | 7 2 7 2 | 6 2 5 2 | 3+4 2 |
| | 9 2 9 2 | 7 2 6 2 | 5+6 2 |
| | 6 2 6 2 | 6 2 5 2 | 3+4 2 |
| | 8 2 8 2 | 7 2 6 2 | 5+6 2 |
| | 7 2 7 2 | 6 2 5 2 | 3+4 2 |
| | 9 2 9 2 | 7 2 6 2 | 5+6 2 |
| | 6 2 6 2 | 6 2 5 2 | 3+4 2 |
| | 8 2 8 2 | 7 2 6 2 | 5+6 2 |
| | 7 2 7 2 | 6 2 5 2 | 3+4 2 |
| | 9 2 9 2 | 7 2 6 2 | 5+6 2 |
| | 7 2 7 2 | 7 2 5 2 | 5+6 2 |
| | 9 2 9 2 | 6 2 6 2 | 3+4 2 |
| | 7 2 7 2 | 7 2 5 2 | 5+6 2 |
| | 7 2 9 2 | 6 2 5 2 | 3+4 2 |
| | 9 2 7 2 | 7 2 6 2 | 5+6 2 |
| | 9 2 9 2 | 6 2 5 2 | 3+4 2 |
| | 6 2 6 2 | 7 2 6 2 | 5+6 2 |
| | 8 2 8 2 | 6 2 6 2 | 3+4 2 |
| | 7 2 7 2 | 7 2 5 2 | 3+4 2 |
| | 9 2 9 2 | 6 2 6 2 | 5+6 2 |
| | 7 2 7 2 | 6 2 5 2 | 3+4 2 |
| | 8 2 6 2 | 7 2 6 2 | 5+6 2 |
| | 9 2 7 2 | 6 2 5 2 | 3+4 2 |
| | 6 2 8 2 | 7 2 6 2 | 5+6 2 |
| | 7 2 9 2 | 6 2 6 2 | 3+4 2 |
| | 9 2 9 2 | 7 2 6 2 | 5+6 2 |
| | 8 2 6 2 | 6 2 5 2 | 3+4 2 |
| | 9 2 7 2 | 7 2 6 2 | 5+6 2 |
| | 7 2 7 2 | 7 2 5 2 | 3+4 2 |
| | 7 2 7 2 | 6 2 6 2 | 5+6 2 |
| | 6 2 6 2 | 6 2 5 2 | 5+6 2 |
| | 6 2 8 2 | 7 2 6 2 | 5+6 2 |
| | 7 2 9 2 | 6 2 5 2 | 3+4 2 |
| | 6 2 6 2 | 7 2 6 2 | 5+6 2 |

Continued on next column

**TABLE LXIV– continued**

| TQC | ICSD | NoE | SF |
|---|---|---|---|
| | 7 2 7 2 | 7 2 5 2 | 3+4 2 |
| | 6 2 8 2 | 6 2 6 2 | 5+6 2 |
| | 8 2 6 2 | 7 2 6 2 | 3+4 2 |
| | 6 2 8 2 | 7 2 5 2 | 3+4 2 |
| | 8 2 6 2 | 6 2 6 2 | 5+6 2 |
| | 6 2 8 2 | 7 2 6 2 | 3+4 2 |
| | 7 2 9 2 | 6 2 6 2 | 5+6 2 |
| | 8 2 6 2 | 6 2 5 2 | 3+4 2 |
| | 9 2 7 2 | 7 2 6 2 | 5+6 2 |
| | 6 2 6 2 | 7 2 5 2 | 3+4 2 |
| | 6 2 8 2 | 6 2 6 2 | 3+4 2 |
| | 7 2 7 2 | 7 2 5 2 | 5+6 2 |
| | 6 2 6 2 | 6 2 6 2 | 3+4 2 |
| | 9 2 9 2 | 6 2 5 2 | 5+6 2 |
| | 7 2 8 2 | 7 2 5 2 | 3+4 2 |
| | 6 2 6 2 | 7 2 6 2 | 3+4 2 |
| | 8 2 9 2 | 6 2 6 2 | 5+6 2 |
| | 9 2 7 2 | 6 2 5 2 | 3+4 2 |
| | 6 2 6 2 | 7 2 6 2 | 5+6 2 |
| | 8 2 8 2 | 7 2 6 2 | 5+6 2 |
| | 9 2 7 2 | 6 2 6 2 | 3+4 2 |
| | 7 2 9 2 | 7 2 6 2 | 5+6 2 |
| | 7 2 7 2 | 6 2 5 2 | 5+6 2 |
| | 6 2 8 2 | 7 2 6 2 | 3+4 2 |
| | 7 2 9 2 | 6 2 6 2 | 5+6 2 |
| | 7 2 7 2 | 7 2 5 2 | 3+4 2 |
| | 9 2 7 2 | 6 2 5 2 | 5+6 2 |
| | 8 2 9 2 | 7 2 6 2 | 3+4 2 |
| | 6 2 8 2 | 6 2 5 2 | 5+6 2 |
| | 9 2 6 2 | 7 2 6 2 | 3+4 2 |
| 139.1.151 | 59649 | 34 | Ca1Si2Zn2 |
| | -GM -M | -P -X | -N |
| | 7 2 9 2 | 7 2 5 2 | 5+6 2 |
| | 9 2 7 2 | 6 2 6 2 | 3+4 2 |
| | 7 2 7 2 | 7 2 6 2 | 5+6 2 |
| | 6 2 8 2 | 6 2 5 2 | 3+4 2 |
| | 8 2 6 2 | 7 2 5 2 | 5+6 2 |
| | 9 2 9 2 | 6 2 6 2 | 3+4 2 |
| | 7 2 7 2 | 7 2 5 2 | 5+6 2 |
| | 6 2 8 2 | 6 2 6 2 | 3+4 2 |
| | 8 2 9 2 | 6 2 6 2 | 3+4 2 |
| | 6 2 6 2 | 7 2 5 2 | 5+6 2 |
| | 9 2 8 2 | 7 2 5 2 | 5+6 2 |
| | 9 2 7 2 | 6 2 6 2 | 3+4 2 |
| | 7 2 8 2 | 7 2 6 2 | 5+6 2 |
| | 6 2 9 2 | 6 2 5 2 | 3+4 2 |
| | 7 2 7 2 | 7 2 5 2 | 3+4 2 |
| | 9 2 6 2 | 6 2 6 2 | 5+6 2 |
| | 8 2 7 2 | 7 2 5 2 | 3+4 2 |
| 139.1.152 | 425469 | 28 | Ca1Rh2Si2 |
| | -GM -M | -P -X | -N |
| | 7 2 7 2 | 7 2 5 2 | 5+6 2 |
| | 9 2 9 2 | 6 2 6 2 | 3+4 2 |
| | 7 2 7 2 | 7 2 6 2 | 3+4 2 |
| | 8 2 8 2 | 6 2 5 2 | 5+6 2 |
| | 6 2 6 2 | 6 2 6 2 | 3+4 2 |
| | 9 2 7 2 | 7 2 5 2 | 5+6 2 |
| | 9 2 9 2 | 7 2 6 2 | 5+6 2 |
| | 7 2 9 2 | 6 2 5 2 | 3+4 2 |
| | 6 2 8 2 | 7 2 5 2 | 3+4 2 |
| | 8 2 6 2 | 6 2 6 2 | 5+6 2 |

Continued on next column

**TABLE LXIV– continued**

| TQC | ICSD | NoE | SF |
|---|---|---|---|
| | 7 2 8 2 | 7 2 5 2 | 5+6 2 |
| | 6 2 8 2 | 6 2 5 2 | 3+4 2 |
| | 6 2 9 2 | 7 2 6 2 | 5+6 2 |
| | 7 2 7 2 | 7 2 6 2 | 3+4 2 |
| 139.1.153 | 50185 | 30 | Ca1P2Rh2 |
| | -GM -M | -P -X | -N |
| | 7 2 7 2 | 7 2 5 2 | 5+6 2 |
| | 9 2 9 2 | 6 2 6 2 | 3+4 2 |
| | 7 2 7 2 | 7 2 6 2 | 3+4 2 |
| | 8 2 6 2 | 6 2 5 2 | 5+6 2 |
| | 9 2 7 2 | 7 2 5 2 | 3+4 2 |
| | 6 2 8 2 | 6 2 6 2 | 5+6 2 |
| | 9 2 9 2 | 7 2 5 2 | 5+6 2 |
| | 7 2 8 2 | 6 2 6 2 | 3+4 2 |
| | 6 2 9 2 | 7 2 5 2 | 3+4 2 |
| | 7 2 6 2 | 6 2 5 2 | 5+6 2 |
| | 8 2 8 2 | 7 2 6 2 | 5+6 2 |
| | 6 2 9 2 | 6 2 6 2 | 3+4 2 |
| | 7 2 8 2 | 7 2 5 2 | 5+6 2 |
| | 6 2 7 2 | 7 2 6 2 | 3+4 2 |
| | 9 2 7 2 | 6 2 5 2 | 3+4 2 |
| 139.1.154 | 36371 | 32 | Ca1P2Pd2 |
| | -GM -M | -P -X | -N |
| | 7 2 7 2 | 7 2 5 2 | 5+6 2 |
| | 9 2 9 2 | 6 2 6 2 | 3+4 2 |
| | 7 2 7 2 | 7 2 6 2 | 3+4 2 |
| | 8 2 6 2 | 6 2 5 2 | 5+6 2 |
| | 9 2 7 2 | 7 2 5 2 | 3+4 2 |
| | 6 2 8 2 | 6 2 6 2 | 5+6 2 |
| | 9 2 9 2 | 7 2 6 2 | 5+6 2 |
| | 7 2 8 2 | 6 2 5 2 | 3+4 2 |
| | 6 2 9 2 | 7 2 5 2 | 5+6 2 |
| | 7 2 6 2 | 6 2 5 2 | 3+4 2 |
| | 8 2 8 2 | 7 2 6 2 | 5+6 2 |
| | 6 2 9 2 | 6 2 6 2 | 3+4 2 |
| | 6 2 8 2 | 7 2 5 2 | 5+6 2 |
| | 7 2 7 2 | 7 2 6 2 | 3+4 2 |
| | 9 2 7 2 | 6 2 5 2 | 3+4 2 |
| | 8 2 6 2 | 6 2 6 2 | 5+6 2 |
| 139.1.155 | 30835 | 22 | Ca1P2Pd2 |
| | -GM -M | -P -X | -N |
| | 7 2 9 2 | 7 2 5 2 | 3+4 2 |
| | 6 2 7 2 | 6 2 5 2 | 5+6 2 |
| | 9 2 8 2 | 6 2 6 2 | 3+4 2 |
| | 7 2 6 2 | 7 2 6 2 | 5+6 2 |
| | 8 2 7 2 | 7 2 6 2 | 3+4 2 |
| | 9 2 9 2 | 6 2 5 2 | 5+6 2 |
| | 8 2 6 2 | 6 2 5 2 | 3+4 2 |
| | 7 2 7 2 | 7 2 6 2 | 5+6 2 |
| | 6 2 9 2 | 7 2 5 2 | 3+4 2 |
| | 6 2 8 2 | 6 2 6 2 | 5+6 2 |
| | 9 2 8 2 | 7 2 5 2 | 3+4 2 |
| 139.1.156 | 10461 | 32 | Ca1Ni2P2 |
| | -GM -M | -P -X | -N |
| | 7 2 7 2 | 7 2 5 2 | 5+6 2 |
| | 9 2 9 2 | 6 2 6 2 | 3+4 2 |
| | 7 2 7 2 | 6 2 6 2 | 5+6 2 |
| | 8 2 9 2 | 7 2 5 2 | 3+4 2 |
| | 9 2 8 2 | 7 2 5 2 | 3+4 2 |
| | 7 2 6 2 | 6 2 6 2 | 5+6 2 |
| | 6 2 7 2 | 7 2 5 2 | 5+6 2 |

Continued on next column

**TABLE LXIV– continued**

| TQC | ICSD | NoE | SF |
|---|---|---|---|
| | 9 2 8 2 | 6 2 5 2 | 3+4 2 |
| | 6 2 9 2 | 7 2 6 2 | 3+4 2 |
| | 7 2 6 2 | 7 2 6 2 | 5+6 2 |
| | 8 2 8 2 | 6 2 5 2 | 3+4 2 |
| | 6 2 9 2 | 6 2 6 2 | 5+6 2 |
| | 7 2 8 2 | 7 2 5 2 | 5+6 2 |
| | 6 2 7 2 | 7 2 6 2 | 3+4 2 |
| | 9 2 7 2 | 6 2 5 2 | 3+4 2 |
| | 8 2 6 2 | 6 2 6 2 | 5+6 2 |
| 139.1.157 | 414231 | 20 | Ca1In1 |
| | -GM -M | -P -X | -N |
| | 7 2 7 2 | 6 2 5 2 | 3+4 2 |
| | 8 2 8 2 | 6 2 6 2 | 5+6 2 |
| | 9 2 9 2 | 7 2 6 2 | 5+6 2 |
| | 6 2 9 2 | 7 2 5 2 | 3+4 2 |
| | 7 2 6 2 | 7 2 6 2 | 3+4 2 |
| | 6 2 7 2 | 7 2 5 2 | 5+6 2 |
| | 7 2 6 2 | 6 2 5 2 | 5+6 2 |
| | 6 2 8 2 | 6 2 5 2 | 3+4 2 |
| | 8 2 9 2 | 7 2 6 2 | 3+4 2 |
| | 9 2 8 2 | 6 2 6 2 | 3+4 2 |
| 139.1.158 | 52761 | 26 | Ca1Ge2Ru2 |
| | -GM -M | -P -X | -N |
| | 7 2 7 2 | 7 2 5 2 | 5+6 2 |
| | 9 2 9 2 | 6 2 6 2 | 3+4 2 |
| | 7 2 7 2 | 6 2 6 2 | 3+4 2 |
| | 8 2 6 2 | 7 2 5 2 | 5+6 2 |
| | 9 2 7 2 | 6 2 6 2 | 3+4 2 |
| | 6 2 8 2 | 7 2 5 2 | 5+6 2 |
| | 9 2 9 2 | 6 2 6 2 | 5+6 2 |
| | 7 2 9 2 | 7 2 5 2 | 3+4 2 |
| | 6 2 8 2 | 7 2 5 2 | 3+4 2 |
| | 8 2 6 2 | 7 2 6 2 | 5+6 2 |
| | 7 2 8 2 | 6 2 5 2 | 3+4 2 |
| | 6 2 9 2 | 6 2 6 2 | 5+6 2 |
| | 6 2 8 2 | 7 2 5 2 | 5+6 2 |
| 139.1.159 | 52759 | 30 | Ca1Ge2Pd2 |
| | -GM -M | -P -X | -N |
| | 7 2 7 2 | 7 2 5 2 | 5+6 2 |
| | 9 2 9 2 | 6 2 6 2 | 3+4 2 |
| | 7 2 7 2 | 7 2 6 2 | 3+4 2 |
| | 8 2 6 2 | 6 2 5 2 | 5+6 2 |
| | 9 2 8 2 | 6 2 6 2 | 3+4 2 |
| | 6 2 7 2 | 7 2 5 2 | 5+6 2 |
| | 7 2 9 2 | 6 2 6 2 | 5+6 2 |
| | 9 2 9 2 | 7 2 5 2 | 3+4 2 |
| | 6 2 8 2 | 7 2 5 2 | 5+6 2 |
| | 8 2 6 2 | 6 2 6 2 | 3+4 2 |
| | 6 2 8 2 | 6 2 5 2 | 3+4 2 |
| | 7 2 7 2 | 7 2 5 2 | 5+6 2 |
| | 9 2 9 2 | 7 2 6 2 | 5+6 2 |
| | 6 2 8 2 | 7 2 6 2 | 3+4 2 |
| | 7 2 7 2 | 6 2 6 2 | 3+4 2 |
| 139.1.160 | 81755 | 30 | Ca1Ge2Ni2 |
| | -GM -M | -P -X | -N |
| | 7 2 9 2 | 7 2 5 2 | 5+6 2 |
| | 9 2 7 2 | 6 2 6 2 | 3+4 2 |
| | 7 2 7 2 | 6 2 6 2 | 3+4 2 |
| | 8 2 6 2 | 7 2 5 2 | 5+6 2 |
| | 9 2 7 2 | 7 2 5 2 | 3+4 2 |
| | 6 2 8 2 | 6 2 6 2 | 5+6 2 |

Continued on next column

**TABLE LXIV– continued**

| TQC | ICSD | NoE | SF |
|---|---|---|---|
| | 9 2 9 2 | 7 2 5 2 | 5+6 2 |
| | 7 2 9 2 | 6 2 5 2 | 3+4 2 |
| | 6 2 8 2 | 7 2 6 2 | 3+4 2 |
| | 8 2 6 2 | 7 2 6 2 | 5+6 2 |
| | 6 2 8 2 | 6 2 5 2 | 3+4 2 |
| | 7 2 9 2 | 6 2 6 2 | 5+6 2 |
| | 9 2 7 2 | 7 2 5 2 | 5+6 2 |
| | 7 2 8 2 | 7 2 6 2 | 3+4 2 |
| | 6 2 7 2 | 6 2 5 2 | 3+4 2 |
| 139.1.161 | 408 | 30 | Ca1Ge2Ni2 |
| | -GM -M | -P -X | -N |
| | 7 2 9 2 | 7 2 5 2 | 5+6 2 |
| | 9 2 7 2 | 6 2 6 2 | 3+4 2 |
| | 7 2 7 2 | 6 2 6 2 | 3+4 2 |
| | 8 2 6 2 | 7 2 5 2 | 5+6 2 |
| | 9 2 7 2 | 7 2 5 2 | 3+4 2 |
| | 6 2 8 2 | 6 2 6 2 | 5+6 2 |
| | 9 2 9 2 | 7 2 5 2 | 5+6 2 |
| | 7 2 9 2 | 6 2 5 2 | 3+4 2 |
| | 6 2 8 2 | 7 2 6 2 | 3+4 2 |
| | 8 2 6 2 | 7 2 6 2 | 5+6 2 |
| | 6 2 8 2 | 6 2 5 2 | 3+4 2 |
| | 7 2 9 2 | 6 2 6 2 | 5+6 2 |
| | 9 2 7 2 | 7 2 5 2 | 5+6 2 |
| | 7 2 8 2 | 7 2 6 2 | 3+4 2 |
| | 6 2 7 2 | 6 2 5 2 | 3+4 2 |
| 139.1.162 | 81743 | 24 | Ca1Ge2Mn2 |
| | -GM -M | -P -X | -N |
| | 7 2 7 2 | 7 2 5 2 | 5+6 2 |
| | 9 2 9 2 | 6 2 6 2 | 3+4 2 |
| | 7 2 7 2 | 6 2 6 2 | 3+4 2 |
| | 9 2 7 2 | 7 2 5 2 | 5+6 2 |
| | 8 2 6 2 | 7 2 6 2 | 3+4 2 |
| | 9 2 9 2 | 6 2 5 2 | 5+6 2 |
| | 6 2 8 2 | 7 2 5 2 | 5+6 2 |
| | 7 2 9 2 | 6 2 5 2 | 3+4 2 |
| | 6 2 8 2 | 7 2 6 2 | 3+4 2 |
| | 7 2 6 2 | 7 2 6 2 | 5+6 2 |
| | 8 2 8 2 | 6 2 5 2 | 5+6 2 |
| | 6 2 9 2 | 6 2 6 2 | 3+4 2 |
| 139.1.163 | 402 | 24 | Ca1Ge2Mn2 |
| | -GM -M | -P -X | -N |
| | 7 2 7 2 | 7 2 5 2 | 5+6 2 |
| | 9 2 9 2 | 6 2 6 2 | 3+4 2 |
| | 7 2 7 2 | 6 2 6 2 | 3+4 2 |
| | 9 2 7 2 | 7 2 5 2 | 5+6 2 |
| | 8 2 6 2 | 7 2 6 2 | 3+4 2 |
| | 9 2 9 2 | 6 2 5 2 | 5+6 2 |
| | 6 2 8 2 | 7 2 5 2 | 5+6 2 |
| | 7 2 9 2 | 6 2 5 2 | 3+4 2 |
| | 6 2 8 2 | 7 2 6 2 | 3+4 2 |
| | 7 2 6 2 | 7 2 6 2 | 5+6 2 |
| | 8 2 8 2 | 6 2 5 2 | 5+6 2 |
| | 6 2 9 2 | 6 2 6 2 | 3+4 2 |
| 139.1.164 | 52755 | 28 | Ca1Ge2Ir2 |
| | -GM -M | -P -X | -N |
| | 7 2 9 2 | 7 2 5 2 | 5+6 2 |
| | 9 2 7 2 | 6 2 6 2 | 3+4 2 |
| | 7 2 7 2 | 7 2 6 2 | 3+4 2 |
| | 8 2 8 2 | 6 2 5 2 | 5+6 2 |
| | 6 2 6 2 | 6 2 6 2 | 3+4 2 |

Continued on next column

**TABLE LXIV– continued**

| TQC | ICSD | NoE | SF |
|---|---|---|---|
| | 9 2 7 2 | 7 2 5 2 | 5+6 2 |
| | 9 2 9 2 | 6 2 6 2 | 5+6 2 |
| | 7 2 8 2 | 7 2 5 2 | 3+4 2 |
| | 6 2 9 2 | 7 2 5 2 | 5+6 2 |
| | 8 2 6 2 | 6 2 5 2 | 3+4 2 |
| | 7 2 8 2 | 7 2 6 2 | 3+4 2 |
| | 6 2 9 2 | 6 2 6 2 | 5+6 2 |
| | 6 2 8 2 | 7 2 5 2 | 5+6 2 |
| | 7 2 7 2 | 7 2 6 2 | 3+4 2 |
| 139.1.165 | 10463 | 28 | Ca1Fe2P2 |
| | -GM -M | -P -X | -N |
| | 7 2 9 2 | 7 2 5 2 | 5+6 2 |
| | 9 2 7 2 | 6 2 6 2 | 3+4 2 |
| | 7 2 7 2 | 6 2 6 2 | 3+4 2 |
| | 9 2 6 2 | 7 2 5 2 | 5+6 2 |
| | 8 2 7 2 | 7 2 5 2 | 3+4 2 |
| | 9 2 9 2 | 6 2 6 2 | 5+6 2 |
| | 6 2 8 2 | 7 2 5 2 | 5+6 2 |
| | 7 2 9 2 | 6 2 5 2 | 3+4 2 |
| | 6 2 8 2 | 7 2 6 2 | 3+4 2 |
| | 7 2 6 2 | 7 2 6 2 | 5+6 2 |
| | 8 2 8 2 | 6 2 5 2 | 3+4 2 |
| | 6 2 9 2 | 6 2 6 2 | 5+6 2 |
| | 7 2 8 2 | 7 2 5 2 | 5+6 2 |
| | 6 2 7 2 | 7 2 5 2 | 3+4 2 |
| 139.1.166 | 41756 | 34 | Ca1Cu2P2 |
| | -GM -M | -P -X | -N |
| | 7 2 7 2 | 7 2 5 2 | 5+6 2 |
| | 9 2 9 2 | 6 2 6 2 | 3+4 2 |
| | 7 2 7 2 | 7 2 6 2 | 5+6 2 |
| | 8 2 6 2 | 6 2 5 2 | 3+4 2 |
| | 9 2 7 2 | 7 2 5 2 | 3+4 2 |
| | 6 2 9 2 | 6 2 6 2 | 5+6 2 |
| | 9 2 8 2 | 7 2 5 2 | 5+6 2 |
| | 7 2 8 2 | 6 2 5 2 | 3+4 2 |
| | 6 2 9 2 | 7 2 6 2 | 5+6 2 |
| | 7 2 6 2 | 6 2 6 2 | 3+4 2 |
| | 8 2 8 2 | 7 2 5 2 | 3+4 2 |
| | 6 2 9 2 | 6 2 6 2 | 5+6 2 |
| | 9 2 7 2 | 7 2 5 2 | 5+6 2 |
| | 7 2 8 2 | 7 2 6 2 | 3+4 2 |
| | 6 2 7 2 | 6 2 6 2 | 3+4 2 |
| | 8 2 6 2 | 6 2 5 2 | 5+6 2 |
| | 9 2 9 2 | 7 2 5 2 | 5+6 2 |
| 139.1.167 | 420974 | 28 | Ca1Co2Si2 |
| | -GM -M | -P -X | -N |
| | 7 2 9 2 | 7 2 5 2 | 5+6 2 |
| | 9 2 7 2 | 6 2 6 2 | 3+4 2 |
| | 7 2 7 2 | 7 2 6 2 | 3+4 2 |
| | 8 2 6 2 | 6 2 5 2 | 5+6 2 |
| | 9 2 7 2 | 7 2 5 2 | 3+4 2 |
| | 6 2 8 2 | 6 2 6 2 | 5+6 2 |
| | 9 2 9 2 | 6 2 5 2 | 5+6 2 |
| | 7 2 9 2 | 7 2 5 2 | 3+4 2 |
| | 6 2 8 2 | 7 2 6 2 | 3+4 2 |
| | 8 2 6 2 | 7 2 6 2 | 5+6 2 |
| | 7 2 8 2 | 6 2 5 2 | 3+4 2 |
| | 6 2 9 2 | 6 2 6 2 | 5+6 2 |
| | 7 2 8 2 | 7 2 5 2 | 5+6 2 |
| | 6 2 7 2 | 7 2 6 2 | 3+4 2 |
| 139.1.168 | 85892 | 30 | Ca1Co2P2 |

Continued on next column

**TABLE LXIV– continued**

| TQC | ICSD | NoE | SF |
|---|---|---|---|
| | -GM -M | -P -X | -N |
| | 7 2 7 2 | 7 2 5 2 | 5+6 2 |
| | 9 2 9 2 | 6 2 6 2 | 3+4 2 |
| | 7 2 7 2 | 6 2 6 2 | 3+4 2 |
| | 8 2 6 2 | 7 2 5 2 | 5+6 2 |
| | 9 2 7 2 | 7 2 5 2 | 3+4 2 |
| | 9 2 9 2 | 6 2 6 2 | 5+6 2 |
| | 6 2 8 2 | 7 2 5 2 | 5+6 2 |
| | 7 2 9 2 | 6 2 5 2 | 3+4 2 |
| | 6 2 8 2 | 7 2 6 2 | 3+4 2 |
| | 7 2 6 2 | 7 2 6 2 | 5+6 2 |
| | 8 2 8 2 | 6 2 5 2 | 5+6 2 |
| | 6 2 9 2 | 6 2 6 2 | 3+4 2 |
| | 7 2 8 2 | 7 2 5 2 | 5+6 2 |
| | 6 2 7 2 | 7 2 6 2 | 3+4 2 |
| | 9 2 7 2 | 6 2 5 2 | 3+4 2 |
| 139.1.169 | 85891 | 30 | Ca1Co2P2 |
| | -GM -M | -P -X | -N |
| | 7 2 7 2 | 7 2 5 2 | 5+6 2 |
| | 9 2 9 2 | 6 2 6 2 | 3+4 2 |
| | 7 2 7 2 | 6 2 6 2 | 3+4 2 |
| | 8 2 6 2 | 7 2 5 2 | 5+6 2 |
| | 9 2 7 2 | 7 2 5 2 | 3+4 2 |
| | 9 2 9 2 | 6 2 6 2 | 5+6 2 |
| | 6 2 8 2 | 7 2 5 2 | 5+6 2 |
| | 7 2 9 2 | 6 2 5 2 | 3+4 2 |
| | 6 2 8 2 | 7 2 6 2 | 3+4 2 |
| | 7 2 6 2 | 7 2 6 2 | 5+6 2 |
| | 8 2 8 2 | 6 2 5 2 | 5+6 2 |
| | 6 2 9 2 | 6 2 6 2 | 3+4 2 |
| | 7 2 8 2 | 7 2 5 2 | 5+6 2 |
| | 6 2 7 2 | 7 2 6 2 | 3+4 2 |
| | 9 2 7 2 | 6 2 5 2 | 3+4 2 |
| 139.1.170 | 85890 | 30 | Ca1Co2P2 |
| | -GM -M | -P -X | -N |
| | 7 2 7 2 | 7 2 5 2 | 5+6 2 |
| | 9 2 9 2 | 6 2 6 2 | 3+4 2 |
| | 7 2 7 2 | 6 2 6 2 | 3+4 2 |
| | 8 2 6 2 | 7 2 5 2 | 5+6 2 |
| | 9 2 7 2 | 7 2 5 2 | 3+4 2 |
| | 9 2 9 2 | 6 2 6 2 | 5+6 2 |
| | 6 2 8 2 | 7 2 5 2 | 5+6 2 |
| | 7 2 9 2 | 6 2 5 2 | 3+4 2 |
| | 6 2 8 2 | 7 2 6 2 | 3+4 2 |
| | 7 2 6 2 | 7 2 6 2 | 5+6 2 |
| | 8 2 8 2 | 6 2 5 2 | 5+6 2 |
| | 6 2 9 2 | 6 2 6 2 | 3+4 2 |
| | 7 2 8 2 | 7 2 5 2 | 5+6 2 |
| | 6 2 7 2 | 7 2 6 2 | 3+4 2 |
| | 9 2 7 2 | 6 2 5 2 | 3+4 2 |
| 139.1.171 | 81751 | 28 | Ca1Co2Ge2 |
| | -GM -M | -P -X | -N |
| | 7 2 9 2 | 7 2 5 2 | 5+6 2 |
| | 9 2 7 2 | 6 2 6 2 | 3+4 2 |
| | 7 2 7 2 | 6 2 6 2 | 3+4 2 |
| | 9 2 6 2 | 7 2 5 2 | 5+6 2 |
| | 8 2 7 2 | 7 2 5 2 | 3+4 2 |
| | 9 2 8 2 | 6 2 6 2 | 5+6 2 |
| | 6 2 9 2 | 7 2 5 2 | 5+6 2 |
| | 7 2 9 2 | 6 2 5 2 | 3+4 2 |
| | 6 2 8 2 | 7 2 6 2 | 3+4 2 |
| | 8 2 6 2 | 7 2 6 2 | 5+6 2 |
| | 7 2 8 2 | 6 2 5 2 | 5+6 2 |
| | 6 2 9 2 | 6 2 6 2 | 3+4 2 |
| | 7 2 8 2 | 7 2 5 2 | 5+6 2 |
| | 6 2 7 2 | 7 2 6 2 | 3+4 2 |
| 139.1.172 | 406 | 28 | Ca1Co2Ge2 |
| | -GM -M | -P -X | -N |
| | 7 2 9 2 | 7 2 5 2 | 5+6 2 |
| | 9 2 7 2 | 6 2 6 2 | 3+4 2 |
| | 7 2 7 2 | 6 2 6 2 | 3+4 2 |
| | 9 2 6 2 | 7 2 5 2 | 5+6 2 |
| | 8 2 7 2 | 7 2 5 2 | 3+4 2 |
| | 9 2 8 2 | 6 2 6 2 | 5+6 2 |
| | 6 2 9 2 | 7 2 5 2 | 5+6 2 |
| | 7 2 9 2 | 6 2 5 2 | 3+4 2 |
| | 6 2 8 2 | 7 2 6 2 | 3+4 2 |
| | 8 2 6 2 | 7 2 6 2 | 5+6 2 |
| | 7 2 8 2 | 6 2 5 2 | 5+6 2 |
| | 6 2 9 2 | 6 2 6 2 | 3+4 2 |
| | 7 2 8 2 | 7 2 5 2 | 5+6 2 |
| | 6 2 7 2 | 7 2 6 2 | 3+4 2 |
| 139.1.173 | 74670 | 32 | C2Yb1 |
| | -GM -M | -P -X | -N |
| | 7 2 7 2 | 6 2 5 2 | 3+4 2 |
| | 9 2 9 2 | 7 2 6 2 | 5+6 2 |
| | 8 2 9 2 | 6 2 6 2 | 5+6 2 |
| | 9 2 8 2 | 7 2 6 2 | 5+6 2 |
| | 7 2 7 2 | 7 2 5 2 | 5+6 2 |
| | 7 2 9 2 | 6 2 6 2 | 3+4 2 |
| | 9 2 9 2 | 7 2 6 2 | 3+4 2 |
| | 8 2 8 2 | 6 2 5 2 | 3+4 2 |
| | 9 2 7 2 | 7 2 6 2 | 5+6 2 |
| | 8 2 9 2 | 6 2 5 2 | 5+6 2 |
| | 9 2 8 2 | 6 2 6 2 | 5+6 2 |
| | 8 2 8 2 | 7 2 6 2 | 5+6 2 |
| | 8 2 9 2 | 6 2 6 2 | 5+6 2 |
| | 8 2 8 2 | 7 2 6 2 | 5+6 2 |
| | 9 2 8 2 | 7 2 6 2 | 5+6 2 |
| | 9 2 9 2 | 6 2 6 2 | 5+6 2 |
| 139.1.174 | 23686 | 32 | C2Yb1 |
| | -GM -M | -P -X | -N |
| | 7 2 7 2 | 6 2 5 2 | 3+4 2 |
| | 9 2 9 2 | 7 2 6 2 | 5+6 2 |
| | 8 2 9 2 | 6 2 6 2 | 5+6 2 |
| | 9 2 8 2 | 7 2 6 2 | 5+6 2 |
| | 7 2 7 2 | 7 2 5 2 | 5+6 2 |
| | 7 2 9 2 | 6 2 6 2 | 3+4 2 |
| | 9 2 9 2 | 7 2 6 2 | 3+4 2 |
| | 8 2 8 2 | 6 2 5 2 | 3+4 2 |
| | 9 2 7 2 | 7 2 6 2 | 5+6 2 |
| | 8 2 9 2 | 6 2 5 2 | 5+6 2 |
| | 9 2 8 2 | 6 2 6 2 | 5+6 2 |
| | 8 2 8 2 | 7 2 6 2 | 5+6 2 |
| | 8 2 9 2 | 6 2 6 2 | 5+6 2 |
| | 8 2 8 2 | 7 2 6 2 | 5+6 2 |
| | 9 2 8 2 | 7 2 6 2 | 5+6 2 |
| | 9 2 9 2 | 6 2 6 2 | 5+6 2 |
| 139.1.175 | 26478 | 22 | C2U1 |
| | -GM -M | -P -X | -N |
| | 7 2 7 2 | 6 2 5 2 | 3+4 2 |
| | 9 2 9 2 | 7 2 6 2 | 5+6 2 |

Continued on next column

**TABLE LXIV– continued**

| TQC | ICSD NoE | SF |
|---|---|---|
| | 8 2 9 2 | 6 2 6 2 5+6 2 |
| | 9 2 8 2 | 7 2 6 2 5+6 2 |
| | 7 2 7 2 | 7 2 5 2 5+6 2 |
| | 7 2 9 2 | 6 2 6 2 3+4 2 |
| | 9 2 9 2 | 7 2 5 2 3+4 2 |
| | 9 2 8 2 | 6 2 5 2 3+4 2 |
| | 8 2 7 2 | 7 2 6 2 5+6 2 |
| | 7 2 6 2 | 7 2 6 2 5+6 2 |
| | 6 2 9 2 | 6 2 6 2 5+6 2 |
| 139.1.176 | 202305 22 | C2U1 |
| | -GM -M | -P -X -N |
| | 7 2 7 2 | 6 2 5 2 3+4 2 |
| | 9 2 9 2 | 7 2 6 2 5+6 2 |
| | 8 2 9 2 | 6 2 6 2 5+6 2 |
| | 9 2 8 2 | 7 2 6 2 5+6 2 |
| | 7 2 7 2 | 7 2 5 2 5+6 2 |
| | 7 2 9 2 | 6 2 6 2 3+4 2 |
| | 9 2 9 2 | 7 2 5 2 3+4 2 |
| | 9 2 8 2 | 6 2 5 2 3+4 2 |
| | 8 2 7 2 | 7 2 6 2 5+6 2 |
| | 7 2 6 2 | 7 2 6 2 5+6 2 |
| | 6 2 9 2 | 6 2 6 2 5+6 2 |
| 139.1.177 | 63560 22 | C2Nd1 |
| | -GM -M | -P -X -N |
| | 7 2 7 2 | 6 2 5 2 3+4 2 |
| | 9 2 9 2 | 7 2 6 2 5+6 2 |
| | 8 2 9 2 | 6 2 6 2 5+6 2 |
| | 9 2 8 2 | 7 2 6 2 5+6 2 |
| | 7 2 7 2 | 7 2 5 2 5+6 2 |
| | 7 2 9 2 | 6 2 6 2 3+4 2 |
| | 9 2 9 2 | 7 2 6 2 3+4 2 |
| | 9 2 8 2 | 6 2 5 2 3+4 2 |
| | 8 2 7 2 | 7 2 6 2 5+6 2 |
| | 8 2 6 2 | 7 2 5 2 5+6 2 |
| | 7 2 9 2 | 6 2 6 2 5+6 2 |
| 139.1.178 | 69888 30 | C2Er1 |
| | -GM -M | -P -X -N |
| | 7 2 7 2 | 6 2 5 2 3+4 2 |
| | 9 2 9 2 | 7 2 6 2 5+6 2 |
| | 8 2 9 2 | 6 2 6 2 5+6 2 |
| | 9 2 8 2 | 7 2 6 2 5+6 2 |
| | 7 2 7 2 | 7 2 5 2 5+6 2 |
| | 7 2 9 2 | 6 2 6 2 3+4 2 |
| | 9 2 9 2 | 7 2 6 2 3+4 2 |
| | 9 2 8 2 | 6 2 5 2 3+4 2 |
| | 8 2 7 2 | 7 2 6 2 5+6 2 |
| | 8 2 9 2 | 6 2 5 2 5+6 2 |
| | 9 2 8 2 | 6 2 6 2 5+6 2 |
| | 8 2 8 2 | 7 2 6 2 5+6 2 |
| | 8 2 9 2 | 6 2 6 2 5+6 2 |
| | 8 2 8 2 | 7 2 6 2 5+6 2 |
| | 9 2 8 2 | 7 2 6 2 5+6 2 |
| 139.1.179 | 23690 30 | C2Er1 |
| | -GM -M | -P -X -N |
| | 7 2 7 2 | 6 2 5 2 3+4 2 |
| | 9 2 9 2 | 7 2 6 2 5+6 2 |
| | 8 2 9 2 | 6 2 6 2 5+6 2 |
| | 9 2 8 2 | 7 2 6 2 5+6 2 |
| | 7 2 7 2 | 7 2 5 2 5+6 2 |
| | 7 2 9 2 | 6 2 6 2 3+4 2 |
| | 9 2 9 2 | 7 2 6 2 3+4 2 |

Continued on next column

**TABLE LXIV– continued**

| TQC | ICSD NoE | SF |
|---|---|---|
| | 9 2 8 2 | 6 2 5 2 3+4 2 |
| | 8 2 7 2 | 7 2 6 2 5+6 2 |
| | 8 2 9 2 | 6 2 5 2 5+6 2 |
| | 9 2 8 2 | 6 2 6 2 5+6 2 |
| | 8 2 8 2 | 7 2 6 2 5+6 2 |
| | 8 2 9 2 | 6 2 6 2 5+6 2 |
| | 8 2 8 2 | 7 2 6 2 5+6 2 |
| | 9 2 8 2 | 7 2 6 2 5+6 2 |
| 139.1.180 | 23689 30 | C2Er1 |
| | -GM -M | -P -X -N |
| | 7 2 7 2 | 6 2 5 2 3+4 2 |
| | 9 2 9 2 | 7 2 6 2 5+6 2 |
| | 8 2 9 2 | 6 2 6 2 5+6 2 |
| | 9 2 8 2 | 7 2 6 2 5+6 2 |
| | 7 2 7 2 | 7 2 5 2 5+6 2 |
| | 7 2 9 2 | 6 2 6 2 3+4 2 |
| | 9 2 9 2 | 7 2 6 2 3+4 2 |
| | 9 2 8 2 | 6 2 5 2 3+4 2 |
| | 8 2 7 2 | 7 2 6 2 5+6 2 |
| | 8 2 9 2 | 6 2 5 2 5+6 2 |
| | 9 2 8 2 | 6 2 6 2 5+6 2 |
| | 8 2 8 2 | 7 2 6 2 5+6 2 |
| | 8 2 9 2 | 6 2 6 2 5+6 2 |
| | 8 2 8 2 | 7 2 6 2 5+6 2 |
| | 9 2 8 2 | 7 2 6 2 5+6 2 |
| 139.1.181 | 380471 88 | C2Cl1N2Ni1Rb1 |
| | -GM -M | -P -X -N |
| | 7 2 9 2 | 6 2 5 2 3+4 2 |
| | 8 2 6 2 | 7 2 6 2 5+6 2 |
| | 7 2 7 2 | 7 2 5 2 5+6 2 |
| | 8 2 8 2 | 6 2 5 2 3+4 2 |
| | 9 2 9 2 | 6 2 6 2 3+4 2 |
| | 6 2 6 2 | 7 2 6 2 5+6 2 |
| | 7 2 7 2 | 7 2 5 2 5+6 2 |
| | 9 2 9 2 | 6 2 6 2 3+4 2 |
| | 6 2 8 2 | 6 2 6 2 3+4 2 |
| | 9 2 7 2 | 7 2 5 2 5+6 2 |
| | 6 2 8 2 | 6 2 6 2 3+4 2 |
| | 9 2 7 2 | 7 2 5 2 5+6 2 |
| | 7 2 9 2 | 6 2 5 2 3+4 2 |
| | 8 2 6 2 | 7 2 6 2 5+6 2 |
| | 7 2 7 2 | 6 2 5 2 3+4 2 |
| | 9 2 9 2 | 7 2 5 2 5+6 2 |
| | 8 2 8 2 | 7 2 6 2 3+4 2 |
| | 6 2 6 2 | 6 2 6 2 5+6 2 |
| | 7 2 7 2 | 7 2 5 2 3+4 2 |
| | 6 2 6 2 | 6 2 5 2 5+6 2 |
| | 6 2 6 2 | 7 2 6 2 3+4 2 |
| | 9 2 9 2 | 6 2 6 2 3+4 2 |
| | 6 2 7 2 | 7 2 5 2 5+6 2 |
| | 7 2 6 2 | 6 2 5 2 5+6 2 |
| | 9 2 9 2 | 7 2 6 2 3+4 2 |
| | 8 2 8 2 | 7 2 5 2 3+4 2 |
| | 8 2 8 2 | 6 2 5 2 5+6 2 |
| | 9 2 9 2 | 6 2 6 2 5+6 2 |
| | 8 2 8 2 | 7 2 5 2 3+4 2 |
| | 7 2 7 2 | 6 2 6 2 5+6 2 |
| | 7 2 7 2 | 7 2 6 2 5+6 2 |
| | 6 2 9 2 | 6 2 5 2 3+4 2 |
| | 9 2 6 2 | 7 2 6 2 3+4 2 |
| | 7 2 7 2 | 6 2 5 2 5+6 2 |

Continued on next column

**TABLE LXIV– continued**

| TQC | ICSD | NoE | SF |
|---|---|---|---|
| | 6 2 6 2 | 6 2 6 2 | 3+4 2 |
| | 9 2 9 2 | 7 2 5 2 | 5+6 2 |
| | 8 2 8 2 | 7 2 6 2 | 3+4 2 |
| | 7 2 7 2 | 6 2 5 2 | 3+4 2 |
| | 6 2 6 2 | 7 2 5 2 | 3+4 2 |
| | 6 2 6 2 | 6 2 5 2 | 5+6 2 |
| | 7 2 7 2 | 7 2 5 2 | 5+6 2 |
| | 7 2 7 2 | 6 2 5 2 | 5+6 2 |
| | 6 2 6 2 | 6 2 5 2 | 3+4 2 |
| | 6 2 6 2 | 7 2 5 2 | 5+6 2 |
| 139.1.182 | 380469 | 88 | C2Cl1Cs1N2Ni1 |
| | -GM -M | -P -X | -N |
| | 7 2 9 2 | 6 2 6 2 | 3+4 2 |
| | 8 2 6 2 | 7 2 5 2 | 5+6 2 |
| | 7 2 7 2 | 7 2 5 2 | 5+6 2 |
| | 8 2 8 2 | 6 2 5 2 | 3+4 2 |
| | 9 2 9 2 | 6 2 6 2 | 3+4 2 |
| | 6 2 6 2 | 7 2 6 2 | 5+6 2 |
| | 7 2 7 2 | 7 2 5 2 | 5+6 2 |
| | 9 2 9 2 | 6 2 6 2 | 3+4 2 |
| | 6 2 8 2 | 6 2 6 2 | 3+4 2 |
| | 7 2 7 2 | 7 2 5 2 | 5+6 2 |
| | 9 2 7 2 | 6 2 5 2 | 3+4 2 |
| | 6 2 9 2 | 7 2 5 2 | 5+6 2 |
| | 7 2 7 2 | 7 2 6 2 | 5+6 2 |
| | 8 2 8 2 | 6 2 5 2 | 3+4 2 |
| | 9 2 6 2 | 7 2 6 2 | 5+6 2 |
| | 9 2 9 2 | 6 2 5 2 | 3+4 2 |
| | 6 2 8 2 | 6 2 6 2 | 3+4 2 |
| | 8 2 6 2 | 7 2 6 2 | 5+6 2 |
| | 7 2 7 2 | 7 2 5 2 | 3+4 2 |
| | 6 2 6 2 | 6 2 5 2 | 5+6 2 |
| | 6 2 6 2 | 7 2 6 2 | 3+4 2 |
| | 9 2 9 2 | 6 2 6 2 | 3+4 2 |
| | 6 2 7 2 | 7 2 5 2 | 5+6 2 |
| | 7 2 6 2 | 6 2 5 2 | 5+6 2 |
| | 9 2 9 2 | 7 2 6 2 | 3+4 2 |
| | 8 2 8 2 | 7 2 5 2 | 3+4 2 |
| | 8 2 8 2 | 6 2 5 2 | 5+6 2 |
| | 9 2 9 2 | 6 2 6 2 | 5+6 2 |
| | 8 2 8 2 | 7 2 6 2 | 3+4 2 |
| | 7 2 7 2 | 6 2 5 2 | 5+6 2 |
| | 7 2 7 2 | 7 2 6 2 | 5+6 2 |
| | 6 2 9 2 | 6 2 5 2 | 3+4 2 |
| | 9 2 6 2 | 7 2 6 2 | 5+6 2 |
| | 7 2 7 2 | 6 2 5 2 | 3+4 2 |
| | 6 2 6 2 | 6 2 6 2 | 3+4 2 |
| | 9 2 9 2 | 7 2 6 2 | 5+6 2 |
| | 8 2 8 2 | 7 2 5 2 | 3+4 2 |
| | 7 2 7 2 | 6 2 5 2 | 3+4 2 |
| | 6 2 6 2 | 7 2 5 2 | 3+4 2 |
| | 6 2 6 2 | 6 2 5 2 | 5+6 2 |
| | 7 2 7 2 | 7 2 5 2 | 5+6 2 |
| | 7 2 7 2 | 6 2 5 2 | 5+6 2 |
| | 6 2 6 2 | 6 2 5 2 | 3+4 2 |
| | 6 2 6 2 | 7 2 5 2 | 5+6 2 |
| 139.1.183 | 380470 | 88 | C2Br1N2Ni1Rb1 |
| | -GM -M | -P -X | -N |
| | 7 2 9 2 | 6 2 5 2 | 3+4 2 |
| | 8 2 6 2 | 7 2 6 2 | 5+6 2 |
| | 7 2 7 2 | 7 2 5 2 | 5+6 2 |

Continued on next column

**TABLE LXIV– continued**

| TQC | ICSD | NoE | SF |
|---|---|---|---|
| | 8 2 8 2 | 6 2 5 2 | 3+4 2 |
| | 9 2 9 2 | 6 2 6 2 | 3+4 2 |
| | 6 2 6 2 | 7 2 6 2 | 5+6 2 |
| | 7 2 7 2 | 7 2 5 2 | 5+6 2 |
| | 9 2 9 2 | 6 2 6 2 | 3+4 2 |
| | 6 2 8 2 | 6 2 6 2 | 3+4 2 |
| | 9 2 7 2 | 7 2 5 2 | 5+6 2 |
| | 6 2 8 2 | 6 2 6 2 | 3+4 2 |
| | 7 2 9 2 | 7 2 5 2 | 5+6 2 |
| | 9 2 7 2 | 6 2 5 2 | 3+4 2 |
| | 8 2 6 2 | 7 2 6 2 | 5+6 2 |
| | 7 2 7 2 | 6 2 5 2 | 3+4 2 |
| | 9 2 9 2 | 7 2 5 2 | 5+6 2 |
| | 8 2 8 2 | 6 2 6 2 | 3+4 2 |
| | 6 2 6 2 | 7 2 6 2 | 5+6 2 |
| | 7 2 7 2 | 7 2 5 2 | 3+4 2 |
| | 6 2 6 2 | 6 2 5 2 | 5+6 2 |
| | 6 2 6 2 | 7 2 6 2 | 3+4 2 |
| | 9 2 9 2 | 6 2 6 2 | 3+4 2 |
| | 7 2 7 2 | 7 2 5 2 | 5+6 2 |
| | 6 2 6 2 | 7 2 5 2 | 5+6 2 |
| | 9 2 9 2 | 6 2 6 2 | 3+4 2 |
| | 8 2 8 2 | 7 2 5 2 | 3+4 2 |
| | 8 2 8 2 | 6 2 5 2 | 5+6 2 |
| | 9 2 9 2 | 6 2 6 2 | 5+6 2 |
| | 8 2 8 2 | 7 2 6 2 | 3+4 2 |
| | 7 2 7 2 | 6 2 6 2 | 5+6 2 |
| | 7 2 7 2 | 7 2 5 2 | 5+6 2 |
| | 6 2 9 2 | 6 2 5 2 | 3+4 2 |
| | 9 2 6 2 | 7 2 6 2 | 3+4 2 |
| | 7 2 7 2 | 6 2 5 2 | 5+6 2 |
| | 6 2 6 2 | 7 2 5 2 | 3+4 2 |
| | 7 2 9 2 | 6 2 6 2 | 5+6 2 |
| | 9 2 8 2 | 7 2 6 2 | 3+4 2 |
| | 6 2 7 2 | 6 2 5 2 | 3+4 2 |
| | 8 2 6 2 | 7 2 5 2 | 3+4 2 |
| | 6 2 6 2 | 6 2 5 2 | 5+6 2 |
| | 7 2 7 2 | 7 2 5 2 | 5+6 2 |
| | 7 2 7 2 | 6 2 5 2 | 5+6 2 |
| | 6 2 6 2 | 6 2 5 2 | 3+4 2 |
| | 6 2 6 2 | 7 2 5 2 | 5+6 2 |
| 139.1.184 | 380468 | 88 | C2Br1Cs1N2Ni1 |
| | -GM -M | -P -X | -N |
| | 7 2 9 2 | 6 2 6 2 | 3+4 2 |
| | 8 2 6 2 | 7 2 5 2 | 5+6 2 |
| | 7 2 7 2 | 7 2 5 2 | 5+6 2 |
| | 8 2 8 2 | 6 2 5 2 | 3+4 2 |
| | 9 2 9 2 | 6 2 6 2 | 3+4 2 |
| | 6 2 6 2 | 7 2 6 2 | 5+6 2 |
| | 7 2 7 2 | 7 2 5 2 | 5+6 2 |
| | 9 2 9 2 | 6 2 6 2 | 3+4 2 |
| | 7 2 7 2 | 6 2 5 2 | 3+4 2 |
| | 6 2 8 2 | 7 2 6 2 | 5+6 2 |
| | 9 2 7 2 | 6 2 5 2 | 3+4 2 |
| | 7 2 9 2 | 7 2 5 2 | 5+6 2 |
| | 6 2 7 2 | 7 2 6 2 | 5+6 2 |
| | 8 2 8 2 | 6 2 5 2 | 3+4 2 |
| | 9 2 6 2 | 7 2 6 2 | 5+6 2 |
| | 9 2 9 2 | 6 2 5 2 | 3+4 2 |
| | 6 2 8 2 | 7 2 6 2 | 3+4 2 |
| | 8 2 6 2 | 6 2 6 2 | 5+6 2 |

Continued on next column

**TABLE LXIV– continued**

| TQC | ICSD | NoE | SF |
|---|---|---|---|
| | 7 2 7 2 | 7 2 5 2 | 3+4 2 |
| | 6 2 6 2 | 6 2 5 2 | 3+4 2 |
| | 6 2 6 2 | 7 2 6 2 | 5+6 2 |
| | 9 2 9 2 | 6 2 6 2 | 3+4 2 |
| | 9 2 7 2 | 7 2 5 2 | 5+6 2 |
| | 6 2 6 2 | 7 2 5 2 | 5+6 2 |
| | 8 2 9 2 | 6 2 6 2 | 3+4 2 |
| | 7 2 8 2 | 7 2 5 2 | 3+4 2 |
| | 8 2 8 2 | 6 2 5 2 | 5+6 2 |
| | 9 2 9 2 | 6 2 6 2 | 5+6 2 |
| | 8 2 8 2 | 7 2 6 2 | 3+4 2 |
| | 7 2 7 2 | 6 2 5 2 | 5+6 2 |
| | 7 2 7 2 | 7 2 6 2 | 5+6 2 |
| | 6 2 9 2 | 6 2 5 2 | 3+4 2 |
| | 9 2 6 2 | 7 2 6 2 | 3+4 2 |
| | 7 2 7 2 | 6 2 5 2 | 5+6 2 |
| | 6 2 6 2 | 7 2 5 2 | 3+4 2 |
| | 9 2 9 2 | 6 2 6 2 | 5+6 2 |
| | 7 2 8 2 | 7 2 6 2 | 3+4 2 |
| | 6 2 7 2 | 6 2 5 2 | 3+4 2 |
| | 8 2 6 2 | 7 2 5 2 | 3+4 2 |
| | 7 2 7 2 | 6 2 5 2 | 5+6 2 |
| | 6 2 6 2 | 7 2 5 2 | 5+6 2 |
| | 7 2 7 2 | 6 2 5 2 | 5+6 2 |
| | 6 2 6 2 | 6 2 5 2 | 3+4 2 |
| | 6 2 6 2 | 7 2 5 2 | 5+6 2 |
| 139.1.185 | 79563 | 48 | C1B2Gd1Ni2 |
| | -GM -M | -P -X | -N |
| | 7 2 7 2 | 6 2 5 2 | 3+4 2 |
| | 9 2 9 2 | 7 2 6 2 | 5+6 2 |
| | 9 2 9 2 | 6 2 6 2 | 5+6 2 |
| | 8 2 8 2 | 7 2 6 2 | 5+6 2 |
| | 7 2 7 2 | 7 2 5 2 | 5+6 2 |
| | 7 2 9 2 | 6 2 6 2 | 3+4 2 |
| | 9 2 9 2 | 7 2 5 2 | 5+6 2 |
| | 9 2 7 2 | 6 2 6 2 | 3+4 2 |
| | 9 2 6 2 | 7 2 5 2 | 3+4 2 |
| | 8 2 7 2 | 6 2 6 2 | 3+4 2 |
| | 6 2 8 2 | 7 2 6 2 | 5+6 2 |
| | 7 2 9 2 | 6 2 5 2 | 5+6 2 |
| | 7 2 8 2 | 6 2 5 2 | 3+4 2 |
| | 6 2 6 2 | 7 2 6 2 | 5+6 2 |
| | 8 2 7 2 | 6 2 5 2 | 3+4 2 |
| | 8 2 9 2 | 7 2 6 2 | 5+6 2 |
| | 9 2 8 2 | 7 2 6 2 | 3+4 2 |
| | 6 2 9 2 | 6 2 5 2 | 5+6 2 |
| | 7 2 8 2 | 6 2 6 2 | 3+4 2 |
| | 6 2 8 2 | 7 2 5 2 | 3+4 2 |
| | 8 2 9 2 | 7 2 6 2 | 5+6 2 |
| | 9 2 7 2 | 6 2 6 2 | 5+6 2 |
| | 9 2 8 2 | 6 2 6 2 | 5+6 2 |
| | 8 2 8 2 | 6 2 6 2 | 5+6 2 |
| 139.1.186 | 56651 | 52 | C1B2Er1Ni2 |
| | -GM -M | -P -X | -N |
| | 7 2 7 2 | 6 2 5 2 | 3+4 2 |
| | 9 2 9 2 | 7 2 6 2 | 5+6 2 |
| | 9 2 9 2 | 6 2 6 2 | 5+6 2 |
| | 8 2 8 2 | 7 2 6 2 | 5+6 2 |
| | 7 2 7 2 | 7 2 5 2 | 5+6 2 |
| | 7 2 9 2 | 6 2 6 2 | 3+4 2 |
| | 9 2 9 2 | 7 2 6 2 | 3+4 2 |

Continued on next column

**TABLE LXIV– continued**

| TQC | ICSD | NoE | SF |
|---|---|---|---|
| | 9 2 7 2 | 6 2 5 2 | 5+6 2 |
| | 9 2 6 2 | 7 2 5 2 | 3+4 2 |
| | 8 2 7 2 | 6 2 6 2 | 3+4 2 |
| | 6 2 8 2 | 7 2 6 2 | 5+6 2 |
| | 7 2 9 2 | 6 2 6 2 | 5+6 2 |
| | 7 2 8 2 | 6 2 5 2 | 3+4 2 |
| | 6 2 7 2 | 7 2 5 2 | 5+6 2 |
| | 8 2 6 2 | 6 2 5 2 | 3+4 2 |
| | 8 2 9 2 | 7 2 6 2 | 5+6 2 |
| | 9 2 8 2 | 7 2 6 2 | 3+4 2 |
| | 8 2 9 2 | 6 2 5 2 | 5+6 2 |
| | 9 2 8 2 | 6 2 6 2 | 3+4 2 |
| | 6 2 9 2 | 7 2 5 2 | 3+4 2 |
| | 8 2 8 2 | 7 2 6 2 | 5+6 2 |
| | 9 2 8 2 | 6 2 6 2 | 5+6 2 |
| | 7 2 7 2 | 6 2 6 2 | 5+6 2 |
| | 6 2 8 2 | 6 2 6 2 | 5+6 2 |
| | 9 2 9 2 | 7 2 6 2 | 5+6 2 |
| | 8 2 8 2 | 7 2 6 2 | 5+6 2 |
| 139.1.187 | 56596 | 50 | C1B2Dy1Ni2 |
| | -GM -M | -P -X | -N |
| | 7 2 7 2 | 7 2 5 2 | 5+6 2 |
| | 9 2 9 2 | 6 2 6 2 | 3+4 2 |
| | 9 2 9 2 | 7 2 6 2 | 3+4 2 |
| | 8 2 8 2 | 6 2 6 2 | 3+4 2 |
| | 7 2 7 2 | 6 2 5 2 | 3+4 2 |
| | 7 2 9 2 | 7 2 6 2 | 5+6 2 |
| | 9 2 9 2 | 6 2 6 2 | 3+4 2 |
| | 9 2 7 2 | 7 2 5 2 | 5+6 2 |
| | 9 2 6 2 | 6 2 5 2 | 5+6 2 |
| | 8 2 7 2 | 7 2 6 2 | 5+6 2 |
| | 6 2 8 2 | 6 2 6 2 | 3+4 2 |
| | 7 2 9 2 | 7 2 6 2 | 3+4 2 |
| | 7 2 8 2 | 7 2 5 2 | 5+6 2 |
| | 6 2 7 2 | 6 2 5 2 | 3+4 2 |
| | 8 2 6 2 | 7 2 5 2 | 5+6 2 |
| | 8 2 9 2 | 6 2 6 2 | 3+4 2 |
| | 9 2 8 2 | 6 2 6 2 | 5+6 2 |
| | 6 2 9 2 | 7 2 5 2 | 3+4 2 |
| | 8 2 8 2 | 6 2 6 2 | 5+6 2 |
| | 9 2 8 2 | 7 2 5 2 | 5+6 2 |
| | 9 2 9 2 | 6 2 6 2 | 3+4 2 |
| | 7 2 8 2 | 7 2 6 2 | 3+4 2 |
| | 6 2 7 2 | 7 2 6 2 | 3+4 2 |
| | 8 2 8 2 | 7 2 6 2 | 3+4 2 |
| | 9 2 9 2 | 6 2 6 2 | 3+4 2 |
| 139.1.188 | 56595 | 50 | C1B2Dy1Ni2 |
| | -GM -M | -P -X | -N |
| | 7 2 7 2 | 7 2 5 2 | 5+6 2 |
| | 9 2 9 2 | 6 2 6 2 | 3+4 2 |
| | 9 2 9 2 | 7 2 6 2 | 3+4 2 |
| | 8 2 8 2 | 6 2 6 2 | 3+4 2 |
| | 7 2 7 2 | 6 2 5 2 | 3+4 2 |
| | 7 2 9 2 | 7 2 6 2 | 5+6 2 |
| | 9 2 9 2 | 6 2 6 2 | 3+4 2 |
| | 9 2 7 2 | 7 2 5 2 | 5+6 2 |
| | 9 2 6 2 | 6 2 5 2 | 5+6 2 |
| | 8 2 7 2 | 7 2 6 2 | 5+6 2 |
| | 6 2 8 2 | 6 2 6 2 | 3+4 2 |
| | 7 2 9 2 | 7 2 6 2 | 3+4 2 |
| | 7 2 8 2 | 7 2 5 2 | 5+6 2 |

Continued on next column

**TABLE LXIV– continued**

| TQC | ICSD | NoE | SF |
|---|---|---|---|
| | 6 2 7 2 | 6 2 5 2 | 3+4 2 |
| | 8 2 6 2 | 7 2 5 2 | 5+6 2 |
| | 8 2 9 2 | 6 2 6 2 | 3+4 2 |
| | 9 2 8 2 | 6 2 6 2 | 5+6 2 |
| | 6 2 9 2 | 7 2 5 2 | 3+4 2 |
| | 8 2 8 2 | 6 2 6 2 | 5+6 2 |
| | 9 2 8 2 | 7 2 5 2 | 5+6 2 |
| | 9 2 9 2 | 6 2 6 2 | 3+4 2 |
| | 7 2 8 2 | 7 2 6 2 | 3+4 2 |
| | 6 2 7 2 | 7 2 6 2 | 3+4 2 |
| | 8 2 8 2 | 7 2 6 2 | 3+4 2 |
| | 9 2 9 2 | 6 2 6 2 | 3+4 2 |
| 139.1.189 | 56594 | 50 | C1B2Dy1Ni2 |
| | -GM -M | -P -X | -N |
| | 7 2 7 2 | 7 2 5 2 | 5+6 2 |
| | 9 2 9 2 | 6 2 6 2 | 3+4 2 |
| | 9 2 9 2 | 7 2 6 2 | 3+4 2 |
| | 8 2 8 2 | 6 2 6 2 | 3+4 2 |
| | 7 2 7 2 | 6 2 5 2 | 3+4 2 |
| | 7 2 9 2 | 7 2 6 2 | 5+6 2 |
| | 9 2 9 2 | 6 2 6 2 | 5+6 2 |
| | 9 2 7 2 | 7 2 5 2 | 3+4 2 |
| | 9 2 6 2 | 6 2 5 2 | 5+6 2 |
| | 8 2 7 2 | 7 2 6 2 | 5+6 2 |
| | 6 2 8 2 | 6 2 6 2 | 3+4 2 |
| | 7 2 9 2 | 7 2 6 2 | 3+4 2 |
| | 7 2 8 2 | 7 2 5 2 | 5+6 2 |
| | 6 2 7 2 | 6 2 5 2 | 3+4 2 |
| | 8 2 6 2 | 7 2 5 2 | 5+6 2 |
| | 8 2 9 2 | 6 2 6 2 | 3+4 2 |
| | 9 2 8 2 | 6 2 6 2 | 5+6 2 |
| | 6 2 9 2 | 7 2 5 2 | 3+4 2 |
| | 8 2 8 2 | 6 2 6 2 | 5+6 2 |
| | 9 2 8 2 | 7 2 5 2 | 5+6 2 |
| | 9 2 9 2 | 6 2 6 2 | 3+4 2 |
| | 7 2 8 2 | 7 2 6 2 | 3+4 2 |
| | 6 2 7 2 | 7 2 6 2 | 3+4 2 |
| | 8 2 8 2 | 7 2 6 2 | 3+4 2 |
| | 9 2 9 2 | 6 2 6 2 | 3+4 2 |
| 139.1.190 | 56593 | 50 | C1B2Dy1Ni2 |
| | -GM -M | -P -X | -N |
| | 7 2 7 2 | 7 2 5 2 | 5+6 2 |
| | 9 2 9 2 | 6 2 6 2 | 3+4 2 |
| | 9 2 9 2 | 7 2 6 2 | 3+4 2 |
| | 8 2 8 2 | 6 2 6 2 | 3+4 2 |
| | 7 2 7 2 | 6 2 5 2 | 3+4 2 |
| | 7 2 9 2 | 7 2 6 2 | 5+6 2 |
| | 9 2 9 2 | 6 2 6 2 | 5+6 2 |
| | 9 2 7 2 | 7 2 5 2 | 3+4 2 |
| | 9 2 6 2 | 6 2 5 2 | 5+6 2 |
| | 8 2 7 2 | 7 2 6 2 | 5+6 2 |
| | 6 2 8 2 | 6 2 6 2 | 3+4 2 |
| | 7 2 9 2 | 7 2 6 2 | 3+4 2 |
| | 7 2 8 2 | 7 2 5 2 | 5+6 2 |
| | 6 2 7 2 | 6 2 5 2 | 3+4 2 |
| | 8 2 6 2 | 7 2 5 2 | 5+6 2 |
| | 8 2 9 2 | 6 2 6 2 | 3+4 2 |
| | 9 2 8 2 | 6 2 6 2 | 5+6 2 |
| | 6 2 9 2 | 7 2 5 2 | 3+4 2 |
| | 8 2 8 2 | 6 2 6 2 | 5+6 2 |
| | 9 2 8 2 | 7 2 5 2 | 5+6 2 |

Continued on next column

**TABLE LXIV– continued**

| TQC | ICSD | NoE | SF |
|---|---|---|---|
| | 9 2 9 2 | 6 2 6 2 | 3+4 2 |
| | 7 2 8 2 | 7 2 6 2 | 3+4 2 |
| | 6 2 7 2 | 7 2 6 2 | 3+4 2 |
| | 8 2 8 2 | 7 2 6 2 | 3+4 2 |
| | 9 2 9 2 | 6 2 6 2 | 3+4 2 |
| 139.1.191 | 95097 | 40 | C1B2Ce1Rh2 |
| | -GM -M | -P -X | -N |
| | 7 2 7 2 | 7 2 5 2 | 5+6 2 |
| | 9 2 9 2 | 6 2 6 2 | 3+4 2 |
| | 9 2 9 2 | 7 2 6 2 | 3+4 2 |
| | 8 2 8 2 | 6 2 6 2 | 3+4 2 |
| | 7 2 7 2 | 6 2 5 2 | 3+4 2 |
| | 9 2 9 2 | 7 2 6 2 | 5+6 2 |
| | 7 2 7 2 | 6 2 5 2 | 3+4 2 |
| | 9 2 9 2 | 7 2 6 2 | 5+6 2 |
| | 9 2 6 2 | 6 2 5 2 | 5+6 2 |
| | 8 2 7 2 | 7 2 6 2 | 5+6 2 |
| | 6 2 8 2 | 6 2 6 2 | 3+4 2 |
| | 7 2 9 2 | 7 2 5 2 | 3+4 2 |
| | 6 2 8 2 | 6 2 6 2 | 5+6 2 |
| | 7 2 6 2 | 7 2 5 2 | 3+4 2 |
| | 8 2 9 2 | 7 2 5 2 | 5+6 2 |
| | 8 2 9 2 | 6 2 6 2 | 3+4 2 |
| | 9 2 8 2 | 6 2 6 2 | 5+6 2 |
| | 6 2 7 2 | 7 2 5 2 | 3+4 2 |
| | 6 2 8 2 | 7 2 6 2 | 5+6 2 |
| | 7 2 8 2 | 6 2 5 2 | 5+6 2 |
| 139.1.192 | 95098 | 40 | C1B2Ce1Ir2 |
| | -GM -M | -P -X | -N |
| | 7 2 7 2 | 7 2 5 2 | 5+6 2 |
| | 9 2 9 2 | 6 2 6 2 | 3+4 2 |
| | 9 2 9 2 | 7 2 6 2 | 3+4 2 |
| | 8 2 8 2 | 6 2 6 2 | 3+4 2 |
| | 7 2 7 2 | 6 2 5 2 | 3+4 2 |
| | 7 2 9 2 | 7 2 6 2 | 5+6 2 |
| | 9 2 7 2 | 6 2 5 2 | 3+4 2 |
| | 9 2 9 2 | 7 2 6 2 | 5+6 2 |
| | 6 2 6 2 | 6 2 5 2 | 5+6 2 |
| | 8 2 8 2 | 7 2 6 2 | 5+6 2 |
| | 9 2 7 2 | 7 2 6 2 | 3+4 2 |
| | 7 2 9 2 | 6 2 5 2 | 3+4 2 |
| | 6 2 8 2 | 6 2 5 2 | 5+6 2 |
| | 7 2 6 2 | 7 2 6 2 | 3+4 2 |
| | 8 2 9 2 | 7 2 5 2 | 5+6 2 |
| | 8 2 7 2 | 6 2 6 2 | 3+4 2 |
| | 9 2 9 2 | 6 2 6 2 | 5+6 2 |
| | 6 2 8 2 | 7 2 5 2 | 3+4 2 |
| | 6 2 8 2 | 7 2 6 2 | 5+6 2 |
| | 7 2 8 2 | 6 2 5 2 | 5+6 2 |
| 139.1.193 | 32544 | 90 | Br7Mn2Rb3 |
| | -GM -M | -P -X | -N |
| | 7 2 9 2 | 6 2 5 2 | 3+4 2 |
| | 9 2 7 2 | 7 2 6 2 | 5+6 2 |
| | 7 2 7 2 | 6 2 5 2 | 3+4 2 |
| | 7 2 7 2 | 6 2 6 2 | 5+6 2 |
| | 9 2 9 2 | 7 2 6 2 | 3+4 2 |
| | 7 2 7 2 | 6 2 5 2 | 3+4 2 |
| | 9 2 9 2 | 7 2 5 2 | 5+6 2 |
| | 6 2 6 2 | 7 2 5 2 | 5+6 2 |
| | 7 2 8 2 | 6 2 6 2 | 3+4 2 |
| | 8 2 7 2 | 7 2 5 2 | 5+6 2 |

Continued on next column

**TABLE LXIV– continued**

| TQC | ICSD | NoE | SF |
|---|---|---|---|
| | 7 2 9 2 | 6 2 5 2 | 3+4 2 |
| | 9 2 7 2 | 7 2 6 2 | 5+6 2 |
| | 9 2 9 2 | 7 2 6 2 | 5+6 2 |
| | 7 2 9 2 | 6 2 5 2 | 3+4 2 |
| | 6 2 8 2 | 7 2 6 2 | 5+6 2 |
| | 8 2 6 2 | 6 2 5 2 | 3+4 2 |
| | 9 2 7 2 | 7 2 6 2 | 5+6 2 |
| | 8 2 8 2 | 6 2 6 2 | 5+6 2 |
| | 9 2 9 2 | 7 2 6 2 | 5+6 2 |
| | 8 2 9 2 | 6 2 6 2 | 3+4 2 |
| | 9 2 8 2 | 7 2 5 2 | 5+6 2 |
| | 6 2 6 2 | 6 2 5 2 | 3+4 2 |
| | 9 2 9 2 | 7 2 6 2 | 3+4 2 |
| | 7 2 7 2 | 6 2 6 2 | 5+6 2 |
| | 7 2 7 2 | 7 2 6 2 | 3+4 2 |
| | 9 2 9 2 | 7 2 5 2 | 5+6 2 |
| | 7 2 9 2 | 6 2 6 2 | 3+4 2 |
| | 9 2 7 2 | 6 2 5 2 | 5+6 2 |
| | 7 2 8 2 | 7 2 6 2 | 3+4 2 |
| | 8 2 7 2 | 6 2 5 2 | 3+4 2 |
| | 6 2 6 2 | 6 2 6 2 | 5+6 2 |
| | 8 2 9 2 | 7 2 5 2 | 3+4 2 |
| | 6 2 8 2 | 6 2 6 2 | 5+6 2 |
| | 9 2 9 2 | 7 2 6 2 | 5+6 2 |
| | 9 2 6 2 | 7 2 5 2 | 3+4 2 |
| | 8 2 8 2 | 6 2 5 2 | 5+6 2 |
| | 7 2 7 2 | 6 2 6 2 | 3+4 2 |
| | 6 2 9 2 | 7 2 6 2 | 5+6 2 |
| | 9 2 6 2 | 7 2 5 2 | 3+4 2 |
| | 8 2 8 2 | 6 2 6 2 | 3+4 2 |
| | 6 2 6 2 | 7 2 5 2 | 5+6 2 |
| | 8 2 8 2 | 6 2 5 2 | 3+4 2 |
| | 7 2 7 2 | 6 2 6 2 | 5+6 2 |
| | 6 2 6 2 | 7 2 6 2 | 3+4 2 |
| | 8 2 8 2 | 6 2 5 2 | 5+6 2 |
| 139.1.194 | 99944 | 90 | Br2Fe2O5Sr3 |
| | -GM -M | -P -X | -N |
| | 7 2 7 2 | 6 2 5 2 | 3+4 2 |
| | 7 2 7 2 | 6 2 5 2 | 3+4 2 |
| | 9 2 9 2 | 7 2 6 2 | 5+6 2 |
| | 7 2 7 2 | 7 2 5 2 | 5+6 2 |
| | 9 2 9 2 | 7 2 6 2 | 3+4 2 |
| | 7 2 7 2 | 6 2 5 2 | 5+6 2 |
| | 6 2 6 2 | 6 2 6 2 | 5+6 2 |
| | 8 2 8 2 | 7 2 5 2 | 3+4 2 |
| | 9 2 9 2 | 7 2 6 2 | 5+6 2 |
| | 7 2 9 2 | 6 2 5 2 | 5+6 2 |
| | 9 2 7 2 | 7 2 6 2 | 3+4 2 |
| | 8 2 8 2 | 6 2 6 2 | 5+6 2 |
| | 9 2 9 2 | 7 2 6 2 | 5+6 2 |
| | 7 2 9 2 | 6 2 5 2 | 3+4 2 |
| | 6 2 6 2 | 7 2 6 2 | 5+6 2 |
| | 8 2 8 2 | 7 2 5 2 | 3+4 2 |
| | 9 2 7 2 | 6 2 6 2 | 5+6 2 |
| | 7 2 9 2 | 6 2 5 2 | 3+4 2 |
| | 9 2 7 2 | 7 2 6 2 | 5+6 2 |
| | 9 2 9 2 | 6 2 6 2 | 3+4 2 |
| | 7 2 9 2 | 7 2 5 2 | 5+6 2 |
| | 9 2 9 2 | 6 2 6 2 | 3+4 2 |
| | 7 2 7 2 | 6 2 5 2 | 3+4 2 |
| | 6 2 6 2 | 6 2 5 2 | 3+4 2 |

Continued on next column

**TABLE LXIV– continued**

| TQC | ICSD | NoE | SF |
|---|---|---|---|
| | 8 2 8 2 | 7 2 6 2 | 5+6 2 |
| | 9 2 7 2 | 7 2 6 2 | 3+4 2 |
| | 9 2 9 2 | 7 2 6 2 | 5+6 2 |
| | 8 2 8 2 | 6 2 6 2 | 3+4 2 |
| | 7 2 6 2 | 6 2 5 2 | 3+4 2 |
| | 6 2 9 2 | 7 2 6 2 | 5+6 2 |
| | 8 2 8 2 | 6 2 5 2 | 5+6 2 |
| | 7 2 9 2 | 7 2 5 2 | 3+4 2 |
| | 6 2 8 2 | 7 2 5 2 | 5+6 2 |
| | 9 2 7 2 | 6 2 6 2 | 3+4 2 |
| | 8 2 6 2 | 7 2 5 2 | 5+6 2 |
| | 7 2 7 2 | 6 2 6 2 | 3+4 2 |
| | 9 2 9 2 | 7 2 6 2 | 5+6 2 |
| | 6 2 8 2 | 6 2 6 2 | 3+4 2 |
| | 9 2 7 2 | 7 2 5 2 | 5+6 2 |
| | 8 2 6 2 | 6 2 6 2 | 3+4 2 |
| | 6 2 6 2 | 7 2 5 2 | 5+6 2 |
| | 8 2 8 2 | 6 2 5 2 | 5+6 2 |
| | 7 2 7 2 | 7 2 5 2 | 5+6 2 |
| | 6 2 6 2 | 6 2 5 2 | 3+4 2 |
| | 7 2 7 2 | 6 2 6 2 | 5+6 2 |
| 139.1.195 | 99943 | 90 | Br2Fe2O5Sr3 |
| | -GM -M | -P -X | -N |
| | 7 2 7 2 | 6 2 5 2 | 3+4 2 |
| | 7 2 7 2 | 6 2 5 2 | 3+4 2 |
| | 9 2 9 2 | 7 2 6 2 | 5+6 2 |
| | 7 2 7 2 | 7 2 5 2 | 5+6 2 |
| | 9 2 9 2 | 7 2 6 2 | 3+4 2 |
| | 7 2 7 2 | 6 2 5 2 | 5+6 2 |
| | 6 2 6 2 | 6 2 6 2 | 5+6 2 |
| | 8 2 8 2 | 7 2 5 2 | 3+4 2 |
| | 9 2 9 2 | 7 2 6 2 | 5+6 2 |
| | 7 2 9 2 | 6 2 5 2 | 5+6 2 |
| | 9 2 7 2 | 7 2 6 2 | 3+4 2 |
| | 8 2 8 2 | 6 2 6 2 | 5+6 2 |
| | 9 2 9 2 | 7 2 6 2 | 5+6 2 |
| | 7 2 9 2 | 6 2 5 2 | 3+4 2 |
| | 6 2 6 2 | 7 2 6 2 | 5+6 2 |
| | 8 2 8 2 | 7 2 5 2 | 3+4 2 |
| | 9 2 7 2 | 6 2 6 2 | 5+6 2 |
| | 7 2 9 2 | 6 2 5 2 | 3+4 2 |
| | 9 2 7 2 | 7 2 6 2 | 5+6 2 |
| | 9 2 9 2 | 6 2 6 2 | 3+4 2 |
| | 7 2 9 2 | 7 2 5 2 | 5+6 2 |
| | 9 2 9 2 | 6 2 6 2 | 3+4 2 |
| | 7 2 7 2 | 6 2 5 2 | 3+4 2 |
| | 6 2 6 2 | 6 2 5 2 | 3+4 2 |
| | 8 2 8 2 | 7 2 6 2 | 5+6 2 |
| | 9 2 7 2 | 7 2 6 2 | 3+4 2 |
| | 9 2 9 2 | 7 2 6 2 | 5+6 2 |
| | 8 2 8 2 | 6 2 6 2 | 3+4 2 |
| | 7 2 6 2 | 6 2 5 2 | 3+4 2 |
| | 6 2 9 2 | 7 2 6 2 | 5+6 2 |
| | 8 2 8 2 | 6 2 5 2 | 5+6 2 |
| | 7 2 9 2 | 7 2 5 2 | 3+4 2 |
| | 6 2 8 2 | 7 2 5 2 | 5+6 2 |
| | 9 2 7 2 | 6 2 6 2 | 3+4 2 |
| | 8 2 6 2 | 7 2 5 2 | 5+6 2 |
| | 7 2 7 2 | 6 2 6 2 | 3+4 2 |
| | 9 2 9 2 | 7 2 6 2 | 5+6 2 |
| | 6 2 7 2 | 6 2 6 2 | 3+4 2 |

Continued on next column

**TABLE LXIV– continued**

| TQC | ICSD | NoE | SF |
|---|---|---|---|
| | 9 2 8 2 | 7 2 5 2 | 5+6 2 |
| | 8 2 6 2 | 6 2 6 2 | 3+4 2 |
| | 6 2 6 2 | 7 2 5 2 | 5+6 2 |
| | 8 2 8 2 | 6 2 5 2 | 5+6 2 |
| | 7 2 7 2 | 7 2 5 2 | 5+6 2 |
| | 6 2 6 2 | 6 2 5 2 | 3+4 2 |
| | 7 2 7 2 | 6 2 6 2 | 5+6 2 |
| 139.1.196 | 69872 | 90 | Br2Fe2O5Sr3 |
| | -GM -M | -P -X | -N |
| | 7 2 7 2 | 6 2 5 2 | 3+4 2 |
| | 7 2 7 2 | 6 2 5 2 | 3+4 2 |
| | 9 2 9 2 | 7 2 6 2 | 5+6 2 |
| | 7 2 7 2 | 7 2 5 2 | 5+6 2 |
| | 9 2 9 2 | 7 2 6 2 | 3+4 2 |
| | 7 2 7 2 | 6 2 5 2 | 5+6 2 |
| | 6 2 6 2 | 6 2 6 2 | 5+6 2 |
| | 8 2 8 2 | 7 2 5 2 | 3+4 2 |
| | 9 2 9 2 | 7 2 6 2 | 5+6 2 |
| | 9 2 9 2 | 6 2 5 2 | 5+6 2 |
| | 7 2 7 2 | 7 2 6 2 | 3+4 2 |
| | 8 2 8 2 | 6 2 6 2 | 5+6 2 |
| | 9 2 9 2 | 7 2 6 2 | 5+6 2 |
| | 7 2 9 2 | 6 2 5 2 | 3+4 2 |
| | 6 2 6 2 | 7 2 6 2 | 5+6 2 |
| | 8 2 8 2 | 7 2 5 2 | 3+4 2 |
| | 9 2 7 2 | 6 2 6 2 | 5+6 2 |
| | 7 2 9 2 | 6 2 5 2 | 3+4 2 |
| | 9 2 7 2 | 7 2 6 2 | 5+6 2 |
| | 9 2 9 2 | 6 2 6 2 | 3+4 2 |
| | 7 2 9 2 | 7 2 5 2 | 5+6 2 |
| | 9 2 9 2 | 6 2 6 2 | 3+4 2 |
| | 7 2 7 2 | 6 2 5 2 | 3+4 2 |
| | 6 2 8 2 | 6 2 5 2 | 3+4 2 |
| | 8 2 6 2 | 7 2 6 2 | 5+6 2 |
| | 9 2 7 2 | 7 2 6 2 | 3+4 2 |
| | 9 2 9 2 | 7 2 6 2 | 5+6 2 |
| | 8 2 8 2 | 6 2 6 2 | 3+4 2 |
| | 7 2 6 2 | 6 2 5 2 | 3+4 2 |
| | 6 2 9 2 | 7 2 6 2 | 5+6 2 |
| | 8 2 8 2 | 6 2 5 2 | 5+6 2 |
| | 7 2 9 2 | 7 2 5 2 | 3+4 2 |
| | 6 2 8 2 | 7 2 5 2 | 5+6 2 |
| | 9 2 7 2 | 6 2 6 2 | 3+4 2 |
| | 8 2 6 2 | 7 2 5 2 | 5+6 2 |
| | 7 2 7 2 | 6 2 6 2 | 3+4 2 |
| | 9 2 9 2 | 7 2 6 2 | 5+6 2 |
| | 6 2 7 2 | 6 2 6 2 | 3+4 2 |
| | 9 2 8 2 | 7 2 5 2 | 5+6 2 |
| | 8 2 6 2 | 6 2 6 2 | 3+4 2 |
| | 6 2 6 2 | 7 2 5 2 | 5+6 2 |
| | 8 2 8 2 | 6 2 5 2 | 5+6 2 |
| | 7 2 7 2 | 7 2 5 2 | 5+6 2 |
| | 6 2 6 2 | 6 2 5 2 | 3+4 2 |
| | 7 2 7 2 | 6 2 6 2 | 5+6 2 |
| 139.1.197 | 430065 | 58 | Bi2F2O1Sr2Ti2 |
| | -GM -M | -P -X | -N |
| | 7 2 9 2 | 6 2 5 2 | 3+4 2 |
| | 9 2 7 2 | 7 2 6 2 | 5+6 2 |
| | 7 2 9 2 | 6 2 5 2 | 3+4 2 |
| | 8 2 6 2 | 7 2 6 2 | 5+6 2 |
| | 7 2 7 2 | 7 2 5 2 | 5+6 2 |

Continued on next column

**TABLE LXIV– continued**

| TQC | ICSD | NoE | SF |
|---|---|---|---|
| | 7 2 9 2 | 6 2 5 2 | 3+4 2 |
| | 9 2 7 2 | 7 2 6 2 | 5+6 2 |
| | 6 2 8 2 | 6 2 5 2 | 3+4 2 |
| | 9 2 7 2 | 7 2 6 2 | 5+6 2 |
| | 8 2 6 2 | 6 2 5 2 | 3+4 2 |
| | 7 2 9 2 | 7 2 6 2 | 5+6 2 |
| | 7 2 7 2 | 6 2 5 2 | 3+4 2 |
| | 9 2 9 2 | 7 2 6 2 | 5+6 2 |
| | 6 2 7 2 | 6 2 5 2 | 3+4 2 |
| | 7 2 8 2 | 7 2 6 2 | 5+6 2 |
| | 9 2 9 2 | 7 2 6 2 | 3+4 2 |
| | 6 2 8 2 | 6 2 6 2 | 3+4 2 |
| | 9 2 7 2 | 6 2 5 2 | 5+6 2 |
| | 8 2 6 2 | 7 2 6 2 | 3+4 2 |
| | 9 2 9 2 | 6 2 5 2 | 5+6 2 |
| | 8 2 8 2 | 7 2 6 2 | 3+4 2 |
| | 9 2 9 2 | 6 2 6 2 | 3+4 2 |
| | 7 2 7 2 | 7 2 6 2 | 5+6 2 |
| | 7 2 9 2 | 6 2 5 2 | 3+4 2 |
| | 9 2 7 2 | 6 2 6 2 | 3+4 2 |
| | 6 2 6 2 | 7 2 5 2 | 5+6 2 |
| | 6 2 6 2 | 7 2 6 2 | 3+4 2 |
| | 7 2 7 2 | 6 2 5 2 | 5+6 2 |
| | 9 2 9 2 | 6 2 6 2 | 3+4 2 |
| 139.1.198 | 430064 | 58 | Bi2F2O1Sr2Ti2 |
| | -GM -M | -P -X | -N |
| | 7 2 9 2 | 6 2 5 2 | 3+4 2 |
| | 9 2 7 2 | 7 2 6 2 | 5+6 2 |
| | 7 2 9 2 | 6 2 5 2 | 3+4 2 |
| | 8 2 6 2 | 7 2 6 2 | 5+6 2 |
| | 7 2 7 2 | 7 2 5 2 | 5+6 2 |
| | 7 2 9 2 | 6 2 5 2 | 3+4 2 |
| | 9 2 7 2 | 7 2 6 2 | 5+6 2 |
| | 6 2 8 2 | 6 2 5 2 | 3+4 2 |
| | 9 2 7 2 | 7 2 6 2 | 5+6 2 |
| | 8 2 6 2 | 6 2 5 2 | 3+4 2 |
| | 7 2 9 2 | 7 2 6 2 | 5+6 2 |
| | 7 2 7 2 | 6 2 5 2 | 3+4 2 |
| | 9 2 9 2 | 7 2 6 2 | 5+6 2 |
| | 6 2 7 2 | 6 2 5 2 | 3+4 2 |
| | 7 2 8 2 | 7 2 6 2 | 5+6 2 |
| | 9 2 9 2 | 7 2 6 2 | 3+4 2 |
| | 6 2 8 2 | 6 2 6 2 | 3+4 2 |
| | 9 2 7 2 | 6 2 5 2 | 5+6 2 |
| | 8 2 6 2 | 7 2 6 2 | 3+4 2 |
| | 9 2 9 2 | 6 2 5 2 | 5+6 2 |
| | 8 2 8 2 | 7 2 6 2 | 3+4 2 |
| | 9 2 9 2 | 6 2 6 2 | 3+4 2 |
| | 7 2 7 2 | 7 2 6 2 | 5+6 2 |
| | 7 2 9 2 | 6 2 5 2 | 3+4 2 |
| | 9 2 7 2 | 6 2 6 2 | 3+4 2 |
| | 6 2 6 2 | 7 2 5 2 | 5+6 2 |
| | 6 2 6 2 | 7 2 6 2 | 3+4 2 |
| | 7 2 7 2 | 6 2 5 2 | 5+6 2 |
| | 9 2 9 2 | 6 2 6 2 | 3+4 2 |
| 139.1.199 | 402954 | 84 | Bi2Eu4O1 |
| | -GM -M | -P -X | -N |
| | 7 2 9 2 | 6 2 5 2 | 3+4 2 |
| | 9 2 7 2 | 7 2 6 2 | 5+6 2 |
| | 7 2 7 2 | 6 2 6 2 | 3+4 2 |
| | 6 2 6 2 | 7 2 6 2 | 5+6 2 |

Continued on next column

**TABLE LXIV– continued**

| TQC | ICSD | NoE | SF |
|---|---|---|---|
| | 7 2 9 2 | 6 2 5 2 | 3+4 2 |
| | 9 2 7 2 | 7 2 6 2 | 5+6 2 |
| | 8 2 8 2 | 6 2 5 2 | 3+4 2 |
| | 9 2 9 2 | 7 2 5 2 | 5+6 2 |
| | 7 2 7 2 | 6 2 5 2 | 3+4 2 |
| | 9 2 9 2 | 7 2 5 2 | 5+6 2 |
| | 9 2 9 2 | 6 2 6 2 | 3+4 2 |
| | 6 2 8 2 | 7 2 6 2 | 5+6 2 |
| | 8 2 6 2 | 6 2 5 2 | 3+4 2 |
| | 8 2 8 2 | 7 2 5 2 | 5+6 2 |
| | 9 2 9 2 | 6 2 5 2 | 3+4 2 |
| | 7 2 7 2 | 7 2 5 2 | 5+6 2 |
| | 8 2 8 2 | 6 2 5 2 | 3+4 2 |
| | 7 2 7 2 | 6 2 6 2 | 5+6 2 |
| | 9 2 9 2 | 7 2 5 2 | 3+4 2 |
| | 9 2 9 2 | 6 2 6 2 | 5+6 2 |
| | 7 2 8 2 | 7 2 6 2 | 5+6 2 |
| | 8 2 9 2 | 7 2 6 2 | 5+6 2 |
| | 9 2 7 2 | 6 2 6 2 | 3+4 2 |
| | 9 2 9 2 | 7 2 5 2 | 5+6 2 |
| | 7 2 7 2 | 7 2 6 2 | 5+6 2 |
| | 6 2 6 2 | 6 2 6 2 | 3+4 2 |
| | 9 2 9 2 | 6 2 5 2 | 5+6 2 |
| | 8 2 8 2 | 7 2 5 2 | 3+4 2 |
| | 6 2 6 2 | 7 2 5 2 | 3+4 2 |
| | 8 2 7 2 | 7 2 5 2 | 5+6 2 |
| | 7 2 6 2 | 6 2 5 2 | 3+4 2 |
| | 9 2 8 2 | 6 2 6 2 | 3+4 2 |
| | 8 2 8 2 | 6 2 5 2 | 5+6 2 |
| | 8 2 9 2 | 7 2 5 2 | 5+6 2 |
| | 6 2 8 2 | 7 2 6 2 | 3+4 2 |
| | 9 2 9 2 | 6 2 5 2 | 3+4 2 |
| | 9 2 6 2 | 7 2 6 2 | 5+6 2 |
| | 8 2 9 2 | 6 2 5 2 | 5+6 2 |
| | 8 2 8 2 | 6 2 5 2 | 3+4 2 |
| | 6 2 9 2 | 7 2 5 2 | 5+6 2 |
| | 9 2 7 2 | 6 2 6 2 | 3+4 2 |
| | 7 2 6 2 | 7 2 5 2 | 5+6 2 |
| 139.1.200 | 41836 | 50 | Bi1Sr2 |
| | -GM -M | -P -X | -N |
| | 7 2 9 2 | 6 2 5 2 | 3+4 2 |
| | 9 2 7 2 | 7 2 6 2 | 5+6 2 |
| | 7 2 7 2 | 6 2 6 2 | 3+4 2 |
| | 6 2 6 2 | 7 2 6 2 | 5+6 2 |
| | 7 2 9 2 | 7 2 5 2 | 3+4 2 |
| | 9 2 7 2 | 6 2 6 2 | 5+6 2 |
| | 8 2 8 2 | 6 2 5 2 | 5+6 2 |
| | 9 2 9 2 | 7 2 5 2 | 3+4 2 |
| | 9 2 9 2 | 6 2 5 2 | 5+6 2 |
| | 7 2 9 2 | 7 2 6 2 | 3+4 2 |
| | 6 2 8 2 | 6 2 5 2 | 3+4 2 |
| | 9 2 7 2 | 7 2 6 2 | 5+6 2 |
| | 8 2 6 2 | 6 2 5 2 | 3+4 2 |
| | 8 2 8 2 | 7 2 5 2 | 5+6 2 |
| | 9 2 9 2 | 7 2 5 2 | 5+6 2 |
| | 8 2 8 2 | 6 2 5 2 | 3+4 2 |
| | 7 2 7 2 | 7 2 6 2 | 3+4 2 |
| | 9 2 9 2 | 6 2 5 2 | 5+6 2 |
| | 7 2 7 2 | 7 2 6 2 | 5+6 2 |
| | 9 2 9 2 | 6 2 5 2 | 3+4 2 |
| | 7 2 7 2 | 6 2 6 2 | 3+4 2 |

Continued on next column

**TABLE LXIV– continued**

| TQC | ICSD | NoE | SF |
|---|---|---|---|
| | 7 2 7 2 | 7 2 6 2 | 5+6 2 |
| | 6 2 9 2 | 7 2 5 2 | 3+4 2 |
| | 9 2 6 2 | 6 2 5 2 | 5+6 2 |
| | 8 2 8 2 | 6 2 6 2 | 5+6 2 |
| 139.1.201 | 107609 | 22 | Be1Pd2 |
| | -GM -M | -P -X | -N |
| | 7 2 7 2 | 6 2 5 2 | 3+4 2 |
| | 9 2 9 2 | 7 2 6 2 | 5+6 2 |
| | 8 2 6 2 | 6 2 6 2 | 3+4 2 |
| | 6 2 7 2 | 7 2 6 2 | 5+6 2 |
| | 7 2 8 2 | 7 2 5 2 | 5+6 2 |
| | 7 2 9 2 | 6 2 5 2 | 5+6 2 |
| | 6 2 8 2 | 7 2 5 2 | 3+4 2 |
| | 9 2 9 2 | 6 2 6 2 | 3+4 2 |
| | 8 2 8 2 | 6 2 5 2 | 3+4 2 |
| | 6 2 7 2 | 7 2 6 2 | 5+6 2 |
| | 7 2 6 2 | 6 2 6 2 | 5+6 2 |
| 139.1.202 | 58717 | 30 | Be12Mo1 |
| | -GM -M | -P -X | -N |
| | 7 2 6 2 | 7 2 6 2 | 3+4 2 |
| | 6 2 9 2 | 6 2 5 2 | 5+6 2 |
| | 9 2 8 2 | 7 2 5 2 | 3+4 2 |
| | 8 2 7 2 | 6 2 5 2 | 5+6 2 |
| | 7 2 7 2 | 6 2 6 2 | 5+6 2 |
| | 8 2 9 2 | 7 2 6 2 | 3+4 2 |
| | 9 2 6 2 | 6 2 5 2 | 3+4 2 |
| | 8 2 6 2 | 7 2 5 2 | 5+6 2 |
| | 7 2 7 2 | 7 2 6 2 | 3+4 2 |
| | 6 2 6 2 | 7 2 5 2 | 5+6 2 |
| | 7 2 6 2 | 6 2 6 2 | 3+4 2 |
| | 6 2 7 2 | 6 2 5 2 | 3+4 2 |
| | 9 2 7 2 | 7 2 6 2 | 5+6 2 |
| | 8 2 9 2 | 6 2 5 2 | 3+4 2 |
| | 9 2 8 2 | 7 2 5 2 | 5+6 2 |
| 139.1.203 | 237677 | 112 | Ba6Ge32Rh8 |
| | -GM -M | -P -X | -N |
| | 7 2 7 2 | 7 2 5 2 | 5+6 2 |
| | 9 2 9 2 | 6 2 6 2 | 3+4 2 |
| | 7 2 7 2 | 7 2 5 2 | 5+6 2 |
| | 9 2 7 2 | 6 2 6 2 | 3+4 2 |
| | 8 2 8 2 | 7 2 6 2 | 5+6 2 |
| | 9 2 9 2 | 6 2 6 2 | 3+4 2 |
| | 6 2 9 2 | 7 2 5 2 | 3+4 2 |
| | 7 2 6 2 | 6 2 5 2 | 5+6 2 |
| | 7 2 7 2 | 7 2 5 2 | 3+4 2 |
| | 8 2 7 2 | 6 2 6 2 | 3+4 2 |
| | 7 2 6 2 | 7 2 5 2 | 5+6 2 |
| | 6 2 9 2 | 6 2 6 2 | 5+6 2 |
| | 8 2 8 2 | 6 2 5 2 | 3+4 2 |
| | 9 2 9 2 | 7 2 6 2 | 5+6 2 |
| | 7 2 6 2 | 7 2 6 2 | 3+4 2 |
| | 6 2 6 2 | 6 2 5 2 | 3+4 2 |
| | 8 2 8 2 | 6 2 5 2 | 5+6 2 |
| | 7 2 9 2 | 6 2 6 2 | 5+6 2 |
| | 9 2 7 2 | 7 2 5 2 | 3+4 2 |
| | 9 2 8 2 | 7 2 6 2 | 5+6 2 |
| | 7 2 7 2 | 6 2 6 2 | 3+4 2 |
| | 8 2 7 2 | 6 2 5 2 | 5+6 2 |
| | 6 2 7 2 | 7 2 6 2 | 5+6 2 |
| | 9 2 6 2 | 6 2 5 2 | 3+4 2 |
| | 8 2 9 2 | 7 2 5 2 | 5+6 2 |

Continued on next column

**TABLE LXIV– continued**

| TQC | ICSD | NoE | SF |
|---|---|---|---|
| | 7 2 8 2 | 6 2 6 2 | 3+4 2 |
| | 6 2 9 2 | 6 2 5 2 | 3+4 2 |
| | 6 2 6 2 | 7 2 6 2 | 5+6 2 |
| | 7 2 6 2 | 6 2 5 2 | 3+4 2 |
| | 7 2 7 2 | 7 2 6 2 | 3+4 2 |
| | 9 2 9 2 | 7 2 5 2 | 5+6 2 |
| | 9 2 7 2 | 7 2 6 2 | 3+4 2 |
| | 8 2 8 2 | 6 2 6 2 | 5+6 2 |
| | 6 2 9 2 | 6 2 6 2 | 5+6 2 |
| | 7 2 9 2 | 7 2 5 2 | 5+6 2 |
| | 6 2 8 2 | 6 2 5 2 | 5+6 2 |
| | 7 2 9 2 | 7 2 5 2 | 3+4 2 |
| | 9 2 8 2 | 7 2 6 2 | 3+4 2 |
| | 8 2 6 2 | 6 2 5 2 | 5+6 2 |
| | 6 2 6 2 | 6 2 5 2 | 3+4 2 |
| | 7 2 8 2 | 7 2 6 2 | 5+6 2 |
| | 9 2 7 2 | 7 2 5 2 | 5+6 2 |
| | 6 2 9 2 | 6 2 5 2 | 3+4 2 |
| | 8 2 6 2 | 7 2 6 2 | 3+4 2 |
| | 7 2 7 2 | 6 2 6 2 | 3+4 2 |
| | 8 2 8 2 | 6 2 5 2 | 5+6 2 |
| | 6 2 7 2 | 7 2 6 2 | 3+4 2 |
| | 9 2 6 2 | 7 2 6 2 | 3+4 2 |
| | 8 2 8 2 | 6 2 6 2 | 5+6 2 |
| | 9 2 9 2 | 7 2 6 2 | 5+6 2 |
| | 9 2 6 2 | 7 2 5 2 | 3+4 2 |
| | 8 2 7 2 | 6 2 6 2 | 5+6 2 |
| | 6 2 6 2 | 7 2 5 2 | 5+6 2 |
| | 9 2 9 2 | 6 2 5 2 | 3+4 2 |
| | 8 2 7 2 | 7 2 5 2 | 3+4 2 |
| | 7 2 8 2 | 6 2 6 2 | 5+6 2 |
| 139.1.204 | 427108 | 130 | Ba3Ge16Ir4 |
| | -GM -M | -P -X | -N |
| | 7 2 7 2 | 6 2 5 2 | 3+4 2 |
| | 7 2 9 2 | 6 2 5 2 | 3+4 2 |
| | 8 2 6 2 | 7 2 6 2 | 5+6 2 |
| | 9 2 9 2 | 7 2 6 2 | 5+6 2 |
| | 6 2 8 2 | 6 2 5 2 | 3+4 2 |
| | 9 2 7 2 | 7 2 6 2 | 5+6 2 |
| | 9 2 7 2 | 6 2 6 2 | 5+6 2 |
| | 7 2 9 2 | 7 2 5 2 | 5+6 2 |
| | 8 2 9 2 | 6 2 6 2 | 3+4 2 |
| | 9 2 8 2 | 7 2 6 2 | 5+6 2 |
| | 6 2 8 2 | 6 2 5 2 | 3+4 2 |
| | 8 2 6 2 | 7 2 6 2 | 5+6 2 |
| | 7 2 7 2 | 7 2 5 2 | 5+6 2 |
| | 7 2 9 2 | 6 2 5 2 | 3+4 2 |
| | 9 2 7 2 | 7 2 6 2 | 5+6 2 |
| | 9 2 7 2 | 7 2 6 2 | 3+4 2 |
| | 8 2 6 2 | 6 2 6 2 | 5+6 2 |
| | 9 2 8 2 | 6 2 6 2 | 3+4 2 |
| | 6 2 7 2 | 7 2 5 2 | 3+4 2 |
| | 6 2 9 2 | 7 2 5 2 | 5+6 2 |
| | 7 2 6 2 | 6 2 5 2 | 3+4 2 |
| | 7 2 9 2 | 6 2 6 2 | 5+6 2 |
| | 8 2 9 2 | 7 2 6 2 | 3+4 2 |
| | 9 2 8 2 | 6 2 6 2 | 5+6 2 |
| | 7 2 9 2 | 7 2 5 2 | 3+4 2 |
| | 6 2 6 2 | 6 2 5 2 | 5+6 2 |
| | 8 2 8 2 | 6 2 5 2 | 5+6 2 |
| | 9 2 7 2 | 7 2 6 2 | 3+4 2 |

Continued on next column

**TABLE LXIV– continued**

| TQC | ICSD | NoE | SF |
|---|---|---|---|
| | 8 2 8 2 | 7 2 6 2 | 5+6 2 |
| | 7 2 7 2 | 6 2 5 2 | 3+4 2 |
| | 9 2 6 2 | 6 2 6 2 | 3+4 2 |
| | 6 2 7 2 | 6 2 5 2 | 5+6 2 |
| | 7 2 9 2 | 7 2 6 2 | 3+4 2 |
| | 9 2 7 2 | 6 2 5 2 | 5+6 2 |
| | 6 2 8 2 | 7 2 5 2 | 5+6 2 |
| | 8 2 9 2 | 6 2 5 2 | 3+4 2 |
| | 7 2 6 2 | 7 2 6 2 | 3+4 2 |
| | 6 2 7 2 | 6 2 6 2 | 5+6 2 |
| | 9 2 6 2 | 7 2 5 2 | 3+4 2 |
| | 7 2 9 2 | 6 2 6 2 | 3+4 2 |
| | 7 2 7 2 | 7 2 5 2 | 5+6 2 |
| | 6 2 9 2 | 6 2 6 2 | 5+6 2 |
| | 8 2 6 2 | 7 2 6 2 | 3+4 2 |
| | 9 2 8 2 | 7 2 5 2 | 5+6 2 |
| | 7 2 6 2 | 6 2 6 2 | 3+4 2 |
| | 6 2 9 2 | 7 2 5 2 | 5+6 2 |
| | 8 2 8 2 | 6 2 5 2 | 5+6 2 |
| | 9 2 8 2 | 7 2 6 2 | 3+4 2 |
| | 8 2 6 2 | 6 2 5 2 | 5+6 2 |
| | 9 2 7 2 | 7 2 6 2 | 3+4 2 |
| | 6 2 9 2 | 6 2 5 2 | 5+6 2 |
| | 7 2 8 2 | 7 2 6 2 | 3+4 2 |
| | 6 2 6 2 | 6 2 5 2 | 3+4 2 |
| | 8 2 7 2 | 7 2 6 2 | 5+6 2 |
| | 7 2 9 2 | 7 2 5 2 | 3+4 2 |
| | 8 2 8 2 | 6 2 6 2 | 5+6 2 |
| | 9 2 7 2 | 7 2 5 2 | 3+4 2 |
| | 6 2 6 2 | 6 2 5 2 | 5+6 2 |
| | 9 2 7 2 | 7 2 5 2 | 5+6 2 |
| | 8 2 8 2 | 7 2 6 2 | 3+4 2 |
| | 6 2 9 2 | 7 2 6 2 | 5+6 2 |
| | 9 2 6 2 | 6 2 6 2 | 3+4 2 |
| | 7 2 9 2 | 6 2 6 2 | 3+4 2 |
| | 9 2 7 2 | 7 2 5 2 | 5+6 2 |
| | 8 2 6 2 | 6 2 6 2 | 3+4 2 |
| 139.1.205 | 194235 | 130 | Ba3Ge16Ir4 |
| | -GM -M | -P -X | -N |
| | 7 2 7 2 | 6 2 5 2 | 3+4 2 |
| | 7 2 9 2 | 6 2 5 2 | 3+4 2 |
| | 8 2 6 2 | 7 2 6 2 | 5+6 2 |
| | 9 2 9 2 | 7 2 6 2 | 5+6 2 |
| | 6 2 8 2 | 6 2 5 2 | 3+4 2 |
| | 9 2 7 2 | 7 2 6 2 | 5+6 2 |
| | 9 2 7 2 | 6 2 6 2 | 5+6 2 |
| | 7 2 9 2 | 7 2 5 2 | 5+6 2 |
| | 8 2 9 2 | 6 2 6 2 | 3+4 2 |
| | 9 2 8 2 | 7 2 6 2 | 5+6 2 |
| | 6 2 8 2 | 6 2 5 2 | 3+4 2 |
| | 8 2 6 2 | 7 2 6 2 | 5+6 2 |
| | 7 2 7 2 | 7 2 5 2 | 5+6 2 |
| | 7 2 9 2 | 6 2 5 2 | 3+4 2 |
| | 9 2 7 2 | 7 2 6 2 | 5+6 2 |
| | 9 2 7 2 | 7 2 6 2 | 3+4 2 |
| | 8 2 8 2 | 6 2 6 2 | 5+6 2 |
| | 9 2 6 2 | 6 2 6 2 | 3+4 2 |
| | 6 2 9 2 | 7 2 5 2 | 3+4 2 |
| | 7 2 7 2 | 7 2 5 2 | 3+4 2 |
| | 6 2 9 2 | 6 2 5 2 | 5+6 2 |
| | 7 2 6 2 | 6 2 6 2 | 5+6 2 |

Continued on next column

**TABLE LXIV– continued**

| TQC | ICSD | NoE | SF |
|---|---|---|---|
| | 8 2 9 2 | 7 2 6 2 | 3+4 2 |
| | 9 2 8 2 | 6 2 6 2 | 5+6 2 |
| | 7 2 9 2 | 7 2 5 2 | 3+4 2 |
| | 6 2 6 2 | 6 2 5 2 | 5+6 2 |
| | 8 2 8 2 | 6 2 5 2 | 5+6 2 |
| | 9 2 7 2 | 7 2 6 2 | 3+4 2 |
| | 8 2 8 2 | 7 2 6 2 | 5+6 2 |
| | 7 2 7 2 | 6 2 5 2 | 3+4 2 |
| | 9 2 6 2 | 6 2 6 2 | 3+4 2 |
| | 6 2 7 2 | 6 2 5 2 | 5+6 2 |
| | 7 2 9 2 | 7 2 6 2 | 3+4 2 |
| | 9 2 7 2 | 6 2 5 2 | 5+6 2 |
| | 6 2 8 2 | 7 2 5 2 | 5+6 2 |
| | 8 2 9 2 | 6 2 5 2 | 3+4 2 |
| | 7 2 6 2 | 7 2 6 2 | 3+4 2 |
| | 6 2 7 2 | 7 2 5 2 | 5+6 2 |
| | 9 2 6 2 | 6 2 6 2 | 3+4 2 |
| | 7 2 9 2 | 6 2 6 2 | 3+4 2 |
| | 7 2 7 2 | 7 2 5 2 | 5+6 2 |
| | 6 2 9 2 | 6 2 6 2 | 5+6 2 |
| | 8 2 6 2 | 7 2 6 2 | 3+4 2 |
| | 7 2 8 2 | 7 2 5 2 | 5+6 2 |
| | 9 2 6 2 | 6 2 6 2 | 3+4 2 |
| | 6 2 9 2 | 7 2 5 2 | 5+6 2 |
| | 8 2 8 2 | 6 2 5 2 | 5+6 2 |
| | 9 2 8 2 | 7 2 6 2 | 3+4 2 |
| | 8 2 6 2 | 6 2 5 2 | 5+6 2 |
| | 9 2 7 2 | 7 2 6 2 | 3+4 2 |
| | 6 2 9 2 | 6 2 5 2 | 5+6 2 |
| | 7 2 8 2 | 7 2 6 2 | 3+4 2 |
| | 6 2 6 2 | 6 2 5 2 | 3+4 2 |
| | 7 2 7 2 | 7 2 6 2 | 5+6 2 |
| | 8 2 9 2 | 7 2 5 2 | 3+4 2 |
| | 8 2 8 2 | 6 2 6 2 | 5+6 2 |
| | 9 2 7 2 | 7 2 5 2 | 3+4 2 |
| | 6 2 6 2 | 6 2 5 2 | 5+6 2 |
| | 9 2 7 2 | 7 2 5 2 | 3+4 2 |
| | 8 2 8 2 | 7 2 6 2 | 5+6 2 |
| | 6 2 9 2 | 6 2 6 2 | 5+6 2 |
| | 9 2 6 2 | 7 2 6 2 | 3+4 2 |
| | 7 2 9 2 | 6 2 6 2 | 3+4 2 |
| | 9 2 7 2 | 7 2 5 2 | 5+6 2 |
| | 8 2 6 2 | 6 2 6 2 | 3+4 2 |
| 139.1.206 | 189363 | 130 | Ba3Ge16Ir4 |
| | -GM -M | -P -X | -N |
| | 7 2 7 2 | 6 2 5 2 | 3+4 2 |
| | 7 2 9 2 | 6 2 5 2 | 3+4 2 |
| | 8 2 6 2 | 7 2 6 2 | 5+6 2 |
| | 9 2 9 2 | 7 2 6 2 | 5+6 2 |
| | 6 2 8 2 | 6 2 5 2 | 3+4 2 |
| | 9 2 7 2 | 7 2 6 2 | 5+6 2 |
| | 9 2 7 2 | 6 2 6 2 | 5+6 2 |
| | 7 2 9 2 | 7 2 5 2 | 5+6 2 |
| | 8 2 9 2 | 6 2 6 2 | 3+4 2 |
| | 9 2 8 2 | 7 2 6 2 | 5+6 2 |
| | 6 2 8 2 | 6 2 5 2 | 3+4 2 |
| | 8 2 6 2 | 7 2 6 2 | 5+6 2 |
| | 7 2 7 2 | 7 2 5 2 | 5+6 2 |
| | 7 2 9 2 | 6 2 6 2 | 3+4 2 |
| | 9 2 7 2 | 7 2 6 2 | 3+4 2 |
| | 9 2 7 2 | 6 2 5 2 | 5+6 2 |

Continued on next column

**TABLE LXIV– continued**

| TQC | ICSD | NoE | SF |
|---|---|---|---|
| | 8 2 8 2 | 6 2 6 2 | 5+6 2 |
| | 9 2 6 2 | 7 2 6 2 | 3+4 2 |
| | 6 2 9 2 | 7 2 5 2 | 3+4 2 |
| | 7 2 7 2 | 6 2 5 2 | 5+6 2 |
| | 6 2 9 2 | 7 2 5 2 | 5+6 2 |
| | 7 2 6 2 | 6 2 6 2 | 3+4 2 |
| | 7 2 9 2 | 7 2 6 2 | 3+4 2 |
| | 8 2 8 2 | 6 2 6 2 | 5+6 2 |
| | 9 2 9 2 | 7 2 5 2 | 3+4 2 |
| | 6 2 6 2 | 6 2 5 2 | 5+6 2 |
| | 8 2 8 2 | 6 2 5 2 | 5+6 2 |
| | 9 2 7 2 | 7 2 6 2 | 3+4 2 |
| | 8 2 8 2 | 7 2 6 2 | 5+6 2 |
| | 7 2 7 2 | 6 2 5 2 | 3+4 2 |
| | 9 2 6 2 | 6 2 6 2 | 3+4 2 |
| | 6 2 7 2 | 7 2 5 2 | 5+6 2 |
| | 7 2 9 2 | 6 2 6 2 | 5+6 2 |
| | 9 2 7 2 | 6 2 5 2 | 3+4 2 |
| | 6 2 9 2 | 6 2 5 2 | 5+6 2 |
| | 8 2 8 2 | 7 2 5 2 | 3+4 2 |
| | 7 2 6 2 | 7 2 6 2 | 3+4 2 |
| | 6 2 7 2 | 6 2 6 2 | 5+6 2 |
| | 9 2 6 2 | 7 2 5 2 | 3+4 2 |
| | 7 2 9 2 | 6 2 6 2 | 3+4 2 |
| | 7 2 7 2 | 7 2 5 2 | 5+6 2 |
| | 6 2 9 2 | 7 2 6 2 | 5+6 2 |
| | 7 2 6 2 | 6 2 6 2 | 5+6 2 |
| | 8 2 8 2 | 7 2 5 2 | 3+4 2 |
| | 9 2 6 2 | 6 2 6 2 | 3+4 2 |
| | 6 2 9 2 | 7 2 5 2 | 5+6 2 |
| | 8 2 8 2 | 6 2 5 2 | 5+6 2 |
| | 9 2 8 2 | 7 2 6 2 | 3+4 2 |
| | 8 2 6 2 | 6 2 5 2 | 5+6 2 |
| | 9 2 7 2 | 7 2 6 2 | 3+4 2 |
| | 6 2 9 2 | 7 2 5 2 | 5+6 2 |
| | 7 2 8 2 | 6 2 5 2 | 3+4 2 |
| | 6 2 6 2 | 6 2 6 2 | 3+4 2 |
| | 8 2 7 2 | 7 2 6 2 | 5+6 2 |
| | 7 2 9 2 | 7 2 5 2 | 3+4 2 |
| | 8 2 8 2 | 6 2 6 2 | 5+6 2 |
| | 9 2 7 2 | 7 2 5 2 | 3+4 2 |
| | 6 2 6 2 | 6 2 5 2 | 5+6 2 |
| | 9 2 7 2 | 7 2 6 2 | 5+6 2 |
| | 8 2 8 2 | 7 2 6 2 | 3+4 2 |
| | 6 2 9 2 | 7 2 5 2 | 5+6 2 |
| | 9 2 6 2 | 6 2 6 2 | 3+4 2 |
| | 7 2 9 2 | 6 2 6 2 | 3+4 2 |
| | 9 2 7 2 | 7 2 5 2 | 5+6 2 |
| | 8 2 6 2 | 6 2 6 2 | 3+4 2 |
| 139.1.207 | 41837 | 50 | Ba2Sb1 |
| | -GM -M | -P -X | -N |
| | 7 2 7 2 | 6 2 5 2 | 5+6 2 |
| | 9 2 9 2 | 7 2 6 2 | 3+4 2 |
| | 7 2 7 2 | 6 2 6 2 | 3+4 2 |
| | 6 2 6 2 | 7 2 6 2 | 5+6 2 |
| | 8 2 8 2 | 7 2 5 2 | 3+4 2 |
| | 7 2 9 2 | 6 2 5 2 | 5+6 2 |
| | 9 2 7 2 | 6 2 6 2 | 5+6 2 |
| | 9 2 9 2 | 7 2 5 2 | 3+4 2 |
| | 9 2 9 2 | 6 2 5 2 | 5+6 2 |
| | 9 2 9 2 | 7 2 5 2 | 3+4 2 |

Continued on next column

**TABLE LXIV– continued**

| TQC | ICSD | NoE | SF |
|---|---|---|---|
| | 7 2 8 2 | 6 2 6 2 | 5+6 2 |
| | 6 2 7 2 | 7 2 5 2 | 3+4 2 |
| | 8 2 6 2 | 7 2 6 2 | 3+4 2 |
| | 8 2 8 2 | 6 2 5 2 | 5+6 2 |
| | 9 2 9 2 | 7 2 5 2 | 5+6 2 |
| | 8 2 8 2 | 6 2 5 2 | 3+4 2 |
| | 7 2 7 2 | 7 2 6 2 | 3+4 2 |
| | 9 2 9 2 | 6 2 5 2 | 5+6 2 |
| | 7 2 7 2 | 7 2 6 2 | 5+6 2 |
| | 7 2 7 2 | 6 2 6 2 | 3+4 2 |
| | 7 2 9 2 | 7 2 6 2 | 3+4 2 |
| | 6 2 6 2 | 6 2 5 2 | 3+4 2 |
| | 9 2 7 2 | 6 2 5 2 | 5+6 2 |
| | 9 2 9 2 | 7 2 5 2 | 5+6 2 |
| | 8 2 8 2 | 6 2 6 2 | 5+6 2 |
| 139.1.208 | 30084 | 32 | Ba2Hg1 |
| | -GM -M | -P -X | -N |
| | 7 2 7 2 | 7 2 5 2 | 5+6 2 |
| | 9 2 9 2 | 6 2 6 2 | 3+4 2 |
| | 7 2 7 2 | 7 2 5 2 | 5+6 2 |
| | 9 2 9 2 | 6 2 6 2 | 3+4 2 |
| | 7 2 7 2 | 7 2 6 2 | 5+6 2 |
| | 6 2 9 2 | 6 2 5 2 | 3+4 2 |
| | 8 2 8 2 | 7 2 5 2 | 5+6 2 |
| | 9 2 6 2 | 6 2 6 2 | 3+4 2 |
| | 6 2 6 2 | 6 2 5 2 | 3+4 2 |
| | 7 2 7 2 | 7 2 5 2 | 3+4 2 |
| | 6 2 6 2 | 7 2 5 2 | 3+4 2 |
| | 7 2 7 2 | 6 2 5 2 | 3+4 2 |
| | 6 2 6 2 | 7 2 5 2 | 3+4 2 |
| | 7 2 7 2 | 6 2 5 2 | 3+4 2 |
| | 9 2 9 2 | 7 2 6 2 | 5+6 2 |
| | 7 2 7 2 | 7 2 6 2 | 3+4 2 |
| 139.1.209 | 183149 | 66 | Ba2F2Mn2O1Se2 |
| | -GM -M | -P -X | -N |
| | 7 2 9 2 | 6 2 5 2 | 3+4 2 |
| | 9 2 7 2 | 7 2 6 2 | 5+6 2 |
| | 7 2 9 2 | 6 2 5 2 | 3+4 2 |
| | 8 2 6 2 | 7 2 6 2 | 5+6 2 |
| | 7 2 7 2 | 7 2 5 2 | 5+6 2 |
| | 7 2 9 2 | 6 2 6 2 | 3+4 2 |
| | 9 2 7 2 | 7 2 5 2 | 5+6 2 |
| | 7 2 7 2 | 6 2 5 2 | 3+4 2 |
| | 9 2 9 2 | 7 2 6 2 | 5+6 2 |
| | 6 2 8 2 | 6 2 6 2 | 3+4 2 |
| | 8 2 6 2 | 7 2 5 2 | 5+6 2 |
| | 7 2 7 2 | 6 2 6 2 | 3+4 2 |
| | 9 2 9 2 | 7 2 5 2 | 5+6 2 |
| | 6 2 7 2 | 7 2 6 2 | 3+4 2 |
| | 9 2 8 2 | 6 2 6 2 | 5+6 2 |
| | 7 2 9 2 | 6 2 6 2 | 3+4 2 |
| | 6 2 8 2 | 7 2 5 2 | 3+4 2 |
| | 9 2 7 2 | 6 2 5 2 | 5+6 2 |
| | 8 2 6 2 | 7 2 6 2 | 3+4 2 |
| | 7 2 9 2 | 6 2 5 2 | 5+6 2 |
| | 9 2 8 2 | 7 2 6 2 | 3+4 2 |
| | 8 2 7 2 | 6 2 6 2 | 3+4 2 |
| | 9 2 9 2 | 6 2 5 2 | 5+6 2 |
| | 7 2 7 2 | 7 2 6 2 | 3+4 2 |
| | 6 2 6 2 | 6 2 6 2 | 3+4 2 |
| | 9 2 9 2 | 6 2 5 2 | 5+6 2 |

Continued on next column

**TABLE LXIV– continued**

| TQC | ICSD | NoE | SF |
|---|---|---|---|
| | 8 2 9 2 | 7 2 5 2 | 3+4 2 |
| | 9 2 8 2 | 7 2 6 2 | 5+6 2 |
| | 6 2 6 2 | 6 2 6 2 | 5+6 2 |
| | 7 2 7 2 | 7 2 6 2 | 3+4 2 |
| | 6 2 6 2 | 6 2 6 2 | 5+6 2 |
| | 7 2 7 2 | 7 2 6 2 | 5+6 2 |
| | 6 2 6 2 | 7 2 6 2 | 3+4 2 |
| 139.1.210 | 71535 | 130 | Ba2Cu2O12Si4 |
| | -GM -M | -P -X | -N |
| | 7 2 9 2 | 6 2 6 2 | 3+4 2 |
| | 8 2 6 2 | 7 2 5 2 | 5+6 2 |
| | 7 2 7 2 | 6 2 5 2 | 3+4 2 |
| | 8 2 8 2 | 6 2 6 2 | 5+6 2 |
| | 9 2 9 2 | 7 2 6 2 | 5+6 2 |
| | 6 2 6 2 | 7 2 5 2 | 3+4 2 |
| | 6 2 6 2 | 7 2 5 2 | 3+4 2 |
| | 7 2 7 2 | 6 2 6 2 | 5+6 2 |
| | 8 2 8 2 | 7 2 6 2 | 3+4 2 |
| | 9 2 9 2 | 6 2 6 2 | 3+4 2 |
| | 9 2 9 2 | 6 2 5 2 | 5+6 2 |
| | 6 2 7 2 | 7 2 5 2 | 5+6 2 |
| | 7 2 6 2 | 6 2 5 2 | 5+6 2 |
| | 8 2 8 2 | 7 2 6 2 | 3+4 2 |
| | 9 2 8 2 | 7 2 5 2 | 3+4 2 |
| | 6 2 7 2 | 6 2 6 2 | 5+6 2 |
| | 6 2 8 2 | 6 2 6 2 | 3+4 2 |
| | 8 2 6 2 | 7 2 5 2 | 5+6 2 |
| | 9 2 7 2 | 7 2 5 2 | 3+4 2 |
| | 7 2 9 2 | 6 2 6 2 | 5+6 2 |
| | 6 2 6 2 | 7 2 5 2 | 3+4 2 |
| | 9 2 8 2 | 7 2 6 2 | 5+6 2 |
| | 8 2 9 2 | 6 2 6 2 | 5+6 2 |
| | 7 2 7 2 | 6 2 5 2 | 3+4 2 |
| | 9 2 9 2 | 7 2 6 2 | 3+4 2 |
| | 7 2 7 2 | 6 2 6 2 | 5+6 2 |
| | 7 2 7 2 | 6 2 5 2 | 5+6 2 |
| | 7 2 7 2 | 6 2 6 2 | 3+4 2 |
| | 6 2 6 2 | 7 2 5 2 | 3+4 2 |
| | 8 2 8 2 | 7 2 5 2 | 5+6 2 |
| | 9 2 9 2 | 7 2 6 2 | 3+4 2 |
| | 8 2 8 2 | 6 2 5 2 | 5+6 2 |
| | 7 2 6 2 | 6 2 5 2 | 3+4 2 |
| | 6 2 6 2 | 7 2 5 2 | 5+6 2 |
| | 8 2 7 2 | 6 2 6 2 | 5+6 2 |
| | 6 2 8 2 | 7 2 5 2 | 3+4 2 |
| | 6 2 9 2 | 6 2 6 2 | 3+4 2 |
| | 9 2 8 2 | 6 2 5 2 | 5+6 2 |
| | 8 2 9 2 | 7 2 6 2 | 5+6 2 |
| | 9 2 6 2 | 7 2 6 2 | 3+4 2 |
| | 6 2 7 2 | 6 2 5 2 | 3+4 2 |
| | 7 2 6 2 | 6 2 6 2 | 5+6 2 |
| | 9 2 9 2 | 6 2 6 2 | 5+6 2 |
| | 8 2 8 2 | 7 2 5 2 | 3+4 2 |
| | 8 2 9 2 | 7 2 6 2 | 5+6 2 |
| | 6 2 8 2 | 7 2 5 2 | 3+4 2 |
| | 7 2 8 2 | 7 2 5 2 | 5+6 2 |
| | 6 2 6 2 | 6 2 6 2 | 3+4 2 |
| | 9 2 7 2 | 6 2 5 2 | 3+4 2 |
| | 7 2 9 2 | 6 2 6 2 | 5+6 2 |
| | 6 2 8 2 | 7 2 6 2 | 5+6 2 |
| | 7 2 6 2 | 7 2 6 2 | 3+4 2 |

Continued on next column

**TABLE LXIV– continued**

| TQC | ICSD | NoE | SF |
|---|---|---|---|
| | 9 2 7 2 | 7 2 5 2 | 5+6 2 |
| | 8 2 7 2 | 6 2 6 2 | 5+6 2 |
| | 8 2 6 2 | 7 2 5 2 | 3+4 2 |
| | 9 2 7 2 | 6 2 5 2 | 3+4 2 |
| | 7 2 9 2 | 7 2 6 2 | 3+4 2 |
| | 9 2 9 2 | 7 2 5 2 | 5+6 2 |
| | 8 2 8 2 | 6 2 5 2 | 3+4 2 |
| | 7 2 6 2 | 7 2 5 2 | 5+6 2 |
| | 6 2 7 2 | 6 2 6 2 | 5+6 2 |
| | 9 2 9 2 | 7 2 6 2 | 3+4 2 |
| | 6 2 6 2 | 6 2 6 2 | 5+6 2 |
| | 8 2 8 2 | 6 2 5 2 | 3+4 2 |
| | 8 2 8 2 | 6 2 5 2 | 5+6 2 |
| 139.1.211 | 78593 | 98 | Ba2Ca1Cu2O8Tl2 |
| | -GM -M | -P -X | -N |
| | 7 2 7 2 | 6 2 5 2 | 3+4 2 |
| | 9 2 9 2 | 7 2 6 2 | 5+6 2 |
| | 7 2 9 2 | 6 2 5 2 | 3+4 2 |
| | 9 2 7 2 | 7 2 6 2 | 5+6 2 |
| | 7 2 7 2 | 6 2 5 2 | 3+4 2 |
| | 9 2 9 2 | 7 2 6 2 | 5+6 2 |
| | 7 2 9 2 | 6 2 5 2 | 3+4 2 |
| | 9 2 7 2 | 7 2 5 2 | 5+6 2 |
| | 6 2 6 2 | 7 2 6 2 | 3+4 2 |
| | 8 2 8 2 | 6 2 6 2 | 5+6 2 |
| | 7 2 7 2 | 6 2 5 2 | 3+4 2 |
| | 9 2 9 2 | 7 2 6 2 | 5+6 2 |
| | 7 2 7 2 | 6 2 6 2 | 3+4 2 |
| | 9 2 9 2 | 7 2 5 2 | 5+6 2 |
| | 6 2 6 2 | 6 2 5 2 | 3+4 2 |
| | 8 2 8 2 | 7 2 6 2 | 5+6 2 |
| | 9 2 7 2 | 7 2 5 2 | 3+4 2 |
| | 7 2 9 2 | 6 2 6 2 | 5+6 2 |
| | 7 2 9 2 | 7 2 5 2 | 5+6 2 |
| | 7 2 9 2 | 6 2 6 2 | 3+4 2 |
| | 6 2 7 2 | 6 2 5 2 | 3+4 2 |
| | 9 2 8 2 | 7 2 6 2 | 5+6 2 |
| | 7 2 7 2 | 6 2 5 2 | 3+4 2 |
| | 9 2 6 2 | 7 2 6 2 | 5+6 2 |
| | 6 2 9 2 | 6 2 5 2 | 3+4 2 |
| | 8 2 8 2 | 7 2 6 2 | 3+4 2 |
| | 7 2 7 2 | 6 2 5 2 | 5+6 2 |
| | 9 2 7 2 | 7 2 6 2 | 5+6 2 |
| | 8 2 6 2 | 6 2 6 2 | 3+4 2 |
| | 9 2 7 2 | 7 2 5 2 | 3+4 2 |
| | 6 2 6 2 | 6 2 6 2 | 5+6 2 |
| | 8 2 8 2 | 7 2 6 2 | 5+6 2 |
| | 6 2 9 2 | 7 2 5 2 | 3+4 2 |
| | 7 2 6 2 | 6 2 5 2 | 5+6 2 |
| | 8 2 6 2 | 7 2 6 2 | 5+6 2 |
| | 9 2 9 2 | 6 2 5 2 | 3+4 2 |
| | 7 2 8 2 | 7 2 5 2 | 5+6 2 |
| | 6 2 9 2 | 6 2 6 2 | 3+4 2 |
| | 8 2 7 2 | 7 2 5 2 | 3+4 2 |
| | 7 2 8 2 | 6 2 6 2 | 5+6 2 |
| | 6 2 6 2 | 6 2 5 2 | 3+4 2 |
| | 9 2 7 2 | 7 2 6 2 | 5+6 2 |
| | 9 2 9 2 | 6 2 6 2 | 5+6 2 |
| | 8 2 8 2 | 7 2 6 2 | 3+4 2 |
| | 6 2 6 2 | 7 2 5 2 | 3+4 2 |
| | 8 2 8 2 | 7 2 5 2 | 3+4 2 |

Continued on next column

**TABLE LXIV– continued**

| TQC | ICSD | NoE | SF |
|---|---|---|---|
| | 7 2 7 2 | 6 2 5 2 | 5+6 2 |
| | 6 2 6 2 | 6 2 6 2 | 5+6 2 |
| | 9 2 9 2 | 7 2 5 2 | 5+6 2 |
| 139.1.212 | 78592 | 98 | Ba2Ca1Cu2O8Tl2 |
| | -GM -M | -P -X | -N |
| | 7 2 7 2 | 6 2 5 2 | 3+4 2 |
| | 9 2 9 2 | 7 2 6 2 | 5+6 2 |
| | 7 2 9 2 | 6 2 5 2 | 3+4 2 |
| | 9 2 7 2 | 7 2 6 2 | 5+6 2 |
| | 7 2 7 2 | 6 2 5 2 | 3+4 2 |
| | 7 2 9 2 | 7 2 6 2 | 5+6 2 |
| | 9 2 9 2 | 6 2 5 2 | 3+4 2 |
| | 9 2 7 2 | 7 2 5 2 | 5+6 2 |
| | 6 2 6 2 | 7 2 6 2 | 3+4 2 |
| | 8 2 8 2 | 6 2 6 2 | 5+6 2 |
| | 7 2 7 2 | 6 2 5 2 | 3+4 2 |
| | 9 2 9 2 | 7 2 6 2 | 5+6 2 |
| | 7 2 7 2 | 6 2 6 2 | 3+4 2 |
| | 9 2 9 2 | 7 2 5 2 | 5+6 2 |
| | 6 2 6 2 | 6 2 5 2 | 3+4 2 |
| | 8 2 8 2 | 7 2 6 2 | 5+6 2 |
| | 9 2 7 2 | 7 2 5 2 | 3+4 2 |
| | 7 2 9 2 | 6 2 6 2 | 5+6 2 |
| | 7 2 9 2 | 7 2 5 2 | 5+6 2 |
| | 7 2 9 2 | 6 2 6 2 | 3+4 2 |
| | 6 2 7 2 | 6 2 5 2 | 3+4 2 |
| | 9 2 8 2 | 7 2 6 2 | 5+6 2 |
| | 7 2 7 2 | 6 2 5 2 | 3+4 2 |
| | 9 2 6 2 | 7 2 6 2 | 5+6 2 |
| | 6 2 9 2 | 6 2 5 2 | 3+4 2 |
| | 8 2 8 2 | 7 2 6 2 | 3+4 2 |
| | 7 2 7 2 | 6 2 5 2 | 5+6 2 |
| | 9 2 7 2 | 7 2 6 2 | 5+6 2 |
| | 8 2 6 2 | 6 2 6 2 | 3+4 2 |
| | 9 2 7 2 | 7 2 5 2 | 3+4 2 |
| | 6 2 6 2 | 6 2 6 2 | 5+6 2 |
| | 8 2 9 2 | 7 2 6 2 | 5+6 2 |
| | 7 2 6 2 | 7 2 5 2 | 3+4 2 |
| | 8 2 8 2 | 6 2 5 2 | 5+6 2 |
| | 6 2 6 2 | 7 2 6 2 | 5+6 2 |
| | 9 2 9 2 | 6 2 5 2 | 3+4 2 |
| | 6 2 8 2 | 7 2 5 2 | 5+6 2 |
| | 7 2 9 2 | 6 2 6 2 | 3+4 2 |
| | 8 2 7 2 | 7 2 5 2 | 3+4 2 |
| | 7 2 8 2 | 6 2 6 2 | 5+6 2 |
| | 6 2 6 2 | 6 2 5 2 | 3+4 2 |
| | 9 2 7 2 | 7 2 6 2 | 5+6 2 |
| | 9 2 9 2 | 6 2 6 2 | 5+6 2 |
| | 8 2 8 2 | 7 2 6 2 | 3+4 2 |
| | 6 2 6 2 | 7 2 5 2 | 3+4 2 |
| | 8 2 8 2 | 7 2 5 2 | 3+4 2 |
| | 7 2 7 2 | 6 2 5 2 | 5+6 2 |
| | 6 2 6 2 | 6 2 6 2 | 5+6 2 |
| | 9 2 9 2 | 7 2 5 2 | 5+6 2 |
| 139.1.213 | 2141 | 50 | Ba2Bi1 |
| | -GM -M | -P -X | -N |
| | 7 2 7 2 | 6 2 5 2 | 5+6 2 |
| | 9 2 9 2 | 7 2 6 2 | 3+4 2 |
| | 7 2 7 2 | 6 2 6 2 | 3+4 2 |
| | 6 2 6 2 | 7 2 6 2 | 5+6 2 |
| | 8 2 8 2 | 7 2 5 2 | 3+4 2 |

Continued on next column

**TABLE LXIV– continued**

| TQC | ICSD | NoE | SF |
|---|---|---|---|
| | 7 2 9 2 | 6 2 5 2 | 5+6 2 |
| | 9 2 7 2 | 6 2 6 2 | 5+6 2 |
| | 9 2 9 2 | 7 2 5 2 | 3+4 2 |
| | 9 2 9 2 | 6 2 5 2 | 5+6 2 |
| | 9 2 9 2 | 7 2 5 2 | 3+4 2 |
| | 7 2 8 2 | 6 2 6 2 | 5+6 2 |
| | 6 2 7 2 | 7 2 5 2 | 3+4 2 |
| | 8 2 6 2 | 7 2 6 2 | 3+4 2 |
| | 8 2 8 2 | 6 2 5 2 | 5+6 2 |
| | 9 2 9 2 | 7 2 5 2 | 5+6 2 |
| | 8 2 8 2 | 6 2 5 2 | 3+4 2 |
| | 7 2 7 2 | 7 2 6 2 | 3+4 2 |
| | 9 2 9 2 | 6 2 5 2 | 5+6 2 |
| | 7 2 7 2 | 7 2 6 2 | 5+6 2 |
| | 9 2 9 2 | 6 2 5 2 | 3+4 2 |
| | 7 2 7 2 | 6 2 6 2 | 3+4 2 |
| | 7 2 7 2 | 7 2 6 2 | 5+6 2 |
| | 6 2 6 2 | 6 2 5 2 | 3+4 2 |
| | 9 2 9 2 | 7 2 5 2 | 5+6 2 |
| | 8 2 6 2 | 6 2 6 2 | 5+6 2 |
| 139.1.214 | 61197 | 40 | Ba1Pd2Sb2 |
| | -GM -M | -P -X | -N |
| | 7 2 7 2 | 6 2 5 2 | 3+4 2 |
| | 9 2 9 2 | 7 2 6 2 | 5+6 2 |
| | 9 2 9 2 | 6 2 6 2 | 5+6 2 |
| | 8 2 8 2 | 7 2 6 2 | 5+6 2 |
| | 7 2 7 2 | 7 2 5 2 | 5+6 2 |
| | 9 2 9 2 | 6 2 6 2 | 3+4 2 |
| | 7 2 7 2 | 7 2 5 2 | 3+4 2 |
| | 9 2 6 2 | 6 2 6 2 | 5+6 2 |
| | 8 2 7 2 | 7 2 6 2 | 3+4 2 |
| | 6 2 8 2 | 6 2 5 2 | 5+6 2 |
| | 7 2 9 2 | 7 2 5 2 | 5+6 2 |
| | 6 2 9 2 | 6 2 5 2 | 3+4 2 |
| | 9 2 8 2 | 7 2 6 2 | 5+6 2 |
| | 8 2 6 2 | 6 2 5 2 | 3+4 2 |
| | 6 2 8 2 | 7 2 6 2 | 3+4 2 |
| | 7 2 7 2 | 6 2 5 2 | 5+6 2 |
| | 7 2 9 2 | 7 2 6 2 | 5+6 2 |
| | 9 2 8 2 | 7 2 6 2 | 3+4 2 |
| | 6 2 7 2 | 6 2 6 2 | 3+4 2 |
| | 6 2 6 2 | 6 2 5 2 | 5+6 2 |
| 139.1.215 | 50188 | 38 | Ba1P2Rh2 |
| | -GM -M | -P -X | -N |
| | 7 2 7 2 | 6 2 5 2 | 3+4 2 |
| | 9 2 9 2 | 7 2 6 2 | 5+6 2 |
| | 9 2 9 2 | 6 2 6 2 | 5+6 2 |
| | 7 2 8 2 | 7 2 6 2 | 5+6 2 |
| | 8 2 7 2 | 7 2 5 2 | 3+4 2 |
| | 9 2 9 2 | 6 2 6 2 | 5+6 2 |
| | 7 2 7 2 | 7 2 5 2 | 3+4 2 |
| | 8 2 6 2 | 6 2 6 2 | 3+4 2 |
| | 9 2 7 2 | 7 2 5 2 | 5+6 2 |
| | 9 2 9 2 | 6 2 6 2 | 5+6 2 |
| | 6 2 8 2 | 7 2 5 2 | 3+4 2 |
| | 7 2 9 2 | 6 2 5 2 | 5+6 2 |
| | 6 2 8 2 | 7 2 6 2 | 3+4 2 |
| | 8 2 6 2 | 6 2 5 2 | 5+6 2 |
| | 7 2 8 2 | 7 2 6 2 | 3+4 2 |
| | 6 2 7 2 | 6 2 5 2 | 5+6 2 |
| | 7 2 9 2 | 7 2 6 2 | 5+6 2 |

Continued on next column

**TABLE LXIV– continued**

| TQC | ICSD | NoE | SF |
|---|---|---|---|
| | 6 2 8 2 | 6 2 6 2 | 3+4 2 |
| | 9 2 9 2 | 7 2 5 2 | 3+4 2 |
| 139.1.216 | 419531 | 34 | Ba1Mn2Sb2 |
| | -GM -M | -P -X | -N |
| | 7 2 7 2 | 6 2 5 2 | 3+4 2 |
| | 9 2 9 2 | 7 2 6 2 | 5+6 2 |
| | 9 2 9 2 | 6 2 6 2 | 5+6 2 |
| | 8 2 8 2 | 7 2 6 2 | 5+6 2 |
| | 7 2 9 2 | 7 2 5 2 | 5+6 2 |
| | 9 2 7 2 | 6 2 6 2 | 3+4 2 |
| | 7 2 7 2 | 7 2 5 2 | 3+4 2 |
| | 9 2 9 2 | 6 2 6 2 | 5+6 2 |
| | 9 2 7 2 | 7 2 5 2 | 5+6 2 |
| | 8 2 6 2 | 6 2 5 2 | 3+4 2 |
| | 7 2 9 2 | 7 2 6 2 | 3+4 2 |
| | 6 2 8 2 | 6 2 5 2 | 5+6 2 |
| | 6 2 8 2 | 6 2 5 2 | 3+4 2 |
| | 7 2 7 2 | 7 2 6 2 | 5+6 2 |
| | 8 2 6 2 | 6 2 6 2 | 5+6 2 |
| | 7 2 9 2 | 7 2 6 2 | 3+4 2 |
| | 6 2 8 2 | 7 2 5 2 | 3+4 2 |
| 139.1.217 | 32019 | 34 | Ba1Mn2Sb2 |
| | -GM -M | -P -X | -N |
| | 7 2 7 2 | 6 2 5 2 | 3+4 2 |
| | 9 2 9 2 | 7 2 6 2 | 5+6 2 |
| | 9 2 9 2 | 6 2 6 2 | 5+6 2 |
| | 8 2 8 2 | 7 2 6 2 | 5+6 2 |
| | 7 2 9 2 | 7 2 5 2 | 5+6 2 |
| | 9 2 7 2 | 6 2 6 2 | 3+4 2 |
| | 7 2 7 2 | 7 2 5 2 | 3+4 2 |
| | 9 2 9 2 | 6 2 6 2 | 5+6 2 |
| | 9 2 7 2 | 7 2 5 2 | 5+6 2 |
| | 8 2 6 2 | 6 2 5 2 | 3+4 2 |
| | 7 2 9 2 | 7 2 6 2 | 3+4 2 |
| | 6 2 8 2 | 6 2 5 2 | 5+6 2 |
| | 6 2 8 2 | 6 2 5 2 | 3+4 2 |
| | 7 2 7 2 | 7 2 6 2 | 5+6 2 |
| | 8 2 6 2 | 6 2 6 2 | 5+6 2 |
| | 7 2 9 2 | 7 2 6 2 | 3+4 2 |
| | 6 2 8 2 | 7 2 5 2 | 3+4 2 |
| 139.1.218 | 77668 | 34 | Ba1Mn2P2 |
| | -GM -M | -P -X | -N |
| | 7 2 7 2 | 6 2 5 2 | 3+4 2 |
| | 9 2 9 2 | 7 2 6 2 | 5+6 2 |
| | 9 2 9 2 | 6 2 6 2 | 5+6 2 |
| | 8 2 8 2 | 7 2 6 2 | 5+6 2 |
| | 7 2 9 2 | 7 2 5 2 | 3+4 2 |
| | 9 2 7 2 | 6 2 6 2 | 5+6 2 |
| | 7 2 7 2 | 7 2 5 2 | 3+4 2 |
| | 9 2 9 2 | 6 2 6 2 | 5+6 2 |
| | 9 2 7 2 | 7 2 5 2 | 5+6 2 |
| | 8 2 6 2 | 7 2 5 2 | 3+4 2 |
| | 7 2 9 2 | 6 2 6 2 | 3+4 2 |
| | 6 2 8 2 | 6 2 6 2 | 5+6 2 |
| | 6 2 8 2 | 6 2 5 2 | 3+4 2 |
| | 7 2 6 2 | 6 2 5 2 | 5+6 2 |
| | 8 2 7 2 | 7 2 6 2 | 5+6 2 |
| | 6 2 9 2 | 7 2 6 2 | 3+4 2 |
| | 7 2 8 2 | 7 2 5 2 | 3+4 2 |
| 139.1.219 | 10469 | 34 | Ba1Mn2P2 |
| | -GM -M | -P -X | -N |

Continued on next column

**TABLE LXIV– continued**

| TQC | ICSD | NoE | SF |
|---|---|---|---|
| | 7 2 7 2 | 6 2 5 2 | 3+4 2 |
| | 9 2 9 2 | 7 2 6 2 | 5+6 2 |
| | 9 2 9 2 | 6 2 6 2 | 5+6 2 |
| | 8 2 8 2 | 7 2 6 2 | 5+6 2 |
| | 7 2 9 2 | 7 2 5 2 | 3+4 2 |
| | 9 2 7 2 | 6 2 6 2 | 5+6 2 |
| | 7 2 7 2 | 7 2 5 2 | 3+4 2 |
| | 9 2 9 2 | 6 2 6 2 | 5+6 2 |
| | 9 2 7 2 | 7 2 5 2 | 5+6 2 |
| | 8 2 6 2 | 7 2 5 2 | 3+4 2 |
| | 7 2 9 2 | 6 2 6 2 | 3+4 2 |
| | 6 2 8 2 | 6 2 6 2 | 5+6 2 |
| | 6 2 8 2 | 6 2 5 2 | 3+4 2 |
| | 7 2 6 2 | 6 2 5 2 | 5+6 2 |
| | 8 2 7 2 | 7 2 6 2 | 5+6 2 |
| | 6 2 9 2 | 7 2 6 2 | 3+4 2 |
| | 7 2 8 2 | 7 2 5 2 | 3+4 2 |
| 139.1.220 | 100024 | 54 | Ba1Mn1Sb2 |
| | -GM -M | -P -X | -N |
| | 7 2 7 2 | 6 2 5 2 | 3+4 2 |
| | 9 2 9 2 | 7 2 6 2 | 5+6 2 |
| | 7 2 7 2 | 6 2 5 2 | 3+4 2 |
| | 9 2 9 2 | 7 2 6 2 | 5+6 2 |
| | 7 2 7 2 | 6 2 6 2 | 3+4 2 |
| | 9 2 9 2 | 7 2 5 2 | 5+6 2 |
| | 8 2 8 2 | 6 2 5 2 | 3+4 2 |
| | 6 2 6 2 | 7 2 6 2 | 5+6 2 |
| | 7 2 7 2 | 6 2 5 2 | 5+6 2 |
| | 7 2 9 2 | 7 2 6 2 | 3+4 2 |
| | 9 2 7 2 | 7 2 6 2 | 3+4 2 |
| | 6 2 6 2 | 6 2 6 2 | 5+6 2 |
| | 7 2 9 2 | 6 2 5 2 | 5+6 2 |
| | 9 2 8 2 | 7 2 5 2 | 3+4 2 |
| | 8 2 7 2 | 7 2 5 2 | 5+6 2 |
| | 9 2 9 2 | 6 2 6 2 | 3+4 2 |
| | 9 2 9 2 | 6 2 5 2 | 5+6 2 |
| | 8 2 7 2 | 7 2 6 2 | 3+4 2 |
| | 9 2 6 2 | 6 2 6 2 | 3+4 2 |
| | 7 2 9 2 | 7 2 5 2 | 5+6 2 |
| | 6 2 8 2 | 6 2 5 2 | 5+6 2 |
| | 6 2 8 2 | 7 2 5 2 | 3+4 2 |
| | 8 2 6 2 | 7 2 6 2 | 3+4 2 |
| | 7 2 9 2 | 6 2 5 2 | 5+6 2 |
| | 8 2 8 2 | 6 2 6 2 | 5+6 2 |
| | 6 2 7 2 | 7 2 5 2 | 3+4 2 |
| | 7 2 8 2 | 6 2 6 2 | 3+4 2 |
| 139.1.221 | 10468 | 36 | Ba1Fe2P2 |
| | -GM -M | -P -X | -N |
| | 7 2 7 2 | 6 2 5 2 | 3+4 2 |
| | 9 2 9 2 | 7 2 6 2 | 5+6 2 |
| | 9 2 9 2 | 6 2 6 2 | 5+6 2 |
| | 8 2 8 2 | 7 2 6 2 | 5+6 2 |
| | 7 2 9 2 | 7 2 5 2 | 3+4 2 |
| | 9 2 7 2 | 6 2 6 2 | 5+6 2 |
| | 7 2 7 2 | 7 2 5 2 | 3+4 2 |
| | 9 2 9 2 | 6 2 6 2 | 5+6 2 |
| | 9 2 6 2 | 7 2 5 2 | 3+4 2 |
| | 8 2 7 2 | 7 2 5 2 | 3+4 2 |
| | 6 2 9 2 | 6 2 6 2 | 5+6 2 |
| | 7 2 8 2 | 6 2 5 2 | 5+6 2 |
| | 6 2 8 2 | 7 2 6 2 | 3+4 2 |

Continued on next column

**TABLE LXIV– continued**

| TQC | ICSD | NoE | SF |
|---|---|---|---|
| | 7 2 6 2 | 6 2 5 2 | 5+6 2 |
| | 8 2 7 2 | 7 2 6 2 | 3+4 2 |
| | 6 2 8 2 | 6 2 6 2 | 5+6 2 |
| | 7 2 9 2 | 7 2 5 2 | 3+4 2 |
| | 6 2 8 2 | 6 2 6 2 | 5+6 2 |
| 139.1.222 | 420975 | 36 | Ba1Co2Ge2 |
| | -GM -M | -P -X | -N |
| | 7 2 7 2 | 6 2 5 2 | 3+4 2 |
| | 9 2 9 2 | 7 2 6 2 | 5+6 2 |
| | 9 2 9 2 | 6 2 6 2 | 5+6 2 |
| | 8 2 8 2 | 7 2 6 2 | 5+6 2 |
| | 7 2 9 2 | 7 2 5 2 | 3+4 2 |
| | 9 2 7 2 | 6 2 6 2 | 5+6 2 |
| | 7 2 7 2 | 7 2 5 2 | 3+4 2 |
| | 9 2 7 2 | 6 2 6 2 | 5+6 2 |
| | 8 2 6 2 | 7 2 5 2 | 3+4 2 |
| | 9 2 9 2 | 7 2 5 2 | 5+6 2 |
| | 6 2 8 2 | 6 2 6 2 | 3+4 2 |
| | 7 2 9 2 | 6 2 5 2 | 5+6 2 |
| | 6 2 8 2 | 6 2 5 2 | 3+4 2 |
| | 8 2 6 2 | 7 2 6 2 | 5+6 2 |
| | 7 2 8 2 | 7 2 6 2 | 3+4 2 |
| | 6 2 7 2 | 7 2 5 2 | 5+6 2 |
| | 7 2 9 2 | 6 2 5 2 | 5+6 2 |
| | 6 2 8 2 | 7 2 6 2 | 3+4 2 |
| 139.1.223 | 174263 | 42 | Ba1Cd2Ge2 |
| | -GM -M | -P -X | -N |
| | 7 2 7 2 | 6 2 5 2 | 3+4 2 |
| | 9 2 9 2 | 7 2 6 2 | 5+6 2 |
| | 9 2 9 2 | 6 2 6 2 | 5+6 2 |
| | 8 2 8 2 | 7 2 6 2 | 5+6 2 |
| | 7 2 7 2 | 7 2 5 2 | 5+6 2 |
| | 9 2 9 2 | 6 2 6 2 | 3+4 2 |
| | 6 2 8 2 | 6 2 5 2 | 3+4 2 |
| | 7 2 6 2 | 7 2 6 2 | 5+6 2 |
| | 8 2 7 2 | 7 2 5 2 | 5+6 2 |
| | 9 2 9 2 | 6 2 6 2 | 3+4 2 |
| | 6 2 8 2 | 7 2 5 2 | 3+4 2 |
| | 7 2 7 2 | 6 2 6 2 | 3+4 2 |
| | 8 2 6 2 | 6 2 5 2 | 5+6 2 |
| | 6 2 8 2 | 7 2 6 2 | 5+6 2 |
| | 9 2 9 2 | 7 2 5 2 | 5+6 2 |
| | 9 2 7 2 | 6 2 6 2 | 3+4 2 |
| | 7 2 9 2 | 7 2 5 2 | 5+6 2 |
| | 6 2 8 2 | 6 2 6 2 | 3+4 2 |
| | 7 2 7 2 | 7 2 5 2 | 3+4 2 |
| | 9 2 6 2 | 7 2 6 2 | 5+6 2 |
| | 8 2 9 2 | 6 2 5 2 | 3+4 2 |
| 139.1.224 | 71642 | 36 | B2Ir2Zn1 |
| | -GM -M | -P -X | -N |
| | 7 2 9 2 | 7 2 5 2 | 5+6 2 |
| | 9 2 7 2 | 6 2 5 2 | 3+4 2 |
| | 7 2 6 2 | 6 2 6 2 | 3+4 2 |
| | 6 2 7 2 | 7 2 6 2 | 3+4 2 |
| | 7 2 7 2 | 6 2 5 2 | 3+4 2 |
| | 7 2 6 2 | 7 2 5 2 | 3+4 2 |
| | 6 2 6 2 | 7 2 5 2 | 3+4 2 |
| | 6 2 9 2 | 6 2 5 2 | 5+6 2 |
| | 8 2 8 2 | 7 2 5 2 | 3+4 2 |
| | 6 2 6 2 | 7 2 6 2 | 3+4 2 |
| | 7 2 9 2 | 6 2 5 2 | 5+6 2 |

Continued on next column

**TABLE LXIV– continued**

| TQC | ICSD | NoE | SF |
|---|---|---|---|
| | 6 2 8 2 | 6 2 5 2 | 5+6 2 |
| | 7 2 7 2 | 7 2 6 2 | 3+4 2 |
| | 9 2 7 2 | 6 2 6 2 | 5+6 2 |
| | 8 2 7 2 | 7 2 6 2 | 3+4 2 |
| | 9 2 6 2 | 7 2 5 2 | 5+6 2 |
| | 6 2 8 2 | 6 2 5 2 | 3+4 2 |
| | 8 2 6 2 | 7 2 6 2 | 5+6 2 |
| 139.1.225 | 426150 | 80 | Au4Ce1Zn2 |
| | -GM -M | -P -X | -N |
| | 7 2 7 2 | 6 2 5 2 | 3+4 2 |
| | 9 2 9 2 | 7 2 6 2 | 5+6 2 |
| | 8 2 8 2 | 6 2 6 2 | 5+6 2 |
| | 9 2 9 2 | 7 2 6 2 | 5+6 2 |
| | 7 2 6 2 | 7 2 6 2 | 5+6 2 |
| | 7 2 9 2 | 7 2 5 2 | 3+4 2 |
| | 8 2 8 2 | 6 2 6 2 | 3+4 2 |
| | 9 2 7 2 | 6 2 5 2 | 5+6 2 |
| | 6 2 7 2 | 7 2 5 2 | 3+4 2 |
| | 8 2 8 2 | 6 2 6 2 | 5+6 2 |
| | 9 2 8 2 | 7 2 5 2 | 3+4 2 |
| | 6 2 6 2 | 6 2 6 2 | 5+6 2 |
| | 9 2 9 2 | 6 2 5 2 | 3+4 2 |
| | 7 2 7 2 | 7 2 5 2 | 5+6 2 |
| | 6 2 7 2 | 7 2 6 2 | 3+4 2 |
| | 7 2 9 2 | 6 2 6 2 | 5+6 2 |
| | 6 2 6 2 | 7 2 5 2 | 3+4 2 |
| | 8 2 8 2 | 6 2 5 2 | 5+6 2 |
| | 7 2 6 2 | 7 2 6 2 | 3+4 2 |
| | 8 2 7 2 | 6 2 6 2 | 3+4 2 |
| | 6 2 9 2 | 7 2 6 2 | 5+6 2 |
| | 9 2 8 2 | 6 2 5 2 | 5+6 2 |
| | 9 2 6 2 | 7 2 5 2 | 5+6 2 |
| | 6 2 7 2 | 6 2 6 2 | 3+4 2 |
| | 7 2 9 2 | 7 2 6 2 | 5+6 2 |
| | 9 2 8 2 | 7 2 5 2 | 3+4 2 |
| | 8 2 7 2 | 6 2 5 2 | 3+4 2 |
| | 9 2 6 2 | 7 2 6 2 | 5+6 2 |
| | 7 2 8 2 | 6 2 6 2 | 3+4 2 |
| | 8 2 9 2 | 6 2 5 2 | 3+4 2 |
| | 6 2 6 2 | 6 2 6 2 | 3+4 2 |
| | 6 2 8 2 | 6 2 5 2 | 5+6 2 |
| | 9 2 7 2 | 7 2 5 2 | 5+6 2 |
| | 8 2 7 2 | 7 2 6 2 | 5+6 2 |
| | 7 2 9 2 | 7 2 6 2 | 3+4 2 |
| | 8 2 6 2 | 6 2 5 2 | 5+6 2 |
| | 6 2 9 2 | 6 2 5 2 | 3+4 2 |
| | 9 2 8 2 | 7 2 6 2 | 5+6 2 |
| | 6 2 9 2 | 7 2 5 2 | 5+6 2 |
| | 7 2 8 2 | 6 2 6 2 | 3+4 2 |
| 139.1.226 | 58388 | 46 | Au2Yb1 |
| | -GM -M | -P -X | -N |
| | 7 2 7 2 | 6 2 5 2 | 3+4 2 |
| | 9 2 9 2 | 7 2 6 2 | 5+6 2 |
| | 9 2 9 2 | 6 2 6 2 | 5+6 2 |
| | 8 2 8 2 | 7 2 6 2 | 5+6 2 |
| | 7 2 9 2 | 7 2 5 2 | 3+4 2 |
| | 8 2 7 2 | 6 2 6 2 | 5+6 2 |
| | 6 2 6 2 | 6 2 5 2 | 3+4 2 |
| | 7 2 8 2 | 7 2 6 2 | 5+6 2 |
| | 9 2 9 2 | 7 2 6 2 | 5+6 2 |
| | 6 2 7 2 | 7 2 5 2 | 3+4 2 |

Continued on next column

**TABLE LXIV– continued**

| TQC | ICSD | NoE | SF |
|---|---|---|---|
| | 7 2 8 2 | 6 2 5 2 | 3+4 2 |
| | 9 2 7 2 | 6 2 6 2 | 3+4 2 |
| | 8 2 6 2 | 7 2 5 2 | 5+6 2 |
| | 6 2 8 2 | 6 2 6 2 | 5+6 2 |
| | 8 2 6 2 | 7 2 5 2 | 3+4 2 |
| | 9 2 9 2 | 6 2 6 2 | 5+6 2 |
| | 8 2 8 2 | 6 2 6 2 | 5+6 2 |
| | 8 2 9 2 | 7 2 6 2 | 5+6 2 |
| | 9 2 8 2 | 7 2 5 2 | 5+6 2 |
| | 7 2 9 2 | 6 2 6 2 | 5+6 2 |
| | 8 2 9 2 | 6 2 6 2 | 5+6 2 |
| | 8 2 8 2 | 7 2 6 2 | 5+6 2 |
| | 9 2 8 2 | 6 2 6 2 | 5+6 2 |
| 139.1.227 | 413 | 40 | Au2Si2Sr1 |
| | -GM -M | -P -X | -N |
| | 7 2 7 2 | 6 2 5 2 | 3+4 2 |
| | 9 2 9 2 | 7 2 6 2 | 5+6 2 |
| | 9 2 9 2 | 6 2 6 2 | 5+6 2 |
| | 8 2 8 2 | 7 2 6 2 | 5+6 2 |
| | 7 2 9 2 | 7 2 5 2 | 5+6 2 |
| | 9 2 7 2 | 6 2 6 2 | 3+4 2 |
| | 6 2 8 2 | 7 2 5 2 | 3+4 2 |
| | 7 2 6 2 | 6 2 6 2 | 5+6 2 |
| | 8 2 7 2 | 6 2 5 2 | 3+4 2 |
| | 9 2 9 2 | 7 2 6 2 | 5+6 2 |
| | 7 2 7 2 | 7 2 5 2 | 3+4 2 |
| | 6 2 8 2 | 6 2 6 2 | 5+6 2 |
| | 8 2 9 2 | 6 2 5 2 | 5+6 2 |
| | 6 2 6 2 | 7 2 6 2 | 3+4 2 |
| | 9 2 8 2 | 6 2 5 2 | 3+4 2 |
| | 9 2 7 2 | 7 2 6 2 | 5+6 2 |
| | 7 2 8 2 | 7 2 5 2 | 5+6 2 |
| | 6 2 9 2 | 7 2 6 2 | 3+4 2 |
| | 7 2 6 2 | 6 2 5 2 | 3+4 2 |
| | 8 2 7 2 | 6 2 6 2 | 5+6 2 |
| 139.1.228 | 37453 | 82 | Au2Cl6Cs2 |
| | -GM -M | -P -X | -N |
| | 7 2 9 2 | 6 2 6 2 | 3+4 2 |
| | 8 2 6 2 | 7 2 5 2 | 5+6 2 |
| | 7 2 7 2 | 6 2 5 2 | 3+4 2 |
| | 7 2 7 2 | 7 2 5 2 | 5+6 2 |
| | 9 2 9 2 | 7 2 6 2 | 5+6 2 |
| | 9 2 9 2 | 6 2 5 2 | 3+4 2 |
| | 8 2 8 2 | 6 2 6 2 | 5+6 2 |
| | 6 2 6 2 | 7 2 6 2 | 3+4 2 |
| | 6 2 8 2 | 6 2 5 2 | 3+4 2 |
| | 9 2 7 2 | 7 2 6 2 | 5+6 2 |
| | 7 2 7 2 | 6 2 5 2 | 3+4 2 |
| | 7 2 9 2 | 6 2 6 2 | 3+4 2 |
| | 6 2 7 2 | 7 2 5 2 | 5+6 2 |
| | 8 2 6 2 | 7 2 5 2 | 5+6 2 |
| | 7 2 8 2 | 6 2 6 2 | 3+4 2 |
| | 9 2 7 2 | 6 2 5 2 | 5+6 2 |
| | 6 2 6 2 | 7 2 5 2 | 5+6 2 |
| | 7 2 7 2 | 6 2 5 2 | 3+4 2 |
| | 6 2 6 2 | 7 2 5 2 | 3+4 2 |
| | 6 2 6 2 | 7 2 5 2 | 3+4 2 |
| | 9 2 9 2 | 6 2 5 2 | 3+4 2 |
| | 7 2 7 2 | 6 2 6 2 | 5+6 2 |
| | 6 2 8 2 | 7 2 6 2 | 5+6 2 |
| | 8 2 6 2 | 7 2 5 2 | 5+6 2 |

Continued on next column

**TABLE LXIV– continued**

| TQC | ICSD | NoE | SF |
|---|---|---|---|
| | 9 2 9 2 | 7 2 6 2 | 3+4 2 |
| | 6 2 6 2 | 6 2 5 2 | 5+6 2 |
| | 6 2 9 2 | 7 2 6 2 | 5+6 2 |
| | 7 2 6 2 | 6 2 5 2 | 5+6 2 |
| | 9 2 8 2 | 6 2 5 2 | 3+4 2 |
| | 8 2 9 2 | 7 2 6 2 | 3+4 2 |
| | 8 2 8 2 | 6 2 6 2 | 3+4 2 |
| | 9 2 7 2 | 7 2 6 2 | 5+6 2 |
| | 9 2 9 2 | 7 2 6 2 | 3+4 2 |
| | 8 2 8 2 | 6 2 6 2 | 5+6 2 |
| | 6 2 6 2 | 7 2 5 2 | 5+6 2 |
| | 7 2 7 2 | 6 2 5 2 | 5+6 2 |
| | 7 2 7 2 | 7 2 5 2 | 3+4 2 |
| | 6 2 7 2 | 6 2 5 2 | 5+6 2 |
| | 7 2 6 2 | 6 2 5 2 | 3+4 2 |
| | 6 2 7 2 | 7 2 5 2 | 3+4 2 |
| | 7 2 6 2 | 7 2 5 2 | 5+6 2 |
| 139.1.229 | 418530 | 42 | Au2Ce1Si2 |
| | -GM -M | -P -X | -N |
| | 7 2 7 2 | 7 2 5 2 | 5+6 2 |
| | 9 2 9 2 | 6 2 6 2 | 3+4 2 |
| | 9 2 9 2 | 7 2 6 2 | 3+4 2 |
| | 8 2 8 2 | 6 2 6 2 | 3+4 2 |
| | 7 2 7 2 | 6 2 5 2 | 3+4 2 |
| | 9 2 9 2 | 7 2 6 2 | 5+6 2 |
| | 7 2 7 2 | 6 2 5 2 | 5+6 2 |
| | 6 2 6 2 | 7 2 6 2 | 3+4 2 |
| | 8 2 8 2 | 6 2 5 2 | 5+6 2 |
| | 9 2 9 2 | 7 2 6 2 | 3+4 2 |
| | 7 2 7 2 | 6 2 5 2 | 5+6 2 |
| | 6 2 6 2 | 7 2 6 2 | 3+4 2 |
| | 8 2 9 2 | 6 2 5 2 | 3+4 2 |
| | 9 2 8 2 | 7 2 6 2 | 5+6 2 |
| | 6 2 8 2 | 7 2 5 2 | 5+6 2 |
| | 7 2 7 2 | 6 2 6 2 | 3+4 2 |
| | 9 2 8 2 | 6 2 5 2 | 3+4 2 |
| | 6 2 9 2 | 6 2 6 2 | 5+6 2 |
| | 7 2 7 2 | 7 2 5 2 | 5+6 2 |
| | 6 2 6 2 | 6 2 5 2 | 3+4 2 |
| | 8 2 8 2 | 7 2 6 2 | 3+4 2 |
| 139.1.230 | 412 | 32 | Au2Ca1Si2 |
| | -GM -M | -P -X | -N |
| | 7 2 9 2 | 7 2 5 2 | 5+6 2 |
| | 9 2 7 2 | 6 2 6 2 | 3+4 2 |
| | 6 2 8 2 | 7 2 6 2 | 3+4 2 |
| | 7 2 6 2 | 6 2 5 2 | 5+6 2 |
| | 8 2 7 2 | 6 2 6 2 | 5+6 2 |
| | 9 2 9 2 | 7 2 5 2 | 3+4 2 |
| | 7 2 7 2 | 7 2 5 2 | 3+4 2 |
| | 6 2 8 2 | 6 2 6 2 | 5+6 2 |
| | 8 2 9 2 | 6 2 5 2 | 5+6 2 |
| | 6 2 6 2 | 7 2 6 2 | 3+4 2 |
| | 9 2 8 2 | 6 2 6 2 | 3+4 2 |
| | 9 2 7 2 | 7 2 5 2 | 5+6 2 |
| | 7 2 8 2 | 7 2 5 2 | 5+6 2 |
| | 6 2 9 2 | 7 2 6 2 | 3+4 2 |
| | 7 2 6 2 | 6 2 5 2 | 3+4 2 |
| | 8 2 7 2 | 6 2 6 2 | 5+6 2 |
| 139.1.231 | 263018 | 70 | As4Ba2Fe2O1Ti2 |
| | -GM -M | -P -X | -N |
| | 7 2 7 2 | 6 2 5 2 | 3+4 2 |
| | 9 2 9 2 | 7 2 6 2 | 5+6 2 |
| | 7 2 7 2 | 6 2 5 2 | 3+4 2 |
| | 7 2 9 2 | 7 2 6 2 | 5+6 2 |
| | 9 2 7 2 | 6 2 5 2 | 3+4 2 |
| | 7 2 9 2 | 6 2 5 2 | 5+6 2 |
| | 9 2 7 2 | 7 2 6 2 | 3+4 2 |
| | 6 2 6 2 | 7 2 6 2 | 5+6 2 |
| | 8 2 8 2 | 6 2 5 2 | 3+4 2 |
| | 7 2 9 2 | 7 2 5 2 | 5+6 2 |
| | 7 2 7 2 | 6 2 6 2 | 3+4 2 |
| | 9 2 7 2 | 7 2 5 2 | 5+6 2 |
| | 9 2 9 2 | 6 2 6 2 | 3+4 2 |
| | 8 2 8 2 | 7 2 6 2 | 5+6 2 |
| | 9 2 9 2 | 6 2 6 2 | 5+6 2 |
| | 7 2 9 2 | 7 2 6 2 | 5+6 2 |
| | 9 2 7 2 | 6 2 6 2 | 3+4 2 |
| | 7 2 9 2 | 7 2 5 2 | 5+6 2 |
| | 9 2 7 2 | 6 2 6 2 | 3+4 2 |
| | 9 2 7 2 | 7 2 6 2 | 5+6 2 |
| | 8 2 6 2 | 7 2 5 2 | 3+4 2 |
| | 7 2 7 2 | 6 2 6 2 | 5+6 2 |
| | 6 2 6 2 | 7 2 5 2 | 5+6 2 |
| | 7 2 9 2 | 6 2 6 2 | 3+4 2 |
| | 6 2 8 2 | 7 2 5 2 | 5+6 2 |
| | 6 2 8 2 | 6 2 5 2 | 5+6 2 |
| | 6 2 6 2 | 7 2 5 2 | 3+4 2 |
| | 9 2 9 2 | 6 2 6 2 | 3+4 2 |
| | 9 2 7 2 | 7 2 5 2 | 3+4 2 |
| | 7 2 9 2 | 6 2 6 2 | 5+6 2 |
| | 8 2 6 2 | 6 2 6 2 | 5+6 2 |
| | 8 2 8 2 | 7 2 5 2 | 3+4 2 |
| | 6 2 8 2 | 6 2 6 2 | 5+6 2 |
| | 6 2 6 2 | 7 2 5 2 | 3+4 2 |
| | 7 2 9 2 | 7 2 6 2 | 3+4 2 |
| 139.1.232 | 417002 | 38 | As2Rh2Sr1 |
| | -GM -M | -P -X | -N |
| | 7 2 7 2 | 6 2 5 2 | 3+4 2 |
| | 9 2 9 2 | 7 2 6 2 | 5+6 2 |
| | 9 2 9 2 | 6 2 6 2 | 5+6 2 |
| | 8 2 8 2 | 7 2 6 2 | 5+6 2 |
| | 7 2 9 2 | 7 2 5 2 | 5+6 2 |
| | 9 2 7 2 | 6 2 6 2 | 3+4 2 |
| | 7 2 7 2 | 7 2 5 2 | 3+4 2 |
| | 9 2 6 2 | 6 2 6 2 | 5+6 2 |
| | 8 2 7 2 | 7 2 5 2 | 3+4 2 |
| | 9 2 9 2 | 6 2 6 2 | 5+6 2 |
| | 6 2 8 2 | 7 2 5 2 | 5+6 2 |
| | 7 2 9 2 | 6 2 5 2 | 3+4 2 |
| | 6 2 8 2 | 7 2 6 2 | 3+4 2 |
| | 8 2 6 2 | 6 2 5 2 | 5+6 2 |
| | 7 2 8 2 | 7 2 6 2 | 5+6 2 |
| | 6 2 9 2 | 6 2 5 2 | 3+4 2 |
| | 7 2 8 2 | 7 2 6 2 | 5+6 2 |
| | 6 2 7 2 | 7 2 5 2 | 3+4 2 |
| | 9 2 7 2 | 6 2 6 2 | 3+4 2 |
| 139.1.233 | 47170 | 40 | As2Pd2Sr1 |
| | -GM -M | -P -X | -N |
| | 7 2 7 2 | 6 2 5 2 | 3+4 2 |
| | 9 2 9 2 | 7 2 6 2 | 5+6 2 |
| | 9 2 9 2 | 6 2 6 2 | 5+6 2 |
| | 8 2 8 2 | 7 2 6 2 | 5+6 2 |

Continued on next column

**TABLE LXIV– continued**

| TQC | ICSD | NoE | SF |
|---|---|---|---|
| | 7 2 7 2 | 7 2 5 2 | 5+6 2 |
| | 9 2 9 2 | 6 2 6 2 | 3+4 2 |
| | 7 2 7 2 | 7 2 6 2 | 3+4 2 |
| | 8 2 6 2 | 6 2 5 2 | 5+6 2 |
| | 9 2 7 2 | 7 2 5 2 | 3+4 2 |
| | 6 2 8 2 | 6 2 6 2 | 5+6 2 |
| | 9 2 9 2 | 7 2 5 2 | 5+6 2 |
| | 7 2 9 2 | 6 2 5 2 | 3+4 2 |
| | 6 2 8 2 | 7 2 6 2 | 5+6 2 |
| | 8 2 6 2 | 6 2 5 2 | 3+4 2 |
| | 7 2 8 2 | 7 2 6 2 | 3+4 2 |
| | 6 2 9 2 | 6 2 5 2 | 5+6 2 |
| | 7 2 8 2 | 7 2 6 2 | 5+6 2 |
| | 6 2 7 2 | 7 2 6 2 | 3+4 2 |
| | 9 2 7 2 | 6 2 5 2 | 3+4 2 |
| | 9 2 9 2 | 6 2 6 2 | 5+6 2 |
| 139.1.234 | 36374 | 40 | As2Pd2Sr1 |
| | -GM -M | -P -X | -N |
| | 7 2 7 2 | 6 2 5 2 | 3+4 2 |
| | 9 2 9 2 | 7 2 6 2 | 5+6 2 |
| | 9 2 9 2 | 6 2 6 2 | 5+6 2 |
| | 8 2 8 2 | 7 2 6 2 | 5+6 2 |
| | 7 2 7 2 | 7 2 5 2 | 5+6 2 |
| | 9 2 9 2 | 6 2 6 2 | 3+4 2 |
| | 7 2 7 2 | 7 2 6 2 | 3+4 2 |
| | 8 2 6 2 | 6 2 5 2 | 5+6 2 |
| | 9 2 7 2 | 7 2 5 2 | 3+4 2 |
| | 6 2 8 2 | 6 2 6 2 | 5+6 2 |
| | 9 2 9 2 | 7 2 5 2 | 5+6 2 |
| | 7 2 9 2 | 6 2 5 2 | 3+4 2 |
| | 6 2 8 2 | 7 2 6 2 | 5+6 2 |
| | 8 2 6 2 | 6 2 5 2 | 3+4 2 |
| | 7 2 8 2 | 7 2 6 2 | 3+4 2 |
| | 6 2 9 2 | 6 2 5 2 | 5+6 2 |
| | 7 2 8 2 | 7 2 6 2 | 5+6 2 |
| | 6 2 7 2 | 7 2 6 2 | 3+4 2 |
| | 9 2 7 2 | 6 2 5 2 | 3+4 2 |
| | 9 2 9 2 | 6 2 6 2 | 5+6 2 |
| 139.1.235 | 402951 | 112 | As2O1Yb4 |
| | -GM -M | -P -X | -N |
| | 7 2 9 2 | 6 2 5 2 | 3+4 2 |
| | 9 2 7 2 | 7 2 6 2 | 5+6 2 |
| | 7 2 7 2 | 6 2 6 2 | 3+4 2 |
| | 6 2 6 2 | 7 2 6 2 | 5+6 2 |
| | 7 2 9 2 | 6 2 5 2 | 3+4 2 |
| | 9 2 7 2 | 7 2 6 2 | 5+6 2 |
| | 8 2 8 2 | 6 2 5 2 | 3+4 2 |
| | 9 2 9 2 | 7 2 5 2 | 5+6 2 |
| | 9 2 9 2 | 6 2 5 2 | 3+4 2 |
| | 7 2 9 2 | 7 2 5 2 | 5+6 2 |
| | 9 2 7 2 | 6 2 6 2 | 3+4 2 |
| | 6 2 8 2 | 7 2 6 2 | 5+6 2 |
| | 8 2 6 2 | 6 2 5 2 | 3+4 2 |
| | 8 2 8 2 | 7 2 5 2 | 5+6 2 |
| | 9 2 9 2 | 6 2 5 2 | 3+4 2 |
| | 8 2 8 2 | 7 2 5 2 | 5+6 2 |
| | 7 2 7 2 | 6 2 5 2 | 3+4 2 |
| | 7 2 7 2 | 6 2 6 2 | 5+6 2 |
| | 9 2 9 2 | 7 2 5 2 | 3+4 2 |
| | 9 2 9 2 | 7 2 6 2 | 5+6 2 |
| | 7 2 8 2 | 6 2 6 2 | 5+6 2 |

Continued on next column

| TQC | ICSD | NoE | SF |
|---|---|---|---|
| | 8 2 9 2 | 7 2 6 2 | 5+6 2 |
| | 9 2 7 2 | 7 2 6 2 | 3+4 2 |
| | 7 2 9 2 | 6 2 6 2 | 5+6 2 |
| | 6 2 7 2 | 6 2 5 2 | 5+6 2 |
| | 9 2 6 2 | 7 2 5 2 | 5+6 2 |
| | 8 2 9 2 | 7 2 5 2 | 3+4 2 |
| | 9 2 8 2 | 6 2 6 2 | 3+4 2 |
| | 6 2 7 2 | 6 2 5 2 | 3+4 2 |
| | 7 2 6 2 | 7 2 5 2 | 3+4 2 |
| | 8 2 8 2 | 6 2 5 2 | 5+6 2 |
| | 9 2 9 2 | 7 2 6 2 | 5+6 2 |
| | 8 2 8 2 | 7 2 6 2 | 3+4 2 |
| | 8 2 8 2 | 6 2 5 2 | 5+6 2 |
| | 9 2 9 2 | 7 2 6 2 | 3+4 2 |
| | 6 2 6 2 | 6 2 5 2 | 5+6 2 |
| | 9 2 9 2 | 7 2 5 2 | 3+4 2 |
| | 8 2 8 2 | 6 2 5 2 | 5+6 2 |
| | 9 2 9 2 | 6 2 5 2 | 3+4 2 |
| | 8 2 8 2 | 7 2 5 2 | 5+6 2 |
| | 7 2 6 2 | 6 2 5 2 | 3+4 2 |
| | 6 2 7 2 | 7 2 5 2 | 5+6 2 |
| | 8 2 6 2 | 6 2 6 2 | 3+4 2 |
| | 6 2 6 2 | 7 2 5 2 | 3+4 2 |
| | 9 2 8 2 | 7 2 6 2 | 5+6 2 |
| | 8 2 9 2 | 6 2 5 2 | 3+4 2 |
| | 8 2 8 2 | 7 2 6 2 | 5+6 2 |
| | 7 2 8 2 | 6 2 5 2 | 5+6 2 |
| | 8 2 8 2 | 6 2 5 2 | 3+4 2 |
| | 9 2 7 2 | 7 2 6 2 | 5+6 2 |
| | 7 2 9 2 | 6 2 5 2 | 3+4 2 |
| | 9 2 9 2 | 7 2 5 2 | 5+6 2 |
| | 8 2 9 2 | 7 2 5 2 | 3+4 2 |
| | 9 2 8 2 | 6 2 5 2 | 3+4 2 |
| | 8 2 7 2 | 6 2 5 2 | 5+6 2 |
| | 9 2 9 2 | 7 2 5 2 | 5+6 2 |
| 139.1.236 | 167013 | 58 | As2F2O1Sr2Ti2 |
| | -GM -M | -P -X | -N |
| | 7 2 9 2 | 6 2 5 2 | 3+4 2 |
| | 9 2 7 2 | 7 2 6 2 | 5+6 2 |
| | 7 2 9 2 | 6 2 5 2 | 3+4 2 |
| | 8 2 6 2 | 7 2 6 2 | 5+6 2 |
| | 7 2 7 2 | 7 2 5 2 | 5+6 2 |
| | 7 2 9 2 | 6 2 6 2 | 3+4 2 |
| | 9 2 7 2 | 7 2 5 2 | 5+6 2 |
| | 6 2 8 2 | 6 2 6 2 | 3+4 2 |
| | 9 2 7 2 | 7 2 5 2 | 5+6 2 |
| | 8 2 6 2 | 6 2 5 2 | 3+4 2 |
| | 7 2 9 2 | 7 2 6 2 | 5+6 2 |
| | 7 2 7 2 | 6 2 5 2 | 3+4 2 |
| | 9 2 9 2 | 7 2 6 2 | 5+6 2 |
| | 7 2 7 2 | 6 2 6 2 | 3+4 2 |
| | 6 2 9 2 | 7 2 6 2 | 3+4 2 |
| | 9 2 8 2 | 6 2 6 2 | 5+6 2 |
| | 6 2 8 2 | 7 2 5 2 | 3+4 2 |
| | 9 2 7 2 | 6 2 6 2 | 5+6 2 |
| | 8 2 6 2 | 6 2 5 2 | 3+4 2 |
| | 9 2 8 2 | 7 2 6 2 | 5+6 2 |
| | 8 2 9 2 | 6 2 5 2 | 3+4 2 |
| | 9 2 9 2 | 7 2 6 2 | 3+4 2 |
| | 7 2 7 2 | 7 2 6 2 | 5+6 2 |
| | 7 2 7 2 | 6 2 6 2 | 3+4 2 |

Continued on next column

**TABLE LXIV– continued**

| TQC | ICSD | NoE | SF |
|---|---|---|---|
| | 6 2 6 2 | 6 2 5 2 | 3+4 2 |
| | 6 2 6 2 | 7 2 5 2 | 3+4 2 |
| | 9 2 9 2 | 6 2 5 2 | 5+6 2 |
| | 9 2 9 2 | 7 2 6 2 | 5+6 2 |
| | 8 2 8 2 | 6 2 6 2 | 5+6 2 |
| 139.1.237 | 78756 | 42 | As2Cu2Sr1 |
| | -GM -M | -P -X | -N |
| | 7 2 7 2 | 6 2 5 2 | 3+4 2 |
| | 9 2 9 2 | 7 2 6 2 | 5+6 2 |
| | 9 2 9 2 | 6 2 6 2 | 5+6 2 |
| | 8 2 8 2 | 7 2 6 2 | 5+6 2 |
| | 7 2 7 2 | 7 2 5 2 | 5+6 2 |
| | 9 2 9 2 | 6 2 6 2 | 3+4 2 |
| | 7 2 7 2 | 7 2 6 2 | 3+4 2 |
| | 9 2 6 2 | 6 2 5 2 | 5+6 2 |
| | 8 2 7 2 | 7 2 5 2 | 3+4 2 |
| | 6 2 9 2 | 6 2 6 2 | 5+6 2 |
| | 9 2 8 2 | 7 2 5 2 | 5+6 2 |
| | 7 2 8 2 | 6 2 5 2 | 3+4 2 |
| | 6 2 9 2 | 7 2 6 2 | 5+6 2 |
| | 8 2 6 2 | 6 2 5 2 | 3+4 2 |
| | 7 2 8 2 | 7 2 6 2 | 3+4 2 |
| | 6 2 7 2 | 6 2 5 2 | 5+6 2 |
| | 9 2 9 2 | 7 2 6 2 | 5+6 2 |
| | 7 2 8 2 | 7 2 6 2 | 3+4 2 |
| | 6 2 7 2 | 6 2 6 2 | 3+4 2 |
| | 9 2 9 2 | 6 2 5 2 | 5+6 2 |
| | 8 2 6 2 | 7 2 5 2 | 5+6 2 |
| 139.1.238 | 36372 | 32 | As2Ca1Pd2 |
| | -GM -M | -P -X | -N |
| | 7 2 7 2 | 7 2 5 2 | 5+6 2 |
| | 9 2 9 2 | 6 2 6 2 | 3+4 2 |
| | 7 2 7 2 | 6 2 6 2 | 3+4 2 |
| | 8 2 6 2 | 7 2 5 2 | 5+6 2 |
| | 9 2 7 2 | 7 2 5 2 | 3+4 2 |
| | 6 2 8 2 | 6 2 6 2 | 5+6 2 |
| | 9 2 9 2 | 7 2 5 2 | 5+6 2 |
| | 7 2 9 2 | 6 2 6 2 | 3+4 2 |
| | 6 2 8 2 | 7 2 5 2 | 5+6 2 |
| | 8 2 6 2 | 6 2 5 2 | 3+4 2 |
| | 7 2 8 2 | 7 2 6 2 | 3+4 2 |
| | 6 2 9 2 | 6 2 6 2 | 5+6 2 |
| | 7 2 8 2 | 7 2 5 2 | 5+6 2 |
| | 6 2 7 2 | 7 2 6 2 | 3+4 2 |
| | 9 2 7 2 | 6 2 5 2 | 3+4 2 |
| | 9 2 9 2 | 6 2 6 2 | 5+6 2 |
| 139.1.239 | 416983 | 38 | As2Ba1Rh2 |
| | -GM -M | -P -X | -N |
| | 7 2 7 2 | 6 2 5 2 | 3+4 2 |
| | 9 2 9 2 | 7 2 6 2 | 5+6 2 |
| | 9 2 9 2 | 6 2 6 2 | 5+6 2 |
| | 7 2 8 2 | 7 2 6 2 | 5+6 2 |
| | 8 2 7 2 | 7 2 5 2 | 3+4 2 |
| | 9 2 9 2 | 6 2 6 2 | 5+6 2 |
| | 7 2 7 2 | 7 2 5 2 | 3+4 2 |
| | 9 2 6 2 | 6 2 6 2 | 5+6 2 |
| | 8 2 7 2 | 7 2 5 2 | 3+4 2 |
| | 9 2 9 2 | 7 2 5 2 | 5+6 2 |
| | 6 2 8 2 | 6 2 6 2 | 5+6 2 |
| | 7 2 9 2 | 6 2 5 2 | 3+4 2 |
| | 6 2 8 2 | 7 2 6 2 | 3+4 2 |

Continued on next column

**TABLE LXIV– continued**

| TQC | ICSD | NoE | SF |
|---|---|---|---|
| | 7 2 6 2 | 6 2 5 2 | 5+6 2 |
| | 8 2 8 2 | 7 2 6 2 | 3+4 2 |
| | 6 2 7 2 | 6 2 5 2 | 5+6 2 |
| | 7 2 9 2 | 7 2 6 2 | 5+6 2 |
| | 6 2 9 2 | 6 2 5 2 | 3+4 2 |
| | 9 2 8 2 | 6 2 6 2 | 3+4 2 |
| 139.1.240 | 41794 | 34 | As2Ba1Mn2 |
| | -GM -M | -P -X | -N |
| | 7 2 7 2 | 6 2 5 2 | 3+4 2 |
| | 9 2 9 2 | 7 2 6 2 | 5+6 2 |
| | 9 2 9 2 | 6 2 6 2 | 5+6 2 |
| | 8 2 8 2 | 7 2 6 2 | 5+6 2 |
| | 7 2 9 2 | 7 2 5 2 | 3+4 2 |
| | 9 2 7 2 | 6 2 6 2 | 5+6 2 |
| | 7 2 7 2 | 7 2 5 2 | 3+4 2 |
| | 9 2 9 2 | 6 2 6 2 | 5+6 2 |
| | 9 2 7 2 | 7 2 5 2 | 5+6 2 |
| | 8 2 6 2 | 7 2 5 2 | 3+4 2 |
| | 7 2 9 2 | 6 2 6 2 | 3+4 2 |
| | 6 2 8 2 | 6 2 6 2 | 5+6 2 |
| | 6 2 8 2 | 6 2 5 2 | 3+4 2 |
| | 7 2 7 2 | 6 2 5 2 | 5+6 2 |
| | 8 2 6 2 | 7 2 6 2 | 5+6 2 |
| | 7 2 9 2 | 7 2 6 2 | 3+4 2 |
| | 6 2 8 2 | 7 2 5 2 | 3+4 2 |
| 139.1.241 | 236307 | 42 | As2Ba1Cu2 |
| | -GM -M | -P -X | -N |
| | 7 2 7 2 | 6 2 5 2 | 3+4 2 |
| | 9 2 9 2 | 7 2 6 2 | 5+6 2 |
| | 7 2 9 2 | 6 2 6 2 | 5+6 2 |
| | 9 2 8 2 | 7 2 6 2 | 5+6 2 |
| | 8 2 7 2 | 7 2 5 2 | 5+6 2 |
| | 9 2 9 2 | 6 2 6 2 | 3+4 2 |
| | 7 2 7 2 | 7 2 5 2 | 3+4 2 |
| | 9 2 6 2 | 6 2 6 2 | 5+6 2 |
| | 8 2 7 2 | 7 2 5 2 | 3+4 2 |
| | 6 2 9 2 | 7 2 6 2 | 5+6 2 |
| | 7 2 8 2 | 6 2 5 2 | 5+6 2 |
| | 6 2 8 2 | 6 2 5 2 | 3+4 2 |
| | 9 2 9 2 | 7 2 6 2 | 5+6 2 |
| | 8 2 6 2 | 6 2 5 2 | 3+4 2 |
| | 7 2 8 2 | 7 2 6 2 | 3+4 2 |
| | 6 2 7 2 | 6 2 5 2 | 5+6 2 |
| | 9 2 9 2 | 7 2 6 2 | 5+6 2 |
| | 7 2 8 2 | 7 2 6 2 | 3+4 2 |
| | 6 2 7 2 | 6 2 5 2 | 3+4 2 |
| | 9 2 6 2 | 7 2 5 2 | 5+6 2 |
| | 8 2 9 2 | 6 2 6 2 | 5+6 2 |
| 139.1.242 | 92989 | 26 | As1Ti1Zr1 |
| | -GM -M | -P -X | -N |
| | 7 2 7 2 | 7 2 6 2 | 3+4 2 |
| | 9 2 9 2 | 6 2 5 2 | 5+6 2 |
| | 7 2 7 2 | 7 2 6 2 | 5+6 2 |
| | 9 2 9 2 | 6 2 6 2 | 3+4 2 |
| | 7 2 7 2 | 6 2 5 2 | 3+4 2 |
| | 6 2 6 2 | 7 2 6 2 | 3+4 2 |
| | 7 2 7 2 | 7 2 5 2 | 5+6 2 |
| | 8 2 6 2 | 6 2 5 2 | 5+6 2 |
| | 9 2 9 2 | 6 2 6 2 | 5+6 2 |
| | 8 2 8 2 | 6 2 6 2 | 3+4 2 |
| | 6 2 6 2 | 7 2 5 2 | 3+4 2 |

Continued on next column

**TABLE LXIV– continued**

| TQC | ICSD | NoE | SF |
|---|---|---|---|
| | 6 2 7 2 | 6 2 6 2 5+6 2 | |
| | 7 2 8 2 | 7 2 6 2 3+4 2 | |
| 139.1.243 | 42358 | 50 | As1Ba2 |
| | -GM -M | -P -X | -N |
| | 7 2 7 2 | 6 2 5 2 5+6 2 | |
| | 9 2 9 2 | 7 2 6 2 3+4 2 | |
| | 7 2 7 2 | 6 2 6 2 3+4 2 | |
| | 6 2 6 2 | 7 2 6 2 5+6 2 | |
| | 8 2 8 2 | 7 2 5 2 3+4 2 | |
| | 7 2 9 2 | 6 2 5 2 5+6 2 | |
| | 9 2 7 2 | 6 2 6 2 5+6 2 | |
| | 9 2 9 2 | 7 2 5 2 3+4 2 | |
| | 9 2 9 2 | 6 2 5 2 3+4 2 | |
| | 9 2 9 2 | 7 2 5 2 5+6 2 | |
| | 7 2 8 2 | 6 2 6 2 5+6 2 | |
| | 6 2 7 2 | 7 2 5 2 3+4 2 | |
| | 8 2 6 2 | 7 2 6 2 3+4 2 | |
| | 8 2 8 2 | 6 2 5 2 5+6 2 | |
| | 9 2 9 2 | 6 2 5 2 5+6 2 | |
| | 8 2 8 2 | 7 2 5 2 3+4 2 | |
| | 7 2 7 2 | 7 2 6 2 5+6 2 | |
| | 9 2 9 2 | 6 2 5 2 3+4 2 | |
| | 7 2 7 2 | 7 2 6 2 5+6 2 | |
| | 7 2 7 2 | 6 2 6 2 3+4 2 | |
| | 6 2 6 2 | 7 2 6 2 3+4 2 | |
| | 7 2 9 2 | 6 2 5 2 3+4 2 | |
| | 9 2 7 2 | 6 2 5 2 5+6 2 | |
| | 9 2 9 2 | 7 2 5 2 5+6 2 | |
| | 8 2 8 2 | 6 2 6 2 5+6 2 | |
| 139.1.244 | 156967 | 66 | Al8Cr4Gd1 |
| | -GM -M | -P -X | -N |
| | 7 2 7 2 | 6 2 5 2 3+4 2 | |
| | 9 2 9 2 | 7 2 6 2 5+6 2 | |
| | 9 2 9 2 | 7 2 6 2 5+6 2 | |
| | 8 2 8 2 | 6 2 6 2 5+6 2 | |
| | 7 2 6 2 | 7 2 6 2 5+6 2 | |
| | 7 2 9 2 | 6 2 5 2 3+4 2 | |
| | 9 2 8 2 | 7 2 5 2 5+6 2 | |
| | 8 2 7 2 | 6 2 6 2 3+4 2 | |
| | 6 2 9 2 | 6 2 6 2 3+4 2 | |
| | 6 2 7 2 | 7 2 5 2 5+6 2 | |
| | 6 2 8 2 | 7 2 6 2 3+4 2 | |
| | 8 2 9 2 | 6 2 5 2 5+6 2 | |
| | 7 2 9 2 | 7 2 5 2 3+4 2 | |
| | 7 2 8 2 | 7 2 5 2 5+6 2 | |
| | 6 2 9 2 | 6 2 6 2 3+4 2 | |
| | 9 2 8 2 | 6 2 6 2 5+6 2 | |
| | 8 2 7 2 | 7 2 5 2 3+4 2 | |
| | 7 2 6 2 | 6 2 6 2 3+4 2 | |
| | 7 2 6 2 | 6 2 6 2 5+6 2 | |
| | 6 2 7 2 | 7 2 6 2 5+6 2 | |
| | 9 2 8 2 | 7 2 5 2 5+6 2 | |
| | 6 2 9 2 | 6 2 5 2 3+4 2 | |
| | 8 2 8 2 | 6 2 5 2 5+6 2 | |
| | 6 2 6 2 | 7 2 6 2 3+4 2 | |
| | 9 2 7 2 | 7 2 6 2 3+4 2 | |
| | 7 2 8 2 | 6 2 5 2 5+6 2 | |
| | 8 2 9 2 | 7 2 5 2 5+6 2 | |
| | 8 2 9 2 | 6 2 6 2 5+6 2 | |
| | 9 2 8 2 | 6 2 6 2 5+6 2 | |
| | 7 2 9 2 | 7 2 6 2 5+6 2 | |

Continued on next column

**TABLE LXIV– continued**

| TQC | ICSD | NoE | SF |
|---|---|---|---|
| | 6 2 8 2 | 6 2 6 2 3+4 2 | |
| | 7 2 8 2 | 7 2 6 2 3+4 2 | |
| | 6 2 8 2 | 7 2 6 2 5+6 2 | |
| 139.1.245 | 57545 | 54 | Al8Ca1Mn4 |
| | -GM -M | -P -X | -N |
| | 7 2 6 2 | 6 2 6 2 5+6 2 | |
| | 8 2 8 2 | 7 2 5 2 3+4 2 | |
| | 9 2 9 2 | 7 2 5 2 5+6 2 | |
| | 7 2 7 2 | 6 2 6 2 3+4 2 | |
| | 6 2 9 2 | 6 2 6 2 5+6 2 | |
| | 6 2 7 2 | 7 2 5 2 3+4 2 | |
| | 6 2 8 2 | 6 2 6 2 3+4 2 | |
| | 8 2 9 2 | 7 2 5 2 5+6 2 | |
| | 7 2 9 2 | 7 2 5 2 5+6 2 | |
| | 6 2 8 2 | 7 2 5 2 3+4 2 | |
| | 7 2 8 2 | 6 2 6 2 3+4 2 | |
| | 9 2 9 2 | 6 2 6 2 5+6 2 | |
| | 8 2 8 2 | 6 2 5 2 3+4 2 | |
| | 7 2 9 2 | 7 2 6 2 3+4 2 | |
| | 7 2 8 2 | 7 2 6 2 5+6 2 | |
| | 6 2 6 2 | 6 2 5 2 5+6 2 | |
| | 9 2 7 2 | 6 2 6 2 3+4 2 | |
| | 8 2 7 2 | 7 2 5 2 3+4 2 | |
| | 6 2 6 2 | 7 2 6 2 5+6 2 | |
| | 9 2 7 2 | 6 2 5 2 5+6 2 | |
| | 6 2 6 2 | 6 2 6 2 3+4 2 | |
| | 7 2 9 2 | 7 2 5 2 5+6 2 | |
| | 6 2 8 2 | 7 2 5 2 3+4 2 | |
| | 7 2 9 2 | 6 2 6 2 5+6 2 | |
| | 7 2 8 2 | 7 2 6 2 5+6 2 | |
| | 7 2 9 2 | 6 2 5 2 3+4 2 | |
| | 6 2 8 2 | 7 2 6 2 5+6 2 | |
| 139.1.246 | 57539 | 70 | Al8Ca1Cu4 |
| | -GM -M | -P -X | -N |
| | 7 2 6 2 | 6 2 6 2 5+6 2 | |
| | 8 2 8 2 | 7 2 5 2 3+4 2 | |
| | 9 2 9 2 | 7 2 5 2 3+4 2 | |
| | 6 2 7 2 | 6 2 6 2 5+6 2 | |
| | 7 2 9 2 | 6 2 6 2 5+6 2 | |
| | 6 2 8 2 | 7 2 5 2 3+4 2 | |
| | 6 2 9 2 | 7 2 6 2 3+4 2 | |
| | 7 2 7 2 | 6 2 5 2 5+6 2 | |
| | 7 2 9 2 | 7 2 5 2 5+6 2 | |
| | 7 2 8 2 | 6 2 5 2 3+4 2 | |
| | 6 2 8 2 | 7 2 6 2 3+4 2 | |
| | 7 2 9 2 | 6 2 6 2 5+6 2 | |
| | 6 2 8 2 | 7 2 5 2 3+4 2 | |
| | 6 2 8 2 | 6 2 6 2 5+6 2 | |
| | 7 2 9 2 | 7 2 6 2 3+4 2 | |
| | 6 2 9 2 | 6 2 5 2 5+6 2 | |
| | 9 2 8 2 | 6 2 6 2 3+4 2 | |
| | 8 2 9 2 | 7 2 5 2 5+6 2 | |
| | 6 2 8 2 | 7 2 6 2 3+4 2 | |
| | 8 2 9 2 | 6 2 5 2 5+6 2 | |
| | 7 2 8 2 | 6 2 5 2 3+4 2 | |
| | 7 2 8 2 | 7 2 6 2 5+6 2 | |
| | 6 2 9 2 | 6 2 5 2 3+4 2 | |
| | 7 2 9 2 | 7 2 6 2 5+6 2 | |
| | 6 2 8 2 | 6 2 6 2 5+6 2 | |
| | 7 2 6 2 | 7 2 5 2 3+4 2 | |
| | 6 2 7 2 | 6 2 6 2 5+6 2 | |

Continued on next column

**TABLE LXIV– continued**

| TQC | ICSD | NoE | SF |
|---|---|---|---|
| | 7 2 6 2 | 7 2 5 2 | 3+4 2 |
| | 9 2 9 2 | 7 2 5 2 | 3+4 2 |
| | 6 2 7 2 | 7 2 6 2 | 5+6 2 |
| | 8 2 7 2 | 6 2 6 2 | 5+6 2 |
| | 7 2 6 2 | 6 2 5 2 | 3+4 2 |
| | 9 2 6 2 | 6 2 6 2 | 3+4 2 |
| | 7 2 9 2 | 7 2 6 2 | 5+6 2 |
| | 6 2 8 2 | 7 2 5 2 | 5+6 2 |
| 139.1.247 | 95799 | 68 | Al4Ge2Ni1Tb2 |
| | -GM -M | -P -X | -N |
| | 7 2 9 2 | 6 2 5 2 | 3+4 2 |
| | 9 2 7 2 | 7 2 6 2 | 5+6 2 |
| | 7 2 9 2 | 6 2 5 2 | 3+4 2 |
| | 9 2 7 2 | 7 2 6 2 | 5+6 2 |
| | 7 2 9 2 | 6 2 5 2 | 3+4 2 |
| | 6 2 8 2 | 7 2 6 2 | 5+6 2 |
| | 9 2 7 2 | 6 2 5 2 | 3+4 2 |
| | 8 2 6 2 | 7 2 6 2 | 5+6 2 |
| | 7 2 7 2 | 7 2 5 2 | 5+6 2 |
| | 9 2 9 2 | 6 2 6 2 | 3+4 2 |
| | 7 2 7 2 | 7 2 6 2 | 3+4 2 |
| | 9 2 9 2 | 6 2 6 2 | 5+6 2 |
| | 6 2 6 2 | 6 2 5 2 | 5+6 2 |
| | 7 2 7 2 | 6 2 5 2 | 3+4 2 |
| | 7 2 7 2 | 7 2 5 2 | 3+4 2 |
| | 6 2 6 2 | 7 2 5 2 | 3+4 2 |
| | 6 2 6 2 | 7 2 6 2 | 3+4 2 |
| | 9 2 9 2 | 6 2 5 2 | 3+4 2 |
| | 8 2 8 2 | 7 2 6 2 | 3+4 2 |
| | 7 2 7 2 | 6 2 5 2 | 5+6 2 |
| | 8 2 6 2 | 7 2 6 2 | 3+4 2 |
| | 6 2 9 2 | 6 2 6 2 | 5+6 2 |
| | 7 2 7 2 | 7 2 5 2 | 5+6 2 |
| | 6 2 6 2 | 6 2 5 2 | 3+4 2 |
| | 9 2 8 2 | 7 2 5 2 | 5+6 2 |
| | 8 2 8 2 | 6 2 5 2 | 3+4 2 |
| | 8 2 6 2 | 7 2 6 2 | 5+6 2 |
| | 7 2 7 2 | 7 2 5 2 | 3+4 2 |
| | 6 2 9 2 | 6 2 5 2 | 3+4 2 |
| | 6 2 8 2 | 7 2 6 2 | 5+6 2 |
| | 9 2 6 2 | 6 2 6 2 | 5+6 2 |
| | 9 2 9 2 | 6 2 5 2 | 3+4 2 |
| | 8 2 6 2 | 7 2 6 2 | 5+6 2 |
| | 8 2 8 2 | 7 2 5 2 | 3+4 2 |
| 139.1.248 | 168816 | 70 | Al2Pd5U1 |
| | -GM -M | -P -X | -N |
| | 7 2 7 2 | 6 2 5 2 | 3+4 2 |
| | 9 2 9 2 | 7 2 6 2 | 5+6 2 |
| | 9 2 9 2 | 6 2 6 2 | 5+6 2 |
| | 8 2 8 2 | 7 2 6 2 | 5+6 2 |
| | 7 2 9 2 | 7 2 5 2 | 3+4 2 |
| | 9 2 7 2 | 6 2 6 2 | 5+6 2 |
| | 7 2 7 2 | 7 2 5 2 | 5+6 2 |
| | 6 2 6 2 | 6 2 6 2 | 3+4 2 |
| | 9 2 7 2 | 7 2 5 2 | 5+6 2 |
| | 7 2 9 2 | 7 2 5 2 | 3+4 2 |
| | 7 2 8 2 | 6 2 6 2 | 3+4 2 |
| | 6 2 9 2 | 6 2 5 2 | 5+6 2 |
| | 6 2 6 2 | 6 2 6 2 | 5+6 2 |
| | 7 2 8 2 | 7 2 5 2 | 3+4 2 |
| | 8 2 6 2 | 6 2 6 2 | 5+6 2 |

Continued on next column

**TABLE LXIV– continued**

| TQC | ICSD | NoE | SF |
|---|---|---|---|
| | 8 2 7 2 | 7 2 5 2 | 5+6 2 |
| | 6 2 9 2 | 6 2 5 2 | 3+4 2 |
| | 9 2 6 2 | 6 2 5 2 | 3+4 2 |
| | 8 2 7 2 | 7 2 5 2 | 3+4 2 |
| | 7 2 6 2 | 7 2 6 2 | 5+6 2 |
| | 9 2 6 2 | 6 2 5 2 | 5+6 2 |
| | 6 2 9 2 | 6 2 6 2 | 3+4 2 |
| | 8 2 8 2 | 7 2 5 2 | 5+6 2 |
| | 7 2 8 2 | 7 2 5 2 | 3+4 2 |
| | 6 2 7 2 | 7 2 6 2 | 5+6 2 |
| | 7 2 6 2 | 6 2 5 2 | 3+4 2 |
| | 6 2 7 2 | 6 2 6 2 | 5+6 2 |
| | 8 2 8 2 | 7 2 6 2 | 5+6 2 |
| | 9 2 6 2 | 6 2 6 2 | 3+4 2 |
| | 6 2 7 2 | 7 2 5 2 | 5+6 2 |
| | 9 2 9 2 | 6 2 5 2 | 5+6 2 |
| | 7 2 9 2 | 6 2 5 2 | 5+6 2 |
| | 8 2 8 2 | 7 2 6 2 | 5+6 2 |
| | 8 2 9 2 | 6 2 6 2 | 3+4 2 |
| | 9 2 8 2 | 7 2 5 2 | 5+6 2 |
| 139.1.249 | 161313 | 70 | Al2Pd5U1 |
| | -GM -M | -P -X | -N |
| | 7 2 7 2 | 6 2 5 2 | 3+4 2 |
| | 9 2 9 2 | 7 2 6 2 | 5+6 2 |
| | 9 2 9 2 | 6 2 6 2 | 5+6 2 |
| | 8 2 8 2 | 7 2 6 2 | 5+6 2 |
| | 7 2 9 2 | 7 2 5 2 | 3+4 2 |
| | 9 2 7 2 | 6 2 6 2 | 5+6 2 |
| | 7 2 7 2 | 7 2 5 2 | 5+6 2 |
| | 6 2 6 2 | 6 2 6 2 | 3+4 2 |
| | 9 2 7 2 | 7 2 5 2 | 5+6 2 |
| | 7 2 9 2 | 7 2 5 2 | 3+4 2 |
| | 7 2 8 2 | 6 2 6 2 | 3+4 2 |
| | 6 2 9 2 | 6 2 5 2 | 5+6 2 |
| | 6 2 6 2 | 6 2 6 2 | 5+6 2 |
| | 7 2 8 2 | 7 2 5 2 | 3+4 2 |
| | 8 2 6 2 | 6 2 6 2 | 5+6 2 |
| | 8 2 7 2 | 7 2 5 2 | 5+6 2 |
| | 6 2 9 2 | 6 2 5 2 | 3+4 2 |
| | 9 2 6 2 | 6 2 5 2 | 3+4 2 |
| | 8 2 7 2 | 7 2 5 2 | 3+4 2 |
| | 7 2 6 2 | 7 2 6 2 | 5+6 2 |
| | 9 2 6 2 | 6 2 5 2 | 5+6 2 |
| | 6 2 9 2 | 6 2 6 2 | 3+4 2 |
| | 8 2 8 2 | 7 2 5 2 | 5+6 2 |
| | 7 2 8 2 | 7 2 5 2 | 3+4 2 |
| | 6 2 7 2 | 7 2 6 2 | 5+6 2 |
| | 7 2 6 2 | 6 2 5 2 | 3+4 2 |
| | 6 2 7 2 | 6 2 6 2 | 5+6 2 |
| | 8 2 8 2 | 7 2 6 2 | 5+6 2 |
| | 9 2 6 2 | 6 2 6 2 | 3+4 2 |
| | 6 2 7 2 | 7 2 5 2 | 5+6 2 |
| | 9 2 9 2 | 6 2 5 2 | 5+6 2 |
| | 7 2 9 2 | 6 2 5 2 | 5+6 2 |
| | 8 2 8 2 | 7 2 6 2 | 5+6 2 |
| | 8 2 9 2 | 6 2 6 2 | 3+4 2 |
| | 9 2 8 2 | 7 2 5 2 | 5+6 2 |
| 139.1.250 | 166270 | 72 | Al2Pd5Pu1 |
| | -GM -M | -P -X | -N |
| | 7 2 7 2 | 6 2 5 2 | 3+4 2 |
| | 9 2 9 2 | 7 2 6 2 | 5+6 2 |

Continued on next column

**TABLE LXIV– continued**

| TQC | ICSD | NoE | SF |
|---|---|---|---|
| | 9 2 9 2 | 6 2 6 2 | 5+6 2 |
| | 8 2 8 2 | 7 2 6 2 | 5+6 2 |
| | 7 2 9 2 | 7 2 5 2 | 3+4 2 |
| | 9 2 7 2 | 6 2 6 2 | 5+6 2 |
| | 9 2 7 2 | 7 2 5 2 | 5+6 2 |
| | 7 2 8 2 | 6 2 6 2 | 3+4 2 |
| | 6 2 9 2 | 7 2 5 2 | 5+6 2 |
| | 7 2 6 2 | 6 2 5 2 | 3+4 2 |
| | 7 2 7 2 | 7 2 6 2 | 5+6 2 |
| | 6 2 9 2 | 6 2 6 2 | 3+4 2 |
| | 6 2 6 2 | 7 2 5 2 | 5+6 2 |
| | 7 2 8 2 | 6 2 5 2 | 3+4 2 |
| | 8 2 9 2 | 7 2 6 2 | 5+6 2 |
| | 8 2 6 2 | 6 2 5 2 | 5+6 2 |
| | 6 2 7 2 | 6 2 5 2 | 3+4 2 |
| | 9 2 6 2 | 7 2 5 2 | 3+4 2 |
| | 8 2 7 2 | 6 2 6 2 | 5+6 2 |
| | 7 2 6 2 | 7 2 5 2 | 3+4 2 |
| | 9 2 6 2 | 6 2 5 2 | 5+6 2 |
| | 6 2 9 2 | 6 2 6 2 | 3+4 2 |
| | 8 2 8 2 | 7 2 5 2 | 5+6 2 |
| | 7 2 7 2 | 7 2 5 2 | 3+4 2 |
| | 6 2 8 2 | 7 2 6 2 | 5+6 2 |
| | 7 2 7 2 | 6 2 5 2 | 3+4 2 |
| | 6 2 6 2 | 6 2 6 2 | 5+6 2 |
| | 8 2 8 2 | 7 2 6 2 | 5+6 2 |
| | 9 2 6 2 | 6 2 6 2 | 3+4 2 |
| | 6 2 7 2 | 7 2 5 2 | 5+6 2 |
| | 9 2 9 2 | 6 2 5 2 | 5+6 2 |
| | 7 2 9 2 | 6 2 5 2 | 5+6 2 |
| | 8 2 8 2 | 7 2 6 2 | 5+6 2 |
| | 8 2 9 2 | 7 2 6 2 | 3+4 2 |
| | 9 2 8 2 | 6 2 5 2 | 5+6 2 |
| | 6 2 8 2 | 6 2 6 2 | 5+6 2 |
| 139.1.251 | 58252 | 44 | Ag2Er1 |
| | -GM -M | -P -X | -N |
| | 7 2 7 2 | 6 2 5 2 | 3+4 2 |
| | 9 2 9 2 | 7 2 6 2 | 5+6 2 |
| | 9 2 9 2 | 6 2 6 2 | 5+6 2 |
| | 8 2 8 2 | 7 2 6 2 | 5+6 2 |
| | 7 2 7 2 | 6 2 5 2 | 3+4 2 |
| | 8 2 6 2 | 7 2 6 2 | 3+4 2 |
| | 6 2 9 2 | 6 2 5 2 | 5+6 2 |
| | 7 2 8 2 | 7 2 6 2 | 5+6 2 |
| | 9 2 9 2 | 7 2 5 2 | 5+6 2 |
| | 6 2 8 2 | 7 2 6 2 | 3+4 2 |
| | 7 2 7 2 | 6 2 5 2 | 3+4 2 |
| | 9 2 7 2 | 6 2 6 2 | 5+6 2 |
| | 6 2 8 2 | 7 2 5 2 | 3+4 2 |
| | 8 2 6 2 | 6 2 6 2 | 5+6 2 |
| | 8 2 6 2 | 7 2 5 2 | 3+4 2 |
| | 9 2 9 2 | 6 2 6 2 | 5+6 2 |
| | 8 2 9 2 | 7 2 6 2 | 5+6 2 |
| | 8 2 8 2 | 6 2 6 2 | 5+6 2 |
| | 7 2 8 2 | 7 2 5 2 | 5+6 2 |
| | 9 2 9 2 | 6 2 6 2 | 5+6 2 |
| | 8 2 9 2 | 6 2 6 2 | 5+6 2 |
| | 9 2 8 2 | 7 2 6 2 | 5+6 2 |

The End

TABLE LXV: The computed irreducible representations for the valence bands for the compounds of SEBR and NLC in space group 140. See the list of tables.

| TQC | ICSD | NoE | SF |
|---|---|---|---|
| 140.1.1 | 86940 | 72 | O3Ru1Sr1 |
| | -GM -M | -P -X | -N |
| | 7 2 9 2 6+7 4 | 6 2 | 2+2 4 |
| | 8 2 6 2 6+7 4 | 5 2 | 2+2 4 |
| | 7 2 7 2 6+7 4 | 5 2 | 2+2 4 |
| | 6 2 7 2 6+7 4 | 5 2 | 2+2 4 |
| | 7 2 8 2 6+7 4 | 6 2 | 2+2 4 |
| | 9 2 6 2 6+7 4 | 6 2 | 2+2 4 |
| | 9 2 9 2 6+7 4 | 5 2 | 2+2 4 |
| | 8 2 9 2 6+7 4 | 6 2 | 2+2 4 |
| | 6 2 7 2 6+7 4 | 5 2 | 2+2 4 |
| | 9 2 8 2 6+7 4 | 6 2 | 2+2 4 |
| | 7 2 6 2 6+7 4 | 5 2 | 2+2 4 |
| | 6 2 8 2 6+7 4 | 6 2 | 2+2 4 |
| | 9 2 9 2 6+7 4 | 6 2 | 2+2 4 |
| | 8 2 7 2 6+7 4 | 5 2 | 2+2 4 |
| | 7 2 7 2 6+7 4 | 5 2 | 2+2 4 |
| | 6 2 6 2 6+7 4 | 5 2 | 2+2 4 |
| | 7 2 7 2 6+7 4 | 5 2 | 2+2 4 |
| | 6 2 6 2 6+7 4 | 5 2 | 2+2 4 |
| | 7 2 7 2 | 5 2 | |
| | 6 2 6 2 | 5 2 | |
| | 9 2 7 2 | 5 2 | |
| | 9 2 6 2 | 6 2 | |
| | 8 2 9 2 | 5 2 | |
| | 9 2 8 2 | 6 2 | |
| | 8 2 9 2 | 6 2 | |
| | 8 2 9 2 | 6 2 | |
| | 7 2 8 2 | 6 2 | |
| | 6 2 8 2 | 6 2 | |
| | 9 2 9 2 | 6 2 | |
| | 9 2 9 2 | 6 2 | |
| | 8 2 8 2 | 6 2 | |
| | 6 2 6 2 | 5 2 | |
| | 7 2 7 2 | 5 2 | |
| | 7 2 6 2 | 5 2 | |
| | 6 2 7 2 | 5 2 | |
| | 6 2 6 2 | 5 2 | |
| 140.1.2 | 174206 | 68 | Mo1O3Sr1 |
| | -GM -M | -P -X | -N |
| | 7 2 9 2 6+7 4 | 6 2 | 2+2 4 |
| | 8 2 6 2 6+7 4 | 5 2 | 2+2 4 |
| | 7 2 7 2 6+7 4 | 5 2 | 2+2 4 |
| | 6 2 7 2 6+7 4 | 5 2 | 2+2 4 |
| | 7 2 6 2 6+7 4 | 5 2 | 2+2 4 |
| | 9 2 8 2 6+7 4 | 6 2 | 2+2 4 |
| | 8 2 9 2 6+7 4 | 6 2 | 2+2 4 |
| | 9 2 9 2 6+7 4 | 6 2 | 2+2 4 |
| | 6 2 7 2 6+7 4 | 5 2 | 2+2 4 |
| | 9 2 8 2 6+7 4 | 6 2 | 2+2 4 |
| | 7 2 6 2 6+7 4 | 5 2 | 2+2 4 |
| | 6 2 8 2 6+7 4 | 6 2 | 2+2 4 |
| | 9 2 9 2 6+7 4 | 6 2 | 2+2 4 |
| | 8 2 7 2 6+7 4 | 5 2 | 2+2 4 |
| | 6 2 6 2 6+7 4 | 5 2 | 2+2 4 |
| | 7 2 7 2 6+7 4 | 5 2 | 2+2 4 |
| | 7 2 7 2 6+7 4 | 5 2 | 2+2 4 |
| | 6 2 6 2 | 5 2 | |

Continued on next column

**TABLE LXV– continued**

| TQC | ICSD | | NoE | SF | |
|---|---|---|---|---|---|
| | 6 2 | 7 2 | | 5 2 | |
| | 7 2 | 6 2 | | 5 2 | |
| | 9 2 | 7 2 | | 5 2 | |
| | 8 2 | 6 2 | | 5 2 | |
| | 9 2 | 9 2 | | 6 2 | |
| | 8 2 | 8 2 | | 6 2 | |
| | 9 2 | 9 2 | | 6 2 | |
| | 8 2 | 8 2 | | 6 2 | |
| | 9 2 | 9 2 | | 6 2 | |
| | 8 2 | 8 2 | | 6 2 | |
| | 9 2 | 9 2 | | 6 2 | |
| | 7 2 | 9 2 | | 6 2 | |
| | 7 2 | 8 2 | | 6 2 | |
| | 6 2 | 7 2 | | 5 2 | |
| | 6 2 | 6 2 | | 5 2 | |
| | 7 2 | 7 2 | | 5 2 | |
| 140.1.3 | 245120 | | 28 | Mg2Pt1 | |
| | -GM | -M | -P | -X | -N |
| | 7 2 | 7 2 | 6+7 4 | 5 2 | 2+2 4 |
| | 9 2 | 9 2 | 6+7 4 | 6 2 | 2+2 4 |
| | 7 2 | 6 2 | 6+7 4 | 6 2 | 2+2 4 |
| | 8 2 | 8 2 | 6+7 4 | 5 2 | 2+2 4 |
| | 6 2 | 9 2 | 6+7 4 | 6 2 | 2+2 4 |
| | 8 2 | 7 2 | 6+7 4 | 5 2 | 2+2 4 |
| | 6 2 | 9 2 | 6+7 4 | 6 2 | 2+2 4 |
| | 9 2 | 8 2 | | 5 2 | |
| | 8 2 | 6 2 | | 6 2 | |
| | 7 2 | 8 2 | | 6 2 | |
| | 6 2 | 8 2 | | 5 2 | |
| | 9 2 | 7 2 | | 5 2 | |
| | 7 2 | 7 2 | | 5 2 | |
| | 8 2 | 6 2 | | 5 2 | |
| 140.1.4 | 88093 | | 32 | Cr1Sb2 | |
| | -GM | -M | -P | -X | -N |
| | 7 2 | 6 2 | 6+7 4 | 6 2 | 2+2 4 |
| | 6 2 | 8 2 | 6+7 4 | 5 2 | 2+2 4 |
| | 9 2 | 9 2 | 6+7 4 | 6 2 | 2+2 4 |
| | 8 2 | 7 2 | 6+7 4 | 5 2 | 2+2 4 |
| | 6 2 | 9 2 | 6+7 4 | 5 2 | 2+2 4 |
| | 8 2 | 7 2 | 6+7 4 | 6 2 | 2+2 4 |
| | 9 2 | 8 2 | 6+7 4 | 5 2 | 2+2 4 |
| | 7 2 | 9 2 | 6+7 4 | 6 2 | 2+2 4 |
| | 8 2 | 6 2 | | 5 2 | |
| | 6 2 | 8 2 | | 6 2 | |
| | 7 2 | 7 2 | | 6 2 | |
| | 7 2 | 6 2 | | 6 2 | |
| | 6 2 | 7 2 | | 5 2 | |
| | 7 2 | 8 2 | | 6 2 | |
| | 9 2 | 6 2 | | 6 2 | |
| | 8 2 | 7 2 | | 5 2 | |
| 140.1.5 | 86811 | | 32 | Cr1Sb2 | |
| | -GM | -M | -P | -X | -N |
| | 7 2 | 6 2 | 6+7 4 | 6 2 | 2+2 4 |
| | 6 2 | 8 2 | 6+7 4 | 5 2 | 2+2 4 |
| | 9 2 | 9 2 | 6+7 4 | 6 2 | 2+2 4 |
| | 8 2 | 7 2 | 6+7 4 | 5 2 | 2+2 4 |
| | 6 2 | 9 2 | 6+7 4 | 5 2 | 2+2 4 |
| | 8 2 | 7 2 | 6+7 4 | 6 2 | 2+2 4 |
| | 9 2 | 8 2 | 6+7 4 | 5 2 | 2+2 4 |
| | 7 2 | 9 2 | 6+7 4 | 6 2 | 2+2 4 |
| | 8 2 | 6 2 | | 5 2 | |

Continued on next column

| TQC | ICSD | | NoE | SF | |
|---|---|---|---|---|---|
| | 6 2 | 8 2 | | 6 2 | |
| | 7 2 | 7 2 | | 6 2 | |
| | 7 2 | 6 2 | | 6 2 | |
| | 6 2 | 7 2 | | 5 2 | |
| | 7 2 | 8 2 | | 6 2 | |
| | 9 2 | 6 2 | | 6 2 | |
| | 8 2 | 7 2 | | 5 2 | |
| 140.1.6 | 281417 | | 104 | Cl1Ga4La3 | |
| | -GM | -M | -P | -X | -N |
| | 7 2 | 7 2 | 6+7 4 | 6 2 | 2+2 4 |
| | 6 2 | 6 2 | 6+7 4 | 6 2 | 2+2 4 |
| | 8 2 | 8 2 | 6+7 4 | 5 2 | 2+2 4 |
| | 9 2 | 9 2 | 6+7 4 | 5 2 | 2+2 4 |
| | 7 2 | 7 2 | 6+7 4 | 5 2 | 2+2 4 |
| | 9 2 | 9 2 | 6+7 4 | 6 2 | 2+2 4 |
| | 7 2 | 7 2 | 6+7 4 | 5 2 | 2+2 4 |
| | 6 2 | 7 2 | 6+7 4 | 5 2 | 2+2 4 |
| | 7 2 | 6 2 | 6+7 4 | 6 2 | 2+2 4 |
| | 8 2 | 8 2 | 6+7 4 | 6 2 | 2+2 4 |
| | 9 2 | 9 2 | 6+7 4 | 5 2 | 2+2 4 |
| | 9 2 | 9 2 | 6+7 4 | 6 2 | 2+2 4 |
| | 7 2 | 7 2 | 6+7 4 | 5 2 | 2+2 4 |
| | 7 2 | 7 2 | 6+7 4 | 5 2 | 2+2 4 |
| | 9 2 | 9 2 | 6+7 4 | 6 2 | 2+2 4 |
| | 6 2 | 6 2 | 6+7 4 | 6 2 | 2+2 4 |
| | 8 2 | 8 2 | 6+7 4 | 6 2 | 2+2 4 |
| | 6 2 | 9 2 | 6+7 4 | 5 2 | 2+2 4 |
| | 7 2 | 7 2 | 6+7 4 | 6 2 | 2+2 4 |
| | 9 2 | 6 2 | 6+7 4 | 5 2 | 2+2 4 |
| | 7 2 | 9 2 | 6+7 4 | 6 2 | 2+2 4 |
| | 8 2 | 8 2 | 6+7 4 | 5 2 | 2+2 4 |
| | 9 2 | 8 2 | 6+7 4 | 5 2 | 2+2 4 |
| | 8 2 | 7 2 | 6+7 4 | 6 2 | 2+2 4 |
| | 6 2 | 6 2 | 6+7 4 | 5 2 | 2+2 4 |
| | 7 2 | 7 2 | 6+7 4 | 5 2 | 2+2 4 |
| | 7 2 | 9 2 | | 6 2 | |
| | 8 2 | 6 2 | | 5 2 | |
| | 9 2 | 9 2 | | 6 2 | |
| | 9 2 | 9 2 | | 5 2 | |
| | 8 2 | 8 2 | | 6 2 | |
| | 9 2 | 9 2 | | 6 2 | |
| | 8 2 | 8 2 | | 6 2 | |
| | 7 2 | 9 2 | | 6 2 | |
| | 6 2 | 8 2 | | 6 2 | |
| | 9 2 | 9 2 | | 6 2 | |
| | 6 2 | 8 2 | | 5 2 | |
| | 8 2 | 7 2 | | 6 2 | |
| | 9 2 | 6 2 | | 5 2 | |
| | 9 2 | 7 2 | | 6 2 | |
| | 9 2 | 9 2 | | 6 2 | |
| | 7 2 | 7 2 | | 5 2 | |
| | 6 2 | 8 2 | | 5 2 | |
| | 8 2 | 8 2 | | 5 2 | |
| | 9 2 | 6 2 | | 6 2 | |
| | 6 2 | 7 2 | | 5 2 | |
| | 7 2 | 6 2 | | 5 2 | |
| | 7 2 | 8 2 | | 5 2 | |
| | 6 2 | 9 2 | | 5 2 | |
| | 8 2 | 6 2 | | 6 2 | |
| | 7 2 | 7 2 | | 6 2 | |

Continued on next column

**TABLE LXV– continued**

| TQC | ICSD | NoE | SF |
|---|---|---|---|
| | 6 2 9 2 | 6 2 | |
| 140.1.7 | 98408 | 236 | Ce1Ni9Si4 |
| | -GM -M | -P -X | -N |
| | 7 2 7 2 | 6+7 4 5 2 | 2+2 4 |
| | 9 2 9 2 | 6+7 4 6 2 | 2+2 4 |
| | 7 2 9 2 | 6+7 4 5 2 | 2+2 4 |
| | 9 2 7 2 | 6+7 4 6 2 | 2+2 4 |
| | 6 2 8 2 | 6+7 4 6 2 | 2+2 4 |
| | 8 2 6 2 | 6+7 4 5 2 | 2+2 4 |
| | 7 2 7 2 | 6+7 4 5 2 | 2+2 4 |
| | 9 2 9 2 | 6+7 4 6 2 | 2+2 4 |
| | 7 2 7 2 | 6+7 4 6 2 | 2+2 4 |
| | 6 2 6 2 | 6+7 4 6 2 | 2+2 4 |
| | 8 2 9 2 | 6+7 4 5 2 | 2+2 4 |
| | 9 2 8 2 | 6+7 4 5 2 | 2+2 4 |
| | 8 2 9 2 | 6+7 4 5 2 | 2+2 4 |
| | 7 2 8 2 | 6+7 4 6 2 | 2+2 4 |
| | 6 2 7 2 | 6+7 4 5 2 | 2+2 4 |
| | 9 2 6 2 | 6+7 4 6 2 | 2+2 4 |
| | 6 2 6 2 | 6+7 4 6 2 | 2+2 4 |
| | 7 2 7 2 | 6+7 4 6 2 | 2+2 4 |
| | 9 2 9 2 | 6+7 4 5 2 | 2+2 4 |
| | 8 2 8 2 | 6+7 4 5 2 | 2+2 4 |
| | 7 2 8 2 | 6+7 4 5 2 | 2+2 4 |
| | 8 2 7 2 | 6+7 4 6 2 | 2+2 4 |
| | 7 2 9 2 | 6+7 4 6 2 | 2+2 4 |
| | 7 2 7 2 | 6+7 4 6 2 | 2+2 4 |
| | 6 2 8 2 | 6+7 4 5 2 | 2+2 4 |
| | 7 2 6 2 | 6+7 4 6 2 | 2+2 4 |
| | 6 2 7 2 | 6+7 4 6 2 | 2+2 4 |
| | 9 2 6 2 | 6+7 4 6 2 | 2+2 4 |
| | 8 2 7 2 | 6+7 4 5 2 | 2+2 4 |
| | 6 2 6 2 | 6+7 4 6 2 | 2+2 4 |
| | 7 2 7 2 | 6+7 4 6 2 | 2+2 4 |
| | 6 2 6 2 | 6+7 4 5 2 | 2+2 4 |
| | 8 2 9 2 | 6+7 4 5 2 | 2+2 4 |
| | 9 2 8 2 | 6+7 4 6 2 | 2+2 4 |
| | 6 2 6 2 | 6+7 4 5 2 | 2+2 4 |
| | 9 2 8 2 | 6+7 4 5 2 | 2+2 4 |
| | 8 2 9 2 | 6+7 4 6 2 | 2+2 4 |
| | 9 2 9 2 | 6+7 4 5 2 | 2+2 4 |
| | 9 2 8 2 | 6+7 4 6 2 | 2+2 4 |
| | 7 2 7 2 | 6+7 4 6 2 | 2+2 4 |
| | 9 2 6 2 | 6+7 4 5 2 | 2+2 4 |
| | 8 2 6 2 | 6+7 4 5 2 | 2+2 4 |
| | 6 2 9 2 | 6+7 4 5 2 | 2+2 4 |
| | 6 2 8 2 | 6+7 4 6 2 | 2+2 4 |
| | 7 2 7 2 | 6+7 4 6 2 | 2+2 4 |
| | 6 2 6 2 | 6+7 4 5 2 | 2+2 4 |
| | 7 2 9 2 | 6+7 4 6 2 | 2+2 4 |
| | 6 2 7 2 | 6+7 4 6 2 | 2+2 4 |
| | 6 2 6 2 | 6+7 4 6 2 | 2+2 4 |
| | 7 2 9 2 | 6+7 4 5 2 | 2+2 4 |
| | 8 2 8 2 | 6+7 4 6 2 | 2+2 4 |
| | 7 2 9 2 | 6+7 4 5 2 | 2+2 4 |
| | 9 2 7 2 | 6+7 4 5 2 | 2+2 4 |
| | 8 2 6 2 | 6+7 4 6 2 | 2+2 4 |
| | 7 2 8 2 | 6+7 4 5 2 | 2+2 4 |
| | 9 2 7 2 | 6+7 4 6 2 | 2+2 4 |
| | 6 2 8 2 | 6+7 4 6 2 | 2+2 4 |
| | 7 2 7 2 | 6+7 4 5 2 | 2+2 4 |

Continued on next column

**TABLE LXV– continued**

| TQC | ICSD | NoE | SF |
|---|---|---|---|
| | 6 2 9 2 | 6+7 4 5 2 | 2+2 4 |
| | 9 2 9 2 | 6 2 | |
| | 8 2 7 2 | 5 2 | |
| | 8 2 8 2 | 5 2 | |
| | 9 2 6 2 | 5 2 | |
| | 8 2 9 2 | 6 2 | |
| | 7 2 8 2 | 5 2 | |
| | 7 2 6 2 | 6 2 | |
| | 7 2 9 2 | 6 2 | |
| | 9 2 7 2 | 6 2 | |
| | 9 2 6 2 | 6 2 | |
| | 6 2 7 2 | 5 2 | |
| | 8 2 9 2 | 6 2 | |
| | 9 2 8 2 | 5 2 | |
| | 8 2 6 2 | 6 2 | |
| | 6 2 9 2 | 5 2 | |
| | 7 2 6 2 | 5 2 | |
| | 6 2 7 2 | 6 2 | |
| | 7 2 7 2 | 6 2 | |
| | 8 2 8 2 | 5 2 | |
| | 6 2 7 2 | 5 2 | |
| | 9 2 9 2 | 6 2 | |
| | 8 2 6 2 | 6 2 | |
| | 6 2 7 2 | 6 2 | |
| | 7 2 6 2 | 5 2 | |
| | 8 2 8 2 | 6 2 | |
| | 7 2 7 2 | 5 2 | |
| | 9 2 9 2 | 6 2 | |
| | 8 2 8 2 | 6 2 | |
| | 6 2 7 2 | 5 2 | |
| | 7 2 6 2 | 5 2 | |
| | 9 2 8 2 | 6 2 | |
| | 6 2 6 2 | 6 2 | |
| | 7 2 9 2 | 5 2 | |
| | 9 2 8 2 | 6 2 | |
| | 6 2 9 2 | 5 2 | |
| | 8 2 8 2 | 5 2 | |
| | 6 2 9 2 | 6 2 | |
| | 7 2 9 2 | 6 2 | |
| | 9 2 7 2 | 5 2 | |
| | 9 2 8 2 | 6 2 | |
| | 7 2 6 2 | 6 2 | |
| | 6 2 7 2 | 5 2 | |
| | 9 2 6 2 | 5 2 | |
| | 8 2 7 2 | 5 2 | |
| | 7 2 6 2 | 5 2 | |
| | 6 2 9 2 | 6 2 | |
| | 6 2 6 2 | 6 2 | |
| | 7 2 9 2 | 5 2 | |
| | 6 2 7 2 | 6 2 | |
| | 8 2 8 2 | 5 2 | |
| | 9 2 6 2 | 5 2 | |
| | 8 2 6 2 | 6 2 | |
| | 7 2 7 2 | 5 2 | |
| | 6 2 6 2 | 6 2 | |
| | 7 2 8 2 | 6 2 | |
| | 8 2 7 2 | 5 2 | |
| | 6 2 6 2 | 6 2 | |
| | 9 2 9 2 | 6 2 | |
| | 6 2 8 2 | 5 2 | |
| 140.1.8 | 106369 | 172 | Cd3Sr5 |

Continued on next column

**TABLE LXV– continued**

| TQC | ICSD | | NoE | | SF |
|---|---|---|---|---|---|
| | -GM | -M | -P | -X | -N |
| | 7 2 | 9 2 | 6+7 4 | 5 2 | 2+2 4 |
| | 6 2 | 7 2 | 6+7 4 | 6 2 | 2+2 4 |
| | 6 2 | 8 2 | 6+7 4 | 5 2 | 2+2 4 |
| | 7 2 | 8 2 | 6+7 4 | 6 2 | 2+2 4 |
| | 8 2 | 9 2 | 6+7 4 | 5 2 | 2+2 4 |
| | 9 2 | 6 2 | 6+7 4 | 6 2 | 2+2 4 |
| | 8 2 | 6 2 | 6+7 4 | 5 2 | 2+2 4 |
| | 9 2 | 7 2 | 6+7 4 | 6 2 | 2+2 4 |
| | 7 2 | 7 2 | 6+7 4 | 5 2 | 2+2 4 |
| | 7 2 | 7 2 | 6+7 4 | 5 2 | 2+2 4 |
| | 6 2 | 8 2 | 6+7 4 | 6 2 | 2+2 4 |
| | 7 2 | 9 2 | 6+7 4 | 5 2 | 2+2 4 |
| | 6 2 | 9 2 | 6+7 4 | 6 2 | 2+2 4 |
| | 7 2 | 8 2 | 6+7 4 | 6 2 | 2+2 4 |
| | 9 2 | 6 2 | 6+7 4 | 5 2 | 2+2 4 |
| | 9 2 | 6 2 | 6+7 4 | 5 2 | 2+2 4 |
| | 8 2 | 7 2 | 6+7 4 | 6 2 | 2+2 4 |
| | 8 2 | 9 2 | 6+7 4 | 5 2 | 2+2 4 |
| | 9 2 | 7 2 | 6+7 4 | 6 2 | 2+2 4 |
| | 9 2 | 9 2 | 6+7 4 | 6 2 | 2+2 4 |
| | 9 2 | 8 2 | 6+7 4 | 6 2 | 2+2 4 |
| | 6 2 | 9 2 | 6+7 4 | 6 2 | 2+2 4 |
| | 7 2 | 9 2 | 6+7 4 | 5 2 | 2+2 4 |
| | 8 2 | 9 2 | 6+7 4 | 6 2 | 2+2 4 |
| | 7 2 | 8 2 | 6+7 4 | 5 2 | 2+2 4 |
| | 6 2 | 6 2 | 6+7 4 | 6 2 | 2+2 4 |
| | 8 2 | 7 2 | 6+7 4 | 6 2 | 2+2 4 |
| | 9 2 | 7 2 | 6+7 4 | 5 2 | 2+2 4 |
| | 7 2 | 8 2 | 6+7 4 | 5 2 | 2+2 4 |
| | 8 2 | 6 2 | 6+7 4 | 6 2 | 2+2 4 |
| | 9 2 | 8 2 | 6+7 4 | 6 2 | 2+2 4 |
| | 7 2 | 9 2 | 6+7 4 | 5 2 | 2+2 4 |
| | 6 2 | 7 2 | 6+7 4 | 5 2 | 2+2 4 |
| | 9 2 | 6 2 | 6+7 4 | 6 2 | 2+2 4 |
| | 6 2 | 7 2 | 6+7 4 | 5 2 | 2+2 4 |
| | 8 2 | 9 2 | 6+7 4 | 5 2 | 2+2 4 |
| | 9 2 | 6 2 | 6+7 4 | 6 2 | 2+2 4 |
| | 8 2 | 8 2 | 6+7 4 | 6 2 | 2+2 4 |
| | 9 2 | 9 2 | 6+7 4 | 6 2 | 2+2 4 |
| | 8 2 | 8 2 | 6+7 4 | 6 2 | 2+2 4 |
| | 6 2 | 6 2 | 6+7 4 | 6 2 | 2+2 4 |
| | 7 2 | 7 2 | 6+7 4 | 6 2 | 2+2 4 |
| | 8 2 | 8 2 | 6+7 4 | 5 2 | 2+2 4 |
| | 9 2 | 9 2 | | 5 2 | |
| | 8 2 | 8 2 | | 5 2 | |
| | 9 2 | 9 2 | | 5 2 | |
| | 7 2 | 7 2 | | 6 2 | |
| | 6 2 | 6 2 | | 6 2 | |
| | 6 2 | 6 2 | | 5 2 | |
| | 8 2 | 8 2 | | 6 2 | |
| | 9 2 | 7 2 | | 5 2 | |
| | 7 2 | 9 2 | | 6 2 | |
| | 9 2 | 9 2 | | 5 2 | |
| | 8 2 | 8 2 | | 5 2 | |
| | 6 2 | 6 2 | | 6 2 | |
| | 7 2 | 7 2 | | 6 2 | |
| | 7 2 | 7 2 | | 6 2 | |
| | 6 2 | 6 2 | | 6 2 | |
| | 9 2 | 9 2 | | 5 2 | |
| | 8 2 | 8 2 | | 5 2 | |

Continued on next column

**TABLE LXV– continued**

| TQC | ICSD | | NoE | | SF |
|---|---|---|---|---|---|
| | 6 2 | 6 2 | | 6 2 | |
| | 7 2 | 7 2 | | 6 2 | |
| | 8 2 | 8 2 | | 5 2 | |
| | 9 2 | 9 2 | | 5 2 | |
| | 6 2 | 6 2 | | 5 2 | |
| | 8 2 | 8 2 | | 6 2 | |
| | 8 2 | 6 2 | | 5 2 | |
| | 6 2 | 8 2 | | 6 2 | |
| | 7 2 | 7 2 | | 5 2 | |
| | 9 2 | 9 2 | | 6 2 | |
| | 7 2 | 7 2 | | 6 2 | |
| | 6 2 | 6 2 | | 6 2 | |
| | 7 2 | 9 2 | | 5 2 | |
| | 9 2 | 7 2 | | 6 2 | |
| | 8 2 | 8 2 | | 5 2 | |
| | 9 2 | 9 2 | | 5 2 | |
| | 9 2 | 7 2 | | 5 2 | |
| | 7 2 | 9 2 | | 6 2 | |
| | 7 2 | 9 2 | | 5 2 | |
| | 6 2 | 7 2 | | 5 2 | |
| | 6 2 | 6 2 | | 5 2 | |
| | 8 2 | 8 2 | | 6 2 | |
| | 9 2 | 7 2 | | 6 2 | |
| | 8 2 | 6 2 | | 5 2 | |
| | 7 2 | 8 2 | | 6 2 | |
| | 7 2 | 7 2 | | 6 2 | |
| 140.1.9 | 106344 | | 92 | | Ca5Hg3 |
| | -GM | -M | -P | -X | -N |
| | 6 2 | 6 2 | 6+7 4 | 6 2 | 2+2 4 |
| | 7 2 | 7 2 | 6+7 4 | 6 2 | 2+2 4 |
| | 8 2 | 8 2 | 6+7 4 | 5 2 | 2+2 4 |
| | 9 2 | 9 2 | 6+7 4 | 5 2 | 2+2 4 |
| | 7 2 | 7 2 | 6+7 4 | 5 2 | 2+2 4 |
| | 9 2 | 9 2 | 6+7 4 | 5 2 | 2+2 4 |
| | 8 2 | 8 2 | 6+7 4 | 6 2 | 2+2 4 |
| | 6 2 | 6 2 | 6+7 4 | 6 2 | 2+2 4 |
| | 6 2 | 6 2 | 6+7 4 | 5 2 | 2+2 4 |
| | 8 2 | 8 2 | 6+7 4 | 6 2 | 2+2 4 |
| | 7 2 | 7 2 | 6+7 4 | 5 2 | 2+2 4 |
| | 9 2 | 9 2 | 6+7 4 | 6 2 | 2+2 4 |
| | 7 2 | 7 2 | 6+7 4 | 5 2 | 2+2 4 |
| | 9 2 | 9 2 | 6+7 4 | 5 2 | 2+2 4 |
| | 6 2 | 6 2 | 6+7 4 | 6 2 | 2+2 4 |
| | 8 2 | 8 2 | 6+7 4 | 6 2 | 2+2 4 |
| | 7 2 | 7 2 | 6+7 4 | 6 2 | 2+2 4 |
| | 6 2 | 6 2 | 6+7 4 | 6 2 | 2+2 4 |
| | 9 2 | 9 2 | 6+7 4 | 5 2 | 2+2 4 |
| | 8 2 | 8 2 | 6+7 4 | 5 2 | 2+2 4 |
| | 7 2 | 8 2 | 6+7 4 | 5 2 | 2+2 4 |
| | 8 2 | 7 2 | 6+7 4 | 5 2 | 2+2 4 |
| | 9 2 | 9 2 | 6+7 4 | 6 2 | 2+2 4 |
| | 6 2 | 6 2 | | 6 2 | |
| | 7 2 | 6 2 | | 6 2 | |
| | 6 2 | 8 2 | | 5 2 | |
| | 8 2 | 6 2 | | 6 2 | |
| | 8 2 | 7 2 | | 5 2 | |
| | 7 2 | 8 2 | | 5 2 | |
| | 9 2 | 9 2 | | 6 2 | |
| | 6 2 | 7 2 | | 6 2 | |
| | 6 2 | 6 2 | | 6 2 | |
| | 7 2 | 9 2 | | 5 2 | |

Continued on next column

**TABLE LXV– continued**

| TQC | ICSD NoE SF |
|---|---|
| | 9 2 7 2 6 2 |
| | 8 2 8 2 5 2 |
| | 9 2 9 2 5 2 |
| | 9 2 9 2 5 2 |
| | 7 2 6 2 6 2 |
| | 7 2 9 2 5 2 |
| | 6 2 7 2 5 2 |
| | 6 2 8 2 6 2 |
| | 8 2 7 2 5 2 |
| | 9 2 7 2 6 2 |
| | 8 2 6 2 6 2 |
| | 7 2 8 2 5 2 |
| | 7 2 7 2 6 2 |
| 140.1.10 | 39599 168 Br5Cd1Cs3 |
| | -GM -M -P -X -N |
| | 7 2 7 2 6+7 4 5 2 2+2 4 |
| | 7 2 7 2 6+7 4 5 2 2+2 4 |
| | 8 2 9 2 6+7 4 6 2 2+2 4 |
| | 9 2 8 2 6+7 4 6 2 2+2 4 |
| | 7 2 9 2 6+7 4 6 2 2+2 4 |
| | 9 2 6 2 6+7 4 5 2 2+2 4 |
| | 6 2 7 2 6+7 4 5 2 2+2 4 |
| | 7 2 7 2 6+7 4 5 2 2+2 4 |
| | 8 2 6 2 6+7 4 5 2 2+2 4 |
| | 6 2 9 2 6+7 4 5 2 2+2 4 |
| | 6 2 8 2 6+7 4 6 2 2+2 4 |
| | 7 2 6 2 6+7 4 6 2 2+2 4 |
| | 9 2 7 2 6+7 4 6 2 2+2 4 |
| | 8 2 8 2 6+7 4 5 2 2+2 4 |
| | 9 2 9 2 6+7 4 5 2 2+2 4 |
| | 7 2 7 2 6+7 4 6 2 2+2 4 |
| | 6 2 9 2 6+7 4 6 2 2+2 4 |
| | 9 2 8 2 6+7 4 6 2 2+2 4 |
| | 8 2 6 2 6+7 4 5 2 2+2 4 |
| | 7 2 7 2 6+7 4 5 2 2+2 4 |
| | 6 2 6 2 6+7 4 6 2 2+2 4 |
| | 8 2 9 2 6+7 4 5 2 2+2 4 |
| | 9 2 8 2 6+7 4 6 2 2+2 4 |
| | 7 2 7 2 6+7 4 5 2 2+2 4 |
| | 9 2 8 2 6+7 4 6 2 2+2 4 |
| | 8 2 9 2 6+7 4 6 2 2+2 4 |
| | 9 2 9 2 6+7 4 6 2 2+2 4 |
| | 9 2 7 2 6+7 4 5 2 2+2 4 |
| | 7 2 6 2 6+7 4 5 2 2+2 4 |
| | 6 2 9 2 6+7 4 6 2 2+2 4 |
| | 7 2 7 2 6+7 4 5 2 2+2 4 |
| | 7 2 9 2 6+7 4 6 2 2+2 4 |
| | 8 2 6 2 6+7 4 6 2 2+2 4 |
| | 6 2 8 2 6+7 4 5 2 2+2 4 |
| | 9 2 7 2 6+7 4 5 2 2+2 4 |
| | 9 2 9 2 6+7 4 6 2 2+2 4 |
| | 8 2 8 2 6+7 4 6 2 2+2 4 |
| | 6 2 9 2 6+7 4 5 2 2+2 4 |
| | 6 2 7 2 6+7 4 5 2 2+2 4 |
| | 9 2 6 2 6+7 4 6 2 2+2 4 |
| | 7 2 8 2 6+7 4 6 2 2+2 4 |
| | 8 2 9 2 6+7 4 6 2 2+2 4 |
| | 7 2 8 2 6 2 |
| | 9 2 6 2 5 2 |
| | 8 2 7 2 5 2 |
| | 9 2 9 2 6 2 |

Continued on next column

**TABLE LXV– continued**

| TQC | ICSD NoE SF |
|---|---|
| | 7 2 9 2 5 2 |
| | 9 2 7 2 6 2 |
| | 6 2 8 2 5 2 |
| | 8 2 6 2 6 2 |
| | 6 2 8 2 5 2 |
| | 7 2 6 2 6 2 |
| | 9 2 7 2 5 2 |
| | 7 2 9 2 6 2 |
| | 9 2 7 2 5 2 |
| | 8 2 8 2 6 2 |
| | 6 2 6 2 6 2 |
| | 8 2 9 2 5 2 |
| | 6 2 9 2 5 2 |
| | 7 2 8 2 6 2 |
| | 8 2 7 2 6 2 |
| | 9 2 6 2 5 2 |
| | 9 2 8 2 6 2 |
| | 7 2 9 2 6 2 |
| | 6 2 7 2 5 2 |
| | 8 2 9 2 6 2 |
| | 9 2 8 2 5 2 |
| | 8 2 6 2 6 2 |
| | 6 2 8 2 5 2 |
| | 9 2 7 2 6 2 |
| | 6 2 7 2 5 2 |
| | 7 2 8 2 6 2 |
| | 9 2 9 2 5 2 |
| | 8 2 6 2 6 2 |
| | 7 2 9 2 6 2 |
| | 8 2 6 2 5 2 |
| | 7 2 7 2 5 2 |
| | 7 2 7 2 6 2 |
| | 6 2 6 2 5 2 |
| | 8 2 8 2 5 2 |
| | 9 2 9 2 5 2 |
| | 7 2 9 2 6 2 |
| | 6 2 7 2 6 2 |
| | 9 2 6 2 5 2 |
| 140.1.11 | 190146 260 Bi1Ho5Ni2 |
| | -GM -M -P -X -N |
| | 7 2 7 2 6+7 4 6 2 2+2 4 |
| | 6 2 9 2 6+7 4 5 2 2+2 4 |
| | 8 2 6 2 6+7 4 5 2 2+2 4 |
| | 9 2 8 2 6+7 4 5 2 2+2 4 |
| | 6 2 8 2 6+7 4 6 2 2+2 4 |
| | 7 2 9 2 6+7 4 5 2 2+2 4 |
| | 8 2 7 2 6+7 4 6 2 2+2 4 |
| | 9 2 6 2 6+7 4 6 2 2+2 4 |
| | 7 2 7 2 6+7 4 5 2 2+2 4 |
| | 7 2 7 2 6+7 4 5 2 2+2 4 |
| | 9 2 9 2 6+7 4 6 2 2+2 4 |
| | 9 2 9 2 6+7 4 6 2 2+2 4 |
| | 6 2 8 2 6+7 4 5 2 2+2 4 |
| | 7 2 7 2 6+7 4 5 2 2+2 4 |
| | 8 2 6 2 6+7 4 6 2 2+2 4 |
| | 9 2 9 2 6+7 4 5 2 2+2 4 |
| | 6 2 6 2 6+7 4 6 2 2+2 4 |
| | 7 2 8 2 6+7 4 6 2 2+2 4 |
| | 8 2 7 2 6+7 4 5 2 2+2 4 |
| | 9 2 9 2 6+7 4 6 2 2+2 4 |
| | 9 2 9 2 6+7 4 6 2 2+2 4 |

Continued on next column

**TABLE LXV– continued**

| TQC | ICSD | NoE | SF |
|---|---|---|---|
| | 8 2 8 2 | 6+7 4 6 2 | 2+2 4 |
| | 8 2 9 2 | 6+7 4 6 2 | 2+2 4 |
| | 6 2 8 2 | 6+7 4 5 2 | 2+2 4 |
| | 7 2 6 2 | 6+7 4 6 2 | 2+2 4 |
| | 9 2 7 2 | 6+7 4 5 2 | 2+2 4 |
| | 7 2 9 2 | 6+7 4 6 2 | 2+2 4 |
| | 8 2 6 2 | 6+7 4 5 2 | 2+2 4 |
| | 6 2 7 2 | 6+7 4 6 2 | 2+2 4 |
| | 9 2 8 2 | 6+7 4 5 2 | 2+2 4 |
| | 7 2 9 2 | 6+7 4 6 2 | 2+2 4 |
| | 9 2 7 2 | 6+7 4 5 2 | 2+2 4 |
| | 6 2 8 2 | 6+7 4 5 2 | 2+2 4 |
| | 7 2 9 2 | 6+7 4 6 2 | 2+2 4 |
| | 8 2 6 2 | 6+7 4 5 2 | 2+2 4 |
| | 6 2 7 2 | 6+7 4 6 2 | 2+2 4 |
| | 9 2 6 2 | 6+7 4 5 2 | 2+2 4 |
| | 8 2 9 2 | 6+7 4 6 2 | 2+2 4 |
| | 9 2 8 2 | 6+7 4 6 2 | 2+2 4 |
| | 8 2 8 2 | 6+7 4 6 2 | 2+2 4 |
| | 7 2 9 2 | 6+7 4 5 2 | 2+2 4 |
| | 9 2 7 2 | 6+7 4 6 2 | 2+2 4 |
| | 7 2 7 2 | 6+7 4 6 2 | 2+2 4 |
| | 7 2 9 2 | 6+7 4 5 2 | 2+2 4 |
| | 9 2 8 2 | 6+7 4 5 2 | 2+2 4 |
| | 6 2 7 2 | 6+7 4 6 2 | 2+2 4 |
| | 6 2 6 2 | 6+7 4 5 2 | 2+2 4 |
| | 7 2 7 2 | 6+7 4 6 2 | 2+2 4 |
| | 9 2 6 2 | 6+7 4 5 2 | 2+2 4 |
| | 8 2 6 2 | 6+7 4 6 2 | 2+2 4 |
| | 6 2 9 2 | 6+7 4 5 2 | 2+2 4 |
| | 7 2 9 2 | 6+7 4 5 2 | 2+2 4 |
| | 8 2 7 2 | 6+7 4 5 2 | 2+2 4 |
| | 9 2 8 2 | 6+7 4 6 2 | 2+2 4 |
| | 7 2 7 2 | 6+7 4 5 2 | 2+2 4 |
| | 8 2 9 2 | 6+7 4 6 2 | 2+2 4 |
| | 9 2 8 2 | 6+7 4 5 2 | 2+2 4 |
| | 8 2 9 2 | 6+7 4 6 2 | 2+2 4 |
| | 9 2 8 2 | 6+7 4 5 2 | 2+2 4 |
| | 8 2 9 2 | 6+7 4 5 2 | 2+2 4 |
| | 6 2 7 2 | 6+7 4 6 2 | 2+2 4 |
| | 9 2 8 2 | 6+7 4 5 2 | 2+2 4 |
| | 6 2 6 2 | 6+7 4 5 2 | 2+2 4 |
| | 7 2 6 2 | 6+7 4 6 2 | 2+2 4 |
| | 6 2 7 2 | 6+7 4 5 2 | 2+2 4 |
| | 9 2 9 2 | 6 2 | |
| | 8 2 6 2 | 6 2 | |
| | 7 2 9 2 | 5 2 | |
| | 7 2 8 2 | 6 2 | |
| | 6 2 7 2 | 6 2 | |
| | 9 2 6 2 | 6 2 | |
| | 8 2 6 2 | 6 2 | |
| | 6 2 8 2 | 5 2 | |
| | 8 2 9 2 | 5 2 | |
| | 7 2 7 2 | 6 2 | |
| | 9 2 6 2 | 6 2 | |
| | 6 2 7 2 | 6 2 | |
| | 8 2 8 2 | 5 2 | |
| | 6 2 8 2 | 6 2 | |
| | 7 2 6 2 | 6 2 | |
| | 8 2 9 2 | 5 2 | |
| | 7 2 8 2 | 6 2 | |

Continued on next column

**TABLE LXV– continued**

| TQC | ICSD | NoE | SF |
|---|---|---|---|
| | 8 2 7 2 | 5 2 | |
| | 9 2 9 2 | 6 2 | |
| | 8 2 6 2 | 5 2 | |
| | 9 2 8 2 | 5 2 | |
| | 6 2 8 2 | 6 2 | |
| | 9 2 8 2 | 6 2 | |
| | 8 2 9 2 | 5 2 | |
| | 9 2 9 2 | 6 2 | |
| | 9 2 6 2 | 5 2 | |
| | 7 2 7 2 | 6 2 | |
| | 6 2 9 2 | 6 2 | |
| | 9 2 9 2 | 5 2 | |
| | 7 2 9 2 | 6 2 | |
| | 8 2 7 2 | 5 2 | |
| | 8 2 8 2 | 5 2 | |
| | 6 2 6 2 | 6 2 | |
| | 7 2 8 2 | 6 2 | |
| | 9 2 7 2 | 6 2 | |
| | 8 2 6 2 | 5 2 | |
| | 6 2 8 2 | 6 2 | |
| | 8 2 8 2 | 6 2 | |
| | 8 2 9 2 | 5 2 | |
| | 7 2 7 2 | 6 2 | |
| | 7 2 6 2 | 5 2 | |
| | 9 2 7 2 | 6 2 | |
| | 7 2 6 2 | 6 2 | |
| | 6 2 7 2 | 6 2 | |
| | 7 2 8 2 | 5 2 | |
| | 9 2 9 2 | 6 2 | |
| | 8 2 7 2 | 5 2 | |
| | 8 2 6 2 | 6 2 | |
| | 9 2 8 2 | 5 2 | |
| | 6 2 9 2 | 6 2 | |
| | 7 2 8 2 | 5 2 | |
| | 8 2 8 2 | 6 2 | |
| | 9 2 6 2 | 6 2 | |
| | 8 2 7 2 | 5 2 | |
| | 6 2 9 2 | 6 2 | |
| | 9 2 8 2 | 5 2 | |
| | 8 2 9 2 | 6 2 | |
| | 9 2 7 2 | 5 2 | |
| | 6 2 8 2 | 5 2 | |
| | 6 2 8 2 | 6 2 | |
| | 9 2 9 2 | 5 2 | |
| | 8 2 8 2 | 6 2 | |
| | 9 2 6 2 | 6 2 | |
| | 8 2 6 2 | 5 2 | |
| | 8 2 9 2 | 6 2 | |
| 140.1.12 | 100084 | 124 | Ba5Sn3 |
| | -GM -M | -P -X | -N |
| | 7 2 7 2 | 6+7 4 5 2 | 2+2 4 |
| | 7 2 7 2 | 6+7 4 5 2 | 2+2 4 |
| | 6 2 9 2 | 6+7 4 6 2 | 2+2 4 |
| | 6 2 8 2 | 6+7 4 5 2 | 2+2 4 |
| | 7 2 8 2 | 6+7 4 6 2 | 2+2 4 |
| | 7 2 9 2 | 6+7 4 6 2 | 2+2 4 |
| | 8 2 6 2 | 6+7 4 5 2 | 2+2 4 |
| | 9 2 6 2 | 6+7 4 5 2 | 2+2 4 |
| | 8 2 7 2 | 6+7 4 6 2 | 2+2 4 |
| | 9 2 7 2 | 6+7 4 5 2 | 2+2 4 |
| | 9 2 9 2 | 6+7 4 6 2 | 2+2 4 |

Continued on next column

**TABLE LXV– continued**

| TQC | ICSD | NoE | SF |
|---|---|---|---|
| | 6 2 9 2 | 6+7 4 | 6 2 2+2 4 |
| | 9 2 8 2 | 6+7 4 | 5 2 2+2 4 |
| | 7 2 9 2 | 6+7 4 | 6 2 2+2 4 |
| | 7 2 8 2 | 6+7 4 | 5 2 2+2 4 |
| | 6 2 6 2 | 6+7 4 | 5 2 2+2 4 |
| | 8 2 6 2 | 6+7 4 | 6 2 2+2 4 |
| | 9 2 7 2 | 6+7 4 | 6 2 2+2 4 |
| | 8 2 9 2 | 6+7 4 | 6 2 2+2 4 |
| | 9 2 7 2 | 6+7 4 | 5 2 2+2 4 |
| | 9 2 8 2 | 6+7 4 | 6 2 2+2 4 |
| | 8 2 9 2 | 6+7 4 | 6 2 2+2 4 |
| | 6 2 9 2 | 6+7 4 | 5 2 2+2 4 |
| | 7 2 8 2 | 6+7 4 | 6 2 2+2 4 |
| | 8 2 6 2 | 6+7 4 | 6 2 2+2 4 |
| | 7 2 9 2 | 6+7 4 | 6 2 2+2 4 |
| | 9 2 7 2 | 6+7 4 | 5 2 2+2 4 |
| | 6 2 7 2 | 6+7 4 | 5 2 2+2 4 |
| | 9 2 6 2 | 6+7 4 | 6 2 2+2 4 |
| | 8 2 8 2 | 6+7 4 | 5 2 2+2 4 |
| | 7 2 9 2 | 6+7 4 | 6 2 2+2 4 |
| | 7 2 8 2 | | 5 2 |
| | 6 2 7 2 | | 6 2 |
| | 9 2 9 2 | | 5 2 |
| | 6 2 6 2 | | 5 2 |
| | 8 2 7 2 | | 6 2 |
| | 9 2 8 2 | | 6 2 |
| | 8 2 9 2 | | 5 2 |
| | 9 2 8 2 | | 6 2 |
| | 8 2 6 2 | | 6 2 |
| | 7 2 7 2 | | 6 2 |
| | 6 2 6 2 | | 6 2 |
| | 7 2 9 2 | | 5 2 |
| | 9 2 7 2 | | 6 2 |
| | 9 2 9 2 | | 5 2 |
| | 8 2 8 2 | | 5 2 |
| | 6 2 6 2 | | 5 2 |
| | 9 2 9 2 | | 5 2 |
| | 7 2 7 2 | | 6 2 |
| | 7 2 9 2 | | 5 2 |
| | 6 2 8 2 | | 5 2 |
| | 7 2 7 2 | | 6 2 |
| | 8 2 7 2 | | 5 2 |
| | 9 2 8 2 | | 6 2 |
| | 8 2 7 2 | | 6 2 |
| | 9 2 6 2 | | 6 2 |
| | 6 2 6 2 | | 6 2 |
| | 8 2 6 2 | | 5 2 |
| | 7 2 9 2 | | 6 2 |
| | 7 2 7 2 | | 5 2 |
| | 9 2 8 2 | | 6 2 |
| | 6 2 9 2 | | 5 2 |
| 140.1.13 | 247208 | 148 | B2Cu2O6Sr1 |
| | -GM -M | -P | -X -N |
| | 7 2 7 2 | 6+7 4 | 5 2 2+2 4 |
| | 9 2 9 2 | 6+7 4 | 6 2 2+2 4 |
| | 7 2 7 2 | 6+7 4 | 6 2 2+2 4 |
| | 6 2 9 2 | 6+7 4 | 6 2 2+2 4 |
| | 8 2 8 2 | 6+7 4 | 5 2 2+2 4 |
| | 9 2 6 2 | 6+7 4 | 5 2 2+2 4 |
| | 6 2 6 2 | 6+7 4 | 6 2 2+2 4 |
| | 7 2 7 2 | 6+7 4 | 6 2 2+2 4 |
| | 9 2 9 2 | 6+7 4 | 5 2 2+2 4 |
| | 8 2 8 2 | 6+7 4 | 5 2 2+2 4 |
| | 7 2 7 2 | 6+7 4 | 6 2 2+2 4 |
| | 9 2 8 2 | 6+7 4 | 5 2 2+2 4 |
| | 8 2 9 2 | 6+7 4 | 5 2 2+2 4 |
| | 6 2 6 2 | 6+7 4 | 6 2 2+2 4 |
| | 7 2 7 2 | 6+7 4 | 5 2 2+2 4 |
| | 9 2 9 2 | 6+7 4 | 6 2 2+2 4 |
| | 7 2 7 2 | 6+7 4 | 5 2 2+2 4 |
| | 6 2 8 2 | 6+7 4 | 5 2 2+2 4 |
| | 8 2 6 2 | 6+7 4 | 6 2 2+2 4 |
| | 9 2 9 2 | 6+7 4 | 6 2 2+2 4 |
| | 7 2 7 2 | 6+7 4 | 6 2 2+2 4 |
| | 6 2 6 2 | 6+7 4 | 6 2 2+2 4 |
| | 7 2 9 2 | 6+7 4 | 5 2 2+2 4 |
| | 8 2 8 2 | 6+7 4 | 5 2 2+2 4 |
| | 9 2 9 2 | 6+7 4 | 5 2 2+2 4 |
| | 9 2 8 2 | 6+7 4 | 6 2 2+2 4 |
| | 8 2 7 2 | 6+7 4 | 6 2 2+2 4 |
| | 6 2 6 2 | 6+7 4 | 6 2 2+2 4 |
| | 7 2 9 2 | 6+7 4 | 5 2 2+2 4 |
| | 9 2 7 2 | 6+7 4 | 6 2 2+2 4 |
| | 6 2 9 2 | 6+7 4 | 5 2 2+2 4 |
| | 8 2 8 2 | 6+7 4 | 6 2 2+2 4 |
| | 7 2 6 2 | 6+7 4 | 5 2 2+2 4 |
| | 6 2 8 2 | 6+7 4 | 5 2 2+2 4 |
| | 8 2 6 2 | 6+7 4 | 5 2 2+2 4 |
| | 6 2 7 2 | 6+7 4 | 6 2 2+2 4 |
| | 6 2 8 2 | 6+7 4 | 5 2 2+2 4 |
| | 9 2 6 2 | | 5 2 |
| | 8 2 9 2 | | 6 2 |
| | 9 2 6 2 | | 6 2 |
| | 7 2 7 2 | | 6 2 |
| | 9 2 9 2 | | 6 2 |
| | 9 2 8 2 | | 5 2 |
| | 7 2 7 2 | | 6 2 |
| | 8 2 9 2 | | 5 2 |
| | 9 2 7 2 | | 5 2 |
| | 7 2 6 2 | | 5 2 |
| | 6 2 8 2 | | 6 2 |
| | 8 2 7 2 | | 6 2 |
| | 8 2 9 2 | | 5 2 |
| | 9 2 8 2 | | 5 2 |
| | 7 2 7 2 | | 6 2 |
| | 8 2 6 2 | | 5 2 |
| | 6 2 6 2 | | 6 2 |
| | 6 2 9 2 | | 5 2 |
| | 7 2 8 2 | | 6 2 |
| | 6 2 6 2 | | 6 2 |
| | 7 2 6 2 | | 6 2 |
| | 7 2 7 2 | | 6 2 |
| | 8 2 8 2 | | 5 2 |
| | 9 2 9 2 | | 5 2 |
| | 6 2 7 2 | | 6 2 |
| | 9 2 6 2 | | 5 2 |
| | 8 2 9 2 | | 5 2 |
| | 8 2 8 2 | | 6 2 |
| | 7 2 7 2 | | 6 2 |
| | 9 2 8 2 | | 5 2 |
| | 8 2 8 2 | | 5 2 |
| | 6 2 9 2 | | 5 2 |

Continued on next column

**TABLE LXV– continued**

| TQC | ICSD | NoE | SF |
|---|---|---|---|
| | 7 2 7 2 | | 6 2 |
| | 6 2 6 2 | | 5 2 |
| | 9 2 9 2 | | 6 2 |
| | 7 2 7 2 | | 5 2 |
| | 9 2 9 2 | | 5 2 |
| 140.1.14 | 247207 | 148 | B2Cu2O6Sr1 |
| | -GM -M | -P | -X -N |
| | 7 2 7 2 | 6+7 4 | 5 2 2+2 4 |
| | 9 2 9 2 | 6+7 4 | 6 2 2+2 4 |
| | 7 2 7 2 | 6+7 4 | 6 2 2+2 4 |
| | 6 2 9 2 | 6+7 4 | 6 2 2+2 4 |
| | 8 2 8 2 | 6+7 4 | 5 2 2+2 4 |
| | 9 2 6 2 | 6+7 4 | 5 2 2+2 4 |
| | 6 2 6 2 | 6+7 4 | 6 2 2+2 4 |
| | 7 2 7 2 | 6+7 4 | 6 2 2+2 4 |
| | 9 2 9 2 | 6+7 4 | 5 2 2+2 4 |
| | 8 2 8 2 | 6+7 4 | 5 2 2+2 4 |
| | 7 2 7 2 | 6+7 4 | 6 2 2+2 4 |
| | 9 2 8 2 | 6+7 4 | 5 2 2+2 4 |
| | 8 2 9 2 | 6+7 4 | 5 2 2+2 4 |
| | 6 2 6 2 | 6+7 4 | 6 2 2+2 4 |
| | 7 2 7 2 | 6+7 4 | 5 2 2+2 4 |
| | 9 2 9 2 | 6+7 4 | 6 2 2+2 4 |
| | 7 2 7 2 | 6+7 4 | 5 2 2+2 4 |
| | 6 2 6 2 | 6+7 4 | 5 2 2+2 4 |
| | 8 2 8 2 | 6+7 4 | 6 2 2+2 4 |
| | 9 2 9 2 | 6+7 4 | 6 2 2+2 4 |
| | 7 2 7 2 | 6+7 4 | 6 2 2+2 4 |
| | 6 2 6 2 | 6+7 4 | 6 2 2+2 4 |
| | 7 2 9 2 | 6+7 4 | 5 2 2+2 4 |
| | 8 2 8 2 | 6+7 4 | 5 2 2+2 4 |
| | 9 2 9 2 | 6+7 4 | 5 2 2+2 4 |
| | 9 2 8 2 | 6+7 4 | 6 2 2+2 4 |
| | 8 2 7 2 | 6+7 4 | 6 2 2+2 4 |
| | 6 2 6 2 | 6+7 4 | 6 2 2+2 4 |
| | 7 2 9 2 | 6+7 4 | 5 2 2+2 4 |
| | 9 2 7 2 | 6+7 4 | 6 2 2+2 4 |
| | 6 2 9 2 | 6+7 4 | 5 2 2+2 4 |
| | 8 2 8 2 | 6+7 4 | 6 2 2+2 4 |
| | 7 2 6 2 | 6+7 4 | 5 2 2+2 4 |
| | 6 2 8 2 | 6+7 4 | 5 2 2+2 4 |
| | 8 2 6 2 | 6+7 4 | 5 2 2+2 4 |
| | 6 2 7 2 | 6+7 4 | 6 2 2+2 4 |
| | 6 2 6 2 | 6+7 4 | 5 2 2+2 4 |
| | 9 2 7 2 | | 5 2 |
| | 8 2 8 2 | | 6 2 |
| | 9 2 9 2 | | 6 2 |
| | 7 2 6 2 | | 6 2 |
| | 9 2 9 2 | | 6 2 |
| | 9 2 8 2 | | 5 2 |
| | 7 2 7 2 | | 6 2 |
| | 8 2 6 2 | | 5 2 |
| | 9 2 7 2 | | 5 2 |
| | 7 2 9 2 | | 5 2 |
| | 8 2 8 2 | | 6 2 |
| | 6 2 7 2 | | 6 2 |
| | 8 2 9 2 | | 5 2 |
| | 9 2 8 2 | | 5 2 |
| | 7 2 7 2 | | 6 2 |
| | 8 2 6 2 | | 5 2 |
| | 6 2 6 2 | | 6 2 |

Continued on next column

**TABLE LXV– continued**

| TQC | ICSD | NoE | SF |
|---|---|---|---|
| | 6 2 9 2 | | 5 2 |
| | 7 2 8 2 | | 6 2 |
| | 6 2 6 2 | | 6 2 |
| | 7 2 6 2 | | 6 2 |
| | 7 2 7 2 | | 6 2 |
| | 8 2 8 2 | | 5 2 |
| | 9 2 9 2 | | 5 2 |
| | 6 2 7 2 | | 6 2 |
| | 9 2 6 2 | | 5 2 |
| | 8 2 9 2 | | 5 2 |
| | 8 2 8 2 | | 6 2 |
| | 7 2 7 2 | | 6 2 |
| | 9 2 8 2 | | 5 2 |
| | 8 2 8 2 | | 5 2 |
| | 6 2 9 2 | | 5 2 |
| | 7 2 7 2 | | 6 2 |
| | 6 2 6 2 | | 5 2 |
| | 9 2 9 2 | | 6 2 |
| | 7 2 7 2 | | 5 2 |
| | 9 2 9 2 | | 5 2 |
| 140.2.1 | 1203 | 124 | Sn3Sr5 |
| | -GM -M | -P | -X -N |
| | 7 2 9 2 | 6+7 4 | 5 2 2+2 4 |
| | 6 2 8 2 | 6+7 4 | 6 2 2+2 4 |
| | 6 2 8 2 | 6+7 4 | 5 2 2+2 4 |
| | 7 2 9 2 | 6+7 4 | 6 2 2+2 4 |
| | 8 2 6 2 | 6+7 4 | 5 2 2+2 4 |
| | 9 2 7 2 | 6+7 4 | 6 2 2+2 4 |
| | 8 2 6 2 | 6+7 4 | 5 2 2+2 4 |
| | 9 2 7 2 | 6+7 4 | 6 2 2+2 4 |
| | 7 2 7 2 | 6+7 4 | 5 2 2+2 4 |
| | 7 2 7 2 | 6+7 4 | 5 2 2+2 4 |
| | 6 2 8 2 | 6+7 4 | 6 2 2+2 4 |
| | 7 2 9 2 | 6+7 4 | 5 2 2+2 4 |
| | 7 2 9 2 | 6+7 4 | 6 2 2+2 4 |
| | 6 2 8 2 | 6+7 4 | 5 2 2+2 4 |
| | 8 2 6 2 | 6+7 4 | 6 2 2+2 4 |
| | 8 2 6 2 | 6+7 4 | 5 2 2+2 4 |
| | 9 2 7 2 | 6+7 4 | 6 2 2+2 4 |
| | 9 2 7 2 | 6+7 4 | 5 2 2+2 4 |
| | 9 2 9 2 | 6+7 4 | 6 2 2+2 4 |
| | 9 2 9 2 | 6+7 4 | 6 2 2+2 4 |
| | 6 2 8 2 | 6+7 4 | 5 2 2+2 4 |
| | 7 2 9 2 | 6+7 4 | 6 2 2+2 4 |
| | 7 2 9 2 | 6+7 4 | 6 2 2+2 4 |
| | 8 2 6 2 | 6+7 4 | 5 2 2+2 4 |
| | 9 2 7 2 | 6+7 4 | 5 2 2+2 4 |
| | 6 2 8 2 | 6+7 4 | 6 2 2+2 4 |
| | 9 2 9 2 | 6+7 4 | 6 2 2+2 4 |
| | 8 2 7 2 | 6+7 4 | 6 2 2+2 4 |
| | 7 2 6 2 | 6+7 4 | 5 2 2+2 4 |
| | 7 2 8 2 | 6+7 4 | 5 2 2+2 4 |
| | 9 2 9 2 | 6+7 4 | 6 2 2+2 4 |
| | 6 2 8 2 | | 6 2 |
| | 8 2 7 2 | | 5 2 |
| | 6 2 6 2 | | 5 2 |
| | 9 2 7 2 | | 6 2 |
| | 8 2 6 2 | | 5 2 |
| | 9 2 9 2 | | 6 2 |
| | 9 2 9 2 | | 6 2 |
| | 8 2 8 2 | | 6 2 |

Continued on next column

**TABLE LXV– continued**

| TQC | ICSD | NoE | SF |
|---|---|---|---|
| | 8 2 8 2 6 2 | | |
| | 7 2 7 2 6 2 | | |
| | 6 2 6 2 6 2 | | |
| | 7 2 9 2 5 2 | | |
| | 9 2 7 2 6 2 | | |
| | 9 2 9 2 5 2 | | |
| | 8 2 8 2 5 2 | | |
| | 9 2 9 2 5 2 | | |
| | 6 2 7 2 6 2 | | |
| | 7 2 6 2 5 2 | | |
| | 7 2 9 2 5 2 | | |
| | 6 2 8 2 5 2 | | |
| | 7 2 7 2 6 2 | | |
| | 8 2 8 2 6 2 | | |
| | 9 2 7 2 6 2 | | |
| | 8 2 7 2 5 2 | | |
| | 9 2 6 2 6 2 | | |
| | 7 2 9 2 5 2 | | |
| | 8 2 8 2 6 2 | | |
| | 9 2 6 2 5 2 | | |
| | 6 2 7 2 6 2 | | |
| | 8 2 9 2 6 2 | | |
| | 7 2 6 2 6 2 | | |
| 140.2.2 | 409797 | 28 | Sb2Ti1 |
| | -GM -M -P -X -N | | |
| | 7 2 6 2 6+7 4 6 2 2+2 4 | | |
| | 6 2 8 2 6+7 4 5 2 2+2 4 | | |
| | 9 2 9 2 6+7 4 6 2 2+2 4 | | |
| | 8 2 7 2 6+7 4 5 2 2+2 4 | | |
| | 6 2 9 2 6+7 4 5 2 2+2 4 | | |
| | 8 2 7 2 6+7 4 6 2 2+2 4 | | |
| | 9 2 8 2 6+7 4 6 2 2+2 4 | | |
| | 7 2 9 2 5 2 | | |
| | 8 2 6 2 5 2 | | |
| | 7 2 8 2 6 2 | | |
| | 6 2 7 2 6 2 | | |
| | 7 2 7 2 6 2 | | |
| | 6 2 6 2 5 2 | | |
| | 7 2 6 2 6 2 | | |
| 140.2.3 | 249576 | 28 | Sb2Ti1 |
| | -GM -M -P -X -N | | |
| | 7 2 6 2 6+7 4 6 2 2+2 4 | | |
| | 6 2 8 2 6+7 4 5 2 2+2 4 | | |
| | 9 2 9 2 6+7 4 6 2 2+2 4 | | |
| | 8 2 7 2 6+7 4 5 2 2+2 4 | | |
| | 6 2 9 2 6+7 4 5 2 2+2 4 | | |
| | 8 2 7 2 6+7 4 6 2 2+2 4 | | |
| | 9 2 8 2 6+7 4 6 2 2+2 4 | | |
| | 7 2 9 2 5 2 | | |
| | 8 2 6 2 5 2 | | |
| | 7 2 8 2 6 2 | | |
| | 6 2 7 2 6 2 | | |
| | 7 2 7 2 6 2 | | |
| | 6 2 6 2 5 2 | | |
| | 7 2 6 2 6 2 | | |
| 140.2.4 | 16582 | 184 | Pu5Si3 |
| | -GM -M -P -X -N | | |
| | 7 2 6 2 6+7 4 6 2 2+2 4 | | |
| | 7 2 8 2 6+7 4 5 2 2+2 4 | | |
| | 8 2 9 2 6+7 4 6 2 2+2 4 | | |
| | 9 2 6 2 6+7 4 5 2 2+2 4 | | |
| | 6 2 7 2 6+7 4 6 2 2+2 4 | | |
| | 6 2 7 2 6+7 4 6 2 2+2 4 | | |
| | 8 2 6 2 6+7 4 5 2 2+2 4 | | |
| | 9 2 8 2 6+7 4 5 2 2+2 4 | | |
| | 7 2 9 2 6+7 4 6 2 2+2 4 | | |
| | 8 2 9 2 6+7 4 5 2 2+2 4 | | |
| | 6 2 7 2 6+7 4 6 2 2+2 4 | | |
| | 9 2 6 2 6+7 4 5 2 2+2 4 | | |
| | 7 2 7 2 6+7 4 6 2 2+2 4 | | |
| | 6 2 8 2 6+7 4 6 2 2+2 4 | | |
| | 8 2 9 2 6+7 4 5 2 2+2 4 | | |
| | 8 2 8 2 6+7 4 5 2 2+2 4 | | |
| | 7 2 9 2 6+7 4 6 2 2+2 4 | | |
| | 6 2 7 2 6+7 4 5 2 2+2 4 | | |
| | 9 2 6 2 6+7 4 6 2 2+2 4 | | |
| | 9 2 8 2 6+7 4 5 2 2+2 4 | | |
| | 6 2 7 2 6+7 4 6 2 2+2 4 | | |
| | 9 2 9 2 6+7 4 5 2 2+2 4 | | |
| | 8 2 6 2 6+7 4 6 2 2+2 4 | | |
| | 7 2 9 2 6+7 4 5 2 2+2 4 | | |
| | 6 2 7 2 6+7 4 5 2 2+2 4 | | |
| | 7 2 8 2 6+7 4 6 2 2+2 4 | | |
| | 6 2 9 2 6+7 4 6 2 2+2 4 | | |
| | 9 2 7 2 6+7 4 5 2 2+2 4 | | |
| | 8 2 6 2 6+7 4 5 2 2+2 4 | | |
| | 8 2 8 2 6+7 4 6 2 2+2 4 | | |
| | 9 2 6 2 6+7 4 6 2 2+2 4 | | |
| | 6 2 7 2 6+7 4 5 2 2+2 4 | | |
| | 7 2 8 2 6+7 4 6 2 2+2 4 | | |
| | 8 2 9 2 6+7 4 6 2 2+2 4 | | |
| | 6 2 7 2 6+7 4 5 2 2+2 4 | | |
| | 8 2 8 2 6+7 4 5 2 2+2 4 | | |
| | 9 2 6 2 6+7 4 5 2 2+2 4 | | |
| | 7 2 9 2 6+7 4 6 2 2+2 4 | | |
| | 7 2 6 2 6+7 4 6 2 2+2 4 | | |
| | 9 2 8 2 6+7 4 5 2 2+2 4 | | |
| | 7 2 7 2 6+7 4 5 2 2+2 4 | | |
| | 7 2 7 2 6+7 4 6 2 2+2 4 | | |
| | 9 2 8 2 6+7 4 5 2 2+2 4 | | |
| | 6 2 9 2 6+7 4 6 2 2+2 4 | | |
| | 8 2 6 2 6+7 4 5 2 2+2 4 | | |
| | 9 2 9 2 6+7 4 6 2 2+2 4 | | |
| | 7 2 7 2 5 2 | | |
| | 8 2 7 2 6 2 | | |
| | 9 2 8 2 6 2 | | |
| | 7 2 9 2 5 2 | | |
| | 6 2 6 2 6 2 | | |
| | 7 2 8 2 5 2 | | |
| | 7 2 9 2 5 2 | | |
| | 6 2 8 2 6 2 | | |
| | 8 2 9 2 5 2 | | |
| | 6 2 7 2 5 2 | | |
| | 8 2 6 2 6 2 | | |
| | 9 2 8 2 5 2 | | |
| | 9 2 7 2 6 2 | | |
| | 8 2 6 2 6 2 | | |
| | 9 2 9 2 5 2 | | |
| | 6 2 6 2 6 2 | | |
| | 6 2 9 2 5 2 | | |
| | 7 2 7 2 6 2 | | |
| | 7 2 6 2 5 2 | | |

Continued on next column

**TABLE LXV– continued**

| TQC | ICSD | | NoE | | SF |
|---|---|---|---|---|---|
| | 9 2 | 8 2 | | 6 2 | |
| | 8 2 | 6 2 | | 5 2 | |
| | 9 2 | 9 2 | | 6 2 | |
| | 6 2 | 7 2 | | 5 2 | |
| | 7 2 | 8 2 | | 5 2 | |
| | 7 2 | 6 2 | | 5 2 | |
| | 8 2 | 8 2 | | 6 2 | |
| | 7 2 | 8 2 | | 5 2 | |
| | 6 2 | 6 2 | | 5 2 | |
| | 8 2 | 9 2 | | 6 2 | |
| | 8 2 | 9 2 | | 6 2 | |
| | 6 2 | 7 2 | | 5 2 | |
| | 7 2 | 7 2 | | 6 2 | |
| | 9 2 | 9 2 | | 5 2 | |
| | 8 2 | 8 2 | | 6 2 | |
| | 6 2 | 6 2 | | 5 2 | |
| | 9 2 | 6 2 | | 6 2 | |
| | 6 2 | 7 2 | | 6 2 | |
| | 7 2 | 8 2 | | 6 2 | |
| | 8 2 | 9 2 | | 5 2 | |
| | 9 2 | 7 2 | | 6 2 | |
| | 6 2 | 9 2 | | 5 2 | |
| | 7 2 | 6 2 | | 5 2 | |
| | 8 2 | 9 2 | | 6 2 | |
| | 9 2 | 8 2 | | 6 2 | |
| | 7 2 | 6 2 | | 5 2 | |
| | 8 2 | 7 2 | | 6 2 | |
| 140.2.5 | 105479 | | 36 | | Ni1Zr2 |
| | -GM | -M | -P | -X | -N |
| | 7 2 | 7 2 | 6+7 4 | 5 2 | 2+2 4 |
| | 9 2 | 9 2 | 6+7 4 | 6 2 | 2+2 4 |
| | 7 2 | 8 2 | 6+7 4 | 6 2 | 2+2 4 |
| | 9 2 | 6 2 | 6+7 4 | 5 2 | 2+2 4 |
| | 8 2 | 7 2 | 6+7 4 | 6 2 | 2+2 4 |
| | 6 2 | 6 2 | 6+7 4 | 5 2 | 2+2 4 |
| | 8 2 | 8 2 | 6+7 4 | 6 2 | 2+2 4 |
| | 8 2 | 9 2 | 6+7 4 | 5 2 | 2+2 4 |
| | 6 2 | 8 2 | 6+7 4 | 6 2 | 2+2 4 |
| | 7 2 | 8 2 | | 6 2 | |
| | 6 2 | 9 2 | | 5 2 | |
| | 7 2 | 7 2 | | 5 2 | |
| | 8 2 | 7 2 | | 6 2 | |
| | 9 2 | 6 2 | | 5 2 | |
| | 8 2 | 9 2 | | 5 2 | |
| | 8 2 | 7 2 | | 5 2 | |
| | 9 2 | 6 2 | | 5 2 | |
| | 9 2 | 6 2 | | 5 2 | |
| 140.2.6 | 260983 | | 188 | | Ni1U6 |
| | -GM | -M | -P | -X | -N |
| | 7 2 | 6 2 | 6+7 4 | 6 2 | 2+2 4 |
| | 6 2 | 8 2 | 6+7 4 | 5 2 | 2+2 4 |
| | 8 2 | 9 2 | 6+7 4 | 6 2 | 2+2 4 |
| | 9 2 | 7 2 | 6+7 4 | 5 2 | 2+2 4 |
| | 7 2 | 6 2 | 6+7 4 | 6 2 | 2+2 4 |
| | 6 2 | 7 2 | 6+7 4 | 5 2 | 2+2 4 |
| | 8 2 | 7 2 | 6+7 4 | 6 2 | 2+2 4 |
| | 9 2 | 8 2 | 6+7 4 | 5 2 | 2+2 4 |
| | 8 2 | 9 2 | 6+7 4 | 6 2 | 2+2 4 |
| | 9 2 | 6 2 | 6+7 4 | 5 2 | 2+2 4 |
| | 6 2 | 8 2 | 6+7 4 | 6 2 | 2+2 4 |
| | 7 2 | 9 2 | 6+7 4 | 5 2 | 2+2 4 |

Continued on next column

**TABLE LXV– continued**

| TQC | ICSD | | NoE | | SF |
|---|---|---|---|---|---|
| | 9 2 | 7 2 | 6+7 4 | 6 2 | 2+2 4 |
| | 8 2 | 6 2 | 6+7 4 | 5 2 | 2+2 4 |
| | 7 2 | 8 2 | 6+7 4 | 6 2 | 2+2 4 |
| | 6 2 | 9 2 | 6+7 4 | 5 2 | 2+2 4 |
| | 6 2 | 6 2 | 6+7 4 | 6 2 | 2+2 4 |
| | 7 2 | 8 2 | 6+7 4 | 6 2 | 2+2 4 |
| | 8 2 | 7 2 | 6+7 4 | 5 2 | 2+2 4 |
| | 9 2 | 9 2 | 6+7 4 | 5 2 | 2+2 4 |
| | 7 2 | 9 2 | 6+7 4 | 6 2 | 2+2 4 |
| | 6 2 | 6 2 | 6+7 4 | 5 2 | 2+2 4 |
| | 8 2 | 7 2 | 6+7 4 | 6 2 | 2+2 4 |
| | 9 2 | 8 2 | 6+7 4 | 5 2 | 2+2 4 |
| | 6 2 | 8 2 | 6+7 4 | 6 2 | 2+2 4 |
| | 9 2 | 6 2 | 6+7 4 | 5 2 | 2+2 4 |
| | 7 2 | 7 2 | 6+7 4 | 6 2 | 2+2 4 |
| | 7 2 | 9 2 | 6+7 4 | 5 2 | 2+2 4 |
| | 6 2 | 7 2 | 6+7 4 | 6 2 | 2+2 4 |
| | 8 2 | 6 2 | 6+7 4 | 6 2 | 2+2 4 |
| | 8 2 | 9 2 | 6+7 4 | 5 2 | 2+2 4 |
| | 9 2 | 9 2 | 6+7 4 | 6 2 | 2+2 4 |
| | 7 2 | 8 2 | 6+7 4 | 5 2 | 2+2 4 |
| | 6 2 | 6 2 | 6+7 4 | 5 2 | 2+2 4 |
| | 8 2 | 7 2 | 6+7 4 | 5 2 | 2+2 4 |
| | 9 2 | 6 2 | 6+7 4 | 6 2 | 2+2 4 |
| | 8 2 | 8 2 | 6+7 4 | 6 2 | 2+2 4 |
| | 7 2 | 7 2 | 6+7 4 | 5 2 | 2+2 4 |
| | 6 2 | 7 2 | 6+7 4 | 6 2 | 2+2 4 |
| | 9 2 | 9 2 | 6+7 4 | 6 2 | 2+2 4 |
| | 7 2 | 6 2 | 6+7 4 | 5 2 | 2+2 4 |
| | 6 2 | 8 2 | 6+7 4 | 6 2 | 2+2 4 |
| | 9 2 | 9 2 | 6+7 4 | 5 2 | 2+2 4 |
| | 8 2 | 7 2 | 6+7 4 | 5 2 | 2+2 4 |
| | 7 2 | 8 2 | 6+7 4 | 6 2 | 2+2 4 |
| | 6 2 | 9 2 | 6+7 4 | 5 2 | 2+2 4 |
| | 9 2 | 8 2 | 6+7 4 | 5 2 | 2+2 4 |
| | 8 2 | 6 2 | | 6 2 | |
| | 7 2 | 9 2 | | 5 2 | |
| | 7 2 | 7 2 | | 6 2 | |
| | 9 2 | 9 2 | | 6 2 | |
| | 8 2 | 8 2 | | 5 2 | |
| | 6 2 | 7 2 | | 6 2 | |
| | 9 2 | 6 2 | | 5 2 | |
| | 8 2 | 8 2 | | 6 2 | |
| | 9 2 | 9 2 | | 5 2 | |
| | 6 2 | 8 2 | | 5 2 | |
| | 6 2 | 6 2 | | 6 2 | |
| | 7 2 | 8 2 | | 5 2 | |
| | 7 2 | 7 2 | | 6 2 | |
| | 8 2 | 6 2 | | 6 2 | |
| | 8 2 | 9 2 | | 5 2 | |
| | 6 2 | 6 2 | | 5 2 | |
| | 9 2 | 8 2 | | 6 2 | |
| | 8 2 | 7 2 | | 6 2 | |
| | 7 2 | 6 2 | | 5 2 | |
| | 6 2 | 9 2 | | 5 2 | |
| | 6 2 | 7 2 | | 5 2 | |
| | 7 2 | 8 2 | | 6 2 | |
| | 8 2 | 8 2 | | 5 2 | |
| | 7 2 | 9 2 | | 5 2 | |
| | 9 2 | 7 2 | | 6 2 | |
| | 6 2 | 6 2 | | 6 2 | |

Continued on next column

**TABLE LXV– continued**

| TQC | ICSD | NoE | SF |
|---|---|---|---|
| | 8 2 7 2 | 6 2 | |
| | 8 2 9 2 | 5 2 | |
| | 7 2 8 2 | 6 2 | |
| | 9 2 8 2 | 5 2 | |
| | 9 2 7 2 | 5 2 | |
| | 8 2 6 2 | 6 2 | |
| | 7 2 8 2 | 5 2 | |
| | 6 2 9 2 | 6 2 | |
| | 9 2 7 2 | 5 2 | |
| | 7 2 6 2 | 6 2 | |
| | 6 2 6 2 | 6 2 | |
| | 9 2 6 2 | 6 2 | |
| | 7 2 7 2 | 5 2 | |
| | 6 2 8 2 | 5 2 | |
| | 7 2 9 2 | 6 2 | |
| | 8 2 9 2 | 5 2 | |
| | 7 2 9 2 | 5 2 | |
| | 7 2 7 2 | 6 2 | |
| | 6 2 8 2 | 5 2 | |
| | 9 2 9 2 | 6 2 | |
| | 8 2 7 2 | 6 2 | |
| 140.2.7 | 409374 | 124 | Ge3Sr5 |
| | -GM -M | -P -X | -N |
| | 7 2 7 2 | 6+7 4 5 2 | 2+2 4 |
| | 6 2 9 2 | 6+7 4 5 2 | 2+2 4 |
| | 6 2 7 2 | 6+7 4 5 2 | 2+2 4 |
| | 7 2 8 2 | 6+7 4 6 2 | 2+2 4 |
| | 7 2 9 2 | 6+7 4 6 2 | 2+2 4 |
| | 9 2 8 2 | 6+7 4 5 2 | 2+2 4 |
| | 8 2 6 2 | 6+7 4 6 2 | 2+2 4 |
| | 8 2 6 2 | 6+7 4 6 2 | 2+2 4 |
| | 9 2 7 2 | 6+7 4 5 2 | 2+2 4 |
| | 7 2 7 2 | 6+7 4 5 2 | 2+2 4 |
| | 9 2 9 2 | 6+7 4 6 2 | 2+2 4 |
| | 9 2 9 2 | 6+7 4 6 2 | 2+2 4 |
| | 6 2 8 2 | 6+7 4 5 2 | 2+2 4 |
| | 6 2 9 2 | 6+7 4 6 2 | 2+2 4 |
| | 7 2 8 2 | 6+7 4 5 2 | 2+2 4 |
| | 8 2 6 2 | 6+7 4 5 2 | 2+2 4 |
| | 9 2 6 2 | 6+7 4 6 2 | 2+2 4 |
| | 7 2 7 2 | 6+7 4 6 2 | 2+2 4 |
| | 8 2 7 2 | 6+7 4 5 2 | 2+2 4 |
| | 9 2 9 2 | 6+7 4 6 2 | 2+2 4 |
| | 9 2 8 2 | 6+7 4 6 2 | 2+2 4 |
| | 8 2 9 2 | 6+7 4 6 2 | 2+2 4 |
| | 6 2 9 2 | 6+7 4 6 2 | 2+2 4 |
| | 9 2 8 2 | 6+7 4 5 2 | 2+2 4 |
| | 8 2 6 2 | 6+7 4 6 2 | 2+2 4 |
| | 7 2 9 2 | 6+7 4 6 2 | 2+2 4 |
| | 7 2 7 2 | 6+7 4 5 2 | 2+2 4 |
| | 6 2 7 2 | 6+7 4 5 2 | 2+2 4 |
| | 9 2 9 2 | 6+7 4 6 2 | 2+2 4 |
| | 7 2 8 2 | 6+7 4 5 2 | 2+2 4 |
| | 8 2 6 2 | 6+7 4 5 2 | 2+2 4 |
| | 7 2 8 2 | 6 2 | |
| | 6 2 7 2 | 5 2 | |
| | 9 2 9 2 | 6 2 | |
| | 8 2 6 2 | 5 2 | |
| | 6 2 7 2 | 6 2 | |
| | 9 2 8 2 | 5 2 | |
| | 9 2 9 2 | 6 2 | |
| | 8 2 6 2 | 6 2 | |
| | 8 2 8 2 | 6 2 | |
| | 7 2 7 2 | 6 2 | |
| | 6 2 6 2 | 6 2 | |
| | 7 2 9 2 | 5 2 | |
| | 9 2 7 2 | 6 2 | |
| | 9 2 9 2 | 5 2 | |
| | 8 2 8 2 | 5 2 | |
| | 9 2 9 2 | 5 2 | |
| | 6 2 7 2 | 5 2 | |
| | 7 2 6 2 | 6 2 | |
| | 7 2 9 2 | 5 2 | |
| | 6 2 8 2 | 5 2 | |
| | 7 2 7 2 | 6 2 | |
| | 9 2 8 2 | 6 2 | |
| | 8 2 7 2 | 6 2 | |
| | 8 2 7 2 | 5 2 | |
| | 9 2 6 2 | 5 2 | |
| | 8 2 9 2 | 6 2 | |
| | 7 2 8 2 | 5 2 | |
| | 9 2 6 2 | 6 2 | |
| | 7 2 7 2 | 6 2 | |
| | 6 2 9 2 | 6 2 | |
| | 8 2 6 2 | 6 2 | |
| 140.2.8 | 420322 | 216 | Cs9O4Sc1 |
| | -GM -M | -P -X | -N |
| | 7 2 7 2 | 6+7 4 5 2 | 2+2 4 |
| | 7 2 7 2 | 6+7 4 5 2 | 2+2 4 |
| | 7 2 6 2 | 6+7 4 6 2 | 2+2 4 |
| | 6 2 7 2 | 6+7 4 6 2 | 2+2 4 |
| | 8 2 8 2 | 6+7 4 5 2 | 2+2 4 |
| | 9 2 9 2 | 6+7 4 5 2 | 2+2 4 |
| | 8 2 9 2 | 6+7 4 5 2 | 2+2 4 |
| | 6 2 6 2 | 6+7 4 6 2 | 2+2 4 |
| | 7 2 7 2 | 6+7 4 6 2 | 2+2 4 |
| | 6 2 7 2 | 6+7 4 5 2 | 2+2 4 |
| | 9 2 8 2 | 6+7 4 6 2 | 2+2 4 |
| | 7 2 6 2 | 6+7 4 6 2 | 2+2 4 |
| | 8 2 8 2 | 6+7 4 5 2 | 2+2 4 |
| | 9 2 9 2 | 6+7 4 5 2 | 2+2 4 |
| | 8 2 8 2 | 6+7 4 5 2 | 2+2 4 |
| | 9 2 9 2 | 6+7 4 5 2 | 2+2 4 |
| | 6 2 7 2 | 6+7 4 6 2 | 2+2 4 |
| | 7 2 6 2 | 6+7 4 6 2 | 2+2 4 |
| | 8 2 9 2 | 6+7 4 5 2 | 2+2 4 |
| | 7 2 6 2 | 6+7 4 6 2 | 2+2 4 |
| | 9 2 8 2 | 6+7 4 5 2 | 2+2 4 |
| | 8 2 8 2 | 6+7 4 5 2 | 2+2 4 |
| | 9 2 9 2 | 6+7 4 5 2 | 2+2 4 |
| | 6 2 6 2 | 6+7 4 6 2 | 2+2 4 |
| | 7 2 7 2 | 6+7 4 6 2 | 2+2 4 |
| | 6 2 7 2 | 6+7 4 6 2 | 2+2 4 |
| | 9 2 9 2 | 6+7 4 6 2 | 2+2 4 |
| | 9 2 9 2 | 6+7 4 6 2 | 2+2 4 |
| | 7 2 6 2 | 6+7 4 6 2 | 2+2 4 |
| | 6 2 7 2 | 6+7 4 6 2 | 2+2 4 |
| | 9 2 8 2 | 6+7 4 5 2 | 2+2 4 |
| | 7 2 6 2 | 6+7 4 6 2 | 2+2 4 |
| | 8 2 9 2 | 6+7 4 5 2 | 2+2 4 |
| | 6 2 7 2 | 6+7 4 6 2 | 2+2 4 |
| | 9 2 8 2 | 6+7 4 5 2 | 2+2 4 |

Continued on next column

**TABLE LXV– continued**

| TQC | ICSD | NoE | SF |
|---|---|---|---|
| | 8 2 9 2 | 6+7 4 5 2 | 2+2 4 |
| | 9 2 8 2 | 6+7 4 5 2 | 2+2 4 |
| | 6 2 6 2 | 6+7 4 6 2 | 2+2 4 |
| | 7 2 7 2 | 6+7 4 6 2 | 2+2 4 |
| | 8 2 9 2 | 6+7 4 5 2 | 2+2 4 |
| | 6 2 7 2 | 6+7 4 6 2 | 2+2 4 |
| | 7 2 6 2 | 6+7 4 6 2 | 2+2 4 |
| | 8 2 9 2 | 6+7 4 5 2 | 2+2 4 |
| | 9 2 8 2 | 6+7 4 5 2 | 2+2 4 |
| | 9 2 9 2 | 6+7 4 6 2 | 2+2 4 |
| | 9 2 9 2 | 6+7 4 6 2 | 2+2 4 |
| | 8 2 8 2 | 6+7 4 6 2 | 2+2 4 |
| | 8 2 8 2 | 6+7 4 6 2 | 2+2 4 |
| | 7 2 6 2 | 6+7 4 6 2 | 2+2 4 |
| | 6 2 7 2 | 6+7 4 6 2 | 2+2 4 |
| | 9 2 8 2 | 6+7 4 5 2 | 2+2 4 |
| | 6 2 7 2 | 6+7 4 6 2 | 2+2 4 |
| | 8 2 9 2 | 6+7 4 5 2 | 2+2 4 |
| | 7 2 6 2 | 6+7 4 6 2 | 2+2 4 |
| | 9 2 8 2 | 5 2 | |
| | 8 2 9 2 | 5 2 | |
| | 9 2 8 2 | 5 2 | |
| | 6 2 6 2 | 6 2 | |
| | 7 2 7 2 | 6 2 | |
| | 8 2 9 2 | 5 2 | |
| | 6 2 7 2 | 6 2 | |
| | 7 2 6 2 | 6 2 | |
| | 9 2 8 2 | 5 2 | |
| | 8 2 9 2 | 5 2 | |
| | 6 2 7 2 | 6 2 | |
| | 9 2 8 2 | 5 2 | |
| | 8 2 9 2 | 5 2 | |
| | 6 2 7 2 | 6 2 | |
| | 7 2 6 2 | 6 2 | |
| | 7 2 6 2 | 6 2 | |
| | 8 2 9 2 | 5 2 | |
| | 9 2 8 2 | 5 2 | |
| | 7 2 6 2 | 6 2 | |
| | 6 2 7 2 | 6 2 | |
| | 8 2 9 2 | 5 2 | |
| | 6 2 7 2 | 6 2 | |
| | 7 2 6 2 | 6 2 | |
| | 9 2 8 2 | 5 2 | |
| | 8 2 9 2 | 5 2 | |
| | 9 2 8 2 | 5 2 | |
| | 7 2 6 2 | 6 2 | |
| | 8 2 9 2 | 5 2 | |
| | 8 2 9 2 | 5 2 | |
| | 9 2 8 2 | 5 2 | |
| | 9 2 9 2 | 5 2 | |
| | 8 2 7 2 | 5 2 | |
| | 6 2 8 2 | 6 2 | |
| | 7 2 6 2 | 6 2 | |
| | 9 2 8 2 | 5 2 | |
| | 6 2 7 2 | 6 2 | |
| | 7 2 7 2 | 6 2 | |
| | 6 2 6 2 | 6 2 | |
| | 6 2 8 2 | 5 2 | |
| | 9 2 7 2 | 6 2 | |
| | 8 2 9 2 | 5 2 | |
| | 6 2 7 2 | 6 2 | |

Continued on next column

**TABLE LXV– continued**

| TQC | ICSD | NoE | SF |
|---|---|---|---|
| | 7 2 6 2 | 6 2 | |
| | 9 2 8 2 | 5 2 | |
| | 8 2 9 2 | 5 2 | |
| | 8 2 9 2 | 5 2 | |
| | 9 2 8 2 | 5 2 | |
| | 7 2 6 2 | 6 2 | |
| | 7 2 6 2 | 6 2 | |
| | 6 2 7 2 | 6 2 | |
| | 7 2 7 2 | 5 2 | |
| | 7 2 7 2 | 6 2 | |
| | 6 2 8 2 | 5 2 | |
| | 9 2 9 2 | 6 2 | |
| 140.2.9 | 419823 | 216 | Cs9In1O4 |
| | -GM -M | -P -X | -N |
| | 7 2 7 2 | 6+7 4 5 2 | 2+2 4 |
| | 7 2 7 2 | 6+7 4 5 2 | 2+2 4 |
| | 7 2 6 2 | 6+7 4 6 2 | 2+2 4 |
| | 6 2 7 2 | 6+7 4 6 2 | 2+2 4 |
| | 8 2 8 2 | 6+7 4 5 2 | 2+2 4 |
| | 9 2 9 2 | 6+7 4 5 2 | 2+2 4 |
| | 8 2 9 2 | 6+7 4 5 2 | 2+2 4 |
| | 6 2 6 2 | 6+7 4 6 2 | 2+2 4 |
| | 7 2 7 2 | 6+7 4 6 2 | 2+2 4 |
| | 6 2 7 2 | 6+7 4 5 2 | 2+2 4 |
| | 9 2 8 2 | 6+7 4 6 2 | 2+2 4 |
| | 7 2 6 2 | 6+7 4 6 2 | 2+2 4 |
| | 8 2 8 2 | 6+7 4 5 2 | 2+2 4 |
| | 9 2 9 2 | 6+7 4 5 2 | 2+2 4 |
| | 8 2 8 2 | 6+7 4 5 2 | 2+2 4 |
| | 9 2 9 2 | 6+7 4 5 2 | 2+2 4 |
| | 6 2 7 2 | 6+7 4 6 2 | 2+2 4 |
| | 7 2 6 2 | 6+7 4 6 2 | 2+2 4 |
| | 8 2 9 2 | 6+7 4 5 2 | 2+2 4 |
| | 7 2 6 2 | 6+7 4 6 2 | 2+2 4 |
| | 9 2 8 2 | 6+7 4 5 2 | 2+2 4 |
| | 6 2 7 2 | 6+7 4 6 2 | 2+2 4 |
| | 9 2 8 2 | 6+7 4 5 2 | 2+2 4 |
| | 6 2 7 2 | 6+7 4 5 2 | 2+2 4 |
| | 8 2 9 2 | 6+7 4 6 2 | 2+2 4 |
| | 7 2 6 2 | 6+7 4 6 2 | 2+2 4 |
| | 9 2 9 2 | 6+7 4 6 2 | 2+2 4 |
| | 9 2 9 2 | 6+7 4 6 2 | 2+2 4 |
| | 7 2 6 2 | 6+7 4 6 2 | 2+2 4 |
| | 6 2 7 2 | 6+7 4 6 2 | 2+2 4 |
| | 9 2 8 2 | 6+7 4 5 2 | 2+2 4 |
| | 7 2 6 2 | 6+7 4 6 2 | 2+2 4 |
| | 8 2 9 2 | 6+7 4 5 2 | 2+2 4 |
| | 6 2 7 2 | 6+7 4 6 2 | 2+2 4 |
| | 9 2 8 2 | 6+7 4 5 2 | 2+2 4 |
| | 8 2 9 2 | 6+7 4 5 2 | 2+2 4 |
| | 9 2 8 2 | 6+7 4 5 2 | 2+2 4 |
| | 6 2 6 2 | 6+7 4 6 2 | 2+2 4 |
| | 7 2 7 2 | 6+7 4 6 2 | 2+2 4 |
| | 8 2 9 2 | 6+7 4 5 2 | 2+2 4 |
| | 6 2 7 2 | 6+7 4 6 2 | 2+2 4 |
| | 7 2 6 2 | 6+7 4 6 2 | 2+2 4 |
| | 8 2 9 2 | 6+7 4 5 2 | 2+2 4 |
| | 9 2 8 2 | 6+7 4 5 2 | 2+2 4 |
| | 9 2 9 2 | 6+7 4 6 2 | 2+2 4 |
| | 9 2 9 2 | 6+7 4 6 2 | 2+2 4 |
| | 8 2 8 2 | 6+7 4 6 2 | 2+2 4 |

Continued on next column

**TABLE LXV– continued**

| TQC | ICSD | NoE | SF |
|---|---|---|---|
| | 8 2 8 2 6+7 4 6 2 2+2 4 | | |
| | 7 2 6 2 6+7 4 6 2 2+2 4 | | |
| | 6 2 7 2 6+7 4 6 2 2+2 4 | | |
| | 6 2 8 2 6+7 4 5 2 2+2 4 | | |
| | 9 2 7 2 6+7 4 6 2 2+2 4 | | |
| | 8 2 9 2 6+7 4 5 2 2+2 4 | | |
| | 7 2 6 2 6+7 4 6 2 2+2 4 | | |
| | 9 2 8 2 5 2 | | |
| | 8 2 9 2 5 2 | | |
| | 9 2 8 2 5 2 | | |
| | 6 2 6 2 6 2 | | |
| | 7 2 7 2 6 2 | | |
| | 8 2 9 2 5 2 | | |
| | 7 2 6 2 6 2 | | |
| | 9 2 8 2 5 2 | | |
| | 6 2 7 2 5 2 | | |
| | 8 2 9 2 6 2 | | |
| | 6 2 7 2 6 2 | | |
| | 8 2 9 2 5 2 | | |
| | 9 2 8 2 5 2 | | |
| | 6 2 7 2 6 2 | | |
| | 7 2 6 2 6 2 | | |
| | 7 2 6 2 6 2 | | |
| | 8 2 9 2 5 2 | | |
| | 9 2 8 2 5 2 | | |
| | 7 2 6 2 6 2 | | |
| | 6 2 7 2 6 2 | | |
| | 8 2 9 2 5 2 | | |
| | 6 2 7 2 6 2 | | |
| | 7 2 6 2 6 2 | | |
| | 9 2 8 2 5 2 | | |
| | 8 2 9 2 5 2 | | |
| | 9 2 8 2 5 2 | | |
| | 7 2 6 2 6 2 | | |
| | 8 2 9 2 5 2 | | |
| | 6 2 7 2 6 2 | | |
| | 9 2 8 2 5 2 | | |
| | 6 2 7 2 6 2 | | |
| | 8 2 9 2 5 2 | | |
| | 9 2 8 2 5 2 | | |
| | 7 2 6 2 6 2 | | |
| | 8 2 9 2 5 2 | | |
| | 9 2 8 2 5 2 | | |
| | 8 2 9 2 5 2 | | |
| | 7 2 6 2 6 2 | | |
| | 8 2 9 2 5 2 | | |
| | 9 2 8 2 5 2 | | |
| | 7 2 6 2 6 2 | | |
| | 6 2 7 2 6 2 | | |
| | 7 2 6 2 6 2 | | |
| | 6 2 7 2 5 2 | | |
| | 9 2 7 2 6 2 | | |
| | 6 2 8 2 6 2 | | |
| | 8 2 9 2 5 2 | | |
| | 9 2 8 2 5 2 | | |
| | 6 2 7 2 6 2 | | |
| | 7 2 6 2 6 2 | | |
| | 7 2 7 2 5 2 | | |
| | 7 2 7 2 6 2 | | |
| | 6 2 8 2 5 2 | | |

Continued on next column

**TABLE LXV– continued**

| TQC | ICSD | NoE | SF |
|---|---|---|---|
| | 9 2 9 2 6 2 | | |
| 140.2.10 | 420320 | 216 | Cs9Ga1O4 |
| | -GM -M -P -X -N | | |
| | 7 2 7 2 6+7 4 5 2 2+2 4 | | |
| | 7 2 7 2 6+7 4 5 2 2+2 4 | | |
| | 7 2 6 2 6+7 4 6 2 2+2 4 | | |
| | 6 2 7 2 6+7 4 6 2 2+2 4 | | |
| | 8 2 8 2 6+7 4 5 2 2+2 4 | | |
| | 9 2 9 2 6+7 4 5 2 2+2 4 | | |
| | 8 2 9 2 6+7 4 5 2 2+2 4 | | |
| | 6 2 6 2 6+7 4 6 2 2+2 4 | | |
| | 7 2 7 2 6+7 4 6 2 2+2 4 | | |
| | 6 2 6 2 6+7 4 5 2 2+2 4 | | |
| | 7 2 7 2 6+7 4 5 2 2+2 4 | | |
| | 8 2 8 2 6+7 4 5 2 2+2 4 | | |
| | 9 2 8 2 6+7 4 6 2 2+2 4 | | |
| | 9 2 9 2 6+7 4 6 2 2+2 4 | | |
| | 6 2 7 2 6+7 4 6 2 2+2 4 | | |
| | 7 2 6 2 6+7 4 6 2 2+2 4 | | |
| | 8 2 8 2 6+7 4 5 2 2+2 4 | | |
| | 9 2 9 2 6+7 4 5 2 2+2 4 | | |
| | 8 2 9 2 6+7 4 5 2 2+2 4 | | |
| | 7 2 6 2 6+7 4 6 2 2+2 4 | | |
| | 9 2 8 2 6+7 4 5 2 2+2 4 | | |
| | 6 2 7 2 6+7 4 6 2 2+2 4 | | |
| | 8 2 8 2 6+7 4 5 2 2+2 4 | | |
| | 9 2 9 2 6+7 4 5 2 2+2 4 | | |
| | 6 2 6 2 6+7 4 6 2 2+2 4 | | |
| | 7 2 7 2 6+7 4 6 2 2+2 4 | | |
| | 9 2 9 2 6+7 4 6 2 2+2 4 | | |
| | 9 2 9 2 6+7 4 6 2 2+2 4 | | |
| | 7 2 6 2 6+7 4 6 2 2+2 4 | | |
| | 6 2 7 2 6+7 4 6 2 2+2 4 | | |
| | 7 2 6 2 6+7 4 6 2 2+2 4 | | |
| | 9 2 8 2 6+7 4 5 2 2+2 4 | | |
| | 8 2 9 2 6+7 4 5 2 2+2 4 | | |
| | 6 2 7 2 6+7 4 6 2 2+2 4 | | |
| | 9 2 8 2 6+7 4 5 2 2+2 4 | | |
| | 8 2 9 2 6+7 4 5 2 2+2 4 | | |
| | 9 2 8 2 6+7 4 5 2 2+2 4 | | |
| | 6 2 6 2 6+7 4 6 2 2+2 4 | | |
| | 7 2 7 2 6+7 4 6 2 2+2 4 | | |
| | 8 2 9 2 6+7 4 5 2 2+2 4 | | |
| | 6 2 7 2 6+7 4 6 2 2+2 4 | | |
| | 7 2 6 2 6+7 4 6 2 2+2 4 | | |
| | 8 2 9 2 6+7 4 5 2 2+2 4 | | |
| | 9 2 8 2 6+7 4 5 2 2+2 4 | | |
| | 9 2 9 2 6+7 4 6 2 2+2 4 | | |
| | 9 2 9 2 6+7 4 6 2 2+2 4 | | |
| | 8 2 8 2 6+7 4 6 2 2+2 4 | | |
| | 8 2 8 2 6+7 4 6 2 2+2 4 | | |
| | 7 2 6 2 6+7 4 6 2 2+2 4 | | |
| | 6 2 7 2 6+7 4 6 2 2+2 4 | | |
| | 9 2 8 2 6+7 4 5 2 2+2 4 | | |
| | 6 2 7 2 6+7 4 6 2 2+2 4 | | |
| | 8 2 9 2 6+7 4 5 2 2+2 4 | | |
| | 9 2 8 2 6+7 4 5 2 2+2 4 | | |
| | 7 2 6 2 6 2 | | |
| | 8 2 9 2 5 2 | | |
| | 7 2 6 2 6 2 | | |
| | 9 2 8 2 5 2 | | |

Continued on next column

**TABLE LXV– continued**

| TQC | ICSD | | NoE | | SF |
|---|---|---|---|---|---|
| | 6 2 | 7 2 | | 6 2 | |
| | 8 2 | 9 2 | | 5 2 | |
| | 6 2 | 7 2 | | 6 2 | |
| | 9 2 | 8 2 | | 5 2 | |
| | 7 2 | 6 2 | | 6 2 | |
| | 6 2 | 7 2 | | 6 2 | |
| | 8 2 | 9 2 | | 5 2 | |
| | 6 2 | 6 2 | | 5 2 | |
| | 7 2 | 9 2 | | 6 2 | |
| | 8 2 | 7 2 | | 6 2 | |
| | 9 2 | 8 2 | | 5 2 | |
| | 8 2 | 9 2 | | 5 2 | |
| | 7 2 | 6 2 | | 6 2 | |
| | 9 2 | 7 2 | | 5 2 | |
| | 6 2 | 8 2 | | 6 2 | |
| | 7 2 | 6 2 | | 6 2 | |
| | 8 2 | 9 2 | | 5 2 | |
| | 7 2 | 6 2 | | 6 2 | |
| | 9 2 | 8 2 | | 5 2 | |
| | 6 2 | 7 2 | | 6 2 | |
| | 8 2 | 9 2 | | 5 2 | |
| | 9 2 | 8 2 | | 5 2 | |
| | 7 2 | 6 2 | | 6 2 | |
| | 8 2 | 9 2 | | 5 2 | |
| | 6 2 | 7 2 | | 6 2 | |
| | 9 2 | 8 2 | | 5 2 | |
| | 8 2 | 9 2 | | 5 2 | |
| | 6 2 | 7 2 | | 6 2 | |
| | 9 2 | 8 2 | | 5 2 | |
| | 7 2 | 6 2 | | 6 2 | |
| | 8 2 | 9 2 | | 5 2 | |
| | 9 2 | 8 2 | | 5 2 | |
| | 7 2 | 6 2 | | 6 2 | |
| | 8 2 | 9 2 | | 5 2 | |
| | 6 2 | 7 2 | | 6 2 | |
| | 9 2 | 7 2 | | 5 2 | |
| | 6 2 | 8 2 | | 6 2 | |
| | 8 2 | 9 2 | | 5 2 | |
| | 9 2 | 8 2 | | 5 2 | |
| | 8 2 | 9 2 | | 5 2 | |
| | 7 2 | 6 2 | | 6 2 | |
| | 9 2 | 8 2 | | 5 2 | |
| | 6 2 | 7 2 | | 6 2 | |
| | 7 2 | 6 2 | | 6 2 | |
| | 7 2 | 6 2 | | 6 2 | |
| | 6 2 | 7 2 | | 6 2 | |
| | 7 2 | 7 2 | | 5 2 | |
| | 7 2 | 8 2 | | 6 2 | |
| | 6 2 | 7 2 | | 5 2 | |
| | 9 2 | 9 2 | | 6 2 | |
| 140.2.11 | 418133 | | 100 | | Ca1Ni4Sn2 |
| | -GM | -M | -P | -X | -N |
| | 7 2 | 7 2 | 6+7 4 | 6 2 | 2+2 4 |
| | 6 2 | 6 2 | 6+7 4 | 5 2 | 2+2 4 |
| | 8 2 | 8 2 | 6+7 4 | 6 2 | 2+2 4 |
| | 9 2 | 9 2 | 6+7 4 | 5 2 | 2+2 4 |
| | 6 2 | 9 2 | 6+7 4 | 5 2 | 2+2 4 |
| | 8 2 | 6 2 | 6+7 4 | 6 2 | 2+2 4 |
| | 7 2 | 7 2 | 6+7 4 | 5 2 | 2+2 4 |
| | 8 2 | 8 2 | 6+7 4 | 6 2 | 2+2 4 |
| | 6 2 | 9 2 | 6+7 4 | 6 2 | 2+2 4 |

Continued on next column

**TABLE LXV– continued**

| TQC | ICSD | | NoE | | SF |
|---|---|---|---|---|---|
| | 9 2 | 7 2 | 6+7 4 | 5 2 | 2+2 4 |
| | 7 2 | 8 2 | 6+7 4 | 5 2 | 2+2 4 |
| | 7 2 | 7 2 | 6+7 4 | 6 2 | 2+2 4 |
| | 9 2 | 6 2 | 6+7 4 | 6 2 | 2+2 4 |
| | 9 2 | 6 2 | 6+7 4 | 5 2 | 2+2 4 |
| | 8 2 | 7 2 | 6+7 4 | 6 2 | 2+2 4 |
| | 8 2 | 9 2 | 6+7 4 | 6 2 | 2+2 4 |
| | 7 2 | 8 2 | 6+7 4 | 6 2 | 2+2 4 |
| | 6 2 | 8 2 | 6+7 4 | 5 2 | 2+2 4 |
| | 6 2 | 9 2 | 6+7 4 | 5 2 | 2+2 4 |
| | 9 2 | 7 2 | 6+7 4 | 6 2 | 2+2 4 |
| | 7 2 | 6 2 | 6+7 4 | 6 2 | 2+2 4 |
| | 8 2 | 6 2 | 6+7 4 | 5 2 | 2+2 4 |
| | 6 2 | 8 2 | 6+7 4 | 6 2 | 2+2 4 |
| | 9 2 | 6 2 | 6+7 4 | 5 2 | 2+2 4 |
| | 9 2 | 7 2 | 6+7 4 | 6 2 | 2+2 4 |
| | 7 2 | 9 2 | | 5 2 | |
| | 8 2 | 9 2 | | 5 2 | |
| | 6 2 | 8 2 | | 6 2 | |
| | 7 2 | 8 2 | | 5 2 | |
| | 9 2 | 6 2 | | 5 2 | |
| | 8 2 | 9 2 | | 6 2 | |
| | 7 2 | 7 2 | | 5 2 | |
| | 6 2 | 7 2 | | 6 2 | |
| | 6 2 | 9 2 | | 6 2 | |
| | 8 2 | 8 2 | | 5 2 | |
| | 7 2 | 6 2 | | 5 2 | |
| | 9 2 | 7 2 | | 6 2 | |
| | 8 2 | 9 2 | | 5 2 | |
| | 7 2 | 8 2 | | 5 2 | |
| | 6 2 | 6 2 | | 6 2 | |
| | 9 2 | 7 2 | | 6 2 | |
| | 9 2 | 6 2 | | 5 2 | |
| | 6 2 | 8 2 | | 6 2 | |
| | 8 2 | 9 2 | | 6 2 | |
| | 6 2 | 9 2 | | 5 2 | |
| | 7 2 | 7 2 | | 5 2 | |
| | 8 2 | 6 2 | | 6 2 | |
| | 8 2 | 8 2 | | 6 2 | |
| | 9 2 | 8 2 | | 5 2 | |
| | 7 2 | 9 2 | | 5 2 | |
| 140.2.12 | 150146 | | 68 | | Bi2Fe1Ti4 |
| | -GM | -M | -P | -X | -N |
| | 7 2 | 6 2 | 6+7 4 | 6 2 | 2+2 4 |
| | 9 2 | 8 2 | 6+7 4 | 5 2 | 2+2 4 |
| | 8 2 | 9 2 | 6+7 4 | 5 2 | 2+2 4 |
| | 6 2 | 7 2 | 6+7 4 | 6 2 | 2+2 4 |
| | 7 2 | 7 2 | 6+7 4 | 5 2 | 2+2 4 |
| | 8 2 | 9 2 | 6+7 4 | 6 2 | 2+2 4 |
| | 9 2 | 7 2 | 6+7 4 | 6 2 | 2+2 4 |
| | 7 2 | 8 2 | 6+7 4 | 5 2 | 2+2 4 |
| | 6 2 | 6 2 | 6+7 4 | 5 2 | 2+2 4 |
| | 7 2 | 9 2 | 6+7 4 | 5 2 | 2+2 4 |
| | 8 2 | 9 2 | 6+7 4 | 6 2 | 2+2 4 |
| | 6 2 | 6 2 | 6+7 4 | 6 2 | 2+2 4 |
| | 9 2 | 8 2 | 6+7 4 | 5 2 | 2+2 4 |
| | 8 2 | 8 2 | 6+7 4 | 6 2 | 2+2 4 |
| | 7 2 | 9 2 | 6+7 4 | 5 2 | 2+2 4 |
| | 9 2 | 9 2 | 6+7 4 | 6 2 | 2+2 4 |
| | 6 2 | 8 2 | 6+7 4 | 6 2 | 2+2 4 |
| | 7 2 | 7 2 | | 5 2 | |

Continued on next column

**TABLE LXV– continued**

| TQC | ICSD | NoE | SF |
|---|---|---|---|
| | 6 2 6 2 | | 6 2 |
| | 6 2 6 2 | | 5 2 |
| | 8 2 7 2 | | 6 2 |
| | 8 2 8 2 | | 6 2 |
| | 8 2 7 2 | | 6 2 |
| | 9 2 8 2 | | 5 2 |
| | 9 2 8 2 | | 5 2 |
| | 7 2 9 2 | | 6 2 |
| | 6 2 6 2 | | 6 2 |
| | 7 2 7 2 | | 5 2 |
| | 7 2 9 2 | | 6 2 |
| | 7 2 7 2 | | 5 2 |
| | 6 2 6 2 | | 5 2 |
| | 8 2 7 2 | | 5 2 |
| | 9 2 8 2 | | 6 2 |
| | 7 2 9 2 | | 5 2 |
| 140.2.13 | 428736 | 168 | Al1Eu3O5 |
| | -GM -M | -P -X | -N |
| | 7 2 7 2 | 6+7 4 5 2 | 2+2 4 |
| | 9 2 9 2 | 6+7 4 6 2 | 2+2 4 |
| | 7 2 7 2 | 6+7 4 6 2 | 2+2 4 |
| | 6 2 6 2 | 6+7 4 6 2 | 2+2 4 |
| | 8 2 8 2 | 6+7 4 5 2 | 2+2 4 |
| | 9 2 9 2 | 6+7 4 5 2 | 2+2 4 |
| | 7 2 9 2 | 6+7 4 5 2 | 2+2 4 |
| | 9 2 7 2 | 6+7 4 6 2 | 2+2 4 |
| | 6 2 6 2 | 6+7 4 5 2 | 2+2 4 |
| | 7 2 7 2 | 6+7 4 6 2 | 2+2 4 |
| | 8 2 8 2 | 6+7 4 6 2 | 2+2 4 |
| | 9 2 9 2 | 6+7 4 5 2 | 2+2 4 |
| | 6 2 6 2 | 6+7 4 5 2 | 2+2 4 |
| | 7 2 9 2 | 6+7 4 6 2 | 2+2 4 |
| | 9 2 9 2 | 6+7 4 5 2 | 2+2 4 |
| | 8 2 8 2 | 6+7 4 6 2 | 2+2 4 |
| | 9 2 7 2 | 6+7 4 5 2 | 2+2 4 |
| | 7 2 7 2 | 6+7 4 6 2 | 2+2 4 |
| | 6 2 8 2 | 6+7 4 5 2 | 2+2 4 |
| | 7 2 6 2 | 6+7 4 6 2 | 2+2 4 |
| | 8 2 7 2 | 6+7 4 5 2 | 2+2 4 |
| | 8 2 9 2 | 6+7 4 6 2 | 2+2 4 |
| | 7 2 6 2 | 6+7 4 5 2 | 2+2 4 |
| | 9 2 8 2 | 6+7 4 6 2 | 2+2 4 |
| | 6 2 7 2 | 6+7 4 5 2 | 2+2 4 |
| | 8 2 7 2 | 6+7 4 5 2 | 2+2 4 |
| | 7 2 8 2 | 6+7 4 6 2 | 2+2 4 |
| | 6 2 6 2 | 6+7 4 6 2 | 2+2 4 |
| | 8 2 6 2 | 6+7 4 6 2 | 2+2 4 |
| | 9 2 8 2 | 6+7 4 6 2 | 2+2 4 |
| | 7 2 9 2 | 6+7 4 5 2 | 2+2 4 |
| | 6 2 9 2 | 6+7 4 5 2 | 2+2 4 |
| | 9 2 7 2 | 6+7 4 5 2 | 2+2 4 |
| | 7 2 7 2 | 6+7 4 5 2 | 2+2 4 |
| | 7 2 6 2 | 6+7 4 6 2 | 2+2 4 |
| | 8 2 9 2 | 6+7 4 5 2 | 2+2 4 |
| | 6 2 8 2 | 6+7 4 5 2 | 2+2 4 |
| | 9 2 8 2 | 6+7 4 5 2 | 2+2 4 |
| | 8 2 9 2 | 6+7 4 6 2 | 2+2 4 |
| | 9 2 9 2 | 6+7 4 5 2 | 2+2 4 |
| | 9 2 7 2 | 6+7 4 6 2 | 2+2 4 |
| | 7 2 6 2 | 6+7 4 6 2 | 2+2 4 |
| | 6 2 8 2 | | 6 2 |

Continued on next column

**TABLE LXV– continued**

| TQC | ICSD | NoE | SF |
|---|---|---|---|
| | 7 2 7 2 | | 6 2 |
| | 6 2 9 2 | | 5 2 |
| | 8 2 7 2 | | 5 2 |
| | 9 2 7 2 | | 6 2 |
| | 8 2 9 2 | | 6 2 |
| | 6 2 8 2 | | 6 2 |
| | 7 2 8 2 | | 5 2 |
| | 6 2 7 2 | | 5 2 |
| | 8 2 6 2 | | 6 2 |
| | 7 2 6 2 | | 5 2 |
| | 6 2 9 2 | | 6 2 |
| | 7 2 8 2 | | 5 2 |
| | 8 2 6 2 | | 5 2 |
| | 9 2 9 2 | | 6 2 |
| | 9 2 8 2 | | 6 2 |
| | 8 2 9 2 | | 5 2 |
| | 9 2 6 2 | | 6 2 |
| | 9 2 7 2 | | 6 2 |
| | 9 2 8 2 | | 6 2 |
| | 9 2 9 2 | | 6 2 |
| | 8 2 9 2 | | 6 2 |
| | 6 2 6 2 | | 5 2 |
| | 8 2 8 2 | | 6 2 |
| | 6 2 7 2 | | 5 2 |
| | 8 2 6 2 | | 5 2 |
| | 7 2 9 2 | | 6 2 |
| | 9 2 6 2 | | 5 2 |
| | 6 2 8 2 | | 5 2 |
| | 6 2 7 2 | | 5 2 |
| | 9 2 6 2 | | 6 2 |
| | 7 2 9 2 | | 6 2 |
| | 8 2 8 2 | | 6 2 |
| | 9 2 6 2 | | 5 2 |
| | 6 2 7 2 | | 5 2 |
| | 7 2 9 2 | | 6 2 |
| | 7 2 8 2 | | 6 2 |
| | 8 2 7 2 | | 5 2 |
| | 9 2 9 2 | | 6 2 |
| | 8 2 8 2 | | 6 2 |
| | 6 2 6 2 | | 6 2 |
| | 9 2 7 2 | | 5 2 |
| 140.2.14 | 420321 | 216 | Al1Cs9O4 |
| | -GM -M | -P -X | -N |
| | 7 2 6 2 | 6+7 4 6 2 | 2+2 4 |
| | 8 2 9 2 | 6+7 4 5 2 | 2+2 4 |
| | 8 2 8 2 | 6+7 4 5 2 | 2+2 4 |
| | 9 2 9 2 | 6+7 4 5 2 | 2+2 4 |
| | 6 2 6 2 | 6+7 4 6 2 | 2+2 4 |
| | 7 2 7 2 | 6+7 4 6 2 | 2+2 4 |
| | 6 2 7 2 | 6+7 4 5 2 | 2+2 4 |
| | 9 2 8 2 | 6+7 4 6 2 | 2+2 4 |
| | 7 2 7 2 | 6+7 4 5 2 | 2+2 4 |
| | 7 2 7 2 | 6+7 4 5 2 | 2+2 4 |
| | 6 2 7 2 | 6+7 4 6 2 | 2+2 4 |
| | 7 2 6 2 | 6+7 4 6 2 | 2+2 4 |
| | 8 2 8 2 | 6+7 4 5 2 | 2+2 4 |
| | 9 2 9 2 | 6+7 4 5 2 | 2+2 4 |
| | 6 2 6 2 | 6+7 4 6 2 | 2+2 4 |
| | 7 2 7 2 | 6+7 4 6 2 | 2+2 4 |
| | 8 2 8 2 | 6+7 4 5 2 | 2+2 4 |
| | 9 2 9 2 | 6+7 4 5 2 | 2+2 4 |

Continued on next column

**TABLE LXV– continued**

| TQC | ICSD NoE SF |
|---|---|
| | 8 2 9 2 6+7 4 5 2 2+2 4 |
| | 7 2 6 2 6+7 4 6 2 2+2 4 |
| | 9 2 8 2 6+7 4 5 2 2+2 4 |
| | 6 2 7 2 6+7 4 5 2 2+2 4 |
| | 8 2 8 2 6+7 4 5 2 2+2 4 |
| | 9 2 9 2 6+7 4 6 2 2+2 4 |
| | 6 2 6 2 6+7 4 6 2 2+2 4 |
| | 7 2 7 2 6+7 4 6 2 2+2 4 |
| | 9 2 9 2 6+7 4 6 2 2+2 4 |
| | 9 2 9 2 6+7 4 6 2 2+2 4 |
| | 9 2 8 2 6+7 4 5 2 2+2 4 |
| | 7 2 6 2 6+7 4 6 2 2+2 4 |
| | 6 2 9 2 6+7 4 5 2 2+2 4 |
| | 8 2 7 2 6+7 4 6 2 2+2 4 |
| | 7 2 6 2 6+7 4 6 2 2+2 4 |
| | 6 2 7 2 6+7 4 6 2 2+2 4 |
| | 9 2 8 2 6+7 4 5 2 2+2 4 |
| | 8 2 9 2 6+7 4 5 2 2+2 4 |
| | 7 2 6 2 6+7 4 6 2 2+2 4 |
| | 9 2 8 2 6+7 4 5 2 2+2 4 |
| | 9 2 9 2 6+7 4 6 2 2+2 4 |
| | 8 2 9 2 6+7 4 6 2 2+2 4 |
| | 8 2 8 2 6+7 4 6 2 2+2 4 |
| | 9 2 8 2 6+7 4 6 2 2+2 4 |
| | 6 2 7 2 6+7 4 6 2 2+2 4 |
| | 8 2 9 2 6+7 4 5 2 2+2 4 |
| | 7 2 6 2 6+7 4 6 2 2+2 4 |
| | 8 2 9 2 6+7 4 5 2 2+2 4 |
| | 6 2 8 2 6+7 4 5 2 2+2 4 |
| | 9 2 7 2 6+7 4 6 2 2+2 4 |
| | 6 2 7 2 6+7 4 6 2 2+2 4 |
| | 9 2 8 2 6+7 4 5 2 2+2 4 |
| | 7 2 6 2 6+7 4 6 2 2+2 4 |
| | 8 2 7 2 6+7 4 5 2 2+2 4 |
| | 6 2 9 2 6+7 4 6 2 2+2 4 |
| | 8 2 9 2 6+7 4 5 2 2+2 4 |
| | 7 2 6 2 6 2 |
| | 9 2 8 2 5 2 |
| | 7 2 6 2 6 2 |
| | 6 2 7 2 6 2 |
| | 6 2 7 2 6 2 |
| | 9 2 8 2 5 2 |
| | 7 2 6 2 6 2 |
| | 8 2 9 2 5 2 |
| | 9 2 8 2 5 2 |
| | 8 2 9 2 5 2 |
| | 8 2 9 2 5 2 |
| | 6 2 7 2 6 2 |
| | 6 2 7 2 5 2 |
| | 9 2 8 2 6 2 |
| | 8 2 9 2 5 2 |
| | 7 2 6 2 6 2 |
| | 9 2 8 2 5 2 |
| | 7 2 6 2 6 2 |
| | 9 2 7 2 5 2 |
| | 6 2 8 2 6 2 |
| | 8 2 9 2 5 2 |
| | 7 2 6 2 6 2 |
| | 9 2 8 2 5 2 |
| | 8 2 9 2 5 2 |
| | 7 2 7 2 6 2 |

Continued on next column

**TABLE LXV– continued**

| TQC | ICSD NoE SF |
|---|---|
| | 6 2 6 2 6 2 |
| | 7 2 6 2 6 2 |
| | 8 2 9 2 5 2 |
| | 6 2 7 2 6 2 |
| | 9 2 8 2 5 2 |
| | 8 2 9 2 5 2 |
| | 6 2 7 2 6 2 |
| | 7 2 8 2 5 2 |
| | 9 2 6 2 6 2 |
| | 8 2 9 2 5 2 |
| | 9 2 8 2 5 2 |
| | 7 2 6 2 6 2 |
| | 6 2 7 2 6 2 |
| | 9 2 9 2 5 2 |
| | 8 2 8 2 5 2 |
| | 6 2 7 2 6 2 |
| | 9 2 8 2 5 2 |
| | 6 2 7 2 6 2 |
| | 7 2 6 2 6 2 |
| | 8 2 9 2 5 2 |
| | 8 2 9 2 5 2 |
| | 7 2 6 2 5 2 |
| | 9 2 8 2 6 2 |
| | 7 2 6 2 6 2 |
| | 6 2 7 2 6 2 |
| | 7 2 7 2 5 2 |
| | 7 2 8 2 6 2 |
| | 6 2 9 2 5 2 |
| | 9 2 7 2 6 2 |

The End

TABLE LXVI: The computed irreducible representations for the valence bands for the compounds of SEBR and NLC in space group 141. See the list of tables.

| TQC | ICSD | NoE | SF | | |
|---|---|---|---|---|---|
| 141.1.1 | 423244 | 64 | Si2Yb1 | | |
| | -GM | -M | -P | -X | -N |
| | 7 2 | 5 4 | 4+5 2 | 3+4 4 | 5+6 2 |
| | 8 2 | 5 4 | 7 2 | 3+4 4 | 3+4 2 |
| | 6 2 | 5 4 | 3+6 2 | 3+4 4 | 5+6 2 |
| | 9 2 | 5 4 | 7 2 | 3+4 4 | 3+4 2 |
| | 6 2 | 5 4 | 7 2 | 3+4 4 | 3+4 2 |
| | 7 2 | 5 4 | 4+5 2 | 3+4 4 | 5+6 2 |
| | 8 2 | 5 4 | 3+6 2 | 3+4 4 | 5+6 2 |
| | 9 2 | 5 4 | 7 2 | 3+4 4 | 3+4 2 |
| | 7 2 | 5 4 | 3+6 2 | 3+4 4 | 3+4 2 |
| | 6 2 | 5 4 | 7 2 | 3+4 4 | 5+6 2 |
| | 8 2 | 5 4 | 7 2 | 3+4 4 | 3+4 2 |
| | 6 2 | 5 4 | 4+5 2 | 3+4 4 | 5+6 2 |
| | 7 2 | 5 4 | 3+6 2 | 3+4 4 | 5+6 2 |
| | 7 2 | 5 4 | 7 2 | 3+4 4 | 3+4 2 |
| | 9 2 | 5 4 | 7 2 | 3+4 4 | 3+4 2 |
| | 8 2 | 5 4 | 4+5 2 | 3+4 4 | 5+6 2 |
| | 9 2 | | 3+6 2 | | 3+4 2 |
| | 8 2 | | 7 2 | | 5+6 2 |
| | 7 2 | | 4+5 2 | | 3+4 2 |
| | 8 2 | | 7 2 | | 3+4 2 |
| | 9 2 | | 7 2 | | 5+6 2 |
| | 6 2 | | 4+5 2 | | 5+6 2 |

Continued on next column

**TABLE LXVI– continued**

| TQC | ICSD | | NoE | | SF |
|---|---|---|---|---|---|
| | 7 2 | | 7 2 | | 3+4 2 |
| | 8 2 | | 3+6 2 | | 5+6 2 |
| | 6 2 | | 4+5 2 | | 3+4 2 |
| | 9 2 | | 7 2 | | 5+6 2 |
| | 7 2 | | 3+6 2 | | 3+4 2 |
| | 6 2 | | 7 2 | | 5+6 2 |
| | 8 2 | | 7 2 | | 3+4 2 |
| | 7 2 | | 4+5 2 | | 5+6 2 |
| | 9 2 | | 7 2 | | 3+4 2 |
| | 6 2 | | 3+6 2 | | 5+6 2 |
| 141.1.2 | 170062 | | 44 | | Sc1Sn2 |
| | -GM | -M | -P | -X | -N |
| | 7 2 | 5 4 | 4+5 2 | 3+4 4 | 5+6 2 |
| | 8 2 | 5 4 | 7 2 | 3+4 4 | 3+4 2 |
| | 6 2 | 5 4 | 3+6 2 | 3+4 4 | 5+6 2 |
| | 7 2 | 5 4 | 7 2 | 3+4 4 | 3+4 2 |
| | 9 2 | 5 4 | 3+6 2 | 3+4 4 | 5+6 2 |
| | 8 2 | 5 4 | 7 2 | 3+4 4 | 3+4 2 |
| | 6 2 | 5 4 | 4+5 2 | 3+4 4 | 5+6 2 |
| | 8 2 | 5 4 | 7 2 | 3+4 4 | 3+4 2 |
| | 7 2 | 5 4 | 7 2 | 3+4 4 | 3+4 2 |
| | 6 2 | 5 4 | 3+6 2 | 3+4 4 | 5+6 2 |
| | 9 2 | 5 4 | 4+5 2 | 3+4 4 | 3+4 2 |
| | 7 2 | | 7 2 | | 5+6 2 |
| | 8 2 | | 7 2 | | 5+6 2 |
| | 9 2 | | 4+5 2 | | 5+6 2 |
| | 6 2 | | 3+6 2 | | 3+4 2 |
| | 7 2 | | 7 2 | | 5+6 2 |
| | 9 2 | | 7 2 | | 3+4 2 |
| | 8 2 | | 3+6 2 | | 5+6 2 |
| | 7 2 | | 4+5 2 | | 3+4 2 |
| | 6 2 | | 7 2 | | 3+4 2 |
| | 9 2 | | 7 2 | | 3+4 2 |
| | 6 2 | | 3+6 2 | | 5+6 2 |
| 141.1.3 | 88368 | | 96 | | O4Tb1V1 |
| | -GM | -M | -P | -X | -N |
| | 7 2 | 5 4 | 4+5 2 | 3+4 4 | 5+6 2 |
| | 8 2 | 5 4 | 7 2 | 3+4 4 | 3+4 2 |
| | 6 2 | 5 4 | 3+6 2 | 3+4 4 | 5+6 2 |
| | 9 2 | 5 4 | 7 2 | 3+4 4 | 3+4 2 |
| | 7 2 | 5 4 | 7 2 | 3+4 4 | 5+6 2 |
| | 6 2 | 5 4 | 3+6 2 | 3+4 4 | 3+4 2 |
| | 9 2 | 5 4 | 4+5 2 | 3+4 4 | 5+6 2 |
| | 8 2 | 5 4 | 7 2 | 3+4 4 | 3+4 2 |
| | 7 2 | 5 4 | 7 2 | 3+4 4 | 5+6 2 |
| | 8 2 | 5 4 | 3+6 2 | 3+4 4 | 3+4 2 |
| | 9 2 | 5 4 | 4+5 2 | 3+4 4 | 5+6 2 |
| | 6 2 | 5 4 | 7 2 | 3+4 4 | 3+4 2 |
| | 8 2 | 5 4 | 7 2 | 3+4 4 | 3+4 2 |
| | 6 2 | 5 4 | 3+6 2 | 3+4 4 | 5+6 2 |
| | 9 2 | 5 4 | 4+5 2 | 3+4 4 | 3+4 2 |
| | 7 2 | 5 4 | 7 2 | 3+4 4 | 5+6 2 |
| | 8 2 | 5 4 | 4+5 2 | 3+4 4 | 5+6 2 |
| | 7 2 | 5 4 | 7 2 | 3+4 4 | 3+4 2 |
| | 6 2 | 5 4 | 3+6 2 | 3+4 4 | 5+6 2 |
| | 9 2 | 5 4 | 7 2 | 3+4 4 | 3+4 2 |
| | 6 2 | 5 4 | 3+6 2 | 3+4 4 | 3+4 2 |
| | 7 2 | 5 4 | 7 2 | 3+4 4 | 5+6 2 |
| | 7 2 | 5 4 | 4+5 2 | 3+4 4 | 5+6 2 |
| | 6 2 | 5 4 | 7 2 | 3+4 4 | 3+4 2 |
| | 7 2 | | 4+5 2 | | 5+6 2 |

Continued on next column

**TABLE LXVI– continued**

| TQC | ICSD | | NoE | | SF |
|---|---|---|---|---|---|
| | 9 2 | | 7 2 | | 5+6 2 |
| | 8 2 | | 7 2 | | 3+4 2 |
| | 9 2 | | 4+5 2 | | 3+4 2 |
| | 6 2 | | 3+6 2 | | 5+6 2 |
| | 8 2 | | 7 2 | | 3+4 2 |
| | 6 2 | | 7 2 | | 5+6 2 |
| | 9 2 | | 3+6 2 | | 3+4 2 |
| | 9 2 | | 4+5 2 | | 3+4 2 |
| | 8 2 | | 7 2 | | 5+6 2 |
| | 9 2 | | 3+6 2 | | 5+6 2 |
| | 6 2 | | 7 2 | | 5+6 2 |
| | 7 2 | | 7 2 | | 3+4 2 |
| | 8 2 | | 4+5 2 | | 3+4 2 |
| | 7 2 | | 7 2 | | 5+6 2 |
| | 8 2 | | 3+6 2 | | 3+4 2 |
| | 7 2 | | 4+5 2 | | 5+6 2 |
| | 8 2 | | 7 2 | | 3+4 2 |
| | 6 2 | | 7 2 | | 5+6 2 |
| | 8 2 | | 3+6 2 | | 3+4 2 |
| | 7 2 | | 4+5 2 | | 5+6 2 |
| | 9 2 | | 7 2 | | 3+4 2 |
| | 6 2 | | 4+5 2 | | 5+6 2 |
| | 7 2 | | 7 2 | | 3+4 2 |
| 141.1.4 | 78078 | | 96 | | O4Tb1V1 |
| | -GM | -M | -P | -X | -N |
| | 7 2 | 5 4 | 4+5 2 | 3+4 4 | 5+6 2 |
| | 8 2 | 5 4 | 7 2 | 3+4 4 | 3+4 2 |
| | 6 2 | 5 4 | 3+6 2 | 3+4 4 | 5+6 2 |
| | 9 2 | 5 4 | 7 2 | 3+4 4 | 3+4 2 |
| | 7 2 | 5 4 | 7 2 | 3+4 4 | 5+6 2 |
| | 6 2 | 5 4 | 3+6 2 | 3+4 4 | 3+4 2 |
| | 9 2 | 5 4 | 4+5 2 | 3+4 4 | 5+6 2 |
| | 8 2 | 5 4 | 7 2 | 3+4 4 | 3+4 2 |
| | 7 2 | 5 4 | 7 2 | 3+4 4 | 5+6 2 |
| | 8 2 | 5 4 | 3+6 2 | 3+4 4 | 3+4 2 |
| | 9 2 | 5 4 | 4+5 2 | 3+4 4 | 5+6 2 |
| | 6 2 | 5 4 | 7 2 | 3+4 4 | 3+4 2 |
| | 8 2 | 5 4 | 7 2 | 3+4 4 | 3+4 2 |
| | 6 2 | 5 4 | 3+6 2 | 3+4 4 | 5+6 2 |
| | 9 2 | 5 4 | 4+5 2 | 3+4 4 | 3+4 2 |
| | 7 2 | 5 4 | 7 2 | 3+4 4 | 5+6 2 |
| | 8 2 | 5 4 | 4+5 2 | 3+4 4 | 5+6 2 |
| | 7 2 | 5 4 | 7 2 | 3+4 4 | 3+4 2 |
| | 6 2 | 5 4 | 3+6 2 | 3+4 4 | 5+6 2 |
| | 9 2 | 5 4 | 7 2 | 3+4 4 | 3+4 2 |
| | 6 2 | 5 4 | 3+6 2 | 3+4 4 | 3+4 2 |
| | 7 2 | 5 4 | 7 2 | 3+4 4 | 5+6 2 |
| | 7 2 | 5 4 | 4+5 2 | 3+4 4 | 5+6 2 |
| | 6 2 | 5 4 | 7 2 | 3+4 4 | 3+4 2 |
| | 9 2 | | 4+5 2 | | 5+6 2 |
| | 7 2 | | 7 2 | | 5+6 2 |
| | 8 2 | | 7 2 | | 3+4 2 |
| | 9 2 | | 4+5 2 | | 3+4 2 |
| | 6 2 | | 3+6 2 | | 5+6 2 |
| | 8 2 | | 7 2 | | 3+4 2 |
| | 6 2 | | 7 2 | | 5+6 2 |
| | 9 2 | | 3+6 2 | | 3+4 2 |
| | 9 2 | | 4+5 2 | | 3+4 2 |
| | 8 2 | | 7 2 | | 5+6 2 |
| | 9 2 | | 3+6 2 | | 5+6 2 |
| | 6 2 | | 7 2 | | 5+6 2 |

Continued on next column

**TABLE LXVI– continued**

| TQC | ICSD | NoE | SF | | |
|---|---|---|---|---|---|
| | 7 2 | | 7 2 | | 3+4 2 |
| | 8 2 | | 4+5 2 | | 3+4 2 |
| | 7 2 | | 7 2 | | 5+6 2 |
| | 8 2 | | 3+6 2 | | 3+4 2 |
| | 7 2 | | 4+5 2 | | 5+6 2 |
| | 8 2 | | 7 2 | | 3+4 2 |
| | 6 2 | | 7 2 | | 5+6 2 |
| | 8 2 | | 3+6 2 | | 3+4 2 |
| | 7 2 | | 4+5 2 | | 5+6 2 |
| | 9 2 | | 7 2 | | 3+4 2 |
| | 6 2 | | 4+5 2 | | 5+6 2 |
| | 7 2 | | 7 2 | | 3+4 2 |
| 141.1.5 | 78076 | 84 | O4Pr1V1 | | |
| | -GM | -M | -P | -X | -N |
| | 7 2 | 5 4 | 3+6 2 | 3+4 4 | 3+4 2 |
| | 8 2 | 5 4 | 7 2 | 3+4 4 | 5+6 2 |
| | 6 2 | 5 4 | 4+5 2 | 3+4 4 | 3+4 2 |
| | 9 2 | 5 4 | 7 2 | 3+4 4 | 5+6 2 |
| | 7 2 | 5 4 | 7 2 | 3+4 4 | 3+4 2 |
| | 7 2 | 5 4 | 3+6 2 | 3+4 4 | 5+6 2 |
| | 6 2 | 5 4 | 4+5 2 | 3+4 4 | 3+4 2 |
| | 9 2 | 5 4 | 7 2 | 3+4 4 | 5+6 2 |
| | 8 2 | 5 4 | 7 2 | 3+4 4 | 3+4 2 |
| | 8 2 | 5 4 | 3+6 2 | 3+4 4 | 5+6 2 |
| | 9 2 | 5 4 | 4+5 2 | 3+4 4 | 5+6 2 |
| | 6 2 | 5 4 | 7 2 | 3+4 4 | 3+4 2 |
| | 8 2 | 5 4 | 7 2 | 3+4 4 | 5+6 2 |
| | 6 2 | 5 4 | 4+5 2 | 3+4 4 | 3+4 2 |
| | 9 2 | 5 4 | 3+6 2 | 3+4 4 | 5+6 2 |
| | 7 2 | 5 4 | 7 2 | 3+4 4 | 3+4 2 |
| | 8 2 | 5 4 | 7 2 | 3+4 4 | 3+4 2 |
| | 7 2 | 5 4 | 4+5 2 | 3+4 4 | 5+6 2 |
| | 6 2 | 5 4 | 3+6 2 | 3+4 4 | 3+4 2 |
| | 9 2 | 5 4 | 7 2 | 3+4 4 | 5+6 2 |
| | 6 2 | 5 4 | 4+5 2 | 3+4 4 | 5+6 2 |
| | 7 2 | | 7 2 | | 3+4 2 |
| | 7 2 | | 3+6 2 | | 5+6 2 |
| | 9 2 | | 7 2 | | 3+4 2 |
| | 8 2 | | 3+6 2 | | 3+4 2 |
| | 6 2 | | 7 2 | | 3+4 2 |
| | 7 2 | | 7 2 | | 5+6 2 |
| | 9 2 | | 3+6 2 | | 5+6 2 |
| | 8 2 | | 4+5 2 | | 5+6 2 |
| | 6 2 | | 7 2 | | 3+4 2 |
| | 6 2 | | 4+5 2 | | 5+6 2 |
| | 9 2 | | 7 2 | | 3+4 2 |
| | 9 2 | | 3+6 2 | | 5+6 2 |
| | 8 2 | | 7 2 | | 3+4 2 |
| | 9 2 | | 4+5 2 | | 3+4 2 |
| | 7 2 | | 7 2 | | 3+4 2 |
| | 7 2 | | 7 2 | | 5+6 2 |
| | 6 2 | | 3+6 2 | | 5+6 2 |
| | 8 2 | | 7 2 | | 3+4 2 |
| | 8 2 | | 4+5 2 | | 5+6 2 |
| | 7 2 | | 3+6 2 | | 3+4 2 |
| | 6 2 | | 7 2 | | 5+6 2 |
| 141.1.6 | 157327 | 84 | O4Pr1V1 | | |
| | -GM | -M | -P | -X | -N |
| | 7 2 | 5 4 | 3+6 2 | 3+4 4 | 3+4 2 |
| | 8 2 | 5 4 | 7 2 | 3+4 4 | 5+6 2 |
| | 6 2 | 5 4 | 4+5 2 | 3+4 4 | 3+4 2 |

Continued on next column

**TABLE LXVI– continued**

| TQC | ICSD | NoE | SF | | |
|---|---|---|---|---|---|
| | 9 2 | 5 4 | 7 2 | 3+4 4 | 5+6 2 |
| | 7 2 | 5 4 | 7 2 | 3+4 4 | 3+4 2 |
| | 7 2 | 5 4 | 3+6 2 | 3+4 4 | 5+6 2 |
| | 9 2 | 5 4 | 4+5 2 | 3+4 4 | 3+4 2 |
| | 6 2 | 5 4 | 7 2 | 3+4 4 | 5+6 2 |
| | 8 2 | 5 4 | 3+6 2 | 3+4 4 | 3+4 2 |
| | 8 2 | 5 4 | 7 2 | 3+4 4 | 5+6 2 |
| | 9 2 | 5 4 | 4+5 2 | 3+4 4 | 5+6 2 |
| | 6 2 | 5 4 | 7 2 | 3+4 4 | 3+4 2 |
| | 8 2 | 5 4 | 7 2 | 3+4 4 | 5+6 2 |
| | 6 2 | 5 4 | 4+5 2 | 3+4 4 | 3+4 2 |
| | 9 2 | 5 4 | 3+6 2 | 3+4 4 | 5+6 2 |
| | 7 2 | 5 4 | 7 2 | 3+4 4 | 3+4 2 |
| | 8 2 | 5 4 | 7 2 | 3+4 4 | 3+4 2 |
| | 6 2 | 5 4 | 3+6 2 | 3+4 4 | 5+6 2 |
| | 9 2 | 5 4 | 4+5 2 | 3+4 4 | 5+6 2 |
| | 7 2 | 5 4 | 7 2 | 3+4 4 | 3+4 2 |
| | 6 2 | 5 4 | 4+5 2 | 3+4 4 | 5+6 2 |
| | 7 2 | | 3+6 2 | | 3+4 2 |
| | 7 2 | | 7 2 | | 5+6 2 |
| | 9 2 | | 7 2 | | 3+4 2 |
| | 8 2 | | 7 2 | | 3+4 2 |
| | 9 2 | | 3+6 2 | | 3+4 2 |
| | 6 2 | | 7 2 | | 5+6 2 |
| | 7 2 | | 4+5 2 | | 5+6 2 |
| | 6 2 | | 3+6 2 | | 5+6 2 |
| | 8 2 | | 7 2 | | 3+4 2 |
| | 6 2 | | 4+5 2 | | 5+6 2 |
| | 9 2 | | 7 2 | | 3+4 2 |
| | 9 2 | | 3+6 2 | | 5+6 2 |
| | 8 2 | | 7 2 | | 3+4 2 |
| | 9 2 | | 4+5 2 | | 3+4 2 |
| | 8 2 | | 7 2 | | 3+4 2 |
| | 7 2 | | 7 2 | | 5+6 2 |
| | 7 2 | | 3+6 2 | | 5+6 2 |
| | 6 2 | | 4+5 2 | | 3+4 2 |
| | 8 2 | | 7 2 | | 5+6 2 |
| | 7 2 | | 3+6 2 | | 3+4 2 |
| | 6 2 | | 7 2 | | 5+6 2 |
| 141.1.7 | 79759 | 104 | O4P1Tm1 | | |
| | -GM | -M | -P | -X | -N |
| | 7 2 | 5 4 | 4+5 2 | 3+4 4 | 5+6 2 |
| | 8 2 | 5 4 | 7 2 | 3+4 4 | 3+4 2 |
| | 6 2 | 5 4 | 3+6 2 | 3+4 4 | 5+6 2 |
| | 9 2 | 5 4 | 7 2 | 3+4 4 | 3+4 2 |
| | 7 2 | 5 4 | 7 2 | 3+4 4 | 3+4 2 |
| | 9 2 | 5 4 | 3+6 2 | 3+4 4 | 5+6 2 |
| | 6 2 | 5 4 | 4+5 2 | 3+4 4 | 5+6 2 |
| | 7 2 | 5 4 | 7 2 | 3+4 4 | 3+4 2 |
| | 8 2 | 5 4 | 3+6 2 | 3+4 4 | 5+6 2 |
| | 8 2 | 5 4 | 7 2 | 3+4 4 | 3+4 2 |
| | 6 2 | 5 4 | 7 2 | 3+4 4 | 5+6 2 |
| | 9 2 | 5 4 | 4+5 2 | 3+4 4 | 3+4 2 |
| | 8 2 | 5 4 | 7 2 | 3+4 4 | 5+6 2 |
| | 9 2 | 5 4 | 4+5 2 | 3+4 4 | 3+4 2 |
| | 6 2 | 5 4 | 3+6 2 | 3+4 4 | 3+4 2 |
| | 7 2 | 5 4 | 7 2 | 3+4 4 | 5+6 2 |
| | 7 2 | 5 4 | 3+6 2 | 3+4 4 | 5+6 2 |
| | 8 2 | 5 4 | 7 2 | 3+4 4 | 3+4 2 |
| | 6 2 | 5 4 | 7 2 | 3+4 4 | 3+4 2 |
| | 6 2 | 5 4 | 4+5 2 | 3+4 4 | 5+6 2 |

Continued on next column

**TABLE LXVI– continued**

| TQC | ICSD | | NoE | SF | |
|---|---|---|---|---|---|
| | 7 2 | 5 4 | 7 2 | 3+4 4 | 3+4 2 |
| | 9 2 | 5 4 | 3+6 2 | 3+4 4 | 3+4 2 |
| | 9 2 | 5 4 | 7 2 | 3+4 4 | 5+6 2 |
| | 8 2 | 5 4 | 4+5 2 | 3+4 4 | 5+6 2 |
| | 8 2 | 5 4 | 3+6 2 | 3+4 4 | 5+6 2 |
| | 9 2 | 5 4 | 7 2 | 3+4 4 | 3+4 2 |
| | 6 2 | | 4+5 2 | | 3+4 2 |
| | 7 2 | | 7 2 | | 5+6 2 |
| | 6 2 | | 7 2 | | 5+6 2 |
| | 7 2 | | 3+6 2 | | 3+4 2 |
| | 8 2 | | 4+5 2 | | 5+6 2 |
| | 9 2 | | 7 2 | | 3+4 2 |
| | 6 2 | | 7 2 | | 5+6 2 |
| | 9 2 | | 4+5 2 | | 3+4 2 |
| | 8 2 | | 7 2 | | 5+6 2 |
| | 7 2 | | 3+6 2 | | 5+6 2 |
| | 9 2 | | 4+5 2 | | 3+4 2 |
| | 6 2 | | 7 2 | | 3+4 2 |
| | 7 2 | | 7 2 | | 5+6 2 |
| | 7 2 | | 3+6 2 | | 3+4 2 |
| | 8 2 | | 4+5 2 | | 5+6 2 |
| | 9 2 | | 7 2 | | 3+4 2 |
| | 8 2 | | 7 2 | | 5+6 2 |
| | 6 2 | | 4+5 2 | | 3+4 2 |
| | 7 2 | | 3+6 2 | | 5+6 2 |
| | 8 2 | | 7 2 | | 3+4 2 |
| | 6 2 | | 7 2 | | 5+6 2 |
| | 8 2 | | 4+5 2 | | 3+4 2 |
| | 7 2 | | 3+6 2 | | 5+6 2 |
| | 6 2 | | 7 2 | | 3+4 2 |
| | 9 2 | | 7 2 | | 5+6 2 |
| | 7 2 | | 3+6 2 | | 3+4 2 |
| 141.1.8 | 36053 | | 104 | O4P1Tm1 | |
| | -GM | -M | -P | -X | -N |
| | 7 2 | 5 4 | 3+6 2 | 3+4 4 | 3+4 2 |
| | 8 2 | 5 4 | 7 2 | 3+4 4 | 5+6 2 |
| | 6 2 | 5 4 | 4+5 2 | 3+4 4 | 3+4 2 |
| | 9 2 | 5 4 | 7 2 | 3+4 4 | 5+6 2 |
| | 7 2 | 5 4 | 7 2 | 3+4 4 | 5+6 2 |
| | 7 2 | 5 4 | 4+5 2 | 3+4 4 | 3+4 2 |
| | 9 2 | 5 4 | 3+6 2 | 3+4 4 | 3+4 2 |
| | 6 2 | 5 4 | 7 2 | 3+4 4 | 5+6 2 |
| | 8 2 | 5 4 | 4+5 2 | 3+4 4 | 3+4 2 |
| | 8 2 | 5 4 | 7 2 | 3+4 4 | 5+6 2 |
| | 6 2 | 5 4 | 7 2 | 3+4 4 | 3+4 2 |
| | 9 2 | 5 4 | 3+6 2 | 3+4 4 | 5+6 2 |
| | 8 2 | 5 4 | 7 2 | 3+4 4 | 3+4 2 |
| | 9 2 | 5 4 | 3+6 2 | 3+4 4 | 5+6 2 |
| | 6 2 | 5 4 | 4+5 2 | 3+4 4 | 5+6 2 |
| | 7 2 | 5 4 | 7 2 | 3+4 4 | 3+4 2 |
| | 7 2 | 5 4 | 4+5 2 | 3+4 4 | 3+4 2 |
| | 8 2 | 5 4 | 7 2 | 3+4 4 | 5+6 2 |
| | 6 2 | 5 4 | 7 2 | 3+4 4 | 5+6 2 |
| | 6 2 | 5 4 | 3+6 2 | 3+4 4 | 3+4 2 |
| | 7 2 | 5 4 | 4+5 2 | 3+4 4 | 5+6 2 |
| | 9 2 | 5 4 | 7 2 | 3+4 4 | 5+6 2 |
| | 9 2 | 5 4 | 7 2 | 3+4 4 | 3+4 2 |
| | 8 2 | 5 4 | 3+6 2 | 3+4 4 | 3+4 2 |
| | 8 2 | 5 4 | 4+5 2 | 3+4 4 | 3+4 2 |
| | 9 2 | 5 4 | 7 2 | 3+4 4 | 5+6 2 |
| | 6 2 | | 3+6 2 | | 5+6 2 |
| | 7 2 | | 7 2 | | 3+4 2 |
| | 6 2 | | 7 2 | | 5+6 2 |
| | 7 2 | | 4+5 2 | | 3+4 2 |
| | 8 2 | | 3+6 2 | | 3+4 2 |
| | 9 2 | | 7 2 | | 5+6 2 |
| | 6 2 | | 7 2 | | 3+4 2 |
| | 9 2 | | 3+6 2 | | 5+6 2 |
| | 8 2 | | 7 2 | | 3+4 2 |
| | 9 2 | | 4+5 2 | | 3+4 2 |
| | 6 2 | | 3+6 2 | | 5+6 2 |
| | 7 2 | | 7 2 | | 5+6 2 |
| | 7 2 | | 7 2 | | 3+4 2 |
| | 7 2 | | 3+6 2 | | 5+6 2 |
| | 8 2 | | 4+5 2 | | 3+4 2 |
| | 9 2 | | 7 2 | | 5+6 2 |
| | 8 2 | | 7 2 | | 3+4 2 |
| | 6 2 | | 3+6 2 | | 5+6 2 |
| | 7 2 | | 4+5 2 | | 3+4 2 |
| | 8 2 | | 7 2 | | 5+6 2 |
| | 6 2 | | 3+6 2 | | 3+4 2 |
| | 8 2 | | 7 2 | | 5+6 2 |
| | 7 2 | | 4+5 2 | | 3+4 2 |
| | 6 2 | | 7 2 | | 5+6 2 |
| | 9 2 | | 7 2 | | 3+4 2 |
| | 7 2 | | 4+5 2 | | 5+6 2 |
| 141.1.9 | 61683 | | 64 | Na1O2Tb1 | |
| | -GM | -M | -P | -X | -N |
| | 7 2 | 5 4 | 4+5 2 | 3+4 4 | 5+6 2 |
| | 8 2 | 5 4 | 7 2 | 3+4 4 | 3+4 2 |
| | 6 2 | 5 4 | 3+6 2 | 3+4 4 | 5+6 2 |
| | 9 2 | 5 4 | 7 2 | 3+4 4 | 3+4 2 |
| | 6 2 | 5 4 | 7 2 | 3+4 4 | 5+6 2 |
| | 7 2 | 5 4 | 4+5 2 | 3+4 4 | 3+4 2 |
| | 9 2 | 5 4 | 3+6 2 | 3+4 4 | 5+6 2 |
| | 8 2 | 5 4 | 7 2 | 3+4 4 | 3+4 2 |
| | 7 2 | 5 4 | 7 2 | 3+4 4 | 3+4 2 |
| | 8 2 | 5 4 | 4+5 2 | 3+4 4 | 5+6 2 |
| | 9 2 | 5 4 | 3+6 2 | 3+4 4 | 3+4 2 |
| | 6 2 | 5 4 | 7 2 | 3+4 4 | 5+6 2 |
| | 7 2 | 5 4 | 3+6 2 | 3+4 4 | 3+4 2 |
| | 6 2 | 5 4 | 7 2 | 3+4 4 | 5+6 2 |
| | 8 2 | 5 4 | 4+5 2 | 3+4 4 | 3+4 2 |
| | 9 2 | 5 4 | 7 2 | 3+4 4 | 5+6 2 |
| | 7 2 | | 3+6 2 | | 3+4 2 |
| | 8 2 | | 7 2 | | 3+4 2 |
| | 6 2 | | 7 2 | | 5+6 2 |
| | 6 2 | | 4+5 2 | | 5+6 2 |
| | 7 2 | | 3+6 2 | | 5+6 2 |
| | 9 2 | | 7 2 | | 5+6 2 |
| | 9 2 | | 7 2 | | 3+4 2 |
| | 8 2 | | 4+5 2 | | 3+4 2 |
| | 6 2 | | 4+5 2 | | 5+6 2 |
| | 8 2 | | 7 2 | | 3+4 2 |
| | 7 2 | | 7 2 | | 5+6 2 |
| | 7 2 | | 4+5 2 | | 3+4 2 |
| | 8 2 | | 3+6 2 | | 5+6 2 |
| | 9 2 | | 7 2 | | 3+4 2 |
| | 6 2 | | 7 2 | | 5+6 2 |
| | 7 2 | | 3+6 2 | | 3+4 2 |
| 141.1.10 | 61682 | | 52 | Na1O2Pr1 | |
| | -GM | -M | -P | -X | -N |

Continued on next column

**TABLE LXVI– continued**

| TQC | ICSD | | NoE | SF | |
|---|---|---|---|---|---|
| | 7 2 | 5 4 | 4+5 2 | 3+4 4 | 5+6 2 |
| | 8 2 | 5 4 | 7 2 | 3+4 4 | 3+4 2 |
| | 6 2 | 5 4 | 3+6 2 | 3+4 4 | 5+6 2 |
| | 9 2 | 5 4 | 7 2 | 3+4 4 | 3+4 2 |
| | 6 2 | 5 4 | 7 2 | 3+4 4 | 5+6 2 |
| | 9 2 | 5 4 | 4+5 2 | 3+4 4 | 3+4 2 |
| | 8 2 | 5 4 | 3+6 2 | 3+4 4 | 5+6 2 |
| | 7 2 | 5 4 | 7 2 | 3+4 4 | 3+4 2 |
| | 7 2 | 5 4 | 7 2 | 3+4 4 | 5+6 2 |
| | 8 2 | 5 4 | 3+6 2 | 3+4 4 | 3+4 2 |
| | 6 2 | 5 4 | 4+5 2 | 3+4 4 | 3+4 2 |
| | 9 2 | 5 4 | 7 2 | 3+4 4 | 5+6 2 |
| | 7 2 | 5 4 | 3+6 2 | 3+4 4 | 3+4 2 |
| | 6 2 | | 7 2 | | 5+6 2 |
| | 9 2 | | 3+6 2 | | 3+4 2 |
| | 8 2 | | 7 2 | | 5+6 2 |
| | 8 2 | | 4+5 2 | | 3+4 2 |
| | 7 2 | | 7 2 | | 3+4 2 |
| | 6 2 | | 7 2 | | 5+6 2 |
| | 9 2 | | 4+5 2 | | 5+6 2 |
| | 6 2 | | 3+6 2 | | 5+6 2 |
| | 7 2 | | 7 2 | | 3+4 2 |
| | 9 2 | | 7 2 | | 5+6 2 |
| | 8 2 | | 4+5 2 | | 3+4 2 |
| | 6 2 | | 4+5 2 | | 3+4 2 |
| | 7 2 | | 7 2 | | 5+6 2 |
| 141.1.11 | 410968 | | 40 | In2Zr1 | |
| | -GM | -M | -P | -X | -N |
| | 7 2 | 5 4 | 3+6 2 | 3+4 4 | 3+4 2 |
| | 8 2 | 5 4 | 7 2 | 3+4 4 | 5+6 2 |
| | 6 2 | 5 4 | 4+5 2 | 3+4 4 | 5+6 2 |
| | 7 2 | 5 4 | 7 2 | 3+4 4 | 3+4 2 |
| | 9 2 | 5 4 | 3+6 2 | 3+4 4 | 3+4 2 |
| | 8 2 | 5 4 | 7 2 | 3+4 4 | 5+6 2 |
| | 6 2 | 5 4 | 7 2 | 3+4 4 | 3+4 2 |
| | 9 2 | 5 4 | 4+5 2 | 3+4 4 | 5+6 2 |
| | 7 2 | 5 4 | 4+5 2 | 3+4 4 | 3+4 2 |
| | 6 2 | 5 4 | 7 2 | 3+4 4 | 5+6 2 |
| | 8 2 | | 3+6 2 | | 3+4 2 |
| | 6 2 | | 7 2 | | 3+4 2 |
| | 9 2 | | 7 2 | | 5+6 2 |
| | 7 2 | | 3+6 2 | | 5+6 2 |
| | 8 2 | | 7 2 | | 3+4 2 |
| | 6 2 | | 4+5 2 | | 3+4 2 |
| | 9 2 | | 4+5 2 | | 5+6 2 |
| | 7 2 | | 7 2 | | 5+6 2 |
| | 8 2 | | 7 2 | | 3+4 2 |
| | 7 2 | | 3+6 2 | | 5+6 2 |
| 141.1.12 | 61422 | | 44 | H1V2 | |
| | -GM | -M | -P | -X | -N |
| | 7 2 | 5 4 | 4+5 2 | 3+4 4 | 5+6 2 |
| | 6 2 | 5 4 | 7 2 | 3+4 4 | 3+4 2 |
| | 7 2 | 5 4 | 3+6 2 | 3+4 4 | 5+6 2 |
| | 6 2 | 5 4 | 7 2 | 3+4 4 | 5+6 2 |
| | 8 2 | 5 4 | 3+6 2 | 3+4 4 | 3+4 2 |
| | 8 2 | 5 4 | 7 2 | 3+4 4 | 5+6 2 |
| | 9 2 | 5 4 | 4+5 2 | 3+4 4 | 3+4 2 |
| | 9 2 | 5 4 | 7 2 | 3+4 4 | 3+4 2 |
| | 7 2 | 5 4 | 3+6 2 | 3+4 4 | 3+4 2 |
| | 7 2 | 5 4 | 7 2 | 3+4 4 | 5+6 2 |
| | 6 2 | 5 4 | 7 2 | 3+4 4 | 5+6 2 |

Continued on next column

**TABLE LXVI– continued**

| TQC | ICSD | | NoE | SF | |
|---|---|---|---|---|---|
| | 9 2 | | 4+5 2 | | 5+6 2 |
| | 7 2 | | 3+6 2 | | 3+4 2 |
| | 8 2 | | 7 2 | | 3+4 2 |
| | 9 2 | | 4+5 2 | | 5+6 2 |
| | 8 2 | | 7 2 | | 3+4 2 |
| | 6 2 | | 3+6 2 | | 5+6 2 |
| | 7 2 | | 7 2 | | 5+6 2 |
| | 6 2 | | 4+5 2 | | 3+4 2 |
| | 7 2 | | 7 2 | | 5+6 2 |
| | 6 2 | | 4+5 2 | | 3+4 2 |
| | 6 2 | | 7 2 | | 3+4 2 |
| 141.1.13 | 60511 | | 44 | H1V2 | |
| | -GM | -M | -P | -X | -N |
| | 7 2 | 5 4 | 4+5 2 | 3+4 4 | 5+6 2 |
| | 6 2 | 5 4 | 7 2 | 3+4 4 | 3+4 2 |
| | 7 2 | 5 4 | 3+6 2 | 3+4 4 | 5+6 2 |
| | 6 2 | 5 4 | 7 2 | 3+4 4 | 5+6 2 |
| | 8 2 | 5 4 | 3+6 2 | 3+4 4 | 3+4 2 |
| | 8 2 | 5 4 | 7 2 | 3+4 4 | 5+6 2 |
| | 9 2 | 5 4 | 4+5 2 | 3+4 4 | 3+4 2 |
| | 9 2 | 5 4 | 7 2 | 3+4 4 | 3+4 2 |
| | 7 2 | 5 4 | 7 2 | 3+4 4 | 3+4 2 |
| | 7 2 | 5 4 | 4+5 2 | 3+4 4 | 5+6 2 |
| | 6 2 | 5 4 | 3+6 2 | 3+4 4 | 5+6 2 |
| | 7 2 | | 7 2 | | 5+6 2 |
| | 9 2 | | 4+5 2 | | 3+4 2 |
| | 8 2 | | 7 2 | | 3+4 2 |
| | 8 2 | | 7 2 | | 3+4 2 |
| | 9 2 | | 3+6 2 | | 3+4 2 |
| | 6 2 | | 4+5 2 | | 5+6 2 |
| | 7 2 | | 7 2 | | 5+6 2 |
| | 6 2 | | 7 2 | | 5+6 2 |
| | 7 2 | | 3+6 2 | | 3+4 2 |
| | 6 2 | | 4+5 2 | | 3+4 2 |
| | 6 2 | | 7 2 | | 5+6 2 |
| 141.1.14 | 103986 | | 36 | Ga2Th1 | |
| | -GM | -M | -P | -X | -N |
| | 7 2 | 5 4 | 4+5 2 | 3+4 4 | 5+6 2 |
| | 8 2 | 5 4 | 7 2 | 3+4 4 | 3+4 2 |
| | 6 2 | 5 4 | 3+6 2 | 3+4 4 | 5+6 2 |
| | 9 2 | 5 4 | 7 2 | 3+4 4 | 3+4 2 |
| | 6 2 | 5 4 | 7 2 | 3+4 4 | 3+4 2 |
| | 7 2 | 5 4 | 4+5 2 | 3+4 4 | 5+6 2 |
| | 8 2 | 5 4 | 3+6 2 | 3+4 4 | 5+6 2 |
| | 9 2 | 5 4 | 7 2 | 3+4 4 | 3+4 2 |
| | 7 2 | 5 4 | 3+6 2 | 3+4 4 | 3+4 2 |
| | 6 2 | | 7 2 | | 5+6 2 |
| | 8 2 | | 7 2 | | 3+4 2 |
| | 9 2 | | 4+5 2 | | 5+6 2 |
| | 6 2 | | 3+6 2 | | 3+4 2 |
| | 7 2 | | 7 2 | | 5+6 2 |
| | 7 2 | | 7 2 | | 3+4 2 |
| | 9 2 | | 4+5 2 | | 5+6 2 |
| | 8 2 | | 3+6 2 | | 3+4 2 |
| | 6 2 | | 7 2 | | 5+6 2 |
| 141.1.15 | 100566 | | 124 | Cu4O3 | |
| | -GM | -M | -P | -X | -N |
| | 7 2 | 5 4 | 3+6 2 | 3+4 4 | 3+4 2 |
| | 6 2 | 5 4 | 7 2 | 3+4 4 | 5+6 2 |
| | 9 2 | 5 4 | 4+5 2 | 3+4 4 | 5+6 2 |
| | 8 2 | 5 4 | 7 2 | 3+4 4 | 3+4 2 |

Continued on next column

**TABLE LXVI– continued**

| TQC | ICSD | | NoE | SF | |
|---|---|---|---|---|---|
| | 8 2 | 5 4 | 3+6 2 | 3+4 4 | 5+6 2 |
| | 7 2 | 5 4 | 7 2 | 3+4 4 | 3+4 2 |
| | 6 2 | 5 4 | 4+5 2 | 3+4 4 | 3+4 2 |
| | 7 2 | 5 4 | 7 2 | 3+4 4 | 3+4 2 |
| | 6 2 | 5 4 | 7 2 | 3+4 4 | 5+6 2 |
| | 7 2 | 5 4 | 3+6 2 | 3+4 4 | 5+6 2 |
| | 7 2 | 5 4 | 4+5 2 | 3+4 4 | 5+6 2 |
| | 6 2 | 5 4 | 7 2 | 3+4 4 | 5+6 2 |
| | 6 2 | 5 4 | 3+6 2 | 3+4 4 | 3+4 2 |
| | 8 2 | 5 4 | 7 2 | 3+4 4 | 3+4 2 |
| | 9 2 | 5 4 | 7 2 | 3+4 4 | 3+4 2 |
| | 7 2 | 5 4 | 3+6 2 | 3+4 4 | 5+6 2 |
| | 6 2 | 5 4 | 4+5 2 | 3+4 4 | 5+6 2 |
| | 9 2 | 5 4 | 7 2 | 3+4 4 | 3+4 2 |
| | 8 2 | 5 4 | 7 2 | 3+4 4 | 3+4 2 |
| | 9 2 | 5 4 | 4+5 2 | 3+4 4 | 3+4 2 |
| | 8 2 | 5 4 | 7 2 | 3+4 4 | 5+6 2 |
| | 7 2 | 5 4 | 3+6 2 | 3+4 4 | 3+4 2 |
| | 9 2 | 5 4 | 4+5 2 | 3+4 4 | 5+6 2 |
| | 8 2 | 5 4 | 7 2 | 3+4 4 | 5+6 2 |
| | 9 2 | 5 4 | 4+5 2 | 3+4 4 | 5+6 2 |
| | 6 2 | 5 4 | 7 2 | 3+4 4 | 3+4 2 |
| | 6 2 | 5 4 | 3+6 2 | 3+4 4 | 5+6 2 |
| | 7 2 | 5 4 | 7 2 | 3+4 4 | 3+4 2 |
| | 6 2 | 5 4 | 4+5 2 | 3+4 4 | 5+6 2 |
| | 7 2 | 5 4 | 7 2 | 3+4 4 | 3+4 2 |
| | 6 2 | 5 4 | 7 2 | 3+4 4 | 3+4 2 |
| | 7 2 | | 7 2 | | 3+4 2 |
| | 7 2 | | 3+6 2 | | 3+4 2 |
| | 6 2 | | 4+5 2 | | 5+6 2 |
| | 7 2 | | 7 2 | | 3+4 2 |
| | 6 2 | | 3+6 2 | | 3+4 2 |
| | 7 2 | | 7 2 | | 5+6 2 |
| | 6 2 | | 4+5 2 | | 3+4 2 |
| | 7 2 | | 3+6 2 | | 5+6 2 |
| | 6 2 | | 7 2 | | 3+4 2 |
| | 7 2 | | 3+6 2 | | 5+6 2 |
| | 7 2 | | 7 2 | | 3+4 2 |
| | 6 2 | | 4+5 2 | | 5+6 2 |
| | 7 2 | | 7 2 | | 3+4 2 |
| | 6 2 | | 3+6 2 | | 5+6 2 |
| | 6 2 | | 7 2 | | 5+6 2 |
| | 6 2 | | 4+5 2 | | 3+4 2 |
| | 7 2 | | 7 2 | | 5+6 2 |
| | 7 2 | | 3+6 2 | | 3+4 2 |
| | 6 2 | | 7 2 | | 5+6 2 |
| | 7 2 | | 4+5 2 | | 3+4 2 |
| | 6 2 | | 7 2 | | 5+6 2 |
| | 7 2 | | 4+5 2 | | 5+6 2 |
| | 6 2 | | 7 2 | | 5+6 2 |
| | 7 2 | | 7 2 | | 3+4 2 |
| | 6 2 | | 7 2 | | 5+6 2 |
| | 7 2 | | 3+6 2 | | 5+6 2 |
| | 6 2 | | 4+5 2 | | 3+4 2 |
| | 7 2 | | 3+6 2 | | 5+6 2 |
| | 7 2 | | 7 2 | | 5+6 2 |
| | 6 2 | | 3+6 2 | | 3+4 2 |
| | 6 2 | | 7 2 | | 5+6 2 |
| 141.1.16 | 35419 | | 124 | Cu2P1Ti1 | |
| | -GM | -M | -P | -X | -N |
| | 7 2 | 5 4 | 3+6 2 | 3+4 4 | 3+4 2 |

Continued on next column

**TABLE LXVI– continued**

| TQC | ICSD | | NoE | SF | |
|---|---|---|---|---|---|
| | 8 2 | 5 4 | 4+5 2 | 3+4 4 | 3+4 2 |
| | 6 2 | 5 4 | 7 2 | 3+4 4 | 5+6 2 |
| | 9 2 | 5 4 | 7 2 | 3+4 4 | 5+6 2 |
| | 7 2 | 5 4 | 4+5 2 | 3+4 4 | 5+6 2 |
| | 6 2 | 5 4 | 7 2 | 3+4 4 | 3+4 2 |
| | 8 2 | 5 4 | 3+6 2 | 3+4 4 | 5+6 2 |
| | 9 2 | 5 4 | 7 2 | 3+4 4 | 3+4 2 |
| | 7 2 | 5 4 | 3+6 2 | 3+4 4 | 3+4 2 |
| | 6 2 | 5 4 | 7 2 | 3+4 4 | 5+6 2 |
| | 8 2 | 5 4 | 4+5 2 | 3+4 4 | 3+4 2 |
| | 9 2 | 5 4 | 7 2 | 3+4 4 | 5+6 2 |
| | 7 2 | 5 4 | 3+6 2 | 3+4 4 | 5+6 2 |
| | 6 2 | 5 4 | 7 2 | 3+4 4 | 3+4 2 |
| | 8 2 | 5 4 | 4+5 2 | 3+4 4 | 5+6 2 |
| | 9 2 | 5 4 | 7 2 | 3+4 4 | 3+4 2 |
| | 8 2 | 5 4 | 4+5 2 | 3+4 4 | 3+4 2 |
| | 9 2 | 5 4 | 7 2 | 3+4 4 | 3+4 2 |
| | 9 2 | 5 4 | 7 2 | 3+4 4 | 5+6 2 |
| | 8 2 | 5 4 | 3+6 2 | 3+4 4 | 5+6 2 |
| | 6 2 | 5 4 | 3+6 2 | 3+4 4 | 3+4 2 |
| | 7 2 | 5 4 | 7 2 | 3+4 4 | 3+4 2 |
| | 6 2 | 5 4 | 7 2 | 3+4 4 | 5+6 2 |
| | 7 2 | 5 4 | 4+5 2 | 3+4 4 | 3+4 2 |
| | 8 2 | 5 4 | 7 2 | 3+4 4 | 3+4 2 |
| | 9 2 | 5 4 | 4+5 2 | 3+4 4 | 5+6 2 |
| | 7 2 | 5 4 | 3+6 2 | 3+4 4 | 5+6 2 |
| | 6 2 | 5 4 | 7 2 | 3+4 4 | 5+6 2 |
| | 7 2 | 5 4 | 7 2 | 3+4 4 | 5+6 2 |
| | 8 2 | 5 4 | 3+6 2 | 3+4 4 | 3+4 2 |
| | 7 2 | 5 4 | 4+5 2 | 3+4 4 | 5+6 2 |
| | 6 2 | | 7 2 | | 5+6 2 |
| | 6 2 | | 4+5 2 | | 3+4 2 |
| | 7 2 | | 7 2 | | 3+4 2 |
| | 6 2 | | 7 2 | | 5+6 2 |
| | 9 2 | | 3+6 2 | | 3+4 2 |
| | 7 2 | | 7 2 | | 3+4 2 |
| | 6 2 | | 3+6 2 | | 5+6 2 |
| | 8 2 | | 4+5 2 | | 5+6 2 |
| | 9 2 | | 7 2 | | 3+4 2 |
| | 9 2 | | 7 2 | | 5+6 2 |
| | 8 2 | | 4+5 2 | | 3+4 2 |
| | 6 2 | | 3+6 2 | | 5+6 2 |
| | 7 2 | | 7 2 | | 5+6 2 |
| | 9 2 | | 4+5 2 | | 3+4 2 |
| | 8 2 | | 7 2 | | 3+4 2 |
| | 8 2 | | 3+6 2 | | 3+4 2 |
| | 9 2 | | 7 2 | | 5+6 2 |
| | 6 2 | | 7 2 | | 5+6 2 |
| | 7 2 | | 3+6 2 | | 3+4 2 |
| | 8 2 | | 4+5 2 | | 3+4 2 |
| | 7 2 | | 7 2 | | 3+4 2 |
| | 8 2 | | 7 2 | | 5+6 2 |
| | 9 2 | | 3+6 2 | | 5+6 2 |
| | 6 2 | | 4+5 2 | | 3+4 2 |
| | 7 2 | | 7 2 | | 5+6 2 |
| | 7 2 | | 4+5 2 | | 5+6 2 |
| | 6 2 | | 7 2 | | 3+4 2 |
| | 8 2 | | 7 2 | | 5+6 2 |
| | 9 2 | | 3+6 2 | | 3+4 2 |
| | 6 2 | | 3+6 2 | | 3+4 2 |
| | 9 2 | | 7 2 | | 5+6 2 |

Continued on next column

**TABLE LXVI– continued**

| TQC | ICSD | NoE | SF | | |
|---|---|---|---|---|---|
| 141.1.17 | 100796 | 100 | Cu2Ge1O4 | | |
| | -GM | -M | -P | -X | -N |
| | 7 2 | 5 4 | 4+5 2 | 3+4 4 | 5+6 2 |
| | 8 2 | 5 4 | 7 2 | 3+4 4 | 3+4 2 |
| | 6 2 | 5 4 | 3+6 2 | 3+4 4 | 3+4 2 |
| | 9 2 | 5 4 | 7 2 | 3+4 4 | 5+6 2 |
| | 8 2 | 5 4 | 7 2 | 3+4 4 | 3+4 2 |
| | 9 2 | 5 4 | 4+5 2 | 3+4 4 | 5+6 2 |
| | 6 2 | 5 4 | 3+6 2 | 3+4 4 | 3+4 2 |
| | 7 2 | 5 4 | 7 2 | 3+4 4 | 5+6 2 |
| | 7 2 | 5 4 | 4+5 2 | 3+4 4 | 3+4 2 |
| | 8 2 | 5 4 | 7 2 | 3+4 4 | 5+6 2 |
| | 6 2 | 5 4 | 3+6 2 | 3+4 4 | 3+4 2 |
| | 6 2 | 5 4 | 7 2 | 3+4 4 | 5+6 2 |
| | 7 2 | 5 4 | 7 2 | 3+4 4 | 3+4 2 |
| | 9 2 | 5 4 | 4+5 2 | 3+4 4 | 5+6 2 |
| | 9 2 | 5 4 | 7 2 | 3+4 4 | 3+4 2 |
| | 8 2 | 5 4 | 3+6 2 | 3+4 4 | 5+6 2 |
| | 7 2 | 5 4 | 4+5 2 | 3+4 4 | 3+4 2 |
| | 6 2 | 5 4 | 7 2 | 3+4 4 | 3+4 2 |
| | 9 2 | 5 4 | 3+6 2 | 3+4 4 | 3+4 2 |
| | 8 2 | 5 4 | 7 2 | 3+4 4 | 5+6 2 |
| | 6 2 | 5 4 | 7 2 | 3+4 4 | 3+4 2 |
| | 7 2 | 5 4 | 3+6 2 | 3+4 4 | 5+6 2 |
| | 6 2 | 5 4 | 7 2 | 3+4 4 | 3+4 2 |
| | 7 2 | 5 4 | 4+5 2 | 3+4 4 | 3+4 2 |
| | 8 2 | 5 4 | 4+5 2 | 3+4 4 | 5+6 2 |
| | 6 2 | | 7 2 | | 5+6 2 |
| | 7 2 | | 7 2 | | 5+6 2 |
| | 7 2 | | 3+6 2 | | 3+4 2 |
| | 8 2 | | 7 2 | | 3+4 2 |
| | 9 2 | | 3+6 2 | | 5+6 2 |
| | 6 2 | | 7 2 | | 5+6 2 |
| | 6 2 | | 4+5 2 | | 5+6 2 |
| | 9 2 | | 3+6 2 | | 3+4 2 |
| | 8 2 | | 7 2 | | 3+4 2 |
| | 9 2 | | 7 2 | | 3+4 2 |
| | 7 2 | | 4+5 2 | | 3+4 2 |
| | 6 2 | | 3+6 2 | | 5+6 2 |
| | 7 2 | | 7 2 | | 3+4 2 |
| | 6 2 | | 4+5 2 | | 3+4 2 |
| | 7 2 | | 7 2 | | 5+6 2 |
| | 7 2 | | 7 2 | | 3+4 2 |
| | 6 2 | | 3+6 2 | | 5+6 2 |
| | 6 2 | | 4+5 2 | | 3+4 2 |
| | 7 2 | | 7 2 | | 3+4 2 |
| | 7 2 | | 3+6 2 | | 3+4 2 |
| | 7 2 | | 7 2 | | 3+4 2 |
| | 6 2 | | 4+5 2 | | 5+6 2 |
| | 6 2 | | 7 2 | | 3+4 2 |
| | 6 2 | | 7 2 | | 3+4 2 |
| | 7 2 | | 4+5 2 | | 3+4 2 |
| 141.1.18 | 84377 | 92 | Cr2Ni1O4 | | |
| | -GM | -M | -P | -X | -N |
| | 7 2 | 5 4 | 4+5 2 | 3+4 4 | 5+6 2 |
| | 6 2 | 5 4 | 7 2 | 3+4 4 | 3+4 2 |
| | 6 2 | 5 4 | 3+6 2 | 3+4 4 | 5+6 2 |
| | 7 2 | 5 4 | 7 2 | 3+4 4 | 5+6 2 |
| | 8 2 | 5 4 | 7 2 | 3+4 4 | 5+6 2 |
| | 9 2 | 5 4 | 3+6 2 | 3+4 4 | 3+4 2 |
| | 8 2 | 5 4 | 4+5 2 | 3+4 4 | 3+4 2 |

Continued on next column

**TABLE LXVI– continued**

| TQC | ICSD | NoE | SF | | |
|---|---|---|---|---|---|
| | 9 2 | 5 4 | 7 2 | 3+4 4 | 3+4 2 |
| | 7 2 | 5 4 | 4+5 2 | 3+4 4 | 5+6 2 |
| | 6 2 | 5 4 | 7 2 | 3+4 4 | 3+4 2 |
| | 7 2 | 5 4 | 3+6 2 | 3+4 4 | 5+6 2 |
| | 6 2 | 5 4 | 7 2 | 3+4 4 | 5+6 2 |
| | 8 2 | 5 4 | 3+6 2 | 3+4 4 | 5+6 2 |
| | 8 2 | 5 4 | 7 2 | 3+4 4 | 3+4 2 |
| | 9 2 | 5 4 | 4+5 2 | 3+4 4 | 3+4 2 |
| | 7 2 | 5 4 | 7 2 | 3+4 4 | 5+6 2 |
| | 6 2 | 5 4 | 4+5 2 | 3+4 4 | 5+6 2 |
| | 6 2 | 5 4 | 7 2 | 3+4 4 | 3+4 2 |
| | 7 2 | 5 4 | 3+6 2 | 3+4 4 | 3+4 2 |
| | 6 2 | 5 4 | 7 2 | 3+4 4 | 3+4 2 |
| | 8 2 | 5 4 | 7 2 | 3+4 4 | 5+6 2 |
| | 9 2 | 5 4 | 3+6 2 | 3+4 4 | 3+4 2 |
| | 7 2 | 5 4 | 4+5 2 | 3+4 4 | 5+6 2 |
| | 6 2 | | 7 2 | | 5+6 2 |
| | 9 2 | | 4+5 2 | | 5+6 2 |
| | 7 2 | | 7 2 | | 3+4 2 |
| | 8 2 | | 3+6 2 | | 3+4 2 |
| | 8 2 | | 7 2 | | 5+6 2 |
| | 9 2 | | 7 2 | | 5+6 2 |
| | 9 2 | | 3+6 2 | | 3+4 2 |
| | 9 2 | | 4+5 2 | | 3+4 2 |
| | 8 2 | | 7 2 | | 3+4 2 |
| | 7 2 | | 7 2 | | 5+6 2 |
| | 6 2 | | 4+5 2 | | 5+6 2 |
| | 7 2 | | 7 2 | | 5+6 2 |
| | 8 2 | | 3+6 2 | | 3+4 2 |
| | 6 2 | | 4+5 2 | | 3+4 2 |
| | 7 2 | | 7 2 | | 5+6 2 |
| | 9 2 | | 3+6 2 | | 5+6 2 |
| | 6 2 | | 7 2 | | 3+4 2 |
| | 7 2 | | 4+5 2 | | 5+6 2 |
| | 6 2 | | 7 2 | | 5+6 2 |
| | 6 2 | | 7 2 | | 3+4 2 |
| | 7 2 | | 3+6 2 | | 3+4 2 |
| | 6 2 | | 7 2 | | 5+6 2 |
| | 9 2 | | 4+5 2 | | 3+4 2 |
| 141.1.19 | 31300 | 92 | Cr2Ni1O4 | | |
| | -GM | -M | -P | -X | -N |
| | 7 2 | 5 4 | 3+6 2 | 3+4 4 | 3+4 2 |
| | 6 2 | 5 4 | 7 2 | 3+4 4 | 5+6 2 |
| | 6 2 | 5 4 | 4+5 2 | 3+4 4 | 3+4 2 |
| | 7 2 | 5 4 | 7 2 | 3+4 4 | 3+4 2 |
| | 8 2 | 5 4 | 3+6 2 | 3+4 4 | 3+4 2 |
| | 8 2 | 5 4 | 7 2 | 3+4 4 | 5+6 2 |
| | 9 2 | 5 4 | 7 2 | 3+4 4 | 5+6 2 |
| | 9 2 | 5 4 | 4+5 2 | 3+4 4 | 5+6 2 |
| | 7 2 | 5 4 | 3+6 2 | 3+4 4 | 3+4 2 |
| | 6 2 | 5 4 | 7 2 | 3+4 4 | 5+6 2 |
| | 7 2 | 5 4 | 4+5 2 | 3+4 4 | 3+4 2 |
| | 6 2 | 5 4 | 7 2 | 3+4 4 | 3+4 2 |
| | 8 2 | 5 4 | 4+5 2 | 3+4 4 | 3+4 2 |
| | 8 2 | 5 4 | 7 2 | 3+4 4 | 5+6 2 |
| | 9 2 | 5 4 | 3+6 2 | 3+4 4 | 5+6 2 |
| | 7 2 | 5 4 | 7 2 | 3+4 4 | 3+4 2 |
| | 6 2 | 5 4 | 3+6 2 | 3+4 4 | 3+4 2 |
| | 6 2 | 5 4 | 7 2 | 3+4 4 | 5+6 2 |
| | 7 2 | 5 4 | 4+5 2 | 3+4 4 | 3+4 2 |
| | 6 2 | 5 4 | 7 2 | 3+4 4 | 5+6 2 |

Continued on next column

**TABLE LXVI– continued**

| TQC | ICSD | | NoE | SF | |
|---|---|---|---|---|---|
| | 7 2 | 5 4 | 3+6 2 | 3+4 4 | 3+4 2 |
| | 6 2 | 5 4 | 7 2 | 3+4 4 | 5+6 2 |
| | 7 2 | 5 4 | 7 2 | 3+4 4 | 3+4 2 |
| | 8 2 | | 4+5 2 | | 5+6 2 |
| | 9 2 | | 3+6 2 | | 3+4 2 |
| | 9 2 | | 7 2 | | 5+6 2 |
| | 8 2 | | 4+5 2 | | 5+6 2 |
| | 8 2 | | 7 2 | | 3+4 2 |
| | 9 2 | | 7 2 | | 3+4 2 |
| | 9 2 | | 4+5 2 | | 5+6 2 |
| | 9 2 | | 7 2 | | 5+6 2 |
| | 8 2 | | 3+6 2 | | 5+6 2 |
| | 7 2 | | 7 2 | | 3+4 2 |
| | 6 2 | | 3+6 2 | | 3+4 2 |
| | 7 2 | | 7 2 | | 5+6 2 |
| | 6 2 | | 4+5 2 | | 3+4 2 |
| | 8 2 | | 3+6 2 | | 5+6 2 |
| | 7 2 | | 7 2 | | 3+4 2 |
| | 9 2 | | 4+5 2 | | 3+4 2 |
| | 6 2 | | 7 2 | | 5+6 2 |
| | 6 2 | | 3+6 2 | | 3+4 2 |
| | 7 2 | | 7 2 | | 5+6 2 |
| | 6 2 | | 7 2 | | 3+4 2 |
| | 7 2 | | 4+5 2 | | 5+6 2 |
| | 6 2 | | 3+6 2 | | 5+6 2 |
| | 9 2 | | 7 2 | | 3+4 2 |
| 141.1.20 | 31299 | | 92 | Cr2Ni1O4 | |
| | -GM | -M | -P | -X | -N |
| | 7 2 | 5 4 | 3+6 2 | 3+4 4 | 3+4 2 |
| | 6 2 | 5 4 | 7 2 | 3+4 4 | 5+6 2 |
| | 6 2 | 5 4 | 4+5 2 | 3+4 4 | 3+4 2 |
| | 7 2 | 5 4 | 7 2 | 3+4 4 | 3+4 2 |
| | 8 2 | 5 4 | 7 2 | 3+4 4 | 3+4 2 |
| | 8 2 | 5 4 | 4+5 2 | 3+4 4 | 5+6 2 |
| | 9 2 | 5 4 | 3+6 2 | 3+4 4 | 5+6 2 |
| | 9 2 | 5 4 | 7 2 | 3+4 4 | 5+6 2 |
| | 7 2 | 5 4 | 3+6 2 | 3+4 4 | 3+4 2 |
| | 6 2 | 5 4 | 7 2 | 3+4 4 | 5+6 2 |
| | 7 2 | 5 4 | 4+5 2 | 3+4 4 | 3+4 2 |
| | 6 2 | 5 4 | 7 2 | 3+4 4 | 3+4 2 |
| | 8 2 | 5 4 | 4+5 2 | 3+4 4 | 3+4 2 |
| | 8 2 | 5 4 | 7 2 | 3+4 4 | 5+6 2 |
| | 9 2 | 5 4 | 3+6 2 | 3+4 4 | 5+6 2 |
| | 7 2 | 5 4 | 7 2 | 3+4 4 | 3+4 2 |
| | 6 2 | 5 4 | 3+6 2 | 3+4 4 | 3+4 2 |
| | 6 2 | 5 4 | 7 2 | 3+4 4 | 5+6 2 |
| | 7 2 | 5 4 | 4+5 2 | 3+4 4 | 3+4 2 |
| | 6 2 | 5 4 | 7 2 | 3+4 4 | 5+6 2 |
| | 7 2 | 5 4 | 3+6 2 | 3+4 4 | 3+4 2 |
| | 6 2 | 5 4 | 7 2 | 3+4 4 | 5+6 2 |
| | 7 2 | 5 4 | 7 2 | 3+4 4 | 3+4 2 |
| | 8 2 | | 4+5 2 | | 5+6 2 |
| | 9 2 | | 3+6 2 | | 3+4 2 |
| | 9 2 | | 7 2 | | 5+6 2 |
| | 8 2 | | 4+5 2 | | 5+6 2 |
| | 8 2 | | 7 2 | | 3+4 2 |
| | 9 2 | | 7 2 | | 3+4 2 |
| | 9 2 | | 4+5 2 | | 5+6 2 |
| | 9 2 | | 3+6 2 | | 5+6 2 |
| | 8 2 | | 7 2 | | 5+6 2 |
| | 7 2 | | 7 2 | | 3+4 2 |

Continued on next column

| TQC | ICSD | | NoE | SF | |
|---|---|---|---|---|---|
| | 6 2 | | 3+6 2 | | 3+4 2 |
| | 7 2 | | 7 2 | | 5+6 2 |
| | 6 2 | | 4+5 2 | | 3+4 2 |
| | 8 2 | | 3+6 2 | | 5+6 2 |
| | 9 2 | | 7 2 | | 3+4 2 |
| | 7 2 | | 4+5 2 | | 3+4 2 |
| | 6 2 | | 7 2 | | 5+6 2 |
| | 6 2 | | 3+6 2 | | 3+4 2 |
| | 7 2 | | 7 2 | | 3+4 2 |
| | 6 2 | | 7 2 | | 5+6 2 |
| | 7 2 | | 4+5 2 | | 5+6 2 |
| | 6 2 | | 7 2 | | 5+6 2 |
| | 9 2 | | 3+6 2 | | 3+4 2 |
| 141.1.21 | 280062 | | 92 | Cr2Ni1O4 | |
| | -GM | -M | -P | -X | -N |
| | 7 2 | 5 4 | 3+6 2 | 3+4 4 | 3+4 2 |
| | 6 2 | 5 4 | 7 2 | 3+4 4 | 5+6 2 |
| | 6 2 | 5 4 | 4+5 2 | 3+4 4 | 3+4 2 |
| | 7 2 | 5 4 | 7 2 | 3+4 4 | 3+4 2 |
| | 8 2 | 5 4 | 7 2 | 3+4 4 | 3+4 2 |
| | 8 2 | 5 4 | 4+5 2 | 3+4 4 | 5+6 2 |
| | 9 2 | 5 4 | 3+6 2 | 3+4 4 | 5+6 2 |
| | 9 2 | 5 4 | 7 2 | 3+4 4 | 5+6 2 |
| | 7 2 | 5 4 | 3+6 2 | 3+4 4 | 3+4 2 |
| | 6 2 | 5 4 | 7 2 | 3+4 4 | 5+6 2 |
| | 7 2 | 5 4 | 4+5 2 | 3+4 4 | 3+4 2 |
| | 6 2 | 5 4 | 7 2 | 3+4 4 | 3+4 2 |
| | 8 2 | 5 4 | 4+5 2 | 3+4 4 | 3+4 2 |
| | 8 2 | 5 4 | 7 2 | 3+4 4 | 5+6 2 |
| | 9 2 | 5 4 | 3+6 2 | 3+4 4 | 5+6 2 |
| | 7 2 | 5 4 | 7 2 | 3+4 4 | 3+4 2 |
| | 6 2 | 5 4 | 3+6 2 | 3+4 4 | 3+4 2 |
| | 6 2 | 5 4 | 7 2 | 3+4 4 | 5+6 2 |
| | 7 2 | 5 4 | 4+5 2 | 3+4 4 | 3+4 2 |
| | 6 2 | 5 4 | 7 2 | 3+4 4 | 5+6 2 |
| | 7 2 | 5 4 | 3+6 2 | 3+4 4 | 3+4 2 |
| | 6 2 | 5 4 | 7 2 | 3+4 4 | 5+6 2 |
| | 7 2 | 5 4 | 7 2 | 3+4 4 | 3+4 2 |
| | 8 2 | | 4+5 2 | | 5+6 2 |
| | 9 2 | | 3+6 2 | | 3+4 2 |
| | 9 2 | | 7 2 | | 5+6 2 |
| | 8 2 | | 4+5 2 | | 5+6 2 |
| | 8 2 | | 7 2 | | 3+4 2 |
| | 9 2 | | 7 2 | | 3+4 2 |
| | 9 2 | | 4+5 2 | | 5+6 2 |
| | 9 2 | | 7 2 | | 5+6 2 |
| | 8 2 | | 3+6 2 | | 5+6 2 |
| | 7 2 | | 7 2 | | 3+4 2 |
| | 6 2 | | 3+6 2 | | 3+4 2 |
| | 7 2 | | 7 2 | | 5+6 2 |
| | 6 2 | | 4+5 2 | | 3+4 2 |
| | 8 2 | | 3+6 2 | | 5+6 2 |
| | 9 2 | | 7 2 | | 3+4 2 |
| | 7 2 | | 4+5 2 | | 3+4 2 |
| | 6 2 | | 7 2 | | 5+6 2 |
| | 6 2 | | 3+6 2 | | 3+4 2 |
| | 7 2 | | 7 2 | | 5+6 2 |
| | 6 2 | | 7 2 | | 3+4 2 |
| | 7 2 | | 4+5 2 | | 5+6 2 |
| | 6 2 | | 7 2 | | 5+6 2 |
| | 9 2 | | 3+6 2 | | 3+4 2 |

Continued on next column

**TABLE LXVI– continued**

| TQC | ICSD | | NoE | | SF |
|---|---|---|---|---|---|
| 141.1.22 | 409800 | | 288 | | Ce1Zn11 |
| | -GM | -M | -P | -X | -N |
| | 7 2 | 5 4 | 4+5 2 | 3+4 4 | 3+4 2 |
| | 8 2 | 5 4 | 7 2 | 3+4 4 | 5+6 2 |
| | 6 2 | 5 4 | 3+6 2 | 3+4 4 | 5+6 2 |
| | 9 2 | 5 4 | 7 2 | 3+4 4 | 3+4 2 |
| | 7 2 | 5 4 | 7 2 | 3+4 4 | 5+6 2 |
| | 8 2 | 5 4 | 4+5 2 | 3+4 4 | 3+4 2 |
| | 6 2 | 5 4 | 3+6 2 | 3+4 4 | 5+6 2 |
| | 9 2 | 5 4 | 7 2 | 3+4 4 | 3+4 2 |
| | 7 2 | 5 4 | 3+6 2 | 3+4 4 | 3+4 2 |
| | 6 2 | 5 4 | 7 2 | 3+4 4 | 5+6 2 |
| | 7 2 | 5 4 | 7 2 | 3+4 4 | 3+4 2 |
| | 9 2 | 5 4 | 4+5 2 | 3+4 4 | 5+6 2 |
| | 7 2 | 5 4 | 4+5 2 | 3+4 4 | 3+4 2 |
| | 6 2 | 5 4 | 7 2 | 3+4 4 | 3+4 2 |
| | 6 2 | 5 4 | 3+6 2 | 3+4 4 | 5+6 2 |
| | 7 2 | 5 4 | 7 2 | 3+4 4 | 3+4 2 |
| | 6 2 | 5 4 | 7 2 | 3+4 4 | 3+4 2 |
| | 6 2 | 5 4 | 7 2 | 3+4 4 | 3+4 2 |
| | 7 2 | 5 4 | 3+6 2 | 3+4 4 | 5+6 2 |
| | 9 2 | 5 4 | 4+5 2 | 3+4 4 | 3+4 2 |
| | 8 2 | 5 4 | 4+5 2 | 3+4 4 | 5+6 2 |
| | 7 2 | 5 4 | 7 2 | 3+4 4 | 5+6 2 |
| | 7 2 | 5 4 | 3+6 2 | 3+4 4 | 3+4 2 |
| | 7 2 | 5 4 | 7 2 | 3+4 4 | 3+4 2 |
| | 8 2 | 5 4 | 3+6 2 | 3+4 4 | 3+4 2 |
| | 6 2 | 5 4 | 7 2 | 3+4 4 | 5+6 2 |
| | 6 2 | 5 4 | 7 2 | 3+4 4 | 5+6 2 |
| | 9 2 | 5 4 | 4+5 2 | 3+4 4 | 3+4 2 |
| | 8 2 | 5 4 | 7 2 | 3+4 4 | 3+4 2 |
| | 7 2 | 5 4 | 3+6 2 | 3+4 4 | 3+4 2 |
| | 6 2 | 5 4 | 4+5 2 | 3+4 4 | 5+6 2 |
| | 8 2 | 5 4 | 7 2 | 3+4 4 | 3+4 2 |
| | 8 2 | 5 4 | 7 2 | 3+4 4 | 5+6 2 |
| | 9 2 | 5 4 | 3+6 2 | 3+4 4 | 5+6 2 |
| | 6 2 | 5 4 | 7 2 | 3+4 4 | 3+4 2 |
| | 7 2 | 5 4 | 4+5 2 | 3+4 4 | 5+6 2 |
| | 7 2 | 5 4 | 3+6 2 | 3+4 4 | 3+4 2 |
| | 6 2 | 5 4 | 7 2 | 3+4 4 | 5+6 2 |
| | 7 2 | 5 4 | 3+6 2 | 3+4 4 | 3+4 2 |
| | 9 2 | 5 4 | 7 2 | 3+4 4 | 5+6 2 |
| | 8 2 | 5 4 | 4+5 2 | 3+4 4 | 3+4 2 |
| | 6 2 | 5 4 | 7 2 | 3+4 4 | 3+4 2 |
| | 7 2 | 5 4 | 7 2 | 3+4 4 | 3+4 2 |
| | 9 2 | 5 4 | 3+6 2 | 3+4 4 | 5+6 2 |
| | 9 2 | 5 4 | 4+5 2 | 3+4 4 | 3+4 2 |
| | 6 2 | 5 4 | 7 2 | 3+4 4 | 5+6 2 |
| | 8 2 | 5 4 | 4+5 2 | 3+4 4 | 3+4 2 |
| | 7 2 | 5 4 | 7 2 | 3+4 4 | 5+6 2 |
| | 9 2 | 5 4 | 4+5 2 | 3+4 4 | 3+4 2 |
| | 6 2 | 5 4 | 3+6 2 | 3+4 4 | 5+6 2 |
| | 8 2 | 5 4 | 7 2 | 3+4 4 | 5+6 2 |
| | 9 2 | 5 4 | 7 2 | 3+4 4 | 3+4 2 |
| | 8 2 | 5 4 | 4+5 2 | 3+4 4 | 5+6 2 |
| | 6 2 | 5 4 | 7 2 | 3+4 4 | 5+6 2 |
| | 9 2 | 5 4 | 7 2 | 3+4 4 | 3+4 2 |
| | 6 2 | 5 4 | 3+6 2 | 3+4 4 | 5+6 2 |
| | 7 2 | 5 4 | 7 2 | 3+4 4 | 3+4 2 |
| | 6 2 | 5 4 | 3+6 2 | 3+4 4 | 3+4 2 |
| | 7 2 | 5 4 | 4+5 2 | 3+4 4 | 3+4 2 |
| | 9 2 | 5 4 | 7 2 | 3+4 4 | 3+4 2 |
| | 8 2 | 5 4 | 7 2 | 3+4 4 | 5+6 2 |
| | 8 2 | 5 4 | 4+5 2 | 3+4 4 | 5+6 2 |
| | 9 2 | 5 4 | 7 2 | 3+4 4 | 5+6 2 |
| | 6 2 | 5 4 | 3+6 2 | 3+4 4 | 3+4 2 |
| | 6 2 | 5 4 | 7 2 | 3+4 4 | 3+4 2 |
| | 8 2 | 5 4 | 7 2 | 3+4 4 | 5+6 2 |
| | 7 2 | 5 4 | 4+5 2 | 3+4 4 | 3+4 2 |
| | 9 2 | 5 4 | 3+6 2 | 3+4 4 | 3+4 2 |
| | 8 2 | 5 4 | 7 2 | 3+4 4 | 5+6 2 |
| | 6 2 | 5 4 | 4+5 2 | 3+4 4 | 5+6 2 |
| | 7 2 | 5 4 | 7 2 | 3+4 4 | 3+4 2 |
| | 9 2 | 5 4 | 3+6 2 | 3+4 4 | 5+6 2 |
| | 7 2 | | 7 2 | | 3+4 2 |
| | 6 2 | | 7 2 | | 5+6 2 |
| | 7 2 | | 4+5 2 | | 3+4 2 |
| | 8 2 | | 3+6 2 | | 5+6 2 |
| | 6 2 | | 7 2 | | 3+4 2 |
| | 9 2 | | 7 2 | | 5+6 2 |
| | 7 2 | | 3+6 2 | | 3+4 2 |
| | 8 2 | | 4+5 2 | | 5+6 2 |
| | 6 2 | | 4+5 2 | | 3+4 2 |
| | 7 2 | | 3+6 2 | | 5+6 2 |
| | 9 2 | | 7 2 | | 5+6 2 |
| | 7 2 | | 7 2 | | 3+4 2 |
| | 9 2 | | 4+5 2 | | 3+4 2 |
| | 8 2 | | 7 2 | | 5+6 2 |
| | 6 2 | | 7 2 | | 5+6 2 |
| | 8 2 | | 3+6 2 | | 3+4 2 |
| | 9 2 | | 7 2 | | 5+6 2 |
| | 6 2 | | 4+5 2 | | 5+6 2 |
| | 8 2 | | 7 2 | | 3+4 2 |
| | 6 2 | | 4+5 2 | | 5+6 2 |
| | 7 2 | | 3+6 2 | | 3+4 2 |
| | 9 2 | | 7 2 | | 5+6 2 |
| | 6 2 | | 3+6 2 | | 3+4 2 |
| | 8 2 | | 4+5 2 | | 3+4 2 |
| | 7 2 | | 7 2 | | 5+6 2 |
| | 9 2 | | 7 2 | | 3+4 2 |
| | 9 2 | | 3+6 2 | | 5+6 2 |
| | 6 2 | | 7 2 | | 3+4 2 |
| | 8 2 | | 3+6 2 | | 5+6 2 |
| | 9 2 | | 7 2 | | 3+4 2 |
| | 7 2 | | 4+5 2 | | 5+6 2 |
| | 7 2 | | 3+6 2 | | 5+6 2 |
| | 8 2 | | 7 2 | | 3+4 2 |
| | 6 2 | | 7 2 | | 3+4 2 |
| | 7 2 | | 4+5 2 | | 5+6 2 |
| | 6 2 | | 7 2 | | 3+4 2 |
| | 9 2 | | 4+5 2 | | 3+4 2 |
| | 6 2 | | 7 2 | | 5+6 2 |
| | 7 2 | | 3+6 2 | | 3+4 2 |
| | 8 2 | | 7 2 | | 3+4 2 |
| | 7 2 | | 4+5 2 | | 5+6 2 |
| | 6 2 | | 7 2 | | 3+4 2 |
| | 9 2 | | 4+5 2 | | 3+4 2 |
| | 7 2 | | 7 2 | | 5+6 2 |
| | 6 2 | | 3+6 2 | | 3+4 2 |
| | 7 2 | | 3+6 2 | | 5+6 2 |
| | 6 2 | | 7 2 | | 3+4 2 |
| | 7 2 | | 7 2 | | 3+4 2 |

Continued on next column

**TABLE LXVI– continued**

| TQC | ICSD | | NoE | | SF |
|---|---|---|---|---|---|
| | 6 2 | | 4+5 2 | | 3+4 2 |
| | 7 2 | | 7 2 | | 5+6 2 |
| | 8 2 | | 3+6 2 | | 5+6 2 |
| | 9 2 | | 7 2 | | 3+4 2 |
| | 8 2 | | 7 2 | | 5+6 2 |
| | 7 2 | | 4+5 2 | | 3+4 2 |
| | 6 2 | | 3+6 2 | | 3+4 2 |
| | 7 2 | | 7 2 | | 5+6 2 |
| | 6 2 | | 7 2 | | 3+4 2 |
| | 9 2 | | 4+5 2 | | 5+6 2 |
| | 6 2 | | 3+6 2 | | 5+6 2 |
| | 9 2 | | 7 2 | | 3+4 2 |
| | 8 2 | | 7 2 | | 5+6 2 |
| | 8 2 | | 4+5 2 | | 3+4 2 |
| | 7 2 | | 3+6 2 | | 5+6 2 |
| | 9 2 | | 7 2 | | 3+4 2 |
| | 9 2 | | 7 2 | | 5+6 2 |
| | 8 2 | | 4+5 2 | | 5+6 2 |
| | 8 2 | | 3+6 2 | | 5+6 2 |
| | 6 2 | | 7 2 | | 5+6 2 |
| | 8 2 | | 4+5 2 | | 3+4 2 |
| | 9 2 | | 7 2 | | 3+4 2 |
| | 7 2 | | 7 2 | | 5+6 2 |
| | 6 2 | | 4+5 2 | | 3+4 2 |
| 141.1.23 | 247146 | | 24 | | Ag1Li1 |
| | -GM | -M | -P | -X | -N |
| | 7 2 | 5 4 | 3+6 2 | 3+4 4 | 3+4 2 |
| | 8 2 | 5 4 | 7 2 | 3+4 4 | 5+6 2 |
| | 7 2 | 5 4 | 4+5 2 | 3+4 4 | 5+6 2 |
| | 6 2 | 5 4 | 7 2 | 3+4 4 | 3+4 2 |
| | 9 2 | 5 4 | 7 2 | 3+4 4 | 3+4 2 |
| | 6 2 | 5 4 | 3+6 2 | 3+4 4 | 5+6 2 |
| | 9 2 | | 7 2 | | 5+6 2 |
| | 8 2 | | 4+5 2 | | 3+4 2 |
| | 9 2 | | 3+6 2 | | 5+6 2 |
| | 6 2 | | 7 2 | | 3+4 2 |
| | 7 2 | | 7 2 | | 5+6 2 |
| | 6 2 | | 4+5 2 | | 3+4 2 |
| 141.1.24 | 247145 | | 24 | | Ag1Li1 |
| | -GM | -M | -P | -X | -N |
| | 7 2 | 5 4 | 3+6 2 | 3+4 4 | 3+4 2 |
| | 8 2 | 5 4 | 7 2 | 3+4 4 | 5+6 2 |
| | 7 2 | 5 4 | 4+5 2 | 3+4 4 | 5+6 2 |
| | 6 2 | 5 4 | 7 2 | 3+4 4 | 3+4 2 |
| | 9 2 | 5 4 | 7 2 | 3+4 4 | 3+4 2 |
| | 6 2 | 5 4 | 3+6 2 | 3+4 4 | 5+6 2 |
| | 9 2 | | 7 2 | | 5+6 2 |
| | 8 2 | | 4+5 2 | | 3+4 2 |
| | 9 2 | | 3+6 2 | | 5+6 2 |
| | 6 2 | | 7 2 | | 3+4 2 |
| | 7 2 | | 7 2 | | 5+6 2 |
| | 6 2 | | 4+5 2 | | 3+4 2 |

The End

TABLE LXVII: The computed irreducible representations for the valence bands for the compounds of SEBR and NLC in space group 147. See the list of tables.

| TQC | ICSD | NoE | SF |
|---|---|---|---|
| 147.1.1 | 166370 | 126 | Ni1Ti1 |

Continued on next column

**TABLE LXVII– continued**

| TQC | ICSD | | NoE | | SF | |
|---|---|---|---|---|---|---|
| | -A | -GM | -H | -K | -L | -M |
| | 5+6 2 | 5+6 2 | 4+4 2 | 4+4 2 | 2+2 2 | 2+2 2 |
| | 8+9 2 | 5+6 2 | 5+6 2 | 5+6 2 | 3+3 2 | 3+3 2 |
| | 5+6 2 | 4+4 2 | 5+6 2 | 5+6 2 | 3+3 2 | 2+2 2 |
| | 4+4 2 | 7+7 2 | 4+4 2 | 4+4 2 | 2+2 2 | 3+3 2 |
| | 8+9 2 | 8+9 2 | 5+6 2 | 5+6 2 | 2+2 2 | 3+3 2 |
| | 7+7 2 | 8+9 2 | 5+6 2 | 5+6 2 | 3+3 2 | 2+2 2 |
| | 5+6 2 | 5+6 2 | 5+6 2 | 5+6 2 | 2+2 2 | 3+3 2 |
| | 8+9 2 | 4+4 2 | 5+6 2 | 4+4 2 | 3+3 2 | 2+2 2 |
| | 8+9 2 | 5+6 2 | 4+4 2 | 5+6 2 | 2+2 2 | 3+3 2 |
| | 4+4 2 | 7+7 2 | 5+6 2 | 5+6 2 | 3+3 2 | 3+3 2 |
| | 5+6 2 | 8+9 2 | 4+4 2 | 4+4 2 | 3+3 2 | 2+2 2 |
| | 7+7 2 | 8+9 2 | 5+6 2 | 5+6 2 | 3+3 2 | 2+2 2 |
| | 8+9 2 | 4+4 2 | 4+4 2 | 5+6 2 | 2+2 2 | 3+3 2 |
| | 5+6 2 | 5+6 2 | 5+6 2 | 4+4 2 | 2+2 2 | 2+2 2 |
| | 4+4 2 | 5+6 2 | 5+6 2 | 5+6 2 | 3+3 2 | 3+3 2 |
| | 5+6 2 | 5+6 2 | 5+6 2 | 5+6 2 | 2+2 2 | 2+2 2 |
| | 7+7 2 | 8+9 2 | 5+6 2 | 5+6 2 | 3+3 2 | 3+3 2 |
| | 8+9 2 | 8+9 2 | 4+4 2 | 4+4 2 | 2+2 2 | 3+3 2 |
| | 7+7 2 | 7+7 2 | 5+6 2 | 5+6 2 | 3+3 2 | 2+2 2 |
| | 8+9 2 | 5+6 2 | 4+4 2 | 4+4 2 | 2+2 2 | 3+3 2 |
| | 8+9 2 | 8+9 2 | 5+6 2 | 5+6 2 | 3+3 2 | 3+3 2 |
| | 4+4 2 | 4+4 2 | 5+6 2 | 5+6 2 | 3+3 2 | 2+2 2 |
| | 5+6 2 | 5+6 2 | 4+4 2 | 5+6 2 | 2+2 2 | 2+2 2 |
| | 5+6 2 | 8+9 2 | 5+6 2 | 4+4 2 | 2+2 2 | 3+3 2 |
| | 8+9 2 | 7+7 2 | 5+6 2 | 4+4 2 | 2+2 2 | 2+2 2 |
| | 5+6 2 | 5+6 2 | 5+6 2 | 5+6 2 | 2+2 2 | 3+3 2 |
| | 7+7 2 | 5+6 2 | 4+4 2 | 5+6 2 | 3+3 2 | 3+3 2 |
| | 4+4 2 | 4+4 2 | 5+6 2 | 5+6 2 | 3+3 2 | 3+3 2 |
| | 8+9 2 | 8+9 2 | 4+4 2 | 4+4 2 | 2+2 2 | 2+2 2 |
| | 5+6 2 | 5+6 2 | 5+6 2 | 5+6 2 | 3+3 2 | 2+2 2 |
| | 8+9 2 | 7+7 2 | 5+6 2 | 5+6 2 | 2+2 2 | 3+3 2 |
| | 7+7 2 | 5+6 2 | 4+4 2 | 5+6 2 | 3+3 2 | 3+3 2 |
| | 8+9 2 | 8+9 2 | 5+6 2 | 4+4 2 | 2+2 2 | 3+3 2 |
| | 5+6 2 | 4+4 2 | 5+6 2 | 5+6 2 | 3+3 2 | 2+2 2 |
| | 4+4 2 | 8+9 2 | 5+6 2 | 4+4 2 | 2+2 2 | 3+3 2 |
| | 8+9 2 | 4+4 2 | 4+4 2 | 5+6 2 | 3+3 2 | 2+2 2 |
| | 7+7 2 | 7+7 2 | 5+6 2 | 5+6 2 | 2+2 2 | 3+3 2 |
| | 5+6 2 | 5+6 2 | 4+4 2 | 4+4 2 | 3+3 2 | 2+2 2 |
| | 8+9 2 | 8+9 2 | 5+6 2 | 5+6 2 | 2+2 2 | 2+2 2 |
| | 7+7 2 | 8+9 2 | 4+4 2 | 4+4 2 | 3+3 2 | 3+3 2 |
| | 8+9 2 | 5+6 2 | 5+6 2 | 5+6 2 | 2+2 2 | 3+3 2 |
| | 4+4 2 | 7+7 2 | 5+6 2 | 5+6 2 | 2+2 2 | 2+2 2 |
| | 5+6 2 | 4+4 2 | 4+4 2 | 5+6 2 | 3+3 2 | 3+3 2 |
| | 5+6 2 | 5+6 2 | 5+6 2 | 4+4 2 | 3+3 2 | 2+2 2 |
| | 5+6 2 | 8+9 2 | 5+6 2 | 5+6 2 | 2+2 2 | 3+3 2 |
| | 4+4 2 | 4+4 2 | 5+6 2 | 5+6 2 | 2+2 2 | 2+2 2 |
| | 8+9 2 | 8+9 2 | 4+4 2 | 4+4 2 | 2+2 2 | 2+2 2 |
| | 5+6 2 | 5+6 2 | 5+6 2 | 5+6 2 | 3+3 2 | 3+3 2 |
| | 8+9 2 | 5+6 2 | 5+6 2 | 5+6 2 | 2+2 2 | 3+3 2 |
| | 5+6 2 | 8+9 2 | 4+4 2 | 4+4 2 | 3+3 2 | 2+2 2 |
| | 4+4 2 | 5+6 2 | 5+6 2 | 5+6 2 | 2+2 2 | 3+3 2 |
| | 5+6 2 | 7+7 2 | 5+6 2 | 5+6 2 | 2+2 2 | 2+2 2 |
| | 7+7 2 | 4+4 2 | 4+4 2 | 5+6 2 | 2+2 2 | 2+2 2 |
| | 8+9 2 | 5+6 2 | 5+6 2 | 4+4 2 | 3+3 2 | 3+3 2 |
| | 7+7 2 | 5+6 2 | 5+6 2 | 5+6 2 | 3+3 2 | 3+3 2 |
| | 8+9 2 | 4+4 2 | 4+4 2 | 4+4 2 | 2+2 2 | 3+3 2 |
| | 5+6 2 | 8+9 2 | 5+6 2 | 5+6 2 | 2+2 2 | 2+2 2 |
| | 8+9 2 | 8+9 2 | 4+4 2 | 5+6 2 | 3+3 2 | 3+3 2 |
| | 5+6 2 | 7+7 2 | 5+6 2 | 4+4 2 | 3+3 2 | 3+3 2 |
| | 4+4 2 | 5+6 2 | 5+6 2 | 5+6 2 | 2+2 2 | 2+2 2 |

Continued on next column

**TABLE LXVII– continued**

| TQC | ICSD | NoE | SF |
|---|---|---|---|
| | 7+7 2 8+9 2 5+6 2 5+6 2 2+2 2 3+3 2 | | |
| | 8+9 2 7+7 2 4+4 2 4+4 2 3+3 2 2+2 2 | | |
| | 8+9 2 5+6 2 5+6 2 5+6 2 2+2 2 2+2 2 | | |
| 147.1.2 | 150943 | 126 | Ni1Ti1 |
| | -A -GM -H -K -L -M | | |
| | 5+6 2 5+6 2 4+4 2 4+4 2 2+2 2 2+2 2 | | |
| | 8+9 2 5+6 2 5+6 2 5+6 2 3+3 2 3+3 2 | | |
| | 8+9 2 7+7 2 5+6 2 5+6 2 2+2 2 2+2 2 | | |
| | 5+6 2 8+9 2 5+6 2 4+4 2 3+3 2 3+3 2 | | |
| | 7+7 2 4+4 2 4+4 2 5+6 2 2+2 2 3+3 2 | | |
| | 4+4 2 8+9 2 5+6 2 5+6 2 3+3 2 3+3 2 | | |
| | 8+9 2 5+6 2 5+6 2 5+6 2 2+2 2 2+2 2 | | |
| | 5+6 2 4+4 2 4+4 2 4+4 2 3+3 2 3+3 2 | | |
| | 4+4 2 5+6 2 5+6 2 5+6 2 3+3 2 3+3 2 | | |
| | 5+6 2 7+7 2 5+6 2 5+6 2 2+2 2 2+2 2 | | |
| | 7+7 2 8+9 2 4+4 2 4+4 2 3+3 2 2+2 2 | | |
| | 8+9 2 4+4 2 5+6 2 5+6 2 2+2 2 3+3 2 | | |
| | 8+9 2 5+6 2 5+6 2 5+6 2 3+3 2 2+2 2 | | |
| | 5+6 2 8+9 2 4+4 2 4+4 2 2+2 2 3+3 2 | | |
| | 4+4 2 5+6 2 5+6 2 5+6 2 3+3 2 3+3 2 | | |
| | 5+6 2 5+6 2 5+6 2 5+6 2 2+2 2 2+2 2 | | |
| | 7+7 2 8+9 2 5+6 2 4+4 2 3+3 2 2+2 2 | | |
| | 8+9 2 8+9 2 4+4 2 5+6 2 2+2 2 3+3 2 | | |
| | 7+7 2 7+7 2 4+4 2 5+6 2 3+3 2 2+2 2 | | |
| | 8+9 2 5+6 2 5+6 2 4+4 2 2+2 2 3+3 2 | | |
| | 8+9 2 8+9 2 5+6 2 5+6 2 2+2 2 3+3 2 | | |
| | 5+6 2 4+4 2 5+6 2 5+6 2 3+3 2 2+2 2 | | |
| | 8+9 2 5+6 2 4+4 2 4+4 2 3+3 2 3+3 2 | | |
| | 4+4 2 5+6 2 5+6 2 5+6 2 2+2 2 2+2 2 | | |
| | 8+9 2 8+9 2 5+6 2 4+4 2 2+2 2 3+3 2 | | |
| | 5+6 2 7+7 2 4+4 2 5+6 2 2+2 2 2+2 2 | | |
| | 7+7 2 4+4 2 5+6 2 5+6 2 3+3 2 3+3 2 | | |
| | 5+6 2 5+6 2 5+6 2 5+6 2 2+2 2 3+3 2 | | |
| | 4+4 2 8+9 2 4+4 2 4+4 2 3+3 2 2+2 2 | | |
| | 5+6 2 5+6 2 5+6 2 5+6 2 3+3 2 3+3 2 | | |
| | 8+9 2 4+4 2 5+6 2 4+4 2 2+2 2 3+3 2 | | |
| | 7+7 2 5+6 2 4+4 2 5+6 2 3+3 2 2+2 2 | | |
| | 8+9 2 7+7 2 5+6 2 5+6 2 2+2 2 3+3 2 | | |
| | 5+6 2 8+9 2 5+6 2 5+6 2 3+3 2 3+3 2 | | |
| | 7+7 2 4+4 2 4+4 2 5+6 2 2+2 2 2+2 2 | | |
| | 4+4 2 8+9 2 5+6 2 4+4 2 2+2 2 3+3 2 | | |
| | 8+9 2 5+6 2 5+6 2 5+6 2 3+3 2 2+2 2 | | |
| | 5+6 2 7+7 2 4+4 2 4+4 2 2+2 2 2+2 2 | | |
| | 8+9 2 8+9 2 5+6 2 5+6 2 3+3 2 3+3 2 | | |
| | 7+7 2 5+6 2 4+4 2 4+4 2 2+2 2 2+2 2 | | |
| | 8+9 2 4+4 2 5+6 2 5+6 2 3+3 2 3+3 2 | | |
| | 4+4 2 8+9 2 5+6 2 5+6 2 2+2 2 2+2 2 | | |
| | 5+6 2 5+6 2 5+6 2 5+6 2 3+3 2 3+3 2 | | |
| | 5+6 2 7+7 2 4+4 2 4+4 2 2+2 2 2+2 2 | | |
| | 8+9 2 8+9 2 5+6 2 5+6 2 2+2 2 3+3 2 | | |
| | 4+4 2 4+4 2 5+6 2 5+6 2 3+3 2 2+2 2 | | |
| | 5+6 2 5+6 2 4+4 2 4+4 2 2+2 2 3+3 2 | | |
| | 5+6 2 5+6 2 5+6 2 5+6 2 2+2 2 2+2 2 | | |
| | 8+9 2 8+9 2 5+6 2 4+4 2 3+3 2 3+3 2 | | |
| | 5+6 2 5+6 2 5+6 2 5+6 2 3+3 2 2+2 2 | | |
| | 4+4 2 4+4 2 4+4 2 5+6 2 2+2 2 3+3 2 | | |
| | 8+9 2 5+6 2 5+6 2 5+6 2 2+2 2 2+2 2 | | |
| | 7+7 2 8+9 2 4+4 2 5+6 2 3+3 2 3+3 2 | | |
| | 5+6 2 7+7 2 5+6 2 4+4 2 2+2 2 2+2 2 | | |
| | 7+7 2 5+6 2 5+6 2 5+6 2 2+2 2 3+3 2 | | |
| | 8+9 2 4+4 2 5+6 2 4+4 2 3+3 2 2+2 2 | | |

Continued on next column

**TABLE LXVII– continued**

| TQC | ICSD | NoE | SF |
|---|---|---|---|
| | 5+6 2 8+9 2 4+4 2 5+6 2 2+2 2 3+3 2 | | |
| | 8+9 2 7+7 2 4+4 2 5+6 2 3+3 2 3+3 2 | | |
| | 5+6 2 8+9 2 5+6 2 5+6 2 3+3 2 3+3 2 | | |
| | 4+4 2 5+6 2 5+6 2 4+4 2 2+2 2 2+2 2 | | |
| | 7+7 2 5+6 2 5+6 2 5+6 2 3+3 2 2+2 2 | | |
| | 8+9 2 4+4 2 5+6 2 5+6 2 2+2 2 3+3 2 | | |
| | 5+6 2 8+9 2 4+4 2 4+4 2 2+2 2 2+2 2 | | |

The End

TABLE LXVIII: The computed irreducible representations for the valence bands for the compounds of SEBR and NLC in space group 148. See the list of tables.

| TQC | ICSD | NoE | SF |
|---|---|---|---|
| 148.1.1 | 54337 | 104 | Pd4Zr3 |
| | -GM -T -F -L | | |
| | 5+6 2 8+9 2 3+3 2 3+3 2 | | |
| | 5+6 2 5+6 2 2+2 2 2+2 2 | | |
| | 4+4 2 4+4 2 3+3 2 3+3 2 | | |
| | 5+6 2 5+6 2 3+3 2 2+2 2 | | |
| | 8+9 2 8+9 2 3+3 2 3+3 2 | | |
| | 8+9 2 7+7 2 2+2 2 2+2 2 | | |
| | 7+7 2 8+9 2 3+3 2 3+3 2 | | |
| | 8+9 2 5+6 2 2+2 2 2+2 2 | | |
| | 4+4 2 4+4 2 3+3 2 2+2 2 | | |
| | 5+6 2 5+6 2 3+3 2 3+3 2 | | |
| | 5+6 2 7+7 2 2+2 2 2+2 2 | | |
| | 4+4 2 4+4 2 3+3 2 3+3 2 | | |
| | 5+6 2 8+9 2 2+2 2 2+2 2 | | |
| | 7+7 2 5+6 2 2+2 2 3+3 2 | | |
| | 8+9 2 8+9 2 3+3 2 3+3 2 | | |
| | 8+9 2 8+9 2 3+3 2 2+2 2 | | |
| | 7+7 2 7+7 2 2+2 2 3+3 2 | | |
| | 5+6 2 5+6 2 3+3 2 3+3 2 | | |
| | 4+4 2 4+4 2 2+2 2 2+2 2 | | |
| | 5+6 2 5+6 2 3+3 2 3+3 2 | | |
| | 8+9 2 7+7 2 3+3 2 2+2 2 | | |
| | 4+4 2 8+9 2 3+3 2 3+3 2 | | |
| | 5+6 2 8+9 2 2+2 2 3+3 2 | | |
| | 8+9 2 5+6 2 3+3 2 2+2 2 | | |
| | 5+6 2 4+4 2 2+2 2 2+2 2 | | |
| | 7+7 2 5+6 2 3+3 2 3+3 2 | | |
| | 5+6 2 5+6 2 3+3 2 3+3 2 | | |
| | 8+9 2 7+7 2 2+2 2 2+2 2 | | |
| | 4+4 2 8+9 2 3+3 2 2+2 2 | | |
| | 5+6 2 5+6 2 2+2 2 3+3 2 | | |
| | 4+4 2 8+9 2 3+3 2 2+2 2 | | |
| | 5+6 2 5+6 2 2+2 2 2+2 2 | | |
| | 7+7 2 4+4 2 2+2 2 3+3 2 | | |
| | 8+9 2 5+6 2 3+3 2 2+2 2 | | |
| | 5+6 2 7+7 2 2+2 2 3+3 2 | | |
| | 4+4 2 8+9 2 3+3 2 2+2 2 | | |
| | 5+6 2 4+4 2 3+3 2 2+2 2 | | |
| | 8+9 2 5+6 2 2+2 2 3+3 2 | | |
| | 5+6 2 8+9 2 2+2 2 3+3 2 | | |
| | 8+9 2 7+7 2 3+3 2 2+2 2 | | |
| | 4+4 2 8+9 2 3+3 2 3+3 2 | | |
| | 7+7 2 8+9 2 3+3 2 2+2 2 | | |
| | 5+6 2 7+7 2 3+3 2 3+3 2 | | |
| | 8+9 2 5+6 2 2+2 2 2+2 2 | | |

Continued on next column

**TABLE LXVIII– continued**

| TQC | ICSD | NoE | SF |
|---|---|---|---|
| | 5+6 2 4+4 2 3+3 2 3+3 2 | | |
| | 4+4 2 8+9 2 3+3 2 2+2 2 | | |
| | 5+6 2 8+9 2 3+3 2 2+2 2 | | |
| | 8+9 2 7+7 2 2+2 2 3+3 2 | | |
| | 4+4 2 8+9 2 2+2 2 2+2 2 | | |
| | 5+6 2 8+9 2 2+2 2 3+3 2 | | |
| | 8+9 2 5+6 2 3+3 2 3+3 2 | | |
| | 7+7 2 4+4 2 2+2 2 3+3 2 | | |
| 148.1.2 | 1096 | 160 | P2Pd15 |
| | -GM -T -F -L | | |
| | 5+6 2 5+6 2 2+2 2 2+2 2 | | |
| | 8+9 2 8+9 2 3+3 2 3+3 2 | | |
| | 5+6 2 8+9 2 3+3 2 3+3 2 | | |
| | 8+9 2 7+7 2 2+2 2 2+2 2 | | |
| | 7+7 2 8+9 2 2+2 2 3+3 2 | | |
| | 4+4 2 4+4 2 3+3 2 2+2 2 | | |
| | 8+9 2 5+6 2 3+3 2 2+2 2 | | |
| | 5+6 2 5+6 2 2+2 2 3+3 2 | | |
| | 5+6 2 8+9 2 2+2 2 2+2 2 | | |
| | 5+6 2 7+7 2 3+3 2 3+3 2 | | |
| | 4+4 2 5+6 2 3+3 2 2+2 2 | | |
| | 8+9 2 8+9 2 3+3 2 2+2 2 | | |
| | 8+9 2 4+4 2 2+2 2 3+3 2 | | |
| | 7+7 2 5+6 2 2+2 2 2+2 2 | | |
| | 5+6 2 8+9 2 3+3 2 3+3 2 | | |
| | 4+4 2 8+9 2 3+3 2 2+2 2 | | |
| | 5+6 2 7+7 2 3+3 2 3+3 2 | | |
| | 8+9 2 5+6 2 2+2 2 2+2 2 | | |
| | 7+7 2 8+9 2 2+2 2 3+3 2 | | |
| | 5+6 2 5+6 2 3+3 2 2+2 2 | | |
| | 5+6 2 4+4 2 3+3 2 2+2 2 | | |
| | 4+4 2 7+7 2 2+2 2 3+3 2 | | |
| | 8+9 2 5+6 2 3+3 2 3+3 2 | | |
| | 7+7 2 8+9 2 2+2 2 2+2 2 | | |
| | 8+9 2 8+9 2 3+3 2 2+2 2 | | |
| | 8+9 2 5+6 2 2+2 2 3+3 2 | | |
| | 5+6 2 8+9 2 2+2 2 2+2 2 | | |
| | 5+6 2 4+4 2 3+3 2 3+3 2 | | |
| | 8+9 2 5+6 2 3+3 2 3+3 2 | | |
| | 7+7 2 7+7 2 2+2 2 2+2 2 | | |
| | 8+9 2 8+9 2 2+2 2 2+2 2 | | |
| | 4+4 2 5+6 2 3+3 2 2+2 2 | | |
| | 5+6 2 7+7 2 3+3 2 3+3 2 | | |
| | 8+9 2 8+9 2 2+2 2 2+2 2 | | |
| | 7+7 2 4+4 2 3+3 2 3+3 2 | | |
| | 4+4 2 5+6 2 3+3 2 3+3 2 | | |
| | 8+9 2 5+6 2 2+2 2 3+3 2 | | |
| | 5+6 2 8+9 2 3+3 2 2+2 2 | | |
| | 4+4 2 7+7 2 2+2 2 2+2 2 | | |
| | 5+6 2 4+4 2 2+2 2 2+2 2 | | |
| | 5+6 2 8+9 2 3+3 2 3+3 2 | | |
| | 7+7 2 4+4 2 2+2 2 2+2 2 | | |
| | 5+6 2 5+6 2 3+3 2 3+3 2 | | |
| | 4+4 2 8+9 2 2+2 2 2+2 2 | | |
| | 8+9 2 7+7 2 3+3 2 3+3 2 | | |
| | 5+6 2 5+6 2 2+2 2 3+3 2 | | |
| | 8+9 2 5+6 2 2+2 2 2+2 2 | | |
| | 5+6 2 4+4 2 2+2 2 3+3 2 | | |
| | 8+9 2 5+6 2 3+3 2 2+2 2 | | |
| | 7+7 2 4+4 2 2+2 2 3+3 2 | | |
| | 4+4 2 7+7 2 3+3 2 3+3 2 | | |

Continued on next column

**TABLE LXVIII– continued**

| TQC | ICSD | NoE | SF |
|---|---|---|---|
| | 5+6 2 5+6 2 3+3 2 3+3 2 | | |
| | 5+6 2 8+9 2 2+2 2 2+2 2 | | |
| | 4+4 2 8+9 2 3+3 2 3+3 2 | | |
| | 5+6 2 5+6 2 3+3 2 2+2 2 | | |
| | 8+9 2 8+9 2 2+2 2 3+3 2 | | |
| | 8+9 2 5+6 2 3+3 2 2+2 2 | | |
| | 7+7 2 7+7 2 2+2 2 2+2 2 | | |
| | 4+4 2 8+9 2 2+2 2 3+3 2 | | |
| | 5+6 2 5+6 2 3+3 2 2+2 2 | | |
| | 7+7 2 4+4 2 2+2 2 3+3 2 | | |
| | 8+9 2 8+9 2 3+3 2 2+2 2 | | |
| | 8+9 2 7+7 2 2+2 2 3+3 2 | | |
| | 5+6 2 5+6 2 3+3 2 2+2 2 | | |
| | 4+4 2 8+9 2 2+2 2 3+3 2 | | |
| | 5+6 2 5+6 2 2+2 2 2+2 2 | | |
| | 7+7 2 7+7 2 3+3 2 3+3 2 | | |
| | 8+9 2 8+9 2 3+3 2 2+2 2 | | |
| | 8+9 2 4+4 2 3+3 2 3+3 2 | | |
| | 4+4 2 8+9 2 2+2 2 2+2 2 | | |
| | 5+6 2 8+9 2 2+2 2 3+3 2 | | |
| | 5+6 2 7+7 2 2+2 2 2+2 2 | | |
| | 5+6 2 5+6 2 3+3 2 2+2 2 | | |
| | 4+4 2 4+4 2 3+3 2 2+2 2 | | |
| | 7+7 2 8+9 2 2+2 2 3+3 2 | | |
| | 8+9 2 5+6 2 3+3 2 2+2 2 | | |
| | 5+6 2 7+7 2 3+3 2 3+3 2 | | |
| | 8+9 2 8+9 2 3+3 2 2+2 2 | | |
| | 4+4 2 8+9 2 2+2 2 3+3 2 | | |
| | 5+6 2 4+4 2 3+3 2 2+2 2 | | |
| 148.1.3 | 37134 | 96 | O12Te1Tl6 |
| | -GM -T -F -L | | |
| | 5+6 2 5+6 2 3+3 2 3+3 2 | | |
| | 8+9 2 8+9 2 2+2 2 2+2 2 | | |
| | 5+6 2 7+7 2 2+2 2 2+2 2 | | |
| | 7+7 2 4+4 2 3+3 2 2+2 2 | | |
| | 8+9 2 5+6 2 2+2 2 3+3 2 | | |
| | 4+4 2 8+9 2 2+2 2 2+2 2 | | |
| | 5+6 2 8+9 2 3+3 2 3+3 2 | | |
| | 8+9 2 5+6 2 2+2 2 2+2 2 | | |
| | 7+7 2 4+4 2 3+3 2 3+3 2 | | |
| | 5+6 2 5+6 2 3+3 2 3+3 2 | | |
| | 4+4 2 7+7 2 2+2 2 2+2 2 | | |
| | 8+9 2 8+9 2 3+3 2 3+3 2 | | |
| | 5+6 2 5+6 2 3+3 2 3+3 2 | | |
| | 8+9 2 8+9 2 2+2 2 2+2 2 | | |
| | 8+9 2 7+7 2 3+3 2 2+2 2 | | |
| | 7+7 2 5+6 2 2+2 2 3+3 2 | | |
| | 5+6 2 4+4 2 3+3 2 3+3 2 | | |
| | 4+4 2 8+9 2 2+2 2 2+2 2 | | |
| | 8+9 2 8+9 2 2+2 2 2+2 2 | | |
| | 7+7 2 5+6 2 3+3 2 3+3 2 | | |
| | 8+9 2 8+9 2 2+2 2 2+2 2 | | |
| | 5+6 2 7+7 2 2+2 2 2+2 2 | | |
| | 5+6 2 5+6 2 2+2 2 3+3 2 | | |
| | 4+4 2 5+6 2 3+3 2 2+2 2 | | |
| | 5+6 2 4+4 2 3+3 2 3+3 2 | | |
| | 5+6 2 8+9 2 3+3 2 3+3 2 | | |
| | 4+4 2 5+6 2 2+2 2 2+2 2 | | |
| | 8+9 2 7+7 2 3+3 2 3+3 2 | | |
| | 7+7 2 8+9 2 2+2 2 2+2 2 | | |
| | 8+9 2 8+9 2 3+3 2 3+3 2 | | |

Continued on next column

**TABLE LXVIII– continued**

| TQC | ICSD NoE SF |
|---|---|
| | 5+6 2 4+4 2 2+2 2 2+2 2 |
| | 7+7 2 5+6 2 2+2 2 3+3 2 |
| | 8+9 2 8+9 2 3+3 2 2+2 2 |
| | 4+4 2 7+7 2 3+3 2 3+3 2 |
| | 5+6 2 4+4 2 2+2 2 2+2 2 |
| | 8+9 2 5+6 2 2+2 2 2+2 2 |
| | 8+9 2 5+6 2 3+3 2 3+3 2 |
| | 5+6 2 8+9 2 2+2 2 2+2 2 |
| | 7+7 2 7+7 2 3+3 2 3+3 2 |
| | 8+9 2 4+4 2 3+3 2 2+2 2 |
| | 4+4 2 5+6 2 2+2 2 3+3 2 |
| | 5+6 2 8+9 2 2+2 2 3+3 2 |
| | 5+6 2 8+9 2 2+2 2 2+2 2 |
| | 8+9 2 7+7 2 3+3 2 2+2 2 |
| | 7+7 2 8+9 2 2+2 2 3+3 2 |
| | 5+6 2 5+6 2 3+3 2 3+3 2 |
| | 4+4 2 4+4 2 3+3 2 2+2 2 |
| | 5+6 2 5+6 2 3+3 2 3+3 2 |
| 148.1.4 | 166373 104 Ni4Ti3 |
| | -GM -T -F -L |
| | 5+6 2 8+9 2 3+3 2 2+2 2 |
| | 5+6 2 5+6 2 2+2 2 3+3 2 |
| | 4+4 2 4+4 2 3+3 2 2+2 2 |
| | 5+6 2 5+6 2 2+2 2 3+3 2 |
| | 8+9 2 7+7 2 3+3 2 2+2 2 |
| | 7+7 2 8+9 2 3+3 2 3+3 2 |
| | 8+9 2 7+7 2 3+3 2 2+2 2 |
| | 4+4 2 8+9 2 2+2 2 2+2 2 |
| | 5+6 2 5+6 2 3+3 2 2+2 2 |
| | 5+6 2 8+9 2 3+3 2 3+3 2 |
| | 4+4 2 4+4 2 2+2 2 2+2 2 |
| | 5+6 2 5+6 2 3+3 2 3+3 2 |
| | 8+9 2 8+9 2 2+2 2 2+2 2 |
| | 8+9 2 7+7 2 2+2 2 3+3 2 |
| | 7+7 2 5+6 2 3+3 2 3+3 2 |
| | 4+4 2 8+9 2 2+2 2 2+2 2 |
| | 5+6 2 7+7 2 3+3 2 3+3 2 |
| | 8+9 2 5+6 2 3+3 2 3+3 2 |
| | 5+6 2 8+9 2 3+3 2 2+2 2 |
| | 7+7 2 4+4 2 2+2 2 2+2 2 |
| | 4+4 2 8+9 2 3+3 2 3+3 2 |
| | 5+6 2 7+7 2 2+2 2 2+2 2 |
| | 8+9 2 8+9 2 3+3 2 2+2 2 |
| | 5+6 2 5+6 2 3+3 2 3+3 2 |
| | 8+9 2 8+9 2 3+3 2 2+2 2 |
| | 7+7 2 5+6 2 2+2 2 3+3 2 |
| | 8+9 2 4+4 2 2+2 2 2+2 2 |
| | 5+6 2 8+9 2 3+3 2 3+3 2 |
| | 5+6 2 7+7 2 2+2 2 2+2 2 |
| | 4+4 2 8+9 2 3+3 2 3+3 2 |
| | 7+7 2 4+4 2 2+2 2 3+3 2 |
| | 8+9 2 5+6 2 2+2 2 3+3 2 |
| | 8+9 2 5+6 2 3+3 2 2+2 2 |
| | 5+6 2 8+9 2 2+2 2 2+2 2 |
| | 4+4 2 5+6 2 3+3 2 2+2 2 |
| | 5+6 2 7+7 2 2+2 2 3+3 2 |
| | 8+9 2 8+9 2 3+3 2 2+2 2 |
| | 4+4 2 4+4 2 2+2 2 3+3 2 |
| | 5+6 2 5+6 2 2+2 2 2+2 2 |
| | 5+6 2 5+6 2 3+3 2 3+3 2 |
| | 7+7 2 4+4 2 3+3 2 2+2 2 |

Continued on next column

**TABLE LXVIII– continued**

| TQC | ICSD NoE SF |
|---|---|
| | 8+9 2 5+6 2 3+3 2 2+2 2 |
| | 5+6 2 5+6 2 3+3 2 3+3 2 |
| | 4+4 2 4+4 2 3+3 2 3+3 2 |
| | 5+6 2 7+7 2 3+3 2 3+3 2 |
| | 4+4 2 8+9 2 3+3 2 3+3 2 |
| | 5+6 2 5+6 2 2+2 2 2+2 2 |
| | 8+9 2 4+4 2 3+3 2 2+2 2 |
| | 8+9 2 5+6 2 2+2 2 3+3 2 |
| | 7+7 2 8+9 2 2+2 2 2+2 2 |
| | 4+4 2 7+7 2 2+2 2 3+3 2 |
| | 5+6 2 8+9 2 2+2 2 2+2 2 |
| 148.1.5 | 166371 104 Ni4Ti3 |
| | -GM -T -F -L |
| | 5+6 2 4+4 2 2+2 2 2+2 2 |
| | 8+9 2 5+6 2 3+3 2 3+3 2 |
| | 5+6 2 7+7 2 2+2 2 3+3 2 |
| | 4+4 2 8+9 2 3+3 2 2+2 2 |
| | 7+7 2 5+6 2 3+3 2 2+2 2 |
| | 8+9 2 8+9 2 2+2 2 3+3 2 |
| | 4+4 2 8+9 2 3+3 2 2+2 2 |
| | 8+9 2 5+6 2 3+3 2 3+3 2 |
| | 5+6 2 4+4 2 2+2 2 2+2 2 |
| | 5+6 2 5+6 2 2+2 2 2+2 2 |
| | 8+9 2 8+9 2 2+2 2 3+3 2 |
| | 7+7 2 7+7 2 3+3 2 3+3 2 |
| | 5+6 2 5+6 2 3+3 2 2+2 2 |
| | 4+4 2 4+4 2 3+3 2 2+2 2 |
| | 5+6 2 5+6 2 2+2 2 3+3 2 |
| | 8+9 2 8+9 2 3+3 2 3+3 2 |
| | 8+9 2 7+7 2 3+3 2 2+2 2 |
| | 7+7 2 5+6 2 2+2 2 3+3 2 |
| | 8+9 2 8+9 2 3+3 2 3+3 2 |
| | 5+6 2 8+9 2 2+2 2 2+2 2 |
| | 8+9 2 7+7 2 3+3 2 3+3 2 |
| | 7+7 2 5+6 2 2+2 2 2+2 2 |
| | 5+6 2 4+4 2 3+3 2 2+2 2 |
| | 5+6 2 8+9 2 2+2 2 3+3 2 |
| | 4+4 2 8+9 2 3+3 2 3+3 2 |
| | 7+7 2 8+9 2 2+2 2 2+2 2 |
| | 8+9 2 7+7 2 2+2 2 2+2 2 |
| | 4+4 2 5+6 2 3+3 2 3+3 2 |
| | 5+6 2 8+9 2 3+3 2 2+2 2 |
| | 8+9 2 4+4 2 2+2 2 3+3 2 |
| | 5+6 2 5+6 2 3+3 2 3+3 2 |
| | 5+6 2 5+6 2 2+2 2 2+2 2 |
| | 4+4 2 4+4 2 2+2 2 2+2 2 |
| | 5+6 2 5+6 2 3+3 2 3+3 2 |
| | 4+4 2 7+7 2 2+2 2 2+2 2 |
| | 5+6 2 8+9 2 3+3 2 3+3 2 |
| | 8+9 2 5+6 2 3+3 2 3+3 2 |
| | 7+7 2 4+4 2 3+3 2 2+2 2 |
| | 5+6 2 5+6 2 2+2 2 2+2 2 |
| | 8+9 2 7+7 2 3+3 2 3+3 2 |
| | 4+4 2 8+9 2 3+3 2 2+2 2 |
| | 5+6 2 8+9 2 3+3 2 3+3 2 |
| | 8+9 2 7+7 2 3+3 2 2+2 2 |
| | 7+7 2 4+4 2 3+3 2 2+2 2 |
| | 4+4 2 5+6 2 2+2 2 3+3 2 |
| | 5+6 2 8+9 2 3+3 2 3+3 2 |
| | 5+6 2 5+6 2 2+2 2 2+2 2 |
| | 5+6 2 4+4 2 3+3 2 3+3 2 |

Continued on next column

**TABLE LXVIII– continued**

| TQC | ICSD NoE SF |
|---|---|
| | 4+4 2 5+6 2 3+3 2 2+2 2 |
| | 8+9 2 8+9 2 2+2 2 3+3 2 |
| | 5+6 2 7+7 2 2+2 2 2+2 2 |
| | 4+4 2 8+9 2 3+3 2 2+2 2 |
| 148.1.6 | 166369 104 Ni4Ti3 |
| | -GM -T -F -L |
| | 5+6 2 4+4 2 2+2 2 3+3 2 |
| | 8+9 2 5+6 2 3+3 2 2+2 2 |
| | 5+6 2 5+6 2 3+3 2 2+2 2 |
| | 7+7 2 7+7 2 2+2 2 3+3 2 |
| | 8+9 2 8+9 2 3+3 2 2+2 2 |
| | 4+4 2 8+9 2 3+3 2 3+3 2 |
| | 5+6 2 5+6 2 2+2 2 3+3 2 |
| | 5+6 2 8+9 2 2+2 2 2+2 2 |
| | 4+4 2 7+7 2 3+3 2 2+2 2 |
| | 8+9 2 4+4 2 2+2 2 2+2 2 |
| | 8+9 2 5+6 2 3+3 2 3+3 2 |
| | 7+7 2 8+9 2 3+3 2 3+3 2 |
| | 5+6 2 8+9 2 2+2 2 2+2 2 |
| | 4+4 2 5+6 2 2+2 2 3+3 2 |
| | 5+6 2 7+7 2 3+3 2 3+3 2 |
| | 8+9 2 8+9 2 3+3 2 2+2 2 |
| | 7+7 2 8+9 2 3+3 2 3+3 2 |
| | 5+6 2 5+6 2 2+2 2 2+2 2 |
| | 8+9 2 4+4 2 3+3 2 3+3 2 |
| | 5+6 2 7+7 2 3+3 2 3+3 2 |
| | 8+9 2 5+6 2 2+2 2 2+2 2 |
| | 5+6 2 8+9 2 3+3 2 2+2 2 |
| | 4+4 2 4+4 2 2+2 2 2+2 2 |
| | 8+9 2 5+6 2 2+2 2 3+3 2 |
| | 7+7 2 8+9 2 3+3 2 2+2 2 |
| | 4+4 2 7+7 2 2+2 2 3+3 2 |
| | 5+6 2 8+9 2 3+3 2 2+2 2 |
| | 7+7 2 4+4 2 2+2 2 3+3 2 |
| | 8+9 2 5+6 2 3+3 2 2+2 2 |
| | 5+6 2 5+6 2 3+3 2 3+3 2 |
| | 4+4 2 7+7 2 2+2 2 3+3 2 |
| | 5+6 2 8+9 2 3+3 2 2+2 2 |
| | 5+6 2 5+6 2 2+2 2 2+2 2 |
| | 8+9 2 8+9 2 3+3 2 3+3 2 |
| | 4+4 2 4+4 2 2+2 2 2+2 2 |
| | 5+6 2 5+6 2 2+2 2 2+2 2 |
| | 7+7 2 8+9 2 3+3 2 3+3 2 |
| | 8+9 2 7+7 2 3+3 2 2+2 2 |
| | 4+4 2 5+6 2 2+2 2 3+3 2 |
| | 8+9 2 4+4 2 3+3 2 2+2 2 |
| | 5+6 2 8+9 2 3+3 2 2+2 2 |
| | 5+6 2 5+6 2 3+3 2 3+3 2 |
| | 4+4 2 4+4 2 3+3 2 2+2 2 |
| | 5+6 2 5+6 2 3+3 2 3+3 2 |
| | 5+6 2 8+9 2 3+3 2 3+3 2 |
| | 5+6 2 7+7 2 3+3 2 3+3 2 |
| | 4+4 2 5+6 2 3+3 2 3+3 2 |
| | 8+9 2 5+6 2 2+2 2 2+2 2 |
| | 7+7 2 4+4 2 2+2 2 3+3 2 |
| | 8+9 2 8+9 2 3+3 2 2+2 2 |
| | 4+4 2 5+6 2 2+2 2 2+2 2 |
| | 5+6 2 7+7 2 2+2 2 2+2 2 |
| 148.1.7 | 171584 64 Ni1O3Ti1 |
| | -GM -T -F -L |
| | 5+6 2 8+9 2 2+2 2 3+3 2 |

Continued on next column

**TABLE LXVIII– continued**

| TQC | ICSD NoE SF |
|---|---|
| | 8+9 2 5+6 2 3+3 2 2+2 2 |
| | 7+7 2 5+6 2 3+3 2 2+2 2 |
| | 8+9 2 4+4 2 2+2 2 3+3 2 |
| | 5+6 2 8+9 2 3+3 2 2+2 2 |
| | 4+4 2 7+7 2 2+2 2 3+3 2 |
| | 5+6 2 8+9 2 2+2 2 2+2 2 |
| | 8+9 2 5+6 2 3+3 2 3+3 2 |
| | 7+7 2 5+6 2 2+2 2 3+3 2 |
| | 8+9 2 4+4 2 3+3 2 2+2 2 |
| | 5+6 2 8+9 2 3+3 2 3+3 2 |
| | 4+4 2 7+7 2 2+2 2 2+2 2 |
| | 5+6 2 8+9 2 2+2 2 3+3 2 |
| | 4+4 2 7+7 2 3+3 2 2+2 2 |
| | 5+6 2 5+6 2 3+3 2 3+3 2 |
| | 7+7 2 8+9 2 2+2 2 2+2 2 |
| | 8+9 2 4+4 2 2+2 2 2+2 2 |
| | 5+6 2 5+6 2 3+3 2 3+3 2 |
| | 5+6 2 8+9 2 2+2 2 3+3 2 |
| | 4+4 2 8+9 2 2+2 2 3+3 2 |
| | 8+9 2 5+6 2 3+3 2 2+2 2 |
| | 8+9 2 7+7 2 2+2 2 2+2 2 |
| | 8+9 2 4+4 2 3+3 2 3+3 2 |
| | 7+7 2 5+6 2 3+3 2 2+2 2 |
| | 4+4 2 5+6 2 2+2 2 3+3 2 |
| | 5+6 2 4+4 2 3+3 2 2+2 2 |
| | 5+6 2 7+7 2 2+2 2 2+2 2 |
| | 8+9 2 8+9 2 3+3 2 3+3 2 |
| | 7+7 2 5+6 2 3+3 2 2+2 2 |
| | 8+9 2 8+9 2 2+2 2 3+3 2 |
| | 5+6 2 4+4 2 2+2 2 2+2 2 |
| | 4+4 2 5+6 2 3+3 2 3+3 2 |
| 148.1.8 | 59375 84 Mo6Te8 |
| | -GM -T -F -L |
| | 5+6 2 8+9 2 2+2 2 3+3 2 |
| | 8+9 2 7+7 2 3+3 2 2+2 2 |
| | 7+7 2 5+6 2 2+2 2 2+2 2 |
| | 8+9 2 4+4 2 3+3 2 3+3 2 |
| | 4+4 2 5+6 2 2+2 2 3+3 2 |
| | 5+6 2 8+9 2 3+3 2 2+2 2 |
| | 8+9 2 8+9 2 3+3 2 2+2 2 |
| | 5+6 2 5+6 2 2+2 2 3+3 2 |
| | 4+4 2 5+6 2 2+2 2 3+3 2 |
| | 5+6 2 8+9 2 3+3 2 2+2 2 |
| | 5+6 2 4+4 2 2+2 2 2+2 2 |
| | 7+7 2 5+6 2 3+3 2 3+3 2 |
| | 8+9 2 5+6 2 3+3 2 3+3 2 |
| | 8+9 2 7+7 2 2+2 2 2+2 2 |
| | 8+9 2 8+9 2 3+3 2 3+3 2 |
| | 5+6 2 8+9 2 3+3 2 2+2 2 |
| | 7+7 2 5+6 2 2+2 2 3+3 2 |
| | 8+9 2 4+4 2 3+3 2 2+2 2 |
| | 5+6 2 8+9 2 2+2 2 2+2 2 |
| | 4+4 2 8+9 2 3+3 2 3+3 2 |
| | 5+6 2 5+6 2 2+2 2 3+3 2 |
| | 4+4 2 7+7 2 3+3 2 3+3 2 |
| | 5+6 2 8+9 2 2+2 2 3+3 2 |
| | 8+9 2 7+7 2 2+2 2 2+2 2 |
| | 7+7 2 5+6 2 3+3 2 2+2 2 |
| | 8+9 2 8+9 2 2+2 2 2+2 2 |
| | 8+9 2 8+9 2 3+3 2 3+3 2 |
| | 5+6 2 7+7 2 2+2 2 2+2 2 |

Continued on next column

**TABLE LXVIII– continued**

| TQC | ICSD | | NoE | SF |
|---|---|---|---|---|
| | 5+6 2 | 5+6 2 | 2+2 2 | 3+3 2 |
| | 4+4 2 | 4+4 2 | 3+3 2 | 3+3 2 |
| | 5+6 2 | 5+6 2 | 2+2 2 | 2+2 2 |
| | 8+9 2 | 7+7 2 | 2+2 2 | 3+3 2 |
| | 7+7 2 | 4+4 2 | 2+2 2 | 2+2 2 |
| | 4+4 2 | 8+9 2 | 3+3 2 | 2+2 2 |
| | 5+6 2 | 5+6 2 | 3+3 2 | 2+2 2 |
| | 5+6 2 | 8+9 2 | 2+2 2 | 3+3 2 |
| | 8+9 2 | 8+9 2 | 3+3 2 | 2+2 2 |
| | 7+7 2 | 5+6 2 | 2+2 2 | 3+3 2 |
| | 8+9 2 | 4+4 2 | 3+3 2 | 2+2 2 |
| | 5+6 2 | 5+6 2 | 2+2 2 | 3+3 2 |
| | 4+4 2 | 7+7 2 | 3+3 2 | 2+2 2 |
| | 8+9 2 | 8+9 2 | 2+2 2 | 3+3 2 |
| 148.1.9 | 46015 | | 108 | Mo6S8Yb1 |
| | -GM | -T | -F | -L |
| | 5+6 2 | 5+6 2 | 3+3 2 | 3+3 2 |
| | 8+9 2 | 8+9 2 | 2+2 2 | 2+2 2 |
| | 8+9 2 | 8+9 2 | 2+2 2 | 2+2 2 |
| | 7+7 2 | 7+7 2 | 2+2 2 | 2+2 2 |
| | 5+6 2 | 5+6 2 | 3+3 2 | 2+2 2 |
| | 8+9 2 | 8+9 2 | 2+2 2 | 3+3 2 |
| | 7+7 2 | 5+6 2 | 2+2 2 | 3+3 2 |
| | 8+9 2 | 4+4 2 | 3+3 2 | 3+3 2 |
| | 4+4 2 | 8+9 2 | 2+2 2 | 2+2 2 |
| | 5+6 2 | 7+7 2 | 3+3 2 | 2+2 2 |
| | 8+9 2 | 5+6 2 | 3+3 2 | 3+3 2 |
| | 5+6 2 | 8+9 2 | 2+2 2 | 2+2 2 |
| | 4+4 2 | 5+6 2 | 3+3 2 | 2+2 2 |
| | 5+6 2 | 8+9 2 | 3+3 2 | 3+3 2 |
| | 5+6 2 | 7+7 2 | 2+2 2 | 2+2 2 |
| | 8+9 2 | 8+9 2 | 3+3 2 | 3+3 2 |
| | 7+7 2 | 8+9 2 | 2+2 2 | 2+2 2 |
| | 8+9 2 | 4+4 2 | 2+2 2 | 3+3 2 |
| | 8+9 2 | 5+6 2 | 3+3 2 | 2+2 2 |
| | 5+6 2 | 7+7 2 | 3+3 2 | 3+3 2 |
| | 8+9 2 | 8+9 2 | 3+3 2 | 2+2 2 |
| | 7+7 2 | 5+6 2 | 2+2 2 | 3+3 2 |
| | 5+6 2 | 5+6 2 | 2+2 2 | 3+3 2 |
| | 4+4 2 | 5+6 2 | 3+3 2 | 2+2 2 |
| | 5+6 2 | 4+4 2 | 2+2 2 | 2+2 2 |
| | 4+4 2 | 8+9 2 | 2+2 2 | 3+3 2 |
| | 5+6 2 | 4+4 2 | 3+3 2 | 2+2 2 |
| | 8+9 2 | 5+6 2 | 2+2 2 | 3+3 2 |
| | 8+9 2 | 8+9 2 | 2+2 2 | 2+2 2 |
| | 7+7 2 | 5+6 2 | 3+3 2 | 3+3 2 |
| | 5+6 2 | 7+7 2 | 3+3 2 | 3+3 2 |
| | 8+9 2 | 5+6 2 | 2+2 2 | 2+2 2 |
| | 5+6 2 | 4+4 2 | 2+2 2 | 2+2 2 |
| | 4+4 2 | 8+9 2 | 3+3 2 | 3+3 2 |
| | 7+7 2 | 8+9 2 | 2+2 2 | 2+2 2 |
| | 8+9 2 | 7+7 2 | 3+3 2 | 3+3 2 |
| | 5+6 2 | 8+9 2 | 2+2 2 | 2+2 2 |
| | 8+9 2 | 4+4 2 | 2+2 2 | 2+2 2 |
| | 7+7 2 | 7+7 2 | 2+2 2 | 2+2 2 |
| | 8+9 2 | 8+9 2 | 3+3 2 | 2+2 2 |
| | 4+4 2 | 8+9 2 | 2+2 2 | 3+3 2 |
| | 5+6 2 | 5+6 2 | 2+2 2 | 2+2 2 |
| | 5+6 2 | 5+6 2 | 3+3 2 | 3+3 2 |
| | 7+7 2 | 8+9 2 | 2+2 2 | 2+2 2 |
| | 8+9 2 | 5+6 2 | 3+3 2 | 3+3 2 |

Continued on next column

**TABLE LXVIII– continued**

| TQC | ICSD | | NoE | SF |
|---|---|---|---|---|
| | 8+9 2 | 7+7 2 | 2+2 2 | 3+3 2 |
| | 8+9 2 | 4+4 2 | 2+2 2 | 2+2 2 |
| | 7+7 2 | 8+9 2 | 2+2 2 | 2+2 2 |
| | 8+9 2 | 5+6 2 | 2+2 2 | 2+2 2 |
| | 8+9 2 | 8+9 2 | 2+2 2 | 2+2 2 |
| | 8+9 2 | 7+7 2 | 2+2 2 | 3+3 2 |
| | 4+4 2 | 8+9 2 | 3+3 2 | 2+2 2 |
| | 5+6 2 | 8+9 2 | 2+2 2 | 2+2 2 |
| | 8+9 2 | 8+9 2 | 3+3 2 | 3+3 2 |
| 148.1.10 | 46014 | | 108 | Mo6S8Yb1 |
| | -GM | -T | -F | -L |
| | 5+6 2 | 5+6 2 | 3+3 2 | 3+3 2 |
| | 8+9 2 | 8+9 2 | 2+2 2 | 2+2 2 |
| | 7+7 2 | 7+7 2 | 2+2 2 | 2+2 2 |
| | 8+9 2 | 8+9 2 | 2+2 2 | 2+2 2 |
| | 5+6 2 | 5+6 2 | 3+3 2 | 2+2 2 |
| | 8+9 2 | 8+9 2 | 2+2 2 | 3+3 2 |
| | 7+7 2 | 5+6 2 | 2+2 2 | 3+3 2 |
| | 8+9 2 | 4+4 2 | 3+3 2 | 3+3 2 |
| | 4+4 2 | 7+7 2 | 2+2 2 | 2+2 2 |
| | 5+6 2 | 8+9 2 | 3+3 2 | 2+2 2 |
| | 8+9 2 | 5+6 2 | 3+3 2 | 3+3 2 |
| | 5+6 2 | 8+9 2 | 2+2 2 | 2+2 2 |
| | 4+4 2 | 5+6 2 | 3+3 2 | 2+2 2 |
| | 5+6 2 | 8+9 2 | 3+3 2 | 3+3 2 |
| | 5+6 2 | 7+7 2 | 2+2 2 | 2+2 2 |
| | 8+9 2 | 8+9 2 | 3+3 2 | 3+3 2 |
| | 7+7 2 | 8+9 2 | 2+2 2 | 2+2 2 |
| | 8+9 2 | 4+4 2 | 2+2 2 | 3+3 2 |
| | 8+9 2 | 5+6 2 | 3+3 2 | 2+2 2 |
| | 8+9 2 | 8+9 2 | 2+2 2 | 3+3 2 |
| | 7+7 2 | 7+7 2 | 3+3 2 | 2+2 2 |
| | 8+9 2 | 8+9 2 | 2+2 2 | 2+2 2 |
| | 5+6 2 | 7+7 2 | 3+3 2 | 2+2 2 |
| | 7+7 2 | 5+6 2 | 2+2 2 | 3+3 2 |
| | 8+9 2 | 8+9 2 | 2+2 2 | 2+2 2 |
| | 5+6 2 | 5+6 2 | 2+2 2 | 3+3 2 |
| | 4+4 2 | 5+6 2 | 3+3 2 | 2+2 2 |
| | 5+6 2 | 4+4 2 | 2+2 2 | 2+2 2 |
| | 4+4 2 | 8+9 2 | 2+2 2 | 3+3 2 |
| | 5+6 2 | 8+9 2 | 2+2 2 | 2+2 2 |
| | 8+9 2 | 8+9 2 | 3+3 2 | 3+3 2 |
| | 8+9 2 | 7+7 2 | 2+2 2 | 2+2 2 |
| | 7+7 2 | 4+4 2 | 2+2 2 | 2+2 2 |
| | 8+9 2 | 5+6 2 | 2+2 2 | 2+2 2 |
| | 8+9 2 | 5+6 2 | 3+3 2 | 2+2 2 |
| | 7+7 2 | 8+9 2 | 2+2 2 | 3+3 2 |
| | 5+6 2 | 5+6 2 | 2+2 2 | 2+2 2 |
| | 8+9 2 | 4+4 2 | 3+3 2 | 3+3 2 |
| | 8+9 2 | 7+7 2 | 2+2 2 | 2+2 2 |
| | 5+6 2 | 8+9 2 | 2+2 2 | 2+2 2 |
| | 4+4 2 | 8+9 2 | 3+3 2 | 3+3 2 |
| | 8+9 2 | 8+9 2 | 2+2 2 | 2+2 2 |
| | 7+7 2 | 7+7 2 | 3+3 2 | 3+3 2 |
| | 5+6 2 | 8+9 2 | 3+3 2 | 2+2 2 |
| | 4+4 2 | 4+4 2 | 2+2 2 | 3+3 2 |
| | 5+6 2 | 5+6 2 | 2+2 2 | 2+2 2 |
| | 5+6 2 | 5+6 2 | 3+3 2 | 3+3 2 |
| | 8+9 2 | 8+9 2 | 3+3 2 | 2+2 2 |
| | 7+7 2 | 5+6 2 | 2+2 2 | 3+3 2 |
| | 8+9 2 | 4+4 2 | 2+2 2 | 3+3 2 |

Continued on next column

**TABLE LXVIII– continued**

| TQC | ICSD | NoE | SF |
|---|---|---|---|
| | 8+9 2 7+7 2 2+2 2 3+3 2 | | |
| | 4+4 2 5+6 2 3+3 2 2+2 2 | | |
| | 5+6 2 8+9 2 2+2 2 2+2 2 | | |
| | 8+9 2 8+9 2 3+3 2 3+3 2 | | |
| 148.1.11 | 46013 | 108 | Mo6S8Yb1 |
| | -GM -T -F -L | | |
| | 5+6 2 5+6 2 3+3 2 3+3 2 | | |
| | 8+9 2 8+9 2 2+2 2 2+2 2 | | |
| | 7+7 2 7+7 2 2+2 2 2+2 2 | | |
| | 8+9 2 8+9 2 2+2 2 2+2 2 | | |
| | 5+6 2 5+6 2 3+3 2 2+2 2 | | |
| | 8+9 2 8+9 2 2+2 2 3+3 2 | | |
| | 7+7 2 5+6 2 2+2 2 3+3 2 | | |
| | 8+9 2 4+4 2 3+3 2 3+3 2 | | |
| | 4+4 2 7+7 2 2+2 2 2+2 2 | | |
| | 5+6 2 8+9 2 3+3 2 2+2 2 | | |
| | 8+9 2 5+6 2 3+3 2 3+3 2 | | |
| | 5+6 2 8+9 2 2+2 2 2+2 2 | | |
| | 4+4 2 5+6 2 3+3 2 3+3 2 | | |
| | 5+6 2 8+9 2 3+3 2 2+2 2 | | |
| | 5+6 2 7+7 2 2+2 2 2+2 2 | | |
| | 8+9 2 8+9 2 3+3 2 3+3 2 | | |
| | 7+7 2 8+9 2 2+2 2 3+3 2 | | |
| | 8+9 2 4+4 2 2+2 2 2+2 2 | | |
| | 8+9 2 5+6 2 3+3 2 3+3 2 | | |
| | 5+6 2 7+7 2 3+3 2 2+2 2 | | |
| | 8+9 2 8+9 2 3+3 2 2+2 2 | | |
| | 7+7 2 5+6 2 2+2 2 3+3 2 | | |
| | 5+6 2 5+6 2 2+2 2 3+3 2 | | |
| | 4+4 2 5+6 2 3+3 2 2+2 2 | | |
| | 5+6 2 4+4 2 2+2 2 2+2 2 | | |
| | 8+9 2 8+9 2 2+2 2 3+3 2 | | |
| | 4+4 2 8+9 2 3+3 2 2+2 2 | | |
| | 5+6 2 7+7 2 2+2 2 3+3 2 | | |
| | 7+7 2 8+9 2 2+2 2 2+2 2 | | |
| | 8+9 2 5+6 2 2+2 2 2+2 2 | | |
| | 8+9 2 4+4 2 2+2 2 2+2 2 | | |
| | 7+7 2 5+6 2 3+3 2 2+2 2 | | |
| | 8+9 2 5+6 2 2+2 2 3+3 2 | | |
| | 5+6 2 4+4 2 3+3 2 3+3 2 | | |
| | 8+9 2 7+7 2 2+2 2 2+2 2 | | |
| | 5+6 2 8+9 2 2+2 2 2+2 2 | | |
| | 8+9 2 8+9 2 3+3 2 3+3 2 | | |
| | 4+4 2 8+9 2 2+2 2 2+2 2 | | |
| | 7+7 2 7+7 2 3+3 2 3+3 2 | | |
| | 8+9 2 8+9 2 2+2 2 2+2 2 | | |
| | 8+9 2 8+9 2 2+2 2 2+2 2 | | |
| | 7+7 2 8+9 2 2+2 2 2+2 2 | | |
| | 5+6 2 7+7 2 2+2 2 2+2 2 | | |
| | 8+9 2 4+4 2 3+3 2 2+2 2 | | |
| | 4+4 2 8+9 2 2+2 2 3+3 2 | | |
| | 5+6 2 5+6 2 2+2 2 2+2 2 | | |
| | 5+6 2 5+6 2 3+3 2 3+3 2 | | |
| | 7+7 2 5+6 2 3+3 2 2+2 2 | | |
| | 8+9 2 8+9 2 2+2 2 3+3 2 | | |
| | 8+9 2 4+4 2 2+2 2 3+3 2 | | |
| | 8+9 2 7+7 2 2+2 2 3+3 2 | | |
| | 4+4 2 5+6 2 3+3 2 2+2 2 | | |
| | 5+6 2 8+9 2 2+2 2 2+2 2 | | |
| | 8+9 2 8+9 2 3+3 2 3+3 2 | | |
| 148.1.12 | 201506 | 108 | Mo6S8Yb1 |

Continued on next column

**TABLE LXVIII– continued**

| TQC | ICSD | NoE | SF |
|---|---|---|---|
| | -GM -T -F -L | | |
| | 5+6 2 5+6 2 3+3 2 3+3 2 | | |
| | 8+9 2 8+9 2 2+2 2 2+2 2 | | |
| | 7+7 2 7+7 2 2+2 2 2+2 2 | | |
| | 8+9 2 8+9 2 2+2 2 2+2 2 | | |
| | 5+6 2 5+6 2 3+3 2 2+2 2 | | |
| | 8+9 2 8+9 2 2+2 2 3+3 2 | | |
| | 7+7 2 5+6 2 2+2 2 3+3 2 | | |
| | 8+9 2 4+4 2 3+3 2 3+3 2 | | |
| | 8+9 2 5+6 2 2+2 2 2+2 2 | | |
| | 4+4 2 7+7 2 3+3 2 2+2 2 | | |
| | 5+6 2 8+9 2 3+3 2 3+3 2 | | |
| | 5+6 2 8+9 2 2+2 2 2+2 2 | | |
| | 4+4 2 5+6 2 3+3 2 3+3 2 | | |
| | 5+6 2 8+9 2 3+3 2 2+2 2 | | |
| | 5+6 2 7+7 2 2+2 2 2+2 2 | | |
| | 8+9 2 8+9 2 3+3 2 3+3 2 | | |
| | 7+7 2 8+9 2 2+2 2 3+3 2 | | |
| | 8+9 2 4+4 2 2+2 2 2+2 2 | | |
| | 8+9 2 5+6 2 3+3 2 3+3 2 | | |
| | 8+9 2 7+7 2 3+3 2 2+2 2 | | |
| | 7+7 2 8+9 2 3+3 2 2+2 2 | | |
| | 5+6 2 5+6 2 2+2 2 2+2 2 | | |
| | 8+9 2 8+9 2 2+2 2 3+3 2 | | |
| | 5+6 2 8+9 2 2+2 2 2+2 2 | | |
| | 8+9 2 7+7 2 2+2 2 2+2 2 | | |
| | 7+7 2 5+6 2 2+2 2 3+3 2 | | |
| | 4+4 2 5+6 2 3+3 2 2+2 2 | | |
| | 5+6 2 4+4 2 2+2 2 2+2 2 | | |
| | 8+9 2 8+9 2 2+2 2 3+3 2 | | |
| | 4+4 2 8+9 2 3+3 2 2+2 2 | | |
| | 5+6 2 7+7 2 2+2 2 3+3 2 | | |
| | 8+9 2 4+4 2 2+2 2 2+2 2 | | |
| | 7+7 2 5+6 2 3+3 2 2+2 2 | | |
| | 8+9 2 8+9 2 2+2 2 2+2 2 | | |
| | 8+9 2 5+6 2 2+2 2 3+3 2 | | |
| | 7+7 2 5+6 2 2+2 2 2+2 2 | | |
| | 8+9 2 4+4 2 2+2 2 2+2 2 | | |
| | 5+6 2 8+9 2 3+3 2 3+3 2 | | |
| | 8+9 2 7+7 2 2+2 2 2+2 2 | | |
| | 5+6 2 8+9 2 2+2 2 2+2 2 | | |
| | 4+4 2 8+9 2 3+3 2 3+3 2 | | |
| | 8+9 2 8+9 2 2+2 2 3+3 2 | | |
| | 7+7 2 7+7 2 3+3 2 2+2 2 | | |
| | 5+6 2 8+9 2 2+2 2 2+2 2 | | |
| | 4+4 2 4+4 2 3+3 2 3+3 2 | | |
| | 5+6 2 5+6 2 2+2 2 2+2 2 | | |
| | 5+6 2 5+6 2 3+3 2 3+3 2 | | |
| | 7+7 2 5+6 2 3+3 2 2+2 2 | | |
| | 8+9 2 8+9 2 2+2 2 3+3 2 | | |
| | 8+9 2 4+4 2 2+2 2 3+3 2 | | |
| | 8+9 2 5+6 2 2+2 2 3+3 2 | | |
| | 4+4 2 7+7 2 3+3 2 2+2 2 | | |
| | 5+6 2 8+9 2 2+2 2 2+2 2 | | |
| | 8+9 2 8+9 2 3+3 2 3+3 2 | | |
| 148.1.13 | 62156 | 94 | Mo6S8Sr1 |
| | -GM -T -F -L | | |
| | 5+6 2 5+6 2 3+3 2 3+3 2 | | |
| | 8+9 2 8+9 2 2+2 2 2+2 2 | | |
| | 8+9 2 8+9 2 2+2 2 2+2 2 | | |
| | 7+7 2 7+7 2 2+2 2 2+2 2 | | |

Continued on next column

**TABLE LXVIII– continued**

| TQC | ICSD | NoE | SF | |
|---|---|---|---|---|
| | 5+6 2 | 8+9 2 | 3+3 2 | 2+2 2 |
| | 8+9 2 | 5+6 2 | 2+2 2 | 3+3 2 |
| | 8+9 2 | 5+6 2 | 2+2 2 | 3+3 2 |
| | 7+7 2 | 4+4 2 | 3+3 2 | 3+3 2 |
| | 4+4 2 | 5+6 2 | 2+2 2 | 2+2 2 |
| | 5+6 2 | 7+7 2 | 3+3 2 | 2+2 2 |
| | 8+9 2 | 8+9 2 | 3+3 2 | 3+3 2 |
| | 5+6 2 | 8+9 2 | 2+2 2 | 2+2 2 |
| | 4+4 2 | 8+9 2 | 3+3 2 | 2+2 2 |
| | 5+6 2 | 5+6 2 | 3+3 2 | 3+3 2 |
| | 5+6 2 | 7+7 2 | 2+2 2 | 2+2 2 |
| | 8+9 2 | 8+9 2 | 3+3 2 | 3+3 2 |
| | 7+7 2 | 8+9 2 | 2+2 2 | 2+2 2 |
| | 8+9 2 | 4+4 2 | 3+3 2 | 3+3 2 |
| | 8+9 2 | 5+6 2 | 2+2 2 | 2+2 2 |
| | 5+6 2 | 7+7 2 | 3+3 2 | 3+3 2 |
| | 7+7 2 | 8+9 2 | 3+3 2 | 2+2 2 |
| | 8+9 2 | 5+6 2 | 2+2 2 | 3+3 2 |
| | 5+6 2 | 5+6 2 | 2+2 2 | 3+3 2 |
| | 4+4 2 | 5+6 2 | 3+3 2 | 2+2 2 |
| | 5+6 2 | 4+4 2 | 2+2 2 | 2+2 2 |
| | 4+4 2 | 8+9 2 | 2+2 2 | 3+3 2 |
| | 5+6 2 | 4+4 2 | 3+3 2 | 3+3 2 |
| | 8+9 2 | 5+6 2 | 2+2 2 | 2+2 2 |
| | 7+7 2 | 8+9 2 | 3+3 2 | 2+2 2 |
| | 8+9 2 | 7+7 2 | 2+2 2 | 3+3 2 |
| | 5+6 2 | 5+6 2 | 2+2 2 | 3+3 2 |
| | 8+9 2 | 5+6 2 | 3+3 2 | 2+2 2 |
| | 5+6 2 | 4+4 2 | 2+2 2 | 2+2 2 |
| | 4+4 2 | 8+9 2 | 3+3 2 | 3+3 2 |
| | 7+7 2 | 8+9 2 | 2+2 2 | 3+3 2 |
| | 8+9 2 | 7+7 2 | 3+3 2 | 2+2 2 |
| | 5+6 2 | 4+4 2 | 2+2 2 | 2+2 2 |
| | 4+4 2 | 8+9 2 | 3+3 2 | 3+3 2 |
| | 5+6 2 | 5+6 2 | 2+2 2 | 2+2 2 |
| | 5+6 2 | 5+6 2 | 3+3 2 | 3+3 2 |
| | 7+7 2 | 5+6 2 | 2+2 2 | 2+2 2 |
| | 8+9 2 | 8+9 2 | 3+3 2 | 3+3 2 |
| | 8+9 2 | 4+4 2 | 2+2 2 | 3+3 2 |
| | 8+9 2 | 5+6 2 | 2+2 2 | 3+3 2 |
| | 4+4 2 | 7+7 2 | 2+2 2 | 2+2 2 |
| | 5+6 2 | 8+9 2 | 3+3 2 | 2+2 2 |
| | 8+9 2 | 8+9 2 | 3+3 2 | 3+3 2 |
| 148.1.14 | 68 | 88 | Mo3Pb0.5Se4 | |
| | -GM | -T | -F | -L |
| | 5+6 2 | 5+6 2 | 3+3 2 | 2+2 2 |
| | 8+9 2 | 8+9 2 | 2+2 2 | 3+3 2 |
| | 8+9 2 | 5+6 2 | 2+2 2 | 3+3 2 |
| | 7+7 2 | 8+9 2 | 3+3 2 | 2+2 2 |
| | 8+9 2 | 5+6 2 | 2+2 2 | 3+3 2 |
| | 4+4 2 | 4+4 2 | 3+3 2 | 2+2 2 |
| | 5+6 2 | 7+7 2 | 3+3 2 | 3+3 2 |
| | 5+6 2 | 8+9 2 | 2+2 2 | 2+2 2 |
| | 5+6 2 | 5+6 2 | 3+3 2 | 3+3 2 |
| | 5+6 2 | 8+9 2 | 2+2 2 | 2+2 2 |
| | 4+4 2 | 7+7 2 | 3+3 2 | 2+2 2 |
| | 8+9 2 | 8+9 2 | 2+2 2 | 3+3 2 |
| | 7+7 2 | 8+9 2 | 2+2 2 | 3+3 2 |
| | 8+9 2 | 4+4 2 | 3+3 2 | 2+2 2 |
| | 8+9 2 | 5+6 2 | 3+3 2 | 2+2 2 |
| | 5+6 2 | 7+7 2 | 3+3 2 | 2+2 2 |

Continued on next column

**TABLE LXVIII– continued**

| TQC | ICSD | NoE | SF | |
|---|---|---|---|---|
| | 8+9 2 | 8+9 2 | 2+2 2 | 3+3 2 |
| | 7+7 2 | 5+6 2 | 3+3 2 | 3+3 2 |
| | 5+6 2 | 5+6 2 | 3+3 2 | 3+3 2 |
| | 5+6 2 | 8+9 2 | 2+2 2 | 2+2 2 |
| | 4+4 2 | 5+6 2 | 2+2 2 | 3+3 2 |
| | 5+6 2 | 5+6 2 | 2+2 2 | 2+2 2 |
| | 5+6 2 | 4+4 2 | 3+3 2 | 2+2 2 |
| | 8+9 2 | 8+9 2 | 2+2 2 | 2+2 2 |
| | 4+4 2 | 5+6 2 | 3+3 2 | 3+3 2 |
| | 7+7 2 | 7+7 2 | 2+2 2 | 3+3 2 |
| | 8+9 2 | 4+4 2 | 3+3 2 | 3+3 2 |
| | 5+6 2 | 5+6 2 | 2+2 2 | 2+2 2 |
| | 8+9 2 | 5+6 2 | 3+3 2 | 3+3 2 |
| | 5+6 2 | 4+4 2 | 2+2 2 | 2+2 2 |
| | 4+4 2 | 8+9 2 | 3+3 2 | 3+3 2 |
| | 8+9 2 | 8+9 2 | 2+2 2 | 2+2 2 |
| | 7+7 2 | 8+9 2 | 3+3 2 | 3+3 2 |
| | 4+4 2 | 7+7 2 | 2+2 2 | 2+2 2 |
| | 5+6 2 | 4+4 2 | 2+2 2 | 3+3 2 |
| | 8+9 2 | 5+6 2 | 3+3 2 | 2+2 2 |
| | 5+6 2 | 5+6 2 | 2+2 2 | 2+2 2 |
| | 7+7 2 | 8+9 2 | 3+3 2 | 3+3 2 |
| | 8+9 2 | 5+6 2 | 2+2 2 | 3+3 2 |
| | 5+6 2 | 8+9 2 | 3+3 2 | 3+3 2 |
| | 8+9 2 | 7+7 2 | 2+2 2 | 2+2 2 |
| | 4+4 2 | 4+4 2 | 2+2 2 | 3+3 2 |
| | 5+6 2 | 5+6 2 | 3+3 2 | 2+2 2 |
| | 8+9 2 | 8+9 2 | 3+3 2 | 3+3 2 |
| 148.1.15 | 89004 | 174 | Mo12Rb2Se14 | |
| | -GM | -T | -F | -L |
| | 5+6 2 | 5+6 2 | 2+2 2 | 2+2 2 |
| | 8+9 2 | 8+9 2 | 3+3 2 | 3+3 2 |
| | 5+6 2 | 5+6 2 | 2+2 2 | 2+2 2 |
| | 8+9 2 | 8+9 2 | 3+3 2 | 3+3 2 |
| | 8+9 2 | 5+6 2 | 2+2 2 | 2+2 2 |
| | 7+7 2 | 4+4 2 | 3+3 2 | 3+3 2 |
| | 5+6 2 | 7+7 2 | 2+2 2 | 2+2 2 |
| | 5+6 2 | 8+9 2 | 3+3 2 | 3+3 2 |
| | 4+4 2 | 5+6 2 | 3+3 2 | 3+3 2 |
| | 7+7 2 | 8+9 2 | 2+2 2 | 2+2 2 |
| | 8+9 2 | 8+9 2 | 3+3 2 | 3+3 2 |
| | 8+9 2 | 7+7 2 | 2+2 2 | 3+3 2 |
| | 5+6 2 | 5+6 2 | 2+2 2 | 2+2 2 |
| | 4+4 2 | 4+4 2 | 3+3 2 | 2+2 2 |
| | 8+9 2 | 8+9 2 | 3+3 2 | 3+3 2 |
| | 5+6 2 | 5+6 2 | 2+2 2 | 2+2 2 |
| | 8+9 2 | 5+6 2 | 3+3 2 | 2+2 2 |
| | 5+6 2 | 8+9 2 | 2+2 2 | 3+3 2 |
| | 5+6 2 | 5+6 2 | 2+2 2 | 2+2 2 |
| | 8+9 2 | 8+9 2 | 3+3 2 | 3+3 2 |
| | 7+7 2 | 4+4 2 | 3+3 2 | 2+2 2 |
| | 4+4 2 | 7+7 2 | 2+2 2 | 3+3 2 |
| | 4+4 2 | 4+4 2 | 2+2 2 | 2+2 2 |
| | 5+6 2 | 5+6 2 | 3+3 2 | 3+3 2 |
| | 5+6 2 | 8+9 2 | 3+3 2 | 3+3 2 |
| | 8+9 2 | 7+7 2 | 2+2 2 | 3+3 2 |
| | 7+7 2 | 8+9 2 | 2+2 2 | 2+2 2 |
| | 8+9 2 | 7+7 2 | 3+3 2 | 2+2 2 |
| | 7+7 2 | 8+9 2 | 3+3 2 | 3+3 2 |
| | 8+9 2 | 5+6 2 | 2+2 2 | 2+2 2 |
| | 5+6 2 | 8+9 2 | 3+3 2 | 2+2 2 |

Continued on next column

**TABLE LXVIII– continued**

| TQC | ICSD | NoE | SF |
|---|---|---|---|
| | 4+4 2 5+6 2 2+2 2 3+3 2 | | |
| | 8+9 2 5+6 2 3+3 2 2+2 2 | | |
| | 5+6 2 8+9 2 3+3 2 3+3 2 | | |
| | 5+6 2 4+4 2 3+3 2 2+2 2 | | |
| | 8+9 2 5+6 2 2+2 2 2+2 2 | | |
| | 5+6 2 5+6 2 3+3 2 3+3 2 | | |
| | 8+9 2 8+9 2 2+2 2 2+2 2 | | |
| | 8+9 2 8+9 2 3+3 2 2+2 2 | | |
| | 7+7 2 7+7 2 2+2 2 3+3 2 | | |
| | 5+6 2 4+4 2 2+2 2 3+3 2 | | |
| | 4+4 2 5+6 2 2+2 2 3+3 2 | | |
| | 5+6 2 5+6 2 3+3 2 3+3 2 | | |
| | 7+7 2 8+9 2 3+3 2 2+2 2 | | |
| | 8+9 2 8+9 2 2+2 2 3+3 2 | | |
| | 5+6 2 7+7 2 2+2 2 2+2 2 | | |
| | 5+6 2 8+9 2 2+2 2 2+2 2 | | |
| | 4+4 2 5+6 2 3+3 2 3+3 2 | | |
| | 8+9 2 4+4 2 3+3 2 3+3 2 | | |
| | 4+4 2 8+9 2 2+2 2 2+2 2 | | |
| | 5+6 2 7+7 2 3+3 2 2+2 2 | | |
| | 8+9 2 5+6 2 2+2 2 3+3 2 | | |
| | 7+7 2 4+4 2 2+2 2 2+2 2 | | |
| | 8+9 2 5+6 2 3+3 2 3+3 2 | | |
| | 5+6 2 8+9 2 2+2 2 2+2 2 | | |
| | 8+9 2 5+6 2 3+3 2 3+3 2 | | |
| | 4+4 2 7+7 2 3+3 2 2+2 2 | | |
| | 5+6 2 5+6 2 2+2 2 3+3 2 | | |
| | 8+9 2 8+9 2 2+2 2 2+2 2 | | |
| | 5+6 2 4+4 2 3+3 2 3+3 2 | | |
| | 7+7 2 5+6 2 2+2 2 3+3 2 | | |
| | 8+9 2 5+6 2 2+2 2 2+2 2 | | |
| | 4+4 2 4+4 2 3+3 2 3+3 2 | | |
| | 8+9 2 8+9 2 3+3 2 2+2 2 | | |
| | 5+6 2 8+9 2 2+2 2 2+2 2 | | |
| | 7+7 2 4+4 2 3+3 2 3+3 2 | | |
| | 4+4 2 7+7 2 2+2 2 2+2 2 | | |
| | 5+6 2 5+6 2 3+3 2 3+3 2 | | |
| | 5+6 2 8+9 2 2+2 2 2+2 2 | | |
| | 8+9 2 5+6 2 3+3 2 2+2 2 | | |
| | 8+9 2 7+7 2 2+2 2 3+3 2 | | |
| | 5+6 2 8+9 2 3+3 2 3+3 2 | | |
| | 7+7 2 5+6 2 3+3 2 2+2 2 | | |
| | 4+4 2 4+4 2 2+2 2 3+3 2 | | |
| | 8+9 2 5+6 2 2+2 2 2+2 2 | | |
| | 5+6 2 8+9 2 3+3 2 2+2 2 | | |
| | 7+7 2 7+7 2 2+2 2 3+3 2 | | |
| | 8+9 2 4+4 2 2+2 2 3+3 2 | | |
| | 4+4 2 5+6 2 3+3 2 2+2 2 | | |
| | 5+6 2 8+9 2 3+3 2 3+3 2 | | |
| | 5+6 2 5+6 2 2+2 2 3+3 2 | | |
| | 8+9 2 7+7 2 2+2 2 2+2 2 | | |
| | 7+7 2 8+9 2 3+3 2 2+2 2 | | |
| | 8+9 2 4+4 2 2+2 2 3+3 2 | | |
| | 4+4 2 5+6 2 3+3 2 3+3 2 | | |
| | 5+6 2 8+9 2 3+3 2 2+2 2 | | |
| | 5+6 2 8+9 2 2+2 2 2+2 2 | | |
| 148.1.16 | 245054 | 126 | Mn3O6Te1 |
| | -GM -T -F -L | | |
| | 5+6 2 5+6 2 3+3 2 3+3 2 | | |
| | 5+6 2 8+9 2 3+3 2 2+2 2 | | |
| | 8+9 2 8+9 2 2+2 2 2+2 2 | | |

Continued on next column

**TABLE LXVIII– continued**

| TQC | ICSD | NoE | SF |
|---|---|---|---|
| | 8+9 2 5+6 2 2+2 2 3+3 2 | | |
| | 7+7 2 4+4 2 2+2 2 2+2 2 | | |
| | 8+9 2 8+9 2 2+2 2 3+3 2 | | |
| | 7+7 2 7+7 2 2+2 2 2+2 2 | | |
| | 8+9 2 5+6 2 2+2 2 3+3 2 | | |
| | 4+4 2 4+4 2 3+3 2 3+3 2 | | |
| | 5+6 2 5+6 2 3+3 2 3+3 2 | | |
| | 4+4 2 7+7 2 3+3 2 2+2 2 | | |
| | 5+6 2 8+9 2 3+3 2 2+2 2 | | |
| | 5+6 2 5+6 2 3+3 2 3+3 2 | | |
| | 5+6 2 8+9 2 3+3 2 2+2 2 | | |
| | 8+9 2 8+9 2 2+2 2 2+2 2 | | |
| | 8+9 2 5+6 2 2+2 2 3+3 2 | | |
| | 7+7 2 8+9 2 2+2 2 2+2 2 | | |
| | 8+9 2 4+4 2 2+2 2 2+2 2 | | |
| | 7+7 2 7+7 2 2+2 2 3+3 2 | | |
| | 8+9 2 5+6 2 2+2 2 3+3 2 | | |
| | 4+4 2 4+4 2 3+3 2 3+3 2 | | |
| | 5+6 2 5+6 2 3+3 2 2+2 2 | | |
| | 4+4 2 5+6 2 3+3 2 3+3 2 | | |
| | 5+6 2 7+7 2 3+3 2 2+2 2 | | |
| | 5+6 2 8+9 2 2+2 2 3+3 2 | | |
| | 8+9 2 8+9 2 3+3 2 2+2 2 | | |
| | 5+6 2 5+6 2 2+2 2 3+3 2 | | |
| | 4+4 2 8+9 2 3+3 2 2+2 2 | | |
| | 5+6 2 8+9 2 2+2 2 2+2 2 | | |
| | 8+9 2 7+7 2 3+3 2 3+3 2 | | |
| | 7+7 2 4+4 2 2+2 2 2+2 2 | | |
| | 4+4 2 5+6 2 3+3 2 3+3 2 | | |
| | 5+6 2 7+7 2 2+2 2 2+2 2 | | |
| | 8+9 2 8+9 2 3+3 2 3+3 2 | | |
| | 7+7 2 5+6 2 2+2 2 2+2 2 | | |
| | 5+6 2 4+4 2 2+2 2 3+3 2 | | |
| | 8+9 2 5+6 2 3+3 2 2+2 2 | | |
| | 8+9 2 8+9 2 2+2 2 3+3 2 | | |
| | 8+9 2 7+7 2 3+3 2 3+3 2 | | |
| | 7+7 2 8+9 2 2+2 2 2+2 2 | | |
| | 4+4 2 5+6 2 2+2 2 3+3 2 | | |
| | 5+6 2 4+4 2 2+2 2 2+2 2 | | |
| | 8+9 2 5+6 2 2+2 2 3+3 2 | | |
| | 5+6 2 5+6 2 3+3 2 2+2 2 | | |
| | 5+6 2 4+4 2 3+3 2 3+3 2 | | |
| | 4+4 2 8+9 2 3+3 2 2+2 2 | | |
| | 8+9 2 7+7 2 3+3 2 3+3 2 | | |
| | 7+7 2 8+9 2 3+3 2 2+2 2 | | |
| | 5+6 2 5+6 2 3+3 2 2+2 2 | | |
| | 8+9 2 8+9 2 2+2 2 3+3 2 | | |
| | 5+6 2 5+6 2 2+2 2 2+2 2 | | |
| | 4+4 2 4+4 2 3+3 2 3+3 2 | | |
| | 5+6 2 8+9 2 3+3 2 3+3 2 | | |
| | 7+7 2 7+7 2 2+2 2 2+2 2 | | |
| | 8+9 2 8+9 2 2+2 2 3+3 2 | | |
| | 4+4 2 5+6 2 2+2 2 3+3 2 | | |
| | 5+6 2 7+7 2 2+2 2 2+2 2 | | |
| | 8+9 2 8+9 2 3+3 2 2+2 2 | | |
| | 8+9 2 5+6 2 3+3 2 3+3 2 | | |
| | 7+7 2 4+4 2 2+2 2 2+2 2 | | |
| | 5+6 2 5+6 2 3+3 2 3+3 2 | | |
| | 5+6 2 8+9 2 2+2 2 2+2 2 | | |
| | 4+4 2 7+7 2 3+3 2 3+3 2 | | |
| 148.1.17 | 183618 | 126 | Mn3O6Te1 |

Continued on next column

**TABLE LXVIII– continued**

| TQC | ICSD NoE SF |
|---|---|
| | -GM -T -F -L |
| | 5+6 2 5+6 2 3+3 2 3+3 2 |
| | 5+6 2 8+9 2 3+3 2 2+2 2 |
| | 8+9 2 8+9 2 2+2 2 2+2 2 |
| | 8+9 2 5+6 2 2+2 2 3+3 2 |
| | 7+7 2 4+4 2 2+2 2 2+2 2 |
| | 8+9 2 8+9 2 2+2 2 3+3 2 |
| | 7+7 2 7+7 2 2+2 2 2+2 2 |
| | 8+9 2 5+6 2 2+2 2 3+3 2 |
| | 4+4 2 4+4 2 3+3 2 3+3 2 |
| | 5+6 2 5+6 2 3+3 2 3+3 2 |
| | 4+4 2 7+7 2 3+3 2 2+2 2 |
| | 5+6 2 8+9 2 3+3 2 2+2 2 |
| | 5+6 2 5+6 2 3+3 2 3+3 2 |
| | 5+6 2 8+9 2 3+3 2 2+2 2 |
| | 8+9 2 8+9 2 2+2 2 2+2 2 |
| | 8+9 2 5+6 2 2+2 2 3+3 2 |
| | 7+7 2 8+9 2 2+2 2 2+2 2 |
| | 8+9 2 4+4 2 2+2 2 2+2 2 |
| | 7+7 2 7+7 2 2+2 2 3+3 2 |
| | 8+9 2 5+6 2 2+2 2 3+3 2 |
| | 4+4 2 4+4 2 3+3 2 3+3 2 |
| | 5+6 2 5+6 2 3+3 2 2+2 2 |
| | 4+4 2 5+6 2 3+3 2 3+3 2 |
| | 5+6 2 7+7 2 3+3 2 2+2 2 |
| | 5+6 2 8+9 2 2+2 2 3+3 2 |
| | 8+9 2 8+9 2 3+3 2 2+2 2 |
| | 5+6 2 5+6 2 2+2 2 3+3 2 |
| | 4+4 2 8+9 2 3+3 2 2+2 2 |
| | 5+6 2 8+9 2 2+2 2 2+2 2 |
| | 8+9 2 7+7 2 3+3 2 3+3 2 |
| | 7+7 2 4+4 2 2+2 2 2+2 2 |
| | 4+4 2 5+6 2 3+3 2 3+3 2 |
| | 5+6 2 7+7 2 2+2 2 2+2 2 |
| | 8+9 2 8+9 2 3+3 2 3+3 2 |
| | 7+7 2 5+6 2 2+2 2 2+2 2 |
| | 5+6 2 4+4 2 2+2 2 3+3 2 |
| | 8+9 2 5+6 2 3+3 2 2+2 2 |
| | 8+9 2 8+9 2 2+2 2 3+3 2 |
| | 8+9 2 7+7 2 2+2 2 3+3 2 |
| | 7+7 2 8+9 2 3+3 2 2+2 2 |
| | 4+4 2 5+6 2 2+2 2 3+3 2 |
| | 5+6 2 4+4 2 2+2 2 2+2 2 |
| | 8+9 2 5+6 2 2+2 2 3+3 2 |
| | 5+6 2 5+6 2 3+3 2 2+2 2 |
| | 5+6 2 4+4 2 3+3 2 3+3 2 |
| | 4+4 2 8+9 2 3+3 2 2+2 2 |
| | 8+9 2 7+7 2 3+3 2 3+3 2 |
| | 7+7 2 8+9 2 3+3 2 2+2 2 |
| | 5+6 2 5+6 2 3+3 2 2+2 2 |
| | 8+9 2 8+9 2 2+2 2 3+3 2 |
| | 5+6 2 5+6 2 2+2 2 2+2 2 |
| | 4+4 2 4+4 2 3+3 2 3+3 2 |
| | 5+6 2 8+9 2 3+3 2 3+3 2 |
| | 7+7 2 7+7 2 2+2 2 2+2 2 |
| | 8+9 2 8+9 2 2+2 2 3+3 2 |
| | 4+4 2 5+6 2 2+2 2 3+3 2 |
| | 5+6 2 7+7 2 2+2 2 2+2 2 |
| | 8+9 2 8+9 2 3+3 2 2+2 2 |
| | 8+9 2 5+6 2 3+3 2 3+3 2 |
| | 7+7 2 4+4 2 2+2 2 2+2 2 |

Continued on next column

**TABLE LXVIII– continued**

| TQC | ICSD NoE SF |
|---|---|
| | 5+6 2 5+6 2 3+3 2 3+3 2 |
| | 5+6 2 8+9 2 2+2 2 2+2 2 |
| | 4+4 2 7+7 2 3+3 2 3+3 2 |
| 148.1.18 | 182436 126 Mn3O6Te1 |
| | -GM -T -F -L |
| | 5+6 2 5+6 2 3+3 2 3+3 2 |
| | 5+6 2 8+9 2 3+3 2 2+2 2 |
| | 8+9 2 8+9 2 2+2 2 2+2 2 |
| | 8+9 2 5+6 2 2+2 2 3+3 2 |
| | 7+7 2 4+4 2 2+2 2 3+3 2 |
| | 8+9 2 8+9 2 2+2 2 2+2 2 |
| | 7+7 2 7+7 2 2+2 2 2+2 2 |
| | 8+9 2 5+6 2 2+2 2 3+3 2 |
| | 4+4 2 4+4 2 3+3 2 3+3 2 |
| | 5+6 2 5+6 2 3+3 2 3+3 2 |
| | 4+4 2 7+7 2 3+3 2 2+2 2 |
| | 5+6 2 8+9 2 3+3 2 2+2 2 |
| | 5+6 2 5+6 2 3+3 2 3+3 2 |
| | 5+6 2 8+9 2 3+3 2 2+2 2 |
| | 8+9 2 8+9 2 2+2 2 2+2 2 |
| | 8+9 2 5+6 2 2+2 2 3+3 2 |
| | 7+7 2 8+9 2 2+2 2 2+2 2 |
| | 8+9 2 4+4 2 2+2 2 2+2 2 |
| | 7+7 2 7+7 2 2+2 2 3+3 2 |
| | 8+9 2 5+6 2 2+2 2 3+3 2 |
| | 4+4 2 4+4 2 3+3 2 3+3 2 |
| | 5+6 2 5+6 2 3+3 2 2+2 2 |
| | 4+4 2 5+6 2 3+3 2 3+3 2 |
| | 5+6 2 7+7 2 3+3 2 2+2 2 |
| | 5+6 2 8+9 2 2+2 2 3+3 2 |
| | 8+9 2 8+9 2 3+3 2 2+2 2 |
| | 5+6 2 5+6 2 2+2 2 3+3 2 |
| | 4+4 2 8+9 2 3+3 2 2+2 2 |
| | 5+6 2 8+9 2 2+2 2 2+2 2 |
| | 8+9 2 7+7 2 3+3 2 3+3 2 |
| | 7+7 2 4+4 2 2+2 2 2+2 2 |
| | 4+4 2 5+6 2 3+3 2 3+3 2 |
| | 5+6 2 7+7 2 2+2 2 2+2 2 |
| | 8+9 2 8+9 2 3+3 2 3+3 2 |
| | 7+7 2 5+6 2 2+2 2 2+2 2 |
| | 5+6 2 4+4 2 2+2 2 3+3 2 |
| | 8+9 2 5+6 2 3+3 2 2+2 2 |
| | 8+9 2 8+9 2 2+2 2 3+3 2 |
| | 8+9 2 7+7 2 3+3 2 3+3 2 |
| | 7+7 2 8+9 2 2+2 2 2+2 2 |
| | 4+4 2 5+6 2 2+2 2 3+3 2 |
| | 5+6 2 4+4 2 2+2 2 2+2 2 |
| | 8+9 2 5+6 2 2+2 2 3+3 2 |
| | 5+6 2 5+6 2 3+3 2 2+2 2 |
| | 5+6 2 4+4 2 3+3 2 3+3 2 |
| | 4+4 2 8+9 2 3+3 2 2+2 2 |
| | 8+9 2 7+7 2 3+3 2 3+3 2 |
| | 7+7 2 8+9 2 3+3 2 2+2 2 |
| | 5+6 2 5+6 2 3+3 2 2+2 2 |
| | 8+9 2 8+9 2 2+2 2 3+3 2 |
| | 5+6 2 5+6 2 3+3 2 2+2 2 |
| | 4+4 2 8+9 2 2+2 2 3+3 2 |
| | 5+6 2 4+4 2 3+3 2 2+2 2 |
| | 7+7 2 7+7 2 2+2 2 3+3 2 |
| | 8+9 2 8+9 2 2+2 2 3+3 2 |
| | 4+4 2 5+6 2 2+2 2 3+3 2 |

Continued on next column

**TABLE LXVIII– continued**

| TQC | ICSD | NoE | SF |
|---|---|---|---|
| | 5+6 2 7+7 2 2+2 2 2+2 2 | | |
| | 8+9 2 8+9 2 3+3 2 2+2 2 | | |
| | 8+9 2 5+6 2 3+3 2 3+3 2 | | |
| | 7+7 2 4+4 2 2+2 2 2+2 2 | | |
| | 5+6 2 5+6 2 3+3 2 3+3 2 | | |
| | 5+6 2 8+9 2 2+2 2 2+2 2 | | |
| | 4+4 2 7+7 2 3+3 2 3+3 2 | | |
| 148.1.19 | 182435 | 126 | Mn3O6Te1 |
| | -GM -T -F -L | | |
| | 5+6 2 5+6 2 3+3 2 3+3 2 | | |
| | 5+6 2 8+9 2 3+3 2 2+2 2 | | |
| | 8+9 2 8+9 2 2+2 2 2+2 2 | | |
| | 8+9 2 5+6 2 2+2 2 3+3 2 | | |
| | 7+7 2 4+4 2 2+2 2 3+3 2 | | |
| | 8+9 2 8+9 2 2+2 2 2+2 2 | | |
| | 7+7 2 7+7 2 2+2 2 2+2 2 | | |
| | 8+9 2 5+6 2 2+2 2 3+3 2 | | |
| | 4+4 2 4+4 2 3+3 2 3+3 2 | | |
| | 5+6 2 5+6 2 3+3 2 2+2 2 | | |
| | 4+4 2 7+7 2 3+3 2 3+3 2 | | |
| | 5+6 2 8+9 2 3+3 2 2+2 2 | | |
| | 5+6 2 5+6 2 3+3 2 3+3 2 | | |
| | 5+6 2 8+9 2 3+3 2 2+2 2 | | |
| | 8+9 2 8+9 2 2+2 2 2+2 2 | | |
| | 8+9 2 5+6 2 2+2 2 3+3 2 | | |
| | 7+7 2 4+4 2 2+2 2 2+2 2 | | |
| | 8+9 2 8+9 2 2+2 2 3+3 2 | | |
| | 7+7 2 7+7 2 2+2 2 2+2 2 | | |
| | 8+9 2 5+6 2 2+2 2 3+3 2 | | |
| | 4+4 2 4+4 2 3+3 2 2+2 2 | | |
| | 5+6 2 5+6 2 3+3 2 3+3 2 | | |
| | 4+4 2 7+7 2 3+3 2 3+3 2 | | |
| | 5+6 2 8+9 2 3+3 2 2+2 2 | | |
| | 5+6 2 5+6 2 2+2 2 3+3 2 | | |
| | 8+9 2 8+9 2 3+3 2 2+2 2 | | |
| | 5+6 2 5+6 2 2+2 2 3+3 2 | | |
| | 4+4 2 8+9 2 3+3 2 2+2 2 | | |
| | 5+6 2 8+9 2 2+2 2 2+2 2 | | |
| | 8+9 2 7+7 2 3+3 2 3+3 2 | | |
| | 7+7 2 4+4 2 2+2 2 2+2 2 | | |
| | 8+9 2 5+6 2 3+3 2 3+3 2 | | |
| | 4+4 2 7+7 2 2+2 2 2+2 2 | | |
| | 7+7 2 8+9 2 3+3 2 3+3 2 | | |
| | 5+6 2 5+6 2 2+2 2 2+2 2 | | |
| | 5+6 2 4+4 2 2+2 2 3+3 2 | | |
| | 8+9 2 5+6 2 3+3 2 2+2 2 | | |
| | 8+9 2 8+9 2 2+2 2 3+3 2 | | |
| | 7+7 2 7+7 2 2+2 2 3+3 2 | | |
| | 8+9 2 8+9 2 3+3 2 2+2 2 | | |
| | 4+4 2 5+6 2 2+2 2 3+3 2 | | |
| | 5+6 2 4+4 2 2+2 2 2+2 2 | | |
| | 8+9 2 5+6 2 2+2 2 3+3 2 | | |
| | 5+6 2 8+9 2 3+3 2 2+2 2 | | |
| | 5+6 2 7+7 2 3+3 2 3+3 2 | | |
| | 4+4 2 5+6 2 3+3 2 2+2 2 | | |
| | 8+9 2 4+4 2 3+3 2 3+3 2 | | |
| | 7+7 2 8+9 2 3+3 2 2+2 2 | | |
| | 5+6 2 5+6 2 3+3 2 2+2 2 | | |
| | 8+9 2 8+9 2 2+2 2 3+3 2 | | |
| | 5+6 2 5+6 2 3+3 2 2+2 2 | | |
| | 5+6 2 4+4 2 2+2 2 3+3 2 | | |
| | 4+4 2 8+9 2 2+2 2 3+3 2 | | |
| | 7+7 2 7+7 2 3+3 2 2+2 2 | | |
| | 8+9 2 8+9 2 2+2 2 3+3 2 | | |
| | 4+4 2 5+6 2 2+2 2 3+3 2 | | |
| | 5+6 2 7+7 2 2+2 2 2+2 2 | | |
| | 8+9 2 8+9 2 3+3 2 2+2 2 | | |
| | 8+9 2 5+6 2 3+3 2 3+3 2 | | |
| | 7+7 2 4+4 2 2+2 2 2+2 2 | | |
| | 5+6 2 5+6 2 3+3 2 3+3 2 | | |
| | 5+6 2 8+9 2 2+2 2 2+2 2 | | |
| | 4+4 2 7+7 2 3+3 2 3+3 2 | | |
| 148.1.20 | 60006 | 58 | Mn1O3Ti1 |
| | -GM -T -F -L | | |
| | 5+6 2 8+9 2 2+2 2 3+3 2 | | |
| | 8+9 2 5+6 2 3+3 2 2+2 2 | | |
| | 7+7 2 5+6 2 2+2 2 2+2 2 | | |
| | 8+9 2 4+4 2 3+3 2 3+3 2 | | |
| | 5+6 2 8+9 2 3+3 2 2+2 2 | | |
| | 4+4 2 7+7 2 2+2 2 3+3 2 | | |
| | 5+6 2 8+9 2 2+2 2 2+2 2 | | |
| | 7+7 2 5+6 2 3+3 2 3+3 2 | | |
| | 8+9 2 8+9 2 2+2 2 3+3 2 | | |
| | 5+6 2 7+7 2 3+3 2 2+2 2 | | |
| | 4+4 2 5+6 2 3+3 2 3+3 2 | | |
| | 8+9 2 4+4 2 2+2 2 3+3 2 | | |
| | 4+4 2 8+9 2 2+2 2 2+2 2 | | |
| | 5+6 2 7+7 2 3+3 2 2+2 2 | | |
| | 5+6 2 5+6 2 3+3 2 3+3 2 | | |
| | 7+7 2 8+9 2 2+2 2 2+2 2 | | |
| | 8+9 2 5+6 2 2+2 2 2+2 2 | | |
| | 5+6 2 4+4 2 3+3 2 3+3 2 | | |
| | 8+9 2 8+9 2 2+2 2 3+3 2 | | |
| | 5+6 2 5+6 2 2+2 2 2+2 2 | | |
| | 4+4 2 8+9 2 3+3 2 3+3 2 | | |
| | 8+9 2 7+7 2 2+2 2 2+2 2 | | |
| | 8+9 2 5+6 2 3+3 2 3+3 2 | | |
| | 7+7 2 4+4 2 3+3 2 2+2 2 | | |
| | 4+4 2 4+4 2 3+3 2 2+2 2 | | |
| | 5+6 2 5+6 2 2+2 2 3+3 2 | | |
| | 5+6 2 7+7 2 2+2 2 2+2 2 | | |
| | 7+7 2 8+9 2 3+3 2 3+3 2 | | |
| | 8+9 2 5+6 2 2+2 2 2+2 2 | | |
| 148.1.21 | 247553 | 58 | Mn1O3Ti1 |
| | -GM -T -F -L | | |
| | 5+6 2 8+9 2 2+2 2 3+3 2 | | |
| | 8+9 2 5+6 2 3+3 2 2+2 2 | | |
| | 7+7 2 5+6 2 2+2 2 2+2 2 | | |
| | 5+6 2 4+4 2 3+3 2 3+3 2 | | |
| | 4+4 2 8+9 2 3+3 2 2+2 2 | | |
| | 8+9 2 7+7 2 2+2 2 3+3 2 | | |
| | 5+6 2 8+9 2 2+2 2 2+2 2 | | |
| | 7+7 2 8+9 2 3+3 2 3+3 2 | | |
| | 8+9 2 7+7 2 2+2 2 3+3 2 | | |
| | 5+6 2 5+6 2 3+3 2 2+2 2 | | |
| | 4+4 2 5+6 2 3+3 2 3+3 2 | | |
| | 8+9 2 4+4 2 2+2 2 2+2 2 | | |
| | 5+6 2 8+9 2 2+2 2 3+3 2 | | |
| | 4+4 2 7+7 2 3+3 2 2+2 2 | | |
| | 5+6 2 8+9 2 3+3 2 3+3 2 | | |
| | 7+7 2 5+6 2 2+2 2 2+2 2 | | |
| | 8+9 2 5+6 2 2+2 2 2+2 2 | | |

Continued on next column

**TABLE LXVIII– continued**

| TQC | ICSD NoE SF |
|---|---|
| | 5+6 2 4+4 2 3+3 2 3+3 2 |
| | 8+9 2 8+9 2 2+2 2 3+3 2 |
| | 5+6 2 8+9 2 2+2 2 3+3 2 |
| | 4+4 2 7+7 2 3+3 2 2+2 2 |
| | 8+9 2 5+6 2 2+2 2 2+2 2 |
| | 8+9 2 5+6 2 3+3 2 3+3 2 |
| | 7+7 2 4+4 2 3+3 2 2+2 2 |
| | 5+6 2 5+6 2 3+3 2 3+3 2 |
| | 4+4 2 4+4 2 2+2 2 2+2 2 |
| | 5+6 2 5+6 2 2+2 2 2+2 2 |
| | 7+7 2 7+7 2 3+3 2 3+3 2 |
| | 8+9 2 8+9 2 3+3 2 2+2 2 |
| 148.1.22 | 171579 58 Mn1O3Ti1 |
| | -GM -T -F -L |
| | 5+6 2 8+9 2 2+2 2 3+3 2 |
| | 8+9 2 5+6 2 3+3 2 2+2 2 |
| | 7+7 2 5+6 2 2+2 2 2+2 2 |
| | 5+6 2 4+4 2 3+3 2 3+3 2 |
| | 4+4 2 8+9 2 3+3 2 2+2 2 |
| | 8+9 2 7+7 2 2+2 2 3+3 2 |
| | 5+6 2 8+9 2 2+2 2 2+2 2 |
| | 7+7 2 5+6 2 3+3 2 3+3 2 |
| | 8+9 2 8+9 2 2+2 2 3+3 2 |
| | 5+6 2 7+7 2 3+3 2 2+2 2 |
| | 4+4 2 5+6 2 3+3 2 3+3 2 |
| | 8+9 2 4+4 2 2+2 2 3+3 2 |
| | 5+6 2 8+9 2 2+2 2 2+2 2 |
| | 4+4 2 7+7 2 3+3 2 2+2 2 |
| | 5+6 2 8+9 2 3+3 2 3+3 2 |
| | 7+7 2 5+6 2 2+2 2 2+2 2 |
| | 8+9 2 5+6 2 2+2 2 2+2 2 |
| | 5+6 2 4+4 2 3+3 2 3+3 2 |
| | 8+9 2 8+9 2 2+2 2 3+3 2 |
| | 5+6 2 5+6 2 2+2 2 3+3 2 |
| | 4+4 2 8+9 2 3+3 2 2+2 2 |
| | 8+9 2 7+7 2 2+2 2 2+2 2 |
| | 8+9 2 5+6 2 3+3 2 3+3 2 |
| | 7+7 2 4+4 2 3+3 2 2+2 2 |
| | 5+6 2 5+6 2 3+3 2 3+3 2 |
| | 4+4 2 4+4 2 2+2 2 2+2 2 |
| | 5+6 2 7+7 2 2+2 2 2+2 2 |
| | 7+7 2 5+6 2 3+3 2 3+3 2 |
| | 8+9 2 8+9 2 2+2 2 2+2 2 |
| 148.1.23 | 73247 58 Ge1Mn1O3 |
| | -GM -T -F -L |
| | 5+6 2 8+9 2 2+2 2 3+3 2 |
| | 8+9 2 5+6 2 3+3 2 2+2 2 |
| | 5+6 2 8+9 2 2+2 2 3+3 2 |
| | 4+4 2 7+7 2 3+3 2 2+2 2 |
| | 8+9 2 5+6 2 2+2 2 3+3 2 |
| | 7+7 2 4+4 2 3+3 2 2+2 2 |
| | 8+9 2 5+6 2 3+3 2 2+2 2 |
| | 5+6 2 8+9 2 2+2 2 3+3 2 |
| | 5+6 2 8+9 2 2+2 2 3+3 2 |
| | 5+6 2 8+9 2 2+2 2 2+2 2 |
| | 4+4 2 7+7 2 3+3 2 2+2 2 |
| | 8+9 2 5+6 2 3+3 2 3+3 2 |
| | 7+7 2 4+4 2 2+2 2 3+3 2 |
| | 5+6 2 5+6 2 3+3 2 2+2 2 |
| | 8+9 2 8+9 2 2+2 2 3+3 2 |
| | 4+4 2 7+7 2 3+3 2 2+2 2 |

Continued on next column

**TABLE LXVIII– continued**

| TQC | ICSD NoE SF |
|---|---|
| | 5+6 2 8+9 2 2+2 2 2+2 2 |
| | 8+9 2 5+6 2 3+3 2 3+3 2 |
| | 7+7 2 4+4 2 3+3 2 3+3 2 |
| | 5+6 2 5+6 2 2+2 2 2+2 2 |
| | 4+4 2 5+6 2 3+3 2 3+3 2 |
| | 8+9 2 4+4 2 2+2 2 2+2 2 |
| | 8+9 2 8+9 2 3+3 2 3+3 2 |
| | 7+7 2 7+7 2 2+2 2 2+2 2 |
| | 5+6 2 5+6 2 3+3 2 3+3 2 |
| | 4+4 2 4+4 2 2+2 2 2+2 2 |
| | 5+6 2 5+6 2 2+2 2 2+2 2 |
| | 7+7 2 8+9 2 3+3 2 3+3 2 |
| | 8+9 2 7+7 2 3+3 2 2+2 2 |
| 148.1.24 | 73246 58 Ge1Mn1O3 |
| | -GM -T -F -L |
| | 5+6 2 8+9 2 2+2 2 3+3 2 |
| | 8+9 2 5+6 2 3+3 2 2+2 2 |
| | 5+6 2 8+9 2 2+2 2 3+3 2 |
| | 4+4 2 7+7 2 3+3 2 2+2 2 |
| | 8+9 2 5+6 2 2+2 2 3+3 2 |
| | 7+7 2 4+4 2 3+3 2 2+2 2 |
| | 8+9 2 5+6 2 3+3 2 2+2 2 |
| | 5+6 2 8+9 2 2+2 2 3+3 2 |
| | 5+6 2 8+9 2 2+2 2 3+3 2 |
| | 5+6 2 8+9 2 2+2 2 2+2 2 |
| | 4+4 2 7+7 2 3+3 2 2+2 2 |
| | 8+9 2 5+6 2 3+3 2 3+3 2 |
| | 7+7 2 4+4 2 2+2 2 3+3 2 |
| | 5+6 2 5+6 2 3+3 2 2+2 2 |
| | 8+9 2 8+9 2 2+2 2 3+3 2 |
| | 4+4 2 7+7 2 3+3 2 2+2 2 |
| | 5+6 2 8+9 2 2+2 2 2+2 2 |
| | 8+9 2 5+6 2 3+3 2 3+3 2 |
| | 7+7 2 4+4 2 3+3 2 3+3 2 |
| | 5+6 2 5+6 2 2+2 2 2+2 2 |
| | 4+4 2 5+6 2 2+2 2 3+3 2 |
| | 8+9 2 4+4 2 3+3 2 2+2 2 |
| | 8+9 2 8+9 2 3+3 2 3+3 2 |
| | 7+7 2 7+7 2 2+2 2 2+2 2 |
| | 5+6 2 5+6 2 3+3 2 3+3 2 |
| | 4+4 2 4+4 2 2+2 2 2+2 2 |
| | 5+6 2 5+6 2 2+2 2 2+2 2 |
| | 7+7 2 8+9 2 3+3 2 3+3 2 |
| | 8+9 2 7+7 2 3+3 2 2+2 2 |
| 148.1.25 | 73245 58 Ge1Mn1O3 |
| | -GM -T -F -L |
| | 5+6 2 8+9 2 2+2 2 3+3 2 |
| | 8+9 2 5+6 2 3+3 2 2+2 2 |
| | 5+6 2 8+9 2 2+2 2 3+3 2 |
| | 4+4 2 7+7 2 3+3 2 2+2 2 |
| | 8+9 2 5+6 2 2+2 2 3+3 2 |
| | 7+7 2 4+4 2 3+3 2 2+2 2 |
| | 8+9 2 5+6 2 3+3 2 2+2 2 |
| | 5+6 2 8+9 2 2+2 2 3+3 2 |
| | 5+6 2 8+9 2 2+2 2 3+3 2 |
| | 5+6 2 8+9 2 2+2 2 2+2 2 |
| | 4+4 2 7+7 2 3+3 2 2+2 2 |
| | 8+9 2 5+6 2 3+3 2 3+3 2 |
| | 7+7 2 4+4 2 2+2 2 3+3 2 |
| | 5+6 2 5+6 2 3+3 2 2+2 2 |
| | 8+9 2 8+9 2 2+2 2 3+3 2 |

Continued on next column

**TABLE LXVIII– continued**

| TQC | ICSD NoE SF |
|---|---|
| | 4+4 2 7+7 2 3+3 2 2+2 2 |
| | 5+6 2 8+9 2 2+2 2 2+2 2 |
| | 8+9 2 5+6 2 3+3 2 3+3 2 |
| | 7+7 2 4+4 2 3+3 2 3+3 2 |
| | 5+6 2 5+6 2 2+2 2 2+2 2 |
| | 4+4 2 5+6 2 2+2 2 3+3 2 |
| | 8+9 2 4+4 2 3+3 2 2+2 2 |
| | 8+9 2 8+9 2 3+3 2 3+3 2 |
| | 7+7 2 7+7 2 2+2 2 2+2 2 |
| | 5+6 2 5+6 2 3+3 2 3+3 2 |
| | 4+4 2 4+4 2 2+2 2 2+2 2 |
| | 5+6 2 5+6 2 2+2 2 2+2 2 |
| | 7+7 2 8+9 2 3+3 2 3+3 2 |
| | 8+9 2 7+7 2 3+3 2 2+2 2 |
| 148.1.26 | 73244 58 Ge1Mn1O3 |
| | -GM -T -F -L |
| | 5+6 2 8+9 2 2+2 2 3+3 2 |
| | 8+9 2 5+6 2 3+3 2 2+2 2 |
| | 5+6 2 8+9 2 2+2 2 3+3 2 |
| | 4+4 2 7+7 2 3+3 2 2+2 2 |
| | 8+9 2 5+6 2 2+2 2 3+3 2 |
| | 7+7 2 4+4 2 3+3 2 2+2 2 |
| | 8+9 2 5+6 2 3+3 2 2+2 2 |
| | 5+6 2 8+9 2 2+2 2 3+3 2 |
| | 5+6 2 8+9 2 2+2 2 3+3 2 |
| | 5+6 2 8+9 2 2+2 2 2+2 2 |
| | 4+4 2 7+7 2 3+3 2 2+2 2 |
| | 8+9 2 5+6 2 3+3 2 3+3 2 |
| | 7+7 2 4+4 2 2+2 2 3+3 2 |
| | 5+6 2 5+6 2 3+3 2 2+2 2 |
| | 8+9 2 8+9 2 2+2 2 2+2 2 |
| | 4+4 2 7+7 2 3+3 2 3+3 2 |
| | 5+6 2 8+9 2 2+2 2 2+2 2 |
| | 8+9 2 5+6 2 3+3 2 3+3 2 |
| | 7+7 2 4+4 2 3+3 2 3+3 2 |
| | 5+6 2 5+6 2 2+2 2 2+2 2 |
| | 4+4 2 5+6 2 2+2 2 3+3 2 |
| | 8+9 2 4+4 2 3+3 2 2+2 2 |
| | 8+9 2 8+9 2 3+3 2 3+3 2 |
| | 7+7 2 7+7 2 2+2 2 2+2 2 |
| | 5+6 2 5+6 2 3+3 2 3+3 2 |
| | 4+4 2 4+4 2 2+2 2 2+2 2 |
| | 5+6 2 5+6 2 2+2 2 2+2 2 |
| | 7+7 2 8+9 2 3+3 2 3+3 2 |
| | 8+9 2 7+7 2 3+3 2 2+2 2 |
| 148.1.27 | 73243 58 Ge1Mn1O3 |
| | -GM -T -F -L |
| | 5+6 2 8+9 2 2+2 2 3+3 2 |
| | 8+9 2 5+6 2 3+3 2 2+2 2 |
| | 5+6 2 8+9 2 2+2 2 3+3 2 |
| | 4+4 2 7+7 2 3+3 2 2+2 2 |
| | 8+9 2 5+6 2 2+2 2 3+3 2 |
| | 7+7 2 4+4 2 3+3 2 2+2 2 |
| | 8+9 2 5+6 2 3+3 2 2+2 2 |
| | 5+6 2 8+9 2 2+2 2 3+3 2 |
| | 5+6 2 8+9 2 2+2 2 3+3 2 |
| | 5+6 2 8+9 2 2+2 2 2+2 2 |
| | 4+4 2 7+7 2 3+3 2 2+2 2 |
| | 8+9 2 5+6 2 3+3 2 3+3 2 |
| | 7+7 2 4+4 2 2+2 2 3+3 2 |
| | 5+6 2 5+6 2 3+3 2 2+2 2 |

Continued on next column

**TABLE LXVIII– continued**

| TQC | ICSD NoE SF |
|---|---|
| | 8+9 2 8+9 2 2+2 2 3+3 2 |
| | 4+4 2 7+7 2 3+3 2 2+2 2 |
| | 5+6 2 8+9 2 2+2 2 2+2 2 |
| | 8+9 2 5+6 2 3+3 2 3+3 2 |
| | 7+7 2 4+4 2 3+3 2 3+3 2 |
| | 5+6 2 5+6 2 2+2 2 2+2 2 |
| | 4+4 2 5+6 2 2+2 2 3+3 2 |
| | 8+9 2 4+4 2 3+3 2 2+2 2 |
| | 8+9 2 8+9 2 3+3 2 3+3 2 |
| | 7+7 2 7+7 2 2+2 2 2+2 2 |
| | 5+6 2 5+6 2 3+3 2 3+3 2 |
| | 4+4 2 4+4 2 2+2 2 2+2 2 |
| | 5+6 2 5+6 2 2+2 2 2+2 2 |
| | 7+7 2 8+9 2 3+3 2 3+3 2 |
| | 8+9 2 7+7 2 3+3 2 2+2 2 |
| 148.1.28 | 69591 58 Ge1Mn1O3 |
| | -GM -T -F -L |
| | 5+6 2 8+9 2 2+2 2 3+3 2 |
| | 8+9 2 5+6 2 3+3 2 2+2 2 |
| | 5+6 2 8+9 2 2+2 2 3+3 2 |
| | 4+4 2 7+7 2 3+3 2 2+2 2 |
| | 8+9 2 5+6 2 2+2 2 3+3 2 |
| | 7+7 2 4+4 2 3+3 2 2+2 2 |
| | 8+9 2 5+6 2 3+3 2 2+2 2 |
| | 5+6 2 8+9 2 2+2 2 3+3 2 |
| | 5+6 2 8+9 2 2+2 2 3+3 2 |
| | 5+6 2 8+9 2 2+2 2 2+2 2 |
| | 4+4 2 7+7 2 3+3 2 2+2 2 |
| | 8+9 2 5+6 2 3+3 2 3+3 2 |
| | 7+7 2 4+4 2 2+2 2 3+3 2 |
| | 5+6 2 5+6 2 3+3 2 2+2 2 |
| | 8+9 2 8+9 2 2+2 2 3+3 2 |
| | 4+4 2 7+7 2 3+3 2 2+2 2 |
| | 5+6 2 8+9 2 2+2 2 2+2 2 |
| | 8+9 2 5+6 2 3+3 2 3+3 2 |
| | 7+7 2 4+4 2 3+3 2 3+3 2 |
| | 5+6 2 5+6 2 2+2 2 2+2 2 |
| | 4+4 2 5+6 2 2+2 2 3+3 2 |
| | 8+9 2 4+4 2 3+3 2 2+2 2 |
| | 8+9 2 8+9 2 3+3 2 3+3 2 |
| | 7+7 2 7+7 2 2+2 2 2+2 2 |
| | 5+6 2 5+6 2 3+3 2 3+3 2 |
| | 4+4 2 4+4 2 2+2 2 2+2 2 |
| | 5+6 2 5+6 2 2+2 2 2+2 2 |
| | 7+7 2 8+9 2 3+3 2 3+3 2 |
| | 8+9 2 7+7 2 3+3 2 2+2 2 |
| 148.1.29 | 247127 198 Cu1O3Si1 |
| | -GM -T -F -L |
| | 5+6 2 5+6 2 3+3 2 3+3 2 |
| | 7+7 2 7+7 2 2+2 2 2+2 2 |
| | 8+9 2 8+9 2 2+2 2 2+2 2 |
| | 4+4 2 4+4 2 3+3 2 3+3 2 |
| | 5+6 2 5+6 2 3+3 2 3+3 2 |
| | 8+9 2 8+9 2 2+2 2 2+2 2 |
| | 5+6 2 8+9 2 2+2 2 2+2 2 |
| | 8+9 2 4+4 2 3+3 2 3+3 2 |
| | 4+4 2 5+6 2 2+2 2 2+2 2 |
| | 5+6 2 7+7 2 2+2 2 3+3 2 |
| | 7+7 2 8+9 2 3+3 2 2+2 2 |
| | 8+9 2 5+6 2 3+3 2 3+3 2 |
| | 5+6 2 4+4 2 3+3 2 2+2 2 |

Continued on next column

**TABLE LXVIII– continued**

| TQC | ICSD | NoE | SF |
|---|---|---|---|
| | 4+4 2 5+6 2 2+2 2 3+3 2 | | |
| | 8+9 2 8+9 2 2+2 2 2+2 2 | | |
| | 5+6 2 5+6 2 3+3 2 3+3 2 | | |
| | 7+7 2 7+7 2 2+2 2 2+2 2 | | |
| | 8+9 2 8+9 2 3+3 2 3+3 2 | | |
| | 8+9 2 8+9 2 2+2 2 2+2 2 | | |
| | 4+4 2 4+4 2 3+3 2 3+3 2 | | |
| | 5+6 2 5+6 2 3+3 2 3+3 2 | | |
| | 5+6 2 5+6 2 2+2 2 2+2 2 | | |
| | 7+7 2 7+7 2 2+2 2 2+2 2 | | |
| | 8+9 2 8+9 2 3+3 2 3+3 2 | | |
| | 4+4 2 4+4 2 2+2 2 2+2 2 | | |
| | 5+6 2 5+6 2 3+3 2 3+3 2 | | |
| | 7+7 2 5+6 2 2+2 2 2+2 2 | | |
| | 8+9 2 8+9 2 3+3 2 3+3 2 | | |
| | 5+6 2 8+9 2 2+2 2 2+2 2 | | |
| | 8+9 2 7+7 2 3+3 2 3+3 2 | | |
| | 4+4 2 8+9 2 2+2 2 2+2 2 | | |
| | 5+6 2 4+4 2 2+2 2 3+3 2 | | |
| | 8+9 2 5+6 2 3+3 2 2+2 2 | | |
| | 8+9 2 5+6 2 3+3 2 3+3 2 | | |
| | 7+7 2 7+7 2 3+3 2 2+2 2 | | |
| | 5+6 2 8+9 2 2+2 2 3+3 2 | | |
| | 4+4 2 7+7 2 2+2 2 3+3 2 | | |
| | 5+6 2 8+9 2 3+3 2 2+2 2 | | |
| | 7+7 2 4+4 2 3+3 2 3+3 2 | | |
| | 8+9 2 5+6 2 2+2 2 2+2 2 | | |
| | 5+6 2 5+6 2 2+2 2 2+2 2 | | |
| | 5+6 2 8+9 2 3+3 2 2+2 2 | | |
| | 8+9 2 5+6 2 2+2 2 3+3 2 | | |
| | 4+4 2 4+4 2 3+3 2 3+3 2 | | |
| | 5+6 2 8+9 2 2+2 2 2+2 2 | | |
| | 7+7 2 5+6 2 3+3 2 3+3 2 | | |
| | 8+9 2 8+9 2 2+2 2 3+3 2 | | |
| | 5+6 2 7+7 2 3+3 2 2+2 2 | | |
| | 5+6 2 5+6 2 3+3 2 3+3 2 | | |
| | 4+4 2 5+6 2 2+2 2 2+2 2 | | |
| | 8+9 2 8+9 2 2+2 2 3+3 2 | | |
| | 8+9 2 8+9 2 3+3 2 2+2 2 | | |
| | 5+6 2 7+7 2 2+2 2 3+3 2 | | |
| | 7+7 2 4+4 2 3+3 2 2+2 2 | | |
| | 8+9 2 5+6 2 3+3 2 3+3 2 | | |
| | 8+9 2 8+9 2 2+2 2 3+3 2 | | |
| | 4+4 2 7+7 2 3+3 2 2+2 2 | | |
| | 5+6 2 5+6 2 2+2 2 2+2 2 | | |
| | 8+9 2 8+9 2 2+2 2 2+2 2 | | |
| | 7+7 2 4+4 2 3+3 2 3+3 2 | | |
| | 7+7 2 5+6 2 3+3 2 2+2 2 | | |
| | 8+9 2 8+9 2 2+2 2 2+2 2 | | |
| | 5+6 2 7+7 2 3+3 2 3+3 2 | | |
| | 4+4 2 5+6 2 2+2 2 3+3 2 | | |
| | 5+6 2 8+9 2 2+2 2 2+2 2 | | |
| | 8+9 2 4+4 2 3+3 2 3+3 2 | | |
| | 7+7 2 7+7 2 2+2 2 2+2 2 | | |
| | 8+9 2 5+6 2 3+3 2 3+3 2 | | |
| | 4+4 2 8+9 2 2+2 2 2+2 2 | | |
| | 5+6 2 4+4 2 3+3 2 3+3 2 | | |
| | 5+6 2 5+6 2 3+3 2 3+3 2 | | |
| | 5+6 2 8+9 2 2+2 2 2+2 2 | | |
| | 8+9 2 4+4 2 2+2 2 2+2 2 | | |
| | 5+6 2 5+6 2 3+3 2 3+3 2 | | |

Continued on next column

**TABLE LXVIII– continued**

| TQC | ICSD | NoE | SF |
|---|---|---|---|
| | 8+9 2 8+9 2 2+2 2 2+2 2 | | |
| | 8+9 2 7+7 2 3+3 2 3+3 2 | | |
| | 7+7 2 5+6 2 2+2 2 2+2 2 | | |
| | 5+6 2 8+9 2 3+3 2 3+3 2 | | |
| | 4+4 2 5+6 2 2+2 2 3+3 2 | | |
| | 8+9 2 4+4 2 3+3 2 2+2 2 | | |
| | 5+6 2 7+7 2 2+2 2 2+2 2 | | |
| | 4+4 2 8+9 2 3+3 2 3+3 2 | | |
| | 8+9 2 4+4 2 2+2 2 2+2 2 | | |
| | 7+7 2 5+6 2 3+3 2 3+3 2 | | |
| | 5+6 2 8+9 2 3+3 2 2+2 2 | | |
| | 7+7 2 5+6 2 2+2 2 3+3 2 | | |
| | 8+9 2 8+9 2 3+3 2 2+2 2 | | |
| | 5+6 2 7+7 2 2+2 2 3+3 2 | | |
| | 4+4 2 8+9 2 3+3 2 2+2 2 | | |
| | 8+9 2 5+6 2 2+2 2 3+3 2 | | |
| | 8+9 2 7+7 2 2+2 2 3+3 2 | | |
| | 5+6 2 8+9 2 3+3 2 2+2 2 | | |
| | 4+4 2 5+6 2 3+3 2 3+3 2 | | |
| | 7+7 2 5+6 2 2+2 2 2+2 2 | | |
| | 8+9 2 4+4 2 3+3 2 2+2 2 | | |
| | 5+6 2 8+9 2 2+2 2 3+3 2 | | |
| | 8+9 2 8+9 2 3+3 2 2+2 2 | | |
| | 7+7 2 5+6 2 2+2 2 3+3 2 | | |
| | 5+6 2 7+7 2 3+3 2 3+3 2 | | |
| 148.1.30 | 84337 | 174 | Cs2Mo12Te14 |
| | -GM | -T | -F -L |
| | 5+6 2 5+6 2 2+2 2 2+2 2 | | |
| | 8+9 2 8+9 2 3+3 2 3+3 2 | | |
| | 5+6 2 8+9 2 2+2 2 3+3 2 | | |
| | 8+9 2 5+6 2 3+3 2 2+2 2 | | |
| | 7+7 2 8+9 2 2+2 2 2+2 2 | | |
| | 8+9 2 7+7 2 3+3 2 3+3 2 | | |
| | 5+6 2 5+6 2 2+2 2 2+2 2 | | |
| | 5+6 2 4+4 2 3+3 2 3+3 2 | | |
| | 4+4 2 5+6 2 2+2 2 2+2 2 | | |
| | 7+7 2 8+9 2 3+3 2 3+3 2 | | |
| | 8+9 2 5+6 2 2+2 2 2+2 2 | | |
| | 8+9 2 4+4 2 3+3 2 3+3 2 | | |
| | 5+6 2 8+9 2 3+3 2 2+2 2 | | |
| | 4+4 2 7+7 2 2+2 2 3+3 2 | | |
| | 8+9 2 5+6 2 2+2 2 2+2 2 | | |
| | 5+6 2 8+9 2 3+3 2 3+3 2 | | |
| | 8+9 2 8+9 2 3+3 2 3+3 2 | | |
| | 5+6 2 5+6 2 2+2 2 2+2 2 | | |
| | 5+6 2 8+9 2 2+2 2 2+2 2 | | |
| | 7+7 2 5+6 2 3+3 2 3+3 2 | | |
| | 8+9 2 7+7 2 3+3 2 3+3 2 | | |
| | 4+4 2 4+4 2 2+2 2 2+2 2 | | |
| | 5+6 2 5+6 2 2+2 2 3+3 2 | | |
| | 8+9 2 7+7 2 3+3 2 2+2 2 | | |
| | 4+4 2 8+9 2 3+3 2 2+2 2 | | |
| | 5+6 2 4+4 2 2+2 2 3+3 2 | | |
| | 7+7 2 5+6 2 3+3 2 3+3 2 | | |
| | 8+9 2 5+6 2 2+2 2 2+2 2 | | |
| | 5+6 2 8+9 2 3+3 2 3+3 2 | | |
| | 4+4 2 4+4 2 2+2 2 2+2 2 | | |
| | 7+7 2 8+9 2 2+2 2 2+2 2 | | |
| | 8+9 2 5+6 2 3+3 2 3+3 2 | | |
| | 5+6 2 8+9 2 3+3 2 3+3 2 | | |
| | 8+9 2 5+6 2 3+3 2 2+2 2 | | |

Continued on next column

**TABLE LXVIII– continued**

| TQC | ICSD | NoE | SF | |
|---|---|---|---|---|
| | 5+6 2 | 7+7 2 | 2+2 2 | 3+3 2 |
| | 8+9 2 | 8+9 2 | 3+3 2 | 3+3 2 |
| | 5+6 2 | 5+6 2 | 2+2 2 | 2+2 2 |
| | 8+9 2 | 8+9 2 | 3+3 2 | 3+3 2 |
| | 8+9 2 | 5+6 2 | 2+2 2 | 2+2 2 |
| | 5+6 2 | 4+4 2 | 3+3 2 | 3+3 2 |
| | 7+7 2 | 8+9 2 | 2+2 2 | 2+2 2 |
| | 4+4 2 | 5+6 2 | 2+2 2 | 2+2 2 |
| | 5+6 2 | 7+7 2 | 3+3 2 | 3+3 2 |
| | 8+9 2 | 8+9 2 | 3+3 2 | 2+2 2 |
| | 7+7 2 | 5+6 2 | 2+2 2 | 3+3 2 |
| | 5+6 2 | 8+9 2 | 2+2 2 | 2+2 2 |
| | 8+9 2 | 4+4 2 | 3+3 2 | 2+2 2 |
| | 5+6 2 | 5+6 2 | 2+2 2 | 3+3 2 |
| | 4+4 2 | 7+7 2 | 3+3 2 | 3+3 2 |
| | 5+6 2 | 5+6 2 | 3+3 2 | 2+2 2 |
| | 4+4 2 | 8+9 2 | 2+2 2 | 3+3 2 |
| | 8+9 2 | 4+4 2 | 2+2 2 | 2+2 2 |
| | 5+6 2 | 8+9 2 | 3+3 2 | 3+3 2 |
| | 7+7 2 | 7+7 2 | 2+2 2 | 2+2 2 |
| | 8+9 2 | 8+9 2 | 3+3 2 | 3+3 2 |
| | 8+9 2 | 5+6 2 | 2+2 2 | 2+2 2 |
| | 5+6 2 | 5+6 2 | 3+3 2 | 3+3 2 |
| | 4+4 2 | 8+9 2 | 2+2 2 | 2+2 2 |
| | 5+6 2 | 4+4 2 | 2+2 2 | 3+3 2 |
| | 8+9 2 | 7+7 2 | 3+3 2 | 2+2 2 |
| | 4+4 2 | 8+9 2 | 2+2 2 | 3+3 2 |
| | 8+9 2 | 8+9 2 | 3+3 2 | 2+2 2 |
| | 5+6 2 | 5+6 2 | 2+2 2 | 2+2 2 |
| | 7+7 2 | 7+7 2 | 3+3 2 | 3+3 2 |
| | 8+9 2 | 5+6 2 | 2+2 2 | 3+3 2 |
| | 7+7 2 | 7+7 2 | 2+2 2 | 3+3 2 |
| | 8+9 2 | 4+4 2 | 3+3 2 | 2+2 2 |
| | 5+6 2 | 8+9 2 | 3+3 2 | 2+2 2 |
| | 4+4 2 | 8+9 2 | 2+2 2 | 3+3 2 |
| | 5+6 2 | 5+6 2 | 3+3 2 | 2+2 2 |
| | 4+4 2 | 5+6 2 | 2+2 2 | 3+3 2 |
| | 5+6 2 | 8+9 2 | 3+3 2 | 3+3 2 |
| | 7+7 2 | 4+4 2 | 2+2 2 | 2+2 2 |
| | 8+9 2 | 7+7 2 | 3+3 2 | 2+2 2 |
| | 5+6 2 | 8+9 2 | 2+2 2 | 3+3 2 |
| | 8+9 2 | 8+9 2 | 2+2 2 | 2+2 2 |
| | 7+7 2 | 5+6 2 | 3+3 2 | 3+3 2 |
| | 8+9 2 | 4+4 2 | 2+2 2 | 2+2 2 |
| | 4+4 2 | 8+9 2 | 3+3 2 | 3+3 2 |
| | 5+6 2 | 7+7 2 | 2+2 2 | 2+2 2 |
| | 5+6 2 | 5+6 2 | 3+3 2 | 2+2 2 |
| | 8+9 2 | 4+4 2 | 2+2 2 | 3+3 2 |
| | 7+7 2 | 5+6 2 | 3+3 2 | 2+2 2 |
| | 8+9 2 | 8+9 2 | 2+2 2 | 3+3 2 |
| | 4+4 2 | 7+7 2 | 3+3 2 | 2+2 2 |
| | 5+6 2 | 5+6 2 | 2+2 2 | 3+3 2 |
| | 5+6 2 | 5+6 2 | 3+3 2 | 3+3 2 |
| 148.1.31 | 40820 | | 174 | Cs2Mo12Se14 |
| | -GM | -T | -F | -L |
| | 5+6 2 | 5+6 2 | 2+2 2 | 2+2 2 |
| | 8+9 2 | 8+9 2 | 3+3 2 | 3+3 2 |
| | 5+6 2 | 8+9 2 | 2+2 2 | 3+3 2 |
| | 8+9 2 | 5+6 2 | 3+3 2 | 2+2 2 |
| | 8+9 2 | 8+9 2 | 2+2 2 | 3+3 2 |
| | 7+7 2 | 7+7 2 | 3+3 2 | 2+2 2 |

Continued on next column

**TABLE LXVIII– continued**

| TQC | ICSD | NoE | SF | |
|---|---|---|---|---|
| | 5+6 2 | 4+4 2 | 2+2 2 | 3+3 2 |
| | 5+6 2 | 5+6 2 | 3+3 2 | 2+2 2 |
| | 4+4 2 | 8+9 2 | 2+2 2 | 2+2 2 |
| | 7+7 2 | 5+6 2 | 3+3 2 | 3+3 2 |
| | 8+9 2 | 4+4 2 | 3+3 2 | 2+2 2 |
| | 8+9 2 | 5+6 2 | 2+2 2 | 2+2 2 |
| | 4+4 2 | 7+7 2 | 3+3 2 | 3+3 2 |
| | 5+6 2 | 8+9 2 | 2+2 2 | 3+3 2 |
| | 8+9 2 | 5+6 2 | 3+3 2 | 2+2 2 |
| | 5+6 2 | 8+9 2 | 2+2 2 | 3+3 2 |
| | 5+6 2 | 5+6 2 | 2+2 2 | 2+2 2 |
| | 8+9 2 | 8+9 2 | 3+3 2 | 3+3 2 |
| | 4+4 2 | 4+4 2 | 2+2 2 | 2+2 2 |
| | 5+6 2 | 5+6 2 | 3+3 2 | 3+3 2 |
| | 7+7 2 | 7+7 2 | 2+2 2 | 2+2 2 |
| | 8+9 2 | 8+9 2 | 3+3 2 | 3+3 2 |
| | 4+4 2 | 5+6 2 | 2+2 2 | 3+3 2 |
| | 5+6 2 | 8+9 2 | 3+3 2 | 2+2 2 |
| | 8+9 2 | 7+7 2 | 3+3 2 | 2+2 2 |
| | 5+6 2 | 4+4 2 | 2+2 2 | 2+2 2 |
| | 7+7 2 | 5+6 2 | 2+2 2 | 3+3 2 |
| | 8+9 2 | 5+6 2 | 3+3 2 | 3+3 2 |
| | 7+7 2 | 8+9 2 | 3+3 2 | 3+3 2 |
| | 8+9 2 | 4+4 2 | 2+2 2 | 2+2 2 |
| | 5+6 2 | 5+6 2 | 3+3 2 | 3+3 2 |
| | 8+9 2 | 8+9 2 | 2+2 2 | 2+2 2 |
| | 4+4 2 | 8+9 2 | 3+3 2 | 3+3 2 |
| | 5+6 2 | 5+6 2 | 3+3 2 | 2+2 2 |
| | 5+6 2 | 7+7 2 | 3+3 2 | 3+3 2 |
| | 8+9 2 | 8+9 2 | 2+2 2 | 3+3 2 |
| | 5+6 2 | 8+9 2 | 3+3 2 | 2+2 2 |
| | 8+9 2 | 5+6 2 | 2+2 2 | 3+3 2 |
| | 8+9 2 | 4+4 2 | 3+3 2 | 3+3 2 |
| | 7+7 2 | 5+6 2 | 2+2 2 | 2+2 2 |
| | 5+6 2 | 7+7 2 | 2+2 2 | 2+2 2 |
| | 4+4 2 | 8+9 2 | 2+2 2 | 2+2 2 |
| | 7+7 2 | 8+9 2 | 3+3 2 | 3+3 2 |
| | 5+6 2 | 5+6 2 | 3+3 2 | 2+2 2 |
| | 8+9 2 | 4+4 2 | 2+2 2 | 2+2 2 |
| | 5+6 2 | 5+6 2 | 2+2 2 | 3+3 2 |
| | 5+6 2 | 8+9 2 | 2+2 2 | 3+3 2 |
| | 4+4 2 | 5+6 2 | 3+3 2 | 2+2 2 |
| | 8+9 2 | 7+7 2 | 3+3 2 | 2+2 2 |
| | 4+4 2 | 5+6 2 | 2+2 2 | 3+3 2 |
| | 5+6 2 | 8+9 2 | 3+3 2 | 3+3 2 |
| | 8+9 2 | 4+4 2 | 2+2 2 | 2+2 2 |
| | 7+7 2 | 7+7 2 | 3+3 2 | 3+3 2 |
| | 8+9 2 | 8+9 2 | 2+2 2 | 2+2 2 |
| | 5+6 2 | 5+6 2 | 2+2 2 | 3+3 2 |
| | 8+9 2 | 8+9 2 | 3+3 2 | 2+2 2 |
| | 4+4 2 | 4+4 2 | 3+3 2 | 3+3 2 |
| | 5+6 2 | 5+6 2 | 2+2 2 | 2+2 2 |
| | 8+9 2 | 8+9 2 | 2+2 2 | 3+3 2 |
| | 5+6 2 | 7+7 2 | 3+3 2 | 2+2 2 |
| | 8+9 2 | 8+9 2 | 2+2 2 | 2+2 2 |
| | 7+7 2 | 8+9 2 | 2+2 2 | 3+3 2 |
| | 4+4 2 | 7+7 2 | 3+3 2 | 2+2 2 |
| | 8+9 2 | 5+6 2 | 3+3 2 | 3+3 2 |
| | 5+6 2 | 5+6 2 | 2+2 2 | 3+3 2 |
| | 7+7 2 | 7+7 2 | 3+3 2 | 3+3 2 |
| | 4+4 2 | 4+4 2 | 2+2 2 | 2+2 2 |

Continued on next column

**TABLE LXVIII– continued**

| TQC | ICSD | | NoE | SF |
|---|---|---|---|---|
| | 5+6 2 | 8+9 2 | 3+3 2 | 2+2 2 |
| | 5+6 2 | 5+6 2 | 2+2 2 | 3+3 2 |
| | 8+9 2 | 8+9 2 | 3+3 2 | 3+3 2 |
| | 5+6 2 | 4+4 2 | 2+2 2 | 2+2 2 |
| | 8+9 2 | 5+6 2 | 3+3 2 | 2+2 2 |
| | 7+7 2 | 8+9 2 | 3+3 2 | 3+3 2 |
| | 4+4 2 | 7+7 2 | 2+2 2 | 2+2 2 |
| | 5+6 2 | 8+9 2 | 2+2 2 | 3+3 2 |
| | 8+9 2 | 5+6 2 | 3+3 2 | 3+3 2 |
| | 7+7 2 | 4+4 2 | 2+2 2 | 2+2 2 |
| | 8+9 2 | 7+7 2 | 2+2 2 | 2+2 2 |
| | 4+4 2 | 8+9 2 | 3+3 2 | 2+2 2 |
| | 5+6 2 | 5+6 2 | 2+2 2 | 3+3 2 |
| | 8+9 2 | 8+9 2 | 3+3 2 | 2+2 2 |
| | 5+6 2 | 4+4 2 | 2+2 2 | 3+3 2 |
| | 7+7 2 | 5+6 2 | 3+3 2 | 3+3 2 |
| | 8+9 2 | 5+6 2 | 2+2 2 | 2+2 2 |
| | 4+4 2 | 7+7 2 | 3+3 2 | 2+2 2 |
| | 5+6 2 | 8+9 2 | 3+3 2 | 3+3 2 |
| | 5+6 2 | 5+6 2 | 2+2 2 | 3+3 2 |
| 148.1.32 | 71020 | | 56 | Cr2Si2Te6 |
| | -GM | -T | -F | -L |
| | 5+6 2 | 5+6 2 | 3+3 2 | 3+3 2 |
| | 8+9 2 | 8+9 2 | 2+2 2 | 2+2 2 |
| | 7+7 2 | 7+7 2 | 2+2 2 | 2+2 2 |
| | 8+9 2 | 8+9 2 | 3+3 2 | 3+3 2 |
| | 5+6 2 | 5+6 2 | 2+2 2 | 3+3 2 |
| | 4+4 2 | 4+4 2 | 3+3 2 | 2+2 2 |
| | 5+6 2 | 5+6 2 | 3+3 2 | 3+3 2 |
| | 8+9 2 | 8+9 2 | 2+2 2 | 2+2 2 |
| | 5+6 2 | 7+7 2 | 3+3 2 | 3+3 2 |
| | 7+7 2 | 8+9 2 | 2+2 2 | 2+2 2 |
| | 8+9 2 | 8+9 2 | 2+2 2 | 2+2 2 |
| | 4+4 2 | 4+4 2 | 3+3 2 | 3+3 2 |
| | 8+9 2 | 5+6 2 | 2+2 2 | 2+2 2 |
| | 5+6 2 | 5+6 2 | 3+3 2 | 3+3 2 |
| | 5+6 2 | 5+6 2 | 2+2 2 | 2+2 2 |
| | 5+6 2 | 8+9 2 | 3+3 2 | 3+3 2 |
| | 7+7 2 | 5+6 2 | 2+2 2 | 3+3 2 |
| | 8+9 2 | 4+4 2 | 2+2 2 | 2+2 2 |
| | 4+4 2 | 7+7 2 | 3+3 2 | 3+3 2 |
| | 5+6 2 | 8+9 2 | 3+3 2 | 2+2 2 |
| | 5+6 2 | 5+6 2 | 2+2 2 | 3+3 2 |
| | 4+4 2 | 5+6 2 | 3+3 2 | 2+2 2 |
| | 8+9 2 | 4+4 2 | 2+2 2 | 3+3 2 |
| | 5+6 2 | 8+9 2 | 3+3 2 | 2+2 2 |
| | 4+4 2 | 7+7 2 | 3+3 2 | 3+3 2 |
| | 7+7 2 | 4+4 2 | 3+3 2 | 3+3 2 |
| | 5+6 2 | 5+6 2 | 2+2 2 | 2+2 2 |
| | 8+9 2 | 8+9 2 | 2+2 2 | 2+2 2 |
| 148.1.33 | 106520 | | 60 | Cr2Se3 |
| | -GM | -T | -F | -L |
| | 5+6 2 | 5+6 2 | 3+3 2 | 2+2 2 |
| | 8+9 2 | 8+9 2 | 2+2 2 | 3+3 2 |
| | 5+6 2 | 8+9 2 | 3+3 2 | 3+3 2 |
| | 4+4 2 | 7+7 2 | 2+2 2 | 2+2 2 |
| | 8+9 2 | 5+6 2 | 2+2 2 | 2+2 2 |
| | 7+7 2 | 4+4 2 | 3+3 2 | 3+3 2 |
| | 5+6 2 | 8+9 2 | 3+3 2 | 3+3 2 |
| | 5+6 2 | 5+6 2 | 3+3 2 | 2+2 2 |
| | 7+7 2 | 7+7 2 | 2+2 2 | 2+2 2 |

Continued on next column

**TABLE LXVIII– continued**

| TQC | ICSD | | NoE | SF |
|---|---|---|---|---|
| | 4+4 2 | 8+9 2 | 3+3 2 | 3+3 2 |
| | 8+9 2 | 4+4 2 | 2+2 2 | 2+2 2 |
| | 5+6 2 | 5+6 2 | 2+2 2 | 3+3 2 |
| | 8+9 2 | 5+6 2 | 3+3 2 | 3+3 2 |
| | 5+6 2 | 8+9 2 | 2+2 2 | 2+2 2 |
| | 4+4 2 | 7+7 2 | 3+3 2 | 2+2 2 |
| | 5+6 2 | 8+9 2 | 2+2 2 | 3+3 2 |
| | 8+9 2 | 5+6 2 | 3+3 2 | 2+2 2 |
| | 5+6 2 | 5+6 2 | 2+2 2 | 3+3 2 |
| | 4+4 2 | 4+4 2 | 3+3 2 | 3+3 2 |
| | 7+7 2 | 5+6 2 | 3+3 2 | 2+2 2 |
| | 8+9 2 | 8+9 2 | 2+2 2 | 3+3 2 |
| | 5+6 2 | 4+4 2 | 3+3 2 | 2+2 2 |
| | 4+4 2 | 8+9 2 | 2+2 2 | 2+2 2 |
| | 5+6 2 | 7+7 2 | 2+2 2 | 3+3 2 |
| | 8+9 2 | 5+6 2 | 3+3 2 | 2+2 2 |
| | 4+4 2 | 4+4 2 | 3+3 2 | 3+3 2 |
| | 5+6 2 | 8+9 2 | 3+3 2 | 3+3 2 |
| | 8+9 2 | 5+6 2 | 2+2 2 | 2+2 2 |
| | 7+7 2 | 7+7 2 | 3+3 2 | 2+2 2 |
| | 5+6 2 | 8+9 2 | 3+3 2 | 3+3 2 |
| 148.1.34 | 79270 | | 56 | Cr2Ge2Te6 |
| | -GM | -T | -F | -L |
| | 5+6 2 | 5+6 2 | 3+3 2 | 3+3 2 |
| | 8+9 2 | 8+9 2 | 2+2 2 | 2+2 2 |
| | 8+9 2 | 5+6 2 | 2+2 2 | 2+2 2 |
| | 7+7 2 | 4+4 2 | 3+3 2 | 3+3 2 |
| | 5+6 2 | 8+9 2 | 2+2 2 | 3+3 2 |
| | 4+4 2 | 7+7 2 | 3+3 2 | 2+2 2 |
| | 5+6 2 | 5+6 2 | 3+3 2 | 3+3 2 |
| | 8+9 2 | 8+9 2 | 2+2 2 | 2+2 2 |
| | 5+6 2 | 7+7 2 | 3+3 2 | 3+3 2 |
| | 7+7 2 | 8+9 2 | 2+2 2 | 2+2 2 |
| | 8+9 2 | 8+9 2 | 2+2 2 | 2+2 2 |
| | 4+4 2 | 4+4 2 | 3+3 2 | 3+3 2 |
| | 8+9 2 | 5+6 2 | 2+2 2 | 2+2 2 |
| | 5+6 2 | 5+6 2 | 3+3 2 | 3+3 2 |
| | 5+6 2 | 5+6 2 | 2+2 2 | 2+2 2 |
| | 5+6 2 | 8+9 2 | 3+3 2 | 3+3 2 |
| | 7+7 2 | 5+6 2 | 2+2 2 | 2+2 2 |
| | 8+9 2 | 4+4 2 | 2+2 2 | 3+3 2 |
| | 4+4 2 | 7+7 2 | 3+3 2 | 3+3 2 |
| | 5+6 2 | 8+9 2 | 3+3 2 | 2+2 2 |
| | 5+6 2 | 5+6 2 | 2+2 2 | 3+3 2 |
| | 4+4 2 | 5+6 2 | 3+3 2 | 2+2 2 |
| | 8+9 2 | 4+4 2 | 2+2 2 | 3+3 2 |
| | 5+6 2 | 8+9 2 | 3+3 2 | 2+2 2 |
| | 4+4 2 | 7+7 2 | 3+3 2 | 3+3 2 |
| | 5+6 2 | 4+4 2 | 3+3 2 | 3+3 2 |
| | 8+9 2 | 5+6 2 | 2+2 2 | 2+2 2 |
| | 7+7 2 | 8+9 2 | 2+2 2 | 2+2 2 |
| 148.1.35 | 79269 | | 56 | Cr2Ge2Te6 |
| | -GM | -T | -F | -L |
| | 5+6 2 | 5+6 2 | 3+3 2 | 3+3 2 |
| | 8+9 2 | 8+9 2 | 2+2 2 | 2+2 2 |
| | 8+9 2 | 5+6 2 | 2+2 2 | 2+2 2 |
| | 7+7 2 | 4+4 2 | 3+3 2 | 3+3 2 |
| | 5+6 2 | 8+9 2 | 2+2 2 | 3+3 2 |
| | 4+4 2 | 7+7 2 | 3+3 2 | 2+2 2 |
| | 5+6 2 | 5+6 2 | 3+3 2 | 3+3 2 |
| | 8+9 2 | 8+9 2 | 2+2 2 | 2+2 2 |

Continued on next column

**TABLE LXVIII– continued**

| TQC | ICSD | | NoE | SF |
|---|---|---|---|---|
| | 5+6 2 | 7+7 2 | 3+3 2 | 3+3 2 |
| | 7+7 2 | 8+9 2 | 2+2 2 | 2+2 2 |
| | 8+9 2 | 8+9 2 | 2+2 2 | 2+2 2 |
| | 4+4 2 | 4+4 2 | 3+3 2 | 3+3 2 |
| | 8+9 2 | 5+6 2 | 2+2 2 | 2+2 2 |
| | 5+6 2 | 5+6 2 | 3+3 2 | 3+3 2 |
| | 5+6 2 | 5+6 2 | 2+2 2 | 2+2 2 |
| | 5+6 2 | 8+9 2 | 3+3 2 | 3+3 2 |
| | 4+4 2 | 5+6 2 | 2+2 2 | 2+2 2 |
| | 7+7 2 | 4+4 2 | 2+2 2 | 3+3 2 |
| | 8+9 2 | 7+7 2 | 3+3 2 | 3+3 2 |
| | 5+6 2 | 8+9 2 | 3+3 2 | 2+2 2 |
| | 5+6 2 | 5+6 2 | 2+2 2 | 2+2 2 |
| | 4+4 2 | 5+6 2 | 2+2 2 | 3+3 2 |
| | 8+9 2 | 4+4 2 | 3+3 2 | 3+3 2 |
| | 5+6 2 | 8+9 2 | 3+3 2 | 2+2 2 |
| | 4+4 2 | 4+4 2 | 3+3 2 | 3+3 2 |
| | 5+6 2 | 5+6 2 | 3+3 2 | 3+3 2 |
| | 7+7 2 | 7+7 2 | 2+2 2 | 2+2 2 |
| | 8+9 2 | 8+9 2 | 2+2 2 | 2+2 2 |
| 148.1.36 | 79268 | | 56 | Cr2Ge2Te6 |
| | -GM | -T | -F | -L |
| | 5+6 2 | 5+6 2 | 3+3 2 | 3+3 2 |
| | 8+9 2 | 8+9 2 | 2+2 2 | 2+2 2 |
| | 8+9 2 | 5+6 2 | 2+2 2 | 2+2 2 |
| | 7+7 2 | 4+4 2 | 3+3 2 | 3+3 2 |
| | 5+6 2 | 8+9 2 | 2+2 2 | 3+3 2 |
| | 4+4 2 | 7+7 2 | 3+3 2 | 2+2 2 |
| | 5+6 2 | 5+6 2 | 3+3 2 | 3+3 2 |
| | 8+9 2 | 8+9 2 | 2+2 2 | 2+2 2 |
| | 5+6 2 | 7+7 2 | 3+3 2 | 3+3 2 |
| | 7+7 2 | 8+9 2 | 2+2 2 | 2+2 2 |
| | 8+9 2 | 8+9 2 | 2+2 2 | 2+2 2 |
| | 4+4 2 | 4+4 2 | 3+3 2 | 3+3 2 |
| | 8+9 2 | 5+6 2 | 2+2 2 | 2+2 2 |
| | 5+6 2 | 5+6 2 | 3+3 2 | 3+3 2 |
| | 5+6 2 | 5+6 2 | 2+2 2 | 2+2 2 |
| | 7+7 2 | 8+9 2 | 3+3 2 | 3+3 2 |
| | 5+6 2 | 5+6 2 | 2+2 2 | 2+2 2 |
| | 8+9 2 | 4+4 2 | 2+2 2 | 3+3 2 |
| | 4+4 2 | 7+7 2 | 3+3 2 | 3+3 2 |
| | 5+6 2 | 8+9 2 | 3+3 2 | 2+2 2 |
| | 5+6 2 | 5+6 2 | 2+2 2 | 3+3 2 |
| | 4+4 2 | 5+6 2 | 3+3 2 | 2+2 2 |
| | 8+9 2 | 4+4 2 | 2+2 2 | 3+3 2 |
| | 5+6 2 | 8+9 2 | 3+3 2 | 2+2 2 |
| | 4+4 2 | 7+7 2 | 3+3 2 | 3+3 2 |
| | 5+6 2 | 4+4 2 | 3+3 2 | 3+3 2 |
| | 8+9 2 | 5+6 2 | 2+2 2 | 2+2 2 |
| | 7+7 2 | 8+9 2 | 2+2 2 | 2+2 2 |
| 148.1.37 | 252042 | | 64 | Cr1Fe1O3 |
| | -GM | -T | -F | -L |
| | 5+6 2 | 5+6 2 | 2+2 2 | 2+2 2 |
| | 8+9 2 | 8+9 2 | 3+3 2 | 3+3 2 |
| | 8+9 2 | 5+6 2 | 3+3 2 | 3+3 2 |
| | 7+7 2 | 4+4 2 | 2+2 2 | 2+2 2 |
| | 5+6 2 | 8+9 2 | 2+2 2 | 2+2 2 |
| | 4+4 2 | 7+7 2 | 3+3 2 | 3+3 2 |
| | 5+6 2 | 8+9 2 | 3+3 2 | 3+3 2 |
| | 8+9 2 | 5+6 2 | 2+2 2 | 2+2 2 |
| | 5+6 2 | 8+9 2 | 2+2 2 | 2+2 2 |

Continued on next column

**TABLE LXVIII– continued**

| TQC | ICSD | | NoE | SF |
|---|---|---|---|---|
| | 4+4 2 | 5+6 2 | 3+3 2 | 3+3 2 |
| | 7+7 2 | 7+7 2 | 3+3 2 | 2+2 2 |
| | 8+9 2 | 4+4 2 | 2+2 2 | 3+3 2 |
| | 5+6 2 | 4+4 2 | 2+2 2 | 2+2 2 |
| | 4+4 2 | 5+6 2 | 3+3 2 | 3+3 2 |
| | 5+6 2 | 5+6 2 | 3+3 2 | 2+2 2 |
| | 7+7 2 | 8+9 2 | 2+2 2 | 3+3 2 |
| | 8+9 2 | 7+7 2 | 2+2 2 | 3+3 2 |
| | 5+6 2 | 8+9 2 | 3+3 2 | 2+2 2 |
| | 8+9 2 | 5+6 2 | 2+2 2 | 2+2 2 |
| | 5+6 2 | 8+9 2 | 2+2 2 | 3+3 2 |
| | 4+4 2 | 5+6 2 | 3+3 2 | 2+2 2 |
| | 8+9 2 | 4+4 2 | 2+2 2 | 3+3 2 |
| | 8+9 2 | 8+9 2 | 3+3 2 | 2+2 2 |
| | 7+7 2 | 7+7 2 | 3+3 2 | 3+3 2 |
| | 5+6 2 | 8+9 2 | 3+3 2 | 2+2 2 |
| | 4+4 2 | 4+4 2 | 2+2 2 | 3+3 2 |
| | 5+6 2 | 7+7 2 | 2+2 2 | 3+3 2 |
| | 8+9 2 | 5+6 2 | 2+2 2 | 2+2 2 |
| | 7+7 2 | 8+9 2 | 3+3 2 | 3+3 2 |
| | 5+6 2 | 5+6 2 | 3+3 2 | 2+2 2 |
| | 4+4 2 | 7+7 2 | 2+2 2 | 3+3 2 |
| | 8+9 2 | 8+9 2 | 3+3 2 | 2+2 2 |
| 148.1.38 | 39426 | | 50 | Cl3Ti1 |
| | -GM | -T | -F | -L |
| | 5+6 2 | 5+6 2 | 2+2 2 | 2+2 2 |
| | 8+9 2 | 8+9 2 | 3+3 2 | 3+3 2 |
| | 5+6 2 | 5+6 2 | 2+2 2 | 3+3 2 |
| | 4+4 2 | 4+4 2 | 3+3 2 | 2+2 2 |
| | 7+7 2 | 7+7 2 | 2+2 2 | 2+2 2 |
| | 8+9 2 | 8+9 2 | 3+3 2 | 3+3 2 |
| | 5+6 2 | 8+9 2 | 3+3 2 | 3+3 2 |
| | 8+9 2 | 5+6 2 | 2+2 2 | 2+2 2 |
| | 4+4 2 | 7+7 2 | 2+2 2 | 2+2 2 |
| | 5+6 2 | 4+4 2 | 3+3 2 | 2+2 2 |
| | 7+7 2 | 8+9 2 | 2+2 2 | 3+3 2 |
| | 8+9 2 | 5+6 2 | 3+3 2 | 2+2 2 |
| | 5+6 2 | 5+6 2 | 2+2 2 | 3+3 2 |
| | 5+6 2 | 5+6 2 | 3+3 2 | 3+3 2 |
| | 4+4 2 | 4+4 2 | 2+2 2 | 2+2 2 |
| | 5+6 2 | 5+6 2 | 3+3 2 | 3+3 2 |
| | 7+7 2 | 7+7 2 | 2+2 2 | 3+3 2 |
| | 8+9 2 | 8+9 2 | 3+3 2 | 2+2 2 |
| | 8+9 2 | 8+9 2 | 2+2 2 | 2+2 2 |
| | 5+6 2 | 8+9 2 | 2+2 2 | 2+2 2 |
| | 4+4 2 | 5+6 2 | 3+3 2 | 3+3 2 |
| | 8+9 2 | 4+4 2 | 3+3 2 | 3+3 2 |
| | 7+7 2 | 8+9 2 | 2+2 2 | 2+2 2 |
| | 8+9 2 | 7+7 2 | 3+3 2 | 3+3 2 |
| | 4+4 2 | 4+4 2 | 2+2 2 | 3+3 2 |
| 148.1.39 | 201976 | | 110 | Cl12N1Sc7 |
| | -GM | -T | -F | -L |
| | 5+6 2 | 8+9 2 | 3+3 2 | 2+2 2 |
| | 4+4 2 | 5+6 2 | 2+2 2 | 3+3 2 |
| | 5+6 2 | 4+4 2 | 2+2 2 | 2+2 2 |
| | 8+9 2 | 5+6 2 | 2+2 2 | 3+3 2 |
| | 7+7 2 | 7+7 2 | 3+3 2 | 3+3 2 |
| | 8+9 2 | 8+9 2 | 3+3 2 | 2+2 2 |
| | 5+6 2 | 7+7 2 | 3+3 2 | 2+2 2 |
| | 4+4 2 | 8+9 2 | 2+2 2 | 3+3 2 |
| | 5+6 2 | 5+6 2 | 2+2 2 | 3+3 2 |

Continued on next column

**TABLE LXVIII– continued**

| TQC | ICSD | NoE | SF |
|---|---|---|---|
| | 7+7 2 4+4 2 3+3 2 2+2 2 | | |
| | 8+9 2 5+6 2 3+3 2 2+2 2 | | |
| | 8+9 2 8+9 2 2+2 2 3+3 2 | | |
| | 5+6 2 5+6 2 3+3 2 3+3 2 | | |
| | 5+6 2 8+9 2 3+3 2 2+2 2 | | |
| | 5+6 2 5+6 2 3+3 2 2+2 2 | | |
| | 4+4 2 8+9 2 2+2 2 2+2 2 | | |
| | 5+6 2 7+7 2 2+2 2 3+3 2 | | |
| | 7+7 2 8+9 2 3+3 2 2+2 2 | | |
| | 8+9 2 4+4 2 2+2 2 3+3 2 | | |
| | 8+9 2 5+6 2 3+3 2 3+3 2 | | |
| | 4+4 2 5+6 2 2+2 2 3+3 2 | | |
| | 5+6 2 8+9 2 3+3 2 3+3 2 | | |
| | 7+7 2 7+7 2 2+2 2 2+2 2 | | |
| | 8+9 2 8+9 2 2+2 2 2+2 2 | | |
| | 8+9 2 5+6 2 3+3 2 3+3 2 | | |
| | 4+4 2 4+4 2 3+3 2 3+3 2 | | |
| | 5+6 2 5+6 2 3+3 2 2+2 2 | | |
| | 5+6 2 4+4 2 2+2 2 2+2 2 | | |
| | 8+9 2 7+7 2 2+2 2 3+3 2 | | |
| | 8+9 2 8+9 2 3+3 2 2+2 2 | | |
| | 7+7 2 5+6 2 2+2 2 3+3 2 | | |
| | 8+9 2 8+9 2 3+3 2 2+2 2 | | |
| | 5+6 2 5+6 2 2+2 2 2+2 2 | | |
| | 4+4 2 4+4 2 2+2 2 3+3 2 | | |
| | 5+6 2 7+7 2 3+3 2 2+2 2 | | |
| | 5+6 2 8+9 2 2+2 2 3+3 2 | | |
| | 4+4 2 5+6 2 3+3 2 2+2 2 | | |
| | 5+6 2 8+9 2 2+2 2 3+3 2 | | |
| | 7+7 2 7+7 2 3+3 2 3+3 2 | | |
| | 8+9 2 8+9 2 3+3 2 3+3 2 | | |
| | 8+9 2 4+4 2 2+2 2 2+2 2 | | |
| | 8+9 2 5+6 2 3+3 2 2+2 2 | | |
| | 7+7 2 5+6 2 2+2 2 3+3 2 | | |
| | 4+4 2 8+9 2 3+3 2 2+2 2 | | |
| | 5+6 2 5+6 2 2+2 2 3+3 2 | | |
| | 8+9 2 8+9 2 3+3 2 3+3 2 | | |
| | 8+9 2 4+4 2 2+2 2 2+2 2 | | |
| | 7+7 2 7+7 2 2+2 2 2+2 2 | | |
| | 5+6 2 7+7 2 2+2 2 2+2 2 | | |
| | 8+9 2 8+9 2 2+2 2 3+3 2 | | |
| | 8+9 2 8+9 2 3+3 2 2+2 2 | | |
| | 7+7 2 5+6 2 2+2 2 2+2 2 | | |
| | 5+6 2 4+4 2 3+3 2 3+3 2 | | |
| | 4+4 2 5+6 2 3+3 2 3+3 2 | | |
| | 5+6 2 5+6 2 3+3 2 2+2 2 | | |
| 148.1.40 | 424476 | 66 | Cl12Co1Sc7 |
| | -GM -T -F -L | | |
| | 5+6 2 5+6 2 3+3 2 3+3 2 | | |
| | 5+6 2 8+9 2 2+2 2 2+2 2 | | |
| | 5+6 2 8+9 2 2+2 2 3+3 2 | | |
| | 4+4 2 7+7 2 3+3 2 2+2 2 | | |
| | 5+6 2 4+4 2 3+3 2 3+3 2 | | |
| | 8+9 2 5+6 2 3+3 2 3+3 2 | | |
| | 5+6 2 5+6 2 3+3 2 3+3 2 | | |
| | 4+4 2 4+4 2 2+2 2 3+3 2 | | |
| | 8+9 2 5+6 2 3+3 2 2+2 2 | | |
| | 7+7 2 5+6 2 3+3 2 2+2 2 | | |
| | 5+6 2 8+9 2 3+3 2 3+3 2 | | |
| | 8+9 2 4+4 2 2+2 2 3+3 2 | | |
| | 7+7 2 5+6 2 3+3 2 2+2 2 | | |

Continued on next column

**TABLE LXVIII– continued**

| TQC | ICSD | NoE | SF |
|---|---|---|---|
| | 8+9 2 8+9 2 2+2 2 3+3 2 | | |
| | 4+4 2 7+7 2 2+2 2 2+2 2 | | |
| | 5+6 2 8+9 2 3+3 2 2+2 2 | | |
| | 4+4 2 7+7 2 2+2 2 2+2 2 | | |
| | 5+6 2 8+9 2 3+3 2 3+3 2 | | |
| | 8+9 2 5+6 2 3+3 2 2+2 2 | | |
| | 5+6 2 4+4 2 2+2 2 3+3 2 | | |
| | 4+4 2 8+9 2 3+3 2 2+2 2 | | |
| | 8+9 2 5+6 2 3+3 2 3+3 2 | | |
| | 5+6 2 5+6 2 2+2 2 2+2 2 | | |
| | 7+7 2 4+4 2 2+2 2 3+3 2 | | |
| | 8+9 2 5+6 2 3+3 2 3+3 2 | | |
| | 8+9 2 8+9 2 2+2 2 3+3 2 | | |
| | 7+7 2 7+7 2 3+3 2 2+2 2 | | |
| | 5+6 2 8+9 2 3+3 2 3+3 2 | | |
| | 4+4 2 5+6 2 3+3 2 2+2 2 | | |
| | 5+6 2 7+7 2 2+2 2 3+3 2 | | |
| | 4+4 2 8+9 2 3+3 2 2+2 2 | | |
| | 8+9 2 4+4 2 2+2 2 3+3 2 | | |
| | 5+6 2 5+6 2 2+2 2 2+2 2 | | |
| 148.1.41 | 183603 | 134 | Ce1Pt5Sb1 |
| | -GM -T -F -L | | |
| | 5+6 2 8+9 2 3+3 2 2+2 2 | | |
| | 5+6 2 5+6 2 3+3 2 3+3 2 | | |
| | 8+9 2 5+6 2 2+2 2 3+3 2 | | |
| | 8+9 2 8+9 2 2+2 2 2+2 2 | | |
| | 8+9 2 5+6 2 2+2 2 3+3 2 | | |
| | 7+7 2 4+4 2 2+2 2 3+3 2 | | |
| | 8+9 2 8+9 2 2+2 2 2+2 2 | | |
| | 7+7 2 7+7 2 2+2 2 2+2 2 | | |
| | 5+6 2 8+9 2 3+3 2 3+3 2 | | |
| | 8+9 2 5+6 2 2+2 2 2+2 2 | | |
| | 5+6 2 5+6 2 2+2 2 2+2 2 | | |
| | 8+9 2 8+9 2 3+3 2 3+3 2 | | |
| | 5+6 2 8+9 2 3+3 2 3+3 2 | | |
| | 8+9 2 5+6 2 2+2 2 2+2 2 | | |
| | 5+6 2 8+9 2 3+3 2 2+2 2 | | |
| | 8+9 2 5+6 2 3+3 2 3+3 2 | | |
| | 7+7 2 4+4 2 2+2 2 2+2 2 | | |
| | 4+4 2 8+9 2 3+3 2 3+3 2 | | |
| | 5+6 2 7+7 2 2+2 2 3+3 2 | | |
| | 5+6 2 5+6 2 2+2 2 2+2 2 | | |
| | 8+9 2 7+7 2 3+3 2 3+3 2 | | |
| | 7+7 2 8+9 2 2+2 2 2+2 2 | | |
| | 4+4 2 5+6 2 3+3 2 3+3 2 | | |
| | 5+6 2 4+4 2 3+3 2 2+2 2 | | |
| | 8+9 2 5+6 2 2+2 2 3+3 2 | | |
| | 5+6 2 8+9 2 3+3 2 2+2 2 | | |
| | 4+4 2 7+7 2 2+2 2 3+3 2 | | |
| | 8+9 2 4+4 2 3+3 2 2+2 2 | | |
| | 7+7 2 5+6 2 2+2 2 2+2 2 | | |
| | 8+9 2 8+9 2 3+3 2 3+3 2 | | |
| | 5+6 2 4+4 2 2+2 2 3+3 2 | | |
| | 4+4 2 5+6 2 3+3 2 2+2 2 | | |
| | 5+6 2 7+7 2 2+2 2 3+3 2 | | |
| | 7+7 2 8+9 2 2+2 2 2+2 2 | | |
| | 8+9 2 5+6 2 3+3 2 2+2 2 | | |
| | 4+4 2 8+9 2 2+2 2 3+3 2 | | |
| | 5+6 2 4+4 2 3+3 2 2+2 2 | | |
| | 7+7 2 7+7 2 2+2 2 3+3 2 | | |
| | 5+6 2 5+6 2 3+3 2 2+2 2 | | |

Continued on next column

**TABLE LXVIII– continued**

| TQC | ICSD NoE SF |
|---|---|
| | 4+4 2 4+4 2 2+2 2 3+3 2 |
| | 8+9 2 8+9 2 3+3 2 3+3 2 |
| | 8+9 2 5+6 2 3+3 2 2+2 2 |
| | 5+6 2 7+7 2 2+2 2 2+2 2 |
| | 7+7 2 8+9 2 3+3 2 3+3 2 |
| | 8+9 2 5+6 2 2+2 2 2+2 2 |
| | 4+4 2 8+9 2 3+3 2 3+3 2 |
| | 5+6 2 7+7 2 2+2 2 2+2 2 |
| | 7+7 2 8+9 2 3+3 2 3+3 2 |
| | 5+6 2 4+4 2 2+2 2 2+2 2 |
| | 8+9 2 5+6 2 2+2 2 3+3 2 |
| | 4+4 2 5+6 2 3+3 2 2+2 2 |
| | 8+9 2 8+9 2 2+2 2 3+3 2 |
| | 5+6 2 7+7 2 3+3 2 2+2 2 |
| | 5+6 2 8+9 2 2+2 2 3+3 2 |
| | 8+9 2 4+4 2 2+2 2 3+3 2 |
| | 7+7 2 5+6 2 3+3 2 2+2 2 |
| | 5+6 2 8+9 2 2+2 2 3+3 2 |
| | 4+4 2 5+6 2 3+3 2 2+2 2 |
| | 8+9 2 7+7 2 3+3 2 3+3 2 |
| | 5+6 2 5+6 2 2+2 2 2+2 2 |
| | 8+9 2 4+4 2 3+3 2 3+3 2 |
| | 7+7 2 5+6 2 2+2 2 3+3 2 |
| | 8+9 2 7+7 2 3+3 2 2+2 2 |
| | 5+6 2 8+9 2 2+2 2 3+3 2 |
| | 8+9 2 8+9 2 2+2 2 2+2 2 |
| | 4+4 2 5+6 2 3+3 2 2+2 2 |
| | 8+9 2 8+9 2 2+2 2 2+2 2 |
| 148.1.42 | 62388 94 Ca1Mo6S8 |
| | -GM -T -F -L |
| | 5+6 2 5+6 2 3+3 2 3+3 2 |
| | 8+9 2 8+9 2 2+2 2 2+2 2 |
| | 7+7 2 7+7 2 2+2 2 2+2 2 |
| | 8+9 2 8+9 2 2+2 2 2+2 2 |
| | 5+6 2 8+9 2 3+3 2 2+2 2 |
| | 8+9 2 5+6 2 2+2 2 3+3 2 |
| | 7+7 2 5+6 2 2+2 2 3+3 2 |
| | 8+9 2 4+4 2 3+3 2 3+3 2 |
| | 4+4 2 7+7 2 2+2 2 2+2 2 |
| | 5+6 2 8+9 2 3+3 2 2+2 2 |
| | 8+9 2 5+6 2 3+3 2 3+3 2 |
| | 5+6 2 8+9 2 2+2 2 2+2 2 |
| | 4+4 2 5+6 2 3+3 2 2+2 2 |
| | 5+6 2 7+7 2 3+3 2 3+3 2 |
| | 5+6 2 8+9 2 2+2 2 2+2 2 |
| | 8+9 2 8+9 2 3+3 2 3+3 2 |
| | 7+7 2 8+9 2 2+2 2 2+2 2 |
| | 8+9 2 4+4 2 2+2 2 3+3 2 |
| | 8+9 2 5+6 2 3+3 2 2+2 2 |
| | 5+6 2 7+7 2 3+3 2 3+3 2 |
| | 7+7 2 8+9 2 3+3 2 2+2 2 |
| | 8+9 2 5+6 2 2+2 2 3+3 2 |
| | 5+6 2 5+6 2 2+2 2 3+3 2 |
| | 4+4 2 5+6 2 3+3 2 2+2 2 |
| | 5+6 2 4+4 2 2+2 2 2+2 2 |
| | 4+4 2 8+9 2 2+2 2 3+3 2 |
| | 5+6 2 4+4 2 3+3 2 2+2 2 |
| | 8+9 2 5+6 2 2+2 2 3+3 2 |
| | 7+7 2 8+9 2 2+2 2 2+2 2 |
| | 8+9 2 7+7 2 3+3 2 3+3 2 |
| | 5+6 2 5+6 2 2+2 2 3+3 2 |

Continued on next column

**TABLE LXVIII– continued**

| TQC | ICSD NoE SF |
|---|---|
| | 8+9 2 5+6 2 3+3 2 2+2 2 |
| | 5+6 2 4+4 2 2+2 2 2+2 2 |
| | 7+7 2 8+9 2 3+3 2 3+3 2 |
| | 8+9 2 8+9 2 2+2 2 3+3 2 |
| | 4+4 2 7+7 2 3+3 2 2+2 2 |
| | 5+6 2 4+4 2 2+2 2 2+2 2 |
| | 4+4 2 8+9 2 3+3 2 3+3 2 |
| | 5+6 2 5+6 2 2+2 2 2+2 2 |
| | 5+6 2 5+6 2 3+3 2 3+3 2 |
| | 7+7 2 5+6 2 2+2 2 2+2 2 |
| | 8+9 2 8+9 2 3+3 2 3+3 2 |
| | 8+9 2 4+4 2 2+2 2 3+3 2 |
| | 8+9 2 5+6 2 2+2 2 3+3 2 |
| | 4+4 2 7+7 2 2+2 2 2+2 2 |
| | 5+6 2 8+9 2 3+3 2 2+2 2 |
| | 8+9 2 8+9 2 3+3 2 3+3 2 |
| 148.1.43 | 39242 50 Br3Ti1 |
| | -GM -T -F -L |
| | 5+6 2 5+6 2 2+2 2 2+2 2 |
| | 8+9 2 8+9 2 3+3 2 3+3 2 |
| | 4+4 2 4+4 2 2+2 2 3+3 2 |
| | 5+6 2 5+6 2 3+3 2 2+2 2 |
| | 7+7 2 7+7 2 2+2 2 2+2 2 |
| | 8+9 2 8+9 2 3+3 2 3+3 2 |
| | 5+6 2 8+9 2 3+3 2 3+3 2 |
| | 8+9 2 5+6 2 2+2 2 2+2 2 |
| | 4+4 2 7+7 2 2+2 2 2+2 2 |
| | 7+7 2 4+4 2 3+3 2 2+2 2 |
| | 5+6 2 5+6 2 2+2 2 3+3 2 |
| | 8+9 2 8+9 2 3+3 2 2+2 2 |
| | 5+6 2 5+6 2 2+2 2 3+3 2 |
| | 5+6 2 5+6 2 3+3 2 3+3 2 |
| | 4+4 2 4+4 2 2+2 2 3+3 2 |
| | 5+6 2 5+6 2 3+3 2 2+2 2 |
| | 8+9 2 8+9 2 2+2 2 3+3 2 |
| | 7+7 2 8+9 2 3+3 2 2+2 2 |
| | 8+9 2 7+7 2 2+2 2 2+2 2 |
| | 8+9 2 8+9 2 3+3 2 3+3 2 |
| | 5+6 2 5+6 2 2+2 2 2+2 2 |
| | 4+4 2 4+4 2 3+3 2 3+3 2 |
| | 8+9 2 8+9 2 2+2 2 2+2 2 |
| | 7+7 2 7+7 2 3+3 2 3+3 2 |
| | 4+4 2 4+4 2 2+2 2 3+3 2 |
| 148.1.44 | 419636 110 Ba6N6O1Re2 |
| | -GM -T -F -L |
| | 5+6 2 8+9 2 3+3 2 2+2 2 |
| | 7+7 2 5+6 2 2+2 2 3+3 2 |
| | 8+9 2 4+4 2 2+2 2 3+3 2 |
| | 4+4 2 5+6 2 3+3 2 2+2 2 |
| | 5+6 2 7+7 2 2+2 2 3+3 2 |
| | 8+9 2 8+9 2 3+3 2 2+2 2 |
| | 5+6 2 8+9 2 3+3 2 2+2 2 |
| | 4+4 2 8+9 2 3+3 2 3+3 2 |
| | 7+7 2 7+7 2 2+2 2 2+2 2 |
| | 5+6 2 4+4 2 2+2 2 3+3 2 |
| | 8+9 2 5+6 2 2+2 2 2+2 2 |
| | 5+6 2 5+6 2 3+3 2 3+3 2 |
| | 8+9 2 8+9 2 3+3 2 2+2 2 |
| | 5+6 2 5+6 2 3+3 2 3+3 2 |
| | 8+9 2 8+9 2 2+2 2 2+2 2 |
| | 5+6 2 5+6 2 2+2 2 3+3 2 |

Continued on next column

**TABLE LXVIII– continued**

| TQC | ICSD NoE SF |
|---|---|
| | 8+9 2 8+9 2 3+3 2 2+2 2 |
| | 7+7 2 4+4 2 2+2 2 2+2 2 |
| | 4+4 2 7+7 2 3+3 2 3+3 2 |
| | 5+6 2 5+6 2 2+2 2 3+3 2 |
| | 8+9 2 5+6 2 2+2 2 2+2 2 |
| | 8+9 2 8+9 2 3+3 2 3+3 2 |
| | 7+7 2 7+7 2 3+3 2 2+2 2 |
| | 4+4 2 4+4 2 2+2 2 3+3 2 |
| | 5+6 2 8+9 2 3+3 2 3+3 2 |
| | 8+9 2 8+9 2 2+2 2 2+2 2 |
| | 5+6 2 5+6 2 3+3 2 2+2 2 |
| | 7+7 2 4+4 2 2+2 2 3+3 2 |
| | 8+9 2 5+6 2 3+3 2 2+2 2 |
| | 4+4 2 7+7 2 3+3 2 3+3 2 |
| | 5+6 2 8+9 2 2+2 2 2+2 2 |
| | 4+4 2 7+7 2 3+3 2 3+3 2 |
| | 5+6 2 4+4 2 2+2 2 2+2 2 |
| | 7+7 2 5+6 2 2+2 2 3+3 2 |
| | 8+9 2 8+9 2 3+3 2 2+2 2 |
| | 4+4 2 8+9 2 2+2 2 3+3 2 |
| | 8+9 2 5+6 2 3+3 2 2+2 2 |
| | 5+6 2 8+9 2 2+2 2 3+3 2 |
| | 5+6 2 7+7 2 3+3 2 2+2 2 |
| | 8+9 2 5+6 2 2+2 2 3+3 2 |
| | 7+7 2 4+4 2 3+3 2 2+2 2 |
| | 5+6 2 8+9 2 2+2 2 3+3 2 |
| | 8+9 2 5+6 2 2+2 2 3+3 2 |
| | 7+7 2 5+6 2 3+3 2 3+3 2 |
| | 8+9 2 8+9 2 2+2 2 2+2 2 |
| | 4+4 2 4+4 2 2+2 2 3+3 2 |
| | 8+9 2 5+6 2 2+2 2 2+2 2 |
| | 5+6 2 5+6 2 3+3 2 3+3 2 |
| | 7+7 2 7+7 2 3+3 2 2+2 2 |
| | 5+6 2 4+4 2 2+2 2 3+3 2 |
| | 8+9 2 8+9 2 2+2 2 2+2 2 |
| | 8+9 2 5+6 2 3+3 2 3+3 2 |
| | 5+6 2 5+6 2 3+3 2 3+3 2 |
| | 8+9 2 8+9 2 2+2 2 2+2 2 |
| | 5+6 2 8+9 2 3+3 2 3+3 2 |
| 148.1.45 | 419467 112 Ba6N6O1Os2 |
| | -GM -T -F -L |
| | 5+6 2 8+9 2 3+3 2 2+2 2 |
| | 7+7 2 5+6 2 2+2 2 3+3 2 |
| | 8+9 2 4+4 2 2+2 2 3+3 2 |
| | 4+4 2 5+6 2 3+3 2 2+2 2 |
| | 5+6 2 7+7 2 2+2 2 3+3 2 |
| | 8+9 2 8+9 2 3+3 2 2+2 2 |
| | 5+6 2 8+9 2 3+3 2 2+2 2 |
| | 5+6 2 8+9 2 3+3 2 3+3 2 |
| | 8+9 2 5+6 2 2+2 2 2+2 2 |
| | 4+4 2 7+7 2 2+2 2 2+2 2 |
| | 7+7 2 4+4 2 3+3 2 3+3 2 |
| | 5+6 2 5+6 2 2+2 2 3+3 2 |
| | 8+9 2 8+9 2 3+3 2 2+2 2 |
| | 5+6 2 5+6 2 2+2 2 3+3 2 |
| | 8+9 2 8+9 2 3+3 2 2+2 2 |
| | 5+6 2 8+9 2 2+2 2 3+3 2 |
| | 8+9 2 5+6 2 3+3 2 2+2 2 |
| | 7+7 2 4+4 2 3+3 2 2+2 2 |
| | 4+4 2 7+7 2 2+2 2 3+3 2 |
| | 5+6 2 5+6 2 2+2 2 3+3 2 |

Continued on next column

**TABLE LXVIII– continued**

| TQC | ICSD NoE SF |
|---|---|
| | 8+9 2 4+4 2 2+2 2 2+2 2 |
| | 8+9 2 5+6 2 3+3 2 3+3 2 |
| | 7+7 2 8+9 2 2+2 2 3+3 2 |
| | 4+4 2 5+6 2 3+3 2 2+2 2 |
| | 5+6 2 7+7 2 2+2 2 2+2 2 |
| | 8+9 2 8+9 2 3+3 2 3+3 2 |
| | 5+6 2 8+9 2 3+3 2 2+2 2 |
| | 7+7 2 7+7 2 2+2 2 3+3 2 |
| | 4+4 2 5+6 2 3+3 2 2+2 2 |
| | 8+9 2 4+4 2 3+3 2 2+2 2 |
| | 5+6 2 8+9 2 2+2 2 3+3 2 |
| | 4+4 2 7+7 2 3+3 2 3+3 2 |
| | 7+7 2 4+4 2 2+2 2 2+2 2 |
| | 5+6 2 8+9 2 2+2 2 3+3 2 |
| | 8+9 2 5+6 2 3+3 2 2+2 2 |
| | 5+6 2 8+9 2 2+2 2 2+2 2 |
| | 8+9 2 5+6 2 3+3 2 3+3 2 |
| | 4+4 2 4+4 2 2+2 2 2+2 2 |
| | 7+7 2 5+6 2 3+3 2 3+3 2 |
| | 5+6 2 7+7 2 3+3 2 3+3 2 |
| | 8+9 2 8+9 2 2+2 2 2+2 2 |
| | 5+6 2 5+6 2 2+2 2 2+2 2 |
| | 8+9 2 8+9 2 3+3 2 3+3 2 |
| | 7+7 2 8+9 2 2+2 2 3+3 2 |
| | 4+4 2 5+6 2 2+2 2 3+3 2 |
| | 8+9 2 7+7 2 2+2 2 3+3 2 |
| | 5+6 2 4+4 2 3+3 2 2+2 2 |
| | 8+9 2 5+6 2 2+2 2 3+3 2 |
| | 8+9 2 5+6 2 2+2 2 2+2 2 |
| | 7+7 2 4+4 2 3+3 2 3+3 2 |
| | 5+6 2 8+9 2 2+2 2 2+2 2 |
| | 5+6 2 5+6 2 3+3 2 3+3 2 |
| | 8+9 2 8+9 2 2+2 2 2+2 2 |
| | 8+9 2 5+6 2 3+3 2 3+3 2 |
| | 5+6 2 8+9 2 3+3 2 2+2 2 |
| | 5+6 2 5+6 2 2+2 2 3+3 2 |
| 148.1.46 | 96087 88 Ba1Ni2O8V2 |
| | -GM -T -F -L |
| | 5+6 2 5+6 2 3+3 2 3+3 2 |
| | 5+6 2 8+9 2 2+2 2 2+2 2 |
| | 8+9 2 5+6 2 3+3 2 3+3 2 |
| | 8+9 2 5+6 2 2+2 2 3+3 2 |
| | 7+7 2 4+4 2 2+2 2 2+2 2 |
| | 5+6 2 8+9 2 3+3 2 3+3 2 |
| | 4+4 2 7+7 2 3+3 2 2+2 2 |
| | 5+6 2 5+6 2 2+2 2 2+2 2 |
| | 8+9 2 8+9 2 3+3 2 3+3 2 |
| | 8+9 2 8+9 2 2+2 2 2+2 2 |
| | 8+9 2 8+9 2 2+2 2 2+2 2 |
| | 7+7 2 7+7 2 2+2 2 2+2 2 |
| | 5+6 2 8+9 2 2+2 2 3+3 2 |
| | 8+9 2 5+6 2 3+3 2 2+2 2 |
| | 7+7 2 4+4 2 2+2 2 3+3 2 |
| | 5+6 2 8+9 2 3+3 2 3+3 2 |
| | 4+4 2 7+7 2 3+3 2 2+2 2 |
| | 8+9 2 5+6 2 2+2 2 2+2 2 |
| | 5+6 2 5+6 2 3+3 2 3+3 2 |
| | 4+4 2 4+4 2 2+2 2 2+2 2 |
| | 8+9 2 8+9 2 2+2 2 3+3 2 |
| | 7+7 2 7+7 2 3+3 2 2+2 2 |
| | 5+6 2 8+9 2 3+3 2 3+3 2 |

Continued on next column

**TABLE LXVIII– continued**

| TQC | ICSD | NoE | SF |
|---|---|---|---|
| | 5+6 2 5+6 2 2+2 2 2+2 2 | | |
| | 4+4 2 5+6 2 3+3 2 3+3 2 | | |
| | 8+9 2 4+4 2 2+2 2 2+2 2 | | |
| | 5+6 2 8+9 2 3+3 2 2+2 2 | | |
| | 8+9 2 8+9 2 2+2 2 3+3 2 | | |
| | 7+7 2 5+6 2 3+3 2 3+3 2 | | |
| | 8+9 2 7+7 2 2+2 2 2+2 2 | | |
| | 5+6 2 8+9 2 3+3 2 2+2 2 | | |
| | 8+9 2 4+4 2 2+2 2 3+3 2 | | |
| | 5+6 2 5+6 2 2+2 2 2+2 2 | | |
| | 4+4 2 5+6 2 3+3 2 3+3 2 | | |
| | 8+9 2 8+9 2 2+2 2 3+3 2 | | |
| | 7+7 2 7+7 2 3+3 2 2+2 2 | | |
| | 4+4 2 7+7 2 2+2 2 3+3 2 | | |
| | 5+6 2 8+9 2 3+3 2 2+2 2 | | |
| | 8+9 2 5+6 2 2+2 2 3+3 2 | | |
| | 5+6 2 8+9 2 3+3 2 2+2 2 | | |
| | 7+7 2 4+4 2 3+3 2 3+3 2 | | |
| | 8+9 2 5+6 2 2+2 2 2+2 2 | | |
| | 5+6 2 8+9 2 3+3 2 3+3 2 | | |
| | 4+4 2 7+7 2 2+2 2 2+2 2 | | |
| 148.1.47 | 96086 | 88 | Ba1Ni2O8V2 |
| | -GM -T -F -L | | |
| | 5+6 2 5+6 2 3+3 2 3+3 2 | | |
| | 5+6 2 8+9 2 2+2 2 2+2 2 | | |
| | 8+9 2 5+6 2 3+3 2 3+3 2 | | |
| | 8+9 2 5+6 2 2+2 2 3+3 2 | | |
| | 7+7 2 4+4 2 2+2 2 2+2 2 | | |
| | 5+6 2 8+9 2 3+3 2 3+3 2 | | |
| | 4+4 2 7+7 2 3+3 2 2+2 2 | | |
| | 5+6 2 5+6 2 2+2 2 2+2 2 | | |
| | 8+9 2 8+9 2 3+3 2 3+3 2 | | |
| | 8+9 2 8+9 2 2+2 2 2+2 2 | | |
| | 8+9 2 8+9 2 2+2 2 2+2 2 | | |
| | 7+7 2 7+7 2 2+2 2 2+2 2 | | |
| | 5+6 2 8+9 2 2+2 2 3+3 2 | | |
| | 8+9 2 5+6 2 3+3 2 2+2 2 | | |
| | 7+7 2 4+4 2 2+2 2 3+3 2 | | |
| | 5+6 2 8+9 2 3+3 2 3+3 2 | | |
| | 4+4 2 7+7 2 3+3 2 2+2 2 | | |
| | 8+9 2 5+6 2 2+2 2 2+2 2 | | |
| | 5+6 2 5+6 2 3+3 2 3+3 2 | | |
| | 4+4 2 4+4 2 2+2 2 2+2 2 | | |
| | 8+9 2 8+9 2 2+2 2 3+3 2 | | |
| | 7+7 2 7+7 2 3+3 2 2+2 2 | | |
| | 5+6 2 8+9 2 3+3 2 3+3 2 | | |
| | 5+6 2 5+6 2 2+2 2 2+2 2 | | |
| | 4+4 2 5+6 2 3+3 2 3+3 2 | | |
| | 8+9 2 4+4 2 2+2 2 2+2 2 | | |
| | 5+6 2 8+9 2 3+3 2 2+2 2 | | |
| | 8+9 2 8+9 2 2+2 2 3+3 2 | | |
| | 7+7 2 5+6 2 3+3 2 3+3 2 | | |
| | 8+9 2 7+7 2 2+2 2 2+2 2 | | |
| | 5+6 2 8+9 2 3+3 2 2+2 2 | | |
| | 8+9 2 4+4 2 2+2 2 3+3 2 | | |
| | 5+6 2 5+6 2 2+2 2 2+2 2 | | |
| | 4+4 2 5+6 2 3+3 2 3+3 2 | | |
| | 8+9 2 8+9 2 2+2 2 3+3 2 | | |
| | 7+7 2 7+7 2 3+3 2 2+2 2 | | |
| | 4+4 2 7+7 2 2+2 2 3+3 2 | | |
| | 5+6 2 8+9 2 3+3 2 2+2 2 | | |

Continued on next column

**TABLE LXVIII– continued**

| TQC | ICSD | NoE | SF |
|---|---|---|---|
| | 8+9 2 5+6 2 2+2 2 3+3 2 | | |
| | 5+6 2 8+9 2 3+3 2 2+2 2 | | |
| | 7+7 2 4+4 2 3+3 2 3+3 2 | | |
| | 8+9 2 5+6 2 2+2 2 2+2 2 | | |
| | 5+6 2 8+9 2 2+2 2 3+3 2 | | |
| | 4+4 2 7+7 2 3+3 2 2+2 2 | | |
| 148.1.48 | 85489 | 94 | Ba1Mo6S8 |
| | -GM -T -F -L | | |
| | 5+6 2 5+6 2 3+3 2 3+3 2 | | |
| | 8+9 2 8+9 2 2+2 2 2+2 2 | | |
| | 5+6 2 8+9 2 2+2 2 2+2 2 | | |
| | 8+9 2 7+7 2 3+3 2 3+3 2 | | |
| | 7+7 2 8+9 2 2+2 2 2+2 2 | | |
| | 8+9 2 5+6 2 3+3 2 3+3 2 | | |
| | 4+4 2 5+6 2 2+2 2 3+3 2 | | |
| | 5+6 2 4+4 2 3+3 2 2+2 2 | | |
| | 8+9 2 5+6 2 2+2 2 3+3 2 | | |
| | 5+6 2 8+9 2 3+3 2 2+2 2 | | |
| | 7+7 2 7+7 2 2+2 2 2+2 2 | | |
| | 8+9 2 8+9 2 2+2 2 2+2 2 | | |
| | 4+4 2 8+9 2 3+3 2 2+2 2 | | |
| | 5+6 2 7+7 2 3+3 2 3+3 2 | | |
| | 5+6 2 8+9 2 2+2 2 2+2 2 | | |
| | 8+9 2 5+6 2 3+3 2 3+3 2 | | |
| | 7+7 2 8+9 2 2+2 2 3+3 2 | | |
| | 8+9 2 4+4 2 3+3 2 2+2 2 | | |
| | 5+6 2 5+6 2 2+2 2 3+3 2 | | |
| | 5+6 2 7+7 2 3+3 2 2+2 2 | | |
| | 8+9 2 8+9 2 3+3 2 2+2 2 | | |
| | 4+4 2 5+6 2 2+2 2 3+3 2 | | |
| | 5+6 2 5+6 2 2+2 2 3+3 2 | | |
| | 8+9 2 4+4 2 3+3 2 2+2 2 | | |
| | 7+7 2 5+6 2 2+2 2 2+2 2 | | |
| | 5+6 2 8+9 2 3+3 2 3+3 2 | | |
| | 4+4 2 5+6 2 2+2 2 3+3 2 | | |
| | 8+9 2 4+4 2 2+2 2 2+2 2 | | |
| | 8+9 2 8+9 2 3+3 2 3+3 2 | | |
| | 7+7 2 7+7 2 2+2 2 2+2 2 | | |
| | 5+6 2 5+6 2 2+2 2 3+3 2 | | |
| | 8+9 2 5+6 2 3+3 2 2+2 2 | | |
| | 5+6 2 4+4 2 3+3 2 2+2 2 | | |
| | 4+4 2 8+9 2 2+2 2 3+3 2 | | |
| | 5+6 2 8+9 2 2+2 2 2+2 2 | | |
| | 8+9 2 7+7 2 3+3 2 3+3 2 | | |
| | 7+7 2 4+4 2 2+2 2 2+2 2 | | |
| | 4+4 2 5+6 2 3+3 2 3+3 2 | | |
| | 5+6 2 8+9 2 2+2 2 2+2 2 | | |
| | 5+6 2 5+6 2 3+3 2 3+3 2 | | |
| | 7+7 2 5+6 2 2+2 2 2+2 2 | | |
| | 8+9 2 8+9 2 3+3 2 3+3 2 | | |
| | 8+9 2 7+7 2 2+2 2 3+3 2 | | |
| | 8+9 2 4+4 2 2+2 2 3+3 2 | | |
| | 4+4 2 5+6 2 2+2 2 2+2 2 | | |
| | 5+6 2 8+9 2 3+3 2 2+2 2 | | |
| | 8+9 2 8+9 2 3+3 2 3+3 2 | | |
| 148.1.49 | 65776 | 94 | Ba1Mo6S8 |
| | -GM -T -F -L | | |
| | 5+6 2 5+6 2 3+3 2 3+3 2 | | |
| | 8+9 2 8+9 2 2+2 2 2+2 2 | | |
| | 5+6 2 8+9 2 3+3 2 2+2 2 | | |
| | 8+9 2 7+7 2 2+2 2 3+3 2 | | |

Continued on next column

**TABLE LXVIII– continued**

| TQC | ICSD | NoE | SF | |
|---|---|---|---|---|
| | 7+7 2 | 8+9 2 | 2+2 2 | 2+2 2 |
| | 8+9 2 | 5+6 2 | 3+3 2 | 3+3 2 |
| | 4+4 2 | 5+6 2 | 2+2 2 | 3+3 2 |
| | 5+6 2 | 4+4 2 | 3+3 2 | 2+2 2 |
| | 8+9 2 | 5+6 2 | 2+2 2 | 3+3 2 |
| | 5+6 2 | 8+9 2 | 3+3 2 | 2+2 2 |
| | 7+7 2 | 7+7 2 | 2+2 2 | 2+2 2 |
| | 8+9 2 | 8+9 2 | 2+2 2 | 2+2 2 |
| | 4+4 2 | 8+9 2 | 3+3 2 | 2+2 2 |
| | 5+6 2 | 7+7 2 | 3+3 2 | 3+3 2 |
| | 5+6 2 | 8+9 2 | 2+2 2 | 2+2 2 |
| | 8+9 2 | 5+6 2 | 3+3 2 | 3+3 2 |
| | 7+7 2 | 8+9 2 | 2+2 2 | 3+3 2 |
| | 8+9 2 | 4+4 2 | 3+3 2 | 2+2 2 |
| | 5+6 2 | 5+6 2 | 2+2 2 | 3+3 2 |
| | 8+9 2 | 7+7 2 | 3+3 2 | 2+2 2 |
| | 5+6 2 | 8+9 2 | 3+3 2 | 2+2 2 |
| | 4+4 2 | 5+6 2 | 2+2 2 | 3+3 2 |
| | 8+9 2 | 5+6 2 | 2+2 2 | 3+3 2 |
| | 5+6 2 | 4+4 2 | 3+3 2 | 2+2 2 |
| | 7+7 2 | 5+6 2 | 2+2 2 | 2+2 2 |
| | 4+4 2 | 8+9 2 | 3+3 2 | 3+3 2 |
| | 5+6 2 | 5+6 2 | 2+2 2 | 3+3 2 |
| | 8+9 2 | 4+4 2 | 2+2 2 | 2+2 2 |
| | 8+9 2 | 8+9 2 | 3+3 2 | 3+3 2 |
| | 7+7 2 | 7+7 2 | 2+2 2 | 2+2 2 |
| | 5+6 2 | 5+6 2 | 2+2 2 | 3+3 2 |
| | 8+9 2 | 5+6 2 | 3+3 2 | 2+2 2 |
| | 5+6 2 | 4+4 2 | 3+3 2 | 2+2 2 |
| | 4+4 2 | 8+9 2 | 2+2 2 | 3+3 2 |
| | 8+9 2 | 8+9 2 | 2+2 2 | 3+3 2 |
| | 7+7 2 | 7+7 2 | 3+3 2 | 2+2 2 |
| | 5+6 2 | 4+4 2 | 2+2 2 | 2+2 2 |
| | 4+4 2 | 5+6 2 | 3+3 2 | 3+3 2 |
| | 5+6 2 | 8+9 2 | 2+2 2 | 2+2 2 |
| | 5+6 2 | 5+6 2 | 3+3 2 | 3+3 2 |
| | 7+7 2 | 5+6 2 | 2+2 2 | 2+2 2 |
| | 8+9 2 | 8+9 2 | 3+3 2 | 3+3 2 |
| | 8+9 2 | 4+4 2 | 2+2 2 | 3+3 2 |
| | 8+9 2 | 7+7 2 | 2+2 2 | 3+3 2 |
| | 4+4 2 | 5+6 2 | 2+2 2 | 2+2 2 |
| | 5+6 2 | 8+9 2 | 3+3 2 | 2+2 2 |
| | 8+9 2 | 8+9 2 | 3+3 2 | 3+3 2 |
| 148.1.50 | 62157 | 94 | Ba1Mo6S8 | |
| | -GM | -T | -F | -L |
| | 5+6 2 | 5+6 2 | 3+3 2 | 3+3 2 |
| | 8+9 2 | 8+9 2 | 2+2 2 | 2+2 2 |
| | 5+6 2 | 8+9 2 | 3+3 2 | 2+2 2 |
| | 8+9 2 | 7+7 2 | 2+2 2 | 3+3 2 |
| | 7+7 2 | 8+9 2 | 2+2 2 | 2+2 2 |
| | 8+9 2 | 5+6 2 | 3+3 2 | 3+3 2 |
| | 4+4 2 | 5+6 2 | 2+2 2 | 3+3 2 |
| | 5+6 2 | 4+4 2 | 3+3 2 | 2+2 2 |
| | 8+9 2 | 5+6 2 | 2+2 2 | 3+3 2 |
| | 5+6 2 | 8+9 2 | 3+3 2 | 2+2 2 |
| | 7+7 2 | 7+7 2 | 2+2 2 | 2+2 2 |
| | 8+9 2 | 8+9 2 | 2+2 2 | 2+2 2 |
| | 4+4 2 | 8+9 2 | 3+3 2 | 2+2 2 |
| | 5+6 2 | 7+7 2 | 3+3 2 | 3+3 2 |
| | 5+6 2 | 8+9 2 | 2+2 2 | 2+2 2 |
| | 8+9 2 | 5+6 2 | 3+3 2 | 3+3 2 |

Continued on next column

**TABLE LXVIII– continued**

| TQC | ICSD | NoE | SF | |
|---|---|---|---|---|
| | 7+7 2 | 8+9 2 | 2+2 2 | 3+3 2 |
| | 8+9 2 | 4+4 2 | 3+3 2 | 2+2 2 |
| | 5+6 2 | 5+6 2 | 2+2 2 | 3+3 2 |
| | 8+9 2 | 7+7 2 | 3+3 2 | 2+2 2 |
| | 5+6 2 | 8+9 2 | 3+3 2 | 2+2 2 |
| | 8+9 2 | 5+6 2 | 2+2 2 | 3+3 2 |
| | 7+7 2 | 5+6 2 | 2+2 2 | 3+3 2 |
| | 4+4 2 | 4+4 2 | 3+3 2 | 2+2 2 |
| | 5+6 2 | 5+6 2 | 2+2 2 | 2+2 2 |
| | 4+4 2 | 8+9 2 | 3+3 2 | 3+3 2 |
| | 5+6 2 | 5+6 2 | 2+2 2 | 3+3 2 |
| | 8+9 2 | 4+4 2 | 2+2 2 | 2+2 2 |
| | 8+9 2 | 8+9 2 | 3+3 2 | 3+3 2 |
| | 7+7 2 | 7+7 2 | 2+2 2 | 2+2 2 |
| | 5+6 2 | 5+6 2 | 2+2 2 | 3+3 2 |
| | 8+9 2 | 5+6 2 | 3+3 2 | 2+2 2 |
| | 5+6 2 | 4+4 2 | 3+3 2 | 2+2 2 |
| | 4+4 2 | 8+9 2 | 2+2 2 | 3+3 2 |
| | 8+9 2 | 8+9 2 | 2+2 2 | 3+3 2 |
| | 7+7 2 | 7+7 2 | 3+3 2 | 2+2 2 |
| | 5+6 2 | 4+4 2 | 2+2 2 | 2+2 2 |
| | 4+4 2 | 8+9 2 | 3+3 2 | 3+3 2 |
| | 5+6 2 | 5+6 2 | 2+2 2 | 2+2 2 |
| | 5+6 2 | 5+6 2 | 3+3 2 | 3+3 2 |
| | 7+7 2 | 5+6 2 | 2+2 2 | 2+2 2 |
| | 8+9 2 | 8+9 2 | 3+3 2 | 3+3 2 |
| | 8+9 2 | 4+4 2 | 2+2 2 | 3+3 2 |
| | 8+9 2 | 7+7 2 | 2+2 2 | 3+3 2 |
| | 4+4 2 | 5+6 2 | 2+2 2 | 2+2 2 |
| | 5+6 2 | 8+9 2 | 3+3 2 | 2+2 2 |
| | 8+9 2 | 8+9 2 | 3+3 2 | 3+3 2 |
| 148.1.51 | 59343 | 86 | Ba1Co2O8P2 | |
| | -GM | -T | -F | -L |
| | 5+6 2 | 5+6 2 | 3+3 2 | 3+3 2 |
| | 5+6 2 | 8+9 2 | 3+3 2 | 2+2 2 |
| | 8+9 2 | 5+6 2 | 2+2 2 | 3+3 2 |
| | 8+9 2 | 5+6 2 | 2+2 2 | 3+3 2 |
| | 7+7 2 | 4+4 2 | 2+2 2 | 2+2 2 |
| | 5+6 2 | 8+9 2 | 3+3 2 | 3+3 2 |
| | 4+4 2 | 7+7 2 | 3+3 2 | 2+2 2 |
| | 5+6 2 | 8+9 2 | 3+3 2 | 2+2 2 |
| | 8+9 2 | 5+6 2 | 2+2 2 | 3+3 2 |
| | 8+9 2 | 8+9 2 | 2+2 2 | 2+2 2 |
| | 8+9 2 | 8+9 2 | 2+2 2 | 2+2 2 |
| | 7+7 2 | 7+7 2 | 2+2 2 | 2+2 2 |
| | 5+6 2 | 8+9 2 | 3+3 2 | 2+2 2 |
| | 8+9 2 | 5+6 2 | 2+2 2 | 3+3 2 |
| | 5+6 2 | 8+9 2 | 3+3 2 | 2+2 2 |
| | 4+4 2 | 7+7 2 | 2+2 2 | 3+3 2 |
| | 8+9 2 | 5+6 2 | 2+2 2 | 3+3 2 |
| | 7+7 2 | 4+4 2 | 3+3 2 | 2+2 2 |
| | 5+6 2 | 8+9 2 | 3+3 2 | 2+2 2 |
| | 8+9 2 | 5+6 2 | 2+2 2 | 3+3 2 |
| | 5+6 2 | 4+4 2 | 2+2 2 | 3+3 2 |
| | 8+9 2 | 5+6 2 | 2+2 2 | 3+3 2 |
| | 7+7 2 | 8+9 2 | 3+3 2 | 2+2 2 |
| | 5+6 2 | 8+9 2 | 3+3 2 | 2+2 2 |
| | 4+4 2 | 7+7 2 | 3+3 2 | 3+3 2 |
| | 8+9 2 | 5+6 2 | 2+2 2 | 2+2 2 |
| | 5+6 2 | 5+6 2 | 3+3 2 | 3+3 2 |
| | 4+4 2 | 4+4 2 | 2+2 2 | 2+2 2 |

Continued on next column

**TABLE LXVIII– continued**

| TQC | ICSD | | NoE | SF |
|---|---|---|---|---|
| | 5+6 2 | 5+6 2 | 3+3 2 | 3+3 2 |
| | 8+9 2 | 8+9 2 | 2+2 2 | 2+2 2 |
| | 7+7 2 | 8+9 2 | 3+3 2 | 2+2 2 |
| | 8+9 2 | 7+7 2 | 2+2 2 | 3+3 2 |
| | 5+6 2 | 5+6 2 | 2+2 2 | 3+3 2 |
| | 4+4 2 | 4+4 2 | 3+3 2 | 2+2 2 |
| | 8+9 2 | 8+9 2 | 2+2 2 | 3+3 2 |
| | 7+7 2 | 7+7 2 | 3+3 2 | 2+2 2 |
| | 4+4 2 | 7+7 2 | 3+3 2 | 2+2 2 |
| | 5+6 2 | 8+9 2 | 2+2 2 | 3+3 2 |
| | 5+6 2 | 8+9 2 | 2+2 2 | 3+3 2 |
| | 8+9 2 | 5+6 2 | 3+3 2 | 2+2 2 |
| | 7+7 2 | 4+4 2 | 3+3 2 | 2+2 2 |
| | 8+9 2 | 5+6 2 | 2+2 2 | 3+3 2 |
| | 5+6 2 | 8+9 2 | 2+2 2 | 3+3 2 |
| 148.1.52 | 237739 | | 190 | Au5Mn1Y3 |
| | -GM | -T | -F | -L |
| | 5+6 2 | 8+9 2 | 2+2 2 | 3+3 2 |
| | 4+4 2 | 7+7 2 | 3+3 2 | 2+2 2 |
| | 5+6 2 | 8+9 2 | 2+2 2 | 3+3 2 |
| | 7+7 2 | 4+4 2 | 2+2 2 | 2+2 2 |
| | 8+9 2 | 5+6 2 | 3+3 2 | 3+3 2 |
| | 8+9 2 | 5+6 2 | 3+3 2 | 2+2 2 |
| | 5+6 2 | 8+9 2 | 3+3 2 | 3+3 2 |
| | 4+4 2 | 7+7 2 | 2+2 2 | 2+2 2 |
| | 5+6 2 | 8+9 2 | 2+2 2 | 3+3 2 |
| | 7+7 2 | 4+4 2 | 3+3 2 | 2+2 2 |
| | 8+9 2 | 5+6 2 | 2+2 2 | 3+3 2 |
| | 8+9 2 | 5+6 2 | 3+3 2 | 2+2 2 |
| | 5+6 2 | 8+9 2 | 3+3 2 | 3+3 2 |
| | 8+9 2 | 5+6 2 | 2+2 2 | 2+2 2 |
| | 5+6 2 | 8+9 2 | 2+2 2 | 3+3 2 |
| | 4+4 2 | 7+7 2 | 3+3 2 | 2+2 2 |
| | 7+7 2 | 5+6 2 | 2+2 2 | 3+3 2 |
| | 5+6 2 | 8+9 2 | 3+3 2 | 2+2 2 |
| | 8+9 2 | 4+4 2 | 2+2 2 | 3+3 2 |
| | 4+4 2 | 7+7 2 | 2+2 2 | 3+3 2 |
| | 5+6 2 | 8+9 2 | 3+3 2 | 2+2 2 |
| | 8+9 2 | 5+6 2 | 3+3 2 | 2+2 2 |
| | 7+7 2 | 4+4 2 | 2+2 2 | 3+3 2 |
| | 8+9 2 | 5+6 2 | 3+3 2 | 2+2 2 |
| | 5+6 2 | 5+6 2 | 2+2 2 | 2+2 2 |
| | 4+4 2 | 8+9 2 | 3+3 2 | 3+3 2 |
| | 8+9 2 | 4+4 2 | 3+3 2 | 2+2 2 |
| | 5+6 2 | 5+6 2 | 3+3 2 | 3+3 2 |
| | 7+7 2 | 7+7 2 | 2+2 2 | 3+3 2 |
| | 5+6 2 | 5+6 2 | 3+3 2 | 2+2 2 |
| | 8+9 2 | 8+9 2 | 2+2 2 | 3+3 2 |
| | 5+6 2 | 8+9 2 | 3+3 2 | 2+2 2 |
| | 4+4 2 | 5+6 2 | 2+2 2 | 2+2 2 |
| | 8+9 2 | 8+9 2 | 2+2 2 | 3+3 2 |
| | 7+7 2 | 7+7 2 | 3+3 2 | 2+2 2 |
| | 5+6 2 | 4+4 2 | 3+3 2 | 3+3 2 |
| | 4+4 2 | 8+9 2 | 2+2 2 | 2+2 2 |
| | 8+9 2 | 5+6 2 | 3+3 2 | 3+3 2 |
| | 5+6 2 | 5+6 2 | 2+2 2 | 2+2 2 |
| | 8+9 2 | 4+4 2 | 3+3 2 | 3+3 2 |
| | 7+7 2 | 7+7 2 | 2+2 2 | 2+2 2 |
| | 4+4 2 | 8+9 2 | 3+3 2 | 3+3 2 |
| | 5+6 2 | 4+4 2 | 3+3 2 | 3+3 2 |
| | 5+6 2 | 5+6 2 | 2+2 2 | 3+3 2 |

Continued on next column

**TABLE LXVIII– continued**

| TQC | ICSD | | NoE | SF |
|---|---|---|---|---|
| | 4+4 2 | 7+7 2 | 3+3 2 | 2+2 2 |
| | 8+9 2 | 8+9 2 | 3+3 2 | 2+2 2 |
| | 5+6 2 | 5+6 2 | 3+3 2 | 2+2 2 |
| | 5+6 2 | 8+9 2 | 2+2 2 | 3+3 2 |
| | 8+9 2 | 5+6 2 | 3+3 2 | 3+3 2 |
| | 4+4 2 | 4+4 2 | 2+2 2 | 2+2 2 |
| | 5+6 2 | 8+9 2 | 3+3 2 | 2+2 2 |
| | 8+9 2 | 5+6 2 | 2+2 2 | 3+3 2 |
| | 7+7 2 | 7+7 2 | 3+3 2 | 3+3 2 |
| | 5+6 2 | 8+9 2 | 3+3 2 | 2+2 2 |
| | 8+9 2 | 8+9 2 | 2+2 2 | 2+2 2 |
| | 7+7 2 | 4+4 2 | 3+3 2 | 3+3 2 |
| | 4+4 2 | 5+6 2 | 2+2 2 | 3+3 2 |
| | 5+6 2 | 7+7 2 | 3+3 2 | 2+2 2 |
| | 8+9 2 | 5+6 2 | 2+2 2 | 3+3 2 |
| | 5+6 2 | 8+9 2 | 3+3 2 | 2+2 2 |
| | 8+9 2 | 5+6 2 | 2+2 2 | 3+3 2 |
| | 4+4 2 | 4+4 2 | 2+2 2 | 2+2 2 |
| | 5+6 2 | 7+7 2 | 3+3 2 | 3+3 2 |
| | 5+6 2 | 5+6 2 | 3+3 2 | 2+2 2 |
| | 8+9 2 | 8+9 2 | 2+2 2 | 3+3 2 |
| | 4+4 2 | 4+4 2 | 3+3 2 | 2+2 2 |
| | 7+7 2 | 8+9 2 | 2+2 2 | 3+3 2 |
| | 5+6 2 | 5+6 2 | 3+3 2 | 2+2 2 |
| | 8+9 2 | 8+9 2 | 2+2 2 | 3+3 2 |
| | 5+6 2 | 7+7 2 | 3+3 2 | 2+2 2 |
| | 4+4 2 | 5+6 2 | 3+3 2 | 3+3 2 |
| | 7+7 2 | 4+4 2 | 2+2 2 | 3+3 2 |
| | 5+6 2 | 5+6 2 | 3+3 2 | 2+2 2 |
| | 4+4 2 | 8+9 2 | 3+3 2 | 3+3 2 |
| | 8+9 2 | 8+9 2 | 2+2 2 | 3+3 2 |
| | 5+6 2 | 7+7 2 | 3+3 2 | 2+2 2 |
| | 7+7 2 | 4+4 2 | 2+2 2 | 2+2 2 |
| | 5+6 2 | 8+9 2 | 3+3 2 | 2+2 2 |
| | 8+9 2 | 5+6 2 | 3+3 2 | 3+3 2 |
| | 8+9 2 | 7+7 2 | 3+3 2 | 3+3 2 |
| | 5+6 2 | 5+6 2 | 2+2 2 | 2+2 2 |
| | 4+4 2 | 8+9 2 | 3+3 2 | 2+2 2 |
| | 5+6 2 | 5+6 2 | 3+3 2 | 2+2 2 |
| | 5+6 2 | 5+6 2 | 3+3 2 | 3+3 2 |
| | 8+9 2 | 4+4 2 | 2+2 2 | 3+3 2 |
| | 7+7 2 | 8+9 2 | 2+2 2 | 2+2 2 |
| | 8+9 2 | 7+7 2 | 3+3 2 | 2+2 2 |
| | 4+4 2 | 5+6 2 | 2+2 2 | 3+3 2 |
| | 5+6 2 | 8+9 2 | 2+2 2 | 2+2 2 |
| | 8+9 2 | 5+6 2 | 2+2 2 | 3+3 2 |
| | 7+7 2 | 4+4 2 | 3+3 2 | 2+2 2 |
| | 5+6 2 | 8+9 2 | 2+2 2 | 3+3 2 |
| | 4+4 2 | 7+7 2 | 3+3 2 | 2+2 2 |
| | 8+9 2 | 8+9 2 | 2+2 2 | 3+3 2 |
| | 8+9 2 | 5+6 2 | 3+3 2 | 2+2 2 |
| 148.1.53 | 91777 | | 68 | Ag1Bi1O3 |
| | -GM | -T | -F | -L |
| | 5+6 2 | 5+6 2 | 2+2 2 | 2+2 2 |
| | 5+6 2 | 8+9 2 | 3+3 2 | 3+3 2 |
| | 8+9 2 | 5+6 2 | 2+2 2 | 2+2 2 |
| | 4+4 2 | 4+4 2 | 3+3 2 | 3+3 2 |
| | 8+9 2 | 8+9 2 | 2+2 2 | 2+2 2 |
| | 7+7 2 | 7+7 2 | 3+3 2 | 3+3 2 |
| | 8+9 2 | 8+9 2 | 3+3 2 | 3+3 2 |
| | 5+6 2 | 5+6 2 | 2+2 2 | 2+2 2 |

Continued on next column

**TABLE LXVIII– continued**

| TQC | ICSD | NoE | SF | |
|---|---|---|---|---|
| | 5+6 2 | 5+6 2 | 2+2 2 | 2+2 2 |
| | 5+6 2 | 5+6 2 | 3+3 2 | 3+3 2 |
| | 4+4 2 | 4+4 2 | 3+3 2 | 2+2 2 |
| | 8+9 2 | 8+9 2 | 2+2 2 | 3+3 2 |
| | 7+7 2 | 7+7 2 | 2+2 2 | 2+2 2 |
| | 7+7 2 | 8+9 2 | 3+3 2 | 3+3 2 |
| | 8+9 2 | 7+7 2 | 2+2 2 | 3+3 2 |
| | 4+4 2 | 8+9 2 | 3+3 2 | 2+2 2 |
| | 5+6 2 | 5+6 2 | 2+2 2 | 3+3 2 |
| | 8+9 2 | 4+4 2 | 3+3 2 | 2+2 2 |
| | 5+6 2 | 5+6 2 | 3+3 2 | 3+3 2 |
| | 8+9 2 | 8+9 2 | 2+2 2 | 2+2 2 |
| | 4+4 2 | 8+9 2 | 3+3 2 | 3+3 2 |
| | 5+6 2 | 7+7 2 | 2+2 2 | 2+2 2 |
| | 8+9 2 | 5+6 2 | 3+3 2 | 2+2 2 |
| | 7+7 2 | 4+4 2 | 2+2 2 | 3+3 2 |
| | 5+6 2 | 5+6 2 | 2+2 2 | 2+2 2 |
| | 7+7 2 | 8+9 2 | 3+3 2 | 3+3 2 |
| | 8+9 2 | 7+7 2 | 2+2 2 | 3+3 2 |
| | 4+4 2 | 8+9 2 | 3+3 2 | 2+2 2 |
| | 5+6 2 | 4+4 2 | 2+2 2 | 3+3 2 |
| | 5+6 2 | 5+6 2 | 3+3 2 | 2+2 2 |
| | 8+9 2 | 5+6 2 | 2+2 2 | 3+3 2 |
| | 4+4 2 | 7+7 2 | 3+3 2 | 2+2 2 |
| | 5+6 2 | 4+4 2 | 2+2 2 | 3+3 2 |
| | 7+7 2 | 8+9 2 | 3+3 2 | 2+2 2 |
| 148.1.54 | 89432 | 68 | Ag1Bi1O3 | |
| | -GM | -T | -F | -L |
| | 5+6 2 | 5+6 2 | 2+2 2 | 2+2 2 |
| | 5+6 2 | 8+9 2 | 3+3 2 | 3+3 2 |
| | 8+9 2 | 5+6 2 | 2+2 2 | 2+2 2 |
| | 4+4 2 | 4+4 2 | 3+3 2 | 3+3 2 |
| | 8+9 2 | 8+9 2 | 2+2 2 | 2+2 2 |
| | 7+7 2 | 7+7 2 | 3+3 2 | 3+3 2 |
| | 8+9 2 | 8+9 2 | 3+3 2 | 3+3 2 |
| | 5+6 2 | 5+6 2 | 2+2 2 | 2+2 2 |
| | 5+6 2 | 5+6 2 | 2+2 2 | 2+2 2 |
| | 5+6 2 | 5+6 2 | 3+3 2 | 3+3 2 |
| | 4+4 2 | 4+4 2 | 3+3 2 | 2+2 2 |
| | 8+9 2 | 8+9 2 | 2+2 2 | 3+3 2 |
| | 7+7 2 | 7+7 2 | 2+2 2 | 2+2 2 |
| | 7+7 2 | 8+9 2 | 3+3 2 | 3+3 2 |
| | 8+9 2 | 7+7 2 | 2+2 2 | 3+3 2 |
| | 4+4 2 | 8+9 2 | 3+3 2 | 2+2 2 |
| | 5+6 2 | 5+6 2 | 2+2 2 | 3+3 2 |
| | 8+9 2 | 4+4 2 | 3+3 2 | 2+2 2 |
| | 5+6 2 | 5+6 2 | 3+3 2 | 3+3 2 |
| | 8+9 2 | 8+9 2 | 2+2 2 | 2+2 2 |
| | 4+4 2 | 8+9 2 | 3+3 2 | 3+3 2 |
| | 5+6 2 | 7+7 2 | 2+2 2 | 2+2 2 |
| | 8+9 2 | 5+6 2 | 3+3 2 | 2+2 2 |
| | 7+7 2 | 4+4 2 | 2+2 2 | 3+3 2 |
| | 5+6 2 | 5+6 2 | 2+2 2 | 2+2 2 |
| | 7+7 2 | 8+9 2 | 3+3 2 | 3+3 2 |
| | 8+9 2 | 7+7 2 | 2+2 2 | 3+3 2 |
| | 4+4 2 | 8+9 2 | 3+3 2 | 2+2 2 |
| | 5+6 2 | 4+4 2 | 2+2 2 | 3+3 2 |
| | 5+6 2 | 5+6 2 | 3+3 2 | 2+2 2 |
| | 8+9 2 | 5+6 2 | 2+2 2 | 3+3 2 |
| | 4+4 2 | 7+7 2 | 3+3 2 | 2+2 2 |
| | 5+6 2 | 4+4 2 | 2+2 2 | 3+3 2 |

Continued on next column

**TABLE LXVIII– continued**

| TQC | ICSD | NoE | SF | |
|---|---|---|---|---|
| | 7+7 2 | 8+9 2 | 3+3 2 | 2+2 2 |

The End

TABLE LXIX: The computed irreducible representations for the valence bands for the compounds of SEBR and NLC in space group 162. See the list of tables.

| TQC | ICSD | | NoE | | SF | |
|---|---|---|---|---|---|---|
| 162.1.1 | 28086 | | 60 | | O6U1V2 | |
| | -A | -GM | -H | -K | -L | -M |
| | 8 2 | 8 2 | 6 2 | 6 2 | 3+4 2 | 3+4 2 |
| | 9 2 | 9 2 | 6 2 | 6 2 | 5+6 2 | 5+6 2 |
| | 9 2 | 9 2 | 4+5 2 | 4+5 2 | 5+6 2 | 5+6 2 |
| | 6+7 2 | 6+7 2 | 6 2 | 6 2 | 5+6 2 | 5+6 2 |
| | 4+5 2 | 8 2 | 6 2 | 6 2 | 3+4 2 | 3+4 2 |
| | 8 2 | 8 2 | 4+5 2 | 4+5 2 | 3+4 2 | 3+4 2 |
| | 8 2 | 4+5 2 | 6 2 | 6 2 | 3+4 2 | 3+4 2 |
| | 9 2 | 9 2 | 6 2 | 6 2 | 5+6 2 | 5+6 2 |
| | 9 2 | 6+7 2 | 4+5 2 | 4+5 2 | 5+6 2 | 5+6 2 |
| | 6+7 2 | 9 2 | 6 2 | 6 2 | 5+6 2 | 5+6 2 |
| | 8 2 | 8 2 | 6 2 | 6 2 | 3+4 2 | 3+4 2 |
| | 4+5 2 | 8 2 | 4+5 2 | 4+5 2 | 3+4 2 | 3+4 2 |
| | 9 2 | 4+5 2 | 6 2 | 6 2 | 3+4 2 | 5+6 2 |
| | 8 2 | 9 2 | 6 2 | 4+5 2 | 5+6 2 | 5+6 2 |
| | 4+5 2 | 4+5 2 | 4+5 2 | 6 2 | 3+4 2 | 3+4 2 |
| | 8 2 | 8 2 | 6 2 | 4+5 2 | 5+6 2 | 5+6 2 |
| | 9 2 | 9 2 | 6 2 | 6 2 | 3+4 2 | 3+4 2 |
| | 6+7 2 | 6+7 2 | 4+5 2 | 6 2 | 5+6 2 | 3+4 2 |
| | 9 2 | 8 2 | 6 2 | 6 2 | 5+6 2 | 5+6 2 |
| | 9 2 | 9 2 | 6 2 | 6 2 | 5+6 2 | 3+4 2 |
| | 8 2 | 8 2 | 6 2 | 6 2 | 3+4 2 | 5+6 2 |
| | 6+7 2 | 6+7 2 | 4+5 2 | 4+5 2 | 5+6 2 | 5+6 2 |
| | 8 2 | 9 2 | 6 2 | 6 2 | 3+4 2 | 3+4 2 |
| | 9 2 | 8 2 | 6 2 | 6 2 | 3+4 2 | 5+6 2 |
| | 8 2 | 4+5 2 | 6 2 | 4+5 2 | 5+6 2 | 3+4 2 |
| | 4+5 2 | 6+7 2 | 4+5 2 | 6 2 | 3+4 2 | 5+6 2 |
| | 9 2 | 9 2 | 6 2 | 6 2 | 5+6 2 | 3+4 2 |
| | 6+7 2 | 9 2 | 4+5 2 | 4+5 2 | 5+6 2 | 5+6 2 |
| | 6+7 2 | 9 2 | 6 2 | 4+5 2 | 5+6 2 | 5+6 2 |
| | 9 2 | 4+5 2 | 4+5 2 | 6 2 | 3+4 2 | 3+4 2 |
| 162.1.2 | 275 | | 62 | | Cr2O6U1 | |
| | -A | -GM | -H | -K | -L | -M |
| | 8 2 | 8 2 | 6 2 | 6 2 | 3+4 2 | 3+4 2 |
| | 9 2 | 9 2 | 6 2 | 6 2 | 5+6 2 | 5+6 2 |
| | 9 2 | 9 2 | 6 2 | 4+5 2 | 5+6 2 | 5+6 2 |
| | 6+7 2 | 6+7 2 | 4+5 2 | 6 2 | 5+6 2 | 5+6 2 |
| | 4+5 2 | 8 2 | 6 2 | 6 2 | 3+4 2 | 3+4 2 |
| | 8 2 | 8 2 | 4+5 2 | 4+5 2 | 3+4 2 | 3+4 2 |
| | 8 2 | 4+5 2 | 6 2 | 6 2 | 3+4 2 | 3+4 2 |
| | 9 2 | 9 2 | 6 2 | 6 2 | 5+6 2 | 5+6 2 |
| | 9 2 | 6+7 2 | 4+5 2 | 4+5 2 | 5+6 2 | 5+6 2 |
| | 6+7 2 | 9 2 | 6 2 | 6 2 | 5+6 2 | 5+6 2 |
| | 8 2 | 8 2 | 6 2 | 6 2 | 3+4 2 | 3+4 2 |
| | 4+5 2 | 8 2 | 4+5 2 | 4+5 2 | 3+4 2 | 3+4 2 |
| | 9 2 | 4+5 2 | 6 2 | 4+5 2 | 3+4 2 | 5+6 2 |
| | 4+5 2 | 9 2 | 6 2 | 6 2 | 5+6 2 | 3+4 2 |
| | 8 2 | 4+5 2 | 4+5 2 | 6 2 | 3+4 2 | 5+6 2 |
| | 6+7 2 | 8 2 | 6 2 | 4+5 2 | 5+6 2 | 3+4 2 |
| | 9 2 | 9 2 | 6 2 | 6 2 | 3+4 2 | 5+6 2 |
| | 8 2 | 6+7 2 | 6 2 | 6 2 | 5+6 2 | 3+4 2 |

Continued on next column

**TABLE LXIX– continued**

| TQC | ICSD | | NoE | | SF | |
|---|---|---|---|---|---|---|
| | 9 2 | 8 2 | 4+5 2 | 6 2 | 5+6 2 | 5+6 2 |
| | 9 2 | 9 2 | 6 2 | 6 2 | 5+6 2 | 3+4 2 |
| | 8 2 | 8 2 | 6 2 | 6 2 | 3+4 2 | 5+6 2 |
| | 6+7 2 | 6+7 2 | 4+5 2 | 4+5 2 | 3+4 2 | 5+6 2 |
| | 8 2 | 9 2 | 6 2 | 6 2 | 5+6 2 | 3+4 2 |
| | 9 2 | 9 2 | 6 2 | 6 2 | 3+4 2 | 5+6 2 |
| | 8 2 | 6+7 2 | 6 2 | 4+5 2 | 5+6 2 | 3+4 2 |
| | 4+5 2 | 8 2 | 4+5 2 | 6 2 | 3+4 2 | 5+6 2 |
| | 9 2 | 4+5 2 | 6 2 | 4+5 2 | 5+6 2 | 3+4 2 |
| | 6+7 2 | 9 2 | 4+5 2 | 6 2 | 5+6 2 | 5+6 2 |
| | 6+7 2 | 4+5 2 | 6 2 | 6 2 | 5+6 2 | 5+6 2 |
| | 9 2 | 8 2 | 4+5 2 | 4+5 2 | 3+4 2 | 3+4 2 |
| | 9 2 | 9 2 | 6 2 | 6 2 | 3+4 2 | 5+6 2 |
| 162.1.3 | 15133 | | 62 | | Cr2O6U1 | |
| | -A | -GM | -H | -K | -L | -M |
| | 8 2 | 8 2 | 6 2 | 6 2 | 3+4 2 | 3+4 2 |
| | 9 2 | 9 2 | 6 2 | 6 2 | 5+6 2 | 5+6 2 |
| | 9 2 | 9 2 | 6 2 | 4+5 2 | 5+6 2 | 5+6 2 |
| | 6+7 2 | 6+7 2 | 4+5 2 | 6 2 | 5+6 2 | 5+6 2 |
| | 8 2 | 8 2 | 6 2 | 6 2 | 3+4 2 | 3+4 2 |
| | 4+5 2 | 8 2 | 4+5 2 | 4+5 2 | 3+4 2 | 3+4 2 |
| | 8 2 | 4+5 2 | 6 2 | 6 2 | 3+4 2 | 3+4 2 |
| | 9 2 | 9 2 | 6 2 | 6 2 | 5+6 2 | 5+6 2 |
| | 9 2 | 6+7 2 | 4+5 2 | 4+5 2 | 5+6 2 | 5+6 2 |
| | 6+7 2 | 9 2 | 6 2 | 6 2 | 5+6 2 | 5+6 2 |
| | 8 2 | 8 2 | 6 2 | 6 2 | 3+4 2 | 3+4 2 |
| | 4+5 2 | 8 2 | 4+5 2 | 4+5 2 | 3+4 2 | 3+4 2 |
| | 9 2 | 4+5 2 | 6 2 | 4+5 2 | 3+4 2 | 5+6 2 |
| | 4+5 2 | 9 2 | 6 2 | 6 2 | 5+6 2 | 3+4 2 |
| | 8 2 | 4+5 2 | 4+5 2 | 6 2 | 3+4 2 | 5+6 2 |
| | 6+7 2 | 8 2 | 6 2 | 4+5 2 | 5+6 2 | 3+4 2 |
| | 9 2 | 9 2 | 6 2 | 6 2 | 3+4 2 | 3+4 2 |
| | 8 2 | 6+7 2 | 4+5 2 | 6 2 | 5+6 2 | 5+6 2 |
| | 9 2 | 8 2 | 6 2 | 6 2 | 5+6 2 | 5+6 2 |
| | 9 2 | 9 2 | 6 2 | 6 2 | 5+6 2 | 3+4 2 |
| | 8 2 | 8 2 | 6 2 | 6 2 | 3+4 2 | 5+6 2 |
| | 6+7 2 | 6+7 2 | 4+5 2 | 4+5 2 | 3+4 2 | 5+6 2 |
| | 8 2 | 9 2 | 6 2 | 6 2 | 5+6 2 | 3+4 2 |
| | 9 2 | 9 2 | 6 2 | 6 2 | 3+4 2 | 5+6 2 |
| | 8 2 | 4+5 2 | 4+5 2 | 4+5 2 | 3+4 2 | 3+4 2 |
| | 4+5 2 | 8 2 | 6 2 | 6 2 | 5+6 2 | 3+4 2 |
| | 9 2 | 6+7 2 | 6 2 | 4+5 2 | 5+6 2 | 5+6 2 |
| | 6+7 2 | 9 2 | 4+5 2 | 6 2 | 5+6 2 | 5+6 2 |
| | 6+7 2 | 4+5 2 | 6 2 | 6 2 | 5+6 2 | 5+6 2 |
| | 9 2 | 8 2 | 4+5 2 | 4+5 2 | 3+4 2 | 3+4 2 |
| | 9 2 | 9 2 | 6 2 | 6 2 | 3+4 2 | 5+6 2 |
| 162.1.4 | 37011 | | 100 | | Ba1Fe4O8Sr1 | |
| | -A | -GM | -H | -K | -L | -M |
| | 9 2 | 8 2 | 6 2 | 6 2 | 5+6 2 | 3+4 2 |
| | 8 2 | 8 2 | 6 2 | 6 2 | 3+4 2 | 3+4 2 |
| | 8 2 | 8 2 | 6 2 | 6 2 | 5+6 2 | 5+6 2 |
| | 9 2 | 9 2 | 4+5 2 | 4+5 2 | 3+4 2 | 3+4 2 |
| | 8 2 | 9 2 | 6 2 | 6 2 | 3+4 2 | 5+6 2 |
| | 9 2 | 8 2 | 4+5 2 | 6 2 | 3+4 2 | 5+6 2 |
| | 8 2 | 9 2 | 6 2 | 4+5 2 | 5+6 2 | 3+4 2 |
| | 4+5 2 | 6+7 2 | 6 2 | 4+5 2 | 3+4 2 | 5+6 2 |
| | 9 2 | 8 2 | 4+5 2 | 6 2 | 5+6 2 | 3+4 2 |
| | 6+7 2 | 4+5 2 | 6 2 | 6 2 | 5+6 2 | 3+4 2 |
| | 8 2 | 9 2 | 6 2 | 6 2 | 3+4 2 | 5+6 2 |
| | 4+5 2 | 6+7 2 | 4+5 2 | 4+5 2 | 3+4 2 | 5+6 2 |
| | 8 2 | 9 2 | 6 2 | 6 2 | 3+4 2 | 5+6 2 |
| | 9 2 | 9 2 | 6 2 | 6 2 | 5+6 2 | 5+6 2 |
| | 6+7 2 | 6+7 2 | 4+5 2 | 4+5 2 | 5+6 2 | 5+6 2 |
| | 9 2 | 9 2 | 6 2 | 6 2 | 5+6 2 | 5+6 2 |
| | 9 2 | 9 2 | 4+5 2 | 4+5 2 | 3+4 2 | 3+4 2 |
| | 8 2 | 8 2 | 6 2 | 6 2 | 5+6 2 | 5+6 2 |
| | 8 2 | 8 2 | 6 2 | 6 2 | 3+4 2 | 3+4 2 |
| | 4+5 2 | 4+5 2 | 4+5 2 | 4+5 2 | 5+6 2 | 5+6 2 |
| | 9 2 | 9 2 | 6 2 | 6 2 | 5+6 2 | 5+6 2 |
| | 6+7 2 | 6+7 2 | 4+5 2 | 6 2 | 3+4 2 | 3+4 2 |
| | 8 2 | 8 2 | 6 2 | 4+5 2 | 5+6 2 | 3+4 2 |
| | 4+5 2 | 4+5 2 | 6 2 | 6 2 | 3+4 2 | 3+4 2 |
| | 8 2 | 9 2 | 6 2 | 6 2 | 5+6 2 | 5+6 2 |
| | 9 2 | 8 2 | 6 2 | 6 2 | 3+4 2 | 3+4 2 |
| | 9 2 | 8 2 | 6 2 | 6 2 | 5+6 2 | 5+6 2 |
| | 6+7 2 | 4+5 2 | 6 2 | 6 2 | 3+4 2 | 5+6 2 |
| | 9 2 | 8 2 | 4+5 2 | 4+5 2 | 5+6 2 | 3+4 2 |
| | 8 2 | 9 2 | 6 2 | 6 2 | 3+4 2 | 3+4 2 |
| | 4+5 2 | 6+7 2 | 4+5 2 | 4+5 2 | 5+6 2 | 5+6 2 |
| | 8 2 | 9 2 | 6 2 | 6 2 | 3+4 2 | 5+6 2 |
| | 9 2 | 8 2 | 4+5 2 | 4+5 2 | 5+6 2 | 5+6 2 |
| | 6+7 2 | 9 2 | 6 2 | 6 2 | 5+6 2 | 3+4 2 |
| | 9 2 | 6+7 2 | 4+5 2 | 4+5 2 | 3+4 2 | 3+4 2 |
| | 8 2 | 4+5 2 | 6 2 | 6 2 | 3+4 2 | 5+6 2 |
| | 8 2 | 8 2 | 6 2 | 6 2 | 3+4 2 | 3+4 2 |
| | 4+5 2 | 9 2 | 4+5 2 | 4+5 2 | 3+4 2 | 3+4 2 |
| | 9 2 | 9 2 | 6 2 | 6 2 | 5+6 2 | 5+6 2 |
| | 6+7 2 | 6+7 2 | 6 2 | 6 2 | 5+6 2 | 5+6 2 |
| | 6+7 2 | 6+7 2 | 6 2 | 6 2 | 5+6 2 | 5+6 2 |
| | 9 2 | 9 2 | 6 2 | 6 2 | 3+4 2 | 3+4 2 |
| | 8 2 | 8 2 | 4+5 2 | 4+5 2 | 5+6 2 | 5+6 2 |
| | 4+5 2 | 4+5 2 | 6 2 | 6 2 | 3+4 2 | 3+4 2 |
| | 9 2 | 9 2 | 6 2 | 6 2 | 3+4 2 | 3+4 2 |
| | 4+5 2 | 4+5 2 | 6 2 | 6 2 | 5+6 2 | 5+6 2 |
| | 8 2 | 8 2 | 6 2 | 4+5 2 | 3+4 2 | 3+4 2 |
| | 6+7 2 | 6+7 2 | 4+5 2 | 6 2 | 5+6 2 | 3+4 2 |
| | 9 2 | 9 2 | 6 2 | 6 2 | 3+4 2 | 5+6 2 |
| | 9 2 | 9 2 | 4+5 2 | 4+5 2 | 5+6 2 | 5+6 2 |
| 162.1.5 | 26557 | | 78 | | Ag6O2 | |
| | -A | -GM | -H | -K | -L | -M |
| | 9 2 | 8 2 | 4+5 2 | 4+5 2 | 3+4 2 | 5+6 2 |
| | 8 2 | 9 2 | 6 2 | 6 2 | 5+6 2 | 3+4 2 |
| | 8 2 | 9 2 | 6 2 | 6 2 | 3+4 2 | 3+4 2 |
| | 6+7 2 | 4+5 2 | 6 2 | 6 2 | 3+4 2 | 5+6 2 |
| | 9 2 | 8 2 | 4+5 2 | 4+5 2 | 5+6 2 | 5+6 2 |
| | 4+5 2 | 8 2 | 6 2 | 6 2 | 5+6 2 | 3+4 2 |
| | 9 2 | 8 2 | 4+5 2 | 4+5 2 | 5+6 2 | 5+6 2 |
| | 8 2 | 8 2 | 6 2 | 6 2 | 5+6 2 | 3+4 2 |
| | 9 2 | 6+7 2 | 6 2 | 6 2 | 3+4 2 | 3+4 2 |
| | 8 2 | 9 2 | 6 2 | 6 2 | 3+4 2 | 3+4 2 |
| | 9 2 | 9 2 | 6 2 | 4+5 2 | 3+4 2 | 3+4 2 |
| | 4+5 2 | 8 2 | 4+5 2 | 6 2 | 5+6 2 | 5+6 2 |
| | 8 2 | 9 2 | 4+5 2 | 6 2 | 3+4 2 | 5+6 2 |
| | 9 2 | 6+7 2 | 6 2 | 4+5 2 | 5+6 2 | 5+6 2 |
| | 6+7 2 | 4+5 2 | 6 2 | 6 2 | 5+6 2 | 3+4 2 |
| | 8 2 | 8 2 | 6 2 | 4+5 2 | 3+4 2 | 5+6 2 |
| | 9 2 | 4+5 2 | 4+5 2 | 6 2 | 5+6 2 | 3+4 2 |
| | 8 2 | 6+7 2 | 6 2 | 6 2 | 3+4 2 | 5+6 2 |
| | 9 2 | 9 2 | 4+5 2 | 6 2 | 5+6 2 | 5+6 2 |
| | 6+7 2 | 9 2 | 6 2 | 6 2 | 3+4 2 | 3+4 2 |
| | 4+5 2 | 8 2 | 6 2 | 4+5 2 | 5+6 2 | 5+6 2 |
| | 8 2 | 8 2 | 6 2 | 4+5 2 | 5+6 2 | 3+4 2 |

Continued on next column

Continued on next column

**TABLE LXIX– continued**

| TQC | ICSD | NoE | SF | | | |
|---|---|---|---|---|---|---|
| | 9 2 | 8 2 | 6 2 | 6 2 | 3+4 2 | 5+6 2 |
| | 8 2 | 4+5 2 | 4+5 2 | 6 2 | 3+4 2 | 3+4 2 |
| | 4+5 2 | 9 2 | 6 2 | 6 2 | 5+6 2 | 5+6 2 |
| | 8 2 | 9 2 | 4+5 2 | 4+5 2 | 5+6 2 | 5+6 2 |
| | 9 2 | 6+7 2 | 6 2 | 6 2 | 3+4 2 | 3+4 2 |
| | 6+7 2 | 9 2 | 4+5 2 | 6 2 | 5+6 2 | 5+6 2 |
| | 9 2 | 8 2 | 6 2 | 4+5 2 | 3+4 2 | 3+4 2 |
| | 8 2 | 6+7 2 | 6 2 | 6 2 | 3+4 2 | 5+6 2 |
| | 6+7 2 | 9 2 | 6 2 | 4+5 2 | 5+6 2 | 3+4 2 |
| | 4+5 2 | 4+5 2 | 4+5 2 | 6 2 | 3+4 2 | 3+4 2 |
| | 8 2 | 8 2 | 6 2 | 6 2 | 3+4 2 | 3+4 2 |
| | 6+7 2 | 4+5 2 | 4+5 2 | 4+5 2 | 5+6 2 | 5+6 2 |
| | 9 2 | 9 2 | 6 2 | 6 2 | 3+4 2 | 5+6 2 |
| | 4+5 2 | 8 2 | 6 2 | 6 2 | 5+6 2 | 5+6 2 |
| | 8 2 | 9 2 | 6 2 | 4+5 2 | 5+6 2 | 3+4 2 |
| | 8 2 | 6+7 2 | 4+5 2 | 6 2 | 3+4 2 | 3+4 2 |
| | 9 2 | 8 2 | 6 2 | 6 2 | 3+4 2 | 5+6 2 |

The End

TABLE LXX: The computed irreducible representations for the valence bands for the compounds of SEBR and NLC in space group 163. See the list of tables.

| TQC | ICSD | NoE | SF | | | |
|---|---|---|---|---|---|---|
| 163.1.1 | 71037 | 112 | Mo1Ni1P8 | | | |
| | -A | -GM | -H | -K | -L | -M |
| | 5+6 4 | 8 2 | 4+5 2 | 4+5 2 | 2+2 4 | 5+6 2 |
| | 5+6 4 | 9 2 | 6 2 | 6 2 | 2+2 4 | 3+4 2 |
| | 4+4 4 | 8 2 | 6 2 | 6 2 | 2+2 4 | 3+4 2 |
| | 5+6 4 | 9 2 | 6 2 | 4+5 2 | 2+2 4 | 5+6 2 |
| | 5+6 4 | 6+7 2 | 4+5 2 | 6 2 | 2+2 4 | 5+6 2 |
| | 5+6 4 | 4+5 2 | 6 2 | 6 2 | 2+2 4 | 3+4 2 |
| | 4+4 4 | 8 2 | 4+5 2 | 4+5 2 | 2+2 4 | 5+6 2 |
| | 5+6 4 | 9 2 | 6 2 | 6 2 | 2+2 4 | 3+4 2 |
| | 5+6 4 | 9 2 | 6 2 | 6 2 | 2+2 4 | 3+4 2 |
| | 4+4 4 | 8 2 | 4+5 2 | 4+5 2 | 2+2 4 | 3+4 2 |
| | 5+6 4 | 8 2 | 6 2 | 6 2 | 2+2 4 | 5+6 2 |
| | 5+6 4 | 4+5 2 | 6 2 | 6 2 | 2+2 4 | 5+6 2 |
| | 4+4 4 | 9 2 | 4+5 2 | 4+5 2 | 2+2 4 | 5+6 2 |
| | 5+6 4 | 6+7 2 | 6 2 | 6 2 | 2+2 4 | 3+4 2 |
| | 5+6 4 | 8 2 | 6 2 | 4+5 2 | 2+2 4 | 3+4 2 |
| | 4+4 4 | 9 2 | 4+5 2 | 6 2 | 2+2 4 | 5+6 2 |
| | 5+6 4 | 8 2 | 6 2 | 4+5 2 | 2+2 4 | 5+6 2 |
| | 5+6 4 | 8 2 | 4+5 2 | 6 2 | 2+2 4 | 3+4 2 |
| | 4+4 4 | 4+5 2 | 6 2 | 6 2 | 2+2 4 | 3+4 2 |
| | 5+6 4 | 9 2 | 6 2 | 6 2 | 2+2 4 | 3+4 2 |
| | 4+4 4 | 9 2 | 4+5 2 | 6 2 | 2+2 4 | 5+6 2 |
| | 5+6 4 | 6+7 2 | 6 2 | 4+5 2 | 2+2 4 | 3+4 2 |
| | 5+6 4 | 8 2 | 4+5 2 | 6 2 | 2+2 4 | 5+6 2 |
| | 5+6 4 | 4+5 2 | 6 2 | 6 2 | 2+2 4 | 3+4 2 |
| | 4+4 4 | 8 2 | 6 2 | 4+5 2 | 2+2 4 | 5+6 2 |
| | 5+6 4 | 9 2 | 6 2 | 6 2 | 2+2 4 | 3+4 2 |
| | 4+4 4 | 6+7 2 | 6 2 | 4+5 2 | 2+2 4 | 5+6 2 |
| | 5+6 4 | 4+5 2 | 4+5 2 | 6 2 | 2+2 4 | 3+4 2 |
| | | 8 2 | 6 2 | 6 2 | | 3+4 2 |
| | | 9 2 | 4+5 2 | 6 2 | | 3+4 2 |
| | | 6+7 2 | 6 2 | 6 2 | | 5+6 2 |
| | | 8 2 | 6 2 | 4+5 2 | | 5+6 2 |
| | | 4+5 2 | 4+5 2 | 4+5 2 | | 5+6 2 |
| | | 8 2 | 6 2 | 6 2 | | 3+4 2 |

Continued on next column

**TABLE LXX– continued**

| TQC | ICSD | NoE | SF | | | |
|---|---|---|---|---|---|---|
| | | 8 2 | 6 2 | 6 2 | | 3+4 2 |
| | | 9 2 | 6 2 | 4+5 2 | | 5+6 2 |
| | | 8 2 | 4+5 2 | 6 2 | | 3+4 2 |
| | | 4+5 2 | 6 2 | 6 2 | | 5+6 2 |
| | | 8 2 | 6 2 | 4+5 2 | | 3+4 2 |
| | | 9 2 | 4+5 2 | 6 2 | | 5+6 2 |
| | | 6+7 2 | 6 2 | 4+5 2 | | 3+4 2 |
| | | 9 2 | 6 2 | 6 2 | | 3+4 2 |
| | | 9 2 | 4+5 2 | 6 2 | | 3+4 2 |
| | | 4+5 2 | 4+5 2 | 6 2 | | 3+4 2 |
| | | 4+5 2 | 6 2 | 4+5 2 | | 3+4 2 |
| | | 8 2 | 6 2 | 6 2 | | 3+4 2 |
| | | 8 2 | 6 2 | 6 2 | | 5+6 2 |
| | | 8 2 | 6 2 | 4+5 2 | | 3+4 2 |
| | | 8 2 | 4+5 2 | 6 2 | | 5+6 2 |
| | | 9 2 | 6 2 | 6 2 | | 3+4 2 |
| | | 6+7 2 | 6 2 | 4+5 2 | | 5+6 2 |
| | | 4+5 2 | 4+5 2 | 6 2 | | 3+4 2 |
| | | 8 2 | 6 2 | 6 2 | | 5+6 2 |
| | | 9 2 | 6 2 | 6 2 | | 5+6 2 |
| | | 6+7 2 | 4+5 2 | 4+5 2 | | 3+4 2 |
| | | 9 2 | 6 2 | 6 2 | | 3+4 2 |
| 163.1.2 | 411131 | 140 | Al1F6Li1Yb1 | | | |
| | -A | -GM | -H | -K | -L | -M |
| | 5+6 4 | 8 2 | 6 2 | 6 2 | 2+2 4 | 3+4 2 |
| | 5+6 4 | 8 2 | 6 2 | 6 2 | 2+2 4 | 3+4 2 |
| | 5+6 4 | 9 2 | 6 2 | 6 2 | 2+2 4 | 5+6 2 |
| | 5+6 4 | 9 2 | 6 2 | 6 2 | 2+2 4 | 5+6 2 |
| | 4+4 4 | 8 2 | 4+5 2 | 4+5 2 | 2+2 4 | 3+4 2 |
| | 5+6 4 | 9 2 | 6 2 | 6 2 | 2+2 4 | 5+6 2 |
| | 4+4 4 | 9 2 | 4+5 2 | 4+5 2 | 2+2 4 | 5+6 2 |
| | 5+6 4 | 6+7 2 | 6 2 | 6 2 | 2+2 4 | 3+4 2 |
| | 4+4 4 | 8 2 | 6 2 | 6 2 | 2+2 4 | 5+6 2 |
| | 5+6 4 | 4+5 2 | 4+5 2 | 4+5 2 | 2+2 4 | 3+4 2 |
| | 5+6 4 | 8 2 | 6 2 | 6 2 | 2+2 4 | 5+6 2 |
| | 5+6 4 | 6+7 2 | 6 2 | 6 2 | 2+2 4 | 3+4 2 |
| | 4+4 4 | 9 2 | 4+5 2 | 4+5 2 | 2+2 4 | 3+4 2 |
| | 5+6 4 | 4+5 2 | 6 2 | 6 2 | 2+2 4 | 3+4 2 |
| | 5+6 4 | 8 2 | 6 2 | 6 2 | 2+2 4 | 5+6 2 |
| | 4+4 4 | 9 2 | 6 2 | 6 2 | 2+2 4 | 5+6 2 |
| | 5+6 4 | 6+7 2 | 4+5 2 | 4+5 2 | 2+2 4 | 5+6 2 |
| | 5+6 4 | 9 2 | 6 2 | 4+5 2 | 2+2 4 | 5+6 2 |
| | 4+4 4 | 6+7 2 | 4+5 2 | 6 2 | 2+2 4 | 5+6 2 |
| | 5+6 4 | 9 2 | 6 2 | 6 2 | 2+2 4 | 5+6 2 |
| | 5+6 4 | 8 2 | 6 2 | 6 2 | 2+2 4 | 3+4 2 |
| | 5+6 4 | 9 2 | 4+5 2 | 4+5 2 | 2+2 4 | 5+6 2 |
| | 4+4 4 | 4+5 2 | 6 2 | 6 2 | 2+2 4 | 3+4 2 |
| | 5+6 4 | 8 2 | 6 2 | 6 2 | 2+2 4 | 3+4 2 |
| | 5+6 4 | 9 2 | 4+5 2 | 4+5 2 | 2+2 4 | 5+6 2 |
| | 4+4 4 | 6+7 2 | 6 2 | 6 2 | 2+2 4 | 3+4 2 |
| | 5+6 4 | 9 2 | 4+5 2 | 4+5 2 | 2+2 4 | 5+6 2 |
| | 4+4 4 | 8 2 | 6 2 | 6 2 | 2+2 4 | 5+6 2 |
| | 5+6 4 | 8 2 | 6 2 | 6 2 | 2+2 4 | 3+4 2 |
| | 4+4 4 | 4+5 2 | 4+5 2 | 4+5 2 | 2+2 4 | 5+6 2 |
| | 5+6 4 | 8 2 | 6 2 | 6 2 | 2+2 4 | 3+4 2 |
| | 5+6 4 | 9 2 | 6 2 | 6 2 | 2+2 4 | 5+6 2 |
| | 4+4 4 | 6+7 2 | 6 2 | 6 2 | 2+2 4 | 5+6 2 |
| | 5+6 4 | 9 2 | 4+5 2 | 4+5 2 | 2+2 4 | 3+4 2 |
| | 5+6 4 | 8 2 | 6 2 | 6 2 | 2+2 4 | 3+4 2 |
| | | 4+5 2 | 6 2 | 4+5 2 | | 3+4 2 |
| | | 8 2 | 4+5 2 | 6 2 | | 5+6 2 |

Continued on next column

**TABLE LXX– continued**

| TQC | ICSD | NoE | | SF |
|---|---|---|---|---|
| | 9 2 | 6 2 | 6 2 | 3+4 2 |
| | 8 2 | 6 2 | 6 2 | 5+6 2 |
| | 6+7 2 | 4+5 2 | 4+5 2 | 5+6 2 |
| | 9 2 | 6 2 | 6 2 | 3+4 2 |
| | 4+5 2 | 6 2 | 6 2 | 5+6 2 |
| | 8 2 | 6 2 | 6 2 | 3+4 2 |
| | 8 2 | 4+5 2 | 6 2 | 3+4 2 |
| | 9 2 | 6 2 | 4+5 2 | 5+6 2 |
| | 6+7 2 | 6 2 | 6 2 | 5+6 2 |
| | 9 2 | 6 2 | 4+5 2 | 3+4 2 |
| | 8 2 | 4+5 2 | 6 2 | 5+6 2 |
| | 4+5 2 | 6 2 | 6 2 | 3+4 2 |
| | 9 2 | 4+5 2 | 4+5 2 | 5+6 2 |
| | 6+7 2 | 6 2 | 4+5 2 | 3+4 2 |
| | 9 2 | 4+5 2 | 6 2 | 3+4 2 |
| | 8 2 | 6 2 | 6 2 | 5+6 2 |
| | 4+5 2 | 4+5 2 | 6 2 | 5+6 2 |
| | 9 2 | 6 2 | 4+5 2 | 5+6 2 |
| | 6+7 2 | 6 2 | 6 2 | 3+4 2 |
| | 9 2 | 6 2 | 6 2 | 5+6 2 |
| | 6+7 2 | 4+5 2 | 4+5 2 | 5+6 2 |
| | 9 2 | 6 2 | 6 2 | 5+6 2 |
| | 9 2 | 6 2 | 6 2 | 5+6 2 |
| | 9 2 | 6 2 | 6 2 | 5+6 2 |
| | 6+7 2 | 4+5 2 | 4+5 2 | 5+6 2 |
| | 8 2 | 6 2 | 6 2 | 5+6 2 |
| | 9 2 | 6 2 | 6 2 | 5+6 2 |
| | 9 2 | 4+5 2 | 4+5 2 | 5+6 2 |
| | 6+7 2 | 6 2 | 6 2 | 5+6 2 |
| | 9 2 | 6 2 | 6 2 | 5+6 2 |
| | 9 2 | 4+5 2 | 4+5 2 | 5+6 2 |
| | 6+7 2 | 6 2 | 6 2 | 5+6 2 |
| | 9 2 | 6 2 | 6 2 | 5+6 2 |

The End

TABLE LXXI: The computed irreducible representations for the valence bands for the compounds of SEBR and NLC in space group 164. See the list of tables.

| TQC | ICSD | | NoE | | SF | |
|---|---|---|---|---|---|---|
| 164.1.1 | 657478 | | 16 | | Te2Zr1 | |
| | -A | -GM | -H | -K | -L | -M |
| | 8 2 | 8 2 | 6 2 | 6 2 | 3+4 2 | 3+4 2 |
| | 9 2 | 9 2 | 4+5 2 | 4+5 2 | 5+6 2 | 5+6 2 |
| | 8 2 | 8 2 | 4+5 2 | 6 2 | 3+4 2 | 3+4 2 |
| | 8 2 | 8 2 | 6 2 | 4+5 2 | 5+6 2 | 5+6 2 |
| | 4+5 2 | 4+5 2 | 6 2 | 6 2 | 3+4 2 | 3+4 2 |
| | 9 2 | 9 2 | 6 2 | 6 2 | 3+4 2 | 5+6 2 |
| | 9 2 | 8 2 | 4+5 2 | 4+5 2 | 5+6 2 | 3+4 2 |
| | 6+7 2 | 4+5 2 | 6 2 | 6 2 | 5+6 2 | 5+6 2 |
| 164.1.2 | 653213 | | 16 | | Te2Zr1 | |
| | -A | -GM | -H | -K | -L | -M |
| | 8 2 | 8 2 | 6 2 | 6 2 | 3+4 2 | 3+4 2 |
| | 9 2 | 9 2 | 4+5 2 | 4+5 2 | 5+6 2 | 5+6 2 |
| | 8 2 | 8 2 | 4+5 2 | 6 2 | 3+4 2 | 3+4 2 |
| | 8 2 | 8 2 | 6 2 | 4+5 2 | 5+6 2 | 5+6 2 |
| | 4+5 2 | 4+5 2 | 6 2 | 6 2 | 3+4 2 | 3+4 2 |
| | 9 2 | 9 2 | 6 2 | 6 2 | 3+4 2 | 5+6 2 |
| | 9 2 | 8 2 | 4+5 2 | 4+5 2 | 5+6 2 | 3+4 2 |
| | 6+7 2 | 4+5 2 | 6 2 | 6 2 | 5+6 2 | 5+6 2 |
| 164.1.3 | 80205 | | 16 | | Si1Te2 | |
| | -A | -GM | -H | -K | -L | -M |
| | 8 2 | 8 2 | 4+5 2 | 4+5 2 | 5+6 2 | 5+6 2 |
| | 9 2 | 9 2 | 6 2 | 6 2 | 3+4 2 | 3+4 2 |
| | 8 2 | 8 2 | 6 2 | 6 2 | 3+4 2 | 3+4 2 |
| | 9 2 | 9 2 | 4+5 2 | 4+5 2 | 5+6 2 | 5+6 2 |
| | 9 2 | 8 2 | 6 2 | 6 2 | 5+6 2 | 5+6 2 |
| | 8 2 | 6+7 2 | 6 2 | 6 2 | 5+6 2 | 5+6 2 |
| | 6+7 2 | 8 2 | 4+5 2 | 4+5 2 | 3+4 2 | 3+4 2 |
| | 4+5 2 | 4+5 2 | 6 2 | 6 2 | 3+4 2 | 3+4 2 |
| 164.1.4 | 651764 | | 58 | | Sb2Yb1Zn2 | |
| | -A | -GM | -H | -K | -L | -M |
| | 8 2 | 8 2 | 6 2 | 6 2 | 3+4 2 | 3+4 2 |
| | 9 2 | 9 2 | 6 2 | 6 2 | 5+6 2 | 5+6 2 |
| | 9 2 | 9 2 | 4+5 2 | 4+5 2 | 5+6 2 | 5+6 2 |
| | 6+7 2 | 6+7 2 | 6 2 | 6 2 | 5+6 2 | 5+6 2 |
| | 9 2 | 8 2 | 6 2 | 6 2 | 3+4 2 | 5+6 2 |
| | 8 2 | 9 2 | 4+5 2 | 4+5 2 | 5+6 2 | 3+4 2 |
| | 4+5 2 | 6+7 2 | 6 2 | 6 2 | 5+6 2 | 3+4 2 |
| | 6+7 2 | 4+5 2 | 6 2 | 6 2 | 3+4 2 | 5+6 2 |
| | 8 2 | 8 2 | 4+5 2 | 4+5 2 | 3+4 2 | 5+6 2 |
| | 9 2 | 9 2 | 6 2 | 6 2 | 5+6 2 | 3+4 2 |
| | 8 2 | 8 2 | 6 2 | 6 2 | 3+4 2 | 5+6 2 |
| | 9 2 | 9 2 | 6 2 | 6 2 | 5+6 2 | 3+4 2 |
| | 6+7 2 | 4+5 2 | 4+5 2 | 4+5 2 | 3+4 2 | 5+6 2 |
| | 9 2 | 8 2 | 6 2 | 6 2 | 5+6 2 | 3+4 2 |
| | 8 2 | 6+7 2 | 6 2 | 6 2 | 3+4 2 | 5+6 2 |
| | 4+5 2 | 9 2 | 4+5 2 | 4+5 2 | 5+6 2 | 3+4 2 |
| | 9 2 | 8 2 | 4+5 2 | 4+5 2 | 5+6 2 | 3+4 2 |
| | 8 2 | 8 2 | 6 2 | 6 2 | 3+4 2 | 5+6 2 |
| | 9 2 | 4+5 2 | 6 2 | 6 2 | 3+4 2 | 5+6 2 |
| | 6+7 2 | 9 2 | 6 2 | 6 2 | 3+4 2 | 5+6 2 |
| | 9 2 | 6+7 2 | 4+5 2 | 4+5 2 | 5+6 2 | 3+4 2 |
| | 9 2 | 9 2 | 6 2 | 6 2 | 5+6 2 | 3+4 2 |
| | 8 2 | 9 2 | 6 2 | 6 2 | 5+6 2 | 5+6 2 |
| | 6+7 2 | 9 2 | 4+5 2 | 4+5 2 | 5+6 2 | 5+6 2 |
| | 4+5 2 | 6+7 2 | 6 2 | 6 2 | 5+6 2 | 5+6 2 |
| | 8 2 | 9 2 | 6 2 | 6 2 | 5+6 2 | 3+4 2 |
| | 9 2 | 9 2 | 4+5 2 | 4+5 2 | 5+6 2 | 5+6 2 |
| | 9 2 | 9 2 | 6 2 | 6 2 | 5+6 2 | 5+6 2 |
| | 6+7 2 | 6+7 2 | 6 2 | 6 2 | 5+6 2 | 5+6 2 |
| 164.1.5 | 419689 | | 58 | | Sb2Yb1Zn2 | |
| | -A | -GM | -H | -K | -L | -M |
| | 8 2 | 8 2 | 6 2 | 6 2 | 3+4 2 | 3+4 2 |
| | 9 2 | 9 2 | 6 2 | 6 2 | 5+6 2 | 5+6 2 |
| | 9 2 | 9 2 | 4+5 2 | 4+5 2 | 5+6 2 | 5+6 2 |
| | 6+7 2 | 6+7 2 | 6 2 | 6 2 | 5+6 2 | 5+6 2 |
| | 9 2 | 8 2 | 6 2 | 6 2 | 3+4 2 | 5+6 2 |
| | 8 2 | 9 2 | 4+5 2 | 4+5 2 | 5+6 2 | 3+4 2 |
| | 4+5 2 | 6+7 2 | 6 2 | 6 2 | 5+6 2 | 3+4 2 |
| | 6+7 2 | 4+5 2 | 6 2 | 6 2 | 3+4 2 | 5+6 2 |
| | 8 2 | 8 2 | 4+5 2 | 4+5 2 | 3+4 2 | 5+6 2 |
| | 9 2 | 9 2 | 6 2 | 6 2 | 5+6 2 | 3+4 2 |
| | 8 2 | 8 2 | 6 2 | 6 2 | 3+4 2 | 3+4 2 |
| | 9 2 | 9 2 | 6 2 | 6 2 | 5+6 2 | 5+6 2 |
| | 6+7 2 | 4+5 2 | 4+5 2 | 4+5 2 | 3+4 2 | 5+6 2 |
| | 9 2 | 8 2 | 6 2 | 6 2 | 5+6 2 | 3+4 2 |
| | 8 2 | 6+7 2 | 6 2 | 6 2 | 3+4 2 | 5+6 2 |
| | 4+5 2 | 9 2 | 4+5 2 | 4+5 2 | 5+6 2 | 3+4 2 |
| | 9 2 | 8 2 | 4+5 2 | 4+5 2 | 5+6 2 | 3+4 2 |
| | 8 2 | 8 2 | 6 2 | 6 2 | 3+4 2 | 5+6 2 |

Continued on next column

**TABLE LXXI– continued**

| TQC | ICSD | | NoE | SF | | |
|---|---|---|---|---|---|---|
| | 6+7 2 | 4+5 2 | 6 2 | 6 2 | 3+4 2 | 5+6 2 |
| | 9 2 | 9 2 | 6 2 | 6 2 | 3+4 2 | 3+4 2 |
| | 9 2 | 6+7 2 | 4+5 2 | 4+5 2 | 5+6 2 | 5+6 2 |
| | 8 2 | 9 2 | 6 2 | 6 2 | 5+6 2 | 3+4 2 |
| | 9 2 | 9 2 | 6 2 | 6 2 | 5+6 2 | 5+6 2 |
| | 4+5 2 | 9 2 | 4+5 2 | 4+5 2 | 5+6 2 | 5+6 2 |
| | 6+7 2 | 6+7 2 | 6 2 | 6 2 | 5+6 2 | 5+6 2 |
| | 8 2 | 9 2 | 6 2 | 6 2 | 5+6 2 | 3+4 2 |
| | 9 2 | 9 2 | 4+5 2 | 4+5 2 | 5+6 2 | 5+6 2 |
| | 9 2 | 9 2 | 6 2 | 6 2 | 5+6 2 | 5+6 2 |
| | 6+7 2 | 6+7 2 | 6 2 | 6 2 | 5+6 2 | 5+6 2 |
| 164.1.6 | 69557 | | 48 | Sb2Te1 | | |
| | -A | -GM | -H | -K | -L | -M |
| | 8 2 | 8 2 | 4+5 2 | 4+5 2 | 5+6 2 | 5+6 2 |
| | 9 2 | 9 2 | 6 2 | 6 2 | 3+4 2 | 3+4 2 |
| | 9 2 | 8 2 | 6 2 | 6 2 | 3+4 2 | 3+4 2 |
| | 8 2 | 8 2 | 4+5 2 | 4+5 2 | 5+6 2 | 5+6 2 |
| | 8 2 | 9 2 | 6 2 | 6 2 | 3+4 2 | 3+4 2 |
| | 9 2 | 9 2 | 6 2 | 6 2 | 5+6 2 | 5+6 2 |
| | 8 2 | 8 2 | 6 2 | 6 2 | 3+4 2 | 3+4 2 |
| | 9 2 | 8 2 | 4+5 2 | 4+5 2 | 3+4 2 | 5+6 2 |
| | 8 2 | 9 2 | 6 2 | 6 2 | 5+6 2 | 3+4 2 |
| | 9 2 | 9 2 | 6 2 | 6 2 | 5+6 2 | 5+6 2 |
| | 8 2 | 8 2 | 6 2 | 6 2 | 3+4 2 | 3+4 2 |
| | 9 2 | 9 2 | 4+5 2 | 4+5 2 | 5+6 2 | 5+6 2 |
| | 9 2 | 8 2 | 6 2 | 6 2 | 5+6 2 | 3+4 2 |
| | 6+7 2 | 6+7 2 | 6 2 | 4+5 2 | 3+4 2 | 5+6 2 |
| | 4+5 2 | 9 2 | 4+5 2 | 6 2 | 5+6 2 | 5+6 2 |
| | 9 2 | 4+5 2 | 4+5 2 | 4+5 2 | 3+4 2 | 3+4 2 |
| | 8 2 | 8 2 | 6 2 | 6 2 | 3+4 2 | 5+6 2 |
| | 8 2 | 9 2 | 6 2 | 6 2 | 3+4 2 | 3+4 2 |
| | 9 2 | 6+7 2 | 6 2 | 6 2 | 5+6 2 | 5+6 2 |
| | 6+7 2 | 4+5 2 | 4+5 2 | 6 2 | 5+6 2 | 3+4 2 |
| | 6+7 2 | 9 2 | 6 2 | 4+5 2 | 3+4 2 | 5+6 2 |
| | 4+5 2 | 6+7 2 | 6 2 | 6 2 | 5+6 2 | 5+6 2 |
| | 9 2 | 8 2 | 6 2 | 6 2 | 5+6 2 | 3+4 2 |
| | 8 2 | 8 2 | 4+5 2 | 4+5 2 | 5+6 2 | 5+6 2 |
| 164.1.7 | 41385 | | 22 | Pt1Te2 | | |
| | -A | -GM | -H | -K | -L | -M |
| | 8 2 | 8 2 | 6 2 | 6 2 | 3+4 2 | 5+6 2 |
| | 9 2 | 9 2 | 4+5 2 | 4+5 2 | 5+6 2 | 3+4 2 |
| | 4+5 2 | 8 2 | 6 2 | 6 2 | 3+4 2 | 3+4 2 |
| | 8 2 | 4+5 2 | 4+5 2 | 6 2 | 5+6 2 | 3+4 2 |
| | 8 2 | 8 2 | 6 2 | 4+5 2 | 3+4 2 | 3+4 2 |
| | 8 2 | 8 2 | 4+5 2 | 4+5 2 | 3+4 2 | 3+4 2 |
| | 4+5 2 | 4+5 2 | 6 2 | 6 2 | 3+4 2 | 5+6 2 |
| | 9 2 | 8 2 | 6 2 | 6 2 | 3+4 2 | 5+6 2 |
| | 9 2 | 9 2 | 6 2 | 6 2 | 3+4 2 | 3+4 2 |
| | 8 2 | 6+7 2 | 4+5 2 | 4+5 2 | 5+6 2 | 5+6 2 |
| | 6+7 2 | 8 2 | 6 2 | 6 2 | 5+6 2 | 3+4 2 |
| 164.1.8 | 41373 | | 22 | Pt1Te2 | | |
| | -A | -GM | -H | -K | -L | -M |
| | 8 2 | 8 2 | 6 2 | 6 2 | 3+4 2 | 5+6 2 |
| | 9 2 | 9 2 | 4+5 2 | 4+5 2 | 5+6 2 | 3+4 2 |
| | 4+5 2 | 8 2 | 6 2 | 6 2 | 3+4 2 | 3+4 2 |
| | 8 2 | 4+5 2 | 4+5 2 | 6 2 | 5+6 2 | 3+4 2 |
| | 8 2 | 8 2 | 6 2 | 4+5 2 | 3+4 2 | 3+4 2 |
| | 8 2 | 8 2 | 4+5 2 | 4+5 2 | 3+4 2 | 3+4 2 |
| | 4+5 2 | 4+5 2 | 6 2 | 6 2 | 3+4 2 | 5+6 2 |
| | 9 2 | 8 2 | 6 2 | 6 2 | 3+4 2 | 5+6 2 |
| | 9 2 | 9 2 | 6 2 | 6 2 | 3+4 2 | 3+4 2 |

Continued on next column

**TABLE LXXI– continued**

| TQC | ICSD | | NoE | SF | | |
|---|---|---|---|---|---|---|
| | 8 2 | 6+7 2 | 4+5 2 | 4+5 2 | 5+6 2 | 5+6 2 |
| | 6+7 2 | 8 2 | 6 2 | 6 2 | 5+6 2 | 3+4 2 |
| 164.1.9 | 83642 | | 22 | Pd1Te2 | | |
| | -A | -GM | -H | -K | -L | -M |
| | 8 2 | 8 2 | 6 2 | 6 2 | 3+4 2 | 5+6 2 |
| | 9 2 | 9 2 | 4+5 2 | 4+5 2 | 5+6 2 | 3+4 2 |
| | 8 2 | 8 2 | 4+5 2 | 6 2 | 3+4 2 | 3+4 2 |
| | 4+5 2 | 8 2 | 6 2 | 4+5 2 | 5+6 2 | 3+4 2 |
| | 8 2 | 4+5 2 | 6 2 | 6 2 | 3+4 2 | 3+4 2 |
| | 8 2 | 8 2 | 6 2 | 4+5 2 | 3+4 2 | 5+6 2 |
| | 4+5 2 | 4+5 2 | 4+5 2 | 6 2 | 3+4 2 | 3+4 2 |
| | 9 2 | 8 2 | 6 2 | 6 2 | 3+4 2 | 3+4 2 |
| | 9 2 | 9 2 | 6 2 | 6 2 | 3+4 2 | 5+6 2 |
| | 8 2 | 6+7 2 | 4+5 2 | 4+5 2 | 5+6 2 | 5+6 2 |
| | 6+7 2 | 8 2 | 6 2 | 6 2 | 5+6 2 | 3+4 2 |
| 164.1.10 | 42555 | | 22 | Pd1Te2 | | |
| | -A | -GM | -H | -K | -L | -M |
| | 8 2 | 8 2 | 6 2 | 6 2 | 3+4 2 | 5+6 2 |
| | 9 2 | 9 2 | 4+5 2 | 4+5 2 | 5+6 2 | 3+4 2 |
| | 8 2 | 8 2 | 4+5 2 | 6 2 | 3+4 2 | 3+4 2 |
| | 4+5 2 | 8 2 | 6 2 | 4+5 2 | 5+6 2 | 3+4 2 |
| | 8 2 | 4+5 2 | 6 2 | 6 2 | 3+4 2 | 3+4 2 |
| | 8 2 | 8 2 | 4+5 2 | 4+5 2 | 3+4 2 | 5+6 2 |
| | 4+5 2 | 4+5 2 | 6 2 | 6 2 | 3+4 2 | 3+4 2 |
| | 9 2 | 8 2 | 6 2 | 6 2 | 3+4 2 | 3+4 2 |
| | 8 2 | 9 2 | 6 2 | 6 2 | 3+4 2 | 5+6 2 |
| | 9 2 | 6+7 2 | 4+5 2 | 4+5 2 | 5+6 2 | 5+6 2 |
| | 6+7 2 | 8 2 | 6 2 | 6 2 | 5+6 2 | 3+4 2 |
| 164.1.11 | 42554 | | 22 | Pd1Te2 | | |
| | -A | -GM | -H | -K | -L | -M |
| | 8 2 | 8 2 | 6 2 | 6 2 | 3+4 2 | 5+6 2 |
| | 9 2 | 9 2 | 4+5 2 | 4+5 2 | 5+6 2 | 3+4 2 |
| | 8 2 | 8 2 | 4+5 2 | 6 2 | 3+4 2 | 3+4 2 |
| | 4+5 2 | 8 2 | 6 2 | 4+5 2 | 5+6 2 | 3+4 2 |
| | 8 2 | 4+5 2 | 6 2 | 6 2 | 3+4 2 | 3+4 2 |
| | 8 2 | 8 2 | 4+5 2 | 4+5 2 | 3+4 2 | 5+6 2 |
| | 4+5 2 | 4+5 2 | 6 2 | 6 2 | 3+4 2 | 3+4 2 |
| | 9 2 | 8 2 | 6 2 | 6 2 | 3+4 2 | 3+4 2 |
| | 9 2 | 9 2 | 6 2 | 4+5 2 | 3+4 2 | 5+6 2 |
| | 8 2 | 6+7 2 | 4+5 2 | 6 2 | 5+6 2 | 5+6 2 |
| | 6+7 2 | 8 2 | 6 2 | 6 2 | 5+6 2 | 3+4 2 |
| 164.1.12 | 41387 | | 22 | Pd1Te2 | | |
| | -A | -GM | -H | -K | -L | -M |
| | 8 2 | 8 2 | 6 2 | 6 2 | 3+4 2 | 5+6 2 |
| | 9 2 | 9 2 | 4+5 2 | 4+5 2 | 5+6 2 | 3+4 2 |
| | 8 2 | 8 2 | 4+5 2 | 6 2 | 3+4 2 | 3+4 2 |
| | 4+5 2 | 8 2 | 6 2 | 4+5 2 | 5+6 2 | 3+4 2 |
| | 8 2 | 4+5 2 | 6 2 | 6 2 | 3+4 2 | 3+4 2 |
| | 8 2 | 8 2 | 4+5 2 | 4+5 2 | 3+4 2 | 5+6 2 |
| | 4+5 2 | 4+5 2 | 6 2 | 6 2 | 3+4 2 | 3+4 2 |
| | 9 2 | 8 2 | 6 2 | 6 2 | 3+4 2 | 3+4 2 |
| | 9 2 | 9 2 | 6 2 | 4+5 2 | 3+4 2 | 5+6 2 |
| | 8 2 | 6+7 2 | 4+5 2 | 6 2 | 5+6 2 | 5+6 2 |
| | 6+7 2 | 8 2 | 6 2 | 6 2 | 5+6 2 | 3+4 2 |
| 164.1.13 | 100271 | | 58 | P2Yb1Zn2 | | |
| | -A | -GM | -H | -K | -L | -M |
| | 8 2 | 8 2 | 6 2 | 6 2 | 3+4 2 | 3+4 2 |
| | 9 2 | 9 2 | 6 2 | 6 2 | 5+6 2 | 5+6 2 |
| | 9 2 | 9 2 | 4+5 2 | 4+5 2 | 5+6 2 | 5+6 2 |
| | 6+7 2 | 6+7 2 | 6 2 | 6 2 | 5+6 2 | 5+6 2 |

Continued on next column

**TABLE LXXI– continued**

| TQC | ICSD | | NoE | SF | | |
|---|---|---|---|---|---|---|
| | 9 2 | 8 2 | 6 2 | 6 2 | 3+4 2 | 5+6 2 |
| | 8 2 | 9 2 | 4+5 2 | 4+5 2 | 5+6 2 | 3+4 2 |
| | 4+5 2 | 6+7 2 | 6 2 | 6 2 | 3+4 2 | 3+4 2 |
| | 6+7 2 | 4+5 2 | 4+5 2 | 4+5 2 | 5+6 2 | 5+6 2 |
| | 8 2 | 9 2 | 6 2 | 6 2 | 3+4 2 | 5+6 2 |
| | 9 2 | 8 2 | 6 2 | 6 2 | 5+6 2 | 3+4 2 |
| | 8 2 | 8 2 | 6 2 | 6 2 | 3+4 2 | 3+4 2 |
| | 9 2 | 9 2 | 4+5 2 | 4+5 2 | 5+6 2 | 5+6 2 |
| | 6+7 2 | 4+5 2 | 6 2 | 6 2 | 3+4 2 | 5+6 2 |
| | 9 2 | 8 2 | 6 2 | 6 2 | 5+6 2 | 3+4 2 |
| | 8 2 | 6+7 2 | 6 2 | 6 2 | 5+6 2 | 3+4 2 |
| | 4+5 2 | 9 2 | 4+5 2 | 4+5 2 | 3+4 2 | 5+6 2 |
| | 9 2 | 8 2 | 6 2 | 4+5 2 | 5+6 2 | 3+4 2 |
| | 8 2 | 8 2 | 4+5 2 | 6 2 | 3+4 2 | 5+6 2 |
| | 6+7 2 | 4+5 2 | 6 2 | 6 2 | 3+4 2 | 5+6 2 |
| | 9 2 | 9 2 | 6 2 | 6 2 | 3+4 2 | 3+4 2 |
| | 4+5 2 | 6+7 2 | 4+5 2 | 4+5 2 | 5+6 2 | 5+6 2 |
| | 8 2 | 9 2 | 6 2 | 6 2 | 5+6 2 | 3+4 2 |
| | 9 2 | 9 2 | 6 2 | 6 2 | 5+6 2 | 5+6 2 |
| | 9 2 | 6+7 2 | 4+5 2 | 4+5 2 | 5+6 2 | 5+6 2 |
| | 6+7 2 | 9 2 | 6 2 | 6 2 | 5+6 2 | 5+6 2 |
| | 9 2 | 9 2 | 6 2 | 6 2 | 5+6 2 | 3+4 2 |
| | 9 2 | 9 2 | 4+5 2 | 4+5 2 | 5+6 2 | 5+6 2 |
| | 6+7 2 | 6+7 2 | 6 2 | 6 2 | 5+6 2 | 5+6 2 |
| | 9 2 | 9 2 | 6 2 | 6 2 | 5+6 2 | 5+6 2 |
| 164.1.14 | 43293 | | 22 | Ni1Te2 | | |
| | -A | -GM | -H | -K | -L | -M |
| | 8 2 | 8 2 | 6 2 | 4+5 2 | 3+4 2 | 5+6 2 |
| | 9 2 | 9 2 | 4+5 2 | 6 2 | 5+6 2 | 3+4 2 |
| | 8 2 | 8 2 | 4+5 2 | 6 2 | 3+4 2 | 3+4 2 |
| | 4+5 2 | 8 2 | 6 2 | 4+5 2 | 5+6 2 | 5+6 2 |
| | 8 2 | 4+5 2 | 6 2 | 6 2 | 3+4 2 | 5+6 2 |
| | 8 2 | 8 2 | 6 2 | 4+5 2 | 3+4 2 | 3+4 2 |
| | 4+5 2 | 4+5 2 | 4+5 2 | 6 2 | 3+4 2 | 3+4 2 |
| | 9 2 | 8 2 | 6 2 | 6 2 | 5+6 2 | 5+6 2 |
| | 8 2 | 9 2 | 6 2 | 6 2 | 3+4 2 | 3+4 2 |
| | 9 2 | 6+7 2 | 4+5 2 | 4+5 2 | 3+4 2 | 3+4 2 |
| | 6+7 2 | 8 2 | 6 2 | 6 2 | 5+6 2 | 3+4 2 |
| 164.1.15 | 42559 | | 22 | Ni1Te2 | | |
| | -A | -GM | -H | -K | -L | -M |
| | 8 2 | 8 2 | 6 2 | 4+5 2 | 5+6 2 | 5+6 2 |
| | 9 2 | 9 2 | 4+5 2 | 6 2 | 3+4 2 | 3+4 2 |
| | 8 2 | 8 2 | 4+5 2 | 6 2 | 3+4 2 | 3+4 2 |
| | 4+5 2 | 8 2 | 6 2 | 4+5 2 | 5+6 2 | 5+6 2 |
| | 8 2 | 4+5 2 | 6 2 | 6 2 | 3+4 2 | 5+6 2 |
| | 8 2 | 8 2 | 6 2 | 4+5 2 | 3+4 2 | 3+4 2 |
| | 4+5 2 | 4+5 2 | 4+5 2 | 6 2 | 5+6 2 | 3+4 2 |
| | 9 2 | 8 2 | 6 2 | 6 2 | 3+4 2 | 5+6 2 |
| | 8 2 | 9 2 | 6 2 | 6 2 | 3+4 2 | 3+4 2 |
| | 9 2 | 6+7 2 | 4+5 2 | 4+5 2 | 3+4 2 | 3+4 2 |
| | 6+7 2 | 8 2 | 6 2 | 6 2 | 5+6 2 | 3+4 2 |
| 164.1.16 | 94478 | | 46 | Nd2S2Te1 | | |
| | -A | -GM | -H | -K | -L | -M |
| | 9 2 | 8 2 | 4+5 2 | 4+5 2 | 3+4 2 | 5+6 2 |
| | 8 2 | 9 2 | 6 2 | 6 2 | 5+6 2 | 3+4 2 |
| | 9 2 | 8 2 | 4+5 2 | 4+5 2 | 3+4 2 | 5+6 2 |
| | 8 2 | 9 2 | 6 2 | 6 2 | 5+6 2 | 3+4 2 |
| | 9 2 | 8 2 | 6 2 | 6 2 | 3+4 2 | 5+6 2 |
| | 8 2 | 4+5 2 | 4+5 2 | 4+5 2 | 5+6 2 | 3+4 2 |
| | 6+7 2 | 6+7 2 | 6 2 | 6 2 | 3+4 2 | 5+6 2 |
| | 4+5 2 | 9 2 | 6 2 | 6 2 | 5+6 2 | 3+4 2 |

Continued on next column

**TABLE LXXI– continued**

| TQC | ICSD | | NoE | SF | | |
|---|---|---|---|---|---|---|
| | 9 2 | 8 2 | 4+5 2 | 4+5 2 | 3+4 2 | 5+6 2 |
| | 8 2 | 9 2 | 6 2 | 6 2 | 5+6 2 | 3+4 2 |
| | 8 2 | 8 2 | 6 2 | 6 2 | 3+4 2 | 3+4 2 |
| | 9 2 | 8 2 | 6 2 | 6 2 | 5+6 2 | 3+4 2 |
| | 9 2 | 8 2 | 4+5 2 | 4+5 2 | 3+4 2 | 5+6 2 |
| | 6+7 2 | 4+5 2 | 6 2 | 6 2 | 3+4 2 | 5+6 2 |
| | 8 2 | 9 2 | 6 2 | 6 2 | 5+6 2 | 5+6 2 |
| | 4+5 2 | 9 2 | 4+5 2 | 4+5 2 | 3+4 2 | 5+6 2 |
| | 9 2 | 6+7 2 | 6 2 | 6 2 | 5+6 2 | 3+4 2 |
| | 8 2 | 9 2 | 6 2 | 6 2 | 5+6 2 | 3+4 2 |
| | 9 2 | 6+7 2 | 6 2 | 4+5 2 | 5+6 2 | 5+6 2 |
| | 6+7 2 | 9 2 | 4+5 2 | 6 2 | 5+6 2 | 5+6 2 |
| | 9 2 | 8 2 | 4+5 2 | 6 2 | 3+4 2 | 3+4 2 |
| | 4+5 2 | 4+5 2 | 6 2 | 4+5 2 | 5+6 2 | 5+6 2 |
| | 8 2 | 8 2 | 6 2 | 6 2 | 3+4 2 | 3+4 2 |
| 164.1.17 | 32514 | | 46 | Nd2O3 | | |
| | -A | -GM | -H | -K | -L | -M |
| | 8 2 | 8 2 | 4+5 2 | 4+5 2 | 5+6 2 | 5+6 2 |
| | 9 2 | 9 2 | 6 2 | 6 2 | 3+4 2 | 3+4 2 |
| | 8 2 | 8 2 | 4+5 2 | 4+5 2 | 5+6 2 | 5+6 2 |
| | 9 2 | 9 2 | 6 2 | 6 2 | 3+4 2 | 3+4 2 |
| | 8 2 | 8 2 | 6 2 | 6 2 | 5+6 2 | 5+6 2 |
| | 9 2 | 9 2 | 6 2 | 4+5 2 | 3+4 2 | 3+4 2 |
| | 9 2 | 8 2 | 4+5 2 | 6 2 | 5+6 2 | 3+4 2 |
| | 4+5 2 | 4+5 2 | 6 2 | 6 2 | 3+4 2 | 5+6 2 |
| | 6+7 2 | 6+7 2 | 6 2 | 6 2 | 5+6 2 | 3+4 2 |
| | 9 2 | 9 2 | 4+5 2 | 4+5 2 | 3+4 2 | 5+6 2 |
| | 8 2 | 8 2 | 6 2 | 6 2 | 5+6 2 | 3+4 2 |
| | 8 2 | 6+7 2 | 4+5 2 | 6 2 | 3+4 2 | 3+4 2 |
| | 4+5 2 | 8 2 | 6 2 | 4+5 2 | 5+6 2 | 5+6 2 |
| | 8 2 | 9 2 | 6 2 | 6 2 | 5+6 2 | 5+6 2 |
| | 9 2 | 4+5 2 | 4+5 2 | 4+5 2 | 3+4 2 | 5+6 2 |
| | 6+7 2 | 8 2 | 6 2 | 6 2 | 5+6 2 | 5+6 2 |
| | 9 2 | 9 2 | 6 2 | 6 2 | 3+4 2 | 3+4 2 |
| | 8 2 | 9 2 | 6 2 | 4+5 2 | 3+4 2 | 3+4 2 |
| | 8 2 | 6+7 2 | 4+5 2 | 6 2 | 3+4 2 | 5+6 2 |
| | 4+5 2 | 9 2 | 6 2 | 6 2 | 3+4 2 | 5+6 2 |
| | 8 2 | 8 2 | 6 2 | 6 2 | 5+6 2 | 3+4 2 |
| | 6+7 2 | 4+5 2 | 4+5 2 | 6 2 | 3+4 2 | 3+4 2 |
| | 9 2 | 8 2 | 6 2 | 4+5 2 | 5+6 2 | 5+6 2 |
| 164.1.18 | 25806 | | 46 | Nd2O2Se1 | | |
| | -A | -GM | -H | -K | -L | -M |
| | 9 2 | 8 2 | 4+5 2 | 4+5 2 | 3+4 2 | 5+6 2 |
| | 8 2 | 9 2 | 6 2 | 6 2 | 5+6 2 | 3+4 2 |
| | 9 2 | 8 2 | 4+5 2 | 4+5 2 | 3+4 2 | 5+6 2 |
| | 8 2 | 9 2 | 6 2 | 6 2 | 5+6 2 | 3+4 2 |
| | 9 2 | 8 2 | 6 2 | 6 2 | 3+4 2 | 5+6 2 |
| | 8 2 | 9 2 | 4+5 2 | 4+5 2 | 5+6 2 | 3+4 2 |
| | 6+7 2 | 4+5 2 | 6 2 | 6 2 | 3+4 2 | 5+6 2 |
| | 8 2 | 6+7 2 | 6 2 | 6 2 | 5+6 2 | 3+4 2 |
| | 4+5 2 | 9 2 | 4+5 2 | 4+5 2 | 5+6 2 | 3+4 2 |
| | 9 2 | 8 2 | 6 2 | 6 2 | 3+4 2 | 5+6 2 |
| | 8 2 | 8 2 | 6 2 | 6 2 | 3+4 2 | 3+4 2 |
| | 9 2 | 8 2 | 6 2 | 6 2 | 5+6 2 | 3+4 2 |
| | 6+7 2 | 4+5 2 | 4+5 2 | 4+5 2 | 3+4 2 | 5+6 2 |
| | 9 2 | 8 2 | 6 2 | 6 2 | 3+4 2 | 5+6 2 |
| | 8 2 | 9 2 | 4+5 2 | 4+5 2 | 5+6 2 | 5+6 2 |
| | 4+5 2 | 6+7 2 | 6 2 | 6 2 | 3+4 2 | 5+6 2 |
| | 8 2 | 9 2 | 6 2 | 6 2 | 5+6 2 | 3+4 2 |
| | 9 2 | 9 2 | 6 2 | 4+5 2 | 5+6 2 | 3+4 2 |
| | 9 2 | 9 2 | 6 2 | 6 2 | 5+6 2 | 5+6 2 |

Continued on next column

**TABLE LXXI– continued**

| TQC | ICSD | NoE | SF | | | |
|---|---|---|---|---|---|---|
| | 6+7 2 | 6+7 2 | 4+5 2 | 6 2 | 5+6 2 | 5+6 2 |
| | 9 2 | 8 2 | 6 2 | 6 2 | 3+4 2 | 3+4 2 |
| | 8 2 | 8 2 | 4+5 2 | 6 2 | 5+6 2 | 3+4 2 |
| | 4+5 2 | 4+5 2 | 6 2 | 4+5 2 | 3+4 2 | 5+6 2 |
| 164.1.19 | 32515 | 46 | Nd2O2S1 | | | |
| | -A | -GM | -H | -K | -L | -M |
| | 9 2 | 8 2 | 4+5 2 | 4+5 2 | 3+4 2 | 5+6 2 |
| | 8 2 | 9 2 | 6 2 | 6 2 | 5+6 2 | 3+4 2 |
| | 9 2 | 8 2 | 4+5 2 | 4+5 2 | 3+4 2 | 5+6 2 |
| | 8 2 | 9 2 | 6 2 | 6 2 | 5+6 2 | 3+4 2 |
| | 9 2 | 8 2 | 6 2 | 6 2 | 3+4 2 | 5+6 2 |
| | 8 2 | 9 2 | 4+5 2 | 4+5 2 | 5+6 2 | 3+4 2 |
| | 8 2 | 4+5 2 | 6 2 | 6 2 | 3+4 2 | 5+6 2 |
| | 6+7 2 | 6+7 2 | 6 2 | 6 2 | 5+6 2 | 3+4 2 |
| | 4+5 2 | 9 2 | 4+5 2 | 4+5 2 | 5+6 2 | 3+4 2 |
| | 9 2 | 8 2 | 6 2 | 6 2 | 3+4 2 | 5+6 2 |
| | 8 2 | 8 2 | 6 2 | 6 2 | 3+4 2 | 3+4 2 |
| | 9 2 | 8 2 | 4+5 2 | 6 2 | 5+6 2 | 3+4 2 |
| | 6+7 2 | 4+5 2 | 6 2 | 4+5 2 | 3+4 2 | 5+6 2 |
| | 9 2 | 8 2 | 6 2 | 6 2 | 3+4 2 | 5+6 2 |
| | 8 2 | 9 2 | 4+5 2 | 4+5 2 | 5+6 2 | 5+6 2 |
| | 4+5 2 | 6+7 2 | 6 2 | 6 2 | 3+4 2 | 5+6 2 |
| | 8 2 | 9 2 | 6 2 | 6 2 | 5+6 2 | 3+4 2 |
| | 9 2 | 9 2 | 6 2 | 4+5 2 | 5+6 2 | 3+4 2 |
| | 9 2 | 9 2 | 4+5 2 | 6 2 | 5+6 2 | 5+6 2 |
| | 6+7 2 | 6+7 2 | 6 2 | 6 2 | 5+6 2 | 5+6 2 |
| | 9 2 | 8 2 | 6 2 | 6 2 | 3+4 2 | 3+4 2 |
| | 8 2 | 8 2 | 4+5 2 | 4+5 2 | 5+6 2 | 3+4 2 |
| | 4+5 2 | 4+5 2 | 6 2 | 6 2 | 3+4 2 | 5+6 2 |
| 164.1.20 | 300245 | 46 | Nd2O2S1 | | | |
| | -A | -GM | -H | -K | -L | -M |
| | 9 2 | 8 2 | 4+5 2 | 4+5 2 | 3+4 2 | 5+6 2 |
| | 8 2 | 9 2 | 6 2 | 6 2 | 5+6 2 | 3+4 2 |
| | 9 2 | 8 2 | 4+5 2 | 4+5 2 | 3+4 2 | 5+6 2 |
| | 8 2 | 9 2 | 6 2 | 6 2 | 5+6 2 | 3+4 2 |
| | 9 2 | 8 2 | 6 2 | 6 2 | 3+4 2 | 5+6 2 |
| | 8 2 | 9 2 | 4+5 2 | 4+5 2 | 5+6 2 | 3+4 2 |
| | 8 2 | 4+5 2 | 6 2 | 6 2 | 3+4 2 | 5+6 2 |
| | 6+7 2 | 6+7 2 | 6 2 | 6 2 | 5+6 2 | 3+4 2 |
| | 4+5 2 | 9 2 | 4+5 2 | 4+5 2 | 5+6 2 | 3+4 2 |
| | 9 2 | 8 2 | 6 2 | 6 2 | 3+4 2 | 5+6 2 |
| | 8 2 | 8 2 | 6 2 | 6 2 | 3+4 2 | 3+4 2 |
| | 9 2 | 8 2 | 4+5 2 | 6 2 | 5+6 2 | 3+4 2 |
| | 6+7 2 | 4+5 2 | 6 2 | 4+5 2 | 3+4 2 | 5+6 2 |
| | 9 2 | 8 2 | 6 2 | 6 2 | 3+4 2 | 5+6 2 |
| | 8 2 | 9 2 | 4+5 2 | 4+5 2 | 5+6 2 | 5+6 2 |
| | 4+5 2 | 6+7 2 | 6 2 | 6 2 | 3+4 2 | 5+6 2 |
| | 9 2 | 9 2 | 6 2 | 6 2 | 5+6 2 | 3+4 2 |
| | 8 2 | 9 2 | 6 2 | 4+5 2 | 5+6 2 | 3+4 2 |
| | 9 2 | 9 2 | 4+5 2 | 6 2 | 5+6 2 | 5+6 2 |
| | 6+7 2 | 6+7 2 | 6 2 | 6 2 | 5+6 2 | 5+6 2 |
| | 9 2 | 8 2 | 6 2 | 6 2 | 3+4 2 | 3+4 2 |
| | 8 2 | 8 2 | 4+5 2 | 4+5 2 | 5+6 2 | 3+4 2 |
| | 4+5 2 | 4+5 2 | 6 2 | 6 2 | 3+4 2 | 5+6 2 |
| 164.1.21 | 380376 | 48 | Mn2Sb2Yb1 | | | |
| | -A | -GM | -H | -K | -L | -M |
| | 8 2 | 8 2 | 6 2 | 6 2 | 3+4 2 | 3+4 2 |
| | 9 2 | 9 2 | 6 2 | 6 2 | 5+6 2 | 5+6 2 |
| | 9 2 | 9 2 | 4+5 2 | 4+5 2 | 5+6 2 | 5+6 2 |
| | 6+7 2 | 6+7 2 | 6 2 | 6 2 | 5+6 2 | 5+6 2 |
| | 8 2 | 8 2 | 6 2 | 6 2 | 3+4 2 | 5+6 2 |

Continued on next column

**TABLE LXXI– continued**

| TQC | ICSD | NoE | SF | | | |
|---|---|---|---|---|---|---|
| | 9 2 | 9 2 | 4+5 2 | 4+5 2 | 5+6 2 | 3+4 2 |
| | 8 2 | 8 2 | 4+5 2 | 6 2 | 3+4 2 | 3+4 2 |
| | 9 2 | 8 2 | 6 2 | 4+5 2 | 5+6 2 | 5+6 2 |
| | 8 2 | 9 2 | 6 2 | 6 2 | 5+6 2 | 5+6 2 |
| | 4+5 2 | 6+7 2 | 6 2 | 6 2 | 3+4 2 | 3+4 2 |
| | 9 2 | 4+5 2 | 4+5 2 | 6 2 | 5+6 2 | 3+4 2 |
| | 6+7 2 | 9 2 | 6 2 | 4+5 2 | 3+4 2 | 5+6 2 |
| | 9 2 | 9 2 | 6 2 | 6 2 | 5+6 2 | 5+6 2 |
| | 9 2 | 6+7 2 | 4+5 2 | 4+5 2 | 5+6 2 | 5+6 2 |
| | 6+7 2 | 9 2 | 6 2 | 6 2 | 5+6 2 | 5+6 2 |
| | 9 2 | 8 2 | 6 2 | 6 2 | 3+4 2 | 3+4 2 |
| | 6+7 2 | 4+5 2 | 6 2 | 6 2 | 5+6 2 | 5+6 2 |
| | 9 2 | 9 2 | 4+5 2 | 6 2 | 3+4 2 | 3+4 2 |
| | 9 2 | 9 2 | 6 2 | 4+5 2 | 5+6 2 | 5+6 2 |
| | 6+7 2 | 6+7 2 | 6 2 | 6 2 | 5+6 2 | 5+6 2 |
| | 9 2 | 9 2 | 6 2 | 6 2 | 5+6 2 | 5+6 2 |
| | 4+5 2 | 8 2 | 4+5 2 | 4+5 2 | 5+6 2 | 5+6 2 |
| | 8 2 | 6+7 2 | 6 2 | 6 2 | 5+6 2 | 3+4 2 |
| | 9 2 | 9 2 | 6 2 | 6 2 | 3+4 2 | 5+6 2 |
| 164.1.22 | 41790 | 34 | Mn2Sb2Sr1 | | | |
| | -A | -GM | -H | -K | -L | -M |
| | 8 2 | 8 2 | 6 2 | 6 2 | 3+4 2 | 3+4 2 |
| | 9 2 | 9 2 | 6 2 | 6 2 | 5+6 2 | 5+6 2 |
| | 9 2 | 9 2 | 4+5 2 | 4+5 2 | 5+6 2 | 5+6 2 |
| | 6+7 2 | 6+7 2 | 6 2 | 6 2 | 5+6 2 | 5+6 2 |
| | 9 2 | 8 2 | 6 2 | 6 2 | 3+4 2 | 5+6 2 |
| | 8 2 | 9 2 | 4+5 2 | 4+5 2 | 5+6 2 | 3+4 2 |
| | 8 2 | 8 2 | 4+5 2 | 4+5 2 | 3+4 2 | 3+4 2 |
| | 9 2 | 9 2 | 6 2 | 6 2 | 5+6 2 | 5+6 2 |
| | 8 2 | 8 2 | 6 2 | 6 2 | 5+6 2 | 3+4 2 |
| | 4+5 2 | 4+5 2 | 6 2 | 6 2 | 3+4 2 | 5+6 2 |
| | 9 2 | 6+7 2 | 4+5 2 | 4+5 2 | 5+6 2 | 3+4 2 |
| | 6+7 2 | 9 2 | 6 2 | 6 2 | 3+4 2 | 5+6 2 |
| | 9 2 | 8 2 | 6 2 | 6 2 | 3+4 2 | 5+6 2 |
| | 6+7 2 | 4+5 2 | 6 2 | 6 2 | 5+6 2 | 3+4 2 |
| | 9 2 | 8 2 | 4+5 2 | 4+5 2 | 5+6 2 | 3+4 2 |
| | 4+5 2 | 6+7 2 | 6 2 | 6 2 | 3+4 2 | 5+6 2 |
| | 8 2 | 9 2 | 6 2 | 6 2 | 3+4 2 | 3+4 2 |
| 164.1.23 | 416094 | 34 | Mn2Sb2Sr1 | | | |
| | -A | -GM | -H | -K | -L | -M |
| | 8 2 | 8 2 | 6 2 | 6 2 | 3+4 2 | 3+4 2 |
| | 9 2 | 9 2 | 6 2 | 6 2 | 5+6 2 | 5+6 2 |
| | 9 2 | 9 2 | 4+5 2 | 4+5 2 | 5+6 2 | 5+6 2 |
| | 6+7 2 | 6+7 2 | 6 2 | 6 2 | 5+6 2 | 5+6 2 |
| | 9 2 | 8 2 | 6 2 | 6 2 | 3+4 2 | 5+6 2 |
| | 8 2 | 9 2 | 4+5 2 | 4+5 2 | 5+6 2 | 3+4 2 |
| | 8 2 | 8 2 | 4+5 2 | 4+5 2 | 3+4 2 | 3+4 2 |
| | 9 2 | 9 2 | 6 2 | 6 2 | 5+6 2 | 5+6 2 |
| | 8 2 | 8 2 | 6 2 | 6 2 | 5+6 2 | 3+4 2 |
| | 4+5 2 | 4+5 2 | 6 2 | 6 2 | 3+4 2 | 5+6 2 |
| | 9 2 | 6+7 2 | 4+5 2 | 4+5 2 | 5+6 2 | 3+4 2 |
| | 6+7 2 | 9 2 | 6 2 | 6 2 | 3+4 2 | 5+6 2 |
| | 9 2 | 8 2 | 6 2 | 6 2 | 3+4 2 | 5+6 2 |
| | 6+7 2 | 4+5 2 | 6 2 | 6 2 | 5+6 2 | 3+4 2 |
| | 9 2 | 8 2 | 4+5 2 | 4+5 2 | 5+6 2 | 3+4 2 |
| | 4+5 2 | 6+7 2 | 6 2 | 6 2 | 3+4 2 | 5+6 2 |
| | 8 2 | 9 2 | 6 2 | 6 2 | 3+4 2 | 3+4 2 |
| 164.1.24 | 77670 | 34 | Mn2P2Sr1 | | | |
| | -A | -GM | -H | -K | -L | -M |
| | 8 2 | 8 2 | 6 2 | 6 2 | 3+4 2 | 3+4 2 |
| | 9 2 | 9 2 | 6 2 | 6 2 | 5+6 2 | 5+6 2 |

Continued on next column

**TABLE LXXI– continued**

| TQC | ICSD | NoE | SF | | | |
|---|---|---|---|---|---|---|
| | 9 2 | 9 2 | 4+5 2 | 4+5 2 | 5+6 2 | 5+6 2 |
| | 6+7 2 | 6+7 2 | 6 2 | 6 2 | 5+6 2 | 5+6 2 |
| | 8 2 | 8 2 | 6 2 | 6 2 | 3+4 2 | 3+4 2 |
| | 9 2 | 9 2 | 4+5 2 | 4+5 2 | 5+6 2 | 5+6 2 |
| | 8 2 | 8 2 | 6 2 | 4+5 2 | 3+4 2 | 3+4 2 |
| | 9 2 | 9 2 | 4+5 2 | 6 2 | 5+6 2 | 5+6 2 |
| | 8 2 | 6+7 2 | 6 2 | 6 2 | 3+4 2 | 3+4 2 |
| | 4+5 2 | 9 2 | 6 2 | 6 2 | 5+6 2 | 3+4 2 |
| | 6+7 2 | 4+5 2 | 4+5 2 | 6 2 | 5+6 2 | 5+6 2 |
| | 9 2 | 8 2 | 6 2 | 4+5 2 | 3+4 2 | 5+6 2 |
| | 9 2 | 8 2 | 6 2 | 6 2 | 3+4 2 | 3+4 2 |
| | 6+7 2 | 4+5 2 | 6 2 | 6 2 | 5+6 2 | 5+6 2 |
| | 9 2 | 8 2 | 4+5 2 | 4+5 2 | 3+4 2 | 3+4 2 |
| | 4+5 2 | 6+7 2 | 6 2 | 6 2 | 5+6 2 | 5+6 2 |
| | 8 2 | 9 2 | 6 2 | 6 2 | 3+4 2 | 3+4 2 |
| 164.1.25 | 77669 | 34 | Mn2P2Sr1 | | | |
| | -A | -GM | -H | -K | -L | -M |
| | 8 2 | 8 2 | 6 2 | 6 2 | 3+4 2 | 3+4 2 |
| | 9 2 | 9 2 | 6 2 | 6 2 | 5+6 2 | 5+6 2 |
| | 9 2 | 9 2 | 4+5 2 | 4+5 2 | 5+6 2 | 5+6 2 |
| | 6+7 2 | 6+7 2 | 6 2 | 6 2 | 5+6 2 | 5+6 2 |
| | 9 2 | 8 2 | 6 2 | 6 2 | 3+4 2 | 3+4 2 |
| | 8 2 | 9 2 | 4+5 2 | 4+5 2 | 5+6 2 | 5+6 2 |
| | 8 2 | 8 2 | 6 2 | 4+5 2 | 3+4 2 | 3+4 2 |
| | 9 2 | 6+7 2 | 4+5 2 | 6 2 | 5+6 2 | 5+6 2 |
| | 4+5 2 | 9 2 | 6 2 | 6 2 | 5+6 2 | 3+4 2 |
| | 8 2 | 9 2 | 6 2 | 6 2 | 3+4 2 | 3+4 2 |
| | 6+7 2 | 4+5 2 | 4+5 2 | 6 2 | 5+6 2 | 5+6 2 |
| | 9 2 | 8 2 | 6 2 | 4+5 2 | 3+4 2 | 5+6 2 |
| | 9 2 | 8 2 | 6 2 | 6 2 | 3+4 2 | 3+4 2 |
| | 6+7 2 | 4+5 2 | 6 2 | 6 2 | 5+6 2 | 5+6 2 |
| | 9 2 | 8 2 | 4+5 2 | 4+5 2 | 5+6 2 | 3+4 2 |
| | 4+5 2 | 6+7 2 | 6 2 | 6 2 | 3+4 2 | 5+6 2 |
| | 8 2 | 9 2 | 6 2 | 6 2 | 3+4 2 | 3+4 2 |
| 164.1.26 | 49019 | 34 | Mn2P2Sr1 | | | |
| | -A | -GM | -H | -K | -L | -M |
| | 8 2 | 8 2 | 6 2 | 6 2 | 3+4 2 | 3+4 2 |
| | 9 2 | 9 2 | 6 2 | 6 2 | 5+6 2 | 5+6 2 |
| | 9 2 | 9 2 | 4+5 2 | 4+5 2 | 5+6 2 | 5+6 2 |
| | 6+7 2 | 6+7 2 | 6 2 | 6 2 | 5+6 2 | 5+6 2 |
| | 9 2 | 8 2 | 6 2 | 6 2 | 3+4 2 | 5+6 2 |
| | 8 2 | 9 2 | 4+5 2 | 4+5 2 | 5+6 2 | 3+4 2 |
| | 8 2 | 8 2 | 6 2 | 4+5 2 | 3+4 2 | 3+4 2 |
| | 9 2 | 6+7 2 | 4+5 2 | 6 2 | 5+6 2 | 5+6 2 |
| | 4+5 2 | 9 2 | 6 2 | 6 2 | 5+6 2 | 3+4 2 |
| | 8 2 | 4+5 2 | 6 2 | 6 2 | 3+4 2 | 3+4 2 |
| | 6+7 2 | 8 2 | 4+5 2 | 6 2 | 5+6 2 | 5+6 2 |
| | 9 2 | 9 2 | 6 2 | 4+5 2 | 3+4 2 | 5+6 2 |
| | 9 2 | 8 2 | 6 2 | 6 2 | 3+4 2 | 3+4 2 |
| | 6+7 2 | 4+5 2 | 6 2 | 6 2 | 5+6 2 | 5+6 2 |
| | 9 2 | 8 2 | 4+5 2 | 4+5 2 | 5+6 2 | 3+4 2 |
| | 4+5 2 | 9 2 | 6 2 | 6 2 | 3+4 2 | 5+6 2 |
| | 8 2 | 6+7 2 | 6 2 | 6 2 | 3+4 2 | 3+4 2 |
| 164.1.27 | 412655 | 38 | Mg2Sb2Yb1 | | | |
| | -A | -GM | -H | -K | -L | -M |
| | 8 2 | 8 2 | 6 2 | 6 2 | 3+4 2 | 3+4 2 |
| | 9 2 | 9 2 | 6 2 | 6 2 | 5+6 2 | 5+6 2 |
| | 9 2 | 9 2 | 4+5 2 | 4+5 2 | 5+6 2 | 5+6 2 |
| | 6+7 2 | 6+7 2 | 6 2 | 6 2 | 5+6 2 | 5+6 2 |
| | 9 2 | 8 2 | 6 2 | 6 2 | 3+4 2 | 5+6 2 |
| | 8 2 | 9 2 | 4+5 2 | 4+5 2 | 5+6 2 | 3+4 2 |

Continued on next column

**TABLE LXXI– continued**

| TQC | ICSD | NoE | SF | | | |
|---|---|---|---|---|---|---|
| | 9 2 | 8 2 | 4+5 2 | 4+5 2 | 3+4 2 | 3+4 2 |
| | 8 2 | 8 2 | 6 2 | 6 2 | 5+6 2 | 5+6 2 |
| | 8 2 | 4+5 2 | 6 2 | 6 2 | 3+4 2 | 5+6 2 |
| | 9 2 | 6+7 2 | 6 2 | 6 2 | 3+4 2 | 3+4 2 |
| | 4+5 2 | 9 2 | 4+5 2 | 4+5 2 | 5+6 2 | 5+6 2 |
| | 6+7 2 | 9 2 | 6 2 | 6 2 | 5+6 2 | 3+4 2 |
| | 9 2 | 9 2 | 6 2 | 6 2 | 5+6 2 | 5+6 2 |
| | 9 2 | 6+7 2 | 4+5 2 | 4+5 2 | 5+6 2 | 5+6 2 |
| | 6+7 2 | 9 2 | 6 2 | 6 2 | 5+6 2 | 5+6 2 |
| | 9 2 | 9 2 | 6 2 | 6 2 | 5+6 2 | 3+4 2 |
| | 9 2 | 6+7 2 | 4+5 2 | 6 2 | 5+6 2 | 5+6 2 |
| | 6+7 2 | 9 2 | 6 2 | 4+5 2 | 5+6 2 | 5+6 2 |
| | 9 2 | 9 2 | 6 2 | 6 2 | 5+6 2 | 5+6 2 |
| 164.1.28 | 191298 | 26 | Li3Pr1Sb2 | | | |
| | -A | -GM | -H | -K | -L | -M |
| | 8 2 | 8 2 | 6 2 | 6 2 | 3+4 2 | 3+4 2 |
| | 9 2 | 9 2 | 6 2 | 6 2 | 5+6 2 | 5+6 2 |
| | 9 2 | 9 2 | 4+5 2 | 4+5 2 | 5+6 2 | 5+6 2 |
| | 6+7 2 | 6+7 2 | 6 2 | 6 2 | 5+6 2 | 5+6 2 |
| | 8 2 | 8 2 | 6 2 | 6 2 | 3+4 2 | 3+4 2 |
| | 9 2 | 9 2 | 4+5 2 | 4+5 2 | 5+6 2 | 5+6 2 |
| | 9 2 | 8 2 | 4+5 2 | 6 2 | 3+4 2 | 3+4 2 |
| | 8 2 | 8 2 | 6 2 | 4+5 2 | 5+6 2 | 5+6 2 |
| | 8 2 | 4+5 2 | 6 2 | 6 2 | 3+4 2 | 5+6 2 |
| | 4+5 2 | 9 2 | 6 2 | 6 2 | 3+4 2 | 3+4 2 |
| | 9 2 | 6+7 2 | 4+5 2 | 4+5 2 | 5+6 2 | 5+6 2 |
| | 6+7 2 | 9 2 | 6 2 | 6 2 | 5+6 2 | 3+4 2 |
| | 9 2 | 9 2 | 6 2 | 6 2 | 3+4 2 | 5+6 2 |
| 164.1.29 | 26018 | 94 | K3Na1O8S2 | | | |
| | -A | -GM | -H | -K | -L | -M |
| | 8 2 | 8 2 | 4+5 2 | 4+5 2 | 3+4 2 | 3+4 2 |
| | 9 2 | 9 2 | 6 2 | 6 2 | 5+6 2 | 5+6 2 |
| | 8 2 | 8 2 | 6 2 | 6 2 | 3+4 2 | 3+4 2 |
| | 9 2 | 8 2 | 4+5 2 | 4+5 2 | 3+4 2 | 5+6 2 |
| | 8 2 | 9 2 | 6 2 | 6 2 | 5+6 2 | 3+4 2 |
| | 8 2 | 8 2 | 6 2 | 6 2 | 5+6 2 | 5+6 2 |
| | 9 2 | 9 2 | 4+5 2 | 4+5 2 | 3+4 2 | 3+4 2 |
| | 4+5 2 | 4+5 2 | 6 2 | 6 2 | 5+6 2 | 5+6 2 |
| | 6+7 2 | 6+7 2 | 6 2 | 6 2 | 3+4 2 | 3+4 2 |
| | 8 2 | 8 2 | 6 2 | 6 2 | 5+6 2 | 5+6 2 |
| | 9 2 | 9 2 | 4+5 2 | 4+5 2 | 3+4 2 | 3+4 2 |
| | 9 2 | 9 2 | 4+5 2 | 4+5 2 | 5+6 2 | 5+6 2 |
| | 8 2 | 8 2 | 6 2 | 6 2 | 3+4 2 | 3+4 2 |
| | 9 2 | 9 2 | 6 2 | 6 2 | 5+6 2 | 5+6 2 |
| | 9 2 | 9 2 | 4+5 2 | 4+5 2 | 3+4 2 | 3+4 2 |
| | 8 2 | 8 2 | 6 2 | 6 2 | 5+6 2 | 5+6 2 |
| | 6+7 2 | 6+7 2 | 6 2 | 6 2 | 3+4 2 | 5+6 2 |
| | 4+5 2 | 4+5 2 | 6 2 | 6 2 | 5+6 2 | 3+4 2 |
| | 9 2 | 9 2 | 6 2 | 6 2 | 5+6 2 | 5+6 2 |
| | 6+7 2 | 6+7 2 | 4+5 2 | 4+5 2 | 5+6 2 | 5+6 2 |
| | 9 2 | 8 2 | 6 2 | 6 2 | 3+4 2 | 5+6 2 |
| | 8 2 | 9 2 | 4+5 2 | 4+5 2 | 5+6 2 | 3+4 2 |
| | 8 2 | 8 2 | 6 2 | 6 2 | 5+6 2 | 3+4 2 |
| | 4+5 2 | 9 2 | 6 2 | 6 2 | 3+4 2 | 5+6 2 |
| | 8 2 | 6+7 2 | 4+5 2 | 4+5 2 | 3+4 2 | 5+6 2 |
| | 9 2 | 8 2 | 6 2 | 6 2 | 5+6 2 | 3+4 2 |
| | 6+7 2 | 4+5 2 | 4+5 2 | 4+5 2 | 5+6 2 | 5+6 2 |
| | 9 2 | 9 2 | 6 2 | 6 2 | 3+4 2 | 3+4 2 |
| | 8 2 | 8 2 | 6 2 | 6 2 | 3+4 2 | 3+4 2 |
| | 4+5 2 | 4+5 2 | 6 2 | 6 2 | 5+6 2 | 5+6 2 |
| | 9 2 | 9 2 | 4+5 2 | 4+5 2 | 3+4 2 | 5+6 2 |

Continued on next column

**TABLE LXXI– continued**

| TQC | ICSD | | NoE | SF | | |
|---|---|---|---|---|---|---|
| | 6+7 2 | 6+7 2 | 6 2 | 6 2 | 5+6 2 | 3+4 2 |
| | 9 2 | 8 2 | 4+5 2 | 6 2 | 3+4 2 | 3+4 2 |
| | 8 2 | 8 2 | 6 2 | 4+5 2 | 5+6 2 | 5+6 2 |
| | 8 2 | 4+5 2 | 6 2 | 6 2 | 5+6 2 | 5+6 2 |
| | 4+5 2 | 9 2 | 4+5 2 | 4+5 2 | 3+4 2 | 5+6 2 |
| | 9 2 | 9 2 | 6 2 | 6 2 | 5+6 2 | 3+4 2 |
| | 6+7 2 | 6+7 2 | 6 2 | 6 2 | 3+4 2 | 3+4 2 |
| | 8 2 | 8 2 | 4+5 2 | 4+5 2 | 5+6 2 | 3+4 2 |
| | 4+5 2 | 4+5 2 | 6 2 | 6 2 | 5+6 2 | 3+4 2 |
| | 9 2 | 9 2 | 6 2 | 6 2 | 3+4 2 | 5+6 2 |
| | 8 2 | 6+7 2 | 4+5 2 | 4+5 2 | 3+4 2 | 5+6 2 |
| | 9 2 | 8 2 | 6 2 | 6 2 | 3+4 2 | 5+6 2 |
| | 6+7 2 | 9 2 | 6 2 | 6 2 | 5+6 2 | 3+4 2 |
| | 8 2 | 8 2 | 6 2 | 6 2 | 5+6 2 | 3+4 2 |
| | 9 2 | 9 2 | 6 2 | 6 2 | 3+4 2 | 5+6 2 |
| | 9 2 | 8 2 | 4+5 2 | 4+5 2 | 3+4 2 | 3+4 2 |
| 164.1.30 | 26014 | | 94 | K3Na1O8S2 | | |
| | -A | -GM | -H | -K | -L | -M |
| | 8 2 | 8 2 | 4+5 2 | 4+5 2 | 3+4 2 | 3+4 2 |
| | 9 2 | 9 2 | 6 2 | 6 2 | 5+6 2 | 5+6 2 |
| | 8 2 | 8 2 | 6 2 | 6 2 | 3+4 2 | 3+4 2 |
| | 9 2 | 8 2 | 4+5 2 | 4+5 2 | 3+4 2 | 5+6 2 |
| | 8 2 | 9 2 | 6 2 | 6 2 | 5+6 2 | 3+4 2 |
| | 8 2 | 8 2 | 6 2 | 6 2 | 5+6 2 | 5+6 2 |
| | 9 2 | 9 2 | 4+5 2 | 4+5 2 | 3+4 2 | 3+4 2 |
| | 4+5 2 | 8 2 | 6 2 | 6 2 | 5+6 2 | 5+6 2 |
| | 6+7 2 | 4+5 2 | 6 2 | 6 2 | 3+4 2 | 3+4 2 |
| | 8 2 | 6+7 2 | 4+5 2 | 4+5 2 | 5+6 2 | 5+6 2 |
| | 9 2 | 9 2 | 6 2 | 6 2 | 3+4 2 | 3+4 2 |
| | 9 2 | 8 2 | 4+5 2 | 4+5 2 | 5+6 2 | 5+6 2 |
| | 8 2 | 9 2 | 6 2 | 6 2 | 3+4 2 | 3+4 2 |
| | 9 2 | 9 2 | 6 2 | 6 2 | 5+6 2 | 5+6 2 |
| | 9 2 | 9 2 | 4+5 2 | 4+5 2 | 3+4 2 | 3+4 2 |
| | 8 2 | 8 2 | 6 2 | 6 2 | 5+6 2 | 5+6 2 |
| | 6+7 2 | 6+7 2 | 6 2 | 6 2 | 3+4 2 | 5+6 2 |
| | 9 2 | 4+5 2 | 6 2 | 6 2 | 5+6 2 | 3+4 2 |
| | 4+5 2 | 9 2 | 4+5 2 | 4+5 2 | 5+6 2 | 5+6 2 |
| | 6+7 2 | 6+7 2 | 6 2 | 6 2 | 5+6 2 | 5+6 2 |
| | 9 2 | 8 2 | 6 2 | 4+5 2 | 3+4 2 | 5+6 2 |
| | 8 2 | 9 2 | 4+5 2 | 6 2 | 5+6 2 | 3+4 2 |
| | 8 2 | 8 2 | 6 2 | 6 2 | 5+6 2 | 3+4 2 |
| | 4+5 2 | 9 2 | 6 2 | 6 2 | 3+4 2 | 5+6 2 |
| | 8 2 | 6+7 2 | 4+5 2 | 4+5 2 | 3+4 2 | 5+6 2 |
| | 9 2 | 8 2 | 6 2 | 6 2 | 5+6 2 | 5+6 2 |
| | 6+7 2 | 4+5 2 | 4+5 2 | 4+5 2 | 5+6 2 | 3+4 2 |
| | 9 2 | 9 2 | 6 2 | 6 2 | 3+4 2 | 3+4 2 |
| | 9 2 | 8 2 | 6 2 | 6 2 | 5+6 2 | 3+4 2 |
| | 6+7 2 | 4+5 2 | 4+5 2 | 6 2 | 3+4 2 | 5+6 2 |
| | 4+5 2 | 9 2 | 6 2 | 4+5 2 | 3+4 2 | 5+6 2 |
| | 8 2 | 6+7 2 | 6 2 | 6 2 | 5+6 2 | 3+4 2 |
| | 9 2 | 8 2 | 4+5 2 | 4+5 2 | 3+4 2 | 5+6 2 |
| | 8 2 | 8 2 | 6 2 | 6 2 | 5+6 2 | 3+4 2 |
| | 8 2 | 4+5 2 | 6 2 | 6 2 | 5+6 2 | 5+6 2 |
| | 4+5 2 | 9 2 | 4+5 2 | 4+5 2 | 3+4 2 | 3+4 2 |
| | 9 2 | 9 2 | 6 2 | 6 2 | 5+6 2 | 5+6 2 |
| | 6+7 2 | 6+7 2 | 6 2 | 6 2 | 3+4 2 | 3+4 2 |
| | 8 2 | 9 2 | 4+5 2 | 4+5 2 | 5+6 2 | 3+4 2 |
| | 4+5 2 | 6+7 2 | 6 2 | 6 2 | 5+6 2 | 3+4 2 |
| | 9 2 | 8 2 | 6 2 | 6 2 | 3+4 2 | 5+6 2 |
| | 9 2 | 4+5 2 | 6 2 | 4+5 2 | 3+4 2 | 5+6 2 |
| | 6+7 2 | 8 2 | 4+5 2 | 6 2 | 3+4 2 | 5+6 2 |

Continued on next column

**TABLE LXXI– continued**

| TQC | ICSD | | NoE | SF | | |
|---|---|---|---|---|---|---|
| | 8 2 | 9 2 | 6 2 | 6 2 | 5+6 2 | 3+4 2 |
| | 8 2 | 8 2 | 6 2 | 6 2 | 5+6 2 | 3+4 2 |
| | 9 2 | 9 2 | 6 2 | 6 2 | 3+4 2 | 3+4 2 |
| | 9 2 | 8 2 | 4+5 2 | 4+5 2 | 3+4 2 | 5+6 2 |
| 164.1.31 | 660273 | | 84 | In2S4Zn1 | | |
| | -A | -GM | -H | -K | -L | -M |
| | 8 2 | 8 2 | 4+5 2 | 4+5 2 | 5+6 2 | 5+6 2 |
| | 9 2 | 9 2 | 6 2 | 6 2 | 3+4 2 | 3+4 2 |
| | 8 2 | 8 2 | 6 2 | 6 2 | 3+4 2 | 3+4 2 |
| | 9 2 | 9 2 | 4+5 2 | 4+5 2 | 5+6 2 | 5+6 2 |
| | 8 2 | 8 2 | 6 2 | 6 2 | 3+4 2 | 3+4 2 |
| | 9 2 | 9 2 | 4+5 2 | 4+5 2 | 5+6 2 | 5+6 2 |
| | 9 2 | 8 2 | 4+5 2 | 4+5 2 | 3+4 2 | 5+6 2 |
| | 8 2 | 9 2 | 6 2 | 6 2 | 5+6 2 | 3+4 2 |
| | 8 2 | 8 2 | 4+5 2 | 4+5 2 | 3+4 2 | 3+4 2 |
| | 9 2 | 9 2 | 6 2 | 6 2 | 5+6 2 | 5+6 2 |
| | 6+7 2 | 4+5 2 | 6 2 | 6 2 | 5+6 2 | 5+6 2 |
| | 4+5 2 | 6+7 2 | 6 2 | 6 2 | 3+4 2 | 3+4 2 |
| | 8 2 | 8 2 | 6 2 | 6 2 | 5+6 2 | 5+6 2 |
| | 9 2 | 9 2 | 6 2 | 6 2 | 3+4 2 | 3+4 2 |
| | 8 2 | 8 2 | 6 2 | 6 2 | 5+6 2 | 5+6 2 |
| | 9 2 | 9 2 | 4+5 2 | 4+5 2 | 3+4 2 | 3+4 2 |
| | 8 2 | 8 2 | 6 2 | 6 2 | 5+6 2 | 5+6 2 |
| | 9 2 | 9 2 | 6 2 | 6 2 | 3+4 2 | 3+4 2 |
| | 6+7 2 | 4+5 2 | 6 2 | 6 2 | 5+6 2 | 5+6 2 |
| | 4+5 2 | 6+7 2 | 4+5 2 | 4+5 2 | 5+6 2 | 5+6 2 |
| | 8 2 | 8 2 | 6 2 | 6 2 | 3+4 2 | 3+4 2 |
| | 9 2 | 9 2 | 6 2 | 6 2 | 3+4 2 | 3+4 2 |
| | 8 2 | 8 2 | 6 2 | 6 2 | 5+6 2 | 5+6 2 |
| | 4+5 2 | 4+5 2 | 4+5 2 | 4+5 2 | 3+4 2 | 3+4 2 |
| | 9 2 | 9 2 | 6 2 | 6 2 | 5+6 2 | 5+6 2 |
| | 6+7 2 | 6+7 2 | 4+5 2 | 4+5 2 | 3+4 2 | 3+4 2 |
| | 8 2 | 8 2 | 6 2 | 6 2 | 5+6 2 | 5+6 2 |
| | 9 2 | 9 2 | 4+5 2 | 4+5 2 | 3+4 2 | 3+4 2 |
| | 8 2 | 8 2 | 4+5 2 | 4+5 2 | 5+6 2 | 3+4 2 |
| | 9 2 | 9 2 | 6 2 | 6 2 | 3+4 2 | 5+6 2 |
| | 4+5 2 | 4+5 2 | 6 2 | 6 2 | 3+4 2 | 3+4 2 |
| | 6+7 2 | 6+7 2 | 6 2 | 6 2 | 5+6 2 | 5+6 2 |
| | 8 2 | 8 2 | 4+5 2 | 4+5 2 | 3+4 2 | 3+4 2 |
| | 9 2 | 8 2 | 6 2 | 6 2 | 5+6 2 | 5+6 2 |
| | 4+5 2 | 9 2 | 4+5 2 | 4+5 2 | 3+4 2 | 5+6 2 |
| | 6+7 2 | 4+5 2 | 6 2 | 6 2 | 5+6 2 | 3+4 2 |
| | 9 2 | 6+7 2 | 6 2 | 6 2 | 3+4 2 | 5+6 2 |
| | 8 2 | 9 2 | 4+5 2 | 4+5 2 | 5+6 2 | 3+4 2 |
| | 9 2 | 8 2 | 6 2 | 6 2 | 3+4 2 | 5+6 2 |
| | 8 2 | 9 2 | 6 2 | 6 2 | 5+6 2 | 3+4 2 |
| | 9 2 | 8 2 | 4+5 2 | 4+5 2 | 3+4 2 | 5+6 2 |
| | 6+7 2 | 4+5 2 | 6 2 | 6 2 | 5+6 2 | 3+4 2 |
| 164.1.32 | 42668 | | 84 | In2S4Zn1 | | |
| | -A | -GM | -H | -K | -L | -M |
| | 8 2 | 8 2 | 4+5 2 | 4+5 2 | 5+6 2 | 5+6 2 |
| | 9 2 | 9 2 | 6 2 | 6 2 | 3+4 2 | 3+4 2 |
| | 8 2 | 8 2 | 6 2 | 6 2 | 3+4 2 | 3+4 2 |
| | 9 2 | 9 2 | 4+5 2 | 4+5 2 | 5+6 2 | 5+6 2 |
| | 8 2 | 8 2 | 6 2 | 6 2 | 3+4 2 | 3+4 2 |
| | 9 2 | 9 2 | 4+5 2 | 4+5 2 | 5+6 2 | 5+6 2 |
| | 9 2 | 8 2 | 4+5 2 | 4+5 2 | 3+4 2 | 5+6 2 |
| | 8 2 | 9 2 | 6 2 | 6 2 | 5+6 2 | 3+4 2 |
| | 8 2 | 8 2 | 4+5 2 | 4+5 2 | 3+4 2 | 3+4 2 |
| | 9 2 | 9 2 | 6 2 | 6 2 | 5+6 2 | 5+6 2 |
| | 6+7 2 | 4+5 2 | 6 2 | 6 2 | 5+6 2 | 5+6 2 |

Continued on next column

**TABLE LXXI– continued**

| TQC | ICSD | | NoE | SF | | |
|---|---|---|---|---|---|---|
| | 4+5 2 | 6+7 2 | 6 2 | 6 2 | 3+4 2 | 3+4 2 |
| | 8 2 | 8 2 | 6 2 | 6 2 | 5+6 2 | 5+6 2 |
| | 9 2 | 9 2 | 6 2 | 6 2 | 3+4 2 | 3+4 2 |
| | 8 2 | 8 2 | 6 2 | 6 2 | 5+6 2 | 5+6 2 |
| | 9 2 | 9 2 | 4+5 2 | 4+5 2 | 3+4 2 | 3+4 2 |
| | 8 2 | 8 2 | 6 2 | 6 2 | 5+6 2 | 5+6 2 |
| | 9 2 | 9 2 | 6 2 | 6 2 | 3+4 2 | 3+4 2 |
| | 6+7 2 | 4+5 2 | 6 2 | 6 2 | 5+6 2 | 5+6 2 |
| | 4+5 2 | 6+7 2 | 4+5 2 | 4+5 2 | 5+6 2 | 5+6 2 |
| | 8 2 | 8 2 | 6 2 | 6 2 | 3+4 2 | 3+4 2 |
| | 9 2 | 9 2 | 6 2 | 6 2 | 3+4 2 | 3+4 2 |
| | 8 2 | 8 2 | 6 2 | 6 2 | 5+6 2 | 5+6 2 |
| | 4+5 2 | 4+5 2 | 4+5 2 | 4+5 2 | 3+4 2 | 3+4 2 |
| | 9 2 | 9 2 | 6 2 | 6 2 | 5+6 2 | 5+6 2 |
| | 6+7 2 | 6+7 2 | 4+5 2 | 4+5 2 | 3+4 2 | 3+4 2 |
| | 8 2 | 8 2 | 6 2 | 6 2 | 5+6 2 | 5+6 2 |
| | 9 2 | 9 2 | 4+5 2 | 4+5 2 | 3+4 2 | 3+4 2 |
| | 8 2 | 8 2 | 4+5 2 | 4+5 2 | 5+6 2 | 3+4 2 |
| | 9 2 | 9 2 | 6 2 | 6 2 | 3+4 2 | 5+6 2 |
| | 4+5 2 | 4+5 2 | 6 2 | 6 2 | 3+4 2 | 3+4 2 |
| | 6+7 2 | 6+7 2 | 6 2 | 6 2 | 5+6 2 | 5+6 2 |
| | 8 2 | 8 2 | 4+5 2 | 4+5 2 | 3+4 2 | 3+4 2 |
| | 9 2 | 8 2 | 6 2 | 6 2 | 5+6 2 | 5+6 2 |
| | 4+5 2 | 9 2 | 4+5 2 | 4+5 2 | 3+4 2 | 5+6 2 |
| | 6+7 2 | 4+5 2 | 6 2 | 6 2 | 5+6 2 | 3+4 2 |
| | 9 2 | 6+7 2 | 6 2 | 6 2 | 3+4 2 | 5+6 2 |
| | 8 2 | 9 2 | 4+5 2 | 4+5 2 | 5+6 2 | 3+4 2 |
| | 9 2 | 8 2 | 6 2 | 6 2 | 3+4 2 | 5+6 2 |
| | 8 2 | 9 2 | 6 2 | 6 2 | 5+6 2 | 3+4 2 |
| | 9 2 | 8 2 | 4+5 2 | 4+5 2 | 3+4 2 | 5+6 2 |
| | 6+7 2 | 4+5 2 | 6 2 | 6 2 | 5+6 2 | 3+4 2 |
| 164.1.33 | 25809 | | 60 | Ho2O2Se1 | | |
| | -A | -GM | -H | -K | -L | -M |
| | 9 2 | 8 2 | 4+5 2 | 4+5 2 | 3+4 2 | 5+6 2 |
| | 8 2 | 9 2 | 6 2 | 6 2 | 5+6 2 | 3+4 2 |
| | 9 2 | 8 2 | 4+5 2 | 4+5 2 | 3+4 2 | 5+6 2 |
| | 8 2 | 9 2 | 6 2 | 6 2 | 5+6 2 | 3+4 2 |
| | 9 2 | 8 2 | 6 2 | 6 2 | 3+4 2 | 5+6 2 |
| | 8 2 | 4+5 2 | 4+5 2 | 4+5 2 | 5+6 2 | 3+4 2 |
| | 6+7 2 | 9 2 | 6 2 | 6 2 | 3+4 2 | 5+6 2 |
| | 4+5 2 | 6+7 2 | 6 2 | 6 2 | 5+6 2 | 3+4 2 |
| | 8 2 | 9 2 | 4+5 2 | 4+5 2 | 5+6 2 | 3+4 2 |
| | 9 2 | 8 2 | 6 2 | 6 2 | 3+4 2 | 5+6 2 |
| | 8 2 | 8 2 | 6 2 | 6 2 | 3+4 2 | 3+4 2 |
| | 9 2 | 8 2 | 6 2 | 6 2 | 5+6 2 | 3+4 2 |
| | 9 2 | 9 2 | 4+5 2 | 4+5 2 | 3+4 2 | 5+6 2 |
| | 6+7 2 | 4+5 2 | 6 2 | 6 2 | 3+4 2 | 5+6 2 |
| | 4+5 2 | 8 2 | 4+5 2 | 6 2 | 5+6 2 | 5+6 2 |
| | 8 2 | 9 2 | 6 2 | 4+5 2 | 3+4 2 | 5+6 2 |
| | 9 2 | 6+7 2 | 6 2 | 6 2 | 5+6 2 | 3+4 2 |
| | 8 2 | 9 2 | 6 2 | 6 2 | 5+6 2 | 3+4 2 |
| | 6+7 2 | 6+7 2 | 6 2 | 4+5 2 | 5+6 2 | 5+6 2 |
| | 9 2 | 9 2 | 4+5 2 | 6 2 | 5+6 2 | 5+6 2 |
| | 8 2 | 8 2 | 6 2 | 6 2 | 3+4 2 | 3+4 2 |
| | 9 2 | 8 2 | 4+5 2 | 4+5 2 | 5+6 2 | 5+6 2 |
| | 4+5 2 | 4+5 2 | 6 2 | 6 2 | 3+4 2 | 3+4 2 |
| | 8 2 | 9 2 | 4+5 2 | 4+5 2 | 3+4 2 | 5+6 2 |
| | 9 2 | 9 2 | 6 2 | 6 2 | 5+6 2 | 3+4 2 |
| | 6+7 2 | 6+7 2 | 6 2 | 6 2 | 5+6 2 | 5+6 2 |
| | 9 2 | 8 2 | 6 2 | 6 2 | 3+4 2 | 3+4 2 |
| | 8 2 | 8 2 | 6 2 | 6 2 | 5+6 2 | 3+4 2 |

Continued on next column

**TABLE LXXI– continued**

| TQC | ICSD | | NoE | SF | | |
|---|---|---|---|---|---|---|
| | 4+5 2 | 4+5 2 | 4+5 2 | 4+5 2 | 3+4 2 | 5+6 2 |
| | 8 2 | 8 2 | 6 2 | 6 2 | 3+4 2 | 5+6 2 |
| 164.1.34 | 185808 | | 100 | Hg1Pt2Se3 | | |
| | -A | -GM | -H | -K | -L | -M |
| | 8 2 | 8 2 | 4+5 2 | 4+5 2 | 5+6 2 | 5+6 2 |
| | 9 2 | 4+5 2 | 6 2 | 6 2 | 3+4 2 | 3+4 2 |
| | 4+5 2 | 8 2 | 6 2 | 6 2 | 5+6 2 | 5+6 2 |
| | 8 2 | 6+7 2 | 6 2 | 6 2 | 3+4 2 | 3+4 2 |
| | 6+7 2 | 9 2 | 6 2 | 4+5 2 | 5+6 2 | 3+4 2 |
| | 9 2 | 9 2 | 4+5 2 | 6 2 | 3+4 2 | 5+6 2 |
| | 8 2 | 8 2 | 6 2 | 6 2 | 5+6 2 | 5+6 2 |
| | 4+5 2 | 4+5 2 | 4+5 2 | 4+5 2 | 3+4 2 | 3+4 2 |
| | 9 2 | 9 2 | 6 2 | 6 2 | 5+6 2 | 5+6 2 |
| | 6+7 2 | 6+7 2 | 6 2 | 6 2 | 3+4 2 | 3+4 2 |
| | 8 2 | 8 2 | 6 2 | 6 2 | 5+6 2 | 3+4 2 |
| | 4+5 2 | 8 2 | 4+5 2 | 6 2 | 3+4 2 | 5+6 2 |
| | 8 2 | 4+5 2 | 6 2 | 4+5 2 | 3+4 2 | 5+6 2 |
| | 8 2 | 8 2 | 4+5 2 | 6 2 | 5+6 2 | 3+4 2 |
| | 4+5 2 | 4+5 2 | 6 2 | 4+5 2 | 5+6 2 | 5+6 2 |
| | 9 2 | 9 2 | 6 2 | 6 2 | 3+4 2 | 3+4 2 |
| | 6+7 2 | 6+7 2 | 4+5 2 | 6 2 | 5+6 2 | 5+6 2 |
| | 8 2 | 8 2 | 6 2 | 4+5 2 | 3+4 2 | 5+6 2 |
| | 9 2 | 9 2 | 6 2 | 6 2 | 5+6 2 | 3+4 2 |
| | 8 2 | 8 2 | 6 2 | 6 2 | 3+4 2 | 3+4 2 |
| | 8 2 | 8 2 | 4+5 2 | 4+5 2 | 5+6 2 | 5+6 2 |
| | 9 2 | 8 2 | 6 2 | 6 2 | 3+4 2 | 5+6 2 |
| | 4+5 2 | 4+5 2 | 6 2 | 4+5 2 | 5+6 2 | 3+4 2 |
| | 8 2 | 9 2 | 4+5 2 | 6 2 | 5+6 2 | 5+6 2 |
| | 9 2 | 8 2 | 6 2 | 6 2 | 3+4 2 | 3+4 2 |
| | 6+7 2 | 8 2 | 4+5 2 | 4+5 2 | 3+4 2 | 3+4 2 |
| | 4+5 2 | 4+5 2 | 6 2 | 6 2 | 5+6 2 | 5+6 2 |
| | 9 2 | 8 2 | 6 2 | 6 2 | 5+6 2 | 5+6 2 |
| | 8 2 | 4+5 2 | 4+5 2 | 4+5 2 | 3+4 2 | 3+4 2 |
| | 8 2 | 9 2 | 6 2 | 6 2 | 3+4 2 | 5+6 2 |
| | 4+5 2 | 6+7 2 | 4+5 2 | 6 2 | 5+6 2 | 3+4 2 |
| | 8 2 | 8 2 | 6 2 | 4+5 2 | 3+4 2 | 3+4 2 |
| | 8 2 | 4+5 2 | 6 2 | 6 2 | 5+6 2 | 5+6 2 |
| | 4+5 2 | 9 2 | 4+5 2 | 4+5 2 | 3+4 2 | 3+4 2 |
| | 8 2 | 8 2 | 6 2 | 6 2 | 5+6 2 | 5+6 2 |
| | 8 2 | 6+7 2 | 6 2 | 6 2 | 3+4 2 | 3+4 2 |
| | 4+5 2 | 8 2 | 6 2 | 6 2 | 5+6 2 | 3+4 2 |
| | 8 2 | 9 2 | 4+5 2 | 6 2 | 3+4 2 | 5+6 2 |
| | 9 2 | 4+5 2 | 6 2 | 4+5 2 | 3+4 2 | 5+6 2 |
| | 8 2 | 8 2 | 6 2 | 6 2 | 5+6 2 | 3+4 2 |
| | 4+5 2 | 9 2 | 6 2 | 4+5 2 | 5+6 2 | 3+4 2 |
| | 8 2 | 4+5 2 | 4+5 2 | 6 2 | 3+4 2 | 5+6 2 |
| | 8 2 | 8 2 | 6 2 | 6 2 | 3+4 2 | 3+4 2 |
| | 6+7 2 | 9 2 | 4+5 2 | 6 2 | 5+6 2 | 5+6 2 |
| | 8 2 | 8 2 | 6 2 | 4+5 2 | 5+6 2 | 5+6 2 |
| | 9 2 | 8 2 | 4+5 2 | 6 2 | 3+4 2 | 3+4 2 |
| | 9 2 | 9 2 | 6 2 | 6 2 | 3+4 2 | 5+6 2 |
| | 4+5 2 | 8 2 | 6 2 | 4+5 2 | 5+6 2 | 3+4 2 |
| | 6+7 2 | 6+7 2 | 6 2 | 6 2 | 3+4 2 | 5+6 2 |
| | 9 2 | 4+5 2 | 6 2 | 6 2 | 5+6 2 | 3+4 2 |
| 164.1.35 | 56153 | | 24 | H2Cd1O2 | | |
| | -A | -GM | -H | -K | -L | -M |
| | 8 2 | 8 2 | 6 2 | 6 2 | 3+4 2 | 5+6 2 |
| | 9 2 | 9 2 | 4+5 2 | 4+5 2 | 5+6 2 | 3+4 2 |
| | 4+5 2 | 4+5 2 | 6 2 | 6 2 | 3+4 2 | 3+4 2 |
| | 8 2 | 8 2 | 4+5 2 | 4+5 2 | 3+4 2 | 3+4 2 |
| | 8 2 | 8 2 | 4+5 2 | 4+5 2 | 3+4 2 | 3+4 2 |

Continued on next column

**TABLE LXXI– continued**

| TQC | ICSD | | NoE | SF | | |
|---|---|---|---|---|---|---|
| | 8 2 | 8 2 | 6 2 | 6 2 | 3+4 2 | 3+4 2 |
| | 4+5 2 | 4+5 2 | 6 2 | 6 2 | 3+4 2 | 3+4 2 |
| | 8 2 | 8 2 | 6 2 | 6 2 | 5+6 2 | 3+4 2 |
| | 9 2 | 9 2 | 6 2 | 6 2 | 3+4 2 | 5+6 2 |
| | 9 2 | 6+7 2 | 4+5 2 | 4+5 2 | 5+6 2 | 5+6 2 |
| | 6+7 2 | 4+5 2 | 6 2 | 6 2 | 5+6 2 | 5+6 2 |
| | 4+5 2 | 8 2 | 4+5 2 | 4+5 2 | 3+4 2 | 3+4 2 |
| 164.1.36 | 41478 | | 10 | H1Si1 | | |
| | -A | -GM | -H | -K | -L | -M |
| | 8 2 | 8 2 | 4+5 2 | 4+5 2 | 5+6 2 | 5+6 2 |
| | 9 2 | 9 2 | 6 2 | 6 2 | 3+4 2 | 3+4 2 |
| | 8 2 | 8 2 | 6 2 | 6 2 | 3+4 2 | 3+4 2 |
| | 4+5 2 | 8 2 | 4+5 2 | 4+5 2 | 5+6 2 | 5+6 2 |
| | 9 2 | 4+5 2 | 6 2 | 6 2 | 5+6 2 | 5+6 2 |
| 164.1.37 | 42876 | | 48 | Ge2Sb2Te5 | | |
| | -A | -GM | -H | -K | -L | -M |
| | 8 2 | 8 2 | 6 2 | 6 2 | 3+4 2 | 3+4 2 |
| | 9 2 | 9 2 | 4+5 2 | 4+5 2 | 5+6 2 | 5+6 2 |
| | 8 2 | 8 2 | 6 2 | 6 2 | 3+4 2 | 3+4 2 |
| | 9 2 | 8 2 | 4+5 2 | 4+5 2 | 3+4 2 | 5+6 2 |
| | 8 2 | 9 2 | 6 2 | 6 2 | 5+6 2 | 3+4 2 |
| | 9 2 | 9 2 | 6 2 | 6 2 | 5+6 2 | 5+6 2 |
| | 8 2 | 8 2 | 4+5 2 | 4+5 2 | 3+4 2 | 3+4 2 |
| | 9 2 | 9 2 | 6 2 | 6 2 | 5+6 2 | 5+6 2 |
| | 8 2 | 8 2 | 6 2 | 6 2 | 3+4 2 | 3+4 2 |
| | 9 2 | 9 2 | 6 2 | 6 2 | 5+6 2 | 5+6 2 |
| | 9 2 | 8 2 | 4+5 2 | 4+5 2 | 3+4 2 | 3+4 2 |
| | 8 2 | 6+7 2 | 6 2 | 6 2 | 5+6 2 | 5+6 2 |
| | 6+7 2 | 9 2 | 4+5 2 | 4+5 2 | 5+6 2 | 5+6 2 |
| | 9 2 | 9 2 | 6 2 | 6 2 | 5+6 2 | 5+6 2 |
| | 4+5 2 | 8 2 | 6 2 | 6 2 | 3+4 2 | 3+4 2 |
| | 9 2 | 4+5 2 | 4+5 2 | 6 2 | 3+4 2 | 3+4 2 |
| | 6+7 2 | 6+7 2 | 6 2 | 4+5 2 | 5+6 2 | 3+4 2 |
| | 8 2 | 8 2 | 6 2 | 6 2 | 5+6 2 | 5+6 2 |
| | 9 2 | 8 2 | 4+5 2 | 6 2 | 3+4 2 | 5+6 2 |
| | 8 2 | 9 2 | 6 2 | 4+5 2 | 3+4 2 | 5+6 2 |
| | 6+7 2 | 9 2 | 6 2 | 6 2 | 3+4 2 | 3+4 2 |
| | 8 2 | 4+5 2 | 4+5 2 | 4+5 2 | 5+6 2 | 5+6 2 |
| | 4+5 2 | 8 2 | 6 2 | 6 2 | 5+6 2 | 3+4 2 |
| | 9 2 | 6+7 2 | 6 2 | 6 2 | 5+6 2 | 5+6 2 |
| 164.1.38 | 59801 | | 44 | Ge2I2La2 | | |
| | -A | -GM | -H | -K | -L | -M |
| | 8 2 | 8 2 | 6 2 | 6 2 | 3+4 2 | 3+4 2 |
| | 9 2 | 9 2 | 6 2 | 6 2 | 5+6 2 | 5+6 2 |
| | 8 2 | 8 2 | 6 2 | 6 2 | 3+4 2 | 3+4 2 |
| | 9 2 | 9 2 | 6 2 | 6 2 | 5+6 2 | 5+6 2 |
| | 8 2 | 8 2 | 4+5 2 | 4+5 2 | 5+6 2 | 5+6 2 |
| | 9 2 | 9 2 | 4+5 2 | 4+5 2 | 3+4 2 | 3+4 2 |
| | 6+7 2 | 6+7 2 | 6 2 | 6 2 | 3+4 2 | 3+4 2 |
| | 4+5 2 | 4+5 2 | 6 2 | 6 2 | 5+6 2 | 5+6 2 |
| | 9 2 | 8 2 | 6 2 | 6 2 | 3+4 2 | 5+6 2 |
| | 8 2 | 9 2 | 4+5 2 | 4+5 2 | 5+6 2 | 3+4 2 |
| | 8 2 | 8 2 | 6 2 | 6 2 | 5+6 2 | 5+6 2 |
| | 9 2 | 9 2 | 4+5 2 | 4+5 2 | 3+4 2 | 3+4 2 |
| | 9 2 | 8 2 | 4+5 2 | 6 2 | 5+6 2 | 3+4 2 |
| | 8 2 | 9 2 | 6 2 | 4+5 2 | 3+4 2 | 5+6 2 |
| | 9 2 | 8 2 | 6 2 | 6 2 | 3+4 2 | 5+6 2 |
| | 8 2 | 4+5 2 | 6 2 | 6 2 | 5+6 2 | 3+4 2 |
| | 4+5 2 | 9 2 | 4+5 2 | 4+5 2 | 5+6 2 | 5+6 2 |
| | 6+7 2 | 6+7 2 | 6 2 | 6 2 | 3+4 2 | 3+4 2 |
| | 8 2 | 8 2 | 6 2 | 6 2 | 3+4 2 | 3+4 2 |

Continued on next column

**TABLE LXXI– continued**

| TQC | ICSD | | NoE | SF | | |
|---|---|---|---|---|---|---|
| | 4+5 2 | 4+5 2 | 6 2 | 6 2 | 5+6 2 | 5+6 2 |
| | 8 2 | 8 2 | 4+5 2 | 4+5 2 | 5+6 2 | 5+6 2 |
| | 9 2 | 9 2 | 4+5 2 | 4+5 2 | 3+4 2 | 3+4 2 |
| 164.1.39 | 42875 | | 66 | Ge1Sb4Te7 | | |
| | -A | -GM | -H | -K | -L | -M |
| | 8 2 | 8 2 | 6 2 | 6 2 | 3+4 2 | 3+4 2 |
| | 9 2 | 8 2 | 4+5 2 | 4+5 2 | 3+4 2 | 5+6 2 |
| | 8 2 | 9 2 | 6 2 | 6 2 | 5+6 2 | 3+4 2 |
| | 9 2 | 9 2 | 4+5 2 | 4+5 2 | 5+6 2 | 5+6 2 |
| | 9 2 | 8 2 | 6 2 | 6 2 | 3+4 2 | 3+4 2 |
| | 8 2 | 9 2 | 6 2 | 6 2 | 5+6 2 | 5+6 2 |
| | 8 2 | 8 2 | 6 2 | 6 2 | 3+4 2 | 3+4 2 |
| | 9 2 | 8 2 | 4+5 2 | 4+5 2 | 5+6 2 | 5+6 2 |
| | 9 2 | 9 2 | 6 2 | 6 2 | 5+6 2 | 3+4 2 |
| | 8 2 | 9 2 | 4+5 2 | 4+5 2 | 3+4 2 | 5+6 2 |
| | 8 2 | 8 2 | 6 2 | 6 2 | 3+4 2 | 3+4 2 |
| | 9 2 | 8 2 | 6 2 | 6 2 | 5+6 2 | 3+4 2 |
| | 9 2 | 9 2 | 6 2 | 6 2 | 5+6 2 | 5+6 2 |
| | 8 2 | 8 2 | 6 2 | 6 2 | 3+4 2 | 5+6 2 |
| | 9 2 | 9 2 | 4+5 2 | 4+5 2 | 5+6 2 | 3+4 2 |
| | 9 2 | 9 2 | 6 2 | 6 2 | 3+4 2 | 5+6 2 |
| | 8 2 | 8 2 | 4+5 2 | 4+5 2 | 5+6 2 | 3+4 2 |
| | 8 2 | 6+7 2 | 6 2 | 6 2 | 3+4 2 | 5+6 2 |
| | 6+7 2 | 9 2 | 6 2 | 6 2 | 5+6 2 | 3+4 2 |
| | 4+5 2 | 8 2 | 4+5 2 | 4+5 2 | 3+4 2 | 5+6 2 |
| | 9 2 | 9 2 | 6 2 | 6 2 | 5+6 2 | 5+6 2 |
| | 6+7 2 | 6+7 2 | 6 2 | 6 2 | 3+4 2 | 5+6 2 |
| | 8 2 | 4+5 2 | 4+5 2 | 4+5 2 | 5+6 2 | 3+4 2 |
| | 9 2 | 8 2 | 6 2 | 6 2 | 3+4 2 | 3+4 2 |
| | 8 2 | 9 2 | 6 2 | 6 2 | 3+4 2 | 5+6 2 |
| | 4+5 2 | 4+5 2 | 6 2 | 6 2 | 5+6 2 | 3+4 2 |
| | 9 2 | 6+7 2 | 4+5 2 | 4+5 2 | 5+6 2 | 5+6 2 |
| | 6+7 2 | 9 2 | 6 2 | 6 2 | 3+4 2 | 3+4 2 |
| | 4+5 2 | 8 2 | 6 2 | 6 2 | 3+4 2 | 5+6 2 |
| | 6+7 2 | 4+5 2 | 4+5 2 | 4+5 2 | 5+6 2 | 5+6 2 |
| | 9 2 | 6+7 2 | 6 2 | 6 2 | 5+6 2 | 3+4 2 |
| | 8 2 | 8 2 | 6 2 | 4+5 2 | 5+6 2 | 5+6 2 |
| | 8 2 | 9 2 | 4+5 2 | 6 2 | 5+6 2 | 3+4 2 |
| 164.1.40 | 25808 | | 54 | Gd2O2Se1 | | |
| | -A | -GM | -H | -K | -L | -M |
| | 9 2 | 8 2 | 4+5 2 | 4+5 2 | 3+4 2 | 5+6 2 |
| | 8 2 | 9 2 | 6 2 | 6 2 | 5+6 2 | 3+4 2 |
| | 9 2 | 8 2 | 4+5 2 | 4+5 2 | 3+4 2 | 5+6 2 |
| | 8 2 | 9 2 | 6 2 | 6 2 | 5+6 2 | 3+4 2 |
| | 9 2 | 8 2 | 6 2 | 6 2 | 3+4 2 | 5+6 2 |
| | 8 2 | 4+5 2 | 4+5 2 | 4+5 2 | 5+6 2 | 3+4 2 |
| | 6+7 2 | 9 2 | 6 2 | 6 2 | 3+4 2 | 5+6 2 |
| | 4+5 2 | 6+7 2 | 6 2 | 6 2 | 5+6 2 | 3+4 2 |
| | 8 2 | 9 2 | 4+5 2 | 4+5 2 | 5+6 2 | 3+4 2 |
| | 9 2 | 8 2 | 6 2 | 6 2 | 3+4 2 | 5+6 2 |
| | 8 2 | 8 2 | 6 2 | 6 2 | 3+4 2 | 3+4 2 |
| | 9 2 | 8 2 | 6 2 | 6 2 | 5+6 2 | 3+4 2 |
| | 6+7 2 | 4+5 2 | 4+5 2 | 4+5 2 | 3+4 2 | 5+6 2 |
| | 9 2 | 8 2 | 6 2 | 6 2 | 3+4 2 | 5+6 2 |
| | 8 2 | 9 2 | 4+5 2 | 4+5 2 | 5+6 2 | 5+6 2 |
| | 4+5 2 | 9 2 | 6 2 | 6 2 | 3+4 2 | 5+6 2 |
| | 8 2 | 6+7 2 | 6 2 | 6 2 | 5+6 2 | 3+4 2 |
| | 9 2 | 9 2 | 6 2 | 4+5 2 | 5+6 2 | 3+4 2 |
| | 9 2 | 9 2 | 6 2 | 6 2 | 5+6 2 | 5+6 2 |
| | 6+7 2 | 6+7 2 | 4+5 2 | 6 2 | 5+6 2 | 5+6 2 |
| | 9 2 | 8 2 | 6 2 | 6 2 | 3+4 2 | 3+4 2 |

Continued on next column

**TABLE LXXI– continued**

| TQC | ICSD | | NoE | SF | | |
|---|---|---|---|---|---|---|
| | 8 2 | 8 2 | 4+5 2 | 4+5 2 | 5+6 2 | 5+6 2 |
| | 4+5 2 | 4+5 2 | 6 2 | 6 2 | 3+4 2 | 3+4 2 |
| | 8 2 | 9 2 | 6 2 | 4+5 2 | 3+4 2 | 5+6 2 |
| | 9 2 | 9 2 | 4+5 2 | 6 2 | 5+6 2 | 3+4 2 |
| | 6+7 2 | 6+7 2 | 6 2 | 6 2 | 5+6 2 | 5+6 2 |
| | 9 2 | 8 2 | 6 2 | 6 2 | 3+4 2 | 3+4 2 |
| 164.1.41 | 91111 | | 54 | Gd2O2S1 | | |
| | -A | -GM | -H | -K | -L | -M |
| | 9 2 | 8 2 | 4+5 2 | 4+5 2 | 3+4 2 | 5+6 2 |
| | 8 2 | 9 2 | 6 2 | 6 2 | 5+6 2 | 3+4 2 |
| | 9 2 | 8 2 | 4+5 2 | 4+5 2 | 3+4 2 | 5+6 2 |
| | 8 2 | 9 2 | 6 2 | 6 2 | 5+6 2 | 3+4 2 |
| | 9 2 | 8 2 | 6 2 | 6 2 | 3+4 2 | 5+6 2 |
| | 8 2 | 4+5 2 | 4+5 2 | 4+5 2 | 5+6 2 | 3+4 2 |
| | 6+7 2 | 9 2 | 6 2 | 6 2 | 3+4 2 | 5+6 2 |
| | 4+5 2 | 6+7 2 | 6 2 | 6 2 | 5+6 2 | 3+4 2 |
| | 8 2 | 9 2 | 4+5 2 | 4+5 2 | 5+6 2 | 3+4 2 |
| | 9 2 | 8 2 | 6 2 | 6 2 | 3+4 2 | 5+6 2 |
| | 8 2 | 8 2 | 6 2 | 6 2 | 3+4 2 | 3+4 2 |
| | 9 2 | 9 2 | 4+5 2 | 6 2 | 5+6 2 | 3+4 2 |
| | 9 2 | 8 2 | 6 2 | 4+5 2 | 3+4 2 | 5+6 2 |
| | 6+7 2 | 8 2 | 6 2 | 6 2 | 5+6 2 | 5+6 2 |
| | 8 2 | 4+5 2 | 6 2 | 4+5 2 | 3+4 2 | 5+6 2 |
| | 4+5 2 | 6+7 2 | 4+5 2 | 6 2 | 3+4 2 | 5+6 2 |
| | 9 2 | 9 2 | 6 2 | 6 2 | 5+6 2 | 3+4 2 |
| | 8 2 | 6+7 2 | 6 2 | 6 2 | 5+6 2 | 5+6 2 |
| | 9 2 | 9 2 | 4+5 2 | 4+5 2 | 5+6 2 | 3+4 2 |
| | 6+7 2 | 9 2 | 6 2 | 6 2 | 5+6 2 | 5+6 2 |
| | 9 2 | 8 2 | 6 2 | 6 2 | 3+4 2 | 3+4 2 |
| | 8 2 | 8 2 | 4+5 2 | 4+5 2 | 5+6 2 | 5+6 2 |
| | 4+5 2 | 4+5 2 | 6 2 | 6 2 | 3+4 2 | 3+4 2 |
| | 8 2 | 9 2 | 6 2 | 4+5 2 | 3+4 2 | 5+6 2 |
| | 9 2 | 9 2 | 4+5 2 | 6 2 | 5+6 2 | 3+4 2 |
| | 6+7 2 | 6+7 2 | 6 2 | 6 2 | 5+6 2 | 5+6 2 |
| | 9 2 | 8 2 | 6 2 | 6 2 | 3+4 2 | 3+4 2 |
| 164.1.42 | 159909 | | 92 | Fe2K3Na1O8 | | |
| | -A | -GM | -H | -K | -L | -M |
| | 9 2 | 8 2 | 6 2 | 6 2 | 5+6 2 | 3+4 2 |
| | 8 2 | 8 2 | 4+5 2 | 4+5 2 | 5+6 2 | 5+6 2 |
| | 9 2 | 9 2 | 6 2 | 6 2 | 3+4 2 | 3+4 2 |
| | 8 2 | 8 2 | 4+5 2 | 4+5 2 | 3+4 2 | 5+6 2 |
| | 9 2 | 9 2 | 6 2 | 6 2 | 5+6 2 | 3+4 2 |
| | 9 2 | 9 2 | 6 2 | 6 2 | 3+4 2 | 3+4 2 |
| | 6+7 2 | 6+7 2 | 6 2 | 6 2 | 5+6 2 | 5+6 2 |
| | 8 2 | 8 2 | 4+5 2 | 4+5 2 | 5+6 2 | 5+6 2 |
| | 4+5 2 | 4+5 2 | 6 2 | 6 2 | 3+4 2 | 3+4 2 |
| | 9 2 | 8 2 | 4+5 2 | 4+5 2 | 3+4 2 | 5+6 2 |
| | 8 2 | 9 2 | 6 2 | 6 2 | 5+6 2 | 3+4 2 |
| | 8 2 | 9 2 | 6 2 | 6 2 | 3+4 2 | 5+6 2 |
| | 8 2 | 8 2 | 6 2 | 6 2 | 5+6 2 | 5+6 2 |
| | 9 2 | 9 2 | 4+5 2 | 4+5 2 | 3+4 2 | 3+4 2 |
| | 4+5 2 | 6+7 2 | 4+5 2 | 4+5 2 | 3+4 2 | 5+6 2 |
| | 8 2 | 4+5 2 | 6 2 | 6 2 | 3+4 2 | 5+6 2 |
| | 4+5 2 | 8 2 | 6 2 | 6 2 | 5+6 2 | 5+6 2 |
| | 6+7 2 | 9 2 | 4+5 2 | 4+5 2 | 3+4 2 | 3+4 2 |
| | 8 2 | 6+7 2 | 6 2 | 6 2 | 5+6 2 | 3+4 2 |
| | 9 2 | 9 2 | 6 2 | 6 2 | 3+4 2 | 5+6 2 |
| | 8 2 | 8 2 | 6 2 | 6 2 | 3+4 2 | 3+4 2 |
| | 9 2 | 4+5 2 | 6 2 | 6 2 | 5+6 2 | 5+6 2 |
| | 6+7 2 | 8 2 | 6 2 | 6 2 | 5+6 2 | 5+6 2 |
| | 9 2 | 6+7 2 | 4+5 2 | 4+5 2 | 3+4 2 | 5+6 2 |

Continued on next column

| TQC | ICSD | | NoE | SF | | |
|---|---|---|---|---|---|---|
| | 4+5 2 | 9 2 | 6 2 | 4+5 2 | 5+6 2 | 3+4 2 |
| | 8 2 | 9 2 | 4+5 2 | 6 2 | 3+4 2 | 3+4 2 |
| | 8 2 | 4+5 2 | 6 2 | 6 2 | 5+6 2 | 3+4 2 |
| | 4+5 2 | 8 2 | 6 2 | 4+5 2 | 3+4 2 | 5+6 2 |
| | 6+7 2 | 9 2 | 4+5 2 | 6 2 | 3+4 2 | 5+6 2 |
| | 9 2 | 6+7 2 | 6 2 | 6 2 | 5+6 2 | 3+4 2 |
| | 8 2 | 8 2 | 4+5 2 | 4+5 2 | 3+4 2 | 5+6 2 |
| | 9 2 | 9 2 | 6 2 | 6 2 | 5+6 2 | 3+4 2 |
| | 9 2 | 9 2 | 6 2 | 6 2 | 5+6 2 | 3+4 2 |
| | 6+7 2 | 8 2 | 6 2 | 4+5 2 | 3+4 2 | 5+6 2 |
| | 8 2 | 8 2 | 4+5 2 | 6 2 | 3+4 2 | 5+6 2 |
| | 9 2 | 4+5 2 | 6 2 | 4+5 2 | 5+6 2 | 3+4 2 |
| | 8 2 | 9 2 | 4+5 2 | 6 2 | 3+4 2 | 5+6 2 |
| | 4+5 2 | 6+7 2 | 6 2 | 6 2 | 5+6 2 | 3+4 2 |
| | 9 2 | 8 2 | 6 2 | 4+5 2 | 3+4 2 | 3+4 2 |
| | 6+7 2 | 4+5 2 | 4+5 2 | 6 2 | 5+6 2 | 5+6 2 |
| | 9 2 | 9 2 | 6 2 | 6 2 | 3+4 2 | 3+4 2 |
| | 8 2 | 9 2 | 4+5 2 | 4+5 2 | 5+6 2 | 5+6 2 |
| | 8 2 | 6+7 2 | 6 2 | 6 2 | 5+6 2 | 5+6 2 |
| | 4+5 2 | 8 2 | 6 2 | 6 2 | 3+4 2 | 3+4 2 |
| | 4+5 2 | 4+5 2 | 6 2 | 6 2 | 5+6 2 | 3+4 2 |
| | 8 2 | 8 2 | 6 2 | 6 2 | 3+4 2 | 5+6 2 |
| 164.1.43 | 155025 | | 104 | Fe2In2Se5 | | |
| | -A | -GM | -H | -K | -L | -M |
| | 8 2 | 8 2 | 6 2 | 6 2 | 5+6 2 | 5+6 2 |
| | 9 2 | 8 2 | 4+5 2 | 4+5 2 | 3+4 2 | 3+4 2 |
| | 9 2 | 9 2 | 6 2 | 6 2 | 3+4 2 | 5+6 2 |
| | 8 2 | 9 2 | 4+5 2 | 4+5 2 | 5+6 2 | 3+4 2 |
| | 8 2 | 8 2 | 6 2 | 6 2 | 5+6 2 | 5+6 2 |
| | 9 2 | 9 2 | 6 2 | 6 2 | 3+4 2 | 3+4 2 |
| | 9 2 | 8 2 | 6 2 | 6 2 | 3+4 2 | 5+6 2 |
| | 8 2 | 9 2 | 6 2 | 6 2 | 5+6 2 | 3+4 2 |
| | 9 2 | 8 2 | 4+5 2 | 4+5 2 | 3+4 2 | 3+4 2 |
| | 8 2 | 9 2 | 6 2 | 6 2 | 5+6 2 | 5+6 2 |
| | 8 2 | 8 2 | 6 2 | 6 2 | 3+4 2 | 3+4 2 |
| | 9 2 | 9 2 | 6 2 | 6 2 | 5+6 2 | 5+6 2 |
| | 9 2 | 8 2 | 6 2 | 6 2 | 5+6 2 | 3+4 2 |
| | 8 2 | 9 2 | 6 2 | 6 2 | 3+4 2 | 5+6 2 |
| | 9 2 | 8 2 | 4+5 2 | 4+5 2 | 5+6 2 | 5+6 2 |
| | 8 2 | 9 2 | 6 2 | 6 2 | 3+4 2 | 3+4 2 |
| | 8 2 | 8 2 | 4+5 2 | 4+5 2 | 3+4 2 | 3+4 2 |
| | 9 2 | 9 2 | 6 2 | 6 2 | 5+6 2 | 5+6 2 |
| | 8 2 | 8 2 | 6 2 | 6 2 | 3+4 2 | 3+4 2 |
| | 4+5 2 | 4+5 2 | 4+5 2 | 4+5 2 | 5+6 2 | 5+6 2 |
| | 6+7 2 | 6+7 2 | 6 2 | 6 2 | 5+6 2 | 5+6 2 |
| | 9 2 | 9 2 | 6 2 | 6 2 | 3+4 2 | 5+6 2 |
| | 8 2 | 8 2 | 4+5 2 | 4+5 2 | 3+4 2 | 3+4 2 |
| | 9 2 | 8 2 | 6 2 | 6 2 | 5+6 2 | 3+4 2 |
| | 8 2 | 9 2 | 6 2 | 6 2 | 3+4 2 | 5+6 2 |
| | 9 2 | 9 2 | 6 2 | 6 2 | 5+6 2 | 3+4 2 |
| | 4+5 2 | 4+5 2 | 4+5 2 | 4+5 2 | 5+6 2 | 5+6 2 |
| | 6+7 2 | 4+5 2 | 6 2 | 6 2 | 3+4 2 | 5+6 2 |
| | 6+7 2 | 6+7 2 | 6 2 | 6 2 | 5+6 2 | 3+4 2 |
| | 4+5 2 | 6+7 2 | 4+5 2 | 4+5 2 | 3+4 2 | 5+6 2 |
| | 9 2 | 8 2 | 6 2 | 6 2 | 5+6 2 | 3+4 2 |
| | 9 2 | 8 2 | 4+5 2 | 4+5 2 | 3+4 2 | 3+4 2 |
| | 8 2 | 9 2 | 6 2 | 6 2 | 3+4 2 | 5+6 2 |
| | 6+7 2 | 4+5 2 | 4+5 2 | 4+5 2 | 5+6 2 | 3+4 2 |
| | 4+5 2 | 6+7 2 | 6 2 | 6 2 | 3+4 2 | 5+6 2 |
| | 8 2 | 9 2 | 6 2 | 6 2 | 5+6 2 | 3+4 2 |
| | 9 2 | 8 2 | 6 2 | 6 2 | 3+4 2 | 5+6 2 |

Continued on next column

**TABLE LXXI– continued**

| TQC | ICSD | | NoE | | SF | |
|---|---|---|---|---|---|---|
| | 8 2 | 9 2 | 4+5 2 | 4+5 2 | 5+6 2 | 3+4 2 |
| | 8 2 | 8 2 | 6 2 | 6 2 | 3+4 2 | 5+6 2 |
| | 6+7 2 | 4+5 2 | 4+5 2 | 4+5 2 | 5+6 2 | 3+4 2 |
| | 4+5 2 | 6+7 2 | 6 2 | 6 2 | 3+4 2 | 5+6 2 |
| | 9 2 | 8 2 | 4+5 2 | 4+5 2 | 5+6 2 | 3+4 2 |
| | 8 2 | 9 2 | 6 2 | 6 2 | 5+6 2 | 5+6 2 |
| | 9 2 | 8 2 | 6 2 | 6 2 | 3+4 2 | 3+4 2 |
| | 4+5 2 | 4+5 2 | 6 2 | 6 2 | 5+6 2 | 5+6 2 |
| | 6+7 2 | 6+7 2 | 4+5 2 | 4+5 2 | 3+4 2 | 3+4 2 |
| | 8 2 | 9 2 | 6 2 | 6 2 | 3+4 2 | 3+4 2 |
| | 9 2 | 8 2 | 4+5 2 | 4+5 2 | 5+6 2 | 5+6 2 |
| | 8 2 | 9 2 | 6 2 | 6 2 | 3+4 2 | 3+4 2 |
| | 6+7 2 | 4+5 2 | 6 2 | 6 2 | 5+6 2 | 5+6 2 |
| | 4+5 2 | 6+7 2 | 6 2 | 6 2 | 3+4 2 | 3+4 2 |
| | 9 2 | 8 2 | 4+5 2 | 4+5 2 | 5+6 2 | 5+6 2 |
| 164.1.44 | 25810 | | 62 | | Er2O2Se1 | |
| | -A | -GM | -H | -K | -L | -M |
| | 9 2 | 8 2 | 4+5 2 | 4+5 2 | 3+4 2 | 5+6 2 |
| | 8 2 | 9 2 | 6 2 | 6 2 | 5+6 2 | 3+4 2 |
| | 9 2 | 8 2 | 4+5 2 | 4+5 2 | 3+4 2 | 5+6 2 |
| | 8 2 | 9 2 | 6 2 | 6 2 | 5+6 2 | 3+4 2 |
| | 9 2 | 8 2 | 6 2 | 6 2 | 3+4 2 | 5+6 2 |
| | 8 2 | 4+5 2 | 4+5 2 | 4+5 2 | 5+6 2 | 3+4 2 |
| | 6+7 2 | 9 2 | 6 2 | 6 2 | 3+4 2 | 5+6 2 |
| | 4+5 2 | 6+7 2 | 6 2 | 6 2 | 5+6 2 | 3+4 2 |
| | 8 2 | 9 2 | 4+5 2 | 4+5 2 | 3+4 2 | 3+4 2 |
| | 9 2 | 8 2 | 6 2 | 6 2 | 5+6 2 | 5+6 2 |
| | 8 2 | 8 2 | 6 2 | 6 2 | 3+4 2 | 3+4 2 |
| | 9 2 | 9 2 | 6 2 | 6 2 | 5+6 2 | 3+4 2 |
| | 9 2 | 8 2 | 4+5 2 | 4+5 2 | 3+4 2 | 5+6 2 |
| | 6+7 2 | 4+5 2 | 6 2 | 6 2 | 5+6 2 | 5+6 2 |
| | 4+5 2 | 8 2 | 6 2 | 6 2 | 3+4 2 | 5+6 2 |
| | 8 2 | 9 2 | 4+5 2 | 4+5 2 | 3+4 2 | 5+6 2 |
| | 9 2 | 6+7 2 | 6 2 | 6 2 | 5+6 2 | 3+4 2 |
| | 8 2 | 9 2 | 6 2 | 6 2 | 5+6 2 | 3+4 2 |
| | 6+7 2 | 6+7 2 | 6 2 | 4+5 2 | 5+6 2 | 5+6 2 |
| | 9 2 | 8 2 | 4+5 2 | 6 2 | 5+6 2 | 5+6 2 |
| | 8 2 | 9 2 | 6 2 | 6 2 | 3+4 2 | 3+4 2 |
| | 9 2 | 8 2 | 4+5 2 | 4+5 2 | 5+6 2 | 5+6 2 |
| | 4+5 2 | 4+5 2 | 6 2 | 6 2 | 3+4 2 | 3+4 2 |
| | 8 2 | 9 2 | 6 2 | 4+5 2 | 3+4 2 | 5+6 2 |
| | 9 2 | 9 2 | 4+5 2 | 6 2 | 5+6 2 | 3+4 2 |
| | 6+7 2 | 6+7 2 | 6 2 | 6 2 | 5+6 2 | 5+6 2 |
| | 9 2 | 8 2 | 6 2 | 6 2 | 3+4 2 | 3+4 2 |
| | 8 2 | 8 2 | 6 2 | 6 2 | 5+6 2 | 3+4 2 |
| | 4+5 2 | 4+5 2 | 4+5 2 | 4+5 2 | 3+4 2 | 5+6 2 |
| | 8 2 | 8 2 | 6 2 | 6 2 | 3+4 2 | 5+6 2 |
| | 8 2 | 9 2 | 4+5 2 | 4+5 2 | 5+6 2 | 3+4 2 |
| 164.1.45 | 35585 | | 36 | | Cu2P2Zr1 | |
| | -A | -GM | -H | -K | -L | -M |
| | 9 2 | 8 2 | 4+5 2 | 6 2 | 3+4 2 | 5+6 2 |
| | 8 2 | 9 2 | 6 2 | 4+5 2 | 5+6 2 | 3+4 2 |
| | 8 2 | 8 2 | 6 2 | 4+5 2 | 3+4 2 | 3+4 2 |
| | 4+5 2 | 4+5 2 | 4+5 2 | 6 2 | 5+6 2 | 5+6 2 |
| | 8 2 | 8 2 | 6 2 | 6 2 | 5+6 2 | 3+4 2 |
| | 6+7 2 | 6+7 2 | 6 2 | 6 2 | 3+4 2 | 3+4 2 |
| | 9 2 | 9 2 | 6 2 | 6 2 | 5+6 2 | 5+6 2 |
| | 9 2 | 9 2 | 4+5 2 | 4+5 2 | 3+4 2 | 5+6 2 |
| | 9 2 | 8 2 | 6 2 | 6 2 | 3+4 2 | 5+6 2 |
| | 6+7 2 | 4+5 2 | 6 2 | 6 2 | 5+6 2 | 3+4 2 |
| | 8 2 | 8 2 | 4+5 2 | 4+5 2 | 3+4 2 | 3+4 2 |

Continued on next column

**TABLE LXXI– continued**

| TQC | ICSD | | NoE | | SF | |
|---|---|---|---|---|---|---|
| | 4+5 2 | 9 2 | 6 2 | 6 2 | 5+6 2 | 5+6 2 |
| | 9 2 | 6+7 2 | 6 2 | 6 2 | 5+6 2 | 3+4 2 |
| | 4+5 2 | 4+5 2 | 6 2 | 6 2 | 3+4 2 | 3+4 2 |
| | 8 2 | 8 2 | 4+5 2 | 4+5 2 | 3+4 2 | 5+6 2 |
| | 8 2 | 8 2 | 6 2 | 6 2 | 3+4 2 | 5+6 2 |
| | 6+7 2 | 8 2 | 4+5 2 | 4+5 2 | 5+6 2 | 5+6 2 |
| | 9 2 | 4+5 2 | 6 2 | 6 2 | 5+6 2 | 3+4 2 |
| 164.1.46 | 407246 | | 46 | | Ce2I2Si2 | |
| | -A | -GM | -H | -K | -L | -M |
| | 8 2 | 8 2 | 6 2 | 6 2 | 3+4 2 | 3+4 2 |
| | 9 2 | 9 2 | 6 2 | 6 2 | 5+6 2 | 5+6 2 |
| | 8 2 | 8 2 | 6 2 | 6 2 | 3+4 2 | 3+4 2 |
| | 9 2 | 9 2 | 6 2 | 6 2 | 5+6 2 | 5+6 2 |
| | 8 2 | 8 2 | 4+5 2 | 4+5 2 | 3+4 2 | 3+4 2 |
| | 9 2 | 9 2 | 6 2 | 6 2 | 5+6 2 | 5+6 2 |
| | 6+7 2 | 6+7 2 | 4+5 2 | 4+5 2 | 3+4 2 | 3+4 2 |
| | 4+5 2 | 4+5 2 | 6 2 | 6 2 | 5+6 2 | 5+6 2 |
| | 9 2 | 8 2 | 6 2 | 6 2 | 3+4 2 | 5+6 2 |
| | 8 2 | 9 2 | 4+5 2 | 4+5 2 | 5+6 2 | 3+4 2 |
| | 8 2 | 8 2 | 6 2 | 6 2 | 5+6 2 | 5+6 2 |
| | 9 2 | 8 2 | 4+5 2 | 4+5 2 | 3+4 2 | 3+4 2 |
| | 8 2 | 9 2 | 6 2 | 6 2 | 5+6 2 | 3+4 2 |
| | 9 2 | 8 2 | 4+5 2 | 4+5 2 | 3+4 2 | 5+6 2 |
| | 8 2 | 9 2 | 6 2 | 6 2 | 5+6 2 | 5+6 2 |
| | 4+5 2 | 4+5 2 | 6 2 | 6 2 | 3+4 2 | 5+6 2 |
| | 9 2 | 8 2 | 6 2 | 6 2 | 5+6 2 | 3+4 2 |
| | 8 2 | 6+7 2 | 4+5 2 | 4+5 2 | 3+4 2 | 5+6 2 |
| | 6+7 2 | 4+5 2 | 6 2 | 6 2 | 5+6 2 | 3+4 2 |
| | 4+5 2 | 9 2 | 6 2 | 6 2 | 3+4 2 | 3+4 2 |
| | 9 2 | 9 2 | 4+5 2 | 4+5 2 | 5+6 2 | 5+6 2 |
| | 9 2 | 8 2 | 6 2 | 6 2 | 3+4 2 | 3+4 2 |
| | 8 2 | 9 2 | 4+5 2 | 4+5 2 | 5+6 2 | 5+6 2 |
| 164.1.47 | 78989 | | 58 | | Cd2Sb2Yb1 | |
| | -A | -GM | -H | -K | -L | -M |
| | 8 2 | 8 2 | 6 2 | 6 2 | 3+4 2 | 3+4 2 |
| | 9 2 | 9 2 | 6 2 | 6 2 | 5+6 2 | 5+6 2 |
| | 9 2 | 9 2 | 4+5 2 | 4+5 2 | 5+6 2 | 5+6 2 |
| | 6+7 2 | 6+7 2 | 6 2 | 6 2 | 5+6 2 | 5+6 2 |
| | 8 2 | 8 2 | 6 2 | 6 2 | 5+6 2 | 3+4 2 |
| | 9 2 | 9 2 | 4+5 2 | 4+5 2 | 3+4 2 | 5+6 2 |
| | 4+5 2 | 4+5 2 | 6 2 | 6 2 | 5+6 2 | 3+4 2 |
| | 6+7 2 | 6+7 2 | 6 2 | 6 2 | 3+4 2 | 5+6 2 |
| | 9 2 | 8 2 | 4+5 2 | 4+5 2 | 3+4 2 | 5+6 2 |
| | 8 2 | 9 2 | 6 2 | 6 2 | 5+6 2 | 3+4 2 |
| | 8 2 | 8 2 | 6 2 | 6 2 | 5+6 2 | 3+4 2 |
| | 9 2 | 4+5 2 | 6 2 | 6 2 | 3+4 2 | 5+6 2 |
| | 6+7 2 | 9 2 | 4+5 2 | 4+5 2 | 3+4 2 | 5+6 2 |
| | 9 2 | 8 2 | 6 2 | 6 2 | 5+6 2 | 3+4 2 |
| | 4+5 2 | 6+7 2 | 6 2 | 6 2 | 3+4 2 | 5+6 2 |
| | 8 2 | 9 2 | 4+5 2 | 4+5 2 | 5+6 2 | 3+4 2 |
| | 9 2 | 8 2 | 4+5 2 | 4+5 2 | 3+4 2 | 3+4 2 |
| | 8 2 | 8 2 | 6 2 | 6 2 | 5+6 2 | 5+6 2 |
| | 8 2 | 4+5 2 | 6 2 | 6 2 | 3+4 2 | 5+6 2 |
| | 6+7 2 | 9 2 | 6 2 | 6 2 | 3+4 2 | 3+4 2 |
| | 9 2 | 6+7 2 | 4+5 2 | 4+5 2 | 5+6 2 | 5+6 2 |
| | 9 2 | 9 2 | 6 2 | 6 2 | 5+6 2 | 3+4 2 |
| | 4+5 2 | 9 2 | 6 2 | 6 2 | 5+6 2 | 5+6 2 |
| | 9 2 | 9 2 | 4+5 2 | 4+5 2 | 5+6 2 | 5+6 2 |
| | 6+7 2 | 6+7 2 | 6 2 | 6 2 | 5+6 2 | 5+6 2 |
| | 8 2 | 9 2 | 6 2 | 6 2 | 5+6 2 | 3+4 2 |
| | 9 2 | 9 2 | 4+5 2 | 6 2 | 5+6 2 | 5+6 2 |

Continued on next column

**TABLE LXXI– continued**

| TQC | ICSD | | NoE | SF | | |
|---|---|---|---|---|---|---|
| | 9 2 | 9 2 | 6 2 | 4+5 2 | 5+6 2 | 5+6 2 |
| | 6+7 2 | 6+7 2 | 6 2 | 6 2 | 5+6 2 | 5+6 2 |
| 164.1.48 | 419690 | | 58 | Cd2Sb2Yb1 | | |
| | -A | -GM | -H | -K | -L | -M |
| | 8 2 | 8 2 | 6 2 | 6 2 | 3+4 2 | 3+4 2 |
| | 9 2 | 9 2 | 6 2 | 6 2 | 5+6 2 | 5+6 2 |
| | 9 2 | 9 2 | 6 2 | 6 2 | 5+6 2 | 5+6 2 |
| | 6+7 2 | 6+7 2 | 4+5 2 | 4+5 2 | 5+6 2 | 5+6 2 |
| | 8 2 | 8 2 | 6 2 | 6 2 | 5+6 2 | 3+4 2 |
| | 9 2 | 9 2 | 4+5 2 | 4+5 2 | 3+4 2 | 5+6 2 |
| | 4+5 2 | 4+5 2 | 6 2 | 6 2 | 5+6 2 | 3+4 2 |
| | 6+7 2 | 6+7 2 | 6 2 | 6 2 | 3+4 2 | 5+6 2 |
| | 9 2 | 8 2 | 4+5 2 | 4+5 2 | 3+4 2 | 5+6 2 |
| | 8 2 | 9 2 | 6 2 | 6 2 | 5+6 2 | 3+4 2 |
| | 8 2 | 8 2 | 6 2 | 6 2 | 5+6 2 | 3+4 2 |
| | 9 2 | 4+5 2 | 6 2 | 6 2 | 3+4 2 | 5+6 2 |
| | 6+7 2 | 9 2 | 4+5 2 | 4+5 2 | 3+4 2 | 5+6 2 |
| | 9 2 | 8 2 | 6 2 | 6 2 | 5+6 2 | 3+4 2 |
| | 4+5 2 | 6+7 2 | 6 2 | 6 2 | 3+4 2 | 5+6 2 |
| | 8 2 | 9 2 | 4+5 2 | 4+5 2 | 5+6 2 | 3+4 2 |
| | 9 2 | 8 2 | 4+5 2 | 4+5 2 | 3+4 2 | 3+4 2 |
| | 8 2 | 8 2 | 6 2 | 6 2 | 5+6 2 | 5+6 2 |
| | 8 2 | 4+5 2 | 6 2 | 6 2 | 3+4 2 | 5+6 2 |
| | 6+7 2 | 9 2 | 6 2 | 6 2 | 3+4 2 | 3+4 2 |
| | 9 2 | 6+7 2 | 4+5 2 | 4+5 2 | 5+6 2 | 5+6 2 |
| | 9 2 | 9 2 | 6 2 | 6 2 | 5+6 2 | 3+4 2 |
| | 4+5 2 | 9 2 | 6 2 | 6 2 | 5+6 2 | 5+6 2 |
| | 9 2 | 9 2 | 4+5 2 | 4+5 2 | 5+6 2 | 5+6 2 |
| | 6+7 2 | 6+7 2 | 6 2 | 6 2 | 5+6 2 | 5+6 2 |
| | 9 2 | 9 2 | 6 2 | 6 2 | 5+6 2 | 3+4 2 |
| | 8 2 | 9 2 | 4+5 2 | 6 2 | 5+6 2 | 5+6 2 |
| | 9 2 | 9 2 | 6 2 | 4+5 2 | 5+6 2 | 5+6 2 |
| | 6+7 2 | 6+7 2 | 6 2 | 6 2 | 5+6 2 | 5+6 2 |
| 164.1.49 | 248526 | | 10 | Ca1Si2 | | |
| | -A | -GM | -H | -K | -L | -M |
| | 9 2 | 8 2 | 4+5 2 | 4+5 2 | 3+4 2 | 5+6 2 |
| | 8 2 | 9 2 | 6 2 | 6 2 | 5+6 2 | 3+4 2 |
| | 9 2 | 8 2 | 6 2 | 6 2 | 3+4 2 | 5+6 2 |
| | 6+7 2 | 4+5 2 | 6 2 | 4+5 2 | 5+6 2 | 3+4 2 |
| | 8 2 | 8 2 | 4+5 2 | 6 2 | 3+4 2 | 3+4 2 |
| 164.1.50 | 248525 | | 10 | Ca1Si2 | | |
| | -A | -GM | -H | -K | -L | -M |
| | 9 2 | 8 2 | 4+5 2 | 4+5 2 | 3+4 2 | 5+6 2 |
| | 8 2 | 9 2 | 6 2 | 6 2 | 5+6 2 | 3+4 2 |
| | 9 2 | 8 2 | 6 2 | 6 2 | 3+4 2 | 5+6 2 |
| | 6+7 2 | 4+5 2 | 6 2 | 4+5 2 | 5+6 2 | 3+4 2 |
| | 8 2 | 8 2 | 4+5 2 | 6 2 | 3+4 2 | 3+4 2 |
| 164.1.51 | 248524 | | 10 | Ca1Si2 | | |
| | -A | -GM | -H | -K | -L | -M |
| | 9 2 | 8 2 | 4+5 2 | 4+5 2 | 3+4 2 | 5+6 2 |
| | 8 2 | 9 2 | 6 2 | 6 2 | 5+6 2 | 3+4 2 |
| | 9 2 | 8 2 | 6 2 | 6 2 | 3+4 2 | 5+6 2 |
| | 6+7 2 | 4+5 2 | 6 2 | 4+5 2 | 5+6 2 | 3+4 2 |
| | 8 2 | 8 2 | 4+5 2 | 6 2 | 3+4 2 | 3+4 2 |
| 164.1.52 | 248523 | | 10 | Ca1Si2 | | |
| | -A | -GM | -H | -K | -L | -M |
| | 9 2 | 8 2 | 4+5 2 | 4+5 2 | 3+4 2 | 5+6 2 |
| | 8 2 | 9 2 | 6 2 | 6 2 | 5+6 2 | 3+4 2 |
| | 9 2 | 8 2 | 6 2 | 6 2 | 5+6 2 | 3+4 2 |
| | 6+7 2 | 8 2 | 6 2 | 4+5 2 | 3+4 2 | 5+6 2 |
| | 8 2 | 4+5 2 | 4+5 2 | 6 2 | 3+4 2 | 3+4 2 |
| 164.1.53 | 248522 | | 10 | Ca1Si2 | | |
| | -A | -GM | -H | -K | -L | -M |
| | 9 2 | 8 2 | 4+5 2 | 4+5 2 | 3+4 2 | 5+6 2 |
| | 8 2 | 9 2 | 6 2 | 6 2 | 5+6 2 | 3+4 2 |
| | 9 2 | 8 2 | 6 2 | 6 2 | 5+6 2 | 3+4 2 |
| | 6+7 2 | 8 2 | 6 2 | 4+5 2 | 3+4 2 | 5+6 2 |
| | 8 2 | 4+5 2 | 4+5 2 | 6 2 | 3+4 2 | 3+4 2 |
| 164.1.54 | 248521 | | 10 | Ca1Si2 | | |
| | -A | -GM | -H | -K | -L | -M |
| | 9 2 | 8 2 | 4+5 2 | 4+5 2 | 3+4 2 | 5+6 2 |
| | 8 2 | 9 2 | 6 2 | 6 2 | 5+6 2 | 3+4 2 |
| | 9 2 | 8 2 | 6 2 | 6 2 | 5+6 2 | 3+4 2 |
| | 6+7 2 | 8 2 | 6 2 | 4+5 2 | 3+4 2 | 5+6 2 |
| | 8 2 | 4+5 2 | 4+5 2 | 6 2 | 3+4 2 | 3+4 2 |
| 164.1.55 | 248520 | | 10 | Ca1Si2 | | |
| | -A | -GM | -H | -K | -L | -M |
| | 9 2 | 8 2 | 4+5 2 | 4+5 2 | 3+4 2 | 5+6 2 |
| | 8 2 | 9 2 | 6 2 | 6 2 | 5+6 2 | 3+4 2 |
| | 9 2 | 8 2 | 6 2 | 6 2 | 5+6 2 | 3+4 2 |
| | 6+7 2 | 8 2 | 6 2 | 4+5 2 | 3+4 2 | 5+6 2 |
| | 8 2 | 4+5 2 | 4+5 2 | 6 2 | 3+4 2 | 5+6 2 |
| 164.1.56 | 248519 | | 10 | Ca1Si2 | | |
| | -A | -GM | -H | -K | -L | -M |
| | 9 2 | 8 2 | 4+5 2 | 4+5 2 | 3+4 2 | 5+6 2 |
| | 8 2 | 9 2 | 6 2 | 6 2 | 5+6 2 | 3+4 2 |
| | 9 2 | 8 2 | 6 2 | 6 2 | 5+6 2 | 3+4 2 |
| | 6+7 2 | 8 2 | 6 2 | 4+5 2 | 3+4 2 | 5+6 2 |
| | 8 2 | 4+5 2 | 4+5 2 | 6 2 | 3+4 2 | 5+6 2 |
| 164.1.57 | 41789 | | 26 | Ca1Mn2Sb2 | | |
| | -A | -GM | -H | -K | -L | -M |
| | 9 2 | 8 2 | 4+5 2 | 4+5 2 | 3+4 2 | 5+6 2 |
| | 8 2 | 9 2 | 6 2 | 6 2 | 5+6 2 | 3+4 2 |
| | 8 2 | 8 2 | 4+5 2 | 4+5 2 | 3+4 2 | 3+4 2 |
| | 9 2 | 9 2 | 6 2 | 6 2 | 5+6 2 | 5+6 2 |
| | 8 2 | 8 2 | 6 2 | 6 2 | 5+6 2 | 5+6 2 |
| | 4+5 2 | 4+5 2 | 6 2 | 6 2 | 3+4 2 | 3+4 2 |
| | 9 2 | 6+7 2 | 4+5 2 | 4+5 2 | 5+6 2 | 3+4 2 |
| | 6+7 2 | 9 2 | 6 2 | 6 2 | 3+4 2 | 5+6 2 |
| | 9 2 | 8 2 | 6 2 | 6 2 | 3+4 2 | 5+6 2 |
| | 6+7 2 | 4+5 2 | 6 2 | 6 2 | 5+6 2 | 3+4 2 |
| | 9 2 | 8 2 | 4+5 2 | 4+5 2 | 5+6 2 | 3+4 2 |
| | 4+5 2 | 9 2 | 6 2 | 6 2 | 3+4 2 | 5+6 2 |
| | 8 2 | 6+7 2 | 6 2 | 6 2 | 3+4 2 | 3+4 2 |
| 164.1.58 | 416093 | | 26 | Ca1Mn2Sb2 | | |
| | -A | -GM | -H | -K | -L | -M |
| | 9 2 | 8 2 | 4+5 2 | 4+5 2 | 3+4 2 | 5+6 2 |
| | 8 2 | 9 2 | 6 2 | 6 2 | 5+6 2 | 3+4 2 |
| | 8 2 | 8 2 | 4+5 2 | 4+5 2 | 3+4 2 | 3+4 2 |
| | 9 2 | 9 2 | 6 2 | 6 2 | 5+6 2 | 5+6 2 |
| | 8 2 | 8 2 | 6 2 | 6 2 | 5+6 2 | 5+6 2 |
| | 4+5 2 | 4+5 2 | 6 2 | 6 2 | 3+4 2 | 3+4 2 |
| | 9 2 | 6+7 2 | 4+5 2 | 4+5 2 | 5+6 2 | 3+4 2 |
| | 6+7 2 | 9 2 | 6 2 | 6 2 | 3+4 2 | 5+6 2 |
| | 9 2 | 8 2 | 6 2 | 6 2 | 3+4 2 | 5+6 2 |
| | 6+7 2 | 4+5 2 | 6 2 | 6 2 | 5+6 2 | 3+4 2 |
| | 9 2 | 8 2 | 4+5 2 | 4+5 2 | 5+6 2 | 3+4 2 |
| | 4+5 2 | 6+7 2 | 6 2 | 6 2 | 3+4 2 | 5+6 2 |
| | 8 2 | 9 2 | 6 2 | 6 2 | 3+4 2 | 3+4 2 |
| 164.1.59 | 163778 | | 26 | Ca1Mn2Sb2 | | |

Continued on next column

**TABLE LXXI– continued**

| TQC | ICSD | | NoE | | SF | |
|---|---|---|---|---|---|---|
| | -A | -GM | -H | -K | -L | -M |
| | 9 2 | 8 2 | 4+5 2 | 4+5 2 | 3+4 2 | 5+6 2 |
| | 8 2 | 9 2 | 6 2 | 6 2 | 5+6 2 | 3+4 2 |
| | 8 2 | 8 2 | 4+5 2 | 4+5 2 | 3+4 2 | 3+4 2 |
| | 9 2 | 9 2 | 6 2 | 6 2 | 5+6 2 | 5+6 2 |
| | 8 2 | 8 2 | 6 2 | 6 2 | 5+6 2 | 5+6 2 |
| | 4+5 2 | 6+7 2 | 6 2 | 6 2 | 3+4 2 | 3+4 2 |
| | 9 2 | 4+5 2 | 4+5 2 | 4+5 2 | 5+6 2 | 3+4 2 |
| | 6+7 2 | 9 2 | 6 2 | 6 2 | 3+4 2 | 5+6 2 |
| | 9 2 | 8 2 | 6 2 | 6 2 | 3+4 2 | 5+6 2 |
| | 6+7 2 | 4+5 2 | 6 2 | 6 2 | 5+6 2 | 3+4 2 |
| | 9 2 | 8 2 | 4+5 2 | 4+5 2 | 5+6 2 | 3+4 2 |
| | 4+5 2 | 6+7 2 | 6 2 | 6 2 | 3+4 2 | 5+6 2 |
| | 8 2 | 9 2 | 6 2 | 6 2 | 3+4 2 | 3+4 2 |
| 164.1.60 | 49017 | | 26 | | Ca1Mn2P2 | |
| | -A | -GM | -H | -K | -L | -M |
| | 9 2 | 8 2 | 4+5 2 | 4+5 2 | 3+4 2 | 5+6 2 |
| | 8 2 | 9 2 | 6 2 | 6 2 | 5+6 2 | 3+4 2 |
| | 8 2 | 8 2 | 4+5 2 | 4+5 2 | 3+4 2 | 3+4 2 |
| | 9 2 | 6+7 2 | 6 2 | 6 2 | 5+6 2 | 5+6 2 |
| | 4+5 2 | 9 2 | 6 2 | 6 2 | 5+6 2 | 3+4 2 |
| | 8 2 | 4+5 2 | 6 2 | 6 2 | 5+6 2 | 5+6 2 |
| | 6+7 2 | 8 2 | 4+5 2 | 6 2 | 3+4 2 | 3+4 2 |
| | 9 2 | 9 2 | 6 2 | 4+5 2 | 3+4 2 | 5+6 2 |
| | 9 2 | 8 2 | 6 2 | 6 2 | 3+4 2 | 5+6 2 |
| | 6+7 2 | 4+5 2 | 6 2 | 6 2 | 5+6 2 | 3+4 2 |
| | 9 2 | 8 2 | 4+5 2 | 4+5 2 | 5+6 2 | 3+4 2 |
| | 4+5 2 | 9 2 | 6 2 | 6 2 | 3+4 2 | 5+6 2 |
| | 8 2 | 6+7 2 | 6 2 | 6 2 | 3+4 2 | 3+4 2 |
| 164.1.61 | 35512 | | 64 | | C6N6Pt1Rb2S6 | |
| | -A | -GM | -H | -K | -L | -M |
| | 9 2 | 8 2 | 4+5 2 | 4+5 2 | 3+4 2 | 5+6 2 |
| | 8 2 | 9 2 | 6 2 | 6 2 | 5+6 2 | 3+4 2 |
| | 8 2 | 8 2 | 6 2 | 6 2 | 3+4 2 | 3+4 2 |
| | 8 2 | 8 2 | 6 2 | 6 2 | 3+4 2 | 3+4 2 |
| | 4+5 2 | 4+5 2 | 4+5 2 | 4+5 2 | 3+4 2 | 3+4 2 |
| | 9 2 | 8 2 | 6 2 | 6 2 | 3+4 2 | 5+6 2 |
| | 8 2 | 9 2 | 4+5 2 | 4+5 2 | 5+6 2 | 3+4 2 |
| | 6+7 2 | 4+5 2 | 6 2 | 6 2 | 3+4 2 | 5+6 2 |
| | 9 2 | 8 2 | 6 2 | 6 2 | 5+6 2 | 3+4 2 |
| | 8 2 | 9 2 | 4+5 2 | 4+5 2 | 3+4 2 | 5+6 2 |
| | 4+5 2 | 6+7 2 | 6 2 | 6 2 | 5+6 2 | 3+4 2 |
| | 9 2 | 9 2 | 6 2 | 6 2 | 5+6 2 | 5+6 2 |
| | 9 2 | 9 2 | 6 2 | 6 2 | 5+6 2 | 5+6 2 |
| | 6+7 2 | 6+7 2 | 4+5 2 | 4+5 2 | 5+6 2 | 5+6 2 |
| | 8 2 | 8 2 | 6 2 | 6 2 | 3+4 2 | 3+4 2 |
| | 4+5 2 | 4+5 2 | 4+5 2 | 4+5 2 | 3+4 2 | 3+4 2 |
| | 8 2 | 8 2 | 6 2 | 6 2 | 3+4 2 | 3+4 2 |
| | 8 2 | 8 2 | 6 2 | 6 2 | 3+4 2 | 3+4 2 |
| | 4+5 2 | 4+5 2 | 4+5 2 | 4+5 2 | 3+4 2 | 3+4 2 |
| | 8 2 | 8 2 | 6 2 | 6 2 | 3+4 2 | 3+4 2 |
| | 9 2 | 9 2 | 6 2 | 6 2 | 5+6 2 | 5+6 2 |
| | 9 2 | 9 2 | 6 2 | 6 2 | 5+6 2 | 5+6 2 |
| | 6+7 2 | 6+7 2 | 4+5 2 | 4+5 2 | 5+6 2 | 5+6 2 |
| | 9 2 | 9 2 | 4+5 2 | 4+5 2 | 5+6 2 | 5+6 2 |
| | 6+7 2 | 6+7 2 | 6 2 | 6 2 | 5+6 2 | 5+6 2 |
| | 9 2 | 9 2 | 6 2 | 6 2 | 5+6 2 | 5+6 2 |
| | 8 2 | 8 2 | 6 2 | 6 2 | 5+6 2 | 5+6 2 |
| | 9 2 | 9 2 | 4+5 2 | 4+5 2 | 3+4 2 | 3+4 2 |
| | 4+5 2 | 4+5 2 | 6 2 | 6 2 | 5+6 2 | 5+6 2 |
| | 8 2 | 8 2 | 6 2 | 6 2 | 3+4 2 | 3+4 2 |

Continued on next column

**TABLE LXXI– continued**

| TQC | ICSD | | NoE | | SF | |
|---|---|---|---|---|---|---|
| | 9 2 | 8 2 | 4+5 2 | 4+5 2 | 3+4 2 | 3+4 2 |
| | 9 2 | 9 2 | 6 2 | 6 2 | 5+6 2 | 5+6 2 |
| 164.1.62 | 35511 | | 64 | | C6K2N6Pt1S6 | |
| | -A | -GM | -H | -K | -L | -M |
| | 9 2 | 8 2 | 4+5 2 | 4+5 2 | 3+4 2 | 5+6 2 |
| | 8 2 | 9 2 | 6 2 | 6 2 | 5+6 2 | 3+4 2 |
| | 9 2 | 8 2 | 6 2 | 6 2 | 3+4 2 | 5+6 2 |
| | 8 2 | 8 2 | 4+5 2 | 4+5 2 | 5+6 2 | 3+4 2 |
| | 8 2 | 9 2 | 6 2 | 6 2 | 3+4 2 | 3+4 2 |
| | 6+7 2 | 4+5 2 | 6 2 | 6 2 | 3+4 2 | 5+6 2 |
| | 4+5 2 | 6+7 2 | 6 2 | 6 2 | 5+6 2 | 3+4 2 |
| | 8 2 | 8 2 | 4+5 2 | 4+5 2 | 3+4 2 | 5+6 2 |
| | 9 2 | 9 2 | 6 2 | 6 2 | 5+6 2 | 3+4 2 |
| | 8 2 | 8 2 | 6 2 | 6 2 | 3+4 2 | 3+4 2 |
| | 4+5 2 | 4+5 2 | 4+5 2 | 4+5 2 | 3+4 2 | 3+4 2 |
| | 9 2 | 9 2 | 6 2 | 6 2 | 5+6 2 | 5+6 2 |
| | 9 2 | 9 2 | 6 2 | 6 2 | 5+6 2 | 5+6 2 |
| | 6+7 2 | 6+7 2 | 4+5 2 | 4+5 2 | 5+6 2 | 5+6 2 |
| | 8 2 | 8 2 | 6 2 | 6 2 | 3+4 2 | 3+4 2 |
| | 4+5 2 | 4+5 2 | 4+5 2 | 4+5 2 | 3+4 2 | 3+4 2 |
| | 8 2 | 8 2 | 6 2 | 6 2 | 3+4 2 | 3+4 2 |
| | 8 2 | 8 2 | 6 2 | 6 2 | 3+4 2 | 3+4 2 |
| | 4+5 2 | 4+5 2 | 4+5 2 | 4+5 2 | 3+4 2 | 3+4 2 |
| | 8 2 | 8 2 | 6 2 | 6 2 | 3+4 2 | 3+4 2 |
| | 9 2 | 9 2 | 6 2 | 6 2 | 5+6 2 | 5+6 2 |
| | 9 2 | 9 2 | 4+5 2 | 4+5 2 | 5+6 2 | 5+6 2 |
| | 6+7 2 | 6+7 2 | 6 2 | 6 2 | 5+6 2 | 5+6 2 |
| | 6+7 2 | 6+7 2 | 4+5 2 | 4+5 2 | 5+6 2 | 5+6 2 |
| | 9 2 | 9 2 | 6 2 | 6 2 | 5+6 2 | 5+6 2 |
| | 9 2 | 8 2 | 6 2 | 6 2 | 5+6 2 | 5+6 2 |
| | 9 2 | 9 2 | 6 2 | 6 2 | 5+6 2 | 5+6 2 |
| | 8 2 | 9 2 | 4+5 2 | 4+5 2 | 3+4 2 | 3+4 2 |
| | 4+5 2 | 4+5 2 | 6 2 | 6 2 | 5+6 2 | 5+6 2 |
| | 8 2 | 8 2 | 6 2 | 6 2 | 3+4 2 | 3+4 2 |
| | 9 2 | 8 2 | 4+5 2 | 4+5 2 | 3+4 2 | 3+4 2 |
| | 9 2 | 9 2 | 6 2 | 6 2 | 5+6 2 | 5+6 2 |
| 164.1.63 | 35510 | | 56 | | C6H8N8Pt1S6 | |
| | -A | -GM | -H | -K | -L | -M |
| | 8 2 | 8 2 | 6 2 | 6 2 | 3+4 2 | 3+4 2 |
| | 8 2 | 8 2 | 6 2 | 6 2 | 3+4 2 | 3+4 2 |
| | 4+5 2 | 4+5 2 | 4+5 2 | 4+5 2 | 3+4 2 | 3+4 2 |
| | 9 2 | 9 2 | 6 2 | 6 2 | 5+6 2 | 5+6 2 |
| | 9 2 | 9 2 | 6 2 | 6 2 | 5+6 2 | 5+6 2 |
| | 6+7 2 | 6+7 2 | 4+5 2 | 4+5 2 | 5+6 2 | 5+6 2 |
| | 9 2 | 8 2 | 4+5 2 | 4+5 2 | 3+4 2 | 5+6 2 |
| | 8 2 | 9 2 | 6 2 | 6 2 | 5+6 2 | 3+4 2 |
| | 8 2 | 8 2 | 6 2 | 6 2 | 3+4 2 | 3+4 2 |
| | 4+5 2 | 4+5 2 | 4+5 2 | 4+5 2 | 3+4 2 | 3+4 2 |
| | 8 2 | 8 2 | 6 2 | 6 2 | 3+4 2 | 3+4 2 |
| | 4+5 2 | 4+5 2 | 6 2 | 6 2 | 3+4 2 | 3+4 2 |
| | 8 2 | 8 2 | 4+5 2 | 4+5 2 | 3+4 2 | 3+4 2 |
| | 8 2 | 8 2 | 6 2 | 6 2 | 3+4 2 | 3+4 2 |
| | 9 2 | 9 2 | 6 2 | 6 2 | 5+6 2 | 5+6 2 |
| | 9 2 | 9 2 | 6 2 | 6 2 | 5+6 2 | 5+6 2 |
| | 6+7 2 | 6+7 2 | 4+5 2 | 4+5 2 | 5+6 2 | 5+6 2 |
| | 9 2 | 9 2 | 4+5 2 | 4+5 2 | 5+6 2 | 5+6 2 |
| | 6+7 2 | 6+7 2 | 6 2 | 6 2 | 5+6 2 | 5+6 2 |
| | 9 2 | 9 2 | 6 2 | 6 2 | 5+6 2 | 5+6 2 |
| | 8 2 | 8 2 | 6 2 | 6 2 | 5+6 2 | 5+6 2 |
| | 9 2 | 9 2 | 4+5 2 | 4+5 2 | 3+4 2 | 3+4 2 |
| | 4+5 2 | 4+5 2 | 6 2 | 6 2 | 5+6 2 | 5+6 2 |

Continued on next column

**TABLE LXXI– continued**

| TQC | ICSD | | NoE | | SF | |
|---|---|---|---|---|---|---|
| | 8 2 | 8 2 | 6 2 | 6 2 | 3+4 2 | 3+4 2 |
| | 9 2 | 8 2 | 6 2 | 6 2 | 3+4 2 | 5+6 2 |
| | 6+7 2 | 4+5 2 | 4+5 2 | 4+5 2 | 5+6 2 | 3+4 2 |
| | 8 2 | 9 2 | 4+5 2 | 6 2 | 5+6 2 | 3+4 2 |
| | 9 2 | 9 2 | 6 2 | 4+5 2 | 3+4 2 | 5+6 2 |
| 164.1.64 | 5005 | | 14 | | C1V2 | |
| | -A | -GM | -H | -K | -L | -M |
| | 8 2 | 8 2 | 6 2 | 6 2 | 3+4 2 | 3+4 2 |
| | 9 2 | 9 2 | 4+5 2 | 4+5 2 | 5+6 2 | 5+6 2 |
| | 6+7 2 | 6+7 2 | 6 2 | 6 2 | 5+6 2 | 5+6 2 |
| | 9 2 | 9 2 | 6 2 | 6 2 | 5+6 2 | 5+6 2 |
| | 8 2 | 8 2 | 6 2 | 6 2 | 3+4 2 | 3+4 2 |
| | 9 2 | 8 2 | 6 2 | 4+5 2 | 5+6 2 | 3+4 2 |
| | 4+5 2 | 4+5 2 | 4+5 2 | 6 2 | 3+4 2 | 5+6 2 |
| 164.1.65 | 409555 | | 14 | | C1Ta2 | |
| | -A | -GM | -H | -K | -L | -M |
| | 8 2 | 8 2 | 6 2 | 6 2 | 3+4 2 | 3+4 2 |
| | 9 2 | 9 2 | 4+5 2 | 4+5 2 | 5+6 2 | 5+6 2 |
| | 6+7 2 | 6+7 2 | 6 2 | 6 2 | 5+6 2 | 5+6 2 |
| | 9 2 | 9 2 | 6 2 | 6 2 | 5+6 2 | 5+6 2 |
| | 8 2 | 8 2 | 6 2 | 6 2 | 3+4 2 | 3+4 2 |
| | 9 2 | 8 2 | 4+5 2 | 4+5 2 | 5+6 2 | 3+4 2 |
| | 4+5 2 | 4+5 2 | 6 2 | 6 2 | 3+4 2 | 5+6 2 |
| 164.1.66 | 95113 | | 42 | | C1Nb2S2 | |
| | -A | -GM | -H | -K | -L | -M |
| | 8 2 | 8 2 | 4+5 2 | 4+5 2 | 5+6 2 | 5+6 2 |
| | 9 2 | 9 2 | 6 2 | 6 2 | 3+4 2 | 3+4 2 |
| | 8 2 | 8 2 | 6 2 | 6 2 | 5+6 2 | 5+6 2 |
| | 9 2 | 9 2 | 4+5 2 | 4+5 2 | 3+4 2 | 3+4 2 |
| | 8 2 | 8 2 | 6 2 | 6 2 | 5+6 2 | 5+6 2 |
| | 4+5 2 | 4+5 2 | 6 2 | 6 2 | 3+4 2 | 3+4 2 |
| | 9 2 | 9 2 | 4+5 2 | 4+5 2 | 5+6 2 | 5+6 2 |
| | 6+7 2 | 6+7 2 | 6 2 | 6 2 | 3+4 2 | 3+4 2 |
| | 9 2 | 8 2 | 4+5 2 | 4+5 2 | 3+4 2 | 3+4 2 |
| | 8 2 | 9 2 | 6 2 | 6 2 | 5+6 2 | 5+6 2 |
| | 8 2 | 8 2 | 6 2 | 6 2 | 3+4 2 | 3+4 2 |
| | 9 2 | 9 2 | 4+5 2 | 4+5 2 | 5+6 2 | 5+6 2 |
| | 6+7 2 | 6+7 2 | 6 2 | 6 2 | 3+4 2 | 3+4 2 |
| | 9 2 | 9 2 | 6 2 | 6 2 | 5+6 2 | 5+6 2 |
| | 8 2 | 8 2 | 6 2 | 6 2 | 5+6 2 | 5+6 2 |
| | 9 2 | 4+5 2 | 4+5 2 | 4+5 2 | 3+4 2 | 5+6 2 |
| | 4+5 2 | 8 2 | 6 2 | 6 2 | 5+6 2 | 3+4 2 |
| | 8 2 | 9 2 | 6 2 | 6 2 | 3+4 2 | 3+4 2 |
| | 8 2 | 8 2 | 4+5 2 | 4+5 2 | 5+6 2 | 3+4 2 |
| | 4+5 2 | 4+5 2 | 6 2 | 6 2 | 3+4 2 | 5+6 2 |
| | 9 2 | 8 2 | 6 2 | 6 2 | 5+6 2 | 5+6 2 |
| 164.1.67 | 95111 | | 42 | | C1Nb2S2 | |
| | -A | -GM | -H | -K | -L | -M |
| | 8 2 | 8 2 | 4+5 2 | 4+5 2 | 5+6 2 | 5+6 2 |
| | 9 2 | 9 2 | 6 2 | 6 2 | 3+4 2 | 3+4 2 |
| | 8 2 | 8 2 | 6 2 | 6 2 | 5+6 2 | 5+6 2 |
| | 9 2 | 9 2 | 4+5 2 | 4+5 2 | 3+4 2 | 3+4 2 |
| | 8 2 | 8 2 | 6 2 | 6 2 | 5+6 2 | 5+6 2 |
| | 4+5 2 | 4+5 2 | 6 2 | 6 2 | 3+4 2 | 3+4 2 |
| | 9 2 | 9 2 | 4+5 2 | 4+5 2 | 5+6 2 | 5+6 2 |
| | 6+7 2 | 6+7 2 | 6 2 | 6 2 | 3+4 2 | 3+4 2 |
| | 9 2 | 8 2 | 4+5 2 | 4+5 2 | 3+4 2 | 3+4 2 |
| | 8 2 | 9 2 | 6 2 | 6 2 | 5+6 2 | 5+6 2 |
| | 8 2 | 8 2 | 6 2 | 6 2 | 3+4 2 | 3+4 2 |
| | 9 2 | 9 2 | 4+5 2 | 4+5 2 | 5+6 2 | 5+6 2 |
| | 6+7 2 | 6+7 2 | 6 2 | 6 2 | 3+4 2 | 3+4 2 |

Continued on next column

**TABLE LXXI– continued**

| TQC | ICSD | | NoE | | SF | |
|---|---|---|---|---|---|---|
| | 9 2 | 9 2 | 6 2 | 6 2 | 5+6 2 | 5+6 2 |
| | 9 2 | 8 2 | 6 2 | 6 2 | 5+6 2 | 5+6 2 |
| | 8 2 | 4+5 2 | 4+5 2 | 4+5 2 | 5+6 2 | 5+6 2 |
| | 4+5 2 | 8 2 | 6 2 | 6 2 | 3+4 2 | 3+4 2 |
| | 8 2 | 9 2 | 6 2 | 6 2 | 5+6 2 | 3+4 2 |
| | 8 2 | 8 2 | 4+5 2 | 4+5 2 | 3+4 2 | 3+4 2 |
| | 4+5 2 | 4+5 2 | 6 2 | 6 2 | 3+4 2 | 5+6 2 |
| | 9 2 | 8 2 | 6 2 | 6 2 | 5+6 2 | 5+6 2 |
| 164.1.68 | 95109 | | 42 | | C1Nb2S2 | |
| | -A | -GM | -H | -K | -L | -M |
| | 8 2 | 8 2 | 4+5 2 | 4+5 2 | 5+6 2 | 5+6 2 |
| | 9 2 | 9 2 | 6 2 | 6 2 | 3+4 2 | 3+4 2 |
| | 8 2 | 8 2 | 6 2 | 6 2 | 5+6 2 | 5+6 2 |
| | 9 2 | 9 2 | 4+5 2 | 4+5 2 | 3+4 2 | 3+4 2 |
| | 8 2 | 8 2 | 6 2 | 6 2 | 5+6 2 | 5+6 2 |
| | 4+5 2 | 4+5 2 | 6 2 | 6 2 | 3+4 2 | 3+4 2 |
| | 9 2 | 9 2 | 4+5 2 | 4+5 2 | 5+6 2 | 5+6 2 |
| | 6+7 2 | 6+7 2 | 6 2 | 6 2 | 3+4 2 | 3+4 2 |
| | 9 2 | 8 2 | 4+5 2 | 4+5 2 | 3+4 2 | 3+4 2 |
| | 8 2 | 9 2 | 6 2 | 6 2 | 5+6 2 | 5+6 2 |
| | 8 2 | 8 2 | 6 2 | 6 2 | 3+4 2 | 3+4 2 |
| | 9 2 | 9 2 | 4+5 2 | 4+5 2 | 5+6 2 | 5+6 2 |
| | 6+7 2 | 6+7 2 | 6 2 | 6 2 | 3+4 2 | 3+4 2 |
| | 9 2 | 9 2 | 6 2 | 6 2 | 5+6 2 | 5+6 2 |
| | 8 2 | 8 2 | 6 2 | 6 2 | 5+6 2 | 5+6 2 |
| | 9 2 | 4+5 2 | 4+5 2 | 4+5 2 | 3+4 2 | 5+6 2 |
| | 4+5 2 | 8 2 | 6 2 | 6 2 | 5+6 2 | 3+4 2 |
| | 8 2 | 9 2 | 6 2 | 6 2 | 3+4 2 | 3+4 2 |
| | 8 2 | 8 2 | 4+5 2 | 4+5 2 | 5+6 2 | 3+4 2 |
| | 4+5 2 | 4+5 2 | 6 2 | 6 2 | 3+4 2 | 5+6 2 |
| | 9 2 | 8 2 | 6 2 | 6 2 | 5+6 2 | 5+6 2 |
| 164.1.69 | 656705 | | 42 | | C1Nb2S2 | |
| | -A | -GM | -H | -K | -L | -M |
| | 8 2 | 8 2 | 4+5 2 | 4+5 2 | 5+6 2 | 5+6 2 |
| | 9 2 | 9 2 | 6 2 | 6 2 | 3+4 2 | 3+4 2 |
| | 8 2 | 8 2 | 6 2 | 6 2 | 5+6 2 | 5+6 2 |
| | 9 2 | 9 2 | 4+5 2 | 4+5 2 | 3+4 2 | 3+4 2 |
| | 8 2 | 8 2 | 6 2 | 6 2 | 5+6 2 | 5+6 2 |
| | 4+5 2 | 4+5 2 | 6 2 | 6 2 | 3+4 2 | 3+4 2 |
| | 9 2 | 9 2 | 4+5 2 | 4+5 2 | 5+6 2 | 5+6 2 |
| | 6+7 2 | 6+7 2 | 6 2 | 6 2 | 3+4 2 | 3+4 2 |
| | 9 2 | 8 2 | 4+5 2 | 4+5 2 | 3+4 2 | 3+4 2 |
| | 8 2 | 9 2 | 6 2 | 6 2 | 5+6 2 | 5+6 2 |
| | 8 2 | 8 2 | 6 2 | 6 2 | 3+4 2 | 3+4 2 |
| | 9 2 | 9 2 | 4+5 2 | 4+5 2 | 5+6 2 | 5+6 2 |
| | 6+7 2 | 6+7 2 | 6 2 | 6 2 | 3+4 2 | 3+4 2 |
| | 9 2 | 9 2 | 6 2 | 6 2 | 5+6 2 | 5+6 2 |
| | 9 2 | 8 2 | 6 2 | 6 2 | 5+6 2 | 5+6 2 |
| | 8 2 | 4+5 2 | 4+5 2 | 4+5 2 | 5+6 2 | 5+6 2 |
| | 4+5 2 | 8 2 | 6 2 | 6 2 | 3+4 2 | 3+4 2 |
| | 8 2 | 9 2 | 6 2 | 6 2 | 5+6 2 | 3+4 2 |
| | 8 2 | 8 2 | 4+5 2 | 4+5 2 | 3+4 2 | 5+6 2 |
| | 4+5 2 | 4+5 2 | 6 2 | 6 2 | 3+4 2 | 3+4 2 |
| | 9 2 | 8 2 | 6 2 | 6 2 | 5+6 2 | 5+6 2 |
| 164.1.70 | 43670 | | 30 | | C1Nb2 | |
| | -A | -GM | -H | -K | -L | -M |
| | 8 2 | 8 2 | 4+5 2 | 4+5 2 | 5+6 2 | 5+6 2 |
| | 9 2 | 9 2 | 6 2 | 6 2 | 3+4 2 | 3+4 2 |
| | 8 2 | 8 2 | 6 2 | 6 2 | 3+4 2 | 5+6 2 |
| | 9 2 | 9 2 | 4+5 2 | 4+5 2 | 5+6 2 | 3+4 2 |
| | 8 2 | 8 2 | 6 2 | 6 2 | 3+4 2 | 5+6 2 |

Continued on next column

**TABLE LXXI– continued**

| TQC | ICSD | | NoE | SF | | |
|---|---|---|---|---|---|---|
| | 9 2 | 4+5 2 | 6 2 | 6 2 | 5+6 2 | 3+4 2 |
| | 6+7 2 | 6+7 2 | 4+5 2 | 4+5 2 | 5+6 2 | 5+6 2 |
| | 4+5 2 | 9 2 | 6 2 | 6 2 | 3+4 2 | 3+4 2 |
| | 8 2 | 8 2 | 6 2 | 6 2 | 3+4 2 | 3+4 2 |
| | 9 2 | 9 2 | 4+5 2 | 4+5 2 | 5+6 2 | 5+6 2 |
| | 6+7 2 | 6+7 2 | 6 2 | 6 2 | 5+6 2 | 5+6 2 |
| | 9 2 | 9 2 | 6 2 | 6 2 | 5+6 2 | 5+6 2 |
| | 8 2 | 8 2 | 6 2 | 6 2 | 3+4 2 | 3+4 2 |
| | 9 2 | 8 2 | 4+5 2 | 4+5 2 | 5+6 2 | 3+4 2 |
| | 4+5 2 | 4+5 2 | 6 2 | 6 2 | 3+4 2 | 5+6 2 |
| 164.1.71 | 82264 | | 52 | C1N2O2Pr2 | | |
| | -A | -GM | -H | -K | -L | -M |
| | 8 2 | 8 2 | 4+5 2 | 4+5 2 | 5+6 2 | 5+6 2 |
| | 9 2 | 9 2 | 6 2 | 6 2 | 3+4 2 | 3+4 2 |
| | 8 2 | 8 2 | 4+5 2 | 4+5 2 | 5+6 2 | 5+6 2 |
| | 9 2 | 9 2 | 6 2 | 6 2 | 3+4 2 | 3+4 2 |
| | 9 2 | 8 2 | 6 2 | 6 2 | 5+6 2 | 3+4 2 |
| | 8 2 | 8 2 | 6 2 | 6 2 | 5+6 2 | 5+6 2 |
| | 9 2 | 9 2 | 4+5 2 | 4+5 2 | 3+4 2 | 3+4 2 |
| | 8 2 | 9 2 | 6 2 | 6 2 | 3+4 2 | 5+6 2 |
| | 4+5 2 | 4+5 2 | 6 2 | 6 2 | 5+6 2 | 5+6 2 |
| | 6+7 2 | 6+7 2 | 6 2 | 6 2 | 3+4 2 | 3+4 2 |
| | 9 2 | 9 2 | 4+5 2 | 4+5 2 | 3+4 2 | 3+4 2 |
| | 8 2 | 8 2 | 6 2 | 6 2 | 5+6 2 | 5+6 2 |
| | 9 2 | 8 2 | 6 2 | 6 2 | 5+6 2 | 3+4 2 |
| | 8 2 | 9 2 | 4+5 2 | 4+5 2 | 3+4 2 | 5+6 2 |
| | 8 2 | 9 2 | 6 2 | 6 2 | 3+4 2 | 5+6 2 |
| | 4+5 2 | 6+7 2 | 6 2 | 6 2 | 3+4 2 | 5+6 2 |
| | 6+7 2 | 8 2 | 6 2 | 4+5 2 | 3+4 2 | 3+4 2 |
| | 9 2 | 4+5 2 | 4+5 2 | 6 2 | 5+6 2 | 5+6 2 |
| | 8 2 | 8 2 | 6 2 | 6 2 | 5+6 2 | 5+6 2 |
| | 8 2 | 9 2 | 4+5 2 | 4+5 2 | 5+6 2 | 3+4 2 |
| | 4+5 2 | 6+7 2 | 6 2 | 6 2 | 3+4 2 | 5+6 2 |
| | 9 2 | 9 2 | 4+5 2 | 6 2 | 3+4 2 | 3+4 2 |
| | 9 2 | 8 2 | 6 2 | 4+5 2 | 5+6 2 | 3+4 2 |
| | 6+7 2 | 4+5 2 | 6 2 | 6 2 | 5+6 2 | 3+4 2 |
| | 8 2 | 8 2 | 6 2 | 6 2 | 3+4 2 | 3+4 2 |
| | 4+5 2 | 6+7 2 | 6 2 | 6 2 | 5+6 2 | 5+6 2 |
| 164.1.72 | 82265 | | 54 | C1N2Nd2O2 | | |
| | -A | -GM | -H | -K | -L | -M |
| | 8 2 | 8 2 | 4+5 2 | 4+5 2 | 5+6 2 | 5+6 2 |
| | 9 2 | 9 2 | 6 2 | 6 2 | 3+4 2 | 3+4 2 |
| | 8 2 | 8 2 | 4+5 2 | 4+5 2 | 5+6 2 | 5+6 2 |
| | 9 2 | 9 2 | 6 2 | 6 2 | 3+4 2 | 3+4 2 |
| | 9 2 | 8 2 | 6 2 | 6 2 | 5+6 2 | 3+4 2 |
| | 8 2 | 8 2 | 6 2 | 6 2 | 5+6 2 | 5+6 2 |
| | 9 2 | 9 2 | 4+5 2 | 4+5 2 | 3+4 2 | 3+4 2 |
| | 8 2 | 9 2 | 6 2 | 6 2 | 3+4 2 | 5+6 2 |
| | 4+5 2 | 4+5 2 | 6 2 | 6 2 | 5+6 2 | 5+6 2 |
| | 6+7 2 | 6+7 2 | 6 2 | 6 2 | 3+4 2 | 3+4 2 |
| | 9 2 | 9 2 | 4+5 2 | 4+5 2 | 3+4 2 | 3+4 2 |
| | 8 2 | 8 2 | 6 2 | 6 2 | 5+6 2 | 5+6 2 |
| | 9 2 | 8 2 | 6 2 | 6 2 | 5+6 2 | 3+4 2 |
| | 8 2 | 9 2 | 4+5 2 | 4+5 2 | 3+4 2 | 5+6 2 |
| | 8 2 | 9 2 | 6 2 | 6 2 | 3+4 2 | 5+6 2 |
| | 4+5 2 | 6+7 2 | 6 2 | 6 2 | 3+4 2 | 5+6 2 |
| | 6+7 2 | 8 2 | 6 2 | 4+5 2 | 3+4 2 | 3+4 2 |
| | 8 2 | 4+5 2 | 4+5 2 | 6 2 | 5+6 2 | 5+6 2 |
| | 9 2 | 8 2 | 6 2 | 6 2 | 5+6 2 | 5+6 2 |
| | 8 2 | 9 2 | 4+5 2 | 4+5 2 | 5+6 2 | 5+6 2 |
| | 4+5 2 | 6+7 2 | 6 2 | 6 2 | 3+4 2 | 3+4 2 |

Continued on next column

**TABLE LXXI– continued**

| TQC | ICSD | | NoE | SF | | |
|---|---|---|---|---|---|---|
| | 9 2 | 9 2 | 6 2 | 6 2 | 3+4 2 | 3+4 2 |
| | 9 2 | 8 2 | 4+5 2 | 4+5 2 | 5+6 2 | 3+4 2 |
| | 6+7 2 | 4+5 2 | 6 2 | 6 2 | 5+6 2 | 3+4 2 |
| | 8 2 | 8 2 | 6 2 | 6 2 | 3+4 2 | 3+4 2 |
| | 4+5 2 | 6+7 2 | 4+5 2 | 4+5 2 | 5+6 2 | 5+6 2 |
| | 8 2 | 9 2 | 6 2 | 6 2 | 3+4 2 | 5+6 2 |
| 164.1.73 | 71625 | | 60 | C1F2Ho2 | | |
| | -A | -GM | -H | -K | -L | -M |
| | 8 2 | 8 2 | 4+5 2 | 4+5 2 | 5+6 2 | 5+6 2 |
| | 9 2 | 9 2 | 6 2 | 6 2 | 3+4 2 | 3+4 2 |
| | 9 2 | 8 2 | 4+5 2 | 4+5 2 | 3+4 2 | 5+6 2 |
| | 8 2 | 9 2 | 6 2 | 6 2 | 5+6 2 | 3+4 2 |
| | 9 2 | 8 2 | 6 2 | 6 2 | 3+4 2 | 5+6 2 |
| | 8 2 | 9 2 | 4+5 2 | 4+5 2 | 5+6 2 | 3+4 2 |
| | 8 2 | 8 2 | 6 2 | 6 2 | 5+6 2 | 5+6 2 |
| | 4+5 2 | 4+5 2 | 6 2 | 6 2 | 3+4 2 | 3+4 2 |
| | 6+7 2 | 6+7 2 | 4+5 2 | 4+5 2 | 5+6 2 | 3+4 2 |
| | 9 2 | 9 2 | 6 2 | 6 2 | 3+4 2 | 5+6 2 |
| | 8 2 | 8 2 | 6 2 | 6 2 | 3+4 2 | 3+4 2 |
| | 9 2 | 8 2 | 6 2 | 6 2 | 5+6 2 | 3+4 2 |
| | 9 2 | 8 2 | 4+5 2 | 4+5 2 | 3+4 2 | 5+6 2 |
| | 6+7 2 | 4+5 2 | 6 2 | 6 2 | 3+4 2 | 5+6 2 |
| | 8 2 | 9 2 | 4+5 2 | 4+5 2 | 5+6 2 | 5+6 2 |
| | 4+5 2 | 9 2 | 6 2 | 6 2 | 3+4 2 | 3+4 2 |
| | 8 2 | 6+7 2 | 6 2 | 6 2 | 5+6 2 | 3+4 2 |
| | 9 2 | 9 2 | 4+5 2 | 4+5 2 | 5+6 2 | 5+6 2 |
| | 6+7 2 | 6+7 2 | 6 2 | 6 2 | 5+6 2 | 5+6 2 |
| | 9 2 | 9 2 | 6 2 | 6 2 | 5+6 2 | 5+6 2 |
| | 9 2 | 8 2 | 6 2 | 6 2 | 3+4 2 | 3+4 2 |
| | 8 2 | 9 2 | 4+5 2 | 4+5 2 | 3+4 2 | 3+4 2 |
| | 4+5 2 | 4+5 2 | 6 2 | 6 2 | 5+6 2 | 5+6 2 |
| | 8 2 | 8 2 | 6 2 | 6 2 | 3+4 2 | 3+4 2 |
| | 9 2 | 9 2 | 4+5 2 | 4+5 2 | 5+6 2 | 5+6 2 |
| | 6+7 2 | 6+7 2 | 6 2 | 6 2 | 5+6 2 | 5+6 2 |
| | 9 2 | 8 2 | 4+5 2 | 4+5 2 | 3+4 2 | 3+4 2 |
| | 8 2 | 9 2 | 6 2 | 6 2 | 5+6 2 | 5+6 2 |
| | 4+5 2 | 4+5 2 | 6 2 | 6 2 | 3+4 2 | 3+4 2 |
| | 8 2 | 8 2 | 6 2 | 6 2 | 5+6 2 | 5+6 2 |
| 164.1.74 | 33912 | | 54 | C1F2Gd2 | | |
| | -A | -GM | -H | -K | -L | -M |
| | 8 2 | 8 2 | 4+5 2 | 4+5 2 | 5+6 2 | 5+6 2 |
| | 9 2 | 9 2 | 6 2 | 6 2 | 3+4 2 | 3+4 2 |
| | 9 2 | 8 2 | 4+5 2 | 4+5 2 | 3+4 2 | 5+6 2 |
| | 8 2 | 9 2 | 6 2 | 6 2 | 5+6 2 | 3+4 2 |
| | 8 2 | 8 2 | 6 2 | 6 2 | 5+6 2 | 5+6 2 |
| | 9 2 | 9 2 | 4+5 2 | 4+5 2 | 3+4 2 | 3+4 2 |
| | 8 2 | 8 2 | 6 2 | 6 2 | 5+6 2 | 5+6 2 |
| | 4+5 2 | 4+5 2 | 6 2 | 6 2 | 3+4 2 | 3+4 2 |
| | 6+7 2 | 6+7 2 | 4+5 2 | 4+5 2 | 5+6 2 | 3+4 2 |
| | 9 2 | 9 2 | 6 2 | 6 2 | 3+4 2 | 5+6 2 |
| | 8 2 | 8 2 | 6 2 | 6 2 | 3+4 2 | 3+4 2 |
| | 9 2 | 8 2 | 6 2 | 6 2 | 5+6 2 | 3+4 2 |
| | 9 2 | 8 2 | 4+5 2 | 4+5 2 | 3+4 2 | 5+6 2 |
| | 6+7 2 | 4+5 2 | 6 2 | 6 2 | 3+4 2 | 5+6 2 |
| | 8 2 | 9 2 | 4+5 2 | 4+5 2 | 5+6 2 | 5+6 2 |
| | 4+5 2 | 9 2 | 6 2 | 6 2 | 3+4 2 | 3+4 2 |
| | 8 2 | 6+7 2 | 6 2 | 6 2 | 5+6 2 | 3+4 2 |
| | 9 2 | 9 2 | 4+5 2 | 4+5 2 | 5+6 2 | 5+6 2 |
| | 6+7 2 | 6+7 2 | 6 2 | 6 2 | 5+6 2 | 5+6 2 |
| | 9 2 | 9 2 | 6 2 | 6 2 | 5+6 2 | 5+6 2 |
| | 9 2 | 8 2 | 6 2 | 6 2 | 3+4 2 | 3+4 2 |

Continued on next column

**TABLE LXXI– continued**

| TQC | ICSD | | NoE | SF | | |
|---|---|---|---|---|---|---|
| | 8 2 | 9 2 | 4+5 2 | 4+5 2 | 3+4 2 | 5+6 2 |
| | 4+5 2 | 4+5 2 | 6 2 | 6 2 | 5+6 2 | 3+4 2 |
| | 8 2 | 8 2 | 6 2 | 6 2 | 3+4 2 | 3+4 2 |
| | 9 2 | 9 2 | 6 2 | 4+5 2 | 5+6 2 | 5+6 2 |
| | 6+7 2 | 6+7 2 | 4+5 2 | 6 2 | 5+6 2 | 5+6 2 |
| | 9 2 | 8 2 | 4+5 2 | 4+5 2 | 3+4 2 | 3+4 2 |
| 164.1.75 | 42753 | | 66 | Bi4Pb1Te7 | | |
| | -A | -GM | -H | -K | -L | -M |
| | 9 2 | 8 2 | 6 2 | 6 2 | 5+6 2 | 3+4 2 |
| | 8 2 | 9 2 | 4+5 2 | 4+5 2 | 3+4 2 | 5+6 2 |
| | 9 2 | 8 2 | 6 2 | 6 2 | 5+6 2 | 3+4 2 |
| | 8 2 | 9 2 | 6 2 | 6 2 | 3+4 2 | 5+6 2 |
| | 9 2 | 8 2 | 6 2 | 6 2 | 5+6 2 | 3+4 2 |
| | 8 2 | 9 2 | 4+5 2 | 4+5 2 | 3+4 2 | 5+6 2 |
| | 9 2 | 8 2 | 6 2 | 6 2 | 5+6 2 | 3+4 2 |
| | 8 2 | 9 2 | 4+5 2 | 4+5 2 | 3+4 2 | 5+6 2 |
| | 9 2 | 8 2 | 6 2 | 6 2 | 5+6 2 | 3+4 2 |
| | 8 2 | 9 2 | 4+5 2 | 4+5 2 | 3+4 2 | 5+6 2 |
| | 9 2 | 8 2 | 6 2 | 6 2 | 5+6 2 | 3+4 2 |
| | 8 2 | 8 2 | 6 2 | 6 2 | 3+4 2 | 3+4 2 |
| | 8 2 | 9 2 | 4+5 2 | 4+5 2 | 3+4 2 | 5+6 2 |
| | 9 2 | 8 2 | 6 2 | 6 2 | 5+6 2 | 3+4 2 |
| | 8 2 | 9 2 | 6 2 | 6 2 | 3+4 2 | 5+6 2 |
| | 9 2 | 9 2 | 6 2 | 6 2 | 5+6 2 | 5+6 2 |
| | 8 2 | 8 2 | 4+5 2 | 4+5 2 | 3+4 2 | 3+4 2 |
| | 8 2 | 9 2 | 6 2 | 4+5 2 | 5+6 2 | 5+6 2 |
| | 9 2 | 8 2 | 4+5 2 | 6 2 | 3+4 2 | 5+6 2 |
| | 9 2 | 6+7 2 | 6 2 | 6 2 | 3+4 2 | 3+4 2 |
| | 4+5 2 | 4+5 2 | 6 2 | 6 2 | 5+6 2 | 5+6 2 |
| | 6+7 2 | 9 2 | 4+5 2 | 4+5 2 | 5+6 2 | 3+4 2 |
| | 8 2 | 8 2 | 6 2 | 6 2 | 3+4 2 | 5+6 2 |
| | 4+5 2 | 6+7 2 | 6 2 | 6 2 | 3+4 2 | 5+6 2 |
| | 9 2 | 9 2 | 4+5 2 | 4+5 2 | 5+6 2 | 3+4 2 |
| | 6+7 2 | 9 2 | 6 2 | 6 2 | 5+6 2 | 5+6 2 |
| | 8 2 | 6+7 2 | 6 2 | 6 2 | 3+4 2 | 3+4 2 |
| | 4+5 2 | 4+5 2 | 6 2 | 4+5 2 | 5+6 2 | 3+4 2 |
| | 8 2 | 8 2 | 4+5 2 | 6 2 | 3+4 2 | 5+6 2 |
| | 9 2 | 8 2 | 6 2 | 6 2 | 5+6 2 | 3+4 2 |
| | 6+7 2 | 4+5 2 | 6 2 | 6 2 | 3+4 2 | 5+6 2 |
| | 9 2 | 6+7 2 | 4+5 2 | 4+5 2 | 3+4 2 | 3+4 2 |
| | 4+5 2 | 9 2 | 6 2 | 6 2 | 3+4 2 | 5+6 2 |
| 164.1.76 | 42707 | | 66 | Bi4Pb1Te7 | | |
| | -A | -GM | -H | -K | -L | -M |
| | 9 2 | 8 2 | 6 2 | 6 2 | 5+6 2 | 3+4 2 |
| | 8 2 | 9 2 | 4+5 2 | 4+5 2 | 3+4 2 | 5+6 2 |
| | 9 2 | 8 2 | 6 2 | 6 2 | 5+6 2 | 3+4 2 |
| | 8 2 | 9 2 | 6 2 | 6 2 | 3+4 2 | 5+6 2 |
| | 9 2 | 8 2 | 4+5 2 | 4+5 2 | 5+6 2 | 3+4 2 |
| | 8 2 | 9 2 | 6 2 | 6 2 | 3+4 2 | 5+6 2 |
| | 9 2 | 8 2 | 6 2 | 6 2 | 5+6 2 | 3+4 2 |
| | 8 2 | 9 2 | 4+5 2 | 4+5 2 | 3+4 2 | 5+6 2 |
| | 9 2 | 8 2 | 6 2 | 6 2 | 5+6 2 | 3+4 2 |
| | 8 2 | 9 2 | 4+5 2 | 4+5 2 | 3+4 2 | 5+6 2 |
| | 9 2 | 8 2 | 6 2 | 6 2 | 5+6 2 | 3+4 2 |
| | 8 2 | 8 2 | 6 2 | 6 2 | 3+4 2 | 3+4 2 |
| | 8 2 | 9 2 | 6 2 | 6 2 | 3+4 2 | 5+6 2 |
| | 9 2 | 8 2 | 4+5 2 | 4+5 2 | 5+6 2 | 3+4 2 |
| | 8 2 | 9 2 | 6 2 | 6 2 | 3+4 2 | 5+6 2 |
| | 9 2 | 8 2 | 6 2 | 6 2 | 5+6 2 | 5+6 2 |
| | 8 2 | 9 2 | 4+5 2 | 4+5 2 | 3+4 2 | 3+4 2 |
| | 9 2 | 9 2 | 6 2 | 6 2 | 5+6 2 | 5+6 2 |

Continued on next column

**TABLE LXXI– continued**

| TQC | ICSD | | NoE | SF | | |
|---|---|---|---|---|---|---|
| | 8 2 | 8 2 | 6 2 | 4+5 2 | 3+4 2 | 5+6 2 |
| | 9 2 | 6+7 2 | 4+5 2 | 6 2 | 3+4 2 | 3+4 2 |
| | 4+5 2 | 4+5 2 | 6 2 | 6 2 | 5+6 2 | 5+6 2 |
| | 6+7 2 | 9 2 | 4+5 2 | 4+5 2 | 5+6 2 | 3+4 2 |
| | 8 2 | 8 2 | 6 2 | 6 2 | 3+4 2 | 5+6 2 |
| | 9 2 | 6+7 2 | 6 2 | 6 2 | 3+4 2 | 3+4 2 |
| | 4+5 2 | 9 2 | 4+5 2 | 4+5 2 | 5+6 2 | 5+6 2 |
| | 6+7 2 | 6+7 2 | 6 2 | 6 2 | 5+6 2 | 5+6 2 |
| | 8 2 | 9 2 | 6 2 | 6 2 | 3+4 2 | 3+4 2 |
| | 4+5 2 | 4+5 2 | 6 2 | 4+5 2 | 5+6 2 | 3+4 2 |
| | 8 2 | 8 2 | 4+5 2 | 6 2 | 3+4 2 | 5+6 2 |
| | 9 2 | 8 2 | 6 2 | 6 2 | 5+6 2 | 3+4 2 |
| | 6+7 2 | 4+5 2 | 6 2 | 6 2 | 3+4 2 | 5+6 2 |
| | 9 2 | 6+7 2 | 4+5 2 | 4+5 2 | 3+4 2 | 3+4 2 |
| | 4+5 2 | 9 2 | 6 2 | 6 2 | 3+4 2 | 5+6 2 |
| 164.1.77 | 250249 | | 66 | Bi4Pb1Te7 | | |
| | -A | -GM | -H | -K | -L | -M |
| | 9 2 | 8 2 | 6 2 | 6 2 | 3+4 2 | 5+6 2 |
| | 8 2 | 9 2 | 6 2 | 6 2 | 5+6 2 | 3+4 2 |
| | 9 2 | 8 2 | 4+5 2 | 4+5 2 | 5+6 2 | 3+4 2 |
| | 8 2 | 9 2 | 6 2 | 6 2 | 3+4 2 | 5+6 2 |
| | 8 2 | 8 2 | 6 2 | 6 2 | 5+6 2 | 3+4 2 |
| | 9 2 | 8 2 | 4+5 2 | 4+5 2 | 5+6 2 | 5+6 2 |
| | 9 2 | 9 2 | 4+5 2 | 4+5 2 | 3+4 2 | 3+4 2 |
| | 8 2 | 9 2 | 6 2 | 6 2 | 3+4 2 | 5+6 2 |
| | 8 2 | 8 2 | 6 2 | 6 2 | 5+6 2 | 3+4 2 |
| | 9 2 | 9 2 | 6 2 | 6 2 | 3+4 2 | 5+6 2 |
| | 9 2 | 8 2 | 4+5 2 | 4+5 2 | 5+6 2 | 3+4 2 |
| | 8 2 | 8 2 | 6 2 | 6 2 | 3+4 2 | 3+4 2 |
| | 8 2 | 9 2 | 6 2 | 6 2 | 3+4 2 | 5+6 2 |
| | 9 2 | 8 2 | 4+5 2 | 4+5 2 | 5+6 2 | 5+6 2 |
| | 8 2 | 9 2 | 6 2 | 6 2 | 3+4 2 | 3+4 2 |
| | 9 2 | 8 2 | 4+5 2 | 4+5 2 | 5+6 2 | 5+6 2 |
| | 8 2 | 9 2 | 6 2 | 6 2 | 3+4 2 | 3+4 2 |
| | 9 2 | 9 2 | 6 2 | 6 2 | 5+6 2 | 5+6 2 |
| | 9 2 | 8 2 | 6 2 | 6 2 | 3+4 2 | 3+4 2 |
| | 8 2 | 9 2 | 4+5 2 | 4+5 2 | 3+4 2 | 5+6 2 |
| | 4+5 2 | 6+7 2 | 6 2 | 6 2 | 5+6 2 | 5+6 2 |
| | 6+7 2 | 4+5 2 | 6 2 | 6 2 | 5+6 2 | 3+4 2 |
| | 8 2 | 8 2 | 4+5 2 | 4+5 2 | 5+6 2 | 3+4 2 |
| | 4+5 2 | 6+7 2 | 6 2 | 6 2 | 3+4 2 | 5+6 2 |
| | 9 2 | 9 2 | 6 2 | 6 2 | 5+6 2 | 5+6 2 |
| | 8 2 | 9 2 | 6 2 | 6 2 | 3+4 2 | 3+4 2 |
| | 6+7 2 | 4+5 2 | 4+5 2 | 4+5 2 | 5+6 2 | 5+6 2 |
| | 4+5 2 | 6+7 2 | 6 2 | 6 2 | 3+4 2 | 5+6 2 |
| | 6+7 2 | 8 2 | 4+5 2 | 4+5 2 | 5+6 2 | 3+4 2 |
| | 8 2 | 6+7 2 | 6 2 | 6 2 | 3+4 2 | 5+6 2 |
| | 4+5 2 | 4+5 2 | 6 2 | 4+5 2 | 3+4 2 | 3+4 2 |
| | 9 2 | 8 2 | 4+5 2 | 6 2 | 3+4 2 | 3+4 2 |
| | 9 2 | 9 2 | 6 2 | 6 2 | 3+4 2 | 5+6 2 |
| 164.1.78 | 42891 | | 66 | Bi4Ge1Te7 | | |
| | -A | -GM | -H | -K | -L | -M |
| | 9 2 | 8 2 | 4+5 2 | 4+5 2 | 3+4 2 | 5+6 2 |
| | 8 2 | 9 2 | 6 2 | 6 2 | 5+6 2 | 3+4 2 |
| | 9 2 | 8 2 | 6 2 | 6 2 | 5+6 2 | 3+4 2 |
| | 8 2 | 9 2 | 6 2 | 6 2 | 3+4 2 | 5+6 2 |
| | 9 2 | 8 2 | 6 2 | 6 2 | 5+6 2 | 3+4 2 |
| | 8 2 | 9 2 | 4+5 2 | 4+5 2 | 3+4 2 | 5+6 2 |
| | 9 2 | 8 2 | 6 2 | 6 2 | 5+6 2 | 3+4 2 |
| | 8 2 | 9 2 | 4+5 2 | 4+5 2 | 3+4 2 | 5+6 2 |
| | 9 2 | 8 2 | 6 2 | 6 2 | 5+6 2 | 3+4 2 |

Continued on next column

**TABLE LXXI– continued**

| TQC | ICSD | | NoE | SF | | |
|---|---|---|---|---|---|---|
| | 8 2 | 9 2 | 4+5 2 | 4+5 2 | 3+4 2 | 5+6 2 |
| | 9 2 | 8 2 | 6 2 | 6 2 | 5+6 2 | 3+4 2 |
| | 8 2 | 8 2 | 6 2 | 6 2 | 3+4 2 | 3+4 2 |
| | 8 2 | 9 2 | 6 2 | 6 2 | 3+4 2 | 5+6 2 |
| | 9 2 | 8 2 | 4+5 2 | 4+5 2 | 5+6 2 | 3+4 2 |
| | 8 2 | 9 2 | 6 2 | 6 2 | 3+4 2 | 5+6 2 |
| | 9 2 | 9 2 | 6 2 | 6 2 | 5+6 2 | 5+6 2 |
| | 8 2 | 8 2 | 4+5 2 | 4+5 2 | 3+4 2 | 3+4 2 |
| | 9 2 | 8 2 | 6 2 | 6 2 | 5+6 2 | 3+4 2 |
| | 8 2 | 9 2 | 6 2 | 6 2 | 3+4 2 | 5+6 2 |
| | 4+5 2 | 6+7 2 | 4+5 2 | 4+5 2 | 3+4 2 | 5+6 2 |
| | 6+7 2 | 4+5 2 | 4+5 2 | 6 2 | 5+6 2 | 5+6 2 |
| | 9 2 | 9 2 | 6 2 | 4+5 2 | 5+6 2 | 3+4 2 |
| | 9 2 | 8 2 | 6 2 | 6 2 | 3+4 2 | 3+4 2 |
| | 8 2 | 9 2 | 6 2 | 6 2 | 5+6 2 | 5+6 2 |
| | 6+7 2 | 9 2 | 4+5 2 | 4+5 2 | 5+6 2 | 5+6 2 |
| | 8 2 | 6+7 2 | 6 2 | 6 2 | 3+4 2 | 5+6 2 |
| | 4+5 2 | 6+7 2 | 6 2 | 6 2 | 3+4 2 | 3+4 2 |
| | 4+5 2 | 4+5 2 | 6 2 | 4+5 2 | 5+6 2 | 3+4 2 |
| | 8 2 | 8 2 | 4+5 2 | 6 2 | 3+4 2 | 5+6 2 |
| | 9 2 | 4+5 2 | 6 2 | 6 2 | 5+6 2 | 3+4 2 |
| | 6+7 2 | 8 2 | 6 2 | 6 2 | 3+4 2 | 5+6 2 |
| | 9 2 | 6+7 2 | 4+5 2 | 4+5 2 | 3+4 2 | 3+4 2 |
| | 4+5 2 | 9 2 | 6 2 | 6 2 | 3+4 2 | 5+6 2 |
| 164.1.79 | 42708 | | 48 | Bi2Pb2Te5 | | |
| | -A | -GM | -H | -K | -L | -M |
| | 8 2 | 8 2 | 6 2 | 6 2 | 3+4 2 | 3+4 2 |
| | 9 2 | 9 2 | 4+5 2 | 4+5 2 | 5+6 2 | 5+6 2 |
| | 9 2 | 8 2 | 6 2 | 6 2 | 3+4 2 | 3+4 2 |
| | 8 2 | 8 2 | 4+5 2 | 4+5 2 | 3+4 2 | 5+6 2 |
| | 8 2 | 9 2 | 6 2 | 6 2 | 5+6 2 | 5+6 2 |
| | 9 2 | 9 2 | 6 2 | 6 2 | 5+6 2 | 3+4 2 |
| | 8 2 | 8 2 | 4+5 2 | 4+5 2 | 3+4 2 | 3+4 2 |
| | 9 2 | 8 2 | 6 2 | 6 2 | 3+4 2 | 5+6 2 |
| | 8 2 | 9 2 | 6 2 | 6 2 | 5+6 2 | 3+4 2 |
| | 9 2 | 9 2 | 4+5 2 | 4+5 2 | 5+6 2 | 5+6 2 |
| | 8 2 | 8 2 | 6 2 | 6 2 | 3+4 2 | 3+4 2 |
| | 9 2 | 8 2 | 6 2 | 6 2 | 5+6 2 | 5+6 2 |
| | 9 2 | 9 2 | 4+5 2 | 4+5 2 | 5+6 2 | 3+4 2 |
| | 6+7 2 | 9 2 | 6 2 | 6 2 | 3+4 2 | 5+6 2 |
| | 9 2 | 6+7 2 | 6 2 | 6 2 | 5+6 2 | 3+4 2 |
| | 4+5 2 | 8 2 | 4+5 2 | 4+5 2 | 3+4 2 | 5+6 2 |
| | 9 2 | 4+5 2 | 6 2 | 6 2 | 3+4 2 | 5+6 2 |
| | 8 2 | 6+7 2 | 6 2 | 6 2 | 5+6 2 | 3+4 2 |
| | 6+7 2 | 9 2 | 4+5 2 | 6 2 | 5+6 2 | 5+6 2 |
| | 8 2 | 4+5 2 | 6 2 | 4+5 2 | 3+4 2 | 5+6 2 |
| | 6+7 2 | 8 2 | 6 2 | 6 2 | 5+6 2 | 5+6 2 |
| | 9 2 | 9 2 | 6 2 | 4+5 2 | 3+4 2 | 3+4 2 |
| | 8 2 | 8 2 | 4+5 2 | 6 2 | 5+6 2 | 5+6 2 |
| | 4+5 2 | 6+7 2 | 6 2 | 6 2 | 3+4 2 | 3+4 2 |
| 164.1.80 | 30372 | | 48 | Bi2Pb2Se5 | | |
| | -A | -GM | -H | -K | -L | -M |
| | 9 2 | 8 2 | 6 2 | 6 2 | 3+4 2 | 5+6 2 |
| | 8 2 | 8 2 | 4+5 2 | 4+5 2 | 3+4 2 | 3+4 2 |
| | 9 2 | 9 2 | 6 2 | 6 2 | 5+6 2 | 5+6 2 |
| | 8 2 | 8 2 | 6 2 | 6 2 | 3+4 2 | 3+4 2 |
| | 8 2 | 9 2 | 4+5 2 | 4+5 2 | 5+6 2 | 3+4 2 |
| | 9 2 | 8 2 | 6 2 | 6 2 | 3+4 2 | 5+6 2 |
| | 8 2 | 9 2 | 6 2 | 6 2 | 5+6 2 | 3+4 2 |
| | 9 2 | 9 2 | 4+5 2 | 4+5 2 | 5+6 2 | 5+6 2 |
| | 8 2 | 8 2 | 6 2 | 6 2 | 3+4 2 | 3+4 2 |

Continued on next column

**TABLE LXXI– continued**

| TQC | ICSD | | NoE | SF | | |
|---|---|---|---|---|---|---|
| | 9 2 | 8 2 | 6 2 | 6 2 | 5+6 2 | 3+4 2 |
| | 8 2 | 9 2 | 4+5 2 | 6 2 | 3+4 2 | 5+6 2 |
| | 9 2 | 8 2 | 6 2 | 6 2 | 5+6 2 | 5+6 2 |
| | 9 2 | 9 2 | 6 2 | 4+5 2 | 3+4 2 | 3+4 2 |
| | 6+7 2 | 4+5 2 | 4+5 2 | 4+5 2 | 5+6 2 | 5+6 2 |
| | 9 2 | 8 2 | 6 2 | 6 2 | 3+4 2 | 5+6 2 |
| | 8 2 | 9 2 | 6 2 | 6 2 | 5+6 2 | 3+4 2 |
| | 9 2 | 6+7 2 | 6 2 | 4+5 2 | 5+6 2 | 5+6 2 |
| | 6+7 2 | 4+5 2 | 4+5 2 | 6 2 | 3+4 2 | 5+6 2 |
| | 4+5 2 | 9 2 | 6 2 | 6 2 | 5+6 2 | 3+4 2 |
| | 6+7 2 | 6+7 2 | 6 2 | 6 2 | 3+4 2 | 5+6 2 |
| | 8 2 | 8 2 | 4+5 2 | 4+5 2 | 3+4 2 | 3+4 2 |
| | 9 2 | 9 2 | 6 2 | 6 2 | 5+6 2 | 5+6 2 |
| | 8 2 | 6+7 2 | 4+5 2 | 4+5 2 | 5+6 2 | 5+6 2 |
| | 4+5 2 | 8 2 | 6 2 | 6 2 | 5+6 2 | 3+4 2 |
| 164.1.81 | 659569 | | 16 | Bi2Mg3 | | |
| | -A | -GM | -H | -K | -L | -M |
| | 8 2 | 8 2 | 4+5 2 | 4+5 2 | 3+4 2 | 5+6 2 |
| | 9 2 | 9 2 | 6 2 | 6 2 | 5+6 2 | 3+4 2 |
| | 9 2 | 8 2 | 4+5 2 | 6 2 | 3+4 2 | 3+4 2 |
| | 8 2 | 8 2 | 6 2 | 4+5 2 | 5+6 2 | 5+6 2 |
| | 9 2 | 9 2 | 6 2 | 6 2 | 3+4 2 | 5+6 2 |
| | 8 2 | 4+5 2 | 6 2 | 6 2 | 5+6 2 | 3+4 2 |
| | 6+7 2 | 8 2 | 4+5 2 | 4+5 2 | 3+4 2 | 5+6 2 |
| | 4+5 2 | 6+7 2 | 6 2 | 6 2 | 5+6 2 | 3+4 2 |
| 164.1.82 | 41791 | | 26 | Bi2Ca1Mn2 | | |
| | -A | -GM | -H | -K | -L | -M |
| | 9 2 | 8 2 | 4+5 2 | 4+5 2 | 3+4 2 | 5+6 2 |
| | 8 2 | 9 2 | 6 2 | 6 2 | 5+6 2 | 3+4 2 |
| | 8 2 | 8 2 | 4+5 2 | 4+5 2 | 3+4 2 | 3+4 2 |
| | 9 2 | 9 2 | 6 2 | 6 2 | 5+6 2 | 5+6 2 |
| | 8 2 | 8 2 | 6 2 | 6 2 | 5+6 2 | 5+6 2 |
| | 4+5 2 | 6+7 2 | 6 2 | 6 2 | 3+4 2 | 3+4 2 |
| | 9 2 | 4+5 2 | 4+5 2 | 4+5 2 | 5+6 2 | 3+4 2 |
| | 6+7 2 | 9 2 | 6 2 | 6 2 | 3+4 2 | 5+6 2 |
| | 9 2 | 8 2 | 6 2 | 6 2 | 3+4 2 | 5+6 2 |
| | 6+7 2 | 4+5 2 | 6 2 | 6 2 | 5+6 2 | 3+4 2 |
| | 9 2 | 8 2 | 4+5 2 | 4+5 2 | 5+6 2 | 3+4 2 |
| | 4+5 2 | 6+7 2 | 6 2 | 6 2 | 3+4 2 | 5+6 2 |
| | 8 2 | 9 2 | 6 2 | 6 2 | 3+4 2 | 3+4 2 |
| 164.1.83 | 30525 | | 66 | Bi1Te1 | | |
| | -A | -GM | -H | -K | -L | -M |
| | 8 2 | 8 2 | 6 2 | 6 2 | 5+6 2 | 5+6 2 |
| | 9 2 | 9 2 | 4+5 2 | 4+5 2 | 3+4 2 | 3+4 2 |
| | 8 2 | 8 2 | 6 2 | 6 2 | 5+6 2 | 5+6 2 |
| | 9 2 | 9 2 | 6 2 | 6 2 | 3+4 2 | 3+4 2 |
| | 9 2 | 8 2 | 4+5 2 | 4+5 2 | 5+6 2 | 5+6 2 |
| | 8 2 | 9 2 | 6 2 | 6 2 | 3+4 2 | 3+4 2 |
| | 9 2 | 8 2 | 6 2 | 6 2 | 3+4 2 | 5+6 2 |
| | 8 2 | 9 2 | 4+5 2 | 4+5 2 | 5+6 2 | 3+4 2 |
| | 9 2 | 8 2 | 6 2 | 6 2 | 3+4 2 | 5+6 2 |
| | 8 2 | 9 2 | 6 2 | 6 2 | 5+6 2 | 3+4 2 |
| | 9 2 | 8 2 | 4+5 2 | 4+5 2 | 3+4 2 | 5+6 2 |
| | 8 2 | 9 2 | 6 2 | 6 2 | 5+6 2 | 3+4 2 |
| | 8 2 | 8 2 | 6 2 | 6 2 | 3+4 2 | 3+4 2 |
| | 9 2 | 9 2 | 4+5 2 | 4+5 2 | 5+6 2 | 5+6 2 |
| | 8 2 | 8 2 | 6 2 | 6 2 | 3+4 2 | 3+4 2 |
| | 9 2 | 9 2 | 6 2 | 6 2 | 5+6 2 | 5+6 2 |
| | 8 2 | 8 2 | 6 2 | 6 2 | 5+6 2 | 3+4 2 |
| | 9 2 | 8 2 | 4+5 2 | 4+5 2 | 3+4 2 | 5+6 2 |
| | 8 2 | 9 2 | 4+5 2 | 4+5 2 | 5+6 2 | 5+6 2 |

Continued on next column

**TABLE LXXI– continued**

| TQC | ICSD | | NoE | SF | | |
|---|---|---|---|---|---|---|
| | 9 2 | 8 2 | 6 2 | 6 2 | 3+4 2 | 3+4 2 |
| | 4+5 2 | 4+5 2 | 6 2 | 6 2 | 5+6 2 | 5+6 2 |
| | 8 2 | 9 2 | 6 2 | 6 2 | 3+4 2 | 3+4 2 |
| | 6+7 2 | 6+7 2 | 6 2 | 6 2 | 5+6 2 | 3+4 2 |
| | 4+5 2 | 4+5 2 | 4+5 2 | 4+5 2 | 5+6 2 | 5+6 2 |
| | 9 2 | 8 2 | 4+5 2 | 4+5 2 | 3+4 2 | 5+6 2 |
| | 6+7 2 | 6+7 2 | 6 2 | 6 2 | 3+4 2 | 3+4 2 |
| | 8 2 | 9 2 | 6 2 | 6 2 | 5+6 2 | 5+6 2 |
| | 9 2 | 8 2 | 6 2 | 4+5 2 | 3+4 2 | 3+4 2 |
| | 4+5 2 | 4+5 2 | 4+5 2 | 6 2 | 5+6 2 | 5+6 2 |
| | 6+7 2 | 6+7 2 | 6 2 | 6 2 | 3+4 2 | 3+4 2 |
| | 8 2 | 9 2 | 6 2 | 6 2 | 3+4 2 | 5+6 2 |
| | 9 2 | 8 2 | 4+5 2 | 4+5 2 | 5+6 2 | 3+4 2 |
| | 9 2 | 9 2 | 6 2 | 6 2 | 3+4 2 | 5+6 2 |
| 164.1.84 | 100654 | | 66 | Bi1Te1 | | |
| | -A | -GM | -H | -K | -L | -M |
| | 9 2 | 8 2 | 6 2 | 6 2 | 3+4 2 | 3+4 2 |
| | 8 2 | 9 2 | 6 2 | 6 2 | 5+6 2 | 5+6 2 |
| | 9 2 | 8 2 | 6 2 | 6 2 | 3+4 2 | 5+6 2 |
| | 8 2 | 9 2 | 4+5 2 | 4+5 2 | 5+6 2 | 3+4 2 |
| | 8 2 | 8 2 | 6 2 | 6 2 | 3+4 2 | 5+6 2 |
| | 9 2 | 9 2 | 6 2 | 6 2 | 5+6 2 | 3+4 2 |
| | 8 2 | 8 2 | 4+5 2 | 4+5 2 | 5+6 2 | 5+6 2 |
| | 9 2 | 9 2 | 6 2 | 6 2 | 3+4 2 | 3+4 2 |
| | 8 2 | 8 2 | 4+5 2 | 4+5 2 | 5+6 2 | 5+6 2 |
| | 9 2 | 9 2 | 6 2 | 6 2 | 3+4 2 | 5+6 2 |
| | 8 2 | 8 2 | 4+5 2 | 4+5 2 | 5+6 2 | 3+4 2 |
| | 9 2 | 9 2 | 6 2 | 6 2 | 3+4 2 | 3+4 2 |
| | 9 2 | 8 2 | 4+5 2 | 4+5 2 | 5+6 2 | 3+4 2 |
| | 8 2 | 9 2 | 6 2 | 6 2 | 3+4 2 | 5+6 2 |
| | 9 2 | 8 2 | 6 2 | 6 2 | 5+6 2 | 3+4 2 |
| | 8 2 | 9 2 | 6 2 | 6 2 | 3+4 2 | 5+6 2 |
| | 8 2 | 8 2 | 4+5 2 | 4+5 2 | 3+4 2 | 3+4 2 |
| | 9 2 | 8 2 | 6 2 | 6 2 | 5+6 2 | 5+6 2 |
| | 9 2 | 8 2 | 6 2 | 6 2 | 3+4 2 | 5+6 2 |
| | 8 2 | 9 2 | 4+5 2 | 4+5 2 | 5+6 2 | 5+6 2 |
| | 6+7 2 | 9 2 | 6 2 | 6 2 | 3+4 2 | 3+4 2 |
| | 4+5 2 | 4+5 2 | 6 2 | 6 2 | 5+6 2 | 3+4 2 |
| | 9 2 | 6+7 2 | 6 2 | 6 2 | 3+4 2 | 3+4 2 |
| | 6+7 2 | 4+5 2 | 4+5 2 | 4+5 2 | 3+4 2 | 5+6 2 |
| | 8 2 | 8 2 | 4+5 2 | 4+5 2 | 5+6 2 | 5+6 2 |
| | 4+5 2 | 6+7 2 | 6 2 | 6 2 | 5+6 2 | 3+4 2 |
| | 9 2 | 9 2 | 6 2 | 6 2 | 3+4 2 | 5+6 2 |
| | 6+7 2 | 8 2 | 4+5 2 | 4+5 2 | 5+6 2 | 3+4 2 |
| | 8 2 | 4+5 2 | 6 2 | 6 2 | 3+4 2 | 5+6 2 |
| | 4+5 2 | 6+7 2 | 6 2 | 6 2 | 5+6 2 | 3+4 2 |
| | 9 2 | 9 2 | 4+5 2 | 6 2 | 5+6 2 | 5+6 2 |
| | 8 2 | 8 2 | 6 2 | 4+5 2 | 3+4 2 | 3+4 2 |
| | 8 2 | 9 2 | 6 2 | 6 2 | 5+6 2 | 5+6 2 |
| 164.1.85 | 99681 | | 170 | Ba6Cl2O12Ru3 | | |
| | -A | -GM | -H | -K | -L | -M |
| | 9 2 | 8 2 | 4+5 2 | 4+5 2 | 3+4 2 | 5+6 2 |
| | 8 2 | 9 2 | 6 2 | 6 2 | 5+6 2 | 3+4 2 |
| | 8 2 | 8 2 | 4+5 2 | 4+5 2 | 5+6 2 | 5+6 2 |
| | 9 2 | 9 2 | 6 2 | 6 2 | 3+4 2 | 3+4 2 |
| | 8 2 | 8 2 | 4+5 2 | 4+5 2 | 5+6 2 | 5+6 2 |
| | 9 2 | 9 2 | 6 2 | 6 2 | 3+4 2 | 3+4 2 |
| | 8 2 | 8 2 | 6 2 | 6 2 | 3+4 2 | 3+4 2 |
| | 9 2 | 9 2 | 6 2 | 6 2 | 5+6 2 | 5+6 2 |
| | 8 2 | 8 2 | 4+5 2 | 4+5 2 | 3+4 2 | 3+4 2 |
| | 4+5 2 | 4+5 2 | 6 2 | 6 2 | 3+4 2 | 3+4 2 |

Continued on next column

**TABLE LXXI– continued**

| TQC | ICSD | | NoE | SF | | |
|---|---|---|---|---|---|---|
| | 8 2 | 8 2 | 6 2 | 6 2 | 5+6 2 | 5+6 2 |
| | 9 2 | 9 2 | 4+5 2 | 4+5 2 | 3+4 2 | 3+4 2 |
| | 6+7 2 | 6+7 2 | 6 2 | 6 2 | 5+6 2 | 5+6 2 |
| | 9 2 | 9 2 | 4+5 2 | 4+5 2 | 5+6 2 | 5+6 2 |
| | 8 2 | 8 2 | 6 2 | 6 2 | 3+4 2 | 3+4 2 |
| | 9 2 | 9 2 | 6 2 | 6 2 | 3+4 2 | 3+4 2 |
| | 4+5 2 | 4+5 2 | 4+5 2 | 4+5 2 | 5+6 2 | 5+6 2 |
| | 6+7 2 | 6+7 2 | 6 2 | 6 2 | 5+6 2 | 5+6 2 |
| | 9 2 | 8 2 | 4+5 2 | 4+5 2 | 3+4 2 | 5+6 2 |
| | 8 2 | 9 2 | 6 2 | 6 2 | 5+6 2 | 3+4 2 |
| | 9 2 | 8 2 | 6 2 | 6 2 | 3+4 2 | 5+6 2 |
| | 8 2 | 9 2 | 4+5 2 | 4+5 2 | 5+6 2 | 3+4 2 |
| | 8 2 | 8 2 | 4+5 2 | 4+5 2 | 3+4 2 | 3+4 2 |
| | 9 2 | 9 2 | 6 2 | 6 2 | 5+6 2 | 5+6 2 |
| | 8 2 | 8 2 | 6 2 | 4+5 2 | 5+6 2 | 5+6 2 |
| | 9 2 | 9 2 | 4+5 2 | 6 2 | 3+4 2 | 3+4 2 |
| | 6+7 2 | 4+5 2 | 6 2 | 6 2 | 3+4 2 | 5+6 2 |
| | 4+5 2 | 6+7 2 | 6 2 | 6 2 | 5+6 2 | 3+4 2 |
| | 9 2 | 8 2 | 6 2 | 6 2 | 3+4 2 | 5+6 2 |
| | 8 2 | 9 2 | 4+5 2 | 4+5 2 | 5+6 2 | 3+4 2 |
| | 4+5 2 | 4+5 2 | 6 2 | 6 2 | 3+4 2 | 3+4 2 |
| | 6+7 2 | 6+7 2 | 4+5 2 | 4+5 2 | 5+6 2 | 5+6 2 |
| | 8 2 | 8 2 | 6 2 | 6 2 | 5+6 2 | 5+6 2 |
| | 9 2 | 9 2 | 6 2 | 6 2 | 3+4 2 | 3+4 2 |
| | 4+5 2 | 4+5 2 | 4+5 2 | 4+5 2 | 5+6 2 | 5+6 2 |
| | 8 2 | 8 2 | 6 2 | 6 2 | 3+4 2 | 3+4 2 |
| | 6+7 2 | 6+7 2 | 6 2 | 6 2 | 5+6 2 | 5+6 2 |
| | 9 2 | 9 2 | 6 2 | 6 2 | 3+4 2 | 3+4 2 |
| | 8 2 | 8 2 | 6 2 | 6 2 | 3+4 2 | 3+4 2 |
| | 4+5 2 | 4+5 2 | 6 2 | 6 2 | 3+4 2 | 3+4 2 |
| | 8 2 | 8 2 | 4+5 2 | 4+5 2 | 3+4 2 | 3+4 2 |
| | 8 2 | 8 2 | 4+5 2 | 4+5 2 | 5+6 2 | 5+6 2 |
| | 9 2 | 9 2 | 6 2 | 6 2 | 3+4 2 | 3+4 2 |
| | 6+7 2 | 6+7 2 | 4+5 2 | 4+5 2 | 3+4 2 | 3+4 2 |
| | 4+5 2 | 4+5 2 | 6 2 | 6 2 | 3+4 2 | 5+6 2 |
| | 8 2 | 8 2 | 6 2 | 6 2 | 5+6 2 | 3+4 2 |
| | 9 2 | 9 2 | 4+5 2 | 4+5 2 | 3+4 2 | 5+6 2 |
| | 9 2 | 8 2 | 6 2 | 6 2 | 5+6 2 | 3+4 2 |
| | 9 2 | 4+5 2 | 4+5 2 | 4+5 2 | 5+6 2 | 5+6 2 |
| | 8 2 | 9 2 | 6 2 | 6 2 | 3+4 2 | 3+4 2 |
| | 4+5 2 | 9 2 | 6 2 | 6 2 | 5+6 2 | 5+6 2 |
| | 6+7 2 | 8 2 | 6 2 | 4+5 2 | 5+6 2 | 3+4 2 |
| | 8 2 | 6+7 2 | 4+5 2 | 6 2 | 3+4 2 | 5+6 2 |
| | 9 2 | 8 2 | 6 2 | 6 2 | 3+4 2 | 5+6 2 |
| | 9 2 | 9 2 | 4+5 2 | 4+5 2 | 5+6 2 | 3+4 2 |
| | 6+7 2 | 4+5 2 | 6 2 | 6 2 | 3+4 2 | 5+6 2 |
| | 8 2 | 6+7 2 | 6 2 | 6 2 | 5+6 2 | 3+4 2 |
| | 4+5 2 | 9 2 | 6 2 | 6 2 | 5+6 2 | 5+6 2 |
| | 8 2 | 8 2 | 6 2 | 6 2 | 3+4 2 | 3+4 2 |
| | 4+5 2 | 9 2 | 4+5 2 | 4+5 2 | 3+4 2 | 3+4 2 |
| | 8 2 | 4+5 2 | 6 2 | 6 2 | 5+6 2 | 5+6 2 |
| | 9 2 | 8 2 | 6 2 | 6 2 | 3+4 2 | 3+4 2 |
| | 8 2 | 8 2 | 4+5 2 | 4+5 2 | 5+6 2 | 5+6 2 |
| | 9 2 | 9 2 | 6 2 | 6 2 | 3+4 2 | 3+4 2 |
| | 6+7 2 | 6+7 2 | 6 2 | 6 2 | 5+6 2 | 5+6 2 |
| | 9 2 | 9 2 | 6 2 | 6 2 | 5+6 2 | 5+6 2 |
| | 6+7 2 | 6+7 2 | 4+5 2 | 4+5 2 | 5+6 2 | 5+6 2 |
| | 8 2 | 8 2 | 6 2 | 6 2 | 3+4 2 | 5+6 2 |
| | 4+5 2 | 4+5 2 | 6 2 | 6 2 | 5+6 2 | 3+4 2 |
| | 9 2 | 9 2 | 4+5 2 | 6 2 | 3+4 2 | 3+4 2 |
| | 9 2 | 9 2 | 6 2 | 4+5 2 | 5+6 2 | 5+6 2 |

Continued on next column

**TABLE LXXI– continued**

| TQC | ICSD | | NoE | SF | | |
|---|---|---|---|---|---|---|
| | 8 2 | 8 2 | 4+5 2 | 4+5 2 | 3+4 2 | 3+4 2 |
| | 9 2 | 9 2 | 6 2 | 6 2 | 3+4 2 | 3+4 2 |
| | 6+7 2 | 8 2 | 6 2 | 6 2 | 5+6 2 | 5+6 2 |
| | 8 2 | 6+7 2 | 6 2 | 6 2 | 5+6 2 | 5+6 2 |
| | 9 2 | 9 2 | 6 2 | 6 2 | 5+6 2 | 5+6 2 |
| | 4+5 2 | 4+5 2 | 4+5 2 | 4+5 2 | 5+6 2 | 5+6 2 |
| | 6+7 2 | 6+7 2 | 6 2 | 6 2 | 3+4 2 | 3+4 2 |
| | 9 2 | 9 2 | 4+5 2 | 4+5 2 | 3+4 2 | 3+4 2 |
| | 8 2 | 8 2 | 6 2 | 6 2 | 5+6 2 | 5+6 2 |
| | 8 2 | 8 2 | 6 2 | 6 2 | 3+4 2 | 3+4 2 |
| | 4+5 2 | 4+5 2 | 4+5 2 | 4+5 2 | 3+4 2 | 3+4 2 |
| | 8 2 | 8 2 | 6 2 | 6 2 | 5+6 2 | 5+6 2 |
| | 9 2 | 9 2 | 6 2 | 6 2 | 3+4 2 | 3+4 2 |
| | 6+7 2 | 6+7 2 | 4+5 2 | 4+5 2 | 3+4 2 | 3+4 2 |
| 164.1.86 | 152482 | | 102 | Ba3Ca1O9Ru2 | | |
| | -A | -GM | -H | -K | -L | -M |
| | 9 2 | 8 2 | 4+5 2 | 4+5 2 | 3+4 2 | 5+6 2 |
| | 8 2 | 9 2 | 6 2 | 6 2 | 5+6 2 | 3+4 2 |
| | 8 2 | 8 2 | 6 2 | 6 2 | 3+4 2 | 3+4 2 |
| | 8 2 | 8 2 | 4+5 2 | 4+5 2 | 5+6 2 | 5+6 2 |
| | 8 2 | 8 2 | 6 2 | 6 2 | 3+4 2 | 3+4 2 |
| | 9 2 | 4+5 2 | 6 2 | 6 2 | 5+6 2 | 5+6 2 |
| | 4+5 2 | 9 2 | 6 2 | 6 2 | 5+6 2 | 3+4 2 |
| | 9 2 | 8 2 | 4+5 2 | 4+5 2 | 3+4 2 | 5+6 2 |
| | 6+7 2 | 8 2 | 6 2 | 6 2 | 5+6 2 | 3+4 2 |
| | 8 2 | 9 2 | 6 2 | 6 2 | 5+6 2 | 3+4 2 |
| | 8 2 | 4+5 2 | 4+5 2 | 4+5 2 | 3+4 2 | 5+6 2 |
| | 4+5 2 | 6+7 2 | 6 2 | 6 2 | 3+4 2 | 5+6 2 |
| | 9 2 | 9 2 | 6 2 | 6 2 | 3+4 2 | 5+6 2 |
| | 8 2 | 8 2 | 4+5 2 | 4+5 2 | 5+6 2 | 3+4 2 |
| | 9 2 | 9 2 | 6 2 | 6 2 | 5+6 2 | 5+6 2 |
| | 6+7 2 | 6+7 2 | 6 2 | 6 2 | 3+4 2 | 5+6 2 |
| | 9 2 | 4+5 2 | 4+5 2 | 4+5 2 | 5+6 2 | 3+4 2 |
| | 4+5 2 | 9 2 | 6 2 | 6 2 | 3+4 2 | 5+6 2 |
| | 6+7 2 | 8 2 | 6 2 | 6 2 | 5+6 2 | 3+4 2 |
| | 8 2 | 6+7 2 | 6 2 | 6 2 | 5+6 2 | 5+6 2 |
| | 9 2 | 9 2 | 4+5 2 | 4+5 2 | 5+6 2 | 5+6 2 |
| | 9 2 | 9 2 | 6 2 | 6 2 | 3+4 2 | 3+4 2 |
| | 6+7 2 | 6+7 2 | 4+5 2 | 6 2 | 5+6 2 | 3+4 2 |
| | 9 2 | 9 2 | 6 2 | 4+5 2 | 3+4 2 | 5+6 2 |
| | 9 2 | 9 2 | 6 2 | 6 2 | 3+4 2 | 3+4 2 |
| | 9 2 | 9 2 | 6 2 | 6 2 | 5+6 2 | 3+4 2 |
| | 6+7 2 | 8 2 | 4+5 2 | 4+5 2 | 3+4 2 | 5+6 2 |
| | 8 2 | 6+7 2 | 6 2 | 6 2 | 5+6 2 | 3+4 2 |
| | 4+5 2 | 4+5 2 | 6 2 | 6 2 | 3+4 2 | 5+6 2 |
| | 8 2 | 8 2 | 4+5 2 | 4+5 2 | 5+6 2 | 5+6 2 |
| | 4+5 2 | 8 2 | 6 2 | 6 2 | 5+6 2 | 3+4 2 |
| | 8 2 | 4+5 2 | 4+5 2 | 4+5 2 | 5+6 2 | 5+6 2 |
| | 9 2 | 8 2 | 6 2 | 6 2 | 3+4 2 | 3+4 2 |
| | 9 2 | 9 2 | 6 2 | 4+5 2 | 3+4 2 | 3+4 2 |
| | 6+7 2 | 6+7 2 | 4+5 2 | 6 2 | 3+4 2 | 5+6 2 |
| | 8 2 | 9 2 | 6 2 | 6 2 | 5+6 2 | 3+4 2 |
| | 9 2 | 9 2 | 6 2 | 6 2 | 3+4 2 | 5+6 2 |
| | 6+7 2 | 6+7 2 | 4+5 2 | 6 2 | 3+4 2 | 3+4 2 |
| | 9 2 | 9 2 | 6 2 | 6 2 | 5+6 2 | 3+4 2 |
| | 9 2 | 8 2 | 6 2 | 4+5 2 | 5+6 2 | 5+6 2 |
| | 8 2 | 4+5 2 | 4+5 2 | 4+5 2 | 5+6 2 | 3+4 2 |
| | 4+5 2 | 9 2 | 6 2 | 6 2 | 3+4 2 | 5+6 2 |
| | 9 2 | 6+7 2 | 6 2 | 6 2 | 3+4 2 | 5+6 2 |
| | 6+7 2 | 9 2 | 4+5 2 | 4+5 2 | 5+6 2 | 3+4 2 |
| | 8 2 | 8 2 | 6 2 | 6 2 | 3+4 2 | 3+4 2 |

Continued on next column

**TABLE LXXI– continued**

| TQC | ICSD | | NoE | SF | | |
|---|---|---|---|---|---|---|
| | 9 2 | 9 2 | 4+5 2 | 4+5 2 | 3+4 2 | 5+6 2 |
| | 9 2 | 9 2 | 6 2 | 6 2 | 5+6 2 | 5+6 2 |
| | 6+7 2 | 6+7 2 | 6 2 | 6 2 | 3+4 2 | 3+4 2 |
| | 8 2 | 8 2 | 6 2 | 6 2 | 5+6 2 | 5+6 2 |
| | 4+5 2 | 4+5 2 | 4+5 2 | 6 2 | 3+4 2 | 3+4 2 |
| | 8 2 | 8 2 | 6 2 | 4+5 2 | 3+4 2 | 5+6 2 |
| 164.1.87 | 409958 | | 18 | Ba1Sn2 | | |
| | -A | -GM | -H | -K | -L | -M |
| | 9 2 | 8 2 | 6 2 | 6 2 | 5+6 2 | 3+4 2 |
| | 8 2 | 9 2 | 6 2 | 6 2 | 3+4 2 | 5+6 2 |
| | 8 2 | 9 2 | 6 2 | 4+5 2 | 3+4 2 | 5+6 2 |
| | 4+5 2 | 6+7 2 | 4+5 2 | 6 2 | 3+4 2 | 5+6 2 |
| | 8 2 | 8 2 | 4+5 2 | 6 2 | 5+6 2 | 5+6 2 |
| | 9 2 | 9 2 | 6 2 | 4+5 2 | 3+4 2 | 3+4 2 |
| | 8 2 | 8 2 | 6 2 | 6 2 | 5+6 2 | 3+4 2 |
| | 4+5 2 | 8 2 | 6 2 | 4+5 2 | 3+4 2 | 5+6 2 |
| | 9 2 | 4+5 2 | 4+5 2 | 6 2 | 5+6 2 | 5+6 2 |
| 164.1.88 | 41107 | | 68 | As4Ge1Te7 | | |
| | -A | -GM | -H | -K | -L | -M |
| | 9 2 | 8 2 | 6 2 | 6 2 | 5+6 2 | 3+4 2 |
| | 8 2 | 9 2 | 6 2 | 6 2 | 3+4 2 | 5+6 2 |
| | 9 2 | 8 2 | 4+5 2 | 4+5 2 | 5+6 2 | 3+4 2 |
| | 8 2 | 8 2 | 6 2 | 6 2 | 5+6 2 | 5+6 2 |
| | 9 2 | 9 2 | 6 2 | 6 2 | 3+4 2 | 3+4 2 |
| | 8 2 | 9 2 | 4+5 2 | 4+5 2 | 3+4 2 | 5+6 2 |
| | 8 2 | 8 2 | 6 2 | 6 2 | 5+6 2 | 3+4 2 |
| | 9 2 | 8 2 | 6 2 | 6 2 | 3+4 2 | 3+4 2 |
| | 9 2 | 9 2 | 4+5 2 | 4+5 2 | 5+6 2 | 5+6 2 |
| | 8 2 | 9 2 | 4+5 2 | 4+5 2 | 3+4 2 | 5+6 2 |
| | 9 2 | 8 2 | 6 2 | 6 2 | 5+6 2 | 3+4 2 |
| | 8 2 | 8 2 | 6 2 | 6 2 | 3+4 2 | 3+4 2 |
| | 8 2 | 9 2 | 6 2 | 6 2 | 3+4 2 | 5+6 2 |
| | 8 2 | 9 2 | 4+5 2 | 4+5 2 | 5+6 2 | 5+6 2 |
| | 9 2 | 8 2 | 6 2 | 6 2 | 3+4 2 | 3+4 2 |
| | 4+5 2 | 6+7 2 | 6 2 | 6 2 | 5+6 2 | 5+6 2 |
| | 8 2 | 9 2 | 6 2 | 6 2 | 3+4 2 | 5+6 2 |
| | 9 2 | 9 2 | 4+5 2 | 4+5 2 | 3+4 2 | 5+6 2 |
| | 8 2 | 8 2 | 6 2 | 6 2 | 5+6 2 | 3+4 2 |
| | 9 2 | 9 2 | 6 2 | 6 2 | 3+4 2 | 5+6 2 |
| | 6+7 2 | 8 2 | 4+5 2 | 4+5 2 | 5+6 2 | 3+4 2 |
| | 4+5 2 | 4+5 2 | 6 2 | 6 2 | 3+4 2 | 3+4 2 |
| | 8 2 | 6+7 2 | 6 2 | 6 2 | 5+6 2 | 5+6 2 |
| | 9 2 | 6+7 2 | 4+5 2 | 4+5 2 | 3+4 2 | 5+6 2 |
| | 6+7 2 | 8 2 | 6 2 | 6 2 | 5+6 2 | 3+4 2 |
| | 4+5 2 | 4+5 2 | 4+5 2 | 4+5 2 | 3+4 2 | 3+4 2 |
| | 9 2 | 9 2 | 6 2 | 6 2 | 5+6 2 | 5+6 2 |
| | 9 2 | 8 2 | 6 2 | 4+5 2 | 5+6 2 | 3+4 2 |
| | 6+7 2 | 4+5 2 | 4+5 2 | 6 2 | 3+4 2 | 5+6 2 |
| | 9 2 | 9 2 | 6 2 | 6 2 | 5+6 2 | 3+4 2 |
| | 8 2 | 9 2 | 6 2 | 6 2 | 5+6 2 | 3+4 2 |
| | 8 2 | 9 2 | 6 2 | 6 2 | 5+6 2 | 5+6 2 |
| | 6+7 2 | 6+7 2 | 4+5 2 | 4+5 2 | 5+6 2 | 5+6 2 |
| | 8 2 | 8 2 | 6 2 | 6 2 | 3+4 2 | 5+6 2 |
| 164.1.89 | 88231 | | 58 | As2Yb1Zn2 | | |
| | -A | -GM | -H | -K | -L | -M |
| | 8 2 | 8 2 | 6 2 | 6 2 | 3+4 2 | 3+4 2 |
| | 9 2 | 9 2 | 6 2 | 6 2 | 5+6 2 | 5+6 2 |
| | 9 2 | 9 2 | 4+5 2 | 4+5 2 | 5+6 2 | 5+6 2 |
| | 6+7 2 | 6+7 2 | 6 2 | 6 2 | 5+6 2 | 5+6 2 |
| | 9 2 | 8 2 | 6 2 | 6 2 | 3+4 2 | 5+6 2 |
| | 8 2 | 9 2 | 4+5 2 | 4+5 2 | 5+6 2 | 3+4 2 |

Continued on next column

**TABLE LXXI– continued**

| TQC | ICSD | | NoE | | SF | |
|---|---|---|---|---|---|---|
| | 4+5 2 | 6+7 2 | 6 2 | 6 2 | 3+4 2 | 3+4 2 |
| | 6+7 2 | 4+5 2 | 4+5 2 | 4+5 2 | 5+6 2 | 5+6 2 |
| | 8 2 | 9 2 | 6 2 | 6 2 | 3+4 2 | 5+6 2 |
| | 9 2 | 8 2 | 6 2 | 6 2 | 5+6 2 | 3+4 2 |
| | 8 2 | 8 2 | 6 2 | 6 2 | 3+4 2 | 3+4 2 |
| | 9 2 | 9 2 | 4+5 2 | 4+5 2 | 5+6 2 | 5+6 2 |
| | 6+7 2 | 4+5 2 | 6 2 | 6 2 | 3+4 2 | 5+6 2 |
| | 9 2 | 8 2 | 6 2 | 6 2 | 5+6 2 | 3+4 2 |
| | 8 2 | 6+7 2 | 6 2 | 6 2 | 5+6 2 | 3+4 2 |
| | 4+5 2 | 9 2 | 4+5 2 | 4+5 2 | 3+4 2 | 5+6 2 |
| | 9 2 | 8 2 | 6 2 | 4+5 2 | 5+6 2 | 3+4 2 |
| | 8 2 | 8 2 | 4+5 2 | 6 2 | 3+4 2 | 5+6 2 |
| | 9 2 | 4+5 2 | 6 2 | 6 2 | 3+4 2 | 5+6 2 |
| | 6+7 2 | 9 2 | 6 2 | 6 2 | 3+4 2 | 3+4 2 |
| | 8 2 | 6+7 2 | 4+5 2 | 4+5 2 | 5+6 2 | 5+6 2 |
| | 4+5 2 | 9 2 | 6 2 | 6 2 | 5+6 2 | 3+4 2 |
| | 9 2 | 9 2 | 6 2 | 6 2 | 5+6 2 | 5+6 2 |
| | 9 2 | 9 2 | 4+5 2 | 4+5 2 | 5+6 2 | 5+6 2 |
| | 6+7 2 | 6+7 2 | 6 2 | 6 2 | 5+6 2 | 5+6 2 |
| | 9 2 | 9 2 | 6 2 | 6 2 | 5+6 2 | 3+4 2 |
| | 8 2 | 9 2 | 4+5 2 | 4+5 2 | 5+6 2 | 5+6 2 |
| | 9 2 | 6+7 2 | 6 2 | 6 2 | 5+6 2 | 5+6 2 |
| | 6+7 2 | 9 2 | 6 2 | 6 2 | 5+6 2 | 5+6 2 |
| 164.1.90 | 39512 | | 36 | | As2Mn2Zn1 | |
| | -A | -GM | -H | -K | -L | -M |
| | 8 2 | 8 2 | 4+5 2 | 4+5 2 | 3+4 2 | 5+6 2 |
| | 9 2 | 9 2 | 6 2 | 6 2 | 5+6 2 | 3+4 2 |
| | 4+5 2 | 4+5 2 | 6 2 | 6 2 | 3+4 2 | 3+4 2 |
| | 8 2 | 8 2 | 4+5 2 | 4+5 2 | 3+4 2 | 3+4 2 |
| | 8 2 | 8 2 | 4+5 2 | 4+5 2 | 3+4 2 | 3+4 2 |
| | 4+5 2 | 4+5 2 | 6 2 | 6 2 | 3+4 2 | 3+4 2 |
| | 8 2 | 8 2 | 6 2 | 6 2 | 3+4 2 | 3+4 2 |
| | 8 2 | 8 2 | 6 2 | 6 2 | 3+4 2 | 3+4 2 |
| | 9 2 | 9 2 | 4+5 2 | 4+5 2 | 5+6 2 | 5+6 2 |
| | 8 2 | 6+7 2 | 6 2 | 6 2 | 5+6 2 | 5+6 2 |
| | 4+5 2 | 8 2 | 6 2 | 6 2 | 5+6 2 | 5+6 2 |
| | 9 2 | 9 2 | 4+5 2 | 4+5 2 | 3+4 2 | 3+4 2 |
| | 6+7 2 | 4+5 2 | 6 2 | 6 2 | 3+4 2 | 3+4 2 |
| | 9 2 | 8 2 | 6 2 | 6 2 | 5+6 2 | 5+6 2 |
| | 6+7 2 | 4+5 2 | 6 2 | 6 2 | 3+4 2 | 3+4 2 |
| | 9 2 | 9 2 | 4+5 2 | 4+5 2 | 3+4 2 | 3+4 2 |
| | 4+5 2 | 6+7 2 | 6 2 | 6 2 | 5+6 2 | 5+6 2 |
| | 8 2 | 8 2 | 6 2 | 6 2 | 3+4 2 | 5+6 2 |
| 164.1.91 | 49020 | | 34 | | As2Mn2Sr1 | |
| | -A | -GM | -H | -K | -L | -M |
| | 8 2 | 8 2 | 6 2 | 6 2 | 3+4 2 | 3+4 2 |
| | 9 2 | 9 2 | 6 2 | 6 2 | 5+6 2 | 5+6 2 |
| | 9 2 | 9 2 | 4+5 2 | 4+5 2 | 5+6 2 | 5+6 2 |
| | 6+7 2 | 6+7 2 | 6 2 | 6 2 | 5+6 2 | 5+6 2 |
| | 9 2 | 8 2 | 6 2 | 6 2 | 3+4 2 | 5+6 2 |
| | 8 2 | 9 2 | 4+5 2 | 4+5 2 | 5+6 2 | 3+4 2 |
| | 8 2 | 8 2 | 4+5 2 | 4+5 2 | 3+4 2 | 3+4 2 |
| | 9 2 | 9 2 | 6 2 | 6 2 | 5+6 2 | 5+6 2 |
| | 8 2 | 8 2 | 6 2 | 6 2 | 5+6 2 | 3+4 2 |
| | 4+5 2 | 6+7 2 | 6 2 | 6 2 | 3+4 2 | 5+6 2 |
| | 9 2 | 4+5 2 | 4+5 2 | 6 2 | 5+6 2 | 3+4 2 |
| | 6+7 2 | 9 2 | 6 2 | 4+5 2 | 3+4 2 | 5+6 2 |
| | 9 2 | 8 2 | 6 2 | 6 2 | 3+4 2 | 5+6 2 |
| | 6+7 2 | 4+5 2 | 6 2 | 6 2 | 5+6 2 | 3+4 2 |
| | 9 2 | 8 2 | 4+5 2 | 4+5 2 | 5+6 2 | 3+4 2 |
| | 4+5 2 | 9 2 | 6 2 | 6 2 | 3+4 2 | 5+6 2 |

Continued on next column

**TABLE LXXI– continued**

| TQC | ICSD | | NoE | | SF | |
|---|---|---|---|---|---|---|
| | 8 2 | 6+7 2 | 6 2 | 6 2 | 3+4 2 | 3+4 2 |
| 164.1.92 | 41793 | | 34 | | As2Mn2Sr1 | |
| | -A | -GM | -H | -K | -L | -M |
| | 8 2 | 8 2 | 6 2 | 6 2 | 3+4 2 | 3+4 2 |
| | 9 2 | 9 2 | 6 2 | 6 2 | 5+6 2 | 5+6 2 |
| | 9 2 | 9 2 | 4+5 2 | 4+5 2 | 5+6 2 | 5+6 2 |
| | 6+7 2 | 6+7 2 | 6 2 | 6 2 | 5+6 2 | 5+6 2 |
| | 9 2 | 8 2 | 6 2 | 6 2 | 3+4 2 | 5+6 2 |
| | 8 2 | 9 2 | 4+5 2 | 4+5 2 | 5+6 2 | 3+4 2 |
| | 8 2 | 8 2 | 4+5 2 | 4+5 2 | 3+4 2 | 3+4 2 |
| | 9 2 | 9 2 | 6 2 | 6 2 | 5+6 2 | 5+6 2 |
| | 8 2 | 8 2 | 6 2 | 6 2 | 5+6 2 | 3+4 2 |
| | 4+5 2 | 6+7 2 | 6 2 | 6 2 | 3+4 2 | 5+6 2 |
| | 9 2 | 4+5 2 | 4+5 2 | 6 2 | 5+6 2 | 3+4 2 |
| | 6+7 2 | 9 2 | 6 2 | 4+5 2 | 3+4 2 | 5+6 2 |
| | 9 2 | 8 2 | 6 2 | 6 2 | 3+4 2 | 3+4 2 |
| | 6+7 2 | 4+5 2 | 6 2 | 6 2 | 5+6 2 | 5+6 2 |
| | 9 2 | 8 2 | 4+5 2 | 4+5 2 | 5+6 2 | 3+4 2 |
| | 4+5 2 | 9 2 | 6 2 | 6 2 | 3+4 2 | 5+6 2 |
| | 8 2 | 6+7 2 | 6 2 | 6 2 | 3+4 2 | 3+4 2 |
| 164.1.93 | 63174 | | 76 | | As2Ge5Te8 | |
| | -A | -GM | -H | -K | -L | -M |
| | 8 2 | 8 2 | 4+5 2 | 4+5 2 | 3+4 2 | 5+6 2 |
| | 9 2 | 9 2 | 6 2 | 6 2 | 5+6 2 | 3+4 2 |
| | 8 2 | 8 2 | 6 2 | 6 2 | 5+6 2 | 5+6 2 |
| | 9 2 | 9 2 | 6 2 | 6 2 | 3+4 2 | 3+4 2 |
| | 8 2 | 8 2 | 6 2 | 6 2 | 3+4 2 | 5+6 2 |
| | 9 2 | 9 2 | 4+5 2 | 4+5 2 | 5+6 2 | 5+6 2 |
| | 8 2 | 8 2 | 4+5 2 | 4+5 2 | 5+6 2 | 3+4 2 |
| | 9 2 | 9 2 | 6 2 | 6 2 | 3+4 2 | 3+4 2 |
| | 8 2 | 8 2 | 4+5 2 | 4+5 2 | 3+4 2 | 3+4 2 |
| | 9 2 | 9 2 | 6 2 | 6 2 | 5+6 2 | 5+6 2 |
| | 8 2 | 8 2 | 6 2 | 6 2 | 3+4 2 | 3+4 2 |
| | 9 2 | 9 2 | 6 2 | 6 2 | 5+6 2 | 5+6 2 |
| | 8 2 | 8 2 | 4+5 2 | 4+5 2 | 3+4 2 | 3+4 2 |
| | 9 2 | 9 2 | 6 2 | 6 2 | 5+6 2 | 5+6 2 |
| | 8 2 | 8 2 | 6 2 | 6 2 | 3+4 2 | 3+4 2 |
| | 9 2 | 8 2 | 6 2 | 6 2 | 5+6 2 | 5+6 2 |
| | 8 2 | 9 2 | 4+5 2 | 6 2 | 3+4 2 | 3+4 2 |
| | 9 2 | 9 2 | 6 2 | 4+5 2 | 5+6 2 | 5+6 2 |
| | 9 2 | 8 2 | 6 2 | 6 2 | 3+4 2 | 3+4 2 |
| | 8 2 | 8 2 | 4+5 2 | 4+5 2 | 5+6 2 | 5+6 2 |
| | 9 2 | 9 2 | 6 2 | 6 2 | 5+6 2 | 5+6 2 |
| | 9 2 | 8 2 | 4+5 2 | 4+5 2 | 3+4 2 | 3+4 2 |
| | 8 2 | 9 2 | 6 2 | 6 2 | 5+6 2 | 5+6 2 |
| | 8 2 | 9 2 | 6 2 | 6 2 | 3+4 2 | 5+6 2 |
| | 9 2 | 8 2 | 4+5 2 | 4+5 2 | 5+6 2 | 3+4 2 |
| | 6+7 2 | 4+5 2 | 6 2 | 6 2 | 3+4 2 | 3+4 2 |
| | 4+5 2 | 6+7 2 | 6 2 | 6 2 | 3+4 2 | 5+6 2 |
| | 8 2 | 6+7 2 | 4+5 2 | 4+5 2 | 5+6 2 | 3+4 2 |
| | 6+7 2 | 4+5 2 | 6 2 | 6 2 | 3+4 2 | 5+6 2 |
| | 6+7 2 | 4+5 2 | 6 2 | 6 2 | 5+6 2 | 5+6 2 |
| | 4+5 2 | 6+7 2 | 4+5 2 | 4+5 2 | 5+6 2 | 3+4 2 |
| | 4+5 2 | 8 2 | 6 2 | 6 2 | 3+4 2 | 5+6 2 |
| | 6+7 2 | 6+7 2 | 6 2 | 6 2 | 5+6 2 | 3+4 2 |
| | 4+5 2 | 9 2 | 6 2 | 4+5 2 | 3+4 2 | 5+6 2 |
| | 9 2 | 4+5 2 | 4+5 2 | 6 2 | 3+4 2 | 3+4 2 |
| | 8 2 | 8 2 | 6 2 | 6 2 | 5+6 2 | 3+4 2 |
| | 9 2 | 9 2 | 4+5 2 | 4+5 2 | 3+4 2 | 5+6 2 |
| | 8 2 | 8 2 | 6 2 | 6 2 | 5+6 2 | 3+4 2 |
| 164.1.94 | 41792 | | 26 | | As2Ca1Mn2 | |

Continued on next column

**TABLE LXXI– continued**

| TQC | ICSD | NoE | SF | | |
|---|---|---|---|---|---|
| | -A | -GM | -H | -K | -L | -M |
| | 9 2 | 8 2 | 4+5 2 | 4+5 2 | 3+4 2 | 5+6 2 |
| | 8 2 | 9 2 | 6 2 | 6 2 | 5+6 2 | 3+4 2 |
| | 8 2 | 8 2 | 4+5 2 | 4+5 2 | 3+4 2 | 3+4 2 |
| | 9 2 | 9 2 | 6 2 | 6 2 | 5+6 2 | 5+6 2 |
| | 8 2 | 8 2 | 6 2 | 6 2 | 5+6 2 | 5+6 2 |
| | 4+5 2 | 6+7 2 | 6 2 | 6 2 | 5+6 2 | 3+4 2 |
| | 9 2 | 4+5 2 | 4+5 2 | 4+5 2 | 3+4 2 | 3+4 2 |
| | 6+7 2 | 9 2 | 6 2 | 6 2 | 3+4 2 | 5+6 2 |
| | 9 2 | 8 2 | 6 2 | 6 2 | 3+4 2 | 5+6 2 |
| | 6+7 2 | 4+5 2 | 6 2 | 6 2 | 5+6 2 | 3+4 2 |
| | 9 2 | 8 2 | 4+5 2 | 4+5 2 | 5+6 2 | 3+4 2 |
| | 4+5 2 | 9 2 | 6 2 | 6 2 | 3+4 2 | 5+6 2 |
| | 8 2 | 6+7 2 | 6 2 | 6 2 | 3+4 2 | 3+4 2 |
| 164.1.95 | 411054 | 38 | Al2Si2Yb1 | | | |
| | -A | -GM | -H | -K | -L | -M |
| | 8 2 | 8 2 | 6 2 | 6 2 | 3+4 2 | 3+4 2 |
| | 9 2 | 9 2 | 6 2 | 6 2 | 5+6 2 | 5+6 2 |
| | 9 2 | 9 2 | 4+5 2 | 4+5 2 | 5+6 2 | 5+6 2 |
| | 6+7 2 | 6+7 2 | 6 2 | 6 2 | 5+6 2 | 5+6 2 |
| | 9 2 | 8 2 | 6 2 | 6 2 | 3+4 2 | 5+6 2 |
| | 8 2 | 9 2 | 4+5 2 | 4+5 2 | 5+6 2 | 3+4 2 |
| | 9 2 | 8 2 | 6 2 | 4+5 2 | 5+6 2 | 3+4 2 |
| | 8 2 | 8 2 | 4+5 2 | 6 2 | 3+4 2 | 5+6 2 |
| | 6+7 2 | 4+5 2 | 6 2 | 6 2 | 3+4 2 | 5+6 2 |
| | 9 2 | 6+7 2 | 6 2 | 6 2 | 3+4 2 | 5+6 2 |
| | 9 2 | 9 2 | 4+5 2 | 4+5 2 | 5+6 2 | 3+4 2 |
| | 9 2 | 9 2 | 6 2 | 6 2 | 5+6 2 | 3+4 2 |
| | 6+7 2 | 9 2 | 6 2 | 6 2 | 5+6 2 | 5+6 2 |
| | 4+5 2 | 6+7 2 | 6 2 | 4+5 2 | 5+6 2 | 5+6 2 |
| | 8 2 | 9 2 | 4+5 2 | 6 2 | 5+6 2 | 5+6 2 |
| | 9 2 | 9 2 | 6 2 | 6 2 | 5+6 2 | 3+4 2 |
| | 9 2 | 9 2 | 4+5 2 | 4+5 2 | 5+6 2 | 5+6 2 |
| | 6+7 2 | 6+7 2 | 6 2 | 6 2 | 5+6 2 | 5+6 2 |
| | 9 2 | 9 2 | 6 2 | 6 2 | 5+6 2 | 5+6 2 |
| 164.1.96 | 411056 | 38 | Al2Ge2Yb1 | | | |
| | -A | -GM | -H | -K | -L | -M |
| | 8 2 | 8 2 | 6 2 | 6 2 | 3+4 2 | 3+4 2 |
| | 9 2 | 9 2 | 6 2 | 6 2 | 5+6 2 | 5+6 2 |
| | 9 2 | 9 2 | 4+5 2 | 4+5 2 | 5+6 2 | 5+6 2 |
| | 6+7 2 | 6+7 2 | 6 2 | 6 2 | 5+6 2 | 5+6 2 |
| | 9 2 | 8 2 | 6 2 | 6 2 | 3+4 2 | 5+6 2 |
| | 8 2 | 9 2 | 4+5 2 | 4+5 2 | 5+6 2 | 3+4 2 |
| | 9 2 | 8 2 | 6 2 | 4+5 2 | 5+6 2 | 3+4 2 |
| | 8 2 | 8 2 | 4+5 2 | 6 2 | 3+4 2 | 5+6 2 |
| | 6+7 2 | 4+5 2 | 6 2 | 6 2 | 3+4 2 | 5+6 2 |
| | 9 2 | 6+7 2 | 6 2 | 6 2 | 3+4 2 | 3+4 2 |
| | 9 2 | 9 2 | 4+5 2 | 4+5 2 | 5+6 2 | 5+6 2 |
| | 9 2 | 9 2 | 6 2 | 6 2 | 5+6 2 | 3+4 2 |
| | 6+7 2 | 9 2 | 6 2 | 6 2 | 5+6 2 | 5+6 2 |
| | 8 2 | 9 2 | 6 2 | 4+5 2 | 5+6 2 | 5+6 2 |
| | 4+5 2 | 6+7 2 | 4+5 2 | 6 2 | 5+6 2 | 5+6 2 |
| | 9 2 | 9 2 | 6 2 | 6 2 | 5+6 2 | 3+4 2 |
| | 9 2 | 9 2 | 4+5 2 | 4+5 2 | 5+6 2 | 5+6 2 |
| | 6+7 2 | 6+7 2 | 6 2 | 6 2 | 5+6 2 | 5+6 2 |
| | 9 2 | 9 2 | 6 2 | 6 2 | 5+6 2 | 5+6 2 |
| 164.1.97 | 409458 | 38 | Al2Ge2Yb1 | | | |
| | -A | -GM | -H | -K | -L | -M |
| | 8 2 | 8 2 | 6 2 | 6 2 | 3+4 2 | 3+4 2 |
| | 9 2 | 9 2 | 6 2 | 6 2 | 5+6 2 | 5+6 2 |
| | 9 2 | 9 2 | 4+5 2 | 4+5 2 | 5+6 2 | 5+6 2 |

Continued on next column

**TABLE LXXI– continued**

| TQC | ICSD | NoE | SF | | |
|---|---|---|---|---|---|
| | 6+7 2 | 6+7 2 | 6 2 | 6 2 | 5+6 2 | 5+6 2 |
| | 9 2 | 8 2 | 6 2 | 6 2 | 3+4 2 | 5+6 2 |
| | 8 2 | 9 2 | 4+5 2 | 4+5 2 | 5+6 2 | 3+4 2 |
| | 9 2 | 8 2 | 6 2 | 4+5 2 | 5+6 2 | 3+4 2 |
| | 8 2 | 8 2 | 4+5 2 | 6 2 | 3+4 2 | 5+6 2 |
| | 6+7 2 | 4+5 2 | 6 2 | 6 2 | 3+4 2 | 5+6 2 |
| | 9 2 | 6+7 2 | 6 2 | 6 2 | 3+4 2 | 3+4 2 |
| | 9 2 | 9 2 | 4+5 2 | 4+5 2 | 5+6 2 | 5+6 2 |
| | 9 2 | 9 2 | 6 2 | 6 2 | 5+6 2 | 3+4 2 |
| | 6+7 2 | 9 2 | 6 2 | 6 2 | 5+6 2 | 5+6 2 |
| | 8 2 | 9 2 | 6 2 | 4+5 2 | 5+6 2 | 5+6 2 |
| | 4+5 2 | 6+7 2 | 4+5 2 | 6 2 | 5+6 2 | 5+6 2 |
| | 9 2 | 9 2 | 6 2 | 6 2 | 5+6 2 | 3+4 2 |
| | 9 2 | 9 2 | 4+5 2 | 4+5 2 | 5+6 2 | 5+6 2 |
| | 6+7 2 | 6+7 2 | 6 2 | 6 2 | 5+6 2 | 5+6 2 |
| | 9 2 | 9 2 | 6 2 | 6 2 | 5+6 2 | 5+6 2 |
| 164.1.98 | 412121 | 28 | Al2Ge2Nd1 | | | |
| | -A | -GM | -H | -K | -L | -M |
| | 8 2 | 8 2 | 6 2 | 6 2 | 3+4 2 | 3+4 2 |
| | 9 2 | 9 2 | 6 2 | 6 2 | 5+6 2 | 5+6 2 |
| | 9 2 | 9 2 | 4+5 2 | 4+5 2 | 5+6 2 | 5+6 2 |
| | 6+7 2 | 6+7 2 | 6 2 | 6 2 | 5+6 2 | 5+6 2 |
| | 9 2 | 8 2 | 6 2 | 6 2 | 3+4 2 | 5+6 2 |
| | 8 2 | 9 2 | 4+5 2 | 4+5 2 | 5+6 2 | 3+4 2 |
| | 9 2 | 8 2 | 6 2 | 4+5 2 | 5+6 2 | 3+4 2 |
| | 8 2 | 8 2 | 4+5 2 | 6 2 | 3+4 2 | 5+6 2 |
| | 9 2 | 4+5 2 | 6 2 | 6 2 | 3+4 2 | 5+6 2 |
| | 6+7 2 | 9 2 | 6 2 | 6 2 | 3+4 2 | 3+4 2 |
| | 8 2 | 6+7 2 | 4+5 2 | 4+5 2 | 5+6 2 | 5+6 2 |
| | 4+5 2 | 9 2 | 6 2 | 6 2 | 5+6 2 | 3+4 2 |
| | 8 2 | 9 2 | 6 2 | 6 2 | 3+4 2 | 3+4 2 |
| | 9 2 | 9 2 | 4+5 2 | 4+5 2 | 5+6 2 | 5+6 2 |
| 164.1.99 | 160311 | 36 | Al2Er1Ge2 | | | |
| | -A | -GM | -H | -K | -L | -M |
| | 8 2 | 8 2 | 6 2 | 6 2 | 3+4 2 | 3+4 2 |
| | 9 2 | 9 2 | 6 2 | 6 2 | 5+6 2 | 5+6 2 |
| | 9 2 | 9 2 | 4+5 2 | 4+5 2 | 5+6 2 | 5+6 2 |
| | 6+7 2 | 6+7 2 | 6 2 | 6 2 | 5+6 2 | 5+6 2 |
| | 9 2 | 8 2 | 6 2 | 6 2 | 3+4 2 | 5+6 2 |
| | 8 2 | 9 2 | 4+5 2 | 4+5 2 | 5+6 2 | 3+4 2 |
| | 9 2 | 8 2 | 6 2 | 4+5 2 | 5+6 2 | 3+4 2 |
| | 8 2 | 8 2 | 4+5 2 | 6 2 | 3+4 2 | 5+6 2 |
| | 6+7 2 | 4+5 2 | 6 2 | 6 2 | 3+4 2 | 5+6 2 |
| | 9 2 | 6+7 2 | 6 2 | 6 2 | 3+4 2 | 5+6 2 |
| | 8 2 | 9 2 | 4+5 2 | 4+5 2 | 5+6 2 | 3+4 2 |
| | 4+5 2 | 9 2 | 6 2 | 6 2 | 5+6 2 | 3+4 2 |
| | 9 2 | 9 2 | 6 2 | 6 2 | 5+6 2 | 5+6 2 |
| | 9 2 | 9 2 | 4+5 2 | 4+5 2 | 5+6 2 | 5+6 2 |
| | 6+7 2 | 6+7 2 | 6 2 | 6 2 | 5+6 2 | 5+6 2 |
| | 9 2 | 9 2 | 6 2 | 6 2 | 3+4 2 | 3+4 2 |
| | 9 2 | 9 2 | 4+5 2 | 4+5 2 | 5+6 2 | 5+6 2 |
| | 6+7 2 | 6+7 2 | 6 2 | 6 2 | 5+6 2 | 5+6 2 |
| 164.1.100 | 50976 | 26 | Al2Ce1Ge2 | | | |
| | -A | -GM | -H | -K | -L | -M |
| | 8 2 | 8 2 | 6 2 | 6 2 | 3+4 2 | 3+4 2 |
| | 9 2 | 9 2 | 6 2 | 6 2 | 5+6 2 | 5+6 2 |
| | 9 2 | 9 2 | 4+5 2 | 4+5 2 | 5+6 2 | 5+6 2 |
| | 6+7 2 | 6+7 2 | 6 2 | 6 2 | 5+6 2 | 5+6 2 |
| | 9 2 | 8 2 | 6 2 | 6 2 | 3+4 2 | 5+6 2 |
| | 8 2 | 9 2 | 4+5 2 | 4+5 2 | 5+6 2 | 3+4 2 |
| | 9 2 | 8 2 | 6 2 | 4+5 2 | 5+6 2 | 3+4 2 |

Continued on next column

**TABLE LXXI– continued**

| TQC | ICSD | | NoE | SF | | |
|---|---|---|---|---|---|---|
| | 8 2 | 8 2 | 4+5 2 | 6 2 | 3+4 2 | 5+6 2 |
| | 9 2 | 4+5 2 | 6 2 | 6 2 | 3+4 2 | 5+6 2 |
| | 6+7 2 | 9 2 | 6 2 | 6 2 | 3+4 2 | 3+4 2 |
| | 8 2 | 9 2 | 4+5 2 | 4+5 2 | 5+6 2 | 5+6 2 |
| | 4+5 2 | 6+7 2 | 6 2 | 6 2 | 5+6 2 | 3+4 2 |
| | 8 2 | 9 2 | 6 2 | 6 2 | 3+4 2 | 3+4 2 |
| 164.1.101 | 94502 | | 44 | Al1La1Si2 | | |
| | -A | -GM | -H | -K | -L | -M |
| | 8 2 | 8 2 | 6 2 | 6 2 | 3+4 2 | 3+4 2 |
| | 9 2 | 9 2 | 6 2 | 6 2 | 5+6 2 | 5+6 2 |
| | 8 2 | 8 2 | 6 2 | 6 2 | 3+4 2 | 3+4 2 |
| | 9 2 | 9 2 | 6 2 | 6 2 | 5+6 2 | 5+6 2 |
| | 8 2 | 8 2 | 4+5 2 | 4+5 2 | 3+4 2 | 3+4 2 |
| | 9 2 | 9 2 | 6 2 | 6 2 | 5+6 2 | 5+6 2 |
| | 6+7 2 | 6+7 2 | 4+5 2 | 4+5 2 | 3+4 2 | 3+4 2 |
| | 4+5 2 | 4+5 2 | 6 2 | 6 2 | 5+6 2 | 5+6 2 |
| | 8 2 | 8 2 | 6 2 | 6 2 | 5+6 2 | 5+6 2 |
| | 9 2 | 8 2 | 4+5 2 | 4+5 2 | 3+4 2 | 5+6 2 |
| | 8 2 | 9 2 | 6 2 | 6 2 | 5+6 2 | 3+4 2 |
| | 9 2 | 8 2 | 4+5 2 | 4+5 2 | 3+4 2 | 3+4 2 |
| | 8 2 | 9 2 | 6 2 | 6 2 | 5+6 2 | 3+4 2 |
| | 4+5 2 | 8 2 | 4+5 2 | 4+5 2 | 3+4 2 | 5+6 2 |
| | 9 2 | 4+5 2 | 6 2 | 6 2 | 5+6 2 | 5+6 2 |
| | 8 2 | 8 2 | 6 2 | 6 2 | 3+4 2 | 5+6 2 |
| | 9 2 | 4+5 2 | 6 2 | 6 2 | 3+4 2 | 5+6 2 |
| | 6+7 2 | 9 2 | 6 2 | 6 2 | 5+6 2 | 3+4 2 |
| | 9 2 | 6+7 2 | 4+5 2 | 4+5 2 | 3+4 2 | 3+4 2 |
| | 8 2 | 9 2 | 6 2 | 6 2 | 3+4 2 | 5+6 2 |
| | 4+5 2 | 9 2 | 4+5 2 | 4+5 2 | 5+6 2 | 3+4 2 |
| | 9 2 | 8 2 | 4+5 2 | 4+5 2 | 5+6 2 | 3+4 2 |
| 164.1.102 | 94501 | | 44 | Al1La1Si2 | | |
| | -A | -GM | -H | -K | -L | -M |
| | 8 2 | 8 2 | 6 2 | 6 2 | 3+4 2 | 3+4 2 |
| | 9 2 | 9 2 | 6 2 | 6 2 | 5+6 2 | 5+6 2 |
| | 8 2 | 8 2 | 6 2 | 6 2 | 3+4 2 | 3+4 2 |
| | 9 2 | 9 2 | 6 2 | 6 2 | 5+6 2 | 5+6 2 |
| | 8 2 | 8 2 | 4+5 2 | 4+5 2 | 3+4 2 | 3+4 2 |
| | 9 2 | 9 2 | 6 2 | 6 2 | 5+6 2 | 5+6 2 |
| | 6+7 2 | 6+7 2 | 4+5 2 | 4+5 2 | 3+4 2 | 3+4 2 |
| | 4+5 2 | 4+5 2 | 6 2 | 6 2 | 5+6 2 | 5+6 2 |
| | 8 2 | 8 2 | 6 2 | 6 2 | 5+6 2 | 5+6 2 |
| | 9 2 | 8 2 | 4+5 2 | 4+5 2 | 3+4 2 | 5+6 2 |
| | 8 2 | 9 2 | 6 2 | 6 2 | 5+6 2 | 3+4 2 |
| | 9 2 | 8 2 | 4+5 2 | 4+5 2 | 3+4 2 | 3+4 2 |
| | 8 2 | 9 2 | 6 2 | 6 2 | 5+6 2 | 3+4 2 |
| | 4+5 2 | 8 2 | 4+5 2 | 4+5 2 | 3+4 2 | 5+6 2 |
| | 9 2 | 4+5 2 | 6 2 | 6 2 | 5+6 2 | 5+6 2 |
| | 8 2 | 8 2 | 6 2 | 6 2 | 3+4 2 | 5+6 2 |
| | 9 2 | 4+5 2 | 6 2 | 6 2 | 3+4 2 | 5+6 2 |
| | 6+7 2 | 9 2 | 6 2 | 6 2 | 5+6 2 | 3+4 2 |
| | 9 2 | 6+7 2 | 4+5 2 | 4+5 2 | 3+4 2 | 3+4 2 |
| | 8 2 | 9 2 | 6 2 | 6 2 | 3+4 2 | 5+6 2 |
| | 4+5 2 | 9 2 | 4+5 2 | 4+5 2 | 5+6 2 | 3+4 2 |
| | 8 2 | 8 2 | 4+5 2 | 4+5 2 | 5+6 2 | 3+4 2 |

The End

TABLE LXXII: The computed irreducible representations for the valence bands for the compounds of SEBR and NLC in space group 165. See the list of tables.

| TQC | ICSD | | NoE | SF | | |
|---|---|---|---|---|---|---|
| 165.1.1 | 154809 | | 84 | H3Y1 | | |
| | -A | -GM | -H | -K | -L | -M |
| | 5+6 4 | 8 2 | 4+4 2 | 4+5 2 | 2+2 4 | 5+6 2 |
| | 4+4 4 | 9 2 | 5+5 2 | 6 2 | 2+2 4 | 3+4 2 |
| | 5+6 4 | 4+5 2 | 6+6 4 | 6 2 | 2+2 4 | 3+4 2 |
| | 5+6 4 | 6+7 2 | 6+6 4 | 4+5 2 | 2+2 4 | 5+6 2 |
| | 4+4 4 | 8 2 | 6+6 4 | 6 2 | 2+2 4 | 5+6 2 |
| | 5+6 4 | 9 2 | 5+5 2 | 6 2 | 2+2 4 | 3+4 2 |
| | 5+6 4 | 8 2 | 6+6 4 | 6 2 | 2+2 4 | 5+6 2 |
| | 5+6 4 | 9 2 | 4+4 2 | 4+5 2 | 2+2 4 | 3+4 2 |
| | 4+4 4 | 4+5 2 | 4+4 2 | 6 2 | 2+2 4 | 3+4 2 |
| | 5+6 4 | 6+7 2 | 5+5 2 | 6 2 | 2+2 4 | 5+6 2 |
| | 5+6 4 | 8 2 | 6+6 4 | 4+5 2 | 2+2 4 | 5+6 2 |
| | 4+4 4 | 9 2 | 6+6 4 | 6 2 | 2+2 4 | 3+4 2 |
| | 5+6 4 | 9 2 | 4+4 2 | 4+5 2 | 2+2 4 | 5+6 2 |
| | 4+4 4 | 8 2 | 5+5 2 | 6 2 | 2+2 4 | 3+4 2 |
| | 5+6 4 | 8 2 | 6+6 4 | 6 2 | 2+2 4 | 5+6 2 |
| | 5+6 4 | 9 2 | 6+6 4 | 4+5 2 | 2+2 4 | 3+4 2 |
| | 5+6 4 | 8 2 | 4+4 2 | 6 2 | 2+2 4 | 5+6 2 |
| | 4+4 4 | 8 2 | 5+5 2 | 6 2 | 2+2 4 | 3+4 2 |
| | 5+6 4 | 4+5 2 | 6+6 4 | 4+5 2 | 2+2 4 | 5+6 2 |
| | 5+6 4 | 6+7 2 | 6+6 4 | 6 2 | 2+2 4 | 3+4 2 |
| | 5+6 4 | 9 2 | 5+5 2 | 6 2 | 2+2 4 | 5+6 2 |
| | | 9 2 | 6+6 4 | 4+5 2 | | 3+4 2 |
| | | 4+5 2 | 4+4 2 | 6 2 | | 5+6 2 |
| | | 6+7 2 | 6+6 4 | 6 2 | | 3+4 2 |
| | | 8 2 | 6+6 4 | 4+5 2 | | 5+6 2 |
| | | 9 2 | 4+4 2 | 6 2 | | 3+4 2 |
| | | 4+5 2 | 6+6 4 | 6 2 | | 5+6 2 |
| | | 6+7 2 | 5+5 2 | 6 2 | | 5+6 2 |
| | | 8 2 | | 4+5 2 | | 3+4 2 |
| | | 9 2 | | 6 2 | | 3+4 2 |
| | | 8 2 | | 4+5 2 | | 3+4 2 |
| | | 8 2 | | 6 2 | | 5+6 2 |
| | | 9 2 | | 6 2 | | 3+4 2 |
| | | 9 2 | | 6 2 | | 3+4 2 |
| | | 8 2 | | 4+5 2 | | 5+6 2 |
| | | 4+5 2 | | 6 2 | | 3+4 2 |
| | | 6+7 2 | | 6 2 | | 5+6 2 |
| | | 9 2 | | 6 2 | | 5+6 2 |
| | | 8 2 | | 4+5 2 | | 5+6 2 |
| | | 8 2 | | 6 2 | | 3+4 2 |
| | | 9 2 | | 4+5 2 | | 3+4 2 |
| | | 8 2 | | 6 2 | | 5+6 2 |
| 165.1.2 | 26776 | | 228 | As1Cu3 | | |
| | -A | -GM | -H | -K | -L | -M |
| | 5+6 4 | 8 2 | 4+4 2 | 4+5 2 | 2+2 4 | 5+6 2 |
| | 4+4 4 | 9 2 | 5+5 2 | 6 2 | 2+2 4 | 3+4 2 |
| | 5+6 4 | 4+5 2 | 6+6 4 | 6 2 | 2+2 4 | 5+6 2 |
| | 4+4 4 | 8 2 | 6+6 4 | 4+5 2 | 2+2 4 | 3+4 2 |
| | 5+6 4 | 6+7 2 | 4+4 2 | 6 2 | 2+2 4 | 5+6 2 |
| | 5+6 4 | 9 2 | 6+6 4 | 6 2 | 2+2 4 | 3+4 2 |
| | 5+6 4 | 8 2 | 5+5 2 | 6 2 | 2+2 4 | 3+4 2 |
| | 5+6 4 | 9 2 | 6+6 4 | 4+5 2 | 2+2 4 | 5+6 2 |
| | 4+4 4 | 6+7 2 | 5+5 2 | 6 2 | 2+2 4 | 5+6 2 |
| | 5+6 4 | 8 2 | 6+6 4 | 6 2 | 2+2 4 | 3+4 2 |
| | 4+4 4 | 9 2 | 4+4 2 | 4+5 2 | 2+2 4 | 5+6 2 |
| | 5+6 4 | 4+5 2 | 6+6 4 | 6 2 | 2+2 4 | 3+4 2 |

Continued on next column

**TABLE LXXII– continued**

| TQC | ICSD | | NoE | | SF | |
|---|---|---|---|---|---|---|
| | 4+4 4 | 8 2 | 6+6 4 | 6 2 | 2+2 4 | 3+4 2 |
| | 5+6 4 | 8 2 | 4+4 2 | 4+5 2 | 2+2 4 | 5+6 2 |
| | 5+6 4 | 9 2 | 6+6 4 | 6 2 | 2+2 4 | 5+6 2 |
| | 4+4 4 | 8 2 | 5+5 2 | 6 2 | 2+2 4 | 3+4 2 |
| | 5+6 4 | 4+5 2 | 6+6 4 | 4+5 2 | 2+2 4 | 5+6 2 |
| | 4+4 4 | 6+7 2 | 5+5 2 | 6 2 | 2+2 4 | 5+6 2 |
| | 5+6 4 | 9 2 | 4+4 2 | 4+5 2 | 2+2 4 | 3+4 2 |
| | 5+6 4 | 9 2 | 6+6 4 | 6 2 | 2+2 4 | 3+4 2 |
| | 4+4 4 | 6+7 2 | 6+6 4 | 6 2 | 2+2 4 | 3+4 2 |
| | 5+6 4 | 9 2 | 5+5 2 | 6 2 | 2+2 4 | 3+4 2 |
| | 5+6 4 | 4+5 2 | 4+4 2 | 4+5 2 | 2+2 4 | 5+6 2 |
| | 5+6 4 | 8 2 | 6+6 4 | 6 2 | 2+2 4 | 5+6 2 |
| | 5+6 4 | 8 2 | 5+5 2 | 4+5 2 | 2+2 4 | 3+4 2 |
| | 4+4 4 | 6+7 2 | 6+6 4 | 6 2 | 2+2 4 | 3+4 2 |
| | 5+6 4 | 9 2 | 4+4 2 | 6 2 | 2+2 4 | 5+6 2 |
| | 4+4 4 | 4+5 2 | 6+6 4 | 4+5 2 | 2+2 4 | 3+4 2 |
| | 5+6 4 | 8 2 | 6+6 4 | 6 2 | 2+2 4 | 5+6 2 |
| | 5+6 4 | 9 2 | 4+4 2 | 6 2 | 2+2 4 | 5+6 2 |
| | 4+4 4 | 6+7 2 | 5+5 2 | 6 2 | 2+2 4 | 5+6 2 |
| | 5+6 4 | 9 2 | 6+6 4 | 6 2 | 2+2 4 | 5+6 2 |
| | 5+6 4 | 8 2 | 6+6 4 | 4+5 2 | 2+2 4 | 3+4 2 |
| | 4+4 4 | 4+5 2 | 4+4 2 | 4+5 2 | 2+2 4 | 3+4 2 |
| | 5+6 4 | 9 2 | 6+6 4 | 6 2 | 2+2 4 | 3+4 2 |
| | 5+6 4 | 9 2 | 5+5 2 | 6 2 | 2+2 4 | 3+4 2 |
| | 4+4 4 | 6+7 2 | 6+6 4 | 6 2 | 2+2 4 | 5+6 2 |
| | 5+6 4 | 6+7 2 | 5+5 2 | 6 2 | 2+2 4 | 3+4 2 |
| | 5+6 4 | 8 2 | 4+4 2 | 4+5 2 | 2+2 4 | 5+6 2 |
| | 4+4 4 | 4+5 2 | 6+6 4 | 6 2 | 2+2 4 | 3+4 2 |
| | 5+6 4 | 9 2 | 6+6 4 | 6 2 | 2+2 4 | 5+6 2 |
| | 4+4 4 | 4+5 2 | 4+4 2 | 4+5 2 | 2+2 4 | 3+4 2 |
| | 5+6 4 | 8 2 | 5+5 2 | 6 2 | 2+2 4 | 5+6 2 |
| | 4+4 4 | 9 2 | 6+6 4 | 4+5 2 | 2+2 4 | 3+4 2 |
| | 5+6 4 | 6+7 2 | 6+6 4 | 6 2 | 2+2 4 | 5+6 2 |
| | 5+6 4 | 8 2 | 5+5 2 | 6 2 | 2+2 4 | 3+4 2 |
| | 5+6 4 | 9 2 | 4+4 2 | 6 2 | 2+2 4 | 5+6 2 |
| | 5+6 4 | 8 2 | 6+6 4 | 6 2 | 2+2 4 | 5+6 2 |
| | 4+4 4 | 9 2 | 6+6 4 | 4+5 2 | 2+2 4 | 3+4 2 |
| | 5+6 4 | 8 2 | 5+5 2 | 6 2 | 2+2 4 | 5+6 2 |
| | 4+4 4 | 4+5 2 | 4+4 2 | 4+5 2 | 2+2 4 | 3+4 2 |
| | 5+6 4 | 8 2 | 6+6 4 | 6 2 | 2+2 4 | 3+4 2 |
| | 5+6 4 | 8 2 | 4+4 2 | 6 2 | 2+2 4 | 5+6 2 |
| | 5+6 4 | 9 2 | 5+5 2 | 4+5 2 | 2+2 4 | 5+6 2 |
| | 4+4 4 | 4+5 2 | 6+6 4 | 6 2 | 2+2 4 | 3+4 2 |
| | 5+6 4 | 6+7 2 | 6+6 4 | 6 2 | 2+2 4 | 3+4 2 |
| | 5+6 4 | 9 2 | 6+6 4 | 4+5 2 | 2+2 4 | 5+6 2 |
| | | 8 2 | 5+5 2 | 6 2 | | 3+4 2 |
| | | 8 2 | 4+4 2 | 6 2 | | 5+6 2 |
| | | 8 2 | 6+6 4 | 4+5 2 | | 5+6 2 |
| | | 9 2 | 5+5 2 | 6 2 | | 3+4 2 |
| | | 4+5 2 | 6+6 4 | 4+5 2 | | 3+4 2 |
| | | 6+7 2 | 4+4 2 | 6 2 | | 5+6 2 |
| | | 8 2 | 6+6 4 | 6 2 | | 5+6 2 |
| | | 8 2 | 5+5 2 | 4+5 2 | | 3+4 2 |
| | | 9 2 | 6+6 4 | 6 2 | | 5+6 2 |
| | | 4+5 2 | 4+4 2 | 6 2 | | 5+6 2 |
| | | 8 2 | 6+6 4 | 4+5 2 | | 3+4 2 |
| | | 9 2 | 6+6 4 | 6 2 | | 5+6 2 |
| | | 6+7 2 | 4+4 2 | 6 2 | | 3+4 2 |
| | | 8 2 | 5+5 2 | 6 2 | | 5+6 2 |
| | | 9 2 | 6+6 4 | 4+5 2 | | 3+4 2 |
| | | 4+5 2 | 4+4 2 | 6 2 | | 5+6 2 |
| | | 9 2 | 5+5 2 | 4+5 2 | | 3+4 2 |
| | | 9 2 | 6+6 4 | 6 2 | | 5+6 2 |
| | | 8 2 | 6+6 4 | 4+5 2 | | 3+4 2 |
| | | 6+7 2 | | 6 2 | | 5+6 2 |
| | | 8 2 | | 4+5 2 | | 3+4 2 |
| | | 4+5 2 | | 6 2 | | 3+4 2 |
| | | 8 2 | | 6 2 | | 5+6 2 |
| | | 4+5 2 | | 6 2 | | 5+6 2 |
| | | 9 2 | | 4+5 2 | | 3+4 2 |
| | | 6+7 2 | | 6 2 | | 5+6 2 |
| | | 8 2 | | 6 2 | | 3+4 2 |
| | | 9 2 | | 4+5 2 | | 5+6 2 |
| | | 8 2 | | 6 2 | | 3+4 2 |
| | | 9 2 | | 6 2 | | 3+4 2 |
| | | 6+7 2 | | 6 2 | | 5+6 2 |
| | | 4+5 2 | | 4+5 2 | | 5+6 2 |
| | | 8 2 | | 4+5 2 | | 3+4 2 |
| | | 9 2 | | 6 2 | | 5+6 2 |
| | | 9 2 | | 6 2 | | 5+6 2 |
| | | 6+7 2 | | 6 2 | | 3+4 2 |
| | | 8 2 | | 4+5 2 | | 3+4 2 |
| | | 8 2 | | 6 2 | | 5+6 2 |
| | | 9 2 | | 6 2 | | 5+6 2 |
| | | 4+5 2 | | 4+5 2 | | 3+4 2 |
| | | 6+7 2 | | 6 2 | | 3+4 2 |
| | | 9 2 | | 6 2 | | 3+4 2 |
| | | 4+5 2 | | 6 2 | | 5+6 2 |
| | | 8 2 | | 4+5 2 | | 3+4 2 |
| | | 6+7 2 | | 6 2 | | 5+6 2 |
| | | 9 2 | | 6 2 | | 5+6 2 |
| | | 8 2 | | 4+5 2 | | 3+4 2 |
| | | 4+5 2 | | 6 2 | | 5+6 2 |
| | | 9 2 | | 6 2 | | 3+4 2 |
| | | 8 2 | | 4+5 2 | | 5+6 2 |
| | | 9 2 | | 6 2 | | 5+6 2 |
| | | 6+7 2 | | 6 2 | | 3+4 2 |
| | | 9 2 | | 4+5 2 | | 5+6 2 |
| | | 8 2 | | 6 2 | | 3+4 2 |
| | | 8 2 | | 4+5 2 | | 5+6 2 |
| | | 9 2 | | 6 2 | | 3+4 2 |
| | | 8 2 | | 6 2 | | 3+4 2 |

The End

TABLE LXXIII: The computed irreducible representations for the valence bands for the compounds of SEBR and NLC in space group 166. See the list of tables.

| TQC | ICSD | | NoE | SF |
|---|---|---|---|---|
| 166.1.1 | 653496 | | 252 | Yb2Zn17 |
| | -GM | -T | -F | -L |
| | 8 2 | 8 2 | 3+4 2 | 3+4 2 |
| | 9 2 | 9 2 | 5+6 2 | 5+6 2 |
| | 8 2 | 8 2 | 3+4 2 | 5+6 2 |
| | 9 2 | 9 2 | 5+6 2 | 3+4 2 |
| | 4+5 2 | 4+5 2 | 5+6 2 | 3+4 2 |
| | 6+7 2 | 6+7 2 | 3+4 2 | 5+6 2 |
| | 8 2 | 9 2 | 3+4 2 | 5+6 2 |
| | 9 2 | 8 2 | 5+6 2 | 3+4 2 |
| | 8 2 | 9 2 | 3+4 2 | 3+4 2 |
| | 8 2 | 9 2 | 5+6 2 | 5+6 2 |

Continued on next column

**TABLE LXXIII– continued**

| TQC | ICSD NoE SF |
|---|---|
| | 8 2 6+7 2 5+6 2 3+4 2 |
| | 4+5 2 9 2 5+6 2 3+4 2 |
| | 8 2 8 2 3+4 2 3+4 2 |
| | 9 2 9 2 3+4 2 5+6 2 |
| | 4+5 2 4+5 2 5+6 2 3+4 2 |
| | 8 2 9 2 5+6 2 5+6 2 |
| | 6+7 2 6+7 2 5+6 2 3+4 2 |
| | 8 2 8 2 3+4 2 5+6 2 |
| | 4+5 2 9 2 3+4 2 5+6 2 |
| | 8 2 6+7 2 5+6 2 3+4 2 |
| | 9 2 9 2 5+6 2 3+4 2 |
| | 8 2 8 2 3+4 2 5+6 2 |
| | 4+5 2 9 2 3+4 2 3+4 2 |
| | 9 2 6+7 2 5+6 2 5+6 2 |
| | 8 2 8 2 5+6 2 3+4 2 |
| | 6+7 2 4+5 2 3+4 2 5+6 2 |
| | 8 2 9 2 5+6 2 3+4 2 |
| | 8 2 6+7 2 5+6 2 3+4 2 |
| | 9 2 9 2 3+4 2 3+4 2 |
| | 4+5 2 8 2 5+6 2 5+6 2 |
| | 9 2 6+7 2 3+4 2 5+6 2 |
| | 8 2 8 2 5+6 2 3+4 2 |
| | 4+5 2 4+5 2 3+4 2 3+4 2 |
| | 9 2 9 2 5+6 2 5+6 2 |
| | 9 2 8 2 3+4 2 5+6 2 |
| | 6+7 2 8 2 3+4 2 3+4 2 |
| | 8 2 9 2 5+6 2 5+6 2 |
| | 8 2 6+7 2 5+6 2 5+6 2 |
| | 4+5 2 9 2 3+4 2 5+6 2 |
| | 6+7 2 4+5 2 5+6 2 3+4 2 |
| | 6+7 2 6+7 2 3+4 2 5+6 2 |
| | 9 2 9 2 3+4 2 3+4 2 |
| | 9 2 8 2 5+6 2 5+6 2 |
| | 4+5 2 4+5 2 3+4 2 3+4 2 |
| | 6+7 2 8 2 5+6 2 3+4 2 |
| | 8 2 9 2 3+4 2 5+6 2 |
| | 8 2 4+5 2 5+6 2 5+6 2 |
| | 9 2 6+7 2 3+4 2 3+4 2 |
| | 6+7 2 8 2 5+6 2 3+4 2 |
| | 8 2 9 2 3+4 2 5+6 2 |
| | 9 2 9 2 5+6 2 3+4 2 |
| | 8 2 8 2 3+4 2 3+4 2 |
| | 9 2 6+7 2 5+6 2 5+6 2 |
| | 8 2 9 2 3+4 2 5+6 2 |
| | 8 2 4+5 2 5+6 2 3+4 2 |
| | 4+5 2 8 2 3+4 2 5+6 2 |
| | 9 2 9 2 5+6 2 3+4 2 |
| | 4+5 2 6+7 2 3+4 2 5+6 2 |
| | 6+7 2 8 2 5+6 2 3+4 2 |
| | 8 2 8 2 5+6 2 3+4 2 |
| | 9 2 9 2 3+4 2 5+6 2 |
| | 6+7 2 8 2 5+6 2 5+6 2 |
| | 8 2 4+5 2 3+4 2 3+4 2 |
| | 8 2 9 2 3+4 2 5+6 2 |
| | 9 2 6+7 2 5+6 2 3+4 2 |
| | 4+5 2 9 2 3+4 2 3+4 2 |
| | 8 2 8 2 5+6 2 5+6 2 |
| | 9 2 9 2 3+4 2 3+4 2 |
| | 6+7 2 4+5 2 5+6 2 5+6 2 |
| | 4+5 2 9 2 3+4 2 3+4 2 |
| | 8 2 6+7 2 5+6 2 5+6 2 |

Continued on next column

**TABLE LXXIII– continued**

| TQC | ICSD NoE SF |
|---|---|
| | 9 2 8 2 3+4 2 3+4 2 |
| | 9 2 4+5 2 5+6 2 5+6 2 |
| | 4+5 2 9 2 5+6 2 3+4 2 |
| | 6+7 2 8 2 5+6 2 5+6 2 |
| | 8 2 8 2 3+4 2 3+4 2 |
| | 8 2 9 2 3+4 2 5+6 2 |
| | 9 2 6+7 2 5+6 2 5+6 2 |
| | 4+5 2 8 2 3+4 2 3+4 2 |
| | 9 2 4+5 2 3+4 2 3+4 2 |
| | 8 2 9 2 5+6 2 5+6 2 |
| | 8 2 8 2 3+4 2 3+4 2 |
| | 9 2 6+7 2 5+6 2 3+4 2 |
| | 8 2 4+5 2 3+4 2 5+6 2 |
| | 9 2 8 2 5+6 2 3+4 2 |
| | 6+7 2 9 2 3+4 2 5+6 2 |
| | 8 2 9 2 5+6 2 5+6 2 |
| | 9 2 9 2 5+6 2 3+4 2 |
| | 4+5 2 6+7 2 3+4 2 5+6 2 |
| | 8 2 8 2 5+6 2 3+4 2 |
| | 4+5 2 9 2 3+4 2 3+4 2 |
| | 8 2 9 2 5+6 2 5+6 2 |
| | 4+5 2 9 2 3+4 2 3+4 2 |
| | 8 2 6+7 2 5+6 2 5+6 2 |
| | 9 2 9 2 5+6 2 3+4 2 |
| | 6+7 2 9 2 5+6 2 3+4 2 |
| | 8 2 6+7 2 3+4 2 5+6 2 |
| | 4+5 2 8 2 5+6 2 5+6 2 |
| | 8 2 4+5 2 3+4 2 3+4 2 |
| | 9 2 8 2 3+4 2 3+4 2 |
| | 6+7 2 9 2 5+6 2 5+6 2 |
| | 9 2 8 2 3+4 2 5+6 2 |
| | 8 2 8 2 5+6 2 3+4 2 |
| | 4+5 2 4+5 2 5+6 2 5+6 2 |
| | 9 2 6+7 2 3+4 2 3+4 2 |
| | 9 2 9 2 3+4 2 5+6 2 |
| | 8 2 9 2 5+6 2 3+4 2 |
| | 8 2 8 2 3+4 2 5+6 2 |
| | 4+5 2 8 2 5+6 2 3+4 2 |
| | 9 2 4+5 2 5+6 2 3+4 2 |
| | 6+7 2 9 2 3+4 2 3+4 2 |
| | 8 2 6+7 2 5+6 2 5+6 2 |
| | 4+5 2 8 2 3+4 2 5+6 2 |
| | 8 2 4+5 2 5+6 2 5+6 2 |
| | 9 2 8 2 3+4 2 5+6 2 |
| | 9 2 4+5 2 5+6 2 3+4 2 |
| | 6+7 2 8 2 3+4 2 3+4 2 |
| | 9 2 9 2 3+4 2 3+4 2 |
| | 8 2 9 2 3+4 2 5+6 2 |
| | 8 2 8 2 5+6 2 3+4 2 |
| | 8 2 9 2 5+6 2 5+6 2 |
| | 9 2 6+7 2 5+6 2 3+4 2 |
| | 4+5 2 8 2 3+4 2 5+6 2 |
| | 8 2 4+5 2 5+6 2 5+6 2 |
| | 6+7 2 8 2 3+4 2 5+6 2 |
| | 9 2 9 2 3+4 2 3+4 2 |
| 166.1.2 | 653487 252 Yb2Zn17 |
| | -GM -T -F -L |
| | 8 2 8 2 3+4 2 3+4 2 |
| | 9 2 9 2 5+6 2 5+6 2 |
| | 8 2 8 2 3+4 2 5+6 2 |
| | 9 2 9 2 5+6 2 3+4 2 |

Continued on next column

**TABLE LXXIII– continued**

| TQC | ICSD | NoE | SF | |
|---|---|---|---|---|
| | 4+5 2 | 4+5 2 | 5+6 2 | 3+4 2 |
| | 6+7 2 | 6+7 2 | 3+4 2 | 5+6 2 |
| | 8 2 | 9 2 | 3+4 2 | 5+6 2 |
| | 9 2 | 8 2 | 5+6 2 | 3+4 2 |
| | 8 2 | 9 2 | 3+4 2 | 3+4 2 |
| | 8 2 | 9 2 | 5+6 2 | 5+6 2 |
| | 8 2 | 6+7 2 | 5+6 2 | 3+4 2 |
| | 4+5 2 | 9 2 | 5+6 2 | 3+4 2 |
| | 8 2 | 8 2 | 3+4 2 | 3+4 2 |
| | 9 2 | 9 2 | 3+4 2 | 5+6 2 |
| | 4+5 2 | 4+5 2 | 5+6 2 | 3+4 2 |
| | 8 2 | 9 2 | 5+6 2 | 5+6 2 |
| | 6+7 2 | 6+7 2 | 5+6 2 | 3+4 2 |
| | 8 2 | 8 2 | 3+4 2 | 5+6 2 |
| | 4+5 2 | 9 2 | 3+4 2 | 5+6 2 |
| | 8 2 | 6+7 2 | 5+6 2 | 3+4 2 |
| | 9 2 | 9 2 | 5+6 2 | 3+4 2 |
| | 8 2 | 8 2 | 3+4 2 | 5+6 2 |
| | 4+5 2 | 9 2 | 3+4 2 | 3+4 2 |
| | 9 2 | 6+7 2 | 5+6 2 | 5+6 2 |
| | 8 2 | 8 2 | 5+6 2 | 3+4 2 |
| | 6+7 2 | 4+5 2 | 3+4 2 | 5+6 2 |
| | 8 2 | 9 2 | 5+6 2 | 3+4 2 |
| | 8 2 | 6+7 2 | 5+6 2 | 3+4 2 |
| | 9 2 | 9 2 | 3+4 2 | 3+4 2 |
| | 4+5 2 | 8 2 | 5+6 2 | 5+6 2 |
| | 9 2 | 6+7 2 | 3+4 2 | 5+6 2 |
| | 8 2 | 8 2 | 5+6 2 | 3+4 2 |
| | 4+5 2 | 4+5 2 | 3+4 2 | 3+4 2 |
| | 9 2 | 9 2 | 5+6 2 | 5+6 2 |
| | 9 2 | 8 2 | 3+4 2 | 5+6 2 |
| | 6+7 2 | 8 2 | 3+4 2 | 3+4 2 |
| | 8 2 | 9 2 | 5+6 2 | 5+6 2 |
| | 8 2 | 6+7 2 | 5+6 2 | 5+6 2 |
| | 4+5 2 | 9 2 | 3+4 2 | 5+6 2 |
| | 6+7 2 | 4+5 2 | 5+6 2 | 3+4 2 |
| | 6+7 2 | 6+7 2 | 3+4 2 | 5+6 2 |
| | 9 2 | 9 2 | 3+4 2 | 3+4 2 |
| | 9 2 | 8 2 | 5+6 2 | 5+6 2 |
| | 4+5 2 | 4+5 2 | 3+4 2 | 3+4 2 |
| | 6+7 2 | 8 2 | 5+6 2 | 3+4 2 |
| | 8 2 | 9 2 | 3+4 2 | 5+6 2 |
| | 8 2 | 4+5 2 | 5+6 2 | 5+6 2 |
| | 9 2 | 6+7 2 | 3+4 2 | 3+4 2 |
| | 6+7 2 | 8 2 | 5+6 2 | 3+4 2 |
| | 8 2 | 9 2 | 3+4 2 | 5+6 2 |
| | 9 2 | 9 2 | 5+6 2 | 3+4 2 |
| | 8 2 | 8 2 | 3+4 2 | 3+4 2 |
| | 9 2 | 6+7 2 | 5+6 2 | 5+6 2 |
| | 8 2 | 9 2 | 3+4 2 | 5+6 2 |
| | 8 2 | 4+5 2 | 5+6 2 | 3+4 2 |
| | 4+5 2 | 8 2 | 3+4 2 | 5+6 2 |
| | 9 2 | 9 2 | 5+6 2 | 3+4 2 |
| | 4+5 2 | 6+7 2 | 3+4 2 | 5+6 2 |
| | 6+7 2 | 8 2 | 5+6 2 | 3+4 2 |
| | 8 2 | 8 2 | 5+6 2 | 3+4 2 |
| | 9 2 | 9 2 | 3+4 2 | 5+6 2 |
| | 6+7 2 | 8 2 | 5+6 2 | 5+6 2 |
| | 8 2 | 4+5 2 | 3+4 2 | 3+4 2 |
| | 8 2 | 9 2 | 3+4 2 | 5+6 2 |
| | 9 2 | 6+7 2 | 5+6 2 | 3+4 2 |

Continued on next column

**TABLE LXXIII– continued**

| TQC | ICSD | NoE | SF | |
|---|---|---|---|---|
| | 4+5 2 | 9 2 | 3+4 2 | 3+4 2 |
| | 8 2 | 8 2 | 5+6 2 | 5+6 2 |
| | 9 2 | 9 2 | 3+4 2 | 3+4 2 |
| | 6+7 2 | 4+5 2 | 5+6 2 | 5+6 2 |
| | 4+5 2 | 9 2 | 3+4 2 | 3+4 2 |
| | 8 2 | 6+7 2 | 5+6 2 | 5+6 2 |
| | 9 2 | 8 2 | 3+4 2 | 3+4 2 |
| | 9 2 | 4+5 2 | 5+6 2 | 5+6 2 |
| | 4+5 2 | 9 2 | 5+6 2 | 3+4 2 |
| | 6+7 2 | 8 2 | 5+6 2 | 5+6 2 |
| | 8 2 | 8 2 | 3+4 2 | 3+4 2 |
| | 8 2 | 9 2 | 3+4 2 | 5+6 2 |
| | 9 2 | 6+7 2 | 5+6 2 | 5+6 2 |
| | 4+5 2 | 8 2 | 3+4 2 | 3+4 2 |
| | 9 2 | 4+5 2 | 3+4 2 | 3+4 2 |
| | 8 2 | 9 2 | 5+6 2 | 5+6 2 |
| | 8 2 | 8 2 | 3+4 2 | 3+4 2 |
| | 9 2 | 6+7 2 | 5+6 2 | 3+4 2 |
| | 8 2 | 4+5 2 | 3+4 2 | 5+6 2 |
| | 9 2 | 8 2 | 5+6 2 | 3+4 2 |
| | 6+7 2 | 9 2 | 3+4 2 | 5+6 2 |
| | 8 2 | 9 2 | 5+6 2 | 5+6 2 |
| | 9 2 | 9 2 | 5+6 2 | 3+4 2 |
| | 4+5 2 | 6+7 2 | 3+4 2 | 5+6 2 |
| | 8 2 | 8 2 | 5+6 2 | 3+4 2 |
| | 4+5 2 | 9 2 | 3+4 2 | 3+4 2 |
| | 8 2 | 9 2 | 5+6 2 | 5+6 2 |
| | 4+5 2 | 9 2 | 3+4 2 | 3+4 2 |
| | 8 2 | 6+7 2 | 5+6 2 | 5+6 2 |
| | 9 2 | 9 2 | 5+6 2 | 3+4 2 |
| | 6+7 2 | 9 2 | 5+6 2 | 3+4 2 |
| | 8 2 | 6+7 2 | 3+4 2 | 5+6 2 |
| | 4+5 2 | 8 2 | 5+6 2 | 5+6 2 |
| | 8 2 | 4+5 2 | 3+4 2 | 3+4 2 |
| | 9 2 | 8 2 | 3+4 2 | 3+4 2 |
| | 6+7 2 | 9 2 | 5+6 2 | 5+6 2 |
| | 9 2 | 8 2 | 3+4 2 | 5+6 2 |
| | 8 2 | 8 2 | 5+6 2 | 3+4 2 |
| | 4+5 2 | 4+5 2 | 5+6 2 | 5+6 2 |
| | 9 2 | 6+7 2 | 3+4 2 | 3+4 2 |
| | 9 2 | 9 2 | 3+4 2 | 5+6 2 |
| | 8 2 | 9 2 | 5+6 2 | 3+4 2 |
| | 8 2 | 8 2 | 3+4 2 | 5+6 2 |
| | 4+5 2 | 8 2 | 5+6 2 | 3+4 2 |
| | 9 2 | 4+5 2 | 5+6 2 | 3+4 2 |
| | 6+7 2 | 9 2 | 3+4 2 | 3+4 2 |
| | 8 2 | 6+7 2 | 5+6 2 | 5+6 2 |
| | 4+5 2 | 8 2 | 3+4 2 | 5+6 2 |
| | 8 2 | 4+5 2 | 5+6 2 | 5+6 2 |
| | 9 2 | 8 2 | 3+4 2 | 5+6 2 |
| | 9 2 | 4+5 2 | 5+6 2 | 3+4 2 |
| | 6+7 2 | 8 2 | 3+4 2 | 3+4 2 |
| | 9 2 | 9 2 | 3+4 2 | 3+4 2 |
| | 8 2 | 9 2 | 3+4 2 | 5+6 2 |
| | 8 2 | 8 2 | 5+6 2 | 3+4 2 |
| | 8 2 | 9 2 | 5+6 2 | 5+6 2 |
| | 9 2 | 6+7 2 | 5+6 2 | 3+4 2 |
| | 4+5 2 | 8 2 | 3+4 2 | 5+6 2 |
| | 8 2 | 4+5 2 | 5+6 2 | 5+6 2 |
| | 6+7 2 | 8 2 | 3+4 2 | 5+6 2 |
| | 9 2 | 9 2 | 3+4 2 | 3+4 2 |

Continued on next column

**TABLE LXXIII– continued**

| TQC | ICSD | NoE | SF | |
|---|---|---|---|---|
| 166.1.3 | 104207 | 252 | Yb2Zn17 | |
| | -GM | -T | -F | -L |
| | 8 2 | 8 2 | 3+4 2 | 3+4 2 |
| | 9 2 | 9 2 | 5+6 2 | 5+6 2 |
| | 8 2 | 8 2 | 3+4 2 | 5+6 2 |
| | 9 2 | 9 2 | 5+6 2 | 3+4 2 |
| | 4+5 2 | 4+5 2 | 5+6 2 | 3+4 2 |
| | 6+7 2 | 6+7 2 | 3+4 2 | 5+6 2 |
| | 8 2 | 9 2 | 3+4 2 | 5+6 2 |
| | 9 2 | 8 2 | 5+6 2 | 3+4 2 |
| | 8 2 | 9 2 | 3+4 2 | 3+4 2 |
| | 8 2 | 9 2 | 5+6 2 | 5+6 2 |
| | 8 2 | 6+7 2 | 5+6 2 | 3+4 2 |
| | 4+5 2 | 9 2 | 5+6 2 | 3+4 2 |
| | 8 2 | 8 2 | 3+4 2 | 3+4 2 |
| | 9 2 | 9 2 | 3+4 2 | 5+6 2 |
| | 4+5 2 | 4+5 2 | 5+6 2 | 3+4 2 |
| | 8 2 | 9 2 | 5+6 2 | 5+6 2 |
| | 6+7 2 | 6+7 2 | 5+6 2 | 3+4 2 |
| | 8 2 | 8 2 | 3+4 2 | 5+6 2 |
| | 4+5 2 | 9 2 | 3+4 2 | 3+4 2 |
| | 8 2 | 6+7 2 | 5+6 2 | 5+6 2 |
| | 9 2 | 9 2 | 5+6 2 | 3+4 2 |
| | 8 2 | 8 2 | 3+4 2 | 5+6 2 |
| | 4+5 2 | 9 2 | 3+4 2 | 3+4 2 |
| | 9 2 | 6+7 2 | 5+6 2 | 5+6 2 |
| | 8 2 | 8 2 | 5+6 2 | 3+4 2 |
| | 6+7 2 | 9 2 | 3+4 2 | 5+6 2 |
| | 8 2 | 4+5 2 | 5+6 2 | 3+4 2 |
| | 8 2 | 6+7 2 | 3+4 2 | 3+4 2 |
| | 9 2 | 9 2 | 5+6 2 | 3+4 2 |
| | 4+5 2 | 8 2 | 5+6 2 | 5+6 2 |
| | 9 2 | 6+7 2 | 3+4 2 | 5+6 2 |
| | 8 2 | 8 2 | 5+6 2 | 3+4 2 |
| | 4+5 2 | 4+5 2 | 5+6 2 | 3+4 2 |
| | 9 2 | 9 2 | 3+4 2 | 5+6 2 |
| | 6+7 2 | 8 2 | 3+4 2 | 5+6 2 |
| | 9 2 | 8 2 | 3+4 2 | 3+4 2 |
| | 8 2 | 9 2 | 5+6 2 | 5+6 2 |
| | 8 2 | 6+7 2 | 3+4 2 | 5+6 2 |
| | 4+5 2 | 9 2 | 5+6 2 | 5+6 2 |
| | 6+7 2 | 4+5 2 | 5+6 2 | 3+4 2 |
| | 6+7 2 | 6+7 2 | 3+4 2 | 5+6 2 |
| | 9 2 | 9 2 | 3+4 2 | 3+4 2 |
| | 9 2 | 8 2 | 5+6 2 | 5+6 2 |
| | 4+5 2 | 4+5 2 | 3+4 2 | 3+4 2 |
| | 6+7 2 | 8 2 | 5+6 2 | 3+4 2 |
| | 8 2 | 9 2 | 3+4 2 | 5+6 2 |
| | 8 2 | 6+7 2 | 5+6 2 | 5+6 2 |
| | 9 2 | 8 2 | 3+4 2 | 3+4 2 |
| | 6+7 2 | 4+5 2 | 5+6 2 | 3+4 2 |
| | 8 2 | 9 2 | 3+4 2 | 5+6 2 |
| | 9 2 | 9 2 | 5+6 2 | 3+4 2 |
| | 8 2 | 8 2 | 3+4 2 | 3+4 2 |
| | 9 2 | 6+7 2 | 5+6 2 | 5+6 2 |
| | 8 2 | 9 2 | 3+4 2 | 5+6 2 |
| | 8 2 | 4+5 2 | 5+6 2 | 3+4 2 |
| | 4+5 2 | 8 2 | 3+4 2 | 5+6 2 |
| | 9 2 | 8 2 | 5+6 2 | 3+4 2 |
| | 4+5 2 | 9 2 | 3+4 2 | 5+6 2 |
| | 8 2 | 6+7 2 | 5+6 2 | 3+4 2 |

Continued on next column

**TABLE LXXIII– continued**

| TQC | ICSD | NoE | SF | |
|---|---|---|---|---|
| | 6+7 2 | 8 2 | 5+6 2 | 3+4 2 |
| | 9 2 | 9 2 | 3+4 2 | 5+6 2 |
| | 6+7 2 | 8 2 | 5+6 2 | 5+6 2 |
| | 8 2 | 9 2 | 3+4 2 | 3+4 2 |
| | 8 2 | 4+5 2 | 3+4 2 | 5+6 2 |
| | 4+5 2 | 6+7 2 | 5+6 2 | 3+4 2 |
| | 9 2 | 9 2 | 3+4 2 | 3+4 2 |
| | 8 2 | 8 2 | 5+6 2 | 5+6 2 |
| | 9 2 | 9 2 | 3+4 2 | 3+4 2 |
| | 6+7 2 | 4+5 2 | 5+6 2 | 5+6 2 |
| | 4+5 2 | 9 2 | 3+4 2 | 3+4 2 |
| | 9 2 | 6+7 2 | 5+6 2 | 5+6 2 |
| | 8 2 | 8 2 | 3+4 2 | 3+4 2 |
| | 9 2 | 4+5 2 | 5+6 2 | 5+6 2 |
| | 6+7 2 | 9 2 | 5+6 2 | 3+4 2 |
| | 4+5 2 | 8 2 | 5+6 2 | 5+6 2 |
| | 8 2 | 8 2 | 3+4 2 | 3+4 2 |
| | 9 2 | 9 2 | 3+4 2 | 5+6 2 |
| | 8 2 | 6+7 2 | 5+6 2 | 5+6 2 |
| | 4+5 2 | 8 2 | 3+4 2 | 3+4 2 |
| | 9 2 | 4+5 2 | 3+4 2 | 3+4 2 |
| | 8 2 | 9 2 | 5+6 2 | 5+6 2 |
| | 8 2 | 8 2 | 3+4 2 | 3+4 2 |
| | 9 2 | 6+7 2 | 5+6 2 | 3+4 2 |
| | 8 2 | 4+5 2 | 3+4 2 | 5+6 2 |
| | 9 2 | 8 2 | 5+6 2 | 3+4 2 |
| | 6+7 2 | 9 2 | 3+4 2 | 5+6 2 |
| | 8 2 | 9 2 | 5+6 2 | 5+6 2 |
| | 9 2 | 9 2 | 5+6 2 | 3+4 2 |
| | 4+5 2 | 6+7 2 | 3+4 2 | 5+6 2 |
| | 8 2 | 8 2 | 5+6 2 | 3+4 2 |
| | 4+5 2 | 9 2 | 3+4 2 | 3+4 2 |
| | 8 2 | 9 2 | 5+6 2 | 5+6 2 |
| | 4+5 2 | 9 2 | 3+4 2 | 3+4 2 |
| | 8 2 | 6+7 2 | 5+6 2 | 5+6 2 |
| | 9 2 | 9 2 | 5+6 2 | 3+4 2 |
| | 6+7 2 | 9 2 | 5+6 2 | 3+4 2 |
| | 8 2 | 6+7 2 | 3+4 2 | 5+6 2 |
| | 4+5 2 | 8 2 | 5+6 2 | 5+6 2 |
| | 8 2 | 4+5 2 | 3+4 2 | 3+4 2 |
| | 9 2 | 8 2 | 3+4 2 | 3+4 2 |
| | 6+7 2 | 9 2 | 5+6 2 | 5+6 2 |
| | 9 2 | 8 2 | 3+4 2 | 5+6 2 |
| | 8 2 | 8 2 | 5+6 2 | 3+4 2 |
| | 4+5 2 | 4+5 2 | 5+6 2 | 5+6 2 |
| | 9 2 | 6+7 2 | 3+4 2 | 3+4 2 |
| | 9 2 | 9 2 | 3+4 2 | 5+6 2 |
| | 8 2 | 9 2 | 5+6 2 | 3+4 2 |
| | 8 2 | 8 2 | 3+4 2 | 5+6 2 |
| | 4+5 2 | 8 2 | 5+6 2 | 3+4 2 |
| | 9 2 | 4+5 2 | 5+6 2 | 3+4 2 |
| | 6+7 2 | 9 2 | 3+4 2 | 3+4 2 |
| | 8 2 | 6+7 2 | 5+6 2 | 5+6 2 |
| | 4+5 2 | 8 2 | 3+4 2 | 5+6 2 |
| | 8 2 | 4+5 2 | 5+6 2 | 5+6 2 |
| | 9 2 | 8 2 | 3+4 2 | 3+4 2 |
| | 9 2 | 4+5 2 | 5+6 2 | 5+6 2 |
| | 6+7 2 | 8 2 | 3+4 2 | 3+4 2 |
| | 9 2 | 9 2 | 3+4 2 | 3+4 2 |
| | 8 2 | 9 2 | 3+4 2 | 5+6 2 |
| | 8 2 | 8 2 | 5+6 2 | 3+4 2 |

Continued on next column

**TABLE LXXIII– continued**

| TQC | ICSD | NoE | SF | |
|---|---|---|---|---|
| | 8 2 | 9 2 | 5+6 2 | 5+6 2 |
| | 9 2 | 6+7 2 | 5+6 2 | 3+4 2 |
| | 4+5 2 | 8 2 | 3+4 2 | 5+6 2 |
| | 8 2 | 4+5 2 | 5+6 2 | 5+6 2 |
| | 6+7 2 | 8 2 | 3+4 2 | 5+6 2 |
| | 9 2 | 9 2 | 3+4 2 | 3+4 2 |
| 166.1.4 | 653467 | 226 | Y2Zn17 | |
| | -GM | -T | -F | -L |
| | 8 2 | 8 2 | 3+4 2 | 3+4 2 |
| | 9 2 | 9 2 | 5+6 2 | 5+6 2 |
| | 8 2 | 9 2 | 3+4 2 | 5+6 2 |
| | 9 2 | 8 2 | 5+6 2 | 3+4 2 |
| | 4+5 2 | 4+5 2 | 5+6 2 | 3+4 2 |
| | 6+7 2 | 6+7 2 | 3+4 2 | 5+6 2 |
| | 8 2 | 9 2 | 3+4 2 | 5+6 2 |
| | 9 2 | 8 2 | 5+6 2 | 3+4 2 |
| | 8 2 | 9 2 | 3+4 2 | 3+4 2 |
| | 8 2 | 9 2 | 5+6 2 | 5+6 2 |
| | 8 2 | 6+7 2 | 5+6 2 | 3+4 2 |
| | 4+5 2 | 9 2 | 5+6 2 | 3+4 2 |
| | 8 2 | 8 2 | 3+4 2 | 3+4 2 |
| | 9 2 | 9 2 | 3+4 2 | 3+4 2 |
| | 4+5 2 | 4+5 2 | 5+6 2 | 5+6 2 |
| | 6+7 2 | 9 2 | 5+6 2 | 5+6 2 |
| | 8 2 | 6+7 2 | 5+6 2 | 3+4 2 |
| | 8 2 | 8 2 | 3+4 2 | 5+6 2 |
| | 4+5 2 | 9 2 | 3+4 2 | 5+6 2 |
| | 8 2 | 6+7 2 | 5+6 2 | 3+4 2 |
| | 9 2 | 9 2 | 5+6 2 | 5+6 2 |
| | 8 2 | 8 2 | 3+4 2 | 3+4 2 |
| | 4+5 2 | 9 2 | 3+4 2 | 3+4 2 |
| | 9 2 | 6+7 2 | 5+6 2 | 5+6 2 |
| | 8 2 | 8 2 | 5+6 2 | 3+4 2 |
| | 6+7 2 | 4+5 2 | 3+4 2 | 5+6 2 |
| | 8 2 | 9 2 | 5+6 2 | 3+4 2 |
| | 8 2 | 6+7 2 | 5+6 2 | 3+4 2 |
| | 9 2 | 9 2 | 3+4 2 | 5+6 2 |
| | 4+5 2 | 8 2 | 5+6 2 | 3+4 2 |
| | 9 2 | 6+7 2 | 3+4 2 | 5+6 2 |
| | 8 2 | 8 2 | 5+6 2 | 3+4 2 |
| | 4+5 2 | 4+5 2 | 3+4 2 | 3+4 2 |
| | 9 2 | 9 2 | 5+6 2 | 5+6 2 |
| | 6+7 2 | 8 2 | 3+4 2 | 5+6 2 |
| | 9 2 | 8 2 | 3+4 2 | 3+4 2 |
| | 8 2 | 9 2 | 5+6 2 | 5+6 2 |
| | 8 2 | 6+7 2 | 5+6 2 | 5+6 2 |
| | 4+5 2 | 9 2 | 3+4 2 | 5+6 2 |
| | 6+7 2 | 4+5 2 | 5+6 2 | 3+4 2 |
| | 6+7 2 | 6+7 2 | 3+4 2 | 5+6 2 |
| | 9 2 | 9 2 | 3+4 2 | 3+4 2 |
| | 4+5 2 | 8 2 | 5+6 2 | 5+6 2 |
| | 9 2 | 4+5 2 | 3+4 2 | 3+4 2 |
| | 6+7 2 | 8 2 | 3+4 2 | 3+4 2 |
| | 8 2 | 9 2 | 5+6 2 | 5+6 2 |
| | 8 2 | 4+5 2 | 5+6 2 | 5+6 2 |
| | 9 2 | 6+7 2 | 3+4 2 | 3+4 2 |
| | 6+7 2 | 8 2 | 5+6 2 | 3+4 2 |
| | 8 2 | 9 2 | 3+4 2 | 5+6 2 |
| | 9 2 | 9 2 | 5+6 2 | 3+4 2 |
| | 8 2 | 8 2 | 3+4 2 | 3+4 2 |
| | 9 2 | 6+7 2 | 5+6 2 | 5+6 2 |

Continued on next column

**TABLE LXXIII– continued**

| TQC | ICSD | NoE | SF | |
|---|---|---|---|---|
| | 8 2 | 9 2 | 3+4 2 | 5+6 2 |
| | 8 2 | 4+5 2 | 5+6 2 | 3+4 2 |
| | 4+5 2 | 8 2 | 3+4 2 | 5+6 2 |
| | 9 2 | 9 2 | 5+6 2 | 3+4 2 |
| | 4+5 2 | 8 2 | 3+4 2 | 3+4 2 |
| | 6+7 2 | 6+7 2 | 5+6 2 | 5+6 2 |
| | 8 2 | 8 2 | 5+6 2 | 3+4 2 |
| | 9 2 | 9 2 | 3+4 2 | 5+6 2 |
| | 6+7 2 | 8 2 | 5+6 2 | 5+6 2 |
| | 8 2 | 4+5 2 | 3+4 2 | 3+4 2 |
| | 8 2 | 9 2 | 3+4 2 | 5+6 2 |
| | 9 2 | 6+7 2 | 5+6 2 | 3+4 2 |
| | 4+5 2 | 8 2 | 3+4 2 | 3+4 2 |
| | 8 2 | 9 2 | 3+4 2 | 5+6 2 |
| | 9 2 | 9 2 | 5+6 2 | 3+4 2 |
| | 6+7 2 | 4+5 2 | 5+6 2 | 5+6 2 |
| | 4+5 2 | 9 2 | 3+4 2 | 3+4 2 |
| | 9 2 | 6+7 2 | 5+6 2 | 5+6 2 |
| | 8 2 | 8 2 | 3+4 2 | 3+4 2 |
| | 9 2 | 9 2 | 5+6 2 | 5+6 2 |
| | 6+7 2 | 4+5 2 | 5+6 2 | 3+4 2 |
| | 4+5 2 | 8 2 | 3+4 2 | 5+6 2 |
| | 8 2 | 8 2 | 5+6 2 | 3+4 2 |
| | 8 2 | 9 2 | 3+4 2 | 5+6 2 |
| | 9 2 | 6+7 2 | 5+6 2 | 3+4 2 |
| | 4+5 2 | 8 2 | 3+4 2 | 5+6 2 |
| | 9 2 | 4+5 2 | 3+4 2 | 3+4 2 |
| | 8 2 | 9 2 | 5+6 2 | 5+6 2 |
| | 8 2 | 8 2 | 3+4 2 | 3+4 2 |
| | 9 2 | 6+7 2 | 5+6 2 | 3+4 2 |
| | 8 2 | 4+5 2 | 3+4 2 | 5+6 2 |
| | 9 2 | 8 2 | 5+6 2 | 3+4 2 |
| | 8 2 | 9 2 | 3+4 2 | 5+6 2 |
| | 6+7 2 | 9 2 | 5+6 2 | 5+6 2 |
| | 4+5 2 | 9 2 | 5+6 2 | 3+4 2 |
| | 9 2 | 6+7 2 | 3+4 2 | 3+4 2 |
| | 8 2 | 8 2 | 5+6 2 | 5+6 2 |
| | 4+5 2 | 9 2 | 3+4 2 | 3+4 2 |
| | 8 2 | 9 2 | 5+6 2 | 5+6 2 |
| | 4+5 2 | 9 2 | 3+4 2 | 3+4 2 |
| | 8 2 | 6+7 2 | 5+6 2 | 5+6 2 |
| | 9 2 | 9 2 | 5+6 2 | 3+4 2 |
| | 6+7 2 | 9 2 | 5+6 2 | 3+4 2 |
| | 8 2 | 6+7 2 | 3+4 2 | 5+6 2 |
| | 4+5 2 | 8 2 | 5+6 2 | 5+6 2 |
| | 8 2 | 4+5 2 | 3+4 2 | 3+4 2 |
| | 9 2 | 8 2 | 5+6 2 | 3+4 2 |
| | 6+7 2 | 9 2 | 3+4 2 | 5+6 2 |
| | 9 2 | 8 2 | 3+4 2 | 5+6 2 |
| | 8 2 | 4+5 2 | 5+6 2 | 3+4 2 |
| | 4+5 2 | 8 2 | 5+6 2 | 5+6 2 |
| | 9 2 | 6+7 2 | 3+4 2 | 3+4 2 |
| | 9 2 | 9 2 | 3+4 2 | 5+6 2 |
| | 8 2 | 8 2 | 5+6 2 | 3+4 2 |
| | 4+5 2 | 4+5 2 | 5+6 2 | 5+6 2 |
| | 8 2 | 4+5 2 | 5+6 2 | 3+4 2 |
| | 8 2 | 8 2 | 3+4 2 | 5+6 2 |
| | 9 2 | 8 2 | 3+4 2 | 5+6 2 |
| | 9 2 | 9 2 | 3+4 2 | 3+4 2 |
| | 6+7 2 | 9 2 | 3+4 2 | 3+4 2 |
| 166.1.5 | 104239 | 226 | Y2Zn17 | |

Continued on next column

**TABLE LXXIII– continued**

| TQC | ICSD | NoE | SF | |
|---|---|---|---|---|
| | -GM | -T | -F | -L |
| | 8 2 | 8 2 | 3+4 2 | 3+4 2 |
| | 9 2 | 9 2 | 5+6 2 | 5+6 2 |
| | 8 2 | 9 2 | 3+4 2 | 5+6 2 |
| | 9 2 | 8 2 | 5+6 2 | 3+4 2 |
| | 4+5 2 | 4+5 2 | 5+6 2 | 3+4 2 |
| | 6+7 2 | 6+7 2 | 3+4 2 | 5+6 2 |
| | 8 2 | 9 2 | 3+4 2 | 5+6 2 |
| | 9 2 | 8 2 | 5+6 2 | 3+4 2 |
| | 8 2 | 9 2 | 3+4 2 | 3+4 2 |
| | 8 2 | 9 2 | 5+6 2 | 5+6 2 |
| | 8 2 | 6+7 2 | 5+6 2 | 3+4 2 |
| | 4+5 2 | 9 2 | 5+6 2 | 3+4 2 |
| | 8 2 | 8 2 | 3+4 2 | 3+4 2 |
| | 9 2 | 4+5 2 | 3+4 2 | 5+6 2 |
| | 4+5 2 | 9 2 | 5+6 2 | 3+4 2 |
| | 6+7 2 | 9 2 | 5+6 2 | 5+6 2 |
| | 8 2 | 6+7 2 | 3+4 2 | 3+4 2 |
| | 8 2 | 8 2 | 5+6 2 | 5+6 2 |
| | 4+5 2 | 9 2 | 3+4 2 | 5+6 2 |
| | 8 2 | 6+7 2 | 5+6 2 | 3+4 2 |
| | 9 2 | 9 2 | 5+6 2 | 5+6 2 |
| | 9 2 | 8 2 | 3+4 2 | 3+4 2 |
| | 8 2 | 9 2 | 3+4 2 | 3+4 2 |
| | 4+5 2 | 6+7 2 | 5+6 2 | 5+6 2 |
| | 6+7 2 | 8 2 | 5+6 2 | 3+4 2 |
| | 8 2 | 4+5 2 | 3+4 2 | 5+6 2 |
| | 8 2 | 9 2 | 5+6 2 | 3+4 2 |
| | 8 2 | 6+7 2 | 3+4 2 | 3+4 2 |
| | 9 2 | 9 2 | 5+6 2 | 5+6 2 |
| | 4+5 2 | 8 2 | 5+6 2 | 3+4 2 |
| | 9 2 | 8 2 | 3+4 2 | 5+6 2 |
| | 8 2 | 6+7 2 | 5+6 2 | 3+4 2 |
| | 9 2 | 4+5 2 | 5+6 2 | 3+4 2 |
| | 4+5 2 | 9 2 | 3+4 2 | 5+6 2 |
| | 6+7 2 | 8 2 | 3+4 2 | 5+6 2 |
| | 9 2 | 8 2 | 3+4 2 | 3+4 2 |
| | 8 2 | 9 2 | 5+6 2 | 5+6 2 |
| | 8 2 | 6+7 2 | 5+6 2 | 5+6 2 |
| | 4+5 2 | 9 2 | 3+4 2 | 5+6 2 |
| | 6+7 2 | 4+5 2 | 5+6 2 | 3+4 2 |
| | 6+7 2 | 6+7 2 | 3+4 2 | 5+6 2 |
| | 9 2 | 9 2 | 3+4 2 | 3+4 2 |
| | 9 2 | 8 2 | 5+6 2 | 5+6 2 |
| | 4+5 2 | 4+5 2 | 3+4 2 | 3+4 2 |
| | 6+7 2 | 8 2 | 3+4 2 | 3+4 2 |
| | 8 2 | 9 2 | 5+6 2 | 5+6 2 |
| | 8 2 | 6+7 2 | 5+6 2 | 5+6 2 |
| | 9 2 | 4+5 2 | 3+4 2 | 3+4 2 |
| | 6+7 2 | 8 2 | 5+6 2 | 3+4 2 |
| | 8 2 | 9 2 | 3+4 2 | 5+6 2 |
| | 9 2 | 9 2 | 5+6 2 | 3+4 2 |
| | 8 2 | 8 2 | 3+4 2 | 3+4 2 |
| | 9 2 | 6+7 2 | 5+6 2 | 5+6 2 |
| | 8 2 | 9 2 | 3+4 2 | 5+6 2 |
| | 4+5 2 | 4+5 2 | 5+6 2 | 3+4 2 |
| | 8 2 | 8 2 | 3+4 2 | 5+6 2 |
| | 9 2 | 8 2 | 5+6 2 | 3+4 2 |
| | 4+5 2 | 9 2 | 5+6 2 | 3+4 2 |
| | 6+7 2 | 6+7 2 | 3+4 2 | 5+6 2 |
| | 8 2 | 8 2 | 5+6 2 | 3+4 2 |
| | 9 2 | 9 2 | 3+4 2 | 5+6 2 |
| | 6+7 2 | 8 2 | 5+6 2 | 5+6 2 |
| | 8 2 | 4+5 2 | 3+4 2 | 3+4 2 |
| | 8 2 | 9 2 | 3+4 2 | 5+6 2 |
| | 4+5 2 | 6+7 2 | 5+6 2 | 3+4 2 |
| | 9 2 | 9 2 | 3+4 2 | 3+4 2 |
| | 8 2 | 8 2 | 3+4 2 | 5+6 2 |
| | 9 2 | 9 2 | 5+6 2 | 3+4 2 |
| | 6+7 2 | 4+5 2 | 5+6 2 | 5+6 2 |
| | 4+5 2 | 9 2 | 3+4 2 | 3+4 2 |
| | 9 2 | 6+7 2 | 5+6 2 | 5+6 2 |
| | 8 2 | 8 2 | 3+4 2 | 3+4 2 |
| | 9 2 | 9 2 | 5+6 2 | 5+6 2 |
| | 6+7 2 | 4+5 2 | 5+6 2 | 3+4 2 |
| | 4+5 2 | 8 2 | 3+4 2 | 5+6 2 |
| | 8 2 | 8 2 | 5+6 2 | 3+4 2 |
| | 8 2 | 9 2 | 3+4 2 | 5+6 2 |
| | 9 2 | 6+7 2 | 5+6 2 | 5+6 2 |
| | 4+5 2 | 8 2 | 3+4 2 | 3+4 2 |
| | 9 2 | 4+5 2 | 5+6 2 | 3+4 2 |
| | 8 2 | 9 2 | 3+4 2 | 5+6 2 |
| | 8 2 | 8 2 | 3+4 2 | 3+4 2 |
| | 9 2 | 6+7 2 | 5+6 2 | 3+4 2 |
| | 8 2 | 4+5 2 | 3+4 2 | 5+6 2 |
| | 9 2 | 8 2 | 5+6 2 | 3+4 2 |
| | 6+7 2 | 9 2 | 3+4 2 | 5+6 2 |
| | 8 2 | 9 2 | 5+6 2 | 5+6 2 |
| | 9 2 | 9 2 | 5+6 2 | 3+4 2 |
| | 4+5 2 | 6+7 2 | 3+4 2 | 3+4 2 |
| | 8 2 | 8 2 | 5+6 2 | 5+6 2 |
| | 4+5 2 | 9 2 | 3+4 2 | 3+4 2 |
| | 8 2 | 9 2 | 5+6 2 | 5+6 2 |
| | 4+5 2 | 9 2 | 5+6 2 | 3+4 2 |
| | 8 2 | 6+7 2 | 3+4 2 | 5+6 2 |
| | 9 2 | 9 2 | 5+6 2 | 3+4 2 |
| | 6+7 2 | 9 2 | 5+6 2 | 3+4 2 |
| | 8 2 | 6+7 2 | 3+4 2 | 5+6 2 |
| | 4+5 2 | 8 2 | 5+6 2 | 5+6 2 |
| | 8 2 | 4+5 2 | 3+4 2 | 3+4 2 |
| | 9 2 | 8 2 | 5+6 2 | 3+4 2 |
| | 6+7 2 | 9 2 | 3+4 2 | 5+6 2 |
| | 9 2 | 8 2 | 3+4 2 | 5+6 2 |
| | 8 2 | 4+5 2 | 5+6 2 | 3+4 2 |
| | 4+5 2 | 8 2 | 5+6 2 | 5+6 2 |
| | 9 2 | 6+7 2 | 3+4 2 | 3+4 2 |
| | 9 2 | 9 2 | 3+4 2 | 5+6 2 |
| | 8 2 | 8 2 | 5+6 2 | 3+4 2 |
| | 4+5 2 | 4+5 2 | 5+6 2 | 5+6 2 |
| | 8 2 | 4+5 2 | 5+6 2 | 3+4 2 |
| | 8 2 | 8 2 | 3+4 2 | 5+6 2 |
| | 9 2 | 8 2 | 3+4 2 | 5+6 2 |
| | 9 2 | 9 2 | 3+4 2 | 3+4 2 |
| | 6+7 2 | 9 2 | 3+4 2 | 3+4 2 |
| 166.1.6 | 653389 | 232 | U2Zn17 | |
| | -GM | -T | -F | -L |
| | 8 2 | 8 2 | 3+4 2 | 3+4 2 |
| | 9 2 | 9 2 | 5+6 2 | 5+6 2 |
| | 8 2 | 9 2 | 3+4 2 | 5+6 2 |
| | 9 2 | 8 2 | 5+6 2 | 3+4 2 |
| | 4+5 2 | 4+5 2 | 5+6 2 | 3+4 2 |
| | 8 2 | 9 2 | 3+4 2 | 5+6 2 |

Continued on next column

**TABLE LXXIII– continued**

| TQC | ICSD | NoE | SF | |
|---|---|---|---|---|
| | 6+7 2 | 6+7 2 | 3+4 2 | 5+6 2 |
| | 9 2 | 8 2 | 5+6 2 | 3+4 2 |
| | 8 2 | 9 2 | 3+4 2 | 3+4 2 |
| | 8 2 | 9 2 | 5+6 2 | 5+6 2 |
| | 8 2 | 6+7 2 | 5+6 2 | 3+4 2 |
| | 4+5 2 | 9 2 | 3+4 2 | 3+4 2 |
| | 8 2 | 9 2 | 5+6 2 | 3+4 2 |
| | 4+5 2 | 8 2 | 3+4 2 | 5+6 2 |
| | 9 2 | 8 2 | 5+6 2 | 3+4 2 |
| | 8 2 | 9 2 | 5+6 2 | 5+6 2 |
| | 6+7 2 | 4+5 2 | 3+4 2 | 3+4 2 |
| | 8 2 | 6+7 2 | 5+6 2 | 5+6 2 |
| | 4+5 2 | 6+7 2 | 3+4 2 | 3+4 2 |
| | 8 2 | 9 2 | 5+6 2 | 5+6 2 |
| | 9 2 | 8 2 | 3+4 2 | 3+4 2 |
| | 9 2 | 9 2 | 5+6 2 | 5+6 2 |
| | 4+5 2 | 9 2 | 3+4 2 | 5+6 2 |
| | 8 2 | 6+7 2 | 5+6 2 | 3+4 2 |
| | 6+7 2 | 8 2 | 5+6 2 | 3+4 2 |
| | 8 2 | 6+7 2 | 3+4 2 | 3+4 2 |
| | 8 2 | 9 2 | 5+6 2 | 5+6 2 |
| | 8 2 | 4+5 2 | 5+6 2 | 3+4 2 |
| | 9 2 | 8 2 | 3+4 2 | 5+6 2 |
| | 9 2 | 9 2 | 5+6 2 | 3+4 2 |
| | 4+5 2 | 6+7 2 | 3+4 2 | 5+6 2 |
| | 9 2 | 4+5 2 | 5+6 2 | 3+4 2 |
| | 8 2 | 8 2 | 5+6 2 | 5+6 2 |
| | 4+5 2 | 9 2 | 3+4 2 | 3+4 2 |
| | 6+7 2 | 8 2 | 3+4 2 | 5+6 2 |
| | 8 2 | 9 2 | 3+4 2 | 5+6 2 |
| | 9 2 | 8 2 | 5+6 2 | 3+4 2 |
| | 8 2 | 6+7 2 | 3+4 2 | 5+6 2 |
| | 4+5 2 | 9 2 | 5+6 2 | 5+6 2 |
| | 6+7 2 | 6+7 2 | 5+6 2 | 3+4 2 |
| | 9 2 | 4+5 2 | 3+4 2 | 5+6 2 |
| | 6+7 2 | 8 2 | 3+4 2 | 3+4 2 |
| | 4+5 2 | 9 2 | 5+6 2 | 5+6 2 |
| | 8 2 | 8 2 | 3+4 2 | 3+4 2 |
| | 9 2 | 4+5 2 | 3+4 2 | 5+6 2 |
| | 6+7 2 | 9 2 | 5+6 2 | 3+4 2 |
| | 8 2 | 8 2 | 5+6 2 | 5+6 2 |
| | 9 2 | 9 2 | 3+4 2 | 3+4 2 |
| | 6+7 2 | 6+7 2 | 3+4 2 | 3+4 2 |
| | 8 2 | 4+5 2 | 5+6 2 | 5+6 2 |
| | 9 2 | 9 2 | 5+6 2 | 3+4 2 |
| | 8 2 | 8 2 | 3+4 2 | 3+4 2 |
| | 8 2 | 9 2 | 5+6 2 | 5+6 2 |
| | 9 2 | 6+7 2 | 3+4 2 | 5+6 2 |
| | 4+5 2 | 8 2 | 5+6 2 | 3+4 2 |
| | 9 2 | 4+5 2 | 3+4 2 | 5+6 2 |
| | 8 2 | 8 2 | 5+6 2 | 3+4 2 |
| | 4+5 2 | 6+7 2 | 5+6 2 | 3+4 2 |
| | 6+7 2 | 9 2 | 3+4 2 | 5+6 2 |
| | 8 2 | 9 2 | 5+6 2 | 3+4 2 |
| | 9 2 | 8 2 | 3+4 2 | 5+6 2 |
| | 4+5 2 | 4+5 2 | 3+4 2 | 5+6 2 |
| | 8 2 | 8 2 | 5+6 2 | 3+4 2 |
| | 8 2 | 9 2 | 3+4 2 | 5+6 2 |
| | 6+7 2 | 6+7 2 | 5+6 2 | 3+4 2 |
| | 8 2 | 9 2 | 3+4 2 | 3+4 2 |
| | 9 2 | 8 2 | 5+6 2 | 5+6 2 |

Continued on next column

**TABLE LXXIII– continued**

| TQC | ICSD | NoE | SF | |
|---|---|---|---|---|
| | 6+7 2 | 9 2 | 3+4 2 | 3+4 2 |
| | 9 2 | 4+5 2 | 5+6 2 | 5+6 2 |
| | 4+5 2 | 9 2 | 3+4 2 | 3+4 2 |
| | 8 2 | 8 2 | 5+6 2 | 5+6 2 |
| | 9 2 | 4+5 2 | 3+4 2 | 3+4 2 |
| | 9 2 | 6+7 2 | 5+6 2 | 5+6 2 |
| | 4+5 2 | 9 2 | 5+6 2 | 3+4 2 |
| | 6+7 2 | 8 2 | 3+4 2 | 3+4 2 |
| | 8 2 | 9 2 | 5+6 2 | 5+6 2 |
| | 9 2 | 8 2 | 3+4 2 | 5+6 2 |
| | 8 2 | 6+7 2 | 5+6 2 | 5+6 2 |
| | 9 2 | 8 2 | 3+4 2 | 3+4 2 |
| | 4+5 2 | 4+5 2 | 3+4 2 | 3+4 2 |
| | 8 2 | 9 2 | 5+6 2 | 5+6 2 |
| | 8 2 | 8 2 | 3+4 2 | 3+4 2 |
| | 9 2 | 6+7 2 | 5+6 2 | 3+4 2 |
| | 8 2 | 4+5 2 | 3+4 2 | 5+6 2 |
| | 9 2 | 8 2 | 5+6 2 | 3+4 2 |
| | 6+7 2 | 9 2 | 3+4 2 | 5+6 2 |
| | 8 2 | 9 2 | 5+6 2 | 5+6 2 |
| | 9 2 | 9 2 | 5+6 2 | 3+4 2 |
| | 4+5 2 | 6+7 2 | 3+4 2 | 3+4 2 |
| | 8 2 | 8 2 | 5+6 2 | 5+6 2 |
| | 4+5 2 | 9 2 | 5+6 2 | 5+6 2 |
| | 8 2 | 9 2 | 3+4 2 | 3+4 2 |
| | 4+5 2 | 6+7 2 | 3+4 2 | 3+4 2 |
| | 8 2 | 9 2 | 5+6 2 | 5+6 2 |
| | 9 2 | 9 2 | 5+6 2 | 3+4 2 |
| | 6+7 2 | 9 2 | 5+6 2 | 3+4 2 |
| | 8 2 | 6+7 2 | 3+4 2 | 5+6 2 |
| | 4+5 2 | 8 2 | 5+6 2 | 5+6 2 |
| | 8 2 | 4+5 2 | 3+4 2 | 3+4 2 |
| | 9 2 | 8 2 | 5+6 2 | 3+4 2 |
| | 9 2 | 9 2 | 3+4 2 | 5+6 2 |
| | 6+7 2 | 8 2 | 3+4 2 | 5+6 2 |
| | 8 2 | 4+5 2 | 5+6 2 | 3+4 2 |
| | 9 2 | 8 2 | 5+6 2 | 5+6 2 |
| | 4+5 2 | 9 2 | 3+4 2 | 3+4 2 |
| | 9 2 | 6+7 2 | 3+4 2 | 3+4 2 |
| | 4+5 2 | 4+5 2 | 5+6 2 | 5+6 2 |
| | 8 2 | 8 2 | 5+6 2 | 5+6 2 |
| | 8 2 | 8 2 | 5+6 2 | 3+4 2 |
| | 8 2 | 4+5 2 | 3+4 2 | 5+6 2 |
| | 9 2 | 9 2 | 3+4 2 | 5+6 2 |
| | 9 2 | 8 2 | 3+4 2 | 3+4 2 |
| | 6+7 2 | 6+7 2 | 3+4 2 | 3+4 2 |
| | 8 2 | 9 2 | 5+6 2 | 5+6 2 |
| | 4+5 2 | 9 2 | 5+6 2 | 3+4 2 |
| | 6+7 2 | 6+7 2 | 5+6 2 | 3+4 2 |
| 166.1.7 | 653384 | 232 | U2Zn17 | |
| | -GM | -T | -F | -L |
| | 8 2 | 8 2 | 3+4 2 | 3+4 2 |
| | 9 2 | 9 2 | 5+6 2 | 5+6 2 |
| | 8 2 | 9 2 | 3+4 2 | 5+6 2 |
| | 9 2 | 8 2 | 5+6 2 | 3+4 2 |
| | 4+5 2 | 4+5 2 | 5+6 2 | 3+4 2 |
| | 8 2 | 9 2 | 3+4 2 | 5+6 2 |
| | 6+7 2 | 6+7 2 | 3+4 2 | 5+6 2 |
| | 9 2 | 8 2 | 5+6 2 | 3+4 2 |
| | 8 2 | 9 2 | 3+4 2 | 3+4 2 |
| | 8 2 | 9 2 | 5+6 2 | 5+6 2 |

Continued on next column

**TABLE LXXIII– continued**

| TQC | ICSD | NoE | SF | |
|---|---|---|---|---|
| | 8 2 | 6+7 2 | 5+6 2 | 3+4 2 |
| | 4+5 2 | 9 2 | 3+4 2 | 3+4 2 |
| | 8 2 | 9 2 | 5+6 2 | 3+4 2 |
| | 9 2 | 8 2 | 3+4 2 | 5+6 2 |
| | 4+5 2 | 4+5 2 | 5+6 2 | 3+4 2 |
| | 6+7 2 | 9 2 | 5+6 2 | 5+6 2 |
| | 8 2 | 8 2 | 5+6 2 | 3+4 2 |
| | 8 2 | 6+7 2 | 3+4 2 | 5+6 2 |
| | 4+5 2 | 9 2 | 5+6 2 | 5+6 2 |
| | 8 2 | 6+7 2 | 3+4 2 | 3+4 2 |
| | 9 2 | 8 2 | 5+6 2 | 3+4 2 |
| | 8 2 | 9 2 | 3+4 2 | 5+6 2 |
| | 9 2 | 9 2 | 3+4 2 | 5+6 2 |
| | 4+5 2 | 6+7 2 | 5+6 2 | 3+4 2 |
| | 6+7 2 | 8 2 | 5+6 2 | 3+4 2 |
| | 8 2 | 4+5 2 | 3+4 2 | 3+4 2 |
| | 8 2 | 9 2 | 5+6 2 | 5+6 2 |
| | 8 2 | 6+7 2 | 5+6 2 | 3+4 2 |
| | 9 2 | 9 2 | 3+4 2 | 5+6 2 |
| | 9 2 | 8 2 | 5+6 2 | 3+4 2 |
| | 4+5 2 | 6+7 2 | 3+4 2 | 5+6 2 |
| | 8 2 | 8 2 | 5+6 2 | 3+4 2 |
| | 9 2 | 4+5 2 | 5+6 2 | 3+4 2 |
| | 4+5 2 | 9 2 | 3+4 2 | 5+6 2 |
| | 6+7 2 | 8 2 | 3+4 2 | 5+6 2 |
| | 8 2 | 9 2 | 3+4 2 | 5+6 2 |
| | 9 2 | 8 2 | 5+6 2 | 3+4 2 |
| | 8 2 | 6+7 2 | 5+6 2 | 5+6 2 |
| | 4+5 2 | 9 2 | 3+4 2 | 5+6 2 |
| | 6+7 2 | 6+7 2 | 5+6 2 | 3+4 2 |
| | 9 2 | 4+5 2 | 3+4 2 | 5+6 2 |
| | 6+7 2 | 8 2 | 3+4 2 | 3+4 2 |
| | 4+5 2 | 9 2 | 5+6 2 | 5+6 2 |
| | 9 2 | 8 2 | 3+4 2 | 3+4 2 |
| | 8 2 | 4+5 2 | 3+4 2 | 3+4 2 |
| | 6+7 2 | 9 2 | 5+6 2 | 5+6 2 |
| | 8 2 | 8 2 | 5+6 2 | 5+6 2 |
| | 9 2 | 9 2 | 3+4 2 | 3+4 2 |
| | 6+7 2 | 6+7 2 | 5+6 2 | 3+4 2 |
| | 8 2 | 4+5 2 | 3+4 2 | 5+6 2 |
| | 9 2 | 9 2 | 5+6 2 | 3+4 2 |
| | 8 2 | 8 2 | 3+4 2 | 3+4 2 |
| | 8 2 | 6+7 2 | 5+6 2 | 5+6 2 |
| | 9 2 | 9 2 | 3+4 2 | 5+6 2 |
| | 4+5 2 | 8 2 | 5+6 2 | 3+4 2 |
| | 9 2 | 4+5 2 | 3+4 2 | 5+6 2 |
| | 8 2 | 6+7 2 | 5+6 2 | 3+4 2 |
| | 4+5 2 | 8 2 | 5+6 2 | 3+4 2 |
| | 6+7 2 | 9 2 | 3+4 2 | 5+6 2 |
| | 8 2 | 9 2 | 5+6 2 | 3+4 2 |
| | 9 2 | 8 2 | 3+4 2 | 5+6 2 |
| | 8 2 | 4+5 2 | 5+6 2 | 5+6 2 |
| | 8 2 | 8 2 | 3+4 2 | 3+4 2 |
| | 4+5 2 | 9 2 | 3+4 2 | 5+6 2 |
| | 6+7 2 | 6+7 2 | 5+6 2 | 3+4 2 |
| | 8 2 | 9 2 | 3+4 2 | 3+4 2 |
| | 9 2 | 8 2 | 5+6 2 | 5+6 2 |
| | 6+7 2 | 9 2 | 3+4 2 | 3+4 2 |
| | 9 2 | 4+5 2 | 5+6 2 | 5+6 2 |
| | 4+5 2 | 9 2 | 3+4 2 | 3+4 2 |
| | 8 2 | 6+7 2 | 5+6 2 | 5+6 2 |
| | 9 2 | 8 2 | 3+4 2 | 3+4 2 |
| | 4+5 2 | 4+5 2 | 5+6 2 | 5+6 2 |
| | 6+7 2 | 9 2 | 5+6 2 | 3+4 2 |
| | 8 2 | 8 2 | 3+4 2 | 3+4 2 |
| | 9 2 | 9 2 | 5+6 2 | 5+6 2 |
| | 9 2 | 8 2 | 3+4 2 | 5+6 2 |
| | 8 2 | 6+7 2 | 5+6 2 | 5+6 2 |
| | 9 2 | 8 2 | 3+4 2 | 3+4 2 |
| | 4+5 2 | 4+5 2 | 5+6 2 | 3+4 2 |
| | 8 2 | 9 2 | 3+4 2 | 5+6 2 |
| | 8 2 | 8 2 | 3+4 2 | 3+4 2 |
| | 9 2 | 6+7 2 | 5+6 2 | 3+4 2 |
| | 8 2 | 4+5 2 | 3+4 2 | 5+6 2 |
| | 9 2 | 8 2 | 5+6 2 | 3+4 2 |
| | 6+7 2 | 9 2 | 3+4 2 | 5+6 2 |
| | 8 2 | 9 2 | 5+6 2 | 5+6 2 |
| | 4+5 2 | 9 2 | 5+6 2 | 3+4 2 |
| | 9 2 | 6+7 2 | 3+4 2 | 3+4 2 |
| | 8 2 | 8 2 | 5+6 2 | 5+6 2 |
| | 4+5 2 | 9 2 | 5+6 2 | 3+4 2 |
| | 8 2 | 9 2 | 3+4 2 | 5+6 2 |
| | 4+5 2 | 6+7 2 | 3+4 2 | 3+4 2 |
| | 8 2 | 9 2 | 5+6 2 | 5+6 2 |
| | 9 2 | 9 2 | 5+6 2 | 3+4 2 |
| | 6+7 2 | 9 2 | 5+6 2 | 3+4 2 |
| | 8 2 | 6+7 2 | 3+4 2 | 5+6 2 |
| | 4+5 2 | 8 2 | 5+6 2 | 5+6 2 |
| | 8 2 | 4+5 2 | 3+4 2 | 3+4 2 |
| | 9 2 | 8 2 | 5+6 2 | 3+4 2 |
| | 9 2 | 9 2 | 3+4 2 | 5+6 2 |
| | 6+7 2 | 8 2 | 3+4 2 | 5+6 2 |
| | 8 2 | 8 2 | 5+6 2 | 3+4 2 |
| | 9 2 | 4+5 2 | 5+6 2 | 5+6 2 |
| | 4+5 2 | 9 2 | 3+4 2 | 3+4 2 |
| | 9 2 | 6+7 2 | 3+4 2 | 3+4 2 |
| | 4+5 2 | 4+5 2 | 5+6 2 | 5+6 2 |
| | 8 2 | 8 2 | 5+6 2 | 5+6 2 |
| | 8 2 | 8 2 | 5+6 2 | 3+4 2 |
| | 8 2 | 4+5 2 | 3+4 2 | 5+6 2 |
| | 9 2 | 9 2 | 3+4 2 | 3+4 2 |
| | 9 2 | 8 2 | 3+4 2 | 5+6 2 |
| | 6+7 2 | 9 2 | 3+4 2 | 3+4 2 |
| | 8 2 | 6+7 2 | 5+6 2 | 5+6 2 |
| | 4+5 2 | 9 2 | 5+6 2 | 3+4 2 |
| | 6+7 2 | 6+7 2 | 5+6 2 | 3+4 2 |
| 166.1.8 | 106208 | 232 | U2Zn17 | |
| | -GM | -T | -F | -L |
| | 8 2 | 8 2 | 3+4 2 | 5+6 2 |
| | 9 2 | 9 2 | 5+6 2 | 3+4 2 |
| | 8 2 | 9 2 | 3+4 2 | 5+6 2 |
| | 9 2 | 8 2 | 5+6 2 | 3+4 2 |
| | 4+5 2 | 4+5 2 | 5+6 2 | 3+4 2 |
| | 8 2 | 9 2 | 3+4 2 | 5+6 2 |
| | 6+7 2 | 6+7 2 | 3+4 2 | 5+6 2 |
| | 9 2 | 8 2 | 5+6 2 | 3+4 2 |
| | 8 2 | 9 2 | 3+4 2 | 3+4 2 |
| | 8 2 | 9 2 | 5+6 2 | 5+6 2 |
| | 8 2 | 6+7 2 | 5+6 2 | 3+4 2 |
| | 4+5 2 | 9 2 | 3+4 2 | 3+4 2 |
| | 8 2 | 9 2 | 5+6 2 | 3+4 2 |
| | 4+5 2 | 8 2 | 3+4 2 | 5+6 2 |

Continued on next column

Continued on next column

**TABLE LXXIII– continued**

| TQC | ICSD | NoE | SF | |
|---|---|---|---|---|
| | 9 2 | 9 2 | 5+6 2 | 3+4 2 |
| | 8 2 | 4+5 2 | 5+6 2 | 5+6 2 |
| | 6+7 2 | 8 2 | 3+4 2 | 3+4 2 |
| | 8 2 | 6+7 2 | 5+6 2 | 5+6 2 |
| | 4+5 2 | 6+7 2 | 3+4 2 | 3+4 2 |
| | 8 2 | 9 2 | 5+6 2 | 5+6 2 |
| | 9 2 | 8 2 | 3+4 2 | 3+4 2 |
| | 9 2 | 9 2 | 5+6 2 | 5+6 2 |
| | 4+5 2 | 9 2 | 3+4 2 | 5+6 2 |
| | 8 2 | 6+7 2 | 5+6 2 | 3+4 2 |
| | 6+7 2 | 8 2 | 5+6 2 | 3+4 2 |
| | 8 2 | 4+5 2 | 3+4 2 | 3+4 2 |
| | 8 2 | 6+7 2 | 5+6 2 | 5+6 2 |
| | 9 2 | 9 2 | 5+6 2 | 3+4 2 |
| | 8 2 | 8 2 | 3+4 2 | 5+6 2 |
| | 9 2 | 9 2 | 5+6 2 | 3+4 2 |
| | 4+5 2 | 6+7 2 | 3+4 2 | 5+6 2 |
| | 9 2 | 4+5 2 | 5+6 2 | 3+4 2 |
| | 8 2 | 8 2 | 5+6 2 | 3+4 2 |
| | 4+5 2 | 8 2 | 3+4 2 | 5+6 2 |
| | 6+7 2 | 9 2 | 3+4 2 | 5+6 2 |
| | 8 2 | 9 2 | 3+4 2 | 5+6 2 |
| | 9 2 | 8 2 | 5+6 2 | 3+4 2 |
| | 8 2 | 6+7 2 | 5+6 2 | 5+6 2 |
| | 4+5 2 | 9 2 | 3+4 2 | 5+6 2 |
| | 6+7 2 | 6+7 2 | 5+6 2 | 3+4 2 |
| | 9 2 | 4+5 2 | 3+4 2 | 5+6 2 |
| | 6+7 2 | 8 2 | 3+4 2 | 3+4 2 |
| | 4+5 2 | 9 2 | 5+6 2 | 5+6 2 |
| | 8 2 | 8 2 | 3+4 2 | 3+4 2 |
| | 9 2 | 4+5 2 | 3+4 2 | 3+4 2 |
| | 6+7 2 | 9 2 | 5+6 2 | 5+6 2 |
| | 8 2 | 8 2 | 5+6 2 | 5+6 2 |
| | 9 2 | 9 2 | 3+4 2 | 3+4 2 |
| | 6+7 2 | 6+7 2 | 3+4 2 | 3+4 2 |
| | 8 2 | 4+5 2 | 5+6 2 | 5+6 2 |
| | 9 2 | 9 2 | 5+6 2 | 3+4 2 |
| | 8 2 | 8 2 | 3+4 2 | 3+4 2 |
| | 8 2 | 6+7 2 | 5+6 2 | 5+6 2 |
| | 9 2 | 9 2 | 3+4 2 | 5+6 2 |
| | 4+5 2 | 8 2 | 5+6 2 | 3+4 2 |
| | 9 2 | 4+5 2 | 3+4 2 | 5+6 2 |
| | 8 2 | 8 2 | 5+6 2 | 3+4 2 |
| | 4+5 2 | 6+7 2 | 5+6 2 | 3+4 2 |
| | 6+7 2 | 9 2 | 3+4 2 | 5+6 2 |
| | 8 2 | 9 2 | 5+6 2 | 3+4 2 |
| | 9 2 | 8 2 | 3+4 2 | 5+6 2 |
| | 4+5 2 | 4+5 2 | 3+4 2 | 5+6 2 |
| | 8 2 | 8 2 | 5+6 2 | 3+4 2 |
| | 8 2 | 9 2 | 3+4 2 | 5+6 2 |
| | 6+7 2 | 6+7 2 | 5+6 2 | 3+4 2 |
| | 8 2 | 9 2 | 3+4 2 | 3+4 2 |
| | 9 2 | 8 2 | 5+6 2 | 5+6 2 |
| | 6+7 2 | 9 2 | 3+4 2 | 3+4 2 |
| | 9 2 | 4+5 2 | 5+6 2 | 5+6 2 |
| | 4+5 2 | 9 2 | 3+4 2 | 3+4 2 |
| | 8 2 | 6+7 2 | 5+6 2 | 5+6 2 |
| | 9 2 | 8 2 | 3+4 2 | 3+4 2 |
| | 4+5 2 | 4+5 2 | 5+6 2 | 5+6 2 |
| | 9 2 | 9 2 | 5+6 2 | 3+4 2 |
| | 6+7 2 | 8 2 | 3+4 2 | 3+4 2 |

Continued on next column

**TABLE LXXIII– continued**

| TQC | ICSD | NoE | SF | |
|---|---|---|---|---|
| | 8 2 | 9 2 | 5+6 2 | 5+6 2 |
| | 9 2 | 8 2 | 3+4 2 | 5+6 2 |
| | 8 2 | 6+7 2 | 5+6 2 | 5+6 2 |
| | 9 2 | 8 2 | 3+4 2 | 3+4 2 |
| | 4+5 2 | 4+5 2 | 5+6 2 | 3+4 2 |
| | 8 2 | 9 2 | 3+4 2 | 5+6 2 |
| | 8 2 | 8 2 | 3+4 2 | 3+4 2 |
| | 9 2 | 6+7 2 | 5+6 2 | 3+4 2 |
| | 8 2 | 4+5 2 | 3+4 2 | 5+6 2 |
| | 9 2 | 8 2 | 5+6 2 | 3+4 2 |
| | 6+7 2 | 9 2 | 3+4 2 | 5+6 2 |
| | 8 2 | 9 2 | 5+6 2 | 5+6 2 |
| | 9 2 | 9 2 | 5+6 2 | 3+4 2 |
| | 4+5 2 | 6+7 2 | 3+4 2 | 3+4 2 |
| | 8 2 | 8 2 | 5+6 2 | 5+6 2 |
| | 4+5 2 | 9 2 | 5+6 2 | 5+6 2 |
| | 8 2 | 9 2 | 3+4 2 | 3+4 2 |
| | 4+5 2 | 6+7 2 | 3+4 2 | 3+4 2 |
| | 8 2 | 9 2 | 5+6 2 | 5+6 2 |
| | 9 2 | 9 2 | 5+6 2 | 3+4 2 |
| | 6+7 2 | 9 2 | 5+6 2 | 3+4 2 |
| | 8 2 | 6+7 2 | 3+4 2 | 5+6 2 |
| | 4+5 2 | 8 2 | 5+6 2 | 5+6 2 |
| | 8 2 | 4+5 2 | 3+4 2 | 3+4 2 |
| | 9 2 | 8 2 | 5+6 2 | 3+4 2 |
| | 9 2 | 9 2 | 3+4 2 | 5+6 2 |
| | 6+7 2 | 8 2 | 3+4 2 | 5+6 2 |
| | 8 2 | 4+5 2 | 5+6 2 | 3+4 2 |
| | 9 2 | 8 2 | 5+6 2 | 5+6 2 |
| | 4+5 2 | 9 2 | 3+4 2 | 3+4 2 |
| | 9 2 | 6+7 2 | 3+4 2 | 3+4 2 |
| | 4+5 2 | 4+5 2 | 5+6 2 | 5+6 2 |
| | 8 2 | 8 2 | 5+6 2 | 5+6 2 |
| | 8 2 | 8 2 | 5+6 2 | 3+4 2 |
| | 8 2 | 4+5 2 | 3+4 2 | 5+6 2 |
| | 9 2 | 9 2 | 3+4 2 | 5+6 2 |
| | 9 2 | 8 2 | 3+4 2 | 3+4 2 |
| | 6+7 2 | 6+7 2 | 3+4 2 | 3+4 2 |
| | 8 2 | 9 2 | 5+6 2 | 5+6 2 |
| | 4+5 2 | 9 2 | 5+6 2 | 3+4 2 |
| | 6+7 2 | 6+7 2 | 5+6 2 | 3+4 2 |
| 166.1.9 | 106207 | 232 | U2Zn17 | |
| | -GM | -T | -F | -L |
| | 8 2 | 8 2 | 3+4 2 | 3+4 2 |
| | 9 2 | 9 2 | 5+6 2 | 5+6 2 |
| | 8 2 | 9 2 | 3+4 2 | 5+6 2 |
| | 9 2 | 8 2 | 5+6 2 | 3+4 2 |
| | 4+5 2 | 4+5 2 | 5+6 2 | 3+4 2 |
| | 8 2 | 9 2 | 3+4 2 | 5+6 2 |
| | 6+7 2 | 6+7 2 | 3+4 2 | 5+6 2 |
| | 9 2 | 8 2 | 5+6 2 | 3+4 2 |
| | 8 2 | 9 2 | 3+4 2 | 3+4 2 |
| | 8 2 | 9 2 | 5+6 2 | 5+6 2 |
| | 8 2 | 6+7 2 | 5+6 2 | 3+4 2 |
| | 4+5 2 | 9 2 | 5+6 2 | 3+4 2 |
| | 8 2 | 9 2 | 3+4 2 | 3+4 2 |
| | 4+5 2 | 8 2 | 3+4 2 | 5+6 2 |
| | 9 2 | 9 2 | 5+6 2 | 3+4 2 |
| | 8 2 | 4+5 2 | 5+6 2 | 5+6 2 |
| | 6+7 2 | 6+7 2 | 5+6 2 | 3+4 2 |
| | 8 2 | 8 2 | 3+4 2 | 5+6 2 |

Continued on next column

**TABLE LXXIII– continued**

| TQC | ICSD | NoE | SF | |
|---|---|---|---|---|
| | 4+5 2 | 6+7 2 | 3+4 2 | 3+4 2 |
| | 8 2 | 9 2 | 5+6 2 | 5+6 2 |
| | 8 2 | 9 2 | 5+6 2 | 3+4 2 |
| | 4+5 2 | 9 2 | 3+4 2 | 5+6 2 |
| | 9 2 | 8 2 | 3+4 2 | 5+6 2 |
| | 9 2 | 6+7 2 | 5+6 2 | 3+4 2 |
| | 8 2 | 8 2 | 5+6 2 | 3+4 2 |
| | 6+7 2 | 9 2 | 5+6 2 | 3+4 2 |
| | 8 2 | 4+5 2 | 3+4 2 | 5+6 2 |
| | 8 2 | 6+7 2 | 5+6 2 | 3+4 2 |
| | 4+5 2 | 8 2 | 3+4 2 | 5+6 2 |
| | 9 2 | 9 2 | 5+6 2 | 3+4 2 |
| | 9 2 | 6+7 2 | 5+6 2 | 5+6 2 |
| | 8 2 | 4+5 2 | 3+4 2 | 3+4 2 |
| | 9 2 | 8 2 | 3+4 2 | 3+4 2 |
| | 4+5 2 | 9 2 | 5+6 2 | 5+6 2 |
| | 6+7 2 | 8 2 | 3+4 2 | 5+6 2 |
| | 8 2 | 9 2 | 3+4 2 | 3+4 2 |
| | 9 2 | 8 2 | 5+6 2 | 5+6 2 |
| | 8 2 | 9 2 | 3+4 2 | 5+6 2 |
| | 4+5 2 | 6+7 2 | 5+6 2 | 5+6 2 |
| | 6+7 2 | 4+5 2 | 5+6 2 | 3+4 2 |
| | 9 2 | 6+7 2 | 3+4 2 | 3+4 2 |
| | 6+7 2 | 8 2 | 3+4 2 | 5+6 2 |
| | 4+5 2 | 9 2 | 5+6 2 | 5+6 2 |
| | 8 2 | 8 2 | 3+4 2 | 3+4 2 |
| | 9 2 | 4+5 2 | 5+6 2 | 3+4 2 |
| | 8 2 | 9 2 | 3+4 2 | 5+6 2 |
| | 6+7 2 | 8 2 | 5+6 2 | 3+4 2 |
| | 9 2 | 9 2 | 3+4 2 | 5+6 2 |
| | 6+7 2 | 6+7 2 | 5+6 2 | 3+4 2 |
| | 8 2 | 4+5 2 | 3+4 2 | 5+6 2 |
| | 9 2 | 9 2 | 5+6 2 | 3+4 2 |
| | 8 2 | 8 2 | 3+4 2 | 3+4 2 |
| | 8 2 | 6+7 2 | 5+6 2 | 5+6 2 |
| | 9 2 | 9 2 | 3+4 2 | 5+6 2 |
| | 4+5 2 | 8 2 | 5+6 2 | 3+4 2 |
| | 9 2 | 4+5 2 | 5+6 2 | 5+6 2 |
| | 8 2 | 8 2 | 3+4 2 | 3+4 2 |
| | 4+5 2 | 6+7 2 | 5+6 2 | 3+4 2 |
| | 6+7 2 | 9 2 | 3+4 2 | 3+4 2 |
| | 8 2 | 9 2 | 5+6 2 | 5+6 2 |
| | 9 2 | 8 2 | 3+4 2 | 5+6 2 |
| | 4+5 2 | 8 2 | 5+6 2 | 5+6 2 |
| | 8 2 | 9 2 | 3+4 2 | 3+4 2 |
| | 8 2 | 4+5 2 | 3+4 2 | 5+6 2 |
| | 9 2 | 6+7 2 | 5+6 2 | 3+4 2 |
| | 6+7 2 | 9 2 | 3+4 2 | 3+4 2 |
| | 8 2 | 8 2 | 5+6 2 | 5+6 2 |
| | 9 2 | 9 2 | 3+4 2 | 3+4 2 |
| | 6+7 2 | 4+5 2 | 3+4 2 | 5+6 2 |
| | 4+5 2 | 9 2 | 5+6 2 | 3+4 2 |
| | 9 2 | 6+7 2 | 5+6 2 | 5+6 2 |
| | 8 2 | 8 2 | 3+4 2 | 3+4 2 |
| | 6+7 2 | 9 2 | 5+6 2 | 5+6 2 |
| | 9 2 | 4+5 2 | 5+6 2 | 3+4 2 |
| | 4+5 2 | 8 2 | 3+4 2 | 3+4 2 |
| | 8 2 | 9 2 | 5+6 2 | 5+6 2 |
| | 9 2 | 8 2 | 3+4 2 | 5+6 2 |
| | 8 2 | 6+7 2 | 5+6 2 | 5+6 2 |
| | 9 2 | 8 2 | 3+4 2 | 3+4 2 |
| | 4+5 2 | 4+5 2 | 3+4 2 | 3+4 2 |
| | 8 2 | 9 2 | 5+6 2 | 5+6 2 |
| | 9 2 | 8 2 | 3+4 2 | 3+4 2 |
| | 8 2 | 4+5 2 | 5+6 2 | 3+4 2 |
| | 8 2 | 6+7 2 | 3+4 2 | 5+6 2 |
| | 9 2 | 8 2 | 5+6 2 | 3+4 2 |
| | 6+7 2 | 9 2 | 3+4 2 | 5+6 2 |
| | 8 2 | 9 2 | 5+6 2 | 5+6 2 |
| | 9 2 | 9 2 | 5+6 2 | 3+4 2 |
| | 4+5 2 | 6+7 2 | 3+4 2 | 3+4 2 |
| | 8 2 | 8 2 | 5+6 2 | 5+6 2 |
| | 4+5 2 | 9 2 | 3+4 2 | 5+6 2 |
| | 8 2 | 9 2 | 5+6 2 | 3+4 2 |
| | 4+5 2 | 6+7 2 | 3+4 2 | 3+4 2 |
| | 8 2 | 9 2 | 5+6 2 | 5+6 2 |
| | 9 2 | 9 2 | 5+6 2 | 3+4 2 |
| | 6+7 2 | 9 2 | 5+6 2 | 3+4 2 |
| | 8 2 | 6+7 2 | 3+4 2 | 5+6 2 |
| | 4+5 2 | 8 2 | 5+6 2 | 5+6 2 |
| | 8 2 | 4+5 2 | 3+4 2 | 3+4 2 |
| | 9 2 | 8 2 | 5+6 2 | 3+4 2 |
| | 9 2 | 9 2 | 3+4 2 | 5+6 2 |
| | 6+7 2 | 8 2 | 3+4 2 | 5+6 2 |
| | 8 2 | 8 2 | 5+6 2 | 3+4 2 |
| | 9 2 | 4+5 2 | 5+6 2 | 5+6 2 |
| | 4+5 2 | 9 2 | 3+4 2 | 3+4 2 |
| | 9 2 | 6+7 2 | 3+4 2 | 3+4 2 |
| | 4+5 2 | 4+5 2 | 5+6 2 | 5+6 2 |
| | 8 2 | 8 2 | 5+6 2 | 5+6 2 |
| | 8 2 | 8 2 | 3+4 2 | 3+4 2 |
| | 8 2 | 4+5 2 | 5+6 2 | 5+6 2 |
| | 9 2 | 9 2 | 3+4 2 | 5+6 2 |
| | 9 2 | 8 2 | 3+4 2 | 3+4 2 |
| | 6+7 2 | 9 2 | 3+4 2 | 3+4 2 |
| | 8 2 | 6+7 2 | 5+6 2 | 5+6 2 |
| | 4+5 2 | 9 2 | 5+6 2 | 3+4 2 |
| | 6+7 2 | 6+7 2 | 5+6 2 | 3+4 2 |
| 166.1.10 | 104248 | 232 | U2Zn17 | |
| | -GM | -T | -F | -L |
| | 8 2 | 8 2 | 3+4 2 | 3+4 2 |
| | 9 2 | 9 2 | 5+6 2 | 5+6 2 |
| | 8 2 | 9 2 | 3+4 2 | 5+6 2 |
| | 9 2 | 8 2 | 5+6 2 | 3+4 2 |
| | 4+5 2 | 4+5 2 | 5+6 2 | 3+4 2 |
| | 8 2 | 9 2 | 3+4 2 | 5+6 2 |
| | 6+7 2 | 6+7 2 | 3+4 2 | 5+6 2 |
| | 9 2 | 8 2 | 5+6 2 | 3+4 2 |
| | 8 2 | 9 2 | 3+4 2 | 3+4 2 |
| | 8 2 | 9 2 | 5+6 2 | 5+6 2 |
| | 8 2 | 6+7 2 | 5+6 2 | 3+4 2 |
| | 4+5 2 | 9 2 | 3+4 2 | 3+4 2 |
| | 8 2 | 9 2 | 5+6 2 | 3+4 2 |
| | 9 2 | 8 2 | 3+4 2 | 5+6 2 |
| | 4+5 2 | 4+5 2 | 5+6 2 | 3+4 2 |
| | 6+7 2 | 9 2 | 5+6 2 | 5+6 2 |
| | 8 2 | 8 2 | 3+4 2 | 3+4 2 |
| | 8 2 | 6+7 2 | 5+6 2 | 5+6 2 |
| | 4+5 2 | 9 2 | 5+6 2 | 5+6 2 |
| | 8 2 | 6+7 2 | 3+4 2 | 3+4 2 |
| | 9 2 | 8 2 | 5+6 2 | 5+6 2 |
| | 8 2 | 9 2 | 3+4 2 | 3+4 2 |

Continued on next column

**TABLE LXXIII– continued**

| TQC | ICSD | NoE | SF | |
|---|---|---|---|---|
| | 9 2 | 9 2 | 3+4 2 | 5+6 2 |
| | 4+5 2 | 8 2 | 5+6 2 | 3+4 2 |
| | 6+7 2 | 6+7 2 | 5+6 2 | 3+4 2 |
| | 8 2 | 4+5 2 | 3+4 2 | 3+4 2 |
| | 8 2 | 9 2 | 5+6 2 | 5+6 2 |
| | 8 2 | 6+7 2 | 5+6 2 | 3+4 2 |
| | 9 2 | 9 2 | 3+4 2 | 5+6 2 |
| | 9 2 | 8 2 | 3+4 2 | 3+4 2 |
| | 4+5 2 | 6+7 2 | 5+6 2 | 5+6 2 |
| | 8 2 | 8 2 | 5+6 2 | 3+4 2 |
| | 9 2 | 4+5 2 | 5+6 2 | 3+4 2 |
| | 4+5 2 | 9 2 | 3+4 2 | 5+6 2 |
| | 6+7 2 | 8 2 | 3+4 2 | 5+6 2 |
| | 8 2 | 9 2 | 3+4 2 | 5+6 2 |
| | 9 2 | 8 2 | 5+6 2 | 3+4 2 |
| | 8 2 | 6+7 2 | 5+6 2 | 5+6 2 |
| | 4+5 2 | 9 2 | 3+4 2 | 5+6 2 |
| | 6+7 2 | 6+7 2 | 5+6 2 | 3+4 2 |
| | 9 2 | 4+5 2 | 3+4 2 | 5+6 2 |
| | 6+7 2 | 8 2 | 3+4 2 | 3+4 2 |
| | 4+5 2 | 9 2 | 5+6 2 | 5+6 2 |
| | 9 2 | 8 2 | 3+4 2 | 3+4 2 |
| | 6+7 2 | 4+5 2 | 3+4 2 | 3+4 2 |
| | 8 2 | 9 2 | 5+6 2 | 5+6 2 |
| | 8 2 | 8 2 | 5+6 2 | 5+6 2 |
| | 9 2 | 9 2 | 3+4 2 | 3+4 2 |
| | 6+7 2 | 6+7 2 | 5+6 2 | 3+4 2 |
| | 8 2 | 4+5 2 | 3+4 2 | 5+6 2 |
| | 9 2 | 9 2 | 5+6 2 | 3+4 2 |
| | 8 2 | 8 2 | 3+4 2 | 3+4 2 |
| | 9 2 | 6+7 2 | 5+6 2 | 5+6 2 |
| | 8 2 | 9 2 | 3+4 2 | 5+6 2 |
| | 4+5 2 | 8 2 | 5+6 2 | 3+4 2 |
| | 9 2 | 4+5 2 | 3+4 2 | 5+6 2 |
| | 8 2 | 6+7 2 | 5+6 2 | 3+4 2 |
| | 4+5 2 | 8 2 | 5+6 2 | 3+4 2 |
| | 6+7 2 | 9 2 | 3+4 2 | 5+6 2 |
| | 8 2 | 9 2 | 5+6 2 | 3+4 2 |
| | 9 2 | 8 2 | 3+4 2 | 5+6 2 |
| | 8 2 | 4+5 2 | 5+6 2 | 5+6 2 |
| | 8 2 | 8 2 | 3+4 2 | 3+4 2 |
| | 4+5 2 | 9 2 | 3+4 2 | 5+6 2 |
| | 6+7 2 | 6+7 2 | 5+6 2 | 3+4 2 |
| | 8 2 | 9 2 | 3+4 2 | 3+4 2 |
| | 9 2 | 8 2 | 5+6 2 | 5+6 2 |
| | 6+7 2 | 9 2 | 3+4 2 | 3+4 2 |
| | 9 2 | 4+5 2 | 5+6 2 | 5+6 2 |
| | 4+5 2 | 9 2 | 3+4 2 | 3+4 2 |
| | 8 2 | 6+7 2 | 5+6 2 | 5+6 2 |
| | 9 2 | 8 2 | 3+4 2 | 3+4 2 |
| | 4+5 2 | 4+5 2 | 5+6 2 | 5+6 2 |
| | 6+7 2 | 9 2 | 5+6 2 | 3+4 2 |
| | 8 2 | 8 2 | 3+4 2 | 5+6 2 |
| | 9 2 | 9 2 | 5+6 2 | 3+4 2 |
| | 8 2 | 8 2 | 3+4 2 | 5+6 2 |
| | 9 2 | 6+7 2 | 5+6 2 | 5+6 2 |
| | 9 2 | 8 2 | 3+4 2 | 3+4 2 |
| | 4+5 2 | 4+5 2 | 5+6 2 | 3+4 2 |
| | 8 2 | 9 2 | 3+4 2 | 5+6 2 |
| | 8 2 | 8 2 | 3+4 2 | 3+4 2 |
| | 9 2 | 6+7 2 | 5+6 2 | 3+4 2 |

Continued on next column

**TABLE LXXIII– continued**

| TQC | ICSD | NoE | SF | |
|---|---|---|---|---|
| | 8 2 | 4+5 2 | 3+4 2 | 5+6 2 |
| | 9 2 | 8 2 | 5+6 2 | 3+4 2 |
| | 8 2 | 9 2 | 3+4 2 | 5+6 2 |
| | 6+7 2 | 9 2 | 5+6 2 | 5+6 2 |
| | 4+5 2 | 9 2 | 5+6 2 | 3+4 2 |
| | 9 2 | 6+7 2 | 3+4 2 | 3+4 2 |
| | 8 2 | 8 2 | 5+6 2 | 5+6 2 |
| | 4+5 2 | 9 2 | 5+6 2 | 3+4 2 |
| | 8 2 | 9 2 | 3+4 2 | 5+6 2 |
| | 4+5 2 | 6+7 2 | 3+4 2 | 3+4 2 |
| | 8 2 | 9 2 | 5+6 2 | 5+6 2 |
| | 9 2 | 9 2 | 5+6 2 | 3+4 2 |
| | 6+7 2 | 9 2 | 5+6 2 | 3+4 2 |
| | 8 2 | 6+7 2 | 3+4 2 | 5+6 2 |
| | 4+5 2 | 8 2 | 5+6 2 | 5+6 2 |
| | 8 2 | 4+5 2 | 3+4 2 | 3+4 2 |
| | 9 2 | 8 2 | 5+6 2 | 3+4 2 |
| | 9 2 | 9 2 | 3+4 2 | 5+6 2 |
| | 6+7 2 | 8 2 | 3+4 2 | 5+6 2 |
| | 8 2 | 8 2 | 5+6 2 | 3+4 2 |
| | 9 2 | 4+5 2 | 5+6 2 | 5+6 2 |
| | 4+5 2 | 9 2 | 3+4 2 | 3+4 2 |
| | 9 2 | 6+7 2 | 3+4 2 | 3+4 2 |
| | 4+5 2 | 4+5 2 | 5+6 2 | 5+6 2 |
| | 8 2 | 8 2 | 5+6 2 | 5+6 2 |
| | 8 2 | 8 2 | 5+6 2 | 3+4 2 |
| | 8 2 | 4+5 2 | 3+4 2 | 5+6 2 |
| | 9 2 | 9 2 | 3+4 2 | 3+4 2 |
| | 9 2 | 8 2 | 3+4 2 | 5+6 2 |
| | 6+7 2 | 9 2 | 3+4 2 | 3+4 2 |
| | 8 2 | 6+7 2 | 5+6 2 | 5+6 2 |
| | 4+5 2 | 9 2 | 5+6 2 | 3+4 2 |
| | 6+7 2 | 6+7 2 | 5+6 2 | 3+4 2 |
| 166.1.11 | 653358 | 250 | Tm2Zn17 | |
| | -GM | -T | -F | -L |
| | 8 2 | 8 2 | 3+4 2 | 3+4 2 |
| | 9 2 | 9 2 | 5+6 2 | 5+6 2 |
| | 8 2 | 8 2 | 3+4 2 | 5+6 2 |
| | 9 2 | 9 2 | 5+6 2 | 3+4 2 |
| | 4+5 2 | 4+5 2 | 5+6 2 | 3+4 2 |
| | 6+7 2 | 6+7 2 | 3+4 2 | 5+6 2 |
| | 8 2 | 9 2 | 3+4 2 | 5+6 2 |
| | 9 2 | 8 2 | 5+6 2 | 3+4 2 |
| | 8 2 | 9 2 | 3+4 2 | 3+4 2 |
| | 8 2 | 9 2 | 5+6 2 | 5+6 2 |
| | 8 2 | 6+7 2 | 5+6 2 | 3+4 2 |
| | 4+5 2 | 9 2 | 5+6 2 | 3+4 2 |
| | 8 2 | 8 2 | 3+4 2 | 3+4 2 |
| | 9 2 | 4+5 2 | 3+4 2 | 3+4 2 |
| | 6+7 2 | 9 2 | 5+6 2 | 5+6 2 |
| | 4+5 2 | 9 2 | 5+6 2 | 5+6 2 |
| | 8 2 | 6+7 2 | 3+4 2 | 3+4 2 |
| | 8 2 | 8 2 | 5+6 2 | 5+6 2 |
| | 4+5 2 | 9 2 | 5+6 2 | 5+6 2 |
| | 8 2 | 6+7 2 | 3+4 2 | 3+4 2 |
| | 9 2 | 9 2 | 5+6 2 | 5+6 2 |
| | 9 2 | 8 2 | 3+4 2 | 3+4 2 |
| | 8 2 | 9 2 | 3+4 2 | 3+4 2 |
| | 4+5 2 | 6+7 2 | 5+6 2 | 5+6 2 |
| | 6+7 2 | 8 2 | 5+6 2 | 3+4 2 |
| | 8 2 | 4+5 2 | 3+4 2 | 5+6 2 |

Continued on next column

**TABLE LXXIII– continued**

| TQC | ICSD | NoE | SF | |
|---|---|---|---|---|
| | 8 2 | 9 2 | 5+6 2 | 3+4 2 |
| | 9 2 | 6+7 2 | 3+4 2 | 3+4 2 |
| | 8 2 | 8 2 | 5+6 2 | 3+4 2 |
| | 4+5 2 | 9 2 | 5+6 2 | 5+6 2 |
| | 9 2 | 6+7 2 | 3+4 2 | 5+6 2 |
| | 8 2 | 8 2 | 5+6 2 | 3+4 2 |
| | 4+5 2 | 4+5 2 | 3+4 2 | 3+4 2 |
| | 9 2 | 9 2 | 5+6 2 | 5+6 2 |
| | 6+7 2 | 8 2 | 3+4 2 | 5+6 2 |
| | 9 2 | 8 2 | 3+4 2 | 3+4 2 |
| | 8 2 | 9 2 | 5+6 2 | 5+6 2 |
| | 8 2 | 6+7 2 | 5+6 2 | 5+6 2 |
| | 4+5 2 | 9 2 | 3+4 2 | 5+6 2 |
| | 6+7 2 | 4+5 2 | 5+6 2 | 3+4 2 |
| | 6+7 2 | 6+7 2 | 3+4 2 | 5+6 2 |
| | 9 2 | 9 2 | 3+4 2 | 3+4 2 |
| | 4+5 2 | 8 2 | 5+6 2 | 5+6 2 |
| | 9 2 | 4+5 2 | 3+4 2 | 3+4 2 |
| | 6+7 2 | 8 2 | 3+4 2 | 3+4 2 |
| | 8 2 | 9 2 | 5+6 2 | 5+6 2 |
| | 8 2 | 4+5 2 | 5+6 2 | 5+6 2 |
| | 9 2 | 6+7 2 | 3+4 2 | 3+4 2 |
| | 6+7 2 | 9 2 | 5+6 2 | 3+4 2 |
| | 8 2 | 8 2 | 3+4 2 | 5+6 2 |
| | 9 2 | 9 2 | 5+6 2 | 3+4 2 |
| | 8 2 | 8 2 | 3+4 2 | 3+4 2 |
| | 9 2 | 6+7 2 | 5+6 2 | 5+6 2 |
| | 8 2 | 9 2 | 3+4 2 | 5+6 2 |
| | 8 2 | 4+5 2 | 5+6 2 | 3+4 2 |
| | 4+5 2 | 8 2 | 3+4 2 | 5+6 2 |
| | 9 2 | 8 2 | 5+6 2 | 3+4 2 |
| | 4+5 2 | 9 2 | 3+4 2 | 5+6 2 |
| | 6+7 2 | 6+7 2 | 5+6 2 | 3+4 2 |
| | 8 2 | 8 2 | 5+6 2 | 3+4 2 |
| | 9 2 | 9 2 | 3+4 2 | 5+6 2 |
| | 6+7 2 | 8 2 | 5+6 2 | 5+6 2 |
| | 8 2 | 4+5 2 | 3+4 2 | 5+6 2 |
| | 8 2 | 9 2 | 3+4 2 | 3+4 2 |
| | 9 2 | 6+7 2 | 5+6 2 | 3+4 2 |
| | 4+5 2 | 9 2 | 3+4 2 | 3+4 2 |
| | 8 2 | 8 2 | 3+4 2 | 5+6 2 |
| | 9 2 | 9 2 | 5+6 2 | 3+4 2 |
| | 6+7 2 | 4+5 2 | 5+6 2 | 5+6 2 |
| | 4+5 2 | 9 2 | 3+4 2 | 3+4 2 |
| | 9 2 | 6+7 2 | 5+6 2 | 5+6 2 |
| | 8 2 | 8 2 | 3+4 2 | 3+4 2 |
| | 9 2 | 4+5 2 | 5+6 2 | 5+6 2 |
| | 4+5 2 | 9 2 | 5+6 2 | 3+4 2 |
| | 6+7 2 | 8 2 | 3+4 2 | 5+6 2 |
| | 8 2 | 8 2 | 5+6 2 | 3+4 2 |
| | 9 2 | 9 2 | 3+4 2 | 5+6 2 |
| | 8 2 | 6+7 2 | 5+6 2 | 5+6 2 |
| | 4+5 2 | 8 2 | 3+4 2 | 3+4 2 |
| | 9 2 | 4+5 2 | 5+6 2 | 3+4 2 |
| | 8 2 | 9 2 | 3+4 2 | 5+6 2 |
| | 8 2 | 8 2 | 3+4 2 | 3+4 2 |
| | 9 2 | 6+7 2 | 5+6 2 | 3+4 2 |
| | 8 2 | 4+5 2 | 3+4 2 | 5+6 2 |
| | 9 2 | 8 2 | 5+6 2 | 3+4 2 |
| | 8 2 | 9 2 | 3+4 2 | 5+6 2 |
| | 6+7 2 | 9 2 | 5+6 2 | 5+6 2 |

Continued on next column

**TABLE LXXIII– continued**

| TQC | ICSD | NoE | SF | |
|---|---|---|---|---|
| | 9 2 | 9 2 | 5+6 2 | 3+4 2 |
| | 4+5 2 | 6+7 2 | 3+4 2 | 5+6 2 |
| | 8 2 | 8 2 | 5+6 2 | 3+4 2 |
| | 4+5 2 | 9 2 | 3+4 2 | 3+4 2 |
| | 8 2 | 9 2 | 5+6 2 | 5+6 2 |
| | 4+5 2 | 9 2 | 3+4 2 | 3+4 2 |
| | 8 2 | 6+7 2 | 5+6 2 | 5+6 2 |
| | 9 2 | 9 2 | 5+6 2 | 3+4 2 |
| | 6+7 2 | 9 2 | 5+6 2 | 3+4 2 |
| | 8 2 | 6+7 2 | 3+4 2 | 5+6 2 |
| | 4+5 2 | 8 2 | 5+6 2 | 5+6 2 |
| | 8 2 | 4+5 2 | 3+4 2 | 3+4 2 |
| | 9 2 | 8 2 | 3+4 2 | 3+4 2 |
| | 6+7 2 | 9 2 | 5+6 2 | 5+6 2 |
| | 9 2 | 8 2 | 3+4 2 | 5+6 2 |
| | 8 2 | 8 2 | 5+6 2 | 3+4 2 |
| | 4+5 2 | 4+5 2 | 5+6 2 | 5+6 2 |
| | 9 2 | 6+7 2 | 3+4 2 | 3+4 2 |
| | 9 2 | 9 2 | 3+4 2 | 5+6 2 |
| | 8 2 | 9 2 | 5+6 2 | 3+4 2 |
| | 8 2 | 8 2 | 5+6 2 | 5+6 2 |
| | 4+5 2 | 8 2 | 3+4 2 | 3+4 2 |
| | 9 2 | 4+5 2 | 5+6 2 | 5+6 2 |
| | 6+7 2 | 9 2 | 3+4 2 | 3+4 2 |
| | 9 2 | 6+7 2 | 5+6 2 | 3+4 2 |
| | 8 2 | 8 2 | 3+4 2 | 5+6 2 |
| | 8 2 | 4+5 2 | 5+6 2 | 5+6 2 |
| | 4+5 2 | 4+5 2 | 5+6 2 | 5+6 2 |
| | 9 2 | 8 2 | 3+4 2 | 3+4 2 |
| | 8 2 | 8 2 | 3+4 2 | 3+4 2 |
| | 6+7 2 | 9 2 | 3+4 2 | 5+6 2 |
| | 9 2 | 9 2 | 3+4 2 | 3+4 2 |
| | 8 2 | 6+7 2 | 5+6 2 | 3+4 2 |
| | 8 2 | 9 2 | 5+6 2 | 5+6 2 |
| | 9 2 | 8 2 | 5+6 2 | 3+4 2 |
| | 4+5 2 | 8 2 | 3+4 2 | 5+6 2 |
| | 8 2 | 9 2 | 5+6 2 | 5+6 2 |
| | 6+7 2 | 4+5 2 | 3+4 2 | 5+6 2 |
| 166.1.12 | 104246 | 250 | Tm2Zn17 | |
| | -GM | -T | -F | -L |
| | 8 2 | 8 2 | 3+4 2 | 3+4 2 |
| | 9 2 | 9 2 | 5+6 2 | 5+6 2 |
| | 8 2 | 8 2 | 3+4 2 | 5+6 2 |
| | 9 2 | 9 2 | 5+6 2 | 3+4 2 |
| | 4+5 2 | 4+5 2 | 5+6 2 | 3+4 2 |
| | 6+7 2 | 6+7 2 | 3+4 2 | 5+6 2 |
| | 8 2 | 9 2 | 3+4 2 | 5+6 2 |
| | 9 2 | 8 2 | 5+6 2 | 3+4 2 |
| | 8 2 | 9 2 | 3+4 2 | 3+4 2 |
| | 8 2 | 9 2 | 5+6 2 | 5+6 2 |
| | 8 2 | 6+7 2 | 5+6 2 | 3+4 2 |
| | 4+5 2 | 9 2 | 5+6 2 | 3+4 2 |
| | 8 2 | 8 2 | 3+4 2 | 3+4 2 |
| | 9 2 | 4+5 2 | 3+4 2 | 5+6 2 |
| | 4+5 2 | 9 2 | 5+6 2 | 3+4 2 |
| | 6+7 2 | 9 2 | 5+6 2 | 5+6 2 |
| | 8 2 | 6+7 2 | 3+4 2 | 3+4 2 |
| | 8 2 | 8 2 | 5+6 2 | 5+6 2 |
| | 4+5 2 | 9 2 | 3+4 2 | 5+6 2 |
| | 8 2 | 6+7 2 | 5+6 2 | 3+4 2 |
| | 9 2 | 9 2 | 5+6 2 | 5+6 2 |

Continued on next column

**TABLE LXXIII– continued**

| TQC | ICSD | NoE | SF | |
|---|---|---|---|---|
| | 9 2 | 8 2 | 3+4 2 | 3+4 2 |
| | 8 2 | 9 2 | 3+4 2 | 3+4 2 |
| | 4+5 2 | 6+7 2 | 5+6 2 | 5+6 2 |
| | 6+7 2 | 8 2 | 5+6 2 | 3+4 2 |
| | 8 2 | 4+5 2 | 3+4 2 | 5+6 2 |
| | 8 2 | 9 2 | 5+6 2 | 3+4 2 |
| | 9 2 | 6+7 2 | 3+4 2 | 3+4 2 |
| | 8 2 | 8 2 | 5+6 2 | 5+6 2 |
| | 4+5 2 | 9 2 | 5+6 2 | 3+4 2 |
| | 9 2 | 6+7 2 | 3+4 2 | 5+6 2 |
| | 8 2 | 8 2 | 5+6 2 | 3+4 2 |
| | 4+5 2 | 4+5 2 | 5+6 2 | 3+4 2 |
| | 9 2 | 9 2 | 3+4 2 | 5+6 2 |
| | 6+7 2 | 8 2 | 3+4 2 | 5+6 2 |
| | 9 2 | 8 2 | 3+4 2 | 3+4 2 |
| | 8 2 | 9 2 | 5+6 2 | 5+6 2 |
| | 8 2 | 6+7 2 | 3+4 2 | 5+6 2 |
| | 4+5 2 | 9 2 | 5+6 2 | 5+6 2 |
| | 6+7 2 | 4+5 2 | 5+6 2 | 3+4 2 |
| | 6+7 2 | 6+7 2 | 3+4 2 | 5+6 2 |
| | 9 2 | 9 2 | 3+4 2 | 3+4 2 |
| | 9 2 | 8 2 | 5+6 2 | 5+6 2 |
| | 4+5 2 | 4+5 2 | 3+4 2 | 3+4 2 |
| | 6+7 2 | 8 2 | 3+4 2 | 3+4 2 |
| | 8 2 | 9 2 | 5+6 2 | 5+6 2 |
| | 8 2 | 4+5 2 | 5+6 2 | 5+6 2 |
| | 9 2 | 9 2 | 3+4 2 | 3+4 2 |
| | 6+7 2 | 6+7 2 | 5+6 2 | 3+4 2 |
| | 8 2 | 8 2 | 3+4 2 | 5+6 2 |
| | 9 2 | 9 2 | 5+6 2 | 3+4 2 |
| | 8 2 | 8 2 | 3+4 2 | 3+4 2 |
| | 9 2 | 6+7 2 | 5+6 2 | 5+6 2 |
| | 8 2 | 9 2 | 3+4 2 | 5+6 2 |
| | 8 2 | 4+5 2 | 5+6 2 | 3+4 2 |
| | 4+5 2 | 8 2 | 3+4 2 | 5+6 2 |
| | 9 2 | 8 2 | 5+6 2 | 3+4 2 |
| | 4+5 2 | 9 2 | 5+6 2 | 5+6 2 |
| | 6+7 2 | 6+7 2 | 3+4 2 | 3+4 2 |
| | 8 2 | 8 2 | 5+6 2 | 3+4 2 |
| | 9 2 | 9 2 | 3+4 2 | 5+6 2 |
| | 6+7 2 | 8 2 | 5+6 2 | 5+6 2 |
| | 8 2 | 9 2 | 3+4 2 | 5+6 2 |
| | 8 2 | 4+5 2 | 3+4 2 | 3+4 2 |
| | 9 2 | 6+7 2 | 5+6 2 | 3+4 2 |
| | 4+5 2 | 9 2 | 3+4 2 | 3+4 2 |
| | 8 2 | 8 2 | 3+4 2 | 5+6 2 |
| | 9 2 | 9 2 | 5+6 2 | 3+4 2 |
| | 6+7 2 | 4+5 2 | 5+6 2 | 5+6 2 |
| | 4+5 2 | 9 2 | 3+4 2 | 3+4 2 |
| | 9 2 | 6+7 2 | 5+6 2 | 5+6 2 |
| | 8 2 | 8 2 | 3+4 2 | 3+4 2 |
| | 9 2 | 4+5 2 | 5+6 2 | 5+6 2 |
| | 4+5 2 | 9 2 | 5+6 2 | 3+4 2 |
| | 6+7 2 | 8 2 | 3+4 2 | 5+6 2 |
| | 8 2 | 8 2 | 5+6 2 | 3+4 2 |
| | 9 2 | 9 2 | 3+4 2 | 5+6 2 |
| | 8 2 | 6+7 2 | 5+6 2 | 5+6 2 |
| | 4+5 2 | 8 2 | 3+4 2 | 3+4 2 |
| | 9 2 | 4+5 2 | 5+6 2 | 3+4 2 |
| | 8 2 | 9 2 | 3+4 2 | 5+6 2 |
| | 8 2 | 8 2 | 3+4 2 | 3+4 2 |

Continued on next column

**TABLE LXXIII– continued**

| TQC | ICSD | NoE | SF | |
|---|---|---|---|---|
| | 9 2 | 6+7 2 | 5+6 2 | 3+4 2 |
| | 8 2 | 4+5 2 | 3+4 2 | 5+6 2 |
| | 9 2 | 8 2 | 5+6 2 | 3+4 2 |
| | 6+7 2 | 9 2 | 3+4 2 | 5+6 2 |
| | 8 2 | 9 2 | 5+6 2 | 5+6 2 |
| | 9 2 | 9 2 | 5+6 2 | 3+4 2 |
| | 4+5 2 | 6+7 2 | 3+4 2 | 3+4 2 |
| | 8 2 | 8 2 | 5+6 2 | 5+6 2 |
| | 4+5 2 | 9 2 | 3+4 2 | 3+4 2 |
| | 8 2 | 9 2 | 5+6 2 | 5+6 2 |
| | 4+5 2 | 9 2 | 3+4 2 | 3+4 2 |
| | 8 2 | 6+7 2 | 5+6 2 | 5+6 2 |
| | 9 2 | 9 2 | 5+6 2 | 3+4 2 |
| | 6+7 2 | 9 2 | 5+6 2 | 3+4 2 |
| | 8 2 | 6+7 2 | 3+4 2 | 5+6 2 |
| | 4+5 2 | 8 2 | 5+6 2 | 5+6 2 |
| | 8 2 | 4+5 2 | 3+4 2 | 3+4 2 |
| | 9 2 | 8 2 | 5+6 2 | 3+4 2 |
| | 6+7 2 | 9 2 | 3+4 2 | 5+6 2 |
| | 9 2 | 8 2 | 3+4 2 | 5+6 2 |
| | 8 2 | 8 2 | 5+6 2 | 3+4 2 |
| | 4+5 2 | 4+5 2 | 5+6 2 | 5+6 2 |
| | 9 2 | 6+7 2 | 3+4 2 | 3+4 2 |
| | 9 2 | 9 2 | 3+4 2 | 5+6 2 |
| | 8 2 | 9 2 | 5+6 2 | 3+4 2 |
| | 8 2 | 8 2 | 5+6 2 | 5+6 2 |
| | 4+5 2 | 8 2 | 3+4 2 | 3+4 2 |
| | 9 2 | 4+5 2 | 5+6 2 | 5+6 2 |
| | 6+7 2 | 9 2 | 3+4 2 | 3+4 2 |
| | 9 2 | 6+7 2 | 5+6 2 | 3+4 2 |
| | 8 2 | 8 2 | 3+4 2 | 5+6 2 |
| | 8 2 | 4+5 2 | 5+6 2 | 5+6 2 |
| | 4+5 2 | 4+5 2 | 5+6 2 | 3+4 2 |
| | 9 2 | 8 2 | 3+4 2 | 5+6 2 |
| | 8 2 | 8 2 | 3+4 2 | 5+6 2 |
| | 6+7 2 | 9 2 | 3+4 2 | 3+4 2 |
| | 9 2 | 9 2 | 3+4 2 | 3+4 2 |
| | 8 2 | 6+7 2 | 5+6 2 | 3+4 2 |
| | 8 2 | 9 2 | 5+6 2 | 5+6 2 |
| | 9 2 | 8 2 | 5+6 2 | 3+4 2 |
| | 4+5 2 | 8 2 | 3+4 2 | 5+6 2 |
| | 8 2 | 9 2 | 5+6 2 | 5+6 2 |
| | 6+7 2 | 4+5 2 | 3+4 2 | 5+6 2 |
| 166.1.13 | 653216 | 54 | Te1Zr3 | |
| | -GM | -T | -F | -L |
| | 8 2 | 9 2 | 5+6 2 | 3+4 2 |
| | 8 2 | 8 2 | 3+4 2 | 5+6 2 |
| | 9 2 | 8 2 | 3+4 2 | 3+4 2 |
| | 8 2 | 9 2 | 5+6 2 | 5+6 2 |
| | 9 2 | 9 2 | 3+4 2 | 3+4 2 |
| | 4+5 2 | 6+7 2 | 5+6 2 | 5+6 2 |
| | 9 2 | 9 2 | 5+6 2 | 3+4 2 |
| | 8 2 | 8 2 | 5+6 2 | 5+6 2 |
| | 6+7 2 | 6+7 2 | 3+4 2 | 5+6 2 |
| | 8 2 | 8 2 | 3+4 2 | 5+6 2 |
| | 9 2 | 9 2 | 5+6 2 | 5+6 2 |
| | 9 2 | 8 2 | 5+6 2 | 3+4 2 |
| | 6+7 2 | 4+5 2 | 3+4 2 | 5+6 2 |
| | 9 2 | 9 2 | 5+6 2 | 3+4 2 |
| | 8 2 | 9 2 | 3+4 2 | 3+4 2 |
| | 4+5 2 | 6+7 2 | 3+4 2 | 5+6 2 |

Continued on next column

**TABLE LXXIII– continued**

| TQC | ICSD | NoE | SF | |
|---|---|---|---|---|
| | 9 2 | 4+5 2 | 3+4 2 | 5+6 2 |
| | 6+7 2 | 8 2 | 3+4 2 | 3+4 2 |
| | 4+5 2 | 8 2 | 5+6 2 | 3+4 2 |
| | 8 2 | 4+5 2 | 3+4 2 | 5+6 2 |
| | 8 2 | 9 2 | 5+6 2 | 5+6 2 |
| | 9 2 | 8 2 | 5+6 2 | 5+6 2 |
| | 8 2 | 9 2 | 3+4 2 | 3+4 2 |
| | 4+5 2 | 6+7 2 | 3+4 2 | 5+6 2 |
| | 8 2 | 8 2 | 3+4 2 | 3+4 2 |
| | 9 2 | 4+5 2 | 3+4 2 | 5+6 2 |
| | 6+7 2 | 8 2 | 5+6 2 | 5+6 2 |
| 166.1.14 | 52499 | 12 | Te1 | |
| | -GM | -T | -F | -L |
| | 8 2 | 9 2 | 5+6 2 | 3+4 2 |
| | 9 2 | 8 2 | 3+4 2 | 5+6 2 |
| | 8 2 | 9 2 | 3+4 2 | 5+6 2 |
| | 8 2 | 6+7 2 | 5+6 2 | 3+4 2 |
| | 4+5 2 | 8 2 | 5+6 2 | 3+4 2 |
| | 9 2 | 8 2 | 3+4 2 | 5+6 2 |
| 166.1.15 | 410956 | 88 | Sn3Sr1 | |
| | -GM | -T | -F | -L |
| | 8 2 | 9 2 | 3+4 2 | 3+4 2 |
| | 9 2 | 8 2 | 5+6 2 | 5+6 2 |
| | 8 2 | 8 2 | 3+4 2 | 3+4 2 |
| | 9 2 | 9 2 | 5+6 2 | 5+6 2 |
| | 8 2 | 9 2 | 5+6 2 | 3+4 2 |
| | 9 2 | 8 2 | 3+4 2 | 5+6 2 |
| | 8 2 | 8 2 | 5+6 2 | 5+6 2 |
| | 9 2 | 9 2 | 3+4 2 | 3+4 2 |
| | 4+5 2 | 6+7 2 | 5+6 2 | 3+4 2 |
| | 8 2 | 9 2 | 3+4 2 | 5+6 2 |
| | 9 2 | 8 2 | 5+6 2 | 3+4 2 |
| | 6+7 2 | 4+5 2 | 3+4 2 | 5+6 2 |
| | 8 2 | 8 2 | 5+6 2 | 5+6 2 |
| | 4+5 2 | 4+5 2 | 3+4 2 | 3+4 2 |
| | 9 2 | 9 2 | 5+6 2 | 5+6 2 |
| | 6+7 2 | 6+7 2 | 3+4 2 | 3+4 2 |
| | 8 2 | 9 2 | 3+4 2 | 5+6 2 |
| | 9 2 | 8 2 | 5+6 2 | 3+4 2 |
| | 8 2 | 8 2 | 3+4 2 | 3+4 2 |
| | 9 2 | 9 2 | 5+6 2 | 5+6 2 |
| | 6+7 2 | 8 2 | 5+6 2 | 3+4 2 |
| | 9 2 | 4+5 2 | 3+4 2 | 5+6 2 |
| | 8 2 | 9 2 | 5+6 2 | 3+4 2 |
| | 4+5 2 | 6+7 2 | 5+6 2 | 5+6 2 |
| | 9 2 | 6+7 2 | 3+4 2 | 5+6 2 |
| | 6+7 2 | 9 2 | 3+4 2 | 3+4 2 |
| | 4+5 2 | 4+5 2 | 5+6 2 | 5+6 2 |
| | 8 2 | 8 2 | 3+4 2 | 3+4 2 |
| | 8 2 | 9 2 | 3+4 2 | 3+4 2 |
| | 8 2 | 8 2 | 5+6 2 | 5+6 2 |
| | 9 2 | 8 2 | 3+4 2 | 5+6 2 |
| | 8 2 | 9 2 | 3+4 2 | 3+4 2 |
| | 4+5 2 | 4+5 2 | 5+6 2 | 5+6 2 |
| | 8 2 | 8 2 | 5+6 2 | 3+4 2 |
| | 9 2 | 6+7 2 | 5+6 2 | 5+6 2 |
| | 6+7 2 | 9 2 | 3+4 2 | 3+4 2 |
| | 9 2 | 8 2 | 5+6 2 | 3+4 2 |
| | 6+7 2 | 9 2 | 3+4 2 | 5+6 2 |
| | 9 2 | 6+7 2 | 3+4 2 | 5+6 2 |
| | 4+5 2 | 9 2 | 5+6 2 | 3+4 2 |

Continued on next column

**TABLE LXXIII– continued**

| TQC | ICSD | NoE | SF | |
|---|---|---|---|---|
| | 8 2 | 4+5 2 | 3+4 2 | 3+4 2 |
| | 9 2 | 8 2 | 5+6 2 | 3+4 2 |
| | 8 2 | 4+5 2 | 3+4 2 | 5+6 2 |
| | 6+7 2 | 8 2 | 5+6 2 | 5+6 2 |
| 166.1.16 | 652055 | 40 | Se1Ti1 | |
| | -GM | -T | -F | -L |
| | 8 2 | 9 2 | 5+6 2 | 3+4 2 |
| | 9 2 | 8 2 | 3+4 2 | 5+6 2 |
| | 8 2 | 8 2 | 5+6 2 | 5+6 2 |
| | 9 2 | 9 2 | 3+4 2 | 3+4 2 |
| | 8 2 | 9 2 | 3+4 2 | 5+6 2 |
| | 8 2 | 8 2 | 3+4 2 | 3+4 2 |
| | 4+5 2 | 9 2 | 5+6 2 | 3+4 2 |
| | 9 2 | 6+7 2 | 5+6 2 | 5+6 2 |
| | 8 2 | 8 2 | 3+4 2 | 3+4 2 |
| | 6+7 2 | 4+5 2 | 5+6 2 | 5+6 2 |
| | 4+5 2 | 8 2 | 3+4 2 | 5+6 2 |
| | 8 2 | 4+5 2 | 3+4 2 | 3+4 2 |
| | 9 2 | 9 2 | 5+6 2 | 3+4 2 |
| | 4+5 2 | 6+7 2 | 5+6 2 | 5+6 2 |
| | 8 2 | 8 2 | 3+4 2 | 5+6 2 |
| | 9 2 | 9 2 | 5+6 2 | 3+4 2 |
| | 6+7 2 | 6+7 2 | 3+4 2 | 5+6 2 |
| | 8 2 | 4+5 2 | 3+4 2 | 3+4 2 |
| | 4+5 2 | 8 2 | 3+4 2 | 3+4 2 |
| | 8 2 | 9 2 | 3+4 2 | 5+6 2 |
| 166.1.17 | 152188 | 58 | Sb8Te3 | |
| | -GM | -T | -F | -L |
| | 8 2 | 8 2 | 5+6 2 | 5+6 2 |
| | 9 2 | 9 2 | 3+4 2 | 3+4 2 |
| | 8 2 | 9 2 | 3+4 2 | 3+4 2 |
| | 9 2 | 8 2 | 3+4 2 | 5+6 2 |
| | 8 2 | 8 2 | 5+6 2 | 3+4 2 |
| | 8 2 | 9 2 | 5+6 2 | 3+4 2 |
| | 9 2 | 9 2 | 5+6 2 | 5+6 2 |
| | 9 2 | 8 2 | 3+4 2 | 3+4 2 |
| | 8 2 | 8 2 | 3+4 2 | 5+6 2 |
| | 8 2 | 9 2 | 5+6 2 | 3+4 2 |
| | 9 2 | 8 2 | 3+4 2 | 5+6 2 |
| | 9 2 | 9 2 | 5+6 2 | 5+6 2 |
| | 8 2 | 8 2 | 3+4 2 | 3+4 2 |
| | 9 2 | 9 2 | 3+4 2 | 5+6 2 |
| | 8 2 | 6+7 2 | 5+6 2 | 5+6 2 |
| | 6+7 2 | 4+5 2 | 5+6 2 | 3+4 2 |
| | 4+5 2 | 9 2 | 3+4 2 | 5+6 2 |
| | 9 2 | 8 2 | 5+6 2 | 3+4 2 |
| | 8 2 | 9 2 | 5+6 2 | 3+4 2 |
| | 9 2 | 8 2 | 3+4 2 | 5+6 2 |
| | 8 2 | 9 2 | 5+6 2 | 3+4 2 |
| | 4+5 2 | 9 2 | 3+4 2 | 5+6 2 |
| | 9 2 | 6+7 2 | 3+4 2 | 3+4 2 |
| | 6+7 2 | 4+5 2 | 5+6 2 | 5+6 2 |
| | 4+5 2 | 6+7 2 | 5+6 2 | 3+4 2 |
| | 8 2 | 6+7 2 | 3+4 2 | 5+6 2 |
| | 6+7 2 | 9 2 | 5+6 2 | 5+6 2 |
| | 9 2 | 8 2 | 5+6 2 | 3+4 2 |
| | 8 2 | 8 2 | 5+6 2 | 5+6 2 |
| 166.1.18 | 152189 | 98 | Sb48Te9 | |
| | -GM | -T | -F | -L |
| | 8 2 | 8 2 | 5+6 2 | 5+6 2 |
| | 9 2 | 9 2 | 3+4 2 | 3+4 2 |

Continued on next column

**TABLE LXXIII– continued**

| TQC | ICSD | NoE | SF | |
|---|---|---|---|---|
| | 8 2 | 9 2 | 3+4 2 | 3+4 2 |
| | 9 2 | 8 2 | 3+4 2 | 5+6 2 |
| | 8 2 | 9 2 | 5+6 2 | 3+4 2 |
| | 9 2 | 8 2 | 5+6 2 | 5+6 2 |
| | 8 2 | 9 2 | 3+4 2 | 3+4 2 |
| | 8 2 | 8 2 | 5+6 2 | 3+4 2 |
| | 9 2 | 8 2 | 3+4 2 | 5+6 2 |
| | 8 2 | 9 2 | 5+6 2 | 3+4 2 |
| | 9 2 | 9 2 | 5+6 2 | 5+6 2 |
| | 9 2 | 8 2 | 3+4 2 | 5+6 2 |
| | 8 2 | 9 2 | 3+4 2 | 3+4 2 |
| | 8 2 | 8 2 | 5+6 2 | 3+4 2 |
| | 9 2 | 8 2 | 3+4 2 | 5+6 2 |
| | 8 2 | 9 2 | 5+6 2 | 3+4 2 |
| | 9 2 | 8 2 | 3+4 2 | 5+6 2 |
| | 8 2 | 9 2 | 5+6 2 | 3+4 2 |
| | 9 2 | 8 2 | 3+4 2 | 5+6 2 |
| | 9 2 | 9 2 | 5+6 2 | 5+6 2 |
| | 8 2 | 8 2 | 3+4 2 | 3+4 2 |
| | 9 2 | 9 2 | 3+4 2 | 5+6 2 |
| | 8 2 | 9 2 | 5+6 2 | 3+4 2 |
| | 6+7 2 | 8 2 | 3+4 2 | 5+6 2 |
| | 9 2 | 6+7 2 | 5+6 2 | 5+6 2 |
| | 4+5 2 | 4+5 2 | 5+6 2 | 3+4 2 |
| | 8 2 | 9 2 | 3+4 2 | 3+4 2 |
| | 9 2 | 8 2 | 5+6 2 | 5+6 2 |
| | 8 2 | 9 2 | 5+6 2 | 3+4 2 |
| | 8 2 | 8 2 | 3+4 2 | 5+6 2 |
| | 9 2 | 9 2 | 3+4 2 | 5+6 2 |
| | 9 2 | 9 2 | 5+6 2 | 3+4 2 |
| | 8 2 | 8 2 | 3+4 2 | 3+4 2 |
| | 8 2 | 9 2 | 5+6 2 | 3+4 2 |
| | 4+5 2 | 6+7 2 | 5+6 2 | 5+6 2 |
| | 9 2 | 4+5 2 | 3+4 2 | 5+6 2 |
| | 6+7 2 | 6+7 2 | 3+4 2 | 3+4 2 |
| | 6+7 2 | 4+5 2 | 5+6 2 | 5+6 2 |
| | 4+5 2 | 9 2 | 5+6 2 | 3+4 2 |
| | 4+5 2 | 6+7 2 | 5+6 2 | 3+4 2 |
| | 6+7 2 | 8 2 | 3+4 2 | 5+6 2 |
| | 9 2 | 6+7 2 | 3+4 2 | 5+6 2 |
| | 4+5 2 | 4+5 2 | 3+4 2 | 3+4 2 |
| | 6+7 2 | 6+7 2 | 5+6 2 | 5+6 2 |
| | 8 2 | 9 2 | 5+6 2 | 5+6 2 |
| | 9 2 | 8 2 | 5+6 2 | 3+4 2 |
| | 8 2 | 8 2 | 5+6 2 | 5+6 2 |
| | 9 2 | 9 2 | 3+4 2 | 3+4 2 |
| | 8 2 | 8 2 | 5+6 2 | 5+6 2 |
| 166.1.19 | 291530 | 28 | Sb3Te4.5 | |
| | -GM | -T | -F | -L |
| | 8 2 | 8 2 | 5+6 2 | 3+4 2 |
| | 9 2 | 9 2 | 3+4 2 | 5+6 2 |
| | 8 2 | 8 2 | 3+4 2 | 3+4 2 |
| | 9 2 | 9 2 | 5+6 2 | 5+6 2 |
| | 8 2 | 8 2 | 3+4 2 | 3+4 2 |
| | 8 2 | 9 2 | 5+6 2 | 5+6 2 |
| | 9 2 | 9 2 | 3+4 2 | 5+6 2 |
| | 8 2 | 8 2 | 5+6 2 | 3+4 2 |
| | 9 2 | 6+7 2 | 5+6 2 | 3+4 2 |
| | 4+5 2 | 9 2 | 3+4 2 | 5+6 2 |
| | 6+7 2 | 4+5 2 | 5+6 2 | 3+4 2 |
| | 9 2 | 9 2 | 3+4 2 | 5+6 2 |
| | 6+7 2 | 6+7 2 | 5+6 2 | 5+6 2 |
| | 8 2 | 8 2 | 5+6 2 | 5+6 2 |
| 166.1.20 | 651728 | 50 | Sb2V3 | |
| | -GM | -T | -F | -L |
| | 8 2 | 8 2 | 5+6 2 | 5+6 2 |
| | 8 2 | 9 2 | 3+4 2 | 3+4 2 |
| | 9 2 | 8 2 | 5+6 2 | 5+6 2 |
| | 9 2 | 9 2 | 3+4 2 | 3+4 2 |
| | 9 2 | 9 2 | 3+4 2 | 3+4 2 |
| | 8 2 | 8 2 | 5+6 2 | 5+6 2 |
| | 4+5 2 | 4+5 2 | 3+4 2 | 5+6 2 |
| | 8 2 | 9 2 | 5+6 2 | 3+4 2 |
| | 8 2 | 6+7 2 | 5+6 2 | 3+4 2 |
| | 9 2 | 8 2 | 3+4 2 | 5+6 2 |
| | 8 2 | 9 2 | 5+6 2 | 3+4 2 |
| | 6+7 2 | 8 2 | 3+4 2 | 5+6 2 |
| | 4+5 2 | 8 2 | 5+6 2 | 5+6 2 |
| | 9 2 | 9 2 | 3+4 2 | 5+6 2 |
| | 6+7 2 | 6+7 2 | 5+6 2 | 3+4 2 |
| | 8 2 | 4+5 2 | 3+4 2 | 5+6 2 |
| | 8 2 | 9 2 | 5+6 2 | 5+6 2 |
| | 4+5 2 | 6+7 2 | 3+4 2 | 3+4 2 |
| | 9 2 | 9 2 | 3+4 2 | 3+4 2 |
| | 8 2 | 8 2 | 5+6 2 | 3+4 2 |
| | 4+5 2 | 8 2 | 3+4 2 | 3+4 2 |
| | 8 2 | 4+5 2 | 5+6 2 | 3+4 2 |
| | 9 2 | 9 2 | 5+6 2 | 5+6 2 |
| | 6+7 2 | 8 2 | 3+4 2 | 5+6 2 |
| | 9 2 | 4+5 2 | 5+6 2 | 3+4 2 |
| 166.1.21 | 41814 | 50 | Sb2V3 | |
| | -GM | -T | -F | -L |
| | 8 2 | 8 2 | 5+6 2 | 5+6 2 |
| | 8 2 | 9 2 | 3+4 2 | 3+4 2 |
| | 9 2 | 8 2 | 5+6 2 | 5+6 2 |
| | 9 2 | 9 2 | 3+4 2 | 3+4 2 |
| | 9 2 | 9 2 | 3+4 2 | 3+4 2 |
| | 8 2 | 8 2 | 5+6 2 | 5+6 2 |
| | 4+5 2 | 4+5 2 | 5+6 2 | 5+6 2 |
| | 8 2 | 9 2 | 3+4 2 | 3+4 2 |
| | 8 2 | 6+7 2 | 5+6 2 | 3+4 2 |
| | 9 2 | 8 2 | 3+4 2 | 5+6 2 |
| | 8 2 | 9 2 | 5+6 2 | 3+4 2 |
| | 6+7 2 | 8 2 | 3+4 2 | 5+6 2 |
| | 4+5 2 | 9 2 | 5+6 2 | 5+6 2 |
| | 9 2 | 8 2 | 3+4 2 | 5+6 2 |
| | 6+7 2 | 6+7 2 | 5+6 2 | 5+6 2 |
| | 8 2 | 4+5 2 | 3+4 2 | 3+4 2 |
| | 4+5 2 | 9 2 | 3+4 2 | 5+6 2 |
| | 8 2 | 9 2 | 5+6 2 | 3+4 2 |
| | 9 2 | 6+7 2 | 3+4 2 | 3+4 2 |
| | 8 2 | 8 2 | 5+6 2 | 3+4 2 |
| | 8 2 | 4+5 2 | 3+4 2 | 3+4 2 |
| | 4+5 2 | 8 2 | 5+6 2 | 3+4 2 |
| | 9 2 | 9 2 | 5+6 2 | 5+6 2 |
| | 6+7 2 | 4+5 2 | 3+4 2 | 3+4 2 |
| | 9 2 | 8 2 | 5+6 2 | 5+6 2 |
| 166.1.22 | 651638 | 28 | Sb2Te3 | |
| | -GM | -T | -F | -L |
| | 8 2 | 9 2 | 5+6 2 | 3+4 2 |
| | 9 2 | 8 2 | 3+4 2 | 5+6 2 |
| | 8 2 | 8 2 | 3+4 2 | 3+4 2 |

Continued on next column

**TABLE LXXIII– continued**

| TQC | ICSD | NoE | SF | |
|---|---|---|---|---|
| | 9 2 | 9 2 | 5+6 2 | 5+6 2 |
| | 8 2 | 8 2 | 3+4 2 | 3+4 2 |
| | 8 2 | 9 2 | 5+6 2 | 5+6 2 |
| | 9 2 | 8 2 | 3+4 2 | 3+4 2 |
| | 8 2 | 9 2 | 5+6 2 | 5+6 2 |
| | 9 2 | 6+7 2 | 5+6 2 | 3+4 2 |
| | 4+5 2 | 4+5 2 | 3+4 2 | 5+6 2 |
| | 6+7 2 | 9 2 | 5+6 2 | 3+4 2 |
| | 9 2 | 9 2 | 3+4 2 | 5+6 2 |
| | 6+7 2 | 6+7 2 | 5+6 2 | 5+6 2 |
| | 8 2 | 8 2 | 5+6 2 | 5+6 2 |
| 166.1.23 | 651635 | 28 | Sb2Te3 | |
| | -GM | -T | -F | -L |
| | 8 2 | 9 2 | 5+6 2 | 3+4 2 |
| | 9 2 | 8 2 | 3+4 2 | 5+6 2 |
| | 8 2 | 8 2 | 3+4 2 | 3+4 2 |
| | 9 2 | 9 2 | 5+6 2 | 5+6 2 |
| | 8 2 | 8 2 | 3+4 2 | 3+4 2 |
| | 8 2 | 9 2 | 5+6 2 | 5+6 2 |
| | 9 2 | 8 2 | 3+4 2 | 3+4 2 |
| | 8 2 | 9 2 | 5+6 2 | 5+6 2 |
| | 9 2 | 6+7 2 | 5+6 2 | 3+4 2 |
| | 4+5 2 | 4+5 2 | 3+4 2 | 5+6 2 |
| | 6+7 2 | 9 2 | 5+6 2 | 3+4 2 |
| | 9 2 | 9 2 | 3+4 2 | 5+6 2 |
| | 6+7 2 | 6+7 2 | 5+6 2 | 5+6 2 |
| | 8 2 | 8 2 | 5+6 2 | 5+6 2 |
| 166.1.24 | 651629 | 28 | Sb2Te3 | |
| | -GM | -T | -F | -L |
| | 8 2 | 8 2 | 5+6 2 | 3+4 2 |
| | 9 2 | 9 2 | 3+4 2 | 5+6 2 |
| | 8 2 | 8 2 | 3+4 2 | 3+4 2 |
| | 9 2 | 9 2 | 5+6 2 | 5+6 2 |
| | 8 2 | 8 2 | 3+4 2 | 3+4 2 |
| | 8 2 | 9 2 | 5+6 2 | 5+6 2 |
| | 9 2 | 9 2 | 3+4 2 | 5+6 2 |
| | 8 2 | 8 2 | 5+6 2 | 3+4 2 |
| | 9 2 | 6+7 2 | 5+6 2 | 3+4 2 |
| | 4+5 2 | 9 2 | 3+4 2 | 5+6 2 |
| | 6+7 2 | 4+5 2 | 5+6 2 | 3+4 2 |
| | 9 2 | 6+7 2 | 3+4 2 | 5+6 2 |
| | 6+7 2 | 9 2 | 5+6 2 | 5+6 2 |
| | 8 2 | 8 2 | 5+6 2 | 5+6 2 |
| 166.1.25 | 5420 | 28 | Sb2Te3 | |
| | -GM | -T | -F | -L |
| | 8 2 | 8 2 | 5+6 2 | 3+4 2 |
| | 9 2 | 9 2 | 3+4 2 | 5+6 2 |
| | 8 2 | 8 2 | 3+4 2 | 3+4 2 |
| | 9 2 | 9 2 | 5+6 2 | 5+6 2 |
| | 8 2 | 8 2 | 3+4 2 | 3+4 2 |
| | 9 2 | 9 2 | 5+6 2 | 5+6 2 |
| | 8 2 | 9 2 | 3+4 2 | 5+6 2 |
| | 8 2 | 8 2 | 5+6 2 | 3+4 2 |
| | 9 2 | 6+7 2 | 5+6 2 | 3+4 2 |
| | 6+7 2 | 9 2 | 3+4 2 | 5+6 2 |
| | 4+5 2 | 4+5 2 | 5+6 2 | 3+4 2 |
| | 9 2 | 6+7 2 | 3+4 2 | 5+6 2 |
| | 6+7 2 | 9 2 | 5+6 2 | 5+6 2 |
| | 8 2 | 8 2 | 5+6 2 | 5+6 2 |
| 166.1.26 | 262171 | 28 | Sb2Te3 | |
| | -GM | -T | -F | -L |
| | 8 2 | 9 2 | 5+6 2 | 3+4 2 |
| | 9 2 | 8 2 | 3+4 2 | 5+6 2 |
| | 8 2 | 8 2 | 3+4 2 | 3+4 2 |
| | 9 2 | 9 2 | 5+6 2 | 5+6 2 |
| | 8 2 | 8 2 | 3+4 2 | 3+4 2 |
| | 8 2 | 9 2 | 5+6 2 | 5+6 2 |
| | 9 2 | 9 2 | 3+4 2 | 3+4 2 |
| | 4+5 2 | 8 2 | 5+6 2 | 5+6 2 |
| | 8 2 | 6+7 2 | 5+6 2 | 3+4 2 |
| | 9 2 | 4+5 2 | 3+4 2 | 5+6 2 |
| | 6+7 2 | 9 2 | 5+6 2 | 3+4 2 |
| | 9 2 | 9 2 | 3+4 2 | 5+6 2 |
| | 6+7 2 | 6+7 2 | 5+6 2 | 5+6 2 |
| | 8 2 | 8 2 | 5+6 2 | 5+6 2 |
| 166.1.27 | 2084 | 28 | Sb2Te3 | |
| | -GM | -T | -F | -L |
| | 8 2 | 8 2 | 5+6 2 | 3+4 2 |
| | 9 2 | 9 2 | 3+4 2 | 5+6 2 |
| | 8 2 | 8 2 | 3+4 2 | 3+4 2 |
| | 9 2 | 9 2 | 5+6 2 | 5+6 2 |
| | 8 2 | 8 2 | 3+4 2 | 3+4 2 |
| | 9 2 | 9 2 | 5+6 2 | 5+6 2 |
| | 8 2 | 9 2 | 3+4 2 | 5+6 2 |
| | 8 2 | 8 2 | 5+6 2 | 3+4 2 |
| | 9 2 | 6+7 2 | 5+6 2 | 3+4 2 |
| | 6+7 2 | 9 2 | 3+4 2 | 5+6 2 |
| | 4+5 2 | 4+5 2 | 5+6 2 | 3+4 2 |
| | 9 2 | 6+7 2 | 3+4 2 | 5+6 2 |
| | 6+7 2 | 9 2 | 5+6 2 | 5+6 2 |
| | 8 2 | 8 2 | 5+6 2 | 5+6 2 |
| 166.1.28 | 20236 | 28 | Sb2Te3 | |
| | -GM | -T | -F | -L |
| | 8 2 | 8 2 | 5+6 2 | 3+4 2 |
| | 9 2 | 9 2 | 3+4 2 | 5+6 2 |
| | 8 2 | 8 2 | 3+4 2 | 3+4 2 |
| | 9 2 | 9 2 | 5+6 2 | 5+6 2 |
| | 8 2 | 8 2 | 3+4 2 | 3+4 2 |
| | 8 2 | 9 2 | 5+6 2 | 5+6 2 |
| | 9 2 | 9 2 | 3+4 2 | 5+6 2 |
| | 8 2 | 8 2 | 5+6 2 | 3+4 2 |
| | 9 2 | 6+7 2 | 5+6 2 | 3+4 2 |
| | 4+5 2 | 9 2 | 3+4 2 | 5+6 2 |
| | 6+7 2 | 4+5 2 | 5+6 2 | 3+4 2 |
| | 9 2 | 6+7 2 | 3+4 2 | 5+6 2 |
| | 6+7 2 | 9 2 | 5+6 2 | 5+6 2 |
| | 8 2 | 8 2 | 5+6 2 | 5+6 2 |
| 166.1.29 | 193347 | 28 | Sb2Te3 | |
| | -GM | -T | -F | -L |
| | 8 2 | 8 2 | 5+6 2 | 3+4 2 |
| | 9 2 | 9 2 | 3+4 2 | 5+6 2 |
| | 8 2 | 8 2 | 3+4 2 | 3+4 2 |
| | 9 2 | 9 2 | 5+6 2 | 5+6 2 |
| | 8 2 | 8 2 | 3+4 2 | 3+4 2 |
| | 9 2 | 9 2 | 5+6 2 | 5+6 2 |
| | 8 2 | 9 2 | 3+4 2 | 5+6 2 |
| | 8 2 | 8 2 | 5+6 2 | 3+4 2 |
| | 9 2 | 6+7 2 | 5+6 2 | 3+4 2 |
| | 6+7 2 | 9 2 | 3+4 2 | 5+6 2 |
| | 4+5 2 | 4+5 2 | 5+6 2 | 3+4 2 |
| | 9 2 | 6+7 2 | 3+4 2 | 5+6 2 |
| | 6+7 2 | 9 2 | 5+6 2 | 5+6 2 |

Continued on next column

**TABLE LXXIII– continued**

| TQC | ICSD | NoE | SF | |
|---|---|---|---|---|
| | 8 2 | 8 2 | 5+6 2 | 5+6 2 |
| 166.1.30 | 193346 | 28 | Sb2Te3 | |
| | -GM | -T | -F | -L |
| | 8 2 | 8 2 | 5+6 2 | 3+4 2 |
| | 9 2 | 9 2 | 3+4 2 | 5+6 2 |
| | 8 2 | 8 2 | 3+4 2 | 3+4 2 |
| | 9 2 | 9 2 | 5+6 2 | 5+6 2 |
| | 8 2 | 8 2 | 3+4 2 | 3+4 2 |
| | 8 2 | 9 2 | 5+6 2 | 5+6 2 |
| | 9 2 | 9 2 | 3+4 2 | 5+6 2 |
| | 8 2 | 8 2 | 5+6 2 | 3+4 2 |
| | 9 2 | 6+7 2 | 5+6 2 | 3+4 2 |
| | 4+5 2 | 9 2 | 3+4 2 | 5+6 2 |
| | 6+7 2 | 4+5 2 | 5+6 2 | 3+4 2 |
| | 9 2 | 9 2 | 3+4 2 | 5+6 2 |
| | 6+7 2 | 6+7 2 | 5+6 2 | 5+6 2 |
| | 8 2 | 8 2 | 5+6 2 | 5+6 2 |
| 166.1.31 | 193345 | 28 | Sb2Te3 | |
| | -GM | -T | -F | -L |
| | 8 2 | 8 2 | 5+6 2 | 3+4 2 |
| | 9 2 | 9 2 | 3+4 2 | 5+6 2 |
| | 8 2 | 8 2 | 3+4 2 | 3+4 2 |
| | 9 2 | 9 2 | 5+6 2 | 5+6 2 |
| | 8 2 | 8 2 | 3+4 2 | 3+4 2 |
| | 9 2 | 9 2 | 5+6 2 | 5+6 2 |
| | 8 2 | 9 2 | 3+4 2 | 5+6 2 |
| | 8 2 | 8 2 | 5+6 2 | 3+4 2 |
| | 9 2 | 6+7 2 | 5+6 2 | 3+4 2 |
| | 6+7 2 | 9 2 | 3+4 2 | 5+6 2 |
| | 4+5 2 | 4+5 2 | 5+6 2 | 3+4 2 |
| | 9 2 | 6+7 2 | 3+4 2 | 5+6 2 |
| | 6+7 2 | 9 2 | 5+6 2 | 5+6 2 |
| | 8 2 | 8 2 | 5+6 2 | 5+6 2 |
| 166.1.32 | 193344 | 28 | Sb2Te3 | |
| | -GM | -T | -F | -L |
| | 8 2 | 8 2 | 5+6 2 | 3+4 2 |
| | 9 2 | 9 2 | 3+4 2 | 5+6 2 |
| | 8 2 | 8 2 | 3+4 2 | 3+4 2 |
| | 9 2 | 9 2 | 5+6 2 | 5+6 2 |
| | 8 2 | 8 2 | 3+4 2 | 3+4 2 |
| | 9 2 | 9 2 | 5+6 2 | 5+6 2 |
| | 8 2 | 9 2 | 3+4 2 | 5+6 2 |
| | 8 2 | 8 2 | 5+6 2 | 3+4 2 |
| | 9 2 | 6+7 2 | 5+6 2 | 3+4 2 |
| | 6+7 2 | 9 2 | 3+4 2 | 5+6 2 |
| | 4+5 2 | 4+5 2 | 5+6 2 | 3+4 2 |
| | 9 2 | 6+7 2 | 3+4 2 | 5+6 2 |
| | 6+7 2 | 9 2 | 5+6 2 | 5+6 2 |
| | 8 2 | 8 2 | 5+6 2 | 5+6 2 |
| 166.1.33 | 193343 | 28 | Sb2Te3 | |
| | -GM | -T | -F | -L |
| | 8 2 | 8 2 | 5+6 2 | 3+4 2 |
| | 9 2 | 9 2 | 3+4 2 | 5+6 2 |
| | 8 2 | 8 2 | 3+4 2 | 3+4 2 |
| | 9 2 | 9 2 | 5+6 2 | 5+6 2 |
| | 8 2 | 8 2 | 3+4 2 | 3+4 2 |
| | 8 2 | 9 2 | 5+6 2 | 5+6 2 |
| | 9 2 | 9 2 | 3+4 2 | 5+6 2 |
| | 8 2 | 8 2 | 5+6 2 | 3+4 2 |
| | 9 2 | 6+7 2 | 5+6 2 | 3+4 2 |
| | 4+5 2 | 9 2 | 3+4 2 | 5+6 2 |

Continued on next column

**TABLE LXXIII– continued**

| TQC | ICSD | NoE | SF | |
|---|---|---|---|---|
| | 6+7 2 | 4+5 2 | 5+6 2 | 3+4 2 |
| | 9 2 | 6+7 2 | 3+4 2 | 5+6 2 |
| | 6+7 2 | 9 2 | 5+6 2 | 5+6 2 |
| | 8 2 | 8 2 | 5+6 2 | 5+6 2 |
| 166.1.34 | 193342 | 28 | Sb2Te3 | |
| | -GM | -T | -F | -L |
| | 8 2 | 8 2 | 5+6 2 | 3+4 2 |
| | 9 2 | 9 2 | 3+4 2 | 5+6 2 |
| | 8 2 | 8 2 | 3+4 2 | 3+4 2 |
| | 9 2 | 9 2 | 5+6 2 | 5+6 2 |
| | 8 2 | 8 2 | 3+4 2 | 3+4 2 |
| | 9 2 | 9 2 | 5+6 2 | 5+6 2 |
| | 8 2 | 9 2 | 3+4 2 | 5+6 2 |
| | 8 2 | 8 2 | 5+6 2 | 3+4 2 |
| | 9 2 | 6+7 2 | 5+6 2 | 3+4 2 |
| | 6+7 2 | 9 2 | 3+4 2 | 5+6 2 |
| | 4+5 2 | 4+5 2 | 5+6 2 | 3+4 2 |
| | 9 2 | 6+7 2 | 3+4 2 | 5+6 2 |
| | 6+7 2 | 9 2 | 5+6 2 | 5+6 2 |
| | 8 2 | 8 2 | 5+6 2 | 5+6 2 |
| 166.1.35 | 193341 | 28 | Sb2Te3 | |
| | -GM | -T | -F | -L |
| | 8 2 | 8 2 | 5+6 2 | 3+4 2 |
| | 9 2 | 9 2 | 3+4 2 | 5+6 2 |
| | 8 2 | 8 2 | 3+4 2 | 3+4 2 |
| | 9 2 | 9 2 | 5+6 2 | 5+6 2 |
| | 8 2 | 8 2 | 3+4 2 | 3+4 2 |
| | 8 2 | 9 2 | 5+6 2 | 5+6 2 |
| | 9 2 | 9 2 | 3+4 2 | 5+6 2 |
| | 8 2 | 8 2 | 5+6 2 | 3+4 2 |
| | 9 2 | 6+7 2 | 5+6 2 | 3+4 2 |
| | 4+5 2 | 9 2 | 3+4 2 | 5+6 2 |
| | 6+7 2 | 4+5 2 | 5+6 2 | 3+4 2 |
| | 9 2 | 9 2 | 3+4 2 | 5+6 2 |
| | 6+7 2 | 6+7 2 | 5+6 2 | 5+6 2 |
| | 8 2 | 8 2 | 5+6 2 | 5+6 2 |
| 166.1.36 | 192780 | 28 | Sb2Te3 | |
| | -GM | -T | -F | -L |
| | 8 2 | 8 2 | 5+6 2 | 3+4 2 |
| | 9 2 | 9 2 | 3+4 2 | 5+6 2 |
| | 8 2 | 8 2 | 3+4 2 | 3+4 2 |
| | 9 2 | 9 2 | 5+6 2 | 5+6 2 |
| | 8 2 | 8 2 | 3+4 2 | 3+4 2 |
| | 9 2 | 9 2 | 5+6 2 | 5+6 2 |
| | 8 2 | 9 2 | 3+4 2 | 5+6 2 |
| | 8 2 | 8 2 | 5+6 2 | 3+4 2 |
| | 9 2 | 6+7 2 | 5+6 2 | 3+4 2 |
| | 6+7 2 | 9 2 | 3+4 2 | 5+6 2 |
| | 4+5 2 | 4+5 2 | 5+6 2 | 3+4 2 |
| | 9 2 | 6+7 2 | 3+4 2 | 5+6 2 |
| | 6+7 2 | 9 2 | 5+6 2 | 5+6 2 |
| | 8 2 | 8 2 | 5+6 2 | 5+6 2 |
| 166.1.37 | 187495 | 28 | Sb2Te3 | |
| | -GM | -T | -F | -L |
| | 8 2 | 8 2 | 3+4 2 | 3+4 2 |
| | 8 2 | 9 2 | 5+6 2 | 3+4 2 |
| | 9 2 | 9 2 | 5+6 2 | 5+6 2 |
| | 9 2 | 8 2 | 3+4 2 | 5+6 2 |
| | 8 2 | 8 2 | 3+4 2 | 3+4 2 |
| | 9 2 | 9 2 | 5+6 2 | 5+6 2 |
| | 6+7 2 | 6+7 2 | 5+6 2 | 5+6 2 |

Continued on next column

**TABLE LXXIII– continued**

| TQC | ICSD | NoE | SF | |
|---|---|---|---|---|
| | 8 2 | 9 2 | 3+4 2 | 5+6 2 |
| | 8 2 | 9 2 | 5+6 2 | 3+4 2 |
| | 4+5 2 | 6+7 2 | 5+6 2 | 3+4 2 |
| | 9 2 | 8 2 | 3+4 2 | 5+6 2 |
| | 8 2 | 8 2 | 5+6 2 | 3+4 2 |
| | 4+5 2 | 4+5 2 | 3+4 2 | 3+4 2 |
| | 9 2 | 9 2 | 3+4 2 | 5+6 2 |
| 166.1.38 | 185952 | 28 | Sb2Te3 | |
| | -GM | -T | -F | -L |
| | 8 2 | 8 2 | 3+4 2 | 3+4 2 |
| | 9 2 | 9 2 | 5+6 2 | 3+4 2 |
| | 8 2 | 8 2 | 3+4 2 | 5+6 2 |
| | 9 2 | 9 2 | 5+6 2 | 5+6 2 |
| | 8 2 | 8 2 | 3+4 2 | 3+4 2 |
| | 9 2 | 9 2 | 5+6 2 | 5+6 2 |
| | 8 2 | 9 2 | 3+4 2 | 5+6 2 |
| | 6+7 2 | 8 2 | 5+6 2 | 3+4 2 |
| | 8 2 | 6+7 2 | 5+6 2 | 5+6 2 |
| | 9 2 | 9 2 | 3+4 2 | 3+4 2 |
| | 4+5 2 | 4+5 2 | 5+6 2 | 3+4 2 |
| | 9 2 | 6+7 2 | 3+4 2 | 5+6 2 |
| | 6+7 2 | 9 2 | 5+6 2 | 5+6 2 |
| | 8 2 | 8 2 | 5+6 2 | 5+6 2 |
| 166.1.39 | 185951 | 28 | Sb2Te3 | |
| | -GM | -T | -F | -L |
| | 8 2 | 9 2 | 5+6 2 | 3+4 2 |
| | 9 2 | 8 2 | 3+4 2 | 5+6 2 |
| | 8 2 | 8 2 | 3+4 2 | 3+4 2 |
| | 9 2 | 9 2 | 5+6 2 | 5+6 2 |
| | 8 2 | 8 2 | 3+4 2 | 3+4 2 |
| | 8 2 | 9 2 | 5+6 2 | 5+6 2 |
| | 9 2 | 8 2 | 3+4 2 | 3+4 2 |
| | 4+5 2 | 9 2 | 5+6 2 | 5+6 2 |
| | 6+7 2 | 6+7 2 | 5+6 2 | 3+4 2 |
| | 9 2 | 4+5 2 | 3+4 2 | 5+6 2 |
| | 8 2 | 9 2 | 5+6 2 | 3+4 2 |
| | 9 2 | 9 2 | 3+4 2 | 5+6 2 |
| | 6+7 2 | 6+7 2 | 5+6 2 | 5+6 2 |
| | 8 2 | 8 2 | 5+6 2 | 5+6 2 |
| 166.1.40 | 185950 | 28 | Sb2Te3 | |
| | -GM | -T | -F | -L |
| | 8 2 | 9 2 | 5+6 2 | 3+4 2 |
| | 9 2 | 8 2 | 3+4 2 | 5+6 2 |
| | 8 2 | 8 2 | 3+4 2 | 3+4 2 |
| | 9 2 | 9 2 | 5+6 2 | 5+6 2 |
| | 8 2 | 8 2 | 3+4 2 | 3+4 2 |
| | 8 2 | 9 2 | 5+6 2 | 5+6 2 |
| | 9 2 | 8 2 | 3+4 2 | 3+4 2 |
| | 4+5 2 | 9 2 | 5+6 2 | 5+6 2 |
| | 6+7 2 | 6+7 2 | 5+6 2 | 3+4 2 |
| | 9 2 | 4+5 2 | 3+4 2 | 5+6 2 |
| | 8 2 | 9 2 | 5+6 2 | 3+4 2 |
| | 9 2 | 9 2 | 3+4 2 | 5+6 2 |
| | 6+7 2 | 6+7 2 | 5+6 2 | 5+6 2 |
| | 8 2 | 8 2 | 5+6 2 | 5+6 2 |
| 166.1.41 | 185949 | 28 | Sb2Te3 | |
| | -GM | -T | -F | -L |
| | 8 2 | 8 2 | 5+6 2 | 3+4 2 |
| | 9 2 | 9 2 | 3+4 2 | 5+6 2 |
| | 8 2 | 8 2 | 3+4 2 | 3+4 2 |
| | 9 2 | 9 2 | 5+6 2 | 5+6 2 |

Continued on next column

**TABLE LXXIII– continued**

| TQC | ICSD | NoE | SF | |
|---|---|---|---|---|
| | 8 2 | 8 2 | 3+4 2 | 3+4 2 |
| | 9 2 | 9 2 | 5+6 2 | 5+6 2 |
| | 8 2 | 8 2 | 3+4 2 | 5+6 2 |
| | 4+5 2 | 9 2 | 5+6 2 | 3+4 2 |
| | 6+7 2 | 6+7 2 | 5+6 2 | 3+4 2 |
| | 8 2 | 4+5 2 | 3+4 2 | 5+6 2 |
| | 9 2 | 9 2 | 5+6 2 | 3+4 2 |
| | 9 2 | 6+7 2 | 3+4 2 | 5+6 2 |
| | 6+7 2 | 9 2 | 5+6 2 | 5+6 2 |
| | 8 2 | 8 2 | 5+6 2 | 5+6 2 |
| 166.1.42 | 185948 | 28 | Sb2Te3 | |
| | -GM | -T | -F | -L |
| | 8 2 | 8 2 | 5+6 2 | 3+4 2 |
| | 9 2 | 9 2 | 3+4 2 | 5+6 2 |
| | 8 2 | 8 2 | 3+4 2 | 3+4 2 |
| | 9 2 | 9 2 | 5+6 2 | 5+6 2 |
| | 8 2 | 8 2 | 3+4 2 | 3+4 2 |
| | 8 2 | 9 2 | 5+6 2 | 5+6 2 |
| | 9 2 | 8 2 | 3+4 2 | 5+6 2 |
| | 4+5 2 | 9 2 | 5+6 2 | 3+4 2 |
| | 6+7 2 | 6+7 2 | 5+6 2 | 3+4 2 |
| | 9 2 | 4+5 2 | 3+4 2 | 5+6 2 |
| | 8 2 | 9 2 | 5+6 2 | 3+4 2 |
| | 9 2 | 9 2 | 3+4 2 | 5+6 2 |
| | 6+7 2 | 6+7 2 | 5+6 2 | 5+6 2 |
| | 8 2 | 8 2 | 5+6 2 | 5+6 2 |
| 166.1.43 | 185947 | 28 | Sb2Te3 | |
| | -GM | -T | -F | -L |
| | 8 2 | 8 2 | 5+6 2 | 3+4 2 |
| | 9 2 | 9 2 | 3+4 2 | 5+6 2 |
| | 8 2 | 8 2 | 3+4 2 | 3+4 2 |
| | 9 2 | 9 2 | 5+6 2 | 5+6 2 |
| | 8 2 | 8 2 | 3+4 2 | 3+4 2 |
| | 8 2 | 9 2 | 5+6 2 | 5+6 2 |
| | 9 2 | 8 2 | 3+4 2 | 3+4 2 |
| | 4+5 2 | 9 2 | 5+6 2 | 5+6 2 |
| | 8 2 | 6+7 2 | 5+6 2 | 3+4 2 |
| | 9 2 | 4+5 2 | 3+4 2 | 5+6 2 |
| | 6+7 2 | 9 2 | 5+6 2 | 3+4 2 |
| | 9 2 | 9 2 | 3+4 2 | 5+6 2 |
| | 6+7 2 | 6+7 2 | 5+6 2 | 5+6 2 |
| | 8 2 | 8 2 | 5+6 2 | 5+6 2 |
| 166.1.44 | 185946 | 28 | Sb2Te3 | |
| | -GM | -T | -F | -L |
| | 8 2 | 8 2 | 5+6 2 | 3+4 2 |
| | 9 2 | 9 2 | 3+4 2 | 5+6 2 |
| | 8 2 | 8 2 | 3+4 2 | 3+4 2 |
| | 9 2 | 9 2 | 5+6 2 | 5+6 2 |
| | 8 2 | 8 2 | 3+4 2 | 3+4 2 |
| | 9 2 | 9 2 | 5+6 2 | 5+6 2 |
| | 8 2 | 9 2 | 3+4 2 | 5+6 2 |
| | 8 2 | 8 2 | 5+6 2 | 3+4 2 |
| | 9 2 | 6+7 2 | 5+6 2 | 3+4 2 |
| | 6+7 2 | 9 2 | 3+4 2 | 5+6 2 |
| | 4+5 2 | 4+5 2 | 5+6 2 | 3+4 2 |
| | 9 2 | 6+7 2 | 3+4 2 | 5+6 2 |
| | 6+7 2 | 9 2 | 5+6 2 | 5+6 2 |
| | 8 2 | 8 2 | 5+6 2 | 5+6 2 |
| 166.1.45 | 185945 | 28 | Sb2Te3 | |
| | -GM | -T | -F | -L |
| | 8 2 | 8 2 | 5+6 2 | 3+4 2 |

Continued on next column

**TABLE LXXIII– continued**

| TQC | ICSD | NoE | SF | |
|---|---|---|---|---|
| | 9 2 | 9 2 | 3+4 2 | 5+6 2 |
| | 8 2 | 8 2 | 3+4 2 | 3+4 2 |
| | 9 2 | 9 2 | 5+6 2 | 5+6 2 |
| | 8 2 | 8 2 | 3+4 2 | 3+4 2 |
| | 9 2 | 9 2 | 5+6 2 | 5+6 2 |
| | 8 2 | 9 2 | 3+4 2 | 5+6 2 |
| | 8 2 | 8 2 | 5+6 2 | 3+4 2 |
| | 9 2 | 6+7 2 | 5+6 2 | 3+4 2 |
| | 6+7 2 | 9 2 | 3+4 2 | 5+6 2 |
| | 4+5 2 | 4+5 2 | 5+6 2 | 3+4 2 |
| | 9 2 | 9 2 | 3+4 2 | 5+6 2 |
| | 6+7 2 | 6+7 2 | 5+6 2 | 5+6 2 |
| | 8 2 | 8 2 | 5+6 2 | 5+6 2 |
| 166.1.46 | 657316 | 38 | Sb2Sn1Te4 | |
| | -GM | -T | -F | -L |
| | 8 2 | 9 2 | 5+6 2 | 3+4 2 |
| | 9 2 | 8 2 | 5+6 2 | 5+6 2 |
| | 8 2 | 9 2 | 3+4 2 | 3+4 2 |
| | 9 2 | 8 2 | 3+4 2 | 5+6 2 |
| | 8 2 | 9 2 | 5+6 2 | 3+4 2 |
| | 9 2 | 8 2 | 3+4 2 | 5+6 2 |
| | 8 2 | 8 2 | 3+4 2 | 3+4 2 |
| | 8 2 | 9 2 | 3+4 2 | 5+6 2 |
| | 9 2 | 9 2 | 5+6 2 | 3+4 2 |
| | 8 2 | 8 2 | 5+6 2 | 5+6 2 |
| | 9 2 | 9 2 | 5+6 2 | 3+4 2 |
| | 8 2 | 6+7 2 | 3+4 2 | 5+6 2 |
| | 4+5 2 | 4+5 2 | 5+6 2 | 3+4 2 |
| | 6+7 2 | 8 2 | 3+4 2 | 5+6 2 |
| | 9 2 | 9 2 | 5+6 2 | 3+4 2 |
| | 4+5 2 | 6+7 2 | 3+4 2 | 5+6 2 |
| | 6+7 2 | 8 2 | 5+6 2 | 3+4 2 |
| | 8 2 | 4+5 2 | 3+4 2 | 5+6 2 |
| | 9 2 | 9 2 | 3+4 2 | 3+4 2 |
| 166.1.47 | 30392 | 38 | Sb2Sn1Te4 | |
| | -GM | -T | -F | -L |
| | 8 2 | 9 2 | 5+6 2 | 3+4 2 |
| | 8 2 | 8 2 | 3+4 2 | 5+6 2 |
| | 9 2 | 9 2 | 5+6 2 | 5+6 2 |
| | 9 2 | 8 2 | 3+4 2 | 3+4 2 |
| | 8 2 | 9 2 | 5+6 2 | 3+4 2 |
| | 9 2 | 8 2 | 3+4 2 | 5+6 2 |
| | 8 2 | 8 2 | 3+4 2 | 3+4 2 |
| | 8 2 | 9 2 | 3+4 2 | 5+6 2 |
| | 9 2 | 9 2 | 5+6 2 | 3+4 2 |
| | 4+5 2 | 6+7 2 | 5+6 2 | 5+6 2 |
| | 8 2 | 8 2 | 5+6 2 | 3+4 2 |
| | 9 2 | 9 2 | 3+4 2 | 3+4 2 |
| | 8 2 | 8 2 | 5+6 2 | 5+6 2 |
| | 9 2 | 4+5 2 | 3+4 2 | 5+6 2 |
| | 6+7 2 | 9 2 | 5+6 2 | 3+4 2 |
| | 4+5 2 | 8 2 | 3+4 2 | 3+4 2 |
| | 6+7 2 | 6+7 2 | 5+6 2 | 5+6 2 |
| | 9 2 | 4+5 2 | 3+4 2 | 5+6 2 |
| | 8 2 | 9 2 | 3+4 2 | 3+4 2 |
| 166.1.48 | 651529 | 28 | Sb2Se2Te1 | |
| | -GM | -T | -F | -L |
| | 8 2 | 8 2 | 5+6 2 | 3+4 2 |
| | 9 2 | 9 2 | 3+4 2 | 5+6 2 |
| | 8 2 | 8 2 | 3+4 2 | 3+4 2 |
| | 9 2 | 9 2 | 5+6 2 | 5+6 2 |
| | 8 2 | 8 2 | 3+4 2 | 3+4 2 |
| | 9 2 | 9 2 | 5+6 2 | 5+6 2 |
| | 8 2 | 9 2 | 5+6 2 | 5+6 2 |
| | 6+7 2 | 6+7 2 | 3+4 2 | 5+6 2 |
| | 9 2 | 8 2 | 5+6 2 | 3+4 2 |
| | 8 2 | 9 2 | 3+4 2 | 3+4 2 |
| | 4+5 2 | 4+5 2 | 3+4 2 | 3+4 2 |
| | 9 2 | 6+7 2 | 5+6 2 | 5+6 2 |
| | 6+7 2 | 9 2 | 5+6 2 | 5+6 2 |
| | 8 2 | 8 2 | 5+6 2 | 5+6 2 |
| 166.1.49 | 52295 | 28 | Sb2Se2Te1 | |
| | -GM | -T | -F | -L |
| | 8 2 | 8 2 | 5+6 2 | 3+4 2 |
| | 9 2 | 9 2 | 3+4 2 | 5+6 2 |
| | 8 2 | 8 2 | 3+4 2 | 3+4 2 |
| | 9 2 | 9 2 | 5+6 2 | 5+6 2 |
| | 8 2 | 8 2 | 3+4 2 | 3+4 2 |
| | 9 2 | 9 2 | 5+6 2 | 5+6 2 |
| | 8 2 | 9 2 | 5+6 2 | 5+6 2 |
| | 9 2 | 6+7 2 | 3+4 2 | 5+6 2 |
| | 6+7 2 | 8 2 | 5+6 2 | 3+4 2 |
| | 8 2 | 9 2 | 3+4 2 | 3+4 2 |
| | 4+5 2 | 4+5 2 | 3+4 2 | 3+4 2 |
| | 9 2 | 6+7 2 | 5+6 2 | 5+6 2 |
| | 6+7 2 | 9 2 | 5+6 2 | 5+6 2 |
| | 8 2 | 8 2 | 5+6 2 | 5+6 2 |
| 166.1.50 | 87416 | 28 | Sb2Se1Te2 | |
| | -GM | -T | -F | -L |
| | 8 2 | 8 2 | 3+4 2 | 5+6 2 |
| | 8 2 | 9 2 | 5+6 2 | 5+6 2 |
| | 9 2 | 9 2 | 3+4 2 | 3+4 2 |
| | 8 2 | 8 2 | 5+6 2 | 3+4 2 |
| | 9 2 | 9 2 | 3+4 2 | 5+6 2 |
| | 8 2 | 8 2 | 3+4 2 | 3+4 2 |
| | 9 2 | 8 2 | 5+6 2 | 5+6 2 |
| | 8 2 | 8 2 | 5+6 2 | 3+4 2 |
| | 9 2 | 4+5 2 | 5+6 2 | 5+6 2 |
| | 4+5 2 | 4+5 2 | 5+6 2 | 3+4 2 |
| | 6+7 2 | 9 2 | 3+4 2 | 5+6 2 |
| | 9 2 | 8 2 | 5+6 2 | 3+4 2 |
| | 6+7 2 | 9 2 | 3+4 2 | 3+4 2 |
| | 8 2 | 6+7 2 | 5+6 2 | 3+4 2 |
| 166.1.51 | 2085 | 28 | Sb2Se1Te2 | |
| | -GM | -T | -F | -L |
| | 8 2 | 8 2 | 3+4 2 | 3+4 2 |
| | 9 2 | 9 2 | 5+6 2 | 3+4 2 |
| | 8 2 | 8 2 | 3+4 2 | 5+6 2 |
| | 9 2 | 9 2 | 5+6 2 | 5+6 2 |
| | 8 2 | 8 2 | 3+4 2 | 3+4 2 |
| | 8 2 | 9 2 | 5+6 2 | 5+6 2 |
| | 9 2 | 9 2 | 3+4 2 | 5+6 2 |
| | 6+7 2 | 6+7 2 | 5+6 2 | 3+4 2 |
| | 9 2 | 8 2 | 5+6 2 | 5+6 2 |
| | 8 2 | 9 2 | 3+4 2 | 3+4 2 |
| | 9 2 | 4+5 2 | 5+6 2 | 3+4 2 |
| | 4+5 2 | 6+7 2 | 3+4 2 | 5+6 2 |
| | 6+7 2 | 9 2 | 5+6 2 | 5+6 2 |
| | 8 2 | 8 2 | 5+6 2 | 5+6 2 |
| 166.1.52 | 651646 | 20 | Sb1Te2Tl1 | |
| | -GM | -T | -F | -L |
| | 8 2 | 9 2 | 5+6 2 | 5+6 2 |

Continued on next column

**TABLE LXXIII– continued**

| TQC | ICSD | NoE | SF |
|---|---|---|---|
| | 9 2 | 8 2 3+4 2 | 3+4 2 |
| | 8 2 | 9 2 3+4 2 | 5+6 2 |
| | 8 2 | 8 2 3+4 2 | 3+4 2 |
| | 9 2 | 8 2 5+6 2 | 3+4 2 |
| | 9 2 | 9 2 5+6 2 | 3+4 2 |
| | 6+7 2 | 4+5 2 5+6 2 | 5+6 2 |
| | 8 2 | 6+7 2 3+4 2 | 3+4 2 |
| | 4+5 2 | 9 2 3+4 2 | 5+6 2 |
| | 9 2 | 8 2 3+4 2 | 5+6 2 |
| 166.1.53 | 15411 | 20 | Sb1Te2Tl1 |
| | -GM | -T -F | -L |
| | 8 2 | 9 2 5+6 2 | 5+6 2 |
| | 9 2 | 8 2 3+4 2 | 3+4 2 |
| | 8 2 | 9 2 3+4 2 | 5+6 2 |
| | 8 2 | 8 2 3+4 2 | 3+4 2 |
| | 9 2 | 8 2 5+6 2 | 3+4 2 |
| | 9 2 | 9 2 5+6 2 | 3+4 2 |
| | 6+7 2 | 4+5 2 5+6 2 | 5+6 2 |
| | 8 2 | 9 2 3+4 2 | 3+4 2 |
| | 4+5 2 | 6+7 2 3+4 2 | 5+6 2 |
| | 9 2 | 8 2 3+4 2 | 5+6 2 |
| 166.1.54 | 651506 | 10 | Sb1 |
| | -GM | -T -F | -L |
| | 8 2 | 9 2 5+6 2 | 3+4 2 |
| | 9 2 | 8 2 3+4 2 | 5+6 2 |
| | 8 2 | 9 2 3+4 2 | 3+4 2 |
| | 8 2 | 6+7 2 5+6 2 | 5+6 2 |
| | 4+5 2 | 8 2 5+6 2 | 3+4 2 |
| 166.1.55 | 651498 | 10 | Sb1 |
| | -GM | -T -F | -L |
| | 8 2 | 9 2 5+6 2 | 3+4 2 |
| | 9 2 | 8 2 3+4 2 | 5+6 2 |
| | 8 2 | 9 2 3+4 2 | 3+4 2 |
| | 8 2 | 6+7 2 5+6 2 | 5+6 2 |
| | 4+5 2 | 8 2 5+6 2 | 3+4 2 |
| 166.1.56 | 651497 | 10 | Sb1 |
| | -GM | -T -F | -L |
| | 8 2 | 9 2 5+6 2 | 3+4 2 |
| | 9 2 | 8 2 3+4 2 | 5+6 2 |
| | 8 2 | 9 2 3+4 2 | 3+4 2 |
| | 8 2 | 6+7 2 5+6 2 | 5+6 2 |
| | 4+5 2 | 8 2 5+6 2 | 3+4 2 |
| 166.1.57 | 651491 | 10 | Sb1 |
| | -GM | -T -F | -L |
| | 8 2 | 9 2 5+6 2 | 3+4 2 |
| | 9 2 | 8 2 3+4 2 | 5+6 2 |
| | 8 2 | 9 2 3+4 2 | 3+4 2 |
| | 8 2 | 6+7 2 5+6 2 | 5+6 2 |
| | 4+5 2 | 8 2 5+6 2 | 3+4 2 |
| 166.1.58 | 651490 | 10 | Sb1 |
| | -GM | -T -F | -L |
| | 8 2 | 9 2 5+6 2 | 3+4 2 |
| | 9 2 | 8 2 3+4 2 | 5+6 2 |
| | 8 2 | 9 2 3+4 2 | 3+4 2 |
| | 8 2 | 6+7 2 5+6 2 | 5+6 2 |
| | 4+5 2 | 8 2 5+6 2 | 3+4 2 |
| 166.1.59 | 651489 | 10 | Sb1 |
| | -GM | -T -F | -L |
| | 8 2 | 9 2 5+6 2 | 3+4 2 |
| | 9 2 | 8 2 3+4 2 | 5+6 2 |

Continued on next column

**TABLE LXXIII– continued**

| TQC | ICSD | NoE | SF |
|---|---|---|---|
| | 8 2 | 9 2 3+4 2 | 3+4 2 |
| | 8 2 | 6+7 2 5+6 2 | 5+6 2 |
| | 4+5 2 | 8 2 5+6 2 | 3+4 2 |
| 166.1.60 | 651488 | 10 | Sb1 |
| | -GM | -T -F | -L |
| | 8 2 | 9 2 5+6 2 | 3+4 2 |
| | 9 2 | 8 2 3+4 2 | 5+6 2 |
| | 8 2 | 9 2 3+4 2 | 3+4 2 |
| | 8 2 | 6+7 2 5+6 2 | 5+6 2 |
| | 4+5 2 | 8 2 5+6 2 | 3+4 2 |
| 166.1.61 | 64697 | 10 | Sb1 |
| | -GM | -T -F | -L |
| | 8 2 | 9 2 5+6 2 | 3+4 2 |
| | 9 2 | 8 2 3+4 2 | 5+6 2 |
| | 8 2 | 9 2 3+4 2 | 3+4 2 |
| | 8 2 | 6+7 2 5+6 2 | 5+6 2 |
| | 4+5 2 | 8 2 5+6 2 | 3+4 2 |
| 166.1.62 | 64696 | 10 | Sb1 |
| | -GM | -T -F | -L |
| | 8 2 | 9 2 5+6 2 | 3+4 2 |
| | 9 2 | 8 2 3+4 2 | 5+6 2 |
| | 8 2 | 9 2 3+4 2 | 3+4 2 |
| | 8 2 | 6+7 2 5+6 2 | 5+6 2 |
| | 4+5 2 | 8 2 5+6 2 | 3+4 2 |
| 166.1.63 | 64695 | 10 | Sb1 |
| | -GM | -T -F | -L |
| | 8 2 | 9 2 5+6 2 | 3+4 2 |
| | 9 2 | 8 2 3+4 2 | 5+6 2 |
| | 8 2 | 9 2 3+4 2 | 3+4 2 |
| | 8 2 | 6+7 2 5+6 2 | 5+6 2 |
| | 4+5 2 | 8 2 5+6 2 | 3+4 2 |
| 166.1.64 | 55402 | 10 | Sb1 |
| | -GM | -T -F | -L |
| | 8 2 | 8 2 5+6 2 | 5+6 2 |
| | 9 2 | 9 2 3+4 2 | 3+4 2 |
| | 8 2 | 8 2 3+4 2 | 5+6 2 |
| | 8 2 | 4+5 2 5+6 2 | 3+4 2 |
| | 4+5 2 | 9 2 5+6 2 | 5+6 2 |
| 166.1.65 | 53795 | 10 | Sb1 |
| | -GM | -T -F | -L |
| | 8 2 | 9 2 5+6 2 | 3+4 2 |
| | 9 2 | 8 2 3+4 2 | 5+6 2 |
| | 8 2 | 9 2 3+4 2 | 3+4 2 |
| | 8 2 | 6+7 2 5+6 2 | 5+6 2 |
| | 4+5 2 | 8 2 5+6 2 | 3+4 2 |
| 166.1.66 | 409754 | 10 | Sb1 |
| | -GM | -T -F | -L |
| | 8 2 | 9 2 5+6 2 | 3+4 2 |
| | 9 2 | 8 2 3+4 2 | 5+6 2 |
| | 8 2 | 9 2 3+4 2 | 3+4 2 |
| | 8 2 | 6+7 2 5+6 2 | 5+6 2 |
| | 4+5 2 | 8 2 5+6 2 | 3+4 2 |
| 166.1.67 | 161493 | 10 | Sb1 |
| | -GM | -T -F | -L |
| | 8 2 | 8 2 5+6 2 | 5+6 2 |
| | 9 2 | 9 2 3+4 2 | 3+4 2 |
| | 8 2 | 8 2 3+4 2 | 5+6 2 |
| | 8 2 | 4+5 2 5+6 2 | 3+4 2 |
| | 4+5 2 | 9 2 5+6 2 | 5+6 2 |
| 166.1.68 | 24757 | 34 | S2Ta1 |

Continued on next column

**TABLE LXXIII– continued**

| TQC | ICSD | NoE | SF | |
|---|---|---|---|---|
| | -GM | -T | -F | -L |
| | 8 2 | 8 2 | 5+6 2 | 5+6 2 |
| | 9 2 | 9 2 | 3+4 2 | 3+4 2 |
| | 8 2 | 9 2 | 5+6 2 | 3+4 2 |
| | 9 2 | 8 2 | 3+4 2 | 5+6 2 |
| | 8 2 | 8 2 | 3+4 2 | 5+6 2 |
| | 9 2 | 9 2 | 5+6 2 | 3+4 2 |
| | 4+5 2 | 4+5 2 | 3+4 2 | 3+4 2 |
| | 6+7 2 | 6+7 2 | 5+6 2 | 5+6 2 |
| | 8 2 | 9 2 | 5+6 2 | 3+4 2 |
| | 9 2 | 8 2 | 3+4 2 | 5+6 2 |
| | 8 2 | 9 2 | 5+6 2 | 5+6 2 |
| | 9 2 | 8 2 | 3+4 2 | 3+4 2 |
| | 8 2 | 9 2 | 5+6 2 | 5+6 2 |
| | 4+5 2 | 8 2 | 3+4 2 | 3+4 2 |
| | 9 2 | 6+7 2 | 5+6 2 | 3+4 2 |
| | 6+7 2 | 4+5 2 | 3+4 2 | 5+6 2 |
| | 8 2 | 9 2 | 5+6 2 | 5+6 2 |
| 166.1.69 | 651179 | 30 | S1Ti1 | |
| | -GM | -T | -F | -L |
| | 8 2 | 8 2 | 3+4 2 | 3+4 2 |
| | 8 2 | 9 2 | 3+4 2 | 5+6 2 |
| | 9 2 | 8 2 | 5+6 2 | 3+4 2 |
| | 9 2 | 9 2 | 5+6 2 | 5+6 2 |
| | 8 2 | 9 2 | 3+4 2 | 5+6 2 |
| | 9 2 | 6+7 2 | 5+6 2 | 3+4 2 |
| | 6+7 2 | 9 2 | 5+6 2 | 5+6 2 |
| | 8 2 | 8 2 | 3+4 2 | 5+6 2 |
| | 4+5 2 | 9 2 | 5+6 2 | 3+4 2 |
| | 9 2 | 6+7 2 | 5+6 2 | 5+6 2 |
| | 9 2 | 8 2 | 3+4 2 | 5+6 2 |
| | 6+7 2 | 4+5 2 | 5+6 2 | 3+4 2 |
| | 8 2 | 4+5 2 | 3+4 2 | 3+4 2 |
| | 4+5 2 | 8 2 | 3+4 2 | 5+6 2 |
| | 8 2 | 8 2 | 3+4 2 | 3+4 2 |
| 166.1.70 | 25561 | 30 | S1Ti1 | |
| | -GM | -T | -F | -L |
| | 8 2 | 8 2 | 3+4 2 | 3+4 2 |
| | 8 2 | 9 2 | 3+4 2 | 5+6 2 |
| | 9 2 | 8 2 | 5+6 2 | 3+4 2 |
| | 9 2 | 9 2 | 5+6 2 | 5+6 2 |
| | 8 2 | 9 2 | 3+4 2 | 5+6 2 |
| | 9 2 | 6+7 2 | 5+6 2 | 3+4 2 |
| | 6+7 2 | 9 2 | 5+6 2 | 5+6 2 |
| | 8 2 | 8 2 | 3+4 2 | 5+6 2 |
| | 4+5 2 | 9 2 | 5+6 2 | 3+4 2 |
| | 9 2 | 6+7 2 | 5+6 2 | 5+6 2 |
| | 9 2 | 8 2 | 3+4 2 | 5+6 2 |
| | 6+7 2 | 4+5 2 | 5+6 2 | 3+4 2 |
| | 8 2 | 4+5 2 | 3+4 2 | 3+4 2 |
| | 4+5 2 | 8 2 | 3+4 2 | 5+6 2 |
| | 8 2 | 8 2 | 3+4 2 | 3+4 2 |
| 166.1.71 | 184500 | 30 | S1Ti1 | |
| | -GM | -T | -F | -L |
| | 8 2 | 8 2 | 3+4 2 | 3+4 2 |
| | 8 2 | 9 2 | 3+4 2 | 5+6 2 |
| | 9 2 | 8 2 | 5+6 2 | 3+4 2 |
| | 9 2 | 9 2 | 5+6 2 | 5+6 2 |
| | 8 2 | 9 2 | 3+4 2 | 5+6 2 |
| | 9 2 | 6+7 2 | 5+6 2 | 3+4 2 |
| | 6+7 2 | 9 2 | 5+6 2 | 5+6 2 |

Continued on next column

**TABLE LXXIII– continued**

| TQC | ICSD | NoE | SF | |
|---|---|---|---|---|
| | 8 2 | 8 2 | 3+4 2 | 5+6 2 |
| | 4+5 2 | 9 2 | 5+6 2 | 3+4 2 |
| | 9 2 | 6+7 2 | 5+6 2 | 5+6 2 |
| | 9 2 | 8 2 | 3+4 2 | 5+6 2 |
| | 6+7 2 | 4+5 2 | 5+6 2 | 3+4 2 |
| | 8 2 | 4+5 2 | 3+4 2 | 3+4 2 |
| | 4+5 2 | 8 2 | 3+4 2 | 5+6 2 |
| | 8 2 | 8 2 | 3+4 2 | 3+4 2 |
| 166.1.72 | 650000 | 236 | Pu2Zn17 | |
| | -GM | -T | -F | -L |
| | 8 2 | 8 2 | 3+4 2 | 3+4 2 |
| | 9 2 | 9 2 | 5+6 2 | 5+6 2 |
| | 8 2 | 9 2 | 3+4 2 | 5+6 2 |
| | 9 2 | 8 2 | 5+6 2 | 3+4 2 |
| | 4+5 2 | 4+5 2 | 5+6 2 | 3+4 2 |
| | 6+7 2 | 9 2 | 3+4 2 | 5+6 2 |
| | 8 2 | 6+7 2 | 3+4 2 | 5+6 2 |
| | 9 2 | 8 2 | 5+6 2 | 3+4 2 |
| | 8 2 | 9 2 | 3+4 2 | 3+4 2 |
| | 8 2 | 9 2 | 5+6 2 | 5+6 2 |
| | 8 2 | 6+7 2 | 5+6 2 | 3+4 2 |
| | 4+5 2 | 9 2 | 3+4 2 | 3+4 2 |
| | 9 2 | 8 2 | 3+4 2 | 3+4 2 |
| | 6+7 2 | 9 2 | 5+6 2 | 3+4 2 |
| | 8 2 | 8 2 | 5+6 2 | 5+6 2 |
| | 8 2 | 9 2 | 5+6 2 | 5+6 2 |
| | 4+5 2 | 4+5 2 | 3+4 2 | 3+4 2 |
| | 9 2 | 6+7 2 | 5+6 2 | 5+6 2 |
| | 8 2 | 9 2 | 5+6 2 | 5+6 2 |
| | 8 2 | 6+7 2 | 5+6 2 | 3+4 2 |
| | 4+5 2 | 8 2 | 3+4 2 | 5+6 2 |
| | 8 2 | 9 2 | 3+4 2 | 3+4 2 |
| | 8 2 | 8 2 | 3+4 2 | 3+4 2 |
| | 9 2 | 6+7 2 | 5+6 2 | 5+6 2 |
| | 4+5 2 | 4+5 2 | 5+6 2 | 3+4 2 |
| | 8 2 | 9 2 | 3+4 2 | 3+4 2 |
| | 6+7 2 | 9 2 | 5+6 2 | 5+6 2 |
| | 4+5 2 | 6+7 2 | 3+4 2 | 3+4 2 |
| | 8 2 | 9 2 | 5+6 2 | 5+6 2 |
| | 9 2 | 8 2 | 3+4 2 | 3+4 2 |
| | 4+5 2 | 8 2 | 5+6 2 | 5+6 2 |
| | 8 2 | 9 2 | 3+4 2 | 3+4 2 |
| | 9 2 | 4+5 2 | 5+6 2 | 5+6 2 |
| | 8 2 | 6+7 2 | 3+4 2 | 3+4 2 |
| | 4+5 2 | 9 2 | 5+6 2 | 5+6 2 |
| | 9 2 | 8 2 | 5+6 2 | 3+4 2 |
| | 6+7 2 | 6+7 2 | 3+4 2 | 5+6 2 |
| | 8 2 | 8 2 | 5+6 2 | 5+6 2 |
| | 9 2 | 9 2 | 5+6 2 | 3+4 2 |
| | 6+7 2 | 9 2 | 3+4 2 | 5+6 2 |
| | 9 2 | 4+5 2 | 3+4 2 | 3+4 2 |
| | 6+7 2 | 8 2 | 3+4 2 | 5+6 2 |
| | 9 2 | 6+7 2 | 5+6 2 | 3+4 2 |
| | 4+5 2 | 8 2 | 3+4 2 | 5+6 2 |
| | 8 2 | 4+5 2 | 5+6 2 | 3+4 2 |
| | 6+7 2 | 4+5 2 | 3+4 2 | 5+6 2 |
| | 8 2 | 6+7 2 | 5+6 2 | 3+4 2 |
| | 9 2 | 8 2 | 3+4 2 | 5+6 2 |
| | 6+7 2 | 9 2 | 5+6 2 | 5+6 2 |
| | 9 2 | 9 2 | 3+4 2 | 3+4 2 |
| | 8 2 | 9 2 | 5+6 2 | 3+4 2 |

Continued on next column

**TABLE LXXIII– continued**

| TQC | ICSD | NoE | SF | |
|---|---|---|---|---|
| | 9 2 | 8 2 | 5+6 2 | 5+6 2 |
| | 8 2 | 9 2 | 3+4 2 | 3+4 2 |
| | 8 2 | 6+7 2 | 3+4 2 | 3+4 2 |
| | 4+5 2 | 6+7 2 | 5+6 2 | 5+6 2 |
| | 8 2 | 8 2 | 3+4 2 | 5+6 2 |
| | 9 2 | 9 2 | 5+6 2 | 3+4 2 |
| | 6+7 2 | 8 2 | 5+6 2 | 5+6 2 |
| | 9 2 | 4+5 2 | 3+4 2 | 3+4 2 |
| | 8 2 | 8 2 | 3+4 2 | 5+6 2 |
| | 4+5 2 | 8 2 | 5+6 2 | 3+4 2 |
| | 8 2 | 9 2 | 3+4 2 | 3+4 2 |
| | 9 2 | 9 2 | 5+6 2 | 5+6 2 |
| | 6+7 2 | 6+7 2 | 3+4 2 | 5+6 2 |
| | 8 2 | 4+5 2 | 5+6 2 | 3+4 2 |
| | 4+5 2 | 8 2 | 5+6 2 | 3+4 2 |
| | 6+7 2 | 9 2 | 3+4 2 | 5+6 2 |
| | 9 2 | 4+5 2 | 3+4 2 | 3+4 2 |
| | 8 2 | 9 2 | 5+6 2 | 5+6 2 |
| | 4+5 2 | 6+7 2 | 3+4 2 | 5+6 2 |
| | 8 2 | 9 2 | 3+4 2 | 3+4 2 |
| | 9 2 | 8 2 | 5+6 2 | 3+4 2 |
| | 8 2 | 9 2 | 5+6 2 | 5+6 2 |
| | 6+7 2 | 4+5 2 | 3+4 2 | 3+4 2 |
| | 9 2 | 8 2 | 5+6 2 | 5+6 2 |
| | 8 2 | 8 2 | 5+6 2 | 3+4 2 |
| | 4+5 2 | 9 2 | 3+4 2 | 5+6 2 |
| | 4+5 2 | 4+5 2 | 5+6 2 | 3+4 2 |
| | 9 2 | 6+7 2 | 3+4 2 | 5+6 2 |
| | 9 2 | 8 2 | 3+4 2 | 3+4 2 |
| | 8 2 | 8 2 | 5+6 2 | 3+4 2 |
| | 8 2 | 9 2 | 3+4 2 | 5+6 2 |
| | 9 2 | 8 2 | 5+6 2 | 3+4 2 |
| | 8 2 | 6+7 2 | 3+4 2 | 5+6 2 |
| | 4+5 2 | 9 2 | 5+6 2 | 3+4 2 |
| | 8 2 | 4+5 2 | 3+4 2 | 5+6 2 |
| | 9 2 | 9 2 | 5+6 2 | 5+6 2 |
| | 6+7 2 | 8 2 | 5+6 2 | 5+6 2 |
| | 8 2 | 6+7 2 | 3+4 2 | 3+4 2 |
| | 9 2 | 9 2 | 5+6 2 | 3+4 2 |
| | 4+5 2 | 9 2 | 3+4 2 | 3+4 2 |
| | 8 2 | 9 2 | 5+6 2 | 5+6 2 |
| | 4+5 2 | 9 2 | 3+4 2 | 3+4 2 |
| | 8 2 | 6+7 2 | 5+6 2 | 5+6 2 |
| | 9 2 | 9 2 | 5+6 2 | 3+4 2 |
| | 6+7 2 | 9 2 | 5+6 2 | 3+4 2 |
| | 8 2 | 6+7 2 | 3+4 2 | 5+6 2 |
| | 4+5 2 | 8 2 | 5+6 2 | 5+6 2 |
| | 9 2 | 4+5 2 | 3+4 2 | 3+4 2 |
| | 8 2 | 8 2 | 3+4 2 | 3+4 2 |
| | 9 2 | 9 2 | 5+6 2 | 5+6 2 |
| | 6+7 2 | 8 2 | 3+4 2 | 5+6 2 |
| | 8 2 | 8 2 | 5+6 2 | 3+4 2 |
| | 9 2 | 4+5 2 | 5+6 2 | 5+6 2 |
| | 4+5 2 | 9 2 | 3+4 2 | 3+4 2 |
| | 9 2 | 6+7 2 | 3+4 2 | 3+4 2 |
| | 4+5 2 | 4+5 2 | 5+6 2 | 5+6 2 |
| | 8 2 | 8 2 | 5+6 2 | 5+6 2 |
| | 8 2 | 9 2 | 3+4 2 | 3+4 2 |
| | 9 2 | 8 2 | 5+6 2 | 5+6 2 |
| | 8 2 | 4+5 2 | 3+4 2 | 3+4 2 |
| | 9 2 | 8 2 | 3+4 2 | 3+4 2 |

Continued on next column

**TABLE LXXIII– continued**

| TQC | ICSD | NoE | SF | |
|---|---|---|---|---|
| | 6+7 2 | 9 2 | 3+4 2 | 5+6 2 |
| | 8 2 | 6+7 2 | 5+6 2 | 3+4 2 |
| | 4+5 2 | 9 2 | 5+6 2 | 5+6 2 |
| | 6+7 2 | 8 2 | 5+6 2 | 3+4 2 |
| | 8 2 | 6+7 2 | 3+4 2 | 5+6 2 |
| | 9 2 | 8 2 | 3+4 2 | 5+6 2 |
| 166.1.73 | 41372 | 54 | Pt3Te4 | |
| | -GM | -T | -F | -L |
| | 8 2 | 8 2 | 3+4 2 | 5+6 2 |
| | 9 2 | 9 2 | 5+6 2 | 3+4 2 |
| | 8 2 | 8 2 | 3+4 2 | 5+6 2 |
| | 9 2 | 9 2 | 5+6 2 | 3+4 2 |
| | 8 2 | 8 2 | 3+4 2 | 3+4 2 |
| | 4+5 2 | 4+5 2 | 5+6 2 | 3+4 2 |
| | 8 2 | 9 2 | 3+4 2 | 5+6 2 |
| | 4+5 2 | 8 2 | 3+4 2 | 3+4 2 |
| | 8 2 | 6+7 2 | 5+6 2 | 5+6 2 |
| | 9 2 | 9 2 | 5+6 2 | 3+4 2 |
| | 6+7 2 | 4+5 2 | 3+4 2 | 3+4 2 |
| | 8 2 | 8 2 | 3+4 2 | 5+6 2 |
| | 9 2 | 8 2 | 3+4 2 | 5+6 2 |
| | 8 2 | 8 2 | 3+4 2 | 3+4 2 |
| | 4+5 2 | 4+5 2 | 5+6 2 | 3+4 2 |
| | 9 2 | 8 2 | 3+4 2 | 5+6 2 |
| | 6+7 2 | 4+5 2 | 3+4 2 | 3+4 2 |
| | 8 2 | 8 2 | 5+6 2 | 5+6 2 |
| | 4+5 2 | 9 2 | 3+4 2 | 3+4 2 |
| | 8 2 | 6+7 2 | 5+6 2 | 5+6 2 |
| | 9 2 | 9 2 | 3+4 2 | 3+4 2 |
| | 8 2 | 9 2 | 5+6 2 | 5+6 2 |
| | 9 2 | 9 2 | 3+4 2 | 3+4 2 |
| | 6+7 2 | 6+7 2 | 5+6 2 | 5+6 2 |
| | 8 2 | 8 2 | 5+6 2 | 3+4 2 |
| | 4+5 2 | 9 2 | 3+4 2 | 3+4 2 |
| | 9 2 | 6+7 2 | 5+6 2 | 5+6 2 |
| 166.1.74 | 41371 | 76 | Pt2Te3 | |
| | -GM | -T | -F | -L |
| | 8 2 | 9 2 | 5+6 2 | 3+4 2 |
| | 9 2 | 8 2 | 3+4 2 | 5+6 2 |
| | 8 2 | 9 2 | 5+6 2 | 3+4 2 |
| | 9 2 | 8 2 | 3+4 2 | 5+6 2 |
| | 8 2 | 9 2 | 3+4 2 | 3+4 2 |
| | 9 2 | 8 2 | 5+6 2 | 5+6 2 |
| | 8 2 | 9 2 | 5+6 2 | 3+4 2 |
| | 4+5 2 | 8 2 | 3+4 2 | 5+6 2 |
| | 6+7 2 | 6+7 2 | 5+6 2 | 5+6 2 |
| | 8 2 | 4+5 2 | 3+4 2 | 5+6 2 |
| | 9 2 | 9 2 | 3+4 2 | 3+4 2 |
| | 9 2 | 8 2 | 5+6 2 | 3+4 2 |
| | 8 2 | 4+5 2 | 3+4 2 | 5+6 2 |
| | 4+5 2 | 8 2 | 5+6 2 | 3+4 2 |
| | 8 2 | 8 2 | 3+4 2 | 5+6 2 |
| | 6+7 2 | 9 2 | 5+6 2 | 5+6 2 |
| | 9 2 | 6+7 2 | 5+6 2 | 3+4 2 |
| | 8 2 | 9 2 | 3+4 2 | 3+4 2 |
| | 9 2 | 8 2 | 5+6 2 | 3+4 2 |
| | 6+7 2 | 9 2 | 3+4 2 | 5+6 2 |
| | 4+5 2 | 4+5 2 | 5+6 2 | 3+4 2 |
| | 9 2 | 6+7 2 | 3+4 2 | 5+6 2 |
| | 9 2 | 9 2 | 5+6 2 | 3+4 2 |
| | 8 2 | 9 2 | 5+6 2 | 3+4 2 |

Continued on next column

**TABLE LXXIII– continued**

| TQC | ICSD | NoE | SF | |
|---|---|---|---|---|
| | 6+7 2 | 6+7 2 | 3+4 2 | 5+6 2 |
| | 9 2 | 9 2 | 3+4 2 | 3+4 2 |
| | 8 2 | 8 2 | 5+6 2 | 3+4 2 |
| | 8 2 | 8 2 | 3+4 2 | 5+6 2 |
| | 4+5 2 | 9 2 | 3+4 2 | 3+4 2 |
| | 9 2 | 4+5 2 | 5+6 2 | 5+6 2 |
| | 8 2 | 8 2 | 3+4 2 | 5+6 2 |
| | 4+5 2 | 6+7 2 | 5+6 2 | 3+4 2 |
| | 8 2 | 8 2 | 3+4 2 | 5+6 2 |
| | 9 2 | 4+5 2 | 5+6 2 | 3+4 2 |
| | 6+7 2 | 8 2 | 3+4 2 | 5+6 2 |
| | 8 2 | 9 2 | 5+6 2 | 5+6 2 |
| | 4+5 2 | 8 2 | 5+6 2 | 3+4 2 |
| | 8 2 | 4+5 2 | 3+4 2 | 5+6 2 |
| 166.1.75 | 41370 | 32 | Pt1Te1 | |
| | -GM | -T | -F | -L |
| | 8 2 | 9 2 | 5+6 2 | 3+4 2 |
| | 9 2 | 8 2 | 3+4 2 | 5+6 2 |
| | 8 2 | 4+5 2 | 5+6 2 | 5+6 2 |
| | 4+5 2 | 8 2 | 3+4 2 | 3+4 2 |
| | 8 2 | 9 2 | 5+6 2 | 5+6 2 |
| | 6+7 2 | 6+7 2 | 3+4 2 | 3+4 2 |
| | 9 2 | 8 2 | 3+4 2 | 3+4 2 |
| | 9 2 | 9 2 | 3+4 2 | 5+6 2 |
| | 9 2 | 9 2 | 5+6 2 | 3+4 2 |
| | 6+7 2 | 6+7 2 | 5+6 2 | 3+4 2 |
| | 8 2 | 8 2 | 3+4 2 | 5+6 2 |
| | 4+5 2 | 4+5 2 | 5+6 2 | 5+6 2 |
| | 8 2 | 9 2 | 3+4 2 | 3+4 2 |
| | 8 2 | 8 2 | 5+6 2 | 3+4 2 |
| | 9 2 | 8 2 | 5+6 2 | 5+6 2 |
| | 4+5 2 | 4+5 2 | 5+6 2 | 5+6 2 |
| 166.1.76 | 649479 | 230 | Pr2Zn17 | |
| | -GM | -T | -F | -L |
| | 8 2 | 9 2 | 3+4 2 | 5+6 2 |
| | 9 2 | 8 2 | 5+6 2 | 3+4 2 |
| | 8 2 | 9 2 | 3+4 2 | 5+6 2 |
| | 9 2 | 8 2 | 5+6 2 | 3+4 2 |
| | 4+5 2 | 4+5 2 | 5+6 2 | 3+4 2 |
| | 8 2 | 9 2 | 3+4 2 | 5+6 2 |
| | 6+7 2 | 6+7 2 | 3+4 2 | 5+6 2 |
| | 9 2 | 8 2 | 5+6 2 | 3+4 2 |
| | 8 2 | 9 2 | 3+4 2 | 3+4 2 |
| | 8 2 | 9 2 | 5+6 2 | 5+6 2 |
| | 8 2 | 6+7 2 | 5+6 2 | 3+4 2 |
| | 4+5 2 | 9 2 | 3+4 2 | 3+4 2 |
| | 8 2 | 9 2 | 5+6 2 | 3+4 2 |
| | 4+5 2 | 9 2 | 3+4 2 | 5+6 2 |
| | 9 2 | 8 2 | 5+6 2 | 3+4 2 |
| | 8 2 | 4+5 2 | 5+6 2 | 5+6 2 |
| | 6+7 2 | 6+7 2 | 3+4 2 | 3+4 2 |
| | 8 2 | 8 2 | 5+6 2 | 5+6 2 |
| | 4+5 2 | 9 2 | 5+6 2 | 3+4 2 |
| | 9 2 | 6+7 2 | 3+4 2 | 5+6 2 |
| | 8 2 | 8 2 | 5+6 2 | 3+4 2 |
| | 8 2 | 9 2 | 3+4 2 | 5+6 2 |
| | 4+5 2 | 9 2 | 3+4 2 | 5+6 2 |
| | 9 2 | 8 2 | 5+6 2 | 3+4 2 |
| | 8 2 | 6+7 2 | 5+6 2 | 3+4 2 |
| | 6+7 2 | 4+5 2 | 3+4 2 | 3+4 2 |
| | 8 2 | 6+7 2 | 5+6 2 | 5+6 2 |

Continued on next column

**TABLE LXXIII– continued**

| TQC | ICSD | NoE | SF | |
|---|---|---|---|---|
| | 8 2 | 9 2 | 5+6 2 | 3+4 2 |
| | 9 2 | 9 2 | 3+4 2 | 5+6 2 |
| | 4+5 2 | 8 2 | 3+4 2 | 3+4 2 |
| | 9 2 | 6+7 2 | 5+6 2 | 5+6 2 |
| | 9 2 | 4+5 2 | 5+6 2 | 3+4 2 |
| | 4+5 2 | 8 2 | 3+4 2 | 3+4 2 |
| | 8 2 | 9 2 | 5+6 2 | 5+6 2 |
| | 6+7 2 | 8 2 | 3+4 2 | 5+6 2 |
| | 8 2 | 9 2 | 3+4 2 | 5+6 2 |
| | 9 2 | 8 2 | 5+6 2 | 3+4 2 |
| | 8 2 | 6+7 2 | 5+6 2 | 5+6 2 |
| | 6+7 2 | 9 2 | 3+4 2 | 5+6 2 |
| | 9 2 | 6+7 2 | 5+6 2 | 3+4 2 |
| | 4+5 2 | 4+5 2 | 3+4 2 | 5+6 2 |
| | 6+7 2 | 9 2 | 3+4 2 | 3+4 2 |
| | 4+5 2 | 8 2 | 5+6 2 | 5+6 2 |
| | 9 2 | 8 2 | 3+4 2 | 3+4 2 |
| | 6+7 2 | 4+5 2 | 3+4 2 | 3+4 2 |
| | 8 2 | 8 2 | 5+6 2 | 5+6 2 |
| | 8 2 | 9 2 | 5+6 2 | 5+6 2 |
| | 6+7 2 | 6+7 2 | 3+4 2 | 3+4 2 |
| | 9 2 | 9 2 | 5+6 2 | 3+4 2 |
| | 8 2 | 4+5 2 | 3+4 2 | 5+6 2 |
| | 9 2 | 9 2 | 5+6 2 | 3+4 2 |
| | 8 2 | 8 2 | 3+4 2 | 3+4 2 |
| | 8 2 | 6+7 2 | 5+6 2 | 5+6 2 |
| | 9 2 | 9 2 | 5+6 2 | 5+6 2 |
| | 4+5 2 | 8 2 | 3+4 2 | 3+4 2 |
| | 8 2 | 6+7 2 | 3+4 2 | 5+6 2 |
| | 9 2 | 4+5 2 | 5+6 2 | 3+4 2 |
| | 4+5 2 | 9 2 | 3+4 2 | 3+4 2 |
| | 6+7 2 | 8 2 | 5+6 2 | 5+6 2 |
| | 8 2 | 8 2 | 5+6 2 | 3+4 2 |
| | 9 2 | 9 2 | 3+4 2 | 5+6 2 |
| | 8 2 | 4+5 2 | 5+6 2 | 3+4 2 |
| | 6+7 2 | 8 2 | 3+4 2 | 5+6 2 |
| | 8 2 | 9 2 | 3+4 2 | 5+6 2 |
| | 4+5 2 | 6+7 2 | 5+6 2 | 3+4 2 |
| | 9 2 | 8 2 | 3+4 2 | 3+4 2 |
| | 8 2 | 9 2 | 5+6 2 | 5+6 2 |
| | 6+7 2 | 9 2 | 3+4 2 | 3+4 2 |
| | 9 2 | 4+5 2 | 5+6 2 | 5+6 2 |
| | 4+5 2 | 6+7 2 | 3+4 2 | 3+4 2 |
| | 8 2 | 9 2 | 5+6 2 | 5+6 2 |
| | 9 2 | 8 2 | 3+4 2 | 3+4 2 |
| | 4+5 2 | 4+5 2 | 5+6 2 | 5+6 2 |
| | 6+7 2 | 9 2 | 5+6 2 | 3+4 2 |
| | 9 2 | 8 2 | 3+4 2 | 5+6 2 |
| | 8 2 | 9 2 | 5+6 2 | 3+4 2 |
| | 8 2 | 8 2 | 3+4 2 | 5+6 2 |
| | 9 2 | 6+7 2 | 5+6 2 | 5+6 2 |
| | 4+5 2 | 8 2 | 3+4 2 | 3+4 2 |
| | 9 2 | 4+5 2 | 3+4 2 | 3+4 2 |
| | 8 2 | 9 2 | 5+6 2 | 5+6 2 |
| | 8 2 | 8 2 | 3+4 2 | 3+4 2 |
| | 9 2 | 6+7 2 | 5+6 2 | 3+4 2 |
| | 8 2 | 4+5 2 | 3+4 2 | 5+6 2 |
| | 9 2 | 8 2 | 5+6 2 | 3+4 2 |
| | 6+7 2 | 9 2 | 3+4 2 | 5+6 2 |
| | 8 2 | 9 2 | 5+6 2 | 5+6 2 |
| | 4+5 2 | 9 2 | 5+6 2 | 3+4 2 |

Continued on next column

**TABLE LXXIII– continued**

| TQC | ICSD | NoE | SF | |
|---|---|---|---|---|
| | 9 2 | 6+7 2 | 3+4 2 | 5+6 2 |
| | 8 2 | 8 2 | 5+6 2 | 3+4 2 |
| | 4+5 2 | 9 2 | 5+6 2 | 3+4 2 |
| | 8 2 | 9 2 | 3+4 2 | 5+6 2 |
| | 4+5 2 | 6+7 2 | 3+4 2 | 3+4 2 |
| | 8 2 | 9 2 | 5+6 2 | 5+6 2 |
| | 9 2 | 9 2 | 5+6 2 | 3+4 2 |
| | 6+7 2 | 9 2 | 5+6 2 | 3+4 2 |
| | 8 2 | 6+7 2 | 3+4 2 | 5+6 2 |
| | 4+5 2 | 8 2 | 5+6 2 | 5+6 2 |
| | 8 2 | 4+5 2 | 3+4 2 | 3+4 2 |
| | 9 2 | 8 2 | 5+6 2 | 3+4 2 |
| | 6+7 2 | 9 2 | 3+4 2 | 5+6 2 |
| | 9 2 | 8 2 | 3+4 2 | 5+6 2 |
| | 8 2 | 8 2 | 5+6 2 | 3+4 2 |
| | 9 2 | 4+5 2 | 5+6 2 | 5+6 2 |
| | 4+5 2 | 6+7 2 | 3+4 2 | 3+4 2 |
| | 9 2 | 9 2 | 3+4 2 | 3+4 2 |
| | 8 2 | 4+5 2 | 5+6 2 | 5+6 2 |
| | 4+5 2 | 8 2 | 5+6 2 | 5+6 2 |
| | 8 2 | 4+5 2 | 5+6 2 | 3+4 2 |
| | 8 2 | 8 2 | 3+4 2 | 5+6 2 |
| | 9 2 | 9 2 | 3+4 2 | 3+4 2 |
| | 9 2 | 8 2 | 3+4 2 | 5+6 2 |
| | 6+7 2 | 6+7 2 | 3+4 2 | 3+4 2 |
| | 8 2 | 9 2 | 5+6 2 | 5+6 2 |
| | 4+5 2 | 9 2 | 5+6 2 | 3+4 2 |
| 166.1.77 | 649469 | 230 | Pr2Zn17 | |
| | -GM | -T | -F | -L |
| | 8 2 | 9 2 | 3+4 2 | 5+6 2 |
| | 9 2 | 8 2 | 5+6 2 | 3+4 2 |
| | 8 2 | 9 2 | 3+4 2 | 5+6 2 |
| | 9 2 | 8 2 | 5+6 2 | 3+4 2 |
| | 4+5 2 | 4+5 2 | 5+6 2 | 3+4 2 |
| | 8 2 | 9 2 | 3+4 2 | 5+6 2 |
| | 6+7 2 | 6+7 2 | 3+4 2 | 5+6 2 |
| | 9 2 | 8 2 | 5+6 2 | 3+4 2 |
| | 8 2 | 9 2 | 3+4 2 | 3+4 2 |
| | 8 2 | 9 2 | 5+6 2 | 5+6 2 |
| | 8 2 | 6+7 2 | 5+6 2 | 3+4 2 |
| | 4+5 2 | 9 2 | 3+4 2 | 3+4 2 |
| | 8 2 | 9 2 | 5+6 2 | 3+4 2 |
| | 4+5 2 | 9 2 | 3+4 2 | 5+6 2 |
| | 9 2 | 8 2 | 5+6 2 | 3+4 2 |
| | 8 2 | 4+5 2 | 5+6 2 | 5+6 2 |
| | 6+7 2 | 6+7 2 | 3+4 2 | 3+4 2 |
| | 8 2 | 8 2 | 5+6 2 | 5+6 2 |
| | 4+5 2 | 9 2 | 5+6 2 | 3+4 2 |
| | 8 2 | 6+7 2 | 3+4 2 | 5+6 2 |
| | 9 2 | 8 2 | 5+6 2 | 3+4 2 |
| | 8 2 | 9 2 | 3+4 2 | 5+6 2 |
| | 4+5 2 | 9 2 | 3+4 2 | 5+6 2 |
| | 9 2 | 8 2 | 5+6 2 | 3+4 2 |
| | 8 2 | 6+7 2 | 5+6 2 | 3+4 2 |
| | 6+7 2 | 4+5 2 | 3+4 2 | 3+4 2 |
| | 8 2 | 6+7 2 | 5+6 2 | 5+6 2 |
| | 8 2 | 9 2 | 5+6 2 | 3+4 2 |
| | 9 2 | 9 2 | 3+4 2 | 5+6 2 |
| | 4+5 2 | 8 2 | 3+4 2 | 3+4 2 |
| | 9 2 | 6+7 2 | 5+6 2 | 5+6 2 |
| | 9 2 | 4+5 2 | 5+6 2 | 3+4 2 |

Continued on next column

**TABLE LXXIII– continued**

| TQC | ICSD | NoE | SF | |
|---|---|---|---|---|
| | 4+5 2 | 8 2 | 3+4 2 | 3+4 2 |
| | 8 2 | 9 2 | 5+6 2 | 5+6 2 |
| | 6+7 2 | 8 2 | 3+4 2 | 5+6 2 |
| | 8 2 | 9 2 | 3+4 2 | 5+6 2 |
| | 9 2 | 8 2 | 5+6 2 | 3+4 2 |
| | 8 2 | 6+7 2 | 5+6 2 | 5+6 2 |
| | 6+7 2 | 9 2 | 3+4 2 | 5+6 2 |
| | 9 2 | 6+7 2 | 5+6 2 | 3+4 2 |
| | 4+5 2 | 4+5 2 | 3+4 2 | 5+6 2 |
| | 6+7 2 | 9 2 | 3+4 2 | 3+4 2 |
| | 4+5 2 | 8 2 | 5+6 2 | 5+6 2 |
| | 9 2 | 8 2 | 3+4 2 | 3+4 2 |
| | 6+7 2 | 4+5 2 | 3+4 2 | 3+4 2 |
| | 8 2 | 8 2 | 5+6 2 | 5+6 2 |
| | 8 2 | 9 2 | 5+6 2 | 5+6 2 |
| | 6+7 2 | 6+7 2 | 3+4 2 | 3+4 2 |
| | 9 2 | 9 2 | 5+6 2 | 3+4 2 |
| | 8 2 | 4+5 2 | 3+4 2 | 5+6 2 |
| | 9 2 | 9 2 | 5+6 2 | 3+4 2 |
| | 8 2 | 8 2 | 3+4 2 | 3+4 2 |
| | 8 2 | 6+7 2 | 5+6 2 | 5+6 2 |
| | 9 2 | 9 2 | 5+6 2 | 5+6 2 |
| | 4+5 2 | 8 2 | 3+4 2 | 3+4 2 |
| | 8 2 | 6+7 2 | 3+4 2 | 5+6 2 |
| | 9 2 | 4+5 2 | 5+6 2 | 3+4 2 |
| | 4+5 2 | 9 2 | 3+4 2 | 3+4 2 |
| | 6+7 2 | 8 2 | 5+6 2 | 5+6 2 |
| | 8 2 | 8 2 | 5+6 2 | 3+4 2 |
| | 9 2 | 9 2 | 3+4 2 | 5+6 2 |
| | 8 2 | 4+5 2 | 5+6 2 | 3+4 2 |
| | 6+7 2 | 8 2 | 3+4 2 | 5+6 2 |
| | 8 2 | 9 2 | 3+4 2 | 5+6 2 |
| | 4+5 2 | 6+7 2 | 5+6 2 | 3+4 2 |
| | 9 2 | 8 2 | 3+4 2 | 3+4 2 |
| | 8 2 | 9 2 | 5+6 2 | 5+6 2 |
| | 6+7 2 | 9 2 | 3+4 2 | 3+4 2 |
| | 9 2 | 4+5 2 | 5+6 2 | 5+6 2 |
| | 4+5 2 | 6+7 2 | 3+4 2 | 3+4 2 |
| | 8 2 | 9 2 | 5+6 2 | 5+6 2 |
| | 9 2 | 8 2 | 3+4 2 | 3+4 2 |
| | 4+5 2 | 4+5 2 | 5+6 2 | 5+6 2 |
| | 6+7 2 | 9 2 | 5+6 2 | 3+4 2 |
| | 9 2 | 8 2 | 3+4 2 | 5+6 2 |
| | 8 2 | 9 2 | 5+6 2 | 3+4 2 |
| | 8 2 | 8 2 | 3+4 2 | 5+6 2 |
| | 9 2 | 6+7 2 | 5+6 2 | 5+6 2 |
| | 4+5 2 | 8 2 | 3+4 2 | 3+4 2 |
| | 9 2 | 4+5 2 | 3+4 2 | 3+4 2 |
| | 8 2 | 9 2 | 5+6 2 | 5+6 2 |
| | 8 2 | 8 2 | 3+4 2 | 3+4 2 |
| | 9 2 | 6+7 2 | 5+6 2 | 3+4 2 |
| | 8 2 | 4+5 2 | 3+4 2 | 5+6 2 |
| | 9 2 | 8 2 | 5+6 2 | 3+4 2 |
| | 6+7 2 | 9 2 | 3+4 2 | 5+6 2 |
| | 8 2 | 9 2 | 5+6 2 | 5+6 2 |
| | 4+5 2 | 9 2 | 5+6 2 | 3+4 2 |
| | 9 2 | 6+7 2 | 3+4 2 | 5+6 2 |
| | 8 2 | 8 2 | 5+6 2 | 3+4 2 |
| | 4+5 2 | 9 2 | 5+6 2 | 3+4 2 |
| | 8 2 | 9 2 | 3+4 2 | 5+6 2 |
| | 4+5 2 | 6+7 2 | 3+4 2 | 3+4 2 |

Continued on next column

**TABLE LXXIII– continued**

| TQC | ICSD | NoE | SF | |
|---|---|---|---|---|
| | 8 2 | 9 2 | 5+6 2 | 5+6 2 |
| | 9 2 | 9 2 | 5+6 2 | 3+4 2 |
| | 6+7 2 | 9 2 | 5+6 2 | 3+4 2 |
| | 8 2 | 6+7 2 | 3+4 2 | 5+6 2 |
| | 4+5 2 | 8 2 | 5+6 2 | 5+6 2 |
| | 8 2 | 4+5 2 | 3+4 2 | 3+4 2 |
| | 9 2 | 8 2 | 5+6 2 | 3+4 2 |
| | 6+7 2 | 9 2 | 3+4 2 | 5+6 2 |
| | 9 2 | 8 2 | 3+4 2 | 5+6 2 |
| | 8 2 | 8 2 | 5+6 2 | 3+4 2 |
| | 9 2 | 4+5 2 | 5+6 2 | 5+6 2 |
| | 4+5 2 | 6+7 2 | 3+4 2 | 3+4 2 |
| | 9 2 | 9 2 | 3+4 2 | 3+4 2 |
| | 8 2 | 4+5 2 | 5+6 2 | 5+6 2 |
| | 4+5 2 | 8 2 | 5+6 2 | 5+6 2 |
| | 8 2 | 4+5 2 | 5+6 2 | 3+4 2 |
| | 8 2 | 8 2 | 3+4 2 | 5+6 2 |
| | 9 2 | 9 2 | 3+4 2 | 3+4 2 |
| | 9 2 | 8 2 | 3+4 2 | 5+6 2 |
| | 6+7 2 | 9 2 | 3+4 2 | 3+4 2 |
| | 8 2 | 6+7 2 | 5+6 2 | 5+6 2 |
| | 4+5 2 | 9 2 | 5+6 2 | 3+4 2 |
| 166.1.78 | 104241 | 230 | Pr2Zn17 | |
| | -GM | -T | -F | -L |
| | 8 2 | 9 2 | 3+4 2 | 5+6 2 |
| | 9 2 | 8 2 | 5+6 2 | 3+4 2 |
| | 8 2 | 9 2 | 3+4 2 | 5+6 2 |
| | 9 2 | 8 2 | 5+6 2 | 3+4 2 |
| | 4+5 2 | 4+5 2 | 5+6 2 | 3+4 2 |
| | 8 2 | 9 2 | 3+4 2 | 5+6 2 |
| | 6+7 2 | 6+7 2 | 3+4 2 | 5+6 2 |
| | 9 2 | 8 2 | 5+6 2 | 3+4 2 |
| | 8 2 | 9 2 | 3+4 2 | 3+4 2 |
| | 8 2 | 9 2 | 5+6 2 | 5+6 2 |
| | 8 2 | 6+7 2 | 5+6 2 | 3+4 2 |
| | 4+5 2 | 9 2 | 3+4 2 | 3+4 2 |
| | 8 2 | 9 2 | 5+6 2 | 3+4 2 |
| | 4+5 2 | 9 2 | 3+4 2 | 5+6 2 |
| | 9 2 | 8 2 | 5+6 2 | 3+4 2 |
| | 8 2 | 4+5 2 | 5+6 2 | 5+6 2 |
| | 6+7 2 | 6+7 2 | 5+6 2 | 3+4 2 |
| | 8 2 | 8 2 | 3+4 2 | 5+6 2 |
| | 4+5 2 | 9 2 | 5+6 2 | 3+4 2 |
| | 8 2 | 6+7 2 | 3+4 2 | 5+6 2 |
| | 9 2 | 8 2 | 5+6 2 | 3+4 2 |
| | 8 2 | 9 2 | 3+4 2 | 5+6 2 |
| | 4+5 2 | 9 2 | 3+4 2 | 5+6 2 |
| | 9 2 | 8 2 | 5+6 2 | 3+4 2 |
| | 8 2 | 6+7 2 | 5+6 2 | 3+4 2 |
| | 6+7 2 | 4+5 2 | 3+4 2 | 3+4 2 |
| | 8 2 | 6+7 2 | 5+6 2 | 5+6 2 |
| | 8 2 | 9 2 | 5+6 2 | 3+4 2 |
| | 9 2 | 9 2 | 3+4 2 | 5+6 2 |
| | 4+5 2 | 8 2 | 3+4 2 | 3+4 2 |
| | 9 2 | 6+7 2 | 5+6 2 | 5+6 2 |
| | 9 2 | 4+5 2 | 5+6 2 | 3+4 2 |
| | 4+5 2 | 8 2 | 3+4 2 | 3+4 2 |
| | 8 2 | 9 2 | 5+6 2 | 5+6 2 |
| | 6+7 2 | 8 2 | 3+4 2 | 5+6 2 |
| | 8 2 | 9 2 | 3+4 2 | 5+6 2 |
| | 9 2 | 8 2 | 5+6 2 | 3+4 2 |

Continued on next column

**TABLE LXXIII– continued**

| TQC | ICSD | NoE | SF | |
|---|---|---|---|---|
| | 8 2 | 6+7 2 | 5+6 2 | 5+6 2 |
| | 6+7 2 | 9 2 | 3+4 2 | 5+6 2 |
| | 9 2 | 6+7 2 | 5+6 2 | 3+4 2 |
| | 4+5 2 | 4+5 2 | 3+4 2 | 5+6 2 |
| | 6+7 2 | 9 2 | 3+4 2 | 3+4 2 |
| | 4+5 2 | 8 2 | 5+6 2 | 5+6 2 |
| | 9 2 | 8 2 | 3+4 2 | 3+4 2 |
| | 6+7 2 | 4+5 2 | 3+4 2 | 3+4 2 |
| | 8 2 | 8 2 | 5+6 2 | 5+6 2 |
| | 8 2 | 9 2 | 5+6 2 | 5+6 2 |
| | 9 2 | 6+7 2 | 3+4 2 | 3+4 2 |
| | 6+7 2 | 9 2 | 5+6 2 | 3+4 2 |
| | 8 2 | 4+5 2 | 3+4 2 | 5+6 2 |
| | 9 2 | 9 2 | 5+6 2 | 3+4 2 |
| | 8 2 | 8 2 | 3+4 2 | 3+4 2 |
| | 8 2 | 6+7 2 | 5+6 2 | 5+6 2 |
| | 9 2 | 9 2 | 5+6 2 | 3+4 2 |
| | 4+5 2 | 8 2 | 3+4 2 | 5+6 2 |
| | 8 2 | 6+7 2 | 3+4 2 | 5+6 2 |
| | 9 2 | 4+5 2 | 5+6 2 | 3+4 2 |
| | 4+5 2 | 9 2 | 3+4 2 | 3+4 2 |
| | 6+7 2 | 8 2 | 5+6 2 | 5+6 2 |
| | 8 2 | 8 2 | 5+6 2 | 3+4 2 |
| | 9 2 | 9 2 | 3+4 2 | 5+6 2 |
| | 8 2 | 4+5 2 | 5+6 2 | 3+4 2 |
| | 6+7 2 | 8 2 | 3+4 2 | 5+6 2 |
| | 8 2 | 9 2 | 3+4 2 | 5+6 2 |
| | 4+5 2 | 6+7 2 | 5+6 2 | 3+4 2 |
| | 9 2 | 8 2 | 3+4 2 | 3+4 2 |
| | 8 2 | 9 2 | 5+6 2 | 5+6 2 |
| | 6+7 2 | 9 2 | 3+4 2 | 3+4 2 |
| | 9 2 | 4+5 2 | 5+6 2 | 5+6 2 |
| | 4+5 2 | 6+7 2 | 3+4 2 | 3+4 2 |
| | 8 2 | 9 2 | 5+6 2 | 5+6 2 |
| | 9 2 | 8 2 | 3+4 2 | 3+4 2 |
| | 4+5 2 | 4+5 2 | 5+6 2 | 5+6 2 |
| | 6+7 2 | 9 2 | 5+6 2 | 3+4 2 |
| | 9 2 | 8 2 | 3+4 2 | 5+6 2 |
| | 8 2 | 9 2 | 5+6 2 | 3+4 2 |
| | 8 2 | 8 2 | 3+4 2 | 5+6 2 |
| | 9 2 | 6+7 2 | 5+6 2 | 5+6 2 |
| | 4+5 2 | 8 2 | 3+4 2 | 3+4 2 |
| | 9 2 | 4+5 2 | 3+4 2 | 3+4 2 |
| | 8 2 | 9 2 | 5+6 2 | 5+6 2 |
| | 8 2 | 8 2 | 3+4 2 | 3+4 2 |
| | 9 2 | 6+7 2 | 5+6 2 | 3+4 2 |
| | 8 2 | 4+5 2 | 3+4 2 | 5+6 2 |
| | 9 2 | 8 2 | 5+6 2 | 3+4 2 |
| | 6+7 2 | 9 2 | 3+4 2 | 5+6 2 |
| | 8 2 | 9 2 | 5+6 2 | 5+6 2 |
| | 4+5 2 | 9 2 | 5+6 2 | 3+4 2 |
| | 9 2 | 6+7 2 | 3+4 2 | 5+6 2 |
| | 8 2 | 8 2 | 5+6 2 | 3+4 2 |
| | 4+5 2 | 9 2 | 5+6 2 | 3+4 2 |
| | 8 2 | 9 2 | 3+4 2 | 5+6 2 |
| | 4+5 2 | 6+7 2 | 3+4 2 | 3+4 2 |
| | 8 2 | 9 2 | 5+6 2 | 5+6 2 |
| | 9 2 | 9 2 | 5+6 2 | 3+4 2 |
| | 6+7 2 | 9 2 | 5+6 2 | 3+4 2 |
| | 8 2 | 6+7 2 | 3+4 2 | 5+6 2 |
| | 4+5 2 | 8 2 | 5+6 2 | 5+6 2 |

Continued on next column

**TABLE LXXIII– continued**

| TQC | ICSD | NoE | SF | |
|---|---|---|---|---|
| | 8 2 | 4+5 2 | 3+4 2 | 3+4 2 |
| | 9 2 | 8 2 | 5+6 2 | 3+4 2 |
| | 6+7 2 | 9 2 | 3+4 2 | 5+6 2 |
| | 9 2 | 8 2 | 3+4 2 | 5+6 2 |
| | 8 2 | 8 2 | 5+6 2 | 3+4 2 |
| | 9 2 | 4+5 2 | 5+6 2 | 5+6 2 |
| | 4+5 2 | 6+7 2 | 3+4 2 | 3+4 2 |
| | 9 2 | 9 2 | 3+4 2 | 3+4 2 |
| | 8 2 | 4+5 2 | 5+6 2 | 5+6 2 |
| | 4+5 2 | 8 2 | 5+6 2 | 5+6 2 |
| | 8 2 | 4+5 2 | 5+6 2 | 3+4 2 |
| | 8 2 | 8 2 | 3+4 2 | 5+6 2 |
| | 9 2 | 9 2 | 3+4 2 | 3+4 2 |
| | 9 2 | 8 2 | 3+4 2 | 5+6 2 |
| | 6+7 2 | 9 2 | 3+4 2 | 3+4 2 |
| | 8 2 | 6+7 2 | 5+6 2 | 5+6 2 |
| | 4+5 2 | 9 2 | 5+6 2 | 3+4 2 |
| 166.1.79 | 54338 | 104 | Pr1 | |
| | -GM | -T | -F | -L |
| | 8 2 | 9 2 | 3+4 2 | 3+4 2 |
| | 9 2 | 8 2 | 5+6 2 | 5+6 2 |
| | 9 2 | 9 2 | 5+6 2 | 5+6 2 |
| | 6+7 2 | 4+5 2 | 3+4 2 | 3+4 2 |
| | 9 2 | 8 2 | 3+4 2 | 3+4 2 |
| | 8 2 | 8 2 | 5+6 2 | 5+6 2 |
| | 4+5 2 | 6+7 2 | 3+4 2 | 5+6 2 |
| | 8 2 | 9 2 | 5+6 2 | 3+4 2 |
| | 8 2 | 9 2 | 5+6 2 | 3+4 2 |
| | 9 2 | 8 2 | 3+4 2 | 5+6 2 |
| | 4+5 2 | 4+5 2 | 3+4 2 | 5+6 2 |
| | 6+7 2 | 6+7 2 | 5+6 2 | 3+4 2 |
| | 8 2 | 9 2 | 5+6 2 | 5+6 2 |
| | 9 2 | 8 2 | 3+4 2 | 3+4 2 |
| | 8 2 | 9 2 | 5+6 2 | 3+4 2 |
| | 9 2 | 8 2 | 3+4 2 | 5+6 2 |
| | 9 2 | 9 2 | 5+6 2 | 5+6 2 |
| | 9 2 | 8 2 | 3+4 2 | 3+4 2 |
| | 6+7 2 | 4+5 2 | 3+4 2 | 3+4 2 |
| | 8 2 | 6+7 2 | 5+6 2 | 5+6 2 |
| | 8 2 | 9 2 | 3+4 2 | 5+6 2 |
| | 4+5 2 | 8 2 | 5+6 2 | 3+4 2 |
| | 8 2 | 8 2 | 5+6 2 | 5+6 2 |
| | 9 2 | 9 2 | 3+4 2 | 3+4 2 |
| | 6+7 2 | 6+7 2 | 3+4 2 | 3+4 2 |
| | 9 2 | 9 2 | 5+6 2 | 5+6 2 |
| | 4+5 2 | 8 2 | 5+6 2 | 3+4 2 |
| | 8 2 | 4+5 2 | 3+4 2 | 5+6 2 |
| | 8 2 | 8 2 | 5+6 2 | 5+6 2 |
| | 4+5 2 | 9 2 | 3+4 2 | 3+4 2 |
| | 9 2 | 4+5 2 | 3+4 2 | 3+4 2 |
| | 6+7 2 | 6+7 2 | 5+6 2 | 5+6 2 |
| | 8 2 | 9 2 | 3+4 2 | 3+4 2 |
| | 8 2 | 8 2 | 3+4 2 | 5+6 2 |
| | 8 2 | 4+5 2 | 5+6 2 | 3+4 2 |
| | 4+5 2 | 8 2 | 3+4 2 | 5+6 2 |
| | 9 2 | 6+7 2 | 5+6 2 | 3+4 2 |
| | 8 2 | 9 2 | 5+6 2 | 3+4 2 |
| | 8 2 | 8 2 | 5+6 2 | 5+6 2 |
| | 4+5 2 | 9 2 | 3+4 2 | 5+6 2 |
| | 6+7 2 | 9 2 | 5+6 2 | 3+4 2 |
| | 9 2 | 8 2 | 3+4 2 | 5+6 2 |
| | 9 2 | 8 2 | 5+6 2 | 5+6 2 |
| | 9 2 | 4+5 2 | 3+4 2 | 3+4 2 |
| | 8 2 | 9 2 | 5+6 2 | 3+4 2 |
| | 8 2 | 6+7 2 | 3+4 2 | 5+6 2 |
| | 8 2 | 8 2 | 3+4 2 | 3+4 2 |
| | 4+5 2 | 9 2 | 5+6 2 | 5+6 2 |
| | 6+7 2 | 8 2 | 3+4 2 | 5+6 2 |
| | 9 2 | 9 2 | 5+6 2 | 3+4 2 |
| | 8 2 | 6+7 2 | 3+4 2 | 3+4 2 |
| | 9 2 | 9 2 | 3+4 2 | 3+4 2 |
| 166.1.80 | 648763 | 48 | Pd3S2Tl2 | |
| | -GM | -T | -F | -L |
| | 8 2 | 9 2 | 5+6 2 | 3+4 2 |
| | 9 2 | 8 2 | 3+4 2 | 5+6 2 |
| | 8 2 | 9 2 | 3+4 2 | 3+4 2 |
| | 8 2 | 8 2 | 3+4 2 | 5+6 2 |
| | 8 2 | 9 2 | 5+6 2 | 3+4 2 |
| | 4+5 2 | 6+7 2 | 5+6 2 | 3+4 2 |
| | 8 2 | 9 2 | 5+6 2 | 3+4 2 |
| | 4+5 2 | 9 2 | 3+4 2 | 5+6 2 |
| | 8 2 | 6+7 2 | 5+6 2 | 5+6 2 |
| | 9 2 | 9 2 | 3+4 2 | 5+6 2 |
| | 8 2 | 8 2 | 3+4 2 | 3+4 2 |
| | 9 2 | 4+5 2 | 5+6 2 | 3+4 2 |
| | 8 2 | 9 2 | 5+6 2 | 5+6 2 |
| | 4+5 2 | 6+7 2 | 5+6 2 | 3+4 2 |
| | 6+7 2 | 9 2 | 5+6 2 | 3+4 2 |
| | 8 2 | 8 2 | 5+6 2 | 5+6 2 |
| | 4+5 2 | 9 2 | 3+4 2 | 3+4 2 |
| | 8 2 | 6+7 2 | 5+6 2 | 3+4 2 |
| | 4+5 2 | 9 2 | 5+6 2 | 3+4 2 |
| | 8 2 | 6+7 2 | 3+4 2 | 5+6 2 |
| | 9 2 | 9 2 | 3+4 2 | 5+6 2 |
| | 8 2 | 9 2 | 3+4 2 | 3+4 2 |
| | 9 2 | 9 2 | 5+6 2 | 5+6 2 |
| | 6+7 2 | 6+7 2 | 5+6 2 | 3+4 2 |
| 166.1.81 | 648762 | 48 | Pd3S2Tl2 | |
| | -GM | -T | -F | -L |
| | 8 2 | 8 2 | 5+6 2 | 5+6 2 |
| | 9 2 | 9 2 | 3+4 2 | 3+4 2 |
| | 8 2 | 8 2 | 3+4 2 | 5+6 2 |
| | 8 2 | 9 2 | 3+4 2 | 3+4 2 |
| | 8 2 | 8 2 | 5+6 2 | 5+6 2 |
| | 4+5 2 | 4+5 2 | 5+6 2 | 5+6 2 |
| | 8 2 | 8 2 | 5+6 2 | 5+6 2 |
| | 4+5 2 | 8 2 | 3+4 2 | 3+4 2 |
| | 8 2 | 4+5 2 | 5+6 2 | 3+4 2 |
| | 9 2 | 8 2 | 3+4 2 | 3+4 2 |
| | 8 2 | 9 2 | 3+4 2 | 5+6 2 |
| | 9 2 | 6+7 2 | 5+6 2 | 5+6 2 |
| | 8 2 | 8 2 | 5+6 2 | 3+4 2 |
| | 4+5 2 | 4+5 2 | 5+6 2 | 5+6 2 |
| | 6+7 2 | 8 2 | 5+6 2 | 5+6 2 |
| | 8 2 | 9 2 | 5+6 2 | 3+4 2 |
| | 4+5 2 | 8 2 | 3+4 2 | 5+6 2 |
| | 8 2 | 4+5 2 | 5+6 2 | 5+6 2 |
| | 4+5 2 | 8 2 | 5+6 2 | 5+6 2 |
| | 8 2 | 4+5 2 | 3+4 2 | 3+4 2 |
| | 9 2 | 8 2 | 3+4 2 | 3+4 2 |
| | 8 2 | 8 2 | 3+4 2 | 5+6 2 |
| | 9 2 | 8 2 | 5+6 2 | 3+4 2 |

Continued on next column

**TABLE LXXIII– continued**

| TQC | ICSD | NoE | SF | |
|---|---|---|---|---|
| | 6+7 2 | 4+5 2 | 5+6 2 | 5+6 2 |
| 166.1.82 | 600019 | 10 | P1 | |
| | -GM | -T | -F | -L |
| | 8 2 | 9 2 | 5+6 2 | 3+4 2 |
| | 8 2 | 8 2 | 3+4 2 | 5+6 2 |
| | 4+5 2 | 9 2 | 3+4 2 | 3+4 2 |
| | 8 2 | 6+7 2 | 5+6 2 | 5+6 2 |
| | 9 2 | 8 2 | 5+6 2 | 3+4 2 |
| 166.1.83 | 53301 | 10 | P1 | |
| | -GM | -T | -F | -L |
| | 8 2 | 9 2 | 5+6 2 | 3+4 2 |
| | 8 2 | 8 2 | 3+4 2 | 5+6 2 |
| | 4+5 2 | 9 2 | 5+6 2 | 3+4 2 |
| | 9 2 | 6+7 2 | 3+4 2 | 5+6 2 |
| | 8 2 | 8 2 | 5+6 2 | 5+6 2 |
| 166.1.84 | 416476 | 72 | O1P1Pr1Zn1 | |
| | -GM | -T | -F | -L |
| | 8 2 | 8 2 | 5+6 2 | 5+6 2 |
| | 9 2 | 9 2 | 3+4 2 | 3+4 2 |
| | 8 2 | 8 2 | 5+6 2 | 5+6 2 |
| | 9 2 | 9 2 | 3+4 2 | 3+4 2 |
| | 8 2 | 8 2 | 5+6 2 | 5+6 2 |
| | 9 2 | 9 2 | 3+4 2 | 3+4 2 |
| | 4+5 2 | 4+5 2 | 5+6 2 | 5+6 2 |
| | 9 2 | 9 2 | 3+4 2 | 3+4 2 |
| | 6+7 2 | 6+7 2 | 3+4 2 | 3+4 2 |
| | 8 2 | 8 2 | 5+6 2 | 5+6 2 |
| | 8 2 | 9 2 | 5+6 2 | 3+4 2 |
| | 9 2 | 8 2 | 3+4 2 | 5+6 2 |
| | 6+7 2 | 4+5 2 | 5+6 2 | 3+4 2 |
| | 4+5 2 | 6+7 2 | 3+4 2 | 5+6 2 |
| | 9 2 | 8 2 | 5+6 2 | 3+4 2 |
| | 8 2 | 9 2 | 3+4 2 | 5+6 2 |
| | 9 2 | 8 2 | 5+6 2 | 3+4 2 |
| | 8 2 | 9 2 | 3+4 2 | 5+6 2 |
| | 4+5 2 | 6+7 2 | 5+6 2 | 3+4 2 |
| | 8 2 | 9 2 | 3+4 2 | 5+6 2 |
| | 9 2 | 4+5 2 | 3+4 2 | 5+6 2 |
| | 6+7 2 | 8 2 | 5+6 2 | 3+4 2 |
| | 8 2 | 9 2 | 3+4 2 | 5+6 2 |
| | 8 2 | 8 2 | 5+6 2 | 3+4 2 |
| | 4+5 2 | 4+5 2 | 3+4 2 | 3+4 2 |
| | 8 2 | 8 2 | 5+6 2 | 5+6 2 |
| | 9 2 | 9 2 | 5+6 2 | 5+6 2 |
| | 6+7 2 | 6+7 2 | 5+6 2 | 5+6 2 |
| | 9 2 | 9 2 | 3+4 2 | 3+4 2 |
| | 9 2 | 8 2 | 3+4 2 | 3+4 2 |
| | 8 2 | 9 2 | 5+6 2 | 3+4 2 |
| | 4+5 2 | 6+7 2 | 5+6 2 | 3+4 2 |
| | 6+7 2 | 8 2 | 3+4 2 | 5+6 2 |
| | 9 2 | 4+5 2 | 3+4 2 | 5+6 2 |
| | 8 2 | 8 2 | 5+6 2 | 3+4 2 |
| | 8 2 | 9 2 | 3+4 2 | 5+6 2 |
| 166.1.85 | 647123 | 236 | Ni7Yb2 | |
| | -GM | -T | -F | -L |
| | 8 2 | 9 2 | 5+6 2 | 3+4 2 |
| | 9 2 | 8 2 | 3+4 2 | 5+6 2 |
| | 8 2 | 8 2 | 3+4 2 | 3+4 2 |
| | 9 2 | 9 2 | 5+6 2 | 5+6 2 |
| | 8 2 | 9 2 | 5+6 2 | 3+4 2 |
| | 8 2 | 8 2 | 3+4 2 | 5+6 2 |

Continued on next column

**TABLE LXXIII– continued**

| TQC | ICSD | NoE | SF | |
|---|---|---|---|---|
| | 9 2 | 9 2 | 3+4 2 | 3+4 2 |
| | 9 2 | 8 2 | 5+6 2 | 5+6 2 |
| | 4+5 2 | 6+7 2 | 5+6 2 | 3+4 2 |
| | 8 2 | 9 2 | 3+4 2 | 5+6 2 |
| | 9 2 | 8 2 | 3+4 2 | 3+4 2 |
| | 8 2 | 6+7 2 | 5+6 2 | 5+6 2 |
| | 6+7 2 | 4+5 2 | 5+6 2 | 3+4 2 |
| | 4+5 2 | 9 2 | 3+4 2 | 5+6 2 |
| | 6+7 2 | 8 2 | 5+6 2 | 3+4 2 |
| | 9 2 | 4+5 2 | 3+4 2 | 5+6 2 |
| | 8 2 | 8 2 | 5+6 2 | 5+6 2 |
| | 8 2 | 9 2 | 3+4 2 | 3+4 2 |
| | 9 2 | 8 2 | 3+4 2 | 5+6 2 |
| | 9 2 | 9 2 | 5+6 2 | 3+4 2 |
| | 8 2 | 8 2 | 5+6 2 | 5+6 2 |
| | 8 2 | 9 2 | 3+4 2 | 3+4 2 |
| | 8 2 | 8 2 | 5+6 2 | 5+6 2 |
| | 8 2 | 9 2 | 3+4 2 | 5+6 2 |
| | 4+5 2 | 9 2 | 5+6 2 | 5+6 2 |
| | 9 2 | 6+7 2 | 3+4 2 | 5+6 2 |
| | 8 2 | 8 2 | 5+6 2 | 3+4 2 |
| | 9 2 | 4+5 2 | 5+6 2 | 5+6 2 |
| | 6+7 2 | 8 2 | 3+4 2 | 3+4 2 |
| | 4+5 2 | 4+5 2 | 3+4 2 | 5+6 2 |
| | 8 2 | 8 2 | 3+4 2 | 3+4 2 |
| | 9 2 | 8 2 | 3+4 2 | 3+4 2 |
| | 8 2 | 8 2 | 5+6 2 | 5+6 2 |
| | 4+5 2 | 4+5 2 | 3+4 2 | 3+4 2 |
| | 8 2 | 8 2 | 5+6 2 | 5+6 2 |
| | 4+5 2 | 9 2 | 3+4 2 | 3+4 2 |
| | 8 2 | 9 2 | 5+6 2 | 5+6 2 |
| | 8 2 | 6+7 2 | 5+6 2 | 5+6 2 |
| | 6+7 2 | 4+5 2 | 5+6 2 | 5+6 2 |
| | 9 2 | 6+7 2 | 3+4 2 | 5+6 2 |
| | 8 2 | 8 2 | 3+4 2 | 3+4 2 |
| | 4+5 2 | 9 2 | 5+6 2 | 3+4 2 |
| | 4+5 2 | 8 2 | 3+4 2 | 5+6 2 |
| | 8 2 | 4+5 2 | 5+6 2 | 5+6 2 |
| | 6+7 2 | 9 2 | 3+4 2 | 3+4 2 |
| | 9 2 | 8 2 | 5+6 2 | 5+6 2 |
| | 9 2 | 4+5 2 | 3+4 2 | 3+4 2 |
| | 8 2 | 8 2 | 3+4 2 | 3+4 2 |
| | 4+5 2 | 9 2 | 3+4 2 | 5+6 2 |
| | 9 2 | 6+7 2 | 5+6 2 | 5+6 2 |
| | 6+7 2 | 6+7 2 | 3+4 2 | 3+4 2 |
| | 9 2 | 9 2 | 5+6 2 | 3+4 2 |
| | 8 2 | 9 2 | 5+6 2 | 5+6 2 |
| | 8 2 | 8 2 | 5+6 2 | 5+6 2 |
| | 4+5 2 | 6+7 2 | 3+4 2 | 3+4 2 |
| | 9 2 | 4+5 2 | 5+6 2 | 5+6 2 |
| | 6+7 2 | 6+7 2 | 3+4 2 | 3+4 2 |
| | 8 2 | 8 2 | 5+6 2 | 5+6 2 |
| | 9 2 | 9 2 | 3+4 2 | 3+4 2 |
| | 4+5 2 | 9 2 | 3+4 2 | 5+6 2 |
| | 6+7 2 | 9 2 | 5+6 2 | 3+4 2 |
| | 8 2 | 6+7 2 | 5+6 2 | 5+6 2 |
| | 8 2 | 9 2 | 5+6 2 | 3+4 2 |
| | 9 2 | 8 2 | 3+4 2 | 3+4 2 |
| | 9 2 | 4+5 2 | 5+6 2 | 5+6 2 |
| | 8 2 | 8 2 | 5+6 2 | 5+6 2 |
| | 9 2 | 9 2 | 3+4 2 | 3+4 2 |

Continued on next column

**TABLE LXXIII– continued**

| TQC | ICSD | NoE | SF | |
|---|---|---|---|---|
| | 4+5 2 | 9 2 | 5+6 2 | 3+4 2 |
| | 6+7 2 | 4+5 2 | 3+4 2 | 3+4 2 |
| | 9 2 | 8 2 | 3+4 2 | 5+6 2 |
| | 8 2 | 6+7 2 | 5+6 2 | 5+6 2 |
| | 6+7 2 | 8 2 | 3+4 2 | 3+4 2 |
| | 8 2 | 4+5 2 | 5+6 2 | 3+4 2 |
| | 9 2 | 8 2 | 3+4 2 | 5+6 2 |
| | 4+5 2 | 9 2 | 5+6 2 | 5+6 2 |
| | 8 2 | 8 2 | 3+4 2 | 3+4 2 |
| | 6+7 2 | 8 2 | 3+4 2 | 5+6 2 |
| | 9 2 | 9 2 | 5+6 2 | 3+4 2 |
| | 4+5 2 | 4+5 2 | 3+4 2 | 5+6 2 |
| | 9 2 | 6+7 2 | 3+4 2 | 3+4 2 |
| | 8 2 | 8 2 | 5+6 2 | 5+6 2 |
| | 8 2 | 9 2 | 5+6 2 | 3+4 2 |
| | 8 2 | 8 2 | 3+4 2 | 3+4 2 |
| | 4+5 2 | 4+5 2 | 5+6 2 | 5+6 2 |
| | 8 2 | 4+5 2 | 5+6 2 | 5+6 2 |
| | 6+7 2 | 6+7 2 | 3+4 2 | 5+6 2 |
| | 9 2 | 8 2 | 5+6 2 | 3+4 2 |
| | 4+5 2 | 9 2 | 3+4 2 | 5+6 2 |
| | 8 2 | 8 2 | 5+6 2 | 3+4 2 |
| | 9 2 | 6+7 2 | 3+4 2 | 5+6 2 |
| | 6+7 2 | 9 2 | 3+4 2 | 3+4 2 |
| | 4+5 2 | 8 2 | 5+6 2 | 5+6 2 |
| | 8 2 | 4+5 2 | 3+4 2 | 5+6 2 |
| | 8 2 | 8 2 | 5+6 2 | 5+6 2 |
| | 4+5 2 | 4+5 2 | 3+4 2 | 3+4 2 |
| | 8 2 | 8 2 | 3+4 2 | 3+4 2 |
| | 8 2 | 9 2 | 5+6 2 | 5+6 2 |
| | 8 2 | 6+7 2 | 5+6 2 | 3+4 2 |
| | 4+5 2 | 8 2 | 5+6 2 | 5+6 2 |
| | 9 2 | 9 2 | 3+4 2 | 5+6 2 |
| | 9 2 | 4+5 2 | 3+4 2 | 3+4 2 |
| | 6+7 2 | 9 2 | 5+6 2 | 5+6 2 |
| | 8 2 | 8 2 | 5+6 2 | 3+4 2 |
| | 9 2 | 6+7 2 | 3+4 2 | 3+4 2 |
| | 8 2 | 9 2 | 5+6 2 | 5+6 2 |
| | 9 2 | 4+5 2 | 5+6 2 | 3+4 2 |
| | 4+5 2 | 8 2 | 3+4 2 | 5+6 2 |
| | 9 2 | 9 2 | 3+4 2 | 3+4 2 |
| | 6+7 2 | 8 2 | 3+4 2 | 5+6 2 |
| | 8 2 | 9 2 | 5+6 2 | 5+6 2 |
| | 9 2 | 8 2 | 3+4 2 | 5+6 2 |
| | 8 2 | 6+7 2 | 5+6 2 | 3+4 2 |
| | 8 2 | 9 2 | 3+4 2 | 3+4 2 |
| | 4+5 2 | 8 2 | 3+4 2 | 5+6 2 |
| | 8 2 | 8 2 | 5+6 2 | 3+4 2 |
| | 9 2 | 4+5 2 | 5+6 2 | 5+6 2 |
| | 8 2 | 8 2 | 3+4 2 | 3+4 2 |
| | 4+5 2 | 9 2 | 3+4 2 | 5+6 2 |
| 166.1.86 | 646928 | 188 | Ni7Th2 | |
| | -GM | -T | -F | -L |
| | 8 2 | 9 2 | 5+6 2 | 3+4 2 |
| | 9 2 | 8 2 | 3+4 2 | 5+6 2 |
| | 8 2 | 8 2 | 3+4 2 | 3+4 2 |
| | 9 2 | 9 2 | 5+6 2 | 5+6 2 |
| | 8 2 | 9 2 | 5+6 2 | 3+4 2 |
| | 9 2 | 8 2 | 3+4 2 | 5+6 2 |
| | 8 2 | 8 2 | 3+4 2 | 3+4 2 |
| | 9 2 | 9 2 | 5+6 2 | 5+6 2 |

Continued on next column

**TABLE LXXIII– continued**

| TQC | ICSD | NoE | SF | |
|---|---|---|---|---|
| | 4+5 2 | 6+7 2 | 5+6 2 | 3+4 2 |
| | 8 2 | 9 2 | 3+4 2 | 5+6 2 |
| | 9 2 | 8 2 | 3+4 2 | 3+4 2 |
| | 8 2 | 6+7 2 | 5+6 2 | 5+6 2 |
| | 6+7 2 | 4+5 2 | 5+6 2 | 3+4 2 |
| | 4+5 2 | 8 2 | 3+4 2 | 5+6 2 |
| | 9 2 | 9 2 | 5+6 2 | 3+4 2 |
| | 6+7 2 | 4+5 2 | 3+4 2 | 5+6 2 |
| | 8 2 | 8 2 | 3+4 2 | 3+4 2 |
| | 9 2 | 9 2 | 5+6 2 | 5+6 2 |
| | 8 2 | 8 2 | 3+4 2 | 3+4 2 |
| | 9 2 | 9 2 | 5+6 2 | 5+6 2 |
| | 8 2 | 8 2 | 5+6 2 | 5+6 2 |
| | 8 2 | 9 2 | 3+4 2 | 5+6 2 |
| | 8 2 | 8 2 | 5+6 2 | 3+4 2 |
| | 8 2 | 9 2 | 5+6 2 | 5+6 2 |
| | 4+5 2 | 6+7 2 | 3+4 2 | 3+4 2 |
| | 9 2 | 8 2 | 3+4 2 | 3+4 2 |
| | 6+7 2 | 4+5 2 | 5+6 2 | 5+6 2 |
| | 9 2 | 9 2 | 3+4 2 | 3+4 2 |
| | 4+5 2 | 8 2 | 3+4 2 | 5+6 2 |
| | 8 2 | 4+5 2 | 5+6 2 | 3+4 2 |
| | 8 2 | 8 2 | 3+4 2 | 5+6 2 |
| | 8 2 | 8 2 | 5+6 2 | 5+6 2 |
| | 4+5 2 | 4+5 2 | 5+6 2 | 3+4 2 |
| | 9 2 | 9 2 | 3+4 2 | 5+6 2 |
| | 9 2 | 6+7 2 | 5+6 2 | 5+6 2 |
| | 6+7 2 | 9 2 | 3+4 2 | 3+4 2 |
| | 8 2 | 8 2 | 5+6 2 | 3+4 2 |
| | 4+5 2 | 6+7 2 | 3+4 2 | 5+6 2 |
| | 6+7 2 | 4+5 2 | 5+6 2 | 3+4 2 |
| | 8 2 | 8 2 | 3+4 2 | 5+6 2 |
| | 9 2 | 9 2 | 3+4 2 | 5+6 2 |
| | 8 2 | 8 2 | 3+4 2 | 5+6 2 |
| | 8 2 | 9 2 | 5+6 2 | 3+4 2 |
| | 4+5 2 | 8 2 | 3+4 2 | 5+6 2 |
| | 9 2 | 6+7 2 | 5+6 2 | 3+4 2 |
| | 8 2 | 8 2 | 5+6 2 | 5+6 2 |
| | 4+5 2 | 4+5 2 | 3+4 2 | 3+4 2 |
| | 9 2 | 8 2 | 3+4 2 | 5+6 2 |
| | 8 2 | 4+5 2 | 5+6 2 | 5+6 2 |
| | 6+7 2 | 9 2 | 5+6 2 | 3+4 2 |
| | 4+5 2 | 6+7 2 | 3+4 2 | 5+6 2 |
| | 9 2 | 9 2 | 5+6 2 | 3+4 2 |
| | 6+7 2 | 9 2 | 3+4 2 | 5+6 2 |
| | 8 2 | 6+7 2 | 3+4 2 | 5+6 2 |
| | 4+5 2 | 8 2 | 3+4 2 | 3+4 2 |
| | 8 2 | 9 2 | 5+6 2 | 3+4 2 |
| | 9 2 | 4+5 2 | 5+6 2 | 5+6 2 |
| | 9 2 | 6+7 2 | 5+6 2 | 5+6 2 |
| | 6+7 2 | 9 2 | 3+4 2 | 3+4 2 |
| | 8 2 | 8 2 | 3+4 2 | 3+4 2 |
| | 4+5 2 | 4+5 2 | 5+6 2 | 5+6 2 |
| | 8 2 | 8 2 | 5+6 2 | 5+6 2 |
| | 9 2 | 9 2 | 5+6 2 | 5+6 2 |
| | 8 2 | 6+7 2 | 5+6 2 | 5+6 2 |
| | 4+5 2 | 8 2 | 3+4 2 | 3+4 2 |
| | 8 2 | 9 2 | 3+4 2 | 5+6 2 |
| | 6+7 2 | 4+5 2 | 3+4 2 | 3+4 2 |
| | 9 2 | 8 2 | 3+4 2 | 3+4 2 |
| | 8 2 | 8 2 | 3+4 2 | 5+6 2 |

Continued on next column

**TABLE LXXIII– continued**

| TQC | ICSD | NoE | SF | |
|---|---|---|---|---|
| | 9 2 | 9 2 | 5+6 2 | 3+4 2 |
| | 4+5 2 | 6+7 2 | 3+4 2 | 3+4 2 |
| | 4+5 2 | 4+5 2 | 3+4 2 | 5+6 2 |
| | 8 2 | 9 2 | 5+6 2 | 3+4 2 |
| | 6+7 2 | 4+5 2 | 5+6 2 | 5+6 2 |
| | 8 2 | 8 2 | 3+4 2 | 5+6 2 |
| | 9 2 | 8 2 | 5+6 2 | 3+4 2 |
| | 8 2 | 6+7 2 | 5+6 2 | 5+6 2 |
| | 8 2 | 8 2 | 5+6 2 | 5+6 2 |
| | 4+5 2 | 9 2 | 3+4 2 | 3+4 2 |
| | 6+7 2 | 4+5 2 | 5+6 2 | 5+6 2 |
| | 9 2 | 8 2 | 5+6 2 | 5+6 2 |
| | 8 2 | 9 2 | 3+4 2 | 3+4 2 |
| | 8 2 | 8 2 | 3+4 2 | 3+4 2 |
| | 9 2 | 8 2 | 3+4 2 | 5+6 2 |
| | 4+5 2 | 6+7 2 | 3+4 2 | 3+4 2 |
| | 6+7 2 | 4+5 2 | 5+6 2 | 5+6 2 |
| | 8 2 | 9 2 | 5+6 2 | 3+4 2 |
| | 4+5 2 | 8 2 | 5+6 2 | 5+6 2 |
| | 8 2 | 4+5 2 | 3+4 2 | 5+6 2 |
| | 4+5 2 | 8 2 | 3+4 2 | 3+4 2 |
| | 8 2 | 9 2 | 5+6 2 | 5+6 2 |
| | 9 2 | 8 2 | 3+4 2 | 5+6 2 |
| | 8 2 | 4+5 2 | 5+6 2 | 3+4 2 |
| | 9 2 | 9 2 | 5+6 2 | 5+6 2 |
| 166.1.87 | 646978 | 110 | Ni3Ti0.67Zr0.33 | |
| | -GM | -T | -F | -L |
| | 8 2 | 8 2 | 3+4 2 | 3+4 2 |
| | 9 2 | 9 2 | 5+6 2 | 5+6 2 |
| | 6+7 2 | 6+7 2 | 5+6 2 | 5+6 2 |
| | 9 2 | 9 2 | 5+6 2 | 5+6 2 |
| | 8 2 | 8 2 | 3+4 2 | 5+6 2 |
| | 8 2 | 9 2 | 5+6 2 | 3+4 2 |
| | 9 2 | 8 2 | 3+4 2 | 3+4 2 |
| | 8 2 | 4+5 2 | 5+6 2 | 5+6 2 |
| | 4+5 2 | 8 2 | 3+4 2 | 3+4 2 |
| | 8 2 | 8 2 | 3+4 2 | 3+4 2 |
| | 4+5 2 | 4+5 2 | 5+6 2 | 5+6 2 |
| | 9 2 | 9 2 | 5+6 2 | 5+6 2 |
| | 6+7 2 | 8 2 | 5+6 2 | 3+4 2 |
| | 8 2 | 6+7 2 | 3+4 2 | 5+6 2 |
| | 8 2 | 9 2 | 5+6 2 | 3+4 2 |
| | 4+5 2 | 8 2 | 3+4 2 | 5+6 2 |
| | 8 2 | 8 2 | 3+4 2 | 3+4 2 |
| | 9 2 | 4+5 2 | 5+6 2 | 3+4 2 |
| | 8 2 | 9 2 | 3+4 2 | 5+6 2 |
| | 8 2 | 6+7 2 | 5+6 2 | 3+4 2 |
| | 4+5 2 | 8 2 | 3+4 2 | 5+6 2 |
| | 8 2 | 9 2 | 5+6 2 | 5+6 2 |
| | 9 2 | 4+5 2 | 5+6 2 | 5+6 2 |
| | 6+7 2 | 8 2 | 3+4 2 | 5+6 2 |
| | 9 2 | 9 2 | 5+6 2 | 3+4 2 |
| | 4+5 2 | 8 2 | 3+4 2 | 5+6 2 |
| | 8 2 | 8 2 | 5+6 2 | 3+4 2 |
| | 8 2 | 4+5 2 | 3+4 2 | 3+4 2 |
| | 8 2 | 4+5 2 | 5+6 2 | 5+6 2 |
| | 6+7 2 | 8 2 | 3+4 2 | 3+4 2 |
| | 9 2 | 8 2 | 3+4 2 | 3+4 2 |
| | 4+5 2 | 8 2 | 5+6 2 | 5+6 2 |
| | 8 2 | 6+7 2 | 5+6 2 | 5+6 2 |
| | 8 2 | 9 2 | 5+6 2 | 3+4 2 |

Continued on next column

**TABLE LXXIII– continued**

| TQC | ICSD | NoE | SF | |
|---|---|---|---|---|
| | 9 2 | 8 2 | 3+4 2 | 3+4 2 |
| | 6+7 2 | 4+5 2 | 3+4 2 | 5+6 2 |
| | 9 2 | 6+7 2 | 5+6 2 | 5+6 2 |
| | 4+5 2 | 9 2 | 3+4 2 | 3+4 2 |
| | 9 2 | 8 2 | 5+6 2 | 5+6 2 |
| | 8 2 | 4+5 2 | 3+4 2 | 3+4 2 |
| | 4+5 2 | 9 2 | 5+6 2 | 5+6 2 |
| | 8 2 | 9 2 | 5+6 2 | 3+4 2 |
| | 9 2 | 6+7 2 | 5+6 2 | 5+6 2 |
| | 6+7 2 | 8 2 | 3+4 2 | 3+4 2 |
| | 8 2 | 9 2 | 5+6 2 | 5+6 2 |
| | 4+5 2 | 8 2 | 5+6 2 | 3+4 2 |
| | 8 2 | 4+5 2 | 3+4 2 | 3+4 2 |
| | 8 2 | 8 2 | 3+4 2 | 5+6 2 |
| | 8 2 | 8 2 | 5+6 2 | 5+6 2 |
| | 9 2 | 8 2 | 3+4 2 | 5+6 2 |
| | 4+5 2 | 9 2 | 5+6 2 | 5+6 2 |
| | 8 2 | 4+5 2 | 5+6 2 | 3+4 2 |
| | 8 2 | 8 2 | 3+4 2 | 5+6 2 |
| | 9 2 | 9 2 | 3+4 2 | 5+6 2 |
| | 9 2 | 9 2 | 5+6 2 | 5+6 2 |
| 166.1.88 | 646399 | 48 | Ni3S2Tl2 | |
| | -GM | -T | -F | -L |
| | 8 2 | 8 2 | 5+6 2 | 5+6 2 |
| | 9 2 | 9 2 | 3+4 2 | 3+4 2 |
| | 8 2 | 9 2 | 3+4 2 | 3+4 2 |
| | 8 2 | 8 2 | 3+4 2 | 5+6 2 |
| | 8 2 | 9 2 | 5+6 2 | 3+4 2 |
| | 8 2 | 6+7 2 | 5+6 2 | 5+6 2 |
| | 4+5 2 | 8 2 | 5+6 2 | 5+6 2 |
| | 9 2 | 4+5 2 | 5+6 2 | 3+4 2 |
| | 9 2 | 9 2 | 3+4 2 | 3+4 2 |
| | 6+7 2 | 9 2 | 3+4 2 | 5+6 2 |
| | 4+5 2 | 8 2 | 3+4 2 | 3+4 2 |
| | 8 2 | 9 2 | 5+6 2 | 5+6 2 |
| | 8 2 | 6+7 2 | 5+6 2 | 3+4 2 |
| | 9 2 | 9 2 | 5+6 2 | 3+4 2 |
| | 8 2 | 6+7 2 | 5+6 2 | 3+4 2 |
| | 4+5 2 | 9 2 | 5+6 2 | 5+6 2 |
| | 8 2 | 9 2 | 3+4 2 | 3+4 2 |
| | 4+5 2 | 6+7 2 | 5+6 2 | 3+4 2 |
| | 8 2 | 9 2 | 3+4 2 | 5+6 2 |
| | 9 2 | 6+7 2 | 5+6 2 | 5+6 2 |
| | 4+5 2 | 9 2 | 3+4 2 | 3+4 2 |
| | 8 2 | 9 2 | 5+6 2 | 3+4 2 |
| | 6+7 2 | 9 2 | 3+4 2 | 5+6 2 |
| | 8 2 | 6+7 2 | 5+6 2 | 3+4 2 |
| 166.1.89 | 646398 | 48 | Ni3S2Tl2 | |
| | -GM | -T | -F | -L |
| | 8 2 | 9 2 | 5+6 2 | 3+4 2 |
| | 9 2 | 8 2 | 3+4 2 | 5+6 2 |
| | 8 2 | 8 2 | 3+4 2 | 5+6 2 |
| | 8 2 | 9 2 | 3+4 2 | 3+4 2 |
| | 8 2 | 8 2 | 5+6 2 | 5+6 2 |
| | 8 2 | 4+5 2 | 5+6 2 | 3+4 2 |
| | 4+5 2 | 9 2 | 5+6 2 | 3+4 2 |
| | 9 2 | 6+7 2 | 5+6 2 | 5+6 2 |
| | 9 2 | 8 2 | 3+4 2 | 5+6 2 |
| | 6+7 2 | 8 2 | 3+4 2 | 3+4 2 |
| | 4+5 2 | 9 2 | 3+4 2 | 5+6 2 |
| | 8 2 | 4+5 2 | 5+6 2 | 3+4 2 |

Continued on next column

**TABLE LXXIII– continued**

| TQC | ICSD | NoE | SF | |
|---|---|---|---|---|
| | 8 2 | 8 2 | 5+6 2 | 5+6 2 |
| | 9 2 | 8 2 | 5+6 2 | 5+6 2 |
| | 8 2 | 4+5 2 | 5+6 2 | 5+6 2 |
| | 4+5 2 | 8 2 | 5+6 2 | 3+4 2 |
| | 8 2 | 8 2 | 3+4 2 | 5+6 2 |
| | 4+5 2 | 4+5 2 | 5+6 2 | 5+6 2 |
| | 8 2 | 8 2 | 3+4 2 | 3+4 2 |
| | 9 2 | 4+5 2 | 5+6 2 | 3+4 2 |
| | 4+5 2 | 8 2 | 3+4 2 | 5+6 2 |
| | 8 2 | 8 2 | 5+6 2 | 5+6 2 |
| | 6+7 2 | 8 2 | 3+4 2 | 3+4 2 |
| | 8 2 | 4+5 2 | 5+6 2 | 5+6 2 |
| 166.1.90 | 422328 | 48 | Ni3S2Tl2 | |
| | -GM | -T | -F | -L |
| | 8 2 | 8 2 | 5+6 2 | 5+6 2 |
| | 9 2 | 9 2 | 3+4 2 | 3+4 2 |
| | 8 2 | 9 2 | 3+4 2 | 3+4 2 |
| | 8 2 | 8 2 | 3+4 2 | 5+6 2 |
| | 8 2 | 9 2 | 5+6 2 | 3+4 2 |
| | 8 2 | 6+7 2 | 5+6 2 | 5+6 2 |
| | 4+5 2 | 8 2 | 5+6 2 | 5+6 2 |
| | 9 2 | 9 2 | 5+6 2 | 3+4 2 |
| | 9 2 | 4+5 2 | 3+4 2 | 3+4 2 |
| | 6+7 2 | 9 2 | 3+4 2 | 5+6 2 |
| | 4+5 2 | 8 2 | 3+4 2 | 3+4 2 |
| | 8 2 | 9 2 | 5+6 2 | 5+6 2 |
| | 8 2 | 6+7 2 | 5+6 2 | 3+4 2 |
| | 9 2 | 9 2 | 5+6 2 | 3+4 2 |
| | 8 2 | 6+7 2 | 5+6 2 | 3+4 2 |
| | 4+5 2 | 9 2 | 5+6 2 | 5+6 2 |
| | 8 2 | 9 2 | 3+4 2 | 3+4 2 |
| | 4+5 2 | 6+7 2 | 5+6 2 | 3+4 2 |
| | 8 2 | 9 2 | 3+4 2 | 5+6 2 |
| | 9 2 | 6+7 2 | 5+6 2 | 5+6 2 |
| | 4+5 2 | 9 2 | 3+4 2 | 3+4 2 |
| | 8 2 | 9 2 | 5+6 2 | 3+4 2 |
| | 6+7 2 | 9 2 | 3+4 2 | 5+6 2 |
| | 8 2 | 6+7 2 | 5+6 2 | 3+4 2 |
| 166.1.91 | 108893 | 48 | Ni3S2Tl2 | |
| | -GM | -T | -F | -L |
| | 8 2 | 9 2 | 5+6 2 | 3+4 2 |
| | 9 2 | 8 2 | 3+4 2 | 5+6 2 |
| | 8 2 | 8 2 | 3+4 2 | 5+6 2 |
| | 8 2 | 9 2 | 3+4 2 | 3+4 2 |
| | 8 2 | 8 2 | 5+6 2 | 5+6 2 |
| | 8 2 | 4+5 2 | 5+6 2 | 3+4 2 |
| | 4+5 2 | 9 2 | 5+6 2 | 3+4 2 |
| | 9 2 | 6+7 2 | 5+6 2 | 5+6 2 |
| | 9 2 | 8 2 | 3+4 2 | 5+6 2 |
| | 6+7 2 | 8 2 | 3+4 2 | 3+4 2 |
| | 4+5 2 | 9 2 | 3+4 2 | 5+6 2 |
| | 8 2 | 8 2 | 5+6 2 | 3+4 2 |
| | 8 2 | 4+5 2 | 5+6 2 | 5+6 2 |
| | 9 2 | 8 2 | 5+6 2 | 5+6 2 |
| | 4+5 2 | 4+5 2 | 5+6 2 | 5+6 2 |
| | 8 2 | 8 2 | 5+6 2 | 3+4 2 |
| | 8 2 | 8 2 | 3+4 2 | 5+6 2 |
| | 4+5 2 | 4+5 2 | 5+6 2 | 3+4 2 |
| | 8 2 | 8 2 | 3+4 2 | 5+6 2 |
| | 9 2 | 4+5 2 | 5+6 2 | 3+4 2 |
| | 4+5 2 | 8 2 | 3+4 2 | 5+6 2 |

Continued on next column

**TABLE LXXIII– continued**

| TQC | ICSD | NoE | SF | |
|---|---|---|---|---|
| | 8 2 | 8 2 | 5+6 2 | 5+6 2 |
| | 6+7 2 | 8 2 | 3+4 2 | 3+4 2 |
| | 8 2 | 4+5 2 | 5+6 2 | 5+6 2 |
| 166.1.92 | 654948 | 50 | Ni3S2Sn2 | |
| | -GM | -T | -F | -L |
| | 8 2 | 9 2 | 5+6 2 | 3+4 2 |
| | 9 2 | 8 2 | 3+4 2 | 5+6 2 |
| | 8 2 | 8 2 | 3+4 2 | 5+6 2 |
| | 8 2 | 9 2 | 3+4 2 | 3+4 2 |
| | 8 2 | 8 2 | 5+6 2 | 5+6 2 |
| | 8 2 | 4+5 2 | 5+6 2 | 3+4 2 |
| | 4+5 2 | 9 2 | 5+6 2 | 3+4 2 |
| | 9 2 | 6+7 2 | 3+4 2 | 5+6 2 |
| | 9 2 | 8 2 | 5+6 2 | 5+6 2 |
| | 6+7 2 | 8 2 | 3+4 2 | 3+4 2 |
| | 9 2 | 9 2 | 3+4 2 | 5+6 2 |
| | 8 2 | 4+5 2 | 5+6 2 | 3+4 2 |
| | 4+5 2 | 8 2 | 5+6 2 | 5+6 2 |
| | 9 2 | 8 2 | 5+6 2 | 5+6 2 |
| | 6+7 2 | 4+5 2 | 5+6 2 | 3+4 2 |
| | 8 2 | 8 2 | 5+6 2 | 5+6 2 |
| | 4+5 2 | 8 2 | 3+4 2 | 5+6 2 |
| | 8 2 | 4+5 2 | 5+6 2 | 5+6 2 |
| | 8 2 | 8 2 | 3+4 2 | 3+4 2 |
| | 4+5 2 | 4+5 2 | 5+6 2 | 3+4 2 |
| | 8 2 | 8 2 | 3+4 2 | 5+6 2 |
| | 4+5 2 | 8 2 | 5+6 2 | 5+6 2 |
| | 8 2 | 9 2 | 3+4 2 | 3+4 2 |
| | 8 2 | 6+7 2 | 5+6 2 | 5+6 2 |
| | 8 2 | 8 2 | 3+4 2 | 3+4 2 |
| 166.1.93 | 646380 | 50 | Ni3S2Sn2 | |
| | -GM | -T | -F | -L |
| | 8 2 | 8 2 | 5+6 2 | 5+6 2 |
| | 9 2 | 9 2 | 3+4 2 | 3+4 2 |
| | 8 2 | 9 2 | 3+4 2 | 3+4 2 |
| | 8 2 | 8 2 | 3+4 2 | 5+6 2 |
| | 8 2 | 9 2 | 5+6 2 | 3+4 2 |
| | 8 2 | 6+7 2 | 5+6 2 | 5+6 2 |
| | 4+5 2 | 8 2 | 5+6 2 | 5+6 2 |
| | 9 2 | 4+5 2 | 5+6 2 | 3+4 2 |
| | 9 2 | 9 2 | 3+4 2 | 3+4 2 |
| | 6+7 2 | 9 2 | 3+4 2 | 5+6 2 |
| | 9 2 | 8 2 | 3+4 2 | 3+4 2 |
| | 8 2 | 6+7 2 | 5+6 2 | 5+6 2 |
| | 4+5 2 | 9 2 | 5+6 2 | 3+4 2 |
| | 9 2 | 9 2 | 5+6 2 | 3+4 2 |
| | 6+7 2 | 6+7 2 | 5+6 2 | 5+6 2 |
| | 8 2 | 9 2 | 5+6 2 | 3+4 2 |
| | 4+5 2 | 9 2 | 3+4 2 | 3+4 2 |
| | 8 2 | 6+7 2 | 5+6 2 | 3+4 2 |
| | 8 2 | 9 2 | 3+4 2 | 5+6 2 |
| | 4+5 2 | 6+7 2 | 5+6 2 | 5+6 2 |
| | 8 2 | 9 2 | 3+4 2 | 3+4 2 |
| | 4+5 2 | 9 2 | 3+4 2 | 3+4 2 |
| | 8 2 | 8 2 | 5+6 2 | 5+6 2 |
| | 8 2 | 4+5 2 | 5+6 2 | 3+4 2 |
| | 8 2 | 9 2 | 3+4 2 | 5+6 2 |
| 166.1.94 | 646379 | 50 | Ni3S2Sn2 | |
| | -GM | -T | -F | -L |
| | 8 2 | 9 2 | 5+6 2 | 3+4 2 |
| | 9 2 | 8 2 | 3+4 2 | 5+6 2 |

Continued on next column

**TABLE LXXIII– continued**

| TQC | ICSD | NoE | SF | |
|---|---|---|---|---|
| | 8 2 | 8 2 | 3+4 2 | 5+6 2 |
| | 8 2 | 9 2 | 3+4 2 | 3+4 2 |
| | 8 2 | 8 2 | 5+6 2 | 5+6 2 |
| | 8 2 | 4+5 2 | 5+6 2 | 3+4 2 |
| | 4+5 2 | 9 2 | 5+6 2 | 3+4 2 |
| | 9 2 | 6+7 2 | 5+6 2 | 5+6 2 |
| | 9 2 | 8 2 | 3+4 2 | 5+6 2 |
| | 6+7 2 | 8 2 | 3+4 2 | 3+4 2 |
| | 9 2 | 9 2 | 3+4 2 | 5+6 2 |
| | 8 2 | 4+5 2 | 5+6 2 | 3+4 2 |
| | 4+5 2 | 8 2 | 5+6 2 | 5+6 2 |
| | 9 2 | 8 2 | 5+6 2 | 5+6 2 |
| | 6+7 2 | 4+5 2 | 5+6 2 | 3+4 2 |
| | 8 2 | 8 2 | 5+6 2 | 5+6 2 |
| | 4+5 2 | 8 2 | 3+4 2 | 5+6 2 |
| | 8 2 | 4+5 2 | 5+6 2 | 5+6 2 |
| | 8 2 | 8 2 | 3+4 2 | 3+4 2 |
| | 4+5 2 | 4+5 2 | 5+6 2 | 3+4 2 |
| | 8 2 | 8 2 | 3+4 2 | 5+6 2 |
| | 4+5 2 | 8 2 | 3+4 2 | 5+6 2 |
| | 8 2 | 9 2 | 5+6 2 | 3+4 2 |
| | 8 2 | 6+7 2 | 5+6 2 | 5+6 2 |
| | 8 2 | 8 2 | 3+4 2 | 3+4 2 |
| 166.1.95 | 402458 | 50 | Ni3S2Sn2 | |
| | -GM | -T | -F | -L |
| | 8 2 | 8 2 | 5+6 2 | 5+6 2 |
| | 9 2 | 9 2 | 3+4 2 | 3+4 2 |
| | 8 2 | 9 2 | 3+4 2 | 3+4 2 |
| | 8 2 | 8 2 | 3+4 2 | 5+6 2 |
| | 8 2 | 9 2 | 5+6 2 | 3+4 2 |
| | 8 2 | 6+7 2 | 5+6 2 | 5+6 2 |
| | 4+5 2 | 8 2 | 5+6 2 | 5+6 2 |
| | 9 2 | 4+5 2 | 3+4 2 | 3+4 2 |
| | 9 2 | 9 2 | 5+6 2 | 3+4 2 |
| | 6+7 2 | 9 2 | 3+4 2 | 5+6 2 |
| | 9 2 | 8 2 | 3+4 2 | 3+4 2 |
| | 8 2 | 6+7 2 | 5+6 2 | 5+6 2 |
| | 4+5 2 | 9 2 | 5+6 2 | 3+4 2 |
| | 9 2 | 9 2 | 5+6 2 | 3+4 2 |
| | 6+7 2 | 6+7 2 | 5+6 2 | 5+6 2 |
| | 8 2 | 9 2 | 5+6 2 | 3+4 2 |
| | 4+5 2 | 9 2 | 3+4 2 | 3+4 2 |
| | 8 2 | 6+7 2 | 5+6 2 | 3+4 2 |
| | 8 2 | 9 2 | 3+4 2 | 5+6 2 |
| | 4+5 2 | 6+7 2 | 5+6 2 | 5+6 2 |
| | 8 2 | 9 2 | 3+4 2 | 3+4 2 |
| | 4+5 2 | 9 2 | 3+4 2 | 3+4 2 |
| | 8 2 | 8 2 | 5+6 2 | 5+6 2 |
| | 8 2 | 4+5 2 | 5+6 2 | 3+4 2 |
| | 8 2 | 9 2 | 3+4 2 | 5+6 2 |
| 166.1.96 | 100217 | 50 | Ni3S2Sn2 | |
| | -GM | -T | -F | -L |
| | 8 2 | 9 2 | 5+6 2 | 3+4 2 |
| | 9 2 | 8 2 | 3+4 2 | 5+6 2 |
| | 8 2 | 8 2 | 3+4 2 | 5+6 2 |
| | 8 2 | 9 2 | 3+4 2 | 3+4 2 |
| | 8 2 | 8 2 | 5+6 2 | 5+6 2 |
| | 8 2 | 4+5 2 | 5+6 2 | 3+4 2 |
| | 4+5 2 | 9 2 | 5+6 2 | 3+4 2 |
| | 9 2 | 6+7 2 | 3+4 2 | 5+6 2 |
| | 9 2 | 8 2 | 5+6 2 | 5+6 2 |

Continued on next column

**TABLE LXXIII– continued**

| TQC | ICSD | NoE | SF | |
|---|---|---|---|---|
| | 6+7 2 | 8 2 | 3+4 2 | 3+4 2 |
| | 9 2 | 9 2 | 3+4 2 | 5+6 2 |
| | 8 2 | 4+5 2 | 5+6 2 | 3+4 2 |
| | 4+5 2 | 8 2 | 5+6 2 | 5+6 2 |
| | 9 2 | 8 2 | 5+6 2 | 5+6 2 |
| | 6+7 2 | 4+5 2 | 5+6 2 | 3+4 2 |
| | 8 2 | 8 2 | 5+6 2 | 5+6 2 |
| | 4+5 2 | 8 2 | 3+4 2 | 5+6 2 |
| | 8 2 | 4+5 2 | 5+6 2 | 5+6 2 |
| | 8 2 | 8 2 | 3+4 2 | 3+4 2 |
| | 4+5 2 | 4+5 2 | 5+6 2 | 3+4 2 |
| | 8 2 | 8 2 | 3+4 2 | 5+6 2 |
| | 4+5 2 | 8 2 | 5+6 2 | 5+6 2 |
| | 8 2 | 9 2 | 3+4 2 | 3+4 2 |
| | 8 2 | 8 2 | 5+6 2 | 5+6 2 |
| | 8 2 | 6+7 2 | 3+4 2 | 3+4 2 |
| 166.1.97 | 646211 | 50 | Ni3Pb2Se2 | |
| | -GM | -T | -F | -L |
| | 8 2 | 8 2 | 5+6 2 | 5+6 2 |
| | 9 2 | 9 2 | 3+4 2 | 3+4 2 |
| | 8 2 | 9 2 | 3+4 2 | 3+4 2 |
| | 8 2 | 8 2 | 3+4 2 | 5+6 2 |
| | 8 2 | 9 2 | 5+6 2 | 3+4 2 |
| | 8 2 | 6+7 2 | 5+6 2 | 5+6 2 |
| | 4+5 2 | 8 2 | 5+6 2 | 5+6 2 |
| | 9 2 | 9 2 | 3+4 2 | 3+4 2 |
| | 9 2 | 4+5 2 | 5+6 2 | 3+4 2 |
| | 6+7 2 | 8 2 | 3+4 2 | 5+6 2 |
| | 9 2 | 9 2 | 3+4 2 | 3+4 2 |
| | 8 2 | 9 2 | 5+6 2 | 5+6 2 |
| | 4+5 2 | 6+7 2 | 5+6 2 | 3+4 2 |
| | 8 2 | 9 2 | 5+6 2 | 3+4 2 |
| | 9 2 | 6+7 2 | 5+6 2 | 3+4 2 |
| | 6+7 2 | 9 2 | 5+6 2 | 5+6 2 |
| | 4+5 2 | 9 2 | 3+4 2 | 3+4 2 |
| | 8 2 | 6+7 2 | 5+6 2 | 3+4 2 |
| | 8 2 | 9 2 | 3+4 2 | 5+6 2 |
| | 4+5 2 | 6+7 2 | 5+6 2 | 5+6 2 |
| | 8 2 | 9 2 | 3+4 2 | 3+4 2 |
| | 4+5 2 | 8 2 | 3+4 2 | 3+4 2 |
| | 8 2 | 9 2 | 5+6 2 | 5+6 2 |
| | 8 2 | 9 2 | 5+6 2 | 3+4 2 |
| | 8 2 | 6+7 2 | 3+4 2 | 3+4 2 |
| 166.1.98 | 646210 | 50 | Ni3Pb2Se2 | |
| | -GM | -T | -F | -L |
| | 8 2 | 8 2 | 5+6 2 | 5+6 2 |
| | 9 2 | 9 2 | 3+4 2 | 3+4 2 |
| | 8 2 | 9 2 | 3+4 2 | 3+4 2 |
| | 8 2 | 8 2 | 3+4 2 | 5+6 2 |
| | 8 2 | 9 2 | 5+6 2 | 3+4 2 |
| | 8 2 | 6+7 2 | 5+6 2 | 5+6 2 |
| | 4+5 2 | 8 2 | 5+6 2 | 5+6 2 |
| | 9 2 | 9 2 | 3+4 2 | 3+4 2 |
| | 9 2 | 4+5 2 | 5+6 2 | 3+4 2 |
| | 6+7 2 | 8 2 | 3+4 2 | 5+6 2 |
| | 9 2 | 9 2 | 3+4 2 | 3+4 2 |
| | 8 2 | 9 2 | 5+6 2 | 5+6 2 |
| | 4+5 2 | 6+7 2 | 5+6 2 | 3+4 2 |
| | 8 2 | 9 2 | 5+6 2 | 3+4 2 |
| | 9 2 | 6+7 2 | 5+6 2 | 3+4 2 |
| | 6+7 2 | 9 2 | 5+6 2 | 5+6 2 |

Continued on next column

**TABLE LXXIII– continued**

| TQC | ICSD | NoE | SF | |
|---|---|---|---|---|
| | 4+5 2 | 9 2 | 3+4 2 | 3+4 2 |
| | 8 2 | 6+7 2 | 5+6 2 | 3+4 2 |
| | 8 2 | 9 2 | 3+4 2 | 5+6 2 |
| | 4+5 2 | 6+7 2 | 5+6 2 | 5+6 2 |
| | 8 2 | 9 2 | 3+4 2 | 3+4 2 |
| | 4+5 2 | 8 2 | 3+4 2 | 3+4 2 |
| | 8 2 | 9 2 | 5+6 2 | 5+6 2 |
| | 8 2 | 9 2 | 5+6 2 | 3+4 2 |
| | 8 2 | 6+7 2 | 3+4 2 | 3+4 2 |
| 166.1.99 | 402453 | 50 | Ni3Pb2Se2 | |
| | -GM | -T | -F | -L |
| | 8 2 | 8 2 | 5+6 2 | 5+6 2 |
| | 9 2 | 9 2 | 3+4 2 | 3+4 2 |
| | 8 2 | 9 2 | 3+4 2 | 3+4 2 |
| | 8 2 | 8 2 | 3+4 2 | 5+6 2 |
| | 8 2 | 9 2 | 5+6 2 | 3+4 2 |
| | 8 2 | 6+7 2 | 5+6 2 | 5+6 2 |
| | 4+5 2 | 8 2 | 5+6 2 | 5+6 2 |
| | 9 2 | 9 2 | 3+4 2 | 3+4 2 |
| | 9 2 | 4+5 2 | 5+6 2 | 3+4 2 |
| | 6+7 2 | 8 2 | 3+4 2 | 5+6 2 |
| | 9 2 | 9 2 | 3+4 2 | 3+4 2 |
| | 8 2 | 9 2 | 5+6 2 | 5+6 2 |
| | 4+5 2 | 6+7 2 | 5+6 2 | 3+4 2 |
| | 8 2 | 9 2 | 5+6 2 | 3+4 2 |
| | 9 2 | 6+7 2 | 5+6 2 | 3+4 2 |
| | 6+7 2 | 9 2 | 5+6 2 | 5+6 2 |
| | 4+5 2 | 9 2 | 3+4 2 | 3+4 2 |
| | 8 2 | 6+7 2 | 5+6 2 | 3+4 2 |
| | 8 2 | 9 2 | 3+4 2 | 5+6 2 |
| | 4+5 2 | 6+7 2 | 5+6 2 | 5+6 2 |
| | 8 2 | 9 2 | 3+4 2 | 3+4 2 |
| | 4+5 2 | 8 2 | 3+4 2 | 3+4 2 |
| | 8 2 | 9 2 | 5+6 2 | 5+6 2 |
| | 8 2 | 9 2 | 5+6 2 | 3+4 2 |
| | 8 2 | 6+7 2 | 3+4 2 | 3+4 2 |
| 166.1.100 | 108892 | 50 | Ni3Pb2Se2 | |
| | -GM | -T | -F | -L |
| | 8 2 | 9 2 | 5+6 2 | 3+4 2 |
| | 9 2 | 8 2 | 3+4 2 | 5+6 2 |
| | 8 2 | 8 2 | 3+4 2 | 5+6 2 |
| | 8 2 | 9 2 | 3+4 2 | 3+4 2 |
| | 8 2 | 8 2 | 5+6 2 | 5+6 2 |
| | 8 2 | 4+5 2 | 5+6 2 | 3+4 2 |
| | 4+5 2 | 9 2 | 5+6 2 | 3+4 2 |
| | 9 2 | 8 2 | 3+4 2 | 5+6 2 |
| | 9 2 | 6+7 2 | 5+6 2 | 5+6 2 |
| | 6+7 2 | 9 2 | 3+4 2 | 3+4 2 |
| | 9 2 | 8 2 | 3+4 2 | 5+6 2 |
| | 8 2 | 8 2 | 5+6 2 | 3+4 2 |
| | 4+5 2 | 4+5 2 | 5+6 2 | 5+6 2 |
| | 8 2 | 8 2 | 5+6 2 | 5+6 2 |
| | 9 2 | 4+5 2 | 5+6 2 | 5+6 2 |
| | 6+7 2 | 8 2 | 5+6 2 | 3+4 2 |
| | 4+5 2 | 8 2 | 3+4 2 | 5+6 2 |
| | 8 2 | 4+5 2 | 5+6 2 | 3+4 2 |
| | 8 2 | 8 2 | 3+4 2 | 5+6 2 |
| | 4+5 2 | 4+5 2 | 5+6 2 | 3+4 2 |
| | 8 2 | 8 2 | 3+4 2 | 5+6 2 |
| | 4+5 2 | 9 2 | 5+6 2 | 5+6 2 |
| | 8 2 | 8 2 | 3+4 2 | 3+4 2 |

Continued on next column

**TABLE LXXIII– continued**

| TQC | ICSD | NoE | SF | |
|---|---|---|---|---|
| | 8 2 | 8 2 | 5+6 2 | 5+6 2 |
| | 8 2 | 4+5 2 | 3+4 2 | 5+6 2 |
| 166.1.101 | 646207 | 50 | Ni3Pb2S2 | |
| | -GM | -T | -F | -L |
| | 8 2 | 9 2 | 5+6 2 | 3+4 2 |
| | 9 2 | 8 2 | 3+4 2 | 5+6 2 |
| | 8 2 | 9 2 | 3+4 2 | 3+4 2 |
| | 8 2 | 8 2 | 3+4 2 | 5+6 2 |
| | 8 2 | 9 2 | 5+6 2 | 3+4 2 |
| | 8 2 | 6+7 2 | 5+6 2 | 3+4 2 |
| | 4+5 2 | 9 2 | 5+6 2 | 5+6 2 |
| | 9 2 | 8 2 | 3+4 2 | 5+6 2 |
| | 9 2 | 4+5 2 | 5+6 2 | 3+4 2 |
| | 6+7 2 | 8 2 | 3+4 2 | 5+6 2 |
| | 9 2 | 9 2 | 3+4 2 | 3+4 2 |
| | 8 2 | 9 2 | 5+6 2 | 5+6 2 |
| | 4+5 2 | 6+7 2 | 5+6 2 | 3+4 2 |
| | 9 2 | 9 2 | 5+6 2 | 3+4 2 |
| | 8 2 | 6+7 2 | 5+6 2 | 5+6 2 |
| | 6+7 2 | 9 2 | 5+6 2 | 3+4 2 |
| | 8 2 | 9 2 | 3+4 2 | 3+4 2 |
| | 4+5 2 | 6+7 2 | 5+6 2 | 3+4 2 |
| | 8 2 | 9 2 | 5+6 2 | 5+6 2 |
| | 4+5 2 | 6+7 2 | 3+4 2 | 5+6 2 |
| | 8 2 | 9 2 | 3+4 2 | 3+4 2 |
| | 4+5 2 | 9 2 | 3+4 2 | 3+4 2 |
| | 8 2 | 8 2 | 5+6 2 | 5+6 2 |
| | 8 2 | 9 2 | 5+6 2 | 3+4 2 |
| | 8 2 | 6+7 2 | 3+4 2 | 3+4 2 |
| 166.1.102 | 646206 | 50 | Ni3Pb2S2 | |
| | -GM | -T | -F | -L |
| | 8 2 | 9 2 | 5+6 2 | 3+4 2 |
| | 9 2 | 8 2 | 3+4 2 | 5+6 2 |
| | 8 2 | 9 2 | 3+4 2 | 3+4 2 |
| | 8 2 | 8 2 | 3+4 2 | 5+6 2 |
| | 8 2 | 9 2 | 5+6 2 | 3+4 2 |
| | 8 2 | 6+7 2 | 5+6 2 | 3+4 2 |
| | 4+5 2 | 9 2 | 5+6 2 | 5+6 2 |
| | 9 2 | 8 2 | 3+4 2 | 5+6 2 |
| | 9 2 | 4+5 2 | 5+6 2 | 3+4 2 |
| | 6+7 2 | 9 2 | 3+4 2 | 5+6 2 |
| | 9 2 | 8 2 | 3+4 2 | 3+4 2 |
| | 8 2 | 9 2 | 5+6 2 | 5+6 2 |
| | 4+5 2 | 6+7 2 | 5+6 2 | 3+4 2 |
| | 9 2 | 9 2 | 5+6 2 | 3+4 2 |
| | 8 2 | 6+7 2 | 5+6 2 | 5+6 2 |
| | 6+7 2 | 9 2 | 5+6 2 | 3+4 2 |
| | 8 2 | 9 2 | 3+4 2 | 3+4 2 |
| | 4+5 2 | 6+7 2 | 5+6 2 | 3+4 2 |
| | 8 2 | 9 2 | 5+6 2 | 5+6 2 |
| | 4+5 2 | 6+7 2 | 3+4 2 | 5+6 2 |
| | 8 2 | 9 2 | 3+4 2 | 3+4 2 |
| | 4+5 2 | 9 2 | 3+4 2 | 3+4 2 |
| | 8 2 | 8 2 | 5+6 2 | 5+6 2 |
| | 8 2 | 9 2 | 5+6 2 | 3+4 2 |
| | 8 2 | 6+7 2 | 3+4 2 | 3+4 2 |
| 166.1.103 | 417632 | 50 | Ni3Pb2S2 | |
| | -GM | -T | -F | -L |
| | 8 2 | 9 2 | 5+6 2 | 3+4 2 |
| | 9 2 | 8 2 | 3+4 2 | 5+6 2 |
| | 8 2 | 9 2 | 3+4 2 | 3+4 2 |

Continued on next column

**TABLE LXXIII– continued**

| TQC | ICSD | NoE | SF | |
|---|---|---|---|---|
| | 8 2 | 8 2 | 3+4 2 | 5+6 2 |
| | 8 2 | 9 2 | 5+6 2 | 3+4 2 |
| | 8 2 | 6+7 2 | 5+6 2 | 3+4 2 |
| | 4+5 2 | 9 2 | 5+6 2 | 5+6 2 |
| | 9 2 | 8 2 | 3+4 2 | 5+6 2 |
| | 9 2 | 4+5 2 | 5+6 2 | 3+4 2 |
| | 6+7 2 | 8 2 | 3+4 2 | 5+6 2 |
| | 9 2 | 9 2 | 3+4 2 | 3+4 2 |
| | 8 2 | 9 2 | 5+6 2 | 5+6 2 |
| | 4+5 2 | 6+7 2 | 5+6 2 | 3+4 2 |
| | 9 2 | 9 2 | 5+6 2 | 3+4 2 |
| | 8 2 | 6+7 2 | 5+6 2 | 5+6 2 |
| | 6+7 2 | 9 2 | 5+6 2 | 3+4 2 |
| | 8 2 | 9 2 | 3+4 2 | 3+4 2 |
| | 4+5 2 | 6+7 2 | 5+6 2 | 3+4 2 |
| | 8 2 | 9 2 | 5+6 2 | 5+6 2 |
| | 4+5 2 | 6+7 2 | 3+4 2 | 5+6 2 |
| | 8 2 | 9 2 | 3+4 2 | 3+4 2 |
| | 4+5 2 | 9 2 | 3+4 2 | 3+4 2 |
| | 8 2 | 8 2 | 5+6 2 | 5+6 2 |
| | 8 2 | 9 2 | 5+6 2 | 3+4 2 |
| | 8 2 | 6+7 2 | 3+4 2 | 3+4 2 |
| 166.1.104 | 24176 | 50 | Ni3Pb2S2 | |
| | -GM | -T | -F | -L |
| | 8 2 | 8 2 | 5+6 2 | 5+6 2 |
| | 9 2 | 9 2 | 3+4 2 | 3+4 2 |
| | 8 2 | 9 2 | 3+4 2 | 3+4 2 |
| | 8 2 | 8 2 | 3+4 2 | 5+6 2 |
| | 8 2 | 9 2 | 5+6 2 | 3+4 2 |
| | 8 2 | 6+7 2 | 5+6 2 | 5+6 2 |
| | 4+5 2 | 8 2 | 5+6 2 | 5+6 2 |
| | 9 2 | 9 2 | 3+4 2 | 3+4 2 |
| | 9 2 | 4+5 2 | 5+6 2 | 3+4 2 |
| | 6+7 2 | 8 2 | 3+4 2 | 5+6 2 |
| | 9 2 | 9 2 | 3+4 2 | 3+4 2 |
| | 4+5 2 | 9 2 | 5+6 2 | 5+6 2 |
| | 8 2 | 6+7 2 | 5+6 2 | 3+4 2 |
| | 8 2 | 9 2 | 5+6 2 | 3+4 2 |
| | 9 2 | 6+7 2 | 5+6 2 | 3+4 2 |
| | 6+7 2 | 9 2 | 5+6 2 | 5+6 2 |
| | 4+5 2 | 9 2 | 3+4 2 | 3+4 2 |
| | 8 2 | 6+7 2 | 5+6 2 | 3+4 2 |
| | 8 2 | 9 2 | 3+4 2 | 5+6 2 |
| | 4+5 2 | 6+7 2 | 5+6 2 | 5+6 2 |
| | 8 2 | 9 2 | 3+4 2 | 3+4 2 |
| | 4+5 2 | 8 2 | 5+6 2 | 3+4 2 |
| | 8 2 | 9 2 | 3+4 2 | 5+6 2 |
| | 8 2 | 9 2 | 5+6 2 | 3+4 2 |
| | 8 2 | 6+7 2 | 3+4 2 | 3+4 2 |
| 166.1.105 | 108895 | 50 | Ni3Pb2S2 | |
| | -GM | -T | -F | -L |
| | 8 2 | 9 2 | 5+6 2 | 3+4 2 |
| | 9 2 | 8 2 | 3+4 2 | 5+6 2 |
| | 8 2 | 8 2 | 3+4 2 | 5+6 2 |
| | 8 2 | 9 2 | 3+4 2 | 3+4 2 |
| | 8 2 | 8 2 | 5+6 2 | 5+6 2 |
| | 8 2 | 4+5 2 | 5+6 2 | 3+4 2 |
| | 4+5 2 | 9 2 | 5+6 2 | 3+4 2 |
| | 9 2 | 8 2 | 3+4 2 | 5+6 2 |
| | 9 2 | 6+7 2 | 5+6 2 | 5+6 2 |
| | 6+7 2 | 9 2 | 3+4 2 | 3+4 2 |

Continued on next column

**TABLE LXXIII– continued**

| TQC | ICSD | NoE | SF | |
|---|---|---|---|---|
| | 9 2 | 8 2 | 3+4 2 | 5+6 2 |
| | 4+5 2 | 8 2 | 5+6 2 | 3+4 2 |
| | 8 2 | 4+5 2 | 5+6 2 | 5+6 2 |
| | 8 2 | 8 2 | 5+6 2 | 5+6 2 |
| | 9 2 | 4+5 2 | 5+6 2 | 3+4 2 |
| | 6+7 2 | 8 2 | 5+6 2 | 5+6 2 |
| | 4+5 2 | 8 2 | 3+4 2 | 5+6 2 |
| | 8 2 | 4+5 2 | 5+6 2 | 5+6 2 |
| | 8 2 | 8 2 | 3+4 2 | 3+4 2 |
| | 4+5 2 | 4+5 2 | 5+6 2 | 3+4 2 |
| | 8 2 | 8 2 | 3+4 2 | 5+6 2 |
| | 4+5 2 | 9 2 | 5+6 2 | 5+6 2 |
| | 8 2 | 8 2 | 3+4 2 | 3+4 2 |
| | 8 2 | 8 2 | 5+6 2 | 5+6 2 |
| | 8 2 | 4+5 2 | 3+4 2 | 5+6 2 |
| 166.1.106 | 646072 | 232 | Nd2Zn17 | |
| | -GM | -T | -F | -L |
| | 8 2 | 8 2 | 3+4 2 | 5+6 2 |
| | 9 2 | 9 2 | 5+6 2 | 3+4 2 |
| | 8 2 | 9 2 | 3+4 2 | 5+6 2 |
| | 9 2 | 8 2 | 5+6 2 | 3+4 2 |
| | 4+5 2 | 4+5 2 | 5+6 2 | 3+4 2 |
| | 6+7 2 | 9 2 | 3+4 2 | 5+6 2 |
| | 8 2 | 6+7 2 | 3+4 2 | 5+6 2 |
| | 9 2 | 8 2 | 5+6 2 | 3+4 2 |
| | 8 2 | 9 2 | 3+4 2 | 3+4 2 |
| | 8 2 | 9 2 | 5+6 2 | 5+6 2 |
| | 8 2 | 6+7 2 | 5+6 2 | 3+4 2 |
| | 4+5 2 | 9 2 | 5+6 2 | 3+4 2 |
| | 8 2 | 9 2 | 3+4 2 | 3+4 2 |
| | 4+5 2 | 8 2 | 3+4 2 | 5+6 2 |
| | 9 2 | 9 2 | 5+6 2 | 3+4 2 |
| | 8 2 | 4+5 2 | 5+6 2 | 5+6 2 |
| | 6+7 2 | 6+7 2 | 5+6 2 | 3+4 2 |
| | 8 2 | 8 2 | 3+4 2 | 5+6 2 |
| | 4+5 2 | 6+7 2 | 5+6 2 | 3+4 2 |
| | 8 2 | 9 2 | 3+4 2 | 5+6 2 |
| | 9 2 | 8 2 | 5+6 2 | 3+4 2 |
| | 8 2 | 9 2 | 3+4 2 | 5+6 2 |
| | 4+5 2 | 9 2 | 3+4 2 | 5+6 2 |
| | 9 2 | 6+7 2 | 5+6 2 | 3+4 2 |
| | 8 2 | 8 2 | 5+6 2 | 3+4 2 |
| | 6+7 2 | 4+5 2 | 3+4 2 | 3+4 2 |
| | 8 2 | 6+7 2 | 5+6 2 | 5+6 2 |
| | 8 2 | 9 2 | 5+6 2 | 3+4 2 |
| | 9 2 | 9 2 | 3+4 2 | 5+6 2 |
| | 4+5 2 | 8 2 | 3+4 2 | 3+4 2 |
| | 9 2 | 6+7 2 | 5+6 2 | 5+6 2 |
| | 9 2 | 4+5 2 | 5+6 2 | 3+4 2 |
| | 4+5 2 | 8 2 | 5+6 2 | 3+4 2 |
| | 8 2 | 9 2 | 3+4 2 | 5+6 2 |
| | 6+7 2 | 8 2 | 3+4 2 | 5+6 2 |
| | 8 2 | 9 2 | 3+4 2 | 5+6 2 |
| | 9 2 | 8 2 | 5+6 2 | 3+4 2 |
| | 8 2 | 6+7 2 | 5+6 2 | 5+6 2 |
| | 6+7 2 | 9 2 | 3+4 2 | 5+6 2 |
| | 9 2 | 6+7 2 | 5+6 2 | 3+4 2 |
| | 4+5 2 | 4+5 2 | 3+4 2 | 5+6 2 |
| | 6+7 2 | 9 2 | 3+4 2 | 3+4 2 |
| | 4+5 2 | 8 2 | 5+6 2 | 5+6 2 |
| | 9 2 | 8 2 | 3+4 2 | 3+4 2 |

Continued on next column

**TABLE LXXIII– continued**

| TQC | ICSD | NoE | SF | |
|---|---|---|---|---|
| | 6+7 2 | 4+5 2 | 3+4 2 | 3+4 2 |
| | 8 2 | 8 2 | 5+6 2 | 5+6 2 |
| | 8 2 | 9 2 | 5+6 2 | 5+6 2 |
| | 9 2 | 6+7 2 | 3+4 2 | 3+4 2 |
| | 6+7 2 | 9 2 | 5+6 2 | 3+4 2 |
| | 8 2 | 4+5 2 | 3+4 2 | 5+6 2 |
| | 9 2 | 9 2 | 5+6 2 | 3+4 2 |
| | 8 2 | 8 2 | 3+4 2 | 3+4 2 |
| | 8 2 | 6+7 2 | 5+6 2 | 5+6 2 |
| | 9 2 | 9 2 | 5+6 2 | 5+6 2 |
| | 4+5 2 | 8 2 | 3+4 2 | 3+4 2 |
| | 8 2 | 4+5 2 | 3+4 2 | 5+6 2 |
| | 9 2 | 6+7 2 | 5+6 2 | 3+4 2 |
| | 4+5 2 | 9 2 | 3+4 2 | 3+4 2 |
| | 6+7 2 | 8 2 | 5+6 2 | 5+6 2 |
| | 8 2 | 8 2 | 5+6 2 | 3+4 2 |
| | 9 2 | 9 2 | 3+4 2 | 5+6 2 |
| | 8 2 | 4+5 2 | 5+6 2 | 5+6 2 |
| | 6+7 2 | 8 2 | 3+4 2 | 3+4 2 |
| | 8 2 | 9 2 | 3+4 2 | 5+6 2 |
| | 4+5 2 | 6+7 2 | 5+6 2 | 3+4 2 |
| | 9 2 | 8 2 | 3+4 2 | 3+4 2 |
| | 8 2 | 9 2 | 5+6 2 | 5+6 2 |
| | 6+7 2 | 9 2 | 3+4 2 | 3+4 2 |
| | 9 2 | 4+5 2 | 5+6 2 | 5+6 2 |
| | 4+5 2 | 6+7 2 | 3+4 2 | 3+4 2 |
| | 8 2 | 9 2 | 5+6 2 | 5+6 2 |
| | 9 2 | 8 2 | 3+4 2 | 3+4 2 |
| | 4+5 2 | 4+5 2 | 5+6 2 | 5+6 2 |
| | 6+7 2 | 9 2 | 5+6 2 | 3+4 2 |
| | 9 2 | 8 2 | 3+4 2 | 5+6 2 |
| | 8 2 | 9 2 | 5+6 2 | 3+4 2 |
| | 8 2 | 8 2 | 3+4 2 | 5+6 2 |
| | 9 2 | 6+7 2 | 5+6 2 | 5+6 2 |
| | 9 2 | 8 2 | 3+4 2 | 3+4 2 |
| | 4+5 2 | 4+5 2 | 3+4 2 | 3+4 2 |
| | 8 2 | 9 2 | 5+6 2 | 5+6 2 |
| | 8 2 | 8 2 | 3+4 2 | 3+4 2 |
| | 9 2 | 6+7 2 | 5+6 2 | 3+4 2 |
| | 8 2 | 4+5 2 | 3+4 2 | 5+6 2 |
| | 9 2 | 8 2 | 5+6 2 | 3+4 2 |
| | 6+7 2 | 9 2 | 3+4 2 | 5+6 2 |
| | 8 2 | 9 2 | 5+6 2 | 5+6 2 |
| | 4+5 2 | 9 2 | 5+6 2 | 3+4 2 |
| | 9 2 | 6+7 2 | 3+4 2 | 3+4 2 |
| | 8 2 | 8 2 | 5+6 2 | 5+6 2 |
| | 4+5 2 | 9 2 | 5+6 2 | 3+4 2 |
| | 8 2 | 9 2 | 3+4 2 | 5+6 2 |
| | 4+5 2 | 6+7 2 | 3+4 2 | 3+4 2 |
| | 8 2 | 9 2 | 5+6 2 | 5+6 2 |
| | 9 2 | 9 2 | 5+6 2 | 3+4 2 |
| | 6+7 2 | 9 2 | 5+6 2 | 3+4 2 |
| | 8 2 | 6+7 2 | 3+4 2 | 5+6 2 |
| | 4+5 2 | 8 2 | 5+6 2 | 5+6 2 |
| | 8 2 | 4+5 2 | 3+4 2 | 3+4 2 |
| | 9 2 | 8 2 | 5+6 2 | 3+4 2 |
| | 6+7 2 | 9 2 | 3+4 2 | 5+6 2 |
| | 9 2 | 8 2 | 3+4 2 | 5+6 2 |
| | 8 2 | 8 2 | 5+6 2 | 3+4 2 |
| | 9 2 | 4+5 2 | 5+6 2 | 5+6 2 |
| | 4+5 2 | 6+7 2 | 3+4 2 | 3+4 2 |

Continued on next column

**TABLE LXXIII– continued**

| TQC | ICSD | NoE | SF | |
|---|---|---|---|---|
| | 9 2 | 9 2 | 3+4 2 | 3+4 2 |
| | 8 2 | 4+5 2 | 5+6 2 | 5+6 2 |
| | 4+5 2 | 8 2 | 5+6 2 | 5+6 2 |
| | 8 2 | 4+5 2 | 5+6 2 | 3+4 2 |
| | 8 2 | 8 2 | 3+4 2 | 5+6 2 |
| | 9 2 | 9 2 | 3+4 2 | 3+4 2 |
| | 9 2 | 8 2 | 3+4 2 | 5+6 2 |
| | 6+7 2 | 9 2 | 3+4 2 | 3+4 2 |
| | 8 2 | 6+7 2 | 5+6 2 | 5+6 2 |
| | 4+5 2 | 9 2 | 5+6 2 | 3+4 2 |
| | 6+7 2 | 6+7 2 | 5+6 2 | 3+4 2 |
| 166.1.107 | 601895 | 232 | Nd2Zn17 | |
| | -GM | -T | -F | -L |
| | 8 2 | 8 2 | 3+4 2 | 5+6 2 |
| | 9 2 | 9 2 | 5+6 2 | 3+4 2 |
| | 8 2 | 9 2 | 3+4 2 | 5+6 2 |
| | 9 2 | 8 2 | 5+6 2 | 3+4 2 |
| | 4+5 2 | 4+5 2 | 5+6 2 | 3+4 2 |
| | 6+7 2 | 9 2 | 3+4 2 | 5+6 2 |
| | 8 2 | 6+7 2 | 3+4 2 | 5+6 2 |
| | 9 2 | 8 2 | 5+6 2 | 3+4 2 |
| | 8 2 | 9 2 | 3+4 2 | 3+4 2 |
| | 8 2 | 9 2 | 5+6 2 | 5+6 2 |
| | 8 2 | 6+7 2 | 5+6 2 | 3+4 2 |
| | 4+5 2 | 9 2 | 5+6 2 | 3+4 2 |
| | 8 2 | 9 2 | 3+4 2 | 3+4 2 |
| | 4+5 2 | 8 2 | 3+4 2 | 5+6 2 |
| | 9 2 | 9 2 | 5+6 2 | 3+4 2 |
| | 8 2 | 4+5 2 | 5+6 2 | 5+6 2 |
| | 6+7 2 | 6+7 2 | 5+6 2 | 3+4 2 |
| | 8 2 | 8 2 | 3+4 2 | 5+6 2 |
| | 4+5 2 | 6+7 2 | 5+6 2 | 3+4 2 |
| | 8 2 | 9 2 | 3+4 2 | 5+6 2 |
| | 9 2 | 8 2 | 5+6 2 | 3+4 2 |
| | 8 2 | 9 2 | 3+4 2 | 5+6 2 |
| | 4+5 2 | 9 2 | 3+4 2 | 5+6 2 |
| | 9 2 | 6+7 2 | 5+6 2 | 3+4 2 |
| | 8 2 | 8 2 | 5+6 2 | 3+4 2 |
| | 6+7 2 | 4+5 2 | 3+4 2 | 3+4 2 |
| | 8 2 | 6+7 2 | 5+6 2 | 5+6 2 |
| | 8 2 | 9 2 | 5+6 2 | 3+4 2 |
| | 9 2 | 9 2 | 3+4 2 | 5+6 2 |
| | 4+5 2 | 8 2 | 3+4 2 | 3+4 2 |
| | 9 2 | 6+7 2 | 5+6 2 | 5+6 2 |
| | 9 2 | 8 2 | 5+6 2 | 3+4 2 |
| | 4+5 2 | 4+5 2 | 3+4 2 | 3+4 2 |
| | 8 2 | 9 2 | 5+6 2 | 5+6 2 |
| | 8 2 | 8 2 | 3+4 2 | 5+6 2 |
| | 6+7 2 | 9 2 | 3+4 2 | 5+6 2 |
| | 9 2 | 8 2 | 5+6 2 | 3+4 2 |
| | 8 2 | 6+7 2 | 5+6 2 | 5+6 2 |
| | 6+7 2 | 9 2 | 3+4 2 | 5+6 2 |
| | 9 2 | 4+5 2 | 5+6 2 | 3+4 2 |
| | 4+5 2 | 6+7 2 | 3+4 2 | 5+6 2 |
| | 6+7 2 | 8 2 | 3+4 2 | 3+4 2 |
| | 9 2 | 9 2 | 5+6 2 | 5+6 2 |
| | 4+5 2 | 8 2 | 3+4 2 | 3+4 2 |
| | 6+7 2 | 4+5 2 | 3+4 2 | 3+4 2 |
| | 8 2 | 8 2 | 5+6 2 | 5+6 2 |
| | 8 2 | 9 2 | 5+6 2 | 5+6 2 |
| | 9 2 | 6+7 2 | 3+4 2 | 3+4 2 |

Continued on next column

**TABLE LXXIII– continued**

| TQC | ICSD | NoE | SF | |
|---|---|---|---|---|
| | 6+7 2 | 9 2 | 5+6 2 | 3+4 2 |
| | 8 2 | 4+5 2 | 3+4 2 | 5+6 2 |
| | 9 2 | 9 2 | 5+6 2 | 3+4 2 |
| | 8 2 | 8 2 | 3+4 2 | 3+4 2 |
| | 8 2 | 6+7 2 | 5+6 2 | 5+6 2 |
| | 9 2 | 9 2 | 5+6 2 | 3+4 2 |
| | 4+5 2 | 8 2 | 3+4 2 | 5+6 2 |
| | 8 2 | 4+5 2 | 3+4 2 | 5+6 2 |
| | 9 2 | 6+7 2 | 5+6 2 | 3+4 2 |
| | 4+5 2 | 9 2 | 3+4 2 | 3+4 2 |
| | 6+7 2 | 8 2 | 5+6 2 | 5+6 2 |
| | 8 2 | 8 2 | 5+6 2 | 3+4 2 |
| | 9 2 | 9 2 | 3+4 2 | 5+6 2 |
| | 8 2 | 4+5 2 | 5+6 2 | 5+6 2 |
| | 6+7 2 | 8 2 | 3+4 2 | 3+4 2 |
| | 8 2 | 9 2 | 3+4 2 | 5+6 2 |
| | 4+5 2 | 6+7 2 | 5+6 2 | 3+4 2 |
| | 9 2 | 8 2 | 3+4 2 | 3+4 2 |
| | 8 2 | 9 2 | 5+6 2 | 5+6 2 |
| | 6+7 2 | 9 2 | 3+4 2 | 3+4 2 |
| | 9 2 | 4+5 2 | 5+6 2 | 5+6 2 |
| | 4+5 2 | 6+7 2 | 3+4 2 | 3+4 2 |
| | 8 2 | 9 2 | 5+6 2 | 5+6 2 |
| | 9 2 | 8 2 | 3+4 2 | 3+4 2 |
| | 4+5 2 | 4+5 2 | 5+6 2 | 5+6 2 |
| | 6+7 2 | 9 2 | 5+6 2 | 3+4 2 |
| | 9 2 | 8 2 | 3+4 2 | 5+6 2 |
| | 8 2 | 9 2 | 5+6 2 | 3+4 2 |
| | 8 2 | 8 2 | 3+4 2 | 5+6 2 |
| | 9 2 | 6+7 2 | 5+6 2 | 5+6 2 |
| | 4+5 2 | 8 2 | 3+4 2 | 3+4 2 |
| | 9 2 | 4+5 2 | 3+4 2 | 3+4 2 |
| | 8 2 | 9 2 | 5+6 2 | 5+6 2 |
| | 8 2 | 8 2 | 3+4 2 | 3+4 2 |
| | 9 2 | 6+7 2 | 5+6 2 | 3+4 2 |
| | 8 2 | 4+5 2 | 3+4 2 | 5+6 2 |
| | 9 2 | 8 2 | 5+6 2 | 3+4 2 |
| | 6+7 2 | 9 2 | 3+4 2 | 5+6 2 |
| | 8 2 | 9 2 | 5+6 2 | 5+6 2 |
| | 4+5 2 | 9 2 | 5+6 2 | 3+4 2 |
| | 9 2 | 6+7 2 | 3+4 2 | 3+4 2 |
| | 8 2 | 8 2 | 5+6 2 | 5+6 2 |
| | 4+5 2 | 9 2 | 5+6 2 | 3+4 2 |
| | 8 2 | 9 2 | 3+4 2 | 5+6 2 |
| | 4+5 2 | 6+7 2 | 3+4 2 | 3+4 2 |
| | 8 2 | 9 2 | 5+6 2 | 5+6 2 |
| | 9 2 | 9 2 | 5+6 2 | 3+4 2 |
| | 6+7 2 | 9 2 | 5+6 2 | 3+4 2 |
| | 8 2 | 6+7 2 | 3+4 2 | 5+6 2 |
| | 4+5 2 | 8 2 | 5+6 2 | 5+6 2 |
| | 8 2 | 4+5 2 | 3+4 2 | 3+4 2 |
| | 9 2 | 8 2 | 5+6 2 | 3+4 2 |
| | 6+7 2 | 9 2 | 3+4 2 | 5+6 2 |
| | 9 2 | 8 2 | 3+4 2 | 5+6 2 |
| | 8 2 | 8 2 | 5+6 2 | 3+4 2 |
| | 9 2 | 4+5 2 | 5+6 2 | 5+6 2 |
| | 4+5 2 | 6+7 2 | 3+4 2 | 3+4 2 |
| | 9 2 | 9 2 | 3+4 2 | 3+4 2 |
| | 8 2 | 4+5 2 | 5+6 2 | 5+6 2 |
| | 4+5 2 | 8 2 | 5+6 2 | 5+6 2 |
| | 8 2 | 4+5 2 | 5+6 2 | 3+4 2 |
| | 8 2 | 8 2 | 3+4 2 | 5+6 2 |
| | 9 2 | 9 2 | 3+4 2 | 3+4 2 |
| | 9 2 | 8 2 | 3+4 2 | 5+6 2 |
| | 6+7 2 | 9 2 | 3+4 2 | 3+4 2 |
| | 8 2 | 6+7 2 | 5+6 2 | 5+6 2 |
| | 4+5 2 | 9 2 | 5+6 2 | 3+4 2 |
| | 6+7 2 | 6+7 2 | 5+6 2 | 3+4 2 |
| 166.1.108 | 104242 | 232 | Nd2Zn17 | |
| | -GM | -T | -F | -L |
| | 8 2 | 8 2 | 3+4 2 | 5+6 2 |
| | 9 2 | 9 2 | 5+6 2 | 3+4 2 |
| | 8 2 | 9 2 | 3+4 2 | 5+6 2 |
| | 9 2 | 8 2 | 5+6 2 | 3+4 2 |
| | 4+5 2 | 4+5 2 | 5+6 2 | 3+4 2 |
| | 6+7 2 | 9 2 | 3+4 2 | 5+6 2 |
| | 8 2 | 6+7 2 | 3+4 2 | 5+6 2 |
| | 9 2 | 8 2 | 5+6 2 | 3+4 2 |
| | 8 2 | 9 2 | 3+4 2 | 3+4 2 |
| | 8 2 | 9 2 | 5+6 2 | 5+6 2 |
| | 8 2 | 6+7 2 | 5+6 2 | 3+4 2 |
| | 4+5 2 | 9 2 | 3+4 2 | 3+4 2 |
| | 8 2 | 9 2 | 5+6 2 | 3+4 2 |
| | 4+5 2 | 8 2 | 3+4 2 | 5+6 2 |
| | 9 2 | 9 2 | 5+6 2 | 3+4 2 |
| | 8 2 | 4+5 2 | 5+6 2 | 5+6 2 |
| | 6+7 2 | 6+7 2 | 5+6 2 | 3+4 2 |
| | 8 2 | 8 2 | 3+4 2 | 5+6 2 |
| | 4+5 2 | 6+7 2 | 5+6 2 | 3+4 2 |
| | 8 2 | 9 2 | 3+4 2 | 5+6 2 |
| | 9 2 | 8 2 | 5+6 2 | 3+4 2 |
| | 8 2 | 9 2 | 3+4 2 | 5+6 2 |
| | 4+5 2 | 9 2 | 3+4 2 | 5+6 2 |
| | 9 2 | 6+7 2 | 5+6 2 | 3+4 2 |
| | 8 2 | 8 2 | 5+6 2 | 3+4 2 |
| | 6+7 2 | 4+5 2 | 3+4 2 | 3+4 2 |
| | 8 2 | 6+7 2 | 5+6 2 | 5+6 2 |
| | 8 2 | 9 2 | 5+6 2 | 3+4 2 |
| | 9 2 | 9 2 | 3+4 2 | 5+6 2 |
| | 4+5 2 | 8 2 | 3+4 2 | 3+4 2 |
| | 9 2 | 6+7 2 | 5+6 2 | 5+6 2 |
| | 9 2 | 8 2 | 5+6 2 | 3+4 2 |
| | 4+5 2 | 4+5 2 | 3+4 2 | 3+4 2 |
| | 8 2 | 9 2 | 5+6 2 | 5+6 2 |
| | 6+7 2 | 8 2 | 3+4 2 | 5+6 2 |
| | 8 2 | 9 2 | 3+4 2 | 5+6 2 |
| | 9 2 | 8 2 | 5+6 2 | 3+4 2 |
| | 8 2 | 6+7 2 | 5+6 2 | 5+6 2 |
| | 6+7 2 | 9 2 | 3+4 2 | 5+6 2 |
| | 9 2 | 6+7 2 | 5+6 2 | 3+4 2 |
| | 4+5 2 | 4+5 2 | 3+4 2 | 5+6 2 |
| | 6+7 2 | 9 2 | 3+4 2 | 3+4 2 |
| | 4+5 2 | 8 2 | 5+6 2 | 5+6 2 |
| | 9 2 | 8 2 | 3+4 2 | 3+4 2 |
| | 6+7 2 | 4+5 2 | 3+4 2 | 3+4 2 |
| | 8 2 | 8 2 | 5+6 2 | 5+6 2 |
| | 8 2 | 9 2 | 5+6 2 | 5+6 2 |
| | 9 2 | 6+7 2 | 3+4 2 | 3+4 2 |
| | 6+7 2 | 9 2 | 5+6 2 | 3+4 2 |
| | 8 2 | 4+5 2 | 3+4 2 | 5+6 2 |
| | 9 2 | 9 2 | 5+6 2 | 3+4 2 |
| | 8 2 | 8 2 | 3+4 2 | 3+4 2 |

Continued on next column

**TABLE LXXIII– continued**

| TQC | ICSD | NoE | SF | |
|---|---|---|---|---|
| | 8 2 | 6+7 2 | 5+6 2 | 5+6 2 |
| | 9 2 | 9 2 | 5+6 2 | 5+6 2 |
| | 4+5 2 | 8 2 | 3+4 2 | 3+4 2 |
| | 8 2 | 4+5 2 | 3+4 2 | 5+6 2 |
| | 9 2 | 6+7 2 | 5+6 2 | 3+4 2 |
| | 4+5 2 | 9 2 | 3+4 2 | 3+4 2 |
| | 6+7 2 | 8 2 | 5+6 2 | 5+6 2 |
| | 8 2 | 8 2 | 5+6 2 | 3+4 2 |
| | 9 2 | 9 2 | 3+4 2 | 5+6 2 |
| | 8 2 | 4+5 2 | 5+6 2 | 5+6 2 |
| | 6+7 2 | 8 2 | 3+4 2 | 3+4 2 |
| | 8 2 | 9 2 | 3+4 2 | 5+6 2 |
| | 4+5 2 | 6+7 2 | 5+6 2 | 3+4 2 |
| | 9 2 | 8 2 | 3+4 2 | 3+4 2 |
| | 8 2 | 9 2 | 5+6 2 | 5+6 2 |
| | 6+7 2 | 9 2 | 3+4 2 | 3+4 2 |
| | 9 2 | 4+5 2 | 5+6 2 | 5+6 2 |
| | 4+5 2 | 6+7 2 | 3+4 2 | 3+4 2 |
| | 8 2 | 9 2 | 5+6 2 | 5+6 2 |
| | 9 2 | 8 2 | 3+4 2 | 3+4 2 |
| | 4+5 2 | 9 2 | 5+6 2 | 5+6 2 |
| | 6+7 2 | 4+5 2 | 5+6 2 | 3+4 2 |
| | 9 2 | 8 2 | 3+4 2 | 5+6 2 |
| | 8 2 | 9 2 | 5+6 2 | 3+4 2 |
| | 8 2 | 8 2 | 3+4 2 | 5+6 2 |
| | 9 2 | 6+7 2 | 5+6 2 | 5+6 2 |
| | 4+5 2 | 8 2 | 3+4 2 | 3+4 2 |
| | 9 2 | 4+5 2 | 3+4 2 | 3+4 2 |
| | 8 2 | 9 2 | 5+6 2 | 5+6 2 |
| | 8 2 | 8 2 | 3+4 2 | 3+4 2 |
| | 9 2 | 6+7 2 | 5+6 2 | 3+4 2 |
| | 8 2 | 4+5 2 | 3+4 2 | 5+6 2 |
| | 9 2 | 8 2 | 5+6 2 | 3+4 2 |
| | 6+7 2 | 9 2 | 3+4 2 | 5+6 2 |
| | 8 2 | 9 2 | 5+6 2 | 5+6 2 |
| | 4+5 2 | 9 2 | 5+6 2 | 3+4 2 |
| | 9 2 | 6+7 2 | 3+4 2 | 5+6 2 |
| | 8 2 | 8 2 | 5+6 2 | 3+4 2 |
| | 4+5 2 | 9 2 | 5+6 2 | 3+4 2 |
| | 8 2 | 9 2 | 3+4 2 | 5+6 2 |
| | 4+5 2 | 6+7 2 | 3+4 2 | 3+4 2 |
| | 8 2 | 9 2 | 5+6 2 | 5+6 2 |
| | 9 2 | 9 2 | 5+6 2 | 3+4 2 |
| | 6+7 2 | 9 2 | 5+6 2 | 3+4 2 |
| | 8 2 | 6+7 2 | 3+4 2 | 5+6 2 |
| | 4+5 2 | 8 2 | 5+6 2 | 5+6 2 |
| | 8 2 | 4+5 2 | 3+4 2 | 3+4 2 |
| | 9 2 | 8 2 | 5+6 2 | 3+4 2 |
| | 6+7 2 | 9 2 | 3+4 2 | 5+6 2 |
| | 9 2 | 8 2 | 3+4 2 | 5+6 2 |
| | 8 2 | 8 2 | 5+6 2 | 3+4 2 |
| | 9 2 | 4+5 2 | 5+6 2 | 5+6 2 |
| | 4+5 2 | 6+7 2 | 3+4 2 | 3+4 2 |
| | 9 2 | 9 2 | 3+4 2 | 3+4 2 |
| | 8 2 | 4+5 2 | 5+6 2 | 5+6 2 |
| | 4+5 2 | 8 2 | 5+6 2 | 5+6 2 |
| | 8 2 | 4+5 2 | 5+6 2 | 3+4 2 |
| | 8 2 | 8 2 | 3+4 2 | 5+6 2 |
| | 9 2 | 9 2 | 3+4 2 | 3+4 2 |
| | 9 2 | 8 2 | 3+4 2 | 5+6 2 |
| | 6+7 2 | 9 2 | 3+4 2 | 3+4 2 |

Continued on next column

**TABLE LXXIII– continued**

| TQC | ICSD | NoE | SF | |
|---|---|---|---|---|
| | 8 2 | 6+7 2 | 5+6 2 | 5+6 2 |
| | 4+5 2 | 9 2 | 5+6 2 | 3+4 2 |
| | 6+7 2 | 6+7 2 | 5+6 2 | 3+4 2 |
| 166.1.109 | 645562 | 162 | Nb6Zn7 | |
| | -GM | -T | -F | -L |
| | 8 2 | 9 2 | 5+6 2 | 3+4 2 |
| | 9 2 | 8 2 | 3+4 2 | 5+6 2 |
| | 8 2 | 8 2 | 3+4 2 | 5+6 2 |
| | 9 2 | 9 2 | 5+6 2 | 3+4 2 |
| | 8 2 | 8 2 | 3+4 2 | 5+6 2 |
| | 9 2 | 9 2 | 5+6 2 | 3+4 2 |
| | 8 2 | 9 2 | 5+6 2 | 3+4 2 |
| | 8 2 | 8 2 | 3+4 2 | 5+6 2 |
| | 9 2 | 9 2 | 3+4 2 | 5+6 2 |
| | 8 2 | 8 2 | 5+6 2 | 3+4 2 |
| | 9 2 | 8 2 | 5+6 2 | 5+6 2 |
| | 9 2 | 9 2 | 3+4 2 | 3+4 2 |
| | 4+5 2 | 6+7 2 | 5+6 2 | 3+4 2 |
| | 8 2 | 9 2 | 3+4 2 | 5+6 2 |
| | 4+5 2 | 4+5 2 | 3+4 2 | 5+6 2 |
| | 8 2 | 8 2 | 5+6 2 | 3+4 2 |
| | 6+7 2 | 9 2 | 5+6 2 | 5+6 2 |
| | 9 2 | 4+5 2 | 3+4 2 | 3+4 2 |
| | 9 2 | 8 2 | 5+6 2 | 3+4 2 |
| | 4+5 2 | 6+7 2 | 5+6 2 | 5+6 2 |
| | 8 2 | 9 2 | 3+4 2 | 5+6 2 |
| | 6+7 2 | 6+7 2 | 3+4 2 | 3+4 2 |
| | 6+7 2 | 4+5 2 | 3+4 2 | 5+6 2 |
| | 9 2 | 8 2 | 5+6 2 | 3+4 2 |
| | 8 2 | 8 2 | 3+4 2 | 3+4 2 |
| | 9 2 | 9 2 | 5+6 2 | 5+6 2 |
| | 8 2 | 8 2 | 3+4 2 | 3+4 2 |
| | 8 2 | 8 2 | 3+4 2 | 3+4 2 |
| | 8 2 | 4+5 2 | 5+6 2 | 5+6 2 |
| | 4+5 2 | 8 2 | 3+4 2 | 3+4 2 |
| | 9 2 | 9 2 | 3+4 2 | 3+4 2 |
| | 6+7 2 | 6+7 2 | 5+6 2 | 5+6 2 |
| | 9 2 | 9 2 | 3+4 2 | 3+4 2 |
| | 9 2 | 9 2 | 3+4 2 | 3+4 2 |
| | 8 2 | 8 2 | 5+6 2 | 5+6 2 |
| | 4+5 2 | 4+5 2 | 5+6 2 | 5+6 2 |
| | 8 2 | 8 2 | 3+4 2 | 3+4 2 |
| | 6+7 2 | 6+7 2 | 5+6 2 | 5+6 2 |
| | 4+5 2 | 4+5 2 | 3+4 2 | 3+4 2 |
| | 9 2 | 9 2 | 5+6 2 | 5+6 2 |
| | 8 2 | 8 2 | 3+4 2 | 3+4 2 |
| | 9 2 | 9 2 | 3+4 2 | 3+4 2 |
| | 8 2 | 8 2 | 5+6 2 | 5+6 2 |
| | 6+7 2 | 9 2 | 5+6 2 | 5+6 2 |
| | 4+5 2 | 4+5 2 | 3+4 2 | 3+4 2 |
| | 9 2 | 6+7 2 | 5+6 2 | 5+6 2 |
| | 4+5 2 | 4+5 2 | 3+4 2 | 3+4 2 |
| | 6+7 2 | 8 2 | 5+6 2 | 5+6 2 |
| | 8 2 | 6+7 2 | 5+6 2 | 5+6 2 |
| | 9 2 | 8 2 | 3+4 2 | 3+4 2 |
| | 8 2 | 9 2 | 3+4 2 | 3+4 2 |
| | 8 2 | 8 2 | 5+6 2 | 5+6 2 |
| | 8 2 | 9 2 | 3+4 2 | 3+4 2 |
| | 9 2 | 8 2 | 3+4 2 | 3+4 2 |
| | 4+5 2 | 9 2 | 3+4 2 | 5+6 2 |
| | 9 2 | 4+5 2 | 5+6 2 | 3+4 2 |

Continued on next column

**TABLE LXXIII– continued**

| TQC | ICSD | NoE | SF | |
|---|---|---|---|---|
| | 6+7 2 | 6+7 2 | 3+4 2 | 3+4 2 |
| | 8 2 | 8 2 | 5+6 2 | 5+6 2 |
| | 4+5 2 | 4+5 2 | 3+4 2 | 3+4 2 |
| | 9 2 | 9 2 | 5+6 2 | 5+6 2 |
| | 8 2 | 8 2 | 3+4 2 | 3+4 2 |
| | 8 2 | 8 2 | 5+6 2 | 5+6 2 |
| | 9 2 | 9 2 | 3+4 2 | 3+4 2 |
| | 6+7 2 | 8 2 | 3+4 2 | 3+4 2 |
| | 9 2 | 4+5 2 | 3+4 2 | 5+6 2 |
| | 8 2 | 9 2 | 5+6 2 | 3+4 2 |
| | 9 2 | 6+7 2 | 3+4 2 | 3+4 2 |
| | 6+7 2 | 8 2 | 5+6 2 | 5+6 2 |
| | 8 2 | 9 2 | 5+6 2 | 5+6 2 |
| | 9 2 | 9 2 | 5+6 2 | 3+4 2 |
| | 6+7 2 | 6+7 2 | 3+4 2 | 5+6 2 |
| | 9 2 | 8 2 | 3+4 2 | 5+6 2 |
| | 8 2 | 4+5 2 | 5+6 2 | 3+4 2 |
| | 4+5 2 | 9 2 | 3+4 2 | 5+6 2 |
| | 9 2 | 9 2 | 5+6 2 | 5+6 2 |
| | 8 2 | 6+7 2 | 5+6 2 | 5+6 2 |
| | 6+7 2 | 8 2 | 3+4 2 | 3+4 2 |
| | 4+5 2 | 4+5 2 | 5+6 2 | 3+4 2 |
| | 8 2 | 8 2 | 3+4 2 | 3+4 2 |
| | 8 2 | 4+5 2 | 5+6 2 | 3+4 2 |
| | 4+5 2 | 8 2 | 3+4 2 | 5+6 2 |
| 166.1.110 | 30364 | 16 | N2W1 | |
| | -GM | -T | -F | -L |
| | 8 2 | 9 2 | 5+6 2 | 3+4 2 |
| | 9 2 | 8 2 | 3+4 2 | 5+6 2 |
| | 8 2 | 8 2 | 5+6 2 | 3+4 2 |
| | 8 2 | 9 2 | 3+4 2 | 5+6 2 |
| | 4+5 2 | 6+7 2 | 3+4 2 | 3+4 2 |
| | 9 2 | 8 2 | 5+6 2 | 3+4 2 |
| | 8 2 | 4+5 2 | 3+4 2 | 5+6 2 |
| | 4+5 2 | 8 2 | 3+4 2 | 3+4 2 |
| 166.1.111 | 44926 | 38 | Mg2Ni3Si1 | |
| | -GM | -T | -F | -L |
| | 8 2 | 9 2 | 3+4 2 | 5+6 2 |
| | 8 2 | 8 2 | 5+6 2 | 3+4 2 |
| | 8 2 | 8 2 | 5+6 2 | 3+4 2 |
| | 4+5 2 | 4+5 2 | 5+6 2 | 5+6 2 |
| | 8 2 | 8 2 | 3+4 2 | 3+4 2 |
| | 8 2 | 8 2 | 5+6 2 | 5+6 2 |
| | 4+5 2 | 4+5 2 | 3+4 2 | 5+6 2 |
| | 8 2 | 8 2 | 5+6 2 | 3+4 2 |
| | 9 2 | 9 2 | 5+6 2 | 5+6 2 |
| | 6+7 2 | 8 2 | 3+4 2 | 3+4 2 |
| | 9 2 | 4+5 2 | 5+6 2 | 5+6 2 |
| | 4+5 2 | 8 2 | 3+4 2 | 3+4 2 |
| | 8 2 | 8 2 | 5+6 2 | 5+6 2 |
| | 8 2 | 4+5 2 | 3+4 2 | 5+6 2 |
| | 8 2 | 8 2 | 5+6 2 | 5+6 2 |
| | 4+5 2 | 4+5 2 | 3+4 2 | 5+6 2 |
| | 4+5 2 | 8 2 | 5+6 2 | 5+6 2 |
| | 8 2 | 6+7 2 | 5+6 2 | 3+4 2 |
| | 8 2 | 9 2 | 3+4 2 | 3+4 2 |
| 166.1.112 | 55617 | 108 | Mg2Nd1Ni9 | |
| | -GM | -T | -F | -L |
| | 8 2 | 8 2 | 3+4 2 | 3+4 2 |
| | 9 2 | 9 2 | 5+6 2 | 5+6 2 |
| | 6+7 2 | 6+7 2 | 5+6 2 | 5+6 2 |

Continued on next column

**TABLE LXXIII– continued**

| TQC | ICSD | NoE | SF | |
|---|---|---|---|---|
| | 9 2 | 9 2 | 5+6 2 | 5+6 2 |
| | 8 2 | 9 2 | 3+4 2 | 5+6 2 |
| | 9 2 | 8 2 | 5+6 2 | 3+4 2 |
| | 8 2 | 9 2 | 3+4 2 | 3+4 2 |
| | 8 2 | 8 2 | 5+6 2 | 5+6 2 |
| | 9 2 | 9 2 | 3+4 2 | 3+4 2 |
| | 8 2 | 8 2 | 3+4 2 | 5+6 2 |
| | 8 2 | 9 2 | 3+4 2 | 3+4 2 |
| | 4+5 2 | 6+7 2 | 3+4 2 | 5+6 2 |
| | 9 2 | 8 2 | 5+6 2 | 5+6 2 |
| | 6+7 2 | 4+5 2 | 3+4 2 | 3+4 2 |
| | 9 2 | 8 2 | 5+6 2 | 5+6 2 |
| | 8 2 | 6+7 2 | 5+6 2 | 3+4 2 |
| | 4+5 2 | 9 2 | 3+4 2 | 5+6 2 |
| | 8 2 | 4+5 2 | 5+6 2 | 3+4 2 |
| | 4+5 2 | 8 2 | 3+4 2 | 5+6 2 |
| | 8 2 | 9 2 | 5+6 2 | 5+6 2 |
| | 6+7 2 | 4+5 2 | 3+4 2 | 5+6 2 |
| | 9 2 | 8 2 | 3+4 2 | 3+4 2 |
| | 9 2 | 9 2 | 5+6 2 | 5+6 2 |
| | 6+7 2 | 6+7 2 | 5+6 2 | 3+4 2 |
| | 4+5 2 | 9 2 | 5+6 2 | 3+4 2 |
| | 8 2 | 9 2 | 3+4 2 | 3+4 2 |
| | 9 2 | 6+7 2 | 3+4 2 | 5+6 2 |
| | 8 2 | 8 2 | 3+4 2 | 3+4 2 |
| | 4+5 2 | 6+7 2 | 5+6 2 | 5+6 2 |
| | 8 2 | 9 2 | 3+4 2 | 5+6 2 |
| | 6+7 2 | 4+5 2 | 5+6 2 | 3+4 2 |
| | 9 2 | 8 2 | 5+6 2 | 5+6 2 |
| | 8 2 | 9 2 | 3+4 2 | 3+4 2 |
| | 9 2 | 4+5 2 | 5+6 2 | 5+6 2 |
| | 4+5 2 | 8 2 | 3+4 2 | 3+4 2 |
| | 8 2 | 6+7 2 | 5+6 2 | 5+6 2 |
| | 9 2 | 9 2 | 3+4 2 | 3+4 2 |
| | 6+7 2 | 8 2 | 3+4 2 | 5+6 2 |
| | 9 2 | 9 2 | 5+6 2 | 3+4 2 |
| | 4+5 2 | 8 2 | 5+6 2 | 5+6 2 |
| | 8 2 | 4+5 2 | 3+4 2 | 3+4 2 |
| | 8 2 | 6+7 2 | 5+6 2 | 5+6 2 |
| | 6+7 2 | 9 2 | 3+4 2 | 5+6 2 |
| | 9 2 | 8 2 | 5+6 2 | 3+4 2 |
| | 4+5 2 | 9 2 | 3+4 2 | 3+4 2 |
| | 9 2 | 9 2 | 3+4 2 | 5+6 2 |
| | 8 2 | 8 2 | 5+6 2 | 5+6 2 |
| | 6+7 2 | 4+5 2 | 5+6 2 | 5+6 2 |
| | 9 2 | 6+7 2 | 3+4 2 | 3+4 2 |
| | 8 2 | 9 2 | 3+4 2 | 5+6 2 |
| | 4+5 2 | 6+7 2 | 3+4 2 | 5+6 2 |
| | 9 2 | 9 2 | 5+6 2 | 3+4 2 |
| | 6+7 2 | 9 2 | 5+6 2 | 5+6 2 |
| | 9 2 | 6+7 2 | 5+6 2 | 5+6 2 |
| 166.1.113 | 41369 | 20 | Li8Pb3 | |
| | -GM | -T | -F | -L |
| | 8 2 | 9 2 | 3+4 2 | 5+6 2 |
| | 8 2 | 8 2 | 3+4 2 | 3+4 2 |
| | 9 2 | 8 2 | 5+6 2 | 3+4 2 |
| | 8 2 | 9 2 | 5+6 2 | 3+4 2 |
| | 9 2 | 8 2 | 5+6 2 | 5+6 2 |
| | 9 2 | 9 2 | 3+4 2 | 5+6 2 |
| | 6+7 2 | 9 2 | 3+4 2 | 5+6 2 |
| | 9 2 | 4+5 2 | 5+6 2 | 3+4 2 |

Continued on next column

**TABLE LXXIII– continued**

| TQC | ICSD | NoE | SF | |
|---|---|---|---|---|
| | 8 2 | 8 2 | 5+6 2 | 5+6 2 |
| | 6+7 2 | 6+7 2 | 5+6 2 | 5+6 2 |
| 166.1.114 | 191137 | 146 | La3Mn2O14Sb3 | |
| | -GM | -T | -F | -L |
| | 8 2 | 8 2 | 5+6 2 | 5+6 2 |
| | 4+5 2 | 4+5 2 | 3+4 2 | 3+4 2 |
| | 8 2 | 8 2 | 5+6 2 | 5+6 2 |
| | 8 2 | 9 2 | 5+6 2 | 3+4 2 |
| | 9 2 | 6+7 2 | 3+4 2 | 5+6 2 |
| | 6+7 2 | 9 2 | 5+6 2 | 3+4 2 |
| | 9 2 | 9 2 | 3+4 2 | 5+6 2 |
| | 9 2 | 4+5 2 | 3+4 2 | 3+4 2 |
| | 4+5 2 | 8 2 | 3+4 2 | 3+4 2 |
| | 8 2 | 8 2 | 3+4 2 | 5+6 2 |
| | 6+7 2 | 8 2 | 5+6 2 | 5+6 2 |
| | 9 2 | 6+7 2 | 3+4 2 | 5+6 2 |
| | 9 2 | 4+5 2 | 5+6 2 | 5+6 2 |
| | 8 2 | 8 2 | 5+6 2 | 3+4 2 |
| | 8 2 | 9 2 | 5+6 2 | 3+4 2 |
| | 4+5 2 | 8 2 | 3+4 2 | 3+4 2 |
| | 8 2 | 9 2 | 5+6 2 | 5+6 2 |
| | 6+7 2 | 6+7 2 | 3+4 2 | 3+4 2 |
| | 9 2 | 9 2 | 5+6 2 | 5+6 2 |
| | 9 2 | 9 2 | 3+4 2 | 3+4 2 |
| | 9 2 | 9 2 | 3+4 2 | 3+4 2 |
| | 6+7 2 | 6+7 2 | 5+6 2 | 3+4 2 |
| | 9 2 | 9 2 | 3+4 2 | 5+6 2 |
| | 9 2 | 9 2 | 3+4 2 | 3+4 2 |
| | 6+7 2 | 6+7 2 | 5+6 2 | 5+6 2 |
| | 9 2 | 9 2 | 3+4 2 | 3+4 2 |
| | 4+5 2 | 6+7 2 | 5+6 2 | 3+4 2 |
| | 8 2 | 9 2 | 3+4 2 | 5+6 2 |
| | 9 2 | 8 2 | 5+6 2 | 3+4 2 |
| | 8 2 | 8 2 | 3+4 2 | 5+6 2 |
| | 9 2 | 9 2 | 3+4 2 | 5+6 2 |
| | 6+7 2 | 4+5 2 | 5+6 2 | 3+4 2 |
| | 9 2 | 8 2 | 5+6 2 | 5+6 2 |
| | 9 2 | 8 2 | 3+4 2 | 5+6 2 |
| | 6+7 2 | 4+5 2 | 3+4 2 | 3+4 2 |
| | 8 2 | 8 2 | 5+6 2 | 3+4 2 |
| | 8 2 | 9 2 | 3+4 2 | 5+6 2 |
| | 9 2 | 4+5 2 | 3+4 2 | 5+6 2 |
| | 4+5 2 | 8 2 | 3+4 2 | 3+4 2 |
| | 8 2 | 9 2 | 5+6 2 | 5+6 2 |
| | 8 2 | 6+7 2 | 5+6 2 | 3+4 2 |
| | 4+5 2 | 4+5 2 | 5+6 2 | 5+6 2 |
| | 8 2 | 8 2 | 5+6 2 | 5+6 2 |
| | 8 2 | 8 2 | 3+4 2 | 3+4 2 |
| | 4+5 2 | 9 2 | 3+4 2 | 5+6 2 |
| | 6+7 2 | 8 2 | 5+6 2 | 3+4 2 |
| | 9 2 | 4+5 2 | 5+6 2 | 3+4 2 |
| | 8 2 | 9 2 | 3+4 2 | 3+4 2 |
| | 8 2 | 8 2 | 3+4 2 | 5+6 2 |
| | 4+5 2 | 6+7 2 | 5+6 2 | 5+6 2 |
| | 9 2 | 9 2 | 3+4 2 | 3+4 2 |
| | 6+7 2 | 8 2 | 5+6 2 | 5+6 2 |
| | 4+5 2 | 9 2 | 3+4 2 | 5+6 2 |
| | 8 2 | 4+5 2 | 3+4 2 | 3+4 2 |
| | 9 2 | 8 2 | 5+6 2 | 3+4 2 |
| | 9 2 | 8 2 | 3+4 2 | 5+6 2 |
| | 8 2 | 4+5 2 | 5+6 2 | 5+6 2 |

Continued on next column

**TABLE LXXIII– continued**

| TQC | ICSD | NoE | SF | |
|---|---|---|---|---|
| | 8 2 | 6+7 2 | 5+6 2 | 3+4 2 |
| | 6+7 2 | 9 2 | 5+6 2 | 5+6 2 |
| | 9 2 | 8 2 | 3+4 2 | 3+4 2 |
| | 9 2 | 6+7 2 | 3+4 2 | 3+4 2 |
| | 4+5 2 | 9 2 | 3+4 2 | 5+6 2 |
| | 8 2 | 9 2 | 5+6 2 | 3+4 2 |
| | 9 2 | 6+7 2 | 5+6 2 | 5+6 2 |
| | 9 2 | 9 2 | 3+4 2 | 3+4 2 |
| | 6+7 2 | 9 2 | 5+6 2 | 3+4 2 |
| | 6+7 2 | 6+7 2 | 3+4 2 | 5+6 2 |
| | 9 2 | 9 2 | 5+6 2 | 3+4 2 |
| | 8 2 | 8 2 | 3+4 2 | 3+4 2 |
| | 8 2 | 4+5 2 | 3+4 2 | 5+6 2 |
| | 4+5 2 | 8 2 | 3+4 2 | 5+6 2 |
| | 4+5 2 | 6+7 2 | 3+4 2 | 3+4 2 |
| | 8 2 | 9 2 | 3+4 2 | 5+6 2 |
| 166.1.115 | 641738 | 114 | La1Rh3 | |
| | -GM | -T | -F | -L |
| | 8 2 | 9 2 | 3+4 2 | 3+4 2 |
| | 8 2 | 8 2 | 5+6 2 | 3+4 2 |
| | 9 2 | 8 2 | 3+4 2 | 5+6 2 |
| | 8 2 | 9 2 | 5+6 2 | 3+4 2 |
| | 9 2 | 9 2 | 5+6 2 | 5+6 2 |
| | 9 2 | 8 2 | 3+4 2 | 5+6 2 |
| | 4+5 2 | 6+7 2 | 5+6 2 | 3+4 2 |
| | 8 2 | 9 2 | 3+4 2 | 5+6 2 |
| | 9 2 | 6+7 2 | 5+6 2 | 5+6 2 |
| | 6+7 2 | 9 2 | 5+6 2 | 5+6 2 |
| | 9 2 | 8 2 | 5+6 2 | 3+4 2 |
| | 6+7 2 | 4+5 2 | 3+4 2 | 5+6 2 |
| | 8 2 | 8 2 | 3+4 2 | 5+6 2 |
| | 9 2 | 9 2 | 5+6 2 | 3+4 2 |
| | 8 2 | 9 2 | 3+4 2 | 3+4 2 |
| | 9 2 | 8 2 | 3+4 2 | 5+6 2 |
| | 8 2 | 9 2 | 3+4 2 | 3+4 2 |
| | 9 2 | 6+7 2 | 5+6 2 | 3+4 2 |
| | 8 2 | 8 2 | 3+4 2 | 5+6 2 |
| | 6+7 2 | 9 2 | 3+4 2 | 5+6 2 |
| | 4+5 2 | 4+5 2 | 5+6 2 | 5+6 2 |
| | 4+5 2 | 8 2 | 3+4 2 | 5+6 2 |
| | 8 2 | 4+5 2 | 5+6 2 | 3+4 2 |
| | 8 2 | 8 2 | 3+4 2 | 3+4 2 |
| | 4+5 2 | 8 2 | 5+6 2 | 5+6 2 |
| | 8 2 | 9 2 | 3+4 2 | 5+6 2 |
| | 9 2 | 6+7 2 | 5+6 2 | 3+4 2 |
| | 9 2 | 4+5 2 | 3+4 2 | 5+6 2 |
| | 6+7 2 | 8 2 | 5+6 2 | 5+6 2 |
| | 8 2 | 9 2 | 5+6 2 | 3+4 2 |
| | 9 2 | 6+7 2 | 3+4 2 | 5+6 2 |
| | 6+7 2 | 9 2 | 3+4 2 | 5+6 2 |
| | 9 2 | 9 2 | 5+6 2 | 3+4 2 |
| | 4+5 2 | 6+7 2 | 3+4 2 | 3+4 2 |
| | 8 2 | 9 2 | 5+6 2 | 5+6 2 |
| | 8 2 | 8 2 | 3+4 2 | 5+6 2 |
| | 4+5 2 | 6+7 2 | 5+6 2 | 3+4 2 |
| | 8 2 | 9 2 | 3+4 2 | 3+4 2 |
| | 6+7 2 | 4+5 2 | 3+4 2 | 3+4 2 |
| | 9 2 | 8 2 | 5+6 2 | 5+6 2 |
| | 8 2 | 8 2 | 5+6 2 | 3+4 2 |
| | 6+7 2 | 4+5 2 | 5+6 2 | 5+6 2 |
| | 9 2 | 6+7 2 | 3+4 2 | 5+6 2 |

Continued on next column

**TABLE LXXIII– continued**

| TQC | ICSD | NoE | SF | |
|---|---|---|---|---|
| | 9 2 | 9 2 | 5+6 2 | 3+4 2 |
| | 4+5 2 | 9 2 | 3+4 2 | 3+4 2 |
| | 8 2 | 8 2 | 3+4 2 | 5+6 2 |
| | 9 2 | 9 2 | 5+6 2 | 3+4 2 |
| | 8 2 | 6+7 2 | 3+4 2 | 5+6 2 |
| | 4+5 2 | 4+5 2 | 5+6 2 | 5+6 2 |
| | 8 2 | 9 2 | 3+4 2 | 3+4 2 |
| | 6+7 2 | 8 2 | 5+6 2 | 5+6 2 |
| | 9 2 | 8 2 | 3+4 2 | 5+6 2 |
| | 4+5 2 | 9 2 | 3+4 2 | 3+4 2 |
| | 8 2 | 8 2 | 5+6 2 | 3+4 2 |
| | 9 2 | 6+7 2 | 5+6 2 | 5+6 2 |
| | 6+7 2 | 4+5 2 | 3+4 2 | 5+6 2 |
| | 8 2 | 8 2 | 5+6 2 | 3+4 2 |
| 166.1.116 | 641724 | 114 | La1Rh3 | |
| | -GM | -T | -F | -L |
| | 8 2 | 8 2 | 3+4 2 | 3+4 2 |
| | 8 2 | 9 2 | 5+6 2 | 3+4 2 |
| | 9 2 | 8 2 | 3+4 2 | 5+6 2 |
| | 8 2 | 9 2 | 5+6 2 | 3+4 2 |
| | 9 2 | 9 2 | 5+6 2 | 5+6 2 |
| | 9 2 | 8 2 | 3+4 2 | 5+6 2 |
| | 4+5 2 | 6+7 2 | 5+6 2 | 3+4 2 |
| | 8 2 | 9 2 | 3+4 2 | 5+6 2 |
| | 9 2 | 6+7 2 | 5+6 2 | 5+6 2 |
| | 6+7 2 | 9 2 | 5+6 2 | 5+6 2 |
| | 9 2 | 8 2 | 5+6 2 | 3+4 2 |
| | 6+7 2 | 4+5 2 | 3+4 2 | 5+6 2 |
| | 8 2 | 8 2 | 3+4 2 | 5+6 2 |
| | 9 2 | 9 2 | 5+6 2 | 3+4 2 |
| | 8 2 | 9 2 | 3+4 2 | 3+4 2 |
| | 9 2 | 8 2 | 3+4 2 | 5+6 2 |
| | 8 2 | 9 2 | 3+4 2 | 3+4 2 |
| | 9 2 | 6+7 2 | 5+6 2 | 3+4 2 |
| | 8 2 | 8 2 | 3+4 2 | 5+6 2 |
| | 6+7 2 | 9 2 | 3+4 2 | 5+6 2 |
| | 4+5 2 | 4+5 2 | 5+6 2 | 5+6 2 |
| | 4+5 2 | 8 2 | 3+4 2 | 5+6 2 |
| | 8 2 | 4+5 2 | 5+6 2 | 3+4 2 |
| | 8 2 | 8 2 | 3+4 2 | 3+4 2 |
| | 4+5 2 | 8 2 | 5+6 2 | 5+6 2 |
| | 8 2 | 9 2 | 3+4 2 | 5+6 2 |
| | 9 2 | 6+7 2 | 5+6 2 | 3+4 2 |
| | 9 2 | 4+5 2 | 3+4 2 | 5+6 2 |
| | 6+7 2 | 9 2 | 5+6 2 | 5+6 2 |
| | 8 2 | 8 2 | 5+6 2 | 3+4 2 |
| | 9 2 | 6+7 2 | 3+4 2 | 5+6 2 |
| | 6+7 2 | 9 2 | 3+4 2 | 5+6 2 |
| | 9 2 | 9 2 | 5+6 2 | 3+4 2 |
| | 4+5 2 | 6+7 2 | 3+4 2 | 3+4 2 |
| | 8 2 | 9 2 | 5+6 2 | 5+6 2 |
| | 8 2 | 8 2 | 3+4 2 | 5+6 2 |
| | 4+5 2 | 6+7 2 | 5+6 2 | 3+4 2 |
| | 8 2 | 9 2 | 3+4 2 | 3+4 2 |
| | 6+7 2 | 4+5 2 | 3+4 2 | 3+4 2 |
| | 9 2 | 8 2 | 5+6 2 | 5+6 2 |
| | 8 2 | 8 2 | 5+6 2 | 3+4 2 |
| | 6+7 2 | 4+5 2 | 5+6 2 | 5+6 2 |
| | 9 2 | 6+7 2 | 3+4 2 | 5+6 2 |
| | 9 2 | 9 2 | 5+6 2 | 3+4 2 |
| | 4+5 2 | 9 2 | 3+4 2 | 3+4 2 |

Continued on next column

**TABLE LXXIII– continued**

| TQC | ICSD | NoE | SF | |
|---|---|---|---|---|
| | 8 2 | 8 2 | 3+4 2 | 5+6 2 |
| | 9 2 | 9 2 | 5+6 2 | 3+4 2 |
| | 8 2 | 6+7 2 | 3+4 2 | 5+6 2 |
| | 4+5 2 | 4+5 2 | 5+6 2 | 5+6 2 |
| | 8 2 | 9 2 | 3+4 2 | 3+4 2 |
| | 6+7 2 | 8 2 | 5+6 2 | 5+6 2 |
| | 9 2 | 8 2 | 3+4 2 | 5+6 2 |
| | 4+5 2 | 9 2 | 3+4 2 | 3+4 2 |
| | 8 2 | 8 2 | 5+6 2 | 3+4 2 |
| | 9 2 | 4+5 2 | 5+6 2 | 5+6 2 |
| | 6+7 2 | 6+7 2 | 3+4 2 | 5+6 2 |
| | 8 2 | 8 2 | 5+6 2 | 3+4 2 |
| 166.1.117 | 640862 | 182 | Ir7Nd2 | |
| | -GM | -T | -F | -L |
| | 8 2 | 8 2 | 3+4 2 | 3+4 2 |
| | 9 2 | 9 2 | 5+6 2 | 5+6 2 |
| | 8 2 | 9 2 | 5+6 2 | 3+4 2 |
| | 9 2 | 8 2 | 3+4 2 | 5+6 2 |
| | 8 2 | 9 2 | 5+6 2 | 3+4 2 |
| | 8 2 | 8 2 | 3+4 2 | 5+6 2 |
| | 9 2 | 9 2 | 5+6 2 | 3+4 2 |
| | 9 2 | 8 2 | 3+4 2 | 5+6 2 |
| | 4+5 2 | 6+7 2 | 5+6 2 | 3+4 2 |
| | 8 2 | 9 2 | 3+4 2 | 5+6 2 |
| | 9 2 | 8 2 | 5+6 2 | 5+6 2 |
| | 6+7 2 | 4+5 2 | 3+4 2 | 3+4 2 |
| | 4+5 2 | 6+7 2 | 3+4 2 | 3+4 2 |
| | 8 2 | 9 2 | 5+6 2 | 5+6 2 |
| | 9 2 | 8 2 | 5+6 2 | 3+4 2 |
| | 6+7 2 | 4+5 2 | 3+4 2 | 5+6 2 |
| | 8 2 | 8 2 | 5+6 2 | 5+6 2 |
| | 8 2 | 9 2 | 3+4 2 | 5+6 2 |
| | 9 2 | 8 2 | 3+4 2 | 3+4 2 |
| | 8 2 | 8 2 | 5+6 2 | 5+6 2 |
| | 8 2 | 8 2 | 5+6 2 | 3+4 2 |
| | 8 2 | 9 2 | 3+4 2 | 5+6 2 |
| | 9 2 | 9 2 | 5+6 2 | 5+6 2 |
| | 4+5 2 | 8 2 | 5+6 2 | 5+6 2 |
| | 9 2 | 9 2 | 3+4 2 | 5+6 2 |
| | 8 2 | 4+5 2 | 3+4 2 | 3+4 2 |
| | 8 2 | 6+7 2 | 5+6 2 | 3+4 2 |
| | 9 2 | 8 2 | 3+4 2 | 3+4 2 |
| | 4+5 2 | 4+5 2 | 3+4 2 | 5+6 2 |
| | 6+7 2 | 9 2 | 5+6 2 | 3+4 2 |
| | 8 2 | 8 2 | 3+4 2 | 5+6 2 |
| | 4+5 2 | 4+5 2 | 5+6 2 | 5+6 2 |
| | 8 2 | 8 2 | 5+6 2 | 5+6 2 |
| | 4+5 2 | 6+7 2 | 3+4 2 | 3+4 2 |
| | 8 2 | 8 2 | 5+6 2 | 3+4 2 |
| | 8 2 | 4+5 2 | 3+4 2 | 5+6 2 |
| | 9 2 | 9 2 | 5+6 2 | 5+6 2 |
| | 8 2 | 8 2 | 3+4 2 | 5+6 2 |
| | 4+5 2 | 6+7 2 | 5+6 2 | 3+4 2 |
| | 6+7 2 | 8 2 | 3+4 2 | 5+6 2 |
| | 9 2 | 9 2 | 3+4 2 | 3+4 2 |
| | 9 2 | 9 2 | 5+6 2 | 3+4 2 |
| | 6+7 2 | 4+5 2 | 5+6 2 | 5+6 2 |
| | 4+5 2 | 8 2 | 3+4 2 | 3+4 2 |
| | 8 2 | 4+5 2 | 3+4 2 | 3+4 2 |
| | 8 2 | 8 2 | 5+6 2 | 5+6 2 |
| | 9 2 | 9 2 | 3+4 2 | 3+4 2 |

Continued on next column

**TABLE LXXIII– continued**

| TQC | ICSD | NoE | SF | |
|---|---|---|---|---|
| | 4+5 2 | 8 2 | 3+4 2 | 5+6 2 |
| | 8 2 | 6+7 2 | 5+6 2 | 5+6 2 |
| | 6+7 2 | 9 2 | 3+4 2 | 3+4 2 |
| | 9 2 | 6+7 2 | 5+6 2 | 5+6 2 |
| | 8 2 | 9 2 | 3+4 2 | 3+4 2 |
| | 9 2 | 9 2 | 5+6 2 | 5+6 2 |
| | 4+5 2 | 8 2 | 3+4 2 | 3+4 2 |
| | 6+7 2 | 6+7 2 | 5+6 2 | 5+6 2 |
| | 9 2 | 6+7 2 | 5+6 2 | 5+6 2 |
| | 8 2 | 9 2 | 3+4 2 | 3+4 2 |
| | 9 2 | 4+5 2 | 5+6 2 | 5+6 2 |
| | 8 2 | 9 2 | 3+4 2 | 3+4 2 |
| | 6+7 2 | 4+5 2 | 5+6 2 | 5+6 2 |
| | 8 2 | 8 2 | 5+6 2 | 5+6 2 |
| | 4+5 2 | 8 2 | 3+4 2 | 3+4 2 |
| | 9 2 | 6+7 2 | 5+6 2 | 3+4 2 |
| | 6+7 2 | 9 2 | 3+4 2 | 3+4 2 |
| | 4+5 2 | 4+5 2 | 3+4 2 | 5+6 2 |
| | 8 2 | 8 2 | 5+6 2 | 3+4 2 |
| | 9 2 | 8 2 | 3+4 2 | 5+6 2 |
| | 4+5 2 | 4+5 2 | 3+4 2 | 3+4 2 |
| | 9 2 | 9 2 | 5+6 2 | 5+6 2 |
| | 8 2 | 8 2 | 5+6 2 | 3+4 2 |
| | 6+7 2 | 8 2 | 3+4 2 | 3+4 2 |
| | 8 2 | 9 2 | 5+6 2 | 5+6 2 |
| | 8 2 | 8 2 | 3+4 2 | 5+6 2 |
| | 4+5 2 | 4+5 2 | 5+6 2 | 5+6 2 |
| | 8 2 | 6+7 2 | 5+6 2 | 3+4 2 |
| | 9 2 | 8 2 | 3+4 2 | 5+6 2 |
| | 8 2 | 8 2 | 5+6 2 | 3+4 2 |
| | 8 2 | 9 2 | 3+4 2 | 5+6 2 |
| | 9 2 | 9 2 | 5+6 2 | 3+4 2 |
| | 4+5 2 | 6+7 2 | 3+4 2 | 5+6 2 |
| | 9 2 | 9 2 | 3+4 2 | 3+4 2 |
| | 8 2 | 8 2 | 5+6 2 | 5+6 2 |
| | 6+7 2 | 4+5 2 | 3+4 2 | 3+4 2 |
| | 8 2 | 9 2 | 5+6 2 | 5+6 2 |
| | 8 2 | 8 2 | 3+4 2 | 5+6 2 |
| | 9 2 | 8 2 | 5+6 2 | 3+4 2 |
| | 4+5 2 | 9 2 | 3+4 2 | 5+6 2 |
| | 6+7 2 | 8 2 | 5+6 2 | 3+4 2 |
| | 9 2 | 4+5 2 | 3+4 2 | 3+4 2 |
| | 8 2 | 6+7 2 | 5+6 2 | 5+6 2 |
| | 9 2 | 9 2 | 5+6 2 | 3+4 2 |
| 166.1.118 | 657607 | 24 | In2Te3 | |
| | -GM | -T | -F | -L |
| | 8 2 | 9 2 | 3+4 2 | 3+4 2 |
| | 8 2 | 8 2 | 5+6 2 | 3+4 2 |
| | 9 2 | 8 2 | 3+4 2 | 5+6 2 |
| | 9 2 | 9 2 | 5+6 2 | 5+6 2 |
| | 8 2 | 8 2 | 3+4 2 | 3+4 2 |
| | 8 2 | 9 2 | 5+6 2 | 5+6 2 |
| | 9 2 | 9 2 | 5+6 2 | 5+6 2 |
| | 4+5 2 | 9 2 | 3+4 2 | 3+4 2 |
| | 6+7 2 | 6+7 2 | 5+6 2 | 3+4 2 |
| | 9 2 | 8 2 | 3+4 2 | 5+6 2 |
| | 6+7 2 | 6+7 2 | 5+6 2 | 5+6 2 |
| | 9 2 | 4+5 2 | 5+6 2 | 3+4 2 |
| 166.1.119 | 640609 | 24 | In2Te3 | |
| | -GM | -T | -F | -L |
| | 8 2 | 8 2 | 3+4 2 | 3+4 2 |

Continued on next column

**TABLE LXXIII– continued**

| TQC | ICSD | NoE | SF | |
|---|---|---|---|---|
| | 8 2 | 9 2 | 5+6 2 | 3+4 2 |
| | 9 2 | 8 2 | 3+4 2 | 5+6 2 |
| | 9 2 | 9 2 | 5+6 2 | 5+6 2 |
| | 8 2 | 8 2 | 3+4 2 | 3+4 2 |
| | 9 2 | 9 2 | 5+6 2 | 5+6 2 |
| | 8 2 | 9 2 | 5+6 2 | 5+6 2 |
| | 9 2 | 9 2 | 3+4 2 | 3+4 2 |
| | 6+7 2 | 8 2 | 5+6 2 | 5+6 2 |
| | 4+5 2 | 6+7 2 | 5+6 2 | 3+4 2 |
| | 9 2 | 6+7 2 | 3+4 2 | 5+6 2 |
| | 6+7 2 | 4+5 2 | 5+6 2 | 3+4 2 |
| 166.1.120 | 640141 | 48 | In2Ni3Se2 | |
| | -GM | -T | -F | -L |
| | 8 2 | 8 2 | 5+6 2 | 5+6 2 |
| | 9 2 | 9 2 | 3+4 2 | 3+4 2 |
| | 8 2 | 9 2 | 3+4 2 | 3+4 2 |
| | 8 2 | 8 2 | 5+6 2 | 5+6 2 |
| | 8 2 | 9 2 | 5+6 2 | 3+4 2 |
| | 4+5 2 | 6+7 2 | 3+4 2 | 5+6 2 |
| | 8 2 | 8 2 | 5+6 2 | 5+6 2 |
| | 9 2 | 4+5 2 | 5+6 2 | 3+4 2 |
| | 9 2 | 9 2 | 3+4 2 | 3+4 2 |
| | 6+7 2 | 9 2 | 3+4 2 | 5+6 2 |
| | 8 2 | 6+7 2 | 3+4 2 | 3+4 2 |
| | 4+5 2 | 9 2 | 5+6 2 | 3+4 2 |
| | 8 2 | 8 2 | 5+6 2 | 3+4 2 |
| | 4+5 2 | 9 2 | 5+6 2 | 5+6 2 |
| | 8 2 | 6+7 2 | 5+6 2 | 3+4 2 |
| | 9 2 | 9 2 | 5+6 2 | 5+6 2 |
| | 8 2 | 9 2 | 3+4 2 | 3+4 2 |
| | 4+5 2 | 6+7 2 | 5+6 2 | 3+4 2 |
| | 9 2 | 9 2 | 3+4 2 | 5+6 2 |
| | 6+7 2 | 6+7 2 | 5+6 2 | 5+6 2 |
| | 8 2 | 9 2 | 3+4 2 | 3+4 2 |
| | 4+5 2 | 9 2 | 3+4 2 | 3+4 2 |
| | 8 2 | 9 2 | 5+6 2 | 5+6 2 |
| | 8 2 | 6+7 2 | 5+6 2 | 3+4 2 |
| 166.1.121 | 640140 | 48 | In2Ni3Se2 | |
| | -GM | -T | -F | -L |
| | 8 2 | 9 2 | 5+6 2 | 3+4 2 |
| | 9 2 | 8 2 | 3+4 2 | 5+6 2 |
| | 8 2 | 8 2 | 3+4 2 | 5+6 2 |
| | 8 2 | 9 2 | 5+6 2 | 3+4 2 |
| | 8 2 | 8 2 | 5+6 2 | 5+6 2 |
| | 4+5 2 | 4+5 2 | 3+4 2 | 3+4 2 |
| | 8 2 | 9 2 | 5+6 2 | 3+4 2 |
| | 9 2 | 6+7 2 | 5+6 2 | 5+6 2 |
| | 9 2 | 8 2 | 3+4 2 | 5+6 2 |
| | 6+7 2 | 8 2 | 3+4 2 | 3+4 2 |
| | 8 2 | 4+5 2 | 3+4 2 | 5+6 2 |
| | 4+5 2 | 8 2 | 5+6 2 | 5+6 2 |
| | 8 2 | 9 2 | 5+6 2 | 5+6 2 |
| | 4+5 2 | 8 2 | 5+6 2 | 3+4 2 |
| | 8 2 | 4+5 2 | 5+6 2 | 5+6 2 |
| | 9 2 | 8 2 | 5+6 2 | 3+4 2 |
| | 8 2 | 8 2 | 3+4 2 | 5+6 2 |
| | 4+5 2 | 4+5 2 | 5+6 2 | 5+6 2 |
| | 9 2 | 8 2 | 3+4 2 | 3+4 2 |
| | 8 2 | 4+5 2 | 5+6 2 | 3+4 2 |
| | 6+7 2 | 8 2 | 3+4 2 | 5+6 2 |
| | 4+5 2 | 8 2 | 3+4 2 | 5+6 2 |

Continued on next column

**TABLE LXXIII– continued**

| TQC | ICSD | NoE | SF | |
|---|---|---|---|---|
| | 8 2 | 8 2 | 5+6 2 | 3+4 2 |
| | 8 2 | 4+5 2 | 5+6 2 | 5+6 2 |
| 166.1.122 | 640135 | 48 | In2Ni3S2 | |
| | -GM | -T | -F | -L |
| | 8 2 | 8 2 | 5+6 2 | 5+6 2 |
| | 9 2 | 9 2 | 3+4 2 | 3+4 2 |
| | 8 2 | 9 2 | 3+4 2 | 3+4 2 |
| | 8 2 | 8 2 | 5+6 2 | 5+6 2 |
| | 8 2 | 9 2 | 5+6 2 | 3+4 2 |
| | 4+5 2 | 6+7 2 | 3+4 2 | 5+6 2 |
| | 8 2 | 8 2 | 5+6 2 | 3+4 2 |
| | 9 2 | 4+5 2 | 5+6 2 | 5+6 2 |
| | 9 2 | 9 2 | 3+4 2 | 3+4 2 |
| | 6+7 2 | 9 2 | 3+4 2 | 5+6 2 |
| | 8 2 | 6+7 2 | 3+4 2 | 3+4 2 |
| | 4+5 2 | 8 2 | 5+6 2 | 3+4 2 |
| | 8 2 | 9 2 | 5+6 2 | 5+6 2 |
| | 4+5 2 | 9 2 | 5+6 2 | 3+4 2 |
| | 8 2 | 6+7 2 | 5+6 2 | 3+4 2 |
| | 9 2 | 9 2 | 5+6 2 | 5+6 2 |
| | 8 2 | 9 2 | 3+4 2 | 3+4 2 |
| | 4+5 2 | 6+7 2 | 5+6 2 | 3+4 2 |
| | 9 2 | 9 2 | 3+4 2 | 5+6 2 |
| | 8 2 | 6+7 2 | 5+6 2 | 5+6 2 |
| | 6+7 2 | 9 2 | 3+4 2 | 3+4 2 |
| | 4+5 2 | 9 2 | 5+6 2 | 3+4 2 |
| | 8 2 | 9 2 | 3+4 2 | 5+6 2 |
| | 8 2 | 6+7 2 | 5+6 2 | 3+4 2 |
| 166.1.123 | 640134 | 48 | In2Ni3S2 | |
| | -GM | -T | -F | -L |
| | 8 2 | 9 2 | 5+6 2 | 3+4 2 |
| | 9 2 | 8 2 | 3+4 2 | 5+6 2 |
| | 8 2 | 8 2 | 3+4 2 | 5+6 2 |
| | 8 2 | 9 2 | 5+6 2 | 3+4 2 |
| | 8 2 | 8 2 | 5+6 2 | 5+6 2 |
| | 4+5 2 | 4+5 2 | 3+4 2 | 3+4 2 |
| | 8 2 | 9 2 | 5+6 2 | 5+6 2 |
| | 9 2 | 6+7 2 | 5+6 2 | 3+4 2 |
| | 9 2 | 8 2 | 3+4 2 | 5+6 2 |
| | 6+7 2 | 8 2 | 3+4 2 | 3+4 2 |
| | 8 2 | 4+5 2 | 3+4 2 | 5+6 2 |
| | 4+5 2 | 8 2 | 5+6 2 | 5+6 2 |
| | 8 2 | 9 2 | 5+6 2 | 3+4 2 |
| | 4+5 2 | 8 2 | 5+6 2 | 5+6 2 |
| | 8 2 | 4+5 2 | 5+6 2 | 5+6 2 |
| | 9 2 | 8 2 | 5+6 2 | 3+4 2 |
| | 8 2 | 8 2 | 3+4 2 | 5+6 2 |
| | 4+5 2 | 4+5 2 | 5+6 2 | 5+6 2 |
| | 9 2 | 8 2 | 3+4 2 | 3+4 2 |
| | 8 2 | 4+5 2 | 5+6 2 | 3+4 2 |
| | 6+7 2 | 8 2 | 3+4 2 | 5+6 2 |
| | 4+5 2 | 8 2 | 5+6 2 | 5+6 2 |
| | 8 2 | 8 2 | 3+4 2 | 3+4 2 |
| | 8 2 | 4+5 2 | 5+6 2 | 5+6 2 |
| 166.1.124 | 415258 | 48 | In2Ni3S2 | |
| | -GM | -T | -F | -L |
| | 8 2 | 8 2 | 5+6 2 | 5+6 2 |
| | 9 2 | 9 2 | 3+4 2 | 3+4 2 |
| | 8 2 | 9 2 | 3+4 2 | 3+4 2 |
| | 8 2 | 8 2 | 5+6 2 | 5+6 2 |
| | 8 2 | 9 2 | 5+6 2 | 3+4 2 |

Continued on next column

**TABLE LXXIII– continued**

| TQC | ICSD | NoE | SF | |
|---|---|---|---|---|
| | 4+5 2 | 6+7 2 | 3+4 2 | 5+6 2 |
| | 8 2 | 8 2 | 5+6 2 | 3+4 2 |
| | 9 2 | 4+5 2 | 5+6 2 | 5+6 2 |
| | 9 2 | 9 2 | 3+4 2 | 3+4 2 |
| | 6+7 2 | 9 2 | 3+4 2 | 5+6 2 |
| | 8 2 | 6+7 2 | 3+4 2 | 3+4 2 |
| | 4+5 2 | 9 2 | 5+6 2 | 3+4 2 |
| | 8 2 | 8 2 | 5+6 2 | 5+6 2 |
| | 4+5 2 | 9 2 | 5+6 2 | 3+4 2 |
| | 8 2 | 6+7 2 | 5+6 2 | 3+4 2 |
| | 9 2 | 9 2 | 5+6 2 | 5+6 2 |
| | 8 2 | 9 2 | 3+4 2 | 3+4 2 |
| | 4+5 2 | 6+7 2 | 5+6 2 | 3+4 2 |
| | 9 2 | 9 2 | 3+4 2 | 5+6 2 |
| | 8 2 | 6+7 2 | 5+6 2 | 5+6 2 |
| | 6+7 2 | 9 2 | 3+4 2 | 3+4 2 |
| | 4+5 2 | 9 2 | 5+6 2 | 3+4 2 |
| | 8 2 | 9 2 | 3+4 2 | 5+6 2 |
| | 8 2 | 6+7 2 | 5+6 2 | 3+4 2 |
| 166.1.125 | 108894 | 48 | In2Ni3S2 | |
| | -GM | -T | -F | -L |
| | 8 2 | 9 2 | 5+6 2 | 3+4 2 |
| | 9 2 | 8 2 | 3+4 2 | 5+6 2 |
| | 8 2 | 8 2 | 3+4 2 | 5+6 2 |
| | 8 2 | 9 2 | 5+6 2 | 3+4 2 |
| | 8 2 | 8 2 | 5+6 2 | 5+6 2 |
| | 4+5 2 | 4+5 2 | 3+4 2 | 3+4 2 |
| | 8 2 | 9 2 | 5+6 2 | 3+4 2 |
| | 9 2 | 6+7 2 | 5+6 2 | 5+6 2 |
| | 9 2 | 8 2 | 3+4 2 | 5+6 2 |
| | 6+7 2 | 8 2 | 3+4 2 | 3+4 2 |
| | 8 2 | 9 2 | 3+4 2 | 5+6 2 |
| | 4+5 2 | 4+5 2 | 5+6 2 | 5+6 2 |
| | 8 2 | 8 2 | 5+6 2 | 3+4 2 |
| | 4+5 2 | 8 2 | 5+6 2 | 5+6 2 |
| | 8 2 | 4+5 2 | 5+6 2 | 5+6 2 |
| | 9 2 | 8 2 | 5+6 2 | 3+4 2 |
| | 8 2 | 8 2 | 3+4 2 | 5+6 2 |
| | 4+5 2 | 4+5 2 | 5+6 2 | 5+6 2 |
| | 9 2 | 8 2 | 3+4 2 | 3+4 2 |
| | 8 2 | 4+5 2 | 5+6 2 | 3+4 2 |
| | 6+7 2 | 8 2 | 3+4 2 | 5+6 2 |
| | 4+5 2 | 8 2 | 5+6 2 | 5+6 2 |
| | 8 2 | 8 2 | 3+4 2 | 3+4 2 |
| | 8 2 | 4+5 2 | 5+6 2 | 5+6 2 |
| 166.1.126 | 172602 | 32 | I1N1Sr2 | |
| | -GM | -T | -F | -L |
| | 8 2 | 8 2 | 5+6 2 | 5+6 2 |
| | 9 2 | 9 2 | 3+4 2 | 3+4 2 |
| | 8 2 | 8 2 | 3+4 2 | 3+4 2 |
| | 9 2 | 9 2 | 5+6 2 | 5+6 2 |
| | 8 2 | 8 2 | 5+6 2 | 5+6 2 |
| | 4+5 2 | 4+5 2 | 3+4 2 | 3+4 2 |
| | 6+7 2 | 6+7 2 | 5+6 2 | 5+6 2 |
| | 9 2 | 9 2 | 3+4 2 | 3+4 2 |
| | 8 2 | 8 2 | 3+4 2 | 3+4 2 |
| | 8 2 | 9 2 | 3+4 2 | 5+6 2 |
| | 9 2 | 9 2 | 5+6 2 | 5+6 2 |
| | 9 2 | 9 2 | 5+6 2 | 5+6 2 |
| | 6+7 2 | 6+7 2 | 5+6 2 | 5+6 2 |
| | 9 2 | 8 2 | 3+4 2 | 3+4 2 |

Continued on next column

**TABLE LXXIII– continued**

| TQC | ICSD | NoE | SF | |
|---|---|---|---|---|
| | 9 2 | 8 2 | 5+6 2 | 3+4 2 |
| | 6+7 2 | 4+5 2 | 5+6 2 | 3+4 2 |
| 166.1.127 | 654919 | 12 | Hg1 | |
| | -GM | -T | -F | -L |
| | 8 2 | 8 2 | 3+4 2 | 3+4 2 |
| | 8 2 | 4+5 2 | 3+4 2 | 3+4 2 |
| | 4+5 2 | 8 2 | 3+4 2 | 3+4 2 |
| | 8 2 | 8 2 | 3+4 2 | 3+4 2 |
| | 4+5 2 | 4+5 2 | 3+4 2 | 3+4 2 |
| | 8 2 | 9 2 | 5+6 2 | 5+6 2 |
| 166.1.128 | 53771 | 12 | Hg1 | |
| | -GM | -T | -F | -L |
| | 8 2 | 8 2 | 3+4 2 | 3+4 2 |
| | 8 2 | 4+5 2 | 3+4 2 | 3+4 2 |
| | 4+5 2 | 8 2 | 3+4 2 | 3+4 2 |
| | 8 2 | 8 2 | 3+4 2 | 3+4 2 |
| | 4+5 2 | 4+5 2 | 3+4 2 | 3+4 2 |
| | 8 2 | 9 2 | 5+6 2 | 3+4 2 |
| 166.1.129 | 174006 | 12 | Hg1 | |
| | -GM | -T | -F | -L |
| | 8 2 | 8 2 | 3+4 2 | 3+4 2 |
| | 8 2 | 4+5 2 | 3+4 2 | 3+4 2 |
| | 4+5 2 | 8 2 | 3+4 2 | 3+4 2 |
| | 8 2 | 8 2 | 3+4 2 | 3+4 2 |
| | 4+5 2 | 4+5 2 | 3+4 2 | 3+4 2 |
| | 8 2 | 9 2 | 5+6 2 | 5+6 2 |
| 166.1.130 | 174005 | 12 | Hg1 | |
| | -GM | -T | -F | -L |
| | 8 2 | 8 2 | 3+4 2 | 3+4 2 |
| | 8 2 | 4+5 2 | 3+4 2 | 3+4 2 |
| | 4+5 2 | 8 2 | 3+4 2 | 3+4 2 |
| | 8 2 | 8 2 | 3+4 2 | 3+4 2 |
| | 4+5 2 | 4+5 2 | 3+4 2 | 3+4 2 |
| | 8 2 | 9 2 | 5+6 2 | 5+6 2 |
| 166.1.131 | 104296 | 12 | Hg1 | |
| | -GM | -T | -F | -L |
| | 8 2 | 8 2 | 3+4 2 | 3+4 2 |
| | 8 2 | 4+5 2 | 3+4 2 | 3+4 2 |
| | 4+5 2 | 8 2 | 3+4 2 | 3+4 2 |
| | 8 2 | 8 2 | 3+4 2 | 3+4 2 |
| | 4+5 2 | 4+5 2 | 3+4 2 | 3+4 2 |
| | 8 2 | 9 2 | 5+6 2 | 5+6 2 |
| 166.1.132 | 638697 | 102 | Hf1Ni3 | |
| | -GM | -T | -F | -L |
| | 8 2 | 9 2 | 5+6 2 | 3+4 2 |
| | 8 2 | 8 2 | 3+4 2 | 5+6 2 |
| | 9 2 | 4+5 2 | 3+4 2 | 3+4 2 |
| | 4+5 2 | 8 2 | 5+6 2 | 5+6 2 |
| | 8 2 | 8 2 | 3+4 2 | 5+6 2 |
| | 9 2 | 8 2 | 5+6 2 | 5+6 2 |
| | 6+7 2 | 4+5 2 | 3+4 2 | 3+4 2 |
| | 8 2 | 9 2 | 5+6 2 | 3+4 2 |
| | 4+5 2 | 6+7 2 | 5+6 2 | 3+4 2 |
| | 8 2 | 8 2 | 3+4 2 | 5+6 2 |
| | 8 2 | 9 2 | 3+4 2 | 3+4 2 |
| | 9 2 | 8 2 | 5+6 2 | 5+6 2 |
| | 4+5 2 | 9 2 | 3+4 2 | 3+4 2 |
| | 8 2 | 6+7 2 | 3+4 2 | 5+6 2 |
| | 8 2 | 8 2 | 5+6 2 | 5+6 2 |
| | 8 2 | 4+5 2 | 3+4 2 | 3+4 2 |

Continued on next column

| TQC | ICSD | NoE | SF | |
|---|---|---|---|---|
| | 8 2 | 9 2 | 5+6 2 | 3+4 2 |
| | 4+5 2 | 8 2 | 5+6 2 | 5+6 2 |
| | 9 2 | 9 2 | 3+4 2 | 5+6 2 |
| | 6+7 2 | 8 2 | 5+6 2 | 3+4 2 |
| | 9 2 | 4+5 2 | 3+4 2 | 5+6 2 |
| | 4+5 2 | 8 2 | 5+6 2 | 3+4 2 |
| | 8 2 | 8 2 | 5+6 2 | 5+6 2 |
| | 6+7 2 | 4+5 2 | 3+4 2 | 5+6 2 |
| | 8 2 | 8 2 | 3+4 2 | 3+4 2 |
| | 8 2 | 4+5 2 | 5+6 2 | 5+6 2 |
| | 9 2 | 6+7 2 | 5+6 2 | 3+4 2 |
| | 9 2 | 9 2 | 3+4 2 | 5+6 2 |
| | 4+5 2 | 8 2 | 5+6 2 | 3+4 2 |
| | 8 2 | 8 2 | 5+6 2 | 3+4 2 |
| | 9 2 | 4+5 2 | 5+6 2 | 3+4 2 |
| | 8 2 | 8 2 | 3+4 2 | 5+6 2 |
| | 6+7 2 | 9 2 | 3+4 2 | 3+4 2 |
| | 9 2 | 6+7 2 | 5+6 2 | 5+6 2 |
| | 4+5 2 | 8 2 | 3+4 2 | 5+6 2 |
| | 8 2 | 4+5 2 | 5+6 2 | 3+4 2 |
| | 9 2 | 9 2 | 5+6 2 | 5+6 2 |
| | 8 2 | 9 2 | 3+4 2 | 3+4 2 |
| | 4+5 2 | 6+7 2 | 3+4 2 | 5+6 2 |
| | 6+7 2 | 9 2 | 5+6 2 | 3+4 2 |
| | 8 2 | 8 2 | 5+6 2 | 5+6 2 |
| | 4+5 2 | 8 2 | 3+4 2 | 3+4 2 |
| | 8 2 | 4+5 2 | 5+6 2 | 3+4 2 |
| | 8 2 | 8 2 | 5+6 2 | 5+6 2 |
| | 9 2 | 8 2 | 3+4 2 | 5+6 2 |
| | 8 2 | 9 2 | 3+4 2 | 5+6 2 |
| | 8 2 | 8 2 | 5+6 2 | 3+4 2 |
| | 4+5 2 | 8 2 | 5+6 2 | 5+6 2 |
| | 8 2 | 4+5 2 | 3+4 2 | 5+6 2 |
| | 9 2 | 9 2 | 3+4 2 | 5+6 2 |
| | 9 2 | 9 2 | 5+6 2 | 5+6 2 |
| 166.1.133 | 2414 | 102 | Hf1Ni3 | |
| | -GM | -T | -F | -L |
| | 8 2 | 9 2 | 5+6 2 | 3+4 2 |
| | 8 2 | 8 2 | 3+4 2 | 5+6 2 |
| | 9 2 | 4+5 2 | 3+4 2 | 3+4 2 |
| | 4+5 2 | 8 2 | 5+6 2 | 5+6 2 |
| | 8 2 | 8 2 | 3+4 2 | 5+6 2 |
| | 9 2 | 8 2 | 5+6 2 | 5+6 2 |
| | 6+7 2 | 4+5 2 | 3+4 2 | 3+4 2 |
| | 8 2 | 9 2 | 5+6 2 | 3+4 2 |
| | 4+5 2 | 6+7 2 | 5+6 2 | 3+4 2 |
| | 8 2 | 8 2 | 3+4 2 | 5+6 2 |
| | 8 2 | 9 2 | 3+4 2 | 3+4 2 |
| | 9 2 | 8 2 | 5+6 2 | 5+6 2 |
| | 4+5 2 | 9 2 | 3+4 2 | 3+4 2 |
| | 8 2 | 6+7 2 | 3+4 2 | 5+6 2 |
| | 8 2 | 8 2 | 5+6 2 | 5+6 2 |
| | 8 2 | 4+5 2 | 3+4 2 | 3+4 2 |
| | 8 2 | 9 2 | 5+6 2 | 3+4 2 |
| | 4+5 2 | 8 2 | 5+6 2 | 5+6 2 |
| | 9 2 | 9 2 | 3+4 2 | 5+6 2 |
| | 6+7 2 | 8 2 | 5+6 2 | 3+4 2 |
| | 9 2 | 4+5 2 | 3+4 2 | 5+6 2 |
| | 4+5 2 | 8 2 | 5+6 2 | 3+4 2 |
| | 8 2 | 4+5 2 | 5+6 2 | 5+6 2 |
| | 6+7 2 | 8 2 | 3+4 2 | 5+6 2 |

Continued on next column

**TABLE LXXIII– continued**

| TQC | ICSD | NoE | SF | |
|---|---|---|---|---|
| | 8 2 | 8 2 | 3+4 2 | 3+4 2 |
| | 8 2 | 4+5 2 | 5+6 2 | 5+6 2 |
| | 9 2 | 6+7 2 | 5+6 2 | 3+4 2 |
| | 9 2 | 9 2 | 3+4 2 | 5+6 2 |
| | 4+5 2 | 8 2 | 5+6 2 | 3+4 2 |
| | 8 2 | 8 2 | 5+6 2 | 3+4 2 |
| | 9 2 | 4+5 2 | 5+6 2 | 3+4 2 |
| | 8 2 | 8 2 | 3+4 2 | 5+6 2 |
| | 6+7 2 | 9 2 | 3+4 2 | 3+4 2 |
| | 9 2 | 6+7 2 | 5+6 2 | 5+6 2 |
| | 4+5 2 | 8 2 | 3+4 2 | 5+6 2 |
| | 8 2 | 4+5 2 | 5+6 2 | 5+6 2 |
| | 9 2 | 9 2 | 5+6 2 | 3+4 2 |
| | 8 2 | 9 2 | 3+4 2 | 3+4 2 |
| | 6+7 2 | 6+7 2 | 3+4 2 | 5+6 2 |
| | 4+5 2 | 9 2 | 5+6 2 | 3+4 2 |
| | 8 2 | 8 2 | 5+6 2 | 5+6 2 |
| | 4+5 2 | 8 2 | 3+4 2 | 3+4 2 |
| | 8 2 | 4+5 2 | 5+6 2 | 3+4 2 |
| | 8 2 | 8 2 | 5+6 2 | 5+6 2 |
| | 9 2 | 8 2 | 3+4 2 | 5+6 2 |
| | 8 2 | 9 2 | 3+4 2 | 5+6 2 |
| | 8 2 | 8 2 | 5+6 2 | 3+4 2 |
| | 4+5 2 | 8 2 | 5+6 2 | 5+6 2 |
| | 8 2 | 4+5 2 | 3+4 2 | 5+6 2 |
| | 9 2 | 9 2 | 3+4 2 | 5+6 2 |
| | 9 2 | 9 2 | 5+6 2 | 5+6 2 |
| 166.1.134 | 171105 | 126 | H9Fe3O15S2 | |
| | -GM | -T | -F | -L |
| | 8 2 | 8 2 | 5+6 2 | 5+6 2 |
| | 9 2 | 9 2 | 3+4 2 | 3+4 2 |
| | 8 2 | 9 2 | 3+4 2 | 3+4 2 |
| | 9 2 | 9 2 | 5+6 2 | 5+6 2 |
| | 8 2 | 8 2 | 5+6 2 | 3+4 2 |
| | 4+5 2 | 6+7 2 | 3+4 2 | 3+4 2 |
| | 9 2 | 8 2 | 5+6 2 | 5+6 2 |
| | 6+7 2 | 4+5 2 | 3+4 2 | 5+6 2 |
| | 8 2 | 9 2 | 5+6 2 | 3+4 2 |
| | 9 2 | 8 2 | 3+4 2 | 5+6 2 |
| | 4+5 2 | 6+7 2 | 5+6 2 | 3+4 2 |
| | 8 2 | 9 2 | 3+4 2 | 5+6 2 |
| | 6+7 2 | 4+5 2 | 3+4 2 | 5+6 2 |
| | 9 2 | 8 2 | 5+6 2 | 3+4 2 |
| | 8 2 | 8 2 | 3+4 2 | 3+4 2 |
| | 8 2 | 9 2 | 5+6 2 | 3+4 2 |
| | 9 2 | 8 2 | 3+4 2 | 5+6 2 |
| | 8 2 | 9 2 | 5+6 2 | 3+4 2 |
| | 4+5 2 | 6+7 2 | 3+4 2 | 5+6 2 |
| | 9 2 | 8 2 | 5+6 2 | 3+4 2 |
| | 6+7 2 | 4+5 2 | 3+4 2 | 5+6 2 |
| | 8 2 | 9 2 | 5+6 2 | 5+6 2 |
| | 9 2 | 8 2 | 3+4 2 | 3+4 2 |
| | 4+5 2 | 6+7 2 | 5+6 2 | 3+4 2 |
| | 8 2 | 9 2 | 5+6 2 | 3+4 2 |
| | 8 2 | 9 2 | 5+6 2 | 5+6 2 |
| | 8 2 | 6+7 2 | 3+4 2 | 3+4 2 |
| | 4+5 2 | 9 2 | 3+4 2 | 3+4 2 |
| | 8 2 | 9 2 | 5+6 2 | 5+6 2 |
| | 9 2 | 9 2 | 3+4 2 | 5+6 2 |
| | 8 2 | 8 2 | 5+6 2 | 3+4 2 |
| | 6+7 2 | 4+5 2 | 3+4 2 | 5+6 2 |

Continued on next column

**TABLE LXXIII– continued**

| TQC | ICSD | NoE | SF | |
|---|---|---|---|---|
| | 9 2 | 8 2 | 5+6 2 | 3+4 2 |
| | 9 2 | 4+5 2 | 5+6 2 | 5+6 2 |
| | 6+7 2 | 8 2 | 3+4 2 | 3+4 2 |
| | 9 2 | 8 2 | 5+6 2 | 5+6 2 |
| | 8 2 | 9 2 | 3+4 2 | 5+6 2 |
| | 4+5 2 | 6+7 2 | 3+4 2 | 3+4 2 |
| | 8 2 | 9 2 | 3+4 2 | 3+4 2 |
| | 4+5 2 | 6+7 2 | 5+6 2 | 5+6 2 |
| | 8 2 | 9 2 | 5+6 2 | 5+6 2 |
| | 9 2 | 8 2 | 3+4 2 | 3+4 2 |
| | 6+7 2 | 4+5 2 | 5+6 2 | 3+4 2 |
| | 9 2 | 9 2 | 5+6 2 | 3+4 2 |
| | 8 2 | 9 2 | 3+4 2 | 5+6 2 |
| | 4+5 2 | 6+7 2 | 3+4 2 | 5+6 2 |
| | 8 2 | 8 2 | 5+6 2 | 3+4 2 |
| | 8 2 | 9 2 | 3+4 2 | 3+4 2 |
| | 4+5 2 | 6+7 2 | 5+6 2 | 5+6 2 |
| | 9 2 | 8 2 | 5+6 2 | 5+6 2 |
| | 9 2 | 8 2 | 3+4 2 | 3+4 2 |
| | 6+7 2 | 4+5 2 | 5+6 2 | 5+6 2 |
| | 9 2 | 8 2 | 3+4 2 | 3+4 2 |
| | 6+7 2 | 4+5 2 | 5+6 2 | 5+6 2 |
| | 9 2 | 8 2 | 3+4 2 | 3+4 2 |
| | 8 2 | 9 2 | 3+4 2 | 5+6 2 |
| | 4+5 2 | 6+7 2 | 5+6 2 | 5+6 2 |
| | 9 2 | 8 2 | 5+6 2 | 5+6 2 |
| | 6+7 2 | 4+5 2 | 3+4 2 | 3+4 2 |
| | 8 2 | 9 2 | 5+6 2 | 3+4 2 |
| | 4+5 2 | 6+7 2 | 5+6 2 | 5+6 2 |
| | 8 2 | 9 2 | 5+6 2 | 5+6 2 |
| | 9 2 | 9 2 | 3+4 2 | 3+4 2 |
| 166.1.135 | 15888 | 74 | H8N2O8Pb1S2 | |
| | -GM | -T | -F | -L |
| | 8 2 | 8 2 | 5+6 2 | 5+6 2 |
| | 9 2 | 9 2 | 3+4 2 | 3+4 2 |
| | 9 2 | 9 2 | 3+4 2 | 3+4 2 |
| | 6+7 2 | 6+7 2 | 5+6 2 | 5+6 2 |
| | 8 2 | 8 2 | 5+6 2 | 5+6 2 |
| | 4+5 2 | 4+5 2 | 3+4 2 | 3+4 2 |
| | 9 2 | 9 2 | 3+4 2 | 3+4 2 |
| | 8 2 | 8 2 | 5+6 2 | 5+6 2 |
| | 8 2 | 9 2 | 5+6 2 | 3+4 2 |
| | 9 2 | 8 2 | 3+4 2 | 5+6 2 |
| | 8 2 | 8 2 | 5+6 2 | 5+6 2 |
| | 9 2 | 9 2 | 3+4 2 | 3+4 2 |
| | 8 2 | 8 2 | 3+4 2 | 3+4 2 |
| | 9 2 | 9 2 | 5+6 2 | 5+6 2 |
| | 8 2 | 8 2 | 5+6 2 | 5+6 2 |
| | 4+5 2 | 4+5 2 | 3+4 2 | 5+6 2 |
| | 9 2 | 9 2 | 5+6 2 | 3+4 2 |
| | 6+7 2 | 6+7 2 | 3+4 2 | 3+4 2 |
| | 8 2 | 8 2 | 3+4 2 | 3+4 2 |
| | 9 2 | 9 2 | 5+6 2 | 5+6 2 |
| | 6+7 2 | 6+7 2 | 5+6 2 | 5+6 2 |
| | 8 2 | 4+5 2 | 3+4 2 | 5+6 2 |
| | 4+5 2 | 8 2 | 5+6 2 | 3+4 2 |
| | 8 2 | 8 2 | 5+6 2 | 3+4 2 |
| | 4+5 2 | 4+5 2 | 3+4 2 | 5+6 2 |
| | 9 2 | 9 2 | 3+4 2 | 5+6 2 |
| | 9 2 | 9 2 | 5+6 2 | 3+4 2 |
| | 6+7 2 | 6+7 2 | 3+4 2 | 3+4 2 |

Continued on next column

**TABLE LXXIII– continued**

| TQC | ICSD | | NoE | SF |
|---|---|---|---|---|
| | 8 2 | 8 2 | 3+4 2 | 3+4 2 |
| | 8 2 | 9 2 | 5+6 2 | 3+4 2 |
| | 4+5 2 | 6+7 2 | 5+6 2 | 5+6 2 |
| | 9 2 | 8 2 | 3+4 2 | 5+6 2 |
| | 6+7 2 | 4+5 2 | 5+6 2 | 3+4 2 |
| | 9 2 | 9 2 | 3+4 2 | 5+6 2 |
| | 8 2 | 8 2 | 3+4 2 | 3+4 2 |
| | 8 2 | 9 2 | 5+6 2 | 5+6 2 |
| | 8 2 | 8 2 | 5+6 2 | 3+4 2 |
| 166.1.136 | 195640 | | 28 | H6Mg1Nb2 |
| | -GM | -T | -F | -L |
| | 8 2 | 9 2 | 5+6 2 | 3+4 2 |
| | 9 2 | 8 2 | 3+4 2 | 5+6 2 |
| | 8 2 | 9 2 | 5+6 2 | 3+4 2 |
| | 9 2 | 8 2 | 3+4 2 | 5+6 2 |
| | 8 2 | 9 2 | 5+6 2 | 3+4 2 |
| | 9 2 | 8 2 | 3+4 2 | 5+6 2 |
| | 4+5 2 | 6+7 2 | 5+6 2 | 3+4 2 |
| | 6+7 2 | 4+5 2 | 3+4 2 | 5+6 2 |
| | 8 2 | 8 2 | 3+4 2 | 5+6 2 |
| | 9 2 | 9 2 | 5+6 2 | 3+4 2 |
| | 8 2 | 8 2 | 3+4 2 | 5+6 2 |
| | 9 2 | 9 2 | 5+6 2 | 3+4 2 |
| | 6+7 2 | 4+5 2 | 3+4 2 | 5+6 2 |
| | 9 2 | 8 2 | 5+6 2 | 3+4 2 |
| 166.1.137 | 67682 | | 124 | H6Fe3O14Pb1S2 |
| | -GM | -T | -F | -L |
| | 8 2 | 8 2 | 5+6 2 | 5+6 2 |
| | 9 2 | 9 2 | 3+4 2 | 3+4 2 |
| | 8 2 | 9 2 | 3+4 2 | 3+4 2 |
| | 4+5 2 | 6+7 2 | 5+6 2 | 5+6 2 |
| | 9 2 | 9 2 | 5+6 2 | 3+4 2 |
| | 6+7 2 | 8 2 | 3+4 2 | 5+6 2 |
| | 8 2 | 4+5 2 | 5+6 2 | 3+4 2 |
| | 9 2 | 8 2 | 3+4 2 | 5+6 2 |
| | 8 2 | 9 2 | 5+6 2 | 3+4 2 |
| | 9 2 | 8 2 | 3+4 2 | 5+6 2 |
| | 4+5 2 | 6+7 2 | 5+6 2 | 3+4 2 |
| | 8 2 | 9 2 | 3+4 2 | 5+6 2 |
| | 6+7 2 | 4+5 2 | 5+6 2 | 3+4 2 |
| | 9 2 | 8 2 | 3+4 2 | 5+6 2 |
| | 8 2 | 8 2 | 5+6 2 | 5+6 2 |
| | 9 2 | 9 2 | 3+4 2 | 3+4 2 |
| | 8 2 | 9 2 | 3+4 2 | 3+4 2 |
| | 4+5 2 | 6+7 2 | 5+6 2 | 3+4 2 |
| | 8 2 | 8 2 | 5+6 2 | 5+6 2 |
| | 9 2 | 4+5 2 | 3+4 2 | 5+6 2 |
| | 6+7 2 | 8 2 | 3+4 2 | 3+4 2 |
| | 9 2 | 9 2 | 5+6 2 | 5+6 2 |
| | 8 2 | 8 2 | 3+4 2 | 3+4 2 |
| | 4+5 2 | 6+7 2 | 5+6 2 | 3+4 2 |
| | 8 2 | 9 2 | 5+6 2 | 5+6 2 |
| | 8 2 | 9 2 | 5+6 2 | 3+4 2 |
| | 8 2 | 9 2 | 3+4 2 | 3+4 2 |
| | 4+5 2 | 9 2 | 3+4 2 | 5+6 2 |
| | 8 2 | 6+7 2 | 5+6 2 | 3+4 2 |
| | 9 2 | 8 2 | 5+6 2 | 5+6 2 |
| | 9 2 | 9 2 | 5+6 2 | 3+4 2 |
| | 6+7 2 | 4+5 2 | 3+4 2 | 5+6 2 |
| | 9 2 | 8 2 | 3+4 2 | 5+6 2 |
| | 8 2 | 4+5 2 | 5+6 2 | 3+4 2 |

Continued on next column

**TABLE LXXIII– continued**

| TQC | ICSD | | NoE | SF |
|---|---|---|---|---|
| | 6+7 2 | 8 2 | 3+4 2 | 3+4 2 |
| | 8 2 | 9 2 | 5+6 2 | 5+6 2 |
| | 4+5 2 | 8 2 | 3+4 2 | 5+6 2 |
| | 9 2 | 9 2 | 3+4 2 | 5+6 2 |
| | 9 2 | 6+7 2 | 5+6 2 | 5+6 2 |
| | 4+5 2 | 6+7 2 | 5+6 2 | 3+4 2 |
| | 8 2 | 9 2 | 3+4 2 | 3+4 2 |
| | 8 2 | 8 2 | 5+6 2 | 3+4 2 |
| | 9 2 | 4+5 2 | 3+4 2 | 3+4 2 |
| | 6+7 2 | 8 2 | 3+4 2 | 5+6 2 |
| | 4+5 2 | 6+7 2 | 5+6 2 | 3+4 2 |
| | 8 2 | 9 2 | 5+6 2 | 5+6 2 |
| | 8 2 | 9 2 | 3+4 2 | 3+4 2 |
| | 9 2 | 9 2 | 3+4 2 | 5+6 2 |
| | 9 2 | 6+7 2 | 5+6 2 | 3+4 2 |
| | 6+7 2 | 8 2 | 5+6 2 | 5+6 2 |
| | 8 2 | 4+5 2 | 5+6 2 | 5+6 2 |
| | 4+5 2 | 8 2 | 3+4 2 | 5+6 2 |
| | 9 2 | 4+5 2 | 3+4 2 | 3+4 2 |
| | 6+7 2 | 8 2 | 5+6 2 | 3+4 2 |
| | 9 2 | 9 2 | 3+4 2 | 3+4 2 |
| | 8 2 | 6+7 2 | 3+4 2 | 5+6 2 |
| | 4+5 2 | 8 2 | 3+4 2 | 5+6 2 |
| | 9 2 | 8 2 | 5+6 2 | 3+4 2 |
| | 6+7 2 | 4+5 2 | 5+6 2 | 3+4 2 |
| | 8 2 | 9 2 | 5+6 2 | 5+6 2 |
| | 4+5 2 | 6+7 2 | 5+6 2 | 3+4 2 |
| | 8 2 | 9 2 | 3+4 2 | 5+6 2 |
| 166.1.138 | 44626 | | 106 | H6Al3Na1O14S2 |
| | -GM | -T | -F | -L |
| | 8 2 | 8 2 | 5+6 2 | 5+6 2 |
| | 9 2 | 9 2 | 3+4 2 | 3+4 2 |
| | 8 2 | 9 2 | 3+4 2 | 3+4 2 |
| | 4+5 2 | 6+7 2 | 5+6 2 | 5+6 2 |
| | 9 2 | 8 2 | 5+6 2 | 3+4 2 |
| | 6+7 2 | 4+5 2 | 3+4 2 | 5+6 2 |
| | 9 2 | 9 2 | 3+4 2 | 3+4 2 |
| | 8 2 | 8 2 | 5+6 2 | 5+6 2 |
| | 8 2 | 9 2 | 5+6 2 | 5+6 2 |
| | 6+7 2 | 4+5 2 | 3+4 2 | 3+4 2 |
| | 9 2 | 8 2 | 5+6 2 | 3+4 2 |
| | 4+5 2 | 6+7 2 | 3+4 2 | 5+6 2 |
| | 8 2 | 9 2 | 3+4 2 | 5+6 2 |
| | 9 2 | 8 2 | 5+6 2 | 3+4 2 |
| | 8 2 | 8 2 | 3+4 2 | 5+6 2 |
| | 9 2 | 9 2 | 5+6 2 | 3+4 2 |
| | 8 2 | 9 2 | 5+6 2 | 3+4 2 |
| | 4+5 2 | 6+7 2 | 3+4 2 | 5+6 2 |
| | 9 2 | 8 2 | 5+6 2 | 3+4 2 |
| | 6+7 2 | 4+5 2 | 3+4 2 | 5+6 2 |
| | 8 2 | 8 2 | 3+4 2 | 5+6 2 |
| | 9 2 | 9 2 | 5+6 2 | 3+4 2 |
| | 4+5 2 | 6+7 2 | 5+6 2 | 5+6 2 |
| | 8 2 | 9 2 | 3+4 2 | 3+4 2 |
| | 8 2 | 8 2 | 5+6 2 | 5+6 2 |
| | 9 2 | 8 2 | 3+4 2 | 3+4 2 |
| | 9 2 | 4+5 2 | 3+4 2 | 3+4 2 |
| | 6+7 2 | 9 2 | 5+6 2 | 5+6 2 |
| | 9 2 | 8 2 | 5+6 2 | 3+4 2 |
| | 8 2 | 4+5 2 | 3+4 2 | 5+6 2 |
| | 4+5 2 | 8 2 | 3+4 2 | 5+6 2 |

Continued on next column

**TABLE LXXIII– continued**

| TQC | ICSD | NoE | SF | |
|---|---|---|---|---|
| | 6+7 2 | 9 2 | 5+6 2 | 3+4 2 |
| | 9 2 | 6+7 2 | 5+6 2 | 5+6 2 |
| | 4+5 2 | 9 2 | 3+4 2 | 3+4 2 |
| | 8 2 | 9 2 | 3+4 2 | 5+6 2 |
| | 9 2 | 6+7 2 | 5+6 2 | 3+4 2 |
| | 6+7 2 | 4+5 2 | 3+4 2 | 3+4 2 |
| | 8 2 | 8 2 | 5+6 2 | 5+6 2 |
| | 9 2 | 8 2 | 5+6 2 | 3+4 2 |
| | 8 2 | 9 2 | 3+4 2 | 5+6 2 |
| | 4+5 2 | 9 2 | 3+4 2 | 3+4 2 |
| | 9 2 | 6+7 2 | 5+6 2 | 5+6 2 |
| | 6+7 2 | 8 2 | 3+4 2 | 3+4 2 |
| | 8 2 | 4+5 2 | 5+6 2 | 3+4 2 |
| | 9 2 | 8 2 | 3+4 2 | 5+6 2 |
| | 8 2 | 9 2 | 3+4 2 | 5+6 2 |
| | 4+5 2 | 8 2 | 5+6 2 | 3+4 2 |
| | 9 2 | 6+7 2 | 5+6 2 | 5+6 2 |
| | 8 2 | 9 2 | 5+6 2 | 5+6 2 |
| | 8 2 | 4+5 2 | 3+4 2 | 3+4 2 |
| | 9 2 | 8 2 | 5+6 2 | 3+4 2 |
| | 6+7 2 | 9 2 | 3+4 2 | 5+6 2 |
| | 9 2 | 9 2 | 3+4 2 | 5+6 2 |
| 166.1.139 | 24344 | 80 | H4Cu1N1O3S1 | |
| | -GM | -T | -F | -L |
| | 8 2 | 9 2 | 5+6 2 | 3+4 2 |
| | 9 2 | 8 2 | 3+4 2 | 5+6 2 |
| | 8 2 | 9 2 | 3+4 2 | 5+6 2 |
| | 4+5 2 | 6+7 2 | 5+6 2 | 3+4 2 |
| | 9 2 | 8 2 | 5+6 2 | 3+4 2 |
| | 6+7 2 | 4+5 2 | 3+4 2 | 5+6 2 |
| | 8 2 | 8 2 | 5+6 2 | 5+6 2 |
| | 9 2 | 9 2 | 3+4 2 | 3+4 2 |
| | 9 2 | 8 2 | 3+4 2 | 5+6 2 |
| | 8 2 | 9 2 | 5+6 2 | 3+4 2 |
| | 8 2 | 9 2 | 5+6 2 | 3+4 2 |
| | 9 2 | 8 2 | 3+4 2 | 5+6 2 |
| | 8 2 | 9 2 | 3+4 2 | 5+6 2 |
| | 4+5 2 | 6+7 2 | 5+6 2 | 3+4 2 |
| | 9 2 | 8 2 | 5+6 2 | 3+4 2 |
| | 6+7 2 | 4+5 2 | 3+4 2 | 5+6 2 |
| | 9 2 | 6+7 2 | 3+4 2 | 5+6 2 |
| | 6+7 2 | 9 2 | 5+6 2 | 3+4 2 |
| | 4+5 2 | 8 2 | 5+6 2 | 5+6 2 |
| | 8 2 | 4+5 2 | 3+4 2 | 3+4 2 |
| | 6+7 2 | 4+5 2 | 5+6 2 | 3+4 2 |
| | 9 2 | 8 2 | 3+4 2 | 5+6 2 |
| | 4+5 2 | 6+7 2 | 3+4 2 | 5+6 2 |
| | 8 2 | 9 2 | 5+6 2 | 3+4 2 |
| | 8 2 | 9 2 | 3+4 2 | 3+4 2 |
| | 9 2 | 8 2 | 5+6 2 | 5+6 2 |
| | 8 2 | 9 2 | 5+6 2 | 5+6 2 |
| | 4+5 2 | 6+7 2 | 3+4 2 | 3+4 2 |
| | 6+7 2 | 4+5 2 | 3+4 2 | 5+6 2 |
| | 9 2 | 8 2 | 5+6 2 | 3+4 2 |
| | 9 2 | 8 2 | 5+6 2 | 3+4 2 |
| | 9 2 | 8 2 | 3+4 2 | 5+6 2 |
| | 8 2 | 9 2 | 5+6 2 | 3+4 2 |
| | 4+5 2 | 6+7 2 | 3+4 2 | 5+6 2 |
| | 8 2 | 9 2 | 5+6 2 | 3+4 2 |
| | 8 2 | 4+5 2 | 3+4 2 | 3+4 2 |
| | 6+7 2 | 8 2 | 5+6 2 | 5+6 2 |

Continued on next column

**TABLE LXXIII– continued**

| TQC | ICSD | NoE | SF | |
|---|---|---|---|---|
| | 8 2 | 9 2 | 3+4 2 | 5+6 2 |
| | 4+5 2 | 8 2 | 3+4 2 | 5+6 2 |
| | 9 2 | 4+5 2 | 5+6 2 | 5+6 2 |
| 166.1.140 | 623755 | 114 | H4Co3Y1 | |
| | -GM | -T | -F | -L |
| | 8 2 | 8 2 | 3+4 2 | 3+4 2 |
| | 8 2 | 9 2 | 5+6 2 | 3+4 2 |
| | 9 2 | 8 2 | 3+4 2 | 5+6 2 |
| | 9 2 | 9 2 | 5+6 2 | 5+6 2 |
| | 8 2 | 9 2 | 5+6 2 | 3+4 2 |
| | 9 2 | 8 2 | 3+4 2 | 5+6 2 |
| | 9 2 | 9 2 | 5+6 2 | 5+6 2 |
| | 6+7 2 | 6+7 2 | 5+6 2 | 5+6 2 |
| | 4+5 2 | 6+7 2 | 5+6 2 | 3+4 2 |
| | 8 2 | 9 2 | 3+4 2 | 5+6 2 |
| | 9 2 | 8 2 | 5+6 2 | 3+4 2 |
| | 6+7 2 | 4+5 2 | 3+4 2 | 5+6 2 |
| | 8 2 | 8 2 | 5+6 2 | 5+6 2 |
| | 9 2 | 9 2 | 3+4 2 | 3+4 2 |
| | 8 2 | 8 2 | 3+4 2 | 3+4 2 |
| | 8 2 | 9 2 | 5+6 2 | 5+6 2 |
| | 9 2 | 9 2 | 3+4 2 | 3+4 2 |
| | 9 2 | 9 2 | 3+4 2 | 5+6 2 |
| | 8 2 | 6+7 2 | 3+4 2 | 3+4 2 |
| | 6+7 2 | 8 2 | 5+6 2 | 3+4 2 |
| | 4+5 2 | 4+5 2 | 3+4 2 | 5+6 2 |
| | 8 2 | 8 2 | 3+4 2 | 5+6 2 |
| | 9 2 | 8 2 | 5+6 2 | 5+6 2 |
| | 8 2 | 9 2 | 5+6 2 | 3+4 2 |
| | 4+5 2 | 4+5 2 | 3+4 2 | 5+6 2 |
| | 8 2 | 8 2 | 5+6 2 | 3+4 2 |
| | 4+5 2 | 6+7 2 | 5+6 2 | 5+6 2 |
| | 8 2 | 9 2 | 3+4 2 | 5+6 2 |
| | 6+7 2 | 4+5 2 | 3+4 2 | 5+6 2 |
| | 9 2 | 8 2 | 5+6 2 | 3+4 2 |
| | 9 2 | 6+7 2 | 3+4 2 | 5+6 2 |
| | 6+7 2 | 9 2 | 5+6 2 | 5+6 2 |
| | 9 2 | 9 2 | 5+6 2 | 3+4 2 |
| | 4+5 2 | 6+7 2 | 3+4 2 | 5+6 2 |
| | 8 2 | 9 2 | 3+4 2 | 3+4 2 |
| | 8 2 | 8 2 | 5+6 2 | 5+6 2 |
| | 4+5 2 | 6+7 2 | 3+4 2 | 3+4 2 |
| | 8 2 | 9 2 | 3+4 2 | 3+4 2 |
| | 9 2 | 8 2 | 5+6 2 | 3+4 2 |
| | 6+7 2 | 4+5 2 | 3+4 2 | 5+6 2 |
| | 6+7 2 | 8 2 | 5+6 2 | 3+4 2 |
| | 9 2 | 9 2 | 5+6 2 | 5+6 2 |
| | 9 2 | 4+5 2 | 3+4 2 | 5+6 2 |
| | 8 2 | 8 2 | 5+6 2 | 3+4 2 |
| | 4+5 2 | 6+7 2 | 3+4 2 | 5+6 2 |
| | 8 2 | 9 2 | 3+4 2 | 3+4 2 |
| | 4+5 2 | 6+7 2 | 5+6 2 | 3+4 2 |
| | 8 2 | 8 2 | 3+4 2 | 3+4 2 |
| | 9 2 | 9 2 | 5+6 2 | 5+6 2 |
| | 8 2 | 4+5 2 | 5+6 2 | 5+6 2 |
| | 6+7 2 | 9 2 | 3+4 2 | 3+4 2 |
| | 9 2 | 8 2 | 5+6 2 | 5+6 2 |
| | 4+5 2 | 8 2 | 3+4 2 | 3+4 2 |
| | 9 2 | 9 2 | 3+4 2 | 5+6 2 |
| | 6+7 2 | 4+5 2 | 5+6 2 | 5+6 2 |
| | 8 2 | 6+7 2 | 3+4 2 | 5+6 2 |

Continued on next column

**TABLE LXXIII– continued**

| TQC | ICSD | NoE | SF | |
|---|---|---|---|---|
| | 8 2 | 9 2 | 5+6 2 | 3+4 2 |
| 166.1.141 | 623744 | 114 | H4Co3Y1 | |
| | -GM | -T | -F | -L |
| | 8 2 | 8 2 | 3+4 2 | 3+4 2 |
| | 8 2 | 9 2 | 5+6 2 | 3+4 2 |
| | 9 2 | 8 2 | 3+4 2 | 5+6 2 |
| | 9 2 | 9 2 | 5+6 2 | 5+6 2 |
| | 8 2 | 9 2 | 5+6 2 | 3+4 2 |
| | 9 2 | 8 2 | 3+4 2 | 5+6 2 |
| | 9 2 | 9 2 | 5+6 2 | 5+6 2 |
| | 6+7 2 | 6+7 2 | 5+6 2 | 5+6 2 |
| | 4+5 2 | 6+7 2 | 5+6 2 | 3+4 2 |
| | 8 2 | 9 2 | 3+4 2 | 5+6 2 |
| | 9 2 | 8 2 | 5+6 2 | 3+4 2 |
| | 6+7 2 | 4+5 2 | 3+4 2 | 5+6 2 |
| | 8 2 | 8 2 | 5+6 2 | 5+6 2 |
| | 9 2 | 9 2 | 3+4 2 | 3+4 2 |
| | 8 2 | 8 2 | 3+4 2 | 3+4 2 |
| | 8 2 | 9 2 | 5+6 2 | 5+6 2 |
| | 9 2 | 9 2 | 3+4 2 | 3+4 2 |
| | 9 2 | 9 2 | 3+4 2 | 5+6 2 |
| | 8 2 | 6+7 2 | 3+4 2 | 3+4 2 |
| | 6+7 2 | 8 2 | 5+6 2 | 3+4 2 |
| | 4+5 2 | 4+5 2 | 3+4 2 | 5+6 2 |
| | 8 2 | 8 2 | 3+4 2 | 5+6 2 |
| | 9 2 | 8 2 | 5+6 2 | 5+6 2 |
| | 8 2 | 9 2 | 5+6 2 | 3+4 2 |
| | 4+5 2 | 4+5 2 | 3+4 2 | 5+6 2 |
| | 8 2 | 8 2 | 5+6 2 | 3+4 2 |
| | 4+5 2 | 6+7 2 | 5+6 2 | 5+6 2 |
| | 8 2 | 9 2 | 3+4 2 | 5+6 2 |
| | 6+7 2 | 4+5 2 | 3+4 2 | 5+6 2 |
| | 9 2 | 8 2 | 5+6 2 | 3+4 2 |
| | 9 2 | 6+7 2 | 3+4 2 | 5+6 2 |
| | 6+7 2 | 9 2 | 5+6 2 | 5+6 2 |
| | 9 2 | 9 2 | 5+6 2 | 3+4 2 |
| | 4+5 2 | 6+7 2 | 3+4 2 | 5+6 2 |
| | 8 2 | 9 2 | 3+4 2 | 3+4 2 |
| | 8 2 | 8 2 | 5+6 2 | 5+6 2 |
| | 4+5 2 | 6+7 2 | 3+4 2 | 3+4 2 |
| | 8 2 | 9 2 | 3+4 2 | 3+4 2 |
| | 9 2 | 8 2 | 5+6 2 | 3+4 2 |
| | 6+7 2 | 4+5 2 | 3+4 2 | 5+6 2 |
| | 6+7 2 | 8 2 | 5+6 2 | 5+6 2 |
| | 9 2 | 9 2 | 3+4 2 | 3+4 2 |
| | 9 2 | 4+5 2 | 5+6 2 | 5+6 2 |
| | 8 2 | 8 2 | 5+6 2 | 3+4 2 |
| | 4+5 2 | 6+7 2 | 3+4 2 | 5+6 2 |
| | 8 2 | 9 2 | 5+6 2 | 3+4 2 |
| | 4+5 2 | 6+7 2 | 3+4 2 | 3+4 2 |
| | 8 2 | 8 2 | 3+4 2 | 3+4 2 |
| | 9 2 | 9 2 | 5+6 2 | 5+6 2 |
| | 8 2 | 4+5 2 | 5+6 2 | 5+6 2 |
| | 6+7 2 | 9 2 | 3+4 2 | 3+4 2 |
| | 9 2 | 8 2 | 5+6 2 | 5+6 2 |
| | 4+5 2 | 8 2 | 3+4 2 | 3+4 2 |
| | 9 2 | 9 2 | 3+4 2 | 5+6 2 |
| | 6+7 2 | 4+5 2 | 5+6 2 | 5+6 2 |
| | 8 2 | 6+7 2 | 3+4 2 | 5+6 2 |
| | 8 2 | 9 2 | 5+6 2 | 3+4 2 |
| 166.1.142 | 604264 | 114 | H4Co3Y1 | |

Continued on next column

**TABLE LXXIII– continued**

| TQC | ICSD | NoE | SF | |
|---|---|---|---|---|
| | -GM | -T | -F | -L |
| | 8 2 | 8 2 | 3+4 2 | 3+4 2 |
| | 8 2 | 9 2 | 5+6 2 | 3+4 2 |
| | 9 2 | 8 2 | 3+4 2 | 5+6 2 |
| | 9 2 | 9 2 | 5+6 2 | 5+6 2 |
| | 8 2 | 9 2 | 5+6 2 | 3+4 2 |
| | 9 2 | 8 2 | 3+4 2 | 5+6 2 |
| | 9 2 | 9 2 | 5+6 2 | 5+6 2 |
| | 6+7 2 | 6+7 2 | 5+6 2 | 5+6 2 |
| | 4+5 2 | 6+7 2 | 5+6 2 | 3+4 2 |
| | 8 2 | 9 2 | 3+4 2 | 5+6 2 |
| | 9 2 | 8 2 | 5+6 2 | 3+4 2 |
| | 6+7 2 | 4+5 2 | 3+4 2 | 5+6 2 |
| | 8 2 | 8 2 | 5+6 2 | 5+6 2 |
| | 9 2 | 9 2 | 3+4 2 | 3+4 2 |
| | 8 2 | 8 2 | 3+4 2 | 3+4 2 |
| | 8 2 | 9 2 | 5+6 2 | 5+6 2 |
| | 9 2 | 9 2 | 3+4 2 | 3+4 2 |
| | 9 2 | 9 2 | 3+4 2 | 5+6 2 |
| | 8 2 | 6+7 2 | 3+4 2 | 3+4 2 |
| | 6+7 2 | 8 2 | 5+6 2 | 3+4 2 |
| | 4+5 2 | 4+5 2 | 3+4 2 | 5+6 2 |
| | 8 2 | 8 2 | 5+6 2 | 5+6 2 |
| | 9 2 | 8 2 | 3+4 2 | 5+6 2 |
| | 8 2 | 9 2 | 5+6 2 | 3+4 2 |
| | 4+5 2 | 4+5 2 | 3+4 2 | 5+6 2 |
| | 8 2 | 8 2 | 5+6 2 | 3+4 2 |
| | 4+5 2 | 6+7 2 | 5+6 2 | 5+6 2 |
| | 8 2 | 9 2 | 3+4 2 | 5+6 2 |
| | 6+7 2 | 4+5 2 | 3+4 2 | 5+6 2 |
| | 9 2 | 8 2 | 5+6 2 | 3+4 2 |
| | 9 2 | 6+7 2 | 3+4 2 | 5+6 2 |
| | 9 2 | 9 2 | 5+6 2 | 5+6 2 |
| | 6+7 2 | 9 2 | 3+4 2 | 3+4 2 |
| | 4+5 2 | 6+7 2 | 5+6 2 | 5+6 2 |
| | 8 2 | 9 2 | 3+4 2 | 3+4 2 |
| | 8 2 | 8 2 | 5+6 2 | 5+6 2 |
| | 4+5 2 | 6+7 2 | 3+4 2 | 3+4 2 |
| | 8 2 | 9 2 | 3+4 2 | 3+4 2 |
| | 9 2 | 8 2 | 5+6 2 | 3+4 2 |
| | 6+7 2 | 4+5 2 | 3+4 2 | 5+6 2 |
| | 6+7 2 | 8 2 | 5+6 2 | 5+6 2 |
| | 9 2 | 9 2 | 3+4 2 | 3+4 2 |
| | 9 2 | 4+5 2 | 5+6 2 | 5+6 2 |
| | 8 2 | 8 2 | 5+6 2 | 3+4 2 |
| | 4+5 2 | 6+7 2 | 3+4 2 | 5+6 2 |
| | 8 2 | 9 2 | 5+6 2 | 3+4 2 |
| | 4+5 2 | 6+7 2 | 3+4 2 | 3+4 2 |
| | 8 2 | 8 2 | 3+4 2 | 3+4 2 |
| | 9 2 | 9 2 | 5+6 2 | 5+6 2 |
| | 8 2 | 4+5 2 | 5+6 2 | 5+6 2 |
| | 6+7 2 | 9 2 | 3+4 2 | 3+4 2 |
| | 9 2 | 8 2 | 5+6 2 | 5+6 2 |
| | 4+5 2 | 9 2 | 3+4 2 | 3+4 2 |
| | 9 2 | 8 2 | 3+4 2 | 5+6 2 |
| | 6+7 2 | 4+5 2 | 5+6 2 | 5+6 2 |
| | 8 2 | 6+7 2 | 3+4 2 | 5+6 2 |
| | 8 2 | 9 2 | 5+6 2 | 3+4 2 |
| 166.1.143 | 602920 | 114 | H4Co3Y1 | |
| | -GM | -T | -F | -L |
| | 8 2 | 8 2 | 3+4 2 | 3+4 2 |

Continued on next column

**TABLE LXXIII– continued**

| TQC | ICSD | NoE | SF | |
|---|---|---|---|---|
| | 8 2 | 9 2 | 5+6 2 | 3+4 2 |
| | 9 2 | 8 2 | 3+4 2 | 5+6 2 |
| | 9 2 | 9 2 | 5+6 2 | 5+6 2 |
| | 8 2 | 9 2 | 5+6 2 | 3+4 2 |
| | 9 2 | 8 2 | 3+4 2 | 5+6 2 |
| | 9 2 | 9 2 | 5+6 2 | 5+6 2 |
| | 6+7 2 | 6+7 2 | 5+6 2 | 5+6 2 |
| | 4+5 2 | 6+7 2 | 5+6 2 | 3+4 2 |
| | 8 2 | 9 2 | 3+4 2 | 5+6 2 |
| | 9 2 | 8 2 | 5+6 2 | 3+4 2 |
| | 6+7 2 | 4+5 2 | 3+4 2 | 5+6 2 |
| | 8 2 | 8 2 | 5+6 2 | 5+6 2 |
| | 9 2 | 9 2 | 3+4 2 | 3+4 2 |
| | 8 2 | 8 2 | 3+4 2 | 3+4 2 |
| | 8 2 | 9 2 | 5+6 2 | 5+6 2 |
| | 9 2 | 9 2 | 3+4 2 | 3+4 2 |
| | 9 2 | 9 2 | 3+4 2 | 5+6 2 |
| | 8 2 | 6+7 2 | 3+4 2 | 3+4 2 |
| | 6+7 2 | 8 2 | 5+6 2 | 3+4 2 |
| | 4+5 2 | 4+5 2 | 3+4 2 | 5+6 2 |
| | 8 2 | 8 2 | 5+6 2 | 5+6 2 |
| | 9 2 | 8 2 | 3+4 2 | 5+6 2 |
| | 8 2 | 9 2 | 5+6 2 | 3+4 2 |
| | 4+5 2 | 4+5 2 | 3+4 2 | 5+6 2 |
| | 8 2 | 8 2 | 5+6 2 | 3+4 2 |
| | 4+5 2 | 6+7 2 | 5+6 2 | 5+6 2 |
| | 8 2 | 9 2 | 3+4 2 | 5+6 2 |
| | 6+7 2 | 4+5 2 | 3+4 2 | 5+6 2 |
| | 9 2 | 8 2 | 5+6 2 | 3+4 2 |
| | 9 2 | 6+7 2 | 3+4 2 | 5+6 2 |
| | 9 2 | 9 2 | 5+6 2 | 5+6 2 |
| | 6+7 2 | 9 2 | 3+4 2 | 3+4 2 |
| | 4+5 2 | 6+7 2 | 5+6 2 | 5+6 2 |
| | 8 2 | 9 2 | 3+4 2 | 3+4 2 |
| | 8 2 | 8 2 | 5+6 2 | 5+6 2 |
| | 4+5 2 | 6+7 2 | 3+4 2 | 3+4 2 |
| | 8 2 | 9 2 | 3+4 2 | 3+4 2 |
| | 9 2 | 8 2 | 5+6 2 | 3+4 2 |
| | 6+7 2 | 4+5 2 | 3+4 2 | 5+6 2 |
| | 6+7 2 | 9 2 | 5+6 2 | 5+6 2 |
| | 9 2 | 8 2 | 3+4 2 | 3+4 2 |
| | 9 2 | 4+5 2 | 5+6 2 | 5+6 2 |
| | 8 2 | 8 2 | 5+6 2 | 3+4 2 |
| | 4+5 2 | 6+7 2 | 3+4 2 | 5+6 2 |
| | 8 2 | 9 2 | 5+6 2 | 3+4 2 |
| | 4+5 2 | 6+7 2 | 3+4 2 | 3+4 2 |
| | 8 2 | 8 2 | 3+4 2 | 3+4 2 |
| | 9 2 | 9 2 | 5+6 2 | 5+6 2 |
| | 8 2 | 4+5 2 | 5+6 2 | 5+6 2 |
| | 6+7 2 | 9 2 | 3+4 2 | 3+4 2 |
| | 9 2 | 8 2 | 5+6 2 | 5+6 2 |
| | 4+5 2 | 9 2 | 3+4 2 | 3+4 2 |
| | 9 2 | 8 2 | 3+4 2 | 5+6 2 |
| | 6+7 2 | 4+5 2 | 5+6 2 | 5+6 2 |
| | 8 2 | 6+7 2 | 3+4 2 | 5+6 2 |
| | 8 2 | 9 2 | 5+6 2 | 3+4 2 |
| 166.1.144 | 623723 | | 120 | H4Co3Pr1 |
| | -GM | -T | -F | -L |
| | 8 2 | 8 2 | 3+4 2 | 3+4 2 |
| | 8 2 | 9 2 | 5+6 2 | 3+4 2 |
| | 9 2 | 8 2 | 3+4 2 | 5+6 2 |

Continued on next column

**TABLE LXXIII– continued**

| TQC | ICSD | NoE | SF | |
|---|---|---|---|---|
| | 8 2 | 9 2 | 5+6 2 | 5+6 2 |
| | 9 2 | 9 2 | 5+6 2 | 3+4 2 |
| | 9 2 | 8 2 | 3+4 2 | 5+6 2 |
| | 4+5 2 | 6+7 2 | 5+6 2 | 3+4 2 |
| | 8 2 | 9 2 | 3+4 2 | 5+6 2 |
| | 6+7 2 | 6+7 2 | 5+6 2 | 5+6 2 |
| | 9 2 | 9 2 | 5+6 2 | 5+6 2 |
| | 9 2 | 8 2 | 5+6 2 | 3+4 2 |
| | 6+7 2 | 4+5 2 | 3+4 2 | 5+6 2 |
| | 8 2 | 8 2 | 5+6 2 | 5+6 2 |
| | 9 2 | 9 2 | 3+4 2 | 3+4 2 |
| | 8 2 | 8 2 | 3+4 2 | 3+4 2 |
| | 9 2 | 9 2 | 3+4 2 | 5+6 2 |
| | 8 2 | 6+7 2 | 5+6 2 | 5+6 2 |
| | 6+7 2 | 9 2 | 3+4 2 | 3+4 2 |
| | 4+5 2 | 9 2 | 5+6 2 | 3+4 2 |
| | 9 2 | 4+5 2 | 3+4 2 | 5+6 2 |
| | 8 2 | 8 2 | 5+6 2 | 3+4 2 |
| | 8 2 | 8 2 | 3+4 2 | 5+6 2 |
| | 4+5 2 | 4+5 2 | 3+4 2 | 3+4 2 |
| | 8 2 | 8 2 | 5+6 2 | 5+6 2 |
| | 4+5 2 | 8 2 | 5+6 2 | 5+6 2 |
| | 8 2 | 4+5 2 | 3+4 2 | 5+6 2 |
| | 9 2 | 9 2 | 5+6 2 | 3+4 2 |
| | 6+7 2 | 6+7 2 | 5+6 2 | 3+4 2 |
| | 9 2 | 8 2 | 3+4 2 | 5+6 2 |
| | 6+7 2 | 6+7 2 | 5+6 2 | 3+4 2 |
| | 9 2 | 9 2 | 3+4 2 | 5+6 2 |
| | 8 2 | 9 2 | 3+4 2 | 5+6 2 |
| | 9 2 | 9 2 | 5+6 2 | 3+4 2 |
| | 8 2 | 8 2 | 3+4 2 | 5+6 2 |
| | 4+5 2 | 6+7 2 | 5+6 2 | 3+4 2 |
| | 6+7 2 | 9 2 | 3+4 2 | 5+6 2 |
| | 8 2 | 8 2 | 5+6 2 | 3+4 2 |
| | 8 2 | 4+5 2 | 5+6 2 | 5+6 2 |
| | 9 2 | 6+7 2 | 3+4 2 | 3+4 2 |
| | 4+5 2 | 9 2 | 3+4 2 | 3+4 2 |
| | 6+7 2 | 6+7 2 | 3+4 2 | 5+6 2 |
| | 9 2 | 4+5 2 | 5+6 2 | 5+6 2 |
| | 4+5 2 | 9 2 | 5+6 2 | 3+4 2 |
| | 8 2 | 4+5 2 | 3+4 2 | 5+6 2 |
| | 9 2 | 8 2 | 3+4 2 | 5+6 2 |
| | 9 2 | 6+7 2 | 5+6 2 | 3+4 2 |
| | 8 2 | 8 2 | 5+6 2 | 3+4 2 |
| | 4+5 2 | 9 2 | 3+4 2 | 5+6 2 |
| | 8 2 | 9 2 | 3+4 2 | 3+4 2 |
| | 6+7 2 | 8 2 | 5+6 2 | 5+6 2 |
| | 8 2 | 9 2 | 3+4 2 | 3+4 2 |
| | 4+5 2 | 4+5 2 | 5+6 2 | 5+6 2 |
| | 9 2 | 6+7 2 | 3+4 2 | 3+4 2 |
| | 6+7 2 | 8 2 | 3+4 2 | 5+6 2 |
| | 9 2 | 9 2 | 5+6 2 | 5+6 2 |
| | 8 2 | 6+7 2 | 3+4 2 | 3+4 2 |
| | 4+5 2 | 9 2 | 3+4 2 | 5+6 2 |
| | 8 2 | 8 2 | 5+6 2 | 5+6 2 |
| | 9 2 | 8 2 | 3+4 2 | 5+6 2 |
| | 8 2 | 9 2 | 5+6 2 | 3+4 2 |
| 166.1.145 | 623698 | | 144 | H4Co3Ho1 |
| | -GM | -T | -F | -L |
| | 8 2 | 8 2 | 3+4 2 | 3+4 2 |
| | 8 2 | 9 2 | 5+6 2 | 3+4 2 |

Continued on next column

**TABLE LXXIII– continued**

| TQC | ICSD | NoE | SF | |
|---|---|---|---|---|
| | 9 2 | 8 2 | 3+4 2 | 5+6 2 |
| | 9 2 | 9 2 | 5+6 2 | 5+6 2 |
| | 8 2 | 9 2 | 5+6 2 | 3+4 2 |
| | 9 2 | 8 2 | 3+4 2 | 5+6 2 |
| | 6+7 2 | 6+7 2 | 5+6 2 | 5+6 2 |
| | 9 2 | 9 2 | 5+6 2 | 5+6 2 |
| | 8 2 | 9 2 | 5+6 2 | 3+4 2 |
| | 9 2 | 8 2 | 3+4 2 | 5+6 2 |
| | 4+5 2 | 6+7 2 | 5+6 2 | 3+4 2 |
| | 6+7 2 | 4+5 2 | 3+4 2 | 5+6 2 |
| | 8 2 | 8 2 | 5+6 2 | 5+6 2 |
| | 9 2 | 9 2 | 5+6 2 | 5+6 2 |
| | 9 2 | 9 2 | 3+4 2 | 3+4 2 |
| | 8 2 | 6+7 2 | 5+6 2 | 5+6 2 |
| | 6+7 2 | 9 2 | 3+4 2 | 3+4 2 |
| | 9 2 | 9 2 | 5+6 2 | 5+6 2 |
| | 9 2 | 8 2 | 5+6 2 | 5+6 2 |
| | 9 2 | 9 2 | 5+6 2 | 5+6 2 |
| | 9 2 | 9 2 | 5+6 2 | 5+6 2 |
| | 8 2 | 6+7 2 | 3+4 2 | 3+4 2 |
| | 6+7 2 | 9 2 | 5+6 2 | 5+6 2 |
| | 9 2 | 9 2 | 3+4 2 | 5+6 2 |
| | 4+5 2 | 6+7 2 | 5+6 2 | 3+4 2 |
| | 8 2 | 4+5 2 | 3+4 2 | 5+6 2 |
| | 6+7 2 | 9 2 | 5+6 2 | 3+4 2 |
| | 8 2 | 8 2 | 3+4 2 | 3+4 2 |
| | 9 2 | 8 2 | 5+6 2 | 5+6 2 |
| | 4+5 2 | 8 2 | 5+6 2 | 5+6 2 |
| | 8 2 | 4+5 2 | 3+4 2 | 5+6 2 |
| | 9 2 | 8 2 | 5+6 2 | 3+4 2 |
| | 4+5 2 | 4+5 2 | 3+4 2 | 5+6 2 |
| | 8 2 | 9 2 | 5+6 2 | 5+6 2 |
| | 6+7 2 | 6+7 2 | 5+6 2 | 3+4 2 |
| | 9 2 | 8 2 | 3+4 2 | 5+6 2 |
| | 8 2 | 9 2 | 3+4 2 | 3+4 2 |
| | 6+7 2 | 6+7 2 | 5+6 2 | 5+6 2 |
| | 9 2 | 9 2 | 5+6 2 | 3+4 2 |
| | 9 2 | 9 2 | 3+4 2 | 5+6 2 |
| | 8 2 | 6+7 2 | 5+6 2 | 5+6 2 |
| | 4+5 2 | 9 2 | 3+4 2 | 3+4 2 |
| | 6+7 2 | 8 2 | 3+4 2 | 5+6 2 |
| | 8 2 | 4+5 2 | 3+4 2 | 3+4 2 |
| | 9 2 | 6+7 2 | 3+4 2 | 5+6 2 |
| | 9 2 | 9 2 | 5+6 2 | 3+4 2 |
| | 8 2 | 8 2 | 5+6 2 | 5+6 2 |
| | 4+5 2 | 4+5 2 | 3+4 2 | 5+6 2 |
| | 8 2 | 8 2 | 3+4 2 | 3+4 2 |
| | 4+5 2 | 9 2 | 5+6 2 | 3+4 2 |
| | 6+7 2 | 8 2 | 5+6 2 | 5+6 2 |
| | 9 2 | 6+7 2 | 5+6 2 | 3+4 2 |
| | 8 2 | 4+5 2 | 3+4 2 | 5+6 2 |
| | 9 2 | 9 2 | 3+4 2 | 3+4 2 |
| | 8 2 | 8 2 | 5+6 2 | 3+4 2 |
| | 6+7 2 | 9 2 | 5+6 2 | 5+6 2 |
| | 4+5 2 | 6+7 2 | 3+4 2 | 3+4 2 |
| | 9 2 | 8 2 | 5+6 2 | 5+6 2 |
| | 6+7 2 | 4+5 2 | 3+4 2 | 5+6 2 |
| | 8 2 | 9 2 | 5+6 2 | 3+4 2 |
| | 9 2 | 9 2 | 3+4 2 | 5+6 2 |
| | 4+5 2 | 8 2 | 5+6 2 | 5+6 2 |
| | 8 2 | 6+7 2 | 3+4 2 | 3+4 2 |

Continued on next column

**TABLE LXXIII– continued**

| TQC | ICSD | NoE | SF | |
|---|---|---|---|---|
| | 4+5 2 | 9 2 | 3+4 2 | 5+6 2 |
| | 8 2 | 6+7 2 | 3+4 2 | 5+6 2 |
| | 9 2 | 9 2 | 5+6 2 | 3+4 2 |
| | 8 2 | 4+5 2 | 3+4 2 | 5+6 2 |
| | 9 2 | 8 2 | 3+4 2 | 5+6 2 |
| | 4+5 2 | 9 2 | 3+4 2 | 3+4 2 |
| | 8 2 | 8 2 | 5+6 2 | 5+6 2 |
| | 8 2 | 8 2 | 3+4 2 | 3+4 2 |
| | 9 2 | 8 2 | 5+6 2 | 5+6 2 |
| 166.1.146 | 602925 | 144 | H4Co3Ho1 | |
| | -GM | -T | -F | -L |
| | 8 2 | 8 2 | 3+4 2 | 3+4 2 |
| | 8 2 | 9 2 | 5+6 2 | 3+4 2 |
| | 9 2 | 8 2 | 3+4 2 | 5+6 2 |
| | 9 2 | 9 2 | 5+6 2 | 5+6 2 |
| | 8 2 | 9 2 | 5+6 2 | 3+4 2 |
| | 9 2 | 8 2 | 3+4 2 | 5+6 2 |
| | 4+5 2 | 6+7 2 | 5+6 2 | 3+4 2 |
| | 9 2 | 9 2 | 3+4 2 | 5+6 2 |
| | 8 2 | 6+7 2 | 5+6 2 | 5+6 2 |
| | 6+7 2 | 9 2 | 5+6 2 | 5+6 2 |
| | 9 2 | 8 2 | 5+6 2 | 3+4 2 |
| | 6+7 2 | 4+5 2 | 3+4 2 | 5+6 2 |
| | 8 2 | 8 2 | 5+6 2 | 5+6 2 |
| | 9 2 | 9 2 | 3+4 2 | 3+4 2 |
| | 8 2 | 8 2 | 3+4 2 | 3+4 2 |
| | 8 2 | 9 2 | 5+6 2 | 5+6 2 |
| | 9 2 | 9 2 | 3+4 2 | 3+4 2 |
| | 9 2 | 9 2 | 3+4 2 | 5+6 2 |
| | 6+7 2 | 6+7 2 | 3+4 2 | 3+4 2 |
| | 8 2 | 8 2 | 5+6 2 | 3+4 2 |
| | 4+5 2 | 4+5 2 | 3+4 2 | 5+6 2 |
| | 8 2 | 8 2 | 5+6 2 | 5+6 2 |
| | 9 2 | 8 2 | 3+4 2 | 5+6 2 |
| | 8 2 | 9 2 | 5+6 2 | 3+4 2 |
| | 8 2 | 4+5 2 | 3+4 2 | 5+6 2 |
| | 4+5 2 | 8 2 | 5+6 2 | 3+4 2 |
| | 4+5 2 | 4+5 2 | 5+6 2 | 5+6 2 |
| | 8 2 | 9 2 | 3+4 2 | 5+6 2 |
| | 9 2 | 6+7 2 | 3+4 2 | 5+6 2 |
| | 6+7 2 | 8 2 | 5+6 2 | 3+4 2 |
| | 9 2 | 9 2 | 3+4 2 | 5+6 2 |
| | 6+7 2 | 6+7 2 | 5+6 2 | 5+6 2 |
| | 9 2 | 9 2 | 3+4 2 | 3+4 2 |
| | 9 2 | 6+7 2 | 5+6 2 | 5+6 2 |
| | 4+5 2 | 9 2 | 3+4 2 | 5+6 2 |
| | 8 2 | 8 2 | 5+6 2 | 3+4 2 |
| | 9 2 | 9 2 | 3+4 2 | 3+4 2 |
| | 9 2 | 8 2 | 3+4 2 | 5+6 2 |
| | 8 2 | 9 2 | 5+6 2 | 5+6 2 |
| | 6+7 2 | 9 2 | 5+6 2 | 3+4 2 |
| | 4+5 2 | 4+5 2 | 3+4 2 | 3+4 2 |
| | 8 2 | 6+7 2 | 5+6 2 | 5+6 2 |
| | 6+7 2 | 9 2 | 3+4 2 | 3+4 2 |
| | 9 2 | 6+7 2 | 5+6 2 | 3+4 2 |
| | 8 2 | 8 2 | 5+6 2 | 5+6 2 |
| | 4+5 2 | 9 2 | 5+6 2 | 5+6 2 |
| | 8 2 | 6+7 2 | 3+4 2 | 3+4 2 |
| | 9 2 | 8 2 | 5+6 2 | 5+6 2 |
| | 6+7 2 | 4+5 2 | 5+6 2 | 3+4 2 |
| | 9 2 | 8 2 | 3+4 2 | 5+6 2 |

Continued on next column

**TABLE LXXIII– continued**

| TQC | ICSD | NoE | SF | |
|---|---|---|---|---|
| | 6+7 2 | 9 2 | 3+4 2 | 5+6 2 |
| | 9 2 | 4+5 2 | 5+6 2 | 5+6 2 |
| | 8 2 | 8 2 | 5+6 2 | 3+4 2 |
| | 4+5 2 | 6+7 2 | 3+4 2 | 5+6 2 |
| | 8 2 | 9 2 | 3+4 2 | 3+4 2 |
| | 9 2 | 8 2 | 5+6 2 | 3+4 2 |
| | 8 2 | 9 2 | 3+4 2 | 3+4 2 |
| | 4+5 2 | 9 2 | 5+6 2 | 5+6 2 |
| | 8 2 | 6+7 2 | 5+6 2 | 5+6 2 |
| | 6+7 2 | 4+5 2 | 3+4 2 | 5+6 2 |
| | 9 2 | 8 2 | 5+6 2 | 3+4 2 |
| | 9 2 | 9 2 | 3+4 2 | 3+4 2 |
| | 4+5 2 | 9 2 | 5+6 2 | 5+6 2 |
| | 8 2 | 6+7 2 | 3+4 2 | 5+6 2 |
| | 9 2 | 9 2 | 5+6 2 | 5+6 2 |
| | 6+7 2 | 8 2 | 3+4 2 | 5+6 2 |
| | 8 2 | 4+5 2 | 3+4 2 | 3+4 2 |
| | 9 2 | 6+7 2 | 5+6 2 | 5+6 2 |
| | 6+7 2 | 9 2 | 5+6 2 | 3+4 2 |
| | 9 2 | 8 2 | 3+4 2 | 3+4 2 |
| | 9 2 | 9 2 | 5+6 2 | 5+6 2 |
| | 4+5 2 | 4+5 2 | 5+6 2 | 3+4 2 |
| 166.1.147 | 623752 | 114 | H3Co3Y1 | |
| | -GM | -T | -F | -L |
| | 8 2 | 8 2 | 3+4 2 | 3+4 2 |
| | 8 2 | 9 2 | 5+6 2 | 3+4 2 |
| | 9 2 | 8 2 | 3+4 2 | 5+6 2 |
| | 9 2 | 9 2 | 5+6 2 | 5+6 2 |
| | 8 2 | 9 2 | 5+6 2 | 3+4 2 |
| | 9 2 | 8 2 | 3+4 2 | 5+6 2 |
| | 9 2 | 9 2 | 5+6 2 | 5+6 2 |
| | 4+5 2 | 6+7 2 | 5+6 2 | 5+6 2 |
| | 6+7 2 | 6+7 2 | 5+6 2 | 3+4 2 |
| | 8 2 | 9 2 | 3+4 2 | 5+6 2 |
| | 9 2 | 8 2 | 5+6 2 | 3+4 2 |
| | 6+7 2 | 4+5 2 | 3+4 2 | 5+6 2 |
| | 8 2 | 8 2 | 5+6 2 | 5+6 2 |
| | 9 2 | 9 2 | 3+4 2 | 3+4 2 |
| | 8 2 | 8 2 | 3+4 2 | 3+4 2 |
| | 8 2 | 9 2 | 5+6 2 | 5+6 2 |
| | 9 2 | 9 2 | 3+4 2 | 3+4 2 |
| | 9 2 | 9 2 | 3+4 2 | 5+6 2 |
| | 8 2 | 6+7 2 | 3+4 2 | 3+4 2 |
| | 6+7 2 | 8 2 | 5+6 2 | 3+4 2 |
| | 4+5 2 | 4+5 2 | 3+4 2 | 5+6 2 |
| | 8 2 | 8 2 | 3+4 2 | 5+6 2 |
| | 9 2 | 8 2 | 5+6 2 | 5+6 2 |
| | 8 2 | 9 2 | 5+6 2 | 3+4 2 |
| | 4+5 2 | 4+5 2 | 3+4 2 | 5+6 2 |
| | 8 2 | 8 2 | 5+6 2 | 3+4 2 |
| | 4+5 2 | 6+7 2 | 5+6 2 | 5+6 2 |
| | 8 2 | 9 2 | 3+4 2 | 5+6 2 |
| | 6+7 2 | 4+5 2 | 3+4 2 | 5+6 2 |
| | 9 2 | 8 2 | 5+6 2 | 3+4 2 |
| | 9 2 | 6+7 2 | 3+4 2 | 5+6 2 |
| | 9 2 | 9 2 | 5+6 2 | 5+6 2 |
| | 6+7 2 | 9 2 | 3+4 2 | 3+4 2 |
| | 4+5 2 | 6+7 2 | 5+6 2 | 5+6 2 |
| | 8 2 | 9 2 | 3+4 2 | 3+4 2 |
| | 8 2 | 8 2 | 5+6 2 | 5+6 2 |
| | 4+5 2 | 6+7 2 | 3+4 2 | 3+4 2 |

Continued on next column

**TABLE LXXIII– continued**

| TQC | ICSD | NoE | SF | |
|---|---|---|---|---|
| | 8 2 | 9 2 | 3+4 2 | 3+4 2 |
| | 9 2 | 8 2 | 5+6 2 | 3+4 2 |
| | 6+7 2 | 4+5 2 | 3+4 2 | 5+6 2 |
| | 6+7 2 | 8 2 | 5+6 2 | 3+4 2 |
| | 9 2 | 9 2 | 3+4 2 | 5+6 2 |
| | 9 2 | 4+5 2 | 5+6 2 | 5+6 2 |
| | 8 2 | 8 2 | 5+6 2 | 3+4 2 |
| | 4+5 2 | 6+7 2 | 3+4 2 | 5+6 2 |
| | 8 2 | 9 2 | 5+6 2 | 3+4 2 |
| | 4+5 2 | 6+7 2 | 3+4 2 | 3+4 2 |
| | 8 2 | 8 2 | 3+4 2 | 3+4 2 |
| | 9 2 | 9 2 | 5+6 2 | 5+6 2 |
| | 8 2 | 4+5 2 | 5+6 2 | 5+6 2 |
| | 6+7 2 | 9 2 | 3+4 2 | 3+4 2 |
| | 9 2 | 8 2 | 5+6 2 | 5+6 2 |
| | 4+5 2 | 8 2 | 3+4 2 | 3+4 2 |
| | 9 2 | 9 2 | 3+4 2 | 5+6 2 |
| | 6+7 2 | 4+5 2 | 5+6 2 | 5+6 2 |
| | 8 2 | 6+7 2 | 3+4 2 | 5+6 2 |
| | 8 2 | 9 2 | 5+6 2 | 3+4 2 |
| 166.1.148 | 24685 | 86 | H3Cl1Co2O3 | |
| | -GM | -T | -F | -L |
| | 8 2 | 8 2 | 3+4 2 | 5+6 2 |
| | 9 2 | 9 2 | 5+6 2 | 3+4 2 |
| | 4+5 2 | 4+5 2 | 5+6 2 | 3+4 2 |
| | 8 2 | 8 2 | 3+4 2 | 5+6 2 |
| | 6+7 2 | 6+7 2 | 5+6 2 | 3+4 2 |
| | 9 2 | 9 2 | 3+4 2 | 5+6 2 |
| | 8 2 | 9 2 | 5+6 2 | 5+6 2 |
| | 9 2 | 8 2 | 3+4 2 | 3+4 2 |
| | 8 2 | 8 2 | 5+6 2 | 5+6 2 |
| | 8 2 | 4+5 2 | 5+6 2 | 5+6 2 |
| | 4+5 2 | 8 2 | 3+4 2 | 3+4 2 |
| | 8 2 | 9 2 | 3+4 2 | 5+6 2 |
| | 9 2 | 6+7 2 | 3+4 2 | 3+4 2 |
| | 9 2 | 8 2 | 5+6 2 | 5+6 2 |
| | 6+7 2 | 8 2 | 3+4 2 | 5+6 2 |
| | 8 2 | 4+5 2 | 3+4 2 | 5+6 2 |
| | 4+5 2 | 9 2 | 5+6 2 | 3+4 2 |
| | 8 2 | 8 2 | 5+6 2 | 3+4 2 |
| | 8 2 | 9 2 | 5+6 2 | 3+4 2 |
| | 4+5 2 | 8 2 | 5+6 2 | 5+6 2 |
| | 8 2 | 4+5 2 | 3+4 2 | 5+6 2 |
| | 9 2 | 9 2 | 3+4 2 | 5+6 2 |
| | 6+7 2 | 6+7 2 | 3+4 2 | 3+4 2 |
| | 4+5 2 | 9 2 | 5+6 2 | 5+6 2 |
| | 8 2 | 6+7 2 | 3+4 2 | 3+4 2 |
| | 9 2 | 8 2 | 5+6 2 | 3+4 2 |
| | 9 2 | 9 2 | 3+4 2 | 5+6 2 |
| | 9 2 | 4+5 2 | 5+6 2 | 5+6 2 |
| | 6+7 2 | 9 2 | 5+6 2 | 3+4 2 |
| | 4+5 2 | 8 2 | 3+4 2 | 3+4 2 |
| | 8 2 | 9 2 | 3+4 2 | 3+4 2 |
| | 8 2 | 6+7 2 | 5+6 2 | 5+6 2 |
| | 9 2 | 8 2 | 5+6 2 | 5+6 2 |
| | 8 2 | 4+5 2 | 3+4 2 | 3+4 2 |
| | 9 2 | 8 2 | 5+6 2 | 5+6 2 |
| | 4+5 2 | 8 2 | 3+4 2 | 5+6 2 |
| | 6+7 2 | 4+5 2 | 3+4 2 | 5+6 2 |
| | 8 2 | 8 2 | 5+6 2 | 5+6 2 |
| | 4+5 2 | 9 2 | 5+6 2 | 3+4 2 |

Continued on next column

**TABLE LXXIII– continued**

| TQC | ICSD | NoE | SF | |
|---|---|---|---|---|
| | 8 2 | 6+7 2 | 3+4 2 | 3+4 2 |
| | 4+5 2 | 9 2 | 5+6 2 | 5+6 2 |
| | 8 2 | 8 2 | 3+4 2 | 5+6 2 |
| | 8 2 | 4+5 2 | 3+4 2 | 3+4 2 |
| 166.1.149 | 155534 | 86 | H3Cl1Co2O3 | |
| | -GM | -T | -F | -L |
| | 8 2 | 8 2 | 3+4 2 | 5+6 2 |
| | 9 2 | 9 2 | 5+6 2 | 3+4 2 |
| | 4+5 2 | 4+5 2 | 5+6 2 | 3+4 2 |
| | 8 2 | 8 2 | 3+4 2 | 5+6 2 |
| | 6+7 2 | 6+7 2 | 5+6 2 | 3+4 2 |
| | 9 2 | 9 2 | 3+4 2 | 5+6 2 |
| | 8 2 | 9 2 | 5+6 2 | 3+4 2 |
| | 9 2 | 8 2 | 3+4 2 | 5+6 2 |
| | 8 2 | 8 2 | 5+6 2 | 5+6 2 |
| | 4+5 2 | 4+5 2 | 5+6 2 | 5+6 2 |
| | 8 2 | 8 2 | 3+4 2 | 5+6 2 |
| | 8 2 | 8 2 | 3+4 2 | 3+4 2 |
| | 9 2 | 9 2 | 3+4 2 | 3+4 2 |
| | 9 2 | 6+7 2 | 5+6 2 | 5+6 2 |
| | 8 2 | 8 2 | 3+4 2 | 3+4 2 |
| | 6+7 2 | 9 2 | 3+4 2 | 5+6 2 |
| | 4+5 2 | 9 2 | 5+6 2 | 5+6 2 |
| | 8 2 | 8 2 | 5+6 2 | 3+4 2 |
| | 8 2 | 4+5 2 | 5+6 2 | 5+6 2 |
| | 4+5 2 | 8 2 | 3+4 2 | 3+4 2 |
| | 8 2 | 4+5 2 | 5+6 2 | 5+6 2 |
| | 9 2 | 6+7 2 | 3+4 2 | 3+4 2 |
| | 6+7 2 | 9 2 | 3+4 2 | 5+6 2 |
| | 4+5 2 | 8 2 | 5+6 2 | 5+6 2 |
| | 8 2 | 9 2 | 3+4 2 | 3+4 2 |
| | 9 2 | 9 2 | 5+6 2 | 3+4 2 |
| | 9 2 | 6+7 2 | 3+4 2 | 5+6 2 |
| | 9 2 | 9 2 | 5+6 2 | 5+6 2 |
| | 6+7 2 | 4+5 2 | 5+6 2 | 3+4 2 |
| | 4+5 2 | 8 2 | 3+4 2 | 3+4 2 |
| | 8 2 | 9 2 | 3+4 2 | 3+4 2 |
| | 9 2 | 6+7 2 | 5+6 2 | 5+6 2 |
| | 8 2 | 8 2 | 5+6 2 | 5+6 2 |
| | 8 2 | 4+5 2 | 3+4 2 | 3+4 2 |
| | 4+5 2 | 8 2 | 5+6 2 | 5+6 2 |
| | 8 2 | 8 2 | 3+4 2 | 5+6 2 |
| | 9 2 | 4+5 2 | 3+4 2 | 5+6 2 |
| | 6+7 2 | 8 2 | 5+6 2 | 5+6 2 |
| | 4+5 2 | 9 2 | 5+6 2 | 3+4 2 |
| | 8 2 | 6+7 2 | 3+4 2 | 3+4 2 |
| | 4+5 2 | 9 2 | 5+6 2 | 5+6 2 |
| | 8 2 | 8 2 | 3+4 2 | 5+6 2 |
| | 8 2 | 4+5 2 | 3+4 2 | 3+4 2 |
| 166.1.150 | 6016 | 44 | H2N2Th1 | |
| | -GM | -T | -F | -L |
| | 8 2 | 9 2 | 5+6 2 | 3+4 2 |
| | 9 2 | 8 2 | 3+4 2 | 5+6 2 |
| | 8 2 | 9 2 | 5+6 2 | 3+4 2 |
| | 9 2 | 8 2 | 3+4 2 | 5+6 2 |
| | 8 2 | 9 2 | 5+6 2 | 3+4 2 |
| | 4+5 2 | 6+7 2 | 3+4 2 | 5+6 2 |
| | 9 2 | 8 2 | 5+6 2 | 3+4 2 |
| | 6+7 2 | 4+5 2 | 3+4 2 | 5+6 2 |
| | 9 2 | 8 2 | 5+6 2 | 5+6 2 |
| | 8 2 | 9 2 | 3+4 2 | 3+4 2 |

Continued on next column

**TABLE LXXIII– continued**

| TQC | ICSD | NoE | SF | |
|---|---|---|---|---|
| | 9 2 | 8 2 | 3+4 2 | 5+6 2 |
| | 8 2 | 9 2 | 5+6 2 | 3+4 2 |
| | 8 2 | 9 2 | 3+4 2 | 5+6 2 |
| | 6+7 2 | 4+5 2 | 5+6 2 | 3+4 2 |
| | 4+5 2 | 6+7 2 | 5+6 2 | 3+4 2 |
| | 9 2 | 8 2 | 5+6 2 | 3+4 2 |
| | 8 2 | 9 2 | 3+4 2 | 5+6 2 |
| | 9 2 | 8 2 | 3+4 2 | 5+6 2 |
| | 8 2 | 8 2 | 5+6 2 | 3+4 2 |
| | 8 2 | 9 2 | 3+4 2 | 5+6 2 |
| | 4+5 2 | 8 2 | 5+6 2 | 5+6 2 |
| | 9 2 | 4+5 2 | 3+4 2 | 3+4 2 |
| 166.1.151 | 423880 | 108 | H2Mo1O4Rb2S1 | |
| | -GM | -T | -F | -L |
| | 8 2 | 8 2 | 5+6 2 | 5+6 2 |
| | 9 2 | 9 2 | 3+4 2 | 3+4 2 |
| | 8 2 | 8 2 | 3+4 2 | 3+4 2 |
| | 9 2 | 9 2 | 5+6 2 | 5+6 2 |
| | 8 2 | 8 2 | 3+4 2 | 3+4 2 |
| | 9 2 | 9 2 | 5+6 2 | 5+6 2 |
| | 9 2 | 9 2 | 5+6 2 | 5+6 2 |
| | 6+7 2 | 6+7 2 | 3+4 2 | 3+4 2 |
| | 8 2 | 8 2 | 5+6 2 | 5+6 2 |
| | 4+5 2 | 4+5 2 | 3+4 2 | 3+4 2 |
| | 8 2 | 9 2 | 3+4 2 | 5+6 2 |
| | 9 2 | 8 2 | 5+6 2 | 3+4 2 |
| | 8 2 | 9 2 | 5+6 2 | 3+4 2 |
| | 9 2 | 8 2 | 3+4 2 | 5+6 2 |
| | 8 2 | 8 2 | 5+6 2 | 5+6 2 |
| | 9 2 | 9 2 | 3+4 2 | 3+4 2 |
| | 8 2 | 8 2 | 5+6 2 | 5+6 2 |
| | 9 2 | 9 2 | 3+4 2 | 3+4 2 |
| | 4+5 2 | 4+5 2 | 5+6 2 | 5+6 2 |
| | 8 2 | 8 2 | 3+4 2 | 3+4 2 |
| | 6+7 2 | 6+7 2 | 5+6 2 | 5+6 2 |
| | 9 2 | 9 2 | 3+4 2 | 3+4 2 |
| | 4+5 2 | 4+5 2 | 5+6 2 | 5+6 2 |
| | 8 2 | 8 2 | 3+4 2 | 3+4 2 |
| | 6+7 2 | 6+7 2 | 5+6 2 | 5+6 2 |
| | 9 2 | 9 2 | 3+4 2 | 3+4 2 |
| | 6+7 2 | 6+7 2 | 5+6 2 | 5+6 2 |
| | 9 2 | 9 2 | 3+4 2 | 3+4 2 |
| | 4+5 2 | 4+5 2 | 3+4 2 | 3+4 2 |
| | 8 2 | 8 2 | 5+6 2 | 5+6 2 |
| | 8 2 | 8 2 | 3+4 2 | 3+4 2 |
| | 9 2 | 9 2 | 5+6 2 | 5+6 2 |
| | 6+7 2 | 6+7 2 | 3+4 2 | 3+4 2 |
| | 8 2 | 8 2 | 5+6 2 | 5+6 2 |
| | 4+5 2 | 4+5 2 | 3+4 2 | 3+4 2 |
| | 9 2 | 9 2 | 5+6 2 | 5+6 2 |
| | 8 2 | 8 2 | 3+4 2 | 3+4 2 |
| | 9 2 | 9 2 | 5+6 2 | 5+6 2 |
| | 9 2 | 9 2 | 3+4 2 | 3+4 2 |
| | 8 2 | 8 2 | 3+4 2 | 3+4 2 |
| | 6+7 2 | 6+7 2 | 5+6 2 | 3+4 2 |
| | 8 2 | 8 2 | 3+4 2 | 5+6 2 |
| | 4+5 2 | 4+5 2 | 5+6 2 | 5+6 2 |
| | 9 2 | 9 2 | 5+6 2 | 5+6 2 |
| | 9 2 | 9 2 | 5+6 2 | 5+6 2 |
| | 8 2 | 8 2 | 3+4 2 | 3+4 2 |
| | 8 2 | 9 2 | 5+6 2 | 3+4 2 |

Continued on next column

**TABLE LXXIII– continued**

| TQC | ICSD | NoE | SF | |
|---|---|---|---|---|
| | 9 2 | 6+7 2 | 3+4 2 | 5+6 2 |
| | 4+5 2 | 8 2 | 3+4 2 | 5+6 2 |
| | 6+7 2 | 4+5 2 | 5+6 2 | 3+4 2 |
| | 8 2 | 9 2 | 3+4 2 | 5+6 2 |
| | 8 2 | 9 2 | 5+6 2 | 5+6 2 |
| | 9 2 | 8 2 | 3+4 2 | 3+4 2 |
| | 4+5 2 | 6+7 2 | 5+6 2 | 3+4 2 |
| 166.1.152 | 623746 | 114 | H1Co3Y1 | |
| | -GM | -T | -F | -L |
| | 8 2 | 8 2 | 3+4 2 | 3+4 2 |
| | 8 2 | 9 2 | 5+6 2 | 3+4 2 |
| | 9 2 | 8 2 | 3+4 2 | 5+6 2 |
| | 8 2 | 9 2 | 5+6 2 | 5+6 2 |
| | 9 2 | 9 2 | 5+6 2 | 3+4 2 |
| | 9 2 | 8 2 | 3+4 2 | 5+6 2 |
| | 4+5 2 | 6+7 2 | 5+6 2 | 3+4 2 |
| | 9 2 | 9 2 | 5+6 2 | 5+6 2 |
| | 8 2 | 6+7 2 | 5+6 2 | 5+6 2 |
| | 6+7 2 | 9 2 | 3+4 2 | 5+6 2 |
| | 9 2 | 8 2 | 5+6 2 | 3+4 2 |
| | 6+7 2 | 4+5 2 | 3+4 2 | 5+6 2 |
| | 8 2 | 8 2 | 5+6 2 | 5+6 2 |
| | 9 2 | 9 2 | 3+4 2 | 3+4 2 |
| | 8 2 | 8 2 | 3+4 2 | 3+4 2 |
| | 8 2 | 9 2 | 5+6 2 | 5+6 2 |
| | 9 2 | 9 2 | 3+4 2 | 3+4 2 |
| | 9 2 | 9 2 | 3+4 2 | 5+6 2 |
| | 8 2 | 6+7 2 | 3+4 2 | 3+4 2 |
| | 6+7 2 | 8 2 | 5+6 2 | 3+4 2 |
| | 4+5 2 | 4+5 2 | 3+4 2 | 5+6 2 |
| | 8 2 | 8 2 | 3+4 2 | 5+6 2 |
| | 9 2 | 8 2 | 5+6 2 | 3+4 2 |
| | 8 2 | 9 2 | 5+6 2 | 5+6 2 |
| | 4+5 2 | 4+5 2 | 3+4 2 | 5+6 2 |
| | 8 2 | 8 2 | 5+6 2 | 3+4 2 |
| | 4+5 2 | 4+5 2 | 5+6 2 | 5+6 2 |
| | 8 2 | 6+7 2 | 3+4 2 | 5+6 2 |
| | 6+7 2 | 8 2 | 3+4 2 | 5+6 2 |
| | 9 2 | 9 2 | 5+6 2 | 3+4 2 |
| | 9 2 | 9 2 | 3+4 2 | 5+6 2 |
| | 9 2 | 6+7 2 | 5+6 2 | 5+6 2 |
| | 6+7 2 | 9 2 | 3+4 2 | 3+4 2 |
| | 4+5 2 | 8 2 | 3+4 2 | 5+6 2 |
| | 8 2 | 6+7 2 | 5+6 2 | 3+4 2 |
| | 8 2 | 9 2 | 5+6 2 | 5+6 2 |
| | 4+5 2 | 6+7 2 | 3+4 2 | 3+4 2 |
| | 9 2 | 8 2 | 3+4 2 | 3+4 2 |
| | 8 2 | 9 2 | 5+6 2 | 3+4 2 |
| | 6+7 2 | 4+5 2 | 3+4 2 | 3+4 2 |
| | 6+7 2 | 8 2 | 5+6 2 | 5+6 2 |
| | 9 2 | 9 2 | 3+4 2 | 5+6 2 |
| | 8 2 | 4+5 2 | 5+6 2 | 5+6 2 |
| | 9 2 | 8 2 | 5+6 2 | 5+6 2 |
| | 4+5 2 | 6+7 2 | 3+4 2 | 3+4 2 |
| | 8 2 | 9 2 | 5+6 2 | 3+4 2 |
| | 4+5 2 | 6+7 2 | 3+4 2 | 3+4 2 |
| | 8 2 | 9 2 | 3+4 2 | 5+6 2 |
| | 9 2 | 8 2 | 5+6 2 | 3+4 2 |
| | 8 2 | 4+5 2 | 5+6 2 | 5+6 2 |
| | 6+7 2 | 9 2 | 3+4 2 | 3+4 2 |
| | 4+5 2 | 8 2 | 3+4 2 | 5+6 2 |

Continued on next column

**TABLE LXXIII– continued**

| TQC | ICSD | NoE | SF | |
|---|---|---|---|---|
| | 9 2 | 9 2 | 5+6 2 | 5+6 2 |
| | 9 2 | 8 2 | 3+4 2 | 3+4 2 |
| | 8 2 | 9 2 | 5+6 2 | 5+6 2 |
| | 6+7 2 | 6+7 2 | 3+4 2 | 5+6 2 |
| | 8 2 | 4+5 2 | 5+6 2 | 3+4 2 |
| 166.1.153 | 40981 | 22 | H1Cl1Sc1 | |
| | -GM | -T | -F | -L |
| | 8 2 | 8 2 | 5+6 2 | 5+6 2 |
| | 9 2 | 9 2 | 3+4 2 | 3+4 2 |
| | 8 2 | 9 2 | 5+6 2 | 3+4 2 |
| | 9 2 | 8 2 | 3+4 2 | 5+6 2 |
| | 8 2 | 8 2 | 5+6 2 | 3+4 2 |
| | 8 2 | 9 2 | 3+4 2 | 5+6 2 |
| | 4+5 2 | 4+5 2 | 5+6 2 | 5+6 2 |
| | 9 2 | 6+7 2 | 3+4 2 | 3+4 2 |
| | 6+7 2 | 9 2 | 5+6 2 | 5+6 2 |
| | 9 2 | 8 2 | 3+4 2 | 3+4 2 |
| | 6+7 2 | 4+5 2 | 3+4 2 | 5+6 2 |
| 166.1.154 | 62226 | 66 | H1Cl1Lu1 | |
| | -GM | -T | -F | -L |
| | 8 2 | 9 2 | 5+6 2 | 3+4 2 |
| | 9 2 | 8 2 | 3+4 2 | 5+6 2 |
| | 8 2 | 9 2 | 5+6 2 | 3+4 2 |
| | 9 2 | 8 2 | 3+4 2 | 5+6 2 |
| | 8 2 | 9 2 | 5+6 2 | 3+4 2 |
| | 9 2 | 8 2 | 3+4 2 | 5+6 2 |
| | 4+5 2 | 6+7 2 | 5+6 2 | 3+4 2 |
| | 6+7 2 | 4+5 2 | 3+4 2 | 5+6 2 |
| | 8 2 | 8 2 | 5+6 2 | 5+6 2 |
| | 9 2 | 9 2 | 3+4 2 | 3+4 2 |
| | 8 2 | 9 2 | 5+6 2 | 3+4 2 |
| | 9 2 | 8 2 | 3+4 2 | 5+6 2 |
| | 8 2 | 8 2 | 3+4 2 | 5+6 2 |
| | 9 2 | 9 2 | 5+6 2 | 3+4 2 |
| | 4+5 2 | 4+5 2 | 5+6 2 | 3+4 2 |
| | 6+7 2 | 6+7 2 | 3+4 2 | 5+6 2 |
| | 8 2 | 8 2 | 3+4 2 | 3+4 2 |
| | 9 2 | 9 2 | 5+6 2 | 5+6 2 |
| | 8 2 | 8 2 | 3+4 2 | 3+4 2 |
| | 9 2 | 9 2 | 5+6 2 | 5+6 2 |
| | 8 2 | 8 2 | 5+6 2 | 5+6 2 |
| | 4+5 2 | 4+5 2 | 3+4 2 | 3+4 2 |
| | 6+7 2 | 6+7 2 | 5+6 2 | 5+6 2 |
| | 9 2 | 9 2 | 3+4 2 | 3+4 2 |
| | 8 2 | 9 2 | 5+6 2 | 3+4 2 |
| | 9 2 | 8 2 | 3+4 2 | 5+6 2 |
| | 8 2 | 6+7 2 | 5+6 2 | 3+4 2 |
| | 4+5 2 | 4+5 2 | 3+4 2 | 5+6 2 |
| | 8 2 | 8 2 | 5+6 2 | 5+6 2 |
| | 6+7 2 | 9 2 | 3+4 2 | 3+4 2 |
| | 9 2 | 9 2 | 5+6 2 | 5+6 2 |
| | 9 2 | 8 2 | 3+4 2 | 3+4 2 |
| | 6+7 2 | 4+5 2 | 3+4 2 | 5+6 2 |
| 166.1.155 | 911 | 52 | H12K2N6Sn1 | |
| | -GM | -T | -F | -L |
| | 8 2 | 9 2 | 5+6 2 | 3+4 2 |
| | 9 2 | 8 2 | 3+4 2 | 5+6 2 |
| | 8 2 | 8 2 | 3+4 2 | 3+4 2 |
| | 9 2 | 9 2 | 5+6 2 | 5+6 2 |
| | 8 2 | 9 2 | 5+6 2 | 5+6 2 |
| | 9 2 | 6+7 2 | 3+4 2 | 3+4 2 |

Continued on next column

**TABLE LXXIII– continued**

| TQC | ICSD | NoE | SF | |
|---|---|---|---|---|
| | 4+5 2 | 8 2 | 5+6 2 | 5+6 2 |
| | 6+7 2 | 4+5 2 | 3+4 2 | 3+4 2 |
| | 8 2 | 9 2 | 5+6 2 | 3+4 2 |
| | 9 2 | 8 2 | 3+4 2 | 5+6 2 |
| | 6+7 2 | 6+7 2 | 5+6 2 | 5+6 2 |
| | 9 2 | 9 2 | 5+6 2 | 5+6 2 |
| | 8 2 | 8 2 | 3+4 2 | 3+4 2 |
| | 4+5 2 | 4+5 2 | 3+4 2 | 3+4 2 |
| | 8 2 | 8 2 | 3+4 2 | 3+4 2 |
| | 9 2 | 9 2 | 5+6 2 | 5+6 2 |
| | 9 2 | 9 2 | 5+6 2 | 5+6 2 |
| | 6+7 2 | 6+7 2 | 5+6 2 | 5+6 2 |
| | 8 2 | 8 2 | 3+4 2 | 3+4 2 |
| | 4+5 2 | 4+5 2 | 3+4 2 | 3+4 2 |
| | 8 2 | 8 2 | 3+4 2 | 3+4 2 |
| | 6+7 2 | 9 2 | 5+6 2 | 5+6 2 |
| | 9 2 | 9 2 | 5+6 2 | 5+6 2 |
| | 9 2 | 6+7 2 | 3+4 2 | 5+6 2 |
| | 4+5 2 | 4+5 2 | 5+6 2 | 5+6 2 |
| | 8 2 | 8 2 | 5+6 2 | 3+4 2 |
| 166.1.156 | 159891 | 114 | H1.24Co3Y1 | |
| | -GM | -T | -F | -L |
| | 8 2 | 9 2 | 5+6 2 | 3+4 2 |
| | 9 2 | 8 2 | 3+4 2 | 5+6 2 |
| | 8 2 | 8 2 | 3+4 2 | 3+4 2 |
| | 8 2 | 9 2 | 5+6 2 | 3+4 2 |
| | 9 2 | 8 2 | 3+4 2 | 5+6 2 |
| | 9 2 | 9 2 | 5+6 2 | 5+6 2 |
| | 4+5 2 | 9 2 | 5+6 2 | 3+4 2 |
| | 8 2 | 6+7 2 | 3+4 2 | 5+6 2 |
| | 9 2 | 8 2 | 5+6 2 | 3+4 2 |
| | 6+7 2 | 4+5 2 | 3+4 2 | 5+6 2 |
| | 6+7 2 | 9 2 | 5+6 2 | 5+6 2 |
| | 9 2 | 6+7 2 | 5+6 2 | 5+6 2 |
| | 8 2 | 8 2 | 5+6 2 | 5+6 2 |
| | 9 2 | 9 2 | 3+4 2 | 3+4 2 |
| | 8 2 | 8 2 | 3+4 2 | 3+4 2 |
| | 8 2 | 9 2 | 5+6 2 | 5+6 2 |
| | 9 2 | 9 2 | 3+4 2 | 3+4 2 |
| | 9 2 | 9 2 | 3+4 2 | 3+4 2 |
| | 6+7 2 | 6+7 2 | 3+4 2 | 5+6 2 |
| | 8 2 | 8 2 | 5+6 2 | 3+4 2 |
| | 4+5 2 | 4+5 2 | 3+4 2 | 5+6 2 |
| | 8 2 | 8 2 | 5+6 2 | 3+4 2 |
| | 9 2 | 4+5 2 | 5+6 2 | 5+6 2 |
| | 4+5 2 | 8 2 | 3+4 2 | 5+6 2 |
| | 8 2 | 8 2 | 3+4 2 | 3+4 2 |
| | 8 2 | 9 2 | 5+6 2 | 5+6 2 |
| | 4+5 2 | 4+5 2 | 5+6 2 | 5+6 2 |
| | 8 2 | 8 2 | 5+6 2 | 3+4 2 |
| | 6+7 2 | 6+7 2 | 3+4 2 | 5+6 2 |
| | 9 2 | 9 2 | 5+6 2 | 5+6 2 |
| | 9 2 | 9 2 | 3+4 2 | 5+6 2 |
| | 9 2 | 6+7 2 | 3+4 2 | 3+4 2 |
| | 6+7 2 | 8 2 | 5+6 2 | 3+4 2 |
| | 4+5 2 | 9 2 | 3+4 2 | 5+6 2 |
| | 8 2 | 6+7 2 | 3+4 2 | 3+4 2 |
| | 8 2 | 9 2 | 5+6 2 | 5+6 2 |
| | 4+5 2 | 6+7 2 | 5+6 2 | 3+4 2 |
| | 6+7 2 | 9 2 | 3+4 2 | 5+6 2 |
| | 8 2 | 8 2 | 3+4 2 | 5+6 2 |

Continued on next column

| TQC | ICSD | NoE | SF | |
|---|---|---|---|---|
| | 9 2 | 4+5 2 | 5+6 2 | 3+4 2 |
| | 9 2 | 8 2 | 3+4 2 | 3+4 2 |
| | 6+7 2 | 4+5 2 | 5+6 2 | 5+6 2 |
| | 9 2 | 9 2 | 5+6 2 | 5+6 2 |
| | 8 2 | 8 2 | 3+4 2 | 3+4 2 |
| | 4+5 2 | 6+7 2 | 3+4 2 | 3+4 2 |
| | 8 2 | 9 2 | 5+6 2 | 5+6 2 |
| | 4+5 2 | 9 2 | 3+4 2 | 5+6 2 |
| | 8 2 | 8 2 | 3+4 2 | 5+6 2 |
| | 9 2 | 6+7 2 | 3+4 2 | 3+4 2 |
| | 8 2 | 4+5 2 | 5+6 2 | 3+4 2 |
| | 6+7 2 | 9 2 | 5+6 2 | 3+4 2 |
| | 9 2 | 8 2 | 5+6 2 | 5+6 2 |
| | 4+5 2 | 9 2 | 3+4 2 | 5+6 2 |
| | 8 2 | 6+7 2 | 3+4 2 | 3+4 2 |
| | 9 2 | 8 2 | 5+6 2 | 5+6 2 |
| | 6+7 2 | 4+5 2 | 3+4 2 | 5+6 2 |
| | 8 2 | 9 2 | 5+6 2 | 5+6 2 |
| 166.1.157 | 94220 | 216 | Ge2Nd2Zn15 | |
| | -GM | -T | -F | -L |
| | 8 2 | 9 2 | 3+4 2 | 5+6 2 |
| | 9 2 | 8 2 | 5+6 2 | 3+4 2 |
| | 8 2 | 9 2 | 3+4 2 | 5+6 2 |
| | 9 2 | 8 2 | 5+6 2 | 3+4 2 |
| | 8 2 | 4+5 2 | 3+4 2 | 5+6 2 |
| | 4+5 2 | 9 2 | 5+6 2 | 3+4 2 |
| | 6+7 2 | 6+7 2 | 3+4 2 | 5+6 2 |
| | 9 2 | 8 2 | 5+6 2 | 3+4 2 |
| | 8 2 | 8 2 | 3+4 2 | 3+4 2 |
| | 8 2 | 9 2 | 5+6 2 | 5+6 2 |
| | 9 2 | 9 2 | 5+6 2 | 5+6 2 |
| | 8 2 | 6+7 2 | 5+6 2 | 3+4 2 |
| | 4+5 2 | 9 2 | 3+4 2 | 3+4 2 |
| | 8 2 | 9 2 | 5+6 2 | 3+4 2 |
| | 4+5 2 | 9 2 | 3+4 2 | 3+4 2 |
| | 8 2 | 8 2 | 3+4 2 | 5+6 2 |
| | 9 2 | 6+7 2 | 5+6 2 | 3+4 2 |
| | 6+7 2 | 4+5 2 | 5+6 2 | 5+6 2 |
| | 8 2 | 6+7 2 | 5+6 2 | 3+4 2 |
| | 4+5 2 | 9 2 | 3+4 2 | 5+6 2 |
| | 8 2 | 9 2 | 3+4 2 | 3+4 2 |
| | 8 2 | 9 2 | 5+6 2 | 3+4 2 |
| | 8 2 | 6+7 2 | 5+6 2 | 5+6 2 |
| | 4+5 2 | 8 2 | 3+4 2 | 5+6 2 |
| | 8 2 | 9 2 | 5+6 2 | 3+4 2 |
| | 9 2 | 9 2 | 3+4 2 | 3+4 2 |
| | 6+7 2 | 8 2 | 5+6 2 | 5+6 2 |
| | 8 2 | 4+5 2 | 5+6 2 | 3+4 2 |
| | 9 2 | 6+7 2 | 3+4 2 | 3+4 2 |
| | 4+5 2 | 8 2 | 5+6 2 | 5+6 2 |
| | 8 2 | 9 2 | 3+4 2 | 5+6 2 |
| | 9 2 | 6+7 2 | 5+6 2 | 3+4 2 |
| | 9 2 | 8 2 | 5+6 2 | 3+4 2 |
| | 8 2 | 8 2 | 3+4 2 | 5+6 2 |
| | 4+5 2 | 4+5 2 | 3+4 2 | 5+6 2 |
| | 6+7 2 | 9 2 | 5+6 2 | 5+6 2 |
| | 9 2 | 8 2 | 3+4 2 | 3+4 2 |
| | 8 2 | 6+7 2 | 5+6 2 | 5+6 2 |
| | 4+5 2 | 9 2 | 3+4 2 | 3+4 2 |
| | 6+7 2 | 4+5 2 | 3+4 2 | 5+6 2 |
| | 9 2 | 8 2 | 5+6 2 | 5+6 2 |

Continued on next column

**TABLE LXXIII– continued**

| TQC | ICSD | NoE | SF | |
|---|---|---|---|---|
| | 6+7 2 | 9 2 | 5+6 2 | 3+4 2 |
| | 8 2 | 9 2 | 3+4 2 | 3+4 2 |
| | 9 2 | 8 2 | 3+4 2 | 5+6 2 |
| | 8 2 | 6+7 2 | 5+6 2 | 3+4 2 |
| | 6+7 2 | 9 2 | 3+4 2 | 5+6 2 |
| | 4+5 2 | 4+5 2 | 5+6 2 | 3+4 2 |
| | 9 2 | 6+7 2 | 3+4 2 | 5+6 2 |
| | 8 2 | 9 2 | 5+6 2 | 3+4 2 |
| | 8 2 | 8 2 | 3+4 2 | 5+6 2 |
| | 9 2 | 4+5 2 | 5+6 2 | 3+4 2 |
| | 9 2 | 6+7 2 | 5+6 2 | 5+6 2 |
| | 8 2 | 9 2 | 5+6 2 | 3+4 2 |
| | 6+7 2 | 8 2 | 3+4 2 | 5+6 2 |
| | 4+5 2 | 9 2 | 3+4 2 | 3+4 2 |
| | 9 2 | 8 2 | 5+6 2 | 3+4 2 |
| | 8 2 | 9 2 | 5+6 2 | 5+6 2 |
| | 6+7 2 | 4+5 2 | 3+4 2 | 5+6 2 |
| | 9 2 | 6+7 2 | 5+6 2 | 3+4 2 |
| | 4+5 2 | 8 2 | 3+4 2 | 3+4 2 |
| | 8 2 | 9 2 | 3+4 2 | 5+6 2 |
| | 9 2 | 8 2 | 5+6 2 | 3+4 2 |
| | 8 2 | 9 2 | 3+4 2 | 5+6 2 |
| | 4+5 2 | 6+7 2 | 3+4 2 | 3+4 2 |
| | 8 2 | 4+5 2 | 5+6 2 | 5+6 2 |
| | 9 2 | 8 2 | 5+6 2 | 3+4 2 |
| | 6+7 2 | 9 2 | 3+4 2 | 3+4 2 |
| | 9 2 | 4+5 2 | 5+6 2 | 5+6 2 |
| | 4+5 2 | 9 2 | 3+4 2 | 3+4 2 |
| | 8 2 | 8 2 | 5+6 2 | 5+6 2 |
| | 8 2 | 9 2 | 3+4 2 | 5+6 2 |
| | 8 2 | 8 2 | 5+6 2 | 3+4 2 |
| | 9 2 | 6+7 2 | 3+4 2 | 3+4 2 |
| | 4+5 2 | 9 2 | 5+6 2 | 5+6 2 |
| | 8 2 | 8 2 | 3+4 2 | 5+6 2 |
| | 6+7 2 | 4+5 2 | 5+6 2 | 3+4 2 |
| | 9 2 | 9 2 | 3+4 2 | 5+6 2 |
| | 8 2 | 6+7 2 | 5+6 2 | 5+6 2 |
| | 4+5 2 | 8 2 | 3+4 2 | 3+4 2 |
| | 9 2 | 9 2 | 5+6 2 | 3+4 2 |
| | 8 2 | 8 2 | 5+6 2 | 3+4 2 |
| | 8 2 | 9 2 | 5+6 2 | 5+6 2 |
| | 4+5 2 | 9 2 | 3+4 2 | 3+4 2 |
| | 8 2 | 6+7 2 | 3+4 2 | 5+6 2 |
| | 9 2 | 9 2 | 5+6 2 | 3+4 2 |
| | 8 2 | 6+7 2 | 5+6 2 | 5+6 2 |
| | 9 2 | 9 2 | 3+4 2 | 3+4 2 |
| | 6+7 2 | 8 2 | 5+6 2 | 3+4 2 |
| | 8 2 | 8 2 | 3+4 2 | 5+6 2 |
| | 4+5 2 | 4+5 2 | 3+4 2 | 3+4 2 |
| | 6+7 2 | 9 2 | 5+6 2 | 5+6 2 |
| | 9 2 | 8 2 | 3+4 2 | 5+6 2 |
| | 8 2 | 9 2 | 5+6 2 | 3+4 2 |
| | 4+5 2 | 8 2 | 5+6 2 | 5+6 2 |
| | 8 2 | 4+5 2 | 3+4 2 | 3+4 2 |
| | 9 2 | 8 2 | 3+4 2 | 5+6 2 |
| | 9 2 | 6+7 2 | 3+4 2 | 3+4 2 |
| | 8 2 | 9 2 | 5+6 2 | 5+6 2 |
| | 4+5 2 | 8 2 | 5+6 2 | 3+4 2 |
| | 9 2 | 4+5 2 | 5+6 2 | 3+4 2 |
| | 8 2 | 8 2 | 3+4 2 | 5+6 2 |
| | 6+7 2 | 4+5 2 | 3+4 2 | 3+4 2 |
| | 9 2 | 9 2 | 3+4 2 | 5+6 2 |
| | 8 2 | 6+7 2 | 5+6 2 | 5+6 2 |
| | 4+5 2 | 8 2 | 5+6 2 | 3+4 2 |
| | 9 2 | 9 2 | 3+4 2 | 3+4 2 |
| | 6+7 2 | 6+7 2 | 5+6 2 | 5+6 2 |
| | 8 2 | 9 2 | 5+6 2 | 5+6 2 |
| 166.1.158 | 99161 | 16 | Ge2Li4Zr1 | |
| | -GM | -T | -F | -L |
| | 8 2 | 8 2 | 5+6 2 | 5+6 2 |
| | 9 2 | 9 2 | 3+4 2 | 3+4 2 |
| | 8 2 | 9 2 | 3+4 2 | 3+4 2 |
| | 8 2 | 8 2 | 5+6 2 | 5+6 2 |
| | 4+5 2 | 4+5 2 | 5+6 2 | 5+6 2 |
| | 8 2 | 9 2 | 3+4 2 | 5+6 2 |
| | 9 2 | 6+7 2 | 3+4 2 | 3+4 2 |
| | 4+5 2 | 9 2 | 3+4 2 | 5+6 2 |
| 166.1.159 | 166193 | 58 | Gd1Sn4Ti6 | |
| | -GM | -T | -F | -L |
| | 8 2 | 8 2 | 3+4 2 | 3+4 2 |
| | 9 2 | 9 2 | 5+6 2 | 5+6 2 |
| | 9 2 | 9 2 | 5+6 2 | 5+6 2 |
| | 6+7 2 | 6+7 2 | 5+6 2 | 5+6 2 |
| | 8 2 | 8 2 | 5+6 2 | 3+4 2 |
| | 8 2 | 9 2 | 3+4 2 | 5+6 2 |
| | 9 2 | 8 2 | 5+6 2 | 5+6 2 |
| | 9 2 | 9 2 | 3+4 2 | 3+4 2 |
| | 9 2 | 9 2 | 3+4 2 | 3+4 2 |
| | 8 2 | 8 2 | 5+6 2 | 5+6 2 |
| | 4+5 2 | 4+5 2 | 3+4 2 | 3+4 2 |
| | 8 2 | 9 2 | 5+6 2 | 5+6 2 |
| | 8 2 | 8 2 | 5+6 2 | 3+4 2 |
| | 4+5 2 | 6+7 2 | 5+6 2 | 5+6 2 |
| | 8 2 | 9 2 | 3+4 2 | 3+4 2 |
| | 9 2 | 8 2 | 3+4 2 | 5+6 2 |
| | 6+7 2 | 8 2 | 5+6 2 | 3+4 2 |
| | 8 2 | 4+5 2 | 3+4 2 | 5+6 2 |
| | 9 2 | 9 2 | 3+4 2 | 5+6 2 |
| | 9 2 | 9 2 | 5+6 2 | 5+6 2 |
| | 6+7 2 | 6+7 2 | 5+6 2 | 5+6 2 |
| | 8 2 | 9 2 | 5+6 2 | 5+6 2 |
| | 9 2 | 9 2 | 5+6 2 | 5+6 2 |
| | 4+5 2 | 6+7 2 | 3+4 2 | 3+4 2 |
| | 6+7 2 | 9 2 | 3+4 2 | 3+4 2 |
| | 9 2 | 6+7 2 | 5+6 2 | 3+4 2 |
| | 8 2 | 8 2 | 3+4 2 | 5+6 2 |
| | 4+5 2 | 4+5 2 | 5+6 2 | 3+4 2 |
| | 8 2 | 9 2 | 5+6 2 | 5+6 2 |
| 166.1.160 | 418524 | 82 | Gd1O1P1Zn1 | |
| | -GM | -T | -F | -L |
| | 8 2 | 8 2 | 5+6 2 | 5+6 2 |
| | 9 2 | 9 2 | 3+4 2 | 3+4 2 |
| | 8 2 | 8 2 | 5+6 2 | 5+6 2 |
| | 9 2 | 9 2 | 3+4 2 | 3+4 2 |
| | 8 2 | 8 2 | 5+6 2 | 5+6 2 |
| | 9 2 | 9 2 | 3+4 2 | 3+4 2 |
| | 4+5 2 | 4+5 2 | 5+6 2 | 5+6 2 |
| | 6+7 2 | 6+7 2 | 3+4 2 | 3+4 2 |
| | 9 2 | 9 2 | 3+4 2 | 3+4 2 |
| | 8 2 | 8 2 | 5+6 2 | 5+6 2 |
| | 8 2 | 9 2 | 5+6 2 | 3+4 2 |
| | 9 2 | 8 2 | 3+4 2 | 5+6 2 |

Continued on next column

Continued on next column

**TABLE LXXIII– continued**

| TQC | ICSD | NoE | SF | |
|---|---|---|---|---|
| | 6+7 2 | 4+5 2 | 5+6 2 | 3+4 2 |
| | 4+5 2 | 6+7 2 | 3+4 2 | 5+6 2 |
| | 9 2 | 8 2 | 5+6 2 | 3+4 2 |
| | 8 2 | 9 2 | 3+4 2 | 5+6 2 |
| | 9 2 | 8 2 | 5+6 2 | 3+4 2 |
| | 8 2 | 9 2 | 3+4 2 | 5+6 2 |
| | 4+5 2 | 6+7 2 | 5+6 2 | 3+4 2 |
| | 8 2 | 9 2 | 3+4 2 | 5+6 2 |
| | 9 2 | 8 2 | 3+4 2 | 5+6 2 |
| | 6+7 2 | 4+5 2 | 5+6 2 | 3+4 2 |
| | 8 2 | 9 2 | 3+4 2 | 5+6 2 |
| | 8 2 | 8 2 | 5+6 2 | 3+4 2 |
| | 4+5 2 | 4+5 2 | 3+4 2 | 3+4 2 |
| | 8 2 | 8 2 | 5+6 2 | 5+6 2 |
| | 9 2 | 9 2 | 5+6 2 | 5+6 2 |
| | 6+7 2 | 6+7 2 | 5+6 2 | 5+6 2 |
| | 9 2 | 9 2 | 3+4 2 | 3+4 2 |
| | 9 2 | 8 2 | 3+4 2 | 3+4 2 |
| | 4+5 2 | 6+7 2 | 5+6 2 | 3+4 2 |
| | 8 2 | 9 2 | 5+6 2 | 3+4 2 |
| | 6+7 2 | 4+5 2 | 3+4 2 | 5+6 2 |
| | 8 2 | 8 2 | 3+4 2 | 5+6 2 |
| | 9 2 | 8 2 | 5+6 2 | 3+4 2 |
| | 8 2 | 9 2 | 3+4 2 | 5+6 2 |
| | 9 2 | 9 2 | 5+6 2 | 3+4 2 |
| | 4+5 2 | 6+7 2 | 3+4 2 | 5+6 2 |
| | 9 2 | 8 2 | 5+6 2 | 3+4 2 |
| | 6+7 2 | 4+5 2 | 3+4 2 | 5+6 2 |
| | 8 2 | 8 2 | 5+6 2 | 3+4 2 |
| 166.1.161 | 55619 | 112 | Gd1Mg2Ni9 | |
| | -GM | -T | -F | -L |
| | 8 2 | 8 2 | 3+4 2 | 3+4 2 |
| | 9 2 | 9 2 | 5+6 2 | 5+6 2 |
| | 6+7 2 | 6+7 2 | 5+6 2 | 5+6 2 |
| | 9 2 | 9 2 | 5+6 2 | 5+6 2 |
| | 8 2 | 9 2 | 3+4 2 | 5+6 2 |
| | 9 2 | 8 2 | 5+6 2 | 3+4 2 |
| | 8 2 | 9 2 | 3+4 2 | 3+4 2 |
| | 8 2 | 8 2 | 5+6 2 | 5+6 2 |
| | 9 2 | 9 2 | 3+4 2 | 3+4 2 |
| | 8 2 | 8 2 | 3+4 2 | 5+6 2 |
| | 8 2 | 9 2 | 3+4 2 | 3+4 2 |
| | 9 2 | 6+7 2 | 5+6 2 | 5+6 2 |
| | 4+5 2 | 8 2 | 3+4 2 | 5+6 2 |
| | 9 2 | 4+5 2 | 3+4 2 | 3+4 2 |
| | 6+7 2 | 8 2 | 5+6 2 | 5+6 2 |
| | 8 2 | 6+7 2 | 5+6 2 | 3+4 2 |
| | 4+5 2 | 9 2 | 3+4 2 | 5+6 2 |
| | 8 2 | 9 2 | 5+6 2 | 3+4 2 |
| | 4+5 2 | 4+5 2 | 3+4 2 | 5+6 2 |
| | 8 2 | 8 2 | 5+6 2 | 5+6 2 |
| | 6+7 2 | 4+5 2 | 3+4 2 | 5+6 2 |
| | 9 2 | 8 2 | 3+4 2 | 3+4 2 |
| | 9 2 | 6+7 2 | 5+6 2 | 3+4 2 |
| | 6+7 2 | 9 2 | 5+6 2 | 5+6 2 |
| | 4+5 2 | 9 2 | 5+6 2 | 3+4 2 |
| | 8 2 | 6+7 2 | 3+4 2 | 5+6 2 |
| | 9 2 | 9 2 | 3+4 2 | 3+4 2 |
| | 8 2 | 6+7 2 | 5+6 2 | 3+4 2 |
| | 4+5 2 | 9 2 | 3+4 2 | 5+6 2 |
| | 8 2 | 8 2 | 5+6 2 | 5+6 2 |

Continued on next column

**TABLE LXXIII– continued**

| TQC | ICSD | NoE | SF | |
|---|---|---|---|---|
| | 6+7 2 | 8 2 | 3+4 2 | 3+4 2 |
| | 8 2 | 4+5 2 | 5+6 2 | 5+6 2 |
| | 9 2 | 9 2 | 3+4 2 | 5+6 2 |
| | 9 2 | 4+5 2 | 5+6 2 | 3+4 2 |
| | 4+5 2 | 8 2 | 5+6 2 | 3+4 2 |
| | 8 2 | 6+7 2 | 3+4 2 | 5+6 2 |
| | 9 2 | 9 2 | 3+4 2 | 3+4 2 |
| | 6+7 2 | 9 2 | 3+4 2 | 5+6 2 |
| | 9 2 | 6+7 2 | 5+6 2 | 3+4 2 |
| | 8 2 | 8 2 | 5+6 2 | 5+6 2 |
| | 9 2 | 8 2 | 3+4 2 | 5+6 2 |
| | 4+5 2 | 4+5 2 | 5+6 2 | 5+6 2 |
| | 8 2 | 9 2 | 5+6 2 | 3+4 2 |
| | 9 2 | 8 2 | 3+4 2 | 3+4 2 |
| | 6+7 2 | 9 2 | 5+6 2 | 5+6 2 |
| | 4+5 2 | 9 2 | 3+4 2 | 3+4 2 |
| | 9 2 | 6+7 2 | 3+4 2 | 5+6 2 |
| | 6+7 2 | 8 2 | 5+6 2 | 5+6 2 |
| | 9 2 | 9 2 | 5+6 2 | 5+6 2 |
| | 8 2 | 4+5 2 | 3+4 2 | 3+4 2 |
| | 6+7 2 | 6+7 2 | 5+6 2 | 5+6 2 |
| | 8 2 | 9 2 | 3+4 2 | 5+6 2 |
| | 4+5 2 | 9 2 | 5+6 2 | 5+6 2 |
| | 9 2 | 6+7 2 | 3+4 2 | 3+4 2 |
| | 9 2 | 9 2 | 5+6 2 | 5+6 2 |
| | 6+7 2 | 6+7 2 | 5+6 2 | 5+6 2 |
| 166.1.162 | 635712 | 54 | Gd1 | |
| | -GM | -T | -F | -L |
| | 8 2 | 9 2 | 5+6 2 | 3+4 2 |
| | 9 2 | 8 2 | 3+4 2 | 5+6 2 |
| | 8 2 | 8 2 | 3+4 2 | 3+4 2 |
| | 8 2 | 9 2 | 5+6 2 | 3+4 2 |
| | 9 2 | 8 2 | 3+4 2 | 5+6 2 |
| | 9 2 | 9 2 | 5+6 2 | 5+6 2 |
| | 8 2 | 9 2 | 5+6 2 | 5+6 2 |
| | 9 2 | 8 2 | 5+6 2 | 3+4 2 |
| | 4+5 2 | 9 2 | 3+4 2 | 5+6 2 |
| | 6+7 2 | 6+7 2 | 5+6 2 | 3+4 2 |
| | 6+7 2 | 4+5 2 | 3+4 2 | 5+6 2 |
| | 9 2 | 6+7 2 | 5+6 2 | 5+6 2 |
| | 8 2 | 9 2 | 3+4 2 | 3+4 2 |
| | 9 2 | 8 2 | 5+6 2 | 3+4 2 |
| | 8 2 | 8 2 | 3+4 2 | 5+6 2 |
| | 9 2 | 9 2 | 3+4 2 | 5+6 2 |
| | 8 2 | 9 2 | 5+6 2 | 3+4 2 |
| | 9 2 | 8 2 | 3+4 2 | 5+6 2 |
| | 9 2 | 4+5 2 | 3+4 2 | 5+6 2 |
| | 6+7 2 | 8 2 | 5+6 2 | 5+6 2 |
| | 8 2 | 9 2 | 5+6 2 | 3+4 2 |
| | 4+5 2 | 6+7 2 | 5+6 2 | 5+6 2 |
| | 6+7 2 | 6+7 2 | 5+6 2 | 3+4 2 |
| | 9 2 | 9 2 | 5+6 2 | 5+6 2 |
| | 9 2 | 9 2 | 3+4 2 | 5+6 2 |
| | 8 2 | 8 2 | 3+4 2 | 3+4 2 |
| | 9 2 | 9 2 | 5+6 2 | 5+6 2 |
| 166.1.163 | 635711 | 54 | Gd1 | |
| | -GM | -T | -F | -L |
| | 8 2 | 9 2 | 5+6 2 | 3+4 2 |
| | 9 2 | 8 2 | 3+4 2 | 5+6 2 |
| | 8 2 | 8 2 | 3+4 2 | 3+4 2 |
| | 8 2 | 9 2 | 5+6 2 | 3+4 2 |

Continued on next column

**TABLE LXXIII– continued**

| TQC | ICSD | NoE | SF | |
|---|---|---|---|---|
| | 9 2 | 8 2 | 3+4 2 | 5+6 2 |
| | 9 2 | 9 2 | 5+6 2 | 5+6 2 |
| | 8 2 | 9 2 | 5+6 2 | 5+6 2 |
| | 9 2 | 8 2 | 5+6 2 | 3+4 2 |
| | 4+5 2 | 9 2 | 3+4 2 | 5+6 2 |
| | 6+7 2 | 6+7 2 | 5+6 2 | 3+4 2 |
| | 6+7 2 | 4+5 2 | 3+4 2 | 5+6 2 |
| | 9 2 | 6+7 2 | 5+6 2 | 5+6 2 |
| | 8 2 | 9 2 | 3+4 2 | 3+4 2 |
| | 9 2 | 8 2 | 3+4 2 | 3+4 2 |
| | 8 2 | 8 2 | 5+6 2 | 5+6 2 |
| | 9 2 | 9 2 | 3+4 2 | 5+6 2 |
| | 8 2 | 9 2 | 5+6 2 | 3+4 2 |
| | 9 2 | 8 2 | 3+4 2 | 5+6 2 |
| | 9 2 | 8 2 | 5+6 2 | 5+6 2 |
| | 6+7 2 | 9 2 | 3+4 2 | 5+6 2 |
| | 8 2 | 4+5 2 | 5+6 2 | 3+4 2 |
| | 4+5 2 | 6+7 2 | 5+6 2 | 5+6 2 |
| | 6+7 2 | 6+7 2 | 5+6 2 | 3+4 2 |
| | 9 2 | 9 2 | 5+6 2 | 5+6 2 |
| | 9 2 | 9 2 | 3+4 2 | 5+6 2 |
| | 8 2 | 8 2 | 3+4 2 | 3+4 2 |
| | 9 2 | 9 2 | 5+6 2 | 3+4 2 |
| 166.1.164 | 52517 | 54 | Gd1 | |
| | -GM | -T | -F | -L |
| | 8 2 | 9 2 | 5+6 2 | 3+4 2 |
| | 8 2 | 8 2 | 3+4 2 | 3+4 2 |
| | 9 2 | 8 2 | 3+4 2 | 5+6 2 |
| | 8 2 | 9 2 | 5+6 2 | 3+4 2 |
| | 9 2 | 8 2 | 3+4 2 | 5+6 2 |
| | 9 2 | 9 2 | 5+6 2 | 5+6 2 |
| | 8 2 | 9 2 | 5+6 2 | 5+6 2 |
| | 9 2 | 9 2 | 5+6 2 | 3+4 2 |
| | 4+5 2 | 8 2 | 3+4 2 | 5+6 2 |
| | 6+7 2 | 6+7 2 | 5+6 2 | 3+4 2 |
| | 6+7 2 | 4+5 2 | 3+4 2 | 5+6 2 |
| | 9 2 | 6+7 2 | 5+6 2 | 5+6 2 |
| | 8 2 | 9 2 | 3+4 2 | 3+4 2 |
| | 9 2 | 8 2 | 5+6 2 | 3+4 2 |
| | 8 2 | 8 2 | 3+4 2 | 5+6 2 |
| | 9 2 | 9 2 | 3+4 2 | 5+6 2 |
| | 8 2 | 9 2 | 5+6 2 | 3+4 2 |
| | 9 2 | 8 2 | 3+4 2 | 5+6 2 |
| | 9 2 | 4+5 2 | 3+4 2 | 5+6 2 |
| | 6+7 2 | 8 2 | 5+6 2 | 5+6 2 |
| | 8 2 | 9 2 | 5+6 2 | 3+4 2 |
| | 4+5 2 | 6+7 2 | 5+6 2 | 5+6 2 |
| | 6+7 2 | 6+7 2 | 5+6 2 | 3+4 2 |
| | 9 2 | 9 2 | 5+6 2 | 5+6 2 |
| | 9 2 | 9 2 | 3+4 2 | 5+6 2 |
| | 8 2 | 8 2 | 3+4 2 | 3+4 2 |
| | 9 2 | 9 2 | 5+6 2 | 5+6 2 |
| 166.1.165 | 38267 | 60 | Ga7Rb1 | |
| | -GM | -T | -F | -L |
| | 8 2 | 9 2 | 5+6 2 | 3+4 2 |
| | 9 2 | 8 2 | 3+4 2 | 5+6 2 |
| | 8 2 | 9 2 | 5+6 2 | 3+4 2 |
| | 9 2 | 8 2 | 3+4 2 | 5+6 2 |
| | 4+5 2 | 6+7 2 | 5+6 2 | 3+4 2 |
| | 8 2 | 8 2 | 3+4 2 | 5+6 2 |
| | 9 2 | 9 2 | 5+6 2 | 3+4 2 |

Continued on next column

**TABLE LXXIII– continued**

| TQC | ICSD | NoE | SF | |
|---|---|---|---|---|
| | 6+7 2 | 4+5 2 | 3+4 2 | 5+6 2 |
| | 8 2 | 8 2 | 3+4 2 | 3+4 2 |
| | 9 2 | 9 2 | 5+6 2 | 5+6 2 |
| | 8 2 | 6+7 2 | 5+6 2 | 5+6 2 |
| | 6+7 2 | 9 2 | 5+6 2 | 5+6 2 |
| | 9 2 | 8 2 | 3+4 2 | 3+4 2 |
| | 9 2 | 9 2 | 3+4 2 | 5+6 2 |
| | 8 2 | 4+5 2 | 5+6 2 | 3+4 2 |
| | 4+5 2 | 8 2 | 3+4 2 | 3+4 2 |
| | 8 2 | 9 2 | 3+4 2 | 3+4 2 |
| | 8 2 | 8 2 | 3+4 2 | 5+6 2 |
| | 4+5 2 | 8 2 | 5+6 2 | 3+4 2 |
| | 9 2 | 4+5 2 | 3+4 2 | 3+4 2 |
| | 9 2 | 9 2 | 5+6 2 | 5+6 2 |
| | 6+7 2 | 6+7 2 | 5+6 2 | 5+6 2 |
| | 8 2 | 8 2 | 5+6 2 | 5+6 2 |
| | 9 2 | 9 2 | 5+6 2 | 3+4 2 |
| | 8 2 | 8 2 | 3+4 2 | 5+6 2 |
| | 6+7 2 | 4+5 2 | 3+4 2 | 5+6 2 |
| | 9 2 | 6+7 2 | 3+4 2 | 3+4 2 |
| | 4+5 2 | 9 2 | 5+6 2 | 3+4 2 |
| | 8 2 | 9 2 | 5+6 2 | 5+6 2 |
| | 9 2 | 8 2 | 3+4 2 | 5+6 2 |
| 166.1.166 | 102891 | 60 | Ga7Rb1 | |
| | -GM | -T | -F | -L |
| | 8 2 | 9 2 | 5+6 2 | 3+4 2 |
| | 9 2 | 8 2 | 3+4 2 | 5+6 2 |
| | 8 2 | 9 2 | 5+6 2 | 3+4 2 |
| | 9 2 | 8 2 | 3+4 2 | 5+6 2 |
| | 4+5 2 | 6+7 2 | 5+6 2 | 3+4 2 |
| | 8 2 | 8 2 | 3+4 2 | 5+6 2 |
| | 9 2 | 9 2 | 5+6 2 | 3+4 2 |
| | 6+7 2 | 4+5 2 | 3+4 2 | 5+6 2 |
| | 8 2 | 8 2 | 3+4 2 | 3+4 2 |
| | 9 2 | 9 2 | 5+6 2 | 5+6 2 |
| | 8 2 | 6+7 2 | 5+6 2 | 5+6 2 |
| | 6+7 2 | 9 2 | 5+6 2 | 5+6 2 |
| | 9 2 | 8 2 | 3+4 2 | 3+4 2 |
| | 9 2 | 9 2 | 3+4 2 | 5+6 2 |
| | 8 2 | 4+5 2 | 5+6 2 | 3+4 2 |
| | 4+5 2 | 8 2 | 3+4 2 | 3+4 2 |
| | 8 2 | 9 2 | 3+4 2 | 3+4 2 |
| | 8 2 | 8 2 | 3+4 2 | 5+6 2 |
| | 4+5 2 | 8 2 | 3+4 2 | 3+4 2 |
| | 9 2 | 4+5 2 | 5+6 2 | 3+4 2 |
| | 9 2 | 9 2 | 5+6 2 | 5+6 2 |
| | 6+7 2 | 6+7 2 | 5+6 2 | 5+6 2 |
| | 8 2 | 8 2 | 5+6 2 | 5+6 2 |
| | 9 2 | 9 2 | 5+6 2 | 3+4 2 |
| | 8 2 | 8 2 | 3+4 2 | 5+6 2 |
| | 6+7 2 | 4+5 2 | 3+4 2 | 5+6 2 |
| | 9 2 | 6+7 2 | 3+4 2 | 3+4 2 |
| | 4+5 2 | 9 2 | 5+6 2 | 3+4 2 |
| | 8 2 | 9 2 | 5+6 2 | 5+6 2 |
| | 9 2 | 8 2 | 3+4 2 | 5+6 2 |
| 166.1.167 | 602235 | 150 | Ga3Ho1 | |
| | -GM | -T | -F | -L |
| | 8 2 | 8 2 | 3+4 2 | 3+4 2 |
| | 8 2 | 9 2 | 3+4 2 | 3+4 2 |
| | 9 2 | 8 2 | 5+6 2 | 5+6 2 |
| | 8 2 | 8 2 | 3+4 2 | 3+4 2 |

Continued on next column

**TABLE LXXIII– continued**

| TQC | ICSD | NoE | SF | |
|---|---|---|---|---|
| | 9 2 | 9 2 | 5+6 2 | 5+6 2 |
| | 9 2 | 9 2 | 5+6 2 | 5+6 2 |
| | 8 2 | 9 2 | 5+6 2 | 3+4 2 |
| | 9 2 | 8 2 | 3+4 2 | 5+6 2 |
| | 8 2 | 8 2 | 3+4 2 | 3+4 2 |
| | 9 2 | 9 2 | 5+6 2 | 5+6 2 |
| | 9 2 | 9 2 | 5+6 2 | 5+6 2 |
| | 6+7 2 | 6+7 2 | 5+6 2 | 5+6 2 |
| | 8 2 | 9 2 | 5+6 2 | 3+4 2 |
| | 4+5 2 | 6+7 2 | 3+4 2 | 5+6 2 |
| | 9 2 | 8 2 | 3+4 2 | 3+4 2 |
| | 6+7 2 | 4+5 2 | 5+6 2 | 5+6 2 |
| | 8 2 | 8 2 | 3+4 2 | 3+4 2 |
| | 9 2 | 9 2 | 5+6 2 | 5+6 2 |
| | 4+5 2 | 4+5 2 | 3+4 2 | 3+4 2 |
| | 6+7 2 | 6+7 2 | 5+6 2 | 5+6 2 |
| | 8 2 | 9 2 | 3+4 2 | 5+6 2 |
| | 8 2 | 8 2 | 5+6 2 | 3+4 2 |
| | 9 2 | 9 2 | 3+4 2 | 5+6 2 |
| | 8 2 | 8 2 | 5+6 2 | 3+4 2 |
| | 9 2 | 8 2 | 3+4 2 | 5+6 2 |
| | 8 2 | 4+5 2 | 5+6 2 | 3+4 2 |
| | 9 2 | 8 2 | 3+4 2 | 5+6 2 |
| | 4+5 2 | 8 2 | 5+6 2 | 3+4 2 |
| | 6+7 2 | 4+5 2 | 5+6 2 | 5+6 2 |
| | 9 2 | 9 2 | 5+6 2 | 3+4 2 |
| | 6+7 2 | 6+7 2 | 3+4 2 | 3+4 2 |
| | 8 2 | 8 2 | 3+4 2 | 3+4 2 |
| | 4+5 2 | 4+5 2 | 5+6 2 | 5+6 2 |
| | 8 2 | 9 2 | 3+4 2 | 5+6 2 |
| | 8 2 | 6+7 2 | 5+6 2 | 5+6 2 |
| | 9 2 | 9 2 | 3+4 2 | 3+4 2 |
| | 4+5 2 | 9 2 | 5+6 2 | 3+4 2 |
| | 8 2 | 8 2 | 3+4 2 | 5+6 2 |
| | 9 2 | 9 2 | 3+4 2 | 5+6 2 |
| | 9 2 | 9 2 | 5+6 2 | 3+4 2 |
| | 9 2 | 6+7 2 | 3+4 2 | 3+4 2 |
| | 6+7 2 | 9 2 | 3+4 2 | 3+4 2 |
| | 8 2 | 8 2 | 5+6 2 | 5+6 2 |
| | 9 2 | 9 2 | 5+6 2 | 5+6 2 |
| | 4+5 2 | 6+7 2 | 3+4 2 | 3+4 2 |
| | 6+7 2 | 4+5 2 | 5+6 2 | 5+6 2 |
| | 8 2 | 8 2 | 5+6 2 | 3+4 2 |
| | 9 2 | 6+7 2 | 3+4 2 | 5+6 2 |
| | 6+7 2 | 9 2 | 5+6 2 | 5+6 2 |
| | 4+5 2 | 8 2 | 5+6 2 | 5+6 2 |
| | 9 2 | 4+5 2 | 3+4 2 | 3+4 2 |
| | 9 2 | 9 2 | 5+6 2 | 5+6 2 |
| | 8 2 | 8 2 | 5+6 2 | 3+4 2 |
| | 8 2 | 8 2 | 3+4 2 | 5+6 2 |
| | 9 2 | 9 2 | 3+4 2 | 3+4 2 |
| | 8 2 | 6+7 2 | 5+6 2 | 3+4 2 |
| | 4+5 2 | 9 2 | 5+6 2 | 5+6 2 |
| | 6+7 2 | 9 2 | 3+4 2 | 3+4 2 |
| | 9 2 | 9 2 | 5+6 2 | 5+6 2 |
| | 9 2 | 6+7 2 | 3+4 2 | 5+6 2 |
| | 6+7 2 | 8 2 | 5+6 2 | 3+4 2 |
| | 8 2 | 4+5 2 | 3+4 2 | 3+4 2 |
| | 9 2 | 9 2 | 3+4 2 | 5+6 2 |
| | 9 2 | 6+7 2 | 5+6 2 | 5+6 2 |
| | 6+7 2 | 8 2 | 5+6 2 | 5+6 2 |
| | 8 2 | 4+5 2 | 3+4 2 | 3+4 2 |
| | 4+5 2 | 8 2 | 5+6 2 | 5+6 2 |
| | 9 2 | 9 2 | 5+6 2 | 3+4 2 |
| | 8 2 | 9 2 | 3+4 2 | 5+6 2 |
| | 9 2 | 8 2 | 5+6 2 | 3+4 2 |
| | 6+7 2 | 6+7 2 | 3+4 2 | 5+6 2 |
| | 9 2 | 9 2 | 5+6 2 | 5+6 2 |
| | 8 2 | 9 2 | 5+6 2 | 3+4 2 |
| | 4+5 2 | 6+7 2 | 3+4 2 | 3+4 2 |
| | 8 2 | 8 2 | 5+6 2 | 5+6 2 |
| 166.1.168 | 657608 | 24 | Ga2Te3 | |
| | -GM | -T | -F | -L |
| | 8 2 | 8 2 | 3+4 2 | 3+4 2 |
| | 9 2 | 9 2 | 5+6 2 | 5+6 2 |
| | 8 2 | 8 2 | 3+4 2 | 3+4 2 |
| | 9 2 | 9 2 | 5+6 2 | 5+6 2 |
| | 8 2 | 9 2 | 3+4 2 | 3+4 2 |
| | 9 2 | 8 2 | 5+6 2 | 5+6 2 |
| | 6+7 2 | 6+7 2 | 5+6 2 | 5+6 2 |
| | 8 2 | 8 2 | 3+4 2 | 3+4 2 |
| | 4+5 2 | 4+5 2 | 5+6 2 | 3+4 2 |
| | 9 2 | 9 2 | 3+4 2 | 5+6 2 |
| | 9 2 | 9 2 | 5+6 2 | 3+4 2 |
| | 6+7 2 | 6+7 2 | 5+6 2 | 5+6 2 |
| 166.1.169 | 635031 | 16 | Ga1P1 | |
| | -GM | -T | -F | -L |
| | 8 2 | 8 2 | 3+4 2 | 3+4 2 |
| | 9 2 | 9 2 | 5+6 2 | 5+6 2 |
| | 8 2 | 8 2 | 5+6 2 | 5+6 2 |
| | 9 2 | 9 2 | 3+4 2 | 3+4 2 |
| | 6+7 2 | 9 2 | 5+6 2 | 5+6 2 |
| | 8 2 | 6+7 2 | 3+4 2 | 3+4 2 |
| | 4+5 2 | 8 2 | 5+6 2 | 3+4 2 |
| | 8 2 | 4+5 2 | 3+4 2 | 5+6 2 |
| 166.1.170 | 96491 | 130 | Ga1Mn1 | |
| | -GM | -T | -F | -L |
| | 8 2 | 4+5 2 | 5+6 2 | 5+6 2 |
| | 4+5 2 | 8 2 | 3+4 2 | 3+4 2 |
| | 8 2 | 6+7 2 | 5+6 2 | 5+6 2 |
| | 9 2 | 9 2 | 3+4 2 | 3+4 2 |
| | 6+7 2 | 9 2 | 3+4 2 | 3+4 2 |
| | 9 2 | 8 2 | 5+6 2 | 5+6 2 |
| | 8 2 | 8 2 | 5+6 2 | 3+4 2 |
| | 8 2 | 9 2 | 3+4 2 | 3+4 2 |
| | 4+5 2 | 4+5 2 | 3+4 2 | 5+6 2 |
| | 8 2 | 8 2 | 5+6 2 | 5+6 2 |
| | 6+7 2 | 9 2 | 5+6 2 | 3+4 2 |
| | 9 2 | 6+7 2 | 3+4 2 | 3+4 2 |
| | 9 2 | 8 2 | 3+4 2 | 5+6 2 |
| | 8 2 | 8 2 | 5+6 2 | 5+6 2 |
| | 9 2 | 9 2 | 3+4 2 | 5+6 2 |
| | 6+7 2 | 9 2 | 5+6 2 | 5+6 2 |
| | 8 2 | 6+7 2 | 5+6 2 | 3+4 2 |
| | 4+5 2 | 9 2 | 3+4 2 | 5+6 2 |
| | 9 2 | 8 2 | 3+4 2 | 3+4 2 |
| | 8 2 | 6+7 2 | 3+4 2 | 5+6 2 |
| | 4+5 2 | 9 2 | 5+6 2 | 3+4 2 |
| | 8 2 | 4+5 2 | 3+4 2 | 3+4 2 |
| | 4+5 2 | 8 2 | 5+6 2 | 5+6 2 |
| | 8 2 | 9 2 | 5+6 2 | 5+6 2 |
| | 9 2 | 9 2 | 3+4 2 | 5+6 2 |

Continued on next column

**TABLE LXXIII– continued**

| TQC | ICSD | NoE | SF | |
|---|---|---|---|---|
| | 6+7 2 | 8 2 | 5+6 2 | 5+6 2 |
| | 9 2 | 4+5 2 | 3+4 2 | 3+4 2 |
| | 9 2 | 6+7 2 | 5+6 2 | 3+4 2 |
| | 6+7 2 | 9 2 | 5+6 2 | 5+6 2 |
| | 9 2 | 8 2 | 3+4 2 | 3+4 2 |
| | 8 2 | 6+7 2 | 3+4 2 | 3+4 2 |
| | 8 2 | 9 2 | 5+6 2 | 5+6 2 |
| | 9 2 | 8 2 | 3+4 2 | 5+6 2 |
| | 4+5 2 | 9 2 | 5+6 2 | 3+4 2 |
| | 8 2 | 4+5 2 | 3+4 2 | 5+6 2 |
| | 8 2 | 9 2 | 5+6 2 | 3+4 2 |
| | 8 2 | 4+5 2 | 5+6 2 | 5+6 2 |
| | 4+5 2 | 8 2 | 3+4 2 | 3+4 2 |
| | 9 2 | 8 2 | 5+6 2 | 5+6 2 |
| | 6+7 2 | 8 2 | 5+6 2 | 3+4 2 |
| | 9 2 | 9 2 | 3+4 2 | 5+6 2 |
| | 4+5 2 | 6+7 2 | 5+6 2 | 5+6 2 |
| | 8 2 | 9 2 | 3+4 2 | 5+6 2 |
| | 4+5 2 | 9 2 | 3+4 2 | 3+4 2 |
| | 8 2 | 6+7 2 | 3+4 2 | 3+4 2 |
| | 6+7 2 | 4+5 2 | 3+4 2 | 3+4 2 |
| | 9 2 | 6+7 2 | 5+6 2 | 5+6 2 |
| | 8 2 | 8 2 | 5+6 2 | 3+4 2 |
| | 8 2 | 9 2 | 3+4 2 | 3+4 2 |
| | 4+5 2 | 9 2 | 5+6 2 | 5+6 2 |
| | 8 2 | 8 2 | 3+4 2 | 5+6 2 |
| | 9 2 | 8 2 | 5+6 2 | 3+4 2 |
| | 6+7 2 | 4+5 2 | 3+4 2 | 3+4 2 |
| | 9 2 | 8 2 | 5+6 2 | 5+6 2 |
| | 9 2 | 4+5 2 | 3+4 2 | 3+4 2 |
| | 6+7 2 | 8 2 | 5+6 2 | 5+6 2 |
| | 9 2 | 9 2 | 5+6 2 | 5+6 2 |
| | 6+7 2 | 6+7 2 | 3+4 2 | 5+6 2 |
| | 9 2 | 9 2 | 3+4 2 | 5+6 2 |
| | 9 2 | 9 2 | 5+6 2 | 3+4 2 |
| | 4+5 2 | 6+7 2 | 5+6 2 | 3+4 2 |
| | 8 2 | 8 2 | 3+4 2 | 3+4 2 |
| | 8 2 | 4+5 2 | 3+4 2 | 5+6 2 |
| | 9 2 | 9 2 | 3+4 2 | 5+6 2 |
| | 4+5 2 | 6+7 2 | 3+4 2 | 3+4 2 |
| 166.1.171 | 35386 | 26 | Ga1Ge1Te1 | |
| | -GM | -T | -F | -L |
| | 8 2 | 9 2 | 5+6 2 | 3+4 2 |
| | 9 2 | 8 2 | 3+4 2 | 5+6 2 |
| | 8 2 | 8 2 | 5+6 2 | 5+6 2 |
| | 9 2 | 9 2 | 3+4 2 | 3+4 2 |
| | 8 2 | 8 2 | 3+4 2 | 5+6 2 |
| | 9 2 | 9 2 | 5+6 2 | 3+4 2 |
| | 8 2 | 9 2 | 5+6 2 | 3+4 2 |
| | 9 2 | 8 2 | 3+4 2 | 5+6 2 |
| | 4+5 2 | 4+5 2 | 5+6 2 | 3+4 2 |
| | 8 2 | 6+7 2 | 3+4 2 | 5+6 2 |
| | 6+7 2 | 8 2 | 3+4 2 | 3+4 2 |
| | 8 2 | 4+5 2 | 5+6 2 | 5+6 2 |
| | 4+5 2 | 9 2 | 5+6 2 | 5+6 2 |
| 166.1.172 | 2513 | 66 | Ga1 | |
| | -GM | -T | -F | -L |
| | 8 2 | 9 2 | 3+4 2 | 5+6 2 |
| | 8 2 | 8 2 | 5+6 2 | 3+4 2 |
| | 4+5 2 | 4+5 2 | 5+6 2 | 5+6 2 |
| | 8 2 | 8 2 | 3+4 2 | 5+6 2 |

Continued on next column

**TABLE LXXIII– continued**

| TQC | ICSD | NoE | SF | |
|---|---|---|---|---|
| | 6+7 2 | 9 2 | 5+6 2 | 3+4 2 |
| | 9 2 | 6+7 2 | 3+4 2 | 3+4 2 |
| | 9 2 | 9 2 | 3+4 2 | 3+4 2 |
| | 9 2 | 8 2 | 3+4 2 | 5+6 2 |
| | 8 2 | 8 2 | 5+6 2 | 3+4 2 |
| | 4+5 2 | 4+5 2 | 5+6 2 | 3+4 2 |
| | 8 2 | 8 2 | 5+6 2 | 5+6 2 |
| | 6+7 2 | 6+7 2 | 3+4 2 | 5+6 2 |
| | 9 2 | 9 2 | 3+4 2 | 3+4 2 |
| | 8 2 | 9 2 | 5+6 2 | 5+6 2 |
| | 9 2 | 6+7 2 | 3+4 2 | 3+4 2 |
| | 4+5 2 | 9 2 | 5+6 2 | 5+6 2 |
| | 8 2 | 4+5 2 | 3+4 2 | 5+6 2 |
| | 6+7 2 | 8 2 | 5+6 2 | 3+4 2 |
| | 9 2 | 8 2 | 5+6 2 | 5+6 2 |
| | 9 2 | 4+5 2 | 3+4 2 | 3+4 2 |
| | 8 2 | 8 2 | 5+6 2 | 5+6 2 |
| | 9 2 | 9 2 | 3+4 2 | 3+4 2 |
| | 9 2 | 6+7 2 | 3+4 2 | 5+6 2 |
| | 6+7 2 | 9 2 | 5+6 2 | 3+4 2 |
| | 8 2 | 9 2 | 3+4 2 | 3+4 2 |
| | 8 2 | 6+7 2 | 5+6 2 | 5+6 2 |
| | 4+5 2 | 9 2 | 3+4 2 | 3+4 2 |
| | 6+7 2 | 8 2 | 5+6 2 | 3+4 2 |
| | 9 2 | 9 2 | 5+6 2 | 5+6 2 |
| | 8 2 | 4+5 2 | 3+4 2 | 3+4 2 |
| | 9 2 | 8 2 | 3+4 2 | 5+6 2 |
| | 6+7 2 | 9 2 | 5+6 2 | 5+6 2 |
| | 9 2 | 8 2 | 5+6 2 | 3+4 2 |
| 166.1.173 | 634071 | 92 | Fe7W6 | |
| | -GM | -T | -F | -L |
| | 8 2 | 9 2 | 5+6 2 | 5+6 2 |
| | 8 2 | 8 2 | 3+4 2 | 3+4 2 |
| | 9 2 | 9 2 | 5+6 2 | 3+4 2 |
| | 8 2 | 8 2 | 3+4 2 | 5+6 2 |
| | 6+7 2 | 4+5 2 | 5+6 2 | 3+4 2 |
| | 9 2 | 8 2 | 3+4 2 | 5+6 2 |
| | 9 2 | 8 2 | 3+4 2 | 3+4 2 |
| | 8 2 | 9 2 | 5+6 2 | 5+6 2 |
| | 4+5 2 | 6+7 2 | 3+4 2 | 3+4 2 |
| | 8 2 | 9 2 | 3+4 2 | 3+4 2 |
| | 9 2 | 8 2 | 5+6 2 | 5+6 2 |
| | 9 2 | 8 2 | 3+4 2 | 3+4 2 |
| | 6+7 2 | 4+5 2 | 3+4 2 | 3+4 2 |
| | 8 2 | 9 2 | 5+6 2 | 5+6 2 |
| | 8 2 | 6+7 2 | 5+6 2 | 5+6 2 |
| | 4+5 2 | 9 2 | 3+4 2 | 3+4 2 |
| | 8 2 | 8 2 | 3+4 2 | 5+6 2 |
| | 9 2 | 8 2 | 3+4 2 | 3+4 2 |
| | 6+7 2 | 9 2 | 5+6 2 | 5+6 2 |
| | 9 2 | 6+7 2 | 5+6 2 | 5+6 2 |
| | 4+5 2 | 4+5 2 | 5+6 2 | 3+4 2 |
| | 8 2 | 9 2 | 3+4 2 | 3+4 2 |
| | 9 2 | 8 2 | 3+4 2 | 5+6 2 |
| | 6+7 2 | 4+5 2 | 5+6 2 | 3+4 2 |
| | 9 2 | 8 2 | 3+4 2 | 3+4 2 |
| | 6+7 2 | 4+5 2 | 5+6 2 | 5+6 2 |
| | 8 2 | 9 2 | 3+4 2 | 3+4 2 |
| | 4+5 2 | 6+7 2 | 3+4 2 | 5+6 2 |
| | 8 2 | 8 2 | 5+6 2 | 3+4 2 |
| | 9 2 | 9 2 | 5+6 2 | 3+4 2 |

Continued on next column

**TABLE LXXIII– continued**

| TQC | ICSD | NoE | SF | |
|---|---|---|---|---|
| | 4+5 2 | 9 2 | 3+4 2 | 5+6 2 |
| | 8 2 | 8 2 | 5+6 2 | 5+6 2 |
| | 9 2 | 6+7 2 | 3+4 2 | 5+6 2 |
| | 6+7 2 | 4+5 2 | 5+6 2 | 3+4 2 |
| | 4+5 2 | 8 2 | 3+4 2 | 5+6 2 |
| | 8 2 | 9 2 | 5+6 2 | 3+4 2 |
| | 8 2 | 4+5 2 | 3+4 2 | 3+4 2 |
| | 6+7 2 | 6+7 2 | 5+6 2 | 3+4 2 |
| | 4+5 2 | 8 2 | 5+6 2 | 5+6 2 |
| | 9 2 | 9 2 | 3+4 2 | 5+6 2 |
| | 9 2 | 4+5 2 | 5+6 2 | 3+4 2 |
| | 8 2 | 8 2 | 3+4 2 | 3+4 2 |
| | 4+5 2 | 8 2 | 3+4 2 | 5+6 2 |
| | 8 2 | 9 2 | 5+6 2 | 5+6 2 |
| | 9 2 | 4+5 2 | 3+4 2 | 3+4 2 |
| | 8 2 | 8 2 | 5+6 2 | 5+6 2 |
| 166.1.174 | 634063 | 92 | Fe7W6 | |
| | -GM | -T | -F | -L |
| | 8 2 | 9 2 | 5+6 2 | 5+6 2 |
| | 8 2 | 8 2 | 3+4 2 | 3+4 2 |
| | 9 2 | 9 2 | 5+6 2 | 3+4 2 |
| | 8 2 | 8 2 | 3+4 2 | 5+6 2 |
| | 6+7 2 | 4+5 2 | 5+6 2 | 3+4 2 |
| | 9 2 | 8 2 | 3+4 2 | 5+6 2 |
| | 9 2 | 8 2 | 3+4 2 | 3+4 2 |
| | 8 2 | 9 2 | 5+6 2 | 5+6 2 |
| | 4+5 2 | 6+7 2 | 3+4 2 | 3+4 2 |
| | 8 2 | 9 2 | 3+4 2 | 3+4 2 |
| | 9 2 | 8 2 | 5+6 2 | 5+6 2 |
| | 9 2 | 8 2 | 3+4 2 | 3+4 2 |
| | 6+7 2 | 4+5 2 | 3+4 2 | 3+4 2 |
| | 8 2 | 9 2 | 5+6 2 | 5+6 2 |
| | 8 2 | 6+7 2 | 5+6 2 | 5+6 2 |
| | 4+5 2 | 9 2 | 3+4 2 | 3+4 2 |
| | 8 2 | 8 2 | 3+4 2 | 5+6 2 |
| | 9 2 | 8 2 | 5+6 2 | 3+4 2 |
| | 6+7 2 | 6+7 2 | 3+4 2 | 5+6 2 |
| | 9 2 | 9 2 | 5+6 2 | 5+6 2 |
| | 4+5 2 | 4+5 2 | 3+4 2 | 3+4 2 |
| | 8 2 | 9 2 | 5+6 2 | 3+4 2 |
| | 9 2 | 8 2 | 3+4 2 | 5+6 2 |
| | 6+7 2 | 4+5 2 | 5+6 2 | 3+4 2 |
| | 9 2 | 8 2 | 3+4 2 | 5+6 2 |
| | 6+7 2 | 4+5 2 | 5+6 2 | 3+4 2 |
| | 8 2 | 9 2 | 3+4 2 | 3+4 2 |
| | 4+5 2 | 6+7 2 | 3+4 2 | 5+6 2 |
| | 8 2 | 8 2 | 5+6 2 | 3+4 2 |
| | 9 2 | 9 2 | 3+4 2 | 3+4 2 |
| | 4+5 2 | 9 2 | 5+6 2 | 5+6 2 |
| | 8 2 | 8 2 | 5+6 2 | 5+6 2 |
| | 6+7 2 | 6+7 2 | 3+4 2 | 5+6 2 |
| | 9 2 | 8 2 | 5+6 2 | 5+6 2 |
| | 4+5 2 | 4+5 2 | 3+4 2 | 3+4 2 |
| | 8 2 | 9 2 | 5+6 2 | 3+4 2 |
| | 8 2 | 4+5 2 | 3+4 2 | 3+4 2 |
| | 6+7 2 | 6+7 2 | 5+6 2 | 3+4 2 |
| | 4+5 2 | 8 2 | 5+6 2 | 5+6 2 |
| | 9 2 | 9 2 | 3+4 2 | 5+6 2 |
| | 9 2 | 4+5 2 | 5+6 2 | 3+4 2 |
| | 8 2 | 8 2 | 3+4 2 | 3+4 2 |
| | 4+5 2 | 8 2 | 3+4 2 | 5+6 2 |

Continued on next column

**TABLE LXXIII– continued**

| TQC | ICSD | NoE | SF | |
|---|---|---|---|---|
| | 8 2 | 9 2 | 5+6 2 | 5+6 2 |
| | 9 2 | 8 2 | 3+4 2 | 3+4 2 |
| | 8 2 | 4+5 2 | 5+6 2 | 5+6 2 |
| 166.1.175 | 634060 | 92 | Fe7W6 | |
| | -GM | -T | -F | -L |
| | 8 2 | 9 2 | 5+6 2 | 5+6 2 |
| | 8 2 | 8 2 | 3+4 2 | 3+4 2 |
| | 9 2 | 9 2 | 5+6 2 | 3+4 2 |
| | 8 2 | 8 2 | 3+4 2 | 5+6 2 |
| | 6+7 2 | 4+5 2 | 5+6 2 | 3+4 2 |
| | 9 2 | 8 2 | 3+4 2 | 5+6 2 |
| | 9 2 | 8 2 | 3+4 2 | 3+4 2 |
| | 8 2 | 9 2 | 5+6 2 | 5+6 2 |
| | 4+5 2 | 6+7 2 | 3+4 2 | 3+4 2 |
| | 8 2 | 9 2 | 3+4 2 | 3+4 2 |
| | 9 2 | 8 2 | 5+6 2 | 5+6 2 |
| | 9 2 | 8 2 | 3+4 2 | 3+4 2 |
| | 6+7 2 | 4+5 2 | 3+4 2 | 3+4 2 |
| | 8 2 | 9 2 | 5+6 2 | 5+6 2 |
| | 8 2 | 6+7 2 | 5+6 2 | 5+6 2 |
| | 4+5 2 | 9 2 | 3+4 2 | 3+4 2 |
| | 8 2 | 8 2 | 3+4 2 | 5+6 2 |
| | 9 2 | 8 2 | 3+4 2 | 3+4 2 |
| | 6+7 2 | 9 2 | 5+6 2 | 5+6 2 |
| | 9 2 | 6+7 2 | 5+6 2 | 5+6 2 |
| | 8 2 | 4+5 2 | 5+6 2 | 3+4 2 |
| | 4+5 2 | 9 2 | 3+4 2 | 3+4 2 |
| | 9 2 | 8 2 | 3+4 2 | 5+6 2 |
| | 6+7 2 | 4+5 2 | 5+6 2 | 3+4 2 |
| | 9 2 | 8 2 | 3+4 2 | 5+6 2 |
| | 6+7 2 | 4+5 2 | 5+6 2 | 3+4 2 |
| | 8 2 | 9 2 | 3+4 2 | 3+4 2 |
| | 4+5 2 | 6+7 2 | 3+4 2 | 5+6 2 |
| | 8 2 | 8 2 | 5+6 2 | 3+4 2 |
| | 9 2 | 9 2 | 3+4 2 | 3+4 2 |
| | 4+5 2 | 9 2 | 5+6 2 | 5+6 2 |
| | 8 2 | 8 2 | 5+6 2 | 5+6 2 |
| | 9 2 | 6+7 2 | 3+4 2 | 5+6 2 |
| | 6+7 2 | 8 2 | 5+6 2 | 5+6 2 |
| | 8 2 | 4+5 2 | 3+4 2 | 3+4 2 |
| | 4+5 2 | 9 2 | 5+6 2 | 3+4 2 |
| | 8 2 | 4+5 2 | 3+4 2 | 3+4 2 |
| | 6+7 2 | 6+7 2 | 5+6 2 | 3+4 2 |
| | 4+5 2 | 8 2 | 5+6 2 | 5+6 2 |
| | 9 2 | 9 2 | 3+4 2 | 5+6 2 |
| | 9 2 | 4+5 2 | 5+6 2 | 3+4 2 |
| | 8 2 | 8 2 | 3+4 2 | 3+4 2 |
| | 4+5 2 | 8 2 | 3+4 2 | 5+6 2 |
| | 8 2 | 9 2 | 5+6 2 | 5+6 2 |
| | 9 2 | 8 2 | 3+4 2 | 3+4 2 |
| | 8 2 | 4+5 2 | 5+6 2 | 5+6 2 |
| 166.1.176 | 103694 | 92 | Fe7W6 | |
| | -GM | -T | -F | -L |
| | 8 2 | 9 2 | 5+6 2 | 5+6 2 |
| | 8 2 | 8 2 | 3+4 2 | 3+4 2 |
| | 9 2 | 9 2 | 5+6 2 | 3+4 2 |
| | 8 2 | 8 2 | 3+4 2 | 5+6 2 |
| | 6+7 2 | 4+5 2 | 5+6 2 | 3+4 2 |
| | 9 2 | 8 2 | 3+4 2 | 5+6 2 |
| | 9 2 | 8 2 | 3+4 2 | 3+4 2 |
| | 8 2 | 9 2 | 5+6 2 | 5+6 2 |

Continued on next column

**TABLE LXXIII– continued**

| TQC | ICSD | NoE | SF | |
|---|---|---|---|---|
| | 4+5 2 | 6+7 2 | 3+4 2 | 3+4 2 |
| | 8 2 | 9 2 | 3+4 2 | 3+4 2 |
| | 9 2 | 8 2 | 5+6 2 | 5+6 2 |
| | 9 2 | 8 2 | 3+4 2 | 3+4 2 |
| | 6+7 2 | 4+5 2 | 3+4 2 | 3+4 2 |
| | 8 2 | 9 2 | 5+6 2 | 5+6 2 |
| | 8 2 | 6+7 2 | 5+6 2 | 5+6 2 |
| | 4+5 2 | 9 2 | 3+4 2 | 3+4 2 |
| | 8 2 | 8 2 | 3+4 2 | 5+6 2 |
| | 9 2 | 8 2 | 3+4 2 | 3+4 2 |
| | 6+7 2 | 9 2 | 5+6 2 | 5+6 2 |
| | 9 2 | 6+7 2 | 5+6 2 | 5+6 2 |
| | 8 2 | 4+5 2 | 5+6 2 | 3+4 2 |
| | 4+5 2 | 9 2 | 3+4 2 | 3+4 2 |
| | 9 2 | 8 2 | 3+4 2 | 5+6 2 |
| | 6+7 2 | 4+5 2 | 5+6 2 | 3+4 2 |
| | 9 2 | 8 2 | 3+4 2 | 3+4 2 |
| | 6+7 2 | 4+5 2 | 5+6 2 | 5+6 2 |
| | 8 2 | 9 2 | 3+4 2 | 3+4 2 |
| | 4+5 2 | 6+7 2 | 3+4 2 | 5+6 2 |
| | 8 2 | 8 2 | 5+6 2 | 3+4 2 |
| | 9 2 | 9 2 | 3+4 2 | 3+4 2 |
| | 4+5 2 | 9 2 | 5+6 2 | 5+6 2 |
| | 8 2 | 8 2 | 5+6 2 | 5+6 2 |
| | 9 2 | 6+7 2 | 3+4 2 | 5+6 2 |
| | 6+7 2 | 4+5 2 | 5+6 2 | 3+4 2 |
| | 4+5 2 | 8 2 | 3+4 2 | 5+6 2 |
| | 8 2 | 9 2 | 5+6 2 | 3+4 2 |
| | 8 2 | 4+5 2 | 3+4 2 | 3+4 2 |
| | 6+7 2 | 6+7 2 | 5+6 2 | 3+4 2 |
| | 4+5 2 | 8 2 | 5+6 2 | 5+6 2 |
| | 9 2 | 9 2 | 3+4 2 | 5+6 2 |
| | 9 2 | 4+5 2 | 5+6 2 | 3+4 2 |
| | 8 2 | 8 2 | 3+4 2 | 3+4 2 |
| | 4+5 2 | 8 2 | 3+4 2 | 5+6 2 |
| | 8 2 | 9 2 | 5+6 2 | 5+6 2 |
| | 9 2 | 4+5 2 | 3+4 2 | 3+4 2 |
| | 8 2 | 8 2 | 5+6 2 | 5+6 2 |
| 166.1.177 | 632784 | 134 | Fe7Nb6 | |
| | -GM | -T | -F | -L |
| | 8 2 | 8 2 | 5+6 2 | 5+6 2 |
| | 8 2 | 9 2 | 3+4 2 | 3+4 2 |
| | 9 2 | 9 2 | 5+6 2 | 3+4 2 |
| | 8 2 | 9 2 | 3+4 2 | 5+6 2 |
| | 9 2 | 8 2 | 3+4 2 | 5+6 2 |
| | 9 2 | 8 2 | 5+6 2 | 3+4 2 |
| | 8 2 | 8 2 | 5+6 2 | 5+6 2 |
| | 8 2 | 9 2 | 3+4 2 | 3+4 2 |
| | 9 2 | 9 2 | 5+6 2 | 3+4 2 |
| | 8 2 | 8 2 | 3+4 2 | 5+6 2 |
| | 9 2 | 9 2 | 3+4 2 | 5+6 2 |
| | 9 2 | 8 2 | 5+6 2 | 3+4 2 |
| | 4+5 2 | 4+5 2 | 5+6 2 | 5+6 2 |
| | 4+5 2 | 6+7 2 | 3+4 2 | 3+4 2 |
| | 8 2 | 8 2 | 5+6 2 | 3+4 2 |
| | 9 2 | 9 2 | 3+4 2 | 5+6 2 |
| | 8 2 | 9 2 | 3+4 2 | 5+6 2 |
| | 6+7 2 | 4+5 2 | 5+6 2 | 3+4 2 |
| | 8 2 | 8 2 | 5+6 2 | 5+6 2 |
| | 9 2 | 9 2 | 3+4 2 | 3+4 2 |
| | 4+5 2 | 6+7 2 | 5+6 2 | 5+6 2 |

Continued on next column

**TABLE LXXIII– continued**

| TQC | ICSD | NoE | SF | |
|---|---|---|---|---|
| | 6+7 2 | 6+7 2 | 3+4 2 | 3+4 2 |
| | 9 2 | 8 2 | 5+6 2 | 3+4 2 |
| | 6+7 2 | 4+5 2 | 3+4 2 | 5+6 2 |
| | 8 2 | 8 2 | 3+4 2 | 3+4 2 |
| | 9 2 | 9 2 | 5+6 2 | 5+6 2 |
| | 8 2 | 8 2 | 5+6 2 | 5+6 2 |
| | 8 2 | 9 2 | 3+4 2 | 3+4 2 |
| | 9 2 | 8 2 | 3+4 2 | 3+4 2 |
| | 6+7 2 | 4+5 2 | 3+4 2 | 3+4 2 |
| | 9 2 | 8 2 | 5+6 2 | 3+4 2 |
| | 8 2 | 9 2 | 3+4 2 | 5+6 2 |
| | 4+5 2 | 6+7 2 | 5+6 2 | 5+6 2 |
| | 8 2 | 9 2 | 3+4 2 | 3+4 2 |
| | 9 2 | 9 2 | 5+6 2 | 3+4 2 |
| | 6+7 2 | 6+7 2 | 3+4 2 | 5+6 2 |
| | 9 2 | 8 2 | 3+4 2 | 5+6 2 |
| | 8 2 | 8 2 | 3+4 2 | 3+4 2 |
| | 8 2 | 4+5 2 | 5+6 2 | 3+4 2 |
| | 4+5 2 | 9 2 | 5+6 2 | 5+6 2 |
| | 9 2 | 8 2 | 5+6 2 | 3+4 2 |
| | 6+7 2 | 4+5 2 | 3+4 2 | 3+4 2 |
| | 8 2 | 9 2 | 3+4 2 | 5+6 2 |
| | 4+5 2 | 6+7 2 | 5+6 2 | 5+6 2 |
| | 8 2 | 9 2 | 3+4 2 | 5+6 2 |
| | 9 2 | 8 2 | 5+6 2 | 5+6 2 |
| | 9 2 | 8 2 | 3+4 2 | 3+4 2 |
| | 6+7 2 | 4+5 2 | 3+4 2 | 5+6 2 |
| | 9 2 | 8 2 | 5+6 2 | 3+4 2 |
| | 6+7 2 | 4+5 2 | 5+6 2 | 3+4 2 |
| | 4+5 2 | 8 2 | 3+4 2 | 5+6 2 |
| | 8 2 | 6+7 2 | 5+6 2 | 3+4 2 |
| | 4+5 2 | 9 2 | 3+4 2 | 3+4 2 |
| | 8 2 | 4+5 2 | 5+6 2 | 5+6 2 |
| | 8 2 | 8 2 | 3+4 2 | 3+4 2 |
| | 9 2 | 8 2 | 5+6 2 | 5+6 2 |
| | 6+7 2 | 9 2 | 5+6 2 | 5+6 2 |
| | 9 2 | 6+7 2 | 3+4 2 | 3+4 2 |
| | 4+5 2 | 9 2 | 3+4 2 | 5+6 2 |
| | 6+7 2 | 4+5 2 | 5+6 2 | 3+4 2 |
| | 8 2 | 8 2 | 3+4 2 | 3+4 2 |
| | 9 2 | 6+7 2 | 5+6 2 | 5+6 2 |
| | 8 2 | 9 2 | 5+6 2 | 3+4 2 |
| | 4+5 2 | 8 2 | 3+4 2 | 5+6 2 |
| | 8 2 | 9 2 | 5+6 2 | 3+4 2 |
| | 9 2 | 4+5 2 | 5+6 2 | 5+6 2 |
| | 8 2 | 8 2 | 3+4 2 | 3+4 2 |
| 166.1.178 | 632622 | 92 | Fe7Mo6 | |
| | -GM | -T | -F | -L |
| | 8 2 | 8 2 | 3+4 2 | 3+4 2 |
| | 9 2 | 9 2 | 5+6 2 | 5+6 2 |
| | 8 2 | 9 2 | 5+6 2 | 5+6 2 |
| | 8 2 | 4+5 2 | 3+4 2 | 3+4 2 |
| | 6+7 2 | 8 2 | 3+4 2 | 3+4 2 |
| | 9 2 | 8 2 | 5+6 2 | 5+6 2 |
| | 8 2 | 8 2 | 3+4 2 | 3+4 2 |
| | 4+5 2 | 9 2 | 3+4 2 | 3+4 2 |
| | 9 2 | 6+7 2 | 5+6 2 | 5+6 2 |
| | 8 2 | 9 2 | 3+4 2 | 3+4 2 |
| | 9 2 | 8 2 | 5+6 2 | 5+6 2 |
| | 9 2 | 4+5 2 | 3+4 2 | 3+4 2 |
| | 6+7 2 | 8 2 | 3+4 2 | 5+6 2 |

Continued on next column

**TABLE LXXIII– continued**

| TQC | ICSD | NoE | SF | |
|---|---|---|---|---|
| | 8 2 | 9 2 | 5+6 2 | 3+4 2 |
| | 8 2 | 6+7 2 | 3+4 2 | 5+6 2 |
| | 4+5 2 | 9 2 | 5+6 2 | 3+4 2 |
| | 8 2 | 8 2 | 3+4 2 | 3+4 2 |
| | 4+5 2 | 4+5 2 | 5+6 2 | 5+6 2 |
| | 6+7 2 | 6+7 2 | 3+4 2 | 5+6 2 |
| | 9 2 | 9 2 | 5+6 2 | 5+6 2 |
| | 9 2 | 9 2 | 3+4 2 | 3+4 2 |
| | 8 2 | 8 2 | 5+6 2 | 5+6 2 |
| | 9 2 | 4+5 2 | 3+4 2 | 3+4 2 |
| | 6+7 2 | 8 2 | 5+6 2 | 3+4 2 |
| | 6+7 2 | 8 2 | 3+4 2 | 3+4 2 |
| | 9 2 | 4+5 2 | 5+6 2 | 5+6 2 |
| | 4+5 2 | 6+7 2 | 3+4 2 | 3+4 2 |
| | 8 2 | 9 2 | 3+4 2 | 5+6 2 |
| | 8 2 | 8 2 | 5+6 2 | 3+4 2 |
| | 4+5 2 | 9 2 | 3+4 2 | 5+6 2 |
| | 8 2 | 6+7 2 | 5+6 2 | 3+4 2 |
| | 9 2 | 9 2 | 5+6 2 | 5+6 2 |
| | 6+7 2 | 8 2 | 3+4 2 | 5+6 2 |
| | 9 2 | 4+5 2 | 5+6 2 | 3+4 2 |
| | 4+5 2 | 8 2 | 5+6 2 | 5+6 2 |
| | 8 2 | 9 2 | 3+4 2 | 3+4 2 |
| | 8 2 | 4+5 2 | 3+4 2 | 3+4 2 |
| | 4+5 2 | 6+7 2 | 5+6 2 | 3+4 2 |
| | 6+7 2 | 8 2 | 5+6 2 | 5+6 2 |
| | 8 2 | 9 2 | 5+6 2 | 5+6 2 |
| | 9 2 | 8 2 | 3+4 2 | 3+4 2 |
| | 9 2 | 4+5 2 | 3+4 2 | 3+4 2 |
| | 4+5 2 | 4+5 2 | 5+6 2 | 5+6 2 |
| | 8 2 | 8 2 | 3+4 2 | 5+6 2 |
| | 9 2 | 9 2 | 3+4 2 | 3+4 2 |
| | 8 2 | 8 2 | 5+6 2 | 5+6 2 |
| 166.1.179 | 632620 | 92 | Fe7Mo6 | |
| | -GM | -T | -F | -L |
| | 8 2 | 8 2 | 3+4 2 | 3+4 2 |
| | 9 2 | 9 2 | 5+6 2 | 5+6 2 |
| | 8 2 | 9 2 | 5+6 2 | 5+6 2 |
| | 8 2 | 4+5 2 | 3+4 2 | 3+4 2 |
| | 6+7 2 | 8 2 | 3+4 2 | 3+4 2 |
| | 9 2 | 8 2 | 5+6 2 | 5+6 2 |
| | 4+5 2 | 8 2 | 3+4 2 | 3+4 2 |
| | 8 2 | 6+7 2 | 3+4 2 | 3+4 2 |
| | 9 2 | 9 2 | 5+6 2 | 5+6 2 |
| | 8 2 | 9 2 | 3+4 2 | 3+4 2 |
| | 9 2 | 8 2 | 5+6 2 | 5+6 2 |
| | 6+7 2 | 4+5 2 | 3+4 2 | 3+4 2 |
| | 9 2 | 8 2 | 3+4 2 | 5+6 2 |
| | 8 2 | 6+7 2 | 5+6 2 | 3+4 2 |
| | 4+5 2 | 9 2 | 5+6 2 | 5+6 2 |
| | 8 2 | 9 2 | 3+4 2 | 3+4 2 |
| | 8 2 | 4+5 2 | 3+4 2 | 3+4 2 |
| | 4+5 2 | 8 2 | 5+6 2 | 5+6 2 |
| | 9 2 | 6+7 2 | 3+4 2 | 5+6 2 |
| | 6+7 2 | 9 2 | 5+6 2 | 5+6 2 |
| | 8 2 | 9 2 | 5+6 2 | 3+4 2 |
| | 9 2 | 4+5 2 | 3+4 2 | 5+6 2 |
| | 6+7 2 | 8 2 | 3+4 2 | 3+4 2 |
| | 9 2 | 4+5 2 | 5+6 2 | 3+4 2 |
| | 6+7 2 | 8 2 | 3+4 2 | 3+4 2 |
| | 9 2 | 8 2 | 5+6 2 | 5+6 2 |
| | 4+5 2 | 6+7 2 | 3+4 2 | 3+4 2 |
| | 8 2 | 9 2 | 3+4 2 | 5+6 2 |
| | 8 2 | 8 2 | 5+6 2 | 3+4 2 |
| | 4+5 2 | 9 2 | 3+4 2 | 5+6 2 |
| | 9 2 | 6+7 2 | 5+6 2 | 3+4 2 |
| | 8 2 | 9 2 | 5+6 2 | 5+6 2 |
| | 6+7 2 | 4+5 2 | 5+6 2 | 3+4 2 |
| | 9 2 | 8 2 | 3+4 2 | 5+6 2 |
| | 4+5 2 | 8 2 | 5+6 2 | 5+6 2 |
| | 8 2 | 9 2 | 3+4 2 | 3+4 2 |
| | 6+7 2 | 4+5 2 | 3+4 2 | 3+4 2 |
| | 8 2 | 6+7 2 | 5+6 2 | 3+4 2 |
| | 4+5 2 | 8 2 | 5+6 2 | 5+6 2 |
| | 8 2 | 9 2 | 5+6 2 | 5+6 2 |
| | 9 2 | 8 2 | 3+4 2 | 3+4 2 |
| | 9 2 | 4+5 2 | 3+4 2 | 3+4 2 |
| | 4+5 2 | 4+5 2 | 5+6 2 | 5+6 2 |
| | 8 2 | 8 2 | 3+4 2 | 5+6 2 |
| | 9 2 | 9 2 | 3+4 2 | 3+4 2 |
| | 8 2 | 8 2 | 5+6 2 | 5+6 2 |
| 166.1.180 | 41380 | 138 | Fe6Ge5Li1 | |
| | -GM | -T | -F | -L |
| | 8 2 | 8 2 | 3+4 2 | 3+4 2 |
| | 8 2 | 9 2 | 5+6 2 | 5+6 2 |
| | 9 2 | 8 2 | 3+4 2 | 3+4 2 |
| | 8 2 | 9 2 | 5+6 2 | 5+6 2 |
| | 9 2 | 9 2 | 5+6 2 | 3+4 2 |
| | 9 2 | 8 2 | 3+4 2 | 5+6 2 |
| | 8 2 | 8 2 | 5+6 2 | 5+6 2 |
| | 9 2 | 9 2 | 3+4 2 | 3+4 2 |
| | 8 2 | 8 2 | 3+4 2 | 3+4 2 |
| | 9 2 | 9 2 | 5+6 2 | 5+6 2 |
| | 9 2 | 8 2 | 3+4 2 | 3+4 2 |
| | 8 2 | 9 2 | 5+6 2 | 5+6 2 |
| | 8 2 | 8 2 | 3+4 2 | 5+6 2 |
| | 9 2 | 4+5 2 | 5+6 2 | 3+4 2 |
| | 4+5 2 | 9 2 | 5+6 2 | 3+4 2 |
| | 8 2 | 9 2 | 3+4 2 | 5+6 2 |
| | 9 2 | 9 2 | 3+4 2 | 5+6 2 |
| | 6+7 2 | 6+7 2 | 3+4 2 | 3+4 2 |
| | 8 2 | 6+7 2 | 5+6 2 | 3+4 2 |
| | 9 2 | 8 2 | 5+6 2 | 5+6 2 |
| | 6+7 2 | 8 2 | 5+6 2 | 5+6 2 |
| | 8 2 | 9 2 | 5+6 2 | 3+4 2 |
| | 4+5 2 | 8 2 | 3+4 2 | 5+6 2 |
| | 9 2 | 4+5 2 | 5+6 2 | 3+4 2 |
| | 8 2 | 9 2 | 3+4 2 | 3+4 2 |
| | 8 2 | 6+7 2 | 5+6 2 | 5+6 2 |
| | 4+5 2 | 9 2 | 3+4 2 | 5+6 2 |
| | 9 2 | 8 2 | 3+4 2 | 3+4 2 |
| | 6+7 2 | 8 2 | 5+6 2 | 5+6 2 |
| | 8 2 | 4+5 2 | 3+4 2 | 5+6 2 |
| | 8 2 | 8 2 | 3+4 2 | 3+4 2 |
| | 4+5 2 | 9 2 | 5+6 2 | 3+4 2 |
| | 8 2 | 6+7 2 | 3+4 2 | 5+6 2 |
| | 9 2 | 9 2 | 3+4 2 | 3+4 2 |
| | 6+7 2 | 6+7 2 | 5+6 2 | 5+6 2 |
| | 8 2 | 9 2 | 3+4 2 | 3+4 2 |
| | 4+5 2 | 6+7 2 | 5+6 2 | 3+4 2 |
| | 9 2 | 8 2 | 5+6 2 | 5+6 2 |
| | 8 2 | 9 2 | 5+6 2 | 5+6 2 |

Continued on next column

**TABLE LXXIII– continued**

| TQC | ICSD | NoE | SF | |
|---|---|---|---|---|
| | 4+5 2 | 4+5 2 | 3+4 2 | 5+6 2 |
| | 9 2 | 8 2 | 3+4 2 | 3+4 2 |
| | 6+7 2 | 4+5 2 | 5+6 2 | 3+4 2 |
| | 9 2 | 9 2 | 5+6 2 | 3+4 2 |
| | 8 2 | 8 2 | 3+4 2 | 5+6 2 |
| | 8 2 | 8 2 | 5+6 2 | 3+4 2 |
| | 4+5 2 | 4+5 2 | 5+6 2 | 5+6 2 |
| | 9 2 | 8 2 | 3+4 2 | 3+4 2 |
| | 6+7 2 | 4+5 2 | 5+6 2 | 3+4 2 |
| | 8 2 | 9 2 | 3+4 2 | 3+4 2 |
| | 9 2 | 8 2 | 3+4 2 | 5+6 2 |
| | 6+7 2 | 6+7 2 | 5+6 2 | 5+6 2 |
| | 9 2 | 9 2 | 3+4 2 | 5+6 2 |
| | 8 2 | 8 2 | 3+4 2 | 3+4 2 |
| | 4+5 2 | 4+5 2 | 5+6 2 | 3+4 2 |
| | 9 2 | 8 2 | 3+4 2 | 5+6 2 |
| | 6+7 2 | 6+7 2 | 5+6 2 | 5+6 2 |
| | 4+5 2 | 9 2 | 5+6 2 | 3+4 2 |
| | 9 2 | 4+5 2 | 3+4 2 | 3+4 2 |
| | 8 2 | 8 2 | 5+6 2 | 5+6 2 |
| | 6+7 2 | 6+7 2 | 3+4 2 | 3+4 2 |
| | 4+5 2 | 9 2 | 3+4 2 | 5+6 2 |
| | 9 2 | 4+5 2 | 3+4 2 | 5+6 2 |
| | 8 2 | 8 2 | 5+6 2 | 3+4 2 |
| | 9 2 | 6+7 2 | 3+4 2 | 5+6 2 |
| | 6+7 2 | 9 2 | 5+6 2 | 3+4 2 |
| | 9 2 | 9 2 | 5+6 2 | 5+6 2 |
| | 9 2 | 8 2 | 5+6 2 | 5+6 2 |
| | 8 2 | 9 2 | 3+4 2 | 3+4 2 |
| | 6+7 2 | 6+7 2 | 3+4 2 | 3+4 2 |
| 166.1.181 | 633920 | 108 | Fe3Th1 | |
| | -GM | -T | -F | -L |
| | 8 2 | 9 2 | 3+4 2 | 3+4 2 |
| | 8 2 | 8 2 | 5+6 2 | 3+4 2 |
| | 9 2 | 8 2 | 3+4 2 | 5+6 2 |
| | 8 2 | 9 2 | 5+6 2 | 3+4 2 |
| | 9 2 | 9 2 | 5+6 2 | 5+6 2 |
| | 9 2 | 8 2 | 3+4 2 | 5+6 2 |
| | 4+5 2 | 6+7 2 | 5+6 2 | 3+4 2 |
| | 8 2 | 9 2 | 3+4 2 | 5+6 2 |
| | 9 2 | 6+7 2 | 5+6 2 | 5+6 2 |
| | 6+7 2 | 8 2 | 5+6 2 | 5+6 2 |
| | 9 2 | 9 2 | 5+6 2 | 3+4 2 |
| | 6+7 2 | 4+5 2 | 3+4 2 | 5+6 2 |
| | 8 2 | 8 2 | 5+6 2 | 3+4 2 |
| | 9 2 | 9 2 | 3+4 2 | 5+6 2 |
| | 8 2 | 8 2 | 3+4 2 | 3+4 2 |
| | 8 2 | 9 2 | 3+4 2 | 5+6 2 |
| | 9 2 | 9 2 | 5+6 2 | 3+4 2 |
| | 9 2 | 9 2 | 3+4 2 | 3+4 2 |
| | 6+7 2 | 4+5 2 | 3+4 2 | 3+4 2 |
| | 4+5 2 | 6+7 2 | 5+6 2 | 5+6 2 |
| | 8 2 | 8 2 | 3+4 2 | 3+4 2 |
| | 8 2 | 8 2 | 5+6 2 | 5+6 2 |
| | 8 2 | 8 2 | 3+4 2 | 5+6 2 |
| | 4+5 2 | 4+5 2 | 5+6 2 | 5+6 2 |
| | 9 2 | 8 2 | 5+6 2 | 5+6 2 |
| | 8 2 | 4+5 2 | 3+4 2 | 5+6 2 |
| | 4+5 2 | 8 2 | 5+6 2 | 3+4 2 |
| | 9 2 | 9 2 | 3+4 2 | 5+6 2 |
| | 8 2 | 6+7 2 | 5+6 2 | 3+4 2 |
| | 6+7 2 | 9 2 | 3+4 2 | 3+4 2 |
| | 9 2 | 9 2 | 3+4 2 | 5+6 2 |
| | 9 2 | 4+5 2 | 5+6 2 | 5+6 2 |
| | 6+7 2 | 6+7 2 | 3+4 2 | 3+4 2 |
| | 8 2 | 8 2 | 5+6 2 | 3+4 2 |
| | 4+5 2 | 9 2 | 3+4 2 | 5+6 2 |
| | 9 2 | 6+7 2 | 3+4 2 | 3+4 2 |
| | 6+7 2 | 8 2 | 5+6 2 | 5+6 2 |
| | 8 2 | 9 2 | 5+6 2 | 3+4 2 |
| | 8 2 | 9 2 | 3+4 2 | 5+6 2 |
| | 9 2 | 9 2 | 3+4 2 | 3+4 2 |
| | 9 2 | 4+5 2 | 5+6 2 | 5+6 2 |
| | 4+5 2 | 8 2 | 3+4 2 | 5+6 2 |
| | 8 2 | 6+7 2 | 5+6 2 | 3+4 2 |
| | 6+7 2 | 8 2 | 3+4 2 | 5+6 2 |
| | 4+5 2 | 6+7 2 | 5+6 2 | 3+4 2 |
| | 8 2 | 9 2 | 3+4 2 | 5+6 2 |
| | 4+5 2 | 8 2 | 5+6 2 | 3+4 2 |
| | 8 2 | 6+7 2 | 3+4 2 | 3+4 2 |
| | 9 2 | 9 2 | 5+6 2 | 5+6 2 |
| | 6+7 2 | 4+5 2 | 3+4 2 | 5+6 2 |
| | 9 2 | 8 2 | 5+6 2 | 5+6 2 |
| | 8 2 | 9 2 | 3+4 2 | 3+4 2 |
| | 4+5 2 | 9 2 | 5+6 2 | 5+6 2 |
| | 8 2 | 4+5 2 | 3+4 2 | 5+6 2 |
| 166.1.182 | 633912 | 108 | Fe3Th1 | |
| | -GM | -T | -F | -L |
| | 8 2 | 9 2 | 3+4 2 | 3+4 2 |
| | 8 2 | 8 2 | 5+6 2 | 3+4 2 |
| | 9 2 | 8 2 | 3+4 2 | 5+6 2 |
| | 8 2 | 9 2 | 5+6 2 | 3+4 2 |
| | 9 2 | 9 2 | 5+6 2 | 5+6 2 |
| | 9 2 | 8 2 | 3+4 2 | 5+6 2 |
| | 4+5 2 | 6+7 2 | 5+6 2 | 3+4 2 |
| | 8 2 | 9 2 | 3+4 2 | 5+6 2 |
| | 9 2 | 6+7 2 | 5+6 2 | 5+6 2 |
| | 6+7 2 | 8 2 | 5+6 2 | 5+6 2 |
| | 9 2 | 9 2 | 5+6 2 | 3+4 2 |
| | 6+7 2 | 4+5 2 | 3+4 2 | 5+6 2 |
| | 8 2 | 8 2 | 5+6 2 | 3+4 2 |
| | 9 2 | 9 2 | 3+4 2 | 5+6 2 |
| | 8 2 | 8 2 | 3+4 2 | 3+4 2 |
| | 8 2 | 9 2 | 3+4 2 | 5+6 2 |
| | 9 2 | 9 2 | 5+6 2 | 3+4 2 |
| | 9 2 | 9 2 | 3+4 2 | 3+4 2 |
| | 6+7 2 | 4+5 2 | 3+4 2 | 3+4 2 |
| | 4+5 2 | 6+7 2 | 5+6 2 | 5+6 2 |
| | 8 2 | 8 2 | 3+4 2 | 3+4 2 |
| | 8 2 | 8 2 | 5+6 2 | 5+6 2 |
| | 8 2 | 8 2 | 3+4 2 | 5+6 2 |
| | 4+5 2 | 4+5 2 | 5+6 2 | 5+6 2 |
| | 9 2 | 8 2 | 5+6 2 | 5+6 2 |
| | 8 2 | 4+5 2 | 3+4 2 | 5+6 2 |
| | 4+5 2 | 8 2 | 5+6 2 | 3+4 2 |
| | 9 2 | 9 2 | 3+4 2 | 5+6 2 |
| | 8 2 | 6+7 2 | 5+6 2 | 3+4 2 |
| | 6+7 2 | 9 2 | 3+4 2 | 3+4 2 |
| | 9 2 | 9 2 | 3+4 2 | 5+6 2 |
| | 9 2 | 4+5 2 | 5+6 2 | 5+6 2 |
| | 6+7 2 | 6+7 2 | 3+4 2 | 3+4 2 |
| | 8 2 | 8 2 | 5+6 2 | 3+4 2 |

Continued on next column

**TABLE LXXIII– continued**

| TQC | ICSD | NoE | SF |
|---|---|---|---|
| | 4+5 2 9 2 3+4 2 5+6 2 | | |
| | 9 2 6+7 2 3+4 2 3+4 2 | | |
| | 6+7 2 8 2 5+6 2 5+6 2 | | |
| | 8 2 9 2 5+6 2 3+4 2 | | |
| | 8 2 9 2 3+4 2 5+6 2 | | |
| | 9 2 9 2 3+4 2 3+4 2 | | |
| | 9 2 4+5 2 5+6 2 5+6 2 | | |
| | 4+5 2 8 2 3+4 2 5+6 2 | | |
| | 8 2 6+7 2 5+6 2 3+4 2 | | |
| | 6+7 2 8 2 3+4 2 5+6 2 | | |
| | 4+5 2 6+7 2 5+6 2 3+4 2 | | |
| | 8 2 9 2 3+4 2 5+6 2 | | |
| | 4+5 2 8 2 5+6 2 3+4 2 | | |
| | 8 2 6+7 2 3+4 2 3+4 2 | | |
| | 9 2 9 2 5+6 2 5+6 2 | | |
| | 6+7 2 4+5 2 3+4 2 5+6 2 | | |
| | 9 2 8 2 5+6 2 5+6 2 | | |
| | 8 2 9 2 3+4 2 3+4 2 | | |
| | 4+5 2 9 2 5+6 2 5+6 2 | | |
| | 8 2 4+5 2 3+4 2 5+6 2 | | |
| 166.1.183 | 633900 | 108 | Fe3Th1 |
| | -GM -T -F -L | | |
| | 8 2 9 2 3+4 2 3+4 2 | | |
| | 8 2 8 2 5+6 2 3+4 2 | | |
| | 9 2 8 2 3+4 2 5+6 2 | | |
| | 8 2 9 2 5+6 2 3+4 2 | | |
| | 9 2 9 2 5+6 2 5+6 2 | | |
| | 9 2 8 2 3+4 2 5+6 2 | | |
| | 4+5 2 6+7 2 5+6 2 3+4 2 | | |
| | 8 2 9 2 3+4 2 5+6 2 | | |
| | 9 2 6+7 2 5+6 2 5+6 2 | | |
| | 6+7 2 8 2 5+6 2 5+6 2 | | |
| | 9 2 9 2 5+6 2 3+4 2 | | |
| | 6+7 2 4+5 2 3+4 2 5+6 2 | | |
| | 8 2 8 2 5+6 2 3+4 2 | | |
| | 9 2 9 2 3+4 2 5+6 2 | | |
| | 8 2 8 2 3+4 2 3+4 2 | | |
| | 8 2 9 2 3+4 2 5+6 2 | | |
| | 9 2 9 2 5+6 2 3+4 2 | | |
| | 9 2 9 2 3+4 2 3+4 2 | | |
| | 6+7 2 4+5 2 3+4 2 3+4 2 | | |
| | 4+5 2 6+7 2 5+6 2 5+6 2 | | |
| | 8 2 8 2 3+4 2 3+4 2 | | |
| | 8 2 8 2 5+6 2 5+6 2 | | |
| | 8 2 8 2 3+4 2 5+6 2 | | |
| | 4+5 2 4+5 2 5+6 2 5+6 2 | | |
| | 9 2 8 2 5+6 2 5+6 2 | | |
| | 8 2 4+5 2 3+4 2 5+6 2 | | |
| | 4+5 2 8 2 5+6 2 3+4 2 | | |
| | 9 2 9 2 3+4 2 5+6 2 | | |
| | 8 2 6+7 2 5+6 2 3+4 2 | | |
| | 6+7 2 9 2 3+4 2 3+4 2 | | |
| | 9 2 9 2 3+4 2 5+6 2 | | |
| | 9 2 4+5 2 5+6 2 5+6 2 | | |
| | 6+7 2 6+7 2 3+4 2 3+4 2 | | |
| | 8 2 8 2 5+6 2 3+4 2 | | |
| | 4+5 2 9 2 3+4 2 5+6 2 | | |
| | 9 2 6+7 2 3+4 2 3+4 2 | | |
| | 6+7 2 8 2 5+6 2 5+6 2 | | |
| | 8 2 9 2 5+6 2 3+4 2 | | |
| | 8 2 9 2 3+4 2 5+6 2 | | |
| | 9 2 9 2 3+4 2 3+4 2 | | |
| | 9 2 4+5 2 5+6 2 5+6 2 | | |
| | 4+5 2 8 2 3+4 2 5+6 2 | | |
| | 8 2 6+7 2 5+6 2 3+4 2 | | |
| | 6+7 2 8 2 3+4 2 5+6 2 | | |
| | 4+5 2 6+7 2 5+6 2 3+4 2 | | |
| | 8 2 9 2 3+4 2 5+6 2 | | |
| | 4+5 2 8 2 5+6 2 3+4 2 | | |
| | 8 2 6+7 2 3+4 2 3+4 2 | | |
| | 9 2 9 2 5+6 2 5+6 2 | | |
| | 6+7 2 4+5 2 3+4 2 5+6 2 | | |
| | 9 2 8 2 5+6 2 5+6 2 | | |
| | 8 2 9 2 3+4 2 3+4 2 | | |
| | 4+5 2 9 2 5+6 2 5+6 2 | | |
| | 8 2 4+5 2 3+4 2 5+6 2 | | |
| 166.1.184 | 633896 | 108 | Fe3Th1 |
| | -GM -T -F -L | | |
| | 8 2 9 2 3+4 2 3+4 2 | | |
| | 8 2 8 2 5+6 2 3+4 2 | | |
| | 9 2 8 2 3+4 2 5+6 2 | | |
| | 8 2 9 2 5+6 2 3+4 2 | | |
| | 9 2 9 2 5+6 2 5+6 2 | | |
| | 9 2 8 2 3+4 2 5+6 2 | | |
| | 4+5 2 6+7 2 5+6 2 3+4 2 | | |
| | 8 2 9 2 3+4 2 5+6 2 | | |
| | 9 2 6+7 2 5+6 2 5+6 2 | | |
| | 6+7 2 8 2 5+6 2 5+6 2 | | |
| | 9 2 9 2 5+6 2 3+4 2 | | |
| | 6+7 2 4+5 2 3+4 2 5+6 2 | | |
| | 8 2 8 2 5+6 2 3+4 2 | | |
| | 9 2 9 2 3+4 2 5+6 2 | | |
| | 8 2 8 2 3+4 2 3+4 2 | | |
| | 8 2 9 2 3+4 2 5+6 2 | | |
| | 9 2 9 2 5+6 2 3+4 2 | | |
| | 9 2 9 2 3+4 2 3+4 2 | | |
| | 6+7 2 4+5 2 3+4 2 3+4 2 | | |
| | 4+5 2 6+7 2 5+6 2 5+6 2 | | |
| | 8 2 8 2 3+4 2 3+4 2 | | |
| | 8 2 8 2 5+6 2 5+6 2 | | |
| | 8 2 8 2 3+4 2 5+6 2 | | |
| | 4+5 2 4+5 2 5+6 2 5+6 2 | | |
| | 9 2 8 2 5+6 2 5+6 2 | | |
| | 8 2 4+5 2 3+4 2 3+4 2 | | |
| | 4+5 2 8 2 5+6 2 5+6 2 | | |
| | 9 2 9 2 3+4 2 5+6 2 | | |
| | 8 2 6+7 2 5+6 2 3+4 2 | | |
| | 6+7 2 9 2 3+4 2 3+4 2 | | |
| | 9 2 9 2 3+4 2 5+6 2 | | |
| | 9 2 4+5 2 5+6 2 5+6 2 | | |
| | 6+7 2 6+7 2 3+4 2 3+4 2 | | |
| | 8 2 8 2 5+6 2 3+4 2 | | |
| | 4+5 2 9 2 3+4 2 5+6 2 | | |
| | 9 2 6+7 2 3+4 2 3+4 2 | | |
| | 6+7 2 8 2 5+6 2 5+6 2 | | |
| | 8 2 9 2 5+6 2 3+4 2 | | |
| | 8 2 9 2 3+4 2 5+6 2 | | |
| | 9 2 9 2 3+4 2 3+4 2 | | |
| | 9 2 4+5 2 5+6 2 5+6 2 | | |
| | 4+5 2 8 2 3+4 2 5+6 2 | | |
| | 8 2 6+7 2 5+6 2 3+4 2 | | |
| | 6+7 2 8 2 3+4 2 5+6 2 | | |

Continued on next column

Continued on next column

**TABLE LXXIII– continued**

| TQC | ICSD | | NoE | SF |
|---|---|---|---|---|
| | 4+5 2 | 6+7 2 | 5+6 2 | 3+4 2 |
| | 8 2 | 9 2 | 3+4 2 | 5+6 2 |
| | 4+5 2 | 8 2 | 5+6 2 | 3+4 2 |
| | 8 2 | 6+7 2 | 3+4 2 | 3+4 2 |
| | 9 2 | 9 2 | 5+6 2 | 5+6 2 |
| | 6+7 2 | 4+5 2 | 3+4 2 | 5+6 2 |
| | 9 2 | 8 2 | 5+6 2 | 5+6 2 |
| | 8 2 | 9 2 | 3+4 2 | 3+4 2 |
| | 4+5 2 | 9 2 | 5+6 2 | 5+6 2 |
| | 8 2 | 4+5 2 | 3+4 2 | 3+4 2 |
| 166.1.185 | 633886 | | 108 | Fe3Th1 |
| | -GM | -T | -F | -L |
| | 8 2 | 9 2 | 3+4 2 | 3+4 2 |
| | 8 2 | 8 2 | 5+6 2 | 3+4 2 |
| | 9 2 | 8 2 | 3+4 2 | 5+6 2 |
| | 8 2 | 9 2 | 5+6 2 | 3+4 2 |
| | 9 2 | 9 2 | 5+6 2 | 5+6 2 |
| | 9 2 | 8 2 | 3+4 2 | 5+6 2 |
| | 4+5 2 | 6+7 2 | 5+6 2 | 3+4 2 |
| | 9 2 | 9 2 | 3+4 2 | 5+6 2 |
| | 8 2 | 6+7 2 | 5+6 2 | 5+6 2 |
| | 6+7 2 | 8 2 | 5+6 2 | 5+6 2 |
| | 9 2 | 9 2 | 5+6 2 | 3+4 2 |
| | 6+7 2 | 4+5 2 | 3+4 2 | 5+6 2 |
| | 8 2 | 8 2 | 5+6 2 | 3+4 2 |
| | 9 2 | 9 2 | 3+4 2 | 5+6 2 |
| | 8 2 | 8 2 | 3+4 2 | 3+4 2 |
| | 8 2 | 9 2 | 3+4 2 | 5+6 2 |
| | 9 2 | 9 2 | 5+6 2 | 3+4 2 |
| | 9 2 | 4+5 2 | 3+4 2 | 3+4 2 |
| | 6+7 2 | 9 2 | 3+4 2 | 3+4 2 |
| | 4+5 2 | 8 2 | 5+6 2 | 5+6 2 |
| | 8 2 | 6+7 2 | 3+4 2 | 3+4 2 |
| | 8 2 | 8 2 | 5+6 2 | 5+6 2 |
| | 8 2 | 8 2 | 3+4 2 | 5+6 2 |
| | 4+5 2 | 4+5 2 | 5+6 2 | 5+6 2 |
| | 9 2 | 8 2 | 5+6 2 | 5+6 2 |
| | 4+5 2 | 4+5 2 | 3+4 2 | 5+6 2 |
| | 8 2 | 8 2 | 5+6 2 | 3+4 2 |
| | 9 2 | 9 2 | 3+4 2 | 5+6 2 |
| | 6+7 2 | 6+7 2 | 5+6 2 | 3+4 2 |
| | 8 2 | 9 2 | 3+4 2 | 3+4 2 |
| | 9 2 | 9 2 | 3+4 2 | 5+6 2 |
| | 9 2 | 4+5 2 | 5+6 2 | 5+6 2 |
| | 6+7 2 | 6+7 2 | 3+4 2 | 3+4 2 |
| | 8 2 | 8 2 | 5+6 2 | 3+4 2 |
| | 4+5 2 | 9 2 | 3+4 2 | 5+6 2 |
| | 9 2 | 6+7 2 | 3+4 2 | 3+4 2 |
| | 6+7 2 | 9 2 | 5+6 2 | 5+6 2 |
| | 8 2 | 8 2 | 3+4 2 | 3+4 2 |
| | 8 2 | 9 2 | 5+6 2 | 5+6 2 |
| | 9 2 | 4+5 2 | 3+4 2 | 3+4 2 |
| | 9 2 | 9 2 | 5+6 2 | 5+6 2 |
| | 4+5 2 | 8 2 | 3+4 2 | 5+6 2 |
| | 6+7 2 | 6+7 2 | 5+6 2 | 5+6 2 |
| | 8 2 | 8 2 | 3+4 2 | 3+4 2 |
| | 4+5 2 | 6+7 2 | 5+6 2 | 3+4 2 |
| | 8 2 | 9 2 | 3+4 2 | 5+6 2 |
| | 4+5 2 | 8 2 | 5+6 2 | 3+4 2 |
| | 8 2 | 6+7 2 | 3+4 2 | 3+4 2 |
| | 9 2 | 9 2 | 5+6 2 | 5+6 2 |

Continued on next column

**TABLE LXXIII– continued**

| TQC | ICSD | | NoE | SF |
|---|---|---|---|---|
| | 6+7 2 | 4+5 2 | 3+4 2 | 5+6 2 |
| | 9 2 | 8 2 | 5+6 2 | 5+6 2 |
| | 8 2 | 9 2 | 3+4 2 | 3+4 2 |
| | 4+5 2 | 9 2 | 5+6 2 | 5+6 2 |
| | 8 2 | 4+5 2 | 3+4 2 | 5+6 2 |
| 166.1.186 | 103654 | | 108 | Fe3Th1 |
| | -GM | -T | -F | -L |
| | 8 2 | 9 2 | 3+4 2 | 3+4 2 |
| | 8 2 | 8 2 | 5+6 2 | 3+4 2 |
| | 9 2 | 8 2 | 3+4 2 | 5+6 2 |
| | 8 2 | 9 2 | 5+6 2 | 3+4 2 |
| | 9 2 | 9 2 | 5+6 2 | 5+6 2 |
| | 9 2 | 8 2 | 3+4 2 | 5+6 2 |
| | 4+5 2 | 6+7 2 | 5+6 2 | 3+4 2 |
| | 8 2 | 9 2 | 3+4 2 | 5+6 2 |
| | 9 2 | 6+7 2 | 5+6 2 | 5+6 2 |
| | 6+7 2 | 8 2 | 5+6 2 | 5+6 2 |
| | 9 2 | 9 2 | 5+6 2 | 3+4 2 |
| | 6+7 2 | 4+5 2 | 3+4 2 | 5+6 2 |
| | 8 2 | 8 2 | 5+6 2 | 3+4 2 |
| | 9 2 | 9 2 | 3+4 2 | 5+6 2 |
| | 8 2 | 8 2 | 3+4 2 | 3+4 2 |
| | 8 2 | 9 2 | 3+4 2 | 5+6 2 |
| | 9 2 | 9 2 | 5+6 2 | 3+4 2 |
| | 9 2 | 8 2 | 3+4 2 | 3+4 2 |
| | 6+7 2 | 9 2 | 3+4 2 | 3+4 2 |
| | 8 2 | 6+7 2 | 5+6 2 | 5+6 2 |
| | 4+5 2 | 4+5 2 | 3+4 2 | 5+6 2 |
| | 8 2 | 8 2 | 5+6 2 | 3+4 2 |
| | 8 2 | 8 2 | 3+4 2 | 5+6 2 |
| | 4+5 2 | 4+5 2 | 5+6 2 | 5+6 2 |
| | 9 2 | 8 2 | 5+6 2 | 5+6 2 |
| | 8 2 | 4+5 2 | 3+4 2 | 5+6 2 |
| | 4+5 2 | 8 2 | 5+6 2 | 3+4 2 |
| | 9 2 | 9 2 | 3+4 2 | 5+6 2 |
| | 6+7 2 | 6+7 2 | 5+6 2 | 3+4 2 |
| | 8 2 | 9 2 | 3+4 2 | 5+6 2 |
| | 9 2 | 9 2 | 3+4 2 | 3+4 2 |
| | 9 2 | 4+5 2 | 5+6 2 | 5+6 2 |
| | 6+7 2 | 8 2 | 3+4 2 | 3+4 2 |
| | 8 2 | 6+7 2 | 5+6 2 | 3+4 2 |
| | 4+5 2 | 9 2 | 3+4 2 | 5+6 2 |
| | 9 2 | 6+7 2 | 3+4 2 | 5+6 2 |
| | 6+7 2 | 8 2 | 5+6 2 | 3+4 2 |
| | 8 2 | 9 2 | 3+4 2 | 3+4 2 |
| | 8 2 | 9 2 | 5+6 2 | 5+6 2 |
| | 9 2 | 9 2 | 3+4 2 | 3+4 2 |
| | 8 2 | 4+5 2 | 5+6 2 | 5+6 2 |
| | 9 2 | 8 2 | 3+4 2 | 5+6 2 |
| | 4+5 2 | 6+7 2 | 5+6 2 | 3+4 2 |
| | 6+7 2 | 8 2 | 3+4 2 | 5+6 2 |
| | 4+5 2 | 6+7 2 | 5+6 2 | 3+4 2 |
| | 8 2 | 9 2 | 3+4 2 | 5+6 2 |
| | 4+5 2 | 8 2 | 5+6 2 | 3+4 2 |
| | 8 2 | 6+7 2 | 3+4 2 | 3+4 2 |
| | 9 2 | 9 2 | 3+4 2 | 5+6 2 |
| | 6+7 2 | 4+5 2 | 5+6 2 | 5+6 2 |
| | 9 2 | 8 2 | 5+6 2 | 5+6 2 |
| | 8 2 | 9 2 | 3+4 2 | 5+6 2 |
| | 4+5 2 | 9 2 | 5+6 2 | 3+4 2 |
| | 8 2 | 4+5 2 | 3+4 2 | 3+4 2 |

Continued on next column

**TABLE LXXIII– continued**

| TQC | ICSD | NoE | SF | |
|---|---|---|---|---|
| 166.1.187 | 71 | 64 | Fe3Sn2 | |
| | -GM | -T | -F | -L |
| | 8 2 | 8 2 | 5+6 2 | 5+6 2 |
| | 8 2 | 9 2 | 3+4 2 | 3+4 2 |
| | 9 2 | 8 2 | 5+6 2 | 5+6 2 |
| | 9 2 | 9 2 | 3+4 2 | 3+4 2 |
| | 9 2 | 9 2 | 3+4 2 | 3+4 2 |
| | 8 2 | 8 2 | 5+6 2 | 5+6 2 |
| | 8 2 | 4+5 2 | 5+6 2 | 5+6 2 |
| | 4+5 2 | 8 2 | 3+4 2 | 3+4 2 |
| | 8 2 | 9 2 | 5+6 2 | 5+6 2 |
| | 9 2 | 6+7 2 | 3+4 2 | 3+4 2 |
| | 6+7 2 | 8 2 | 3+4 2 | 5+6 2 |
| | 9 2 | 9 2 | 3+4 2 | 3+4 2 |
| | 6+7 2 | 9 2 | 5+6 2 | 5+6 2 |
| | 8 2 | 6+7 2 | 5+6 2 | 5+6 2 |
| | 8 2 | 9 2 | 5+6 2 | 5+6 2 |
| | 4+5 2 | 6+7 2 | 3+4 2 | 3+4 2 |
| | 4+5 2 | 9 2 | 5+6 2 | 3+4 2 |
| | 9 2 | 8 2 | 3+4 2 | 3+4 2 |
| | 8 2 | 4+5 2 | 5+6 2 | 5+6 2 |
| | 8 2 | 8 2 | 3+4 2 | 3+4 2 |
| | 4+5 2 | 4+5 2 | 5+6 2 | 3+4 2 |
| | 6+7 2 | 9 2 | 3+4 2 | 3+4 2 |
| | 9 2 | 8 2 | 5+6 2 | 5+6 2 |
| | 8 2 | 4+5 2 | 5+6 2 | 5+6 2 |
| | 9 2 | 8 2 | 3+4 2 | 3+4 2 |
| | 6+7 2 | 8 2 | 3+4 2 | 5+6 2 |
| | 9 2 | 6+7 2 | 5+6 2 | 3+4 2 |
| | 4+5 2 | 9 2 | 5+6 2 | 5+6 2 |
| | 8 2 | 4+5 2 | 5+6 2 | 3+4 2 |
| | 9 2 | 8 2 | 3+4 2 | 3+4 2 |
| | 9 2 | 8 2 | 3+4 2 | 5+6 2 |
| | 6+7 2 | 4+5 2 | 5+6 2 | 3+4 2 |
| 166.1.188 | 631893 | 126 | Fe3Gd1 | |
| | -GM | -T | -F | -L |
| | 8 2 | 8 2 | 3+4 2 | 3+4 2 |
| | 8 2 | 9 2 | 5+6 2 | 3+4 2 |
| | 9 2 | 8 2 | 3+4 2 | 5+6 2 |
| | 8 2 | 9 2 | 5+6 2 | 5+6 2 |
| | 9 2 | 9 2 | 5+6 2 | 3+4 2 |
| | 9 2 | 8 2 | 3+4 2 | 5+6 2 |
| | 4+5 2 | 6+7 2 | 5+6 2 | 3+4 2 |
| | 9 2 | 9 2 | 3+4 2 | 5+6 2 |
| | 8 2 | 6+7 2 | 5+6 2 | 5+6 2 |
| | 6+7 2 | 9 2 | 5+6 2 | 5+6 2 |
| | 9 2 | 8 2 | 5+6 2 | 3+4 2 |
| | 6+7 2 | 4+5 2 | 3+4 2 | 5+6 2 |
| | 8 2 | 8 2 | 5+6 2 | 5+6 2 |
| | 9 2 | 9 2 | 3+4 2 | 3+4 2 |
| | 8 2 | 8 2 | 3+4 2 | 3+4 2 |
| | 8 2 | 9 2 | 5+6 2 | 5+6 2 |
| | 9 2 | 9 2 | 3+4 2 | 3+4 2 |
| | 9 2 | 9 2 | 3+4 2 | 5+6 2 |
| | 6+7 2 | 6+7 2 | 3+4 2 | 3+4 2 |
| | 8 2 | 8 2 | 5+6 2 | 3+4 2 |
| | 4+5 2 | 4+5 2 | 3+4 2 | 5+6 2 |
| | 8 2 | 8 2 | 3+4 2 | 5+6 2 |
| | 9 2 | 8 2 | 5+6 2 | 3+4 2 |
| | 8 2 | 4+5 2 | 5+6 2 | 5+6 2 |
| | 4+5 2 | 8 2 | 5+6 2 | 5+6 2 |

Continued on next column

**TABLE LXXIII– continued**

| TQC | ICSD | NoE | SF | |
|---|---|---|---|---|
| | 8 2 | 4+5 2 | 3+4 2 | 3+4 2 |
| | 4+5 2 | 9 2 | 5+6 2 | 5+6 2 |
| | 8 2 | 8 2 | 3+4 2 | 5+6 2 |
| | 9 2 | 9 2 | 5+6 2 | 5+6 2 |
| | 6+7 2 | 6+7 2 | 3+4 2 | 5+6 2 |
| | 9 2 | 9 2 | 3+4 2 | 3+4 2 |
| | 6+7 2 | 6+7 2 | 3+4 2 | 5+6 2 |
| | 9 2 | 9 2 | 3+4 2 | 5+6 2 |
| | 4+5 2 | 6+7 2 | 5+6 2 | 3+4 2 |
| | 8 2 | 9 2 | 3+4 2 | 3+4 2 |
| | 8 2 | 8 2 | 5+6 2 | 3+4 2 |
| | 8 2 | 9 2 | 3+4 2 | 5+6 2 |
| | 9 2 | 4+5 2 | 5+6 2 | 3+4 2 |
| | 6+7 2 | 6+7 2 | 3+4 2 | 5+6 2 |
| | 4+5 2 | 8 2 | 5+6 2 | 3+4 2 |
| | 9 2 | 9 2 | 5+6 2 | 3+4 2 |
| | 9 2 | 8 2 | 3+4 2 | 5+6 2 |
| | 6+7 2 | 9 2 | 5+6 2 | 5+6 2 |
| | 8 2 | 6+7 2 | 3+4 2 | 5+6 2 |
| | 9 2 | 9 2 | 5+6 2 | 3+4 2 |
| | 4+5 2 | 4+5 2 | 5+6 2 | 3+4 2 |
| | 9 2 | 9 2 | 3+4 2 | 5+6 2 |
| | 8 2 | 8 2 | 5+6 2 | 3+4 2 |
| | 6+7 2 | 6+7 2 | 5+6 2 | 5+6 2 |
| | 9 2 | 8 2 | 5+6 2 | 3+4 2 |
| | 6+7 2 | 9 2 | 3+4 2 | 5+6 2 |
| | 9 2 | 8 2 | 5+6 2 | 5+6 2 |
| | 8 2 | 4+5 2 | 3+4 2 | 3+4 2 |
| | 4+5 2 | 6+7 2 | 5+6 2 | 5+6 2 |
| | 8 2 | 8 2 | 3+4 2 | 5+6 2 |
| | 9 2 | 9 2 | 5+6 2 | 3+4 2 |
| | 8 2 | 9 2 | 3+4 2 | 3+4 2 |
| | 4+5 2 | 9 2 | 5+6 2 | 5+6 2 |
| | 8 2 | 4+5 2 | 5+6 2 | 5+6 2 |
| | 9 2 | 8 2 | 3+4 2 | 3+4 2 |
| | 6+7 2 | 6+7 2 | 5+6 2 | 5+6 2 |
| | 8 2 | 9 2 | 5+6 2 | 5+6 2 |
| | 9 2 | 9 2 | 3+4 2 | 5+6 2 |
| 166.1.189 | 631888 | 126 | Fe3Gd1 | |
| | -GM | -T | -F | -L |
| | 8 2 | 8 2 | 3+4 2 | 3+4 2 |
| | 8 2 | 9 2 | 5+6 2 | 3+4 2 |
| | 9 2 | 8 2 | 3+4 2 | 5+6 2 |
| | 8 2 | 9 2 | 5+6 2 | 5+6 2 |
| | 9 2 | 9 2 | 5+6 2 | 3+4 2 |
| | 9 2 | 8 2 | 3+4 2 | 5+6 2 |
| | 4+5 2 | 6+7 2 | 5+6 2 | 3+4 2 |
| | 9 2 | 9 2 | 3+4 2 | 5+6 2 |
| | 8 2 | 6+7 2 | 5+6 2 | 5+6 2 |
| | 6+7 2 | 9 2 | 5+6 2 | 5+6 2 |
| | 9 2 | 8 2 | 5+6 2 | 3+4 2 |
| | 6+7 2 | 4+5 2 | 3+4 2 | 5+6 2 |
| | 8 2 | 8 2 | 5+6 2 | 5+6 2 |
| | 9 2 | 9 2 | 3+4 2 | 3+4 2 |
| | 8 2 | 8 2 | 3+4 2 | 3+4 2 |
| | 8 2 | 9 2 | 5+6 2 | 5+6 2 |
| | 9 2 | 9 2 | 3+4 2 | 3+4 2 |
| | 9 2 | 9 2 | 3+4 2 | 5+6 2 |
| | 6+7 2 | 6+7 2 | 3+4 2 | 3+4 2 |
| | 8 2 | 8 2 | 5+6 2 | 3+4 2 |
| | 4+5 2 | 4+5 2 | 3+4 2 | 5+6 2 |

Continued on next column

**TABLE LXXIII– continued**

| TQC | ICSD | NoE | SF | |
|---|---|---|---|---|
| | 8 2 | 8 2 | 3+4 2 | 5+6 2 |
| | 9 2 | 8 2 | 5+6 2 | 3+4 2 |
| | 8 2 | 4+5 2 | 5+6 2 | 5+6 2 |
| | 4+5 2 | 8 2 | 5+6 2 | 5+6 2 |
| | 8 2 | 4+5 2 | 3+4 2 | 3+4 2 |
| | 4+5 2 | 9 2 | 5+6 2 | 5+6 2 |
| | 8 2 | 8 2 | 3+4 2 | 5+6 2 |
| | 9 2 | 9 2 | 5+6 2 | 5+6 2 |
| | 6+7 2 | 6+7 2 | 3+4 2 | 5+6 2 |
| | 9 2 | 9 2 | 3+4 2 | 3+4 2 |
| | 6+7 2 | 6+7 2 | 3+4 2 | 5+6 2 |
| | 9 2 | 9 2 | 3+4 2 | 5+6 2 |
| | 4+5 2 | 6+7 2 | 5+6 2 | 3+4 2 |
| | 8 2 | 9 2 | 3+4 2 | 3+4 2 |
| | 8 2 | 8 2 | 5+6 2 | 3+4 2 |
| | 8 2 | 9 2 | 3+4 2 | 5+6 2 |
| | 9 2 | 4+5 2 | 5+6 2 | 3+4 2 |
| | 6+7 2 | 6+7 2 | 5+6 2 | 5+6 2 |
| | 4+5 2 | 8 2 | 3+4 2 | 3+4 2 |
| | 9 2 | 9 2 | 5+6 2 | 3+4 2 |
| | 9 2 | 8 2 | 3+4 2 | 5+6 2 |
| | 6+7 2 | 9 2 | 5+6 2 | 5+6 2 |
| | 8 2 | 6+7 2 | 3+4 2 | 5+6 2 |
| | 9 2 | 9 2 | 5+6 2 | 3+4 2 |
| | 4+5 2 | 4+5 2 | 5+6 2 | 3+4 2 |
| | 9 2 | 9 2 | 3+4 2 | 5+6 2 |
| | 8 2 | 8 2 | 5+6 2 | 3+4 2 |
| | 6+7 2 | 6+7 2 | 5+6 2 | 5+6 2 |
| | 9 2 | 8 2 | 5+6 2 | 3+4 2 |
| | 6+7 2 | 9 2 | 3+4 2 | 5+6 2 |
| | 9 2 | 8 2 | 5+6 2 | 5+6 2 |
| | 8 2 | 4+5 2 | 3+4 2 | 3+4 2 |
| | 4+5 2 | 6+7 2 | 5+6 2 | 5+6 2 |
| | 8 2 | 8 2 | 3+4 2 | 5+6 2 |
| | 9 2 | 9 2 | 5+6 2 | 3+4 2 |
| | 8 2 | 9 2 | 3+4 2 | 3+4 2 |
| | 4+5 2 | 9 2 | 5+6 2 | 5+6 2 |
| | 8 2 | 4+5 2 | 5+6 2 | 5+6 2 |
| | 9 2 | 8 2 | 3+4 2 | 3+4 2 |
| | 6+7 2 | 6+7 2 | 5+6 2 | 5+6 2 |
| | 8 2 | 9 2 | 5+6 2 | 5+6 2 |
| | 9 2 | 9 2 | 3+4 2 | 5+6 2 |
| 166.1.190 | 631864 | 126 | Fe3Gd1 | |
| | -GM | -T | -F | -L |
| | 8 2 | 8 2 | 3+4 2 | 3+4 2 |
| | 8 2 | 9 2 | 5+6 2 | 3+4 2 |
| | 9 2 | 8 2 | 3+4 2 | 5+6 2 |
| | 8 2 | 9 2 | 5+6 2 | 5+6 2 |
| | 9 2 | 9 2 | 5+6 2 | 3+4 2 |
| | 9 2 | 8 2 | 3+4 2 | 5+6 2 |
| | 4+5 2 | 6+7 2 | 5+6 2 | 3+4 2 |
| | 9 2 | 9 2 | 3+4 2 | 5+6 2 |
| | 8 2 | 6+7 2 | 5+6 2 | 5+6 2 |
| | 6+7 2 | 9 2 | 5+6 2 | 5+6 2 |
| | 9 2 | 8 2 | 5+6 2 | 3+4 2 |
| | 6+7 2 | 4+5 2 | 3+4 2 | 5+6 2 |
| | 8 2 | 8 2 | 5+6 2 | 5+6 2 |
| | 9 2 | 9 2 | 3+4 2 | 3+4 2 |
| | 8 2 | 8 2 | 3+4 2 | 3+4 2 |
| | 8 2 | 9 2 | 5+6 2 | 5+6 2 |
| | 9 2 | 9 2 | 3+4 2 | 3+4 2 |

Continued on next column

**TABLE LXXIII– continued**

| TQC | ICSD | NoE | SF | |
|---|---|---|---|---|
| | 9 2 | 9 2 | 3+4 2 | 5+6 2 |
| | 6+7 2 | 6+7 2 | 3+4 2 | 3+4 2 |
| | 8 2 | 8 2 | 5+6 2 | 3+4 2 |
| | 4+5 2 | 4+5 2 | 3+4 2 | 5+6 2 |
| | 8 2 | 8 2 | 3+4 2 | 5+6 2 |
| | 9 2 | 8 2 | 5+6 2 | 3+4 2 |
| | 8 2 | 4+5 2 | 5+6 2 | 5+6 2 |
| | 4+5 2 | 8 2 | 5+6 2 | 5+6 2 |
| | 8 2 | 4+5 2 | 3+4 2 | 3+4 2 |
| | 4+5 2 | 9 2 | 5+6 2 | 5+6 2 |
| | 8 2 | 8 2 | 3+4 2 | 5+6 2 |
| | 9 2 | 9 2 | 5+6 2 | 5+6 2 |
| | 6+7 2 | 6+7 2 | 3+4 2 | 5+6 2 |
| | 9 2 | 9 2 | 3+4 2 | 3+4 2 |
| | 6+7 2 | 6+7 2 | 3+4 2 | 5+6 2 |
| | 9 2 | 9 2 | 3+4 2 | 5+6 2 |
| | 4+5 2 | 6+7 2 | 5+6 2 | 3+4 2 |
| | 8 2 | 9 2 | 3+4 2 | 3+4 2 |
| | 8 2 | 8 2 | 5+6 2 | 3+4 2 |
| | 8 2 | 9 2 | 3+4 2 | 5+6 2 |
| | 9 2 | 4+5 2 | 5+6 2 | 3+4 2 |
| | 6+7 2 | 6+7 2 | 5+6 2 | 5+6 2 |
| | 4+5 2 | 8 2 | 3+4 2 | 3+4 2 |
| | 9 2 | 9 2 | 5+6 2 | 3+4 2 |
| | 9 2 | 8 2 | 3+4 2 | 5+6 2 |
| | 6+7 2 | 9 2 | 5+6 2 | 5+6 2 |
| | 8 2 | 6+7 2 | 3+4 2 | 5+6 2 |
| | 9 2 | 9 2 | 5+6 2 | 3+4 2 |
| | 4+5 2 | 4+5 2 | 5+6 2 | 3+4 2 |
| | 9 2 | 9 2 | 3+4 2 | 5+6 2 |
| | 8 2 | 8 2 | 5+6 2 | 3+4 2 |
| | 6+7 2 | 6+7 2 | 5+6 2 | 5+6 2 |
| | 9 2 | 8 2 | 5+6 2 | 3+4 2 |
| | 6+7 2 | 9 2 | 3+4 2 | 5+6 2 |
| | 9 2 | 8 2 | 5+6 2 | 5+6 2 |
| | 8 2 | 4+5 2 | 3+4 2 | 3+4 2 |
| | 4+5 2 | 6+7 2 | 5+6 2 | 5+6 2 |
| | 8 2 | 8 2 | 3+4 2 | 5+6 2 |
| | 9 2 | 9 2 | 5+6 2 | 3+4 2 |
| | 8 2 | 9 2 | 3+4 2 | 3+4 2 |
| | 4+5 2 | 9 2 | 5+6 2 | 5+6 2 |
| | 8 2 | 4+5 2 | 5+6 2 | 5+6 2 |
| | 9 2 | 8 2 | 3+4 2 | 3+4 2 |
| | 6+7 2 | 6+7 2 | 5+6 2 | 5+6 2 |
| | 8 2 | 9 2 | 5+6 2 | 5+6 2 |
| | 9 2 | 9 2 | 3+4 2 | 5+6 2 |
| 166.1.191 | 103479 | 126 | Fe3Gd1 | |
| | -GM | -T | -F | -L |
| | 8 2 | 8 2 | 3+4 2 | 3+4 2 |
| | 8 2 | 9 2 | 5+6 2 | 3+4 2 |
| | 9 2 | 8 2 | 3+4 2 | 5+6 2 |
| | 8 2 | 9 2 | 5+6 2 | 5+6 2 |
| | 9 2 | 9 2 | 5+6 2 | 3+4 2 |
| | 9 2 | 8 2 | 3+4 2 | 5+6 2 |
| | 4+5 2 | 6+7 2 | 5+6 2 | 3+4 2 |
| | 9 2 | 9 2 | 3+4 2 | 5+6 2 |
| | 8 2 | 6+7 2 | 5+6 2 | 5+6 2 |
| | 6+7 2 | 9 2 | 5+6 2 | 5+6 2 |
| | 9 2 | 8 2 | 5+6 2 | 3+4 2 |
| | 6+7 2 | 4+5 2 | 3+4 2 | 5+6 2 |
| | 8 2 | 8 2 | 5+6 2 | 5+6 2 |

Continued on next column

**TABLE LXXIII– continued**

| TQC | ICSD | NoE | SF | |
|---|---|---|---|---|
| | 9 2 | 9 2 | 3+4 2 | 3+4 2 |
| | 8 2 | 8 2 | 3+4 2 | 3+4 2 |
| | 8 2 | 9 2 | 5+6 2 | 5+6 2 |
| | 9 2 | 9 2 | 3+4 2 | 3+4 2 |
| | 9 2 | 9 2 | 3+4 2 | 5+6 2 |
| | 6+7 2 | 6+7 2 | 3+4 2 | 3+4 2 |
| | 8 2 | 8 2 | 5+6 2 | 3+4 2 |
| | 4+5 2 | 4+5 2 | 3+4 2 | 5+6 2 |
| | 8 2 | 8 2 | 3+4 2 | 5+6 2 |
| | 9 2 | 8 2 | 5+6 2 | 3+4 2 |
| | 8 2 | 4+5 2 | 5+6 2 | 5+6 2 |
| | 4+5 2 | 8 2 | 5+6 2 | 3+4 2 |
| | 8 2 | 4+5 2 | 3+4 2 | 5+6 2 |
| | 4+5 2 | 9 2 | 5+6 2 | 5+6 2 |
| | 8 2 | 8 2 | 3+4 2 | 5+6 2 |
| | 9 2 | 9 2 | 5+6 2 | 5+6 2 |
| | 6+7 2 | 6+7 2 | 3+4 2 | 5+6 2 |
| | 9 2 | 9 2 | 3+4 2 | 3+4 2 |
| | 6+7 2 | 6+7 2 | 3+4 2 | 5+6 2 |
| | 9 2 | 9 2 | 3+4 2 | 5+6 2 |
| | 4+5 2 | 6+7 2 | 5+6 2 | 3+4 2 |
| | 8 2 | 9 2 | 3+4 2 | 3+4 2 |
| | 8 2 | 8 2 | 5+6 2 | 3+4 2 |
| | 8 2 | 9 2 | 3+4 2 | 5+6 2 |
| | 9 2 | 4+5 2 | 5+6 2 | 3+4 2 |
| | 6+7 2 | 6+7 2 | 3+4 2 | 5+6 2 |
| | 4+5 2 | 8 2 | 5+6 2 | 3+4 2 |
| | 9 2 | 9 2 | 5+6 2 | 3+4 2 |
| | 9 2 | 8 2 | 3+4 2 | 5+6 2 |
| | 6+7 2 | 9 2 | 5+6 2 | 5+6 2 |
| | 8 2 | 6+7 2 | 3+4 2 | 5+6 2 |
| | 9 2 | 9 2 | 5+6 2 | 3+4 2 |
| | 4+5 2 | 4+5 2 | 5+6 2 | 3+4 2 |
| | 9 2 | 9 2 | 3+4 2 | 5+6 2 |
| | 8 2 | 8 2 | 5+6 2 | 3+4 2 |
| | 6+7 2 | 6+7 2 | 5+6 2 | 5+6 2 |
| | 9 2 | 8 2 | 5+6 2 | 3+4 2 |
| | 6+7 2 | 9 2 | 3+4 2 | 5+6 2 |
| | 9 2 | 8 2 | 5+6 2 | 5+6 2 |
| | 8 2 | 4+5 2 | 3+4 2 | 3+4 2 |
| | 8 2 | 6+7 2 | 5+6 2 | 5+6 2 |
| | 4+5 2 | 8 2 | 3+4 2 | 5+6 2 |
| | 9 2 | 9 2 | 5+6 2 | 3+4 2 |
| | 8 2 | 9 2 | 3+4 2 | 3+4 2 |
| | 4+5 2 | 9 2 | 5+6 2 | 5+6 2 |
| | 8 2 | 4+5 2 | 5+6 2 | 5+6 2 |
| | 9 2 | 8 2 | 3+4 2 | 3+4 2 |
| | 6+7 2 | 6+7 2 | 5+6 2 | 5+6 2 |
| | 8 2 | 9 2 | 5+6 2 | 5+6 2 |
| | 9 2 | 9 2 | 3+4 2 | 5+6 2 |
| 166.1.192 | 4192 | 64 | Fe2O4Yb1 | |
| | -GM | -T | -F | -L |
| | 8 2 | 8 2 | 3+4 2 | 3+4 2 |
| | 9 2 | 9 2 | 5+6 2 | 5+6 2 |
| | 9 2 | 9 2 | 5+6 2 | 5+6 2 |
| | 6+7 2 | 6+7 2 | 5+6 2 | 5+6 2 |
| | 8 2 | 9 2 | 5+6 2 | 3+4 2 |
| | 9 2 | 8 2 | 3+4 2 | 5+6 2 |
| | 8 2 | 8 2 | 3+4 2 | 3+4 2 |
| | 9 2 | 9 2 | 5+6 2 | 5+6 2 |
| | 8 2 | 8 2 | 3+4 2 | 5+6 2 |

Continued on next column

**TABLE LXXIII– continued**

| TQC | ICSD | NoE | SF | |
|---|---|---|---|---|
| | 8 2 | 9 2 | 5+6 2 | 3+4 2 |
| | 4+5 2 | 9 2 | 3+4 2 | 3+4 2 |
| | 8 2 | 6+7 2 | 5+6 2 | 3+4 2 |
| | 6+7 2 | 4+5 2 | 5+6 2 | 5+6 2 |
| | 9 2 | 8 2 | 3+4 2 | 3+4 2 |
| | 9 2 | 4+5 2 | 3+4 2 | 5+6 2 |
| | 4+5 2 | 8 2 | 5+6 2 | 3+4 2 |
| | 8 2 | 9 2 | 3+4 2 | 5+6 2 |
| | 6+7 2 | 8 2 | 5+6 2 | 3+4 2 |
| | 9 2 | 6+7 2 | 5+6 2 | 5+6 2 |
| | 9 2 | 9 2 | 3+4 2 | 5+6 2 |
| | 9 2 | 9 2 | 5+6 2 | 5+6 2 |
| | 9 2 | 9 2 | 5+6 2 | 5+6 2 |
| | 6+7 2 | 6+7 2 | 5+6 2 | 5+6 2 |
| | 8 2 | 9 2 | 3+4 2 | 5+6 2 |
| | 4+5 2 | 6+7 2 | 5+6 2 | 3+4 2 |
| | 9 2 | 8 2 | 5+6 2 | 3+4 2 |
| | 6+7 2 | 4+5 2 | 5+6 2 | 5+6 2 |
| | 9 2 | 9 2 | 3+4 2 | 5+6 2 |
| | 9 2 | 9 2 | 5+6 2 | 5+6 2 |
| | 6+7 2 | 6+7 2 | 3+4 2 | 5+6 2 |
| | 9 2 | 9 2 | 5+6 2 | 3+4 2 |
| | 9 2 | 8 2 | 5+6 2 | 5+6 2 |
| 166.1.193 | 195811 | 64 | Fe2O4Yb1 | |
| | -GM | -T | -F | -L |
| | 8 2 | 8 2 | 3+4 2 | 3+4 2 |
| | 9 2 | 9 2 | 5+6 2 | 5+6 2 |
| | 9 2 | 9 2 | 5+6 2 | 5+6 2 |
| | 6+7 2 | 6+7 2 | 5+6 2 | 5+6 2 |
| | 8 2 | 9 2 | 5+6 2 | 3+4 2 |
| | 9 2 | 8 2 | 3+4 2 | 5+6 2 |
| | 8 2 | 8 2 | 3+4 2 | 3+4 2 |
| | 9 2 | 9 2 | 5+6 2 | 5+6 2 |
| | 8 2 | 9 2 | 3+4 2 | 5+6 2 |
| | 8 2 | 6+7 2 | 5+6 2 | 3+4 2 |
| | 4+5 2 | 8 2 | 3+4 2 | 3+4 2 |
| | 8 2 | 9 2 | 5+6 2 | 3+4 2 |
| | 6+7 2 | 4+5 2 | 5+6 2 | 5+6 2 |
| | 9 2 | 8 2 | 3+4 2 | 3+4 2 |
| | 9 2 | 4+5 2 | 3+4 2 | 5+6 2 |
| | 4+5 2 | 8 2 | 5+6 2 | 5+6 2 |
| | 8 2 | 9 2 | 3+4 2 | 3+4 2 |
| | 6+7 2 | 8 2 | 5+6 2 | 3+4 2 |
| | 9 2 | 6+7 2 | 5+6 2 | 5+6 2 |
| | 9 2 | 9 2 | 3+4 2 | 5+6 2 |
| | 9 2 | 9 2 | 5+6 2 | 5+6 2 |
| | 9 2 | 9 2 | 5+6 2 | 5+6 2 |
| | 6+7 2 | 6+7 2 | 5+6 2 | 5+6 2 |
| | 8 2 | 9 2 | 3+4 2 | 5+6 2 |
| | 4+5 2 | 6+7 2 | 5+6 2 | 3+4 2 |
| | 9 2 | 8 2 | 5+6 2 | 3+4 2 |
| | 6+7 2 | 4+5 2 | 5+6 2 | 5+6 2 |
| | 9 2 | 9 2 | 3+4 2 | 5+6 2 |
| | 9 2 | 9 2 | 5+6 2 | 5+6 2 |
| | 6+7 2 | 6+7 2 | 3+4 2 | 5+6 2 |
| | 9 2 | 8 2 | 5+6 2 | 3+4 2 |
| | 9 2 | 9 2 | 5+6 2 | 5+6 2 |
| 166.1.194 | 195810 | 64 | Fe2O4Yb1 | |
| | -GM | -T | -F | -L |
| | 8 2 | 8 2 | 3+4 2 | 3+4 2 |
| | 9 2 | 9 2 | 5+6 2 | 5+6 2 |

Continued on next column

**TABLE LXXIII– continued**

| TQC | ICSD | NoE | SF |
|---|---|---|---|
| | 9 2 | 9 2 5+6 2 5+6 2 | |
| | 6+7 2 6+7 2 5+6 2 5+6 2 | | |
| | 8 2 | 9 2 5+6 2 3+4 2 | |
| | 9 2 | 8 2 3+4 2 5+6 2 | |
| | 8 2 | 8 2 3+4 2 3+4 2 | |
| | 9 2 | 9 2 5+6 2 5+6 2 | |
| | 8 2 | 9 2 3+4 2 5+6 2 | |
| | 8 2 | 9 2 5+6 2 3+4 2 | |
| | 4+5 2 6+7 2 5+6 2 3+4 2 | | |
| | 8 2 | 8 2 3+4 2 3+4 2 | |
| | 6+7 2 4+5 2 5+6 2 5+6 2 | | |
| | 9 2 | 8 2 3+4 2 3+4 2 | |
| | 9 2 4+5 2 3+4 2 5+6 2 | | |
| | 4+5 2 | 8 2 5+6 2 3+4 2 | |
| | 8 2 | 9 2 3+4 2 5+6 2 | |
| | 6+7 2 | 8 2 5+6 2 3+4 2 | |
| | 9 2 6+7 2 5+6 2 5+6 2 | | |
| | 9 2 | 9 2 3+4 2 5+6 2 | |
| | 9 2 | 9 2 5+6 2 5+6 2 | |
| | 9 2 | 9 2 5+6 2 5+6 2 | |
| | 6+7 2 6+7 2 5+6 2 5+6 2 | | |
| | 8 2 | 9 2 3+4 2 5+6 2 | |
| | 4+5 2 6+7 2 5+6 2 3+4 2 | | |
| | 9 2 | 8 2 5+6 2 3+4 2 | |
| | 6+7 2 4+5 2 5+6 2 5+6 2 | | |
| | 9 2 | 9 2 3+4 2 5+6 2 | |
| | 9 2 | 9 2 5+6 2 5+6 2 | |
| | 6+7 2 6+7 2 3+4 2 5+6 2 | | |
| | 9 2 | 8 2 5+6 2 3+4 2 | |
| | 9 2 | 9 2 5+6 2 5+6 2 | |
| 166.1.195 | 195809 | 64 | Fe2O4Yb1 |
| | -GM -T -F -L | | |
| | 8 2 | 8 2 3+4 2 3+4 2 | |
| | 9 2 | 9 2 5+6 2 5+6 2 | |
| | 9 2 | 9 2 5+6 2 5+6 2 | |
| | 6+7 2 6+7 2 5+6 2 5+6 2 | | |
| | 8 2 | 9 2 5+6 2 3+4 2 | |
| | 9 2 | 8 2 3+4 2 5+6 2 | |
| | 8 2 | 8 2 3+4 2 3+4 2 | |
| | 9 2 | 9 2 5+6 2 5+6 2 | |
| | 8 2 6+7 2 3+4 2 5+6 2 | | |
| | 4+5 2 | 9 2 5+6 2 3+4 2 | |
| | 8 2 | 9 2 5+6 2 3+4 2 | |
| | 6+7 2 | 8 2 3+4 2 3+4 2 | |
| | 9 2 4+5 2 5+6 2 5+6 2 | | |
| | 8 2 | 8 2 3+4 2 3+4 2 | |
| | 9 2 4+5 2 3+4 2 5+6 2 | | |
| | 4+5 2 | 8 2 5+6 2 5+6 2 | |
| | 8 2 | 9 2 3+4 2 3+4 2 | |
| | 9 2 | 8 2 5+6 2 3+4 2 | |
| | 6+7 2 6+7 2 5+6 2 5+6 2 | | |
| | 9 2 | 9 2 3+4 2 5+6 2 | |
| | 9 2 | 9 2 5+6 2 5+6 2 | |
| | 9 2 | 9 2 5+6 2 5+6 2 | |
| | 6+7 2 6+7 2 5+6 2 5+6 2 | | |
| | 8 2 | 9 2 3+4 2 5+6 2 | |
| | 4+5 2 | 8 2 5+6 2 3+4 2 | |
| | 9 2 4+5 2 5+6 2 3+4 2 | | |
| | 6+7 2 6+7 2 3+4 2 5+6 2 | | |
| | 9 2 | 9 2 3+4 2 5+6 2 | |
| | 9 2 | 8 2 5+6 2 3+4 2 | |

Continued on next column

**TABLE LXXIII– continued**

| TQC | ICSD | NoE | SF |
|---|---|---|---|
| | 8 2 | 9 2 5+6 2 3+4 2 | |
| | 6+7 2 6+7 2 5+6 2 5+6 2 | | |
| | 9 2 | 9 2 3+4 2 5+6 2 | |
| 166.1.196 | 29011 | 46 | Fe1O2Tl1 |
| | -GM -T -F -L | | |
| | 8 2 | 8 2 5+6 2 5+6 2 | |
| | 8 2 | 8 2 3+4 2 3+4 2 | |
| | 4+5 2 4+5 2 5+6 2 5+6 2 | | |
| | 8 2 | 9 2 3+4 2 5+6 2 | |
| | 9 2 | 9 2 3+4 2 3+4 2 | |
| | 8 2 | 9 2 5+6 2 3+4 2 | |
| | 9 2 6+7 2 5+6 2 5+6 2 | | |
| | 6+7 2 | 9 2 3+4 2 3+4 2 | |
| | 9 2 | 9 2 3+4 2 3+4 2 | |
| | 6+7 2 6+7 2 5+6 2 3+4 2 | | |
| | 9 2 | 8 2 3+4 2 3+4 2 | |
| | 9 2 | 8 2 5+6 2 5+6 2 | |
| | 9 2 | 9 2 3+4 2 5+6 2 | |
| | 6+7 2 | 9 2 3+4 2 5+6 2 | |
| | 9 2 6+7 2 5+6 2 3+4 2 | | |
| | 9 2 | 8 2 5+6 2 3+4 2 | |
| | 9 2 4+5 2 3+4 2 3+4 2 | | |
| | 6+7 2 | 9 2 5+6 2 3+4 2 | |
| | 4+5 2 4+5 2 5+6 2 5+6 2 | | |
| | 8 2 | 8 2 3+4 2 5+6 2 | |
| | 8 2 | 8 2 5+6 2 3+4 2 | |
| | 8 2 4+5 2 3+4 2 3+4 2 | | |
| | 4+5 2 | 9 2 5+6 2 5+6 2 | |
| 166.1.197 | 631854 | 148 | Fe15Ga2Y2 |
| | -GM -T -F -L | | |
| | 8 2 | 8 2 3+4 2 3+4 2 | |
| | 9 2 | 9 2 5+6 2 5+6 2 | |
| | 8 2 | 9 2 3+4 2 5+6 2 | |
| | 9 2 | 8 2 5+6 2 3+4 2 | |
| | 8 2 | 9 2 3+4 2 5+6 2 | |
| | 9 2 | 8 2 5+6 2 3+4 2 | |
| | 4+5 2 4+5 2 5+6 2 3+4 2 | | |
| | 6+7 2 6+7 2 3+4 2 5+6 2 | | |
| | 8 2 | 8 2 3+4 2 3+4 2 | |
| | 9 2 | 9 2 5+6 2 5+6 2 | |
| | 8 2 | 9 2 5+6 2 5+6 2 | |
| | 9 2 6+7 2 5+6 2 3+4 2 | | |
| | 6+7 2 | 8 2 3+4 2 3+4 2 | |
| | 8 2 | 8 2 3+4 2 5+6 2 | |
| | 4+5 2 4+5 2 5+6 2 3+4 2 | | |
| | 8 2 | 9 2 3+4 2 3+4 2 | |
| | 8 2 | 9 2 5+6 2 3+4 2 | |
| | 4+5 2 | 9 2 3+4 2 5+6 2 | |
| | 8 2 6+7 2 5+6 2 3+4 2 | | |
| | 8 2 | 9 2 3+4 2 5+6 2 | |
| | 6+7 2 | 8 2 5+6 2 5+6 2 | |
| | 9 2 6+7 2 3+4 2 3+4 2 | | |
| | 9 2 | 9 2 5+6 2 3+4 2 | |
| | 8 2 | 9 2 5+6 2 3+4 2 | |
| | 4+5 2 6+7 2 3+4 2 5+6 2 | | |
| | 8 2 4+5 2 5+6 2 5+6 2 | | |
| | 4+5 2 | 8 2 3+4 2 5+6 2 | |
| | 8 2 | 8 2 5+6 2 3+4 2 | |
| | 4+5 2 | 9 2 5+6 2 5+6 2 | |
| | 8 2 4+5 2 3+4 2 3+4 2 | | |
| | 9 2 6+7 2 5+6 2 3+4 2 | | |

Continued on next column

**TABLE LXXIII– continued**

| TQC | ICSD | NoE | SF | |
|---|---|---|---|---|
| | 6+7 2 | 9 2 | 5+6 2 | 5+6 2 |
| | 8 2 | 8 2 | 3+4 2 | 5+6 2 |
| | 9 2 | 8 2 | 3+4 2 | 3+4 2 |
| | 6+7 2 | 4+5 2 | 5+6 2 | 3+4 2 |
| | 8 2 | 9 2 | 5+6 2 | 5+6 2 |
| | 4+5 2 | 9 2 | 3+4 2 | 5+6 2 |
| | 9 2 | 6+7 2 | 3+4 2 | 5+6 2 |
| | 6+7 2 | 9 2 | 5+6 2 | 5+6 2 |
| | 8 2 | 9 2 | 3+4 2 | 3+4 2 |
| | 9 2 | 6+7 2 | 3+4 2 | 3+4 2 |
| | 4+5 2 | 8 2 | 5+6 2 | 3+4 2 |
| | 8 2 | 8 2 | 3+4 2 | 5+6 2 |
| | 9 2 | 8 2 | 5+6 2 | 3+4 2 |
| | 4+5 2 | 4+5 2 | 3+4 2 | 3+4 2 |
| | 9 2 | 8 2 | 5+6 2 | 5+6 2 |
| | 8 2 | 9 2 | 3+4 2 | 5+6 2 |
| | 6+7 2 | 4+5 2 | 5+6 2 | 3+4 2 |
| | 9 2 | 9 2 | 3+4 2 | 3+4 2 |
| | 8 2 | 6+7 2 | 5+6 2 | 3+4 2 |
| | 4+5 2 | 8 2 | 3+4 2 | 5+6 2 |
| | 9 2 | 8 2 | 5+6 2 | 5+6 2 |
| | 8 2 | 6+7 2 | 5+6 2 | 5+6 2 |
| | 8 2 | 9 2 | 3+4 2 | 3+4 2 |
| | 6+7 2 | 9 2 | 3+4 2 | 5+6 2 |
| | 9 2 | 8 2 | 5+6 2 | 3+4 2 |
| | 9 2 | 4+5 2 | 5+6 2 | 5+6 2 |
| | 6+7 2 | 9 2 | 3+4 2 | 3+4 2 |
| | 8 2 | 8 2 | 5+6 2 | 5+6 2 |
| | 9 2 | 4+5 2 | 3+4 2 | 3+4 2 |
| | 9 2 | 9 2 | 5+6 2 | 5+6 2 |
| | 8 2 | 6+7 2 | 5+6 2 | 3+4 2 |
| | 6+7 2 | 8 2 | 3+4 2 | 5+6 2 |
| | 4+5 2 | 8 2 | 5+6 2 | 3+4 2 |
| | 8 2 | 4+5 2 | 3+4 2 | 3+4 2 |
| | 4+5 2 | 9 2 | 3+4 2 | 5+6 2 |
| | 9 2 | 6+7 2 | 3+4 2 | 3+4 2 |
| | 9 2 | 6+7 2 | 3+4 2 | 5+6 2 |
| | 6+7 2 | 9 2 | 5+6 2 | 5+6 2 |
| | 8 2 | 9 2 | 5+6 2 | 3+4 2 |
| | 9 2 | 9 2 | 5+6 2 | 3+4 2 |
| | 4+5 2 | 8 2 | 3+4 2 | 5+6 2 |
| | 8 2 | 8 2 | 5+6 2 | 5+6 2 |
| | 8 2 | 4+5 2 | 3+4 2 | 3+4 2 |
| 166.1.198 | 95660 | 54 | F1Nd1O1 | |
| | -GM | -T | -F | -L |
| | 8 2 | 9 2 | 5+6 2 | 3+4 2 |
| | 9 2 | 8 2 | 3+4 2 | 5+6 2 |
| | 8 2 | 8 2 | 5+6 2 | 5+6 2 |
| | 9 2 | 9 2 | 3+4 2 | 3+4 2 |
| | 8 2 | 9 2 | 5+6 2 | 3+4 2 |
| | 9 2 | 8 2 | 3+4 2 | 5+6 2 |
| | 8 2 | 9 2 | 5+6 2 | 3+4 2 |
| | 9 2 | 8 2 | 3+4 2 | 5+6 2 |
| | 4+5 2 | 6+7 2 | 5+6 2 | 3+4 2 |
| | 6+7 2 | 4+5 2 | 3+4 2 | 5+6 2 |
| | 9 2 | 8 2 | 3+4 2 | 5+6 2 |
| | 8 2 | 9 2 | 5+6 2 | 3+4 2 |
| | 8 2 | 8 2 | 3+4 2 | 3+4 2 |
| | 8 2 | 8 2 | 5+6 2 | 5+6 2 |
| | 4+5 2 | 4+5 2 | 5+6 2 | 5+6 2 |
| | 9 2 | 9 2 | 5+6 2 | 5+6 2 |

Continued on next column

**TABLE LXXIII– continued**

| TQC | ICSD | NoE | SF | |
|---|---|---|---|---|
| | 6+7 2 | 6+7 2 | 3+4 2 | 3+4 2 |
| | 9 2 | 9 2 | 3+4 2 | 3+4 2 |
| | 8 2 | 9 2 | 3+4 2 | 5+6 2 |
| | 8 2 | 9 2 | 5+6 2 | 3+4 2 |
| | 4+5 2 | 6+7 2 | 5+6 2 | 3+4 2 |
| | 9 2 | 8 2 | 5+6 2 | 3+4 2 |
| | 6+7 2 | 4+5 2 | 3+4 2 | 5+6 2 |
| | 9 2 | 8 2 | 3+4 2 | 5+6 2 |
| | 8 2 | 9 2 | 3+4 2 | 5+6 2 |
| | 9 2 | 9 2 | 5+6 2 | 3+4 2 |
| | 4+5 2 | 6+7 2 | 5+6 2 | 3+4 2 |
| 166.1.199 | 95659 | 54 | F1Nd1O1 | |
| | -GM | -T | -F | -L |
| | 8 2 | 9 2 | 5+6 2 | 3+4 2 |
| | 9 2 | 8 2 | 3+4 2 | 5+6 2 |
| | 8 2 | 8 2 | 5+6 2 | 5+6 2 |
| | 9 2 | 9 2 | 3+4 2 | 3+4 2 |
| | 8 2 | 9 2 | 5+6 2 | 3+4 2 |
| | 9 2 | 8 2 | 3+4 2 | 5+6 2 |
| | 8 2 | 9 2 | 5+6 2 | 3+4 2 |
| | 9 2 | 8 2 | 3+4 2 | 5+6 2 |
| | 4+5 2 | 6+7 2 | 5+6 2 | 3+4 2 |
| | 6+7 2 | 4+5 2 | 3+4 2 | 5+6 2 |
| | 9 2 | 8 2 | 3+4 2 | 5+6 2 |
| | 8 2 | 9 2 | 5+6 2 | 3+4 2 |
| | 8 2 | 8 2 | 3+4 2 | 3+4 2 |
| | 8 2 | 8 2 | 5+6 2 | 5+6 2 |
| | 4+5 2 | 4+5 2 | 5+6 2 | 5+6 2 |
| | 9 2 | 9 2 | 5+6 2 | 5+6 2 |
| | 6+7 2 | 6+7 2 | 3+4 2 | 3+4 2 |
| | 9 2 | 9 2 | 3+4 2 | 3+4 2 |
| | 8 2 | 9 2 | 3+4 2 | 5+6 2 |
| | 8 2 | 9 2 | 5+6 2 | 3+4 2 |
| | 4+5 2 | 6+7 2 | 5+6 2 | 3+4 2 |
| | 9 2 | 8 2 | 5+6 2 | 3+4 2 |
| | 6+7 2 | 4+5 2 | 3+4 2 | 5+6 2 |
| | 9 2 | 8 2 | 3+4 2 | 5+6 2 |
| | 8 2 | 9 2 | 3+4 2 | 5+6 2 |
| | 9 2 | 9 2 | 5+6 2 | 3+4 2 |
| | 4+5 2 | 6+7 2 | 5+6 2 | 3+4 2 |
| 166.1.200 | 95658 | 54 | F1Nd1O1 | |
| | -GM | -T | -F | -L |
| | 8 2 | 9 2 | 5+6 2 | 3+4 2 |
| | 9 2 | 8 2 | 3+4 2 | 5+6 2 |
| | 8 2 | 8 2 | 5+6 2 | 5+6 2 |
| | 9 2 | 9 2 | 3+4 2 | 3+4 2 |
| | 8 2 | 9 2 | 5+6 2 | 3+4 2 |
| | 9 2 | 8 2 | 3+4 2 | 5+6 2 |
| | 8 2 | 9 2 | 5+6 2 | 3+4 2 |
| | 9 2 | 8 2 | 3+4 2 | 5+6 2 |
| | 4+5 2 | 6+7 2 | 5+6 2 | 3+4 2 |
| | 6+7 2 | 4+5 2 | 3+4 2 | 5+6 2 |
| | 9 2 | 8 2 | 3+4 2 | 5+6 2 |
| | 8 2 | 9 2 | 5+6 2 | 3+4 2 |
| | 8 2 | 8 2 | 3+4 2 | 3+4 2 |
| | 8 2 | 8 2 | 5+6 2 | 5+6 2 |
| | 4+5 2 | 4+5 2 | 5+6 2 | 5+6 2 |
| | 9 2 | 9 2 | 5+6 2 | 5+6 2 |
| | 6+7 2 | 6+7 2 | 3+4 2 | 3+4 2 |
| | 9 2 | 9 2 | 3+4 2 | 3+4 2 |
| | 8 2 | 9 2 | 3+4 2 | 5+6 2 |

Continued on next column

**TABLE LXXIII– continued**

| TQC | ICSD | NoE | SF | |
|---|---|---|---|---|
| | 8 2 | 9 2 | 5+6 2 | 3+4 2 |
| | 4+5 2 | 6+7 2 | 5+6 2 | 3+4 2 |
| | 9 2 | 8 2 | 5+6 2 | 3+4 2 |
| | 6+7 2 | 4+5 2 | 3+4 2 | 5+6 2 |
| | 9 2 | 8 2 | 3+4 2 | 5+6 2 |
| | 8 2 | 9 2 | 3+4 2 | 5+6 2 |
| | 9 2 | 9 2 | 5+6 2 | 3+4 2 |
| | 4+5 2 | 6+7 2 | 5+6 2 | 3+4 2 |
| 166.1.201 | 95657 | 54 | F1Nd1O1 | |
| | -GM | -T | -F | -L |
| | 8 2 | 9 2 | 5+6 2 | 3+4 2 |
| | 9 2 | 8 2 | 3+4 2 | 5+6 2 |
| | 8 2 | 8 2 | 5+6 2 | 5+6 2 |
| | 9 2 | 9 2 | 3+4 2 | 3+4 2 |
| | 8 2 | 9 2 | 5+6 2 | 3+4 2 |
| | 9 2 | 8 2 | 3+4 2 | 5+6 2 |
| | 8 2 | 9 2 | 5+6 2 | 3+4 2 |
| | 9 2 | 8 2 | 3+4 2 | 5+6 2 |
| | 4+5 2 | 6+7 2 | 5+6 2 | 3+4 2 |
| | 6+7 2 | 4+5 2 | 3+4 2 | 5+6 2 |
| | 9 2 | 8 2 | 3+4 2 | 5+6 2 |
| | 8 2 | 9 2 | 5+6 2 | 3+4 2 |
| | 8 2 | 8 2 | 3+4 2 | 3+4 2 |
| | 8 2 | 8 2 | 5+6 2 | 5+6 2 |
| | 4+5 2 | 4+5 2 | 5+6 2 | 5+6 2 |
| | 9 2 | 9 2 | 5+6 2 | 5+6 2 |
| | 6+7 2 | 6+7 2 | 3+4 2 | 3+4 2 |
| | 9 2 | 9 2 | 3+4 2 | 3+4 2 |
| | 8 2 | 9 2 | 3+4 2 | 5+6 2 |
| | 8 2 | 9 2 | 5+6 2 | 3+4 2 |
| | 4+5 2 | 6+7 2 | 5+6 2 | 3+4 2 |
| | 9 2 | 8 2 | 5+6 2 | 3+4 2 |
| | 6+7 2 | 4+5 2 | 3+4 2 | 5+6 2 |
| | 9 2 | 8 2 | 3+4 2 | 5+6 2 |
| | 8 2 | 9 2 | 3+4 2 | 5+6 2 |
| | 9 2 | 9 2 | 5+6 2 | 3+4 2 |
| | 4+5 2 | 6+7 2 | 5+6 2 | 3+4 2 |
| 166.1.202 | 95656 | 54 | F1Nd1O1 | |
| | -GM | -T | -F | -L |
| | 8 2 | 9 2 | 5+6 2 | 3+4 2 |
| | 9 2 | 8 2 | 3+4 2 | 5+6 2 |
| | 8 2 | 8 2 | 5+6 2 | 5+6 2 |
| | 9 2 | 9 2 | 3+4 2 | 3+4 2 |
| | 8 2 | 9 2 | 5+6 2 | 3+4 2 |
| | 9 2 | 8 2 | 3+4 2 | 5+6 2 |
| | 8 2 | 9 2 | 5+6 2 | 3+4 2 |
| | 9 2 | 8 2 | 3+4 2 | 5+6 2 |
| | 4+5 2 | 6+7 2 | 5+6 2 | 3+4 2 |
| | 6+7 2 | 4+5 2 | 3+4 2 | 5+6 2 |
| | 9 2 | 8 2 | 3+4 2 | 5+6 2 |
| | 8 2 | 9 2 | 5+6 2 | 3+4 2 |
| | 8 2 | 8 2 | 3+4 2 | 3+4 2 |
| | 8 2 | 8 2 | 5+6 2 | 5+6 2 |
| | 4+5 2 | 4+5 2 | 5+6 2 | 5+6 2 |
| | 9 2 | 9 2 | 5+6 2 | 5+6 2 |
| | 6+7 2 | 6+7 2 | 3+4 2 | 3+4 2 |
| | 9 2 | 9 2 | 3+4 2 | 3+4 2 |
| | 8 2 | 9 2 | 3+4 2 | 5+6 2 |
| | 8 2 | 9 2 | 5+6 2 | 3+4 2 |
| | 4+5 2 | 6+7 2 | 5+6 2 | 3+4 2 |
| | 9 2 | 8 2 | 5+6 2 | 3+4 2 |

Continued on next column

**TABLE LXXIII– continued**

| TQC | ICSD | NoE | SF | |
|---|---|---|---|---|
| | 6+7 2 | 4+5 2 | 3+4 2 | 5+6 2 |
| | 9 2 | 8 2 | 3+4 2 | 5+6 2 |
| | 8 2 | 9 2 | 3+4 2 | 5+6 2 |
| | 9 2 | 9 2 | 5+6 2 | 3+4 2 |
| | 4+5 2 | 6+7 2 | 5+6 2 | 3+4 2 |
| 166.1.203 | 95655 | 54 | F1Nd1O1 | |
| | -GM | -T | -F | -L |
| | 8 2 | 9 2 | 5+6 2 | 3+4 2 |
| | 9 2 | 8 2 | 3+4 2 | 5+6 2 |
| | 8 2 | 8 2 | 5+6 2 | 5+6 2 |
| | 9 2 | 9 2 | 3+4 2 | 3+4 2 |
| | 8 2 | 9 2 | 5+6 2 | 3+4 2 |
| | 9 2 | 8 2 | 3+4 2 | 5+6 2 |
| | 8 2 | 9 2 | 5+6 2 | 3+4 2 |
| | 9 2 | 8 2 | 3+4 2 | 5+6 2 |
| | 4+5 2 | 6+7 2 | 5+6 2 | 3+4 2 |
| | 6+7 2 | 4+5 2 | 3+4 2 | 5+6 2 |
| | 9 2 | 8 2 | 3+4 2 | 5+6 2 |
| | 8 2 | 9 2 | 5+6 2 | 3+4 2 |
| | 8 2 | 8 2 | 3+4 2 | 3+4 2 |
| | 8 2 | 8 2 | 5+6 2 | 5+6 2 |
| | 4+5 2 | 4+5 2 | 5+6 2 | 5+6 2 |
| | 9 2 | 9 2 | 5+6 2 | 5+6 2 |
| | 6+7 2 | 6+7 2 | 3+4 2 | 3+4 2 |
| | 9 2 | 9 2 | 3+4 2 | 3+4 2 |
| | 8 2 | 9 2 | 3+4 2 | 5+6 2 |
| | 8 2 | 9 2 | 5+6 2 | 3+4 2 |
| | 4+5 2 | 6+7 2 | 5+6 2 | 3+4 2 |
| | 9 2 | 8 2 | 5+6 2 | 3+4 2 |
| | 6+7 2 | 4+5 2 | 3+4 2 | 5+6 2 |
| | 9 2 | 8 2 | 3+4 2 | 5+6 2 |
| | 8 2 | 9 2 | 3+4 2 | 5+6 2 |
| | 9 2 | 9 2 | 5+6 2 | 3+4 2 |
| | 4+5 2 | 6+7 2 | 5+6 2 | 3+4 2 |
| 166.1.204 | 184005 | 54 | F1Nd1O1 | |
| | -GM | -T | -F | -L |
| | 8 2 | 8 2 | 5+6 2 | 5+6 2 |
| | 9 2 | 9 2 | 3+4 2 | 3+4 2 |
| | 8 2 | 9 2 | 5+6 2 | 3+4 2 |
| | 9 2 | 8 2 | 3+4 2 | 5+6 2 |
| | 8 2 | 8 2 | 5+6 2 | 5+6 2 |
| | 9 2 | 9 2 | 3+4 2 | 3+4 2 |
| | 8 2 | 8 2 | 5+6 2 | 5+6 2 |
| | 9 2 | 9 2 | 3+4 2 | 3+4 2 |
| | 4+5 2 | 4+5 2 | 5+6 2 | 5+6 2 |
| | 6+7 2 | 6+7 2 | 3+4 2 | 3+4 2 |
| | 9 2 | 9 2 | 3+4 2 | 3+4 2 |
| | 8 2 | 8 2 | 5+6 2 | 5+6 2 |
| | 8 2 | 9 2 | 3+4 2 | 5+6 2 |
| | 8 2 | 9 2 | 5+6 2 | 3+4 2 |
| | 4+5 2 | 6+7 2 | 5+6 2 | 3+4 2 |
| | 9 2 | 8 2 | 5+6 2 | 3+4 2 |
| | 6+7 2 | 4+5 2 | 3+4 2 | 5+6 2 |
| | 9 2 | 8 2 | 3+4 2 | 5+6 2 |
| | 8 2 | 8 2 | 3+4 2 | 3+4 2 |
| | 8 2 | 8 2 | 5+6 2 | 5+6 2 |
| | 4+5 2 | 4+5 2 | 5+6 2 | 5+6 2 |
| | 9 2 | 9 2 | 5+6 2 | 5+6 2 |
| | 6+7 2 | 6+7 2 | 3+4 2 | 3+4 2 |
| | 9 2 | 9 2 | 3+4 2 | 3+4 2 |
| | 8 2 | 8 2 | 3+4 2 | 3+4 2 |

Continued on next column

**TABLE LXXIII– continued**

| TQC | ICSD | | NoE | SF |
|---|---|---|---|---|
| | 9 2 | 8 2 | 5+6 2 | 5+6 2 |
| | 4+5 2 | 4+5 2 | 5+6 2 | 5+6 2 |
| 166.1.205 | 184011 | | 68 | F1Ho1O1 |
| | -GM | -T | -F | -L |
| | 8 2 | 8 2 | 5+6 2 | 5+6 2 |
| | 9 2 | 9 2 | 3+4 2 | 3+4 2 |
| | 8 2 | 9 2 | 5+6 2 | 3+4 2 |
| | 9 2 | 8 2 | 3+4 2 | 5+6 2 |
| | 8 2 | 9 2 | 5+6 2 | 3+4 2 |
| | 9 2 | 8 2 | 3+4 2 | 5+6 2 |
| | 8 2 | 8 2 | 5+6 2 | 5+6 2 |
| | 4+5 2 | 4+5 2 | 3+4 2 | 3+4 2 |
| | 6+7 2 | 6+7 2 | 5+6 2 | 5+6 2 |
| | 9 2 | 9 2 | 3+4 2 | 3+4 2 |
| | 9 2 | 9 2 | 3+4 2 | 3+4 2 |
| | 8 2 | 8 2 | 5+6 2 | 5+6 2 |
| | 8 2 | 9 2 | 3+4 2 | 5+6 2 |
| | 8 2 | 9 2 | 5+6 2 | 3+4 2 |
| | 4+5 2 | 6+7 2 | 5+6 2 | 3+4 2 |
| | 9 2 | 8 2 | 5+6 2 | 3+4 2 |
| | 6+7 2 | 4+5 2 | 3+4 2 | 5+6 2 |
| | 9 2 | 8 2 | 3+4 2 | 5+6 2 |
| | 8 2 | 8 2 | 3+4 2 | 3+4 2 |
| | 8 2 | 8 2 | 5+6 2 | 5+6 2 |
| | 4+5 2 | 4+5 2 | 5+6 2 | 5+6 2 |
| | 9 2 | 9 2 | 5+6 2 | 5+6 2 |
| | 6+7 2 | 6+7 2 | 3+4 2 | 3+4 2 |
| | 9 2 | 9 2 | 3+4 2 | 3+4 2 |
| | 8 2 | 8 2 | 5+6 2 | 5+6 2 |
| | 9 2 | 9 2 | 3+4 2 | 3+4 2 |
| | 8 2 | 8 2 | 5+6 2 | 5+6 2 |
| | 4+5 2 | 4+5 2 | 3+4 2 | 3+4 2 |
| | 6+7 2 | 6+7 2 | 5+6 2 | 5+6 2 |
| | 9 2 | 9 2 | 3+4 2 | 3+4 2 |
| | 8 2 | 8 2 | 3+4 2 | 3+4 2 |
| | 9 2 | 4+5 2 | 5+6 2 | 5+6 2 |
| | 4+5 2 | 8 2 | 5+6 2 | 5+6 2 |
| | 8 2 | 9 2 | 3+4 2 | 3+4 2 |
| 166.1.206 | 202068 | | 46 | Eu2I1P1 |
| | -GM | -T | -F | -L |
| | 8 2 | 9 2 | 5+6 2 | 3+4 2 |
| | 9 2 | 8 2 | 3+4 2 | 5+6 2 |
| | 8 2 | 9 2 | 5+6 2 | 3+4 2 |
| | 9 2 | 8 2 | 3+4 2 | 5+6 2 |
| | 8 2 | 9 2 | 3+4 2 | 5+6 2 |
| | 4+5 2 | 8 2 | 5+6 2 | 3+4 2 |
| | 9 2 | 6+7 2 | 5+6 2 | 3+4 2 |
| | 6+7 2 | 4+5 2 | 3+4 2 | 5+6 2 |
| | 8 2 | 8 2 | 3+4 2 | 3+4 2 |
| | 8 2 | 9 2 | 3+4 2 | 5+6 2 |
| | 9 2 | 9 2 | 5+6 2 | 5+6 2 |
| | 9 2 | 9 2 | 5+6 2 | 5+6 2 |
| | 6+7 2 | 6+7 2 | 5+6 2 | 5+6 2 |
| | 9 2 | 8 2 | 5+6 2 | 3+4 2 |
| | 9 2 | 8 2 | 5+6 2 | 3+4 2 |
| | 6+7 2 | 4+5 2 | 5+6 2 | 3+4 2 |
| | 8 2 | 8 2 | 3+4 2 | 3+4 2 |
| | 4+5 2 | 9 2 | 3+4 2 | 5+6 2 |
| | 8 2 | 8 2 | 3+4 2 | 5+6 2 |
| | 9 2 | 6+7 2 | 5+6 2 | 3+4 2 |
| | 9 2 | 4+5 2 | 3+4 2 | 5+6 2 |

Continued on next column

**TABLE LXXIII– continued**

| TQC | ICSD | | NoE | SF |
|---|---|---|---|---|
| | 6+7 2 | 9 2 | 5+6 2 | 3+4 2 |
| | 8 2 | 8 2 | 5+6 2 | 5+6 2 |
| 166.1.207 | 67700 | | 62 | Er1Fe2O4 |
| | -GM | -T | -F | -L |
| | 8 2 | 8 2 | 3+4 2 | 3+4 2 |
| | 9 2 | 9 2 | 5+6 2 | 5+6 2 |
| | 9 2 | 9 2 | 5+6 2 | 5+6 2 |
| | 6+7 2 | 6+7 2 | 5+6 2 | 5+6 2 |
| | 8 2 | 9 2 | 5+6 2 | 3+4 2 |
| | 9 2 | 8 2 | 3+4 2 | 5+6 2 |
| | 8 2 | 8 2 | 3+4 2 | 3+4 2 |
| | 9 2 | 9 2 | 5+6 2 | 5+6 2 |
| | 8 2 | 8 2 | 3+4 2 | 5+6 2 |
| | 8 2 | 9 2 | 5+6 2 | 3+4 2 |
| | 4+5 2 | 6+7 2 | 3+4 2 | 3+4 2 |
| | 8 2 | 9 2 | 5+6 2 | 3+4 2 |
| | 6+7 2 | 4+5 2 | 3+4 2 | 5+6 2 |
| | 9 2 | 8 2 | 5+6 2 | 3+4 2 |
| | 9 2 | 4+5 2 | 5+6 2 | 5+6 2 |
| | 4+5 2 | 8 2 | 3+4 2 | 3+4 2 |
| | 8 2 | 9 2 | 3+4 2 | 3+4 2 |
| | 9 2 | 9 2 | 5+6 2 | 5+6 2 |
| | 6+7 2 | 6+7 2 | 5+6 2 | 5+6 2 |
| | 9 2 | 8 2 | 3+4 2 | 5+6 2 |
| | 9 2 | 9 2 | 5+6 2 | 5+6 2 |
| | 9 2 | 9 2 | 5+6 2 | 5+6 2 |
| | 6+7 2 | 6+7 2 | 5+6 2 | 5+6 2 |
| | 8 2 | 9 2 | 3+4 2 | 5+6 2 |
| | 4+5 2 | 8 2 | 5+6 2 | 3+4 2 |
| | 9 2 | 6+7 2 | 5+6 2 | 3+4 2 |
| | 6+7 2 | 4+5 2 | 3+4 2 | 5+6 2 |
| | 9 2 | 9 2 | 5+6 2 | 5+6 2 |
| | 9 2 | 9 2 | 3+4 2 | 5+6 2 |
| | 6+7 2 | 6+7 2 | 5+6 2 | 3+4 2 |
| | 9 2 | 8 2 | 5+6 2 | 5+6 2 |
| 166.1.208 | 184012 | | 70 | Er1F1O1 |
| | -GM | -T | -F | -L |
| | 8 2 | 8 2 | 5+6 2 | 5+6 2 |
| | 9 2 | 9 2 | 3+4 2 | 3+4 2 |
| | 8 2 | 9 2 | 5+6 2 | 3+4 2 |
| | 9 2 | 8 2 | 3+4 2 | 5+6 2 |
| | 8 2 | 9 2 | 5+6 2 | 3+4 2 |
| | 9 2 | 8 2 | 3+4 2 | 5+6 2 |
| | 8 2 | 8 2 | 5+6 2 | 5+6 2 |
| | 4+5 2 | 4+5 2 | 3+4 2 | 3+4 2 |
| | 6+7 2 | 6+7 2 | 5+6 2 | 5+6 2 |
| | 9 2 | 9 2 | 3+4 2 | 3+4 2 |
| | 9 2 | 9 2 | 3+4 2 | 5+6 2 |
| | 8 2 | 8 2 | 5+6 2 | 3+4 2 |
| | 8 2 | 9 2 | 3+4 2 | 5+6 2 |
| | 8 2 | 6+7 2 | 5+6 2 | 3+4 2 |
| | 4+5 2 | 9 2 | 5+6 2 | 3+4 2 |
| | 9 2 | 8 2 | 5+6 2 | 3+4 2 |
| | 6+7 2 | 4+5 2 | 3+4 2 | 5+6 2 |
| | 9 2 | 8 2 | 3+4 2 | 5+6 2 |
| | 8 2 | 8 2 | 3+4 2 | 3+4 2 |
| | 8 2 | 8 2 | 5+6 2 | 5+6 2 |
| | 4+5 2 | 4+5 2 | 5+6 2 | 5+6 2 |
| | 9 2 | 9 2 | 5+6 2 | 5+6 2 |
| | 6+7 2 | 6+7 2 | 3+4 2 | 3+4 2 |
| | 9 2 | 9 2 | 3+4 2 | 3+4 2 |

Continued on next column

**TABLE LXXIII– continued**

| TQC | ICSD | | NoE | SF |
|---|---|---|---|---|
| | 8 2 | 8 2 | 5+6 2 | 5+6 2 |
| | 9 2 | 9 2 | 3+4 2 | 3+4 2 |
| | 8 2 | 8 2 | 5+6 2 | 5+6 2 |
| | 4+5 2 | 4+5 2 | 3+4 2 | 3+4 2 |
| | 6+7 2 | 6+7 2 | 5+6 2 | 5+6 2 |
| | 9 2 | 9 2 | 3+4 2 | 3+4 2 |
| | 8 2 | 8 2 | 3+4 2 | 3+4 2 |
| | 9 2 | 4+5 2 | 5+6 2 | 5+6 2 |
| | 4+5 2 | 9 2 | 5+6 2 | 5+6 2 |
| | 8 2 | 8 2 | 3+4 2 | 3+4 2 |
| | 6+7 2 | 6+7 2 | 5+6 2 | 5+6 2 |
| 166.1.209 | 630377 | | 244 | Dy2Zn17 |
| | -GM | -T | -F | -L |
| | 8 2 | 8 2 | 3+4 2 | 3+4 2 |
| | 9 2 | 9 2 | 5+6 2 | 5+6 2 |
| | 8 2 | 9 2 | 3+4 2 | 5+6 2 |
| | 9 2 | 8 2 | 5+6 2 | 3+4 2 |
| | 8 2 | 4+5 2 | 3+4 2 | 5+6 2 |
| | 4+5 2 | 9 2 | 5+6 2 | 3+4 2 |
| | 6+7 2 | 6+7 2 | 3+4 2 | 5+6 2 |
| | 9 2 | 8 2 | 5+6 2 | 3+4 2 |
| | 8 2 | 9 2 | 3+4 2 | 3+4 2 |
| | 8 2 | 9 2 | 5+6 2 | 5+6 2 |
| | 8 2 | 6+7 2 | 5+6 2 | 3+4 2 |
| | 4+5 2 | 9 2 | 3+4 2 | 3+4 2 |
| | 8 2 | 8 2 | 5+6 2 | 3+4 2 |
| | 9 2 | 4+5 2 | 3+4 2 | 3+4 2 |
| | 6+7 2 | 9 2 | 5+6 2 | 5+6 2 |
| | 4+5 2 | 9 2 | 5+6 2 | 5+6 2 |
| | 8 2 | 8 2 | 3+4 2 | 3+4 2 |
| | 8 2 | 6+7 2 | 5+6 2 | 5+6 2 |
| | 4+5 2 | 9 2 | 5+6 2 | 5+6 2 |
| | 8 2 | 6+7 2 | 3+4 2 | 3+4 2 |
| | 9 2 | 9 2 | 5+6 2 | 5+6 2 |
| | 9 2 | 8 2 | 3+4 2 | 3+4 2 |
| | 8 2 | 9 2 | 3+4 2 | 5+6 2 |
| | 4+5 2 | 6+7 2 | 5+6 2 | 3+4 2 |
| | 6+7 2 | 8 2 | 5+6 2 | 3+4 2 |
| | 8 2 | 4+5 2 | 3+4 2 | 5+6 2 |
| | 8 2 | 9 2 | 5+6 2 | 3+4 2 |
| | 8 2 | 6+7 2 | 5+6 2 | 3+4 2 |
| | 9 2 | 9 2 | 3+4 2 | 5+6 2 |
| | 4+5 2 | 8 2 | 5+6 2 | 3+4 2 |
| | 9 2 | 6+7 2 | 3+4 2 | 5+6 2 |
| | 8 2 | 8 2 | 5+6 2 | 3+4 2 |
| | 4+5 2 | 4+5 2 | 5+6 2 | 3+4 2 |
| | 9 2 | 9 2 | 3+4 2 | 5+6 2 |
| | 6+7 2 | 8 2 | 3+4 2 | 5+6 2 |
| | 9 2 | 9 2 | 3+4 2 | 5+6 2 |
| | 8 2 | 8 2 | 5+6 2 | 3+4 2 |
| | 8 2 | 6+7 2 | 5+6 2 | 5+6 2 |
| | 4+5 2 | 9 2 | 3+4 2 | 5+6 2 |
| | 6+7 2 | 4+5 2 | 5+6 2 | 3+4 2 |
| | 9 2 | 6+7 2 | 3+4 2 | 3+4 2 |
| | 6+7 2 | 9 2 | 3+4 2 | 5+6 2 |
| | 4+5 2 | 8 2 | 5+6 2 | 5+6 2 |
| | 9 2 | 8 2 | 3+4 2 | 3+4 2 |
| | 6+7 2 | 4+5 2 | 3+4 2 | 3+4 2 |
| | 8 2 | 9 2 | 5+6 2 | 5+6 2 |
| | 8 2 | 8 2 | 5+6 2 | 5+6 2 |
| | 9 2 | 4+5 2 | 3+4 2 | 3+4 2 |

Continued on next column

**TABLE LXXIII– continued**

| TQC | ICSD | | NoE | SF |
|---|---|---|---|---|
| | 6+7 2 | 6+7 2 | 5+6 2 | 3+4 2 |
| | 8 2 | 9 2 | 3+4 2 | 5+6 2 |
| | 9 2 | 9 2 | 5+6 2 | 3+4 2 |
| | 8 2 | 8 2 | 3+4 2 | 3+4 2 |
| | 9 2 | 9 2 | 5+6 2 | 5+6 2 |
| | 8 2 | 6+7 2 | 3+4 2 | 5+6 2 |
| | 4+5 2 | 4+5 2 | 5+6 2 | 3+4 2 |
| | 8 2 | 8 2 | 3+4 2 | 5+6 2 |
| | 9 2 | 8 2 | 5+6 2 | 3+4 2 |
| | 4+5 2 | 9 2 | 3+4 2 | 5+6 2 |
| | 6+7 2 | 6+7 2 | 5+6 2 | 3+4 2 |
| | 8 2 | 8 2 | 5+6 2 | 3+4 2 |
| | 9 2 | 9 2 | 3+4 2 | 5+6 2 |
| | 6+7 2 | 8 2 | 5+6 2 | 5+6 2 |
| | 8 2 | 4+5 2 | 3+4 2 | 3+4 2 |
| | 8 2 | 9 2 | 3+4 2 | 5+6 2 |
| | 9 2 | 6+7 2 | 5+6 2 | 3+4 2 |
| | 4+5 2 | 9 2 | 3+4 2 | 5+6 2 |
| | 8 2 | 8 2 | 3+4 2 | 3+4 2 |
| | 9 2 | 9 2 | 5+6 2 | 3+4 2 |
| | 6+7 2 | 4+5 2 | 5+6 2 | 5+6 2 |
| | 4+5 2 | 8 2 | 3+4 2 | 3+4 2 |
| | 8 2 | 9 2 | 5+6 2 | 5+6 2 |
| | 9 2 | 6+7 2 | 3+4 2 | 3+4 2 |
| | 9 2 | 4+5 2 | 5+6 2 | 5+6 2 |
| | 4+5 2 | 9 2 | 5+6 2 | 3+4 2 |
| | 6+7 2 | 8 2 | 3+4 2 | 5+6 2 |
| | 8 2 | 8 2 | 5+6 2 | 3+4 2 |
| | 9 2 | 9 2 | 3+4 2 | 5+6 2 |
| | 8 2 | 6+7 2 | 5+6 2 | 5+6 2 |
| | 4+5 2 | 8 2 | 3+4 2 | 3+4 2 |
| | 9 2 | 4+5 2 | 3+4 2 | 3+4 2 |
| | 8 2 | 9 2 | 5+6 2 | 5+6 2 |
| | 8 2 | 8 2 | 3+4 2 | 3+4 2 |
| | 9 2 | 6+7 2 | 5+6 2 | 3+4 2 |
| | 9 2 | 4+5 2 | 3+4 2 | 5+6 2 |
| | 8 2 | 8 2 | 5+6 2 | 3+4 2 |
| | 6+7 2 | 9 2 | 3+4 2 | 5+6 2 |
| | 8 2 | 9 2 | 5+6 2 | 5+6 2 |
| | 9 2 | 9 2 | 5+6 2 | 3+4 2 |
| | 4+5 2 | 6+7 2 | 3+4 2 | 5+6 2 |
| | 8 2 | 8 2 | 5+6 2 | 3+4 2 |
| | 4+5 2 | 9 2 | 5+6 2 | 3+4 2 |
| | 8 2 | 9 2 | 3+4 2 | 5+6 2 |
| | 4+5 2 | 9 2 | 3+4 2 | 3+4 2 |
| | 8 2 | 6+7 2 | 5+6 2 | 5+6 2 |
| | 9 2 | 9 2 | 5+6 2 | 3+4 2 |
| | 6+7 2 | 9 2 | 5+6 2 | 3+4 2 |
| | 8 2 | 6+7 2 | 3+4 2 | 5+6 2 |
| | 4+5 2 | 8 2 | 5+6 2 | 5+6 2 |
| | 8 2 | 4+5 2 | 3+4 2 | 3+4 2 |
| | 9 2 | 8 2 | 5+6 2 | 3+4 2 |
| | 6+7 2 | 9 2 | 3+4 2 | 5+6 2 |
| | 9 2 | 8 2 | 3+4 2 | 5+6 2 |
| | 8 2 | 8 2 | 5+6 2 | 3+4 2 |
| | 4+5 2 | 4+5 2 | 5+6 2 | 5+6 2 |
| | 9 2 | 6+7 2 | 3+4 2 | 3+4 2 |
| | 9 2 | 9 2 | 3+4 2 | 5+6 2 |
| | 8 2 | 9 2 | 5+6 2 | 3+4 2 |
| | 4+5 2 | 8 2 | 5+6 2 | 5+6 2 |
| | 8 2 | 4+5 2 | 3+4 2 | 3+4 2 |

Continued on next column

**TABLE LXXIII– continued**

| TQC | ICSD | NoE | SF | |
|---|---|---|---|---|
| | 9 2 | 6+7 2 | 5+6 2 | 5+6 2 |
| | 9 2 | 9 2 | 3+4 2 | 3+4 2 |
| | 8 2 | 8 2 | 5+6 2 | 3+4 2 |
| | 6+7 2 | 8 2 | 3+4 2 | 5+6 2 |
| | 8 2 | 4+5 2 | 5+6 2 | 5+6 2 |
| | 4+5 2 | 4+5 2 | 5+6 2 | 3+4 2 |
| | 8 2 | 8 2 | 3+4 2 | 5+6 2 |
| | 6+7 2 | 8 2 | 3+4 2 | 5+6 2 |
| | 9 2 | 9 2 | 3+4 2 | 3+4 2 |
| | 9 2 | 9 2 | 3+4 2 | 3+4 2 |
| | 8 2 | 6+7 2 | 5+6 2 | 3+4 2 |
| | 8 2 | 8 2 | 5+6 2 | 5+6 2 |
| | 4+5 2 | 9 2 | 3+4 2 | 3+4 2 |
| 166.1.210 | 629979 | 220 | Dy2Ni7 | |
| | -GM | -T | -F | -L |
| | 8 2 | 9 2 | 3+4 2 | 3+4 2 |
| | 8 2 | 8 2 | 5+6 2 | 5+6 2 |
| | 9 2 | 9 2 | 5+6 2 | 3+4 2 |
| | 9 2 | 8 2 | 3+4 2 | 5+6 2 |
| | 8 2 | 9 2 | 5+6 2 | 3+4 2 |
| | 8 2 | 8 2 | 3+4 2 | 5+6 2 |
| | 9 2 | 9 2 | 5+6 2 | 3+4 2 |
| | 9 2 | 8 2 | 3+4 2 | 5+6 2 |
| | 4+5 2 | 6+7 2 | 5+6 2 | 3+4 2 |
| | 8 2 | 9 2 | 3+4 2 | 5+6 2 |
| | 9 2 | 8 2 | 3+4 2 | 3+4 2 |
| | 8 2 | 6+7 2 | 5+6 2 | 5+6 2 |
| | 6+7 2 | 4+5 2 | 5+6 2 | 3+4 2 |
| | 4+5 2 | 9 2 | 3+4 2 | 5+6 2 |
| | 9 2 | 8 2 | 5+6 2 | 3+4 2 |
| | 6+7 2 | 4+5 2 | 3+4 2 | 5+6 2 |
| | 8 2 | 8 2 | 5+6 2 | 5+6 2 |
| | 9 2 | 9 2 | 3+4 2 | 3+4 2 |
| | 8 2 | 8 2 | 3+4 2 | 3+4 2 |
| | 9 2 | 9 2 | 5+6 2 | 5+6 2 |
| | 8 2 | 8 2 | 5+6 2 | 5+6 2 |
| | 8 2 | 9 2 | 3+4 2 | 3+4 2 |
| | 8 2 | 8 2 | 5+6 2 | 5+6 2 |
| | 8 2 | 9 2 | 3+4 2 | 5+6 2 |
| | 4+5 2 | 9 2 | 5+6 2 | 5+6 2 |
| | 9 2 | 6+7 2 | 3+4 2 | 3+4 2 |
| | 9 2 | 8 2 | 3+4 2 | 5+6 2 |
| | 6+7 2 | 4+5 2 | 5+6 2 | 3+4 2 |
| | 8 2 | 8 2 | 5+6 2 | 5+6 2 |
| | 8 2 | 4+5 2 | 3+4 2 | 3+4 2 |
| | 4+5 2 | 8 2 | 3+4 2 | 5+6 2 |
| | 9 2 | 8 2 | 3+4 2 | 3+4 2 |
| | 8 2 | 4+5 2 | 5+6 2 | 5+6 2 |
| | 4+5 2 | 8 2 | 3+4 2 | 3+4 2 |
| | 8 2 | 8 2 | 5+6 2 | 3+4 2 |
| | 4+5 2 | 9 2 | 5+6 2 | 5+6 2 |
| | 6+7 2 | 9 2 | 3+4 2 | 5+6 2 |
| | 9 2 | 6+7 2 | 5+6 2 | 5+6 2 |
| | 8 2 | 4+5 2 | 5+6 2 | 5+6 2 |
| | 8 2 | 6+7 2 | 3+4 2 | 5+6 2 |
| | 8 2 | 8 2 | 3+4 2 | 3+4 2 |
| | 4+5 2 | 9 2 | 3+4 2 | 3+4 2 |
| | 6+7 2 | 8 2 | 5+6 2 | 5+6 2 |
| | 9 2 | 4+5 2 | 3+4 2 | 3+4 2 |
| | 4+5 2 | 9 2 | 5+6 2 | 5+6 2 |
| | 8 2 | 8 2 | 5+6 2 | 5+6 2 |

Continued on next column

**TABLE LXXIII– continued**

| TQC | ICSD | NoE | SF | |
|---|---|---|---|---|
| | 9 2 | 4+5 2 | 3+4 2 | 3+4 2 |
| | 8 2 | 8 2 | 3+4 2 | 5+6 2 |
| | 4+5 2 | 9 2 | 3+4 2 | 5+6 2 |
| | 9 2 | 6+7 2 | 3+4 2 | 3+4 2 |
| | 6+7 2 | 6+7 2 | 5+6 2 | 5+6 2 |
| | 9 2 | 9 2 | 5+6 2 | 3+4 2 |
| | 8 2 | 9 2 | 5+6 2 | 3+4 2 |
| | 8 2 | 6+7 2 | 5+6 2 | 5+6 2 |
| | 9 2 | 6+7 2 | 3+4 2 | 3+4 2 |
| | 4+5 2 | 9 2 | 5+6 2 | 5+6 2 |
| | 6+7 2 | 9 2 | 3+4 2 | 3+4 2 |
| | 8 2 | 8 2 | 3+4 2 | 5+6 2 |
| | 9 2 | 4+5 2 | 5+6 2 | 5+6 2 |
| | 6+7 2 | 8 2 | 5+6 2 | 3+4 2 |
| | 9 2 | 4+5 2 | 3+4 2 | 3+4 2 |
| | 8 2 | 9 2 | 5+6 2 | 3+4 2 |
| | 9 2 | 8 2 | 5+6 2 | 5+6 2 |
| | 4+5 2 | 8 2 | 3+4 2 | 3+4 2 |
| | 9 2 | 6+7 2 | 5+6 2 | 5+6 2 |
| | 6+7 2 | 4+5 2 | 3+4 2 | 5+6 2 |
| | 8 2 | 9 2 | 3+4 2 | 3+4 2 |
| | 6+7 2 | 8 2 | 3+4 2 | 3+4 2 |
| | 4+5 2 | 9 2 | 5+6 2 | 5+6 2 |
| | 8 2 | 8 2 | 3+4 2 | 3+4 2 |
| | 8 2 | 4+5 2 | 5+6 2 | 5+6 2 |
| | 4+5 2 | 9 2 | 5+6 2 | 3+4 2 |
| | 8 2 | 9 2 | 3+4 2 | 5+6 2 |
| | 8 2 | 8 2 | 5+6 2 | 5+6 2 |
| | 9 2 | 9 2 | 5+6 2 | 3+4 2 |
| | 8 2 | 8 2 | 3+4 2 | 3+4 2 |
| | 9 2 | 6+7 2 | 5+6 2 | 5+6 2 |
| | 4+5 2 | 8 2 | 5+6 2 | 3+4 2 |
| | 9 2 | 4+5 2 | 3+4 2 | 5+6 2 |
| | 9 2 | 8 2 | 3+4 2 | 3+4 2 |
| | 6+7 2 | 6+7 2 | 5+6 2 | 5+6 2 |
| | 8 2 | 9 2 | 3+4 2 | 3+4 2 |
| | 8 2 | 6+7 2 | 3+4 2 | 5+6 2 |
| | 4+5 2 | 8 2 | 5+6 2 | 3+4 2 |
| | 8 2 | 9 2 | 3+4 2 | 5+6 2 |
| | 9 2 | 4+5 2 | 3+4 2 | 3+4 2 |
| | 4+5 2 | 4+5 2 | 5+6 2 | 3+4 2 |
| | 6+7 2 | 8 2 | 5+6 2 | 5+6 2 |
| | 8 2 | 9 2 | 3+4 2 | 5+6 2 |
| | 9 2 | 6+7 2 | 5+6 2 | 3+4 2 |
| | 6+7 2 | 8 2 | 3+4 2 | 5+6 2 |
| | 4+5 2 | 8 2 | 3+4 2 | 5+6 2 |
| | 8 2 | 4+5 2 | 5+6 2 | 5+6 2 |
| | 4+5 2 | 8 2 | 3+4 2 | 5+6 2 |
| | 8 2 | 4+5 2 | 5+6 2 | 3+4 2 |
| | 8 2 | 8 2 | 5+6 2 | 3+4 2 |
| | 8 2 | 6+7 2 | 3+4 2 | 5+6 2 |
| | 4+5 2 | 9 2 | 5+6 2 | 3+4 2 |
| | 9 2 | 9 2 | 5+6 2 | 5+6 2 |
| | 9 2 | 8 2 | 3+4 2 | 5+6 2 |
| | 8 2 | 4+5 2 | 3+4 2 | 3+4 2 |
| | 6+7 2 | 9 2 | 5+6 2 | 5+6 2 |
| | 9 2 | 8 2 | 5+6 2 | 3+4 2 |
| | 6+7 2 | 6+7 2 | 3+4 2 | 3+4 2 |
| | 8 2 | 9 2 | 5+6 2 | 5+6 2 |
| | 9 2 | 9 2 | 3+4 2 | 3+4 2 |
| | 8 2 | 4+5 2 | 5+6 2 | 5+6 2 |

Continued on next column

**TABLE LXXIII– continued**

| TQC | ICSD | NoE | SF |
|---|---|---|---|
| | 4+5 2 8 2 3+4 2 3+4 2 | | |
| | 9 2 8 2 5+6 2 3+4 2 | | |
| | 8 2 9 2 3+4 2 5+6 2 | | |
| 166.1.211 | 262377 | 220 | Dy2Ni7 |
| | -GM -T -F -L | | |
| | 8 2 9 2 3+4 2 3+4 2 | | |
| | 8 2 8 2 5+6 2 3+4 2 | | |
| | 9 2 9 2 5+6 2 5+6 2 | | |
| | 9 2 8 2 3+4 2 5+6 2 | | |
| | 8 2 9 2 5+6 2 3+4 2 | | |
| | 8 2 8 2 3+4 2 3+4 2 | | |
| | 9 2 9 2 5+6 2 5+6 2 | | |
| | 9 2 8 2 3+4 2 5+6 2 | | |
| | 4+5 2 6+7 2 5+6 2 3+4 2 | | |
| | 8 2 9 2 3+4 2 5+6 2 | | |
| | 8 2 8 2 3+4 2 3+4 2 | | |
| | 9 2 6+7 2 5+6 2 5+6 2 | | |
| | 6+7 2 4+5 2 3+4 2 3+4 2 | | |
| | 4+5 2 9 2 5+6 2 5+6 2 | | |
| | 9 2 8 2 5+6 2 3+4 2 | | |
| | 6+7 2 4+5 2 3+4 2 5+6 2 | | |
| | 8 2 8 2 5+6 2 5+6 2 | | |
| | 9 2 9 2 3+4 2 3+4 2 | | |
| | 8 2 8 2 3+4 2 3+4 2 | | |
| | 9 2 9 2 5+6 2 5+6 2 | | |
| | 8 2 8 2 5+6 2 5+6 2 | | |
| | 8 2 8 2 3+4 2 3+4 2 | | |
| | 8 2 9 2 5+6 2 5+6 2 | | |
| | 8 2 9 2 5+6 2 5+6 2 | | |
| | 4+5 2 9 2 3+4 2 5+6 2 | | |
| | 9 2 8 2 3+4 2 3+4 2 | | |
| | 9 2 6+7 2 3+4 2 5+6 2 | | |
| | 8 2 4+5 2 5+6 2 3+4 2 | | |
| | 6+7 2 8 2 5+6 2 5+6 2 | | |
| | 4+5 2 4+5 2 3+4 2 3+4 2 | | |
| | 8 2 8 2 3+4 2 3+4 2 | | |
| | 9 2 8 2 5+6 2 5+6 2 | | |
| | 8 2 4+5 2 3+4 2 3+4 2 | | |
| | 4+5 2 8 2 3+4 2 5+6 2 | | |
| | 8 2 8 2 5+6 2 3+4 2 | | |
| | 4+5 2 9 2 5+6 2 5+6 2 | | |
| | 6+7 2 9 2 5+6 2 5+6 2 | | |
| | 9 2 6+7 2 5+6 2 5+6 2 | | |
| | 8 2 4+5 2 3+4 2 5+6 2 | | |
| | 8 2 8 2 3+4 2 5+6 2 | | |
| | 8 2 6+7 2 5+6 2 3+4 2 | | |
| | 4+5 2 9 2 3+4 2 3+4 2 | | |
| | 6+7 2 8 2 5+6 2 5+6 2 | | |
| | 9 2 4+5 2 3+4 2 3+4 2 | | |
| | 8 2 8 2 3+4 2 5+6 2 | | |
| | 4+5 2 4+5 2 5+6 2 3+4 2 | | |
| | 9 2 9 2 3+4 2 5+6 2 | | |
| | 8 2 8 2 3+4 2 3+4 2 | | |
| | 4+5 2 9 2 3+4 2 5+6 2 | | |
| | 9 2 6+7 2 3+4 2 5+6 2 | | |
| | 6+7 2 6+7 2 5+6 2 3+4 2 | | |
| | 9 2 9 2 5+6 2 5+6 2 | | |
| | 8 2 9 2 5+6 2 3+4 2 | | |
| | 9 2 6+7 2 3+4 2 5+6 2 | | |
| | 6+7 2 6+7 2 5+6 2 3+4 2 | | |
| | 8 2 9 2 5+6 2 5+6 2 | | |

Continued on next column

**TABLE LXXIII– continued**

| TQC | ICSD | NoE | SF |
|---|---|---|---|
| | 4+5 2 8 2 5+6 2 3+4 2 | | |
| | 8 2 9 2 3+4 2 5+6 2 | | |
| | 9 2 4+5 2 3+4 2 5+6 2 | | |
| | 6+7 2 8 2 5+6 2 3+4 2 | | |
| | 9 2 4+5 2 5+6 2 3+4 2 | | |
| | 8 2 9 2 3+4 2 5+6 2 | | |
| | 4+5 2 8 2 5+6 2 3+4 2 | | |
| | 9 2 8 2 3+4 2 5+6 2 | | |
| | 8 2 6+7 2 5+6 2 5+6 2 | | |
| | 9 2 4+5 2 3+4 2 3+4 2 | | |
| | 6+7 2 8 2 3+4 2 3+4 2 | | |
| | 6+7 2 9 2 3+4 2 3+4 2 | | |
| | 4+5 2 4+5 2 5+6 2 5+6 2 | | |
| | 8 2 8 2 3+4 2 3+4 2 | | |
| | 4+5 2 9 2 5+6 2 5+6 2 | | |
| | 8 2 9 2 5+6 2 3+4 2 | | |
| | 8 2 8 2 3+4 2 5+6 2 | | |
| | 9 2 4+5 2 5+6 2 5+6 2 | | |
| | 8 2 8 2 3+4 2 3+4 2 | | |
| | 4+5 2 6+7 2 5+6 2 3+4 2 | | |
| | 8 2 9 2 5+6 2 5+6 2 | | |
| | 9 2 8 2 5+6 2 5+6 2 | | |
| | 9 2 9 2 3+4 2 3+4 2 | | |
| | 6+7 2 8 2 3+4 2 3+4 2 | | |
| | 9 2 6+7 2 5+6 2 5+6 2 | | |
| | 8 2 8 2 3+4 2 5+6 2 | | |
| | 8 2 9 2 3+4 2 3+4 2 | | |
| | 4+5 2 6+7 2 5+6 2 3+4 2 | | |
| | 8 2 4+5 2 3+4 2 5+6 2 | | |
| | 8 2 9 2 5+6 2 3+4 2 | | |
| | 4+5 2 8 2 3+4 2 5+6 2 | | |
| | 9 2 4+5 2 5+6 2 3+4 2 | | |
| | 6+7 2 9 2 3+4 2 5+6 2 | | |
| | 9 2 6+7 2 5+6 2 5+6 2 | | |
| | 6+7 2 8 2 3+4 2 3+4 2 | | |
| | 4+5 2 8 2 5+6 2 5+6 2 | | |
| | 8 2 4+5 2 3+4 2 5+6 2 | | |
| | 8 2 8 2 5+6 2 5+6 2 | | |
| | 4+5 2 4+5 2 3+4 2 3+4 2 | | |
| | 8 2 8 2 5+6 2 3+4 2 | | |
| | 8 2 6+7 2 3+4 2 5+6 2 | | |
| | 4+5 2 9 2 5+6 2 5+6 2 | | |
| | 9 2 9 2 5+6 2 3+4 2 | | |
| | 9 2 8 2 3+4 2 5+6 2 | | |
| | 8 2 4+5 2 3+4 2 5+6 2 | | |
| | 6+7 2 9 2 5+6 2 3+4 2 | | |
| | 6+7 2 8 2 3+4 2 3+4 2 | | |
| | 9 2 9 2 5+6 2 3+4 2 | | |
| | 9 2 6+7 2 5+6 2 5+6 2 | | |
| | 8 2 9 2 3+4 2 3+4 2 | | |
| | 8 2 4+5 2 5+6 2 5+6 2 | | |
| | 9 2 8 2 3+4 2 3+4 2 | | |
| | 9 2 9 2 5+6 2 3+4 2 | | |
| | 4+5 2 8 2 3+4 2 5+6 2 | | |
| 166.1.212 | 630007 | 150 | Dy1Ni3 |
| | -GM -T -F -L | | |
| | 8 2 9 2 3+4 2 3+4 2 | | |
| | 8 2 8 2 5+6 2 3+4 2 | | |
| | 9 2 8 2 3+4 2 5+6 2 | | |
| | 8 2 9 2 5+6 2 3+4 2 | | |
| | 9 2 9 2 5+6 2 5+6 2 | | |

Continued on next column

**TABLE LXXIII– continued**

| TQC | ICSD | | NoE | SF |
|---|---|---|---|---|
| | 9 2 | 8 2 | 3+4 2 | 5+6 2 |
| | 4+5 2 | 6+7 2 | 5+6 2 | 3+4 2 |
| | 8 2 | 9 2 | 3+4 2 | 5+6 2 |
| | 9 2 | 6+7 2 | 5+6 2 | 5+6 2 |
| | 6+7 2 | 9 2 | 5+6 2 | 5+6 2 |
| | 9 2 | 8 2 | 5+6 2 | 3+4 2 |
| | 6+7 2 | 4+5 2 | 3+4 2 | 5+6 2 |
| | 8 2 | 8 2 | 5+6 2 | 5+6 2 |
| | 9 2 | 9 2 | 3+4 2 | 3+4 2 |
| | 8 2 | 8 2 | 3+4 2 | 3+4 2 |
| | 8 2 | 9 2 | 5+6 2 | 5+6 2 |
| | 9 2 | 9 2 | 3+4 2 | 3+4 2 |
| | 8 2 | 9 2 | 3+4 2 | 5+6 2 |
| | 9 2 | 6+7 2 | 3+4 2 | 3+4 2 |
| | 4+5 2 | 8 2 | 5+6 2 | 5+6 2 |
| | 6+7 2 | 4+5 2 | 3+4 2 | 3+4 2 |
| | 8 2 | 8 2 | 5+6 2 | 5+6 2 |
| | 9 2 | 8 2 | 3+4 2 | 3+4 2 |
| | 8 2 | 4+5 2 | 5+6 2 | 5+6 2 |
| | 4+5 2 | 8 2 | 3+4 2 | 5+6 2 |
| | 8 2 | 9 2 | 5+6 2 | 3+4 2 |
| | 4+5 2 | 4+5 2 | 3+4 2 | 5+6 2 |
| | 8 2 | 8 2 | 5+6 2 | 5+6 2 |
| | 9 2 | 9 2 | 3+4 2 | 5+6 2 |
| | 6+7 2 | 6+7 2 | 5+6 2 | 5+6 2 |
| | 9 2 | 9 2 | 3+4 2 | 3+4 2 |
| | 6+7 2 | 6+7 2 | 5+6 2 | 5+6 2 |
| | 9 2 | 9 2 | 3+4 2 | 5+6 2 |
| | 4+5 2 | 6+7 2 | 3+4 2 | 3+4 2 |
| | 8 2 | 9 2 | 3+4 2 | 5+6 2 |
| | 8 2 | 8 2 | 5+6 2 | 3+4 2 |
| | 4+5 2 | 6+7 2 | 3+4 2 | 3+4 2 |
| | 8 2 | 9 2 | 5+6 2 | 3+4 2 |
| | 9 2 | 4+5 2 | 5+6 2 | 5+6 2 |
| | 6+7 2 | 8 2 | 3+4 2 | 3+4 2 |
| | 9 2 | 9 2 | 5+6 2 | 3+4 2 |
| | 6+7 2 | 8 2 | 5+6 2 | 5+6 2 |
| | 9 2 | 6+7 2 | 5+6 2 | 5+6 2 |
| | 8 2 | 8 2 | 3+4 2 | 5+6 2 |
| | 9 2 | 4+5 2 | 3+4 2 | 3+4 2 |
| | 4+5 2 | 8 2 | 3+4 2 | 3+4 2 |
| | 6+7 2 | 9 2 | 5+6 2 | 5+6 2 |
| | 8 2 | 4+5 2 | 3+4 2 | 5+6 2 |
| | 8 2 | 6+7 2 | 5+6 2 | 3+4 2 |
| | 9 2 | 9 2 | 5+6 2 | 5+6 2 |
| | 4+5 2 | 9 2 | 3+4 2 | 3+4 2 |
| | 8 2 | 9 2 | 5+6 2 | 3+4 2 |
| | 9 2 | 8 2 | 5+6 2 | 5+6 2 |
| | 8 2 | 6+7 2 | 5+6 2 | 3+4 2 |
| | 9 2 | 9 2 | 3+4 2 | 5+6 2 |
| | 6+7 2 | 8 2 | 5+6 2 | 5+6 2 |
| | 9 2 | 9 2 | 3+4 2 | 5+6 2 |
| | 4+5 2 | 9 2 | 5+6 2 | 3+4 2 |
| | 8 2 | 4+5 2 | 3+4 2 | 5+6 2 |
| | 4+5 2 | 8 2 | 5+6 2 | 3+4 2 |
| | 9 2 | 6+7 2 | 5+6 2 | 5+6 2 |
| | 6+7 2 | 9 2 | 3+4 2 | 5+6 2 |
| | 9 2 | 6+7 2 | 3+4 2 | 5+6 2 |
| | 8 2 | 8 2 | 5+6 2 | 3+4 2 |
| | 6+7 2 | 4+5 2 | 3+4 2 | 5+6 2 |
| | 8 2 | 9 2 | 3+4 2 | 3+4 2 |
| | 4+5 2 | 9 2 | 5+6 2 | 5+6 2 |
| | 9 2 | 6+7 2 | 3+4 2 | 5+6 2 |
| | 9 2 | 9 2 | 5+6 2 | 3+4 2 |
| | 8 2 | 4+5 2 | 5+6 2 | 3+4 2 |
| | 6+7 2 | 8 2 | 5+6 2 | 5+6 2 |
| | 9 2 | 8 2 | 3+4 2 | 5+6 2 |
| | 8 2 | 6+7 2 | 5+6 2 | 3+4 2 |
| | 9 2 | 9 2 | 3+4 2 | 3+4 2 |
| | 6+7 2 | 9 2 | 5+6 2 | 5+6 2 |
| 166.1.213 | 629986 | | 150 | Dy1Ni3 |
| | -GM | -T | -F | -L |
| | 8 2 | 9 2 | 3+4 2 | 3+4 2 |
| | 8 2 | 8 2 | 5+6 2 | 3+4 2 |
| | 9 2 | 8 2 | 3+4 2 | 5+6 2 |
| | 8 2 | 9 2 | 5+6 2 | 3+4 2 |
| | 9 2 | 9 2 | 5+6 2 | 5+6 2 |
| | 9 2 | 8 2 | 3+4 2 | 5+6 2 |
| | 4+5 2 | 6+7 2 | 5+6 2 | 3+4 2 |
| | 8 2 | 9 2 | 3+4 2 | 5+6 2 |
| | 9 2 | 6+7 2 | 5+6 2 | 5+6 2 |
| | 6+7 2 | 9 2 | 5+6 2 | 5+6 2 |
| | 9 2 | 8 2 | 5+6 2 | 3+4 2 |
| | 6+7 2 | 4+5 2 | 3+4 2 | 5+6 2 |
| | 8 2 | 8 2 | 5+6 2 | 5+6 2 |
| | 9 2 | 9 2 | 3+4 2 | 3+4 2 |
| | 8 2 | 8 2 | 3+4 2 | 3+4 2 |
| | 8 2 | 9 2 | 5+6 2 | 5+6 2 |
| | 9 2 | 9 2 | 3+4 2 | 3+4 2 |
| | 8 2 | 9 2 | 3+4 2 | 5+6 2 |
| | 9 2 | 6+7 2 | 3+4 2 | 3+4 2 |
| | 4+5 2 | 8 2 | 5+6 2 | 5+6 2 |
| | 6+7 2 | 4+5 2 | 3+4 2 | 3+4 2 |
| | 8 2 | 8 2 | 5+6 2 | 5+6 2 |
| | 9 2 | 8 2 | 3+4 2 | 3+4 2 |
| | 8 2 | 4+5 2 | 5+6 2 | 5+6 2 |
| | 4+5 2 | 8 2 | 3+4 2 | 5+6 2 |
| | 8 2 | 9 2 | 5+6 2 | 3+4 2 |
| | 4+5 2 | 4+5 2 | 3+4 2 | 5+6 2 |
| | 8 2 | 8 2 | 5+6 2 | 5+6 2 |
| | 9 2 | 9 2 | 3+4 2 | 5+6 2 |
| | 6+7 2 | 6+7 2 | 5+6 2 | 5+6 2 |
| | 9 2 | 9 2 | 3+4 2 | 3+4 2 |
| | 6+7 2 | 6+7 2 | 5+6 2 | 5+6 2 |
| | 9 2 | 9 2 | 3+4 2 | 5+6 2 |
| | 4+5 2 | 6+7 2 | 3+4 2 | 3+4 2 |
| | 8 2 | 9 2 | 3+4 2 | 5+6 2 |
| | 8 2 | 8 2 | 5+6 2 | 3+4 2 |
| | 4+5 2 | 6+7 2 | 3+4 2 | 3+4 2 |
| | 8 2 | 9 2 | 5+6 2 | 3+4 2 |
| | 9 2 | 4+5 2 | 5+6 2 | 5+6 2 |
| | 6+7 2 | 8 2 | 3+4 2 | 3+4 2 |
| | 9 2 | 9 2 | 5+6 2 | 3+4 2 |
| | 6+7 2 | 8 2 | 5+6 2 | 5+6 2 |
| | 9 2 | 6+7 2 | 5+6 2 | 5+6 2 |
| | 8 2 | 8 2 | 3+4 2 | 5+6 2 |
| | 9 2 | 4+5 2 | 3+4 2 | 3+4 2 |
| | 4+5 2 | 8 2 | 3+4 2 | 3+4 2 |
| | 6+7 2 | 9 2 | 5+6 2 | 5+6 2 |
| | 8 2 | 4+5 2 | 3+4 2 | 5+6 2 |
| | 8 2 | 6+7 2 | 5+6 2 | 3+4 2 |
| | 9 2 | 9 2 | 5+6 2 | 5+6 2 |

Continued on next column

**TABLE LXXIII– continued**

| TQC | ICSD | NoE | SF | |
|---|---|---|---|---|
| | 4+5 2 | 9 2 | 3+4 2 | 3+4 2 |
| | 8 2 | 9 2 | 5+6 2 | 3+4 2 |
| | 9 2 | 8 2 | 5+6 2 | 5+6 2 |
| | 8 2 | 6+7 2 | 5+6 2 | 3+4 2 |
| | 9 2 | 9 2 | 3+4 2 | 5+6 2 |
| | 6+7 2 | 8 2 | 5+6 2 | 5+6 2 |
| | 9 2 | 9 2 | 3+4 2 | 5+6 2 |
| | 4+5 2 | 9 2 | 5+6 2 | 3+4 2 |
| | 8 2 | 4+5 2 | 3+4 2 | 5+6 2 |
| | 4+5 2 | 8 2 | 5+6 2 | 3+4 2 |
| | 9 2 | 6+7 2 | 5+6 2 | 5+6 2 |
| | 6+7 2 | 9 2 | 3+4 2 | 5+6 2 |
| | 9 2 | 6+7 2 | 3+4 2 | 5+6 2 |
| | 8 2 | 8 2 | 5+6 2 | 3+4 2 |
| | 6+7 2 | 4+5 2 | 3+4 2 | 5+6 2 |
| | 8 2 | 9 2 | 3+4 2 | 3+4 2 |
| | 4+5 2 | 9 2 | 5+6 2 | 5+6 2 |
| | 9 2 | 6+7 2 | 3+4 2 | 5+6 2 |
| | 9 2 | 9 2 | 5+6 2 | 3+4 2 |
| | 8 2 | 4+5 2 | 5+6 2 | 3+4 2 |
| | 6+7 2 | 8 2 | 5+6 2 | 5+6 2 |
| | 9 2 | 8 2 | 3+4 2 | 5+6 2 |
| | 8 2 | 6+7 2 | 5+6 2 | 3+4 2 |
| | 9 2 | 9 2 | 3+4 2 | 3+4 2 |
| | 6+7 2 | 9 2 | 5+6 2 | 5+6 2 |
| 166.1.214 | 629980 | 150 | Dy1Ni3 | |
| | -GM | -T | -F | -L |
| | 8 2 | 9 2 | 3+4 2 | 3+4 2 |
| | 8 2 | 8 2 | 5+6 2 | 3+4 2 |
| | 9 2 | 8 2 | 3+4 2 | 5+6 2 |
| | 8 2 | 9 2 | 5+6 2 | 3+4 2 |
| | 9 2 | 9 2 | 5+6 2 | 5+6 2 |
| | 9 2 | 8 2 | 3+4 2 | 5+6 2 |
| | 4+5 2 | 6+7 2 | 5+6 2 | 3+4 2 |
| | 8 2 | 9 2 | 3+4 2 | 5+6 2 |
| | 9 2 | 6+7 2 | 5+6 2 | 5+6 2 |
| | 6+7 2 | 9 2 | 5+6 2 | 5+6 2 |
| | 9 2 | 8 2 | 5+6 2 | 3+4 2 |
| | 6+7 2 | 4+5 2 | 3+4 2 | 5+6 2 |
| | 8 2 | 8 2 | 5+6 2 | 5+6 2 |
| | 9 2 | 9 2 | 3+4 2 | 3+4 2 |
| | 8 2 | 8 2 | 3+4 2 | 3+4 2 |
| | 8 2 | 9 2 | 5+6 2 | 5+6 2 |
| | 9 2 | 9 2 | 3+4 2 | 3+4 2 |
| | 8 2 | 9 2 | 3+4 2 | 5+6 2 |
| | 9 2 | 6+7 2 | 3+4 2 | 3+4 2 |
| | 4+5 2 | 8 2 | 5+6 2 | 5+6 2 |
| | 6+7 2 | 4+5 2 | 3+4 2 | 3+4 2 |
| | 8 2 | 8 2 | 5+6 2 | 5+6 2 |
| | 9 2 | 8 2 | 3+4 2 | 3+4 2 |
| | 8 2 | 4+5 2 | 5+6 2 | 5+6 2 |
| | 4+5 2 | 8 2 | 3+4 2 | 5+6 2 |
| | 8 2 | 9 2 | 5+6 2 | 3+4 2 |
| | 4+5 2 | 4+5 2 | 3+4 2 | 5+6 2 |
| | 8 2 | 8 2 | 5+6 2 | 5+6 2 |
| | 9 2 | 9 2 | 3+4 2 | 5+6 2 |
| | 6+7 2 | 6+7 2 | 5+6 2 | 5+6 2 |
| | 9 2 | 9 2 | 3+4 2 | 3+4 2 |
| | 6+7 2 | 6+7 2 | 3+4 2 | 5+6 2 |
| | 9 2 | 9 2 | 5+6 2 | 5+6 2 |
| | 4+5 2 | 6+7 2 | 3+4 2 | 3+4 2 |

Continued on next column

**TABLE LXXIII– continued**

| TQC | ICSD | NoE | SF | |
|---|---|---|---|---|
| | 8 2 | 9 2 | 3+4 2 | 5+6 2 |
| | 8 2 | 8 2 | 5+6 2 | 3+4 2 |
| | 4+5 2 | 6+7 2 | 3+4 2 | 3+4 2 |
| | 8 2 | 9 2 | 5+6 2 | 3+4 2 |
| | 9 2 | 4+5 2 | 5+6 2 | 5+6 2 |
| | 6+7 2 | 8 2 | 3+4 2 | 3+4 2 |
| | 9 2 | 9 2 | 5+6 2 | 3+4 2 |
| | 6+7 2 | 8 2 | 5+6 2 | 5+6 2 |
| | 9 2 | 6+7 2 | 5+6 2 | 5+6 2 |
| | 8 2 | 4+5 2 | 3+4 2 | 5+6 2 |
| | 9 2 | 8 2 | 3+4 2 | 3+4 2 |
| | 4+5 2 | 8 2 | 3+4 2 | 3+4 2 |
| | 6+7 2 | 9 2 | 5+6 2 | 5+6 2 |
| | 8 2 | 4+5 2 | 3+4 2 | 5+6 2 |
| | 8 2 | 6+7 2 | 5+6 2 | 3+4 2 |
| | 9 2 | 9 2 | 5+6 2 | 3+4 2 |
| | 4+5 2 | 9 2 | 3+4 2 | 5+6 2 |
| | 8 2 | 9 2 | 5+6 2 | 3+4 2 |
| | 9 2 | 8 2 | 5+6 2 | 5+6 2 |
| | 8 2 | 9 2 | 5+6 2 | 3+4 2 |
| | 9 2 | 6+7 2 | 3+4 2 | 5+6 2 |
| | 6+7 2 | 8 2 | 5+6 2 | 5+6 2 |
| | 9 2 | 9 2 | 3+4 2 | 5+6 2 |
| | 4+5 2 | 9 2 | 5+6 2 | 3+4 2 |
| | 8 2 | 4+5 2 | 3+4 2 | 3+4 2 |
| | 9 2 | 8 2 | 5+6 2 | 5+6 2 |
| | 4+5 2 | 6+7 2 | 5+6 2 | 5+6 2 |
| | 6+7 2 | 9 2 | 3+4 2 | 5+6 2 |
| | 9 2 | 6+7 2 | 3+4 2 | 5+6 2 |
| | 8 2 | 8 2 | 5+6 2 | 3+4 2 |
| | 6+7 2 | 4+5 2 | 3+4 2 | 5+6 2 |
| | 8 2 | 9 2 | 3+4 2 | 3+4 2 |
| | 4+5 2 | 9 2 | 5+6 2 | 5+6 2 |
| | 9 2 | 6+7 2 | 3+4 2 | 5+6 2 |
| | 8 2 | 9 2 | 5+6 2 | 3+4 2 |
| | 9 2 | 8 2 | 5+6 2 | 3+4 2 |
| | 6+7 2 | 4+5 2 | 5+6 2 | 5+6 2 |
| | 9 2 | 8 2 | 3+4 2 | 5+6 2 |
| | 8 2 | 6+7 2 | 5+6 2 | 3+4 2 |
| | 9 2 | 9 2 | 3+4 2 | 3+4 2 |
| | 6+7 2 | 9 2 | 5+6 2 | 5+6 2 |
| 166.1.215 | 629598 | 132 | Dy1Fe3 | |
| | -GM | -T | -F | -L |
| | 8 2 | 8 2 | 3+4 2 | 3+4 2 |
| | 8 2 | 9 2 | 5+6 2 | 3+4 2 |
| | 9 2 | 8 2 | 3+4 2 | 5+6 2 |
| | 8 2 | 9 2 | 5+6 2 | 3+4 2 |
| | 9 2 | 9 2 | 5+6 2 | 5+6 2 |
| | 9 2 | 8 2 | 3+4 2 | 5+6 2 |
| | 4+5 2 | 6+7 2 | 5+6 2 | 3+4 2 |
| | 8 2 | 9 2 | 3+4 2 | 5+6 2 |
| | 9 2 | 6+7 2 | 5+6 2 | 5+6 2 |
| | 6+7 2 | 9 2 | 5+6 2 | 5+6 2 |
| | 9 2 | 8 2 | 5+6 2 | 3+4 2 |
| | 6+7 2 | 4+5 2 | 3+4 2 | 5+6 2 |
| | 8 2 | 8 2 | 5+6 2 | 5+6 2 |
| | 9 2 | 9 2 | 3+4 2 | 3+4 2 |
| | 8 2 | 8 2 | 3+4 2 | 3+4 2 |
| | 8 2 | 9 2 | 5+6 2 | 5+6 2 |
| | 9 2 | 9 2 | 3+4 2 | 3+4 2 |
| | 9 2 | 9 2 | 3+4 2 | 5+6 2 |

Continued on next column

**TABLE LXXIII– continued**

| TQC | ICSD | | NoE | SF |
|---|---|---|---|---|
| | 6+7 2 | 6+7 2 | 3+4 2 | 3+4 2 |
| | 8 2 | 8 2 | 5+6 2 | 5+6 2 |
| | 8 2 | 4+5 2 | 3+4 2 | 3+4 2 |
| | 4+5 2 | 8 2 | 3+4 2 | 5+6 2 |
| | 9 2 | 4+5 2 | 5+6 2 | 3+4 2 |
| | 8 2 | 8 2 | 5+6 2 | 5+6 2 |
| | 4+5 2 | 8 2 | 3+4 2 | 3+4 2 |
| | 8 2 | 9 2 | 5+6 2 | 5+6 2 |
| | 4+5 2 | 4+5 2 | 3+4 2 | 5+6 2 |
| | 8 2 | 8 2 | 5+6 2 | 5+6 2 |
| | 9 2 | 9 2 | 5+6 2 | 5+6 2 |
| | 6+7 2 | 6+7 2 | 3+4 2 | 5+6 2 |
| | 9 2 | 9 2 | 3+4 2 | 5+6 2 |
| | 6+7 2 | 6+7 2 | 3+4 2 | 3+4 2 |
| | 9 2 | 9 2 | 5+6 2 | 5+6 2 |
| | 4+5 2 | 6+7 2 | 3+4 2 | 3+4 2 |
| | 8 2 | 9 2 | 3+4 2 | 3+4 2 |
| | 8 2 | 8 2 | 5+6 2 | 5+6 2 |
| | 9 2 | 9 2 | 5+6 2 | 5+6 2 |
| | 4+5 2 | 4+5 2 | 3+4 2 | 3+4 2 |
| | 6+7 2 | 8 2 | 3+4 2 | 3+4 2 |
| | 8 2 | 9 2 | 5+6 2 | 5+6 2 |
| | 9 2 | 6+7 2 | 5+6 2 | 3+4 2 |
| | 9 2 | 9 2 | 5+6 2 | 3+4 2 |
| | 8 2 | 9 2 | 5+6 2 | 5+6 2 |
| | 6+7 2 | 6+7 2 | 3+4 2 | 5+6 2 |
| | 9 2 | 9 2 | 3+4 2 | 3+4 2 |
| | 4+5 2 | 6+7 2 | 5+6 2 | 3+4 2 |
| | 8 2 | 8 2 | 5+6 2 | 5+6 2 |
| | 9 2 | 4+5 2 | 5+6 2 | 5+6 2 |
| | 6+7 2 | 8 2 | 3+4 2 | 3+4 2 |
| | 9 2 | 8 2 | 5+6 2 | 5+6 2 |
| | 6+7 2 | 4+5 2 | 5+6 2 | 5+6 2 |
| | 9 2 | 9 2 | 3+4 2 | 3+4 2 |
| | 8 2 | 6+7 2 | 5+6 2 | 5+6 2 |
| | 8 2 | 8 2 | 3+4 2 | 5+6 2 |
| | 4+5 2 | 9 2 | 3+4 2 | 3+4 2 |
| | 9 2 | 8 2 | 5+6 2 | 3+4 2 |
| | 4+5 2 | 9 2 | 3+4 2 | 3+4 2 |
| | 8 2 | 9 2 | 5+6 2 | 5+6 2 |
| | 9 2 | 9 2 | 5+6 2 | 5+6 2 |
| | 8 2 | 4+5 2 | 3+4 2 | 3+4 2 |
| | 6+7 2 | 8 2 | 5+6 2 | 5+6 2 |
| | 8 2 | 9 2 | 5+6 2 | 5+6 2 |
| | 9 2 | 6+7 2 | 5+6 2 | 3+4 2 |
| | 9 2 | 6+7 2 | 3+4 2 | 5+6 2 |
| | 4+5 2 | 9 2 | 3+4 2 | 5+6 2 |
| | 6+7 2 | 4+5 2 | 5+6 2 | 3+4 2 |
| 166.1.216 | 184010 | | 66 | Dy1F1O1 |
| | -GM | -T | -F | -L |
| | 8 2 | 8 2 | 5+6 2 | 5+6 2 |
| | 9 2 | 9 2 | 3+4 2 | 3+4 2 |
| | 8 2 | 9 2 | 5+6 2 | 3+4 2 |
| | 9 2 | 8 2 | 3+4 2 | 5+6 2 |
| | 8 2 | 9 2 | 5+6 2 | 3+4 2 |
| | 9 2 | 8 2 | 3+4 2 | 5+6 2 |
| | 8 2 | 8 2 | 5+6 2 | 5+6 2 |
| | 4+5 2 | 4+5 2 | 3+4 2 | 3+4 2 |
| | 6+7 2 | 6+7 2 | 5+6 2 | 5+6 2 |
| | 9 2 | 9 2 | 3+4 2 | 3+4 2 |
| | 9 2 | 9 2 | 3+4 2 | 3+4 2 |

Continued on next column

**TABLE LXXIII– continued**

| TQC | ICSD | | NoE | SF |
|---|---|---|---|---|
| | 8 2 | 8 2 | 5+6 2 | 5+6 2 |
| | 8 2 | 9 2 | 3+4 2 | 5+6 2 |
| | 8 2 | 6+7 2 | 5+6 2 | 3+4 2 |
| | 4+5 2 | 9 2 | 5+6 2 | 3+4 2 |
| | 9 2 | 8 2 | 5+6 2 | 3+4 2 |
| | 6+7 2 | 4+5 2 | 3+4 2 | 5+6 2 |
| | 9 2 | 8 2 | 3+4 2 | 5+6 2 |
| | 8 2 | 8 2 | 3+4 2 | 3+4 2 |
| | 8 2 | 8 2 | 5+6 2 | 5+6 2 |
| | 4+5 2 | 4+5 2 | 5+6 2 | 5+6 2 |
| | 9 2 | 9 2 | 5+6 2 | 5+6 2 |
| | 6+7 2 | 6+7 2 | 3+4 2 | 3+4 2 |
| | 9 2 | 9 2 | 3+4 2 | 3+4 2 |
| | 8 2 | 8 2 | 3+4 2 | 5+6 2 |
| | 9 2 | 9 2 | 5+6 2 | 3+4 2 |
| | 8 2 | 8 2 | 5+6 2 | 5+6 2 |
| | 4+5 2 | 4+5 2 | 3+4 2 | 3+4 2 |
| | 6+7 2 | 6+7 2 | 5+6 2 | 5+6 2 |
| | 9 2 | 9 2 | 3+4 2 | 3+4 2 |
| | 8 2 | 8 2 | 3+4 2 | 3+4 2 |
| | 9 2 | 4+5 2 | 5+6 2 | 5+6 2 |
| | 4+5 2 | 9 2 | 5+6 2 | 5+6 2 |
| 166.1.217 | 629540 | | 60 | Dy1 |
| | -GM | -T | -F | -L |
| | 8 2 | 9 2 | 5+6 2 | 3+4 2 |
| | 9 2 | 8 2 | 3+4 2 | 5+6 2 |
| | 8 2 | 8 2 | 3+4 2 | 3+4 2 |
| | 8 2 | 9 2 | 5+6 2 | 5+6 2 |
| | 9 2 | 8 2 | 3+4 2 | 3+4 2 |
| | 9 2 | 9 2 | 5+6 2 | 5+6 2 |
| | 8 2 | 9 2 | 5+6 2 | 5+6 2 |
| | 9 2 | 8 2 | 5+6 2 | 3+4 2 |
| | 4+5 2 | 9 2 | 3+4 2 | 5+6 2 |
| | 6+7 2 | 6+7 2 | 5+6 2 | 3+4 2 |
| | 6+7 2 | 4+5 2 | 3+4 2 | 5+6 2 |
| | 9 2 | 6+7 2 | 5+6 2 | 5+6 2 |
| | 8 2 | 9 2 | 3+4 2 | 3+4 2 |
| | 9 2 | 8 2 | 3+4 2 | 3+4 2 |
| | 8 2 | 8 2 | 5+6 2 | 5+6 2 |
| | 9 2 | 9 2 | 3+4 2 | 5+6 2 |
| | 8 2 | 9 2 | 5+6 2 | 3+4 2 |
| | 9 2 | 8 2 | 3+4 2 | 5+6 2 |
| | 9 2 | 8 2 | 5+6 2 | 5+6 2 |
| | 6+7 2 | 9 2 | 3+4 2 | 5+6 2 |
| | 8 2 | 4+5 2 | 5+6 2 | 3+4 2 |
| | 4+5 2 | 6+7 2 | 5+6 2 | 5+6 2 |
| | 6+7 2 | 6+7 2 | 5+6 2 | 3+4 2 |
| | 9 2 | 9 2 | 5+6 2 | 5+6 2 |
| | 9 2 | 9 2 | 3+4 2 | 5+6 2 |
| | 8 2 | 8 2 | 3+4 2 | 3+4 2 |
| | 9 2 | 9 2 | 5+6 2 | 3+4 2 |
| | 8 2 | 8 2 | 5+6 2 | 5+6 2 |
| | 9 2 | 9 2 | 5+6 2 | 5+6 2 |
| | 9 2 | 9 2 | 3+4 2 | 5+6 2 |
| 166.1.218 | 629537 | | 60 | Dy1 |
| | -GM | -T | -F | -L |
| | 8 2 | 9 2 | 5+6 2 | 3+4 2 |
| | 9 2 | 8 2 | 3+4 2 | 5+6 2 |
| | 8 2 | 8 2 | 3+4 2 | 3+4 2 |
| | 8 2 | 9 2 | 5+6 2 | 3+4 2 |
| | 9 2 | 8 2 | 3+4 2 | 5+6 2 |

Continued on next column

**TABLE LXXIII– continued**

| TQC | ICSD | NoE | SF | |
|---|---|---|---|---|
| | 9 2 | 9 2 | 5+6 2 | 5+6 2 |
| | 8 2 | 9 2 | 5+6 2 | 5+6 2 |
| | 9 2 | 8 2 | 5+6 2 | 3+4 2 |
| | 4+5 2 | 9 2 | 3+4 2 | 5+6 2 |
| | 6+7 2 | 6+7 2 | 5+6 2 | 3+4 2 |
| | 6+7 2 | 4+5 2 | 3+4 2 | 5+6 2 |
| | 9 2 | 6+7 2 | 5+6 2 | 5+6 2 |
| | 8 2 | 9 2 | 3+4 2 | 3+4 2 |
| | 9 2 | 8 2 | 3+4 2 | 3+4 2 |
| | 8 2 | 8 2 | 5+6 2 | 5+6 2 |
| | 9 2 | 9 2 | 3+4 2 | 5+6 2 |
| | 8 2 | 9 2 | 5+6 2 | 3+4 2 |
| | 9 2 | 8 2 | 3+4 2 | 5+6 2 |
| | 9 2 | 8 2 | 5+6 2 | 5+6 2 |
| | 6+7 2 | 9 2 | 3+4 2 | 5+6 2 |
| | 4+5 2 | 4+5 2 | 5+6 2 | 3+4 2 |
| | 8 2 | 6+7 2 | 5+6 2 | 5+6 2 |
| | 6+7 2 | 6+7 2 | 5+6 2 | 3+4 2 |
| | 9 2 | 9 2 | 5+6 2 | 5+6 2 |
| | 9 2 | 9 2 | 3+4 2 | 5+6 2 |
| | 8 2 | 8 2 | 3+4 2 | 3+4 2 |
| | 9 2 | 9 2 | 5+6 2 | 5+6 2 |
| | 8 2 | 8 2 | 5+6 2 | 3+4 2 |
| | 9 2 | 9 2 | 5+6 2 | 5+6 2 |
| | 9 2 | 9 2 | 3+4 2 | 5+6 2 |
| 166.1.219 | 52518 | 60 | Dy1 | |
| | -GM | -T | -F | -L |
| | 8 2 | 8 2 | 3+4 2 | 3+4 2 |
| | 8 2 | 9 2 | 5+6 2 | 3+4 2 |
| | 9 2 | 8 2 | 3+4 2 | 5+6 2 |
| | 8 2 | 9 2 | 5+6 2 | 3+4 2 |
| | 9 2 | 9 2 | 5+6 2 | 5+6 2 |
| | 9 2 | 8 2 | 3+4 2 | 5+6 2 |
| | 8 2 | 9 2 | 5+6 2 | 5+6 2 |
| | 9 2 | 9 2 | 5+6 2 | 3+4 2 |
| | 4+5 2 | 8 2 | 3+4 2 | 5+6 2 |
| | 6+7 2 | 6+7 2 | 5+6 2 | 3+4 2 |
| | 6+7 2 | 6+7 2 | 3+4 2 | 5+6 2 |
| | 9 2 | 4+5 2 | 5+6 2 | 5+6 2 |
| | 8 2 | 9 2 | 3+4 2 | 3+4 2 |
| | 9 2 | 8 2 | 5+6 2 | 3+4 2 |
| | 8 2 | 8 2 | 3+4 2 | 5+6 2 |
| | 9 2 | 9 2 | 3+4 2 | 5+6 2 |
| | 8 2 | 9 2 | 5+6 2 | 3+4 2 |
| | 9 2 | 8 2 | 3+4 2 | 5+6 2 |
| | 6+7 2 | 4+5 2 | 5+6 2 | 5+6 2 |
| | 9 2 | 9 2 | 5+6 2 | 5+6 2 |
| | 4+5 2 | 6+7 2 | 5+6 2 | 3+4 2 |
| | 8 2 | 8 2 | 5+6 2 | 5+6 2 |
| | 6+7 2 | 6+7 2 | 3+4 2 | 3+4 2 |
| | 9 2 | 9 2 | 5+6 2 | 5+6 2 |
| | 9 2 | 9 2 | 3+4 2 | 5+6 2 |
| | 8 2 | 8 2 | 3+4 2 | 3+4 2 |
| | 9 2 | 9 2 | 5+6 2 | 5+6 2 |
| | 8 2 | 8 2 | 5+6 2 | 3+4 2 |
| | 9 2 | 9 2 | 3+4 2 | 5+6 2 |
| | 9 2 | 9 2 | 5+6 2 | 5+6 2 |
| 166.1.220 | 190938 | 26 | Cu1N2Ta1 | |
| | -GM | -T | -F | -L |
| | 8 2 | 9 2 | 3+4 2 | 5+6 2 |
| | 9 2 | 8 2 | 5+6 2 | 3+4 2 |

Continued on next column

**TABLE LXXIII– continued**

| TQC | ICSD | NoE | SF | |
|---|---|---|---|---|
| | 4+5 2 | 6+7 2 | 3+4 2 | 5+6 2 |
| | 8 2 | 9 2 | 3+4 2 | 5+6 2 |
| | 8 2 | 8 2 | 5+6 2 | 3+4 2 |
| | 9 2 | 9 2 | 5+6 2 | 3+4 2 |
| | 8 2 | 8 2 | 3+4 2 | 5+6 2 |
| | 4+5 2 | 4+5 2 | 3+4 2 | 3+4 2 |
| | 8 2 | 8 2 | 3+4 2 | 3+4 2 |
| | 8 2 | 8 2 | 3+4 2 | 3+4 2 |
| | 4+5 2 | 4+5 2 | 3+4 2 | 3+4 2 |
| | 8 2 | 8 2 | 3+4 2 | 3+4 2 |
| | 9 2 | 8 2 | 3+4 2 | 5+6 2 |
| 166.1.221 | 80259 | 36 | Cu1I1 | |
| | -GM | -T | -F | -L |
| | 8 2 | 9 2 | 3+4 2 | 5+6 2 |
| | 8 2 | 8 2 | 3+4 2 | 3+4 2 |
| | 9 2 | 8 2 | 5+6 2 | 3+4 2 |
| | 6+7 2 | 4+5 2 | 5+6 2 | 3+4 2 |
| | 9 2 | 8 2 | 5+6 2 | 3+4 2 |
| | 4+5 2 | 9 2 | 3+4 2 | 5+6 2 |
| | 8 2 | 6+7 2 | 5+6 2 | 3+4 2 |
| | 9 2 | 9 2 | 5+6 2 | 5+6 2 |
| | 4+5 2 | 6+7 2 | 3+4 2 | 5+6 2 |
| | 8 2 | 9 2 | 3+4 2 | 5+6 2 |
| | 9 2 | 8 2 | 5+6 2 | 3+4 2 |
| | 6+7 2 | 4+5 2 | 3+4 2 | 5+6 2 |
| | 8 2 | 8 2 | 3+4 2 | 3+4 2 |
| | 9 2 | 9 2 | 5+6 2 | 5+6 2 |
| | 9 2 | 8 2 | 5+6 2 | 3+4 2 |
| | 8 2 | 4+5 2 | 5+6 2 | 3+4 2 |
| | 6+7 2 | 9 2 | 5+6 2 | 5+6 2 |
| | 9 2 | 9 2 | 5+6 2 | 5+6 2 |
| 166.1.222 | 80257 | 36 | Cu1I1 | |
| | -GM | -T | -F | -L |
| | 8 2 | 9 2 | 3+4 2 | 5+6 2 |
| | 8 2 | 8 2 | 3+4 2 | 3+4 2 |
| | 9 2 | 8 2 | 5+6 2 | 3+4 2 |
| | 6+7 2 | 4+5 2 | 5+6 2 | 3+4 2 |
| | 9 2 | 8 2 | 5+6 2 | 3+4 2 |
| | 9 2 | 9 2 | 3+4 2 | 5+6 2 |
| | 4+5 2 | 6+7 2 | 5+6 2 | 5+6 2 |
| | 8 2 | 9 2 | 5+6 2 | 3+4 2 |
| | 4+5 2 | 6+7 2 | 3+4 2 | 5+6 2 |
| | 8 2 | 9 2 | 3+4 2 | 5+6 2 |
| | 9 2 | 8 2 | 5+6 2 | 3+4 2 |
| | 6+7 2 | 4+5 2 | 3+4 2 | 5+6 2 |
| | 8 2 | 8 2 | 3+4 2 | 3+4 2 |
| | 9 2 | 9 2 | 5+6 2 | 5+6 2 |
| | 6+7 2 | 8 2 | 5+6 2 | 3+4 2 |
| | 9 2 | 4+5 2 | 5+6 2 | 3+4 2 |
| | 8 2 | 9 2 | 5+6 2 | 5+6 2 |
| | 9 2 | 9 2 | 5+6 2 | 5+6 2 |
| 166.1.223 | 80253 | 36 | Cu1I1 | |
| | -GM | -T | -F | -L |
| | 8 2 | 9 2 | 3+4 2 | 5+6 2 |
| | 8 2 | 8 2 | 3+4 2 | 3+4 2 |
| | 9 2 | 8 2 | 5+6 2 | 3+4 2 |
| | 6+7 2 | 4+5 2 | 5+6 2 | 3+4 2 |
| | 9 2 | 8 2 | 5+6 2 | 3+4 2 |
| | 9 2 | 9 2 | 5+6 2 | 5+6 2 |
| | 4+5 2 | 6+7 2 | 3+4 2 | 5+6 2 |
| | 8 2 | 9 2 | 5+6 2 | 3+4 2 |

Continued on next column

**TABLE LXXIII– continued**

| TQC | ICSD | NoE | SF |
|---|---|---|---|
| | 4+5 2 6+7 2 3+4 2 5+6 2 | | |
| | 8 2 9 2 3+4 2 5+6 2 | | |
| | 9 2 8 2 5+6 2 3+4 2 | | |
| | 6+7 2 4+5 2 3+4 2 5+6 2 | | |
| | 8 2 8 2 3+4 2 3+4 2 | | |
| | 9 2 9 2 5+6 2 5+6 2 | | |
| | 6+7 2 8 2 5+6 2 3+4 2 | | |
| | 9 2 4+5 2 5+6 2 5+6 2 | | |
| | 8 2 9 2 5+6 2 3+4 2 | | |
| | 9 2 9 2 5+6 2 5+6 2 | | |
| 166.1.224 | 80252 | 36 | Cu1I1 |
| | -GM -T -F -L | | |
| | 8 2 9 2 3+4 2 5+6 2 | | |
| | 8 2 8 2 3+4 2 3+4 2 | | |
| | 9 2 8 2 5+6 2 3+4 2 | | |
| | 6+7 2 4+5 2 5+6 2 3+4 2 | | |
| | 9 2 8 2 5+6 2 3+4 2 | | |
| | 9 2 9 2 5+6 2 5+6 2 | | |
| | 4+5 2 6+7 2 3+4 2 5+6 2 | | |
| | 8 2 9 2 5+6 2 3+4 2 | | |
| | 4+5 2 6+7 2 3+4 2 5+6 2 | | |
| | 8 2 9 2 3+4 2 5+6 2 | | |
| | 9 2 8 2 5+6 2 3+4 2 | | |
| | 6+7 2 4+5 2 3+4 2 5+6 2 | | |
| | 8 2 8 2 3+4 2 3+4 2 | | |
| | 9 2 9 2 5+6 2 5+6 2 | | |
| | 6+7 2 8 2 5+6 2 3+4 2 | | |
| | 9 2 4+5 2 5+6 2 5+6 2 | | |
| | 8 2 9 2 5+6 2 3+4 2 | | |
| | 9 2 9 2 5+6 2 5+6 2 | | |
| 166.1.225 | 410393 | 120 | Cs9F27Ni9 |
| | -GM -T -F -L | | |
| | 8 2 9 2 3+4 2 5+6 2 | | |
| | 9 2 8 2 5+6 2 3+4 2 | | |
| | 8 2 8 2 5+6 2 3+4 2 | | |
| | 4+5 2 8 2 3+4 2 5+6 2 | | |
| | 8 2 4+5 2 5+6 2 5+6 2 | | |
| | 6+7 2 6+7 2 5+6 2 3+4 2 | | |
| | 9 2 9 2 3+4 2 5+6 2 | | |
| | 4+5 2 4+5 2 3+4 2 3+4 2 | | |
| | 8 2 8 2 5+6 2 5+6 2 | | |
| | 8 2 9 2 5+6 2 3+4 2 | | |
| | 9 2 8 2 3+4 2 5+6 2 | | |
| | 8 2 8 2 3+4 2 3+4 2 | | |
| | 8 2 9 2 5+6 2 3+4 2 | | |
| | 9 2 9 2 3+4 2 5+6 2 | | |
| | 9 2 8 2 5+6 2 5+6 2 | | |
| | 4+5 2 6+7 2 5+6 2 3+4 2 | | |
| | 6+7 2 9 2 3+4 2 5+6 2 | | |
| | 9 2 9 2 5+6 2 5+6 2 | | |
| | 8 2 6+7 2 3+4 2 3+4 2 | | |
| | 9 2 4+5 2 5+6 2 5+6 2 | | |
| | 6+7 2 8 2 5+6 2 5+6 2 | | |
| | 8 2 9 2 3+4 2 3+4 2 | | |
| | 9 2 9 2 3+4 2 5+6 2 | | |
| | 9 2 8 2 5+6 2 5+6 2 | | |
| | 6+7 2 9 2 3+4 2 3+4 2 | | |
| | 9 2 6+7 2 5+6 2 3+4 2 | | |
| | 8 2 9 2 3+4 2 5+6 2 | | |
| | 4+5 2 8 2 3+4 2 3+4 2 | | |
| | 8 2 6+7 2 3+4 2 5+6 2 | | |

Continued on next column

**TABLE LXXIII– continued**

| TQC | ICSD | NoE | SF |
|---|---|---|---|
| | 9 2 4+5 2 3+4 2 3+4 2 | | |
| | 6+7 2 9 2 3+4 2 5+6 2 | | |
| | 8 2 9 2 5+6 2 5+6 2 | | |
| | 9 2 8 2 5+6 2 5+6 2 | | |
| | 4+5 2 8 2 5+6 2 3+4 2 | | |
| | 9 2 6+7 2 5+6 2 3+4 2 | | |
| | 6+7 2 4+5 2 3+4 2 3+4 2 | | |
| | 9 2 9 2 5+6 2 3+4 2 | | |
| | 6+7 2 9 2 3+4 2 3+4 2 | | |
| | 8 2 6+7 2 3+4 2 5+6 2 | | |
| | 9 2 9 2 5+6 2 3+4 2 | | |
| | 4+5 2 9 2 3+4 2 5+6 2 | | |
| | 9 2 6+7 2 5+6 2 5+6 2 | | |
| | 9 2 8 2 3+4 2 3+4 2 | | |
| | 6+7 2 8 2 3+4 2 3+4 2 | | |
| | 9 2 9 2 5+6 2 5+6 2 | | |
| | 8 2 4+5 2 5+6 2 3+4 2 | | |
| | 9 2 6+7 2 5+6 2 5+6 2 | | |
| | 6+7 2 9 2 3+4 2 3+4 2 | | |
| | 8 2 9 2 3+4 2 5+6 2 | | |
| | 4+5 2 6+7 2 3+4 2 5+6 2 | | |
| | 8 2 4+5 2 5+6 2 5+6 2 | | |
| | 4+5 2 8 2 3+4 2 5+6 2 | | |
| | 8 2 9 2 5+6 2 3+4 2 | | |
| | 8 2 8 2 3+4 2 3+4 2 | | |
| | 9 2 6+7 2 5+6 2 5+6 2 | | |
| | 6+7 2 9 2 3+4 2 5+6 2 | | |
| | 9 2 9 2 3+4 2 3+4 2 | | |
| | 8 2 8 2 5+6 2 5+6 2 | | |
| | 4+5 2 4+5 2 3+4 2 3+4 2 | | |
| | 8 2 9 2 3+4 2 5+6 2 | | |
| 166.1.226 | 78928 | 146 | Cs4F12Fe2K1Li1 |
| | -GM -T -F -L | | |
| | 8 2 8 2 3+4 2 3+4 2 | | |
| | 8 2 9 2 3+4 2 5+6 2 | | |
| | 9 2 8 2 5+6 2 3+4 2 | | |
| | 8 2 9 2 3+4 2 5+6 2 | | |
| | 9 2 8 2 5+6 2 5+6 2 | | |
| | 9 2 9 2 5+6 2 3+4 2 | | |
| | 6+7 2 6+7 2 5+6 2 5+6 2 | | |
| | 4+5 2 8 2 3+4 2 3+4 2 | | |
| | 8 2 4+5 2 3+4 2 3+4 2 | | |
| | 6+7 2 4+5 2 5+6 2 3+4 2 | | |
| | 9 2 8 2 5+6 2 3+4 2 | | |
| | 4+5 2 6+7 2 3+4 2 5+6 2 | | |
| | 8 2 9 2 3+4 2 5+6 2 | | |
| | 8 2 9 2 5+6 2 3+4 2 | | |
| | 9 2 8 2 3+4 2 5+6 2 | | |
| | 9 2 9 2 3+4 2 3+4 2 | | |
| | 8 2 8 2 5+6 2 5+6 2 | | |
| | 9 2 9 2 5+6 2 5+6 2 | | |
| | 6+7 2 6+7 2 5+6 2 5+6 2 | | |
| | 9 2 9 2 5+6 2 5+6 2 | | |
| | 8 2 9 2 5+6 2 3+4 2 | | |
| | 9 2 8 2 3+4 2 5+6 2 | | |
| | 8 2 9 2 5+6 2 5+6 2 | | |
| | 9 2 8 2 3+4 2 3+4 2 | | |
| | 4+5 2 6+7 2 5+6 2 3+4 2 | | |
| | 8 2 4+5 2 3+4 2 5+6 2 | | |
| | 6+7 2 8 2 5+6 2 3+4 2 | | |
| | 4+5 2 9 2 3+4 2 5+6 2 | | |

Continued on next column

**TABLE LXXIII– continued**

| TQC | ICSD | NoE | SF | |
|---|---|---|---|---|
| | 9 2 | 6+7 2 | 5+6 2 | 5+6 2 |
| | 8 2 | 4+5 2 | 3+4 2 | 3+4 2 |
| | 6+7 2 | 9 2 | 5+6 2 | 3+4 2 |
| | 9 2 | 8 2 | 3+4 2 | 5+6 2 |
| | 8 2 | 8 2 | 3+4 2 | 3+4 2 |
| | 9 2 | 9 2 | 5+6 2 | 5+6 2 |
| | 8 2 | 8 2 | 3+4 2 | 3+4 2 |
| | 4+5 2 | 4+5 2 | 3+4 2 | 3+4 2 |
| | 9 2 | 9 2 | 5+6 2 | 5+6 2 |
| | 6+7 2 | 6+7 2 | 5+6 2 | 5+6 2 |
| | 8 2 | 8 2 | 3+4 2 | 3+4 2 |
| | 9 2 | 9 2 | 5+6 2 | 5+6 2 |
| | 9 2 | 8 2 | 3+4 2 | 3+4 2 |
| | 9 2 | 4+5 2 | 5+6 2 | 5+6 2 |
| | 8 2 | 9 2 | 3+4 2 | 3+4 2 |
| | 6+7 2 | 8 2 | 3+4 2 | 3+4 2 |
| | 8 2 | 9 2 | 5+6 2 | 5+6 2 |
| | 4+5 2 | 6+7 2 | 3+4 2 | 5+6 2 |
| | 9 2 | 8 2 | 5+6 2 | 5+6 2 |
| | 6+7 2 | 9 2 | 5+6 2 | 3+4 2 |
| | 8 2 | 4+5 2 | 5+6 2 | 3+4 2 |
| | 4+5 2 | 6+7 2 | 3+4 2 | 5+6 2 |
| | 8 2 | 8 2 | 5+6 2 | 3+4 2 |
| | 6+7 2 | 4+5 2 | 5+6 2 | 3+4 2 |
| | 9 2 | 9 2 | 3+4 2 | 5+6 2 |
| | 4+5 2 | 8 2 | 3+4 2 | 3+4 2 |
| | 8 2 | 6+7 2 | 5+6 2 | 5+6 2 |
| | 9 2 | 9 2 | 3+4 2 | 3+4 2 |
| | 8 2 | 9 2 | 5+6 2 | 5+6 2 |
| | 9 2 | 8 2 | 3+4 2 | 5+6 2 |
| | 8 2 | 8 2 | 3+4 2 | 3+4 2 |
| | 4+5 2 | 9 2 | 5+6 2 | 5+6 2 |
| | 9 2 | 4+5 2 | 3+4 2 | 5+6 2 |
| | 6+7 2 | 6+7 2 | 5+6 2 | 3+4 2 |
| | 9 2 | 9 2 | 3+4 2 | 3+4 2 |
| | 8 2 | 6+7 2 | 5+6 2 | 5+6 2 |
| | 9 2 | 8 2 | 3+4 2 | 5+6 2 |
| | 4+5 2 | 8 2 | 5+6 2 | 3+4 2 |
| | 6+7 2 | 4+5 2 | 5+6 2 | 3+4 2 |
| | 8 2 | 9 2 | 3+4 2 | 5+6 2 |
| | 8 2 | 9 2 | 3+4 2 | 3+4 2 |
| | 6+7 2 | 4+5 2 | 5+6 2 | 5+6 2 |
| | 9 2 | 8 2 | 5+6 2 | 5+6 2 |
| | 4+5 2 | 6+7 2 | 3+4 2 | 3+4 2 |
| | 8 2 | 8 2 | 5+6 2 | 5+6 2 |
| 166.1.227 | 42152 | 142 | Cs2F6Na1Ni1 | |
| | -GM | -T | -F | -L |
| | 8 2 | 9 2 | 3+4 2 | 5+6 2 |
| | 8 2 | 8 2 | 3+4 2 | 3+4 2 |
| | 9 2 | 8 2 | 5+6 2 | 3+4 2 |
| | 9 2 | 9 2 | 5+6 2 | 5+6 2 |
| | 6+7 2 | 6+7 2 | 5+6 2 | 3+4 2 |
| | 9 2 | 9 2 | 5+6 2 | 5+6 2 |
| | 9 2 | 8 2 | 5+6 2 | 3+4 2 |
| | 6+7 2 | 4+5 2 | 5+6 2 | 5+6 2 |
| | 4+5 2 | 6+7 2 | 3+4 2 | 5+6 2 |
| | 8 2 | 9 2 | 3+4 2 | 3+4 2 |
| | 4+5 2 | 4+5 2 | 3+4 2 | 5+6 2 |
| | 8 2 | 8 2 | 3+4 2 | 3+4 2 |
| | 9 2 | 9 2 | 3+4 2 | 3+4 2 |
| | 8 2 | 8 2 | 5+6 2 | 5+6 2 |
| | 8 2 | 9 2 | 3+4 2 | 3+4 2 |
| | 9 2 | 8 2 | 5+6 2 | 5+6 2 |
| | 8 2 | 8 2 | 3+4 2 | 5+6 2 |
| | 9 2 | 9 2 | 5+6 2 | 3+4 2 |
| | 8 2 | 9 2 | 5+6 2 | 3+4 2 |
| | 9 2 | 8 2 | 3+4 2 | 5+6 2 |
| | 4+5 2 | 4+5 2 | 3+4 2 | 5+6 2 |
| | 8 2 | 8 2 | 5+6 2 | 3+4 2 |
| | 6+7 2 | 6+7 2 | 3+4 2 | 5+6 2 |
| | 4+5 2 | 9 2 | 5+6 2 | 3+4 2 |
| | 9 2 | 6+7 2 | 5+6 2 | 3+4 2 |
| | 8 2 | 9 2 | 3+4 2 | 5+6 2 |
| | 6+7 2 | 4+5 2 | 3+4 2 | 3+4 2 |
| | 9 2 | 8 2 | 5+6 2 | 5+6 2 |
| | 8 2 | 8 2 | 3+4 2 | 3+4 2 |
| | 8 2 | 9 2 | 3+4 2 | 5+6 2 |
| | 4+5 2 | 8 2 | 3+4 2 | 3+4 2 |
| | 8 2 | 4+5 2 | 3+4 2 | 3+4 2 |
| | 8 2 | 9 2 | 3+4 2 | 5+6 2 |
| | 4+5 2 | 6+7 2 | 3+4 2 | 5+6 2 |
| | 9 2 | 9 2 | 3+4 2 | 5+6 2 |
| | 8 2 | 8 2 | 5+6 2 | 3+4 2 |
| | 9 2 | 9 2 | 3+4 2 | 5+6 2 |
| | 9 2 | 8 2 | 5+6 2 | 3+4 2 |
| | 6+7 2 | 6+7 2 | 5+6 2 | 5+6 2 |
| | 8 2 | 4+5 2 | 3+4 2 | 5+6 2 |
| | 9 2 | 9 2 | 5+6 2 | 5+6 2 |
| | 8 2 | 8 2 | 5+6 2 | 3+4 2 |
| | 6+7 2 | 9 2 | 3+4 2 | 3+4 2 |
| | 4+5 2 | 8 2 | 3+4 2 | 3+4 2 |
| | 8 2 | 4+5 2 | 5+6 2 | 3+4 2 |
| | 4+5 2 | 6+7 2 | 3+4 2 | 5+6 2 |
| | 9 2 | 4+5 2 | 5+6 2 | 3+4 2 |
| | 6+7 2 | 8 2 | 5+6 2 | 5+6 2 |
| | 9 2 | 8 2 | 5+6 2 | 3+4 2 |
| | 9 2 | 9 2 | 5+6 2 | 5+6 2 |
| | 6+7 2 | 6+7 2 | 5+6 2 | 3+4 2 |
| | 9 2 | 9 2 | 5+6 2 | 5+6 2 |
| | 9 2 | 9 2 | 5+6 2 | 5+6 2 |
| | 9 2 | 8 2 | 5+6 2 | 3+4 2 |
| | 9 2 | 9 2 | 5+6 2 | 3+4 2 |
| | 6+7 2 | 8 2 | 5+6 2 | 5+6 2 |
| | 8 2 | 6+7 2 | 3+4 2 | 5+6 2 |
| | 9 2 | 9 2 | 5+6 2 | 3+4 2 |
| | 4+5 2 | 6+7 2 | 3+4 2 | 5+6 2 |
| | 8 2 | 4+5 2 | 3+4 2 | 3+4 2 |
| | 6+7 2 | 8 2 | 3+4 2 | 5+6 2 |
| | 8 2 | 9 2 | 5+6 2 | 3+4 2 |
| | 4+5 2 | 8 2 | 3+4 2 | 5+6 2 |
| | 8 2 | 4+5 2 | 3+4 2 | 3+4 2 |
| | 8 2 | 9 2 | 3+4 2 | 5+6 2 |
| | 8 2 | 8 2 | 3+4 2 | 3+4 2 |
| | 4+5 2 | 4+5 2 | 3+4 2 | 5+6 2 |
| | 4+5 2 | 6+7 2 | 3+4 2 | 3+4 2 |
| | 8 2 | 9 2 | 3+4 2 | 5+6 2 |
| | 8 2 | 8 2 | 3+4 2 | 3+4 2 |
| | 8 2 | 9 2 | 3+4 2 | 3+4 2 |
| 166.1.228 | 16255 | 138 | Cs2F6Fe1Na1 | |
| | -GM | -T | -F | -L |
| | 8 2 | 9 2 | 3+4 2 | 5+6 2 |
| | 8 2 | 8 2 | 5+6 2 | 3+4 2 |

Continued on next column

**TABLE LXXIII– continued**

| TQC | ICSD | NoE | SF | |
|---|---|---|---|---|
| | 9 2 | 8 2 | 3+4 2 | 3+4 2 |
| | 6+7 2 | 4+5 2 | 5+6 2 | 3+4 2 |
| | 4+5 2 | 6+7 2 | 3+4 2 | 5+6 2 |
| | 8 2 | 9 2 | 3+4 2 | 5+6 2 |
| | 9 2 | 8 2 | 5+6 2 | 3+4 2 |
| | 9 2 | 9 2 | 5+6 2 | 5+6 2 |
| | 6+7 2 | 6+7 2 | 5+6 2 | 5+6 2 |
| | 4+5 2 | 4+5 2 | 3+4 2 | 3+4 2 |
| | 8 2 | 8 2 | 3+4 2 | 3+4 2 |
| | 9 2 | 9 2 | 5+6 2 | 5+6 2 |
| | 9 2 | 9 2 | 3+4 2 | 3+4 2 |
| | 8 2 | 8 2 | 5+6 2 | 5+6 2 |
| | 9 2 | 8 2 | 3+4 2 | 5+6 2 |
| | 8 2 | 9 2 | 5+6 2 | 3+4 2 |
| | 8 2 | 9 2 | 3+4 2 | 5+6 2 |
| | 9 2 | 8 2 | 5+6 2 | 3+4 2 |
| | 8 2 | 9 2 | 5+6 2 | 3+4 2 |
| | 9 2 | 8 2 | 3+4 2 | 5+6 2 |
| | 4+5 2 | 6+7 2 | 3+4 2 | 5+6 2 |
| | 8 2 | 4+5 2 | 5+6 2 | 3+4 2 |
| | 4+5 2 | 9 2 | 5+6 2 | 3+4 2 |
| | 6+7 2 | 8 2 | 3+4 2 | 5+6 2 |
| | 8 2 | 9 2 | 5+6 2 | 3+4 2 |
| | 9 2 | 6+7 2 | 3+4 2 | 5+6 2 |
| | 6+7 2 | 4+5 2 | 5+6 2 | 3+4 2 |
| | 9 2 | 8 2 | 3+4 2 | 5+6 2 |
| | 8 2 | 8 2 | 3+4 2 | 5+6 2 |
| | 8 2 | 9 2 | 3+4 2 | 3+4 2 |
| | 8 2 | 9 2 | 3+4 2 | 3+4 2 |
| | 4+5 2 | 6+7 2 | 3+4 2 | 5+6 2 |
| | 4+5 2 | 4+5 2 | 3+4 2 | 5+6 2 |
| | 8 2 | 8 2 | 3+4 2 | 3+4 2 |
| | 9 2 | 8 2 | 3+4 2 | 5+6 2 |
| | 8 2 | 8 2 | 5+6 2 | 3+4 2 |
| | 6+7 2 | 9 2 | 5+6 2 | 5+6 2 |
| | 9 2 | 4+5 2 | 3+4 2 | 3+4 2 |
| | 9 2 | 9 2 | 5+6 2 | 3+4 2 |
| | 6+7 2 | 6+7 2 | 5+6 2 | 5+6 2 |
| | 9 2 | 9 2 | 5+6 2 | 5+6 2 |
| | 8 2 | 8 2 | 3+4 2 | 5+6 2 |
| | 8 2 | 9 2 | 3+4 2 | 3+4 2 |
| | 4+5 2 | 4+5 2 | 3+4 2 | 3+4 2 |
| | 4+5 2 | 6+7 2 | 5+6 2 | 3+4 2 |
| | 8 2 | 8 2 | 3+4 2 | 5+6 2 |
| | 6+7 2 | 8 2 | 5+6 2 | 3+4 2 |
| | 9 2 | 4+5 2 | 5+6 2 | 3+4 2 |
| | 9 2 | 8 2 | 5+6 2 | 3+4 2 |
| | 9 2 | 8 2 | 5+6 2 | 5+6 2 |
| | 9 2 | 9 2 | 5+6 2 | 5+6 2 |
| | 9 2 | 9 2 | 5+6 2 | 3+4 2 |
| | 9 2 | 8 2 | 5+6 2 | 3+4 2 |
| | 6+7 2 | 6+7 2 | 5+6 2 | 5+6 2 |
| | 8 2 | 9 2 | 3+4 2 | 5+6 2 |
| | 4+5 2 | 6+7 2 | 5+6 2 | 3+4 2 |
| | 6+7 2 | 4+5 2 | 3+4 2 | 5+6 2 |
| | 8 2 | 9 2 | 3+4 2 | 5+6 2 |
| | 9 2 | 9 2 | 5+6 2 | 5+6 2 |
| | 9 2 | 9 2 | 5+6 2 | 5+6 2 |
| | 6+7 2 | 6+7 2 | 5+6 2 | 5+6 2 |
| | 8 2 | 8 2 | 3+4 2 | 3+4 2 |
| | 4+5 2 | 4+5 2 | 3+4 2 | 3+4 2 |
| | 8 2 | 8 2 | 3+4 2 | 3+4 2 |
| | 8 2 | 8 2 | 3+4 2 | 5+6 2 |
| | 8 2 | 9 2 | 3+4 2 | 3+4 2 |
| | 4+5 2 | 4+5 2 | 3+4 2 | 5+6 2 |
| | 4+5 2 | 6+7 2 | 3+4 2 | 3+4 2 |
| | 8 2 | 8 2 | 3+4 2 | 5+6 2 |
| 166.1.229 | 102864 | 60 | Cs1Ga7 | |
| | -GM | -T | -F | -L |
| | 8 2 | 9 2 | 5+6 2 | 3+4 2 |
| | 9 2 | 8 2 | 3+4 2 | 5+6 2 |
| | 8 2 | 8 2 | 3+4 2 | 3+4 2 |
| | 8 2 | 9 2 | 5+6 2 | 5+6 2 |
| | 9 2 | 8 2 | 3+4 2 | 3+4 2 |
| | 9 2 | 6+7 2 | 5+6 2 | 5+6 2 |
| | 4+5 2 | 9 2 | 5+6 2 | 3+4 2 |
| | 8 2 | 9 2 | 3+4 2 | 5+6 2 |
| | 6+7 2 | 8 2 | 5+6 2 | 3+4 2 |
| | 9 2 | 4+5 2 | 3+4 2 | 5+6 2 |
| | 8 2 | 6+7 2 | 5+6 2 | 5+6 2 |
| | 6+7 2 | 9 2 | 5+6 2 | 5+6 2 |
| | 9 2 | 9 2 | 3+4 2 | 3+4 2 |
| | 9 2 | 8 2 | 3+4 2 | 5+6 2 |
| | 8 2 | 9 2 | 5+6 2 | 3+4 2 |
| | 4+5 2 | 8 2 | 3+4 2 | 3+4 2 |
| | 8 2 | 4+5 2 | 3+4 2 | 3+4 2 |
| | 8 2 | 8 2 | 3+4 2 | 5+6 2 |
| | 4+5 2 | 8 2 | 5+6 2 | 3+4 2 |
| | 9 2 | 4+5 2 | 5+6 2 | 5+6 2 |
| | 9 2 | 9 2 | 3+4 2 | 3+4 2 |
| | 6+7 2 | 6+7 2 | 5+6 2 | 5+6 2 |
| | 8 2 | 9 2 | 5+6 2 | 5+6 2 |
| | 9 2 | 8 2 | 5+6 2 | 5+6 2 |
| | 8 2 | 8 2 | 3+4 2 | 3+4 2 |
| | 6+7 2 | 4+5 2 | 3+4 2 | 5+6 2 |
| | 9 2 | 9 2 | 3+4 2 | 3+4 2 |
| | 4+5 2 | 6+7 2 | 5+6 2 | 3+4 2 |
| | 8 2 | 9 2 | 5+6 2 | 5+6 2 |
| | 9 2 | 8 2 | 3+4 2 | 5+6 2 |
| 166.1.230 | 625563 | 170 | Co7Y2 | |
| | -GM | -T | -F | -L |
| | 8 2 | 8 2 | 3+4 2 | 3+4 2 |
| | 9 2 | 9 2 | 5+6 2 | 5+6 2 |
| | 8 2 | 9 2 | 5+6 2 | 3+4 2 |
| | 9 2 | 8 2 | 3+4 2 | 5+6 2 |
| | 8 2 | 8 2 | 3+4 2 | 3+4 2 |
| | 9 2 | 9 2 | 5+6 2 | 5+6 2 |
| | 8 2 | 9 2 | 5+6 2 | 3+4 2 |
| | 9 2 | 8 2 | 3+4 2 | 5+6 2 |
| | 8 2 | 8 2 | 5+6 2 | 3+4 2 |
| | 4+5 2 | 9 2 | 3+4 2 | 5+6 2 |
| | 9 2 | 6+7 2 | 5+6 2 | 5+6 2 |
| | 6+7 2 | 4+5 2 | 3+4 2 | 3+4 2 |
| | 4+5 2 | 6+7 2 | 5+6 2 | 3+4 2 |
| | 8 2 | 9 2 | 3+4 2 | 5+6 2 |
| | 9 2 | 8 2 | 5+6 2 | 3+4 2 |
| | 6+7 2 | 4+5 2 | 3+4 2 | 5+6 2 |
| | 8 2 | 8 2 | 5+6 2 | 5+6 2 |
| | 9 2 | 9 2 | 3+4 2 | 3+4 2 |
| | 8 2 | 8 2 | 3+4 2 | 3+4 2 |
| | 9 2 | 9 2 | 5+6 2 | 5+6 2 |
| | 8 2 | 8 2 | 5+6 2 | 5+6 2 |

Continued on next column

**TABLE LXXIII– continued**

| TQC | ICSD | NoE | SF | |
|---|---|---|---|---|
| | 8 2 | 8 2 | 3+4 2 | 3+4 2 |
| | 8 2 | 9 2 | 5+6 2 | 5+6 2 |
| | 8 2 | 9 2 | 5+6 2 | 5+6 2 |
| | 4+5 2 | 9 2 | 3+4 2 | 5+6 2 |
| | 9 2 | 6+7 2 | 3+4 2 | 5+6 2 |
| | 9 2 | 8 2 | 5+6 2 | 3+4 2 |
| | 6+7 2 | 4+5 2 | 3+4 2 | 3+4 2 |
| | 8 2 | 8 2 | 5+6 2 | 5+6 2 |
| | 4+5 2 | 4+5 2 | 3+4 2 | 3+4 2 |
| | 8 2 | 8 2 | 3+4 2 | 3+4 2 |
| | 9 2 | 4+5 2 | 3+4 2 | 5+6 2 |
| | 8 2 | 8 2 | 5+6 2 | 5+6 2 |
| | 4+5 2 | 8 2 | 3+4 2 | 3+4 2 |
| | 8 2 | 8 2 | 5+6 2 | 3+4 2 |
| | 4+5 2 | 9 2 | 5+6 2 | 5+6 2 |
| | 6+7 2 | 6+7 2 | 5+6 2 | 5+6 2 |
| | 8 2 | 9 2 | 5+6 2 | 5+6 2 |
| | 8 2 | 4+5 2 | 3+4 2 | 5+6 2 |
| | 9 2 | 8 2 | 3+4 2 | 5+6 2 |
| | 4+5 2 | 6+7 2 | 3+4 2 | 3+4 2 |
| | 8 2 | 4+5 2 | 5+6 2 | 3+4 2 |
| | 6+7 2 | 9 2 | 5+6 2 | 5+6 2 |
| | 9 2 | 8 2 | 3+4 2 | 5+6 2 |
| | 4+5 2 | 9 2 | 3+4 2 | 3+4 2 |
| | 8 2 | 8 2 | 5+6 2 | 5+6 2 |
| | 8 2 | 4+5 2 | 3+4 2 | 3+4 2 |
| | 9 2 | 8 2 | 3+4 2 | 5+6 2 |
| | 4+5 2 | 9 2 | 3+4 2 | 3+4 2 |
| | 9 2 | 6+7 2 | 3+4 2 | 5+6 2 |
| | 6+7 2 | 6+7 2 | 5+6 2 | 3+4 2 |
| | 9 2 | 9 2 | 5+6 2 | 3+4 2 |
| | 6+7 2 | 6+7 2 | 3+4 2 | 5+6 2 |
| | 9 2 | 9 2 | 5+6 2 | 5+6 2 |
| | 8 2 | 6+7 2 | 5+6 2 | 3+4 2 |
| | 4+5 2 | 9 2 | 5+6 2 | 5+6 2 |
| | 8 2 | 9 2 | 3+4 2 | 3+4 2 |
| | 8 2 | 4+5 2 | 5+6 2 | 5+6 2 |
| | 9 2 | 8 2 | 3+4 2 | 3+4 2 |
| | 6+7 2 | 4+5 2 | 5+6 2 | 5+6 2 |
| | 9 2 | 8 2 | 3+4 2 | 3+4 2 |
| | 8 2 | 8 2 | 5+6 2 | 3+4 2 |
| | 6+7 2 | 9 2 | 5+6 2 | 5+6 2 |
| | 9 2 | 8 2 | 5+6 2 | 3+4 2 |
| | 4+5 2 | 4+5 2 | 3+4 2 | 5+6 2 |
| | 8 2 | 8 2 | 3+4 2 | 3+4 2 |
| | 4+5 2 | 6+7 2 | 3+4 2 | 5+6 2 |
| | 8 2 | 9 2 | 3+4 2 | 3+4 2 |
| | 4+5 2 | 8 2 | 5+6 2 | 5+6 2 |
| | 8 2 | 4+5 2 | 5+6 2 | 3+4 2 |
| | 9 2 | 8 2 | 3+4 2 | 3+4 2 |
| | 8 2 | 9 2 | 5+6 2 | 5+6 2 |
| | 4+5 2 | 4+5 2 | 5+6 2 | 5+6 2 |
| | 6+7 2 | 6+7 2 | 5+6 2 | 5+6 2 |
| | 8 2 | 8 2 | 3+4 2 | 3+4 2 |
| | 8 2 | 9 2 | 3+4 2 | 5+6 2 |
| | 9 2 | 8 2 | 5+6 2 | 3+4 2 |
| | 9 2 | 8 2 | 3+4 2 | 5+6 2 |
| | 6+7 2 | 6+7 2 | 5+6 2 | 3+4 2 |
| | 4+5 2 | 8 2 | 3+4 2 | 5+6 2 |
| | 8 2 | 9 2 | 3+4 2 | 5+6 2 |
| | 8 2 | 4+5 2 | 5+6 2 | 3+4 2 |

Continued on next column

**TABLE LXXIII– continued**

| TQC | ICSD | NoE | SF | |
|---|---|---|---|---|
| | 4+5 2 | 4+5 2 | 5+6 2 | 5+6 2 |
| | 8 2 | 8 2 | 3+4 2 | 3+4 2 |
| | 6+7 2 | 6+7 2 | 3+4 2 | 5+6 2 |
| 166.1.231 | 625438 | 174 | Co7Th2 | |
| | -GM | -T | -F | -L |
| | 8 2 | 8 2 | 3+4 2 | 3+4 2 |
| | 9 2 | 9 2 | 5+6 2 | 5+6 2 |
| | 8 2 | 9 2 | 5+6 2 | 3+4 2 |
| | 9 2 | 8 2 | 3+4 2 | 5+6 2 |
| | 8 2 | 9 2 | 5+6 2 | 3+4 2 |
| | 8 2 | 8 2 | 3+4 2 | 5+6 2 |
| | 9 2 | 9 2 | 5+6 2 | 3+4 2 |
| | 9 2 | 8 2 | 3+4 2 | 5+6 2 |
| | 4+5 2 | 6+7 2 | 5+6 2 | 3+4 2 |
| | 8 2 | 9 2 | 3+4 2 | 5+6 2 |
| | 9 2 | 8 2 | 3+4 2 | 3+4 2 |
| | 6+7 2 | 6+7 2 | 5+6 2 | 5+6 2 |
| | 8 2 | 4+5 2 | 5+6 2 | 3+4 2 |
| | 4+5 2 | 9 2 | 3+4 2 | 5+6 2 |
| | 9 2 | 8 2 | 5+6 2 | 3+4 2 |
| | 6+7 2 | 4+5 2 | 3+4 2 | 5+6 2 |
| | 8 2 | 8 2 | 3+4 2 | 3+4 2 |
| | 9 2 | 9 2 | 5+6 2 | 5+6 2 |
| | 8 2 | 8 2 | 3+4 2 | 5+6 2 |
| | 9 2 | 9 2 | 5+6 2 | 3+4 2 |
| | 8 2 | 8 2 | 5+6 2 | 5+6 2 |
| | 8 2 | 6+7 2 | 3+4 2 | 5+6 2 |
| | 4+5 2 | 9 2 | 5+6 2 | 3+4 2 |
| | 8 2 | 4+5 2 | 5+6 2 | 5+6 2 |
| | 6+7 2 | 8 2 | 3+4 2 | 5+6 2 |
| | 9 2 | 8 2 | 3+4 2 | 3+4 2 |
| | 8 2 | 9 2 | 3+4 2 | 3+4 2 |
| | 4+5 2 | 9 2 | 5+6 2 | 5+6 2 |
| | 8 2 | 4+5 2 | 3+4 2 | 5+6 2 |
| | 8 2 | 8 2 | 5+6 2 | 5+6 2 |
| | 9 2 | 8 2 | 3+4 2 | 3+4 2 |
| | 8 2 | 8 2 | 5+6 2 | 3+4 2 |
| | 4+5 2 | 4+5 2 | 3+4 2 | 5+6 2 |
| | 8 2 | 8 2 | 5+6 2 | 5+6 2 |
| | 9 2 | 9 2 | 5+6 2 | 5+6 2 |
| | 9 2 | 6+7 2 | 3+4 2 | 3+4 2 |
| | 6+7 2 | 4+5 2 | 5+6 2 | 5+6 2 |
| | 4+5 2 | 9 2 | 3+4 2 | 3+4 2 |
| | 8 2 | 8 2 | 5+6 2 | 3+4 2 |
| | 6+7 2 | 6+7 2 | 3+4 2 | 5+6 2 |
| | 9 2 | 9 2 | 3+4 2 | 5+6 2 |
| | 4+5 2 | 8 2 | 5+6 2 | 3+4 2 |
| | 8 2 | 8 2 | 3+4 2 | 5+6 2 |
| | 8 2 | 4+5 2 | 3+4 2 | 3+4 2 |
| | 4+5 2 | 8 2 | 5+6 2 | 5+6 2 |
| | 8 2 | 9 2 | 5+6 2 | 3+4 2 |
| | 9 2 | 6+7 2 | 3+4 2 | 5+6 2 |
| | 9 2 | 8 2 | 5+6 2 | 3+4 2 |
| | 6+7 2 | 4+5 2 | 3+4 2 | 5+6 2 |
| | 8 2 | 9 2 | 5+6 2 | 5+6 2 |
| | 4+5 2 | 6+7 2 | 3+4 2 | 3+4 2 |
| | 9 2 | 9 2 | 5+6 2 | 5+6 2 |
| | 6+7 2 | 4+5 2 | 5+6 2 | 5+6 2 |
| | 9 2 | 8 2 | 3+4 2 | 3+4 2 |
| | 8 2 | 9 2 | 5+6 2 | 3+4 2 |
| | 8 2 | 9 2 | 3+4 2 | 3+4 2 |

Continued on next column

**TABLE LXXIII– continued**

| TQC | ICSD | NoE | SF | |
|---|---|---|---|---|
| | 9 2 | 6+7 2 | 3+4 2 | 5+6 2 |
| | 9 2 | 8 2 | 5+6 2 | 3+4 2 |
| | 8 2 | 9 2 | 5+6 2 | 5+6 2 |
| | 4+5 2 | 8 2 | 3+4 2 | 5+6 2 |
| | 6+7 2 | 4+5 2 | 5+6 2 | 3+4 2 |
| | 8 2 | 6+7 2 | 3+4 2 | 5+6 2 |
| | 4+5 2 | 9 2 | 3+4 2 | 3+4 2 |
| | 8 2 | 8 2 | 5+6 2 | 5+6 2 |
| | 4+5 2 | 6+7 2 | 3+4 2 | 3+4 2 |
| | 9 2 | 8 2 | 5+6 2 | 3+4 2 |
| | 9 2 | 4+5 2 | 3+4 2 | 5+6 2 |
| | 6+7 2 | 9 2 | 3+4 2 | 3+4 2 |
| | 8 2 | 8 2 | 5+6 2 | 5+6 2 |
| | 4+5 2 | 4+5 2 | 5+6 2 | 3+4 2 |
| | 6+7 2 | 9 2 | 3+4 2 | 3+4 2 |
| | 8 2 | 8 2 | 5+6 2 | 5+6 2 |
| | 8 2 | 8 2 | 3+4 2 | 3+4 2 |
| | 4+5 2 | 4+5 2 | 5+6 2 | 5+6 2 |
| | 9 2 | 8 2 | 3+4 2 | 5+6 2 |
| | 8 2 | 8 2 | 5+6 2 | 3+4 2 |
| | 8 2 | 6+7 2 | 3+4 2 | 5+6 2 |
| | 8 2 | 9 2 | 5+6 2 | 5+6 2 |
| | 9 2 | 6+7 2 | 5+6 2 | 3+4 2 |
| | 6+7 2 | 9 2 | 3+4 2 | 5+6 2 |
| | 4+5 2 | 4+5 2 | 3+4 2 | 5+6 2 |
| | 6+7 2 | 8 2 | 5+6 2 | 3+4 2 |
| | 8 2 | 6+7 2 | 3+4 2 | 5+6 2 |
| | 8 2 | 9 2 | 3+4 2 | 3+4 2 |
| | 4+5 2 | 8 2 | 5+6 2 | 3+4 2 |
| | 9 2 | 4+5 2 | 5+6 2 | 5+6 2 |
| | 8 2 | 9 2 | 5+6 2 | 5+6 2 |
| 166.1.232 | 625416 | 174 | Co7Th2 | |
| | -GM | -T | -F | -L |
| | 8 2 | 8 2 | 3+4 2 | 3+4 2 |
| | 9 2 | 9 2 | 5+6 2 | 5+6 2 |
| | 8 2 | 9 2 | 5+6 2 | 3+4 2 |
| | 9 2 | 8 2 | 3+4 2 | 5+6 2 |
| | 8 2 | 9 2 | 5+6 2 | 3+4 2 |
| | 8 2 | 8 2 | 3+4 2 | 5+6 2 |
| | 9 2 | 9 2 | 5+6 2 | 3+4 2 |
| | 9 2 | 8 2 | 3+4 2 | 5+6 2 |
| | 4+5 2 | 6+7 2 | 5+6 2 | 3+4 2 |
| | 8 2 | 9 2 | 3+4 2 | 5+6 2 |
| | 9 2 | 8 2 | 3+4 2 | 3+4 2 |
| | 6+7 2 | 6+7 2 | 5+6 2 | 5+6 2 |
| | 8 2 | 4+5 2 | 5+6 2 | 3+4 2 |
| | 4+5 2 | 9 2 | 3+4 2 | 5+6 2 |
| | 9 2 | 8 2 | 5+6 2 | 3+4 2 |
| | 6+7 2 | 4+5 2 | 3+4 2 | 5+6 2 |
| | 8 2 | 8 2 | 3+4 2 | 3+4 2 |
| | 9 2 | 9 2 | 5+6 2 | 5+6 2 |
| | 8 2 | 8 2 | 3+4 2 | 5+6 2 |
| | 9 2 | 9 2 | 5+6 2 | 3+4 2 |
| | 8 2 | 8 2 | 5+6 2 | 5+6 2 |
| | 8 2 | 6+7 2 | 3+4 2 | 5+6 2 |
| | 4+5 2 | 9 2 | 5+6 2 | 3+4 2 |
| | 8 2 | 4+5 2 | 5+6 2 | 5+6 2 |
| | 6+7 2 | 8 2 | 3+4 2 | 5+6 2 |
| | 9 2 | 8 2 | 3+4 2 | 3+4 2 |
| | 8 2 | 9 2 | 3+4 2 | 3+4 2 |
| | 4+5 2 | 4+5 2 | 5+6 2 | 5+6 2 |

Continued on next column

**TABLE LXXIII– continued**

| TQC | ICSD | NoE | SF | |
|---|---|---|---|---|
| | 8 2 | 9 2 | 3+4 2 | 5+6 2 |
| | 8 2 | 8 2 | 5+6 2 | 5+6 2 |
| | 9 2 | 8 2 | 3+4 2 | 3+4 2 |
| | 8 2 | 8 2 | 5+6 2 | 3+4 2 |
| | 4+5 2 | 4+5 2 | 3+4 2 | 5+6 2 |
| | 8 2 | 8 2 | 5+6 2 | 5+6 2 |
| | 9 2 | 9 2 | 5+6 2 | 5+6 2 |
| | 9 2 | 6+7 2 | 3+4 2 | 3+4 2 |
| | 6+7 2 | 4+5 2 | 5+6 2 | 5+6 2 |
| | 4+5 2 | 8 2 | 3+4 2 | 3+4 2 |
| | 8 2 | 9 2 | 5+6 2 | 3+4 2 |
| | 6+7 2 | 6+7 2 | 3+4 2 | 5+6 2 |
| | 9 2 | 9 2 | 3+4 2 | 5+6 2 |
| | 4+5 2 | 8 2 | 5+6 2 | 3+4 2 |
| | 8 2 | 8 2 | 3+4 2 | 5+6 2 |
| | 8 2 | 4+5 2 | 5+6 2 | 3+4 2 |
| | 4+5 2 | 8 2 | 3+4 2 | 5+6 2 |
| | 8 2 | 9 2 | 5+6 2 | 3+4 2 |
| | 9 2 | 6+7 2 | 3+4 2 | 5+6 2 |
| | 9 2 | 8 2 | 5+6 2 | 3+4 2 |
| | 6+7 2 | 4+5 2 | 3+4 2 | 5+6 2 |
| | 8 2 | 9 2 | 5+6 2 | 5+6 2 |
| | 4+5 2 | 6+7 2 | 3+4 2 | 3+4 2 |
| | 9 2 | 9 2 | 5+6 2 | 5+6 2 |
| | 6+7 2 | 4+5 2 | 5+6 2 | 5+6 2 |
| | 9 2 | 8 2 | 3+4 2 | 3+4 2 |
| | 8 2 | 9 2 | 5+6 2 | 3+4 2 |
| | 9 2 | 9 2 | 3+4 2 | 3+4 2 |
| | 8 2 | 6+7 2 | 3+4 2 | 5+6 2 |
| | 9 2 | 8 2 | 5+6 2 | 3+4 2 |
| | 8 2 | 9 2 | 5+6 2 | 5+6 2 |
| | 4+5 2 | 8 2 | 3+4 2 | 5+6 2 |
| | 6+7 2 | 4+5 2 | 5+6 2 | 3+4 2 |
| | 8 2 | 6+7 2 | 3+4 2 | 5+6 2 |
| | 4+5 2 | 9 2 | 3+4 2 | 3+4 2 |
| | 8 2 | 8 2 | 5+6 2 | 5+6 2 |
| | 4+5 2 | 6+7 2 | 3+4 2 | 3+4 2 |
| | 9 2 | 8 2 | 5+6 2 | 3+4 2 |
| | 9 2 | 4+5 2 | 3+4 2 | 5+6 2 |
| | 6+7 2 | 9 2 | 3+4 2 | 3+4 2 |
| | 8 2 | 8 2 | 5+6 2 | 5+6 2 |
| | 4+5 2 | 4+5 2 | 5+6 2 | 3+4 2 |
| | 6+7 2 | 9 2 | 3+4 2 | 3+4 2 |
| | 8 2 | 8 2 | 5+6 2 | 5+6 2 |
| | 8 2 | 8 2 | 3+4 2 | 3+4 2 |
| | 4+5 2 | 4+5 2 | 5+6 2 | 5+6 2 |
| | 9 2 | 8 2 | 3+4 2 | 5+6 2 |
| | 8 2 | 8 2 | 5+6 2 | 3+4 2 |
| | 8 2 | 6+7 2 | 3+4 2 | 5+6 2 |
| | 9 2 | 9 2 | 5+6 2 | 5+6 2 |
| | 8 2 | 6+7 2 | 5+6 2 | 3+4 2 |
| | 6+7 2 | 9 2 | 3+4 2 | 5+6 2 |
| | 4+5 2 | 4+5 2 | 3+4 2 | 5+6 2 |
| | 6+7 2 | 8 2 | 5+6 2 | 3+4 2 |
| | 8 2 | 6+7 2 | 3+4 2 | 5+6 2 |
| | 8 2 | 9 2 | 3+4 2 | 3+4 2 |
| | 4+5 2 | 8 2 | 5+6 2 | 3+4 2 |
| | 9 2 | 4+5 2 | 5+6 2 | 5+6 2 |
| | 8 2 | 9 2 | 5+6 2 | 5+6 2 |
| 166.1.233 | 184367 | 178 | Co7Pr2 | |
| | -GM | -T | -F | -L |

Continued on next column

**TABLE LXXIII– continued**

| TQC | ICSD | NoE | SF | |
|---|---|---|---|---|
| | 8 2 | 9 2 | 3+4 2 | 3+4 2 |
| | 9 2 | 8 2 | 5+6 2 | 5+6 2 |
| | 8 2 | 9 2 | 5+6 2 | 3+4 2 |
| | 9 2 | 8 2 | 3+4 2 | 5+6 2 |
| | 8 2 | 9 2 | 5+6 2 | 3+4 2 |
| | 8 2 | 8 2 | 3+4 2 | 5+6 2 |
| | 9 2 | 9 2 | 5+6 2 | 3+4 2 |
| | 9 2 | 8 2 | 3+4 2 | 5+6 2 |
| | 4+5 2 | 6+7 2 | 5+6 2 | 3+4 2 |
| | 8 2 | 9 2 | 3+4 2 | 5+6 2 |
| | 9 2 | 8 2 | 5+6 2 | 3+4 2 |
| | 6+7 2 | 6+7 2 | 3+4 2 | 5+6 2 |
| | 4+5 2 | 4+5 2 | 5+6 2 | 3+4 2 |
| | 8 2 | 9 2 | 3+4 2 | 5+6 2 |
| | 9 2 | 8 2 | 5+6 2 | 3+4 2 |
| | 6+7 2 | 4+5 2 | 3+4 2 | 5+6 2 |
| | 8 2 | 8 2 | 5+6 2 | 5+6 2 |
| | 8 2 | 9 2 | 3+4 2 | 3+4 2 |
| | 9 2 | 8 2 | 3+4 2 | 5+6 2 |
| | 8 2 | 9 2 | 5+6 2 | 3+4 2 |
| | 4+5 2 | 8 2 | 5+6 2 | 5+6 2 |
| | 8 2 | 4+5 2 | 3+4 2 | 5+6 2 |
| | 9 2 | 8 2 | 5+6 2 | 5+6 2 |
| | 8 2 | 8 2 | 5+6 2 | 5+6 2 |
| | 8 2 | 9 2 | 5+6 2 | 3+4 2 |
| | 4+5 2 | 6+7 2 | 3+4 2 | 3+4 2 |
| | 8 2 | 9 2 | 3+4 2 | 5+6 2 |
| | 4+5 2 | 4+5 2 | 3+4 2 | 3+4 2 |
| | 9 2 | 8 2 | 5+6 2 | 3+4 2 |
| | 8 2 | 9 2 | 3+4 2 | 5+6 2 |
| | 6+7 2 | 4+5 2 | 5+6 2 | 5+6 2 |
| | 8 2 | 8 2 | 3+4 2 | 3+4 2 |
| | 9 2 | 8 2 | 5+6 2 | 5+6 2 |
| | 8 2 | 4+5 2 | 3+4 2 | 5+6 2 |
| | 4+5 2 | 8 2 | 3+4 2 | 5+6 2 |
| | 8 2 | 8 2 | 5+6 2 | 3+4 2 |
| | 9 2 | 9 2 | 5+6 2 | 5+6 2 |
| | 6+7 2 | 6+7 2 | 3+4 2 | 3+4 2 |
| | 9 2 | 9 2 | 5+6 2 | 3+4 2 |
| | 8 2 | 8 2 | 3+4 2 | 5+6 2 |
| | 4+5 2 | 6+7 2 | 3+4 2 | 3+4 2 |
| | 8 2 | 8 2 | 5+6 2 | 5+6 2 |
| | 4+5 2 | 9 2 | 3+4 2 | 5+6 2 |
| | 9 2 | 4+5 2 | 3+4 2 | 5+6 2 |
| | 6+7 2 | 8 2 | 3+4 2 | 3+4 2 |
| | 8 2 | 4+5 2 | 5+6 2 | 5+6 2 |
| | 9 2 | 9 2 | 5+6 2 | 3+4 2 |
| | 6+7 2 | 8 2 | 5+6 2 | 5+6 2 |
| | 9 2 | 9 2 | 3+4 2 | 5+6 2 |
| | 8 2 | 6+7 2 | 5+6 2 | 3+4 2 |
| | 9 2 | 6+7 2 | 3+4 2 | 3+4 2 |
| | 4+5 2 | 4+5 2 | 3+4 2 | 5+6 2 |
| | 6+7 2 | 8 2 | 5+6 2 | 5+6 2 |
| | 8 2 | 9 2 | 3+4 2 | 3+4 2 |
| | 9 2 | 9 2 | 5+6 2 | 3+4 2 |
| | 4+5 2 | 6+7 2 | 5+6 2 | 5+6 2 |
| | 8 2 | 9 2 | 3+4 2 | 5+6 2 |
| | 8 2 | 8 2 | 5+6 2 | 3+4 2 |
| | 6+7 2 | 6+7 2 | 5+6 2 | 3+4 2 |
| | 9 2 | 4+5 2 | 3+4 2 | 5+6 2 |
| | 9 2 | 9 2 | 5+6 2 | 3+4 2 |

Continued on next column

**TABLE LXXIII– continued**

| TQC | ICSD | NoE | SF | |
|---|---|---|---|---|
| | 9 2 | 9 2 | 3+4 2 | 3+4 2 |
| | 6+7 2 | 8 2 | 3+4 2 | 5+6 2 |
| | 4+5 2 | 8 2 | 3+4 2 | 5+6 2 |
| | 8 2 | 4+5 2 | 5+6 2 | 3+4 2 |
| | 8 2 | 6+7 2 | 3+4 2 | 3+4 2 |
| | 4+5 2 | 8 2 | 5+6 2 | 5+6 2 |
| | 9 2 | 8 2 | 3+4 2 | 3+4 2 |
| | 4+5 2 | 9 2 | 5+6 2 | 5+6 2 |
| | 8 2 | 8 2 | 5+6 2 | 3+4 2 |
| | 8 2 | 4+5 2 | 5+6 2 | 5+6 2 |
| | 6+7 2 | 6+7 2 | 3+4 2 | 5+6 2 |
| | 8 2 | 9 2 | 3+4 2 | 3+4 2 |
| | 8 2 | 9 2 | 5+6 2 | 3+4 2 |
| | 9 2 | 8 2 | 3+4 2 | 5+6 2 |
| | 4+5 2 | 6+7 2 | 5+6 2 | 3+4 2 |
| | 8 2 | 9 2 | 3+4 2 | 5+6 2 |
| | 8 2 | 4+5 2 | 5+6 2 | 3+4 2 |
| | 9 2 | 8 2 | 3+4 2 | 5+6 2 |
| | 6+7 2 | 9 2 | 5+6 2 | 5+6 2 |
| | 4+5 2 | 4+5 2 | 3+4 2 | 3+4 2 |
| | 8 2 | 8 2 | 5+6 2 | 5+6 2 |
| | 9 2 | 8 2 | 3+4 2 | 5+6 2 |
| | 4+5 2 | 6+7 2 | 5+6 2 | 3+4 2 |
| | 8 2 | 8 2 | 3+4 2 | 5+6 2 |
| | 9 2 | 4+5 2 | 3+4 2 | 3+4 2 |
| | 6+7 2 | 9 2 | 5+6 2 | 5+6 2 |
| | 8 2 | 8 2 | 3+4 2 | 3+4 2 |
| | 8 2 | 9 2 | 5+6 2 | 5+6 2 |
| 166.1.234 | 623985 | 170 | Co7La2 | |
| | -GM | -T | -F | -L |
| | 8 2 | 9 2 | 3+4 2 | 3+4 2 |
| | 9 2 | 8 2 | 5+6 2 | 5+6 2 |
| | 8 2 | 9 2 | 5+6 2 | 3+4 2 |
| | 9 2 | 8 2 | 3+4 2 | 5+6 2 |
| | 8 2 | 9 2 | 5+6 2 | 3+4 2 |
| | 8 2 | 8 2 | 3+4 2 | 5+6 2 |
| | 9 2 | 9 2 | 5+6 2 | 3+4 2 |
| | 9 2 | 8 2 | 3+4 2 | 5+6 2 |
| | 4+5 2 | 6+7 2 | 5+6 2 | 3+4 2 |
| | 8 2 | 9 2 | 3+4 2 | 5+6 2 |
| | 9 2 | 8 2 | 5+6 2 | 3+4 2 |
| | 6+7 2 | 6+7 2 | 3+4 2 | 5+6 2 |
| | 4+5 2 | 4+5 2 | 5+6 2 | 3+4 2 |
| | 8 2 | 9 2 | 3+4 2 | 5+6 2 |
| | 9 2 | 8 2 | 5+6 2 | 3+4 2 |
| | 6+7 2 | 4+5 2 | 3+4 2 | 5+6 2 |
| | 8 2 | 8 2 | 5+6 2 | 5+6 2 |
| | 9 2 | 9 2 | 3+4 2 | 3+4 2 |
| | 8 2 | 8 2 | 3+4 2 | 5+6 2 |
| | 9 2 | 9 2 | 5+6 2 | 3+4 2 |
| | 8 2 | 8 2 | 5+6 2 | 5+6 2 |
| | 8 2 | 8 2 | 3+4 2 | 5+6 2 |
| | 8 2 | 9 2 | 5+6 2 | 5+6 2 |
| | 4+5 2 | 6+7 2 | 3+4 2 | 3+4 2 |
| | 9 2 | 9 2 | 5+6 2 | 5+6 2 |
| | 6+7 2 | 8 2 | 3+4 2 | 3+4 2 |
| | 8 2 | 4+5 2 | 3+4 2 | 3+4 2 |
| | 4+5 2 | 8 2 | 5+6 2 | 5+6 2 |
| | 8 2 | 4+5 2 | 3+4 2 | 5+6 2 |
| | 9 2 | 9 2 | 3+4 2 | 3+4 2 |
| | 8 2 | 8 2 | 5+6 2 | 5+6 2 |

Continued on next column

**TABLE LXXIII– continued**

| TQC | ICSD | NoE | SF | |
|---|---|---|---|---|
| | 4+5 2 | 4+5 2 | 5+6 2 | 3+4 2 |
| | 8 2 | 8 2 | 5+6 2 | 5+6 2 |
| | 8 2 | 8 2 | 5+6 2 | 5+6 2 |
| | 4+5 2 | 9 2 | 3+4 2 | 3+4 2 |
| | 8 2 | 6+7 2 | 3+4 2 | 5+6 2 |
| | 6+7 2 | 4+5 2 | 5+6 2 | 5+6 2 |
| | 9 2 | 8 2 | 3+4 2 | 5+6 2 |
| | 9 2 | 9 2 | 5+6 2 | 3+4 2 |
| | 4+5 2 | 6+7 2 | 3+4 2 | 5+6 2 |
| | 8 2 | 9 2 | 3+4 2 | 3+4 2 |
| | 6+7 2 | 8 2 | 5+6 2 | 5+6 2 |
| | 9 2 | 8 2 | 3+4 2 | 3+4 2 |
| | 8 2 | 4+5 2 | 3+4 2 | 5+6 2 |
| | 4+5 2 | 8 2 | 3+4 2 | 3+4 2 |
| | 8 2 | 4+5 2 | 5+6 2 | 3+4 2 |
| | 9 2 | 9 2 | 5+6 2 | 3+4 2 |
| | 8 2 | 6+7 2 | 5+6 2 | 5+6 2 |
| | 6+7 2 | 8 2 | 3+4 2 | 5+6 2 |
| | 4+5 2 | 9 2 | 3+4 2 | 3+4 2 |
| | 9 2 | 6+7 2 | 5+6 2 | 5+6 2 |
| | 9 2 | 9 2 | 3+4 2 | 3+4 2 |
| | 8 2 | 9 2 | 5+6 2 | 5+6 2 |
| | 6+7 2 | 4+5 2 | 3+4 2 | 3+4 2 |
| | 9 2 | 6+7 2 | 5+6 2 | 5+6 2 |
| | 8 2 | 8 2 | 5+6 2 | 5+6 2 |
| | 9 2 | 9 2 | 5+6 2 | 5+6 2 |
| | 4+5 2 | 8 2 | 3+4 2 | 3+4 2 |
| | 6+7 2 | 4+5 2 | 5+6 2 | 3+4 2 |
| | 9 2 | 6+7 2 | 3+4 2 | 5+6 2 |
| | 8 2 | 9 2 | 5+6 2 | 3+4 2 |
| | 8 2 | 8 2 | 3+4 2 | 5+6 2 |
| | 4+5 2 | 9 2 | 3+4 2 | 3+4 2 |
| | 8 2 | 8 2 | 5+6 2 | 5+6 2 |
| | 4+5 2 | 4+5 2 | 3+4 2 | 3+4 2 |
| | 6+7 2 | 6+7 2 | 5+6 2 | 3+4 2 |
| | 9 2 | 8 2 | 3+4 2 | 5+6 2 |
| | 8 2 | 4+5 2 | 3+4 2 | 5+6 2 |
| | 4+5 2 | 8 2 | 5+6 2 | 3+4 2 |
| | 9 2 | 9 2 | 3+4 2 | 3+4 2 |
| | 4+5 2 | 8 2 | 5+6 2 | 3+4 2 |
| | 8 2 | 4+5 2 | 5+6 2 | 5+6 2 |
| | 8 2 | 9 2 | 5+6 2 | 5+6 2 |
| | 6+7 2 | 8 2 | 3+4 2 | 3+4 2 |
| | 8 2 | 8 2 | 5+6 2 | 5+6 2 |
| | 8 2 | 6+7 2 | 3+4 2 | 3+4 2 |
| | 9 2 | 9 2 | 3+4 2 | 5+6 2 |
| | 9 2 | 6+7 2 | 5+6 2 | 5+6 2 |
| | 6+7 2 | 8 2 | 3+4 2 | 5+6 2 |
| | 4+5 2 | 9 2 | 5+6 2 | 3+4 2 |
| | 8 2 | 4+5 2 | 3+4 2 | 3+4 2 |
| | 8 2 | 8 2 | 5+6 2 | 5+6 2 |
| | 4+5 2 | 8 2 | 3+4 2 | 5+6 2 |
| | 8 2 | 4+5 2 | 5+6 2 | 3+4 2 |
| | 6+7 2 | 6+7 2 | 3+4 2 | 5+6 2 |
| 166.1.235 | 623978 | 170 | Co7La2 | |
| | -GM | -T | -F | -L |
| | 8 2 | 9 2 | 3+4 2 | 3+4 2 |
| | 9 2 | 8 2 | 5+6 2 | 5+6 2 |
| | 8 2 | 9 2 | 5+6 2 | 3+4 2 |
| | 9 2 | 8 2 | 3+4 2 | 5+6 2 |
| | 8 2 | 9 2 | 5+6 2 | 3+4 2 |

Continued on next column

**TABLE LXXIII– continued**

| TQC | ICSD | NoE | SF | |
|---|---|---|---|---|
| | 8 2 | 8 2 | 3+4 2 | 5+6 2 |
| | 9 2 | 9 2 | 5+6 2 | 3+4 2 |
| | 9 2 | 8 2 | 3+4 2 | 5+6 2 |
| | 4+5 2 | 6+7 2 | 5+6 2 | 3+4 2 |
| | 8 2 | 9 2 | 3+4 2 | 5+6 2 |
| | 9 2 | 8 2 | 5+6 2 | 3+4 2 |
| | 6+7 2 | 6+7 2 | 3+4 2 | 5+6 2 |
| | 4+5 2 | 4+5 2 | 5+6 2 | 3+4 2 |
| | 8 2 | 9 2 | 3+4 2 | 5+6 2 |
| | 9 2 | 8 2 | 5+6 2 | 3+4 2 |
| | 6+7 2 | 4+5 2 | 3+4 2 | 5+6 2 |
| | 8 2 | 8 2 | 5+6 2 | 5+6 2 |
| | 9 2 | 9 2 | 3+4 2 | 3+4 2 |
| | 8 2 | 8 2 | 3+4 2 | 5+6 2 |
| | 8 2 | 9 2 | 5+6 2 | 3+4 2 |
| | 9 2 | 8 2 | 5+6 2 | 5+6 2 |
| | 8 2 | 8 2 | 3+4 2 | 5+6 2 |
| | 8 2 | 9 2 | 5+6 2 | 5+6 2 |
| | 4+5 2 | 6+7 2 | 5+6 2 | 3+4 2 |
| | 9 2 | 9 2 | 3+4 2 | 5+6 2 |
| | 6+7 2 | 8 2 | 3+4 2 | 3+4 2 |
| | 4+5 2 | 4+5 2 | 5+6 2 | 3+4 2 |
| | 8 2 | 8 2 | 3+4 2 | 5+6 2 |
| | 8 2 | 4+5 2 | 3+4 2 | 3+4 2 |
| | 9 2 | 9 2 | 5+6 2 | 5+6 2 |
| | 8 2 | 8 2 | 3+4 2 | 5+6 2 |
| | 4+5 2 | 4+5 2 | 5+6 2 | 3+4 2 |
| | 8 2 | 8 2 | 5+6 2 | 5+6 2 |
| | 8 2 | 8 2 | 3+4 2 | 5+6 2 |
| | 4+5 2 | 9 2 | 5+6 2 | 3+4 2 |
| | 8 2 | 6+7 2 | 3+4 2 | 5+6 2 |
| | 6+7 2 | 4+5 2 | 5+6 2 | 5+6 2 |
| | 9 2 | 8 2 | 3+4 2 | 5+6 2 |
| | 9 2 | 9 2 | 5+6 2 | 3+4 2 |
| | 6+7 2 | 6+7 2 | 3+4 2 | 5+6 2 |
| | 4+5 2 | 9 2 | 3+4 2 | 3+4 2 |
| | 8 2 | 8 2 | 5+6 2 | 5+6 2 |
| | 9 2 | 8 2 | 3+4 2 | 3+4 2 |
| | 8 2 | 4+5 2 | 3+4 2 | 3+4 2 |
| | 4+5 2 | 8 2 | 3+4 2 | 5+6 2 |
| | 8 2 | 4+5 2 | 5+6 2 | 3+4 2 |
| | 9 2 | 9 2 | 5+6 2 | 5+6 2 |
| | 8 2 | 8 2 | 5+6 2 | 5+6 2 |
| | 6+7 2 | 6+7 2 | 3+4 2 | 3+4 2 |
| | 4+5 2 | 9 2 | 3+4 2 | 3+4 2 |
| | 9 2 | 6+7 2 | 5+6 2 | 5+6 2 |
| | 9 2 | 9 2 | 3+4 2 | 5+6 2 |
| | 8 2 | 9 2 | 5+6 2 | 3+4 2 |
| | 9 2 | 4+5 2 | 3+4 2 | 3+4 2 |
| | 6+7 2 | 6+7 2 | 5+6 2 | 5+6 2 |
| | 8 2 | 8 2 | 5+6 2 | 5+6 2 |
| | 9 2 | 9 2 | 5+6 2 | 5+6 2 |
| | 4+5 2 | 8 2 | 3+4 2 | 3+4 2 |
| | 6+7 2 | 4+5 2 | 5+6 2 | 3+4 2 |
| | 9 2 | 6+7 2 | 3+4 2 | 5+6 2 |
| | 8 2 | 8 2 | 5+6 2 | 3+4 2 |
| | 8 2 | 9 2 | 3+4 2 | 5+6 2 |
| | 4+5 2 | 9 2 | 3+4 2 | 3+4 2 |
| | 8 2 | 4+5 2 | 5+6 2 | 5+6 2 |
| | 4+5 2 | 6+7 2 | 3+4 2 | 3+4 2 |
| | 6+7 2 | 8 2 | 5+6 2 | 3+4 2 |

Continued on next column

**TABLE LXXIII– continued**

| TQC | ICSD | NoE | SF | |
|---|---|---|---|---|
| | 9 2 | 8 2 | 3+4 2 | 5+6 2 |
| | 4+5 2 | 4+5 2 | 3+4 2 | 5+6 2 |
| | 8 2 | 9 2 | 5+6 2 | 3+4 2 |
| | 9 2 | 8 2 | 3+4 2 | 3+4 2 |
| | 8 2 | 8 2 | 5+6 2 | 3+4 2 |
| | 6+7 2 | 4+5 2 | 5+6 2 | 5+6 2 |
| | 8 2 | 8 2 | 3+4 2 | 5+6 2 |
| | 4+5 2 | 9 2 | 5+6 2 | 3+4 2 |
| | 8 2 | 8 2 | 5+6 2 | 5+6 2 |
| | 8 2 | 6+7 2 | 3+4 2 | 5+6 2 |
| | 9 2 | 9 2 | 5+6 2 | 3+4 2 |
| | 9 2 | 6+7 2 | 3+4 2 | 5+6 2 |
| | 4+5 2 | 8 2 | 3+4 2 | 5+6 2 |
| | 6+7 2 | 9 2 | 5+6 2 | 3+4 2 |
| | 8 2 | 4+5 2 | 3+4 2 | 3+4 2 |
| | 8 2 | 8 2 | 5+6 2 | 5+6 2 |
| | 4+5 2 | 4+5 2 | 3+4 2 | 5+6 2 |
| | 8 2 | 8 2 | 5+6 2 | 3+4 2 |
| | 6+7 2 | 6+7 2 | 3+4 2 | 5+6 2 |
| 166.1.236 | 623284 | 198 | Co7Gd2 | |
| | -GM | -T | -F | -L |
| | 8 2 | 8 2 | 3+4 2 | 3+4 2 |
| | 9 2 | 9 2 | 5+6 2 | 5+6 2 |
| | 8 2 | 9 2 | 5+6 2 | 3+4 2 |
| | 9 2 | 8 2 | 3+4 2 | 5+6 2 |
| | 8 2 | 9 2 | 5+6 2 | 3+4 2 |
| | 8 2 | 8 2 | 3+4 2 | 5+6 2 |
| | 9 2 | 9 2 | 5+6 2 | 3+4 2 |
| | 9 2 | 8 2 | 3+4 2 | 5+6 2 |
| | 4+5 2 | 6+7 2 | 5+6 2 | 3+4 2 |
| | 8 2 | 9 2 | 3+4 2 | 5+6 2 |
| | 9 2 | 8 2 | 3+4 2 | 3+4 2 |
| | 6+7 2 | 6+7 2 | 5+6 2 | 5+6 2 |
| | 4+5 2 | 4+5 2 | 5+6 2 | 3+4 2 |
| | 8 2 | 9 2 | 3+4 2 | 5+6 2 |
| | 9 2 | 8 2 | 5+6 2 | 3+4 2 |
| | 6+7 2 | 4+5 2 | 3+4 2 | 5+6 2 |
| | 8 2 | 8 2 | 5+6 2 | 5+6 2 |
| | 9 2 | 9 2 | 3+4 2 | 3+4 2 |
| | 8 2 | 8 2 | 3+4 2 | 5+6 2 |
| | 9 2 | 9 2 | 5+6 2 | 3+4 2 |
| | 8 2 | 8 2 | 5+6 2 | 5+6 2 |
| | 8 2 | 8 2 | 3+4 2 | 5+6 2 |
| | 8 2 | 9 2 | 5+6 2 | 3+4 2 |
| | 8 2 | 9 2 | 5+6 2 | 5+6 2 |
| | 4+5 2 | 9 2 | 3+4 2 | 5+6 2 |
| | 8 2 | 6+7 2 | 5+6 2 | 5+6 2 |
| | 4+5 2 | 8 2 | 3+4 2 | 3+4 2 |
| | 9 2 | 4+5 2 | 3+4 2 | 5+6 2 |
| | 6+7 2 | 8 2 | 5+6 2 | 3+4 2 |
| | 9 2 | 4+5 2 | 3+4 2 | 3+4 2 |
| | 8 2 | 8 2 | 3+4 2 | 3+4 2 |
| | 4+5 2 | 4+5 2 | 3+4 2 | 5+6 2 |
| | 8 2 | 8 2 | 5+6 2 | 5+6 2 |
| | 8 2 | 8 2 | 5+6 2 | 5+6 2 |
| | 9 2 | 8 2 | 3+4 2 | 3+4 2 |
| | 4+5 2 | 9 2 | 5+6 2 | 3+4 2 |
| | 8 2 | 6+7 2 | 5+6 2 | 5+6 2 |
| | 6+7 2 | 4+5 2 | 3+4 2 | 5+6 2 |
| | 9 2 | 9 2 | 5+6 2 | 5+6 2 |
| | 8 2 | 8 2 | 3+4 2 | 5+6 2 |

Continued on next column

**TABLE LXXIII– continued**

| TQC | ICSD | NoE | SF | |
|---|---|---|---|---|
| | 4+5 2 | 6+7 2 | 3+4 2 | 3+4 2 |
| | 8 2 | 9 2 | 5+6 2 | 3+4 2 |
| | 6+7 2 | 8 2 | 3+4 2 | 5+6 2 |
| | 4+5 2 | 4+5 2 | 5+6 2 | 3+4 2 |
| | 8 2 | 8 2 | 3+4 2 | 5+6 2 |
| | 9 2 | 4+5 2 | 3+4 2 | 3+4 2 |
| | 9 2 | 9 2 | 5+6 2 | 5+6 2 |
| | 8 2 | 8 2 | 3+4 2 | 3+4 2 |
| | 6+7 2 | 9 2 | 3+4 2 | 5+6 2 |
| | 4+5 2 | 6+7 2 | 3+4 2 | 5+6 2 |
| | 9 2 | 6+7 2 | 5+6 2 | 3+4 2 |
| | 9 2 | 9 2 | 5+6 2 | 3+4 2 |
| | 8 2 | 9 2 | 5+6 2 | 5+6 2 |
| | 9 2 | 6+7 2 | 5+6 2 | 5+6 2 |
| | 8 2 | 9 2 | 3+4 2 | 3+4 2 |
| | 6+7 2 | 6+7 2 | 5+6 2 | 5+6 2 |
| | 4+5 2 | 9 2 | 5+6 2 | 5+6 2 |
| | 8 2 | 8 2 | 3+4 2 | 3+4 2 |
| | 9 2 | 4+5 2 | 5+6 2 | 5+6 2 |
| | 6+7 2 | 4+5 2 | 3+4 2 | 3+4 2 |
| | 9 2 | 8 2 | 5+6 2 | 3+4 2 |
| | 9 2 | 8 2 | 3+4 2 | 5+6 2 |
| | 8 2 | 8 2 | 5+6 2 | 3+4 2 |
| | 4+5 2 | 9 2 | 3+4 2 | 3+4 2 |
| | 8 2 | 6+7 2 | 3+4 2 | 5+6 2 |
| | 6+7 2 | 4+5 2 | 5+6 2 | 5+6 2 |
| | 9 2 | 9 2 | 3+4 2 | 3+4 2 |
| | 6+7 2 | 8 2 | 3+4 2 | 3+4 2 |
| | 8 2 | 9 2 | 5+6 2 | 3+4 2 |
| | 8 2 | 9 2 | 5+6 2 | 5+6 2 |
| | 4+5 2 | 4+5 2 | 5+6 2 | 3+4 2 |
| | 9 2 | 8 2 | 3+4 2 | 5+6 2 |
| | 4+5 2 | 8 2 | 5+6 2 | 5+6 2 |
| | 8 2 | 8 2 | 3+4 2 | 3+4 2 |
| | 8 2 | 9 2 | 5+6 2 | 5+6 2 |
| | 9 2 | 6+7 2 | 3+4 2 | 3+4 2 |
| | 4+5 2 | 9 2 | 5+6 2 | 5+6 2 |
| | 8 2 | 4+5 2 | 3+4 2 | 3+4 2 |
| | 9 2 | 8 2 | 5+6 2 | 5+6 2 |
| | 6+7 2 | 8 2 | 3+4 2 | 3+4 2 |
| | 9 2 | 6+7 2 | 5+6 2 | 5+6 2 |
| | 8 2 | 8 2 | 3+4 2 | 3+4 2 |
| | 8 2 | 4+5 2 | 5+6 2 | 3+4 2 |
| | 4+5 2 | 8 2 | 3+4 2 | 5+6 2 |
| | 8 2 | 9 2 | 5+6 2 | 3+4 2 |
| | 8 2 | 6+7 2 | 3+4 2 | 5+6 2 |
| | 4+5 2 | 9 2 | 3+4 2 | 3+4 2 |
| | 9 2 | 4+5 2 | 5+6 2 | 5+6 2 |
| | 6+7 2 | 8 2 | 5+6 2 | 5+6 2 |
| | 9 2 | 9 2 | 3+4 2 | 3+4 2 |
| | 6+7 2 | 6+7 2 | 3+4 2 | 5+6 2 |
| | 4+5 2 | 8 2 | 5+6 2 | 5+6 2 |
| | 8 2 | 8 2 | 3+4 2 | 5+6 2 |
| | 8 2 | 4+5 2 | 5+6 2 | 3+4 2 |
| | 8 2 | 4+5 2 | 3+4 2 | 5+6 2 |
| | 4+5 2 | 8 2 | 5+6 2 | 3+4 2 |
| | 9 2 | 9 2 | 3+4 2 | 5+6 2 |
| | 8 2 | 9 2 | 5+6 2 | 3+4 2 |
| | 9 2 | 9 2 | 5+6 2 | 5+6 2 |
| 166.1.237 | 658884 | 206 | Co7Dy2 | |
| | -GM | -T | -F | -L |

Continued on next column

**TABLE LXXIII– continued**

| TQC | ICSD | NoE | SF | |
|---|---|---|---|---|
| | 8 2 | 8 2 | 3+4 2 | 3+4 2 |
| | 9 2 | 9 2 | 5+6 2 | 5+6 2 |
| | 8 2 | 9 2 | 5+6 2 | 3+4 2 |
| | 9 2 | 8 2 | 3+4 2 | 5+6 2 |
| | 8 2 | 9 2 | 3+4 2 | 3+4 2 |
| | 8 2 | 8 2 | 5+6 2 | 5+6 2 |
| | 9 2 | 9 2 | 5+6 2 | 3+4 2 |
| | 9 2 | 8 2 | 3+4 2 | 5+6 2 |
| | 4+5 2 | 6+7 2 | 5+6 2 | 3+4 2 |
| | 8 2 | 9 2 | 3+4 2 | 5+6 2 |
| | 9 2 | 8 2 | 5+6 2 | 3+4 2 |
| | 6+7 2 | 6+7 2 | 3+4 2 | 5+6 2 |
| | 4+5 2 | 4+5 2 | 3+4 2 | 3+4 2 |
| | 8 2 | 9 2 | 5+6 2 | 5+6 2 |
| | 9 2 | 8 2 | 5+6 2 | 3+4 2 |
| | 6+7 2 | 4+5 2 | 3+4 2 | 5+6 2 |
| | 8 2 | 8 2 | 5+6 2 | 5+6 2 |
| | 9 2 | 9 2 | 3+4 2 | 3+4 2 |
| | 8 2 | 8 2 | 3+4 2 | 5+6 2 |
| | 9 2 | 9 2 | 5+6 2 | 3+4 2 |
| | 8 2 | 8 2 | 5+6 2 | 5+6 2 |
| | 8 2 | 8 2 | 3+4 2 | 3+4 2 |
| | 8 2 | 9 2 | 5+6 2 | 5+6 2 |
| | 8 2 | 9 2 | 5+6 2 | 5+6 2 |
| | 4+5 2 | 9 2 | 3+4 2 | 5+6 2 |
| | 9 2 | 6+7 2 | 5+6 2 | 5+6 2 |
| | 8 2 | 8 2 | 3+4 2 | 3+4 2 |
| | 4+5 2 | 4+5 2 | 3+4 2 | 5+6 2 |
| | 9 2 | 8 2 | 5+6 2 | 3+4 2 |
| | 6+7 2 | 4+5 2 | 3+4 2 | 3+4 2 |
| | 8 2 | 8 2 | 3+4 2 | 3+4 2 |
| | 8 2 | 8 2 | 3+4 2 | 5+6 2 |
| | 4+5 2 | 4+5 2 | 5+6 2 | 5+6 2 |
| | 8 2 | 8 2 | 3+4 2 | 3+4 2 |
| | 9 2 | 8 2 | 5+6 2 | 5+6 2 |
| | 4+5 2 | 9 2 | 5+6 2 | 3+4 2 |
| | 8 2 | 6+7 2 | 5+6 2 | 5+6 2 |
| | 9 2 | 9 2 | 3+4 2 | 5+6 2 |
| | 6+7 2 | 4+5 2 | 5+6 2 | 5+6 2 |
| | 8 2 | 8 2 | 3+4 2 | 5+6 2 |
| | 4+5 2 | 6+7 2 | 3+4 2 | 3+4 2 |
| | 8 2 | 9 2 | 5+6 2 | 3+4 2 |
| | 6+7 2 | 8 2 | 5+6 2 | 5+6 2 |
| | 4+5 2 | 4+5 2 | 3+4 2 | 3+4 2 |
| | 8 2 | 9 2 | 3+4 2 | 5+6 2 |
| | 9 2 | 8 2 | 3+4 2 | 5+6 2 |
| | 9 2 | 4+5 2 | 5+6 2 | 3+4 2 |
| | 8 2 | 8 2 | 3+4 2 | 3+4 2 |
| | 6+7 2 | 9 2 | 3+4 2 | 5+6 2 |
| | 4+5 2 | 6+7 2 | 3+4 2 | 5+6 2 |
| | 9 2 | 6+7 2 | 5+6 2 | 3+4 2 |
| | 9 2 | 9 2 | 5+6 2 | 3+4 2 |
| | 9 2 | 9 2 | 5+6 2 | 5+6 2 |
| | 8 2 | 6+7 2 | 5+6 2 | 5+6 2 |
| | 8 2 | 9 2 | 3+4 2 | 3+4 2 |
| | 6+7 2 | 6+7 2 | 5+6 2 | 5+6 2 |
| | 4+5 2 | 9 2 | 5+6 2 | 5+6 2 |
| | 8 2 | 8 2 | 3+4 2 | 3+4 2 |
| | 9 2 | 4+5 2 | 3+4 2 | 5+6 2 |
| | 9 2 | 8 2 | 5+6 2 | 3+4 2 |
| | 6+7 2 | 9 2 | 5+6 2 | 3+4 2 |

Continued on next column

**TABLE LXXIII– continued**

| TQC | ICSD | NoE | SF | |
|---|---|---|---|---|
| | 8 2 | 8 2 | 3+4 2 | 5+6 2 |
| | 8 2 | 4+5 2 | 5+6 2 | 3+4 2 |
| | 9 2 | 9 2 | 3+4 2 | 3+4 2 |
| | 4+5 2 | 6+7 2 | 3+4 2 | 5+6 2 |
| | 9 2 | 8 2 | 5+6 2 | 5+6 2 |
| | 4+5 2 | 9 2 | 3+4 2 | 3+4 2 |
| | 8 2 | 9 2 | 5+6 2 | 3+4 2 |
| | 6+7 2 | 8 2 | 3+4 2 | 3+4 2 |
| | 9 2 | 4+5 2 | 5+6 2 | 3+4 2 |
| | 8 2 | 6+7 2 | 3+4 2 | 5+6 2 |
| | 6+7 2 | 8 2 | 5+6 2 | 3+4 2 |
| | 8 2 | 4+5 2 | 3+4 2 | 5+6 2 |
| | 9 2 | 8 2 | 5+6 2 | 5+6 2 |
| | 6+7 2 | 9 2 | 3+4 2 | 3+4 2 |
| | 4+5 2 | 8 2 | 5+6 2 | 5+6 2 |
| | 9 2 | 4+5 2 | 3+4 2 | 3+4 2 |
| | 8 2 | 8 2 | 5+6 2 | 5+6 2 |
| | 4+5 2 | 9 2 | 3+4 2 | 5+6 2 |
| | 8 2 | 8 2 | 5+6 2 | 3+4 2 |
| | 9 2 | 6+7 2 | 3+4 2 | 5+6 2 |
| | 8 2 | 8 2 | 5+6 2 | 3+4 2 |
| | 8 2 | 4+5 2 | 3+4 2 | 3+4 2 |
| | 4+5 2 | 9 2 | 5+6 2 | 5+6 2 |
| | 8 2 | 8 2 | 5+6 2 | 5+6 2 |
| | 8 2 | 4+5 2 | 5+6 2 | 5+6 2 |
| | 4+5 2 | 6+7 2 | 3+4 2 | 3+4 2 |
| | 9 2 | 8 2 | 3+4 2 | 3+4 2 |
| | 6+7 2 | 9 2 | 5+6 2 | 5+6 2 |
| | 9 2 | 8 2 | 3+4 2 | 5+6 2 |
| | 4+5 2 | 9 2 | 3+4 2 | 3+4 2 |
| | 6+7 2 | 6+7 2 | 5+6 2 | 3+4 2 |
| | 8 2 | 9 2 | 3+4 2 | 5+6 2 |
| | 9 2 | 4+5 2 | 5+6 2 | 5+6 2 |
| | 8 2 | 8 2 | 3+4 2 | 5+6 2 |
| | 8 2 | 9 2 | 5+6 2 | 3+4 2 |
| | 9 2 | 8 2 | 3+4 2 | 5+6 2 |
| | 4+5 2 | 4+5 2 | 5+6 2 | 3+4 2 |
| | 8 2 | 6+7 2 | 3+4 2 | 5+6 2 |
| | 6+7 2 | 8 2 | 5+6 2 | 3+4 2 |
| | 8 2 | 8 2 | 5+6 2 | 5+6 2 |
| | 9 2 | 9 2 | 3+4 2 | 3+4 2 |
| | 9 2 | 9 2 | 5+6 2 | 3+4 2 |
| 166.1.238 | 656092 | 114 | Co3Y1 | |
| | -GM | -T | -F | -L |
| | 8 2 | 8 2 | 3+4 2 | 3+4 2 |
| | 8 2 | 9 2 | 5+6 2 | 3+4 2 |
| | 9 2 | 8 2 | 3+4 2 | 5+6 2 |
| | 8 2 | 9 2 | 5+6 2 | 5+6 2 |
| | 9 2 | 9 2 | 5+6 2 | 3+4 2 |
| | 9 2 | 8 2 | 3+4 2 | 5+6 2 |
| | 4+5 2 | 6+7 2 | 5+6 2 | 3+4 2 |
| | 9 2 | 9 2 | 5+6 2 | 5+6 2 |
| | 8 2 | 6+7 2 | 3+4 2 | 5+6 2 |
| | 6+7 2 | 9 2 | 5+6 2 | 5+6 2 |
| | 9 2 | 8 2 | 5+6 2 | 3+4 2 |
| | 6+7 2 | 4+5 2 | 3+4 2 | 5+6 2 |
| | 8 2 | 8 2 | 5+6 2 | 5+6 2 |
| | 9 2 | 9 2 | 3+4 2 | 3+4 2 |
| | 8 2 | 8 2 | 3+4 2 | 3+4 2 |
| | 8 2 | 9 2 | 5+6 2 | 5+6 2 |
| | 9 2 | 9 2 | 3+4 2 | 3+4 2 |

Continued on next column

**TABLE LXXIII– continued**

| TQC | ICSD | NoE | SF | |
|---|---|---|---|---|
| | 9 2 | 9 2 | 3+4 2 | 3+4 2 |
| | 6+7 2 | 6+7 2 | 3+4 2 | 5+6 2 |
| | 8 2 | 8 2 | 5+6 2 | 3+4 2 |
| | 4+5 2 | 4+5 2 | 3+4 2 | 5+6 2 |
| | 8 2 | 8 2 | 3+4 2 | 5+6 2 |
| | 9 2 | 4+5 2 | 5+6 2 | 3+4 2 |
| | 4+5 2 | 8 2 | 5+6 2 | 5+6 2 |
| | 8 2 | 8 2 | 3+4 2 | 3+4 2 |
| | 8 2 | 9 2 | 5+6 2 | 5+6 2 |
| | 4+5 2 | 4+5 2 | 5+6 2 | 5+6 2 |
| | 8 2 | 8 2 | 5+6 2 | 5+6 2 |
| | 6+7 2 | 6+7 2 | 3+4 2 | 5+6 2 |
| | 9 2 | 9 2 | 3+4 2 | 5+6 2 |
| | 9 2 | 9 2 | 3+4 2 | 3+4 2 |
| | 6+7 2 | 6+7 2 | 5+6 2 | 5+6 2 |
| | 9 2 | 9 2 | 3+4 2 | 3+4 2 |
| | 4+5 2 | 6+7 2 | 5+6 2 | 5+6 2 |
| | 8 2 | 9 2 | 3+4 2 | 3+4 2 |
| | 8 2 | 8 2 | 3+4 2 | 5+6 2 |
| | 4+5 2 | 6+7 2 | 3+4 2 | 3+4 2 |
| | 9 2 | 9 2 | 5+6 2 | 3+4 2 |
| | 6+7 2 | 8 2 | 5+6 2 | 3+4 2 |
| | 8 2 | 4+5 2 | 3+4 2 | 5+6 2 |
| | 6+7 2 | 8 2 | 5+6 2 | 3+4 2 |
| | 9 2 | 4+5 2 | 5+6 2 | 5+6 2 |
| | 9 2 | 8 2 | 3+4 2 | 5+6 2 |
| | 8 2 | 9 2 | 5+6 2 | 5+6 2 |
| | 4+5 2 | 6+7 2 | 3+4 2 | 3+4 2 |
| | 8 2 | 9 2 | 3+4 2 | 3+4 2 |
| | 4+5 2 | 8 2 | 5+6 2 | 3+4 2 |
| | 8 2 | 9 2 | 3+4 2 | 5+6 2 |
| | 9 2 | 6+7 2 | 5+6 2 | 3+4 2 |
| | 8 2 | 4+5 2 | 5+6 2 | 5+6 2 |
| | 6+7 2 | 9 2 | 3+4 2 | 3+4 2 |
| | 9 2 | 8 2 | 5+6 2 | 5+6 2 |
| | 4+5 2 | 8 2 | 3+4 2 | 3+4 2 |
| | 9 2 | 4+5 2 | 5+6 2 | 5+6 2 |
| | 6+7 2 | 9 2 | 3+4 2 | 5+6 2 |
| | 8 2 | 6+7 2 | 5+6 2 | 5+6 2 |
| | 8 2 | 9 2 | 3+4 2 | 3+4 2 |
| 166.1.239 | 656079 | 114 | Co3Y1 | |
| | -GM | -T | -F | -L |
| | 8 2 | 8 2 | 3+4 2 | 3+4 2 |
| | 8 2 | 9 2 | 5+6 2 | 3+4 2 |
| | 9 2 | 8 2 | 3+4 2 | 5+6 2 |
| | 9 2 | 9 2 | 5+6 2 | 5+6 2 |
| | 8 2 | 9 2 | 5+6 2 | 3+4 2 |
| | 9 2 | 8 2 | 3+4 2 | 5+6 2 |
| | 9 2 | 9 2 | 5+6 2 | 5+6 2 |
| | 4+5 2 | 6+7 2 | 5+6 2 | 3+4 2 |
| | 6+7 2 | 6+7 2 | 5+6 2 | 5+6 2 |
| | 8 2 | 9 2 | 3+4 2 | 5+6 2 |
| | 9 2 | 8 2 | 5+6 2 | 3+4 2 |
| | 6+7 2 | 4+5 2 | 3+4 2 | 5+6 2 |
| | 8 2 | 8 2 | 5+6 2 | 5+6 2 |
| | 9 2 | 9 2 | 3+4 2 | 3+4 2 |
| | 8 2 | 8 2 | 3+4 2 | 3+4 2 |
| | 8 2 | 9 2 | 5+6 2 | 5+6 2 |
| | 9 2 | 9 2 | 3+4 2 | 3+4 2 |
| | 9 2 | 9 2 | 3+4 2 | 5+6 2 |
| | 8 2 | 6+7 2 | 3+4 2 | 3+4 2 |

Continued on next column

**TABLE LXXIII– continued**

| TQC | ICSD | NoE | SF | |
|---|---|---|---|---|
| | 4+5 2 | 8 2 | 5+6 2 | 3+4 2 |
| | 6+7 2 | 4+5 2 | 3+4 2 | 5+6 2 |
| | 8 2 | 8 2 | 3+4 2 | 5+6 2 |
| | 9 2 | 8 2 | 5+6 2 | 5+6 2 |
| | 4+5 2 | 4+5 2 | 5+6 2 | 3+4 2 |
| | 8 2 | 8 2 | 3+4 2 | 5+6 2 |
| | 8 2 | 9 2 | 5+6 2 | 3+4 2 |
| | 4+5 2 | 6+7 2 | 3+4 2 | 5+6 2 |
| | 8 2 | 9 2 | 5+6 2 | 5+6 2 |
| | 6+7 2 | 4+5 2 | 3+4 2 | 5+6 2 |
| | 9 2 | 8 2 | 5+6 2 | 5+6 2 |
| | 9 2 | 6+7 2 | 3+4 2 | 3+4 2 |
| | 6+7 2 | 9 2 | 5+6 2 | 5+6 2 |
| | 9 2 | 9 2 | 3+4 2 | 5+6 2 |
| | 4+5 2 | 6+7 2 | 5+6 2 | 3+4 2 |
| | 8 2 | 9 2 | 3+4 2 | 5+6 2 |
| | 8 2 | 8 2 | 3+4 2 | 3+4 2 |
| | 4+5 2 | 6+7 2 | 5+6 2 | 3+4 2 |
| | 9 2 | 9 2 | 3+4 2 | 3+4 2 |
| | 6+7 2 | 8 2 | 5+6 2 | 3+4 2 |
| | 8 2 | 4+5 2 | 3+4 2 | 5+6 2 |
| | 6+7 2 | 8 2 | 5+6 2 | 3+4 2 |
| | 9 2 | 4+5 2 | 5+6 2 | 5+6 2 |
| | 9 2 | 8 2 | 3+4 2 | 5+6 2 |
| | 8 2 | 9 2 | 5+6 2 | 5+6 2 |
| | 4+5 2 | 6+7 2 | 3+4 2 | 3+4 2 |
| | 8 2 | 9 2 | 3+4 2 | 3+4 2 |
| | 4+5 2 | 8 2 | 5+6 2 | 3+4 2 |
| | 8 2 | 6+7 2 | 3+4 2 | 5+6 2 |
| | 9 2 | 9 2 | 5+6 2 | 3+4 2 |
| | 8 2 | 4+5 2 | 5+6 2 | 5+6 2 |
| | 6+7 2 | 9 2 | 3+4 2 | 3+4 2 |
| | 9 2 | 8 2 | 5+6 2 | 5+6 2 |
| | 4+5 2 | 8 2 | 3+4 2 | 3+4 2 |
| | 9 2 | 4+5 2 | 5+6 2 | 5+6 2 |
| | 6+7 2 | 9 2 | 3+4 2 | 5+6 2 |
| | 8 2 | 6+7 2 | 5+6 2 | 5+6 2 |
| | 8 2 | 9 2 | 3+4 2 | 3+4 2 |
| 166.1.240 | 625636 | 114 | Co3Y1 | |
| | -GM | -T | -F | -L |
| | 8 2 | 8 2 | 3+4 2 | 3+4 2 |
| | 8 2 | 9 2 | 5+6 2 | 3+4 2 |
| | 9 2 | 8 2 | 3+4 2 | 5+6 2 |
| | 8 2 | 9 2 | 5+6 2 | 5+6 2 |
| | 9 2 | 9 2 | 5+6 2 | 3+4 2 |
| | 9 2 | 8 2 | 3+4 2 | 5+6 2 |
| | 4+5 2 | 6+7 2 | 5+6 2 | 3+4 2 |
| | 9 2 | 9 2 | 5+6 2 | 5+6 2 |
| | 8 2 | 6+7 2 | 3+4 2 | 5+6 2 |
| | 6+7 2 | 9 2 | 5+6 2 | 5+6 2 |
| | 9 2 | 8 2 | 5+6 2 | 3+4 2 |
| | 6+7 2 | 4+5 2 | 3+4 2 | 5+6 2 |
| | 8 2 | 8 2 | 5+6 2 | 5+6 2 |
| | 9 2 | 9 2 | 3+4 2 | 3+4 2 |
| | 8 2 | 8 2 | 3+4 2 | 3+4 2 |
| | 8 2 | 9 2 | 5+6 2 | 5+6 2 |
| | 9 2 | 9 2 | 3+4 2 | 3+4 2 |
| | 9 2 | 9 2 | 3+4 2 | 3+4 2 |
| | 6+7 2 | 6+7 2 | 3+4 2 | 5+6 2 |
| | 8 2 | 8 2 | 5+6 2 | 3+4 2 |
| | 4+5 2 | 4+5 2 | 3+4 2 | 5+6 2 |

Continued on next column

**TABLE LXXIII– continued**

| TQC | ICSD | NoE | SF | |
|---|---|---|---|---|
| | 8 2 | 8 2 | 3+4 2 | 5+6 2 |
| | 9 2 | 4+5 2 | 5+6 2 | 3+4 2 |
| | 4+5 2 | 8 2 | 5+6 2 | 5+6 2 |
| | 8 2 | 8 2 | 3+4 2 | 5+6 2 |
| | 8 2 | 9 2 | 5+6 2 | 3+4 2 |
| | 4+5 2 | 4+5 2 | 5+6 2 | 5+6 2 |
| | 8 2 | 8 2 | 5+6 2 | 5+6 2 |
| | 6+7 2 | 6+7 2 | 3+4 2 | 5+6 2 |
| | 9 2 | 9 2 | 3+4 2 | 5+6 2 |
| | 9 2 | 9 2 | 3+4 2 | 3+4 2 |
| | 6+7 2 | 6+7 2 | 5+6 2 | 5+6 2 |
| | 9 2 | 9 2 | 3+4 2 | 3+4 2 |
| | 4+5 2 | 6+7 2 | 5+6 2 | 5+6 2 |
| | 8 2 | 9 2 | 3+4 2 | 3+4 2 |
| | 8 2 | 8 2 | 3+4 2 | 5+6 2 |
| | 4+5 2 | 6+7 2 | 5+6 2 | 3+4 2 |
| | 9 2 | 9 2 | 3+4 2 | 3+4 2 |
| | 6+7 2 | 8 2 | 5+6 2 | 3+4 2 |
| | 8 2 | 4+5 2 | 3+4 2 | 5+6 2 |
| | 6+7 2 | 8 2 | 5+6 2 | 3+4 2 |
| | 9 2 | 4+5 2 | 5+6 2 | 5+6 2 |
| | 9 2 | 8 2 | 3+4 2 | 5+6 2 |
| | 8 2 | 9 2 | 5+6 2 | 5+6 2 |
| | 4+5 2 | 6+7 2 | 3+4 2 | 3+4 2 |
| | 8 2 | 9 2 | 3+4 2 | 3+4 2 |
| | 4+5 2 | 8 2 | 5+6 2 | 3+4 2 |
| | 8 2 | 9 2 | 3+4 2 | 3+4 2 |
| | 9 2 | 6+7 2 | 5+6 2 | 5+6 2 |
| | 8 2 | 4+5 2 | 5+6 2 | 5+6 2 |
| | 6+7 2 | 9 2 | 3+4 2 | 3+4 2 |
| | 9 2 | 8 2 | 5+6 2 | 5+6 2 |
| | 4+5 2 | 8 2 | 3+4 2 | 3+4 2 |
| | 9 2 | 4+5 2 | 5+6 2 | 5+6 2 |
| | 6+7 2 | 9 2 | 3+4 2 | 5+6 2 |
| | 8 2 | 6+7 2 | 5+6 2 | 5+6 2 |
| | 8 2 | 9 2 | 3+4 2 | 3+4 2 |
| 166.1.241 | 625625 | 114 | Co3Y1 | |
| | -GM | -T | -F | -L |
| | 8 2 | 8 2 | 3+4 2 | 3+4 2 |
| | 8 2 | 9 2 | 5+6 2 | 3+4 2 |
| | 9 2 | 8 2 | 3+4 2 | 5+6 2 |
| | 8 2 | 9 2 | 5+6 2 | 5+6 2 |
| | 9 2 | 9 2 | 5+6 2 | 3+4 2 |
| | 9 2 | 8 2 | 3+4 2 | 5+6 2 |
| | 4+5 2 | 6+7 2 | 5+6 2 | 3+4 2 |
| | 9 2 | 9 2 | 5+6 2 | 5+6 2 |
| | 8 2 | 6+7 2 | 3+4 2 | 5+6 2 |
| | 6+7 2 | 9 2 | 5+6 2 | 5+6 2 |
| | 9 2 | 8 2 | 5+6 2 | 3+4 2 |
| | 6+7 2 | 4+5 2 | 3+4 2 | 5+6 2 |
| | 8 2 | 8 2 | 5+6 2 | 5+6 2 |
| | 9 2 | 9 2 | 3+4 2 | 3+4 2 |
| | 8 2 | 8 2 | 3+4 2 | 3+4 2 |
| | 8 2 | 9 2 | 5+6 2 | 5+6 2 |
| | 9 2 | 9 2 | 3+4 2 | 3+4 2 |
| | 9 2 | 9 2 | 3+4 2 | 3+4 2 |
| | 6+7 2 | 6+7 2 | 3+4 2 | 5+6 2 |
| | 8 2 | 8 2 | 5+6 2 | 3+4 2 |
| | 4+5 2 | 4+5 2 | 3+4 2 | 5+6 2 |
| | 8 2 | 8 2 | 3+4 2 | 5+6 2 |
| | 9 2 | 4+5 2 | 5+6 2 | 3+4 2 |

Continued on next column

| TQC | ICSD | NoE | SF | |
|---|---|---|---|---|
| | 4+5 2 | 8 2 | 5+6 2 | 5+6 2 |
| | 8 2 | 8 2 | 3+4 2 | 5+6 2 |
| | 8 2 | 9 2 | 5+6 2 | 3+4 2 |
| | 4+5 2 | 4+5 2 | 5+6 2 | 5+6 2 |
| | 8 2 | 8 2 | 5+6 2 | 5+6 2 |
| | 6+7 2 | 6+7 2 | 3+4 2 | 5+6 2 |
| | 9 2 | 9 2 | 3+4 2 | 5+6 2 |
| | 9 2 | 9 2 | 3+4 2 | 3+4 2 |
| | 6+7 2 | 6+7 2 | 5+6 2 | 5+6 2 |
| | 9 2 | 9 2 | 3+4 2 | 3+4 2 |
| | 4+5 2 | 6+7 2 | 5+6 2 | 5+6 2 |
| | 8 2 | 9 2 | 3+4 2 | 3+4 2 |
| | 8 2 | 8 2 | 3+4 2 | 5+6 2 |
| | 4+5 2 | 6+7 2 | 5+6 2 | 3+4 2 |
| | 9 2 | 9 2 | 3+4 2 | 3+4 2 |
| | 6+7 2 | 8 2 | 5+6 2 | 3+4 2 |
| | 8 2 | 4+5 2 | 3+4 2 | 5+6 2 |
| | 6+7 2 | 8 2 | 5+6 2 | 3+4 2 |
| | 9 2 | 4+5 2 | 5+6 2 | 5+6 2 |
| | 9 2 | 8 2 | 3+4 2 | 5+6 2 |
| | 8 2 | 9 2 | 5+6 2 | 5+6 2 |
| | 4+5 2 | 6+7 2 | 3+4 2 | 3+4 2 |
| | 8 2 | 9 2 | 3+4 2 | 3+4 2 |
| | 4+5 2 | 8 2 | 5+6 2 | 3+4 2 |
| | 8 2 | 9 2 | 3+4 2 | 3+4 2 |
| | 9 2 | 6+7 2 | 5+6 2 | 5+6 2 |
| | 8 2 | 4+5 2 | 5+6 2 | 5+6 2 |
| | 6+7 2 | 9 2 | 3+4 2 | 3+4 2 |
| | 9 2 | 8 2 | 5+6 2 | 5+6 2 |
| | 4+5 2 | 8 2 | 3+4 2 | 3+4 2 |
| | 9 2 | 4+5 2 | 5+6 2 | 5+6 2 |
| | 6+7 2 | 9 2 | 3+4 2 | 5+6 2 |
| | 8 2 | 6+7 2 | 5+6 2 | 5+6 2 |
| | 8 2 | 9 2 | 3+4 2 | 3+4 2 |
| 166.1.242 | 625605 | 114 | Co3Y1 | |
| | -GM | -T | -F | -L |
| | 8 2 | 8 2 | 3+4 2 | 3+4 2 |
| | 8 2 | 9 2 | 5+6 2 | 3+4 2 |
| | 9 2 | 8 2 | 3+4 2 | 5+6 2 |
| | 8 2 | 9 2 | 5+6 2 | 5+6 2 |
| | 9 2 | 9 2 | 5+6 2 | 3+4 2 |
| | 9 2 | 8 2 | 3+4 2 | 5+6 2 |
| | 4+5 2 | 6+7 2 | 5+6 2 | 3+4 2 |
| | 9 2 | 9 2 | 5+6 2 | 5+6 2 |
| | 8 2 | 6+7 2 | 3+4 2 | 5+6 2 |
| | 6+7 2 | 9 2 | 5+6 2 | 5+6 2 |
| | 9 2 | 8 2 | 5+6 2 | 3+4 2 |
| | 6+7 2 | 4+5 2 | 3+4 2 | 5+6 2 |
| | 8 2 | 8 2 | 5+6 2 | 5+6 2 |
| | 9 2 | 9 2 | 3+4 2 | 3+4 2 |
| | 8 2 | 8 2 | 3+4 2 | 3+4 2 |
| | 8 2 | 9 2 | 5+6 2 | 5+6 2 |
| | 9 2 | 9 2 | 3+4 2 | 3+4 2 |
| | 9 2 | 9 2 | 3+4 2 | 3+4 2 |
| | 6+7 2 | 6+7 2 | 3+4 2 | 5+6 2 |
| | 8 2 | 8 2 | 5+6 2 | 3+4 2 |
| | 4+5 2 | 4+5 2 | 3+4 2 | 5+6 2 |
| | 8 2 | 8 2 | 3+4 2 | 5+6 2 |
| | 9 2 | 4+5 2 | 5+6 2 | 3+4 2 |
| | 4+5 2 | 8 2 | 5+6 2 | 5+6 2 |
| | 8 2 | 8 2 | 3+4 2 | 5+6 2 |

Continued on next column

**TABLE LXXIII– continued**

| TQC | ICSD | | NoE | SF |
|---|---|---|---|---|
| | 8 2 | 9 2 | 5+6 2 | 3+4 2 |
| | 4+5 2 | 4+5 2 | 5+6 2 | 5+6 2 |
| | 8 2 | 8 2 | 5+6 2 | 5+6 2 |
| | 6+7 2 | 6+7 2 | 3+4 2 | 5+6 2 |
| | 9 2 | 9 2 | 3+4 2 | 5+6 2 |
| | 9 2 | 9 2 | 3+4 2 | 3+4 2 |
| | 6+7 2 | 6+7 2 | 5+6 2 | 5+6 2 |
| | 9 2 | 9 2 | 3+4 2 | 3+4 2 |
| | 4+5 2 | 6+7 2 | 5+6 2 | 5+6 2 |
| | 8 2 | 9 2 | 3+4 2 | 3+4 2 |
| | 8 2 | 8 2 | 3+4 2 | 5+6 2 |
| | 4+5 2 | 6+7 2 | 5+6 2 | 3+4 2 |
| | 9 2 | 9 2 | 3+4 2 | 3+4 2 |
| | 6+7 2 | 8 2 | 5+6 2 | 3+4 2 |
| | 8 2 | 4+5 2 | 3+4 2 | 5+6 2 |
| | 6+7 2 | 8 2 | 5+6 2 | 3+4 2 |
| | 9 2 | 4+5 2 | 5+6 2 | 5+6 2 |
| | 9 2 | 8 2 | 3+4 2 | 5+6 2 |
| | 8 2 | 9 2 | 5+6 2 | 5+6 2 |
| | 4+5 2 | 6+7 2 | 3+4 2 | 3+4 2 |
| | 8 2 | 9 2 | 3+4 2 | 3+4 2 |
| | 4+5 2 | 8 2 | 5+6 2 | 3+4 2 |
| | 8 2 | 9 2 | 3+4 2 | 3+4 2 |
| | 9 2 | 6+7 2 | 5+6 2 | 5+6 2 |
| | 8 2 | 4+5 2 | 5+6 2 | 5+6 2 |
| | 6+7 2 | 9 2 | 3+4 2 | 3+4 2 |
| | 9 2 | 8 2 | 5+6 2 | 5+6 2 |
| | 4+5 2 | 8 2 | 3+4 2 | 3+4 2 |
| | 9 2 | 4+5 2 | 5+6 2 | 5+6 2 |
| | 6+7 2 | 9 2 | 3+4 2 | 5+6 2 |
| | 8 2 | 6+7 2 | 5+6 2 | 5+6 2 |
| | 8 2 | 9 2 | 3+4 2 | 3+4 2 |
| 166.1.243 | 625600 | | 114 | Co3Y1 |
| | -GM | -T | -F | -L |
| | 8 2 | 8 2 | 3+4 2 | 3+4 2 |
| | 8 2 | 9 2 | 5+6 2 | 3+4 2 |
| | 9 2 | 8 2 | 3+4 2 | 5+6 2 |
| | 8 2 | 9 2 | 5+6 2 | 5+6 2 |
| | 9 2 | 9 2 | 5+6 2 | 3+4 2 |
| | 9 2 | 8 2 | 3+4 2 | 5+6 2 |
| | 4+5 2 | 6+7 2 | 5+6 2 | 3+4 2 |
| | 9 2 | 9 2 | 5+6 2 | 5+6 2 |
| | 8 2 | 6+7 2 | 3+4 2 | 5+6 2 |
| | 6+7 2 | 9 2 | 5+6 2 | 5+6 2 |
| | 9 2 | 8 2 | 5+6 2 | 3+4 2 |
| | 6+7 2 | 4+5 2 | 3+4 2 | 5+6 2 |
| | 8 2 | 8 2 | 5+6 2 | 5+6 2 |
| | 9 2 | 9 2 | 3+4 2 | 3+4 2 |
| | 8 2 | 8 2 | 3+4 2 | 3+4 2 |
| | 8 2 | 9 2 | 5+6 2 | 5+6 2 |
| | 9 2 | 9 2 | 3+4 2 | 3+4 2 |
| | 9 2 | 9 2 | 3+4 2 | 3+4 2 |
| | 6+7 2 | 6+7 2 | 3+4 2 | 5+6 2 |
| | 8 2 | 8 2 | 5+6 2 | 3+4 2 |
| | 4+5 2 | 4+5 2 | 3+4 2 | 5+6 2 |
| | 8 2 | 8 2 | 3+4 2 | 5+6 2 |
| | 9 2 | 4+5 2 | 5+6 2 | 3+4 2 |
| | 4+5 2 | 8 2 | 5+6 2 | 5+6 2 |
| | 8 2 | 8 2 | 3+4 2 | 5+6 2 |
| | 8 2 | 9 2 | 5+6 2 | 3+4 2 |
| | 4+5 2 | 4+5 2 | 5+6 2 | 5+6 2 |
| | 8 2 | 8 2 | 5+6 2 | 5+6 2 |
| | 6+7 2 | 6+7 2 | 3+4 2 | 5+6 2 |
| | 9 2 | 9 2 | 3+4 2 | 5+6 2 |
| | 9 2 | 9 2 | 3+4 2 | 3+4 2 |
| | 6+7 2 | 6+7 2 | 5+6 2 | 5+6 2 |
| | 9 2 | 9 2 | 3+4 2 | 3+4 2 |
| | 4+5 2 | 6+7 2 | 5+6 2 | 5+6 2 |
| | 8 2 | 9 2 | 3+4 2 | 3+4 2 |
| | 8 2 | 8 2 | 3+4 2 | 5+6 2 |
| | 4+5 2 | 6+7 2 | 5+6 2 | 3+4 2 |
| | 9 2 | 9 2 | 3+4 2 | 3+4 2 |
| | 6+7 2 | 8 2 | 5+6 2 | 3+4 2 |
| | 8 2 | 4+5 2 | 3+4 2 | 5+6 2 |
| | 6+7 2 | 8 2 | 5+6 2 | 3+4 2 |
| | 9 2 | 4+5 2 | 5+6 2 | 5+6 2 |
| | 9 2 | 8 2 | 3+4 2 | 5+6 2 |
| | 8 2 | 9 2 | 5+6 2 | 5+6 2 |
| | 4+5 2 | 6+7 2 | 3+4 2 | 3+4 2 |
| | 8 2 | 9 2 | 3+4 2 | 3+4 2 |
| | 4+5 2 | 8 2 | 5+6 2 | 3+4 2 |
| | 8 2 | 9 2 | 3+4 2 | 3+4 2 |
| | 9 2 | 6+7 2 | 5+6 2 | 5+6 2 |
| | 8 2 | 4+5 2 | 5+6 2 | 5+6 2 |
| | 6+7 2 | 9 2 | 3+4 2 | 3+4 2 |
| | 9 2 | 8 2 | 5+6 2 | 5+6 2 |
| | 4+5 2 | 8 2 | 3+4 2 | 3+4 2 |
| | 9 2 | 4+5 2 | 5+6 2 | 5+6 2 |
| | 6+7 2 | 9 2 | 3+4 2 | 5+6 2 |
| | 8 2 | 6+7 2 | 5+6 2 | 5+6 2 |
| | 8 2 | 9 2 | 3+4 2 | 3+4 2 |
| 166.1.244 | 625590 | | 114 | Co3Y1 |
| | -GM | -T | -F | -L |
| | 8 2 | 8 2 | 3+4 2 | 3+4 2 |
| | 8 2 | 9 2 | 5+6 2 | 3+4 2 |
| | 9 2 | 8 2 | 3+4 2 | 5+6 2 |
| | 8 2 | 9 2 | 5+6 2 | 5+6 2 |
| | 9 2 | 9 2 | 5+6 2 | 3+4 2 |
| | 9 2 | 8 2 | 3+4 2 | 5+6 2 |
| | 4+5 2 | 6+7 2 | 5+6 2 | 3+4 2 |
| | 9 2 | 9 2 | 5+6 2 | 5+6 2 |
| | 8 2 | 6+7 2 | 3+4 2 | 5+6 2 |
| | 6+7 2 | 9 2 | 5+6 2 | 5+6 2 |
| | 9 2 | 8 2 | 5+6 2 | 3+4 2 |
| | 6+7 2 | 4+5 2 | 3+4 2 | 5+6 2 |
| | 8 2 | 8 2 | 5+6 2 | 5+6 2 |
| | 9 2 | 9 2 | 3+4 2 | 3+4 2 |
| | 8 2 | 8 2 | 3+4 2 | 3+4 2 |
| | 8 2 | 9 2 | 5+6 2 | 5+6 2 |
| | 9 2 | 9 2 | 3+4 2 | 3+4 2 |
| | 9 2 | 9 2 | 3+4 2 | 3+4 2 |
| | 6+7 2 | 6+7 2 | 3+4 2 | 5+6 2 |
| | 8 2 | 8 2 | 5+6 2 | 3+4 2 |
| | 4+5 2 | 4+5 2 | 3+4 2 | 5+6 2 |
| | 8 2 | 8 2 | 3+4 2 | 5+6 2 |
| | 9 2 | 4+5 2 | 5+6 2 | 3+4 2 |
| | 4+5 2 | 8 2 | 5+6 2 | 5+6 2 |
| | 8 2 | 8 2 | 3+4 2 | 3+4 2 |
| | 8 2 | 9 2 | 5+6 2 | 5+6 2 |
| | 4+5 2 | 4+5 2 | 5+6 2 | 5+6 2 |
| | 8 2 | 8 2 | 5+6 2 | 5+6 2 |
| | 6+7 2 | 6+7 2 | 3+4 2 | 5+6 2 |

Continued on next column

**TABLE LXXIII– continued**

| TQC | ICSD | NoE | SF | |
|---|---|---|---|---|
| | 9 2 | 9 2 | 3+4 2 | 5+6 2 |
| | 9 2 | 9 2 | 3+4 2 | 3+4 2 |
| | 6+7 2 | 6+7 2 | 5+6 2 | 5+6 2 |
| | 9 2 | 9 2 | 3+4 2 | 3+4 2 |
| | 4+5 2 | 6+7 2 | 5+6 2 | 5+6 2 |
| | 8 2 | 9 2 | 3+4 2 | 3+4 2 |
| | 8 2 | 8 2 | 3+4 2 | 5+6 2 |
| | 4+5 2 | 6+7 2 | 5+6 2 | 3+4 2 |
| | 9 2 | 9 2 | 3+4 2 | 3+4 2 |
| | 6+7 2 | 8 2 | 5+6 2 | 3+4 2 |
| | 8 2 | 4+5 2 | 3+4 2 | 5+6 2 |
| | 6+7 2 | 8 2 | 5+6 2 | 3+4 2 |
| | 9 2 | 4+5 2 | 5+6 2 | 5+6 2 |
| | 9 2 | 8 2 | 3+4 2 | 5+6 2 |
| | 8 2 | 9 2 | 5+6 2 | 5+6 2 |
| | 4+5 2 | 6+7 2 | 3+4 2 | 3+4 2 |
| | 8 2 | 9 2 | 3+4 2 | 3+4 2 |
| | 4+5 2 | 8 2 | 5+6 2 | 3+4 2 |
| | 8 2 | 9 2 | 3+4 2 | 5+6 2 |
| | 9 2 | 6+7 2 | 5+6 2 | 3+4 2 |
| | 8 2 | 4+5 2 | 5+6 2 | 5+6 2 |
| | 6+7 2 | 9 2 | 3+4 2 | 3+4 2 |
| | 9 2 | 8 2 | 5+6 2 | 5+6 2 |
| | 4+5 2 | 8 2 | 3+4 2 | 3+4 2 |
| | 9 2 | 4+5 2 | 5+6 2 | 5+6 2 |
| | 6+7 2 | 9 2 | 3+4 2 | 5+6 2 |
| | 8 2 | 6+7 2 | 5+6 2 | 5+6 2 |
| | 8 2 | 9 2 | 3+4 2 | 3+4 2 |
| 166.1.245 | 625580 | 114 | Co3Y1 | |
| | -GM | -T | -F | -L |
| | 8 2 | 8 2 | 3+4 2 | 3+4 2 |
| | 8 2 | 9 2 | 5+6 2 | 3+4 2 |
| | 9 2 | 8 2 | 3+4 2 | 5+6 2 |
| | 8 2 | 9 2 | 5+6 2 | 5+6 2 |
| | 9 2 | 9 2 | 5+6 2 | 3+4 2 |
| | 9 2 | 8 2 | 3+4 2 | 5+6 2 |
| | 4+5 2 | 6+7 2 | 5+6 2 | 3+4 2 |
| | 9 2 | 9 2 | 5+6 2 | 5+6 2 |
| | 8 2 | 6+7 2 | 3+4 2 | 5+6 2 |
| | 6+7 2 | 9 2 | 5+6 2 | 5+6 2 |
| | 9 2 | 8 2 | 5+6 2 | 3+4 2 |
| | 6+7 2 | 4+5 2 | 3+4 2 | 5+6 2 |
| | 8 2 | 8 2 | 5+6 2 | 5+6 2 |
| | 9 2 | 9 2 | 3+4 2 | 3+4 2 |
| | 8 2 | 8 2 | 3+4 2 | 3+4 2 |
| | 8 2 | 9 2 | 5+6 2 | 5+6 2 |
| | 9 2 | 9 2 | 3+4 2 | 3+4 2 |
| | 9 2 | 9 2 | 3+4 2 | 3+4 2 |
| | 6+7 2 | 6+7 2 | 3+4 2 | 5+6 2 |
| | 8 2 | 8 2 | 5+6 2 | 3+4 2 |
| | 4+5 2 | 4+5 2 | 3+4 2 | 5+6 2 |
| | 8 2 | 8 2 | 3+4 2 | 5+6 2 |
| | 9 2 | 4+5 2 | 5+6 2 | 3+4 2 |
| | 4+5 2 | 8 2 | 5+6 2 | 5+6 2 |
| | 8 2 | 8 2 | 3+4 2 | 5+6 2 |
| | 8 2 | 9 2 | 5+6 2 | 3+4 2 |
| | 4+5 2 | 4+5 2 | 5+6 2 | 5+6 2 |
| | 8 2 | 8 2 | 5+6 2 | 5+6 2 |
| | 6+7 2 | 6+7 2 | 3+4 2 | 5+6 2 |
| | 9 2 | 9 2 | 3+4 2 | 5+6 2 |
| | 9 2 | 9 2 | 3+4 2 | 3+4 2 |

Continued on next column

**TABLE LXXIII– continued**

| TQC | ICSD | NoE | SF | |
|---|---|---|---|---|
| | 6+7 2 | 6+7 2 | 5+6 2 | 5+6 2 |
| | 9 2 | 9 2 | 3+4 2 | 3+4 2 |
| | 4+5 2 | 6+7 2 | 5+6 2 | 5+6 2 |
| | 8 2 | 9 2 | 3+4 2 | 3+4 2 |
| | 8 2 | 8 2 | 3+4 2 | 5+6 2 |
| | 4+5 2 | 6+7 2 | 5+6 2 | 3+4 2 |
| | 9 2 | 9 2 | 3+4 2 | 3+4 2 |
| | 6+7 2 | 8 2 | 5+6 2 | 3+4 2 |
| | 8 2 | 4+5 2 | 3+4 2 | 5+6 2 |
| | 6+7 2 | 8 2 | 5+6 2 | 3+4 2 |
| | 9 2 | 4+5 2 | 5+6 2 | 5+6 2 |
| | 9 2 | 8 2 | 3+4 2 | 5+6 2 |
| | 8 2 | 9 2 | 5+6 2 | 5+6 2 |
| | 4+5 2 | 6+7 2 | 3+4 2 | 3+4 2 |
| | 8 2 | 9 2 | 3+4 2 | 3+4 2 |
| | 4+5 2 | 8 2 | 5+6 2 | 3+4 2 |
| | 8 2 | 9 2 | 3+4 2 | 3+4 2 |
| | 9 2 | 6+7 2 | 5+6 2 | 5+6 2 |
| | 8 2 | 4+5 2 | 5+6 2 | 5+6 2 |
| | 6+7 2 | 9 2 | 3+4 2 | 3+4 2 |
| | 9 2 | 8 2 | 5+6 2 | 5+6 2 |
| | 4+5 2 | 8 2 | 3+4 2 | 3+4 2 |
| | 9 2 | 4+5 2 | 5+6 2 | 5+6 2 |
| | 6+7 2 | 9 2 | 3+4 2 | 5+6 2 |
| | 8 2 | 6+7 2 | 5+6 2 | 5+6 2 |
| | 8 2 | 9 2 | 3+4 2 | 3+4 2 |
| 166.1.246 | 625572 | 114 | Co3Y1 | |
| | -GM | -T | -F | -L |
| | 8 2 | 8 2 | 3+4 2 | 3+4 2 |
| | 8 2 | 9 2 | 5+6 2 | 3+4 2 |
| | 9 2 | 8 2 | 3+4 2 | 5+6 2 |
| | 8 2 | 9 2 | 5+6 2 | 5+6 2 |
| | 9 2 | 9 2 | 5+6 2 | 3+4 2 |
| | 9 2 | 8 2 | 3+4 2 | 5+6 2 |
| | 4+5 2 | 6+7 2 | 5+6 2 | 3+4 2 |
| | 9 2 | 9 2 | 5+6 2 | 5+6 2 |
| | 8 2 | 6+7 2 | 3+4 2 | 5+6 2 |
| | 6+7 2 | 9 2 | 5+6 2 | 5+6 2 |
| | 9 2 | 8 2 | 5+6 2 | 3+4 2 |
| | 6+7 2 | 4+5 2 | 3+4 2 | 5+6 2 |
| | 8 2 | 8 2 | 5+6 2 | 5+6 2 |
| | 9 2 | 9 2 | 3+4 2 | 3+4 2 |
| | 8 2 | 8 2 | 3+4 2 | 3+4 2 |
| | 8 2 | 9 2 | 5+6 2 | 5+6 2 |
| | 9 2 | 9 2 | 3+4 2 | 3+4 2 |
| | 9 2 | 9 2 | 3+4 2 | 3+4 2 |
| | 6+7 2 | 6+7 2 | 3+4 2 | 5+6 2 |
| | 8 2 | 8 2 | 5+6 2 | 3+4 2 |
| | 4+5 2 | 4+5 2 | 3+4 2 | 5+6 2 |
| | 8 2 | 8 2 | 3+4 2 | 5+6 2 |
| | 9 2 | 4+5 2 | 5+6 2 | 3+4 2 |
| | 4+5 2 | 8 2 | 5+6 2 | 5+6 2 |
| | 8 2 | 8 2 | 3+4 2 | 5+6 2 |
| | 8 2 | 9 2 | 5+6 2 | 3+4 2 |
| | 4+5 2 | 4+5 2 | 5+6 2 | 5+6 2 |
| | 8 2 | 8 2 | 5+6 2 | 5+6 2 |
| | 6+7 2 | 6+7 2 | 3+4 2 | 5+6 2 |
| | 9 2 | 9 2 | 3+4 2 | 5+6 2 |
| | 9 2 | 9 2 | 3+4 2 | 3+4 2 |
| | 6+7 2 | 6+7 2 | 5+6 2 | 5+6 2 |
| | 9 2 | 9 2 | 3+4 2 | 3+4 2 |

Continued on next column

**TABLE LXXIII– continued**

| TQC | ICSD | | NoE | SF |
|---|---|---|---|---|
| | 4+5 2 | 6+7 2 | 5+6 2 | 5+6 2 |
| | 8 2 | 9 2 | 3+4 2 | 3+4 2 |
| | 8 2 | 8 2 | 3+4 2 | 5+6 2 |
| | 4+5 2 | 6+7 2 | 5+6 2 | 3+4 2 |
| | 9 2 | 9 2 | 3+4 2 | 3+4 2 |
| | 6+7 2 | 8 2 | 5+6 2 | 3+4 2 |
| | 8 2 | 4+5 2 | 3+4 2 | 5+6 2 |
| | 6+7 2 | 8 2 | 5+6 2 | 3+4 2 |
| | 9 2 | 4+5 2 | 5+6 2 | 5+6 2 |
| | 9 2 | 8 2 | 3+4 2 | 5+6 2 |
| | 8 2 | 9 2 | 5+6 2 | 5+6 2 |
| | 4+5 2 | 6+7 2 | 3+4 2 | 3+4 2 |
| | 8 2 | 9 2 | 3+4 2 | 3+4 2 |
| | 4+5 2 | 8 2 | 5+6 2 | 3+4 2 |
| | 8 2 | 9 2 | 3+4 2 | 3+4 2 |
| | 9 2 | 6+7 2 | 5+6 2 | 5+6 2 |
| | 8 2 | 4+5 2 | 5+6 2 | 5+6 2 |
| | 6+7 2 | 9 2 | 3+4 2 | 3+4 2 |
| | 9 2 | 8 2 | 5+6 2 | 5+6 2 |
| | 4+5 2 | 8 2 | 3+4 2 | 3+4 2 |
| | 9 2 | 4+5 2 | 5+6 2 | 5+6 2 |
| | 6+7 2 | 9 2 | 3+4 2 | 5+6 2 |
| | 8 2 | 6+7 2 | 5+6 2 | 5+6 2 |
| | 8 2 | 9 2 | 3+4 2 | 3+4 2 |
| 166.1.247 | 625564 | | 114 | Co3Y1 |
| | -GM | -T | -F | -L |
| | 8 2 | 8 2 | 3+4 2 | 3+4 2 |
| | 8 2 | 9 2 | 5+6 2 | 3+4 2 |
| | 9 2 | 8 2 | 3+4 2 | 5+6 2 |
| | 8 2 | 9 2 | 5+6 2 | 5+6 2 |
| | 9 2 | 9 2 | 5+6 2 | 3+4 2 |
| | 9 2 | 8 2 | 3+4 2 | 5+6 2 |
| | 4+5 2 | 6+7 2 | 5+6 2 | 3+4 2 |
| | 9 2 | 9 2 | 5+6 2 | 5+6 2 |
| | 8 2 | 6+7 2 | 3+4 2 | 5+6 2 |
| | 6+7 2 | 9 2 | 5+6 2 | 5+6 2 |
| | 9 2 | 8 2 | 5+6 2 | 3+4 2 |
| | 6+7 2 | 4+5 2 | 3+4 2 | 5+6 2 |
| | 8 2 | 8 2 | 5+6 2 | 5+6 2 |
| | 9 2 | 9 2 | 3+4 2 | 3+4 2 |
| | 8 2 | 8 2 | 3+4 2 | 3+4 2 |
| | 8 2 | 9 2 | 5+6 2 | 5+6 2 |
| | 9 2 | 9 2 | 3+4 2 | 3+4 2 |
| | 9 2 | 9 2 | 3+4 2 | 3+4 2 |
| | 6+7 2 | 6+7 2 | 3+4 2 | 5+6 2 |
| | 8 2 | 8 2 | 5+6 2 | 3+4 2 |
| | 4+5 2 | 4+5 2 | 3+4 2 | 5+6 2 |
| | 8 2 | 8 2 | 3+4 2 | 5+6 2 |
| | 9 2 | 4+5 2 | 5+6 2 | 3+4 2 |
| | 4+5 2 | 8 2 | 5+6 2 | 5+6 2 |
| | 8 2 | 8 2 | 3+4 2 | 5+6 2 |
| | 8 2 | 9 2 | 5+6 2 | 3+4 2 |
| | 4+5 2 | 4+5 2 | 5+6 2 | 5+6 2 |
| | 8 2 | 8 2 | 5+6 2 | 5+6 2 |
| | 6+7 2 | 6+7 2 | 3+4 2 | 5+6 2 |
| | 9 2 | 9 2 | 3+4 2 | 5+6 2 |
| | 9 2 | 9 2 | 3+4 2 | 3+4 2 |
| | 6+7 2 | 6+7 2 | 5+6 2 | 5+6 2 |
| | 9 2 | 9 2 | 3+4 2 | 3+4 2 |
| | 4+5 2 | 6+7 2 | 5+6 2 | 5+6 2 |
| | 8 2 | 9 2 | 3+4 2 | 3+4 2 |

Continued on next column

**TABLE LXXIII– continued**

| TQC | ICSD | | NoE | SF |
|---|---|---|---|---|
| | 8 2 | 8 2 | 3+4 2 | 5+6 2 |
| | 4+5 2 | 6+7 2 | 5+6 2 | 3+4 2 |
| | 9 2 | 9 2 | 3+4 2 | 3+4 2 |
| | 6+7 2 | 8 2 | 5+6 2 | 3+4 2 |
| | 8 2 | 4+5 2 | 3+4 2 | 5+6 2 |
| | 6+7 2 | 8 2 | 5+6 2 | 3+4 2 |
| | 9 2 | 4+5 2 | 5+6 2 | 5+6 2 |
| | 9 2 | 8 2 | 3+4 2 | 5+6 2 |
| | 8 2 | 9 2 | 5+6 2 | 5+6 2 |
| | 4+5 2 | 6+7 2 | 3+4 2 | 3+4 2 |
| | 8 2 | 9 2 | 3+4 2 | 3+4 2 |
| | 4+5 2 | 8 2 | 5+6 2 | 3+4 2 |
| | 8 2 | 9 2 | 3+4 2 | 3+4 2 |
| | 9 2 | 6+7 2 | 5+6 2 | 5+6 2 |
| | 8 2 | 4+5 2 | 5+6 2 | 5+6 2 |
| | 6+7 2 | 9 2 | 3+4 2 | 3+4 2 |
| | 9 2 | 8 2 | 5+6 2 | 5+6 2 |
| | 4+5 2 | 8 2 | 3+4 2 | 3+4 2 |
| | 9 2 | 4+5 2 | 5+6 2 | 5+6 2 |
| | 6+7 2 | 9 2 | 3+4 2 | 5+6 2 |
| | 8 2 | 6+7 2 | 5+6 2 | 5+6 2 |
| | 8 2 | 9 2 | 3+4 2 | 3+4 2 |
| 166.1.248 | 603098 | | 114 | Co3Y1 |
| | -GM | -T | -F | -L |
| | 8 2 | 8 2 | 3+4 2 | 3+4 2 |
| | 8 2 | 9 2 | 5+6 2 | 3+4 2 |
| | 9 2 | 8 2 | 3+4 2 | 5+6 2 |
| | 8 2 | 9 2 | 5+6 2 | 5+6 2 |
| | 9 2 | 9 2 | 5+6 2 | 3+4 2 |
| | 9 2 | 8 2 | 3+4 2 | 5+6 2 |
| | 4+5 2 | 6+7 2 | 5+6 2 | 3+4 2 |
| | 9 2 | 9 2 | 5+6 2 | 5+6 2 |
| | 6+7 2 | 6+7 2 | 5+6 2 | 5+6 2 |
| | 8 2 | 9 2 | 3+4 2 | 5+6 2 |
| | 9 2 | 8 2 | 5+6 2 | 3+4 2 |
| | 6+7 2 | 4+5 2 | 3+4 2 | 5+6 2 |
| | 8 2 | 8 2 | 5+6 2 | 5+6 2 |
| | 9 2 | 9 2 | 3+4 2 | 3+4 2 |
| | 8 2 | 8 2 | 3+4 2 | 3+4 2 |
| | 8 2 | 9 2 | 5+6 2 | 5+6 2 |
| | 9 2 | 9 2 | 3+4 2 | 3+4 2 |
| | 9 2 | 9 2 | 3+4 2 | 3+4 2 |
| | 6+7 2 | 6+7 2 | 3+4 2 | 5+6 2 |
| | 8 2 | 8 2 | 5+6 2 | 3+4 2 |
| | 4+5 2 | 4+5 2 | 3+4 2 | 5+6 2 |
| | 8 2 | 8 2 | 3+4 2 | 5+6 2 |
| | 9 2 | 8 2 | 5+6 2 | 5+6 2 |
| | 4+5 2 | 4+5 2 | 5+6 2 | 3+4 2 |
| | 8 2 | 8 2 | 3+4 2 | 5+6 2 |
| | 8 2 | 9 2 | 5+6 2 | 3+4 2 |
| | 4+5 2 | 4+5 2 | 3+4 2 | 5+6 2 |
| | 8 2 | 6+7 2 | 5+6 2 | 5+6 2 |
| | 6+7 2 | 9 2 | 5+6 2 | 5+6 2 |
| | 9 2 | 8 2 | 3+4 2 | 5+6 2 |
| | 9 2 | 6+7 2 | 3+4 2 | 3+4 2 |
| | 6+7 2 | 9 2 | 5+6 2 | 5+6 2 |
| | 9 2 | 9 2 | 3+4 2 | 5+6 2 |
| | 4+5 2 | 6+7 2 | 5+6 2 | 3+4 2 |
| | 8 2 | 9 2 | 3+4 2 | 3+4 2 |
| | 8 2 | 8 2 | 3+4 2 | 5+6 2 |
| | 4+5 2 | 6+7 2 | 5+6 2 | 3+4 2 |

Continued on next column

**TABLE LXXIII– continued**

| TQC | ICSD | NoE | SF | |
|---|---|---|---|---|
| | 9 2 | 9 2 | 3+4 2 | 3+4 2 |
| | 6+7 2 | 8 2 | 5+6 2 | 3+4 2 |
| | 8 2 | 4+5 2 | 3+4 2 | 5+6 2 |
| | 6+7 2 | 8 2 | 5+6 2 | 3+4 2 |
| | 9 2 | 4+5 2 | 5+6 2 | 5+6 2 |
| | 9 2 | 8 2 | 3+4 2 | 5+6 2 |
| | 8 2 | 9 2 | 5+6 2 | 5+6 2 |
| | 4+5 2 | 6+7 2 | 3+4 2 | 3+4 2 |
| | 8 2 | 9 2 | 3+4 2 | 3+4 2 |
| | 4+5 2 | 8 2 | 5+6 2 | 3+4 2 |
| | 8 2 | 9 2 | 3+4 2 | 5+6 2 |
| | 9 2 | 6+7 2 | 5+6 2 | 3+4 2 |
| | 8 2 | 4+5 2 | 5+6 2 | 5+6 2 |
| | 6+7 2 | 9 2 | 3+4 2 | 3+4 2 |
| | 9 2 | 8 2 | 5+6 2 | 5+6 2 |
| | 4+5 2 | 8 2 | 3+4 2 | 3+4 2 |
| | 9 2 | 4+5 2 | 3+4 2 | 5+6 2 |
| | 6+7 2 | 9 2 | 5+6 2 | 5+6 2 |
| | 8 2 | 6+7 2 | 5+6 2 | 5+6 2 |
| | 8 2 | 9 2 | 3+4 2 | 3+4 2 |
| 166.1.249 | 159894 | 114 | Co3Y1 | |
| | -GM | -T | -F | -L |
| | 8 2 | 8 2 | 3+4 2 | 3+4 2 |
| | 8 2 | 9 2 | 5+6 2 | 3+4 2 |
| | 9 2 | 8 2 | 3+4 2 | 5+6 2 |
| | 8 2 | 9 2 | 5+6 2 | 5+6 2 |
| | 9 2 | 9 2 | 5+6 2 | 3+4 2 |
| | 9 2 | 8 2 | 3+4 2 | 5+6 2 |
| | 4+5 2 | 6+7 2 | 5+6 2 | 3+4 2 |
| | 9 2 | 9 2 | 5+6 2 | 5+6 2 |
| | 6+7 2 | 6+7 2 | 5+6 2 | 5+6 2 |
| | 8 2 | 9 2 | 3+4 2 | 5+6 2 |
| | 9 2 | 8 2 | 5+6 2 | 3+4 2 |
| | 6+7 2 | 4+5 2 | 3+4 2 | 5+6 2 |
| | 8 2 | 8 2 | 5+6 2 | 5+6 2 |
| | 9 2 | 9 2 | 3+4 2 | 3+4 2 |
| | 8 2 | 8 2 | 3+4 2 | 3+4 2 |
| | 8 2 | 9 2 | 5+6 2 | 5+6 2 |
| | 9 2 | 9 2 | 3+4 2 | 3+4 2 |
| | 9 2 | 9 2 | 3+4 2 | 5+6 2 |
| | 8 2 | 6+7 2 | 3+4 2 | 3+4 2 |
| | 6+7 2 | 8 2 | 5+6 2 | 3+4 2 |
| | 4+5 2 | 4+5 2 | 3+4 2 | 5+6 2 |
| | 8 2 | 8 2 | 3+4 2 | 5+6 2 |
| | 9 2 | 8 2 | 5+6 2 | 5+6 2 |
| | 4+5 2 | 4+5 2 | 5+6 2 | 3+4 2 |
| | 8 2 | 8 2 | 3+4 2 | 5+6 2 |
| | 8 2 | 9 2 | 5+6 2 | 3+4 2 |
| | 4+5 2 | 6+7 2 | 3+4 2 | 5+6 2 |
| | 8 2 | 4+5 2 | 5+6 2 | 5+6 2 |
| | 6+7 2 | 9 2 | 3+4 2 | 5+6 2 |
| | 9 2 | 8 2 | 5+6 2 | 5+6 2 |
| | 9 2 | 6+7 2 | 3+4 2 | 3+4 2 |
| | 6+7 2 | 9 2 | 5+6 2 | 5+6 2 |
| | 9 2 | 9 2 | 3+4 2 | 5+6 2 |
| | 4+5 2 | 6+7 2 | 3+4 2 | 3+4 2 |
| | 8 2 | 9 2 | 5+6 2 | 5+6 2 |
| | 8 2 | 8 2 | 5+6 2 | 3+4 2 |
| | 4+5 2 | 6+7 2 | 3+4 2 | 3+4 2 |
| | 9 2 | 9 2 | 3+4 2 | 3+4 2 |
| | 6+7 2 | 8 2 | 5+6 2 | 3+4 2 |

Continued on next column

**TABLE LXXIII– continued**

| TQC | ICSD | NoE | SF | |
|---|---|---|---|---|
| | 8 2 | 4+5 2 | 3+4 2 | 5+6 2 |
| | 6+7 2 | 8 2 | 5+6 2 | 3+4 2 |
| | 9 2 | 4+5 2 | 5+6 2 | 5+6 2 |
| | 9 2 | 8 2 | 3+4 2 | 5+6 2 |
| | 8 2 | 9 2 | 5+6 2 | 5+6 2 |
| | 4+5 2 | 6+7 2 | 3+4 2 | 3+4 2 |
| | 8 2 | 9 2 | 3+4 2 | 3+4 2 |
| | 4+5 2 | 8 2 | 5+6 2 | 3+4 2 |
| | 8 2 | 9 2 | 3+4 2 | 5+6 2 |
| | 9 2 | 6+7 2 | 5+6 2 | 3+4 2 |
| | 8 2 | 4+5 2 | 5+6 2 | 5+6 2 |
| | 6+7 2 | 9 2 | 3+4 2 | 3+4 2 |
| | 9 2 | 8 2 | 5+6 2 | 5+6 2 |
| | 9 2 | 8 2 | 3+4 2 | 3+4 2 |
| | 4+5 2 | 4+5 2 | 3+4 2 | 5+6 2 |
| | 6+7 2 | 9 2 | 5+6 2 | 5+6 2 |
| | 8 2 | 6+7 2 | 5+6 2 | 5+6 2 |
| | 8 2 | 9 2 | 3+4 2 | 3+4 2 |
| 166.1.250 | 159890 | 114 | Co3Y1 | |
| | -GM | -T | -F | -L |
| | 8 2 | 8 2 | 3+4 2 | 3+4 2 |
| | 8 2 | 9 2 | 5+6 2 | 3+4 2 |
| | 9 2 | 8 2 | 3+4 2 | 5+6 2 |
| | 8 2 | 9 2 | 5+6 2 | 5+6 2 |
| | 9 2 | 9 2 | 5+6 2 | 3+4 2 |
| | 9 2 | 8 2 | 3+4 2 | 5+6 2 |
| | 4+5 2 | 6+7 2 | 5+6 2 | 3+4 2 |
| | 9 2 | 9 2 | 5+6 2 | 5+6 2 |
| | 6+7 2 | 6+7 2 | 5+6 2 | 5+6 2 |
| | 8 2 | 9 2 | 3+4 2 | 5+6 2 |
| | 9 2 | 8 2 | 5+6 2 | 3+4 2 |
| | 6+7 2 | 4+5 2 | 3+4 2 | 5+6 2 |
| | 8 2 | 8 2 | 5+6 2 | 5+6 2 |
| | 9 2 | 9 2 | 3+4 2 | 3+4 2 |
| | 8 2 | 8 2 | 3+4 2 | 3+4 2 |
| | 8 2 | 9 2 | 5+6 2 | 5+6 2 |
| | 9 2 | 9 2 | 3+4 2 | 3+4 2 |
| | 9 2 | 9 2 | 3+4 2 | 3+4 2 |
| | 6+7 2 | 6+7 2 | 3+4 2 | 5+6 2 |
| | 8 2 | 8 2 | 5+6 2 | 3+4 2 |
| | 4+5 2 | 4+5 2 | 3+4 2 | 5+6 2 |
| | 8 2 | 8 2 | 3+4 2 | 5+6 2 |
| | 9 2 | 8 2 | 5+6 2 | 5+6 2 |
| | 4+5 2 | 4+5 2 | 5+6 2 | 3+4 2 |
| | 8 2 | 8 2 | 3+4 2 | 5+6 2 |
| | 8 2 | 9 2 | 5+6 2 | 3+4 2 |
| | 4+5 2 | 4+5 2 | 3+4 2 | 5+6 2 |
| | 8 2 | 6+7 2 | 5+6 2 | 5+6 2 |
| | 6+7 2 | 8 2 | 5+6 2 | 5+6 2 |
| | 9 2 | 9 2 | 3+4 2 | 5+6 2 |
| | 9 2 | 6+7 2 | 3+4 2 | 3+4 2 |
| | 6+7 2 | 9 2 | 5+6 2 | 5+6 2 |
| | 9 2 | 9 2 | 3+4 2 | 5+6 2 |
| | 4+5 2 | 6+7 2 | 5+6 2 | 3+4 2 |
| | 8 2 | 9 2 | 3+4 2 | 3+4 2 |
| | 8 2 | 8 2 | 3+4 2 | 5+6 2 |
| | 4+5 2 | 6+7 2 | 5+6 2 | 3+4 2 |
| | 9 2 | 9 2 | 3+4 2 | 3+4 2 |
| | 6+7 2 | 8 2 | 5+6 2 | 3+4 2 |
| | 8 2 | 4+5 2 | 3+4 2 | 5+6 2 |
| | 6+7 2 | 8 2 | 5+6 2 | 3+4 2 |

Continued on next column

**TABLE LXXIII– continued**

| TQC | ICSD | NoE | SF | |
|---|---|---|---|---|
| | 9 2 | 4+5 2 | 5+6 2 | 5+6 2 |
| | 9 2 | 8 2 | 3+4 2 | 5+6 2 |
| | 8 2 | 9 2 | 5+6 2 | 5+6 2 |
| | 4+5 2 | 6+7 2 | 3+4 2 | 3+4 2 |
| | 8 2 | 9 2 | 3+4 2 | 3+4 2 |
| | 4+5 2 | 8 2 | 5+6 2 | 3+4 2 |
| | 8 2 | 9 2 | 3+4 2 | 5+6 2 |
| | 9 2 | 6+7 2 | 5+6 2 | 3+4 2 |
| | 8 2 | 4+5 2 | 5+6 2 | 5+6 2 |
| | 6+7 2 | 9 2 | 3+4 2 | 3+4 2 |
| | 9 2 | 8 2 | 5+6 2 | 5+6 2 |
| | 4+5 2 | 8 2 | 3+4 2 | 3+4 2 |
| | 9 2 | 4+5 2 | 3+4 2 | 5+6 2 |
| | 6+7 2 | 9 2 | 5+6 2 | 5+6 2 |
| | 8 2 | 6+7 2 | 5+6 2 | 5+6 2 |
| | 8 2 | 9 2 | 3+4 2 | 3+4 2 |
| 166.1.251 | 102729 | 114 | Co3Y1 | |
| | -GM | -T | -F | -L |
| | 8 2 | 8 2 | 3+4 2 | 3+4 2 |
| | 8 2 | 9 2 | 5+6 2 | 3+4 2 |
| | 9 2 | 8 2 | 3+4 2 | 5+6 2 |
| | 8 2 | 9 2 | 5+6 2 | 5+6 2 |
| | 9 2 | 9 2 | 5+6 2 | 3+4 2 |
| | 9 2 | 8 2 | 3+4 2 | 5+6 2 |
| | 4+5 2 | 9 2 | 5+6 2 | 3+4 2 |
| | 9 2 | 6+7 2 | 5+6 2 | 5+6 2 |
| | 8 2 | 6+7 2 | 5+6 2 | 5+6 2 |
| | 6+7 2 | 9 2 | 3+4 2 | 5+6 2 |
| | 9 2 | 8 2 | 5+6 2 | 3+4 2 |
| | 6+7 2 | 4+5 2 | 3+4 2 | 5+6 2 |
| | 8 2 | 8 2 | 5+6 2 | 5+6 2 |
| | 9 2 | 9 2 | 3+4 2 | 3+4 2 |
| | 8 2 | 8 2 | 3+4 2 | 3+4 2 |
| | 8 2 | 9 2 | 5+6 2 | 5+6 2 |
| | 9 2 | 9 2 | 3+4 2 | 3+4 2 |
| | 9 2 | 9 2 | 3+4 2 | 3+4 2 |
| | 6+7 2 | 6+7 2 | 3+4 2 | 5+6 2 |
| | 8 2 | 8 2 | 5+6 2 | 3+4 2 |
| | 4+5 2 | 4+5 2 | 3+4 2 | 5+6 2 |
| | 8 2 | 8 2 | 3+4 2 | 3+4 2 |
| | 9 2 | 8 2 | 5+6 2 | 5+6 2 |
| | 4+5 2 | 4+5 2 | 5+6 2 | 5+6 2 |
| | 8 2 | 8 2 | 3+4 2 | 5+6 2 |
| | 8 2 | 9 2 | 5+6 2 | 3+4 2 |
| | 4+5 2 | 4+5 2 | 5+6 2 | 5+6 2 |
| | 8 2 | 8 2 | 5+6 2 | 5+6 2 |
| | 6+7 2 | 6+7 2 | 3+4 2 | 3+4 2 |
| | 9 2 | 9 2 | 3+4 2 | 5+6 2 |
| | 9 2 | 9 2 | 3+4 2 | 5+6 2 |
| | 6+7 2 | 6+7 2 | 3+4 2 | 5+6 2 |
| | 9 2 | 9 2 | 5+6 2 | 3+4 2 |
| | 4+5 2 | 6+7 2 | 3+4 2 | 5+6 2 |
| | 8 2 | 9 2 | 5+6 2 | 3+4 2 |
| | 8 2 | 8 2 | 3+4 2 | 3+4 2 |
| | 4+5 2 | 6+7 2 | 3+4 2 | 5+6 2 |
| | 9 2 | 9 2 | 5+6 2 | 3+4 2 |
| | 6+7 2 | 4+5 2 | 5+6 2 | 5+6 2 |
| | 8 2 | 8 2 | 5+6 2 | 3+4 2 |
| | 6+7 2 | 8 2 | 3+4 2 | 3+4 2 |
| | 9 2 | 4+5 2 | 3+4 2 | 5+6 2 |
| | 9 2 | 8 2 | 5+6 2 | 5+6 2 |

Continued on next column

**TABLE LXXIII– continued**

| TQC | ICSD | NoE | SF | |
|---|---|---|---|---|
| | 8 2 | 9 2 | 5+6 2 | 5+6 2 |
| | 4+5 2 | 6+7 2 | 3+4 2 | 3+4 2 |
| | 8 2 | 9 2 | 3+4 2 | 3+4 2 |
| | 4+5 2 | 6+7 2 | 5+6 2 | 3+4 2 |
| | 8 2 | 8 2 | 3+4 2 | 3+4 2 |
| | 9 2 | 9 2 | 5+6 2 | 5+6 2 |
| | 8 2 | 4+5 2 | 5+6 2 | 5+6 2 |
| | 6+7 2 | 9 2 | 3+4 2 | 5+6 2 |
| | 9 2 | 8 2 | 5+6 2 | 3+4 2 |
| | 4+5 2 | 8 2 | 3+4 2 | 3+4 2 |
| | 9 2 | 4+5 2 | 5+6 2 | 5+6 2 |
| | 6+7 2 | 9 2 | 3+4 2 | 5+6 2 |
| | 8 2 | 6+7 2 | 5+6 2 | 5+6 2 |
| | 8 2 | 9 2 | 3+4 2 | 3+4 2 |
| 166.1.252 | 625506 | 150 | Co3Tm1 | |
| | -GM | -T | -F | -L |
| | 8 2 | 8 2 | 3+4 2 | 3+4 2 |
| | 8 2 | 9 2 | 5+6 2 | 3+4 2 |
| | 9 2 | 8 2 | 3+4 2 | 5+6 2 |
| | 8 2 | 9 2 | 5+6 2 | 5+6 2 |
| | 9 2 | 9 2 | 5+6 2 | 3+4 2 |
| | 9 2 | 8 2 | 3+4 2 | 5+6 2 |
| | 4+5 2 | 6+7 2 | 5+6 2 | 3+4 2 |
| | 9 2 | 9 2 | 3+4 2 | 5+6 2 |
| | 8 2 | 6+7 2 | 5+6 2 | 5+6 2 |
| | 6+7 2 | 9 2 | 5+6 2 | 5+6 2 |
| | 9 2 | 8 2 | 5+6 2 | 3+4 2 |
| | 6+7 2 | 4+5 2 | 3+4 2 | 5+6 2 |
| | 8 2 | 8 2 | 5+6 2 | 5+6 2 |
| | 9 2 | 9 2 | 3+4 2 | 3+4 2 |
| | 8 2 | 8 2 | 3+4 2 | 3+4 2 |
| | 8 2 | 9 2 | 5+6 2 | 5+6 2 |
| | 9 2 | 9 2 | 3+4 2 | 3+4 2 |
| | 9 2 | 9 2 | 3+4 2 | 5+6 2 |
| | 6+7 2 | 6+7 2 | 3+4 2 | 3+4 2 |
| | 8 2 | 8 2 | 5+6 2 | 3+4 2 |
| | 4+5 2 | 4+5 2 | 3+4 2 | 5+6 2 |
| | 8 2 | 8 2 | 3+4 2 | 5+6 2 |
| | 9 2 | 8 2 | 5+6 2 | 3+4 2 |
| | 8 2 | 4+5 2 | 5+6 2 | 5+6 2 |
| | 4+5 2 | 8 2 | 3+4 2 | 5+6 2 |
| | 8 2 | 9 2 | 5+6 2 | 3+4 2 |
| | 4+5 2 | 4+5 2 | 3+4 2 | 5+6 2 |
| | 8 2 | 9 2 | 5+6 2 | 5+6 2 |
| | 9 2 | 8 2 | 5+6 2 | 5+6 2 |
| | 6+7 2 | 6+7 2 | 3+4 2 | 5+6 2 |
| | 6+7 2 | 9 2 | 3+4 2 | 3+4 2 |
| | 9 2 | 6+7 2 | 5+6 2 | 5+6 2 |
| | 9 2 | 9 2 | 3+4 2 | 5+6 2 |
| | 9 2 | 6+7 2 | 3+4 2 | 3+4 2 |
| | 4+5 2 | 9 2 | 5+6 2 | 5+6 2 |
| | 8 2 | 8 2 | 5+6 2 | 3+4 2 |
| | 9 2 | 9 2 | 3+4 2 | 3+4 2 |
| | 8 2 | 8 2 | 3+4 2 | 5+6 2 |
| | 9 2 | 9 2 | 5+6 2 | 3+4 2 |
| | 8 2 | 9 2 | 3+4 2 | 5+6 2 |
| | 4+5 2 | 4+5 2 | 5+6 2 | 3+4 2 |
| | 6+7 2 | 6+7 2 | 5+6 2 | 3+4 2 |
| | 6+7 2 | 9 2 | 3+4 2 | 5+6 2 |
| | 9 2 | 6+7 2 | 5+6 2 | 3+4 2 |
| | 8 2 | 8 2 | 5+6 2 | 5+6 2 |

Continued on next column

**TABLE LXXIII– continued**

| TQC | ICSD | NoE | SF | |
|---|---|---|---|---|
| | 4+5 2 | 9 2 | 5+6 2 | 5+6 2 |
| | 8 2 | 6+7 2 | 3+4 2 | 5+6 2 |
| | 9 2 | 8 2 | 5+6 2 | 3+4 2 |
| | 6+7 2 | 4+5 2 | 3+4 2 | 3+4 2 |
| | 9 2 | 8 2 | 5+6 2 | 5+6 2 |
| | 6+7 2 | 4+5 2 | 5+6 2 | 5+6 2 |
| | 9 2 | 8 2 | 3+4 2 | 5+6 2 |
| | 8 2 | 9 2 | 5+6 2 | 3+4 2 |
| | 4+5 2 | 6+7 2 | 3+4 2 | 5+6 2 |
| | 8 2 | 9 2 | 3+4 2 | 3+4 2 |
| | 9 2 | 8 2 | 5+6 2 | 3+4 2 |
| | 8 2 | 9 2 | 3+4 2 | 3+4 2 |
| | 4+5 2 | 9 2 | 5+6 2 | 5+6 2 |
| | 8 2 | 8 2 | 5+6 2 | 5+6 2 |
| | 9 2 | 6+7 2 | 3+4 2 | 3+4 2 |
| | 6+7 2 | 4+5 2 | 5+6 2 | 5+6 2 |
| | 9 2 | 9 2 | 5+6 2 | 5+6 2 |
| | 8 2 | 9 2 | 3+4 2 | 3+4 2 |
| | 4+5 2 | 6+7 2 | 5+6 2 | 5+6 2 |
| | 9 2 | 9 2 | 3+4 2 | 5+6 2 |
| | 6+7 2 | 8 2 | 3+4 2 | 5+6 2 |
| | 9 2 | 4+5 2 | 5+6 2 | 3+4 2 |
| | 8 2 | 9 2 | 5+6 2 | 3+4 2 |
| | 9 2 | 6+7 2 | 3+4 2 | 5+6 2 |
| | 6+7 2 | 8 2 | 5+6 2 | 5+6 2 |
| | 9 2 | 9 2 | 5+6 2 | 3+4 2 |
| | 4+5 2 | 8 2 | 3+4 2 | 5+6 2 |
| | 8 2 | 4+5 2 | 5+6 2 | 5+6 2 |
| | 9 2 | 9 2 | 3+4 2 | 3+4 2 |
| | 8 2 | 8 2 | 5+6 2 | 5+6 2 |
| 166.1.253 | 625499 | 150 | Co3Tm1 | |
| | -GM | -T | -F | -L |
| | 8 2 | 8 2 | 3+4 2 | 3+4 2 |
| | 8 2 | 9 2 | 5+6 2 | 3+4 2 |
| | 9 2 | 8 2 | 3+4 2 | 5+6 2 |
| | 8 2 | 9 2 | 5+6 2 | 5+6 2 |
| | 9 2 | 9 2 | 5+6 2 | 3+4 2 |
| | 9 2 | 8 2 | 3+4 2 | 5+6 2 |
| | 4+5 2 | 6+7 2 | 5+6 2 | 3+4 2 |
| | 9 2 | 9 2 | 3+4 2 | 5+6 2 |
| | 8 2 | 6+7 2 | 5+6 2 | 5+6 2 |
| | 6+7 2 | 9 2 | 5+6 2 | 5+6 2 |
| | 9 2 | 8 2 | 5+6 2 | 3+4 2 |
| | 6+7 2 | 4+5 2 | 3+4 2 | 5+6 2 |
| | 8 2 | 8 2 | 5+6 2 | 5+6 2 |
| | 9 2 | 9 2 | 3+4 2 | 3+4 2 |
| | 8 2 | 8 2 | 3+4 2 | 3+4 2 |
| | 8 2 | 9 2 | 5+6 2 | 5+6 2 |
| | 9 2 | 9 2 | 3+4 2 | 3+4 2 |
| | 9 2 | 9 2 | 3+4 2 | 5+6 2 |
| | 6+7 2 | 6+7 2 | 3+4 2 | 3+4 2 |
| | 8 2 | 8 2 | 5+6 2 | 3+4 2 |
| | 4+5 2 | 4+5 2 | 3+4 2 | 5+6 2 |
| | 8 2 | 8 2 | 3+4 2 | 5+6 2 |
| | 9 2 | 8 2 | 5+6 2 | 3+4 2 |
| | 8 2 | 4+5 2 | 5+6 2 | 5+6 2 |
| | 4+5 2 | 8 2 | 3+4 2 | 5+6 2 |
| | 8 2 | 9 2 | 5+6 2 | 3+4 2 |
| | 4+5 2 | 4+5 2 | 3+4 2 | 5+6 2 |
| | 8 2 | 9 2 | 5+6 2 | 5+6 2 |
| | 9 2 | 8 2 | 5+6 2 | 5+6 2 |

Continued on next column

**TABLE LXXIII– continued**

| TQC | ICSD | NoE | SF | |
|---|---|---|---|---|
| | 6+7 2 | 6+7 2 | 3+4 2 | 5+6 2 |
| | 6+7 2 | 9 2 | 3+4 2 | 3+4 2 |
| | 9 2 | 6+7 2 | 5+6 2 | 5+6 2 |
| | 9 2 | 9 2 | 3+4 2 | 5+6 2 |
| | 9 2 | 6+7 2 | 3+4 2 | 3+4 2 |
| | 4+5 2 | 9 2 | 5+6 2 | 5+6 2 |
| | 8 2 | 8 2 | 5+6 2 | 3+4 2 |
| | 9 2 | 9 2 | 3+4 2 | 3+4 2 |
| | 8 2 | 9 2 | 3+4 2 | 5+6 2 |
| | 9 2 | 8 2 | 5+6 2 | 3+4 2 |
| | 8 2 | 9 2 | 3+4 2 | 5+6 2 |
| | 4+5 2 | 4+5 2 | 5+6 2 | 3+4 2 |
| | 6+7 2 | 6+7 2 | 5+6 2 | 3+4 2 |
| | 6+7 2 | 9 2 | 3+4 2 | 5+6 2 |
| | 9 2 | 6+7 2 | 5+6 2 | 5+6 2 |
| | 8 2 | 8 2 | 5+6 2 | 3+4 2 |
| | 4+5 2 | 9 2 | 5+6 2 | 5+6 2 |
| | 8 2 | 6+7 2 | 3+4 2 | 3+4 2 |
| | 9 2 | 8 2 | 5+6 2 | 5+6 2 |
| | 6+7 2 | 4+5 2 | 3+4 2 | 3+4 2 |
| | 9 2 | 8 2 | 5+6 2 | 5+6 2 |
| | 6+7 2 | 4+5 2 | 5+6 2 | 5+6 2 |
| | 9 2 | 8 2 | 3+4 2 | 5+6 2 |
| | 8 2 | 9 2 | 5+6 2 | 3+4 2 |
| | 4+5 2 | 6+7 2 | 3+4 2 | 5+6 2 |
| | 8 2 | 9 2 | 3+4 2 | 3+4 2 |
| | 9 2 | 8 2 | 5+6 2 | 3+4 2 |
| | 8 2 | 9 2 | 3+4 2 | 3+4 2 |
| | 4+5 2 | 9 2 | 5+6 2 | 5+6 2 |
| | 8 2 | 8 2 | 5+6 2 | 5+6 2 |
| | 9 2 | 6+7 2 | 3+4 2 | 3+4 2 |
| | 6+7 2 | 4+5 2 | 5+6 2 | 5+6 2 |
| | 9 2 | 9 2 | 5+6 2 | 5+6 2 |
| | 8 2 | 9 2 | 3+4 2 | 3+4 2 |
| | 4+5 2 | 6+7 2 | 5+6 2 | 5+6 2 |
| | 9 2 | 9 2 | 3+4 2 | 5+6 2 |
| | 6+7 2 | 8 2 | 3+4 2 | 5+6 2 |
| | 9 2 | 4+5 2 | 5+6 2 | 3+4 2 |
| | 8 2 | 9 2 | 5+6 2 | 3+4 2 |
| | 9 2 | 6+7 2 | 3+4 2 | 5+6 2 |
| | 6+7 2 | 8 2 | 5+6 2 | 5+6 2 |
| | 9 2 | 9 2 | 5+6 2 | 3+4 2 |
| | 4+5 2 | 4+5 2 | 3+4 2 | 5+6 2 |
| | 8 2 | 8 2 | 5+6 2 | 5+6 2 |
| | 9 2 | 9 2 | 3+4 2 | 3+4 2 |
| | 8 2 | 8 2 | 5+6 2 | 5+6 2 |
| 166.1.254 | 24581 | 96 | Co3Ta1 | |
| | -GM | -T | -F | -L |
| | 8 2 | 8 2 | 5+6 2 | 3+4 2 |
| | 8 2 | 9 2 | 3+4 2 | 5+6 2 |
| | 9 2 | 4+5 2 | 3+4 2 | 3+4 2 |
| | 4+5 2 | 8 2 | 5+6 2 | 5+6 2 |
| | 8 2 | 8 2 | 3+4 2 | 5+6 2 |
| | 4+5 2 | 4+5 2 | 3+4 2 | 3+4 2 |
| | 8 2 | 8 2 | 5+6 2 | 3+4 2 |
| | 9 2 | 6+7 2 | 5+6 2 | 5+6 2 |
| | 6+7 2 | 9 2 | 5+6 2 | 3+4 2 |
| | 8 2 | 8 2 | 3+4 2 | 3+4 2 |
| | 8 2 | 8 2 | 3+4 2 | 5+6 2 |
| | 9 2 | 9 2 | 5+6 2 | 3+4 2 |
| | 4+5 2 | 9 2 | 5+6 2 | 5+6 2 |

Continued on next column

**TABLE LXXIII– continued**

| TQC | ICSD | NoE | SF | |
|---|---|---|---|---|
| | 8 2 | 6+7 2 | 3+4 2 | 3+4 2 |
| | 8 2 | 8 2 | 5+6 2 | 5+6 2 |
| | 8 2 | 4+5 2 | 3+4 2 | 3+4 2 |
| | 4+5 2 | 8 2 | 3+4 2 | 5+6 2 |
| | 8 2 | 9 2 | 5+6 2 | 5+6 2 |
| | 9 2 | 4+5 2 | 3+4 2 | 5+6 2 |
| | 6+7 2 | 8 2 | 5+6 2 | 3+4 2 |
| | 9 2 | 9 2 | 5+6 2 | 5+6 2 |
| | 4+5 2 | 8 2 | 3+4 2 | 3+4 2 |
| | 8 2 | 8 2 | 3+4 2 | 5+6 2 |
| | 8 2 | 4+5 2 | 5+6 2 | 3+4 2 |
| | 9 2 | 9 2 | 5+6 2 | 5+6 2 |
| | 6+7 2 | 8 2 | 3+4 2 | 3+4 2 |
| | 8 2 | 8 2 | 3+4 2 | 5+6 2 |
| | 9 2 | 4+5 2 | 5+6 2 | 3+4 2 |
| | 8 2 | 8 2 | 5+6 2 | 5+6 2 |
| | 4+5 2 | 6+7 2 | 3+4 2 | 3+4 2 |
| | 9 2 | 4+5 2 | 5+6 2 | 3+4 2 |
| | 8 2 | 9 2 | 3+4 2 | 3+4 2 |
| | 8 2 | 8 2 | 5+6 2 | 5+6 2 |
| | 9 2 | 6+7 2 | 3+4 2 | 5+6 2 |
| | 4+5 2 | 9 2 | 5+6 2 | 5+6 2 |
| | 6+7 2 | 8 2 | 3+4 2 | 5+6 2 |
| | 8 2 | 4+5 2 | 5+6 2 | 3+4 2 |
| | 4+5 2 | 9 2 | 3+4 2 | 3+4 2 |
| | 9 2 | 6+7 2 | 5+6 2 | 5+6 2 |
| | 6+7 2 | 8 2 | 3+4 2 | 3+4 2 |
| | 8 2 | 9 2 | 5+6 2 | 5+6 2 |
| | 8 2 | 8 2 | 5+6 2 | 3+4 2 |
| | 4+5 2 | 4+5 2 | 3+4 2 | 5+6 2 |
| | 9 2 | 8 2 | 5+6 2 | 3+4 2 |
| | 8 2 | 8 2 | 3+4 2 | 5+6 2 |
| | 8 2 | 8 2 | 5+6 2 | 5+6 2 |
| | 9 2 | 9 2 | 3+4 2 | 5+6 2 |
| | 4+5 2 | 4+5 2 | 5+6 2 | 5+6 2 |
| 166.1.255 | 624740 | 120 | Co3Pr1 | |
| | -GM | -T | -F | -L |
| | 8 2 | 9 2 | 5+6 2 | 3+4 2 |
| | 8 2 | 8 2 | 3+4 2 | 3+4 2 |
| | 9 2 | 8 2 | 3+4 2 | 5+6 2 |
| | 8 2 | 9 2 | 5+6 2 | 3+4 2 |
| | 9 2 | 9 2 | 3+4 2 | 5+6 2 |
| | 9 2 | 8 2 | 5+6 2 | 5+6 2 |
| | 4+5 2 | 6+7 2 | 5+6 2 | 3+4 2 |
| | 8 2 | 9 2 | 3+4 2 | 5+6 2 |
| | 9 2 | 6+7 2 | 5+6 2 | 5+6 2 |
| | 6+7 2 | 8 2 | 5+6 2 | 5+6 2 |
| | 9 2 | 9 2 | 5+6 2 | 3+4 2 |
| | 6+7 2 | 4+5 2 | 3+4 2 | 5+6 2 |
| | 8 2 | 8 2 | 5+6 2 | 5+6 2 |
| | 9 2 | 9 2 | 3+4 2 | 3+4 2 |
| | 8 2 | 8 2 | 3+4 2 | 3+4 2 |
| | 8 2 | 9 2 | 3+4 2 | 5+6 2 |
| | 9 2 | 9 2 | 5+6 2 | 3+4 2 |
| | 8 2 | 6+7 2 | 3+4 2 | 3+4 2 |
| | 4+5 2 | 8 2 | 3+4 2 | 5+6 2 |
| | 9 2 | 9 2 | 5+6 2 | 5+6 2 |
| | 6+7 2 | 4+5 2 | 3+4 2 | 5+6 2 |
| | 8 2 | 8 2 | 5+6 2 | 3+4 2 |
| | 8 2 | 4+5 2 | 3+4 2 | 5+6 2 |
| | 4+5 2 | 8 2 | 5+6 2 | 3+4 2 |

Continued on next column

**TABLE LXXIII– continued**

| TQC | ICSD | NoE | SF | |
|---|---|---|---|---|
| | 9 2 | 8 2 | 3+4 2 | 3+4 2 |
| | 4+5 2 | 9 2 | 5+6 2 | 5+6 2 |
| | 8 2 | 4+5 2 | 3+4 2 | 5+6 2 |
| | 8 2 | 8 2 | 3+4 2 | 5+6 2 |
| | 6+7 2 | 9 2 | 3+4 2 | 5+6 2 |
| | 9 2 | 6+7 2 | 5+6 2 | 3+4 2 |
| | 9 2 | 9 2 | 3+4 2 | 5+6 2 |
| | 6+7 2 | 6+7 2 | 5+6 2 | 5+6 2 |
| | 9 2 | 9 2 | 5+6 2 | 3+4 2 |
| | 4+5 2 | 6+7 2 | 3+4 2 | 5+6 2 |
| | 8 2 | 9 2 | 5+6 2 | 5+6 2 |
| | 9 2 | 4+5 2 | 3+4 2 | 3+4 2 |
| | 6+7 2 | 8 2 | 3+4 2 | 3+4 2 |
| | 8 2 | 8 2 | 5+6 2 | 3+4 2 |
| | 4+5 2 | 6+7 2 | 3+4 2 | 3+4 2 |
| | 8 2 | 9 2 | 5+6 2 | 5+6 2 |
| | 9 2 | 8 2 | 5+6 2 | 5+6 2 |
| | 6+7 2 | 9 2 | 3+4 2 | 3+4 2 |
| | 9 2 | 8 2 | 5+6 2 | 5+6 2 |
| | 8 2 | 4+5 2 | 5+6 2 | 5+6 2 |
| | 9 2 | 6+7 2 | 3+4 2 | 3+4 2 |
| | 4+5 2 | 8 2 | 3+4 2 | 3+4 2 |
| | 8 2 | 9 2 | 5+6 2 | 5+6 2 |
| | 6+7 2 | 6+7 2 | 3+4 2 | 3+4 2 |
| | 4+5 2 | 4+5 2 | 5+6 2 | 3+4 2 |
| | 8 2 | 9 2 | 5+6 2 | 5+6 2 |
| | 8 2 | 9 2 | 3+4 2 | 3+4 2 |
| | 9 2 | 8 2 | 5+6 2 | 5+6 2 |
| | 4+5 2 | 8 2 | 3+4 2 | 3+4 2 |
| | 8 2 | 4+5 2 | 5+6 2 | 5+6 2 |
| | 9 2 | 9 2 | 3+4 2 | 5+6 2 |
| | 6+7 2 | 9 2 | 5+6 2 | 5+6 2 |
| | 9 2 | 6+7 2 | 3+4 2 | 3+4 2 |
| | 8 2 | 9 2 | 3+4 2 | 5+6 2 |
| | 4+5 2 | 9 2 | 3+4 2 | 3+4 2 |
| | 9 2 | 6+7 2 | 5+6 2 | 5+6 2 |
| 166.1.256 | 624707 | 120 | Co3Pr1 | |
| | -GM | -T | -F | -L |
| | 8 2 | 9 2 | 5+6 2 | 3+4 2 |
| | 8 2 | 8 2 | 3+4 2 | 3+4 2 |
| | 9 2 | 8 2 | 3+4 2 | 5+6 2 |
| | 8 2 | 9 2 | 5+6 2 | 3+4 2 |
| | 9 2 | 9 2 | 3+4 2 | 5+6 2 |
| | 9 2 | 8 2 | 5+6 2 | 5+6 2 |
| | 4+5 2 | 6+7 2 | 5+6 2 | 3+4 2 |
| | 8 2 | 9 2 | 3+4 2 | 5+6 2 |
| | 9 2 | 6+7 2 | 5+6 2 | 5+6 2 |
| | 6+7 2 | 9 2 | 5+6 2 | 5+6 2 |
| | 9 2 | 8 2 | 5+6 2 | 3+4 2 |
| | 6+7 2 | 4+5 2 | 3+4 2 | 5+6 2 |
| | 8 2 | 8 2 | 5+6 2 | 5+6 2 |
| | 9 2 | 9 2 | 3+4 2 | 3+4 2 |
| | 8 2 | 8 2 | 3+4 2 | 3+4 2 |
| | 8 2 | 9 2 | 3+4 2 | 5+6 2 |
| | 9 2 | 9 2 | 5+6 2 | 3+4 2 |
| | 8 2 | 6+7 2 | 3+4 2 | 3+4 2 |
| | 4+5 2 | 9 2 | 3+4 2 | 5+6 2 |
| | 9 2 | 8 2 | 5+6 2 | 5+6 2 |
| | 6+7 2 | 4+5 2 | 3+4 2 | 5+6 2 |
| | 8 2 | 8 2 | 5+6 2 | 3+4 2 |
| | 4+5 2 | 4+5 2 | 3+4 2 | 5+6 2 |

Continued on next column

**TABLE LXXIII– continued**

| TQC | ICSD | NoE | SF | |
|---|---|---|---|---|
| | 8 2 | 8 2 | 5+6 2 | 3+4 2 |
| | 9 2 | 8 2 | 3+4 2 | 3+4 2 |
| | 4+5 2 | 9 2 | 5+6 2 | 5+6 2 |
| | 8 2 | 4+5 2 | 3+4 2 | 5+6 2 |
| | 8 2 | 8 2 | 3+4 2 | 5+6 2 |
| | 6+7 2 | 6+7 2 | 3+4 2 | 5+6 2 |
| | 9 2 | 9 2 | 5+6 2 | 3+4 2 |
| | 9 2 | 9 2 | 3+4 2 | 5+6 2 |
| | 6+7 2 | 6+7 2 | 5+6 2 | 5+6 2 |
| | 9 2 | 9 2 | 5+6 2 | 3+4 2 |
| | 4+5 2 | 6+7 2 | 3+4 2 | 5+6 2 |
| | 8 2 | 9 2 | 5+6 2 | 5+6 2 |
| | 9 2 | 8 2 | 3+4 2 | 3+4 2 |
| | 6+7 2 | 4+5 2 | 5+6 2 | 3+4 2 |
| | 8 2 | 8 2 | 3+4 2 | 3+4 2 |
| | 4+5 2 | 6+7 2 | 3+4 2 | 3+4 2 |
| | 8 2 | 9 2 | 5+6 2 | 5+6 2 |
| | 9 2 | 8 2 | 5+6 2 | 5+6 2 |
| | 6+7 2 | 9 2 | 3+4 2 | 3+4 2 |
| | 9 2 | 8 2 | 5+6 2 | 5+6 2 |
| | 8 2 | 4+5 2 | 5+6 2 | 5+6 2 |
| | 9 2 | 6+7 2 | 3+4 2 | 3+4 2 |
| | 4+5 2 | 8 2 | 3+4 2 | 3+4 2 |
| | 8 2 | 9 2 | 5+6 2 | 5+6 2 |
| | 6+7 2 | 6+7 2 | 3+4 2 | 3+4 2 |
| | 4+5 2 | 4+5 2 | 5+6 2 | 5+6 2 |
| | 8 2 | 9 2 | 5+6 2 | 3+4 2 |
| | 8 2 | 9 2 | 3+4 2 | 3+4 2 |
| | 9 2 | 8 2 | 5+6 2 | 5+6 2 |
| | 4+5 2 | 8 2 | 3+4 2 | 3+4 2 |
| | 8 2 | 4+5 2 | 5+6 2 | 5+6 2 |
| | 9 2 | 9 2 | 3+4 2 | 5+6 2 |
| | 6+7 2 | 9 2 | 5+6 2 | 5+6 2 |
| | 9 2 | 6+7 2 | 3+4 2 | 3+4 2 |
| | 8 2 | 9 2 | 3+4 2 | 5+6 2 |
| | 4+5 2 | 9 2 | 3+4 2 | 3+4 2 |
| | 9 2 | 6+7 2 | 5+6 2 | 5+6 2 |
| 166.1.257 | 624697 | 120 | Co3Pr1 | |
| | -GM | -T | -F | -L |
| | 8 2 | 9 2 | 5+6 2 | 3+4 2 |
| | 8 2 | 8 2 | 3+4 2 | 3+4 2 |
| | 9 2 | 8 2 | 3+4 2 | 5+6 2 |
| | 8 2 | 9 2 | 5+6 2 | 3+4 2 |
| | 9 2 | 9 2 | 3+4 2 | 5+6 2 |
| | 9 2 | 8 2 | 5+6 2 | 5+6 2 |
| | 4+5 2 | 6+7 2 | 5+6 2 | 3+4 2 |
| | 8 2 | 9 2 | 3+4 2 | 5+6 2 |
| | 9 2 | 6+7 2 | 5+6 2 | 5+6 2 |
| | 6+7 2 | 9 2 | 5+6 2 | 5+6 2 |
| | 9 2 | 8 2 | 5+6 2 | 3+4 2 |
| | 6+7 2 | 4+5 2 | 3+4 2 | 5+6 2 |
| | 8 2 | 8 2 | 5+6 2 | 5+6 2 |
| | 9 2 | 9 2 | 3+4 2 | 3+4 2 |
| | 8 2 | 8 2 | 3+4 2 | 3+4 2 |
| | 8 2 | 9 2 | 3+4 2 | 5+6 2 |
| | 9 2 | 9 2 | 5+6 2 | 3+4 2 |
| | 8 2 | 6+7 2 | 3+4 2 | 3+4 2 |
| | 4+5 2 | 9 2 | 3+4 2 | 5+6 2 |
| | 9 2 | 8 2 | 5+6 2 | 5+6 2 |
| | 6+7 2 | 4+5 2 | 3+4 2 | 5+6 2 |
| | 8 2 | 8 2 | 5+6 2 | 3+4 2 |

Continued on next column

**TABLE LXXIII– continued**

| TQC | ICSD | NoE | SF | |
|---|---|---|---|---|
| | 4+5 2 | 4+5 2 | 3+4 2 | 5+6 2 |
| | 8 2 | 8 2 | 5+6 2 | 3+4 2 |
| | 9 2 | 8 2 | 3+4 2 | 3+4 2 |
| | 4+5 2 | 9 2 | 5+6 2 | 5+6 2 |
| | 8 2 | 4+5 2 | 3+4 2 | 5+6 2 |
| | 8 2 | 8 2 | 3+4 2 | 5+6 2 |
| | 6+7 2 | 6+7 2 | 3+4 2 | 5+6 2 |
| | 9 2 | 9 2 | 5+6 2 | 3+4 2 |
| | 9 2 | 9 2 | 3+4 2 | 5+6 2 |
| | 6+7 2 | 6+7 2 | 5+6 2 | 5+6 2 |
| | 9 2 | 9 2 | 5+6 2 | 3+4 2 |
| | 4+5 2 | 6+7 2 | 3+4 2 | 5+6 2 |
| | 8 2 | 9 2 | 5+6 2 | 5+6 2 |
| | 9 2 | 8 2 | 3+4 2 | 3+4 2 |
| | 6+7 2 | 4+5 2 | 5+6 2 | 3+4 2 |
| | 8 2 | 8 2 | 3+4 2 | 3+4 2 |
| | 4+5 2 | 6+7 2 | 3+4 2 | 3+4 2 |
| | 8 2 | 9 2 | 5+6 2 | 5+6 2 |
| | 9 2 | 8 2 | 5+6 2 | 5+6 2 |
| | 6+7 2 | 9 2 | 3+4 2 | 3+4 2 |
| | 9 2 | 8 2 | 5+6 2 | 5+6 2 |
| | 8 2 | 4+5 2 | 5+6 2 | 5+6 2 |
| | 9 2 | 6+7 2 | 3+4 2 | 3+4 2 |
| | 4+5 2 | 8 2 | 3+4 2 | 3+4 2 |
| | 8 2 | 9 2 | 5+6 2 | 5+6 2 |
| | 6+7 2 | 6+7 2 | 3+4 2 | 3+4 2 |
| | 4+5 2 | 4+5 2 | 5+6 2 | 5+6 2 |
| | 8 2 | 9 2 | 5+6 2 | 3+4 2 |
| | 8 2 | 9 2 | 3+4 2 | 3+4 2 |
| | 9 2 | 8 2 | 5+6 2 | 5+6 2 |
| | 4+5 2 | 8 2 | 3+4 2 | 3+4 2 |
| | 8 2 | 4+5 2 | 5+6 2 | 5+6 2 |
| | 9 2 | 9 2 | 3+4 2 | 5+6 2 |
| | 6+7 2 | 9 2 | 5+6 2 | 5+6 2 |
| | 9 2 | 6+7 2 | 3+4 2 | 3+4 2 |
| | 8 2 | 9 2 | 3+4 2 | 5+6 2 |
| | 4+5 2 | 9 2 | 3+4 2 | 3+4 2 |
| | 9 2 | 6+7 2 | 5+6 2 | 5+6 2 |
| 166.1.258 | 108889 | 120 | Co3Pr1 | |
| | -GM | -T | -F | -L |
| | 8 2 | 9 2 | 5+6 2 | 3+4 2 |
| | 8 2 | 8 2 | 3+4 2 | 3+4 2 |
| | 9 2 | 8 2 | 3+4 2 | 5+6 2 |
| | 8 2 | 9 2 | 5+6 2 | 3+4 2 |
| | 9 2 | 9 2 | 3+4 2 | 5+6 2 |
| | 9 2 | 8 2 | 5+6 2 | 5+6 2 |
| | 4+5 2 | 6+7 2 | 5+6 2 | 3+4 2 |
| | 8 2 | 9 2 | 3+4 2 | 5+6 2 |
| | 9 2 | 6+7 2 | 5+6 2 | 5+6 2 |
| | 6+7 2 | 9 2 | 5+6 2 | 5+6 2 |
| | 9 2 | 8 2 | 5+6 2 | 3+4 2 |
| | 6+7 2 | 4+5 2 | 3+4 2 | 5+6 2 |
| | 8 2 | 8 2 | 5+6 2 | 5+6 2 |
| | 9 2 | 9 2 | 3+4 2 | 3+4 2 |
| | 8 2 | 8 2 | 3+4 2 | 3+4 2 |
| | 8 2 | 9 2 | 3+4 2 | 5+6 2 |
| | 9 2 | 9 2 | 5+6 2 | 3+4 2 |
| | 8 2 | 6+7 2 | 3+4 2 | 3+4 2 |
| | 4+5 2 | 9 2 | 3+4 2 | 5+6 2 |
| | 9 2 | 8 2 | 5+6 2 | 5+6 2 |
| | 6+7 2 | 4+5 2 | 3+4 2 | 5+6 2 |

Continued on next column

**TABLE LXXIII– continued**

| TQC | ICSD | NoE | SF | |
|---|---|---|---|---|
| | 8 2 | 8 2 | 5+6 2 | 3+4 2 |
| | 4+5 2 | 4+5 2 | 3+4 2 | 5+6 2 |
| | 8 2 | 8 2 | 5+6 2 | 3+4 2 |
| | 9 2 | 8 2 | 3+4 2 | 3+4 2 |
| | 4+5 2 | 9 2 | 5+6 2 | 5+6 2 |
| | 8 2 | 4+5 2 | 3+4 2 | 5+6 2 |
| | 8 2 | 8 2 | 3+4 2 | 5+6 2 |
| | 6+7 2 | 6+7 2 | 3+4 2 | 5+6 2 |
| | 9 2 | 9 2 | 5+6 2 | 3+4 2 |
| | 9 2 | 9 2 | 3+4 2 | 5+6 2 |
| | 6+7 2 | 6+7 2 | 5+6 2 | 5+6 2 |
| | 9 2 | 9 2 | 5+6 2 | 3+4 2 |
| | 4+5 2 | 6+7 2 | 3+4 2 | 5+6 2 |
| | 8 2 | 9 2 | 5+6 2 | 5+6 2 |
| | 9 2 | 8 2 | 3+4 2 | 3+4 2 |
| | 6+7 2 | 4+5 2 | 5+6 2 | 3+4 2 |
| | 8 2 | 8 2 | 3+4 2 | 3+4 2 |
| | 4+5 2 | 6+7 2 | 3+4 2 | 3+4 2 |
| | 8 2 | 9 2 | 5+6 2 | 5+6 2 |
| | 9 2 | 8 2 | 5+6 2 | 5+6 2 |
| | 6+7 2 | 9 2 | 3+4 2 | 3+4 2 |
| | 9 2 | 8 2 | 5+6 2 | 5+6 2 |
| | 8 2 | 4+5 2 | 5+6 2 | 5+6 2 |
| | 9 2 | 6+7 2 | 3+4 2 | 3+4 2 |
| | 4+5 2 | 8 2 | 3+4 2 | 3+4 2 |
| | 8 2 | 9 2 | 5+6 2 | 5+6 2 |
| | 6+7 2 | 6+7 2 | 3+4 2 | 3+4 2 |
| | 4+5 2 | 4+5 2 | 5+6 2 | 5+6 2 |
| | 8 2 | 9 2 | 5+6 2 | 3+4 2 |
| | 8 2 | 9 2 | 3+4 2 | 3+4 2 |
| | 9 2 | 8 2 | 5+6 2 | 5+6 2 |
| | 4+5 2 | 8 2 | 3+4 2 | 3+4 2 |
| | 8 2 | 4+5 2 | 5+6 2 | 5+6 2 |
| | 9 2 | 9 2 | 3+4 2 | 5+6 2 |
| | 6+7 2 | 9 2 | 5+6 2 | 5+6 2 |
| | 9 2 | 6+7 2 | 3+4 2 | 3+4 2 |
| | 8 2 | 9 2 | 3+4 2 | 5+6 2 |
| | 4+5 2 | 9 2 | 3+4 2 | 3+4 2 |
| | 9 2 | 6+7 2 | 5+6 2 | 5+6 2 |
| 166.1.259 | 624054 | 156 | Co3Lu1 | |
| | -GM | -T | -F | -L |
| | 8 2 | 8 2 | 3+4 2 | 3+4 2 |
| | 8 2 | 9 2 | 5+6 2 | 3+4 2 |
| | 9 2 | 8 2 | 3+4 2 | 5+6 2 |
| | 9 2 | 9 2 | 5+6 2 | 5+6 2 |
| | 8 2 | 9 2 | 5+6 2 | 3+4 2 |
| | 9 2 | 8 2 | 3+4 2 | 5+6 2 |
| | 9 2 | 9 2 | 5+6 2 | 5+6 2 |
| | 4+5 2 | 6+7 2 | 5+6 2 | 3+4 2 |
| | 6+7 2 | 6+7 2 | 5+6 2 | 5+6 2 |
| | 8 2 | 9 2 | 3+4 2 | 5+6 2 |
| | 9 2 | 8 2 | 5+6 2 | 3+4 2 |
| | 6+7 2 | 4+5 2 | 3+4 2 | 5+6 2 |
| | 8 2 | 8 2 | 5+6 2 | 5+6 2 |
| | 9 2 | 9 2 | 5+6 2 | 5+6 2 |
| | 6+7 2 | 9 2 | 5+6 2 | 5+6 2 |
| | 9 2 | 6+7 2 | 5+6 2 | 5+6 2 |
| | 9 2 | 9 2 | 3+4 2 | 3+4 2 |
| | 8 2 | 8 2 | 3+4 2 | 5+6 2 |
| | 4+5 2 | 9 2 | 5+6 2 | 3+4 2 |
| | 6+7 2 | 4+5 2 | 3+4 2 | 5+6 2 |

Continued on next column

**TABLE LXXIII– continued**

| TQC | ICSD | NoE | SF | |
|---|---|---|---|---|
| | 9 2 | 6+7 2 | 3+4 2 | 3+4 2 |
| | 8 2 | 9 2 | 5+6 2 | 5+6 2 |
| | 9 2 | 8 2 | 5+6 2 | 3+4 2 |
| | 8 2 | 8 2 | 3+4 2 | 3+4 2 |
| | 8 2 | 9 2 | 5+6 2 | 5+6 2 |
| | 9 2 | 9 2 | 5+6 2 | 5+6 2 |
| | 6+7 2 | 6+7 2 | 5+6 2 | 5+6 2 |
| | 9 2 | 9 2 | 5+6 2 | 5+6 2 |
| | 9 2 | 9 2 | 5+6 2 | 5+6 2 |
| | 8 2 | 9 2 | 3+4 2 | 3+4 2 |
| | 9 2 | 8 2 | 3+4 2 | 5+6 2 |
| | 6+7 2 | 4+5 2 | 3+4 2 | 5+6 2 |
| | 9 2 | 9 2 | 5+6 2 | 3+4 2 |
| | 4+5 2 | 8 2 | 5+6 2 | 3+4 2 |
| | 8 2 | 6+7 2 | 3+4 2 | 5+6 2 |
| | 8 2 | 8 2 | 5+6 2 | 3+4 2 |
| | 9 2 | 9 2 | 5+6 2 | 5+6 2 |
| | 9 2 | 9 2 | 3+4 2 | 3+4 2 |
| | 6+7 2 | 6+7 2 | 3+4 2 | 3+4 2 |
| | 4+5 2 | 9 2 | 3+4 2 | 5+6 2 |
| | 9 2 | 8 2 | 3+4 2 | 5+6 2 |
| | 8 2 | 8 2 | 5+6 2 | 3+4 2 |
| | 8 2 | 4+5 2 | 3+4 2 | 3+4 2 |
| | 9 2 | 8 2 | 5+6 2 | 5+6 2 |
| | 8 2 | 9 2 | 5+6 2 | 5+6 2 |
| | 4+5 2 | 4+5 2 | 3+4 2 | 5+6 2 |
| | 8 2 | 8 2 | 5+6 2 | 3+4 2 |
| | 4+5 2 | 9 2 | 5+6 2 | 5+6 2 |
| | 8 2 | 6+7 2 | 3+4 2 | 5+6 2 |
| | 9 2 | 4+5 2 | 5+6 2 | 5+6 2 |
| | 6+7 2 | 8 2 | 3+4 2 | 5+6 2 |
| | 9 2 | 9 2 | 3+4 2 | 3+4 2 |
| | 6+7 2 | 6+7 2 | 5+6 2 | 5+6 2 |
| | 9 2 | 9 2 | 3+4 2 | 5+6 2 |
| | 4+5 2 | 9 2 | 5+6 2 | 3+4 2 |
| | 8 2 | 6+7 2 | 3+4 2 | 3+4 2 |
| | 8 2 | 8 2 | 3+4 2 | 5+6 2 |
| | 4+5 2 | 9 2 | 5+6 2 | 3+4 2 |
| | 8 2 | 6+7 2 | 3+4 2 | 3+4 2 |
| | 9 2 | 8 2 | 5+6 2 | 3+4 2 |
| | 6+7 2 | 4+5 2 | 3+4 2 | 5+6 2 |
| | 9 2 | 8 2 | 5+6 2 | 3+4 2 |
| | 6+7 2 | 4+5 2 | 5+6 2 | 5+6 2 |
| | 9 2 | 8 2 | 5+6 2 | 5+6 2 |
| | 8 2 | 9 2 | 3+4 2 | 3+4 2 |
| | 4+5 2 | 6+7 2 | 3+4 2 | 5+6 2 |
| | 8 2 | 9 2 | 3+4 2 | 3+4 2 |
| | 4+5 2 | 6+7 2 | 5+6 2 | 3+4 2 |
| | 8 2 | 8 2 | 3+4 2 | 3+4 2 |
| | 9 2 | 9 2 | 5+6 2 | 5+6 2 |
| | 8 2 | 4+5 2 | 5+6 2 | 5+6 2 |
| | 6+7 2 | 9 2 | 3+4 2 | 3+4 2 |
| | 9 2 | 8 2 | 5+6 2 | 5+6 2 |
| | 9 2 | 8 2 | 3+4 2 | 3+4 2 |
| | 4+5 2 | 4+5 2 | 3+4 2 | 5+6 2 |
| | 6+7 2 | 9 2 | 5+6 2 | 5+6 2 |
| | 8 2 | 6+7 2 | 5+6 2 | 5+6 2 |
| | 8 2 | 9 2 | 3+4 2 | 3+4 2 |
| 166.1.260 | 624039 | 156 | Co3Lu1 | |
| | -GM | -T | -F | -L |
| | 8 2 | 8 2 | 3+4 2 | 3+4 2 |

Continued on next column

**TABLE LXXIII– continued**

| TQC | ICSD | NoE | SF |
|---|---|---|---|
| | 8 2 | 9 2 5+6 2 | 3+4 2 |
| | 9 2 | 8 2 3+4 2 | 5+6 2 |
| | 9 2 | 9 2 5+6 2 | 5+6 2 |
| | 8 2 | 9 2 5+6 2 | 3+4 2 |
| | 9 2 | 8 2 3+4 2 | 5+6 2 |
| | 9 2 | 9 2 5+6 2 | 5+6 2 |
| | 4+5 2 | 6+7 2 5+6 2 | 3+4 2 |
| | 6+7 2 | 6+7 2 5+6 2 | 5+6 2 |
| | 8 2 | 9 2 3+4 2 | 5+6 2 |
| | 9 2 | 8 2 5+6 2 | 3+4 2 |
| | 6+7 2 | 4+5 2 3+4 2 | 5+6 2 |
| | 8 2 | 8 2 5+6 2 | 5+6 2 |
| | 9 2 | 9 2 5+6 2 | 5+6 2 |
| | 6+7 2 | 9 2 5+6 2 | 5+6 2 |
| | 9 2 | 6+7 2 5+6 2 | 5+6 2 |
| | 9 2 | 9 2 3+4 2 | 3+4 2 |
| | 8 2 | 8 2 3+4 2 | 5+6 2 |
| | 4+5 2 | 9 2 5+6 2 | 3+4 2 |
| | 6+7 2 | 4+5 2 3+4 2 | 5+6 2 |
| | 9 2 | 6+7 2 5+6 2 | 3+4 2 |
| | 8 2 | 9 2 3+4 2 | 5+6 2 |
| | 9 2 | 8 2 5+6 2 | 3+4 2 |
| | 8 2 | 8 2 3+4 2 | 3+4 2 |
| | 8 2 | 9 2 5+6 2 | 5+6 2 |
| | 9 2 | 9 2 5+6 2 | 5+6 2 |
| | 6+7 2 | 6+7 2 5+6 2 | 5+6 2 |
| | 9 2 | 9 2 5+6 2 | 5+6 2 |
| | 9 2 | 9 2 5+6 2 | 5+6 2 |
| | 8 2 | 9 2 3+4 2 | 5+6 2 |
| | 9 2 | 8 2 3+4 2 | 3+4 2 |
| | 6+7 2 | 4+5 2 3+4 2 | 5+6 2 |
| | 9 2 | 9 2 5+6 2 | 3+4 2 |
| | 4+5 2 | 8 2 5+6 2 | 3+4 2 |
| | 8 2 | 6+7 2 5+6 2 | 5+6 2 |
| | 8 2 | 8 2 3+4 2 | 3+4 2 |
| | 9 2 | 9 2 5+6 2 | 5+6 2 |
| | 9 2 | 9 2 3+4 2 | 3+4 2 |
| | 6+7 2 | 6+7 2 3+4 2 | 3+4 2 |
| | 4+5 2 | 9 2 3+4 2 | 5+6 2 |
| | 9 2 | 8 2 3+4 2 | 5+6 2 |
| | 8 2 | 8 2 5+6 2 | 3+4 2 |
| | 8 2 | 4+5 2 3+4 2 | 3+4 2 |
| | 9 2 | 8 2 5+6 2 | 5+6 2 |
| | 8 2 | 9 2 5+6 2 | 5+6 2 |
| | 4+5 2 | 4+5 2 3+4 2 | 5+6 2 |
| | 8 2 | 8 2 5+6 2 | 3+4 2 |
| | 4+5 2 | 9 2 3+4 2 | 5+6 2 |
| | 8 2 | 6+7 2 5+6 2 | 5+6 2 |
| | 9 2 | 4+5 2 5+6 2 | 5+6 2 |
| | 6+7 2 | 8 2 3+4 2 | 5+6 2 |
| | 9 2 | 9 2 3+4 2 | 3+4 2 |
| | 6+7 2 | 6+7 2 5+6 2 | 5+6 2 |
| | 9 2 | 9 2 3+4 2 | 5+6 2 |
| | 4+5 2 | 9 2 5+6 2 | 3+4 2 |
| | 8 2 | 6+7 2 3+4 2 | 3+4 2 |
| | 8 2 | 8 2 3+4 2 | 5+6 2 |
| | 4+5 2 | 9 2 5+6 2 | 3+4 2 |
| | 8 2 | 6+7 2 3+4 2 | 3+4 2 |
| | 9 2 | 8 2 5+6 2 | 3+4 2 |
| | 6+7 2 | 4+5 2 3+4 2 | 5+6 2 |
| | 9 2 | 8 2 5+6 2 | 3+4 2 |

Continued on next column

**TABLE LXXIII– continued**

| TQC | ICSD | NoE | SF |
|---|---|---|---|
| | 6+7 2 | 4+5 2 5+6 2 | 5+6 2 |
| | 9 2 | 8 2 5+6 2 | 5+6 2 |
| | 8 2 | 9 2 3+4 2 | 3+4 2 |
| | 4+5 2 | 6+7 2 3+4 2 | 5+6 2 |
| | 8 2 | 9 2 3+4 2 | 3+4 2 |
| | 4+5 2 | 6+7 2 5+6 2 | 3+4 2 |
| | 8 2 | 8 2 3+4 2 | 3+4 2 |
| | 9 2 | 9 2 5+6 2 | 5+6 2 |
| | 8 2 | 4+5 2 5+6 2 | 5+6 2 |
| | 6+7 2 | 9 2 3+4 2 | 3+4 2 |
| | 9 2 | 8 2 5+6 2 | 5+6 2 |
| | 9 2 | 8 2 3+4 2 | 3+4 2 |
| | 4+5 2 | 4+5 2 3+4 2 | 5+6 2 |
| | 6+7 2 | 9 2 5+6 2 | 5+6 2 |
| | 8 2 | 6+7 2 5+6 2 | 5+6 2 |
| | 8 2 | 9 2 3+4 2 | 3+4 2 |
| 166.1.261 | 246307 | 14 | Co1Li1O2 |
| | -GM | -T -F | -L |
| | 8 2 | 9 2 5+6 2 | 3+4 2 |
| | 9 2 | 8 2 3+4 2 | 5+6 2 |
| | 8 2 | 9 2 5+6 2 | 3+4 2 |
| | 8 2 | 8 2 3+4 2 | 5+6 2 |
| | 4+5 2 | 9 2 5+6 2 | 3+4 2 |
| | 9 2 | 6+7 2 5+6 2 | 5+6 2 |
| | 9 2 | 8 2 3+4 2 | 3+4 2 |
| 166.1.262 | 657284 | 204 | Co17Gd2N3 |
| | -GM | -T -F | -L |
| | 8 2 | 9 2 3+4 2 | 5+6 2 |
| | 9 2 | 8 2 5+6 2 | 3+4 2 |
| | 8 2 | 8 2 3+4 2 | 5+6 2 |
| | 9 2 | 9 2 5+6 2 | 3+4 2 |
| | 4+5 2 | 4+5 2 5+6 2 | 5+6 2 |
| | 8 2 | 6+7 2 3+4 2 | 3+4 2 |
| | 6+7 2 | 8 2 3+4 2 | 5+6 2 |
| | 9 2 | 9 2 5+6 2 | 3+4 2 |
| | 8 2 | 8 2 5+6 2 | 5+6 2 |
| | 8 2 | 8 2 3+4 2 | 3+4 2 |
| | 4+5 2 | 4+5 2 5+6 2 | 5+6 2 |
| | 8 2 | 9 2 3+4 2 | 3+4 2 |
| | 9 2 | 6+7 2 5+6 2 | 3+4 2 |
| | 6+7 2 | 9 2 5+6 2 | 5+6 2 |
| | 6+7 2 | 9 2 3+4 2 | 5+6 2 |
| | 9 2 | 6+7 2 3+4 2 | 3+4 2 |
| | 9 2 | 9 2 3+4 2 | 3+4 2 |
| | 9 2 | 9 2 3+4 2 | 5+6 2 |
| | 9 2 | 8 2 5+6 2 | 3+4 2 |
| | 9 2 | 9 2 5+6 2 | 3+4 2 |
| | 6+7 2 | 9 2 5+6 2 | 5+6 2 |
| | 8 2 | 6+7 2 3+4 2 | 3+4 2 |
| | 4+5 2 | 6+7 2 3+4 2 | 3+4 2 |
| | 8 2 | 9 2 3+4 2 | 3+4 2 |
| | 8 2 | 4+5 2 5+6 2 | 5+6 2 |
| | 6+7 2 | 8 2 3+4 2 | 5+6 2 |
| | 9 2 | 8 2 5+6 2 | 3+4 2 |
| | 4+5 2 | 9 2 5+6 2 | 3+4 2 |
| | 8 2 | 9 2 3+4 2 | 5+6 2 |
| | 9 2 | 6+7 2 3+4 2 | 3+4 2 |
| | 8 2 | 9 2 5+6 2 | 5+6 2 |
| | 4+5 2 | 6+7 2 5+6 2 | 5+6 2 |
| | 8 2 | 9 2 3+4 2 | 3+4 2 |
| | 8 2 | 4+5 2 5+6 2 | 5+6 2 |

Continued on next column

**TABLE LXXIII– continued**

| TQC | ICSD | NoE | SF | |
|---|---|---|---|---|
| | 6+7 2 | 8 2 | 5+6 2 | 3+4 2 |
| | 9 2 | 4+5 2 | 5+6 2 | 5+6 2 |
| | 8 2 | 8 2 | 3+4 2 | 3+4 2 |
| | 4+5 2 | 8 2 | 3+4 2 | 3+4 2 |
| | 8 2 | 6+7 2 | 5+6 2 | 5+6 2 |
| | 4+5 2 | 9 2 | 3+4 2 | 3+4 2 |
| | 8 2 | 6+7 2 | 5+6 2 | 5+6 2 |
| | 6+7 2 | 4+5 2 | 3+4 2 | 5+6 2 |
| | 9 2 | 9 2 | 3+4 2 | 5+6 2 |
| | 6+7 2 | 8 2 | 5+6 2 | 3+4 2 |
| | 9 2 | 9 2 | 5+6 2 | 5+6 2 |
| | 4+5 2 | 9 2 | 3+4 2 | 5+6 2 |
| | 8 2 | 8 2 | 5+6 2 | 3+4 2 |
| | 8 2 | 6+7 2 | 5+6 2 | 5+6 2 |
| | 9 2 | 9 2 | 5+6 2 | 3+4 2 |
| | 6+7 2 | 6+7 2 | 3+4 2 | 3+4 2 |
| | 8 2 | 9 2 | 5+6 2 | 5+6 2 |
| | 9 2 | 8 2 | 3+4 2 | 3+4 2 |
| | 4+5 2 | 9 2 | 3+4 2 | 3+4 2 |
| | 8 2 | 4+5 2 | 5+6 2 | 3+4 2 |
| | 6+7 2 | 8 2 | 3+4 2 | 5+6 2 |
| | 9 2 | 9 2 | 5+6 2 | 5+6 2 |
| | 8 2 | 8 2 | 3+4 2 | 3+4 2 |
| | 4+5 2 | 6+7 2 | 3+4 2 | 3+4 2 |
| | 9 2 | 9 2 | 3+4 2 | 5+6 2 |
| | 9 2 | 4+5 2 | 5+6 2 | 3+4 2 |
| | 6+7 2 | 8 2 | 3+4 2 | 5+6 2 |
| | 9 2 | 8 2 | 5+6 2 | 3+4 2 |
| | 8 2 | 4+5 2 | 3+4 2 | 5+6 2 |
| | 9 2 | 8 2 | 5+6 2 | 5+6 2 |
| | 4+5 2 | 8 2 | 3+4 2 | 3+4 2 |
| | 8 2 | 9 2 | 5+6 2 | 5+6 2 |
| | 8 2 | 4+5 2 | 5+6 2 | 3+4 2 |
| | 8 2 | 8 2 | 3+4 2 | 5+6 2 |
| | 4+5 2 | 9 2 | 5+6 2 | 3+4 2 |
| | 8 2 | 8 2 | 3+4 2 | 5+6 2 |
| | 9 2 | 4+5 2 | 3+4 2 | 5+6 2 |
| | 6+7 2 | 8 2 | 5+6 2 | 3+4 2 |
| | 9 2 | 6+7 2 | 5+6 2 | 3+4 2 |
| | 9 2 | 9 2 | 3+4 2 | 5+6 2 |
| | 8 2 | 8 2 | 5+6 2 | 3+4 2 |
| | 4+5 2 | 4+5 2 | 3+4 2 | 5+6 2 |
| | 8 2 | 6+7 2 | 3+4 2 | 3+4 2 |
| | 8 2 | 9 2 | 5+6 2 | 5+6 2 |
| | 9 2 | 8 2 | 3+4 2 | 3+4 2 |
| | 6+7 2 | 9 2 | 5+6 2 | 5+6 2 |
| | 9 2 | 6+7 2 | 5+6 2 | 3+4 2 |
| | 4+5 2 | 4+5 2 | 3+4 2 | 5+6 2 |
| | 8 2 | 8 2 | 5+6 2 | 3+4 2 |
| | 4+5 2 | 9 2 | 3+4 2 | 5+6 2 |
| | 6+7 2 | 8 2 | 5+6 2 | 3+4 2 |
| | 8 2 | 6+7 2 | 3+4 2 | 5+6 2 |
| | 9 2 | 9 2 | 5+6 2 | 3+4 2 |
| | 8 2 | 8 2 | 5+6 2 | 3+4 2 |
| | 8 2 | 6+7 2 | 3+4 2 | 5+6 2 |
| | 4+5 2 | 9 2 | 5+6 2 | 3+4 2 |
| | 9 2 | 9 2 | 3+4 2 | 5+6 2 |
| | 8 2 | 4+5 2 | 5+6 2 | 3+4 2 |
| | 8 2 | 8 2 | 3+4 2 | 3+4 2 |
| | 4+5 2 | 4+5 2 | 5+6 2 | 5+6 2 |
| | 6+7 2 | 6+7 2 | 5+6 2 | 3+4 2 |

Continued on next column

**TABLE LXXIII– continued**

| TQC | ICSD | NoE | SF | |
|---|---|---|---|---|
| | 8 2 | 8 2 | 3+4 2 | 5+6 2 |
| | 9 2 | 9 2 | 3+4 2 | 3+4 2 |
| | 9 2 | 9 2 | 5+6 2 | 5+6 2 |
| | 8 2 | 8 2 | 5+6 2 | 3+4 2 |
| | 4+5 2 | 6+7 2 | 3+4 2 | 3+4 2 |
| | 9 2 | 9 2 | 5+6 2 | 5+6 2 |
| | 6+7 2 | 9 2 | 3+4 2 | 5+6 2 |
| 166.1.263 | 79717 | 198 | Co10Ge3O16 | |
| | -GM | -T | -F | -L |
| | 8 2 | 8 2 | 3+4 2 | 3+4 2 |
| | 8 2 | 9 2 | 5+6 2 | 3+4 2 |
| | 9 2 | 8 2 | 3+4 2 | 5+6 2 |
| | 9 2 | 9 2 | 5+6 2 | 5+6 2 |
| | 9 2 | 8 2 | 3+4 2 | 5+6 2 |
| | 6+7 2 | 9 2 | 5+6 2 | 5+6 2 |
| | 9 2 | 6+7 2 | 5+6 2 | 5+6 2 |
| | 9 2 | 8 2 | 3+4 2 | 3+4 2 |
| | 6+7 2 | 4+5 2 | 5+6 2 | 3+4 2 |
| | 8 2 | 9 2 | 5+6 2 | 5+6 2 |
| | 4+5 2 | 6+7 2 | 3+4 2 | 5+6 2 |
| | 8 2 | 9 2 | 5+6 2 | 3+4 2 |
| | 8 2 | 8 2 | 3+4 2 | 3+4 2 |
| | 9 2 | 9 2 | 5+6 2 | 5+6 2 |
| | 4+5 2 | 4+5 2 | 3+4 2 | 3+4 2 |
| | 8 2 | 8 2 | 3+4 2 | 3+4 2 |
| | 8 2 | 9 2 | 3+4 2 | 5+6 2 |
| | 9 2 | 8 2 | 5+6 2 | 3+4 2 |
| | 8 2 | 8 2 | 3+4 2 | 3+4 2 |
| | 8 2 | 9 2 | 5+6 2 | 5+6 2 |
| | 4+5 2 | 8 2 | 3+4 2 | 3+4 2 |
| | 8 2 | 6+7 2 | 5+6 2 | 5+6 2 |
| | 9 2 | 4+5 2 | 5+6 2 | 5+6 2 |
| | 9 2 | 8 2 | 3+4 2 | 5+6 2 |
| | 6+7 2 | 9 2 | 5+6 2 | 3+4 2 |
| | 8 2 | 9 2 | 3+4 2 | 3+4 2 |
| | 4+5 2 | 9 2 | 5+6 2 | 3+4 2 |
| | 9 2 | 6+7 2 | 3+4 2 | 5+6 2 |
| | 8 2 | 8 2 | 5+6 2 | 5+6 2 |
| | 9 2 | 4+5 2 | 3+4 2 | 3+4 2 |
| | 6+7 2 | 8 2 | 5+6 2 | 3+4 2 |
| | 9 2 | 8 2 | 3+4 2 | 5+6 2 |
| | 6+7 2 | 4+5 2 | 5+6 2 | 5+6 2 |
| | 8 2 | 8 2 | 3+4 2 | 5+6 2 |
| | 9 2 | 9 2 | 3+4 2 | 5+6 2 |
| | 8 2 | 6+7 2 | 5+6 2 | 5+6 2 |
| | 4+5 2 | 9 2 | 3+4 2 | 5+6 2 |
| | 9 2 | 4+5 2 | 5+6 2 | 3+4 2 |
| | 6+7 2 | 8 2 | 3+4 2 | 3+4 2 |
| | 4+5 2 | 9 2 | 5+6 2 | 3+4 2 |
| | 8 2 | 6+7 2 | 5+6 2 | 3+4 2 |
| | 8 2 | 9 2 | 5+6 2 | 5+6 2 |
| | 8 2 | 8 2 | 3+4 2 | 3+4 2 |
| | 9 2 | 4+5 2 | 3+4 2 | 5+6 2 |
| | 4+5 2 | 8 2 | 5+6 2 | 5+6 2 |
| | 8 2 | 9 2 | 5+6 2 | 5+6 2 |
| | 9 2 | 6+7 2 | 5+6 2 | 3+4 2 |
| | 6+7 2 | 9 2 | 3+4 2 | 3+4 2 |
| | 9 2 | 6+7 2 | 5+6 2 | 3+4 2 |
| | 4+5 2 | 9 2 | 3+4 2 | 3+4 2 |
| | 8 2 | 8 2 | 3+4 2 | 5+6 2 |
| | 9 2 | 9 2 | 3+4 2 | 5+6 2 |

Continued on next column

**TABLE LXXIII– continued**

| TQC | ICSD | NoE | SF | |
|---|---|---|---|---|
| | 6+7 2 | 4+5 2 | 5+6 2 | 3+4 2 |
| | 9 2 | 8 2 | 3+4 2 | 5+6 2 |
| | 9 2 | 8 2 | 3+4 2 | 5+6 2 |
| | 6+7 2 | 9 2 | 5+6 2 | 3+4 2 |
| | 8 2 | 6+7 2 | 5+6 2 | 5+6 2 |
| | 8 2 | 9 2 | 3+4 2 | 5+6 2 |
| | 4+5 2 | 4+5 2 | 5+6 2 | 3+4 2 |
| | 4+5 2 | 8 2 | 3+4 2 | 3+4 2 |
| | 8 2 | 8 2 | 3+4 2 | 5+6 2 |
| | 9 2 | 4+5 2 | 5+6 2 | 3+4 2 |
| | 9 2 | 9 2 | 5+6 2 | 3+4 2 |
| | 6+7 2 | 6+7 2 | 3+4 2 | 3+4 2 |
| | 8 2 | 8 2 | 5+6 2 | 5+6 2 |
| | 8 2 | 4+5 2 | 3+4 2 | 5+6 2 |
| | 4+5 2 | 8 2 | 3+4 2 | 3+4 2 |
| | 8 2 | 9 2 | 3+4 2 | 5+6 2 |
| | 9 2 | 6+7 2 | 3+4 2 | 5+6 2 |
| | 9 2 | 9 2 | 5+6 2 | 3+4 2 |
| | 4+5 2 | 8 2 | 5+6 2 | 5+6 2 |
| | 6+7 2 | 9 2 | 3+4 2 | 5+6 2 |
| | 8 2 | 6+7 2 | 3+4 2 | 3+4 2 |
| | 8 2 | 9 2 | 5+6 2 | 3+4 2 |
| | 9 2 | 8 2 | 5+6 2 | 3+4 2 |
| | 4+5 2 | 4+5 2 | 5+6 2 | 5+6 2 |
| | 6+7 2 | 8 2 | 3+4 2 | 3+4 2 |
| | 8 2 | 4+5 2 | 5+6 2 | 5+6 2 |
| | 8 2 | 8 2 | 3+4 2 | 5+6 2 |
| | 9 2 | 8 2 | 5+6 2 | 3+4 2 |
| | 4+5 2 | 4+5 2 | 5+6 2 | 5+6 2 |
| | 8 2 | 8 2 | 3+4 2 | 5+6 2 |
| | 4+5 2 | 9 2 | 5+6 2 | 3+4 2 |
| | 8 2 | 6+7 2 | 3+4 2 | 5+6 2 |
| | 8 2 | 8 2 | 5+6 2 | 3+4 2 |
| | 9 2 | 4+5 2 | 3+4 2 | 5+6 2 |
| | 8 2 | 8 2 | 5+6 2 | 3+4 2 |
| | 4+5 2 | 9 2 | 5+6 2 | 5+6 2 |
| | 8 2 | 9 2 | 3+4 2 | 5+6 2 |
| | 9 2 | 8 2 | 3+4 2 | 5+6 2 |
| | 8 2 | 9 2 | 3+4 2 | 3+4 2 |
| | 8 2 | 8 2 | 5+6 2 | 3+4 2 |
| | 6+7 2 | 6+7 2 | 5+6 2 | 5+6 2 |
| | 4+5 2 | 4+5 2 | 3+4 2 | 5+6 2 |
| | 8 2 | 8 2 | 3+4 2 | 5+6 2 |
| | 8 2 | 8 2 | 5+6 2 | 5+6 2 |
| | 4+5 2 | 4+5 2 | 5+6 2 | 3+4 2 |
| | 8 2 | 8 2 | 3+4 2 | 5+6 2 |
| | 4+5 2 | 4+5 2 | 3+4 2 | 3+4 2 |
| 166.1.264 | 401026 | 70 | Cl8Na2Ti3 | |
| | -GM | -T | -F | -L |
| | 8 2 | 8 2 | 5+6 2 | 5+6 2 |
| | 9 2 | 9 2 | 3+4 2 | 3+4 2 |
| | 8 2 | 9 2 | 3+4 2 | 3+4 2 |
| | 9 2 | 6+7 2 | 5+6 2 | 5+6 2 |
| | 4+5 2 | 9 2 | 3+4 2 | 5+6 2 |
| | 8 2 | 8 2 | 5+6 2 | 5+6 2 |
| | 6+7 2 | 4+5 2 | 5+6 2 | 3+4 2 |
| | 9 2 | 8 2 | 3+4 2 | 3+4 2 |
| | 8 2 | 8 2 | 5+6 2 | 5+6 2 |
| | 4+5 2 | 4+5 2 | 5+6 2 | 3+4 2 |
| | 8 2 | 8 2 | 3+4 2 | 5+6 2 |
| | 9 2 | 9 2 | 5+6 2 | 5+6 2 |

Continued on next column

**TABLE LXXIII– continued**

| TQC | ICSD | NoE | SF | |
|---|---|---|---|---|
| | 8 2 | 9 2 | 3+4 2 | 3+4 2 |
| | 9 2 | 6+7 2 | 5+6 2 | 3+4 2 |
| | 6+7 2 | 8 2 | 3+4 2 | 5+6 2 |
| | 8 2 | 6+7 2 | 3+4 2 | 3+4 2 |
| | 4+5 2 | 9 2 | 5+6 2 | 3+4 2 |
| | 8 2 | 8 2 | 3+4 2 | 5+6 2 |
| | 8 2 | 4+5 2 | 5+6 2 | 5+6 2 |
| | 8 2 | 8 2 | 5+6 2 | 3+4 2 |
| | 4+5 2 | 9 2 | 3+4 2 | 5+6 2 |
| | 6+7 2 | 8 2 | 3+4 2 | 5+6 2 |
| | 9 2 | 4+5 2 | 5+6 2 | 3+4 2 |
| | 9 2 | 9 2 | 5+6 2 | 3+4 2 |
| | 9 2 | 8 2 | 3+4 2 | 5+6 2 |
| | 9 2 | 9 2 | 3+4 2 | 5+6 2 |
| | 6+7 2 | 9 2 | 5+6 2 | 3+4 2 |
| | 8 2 | 6+7 2 | 3+4 2 | 3+4 2 |
| | 4+5 2 | 8 2 | 3+4 2 | 5+6 2 |
| | 9 2 | 4+5 2 | 5+6 2 | 3+4 2 |
| | 9 2 | 9 2 | 5+6 2 | 5+6 2 |
| | 6+7 2 | 6+7 2 | 3+4 2 | 3+4 2 |
| | 8 2 | 8 2 | 5+6 2 | 5+6 2 |
| | 4+5 2 | 4+5 2 | 5+6 2 | 5+6 2 |
| | 8 2 | 8 2 | 3+4 2 | 5+6 2 |
| 166.1.265 | 869 | 22 | Cl1Zr1 | |
| | -GM | -T | -F | -L |
| | 8 2 | 8 2 | 5+6 2 | 5+6 2 |
| | 9 2 | 9 2 | 3+4 2 | 3+4 2 |
| | 8 2 | 8 2 | 3+4 2 | 3+4 2 |
| | 9 2 | 9 2 | 5+6 2 | 5+6 2 |
| | 8 2 | 9 2 | 5+6 2 | 5+6 2 |
| | 4+5 2 | 8 2 | 3+4 2 | 3+4 2 |
| | 9 2 | 6+7 2 | 5+6 2 | 5+6 2 |
| | 6+7 2 | 4+5 2 | 3+4 2 | 3+4 2 |
| | 8 2 | 9 2 | 3+4 2 | 5+6 2 |
| | 9 2 | 8 2 | 5+6 2 | 5+6 2 |
| | 6+7 2 | 4+5 2 | 3+4 2 | 3+4 2 |
| 166.1.266 | 20145 | 22 | Cl1Zr1 | |
| | -GM | -T | -F | -L |
| | 8 2 | 8 2 | 5+6 2 | 5+6 2 |
| | 9 2 | 9 2 | 3+4 2 | 3+4 2 |
| | 8 2 | 8 2 | 3+4 2 | 3+4 2 |
| | 9 2 | 9 2 | 5+6 2 | 5+6 2 |
| | 8 2 | 8 2 | 3+4 2 | 5+6 2 |
| | 8 2 | 8 2 | 5+6 2 | 3+4 2 |
| | 4+5 2 | 4+5 2 | 5+6 2 | 5+6 2 |
| | 9 2 | 9 2 | 3+4 2 | 5+6 2 |
| | 9 2 | 9 2 | 5+6 2 | 3+4 2 |
| | 6+7 2 | 6+7 2 | 3+4 2 | 5+6 2 |
| | 6+7 2 | 4+5 2 | 3+4 2 | 3+4 2 |
| 166.1.267 | 6077 | 74 | Cl1O1Yb1 | |
| | -GM | -T | -F | -L |
| | 8 2 | 8 2 | 5+6 2 | 5+6 2 |
| | 9 2 | 9 2 | 3+4 2 | 3+4 2 |
| | 8 2 | 8 2 | 5+6 2 | 5+6 2 |
| | 9 2 | 9 2 | 3+4 2 | 3+4 2 |
| | 8 2 | 8 2 | 5+6 2 | 5+6 2 |
| | 4+5 2 | 4+5 2 | 3+4 2 | 3+4 2 |
| | 9 2 | 9 2 | 5+6 2 | 5+6 2 |
| | 6+7 2 | 6+7 2 | 3+4 2 | 3+4 2 |
| | 8 2 | 8 2 | 5+6 2 | 5+6 2 |
| | 9 2 | 9 2 | 3+4 2 | 3+4 2 |

Continued on next column

**TABLE LXXIII– continued**

| TQC | ICSD | NoE | SF | |
|---|---|---|---|---|
| | 8 2 | 9 2 | 5+6 2 | 3+4 2 |
| | 9 2 | 8 2 | 3+4 2 | 5+6 2 |
| | 8 2 | 8 2 | 3+4 2 | 3+4 2 |
| | 8 2 | 9 2 | 5+6 2 | 5+6 2 |
| | 4+5 2 | 4+5 2 | 5+6 2 | 5+6 2 |
| | 9 2 | 8 2 | 3+4 2 | 3+4 2 |
| | 6+7 2 | 6+7 2 | 5+6 2 | 5+6 2 |
| | 9 2 | 9 2 | 5+6 2 | 5+6 2 |
| | 8 2 | 8 2 | 3+4 2 | 3+4 2 |
| | 4+5 2 | 9 2 | 3+4 2 | 3+4 2 |
| | 9 2 | 4+5 2 | 5+6 2 | 5+6 2 |
| | 6+7 2 | 6+7 2 | 3+4 2 | 3+4 2 |
| | 8 2 | 8 2 | 5+6 2 | 3+4 2 |
| | 8 2 | 9 2 | 5+6 2 | 5+6 2 |
| | 9 2 | 8 2 | 3+4 2 | 5+6 2 |
| | 9 2 | 9 2 | 5+6 2 | 3+4 2 |
| | 8 2 | 8 2 | 3+4 2 | 5+6 2 |
| | 4+5 2 | 9 2 | 5+6 2 | 3+4 2 |
| | 9 2 | 4+5 2 | 3+4 2 | 5+6 2 |
| | 6+7 2 | 6+7 2 | 3+4 2 | 3+4 2 |
| | 8 2 | 8 2 | 3+4 2 | 3+4 2 |
| | 8 2 | 9 2 | 5+6 2 | 5+6 2 |
| | 9 2 | 8 2 | 5+6 2 | 5+6 2 |
| | 4+5 2 | 6+7 2 | 5+6 2 | 3+4 2 |
| | 6+7 2 | 4+5 2 | 3+4 2 | 5+6 2 |
| | 8 2 | 9 2 | 3+4 2 | 3+4 2 |
| | 9 2 | 8 2 | 5+6 2 | 5+6 2 |
| 166.1.268 | 24410 | 36 | Cl1La1 | |
| | -GM | -T | -F | -L |
| | 8 2 | 9 2 | 5+6 2 | 3+4 2 |
| | 9 2 | 8 2 | 3+4 2 | 5+6 2 |
| | 8 2 | 9 2 | 5+6 2 | 3+4 2 |
| | 9 2 | 8 2 | 3+4 2 | 5+6 2 |
| | 8 2 | 8 2 | 5+6 2 | 5+6 2 |
| | 9 2 | 9 2 | 3+4 2 | 3+4 2 |
| | 4+5 2 | 6+7 2 | 5+6 2 | 3+4 2 |
| | 6+7 2 | 4+5 2 | 3+4 2 | 5+6 2 |
| | 8 2 | 9 2 | 5+6 2 | 3+4 2 |
| | 9 2 | 8 2 | 3+4 2 | 5+6 2 |
| | 8 2 | 8 2 | 3+4 2 | 3+4 2 |
| | 8 2 | 8 2 | 5+6 2 | 5+6 2 |
| | 4+5 2 | 9 2 | 5+6 2 | 5+6 2 |
| | 9 2 | 4+5 2 | 5+6 2 | 5+6 2 |
| | 9 2 | 9 2 | 3+4 2 | 3+4 2 |
| | 6+7 2 | 6+7 2 | 3+4 2 | 3+4 2 |
| | 8 2 | 9 2 | 3+4 2 | 5+6 2 |
| | 9 2 | 8 2 | 3+4 2 | 3+4 2 |
| 166.1.269 | 154513 | 206 | Ce2Cu5Zn2 | |
| | -GM | -T | -F | -L |
| | 8 2 | 9 2 | 5+6 2 | 3+4 2 |
| | 8 2 | 8 2 | 3+4 2 | 3+4 2 |
| | 9 2 | 9 2 | 5+6 2 | 5+6 2 |
| | 9 2 | 8 2 | 3+4 2 | 5+6 2 |
| | 8 2 | 9 2 | 5+6 2 | 3+4 2 |
| | 8 2 | 8 2 | 3+4 2 | 5+6 2 |
| | 9 2 | 9 2 | 3+4 2 | 3+4 2 |
| | 9 2 | 8 2 | 5+6 2 | 5+6 2 |
| | 4+5 2 | 6+7 2 | 5+6 2 | 3+4 2 |
| | 8 2 | 9 2 | 3+4 2 | 5+6 2 |
| | 6+7 2 | 8 2 | 5+6 2 | 5+6 2 |
| | 8 2 | 6+7 2 | 3+4 2 | 3+4 2 |

Continued on next column

**TABLE LXXIII– continued**

| TQC | ICSD | NoE | SF | |
|---|---|---|---|---|
| | 9 2 | 4+5 2 | 3+4 2 | 3+4 2 |
| | 4+5 2 | 8 2 | 5+6 2 | 5+6 2 |
| | 6+7 2 | 9 2 | 5+6 2 | 3+4 2 |
| | 9 2 | 4+5 2 | 3+4 2 | 5+6 2 |
| | 8 2 | 8 2 | 5+6 2 | 5+6 2 |
| | 8 2 | 8 2 | 5+6 2 | 5+6 2 |
| | 8 2 | 4+5 2 | 5+6 2 | 5+6 2 |
| | 4+5 2 | 8 2 | 3+4 2 | 3+4 2 |
| | 8 2 | 8 2 | 3+4 2 | 3+4 2 |
| | 8 2 | 9 2 | 5+6 2 | 5+6 2 |
| | 4+5 2 | 8 2 | 3+4 2 | 5+6 2 |
| | 8 2 | 4+5 2 | 3+4 2 | 5+6 2 |
| | 8 2 | 9 2 | 5+6 2 | 5+6 2 |
| | 4+5 2 | 8 2 | 3+4 2 | 3+4 2 |
| | 8 2 | 6+7 2 | 3+4 2 | 3+4 2 |
| | 8 2 | 8 2 | 5+6 2 | 5+6 2 |
| | 9 2 | 4+5 2 | 3+4 2 | 5+6 2 |
| | 4+5 2 | 6+7 2 | 3+4 2 | 3+4 2 |
| | 4+5 2 | 9 2 | 3+4 2 | 5+6 2 |
| | 8 2 | 9 2 | 5+6 2 | 5+6 2 |
| | 8 2 | 8 2 | 3+4 2 | 5+6 2 |
| | 4+5 2 | 4+5 2 | 5+6 2 | 5+6 2 |
| | 8 2 | 8 2 | 5+6 2 | 5+6 2 |
| | 8 2 | 8 2 | 5+6 2 | 5+6 2 |
| | 8 2 | 4+5 2 | 5+6 2 | 5+6 2 |
| | 4+5 2 | 8 2 | 3+4 2 | 3+4 2 |
| | 8 2 | 8 2 | 3+4 2 | 5+6 2 |
| | 9 2 | 9 2 | 5+6 2 | 3+4 2 |
| | 8 2 | 9 2 | 3+4 2 | 5+6 2 |
| | 9 2 | 8 2 | 5+6 2 | 5+6 2 |
| | 9 2 | 4+5 2 | 5+6 2 | 3+4 2 |
| | 6+7 2 | 9 2 | 5+6 2 | 5+6 2 |
| | 8 2 | 8 2 | 3+4 2 | 3+4 2 |
| | 4+5 2 | 9 2 | 3+4 2 | 3+4 2 |
| | 8 2 | 6+7 2 | 3+4 2 | 3+4 2 |
| | 9 2 | 8 2 | 3+4 2 | 5+6 2 |
| | 8 2 | 8 2 | 5+6 2 | 3+4 2 |
| | 6+7 2 | 9 2 | 3+4 2 | 5+6 2 |
| | 9 2 | 6+7 2 | 5+6 2 | 3+4 2 |
| | 8 2 | 9 2 | 3+4 2 | 5+6 2 |
| | 6+7 2 | 4+5 2 | 5+6 2 | 3+4 2 |
| | 9 2 | 8 2 | 5+6 2 | 5+6 2 |
| | 4+5 2 | 4+5 2 | 3+4 2 | 3+4 2 |
| | 4+5 2 | 8 2 | 5+6 2 | 5+6 2 |
| | 8 2 | 6+7 2 | 3+4 2 | 3+4 2 |
| | 8 2 | 9 2 | 5+6 2 | 5+6 2 |
| | 9 2 | 6+7 2 | 3+4 2 | 3+4 2 |
| | 6+7 2 | 8 2 | 5+6 2 | 5+6 2 |
| | 9 2 | 4+5 2 | 5+6 2 | 3+4 2 |
| | 4+5 2 | 9 2 | 3+4 2 | 5+6 2 |
| | 8 2 | 9 2 | 5+6 2 | 3+4 2 |
| | 9 2 | 9 2 | 3+4 2 | 5+6 2 |
| | 6+7 2 | 8 2 | 5+6 2 | 3+4 2 |
| | 9 2 | 6+7 2 | 3+4 2 | 5+6 2 |
| | 8 2 | 9 2 | 5+6 2 | 3+4 2 |
| | 9 2 | 4+5 2 | 5+6 2 | 3+4 2 |
| | 4+5 2 | 8 2 | 3+4 2 | 5+6 2 |
| | 6+7 2 | 4+5 2 | 3+4 2 | 5+6 2 |
| | 8 2 | 8 2 | 5+6 2 | 5+6 2 |
| | 9 2 | 8 2 | 3+4 2 | 3+4 2 |
| | 6+7 2 | 6+7 2 | 3+4 2 | 3+4 2 |

Continued on next column

**TABLE LXXIII– continued**

| TQC | ICSD | NoE | SF | |
|---|---|---|---|---|
| | 4+5 2 | 8 2 | 5+6 2 | 5+6 2 |
| | 9 2 | 9 2 | 5+6 2 | 5+6 2 |
| | 8 2 | 4+5 2 | 3+4 2 | 3+4 2 |
| | 9 2 | 6+7 2 | 3+4 2 | 5+6 2 |
| | 6+7 2 | 9 2 | 5+6 2 | 3+4 2 |
| | 8 2 | 8 2 | 3+4 2 | 5+6 2 |
| | 4+5 2 | 4+5 2 | 5+6 2 | 3+4 2 |
| | 8 2 | 9 2 | 5+6 2 | 5+6 2 |
| | 9 2 | 8 2 | 3+4 2 | 3+4 2 |
| | 4+5 2 | 6+7 2 | 5+6 2 | 3+4 2 |
| | 8 2 | 9 2 | 3+4 2 | 5+6 2 |
| | 8 2 | 9 2 | 5+6 2 | 5+6 2 |
| | 9 2 | 6+7 2 | 3+4 2 | 5+6 2 |
| | 6+7 2 | 8 2 | 3+4 2 | 3+4 2 |
| | 8 2 | 4+5 2 | 5+6 2 | 3+4 2 |
| | 4+5 2 | 8 2 | 5+6 2 | 3+4 2 |
| | 8 2 | 9 2 | 3+4 2 | 5+6 2 |
| | 9 2 | 8 2 | 3+4 2 | 3+4 2 |
| | 9 2 | 9 2 | 5+6 2 | 5+6 2 |
| | 6+7 2 | 4+5 2 | 3+4 2 | 5+6 2 |
| | 9 2 | 9 2 | 5+6 2 | 3+4 2 |
| | 8 2 | 8 2 | 3+4 2 | 3+4 2 |
| | 4+5 2 | 6+7 2 | 5+6 2 | 5+6 2 |
| | 9 2 | 9 2 | 3+4 2 | 5+6 2 |
| | 8 2 | 8 2 | 5+6 2 | 3+4 2 |
| | 9 2 | 8 2 | 5+6 2 | 3+4 2 |
| | 9 2 | 4+5 2 | 5+6 2 | 5+6 2 |
| | 6+7 2 | 9 2 | 3+4 2 | 5+6 2 |
| | 9 2 | 6+7 2 | 5+6 2 | 3+4 2 |
| | 6+7 2 | 9 2 | 3+4 2 | 3+4 2 |
| 166.1.270 | 620668 | 174 | Ce2Co7 | |
| | -GM | -T | -F | -L |
| | 8 2 | 9 2 | 5+6 2 | 3+4 2 |
| | 8 2 | 8 2 | 3+4 2 | 5+6 2 |
| | 9 2 | 9 2 | 5+6 2 | 3+4 2 |
| | 9 2 | 8 2 | 3+4 2 | 5+6 2 |
| | 8 2 | 9 2 | 5+6 2 | 3+4 2 |
| | 8 2 | 8 2 | 3+4 2 | 5+6 2 |
| | 9 2 | 9 2 | 3+4 2 | 3+4 2 |
| | 9 2 | 8 2 | 5+6 2 | 5+6 2 |
| | 4+5 2 | 6+7 2 | 5+6 2 | 3+4 2 |
| | 8 2 | 9 2 | 3+4 2 | 5+6 2 |
| | 9 2 | 8 2 | 3+4 2 | 3+4 2 |
| | 6+7 2 | 6+7 2 | 5+6 2 | 5+6 2 |
| | 8 2 | 4+5 2 | 3+4 2 | 3+4 2 |
| | 4+5 2 | 9 2 | 5+6 2 | 5+6 2 |
| | 9 2 | 8 2 | 5+6 2 | 3+4 2 |
| | 6+7 2 | 4+5 2 | 3+4 2 | 5+6 2 |
| | 8 2 | 8 2 | 5+6 2 | 5+6 2 |
| | 9 2 | 9 2 | 3+4 2 | 3+4 2 |
| | 8 2 | 8 2 | 3+4 2 | 5+6 2 |
| | 9 2 | 9 2 | 5+6 2 | 3+4 2 |
| | 8 2 | 8 2 | 5+6 2 | 5+6 2 |
| | 8 2 | 8 2 | 3+4 2 | 5+6 2 |
| | 8 2 | 9 2 | 5+6 2 | 5+6 2 |
| | 4+5 2 | 6+7 2 | 5+6 2 | 3+4 2 |
| | 9 2 | 9 2 | 3+4 2 | 5+6 2 |
| | 6+7 2 | 8 2 | 3+4 2 | 3+4 2 |
| | 8 2 | 4+5 2 | 3+4 2 | 5+6 2 |
| | 8 2 | 8 2 | 5+6 2 | 3+4 2 |
| | 4+5 2 | 4+5 2 | 3+4 2 | 3+4 2 |
| | 9 2 | 9 2 | 3+4 2 | 5+6 2 |
| | 8 2 | 8 2 | 5+6 2 | 5+6 2 |
| | 4+5 2 | 4+5 2 | 5+6 2 | 3+4 2 |
| | 8 2 | 8 2 | 5+6 2 | 5+6 2 |
| | 8 2 | 8 2 | 3+4 2 | 5+6 2 |
| | 9 2 | 9 2 | 3+4 2 | 3+4 2 |
| | 4+5 2 | 6+7 2 | 5+6 2 | 5+6 2 |
| | 6+7 2 | 4+5 2 | 5+6 2 | 5+6 2 |
| | 9 2 | 8 2 | 5+6 2 | 3+4 2 |
| | 8 2 | 9 2 | 3+4 2 | 5+6 2 |
| | 6+7 2 | 6+7 2 | 3+4 2 | 5+6 2 |
| | 4+5 2 | 9 2 | 3+4 2 | 3+4 2 |
| | 8 2 | 8 2 | 3+4 2 | 5+6 2 |
| | 9 2 | 8 2 | 5+6 2 | 3+4 2 |
| | 8 2 | 4+5 2 | 3+4 2 | 5+6 2 |
| | 4+5 2 | 8 2 | 5+6 2 | 3+4 2 |
| | 8 2 | 8 2 | 5+6 2 | 3+4 2 |
| | 9 2 | 4+5 2 | 3+4 2 | 5+6 2 |
| | 9 2 | 9 2 | 5+6 2 | 3+4 2 |
| | 6+7 2 | 6+7 2 | 3+4 2 | 5+6 2 |
| | 8 2 | 9 2 | 3+4 2 | 5+6 2 |
| | 4+5 2 | 6+7 2 | 5+6 2 | 3+4 2 |
| | 9 2 | 9 2 | 3+4 2 | 3+4 2 |
| | 8 2 | 9 2 | 5+6 2 | 5+6 2 |
| | 6+7 2 | 4+5 2 | 5+6 2 | 5+6 2 |
| | 9 2 | 8 2 | 5+6 2 | 3+4 2 |
| | 9 2 | 9 2 | 3+4 2 | 5+6 2 |
| | 8 2 | 6+7 2 | 5+6 2 | 5+6 2 |
| | 4+5 2 | 8 2 | 3+4 2 | 3+4 2 |
| | 8 2 | 6+7 2 | 3+4 2 | 3+4 2 |
| | 9 2 | 9 2 | 5+6 2 | 5+6 2 |
| | 6+7 2 | 4+5 2 | 5+6 2 | 5+6 2 |
| | 8 2 | 8 2 | 3+4 2 | 3+4 2 |
| | 4+5 2 | 9 2 | 3+4 2 | 3+4 2 |
| | 6+7 2 | 8 2 | 5+6 2 | 3+4 2 |
| | 8 2 | 6+7 2 | 3+4 2 | 5+6 2 |
| | 4+5 2 | 8 2 | 5+6 2 | 3+4 2 |
| | 9 2 | 4+5 2 | 3+4 2 | 5+6 2 |
| | 9 2 | 9 2 | 3+4 2 | 3+4 2 |
| | 8 2 | 4+5 2 | 5+6 2 | 5+6 2 |
| | 6+7 2 | 8 2 | 5+6 2 | 3+4 2 |
| | 4+5 2 | 9 2 | 3+4 2 | 3+4 2 |
| | 8 2 | 8 2 | 5+6 2 | 5+6 2 |
| | 8 2 | 8 2 | 3+4 2 | 5+6 2 |
| | 4+5 2 | 4+5 2 | 5+6 2 | 3+4 2 |
| | 9 2 | 8 2 | 5+6 2 | 5+6 2 |
| | 8 2 | 8 2 | 3+4 2 | 5+6 2 |
| | 8 2 | 6+7 2 | 3+4 2 | 3+4 2 |
| | 8 2 | 9 2 | 5+6 2 | 5+6 2 |
| | 9 2 | 6+7 2 | 3+4 2 | 3+4 2 |
| | 4+5 2 | 9 2 | 5+6 2 | 5+6 2 |
| | 6+7 2 | 4+5 2 | 3+4 2 | 5+6 2 |
| | 8 2 | 8 2 | 5+6 2 | 3+4 2 |
| | 8 2 | 8 2 | 3+4 2 | 3+4 2 |
| | 4+5 2 | 4+5 2 | 5+6 2 | 5+6 2 |
| | 6+7 2 | 9 2 | 3+4 2 | 3+4 2 |
| | 9 2 | 6+7 2 | 5+6 2 | 5+6 2 |
| | 9 2 | 9 2 | 3+4 2 | 5+6 2 |
| 166.1.271 | 620636 | 174 | Ce2Co7 | |
| | -GM | -T | -F | -L |
| | 8 2 | 9 2 | 5+6 2 | 3+4 2 |

Continued on next column

**TABLE LXXIII– continued**

| TQC | ICSD | NoE | SF | |
|---|---|---|---|---|
| | 8 2 | 8 2 | 3+4 2 | 5+6 2 |
| | 9 2 | 9 2 | 5+6 2 | 3+4 2 |
| | 9 2 | 8 2 | 3+4 2 | 5+6 2 |
| | 8 2 | 9 2 | 5+6 2 | 3+4 2 |
| | 8 2 | 8 2 | 3+4 2 | 5+6 2 |
| | 9 2 | 9 2 | 3+4 2 | 3+4 2 |
| | 9 2 | 8 2 | 5+6 2 | 5+6 2 |
| | 4+5 2 | 6+7 2 | 5+6 2 | 3+4 2 |
| | 8 2 | 9 2 | 3+4 2 | 5+6 2 |
| | 9 2 | 8 2 | 3+4 2 | 3+4 2 |
| | 6+7 2 | 6+7 2 | 5+6 2 | 5+6 2 |
| | 8 2 | 4+5 2 | 5+6 2 | 3+4 2 |
| | 4+5 2 | 9 2 | 3+4 2 | 5+6 2 |
| | 9 2 | 8 2 | 5+6 2 | 3+4 2 |
| | 6+7 2 | 4+5 2 | 3+4 2 | 5+6 2 |
| | 8 2 | 8 2 | 5+6 2 | 5+6 2 |
| | 9 2 | 9 2 | 3+4 2 | 3+4 2 |
| | 8 2 | 8 2 | 3+4 2 | 5+6 2 |
| | 9 2 | 9 2 | 5+6 2 | 3+4 2 |
| | 8 2 | 8 2 | 5+6 2 | 5+6 2 |
| | 8 2 | 8 2 | 3+4 2 | 5+6 2 |
| | 8 2 | 9 2 | 5+6 2 | 5+6 2 |
| | 4+5 2 | 6+7 2 | 5+6 2 | 3+4 2 |
| | 9 2 | 9 2 | 3+4 2 | 5+6 2 |
| | 6+7 2 | 8 2 | 3+4 2 | 3+4 2 |
| | 8 2 | 4+5 2 | 3+4 2 | 3+4 2 |
| | 8 2 | 8 2 | 5+6 2 | 5+6 2 |
| | 4+5 2 | 4+5 2 | 3+4 2 | 5+6 2 |
| | 9 2 | 9 2 | 3+4 2 | 3+4 2 |
| | 8 2 | 8 2 | 5+6 2 | 5+6 2 |
| | 4+5 2 | 4+5 2 | 5+6 2 | 3+4 2 |
| | 8 2 | 8 2 | 5+6 2 | 5+6 2 |
| | 8 2 | 8 2 | 3+4 2 | 5+6 2 |
| | 9 2 | 9 2 | 3+4 2 | 3+4 2 |
| | 4+5 2 | 6+7 2 | 5+6 2 | 5+6 2 |
| | 6+7 2 | 4+5 2 | 5+6 2 | 5+6 2 |
| | 9 2 | 8 2 | 5+6 2 | 3+4 2 |
| | 8 2 | 9 2 | 3+4 2 | 5+6 2 |
| | 4+5 2 | 6+7 2 | 3+4 2 | 5+6 2 |
| | 8 2 | 9 2 | 3+4 2 | 3+4 2 |
| | 6+7 2 | 8 2 | 3+4 2 | 5+6 2 |
| | 9 2 | 8 2 | 5+6 2 | 3+4 2 |
| | 8 2 | 4+5 2 | 3+4 2 | 5+6 2 |
| | 4+5 2 | 8 2 | 5+6 2 | 3+4 2 |
| | 8 2 | 8 2 | 5+6 2 | 3+4 2 |
| | 9 2 | 4+5 2 | 3+4 2 | 3+4 2 |
| | 9 2 | 9 2 | 5+6 2 | 5+6 2 |
| | 6+7 2 | 6+7 2 | 3+4 2 | 5+6 2 |
| | 8 2 | 9 2 | 3+4 2 | 5+6 2 |
| | 4+5 2 | 6+7 2 | 5+6 2 | 3+4 2 |
| | 9 2 | 9 2 | 3+4 2 | 3+4 2 |
| | 8 2 | 9 2 | 5+6 2 | 5+6 2 |
| | 6+7 2 | 4+5 2 | 5+6 2 | 5+6 2 |
| | 9 2 | 8 2 | 3+4 2 | 3+4 2 |
| | 9 2 | 6+7 2 | 5+6 2 | 5+6 2 |
| | 8 2 | 9 2 | 5+6 2 | 5+6 2 |
| | 4+5 2 | 8 2 | 3+4 2 | 3+4 2 |
| | 9 2 | 6+7 2 | 3+4 2 | 3+4 2 |
| | 8 2 | 4+5 2 | 5+6 2 | 5+6 2 |
| | 6+7 2 | 9 2 | 5+6 2 | 5+6 2 |
| | 8 2 | 8 2 | 3+4 2 | 3+4 2 |

Continued on next column

**TABLE LXXIII– continued**

| TQC | ICSD | NoE | SF | |
|---|---|---|---|---|
| | 4+5 2 | 9 2 | 3+4 2 | 3+4 2 |
| | 6+7 2 | 8 2 | 5+6 2 | 5+6 2 |
| | 4+5 2 | 6+7 2 | 3+4 2 | 3+4 2 |
| | 8 2 | 8 2 | 5+6 2 | 3+4 2 |
| | 9 2 | 4+5 2 | 3+4 2 | 5+6 2 |
| | 9 2 | 8 2 | 3+4 2 | 3+4 2 |
| | 8 2 | 9 2 | 5+6 2 | 5+6 2 |
| | 6+7 2 | 4+5 2 | 5+6 2 | 3+4 2 |
| | 4+5 2 | 9 2 | 3+4 2 | 3+4 2 |
| | 8 2 | 8 2 | 5+6 2 | 5+6 2 |
| | 8 2 | 8 2 | 3+4 2 | 5+6 2 |
| | 4+5 2 | 4+5 2 | 5+6 2 | 3+4 2 |
| | 9 2 | 8 2 | 5+6 2 | 5+6 2 |
| | 8 2 | 8 2 | 3+4 2 | 5+6 2 |
| | 8 2 | 6+7 2 | 3+4 2 | 3+4 2 |
| | 8 2 | 9 2 | 5+6 2 | 5+6 2 |
| | 9 2 | 6+7 2 | 5+6 2 | 3+4 2 |
| | 4+5 2 | 9 2 | 3+4 2 | 5+6 2 |
| | 6+7 2 | 4+5 2 | 3+4 2 | 3+4 2 |
| | 8 2 | 8 2 | 5+6 2 | 5+6 2 |
| | 8 2 | 8 2 | 3+4 2 | 3+4 2 |
| | 4+5 2 | 4+5 2 | 5+6 2 | 5+6 2 |
| | 6+7 2 | 6+7 2 | 3+4 2 | 3+4 2 |
| | 9 2 | 9 2 | 5+6 2 | 5+6 2 |
| | 9 2 | 9 2 | 3+4 2 | 5+6 2 |
| 166.1.272 | 603443 | 192 | Ce2Co5Fe12N3 | |
| | -GM | -T | -F | -L |
| | 8 2 | 8 2 | 3+4 2 | 5+6 2 |
| | 9 2 | 9 2 | 5+6 2 | 3+4 2 |
| | 8 2 | 8 2 | 5+6 2 | 5+6 2 |
| | 9 2 | 9 2 | 3+4 2 | 3+4 2 |
| | 4+5 2 | 4+5 2 | 5+6 2 | 5+6 2 |
| | 6+7 2 | 6+7 2 | 3+4 2 | 3+4 2 |
| | 8 2 | 9 2 | 3+4 2 | 5+6 2 |
| | 9 2 | 8 2 | 5+6 2 | 3+4 2 |
| | 8 2 | 8 2 | 3+4 2 | 3+4 2 |
| | 4+5 2 | 4+5 2 | 5+6 2 | 5+6 2 |
| | 8 2 | 8 2 | 5+6 2 | 3+4 2 |
| | 8 2 | 9 2 | 5+6 2 | 5+6 2 |
| | 8 2 | 8 2 | 3+4 2 | 5+6 2 |
| | 4+5 2 | 9 2 | 3+4 2 | 5+6 2 |
| | 8 2 | 4+5 2 | 5+6 2 | 5+6 2 |
| | 4+5 2 | 6+7 2 | 5+6 2 | 3+4 2 |
| | 8 2 | 8 2 | 5+6 2 | 5+6 2 |
| | 8 2 | 4+5 2 | 3+4 2 | 3+4 2 |
| | 4+5 2 | 8 2 | 5+6 2 | 5+6 2 |
| | 9 2 | 8 2 | 3+4 2 | 5+6 2 |
| | 6+7 2 | 4+5 2 | 5+6 2 | 3+4 2 |
| | 8 2 | 8 2 | 5+6 2 | 3+4 2 |
| | 8 2 | 8 2 | 5+6 2 | 5+6 2 |
| | 4+5 2 | 8 2 | 3+4 2 | 3+4 2 |
| | 8 2 | 4+5 2 | 5+6 2 | 5+6 2 |
| | 8 2 | 8 2 | 5+6 2 | 3+4 2 |
| | 8 2 | 8 2 | 3+4 2 | 5+6 2 |
| | 9 2 | 8 2 | 3+4 2 | 5+6 2 |
| | 8 2 | 9 2 | 5+6 2 | 3+4 2 |
| | 9 2 | 9 2 | 3+4 2 | 5+6 2 |
| | 6+7 2 | 9 2 | 5+6 2 | 3+4 2 |
| | 8 2 | 6+7 2 | 5+6 2 | 3+4 2 |
| | 4+5 2 | 9 2 | 3+4 2 | 5+6 2 |
| | 9 2 | 8 2 | 3+4 2 | 3+4 2 |

Continued on next column

**TABLE LXXIII– continued**

| TQC | ICSD | NoE | SF | |
|---|---|---|---|---|
| | 8 2 | 9 2 | 3+4 2 | 3+4 2 |
| | 4+5 2 | 6+7 2 | 5+6 2 | 5+6 2 |
| | 8 2 | 4+5 2 | 5+6 2 | 5+6 2 |
| | 4+5 2 | 9 2 | 5+6 2 | 3+4 2 |
| | 8 2 | 6+7 2 | 3+4 2 | 5+6 2 |
| | 6+7 2 | 6+7 2 | 5+6 2 | 3+4 2 |
| | 9 2 | 9 2 | 3+4 2 | 5+6 2 |
| | 9 2 | 8 2 | 5+6 2 | 5+6 2 |
| | 6+7 2 | 9 2 | 3+4 2 | 3+4 2 |
| | 8 2 | 6+7 2 | 3+4 2 | 3+4 2 |
| | 6+7 2 | 9 2 | 5+6 2 | 3+4 2 |
| | 9 2 | 8 2 | 3+4 2 | 5+6 2 |
| | 8 2 | 9 2 | 5+6 2 | 3+4 2 |
| | 8 2 | 4+5 2 | 5+6 2 | 5+6 2 |
| | 4+5 2 | 6+7 2 | 5+6 2 | 5+6 2 |
| | 9 2 | 9 2 | 5+6 2 | 3+4 2 |
| | 6+7 2 | 9 2 | 3+4 2 | 3+4 2 |
| | 8 2 | 4+5 2 | 3+4 2 | 5+6 2 |
| | 9 2 | 8 2 | 5+6 2 | 3+4 2 |
| | 4+5 2 | 9 2 | 3+4 2 | 3+4 2 |
| | 8 2 | 6+7 2 | 5+6 2 | 5+6 2 |
| | 9 2 | 9 2 | 3+4 2 | 5+6 2 |
| | 9 2 | 6+7 2 | 3+4 2 | 3+4 2 |
| | 6+7 2 | 9 2 | 5+6 2 | 3+4 2 |
| | 4+5 2 | 8 2 | 5+6 2 | 5+6 2 |
| | 9 2 | 4+5 2 | 3+4 2 | 5+6 2 |
| | 8 2 | 8 2 | 3+4 2 | 5+6 2 |
| | 8 2 | 8 2 | 5+6 2 | 5+6 2 |
| | 9 2 | 8 2 | 3+4 2 | 3+4 2 |
| | 6+7 2 | 9 2 | 3+4 2 | 3+4 2 |
| | 9 2 | 6+7 2 | 5+6 2 | 3+4 2 |
| | 8 2 | 9 2 | 3+4 2 | 3+4 2 |
| | 4+5 2 | 9 2 | 5+6 2 | 3+4 2 |
| | 8 2 | 4+5 2 | 3+4 2 | 5+6 2 |
| | 8 2 | 8 2 | 3+4 2 | 5+6 2 |
| | 4+5 2 | 4+5 2 | 5+6 2 | 3+4 2 |
| | 9 2 | 8 2 | 5+6 2 | 5+6 2 |
| | 9 2 | 8 2 | 5+6 2 | 5+6 2 |
| | 6+7 2 | 6+7 2 | 3+4 2 | 3+4 2 |
| | 9 2 | 9 2 | 5+6 2 | 3+4 2 |
| | 6+7 2 | 4+5 2 | 3+4 2 | 5+6 2 |
| | 8 2 | 8 2 | 5+6 2 | 5+6 2 |
| | 9 2 | 9 2 | 3+4 2 | 3+4 2 |
| | 4+5 2 | 6+7 2 | 3+4 2 | 3+4 2 |
| | 8 2 | 9 2 | 5+6 2 | 5+6 2 |
| | 8 2 | 8 2 | 3+4 2 | 3+4 2 |
| | 9 2 | 4+5 2 | 5+6 2 | 5+6 2 |
| | 6+7 2 | 9 2 | 5+6 2 | 3+4 2 |
| | 4+5 2 | 8 2 | 3+4 2 | 3+4 2 |
| | 9 2 | 9 2 | 3+4 2 | 5+6 2 |
| | 8 2 | 6+7 2 | 5+6 2 | 5+6 2 |
| | 4+5 2 | 9 2 | 5+6 2 | 5+6 2 |
| | 9 2 | 6+7 2 | 3+4 2 | 3+4 2 |
| | 8 2 | 9 2 | 3+4 2 | 3+4 2 |
| | 8 2 | 8 2 | 5+6 2 | 3+4 2 |
| | 8 2 | 6+7 2 | 3+4 2 | 3+4 2 |
| | 6+7 2 | 9 2 | 5+6 2 | 5+6 2 |
| | 9 2 | 8 2 | 3+4 2 | 3+4 2 |
| | 4+5 2 | 4+5 2 | 5+6 2 | 5+6 2 |
| | 8 2 | 8 2 | 5+6 2 | 5+6 2 |
| | 8 2 | 9 2 | 3+4 2 | 3+4 2 |

Continued on next column

**TABLE LXXIII– continued**

| TQC | ICSD | NoE | SF | |
|---|---|---|---|---|
| | 4+5 2 | 4+5 2 | 3+4 2 | 3+4 2 |
| 166.1.273 | 201355 | 162 | Cd1Cs4F12Ni3 | |
| | -GM | -T | -F | -L |
| | 8 2 | 9 2 | 3+4 2 | 5+6 2 |
| | 9 2 | 8 2 | 5+6 2 | 3+4 2 |
| | 6+7 2 | 4+5 2 | 5+6 2 | 3+4 2 |
| | 9 2 | 8 2 | 3+4 2 | 3+4 2 |
| | 4+5 2 | 6+7 2 | 5+6 2 | 5+6 2 |
| | 8 2 | 9 2 | 3+4 2 | 5+6 2 |
| | 8 2 | 8 2 | 3+4 2 | 3+4 2 |
| | 9 2 | 9 2 | 5+6 2 | 5+6 2 |
| | 9 2 | 9 2 | 3+4 2 | 3+4 2 |
| | 6+7 2 | 6+7 2 | 5+6 2 | 5+6 2 |
| | 4+5 2 | 4+5 2 | 3+4 2 | 3+4 2 |
| | 8 2 | 8 2 | 5+6 2 | 5+6 2 |
| | 8 2 | 9 2 | 5+6 2 | 3+4 2 |
| | 9 2 | 8 2 | 3+4 2 | 5+6 2 |
| | 9 2 | 9 2 | 3+4 2 | 3+4 2 |
| | 8 2 | 8 2 | 5+6 2 | 5+6 2 |
| | 8 2 | 9 2 | 5+6 2 | 3+4 2 |
| | 9 2 | 8 2 | 3+4 2 | 5+6 2 |
| | 8 2 | 8 2 | 3+4 2 | 3+4 2 |
| | 9 2 | 9 2 | 5+6 2 | 5+6 2 |
| | 4+5 2 | 4+5 2 | 3+4 2 | 3+4 2 |
| | 8 2 | 8 2 | 3+4 2 | 3+4 2 |
| | 4+5 2 | 4+5 2 | 3+4 2 | 3+4 2 |
| | 8 2 | 8 2 | 3+4 2 | 3+4 2 |
| | 4+5 2 | 6+7 2 | 5+6 2 | 3+4 2 |
| | 8 2 | 9 2 | 3+4 2 | 5+6 2 |
| | 6+7 2 | 4+5 2 | 3+4 2 | 3+4 2 |
| | 9 2 | 8 2 | 5+6 2 | 5+6 2 |
| | 8 2 | 8 2 | 3+4 2 | 5+6 2 |
| | 8 2 | 9 2 | 5+6 2 | 3+4 2 |
| | 6+7 2 | 6+7 2 | 3+4 2 | 5+6 2 |
| | 8 2 | 8 2 | 5+6 2 | 3+4 2 |
| | 4+5 2 | 4+5 2 | 3+4 2 | 3+4 2 |
| | 9 2 | 8 2 | 3+4 2 | 3+4 2 |
| | 9 2 | 9 2 | 3+4 2 | 5+6 2 |
| | 8 2 | 9 2 | 5+6 2 | 5+6 2 |
| | 9 2 | 8 2 | 3+4 2 | 3+4 2 |
| | 6+7 2 | 9 2 | 5+6 2 | 5+6 2 |
| | 9 2 | 6+7 2 | 5+6 2 | 5+6 2 |
| | 8 2 | 9 2 | 3+4 2 | 5+6 2 |
| | 8 2 | 9 2 | 5+6 2 | 5+6 2 |
| | 4+5 2 | 6+7 2 | 3+4 2 | 3+4 2 |
| | 9 2 | 8 2 | 5+6 2 | 5+6 2 |
| | 6+7 2 | 4+5 2 | 3+4 2 | 5+6 2 |
| | 8 2 | 9 2 | 3+4 2 | 5+6 2 |
| | 9 2 | 8 2 | 5+6 2 | 3+4 2 |
| | 4+5 2 | 6+7 2 | 5+6 2 | 3+4 2 |
| | 9 2 | 4+5 2 | 5+6 2 | 3+4 2 |
| | 6+7 2 | 8 2 | 5+6 2 | 3+4 2 |
| | 8 2 | 9 2 | 3+4 2 | 5+6 2 |
| | 4+5 2 | 6+7 2 | 3+4 2 | 5+6 2 |
| | 9 2 | 8 2 | 5+6 2 | 3+4 2 |
| | 8 2 | 9 2 | 5+6 2 | 5+6 2 |
| | 6+7 2 | 4+5 2 | 5+6 2 | 3+4 2 |
| | 9 2 | 8 2 | 3+4 2 | 3+4 2 |
| | 8 2 | 8 2 | 3+4 2 | 3+4 2 |
| | 9 2 | 9 2 | 5+6 2 | 3+4 2 |
| | 8 2 | 6+7 2 | 3+4 2 | 5+6 2 |

Continued on next column

**TABLE LXXIII– continued**

| TQC | ICSD | NoE | SF | |
|---|---|---|---|---|
| | 9 2 | 9 2 | 5+6 2 | 5+6 2 |
| | 4+5 2 | 8 2 | 3+4 2 | 3+4 2 |
| | 6+7 2 | 4+5 2 | 5+6 2 | 5+6 2 |
| | 9 2 | 9 2 | 3+4 2 | 5+6 2 |
| | 4+5 2 | 9 2 | 5+6 2 | 3+4 2 |
| | 8 2 | 8 2 | 5+6 2 | 5+6 2 |
| | 9 2 | 4+5 2 | 3+4 2 | 3+4 2 |
| | 8 2 | 6+7 2 | 3+4 2 | 5+6 2 |
| | 6+7 2 | 8 2 | 5+6 2 | 3+4 2 |
| | 8 2 | 8 2 | 3+4 2 | 3+4 2 |
| | 4+5 2 | 4+5 2 | 3+4 2 | 3+4 2 |
| | 8 2 | 9 2 | 3+4 2 | 5+6 2 |
| | 4+5 2 | 6+7 2 | 3+4 2 | 5+6 2 |
| | 8 2 | 9 2 | 3+4 2 | 5+6 2 |
| | 9 2 | 8 2 | 5+6 2 | 3+4 2 |
| | 6+7 2 | 4+5 2 | 5+6 2 | 5+6 2 |
| | 4+5 2 | 6+7 2 | 3+4 2 | 3+4 2 |
| | 8 2 | 9 2 | 3+4 2 | 5+6 2 |
| | 9 2 | 8 2 | 5+6 2 | 3+4 2 |
| | 8 2 | 9 2 | 3+4 2 | 5+6 2 |
| | 9 2 | 6+7 2 | 3+4 2 | 3+4 2 |
| | 6+7 2 | 9 2 | 5+6 2 | 5+6 2 |
| | 8 2 | 9 2 | 5+6 2 | 5+6 2 |
| 166.1.274 | 166533 | 16 | Ca2I1P1 | |
| | -GM | -T | -F | -L |
| | 8 2 | 8 2 | 3+4 2 | 3+4 2 |
| | 8 2 | 9 2 | 3+4 2 | 5+6 2 |
| | 9 2 | 9 2 | 5+6 2 | 5+6 2 |
| | 9 2 | 9 2 | 5+6 2 | 5+6 2 |
| | 6+7 2 | 6+7 2 | 5+6 2 | 5+6 2 |
| | 9 2 | 8 2 | 5+6 2 | 3+4 2 |
| | 9 2 | 8 2 | 5+6 2 | 3+4 2 |
| | 6+7 2 | 4+5 2 | 3+4 2 | 3+4 2 |
| 166.1.275 | 424597 | 38 | Ca2Ge1Pd3 | |
| | -GM | -T | -F | -L |
| | 8 2 | 9 2 | 3+4 2 | 5+6 2 |
| | 8 2 | 8 2 | 5+6 2 | 3+4 2 |
| | 8 2 | 8 2 | 5+6 2 | 5+6 2 |
| | 8 2 | 4+5 2 | 5+6 2 | 5+6 2 |
| | 4+5 2 | 8 2 | 5+6 2 | 3+4 2 |
| | 8 2 | 8 2 | 5+6 2 | 5+6 2 |
| | 4+5 2 | 4+5 2 | 3+4 2 | 3+4 2 |
| | 8 2 | 8 2 | 3+4 2 | 3+4 2 |
| | 8 2 | 4+5 2 | 5+6 2 | 5+6 2 |
| | 4+5 2 | 8 2 | 5+6 2 | 3+4 2 |
| | 8 2 | 8 2 | 3+4 2 | 5+6 2 |
| | 4+5 2 | 8 2 | 5+6 2 | 5+6 2 |
| | 8 2 | 4+5 2 | 3+4 2 | 3+4 2 |
| | 9 2 | 8 2 | 3+4 2 | 5+6 2 |
| | 6+7 2 | 8 2 | 5+6 2 | 5+6 2 |
| | 8 2 | 4+5 2 | 5+6 2 | 5+6 2 |
| | 4+5 2 | 8 2 | 5+6 2 | 5+6 2 |
| | 9 2 | 9 2 | 3+4 2 | 3+4 2 |
| | 8 2 | 6+7 2 | 5+6 2 | 3+4 2 |
| 166.1.276 | 52789 | 36 | Ca1Si2 | |
| | -GM | -T | -F | -L |
| | 8 2 | 8 2 | 3+4 2 | 3+4 2 |
| | 9 2 | 9 2 | 5+6 2 | 5+6 2 |
| | 8 2 | 8 2 | 3+4 2 | 3+4 2 |
| | 9 2 | 9 2 | 5+6 2 | 5+6 2 |
| | 8 2 | 8 2 | 3+4 2 | 3+4 2 |

Continued on next column

**TABLE LXXIII– continued**

| TQC | ICSD | NoE | SF | |
|---|---|---|---|---|
| | 9 2 | 9 2 | 5+6 2 | 5+6 2 |
| | 4+5 2 | 4+5 2 | 3+4 2 | 3+4 2 |
| | 6+7 2 | 6+7 2 | 5+6 2 | 5+6 2 |
| | 8 2 | 8 2 | 5+6 2 | 5+6 2 |
| | 8 2 | 9 2 | 5+6 2 | 3+4 2 |
| | 9 2 | 9 2 | 3+4 2 | 3+4 2 |
| | 9 2 | 8 2 | 3+4 2 | 5+6 2 |
| | 8 2 | 9 2 | 3+4 2 | 3+4 2 |
| | 8 2 | 6+7 2 | 3+4 2 | 5+6 2 |
| | 4+5 2 | 8 2 | 5+6 2 | 5+6 2 |
| | 8 2 | 4+5 2 | 5+6 2 | 3+4 2 |
| | 4+5 2 | 9 2 | 5+6 2 | 3+4 2 |
| | 9 2 | 8 2 | 5+6 2 | 5+6 2 |
| 166.1.277 | 409057 | 10 | Ca1Si2 | |
| | -GM | -T | -F | -L |
| | 8 2 | 9 2 | 5+6 2 | 3+4 2 |
| | 9 2 | 8 2 | 3+4 2 | 5+6 2 |
| | 8 2 | 9 2 | 3+4 2 | 5+6 2 |
| | 8 2 | 6+7 2 | 5+6 2 | 3+4 2 |
| | 4+5 2 | 8 2 | 5+6 2 | 3+4 2 |
| 166.1.278 | 32006 | 20 | Ca1Si2 | |
| | -GM | -T | -F | -L |
| | 8 2 | 8 2 | 5+6 2 | 5+6 2 |
| | 8 2 | 9 2 | 5+6 2 | 3+4 2 |
| | 9 2 | 9 2 | 3+4 2 | 3+4 2 |
| | 9 2 | 8 2 | 3+4 2 | 5+6 2 |
| | 8 2 | 8 2 | 3+4 2 | 3+4 2 |
| | 8 2 | 4+5 2 | 3+4 2 | 5+6 2 |
| | 4+5 2 | 9 2 | 5+6 2 | 5+6 2 |
| | 8 2 | 6+7 2 | 5+6 2 | 3+4 2 |
| | 4+5 2 | 9 2 | 5+6 2 | 3+4 2 |
| | 9 2 | 8 2 | 5+6 2 | 5+6 2 |
| 166.1.279 | 248518 | 20 | Ca1Si2 | |
| | -GM | -T | -F | -L |
| | 8 2 | 9 2 | 5+6 2 | 3+4 2 |
| | 8 2 | 8 2 | 5+6 2 | 5+6 2 |
| | 9 2 | 8 2 | 3+4 2 | 5+6 2 |
| | 9 2 | 9 2 | 3+4 2 | 3+4 2 |
| | 8 2 | 9 2 | 3+4 2 | 5+6 2 |
| | 8 2 | 9 2 | 5+6 2 | 3+4 2 |
| | 4+5 2 | 6+7 2 | 3+4 2 | 3+4 2 |
| | 8 2 | 8 2 | 5+6 2 | 5+6 2 |
| | 4+5 2 | 4+5 2 | 5+6 2 | 3+4 2 |
| | 9 2 | 8 2 | 5+6 2 | 5+6 2 |
| 166.1.280 | 248517 | 20 | Ca1Si2 | |
| | -GM | -T | -F | -L |
| | 8 2 | 8 2 | 5+6 2 | 5+6 2 |
| | 8 2 | 9 2 | 5+6 2 | 3+4 2 |
| | 9 2 | 8 2 | 3+4 2 | 3+4 2 |
| | 9 2 | 9 2 | 3+4 2 | 5+6 2 |
| | 8 2 | 8 2 | 5+6 2 | 5+6 2 |
| | 8 2 | 4+5 2 | 3+4 2 | 5+6 2 |
| | 4+5 2 | 9 2 | 3+4 2 | 3+4 2 |
| | 8 2 | 9 2 | 5+6 2 | 3+4 2 |
| | 4+5 2 | 6+7 2 | 5+6 2 | 3+4 2 |
| | 9 2 | 8 2 | 5+6 2 | 5+6 2 |
| 166.1.281 | 248516 | 20 | Ca1Si2 | |
| | -GM | -T | -F | -L |
| | 8 2 | 8 2 | 5+6 2 | 5+6 2 |
| | 8 2 | 9 2 | 5+6 2 | 3+4 2 |

Continued on next column

**TABLE LXXIII– continued**

| TQC | ICSD | NoE | SF | |
|---|---|---|---|---|
| | 9 2 | 8 2 | 3+4 2 | 3+4 2 |
| | 9 2 | 9 2 | 3+4 2 | 5+6 2 |
| | 8 2 | 8 2 | 5+6 2 | 5+6 2 |
| | 8 2 | 4+5 2 | 3+4 2 | 5+6 2 |
| | 4+5 2 | 9 2 | 3+4 2 | 3+4 2 |
| | 8 2 | 9 2 | 5+6 2 | 3+4 2 |
| | 4+5 2 | 6+7 2 | 5+6 2 | 3+4 2 |
| | 9 2 | 8 2 | 5+6 2 | 5+6 2 |
| 166.1.282 | 248515 | 20 | Ca1Si2 | |
| | -GM | -T | -F | -L |
| | 8 2 | 8 2 | 5+6 2 | 5+6 2 |
| | 8 2 | 9 2 | 5+6 2 | 3+4 2 |
| | 9 2 | 8 2 | 3+4 2 | 3+4 2 |
| | 9 2 | 9 2 | 3+4 2 | 5+6 2 |
| | 8 2 | 8 2 | 5+6 2 | 5+6 2 |
| | 8 2 | 4+5 2 | 3+4 2 | 5+6 2 |
| | 4+5 2 | 9 2 | 3+4 2 | 3+4 2 |
| | 8 2 | 9 2 | 5+6 2 | 3+4 2 |
| | 4+5 2 | 6+7 2 | 5+6 2 | 3+4 2 |
| | 9 2 | 8 2 | 5+6 2 | 5+6 2 |
| 166.1.283 | 193539 | 10 | Ca1Si2 | |
| | -GM | -T | -F | -L |
| | 8 2 | 9 2 | 5+6 2 | 3+4 2 |
| | 9 2 | 8 2 | 3+4 2 | 5+6 2 |
| | 8 2 | 9 2 | 3+4 2 | 5+6 2 |
| | 8 2 | 6+7 2 | 5+6 2 | 3+4 2 |
| | 4+5 2 | 8 2 | 5+6 2 | 3+4 2 |
| 166.1.284 | 15632 | 10 | Ca1Si2 | |
| | -GM | -T | -F | -L |
| | 8 2 | 9 2 | 5+6 2 | 3+4 2 |
| | 9 2 | 8 2 | 3+4 2 | 5+6 2 |
| | 8 2 | 9 2 | 3+4 2 | 5+6 2 |
| | 8 2 | 6+7 2 | 5+6 2 | 3+4 2 |
| | 4+5 2 | 8 2 | 5+6 2 | 3+4 2 |
| 166.1.285 | 77316 | 36 | Ca1Ge2 | |
| | -GM | -T | -F | -L |
| | 8 2 | 8 2 | 3+4 2 | 3+4 2 |
| | 9 2 | 9 2 | 5+6 2 | 5+6 2 |
| | 8 2 | 8 2 | 3+4 2 | 3+4 2 |
| | 9 2 | 9 2 | 5+6 2 | 5+6 2 |
| | 8 2 | 8 2 | 3+4 2 | 3+4 2 |
| | 9 2 | 9 2 | 5+6 2 | 5+6 2 |
| | 4+5 2 | 4+5 2 | 3+4 2 | 3+4 2 |
| | 6+7 2 | 6+7 2 | 5+6 2 | 5+6 2 |
| | 8 2 | 8 2 | 5+6 2 | 5+6 2 |
| | 8 2 | 9 2 | 5+6 2 | 3+4 2 |
| | 9 2 | 9 2 | 3+4 2 | 3+4 2 |
| | 9 2 | 8 2 | 3+4 2 | 5+6 2 |
| | 8 2 | 9 2 | 3+4 2 | 3+4 2 |
| | 8 2 | 8 2 | 3+4 2 | 5+6 2 |
| | 4+5 2 | 9 2 | 5+6 2 | 5+6 2 |
| | 8 2 | 6+7 2 | 5+6 2 | 3+4 2 |
| | 4+5 2 | 4+5 2 | 5+6 2 | 3+4 2 |
| | 9 2 | 8 2 | 5+6 2 | 5+6 2 |
| 166.1.286 | 619304 | 20 | Ca1Ge2 | |
| | -GM | -T | -F | -L |
| | 8 2 | 8 2 | 5+6 2 | 5+6 2 |
| | 8 2 | 9 2 | 5+6 2 | 3+4 2 |
| | 9 2 | 9 2 | 3+4 2 | 3+4 2 |
| | 9 2 | 8 2 | 3+4 2 | 5+6 2 |

Continued on next column

**TABLE LXXIII– continued**

| TQC | ICSD | NoE | SF | |
|---|---|---|---|---|
| | 8 2 | 9 2 | 3+4 2 | 3+4 2 |
| | 8 2 | 8 2 | 3+4 2 | 5+6 2 |
| | 4+5 2 | 9 2 | 5+6 2 | 5+6 2 |
| | 8 2 | 4+5 2 | 5+6 2 | 3+4 2 |
| | 4+5 2 | 8 2 | 5+6 2 | 3+4 2 |
| | 9 2 | 6+7 2 | 5+6 2 | 5+6 2 |
| 166.1.287 | 245612 | 20 | Ca1Ge2 | |
| | -GM | -T | -F | -L |
| | 8 2 | 8 2 | 5+6 2 | 5+6 2 |
| | 8 2 | 9 2 | 5+6 2 | 3+4 2 |
| | 9 2 | 9 2 | 3+4 2 | 3+4 2 |
| | 9 2 | 8 2 | 3+4 2 | 5+6 2 |
| | 8 2 | 8 2 | 3+4 2 | 3+4 2 |
| | 8 2 | 9 2 | 3+4 2 | 5+6 2 |
| | 4+5 2 | 4+5 2 | 5+6 2 | 5+6 2 |
| | 8 2 | 8 2 | 5+6 2 | 3+4 2 |
| | 4+5 2 | 9 2 | 5+6 2 | 3+4 2 |
| | 9 2 | 6+7 2 | 5+6 2 | 5+6 2 |
| 166.1.288 | 110107 | 20 | Ca1Ge2 | |
| | -GM | -T | -F | -L |
| | 8 2 | 8 2 | 5+6 2 | 5+6 2 |
| | 8 2 | 9 2 | 5+6 2 | 3+4 2 |
| | 9 2 | 9 2 | 3+4 2 | 3+4 2 |
| | 9 2 | 8 2 | 3+4 2 | 5+6 2 |
| | 8 2 | 9 2 | 3+4 2 | 3+4 2 |
| | 8 2 | 8 2 | 3+4 2 | 5+6 2 |
| | 4+5 2 | 9 2 | 5+6 2 | 5+6 2 |
| | 8 2 | 4+5 2 | 5+6 2 | 3+4 2 |
| | 4+5 2 | 6+7 2 | 5+6 2 | 3+4 2 |
| | 9 2 | 8 2 | 5+6 2 | 5+6 2 |
| 166.1.289 | 20822 | 52 | C5Ti8 | |
| | -GM | -T | -F | -L |
| | 8 2 | 9 2 | 3+4 2 | 3+4 2 |
| | 8 2 | 8 2 | 5+6 2 | 5+6 2 |
| | 4+5 2 | 6+7 2 | 5+6 2 | 3+4 2 |
| | 8 2 | 9 2 | 3+4 2 | 5+6 2 |
| | 8 2 | 9 2 | 3+4 2 | 3+4 2 |
| | 9 2 | 8 2 | 5+6 2 | 5+6 2 |
| | 6+7 2 | 4+5 2 | 5+6 2 | 5+6 2 |
| | 9 2 | 8 2 | 3+4 2 | 3+4 2 |
| | 9 2 | 8 2 | 5+6 2 | 5+6 2 |
| | 6+7 2 | 6+7 2 | 3+4 2 | 3+4 2 |
| | 9 2 | 9 2 | 3+4 2 | 5+6 2 |
| | 9 2 | 4+5 2 | 3+4 2 | 3+4 2 |
| | 6+7 2 | 8 2 | 5+6 2 | 5+6 2 |
| | 9 2 | 8 2 | 5+6 2 | 3+4 2 |
| | 9 2 | 9 2 | 3+4 2 | 3+4 2 |
| | 6+7 2 | 4+5 2 | 5+6 2 | 5+6 2 |
| | 9 2 | 8 2 | 5+6 2 | 5+6 2 |
| | 8 2 | 8 2 | 3+4 2 | 3+4 2 |
| | 8 2 | 4+5 2 | 5+6 2 | 5+6 2 |
| | 4+5 2 | 8 2 | 5+6 2 | 5+6 2 |
| | 8 2 | 8 2 | 5+6 2 | 3+4 2 |
| | 9 2 | 9 2 | 3+4 2 | 5+6 2 |
| | 6+7 2 | 4+5 2 | 5+6 2 | 5+6 2 |
| | 9 2 | 8 2 | 3+4 2 | 3+4 2 |
| | 4+5 2 | 6+7 2 | 5+6 2 | 3+4 2 |
| | 8 2 | 9 2 | 3+4 2 | 5+6 2 |
| 166.1.290 | 657171 | 186 | C3Fe17Tb2 | |
| | -GM | -T | -F | -L |
| | 8 2 | 9 2 | 3+4 2 | 5+6 2 |

Continued on next column

**TABLE LXXIII– continued**

| TQC | ICSD | NoE | SF | |
|---|---|---|---|---|
| | 9 2 | 8 2 | 5+6 2 | 3+4 2 |
| | 8 2 | 8 2 | 3+4 2 | 5+6 2 |
| | 9 2 | 9 2 | 5+6 2 | 3+4 2 |
| | 4+5 2 | 4+5 2 | 5+6 2 | 5+6 2 |
| | 6+7 2 | 6+7 2 | 3+4 2 | 3+4 2 |
| | 8 2 | 9 2 | 3+4 2 | 5+6 2 |
| | 9 2 | 8 2 | 5+6 2 | 3+4 2 |
| | 8 2 | 8 2 | 5+6 2 | 5+6 2 |
| | 8 2 | 8 2 | 3+4 2 | 3+4 2 |
| | 4+5 2 | 4+5 2 | 5+6 2 | 5+6 2 |
| | 8 2 | 9 2 | 3+4 2 | 3+4 2 |
| | 9 2 | 9 2 | 5+6 2 | 3+4 2 |
| | 6+7 2 | 6+7 2 | 5+6 2 | 5+6 2 |
| | 6+7 2 | 9 2 | 3+4 2 | 5+6 2 |
| | 9 2 | 8 2 | 3+4 2 | 3+4 2 |
| | 9 2 | 9 2 | 3+4 2 | 3+4 2 |
| | 9 2 | 6+7 2 | 3+4 2 | 5+6 2 |
| | 9 2 | 9 2 | 5+6 2 | 3+4 2 |
| | 8 2 | 9 2 | 5+6 2 | 3+4 2 |
| | 8 2 | 9 2 | 5+6 2 | 5+6 2 |
| | 4+5 2 | 6+7 2 | 3+4 2 | 3+4 2 |
| | 9 2 | 9 2 | 3+4 2 | 3+4 2 |
| | 6+7 2 | 6+7 2 | 3+4 2 | 3+4 2 |
| | 8 2 | 8 2 | 5+6 2 | 5+6 2 |
| | 8 2 | 4+5 2 | 3+4 2 | 3+4 2 |
| | 4+5 2 | 8 2 | 5+6 2 | 3+4 2 |
| | 8 2 | 9 2 | 5+6 2 | 5+6 2 |
| | 9 2 | 9 2 | 5+6 2 | 5+6 2 |
| | 6+7 2 | 9 2 | 3+4 2 | 3+4 2 |
| | 8 2 | 6+7 2 | 3+4 2 | 5+6 2 |
| | 4+5 2 | 9 2 | 5+6 2 | 5+6 2 |
| | 8 2 | 6+7 2 | 3+4 2 | 5+6 2 |
| | 9 2 | 8 2 | 5+6 2 | 3+4 2 |
| | 8 2 | 4+5 2 | 5+6 2 | 5+6 2 |
| | 9 2 | 8 2 | 5+6 2 | 3+4 2 |
| | 8 2 | 4+5 2 | 3+4 2 | 3+4 2 |
| | 4+5 2 | 8 2 | 3+4 2 | 3+4 2 |
| | 6+7 2 | 9 2 | 5+6 2 | 3+4 2 |
| | 4+5 2 | 6+7 2 | 3+4 2 | 5+6 2 |
| | 8 2 | 9 2 | 5+6 2 | 5+6 2 |
| | 9 2 | 6+7 2 | 3+4 2 | 5+6 2 |
| | 6+7 2 | 9 2 | 5+6 2 | 5+6 2 |
| | 9 2 | 4+5 2 | 3+4 2 | 3+4 2 |
| | 8 2 | 8 2 | 5+6 2 | 5+6 2 |
| | 6+7 2 | 9 2 | 5+6 2 | 3+4 2 |
| | 4+5 2 | 6+7 2 | 3+4 2 | 3+4 2 |
| | 8 2 | 9 2 | 5+6 2 | 5+6 2 |
| | 9 2 | 8 2 | 5+6 2 | 3+4 2 |
| | 8 2 | 9 2 | 3+4 2 | 5+6 2 |
| | 4+5 2 | 6+7 2 | 5+6 2 | 5+6 2 |
| | 8 2 | 9 2 | 3+4 2 | 3+4 2 |
| | 9 2 | 8 2 | 5+6 2 | 3+4 2 |
| | 9 2 | 4+5 2 | 3+4 2 | 5+6 2 |
| | 8 2 | 8 2 | 5+6 2 | 3+4 2 |
| | 6+7 2 | 8 2 | 3+4 2 | 5+6 2 |
| | 4+5 2 | 9 2 | 3+4 2 | 3+4 2 |
| | 6+7 2 | 6+7 2 | 3+4 2 | 3+4 2 |
| | 9 2 | 9 2 | 3+4 2 | 5+6 2 |
| | 9 2 | 4+5 2 | 5+6 2 | 3+4 2 |
| | 6+7 2 | 8 2 | 3+4 2 | 5+6 2 |
| | 9 2 | 8 2 | 5+6 2 | 3+4 2 |

Continued on next column

**TABLE LXXIII– continued**

| TQC | ICSD | NoE | SF | |
|---|---|---|---|---|
| | 9 2 | 8 2 | 3+4 2 | 5+6 2 |
| | 8 2 | 4+5 2 | 5+6 2 | 5+6 2 |
| | 8 2 | 8 2 | 3+4 2 | 5+6 2 |
| | 8 2 | 9 2 | 5+6 2 | 3+4 2 |
| | 4+5 2 | 4+5 2 | 5+6 2 | 3+4 2 |
| | 8 2 | 8 2 | 5+6 2 | 3+4 2 |
| | 4+5 2 | 9 2 | 3+4 2 | 5+6 2 |
| | 8 2 | 8 2 | 3+4 2 | 3+4 2 |
| | 9 2 | 6+7 2 | 5+6 2 | 5+6 2 |
| | 9 2 | 9 2 | 5+6 2 | 5+6 2 |
| | 6+7 2 | 4+5 2 | 3+4 2 | 3+4 2 |
| | 8 2 | 8 2 | 5+6 2 | 5+6 2 |
| | 8 2 | 8 2 | 3+4 2 | 5+6 2 |
| | 9 2 | 4+5 2 | 3+4 2 | 3+4 2 |
| | 6+7 2 | 9 2 | 5+6 2 | 5+6 2 |
| | 4+5 2 | 6+7 2 | 3+4 2 | 3+4 2 |
| | 9 2 | 8 2 | 5+6 2 | 3+4 2 |
| | 8 2 | 9 2 | 5+6 2 | 5+6 2 |
| | 4+5 2 | 9 2 | 3+4 2 | 5+6 2 |
| | 8 2 | 6+7 2 | 3+4 2 | 3+4 2 |
| | 9 2 | 8 2 | 5+6 2 | 5+6 2 |
| | 8 2 | 4+5 2 | 5+6 2 | 3+4 2 |
| | 4+5 2 | 9 2 | 3+4 2 | 3+4 2 |
| | 8 2 | 6+7 2 | 5+6 2 | 3+4 2 |
| | 8 2 | 8 2 | 5+6 2 | 5+6 2 |
| | 9 2 | 8 2 | 3+4 2 | 5+6 2 |
| | 4+5 2 | 4+5 2 | 3+4 2 | 3+4 2 |
| | 6+7 2 | 9 2 | 5+6 2 | 3+4 2 |
| | 8 2 | 9 2 | 3+4 2 | 5+6 2 |
| | 9 2 | 6+7 2 | 5+6 2 | 3+4 2 |
| | 6+7 2 | 8 2 | 5+6 2 | 5+6 2 |
| 166.1.291 | 659087 | 192 | C3Er2Fe17 | |
| | -GM | -T | -F | -L |
| | 8 2 | 8 2 | 3+4 2 | 3+4 2 |
| | 9 2 | 9 2 | 5+6 2 | 5+6 2 |
| | 8 2 | 8 2 | 3+4 2 | 5+6 2 |
| | 9 2 | 9 2 | 5+6 2 | 3+4 2 |
| | 4+5 2 | 4+5 2 | 5+6 2 | 5+6 2 |
| | 6+7 2 | 6+7 2 | 3+4 2 | 3+4 2 |
| | 8 2 | 9 2 | 3+4 2 | 5+6 2 |
| | 9 2 | 8 2 | 5+6 2 | 3+4 2 |
| | 8 2 | 8 2 | 5+6 2 | 5+6 2 |
| | 8 2 | 8 2 | 3+4 2 | 3+4 2 |
| | 4+5 2 | 4+5 2 | 5+6 2 | 5+6 2 |
| | 8 2 | 9 2 | 3+4 2 | 3+4 2 |
| | 9 2 | 9 2 | 5+6 2 | 3+4 2 |
| | 6+7 2 | 6+7 2 | 5+6 2 | 5+6 2 |
| | 9 2 | 9 2 | 3+4 2 | 5+6 2 |
| | 6+7 2 | 9 2 | 3+4 2 | 3+4 2 |
| | 9 2 | 6+7 2 | 3+4 2 | 3+4 2 |
| | 9 2 | 9 2 | 3+4 2 | 3+4 2 |
| | 9 2 | 8 2 | 5+6 2 | 5+6 2 |
| | 9 2 | 9 2 | 5+6 2 | 3+4 2 |
| | 8 2 | 9 2 | 5+6 2 | 5+6 2 |
| | 8 2 | 6+7 2 | 3+4 2 | 3+4 2 |
| | 4+5 2 | 9 2 | 3+4 2 | 3+4 2 |
| | 6+7 2 | 6+7 2 | 3+4 2 | 3+4 2 |
| | 8 2 | 8 2 | 3+4 2 | 5+6 2 |
| | 8 2 | 4+5 2 | 5+6 2 | 3+4 2 |
| | 9 2 | 8 2 | 5+6 2 | 3+4 2 |
| | 8 2 | 9 2 | 5+6 2 | 5+6 2 |

Continued on next column

**TABLE LXXIII– continued**

| TQC | ICSD | | NoE | SF |
|---|---|---|---|---|
| | 4+5 2 | 9 2 | 5+6 2 | 3+4 2 |
| | 6+7 2 | 6+7 2 | 3+4 2 | 5+6 2 |
| | 8 2 | 9 2 | 3+4 2 | 5+6 2 |
| | 4+5 2 | 9 2 | 5+6 2 | 5+6 2 |
| | 8 2 | 6+7 2 | 5+6 2 | 5+6 2 |
| | 9 2 | 8 2 | 3+4 2 | 3+4 2 |
| | 8 2 | 4+5 2 | 5+6 2 | 5+6 2 |
| | 9 2 | 8 2 | 5+6 2 | 3+4 2 |
| | 6+7 2 | 8 2 | 3+4 2 | 3+4 2 |
| | 4+5 2 | 4+5 2 | 3+4 2 | 3+4 2 |
| | 8 2 | 9 2 | 3+4 2 | 5+6 2 |
| | 4+5 2 | 6+7 2 | 5+6 2 | 3+4 2 |
| | 8 2 | 9 2 | 5+6 2 | 5+6 2 |
| | 9 2 | 6+7 2 | 3+4 2 | 5+6 2 |
| | 6+7 2 | 9 2 | 5+6 2 | 5+6 2 |
| | 9 2 | 4+5 2 | 5+6 2 | 3+4 2 |
| | 9 2 | 8 2 | 3+4 2 | 3+4 2 |
| | 6+7 2 | 9 2 | 5+6 2 | 5+6 2 |
| | 8 2 | 8 2 | 3+4 2 | 3+4 2 |
| | 4+5 2 | 9 2 | 5+6 2 | 5+6 2 |
| | 8 2 | 6+7 2 | 5+6 2 | 3+4 2 |
| | 4+5 2 | 6+7 2 | 3+4 2 | 5+6 2 |
| | 8 2 | 9 2 | 3+4 2 | 3+4 2 |
| | 8 2 | 9 2 | 5+6 2 | 5+6 2 |
| | 8 2 | 8 2 | 5+6 2 | 3+4 2 |
| | 9 2 | 8 2 | 3+4 2 | 3+4 2 |
| | 4+5 2 | 4+5 2 | 3+4 2 | 5+6 2 |
| | 9 2 | 8 2 | 3+4 2 | 5+6 2 |
| | 6+7 2 | 9 2 | 5+6 2 | 3+4 2 |
| | 9 2 | 6+7 2 | 3+4 2 | 3+4 2 |
| | 6+7 2 | 9 2 | 3+4 2 | 5+6 2 |
| | 9 2 | 4+5 2 | 5+6 2 | 3+4 2 |
| | 6+7 2 | 8 2 | 3+4 2 | 3+4 2 |
| | 9 2 | 9 2 | 3+4 2 | 5+6 2 |
| | 8 2 | 8 2 | 5+6 2 | 5+6 2 |
| | 4+5 2 | 4+5 2 | 5+6 2 | 5+6 2 |
| | 9 2 | 9 2 | 3+4 2 | 3+4 2 |
| | 8 2 | 6+7 2 | 5+6 2 | 5+6 2 |
| | 9 2 | 8 2 | 5+6 2 | 3+4 2 |
| | 8 2 | 8 2 | 3+4 2 | 5+6 2 |
| | 6+7 2 | 9 2 | 3+4 2 | 3+4 2 |
| | 8 2 | 8 2 | 5+6 2 | 5+6 2 |
| | 9 2 | 4+5 2 | 3+4 2 | 3+4 2 |
| | 8 2 | 8 2 | 5+6 2 | 5+6 2 |
| | 4+5 2 | 8 2 | 5+6 2 | 3+4 2 |
| | 8 2 | 4+5 2 | 5+6 2 | 5+6 2 |
| | 8 2 | 8 2 | 3+4 2 | 3+4 2 |
| | 4+5 2 | 4+5 2 | 3+4 2 | 5+6 2 |
| | 9 2 | 9 2 | 3+4 2 | 5+6 2 |
| | 6+7 2 | 8 2 | 5+6 2 | 3+4 2 |
| | 8 2 | 6+7 2 | 5+6 2 | 5+6 2 |
| | 9 2 | 9 2 | 5+6 2 | 3+4 2 |
| | 4+5 2 | 8 2 | 3+4 2 | 3+4 2 |
| | 8 2 | 6+7 2 | 5+6 2 | 5+6 2 |
| | 8 2 | 8 2 | 3+4 2 | 5+6 2 |
| | 9 2 | 9 2 | 5+6 2 | 3+4 2 |
| | 8 2 | 4+5 2 | 5+6 2 | 3+4 2 |
| | 4+5 2 | 9 2 | 3+4 2 | 3+4 2 |
| | 9 2 | 6+7 2 | 5+6 2 | 5+6 2 |
| | 4+5 2 | 8 2 | 3+4 2 | 3+4 2 |
| | 9 2 | 9 2 | 5+6 2 | 5+6 2 |

Continued on next column

**TABLE LXXIII– continued**

| TQC | ICSD | | NoE | SF |
|---|---|---|---|---|
| | 8 2 | 6+7 2 | 3+4 2 | 3+4 2 |
| | 6+7 2 | 9 2 | 5+6 2 | 5+6 2 |
| | 9 2 | 4+5 2 | 3+4 2 | 5+6 2 |
| | 8 2 | 9 2 | 3+4 2 | 3+4 2 |
| | 8 2 | 8 2 | 5+6 2 | 5+6 2 |
| | 4+5 2 | 8 2 | 3+4 2 | 3+4 2 |
| | 6+7 2 | 4+5 2 | 5+6 2 | 5+6 2 |
| 166.1.292 | 41260 | | 40 | C3Al6N2 |
| | -GM | -T | -F | -L |
| | 8 2 | 8 2 | 5+6 2 | 3+4 2 |
| | 9 2 | 9 2 | 3+4 2 | 5+6 2 |
| | 8 2 | 8 2 | 3+4 2 | 3+4 2 |
| | 8 2 | 9 2 | 5+6 2 | 3+4 2 |
| | 9 2 | 8 2 | 3+4 2 | 5+6 2 |
| | 9 2 | 9 2 | 5+6 2 | 5+6 2 |
| | 8 2 | 8 2 | 5+6 2 | 5+6 2 |
| | 9 2 | 9 2 | 3+4 2 | 3+4 2 |
| | 8 2 | 9 2 | 3+4 2 | 5+6 2 |
| | 9 2 | 6+7 2 | 5+6 2 | 3+4 2 |
| | 6+7 2 | 8 2 | 5+6 2 | 5+6 2 |
| | 8 2 | 4+5 2 | 5+6 2 | 5+6 2 |
| | 4+5 2 | 9 2 | 3+4 2 | 3+4 2 |
| | 9 2 | 6+7 2 | 3+4 2 | 3+4 2 |
| | 6+7 2 | 9 2 | 5+6 2 | 5+6 2 |
| | 9 2 | 8 2 | 5+6 2 | 5+6 2 |
| | 8 2 | 9 2 | 5+6 2 | 3+4 2 |
| | 4+5 2 | 6+7 2 | 3+4 2 | 3+4 2 |
| | 9 2 | 8 2 | 5+6 2 | 5+6 2 |
| | 6+7 2 | 4+5 2 | 3+4 2 | 3+4 2 |
| 166.1.293 | 74178 | | 146 | C2Ba1Ce1F1O6 |
| | -GM | -T | -F | -L |
| | 8 2 | 8 2 | 3+4 2 | 3+4 2 |
| | 9 2 | 9 2 | 5+6 2 | 5+6 2 |
| | 8 2 | 8 2 | 3+4 2 | 3+4 2 |
| | 8 2 | 9 2 | 3+4 2 | 5+6 2 |
| | 9 2 | 8 2 | 5+6 2 | 3+4 2 |
| | 8 2 | 9 2 | 3+4 2 | 5+6 2 |
| | 9 2 | 9 2 | 5+6 2 | 5+6 2 |
| | 8 2 | 8 2 | 3+4 2 | 3+4 2 |
| | 8 2 | 9 2 | 5+6 2 | 3+4 2 |
| | 9 2 | 8 2 | 3+4 2 | 5+6 2 |
| | 9 2 | 8 2 | 5+6 2 | 3+4 2 |
| | 8 2 | 9 2 | 3+4 2 | 5+6 2 |
| | 6+7 2 | 4+5 2 | 5+6 2 | 3+4 2 |
| | 4+5 2 | 6+7 2 | 3+4 2 | 5+6 2 |
| | 8 2 | 8 2 | 5+6 2 | 5+6 2 |
| | 4+5 2 | 4+5 2 | 3+4 2 | 3+4 2 |
| | 9 2 | 9 2 | 5+6 2 | 5+6 2 |
| | 6+7 2 | 6+7 2 | 3+4 2 | 3+4 2 |
| | 8 2 | 8 2 | 5+6 2 | 3+4 2 |
| | 9 2 | 9 2 | 3+4 2 | 5+6 2 |
| | 4+5 2 | 4+5 2 | 5+6 2 | 3+4 2 |
| | 6+7 2 | 6+7 2 | 3+4 2 | 5+6 2 |
| | 8 2 | 9 2 | 3+4 2 | 5+6 2 |
| | 9 2 | 8 2 | 5+6 2 | 3+4 2 |
| | 9 2 | 9 2 | 5+6 2 | 5+6 2 |
| | 9 2 | 8 2 | 5+6 2 | 3+4 2 |
| | 9 2 | 8 2 | 5+6 2 | 3+4 2 |
| | 9 2 | 9 2 | 5+6 2 | 5+6 2 |
| | 6+7 2 | 6+7 2 | 5+6 2 | 5+6 2 |
| | 6+7 2 | 4+5 2 | 5+6 2 | 3+4 2 |

Continued on next column

**TABLE LXXIII– continued**

| TQC | ICSD | NoE | SF | |
|---|---|---|---|---|
| | 8 2 | 9 2 | 3+4 2 | 5+6 2 |
| | 9 2 | 8 2 | 5+6 2 | 3+4 2 |
| | 8 2 | 8 2 | 3+4 2 | 3+4 2 |
| | 9 2 | 9 2 | 5+6 2 | 5+6 2 |
| | 8 2 | 9 2 | 3+4 2 | 5+6 2 |
| | 8 2 | 8 2 | 5+6 2 | 3+4 2 |
| | 4+5 2 | 9 2 | 3+4 2 | 5+6 2 |
| | 9 2 | 4+5 2 | 5+6 2 | 3+4 2 |
| | 6+7 2 | 6+7 2 | 3+4 2 | 5+6 2 |
| | 9 2 | 8 2 | 3+4 2 | 5+6 2 |
| | 8 2 | 8 2 | 5+6 2 | 3+4 2 |
| | 4+5 2 | 8 2 | 5+6 2 | 3+4 2 |
| | 8 2 | 4+5 2 | 3+4 2 | 3+4 2 |
| | 9 2 | 9 2 | 3+4 2 | 3+4 2 |
| | 6+7 2 | 6+7 2 | 5+6 2 | 5+6 2 |
| | 9 2 | 9 2 | 5+6 2 | 5+6 2 |
| | 8 2 | 9 2 | 3+4 2 | 5+6 2 |
| | 4+5 2 | 6+7 2 | 5+6 2 | 3+4 2 |
| | 8 2 | 9 2 | 3+4 2 | 5+6 2 |
| | 4+5 2 | 8 2 | 3+4 2 | 3+4 2 |
| | 8 2 | 4+5 2 | 5+6 2 | 5+6 2 |
| | 9 2 | 9 2 | 3+4 2 | 3+4 2 |
| | 6+7 2 | 6+7 2 | 5+6 2 | 5+6 2 |
| | 9 2 | 8 2 | 3+4 2 | 3+4 2 |
| | 9 2 | 8 2 | 5+6 2 | 3+4 2 |
| | 6+7 2 | 4+5 2 | 3+4 2 | 3+4 2 |
| | 8 2 | 8 2 | 5+6 2 | 5+6 2 |
| | 4+5 2 | 4+5 2 | 5+6 2 | 5+6 2 |
| | 8 2 | 9 2 | 3+4 2 | 3+4 2 |
| | 4+5 2 | 6+7 2 | 3+4 2 | 5+6 2 |
| | 9 2 | 8 2 | 5+6 2 | 3+4 2 |
| | 6+7 2 | 4+5 2 | 3+4 2 | 5+6 2 |
| | 8 2 | 8 2 | 5+6 2 | 5+6 2 |
| | 4+5 2 | 4+5 2 | 5+6 2 | 3+4 2 |
| | 9 2 | 9 2 | 3+4 2 | 3+4 2 |
| | 6+7 2 | 6+7 2 | 5+6 2 | 5+6 2 |
| | 9 2 | 9 2 | 3+4 2 | 3+4 2 |
| | 6+7 2 | 6+7 2 | 5+6 2 | 5+6 2 |
| | 9 2 | 8 2 | 5+6 2 | 5+6 2 |
| | 8 2 | 9 2 | 3+4 2 | 3+4 2 |
| | 9 2 | 9 2 | 3+4 2 | 3+4 2 |
| | 8 2 | 8 2 | 5+6 2 | 5+6 2 |
| | 8 2 | 9 2 | 3+4 2 | 3+4 2 |
| 166.1.294 | 96188 | 26 | C1Y2 | |
| | -GM | -T | -F | -L |
| | 8 2 | 8 2 | 5+6 2 | 5+6 2 |
| | 9 2 | 9 2 | 3+4 2 | 3+4 2 |
| | 8 2 | 8 2 | 5+6 2 | 5+6 2 |
| | 9 2 | 9 2 | 3+4 2 | 3+4 2 |
| | 8 2 | 8 2 | 5+6 2 | 3+4 2 |
| | 4+5 2 | 9 2 | 3+4 2 | 5+6 2 |
| | 6+7 2 | 4+5 2 | 5+6 2 | 5+6 2 |
| | 9 2 | 6+7 2 | 3+4 2 | 3+4 2 |
| | 8 2 | 8 2 | 3+4 2 | 3+4 2 |
| | 9 2 | 9 2 | 5+6 2 | 5+6 2 |
| | 8 2 | 6+7 2 | 5+6 2 | 5+6 2 |
| | 6+7 2 | 9 2 | 5+6 2 | 5+6 2 |
| | 9 2 | 9 2 | 3+4 2 | 3+4 2 |
| 166.1.295 | 22283 | 26 | C1Y2 | |
| | -GM | -T | -F | -L |
| | 8 2 | 8 2 | 5+6 2 | 5+6 2 |

Continued on next column

**TABLE LXXIII– continued**

| TQC | ICSD | NoE | SF | |
|---|---|---|---|---|
| | 9 2 | 9 2 | 3+4 2 | 3+4 2 |
| | 8 2 | 8 2 | 5+6 2 | 5+6 2 |
| | 9 2 | 9 2 | 3+4 2 | 3+4 2 |
| | 8 2 | 8 2 | 5+6 2 | 3+4 2 |
| | 4+5 2 | 9 2 | 3+4 2 | 5+6 2 |
| | 6+7 2 | 4+5 2 | 5+6 2 | 5+6 2 |
| | 9 2 | 6+7 2 | 3+4 2 | 3+4 2 |
| | 8 2 | 8 2 | 3+4 2 | 3+4 2 |
| | 9 2 | 9 2 | 5+6 2 | 5+6 2 |
| | 8 2 | 6+7 2 | 5+6 2 | 5+6 2 |
| | 6+7 2 | 9 2 | 5+6 2 | 5+6 2 |
| | 9 2 | 9 2 | 3+4 2 | 3+4 2 |
| 166.1.296 | 42762 | 42 | C1Tb2 | |
| | -GM | -T | -F | -L |
| | 8 2 | 8 2 | 5+6 2 | 5+6 2 |
| | 9 2 | 9 2 | 3+4 2 | 3+4 2 |
| | 8 2 | 8 2 | 5+6 2 | 5+6 2 |
| | 9 2 | 9 2 | 3+4 2 | 3+4 2 |
| | 8 2 | 8 2 | 3+4 2 | 3+4 2 |
| | 4+5 2 | 9 2 | 5+6 2 | 5+6 2 |
| | 9 2 | 4+5 2 | 5+6 2 | 5+6 2 |
| | 6+7 2 | 6+7 2 | 3+4 2 | 3+4 2 |
| | 8 2 | 8 2 | 3+4 2 | 3+4 2 |
| | 9 2 | 9 2 | 5+6 2 | 5+6 2 |
| | 8 2 | 6+7 2 | 5+6 2 | 5+6 2 |
| | 6+7 2 | 9 2 | 5+6 2 | 5+6 2 |
| | 9 2 | 9 2 | 3+4 2 | 3+4 2 |
| | 9 2 | 8 2 | 3+4 2 | 3+4 2 |
| | 4+5 2 | 9 2 | 3+4 2 | 5+6 2 |
| | 8 2 | 8 2 | 5+6 2 | 3+4 2 |
| | 9 2 | 4+5 2 | 5+6 2 | 5+6 2 |
| | 8 2 | 6+7 2 | 5+6 2 | 5+6 2 |
| | 6+7 2 | 9 2 | 3+4 2 | 3+4 2 |
| | 8 2 | 8 2 | 3+4 2 | 5+6 2 |
| | 9 2 | 8 2 | 3+4 2 | 3+4 2 |
| 166.1.297 | 42761 | 42 | C1Tb2 | |
| | -GM | -T | -F | -L |
| | 8 2 | 8 2 | 5+6 2 | 5+6 2 |
| | 9 2 | 9 2 | 3+4 2 | 3+4 2 |
| | 8 2 | 8 2 | 5+6 2 | 5+6 2 |
| | 9 2 | 9 2 | 3+4 2 | 3+4 2 |
| | 8 2 | 8 2 | 3+4 2 | 3+4 2 |
| | 4+5 2 | 9 2 | 5+6 2 | 5+6 2 |
| | 6+7 2 | 4+5 2 | 5+6 2 | 5+6 2 |
| | 9 2 | 6+7 2 | 3+4 2 | 3+4 2 |
| | 8 2 | 8 2 | 3+4 2 | 3+4 2 |
| | 9 2 | 9 2 | 5+6 2 | 5+6 2 |
| | 8 2 | 6+7 2 | 5+6 2 | 5+6 2 |
| | 6+7 2 | 9 2 | 5+6 2 | 5+6 2 |
| | 9 2 | 9 2 | 3+4 2 | 3+4 2 |
| | 9 2 | 8 2 | 3+4 2 | 3+4 2 |
| | 4+5 2 | 9 2 | 3+4 2 | 5+6 2 |
| | 8 2 | 8 2 | 5+6 2 | 3+4 2 |
| | 9 2 | 4+5 2 | 5+6 2 | 5+6 2 |
| | 8 2 | 6+7 2 | 5+6 2 | 5+6 2 |
| | 6+7 2 | 9 2 | 3+4 2 | 3+4 2 |
| | 8 2 | 8 2 | 3+4 2 | 5+6 2 |
| | 9 2 | 8 2 | 3+4 2 | 3+4 2 |
| 166.1.298 | 108187 | 42 | C1Tb2 | |
| | -GM | -T | -F | -L |
| | 8 2 | 8 2 | 5+6 2 | 5+6 2 |

Continued on next column

**TABLE LXXIII– continued**

| TQC | ICSD | NoE | SF | |
|---|---|---|---|---|
| | 9 2 | 9 2 | 3+4 2 | 3+4 2 |
| | 8 2 | 8 2 | 5+6 2 | 5+6 2 |
| | 9 2 | 9 2 | 3+4 2 | 3+4 2 |
| | 8 2 | 8 2 | 3+4 2 | 3+4 2 |
| | 4+5 2 | 9 2 | 5+6 2 | 5+6 2 |
| | 6+7 2 | 4+5 2 | 5+6 2 | 5+6 2 |
| | 9 2 | 6+7 2 | 3+4 2 | 3+4 2 |
| | 8 2 | 8 2 | 3+4 2 | 3+4 2 |
| | 9 2 | 9 2 | 5+6 2 | 5+6 2 |
| | 8 2 | 6+7 2 | 5+6 2 | 5+6 2 |
| | 6+7 2 | 9 2 | 5+6 2 | 5+6 2 |
| | 9 2 | 9 2 | 3+4 2 | 3+4 2 |
| | 9 2 | 8 2 | 3+4 2 | 3+4 2 |
| | 4+5 2 | 9 2 | 3+4 2 | 5+6 2 |
| | 8 2 | 8 2 | 5+6 2 | 3+4 2 |
| | 9 2 | 4+5 2 | 5+6 2 | 5+6 2 |
| | 8 2 | 6+7 2 | 5+6 2 | 5+6 2 |
| | 6+7 2 | 9 2 | 3+4 2 | 3+4 2 |
| | 8 2 | 8 2 | 3+4 2 | 5+6 2 |
| | 9 2 | 8 2 | 3+4 2 | 3+4 2 |
| 166.1.299 | 23791 | 26 | C1S2Ta2 | |
| | -GM | -T | -F | -L |
| | 8 2 | 8 2 | 3+4 2 | 3+4 2 |
| | 9 2 | 9 2 | 5+6 2 | 5+6 2 |
| | 8 2 | 8 2 | 3+4 2 | 3+4 2 |
| | 9 2 | 9 2 | 5+6 2 | 5+6 2 |
| | 8 2 | 6+7 2 | 3+4 2 | 3+4 2 |
| | 6+7 2 | 9 2 | 5+6 2 | 5+6 2 |
| | 9 2 | 9 2 | 5+6 2 | 5+6 2 |
| | 4+5 2 | 8 2 | 5+6 2 | 5+6 2 |
| | 8 2 | 4+5 2 | 3+4 2 | 3+4 2 |
| | 9 2 | 8 2 | 5+6 2 | 3+4 2 |
| | 8 2 | 8 2 | 3+4 2 | 3+4 2 |
| | 4+5 2 | 4+5 2 | 3+4 2 | 5+6 2 |
| | 8 2 | 9 2 | 5+6 2 | 5+6 2 |
| 166.1.300 | 95112 | 42 | C1Nb2S2 | |
| | -GM | -T | -F | -L |
| | 8 2 | 8 2 | 5+6 2 | 5+6 2 |
| | 9 2 | 9 2 | 3+4 2 | 3+4 2 |
| | 8 2 | 8 2 | 5+6 2 | 5+6 2 |
| | 9 2 | 9 2 | 3+4 2 | 3+4 2 |
| | 8 2 | 8 2 | 5+6 2 | 5+6 2 |
| | 4+5 2 | 4+5 2 | 3+4 2 | 3+4 2 |
| | 9 2 | 9 2 | 5+6 2 | 5+6 2 |
| | 6+7 2 | 6+7 2 | 3+4 2 | 3+4 2 |
| | 8 2 | 8 2 | 3+4 2 | 3+4 2 |
| | 9 2 | 9 2 | 5+6 2 | 5+6 2 |
| | 8 2 | 8 2 | 3+4 2 | 3+4 2 |
| | 9 2 | 9 2 | 5+6 2 | 5+6 2 |
| | 6+7 2 | 6+7 2 | 3+4 2 | 3+4 2 |
| | 9 2 | 9 2 | 5+6 2 | 5+6 2 |
| | 8 2 | 8 2 | 5+6 2 | 5+6 2 |
| | 4+5 2 | 4+5 2 | 3+4 2 | 3+4 2 |
| | 8 2 | 8 2 | 5+6 2 | 5+6 2 |
| | 8 2 | 9 2 | 3+4 2 | 3+4 2 |
| | 4+5 2 | 8 2 | 5+6 2 | 3+4 2 |
| | 9 2 | 4+5 2 | 5+6 2 | 5+6 2 |
| | 9 2 | 9 2 | 3+4 2 | 5+6 2 |
| 166.1.301 | 95110 | 42 | C1Nb2S2 | |
| | -GM | -T | -F | -L |
| | 8 2 | 8 2 | 5+6 2 | 5+6 2 |

Continued on next column

**TABLE LXXIII– continued**

| TQC | ICSD | NoE | SF | |
|---|---|---|---|---|
| | 9 2 | 9 2 | 3+4 2 | 3+4 2 |
| | 8 2 | 8 2 | 5+6 2 | 5+6 2 |
| | 9 2 | 9 2 | 3+4 2 | 3+4 2 |
| | 8 2 | 8 2 | 5+6 2 | 5+6 2 |
| | 4+5 2 | 4+5 2 | 3+4 2 | 3+4 2 |
| | 6+7 2 | 6+7 2 | 5+6 2 | 5+6 2 |
| | 9 2 | 9 2 | 3+4 2 | 3+4 2 |
| | 8 2 | 8 2 | 3+4 2 | 3+4 2 |
| | 9 2 | 9 2 | 5+6 2 | 5+6 2 |
| | 8 2 | 8 2 | 3+4 2 | 3+4 2 |
| | 9 2 | 9 2 | 5+6 2 | 5+6 2 |
| | 6+7 2 | 6+7 2 | 3+4 2 | 3+4 2 |
| | 9 2 | 9 2 | 5+6 2 | 5+6 2 |
| | 8 2 | 8 2 | 3+4 2 | 5+6 2 |
| | 4+5 2 | 9 2 | 5+6 2 | 3+4 2 |
| | 8 2 | 4+5 2 | 5+6 2 | 5+6 2 |
| | 8 2 | 8 2 | 5+6 2 | 3+4 2 |
| | 4+5 2 | 8 2 | 3+4 2 | 3+4 2 |
| | 9 2 | 4+5 2 | 3+4 2 | 5+6 2 |
| | 9 2 | 9 2 | 5+6 2 | 5+6 2 |
| 166.1.302 | 53780 | 8 | C1 | |
| | -GM | -T | -F | -L |
| | 8 2 | 9 2 | 5+6 2 | 3+4 2 |
| | 9 2 | 8 2 | 3+4 2 | 5+6 2 |
| | 8 2 | 9 2 | 5+6 2 | 3+4 2 |
| | 4+5 2 | 6+7 2 | 3+4 2 | 5+6 2 |
| 166.1.303 | 31829 | 8 | C1 | |
| | -GM | -T | -F | -L |
| | 8 2 | 9 2 | 5+6 2 | 3+4 2 |
| | 9 2 | 8 2 | 3+4 2 | 5+6 2 |
| | 8 2 | 9 2 | 5+6 2 | 3+4 2 |
| | 4+5 2 | 6+7 2 | 3+4 2 | 5+6 2 |
| 166.1.304 | 29123 | 8 | C1 | |
| | -GM | -T | -F | -L |
| | 8 2 | 9 2 | 5+6 2 | 3+4 2 |
| | 9 2 | 8 2 | 3+4 2 | 5+6 2 |
| | 8 2 | 9 2 | 5+6 2 | 3+4 2 |
| | 4+5 2 | 6+7 2 | 3+4 2 | 5+6 2 |
| 166.1.305 | 602542 | 72 | C1B2U1 | |
| | -GM | -T | -F | -L |
| | 8 2 | 8 2 | 3+4 2 | 3+4 2 |
| | 8 2 | 8 2 | 5+6 2 | 5+6 2 |
| | 9 2 | 9 2 | 3+4 2 | 3+4 2 |
| | 9 2 | 9 2 | 5+6 2 | 5+6 2 |
| | 8 2 | 8 2 | 3+4 2 | 3+4 2 |
| | 9 2 | 9 2 | 5+6 2 | 5+6 2 |
| | 8 2 | 9 2 | 5+6 2 | 5+6 2 |
| | 9 2 | 6+7 2 | 5+6 2 | 3+4 2 |
| | 4+5 2 | 4+5 2 | 3+4 2 | 5+6 2 |
| | 6+7 2 | 8 2 | 3+4 2 | 3+4 2 |
| | 9 2 | 9 2 | 5+6 2 | 5+6 2 |
| | 6+7 2 | 6+7 2 | 5+6 2 | 5+6 2 |
| | 8 2 | 9 2 | 3+4 2 | 3+4 2 |
| | 8 2 | 9 2 | 5+6 2 | 3+4 2 |
| | 4+5 2 | 6+7 2 | 5+6 2 | 5+6 2 |
| | 9 2 | 8 2 | 3+4 2 | 5+6 2 |
| | 6+7 2 | 4+5 2 | 5+6 2 | 3+4 2 |
| | 8 2 | 9 2 | 3+4 2 | 5+6 2 |
| | 9 2 | 8 2 | 3+4 2 | 5+6 2 |
| | 9 2 | 8 2 | 5+6 2 | 3+4 2 |
| | 9 2 | 8 2 | 3+4 2 | 3+4 2 |

Continued on next column

**TABLE LXXIII– continued**

| TQC | ICSD | NoE | SF | |
|---|---|---|---|---|
| | 6+7 2 | 8 2 | 5+6 2 | 3+4 2 |
| | 4+5 2 | 4+5 2 | 5+6 2 | 5+6 2 |
| | 8 2 | 6+7 2 | 3+4 2 | 3+4 2 |
| | 9 2 | 8 2 | 3+4 2 | 5+6 2 |
| | 9 2 | 9 2 | 3+4 2 | 5+6 2 |
| | 6+7 2 | 4+5 2 | 5+6 2 | 5+6 2 |
| | 8 2 | 8 2 | 3+4 2 | 3+4 2 |
| | 9 2 | 4+5 2 | 3+4 2 | 5+6 2 |
| | 6+7 2 | 9 2 | 5+6 2 | 3+4 2 |
| | 4+5 2 | 6+7 2 | 3+4 2 | 5+6 2 |
| | 8 2 | 8 2 | 5+6 2 | 5+6 2 |
| | 9 2 | 9 2 | 5+6 2 | 3+4 2 |
| | 6+7 2 | 9 2 | 3+4 2 | 3+4 2 |
| | 9 2 | 9 2 | 5+6 2 | 5+6 2 |
| | 8 2 | 6+7 2 | 5+6 2 | 5+6 2 |
| 166.1.306 | 44142 | 72 | C1B2U1 | |
| | -GM | -T | -F | -L |
| | 8 2 | 8 2 | 3+4 2 | 3+4 2 |
| | 8 2 | 8 2 | 5+6 2 | 5+6 2 |
| | 9 2 | 9 2 | 3+4 2 | 3+4 2 |
| | 9 2 | 9 2 | 5+6 2 | 5+6 2 |
| | 8 2 | 8 2 | 3+4 2 | 3+4 2 |
| | 9 2 | 9 2 | 5+6 2 | 5+6 2 |
| | 8 2 | 9 2 | 5+6 2 | 5+6 2 |
| | 9 2 | 6+7 2 | 5+6 2 | 3+4 2 |
| | 4+5 2 | 4+5 2 | 3+4 2 | 5+6 2 |
| | 6+7 2 | 8 2 | 3+4 2 | 3+4 2 |
| | 9 2 | 9 2 | 5+6 2 | 5+6 2 |
| | 6+7 2 | 6+7 2 | 5+6 2 | 5+6 2 |
| | 8 2 | 9 2 | 3+4 2 | 3+4 2 |
| | 8 2 | 9 2 | 5+6 2 | 3+4 2 |
| | 4+5 2 | 6+7 2 | 5+6 2 | 5+6 2 |
| | 9 2 | 8 2 | 3+4 2 | 5+6 2 |
| | 6+7 2 | 4+5 2 | 5+6 2 | 3+4 2 |
| | 8 2 | 9 2 | 3+4 2 | 5+6 2 |
| | 9 2 | 8 2 | 3+4 2 | 5+6 2 |
| | 9 2 | 8 2 | 5+6 2 | 3+4 2 |
| | 9 2 | 8 2 | 3+4 2 | 3+4 2 |
| | 6+7 2 | 8 2 | 5+6 2 | 3+4 2 |
| | 4+5 2 | 4+5 2 | 5+6 2 | 5+6 2 |
| | 8 2 | 8 2 | 3+4 2 | 3+4 2 |
| | 9 2 | 6+7 2 | 3+4 2 | 5+6 2 |
| | 9 2 | 9 2 | 3+4 2 | 5+6 2 |
| | 6+7 2 | 4+5 2 | 5+6 2 | 5+6 2 |
| | 8 2 | 8 2 | 3+4 2 | 3+4 2 |
| | 9 2 | 4+5 2 | 3+4 2 | 5+6 2 |
| | 6+7 2 | 9 2 | 5+6 2 | 3+4 2 |
| | 4+5 2 | 6+7 2 | 3+4 2 | 5+6 2 |
| | 8 2 | 8 2 | 5+6 2 | 5+6 2 |
| | 9 2 | 9 2 | 5+6 2 | 3+4 2 |
| | 6+7 2 | 9 2 | 3+4 2 | 3+4 2 |
| | 9 2 | 9 2 | 5+6 2 | 5+6 2 |
| | 8 2 | 6+7 2 | 5+6 2 | 5+6 2 |
| 166.1.307 | 238975 | 78 | C1B2Pu1 | |
| | -GM | -T | -F | -L |
| | 8 2 | 8 2 | 3+4 2 | 3+4 2 |
| | 8 2 | 8 2 | 5+6 2 | 5+6 2 |
| | 9 2 | 9 2 | 3+4 2 | 3+4 2 |
| | 9 2 | 9 2 | 5+6 2 | 5+6 2 |
| | 8 2 | 8 2 | 5+6 2 | 3+4 2 |
| | 9 2 | 9 2 | 3+4 2 | 5+6 2 |

Continued on next column

**TABLE LXXIII– continued**

| TQC | ICSD | NoE | SF | |
|---|---|---|---|---|
| | 8 2 | 9 2 | 5+6 2 | 5+6 2 |
| | 9 2 | 6+7 2 | 5+6 2 | 3+4 2 |
| | 4+5 2 | 4+5 2 | 3+4 2 | 5+6 2 |
| | 6+7 2 | 8 2 | 3+4 2 | 3+4 2 |
| | 9 2 | 9 2 | 5+6 2 | 5+6 2 |
| | 6+7 2 | 6+7 2 | 5+6 2 | 5+6 2 |
| | 8 2 | 9 2 | 3+4 2 | 3+4 2 |
| | 8 2 | 9 2 | 5+6 2 | 3+4 2 |
| | 4+5 2 | 6+7 2 | 5+6 2 | 5+6 2 |
| | 9 2 | 8 2 | 3+4 2 | 5+6 2 |
| | 6+7 2 | 4+5 2 | 5+6 2 | 3+4 2 |
| | 8 2 | 9 2 | 3+4 2 | 5+6 2 |
| | 9 2 | 8 2 | 3+4 2 | 5+6 2 |
| | 9 2 | 8 2 | 5+6 2 | 3+4 2 |
| | 9 2 | 8 2 | 3+4 2 | 3+4 2 |
| | 6+7 2 | 8 2 | 5+6 2 | 5+6 2 |
| | 4+5 2 | 4+5 2 | 5+6 2 | 3+4 2 |
| | 8 2 | 6+7 2 | 3+4 2 | 3+4 2 |
| | 9 2 | 9 2 | 3+4 2 | 5+6 2 |
| | 9 2 | 8 2 | 5+6 2 | 5+6 2 |
| | 6+7 2 | 4+5 2 | 3+4 2 | 3+4 2 |
| | 8 2 | 8 2 | 3+4 2 | 5+6 2 |
| | 9 2 | 4+5 2 | 5+6 2 | 5+6 2 |
| | 6+7 2 | 9 2 | 3+4 2 | 3+4 2 |
| | 4+5 2 | 6+7 2 | 3+4 2 | 5+6 2 |
| | 8 2 | 9 2 | 5+6 2 | 5+6 2 |
| | 9 2 | 8 2 | 5+6 2 | 3+4 2 |
| | 6+7 2 | 9 2 | 5+6 2 | 5+6 2 |
| | 8 2 | 9 2 | 3+4 2 | 5+6 2 |
| | 4+5 2 | 6+7 2 | 5+6 2 | 3+4 2 |
| | 9 2 | 8 2 | 5+6 2 | 5+6 2 |
| | 8 2 | 9 2 | 3+4 2 | 5+6 2 |
| | 9 2 | 4+5 2 | 5+6 2 | 3+4 2 |
| 166.1.308 | 1168 | 22 | Br1Zr1 | |
| | -GM | -T | -F | -L |
| | 8 2 | 8 2 | 5+6 2 | 5+6 2 |
| | 9 2 | 9 2 | 3+4 2 | 3+4 2 |
| | 8 2 | 8 2 | 3+4 2 | 3+4 2 |
| | 9 2 | 9 2 | 5+6 2 | 5+6 2 |
| | 8 2 | 9 2 | 5+6 2 | 5+6 2 |
| | 9 2 | 8 2 | 3+4 2 | 3+4 2 |
| | 4+5 2 | 6+7 2 | 5+6 2 | 5+6 2 |
| | 6+7 2 | 4+5 2 | 3+4 2 | 3+4 2 |
| | 8 2 | 9 2 | 3+4 2 | 5+6 2 |
| | 9 2 | 8 2 | 3+4 2 | 3+4 2 |
| | 6+7 2 | 4+5 2 | 5+6 2 | 5+6 2 |
| 166.1.309 | 23354 | 36 | Br1La1 | |
| | -GM | -T | -F | -L |
| | 8 2 | 9 2 | 5+6 2 | 3+4 2 |
| | 9 2 | 8 2 | 3+4 2 | 5+6 2 |
| | 8 2 | 9 2 | 5+6 2 | 3+4 2 |
| | 9 2 | 8 2 | 3+4 2 | 5+6 2 |
| | 8 2 | 9 2 | 5+6 2 | 3+4 2 |
| | 9 2 | 8 2 | 3+4 2 | 5+6 2 |
| | 4+5 2 | 6+7 2 | 5+6 2 | 3+4 2 |
| | 6+7 2 | 4+5 2 | 3+4 2 | 5+6 2 |
| | 8 2 | 9 2 | 5+6 2 | 3+4 2 |
| | 9 2 | 8 2 | 3+4 2 | 5+6 2 |
| | 8 2 | 8 2 | 3+4 2 | 5+6 2 |
| | 9 2 | 9 2 | 5+6 2 | 3+4 2 |
| | 8 2 | 8 2 | 5+6 2 | 5+6 2 |

Continued on next column

**TABLE LXXIII– continued**

| TQC | ICSD | NoE | SF |
|---|---|---|---|
| | 9 2 9 2 3+4 2 3+4 2 | | |
| | 4+5 2 6+7 2 5+6 2 5+6 2 | | |
| | 6+7 2 4+5 2 3+4 2 3+4 2 | | |
| | 8 2 9 2 3+4 2 5+6 2 | | |
| | 9 2 8 2 3+4 2 5+6 2 | | |
| 166.1.310 | 202067 | 46 | Br1Eu2P1 |
| | -GM -T -F -L | | |
| | 8 2 9 2 5+6 2 3+4 2 | | |
| | 9 2 8 2 3+4 2 5+6 2 | | |
| | 8 2 9 2 5+6 2 3+4 2 | | |
| | 9 2 8 2 3+4 2 5+6 2 | | |
| | 8 2 9 2 3+4 2 3+4 2 | | |
| | 4+5 2 8 2 5+6 2 5+6 2 | | |
| | 9 2 6+7 2 5+6 2 3+4 2 | | |
| | 6+7 2 4+5 2 3+4 2 5+6 2 | | |
| | 8 2 8 2 3+4 2 3+4 2 | | |
| | 8 2 9 2 3+4 2 5+6 2 | | |
| | 9 2 9 2 5+6 2 5+6 2 | | |
| | 9 2 9 2 5+6 2 5+6 2 | | |
| | 6+7 2 6+7 2 5+6 2 5+6 2 | | |
| | 9 2 8 2 5+6 2 3+4 2 | | |
| | 9 2 8 2 5+6 2 3+4 2 | | |
| | 6+7 2 4+5 2 5+6 2 3+4 2 | | |
| | 8 2 8 2 3+4 2 3+4 2 | | |
| | 4+5 2 9 2 3+4 2 5+6 2 | | |
| | 8 2 8 2 3+4 2 5+6 2 | | |
| | 9 2 6+7 2 5+6 2 3+4 2 | | |
| | 9 2 9 2 3+4 2 5+6 2 | | |
| | 6+7 2 4+5 2 5+6 2 3+4 2 | | |
| | 8 2 8 2 5+6 2 5+6 2 | | |
| 166.1.311 | 95551 | 94 | Bi6Pb1Te10 |
| | -GM -T -F -L | | |
| | 8 2 9 2 5+6 2 3+4 2 | | |
| | 9 2 8 2 3+4 2 5+6 2 | | |
| | 8 2 8 2 5+6 2 5+6 2 | | |
| | 9 2 9 2 3+4 2 3+4 2 | | |
| | 8 2 9 2 5+6 2 3+4 2 | | |
| | 9 2 8 2 3+4 2 5+6 2 | | |
| | 8 2 8 2 5+6 2 3+4 2 | | |
| | 8 2 9 2 3+4 2 5+6 2 | | |
| | 9 2 8 2 5+6 2 5+6 2 | | |
| | 9 2 9 2 3+4 2 3+4 2 | | |
| | 8 2 9 2 5+6 2 3+4 2 | | |
| | 9 2 8 2 3+4 2 5+6 2 | | |
| | 8 2 8 2 5+6 2 3+4 2 | | |
| | 9 2 9 2 3+4 2 5+6 2 | | |
| | 8 2 9 2 5+6 2 3+4 2 | | |
| | 9 2 8 2 3+4 2 5+6 2 | | |
| | 8 2 8 2 3+4 2 3+4 2 | | |
| | 8 2 9 2 3+4 2 5+6 2 | | |
| | 9 2 8 2 5+6 2 3+4 2 | | |
| | 8 2 9 2 5+6 2 5+6 2 | | |
| | 9 2 8 2 3+4 2 3+4 2 | | |
| | 8 2 9 2 5+6 2 5+6 2 | | |
| | 9 2 8 2 3+4 2 3+4 2 | | |
| | 8 2 9 2 5+6 2 5+6 2 | | |
| | 9 2 9 2 3+4 2 3+4 2 | | |
| | 8 2 8 2 5+6 2 5+6 2 | | |
| | 9 2 9 2 3+4 2 3+4 2 | | |
| | 8 2 9 2 5+6 2 3+4 2 | | |
| | 9 2 8 2 5+6 2 5+6 2 | | |

Continued on next column

**TABLE LXXIII– continued**

| TQC | ICSD | NoE | SF |
|---|---|---|---|
| | 4+5 2 6+7 2 3+4 2 5+6 2 | | |
| | 6+7 2 4+5 2 5+6 2 3+4 2 | | |
| | 4+5 2 6+7 2 5+6 2 5+6 2 | | |
| | 8 2 8 2 3+4 2 3+4 2 | | |
| | 6+7 2 4+5 2 3+4 2 5+6 2 | | |
| | 9 2 9 2 5+6 2 5+6 2 | | |
| | 8 2 8 2 5+6 2 3+4 2 | | |
| | 9 2 9 2 3+4 2 5+6 2 | | |
| | 4+5 2 6+7 2 5+6 2 3+4 2 | | |
| | 6+7 2 4+5 2 3+4 2 5+6 2 | | |
| | 6+7 2 6+7 2 5+6 2 3+4 2 | | |
| | 4+5 2 8 2 3+4 2 5+6 2 | | |
| | 8 2 4+5 2 3+4 2 3+4 2 | | |
| | 6+7 2 6+7 2 5+6 2 5+6 2 | | |
| | 4+5 2 4+5 2 3+4 2 3+4 2 | | |
| | 8 2 9 2 5+6 2 3+4 2 | | |
| | 9 2 8 2 3+4 2 5+6 2 | | |
| | 9 2 9 2 3+4 2 3+4 2 | | |
| 166.1.312 | 20319 | 38 | Bi4Se3 |
| | -GM -T -F -L | | |
| | 8 2 8 2 5+6 2 5+6 2 | | |
| | 9 2 9 2 3+4 2 3+4 2 | | |
| | 8 2 8 2 3+4 2 3+4 2 | | |
| | 8 2 9 2 5+6 2 5+6 2 | | |
| | 9 2 9 2 5+6 2 3+4 2 | | |
| | 8 2 8 2 3+4 2 3+4 2 | | |
| | 9 2 8 2 3+4 2 5+6 2 | | |
| | 9 2 9 2 5+6 2 5+6 2 | | |
| | 8 2 8 2 3+4 2 3+4 2 | | |
| | 8 2 9 2 5+6 2 5+6 2 | | |
| | 6+7 2 6+7 2 5+6 2 5+6 2 | | |
| | 9 2 9 2 3+4 2 3+4 2 | | |
| | 4+5 2 4+5 2 5+6 2 5+6 2 | | |
| | 9 2 9 2 3+4 2 3+4 2 | | |
| | 6+7 2 6+7 2 3+4 2 5+6 2 | | |
| | 9 2 8 2 5+6 2 3+4 2 | | |
| | 8 2 9 2 5+6 2 5+6 2 | | |
| | 8 2 8 2 5+6 2 5+6 2 | | |
| | 9 2 9 2 5+6 2 3+4 2 | | |
| 166.1.313 | 74348 | 28 | Bi2Te3 |
| | -GM -T -F -L | | |
| | 8 2 8 2 5+6 2 3+4 2 | | |
| | 9 2 9 2 3+4 2 5+6 2 | | |
| | 8 2 8 2 3+4 2 3+4 2 | | |
| | 9 2 9 2 5+6 2 5+6 2 | | |
| | 8 2 8 2 3+4 2 3+4 2 | | |
| | 9 2 9 2 5+6 2 5+6 2 | | |
| | 8 2 8 2 3+4 2 5+6 2 | | |
| | 8 2 9 2 5+6 2 3+4 2 | | |
| | 9 2 9 2 5+6 2 3+4 2 | | |
| | 4+5 2 6+7 2 3+4 2 5+6 2 | | |
| | 6+7 2 9 2 5+6 2 3+4 2 | | |
| | 9 2 4+5 2 3+4 2 5+6 2 | | |
| | 6+7 2 6+7 2 5+6 2 5+6 2 | | |
| | 8 2 8 2 5+6 2 5+6 2 | | |
| 166.1.314 | 658764 | 28 | Bi2Te3 |
| | -GM -T -F -L | | |
| | 8 2 8 2 3+4 2 3+4 2 | | |
| | 9 2 9 2 5+6 2 5+6 2 | | |
| | 8 2 8 2 3+4 2 3+4 2 | | |
| | 9 2 9 2 5+6 2 5+6 2 | | |

Continued on next column

**TABLE LXXIII– continued**

| TQC | ICSD | | NoE | SF |
|---|---|---|---|---|
| | 8 2 | 8 2 | 3+4 2 | 3+4 2 |
| | 9 2 | 9 2 | 5+6 2 | 5+6 2 |
| | 8 2 | 9 2 | 3+4 2 | 5+6 2 |
| | 8 2 | 8 2 | 5+6 2 | 3+4 2 |
| | 9 2 | 9 2 | 5+6 2 | 3+4 2 |
| | 6+7 2 | 6+7 2 | 3+4 2 | 5+6 2 |
| | 9 2 | 9 2 | 5+6 2 | 5+6 2 |
| | 4+5 2 | 6+7 2 | 5+6 2 | 3+4 2 |
| | 8 2 | 4+5 2 | 3+4 2 | 5+6 2 |
| | 6+7 2 | 8 2 | 5+6 2 | 5+6 2 |
| 166.1.315 | 44983 | | 28 | Bi2Te3 |
| | -GM | -T | -F | -L |
| | 8 2 | 8 2 | 5+6 2 | 3+4 2 |
| | 9 2 | 9 2 | 3+4 2 | 5+6 2 |
| | 8 2 | 8 2 | 3+4 2 | 3+4 2 |
| | 9 2 | 9 2 | 5+6 2 | 5+6 2 |
| | 8 2 | 8 2 | 3+4 2 | 3+4 2 |
| | 9 2 | 9 2 | 5+6 2 | 5+6 2 |
| | 8 2 | 8 2 | 3+4 2 | 5+6 2 |
| | 8 2 | 9 2 | 5+6 2 | 3+4 2 |
| | 9 2 | 9 2 | 5+6 2 | 3+4 2 |
| | 4+5 2 | 6+7 2 | 3+4 2 | 5+6 2 |
| | 6+7 2 | 4+5 2 | 5+6 2 | 3+4 2 |
| | 9 2 | 9 2 | 3+4 2 | 5+6 2 |
| | 6+7 2 | 6+7 2 | 5+6 2 | 5+6 2 |
| | 8 2 | 8 2 | 5+6 2 | 5+6 2 |
| 166.1.316 | 42546 | | 28 | Bi2Te3 |
| | -GM | -T | -F | -L |
| | 8 2 | 8 2 | 5+6 2 | 3+4 2 |
| | 9 2 | 9 2 | 3+4 2 | 5+6 2 |
| | 8 2 | 8 2 | 3+4 2 | 3+4 2 |
| | 9 2 | 9 2 | 5+6 2 | 5+6 2 |
| | 8 2 | 8 2 | 3+4 2 | 3+4 2 |
| | 9 2 | 9 2 | 5+6 2 | 5+6 2 |
| | 8 2 | 8 2 | 3+4 2 | 5+6 2 |
| | 8 2 | 9 2 | 5+6 2 | 3+4 2 |
| | 9 2 | 9 2 | 5+6 2 | 3+4 2 |
| | 4+5 2 | 6+7 2 | 3+4 2 | 5+6 2 |
| | 6+7 2 | 9 2 | 5+6 2 | 3+4 2 |
| | 9 2 | 4+5 2 | 3+4 2 | 5+6 2 |
| | 6+7 2 | 6+7 2 | 5+6 2 | 5+6 2 |
| | 8 2 | 8 2 | 5+6 2 | 5+6 2 |
| 166.1.317 | 193340 | | 28 | Bi2Te3 |
| | -GM | -T | -F | -L |
| | 8 2 | 8 2 | 5+6 2 | 3+4 2 |
| | 9 2 | 9 2 | 3+4 2 | 5+6 2 |
| | 8 2 | 8 2 | 3+4 2 | 3+4 2 |
| | 9 2 | 9 2 | 5+6 2 | 5+6 2 |
| | 8 2 | 8 2 | 3+4 2 | 3+4 2 |
| | 9 2 | 9 2 | 5+6 2 | 5+6 2 |
| | 8 2 | 8 2 | 3+4 2 | 5+6 2 |
| | 8 2 | 9 2 | 5+6 2 | 3+4 2 |
| | 9 2 | 9 2 | 5+6 2 | 3+4 2 |
| | 4+5 2 | 6+7 2 | 3+4 2 | 5+6 2 |
| | 6+7 2 | 9 2 | 5+6 2 | 3+4 2 |
| | 9 2 | 4+5 2 | 3+4 2 | 5+6 2 |
| | 6+7 2 | 6+7 2 | 5+6 2 | 5+6 2 |
| | 8 2 | 8 2 | 5+6 2 | 5+6 2 |
| 166.1.318 | 193339 | | 28 | Bi2Te3 |
| | -GM | -T | -F | -L |
| | 8 2 | 8 2 | 5+6 2 | 3+4 2 |

Continued on next column

**TABLE LXXIII– continued**

| TQC | ICSD | | NoE | SF |
|---|---|---|---|---|
| | 9 2 | 9 2 | 3+4 2 | 5+6 2 |
| | 8 2 | 8 2 | 3+4 2 | 3+4 2 |
| | 9 2 | 9 2 | 5+6 2 | 5+6 2 |
| | 8 2 | 8 2 | 3+4 2 | 3+4 2 |
| | 9 2 | 9 2 | 5+6 2 | 5+6 2 |
| | 8 2 | 8 2 | 3+4 2 | 5+6 2 |
| | 8 2 | 9 2 | 5+6 2 | 3+4 2 |
| | 9 2 | 9 2 | 5+6 2 | 3+4 2 |
| | 4+5 2 | 6+7 2 | 3+4 2 | 5+6 2 |
| | 6+7 2 | 9 2 | 5+6 2 | 3+4 2 |
| | 9 2 | 4+5 2 | 3+4 2 | 5+6 2 |
| | 6+7 2 | 6+7 2 | 5+6 2 | 5+6 2 |
| | 8 2 | 8 2 | 5+6 2 | 5+6 2 |
| 166.1.319 | 193338 | | 28 | Bi2Te3 |
| | -GM | -T | -F | -L |
| | 8 2 | 8 2 | 5+6 2 | 3+4 2 |
| | 9 2 | 9 2 | 3+4 2 | 5+6 2 |
| | 8 2 | 8 2 | 3+4 2 | 3+4 2 |
| | 9 2 | 9 2 | 5+6 2 | 5+6 2 |
| | 8 2 | 8 2 | 3+4 2 | 3+4 2 |
| | 9 2 | 9 2 | 5+6 2 | 5+6 2 |
| | 8 2 | 8 2 | 3+4 2 | 5+6 2 |
| | 8 2 | 9 2 | 5+6 2 | 3+4 2 |
| | 9 2 | 9 2 | 5+6 2 | 3+4 2 |
| | 4+5 2 | 6+7 2 | 3+4 2 | 5+6 2 |
| | 6+7 2 | 9 2 | 5+6 2 | 3+4 2 |
| | 9 2 | 4+5 2 | 3+4 2 | 5+6 2 |
| | 6+7 2 | 6+7 2 | 5+6 2 | 5+6 2 |
| | 8 2 | 8 2 | 5+6 2 | 5+6 2 |
| 166.1.320 | 193337 | | 28 | Bi2Te3 |
| | -GM | -T | -F | -L |
| | 8 2 | 8 2 | 5+6 2 | 3+4 2 |
| | 9 2 | 9 2 | 3+4 2 | 5+6 2 |
| | 8 2 | 8 2 | 3+4 2 | 3+4 2 |
| | 9 2 | 9 2 | 5+6 2 | 5+6 2 |
| | 8 2 | 8 2 | 3+4 2 | 3+4 2 |
| | 9 2 | 9 2 | 5+6 2 | 5+6 2 |
| | 8 2 | 8 2 | 3+4 2 | 5+6 2 |
| | 8 2 | 9 2 | 5+6 2 | 3+4 2 |
| | 9 2 | 9 2 | 5+6 2 | 3+4 2 |
| | 4+5 2 | 6+7 2 | 3+4 2 | 5+6 2 |
| | 6+7 2 | 9 2 | 5+6 2 | 3+4 2 |
| | 9 2 | 4+5 2 | 3+4 2 | 5+6 2 |
| | 6+7 2 | 6+7 2 | 5+6 2 | 5+6 2 |
| | 8 2 | 8 2 | 5+6 2 | 5+6 2 |
| 166.1.321 | 193336 | | 28 | Bi2Te3 |
| | -GM | -T | -F | -L |
| | 8 2 | 8 2 | 5+6 2 | 3+4 2 |
| | 9 2 | 9 2 | 3+4 2 | 5+6 2 |
| | 8 2 | 8 2 | 3+4 2 | 3+4 2 |
| | 9 2 | 9 2 | 5+6 2 | 5+6 2 |
| | 8 2 | 8 2 | 3+4 2 | 3+4 2 |
| | 9 2 | 9 2 | 5+6 2 | 5+6 2 |
| | 8 2 | 8 2 | 3+4 2 | 5+6 2 |
| | 8 2 | 9 2 | 5+6 2 | 3+4 2 |
| | 9 2 | 9 2 | 5+6 2 | 3+4 2 |
| | 4+5 2 | 6+7 2 | 3+4 2 | 5+6 2 |
| | 6+7 2 | 9 2 | 5+6 2 | 3+4 2 |
| | 9 2 | 4+5 2 | 3+4 2 | 5+6 2 |
| | 6+7 2 | 6+7 2 | 5+6 2 | 5+6 2 |
| | 8 2 | 8 2 | 5+6 2 | 5+6 2 |

Continued on next column

**TABLE LXXIII– continued**

| TQC | ICSD | NoE | SF |
|---|---|---|---|
| 166.1.322 | 193335 | 28 | Bi2Te3 |
| | -GM -T -F -L | | |
| | 8 2 8 2 5+6 2 3+4 2 | | |
| | 9 2 9 2 3+4 2 5+6 2 | | |
| | 8 2 8 2 3+4 2 3+4 2 | | |
| | 9 2 9 2 5+6 2 5+6 2 | | |
| | 8 2 8 2 3+4 2 3+4 2 | | |
| | 9 2 9 2 5+6 2 5+6 2 | | |
| | 8 2 8 2 3+4 2 5+6 2 | | |
| | 8 2 9 2 5+6 2 3+4 2 | | |
| | 9 2 9 2 5+6 2 3+4 2 | | |
| | 4+5 2 6+7 2 3+4 2 5+6 2 | | |
| | 6+7 2 9 2 5+6 2 3+4 2 | | |
| | 9 2 4+5 2 3+4 2 5+6 2 | | |
| | 6+7 2 6+7 2 5+6 2 5+6 2 | | |
| | 8 2 8 2 5+6 2 5+6 2 | | |
| 166.1.323 | 193334 | 28 | Bi2Te3 |
| | -GM -T -F -L | | |
| | 8 2 8 2 5+6 2 3+4 2 | | |
| | 9 2 9 2 3+4 2 5+6 2 | | |
| | 8 2 8 2 3+4 2 3+4 2 | | |
| | 9 2 9 2 5+6 2 5+6 2 | | |
| | 8 2 8 2 3+4 2 3+4 2 | | |
| | 9 2 9 2 5+6 2 5+6 2 | | |
| | 8 2 8 2 3+4 2 5+6 2 | | |
| | 8 2 9 2 5+6 2 3+4 2 | | |
| | 9 2 9 2 5+6 2 3+4 2 | | |
| | 4+5 2 6+7 2 3+4 2 5+6 2 | | |
| | 6+7 2 9 2 5+6 2 3+4 2 | | |
| | 9 2 4+5 2 3+4 2 5+6 2 | | |
| | 6+7 2 6+7 2 5+6 2 5+6 2 | | |
| | 8 2 8 2 5+6 2 5+6 2 | | |
| 166.1.324 | 193333 | 28 | Bi2Te3 |
| | -GM -T -F -L | | |
| | 8 2 8 2 5+6 2 3+4 2 | | |
| | 9 2 9 2 3+4 2 5+6 2 | | |
| | 8 2 8 2 3+4 2 3+4 2 | | |
| | 9 2 9 2 5+6 2 5+6 2 | | |
| | 8 2 8 2 3+4 2 3+4 2 | | |
| | 9 2 9 2 5+6 2 5+6 2 | | |
| | 8 2 8 2 3+4 2 5+6 2 | | |
| | 8 2 9 2 5+6 2 3+4 2 | | |
| | 9 2 9 2 5+6 2 3+4 2 | | |
| | 4+5 2 6+7 2 3+4 2 5+6 2 | | |
| | 6+7 2 9 2 5+6 2 3+4 2 | | |
| | 9 2 4+5 2 3+4 2 5+6 2 | | |
| | 6+7 2 6+7 2 5+6 2 5+6 2 | | |
| | 8 2 8 2 5+6 2 5+6 2 | | |
| 166.1.325 | 193332 | 28 | Bi2Te3 |
| | -GM -T -F -L | | |
| | 8 2 8 2 5+6 2 3+4 2 | | |
| | 9 2 9 2 3+4 2 5+6 2 | | |
| | 8 2 8 2 3+4 2 3+4 2 | | |
| | 9 2 9 2 5+6 2 5+6 2 | | |
| | 8 2 8 2 3+4 2 3+4 2 | | |
| | 9 2 9 2 5+6 2 5+6 2 | | |
| | 8 2 8 2 3+4 2 5+6 2 | | |
| | 8 2 9 2 5+6 2 3+4 2 | | |
| | 9 2 9 2 5+6 2 3+4 2 | | |
| | 4+5 2 6+7 2 3+4 2 5+6 2 | | |
| | 6+7 2 9 2 5+6 2 3+4 2 | | |

Continued on next column

**TABLE LXXIII– continued**

| TQC | ICSD | NoE | SF |
|---|---|---|---|
| | 9 2 4+5 2 3+4 2 5+6 2 | | |
| | 6+7 2 6+7 2 5+6 2 5+6 2 | | |
| | 8 2 8 2 5+6 2 5+6 2 | | |
| 166.1.326 | 193331 | 28 | Bi2Te3 |
| | -GM -T -F -L | | |
| | 8 2 8 2 5+6 2 3+4 2 | | |
| | 9 2 9 2 3+4 2 5+6 2 | | |
| | 8 2 8 2 3+4 2 3+4 2 | | |
| | 9 2 9 2 5+6 2 5+6 2 | | |
| | 8 2 8 2 3+4 2 3+4 2 | | |
| | 9 2 9 2 5+6 2 5+6 2 | | |
| | 8 2 8 2 3+4 2 5+6 2 | | |
| | 8 2 9 2 5+6 2 3+4 2 | | |
| | 9 2 9 2 5+6 2 3+4 2 | | |
| | 4+5 2 6+7 2 3+4 2 5+6 2 | | |
| | 6+7 2 9 2 5+6 2 3+4 2 | | |
| | 9 2 4+5 2 3+4 2 5+6 2 | | |
| | 6+7 2 6+7 2 5+6 2 5+6 2 | | |
| | 8 2 8 2 5+6 2 5+6 2 | | |
| 166.1.327 | 193330 | 28 | Bi2Te3 |
| | -GM -T -F -L | | |
| | 8 2 8 2 5+6 2 3+4 2 | | |
| | 9 2 9 2 3+4 2 5+6 2 | | |
| | 8 2 8 2 3+4 2 3+4 2 | | |
| | 9 2 9 2 5+6 2 5+6 2 | | |
| | 8 2 8 2 3+4 2 3+4 2 | | |
| | 9 2 9 2 5+6 2 5+6 2 | | |
| | 8 2 8 2 3+4 2 5+6 2 | | |
| | 8 2 9 2 5+6 2 3+4 2 | | |
| | 9 2 9 2 5+6 2 3+4 2 | | |
| | 4+5 2 6+7 2 3+4 2 5+6 2 | | |
| | 6+7 2 9 2 5+6 2 3+4 2 | | |
| | 9 2 4+5 2 3+4 2 5+6 2 | | |
| | 6+7 2 6+7 2 5+6 2 5+6 2 | | |
| | 8 2 8 2 5+6 2 5+6 2 | | |
| 166.1.328 | 184631 | 28 | Bi2Te3 |
| | -GM -T -F -L | | |
| | 8 2 8 2 5+6 2 3+4 2 | | |
| | 9 2 9 2 3+4 2 5+6 2 | | |
| | 8 2 8 2 3+4 2 3+4 2 | | |
| | 9 2 9 2 5+6 2 5+6 2 | | |
| | 8 2 8 2 3+4 2 3+4 2 | | |
| | 9 2 9 2 5+6 2 5+6 2 | | |
| | 8 2 8 2 3+4 2 5+6 2 | | |
| | 8 2 9 2 5+6 2 3+4 2 | | |
| | 9 2 9 2 5+6 2 3+4 2 | | |
| | 4+5 2 6+7 2 3+4 2 5+6 2 | | |
| | 6+7 2 9 2 5+6 2 3+4 2 | | |
| | 9 2 4+5 2 3+4 2 5+6 2 | | |
| | 6+7 2 6+7 2 5+6 2 5+6 2 | | |
| | 8 2 8 2 5+6 2 5+6 2 | | |
| 166.1.329 | 158366 | 28 | Bi2Te3 |
| | -GM -T -F -L | | |
| | 8 2 8 2 3+4 2 3+4 2 | | |
| | 9 2 9 2 5+6 2 5+6 2 | | |
| | 8 2 8 2 3+4 2 3+4 2 | | |
| | 9 2 9 2 5+6 2 5+6 2 | | |
| | 8 2 8 2 3+4 2 3+4 2 | | |
| | 9 2 9 2 5+6 2 5+6 2 | | |
| | 8 2 9 2 3+4 2 5+6 2 | | |
| | 8 2 8 2 5+6 2 3+4 2 | | |

Continued on next column

**TABLE LXXIII– continued**

| TQC | ICSD | NoE | SF | |
|---|---|---|---|---|
| | 9 2 | 9 2 | 5+6 2 | 3+4 2 |
| | 6+7 2 | 6+7 2 | 3+4 2 | 5+6 2 |
| | 9 2 | 9 2 | 5+6 2 | 3+4 2 |
| | 4+5 2 | 6+7 2 | 5+6 2 | 5+6 2 |
| | 8 2 | 4+5 2 | 3+4 2 | 5+6 2 |
| | 6+7 2 | 8 2 | 5+6 2 | 5+6 2 |
| 166.1.330 | 15753 | 28 | Bi2Te3 | |
| | -GM | -T | -F | -L |
| | 8 2 | 8 2 | 3+4 2 | 3+4 2 |
| | 9 2 | 9 2 | 5+6 2 | 5+6 2 |
| | 8 2 | 8 2 | 3+4 2 | 3+4 2 |
| | 9 2 | 9 2 | 5+6 2 | 5+6 2 |
| | 8 2 | 8 2 | 3+4 2 | 3+4 2 |
| | 9 2 | 9 2 | 5+6 2 | 5+6 2 |
| | 8 2 | 9 2 | 3+4 2 | 5+6 2 |
| | 8 2 | 8 2 | 5+6 2 | 3+4 2 |
| | 9 2 | 9 2 | 5+6 2 | 3+4 2 |
| | 6+7 2 | 6+7 2 | 3+4 2 | 5+6 2 |
| | 4+5 2 | 9 2 | 5+6 2 | 3+4 2 |
| | 9 2 | 6+7 2 | 5+6 2 | 5+6 2 |
| | 8 2 | 4+5 2 | 3+4 2 | 5+6 2 |
| | 6+7 2 | 8 2 | 5+6 2 | 5+6 2 |
| 166.1.331 | 42545 | 28 | Bi2Se3 | |
| | -GM | -T | -F | -L |
| | 8 2 | 8 2 | 3+4 2 | 3+4 2 |
| | 9 2 | 9 2 | 5+6 2 | 3+4 2 |
| | 8 2 | 8 2 | 3+4 2 | 5+6 2 |
| | 9 2 | 9 2 | 5+6 2 | 5+6 2 |
| | 8 2 | 8 2 | 3+4 2 | 3+4 2 |
| | 9 2 | 9 2 | 5+6 2 | 5+6 2 |
| | 8 2 | 9 2 | 3+4 2 | 5+6 2 |
| | 8 2 | 8 2 | 5+6 2 | 3+4 2 |
| | 9 2 | 6+7 2 | 5+6 2 | 5+6 2 |
| | 6+7 2 | 9 2 | 3+4 2 | 3+4 2 |
| | 4+5 2 | 9 2 | 5+6 2 | 3+4 2 |
| | 9 2 | 6+7 2 | 3+4 2 | 5+6 2 |
| | 6+7 2 | 4+5 2 | 5+6 2 | 5+6 2 |
| | 8 2 | 8 2 | 5+6 2 | 5+6 2 |
| 166.1.332 | 20385 | 28 | Bi2Se3 | |
| | -GM | -T | -F | -L |
| | 8 2 | 8 2 | 3+4 2 | 3+4 2 |
| | 9 2 | 9 2 | 5+6 2 | 3+4 2 |
| | 8 2 | 8 2 | 3+4 2 | 5+6 2 |
| | 9 2 | 9 2 | 5+6 2 | 5+6 2 |
| | 8 2 | 8 2 | 3+4 2 | 3+4 2 |
| | 9 2 | 9 2 | 5+6 2 | 5+6 2 |
| | 8 2 | 9 2 | 3+4 2 | 5+6 2 |
| | 9 2 | 6+7 2 | 5+6 2 | 3+4 2 |
| | 6+7 2 | 8 2 | 5+6 2 | 5+6 2 |
| | 8 2 | 9 2 | 3+4 2 | 3+4 2 |
| | 4+5 2 | 6+7 2 | 5+6 2 | 3+4 2 |
| | 9 2 | 9 2 | 5+6 2 | 5+6 2 |
| | 6+7 2 | 4+5 2 | 3+4 2 | 5+6 2 |
| | 8 2 | 8 2 | 5+6 2 | 5+6 2 |
| 166.1.333 | 165226 | 28 | Bi2Se3 | |
| | -GM | -T | -F | -L |
| | 8 2 | 8 2 | 5+6 2 | 3+4 2 |
| | 9 2 | 9 2 | 3+4 2 | 5+6 2 |
| | 8 2 | 8 2 | 3+4 2 | 3+4 2 |
| | 9 2 | 9 2 | 5+6 2 | 5+6 2 |
| | 8 2 | 8 2 | 3+4 2 | 3+4 2 |

Continued on next column

**TABLE LXXIII– continued**

| TQC | ICSD | NoE | SF | |
|---|---|---|---|---|
| | 9 2 | 9 2 | 5+6 2 | 5+6 2 |
| | 8 2 | 9 2 | 3+4 2 | 5+6 2 |
| | 8 2 | 8 2 | 5+6 2 | 3+4 2 |
| | 9 2 | 6+7 2 | 5+6 2 | 3+4 2 |
| | 4+5 2 | 9 2 | 3+4 2 | 5+6 2 |
| | 6+7 2 | 9 2 | 5+6 2 | 3+4 2 |
| | 9 2 | 4+5 2 | 3+4 2 | 5+6 2 |
| | 6+7 2 | 6+7 2 | 5+6 2 | 5+6 2 |
| | 8 2 | 8 2 | 5+6 2 | 5+6 2 |
| 166.1.334 | 54838 | 28 | Bi2Se2Te1 | |
| | -GM | -T | -F | -L |
| | 8 2 | 8 2 | 5+6 2 | 3+4 2 |
| | 9 2 | 9 2 | 3+4 2 | 5+6 2 |
| | 8 2 | 8 2 | 3+4 2 | 3+4 2 |
| | 9 2 | 9 2 | 5+6 2 | 5+6 2 |
| | 8 2 | 8 2 | 3+4 2 | 3+4 2 |
| | 9 2 | 9 2 | 5+6 2 | 5+6 2 |
| | 8 2 | 9 2 | 5+6 2 | 5+6 2 |
| | 8 2 | 8 2 | 3+4 2 | 3+4 2 |
| | 9 2 | 6+7 2 | 5+6 2 | 5+6 2 |
| | 6+7 2 | 9 2 | 3+4 2 | 3+4 2 |
| | 4+5 2 | 9 2 | 5+6 2 | 3+4 2 |
| | 9 2 | 4+5 2 | 3+4 2 | 5+6 2 |
| | 6+7 2 | 6+7 2 | 5+6 2 | 5+6 2 |
| | 8 2 | 8 2 | 5+6 2 | 5+6 2 |
| 166.1.335 | 43512 | 28 | Bi2Se1Te2 | |
| | -GM | -T | -F | -L |
| | 8 2 | 8 2 | 3+4 2 | 3+4 2 |
| | 9 2 | 9 2 | 5+6 2 | 5+6 2 |
| | 8 2 | 8 2 | 3+4 2 | 3+4 2 |
| | 9 2 | 9 2 | 5+6 2 | 5+6 2 |
| | 8 2 | 8 2 | 3+4 2 | 3+4 2 |
| | 8 2 | 9 2 | 5+6 2 | 5+6 2 |
| | 9 2 | 9 2 | 3+4 2 | 5+6 2 |
| | 8 2 | 8 2 | 5+6 2 | 3+4 2 |
| | 9 2 | 6+7 2 | 5+6 2 | 3+4 2 |
| | 6+7 2 | 9 2 | 3+4 2 | 5+6 2 |
| | 9 2 | 9 2 | 5+6 2 | 3+4 2 |
| | 4+5 2 | 6+7 2 | 3+4 2 | 5+6 2 |
| | 6+7 2 | 4+5 2 | 5+6 2 | 5+6 2 |
| | 8 2 | 8 2 | 5+6 2 | 5+6 2 |
| 166.1.336 | 658633 | 38 | Bi2Ge1Te4 | |
| | -GM | -T | -F | -L |
| | 8 2 | 9 2 | 5+6 2 | 3+4 2 |
| | 9 2 | 8 2 | 3+4 2 | 5+6 2 |
| | 8 2 | 9 2 | 5+6 2 | 3+4 2 |
| | 9 2 | 8 2 | 3+4 2 | 5+6 2 |
| | 8 2 | 9 2 | 5+6 2 | 3+4 2 |
| | 9 2 | 8 2 | 3+4 2 | 5+6 2 |
| | 8 2 | 8 2 | 3+4 2 | 3+4 2 |
| | 8 2 | 9 2 | 3+4 2 | 5+6 2 |
| | 9 2 | 8 2 | 5+6 2 | 3+4 2 |
| | 8 2 | 9 2 | 5+6 2 | 3+4 2 |
| | 9 2 | 9 2 | 5+6 2 | 5+6 2 |
| | 8 2 | 6+7 2 | 3+4 2 | 5+6 2 |
| | 4+5 2 | 9 2 | 5+6 2 | 3+4 2 |
| | 9 2 | 8 2 | 3+4 2 | 5+6 2 |
| | 6+7 2 | 4+5 2 | 5+6 2 | 3+4 2 |
| | 4+5 2 | 6+7 2 | 3+4 2 | 5+6 2 |
| | 8 2 | 8 2 | 5+6 2 | 3+4 2 |
| | 6+7 2 | 4+5 2 | 3+4 2 | 5+6 2 |

Continued on next column

**TABLE LXXIII– continued**

| TQC | ICSD | NoE | SF | |
|---|---|---|---|---|
| | 9 2 | 9 2 | 3+4 2 | 3+4 2 |
| 166.1.337 | 30394 | 38 | Bi2Ge1Te4 | |
| | -GM | -T | -F | -L |
| | 8 2 | 9 2 | 5+6 2 | 5+6 2 |
| | 9 2 | 8 2 | 3+4 2 | 3+4 2 |
| | 8 2 | 8 2 | 5+6 2 | 3+4 2 |
| | 9 2 | 9 2 | 3+4 2 | 5+6 2 |
| | 8 2 | 9 2 | 5+6 2 | 3+4 2 |
| | 9 2 | 8 2 | 3+4 2 | 5+6 2 |
| | 8 2 | 8 2 | 3+4 2 | 3+4 2 |
| | 8 2 | 9 2 | 5+6 2 | 5+6 2 |
| | 9 2 | 8 2 | 3+4 2 | 3+4 2 |
| | 8 2 | 9 2 | 5+6 2 | 3+4 2 |
| | 8 2 | 9 2 | 5+6 2 | 5+6 2 |
| | 9 2 | 6+7 2 | 3+4 2 | 5+6 2 |
| | 4+5 2 | 4+5 2 | 5+6 2 | 3+4 2 |
| | 6+7 2 | 8 2 | 3+4 2 | 5+6 2 |
| | 9 2 | 6+7 2 | 5+6 2 | 3+4 2 |
| | 4+5 2 | 9 2 | 5+6 2 | 5+6 2 |
| | 6+7 2 | 8 2 | 3+4 2 | 3+4 2 |
| | 8 2 | 4+5 2 | 3+4 2 | 5+6 2 |
| | 9 2 | 9 2 | 3+4 2 | 3+4 2 |
| 166.1.338 | 172572 | 20 | Bi1S2Tl1 | |
| | -GM | -T | -F | -L |
| | 8 2 | 8 2 | 5+6 2 | 3+4 2 |
| | 9 2 | 9 2 | 3+4 2 | 5+6 2 |
| | 8 2 | 8 2 | 3+4 2 | 3+4 2 |
| | 8 2 | 9 2 | 5+6 2 | 5+6 2 |
| | 9 2 | 9 2 | 3+4 2 | 5+6 2 |
| | 9 2 | 6+7 2 | 5+6 2 | 5+6 2 |
| | 6+7 2 | 8 2 | 5+6 2 | 3+4 2 |
| | 8 2 | 4+5 2 | 3+4 2 | 5+6 2 |
| | 4+5 2 | 8 2 | 3+4 2 | 3+4 2 |
| | 9 2 | 9 2 | 3+4 2 | 3+4 2 |
| 166.1.339 | 64705 | 10 | Bi1 | |
| | -GM | -T | -F | -L |
| | 8 2 | 9 2 | 5+6 2 | 3+4 2 |
| | 9 2 | 8 2 | 3+4 2 | 5+6 2 |
| | 8 2 | 9 2 | 3+4 2 | 3+4 2 |
| | 8 2 | 8 2 | 5+6 2 | 5+6 2 |
| | 4+5 2 | 6+7 2 | 5+6 2 | 5+6 2 |
| 166.1.340 | 64704 | 10 | Bi1 | |
| | -GM | -T | -F | -L |
| | 8 2 | 9 2 | 5+6 2 | 3+4 2 |
| | 9 2 | 8 2 | 3+4 2 | 5+6 2 |
| | 8 2 | 9 2 | 3+4 2 | 3+4 2 |
| | 8 2 | 8 2 | 5+6 2 | 5+6 2 |
| | 4+5 2 | 6+7 2 | 5+6 2 | 5+6 2 |
| 166.1.341 | 64703 | 10 | Bi1 | |
| | -GM | -T | -F | -L |
| | 8 2 | 9 2 | 5+6 2 | 3+4 2 |
| | 9 2 | 8 2 | 3+4 2 | 5+6 2 |
| | 8 2 | 9 2 | 3+4 2 | 3+4 2 |
| | 8 2 | 8 2 | 5+6 2 | 5+6 2 |
| | 4+5 2 | 6+7 2 | 5+6 2 | 5+6 2 |
| 166.1.342 | 53797 | 10 | Bi1 | |
| | -GM | -T | -F | -L |
| | 8 2 | 9 2 | 5+6 2 | 3+4 2 |
| | 9 2 | 8 2 | 3+4 2 | 5+6 2 |
| | 8 2 | 9 2 | 3+4 2 | 5+6 2 |

Continued on next column

**TABLE LXXIII– continued**

| TQC | ICSD | NoE | SF | |
|---|---|---|---|---|
| | 8 2 | 8 2 | 5+6 2 | 3+4 2 |
| | 4+5 2 | 6+7 2 | 5+6 2 | 5+6 2 |
| 166.1.343 | 246663 | 10 | Bi1 | |
| | -GM | -T | -F | -L |
| | 8 2 | 9 2 | 5+6 2 | 3+4 2 |
| | 9 2 | 8 2 | 3+4 2 | 5+6 2 |
| | 8 2 | 9 2 | 3+4 2 | 3+4 2 |
| | 8 2 | 8 2 | 5+6 2 | 5+6 2 |
| | 4+5 2 | 6+7 2 | 5+6 2 | 5+6 2 |
| 166.1.344 | 242004 | 10 | Bi1 | |
| | -GM | -T | -F | -L |
| | 8 2 | 9 2 | 5+6 2 | 3+4 2 |
| | 9 2 | 8 2 | 3+4 2 | 5+6 2 |
| | 8 2 | 9 2 | 3+4 2 | 3+4 2 |
| | 8 2 | 8 2 | 5+6 2 | 5+6 2 |
| | 4+5 2 | 6+7 2 | 5+6 2 | 5+6 2 |
| 166.1.345 | 242003 | 10 | Bi1 | |
| | -GM | -T | -F | -L |
| | 8 2 | 9 2 | 5+6 2 | 3+4 2 |
| | 9 2 | 8 2 | 3+4 2 | 5+6 2 |
| | 8 2 | 9 2 | 3+4 2 | 3+4 2 |
| | 8 2 | 8 2 | 5+6 2 | 5+6 2 |
| | 4+5 2 | 6+7 2 | 5+6 2 | 5+6 2 |
| 166.1.346 | 242002 | 10 | Bi1 | |
| | -GM | -T | -F | -L |
| | 8 2 | 9 2 | 5+6 2 | 3+4 2 |
| | 9 2 | 8 2 | 3+4 2 | 5+6 2 |
| | 8 2 | 9 2 | 3+4 2 | 3+4 2 |
| | 8 2 | 8 2 | 5+6 2 | 5+6 2 |
| | 4+5 2 | 6+7 2 | 5+6 2 | 5+6 2 |
| 166.1.347 | 242001 | 10 | Bi1 | |
| | -GM | -T | -F | -L |
| | 8 2 | 9 2 | 5+6 2 | 3+4 2 |
| | 9 2 | 8 2 | 3+4 2 | 5+6 2 |
| | 8 2 | 9 2 | 3+4 2 | 3+4 2 |
| | 8 2 | 8 2 | 5+6 2 | 5+6 2 |
| | 4+5 2 | 6+7 2 | 5+6 2 | 5+6 2 |
| 166.1.348 | 242000 | 10 | Bi1 | |
| | -GM | -T | -F | -L |
| | 8 2 | 9 2 | 5+6 2 | 3+4 2 |
| | 9 2 | 8 2 | 3+4 2 | 5+6 2 |
| | 8 2 | 9 2 | 3+4 2 | 3+4 2 |
| | 8 2 | 8 2 | 5+6 2 | 5+6 2 |
| | 4+5 2 | 6+7 2 | 5+6 2 | 5+6 2 |
| 166.1.349 | 241999 | 10 | Bi1 | |
| | -GM | -T | -F | -L |
| | 8 2 | 9 2 | 5+6 2 | 3+4 2 |
| | 9 2 | 8 2 | 3+4 2 | 5+6 2 |
| | 8 2 | 9 2 | 3+4 2 | 3+4 2 |
| | 8 2 | 8 2 | 5+6 2 | 5+6 2 |
| | 4+5 2 | 6+7 2 | 5+6 2 | 5+6 2 |
| 166.1.350 | 241984 | 10 | Bi1 | |
| | -GM | -T | -F | -L |
| | 8 2 | 9 2 | 5+6 2 | 3+4 2 |
| | 9 2 | 8 2 | 3+4 2 | 5+6 2 |
| | 8 2 | 9 2 | 3+4 2 | 3+4 2 |
| | 8 2 | 8 2 | 5+6 2 | 5+6 2 |
| | 4+5 2 | 6+7 2 | 5+6 2 | 5+6 2 |
| 166.1.351 | 241983 | 10 | Bi1 | |
| | -GM | -T | -F | -L |

Continued on next column

**TABLE LXXIII– continued**

| TQC | ICSD | NoE | SF | |
|---|---|---|---|---|
| | 8 2 | 9 2 | 5+6 2 | 3+4 2 |
| | 9 2 | 8 2 | 3+4 2 | 5+6 2 |
| | 8 2 | 9 2 | 3+4 2 | 3+4 2 |
| | 8 2 | 8 2 | 5+6 2 | 5+6 2 |
| | 4+5 2 | 6+7 2 | 5+6 2 | 5+6 2 |
| 166.1.352 | 58724 | 60 | Be17Nb2 | |
| | -GM | -T | -F | -L |
| | 8 2 | 8 2 | 3+4 2 | 3+4 2 |
| | 9 2 | 9 2 | 5+6 2 | 5+6 2 |
| | 8 2 | 8 2 | 3+4 2 | 3+4 2 |
| | 9 2 | 9 2 | 5+6 2 | 5+6 2 |
| | 8 2 | 6+7 2 | 3+4 2 | 3+4 2 |
| | 4+5 2 | 8 2 | 5+6 2 | 5+6 2 |
| | 6+7 2 | 4+5 2 | 3+4 2 | 3+4 2 |
| | 9 2 | 9 2 | 5+6 2 | 5+6 2 |
| | 8 2 | 9 2 | 3+4 2 | 5+6 2 |
| | 9 2 | 8 2 | 5+6 2 | 3+4 2 |
| | 6+7 2 | 8 2 | 5+6 2 | 5+6 2 |
| | 8 2 | 4+5 2 | 3+4 2 | 3+4 2 |
| | 4+5 2 | 9 2 | 5+6 2 | 3+4 2 |
| | 8 2 | 6+7 2 | 3+4 2 | 5+6 2 |
| | 9 2 | 9 2 | 5+6 2 | 5+6 2 |
| | 9 2 | 8 2 | 3+4 2 | 3+4 2 |
| | 6+7 2 | 9 2 | 5+6 2 | 3+4 2 |
| | 8 2 | 6+7 2 | 3+4 2 | 5+6 2 |
| | 4+5 2 | 9 2 | 5+6 2 | 3+4 2 |
| | 9 2 | 9 2 | 3+4 2 | 5+6 2 |
| | 9 2 | 6+7 2 | 5+6 2 | 5+6 2 |
| | 4+5 2 | 8 2 | 5+6 2 | 3+4 2 |
| | 8 2 | 4+5 2 | 3+4 2 | 3+4 2 |
| | 9 2 | 9 2 | 3+4 2 | 3+4 2 |
| | 8 2 | 6+7 2 | 5+6 2 | 3+4 2 |
| | 9 2 | 9 2 | 3+4 2 | 5+6 2 |
| | 6+7 2 | 8 2 | 3+4 2 | 3+4 2 |
| | 8 2 | 4+5 2 | 3+4 2 | 5+6 2 |
| | 9 2 | 8 2 | 5+6 2 | 5+6 2 |
| | 6+7 2 | 9 2 | 5+6 2 | 5+6 2 |
| 166.1.353 | 71755 | 206 | Ba7Cl2O15Ru4 | |
| | -GM | -T | -F | -L |
| | 8 2 | 9 2 | 5+6 2 | 3+4 2 |
| | 9 2 | 8 2 | 3+4 2 | 5+6 2 |
| | 8 2 | 8 2 | 3+4 2 | 3+4 2 |
| | 9 2 | 9 2 | 5+6 2 | 5+6 2 |
| | 8 2 | 8 2 | 3+4 2 | 3+4 2 |
| | 9 2 | 9 2 | 5+6 2 | 5+6 2 |
| | 8 2 | 8 2 | 3+4 2 | 3+4 2 |
| | 8 2 | 8 2 | 5+6 2 | 5+6 2 |
| | 9 2 | 9 2 | 3+4 2 | 3+4 2 |
| | 8 2 | 8 2 | 3+4 2 | 3+4 2 |
| | 4+5 2 | 4+5 2 | 5+6 2 | 5+6 2 |
| | 8 2 | 8 2 | 5+6 2 | 5+6 2 |
| | 9 2 | 9 2 | 5+6 2 | 5+6 2 |
| | 6+7 2 | 6+7 2 | 3+4 2 | 3+4 2 |
| | 8 2 | 8 2 | 3+4 2 | 3+4 2 |
| | 4+5 2 | 4+5 2 | 5+6 2 | 5+6 2 |
| | 9 2 | 9 2 | 3+4 2 | 3+4 2 |
| | 8 2 | 8 2 | 5+6 2 | 5+6 2 |
| | 8 2 | 9 2 | 5+6 2 | 5+6 2 |
| | 9 2 | 8 2 | 3+4 2 | 3+4 2 |
| | 6+7 2 | 6+7 2 | 5+6 2 | 5+6 2 |
| | 4+5 2 | 4+5 2 | 3+4 2 | 3+4 2 |

Continued on next column

**TABLE LXXIII– continued**

| TQC | ICSD | NoE | SF | |
|---|---|---|---|---|
| | 8 2 | 9 2 | 5+6 2 | 3+4 2 |
| | 9 2 | 8 2 | 3+4 2 | 5+6 2 |
| | 8 2 | 9 2 | 5+6 2 | 3+4 2 |
| | 9 2 | 8 2 | 3+4 2 | 5+6 2 |
| | 9 2 | 9 2 | 5+6 2 | 5+6 2 |
| | 8 2 | 8 2 | 3+4 2 | 3+4 2 |
| | 9 2 | 9 2 | 5+6 2 | 5+6 2 |
| | 9 2 | 9 2 | 3+4 2 | 3+4 2 |
| | 8 2 | 8 2 | 5+6 2 | 5+6 2 |
| | 4+5 2 | 6+7 2 | 5+6 2 | 3+4 2 |
| | 6+7 2 | 4+5 2 | 3+4 2 | 5+6 2 |
| | 8 2 | 9 2 | 5+6 2 | 3+4 2 |
| | 9 2 | 8 2 | 3+4 2 | 5+6 2 |
| | 6+7 2 | 6+7 2 | 3+4 2 | 5+6 2 |
| | 4+5 2 | 4+5 2 | 5+6 2 | 3+4 2 |
| | 9 2 | 6+7 2 | 5+6 2 | 5+6 2 |
| | 6+7 2 | 9 2 | 5+6 2 | 5+6 2 |
| | 8 2 | 8 2 | 3+4 2 | 3+4 2 |
| | 9 2 | 9 2 | 5+6 2 | 5+6 2 |
| | 4+5 2 | 4+5 2 | 3+4 2 | 3+4 2 |
| | 6+7 2 | 6+7 2 | 5+6 2 | 5+6 2 |
| | 9 2 | 9 2 | 3+4 2 | 3+4 2 |
| | 8 2 | 8 2 | 5+6 2 | 5+6 2 |
| | 9 2 | 9 2 | 3+4 2 | 3+4 2 |
| | 6+7 2 | 6+7 2 | 5+6 2 | 5+6 2 |
| | 9 2 | 9 2 | 3+4 2 | 3+4 2 |
| | 8 2 | 8 2 | 3+4 2 | 3+4 2 |
| | 9 2 | 9 2 | 3+4 2 | 3+4 2 |
| | 8 2 | 8 2 | 5+6 2 | 5+6 2 |
| | 4+5 2 | 4+5 2 | 5+6 2 | 5+6 2 |
| | 9 2 | 9 2 | 3+4 2 | 5+6 2 |
| | 6+7 2 | 6+7 2 | 5+6 2 | 3+4 2 |
| | 9 2 | 9 2 | 3+4 2 | 3+4 2 |
| | 8 2 | 6+7 2 | 5+6 2 | 5+6 2 |
| | 8 2 | 8 2 | 5+6 2 | 5+6 2 |
| | 6+7 2 | 9 2 | 3+4 2 | 3+4 2 |
| | 4+5 2 | 4+5 2 | 3+4 2 | 3+4 2 |
| | 8 2 | 8 2 | 5+6 2 | 5+6 2 |
| | 4+5 2 | 9 2 | 3+4 2 | 3+4 2 |
| | 9 2 | 6+7 2 | 3+4 2 | 5+6 2 |
| | 9 2 | 8 2 | 5+6 2 | 3+4 2 |
| | 8 2 | 4+5 2 | 5+6 2 | 3+4 2 |
| | 9 2 | 9 2 | 3+4 2 | 5+6 2 |
| | 6+7 2 | 8 2 | 5+6 2 | 5+6 2 |
| | 9 2 | 9 2 | 5+6 2 | 3+4 2 |
| | 6+7 2 | 6+7 2 | 3+4 2 | 3+4 2 |
| | 8 2 | 8 2 | 3+4 2 | 5+6 2 |
| | 4+5 2 | 9 2 | 5+6 2 | 5+6 2 |
| | 9 2 | 4+5 2 | 5+6 2 | 5+6 2 |
| | 9 2 | 8 2 | 3+4 2 | 3+4 2 |
| | 8 2 | 9 2 | 3+4 2 | 3+4 2 |
| | 8 2 | 8 2 | 3+4 2 | 3+4 2 |
| | 9 2 | 4+5 2 | 5+6 2 | 5+6 2 |
| | 4+5 2 | 9 2 | 3+4 2 | 3+4 2 |
| | 6+7 2 | 6+7 2 | 5+6 2 | 5+6 2 |
| | 8 2 | 9 2 | 3+4 2 | 3+4 2 |
| | 9 2 | 8 2 | 3+4 2 | 3+4 2 |
| | 6+7 2 | 6+7 2 | 5+6 2 | 5+6 2 |
| | 9 2 | 9 2 | 5+6 2 | 5+6 2 |
| | 8 2 | 8 2 | 3+4 2 | 3+4 2 |
| | 4+5 2 | 4+5 2 | 5+6 2 | 5+6 2 |

Continued on next column

**TABLE LXXIII– continued**

| TQC | ICSD | NoE | SF | |
|---|---|---|---|---|
| | 9 2 | 9 2 | 3+4 2 | 3+4 2 |
| | 9 2 | 9 2 | 5+6 2 | 5+6 2 |
| | 6+7 2 | 6+7 2 | 5+6 2 | 3+4 2 |
| | 8 2 | 8 2 | 3+4 2 | 5+6 2 |
| | 9 2 | 9 2 | 5+6 2 | 5+6 2 |
| | 9 2 | 9 2 | 5+6 2 | 5+6 2 |
| | 8 2 | 8 2 | 3+4 2 | 3+4 2 |
| | 6+7 2 | 6+7 2 | 3+4 2 | 3+4 2 |
| | 4+5 2 | 4+5 2 | 5+6 2 | 5+6 2 |
| | 9 2 | 9 2 | 3+4 2 | 3+4 2 |
| | 6+7 2 | 6+7 2 | 3+4 2 | 3+4 2 |
| | 8 2 | 8 2 | 3+4 2 | 3+4 2 |
| | 9 2 | 9 2 | 5+6 2 | 5+6 2 |
| | 8 2 | 8 2 | 5+6 2 | 5+6 2 |
| | 4+5 2 | 4+5 2 | 3+4 2 | 3+4 2 |
| | 8 2 | 8 2 | 5+6 2 | 5+6 2 |
| | 9 2 | 9 2 | 3+4 2 | 3+4 2 |
| | 6+7 2 | 6+7 2 | 5+6 2 | 5+6 2 |
| | 9 2 | 9 2 | 5+6 2 | 5+6 2 |
| | 4+5 2 | 4+5 2 | 3+4 2 | 3+4 2 |
| 166.1.354 | 73069 | 206 | Ba7Br2O15Ru4 | |
| | -GM | -T | -F | -L |
| | 8 2 | 9 2 | 3+4 2 | 3+4 2 |
| | 9 2 | 8 2 | 5+6 2 | 5+6 2 |
| | 8 2 | 8 2 | 3+4 2 | 3+4 2 |
| | 9 2 | 9 2 | 5+6 2 | 5+6 2 |
| | 8 2 | 8 2 | 3+4 2 | 3+4 2 |
| | 9 2 | 9 2 | 5+6 2 | 5+6 2 |
| | 8 2 | 8 2 | 3+4 2 | 3+4 2 |
| | 8 2 | 8 2 | 5+6 2 | 5+6 2 |
| | 9 2 | 9 2 | 3+4 2 | 3+4 2 |
| | 8 2 | 8 2 | 5+6 2 | 5+6 2 |
| | 4+5 2 | 4+5 2 | 3+4 2 | 3+4 2 |
| | 8 2 | 8 2 | 5+6 2 | 5+6 2 |
| | 9 2 | 9 2 | 5+6 2 | 5+6 2 |
| | 6+7 2 | 6+7 2 | 3+4 2 | 3+4 2 |
| | 8 2 | 8 2 | 3+4 2 | 3+4 2 |
| | 9 2 | 9 2 | 5+6 2 | 5+6 2 |
| | 4+5 2 | 4+5 2 | 3+4 2 | 3+4 2 |
| | 8 2 | 8 2 | 5+6 2 | 5+6 2 |
| | 9 2 | 9 2 | 3+4 2 | 3+4 2 |
| | 8 2 | 8 2 | 5+6 2 | 5+6 2 |
| | 6+7 2 | 6+7 2 | 5+6 2 | 5+6 2 |
| | 4+5 2 | 4+5 2 | 3+4 2 | 3+4 2 |
| | 8 2 | 9 2 | 5+6 2 | 3+4 2 |
| | 9 2 | 8 2 | 3+4 2 | 5+6 2 |
| | 9 2 | 8 2 | 5+6 2 | 3+4 2 |
| | 8 2 | 9 2 | 3+4 2 | 5+6 2 |
| | 9 2 | 9 2 | 5+6 2 | 5+6 2 |
| | 8 2 | 8 2 | 3+4 2 | 3+4 2 |
| | 9 2 | 9 2 | 5+6 2 | 5+6 2 |
| | 9 2 | 9 2 | 3+4 2 | 3+4 2 |
| | 8 2 | 8 2 | 5+6 2 | 5+6 2 |
| | 4+5 2 | 6+7 2 | 5+6 2 | 3+4 2 |
| | 6+7 2 | 4+5 2 | 3+4 2 | 5+6 2 |
| | 8 2 | 9 2 | 3+4 2 | 5+6 2 |
| | 9 2 | 8 2 | 5+6 2 | 3+4 2 |
| | 6+7 2 | 6+7 2 | 3+4 2 | 5+6 2 |
| | 4+5 2 | 4+5 2 | 5+6 2 | 3+4 2 |
| | 9 2 | 6+7 2 | 5+6 2 | 5+6 2 |
| | 6+7 2 | 9 2 | 3+4 2 | 3+4 2 |
| | 8 2 | 8 2 | 5+6 2 | 5+6 2 |
| | 9 2 | 9 2 | 5+6 2 | 5+6 2 |
| | 4+5 2 | 4+5 2 | 3+4 2 | 3+4 2 |
| | 6+7 2 | 6+7 2 | 5+6 2 | 5+6 2 |
| | 9 2 | 9 2 | 3+4 2 | 3+4 2 |
| | 8 2 | 8 2 | 5+6 2 | 5+6 2 |
| | 9 2 | 9 2 | 3+4 2 | 3+4 2 |
| | 6+7 2 | 6+7 2 | 5+6 2 | 5+6 2 |
| | 9 2 | 9 2 | 3+4 2 | 3+4 2 |
| | 8 2 | 8 2 | 3+4 2 | 3+4 2 |
| | 9 2 | 9 2 | 3+4 2 | 3+4 2 |
| | 8 2 | 8 2 | 5+6 2 | 5+6 2 |
| | 4+5 2 | 4+5 2 | 5+6 2 | 5+6 2 |
| | 9 2 | 9 2 | 3+4 2 | 3+4 2 |
| | 6+7 2 | 6+7 2 | 5+6 2 | 5+6 2 |
| | 9 2 | 9 2 | 3+4 2 | 3+4 2 |
| | 6+7 2 | 6+7 2 | 5+6 2 | 5+6 2 |
| | 8 2 | 8 2 | 5+6 2 | 5+6 2 |
| | 8 2 | 8 2 | 3+4 2 | 3+4 2 |
| | 4+5 2 | 4+5 2 | 3+4 2 | 3+4 2 |
| | 9 2 | 9 2 | 5+6 2 | 5+6 2 |
| | 8 2 | 9 2 | 3+4 2 | 3+4 2 |
| | 9 2 | 9 2 | 5+6 2 | 5+6 2 |
| | 4+5 2 | 8 2 | 3+4 2 | 5+6 2 |
| | 9 2 | 6+7 2 | 5+6 2 | 3+4 2 |
| | 8 2 | 4+5 2 | 3+4 2 | 3+4 2 |
| | 6+7 2 | 9 2 | 5+6 2 | 5+6 2 |
| | 9 2 | 9 2 | 5+6 2 | 3+4 2 |
| | 8 2 | 8 2 | 3+4 2 | 5+6 2 |
| | 4+5 2 | 6+7 2 | 5+6 2 | 3+4 2 |
| | 9 2 | 8 2 | 3+4 2 | 5+6 2 |
| | 6+7 2 | 8 2 | 3+4 2 | 3+4 2 |
| | 8 2 | 4+5 2 | 5+6 2 | 3+4 2 |
| | 4+5 2 | 8 2 | 3+4 2 | 3+4 2 |
| | 8 2 | 9 2 | 5+6 2 | 5+6 2 |
| | 9 2 | 4+5 2 | 3+4 2 | 5+6 2 |
| | 9 2 | 9 2 | 3+4 2 | 3+4 2 |
| | 6+7 2 | 6+7 2 | 5+6 2 | 5+6 2 |
| | 8 2 | 9 2 | 3+4 2 | 3+4 2 |
| | 9 2 | 6+7 2 | 3+4 2 | 3+4 2 |
| | 6+7 2 | 8 2 | 5+6 2 | 5+6 2 |
| | 9 2 | 9 2 | 5+6 2 | 5+6 2 |
| | 8 2 | 8 2 | 3+4 2 | 3+4 2 |
| | 4+5 2 | 4+5 2 | 5+6 2 | 5+6 2 |
| | 9 2 | 9 2 | 3+4 2 | 3+4 2 |
| | 9 2 | 9 2 | 5+6 2 | 5+6 2 |
| | 6+7 2 | 6+7 2 | 5+6 2 | 5+6 2 |
| | 8 2 | 8 2 | 3+4 2 | 3+4 2 |
| | 9 2 | 9 2 | 5+6 2 | 5+6 2 |
| | 8 2 | 8 2 | 5+6 2 | 3+4 2 |
| | 9 2 | 9 2 | 3+4 2 | 5+6 2 |
| | 6+7 2 | 6+7 2 | 3+4 2 | 3+4 2 |
| | 4+5 2 | 4+5 2 | 5+6 2 | 5+6 2 |
| | 9 2 | 9 2 | 3+4 2 | 3+4 2 |
| | 6+7 2 | 6+7 2 | 3+4 2 | 3+4 2 |
| | 8 2 | 8 2 | 3+4 2 | 3+4 2 |
| | 9 2 | 9 2 | 5+6 2 | 5+6 2 |
| | 8 2 | 8 2 | 5+6 2 | 5+6 2 |
| | 4+5 2 | 4+5 2 | 3+4 2 | 3+4 2 |
| | 8 2 | 8 2 | 5+6 2 | 5+6 2 |
| | 9 2 | 9 2 | 3+4 2 | 3+4 2 |

Continued on next column

**TABLE LXXIII– continued**

| TQC | ICSD | NoE | SF | |
|---|---|---|---|---|
| | 6+7 2 | 6+7 2 | 5+6 2 | 5+6 2 |
| | 9 2 | 9 2 | 5+6 2 | 5+6 2 |
| | 4+5 2 | 4+5 2 | 3+4 2 | 3+4 2 |
| 166.1.355 | 99662 | 146 | Ba4Mn3O12Pr1 | |
| | -GM | -T | -F | -L |
| | 8 2 | 8 2 | 3+4 2 | 3+4 2 |
| | 8 2 | 9 2 | 5+6 2 | 5+6 2 |
| | 9 2 | 8 2 | 3+4 2 | 3+4 2 |
| | 8 2 | 9 2 | 5+6 2 | 3+4 2 |
| | 9 2 | 8 2 | 3+4 2 | 5+6 2 |
| | 9 2 | 9 2 | 5+6 2 | 5+6 2 |
| | 9 2 | 9 2 | 5+6 2 | 5+6 2 |
| | 6+7 2 | 6+7 2 | 5+6 2 | 5+6 2 |
| | 8 2 | 8 2 | 3+4 2 | 3+4 2 |
| | 8 2 | 9 2 | 3+4 2 | 5+6 2 |
| | 4+5 2 | 8 2 | 3+4 2 | 3+4 2 |
| | 8 2 | 4+5 2 | 3+4 2 | 3+4 2 |
| | 9 2 | 8 2 | 5+6 2 | 3+4 2 |
| | 9 2 | 8 2 | 3+4 2 | 5+6 2 |
| | 6+7 2 | 4+5 2 | 5+6 2 | 3+4 2 |
| | 8 2 | 9 2 | 5+6 2 | 3+4 2 |
| | 4+5 2 | 6+7 2 | 3+4 2 | 5+6 2 |
| | 9 2 | 9 2 | 5+6 2 | 5+6 2 |
| | 9 2 | 9 2 | 5+6 2 | 5+6 2 |
| | 6+7 2 | 6+7 2 | 5+6 2 | 5+6 2 |
| | 8 2 | 8 2 | 5+6 2 | 5+6 2 |
| | 9 2 | 9 2 | 3+4 2 | 3+4 2 |
| | 8 2 | 9 2 | 5+6 2 | 3+4 2 |
| | 9 2 | 8 2 | 3+4 2 | 5+6 2 |
| | 4+5 2 | 4+5 2 | 5+6 2 | 3+4 2 |
| | 8 2 | 6+7 2 | 3+4 2 | 5+6 2 |
| | 6+7 2 | 8 2 | 5+6 2 | 3+4 2 |
| | 4+5 2 | 9 2 | 3+4 2 | 5+6 2 |
| | 9 2 | 6+7 2 | 5+6 2 | 3+4 2 |
| | 8 2 | 9 2 | 3+4 2 | 5+6 2 |
| | 9 2 | 4+5 2 | 5+6 2 | 3+4 2 |
| | 6+7 2 | 8 2 | 3+4 2 | 5+6 2 |
| | 8 2 | 9 2 | 3+4 2 | 3+4 2 |
| | 8 2 | 8 2 | 3+4 2 | 3+4 2 |
| | 4+5 2 | 4+5 2 | 3+4 2 | 3+4 2 |
| | 8 2 | 8 2 | 3+4 2 | 5+6 2 |
| | 9 2 | 8 2 | 5+6 2 | 3+4 2 |
| | 8 2 | 9 2 | 3+4 2 | 5+6 2 |
| | 9 2 | 9 2 | 3+4 2 | 5+6 2 |
| | 9 2 | 8 2 | 5+6 2 | 5+6 2 |
| | 8 2 | 6+7 2 | 5+6 2 | 5+6 2 |
| | 6+7 2 | 9 2 | 3+4 2 | 3+4 2 |
| | 9 2 | 6+7 2 | 5+6 2 | 3+4 2 |
| | 4+5 2 | 4+5 2 | 5+6 2 | 5+6 2 |
| | 6+7 2 | 8 2 | 3+4 2 | 5+6 2 |
| | 8 2 | 9 2 | 5+6 2 | 3+4 2 |
| | 4+5 2 | 8 2 | 3+4 2 | 5+6 2 |
| | 8 2 | 4+5 2 | 3+4 2 | 5+6 2 |
| | 4+5 2 | 9 2 | 5+6 2 | 5+6 2 |
| | 9 2 | 6+7 2 | 3+4 2 | 3+4 2 |
| | 8 2 | 9 2 | 5+6 2 | 5+6 2 |
| | 9 2 | 6+7 2 | 5+6 2 | 3+4 2 |
| | 6+7 2 | 8 2 | 3+4 2 | 3+4 2 |
| | 9 2 | 4+5 2 | 5+6 2 | 3+4 2 |
| | 6+7 2 | 9 2 | 3+4 2 | 5+6 2 |
| | 9 2 | 9 2 | 3+4 2 | 3+4 2 |

Continued on next column

**TABLE LXXIII– continued**

| TQC | ICSD | NoE | SF | |
|---|---|---|---|---|
| | 8 2 | 9 2 | 5+6 2 | 3+4 2 |
| | 4+5 2 | 6+7 2 | 5+6 2 | 5+6 2 |
| | 9 2 | 8 2 | 5+6 2 | 5+6 2 |
| | 8 2 | 8 2 | 3+4 2 | 5+6 2 |
| | 6+7 2 | 9 2 | 3+4 2 | 3+4 2 |
| | 8 2 | 6+7 2 | 5+6 2 | 3+4 2 |
| | 4+5 2 | 8 2 | 5+6 2 | 5+6 2 |
| | 9 2 | 4+5 2 | 5+6 2 | 5+6 2 |
| | 9 2 | 8 2 | 3+4 2 | 3+4 2 |
| | 9 2 | 8 2 | 3+4 2 | 3+4 2 |
| | 6+7 2 | 4+5 2 | 5+6 2 | 5+6 2 |
| | 8 2 | 9 2 | 5+6 2 | 3+4 2 |
| | 8 2 | 9 2 | 3+4 2 | 5+6 2 |
| | 4+5 2 | 6+7 2 | 3+4 2 | 3+4 2 |
| | 8 2 | 9 2 | 5+6 2 | 5+6 2 |
| | 9 2 | 8 2 | 5+6 2 | 5+6 2 |
| | 6+7 2 | 4+5 2 | 3+4 2 | 5+6 2 |
| 166.1.356 | 174186 | 158 | Ba4Er1O12Ru3 | |
| | -GM | -T | -F | -L |
| | 8 2 | 8 2 | 3+4 2 | 3+4 2 |
| | 8 2 | 9 2 | 5+6 2 | 3+4 2 |
| | 9 2 | 8 2 | 3+4 2 | 5+6 2 |
| | 8 2 | 8 2 | 5+6 2 | 5+6 2 |
| | 9 2 | 9 2 | 3+4 2 | 3+4 2 |
| | 9 2 | 9 2 | 5+6 2 | 5+6 2 |
| | 9 2 | 9 2 | 5+6 2 | 5+6 2 |
| | 6+7 2 | 6+7 2 | 5+6 2 | 5+6 2 |
| | 8 2 | 9 2 | 3+4 2 | 5+6 2 |
| | 9 2 | 8 2 | 3+4 2 | 3+4 2 |
| | 8 2 | 9 2 | 5+6 2 | 5+6 2 |
| | 4+5 2 | 6+7 2 | 3+4 2 | 3+4 2 |
| | 8 2 | 8 2 | 3+4 2 | 5+6 2 |
| | 9 2 | 4+5 2 | 5+6 2 | 3+4 2 |
| | 6+7 2 | 8 2 | 5+6 2 | 3+4 2 |
| | 8 2 | 8 2 | 3+4 2 | 3+4 2 |
| | 4+5 2 | 4+5 2 | 3+4 2 | 3+4 2 |
| | 9 2 | 9 2 | 5+6 2 | 5+6 2 |
| | 9 2 | 9 2 | 5+6 2 | 5+6 2 |
| | 6+7 2 | 6+7 2 | 5+6 2 | 5+6 2 |
| | 8 2 | 9 2 | 5+6 2 | 3+4 2 |
| | 9 2 | 8 2 | 3+4 2 | 5+6 2 |
| | 8 2 | 9 2 | 5+6 2 | 5+6 2 |
| | 9 2 | 8 2 | 3+4 2 | 3+4 2 |
| | 4+5 2 | 6+7 2 | 5+6 2 | 3+4 2 |
| | 8 2 | 4+5 2 | 3+4 2 | 5+6 2 |
| | 6+7 2 | 9 2 | 5+6 2 | 5+6 2 |
| | 4+5 2 | 8 2 | 3+4 2 | 3+4 2 |
| | 9 2 | 6+7 2 | 5+6 2 | 5+6 2 |
| | 8 2 | 4+5 2 | 3+4 2 | 3+4 2 |
| | 6+7 2 | 9 2 | 5+6 2 | 5+6 2 |
| | 9 2 | 8 2 | 3+4 2 | 3+4 2 |
| | 8 2 | 9 2 | 3+4 2 | 5+6 2 |
| | 8 2 | 8 2 | 3+4 2 | 5+6 2 |
| | 4+5 2 | 4+5 2 | 3+4 2 | 3+4 2 |
| | 8 2 | 9 2 | 3+4 2 | 3+4 2 |
| | 9 2 | 6+7 2 | 3+4 2 | 5+6 2 |
| | 4+5 2 | 9 2 | 3+4 2 | 3+4 2 |
| | 6+7 2 | 6+7 2 | 5+6 2 | 5+6 2 |
| | 8 2 | 8 2 | 5+6 2 | 5+6 2 |
| | 9 2 | 9 2 | 5+6 2 | 5+6 2 |
| | 9 2 | 8 2 | 3+4 2 | 3+4 2 |

Continued on next column

**TABLE LXXIII– continued**

| TQC | ICSD | NoE | SF | |
|---|---|---|---|---|
| | 8 2 | 8 2 | 5+6 2 | 5+6 2 |
| | 6+7 2 | 4+5 2 | 5+6 2 | 3+4 2 |
| | 4+5 2 | 9 2 | 3+4 2 | 3+4 2 |
| | 9 2 | 6+7 2 | 3+4 2 | 5+6 2 |
| | 8 2 | 8 2 | 5+6 2 | 5+6 2 |
| | 8 2 | 9 2 | 3+4 2 | 3+4 2 |
| | 8 2 | 8 2 | 3+4 2 | 5+6 2 |
| | 4+5 2 | 4+5 2 | 5+6 2 | 3+4 2 |
| | 9 2 | 6+7 2 | 5+6 2 | 5+6 2 |
| | 9 2 | 9 2 | 3+4 2 | 3+4 2 |
| | 6+7 2 | 8 2 | 5+6 2 | 5+6 2 |
| | 9 2 | 9 2 | 5+6 2 | 3+4 2 |
| | 6+7 2 | 4+5 2 | 3+4 2 | 3+4 2 |
| | 8 2 | 8 2 | 3+4 2 | 3+4 2 |
| | 4+5 2 | 9 2 | 5+6 2 | 3+4 2 |
| | 9 2 | 6+7 2 | 5+6 2 | 5+6 2 |
| | 9 2 | 9 2 | 3+4 2 | 5+6 2 |
| | 6+7 2 | 8 2 | 5+6 2 | 5+6 2 |
| | 8 2 | 8 2 | 3+4 2 | 3+4 2 |
| | 9 2 | 9 2 | 5+6 2 | 3+4 2 |
| | 9 2 | 6+7 2 | 5+6 2 | 5+6 2 |
| | 9 2 | 4+5 2 | 5+6 2 | 3+4 2 |
| | 8 2 | 9 2 | 5+6 2 | 5+6 2 |
| | 6+7 2 | 8 2 | 3+4 2 | 3+4 2 |
| | 4+5 2 | 8 2 | 5+6 2 | 5+6 2 |
| | 8 2 | 4+5 2 | 3+4 2 | 3+4 2 |
| | 8 2 | 9 2 | 3+4 2 | 5+6 2 |
| | 9 2 | 9 2 | 5+6 2 | 5+6 2 |
| | 6+7 2 | 6+7 2 | 5+6 2 | 5+6 2 |
| | 9 2 | 9 2 | 5+6 2 | 5+6 2 |
| | 4+5 2 | 8 2 | 3+4 2 | 3+4 2 |
| | 8 2 | 4+5 2 | 5+6 2 | 5+6 2 |
| | 9 2 | 6+7 2 | 5+6 2 | 5+6 2 |
| | 9 2 | 9 2 | 3+4 2 | 5+6 2 |
| | 6+7 2 | 8 2 | 5+6 2 | 5+6 2 |
| | 4+5 2 | 9 2 | 3+4 2 | 5+6 2 |
| | 9 2 | 6+7 2 | 5+6 2 | 5+6 2 |
| 166.1.357 | 20621 | 42 | Ba3O9Sc4 | |
| | -GM | -T | -F | -L |
| | 8 2 | 9 2 | 3+4 2 | 5+6 2 |
| | 9 2 | 8 2 | 5+6 2 | 3+4 2 |
| | 8 2 | 8 2 | 3+4 2 | 3+4 2 |
| | 9 2 | 9 2 | 3+4 2 | 5+6 2 |
| | 8 2 | 8 2 | 5+6 2 | 5+6 2 |
| | 9 2 | 9 2 | 5+6 2 | 3+4 2 |
| | 8 2 | 9 2 | 5+6 2 | 5+6 2 |
| | 6+7 2 | 6+7 2 | 3+4 2 | 5+6 2 |
| | 9 2 | 4+5 2 | 5+6 2 | 3+4 2 |
| | 4+5 2 | 9 2 | 3+4 2 | 5+6 2 |
| | 6+7 2 | 8 2 | 5+6 2 | 3+4 2 |
| | 9 2 | 6+7 2 | 5+6 2 | 5+6 2 |
| | 8 2 | 9 2 | 3+4 2 | 3+4 2 |
| | 8 2 | 8 2 | 5+6 2 | 5+6 2 |
| | 9 2 | 8 2 | 3+4 2 | 3+4 2 |
| | 8 2 | 4+5 2 | 5+6 2 | 3+4 2 |
| | 4+5 2 | 8 2 | 3+4 2 | 5+6 2 |
| | 9 2 | 4+5 2 | 5+6 2 | 3+4 2 |
| | 8 2 | 8 2 | 3+4 2 | 5+6 2 |
| | 6+7 2 | 9 2 | 3+4 2 | 5+6 2 |
| | 4+5 2 | 6+7 2 | 5+6 2 | 3+4 2 |
| 166.1.358 | 25005 | 144 | Ba2Ni1O6Te1 | |

Continued on next column

**TABLE LXXIII– continued**

| TQC | ICSD | NoE | SF | |
|---|---|---|---|---|
| | -GM | -T | -F | -L |
| | 8 2 | 9 2 | 5+6 2 | 3+4 2 |
| | 9 2 | 8 2 | 3+4 2 | 5+6 2 |
| | 8 2 | 8 2 | 5+6 2 | 5+6 2 |
| | 9 2 | 9 2 | 3+4 2 | 3+4 2 |
| | 8 2 | 9 2 | 3+4 2 | 5+6 2 |
| | 8 2 | 8 2 | 3+4 2 | 3+4 2 |
| | 9 2 | 8 2 | 5+6 2 | 3+4 2 |
| | 9 2 | 8 2 | 5+6 2 | 3+4 2 |
| | 6+7 2 | 4+5 2 | 5+6 2 | 3+4 2 |
| | 9 2 | 9 2 | 5+6 2 | 5+6 2 |
| | 9 2 | 9 2 | 5+6 2 | 5+6 2 |
| | 6+7 2 | 6+7 2 | 5+6 2 | 5+6 2 |
| | 8 2 | 9 2 | 3+4 2 | 5+6 2 |
| | 4+5 2 | 6+7 2 | 3+4 2 | 5+6 2 |
| | 8 2 | 8 2 | 3+4 2 | 3+4 2 |
| | 4+5 2 | 4+5 2 | 3+4 2 | 3+4 2 |
| | 8 2 | 8 2 | 3+4 2 | 3+4 2 |
| | 9 2 | 9 2 | 5+6 2 | 5+6 2 |
| | 8 2 | 9 2 | 5+6 2 | 3+4 2 |
| | 9 2 | 8 2 | 3+4 2 | 5+6 2 |
| | 8 2 | 6+7 2 | 3+4 2 | 3+4 2 |
| | 4+5 2 | 4+5 2 | 5+6 2 | 5+6 2 |
| | 6+7 2 | 8 2 | 3+4 2 | 5+6 2 |
| | 8 2 | 9 2 | 5+6 2 | 3+4 2 |
| | 4+5 2 | 9 2 | 3+4 2 | 3+4 2 |
| | 9 2 | 6+7 2 | 5+6 2 | 5+6 2 |
| | 6+7 2 | 4+5 2 | 5+6 2 | 3+4 2 |
| | 9 2 | 8 2 | 3+4 2 | 5+6 2 |
| | 8 2 | 8 2 | 3+4 2 | 3+4 2 |
| | 8 2 | 9 2 | 3+4 2 | 5+6 2 |
| | 9 2 | 9 2 | 5+6 2 | 5+6 2 |
| | 9 2 | 8 2 | 5+6 2 | 3+4 2 |
| | 9 2 | 9 2 | 5+6 2 | 5+6 2 |
| | 6+7 2 | 6+7 2 | 5+6 2 | 5+6 2 |
| | 9 2 | 8 2 | 5+6 2 | 3+4 2 |
| | 6+7 2 | 4+5 2 | 5+6 2 | 3+4 2 |
| | 8 2 | 9 2 | 3+4 2 | 5+6 2 |
| | 8 2 | 8 2 | 3+4 2 | 3+4 2 |
| | 4+5 2 | 8 2 | 3+4 2 | 5+6 2 |
| | 8 2 | 4+5 2 | 3+4 2 | 3+4 2 |
| | 9 2 | 9 2 | 3+4 2 | 5+6 2 |
| | 8 2 | 8 2 | 5+6 2 | 3+4 2 |
| | 4+5 2 | 6+7 2 | 5+6 2 | 3+4 2 |
| | 9 2 | 4+5 2 | 3+4 2 | 3+4 2 |
| | 6+7 2 | 8 2 | 5+6 2 | 5+6 2 |
| | 8 2 | 4+5 2 | 3+4 2 | 3+4 2 |
| | 4+5 2 | 8 2 | 3+4 2 | 3+4 2 |
| | 8 2 | 9 2 | 5+6 2 | 5+6 2 |
| | 4+5 2 | 6+7 2 | 3+4 2 | 5+6 2 |
| | 9 2 | 9 2 | 5+6 2 | 3+4 2 |
| | 6+7 2 | 6+7 2 | 5+6 2 | 5+6 2 |
| | 9 2 | 9 2 | 3+4 2 | 5+6 2 |
| | 9 2 | 9 2 | 5+6 2 | 3+4 2 |
| | 9 2 | 6+7 2 | 5+6 2 | 3+4 2 |
| | 6+7 2 | 8 2 | 3+4 2 | 5+6 2 |
| | 9 2 | 9 2 | 3+4 2 | 3+4 2 |
| | 9 2 | 8 2 | 5+6 2 | 5+6 2 |
| | 8 2 | 4+5 2 | 3+4 2 | 5+6 2 |
| | 6+7 2 | 9 2 | 5+6 2 | 3+4 2 |
| | 4+5 2 | 6+7 2 | 5+6 2 | 5+6 2 |

Continued on next column

**TABLE LXXIII– continued**

| TQC | ICSD | | NoE | SF |
|---|---|---|---|---|
| | 8 2 | 9 2 | 5+6 2 | 3+4 2 |
| | 8 2 | 8 2 | 3+4 2 | 5+6 2 |
| | 4+5 2 | 4+5 2 | 3+4 2 | 5+6 2 |
| | 8 2 | 8 2 | 3+4 2 | 3+4 2 |
| | 8 2 | 8 2 | 3+4 2 | 3+4 2 |
| | 4+5 2 | 4+5 2 | 5+6 2 | 5+6 2 |
| | 8 2 | 8 2 | 5+6 2 | 5+6 2 |
| | 9 2 | 9 2 | 5+6 2 | 5+6 2 |
| | 6+7 2 | 6+7 2 | 3+4 2 | 3+4 2 |
| | 9 2 | 9 2 | 3+4 2 | 3+4 2 |
| | 9 2 | 9 2 | 3+4 2 | 3+4 2 |
| | 6+7 2 | 6+7 2 | 5+6 2 | 5+6 2 |
| 166.1.359 | 58660 | | 54 | Ba2Mg17 |
| | -GM | -T | -F | -L |
| | 8 2 | 8 2 | 3+4 2 | 3+4 2 |
| | 9 2 | 9 2 | 5+6 2 | 5+6 2 |
| | 8 2 | 9 2 | 3+4 2 | 5+6 2 |
| | 9 2 | 8 2 | 5+6 2 | 3+4 2 |
| | 4+5 2 | 4+5 2 | 5+6 2 | 3+4 2 |
| | 8 2 | 9 2 | 3+4 2 | 5+6 2 |
| | 6+7 2 | 6+7 2 | 3+4 2 | 5+6 2 |
| | 9 2 | 8 2 | 5+6 2 | 3+4 2 |
| | 8 2 | 9 2 | 3+4 2 | 3+4 2 |
| | 9 2 | 8 2 | 5+6 2 | 5+6 2 |
| | 6+7 2 | 9 2 | 5+6 2 | 3+4 2 |
| | 8 2 | 6+7 2 | 3+4 2 | 5+6 2 |
| | 4+5 2 | 8 2 | 5+6 2 | 5+6 2 |
| | 8 2 | 4+5 2 | 3+4 2 | 3+4 2 |
| | 9 2 | 8 2 | 5+6 2 | 3+4 2 |
| | 9 2 | 9 2 | 3+4 2 | 5+6 2 |
| | 6+7 2 | 8 2 | 3+4 2 | 5+6 2 |
| | 8 2 | 4+5 2 | 5+6 2 | 3+4 2 |
| | 4+5 2 | 9 2 | 5+6 2 | 5+6 2 |
| | 9 2 | 6+7 2 | 3+4 2 | 3+4 2 |
| | 4+5 2 | 8 2 | 3+4 2 | 3+4 2 |
| | 8 2 | 4+5 2 | 5+6 2 | 5+6 2 |
| | 8 2 | 8 2 | 5+6 2 | 3+4 2 |
| | 9 2 | 9 2 | 3+4 2 | 5+6 2 |
| | 8 2 | 8 2 | 5+6 2 | 5+6 2 |
| | 9 2 | 4+5 2 | 3+4 2 | 3+4 2 |
| | 6+7 2 | 8 2 | 3+4 2 | 5+6 2 |
| 166.1.360 | 58665 | | 66 | Ba1Pb3 |
| | -GM | -T | -F | -L |
| | 8 2 | 8 2 | 3+4 2 | 3+4 2 |
| | 8 2 | 9 2 | 5+6 2 | 3+4 2 |
| | 9 2 | 8 2 | 3+4 2 | 5+6 2 |
| | 9 2 | 9 2 | 5+6 2 | 5+6 2 |
| | 8 2 | 9 2 | 5+6 2 | 3+4 2 |
| | 9 2 | 8 2 | 3+4 2 | 5+6 2 |
| | 9 2 | 9 2 | 5+6 2 | 5+6 2 |
| | 6+7 2 | 6+7 2 | 5+6 2 | 5+6 2 |
| | 8 2 | 9 2 | 3+4 2 | 3+4 2 |
| | 4+5 2 | 8 2 | 5+6 2 | 5+6 2 |
| | 9 2 | 6+7 2 | 5+6 2 | 3+4 2 |
| | 6+7 2 | 4+5 2 | 3+4 2 | 5+6 2 |
| | 8 2 | 9 2 | 3+4 2 | 5+6 2 |
| | 9 2 | 8 2 | 5+6 2 | 3+4 2 |
| | 8 2 | 8 2 | 5+6 2 | 3+4 2 |
| | 8 2 | 4+5 2 | 3+4 2 | 5+6 2 |
| | 9 2 | 8 2 | 5+6 2 | 3+4 2 |
| | 4+5 2 | 9 2 | 3+4 2 | 5+6 2 |

Continued on next column

**TABLE LXXIII– continued**

| TQC | ICSD | | NoE | SF |
|---|---|---|---|---|
| | 6+7 2 | 4+5 2 | 3+4 2 | 5+6 2 |
| | 8 2 | 8 2 | 5+6 2 | 3+4 2 |
| | 4+5 2 | 6+7 2 | 5+6 2 | 5+6 2 |
| | 8 2 | 9 2 | 3+4 2 | 3+4 2 |
| | 9 2 | 9 2 | 5+6 2 | 5+6 2 |
| | 9 2 | 8 2 | 3+4 2 | 3+4 2 |
| | 6+7 2 | 6+7 2 | 3+4 2 | 5+6 2 |
| | 9 2 | 9 2 | 5+6 2 | 3+4 2 |
| | 6+7 2 | 4+5 2 | 3+4 2 | 3+4 2 |
| | 4+5 2 | 9 2 | 5+6 2 | 3+4 2 |
| | 8 2 | 6+7 2 | 3+4 2 | 3+4 2 |
| | 9 2 | 8 2 | 3+4 2 | 5+6 2 |
| | 8 2 | 9 2 | 5+6 2 | 3+4 2 |
| | 9 2 | 8 2 | 3+4 2 | 5+6 2 |
| | 6+7 2 | 4+5 2 | 5+6 2 | 5+6 2 |
| 166.1.361 | 419973 | | 66 | Ba1Pb3 |
| | -GM | -T | -F | -L |
| | 8 2 | 8 2 | 3+4 2 | 3+4 2 |
| | 8 2 | 9 2 | 5+6 2 | 3+4 2 |
| | 9 2 | 8 2 | 3+4 2 | 5+6 2 |
| | 9 2 | 9 2 | 5+6 2 | 5+6 2 |
| | 8 2 | 9 2 | 5+6 2 | 3+4 2 |
| | 9 2 | 8 2 | 3+4 2 | 5+6 2 |
| | 9 2 | 9 2 | 5+6 2 | 5+6 2 |
| | 6+7 2 | 6+7 2 | 5+6 2 | 5+6 2 |
| | 8 2 | 9 2 | 3+4 2 | 3+4 2 |
| | 4+5 2 | 8 2 | 5+6 2 | 5+6 2 |
| | 6+7 2 | 6+7 2 | 5+6 2 | 3+4 2 |
| | 9 2 | 4+5 2 | 3+4 2 | 5+6 2 |
| | 8 2 | 9 2 | 3+4 2 | 5+6 2 |
| | 9 2 | 8 2 | 5+6 2 | 3+4 2 |
| | 8 2 | 8 2 | 3+4 2 | 3+4 2 |
| | 9 2 | 4+5 2 | 5+6 2 | 5+6 2 |
| | 6+7 2 | 8 2 | 5+6 2 | 3+4 2 |
| | 8 2 | 4+5 2 | 5+6 2 | 5+6 2 |
| | 4+5 2 | 8 2 | 3+4 2 | 3+4 2 |
| | 8 2 | 9 2 | 3+4 2 | 5+6 2 |
| | 4+5 2 | 6+7 2 | 5+6 2 | 5+6 2 |
| | 8 2 | 9 2 | 3+4 2 | 5+6 2 |
| | 9 2 | 9 2 | 3+4 2 | 3+4 2 |
| | 9 2 | 8 2 | 5+6 2 | 3+4 2 |
| | 6+7 2 | 6+7 2 | 3+4 2 | 5+6 2 |
| | 9 2 | 9 2 | 3+4 2 | 3+4 2 |
| | 6+7 2 | 4+5 2 | 5+6 2 | 3+4 2 |
| | 8 2 | 9 2 | 5+6 2 | 5+6 2 |
| | 4+5 2 | 6+7 2 | 3+4 2 | 3+4 2 |
| | 9 2 | 8 2 | 3+4 2 | 3+4 2 |
| | 8 2 | 9 2 | 5+6 2 | 5+6 2 |
| | 9 2 | 4+5 2 | 3+4 2 | 3+4 2 |
| | 6+7 2 | 8 2 | 5+6 2 | 5+6 2 |
| 166.1.362 | 99682 | | 108 | Ba1O3Ru1 |
| | -GM | -T | -F | -L |
| | 8 2 | 8 2 | 5+6 2 | 5+6 2 |
| | 9 2 | 9 2 | 3+4 2 | 3+4 2 |
| | 8 2 | 9 2 | 3+4 2 | 5+6 2 |
| | 8 2 | 8 2 | 3+4 2 | 3+4 2 |
| | 8 2 | 9 2 | 5+6 2 | 5+6 2 |
| | 4+5 2 | 8 2 | 3+4 2 | 3+4 2 |
| | 9 2 | 4+5 2 | 3+4 2 | 3+4 2 |
| | 8 2 | 9 2 | 5+6 2 | 3+4 2 |
| | 8 2 | 6+7 2 | 5+6 2 | 5+6 2 |

Continued on next column

**TABLE LXXIII– continued**

| TQC | ICSD | NoE | SF | |
|---|---|---|---|---|
| | 9 2 | 9 2 | 3+4 2 | 3+4 2 |
| | 4+5 2 | 9 2 | 5+6 2 | 5+6 2 |
| | 6+7 2 | 6+7 2 | 5+6 2 | 5+6 2 |
| | 8 2 | 8 2 | 5+6 2 | 5+6 2 |
| | 9 2 | 8 2 | 3+4 2 | 3+4 2 |
| | 9 2 | 9 2 | 5+6 2 | 3+4 2 |
| | 4+5 2 | 4+5 2 | 5+6 2 | 5+6 2 |
| | 6+7 2 | 8 2 | 3+4 2 | 3+4 2 |
| | 9 2 | 4+5 2 | 5+6 2 | 3+4 2 |
| | 8 2 | 6+7 2 | 3+4 2 | 5+6 2 |
| | 6+7 2 | 8 2 | 5+6 2 | 3+4 2 |
| | 9 2 | 9 2 | 5+6 2 | 3+4 2 |
| | 8 2 | 8 2 | 3+4 2 | 3+4 2 |
| | 9 2 | 8 2 | 3+4 2 | 5+6 2 |
| | 6+7 2 | 4+5 2 | 5+6 2 | 3+4 2 |
| | 9 2 | 8 2 | 3+4 2 | 5+6 2 |
| | 8 2 | 8 2 | 3+4 2 | 3+4 2 |
| | 4+5 2 | 4+5 2 | 3+4 2 | 5+6 2 |
| | 9 2 | 9 2 | 3+4 2 | 3+4 2 |
| | 8 2 | 8 2 | 3+4 2 | 3+4 2 |
| | 9 2 | 4+5 2 | 5+6 2 | 3+4 2 |
| | 4+5 2 | 9 2 | 3+4 2 | 5+6 2 |
| | 8 2 | 6+7 2 | 5+6 2 | 5+6 2 |
| | 6+7 2 | 8 2 | 5+6 2 | 3+4 2 |
| | 9 2 | 9 2 | 3+4 2 | 5+6 2 |
| | 9 2 | 6+7 2 | 5+6 2 | 5+6 2 |
| | 6+7 2 | 8 2 | 3+4 2 | 5+6 2 |
| | 4+5 2 | 4+5 2 | 5+6 2 | 3+4 2 |
| | 8 2 | 8 2 | 3+4 2 | 5+6 2 |
| | 9 2 | 9 2 | 5+6 2 | 3+4 2 |
| | 6+7 2 | 8 2 | 3+4 2 | 5+6 2 |
| | 9 2 | 4+5 2 | 5+6 2 | 3+4 2 |
| | 8 2 | 8 2 | 5+6 2 | 5+6 2 |
| | 9 2 | 9 2 | 3+4 2 | 3+4 2 |
| | 6+7 2 | 6+7 2 | 5+6 2 | 5+6 2 |
| | 9 2 | 9 2 | 5+6 2 | 5+6 2 |
| | 9 2 | 8 2 | 3+4 2 | 3+4 2 |
| | 8 2 | 4+5 2 | 5+6 2 | 5+6 2 |
| | 9 2 | 8 2 | 3+4 2 | 5+6 2 |
| | 6+7 2 | 9 2 | 5+6 2 | 3+4 2 |
| | 4+5 2 | 8 2 | 3+4 2 | 3+4 2 |
| | 8 2 | 6+7 2 | 3+4 2 | 3+4 2 |
| | 4+5 2 | 9 2 | 3+4 2 | 5+6 2 |
| | 8 2 | 4+5 2 | 5+6 2 | 3+4 2 |
| | 9 2 | 8 2 | 5+6 2 | 5+6 2 |
| 166.1.363 | 91075 | | 108 | Ba1O3Ru1 |
| | -GM | -T | -F | -L |
| | 8 2 | 9 2 | 5+6 2 | 3+4 2 |
| | 9 2 | 8 2 | 3+4 2 | 5+6 2 |
| | 8 2 | 8 2 | 3+4 2 | 3+4 2 |
| | 8 2 | 8 2 | 3+4 2 | 5+6 2 |
| | 8 2 | 9 2 | 5+6 2 | 3+4 2 |
| | 4+5 2 | 9 2 | 3+4 2 | 5+6 2 |
| | 9 2 | 6+7 2 | 3+4 2 | 5+6 2 |
| | 8 2 | 8 2 | 5+6 2 | 5+6 2 |
| | 8 2 | 4+5 2 | 5+6 2 | 3+4 2 |
| | 4+5 2 | 8 2 | 3+4 2 | 5+6 2 |
| | 9 2 | 8 2 | 5+6 2 | 3+4 2 |
| | 6+7 2 | 4+5 2 | 5+6 2 | 3+4 2 |
| | 8 2 | 9 2 | 5+6 2 | 3+4 2 |
| | 9 2 | 9 2 | 3+4 2 | 5+6 2 |

Continued on next column

**TABLE LXXIII– continued**

| TQC | ICSD | NoE | SF | |
|---|---|---|---|---|
| | 9 2 | 8 2 | 5+6 2 | 5+6 2 |
| | 4+5 2 | 6+7 2 | 5+6 2 | 3+4 2 |
| | 6+7 2 | 9 2 | 3+4 2 | 5+6 2 |
| | 8 2 | 6+7 2 | 5+6 2 | 5+6 2 |
| | 9 2 | 4+5 2 | 3+4 2 | 3+4 2 |
| | 6+7 2 | 9 2 | 5+6 2 | 5+6 2 |
| | 9 2 | 8 2 | 5+6 2 | 5+6 2 |
| | 8 2 | 9 2 | 3+4 2 | 5+6 2 |
| | 9 2 | 9 2 | 3+4 2 | 3+4 2 |
| | 6+7 2 | 6+7 2 | 5+6 2 | 3+4 2 |
| | 9 2 | 9 2 | 3+4 2 | 5+6 2 |
| | 8 2 | 9 2 | 3+4 2 | 5+6 2 |
| | 4+5 2 | 6+7 2 | 3+4 2 | 3+4 2 |
| | 9 2 | 8 2 | 3+4 2 | 5+6 2 |
| | 8 2 | 9 2 | 3+4 2 | 5+6 2 |
| | 9 2 | 6+7 2 | 5+6 2 | 5+6 2 |
| | 4+5 2 | 8 2 | 3+4 2 | 3+4 2 |
| | 8 2 | 4+5 2 | 5+6 2 | 3+4 2 |
| | 6+7 2 | 9 2 | 3+4 2 | 5+6 2 |
| | 9 2 | 8 2 | 5+6 2 | 3+4 2 |
| | 9 2 | 4+5 2 | 5+6 2 | 3+4 2 |
| | 6+7 2 | 9 2 | 3+4 2 | 3+4 2 |
| | 4+5 2 | 6+7 2 | 5+6 2 | 5+6 2 |
| | 8 2 | 9 2 | 3+4 2 | 3+4 2 |
| | 9 2 | 8 2 | 5+6 2 | 5+6 2 |
| | 6+7 2 | 9 2 | 3+4 2 | 3+4 2 |
| | 9 2 | 6+7 2 | 5+6 2 | 5+6 2 |
| | 8 2 | 9 2 | 5+6 2 | 3+4 2 |
| | 9 2 | 8 2 | 3+4 2 | 5+6 2 |
| | 6+7 2 | 4+5 2 | 5+6 2 | 3+4 2 |
| | 9 2 | 8 2 | 5+6 2 | 3+4 2 |
| | 9 2 | 9 2 | 3+4 2 | 5+6 2 |
| | 8 2 | 6+7 2 | 5+6 2 | 3+4 2 |
| | 9 2 | 9 2 | 3+4 2 | 3+4 2 |
| | 6+7 2 | 8 2 | 5+6 2 | 5+6 2 |
| | 4+5 2 | 9 2 | 3+4 2 | 5+6 2 |
| | 8 2 | 4+5 2 | 3+4 2 | 5+6 2 |
| | 4+5 2 | 8 2 | 3+4 2 | 3+4 2 |
| | 8 2 | 6+7 2 | 5+6 2 | 5+6 2 |
| | 9 2 | 9 2 | 5+6 2 | 3+4 2 |
| 166.1.364 | 51293 | | 108 | Ba1O3Ru1 |
| | -GM | -T | -F | -L |
| | 8 2 | 8 2 | 3+4 2 | 3+4 2 |
| | 8 2 | 9 2 | 5+6 2 | 3+4 2 |
| | 9 2 | 8 2 | 3+4 2 | 5+6 2 |
| | 8 2 | 8 2 | 3+4 2 | 5+6 2 |
| | 8 2 | 9 2 | 5+6 2 | 3+4 2 |
| | 4+5 2 | 8 2 | 3+4 2 | 5+6 2 |
| | 8 2 | 4+5 2 | 5+6 2 | 5+6 2 |
| | 9 2 | 9 2 | 3+4 2 | 5+6 2 |
| | 8 2 | 6+7 2 | 3+4 2 | 5+6 2 |
| | 4+5 2 | 8 2 | 5+6 2 | 3+4 2 |
| | 9 2 | 8 2 | 5+6 2 | 3+4 2 |
| | 6+7 2 | 4+5 2 | 5+6 2 | 3+4 2 |
| | 9 2 | 9 2 | 5+6 2 | 5+6 2 |
| | 8 2 | 9 2 | 5+6 2 | 3+4 2 |
| | 9 2 | 8 2 | 3+4 2 | 5+6 2 |
| | 6+7 2 | 6+7 2 | 5+6 2 | 3+4 2 |
| | 4+5 2 | 9 2 | 5+6 2 | 5+6 2 |
| | 9 2 | 6+7 2 | 3+4 2 | 5+6 2 |
| | 8 2 | 9 2 | 5+6 2 | 5+6 2 |

Continued on next column

**TABLE LXXIII– continued**

| TQC | ICSD | NoE | SF | |
|---|---|---|---|---|
| | 6+7 2 | 4+5 2 | 3+4 2 | 3+4 2 |
| | 9 2 | 8 2 | 5+6 2 | 5+6 2 |
| | 9 2 | 9 2 | 3+4 2 | 5+6 2 |
| | 6+7 2 | 9 2 | 3+4 2 | 3+4 2 |
| | 8 2 | 6+7 2 | 5+6 2 | 3+4 2 |
| | 9 2 | 9 2 | 3+4 2 | 5+6 2 |
| | 8 2 | 9 2 | 3+4 2 | 3+4 2 |
| | 4+5 2 | 6+7 2 | 3+4 2 | 5+6 2 |
| | 9 2 | 8 2 | 3+4 2 | 5+6 2 |
| | 8 2 | 9 2 | 3+4 2 | 5+6 2 |
| | 9 2 | 6+7 2 | 3+4 2 | 5+6 2 |
| | 4+5 2 | 8 2 | 5+6 2 | 3+4 2 |
| | 8 2 | 9 2 | 3+4 2 | 5+6 2 |
| | 6+7 2 | 4+5 2 | 5+6 2 | 3+4 2 |
| | 9 2 | 9 2 | 5+6 2 | 3+4 2 |
| | 9 2 | 6+7 2 | 5+6 2 | 3+4 2 |
| | 6+7 2 | 8 2 | 3+4 2 | 3+4 2 |
| | 4+5 2 | 4+5 2 | 5+6 2 | 5+6 2 |
| | 8 2 | 9 2 | 3+4 2 | 5+6 2 |
| | 9 2 | 8 2 | 5+6 2 | 3+4 2 |
| | 6+7 2 | 9 2 | 5+6 2 | 3+4 2 |
| | 9 2 | 6+7 2 | 3+4 2 | 5+6 2 |
| | 8 2 | 9 2 | 5+6 2 | 3+4 2 |
| | 9 2 | 8 2 | 3+4 2 | 5+6 2 |
| | 6+7 2 | 8 2 | 5+6 2 | 3+4 2 |
| | 9 2 | 4+5 2 | 5+6 2 | 3+4 2 |
| | 9 2 | 9 2 | 3+4 2 | 3+4 2 |
| | 8 2 | 6+7 2 | 5+6 2 | 5+6 2 |
| | 9 2 | 9 2 | 3+4 2 | 3+4 2 |
| | 6+7 2 | 8 2 | 5+6 2 | 5+6 2 |
| | 4+5 2 | 4+5 2 | 3+4 2 | 5+6 2 |
| | 8 2 | 8 2 | 3+4 2 | 5+6 2 |
| | 4+5 2 | 9 2 | 3+4 2 | 3+4 2 |
| | 8 2 | 6+7 2 | 5+6 2 | 3+4 2 |
| | 9 2 | 9 2 | 5+6 2 | 5+6 2 |
| 166.1.365 | 172175 | 108 | Ba1O3Ru1 | |
| | -GM | -T | -F | -L |
| | 8 2 | 8 2 | 3+4 2 | 3+4 2 |
| | 8 2 | 9 2 | 5+6 2 | 3+4 2 |
| | 9 2 | 8 2 | 3+4 2 | 5+6 2 |
| | 8 2 | 8 2 | 5+6 2 | 5+6 2 |
| | 8 2 | 8 2 | 3+4 2 | 3+4 2 |
| | 4+5 2 | 4+5 2 | 5+6 2 | 5+6 2 |
| | 8 2 | 9 2 | 3+4 2 | 5+6 2 |
| | 9 2 | 8 2 | 5+6 2 | 3+4 2 |
| | 8 2 | 9 2 | 3+4 2 | 5+6 2 |
| | 4+5 2 | 6+7 2 | 5+6 2 | 3+4 2 |
| | 9 2 | 8 2 | 3+4 2 | 5+6 2 |
| | 6+7 2 | 4+5 2 | 5+6 2 | 3+4 2 |
| | 9 2 | 9 2 | 5+6 2 | 5+6 2 |
| | 8 2 | 9 2 | 5+6 2 | 3+4 2 |
| | 9 2 | 8 2 | 3+4 2 | 5+6 2 |
| | 6+7 2 | 6+7 2 | 5+6 2 | 5+6 2 |
| | 9 2 | 9 2 | 5+6 2 | 5+6 2 |
| | 4+5 2 | 6+7 2 | 3+4 2 | 3+4 2 |
| | 8 2 | 9 2 | 5+6 2 | 5+6 2 |
| | 9 2 | 4+5 2 | 5+6 2 | 3+4 2 |
| | 6+7 2 | 8 2 | 3+4 2 | 5+6 2 |
| | 9 2 | 9 2 | 3+4 2 | 3+4 2 |
| | 6+7 2 | 6+7 2 | 5+6 2 | 5+6 2 |
| | 9 2 | 9 2 | 3+4 2 | 3+4 2 |
| | 9 2 | 9 2 | 3+4 2 | 3+4 2 |
| | 6+7 2 | 6+7 2 | 3+4 2 | 3+4 2 |
| | 9 2 | 9 2 | 3+4 2 | 5+6 2 |
| | 8 2 | 9 2 | 5+6 2 | 3+4 2 |
| | 9 2 | 6+7 2 | 5+6 2 | 5+6 2 |
| | 6+7 2 | 9 2 | 3+4 2 | 3+4 2 |
| | 9 2 | 8 2 | 3+4 2 | 5+6 2 |
| | 4+5 2 | 9 2 | 3+4 2 | 5+6 2 |
| | 8 2 | 9 2 | 3+4 2 | 5+6 2 |
| | 8 2 | 6+7 2 | 3+4 2 | 5+6 2 |
| | 9 2 | 9 2 | 5+6 2 | 5+6 2 |
| | 8 2 | 8 2 | 3+4 2 | 3+4 2 |
| | 4+5 2 | 4+5 2 | 3+4 2 | 5+6 2 |
| | 4+5 2 | 9 2 | 5+6 2 | 3+4 2 |
| | 8 2 | 6+7 2 | 5+6 2 | 3+4 2 |
| | 9 2 | 8 2 | 5+6 2 | 3+4 2 |
| | 6+7 2 | 4+5 2 | 5+6 2 | 5+6 2 |
| | 8 2 | 8 2 | 5+6 2 | 3+4 2 |
| | 9 2 | 4+5 2 | 3+4 2 | 3+4 2 |
| | 6+7 2 | 8 2 | 5+6 2 | 5+6 2 |
| | 8 2 | 8 2 | 3+4 2 | 5+6 2 |
| | 4+5 2 | 9 2 | 5+6 2 | 3+4 2 |
| | 9 2 | 6+7 2 | 5+6 2 | 5+6 2 |
| | 9 2 | 8 2 | 3+4 2 | 3+4 2 |
| | 8 2 | 8 2 | 5+6 2 | 3+4 2 |
| | 9 2 | 4+5 2 | 3+4 2 | 5+6 2 |
| | 6+7 2 | 9 2 | 5+6 2 | 3+4 2 |
| | 4+5 2 | 9 2 | 5+6 2 | 5+6 2 |
| | 8 2 | 6+7 2 | 3+4 2 | 3+4 2 |
| | 8 2 | 9 2 | 3+4 2 | 5+6 2 |
| 166.1.366 | 110770 | 108 | Ba1O3Ru1 | |
| | -GM | -T | -F | -L |
| | 8 2 | 9 2 | 5+6 2 | 3+4 2 |
| | 9 2 | 8 2 | 3+4 2 | 5+6 2 |
| | 8 2 | 8 2 | 3+4 2 | 3+4 2 |
| | 8 2 | 8 2 | 3+4 2 | 5+6 2 |
| | 8 2 | 9 2 | 5+6 2 | 3+4 2 |
| | 4+5 2 | 9 2 | 3+4 2 | 5+6 2 |
| | 9 2 | 6+7 2 | 3+4 2 | 5+6 2 |
| | 8 2 | 8 2 | 5+6 2 | 5+6 2 |
| | 8 2 | 4+5 2 | 5+6 2 | 3+4 2 |
| | 4+5 2 | 8 2 | 3+4 2 | 5+6 2 |
| | 9 2 | 8 2 | 5+6 2 | 3+4 2 |
| | 6+7 2 | 4+5 2 | 5+6 2 | 3+4 2 |
| | 8 2 | 9 2 | 5+6 2 | 3+4 2 |
| | 9 2 | 9 2 | 3+4 2 | 5+6 2 |
| | 9 2 | 8 2 | 5+6 2 | 5+6 2 |
| | 4+5 2 | 6+7 2 | 5+6 2 | 3+4 2 |
| | 6+7 2 | 9 2 | 3+4 2 | 5+6 2 |
| | 8 2 | 6+7 2 | 5+6 2 | 5+6 2 |
| | 9 2 | 4+5 2 | 3+4 2 | 3+4 2 |
| | 6+7 2 | 9 2 | 5+6 2 | 5+6 2 |
| | 9 2 | 8 2 | 5+6 2 | 5+6 2 |
| | 8 2 | 9 2 | 3+4 2 | 5+6 2 |
| | 9 2 | 9 2 | 3+4 2 | 3+4 2 |
| | 6+7 2 | 6+7 2 | 5+6 2 | 3+4 2 |
| | 9 2 | 9 2 | 3+4 2 | 5+6 2 |
| | 8 2 | 9 2 | 3+4 2 | 5+6 2 |
| | 4+5 2 | 6+7 2 | 3+4 2 | 3+4 2 |
| | 9 2 | 8 2 | 3+4 2 | 5+6 2 |
| | 8 2 | 9 2 | 3+4 2 | 5+6 2 |

Continued on next column

**TABLE LXXIII– continued**

| TQC | ICSD | NoE | SF | |
|---|---|---|---|---|
| | 9 2 | 6+7 2 | 5+6 2 | 5+6 2 |
| | 4+5 2 | 8 2 | 3+4 2 | 3+4 2 |
| | 8 2 | 4+5 2 | 5+6 2 | 3+4 2 |
| | 6+7 2 | 9 2 | 3+4 2 | 5+6 2 |
| | 9 2 | 8 2 | 5+6 2 | 3+4 2 |
| | 9 2 | 4+5 2 | 5+6 2 | 3+4 2 |
| | 6+7 2 | 9 2 | 3+4 2 | 3+4 2 |
| | 4+5 2 | 6+7 2 | 5+6 2 | 5+6 2 |
| | 8 2 | 9 2 | 3+4 2 | 3+4 2 |
| | 9 2 | 8 2 | 5+6 2 | 5+6 2 |
| | 6+7 2 | 9 2 | 3+4 2 | 3+4 2 |
| | 9 2 | 6+7 2 | 5+6 2 | 5+6 2 |
| | 8 2 | 9 2 | 5+6 2 | 3+4 2 |
| | 9 2 | 8 2 | 3+4 2 | 5+6 2 |
| | 6+7 2 | 4+5 2 | 5+6 2 | 3+4 2 |
| | 9 2 | 8 2 | 5+6 2 | 3+4 2 |
| | 9 2 | 9 2 | 3+4 2 | 5+6 2 |
| | 8 2 | 6+7 2 | 5+6 2 | 3+4 2 |
| | 9 2 | 9 2 | 3+4 2 | 3+4 2 |
| | 6+7 2 | 8 2 | 5+6 2 | 5+6 2 |
| | 4+5 2 | 9 2 | 3+4 2 | 5+6 2 |
| | 8 2 | 4+5 2 | 3+4 2 | 5+6 2 |
| | 4+5 2 | 8 2 | 3+4 2 | 3+4 2 |
| | 8 2 | 6+7 2 | 5+6 2 | 5+6 2 |
| | 9 2 | 9 2 | 5+6 2 | 3+4 2 |
| 166.1.367 | 10253 | 108 | Ba1O3Ru1 | |
| | -GM | -T | -F | -L |
| | 8 2 | 9 2 | 5+6 2 | 3+4 2 |
| | 9 2 | 8 2 | 3+4 2 | 5+6 2 |
| | 8 2 | 8 2 | 3+4 2 | 3+4 2 |
| | 8 2 | 8 2 | 3+4 2 | 5+6 2 |
| | 8 2 | 9 2 | 5+6 2 | 3+4 2 |
| | 4+5 2 | 8 2 | 3+4 2 | 5+6 2 |
| | 8 2 | 4+5 2 | 5+6 2 | 5+6 2 |
| | 9 2 | 9 2 | 3+4 2 | 5+6 2 |
| | 8 2 | 6+7 2 | 3+4 2 | 3+4 2 |
| | 4+5 2 | 8 2 | 5+6 2 | 5+6 2 |
| | 9 2 | 8 2 | 5+6 2 | 3+4 2 |
| | 6+7 2 | 4+5 2 | 5+6 2 | 3+4 2 |
| | 9 2 | 9 2 | 5+6 2 | 3+4 2 |
| | 8 2 | 9 2 | 3+4 2 | 5+6 2 |
| | 9 2 | 8 2 | 5+6 2 | 5+6 2 |
| | 6+7 2 | 6+7 2 | 5+6 2 | 3+4 2 |
| | 4+5 2 | 9 2 | 3+4 2 | 5+6 2 |
| | 9 2 | 6+7 2 | 5+6 2 | 5+6 2 |
| | 8 2 | 4+5 2 | 3+4 2 | 5+6 2 |
| | 6+7 2 | 9 2 | 5+6 2 | 3+4 2 |
| | 9 2 | 8 2 | 5+6 2 | 5+6 2 |
| | 9 2 | 9 2 | 3+4 2 | 3+4 2 |
| | 6+7 2 | 9 2 | 3+4 2 | 5+6 2 |
| | 8 2 | 6+7 2 | 5+6 2 | 3+4 2 |
| | 9 2 | 9 2 | 3+4 2 | 5+6 2 |
| | 8 2 | 9 2 | 3+4 2 | 5+6 2 |
| | 4+5 2 | 6+7 2 | 3+4 2 | 3+4 2 |
| | 9 2 | 8 2 | 3+4 2 | 5+6 2 |
| | 8 2 | 9 2 | 3+4 2 | 5+6 2 |
| | 9 2 | 6+7 2 | 3+4 2 | 5+6 2 |
| | 4+5 2 | 8 2 | 5+6 2 | 3+4 2 |
| | 6+7 2 | 4+5 2 | 5+6 2 | 3+4 2 |
| | 8 2 | 9 2 | 3+4 2 | 5+6 2 |
| | 9 2 | 8 2 | 5+6 2 | 3+4 2 |

Continued on next column

**TABLE LXXIII– continued**

| TQC | ICSD | NoE | SF | |
|---|---|---|---|---|
| | 9 2 | 9 2 | 5+6 2 | 3+4 2 |
| | 6+7 2 | 4+5 2 | 3+4 2 | 3+4 2 |
| | 4+5 2 | 6+7 2 | 5+6 2 | 5+6 2 |
| | 8 2 | 9 2 | 3+4 2 | 3+4 2 |
| | 9 2 | 8 2 | 5+6 2 | 5+6 2 |
| | 6+7 2 | 9 2 | 3+4 2 | 3+4 2 |
| | 9 2 | 6+7 2 | 5+6 2 | 5+6 2 |
| | 8 2 | 9 2 | 5+6 2 | 3+4 2 |
| | 9 2 | 8 2 | 3+4 2 | 5+6 2 |
| | 6+7 2 | 4+5 2 | 5+6 2 | 3+4 2 |
| | 9 2 | 8 2 | 5+6 2 | 3+4 2 |
| | 9 2 | 9 2 | 3+4 2 | 3+4 2 |
| | 8 2 | 6+7 2 | 5+6 2 | 5+6 2 |
| | 9 2 | 9 2 | 3+4 2 | 3+4 2 |
| | 6+7 2 | 8 2 | 5+6 2 | 5+6 2 |
| | 4+5 2 | 4+5 2 | 3+4 2 | 5+6 2 |
| | 8 2 | 8 2 | 3+4 2 | 5+6 2 |
| | 4+5 2 | 9 2 | 3+4 2 | 3+4 2 |
| | 8 2 | 6+7 2 | 5+6 2 | 5+6 2 |
| | 9 2 | 9 2 | 5+6 2 | 3+4 2 |
| 166.1.368 | 20898 | 98 | Ba1Ni4O8 | |
| | -GM | -T | -F | -L |
| | 8 2 | 9 2 | 3+4 2 | 5+6 2 |
| | 8 2 | 8 2 | 5+6 2 | 3+4 2 |
| | 9 2 | 6+7 2 | 3+4 2 | 5+6 2 |
| | 9 2 | 9 2 | 5+6 2 | 5+6 2 |
| | 6+7 2 | 8 2 | 3+4 2 | 3+4 2 |
| | 4+5 2 | 4+5 2 | 5+6 2 | 3+4 2 |
| | 8 2 | 9 2 | 3+4 2 | 5+6 2 |
| | 8 2 | 8 2 | 5+6 2 | 3+4 2 |
| | 9 2 | 9 2 | 3+4 2 | 5+6 2 |
| | 9 2 | 8 2 | 5+6 2 | 3+4 2 |
| | 6+7 2 | 8 2 | 5+6 2 | 3+4 2 |
| | 9 2 | 4+5 2 | 5+6 2 | 3+4 2 |
| | 8 2 | 4+5 2 | 3+4 2 | 5+6 2 |
| | 4+5 2 | 8 2 | 3+4 2 | 5+6 2 |
| | 8 2 | 8 2 | 5+6 2 | 3+4 2 |
| | 8 2 | 8 2 | 5+6 2 | 5+6 2 |
| | 4+5 2 | 9 2 | 3+4 2 | 3+4 2 |
| | 8 2 | 6+7 2 | 5+6 2 | 3+4 2 |
| | 9 2 | 4+5 2 | 3+4 2 | 5+6 2 |
| | 6+7 2 | 8 2 | 5+6 2 | 3+4 2 |
| | 9 2 | 9 2 | 3+4 2 | 3+4 2 |
| | 4+5 2 | 8 2 | 5+6 2 | 5+6 2 |
| | 8 2 | 4+5 2 | 5+6 2 | 5+6 2 |
| | 8 2 | 8 2 | 3+4 2 | 3+4 2 |
| | 8 2 | 8 2 | 3+4 2 | 5+6 2 |
| | 4+5 2 | 4+5 2 | 3+4 2 | 5+6 2 |
| | 8 2 | 8 2 | 5+6 2 | 3+4 2 |
| | 9 2 | 9 2 | 5+6 2 | 3+4 2 |
| | 9 2 | 6+7 2 | 3+4 2 | 5+6 2 |
| | 6+7 2 | 9 2 | 5+6 2 | 3+4 2 |
| | 9 2 | 6+7 2 | 5+6 2 | 5+6 2 |
| | 6+7 2 | 9 2 | 3+4 2 | 5+6 2 |
| | 9 2 | 9 2 | 3+4 2 | 3+4 2 |
| | 8 2 | 9 2 | 5+6 2 | 3+4 2 |
| | 4+5 2 | 8 2 | 3+4 2 | 5+6 2 |
| | 4+5 2 | 4+5 2 | 5+6 2 | 3+4 2 |
| | 8 2 | 4+5 2 | 5+6 2 | 5+6 2 |
| | 8 2 | 8 2 | 3+4 2 | 5+6 2 |
| | 9 2 | 8 2 | 5+6 2 | 3+4 2 |

Continued on next column

**TABLE LXXIII– continued**

| TQC | ICSD | | NoE | SF |
|---|---|---|---|---|
| | 8 2 | 8 2 | 3+4 2 | 5+6 2 |
| | 8 2 | 8 2 | 5+6 2 | 3+4 2 |
| | 4+5 2 | 4+5 2 | 3+4 2 | 3+4 2 |
| | 9 2 | 9 2 | 5+6 2 | 5+6 2 |
| | 6+7 2 | 6+7 2 | 3+4 2 | 3+4 2 |
| | 4+5 2 | 4+5 2 | 3+4 2 | 5+6 2 |
| | 8 2 | 8 2 | 5+6 2 | 3+4 2 |
| | 8 2 | 8 2 | 3+4 2 | 5+6 2 |
| | 8 2 | 8 2 | 5+6 2 | 3+4 2 |
| | 8 2 | 8 2 | 3+4 2 | 5+6 2 |
| 166.1.369 | 657452 | | 124 | B6Fe2Gd5 |
| | -GM | -T | -F | -L |
| | 8 2 | 8 2 | 5+6 2 | 5+6 2 |
| | 9 2 | 9 2 | 3+4 2 | 3+4 2 |
| | 8 2 | 8 2 | 3+4 2 | 3+4 2 |
| | 9 2 | 9 2 | 3+4 2 | 5+6 2 |
| | 8 2 | 8 2 | 5+6 2 | 3+4 2 |
| | 9 2 | 9 2 | 3+4 2 | 3+4 2 |
| | 8 2 | 8 2 | 5+6 2 | 5+6 2 |
| | 9 2 | 9 2 | 5+6 2 | 5+6 2 |
| | 8 2 | 8 2 | 3+4 2 | 5+6 2 |
| | 9 2 | 9 2 | 5+6 2 | 3+4 2 |
| | 4+5 2 | 4+5 2 | 5+6 2 | 5+6 2 |
| | 6+7 2 | 6+7 2 | 3+4 2 | 3+4 2 |
| | 9 2 | 9 2 | 5+6 2 | 5+6 2 |
| | 8 2 | 9 2 | 3+4 2 | 3+4 2 |
| | 9 2 | 8 2 | 5+6 2 | 5+6 2 |
| | 8 2 | 6+7 2 | 3+4 2 | 5+6 2 |
| | 6+7 2 | 8 2 | 5+6 2 | 3+4 2 |
| | 9 2 | 9 2 | 5+6 2 | 5+6 2 |
| | 4+5 2 | 4+5 2 | 3+4 2 | 3+4 2 |
| | 6+7 2 | 6+7 2 | 5+6 2 | 5+6 2 |
| | 8 2 | 9 2 | 5+6 2 | 3+4 2 |
| | 8 2 | 9 2 | 3+4 2 | 5+6 2 |
| | 4+5 2 | 6+7 2 | 5+6 2 | 3+4 2 |
| | 9 2 | 8 2 | 3+4 2 | 5+6 2 |
| | 6+7 2 | 4+5 2 | 5+6 2 | 3+4 2 |
| | 9 2 | 8 2 | 3+4 2 | 5+6 2 |
| | 8 2 | 8 2 | 3+4 2 | 5+6 2 |
| | 9 2 | 9 2 | 5+6 2 | 3+4 2 |
| | 8 2 | 6+7 2 | 3+4 2 | 5+6 2 |
| | 4+5 2 | 9 2 | 5+6 2 | 3+4 2 |
| | 8 2 | 8 2 | 5+6 2 | 3+4 2 |
| | 9 2 | 4+5 2 | 3+4 2 | 3+4 2 |
| | 6+7 2 | 9 2 | 5+6 2 | 5+6 2 |
| | 8 2 | 8 2 | 3+4 2 | 5+6 2 |
| | 8 2 | 8 2 | 3+4 2 | 3+4 2 |
| | 4+5 2 | 4+5 2 | 3+4 2 | 5+6 2 |
| | 9 2 | 9 2 | 5+6 2 | 3+4 2 |
| | 4+5 2 | 6+7 2 | 3+4 2 | 3+4 2 |
| | 8 2 | 9 2 | 3+4 2 | 5+6 2 |
| | 6+7 2 | 6+7 2 | 5+6 2 | 3+4 2 |
| | 8 2 | 4+5 2 | 5+6 2 | 5+6 2 |
| | 6+7 2 | 8 2 | 5+6 2 | 3+4 2 |
| | 9 2 | 8 2 | 3+4 2 | 5+6 2 |
| | 9 2 | 8 2 | 5+6 2 | 3+4 2 |
| | 9 2 | 9 2 | 3+4 2 | 5+6 2 |
| | 8 2 | 8 2 | 5+6 2 | 5+6 2 |
| | 9 2 | 4+5 2 | 3+4 2 | 3+4 2 |
| | 4+5 2 | 9 2 | 5+6 2 | 5+6 2 |
| | 8 2 | 9 2 | 5+6 2 | 5+6 2 |

Continued on next column

| TQC | ICSD | | NoE | SF |
|---|---|---|---|---|
| | 6+7 2 | 6+7 2 | 5+6 2 | 3+4 2 |
| | 9 2 | 8 2 | 5+6 2 | 5+6 2 |
| | 8 2 | 9 2 | 3+4 2 | 5+6 2 |
| | 9 2 | 9 2 | 3+4 2 | 3+4 2 |
| | 8 2 | 8 2 | 3+4 2 | 3+4 2 |
| | 6+7 2 | 6+7 2 | 5+6 2 | 5+6 2 |
| | 9 2 | 9 2 | 3+4 2 | 5+6 2 |
| | 4+5 2 | 4+5 2 | 5+6 2 | 3+4 2 |
| | 6+7 2 | 6+7 2 | 5+6 2 | 5+6 2 |
| | 9 2 | 9 2 | 3+4 2 | 3+4 2 |
| | 4+5 2 | 4+5 2 | 5+6 2 | 5+6 2 |
| | 9 2 | 8 2 | 5+6 2 | 3+4 2 |
| | 8 2 | 9 2 | 5+6 2 | 5+6 2 |
| 166.1.370 | 657450 | | 94 | B6Ce5Fe2 |
| | -GM | -T | -F | -L |
| | 8 2 | 8 2 | 5+6 2 | 5+6 2 |
| | 9 2 | 9 2 | 3+4 2 | 3+4 2 |
| | 8 2 | 8 2 | 3+4 2 | 3+4 2 |
| | 9 2 | 9 2 | 5+6 2 | 5+6 2 |
| | 8 2 | 8 2 | 3+4 2 | 3+4 2 |
| | 9 2 | 8 2 | 5+6 2 | 5+6 2 |
| | 8 2 | 9 2 | 3+4 2 | 3+4 2 |
| | 9 2 | 9 2 | 5+6 2 | 5+6 2 |
| | 8 2 | 8 2 | 3+4 2 | 3+4 2 |
| | 9 2 | 9 2 | 5+6 2 | 5+6 2 |
| | 6+7 2 | 4+5 2 | 3+4 2 | 5+6 2 |
| | 4+5 2 | 6+7 2 | 5+6 2 | 3+4 2 |
| | 9 2 | 9 2 | 3+4 2 | 5+6 2 |
| | 8 2 | 9 2 | 5+6 2 | 3+4 2 |
| | 9 2 | 6+7 2 | 5+6 2 | 5+6 2 |
| | 6+7 2 | 8 2 | 5+6 2 | 5+6 2 |
| | 8 2 | 8 2 | 3+4 2 | 3+4 2 |
| | 9 2 | 9 2 | 5+6 2 | 5+6 2 |
| | 4+5 2 | 4+5 2 | 3+4 2 | 3+4 2 |
| | 6+7 2 | 6+7 2 | 5+6 2 | 5+6 2 |
| | 8 2 | 9 2 | 5+6 2 | 3+4 2 |
| | 8 2 | 9 2 | 3+4 2 | 5+6 2 |
| | 4+5 2 | 6+7 2 | 5+6 2 | 3+4 2 |
| | 9 2 | 8 2 | 3+4 2 | 5+6 2 |
| | 6+7 2 | 4+5 2 | 5+6 2 | 3+4 2 |
| | 9 2 | 8 2 | 3+4 2 | 5+6 2 |
| | 8 2 | 9 2 | 3+4 2 | 5+6 2 |
| | 9 2 | 8 2 | 5+6 2 | 3+4 2 |
| | 4+5 2 | 8 2 | 3+4 2 | 5+6 2 |
| | 8 2 | 4+5 2 | 5+6 2 | 3+4 2 |
| | 9 2 | 6+7 2 | 5+6 2 | 3+4 2 |
| | 6+7 2 | 9 2 | 3+4 2 | 5+6 2 |
| | 8 2 | 9 2 | 5+6 2 | 3+4 2 |
| | 8 2 | 8 2 | 3+4 2 | 5+6 2 |
| | 4+5 2 | 4+5 2 | 3+4 2 | 3+4 2 |
| | 8 2 | 8 2 | 3+4 2 | 5+6 2 |
| | 4+5 2 | 9 2 | 3+4 2 | 3+4 2 |
| | 8 2 | 6+7 2 | 5+6 2 | 5+6 2 |
| | 9 2 | 9 2 | 5+6 2 | 3+4 2 |
| | 8 2 | 6+7 2 | 3+4 2 | 3+4 2 |
| | 6+7 2 | 8 2 | 5+6 2 | 5+6 2 |
| | 6+7 2 | 4+5 2 | 5+6 2 | 3+4 2 |
| | 9 2 | 8 2 | 3+4 2 | 3+4 2 |
| | 9 2 | 8 2 | 3+4 2 | 5+6 2 |
| | 8 2 | 4+5 2 | 5+6 2 | 5+6 2 |
| | 4+5 2 | 9 2 | 5+6 2 | 3+4 2 |

Continued on next column

**TABLE LXXIII– continued**

| TQC | ICSD | NoE | SF | |
|---|---|---|---|---|
| | 9 2 | 8 2 | 5+6 2 | 5+6 2 |
| 166.1.371 | 39554 | 24 | B4Mo2 | |
| | -GM | -T | -F | -L |
| | 8 2 | 8 2 | 5+6 2 | 5+6 2 |
| | 8 2 | 9 2 | 3+4 2 | 3+4 2 |
| | 9 2 | 9 2 | 5+6 2 | 3+4 2 |
| | 8 2 | 8 2 | 3+4 2 | 5+6 2 |
| | 4+5 2 | 4+5 2 | 5+6 2 | 5+6 2 |
| | 8 2 | 8 2 | 5+6 2 | 5+6 2 |
| | 9 2 | 8 2 | 3+4 2 | 3+4 2 |
| | 6+7 2 | 6+7 2 | 3+4 2 | 3+4 2 |
| | 9 2 | 9 2 | 5+6 2 | 3+4 2 |
| | 9 2 | 6+7 2 | 3+4 2 | 5+6 2 |
| | 8 2 | 9 2 | 5+6 2 | 3+4 2 |
| | 6+7 2 | 9 2 | 3+4 2 | 5+6 2 |
| 166.1.372 | 191908 | 24 | B4Mo2 | |
| | -GM | -T | -F | -L |
| | 8 2 | 8 2 | 5+6 2 | 5+6 2 |
| | 8 2 | 9 2 | 3+4 2 | 3+4 2 |
| | 9 2 | 9 2 | 5+6 2 | 3+4 2 |
| | 8 2 | 8 2 | 3+4 2 | 5+6 2 |
| | 4+5 2 | 4+5 2 | 5+6 2 | 5+6 2 |
| | 8 2 | 8 2 | 5+6 2 | 5+6 2 |
| | 9 2 | 8 2 | 3+4 2 | 3+4 2 |
| | 6+7 2 | 6+7 2 | 3+4 2 | 3+4 2 |
| | 9 2 | 9 2 | 3+4 2 | 3+4 2 |
| | 9 2 | 6+7 2 | 5+6 2 | 5+6 2 |
| | 8 2 | 9 2 | 5+6 2 | 3+4 2 |
| | 4+5 2 | 9 2 | 3+4 2 | 5+6 2 |
| 166.1.373 | 601082 | 136 | B3Co1Dy2.5 | |
| | -GM | -T | -F | -L |
| | 8 2 | 8 2 | 3+4 2 | 3+4 2 |
| | 8 2 | 8 2 | 5+6 2 | 5+6 2 |
| | 9 2 | 9 2 | 3+4 2 | 3+4 2 |
| | 8 2 | 9 2 | 3+4 2 | 3+4 2 |
| | 9 2 | 8 2 | 5+6 2 | 5+6 2 |
| | 9 2 | 9 2 | 5+6 2 | 5+6 2 |
| | 9 2 | 9 2 | 3+4 2 | 3+4 2 |
| | 8 2 | 8 2 | 5+6 2 | 5+6 2 |
| | 8 2 | 8 2 | 5+6 2 | 5+6 2 |
| | 9 2 | 9 2 | 3+4 2 | 3+4 2 |
| | 6+7 2 | 4+5 2 | 5+6 2 | 5+6 2 |
| | 4+5 2 | 6+7 2 | 3+4 2 | 3+4 2 |
| | 6+7 2 | 9 2 | 5+6 2 | 5+6 2 |
| | 9 2 | 6+7 2 | 3+4 2 | 3+4 2 |
| | 8 2 | 9 2 | 5+6 2 | 5+6 2 |
| | 9 2 | 8 2 | 5+6 2 | 5+6 2 |
| | 8 2 | 9 2 | 3+4 2 | 3+4 2 |
| | 9 2 | 8 2 | 5+6 2 | 5+6 2 |
| | 4+5 2 | 4+5 2 | 3+4 2 | 5+6 2 |
| | 6+7 2 | 6+7 2 | 5+6 2 | 3+4 2 |
| | 8 2 | 9 2 | 5+6 2 | 3+4 2 |
| | 8 2 | 9 2 | 3+4 2 | 5+6 2 |
| | 4+5 2 | 6+7 2 | 5+6 2 | 3+4 2 |
| | 9 2 | 8 2 | 3+4 2 | 5+6 2 |
| | 6+7 2 | 4+5 2 | 5+6 2 | 3+4 2 |
| | 9 2 | 8 2 | 3+4 2 | 5+6 2 |
| | 8 2 | 8 2 | 5+6 2 | 3+4 2 |
| | 9 2 | 9 2 | 3+4 2 | 5+6 2 |
| | 4+5 2 | 6+7 2 | 3+4 2 | 5+6 2 |
| | 8 2 | 9 2 | 5+6 2 | 3+4 2 |

Continued on next column

**TABLE LXXIII– continued**

| TQC | ICSD | NoE | SF | |
|---|---|---|---|---|
| | 8 2 | 8 2 | 5+6 2 | 3+4 2 |
| | 6+7 2 | 4+5 2 | 5+6 2 | 3+4 2 |
| | 9 2 | 8 2 | 3+4 2 | 5+6 2 |
| | 9 2 | 8 2 | 3+4 2 | 5+6 2 |
| | 8 2 | 4+5 2 | 3+4 2 | 5+6 2 |
| | 6+7 2 | 9 2 | 3+4 2 | 3+4 2 |
| | 8 2 | 9 2 | 5+6 2 | 3+4 2 |
| | 4+5 2 | 6+7 2 | 5+6 2 | 5+6 2 |
| | 4+5 2 | 6+7 2 | 3+4 2 | 3+4 2 |
| | 8 2 | 9 2 | 3+4 2 | 5+6 2 |
| | 9 2 | 4+5 2 | 5+6 2 | 3+4 2 |
| | 6+7 2 | 9 2 | 5+6 2 | 3+4 2 |
| | 9 2 | 8 2 | 3+4 2 | 5+6 2 |
| | 8 2 | 8 2 | 5+6 2 | 3+4 2 |
| | 9 2 | 8 2 | 3+4 2 | 5+6 2 |
| | 9 2 | 9 2 | 5+6 2 | 5+6 2 |
| | 8 2 | 8 2 | 5+6 2 | 3+4 2 |
| | 4+5 2 | 4+5 2 | 3+4 2 | 5+6 2 |
| | 6+7 2 | 9 2 | 5+6 2 | 5+6 2 |
| | 9 2 | 6+7 2 | 5+6 2 | 5+6 2 |
| | 9 2 | 9 2 | 3+4 2 | 3+4 2 |
| | 8 2 | 9 2 | 5+6 2 | 5+6 2 |
| | 6+7 2 | 6+7 2 | 3+4 2 | 3+4 2 |
| | 8 2 | 9 2 | 5+6 2 | 5+6 2 |
| | 9 2 | 8 2 | 3+4 2 | 3+4 2 |
| | 8 2 | 9 2 | 3+4 2 | 5+6 2 |
| | 9 2 | 8 2 | 5+6 2 | 3+4 2 |
| | 4+5 2 | 4+5 2 | 5+6 2 | 5+6 2 |
| | 6+7 2 | 6+7 2 | 3+4 2 | 3+4 2 |
| | 4+5 2 | 4+5 2 | 5+6 2 | 5+6 2 |
| | 9 2 | 8 2 | 5+6 2 | 5+6 2 |
| | 8 2 | 8 2 | 3+4 2 | 3+4 2 |
| | 9 2 | 9 2 | 5+6 2 | 5+6 2 |
| | 6+7 2 | 9 2 | 5+6 2 | 5+6 2 |
| | 9 2 | 6+7 2 | 3+4 2 | 3+4 2 |
| | 8 2 | 8 2 | 5+6 2 | 5+6 2 |
| | 8 2 | 9 2 | 3+4 2 | 5+6 2 |
| | 9 2 | 9 2 | 5+6 2 | 3+4 2 |
| 166.1.374 | 44451 | 24 | B2Mo1 | |
| | -GM | -T | -F | -L |
| | 8 2 | 8 2 | 5+6 2 | 5+6 2 |
| | 8 2 | 9 2 | 3+4 2 | 3+4 2 |
| | 9 2 | 9 2 | 5+6 2 | 3+4 2 |
| | 8 2 | 8 2 | 3+4 2 | 5+6 2 |
| | 4+5 2 | 4+5 2 | 5+6 2 | 5+6 2 |
| | 8 2 | 8 2 | 5+6 2 | 5+6 2 |
| | 9 2 | 8 2 | 3+4 2 | 3+4 2 |
| | 6+7 2 | 6+7 2 | 3+4 2 | 3+4 2 |
| | 9 2 | 9 2 | 5+6 2 | 3+4 2 |
| | 9 2 | 6+7 2 | 3+4 2 | 5+6 2 |
| | 8 2 | 9 2 | 5+6 2 | 3+4 2 |
| | 6+7 2 | 9 2 | 3+4 2 | 5+6 2 |
| 166.1.375 | 418398 | 24 | B2Mo1 | |
| | -GM | -T | -F | -L |
| | 8 2 | 8 2 | 5+6 2 | 5+6 2 |
| | 8 2 | 9 2 | 3+4 2 | 3+4 2 |
| | 9 2 | 9 2 | 5+6 2 | 3+4 2 |
| | 8 2 | 8 2 | 3+4 2 | 5+6 2 |
| | 4+5 2 | 4+5 2 | 5+6 2 | 5+6 2 |
| | 8 2 | 8 2 | 5+6 2 | 5+6 2 |
| | 9 2 | 8 2 | 3+4 2 | 3+4 2 |

Continued on next column

**TABLE LXXIII– continued**

| TQC | ICSD | NoE | SF | |
|---|---|---|---|---|
| | 6+7 2 | 6+7 2 | 3+4 2 | 3+4 2 |
| | 9 2 | 9 2 | 3+4 2 | 3+4 2 |
| | 9 2 | 6+7 2 | 5+6 2 | 5+6 2 |
| | 6+7 2 | 9 2 | 5+6 2 | 3+4 2 |
| | 8 2 | 9 2 | 3+4 2 | 5+6 2 |
| 166.1.376 | 418397 | 24 | B2Mo1 | |
| | -GM | -T | -F | -L |
| | 8 2 | 8 2 | 5+6 2 | 5+6 2 |
| | 8 2 | 9 2 | 3+4 2 | 3+4 2 |
| | 9 2 | 9 2 | 5+6 2 | 3+4 2 |
| | 8 2 | 8 2 | 3+4 2 | 5+6 2 |
| | 4+5 2 | 4+5 2 | 5+6 2 | 5+6 2 |
| | 8 2 | 8 2 | 5+6 2 | 5+6 2 |
| | 9 2 | 8 2 | 3+4 2 | 3+4 2 |
| | 6+7 2 | 6+7 2 | 3+4 2 | 3+4 2 |
| | 9 2 | 9 2 | 3+4 2 | 3+4 2 |
| | 9 2 | 6+7 2 | 5+6 2 | 5+6 2 |
| | 6+7 2 | 9 2 | 5+6 2 | 3+4 2 |
| | 8 2 | 9 2 | 3+4 2 | 5+6 2 |
| 166.1.377 | 40907 | 24 | B2Mo1 | |
| | -GM | -T | -F | -L |
| | 8 2 | 8 2 | 5+6 2 | 5+6 2 |
| | 8 2 | 9 2 | 3+4 2 | 3+4 2 |
| | 9 2 | 9 2 | 5+6 2 | 3+4 2 |
| | 8 2 | 8 2 | 3+4 2 | 5+6 2 |
| | 4+5 2 | 4+5 2 | 5+6 2 | 5+6 2 |
| | 8 2 | 8 2 | 5+6 2 | 5+6 2 |
| | 9 2 | 8 2 | 3+4 2 | 3+4 2 |
| | 6+7 2 | 6+7 2 | 3+4 2 | 3+4 2 |
| | 9 2 | 9 2 | 5+6 2 | 3+4 2 |
| | 9 2 | 6+7 2 | 3+4 2 | 5+6 2 |
| | 8 2 | 9 2 | 5+6 2 | 3+4 2 |
| | 6+7 2 | 9 2 | 3+4 2 | 5+6 2 |
| 166.1.378 | 252125 | 24 | B2Mo1 | |
| | -GM | -T | -F | -L |
| | 8 2 | 8 2 | 5+6 2 | 5+6 2 |
| | 8 2 | 9 2 | 3+4 2 | 3+4 2 |
| | 9 2 | 9 2 | 5+6 2 | 3+4 2 |
| | 8 2 | 8 2 | 3+4 2 | 5+6 2 |
| | 4+5 2 | 4+5 2 | 5+6 2 | 5+6 2 |
| | 8 2 | 8 2 | 5+6 2 | 5+6 2 |
| | 9 2 | 8 2 | 3+4 2 | 3+4 2 |
| | 6+7 2 | 6+7 2 | 3+4 2 | 3+4 2 |
| | 9 2 | 9 2 | 5+6 2 | 3+4 2 |
| | 9 2 | 6+7 2 | 3+4 2 | 5+6 2 |
| | 6+7 2 | 9 2 | 5+6 2 | 3+4 2 |
| | 8 2 | 9 2 | 3+4 2 | 5+6 2 |
| 166.1.379 | 97343 | 150 | B2Ce2Ir5 | |
| | -GM | -T | -F | -L |
| | 8 2 | 9 2 | 3+4 2 | 3+4 2 |
| | 8 2 | 8 2 | 5+6 2 | 5+6 2 |
| | 9 2 | 9 2 | 3+4 2 | 3+4 2 |
| | 9 2 | 8 2 | 5+6 2 | 5+6 2 |
| | 8 2 | 9 2 | 3+4 2 | 3+4 2 |
| | 8 2 | 8 2 | 5+6 2 | 5+6 2 |
| | 9 2 | 9 2 | 3+4 2 | 3+4 2 |
| | 9 2 | 8 2 | 5+6 2 | 5+6 2 |
| | 8 2 | 6+7 2 | 3+4 2 | 3+4 2 |
| | 4+5 2 | 8 2 | 5+6 2 | 5+6 2 |
| | 9 2 | 9 2 | 5+6 2 | 3+4 2 |
| | 6+7 2 | 6+7 2 | 3+4 2 | 5+6 2 |

Continued on next column

**TABLE LXXIII– continued**

| TQC | ICSD | NoE | SF | |
|---|---|---|---|---|
| | 4+5 2 | 4+5 2 | 5+6 2 | 3+4 2 |
| | 8 2 | 8 2 | 3+4 2 | 5+6 2 |
| | 6+7 2 | 4+5 2 | 3+4 2 | 3+4 2 |
| | 9 2 | 9 2 | 5+6 2 | 5+6 2 |
| | 8 2 | 8 2 | 5+6 2 | 5+6 2 |
| | 9 2 | 9 2 | 3+4 2 | 3+4 2 |
| | 8 2 | 8 2 | 5+6 2 | 5+6 2 |
| | 8 2 | 8 2 | 3+4 2 | 3+4 2 |
| | 9 2 | 9 2 | 5+6 2 | 5+6 2 |
| | 8 2 | 8 2 | 3+4 2 | 5+6 2 |
| | 4+5 2 | 4+5 2 | 3+4 2 | 3+4 2 |
| | 8 2 | 9 2 | 5+6 2 | 5+6 2 |
| | 9 2 | 9 2 | 5+6 2 | 3+4 2 |
| | 8 2 | 6+7 2 | 3+4 2 | 5+6 2 |
| | 9 2 | 8 2 | 5+6 2 | 3+4 2 |
| | 6+7 2 | 8 2 | 5+6 2 | 5+6 2 |
| | 4+5 2 | 4+5 2 | 5+6 2 | 5+6 2 |
| | 8 2 | 9 2 | 3+4 2 | 5+6 2 |
| | 8 2 | 8 2 | 5+6 2 | 5+6 2 |
| | 4+5 2 | 8 2 | 3+4 2 | 3+4 2 |
| | 8 2 | 4+5 2 | 5+6 2 | 5+6 2 |
| | 8 2 | 8 2 | 3+4 2 | 3+4 2 |
| | 4+5 2 | 4+5 2 | 3+4 2 | 3+4 2 |
| | 8 2 | 8 2 | 3+4 2 | 5+6 2 |
| | 6+7 2 | 6+7 2 | 3+4 2 | 3+4 2 |
| | 9 2 | 9 2 | 5+6 2 | 5+6 2 |
| | 9 2 | 9 2 | 3+4 2 | 3+4 2 |
| | 9 2 | 9 2 | 5+6 2 | 5+6 2 |
| | 8 2 | 6+7 2 | 3+4 2 | 3+4 2 |
| | 8 2 | 8 2 | 5+6 2 | 5+6 2 |
| | 4+5 2 | 9 2 | 3+4 2 | 5+6 2 |
| | 6+7 2 | 4+5 2 | 5+6 2 | 3+4 2 |
| | 9 2 | 8 2 | 3+4 2 | 5+6 2 |
| | 4+5 2 | 9 2 | 5+6 2 | 3+4 2 |
| | 8 2 | 6+7 2 | 3+4 2 | 5+6 2 |
| | 8 2 | 9 2 | 5+6 2 | 3+4 2 |
| | 8 2 | 8 2 | 3+4 2 | 5+6 2 |
| | 4+5 2 | 8 2 | 5+6 2 | 5+6 2 |
| | 9 2 | 6+7 2 | 3+4 2 | 3+4 2 |
| | 4+5 2 | 4+5 2 | 3+4 2 | 3+4 2 |
| | 8 2 | 9 2 | 5+6 2 | 5+6 2 |
| | 6+7 2 | 4+5 2 | 5+6 2 | 3+4 2 |
| | 9 2 | 8 2 | 3+4 2 | 5+6 2 |
| | 4+5 2 | 4+5 2 | 3+4 2 | 3+4 2 |
| | 8 2 | 6+7 2 | 5+6 2 | 5+6 2 |
| | 6+7 2 | 8 2 | 3+4 2 | 3+4 2 |
| | 9 2 | 8 2 | 5+6 2 | 5+6 2 |
| | 9 2 | 8 2 | 5+6 2 | 5+6 2 |
| | 8 2 | 9 2 | 3+4 2 | 5+6 2 |
| | 4+5 2 | 4+5 2 | 5+6 2 | 3+4 2 |
| | 6+7 2 | 9 2 | 3+4 2 | 3+4 2 |
| | 8 2 | 6+7 2 | 5+6 2 | 5+6 2 |
| | 9 2 | 8 2 | 3+4 2 | 3+4 2 |
| | 8 2 | 9 2 | 5+6 2 | 5+6 2 |
| | 9 2 | 8 2 | 5+6 2 | 3+4 2 |
| | 8 2 | 9 2 | 5+6 2 | 5+6 2 |
| | 6+7 2 | 6+7 2 | 3+4 2 | 3+4 2 |
| | 9 2 | 8 2 | 3+4 2 | 3+4 2 |
| | 4+5 2 | 9 2 | 3+4 2 | 5+6 2 |
| | 8 2 | 4+5 2 | 5+6 2 | 5+6 2 |
| | 8 2 | 4+5 2 | 3+4 2 | 5+6 2 |

Continued on next column

**TABLE LXXIII– continued**

| TQC | ICSD | NoE | SF | |
|---|---|---|---|---|
| | 4+5 2 | 8 2 | 5+6 2 | 3+4 2 |
| | 8 2 | 9 2 | 5+6 2 | 5+6 2 |
| 166.1.380 | 409982 | 54 | B1Br1Eu2N2 | |
| | -GM | -T | -F | -L |
| | 8 2 | 8 2 | 3+4 2 | 3+4 2 |
| | 9 2 | 9 2 | 5+6 2 | 5+6 2 |
| | 8 2 | 8 2 | 3+4 2 | 3+4 2 |
| | 9 2 | 9 2 | 5+6 2 | 5+6 2 |
| | 8 2 | 8 2 | 3+4 2 | 3+4 2 |
| | 9 2 | 9 2 | 5+6 2 | 5+6 2 |
| | 4+5 2 | 4+5 2 | 5+6 2 | 3+4 2 |
| | 6+7 2 | 6+7 2 | 3+4 2 | 5+6 2 |
| | 8 2 | 8 2 | 3+4 2 | 3+4 2 |
| | 8 2 | 9 2 | 3+4 2 | 5+6 2 |
| | 9 2 | 8 2 | 5+6 2 | 3+4 2 |
| | 8 2 | 9 2 | 5+6 2 | 5+6 2 |
| | 9 2 | 8 2 | 3+4 2 | 5+6 2 |
| | 9 2 | 9 2 | 5+6 2 | 3+4 2 |
| | 6+7 2 | 8 2 | 5+6 2 | 5+6 2 |
| | 9 2 | 4+5 2 | 5+6 2 | 5+6 2 |
| | 9 2 | 9 2 | 5+6 2 | 3+4 2 |
| | 6+7 2 | 6+7 2 | 5+6 2 | 3+4 2 |
| | 8 2 | 9 2 | 3+4 2 | 5+6 2 |
| | 4+5 2 | 6+7 2 | 3+4 2 | 5+6 2 |
| | 8 2 | 8 2 | 3+4 2 | 3+4 2 |
| | 9 2 | 8 2 | 5+6 2 | 3+4 2 |
| | 8 2 | 9 2 | 5+6 2 | 3+4 2 |
| | 4+5 2 | 4+5 2 | 3+4 2 | 5+6 2 |
| | 8 2 | 9 2 | 3+4 2 | 5+6 2 |
| | 6+7 2 | 6+7 2 | 5+6 2 | 5+6 2 |
| | 9 2 | 9 2 | 3+4 2 | 3+4 2 |
| 166.1.381 | 262037 | 72 | As3Cd4Rb1 | |
| | -GM | -T | -F | -L |
| | 8 2 | 9 2 | 3+4 2 | 5+6 2 |
| | 9 2 | 8 2 | 5+6 2 | 3+4 2 |
| | 8 2 | 8 2 | 5+6 2 | 3+4 2 |
| | 9 2 | 8 2 | 3+4 2 | 3+4 2 |
| | 8 2 | 9 2 | 5+6 2 | 3+4 2 |
| | 6+7 2 | 4+5 2 | 3+4 2 | 5+6 2 |
| | 9 2 | 8 2 | 5+6 2 | 3+4 2 |
| | 6+7 2 | 6+7 2 | 5+6 2 | 5+6 2 |
| | 4+5 2 | 4+5 2 | 3+4 2 | 3+4 2 |
| | 8 2 | 8 2 | 5+6 2 | 5+6 2 |
| | 9 2 | 9 2 | 3+4 2 | 3+4 2 |
| | 4+5 2 | 6+7 2 | 5+6 2 | 5+6 2 |
| | 6+7 2 | 4+5 2 | 3+4 2 | 3+4 2 |
| | 9 2 | 9 2 | 5+6 2 | 5+6 2 |
| | 8 2 | 8 2 | 3+4 2 | 3+4 2 |
| | 9 2 | 9 2 | 5+6 2 | 5+6 2 |
| | 9 2 | 9 2 | 3+4 2 | 3+4 2 |
| | 4+5 2 | 4+5 2 | 5+6 2 | 5+6 2 |
| | 8 2 | 8 2 | 5+6 2 | 5+6 2 |
| | 6+7 2 | 6+7 2 | 3+4 2 | 3+4 2 |
| | 8 2 | 8 2 | 3+4 2 | 5+6 2 |
| | 8 2 | 8 2 | 5+6 2 | 3+4 2 |
| | 9 2 | 9 2 | 3+4 2 | 5+6 2 |
| | 4+5 2 | 4+5 2 | 5+6 2 | 3+4 2 |
| | 6+7 2 | 6+7 2 | 3+4 2 | 3+4 2 |
| | 9 2 | 9 2 | 3+4 2 | 5+6 2 |
| | 8 2 | 8 2 | 5+6 2 | 3+4 2 |
| | 9 2 | 9 2 | 5+6 2 | 5+6 2 |

Continued on next column

**TABLE LXXIII– continued**

| TQC | ICSD | NoE | SF | |
|---|---|---|---|---|
| | 8 2 | 8 2 | 5+6 2 | 5+6 2 |
| | 9 2 | 9 2 | 3+4 2 | 3+4 2 |
| | 6+7 2 | 9 2 | 3+4 2 | 3+4 2 |
| | 8 2 | 6+7 2 | 5+6 2 | 5+6 2 |
| | 8 2 | 8 2 | 5+6 2 | 5+6 2 |
| | 9 2 | 8 2 | 3+4 2 | 3+4 2 |
| | 4+5 2 | 9 2 | 5+6 2 | 5+6 2 |
| | 9 2 | 4+5 2 | 5+6 2 | 5+6 2 |
| 166.1.382 | 262035 | 64 | As3Cd4Na1 | |
| | -GM | -T | -F | -L |
| | 8 2 | 8 2 | 3+4 2 | 3+4 2 |
| | 9 2 | 9 2 | 5+6 2 | 3+4 2 |
| | 8 2 | 8 2 | 3+4 2 | 5+6 2 |
| | 6+7 2 | 6+7 2 | 5+6 2 | 5+6 2 |
| | 4+5 2 | 4+5 2 | 3+4 2 | 3+4 2 |
| | 8 2 | 8 2 | 5+6 2 | 5+6 2 |
| | 9 2 | 9 2 | 3+4 2 | 3+4 2 |
| | 6+7 2 | 6+7 2 | 5+6 2 | 5+6 2 |
| | 4+5 2 | 4+5 2 | 3+4 2 | 3+4 2 |
| | 9 2 | 9 2 | 5+6 2 | 5+6 2 |
| | 8 2 | 8 2 | 3+4 2 | 3+4 2 |
| | 9 2 | 9 2 | 5+6 2 | 5+6 2 |
| | 9 2 | 9 2 | 3+4 2 | 3+4 2 |
| | 8 2 | 4+5 2 | 5+6 2 | 5+6 2 |
| | 4+5 2 | 8 2 | 5+6 2 | 5+6 2 |
| | 8 2 | 6+7 2 | 3+4 2 | 3+4 2 |
| | 6+7 2 | 8 2 | 3+4 2 | 3+4 2 |
| | 8 2 | 9 2 | 5+6 2 | 5+6 2 |
| | 9 2 | 8 2 | 5+6 2 | 5+6 2 |
| | 4+5 2 | 4+5 2 | 3+4 2 | 3+4 2 |
| | 6+7 2 | 6+7 2 | 3+4 2 | 3+4 2 |
| | 8 2 | 9 2 | 5+6 2 | 3+4 2 |
| | 9 2 | 8 2 | 3+4 2 | 5+6 2 |
| | 9 2 | 9 2 | 5+6 2 | 5+6 2 |
| | 8 2 | 8 2 | 5+6 2 | 5+6 2 |
| | 9 2 | 9 2 | 3+4 2 | 3+4 2 |
| | 6+7 2 | 9 2 | 5+6 2 | 3+4 2 |
| | 8 2 | 6+7 2 | 3+4 2 | 5+6 2 |
| | 8 2 | 8 2 | 5+6 2 | 5+6 2 |
| | 4+5 2 | 9 2 | 3+4 2 | 3+4 2 |
| | 9 2 | 4+5 2 | 5+6 2 | 5+6 2 |
| | 9 2 | 8 2 | 5+6 2 | 5+6 2 |
| 166.1.383 | 262032 | 72 | As3Cd4K1 | |
| | -GM | -T | -F | -L |
| | 8 2 | 9 2 | 3+4 2 | 5+6 2 |
| | 9 2 | 8 2 | 5+6 2 | 3+4 2 |
| | 9 2 | 8 2 | 5+6 2 | 3+4 2 |
| | 6+7 2 | 4+5 2 | 5+6 2 | 3+4 2 |
| | 8 2 | 8 2 | 3+4 2 | 3+4 2 |
| | 8 2 | 9 2 | 3+4 2 | 5+6 2 |
| | 9 2 | 8 2 | 5+6 2 | 3+4 2 |
| | 6+7 2 | 6+7 2 | 5+6 2 | 5+6 2 |
| | 4+5 2 | 4+5 2 | 3+4 2 | 3+4 2 |
| | 8 2 | 8 2 | 5+6 2 | 5+6 2 |
| | 9 2 | 9 2 | 3+4 2 | 3+4 2 |
| | 6+7 2 | 6+7 2 | 5+6 2 | 5+6 2 |
| | 4+5 2 | 4+5 2 | 3+4 2 | 3+4 2 |
| | 9 2 | 9 2 | 5+6 2 | 5+6 2 |
| | 8 2 | 8 2 | 3+4 2 | 3+4 2 |
| | 9 2 | 9 2 | 5+6 2 | 5+6 2 |
| | 9 2 | 9 2 | 3+4 2 | 3+4 2 |

Continued on next column

**TABLE LXXIII– continued**

| TQC | ICSD | NoE | SF | |
|---|---|---|---|---|
| | 4+5 2 | 4+5 2 | 5+6 2 | 5+6 2 |
| | 8 2 | 8 2 | 5+6 2 | 5+6 2 |
| | 6+7 2 | 6+7 2 | 3+4 2 | 3+4 2 |
| | 8 2 | 8 2 | 3+4 2 | 5+6 2 |
| | 8 2 | 8 2 | 5+6 2 | 3+4 2 |
| | 9 2 | 9 2 | 5+6 2 | 5+6 2 |
| | 4+5 2 | 4+5 2 | 3+4 2 | 3+4 2 |
| | 6+7 2 | 6+7 2 | 3+4 2 | 3+4 2 |
| | 9 2 | 9 2 | 3+4 2 | 5+6 2 |
| | 8 2 | 8 2 | 5+6 2 | 3+4 2 |
| | 9 2 | 9 2 | 5+6 2 | 5+6 2 |
| | 8 2 | 8 2 | 5+6 2 | 5+6 2 |
| | 9 2 | 9 2 | 3+4 2 | 3+4 2 |
| | 6+7 2 | 9 2 | 3+4 2 | 3+4 2 |
| | 8 2 | 6+7 2 | 5+6 2 | 5+6 2 |
| | 8 2 | 8 2 | 5+6 2 | 5+6 2 |
| | 9 2 | 9 2 | 3+4 2 | 3+4 2 |
| | 4+5 2 | 8 2 | 5+6 2 | 5+6 2 |
| | 9 2 | 4+5 2 | 5+6 2 | 5+6 2 |
| 166.1.384 | 68110 | 28 | As2Te3 | |
| | -GM | -T | -F | -L |
| | 8 2 | 9 2 | 5+6 2 | 3+4 2 |
| | 9 2 | 8 2 | 3+4 2 | 5+6 2 |
| | 8 2 | 8 2 | 3+4 2 | 3+4 2 |
| | 9 2 | 9 2 | 5+6 2 | 5+6 2 |
| | 8 2 | 8 2 | 3+4 2 | 3+4 2 |
| | 8 2 | 9 2 | 5+6 2 | 5+6 2 |
| | 9 2 | 8 2 | 3+4 2 | 3+4 2 |
| | 9 2 | 9 2 | 5+6 2 | 3+4 2 |
| | 8 2 | 6+7 2 | 5+6 2 | 5+6 2 |
| | 4+5 2 | 4+5 2 | 3+4 2 | 5+6 2 |
| | 6+7 2 | 9 2 | 5+6 2 | 3+4 2 |
| | 9 2 | 9 2 | 3+4 2 | 5+6 2 |
| | 6+7 2 | 6+7 2 | 5+6 2 | 5+6 2 |
| | 8 2 | 8 2 | 5+6 2 | 5+6 2 |
| 166.1.385 | 82371 | 28 | As2Sn2Sr1 | |
| | -GM | -T | -F | -L |
| | 8 2 | 8 2 | 3+4 2 | 3+4 2 |
| | 9 2 | 9 2 | 5+6 2 | 5+6 2 |
| | 9 2 | 9 2 | 5+6 2 | 5+6 2 |
| | 6+7 2 | 6+7 2 | 5+6 2 | 5+6 2 |
| | 8 2 | 9 2 | 3+4 2 | 3+4 2 |
| | 9 2 | 8 2 | 5+6 2 | 5+6 2 |
| | 8 2 | 9 2 | 5+6 2 | 3+4 2 |
| | 9 2 | 8 2 | 3+4 2 | 5+6 2 |
| | 8 2 | 8 2 | 3+4 2 | 5+6 2 |
| | 8 2 | 4+5 2 | 5+6 2 | 3+4 2 |
| | 4+5 2 | 9 2 | 5+6 2 | 3+4 2 |
| | 9 2 | 8 2 | 3+4 2 | 3+4 2 |
| | 6+7 2 | 6+7 2 | 5+6 2 | 5+6 2 |
| | 8 2 | 9 2 | 3+4 2 | 5+6 2 |
| 166.1.386 | 68114 | 68 | As2Ge4Te7 | |
| | -GM | -T | -F | -L |
| | 8 2 | 9 2 | 5+6 2 | 3+4 2 |
| | 9 2 | 8 2 | 3+4 2 | 5+6 2 |
| | 8 2 | 8 2 | 5+6 2 | 3+4 2 |
| | 9 2 | 9 2 | 3+4 2 | 5+6 2 |
| | 8 2 | 8 2 | 3+4 2 | 3+4 2 |
| | 9 2 | 9 2 | 5+6 2 | 5+6 2 |
| | 8 2 | 8 2 | 3+4 2 | 3+4 2 |
| | 8 2 | 9 2 | 5+6 2 | 3+4 2 |

Continued on next column

**TABLE LXXIII– continued**

| TQC | ICSD | NoE | SF | |
|---|---|---|---|---|
| | 9 2 | 8 2 | 3+4 2 | 5+6 2 |
| | 9 2 | 9 2 | 5+6 2 | 5+6 2 |
| | 8 2 | 8 2 | 3+4 2 | 3+4 2 |
| | 9 2 | 9 2 | 5+6 2 | 5+6 2 |
| | 8 2 | 8 2 | 3+4 2 | 3+4 2 |
| | 8 2 | 9 2 | 3+4 2 | 5+6 2 |
| | 9 2 | 8 2 | 5+6 2 | 3+4 2 |
| | 8 2 | 9 2 | 5+6 2 | 5+6 2 |
| | 4+5 2 | 6+7 2 | 3+4 2 | 3+4 2 |
| | 6+7 2 | 4+5 2 | 5+6 2 | 3+4 2 |
| | 9 2 | 9 2 | 5+6 2 | 5+6 2 |
| | 8 2 | 8 2 | 3+4 2 | 3+4 2 |
| | 9 2 | 9 2 | 3+4 2 | 5+6 2 |
| | 9 2 | 8 2 | 5+6 2 | 5+6 2 |
| | 8 2 | 9 2 | 5+6 2 | 5+6 2 |
| | 8 2 | 8 2 | 5+6 2 | 3+4 2 |
| | 4+5 2 | 6+7 2 | 3+4 2 | 5+6 2 |
| | 6+7 2 | 4+5 2 | 3+4 2 | 3+4 2 |
| | 9 2 | 9 2 | 5+6 2 | 3+4 2 |
| | 9 2 | 6+7 2 | 3+4 2 | 5+6 2 |
| | 6+7 2 | 9 2 | 5+6 2 | 3+4 2 |
| | 4+5 2 | 4+5 2 | 3+4 2 | 5+6 2 |
| | 6+7 2 | 6+7 2 | 5+6 2 | 5+6 2 |
| | 8 2 | 8 2 | 3+4 2 | 3+4 2 |
| | 9 2 | 8 2 | 5+6 2 | 5+6 2 |
| | 8 2 | 9 2 | 5+6 2 | 5+6 2 |
| 166.1.387 | 68113 | 56 | As2Ge3Te6 | |
| | -GM | -T | -F | -L |
| | 8 2 | 8 2 | 5+6 2 | 3+4 2 |
| | 9 2 | 9 2 | 5+6 2 | 5+6 2 |
| | 8 2 | 8 2 | 3+4 2 | 3+4 2 |
| | 9 2 | 9 2 | 3+4 2 | 5+6 2 |
| | 8 2 | 8 2 | 5+6 2 | 5+6 2 |
| | 9 2 | 9 2 | 3+4 2 | 3+4 2 |
| | 8 2 | 8 2 | 3+4 2 | 3+4 2 |
| | 9 2 | 9 2 | 5+6 2 | 5+6 2 |
| | 8 2 | 8 2 | 3+4 2 | 3+4 2 |
| | 9 2 | 9 2 | 5+6 2 | 5+6 2 |
| | 8 2 | 8 2 | 3+4 2 | 3+4 2 |
| | 8 2 | 9 2 | 5+6 2 | 5+6 2 |
| | 9 2 | 9 2 | 3+4 2 | 3+4 2 |
| | 8 2 | 8 2 | 5+6 2 | 5+6 2 |
| | 9 2 | 9 2 | 3+4 2 | 5+6 2 |
| | 8 2 | 8 2 | 5+6 2 | 3+4 2 |
| | 9 2 | 9 2 | 5+6 2 | 3+4 2 |
| | 9 2 | 8 2 | 3+4 2 | 5+6 2 |
| | 8 2 | 6+7 2 | 5+6 2 | 3+4 2 |
| | 4+5 2 | 6+7 2 | 3+4 2 | 5+6 2 |
| | 6+7 2 | 4+5 2 | 5+6 2 | 5+6 2 |
| | 6+7 2 | 9 2 | 5+6 2 | 3+4 2 |
| | 4+5 2 | 8 2 | 3+4 2 | 5+6 2 |
| | 6+7 2 | 4+5 2 | 5+6 2 | 3+4 2 |
| | 4+5 2 | 6+7 2 | 3+4 2 | 5+6 2 |
| | 8 2 | 4+5 2 | 5+6 2 | 3+4 2 |
| | 9 2 | 9 2 | 3+4 2 | 3+4 2 |
| | 8 2 | 8 2 | 3+4 2 | 5+6 2 |
| 166.1.388 | 422526 | 18 | As2Ca1Ga2 | |
| | -GM | -T | -F | -L |
| | 8 2 | 9 2 | 3+4 2 | 3+4 2 |
| | 9 2 | 8 2 | 5+6 2 | 5+6 2 |
| | 8 2 | 9 2 | 3+4 2 | 5+6 2 |

Continued on next column

**TABLE LXXIII– continued**

| TQC | ICSD | NoE | SF | |
|---|---|---|---|---|
| | 9 2 | 8 2 | 5+6 2 | 3+4 2 |
| | 8 2 | 8 2 | 3+4 2 | 5+6 2 |
| | 8 2 | 4+5 2 | 5+6 2 | 3+4 2 |
| | 4+5 2 | 9 2 | 5+6 2 | 3+4 2 |
| | 9 2 | 6+7 2 | 3+4 2 | 5+6 2 |
| | 6+7 2 | 8 2 | 3+4 2 | 5+6 2 |
| 166.1.389 | 53794 | 10 | As1 | |
| | -GM | -T | -F | -L |
| | 8 2 | 9 2 | 5+6 2 | 3+4 2 |
| | 9 2 | 8 2 | 3+4 2 | 5+6 2 |
| | 8 2 | 9 2 | 3+4 2 | 5+6 2 |
| | 8 2 | 6+7 2 | 5+6 2 | 3+4 2 |
| | 4+5 2 | 8 2 | 5+6 2 | 3+4 2 |
| 166.1.390 | 16518 | 10 | As1 | |
| | -GM | -T | -F | -L |
| | 8 2 | 9 2 | 5+6 2 | 3+4 2 |
| | 9 2 | 8 2 | 3+4 2 | 5+6 2 |
| | 8 2 | 9 2 | 3+4 2 | 3+4 2 |
| | 8 2 | 6+7 2 | 5+6 2 | 5+6 2 |
| | 4+5 2 | 8 2 | 5+6 2 | 3+4 2 |
| 166.1.391 | 16517 | 10 | As1 | |
| | -GM | -T | -F | -L |
| | 8 2 | 9 2 | 5+6 2 | 3+4 2 |
| | 9 2 | 8 2 | 3+4 2 | 5+6 2 |
| | 8 2 | 9 2 | 3+4 2 | 3+4 2 |
| | 8 2 | 6+7 2 | 5+6 2 | 5+6 2 |
| | 4+5 2 | 8 2 | 5+6 2 | 3+4 2 |
| 166.1.392 | 16516 | 10 | As1 | |
| | -GM | -T | -F | -L |
| | 8 2 | 9 2 | 5+6 2 | 3+4 2 |
| | 9 2 | 8 2 | 3+4 2 | 5+6 2 |
| | 8 2 | 9 2 | 3+4 2 | 3+4 2 |
| | 8 2 | 6+7 2 | 5+6 2 | 5+6 2 |
| | 4+5 2 | 8 2 | 5+6 2 | 3+4 2 |
| 166.1.393 | 281656 | 80 | Al6Gd2Pt1Si4 | |
| | -GM | -T | -F | -L |
| | 8 2 | 8 2 | 3+4 2 | 3+4 2 |
| | 9 2 | 9 2 | 5+6 2 | 5+6 2 |
| | 8 2 | 8 2 | 3+4 2 | 3+4 2 |
| | 9 2 | 9 2 | 5+6 2 | 5+6 2 |
| | 8 2 | 8 2 | 3+4 2 | 3+4 2 |
| | 9 2 | 9 2 | 5+6 2 | 5+6 2 |
| | 4+5 2 | 4+5 2 | 3+4 2 | 3+4 2 |
| | 6+7 2 | 6+7 2 | 5+6 2 | 5+6 2 |
| | 8 2 | 8 2 | 3+4 2 | 3+4 2 |
| | 9 2 | 9 2 | 5+6 2 | 5+6 2 |
| | 8 2 | 9 2 | 3+4 2 | 3+4 2 |
| | 8 2 | 8 2 | 5+6 2 | 3+4 2 |
| | 9 2 | 8 2 | 3+4 2 | 3+4 2 |
| | 4+5 2 | 4+5 2 | 5+6 2 | 5+6 2 |
| | 8 2 | 8 2 | 3+4 2 | 5+6 2 |
| | 8 2 | 8 2 | 3+4 2 | 3+4 2 |
| | 9 2 | 9 2 | 5+6 2 | 5+6 2 |
| | 8 2 | 8 2 | 3+4 2 | 3+4 2 |
| | 4+5 2 | 4+5 2 | 3+4 2 | 3+4 2 |
| | 8 2 | 9 2 | 3+4 2 | 3+4 2 |
| | 8 2 | 8 2 | 3+4 2 | 5+6 2 |
| | 9 2 | 9 2 | 3+4 2 | 3+4 2 |
| | 6+7 2 | 6+7 2 | 5+6 2 | 3+4 2 |
| | 9 2 | 4+5 2 | 5+6 2 | 5+6 2 |

Continued on next column

**TABLE LXXIII– continued**

| TQC | ICSD | NoE | SF | |
|---|---|---|---|---|
| | 4+5 2 | 8 2 | 5+6 2 | 5+6 2 |
| | 8 2 | 9 2 | 3+4 2 | 3+4 2 |
| | 4+5 2 | 6+7 2 | 3+4 2 | 3+4 2 |
| | 8 2 | 9 2 | 5+6 2 | 5+6 2 |
| | 9 2 | 8 2 | 5+6 2 | 3+4 2 |
| | 6+7 2 | 4+5 2 | 5+6 2 | 3+4 2 |
| | 9 2 | 8 2 | 5+6 2 | 5+6 2 |
| | 8 2 | 9 2 | 3+4 2 | 5+6 2 |
| | 8 2 | 9 2 | 3+4 2 | 5+6 2 |
| | 4+5 2 | 6+7 2 | 3+4 2 | 3+4 2 |
| | 9 2 | 8 2 | 5+6 2 | 3+4 2 |
| | 9 2 | 8 2 | 3+4 2 | 5+6 2 |
| | 6+7 2 | 4+5 2 | 5+6 2 | 5+6 2 |
| | 9 2 | 9 2 | 5+6 2 | 3+4 2 |
| | 8 2 | 9 2 | 3+4 2 | 5+6 2 |
| | 9 2 | 8 2 | 5+6 2 | 3+4 2 |
| 166.1.394 | 281660 | 84 | Al6Dy2Pt1Si4 | |
| | -GM | -T | -F | -L |
| | 8 2 | 8 2 | 3+4 2 | 3+4 2 |
| | 9 2 | 9 2 | 5+6 2 | 5+6 2 |
| | 8 2 | 8 2 | 3+4 2 | 3+4 2 |
| | 9 2 | 9 2 | 5+6 2 | 5+6 2 |
| | 8 2 | 8 2 | 3+4 2 | 3+4 2 |
| | 9 2 | 9 2 | 5+6 2 | 5+6 2 |
| | 4+5 2 | 4+5 2 | 3+4 2 | 3+4 2 |
| | 6+7 2 | 6+7 2 | 5+6 2 | 5+6 2 |
| | 8 2 | 9 2 | 5+6 2 | 3+4 2 |
| | 8 2 | 8 2 | 3+4 2 | 3+4 2 |
| | 9 2 | 9 2 | 5+6 2 | 5+6 2 |
| | 9 2 | 8 2 | 3+4 2 | 5+6 2 |
| | 8 2 | 8 2 | 3+4 2 | 3+4 2 |
| | 8 2 | 4+5 2 | 3+4 2 | 3+4 2 |
| | 4+5 2 | 9 2 | 5+6 2 | 5+6 2 |
| | 9 2 | 8 2 | 3+4 2 | 5+6 2 |
| | 8 2 | 8 2 | 5+6 2 | 3+4 2 |
| | 8 2 | 8 2 | 3+4 2 | 5+6 2 |
| | 8 2 | 9 2 | 5+6 2 | 3+4 2 |
| | 4+5 2 | 4+5 2 | 3+4 2 | 3+4 2 |
| | 8 2 | 8 2 | 3+4 2 | 3+4 2 |
| | 9 2 | 6+7 2 | 3+4 2 | 3+4 2 |
| | 4+5 2 | 9 2 | 3+4 2 | 3+4 2 |
| | 8 2 | 6+7 2 | 5+6 2 | 3+4 2 |
| | 6+7 2 | 4+5 2 | 5+6 2 | 3+4 2 |
| | 9 2 | 8 2 | 5+6 2 | 5+6 2 |
| | 4+5 2 | 9 2 | 3+4 2 | 5+6 2 |
| | 6+7 2 | 4+5 2 | 5+6 2 | 5+6 2 |
| | 9 2 | 8 2 | 3+4 2 | 3+4 2 |
| | 8 2 | 8 2 | 3+4 2 | 3+4 2 |
| | 9 2 | 9 2 | 3+4 2 | 5+6 2 |
| | 8 2 | 9 2 | 5+6 2 | 5+6 2 |
| | 8 2 | 8 2 | 5+6 2 | 3+4 2 |
| | 9 2 | 6+7 2 | 3+4 2 | 5+6 2 |
| | 4+5 2 | 9 2 | 3+4 2 | 3+4 2 |
| | 6+7 2 | 8 2 | 5+6 2 | 5+6 2 |
| | 9 2 | 4+5 2 | 5+6 2 | 3+4 2 |
| | 8 2 | 8 2 | 3+4 2 | 5+6 2 |
| | 9 2 | 9 2 | 3+4 2 | 5+6 2 |
| | 8 2 | 6+7 2 | 5+6 2 | 5+6 2 |
| | 4+5 2 | 9 2 | 3+4 2 | 3+4 2 |
| | 8 2 | 8 2 | 5+6 2 | 3+4 2 |
| 166.1.395 | 58175 | 84 | Al3Tb1 | |

Continued on next column

**TABLE LXXIII– continued**

| TQC | ICSD | NoE | SF | |
|---|---|---|---|---|
| | -GM | -T | -F | -L |
| | 8 2 | 8 2 | 3+4 2 | 3+4 2 |
| | 8 2 | 9 2 | 3+4 2 | 3+4 2 |
| | 9 2 | 8 2 | 5+6 2 | 5+6 2 |
| | 9 2 | 9 2 | 5+6 2 | 5+6 2 |
| | 8 2 | 9 2 | 5+6 2 | 3+4 2 |
| | 9 2 | 8 2 | 3+4 2 | 5+6 2 |
| | 9 2 | 6+7 2 | 5+6 2 | 5+6 2 |
| | 6+7 2 | 9 2 | 5+6 2 | 5+6 2 |
| | 8 2 | 9 2 | 3+4 2 | 3+4 2 |
| | 4+5 2 | 6+7 2 | 5+6 2 | 5+6 2 |
| | 9 2 | 8 2 | 5+6 2 | 3+4 2 |
| | 6+7 2 | 4+5 2 | 3+4 2 | 5+6 2 |
| | 8 2 | 9 2 | 3+4 2 | 5+6 2 |
| | 9 2 | 8 2 | 5+6 2 | 3+4 2 |
| | 8 2 | 8 2 | 3+4 2 | 3+4 2 |
| | 9 2 | 4+5 2 | 5+6 2 | 5+6 2 |
| | 6+7 2 | 8 2 | 5+6 2 | 5+6 2 |
| | 8 2 | 9 2 | 3+4 2 | 3+4 2 |
| | 4+5 2 | 9 2 | 3+4 2 | 5+6 2 |
| | 8 2 | 6+7 2 | 3+4 2 | 3+4 2 |
| | 8 2 | 4+5 2 | 5+6 2 | 3+4 2 |
| | 9 2 | 8 2 | 5+6 2 | 5+6 2 |
| | 4+5 2 | 9 2 | 5+6 2 | 3+4 2 |
| | 9 2 | 8 2 | 3+4 2 | 5+6 2 |
| | 9 2 | 9 2 | 3+4 2 | 3+4 2 |
| | 6+7 2 | 6+7 2 | 5+6 2 | 3+4 2 |
| | 6+7 2 | 9 2 | 3+4 2 | 5+6 2 |
| | 9 2 | 8 2 | 3+4 2 | 5+6 2 |
| | 8 2 | 4+5 2 | 5+6 2 | 3+4 2 |
| | 9 2 | 9 2 | 5+6 2 | 5+6 2 |
| | 4+5 2 | 6+7 2 | 5+6 2 | 5+6 2 |
| | 9 2 | 8 2 | 5+6 2 | 3+4 2 |
| | 8 2 | 9 2 | 3+4 2 | 5+6 2 |
| | 6+7 2 | 9 2 | 5+6 2 | 3+4 2 |
| | 9 2 | 6+7 2 | 3+4 2 | 5+6 2 |
| | 9 2 | 9 2 | 5+6 2 | 3+4 2 |
| | 6+7 2 | 8 2 | 3+4 2 | 5+6 2 |
| | 8 2 | 6+7 2 | 5+6 2 | 5+6 2 |
| | 4+5 2 | 9 2 | 3+4 2 | 3+4 2 |
| | 8 2 | 4+5 2 | 5+6 2 | 5+6 2 |
| | 9 2 | 8 2 | 5+6 2 | 3+4 2 |
| | 9 2 | 9 2 | 3+4 2 | 5+6 2 |
| 166.1.396 | 420092 | 100 | Al10Ba7 | |
| | -GM | -T | -F | -L |
| | 8 2 | 9 2 | 3+4 2 | 5+6 2 |
| | 8 2 | 8 2 | 5+6 2 | 5+6 2 |
| | 9 2 | 9 2 | 3+4 2 | 3+4 2 |
| | 8 2 | 8 2 | 3+4 2 | 3+4 2 |
| | 9 2 | 9 2 | 5+6 2 | 5+6 2 |
| | 8 2 | 8 2 | 5+6 2 | 3+4 2 |
| | 9 2 | 9 2 | 3+4 2 | 5+6 2 |
| | 9 2 | 8 2 | 5+6 2 | 3+4 2 |
| | 8 2 | 8 2 | 5+6 2 | 5+6 2 |
| | 9 2 | 9 2 | 3+4 2 | 3+4 2 |
| | 8 2 | 8 2 | 3+4 2 | 3+4 2 |
| | 9 2 | 9 2 | 5+6 2 | 5+6 2 |
| | 8 2 | 9 2 | 5+6 2 | 3+4 2 |
| | 9 2 | 8 2 | 3+4 2 | 5+6 2 |
| | 6+7 2 | 4+5 2 | 5+6 2 | 3+4 2 |
| | 4+5 2 | 4+5 2 | 5+6 2 | 5+6 2 |

Continued on next column

**TABLE LXXIII– continued**

| TQC | ICSD | NoE | SF | |
|---|---|---|---|---|
| | 9 2 | 8 2 | 5+6 2 | 3+4 2 |
| | 6+7 2 | 6+7 2 | 3+4 2 | 3+4 2 |
| | 8 2 | 8 2 | 3+4 2 | 3+4 2 |
| | 9 2 | 9 2 | 5+6 2 | 5+6 2 |
| | 8 2 | 8 2 | 3+4 2 | 3+4 2 |
| | 9 2 | 9 2 | 5+6 2 | 5+6 2 |
| | 4+5 2 | 4+5 2 | 5+6 2 | 5+6 2 |
| | 6+7 2 | 6+7 2 | 3+4 2 | 3+4 2 |
| | 4+5 2 | 6+7 2 | 3+4 2 | 5+6 2 |
| | 8 2 | 9 2 | 5+6 2 | 3+4 2 |
| | 6+7 2 | 4+5 2 | 3+4 2 | 5+6 2 |
| | 9 2 | 8 2 | 5+6 2 | 3+4 2 |
| | 8 2 | 9 2 | 5+6 2 | 5+6 2 |
| | 9 2 | 8 2 | 3+4 2 | 3+4 2 |
| | 8 2 | 9 2 | 3+4 2 | 5+6 2 |
| | 9 2 | 8 2 | 5+6 2 | 3+4 2 |
| | 9 2 | 9 2 | 3+4 2 | 5+6 2 |
| | 8 2 | 8 2 | 5+6 2 | 3+4 2 |
| | 8 2 | 9 2 | 3+4 2 | 5+6 2 |
| | 4+5 2 | 6+7 2 | 5+6 2 | 3+4 2 |
| | 9 2 | 8 2 | 3+4 2 | 5+6 2 |
| | 6+7 2 | 4+5 2 | 5+6 2 | 3+4 2 |
| | 8 2 | 9 2 | 5+6 2 | 5+6 2 |
| | 9 2 | 8 2 | 3+4 2 | 3+4 2 |
| | 8 2 | 9 2 | 3+4 2 | 5+6 2 |
| | 9 2 | 6+7 2 | 5+6 2 | 3+4 2 |
| | 4+5 2 | 9 2 | 3+4 2 | 5+6 2 |
| | 6+7 2 | 8 2 | 3+4 2 | 5+6 2 |
| | 8 2 | 6+7 2 | 5+6 2 | 3+4 2 |
| | 4+5 2 | 4+5 2 | 5+6 2 | 3+4 2 |
| | 9 2 | 9 2 | 5+6 2 | 3+4 2 |
| | 6+7 2 | 8 2 | 3+4 2 | 5+6 2 |
| | 8 2 | 4+5 2 | 3+4 2 | 3+4 2 |
| | 9 2 | 8 2 | 3+4 2 | 5+6 2 |
| 166.1.397 | 424311 | 64 | Ag4Sb2Sr1 | |
| | -GM | -T | -F | -L |
| | 8 2 | 8 2 | 3+4 2 | 3+4 2 |
| | 9 2 | 9 2 | 5+6 2 | 5+6 2 |
| | 9 2 | 9 2 | 5+6 2 | 5+6 2 |
| | 6+7 2 | 6+7 2 | 5+6 2 | 5+6 2 |
| | 8 2 | 9 2 | 5+6 2 | 3+4 2 |
| | 9 2 | 8 2 | 3+4 2 | 5+6 2 |
| | 6+7 2 | 4+5 2 | 5+6 2 | 3+4 2 |
| | 9 2 | 8 2 | 3+4 2 | 5+6 2 |
| | 4+5 2 | 6+7 2 | 5+6 2 | 3+4 2 |
| | 8 2 | 8 2 | 3+4 2 | 5+6 2 |
| | 6+7 2 | 4+5 2 | 3+4 2 | 3+4 2 |
| | 4+5 2 | 9 2 | 5+6 2 | 5+6 2 |
| | 8 2 | 6+7 2 | 3+4 2 | 5+6 2 |
| | 9 2 | 8 2 | 5+6 2 | 3+4 2 |
| | 8 2 | 9 2 | 3+4 2 | 5+6 2 |
| | 8 2 | 9 2 | 3+4 2 | 3+4 2 |
| | 9 2 | 9 2 | 5+6 2 | 5+6 2 |
| | 8 2 | 8 2 | 5+6 2 | 3+4 2 |
| | 4+5 2 | 9 2 | 3+4 2 | 5+6 2 |
| | 9 2 | 6+7 2 | 5+6 2 | 3+4 2 |
| | 8 2 | 4+5 2 | 3+4 2 | 5+6 2 |
| | 6+7 2 | 8 2 | 5+6 2 | 3+4 2 |
| | 4+5 2 | 6+7 2 | 5+6 2 | 3+4 2 |
| | 9 2 | 9 2 | 3+4 2 | 5+6 2 |
| | 8 2 | 8 2 | 5+6 2 | 3+4 2 |

Continued on next column

**TABLE LXXIII– continued**

| TQC | ICSD | | NoE | SF |
|---|---|---|---|---|
| | 6+7 2 | 4+5 2 | 3+4 2 | 5+6 2 |
| | 9 2 | 9 2 | 3+4 2 | 5+6 2 |
| | 9 2 | 8 2 | 5+6 2 | 3+4 2 |
| | 8 2 | 8 2 | 5+6 2 | 3+4 2 |
| | 4+5 2 | 4+5 2 | 3+4 2 | 3+4 2 |
| | 8 2 | 9 2 | 5+6 2 | 5+6 2 |
| | 9 2 | 8 2 | 3+4 2 | 5+6 2 |
| 166.1.398 | 416373 | | 44 | Ag2Ni1O2 |
| | -GM | -T | -F | -L |
| | 8 2 | 8 2 | 5+6 2 | 5+6 2 |
| | 9 2 | 9 2 | 3+4 2 | 3+4 2 |
| | 8 2 | 8 2 | 3+4 2 | 3+4 2 |
| | 9 2 | 9 2 | 5+6 2 | 5+6 2 |
| | 4+5 2 | 8 2 | 3+4 2 | 5+6 2 |
| | 8 2 | 4+5 2 | 5+6 2 | 3+4 2 |
| | 8 2 | 8 2 | 3+4 2 | 3+4 2 |
| | 6+7 2 | 6+7 2 | 3+4 2 | 3+4 2 |
| | 4+5 2 | 4+5 2 | 5+6 2 | 5+6 2 |
| | 8 2 | 9 2 | 5+6 2 | 5+6 2 |
| | 9 2 | 9 2 | 3+4 2 | 3+4 2 |
| | 8 2 | 8 2 | 5+6 2 | 5+6 2 |
| | 9 2 | 9 2 | 5+6 2 | 3+4 2 |
| | 8 2 | 6+7 2 | 3+4 2 | 3+4 2 |
| | 4+5 2 | 8 2 | 3+4 2 | 5+6 2 |
| | 9 2 | 4+5 2 | 5+6 2 | 3+4 2 |
| | 6+7 2 | 9 2 | 5+6 2 | 5+6 2 |
| | 8 2 | 9 2 | 3+4 2 | 3+4 2 |
| | 4+5 2 | 6+7 2 | 3+4 2 | 5+6 2 |
| | 9 2 | 8 2 | 3+4 2 | 3+4 2 |
| | 6+7 2 | 4+5 2 | 3+4 2 | 3+4 2 |
| | 8 2 | 8 2 | 3+4 2 | 3+4 2 |
| 166.1.399 | 412279 | | 44 | Ag2Ni1O2 |
| | -GM | -T | -F | -L |
| | 8 2 | 8 2 | 5+6 2 | 5+6 2 |
| | 9 2 | 9 2 | 3+4 2 | 3+4 2 |
| | 8 2 | 8 2 | 3+4 2 | 3+4 2 |
| | 9 2 | 9 2 | 5+6 2 | 5+6 2 |
| | 4+5 2 | 4+5 2 | 3+4 2 | 5+6 2 |
| | 8 2 | 8 2 | 5+6 2 | 3+4 2 |
| | 8 2 | 8 2 | 3+4 2 | 3+4 2 |
| | 6+7 2 | 6+7 2 | 3+4 2 | 3+4 2 |
| | 4+5 2 | 4+5 2 | 5+6 2 | 5+6 2 |
| | 8 2 | 9 2 | 5+6 2 | 5+6 2 |
| | 9 2 | 9 2 | 3+4 2 | 3+4 2 |
| | 8 2 | 8 2 | 5+6 2 | 5+6 2 |
| | 9 2 | 9 2 | 5+6 2 | 3+4 2 |
| | 8 2 | 6+7 2 | 3+4 2 | 3+4 2 |
| | 4+5 2 | 8 2 | 3+4 2 | 5+6 2 |
| | 9 2 | 4+5 2 | 5+6 2 | 3+4 2 |
| | 6+7 2 | 9 2 | 5+6 2 | 5+6 2 |
| | 8 2 | 9 2 | 3+4 2 | 3+4 2 |
| | 4+5 2 | 6+7 2 | 3+4 2 | 5+6 2 |
| | 9 2 | 8 2 | 3+4 2 | 3+4 2 |
| | 6+7 2 | 4+5 2 | 3+4 2 | 3+4 2 |
| | 8 2 | 8 2 | 3+4 2 | 3+4 2 |
| 166.1.400 | 172558 | | 44 | Ag2Ni1O2 |
| | -GM | -T | -F | -L |
| | 8 2 | 8 2 | 5+6 2 | 5+6 2 |
| | 9 2 | 9 2 | 3+4 2 | 3+4 2 |
| | 8 2 | 8 2 | 3+4 2 | 3+4 2 |
| | 9 2 | 9 2 | 5+6 2 | 5+6 2 |

Continued on next column

**TABLE LXXIII– continued**

| TQC | ICSD | | NoE | SF |
|---|---|---|---|---|
| | 4+5 2 | 8 2 | 3+4 2 | 5+6 2 |
| | 8 2 | 4+5 2 | 5+6 2 | 3+4 2 |
| | 8 2 | 8 2 | 3+4 2 | 3+4 2 |
| | 6+7 2 | 6+7 2 | 3+4 2 | 3+4 2 |
| | 4+5 2 | 4+5 2 | 5+6 2 | 5+6 2 |
| | 8 2 | 9 2 | 5+6 2 | 5+6 2 |
| | 9 2 | 9 2 | 3+4 2 | 3+4 2 |
| | 8 2 | 8 2 | 5+6 2 | 5+6 2 |
| | 9 2 | 9 2 | 5+6 2 | 3+4 2 |
| | 8 2 | 6+7 2 | 3+4 2 | 3+4 2 |
| | 4+5 2 | 8 2 | 3+4 2 | 5+6 2 |
| | 9 2 | 4+5 2 | 5+6 2 | 3+4 2 |
| | 6+7 2 | 9 2 | 5+6 2 | 5+6 2 |
| | 8 2 | 9 2 | 3+4 2 | 3+4 2 |
| | 4+5 2 | 8 2 | 3+4 2 | 5+6 2 |
| | 9 2 | 6+7 2 | 3+4 2 | 3+4 2 |
| | 6+7 2 | 4+5 2 | 3+4 2 | 3+4 2 |
| | 8 2 | 8 2 | 3+4 2 | 3+4 2 |
| 166.1.401 | 160582 | | 44 | Ag2Ni1O2 |
| | -GM | -T | -F | -L |
| | 8 2 | 8 2 | 5+6 2 | 5+6 2 |
| | 9 2 | 9 2 | 3+4 2 | 3+4 2 |
| | 8 2 | 8 2 | 3+4 2 | 3+4 2 |
| | 9 2 | 9 2 | 5+6 2 | 5+6 2 |
| | 4+5 2 | 4+5 2 | 3+4 2 | 5+6 2 |
| | 8 2 | 8 2 | 5+6 2 | 3+4 2 |
| | 8 2 | 8 2 | 3+4 2 | 3+4 2 |
| | 6+7 2 | 6+7 2 | 3+4 2 | 3+4 2 |
| | 4+5 2 | 4+5 2 | 5+6 2 | 5+6 2 |
| | 8 2 | 9 2 | 5+6 2 | 5+6 2 |
| | 9 2 | 9 2 | 3+4 2 | 3+4 2 |
| | 8 2 | 8 2 | 5+6 2 | 5+6 2 |
| | 9 2 | 9 2 | 5+6 2 | 3+4 2 |
| | 8 2 | 6+7 2 | 3+4 2 | 3+4 2 |
| | 4+5 2 | 8 2 | 3+4 2 | 5+6 2 |
| | 9 2 | 4+5 2 | 5+6 2 | 3+4 2 |
| | 6+7 2 | 9 2 | 5+6 2 | 5+6 2 |
| | 8 2 | 9 2 | 3+4 2 | 3+4 2 |
| | 4+5 2 | 6+7 2 | 3+4 2 | 5+6 2 |
| | 9 2 | 8 2 | 3+4 2 | 3+4 2 |
| | 6+7 2 | 4+5 2 | 3+4 2 | 3+4 2 |
| | 8 2 | 8 2 | 3+4 2 | 3+4 2 |
| 166.1.402 | 160581 | | 44 | Ag2Ni1O2 |
| | -GM | -T | -F | -L |
| | 8 2 | 8 2 | 5+6 2 | 5+6 2 |
| | 9 2 | 9 2 | 3+4 2 | 3+4 2 |
| | 8 2 | 8 2 | 3+4 2 | 3+4 2 |
| | 9 2 | 9 2 | 5+6 2 | 5+6 2 |
| | 4+5 2 | 4+5 2 | 3+4 2 | 5+6 2 |
| | 8 2 | 8 2 | 5+6 2 | 3+4 2 |
| | 8 2 | 8 2 | 3+4 2 | 3+4 2 |
| | 6+7 2 | 6+7 2 | 3+4 2 | 3+4 2 |
| | 4+5 2 | 4+5 2 | 5+6 2 | 5+6 2 |
| | 8 2 | 9 2 | 5+6 2 | 5+6 2 |
| | 9 2 | 9 2 | 3+4 2 | 3+4 2 |
| | 8 2 | 8 2 | 5+6 2 | 5+6 2 |
| | 9 2 | 9 2 | 5+6 2 | 3+4 2 |
| | 8 2 | 6+7 2 | 3+4 2 | 3+4 2 |
| | 4+5 2 | 8 2 | 3+4 2 | 5+6 2 |
| | 9 2 | 4+5 2 | 5+6 2 | 3+4 2 |
| | 6+7 2 | 9 2 | 5+6 2 | 5+6 2 |

Continued on next column

**TABLE LXXIII– continued**

| TQC | ICSD | | NoE | SF |
|---|---|---|---|---|
| | 8 2 | 9 2 | 3+4 2 | 3+4 2 |
| | 4+5 2 | 6+7 2 | 3+4 2 | 5+6 2 |
| | 9 2 | 8 2 | 3+4 2 | 3+4 2 |
| | 6+7 2 | 4+5 2 | 3+4 2 | 3+4 2 |
| | 8 2 | 8 2 | 3+4 2 | 3+4 2 |
| 166.1.403 | 160580 | | 44 | Ag2Ni1O2 |
| | -GM | -T | -F | -L |
| | 8 2 | 8 2 | 5+6 2 | 5+6 2 |
| | 9 2 | 9 2 | 3+4 2 | 3+4 2 |
| | 8 2 | 8 2 | 3+4 2 | 3+4 2 |
| | 9 2 | 9 2 | 5+6 2 | 5+6 2 |
| | 4+5 2 | 4+5 2 | 3+4 2 | 5+6 2 |
| | 8 2 | 8 2 | 5+6 2 | 3+4 2 |
| | 8 2 | 8 2 | 3+4 2 | 3+4 2 |
| | 6+7 2 | 6+7 2 | 3+4 2 | 3+4 2 |
| | 4+5 2 | 4+5 2 | 5+6 2 | 5+6 2 |
| | 8 2 | 9 2 | 5+6 2 | 5+6 2 |
| | 9 2 | 9 2 | 3+4 2 | 3+4 2 |
| | 8 2 | 8 2 | 5+6 2 | 5+6 2 |
| | 9 2 | 9 2 | 5+6 2 | 3+4 2 |
| | 8 2 | 6+7 2 | 3+4 2 | 3+4 2 |
| | 4+5 2 | 8 2 | 3+4 2 | 5+6 2 |
| | 9 2 | 4+5 2 | 5+6 2 | 3+4 2 |
| | 6+7 2 | 9 2 | 5+6 2 | 5+6 2 |
| | 8 2 | 9 2 | 3+4 2 | 3+4 2 |
| | 4+5 2 | 6+7 2 | 3+4 2 | 5+6 2 |
| | 9 2 | 8 2 | 3+4 2 | 3+4 2 |
| | 6+7 2 | 4+5 2 | 3+4 2 | 3+4 2 |
| | 8 2 | 8 2 | 3+4 2 | 3+4 2 |
| 166.1.404 | 160579 | | 44 | Ag2Ni1O2 |
| | -GM | -T | -F | -L |
| | 8 2 | 8 2 | 5+6 2 | 5+6 2 |
| | 9 2 | 9 2 | 3+4 2 | 3+4 2 |
| | 8 2 | 8 2 | 3+4 2 | 3+4 2 |
| | 9 2 | 9 2 | 5+6 2 | 5+6 2 |
| | 4+5 2 | 4+5 2 | 3+4 2 | 5+6 2 |
| | 8 2 | 8 2 | 5+6 2 | 3+4 2 |
| | 8 2 | 8 2 | 3+4 2 | 3+4 2 |
| | 6+7 2 | 6+7 2 | 3+4 2 | 3+4 2 |
| | 4+5 2 | 4+5 2 | 5+6 2 | 5+6 2 |
| | 8 2 | 9 2 | 5+6 2 | 5+6 2 |
| | 9 2 | 9 2 | 3+4 2 | 3+4 2 |
| | 8 2 | 8 2 | 5+6 2 | 5+6 2 |
| | 9 2 | 9 2 | 5+6 2 | 3+4 2 |
| | 8 2 | 6+7 2 | 3+4 2 | 3+4 2 |
| | 4+5 2 | 8 2 | 3+4 2 | 5+6 2 |
| | 9 2 | 4+5 2 | 5+6 2 | 3+4 2 |
| | 6+7 2 | 9 2 | 5+6 2 | 5+6 2 |
| | 8 2 | 9 2 | 3+4 2 | 3+4 2 |
| | 4+5 2 | 6+7 2 | 3+4 2 | 5+6 2 |
| | 9 2 | 8 2 | 3+4 2 | 3+4 2 |
| | 6+7 2 | 4+5 2 | 3+4 2 | 3+4 2 |
| | 8 2 | 8 2 | 3+4 2 | 3+4 2 |
| 166.1.405 | 160578 | | 44 | Ag2Ni1O2 |
| | -GM | -T | -F | -L |
| | 8 2 | 8 2 | 5+6 2 | 5+6 2 |
| | 9 2 | 9 2 | 3+4 2 | 3+4 2 |
| | 8 2 | 8 2 | 3+4 2 | 3+4 2 |
| | 9 2 | 9 2 | 5+6 2 | 5+6 2 |
| | 4+5 2 | 4+5 2 | 3+4 2 | 5+6 2 |
| | 8 2 | 8 2 | 5+6 2 | 3+4 2 |

Continued on next column

**TABLE LXXIII– continued**

| TQC | ICSD | | NoE | SF |
|---|---|---|---|---|
| | 8 2 | 8 2 | 3+4 2 | 3+4 2 |
| | 6+7 2 | 6+7 2 | 3+4 2 | 3+4 2 |
| | 4+5 2 | 4+5 2 | 5+6 2 | 5+6 2 |
| | 8 2 | 9 2 | 5+6 2 | 5+6 2 |
| | 9 2 | 9 2 | 3+4 2 | 3+4 2 |
| | 8 2 | 8 2 | 5+6 2 | 5+6 2 |
| | 9 2 | 9 2 | 5+6 2 | 3+4 2 |
| | 8 2 | 6+7 2 | 3+4 2 | 3+4 2 |
| | 4+5 2 | 8 2 | 3+4 2 | 5+6 2 |
| | 9 2 | 4+5 2 | 5+6 2 | 3+4 2 |
| | 6+7 2 | 9 2 | 5+6 2 | 5+6 2 |
| | 8 2 | 9 2 | 3+4 2 | 3+4 2 |
| | 4+5 2 | 8 2 | 3+4 2 | 5+6 2 |
| | 9 2 | 6+7 2 | 3+4 2 | 3+4 2 |
| | 6+7 2 | 4+5 2 | 3+4 2 | 3+4 2 |
| | 8 2 | 8 2 | 3+4 2 | 3+4 2 |
| 166.1.406 | 160577 | | 44 | Ag2Ni1O2 |
| | -GM | -T | -F | -L |
| | 8 2 | 8 2 | 5+6 2 | 5+6 2 |
| | 9 2 | 9 2 | 3+4 2 | 3+4 2 |
| | 8 2 | 8 2 | 3+4 2 | 3+4 2 |
| | 9 2 | 9 2 | 5+6 2 | 5+6 2 |
| | 4+5 2 | 8 2 | 3+4 2 | 5+6 2 |
| | 8 2 | 4+5 2 | 5+6 2 | 3+4 2 |
| | 8 2 | 8 2 | 3+4 2 | 3+4 2 |
| | 6+7 2 | 6+7 2 | 3+4 2 | 3+4 2 |
| | 4+5 2 | 4+5 2 | 5+6 2 | 5+6 2 |
| | 8 2 | 9 2 | 5+6 2 | 5+6 2 |
| | 9 2 | 9 2 | 3+4 2 | 3+4 2 |
| | 8 2 | 8 2 | 5+6 2 | 5+6 2 |
| | 9 2 | 9 2 | 5+6 2 | 3+4 2 |
| | 8 2 | 6+7 2 | 3+4 2 | 3+4 2 |
| | 4+5 2 | 8 2 | 3+4 2 | 5+6 2 |
| | 9 2 | 4+5 2 | 5+6 2 | 3+4 2 |
| | 6+7 2 | 9 2 | 5+6 2 | 5+6 2 |
| | 8 2 | 9 2 | 3+4 2 | 3+4 2 |
| | 4+5 2 | 8 2 | 3+4 2 | 5+6 2 |
| | 9 2 | 6+7 2 | 3+4 2 | 3+4 2 |
| | 6+7 2 | 4+5 2 | 3+4 2 | 3+4 2 |
| | 8 2 | 8 2 | 3+4 2 | 3+4 2 |
| 166.1.407 | 160576 | | 44 | Ag2Ni1O2 |
| | -GM | -T | -F | -L |
| | 8 2 | 8 2 | 5+6 2 | 5+6 2 |
| | 9 2 | 9 2 | 3+4 2 | 3+4 2 |
| | 8 2 | 8 2 | 3+4 2 | 3+4 2 |
| | 9 2 | 9 2 | 5+6 2 | 5+6 2 |
| | 4+5 2 | 4+5 2 | 3+4 2 | 5+6 2 |
| | 8 2 | 8 2 | 5+6 2 | 3+4 2 |
| | 8 2 | 8 2 | 3+4 2 | 3+4 2 |
| | 6+7 2 | 6+7 2 | 3+4 2 | 3+4 2 |
| | 4+5 2 | 4+5 2 | 5+6 2 | 5+6 2 |
| | 8 2 | 9 2 | 5+6 2 | 5+6 2 |
| | 9 2 | 9 2 | 3+4 2 | 3+4 2 |
| | 8 2 | 8 2 | 5+6 2 | 5+6 2 |
| | 9 2 | 9 2 | 5+6 2 | 3+4 2 |
| | 8 2 | 6+7 2 | 3+4 2 | 3+4 2 |
| | 4+5 2 | 8 2 | 3+4 2 | 5+6 2 |
| | 9 2 | 4+5 2 | 5+6 2 | 3+4 2 |
| | 6+7 2 | 9 2 | 5+6 2 | 5+6 2 |
| | 8 2 | 9 2 | 3+4 2 | 3+4 2 |
| | 4+5 2 | 8 2 | 3+4 2 | 5+6 2 |

Continued on next column

**TABLE LXXIII– continued**

| TQC | ICSD | NoE | SF | |
|---|---|---|---|---|
| | 9 2 | 6+7 2 | 3+4 2 | 3+4 2 |
| | 6+7 2 | 4+5 2 | 3+4 2 | 3+4 2 |
| | 8 2 | 8 2 | 3+4 2 | 3+4 2 |
| 166.1.408 | 160575 | 44 | Ag2Ni1O2 | |
| | -GM | -T | -F | -L |
| | 8 2 | 8 2 | 5+6 2 | 5+6 2 |
| | 9 2 | 9 2 | 3+4 2 | 3+4 2 |
| | 8 2 | 8 2 | 3+4 2 | 3+4 2 |
| | 9 2 | 9 2 | 5+6 2 | 5+6 2 |
| | 4+5 2 | 4+5 2 | 3+4 2 | 5+6 2 |
| | 8 2 | 8 2 | 5+6 2 | 3+4 2 |
| | 8 2 | 8 2 | 3+4 2 | 3+4 2 |
| | 6+7 2 | 6+7 2 | 3+4 2 | 3+4 2 |
| | 4+5 2 | 4+5 2 | 5+6 2 | 5+6 2 |
| | 8 2 | 9 2 | 5+6 2 | 5+6 2 |
| | 9 2 | 9 2 | 3+4 2 | 3+4 2 |
| | 8 2 | 8 2 | 5+6 2 | 5+6 2 |
| | 9 2 | 9 2 | 5+6 2 | 3+4 2 |
| | 8 2 | 6+7 2 | 3+4 2 | 3+4 2 |
| | 4+5 2 | 8 2 | 3+4 2 | 5+6 2 |
| | 9 2 | 4+5 2 | 5+6 2 | 3+4 2 |
| | 6+7 2 | 9 2 | 5+6 2 | 5+6 2 |
| | 8 2 | 9 2 | 3+4 2 | 3+4 2 |
| | 4+5 2 | 8 2 | 3+4 2 | 5+6 2 |
| | 9 2 | 6+7 2 | 3+4 2 | 3+4 2 |
| | 6+7 2 | 4+5 2 | 3+4 2 | 3+4 2 |
| | 8 2 | 8 2 | 3+4 2 | 3+4 2 |
| 166.1.409 | 160574 | 44 | Ag2Ni1O2 | |
| | -GM | -T | -F | -L |
| | 8 2 | 8 2 | 5+6 2 | 5+6 2 |
| | 9 2 | 9 2 | 3+4 2 | 3+4 2 |
| | 8 2 | 8 2 | 3+4 2 | 3+4 2 |
| | 9 2 | 9 2 | 5+6 2 | 5+6 2 |
| | 4+5 2 | 8 2 | 3+4 2 | 5+6 2 |
| | 8 2 | 4+5 2 | 5+6 2 | 3+4 2 |
| | 8 2 | 8 2 | 3+4 2 | 3+4 2 |
| | 6+7 2 | 6+7 2 | 3+4 2 | 3+4 2 |
| | 4+5 2 | 4+5 2 | 5+6 2 | 5+6 2 |
| | 8 2 | 9 2 | 5+6 2 | 5+6 2 |
| | 9 2 | 9 2 | 3+4 2 | 3+4 2 |
| | 8 2 | 8 2 | 5+6 2 | 5+6 2 |
| | 9 2 | 9 2 | 5+6 2 | 3+4 2 |
| | 8 2 | 6+7 2 | 3+4 2 | 3+4 2 |
| | 4+5 2 | 8 2 | 3+4 2 | 5+6 2 |
| | 9 2 | 4+5 2 | 5+6 2 | 3+4 2 |
| | 6+7 2 | 9 2 | 5+6 2 | 5+6 2 |
| | 8 2 | 9 2 | 3+4 2 | 3+4 2 |
| | 4+5 2 | 8 2 | 3+4 2 | 5+6 2 |
| | 9 2 | 6+7 2 | 3+4 2 | 3+4 2 |
| | 6+7 2 | 4+5 2 | 3+4 2 | 3+4 2 |
| | 8 2 | 8 2 | 3+4 2 | 3+4 2 |
| 166.1.410 | 251620 | 68 | Ag1Ge1Li2 | |
| | -GM | -T | -F | -L |
| | 8 2 | 9 2 | 3+4 2 | 5+6 2 |
| | 8 2 | 8 2 | 3+4 2 | 3+4 2 |
| | 9 2 | 9 2 | 5+6 2 | 5+6 2 |
| | 9 2 | 8 2 | 5+6 2 | 3+4 2 |
| | 4+5 2 | 4+5 2 | 5+6 2 | 3+4 2 |
| | 6+7 2 | 4+5 2 | 3+4 2 | 5+6 2 |
| | 8 2 | 9 2 | 3+4 2 | 5+6 2 |
| | 6+7 2 | 6+7 2 | 5+6 2 | 3+4 2 |

Continued on next column

**TABLE LXXIII– continued**

| TQC | ICSD | NoE | SF | |
|---|---|---|---|---|
| | 4+5 2 | 8 2 | 3+4 2 | 3+4 2 |
| | 9 2 | 6+7 2 | 5+6 2 | 5+6 2 |
| | 8 2 | 9 2 | 5+6 2 | 5+6 2 |
| | 8 2 | 8 2 | 3+4 2 | 3+4 2 |
| | 9 2 | 8 2 | 5+6 2 | 3+4 2 |
| | 9 2 | 8 2 | 3+4 2 | 5+6 2 |
| | 8 2 | 9 2 | 3+4 2 | 3+4 2 |
| | 9 2 | 9 2 | 5+6 2 | 5+6 2 |
| | 8 2 | 9 2 | 5+6 2 | 3+4 2 |
| | 6+7 2 | 6+7 2 | 3+4 2 | 5+6 2 |
| | 9 2 | 9 2 | 5+6 2 | 5+6 2 |
| | 4+5 2 | 8 2 | 3+4 2 | 3+4 2 |
| | 6+7 2 | 4+5 2 | 3+4 2 | 5+6 2 |
| | 8 2 | 6+7 2 | 5+6 2 | 3+4 2 |
| | 4+5 2 | 8 2 | 5+6 2 | 5+6 2 |
| | 9 2 | 4+5 2 | 5+6 2 | 3+4 2 |
| | 8 2 | 8 2 | 3+4 2 | 3+4 2 |
| | 8 2 | 9 2 | 3+4 2 | 5+6 2 |
| | 9 2 | 8 2 | 3+4 2 | 3+4 2 |
| | 9 2 | 4+5 2 | 5+6 2 | 5+6 2 |
| | 6+7 2 | 9 2 | 3+4 2 | 5+6 2 |
| | 8 2 | 8 2 | 5+6 2 | 3+4 2 |
| | 9 2 | 9 2 | 5+6 2 | 3+4 2 |
| | 4+5 2 | 6+7 2 | 3+4 2 | 5+6 2 |
| | 6+7 2 | 8 2 | 5+6 2 | 5+6 2 |
| | 8 2 | 4+5 2 | 3+4 2 | 3+4 2 |

The End

TABLE LXXIV: The computed irreducible representations for the valence bands for the compounds of SEBR and NLC in space group 167. See the list of tables.

| TQC | ICSD | NoE | SF | |
|---|---|---|---|---|
| 167.1.1 | 83385 | 132 | S3Yb2 | |
| | -GM | -T | -F | -L |
| | 8 2 | 5+6 4 | 3+4 2 | 2+2 4 |
| | 8 2 | 5+6 4 | 5+6 2 | 2+2 4 |
| | 9 2 | 5+6 4 | 3+4 2 | 2+2 4 |
| | 9 2 | 5+6 4 | 5+6 2 | 2+2 4 |
| | 8 2 | 4+4 4 | 5+6 2 | 2+2 4 |
| | 8 2 | 5+6 4 | 3+4 2 | 2+2 4 |
| | 9 2 | 5+6 4 | 5+6 2 | 2+2 4 |
| | 9 2 | 4+4 4 | 3+4 2 | 2+2 4 |
| | 4+5 2 | 5+6 4 | 3+4 2 | 2+2 4 |
| | 9 2 | 5+6 4 | 5+6 2 | 2+2 4 |
| | 8 2 | 4+4 4 | 5+6 2 | 2+2 4 |
| | 4+5 2 | 5+6 4 | 3+4 2 | 2+2 4 |
| | 8 2 | 5+6 4 | 5+6 2 | 2+2 4 |
| | 6+7 2 | 5+6 4 | 3+4 2 | 2+2 4 |
| | 6+7 2 | 4+4 4 | 5+6 2 | 2+2 4 |
| | 9 2 | 5+6 4 | 3+4 2 | 2+2 4 |
| | 8 2 | 4+4 4 | 5+6 2 | 2+2 4 |
| | 9 2 | 4+4 4 | 3+4 2 | 2+2 4 |
| | 6+7 2 | 5+6 4 | 5+6 2 | 2+2 4 |
| | 8 2 | 5+6 4 | 3+4 2 | 2+2 4 |
| | 9 2 | 5+6 4 | 3+4 2 | 2+2 4 |
| | 4+5 2 | 4+4 4 | 5+6 2 | 2+2 4 |
| | 8 2 | 5+6 4 | 3+4 2 | 2+2 4 |
| | 9 2 | 5+6 4 | 5+6 2 | 2+2 4 |
| | 9 2 | 5+6 4 | 5+6 2 | 2+2 4 |

Continued on next column

**TABLE LXXIV– continued**

| TQC | ICSD | | NoE | SF |
|---|---|---|---|---|
| | 8 2 | 4+4 4 | 3+4 2 | 2+2 4 |
| | 6+7 2 | 5+6 4 | 3+4 2 | 2+2 4 |
| | 4+5 2 | 4+4 4 | 5+6 2 | 2+2 4 |
| | 8 2 | 5+6 4 | 3+4 2 | 2+2 4 |
| | 8 2 | 5+6 4 | 5+6 2 | 2+2 4 |
| | 4+5 2 | 5+6 4 | 3+4 2 | 2+2 4 |
| | 8 2 | 5+6 4 | 5+6 2 | 2+2 4 |
| | 9 2 | 4+4 4 | 3+4 2 | 2+2 4 |
| | 6+7 2 | | 5+6 2 | |
| | 6+7 2 | | 3+4 2 | |
| | 9 2 | | 5+6 2 | |
| | 4+5 2 | | 3+4 2 | |
| | 8 2 | | 5+6 2 | |
| | 9 2 | | 5+6 2 | |
| | 8 2 | | 3+4 2 | |
| | 4+5 2 | | 3+4 2 | |
| | 9 2 | | 5+6 2 | |
| | 8 2 | | 5+6 2 | |
| | 6+7 2 | | 5+6 2 | |
| | 9 2 | | 3+4 2 | |
| | 8 2 | | 3+4 2 | |
| | 9 2 | | 5+6 2 | |
| | 9 2 | | 3+4 2 | |
| | 8 2 | | 5+6 2 | |
| | 4+5 2 | | 5+6 2 | |
| | 6+7 2 | | 3+4 2 | |
| | 9 2 | | 3+4 2 | |
| | 8 2 | | 5+6 2 | |
| | 8 2 | | 3+4 2 | |
| | 9 2 | | 5+6 2 | |
| | 8 2 | | 3+4 2 | |
| | 4+5 2 | | 5+6 2 | |
| | 9 2 | | 3+4 2 | |
| | 6+7 2 | | 5+6 2 | |
| | 9 2 | | 3+4 2 | |
| | 8 2 | | 3+4 2 | |
| | 9 2 | | 5+6 2 | |
| | 4+5 2 | | 5+6 2 | |
| | 8 2 | | 5+6 2 | |
| | 8 2 | | 3+4 2 | |
| | 6+7 2 | | 5+6 2 | |
| 167.1.2 | 75248 | | 132 | S3Yb2 |
| | -GM | -T | -F | -L |
| | 8 2 | 5+6 4 | 3+4 2 | 2+2 4 |
| | 8 2 | 5+6 4 | 5+6 2 | 2+2 4 |
| | 9 2 | 5+6 4 | 3+4 2 | 2+2 4 |
| | 9 2 | 5+6 4 | 5+6 2 | 2+2 4 |
| | 8 2 | 4+4 4 | 5+6 2 | 2+2 4 |
| | 8 2 | 5+6 4 | 3+4 2 | 2+2 4 |
| | 9 2 | 5+6 4 | 5+6 2 | 2+2 4 |
| | 9 2 | 4+4 4 | 3+4 2 | 2+2 4 |
| | 4+5 2 | 5+6 4 | 3+4 2 | 2+2 4 |
| | 9 2 | 5+6 4 | 5+6 2 | 2+2 4 |
| | 8 2 | 4+4 4 | 5+6 2 | 2+2 4 |
| | 4+5 2 | 5+6 4 | 3+4 2 | 2+2 4 |
| | 8 2 | 5+6 4 | 5+6 2 | 2+2 4 |
| | 6+7 2 | 4+4 4 | 3+4 2 | 2+2 4 |
| | 6+7 2 | 5+6 4 | 5+6 2 | 2+2 4 |
| | 9 2 | 5+6 4 | 3+4 2 | 2+2 4 |
| | 8 2 | 4+4 4 | 5+6 2 | 2+2 4 |
| | 9 2 | 4+4 4 | 3+4 2 | 2+2 4 |

Continued on next column

**TABLE LXXIV– continued**

| TQC | ICSD | | NoE | SF |
|---|---|---|---|---|
| | 6+7 2 | 5+6 4 | 5+6 2 | 2+2 4 |
| | 8 2 | 5+6 4 | 3+4 2 | 2+2 4 |
| | 9 2 | 5+6 4 | 3+4 2 | 2+2 4 |
| | 4+5 2 | 4+4 4 | 5+6 2 | 2+2 4 |
| | 8 2 | 5+6 4 | 3+4 2 | 2+2 4 |
| | 9 2 | 5+6 4 | 5+6 2 | 2+2 4 |
| | 9 2 | 5+6 4 | 5+6 2 | 2+2 4 |
| | 8 2 | 4+4 4 | 3+4 2 | 2+2 4 |
| | 6+7 2 | 5+6 4 | 3+4 2 | 2+2 4 |
| | 4+5 2 | 4+4 4 | 5+6 2 | 2+2 4 |
| | 8 2 | 5+6 4 | 3+4 2 | 2+2 4 |
| | 8 2 | 5+6 4 | 5+6 2 | 2+2 4 |
| | 4+5 2 | 5+6 4 | 3+4 2 | 2+2 4 |
| | 8 2 | 5+6 4 | 5+6 2 | 2+2 4 |
| | 9 2 | 4+4 4 | 5+6 2 | 2+2 4 |
| | 6+7 2 | | 3+4 2 | |
| | 6+7 2 | | 3+4 2 | |
| | 9 2 | | 5+6 2 | |
| | 4+5 2 | | 3+4 2 | |
| | 8 2 | | 5+6 2 | |
| | 9 2 | | 5+6 2 | |
| | 8 2 | | 3+4 2 | |
| | 4+5 2 | | 3+4 2 | |
| | 9 2 | | 5+6 2 | |
| | 8 2 | | 5+6 2 | |
| | 6+7 2 | | 5+6 2 | |
| | 9 2 | | 3+4 2 | |
| | 8 2 | | 3+4 2 | |
| | 9 2 | | 5+6 2 | |
| | 9 2 | | 3+4 2 | |
| | 8 2 | | 5+6 2 | |
| | 4+5 2 | | 5+6 2 | |
| | 6+7 2 | | 3+4 2 | |
| | 9 2 | | 3+4 2 | |
| | 8 2 | | 5+6 2 | |
| | 8 2 | | 3+4 2 | |
| | 9 2 | | 5+6 2 | |
| | 8 2 | | 3+4 2 | |
| | 4+5 2 | | 5+6 2 | |
| | 9 2 | | 3+4 2 | |
| | 6+7 2 | | 5+6 2 | |
| | 9 2 | | 3+4 2 | |
| | 8 2 | | 3+4 2 | |
| | 9 2 | | 5+6 2 | |
| | 4+5 2 | | 5+6 2 | |
| | 8 2 | | 5+6 2 | |
| | 8 2 | | 3+4 2 | |
| | 6+7 2 | | 5+6 2 | |
| 167.1.3 | 95761 | | 56 | O3V2 |
| | -GM | -T | -F | -L |
| | 8 2 | 5+6 4 | 5+6 2 | 2+2 4 |
| | 9 2 | 5+6 4 | 3+4 2 | 2+2 4 |
| | 6+7 2 | 4+4 4 | 5+6 2 | 2+2 4 |
| | 9 2 | 5+6 4 | 3+4 2 | 2+2 4 |
| | 8 2 | 4+4 4 | 3+4 2 | 2+2 4 |
| | 4+5 2 | 5+6 4 | 5+6 2 | 2+2 4 |
| | 8 2 | 5+6 4 | 3+4 2 | 2+2 4 |
| | 6+7 2 | 4+4 4 | 5+6 2 | 2+2 4 |
| | 9 2 | 5+6 4 | 5+6 2 | 2+2 4 |
| | 8 2 | 5+6 4 | 3+4 2 | 2+2 4 |
| | 4+5 2 | 5+6 4 | 3+4 2 | 2+2 4 |

Continued on next column

**TABLE LXXIV– continued**

| TQC | ICSD | NoE | SF | |
|---|---|---|---|---|
| | 9 2 | 4+4 4 | 3+4 2 | 2+2 4 |
| | 8 2 | 4+4 4 | 5+6 2 | 2+2 4 |
| | 4+5 2 | 5+6 4 | 5+6 2 | 2+2 4 |
| | 6+7 2 | | 5+6 2 | |
| | 9 2 | | 3+4 2 | |
| | 8 2 | | 3+4 2 | |
| | 8 2 | | 5+6 2 | |
| | 9 2 | | 5+6 2 | |
| | 8 2 | | 5+6 2 | |
| | 4+5 2 | | 3+4 2 | |
| | 9 2 | | 5+6 2 | |
| | 9 2 | | 3+4 2 | |
| | 6+7 2 | | 3+4 2 | |
| | 8 2 | | 3+4 2 | |
| | 4+5 2 | | 5+6 2 | |
| | 8 2 | | 5+6 2 | |
| | 4+5 2 | | 5+6 2 | |
| 167.1.4 | 95760 | 56 | O3V2 | |
| | -GM | -T | -F | -L |
| | 8 2 | 5+6 4 | 5+6 2 | 2+2 4 |
| | 9 2 | 5+6 4 | 3+4 2 | 2+2 4 |
| | 6+7 2 | 4+4 4 | 5+6 2 | 2+2 4 |
| | 9 2 | 5+6 4 | 3+4 2 | 2+2 4 |
| | 8 2 | 4+4 4 | 3+4 2 | 2+2 4 |
| | 4+5 2 | 5+6 4 | 5+6 2 | 2+2 4 |
| | 8 2 | 5+6 4 | 3+4 2 | 2+2 4 |
| | 6+7 2 | 4+4 4 | 5+6 2 | 2+2 4 |
| | 9 2 | 5+6 4 | 5+6 2 | 2+2 4 |
| | 9 2 | 5+6 4 | 3+4 2 | 2+2 4 |
| | 8 2 | 5+6 4 | 3+4 2 | 2+2 4 |
| | 4+5 2 | 4+4 4 | 5+6 2 | 2+2 4 |
| | 8 2 | 4+4 4 | 3+4 2 | 2+2 4 |
| | 4+5 2 | 5+6 4 | 5+6 2 | 2+2 4 |
| | 6+7 2 | | 5+6 2 | |
| | 9 2 | | 3+4 2 | |
| | 8 2 | | 3+4 2 | |
| | 8 2 | | 5+6 2 | |
| | 9 2 | | 5+6 2 | |
| | 8 2 | | 5+6 2 | |
| | 4+5 2 | | 3+4 2 | |
| | 9 2 | | 5+6 2 | |
| | 9 2 | | 3+4 2 | |
| | 6+7 2 | | 3+4 2 | |
| | 8 2 | | 3+4 2 | |
| | 4+5 2 | | 5+6 2 | |
| | 8 2 | | 5+6 2 | |
| | 4+5 2 | | 5+6 2 | |
| 167.1.5 | 94768 | 56 | O3V2 | |
| | -GM | -T | -F | -L |
| | 8 2 | 5+6 4 | 5+6 2 | 2+2 4 |
| | 9 2 | 5+6 4 | 3+4 2 | 2+2 4 |
| | 6+7 2 | 4+4 4 | 5+6 2 | 2+2 4 |
| | 9 2 | 5+6 4 | 3+4 2 | 2+2 4 |
| | 8 2 | 4+4 4 | 3+4 2 | 2+2 4 |
| | 4+5 2 | 5+6 4 | 5+6 2 | 2+2 4 |
| | 8 2 | 5+6 4 | 3+4 2 | 2+2 4 |
| | 6+7 2 | 4+4 4 | 5+6 2 | 2+2 4 |
| | 9 2 | 5+6 4 | 5+6 2 | 2+2 4 |
| | 9 2 | 5+6 4 | 3+4 2 | 2+2 4 |
| | 8 2 | 5+6 4 | 3+4 2 | 2+2 4 |
| | 4+5 2 | 4+4 4 | 5+6 2 | 2+2 4 |

Continued on next column

**TABLE LXXIV– continued**

| TQC | ICSD | NoE | SF | |
|---|---|---|---|---|
| | 4+5 2 | 4+4 4 | 3+4 2 | 2+2 4 |
| | 8 2 | 5+6 4 | 5+6 2 | 2+2 4 |
| | 6+7 2 | | 5+6 2 | |
| | 9 2 | | 3+4 2 | |
| | 8 2 | | 3+4 2 | |
| | 8 2 | | 5+6 2 | |
| | 9 2 | | 5+6 2 | |
| | 8 2 | | 5+6 2 | |
| | 4+5 2 | | 3+4 2 | |
| | 9 2 | | 5+6 2 | |
| | 9 2 | | 3+4 2 | |
| | 6+7 2 | | 3+4 2 | |
| | 8 2 | | 3+4 2 | |
| | 4+5 2 | | 5+6 2 | |
| | 8 2 | | 5+6 2 | |
| | 6+7 2 | | 5+6 2 | |
| 167.1.6 | 64796 | 56 | O3V2 | |
| | -GM | -T | -F | -L |
| | 8 2 | 5+6 4 | 5+6 2 | 2+2 4 |
| | 9 2 | 5+6 4 | 3+4 2 | 2+2 4 |
| | 6+7 2 | 4+4 4 | 5+6 2 | 2+2 4 |
| | 9 2 | 5+6 4 | 3+4 2 | 2+2 4 |
| | 8 2 | 5+6 4 | 3+4 2 | 2+2 4 |
| | 4+5 2 | 4+4 4 | 5+6 2 | 2+2 4 |
| | 8 2 | 5+6 4 | 3+4 2 | 2+2 4 |
| | 6+7 2 | 4+4 4 | 5+6 2 | 2+2 4 |
| | 9 2 | 5+6 4 | 5+6 2 | 2+2 4 |
| | 9 2 | 5+6 4 | 3+4 2 | 2+2 4 |
| | 8 2 | 5+6 4 | 3+4 2 | 2+2 4 |
| | 4+5 2 | 4+4 4 | 5+6 2 | 2+2 4 |
| | 4+5 2 | 4+4 4 | 3+4 2 | 2+2 4 |
| | 8 2 | 5+6 4 | 5+6 2 | 2+2 4 |
| | 6+7 2 | | 5+6 2 | |
| | 9 2 | | 3+4 2 | |
| | 8 2 | | 5+6 2 | |
| | 8 2 | | 3+4 2 | |
| | 9 2 | | 5+6 2 | |
| | 8 2 | | 5+6 2 | |
| | 4+5 2 | | 3+4 2 | |
| | 9 2 | | 5+6 2 | |
| | 9 2 | | 3+4 2 | |
| | 6+7 2 | | 3+4 2 | |
| | 8 2 | | 3+4 2 | |
| | 4+5 2 | | 5+6 2 | |
| | 8 2 | | 5+6 2 | |
| | 6+7 2 | | 5+6 2 | |
| 167.1.7 | 64786 | 56 | O3V2 | |
| | -GM | -T | -F | -L |
| | 8 2 | 5+6 4 | 5+6 2 | 2+2 4 |
| | 9 2 | 5+6 4 | 3+4 2 | 2+2 4 |
| | 6+7 2 | 4+4 4 | 5+6 2 | 2+2 4 |
| | 9 2 | 5+6 4 | 3+4 2 | 2+2 4 |
| | 8 2 | 5+6 4 | 3+4 2 | 2+2 4 |
| | 4+5 2 | 4+4 4 | 5+6 2 | 2+2 4 |
| | 8 2 | 5+6 4 | 3+4 2 | 2+2 4 |
| | 6+7 2 | 4+4 4 | 5+6 2 | 2+2 4 |
| | 9 2 | 5+6 4 | 5+6 2 | 2+2 4 |
| | 9 2 | 5+6 4 | 3+4 2 | 2+2 4 |
| | 8 2 | 5+6 4 | 3+4 2 | 2+2 4 |
| | 4+5 2 | 4+4 4 | 5+6 2 | 2+2 4 |
| | 4+5 2 | 4+4 4 | 3+4 2 | 2+2 4 |

Continued on next column

**TABLE LXXIV– continued**

| TQC | ICSD | NoE | SF |
|---|---|---|---|
| | 8 2 5+6 4 5+6 2 2+2 4 | | |
| | 6+7 2 5+6 2 | | |
| | 9 2 3+4 2 | | |
| | 8 2 5+6 2 | | |
| | 8 2 3+4 2 | | |
| | 9 2 5+6 2 | | |
| | 8 2 5+6 2 | | |
| | 4+5 2 3+4 2 | | |
| | 9 2 5+6 2 | | |
| | 9 2 3+4 2 | | |
| | 6+7 2 3+4 2 | | |
| | 8 2 3+4 2 | | |
| | 4+5 2 5+6 2 | | |
| | 8 2 5+6 2 | | |
| | 6+7 2 5+6 2 | | |
| 167.1.8 | 25780 | 56 | O3V2 |
| | -GM -T -F -L | | |
| | 8 2 5+6 4 5+6 2 2+2 4 | | |
| | 9 2 5+6 4 3+4 2 2+2 4 | | |
| | 6+7 2 4+4 4 5+6 2 2+2 4 | | |
| | 9 2 5+6 4 3+4 2 2+2 4 | | |
| | 8 2 4+4 4 3+4 2 2+2 4 | | |
| | 4+5 2 5+6 4 5+6 2 2+2 4 | | |
| | 8 2 5+6 4 3+4 2 2+2 4 | | |
| | 6+7 2 4+4 4 5+6 2 2+2 4 | | |
| | 9 2 5+6 4 5+6 2 2+2 4 | | |
| | 9 2 5+6 4 3+4 2 2+2 4 | | |
| | 8 2 5+6 4 3+4 2 2+2 4 | | |
| | 4+5 2 4+4 4 3+4 2 2+2 4 | | |
| | 8 2 4+4 4 5+6 2 2+2 4 | | |
| | 4+5 2 5+6 4 5+6 2 2+2 4 | | |
| | 6+7 2 5+6 2 | | |
| | 9 2 3+4 2 | | |
| | 8 2 3+4 2 | | |
| | 8 2 5+6 2 | | |
| | 9 2 5+6 2 | | |
| | 8 2 5+6 2 | | |
| | 4+5 2 3+4 2 | | |
| | 9 2 5+6 2 | | |
| | 9 2 3+4 2 | | |
| | 6+7 2 3+4 2 | | |
| | 8 2 3+4 2 | | |
| | 4+5 2 5+6 2 | | |
| | 8 2 5+6 2 | | |
| | 4+5 2 5+6 2 | | |
| 167.1.9 | 201109 | 56 | O3V2 |
| | -GM -T -F -L | | |
| | 8 2 5+6 4 5+6 2 2+2 4 | | |
| | 9 2 5+6 4 3+4 2 2+2 4 | | |
| | 6+7 2 4+4 4 5+6 2 2+2 4 | | |
| | 9 2 5+6 4 3+4 2 2+2 4 | | |
| | 8 2 4+4 4 3+4 2 2+2 4 | | |
| | 4+5 2 5+6 4 5+6 2 2+2 4 | | |
| | 8 2 5+6 4 3+4 2 2+2 4 | | |
| | 6+7 2 4+4 4 5+6 2 2+2 4 | | |
| | 9 2 5+6 4 5+6 2 2+2 4 | | |
| | 8 2 5+6 4 3+4 2 2+2 4 | | |
| | 4+5 2 5+6 4 3+4 2 2+2 4 | | |
| | 9 2 4+4 4 3+4 2 2+2 4 | | |
| | 8 2 4+4 4 5+6 2 2+2 4 | | |
| | 4+5 2 5+6 4 5+6 2 2+2 4 | | |

Continued on next column

**TABLE LXXIV– continued**

| TQC | ICSD | NoE | SF |
|---|---|---|---|
| | 6+7 2 5+6 2 | | |
| | 9 2 3+4 2 | | |
| | 8 2 3+4 2 | | |
| | 8 2 5+6 2 | | |
| | 9 2 5+6 2 | | |
| | 8 2 5+6 2 | | |
| | 4+5 2 3+4 2 | | |
| | 9 2 5+6 2 | | |
| | 9 2 3+4 2 | | |
| | 6+7 2 3+4 2 | | |
| | 8 2 3+4 2 | | |
| | 4+5 2 5+6 2 | | |
| | 8 2 5+6 2 | | |
| | 4+5 2 5+6 2 | | |
| 167.1.10 | 201108 | 56 | O3V2 |
| | -GM -T -F -L | | |
| | 8 2 5+6 4 5+6 2 2+2 4 | | |
| | 9 2 5+6 4 3+4 2 2+2 4 | | |
| | 6+7 2 4+4 4 5+6 2 2+2 4 | | |
| | 9 2 5+6 4 3+4 2 2+2 4 | | |
| | 8 2 4+4 4 3+4 2 2+2 4 | | |
| | 4+5 2 5+6 4 5+6 2 2+2 4 | | |
| | 8 2 5+6 4 3+4 2 2+2 4 | | |
| | 6+7 2 4+4 4 5+6 2 2+2 4 | | |
| | 9 2 5+6 4 5+6 2 2+2 4 | | |
| | 9 2 5+6 4 3+4 2 2+2 4 | | |
| | 8 2 5+6 4 3+4 2 2+2 4 | | |
| | 4+5 2 4+4 4 3+4 2 2+2 4 | | |
| | 8 2 4+4 4 5+6 2 2+2 4 | | |
| | 4+5 2 5+6 4 5+6 2 2+2 4 | | |
| | 6+7 2 5+6 2 | | |
| | 9 2 3+4 2 | | |
| | 8 2 3+4 2 | | |
| | 8 2 5+6 2 | | |
| | 9 2 5+6 2 | | |
| | 8 2 5+6 2 | | |
| | 4+5 2 3+4 2 | | |
| | 9 2 5+6 2 | | |
| | 9 2 3+4 2 | | |
| | 6+7 2 3+4 2 | | |
| | 8 2 3+4 2 | | |
| | 4+5 2 5+6 2 | | |
| | 8 2 5+6 2 | | |
| | 6+7 2 5+6 2 | | |
| 167.1.11 | 201107 | 56 | O3V2 |
| | -GM -T -F -L | | |
| | 8 2 5+6 4 5+6 2 2+2 4 | | |
| | 9 2 5+6 4 3+4 2 2+2 4 | | |
| | 6+7 2 4+4 4 5+6 2 2+2 4 | | |
| | 9 2 5+6 4 3+4 2 2+2 4 | | |
| | 8 2 4+4 4 3+4 2 2+2 4 | | |
| | 4+5 2 5+6 4 5+6 2 2+2 4 | | |
| | 8 2 5+6 4 3+4 2 2+2 4 | | |
| | 6+7 2 4+4 4 5+6 2 2+2 4 | | |
| | 9 2 5+6 4 5+6 2 2+2 4 | | |
| | 9 2 5+6 4 3+4 2 2+2 4 | | |
| | 8 2 5+6 4 3+4 2 2+2 4 | | |
| | 4+5 2 4+4 4 3+4 2 2+2 4 | | |
| | 4+5 2 4+4 4 5+6 2 2+2 4 | | |
| | 8 2 5+6 4 5+6 2 2+2 4 | | |
| | 6+7 2 5+6 2 | | |

Continued on next column

**TABLE LXXIV– continued**

| TQC | ICSD | NoE | SF | |
|---|---|---|---|---|
| | 9 2 | | 3+4 2 | |
| | 8 2 | | 3+4 2 | |
| | 8 2 | | 5+6 2 | |
| | 9 2 | | 5+6 2 | |
| | 8 2 | | 5+6 2 | |
| | 4+5 2 | | 3+4 2 | |
| | 9 2 | | 5+6 2 | |
| | 9 2 | | 3+4 2 | |
| | 6+7 2 | | 3+4 2 | |
| | 8 2 | | 3+4 2 | |
| | 4+5 2 | | 5+6 2 | |
| | 8 2 | | 5+6 2 | |
| | 6+7 2 | | 5+6 2 | |
| 167.1.12 | 201106 | 56 | O3V2 | |
| | -GM | -T | -F | -L |
| | 8 2 | 5+6 4 | 5+6 2 | 2+2 4 |
| | 9 2 | 5+6 4 | 3+4 2 | 2+2 4 |
| | 6+7 2 | 4+4 4 | 5+6 2 | 2+2 4 |
| | 9 2 | 5+6 4 | 3+4 2 | 2+2 4 |
| | 8 2 | 5+6 4 | 3+4 2 | 2+2 4 |
| | 4+5 2 | 4+4 4 | 5+6 2 | 2+2 4 |
| | 8 2 | 5+6 4 | 3+4 2 | 2+2 4 |
| | 6+7 2 | 4+4 4 | 5+6 2 | 2+2 4 |
| | 9 2 | 5+6 4 | 5+6 2 | 2+2 4 |
| | 9 2 | 5+6 4 | 3+4 2 | 2+2 4 |
| | 8 2 | 5+6 4 | 3+4 2 | 2+2 4 |
| | 4+5 2 | 4+4 4 | 5+6 2 | 2+2 4 |
| | 4+5 2 | 4+4 4 | 3+4 2 | 2+2 4 |
| | 8 2 | 5+6 4 | 5+6 2 | 2+2 4 |
| | 8 2 | | 5+6 2 | |
| | 6+7 2 | | 3+4 2 | |
| | 9 2 | | 5+6 2 | |
| | 8 2 | | 3+4 2 | |
| | 9 2 | | 5+6 2 | |
| | 8 2 | | 5+6 2 | |
| | 4+5 2 | | 3+4 2 | |
| | 9 2 | | 5+6 2 | |
| | 9 2 | | 3+4 2 | |
| | 6+7 2 | | 3+4 2 | |
| | 8 2 | | 3+4 2 | |
| | 4+5 2 | | 5+6 2 | |
| | 8 2 | | 5+6 2 | |
| | 6+7 2 | | 5+6 2 | |
| 167.1.13 | 1874 | 56 | O3V2 | |
| | -GM | -T | -F | -L |
| | 8 2 | 5+6 4 | 5+6 2 | 2+2 4 |
| | 9 2 | 5+6 4 | 3+4 2 | 2+2 4 |
| | 6+7 2 | 4+4 4 | 5+6 2 | 2+2 4 |
| | 9 2 | 5+6 4 | 3+4 2 | 2+2 4 |
| | 8 2 | 5+6 4 | 3+4 2 | 2+2 4 |
| | 4+5 2 | 4+4 4 | 5+6 2 | 2+2 4 |
| | 8 2 | 5+6 4 | 3+4 2 | 2+2 4 |
| | 6+7 2 | 4+4 4 | 5+6 2 | 2+2 4 |
| | 9 2 | 5+6 4 | 5+6 2 | 2+2 4 |
| | 9 2 | 5+6 4 | 3+4 2 | 2+2 4 |
| | 8 2 | 5+6 4 | 3+4 2 | 2+2 4 |
| | 4+5 2 | 4+4 4 | 5+6 2 | 2+2 4 |
| | 8 2 | 4+4 4 | 3+4 2 | 2+2 4 |
| | 4+5 2 | 5+6 4 | 5+6 2 | 2+2 4 |
| | 8 2 | | 5+6 2 | |
| | 6+7 2 | | 3+4 2 | |

Continued on next column

**TABLE LXXIV– continued**

| TQC | ICSD | NoE | SF | |
|---|---|---|---|---|
| | 9 2 | | 5+6 2 | |
| | 8 2 | | 3+4 2 | |
| | 9 2 | | 5+6 2 | |
| | 8 2 | | 5+6 2 | |
| | 4+5 2 | | 3+4 2 | |
| | 9 2 | | 5+6 2 | |
| | 9 2 | | 3+4 2 | |
| | 6+7 2 | | 3+4 2 | |
| | 8 2 | | 3+4 2 | |
| | 4+5 2 | | 5+6 2 | |
| | 8 2 | | 5+6 2 | |
| | 4+5 2 | | 5+6 2 | |
| 167.1.14 | 1873 | 56 | O3V2 | |
| | -GM | -T | -F | -L |
| | 8 2 | 5+6 4 | 5+6 2 | 2+2 4 |
| | 9 2 | 5+6 4 | 3+4 2 | 2+2 4 |
| | 6+7 2 | 4+4 4 | 5+6 2 | 2+2 4 |
| | 9 2 | 5+6 4 | 3+4 2 | 2+2 4 |
| | 8 2 | 5+6 4 | 3+4 2 | 2+2 4 |
| | 4+5 2 | 4+4 4 | 5+6 2 | 2+2 4 |
| | 8 2 | 5+6 4 | 3+4 2 | 2+2 4 |
| | 6+7 2 | 4+4 4 | 5+6 2 | 2+2 4 |
| | 9 2 | 5+6 4 | 5+6 2 | 2+2 4 |
| | 9 2 | 5+6 4 | 3+4 2 | 2+2 4 |
| | 8 2 | 5+6 4 | 3+4 2 | 2+2 4 |
| | 4+5 2 | 4+4 4 | 5+6 2 | 2+2 4 |
| | 8 2 | 4+4 4 | 3+4 2 | 2+2 4 |
| | 4+5 2 | 5+6 4 | 5+6 2 | 2+2 4 |
| | 8 2 | | 5+6 2 | |
| | 6+7 2 | | 3+4 2 | |
| | 9 2 | | 5+6 2 | |
| | 8 2 | | 3+4 2 | |
| | 9 2 | | 5+6 2 | |
| | 8 2 | | 5+6 2 | |
| | 4+5 2 | | 3+4 2 | |
| | 9 2 | | 5+6 2 | |
| | 9 2 | | 3+4 2 | |
| | 6+7 2 | | 3+4 2 | |
| | 8 2 | | 3+4 2 | |
| | 4+5 2 | | 5+6 2 | |
| | 8 2 | | 5+6 2 | |
| | 4+5 2 | | 5+6 2 | |
| 167.1.15 | 1872 | 56 | O3V2 | |
| | -GM | -T | -F | -L |
| | 8 2 | 5+6 4 | 5+6 2 | 2+2 4 |
| | 9 2 | 5+6 4 | 3+4 2 | 2+2 4 |
| | 6+7 2 | 4+4 4 | 5+6 2 | 2+2 4 |
| | 9 2 | 5+6 4 | 3+4 2 | 2+2 4 |
| | 8 2 | 5+6 4 | 3+4 2 | 2+2 4 |
| | 4+5 2 | 4+4 4 | 5+6 2 | 2+2 4 |
| | 8 2 | 5+6 4 | 3+4 2 | 2+2 4 |
| | 6+7 2 | 4+4 4 | 5+6 2 | 2+2 4 |
| | 9 2 | 5+6 4 | 5+6 2 | 2+2 4 |
| | 9 2 | 5+6 4 | 3+4 2 | 2+2 4 |
| | 8 2 | 5+6 4 | 3+4 2 | 2+2 4 |
| | 4+5 2 | 4+4 4 | 5+6 2 | 2+2 4 |
| | 8 2 | 4+4 4 | 3+4 2 | 2+2 4 |
| | 4+5 2 | 5+6 4 | 5+6 2 | 2+2 4 |
| | 8 2 | | 5+6 2 | |
| | 6+7 2 | | 3+4 2 | |
| | 9 2 | | 5+6 2 | |

Continued on next column

**TABLE LXXIV– continued**

| TQC | ICSD | NoE | SF | |
|---|---|---|---|---|
| | 8 2 | | 3+4 2 | |
| | 9 2 | | 5+6 2 | |
| | 8 2 | | 5+6 2 | |
| | 4+5 2 | | 3+4 2 | |
| | 9 2 | | 5+6 2 | |
| | 9 2 | | 3+4 2 | |
| | 6+7 2 | | 3+4 2 | |
| | 8 2 | | 3+4 2 | |
| | 4+5 2 | | 5+6 2 | |
| | 8 2 | | 5+6 2 | |
| | 4+5 2 | | 5+6 2 | |
| 167.1.16 | 1870 | 56 | O3V2 | |
| | -GM | -T | -F | -L |
| | 8 2 | 5+6 4 | 5+6 2 | 2+2 4 |
| | 9 2 | 5+6 4 | 3+4 2 | 2+2 4 |
| | 6+7 2 | 4+4 4 | 5+6 2 | 2+2 4 |
| | 9 2 | 5+6 4 | 3+4 2 | 2+2 4 |
| | 8 2 | 5+6 4 | 3+4 2 | 2+2 4 |
| | 4+5 2 | 4+4 4 | 5+6 2 | 2+2 4 |
| | 8 2 | 5+6 4 | 3+4 2 | 2+2 4 |
| | 6+7 2 | 4+4 4 | 5+6 2 | 2+2 4 |
| | 9 2 | 5+6 4 | 5+6 2 | 2+2 4 |
| | 9 2 | 5+6 4 | 3+4 2 | 2+2 4 |
| | 8 2 | 5+6 4 | 3+4 2 | 2+2 4 |
| | 4+5 2 | 4+4 4 | 5+6 2 | 2+2 4 |
| | 4+5 2 | 4+4 4 | 3+4 2 | 2+2 4 |
| | 8 2 | 5+6 4 | 5+6 2 | 2+2 4 |
| | 6+7 2 | | 5+6 2 | |
| | 9 2 | | 3+4 2 | |
| | 8 2 | | 5+6 2 | |
| | 8 2 | | 3+4 2 | |
| | 9 2 | | 5+6 2 | |
| | 8 2 | | 5+6 2 | |
| | 4+5 2 | | 3+4 2 | |
| | 9 2 | | 5+6 2 | |
| | 9 2 | | 3+4 2 | |
| | 6+7 2 | | 3+4 2 | |
| | 8 2 | | 3+4 2 | |
| | 4+5 2 | | 5+6 2 | |
| | 8 2 | | 5+6 2 | |
| | 6+7 2 | | 5+6 2 | |
| 167.1.17 | 1869 | 56 | O3V2 | |
| | -GM | -T | -F | -L |
| | 8 2 | 5+6 4 | 5+6 2 | 2+2 4 |
| | 9 2 | 5+6 4 | 3+4 2 | 2+2 4 |
| | 6+7 2 | 4+4 4 | 5+6 2 | 2+2 4 |
| | 9 2 | 5+6 4 | 3+4 2 | 2+2 4 |
| | 8 2 | 4+4 4 | 3+4 2 | 2+2 4 |
| | 4+5 2 | 5+6 4 | 5+6 2 | 2+2 4 |
| | 8 2 | 5+6 4 | 3+4 2 | 2+2 4 |
| | 6+7 2 | 4+4 4 | 5+6 2 | 2+2 4 |
| | 9 2 | 5+6 4 | 5+6 2 | 2+2 4 |
| | 9 2 | 5+6 4 | 3+4 2 | 2+2 4 |
| | 8 2 | 5+6 4 | 3+4 2 | 2+2 4 |
| | 4+5 2 | 4+4 4 | 5+6 2 | 2+2 4 |
| | 4+5 2 | 4+4 4 | 3+4 2 | 2+2 4 |
| | 8 2 | 5+6 4 | 5+6 2 | 2+2 4 |
| | 6+7 2 | | 5+6 2 | |
| | 9 2 | | 3+4 2 | |
| | 8 2 | | 3+4 2 | |
| | 8 2 | | 5+6 2 | |

Continued on next column

**TABLE LXXIV– continued**

| TQC | ICSD | NoE | SF | |
|---|---|---|---|---|
| | 9 2 | | 5+6 2 | |
| | 8 2 | | 5+6 2 | |
| | 4+5 2 | | 3+4 2 | |
| | 9 2 | | 5+6 2 | |
| | 9 2 | | 3+4 2 | |
| | 6+7 2 | | 3+4 2 | |
| | 8 2 | | 3+4 2 | |
| | 4+5 2 | | 5+6 2 | |
| | 8 2 | | 5+6 2 | |
| | 6+7 2 | | 5+6 2 | |
| 167.1.18 | 1473 | 56 | O3V2 | |
| | -GM | -T | -F | -L |
| | 8 2 | 5+6 4 | 5+6 2 | 2+2 4 |
| | 9 2 | 5+6 4 | 3+4 2 | 2+2 4 |
| | 6+7 2 | 4+4 4 | 5+6 2 | 2+2 4 |
| | 9 2 | 5+6 4 | 3+4 2 | 2+2 4 |
| | 8 2 | 4+4 4 | 3+4 2 | 2+2 4 |
| | 4+5 2 | 5+6 4 | 5+6 2 | 2+2 4 |
| | 8 2 | 5+6 4 | 3+4 2 | 2+2 4 |
| | 6+7 2 | 4+4 4 | 5+6 2 | 2+2 4 |
| | 9 2 | 5+6 4 | 5+6 2 | 2+2 4 |
| | 9 2 | 5+6 4 | 3+4 2 | 2+2 4 |
| | 8 2 | 5+6 4 | 3+4 2 | 2+2 4 |
| | 4+5 2 | 4+4 4 | 5+6 2 | 2+2 4 |
| | 4+5 2 | 4+4 4 | 3+4 2 | 2+2 4 |
| | 8 2 | 5+6 4 | 5+6 2 | 2+2 4 |
| | 6+7 2 | | 5+6 2 | |
| | 9 2 | | 3+4 2 | |
| | 8 2 | | 3+4 2 | |
| | 8 2 | | 5+6 2 | |
| | 9 2 | | 5+6 2 | |
| | 8 2 | | 5+6 2 | |
| | 4+5 2 | | 3+4 2 | |
| | 9 2 | | 5+6 2 | |
| | 9 2 | | 3+4 2 | |
| | 6+7 2 | | 3+4 2 | |
| | 8 2 | | 3+4 2 | |
| | 4+5 2 | | 5+6 2 | |
| | 8 2 | | 5+6 2 | |
| | 6+7 2 | | 5+6 2 | |
| 167.1.19 | 167061 | 72 | La1Mn1O3.31 | |
| | -GM | -T | -F | -L |
| | 8 2 | 5+6 4 | 5+6 2 | 2+2 4 |
| | 9 2 | 5+6 4 | 3+4 2 | 2+2 4 |
| | 8 2 | 5+6 4 | 3+4 2 | 2+2 4 |
| | 8 2 | 4+4 4 | 3+4 2 | 2+2 4 |
| | 4+5 2 | 5+6 4 | 5+6 2 | 2+2 4 |
| | 9 2 | 4+4 4 | 5+6 2 | 2+2 4 |
| | 6+7 2 | 5+6 4 | 3+4 2 | 2+2 4 |
| | 9 2 | 4+4 4 | 5+6 2 | 2+2 4 |
| | 8 2 | 5+6 4 | 3+4 2 | 2+2 4 |
| | 9 2 | 5+6 4 | 5+6 2 | 2+2 4 |
| | 8 2 | 5+6 4 | 3+4 2 | 2+2 4 |
| | 4+5 2 | 4+4 4 | 5+6 2 | 2+2 4 |
| | 6+7 2 | 5+6 4 | 5+6 2 | 2+2 4 |
| | 9 2 | 5+6 4 | 3+4 2 | 2+2 4 |
| | 4+5 2 | 5+6 4 | 3+4 2 | 2+2 4 |
| | 8 2 | 4+4 4 | 3+4 2 | 2+2 4 |
| | 8 2 | 5+6 4 | 3+4 2 | 2+2 4 |
| | 4+5 2 | 4+4 4 | 3+4 2 | 2+2 4 |
| | 8 2 | | 3+4 2 | |

Continued on next column

**TABLE LXXIV– continued**

| TQC | ICSD | | NoE | SF |
|---|---|---|---|---|
| | 8 2 | | 3+4 2 | |
| | 4+5 2 | | 3+4 2 | |
| | 8 2 | | 3+4 2 | |
| | 8 2 | | 3+4 2 | |
| | 4+5 2 | | 3+4 2 | |
| | 8 2 | | 5+6 2 | |
| | 9 2 | | 5+6 2 | |
| | 6+7 2 | | 5+6 2 | |
| | 9 2 | | 3+4 2 | |
| | 9 2 | | 3+4 2 | |
| | 8 2 | | 5+6 2 | |
| | 6+7 2 | | 5+6 2 | |
| | 9 2 | | 3+4 2 | |
| | 8 2 | | 3+4 2 | |
| | 4+5 2 | | 5+6 2 | |
| | 9 2 | | 5+6 2 | |
| | 6+7 2 | | 3+4 2 | |
| 167.1.20 | 167062 | | 72 | La1Mn1O3.29 |
| | -GM | -T | -F | -L |
| | 8 2 | 5+6 4 | 5+6 2 | 2+2 4 |
| | 9 2 | 5+6 4 | 3+4 2 | 2+2 4 |
| | 8 2 | 5+6 4 | 3+4 2 | 2+2 4 |
| | 8 2 | 4+4 4 | 3+4 2 | 2+2 4 |
| | 4+5 2 | 5+6 4 | 5+6 2 | 2+2 4 |
| | 9 2 | 4+4 4 | 5+6 2 | 2+2 4 |
| | 6+7 2 | 5+6 4 | 3+4 2 | 2+2 4 |
| | 9 2 | 4+4 4 | 5+6 2 | 2+2 4 |
| | 8 2 | 5+6 4 | 3+4 2 | 2+2 4 |
| | 9 2 | 5+6 4 | 5+6 2 | 2+2 4 |
| | 8 2 | 4+4 4 | 3+4 2 | 2+2 4 |
| | 4+5 2 | 5+6 4 | 5+6 2 | 2+2 4 |
| | 6+7 2 | 5+6 4 | 3+4 2 | 2+2 4 |
| | 9 2 | 5+6 4 | 5+6 2 | 2+2 4 |
| | 4+5 2 | 5+6 4 | 3+4 2 | 2+2 4 |
| | 8 2 | 4+4 4 | 3+4 2 | 2+2 4 |
| | 8 2 | 5+6 4 | 3+4 2 | 2+2 4 |
| | 4+5 2 | 4+4 4 | 3+4 2 | 2+2 4 |
| | 8 2 | | 3+4 2 | |
| | 8 2 | | 3+4 2 | |
| | 4+5 2 | | 3+4 2 | |
| | 8 2 | | 3+4 2 | |
| | 8 2 | | 3+4 2 | |
| | 4+5 2 | | 3+4 2 | |
| | 8 2 | | 5+6 2 | |
| | 9 2 | | 5+6 2 | |
| | 6+7 2 | | 3+4 2 | |
| | 9 2 | | 5+6 2 | |
| | 9 2 | | 3+4 2 | |
| | 8 2 | | 3+4 2 | |
| | 6+7 2 | | 5+6 2 | |
| | 9 2 | | 5+6 2 | |
| | 4+5 2 | | 3+4 2 | |
| | 8 2 | | 5+6 2 | |
| | 9 2 | | 5+6 2 | |
| | 6+7 2 | | 3+4 2 | |
| 167.1.21 | 167065 | | 72 | La1Mn1O3.26 |
| | -GM | -T | -F | -L |
| | 8 2 | 5+6 4 | 5+6 2 | 2+2 4 |
| | 9 2 | 5+6 4 | 3+4 2 | 2+2 4 |
| | 8 2 | 5+6 4 | 3+4 2 | 2+2 4 |
| | 8 2 | 4+4 4 | 3+4 2 | 2+2 4 |
| | 4+5 2 | 5+6 4 | 5+6 2 | 2+2 4 |
| | 9 2 | 4+4 4 | 5+6 2 | 2+2 4 |
| | 6+7 2 | 5+6 4 | 3+4 2 | 2+2 4 |
| | 9 2 | 4+4 4 | 5+6 2 | 2+2 4 |
| | 8 2 | 5+6 4 | 3+4 2 | 2+2 4 |
| | 9 2 | 5+6 4 | 5+6 2 | 2+2 4 |
| | 8 2 | 4+4 4 | 3+4 2 | 2+2 4 |
| | 4+5 2 | 5+6 4 | 5+6 2 | 2+2 4 |
| | 9 2 | 5+6 4 | 3+4 2 | 2+2 4 |
| | 6+7 2 | 5+6 4 | 5+6 2 | 2+2 4 |
| | 4+5 2 | 5+6 4 | 3+4 2 | 2+2 4 |
| | 8 2 | 4+4 4 | 3+4 2 | 2+2 4 |
| | 8 2 | 5+6 4 | 3+4 2 | 2+2 4 |
| | 4+5 2 | 4+4 4 | 3+4 2 | 2+2 4 |
| | 8 2 | | 3+4 2 | |
| | 8 2 | | 3+4 2 | |
| | 4+5 2 | | 3+4 2 | |
| | 8 2 | | 3+4 2 | |
| | 8 2 | | 3+4 2 | |
| | 8 2 | | 3+4 2 | |
| | 4+5 2 | | 5+6 2 | |
| | 9 2 | | 5+6 2 | |
| | 6+7 2 | | 3+4 2 | |
| | 9 2 | | 5+6 2 | |
| | 9 2 | | 3+4 2 | |
| | 8 2 | | 5+6 2 | |
| | 6+7 2 | | 3+4 2 | |
| | 9 2 | | 5+6 2 | |
| | 4+5 2 | | 3+4 2 | |
| | 8 2 | | 5+6 2 | |
| | 9 2 | | 5+6 2 | |
| | 6+7 2 | | 3+4 2 | |
| 167.1.22 | 167064 | | 72 | La1Mn1O3.26 |
| | -GM | -T | -F | -L |
| | 8 2 | 5+6 4 | 5+6 2 | 2+2 4 |
| | 9 2 | 5+6 4 | 3+4 2 | 2+2 4 |
| | 8 2 | 5+6 4 | 3+4 2 | 2+2 4 |
| | 8 2 | 4+4 4 | 3+4 2 | 2+2 4 |
| | 4+5 2 | 5+6 4 | 3+4 2 | 2+2 4 |
| | 9 2 | 4+4 4 | 5+6 2 | 2+2 4 |
| | 6+7 2 | 5+6 4 | 5+6 2 | 2+2 4 |
| | 9 2 | 5+6 4 | 5+6 2 | 2+2 4 |
| | 8 2 | 4+4 4 | 3+4 2 | 2+2 4 |
| | 4+5 2 | 5+6 4 | 3+4 2 | 2+2 4 |
| | 8 2 | 4+4 4 | 3+4 2 | 2+2 4 |
| | 4+5 2 | 5+6 4 | 3+4 2 | 2+2 4 |
| | 8 2 | 5+6 4 | 5+6 2 | 2+2 4 |
| | 9 2 | 5+6 4 | 3+4 2 | 2+2 4 |
| | 4+5 2 | 4+4 4 | 3+4 2 | 2+2 4 |
| | 8 2 | 5+6 4 | 3+4 2 | 2+2 4 |
| | 8 2 | 5+6 4 | 3+4 2 | 2+2 4 |
| | 8 2 | 4+4 4 | 3+4 2 | 2+2 4 |
| | 4+5 2 | | 3+4 2 | |
| | 8 2 | | 3+4 2 | |
| | 6+7 2 | | 5+6 2 | |
| | 9 2 | | 5+6 2 | |
| | 8 2 | | 3+4 2 | |
| | 4+5 2 | | 3+4 2 | |
| | 8 2 | | 3+4 2 | |
| | 9 2 | | 5+6 2 | |
| | 6+7 2 | | 5+6 2 | |

Continued on next column

**TABLE LXXIV– continued**

| TQC | ICSD | | NoE | SF |
|---|---|---|---|---|
| | 9 2 | | 5+6 2 | |
| | 8 2 | | 3+4 2 | |
| | 9 2 | | 3+4 2 | |
| | 6+7 2 | | 5+6 2 | |
| | 9 2 | | 5+6 2 | |
| | 4+5 2 | | 3+4 2 | |
| | 8 2 | | 3+4 2 | |
| | 9 2 | | 5+6 2 | |
| | 6+7 2 | | 5+6 2 | |
| 167.1.23 | 167066 | | 72 | La1Mn1O3.22 |
| | -GM | -T | -F | -L |
| | 8 2 | 5+6 4 | 5+6 2 | 2+2 4 |
| | 9 2 | 5+6 4 | 3+4 2 | 2+2 4 |
| | 8 2 | 5+6 4 | 3+4 2 | 2+2 4 |
| | 8 2 | 4+4 4 | 3+4 2 | 2+2 4 |
| | 4+5 2 | 5+6 4 | 5+6 2 | 2+2 4 |
| | 9 2 | 4+4 4 | 5+6 2 | 2+2 4 |
| | 6+7 2 | 5+6 4 | 3+4 2 | 2+2 4 |
| | 9 2 | 4+4 4 | 5+6 2 | 2+2 4 |
| | 8 2 | 5+6 4 | 3+4 2 | 2+2 4 |
| | 9 2 | 5+6 4 | 5+6 2 | 2+2 4 |
| | 8 2 | 5+6 4 | 3+4 2 | 2+2 4 |
| | 4+5 2 | 4+4 4 | 3+4 2 | 2+2 4 |
| | 4+5 2 | 5+6 4 | 3+4 2 | 2+2 4 |
| | 8 2 | 5+6 4 | 5+6 2 | 2+2 4 |
| | 6+7 2 | 5+6 4 | 5+6 2 | 2+2 4 |
| | 9 2 | 4+4 4 | 3+4 2 | 2+2 4 |
| | 8 2 | 5+6 4 | 3+4 2 | 2+2 4 |
| | 4+5 2 | 4+4 4 | 3+4 2 | 2+2 4 |
| | 8 2 | | 3+4 2 | |
| | 8 2 | | 3+4 2 | |
| | 4+5 2 | | 3+4 2 | |
| | 8 2 | | 3+4 2 | |
| | 8 2 | | 3+4 2 | |
| | 8 2 | | 3+4 2 | |
| | 4+5 2 | | 5+6 2 | |
| | 9 2 | | 5+6 2 | |
| | 6+7 2 | | 3+4 2 | |
| | 9 2 | | 5+6 2 | |
| | 8 2 | | 3+4 2 | |
| | 9 2 | | 3+4 2 | |
| | 6+7 2 | | 5+6 2 | |
| | 9 2 | | 5+6 2 | |
| | 4+5 2 | | 3+4 2 | |
| | 8 2 | | 3+4 2 | |
| | 9 2 | | 5+6 2 | |
| | 6+7 2 | | 5+6 2 | |
| 167.1.24 | 167068 | | 72 | La1Mn1O3.15 |
| | -GM | -T | -F | -L |
| | 8 2 | 5+6 4 | 5+6 2 | 2+2 4 |
| | 9 2 | 5+6 4 | 3+4 2 | 2+2 4 |
| | 8 2 | 5+6 4 | 3+4 2 | 2+2 4 |
| | 8 2 | 4+4 4 | 3+4 2 | 2+2 4 |
| | 4+5 2 | 5+6 4 | 5+6 2 | 2+2 4 |
| | 9 2 | 4+4 4 | 5+6 2 | 2+2 4 |
| | 6+7 2 | 5+6 4 | 3+4 2 | 2+2 4 |
| | 9 2 | 4+4 4 | 5+6 2 | 2+2 4 |
| | 8 2 | 5+6 4 | 3+4 2 | 2+2 4 |
| | 9 2 | 5+6 4 | 5+6 2 | 2+2 4 |
| | 8 2 | 4+4 4 | 3+4 2 | 2+2 4 |
| | 4+5 2 | 5+6 4 | 5+6 2 | 2+2 4 |
| | 9 2 | 5+6 4 | 3+4 2 | 2+2 4 |
| | 6+7 2 | 5+6 4 | 5+6 2 | 2+2 4 |
| | 4+5 2 | 5+6 4 | 3+4 2 | 2+2 4 |
| | 8 2 | 4+4 4 | 3+4 2 | 2+2 4 |
| | 8 2 | 5+6 4 | 3+4 2 | 2+2 4 |
| | 4+5 2 | 4+4 4 | 3+4 2 | 2+2 4 |
| | 8 2 | | 3+4 2 | |
| | 8 2 | | 3+4 2 | |
| | 4+5 2 | | 3+4 2 | |
| | 8 2 | | 3+4 2 | |
| | 8 2 | | 3+4 2 | |
| | 4+5 2 | | 3+4 2 | |
| | 8 2 | | 5+6 2 | |
| | 9 2 | | 5+6 2 | |
| | 6+7 2 | | 3+4 2 | |
| | 9 2 | | 5+6 2 | |
| | 8 2 | | 3+4 2 | |
| | 9 2 | | 3+4 2 | |
| | 6+7 2 | | 5+6 2 | |
| | 9 2 | | 5+6 2 | |
| | 4+5 2 | | 3+4 2 | |
| | 8 2 | | 3+4 2 | |
| | 9 2 | | 5+6 2 | |
| | 6+7 2 | | 5+6 2 | |
| 167.1.25 | 167069 | | 72 | La1Mn1O3.13 |
| | -GM | -T | -F | -L |
| | 8 2 | 5+6 4 | 5+6 2 | 2+2 4 |
| | 9 2 | 5+6 4 | 3+4 2 | 2+2 4 |
| | 8 2 | 5+6 4 | 3+4 2 | 2+2 4 |
| | 8 2 | 4+4 4 | 3+4 2 | 2+2 4 |
| | 4+5 2 | 5+6 4 | 5+6 2 | 2+2 4 |
| | 9 2 | 4+4 4 | 5+6 2 | 2+2 4 |
| | 6+7 2 | 5+6 4 | 3+4 2 | 2+2 4 |
| | 9 2 | 5+6 4 | 5+6 2 | 2+2 4 |
| | 8 2 | 4+4 4 | 3+4 2 | 2+2 4 |
| | 9 2 | 5+6 4 | 5+6 2 | 2+2 4 |
| | 8 2 | 4+4 4 | 3+4 2 | 2+2 4 |
| | 4+5 2 | 5+6 4 | 3+4 2 | 2+2 4 |
| | 9 2 | 5+6 4 | 3+4 2 | 2+2 4 |
| | 4+5 2 | 5+6 4 | 5+6 2 | 2+2 4 |
| | 6+7 2 | 4+4 4 | 5+6 2 | 2+2 4 |
| | 8 2 | 5+6 4 | 3+4 2 | 2+2 4 |
| | 8 2 | 5+6 4 | 3+4 2 | 2+2 4 |
| | 4+5 2 | 4+4 4 | 3+4 2 | 2+2 4 |
| | 8 2 | | 3+4 2 | |
| | 8 2 | | 3+4 2 | |
| | 4+5 2 | | 3+4 2 | |
| | 8 2 | | 3+4 2 | |
| | 8 2 | | 3+4 2 | |
| | 4+5 2 | | 3+4 2 | |
| | 8 2 | | 5+6 2 | |
| | 9 2 | | 3+4 2 | |
| | 6+7 2 | | 5+6 2 | |
| | 9 2 | | 5+6 2 | |
| | 8 2 | | 3+4 2 | |
| | 9 2 | | 3+4 2 | |
| | 4+5 2 | | 5+6 2 | |
| | 6+7 2 | | 5+6 2 | |
| | 8 2 | | 3+4 2 | |
| | 9 2 | | 3+4 2 | |
| | 9 2 | | 5+6 2 | |

Continued on next column

**TABLE LXXIV– continued**

| TQC | ICSD | | NoE | SF |
|---|---|---|---|---|
| | 6+7 2 | | 5+6 2 | |
| 167.1.26 | 96070 | | 68 | Fe2O3 |
| | -GM | -T | -F | -L |
| | 8 2 | 5+6 4 | 5+6 2 | 2+2 4 |
| | 9 2 | 5+6 4 | 3+4 2 | 2+2 4 |
| | 9 2 | 4+4 4 | 3+4 2 | 2+2 4 |
| | 6+7 2 | 5+6 4 | 5+6 2 | 2+2 4 |
| | 8 2 | 5+6 4 | 5+6 2 | 2+2 4 |
| | 4+5 2 | 4+4 4 | 3+4 2 | 2+2 4 |
| | 8 2 | 5+6 4 | 3+4 2 | 2+2 4 |
| | 8 2 | 4+4 4 | 5+6 2 | 2+2 4 |
| | 4+5 2 | 5+6 4 | 5+6 2 | 2+2 4 |
| | 9 2 | 5+6 4 | 3+4 2 | 2+2 4 |
| | 9 2 | 5+6 4 | 3+4 2 | 2+2 4 |
| | 6+7 2 | 4+4 4 | 5+6 2 | 2+2 4 |
| | 8 2 | 4+4 4 | 5+6 2 | 2+2 4 |
| | 4+5 2 | 5+6 4 | 3+4 2 | 2+2 4 |
| | 8 2 | 5+6 4 | 3+4 2 | 2+2 4 |
| | 6+7 2 | 4+4 4 | 5+6 2 | 2+2 4 |
| | 9 2 | 5+6 4 | 5+6 2 | 2+2 4 |
| | 8 2 | | 3+4 2 | |
| | 9 2 | | 5+6 2 | |
| | 8 2 | | 3+4 2 | |
| | 4+5 2 | | 5+6 2 | |
| | 9 2 | | 5+6 2 | |
| | 6+7 2 | | 3+4 2 | |
| | 9 2 | | 3+4 2 | |
| | 4+5 2 | | 3+4 2 | |
| | 8 2 | | 5+6 2 | |
| | 8 2 | | 5+6 2 | |
| | 8 2 | | 5+6 2 | |
| | 4+5 2 | | 3+4 2 | |
| | 9 2 | | 3+4 2 | |
| | 6+7 2 | | 5+6 2 | |
| | 9 2 | | 3+4 2 | |
| | 8 2 | | 5+6 2 | |
| | 6+7 2 | | 3+4 2 | |
| 167.1.27 | 96069 | | 68 | Fe2O3 |
| | -GM | -T | -F | -L |
| | 8 2 | 5+6 4 | 5+6 2 | 2+2 4 |
| | 9 2 | 5+6 4 | 3+4 2 | 2+2 4 |
| | 9 2 | 4+4 4 | 3+4 2 | 2+2 4 |
| | 6+7 2 | 5+6 4 | 5+6 2 | 2+2 4 |
| | 8 2 | 5+6 4 | 5+6 2 | 2+2 4 |
| | 4+5 2 | 4+4 4 | 3+4 2 | 2+2 4 |
| | 8 2 | 5+6 4 | 3+4 2 | 2+2 4 |
| | 8 2 | 4+4 4 | 5+6 2 | 2+2 4 |
| | 4+5 2 | 5+6 4 | 5+6 2 | 2+2 4 |
| | 9 2 | 5+6 4 | 3+4 2 | 2+2 4 |
| | 9 2 | 5+6 4 | 3+4 2 | 2+2 4 |
| | 6+7 2 | 4+4 4 | 5+6 2 | 2+2 4 |
| | 8 2 | 4+4 4 | 5+6 2 | 2+2 4 |
| | 4+5 2 | 5+6 4 | 3+4 2 | 2+2 4 |
| | 8 2 | 5+6 4 | 3+4 2 | 2+2 4 |
| | 6+7 2 | 4+4 4 | 5+6 2 | 2+2 4 |
| | 9 2 | 5+6 4 | 5+6 2 | 2+2 4 |
| | 8 2 | | 3+4 2 | |
| | 9 2 | | 5+6 2 | |
| | 8 2 | | 3+4 2 | |
| | 4+5 2 | | 5+6 2 | |
| | 9 2 | | 5+6 2 | |

Continued on next column

**TABLE LXXIV– continued**

| TQC | ICSD | | NoE | SF |
|---|---|---|---|---|
| | 6+7 2 | | 3+4 2 | |
| | 9 2 | | 3+4 2 | |
| | 8 2 | | 3+4 2 | |
| | 4+5 2 | | 5+6 2 | |
| | 8 2 | | 5+6 2 | |
| | 8 2 | | 5+6 2 | |
| | 4+5 2 | | 3+4 2 | |
| | 9 2 | | 3+4 2 | |
| | 6+7 2 | | 5+6 2 | |
| | 9 2 | | 3+4 2 | |
| | 8 2 | | 5+6 2 | |
| | 6+7 2 | | 3+4 2 | |
| 167.1.28 | 88418 | | 68 | Fe2O3 |
| | -GM | -T | -F | -L |
| | 8 2 | 5+6 4 | 5+6 2 | 2+2 4 |
| | 9 2 | 5+6 4 | 3+4 2 | 2+2 4 |
| | 9 2 | 4+4 4 | 3+4 2 | 2+2 4 |
| | 6+7 2 | 5+6 4 | 5+6 2 | 2+2 4 |
| | 8 2 | 5+6 4 | 5+6 2 | 2+2 4 |
| | 4+5 2 | 4+4 4 | 3+4 2 | 2+2 4 |
| | 8 2 | 5+6 4 | 3+4 2 | 2+2 4 |
| | 9 2 | 4+4 4 | 5+6 2 | 2+2 4 |
| | 8 2 | 5+6 4 | 5+6 2 | 2+2 4 |
| | 4+5 2 | 5+6 4 | 3+4 2 | 2+2 4 |
| | 9 2 | 5+6 4 | 3+4 2 | 2+2 4 |
| | 6+7 2 | 4+4 4 | 5+6 2 | 2+2 4 |
| | 8 2 | 4+4 4 | 5+6 2 | 2+2 4 |
| | 4+5 2 | 5+6 4 | 3+4 2 | 2+2 4 |
| | 8 2 | 5+6 4 | 3+4 2 | 2+2 4 |
| | 6+7 2 | 4+4 4 | 5+6 2 | 2+2 4 |
| | 9 2 | 5+6 4 | 5+6 2 | 2+2 4 |
| | 8 2 | | 3+4 2 | |
| | 9 2 | | 5+6 2 | |
| | 8 2 | | 3+4 2 | |
| | 4+5 2 | | 5+6 2 | |
| | 9 2 | | 5+6 2 | |
| | 6+7 2 | | 3+4 2 | |
| | 9 2 | | 3+4 2 | |
| | 4+5 2 | | 5+6 2 | |
| | 8 2 | | 3+4 2 | |
| | 8 2 | | 5+6 2 | |
| | 9 2 | | 5+6 2 | |
| | 8 2 | | 3+4 2 | |
| | 4+5 2 | | 3+4 2 | |
| | 6+7 2 | | 5+6 2 | |
| | 9 2 | | 3+4 2 | |
| | 8 2 | | 5+6 2 | |
| | 6+7 2 | | 3+4 2 | |
| 167.1.29 | 88417 | | 68 | Fe2O3 |
| | -GM | -T | -F | -L |
| | 8 2 | 5+6 4 | 5+6 2 | 2+2 4 |
| | 9 2 | 5+6 4 | 3+4 2 | 2+2 4 |
| | 9 2 | 4+4 4 | 3+4 2 | 2+2 4 |
| | 6+7 2 | 5+6 4 | 5+6 2 | 2+2 4 |
| | 8 2 | 5+6 4 | 5+6 2 | 2+2 4 |
| | 4+5 2 | 4+4 4 | 3+4 2 | 2+2 4 |
| | 8 2 | 5+6 4 | 3+4 2 | 2+2 4 |
| | 9 2 | 4+4 4 | 5+6 2 | 2+2 4 |
| | 8 2 | 5+6 4 | 5+6 2 | 2+2 4 |
| | 4+5 2 | 5+6 4 | 3+4 2 | 2+2 4 |
| | 9 2 | 5+6 4 | 3+4 2 | 2+2 4 |

Continued on next column

**TABLE LXXIV– continued**

| TQC | ICSD | NoE | SF | |
|---|---|---|---|---|
| | 6+7 2 | 4+4 4 | 5+6 2 | 2+2 4 |
| | 8 2 | 4+4 4 | 5+6 2 | 2+2 4 |
| | 4+5 2 | 5+6 4 | 3+4 2 | 2+2 4 |
| | 8 2 | 5+6 4 | 3+4 2 | 2+2 4 |
| | 6+7 2 | 4+4 4 | 5+6 2 | 2+2 4 |
| | 9 2 | 5+6 4 | 5+6 2 | 2+2 4 |
| | 8 2 | | 3+4 2 | |
| | 9 2 | | 5+6 2 | |
| | 8 2 | | 3+4 2 | |
| | 4+5 2 | | 5+6 2 | |
| | 9 2 | | 5+6 2 | |
| | 6+7 2 | | 3+4 2 | |
| | 9 2 | | 3+4 2 | |
| | 4+5 2 | | 5+6 2 | |
| | 8 2 | | 3+4 2 | |
| | 8 2 | | 5+6 2 | |
| | 9 2 | | 5+6 2 | |
| | 8 2 | | 3+4 2 | |
| | 6+7 2 | | 3+4 2 | |
| | 4+5 2 | | 5+6 2 | |
| | 9 2 | | 3+4 2 | |
| | 8 2 | | 5+6 2 | |
| | 6+7 2 | | 3+4 2 | |
| 167.1.30 | 82904 | 68 | Fe2O3 | |
| | -GM | -T | -F | -L |
| | 8 2 | 5+6 4 | 5+6 2 | 2+2 4 |
| | 9 2 | 5+6 4 | 3+4 2 | 2+2 4 |
| | 9 2 | 4+4 4 | 3+4 2 | 2+2 4 |
| | 6+7 2 | 5+6 4 | 5+6 2 | 2+2 4 |
| | 8 2 | 5+6 4 | 5+6 2 | 2+2 4 |
| | 4+5 2 | 4+4 4 | 3+4 2 | 2+2 4 |
| | 8 2 | 5+6 4 | 3+4 2 | 2+2 4 |
| | 9 2 | 4+4 4 | 5+6 2 | 2+2 4 |
| | 8 2 | 5+6 4 | 5+6 2 | 2+2 4 |
| | 4+5 2 | 5+6 4 | 3+4 2 | 2+2 4 |
| | 9 2 | 5+6 4 | 3+4 2 | 2+2 4 |
| | 6+7 2 | 4+4 4 | 5+6 2 | 2+2 4 |
| | 8 2 | 4+4 4 | 5+6 2 | 2+2 4 |
| | 4+5 2 | 5+6 4 | 3+4 2 | 2+2 4 |
| | 8 2 | 5+6 4 | 3+4 2 | 2+2 4 |
| | 6+7 2 | 4+4 4 | 5+6 2 | 2+2 4 |
| | 9 2 | 5+6 4 | 5+6 2 | 2+2 4 |
| | 8 2 | | 3+4 2 | |
| | 9 2 | | 5+6 2 | |
| | 8 2 | | 3+4 2 | |
| | 4+5 2 | | 5+6 2 | |
| | 9 2 | | 5+6 2 | |
| | 6+7 2 | | 3+4 2 | |
| | 9 2 | | 3+4 2 | |
| | 4+5 2 | | 5+6 2 | |
| | 8 2 | | 3+4 2 | |
| | 8 2 | | 5+6 2 | |
| | 9 2 | | 5+6 2 | |
| | 8 2 | | 3+4 2 | |
| | 4+5 2 | | 3+4 2 | |
| | 6+7 2 | | 5+6 2 | |
| | 9 2 | | 3+4 2 | |
| | 8 2 | | 5+6 2 | |
| | 6+7 2 | | 3+4 2 | |
| 167.1.31 | 82903 | 68 | Fe2O3 | |
| | -GM | -T | -F | -L |

Continued on next column

**TABLE LXXIV– continued**

| TQC | ICSD | NoE | SF | |
|---|---|---|---|---|
| | 8 2 | 5+6 4 | 5+6 2 | 2+2 4 |
| | 9 2 | 5+6 4 | 3+4 2 | 2+2 4 |
| | 9 2 | 4+4 4 | 3+4 2 | 2+2 4 |
| | 6+7 2 | 5+6 4 | 5+6 2 | 2+2 4 |
| | 8 2 | 5+6 4 | 5+6 2 | 2+2 4 |
| | 4+5 2 | 4+4 4 | 3+4 2 | 2+2 4 |
| | 8 2 | 5+6 4 | 3+4 2 | 2+2 4 |
| | 9 2 | 4+4 4 | 5+6 2 | 2+2 4 |
| | 8 2 | 5+6 4 | 5+6 2 | 2+2 4 |
| | 4+5 2 | 5+6 4 | 3+4 2 | 2+2 4 |
| | 9 2 | 5+6 4 | 3+4 2 | 2+2 4 |
| | 6+7 2 | 4+4 4 | 5+6 2 | 2+2 4 |
| | 8 2 | 4+4 4 | 5+6 2 | 2+2 4 |
| | 4+5 2 | 5+6 4 | 3+4 2 | 2+2 4 |
| | 8 2 | 5+6 4 | 3+4 2 | 2+2 4 |
| | 6+7 2 | 4+4 4 | 5+6 2 | 2+2 4 |
| | 9 2 | 5+6 4 | 5+6 2 | 2+2 4 |
| | 8 2 | | 3+4 2 | |
| | 9 2 | | 5+6 2 | |
| | 8 2 | | 3+4 2 | |
| | 4+5 2 | | 5+6 2 | |
| | 9 2 | | 5+6 2 | |
| | 6+7 2 | | 3+4 2 | |
| | 9 2 | | 3+4 2 | |
| | 4+5 2 | | 5+6 2 | |
| | 8 2 | | 3+4 2 | |
| | 8 2 | | 5+6 2 | |
| | 9 2 | | 5+6 2 | |
| | 8 2 | | 3+4 2 | |
| | 4+5 2 | | 3+4 2 | |
| | 6+7 2 | | 5+6 2 | |
| | 9 2 | | 3+4 2 | |
| | 8 2 | | 5+6 2 | |
| | 6+7 2 | | 3+4 2 | |
| 167.1.32 | 82902 | 68 | Fe2O3 | |
| | -GM | -T | -F | -L |
| | 8 2 | 5+6 4 | 5+6 2 | 2+2 4 |
| | 9 2 | 5+6 4 | 3+4 2 | 2+2 4 |
| | 9 2 | 4+4 4 | 3+4 2 | 2+2 4 |
| | 6+7 2 | 5+6 4 | 5+6 2 | 2+2 4 |
| | 8 2 | 5+6 4 | 5+6 2 | 2+2 4 |
| | 4+5 2 | 4+4 4 | 3+4 2 | 2+2 4 |
| | 8 2 | 5+6 4 | 3+4 2 | 2+2 4 |
| | 9 2 | 4+4 4 | 5+6 2 | 2+2 4 |
| | 8 2 | 5+6 4 | 5+6 2 | 2+2 4 |
| | 4+5 2 | 5+6 4 | 3+4 2 | 2+2 4 |
| | 9 2 | 5+6 4 | 3+4 2 | 2+2 4 |
| | 6+7 2 | 4+4 4 | 5+6 2 | 2+2 4 |
| | 8 2 | 4+4 4 | 5+6 2 | 2+2 4 |
| | 4+5 2 | 5+6 4 | 3+4 2 | 2+2 4 |
| | 8 2 | 5+6 4 | 3+4 2 | 2+2 4 |
| | 6+7 2 | 4+4 4 | 5+6 2 | 2+2 4 |
| | 9 2 | 5+6 4 | 5+6 2 | 2+2 4 |
| | 8 2 | | 3+4 2 | |
| | 9 2 | | 5+6 2 | |
| | 8 2 | | 3+4 2 | |
| | 4+5 2 | | 5+6 2 | |
| | 9 2 | | 5+6 2 | |
| | 6+7 2 | | 3+4 2 | |
| | 9 2 | | 3+4 2 | |
| | 4+5 2 | | 5+6 2 | |

Continued on next column

**TABLE LXXIV– continued**

| TQC | ICSD | | NoE | SF |
|---|---|---|---|---|
| | 8 2 | | 3+4 2 | |
| | 8 2 | | 5+6 2 | |
| | 9 2 | | 5+6 2 | |
| | 8 2 | | 3+4 2 | |
| | 4+5 2 | | 3+4 2 | |
| | 6+7 2 | | 5+6 2 | |
| | 9 2 | | 3+4 2 | |
| | 8 2 | | 5+6 2 | |
| | 6+7 2 | | 3+4 2 | |
| 167.1.33 | 82137 | | 68 | Fe2O3 |
| | -GM | -T | -F | -L |
| | 8 2 | 5+6 4 | 5+6 2 | 2+2 4 |
| | 9 2 | 5+6 4 | 3+4 2 | 2+2 4 |
| | 9 2 | 4+4 4 | 3+4 2 | 2+2 4 |
| | 6+7 2 | 5+6 4 | 5+6 2 | 2+2 4 |
| | 8 2 | 5+6 4 | 5+6 2 | 2+2 4 |
| | 4+5 2 | 4+4 4 | 3+4 2 | 2+2 4 |
| | 8 2 | 5+6 4 | 3+4 2 | 2+2 4 |
| | 9 2 | 4+4 4 | 5+6 2 | 2+2 4 |
| | 8 2 | 5+6 4 | 5+6 2 | 2+2 4 |
| | 4+5 2 | 5+6 4 | 3+4 2 | 2+2 4 |
| | 9 2 | 5+6 4 | 3+4 2 | 2+2 4 |
| | 6+7 2 | 4+4 4 | 5+6 2 | 2+2 4 |
| | 8 2 | 4+4 4 | 5+6 2 | 2+2 4 |
| | 4+5 2 | 5+6 4 | 3+4 2 | 2+2 4 |
| | 8 2 | 5+6 4 | 3+4 2 | 2+2 4 |
| | 6+7 2 | 4+4 4 | 5+6 2 | 2+2 4 |
| | 9 2 | 5+6 4 | 5+6 2 | 2+2 4 |
| | 8 2 | | 3+4 2 | |
| | 9 2 | | 5+6 2 | |
| | 8 2 | | 3+4 2 | |
| | 4+5 2 | | 5+6 2 | |
| | 9 2 | | 5+6 2 | |
| | 6+7 2 | | 3+4 2 | |
| | 9 2 | | 3+4 2 | |
| | 4+5 2 | | 5+6 2 | |
| | 8 2 | | 3+4 2 | |
| | 8 2 | | 5+6 2 | |
| | 9 2 | | 5+6 2 | |
| | 6+7 2 | | 3+4 2 | |
| | 4+5 2 | | 3+4 2 | |
| | 8 2 | | 5+6 2 | |
| | 9 2 | | 3+4 2 | |
| | 8 2 | | 3+4 2 | |
| | 6+7 2 | | 5+6 2 | |
| 167.1.34 | 82136 | | 68 | Fe2O3 |
| | -GM | -T | -F | -L |
| | 8 2 | 5+6 4 | 5+6 2 | 2+2 4 |
| | 9 2 | 5+6 4 | 3+4 2 | 2+2 4 |
| | 6+7 2 | 4+4 4 | 3+4 2 | 2+2 4 |
| | 9 2 | 5+6 4 | 5+6 2 | 2+2 4 |
| | 8 2 | 5+6 4 | 5+6 2 | 2+2 4 |
| | 4+5 2 | 4+4 4 | 3+4 2 | 2+2 4 |
| | 8 2 | 5+6 4 | 3+4 2 | 2+2 4 |
| | 8 2 | 4+4 4 | 5+6 2 | 2+2 4 |
| | 4+5 2 | 5+6 4 | 5+6 2 | 2+2 4 |
| | 9 2 | 5+6 4 | 3+4 2 | 2+2 4 |
| | 6+7 2 | 5+6 4 | 3+4 2 | 2+2 4 |
| | 9 2 | 4+4 4 | 5+6 2 | 2+2 4 |
| | 4+5 2 | 4+4 4 | 5+6 2 | 2+2 4 |
| | 8 2 | 5+6 4 | 3+4 2 | 2+2 4 |
| | 8 2 | 5+6 4 | 3+4 2 | 2+2 4 |
| | 6+7 2 | 4+4 4 | 5+6 2 | 2+2 4 |
| | 9 2 | 5+6 4 | 5+6 2 | 2+2 4 |
| | 8 2 | | 3+4 2 | |
| | 9 2 | | 5+6 2 | |
| | 8 2 | | 5+6 2 | |
| | 4+5 2 | | 3+4 2 | |
| | 9 2 | | 5+6 2 | |
| | 6+7 2 | | 3+4 2 | |
| | 9 2 | | 3+4 2 | |
| | 8 2 | | 3+4 2 | |
| | 4+5 2 | | 5+6 2 | |
| | 8 2 | | 5+6 2 | |
| | 8 2 | | 5+6 2 | |
| | 4+5 2 | | 3+4 2 | |
| | 9 2 | | 3+4 2 | |
| | 6+7 2 | | 5+6 2 | |
| | 9 2 | | 3+4 2 | |
| | 8 2 | | 5+6 2 | |
| | 6+7 2 | | 3+4 2 | |
| 167.1.35 | 82135 | | 68 | Fe2O3 |
| | -GM | -T | -F | -L |
| | 8 2 | 5+6 4 | 5+6 2 | 2+2 4 |
| | 9 2 | 5+6 4 | 3+4 2 | 2+2 4 |
| | 8 2 | 4+4 4 | 3+4 2 | 2+2 4 |
| | 4+5 2 | 5+6 4 | 5+6 2 | 2+2 4 |
| | 9 2 | 5+6 4 | 5+6 2 | 2+2 4 |
| | 6+7 2 | 4+4 4 | 3+4 2 | 2+2 4 |
| | 8 2 | 5+6 4 | 3+4 2 | 2+2 4 |
| | 9 2 | 4+4 4 | 5+6 2 | 2+2 4 |
| | 8 2 | 5+6 4 | 5+6 2 | 2+2 4 |
| | 4+5 2 | 5+6 4 | 3+4 2 | 2+2 4 |
| | 9 2 | 5+6 4 | 3+4 2 | 2+2 4 |
| | 6+7 2 | 4+4 4 | 5+6 2 | 2+2 4 |
| | 8 2 | 4+4 4 | 3+4 2 | 2+2 4 |
| | 4+5 2 | 5+6 4 | 5+6 2 | 2+2 4 |
| | 8 2 | 5+6 4 | 3+4 2 | 2+2 4 |
| | 6+7 2 | 4+4 4 | 5+6 2 | 2+2 4 |
| | 9 2 | 5+6 4 | 5+6 2 | 2+2 4 |
| | 8 2 | | 3+4 2 | |
| | 9 2 | | 5+6 2 | |
| | 8 2 | | 3+4 2 | |
| | 4+5 2 | | 5+6 2 | |
| | 9 2 | | 5+6 2 | |
| | 9 2 | | 3+4 2 | |
| | 6+7 2 | | 3+4 2 | |
| | 4+5 2 | | 5+6 2 | |
| | 8 2 | | 3+4 2 | |
| | 8 2 | | 5+6 2 | |
| | 9 2 | | 5+6 2 | |
| | 6+7 2 | | 3+4 2 | |
| | 8 2 | | 3+4 2 | |
| | 4+5 2 | | 5+6 2 | |
| | 9 2 | | 3+4 2 | |
| | 8 2 | | 5+6 2 | |
| | 6+7 2 | | 3+4 2 | |
| 167.1.36 | 82134 | | 68 | Fe2O3 |
| | -GM | -T | -F | -L |
| | 8 2 | 5+6 4 | 5+6 2 | 2+2 4 |
| | 9 2 | 5+6 4 | 3+4 2 | 2+2 4 |
| | 6+7 2 | 4+4 4 | 5+6 2 | 2+2 4 |

Continued on next column

**TABLE LXXIV– continued**

| TQC | ICSD | NoE | SF |
|---|---|---|---|
| | 9 2 5+6 4 3+4 2 2+2 4 | | |
| | 8 2 4+4 4 5+6 2 2+2 4 | | |
| | 4+5 2 5+6 4 3+4 2 2+2 4 | | |
| | 8 2 5+6 4 3+4 2 2+2 4 | | |
| | 8 2 4+4 4 5+6 2 2+2 4 | | |
| | 4+5 2 5+6 4 5+6 2 2+2 4 | | |
| | 9 2 5+6 4 3+4 2 2+2 4 | | |
| | 6+7 2 5+6 4 3+4 2 2+2 4 | | |
| | 9 2 4+4 4 5+6 2 2+2 4 | | |
| | 4+5 2 4+4 4 5+6 2 2+2 4 | | |
| | 8 2 5+6 4 3+4 2 2+2 4 | | |
| | 8 2 5+6 4 3+4 2 2+2 4 | | |
| | 6+7 2 4+4 4 5+6 2 2+2 4 | | |
| | 9 2 5+6 4 3+4 2 2+2 4 | | |
| | 8 2 | 5+6 2 | |
| | 9 2 | 5+6 2 | |
| | 8 2 | 5+6 2 | |
| | 4+5 2 | 3+4 2 | |
| | 9 2 | 3+4 2 | |
| | 6+7 2 | 5+6 2 | |
| | 9 2 | 3+4 2 | |
| | 8 2 | 3+4 2 | |
| | 4+5 2 | 5+6 2 | |
| | 8 2 | 5+6 2 | |
| | 8 2 | 5+6 2 | |
| | 4+5 2 | 3+4 2 | |
| | 9 2 | 5+6 2 | |
| | 6+7 2 | 3+4 2 | |
| | 9 2 | 3+4 2 | |
| | 8 2 | 5+6 2 | |
| | 6+7 2 | 3+4 2 | |
| 167.1.37 | 81248 | 68 | Fe2O3 |
| | -GM -T | -F | -L |
| | 8 2 5+6 4 5+6 2 2+2 4 | | |
| | 9 2 5+6 4 3+4 2 2+2 4 | | |
| | 9 2 4+4 4 3+4 2 2+2 4 | | |
| | 6+7 2 5+6 4 5+6 2 2+2 4 | | |
| | 8 2 5+6 4 5+6 2 2+2 4 | | |
| | 4+5 2 4+4 4 3+4 2 2+2 4 | | |
| | 8 2 5+6 4 3+4 2 2+2 4 | | |
| | 9 2 4+4 4 5+6 2 2+2 4 | | |
| | 8 2 5+6 4 5+6 2 2+2 4 | | |
| | 4+5 2 5+6 4 3+4 2 2+2 4 | | |
| | 9 2 5+6 4 3+4 2 2+2 4 | | |
| | 6+7 2 4+4 4 5+6 2 2+2 4 | | |
| | 8 2 4+4 4 5+6 2 2+2 4 | | |
| | 4+5 2 5+6 4 3+4 2 2+2 4 | | |
| | 8 2 5+6 4 3+4 2 2+2 4 | | |
| | 6+7 2 4+4 4 5+6 2 2+2 4 | | |
| | 9 2 5+6 4 5+6 2 2+2 4 | | |
| | 8 2 | 3+4 2 | |
| | 9 2 | 5+6 2 | |
| | 8 2 | 3+4 2 | |
| | 4+5 2 | 5+6 2 | |
| | 9 2 | 5+6 2 | |
| | 6+7 2 | 3+4 2 | |
| | 9 2 | 3+4 2 | |
| | 4+5 2 | 5+6 2 | |
| | 8 2 | 3+4 2 | |
| | 8 2 | 5+6 2 | |
| | 9 2 | 5+6 2 | |
| | 8 2 | 3+4 2 | |
| | 4+5 2 | 3+4 2 | |
| | 6+7 2 | 5+6 2 | |
| | 9 2 | 3+4 2 | |
| | 8 2 | 5+6 2 | |
| | 6+7 2 | 3+4 2 | |
| 167.1.38 | 71194 | 68 | Fe2O3 |
| | -GM -T | -F | -L |
| | 8 2 5+6 4 5+6 2 2+2 4 | | |
| | 9 2 5+6 4 3+4 2 2+2 4 | | |
| | 9 2 4+4 4 3+4 2 2+2 4 | | |
| | 6+7 2 5+6 4 5+6 2 2+2 4 | | |
| | 8 2 5+6 4 5+6 2 2+2 4 | | |
| | 4+5 2 4+4 4 3+4 2 2+2 4 | | |
| | 8 2 5+6 4 3+4 2 2+2 4 | | |
| | 9 2 4+4 4 5+6 2 2+2 4 | | |
| | 8 2 5+6 4 5+6 2 2+2 4 | | |
| | 4+5 2 5+6 4 3+4 2 2+2 4 | | |
| | 9 2 5+6 4 3+4 2 2+2 4 | | |
| | 6+7 2 4+4 4 5+6 2 2+2 4 | | |
| | 8 2 4+4 4 5+6 2 2+2 4 | | |
| | 4+5 2 5+6 4 3+4 2 2+2 4 | | |
| | 8 2 5+6 4 3+4 2 2+2 4 | | |
| | 6+7 2 4+4 4 5+6 2 2+2 4 | | |
| | 9 2 5+6 4 5+6 2 2+2 4 | | |
| | 8 2 | 3+4 2 | |
| | 9 2 | 5+6 2 | |
| | 8 2 | 3+4 2 | |
| | 4+5 2 | 5+6 2 | |
| | 9 2 | 5+6 2 | |
| | 6+7 2 | 3+4 2 | |
| | 9 2 | 3+4 2 | |
| | 4+5 2 | 5+6 2 | |
| | 8 2 | 3+4 2 | |
| | 8 2 | 5+6 2 | |
| | 9 2 | 5+6 2 | |
| | 8 2 | 3+4 2 | |
| | 4+5 2 | 3+4 2 | |
| | 6+7 2 | 5+6 2 | |
| | 9 2 | 3+4 2 | |
| | 8 2 | 5+6 2 | |
| | 6+7 2 | 3+4 2 | |
| 167.1.39 | 66756 | 68 | Fe2O3 |
| | -GM -T | -F | -L |
| | 8 2 5+6 4 5+6 2 2+2 4 | | |
| | 9 2 5+6 4 3+4 2 2+2 4 | | |
| | 9 2 4+4 4 3+4 2 2+2 4 | | |
| | 6+7 2 5+6 4 5+6 2 2+2 4 | | |
| | 8 2 5+6 4 5+6 2 2+2 4 | | |
| | 4+5 2 4+4 4 3+4 2 2+2 4 | | |
| | 8 2 5+6 4 3+4 2 2+2 4 | | |
| | 8 2 4+4 4 5+6 2 2+2 4 | | |
| | 9 2 5+6 4 5+6 2 2+2 4 | | |
| | 4+5 2 5+6 4 3+4 2 2+2 4 | | |
| | 9 2 5+6 4 3+4 2 2+2 4 | | |
| | 6+7 2 4+4 4 5+6 2 2+2 4 | | |
| | 8 2 4+4 4 5+6 2 2+2 4 | | |
| | 4+5 2 5+6 4 3+4 2 2+2 4 | | |
| | 8 2 5+6 4 3+4 2 2+2 4 | | |
| | 6+7 2 4+4 4 5+6 2 2+2 4 | | |
| | 9 2 5+6 4 5+6 2 2+2 4 | | |

Continued on next column

**TABLE LXXIV– continued**

| TQC | ICSD | | NoE | SF |
|---|---|---|---|---|
| | 8 2 | | 3+4 2 | |
| | 9 2 | | 5+6 2 | |
| | 8 2 | | 3+4 2 | |
| | 4+5 2 | | 5+6 2 | |
| | 9 2 | | 5+6 2 | |
| | 6+7 2 | | 3+4 2 | |
| | 9 2 | | 3+4 2 | |
| | 4+5 2 | | 5+6 2 | |
| | 8 2 | | 3+4 2 | |
| | 8 2 | | 5+6 2 | |
| | 8 2 | | 5+6 2 | |
| | 9 2 | | 3+4 2 | |
| | 4+5 2 | | 3+4 2 | |
| | 6+7 2 | | 5+6 2 | |
| | 9 2 | | 3+4 2 | |
| | 8 2 | | 5+6 2 | |
| | 6+7 2 | | 3+4 2 | |
| 167.1.40 | 64599 | | 68 | Fe2O3 |
| | -GM | -T | -F | -L |
| | 8 2 | 5+6 4 | 5+6 2 | 2+2 4 |
| | 9 2 | 5+6 4 | 3+4 2 | 2+2 4 |
| | 9 2 | 4+4 4 | 3+4 2 | 2+2 4 |
| | 6+7 2 | 5+6 4 | 5+6 2 | 2+2 4 |
| | 8 2 | 5+6 4 | 5+6 2 | 2+2 4 |
| | 4+5 2 | 4+4 4 | 3+4 2 | 2+2 4 |
| | 8 2 | 5+6 4 | 3+4 2 | 2+2 4 |
| | 9 2 | 4+4 4 | 5+6 2 | 2+2 4 |
| | 8 2 | 5+6 4 | 5+6 2 | 2+2 4 |
| | 4+5 2 | 5+6 4 | 3+4 2 | 2+2 4 |
| | 9 2 | 5+6 4 | 3+4 2 | 2+2 4 |
| | 6+7 2 | 4+4 4 | 5+6 2 | 2+2 4 |
| | 8 2 | 4+4 4 | 5+6 2 | 2+2 4 |
| | 4+5 2 | 5+6 4 | 3+4 2 | 2+2 4 |
| | 8 2 | 5+6 4 | 3+4 2 | 2+2 4 |
| | 6+7 2 | 4+4 4 | 5+6 2 | 2+2 4 |
| | 9 2 | 5+6 4 | 5+6 2 | 2+2 4 |
| | 8 2 | | 3+4 2 | |
| | 9 2 | | 5+6 2 | |
| | 8 2 | | 3+4 2 | |
| | 4+5 2 | | 5+6 2 | |
| | 9 2 | | 5+6 2 | |
| | 6+7 2 | | 3+4 2 | |
| | 9 2 | | 3+4 2 | |
| | 4+5 2 | | 5+6 2 | |
| | 8 2 | | 3+4 2 | |
| | 8 2 | | 5+6 2 | |
| | 9 2 | | 5+6 2 | |
| | 8 2 | | 3+4 2 | |
| | 4+5 2 | | 3+4 2 | |
| | 6+7 2 | | 5+6 2 | |
| | 9 2 | | 3+4 2 | |
| | 8 2 | | 5+6 2 | |
| | 6+7 2 | | 3+4 2 | |
| 167.1.41 | 56372 | | 68 | Fe2O3 |
| | -GM | -T | -F | -L |
| | 8 2 | 5+6 4 | 5+6 2 | 2+2 4 |
| | 9 2 | 5+6 4 | 3+4 2 | 2+2 4 |
| | 9 2 | 4+4 4 | 3+4 2 | 2+2 4 |
| | 6+7 2 | 5+6 4 | 5+6 2 | 2+2 4 |
| | 8 2 | 5+6 4 | 5+6 2 | 2+2 4 |
| | 4+5 2 | 4+4 4 | 3+4 2 | 2+2 4 |

Continued on next column

**TABLE LXXIV– continued**

| TQC | ICSD | | NoE | SF |
|---|---|---|---|---|
| | 8 2 | 5+6 4 | 3+4 2 | 2+2 4 |
| | 9 2 | 4+4 4 | 5+6 2 | 2+2 4 |
| | 8 2 | 5+6 4 | 5+6 2 | 2+2 4 |
| | 4+5 2 | 5+6 4 | 3+4 2 | 2+2 4 |
| | 9 2 | 5+6 4 | 3+4 2 | 2+2 4 |
| | 6+7 2 | 4+4 4 | 5+6 2 | 2+2 4 |
| | 8 2 | 4+4 4 | 5+6 2 | 2+2 4 |
| | 4+5 2 | 5+6 4 | 3+4 2 | 2+2 4 |
| | 8 2 | 5+6 4 | 3+4 2 | 2+2 4 |
| | 6+7 2 | 4+4 4 | 5+6 2 | 2+2 4 |
| | 9 2 | 5+6 4 | 5+6 2 | 2+2 4 |
| | 8 2 | | 3+4 2 | |
| | 9 2 | | 5+6 2 | |
| | 8 2 | | 3+4 2 | |
| | 4+5 2 | | 5+6 2 | |
| | 9 2 | | 5+6 2 | |
| | 6+7 2 | | 3+4 2 | |
| | 9 2 | | 3+4 2 | |
| | 4+5 2 | | 5+6 2 | |
| | 8 2 | | 3+4 2 | |
| | 8 2 | | 5+6 2 | |
| | 9 2 | | 5+6 2 | |
| | 8 2 | | 3+4 2 | |
| | 4+5 2 | | 3+4 2 | |
| | 6+7 2 | | 5+6 2 | |
| | 9 2 | | 3+4 2 | |
| | 8 2 | | 5+6 2 | |
| | 6+7 2 | | 3+4 2 | |
| 167.1.42 | 43465 | | 68 | Fe2O3 |
| | -GM | -T | -F | -L |
| | 8 2 | 5+6 4 | 5+6 2 | 2+2 4 |
| | 9 2 | 5+6 4 | 3+4 2 | 2+2 4 |
| | 8 2 | 4+4 4 | 3+4 2 | 2+2 4 |
| | 9 2 | 5+6 4 | 5+6 2 | 2+2 4 |
| | 6+7 2 | 5+6 4 | 5+6 2 | 2+2 4 |
| | 4+5 2 | 4+4 4 | 3+4 2 | 2+2 4 |
| | 8 2 | 5+6 4 | 3+4 2 | 2+2 4 |
| | 9 2 | 4+4 4 | 5+6 2 | 2+2 4 |
| | 8 2 | 5+6 4 | 5+6 2 | 2+2 4 |
| | 4+5 2 | 5+6 4 | 3+4 2 | 2+2 4 |
| | 9 2 | 5+6 4 | 3+4 2 | 2+2 4 |
| | 6+7 2 | 4+4 4 | 5+6 2 | 2+2 4 |
| | 8 2 | 4+4 4 | 3+4 2 | 2+2 4 |
| | 4+5 2 | 5+6 4 | 5+6 2 | 2+2 4 |
| | 8 2 | 5+6 4 | 3+4 2 | 2+2 4 |
| | 6+7 2 | 4+4 4 | 5+6 2 | 2+2 4 |
| | 9 2 | 5+6 4 | 5+6 2 | 2+2 4 |
| | 8 2 | | 3+4 2 | |
| | 9 2 | | 5+6 2 | |
| | 8 2 | | 3+4 2 | |
| | 4+5 2 | | 5+6 2 | |
| | 9 2 | | 5+6 2 | |
| | 6+7 2 | | 3+4 2 | |
| | 9 2 | | 3+4 2 | |
| | 4+5 2 | | 5+6 2 | |
| | 8 2 | | 3+4 2 | |
| | 8 2 | | 5+6 2 | |
| | 9 2 | | 5+6 2 | |
| | 6+7 2 | | 3+4 2 | |
| | 8 2 | | 3+4 2 | |
| | 4+5 2 | | 5+6 2 | |

Continued on next column

**TABLE LXXIV– continued**

| TQC | ICSD | | NoE | SF |
|---|---|---|---|---|
| | 9 2 | | 3+4 2 | |
| | 8 2 | | 5+6 2 | |
| | 6+7 2 | | 3+4 2 | |
| 167.1.43 | 415251 | | 68 | Fe2O3 |
| | -GM | -T | -F | -L |
| | 8 2 | 5+6 4 | 5+6 2 | 2+2 4 |
| | 9 2 | 5+6 4 | 3+4 2 | 2+2 4 |
| | 9 2 | 4+4 4 | 3+4 2 | 2+2 4 |
| | 6+7 2 | 5+6 4 | 5+6 2 | 2+2 4 |
| | 8 2 | 5+6 4 | 5+6 2 | 2+2 4 |
| | 4+5 2 | 4+4 4 | 3+4 2 | 2+2 4 |
| | 8 2 | 5+6 4 | 3+4 2 | 2+2 4 |
| | 9 2 | 4+4 4 | 5+6 2 | 2+2 4 |
| | 8 2 | 5+6 4 | 5+6 2 | 2+2 4 |
| | 4+5 2 | 5+6 4 | 3+4 2 | 2+2 4 |
| | 9 2 | 5+6 4 | 3+4 2 | 2+2 4 |
| | 6+7 2 | 4+4 4 | 5+6 2 | 2+2 4 |
| | 8 2 | 4+4 4 | 5+6 2 | 2+2 4 |
| | 4+5 2 | 5+6 4 | 3+4 2 | 2+2 4 |
| | 8 2 | 5+6 4 | 3+4 2 | 2+2 4 |
| | 6+7 2 | 4+4 4 | 5+6 2 | 2+2 4 |
| | 9 2 | 5+6 4 | 5+6 2 | 2+2 4 |
| | 8 2 | | 3+4 2 | |
| | 9 2 | | 5+6 2 | |
| | 8 2 | | 3+4 2 | |
| | 4+5 2 | | 5+6 2 | |
| | 9 2 | | 5+6 2 | |
| | 6+7 2 | | 3+4 2 | |
| | 9 2 | | 3+4 2 | |
| | 4+5 2 | | 5+6 2 | |
| | 8 2 | | 3+4 2 | |
| | 8 2 | | 5+6 2 | |
| | 9 2 | | 5+6 2 | |
| | 8 2 | | 3+4 2 | |
| | 4+5 2 | | 3+4 2 | |
| | 6+7 2 | | 5+6 2 | |
| | 9 2 | | 3+4 2 | |
| | 8 2 | | 5+6 2 | |
| | 6+7 2 | | 3+4 2 | |
| 167.1.44 | 40142 | | 68 | Fe2O3 |
| | -GM | -T | -F | -L |
| | 8 2 | 5+6 4 | 5+6 2 | 2+2 4 |
| | 9 2 | 5+6 4 | 3+4 2 | 2+2 4 |
| | 9 2 | 4+4 4 | 3+4 2 | 2+2 4 |
| | 6+7 2 | 5+6 4 | 5+6 2 | 2+2 4 |
| | 8 2 | 5+6 4 | 5+6 2 | 2+2 4 |
| | 4+5 2 | 4+4 4 | 3+4 2 | 2+2 4 |
| | 8 2 | 5+6 4 | 3+4 2 | 2+2 4 |
| | 9 2 | 4+4 4 | 5+6 2 | 2+2 4 |
| | 8 2 | 5+6 4 | 5+6 2 | 2+2 4 |
| | 4+5 2 | 5+6 4 | 3+4 2 | 2+2 4 |
| | 9 2 | 5+6 4 | 3+4 2 | 2+2 4 |
| | 6+7 2 | 4+4 4 | 5+6 2 | 2+2 4 |
| | 8 2 | 4+4 4 | 5+6 2 | 2+2 4 |
| | 4+5 2 | 5+6 4 | 3+4 2 | 2+2 4 |
| | 8 2 | 5+6 4 | 3+4 2 | 2+2 4 |
| | 6+7 2 | 4+4 4 | 5+6 2 | 2+2 4 |
| | 9 2 | 5+6 4 | 5+6 2 | 2+2 4 |
| | 8 2 | | 3+4 2 | |
| | 9 2 | | 5+6 2 | |
| | 8 2 | | 3+4 2 | |

Continued on next column

**TABLE LXXIV– continued**

| TQC | ICSD | | NoE | SF |
|---|---|---|---|---|
| | 4+5 2 | | 5+6 2 | |
| | 9 2 | | 5+6 2 | |
| | 6+7 2 | | 3+4 2 | |
| | 9 2 | | 3+4 2 | |
| | 4+5 2 | | 5+6 2 | |
| | 8 2 | | 3+4 2 | |
| | 8 2 | | 5+6 2 | |
| | 9 2 | | 5+6 2 | |
| | 8 2 | | 3+4 2 | |
| | 4+5 2 | | 3+4 2 | |
| | 6+7 2 | | 5+6 2 | |
| | 9 2 | | 3+4 2 | |
| | 8 2 | | 5+6 2 | |
| | 6+7 2 | | 3+4 2 | |
| 167.1.45 | 24791 | | 68 | Fe2O3 |
| | -GM | -T | -F | -L |
| | 8 2 | 5+6 4 | 5+6 2 | 2+2 4 |
| | 9 2 | 5+6 4 | 3+4 2 | 2+2 4 |
| | 8 2 | 4+4 4 | 3+4 2 | 2+2 4 |
| | 4+5 2 | 5+6 4 | 5+6 2 | 2+2 4 |
| | 9 2 | 5+6 4 | 5+6 2 | 2+2 4 |
| | 6+7 2 | 4+4 4 | 3+4 2 | 2+2 4 |
| | 8 2 | 5+6 4 | 3+4 2 | 2+2 4 |
| | 9 2 | 4+4 4 | 5+6 2 | 2+2 4 |
| | 8 2 | 5+6 4 | 5+6 2 | 2+2 4 |
| | 4+5 2 | 5+6 4 | 3+4 2 | 2+2 4 |
| | 9 2 | 5+6 4 | 3+4 2 | 2+2 4 |
| | 6+7 2 | 4+4 4 | 5+6 2 | 2+2 4 |
| | 8 2 | 4+4 4 | 3+4 2 | 2+2 4 |
| | 4+5 2 | 5+6 4 | 5+6 2 | 2+2 4 |
| | 8 2 | 5+6 4 | 3+4 2 | 2+2 4 |
| | 6+7 2 | 4+4 4 | 5+6 2 | 2+2 4 |
| | 9 2 | 5+6 4 | 5+6 2 | 2+2 4 |
| | 8 2 | | 3+4 2 | |
| | 9 2 | | 5+6 2 | |
| | 8 2 | | 5+6 2 | |
| | 4+5 2 | | 3+4 2 | |
| | 9 2 | | 5+6 2 | |
| | 9 2 | | 3+4 2 | |
| | 6+7 2 | | 3+4 2 | |
| | 4+5 2 | | 5+6 2 | |
| | 8 2 | | 3+4 2 | |
| | 8 2 | | 5+6 2 | |
| | 9 2 | | 5+6 2 | |
| | 6+7 2 | | 3+4 2 | |
| | 4+5 2 | | 3+4 2 | |
| | 8 2 | | 5+6 2 | |
| | 9 2 | | 3+4 2 | |
| | 8 2 | | 5+6 2 | |
| | 6+7 2 | | 3+4 2 | |
| 167.1.46 | 245851 | | 68 | Fe2O3 |
| | -GM | -T | -F | -L |
| | 8 2 | 5+6 4 | 5+6 2 | 2+2 4 |
| | 9 2 | 5+6 4 | 3+4 2 | 2+2 4 |
| | 9 2 | 4+4 4 | 3+4 2 | 2+2 4 |
| | 6+7 2 | 5+6 4 | 5+6 2 | 2+2 4 |
| | 8 2 | 5+6 4 | 5+6 2 | 2+2 4 |
| | 4+5 2 | 4+4 4 | 3+4 2 | 2+2 4 |
| | 8 2 | 5+6 4 | 3+4 2 | 2+2 4 |
| | 8 2 | 4+4 4 | 5+6 2 | 2+2 4 |
| | 9 2 | 5+6 4 | 5+6 2 | 2+2 4 |

Continued on next column

**TABLE LXXIV– continued**

| TQC | ICSD | NoE | SF | |
|---|---|---|---|---|
| | 4+5 2 | 5+6 4 | 3+4 2 | 2+2 4 |
| | 9 2 | 5+6 4 | 3+4 2 | 2+2 4 |
| | 6+7 2 | 4+4 4 | 5+6 2 | 2+2 4 |
| | 8 2 | 4+4 4 | 5+6 2 | 2+2 4 |
| | 4+5 2 | 5+6 4 | 3+4 2 | 2+2 4 |
| | 8 2 | 5+6 4 | 3+4 2 | 2+2 4 |
| | 6+7 2 | 4+4 4 | 5+6 2 | 2+2 4 |
| | 9 2 | 5+6 4 | 5+6 2 | 2+2 4 |
| | 8 2 | | 3+4 2 | |
| | 9 2 | | 5+6 2 | |
| | 8 2 | | 3+4 2 | |
| | 4+5 2 | | 5+6 2 | |
| | 9 2 | | 5+6 2 | |
| | 6+7 2 | | 3+4 2 | |
| | 9 2 | | 3+4 2 | |
| | 4+5 2 | | 5+6 2 | |
| | 8 2 | | 3+4 2 | |
| | 8 2 | | 5+6 2 | |
| | 8 2 | | 5+6 2 | |
| | 9 2 | | 3+4 2 | |
| | 4+5 2 | | 3+4 2 | |
| | 6+7 2 | | 5+6 2 | |
| | 9 2 | | 3+4 2 | |
| | 8 2 | | 5+6 2 | |
| | 6+7 2 | | 3+4 2 | |
| 167.1.47 | 24004 | 68 | Fe2O3 | |
| | -GM | -T | -F | -L |
| | 8 2 | 5+6 4 | 5+6 2 | 2+2 4 |
| | 9 2 | 5+6 4 | 3+4 2 | 2+2 4 |
| | 8 2 | 4+4 4 | 3+4 2 | 2+2 4 |
| | 4+5 2 | 5+6 4 | 5+6 2 | 2+2 4 |
| | 9 2 | 5+6 4 | 5+6 2 | 2+2 4 |
| | 6+7 2 | 4+4 4 | 3+4 2 | 2+2 4 |
| | 8 2 | 5+6 4 | 3+4 2 | 2+2 4 |
| | 9 2 | 4+4 4 | 5+6 2 | 2+2 4 |
| | 8 2 | 5+6 4 | 5+6 2 | 2+2 4 |
| | 4+5 2 | 5+6 4 | 3+4 2 | 2+2 4 |
| | 9 2 | 5+6 4 | 3+4 2 | 2+2 4 |
| | 6+7 2 | 4+4 4 | 5+6 2 | 2+2 4 |
| | 8 2 | 4+4 4 | 3+4 2 | 2+2 4 |
| | 4+5 2 | 5+6 4 | 5+6 2 | 2+2 4 |
| | 8 2 | 5+6 4 | 3+4 2 | 2+2 4 |
| | 6+7 2 | 4+4 4 | 5+6 2 | 2+2 4 |
| | 9 2 | 5+6 4 | 5+6 2 | 2+2 4 |
| | 8 2 | | 3+4 2 | |
| | 8 2 | | 5+6 2 | |
| | 4+5 2 | | 3+4 2 | |
| | 9 2 | | 5+6 2 | |
| | 9 2 | | 5+6 2 | |
| | 9 2 | | 3+4 2 | |
| | 6+7 2 | | 3+4 2 | |
| | 4+5 2 | | 5+6 2 | |
| | 8 2 | | 3+4 2 | |
| | 8 2 | | 5+6 2 | |
| | 9 2 | | 5+6 2 | |
| | 6+7 2 | | 3+4 2 | |
| | 4+5 2 | | 3+4 2 | |
| | 8 2 | | 5+6 2 | |
| | 9 2 | | 3+4 2 | |
| | 8 2 | | 3+4 2 | |
| | 6+7 2 | | 5+6 2 | |

Continued on next column

**TABLE LXXIV– continued**

| TQC | ICSD | NoE | SF | |
|---|---|---|---|---|
| 167.1.48 | 22505 | 68 | Fe2O3 | |
| | -GM | -T | -F | -L |
| | 8 2 | 5+6 4 | 5+6 2 | 2+2 4 |
| | 9 2 | 5+6 4 | 3+4 2 | 2+2 4 |
| | 9 2 | 4+4 4 | 3+4 2 | 2+2 4 |
| | 6+7 2 | 5+6 4 | 5+6 2 | 2+2 4 |
| | 8 2 | 5+6 4 | 5+6 2 | 2+2 4 |
| | 4+5 2 | 4+4 4 | 3+4 2 | 2+2 4 |
| | 8 2 | 5+6 4 | 3+4 2 | 2+2 4 |
| | 9 2 | 4+4 4 | 5+6 2 | 2+2 4 |
| | 8 2 | 5+6 4 | 5+6 2 | 2+2 4 |
| | 4+5 2 | 5+6 4 | 3+4 2 | 2+2 4 |
| | 9 2 | 5+6 4 | 3+4 2 | 2+2 4 |
| | 6+7 2 | 4+4 4 | 5+6 2 | 2+2 4 |
| | 8 2 | 4+4 4 | 5+6 2 | 2+2 4 |
| | 4+5 2 | 5+6 4 | 3+4 2 | 2+2 4 |
| | 8 2 | 5+6 4 | 3+4 2 | 2+2 4 |
| | 6+7 2 | 4+4 4 | 5+6 2 | 2+2 4 |
| | 9 2 | 5+6 4 | 5+6 2 | 2+2 4 |
| | 8 2 | | 3+4 2 | |
| | 9 2 | | 5+6 2 | |
| | 8 2 | | 3+4 2 | |
| | 4+5 2 | | 5+6 2 | |
| | 9 2 | | 5+6 2 | |
| | 6+7 2 | | 3+4 2 | |
| | 9 2 | | 3+4 2 | |
| | 4+5 2 | | 5+6 2 | |
| | 8 2 | | 3+4 2 | |
| | 8 2 | | 5+6 2 | |
| | 9 2 | | 5+6 2 | |
| | 8 2 | | 3+4 2 | |
| | 4+5 2 | | 3+4 2 | |
| | 6+7 2 | | 5+6 2 | |
| | 9 2 | | 3+4 2 | |
| | 8 2 | | 5+6 2 | |
| | 6+7 2 | | 3+4 2 | |
| 167.1.49 | 201101 | 68 | Fe2O3 | |
| | -GM | -T | -F | -L |
| | 8 2 | 5+6 4 | 5+6 2 | 2+2 4 |
| | 9 2 | 5+6 4 | 3+4 2 | 2+2 4 |
| | 9 2 | 4+4 4 | 3+4 2 | 2+2 4 |
| | 6+7 2 | 5+6 4 | 5+6 2 | 2+2 4 |
| | 8 2 | 5+6 4 | 5+6 2 | 2+2 4 |
| | 4+5 2 | 4+4 4 | 3+4 2 | 2+2 4 |
| | 8 2 | 5+6 4 | 3+4 2 | 2+2 4 |
| | 8 2 | 4+4 4 | 5+6 2 | 2+2 4 |
| | 4+5 2 | 5+6 4 | 5+6 2 | 2+2 4 |
| | 9 2 | 5+6 4 | 3+4 2 | 2+2 4 |
| | 9 2 | 5+6 4 | 3+4 2 | 2+2 4 |
| | 6+7 2 | 4+4 4 | 5+6 2 | 2+2 4 |
| | 8 2 | 4+4 4 | 5+6 2 | 2+2 4 |
| | 4+5 2 | 5+6 4 | 3+4 2 | 2+2 4 |
| | 8 2 | 5+6 4 | 3+4 2 | 2+2 4 |
| | 6+7 2 | 4+4 4 | 5+6 2 | 2+2 4 |
| | 9 2 | 5+6 4 | 5+6 2 | 2+2 4 |
| | 8 2 | | 3+4 2 | |
| | 9 2 | | 5+6 2 | |
| | 8 2 | | 3+4 2 | |
| | 4+5 2 | | 5+6 2 | |
| | 9 2 | | 5+6 2 | |
| | 6+7 2 | | 3+4 2 | |

Continued on next column

**TABLE LXXIV– continued**

| TQC | ICSD | | NoE | SF |
|---|---|---|---|---|
| | 9 2 | | 3+4 2 | |
| | 8 2 | | 3+4 2 | |
| | 4+5 2 | | 5+6 2 | |
| | 8 2 | | 5+6 2 | |
| | 8 2 | | 5+6 2 | |
| | 4+5 2 | | 3+4 2 | |
| | 9 2 | | 3+4 2 | |
| | 6+7 2 | | 5+6 2 | |
| | 9 2 | | 3+4 2 | |
| | 8 2 | | 5+6 2 | |
| | 6+7 2 | | 3+4 2 | |
| 167.1.50 | 201100 | | 68 | Fe2O3 |
| | -GM | -T | -F | -L |
| | 8 2 | 5+6 4 | 5+6 2 | 2+2 4 |
| | 9 2 | 5+6 4 | 3+4 2 | 2+2 4 |
| | 9 2 | 4+4 4 | 3+4 2 | 2+2 4 |
| | 6+7 2 | 5+6 4 | 5+6 2 | 2+2 4 |
| | 8 2 | 5+6 4 | 5+6 2 | 2+2 4 |
| | 4+5 2 | 4+4 4 | 3+4 2 | 2+2 4 |
| | 8 2 | 5+6 4 | 3+4 2 | 2+2 4 |
| | 9 2 | 4+4 4 | 5+6 2 | 2+2 4 |
| | 8 2 | 5+6 4 | 5+6 2 | 2+2 4 |
| | 4+5 2 | 5+6 4 | 3+4 2 | 2+2 4 |
| | 9 2 | 5+6 4 | 3+4 2 | 2+2 4 |
| | 6+7 2 | 4+4 4 | 5+6 2 | 2+2 4 |
| | 8 2 | 4+4 4 | 5+6 2 | 2+2 4 |
| | 4+5 2 | 5+6 4 | 3+4 2 | 2+2 4 |
| | 8 2 | 5+6 4 | 3+4 2 | 2+2 4 |
| | 6+7 2 | 4+4 4 | 5+6 2 | 2+2 4 |
| | 9 2 | 5+6 4 | 5+6 2 | 2+2 4 |
| | 8 2 | | 3+4 2 | |
| | 9 2 | | 5+6 2 | |
| | 8 2 | | 3+4 2 | |
| | 4+5 2 | | 5+6 2 | |
| | 9 2 | | 5+6 2 | |
| | 6+7 2 | | 3+4 2 | |
| | 9 2 | | 3+4 2 | |
| | 4+5 2 | | 5+6 2 | |
| | 8 2 | | 3+4 2 | |
| | 8 2 | | 5+6 2 | |
| | 9 2 | | 5+6 2 | |
| | 6+7 2 | | 3+4 2 | |
| | 8 2 | | 3+4 2 | |
| | 4+5 2 | | 5+6 2 | |
| | 9 2 | | 3+4 2 | |
| | 8 2 | | 5+6 2 | |
| | 6+7 2 | | 3+4 2 | |
| 167.1.51 | 201099 | | 68 | Fe2O3 |
| | -GM | -T | -F | -L |
| | 8 2 | 5+6 4 | 5+6 2 | 2+2 4 |
| | 9 2 | 5+6 4 | 3+4 2 | 2+2 4 |
| | 9 2 | 4+4 4 | 3+4 2 | 2+2 4 |
| | 6+7 2 | 5+6 4 | 5+6 2 | 2+2 4 |
| | 8 2 | 5+6 4 | 5+6 2 | 2+2 4 |
| | 4+5 2 | 4+4 4 | 3+4 2 | 2+2 4 |
| | 8 2 | 5+6 4 | 3+4 2 | 2+2 4 |
| | 9 2 | 4+4 4 | 5+6 2 | 2+2 4 |
| | 8 2 | 5+6 4 | 5+6 2 | 2+2 4 |
| | 4+5 2 | 5+6 4 | 3+4 2 | 2+2 4 |
| | 9 2 | 5+6 4 | 3+4 2 | 2+2 4 |
| | 6+7 2 | 4+4 4 | 5+6 2 | 2+2 4 |

Continued on next column

**TABLE LXXIV– continued**

| TQC | ICSD | | NoE | SF |
|---|---|---|---|---|
| | 8 2 | 4+4 4 | 5+6 2 | 2+2 4 |
| | 4+5 2 | 5+6 4 | 3+4 2 | 2+2 4 |
| | 8 2 | 5+6 4 | 3+4 2 | 2+2 4 |
| | 6+7 2 | 4+4 4 | 5+6 2 | 2+2 4 |
| | 9 2 | 5+6 4 | 5+6 2 | 2+2 4 |
| | 8 2 | | 3+4 2 | |
| | 9 2 | | 5+6 2 | |
| | 8 2 | | 3+4 2 | |
| | 4+5 2 | | 5+6 2 | |
| | 9 2 | | 5+6 2 | |
| | 6+7 2 | | 3+4 2 | |
| | 9 2 | | 3+4 2 | |
| | 4+5 2 | | 5+6 2 | |
| | 8 2 | | 3+4 2 | |
| | 8 2 | | 5+6 2 | |
| | 9 2 | | 5+6 2 | |
| | 6+7 2 | | 3+4 2 | |
| | 8 2 | | 3+4 2 | |
| | 4+5 2 | | 5+6 2 | |
| | 9 2 | | 3+4 2 | |
| | 8 2 | | 5+6 2 | |
| | 6+7 2 | | 3+4 2 | |
| 167.1.52 | 201098 | | 68 | Fe2O3 |
| | -GM | -T | -F | -L |
| | 8 2 | 5+6 4 | 5+6 2 | 2+2 4 |
| | 9 2 | 5+6 4 | 3+4 2 | 2+2 4 |
| | 9 2 | 4+4 4 | 3+4 2 | 2+2 4 |
| | 6+7 2 | 5+6 4 | 5+6 2 | 2+2 4 |
| | 8 2 | 5+6 4 | 5+6 2 | 2+2 4 |
| | 4+5 2 | 4+4 4 | 3+4 2 | 2+2 4 |
| | 8 2 | 5+6 4 | 3+4 2 | 2+2 4 |
| | 9 2 | 4+4 4 | 5+6 2 | 2+2 4 |
| | 8 2 | 5+6 4 | 5+6 2 | 2+2 4 |
| | 4+5 2 | 5+6 4 | 3+4 2 | 2+2 4 |
| | 9 2 | 5+6 4 | 3+4 2 | 2+2 4 |
| | 6+7 2 | 4+4 4 | 5+6 2 | 2+2 4 |
| | 8 2 | 4+4 4 | 5+6 2 | 2+2 4 |
| | 4+5 2 | 5+6 4 | 3+4 2 | 2+2 4 |
| | 8 2 | 5+6 4 | 3+4 2 | 2+2 4 |
| | 6+7 2 | 4+4 4 | 5+6 2 | 2+2 4 |
| | 9 2 | 5+6 4 | 5+6 2 | 2+2 4 |
| | 8 2 | | 3+4 2 | |
| | 9 2 | | 5+6 2 | |
| | 8 2 | | 3+4 2 | |
| | 4+5 2 | | 5+6 2 | |
| | 9 2 | | 5+6 2 | |
| | 6+7 2 | | 3+4 2 | |
| | 9 2 | | 3+4 2 | |
| | 4+5 2 | | 5+6 2 | |
| | 8 2 | | 3+4 2 | |
| | 8 2 | | 5+6 2 | |
| | 9 2 | | 5+6 2 | |
| | 8 2 | | 3+4 2 | |
| | 4+5 2 | | 3+4 2 | |
| | 6+7 2 | | 5+6 2 | |
| | 9 2 | | 3+4 2 | |
| | 8 2 | | 5+6 2 | |
| | 6+7 2 | | 3+4 2 | |
| 167.1.53 | 201097 | | 68 | Fe2O3 |
| | -GM | -T | -F | -L |
| | 8 2 | 5+6 4 | 5+6 2 | 2+2 4 |

Continued on next column

**TABLE LXXIV– continued**

| TQC | ICSD | NoE | SF | |
|---|---|---|---|---|
| | 9 2 | 5+6 4 | 3+4 2 | 2+2 4 |
| | 9 2 | 4+4 4 | 3+4 2 | 2+2 4 |
| | 6+7 2 | 5+6 4 | 5+6 2 | 2+2 4 |
| | 8 2 | 5+6 4 | 5+6 2 | 2+2 4 |
| | 4+5 2 | 4+4 4 | 3+4 2 | 2+2 4 |
| | 8 2 | 5+6 4 | 3+4 2 | 2+2 4 |
| | 8 2 | 4+4 4 | 5+6 2 | 2+2 4 |
| | 9 2 | 5+6 4 | 5+6 2 | 2+2 4 |
| | 4+5 2 | 5+6 4 | 3+4 2 | 2+2 4 |
| | 9 2 | 5+6 4 | 3+4 2 | 2+2 4 |
| | 6+7 2 | 4+4 4 | 5+6 2 | 2+2 4 |
| | 8 2 | 4+4 4 | 5+6 2 | 2+2 4 |
| | 4+5 2 | 5+6 4 | 3+4 2 | 2+2 4 |
| | 8 2 | 5+6 4 | 3+4 2 | 2+2 4 |
| | 6+7 2 | 4+4 4 | 5+6 2 | 2+2 4 |
| | 9 2 | 5+6 4 | 5+6 2 | 2+2 4 |
| | 8 2 | | 3+4 2 | |
| | 9 2 | | 5+6 2 | |
| | 8 2 | | 3+4 2 | |
| | 4+5 2 | | 5+6 2 | |
| | 9 2 | | 5+6 2 | |
| | 6+7 2 | | 3+4 2 | |
| | 9 2 | | 3+4 2 | |
| | 4+5 2 | | 5+6 2 | |
| | 8 2 | | 3+4 2 | |
| | 8 2 | | 5+6 2 | |
| | 9 2 | | 5+6 2 | |
| | 8 2 | | 3+4 2 | |
| | 4+5 2 | | 3+4 2 | |
| | 6+7 2 | | 5+6 2 | |
| | 9 2 | | 3+4 2 | |
| | 8 2 | | 5+6 2 | |
| | 6+7 2 | | 3+4 2 | |
| 167.1.54 | 201096 | 68 | Fe2O3 | |
| | -GM | -T | -F | -L |
| | 8 2 | 5+6 4 | 5+6 2 | 2+2 4 |
| | 9 2 | 5+6 4 | 3+4 2 | 2+2 4 |
| | 9 2 | 4+4 4 | 3+4 2 | 2+2 4 |
| | 6+7 2 | 5+6 4 | 5+6 2 | 2+2 4 |
| | 8 2 | 5+6 4 | 5+6 2 | 2+2 4 |
| | 4+5 2 | 4+4 4 | 3+4 2 | 2+2 4 |
| | 8 2 | 5+6 4 | 3+4 2 | 2+2 4 |
| | 9 2 | 4+4 4 | 5+6 2 | 2+2 4 |
| | 8 2 | 5+6 4 | 5+6 2 | 2+2 4 |
| | 4+5 2 | 5+6 4 | 3+4 2 | 2+2 4 |
| | 9 2 | 5+6 4 | 3+4 2 | 2+2 4 |
| | 6+7 2 | 4+4 4 | 5+6 2 | 2+2 4 |
| | 8 2 | 4+4 4 | 5+6 2 | 2+2 4 |
| | 4+5 2 | 5+6 4 | 3+4 2 | 2+2 4 |
| | 8 2 | 5+6 4 | 3+4 2 | 2+2 4 |
| | 6+7 2 | 4+4 4 | 5+6 2 | 2+2 4 |
| | 9 2 | 5+6 4 | 5+6 2 | 2+2 4 |
| | 8 2 | | 3+4 2 | |
| | 9 2 | | 5+6 2 | |
| | 8 2 | | 3+4 2 | |
| | 4+5 2 | | 5+6 2 | |
| | 9 2 | | 5+6 2 | |
| | 6+7 2 | | 3+4 2 | |
| | 9 2 | | 3+4 2 | |
| | 4+5 2 | | 5+6 2 | |
| | 8 2 | | 3+4 2 | |

Continued on next column

**TABLE LXXIV– continued**

| TQC | ICSD | NoE | SF | |
|---|---|---|---|---|
| | 8 2 | | 5+6 2 | |
| | 9 2 | | 5+6 2 | |
| | 8 2 | | 3+4 2 | |
| | 4+5 2 | | 3+4 2 | |
| | 6+7 2 | | 5+6 2 | |
| | 9 2 | | 3+4 2 | |
| | 8 2 | | 5+6 2 | |
| | 6+7 2 | | 3+4 2 | |
| 167.1.55 | 184766 | 68 | Fe2O3 | |
| | -GM | -T | -F | -L |
| | 8 2 | 5+6 4 | 5+6 2 | 2+2 4 |
| | 9 2 | 5+6 4 | 3+4 2 | 2+2 4 |
| | 9 2 | 4+4 4 | 3+4 2 | 2+2 4 |
| | 6+7 2 | 5+6 4 | 5+6 2 | 2+2 4 |
| | 8 2 | 5+6 4 | 5+6 2 | 2+2 4 |
| | 4+5 2 | 4+4 4 | 3+4 2 | 2+2 4 |
| | 8 2 | 5+6 4 | 3+4 2 | 2+2 4 |
| | 9 2 | 4+4 4 | 5+6 2 | 2+2 4 |
| | 8 2 | 5+6 4 | 5+6 2 | 2+2 4 |
| | 4+5 2 | 5+6 4 | 3+4 2 | 2+2 4 |
| | 9 2 | 5+6 4 | 3+4 2 | 2+2 4 |
| | 6+7 2 | 4+4 4 | 5+6 2 | 2+2 4 |
| | 8 2 | 4+4 4 | 5+6 2 | 2+2 4 |
| | 4+5 2 | 5+6 4 | 3+4 2 | 2+2 4 |
| | 8 2 | 5+6 4 | 3+4 2 | 2+2 4 |
| | 6+7 2 | 4+4 4 | 5+6 2 | 2+2 4 |
| | 9 2 | 5+6 4 | 5+6 2 | 2+2 4 |
| | 8 2 | | 3+4 2 | |
| | 9 2 | | 5+6 2 | |
| | 8 2 | | 3+4 2 | |
| | 4+5 2 | | 5+6 2 | |
| | 9 2 | | 5+6 2 | |
| | 6+7 2 | | 3+4 2 | |
| | 9 2 | | 3+4 2 | |
| | 4+5 2 | | 5+6 2 | |
| | 8 2 | | 3+4 2 | |
| | 8 2 | | 5+6 2 | |
| | 9 2 | | 5+6 2 | |
| | 8 2 | | 3+4 2 | |
| | 4+5 2 | | 3+4 2 | |
| | 6+7 2 | | 5+6 2 | |
| | 9 2 | | 3+4 2 | |
| | 8 2 | | 5+6 2 | |
| | 6+7 2 | | 3+4 2 | |
| 167.1.56 | 182848 | 68 | Fe2O3 | |
| | -GM | -T | -F | -L |
| | 8 2 | 5+6 4 | 5+6 2 | 2+2 4 |
| | 9 2 | 5+6 4 | 3+4 2 | 2+2 4 |
| | 9 2 | 4+4 4 | 3+4 2 | 2+2 4 |
| | 6+7 2 | 5+6 4 | 5+6 2 | 2+2 4 |
| | 8 2 | 5+6 4 | 5+6 2 | 2+2 4 |
| | 4+5 2 | 4+4 4 | 3+4 2 | 2+2 4 |
| | 8 2 | 5+6 4 | 3+4 2 | 2+2 4 |
| | 8 2 | 4+4 4 | 5+6 2 | 2+2 4 |
| | 4+5 2 | 5+6 4 | 5+6 2 | 2+2 4 |
| | 9 2 | 5+6 4 | 3+4 2 | 2+2 4 |
| | 9 2 | 5+6 4 | 3+4 2 | 2+2 4 |
| | 6+7 2 | 4+4 4 | 5+6 2 | 2+2 4 |
| | 8 2 | 4+4 4 | 5+6 2 | 2+2 4 |
| | 4+5 2 | 5+6 4 | 3+4 2 | 2+2 4 |
| | 8 2 | 5+6 4 | 3+4 2 | 2+2 4 |

Continued on next column

**TABLE LXXIV– continued**

| TQC | ICSD | | NoE | SF |
|---|---|---|---|---|
| | 6+7 2 | 4+4 4 | 5+6 2 | 2+2 4 |
| | 9 2 | 5+6 4 | 5+6 2 | 2+2 4 |
| | 8 2 | | 3+4 2 | |
| | 9 2 | | 5+6 2 | |
| | 8 2 | | 3+4 2 | |
| | 4+5 2 | | 5+6 2 | |
| | 9 2 | | 5+6 2 | |
| | 6+7 2 | | 3+4 2 | |
| | 9 2 | | 3+4 2 | |
| | 4+5 2 | | 3+4 2 | |
| | 8 2 | | 5+6 2 | |
| | 8 2 | | 5+6 2 | |
| | 9 2 | | 5+6 2 | |
| | 8 2 | | 3+4 2 | |
| | 4+5 2 | | 3+4 2 | |
| | 6+7 2 | | 5+6 2 | |
| | 9 2 | | 3+4 2 | |
| | 8 2 | | 5+6 2 | |
| | 6+7 2 | | 3+4 2 | |
| 167.1.57 | 182847 | | 68 | Fe2O3 |
| | -GM | -T | -F | -L |
| | 8 2 | 5+6 4 | 5+6 2 | 2+2 4 |
| | 9 2 | 5+6 4 | 3+4 2 | 2+2 4 |
| | 9 2 | 4+4 4 | 3+4 2 | 2+2 4 |
| | 6+7 2 | 5+6 4 | 5+6 2 | 2+2 4 |
| | 8 2 | 5+6 4 | 5+6 2 | 2+2 4 |
| | 4+5 2 | 4+4 4 | 3+4 2 | 2+2 4 |
| | 8 2 | 5+6 4 | 3+4 2 | 2+2 4 |
| | 8 2 | 4+4 4 | 5+6 2 | 2+2 4 |
| | 4+5 2 | 5+6 4 | 5+6 2 | 2+2 4 |
| | 9 2 | 5+6 4 | 3+4 2 | 2+2 4 |
| | 9 2 | 5+6 4 | 3+4 2 | 2+2 4 |
| | 6+7 2 | 4+4 4 | 5+6 2 | 2+2 4 |
| | 8 2 | 4+4 4 | 5+6 2 | 2+2 4 |
| | 4+5 2 | 5+6 4 | 3+4 2 | 2+2 4 |
| | 8 2 | 5+6 4 | 3+4 2 | 2+2 4 |
| | 6+7 2 | 4+4 4 | 5+6 2 | 2+2 4 |
| | 9 2 | 5+6 4 | 5+6 2 | 2+2 4 |
| | 8 2 | | 3+4 2 | |
| | 9 2 | | 5+6 2 | |
| | 8 2 | | 3+4 2 | |
| | 4+5 2 | | 5+6 2 | |
| | 9 2 | | 5+6 2 | |
| | 6+7 2 | | 3+4 2 | |
| | 9 2 | | 3+4 2 | |
| | 4+5 2 | | 3+4 2 | |
| | 8 2 | | 5+6 2 | |
| | 8 2 | | 5+6 2 | |
| | 9 2 | | 5+6 2 | |
| | 8 2 | | 3+4 2 | |
| | 4+5 2 | | 3+4 2 | |
| | 6+7 2 | | 5+6 2 | |
| | 9 2 | | 3+4 2 | |
| | 8 2 | | 5+6 2 | |
| | 6+7 2 | | 3+4 2 | |
| 167.1.58 | 182846 | | 68 | Fe2O3 |
| | -GM | -T | -F | -L |
| | 8 2 | 5+6 4 | 5+6 2 | 2+2 4 |
| | 9 2 | 5+6 4 | 3+4 2 | 2+2 4 |
| | 9 2 | 4+4 4 | 3+4 2 | 2+2 4 |
| | 6+7 2 | 5+6 4 | 5+6 2 | 2+2 4 |
| | 8 2 | 5+6 4 | 5+6 2 | 2+2 4 |
| | 4+5 2 | 4+4 4 | 3+4 2 | 2+2 4 |
| | 8 2 | 5+6 4 | 3+4 2 | 2+2 4 |
| | 8 2 | 4+4 4 | 5+6 2 | 2+2 4 |
| | 4+5 2 | 5+6 4 | 5+6 2 | 2+2 4 |
| | 9 2 | 5+6 4 | 3+4 2 | 2+2 4 |
| | 9 2 | 5+6 4 | 3+4 2 | 2+2 4 |
| | 6+7 2 | 4+4 4 | 5+6 2 | 2+2 4 |
| | 8 2 | 4+4 4 | 5+6 2 | 2+2 4 |
| | 4+5 2 | 5+6 4 | 3+4 2 | 2+2 4 |
| | 8 2 | 5+6 4 | 3+4 2 | 2+2 4 |
| | 6+7 2 | 4+4 4 | 5+6 2 | 2+2 4 |
| | 9 2 | 5+6 4 | 5+6 2 | 2+2 4 |
| | 8 2 | | 3+4 2 | |
| | 9 2 | | 5+6 2 | |
| | 8 2 | | 3+4 2 | |
| | 4+5 2 | | 5+6 2 | |
| | 9 2 | | 5+6 2 | |
| | 6+7 2 | | 3+4 2 | |
| | 9 2 | | 3+4 2 | |
| | 4+5 2 | | 3+4 2 | |
| | 8 2 | | 5+6 2 | |
| | 8 2 | | 5+6 2 | |
| | 8 2 | | 5+6 2 | |
| | 9 2 | | 3+4 2 | |
| | 4+5 2 | | 3+4 2 | |
| | 6+7 2 | | 5+6 2 | |
| | 9 2 | | 3+4 2 | |
| | 8 2 | | 5+6 2 | |
| | 6+7 2 | | 3+4 2 | |
| 167.1.59 | 182845 | | 68 | Fe2O3 |
| | -GM | -T | -F | -L |
| | 8 2 | 5+6 4 | 5+6 2 | 2+2 4 |
| | 9 2 | 5+6 4 | 3+4 2 | 2+2 4 |
| | 9 2 | 4+4 4 | 3+4 2 | 2+2 4 |
| | 6+7 2 | 5+6 4 | 5+6 2 | 2+2 4 |
| | 8 2 | 5+6 4 | 5+6 2 | 2+2 4 |
| | 4+5 2 | 4+4 4 | 3+4 2 | 2+2 4 |
| | 8 2 | 5+6 4 | 3+4 2 | 2+2 4 |
| | 8 2 | 4+4 4 | 5+6 2 | 2+2 4 |
| | 4+5 2 | 5+6 4 | 5+6 2 | 2+2 4 |
| | 9 2 | 5+6 4 | 3+4 2 | 2+2 4 |
| | 9 2 | 5+6 4 | 3+4 2 | 2+2 4 |
| | 6+7 2 | 4+4 4 | 5+6 2 | 2+2 4 |
| | 8 2 | 4+4 4 | 5+6 2 | 2+2 4 |
| | 4+5 2 | 5+6 4 | 3+4 2 | 2+2 4 |
| | 8 2 | 5+6 4 | 3+4 2 | 2+2 4 |
| | 6+7 2 | 4+4 4 | 5+6 2 | 2+2 4 |
| | 9 2 | 5+6 4 | 5+6 2 | 2+2 4 |
| | 8 2 | | 3+4 2 | |
| | 9 2 | | 5+6 2 | |
| | 8 2 | | 3+4 2 | |
| | 4+5 2 | | 5+6 2 | |
| | 9 2 | | 5+6 2 | |
| | 6+7 2 | | 3+4 2 | |
| | 9 2 | | 3+4 2 | |
| | 4+5 2 | | 5+6 2 | |
| | 8 2 | | 3+4 2 | |
| | 8 2 | | 5+6 2 | |
| | 9 2 | | 5+6 2 | |
| | 8 2 | | 3+4 2 | |

Continued on next column

**TABLE LXXIV– continued**

| TQC | ICSD | | NoE | SF |
|---|---|---|---|---|
| | 4+5 2 | | 3+4 2 | |
| | 6+7 2 | | 5+6 2 | |
| | 9 2 | | 3+4 2 | |
| | 8 2 | | 5+6 2 | |
| | 6+7 2 | | 3+4 2 | |
| 167.1.60 | 182844 | | 68 | Fe2O3 |
| | -GM | -T | -F | -L |
| | 8 2 | 5+6 4 | 5+6 2 | 2+2 4 |
| | 9 2 | 5+6 4 | 3+4 2 | 2+2 4 |
| | 9 2 | 4+4 4 | 3+4 2 | 2+2 4 |
| | 6+7 2 | 5+6 4 | 5+6 2 | 2+2 4 |
| | 8 2 | 5+6 4 | 5+6 2 | 2+2 4 |
| | 4+5 2 | 4+4 4 | 3+4 2 | 2+2 4 |
| | 8 2 | 5+6 4 | 3+4 2 | 2+2 4 |
| | 8 2 | 4+4 4 | 5+6 2 | 2+2 4 |
| | 4+5 2 | 5+6 4 | 5+6 2 | 2+2 4 |
| | 9 2 | 5+6 4 | 3+4 2 | 2+2 4 |
| | 9 2 | 5+6 4 | 3+4 2 | 2+2 4 |
| | 6+7 2 | 4+4 4 | 5+6 2 | 2+2 4 |
| | 8 2 | 4+4 4 | 5+6 2 | 2+2 4 |
| | 4+5 2 | 5+6 4 | 3+4 2 | 2+2 4 |
| | 8 2 | 5+6 4 | 3+4 2 | 2+2 4 |
| | 6+7 2 | 4+4 4 | 5+6 2 | 2+2 4 |
| | 9 2 | 5+6 4 | 5+6 2 | 2+2 4 |
| | 8 2 | | 3+4 2 | |
| | 9 2 | | 5+6 2 | |
| | 8 2 | | 3+4 2 | |
| | 4+5 2 | | 5+6 2 | |
| | 9 2 | | 5+6 2 | |
| | 6+7 2 | | 3+4 2 | |
| | 9 2 | | 3+4 2 | |
| | 4+5 2 | | 5+6 2 | |
| | 8 2 | | 3+4 2 | |
| | 8 2 | | 5+6 2 | |
| | 9 2 | | 5+6 2 | |
| | 8 2 | | 3+4 2 | |
| | 4+5 2 | | 3+4 2 | |
| | 6+7 2 | | 5+6 2 | |
| | 9 2 | | 3+4 2 | |
| | 8 2 | | 5+6 2 | |
| | 6+7 2 | | 3+4 2 | |
| 167.1.61 | 182843 | | 68 | Fe2O3 |
| | -GM | -T | -F | -L |
| | 8 2 | 5+6 4 | 5+6 2 | 2+2 4 |
| | 9 2 | 5+6 4 | 3+4 2 | 2+2 4 |
| | 9 2 | 4+4 4 | 3+4 2 | 2+2 4 |
| | 6+7 2 | 5+6 4 | 5+6 2 | 2+2 4 |
| | 8 2 | 5+6 4 | 5+6 2 | 2+2 4 |
| | 4+5 2 | 4+4 4 | 3+4 2 | 2+2 4 |
| | 8 2 | 5+6 4 | 3+4 2 | 2+2 4 |
| | 8 2 | 4+4 4 | 5+6 2 | 2+2 4 |
| | 4+5 2 | 5+6 4 | 5+6 2 | 2+2 4 |
| | 9 2 | 5+6 4 | 3+4 2 | 2+2 4 |
| | 9 2 | 5+6 4 | 3+4 2 | 2+2 4 |
| | 6+7 2 | 4+4 4 | 5+6 2 | 2+2 4 |
| | 8 2 | 4+4 4 | 5+6 2 | 2+2 4 |
| | 4+5 2 | 5+6 4 | 3+4 2 | 2+2 4 |
| | 8 2 | 5+6 4 | 3+4 2 | 2+2 4 |
| | 6+7 2 | 4+4 4 | 5+6 2 | 2+2 4 |
| | 9 2 | 5+6 4 | 5+6 2 | 2+2 4 |
| | 8 2 | | 3+4 2 | |

Continued on next column

**TABLE LXXIV– continued**

| TQC | ICSD | | NoE | SF |
|---|---|---|---|---|
| | 9 2 | | 5+6 2 | |
| | 8 2 | | 3+4 2 | |
| | 4+5 2 | | 5+6 2 | |
| | 9 2 | | 5+6 2 | |
| | 6+7 2 | | 3+4 2 | |
| | 9 2 | | 3+4 2 | |
| | 8 2 | | 3+4 2 | |
| | 4+5 2 | | 5+6 2 | |
| | 8 2 | | 5+6 2 | |
| | 8 2 | | 5+6 2 | |
| | 4+5 2 | | 3+4 2 | |
| | 9 2 | | 3+4 2 | |
| | 6+7 2 | | 5+6 2 | |
| | 9 2 | | 3+4 2 | |
| | 8 2 | | 5+6 2 | |
| | 6+7 2 | | 3+4 2 | |
| 167.1.62 | 182842 | | 68 | Fe2O3 |
| | -GM | -T | -F | -L |
| | 8 2 | 5+6 4 | 5+6 2 | 2+2 4 |
| | 9 2 | 5+6 4 | 3+4 2 | 2+2 4 |
| | 9 2 | 4+4 4 | 3+4 2 | 2+2 4 |
| | 6+7 2 | 5+6 4 | 5+6 2 | 2+2 4 |
| | 8 2 | 5+6 4 | 5+6 2 | 2+2 4 |
| | 4+5 2 | 4+4 4 | 3+4 2 | 2+2 4 |
| | 8 2 | 5+6 4 | 3+4 2 | 2+2 4 |
| | 8 2 | 4+4 4 | 5+6 2 | 2+2 4 |
| | 4+5 2 | 5+6 4 | 5+6 2 | 2+2 4 |
| | 9 2 | 5+6 4 | 3+4 2 | 2+2 4 |
| | 9 2 | 5+6 4 | 3+4 2 | 2+2 4 |
| | 6+7 2 | 4+4 4 | 5+6 2 | 2+2 4 |
| | 8 2 | 4+4 4 | 5+6 2 | 2+2 4 |
| | 4+5 2 | 5+6 4 | 3+4 2 | 2+2 4 |
| | 8 2 | 5+6 4 | 3+4 2 | 2+2 4 |
| | 6+7 2 | 4+4 4 | 5+6 2 | 2+2 4 |
| | 9 2 | 5+6 4 | 5+6 2 | 2+2 4 |
| | 8 2 | | 3+4 2 | |
| | 9 2 | | 5+6 2 | |
| | 8 2 | | 3+4 2 | |
| | 4+5 2 | | 5+6 2 | |
| | 9 2 | | 5+6 2 | |
| | 6+7 2 | | 3+4 2 | |
| | 9 2 | | 3+4 2 | |
| | 4+5 2 | | 5+6 2 | |
| | 8 2 | | 3+4 2 | |
| | 8 2 | | 5+6 2 | |
| | 9 2 | | 5+6 2 | |
| | 8 2 | | 3+4 2 | |
| | 4+5 2 | | 3+4 2 | |
| | 6+7 2 | | 5+6 2 | |
| | 9 2 | | 3+4 2 | |
| | 8 2 | | 5+6 2 | |
| | 6+7 2 | | 3+4 2 | |
| 167.1.63 | 182841 | | 68 | Fe2O3 |
| | -GM | -T | -F | -L |
| | 8 2 | 5+6 4 | 5+6 2 | 2+2 4 |
| | 9 2 | 5+6 4 | 3+4 2 | 2+2 4 |
| | 9 2 | 4+4 4 | 3+4 2 | 2+2 4 |
| | 6+7 2 | 5+6 4 | 5+6 2 | 2+2 4 |
| | 8 2 | 5+6 4 | 5+6 2 | 2+2 4 |
| | 4+5 2 | 4+4 4 | 3+4 2 | 2+2 4 |
| | 8 2 | 5+6 4 | 3+4 2 | 2+2 4 |

Continued on next column

**TABLE LXXIV– continued**

| TQC | ICSD | NoE | SF |
|---|---|---|---|
| | 8 2 4+4 4 5+6 2 2+2 4 | | |
| | 4+5 2 5+6 4 5+6 2 2+2 4 | | |
| | 9 2 5+6 4 3+4 2 2+2 4 | | |
| | 9 2 5+6 4 3+4 2 2+2 4 | | |
| | 6+7 2 4+4 4 5+6 2 2+2 4 | | |
| | 8 2 4+4 4 5+6 2 2+2 4 | | |
| | 4+5 2 5+6 4 3+4 2 2+2 4 | | |
| | 8 2 5+6 4 3+4 2 2+2 4 | | |
| | 6+7 2 4+4 4 5+6 2 2+2 4 | | |
| | 9 2 5+6 4 5+6 2 2+2 4 | | |
| | 8 2 3+4 2 | | |
| | 9 2 5+6 2 | | |
| | 8 2 3+4 2 | | |
| | 4+5 2 5+6 2 | | |
| | 9 2 5+6 2 | | |
| | 6+7 2 3+4 2 | | |
| | 9 2 3+4 2 | | |
| | 4+5 2 5+6 2 | | |
| | 8 2 3+4 2 | | |
| | 8 2 5+6 2 | | |
| | 8 2 5+6 2 | | |
| | 9 2 3+4 2 | | |
| | 4+5 2 3+4 2 | | |
| | 6+7 2 5+6 2 | | |
| | 9 2 3+4 2 | | |
| | 8 2 5+6 2 | | |
| | 6+7 2 3+4 2 | | |
| 167.1.64 | 182840 | 68 | Fe2O3 |
| | -GM -T -F -L | | |
| | 8 2 5+6 4 5+6 2 2+2 4 | | |
| | 9 2 5+6 4 3+4 2 2+2 4 | | |
| | 9 2 4+4 4 3+4 2 2+2 4 | | |
| | 6+7 2 5+6 4 5+6 2 2+2 4 | | |
| | 8 2 5+6 4 5+6 2 2+2 4 | | |
| | 4+5 2 4+4 4 3+4 2 2+2 4 | | |
| | 8 2 5+6 4 3+4 2 2+2 4 | | |
| | 8 2 4+4 4 5+6 2 2+2 4 | | |
| | 4+5 2 5+6 4 5+6 2 2+2 4 | | |
| | 9 2 5+6 4 3+4 2 2+2 4 | | |
| | 9 2 5+6 4 3+4 2 2+2 4 | | |
| | 6+7 2 4+4 4 5+6 2 2+2 4 | | |
| | 8 2 4+4 4 5+6 2 2+2 4 | | |
| | 4+5 2 5+6 4 3+4 2 2+2 4 | | |
| | 8 2 5+6 4 3+4 2 2+2 4 | | |
| | 6+7 2 4+4 4 5+6 2 2+2 4 | | |
| | 9 2 5+6 4 5+6 2 2+2 4 | | |
| | 8 2 3+4 2 | | |
| | 9 2 5+6 2 | | |
| | 8 2 3+4 2 | | |
| | 4+5 2 5+6 2 | | |
| | 9 2 5+6 2 | | |
| | 6+7 2 3+4 2 | | |
| | 9 2 3+4 2 | | |
| | 4+5 2 5+6 2 | | |
| | 8 2 3+4 2 | | |
| | 8 2 5+6 2 | | |
| | 8 2 5+6 2 | | |
| | 9 2 3+4 2 | | |
| | 4+5 2 3+4 2 | | |
| | 6+7 2 5+6 2 | | |
| | 9 2 3+4 2 | | |

Continued on next column

**TABLE LXXIV– continued**

| TQC | ICSD | NoE | SF |
|---|---|---|---|
| | 8 2 5+6 2 | | |
| | 6+7 2 3+4 2 | | |
| 167.1.65 | 182839 | 68 | Fe2O3 |
| | -GM -T -F -L | | |
| | 8 2 5+6 4 5+6 2 2+2 4 | | |
| | 9 2 5+6 4 3+4 2 2+2 4 | | |
| | 9 2 4+4 4 3+4 2 2+2 4 | | |
| | 6+7 2 5+6 4 5+6 2 2+2 4 | | |
| | 8 2 5+6 4 5+6 2 2+2 4 | | |
| | 4+5 2 4+4 4 3+4 2 2+2 4 | | |
| | 8 2 5+6 4 3+4 2 2+2 4 | | |
| | 9 2 4+4 4 5+6 2 2+2 4 | | |
| | 8 2 5+6 4 5+6 2 2+2 4 | | |
| | 4+5 2 5+6 4 3+4 2 2+2 4 | | |
| | 9 2 5+6 4 3+4 2 2+2 4 | | |
| | 6+7 2 4+4 4 5+6 2 2+2 4 | | |
| | 8 2 4+4 4 5+6 2 2+2 4 | | |
| | 4+5 2 5+6 4 3+4 2 2+2 4 | | |
| | 8 2 5+6 4 3+4 2 2+2 4 | | |
| | 6+7 2 4+4 4 5+6 2 2+2 4 | | |
| | 9 2 5+6 4 5+6 2 2+2 4 | | |
| | 8 2 3+4 2 | | |
| | 9 2 5+6 2 | | |
| | 8 2 3+4 2 | | |
| | 4+5 2 5+6 2 | | |
| | 9 2 5+6 2 | | |
| | 6+7 2 3+4 2 | | |
| | 9 2 3+4 2 | | |
| | 4+5 2 5+6 2 | | |
| | 8 2 3+4 2 | | |
| | 8 2 5+6 2 | | |
| | 9 2 5+6 2 | | |
| | 8 2 3+4 2 | | |
| | 4+5 2 3+4 2 | | |
| | 6+7 2 5+6 2 | | |
| | 9 2 3+4 2 | | |
| | 8 2 5+6 2 | | |
| | 6+7 2 3+4 2 | | |
| 167.1.66 | 173655 | 68 | Fe2O3 |
| | -GM -T -F -L | | |
| | 8 2 5+6 4 5+6 2 2+2 4 | | |
| | 9 2 5+6 4 3+4 2 2+2 4 | | |
| | 9 2 4+4 4 3+4 2 2+2 4 | | |
| | 6+7 2 5+6 4 5+6 2 2+2 4 | | |
| | 8 2 5+6 4 5+6 2 2+2 4 | | |
| | 4+5 2 4+4 4 3+4 2 2+2 4 | | |
| | 8 2 5+6 4 3+4 2 2+2 4 | | |
| | 8 2 4+4 4 5+6 2 2+2 4 | | |
| | 9 2 5+6 4 5+6 2 2+2 4 | | |
| | 4+5 2 5+6 4 3+4 2 2+2 4 | | |
| | 9 2 5+6 4 3+4 2 2+2 4 | | |
| | 6+7 2 4+4 4 5+6 2 2+2 4 | | |
| | 8 2 4+4 4 5+6 2 2+2 4 | | |
| | 4+5 2 5+6 4 3+4 2 2+2 4 | | |
| | 8 2 5+6 4 3+4 2 2+2 4 | | |
| | 6+7 2 4+4 4 5+6 2 2+2 4 | | |
| | 9 2 5+6 4 5+6 2 2+2 4 | | |
| | 8 2 3+4 2 | | |
| | 9 2 5+6 2 | | |
| | 8 2 3+4 2 | | |
| | 4+5 2 5+6 2 | | |

Continued on next column

**TABLE LXXIV– continued**

| TQC | ICSD | | NoE | SF |
|---|---|---|---|---|
| | 9 2 | | 5+6 2 | |
| | 6+7 2 | | 3+4 2 | |
| | 9 2 | | 3+4 2 | |
| | 4+5 2 | | 5+6 2 | |
| | 8 2 | | 3+4 2 | |
| | 8 2 | | 5+6 2 | |
| | 8 2 | | 5+6 2 | |
| | 9 2 | | 3+4 2 | |
| | 4+5 2 | | 3+4 2 | |
| | 6+7 2 | | 5+6 2 | |
| | 9 2 | | 3+4 2 | |
| | 8 2 | | 5+6 2 | |
| | 6+7 2 | | 3+4 2 | |
| 167.1.67 | 173653 | | 68 | Fe2O3 |
| | -GM | -T | -F | -L |
| | 8 2 | 5+6 4 | 5+6 2 | 2+2 4 |
| | 9 2 | 5+6 4 | 3+4 2 | 2+2 4 |
| | 9 2 | 4+4 4 | 3+4 2 | 2+2 4 |
| | 6+7 2 | 5+6 4 | 5+6 2 | 2+2 4 |
| | 8 2 | 5+6 4 | 5+6 2 | 2+2 4 |
| | 4+5 2 | 4+4 4 | 3+4 2 | 2+2 4 |
| | 8 2 | 5+6 4 | 3+4 2 | 2+2 4 |
| | 8 2 | 4+4 4 | 5+6 2 | 2+2 4 |
| | 9 2 | 5+6 4 | 5+6 2 | 2+2 4 |
| | 4+5 2 | 5+6 4 | 3+4 2 | 2+2 4 |
| | 9 2 | 5+6 4 | 3+4 2 | 2+2 4 |
| | 6+7 2 | 4+4 4 | 5+6 2 | 2+2 4 |
| | 8 2 | 4+4 4 | 5+6 2 | 2+2 4 |
| | 4+5 2 | 5+6 4 | 3+4 2 | 2+2 4 |
| | 8 2 | 5+6 4 | 3+4 2 | 2+2 4 |
| | 6+7 2 | 4+4 4 | 5+6 2 | 2+2 4 |
| | 9 2 | 5+6 4 | 5+6 2 | 2+2 4 |
| | 8 2 | | 3+4 2 | |
| | 9 2 | | 5+6 2 | |
| | 8 2 | | 3+4 2 | |
| | 4+5 2 | | 5+6 2 | |
| | 9 2 | | 5+6 2 | |
| | 6+7 2 | | 3+4 2 | |
| | 9 2 | | 3+4 2 | |
| | 4+5 2 | | 5+6 2 | |
| | 8 2 | | 3+4 2 | |
| | 8 2 | | 5+6 2 | |
| | 8 2 | | 5+6 2 | |
| | 9 2 | | 3+4 2 | |
| | 4+5 2 | | 3+4 2 | |
| | 6+7 2 | | 5+6 2 | |
| | 9 2 | | 3+4 2 | |
| | 8 2 | | 5+6 2 | |
| | 6+7 2 | | 3+4 2 | |
| 167.1.68 | 161294 | | 68 | Fe2O3 |
| | -GM | -T | -F | -L |
| | 8 2 | 5+6 4 | 5+6 2 | 2+2 4 |
| | 9 2 | 5+6 4 | 3+4 2 | 2+2 4 |
| | 9 2 | 4+4 4 | 3+4 2 | 2+2 4 |
| | 6+7 2 | 5+6 4 | 5+6 2 | 2+2 4 |
| | 8 2 | 5+6 4 | 5+6 2 | 2+2 4 |
| | 4+5 2 | 4+4 4 | 3+4 2 | 2+2 4 |
| | 8 2 | 5+6 4 | 3+4 2 | 2+2 4 |
| | 8 2 | 4+4 4 | 5+6 2 | 2+2 4 |
| | 4+5 2 | 5+6 4 | 5+6 2 | 2+2 4 |
| | 9 2 | 5+6 4 | 3+4 2 | 2+2 4 |

Continued on next column

**TABLE LXXIV– continued**

| TQC | ICSD | | NoE | SF |
|---|---|---|---|---|
| | 9 2 | 5+6 4 | 3+4 2 | 2+2 4 |
| | 6+7 2 | 4+4 4 | 5+6 2 | 2+2 4 |
| | 8 2 | 4+4 4 | 5+6 2 | 2+2 4 |
| | 4+5 2 | 5+6 4 | 3+4 2 | 2+2 4 |
| | 8 2 | 5+6 4 | 3+4 2 | 2+2 4 |
| | 6+7 2 | 4+4 4 | 5+6 2 | 2+2 4 |
| | 9 2 | 5+6 4 | 5+6 2 | 2+2 4 |
| | 8 2 | | 3+4 2 | |
| | 9 2 | | 5+6 2 | |
| | 8 2 | | 3+4 2 | |
| | 4+5 2 | | 5+6 2 | |
| | 9 2 | | 5+6 2 | |
| | 6+7 2 | | 3+4 2 | |
| | 9 2 | | 3+4 2 | |
| | 8 2 | | 3+4 2 | |
| | 4+5 2 | | 5+6 2 | |
| | 8 2 | | 5+6 2 | |
| | 8 2 | | 5+6 2 | |
| | 4+5 2 | | 3+4 2 | |
| | 9 2 | | 3+4 2 | |
| | 6+7 2 | | 5+6 2 | |
| | 9 2 | | 3+4 2 | |
| | 8 2 | | 5+6 2 | |
| | 6+7 2 | | 3+4 2 | |
| 167.1.69 | 161293 | | 68 | Fe2O3 |
| | -GM | -T | -F | -L |
| | 8 2 | 5+6 4 | 5+6 2 | 2+2 4 |
| | 9 2 | 5+6 4 | 3+4 2 | 2+2 4 |
| | 9 2 | 4+4 4 | 3+4 2 | 2+2 4 |
| | 6+7 2 | 5+6 4 | 5+6 2 | 2+2 4 |
| | 8 2 | 5+6 4 | 5+6 2 | 2+2 4 |
| | 4+5 2 | 4+4 4 | 3+4 2 | 2+2 4 |
| | 8 2 | 5+6 4 | 3+4 2 | 2+2 4 |
| | 8 2 | 4+4 4 | 5+6 2 | 2+2 4 |
| | 4+5 2 | 5+6 4 | 5+6 2 | 2+2 4 |
| | 9 2 | 5+6 4 | 3+4 2 | 2+2 4 |
| | 9 2 | 5+6 4 | 3+4 2 | 2+2 4 |
| | 6+7 2 | 4+4 4 | 5+6 2 | 2+2 4 |
| | 8 2 | 4+4 4 | 5+6 2 | 2+2 4 |
| | 4+5 2 | 5+6 4 | 3+4 2 | 2+2 4 |
| | 8 2 | 5+6 4 | 3+4 2 | 2+2 4 |
| | 6+7 2 | 4+4 4 | 5+6 2 | 2+2 4 |
| | 9 2 | 5+6 4 | 5+6 2 | 2+2 4 |
| | 8 2 | | 3+4 2 | |
| | 9 2 | | 5+6 2 | |
| | 8 2 | | 3+4 2 | |
| | 4+5 2 | | 5+6 2 | |
| | 9 2 | | 5+6 2 | |
| | 6+7 2 | | 3+4 2 | |
| | 9 2 | | 3+4 2 | |
| | 8 2 | | 3+4 2 | |
| | 4+5 2 | | 5+6 2 | |
| | 8 2 | | 5+6 2 | |
| | 8 2 | | 5+6 2 | |
| | 4+5 2 | | 3+4 2 | |
| | 9 2 | | 3+4 2 | |
| | 6+7 2 | | 5+6 2 | |
| | 9 2 | | 3+4 2 | |
| | 8 2 | | 5+6 2 | |
| | 6+7 2 | | 3+4 2 | |
| 167.1.70 | 161292 | | 68 | Fe2O3 |

Continued on next column

**TABLE LXXIV– continued**

| TQC | ICSD | | NoE | SF |
|---|---|---|---|---|
| | -GM | -T | -F | -L |
| | 8 2 | 5+6 4 | 5+6 2 | 2+2 4 |
| | 9 2 | 5+6 4 | 3+4 2 | 2+2 4 |
| | 9 2 | 4+4 4 | 3+4 2 | 2+2 4 |
| | 6+7 2 | 5+6 4 | 5+6 2 | 2+2 4 |
| | 8 2 | 5+6 4 | 5+6 2 | 2+2 4 |
| | 4+5 2 | 4+4 4 | 3+4 2 | 2+2 4 |
| | 8 2 | 5+6 4 | 3+4 2 | 2+2 4 |
| | 8 2 | 4+4 4 | 5+6 2 | 2+2 4 |
| | 4+5 2 | 5+6 4 | 5+6 2 | 2+2 4 |
| | 9 2 | 5+6 4 | 3+4 2 | 2+2 4 |
| | 9 2 | 5+6 4 | 3+4 2 | 2+2 4 |
| | 6+7 2 | 4+4 4 | 5+6 2 | 2+2 4 |
| | 8 2 | 4+4 4 | 5+6 2 | 2+2 4 |
| | 4+5 2 | 5+6 4 | 3+4 2 | 2+2 4 |
| | 8 2 | 5+6 4 | 3+4 2 | 2+2 4 |
| | 6+7 2 | 4+4 4 | 5+6 2 | 2+2 4 |
| | 9 2 | 5+6 4 | 5+6 2 | 2+2 4 |
| | 8 2 | | 3+4 2 | |
| | 9 2 | | 5+6 2 | |
| | 8 2 | | 3+4 2 | |
| | 4+5 2 | | 5+6 2 | |
| | 9 2 | | 5+6 2 | |
| | 6+7 2 | | 3+4 2 | |
| | 9 2 | | 3+4 2 | |
| | 8 2 | | 3+4 2 | |
| | 4+5 2 | | 5+6 2 | |
| | 8 2 | | 5+6 2 | |
| | 8 2 | | 5+6 2 | |
| | 4+5 2 | | 3+4 2 | |
| | 9 2 | | 3+4 2 | |
| | 6+7 2 | | 5+6 2 | |
| | 9 2 | | 3+4 2 | |
| | 8 2 | | 5+6 2 | |
| | 6+7 2 | | 3+4 2 | |
| 167.1.71 | 161291 | | 68 | Fe2O3 |
| | -GM | -T | -F | -L |
| | 8 2 | 5+6 4 | 5+6 2 | 2+2 4 |
| | 9 2 | 5+6 4 | 3+4 2 | 2+2 4 |
| | 9 2 | 4+4 4 | 3+4 2 | 2+2 4 |
| | 6+7 2 | 5+6 4 | 5+6 2 | 2+2 4 |
| | 8 2 | 5+6 4 | 5+6 2 | 2+2 4 |
| | 4+5 2 | 4+4 4 | 3+4 2 | 2+2 4 |
| | 8 2 | 5+6 4 | 3+4 2 | 2+2 4 |
| | 8 2 | 4+4 4 | 5+6 2 | 2+2 4 |
| | 4+5 2 | 5+6 4 | 5+6 2 | 2+2 4 |
| | 9 2 | 5+6 4 | 3+4 2 | 2+2 4 |
| | 9 2 | 5+6 4 | 3+4 2 | 2+2 4 |
| | 6+7 2 | 4+4 4 | 5+6 2 | 2+2 4 |
| | 8 2 | 4+4 4 | 5+6 2 | 2+2 4 |
| | 4+5 2 | 5+6 4 | 3+4 2 | 2+2 4 |
| | 8 2 | 5+6 4 | 3+4 2 | 2+2 4 |
| | 6+7 2 | 4+4 4 | 5+6 2 | 2+2 4 |
| | 9 2 | 5+6 4 | 5+6 2 | 2+2 4 |
| | 8 2 | | 3+4 2 | |
| | 9 2 | | 5+6 2 | |
| | 8 2 | | 3+4 2 | |
| | 4+5 2 | | 5+6 2 | |
| | 9 2 | | 5+6 2 | |
| | 6+7 2 | | 3+4 2 | |
| | 9 2 | | 3+4 2 | |

Continued on next column

**TABLE LXXIV– continued**

| TQC | ICSD | | NoE | SF |
|---|---|---|---|---|
| | 8 2 | | 3+4 2 | |
| | 4+5 2 | | 5+6 2 | |
| | 8 2 | | 5+6 2 | |
| | 8 2 | | 5+6 2 | |
| | 4+5 2 | | 3+4 2 | |
| | 9 2 | | 3+4 2 | |
| | 6+7 2 | | 5+6 2 | |
| | 9 2 | | 3+4 2 | |
| | 8 2 | | 5+6 2 | |
| | 6+7 2 | | 3+4 2 | |
| 167.1.72 | 161290 | | 68 | Fe2O3 |
| | -GM | -T | -F | -L |
| | 8 2 | 5+6 4 | 5+6 2 | 2+2 4 |
| | 9 2 | 5+6 4 | 3+4 2 | 2+2 4 |
| | 9 2 | 4+4 4 | 3+4 2 | 2+2 4 |
| | 6+7 2 | 5+6 4 | 5+6 2 | 2+2 4 |
| | 8 2 | 5+6 4 | 5+6 2 | 2+2 4 |
| | 4+5 2 | 4+4 4 | 3+4 2 | 2+2 4 |
| | 8 2 | 5+6 4 | 3+4 2 | 2+2 4 |
| | 8 2 | 4+4 4 | 5+6 2 | 2+2 4 |
| | 4+5 2 | 5+6 4 | 5+6 2 | 2+2 4 |
| | 9 2 | 5+6 4 | 3+4 2 | 2+2 4 |
| | 9 2 | 5+6 4 | 3+4 2 | 2+2 4 |
| | 6+7 2 | 4+4 4 | 5+6 2 | 2+2 4 |
| | 8 2 | 4+4 4 | 5+6 2 | 2+2 4 |
| | 4+5 2 | 5+6 4 | 3+4 2 | 2+2 4 |
| | 8 2 | 5+6 4 | 3+4 2 | 2+2 4 |
| | 6+7 2 | 4+4 4 | 5+6 2 | 2+2 4 |
| | 9 2 | 5+6 4 | 5+6 2 | 2+2 4 |
| | 8 2 | | 3+4 2 | |
| | 9 2 | | 5+6 2 | |
| | 8 2 | | 3+4 2 | |
| | 4+5 2 | | 5+6 2 | |
| | 9 2 | | 5+6 2 | |
| | 6+7 2 | | 3+4 2 | |
| | 9 2 | | 3+4 2 | |
| | 4+5 2 | | 3+4 2 | |
| | 8 2 | | 5+6 2 | |
| | 8 2 | | 5+6 2 | |
| | 8 2 | | 5+6 2 | |
| | 4+5 2 | | 3+4 2 | |
| | 9 2 | | 3+4 2 | |
| | 6+7 2 | | 5+6 2 | |
| | 9 2 | | 3+4 2 | |
| | 8 2 | | 5+6 2 | |
| | 6+7 2 | | 3+4 2 | |
| 167.1.73 | 161289 | | 68 | Fe2O3 |
| | -GM | -T | -F | -L |
| | 8 2 | 5+6 4 | 5+6 2 | 2+2 4 |
| | 9 2 | 5+6 4 | 3+4 2 | 2+2 4 |
| | 9 2 | 4+4 4 | 3+4 2 | 2+2 4 |
| | 6+7 2 | 5+6 4 | 5+6 2 | 2+2 4 |
| | 8 2 | 5+6 4 | 5+6 2 | 2+2 4 |
| | 4+5 2 | 4+4 4 | 3+4 2 | 2+2 4 |
| | 8 2 | 5+6 4 | 3+4 2 | 2+2 4 |
| | 9 2 | 4+4 4 | 5+6 2 | 2+2 4 |
| | 8 2 | 5+6 4 | 5+6 2 | 2+2 4 |
| | 4+5 2 | 5+6 4 | 3+4 2 | 2+2 4 |
| | 9 2 | 5+6 4 | 3+4 2 | 2+2 4 |
| | 6+7 2 | 4+4 4 | 5+6 2 | 2+2 4 |
| | 8 2 | 4+4 4 | 5+6 2 | 2+2 4 |

Continued on next column

**TABLE LXXIV– continued**

| TQC | ICSD | NoE | SF | |
|---|---|---|---|---|
| | 4+5 2 | 5+6 4 | 3+4 2 | 2+2 4 |
| | 8 2 | 5+6 4 | 3+4 2 | 2+2 4 |
| | 6+7 2 | 4+4 4 | 5+6 2 | 2+2 4 |
| | 9 2 | 5+6 4 | 5+6 2 | 2+2 4 |
| | 8 2 | | 3+4 2 | |
| | 9 2 | | 5+6 2 | |
| | 8 2 | | 3+4 2 | |
| | 4+5 2 | | 5+6 2 | |
| | 9 2 | | 5+6 2 | |
| | 6+7 2 | | 3+4 2 | |
| | 9 2 | | 3+4 2 | |
| | 4+5 2 | | 5+6 2 | |
| | 8 2 | | 3+4 2 | |
| | 8 2 | | 5+6 2 | |
| | 9 2 | | 5+6 2 | |
| | 8 2 | | 3+4 2 | |
| | 4+5 2 | | 3+4 2 | |
| | 6+7 2 | | 5+6 2 | |
| | 9 2 | | 3+4 2 | |
| | 8 2 | | 5+6 2 | |
| | 6+7 2 | | 3+4 2 | |
| 167.1.74 | 161288 | 68 | Fe2O3 | |
| | -GM | -T | -F | -L |
| | 8 2 | 5+6 4 | 5+6 2 | 2+2 4 |
| | 9 2 | 5+6 4 | 3+4 2 | 2+2 4 |
| | 9 2 | 4+4 4 | 3+4 2 | 2+2 4 |
| | 6+7 2 | 5+6 4 | 5+6 2 | 2+2 4 |
| | 8 2 | 5+6 4 | 5+6 2 | 2+2 4 |
| | 4+5 2 | 4+4 4 | 3+4 2 | 2+2 4 |
| | 8 2 | 5+6 4 | 3+4 2 | 2+2 4 |
| | 9 2 | 4+4 4 | 5+6 2 | 2+2 4 |
| | 8 2 | 5+6 4 | 5+6 2 | 2+2 4 |
| | 4+5 2 | 5+6 4 | 3+4 2 | 2+2 4 |
| | 9 2 | 5+6 4 | 3+4 2 | 2+2 4 |
| | 6+7 2 | 4+4 4 | 5+6 2 | 2+2 4 |
| | 8 2 | 4+4 4 | 5+6 2 | 2+2 4 |
| | 4+5 2 | 5+6 4 | 3+4 2 | 2+2 4 |
| | 8 2 | 5+6 4 | 3+4 2 | 2+2 4 |
| | 6+7 2 | 4+4 4 | 5+6 2 | 2+2 4 |
| | 9 2 | 5+6 4 | 5+6 2 | 2+2 4 |
| | 8 2 | | 3+4 2 | |
| | 9 2 | | 5+6 2 | |
| | 8 2 | | 3+4 2 | |
| | 4+5 2 | | 5+6 2 | |
| | 9 2 | | 5+6 2 | |
| | 6+7 2 | | 3+4 2 | |
| | 9 2 | | 3+4 2 | |
| | 4+5 2 | | 5+6 2 | |
| | 8 2 | | 3+4 2 | |
| | 8 2 | | 5+6 2 | |
| | 9 2 | | 5+6 2 | |
| | 8 2 | | 3+4 2 | |
| | 4+5 2 | | 3+4 2 | |
| | 6+7 2 | | 5+6 2 | |
| | 9 2 | | 3+4 2 | |
| | 8 2 | | 5+6 2 | |
| | 6+7 2 | | 3+4 2 | |
| 167.1.75 | 161287 | 68 | Fe2O3 | |
| | -GM | -T | -F | -L |
| | 8 2 | 5+6 4 | 5+6 2 | 2+2 4 |
| | 9 2 | 5+6 4 | 3+4 2 | 2+2 4 |

Continued on next column

**TABLE LXXIV– continued**

| TQC | ICSD | NoE | SF | |
|---|---|---|---|---|
| | 9 2 | 4+4 4 | 3+4 2 | 2+2 4 |
| | 6+7 2 | 5+6 4 | 5+6 2 | 2+2 4 |
| | 8 2 | 5+6 4 | 5+6 2 | 2+2 4 |
| | 4+5 2 | 4+4 4 | 3+4 2 | 2+2 4 |
| | 8 2 | 5+6 4 | 3+4 2 | 2+2 4 |
| | 9 2 | 4+4 4 | 5+6 2 | 2+2 4 |
| | 8 2 | 5+6 4 | 5+6 2 | 2+2 4 |
| | 4+5 2 | 5+6 4 | 3+4 2 | 2+2 4 |
| | 9 2 | 5+6 4 | 3+4 2 | 2+2 4 |
| | 6+7 2 | 4+4 4 | 5+6 2 | 2+2 4 |
| | 8 2 | 4+4 4 | 5+6 2 | 2+2 4 |
| | 4+5 2 | 5+6 4 | 3+4 2 | 2+2 4 |
| | 8 2 | 5+6 4 | 3+4 2 | 2+2 4 |
| | 6+7 2 | 4+4 4 | 5+6 2 | 2+2 4 |
| | 9 2 | 5+6 4 | 5+6 2 | 2+2 4 |
| | 8 2 | | 3+4 2 | |
| | 9 2 | | 5+6 2 | |
| | 8 2 | | 3+4 2 | |
| | 4+5 2 | | 5+6 2 | |
| | 9 2 | | 5+6 2 | |
| | 6+7 2 | | 3+4 2 | |
| | 9 2 | | 3+4 2 | |
| | 4+5 2 | | 5+6 2 | |
| | 8 2 | | 3+4 2 | |
| | 8 2 | | 5+6 2 | |
| | 9 2 | | 5+6 2 | |
| | 8 2 | | 3+4 2 | |
| | 4+5 2 | | 3+4 2 | |
| | 6+7 2 | | 5+6 2 | |
| | 9 2 | | 3+4 2 | |
| | 8 2 | | 5+6 2 | |
| | 6+7 2 | | 3+4 2 | |
| 167.1.76 | 161286 | 68 | Fe2O3 | |
| | -GM | -T | -F | -L |
| | 8 2 | 5+6 4 | 5+6 2 | 2+2 4 |
| | 9 2 | 5+6 4 | 3+4 2 | 2+2 4 |
| | 9 2 | 4+4 4 | 3+4 2 | 2+2 4 |
| | 6+7 2 | 5+6 4 | 5+6 2 | 2+2 4 |
| | 8 2 | 5+6 4 | 5+6 2 | 2+2 4 |
| | 4+5 2 | 4+4 4 | 3+4 2 | 2+2 4 |
| | 8 2 | 5+6 4 | 3+4 2 | 2+2 4 |
| | 9 2 | 4+4 4 | 5+6 2 | 2+2 4 |
| | 8 2 | 5+6 4 | 5+6 2 | 2+2 4 |
| | 4+5 2 | 5+6 4 | 3+4 2 | 2+2 4 |
| | 9 2 | 5+6 4 | 3+4 2 | 2+2 4 |
| | 6+7 2 | 4+4 4 | 5+6 2 | 2+2 4 |
| | 8 2 | 4+4 4 | 5+6 2 | 2+2 4 |
| | 4+5 2 | 5+6 4 | 3+4 2 | 2+2 4 |
| | 8 2 | 5+6 4 | 3+4 2 | 2+2 4 |
| | 6+7 2 | 4+4 4 | 5+6 2 | 2+2 4 |
| | 9 2 | 5+6 4 | 5+6 2 | 2+2 4 |
| | 8 2 | | 3+4 2 | |
| | 9 2 | | 5+6 2 | |
| | 8 2 | | 3+4 2 | |
| | 4+5 2 | | 5+6 2 | |
| | 9 2 | | 5+6 2 | |
| | 6+7 2 | | 3+4 2 | |
| | 9 2 | | 3+4 2 | |
| | 4+5 2 | | 5+6 2 | |
| | 8 2 | | 3+4 2 | |
| | 8 2 | | 5+6 2 | |

Continued on next column

**TABLE LXXIV– continued**

| TQC | ICSD | | NoE | SF |
|---|---|---|---|---|
| | 9 2 | | 5+6 2 | |
| | 8 2 | | 3+4 2 | |
| | 4+5 2 | | 3+4 2 | |
| | 6+7 2 | | 5+6 2 | |
| | 9 2 | | 3+4 2 | |
| | 8 2 | | 5+6 2 | |
| | 6+7 2 | | 3+4 2 | |
| 167.1.77 | 161285 | | 68 | Fe2O3 |
| | -GM | -T | -F | -L |
| | 8 2 | 5+6 4 | 5+6 2 | 2+2 4 |
| | 9 2 | 5+6 4 | 3+4 2 | 2+2 4 |
| | 9 2 | 4+4 4 | 3+4 2 | 2+2 4 |
| | 6+7 2 | 5+6 4 | 5+6 2 | 2+2 4 |
| | 8 2 | 5+6 4 | 5+6 2 | 2+2 4 |
| | 4+5 2 | 4+4 4 | 3+4 2 | 2+2 4 |
| | 8 2 | 5+6 4 | 3+4 2 | 2+2 4 |
| | 9 2 | 4+4 4 | 5+6 2 | 2+2 4 |
| | 8 2 | 5+6 4 | 5+6 2 | 2+2 4 |
| | 4+5 2 | 5+6 4 | 3+4 2 | 2+2 4 |
| | 9 2 | 5+6 4 | 3+4 2 | 2+2 4 |
| | 6+7 2 | 4+4 4 | 5+6 2 | 2+2 4 |
| | 8 2 | 4+4 4 | 5+6 2 | 2+2 4 |
| | 4+5 2 | 5+6 4 | 3+4 2 | 2+2 4 |
| | 8 2 | 5+6 4 | 3+4 2 | 2+2 4 |
| | 6+7 2 | 4+4 4 | 5+6 2 | 2+2 4 |
| | 9 2 | 5+6 4 | 5+6 2 | 2+2 4 |
| | 8 2 | | 3+4 2 | |
| | 9 2 | | 5+6 2 | |
| | 8 2 | | 3+4 2 | |
| | 4+5 2 | | 5+6 2 | |
| | 9 2 | | 5+6 2 | |
| | 6+7 2 | | 3+4 2 | |
| | 9 2 | | 3+4 2 | |
| | 4+5 2 | | 5+6 2 | |
| | 8 2 | | 3+4 2 | |
| | 8 2 | | 5+6 2 | |
| | 9 2 | | 5+6 2 | |
| | 8 2 | | 3+4 2 | |
| | 4+5 2 | | 3+4 2 | |
| | 6+7 2 | | 5+6 2 | |
| | 9 2 | | 3+4 2 | |
| | 8 2 | | 5+6 2 | |
| | 6+7 2 | | 3+4 2 | |
| 167.1.78 | 161284 | | 68 | Fe2O3 |
| | -GM | -T | -F | -L |
| | 8 2 | 5+6 4 | 5+6 2 | 2+2 4 |
| | 9 2 | 5+6 4 | 3+4 2 | 2+2 4 |
| | 9 2 | 4+4 4 | 3+4 2 | 2+2 4 |
| | 6+7 2 | 5+6 4 | 5+6 2 | 2+2 4 |
| | 8 2 | 5+6 4 | 5+6 2 | 2+2 4 |
| | 4+5 2 | 4+4 4 | 3+4 2 | 2+2 4 |
| | 8 2 | 5+6 4 | 3+4 2 | 2+2 4 |
| | 9 2 | 4+4 4 | 5+6 2 | 2+2 4 |
| | 8 2 | 5+6 4 | 5+6 2 | 2+2 4 |
| | 4+5 2 | 5+6 4 | 3+4 2 | 2+2 4 |
| | 9 2 | 5+6 4 | 3+4 2 | 2+2 4 |
| | 6+7 2 | 4+4 4 | 5+6 2 | 2+2 4 |
| | 8 2 | 4+4 4 | 5+6 2 | 2+2 4 |
| | 4+5 2 | 5+6 4 | 3+4 2 | 2+2 4 |
| | 8 2 | 5+6 4 | 3+4 2 | 2+2 4 |
| | 6+7 2 | 4+4 4 | 5+6 2 | 2+2 4 |

Continued on next column

**TABLE LXXIV– continued**

| TQC | ICSD | | NoE | SF |
|---|---|---|---|---|
| | 9 2 | 5+6 4 | 5+6 2 | 2+2 4 |
| | 8 2 | | 3+4 2 | |
| | 9 2 | | 5+6 2 | |
| | 8 2 | | 3+4 2 | |
| | 4+5 2 | | 5+6 2 | |
| | 9 2 | | 5+6 2 | |
| | 6+7 2 | | 3+4 2 | |
| | 9 2 | | 3+4 2 | |
| | 4+5 2 | | 5+6 2 | |
| | 8 2 | | 3+4 2 | |
| | 8 2 | | 5+6 2 | |
| | 9 2 | | 5+6 2 | |
| | 8 2 | | 3+4 2 | |
| | 4+5 2 | | 3+4 2 | |
| | 6+7 2 | | 5+6 2 | |
| | 9 2 | | 3+4 2 | |
| | 8 2 | | 5+6 2 | |
| | 6+7 2 | | 3+4 2 | |
| 167.1.79 | 161283 | | 68 | Fe2O3 |
| | -GM | -T | -F | -L |
| | 8 2 | 5+6 4 | 5+6 2 | 2+2 4 |
| | 9 2 | 5+6 4 | 3+4 2 | 2+2 4 |
| | 9 2 | 4+4 4 | 3+4 2 | 2+2 4 |
| | 6+7 2 | 5+6 4 | 5+6 2 | 2+2 4 |
| | 8 2 | 5+6 4 | 5+6 2 | 2+2 4 |
| | 4+5 2 | 4+4 4 | 3+4 2 | 2+2 4 |
| | 8 2 | 5+6 4 | 3+4 2 | 2+2 4 |
| | 9 2 | 4+4 4 | 5+6 2 | 2+2 4 |
| | 8 2 | 5+6 4 | 5+6 2 | 2+2 4 |
| | 4+5 2 | 5+6 4 | 3+4 2 | 2+2 4 |
| | 9 2 | 5+6 4 | 3+4 2 | 2+2 4 |
| | 6+7 2 | 4+4 4 | 5+6 2 | 2+2 4 |
| | 8 2 | 4+4 4 | 5+6 2 | 2+2 4 |
| | 4+5 2 | 5+6 4 | 3+4 2 | 2+2 4 |
| | 8 2 | 5+6 4 | 3+4 2 | 2+2 4 |
| | 6+7 2 | 4+4 4 | 5+6 2 | 2+2 4 |
| | 9 2 | 5+6 4 | 5+6 2 | 2+2 4 |
| | 8 2 | | 3+4 2 | |
| | 9 2 | | 5+6 2 | |
| | 8 2 | | 3+4 2 | |
| | 4+5 2 | | 5+6 2 | |
| | 9 2 | | 5+6 2 | |
| | 6+7 2 | | 3+4 2 | |
| | 9 2 | | 3+4 2 | |
| | 4+5 2 | | 5+6 2 | |
| | 8 2 | | 3+4 2 | |
| | 8 2 | | 5+6 2 | |
| | 9 2 | | 5+6 2 | |
| | 8 2 | | 3+4 2 | |
| | 4+5 2 | | 3+4 2 | |
| | 6+7 2 | | 5+6 2 | |
| | 9 2 | | 3+4 2 | |
| | 8 2 | | 5+6 2 | |
| | 6+7 2 | | 3+4 2 | |
| 167.1.80 | 15840 | | 68 | Fe2O3 |
| | -GM | -T | -F | -L |
| | 8 2 | 5+6 4 | 5+6 2 | 2+2 4 |
| | 9 2 | 5+6 4 | 3+4 2 | 2+2 4 |
| | 9 2 | 4+4 4 | 3+4 2 | 2+2 4 |
| | 6+7 2 | 5+6 4 | 5+6 2 | 2+2 4 |
| | 8 2 | 5+6 4 | 5+6 2 | 2+2 4 |

Continued on next column

**TABLE LXXIV– continued**

| TQC | ICSD | NoE | SF |
|---|---|---|---|
| | 4+5 2 4+4 4 3+4 2 2+2 4 | | |
| | 8 2 5+6 4 3+4 2 2+2 4 | | |
| | 9 2 4+4 4 5+6 2 2+2 4 | | |
| | 8 2 5+6 4 5+6 2 2+2 4 | | |
| | 4+5 2 5+6 4 3+4 2 2+2 4 | | |
| | 9 2 5+6 4 3+4 2 2+2 4 | | |
| | 6+7 2 4+4 4 5+6 2 2+2 4 | | |
| | 8 2 4+4 4 5+6 2 2+2 4 | | |
| | 4+5 2 5+6 4 3+4 2 2+2 4 | | |
| | 8 2 5+6 4 3+4 2 2+2 4 | | |
| | 6+7 2 4+4 4 5+6 2 2+2 4 | | |
| | 9 2 5+6 4 5+6 2 2+2 4 | | |
| | 8 2 3+4 2 | | |
| | 9 2 5+6 2 | | |
| | 8 2 3+4 2 | | |
| | 4+5 2 5+6 2 | | |
| | 9 2 5+6 2 | | |
| | 6+7 2 3+4 2 | | |
| | 9 2 3+4 2 | | |
| | 4+5 2 5+6 2 | | |
| | 8 2 3+4 2 | | |
| | 8 2 5+6 2 | | |
| | 9 2 5+6 2 | | |
| | 8 2 3+4 2 | | |
| | 4+5 2 3+4 2 | | |
| | 6+7 2 5+6 2 | | |
| | 9 2 3+4 2 | | |
| | 8 2 5+6 2 | | |
| | 6+7 2 3+4 2 | | |
| 167.1.81 | 154191 | 68 | Fe2O3 |
| | -GM -T -F -L | | |
| | 8 2 5+6 4 5+6 2 2+2 4 | | |
| | 9 2 5+6 4 3+4 2 2+2 4 | | |
| | 9 2 4+4 4 3+4 2 2+2 4 | | |
| | 6+7 2 5+6 4 5+6 2 2+2 4 | | |
| | 8 2 5+6 4 5+6 2 2+2 4 | | |
| | 4+5 2 4+4 4 3+4 2 2+2 4 | | |
| | 8 2 5+6 4 3+4 2 2+2 4 | | |
| | 9 2 4+4 4 5+6 2 2+2 4 | | |
| | 8 2 5+6 4 5+6 2 2+2 4 | | |
| | 4+5 2 5+6 4 3+4 2 2+2 4 | | |
| | 9 2 5+6 4 3+4 2 2+2 4 | | |
| | 6+7 2 4+4 4 5+6 2 2+2 4 | | |
| | 8 2 4+4 4 5+6 2 2+2 4 | | |
| | 4+5 2 5+6 4 3+4 2 2+2 4 | | |
| | 8 2 5+6 4 3+4 2 2+2 4 | | |
| | 6+7 2 4+4 4 5+6 2 2+2 4 | | |
| | 9 2 5+6 4 5+6 2 2+2 4 | | |
| | 8 2 3+4 2 | | |
| | 9 2 5+6 2 | | |
| | 8 2 3+4 2 | | |
| | 4+5 2 5+6 2 | | |
| | 9 2 5+6 2 | | |
| | 6+7 2 3+4 2 | | |
| | 9 2 3+4 2 | | |
| | 4+5 2 5+6 2 | | |
| | 8 2 3+4 2 | | |
| | 8 2 5+6 2 | | |
| | 9 2 5+6 2 | | |
| | 8 2 3+4 2 | | |
| | 4+5 2 3+4 2 | | |

Continued on next column

**TABLE LXXIV– continued**

| TQC | ICSD | NoE | SF |
|---|---|---|---|
| | 6+7 2 5+6 2 | | |
| | 9 2 3+4 2 | | |
| | 8 2 5+6 2 | | |
| | 6+7 2 3+4 2 | | |
| 167.1.82 | 154190 | 68 | Fe2O3 |
| | -GM -T -F -L | | |
| | 8 2 5+6 4 5+6 2 2+2 4 | | |
| | 9 2 5+6 4 3+4 2 2+2 4 | | |
| | 9 2 4+4 4 3+4 2 2+2 4 | | |
| | 6+7 2 5+6 4 5+6 2 2+2 4 | | |
| | 8 2 5+6 4 5+6 2 2+2 4 | | |
| | 4+5 2 4+4 4 3+4 2 2+2 4 | | |
| | 8 2 5+6 4 3+4 2 2+2 4 | | |
| | 9 2 4+4 4 5+6 2 2+2 4 | | |
| | 8 2 5+6 4 5+6 2 2+2 4 | | |
| | 4+5 2 5+6 4 3+4 2 2+2 4 | | |
| | 9 2 5+6 4 3+4 2 2+2 4 | | |
| | 6+7 2 4+4 4 5+6 2 2+2 4 | | |
| | 8 2 4+4 4 5+6 2 2+2 4 | | |
| | 4+5 2 5+6 4 3+4 2 2+2 4 | | |
| | 8 2 5+6 4 3+4 2 2+2 4 | | |
| | 6+7 2 4+4 4 5+6 2 2+2 4 | | |
| | 9 2 5+6 4 5+6 2 2+2 4 | | |
| | 8 2 3+4 2 | | |
| | 9 2 5+6 2 | | |
| | 8 2 3+4 2 | | |
| | 4+5 2 5+6 2 | | |
| | 9 2 5+6 2 | | |
| | 6+7 2 3+4 2 | | |
| | 9 2 3+4 2 | | |
| | 4+5 2 5+6 2 | | |
| | 8 2 3+4 2 | | |
| | 8 2 5+6 2 | | |
| | 9 2 5+6 2 | | |
| | 8 2 3+4 2 | | |
| | 4+5 2 3+4 2 | | |
| | 6+7 2 5+6 2 | | |
| | 9 2 3+4 2 | | |
| | 8 2 5+6 2 | | |
| | 6+7 2 3+4 2 | | |
| 167.1.83 | 173470 | 60 | Cr2O3 |
| | -GM -T -F -L | | |
| | 8 2 5+6 4 5+6 2 2+2 4 | | |
| | 9 2 5+6 4 3+4 2 2+2 4 | | |
| | 6+7 2 4+4 4 5+6 2 2+2 4 | | |
| | 9 2 5+6 4 3+4 2 2+2 4 | | |
| | 8 2 4+4 4 3+4 2 2+2 4 | | |
| | 4+5 2 5+6 4 5+6 2 2+2 4 | | |
| | 8 2 5+6 4 3+4 2 2+2 4 | | |
| | 6+7 2 4+4 4 5+6 2 2+2 4 | | |
| | 9 2 5+6 4 5+6 2 2+2 4 | | |
| | 8 2 5+6 4 3+4 2 2+2 4 | | |
| | 4+5 2 5+6 4 3+4 2 2+2 4 | | |
| | 9 2 4+4 4 5+6 2 2+2 4 | | |
| | 8 2 4+4 4 3+4 2 2+2 4 | | |
| | 4+5 2 5+6 4 5+6 2 2+2 4 | | |
| | 6+7 2 5+6 4 5+6 2 2+2 4 | | |
| | 9 2 3+4 2 | | |
| | 8 2 5+6 2 | | |
| | 8 2 3+4 2 | | |
| | 9 2 5+6 2 | | |

Continued on next column

**TABLE LXXIV– continued**

| TQC | ICSD | | NoE | SF |
|---|---|---|---|---|
| | 8 2 | | 5+6 2 | |
| | 4+5 2 | | 5+6 2 | |
| | 9 2 | | 3+4 2 | |
| | 6+7 2 | | 3+4 2 | |
| | 9 2 | | 3+4 2 | |
| | 8 2 | | 3+4 2 | |
| | 4+5 2 | | 5+6 2 | |
| | 8 2 | | 5+6 2 | |
| | 8 2 | | 5+6 2 | |
| | 4+5 2 | | 3+4 2 | |
| | 9 2 | | 5+6 2 | |
| 167.1.84 | 201966 | | 204 | Cr2Cs3I9 |
| | -GM | -T | -F | -L |
| | 8 2 | 4+4 4 | 3+4 2 | 2+2 4 |
| | 4+5 2 | 5+6 4 | 3+4 2 | 2+2 4 |
| | 6+7 2 | 5+6 4 | 3+4 2 | 2+2 4 |
| | 8 2 | 5+6 4 | 5+6 2 | 2+2 4 |
| | 9 2 | 4+4 4 | 5+6 2 | 2+2 4 |
| | 9 2 | 5+6 4 | 5+6 2 | 2+2 4 |
| | 8 2 | 5+6 4 | 3+4 2 | 2+2 4 |
| | 9 2 | 5+6 4 | 5+6 2 | 2+2 4 |
| | 4+5 2 | 4+4 4 | 3+4 2 | 2+2 4 |
| | 8 2 | 5+6 4 | 5+6 2 | 2+2 4 |
| | 6+7 2 | 4+4 4 | 3+4 2 | 2+2 4 |
| | 9 2 | 5+6 4 | 5+6 2 | 2+2 4 |
| | 8 2 | 5+6 4 | 3+4 2 | 2+2 4 |
| | 8 2 | 4+4 4 | 5+6 2 | 2+2 4 |
| | 9 2 | 5+6 4 | 3+4 2 | 2+2 4 |
| | 9 2 | 4+4 4 | 5+6 2 | 2+2 4 |
| | 6+7 2 | 5+6 4 | 5+6 2 | 2+2 4 |
| | 9 2 | 5+6 4 | 3+4 2 | 2+2 4 |
| | 4+5 2 | 4+4 4 | 5+6 2 | 2+2 4 |
| | 8 2 | 5+6 4 | 3+4 2 | 2+2 4 |
| | 6+7 2 | 5+6 4 | 5+6 2 | 2+2 4 |
| | 9 2 | 5+6 4 | 5+6 2 | 2+2 4 |
| | 4+5 2 | 5+6 4 | 3+4 2 | 2+2 4 |
| | 8 2 | 4+4 4 | 3+4 2 | 2+2 4 |
| | 8 2 | 5+6 4 | 5+6 2 | 2+2 4 |
| | 9 2 | 4+4 4 | 3+4 2 | 2+2 4 |
| | 6+7 2 | 5+6 4 | 5+6 2 | 2+2 4 |
| | 8 2 | 5+6 4 | 3+4 2 | 2+2 4 |
| | 4+5 2 | 4+4 4 | 3+4 2 | 2+2 4 |
| | 9 2 | 5+6 4 | 5+6 2 | 2+2 4 |
| | 4+5 2 | 5+6 4 | 5+6 2 | 2+2 4 |
| | 8 2 | 5+6 4 | 3+4 2 | 2+2 4 |
| | 9 2 | 4+4 4 | 5+6 2 | 2+2 4 |
| | 6+7 2 | 5+6 4 | 5+6 2 | 2+2 4 |
| | 8 2 | 4+4 4 | 3+4 2 | 2+2 4 |
| | 6+7 2 | 5+6 4 | 3+4 2 | 2+2 4 |
| | 8 2 | 5+6 4 | 5+6 2 | 2+2 4 |
| | 9 2 | 4+4 4 | 3+4 2 | 2+2 4 |
| | 4+5 2 | 5+6 4 | 3+4 2 | 2+2 4 |
| | 9 2 | 5+6 4 | 5+6 2 | 2+2 4 |
| | 8 2 | 5+6 4 | 5+6 2 | 2+2 4 |
| | 9 2 | 4+4 4 | 3+4 2 | 2+2 4 |
| | 8 2 | 5+6 4 | 5+6 2 | 2+2 4 |
| | 9 2 | 4+4 4 | 3+4 2 | 2+2 4 |
| | 8 2 | 5+6 4 | 3+4 2 | 2+2 4 |
| | 9 2 | 5+6 4 | 5+6 2 | 2+2 4 |
| | 6+7 2 | 5+6 4 | 3+4 2 | 2+2 4 |
| | 4+5 2 | 4+4 4 | 5+6 2 | 2+2 4 |

Continued on next column

**TABLE LXXIV– continued**

| TQC | ICSD | | NoE | SF |
|---|---|---|---|---|
| | 9 2 | 5+6 4 | 3+4 2 | 2+2 4 |
| | 8 2 | 4+4 4 | 5+6 2 | 2+2 4 |
| | 6+7 2 | 5+6 4 | 3+4 2 | 2+2 4 |
| | 9 2 | | 5+6 2 | |
| | 4+5 2 | | 3+4 2 | |
| | 8 2 | | 5+6 2 | |
| | 9 2 | | 5+6 2 | |
| | 8 2 | | 3+4 2 | |
| | 9 2 | | 3+4 2 | |
| | 4+5 2 | | 5+6 2 | |
| | 8 2 | | 3+4 2 | |
| | 6+7 2 | | 5+6 2 | |
| | 8 2 | | 3+4 2 | |
| | 9 2 | | 5+6 2 | |
| | 8 2 | | 3+4 2 | |
| | 6+7 2 | | 5+6 2 | |
| | 4+5 2 | | 3+4 2 | |
| | 9 2 | | 5+6 2 | |
| | 8 2 | | 3+4 2 | |
| | 6+7 2 | | 5+6 2 | |
| | 4+5 2 | | 3+4 2 | |
| | 9 2 | | 5+6 2 | |
| | 8 2 | | 3+4 2 | |
| | 8 2 | | 5+6 2 | |
| | 6+7 2 | | 5+6 2 | |
| | 9 2 | | 3+4 2 | |
| | 9 2 | | 5+6 2 | |
| | 8 2 | | 3+4 2 | |
| | 9 2 | | 5+6 2 | |
| | 9 2 | | 5+6 2 | |
| | 4+5 2 | | 3+4 2 | |
| | 8 2 | | 3+4 2 | |
| | 9 2 | | 5+6 2 | |
| | 8 2 | | 5+6 2 | |
| | 6+7 2 | | 3+4 2 | |
| | 9 2 | | 3+4 2 | |
| | 8 2 | | 5+6 2 | |
| | 4+5 2 | | 3+4 2 | |
| | 9 2 | | 5+6 2 | |
| | 8 2 | | 3+4 2 | |
| | 9 2 | | 5+6 2 | |
| | 4+5 2 | | 3+4 2 | |
| | 6+7 2 | | 3+4 2 | |
| | 8 2 | | 5+6 2 | |
| | 4+5 2 | | 3+4 2 | |
| | 8 2 | | 3+4 2 | |
| | 9 2 | | 5+6 2 | |
| | 6+7 2 | | 5+6 2 | |
| | 8 2 | | 3+4 2 | |
| | 9 2 | | 5+6 2 | |
| | 4+5 2 | | 5+6 2 | |
| | 6+7 2 | | 3+4 2 | |
| | 9 2 | | 5+6 2 | |
| | 8 2 | | 5+6 2 | |
| 167.1.85 | 50809 | | 120 | Ca3Co1O6Rh1 |
| | -GM | -T | -F | -L |
| | 8 2 | 5+6 4 | 3+4 2 | 2+2 4 |
| | 8 2 | 5+6 4 | 3+4 2 | 2+2 4 |
| | 9 2 | 5+6 4 | 5+6 2 | 2+2 4 |
| | 4+5 2 | 4+4 4 | 3+4 2 | 2+2 4 |
| | 8 2 | 5+6 4 | 5+6 2 | 2+2 4 |

Continued on next column

**TABLE LXXIV– continued**

| TQC | ICSD | NoE | SF | |
|---|---|---|---|---|
| | 4+5 2 | 4+4 4 | 3+4 2 | 2+2 4 |
| | 8 2 | 5+6 4 | 3+4 2 | 2+2 4 |
| | 9 2 | 5+6 4 | 5+6 2 | 2+2 4 |
| | 9 2 | 4+4 4 | 3+4 2 | 2+2 4 |
| | 6+7 2 | 4+4 4 | 5+6 2 | 2+2 4 |
| | 9 2 | 5+6 4 | 5+6 2 | 2+2 4 |
| | 6+7 2 | 5+6 4 | 5+6 2 | 2+2 4 |
| | 8 2 | 5+6 4 | 3+4 2 | 2+2 4 |
| | 4+5 2 | 5+6 4 | 3+4 2 | 2+2 4 |
| | 8 2 | 4+4 4 | 3+4 2 | 2+2 4 |
| | 8 2 | 5+6 4 | 3+4 2 | 2+2 4 |
| | 4+5 2 | 4+4 4 | 3+4 2 | 2+2 4 |
| | 8 2 | 5+6 4 | 3+4 2 | 2+2 4 |
| | 9 2 | 5+6 4 | 3+4 2 | 2+2 4 |
| | 8 2 | 4+4 4 | 3+4 2 | 2+2 4 |
| | 4+5 2 | 5+6 4 | 3+4 2 | 2+2 4 |
| | 8 2 | 5+6 4 | 3+4 2 | 2+2 4 |
| | 4+5 2 | 4+4 4 | 3+4 2 | 2+2 4 |
| | 8 2 | 5+6 4 | 5+6 2 | 2+2 4 |
| | 8 2 | 5+6 4 | 3+4 2 | 2+2 4 |
| | 9 2 | 4+4 4 | 5+6 2 | 2+2 4 |
| | 6+7 2 | 5+6 4 | 5+6 2 | 2+2 4 |
| | 9 2 | 5+6 4 | 5+6 2 | 2+2 4 |
| | 6+7 2 | 4+4 4 | 5+6 2 | 2+2 4 |
| | 9 2 | 5+6 4 | 5+6 2 | 2+2 4 |
| | 9 2 | | 5+6 2 | |
| | 6+7 2 | | 5+6 2 | |
| | 9 2 | | 5+6 2 | |
| | 9 2 | | 5+6 2 | |
| | 6+7 2 | | 3+4 2 | |
| | 8 2 | | 5+6 2 | |
| | 4+5 2 | | 5+6 2 | |
| | 9 2 | | 5+6 2 | |
| | 8 2 | | 3+4 2 | |
| | 4+5 2 | | 3+4 2 | |
| | 8 2 | | 5+6 2 | |
| | 8 2 | | 3+4 2 | |
| | 9 2 | | 5+6 2 | |
| | 6+7 2 | | 5+6 2 | |
| | 9 2 | | 3+4 2 | |
| | 9 2 | | 5+6 2 | |
| | 6+7 2 | | 5+6 2 | |
| | 9 2 | | 3+4 2 | |
| | 8 2 | | 3+4 2 | |
| | 4+5 2 | | 3+4 2 | |
| | 8 2 | | 3+4 2 | |
| | 4+5 2 | | 5+6 2 | |
| | 9 2 | | 3+4 2 | |
| | 8 2 | | 3+4 2 | |
| | 6+7 2 | | 3+4 2 | |
| | 8 2 | | 5+6 2 | |
| | 9 2 | | 5+6 2 | |
| | 4+5 2 | | 3+4 2 | |
| | 8 2 | | 3+4 2 | |
| | 8 2 | | 5+6 2 | |
| 167.1.86 | 421391 | 108 | C3Cr2N6 | |
| | -GM | -T | -F | -L |
| | 8 2 | 5+6 4 | 5+6 2 | 2+2 4 |
| | 9 2 | 4+4 4 | 3+4 2 | 2+2 4 |
| | 6+7 2 | 5+6 4 | 3+4 2 | 2+2 4 |
| | 9 2 | 5+6 4 | 5+6 2 | 2+2 4 |

Continued on next column

**TABLE LXXIV– continued**

| TQC | ICSD | NoE | SF | |
|---|---|---|---|---|
| | 4+5 2 | 4+4 4 | 3+4 2 | 2+2 4 |
| | 8 2 | 5+6 4 | 5+6 2 | 2+2 4 |
| | 8 2 | 5+6 4 | 5+6 2 | 2+2 4 |
| | 9 2 | 5+6 4 | 3+4 2 | 2+2 4 |
| | 8 2 | 4+4 4 | 3+4 2 | 2+2 4 |
| | 4+5 2 | 5+6 4 | 5+6 2 | 2+2 4 |
| | 6+7 2 | 4+4 4 | 3+4 2 | 2+2 4 |
| | 9 2 | 5+6 4 | 5+6 2 | 2+2 4 |
| | 8 2 | 5+6 4 | 5+6 2 | 2+2 4 |
| | 8 2 | 4+4 4 | 3+4 2 | 2+2 4 |
| | 8 2 | 5+6 4 | 5+6 2 | 2+2 4 |
| | 4+5 2 | 5+6 4 | 3+4 2 | 2+2 4 |
| | 9 2 | 5+6 4 | 5+6 2 | 2+2 4 |
| | 6+7 2 | 4+4 4 | 3+4 2 | 2+2 4 |
| | 9 2 | 5+6 4 | 5+6 2 | 2+2 4 |
| | 8 2 | 4+4 4 | 5+6 2 | 2+2 4 |
| | 4+5 2 | 5+6 4 | 3+4 2 | 2+2 4 |
| | 8 2 | 5+6 4 | 3+4 2 | 2+2 4 |
| | 6+7 2 | 4+4 4 | 5+6 2 | 2+2 4 |
| | 9 2 | 5+6 4 | 3+4 2 | 2+2 4 |
| | 9 2 | 4+4 4 | 3+4 2 | 2+2 4 |
| | 4+5 2 | 5+6 4 | 5+6 2 | 2+2 4 |
| | 8 2 | 5+6 4 | 5+6 2 | 2+2 4 |
| | 8 2 | | 5+6 2 | |
| | 6+7 2 | | 3+4 2 | |
| | 9 2 | | 3+4 2 | |
| | 9 2 | | 5+6 2 | |
| | 8 2 | | 5+6 2 | |
| | 4+5 2 | | 3+4 2 | |
| | 9 2 | | 3+4 2 | |
| | 9 2 | | 5+6 2 | |
| | 6+7 2 | | 3+4 2 | |
| | 9 2 | | 3+4 2 | |
| | 6+7 2 | | 5+6 2 | |
| | 9 2 | | 5+6 2 | |
| | 8 2 | | 5+6 2 | |
| | 4+5 2 | | 3+4 2 | |
| | 8 2 | | 3+4 2 | |
| | 8 2 | | 3+4 2 | |
| | 4+5 2 | | 5+6 2 | |
| | 8 2 | | 5+6 2 | |
| | 9 2 | | 5+6 2 | |
| | 9 2 | | 3+4 2 | |
| | 6+7 2 | | 3+4 2 | |
| | 8 2 | | 5+6 2 | |
| | 4+5 2 | | 3+4 2 | |
| | 6+7 2 | | 3+4 2 | |
| | 9 2 | | 5+6 2 | |
| | 8 2 | | 5+6 2 | |
| | 8 2 | | 5+6 2 | |
| 167.1.87 | 60625 | 180 | Br6Cd1Rb4 | |
| | -GM | -T | -F | -L |
| | 8 2 | 5+6 4 | 3+4 2 | 2+2 4 |
| | 9 2 | 4+4 4 | 5+6 2 | 2+2 4 |
| | 8 2 | 5+6 4 | 5+6 2 | 2+2 4 |
| | 6+7 2 | 5+6 4 | 3+4 2 | 2+2 4 |
| | 9 2 | 5+6 4 | 3+4 2 | 2+2 4 |
| | 4+5 2 | 4+4 4 | 5+6 2 | 2+2 4 |
| | 8 2 | 5+6 4 | 3+4 2 | 2+2 4 |
| | 9 2 | 4+4 4 | 5+6 2 | 2+2 4 |
| | 6+7 2 | 5+6 4 | 3+4 2 | 2+2 4 |

Continued on next column

**TABLE LXXIV– continued**

| TQC | ICSD | NoE | SF |
|---|---|---|---|
| | 9 2 5+6 4 3+4 2 2+2 4 | | |
| | 4+5 2 5+6 4 5+6 2 2+2 4 | | |
| | 8 2 5+6 4 5+6 2 2+2 4 | | |
| | 8 2 4+4 4 5+6 2 2+2 4 | | |
| | 9 2 5+6 4 3+4 2 2+2 4 | | |
| | 8 2 4+4 4 5+6 2 2+2 4 | | |
| | 9 2 5+6 4 3+4 2 2+2 4 | | |
| | 8 2 4+4 4 5+6 2 2+2 4 | | |
| | 4+5 2 5+6 4 3+4 2 2+2 4 | | |
| | 6+7 2 5+6 4 5+6 2 2+2 4 | | |
| | 9 2 5+6 4 3+4 2 2+2 4 | | |
| | 8 2 4+4 4 3+4 2 2+2 4 | | |
| | 9 2 5+6 4 5+6 2 2+2 4 | | |
| | 8 2 5+6 4 3+4 2 2+2 4 | | |
| | 9 2 4+4 4 5+6 2 2+2 4 | | |
| | 4+5 2 5+6 4 3+4 2 2+2 4 | | |
| | 6+7 2 4+4 4 5+6 2 2+2 4 | | |
| | 4+5 2 5+6 4 5+6 2 2+2 4 | | |
| | 8 2 5+6 4 3+4 2 2+2 4 | | |
| | 9 2 4+4 4 5+6 2 2+2 4 | | |
| | 6+7 2 5+6 4 3+4 2 2+2 4 | | |
| | 9 2 5+6 4 3+4 2 2+2 4 | | |
| | 8 2 5+6 4 5+6 2 2+2 4 | | |
| | 8 2 4+4 4 5+6 2 2+2 4 | | |
| | 9 2 5+6 4 3+4 2 2+2 4 | | |
| | 8 2 5+6 4 5+6 2 2+2 4 | | |
| | 9 2 4+4 4 3+4 2 2+2 4 | | |
| | 4+5 2 5+6 4 3+4 2 2+2 4 | | |
| | 6+7 2 5+6 4 5+6 2 2+2 4 | | |
| | 9 2 4+4 4 3+4 2 2+2 4 | | |
| | 9 2 5+6 4 5+6 2 2+2 4 | | |
| | 6+7 2 4+4 4 3+4 2 2+2 4 | | |
| | 4+5 2 5+6 4 5+6 2 2+2 4 | | |
| | 8 2 5+6 4 3+4 2 2+2 4 | | |
| | 8 2 5+6 4 5+6 2 2+2 4 | | |
| | 8 2 4+4 4 3+4 2 2+2 4 | | |
| | 8 2 3+4 2 | | |
| | 4+5 2 3+4 2 | | |
| | 4+5 2 3+4 2 | | |
| | 8 2 3+4 2 | | |
| | 8 2 3+4 2 | | |
| | 4+5 2 3+4 2 | | |
| | 4+5 2 3+4 2 | | |
| | 8 2 3+4 2 | | |
| | 8 2 3+4 2 | | |
| | 8 2 3+4 2 | | |
| | 8 2 5+6 2 | | |
| | 4+5 2 5+6 2 | | |
| | 6+7 2 3+4 2 | | |
| | 9 2 5+6 2 | | |
| | 9 2 3+4 2 | | |
| | 9 2 3+4 2 | | |
| | 4+5 2 5+6 2 | | |
| | 8 2 3+4 2 | | |
| | 8 2 5+6 2 | | |
| | 6+7 2 3+4 2 | | |
| | 8 2 5+6 2 | | |
| | 9 2 5+6 2 | | |
| | 8 2 3+4 2 | | |
| | 4+5 2 3+4 2 | | |
| | 6+7 2 5+6 2 | | |

Continued on next column

**TABLE LXXIV– continued**

| TQC | ICSD | NoE | SF |
|---|---|---|---|
| | 9 2 3+4 2 | | |
| | 9 2 5+6 2 | | |
| | 6+7 2 3+4 2 | | |
| | 8 2 5+6 2 | | |
| | 9 2 5+6 2 | | |
| | 4+5 2 3+4 2 | | |
| | 8 2 5+6 2 | | |
| | 9 2 3+4 2 | | |
| | 8 2 3+4 2 | | |
| | 8 2 3+4 2 | | |
| | 9 2 5+6 2 | | |
| | 8 2 5+6 2 | | |
| | 4+5 2 3+4 2 | | |
| | 6+7 2 5+6 2 | | |
| | 8 2 5+6 2 | | |
| | 9 2 5+6 2 | | |
| | 4+5 2 3+4 2 | | |
| | 8 2 3+4 2 | | |
| | 6+7 2 3+4 2 | | |
| | 9 2 5+6 2 | | |
| 167.1.88 | 243759 | 200 | Ba3O9Os2 |
| | -GM -T -F -L | | |
| | 8 2 4+4 4 5+6 2 2+2 4 | | |
| | 4+5 2 5+6 4 3+4 2 2+2 4 | | |
| | 6+7 2 5+6 4 5+6 2 2+2 4 | | |
| | 8 2 5+6 4 3+4 2 2+2 4 | | |
| | 9 2 5+6 4 3+4 2 2+2 4 | | |
| | 9 2 5+6 4 5+6 2 2+2 4 | | |
| | 8 2 4+4 4 3+4 2 2+2 4 | | |
| | 9 2 5+6 4 5+6 2 2+2 4 | | |
| | 8 2 4+4 4 3+4 2 2+2 4 | | |
| | 9 2 5+6 4 5+6 2 2+2 4 | | |
| | 6+7 2 5+6 4 5+6 2 2+2 4 | | |
| | 9 2 4+4 4 5+6 2 2+2 4 | | |
| | 8 2 5+6 4 3+4 2 2+2 4 | | |
| | 9 2 4+4 4 5+6 2 2+2 4 | | |
| | 6+7 2 5+6 4 3+4 2 2+2 4 | | |
| | 4+5 2 4+4 4 5+6 2 2+2 4 | | |
| | 8 2 5+6 4 3+4 2 2+2 4 | | |
| | 4+5 2 5+6 4 3+4 2 2+2 4 | | |
| | 8 2 4+4 4 3+4 2 2+2 4 | | |
| | 9 2 5+6 4 5+6 2 2+2 4 | | |
| | 8 2 5+6 4 3+4 2 2+2 4 | | |
| | 9 2 5+6 4 5+6 2 2+2 4 | | |
| | 4+5 2 4+4 4 3+4 2 2+2 4 | | |
| | 6+7 2 5+6 4 5+6 2 2+2 4 | | |
| | 8 2 4+4 4 5+6 2 2+2 4 | | |
| | 6+7 2 5+6 4 3+4 2 2+2 4 | | |
| | 8 2 4+4 4 5+6 2 2+2 4 | | |
| | 9 2 5+6 4 5+6 2 2+2 4 | | |
| | 4+5 2 5+6 4 3+4 2 2+2 4 | | |
| | 9 2 5+6 4 3+4 2 2+2 4 | | |
| | 4+5 2 4+4 4 5+6 2 2+2 4 | | |
| | 8 2 5+6 4 3+4 2 2+2 4 | | |
| | 9 2 5+6 4 5+6 2 2+2 4 | | |
| | 8 2 5+6 4 5+6 2 2+2 4 | | |
| | 6+7 2 4+4 4 3+4 2 2+2 4 | | |
| | 4+5 2 5+6 4 5+6 2 2+2 4 | | |
| | 8 2 5+6 4 3+4 2 2+2 4 | | |
| | 9 2 4+4 4 3+4 2 2+2 4 | | |
| | 9 2 5+6 4 5+6 2 2+2 4 | | |

Continued on next column

**TABLE LXXIV– continued**

| TQC | ICSD | NoE | SF | |
|---|---|---|---|---|
| | 6+7 2 | 5+6 4 | 5+6 2 | 2+2 4 |
| | 8 2 | 4+4 4 | 3+4 2 | 2+2 4 |
| | 9 2 | 5+6 4 | 3+4 2 | 2+2 4 |
| | 6+7 2 | 5+6 4 | 5+6 2 | 2+2 4 |
| | 8 2 | 5+6 4 | 3+4 2 | 2+2 4 |
| | 9 2 | 4+4 4 | 3+4 2 | 2+2 4 |
| | 4+5 2 | 5+6 4 | 3+4 2 | 2+2 4 |
| | 6+7 2 | 4+4 4 | 5+6 2 | 2+2 4 |
| | 8 2 | 5+6 4 | 3+4 2 | 2+2 4 |
| | 9 2 | 5+6 4 | 5+6 2 | 2+2 4 |
| | 9 2 | 4+4 4 | 3+4 2 | 2+2 4 |
| | 8 2 | | 5+6 2 | |
| | 6+7 2 | | 3+4 2 | |
| | 9 2 | | 5+6 2 | |
| | 8 2 | | 5+6 2 | |
| | 4+5 2 | | 3+4 2 | |
| | 8 2 | | 5+6 2 | |
| | 9 2 | | 3+4 2 | |
| | 4+5 2 | | 5+6 2 | |
| | 9 2 | | 5+6 2 | |
| | 8 2 | | 3+4 2 | |
| | 8 2 | | 5+6 2 | |
| | 6+7 2 | | 3+4 2 | |
| | 4+5 2 | | 3+4 2 | |
| | 9 2 | | 5+6 2 | |
| | 8 2 | | 3+4 2 | |
| | 8 2 | | 3+4 2 | |
| | 9 2 | | 5+6 2 | |
| | 9 2 | | 5+6 2 | |
| | 4+5 2 | | 5+6 2 | |
| | 8 2 | | 3+4 2 | |
| | 9 2 | | 3+4 2 | |
| | 6+7 2 | | 5+6 2 | |
| | 8 2 | | 3+4 2 | |
| | 4+5 2 | | 5+6 2 | |
| | 9 2 | | 3+4 2 | |
| | 6+7 2 | | 5+6 2 | |
| | 9 2 | | 5+6 2 | |
| | 8 2 | | 3+4 2 | |
| | 9 2 | | 5+6 2 | |
| | 8 2 | | 5+6 2 | |
| | 8 2 | | 3+4 2 | |
| | 4+5 2 | | 3+4 2 | |
| | 6+7 2 | | 5+6 2 | |
| | 9 2 | | 5+6 2 | |
| | 8 2 | | 3+4 2 | |
| | 4+5 2 | | 3+4 2 | |
| | 9 2 | | 3+4 2 | |
| | 8 2 | | 5+6 2 | |
| | 9 2 | | 5+6 2 | |
| | 6+7 2 | | 5+6 2 | |
| | 9 2 | | 3+4 2 | |
| | 9 2 | | 5+6 2 | |
| | 6+7 2 | | 3+4 2 | |
| | 8 2 | | 3+4 2 | |
| | 4+5 2 | | 3+4 2 | |
| | 8 2 | | 5+6 2 | |
| | 8 2 | | 3+4 2 | |
| | 6+7 2 | | 5+6 2 | |
| | 9 2 | | 5+6 2 | |
| | 6+7 2 | | 3+4 2 | |

Continued on next column

**TABLE LXXIV– continued**

| TQC | ICSD | NoE | SF | |
|---|---|---|---|---|
| 167.1.89 | 97540 | 156 | Ba3Cr2S6 | |
| | -GM | -T | -F | -L |
| | 8 2 | 4+4 4 | 5+6 2 | 2+2 4 |
| | 6+7 2 | 5+6 4 | 3+4 2 | 2+2 4 |
| | 9 2 | 5+6 4 | 3+4 2 | 2+2 4 |
| | 4+5 2 | 5+6 4 | 5+6 2 | 2+2 4 |
| | 8 2 | 4+4 4 | 3+4 2 | 2+2 4 |
| | 9 2 | 5+6 4 | 5+6 2 | 2+2 4 |
| | 8 2 | 5+6 4 | 3+4 2 | 2+2 4 |
| | 4+5 2 | 5+6 4 | 5+6 2 | 2+2 4 |
| | 6+7 2 | 5+6 4 | 3+4 2 | 2+2 4 |
| | 9 2 | 4+4 4 | 5+6 2 | 2+2 4 |
| | 8 2 | 4+4 4 | 5+6 2 | 2+2 4 |
| | 9 2 | 5+6 4 | 3+4 2 | 2+2 4 |
| | 8 2 | 4+4 4 | 3+4 2 | 2+2 4 |
| | 8 2 | 5+6 4 | 3+4 2 | 2+2 4 |
| | 9 2 | 5+6 4 | 5+6 2 | 2+2 4 |
| | 9 2 | 5+6 4 | 5+6 2 | 2+2 4 |
| | 4+5 2 | 4+4 4 | 5+6 2 | 2+2 4 |
| | 6+7 2 | 5+6 4 | 3+4 2 | 2+2 4 |
| | 8 2 | 5+6 4 | 5+6 2 | 2+2 4 |
| | 9 2 | 5+6 4 | 3+4 2 | 2+2 4 |
| | 6+7 2 | 5+6 4 | 5+6 2 | 2+2 4 |
| | 4+5 2 | 4+4 4 | 3+4 2 | 2+2 4 |
| | 8 2 | 5+6 4 | 5+6 2 | 2+2 4 |
| | 9 2 | 5+6 4 | 3+4 2 | 2+2 4 |
| | 8 2 | 4+4 4 | 3+4 2 | 2+2 4 |
| | 9 2 | 4+4 4 | 5+6 2 | 2+2 4 |
| | 8 2 | 5+6 4 | 3+4 2 | 2+2 4 |
| | 4+5 2 | 4+4 4 | 5+6 2 | 2+2 4 |
| | 9 2 | 5+6 4 | 3+4 2 | 2+2 4 |
| | 6+7 2 | 5+6 4 | 5+6 2 | 2+2 4 |
| | 9 2 | 5+6 4 | 3+4 2 | 2+2 4 |
| | 8 2 | 4+4 4 | 5+6 2 | 2+2 4 |
| | 6+7 2 | 5+6 4 | 3+4 2 | 2+2 4 |
| | 4+5 2 | 5+6 4 | 5+6 2 | 2+2 4 |
| | 9 2 | 4+4 4 | 3+4 2 | 2+2 4 |
| | 8 2 | 5+6 4 | 5+6 2 | 2+2 4 |
| | 8 2 | 5+6 4 | 3+4 2 | 2+2 4 |
| | 8 2 | 4+4 4 | 3+4 2 | 2+2 4 |
| | 9 2 | 5+6 4 | 5+6 2 | 2+2 4 |
| | 8 2 | | 3+4 2 | |
| | 4+5 2 | | 3+4 2 | |
| | 6+7 2 | | 5+6 2 | |
| | 9 2 | | 5+6 2 | |
| | 8 2 | | 3+4 2 | |
| | 9 2 | | 5+6 2 | |
| | 4+5 2 | | 3+4 2 | |
| | 8 2 | | 3+4 2 | |
| | 8 2 | | 5+6 2 | |
| | 4+5 2 | | 3+4 2 | |
| | 9 2 | | 5+6 2 | |
| | 6+7 2 | | 3+4 2 | |
| | 9 2 | | 5+6 2 | |
| | 8 2 | | 3+4 2 | |
| | 8 2 | | 5+6 2 | |
| | 4+5 2 | | 5+6 2 | |
| | 6+7 2 | | 5+6 2 | |
| | 9 2 | | 3+4 2 | |
| | 6+7 2 | | 3+4 2 | |
| | 8 2 | | 5+6 2 | |

Continued on next column

**TABLE LXXIV– continued**

| TQC | ICSD | NoE | SF | |
|---|---|---|---|---|
| | 4+5 2 | 5+6 2 | | |
| | 9 2 | 3+4 2 | | |
| | 8 2 | 3+4 2 | | |
| | 8 2 | 5+6 2 | | |
| | 9 2 | 3+4 2 | | |
| | 8 2 | 5+6 2 | | |
| | 4+5 2 | 3+4 2 | | |
| | 9 2 | 3+4 2 | | |
| | 9 2 | 5+6 2 | | |
| | 6+7 2 | 5+6 2 | | |
| | 9 2 | 3+4 2 | | |
| | 6+7 2 | 5+6 2 | | |
| | 9 2 | 5+6 2 | | |
| | 8 2 | 3+4 2 | | |
| | 4+5 2 | 3+4 2 | | |
| | 8 2 | 3+4 2 | | |
| | 4+5 2 | 3+4 2 | | |
| | 8 2 | 5+6 2 | | |
| | 9 2 | 5+6 2 | | |
| 167.1.90 | 402170 | 156 | Au3Ga7Yb1 | |
| | -GM | -T | -F | -L |
| | 8 2 | 5+6 4 | 3+4 2 | 2+2 4 |
| | 8 2 | 5+6 4 | 3+4 2 | 2+2 4 |
| | 9 2 | 5+6 4 | 5+6 2 | 2+2 4 |
| | 9 2 | 4+4 4 | 5+6 2 | 2+2 4 |
| | 9 2 | 5+6 4 | 5+6 2 | 2+2 4 |
| | 9 2 | 5+6 4 | 5+6 2 | 2+2 4 |
| | 6+7 2 | 5+6 4 | 5+6 2 | 2+2 4 |
| | 6+7 2 | 4+4 4 | 5+6 2 | 2+2 4 |
| | 8 2 | 4+4 4 | 5+6 2 | 2+2 4 |
| | 9 2 | 5+6 4 | 3+4 2 | 2+2 4 |
| | 8 2 | 5+6 4 | 5+6 2 | 2+2 4 |
| | 9 2 | 5+6 4 | 3+4 2 | 2+2 4 |
| | 4+5 2 | 4+4 4 | 5+6 2 | 2+2 4 |
| | 8 2 | 5+6 4 | 3+4 2 | 2+2 4 |
| | 9 2 | 5+6 4 | 3+4 2 | 2+2 4 |
| | 9 2 | 4+4 4 | 5+6 2 | 2+2 4 |
| | 6+7 2 | 5+6 4 | 3+4 2 | 2+2 4 |
| | 8 2 | 5+6 4 | 5+6 2 | 2+2 4 |
| | 4+5 2 | 4+4 4 | 3+4 2 | 2+2 4 |
| | 8 2 | 5+6 4 | 3+4 2 | 2+2 4 |
| | 6+7 2 | 5+6 4 | 5+6 2 | 2+2 4 |
| | 4+5 2 | 5+6 4 | 5+6 2 | 2+2 4 |
| | 8 2 | 4+4 4 | 3+4 2 | 2+2 4 |
| | 9 2 | 5+6 4 | 5+6 2 | 2+2 4 |
| | 6+7 2 | 4+4 4 | 3+4 2 | 2+2 4 |
| | 8 2 | 5+6 4 | 5+6 2 | 2+2 4 |
| | 9 2 | 4+4 4 | 5+6 2 | 2+2 4 |
| | 8 2 | 5+6 4 | 3+4 2 | 2+2 4 |
| | 4+5 2 | 5+6 4 | 5+6 2 | 2+2 4 |
| | 9 2 | 5+6 4 | 3+4 2 | 2+2 4 |
| | 9 2 | 4+4 4 | 3+4 2 | 2+2 4 |
| | 8 2 | 5+6 4 | 5+6 2 | 2+2 4 |
| | 8 2 | 5+6 4 | 3+4 2 | 2+2 4 |
| | 6+7 2 | 5+6 4 | 5+6 2 | 2+2 4 |
| | 9 2 | 4+4 4 | 3+4 2 | 2+2 4 |
| | 8 2 | 5+6 4 | 5+6 2 | 2+2 4 |
| | 4+5 2 | 5+6 4 | 3+4 2 | 2+2 4 |
| | 8 2 | 5+6 4 | 5+6 2 | 2+2 4 |
| | 9 2 | 4+4 4 | 5+6 2 | 2+2 4 |
| | 6+7 2 | | 3+4 2 | |

Continued on next column

**TABLE LXXIV– continued**

| TQC | ICSD | NoE | SF | |
|---|---|---|---|---|
| | 8 2 | 3+4 2 | | |
| | 9 2 | 5+6 2 | | |
| | 4+5 2 | 5+6 2 | | |
| | 6+7 2 | 3+4 2 | | |
| | 9 2 | 3+4 2 | | |
| | 9 2 | 5+6 2 | | |
| | 8 2 | 5+6 2 | | |
| | 4+5 2 | 3+4 2 | | |
| | 9 2 | 5+6 2 | | |
| | 8 2 | 3+4 2 | | |
| | 6+7 2 | 5+6 2 | | |
| | 9 2 | 5+6 2 | | |
| | 8 2 | 3+4 2 | | |
| | 8 2 | 5+6 2 | | |
| | 4+5 2 | 3+4 2 | | |
| | 6+7 2 | 3+4 2 | | |
| | 9 2 | 5+6 2 | | |
| | 9 2 | 3+4 2 | | |
| | 9 2 | 5+6 2 | | |
| | 6+7 2 | 5+6 2 | | |
| | 9 2 | 5+6 2 | | |
| | 8 2 | 5+6 2 | | |
| | 6+7 2 | 5+6 2 | | |
| | 9 2 | 5+6 2 | | |
| | 8 2 | 3+4 2 | | |
| | 9 2 | 5+6 2 | | |
| | 9 2 | 3+4 2 | | |
| | 4+5 2 | 3+4 2 | | |
| | 9 2 | 5+6 2 | | |
| | 9 2 | 5+6 2 | | |
| | 6+7 2 | 3+4 2 | | |
| | 9 2 | 5+6 2 | | |
| | 9 2 | 5+6 2 | | |
| | 6+7 2 | 5+6 2 | | |
| | 9 2 | 5+6 2 | | |
| | 9 2 | 5+6 2 | | |
| | 9 2 | 5+6 2 | | |
| | 6+7 2 | 5+6 2 | | |
| 167.1.91 | 391108 | 148 | Al7Au3Dy1 | |
| | -GM | -T | -F | -L |
| | 8 2 | 5+6 4 | 3+4 2 | 2+2 4 |
| | 8 2 | 5+6 4 | 3+4 2 | 2+2 4 |
| | 9 2 | 5+6 4 | 5+6 2 | 2+2 4 |
| | 9 2 | 4+4 4 | 5+6 2 | 2+2 4 |
| | 9 2 | 5+6 4 | 5+6 2 | 2+2 4 |
| | 9 2 | 5+6 4 | 5+6 2 | 2+2 4 |
| | 6+7 2 | 5+6 4 | 5+6 2 | 2+2 4 |
| | 6+7 2 | 4+4 4 | 5+6 2 | 2+2 4 |
| | 8 2 | 4+4 4 | 5+6 2 | 2+2 4 |
| | 4+5 2 | 5+6 4 | 3+4 2 | 2+2 4 |
| | 9 2 | 5+6 4 | 5+6 2 | 2+2 4 |
| | 8 2 | 5+6 4 | 3+4 2 | 2+2 4 |
| | 8 2 | 4+4 4 | 5+6 2 | 2+2 4 |
| | 9 2 | 5+6 4 | 3+4 2 | 2+2 4 |
| | 9 2 | 5+6 4 | 3+4 2 | 2+2 4 |
| | 6+7 2 | 4+4 4 | 5+6 2 | 2+2 4 |
| | 8 2 | 5+6 4 | 3+4 2 | 2+2 4 |
| | 9 2 | 4+4 4 | 5+6 2 | 2+2 4 |
| | 4+5 2 | 5+6 4 | 3+4 2 | 2+2 4 |
| | 8 2 | 5+6 4 | 5+6 2 | 2+2 4 |
| | 4+5 2 | 5+6 4 | 3+4 2 | 2+2 4 |

Continued on next column

**TABLE LXXIV– continued**

| TQC | ICSD | | NoE | SF |
|---|---|---|---|---|
| | 6+7 2 | 4+4 4 | 3+4 2 | 2+2 4 |
| | 8 2 | 5+6 4 | 5+6 2 | 2+2 4 |
| | 9 2 | 5+6 4 | 5+6 2 | 2+2 4 |
| | 9 2 | 4+4 4 | 3+4 2 | 2+2 4 |
| | 8 2 | 5+6 4 | 5+6 2 | 2+2 4 |
| | 6+7 2 | 5+6 4 | 5+6 2 | 2+2 4 |
| | 4+5 2 | 4+4 4 | 3+4 2 | 2+2 4 |
| | 8 2 | 5+6 4 | 3+4 2 | 2+2 4 |
| | 9 2 | 5+6 4 | 5+6 2 | 2+2 4 |
| | 9 2 | 4+4 4 | 3+4 2 | 2+2 4 |
| | 8 2 | 5+6 4 | 3+4 2 | 2+2 4 |
| | 6+7 2 | 5+6 4 | 5+6 2 | 2+2 4 |
| | 8 2 | 5+6 4 | 5+6 2 | 2+2 4 |
| | 9 2 | 4+4 4 | 3+4 2 | 2+2 4 |
| | 8 2 | 5+6 4 | 3+4 2 | 2+2 4 |
| | 4+5 2 | 5+6 4 | 5+6 2 | 2+2 4 |
| | 8 2 | | 5+6 2 | |
| | 6+7 2 | | 3+4 2 | |
| | 9 2 | | 3+4 2 | |
| | 8 2 | | 5+6 2 | |
| | 9 2 | | 5+6 2 | |
| | 4+5 2 | | 3+4 2 | |
| | 9 2 | | 5+6 2 | |
| | 6+7 2 | | 3+4 2 | |
| | 8 2 | | 5+6 2 | |
| | 4+5 2 | | 5+6 2 | |
| | 9 2 | | 3+4 2 | |
| | 9 2 | | 5+6 2 | |
| | 6+7 2 | | 3+4 2 | |
| | 8 2 | | 5+6 2 | |
| | 9 2 | | 5+6 2 | |
| | 8 2 | | 3+4 2 | |
| | 8 2 | | 3+4 2 | |
| | 4+5 2 | | 5+6 2 | |
| | 8 2 | | 3+4 2 | |
| | 6+7 2 | | 5+6 2 | |
| | 9 2 | | 3+4 2 | |
| | 9 2 | | 3+4 2 | |
| | 6+7 2 | | 3+4 2 | |
| | 9 2 | | 5+6 2 | |
| | 9 2 | | 5+6 2 | |
| | 6+7 2 | | 5+6 2 | |
| | 9 2 | | 5+6 2 | |
| | 8 2 | | 5+6 2 | |
| | 9 2 | | 5+6 2 | |
| | 4+5 2 | | 5+6 2 | |
| | 9 2 | | 3+4 2 | |
| | 9 2 | | 5+6 2 | |
| | 9 2 | | 3+4 2 | |
| | 6+7 2 | | 5+6 2 | |
| | 9 2 | | 5+6 2 | |
| | 9 2 | | 5+6 2 | |
| | 9 2 | | 5+6 2 | |
| 167.1.92 | 391101 | | 132 | Al7Au3Ce1 |
| | -GM | -T | -F | -L |
| | 8 2 | 5+6 4 | 3+4 2 | 2+2 4 |
| | 8 2 | 5+6 4 | 3+4 2 | 2+2 4 |
| | 9 2 | 5+6 4 | 5+6 2 | 2+2 4 |
| | 9 2 | 4+4 4 | 5+6 2 | 2+2 4 |
| | 9 2 | 5+6 4 | 5+6 2 | 2+2 4 |
| | 9 2 | 5+6 4 | 5+6 2 | 2+2 4 |
| | 6+7 2 | 5+6 4 | 5+6 2 | 2+2 4 |
| | 6+7 2 | 4+4 4 | 5+6 2 | 2+2 4 |
| | 8 2 | 4+4 4 | 5+6 2 | 2+2 4 |
| | 4+5 2 | 5+6 4 | 3+4 2 | 2+2 4 |
| | 8 2 | 5+6 4 | 3+4 2 | 2+2 4 |
| | 8 2 | 5+6 4 | 5+6 2 | 2+2 4 |
| | 9 2 | 4+4 4 | 5+6 2 | 2+2 4 |
| | 9 2 | 5+6 4 | 3+4 2 | 2+2 4 |
| | 9 2 | 5+6 4 | 3+4 2 | 2+2 4 |
| | 8 2 | 4+4 4 | 5+6 2 | 2+2 4 |
| | 6+7 2 | 5+6 4 | 3+4 2 | 2+2 4 |
| | 4+5 2 | 4+4 4 | 5+6 2 | 2+2 4 |
| | 9 2 | 5+6 4 | 3+4 2 | 2+2 4 |
| | 8 2 | 5+6 4 | 3+4 2 | 2+2 4 |
| | 4+5 2 | 5+6 4 | 5+6 2 | 2+2 4 |
| | 8 2 | 4+4 4 | 3+4 2 | 2+2 4 |
| | 6+7 2 | 5+6 4 | 5+6 2 | 2+2 4 |
| | 9 2 | 5+6 4 | 5+6 2 | 2+2 4 |
| | 9 2 | 4+4 4 | 3+4 2 | 2+2 4 |
| | 8 2 | 5+6 4 | 5+6 2 | 2+2 4 |
| | 6+7 2 | 4+4 4 | 5+6 2 | 2+2 4 |
| | 4+5 2 | 5+6 4 | 3+4 2 | 2+2 4 |
| | 8 2 | 5+6 4 | 3+4 2 | 2+2 4 |
| | 9 2 | 5+6 4 | 5+6 2 | 2+2 4 |
| | 9 2 | 5+6 4 | 3+4 2 | 2+2 4 |
| | 8 2 | 4+4 4 | 3+4 2 | 2+2 4 |
| | 6+7 2 | 5+6 4 | 5+6 2 | 2+2 4 |
| | 8 2 | | 3+4 2 | |
| | 9 2 | | 5+6 2 | |
| | 4+5 2 | | 3+4 2 | |
| | 8 2 | | 5+6 2 | |
| | 8 2 | | 5+6 2 | |
| | 6+7 2 | | 3+4 2 | |
| | 9 2 | | 3+4 2 | |
| | 4+5 2 | | 5+6 2 | |
| | 8 2 | | 5+6 2 | |
| | 9 2 | | 3+4 2 | |
| | 9 2 | | 3+4 2 | |
| | 6+7 2 | | 5+6 2 | |
| | 4+5 2 | | 5+6 2 | |
| | 8 2 | | 5+6 2 | |
| | 9 2 | | 3+4 2 | |
| | 9 2 | | 3+4 2 | |
| | 8 2 | | 5+6 2 | |
| | 6+7 2 | | 5+6 2 | |
| | 9 2 | | 3+4 2 | |
| | 8 2 | | 5+6 2 | |
| | 8 2 | | 3+4 2 | |
| | 8 2 | | 5+6 2 | |
| | 4+5 2 | | 3+4 2 | |
| | 6+7 2 | | 3+4 2 | |
| | 9 2 | | 5+6 2 | |
| | 8 2 | | 3+4 2 | |
| | 4+5 2 | | 3+4 2 | |
| | 9 2 | | 3+4 2 | |
| | 9 2 | | 5+6 2 | |
| | 9 2 | | 3+4 2 | |
| | 6+7 2 | | 5+6 2 | |
| | 9 2 | | 5+6 2 | |
| | 9 2 | | 5+6 2 | |

Continued on next column

The End

TABLE LXXV: The computed irreducible representations for the valence bands for the compounds of SEBR and NLC in space group 176. See the list of tables.

| TQC | ICSD | | NoE | SF | | |
|---|---|---|---|---|---|---|
| 176.1.1 | 80772 | | 104 | Mn1N3Sr3 | | |
| | -A | -GM | -H | -K | -L | -M |
| | 5+6 4 | 10+11 2 | 8+8 2 | 10+11 2 | 2+2 4 | 3+4 2 |
| | 4+4 4 | 13+14 2 | 11+9 2 | 7+8 2 | 2+2 4 | 5+6 2 |
| | 5+6 4 | 16+17 2 | 11+9 2 | 12+9 2 | 2+2 4 | 5+6 2 |
| | 4+4 4 | 7+8 2 | 10+12 2 | 7+8 2 | 2+2 4 | 3+4 2 |
| | 5+6 4 | 12+9 2 | 7+7 2 | 12+9 2 | 2+2 4 | 3+4 2 |
| | 5+6 4 | 15+18 2 | 10+12 2 | 10+11 2 | 2+2 4 | 5+6 2 |
| | 5+6 4 | 12+9 2 | 7+7 2 | 12+9 2 | 2+2 4 | 3+4 2 |
| | 5+6 4 | 13+14 2 | 10+12 2 | 7+8 2 | 2+2 4 | 5+6 2 |
| | 4+4 4 | 15+18 2 | 8+8 2 | 10+11 2 | 2+2 4 | 5+6 2 |
| | 5+6 4 | 7+8 2 | 11+9 2 | 7+8 2 | 2+2 4 | 3+4 2 |
| | 4+4 4 | 10+11 2 | 11+9 2 | 12+9 2 | 2+2 4 | 5+6 2 |
| | 5+6 4 | 16+17 2 | 10+12 2 | 10+11 2 | 2+2 4 | 3+4 2 |
| | 5+6 4 | 12+9 2 | 11+9 2 | 12+9 2 | 2+2 4 | 3+4 2 |
| | 4+4 4 | 15+18 2 | 10+12 2 | 7+8 2 | 2+2 4 | 5+6 2 |
| | 5+6 4 | 16+17 2 | 7+7 2 | 10+11 2 | 2+2 4 | 5+6 2 |
| | 4+4 4 | 12+9 2 | 8+8 2 | 7+8 2 | 2+2 4 | 3+4 2 |
| | 5+6 4 | 10+11 2 | 11+9 2 | 10+11 2 | 2+2 4 | 5+6 2 |
| | 5+6 4 | 13+14 2 | 10+12 2 | 12+9 2 | 2+2 4 | 3+4 2 |
| | 4+4 4 | 7+8 2 | 10+12 2 | 7+8 2 | 2+2 4 | 5+6 2 |
| | 5+6 4 | 15+18 2 | 11+9 2 | 12+9 2 | 2+2 4 | 5+6 2 |
| | 5+6 4 | 7+8 2 | 8+8 2 | 10+11 2 | 2+2 4 | 3+4 2 |
| | 5+6 4 | 10+11 2 | 7+7 2 | 7+8 2 | 2+2 4 | 3+4 2 |
| | 4+4 4 | 13+14 2 | 11+9 2 | 10+11 2 | 2+2 4 | 3+4 2 |
| | 5+6 4 | 16+17 2 | 10+12 2 | 12+9 2 | 2+2 4 | 5+6 2 |
| | 5+6 4 | 10+11 2 | 7+7 2 | 7+8 2 | 2+2 4 | 5+6 2 |
| | 5+6 4 | 15+18 2 | 10+12 2 | 12+9 2 | 2+2 4 | 3+4 2 |
| | | 13+14 2 | 11+9 2 | 10+11 2 | | 3+4 2 |
| | | 7+8 2 | 8+8 2 | 7+8 2 | | 5+6 2 |
| | | 16+17 2 | 10+12 2 | 12+9 2 | | 5+6 2 |
| | | 12+9 2 | 11+9 2 | 10+11 2 | | 3+4 2 |
| | | 7+8 2 | 11+9 2 | 10+11 2 | | 3+4 2 |
| | | 12+9 2 | 8+8 2 | 12+9 2 | | 5+6 2 |
| | | 13+14 2 | 10+12 2 | 7+8 2 | | 5+6 2 |
| | | 16+17 2 | 11+9 2 | 10+11 2 | | 3+4 2 |
| | | 15+18 2 | 7+7 2 | 7+8 2 | | 3+4 2 |
| | | 10+11 2 | 10+12 2 | 12+9 2 | | 5+6 2 |
| | | 10+11 2 | 10+12 2 | 12+9 2 | | 5+6 2 |
| | | 7+8 2 | 7+7 2 | 7+8 2 | | 3+4 2 |
| | | 15+18 2 | 10+12 2 | 10+11 2 | | 5+6 2 |
| | | 12+9 2 | 11+9 2 | 12+9 2 | | 3+4 2 |
| | | 13+14 2 | 8+8 2 | 7+8 2 | | 3+4 2 |
| | | 7+8 2 | 11+9 2 | 10+11 2 | | 5+6 2 |
| | | 16+17 2 | 11+9 2 | 12+9 2 | | 5+6 2 |
| | | 13+14 2 | 8+8 2 | 7+8 2 | | 3+4 2 |
| | | 12+9 2 | 11+9 2 | 10+11 2 | | 5+6 2 |
| | | 16+17 2 | 10+12 2 | 10+11 2 | | 3+4 2 |
| | | 10+11 2 | 10+12 2 | 7+8 2 | | 5+6 2 |
| | | 15+18 2 | 7+7 2 | 12+9 2 | | 3+4 2 |
| | | 15+18 2 | 7+7 2 | 12+9 2 | | 3+4 2 |
| | | 10+11 2 | 10+12 2 | 7+8 2 | | 5+6 2 |
| | | 10+11 2 | 10+12 2 | 10+11 2 | | 3+4 2 |
| | | 15+18 2 | 11+9 2 | 12+9 2 | | 5+6 2 |
| 176.1.2 | 251133 | | 240 | Ba4Nb2O9 | | |
| | -A | -GM | -H | -K | -L | -M |
| | 5+6 4 | 10+11 2 | 8+8 2 | 7+8 2 | 2+2 4 | 3+4 2 |
| | 5+6 4 | 15+18 2 | 11+9 2 | 12+9 2 | 2+2 4 | 5+6 2 |
| | 5+6 4 | 12+9 2 | 7+7 2 | 7+8 2 | 2+2 4 | 3+4 2 |
| | 5+6 4 | 16+17 2 | 10+12 2 | 10+11 2 | 2+2 4 | 5+6 2 |
| | 4+4 4 | 10+11 2 | 8+8 2 | 7+8 2 | 2+2 4 | 3+4 2 |
| | 5+6 4 | 15+18 2 | 11+9 2 | 12+9 2 | 2+2 4 | 5+6 2 |
| | 4+4 4 | 12+9 2 | 7+7 2 | 7+8 2 | 2+2 4 | 3+4 2 |
| | 5+6 4 | 16+17 2 | 10+12 2 | 10+11 2 | 2+2 4 | 5+6 2 |
| | 5+6 4 | 7+8 2 | 11+9 2 | 12+9 2 | 2+2 4 | 3+4 2 |
| | 5+6 4 | 13+14 2 | 10+12 2 | 10+11 2 | 2+2 4 | 5+6 2 |
| | 5+6 4 | 10+11 2 | 8+8 2 | 7+8 2 | 2+2 4 | 3+4 2 |
| | 5+6 4 | 15+18 2 | 11+9 2 | 12+9 2 | 2+2 4 | 5+6 2 |
| | 5+6 4 | 7+8 2 | 11+9 2 | 12+9 2 | 2+2 4 | 3+4 2 |
| | 5+6 4 | 13+14 2 | 10+12 2 | 10+11 2 | 2+2 4 | 5+6 2 |
| | 4+4 4 | 12+9 2 | 7+7 2 | 7+8 2 | 2+2 4 | 3+4 2 |
| | 5+6 4 | 16+17 2 | 10+12 2 | 10+11 2 | 2+2 4 | 5+6 2 |
| | 5+6 4 | 10+11 2 | 11+9 2 | 10+11 2 | 2+2 4 | 3+4 2 |
| | 4+4 4 | 15+18 2 | 10+12 2 | 10+11 2 | 2+2 4 | 5+6 2 |
| | 5+6 4 | 10+11 2 | 7+7 2 | 7+8 2 | 2+2 4 | 5+6 2 |
| | 4+4 4 | 12+9 2 | 10+12 2 | 12+9 2 | 2+2 4 | 3+4 2 |
| | 5+6 4 | 15+18 2 | 8+8 2 | 7+8 2 | 2+2 4 | 5+6 2 |
| | 5+6 4 | 16+17 2 | 11+9 2 | 10+11 2 | 2+2 4 | 3+4 2 |
| | 5+6 4 | 12+9 2 | 11+9 2 | 12+9 2 | 2+2 4 | 3+4 2 |
| | 5+6 4 | 10+11 2 | 10+12 2 | 10+11 2 | 2+2 4 | 3+4 2 |
| | 4+4 4 | 15+18 2 | 11+9 2 | 12+9 2 | 2+2 4 | 5+6 2 |
| | 5+6 4 | 10+11 2 | 8+8 2 | 7+8 2 | 2+2 4 | 3+4 2 |
| | 5+6 4 | 16+17 2 | 7+7 2 | 7+8 2 | 2+2 4 | 5+6 2 |
| | 4+4 4 | 12+9 2 | 10+12 2 | 10+11 2 | 2+2 4 | 3+4 2 |
| | 5+6 4 | 13+14 2 | 10+12 2 | 12+9 2 | 2+2 4 | 5+6 2 |
| | 4+4 4 | 7+8 2 | 11+9 2 | 10+11 2 | 2+2 4 | 3+4 2 |
| | 5+6 4 | 16+17 2 | 10+12 2 | 12+9 2 | 2+2 4 | 5+6 2 |
| | 5+6 4 | 15+18 2 | 11+9 2 | 7+8 2 | 2+2 4 | 5+6 2 |
| | 4+4 4 | 15+18 2 | 7+7 2 | 10+11 2 | 2+2 4 | 5+6 2 |
| | 5+6 4 | 16+17 2 | 10+12 2 | 12+9 2 | 2+2 4 | 5+6 2 |
| | 4+4 4 | 13+14 2 | 11+9 2 | 7+8 2 | 2+2 4 | 3+4 2 |
| | 5+6 4 | 13+14 2 | 8+8 2 | 10+11 2 | 2+2 4 | 5+6 2 |
| | 4+4 4 | 10+11 2 | 7+7 2 | 7+8 2 | 2+2 4 | 3+4 2 |
| | 5+6 4 | 12+9 2 | 10+12 2 | 12+9 2 | 2+2 4 | 5+6 2 |
| | 5+6 4 | 7+8 2 | 8+8 2 | 7+8 2 | 2+2 4 | 3+4 2 |
| | 4+4 4 | 7+8 2 | 11+9 2 | 10+11 2 | 2+2 4 | 3+4 2 |
| | 5+6 4 | 12+9 2 | 11+9 2 | 12+9 2 | 2+2 4 | 3+4 2 |
| | 5+6 4 | 10+11 2 | 10+12 2 | 10+11 2 | 2+2 4 | 3+4 2 |
| | 5+6 4 | 12+9 2 | 10+12 2 | 12+9 2 | 2+2 4 | 5+6 2 |
| | 5+6 4 | 16+17 2 | 11+9 2 | 12+9 2 | 2+2 4 | 3+4 2 |
| | 4+4 4 | 10+11 2 | 10+12 2 | 12+9 2 | 2+2 4 | 5+6 2 |
| | 5+6 4 | 15+18 2 | 7+7 2 | 7+8 2 | 2+2 4 | 3+4 2 |
| | 5+6 4 | 16+17 2 | 8+8 2 | 7+8 2 | 2+2 4 | 5+6 2 |
| | 4+4 4 | 12+9 2 | 11+9 2 | 10+11 2 | 2+2 4 | 3+4 2 |
| | 5+6 4 | 7+8 2 | 7+7 2 | 7+8 2 | 2+2 4 | 5+6 2 |
| | 5+6 4 | 13+14 2 | 8+8 2 | 7+8 2 | 2+2 4 | 3+4 2 |
| | 4+4 4 | 12+9 2 | 10+12 2 | 12+9 2 | 2+2 4 | 5+6 2 |
| | 5+6 4 | 16+17 2 | 11+9 2 | 12+9 2 | 2+2 4 | 3+4 2 |
| | 5+6 4 | 10+11 2 | 10+12 2 | 12+9 2 | 2+2 4 | 5+6 2 |
| | 5+6 4 | 13+14 2 | 7+7 2 | 7+8 2 | 2+2 4 | 5+6 2 |
| | 5+6 4 | 15+18 2 | 11+9 2 | 10+11 2 | 2+2 4 | 3+4 2 |
| | 4+4 4 | 13+14 2 | 8+8 2 | 7+8 2 | 2+2 4 | 3+4 2 |
| | 4+4 4 | 16+17 2 | 11+9 2 | 12+9 2 | 2+2 4 | 5+6 2 |
| | 5+6 4 | 7+8 2 | 10+12 2 | 10+11 2 | 2+2 4 | 5+6 2 |
| | 4+4 4 | 12+9 2 | 11+9 2 | 12+9 2 | 2+2 4 | 3+4 2 |
| | 5+6 4 | 7+8 2 | 10+12 2 | 10+11 2 | 2+2 4 | 3+4 2 |
| | | 15+18 2 | 11+9 2 | 10+11 2 | | 5+6 2 |
| | | 16+17 2 | 10+12 2 | 12+9 2 | | 5+6 2 |
| | | 15+18 2 | 10+12 2 | 10+11 2 | | 5+6 2 |

Continued on next column

**TABLE LXXV– continued**

| TQC | ICSD | NoE | SF | |
|---|---|---|---|---|
| | 16+17 2 | 7+7 2 | 7+8 2 | 5+6 2 |
| | 13+14 2 | 11+9 2 | 7+8 2 | 5+6 2 |
| | 13+14 2 | 8+8 2 | 12+9 2 | 5+6 2 |
| | 12+9 2 | 7+7 2 | 7+8 2 | 3+4 2 |
| | 16+17 2 | 10+12 2 | 12+9 2 | 5+6 2 |
| | 7+8 2 | 10+12 2 | 10+11 2 | 3+4 2 |
| | 13+14 2 | 11+9 2 | 10+11 2 | 5+6 2 |
| | 12+9 2 | 10+12 2 | 10+11 2 | 3+4 2 |
| | 16+17 2 | 11+9 2 | 10+11 2 | 5+6 2 |
| | 7+8 2 | 10+12 2 | 12+9 2 | 3+4 2 |
| | 13+14 2 | 7+7 2 | 7+8 2 | 5+6 2 |
| | 10+11 2 | 10+12 2 | 7+8 2 | 3+4 2 |
| | 7+8 2 | 7+7 2 | 10+11 2 | 3+4 2 |
| | 15+18 2 | 8+8 2 | 7+8 2 | 3+4 2 |
| | 12+9 2 | 11+9 2 | 10+11 2 | 5+6 2 |
| | 16+17 2 | 11+9 2 | 12+9 2 | 5+6 2 |
| | 13+14 2 | 10+12 2 | 12+9 2 | 5+6 2 |
| | 10+11 2 | 11+9 2 | 12+9 2 | 3+4 2 |
| | 15+18 2 | 8+8 2 | 7+8 2 | 3+4 2 |
| | 10+11 2 | 8+8 2 | 7+8 2 | 5+6 2 |
| | 15+18 2 | 11+9 2 | 12+9 2 | 5+6 2 |
| | 10+11 2 | 11+9 2 | 12+9 2 | 3+4 2 |
| | 15+18 2 | 8+8 2 | 7+8 2 | 5+6 2 |
| | 10+11 2 | 7+7 2 | 7+8 2 | 3+4 2 |
| | 7+8 2 | 8+8 2 | 10+11 2 | 3+4 2 |
| | 12+9 2 | 10+12 2 | 12+9 2 | 3+4 2 |
| | 10+11 2 | 11+9 2 | 7+8 2 | 5+6 2 |
| | 12+9 2 | 10+12 2 | 10+11 2 | 3+4 2 |
| | 7+8 2 | 11+9 2 | 10+11 2 | 3+4 2 |
| | 7+8 2 | 7+7 2 | 7+8 2 | 5+6 2 |
| | 15+18 2 | 10+12 2 | 12+9 2 | 5+6 2 |
| | 13+14 2 | 11+9 2 | 12+9 2 | 3+4 2 |
| | 10+11 2 | 8+8 2 | 7+8 2 | 5+6 2 |
| | 12+9 2 | 11+9 2 | 12+9 2 | 3+4 2 |
| | 16+17 2 | 10+12 2 | 12+9 2 | 3+4 2 |
| | 7+8 2 | 11+9 2 | 10+11 2 | 3+4 2 |
| | 16+17 2 | 10+12 2 | 7+8 2 | 5+6 2 |
| | 12+9 2 | 7+7 2 | 10+11 2 | 3+4 2 |
| | 13+14 2 | 10+12 2 | 10+11 2 | 3+4 2 |
| | 15+18 2 | 11+9 2 | 7+8 2 | 5+6 2 |
| | 15+18 2 | 8+8 2 | 10+11 2 | 3+4 2 |
| | 16+17 2 | 11+9 2 | 12+9 2 | 5+6 2 |
| | 10+11 2 | 10+12 2 | 7+8 2 | 3+4 2 |
| | 13+14 2 | 8+8 2 | 12+9 2 | 5+6 2 |
| | 12+9 2 | 11+9 2 | 10+11 2 | 3+4 2 |
| | 7+8 2 | 7+7 2 | 7+8 2 | 3+4 2 |
| | 16+17 2 | 10+12 2 | 10+11 2 | 5+6 2 |
| | 12+9 2 | 10+12 2 | 10+11 2 | 5+6 2 |
| | 10+11 2 | 7+7 2 | 12+9 2 | 5+6 2 |
| | 7+8 2 | 10+12 2 | 7+8 2 | 3+4 2 |
| | 13+14 2 | 11+9 2 | 12+9 2 | 3+4 2 |
| | 16+17 2 | 11+9 2 | 12+9 2 | 5+6 2 |
| | 15+18 2 | 10+12 2 | 10+11 2 | 5+6 2 |
| | 13+14 2 | 8+8 2 | 7+8 2 | 5+6 2 |
| | 13+14 2 | 7+7 2 | 7+8 2 | 5+6 2 |
| | 15+18 2 | 11+9 2 | 10+11 2 | 5+6 2 |
| | 10+11 2 | 10+12 2 | 12+9 2 | 5+6 2 |

The End

TABLE LXXVI: The computed irreducible representations for the valence bands for the compounds of SEBR and NLC in space group 189. See the list of tables.

| TQC | ICSD | NoE | SF | | | | | |
|---|---|---|---|---|---|---|---|---|
| 189.1.1 | 51718 | 92 | Si5Yb3 | | | | | |
| | -A | -GM | -HA | -H | -KA | -K | -L | -M |
| | 9 2 | 9 2 | 7 2 | 7 2 | 7 2 | 7 2 | 5 2 | 5 2 |
| | 7 2 | 7 2 | 8 2 | 8 2 | 8 2 | 8 2 | 5 2 | 5 2 |
| | 8 2 | 8 2 | 9 2 | 9 2 | 9 2 | 9 2 | 5 2 | 5 2 |
| | 7 2 | 7 2 | 8 2 | 8 2 | 8 2 | 8 2 | 5 2 | 5 2 |
| | 9 2 | 9 2 | 7 2 | 7 2 | 7 2 | 7 2 | 5 2 | 5 2 |
| | 8 2 | 8 2 | 9 2 | 9 2 | 9 2 | 9 2 | 5 2 | 5 2 |
| | 9 2 | 9 2 | 7 2 | 7 2 | 7 2 | 7 2 | 5 2 | 5 2 |
| | 8 2 | 8 2 | 8 2 | 8 2 | 8 2 | 8 2 | 5 2 | 5 2 |
| | 7 2 | 7 2 | 9 2 | 9 2 | 9 2 | 9 2 | 5 2 | 5 2 |
| | 8 2 | 9 2 | 7 2 | 7 2 | 7 2 | 7 2 | 5 2 | 5 2 |
| | 9 2 | 8 2 | 9 2 | 9 2 | 9 2 | 9 2 | 5 2 | 5 2 |
| | 7 2 | 7 2 | 8 2 | 8 2 | 8 2 | 8 2 | 5 2 | 5 2 |
| | 8 2 | 9 2 | 7 2 | 7 2 | 7 2 | 7 2 | 5 2 | 5 2 |
| | 8 2 | 9 2 | 9 2 | 9 2 | 8 2 | 8 2 | 5 2 | 5 2 |
| | 7 2 | 7 2 | 8 2 | 8 2 | 9 2 | 9 2 | 5 2 | 5 2 |
| | 9 2 | 8 2 | 7 2 | 7 2 | 8 2 | 8 2 | 5 2 | 5 2 |
| | 8 2 | 9 2 | 9 2 | 9 2 | 7 2 | 7 2 | 5 2 | 5 2 |
| | 9 2 | 7 2 | 8 2 | 8 2 | 9 2 | 9 2 | 5 2 | 5 2 |
| | 7 2 | 8 2 | 7 2 | 7 2 | 9 2 | 9 2 | 5 2 | 5 2 |
| | 9 2 | 8 2 | 8 2 | 8 2 | 7 2 | 7 2 | 5 2 | 5 2 |
| | 9 2 | 8 2 | 9 2 | 9 2 | 9 2 | 9 2 | 5 2 | 5 2 |
| | 7 2 | 9 2 | 8 2 | 8 2 | 8 2 | 8 2 | 5 2 | 5 2 |
| | 7 2 | 7 2 | 7 2 | 7 2 | 7 2 | 7 2 | 5 2 | 5 2 |
| | 8 2 | 9 2 | 8 2 | 8 2 | 8 2 | 8 2 | 5 2 | 5 2 |
| | 9 2 | 7 2 | 9 2 | 9 2 | 9 2 | 9 2 | 5 2 | 5 2 |
| | 8 2 | 7 2 | 9 2 | 9 2 | 7 2 | 7 2 | 5 2 | 5 2 |
| | 9 2 | 8 2 | 7 2 | 7 2 | 8 2 | 8 2 | 5 2 | 5 2 |
| | 8 2 | 9 2 | 8 2 | 8 2 | 9 2 | 9 2 | 5 2 | 5 2 |
| | 7 2 | 8 2 | 7 2 | 7 2 | 7 2 | 7 2 | 5 2 | 5 2 |
| | 7 2 | 8 2 | 9 2 | 9 2 | 9 2 | 9 2 | 5 2 | 5 2 |
| | 9 2 | 7 2 | 8 2 | 8 2 | 8 2 | 8 2 | 5 2 | 5 2 |
| | 8 2 | 9 2 | 9 2 | 9 2 | 7 2 | 7 2 | 5 2 | 5 2 |
| | 9 2 | 9 2 | 7 2 | 7 2 | 8 2 | 8 2 | 5 2 | 5 2 |
| | 9 2 | 8 2 | 7 2 | 7 2 | 7 2 | 7 2 | 5 2 | 5 2 |
| | 7 2 | 7 2 | 8 2 | 8 2 | 8 2 | 8 2 | 5 2 | 5 2 |
| | 8 2 | 9 2 | 7 2 | 7 2 | 9 2 | 9 2 | 5 2 | 5 2 |
| | 8 2 | 7 2 | 9 2 | 9 2 | 7 2 | 7 2 | 5 2 | 5 2 |
| | 7 2 | 8 2 | 8 2 | 8 2 | 9 2 | 9 2 | 5 2 | 5 2 |
| | 9 2 | 9 2 | 9 2 | 9 2 | 8 2 | 8 2 | 5 2 | 5 2 |
| | 8 2 | 7 2 | 7 2 | 7 2 | 9 2 | 9 2 | 5 2 | 5 2 |
| | 7 2 | 9 2 | 8 2 | 8 2 | 7 2 | 7 2 | 5 2 | 5 2 |
| | 9 2 | 8 2 | 9 2 | 9 2 | 8 2 | 8 2 | 5 2 | 5 2 |
| | 7 2 | 7 2 | 8 2 | 8 2 | 7 2 | 7 2 | 5 2 | 5 2 |
| | 8 2 | 8 2 | 7 2 | 7 2 | 8 2 | 8 2 | 5 2 | 5 2 |
| | 9 2 | 9 2 | 9 2 | 9 2 | 9 2 | 9 2 | 5 2 | 5 2 |
| | 9 2 | 8 2 | 8 2 | 8 2 | 9 2 | 9 2 | 5 2 | 5 2 |
| 189.1.2 | 415599 | 92 | Si5Yb3 | | | | | |
| | -A | -GM | -HA | -H | -KA | -K | -L | -M |
| | 9 2 | 9 2 | 7 2 | 7 2 | 7 2 | 7 2 | 5 2 | 5 2 |
| | 7 2 | 7 2 | 8 2 | 8 2 | 8 2 | 8 2 | 5 2 | 5 2 |
| | 8 2 | 8 2 | 9 2 | 9 2 | 9 2 | 9 2 | 5 2 | 5 2 |
| | 7 2 | 7 2 | 8 2 | 8 2 | 8 2 | 8 2 | 5 2 | 5 2 |
| | 9 2 | 9 2 | 7 2 | 7 2 | 7 2 | 7 2 | 5 2 | 5 2 |
| | 8 2 | 8 2 | 9 2 | 9 2 | 9 2 | 9 2 | 5 2 | 5 2 |
| | 9 2 | 9 2 | 7 2 | 7 2 | 7 2 | 7 2 | 5 2 | 5 2 |
| | 8 2 | 8 2 | 8 2 | 8 2 | 8 2 | 8 2 | 5 2 | 5 2 |

Continued on next column

**TABLE LXXVI– continued**

| TQC | ICSD | NoE | | SF | | | | |
|---|---|---|---|---|---|---|---|---|
| | 7 2 | 7 2 | 9 2 | 9 2 | 9 2 | 9 2 | 5 2 | 5 2 |
| | 8 2 | 9 2 | 7 2 | 7 2 | 7 2 | 7 2 | 5 2 | 5 2 |
| | 9 2 | 8 2 | 9 2 | 9 2 | 9 2 | 9 2 | 5 2 | 5 2 |
| | 7 2 | 7 2 | 8 2 | 8 2 | 8 2 | 8 2 | 5 2 | 5 2 |
| | 8 2 | 9 2 | 7 2 | 7 2 | 7 2 | 7 2 | 5 2 | 5 2 |
| | 8 2 | 9 2 | 9 2 | 9 2 | 8 2 | 8 2 | 5 2 | 5 2 |
| | 7 2 | 7 2 | 8 2 | 8 2 | 9 2 | 9 2 | 5 2 | 5 2 |
| | 9 2 | 8 2 | 7 2 | 7 2 | 8 2 | 8 2 | 5 2 | 5 2 |
| | 8 2 | 9 2 | 9 2 | 9 2 | 7 2 | 7 2 | 5 2 | 5 2 |
| | 9 2 | 7 2 | 7 2 | 7 2 | 9 2 | 9 2 | 5 2 | 5 2 |
| | 7 2 | 8 2 | 8 2 | 8 2 | 9 2 | 9 2 | 5 2 | 5 2 |
| | 9 2 | 8 2 | 8 2 | 8 2 | 7 2 | 7 2 | 5 2 | 5 2 |
| | 9 2 | 8 2 | 9 2 | 9 2 | 9 2 | 9 2 | 5 2 | 5 2 |
| | 7 2 | 9 2 | 8 2 | 8 2 | 8 2 | 8 2 | 5 2 | 5 2 |
| | 7 2 | 7 2 | 7 2 | 7 2 | 7 2 | 7 2 | 5 2 | 5 2 |
| | 8 2 | 9 2 | 8 2 | 8 2 | 8 2 | 8 2 | 5 2 | 5 2 |
| | 9 2 | 7 2 | 9 2 | 9 2 | 9 2 | 9 2 | 5 2 | 5 2 |
| | 8 2 | 7 2 | 9 2 | 9 2 | 7 2 | 7 2 | 5 2 | 5 2 |
| | 9 2 | 8 2 | 7 2 | 7 2 | 8 2 | 8 2 | 5 2 | 5 2 |
| | 8 2 | 9 2 | 8 2 | 8 2 | 9 2 | 9 2 | 5 2 | 5 2 |
| | 7 2 | 8 2 | 7 2 | 7 2 | 7 2 | 7 2 | 5 2 | 5 2 |
| | 7 2 | 8 2 | 9 2 | 9 2 | 9 2 | 9 2 | 5 2 | 5 2 |
| | 9 2 | 7 2 | 8 2 | 8 2 | 8 2 | 8 2 | 5 2 | 5 2 |
| | 8 2 | 9 2 | 9 2 | 9 2 | 7 2 | 7 2 | 5 2 | 5 2 |
| | 9 2 | 9 2 | 7 2 | 7 2 | 8 2 | 8 2 | 5 2 | 5 2 |
| | 9 2 | 8 2 | 7 2 | 7 2 | 7 2 | 7 2 | 5 2 | 5 2 |
| | 7 2 | 7 2 | 8 2 | 8 2 | 8 2 | 8 2 | 5 2 | 5 2 |
| | 8 2 | 9 2 | 7 2 | 7 2 | 9 2 | 9 2 | 5 2 | 5 2 |
| | 8 2 | 7 2 | 9 2 | 9 2 | 7 2 | 7 2 | 5 2 | 5 2 |
| | 7 2 | 8 2 | 8 2 | 8 2 | 9 2 | 9 2 | 5 2 | 5 2 |
| | 9 2 | 9 2 | 9 2 | 9 2 | 8 2 | 8 2 | 5 2 | 5 2 |
| | 8 2 | 7 2 | 7 2 | 7 2 | 9 2 | 9 2 | 5 2 | 5 2 |
| | 7 2 | 9 2 | 8 2 | 8 2 | 7 2 | 7 2 | 5 2 | 5 2 |
| | 9 2 | 8 2 | 9 2 | 9 2 | 8 2 | 8 2 | 5 2 | 5 2 |
| | 7 2 | 7 2 | 8 2 | 8 2 | 7 2 | 7 2 | 5 2 | 5 2 |
| | 8 2 | 8 2 | 7 2 | 7 2 | 8 2 | 8 2 | 5 2 | 5 2 |
| | 9 2 | 9 2 | 9 2 | 9 2 | 9 2 | 9 2 | 5 2 | 5 2 |
| | 9 2 | 8 2 | 8 2 | 8 2 | 9 2 | 9 2 | 5 2 | 5 2 |
| 189.1.3 | 408117 | 92 | | Si5Yb3 | | | | |
| | -A | -GM | -HA | -H | -KA | -K | -L | -M |
| | 9 2 | 9 2 | 7 2 | 7 2 | 7 2 | 7 2 | 5 2 | 5 2 |
| | 7 2 | 7 2 | 8 2 | 8 2 | 8 2 | 8 2 | 5 2 | 5 2 |
| | 8 2 | 8 2 | 9 2 | 9 2 | 9 2 | 9 2 | 5 2 | 5 2 |
| | 7 2 | 7 2 | 8 2 | 8 2 | 8 2 | 8 2 | 5 2 | 5 2 |
| | 9 2 | 9 2 | 7 2 | 7 2 | 7 2 | 7 2 | 5 2 | 5 2 |
| | 8 2 | 8 2 | 9 2 | 9 2 | 9 2 | 9 2 | 5 2 | 5 2 |
| | 9 2 | 9 2 | 7 2 | 7 2 | 7 2 | 7 2 | 5 2 | 5 2 |
| | 8 2 | 8 2 | 8 2 | 8 2 | 8 2 | 8 2 | 5 2 | 5 2 |
| | 7 2 | 7 2 | 9 2 | 9 2 | 9 2 | 9 2 | 5 2 | 5 2 |
| | 8 2 | 9 2 | 7 2 | 7 2 | 7 2 | 7 2 | 5 2 | 5 2 |
| | 9 2 | 8 2 | 9 2 | 9 2 | 9 2 | 9 2 | 5 2 | 5 2 |
| | 7 2 | 7 2 | 8 2 | 8 2 | 8 2 | 8 2 | 5 2 | 5 2 |
| | 8 2 | 9 2 | 7 2 | 7 2 | 7 2 | 7 2 | 5 2 | 5 2 |
| | 8 2 | 9 2 | 9 2 | 9 2 | 8 2 | 8 2 | 5 2 | 5 2 |
| | 7 2 | 7 2 | 8 2 | 8 2 | 9 2 | 9 2 | 5 2 | 5 2 |
| | 9 2 | 8 2 | 7 2 | 7 2 | 8 2 | 8 2 | 5 2 | 5 2 |
| | 8 2 | 9 2 | 9 2 | 9 2 | 7 2 | 7 2 | 5 2 | 5 2 |
| | 9 2 | 7 2 | 8 2 | 8 2 | 9 2 | 9 2 | 5 2 | 5 2 |
| | 7 2 | 8 2 | 7 2 | 7 2 | 9 2 | 9 2 | 5 2 | 5 2 |
| | 9 2 | 8 2 | 8 2 | 8 2 | 7 2 | 7 2 | 5 2 | 5 2 |
| | 9 2 | 8 2 | 9 2 | 9 2 | 9 2 | 9 2 | 5 2 | 5 2 |
| | 7 2 | 9 2 | 8 2 | 8 2 | 8 2 | 8 2 | 5 2 | 5 2 |
| | 7 2 | 7 2 | 7 2 | 7 2 | 7 2 | 7 2 | 5 2 | 5 2 |
| | 8 2 | 9 2 | 8 2 | 8 2 | 8 2 | 8 2 | 5 2 | 5 2 |
| | 9 2 | 7 2 | 9 2 | 9 2 | 9 2 | 9 2 | 5 2 | 5 2 |
| | 8 2 | 7 2 | 9 2 | 9 2 | 7 2 | 7 2 | 5 2 | 5 2 |
| | 9 2 | 8 2 | 7 2 | 7 2 | 8 2 | 8 2 | 5 2 | 5 2 |
| | 8 2 | 9 2 | 8 2 | 8 2 | 9 2 | 9 2 | 5 2 | 5 2 |
| | 7 2 | 8 2 | 7 2 | 7 2 | 7 2 | 7 2 | 5 2 | 5 2 |
| | 7 2 | 8 2 | 9 2 | 9 2 | 9 2 | 9 2 | 5 2 | 5 2 |
| | 9 2 | 7 2 | 8 2 | 8 2 | 8 2 | 8 2 | 5 2 | 5 2 |
| | 8 2 | 9 2 | 9 2 | 9 2 | 7 2 | 7 2 | 5 2 | 5 2 |
| | 9 2 | 9 2 | 7 2 | 7 2 | 8 2 | 8 2 | 5 2 | 5 2 |
| | 9 2 | 8 2 | 7 2 | 7 2 | 7 2 | 7 2 | 5 2 | 5 2 |
| | 7 2 | 7 2 | 8 2 | 8 2 | 8 2 | 8 2 | 5 2 | 5 2 |
| | 8 2 | 9 2 | 7 2 | 7 2 | 9 2 | 9 2 | 5 2 | 5 2 |
| | 8 2 | 7 2 | 9 2 | 9 2 | 7 2 | 7 2 | 5 2 | 5 2 |
| | 7 2 | 8 2 | 8 2 | 8 2 | 9 2 | 9 2 | 5 2 | 5 2 |
| | 9 2 | 9 2 | 9 2 | 9 2 | 8 2 | 8 2 | 5 2 | 5 2 |
| | 8 2 | 7 2 | 7 2 | 7 2 | 9 2 | 9 2 | 5 2 | 5 2 |
| | 7 2 | 9 2 | 8 2 | 8 2 | 7 2 | 7 2 | 5 2 | 5 2 |
| | 9 2 | 8 2 | 9 2 | 9 2 | 8 2 | 8 2 | 5 2 | 5 2 |
| | 7 2 | 7 2 | 8 2 | 8 2 | 7 2 | 7 2 | 5 2 | 5 2 |
| | 8 2 | 8 2 | 7 2 | 7 2 | 8 2 | 8 2 | 5 2 | 5 2 |
| | 9 2 | 9 2 | 9 2 | 9 2 | 9 2 | 9 2 | 5 2 | 5 2 |
| | 9 2 | 8 2 | 8 2 | 8 2 | 9 2 | 9 2 | 5 2 | 5 2 |
| 189.1.4 | 410418 | 114 | | Pt1Sn1Yb1 | | | | |
| | -A | -GM | -HA | -H | -KA | -K | -L | -M |
| | 9 2 | 9 2 | 7 2 | 7 2 | 7 2 | 7 2 | 5 2 | 5 2 |
| | 7 2 | 7 2 | 8 2 | 8 2 | 8 2 | 8 2 | 5 2 | 5 2 |
| | 8 2 | 8 2 | 9 2 | 9 2 | 9 2 | 9 2 | 5 2 | 5 2 |
| | 7 2 | 7 2 | 8 2 | 8 2 | 8 2 | 8 2 | 5 2 | 5 2 |
| | 9 2 | 9 2 | 9 2 | 9 2 | 7 2 | 7 2 | 5 2 | 5 2 |
| | 8 2 | 8 2 | 7 2 | 7 2 | 9 2 | 9 2 | 5 2 | 5 2 |
| | 8 2 | 8 2 | 9 2 | 9 2 | 7 2 | 7 2 | 5 2 | 5 2 |
| | 7 2 | 9 2 | 7 2 | 7 2 | 9 2 | 9 2 | 5 2 | 5 2 |
| | 9 2 | 7 2 | 8 2 | 8 2 | 8 2 | 8 2 | 5 2 | 5 2 |
| | 9 2 | 9 2 | 7 2 | 7 2 | 8 2 | 8 2 | 5 2 | 5 2 |
| | 8 2 | 8 2 | 8 2 | 8 2 | 7 2 | 7 2 | 5 2 | 5 2 |
| | 7 2 | 7 2 | 9 2 | 9 2 | 9 2 | 9 2 | 5 2 | 5 2 |
| | 8 2 | 9 2 | 8 2 | 8 2 | 9 2 | 9 2 | 5 2 | 5 2 |
| | 9 2 | 7 2 | 9 2 | 9 2 | 7 2 | 7 2 | 5 2 | 5 2 |
| | 7 2 | 8 2 | 7 2 | 7 2 | 8 2 | 8 2 | 5 2 | 5 2 |
| | 9 2 | 9 2 | 9 2 | 9 2 | 9 2 | 9 2 | 5 2 | 5 2 |
| | 7 2 | 7 2 | 9 2 | 9 2 | 8 2 | 8 2 | 5 2 | 5 2 |
| | 8 2 | 9 2 | 7 2 | 7 2 | 7 2 | 7 2 | 5 2 | 5 2 |
| | 8 2 | 8 2 | 7 2 | 7 2 | 9 2 | 9 2 | 5 2 | 5 2 |
| | 8 2 | 9 2 | 8 2 | 8 2 | 7 2 | 7 2 | 5 2 | 5 2 |
| | 9 2 | 7 2 | 8 2 | 8 2 | 8 2 | 8 2 | 5 2 | 5 2 |
| | 7 2 | 9 2 | 7 2 | 7 2 | 7 2 | 7 2 | 5 2 | 5 2 |
| | 7 2 | 7 2 | 9 2 | 9 2 | 9 2 | 9 2 | 5 2 | 5 2 |
| | 8 2 | 9 2 | 9 2 | 9 2 | 8 2 | 8 2 | 5 2 | 5 2 |
| | 7 2 | 8 2 | 8 2 | 8 2 | 7 2 | 7 2 | 5 2 | 5 2 |
| | 8 2 | 9 2 | 9 2 | 9 2 | 9 2 | 9 2 | 5 2 | 5 2 |
| | 9 2 | 7 2 | 8 2 | 8 2 | 8 2 | 8 2 | 5 2 | 5 2 |
| | 7 2 | 9 2 | 7 2 | 7 2 | 7 2 | 7 2 | 5 2 | 5 2 |
| | 9 2 | 7 2 | 9 2 | 9 2 | 8 2 | 8 2 | 5 2 | 5 2 |
| | 7 2 | 9 2 | 7 2 | 7 2 | 9 2 | 9 2 | 5 2 | 5 2 |
| | 9 2 | 7 2 | 9 2 | 9 2 | 9 2 | 9 2 | 5 2 | 5 2 |
| | 8 2 | 8 2 | 8 2 | 8 2 | 8 2 | 8 2 | 5 2 | 5 2 |
| | 8 2 | 7 2 | 7 2 | 7 2 | 7 2 | 7 2 | 5 2 | 5 2 |
| | 7 2 | 8 2 | 9 2 | 9 2 | 7 2 | 7 2 | 5 2 | 5 2 |

Continued on next column

**TABLE LXXVI– continued**

| TQC | ICSD | | NoE | | SF | | | |
|---|---|---|---|---|---|---|---|---|
| | 9 2 | 8 2 | 8 2 | 8 2 | 8 2 | 8 2 | 5 2 | 5 2 |
| | 8 2 | 9 2 | 7 2 | 7 2 | 9 2 | 9 2 | 5 2 | 5 2 |
| | 7 2 | 7 2 | 8 2 | 8 2 | 7 2 | 7 2 | 5 2 | 5 2 |
| | 9 2 | 8 2 | 7 2 | 7 2 | 9 2 | 9 2 | 5 2 | 5 2 |
| | 7 2 | 9 2 | 9 2 | 9 2 | 8 2 | 8 2 | 5 2 | 5 2 |
| | 9 2 | 7 2 | 7 2 | 7 2 | 9 2 | 9 2 | 5 2 | 5 2 |
| | 8 2 | 9 2 | 9 2 | 9 2 | 8 2 | 8 2 | 5 2 | 5 2 |
| | 8 2 | 8 2 | 8 2 | 8 2 | 7 2 | 7 2 | 5 2 | 5 2 |
| | 9 2 | 7 2 | 9 2 | 9 2 | 7 2 | 7 2 | 5 2 | 5 2 |
| | 8 2 | 8 2 | 7 2 | 7 2 | 8 2 | 8 2 | 5 2 | 5 2 |
| | 7 2 | 8 2 | 8 2 | 8 2 | 9 2 | 9 2 | 5 2 | 5 2 |
| | 9 2 | 9 2 | 8 2 | 8 2 | 8 2 | 8 2 | 5 2 | 5 2 |
| | 7 2 | 7 2 | 7 2 | 7 2 | 9 2 | 9 2 | 5 2 | 5 2 |
| | 9 2 | 8 2 | 8 2 | 8 2 | 7 2 | 7 2 | 5 2 | 5 2 |
| | 9 2 | 9 2 | 7 2 | 7 2 | 8 2 | 8 2 | 5 2 | 5 2 |
| | 8 2 | 7 2 | 9 2 | 9 2 | 9 2 | 9 2 | 5 2 | 5 2 |
| | 7 2 | 7 2 | 8 2 | 8 2 | 7 2 | 7 2 | 5 2 | 5 2 |
| | 8 2 | 8 2 | 9 2 | 9 2 | 9 2 | 9 2 | 5 2 | 5 2 |
| | 7 2 | 9 2 | 7 2 | 7 2 | 7 2 | 7 2 | 5 2 | 5 2 |
| | 9 2 | 7 2 | 9 2 | 9 2 | 8 2 | 8 2 | 5 2 | 5 2 |
| | 7 2 | 9 2 | 7 2 | 7 2 | 8 2 | 8 2 | 5 2 | 5 2 |
| | 8 2 | 8 2 | 9 2 | 9 2 | 7 2 | 7 2 | 5 2 | 5 2 |
| | 9 2 | 8 2 | 8 2 | 8 2 | 9 2 | 9 2 | 5 2 | 5 2 |
| 189.1.5 | 415370 | | 78 | | Pr1Rh1Sn1 | | | |
| | -A | -GM | -HA | -H | -KA | -K | -L | -M |
| | 9 2 | 9 2 | 7 2 | 7 2 | 7 2 | 7 2 | 5 2 | 5 2 |
| | 7 2 | 7 2 | 8 2 | 8 2 | 8 2 | 8 2 | 5 2 | 5 2 |
| | 8 2 | 8 2 | 9 2 | 9 2 | 9 2 | 9 2 | 5 2 | 5 2 |
| | 7 2 | 7 2 | 8 2 | 8 2 | 8 2 | 8 2 | 5 2 | 5 2 |
| | 9 2 | 9 2 | 9 2 | 9 2 | 9 2 | 9 2 | 5 2 | 5 2 |
| | 8 2 | 8 2 | 7 2 | 7 2 | 7 2 | 7 2 | 5 2 | 5 2 |
| | 8 2 | 9 2 | 7 2 | 7 2 | 7 2 | 7 2 | 5 2 | 5 2 |
| | 9 2 | 8 2 | 9 2 | 9 2 | 9 2 | 9 2 | 5 2 | 5 2 |
| | 7 2 | 7 2 | 8 2 | 8 2 | 8 2 | 8 2 | 5 2 | 5 2 |
| | 8 2 | 9 2 | 7 2 | 7 2 | 8 2 | 8 2 | 5 2 | 5 2 |
| | 9 2 | 8 2 | 9 2 | 9 2 | 7 2 | 7 2 | 5 2 | 5 2 |
| | 7 2 | 7 2 | 8 2 | 8 2 | 9 2 | 9 2 | 5 2 | 5 2 |
| | 8 2 | 9 2 | 8 2 | 8 2 | 9 2 | 9 2 | 5 2 | 5 2 |
| | 9 2 | 7 2 | 9 2 | 9 2 | 7 2 | 7 2 | 5 2 | 5 2 |
| | 7 2 | 8 2 | 7 2 | 7 2 | 8 2 | 8 2 | 5 2 | 5 2 |
| | 9 2 | 9 2 | 9 2 | 9 2 | 9 2 | 9 2 | 5 2 | 5 2 |
| | 8 2 | 7 2 | 7 2 | 7 2 | 8 2 | 8 2 | 5 2 | 5 2 |
| | 7 2 | 9 2 | 9 2 | 9 2 | 7 2 | 7 2 | 5 2 | 5 2 |
| | 8 2 | 8 2 | 7 2 | 7 2 | 9 2 | 9 2 | 5 2 | 5 2 |
| | 8 2 | 9 2 | 8 2 | 8 2 | 7 2 | 7 2 | 5 2 | 5 2 |
| | 7 2 | 7 2 | 8 2 | 8 2 | 8 2 | 8 2 | 5 2 | 5 2 |
| | 9 2 | 7 2 | 8 2 | 8 2 | 7 2 | 7 2 | 5 2 | 5 2 |
| | 7 2 | 8 2 | 9 2 | 9 2 | 9 2 | 9 2 | 5 2 | 5 2 |
| | 9 2 | 9 2 | 7 2 | 7 2 | 8 2 | 8 2 | 5 2 | 5 2 |
| | 8 2 | 9 2 | 9 2 | 9 2 | 9 2 | 9 2 | 5 2 | 5 2 |
| | 7 2 | 9 2 | 9 2 | 9 2 | 7 2 | 7 2 | 5 2 | 5 2 |
| | 8 2 | 7 2 | 7 2 | 7 2 | 8 2 | 8 2 | 5 2 | 5 2 |
| | 9 2 | 7 2 | 8 2 | 8 2 | 7 2 | 7 2 | 5 2 | 5 2 |
| | 7 2 | 9 2 | 9 2 | 9 2 | 9 2 | 9 2 | 5 2 | 5 2 |
| | 8 2 | 7 2 | 7 2 | 7 2 | 8 2 | 8 2 | 5 2 | 5 2 |
| | 7 2 | 7 2 | 9 2 | 9 2 | 9 2 | 9 2 | 5 2 | 5 2 |
| | 8 2 | 8 2 | 7 2 | 7 2 | 8 2 | 8 2 | 5 2 | 5 2 |
| | 9 2 | 9 2 | 8 2 | 8 2 | 7 2 | 7 2 | 5 2 | 5 2 |
| | 9 2 | 8 2 | 9 2 | 9 2 | 8 2 | 8 2 | 5 2 | 5 2 |
| | 7 2 | 8 2 | 7 2 | 7 2 | 7 2 | 7 2 | 5 2 | 5 2 |
| | 8 2 | 7 2 | 8 2 | 8 2 | 8 2 | 8 2 | 5 2 | 5 2 |

Continued on next column

**TABLE LXXVI– continued**

| TQC | ICSD | | NoE | | SF | | | |
|---|---|---|---|---|---|---|---|---|
| | 7 2 | 9 2 | 8 2 | 8 2 | 9 2 | 9 2 | 5 2 | 5 2 |
| | 9 2 | 8 2 | 7 2 | 7 2 | 7 2 | 7 2 | 5 2 | 5 2 |
| | 9 2 | 9 2 | 8 2 | 8 2 | 9 2 | 9 2 | 5 2 | 5 2 |
| 189.1.6 | 196273 | | 30 | | O1Ti1 | | | |
| | -A | -GM | -HA | -H | -KA | -K | -L | -M |
| | 9 2 | 9 2 | 8 2 | 8 2 | 7 2 | 7 2 | 5 2 | 5 2 |
| | 8 2 | 8 2 | 7 2 | 7 2 | 8 2 | 8 2 | 5 2 | 5 2 |
| | 7 2 | 7 2 | 9 2 | 9 2 | 9 2 | 9 2 | 5 2 | 5 2 |
| | 8 2 | 9 2 | 7 2 | 7 2 | 8 2 | 8 2 | 5 2 | 5 2 |
| | 7 2 | 8 2 | 9 2 | 9 2 | 7 2 | 7 2 | 5 2 | 5 2 |
| | 8 2 | 7 2 | 8 2 | 8 2 | 9 2 | 9 2 | 5 2 | 5 2 |
| | 9 2 | 9 2 | 7 2 | 7 2 | 8 2 | 8 2 | 5 2 | 5 2 |
| | 7 2 | 7 2 | 8 2 | 8 2 | 7 2 | 7 2 | 5 2 | 5 2 |
| | 9 2 | 9 2 | 9 2 | 9 2 | 9 2 | 9 2 | 5 2 | 5 2 |
| | 9 2 | 8 2 | 9 2 | 9 2 | 7 2 | 7 2 | 5 2 | 5 2 |
| | 8 2 | 7 2 | 7 2 | 7 2 | 8 2 | 8 2 | 5 2 | 5 2 |
| | 7 2 | 8 2 | 8 2 | 8 2 | 9 2 | 9 2 | 5 2 | 5 2 |
| | 9 2 | 9 2 | 9 2 | 9 2 | 9 2 | 9 2 | 5 2 | 5 2 |
| | 7 2 | 7 2 | 7 2 | 7 2 | 8 2 | 8 2 | 5 2 | 5 2 |
| | 8 2 | 9 2 | 8 2 | 8 2 | 7 2 | 7 2 | 5 2 | 5 2 |
| 189.1.7 | 183337 | | 108 | | Nd1Pd1Zn1 | | | |
| | -A | -GM | -HA | -H | -KA | -K | -L | -M |
| | 9 2 | 9 2 | 7 2 | 7 2 | 7 2 | 7 2 | 5 2 | 5 2 |
| | 7 2 | 7 2 | 8 2 | 8 2 | 8 2 | 8 2 | 5 2 | 5 2 |
| | 8 2 | 8 2 | 9 2 | 9 2 | 9 2 | 9 2 | 5 2 | 5 2 |
| | 7 2 | 9 2 | 8 2 | 8 2 | 8 2 | 8 2 | 5 2 | 5 2 |
| | 9 2 | 7 2 | 9 2 | 9 2 | 7 2 | 7 2 | 5 2 | 5 2 |
| | 8 2 | 8 2 | 7 2 | 7 2 | 9 2 | 9 2 | 5 2 | 5 2 |
| | 9 2 | 9 2 | 7 2 | 7 2 | 7 2 | 7 2 | 5 2 | 5 2 |
| | 8 2 | 8 2 | 8 2 | 8 2 | 8 2 | 8 2 | 5 2 | 5 2 |
| | 7 2 | 7 2 | 9 2 | 9 2 | 9 2 | 9 2 | 5 2 | 5 2 |
| | 8 2 | 9 2 | 7 2 | 7 2 | 8 2 | 8 2 | 5 2 | 5 2 |
| | 9 2 | 8 2 | 9 2 | 9 2 | 7 2 | 7 2 | 5 2 | 5 2 |
| | 7 2 | 7 2 | 8 2 | 8 2 | 9 2 | 9 2 | 5 2 | 5 2 |
| | 9 2 | 9 2 | 9 2 | 9 2 | 9 2 | 9 2 | 5 2 | 5 2 |
| | 7 2 | 7 2 | 7 2 | 7 2 | 7 2 | 7 2 | 5 2 | 5 2 |
| | 8 2 | 8 2 | 8 2 | 8 2 | 8 2 | 8 2 | 5 2 | 5 2 |
| | 9 2 | 9 2 | 9 2 | 9 2 | 7 2 | 7 2 | 5 2 | 5 2 |
| | 7 2 | 8 2 | 8 2 | 8 2 | 9 2 | 9 2 | 5 2 | 5 2 |
| | 8 2 | 7 2 | 7 2 | 7 2 | 8 2 | 8 2 | 5 2 | 5 2 |
| | 9 2 | 9 2 | 9 2 | 9 2 | 9 2 | 9 2 | 5 2 | 5 2 |
| | 8 2 | 7 2 | 8 2 | 8 2 | 7 2 | 7 2 | 5 2 | 5 2 |
| | 7 2 | 9 2 | 7 2 | 7 2 | 8 2 | 8 2 | 5 2 | 5 2 |
| | 8 2 | 9 2 | 8 2 | 8 2 | 9 2 | 9 2 | 5 2 | 5 2 |
| | 7 2 | 8 2 | 7 2 | 7 2 | 8 2 | 8 2 | 5 2 | 5 2 |
| | 9 2 | 7 2 | 9 2 | 9 2 | 7 2 | 7 2 | 5 2 | 5 2 |
| | 8 2 | 8 2 | 9 2 | 9 2 | 8 2 | 8 2 | 5 2 | 5 2 |
| | 9 2 | 8 2 | 8 2 | 8 2 | 9 2 | 9 2 | 5 2 | 5 2 |
| | 7 2 | 7 2 | 7 2 | 7 2 | 7 2 | 7 2 | 5 2 | 5 2 |
| | 7 2 | 9 2 | 8 2 | 8 2 | 9 2 | 9 2 | 5 2 | 5 2 |
| | 8 2 | 7 2 | 9 2 | 9 2 | 7 2 | 7 2 | 5 2 | 5 2 |
| | 9 2 | 8 2 | 7 2 | 7 2 | 8 2 | 8 2 | 5 2 | 5 2 |
| | 8 2 | 9 2 | 7 2 | 7 2 | 8 2 | 8 2 | 5 2 | 5 2 |
| | 9 2 | 9 2 | 9 2 | 9 2 | 7 2 | 7 2 | 5 2 | 5 2 |
| | 7 2 | 9 2 | 9 2 | 9 2 | 9 2 | 9 2 | 5 2 | 5 2 |
| | 7 2 | 7 2 | 8 2 | 8 2 | 8 2 | 8 2 | 5 2 | 5 2 |
| | 8 2 | 9 2 | 7 2 | 7 2 | 7 2 | 7 2 | 5 2 | 5 2 |
| | 7 2 | 7 2 | 8 2 | 8 2 | 8 2 | 8 2 | 5 2 | 5 2 |
| | 9 2 | 7 2 | 9 2 | 9 2 | 7 2 | 7 2 | 5 2 | 5 2 |
| | 7 2 | 8 2 | 7 2 | 7 2 | 9 2 | 9 2 | 5 2 | 5 2 |
| | 8 2 | 9 2 | 9 2 | 9 2 | 8 2 | 8 2 | 5 2 | 5 2 |

Continued on next column

**TABLE LXXVI– continued**

| TQC | ICSD | NoE | SF | | |
|---|---|---|---|---|---|
| | 7 2 | 7 2 | 8 2 8 2 | 7 2 | 7 2 5 2 5 2 |
| | 9 2 | 8 2 | 8 2 8 2 | 9 2 | 9 2 5 2 5 2 |
| | 8 2 | 9 2 | 9 2 9 2 | 9 2 | 9 2 5 2 5 2 |
| | 8 2 | 7 2 | 7 2 7 2 | 8 2 | 8 2 5 2 5 2 |
| | 8 2 | 9 2 | 9 2 9 2 | 9 2 | 9 2 5 2 5 2 |
| | 9 2 | 7 2 | 7 2 7 2 | 7 2 | 7 2 5 2 5 2 |
| | 8 2 | 8 2 | 9 2 9 2 | 9 2 | 9 2 5 2 5 2 |
| | 7 2 | 7 2 | 7 2 7 2 | 8 2 | 8 2 5 2 5 2 |
| | 9 2 | 8 2 | 8 2 8 2 | 7 2 | 7 2 5 2 5 2 |
| | 9 2 | 9 2 | 9 2 9 2 | 8 2 | 8 2 5 2 5 2 |
| | 7 2 | 7 2 | 7 2 7 2 | 7 2 | 7 2 5 2 5 2 |
| | 7 2 | 8 2 | 8 2 8 2 | 9 2 | 9 2 5 2 5 2 |
| | 8 2 | 8 2 | 8 2 8 2 | 7 2 | 7 2 5 2 5 2 |
| | 9 2 | 9 2 | 7 2 7 2 | 8 2 | 8 2 5 2 5 2 |
| | 9 2 | 9 2 | 8 2 8 2 | 9 2 | 9 2 5 2 5 2 |
| 189.1.8 | 419191 | 84 | Nd1Pd1Sn1 | | |
| | -A | -GM | -HA -H | -KA | -K -L -M |
| | 9 2 | 9 2 | 7 2 7 2 | 7 2 | 7 2 5 2 5 2 |
| | 7 2 | 7 2 | 8 2 8 2 | 8 2 | 8 2 5 2 5 2 |
| | 8 2 | 8 2 | 9 2 9 2 | 9 2 | 9 2 5 2 5 2 |
| | 7 2 | 7 2 | 8 2 8 2 | 8 2 | 8 2 5 2 5 2 |
| | 9 2 | 9 2 | 9 2 9 2 | 9 2 | 9 2 5 2 5 2 |
| | 8 2 | 8 2 | 7 2 7 2 | 7 2 | 7 2 5 2 5 2 |
| | 8 2 | 8 2 | 7 2 7 2 | 7 2 | 7 2 5 2 5 2 |
| | 9 2 | 9 2 | 9 2 9 2 | 9 2 | 9 2 5 2 5 2 |
| | 7 2 | 7 2 | 8 2 8 2 | 8 2 | 8 2 5 2 5 2 |
| | 8 2 | 9 2 | 7 2 7 2 | 8 2 | 8 2 5 2 5 2 |
| | 9 2 | 8 2 | 9 2 9 2 | 7 2 | 7 2 5 2 5 2 |
| | 7 2 | 7 2 | 8 2 8 2 | 9 2 | 9 2 5 2 5 2 |
| | 8 2 | 9 2 | 8 2 8 2 | 9 2 | 9 2 5 2 5 2 |
| | 9 2 | 7 2 | 9 2 9 2 | 7 2 | 7 2 5 2 5 2 |
| | 7 2 | 8 2 | 7 2 7 2 | 8 2 | 8 2 5 2 5 2 |
| | 9 2 | 9 2 | 9 2 9 2 | 9 2 | 9 2 5 2 5 2 |
| | 7 2 | 9 2 | 7 2 7 2 | 8 2 | 8 2 5 2 5 2 |
| | 8 2 | 9 2 | 9 2 9 2 | 7 2 | 7 2 5 2 5 2 |
| | 8 2 | 7 2 | 7 2 7 2 | 9 2 | 9 2 5 2 5 2 |
| | 7 2 | 7 2 | 8 2 8 2 | 7 2 | 7 2 5 2 5 2 |
| | 9 2 | 8 2 | 8 2 8 2 | 8 2 | 8 2 5 2 5 2 |
| | 7 2 | 7 2 | 7 2 7 2 | 7 2 | 7 2 5 2 5 2 |
| | 8 2 | 9 2 | 9 2 9 2 | 9 2 | 9 2 5 2 5 2 |
| | 8 2 | 8 2 | 8 2 8 2 | 8 2 | 8 2 5 2 5 2 |
| | 7 2 | 9 2 | 9 2 9 2 | 7 2 | 7 2 5 2 5 2 |
| | 9 2 | 7 2 | 9 2 9 2 | 8 2 | 8 2 5 2 5 2 |
| | 8 2 | 7 2 | 8 2 8 2 | 9 2 | 9 2 5 2 5 2 |
| | 7 2 | 9 2 | 7 2 7 2 | 7 2 | 7 2 5 2 5 2 |
| | 9 2 | 9 2 | 9 2 9 2 | 9 2 | 9 2 5 2 5 2 |
| | 8 2 | 7 2 | 9 2 9 2 | 8 2 | 8 2 5 2 5 2 |
| | 7 2 | 7 2 | 7 2 7 2 | 9 2 | 9 2 5 2 5 2 |
| | 8 2 | 8 2 | 7 2 7 2 | 8 2 | 8 2 5 2 5 2 |
| | 9 2 | 8 2 | 8 2 8 2 | 7 2 | 7 2 5 2 5 2 |
| | 7 2 | 9 2 | 9 2 9 2 | 8 2 | 8 2 5 2 5 2 |
| | 9 2 | 8 2 | 7 2 7 2 | 7 2 | 7 2 5 2 5 2 |
| | 8 2 | 7 2 | 8 2 8 2 | 8 2 | 8 2 5 2 5 2 |
| | 7 2 | 8 2 | 8 2 8 2 | 9 2 | 9 2 5 2 5 2 |
| | 9 2 | 9 2 | 8 2 8 2 | 7 2 | 7 2 5 2 5 2 |
| | 9 2 | 9 2 | 7 2 7 2 | 9 2 | 9 2 5 2 5 2 |
| | 7 2 | 8 2 | 9 2 9 2 | 8 2 | 8 2 5 2 5 2 |
| | 8 2 | 7 2 | 7 2 7 2 | 9 2 | 9 2 5 2 5 2 |
| | 8 2 | 9 2 | 9 2 9 2 | 7 2 | 7 2 5 2 5 2 |
| 189.1.9 | 414438 | 48 | Mg1Tl1Y1 | | |
| | -A | -GM | -HA -H | -KA | -K -L -M |

Continued on next column

**TABLE LXXVI– continued**

| TQC | ICSD | NoE | SF | | |
|---|---|---|---|---|---|
| | 9 2 | 9 2 | 7 2 7 2 | 7 2 | 7 2 5 2 5 2 |
| | 7 2 | 7 2 | 8 2 8 2 | 8 2 | 8 2 5 2 5 2 |
| | 8 2 | 8 2 | 9 2 9 2 | 9 2 | 9 2 5 2 5 2 |
| | 9 2 | 7 2 | 8 2 8 2 | 8 2 | 8 2 5 2 5 2 |
| | 7 2 | 9 2 | 7 2 7 2 | 9 2 | 9 2 5 2 5 2 |
| | 8 2 | 8 2 | 9 2 9 2 | 7 2 | 7 2 5 2 5 2 |
| | 8 2 | 8 2 | 9 2 9 2 | 7 2 | 7 2 5 2 5 2 |
| | 9 2 | 9 2 | 7 2 7 2 | 9 2 | 9 2 5 2 5 2 |
| | 7 2 | 7 2 | 8 2 8 2 | 8 2 | 8 2 5 2 5 2 |
| | 9 2 | 9 2 | 7 2 7 2 | 7 2 | 7 2 5 2 5 2 |
| | 8 2 | 8 2 | 8 2 8 2 | 8 2 | 8 2 5 2 5 2 |
| | 7 2 | 7 2 | 9 2 9 2 | 9 2 | 9 2 5 2 5 2 |
| | 8 2 | 9 2 | 9 2 9 2 | 9 2 | 9 2 5 2 5 2 |
| | 9 2 | 9 2 | 7 2 7 2 | 7 2 | 7 2 5 2 5 2 |
| | 8 2 | 9 2 | 9 2 9 2 | 8 2 | 8 2 5 2 5 2 |
| | 8 2 | 8 2 | 8 2 8 2 | 8 2 | 8 2 5 2 5 2 |
| | 9 2 | 7 2 | 8 2 8 2 | 7 2 | 7 2 5 2 5 2 |
| | 9 2 | 8 2 | 7 2 7 2 | 9 2 | 9 2 5 2 5 2 |
| | 7 2 | 7 2 | 7 2 7 2 | 8 2 | 8 2 5 2 5 2 |
| | 9 2 | 8 2 | 9 2 9 2 | 7 2 | 7 2 5 2 5 2 |
| | 8 2 | 8 2 | 8 2 8 2 | 9 2 | 9 2 5 2 5 2 |
| | 7 2 | 9 2 | 8 2 8 2 | 8 2 | 8 2 5 2 5 2 |
| | 7 2 | 8 2 | 7 2 7 2 | 7 2 | 7 2 5 2 5 2 |
| | 9 2 | 7 2 | 8 2 8 2 | 9 2 | 9 2 5 2 5 2 |
| 189.1.10 | 54344 | 90 | Mg1Sn1Yb1 | | |
| | -A | -GM | -HA -H | -KA | -K -L -M |
| | 8 2 | 9 2 | 7 2 7 2 | 7 2 | 7 2 5 2 5 2 |
| | 7 2 | 7 2 | 8 2 8 2 | 8 2 | 8 2 5 2 5 2 |
| | 9 2 | 8 2 | 9 2 9 2 | 9 2 | 9 2 5 2 5 2 |
| | 7 2 | 7 2 | 9 2 9 2 | 8 2 | 8 2 5 2 5 2 |
| | 8 2 | 9 2 | 7 2 7 2 | 7 2 | 7 2 5 2 5 2 |
| | 9 2 | 8 2 | 8 2 8 2 | 9 2 | 9 2 5 2 5 2 |
| | 9 2 | 8 2 | 7 2 7 2 | 7 2 | 7 2 5 2 5 2 |
| | 8 2 | 9 2 | 8 2 8 2 | 9 2 | 9 2 5 2 5 2 |
| | 7 2 | 7 2 | 9 2 9 2 | 8 2 | 8 2 5 2 5 2 |
| | 8 2 | 9 2 | 7 2 7 2 | 7 2 | 7 2 5 2 5 2 |
| | 9 2 | 8 2 | 9 2 9 2 | 8 2 | 8 2 5 2 5 2 |
| | 7 2 | 7 2 | 8 2 8 2 | 9 2 | 9 2 5 2 5 2 |
| | 9 2 | 9 2 | 8 2 8 2 | 9 2 | 9 2 5 2 5 2 |
| | 8 2 | 9 2 | 7 2 7 2 | 7 2 | 7 2 5 2 5 2 |
| | 9 2 | 9 2 | 8 2 8 2 | 8 2 | 8 2 5 2 5 2 |
| | 9 2 | 8 2 | 9 2 9 2 | 9 2 | 9 2 5 2 5 2 |
| | 8 2 | 7 2 | 7 2 7 2 | 8 2 | 8 2 5 2 5 2 |
| | 8 2 | 8 2 | 9 2 9 2 | 7 2 | 7 2 5 2 5 2 |
| | 7 2 | 7 2 | 7 2 7 2 | 8 2 | 8 2 5 2 5 2 |
| | 8 2 | 8 2 | 8 2 8 2 | 7 2 | 7 2 5 2 5 2 |
| | 7 2 | 8 2 | 9 2 9 2 | 9 2 | 9 2 5 2 5 2 |
| | 9 2 | 7 2 | 9 2 9 2 | 7 2 | 7 2 5 2 5 2 |
| | 7 2 | 8 2 | 7 2 7 2 | 8 2 | 8 2 5 2 5 2 |
| | 8 2 | 9 2 | 9 2 9 2 | 9 2 | 9 2 5 2 5 2 |
| | 7 2 | 7 2 | 8 2 8 2 | 7 2 | 7 2 5 2 5 2 |
| | 9 2 | 9 2 | 7 2 7 2 | 8 2 | 8 2 5 2 5 2 |
| | 8 2 | 9 2 | 8 2 8 2 | 9 2 | 9 2 5 2 5 2 |
| | 9 2 | 7 2 | 9 2 9 2 | 8 2 | 8 2 5 2 5 2 |
| | 8 2 | 8 2 | 7 2 7 2 | 7 2 | 7 2 5 2 5 2 |
| | 9 2 | 8 2 | 8 2 8 2 | 8 2 | 8 2 5 2 5 2 |
| | 7 2 | 8 2 | 9 2 9 2 | 9 2 | 9 2 5 2 5 2 |
| | 8 2 | 9 2 | 7 2 7 2 | 7 2 | 7 2 5 2 5 2 |
| | 7 2 | 7 2 | 9 2 9 2 | 9 2 | 9 2 5 2 5 2 |
| | 9 2 | 8 2 | 8 2 8 2 | 9 2 | 9 2 5 2 5 2 |
| | 7 2 | 9 2 | 7 2 7 2 | 8 2 | 8 2 5 2 5 2 |

Continued on next column

**TABLE LXXVI– continued**

| TQC | ICSD | NoE | SF | | | | | |
|---|---|---|---|---|---|---|---|---|
| | 8 2 | 9 2 | 8 2 | 8 2 | 7 2 | 7 2 | 5 2 | 5 2 |
| | 7 2 | 7 2 | 7 2 | 7 2 | 8 2 | 8 2 | 5 2 | 5 2 |
| | 8 2 | 9 2 | 9 2 | 9 2 | 9 2 | 9 2 | 5 2 | 5 2 |
| | 9 2 | 7 2 | 8 2 | 8 2 | 7 2 | 7 2 | 5 2 | 5 2 |
| | 7 2 | 7 2 | 7 2 | 7 2 | 7 2 | 7 2 | 5 2 | 5 2 |
| | 8 2 | 9 2 | 9 2 | 9 2 | 8 2 | 8 2 | 5 2 | 5 2 |
| | 9 2 | 8 2 | 7 2 | 7 2 | 9 2 | 9 2 | 5 2 | 5 2 |
| | 9 2 | 8 2 | 9 2 | 9 2 | 8 2 | 8 2 | 5 2 | 5 2 |
| | 8 2 | 8 2 | 8 2 | 8 2 | 7 2 | 7 2 | 5 2 | 5 2 |
| | 7 2 | 7 2 | 9 2 | 9 2 | 9 2 | 9 2 | 5 2 | 5 2 |
| 189.1.11 | 411152 | 50 | Li5N3Ni3 | | | | | |
| | -A | -GM | -HA | -H | -KA | -K | -L | -M |
| | 8 2 | 9 2 | 7 2 | 7 2 | 7 2 | 7 2 | 5 2 | 5 2 |
| | 7 2 | 7 2 | 9 2 | 9 2 | 8 2 | 8 2 | 5 2 | 5 2 |
| | 9 2 | 8 2 | 8 2 | 8 2 | 9 2 | 9 2 | 5 2 | 5 2 |
| | 9 2 | 9 2 | 7 2 | 7 2 | 8 2 | 8 2 | 5 2 | 5 2 |
| | 7 2 | 7 2 | 8 2 | 8 2 | 7 2 | 7 2 | 5 2 | 5 2 |
| | 8 2 | 8 2 | 9 2 | 9 2 | 9 2 | 9 2 | 5 2 | 5 2 |
| | 8 2 | 9 2 | 9 2 | 9 2 | 9 2 | 9 2 | 5 2 | 5 2 |
| | 7 2 | 9 2 | 7 2 | 7 2 | 7 2 | 7 2 | 5 2 | 5 2 |
| | 9 2 | 8 2 | 8 2 | 8 2 | 9 2 | 9 2 | 5 2 | 5 2 |
| | 8 2 | 7 2 | 8 2 | 8 2 | 8 2 | 8 2 | 5 2 | 5 2 |
| | 9 2 | 9 2 | 7 2 | 7 2 | 7 2 | 7 2 | 5 2 | 5 2 |
| | 7 2 | 8 2 | 9 2 | 9 2 | 8 2 | 8 2 | 5 2 | 5 2 |
| | 9 2 | 7 2 | 8 2 | 8 2 | 9 2 | 9 2 | 5 2 | 5 2 |
| | 8 2 | 8 2 | 7 2 | 7 2 | 7 2 | 7 2 | 5 2 | 5 2 |
| | 9 2 | 7 2 | 9 2 | 9 2 | 8 2 | 8 2 | 5 2 | 5 2 |
| | 7 2 | 9 2 | 7 2 | 7 2 | 7 2 | 7 2 | 5 2 | 5 2 |
| | 7 2 | 8 2 | 8 2 | 8 2 | 9 2 | 9 2 | 5 2 | 5 2 |
| | 9 2 | 9 2 | 9 2 | 9 2 | 8 2 | 8 2 | 5 2 | 5 2 |
| | 8 2 | 7 2 | 9 2 | 9 2 | 8 2 | 8 2 | 5 2 | 5 2 |
| | 7 2 | 7 2 | 8 2 | 8 2 | 7 2 | 7 2 | 5 2 | 5 2 |
| | 8 2 | 8 2 | 7 2 | 7 2 | 9 2 | 9 2 | 5 2 | 5 2 |
| | 8 2 | 7 2 | 9 2 | 9 2 | 8 2 | 8 2 | 5 2 | 5 2 |
| | 7 2 | 9 2 | 8 2 | 8 2 | 7 2 | 7 2 | 5 2 | 5 2 |
| | 9 2 | 8 2 | 7 2 | 7 2 | 9 2 | 9 2 | 5 2 | 5 2 |
| | 8 2 | 8 2 | 8 2 | 8 2 | 9 2 | 9 2 | 5 2 | 5 2 |
| 189.1.12 | 415594 | 72 | La1Rh1Sn1 | | | | | |
| | -A | -GM | -HA | -H | -KA | -K | -L | -M |
| | 9 2 | 9 2 | 7 2 | 7 2 | 7 2 | 7 2 | 5 2 | 5 2 |
| | 7 2 | 7 2 | 8 2 | 8 2 | 8 2 | 8 2 | 5 2 | 5 2 |
| | 8 2 | 8 2 | 9 2 | 9 2 | 9 2 | 9 2 | 5 2 | 5 2 |
| | 7 2 | 9 2 | 8 2 | 8 2 | 8 2 | 8 2 | 5 2 | 5 2 |
| | 9 2 | 7 2 | 9 2 | 9 2 | 9 2 | 9 2 | 5 2 | 5 2 |
| | 8 2 | 8 2 | 7 2 | 7 2 | 7 2 | 7 2 | 5 2 | 5 2 |
| | 9 2 | 8 2 | 7 2 | 7 2 | 7 2 | 7 2 | 5 2 | 5 2 |
| | 8 2 | 9 2 | 9 2 | 9 2 | 9 2 | 9 2 | 5 2 | 5 2 |
| | 7 2 | 7 2 | 8 2 | 8 2 | 8 2 | 8 2 | 5 2 | 5 2 |
| | 8 2 | 9 2 | 7 2 | 7 2 | 8 2 | 8 2 | 5 2 | 5 2 |
| | 9 2 | 8 2 | 9 2 | 9 2 | 7 2 | 7 2 | 5 2 | 5 2 |
| | 7 2 | 7 2 | 8 2 | 8 2 | 9 2 | 9 2 | 5 2 | 5 2 |
| | 8 2 | 9 2 | 8 2 | 8 2 | 9 2 | 9 2 | 5 2 | 5 2 |
| | 9 2 | 7 2 | 9 2 | 9 2 | 7 2 | 7 2 | 5 2 | 5 2 |
| | 7 2 | 8 2 | 7 2 | 7 2 | 8 2 | 8 2 | 5 2 | 5 2 |
| | 9 2 | 9 2 | 7 2 | 7 2 | 9 2 | 9 2 | 5 2 | 5 2 |
| | 8 2 | 9 2 | 9 2 | 9 2 | 8 2 | 8 2 | 5 2 | 5 2 |
| | 8 2 | 7 2 | 9 2 | 9 2 | 7 2 | 7 2 | 5 2 | 5 2 |
| | 7 2 | 9 2 | 7 2 | 7 2 | 9 2 | 9 2 | 5 2 | 5 2 |
| | 8 2 | 8 2 | 8 2 | 8 2 | 7 2 | 7 2 | 5 2 | 5 2 |
| | 7 2 | 7 2 | 8 2 | 8 2 | 8 2 | 8 2 | 5 2 | 5 2 |
| | 9 2 | 7 2 | 8 2 | 8 2 | 7 2 | 7 2 | 5 2 | 5 2 |
| | 7 2 | 8 2 | 9 2 | 9 2 | 9 2 | 9 2 | 5 2 | 5 2 |
| | 9 2 | 9 2 | 7 2 | 7 2 | 8 2 | 8 2 | 5 2 | 5 2 |
| | 8 2 | 9 2 | 9 2 | 9 2 | 7 2 | 7 2 | 5 2 | 5 2 |
| | 7 2 | 9 2 | 9 2 | 9 2 | 8 2 | 8 2 | 5 2 | 5 2 |
| | 9 2 | 7 2 | 7 2 | 7 2 | 9 2 | 9 2 | 5 2 | 5 2 |
| | 8 2 | 7 2 | 8 2 | 8 2 | 7 2 | 7 2 | 5 2 | 5 2 |
| | 7 2 | 9 2 | 9 2 | 9 2 | 9 2 | 9 2 | 5 2 | 5 2 |
| | 8 2 | 7 2 | 7 2 | 7 2 | 8 2 | 8 2 | 5 2 | 5 2 |
| | 7 2 | 7 2 | 9 2 | 9 2 | 9 2 | 9 2 | 5 2 | 5 2 |
| | 8 2 | 8 2 | 7 2 | 7 2 | 8 2 | 8 2 | 5 2 | 5 2 |
| | 9 2 | 8 2 | 8 2 | 8 2 | 7 2 | 7 2 | 5 2 | 5 2 |
| | 9 2 | 9 2 | 9 2 | 9 2 | 8 2 | 8 2 | 5 2 | 5 2 |
| | 7 2 | 8 2 | 7 2 | 7 2 | 7 2 | 7 2 | 5 2 | 5 2 |
| | 8 2 | 8 2 | 8 2 | 8 2 | 8 2 | 8 2 | 5 2 | 5 2 |
| 189.1.13 | 414439 | 48 | La1Mg1Tl1 | | | | | |
| | -A | -GM | -HA | -H | -KA | -K | -L | -M |
| | 9 2 | 9 2 | 7 2 | 7 2 | 7 2 | 7 2 | 5 2 | 5 2 |
| | 7 2 | 7 2 | 8 2 | 8 2 | 8 2 | 8 2 | 5 2 | 5 2 |
| | 8 2 | 8 2 | 9 2 | 9 2 | 9 2 | 9 2 | 5 2 | 5 2 |
| | 9 2 | 9 2 | 8 2 | 8 2 | 8 2 | 8 2 | 5 2 | 5 2 |
| | 7 2 | 7 2 | 7 2 | 7 2 | 9 2 | 9 2 | 5 2 | 5 2 |
| | 8 2 | 8 2 | 9 2 | 9 2 | 7 2 | 7 2 | 5 2 | 5 2 |
| | 8 2 | 9 2 | 7 2 | 7 2 | 7 2 | 7 2 | 5 2 | 5 2 |
| | 9 2 | 8 2 | 9 2 | 9 2 | 9 2 | 9 2 | 5 2 | 5 2 |
| | 7 2 | 7 2 | 8 2 | 8 2 | 8 2 | 8 2 | 5 2 | 5 2 |
| | 9 2 | 9 2 | 7 2 | 7 2 | 7 2 | 7 2 | 5 2 | 5 2 |
| | 8 2 | 8 2 | 8 2 | 8 2 | 8 2 | 8 2 | 5 2 | 5 2 |
| | 7 2 | 7 2 | 9 2 | 9 2 | 9 2 | 9 2 | 5 2 | 5 2 |
| | 9 2 | 9 2 | 9 2 | 9 2 | 9 2 | 9 2 | 5 2 | 5 2 |
| | 8 2 | 9 2 | 7 2 | 7 2 | 7 2 | 7 2 | 5 2 | 5 2 |
| | 8 2 | 9 2 | 9 2 | 9 2 | 8 2 | 8 2 | 5 2 | 5 2 |
| | 8 2 | 8 2 | 8 2 | 8 2 | 8 2 | 8 2 | 5 2 | 5 2 |
| | 9 2 | 7 2 | 8 2 | 8 2 | 7 2 | 7 2 | 5 2 | 5 2 |
| | 7 2 | 8 2 | 7 2 | 7 2 | 9 2 | 9 2 | 5 2 | 5 2 |
| | 9 2 | 7 2 | 7 2 | 7 2 | 8 2 | 8 2 | 5 2 | 5 2 |
| | 9 2 | 8 2 | 9 2 | 9 2 | 7 2 | 7 2 | 5 2 | 5 2 |
| | 8 2 | 9 2 | 8 2 | 8 2 | 9 2 | 9 2 | 5 2 | 5 2 |
| | 7 2 | 8 2 | 8 2 | 8 2 | 8 2 | 8 2 | 5 2 | 5 2 |
| | 7 2 | 8 2 | 7 2 | 7 2 | 7 2 | 7 2 | 5 2 | 5 2 |
| | 9 2 | 7 2 | 8 2 | 8 2 | 9 2 | 9 2 | 5 2 | 5 2 |
| 189.1.14 | 160910 | 48 | In1Mg1Y1 | | | | | |
| | -A | -GM | -HA | -H | -KA | -K | -L | -M |
| | 9 2 | 9 2 | 7 2 | 7 2 | 7 2 | 7 2 | 5 2 | 5 2 |
| | 7 2 | 7 2 | 8 2 | 8 2 | 8 2 | 8 2 | 5 2 | 5 2 |
| | 8 2 | 8 2 | 9 2 | 9 2 | 9 2 | 9 2 | 5 2 | 5 2 |
| | 7 2 | 7 2 | 8 2 | 8 2 | 8 2 | 8 2 | 5 2 | 5 2 |
| | 9 2 | 9 2 | 7 2 | 7 2 | 9 2 | 9 2 | 5 2 | 5 2 |
| | 8 2 | 8 2 | 9 2 | 9 2 | 7 2 | 7 2 | 5 2 | 5 2 |
| | 8 2 | 8 2 | 9 2 | 9 2 | 7 2 | 7 2 | 5 2 | 5 2 |
| | 7 2 | 9 2 | 7 2 | 7 2 | 9 2 | 9 2 | 5 2 | 5 2 |
| | 9 2 | 7 2 | 8 2 | 8 2 | 8 2 | 8 2 | 5 2 | 5 2 |
| | 9 2 | 9 2 | 7 2 | 7 2 | 7 2 | 7 2 | 5 2 | 5 2 |
| | 8 2 | 8 2 | 8 2 | 8 2 | 8 2 | 8 2 | 5 2 | 5 2 |
| | 7 2 | 7 2 | 9 2 | 9 2 | 9 2 | 9 2 | 5 2 | 5 2 |
| | 8 2 | 9 2 | 9 2 | 9 2 | 9 2 | 9 2 | 5 2 | 5 2 |
| | 9 2 | 9 2 | 7 2 | 7 2 | 7 2 | 7 2 | 5 2 | 5 2 |
| | 8 2 | 9 2 | 9 2 | 9 2 | 8 2 | 8 2 | 5 2 | 5 2 |
| | 8 2 | 8 2 | 8 2 | 8 2 | 8 2 | 8 2 | 5 2 | 5 2 |
| | 9 2 | 7 2 | 8 2 | 8 2 | 7 2 | 7 2 | 5 2 | 5 2 |
| | 9 2 | 8 2 | 7 2 | 7 2 | 9 2 | 9 2 | 5 2 | 5 2 |
| | 7 2 | 7 2 | 7 2 | 7 2 | 8 2 | 8 2 | 5 2 | 5 2 |

Continued on next column

**TABLE LXXVI– continued**

| TQC | ICSD | NoE | SF | | | | | |
|---|---|---|---|---|---|---|---|---|
| | 9 2 | 8 2 | 9 2 | 9 2 | 7 2 | 7 2 | 5 2 | 5 2 |
| | 7 2 | 9 2 | 8 2 | 8 2 | 8 2 | 8 2 | 5 2 | 5 2 |
| | 8 2 | 8 2 | 8 2 | 8 2 | 9 2 | 9 2 | 5 2 | 5 2 |
| | 7 2 | 8 2 | 7 2 | 7 2 | 7 2 | 7 2 | 5 2 | 5 2 |
| | 9 2 | 7 2 | 8 2 | 8 2 | 9 2 | 9 2 | 5 2 | 5 2 |
| 189.1.15 | 85966 | 44 | Hf6Ni1Sb2 | | | | | |
| | -A | -GM | -HA | -H | -KA | -K | -L | -M |
| | 8 2 | 9 2 | 7 2 | 7 2 | 8 2 | 8 2 | 5 2 | 5 2 |
| | 8 2 | 9 2 | 9 2 | 9 2 | 7 2 | 7 2 | 5 2 | 5 2 |
| | 9 2 | 8 2 | 7 2 | 7 2 | 9 2 | 9 2 | 5 2 | 5 2 |
| | 9 2 | 7 2 | 8 2 | 8 2 | 7 2 | 7 2 | 5 2 | 5 2 |
| | 7 2 | 9 2 | 8 2 | 8 2 | 8 2 | 8 2 | 5 2 | 5 2 |
| | 9 2 | 8 2 | 7 2 | 7 2 | 9 2 | 9 2 | 5 2 | 5 2 |
| | 9 2 | 7 2 | 9 2 | 9 2 | 9 2 | 9 2 | 5 2 | 5 2 |
| | 9 2 | 9 2 | 9 2 | 9 2 | 9 2 | 9 2 | 5 2 | 5 2 |
| | 7 2 | 7 2 | 8 2 | 8 2 | 7 2 | 7 2 | 5 2 | 5 2 |
| | 7 2 | 8 2 | 7 2 | 7 2 | 8 2 | 8 2 | 5 2 | 5 2 |
| | 8 2 | 9 2 | 9 2 | 9 2 | 9 2 | 9 2 | 5 2 | 5 2 |
| | 7 2 | 7 2 | 7 2 | 7 2 | 7 2 | 7 2 | 5 2 | 5 2 |
| | 9 2 | 8 2 | 8 2 | 8 2 | 7 2 | 7 2 | 5 2 | 5 2 |
| | 8 2 | 7 2 | 9 2 | 9 2 | 8 2 | 8 2 | 5 2 | 5 2 |
| | 9 2 | 9 2 | 8 2 | 8 2 | 9 2 | 9 2 | 5 2 | 5 2 |
| | 7 2 | 8 2 | 8 2 | 8 2 | 7 2 | 7 2 | 5 2 | 5 2 |
| | 8 2 | 8 2 | 9 2 | 9 2 | 9 2 | 9 2 | 5 2 | 5 2 |
| | 8 2 | 8 2 | 7 2 | 7 2 | 7 2 | 7 2 | 5 2 | 5 2 |
| | 7 2 | 7 2 | 9 2 | 9 2 | 8 2 | 8 2 | 5 2 | 5 2 |
| | 8 2 | 7 2 | 7 2 | 7 2 | 8 2 | 8 2 | 5 2 | 5 2 |
| | 9 2 | 9 2 | 9 2 | 9 2 | 9 2 | 9 2 | 5 2 | 5 2 |
| | 7 2 | 8 2 | 8 2 | 8 2 | 8 2 | 8 2 | 5 2 | 5 2 |
| 189.1.16 | 409925 | 54 | Ge1Pd1Ti1 | | | | | |
| | -A | -GM | -HA | -H | -KA | -K | -L | -M |
| | 9 2 | 9 2 | 9 2 | 9 2 | 9 2 | 9 2 | 5 2 | 5 2 |
| | 8 2 | 9 2 | 7 2 | 7 2 | 8 2 | 8 2 | 5 2 | 5 2 |
| | 8 2 | 9 2 | 9 2 | 9 2 | 7 2 | 7 2 | 5 2 | 5 2 |
| | 8 2 | 7 2 | 8 2 | 8 2 | 9 2 | 9 2 | 5 2 | 5 2 |
| | 9 2 | 8 2 | 7 2 | 7 2 | 8 2 | 8 2 | 5 2 | 5 2 |
| | 7 2 | 8 2 | 8 2 | 8 2 | 7 2 | 7 2 | 5 2 | 5 2 |
| | 8 2 | 7 2 | 8 2 | 8 2 | 8 2 | 8 2 | 5 2 | 5 2 |
| | 7 2 | 8 2 | 9 2 | 9 2 | 8 2 | 8 2 | 5 2 | 5 2 |
| | 9 2 | 9 2 | 7 2 | 7 2 | 7 2 | 7 2 | 5 2 | 5 2 |
| | 9 2 | 8 2 | 7 2 | 7 2 | 9 2 | 9 2 | 5 2 | 5 2 |
| | 9 2 | 7 2 | 8 2 | 8 2 | 7 2 | 7 2 | 5 2 | 5 2 |
| | 7 2 | 8 2 | 8 2 | 8 2 | 9 2 | 9 2 | 5 2 | 5 2 |
| | 8 2 | 7 2 | 9 2 | 9 2 | 9 2 | 9 2 | 5 2 | 5 2 |
| | 9 2 | 9 2 | 7 2 | 7 2 | 7 2 | 7 2 | 5 2 | 5 2 |
| | 7 2 | 8 2 | 9 2 | 9 2 | 9 2 | 9 2 | 5 2 | 5 2 |
| | 7 2 | 7 2 | 9 2 | 9 2 | 8 2 | 8 2 | 5 2 | 5 2 |
| | 8 2 | 9 2 | 8 2 | 8 2 | 7 2 | 7 2 | 5 2 | 5 2 |
| | 8 2 | 7 2 | 7 2 | 7 2 | 8 2 | 8 2 | 5 2 | 5 2 |
| | 9 2 | 9 2 | 8 2 | 8 2 | 9 2 | 9 2 | 5 2 | 5 2 |
| | 7 2 | 8 2 | 7 2 | 7 2 | 8 2 | 8 2 | 5 2 | 5 2 |
| | 9 2 | 7 2 | 9 2 | 9 2 | 7 2 | 7 2 | 5 2 | 5 2 |
| | 8 2 | 9 2 | 7 2 | 7 2 | 7 2 | 7 2 | 5 2 | 5 2 |
| | 7 2 | 9 2 | 9 2 | 9 2 | 9 2 | 9 2 | 5 2 | 5 2 |
| | 9 2 | 8 2 | 8 2 | 8 2 | 7 2 | 7 2 | 5 2 | 5 2 |
| | 8 2 | 7 2 | 7 2 | 7 2 | 8 2 | 8 2 | 5 2 | 5 2 |
| | 7 2 | 8 2 | 8 2 | 8 2 | 9 2 | 9 2 | 5 2 | 5 2 |
| | 9 2 | 8 2 | 8 2 | 8 2 | 8 2 | 8 2 | 5 2 | 5 2 |
| 189.1.17 | 59177 | 54 | Ge1Li1Pr1 | | | | | |
| | -A | -GM | -HA | -H | -KA | -K | -L | -M |
| | 9 2 | 9 2 | 7 2 | 7 2 | 7 2 | 7 2 | 5 2 | 5 2 |

Continued on next column

**TABLE LXXVI– continued**

| TQC | ICSD | NoE | SF | | | | | |
|---|---|---|---|---|---|---|---|---|
| | 7 2 | 7 2 | 8 2 | 8 2 | 8 2 | 8 2 | 5 2 | 5 2 |
| | 8 2 | 8 2 | 9 2 | 9 2 | 9 2 | 9 2 | 5 2 | 5 2 |
| | 9 2 | 9 2 | 8 2 | 8 2 | 8 2 | 8 2 | 5 2 | 5 2 |
| | 7 2 | 7 2 | 7 2 | 7 2 | 7 2 | 7 2 | 5 2 | 5 2 |
| | 8 2 | 8 2 | 9 2 | 9 2 | 9 2 | 9 2 | 5 2 | 5 2 |
| | 9 2 | 9 2 | 7 2 | 7 2 | 7 2 | 7 2 | 5 2 | 5 2 |
| | 8 2 | 8 2 | 9 2 | 9 2 | 9 2 | 9 2 | 5 2 | 5 2 |
| | 7 2 | 7 2 | 8 2 | 8 2 | 8 2 | 8 2 | 5 2 | 5 2 |
| | 9 2 | 9 2 | 7 2 | 7 2 | 7 2 | 7 2 | 5 2 | 5 2 |
| | 8 2 | 8 2 | 8 2 | 8 2 | 8 2 | 8 2 | 5 2 | 5 2 |
| | 7 2 | 7 2 | 9 2 | 9 2 | 9 2 | 9 2 | 5 2 | 5 2 |
| | 8 2 | 9 2 | 9 2 | 9 2 | 9 2 | 9 2 | 5 2 | 5 2 |
| | 9 2 | 9 2 | 7 2 | 7 2 | 7 2 | 7 2 | 5 2 | 5 2 |
| | 8 2 | 9 2 | 9 2 | 9 2 | 8 2 | 8 2 | 5 2 | 5 2 |
| | 9 2 | 8 2 | 8 2 | 8 2 | 9 2 | 9 2 | 5 2 | 5 2 |
| | 8 2 | 7 2 | 7 2 | 7 2 | 8 2 | 8 2 | 5 2 | 5 2 |
| | 9 2 | 8 2 | 8 2 | 8 2 | 7 2 | 7 2 | 5 2 | 5 2 |
| | 7 2 | 7 2 | 7 2 | 7 2 | 7 2 | 7 2 | 5 2 | 5 2 |
| | 7 2 | 8 2 | 9 2 | 9 2 | 9 2 | 9 2 | 5 2 | 5 2 |
| | 8 2 | 8 2 | 8 2 | 8 2 | 7 2 | 7 2 | 5 2 | 5 2 |
| | 9 2 | 7 2 | 7 2 | 7 2 | 8 2 | 8 2 | 5 2 | 5 2 |
| | 7 2 | 8 2 | 8 2 | 8 2 | 8 2 | 8 2 | 5 2 | 5 2 |
| | 9 2 | 9 2 | 8 2 | 8 2 | 9 2 | 9 2 | 5 2 | 5 2 |
| | 8 2 | 8 2 | 7 2 | 7 2 | 8 2 | 8 2 | 5 2 | 5 2 |
| | 9 2 | 9 2 | 9 2 | 9 2 | 7 2 | 7 2 | 5 2 | 5 2 |
| | 8 2 | 9 2 | 8 2 | 8 2 | 9 2 | 9 2 | 5 2 | 5 2 |
| 189.1.18 | 240950 | 48 | Ga1Mg1Y1 | | | | | |
| | -A | -GM | -HA | -H | -KA | -K | -L | -M |
| | 8 2 | 9 2 | 7 2 | 7 2 | 7 2 | 7 2 | 5 2 | 5 2 |
| | 7 2 | 7 2 | 9 2 | 9 2 | 8 2 | 8 2 | 5 2 | 5 2 |
| | 9 2 | 8 2 | 8 2 | 8 2 | 9 2 | 9 2 | 5 2 | 5 2 |
| | 7 2 | 7 2 | 9 2 | 9 2 | 8 2 | 8 2 | 5 2 | 5 2 |
| | 8 2 | 9 2 | 8 2 | 8 2 | 9 2 | 9 2 | 5 2 | 5 2 |
| | 9 2 | 8 2 | 7 2 | 7 2 | 7 2 | 7 2 | 5 2 | 5 2 |
| | 9 2 | 8 2 | 7 2 | 7 2 | 7 2 | 7 2 | 5 2 | 5 2 |
| | 8 2 | 9 2 | 8 2 | 8 2 | 9 2 | 9 2 | 5 2 | 5 2 |
| | 7 2 | 7 2 | 9 2 | 9 2 | 8 2 | 8 2 | 5 2 | 5 2 |
| | 8 2 | 9 2 | 7 2 | 7 2 | 7 2 | 7 2 | 5 2 | 5 2 |
| | 9 2 | 8 2 | 9 2 | 9 2 | 8 2 | 8 2 | 5 2 | 5 2 |
| | 7 2 | 7 2 | 8 2 | 8 2 | 9 2 | 9 2 | 5 2 | 5 2 |
| | 8 2 | 9 2 | 8 2 | 8 2 | 9 2 | 9 2 | 5 2 | 5 2 |
| | 9 2 | 9 2 | 7 2 | 7 2 | 7 2 | 7 2 | 5 2 | 5 2 |
| | 9 2 | 9 2 | 8 2 | 8 2 | 8 2 | 8 2 | 5 2 | 5 2 |
| | 9 2 | 7 2 | 9 2 | 9 2 | 8 2 | 8 2 | 5 2 | 5 2 |
| | 8 2 | 8 2 | 7 2 | 7 2 | 7 2 | 7 2 | 5 2 | 5 2 |
| | 8 2 | 8 2 | 9 2 | 9 2 | 9 2 | 9 2 | 5 2 | 5 2 |
| | 7 2 | 7 2 | 7 2 | 7 2 | 8 2 | 8 2 | 5 2 | 5 2 |
| | 7 2 | 8 2 | 8 2 | 8 2 | 7 2 | 7 2 | 5 2 | 5 2 |
| | 8 2 | 9 2 | 9 2 | 9 2 | 8 2 | 8 2 | 5 2 | 5 2 |
| | 9 2 | 8 2 | 9 2 | 9 2 | 9 2 | 9 2 | 5 2 | 5 2 |
| | 7 2 | 8 2 | 7 2 | 7 2 | 7 2 | 7 2 | 5 2 | 5 2 |
| | 8 2 | 7 2 | 9 2 | 9 2 | 9 2 | 9 2 | 5 2 | 5 2 |
| 189.1.19 | 160909 | 48 | Ga1Mg1Y1 | | | | | |
| | -A | -GM | -HA | -H | -KA | -K | -L | -M |
| | 9 2 | 9 2 | 7 2 | 7 2 | 7 2 | 7 2 | 5 2 | 5 2 |
| | 7 2 | 7 2 | 8 2 | 8 2 | 8 2 | 8 2 | 5 2 | 5 2 |
| | 8 2 | 8 2 | 9 2 | 9 2 | 9 2 | 9 2 | 5 2 | 5 2 |
| | 7 2 | 7 2 | 8 2 | 8 2 | 8 2 | 8 2 | 5 2 | 5 2 |
| | 9 2 | 9 2 | 9 2 | 9 2 | 9 2 | 9 2 | 5 2 | 5 2 |
| | 8 2 | 8 2 | 7 2 | 7 2 | 7 2 | 7 2 | 5 2 | 5 2 |
| | 8 2 | 8 2 | 7 2 | 7 2 | 7 2 | 7 2 | 5 2 | 5 2 |

Continued on next column

**TABLE LXXVI– continued**

| TQC | ICSD | NoE | SF | | | | | |
|---|---|---|---|---|---|---|---|---|
| | 9 2 | 9 2 | 9 2 | 9 2 | 9 2 | 9 2 | 5 2 | 5 2 |
| | 7 2 | 7 2 | 8 2 | 8 2 | 8 2 | 8 2 | 5 2 | 5 2 |
| | 9 2 | 9 2 | 7 2 | 7 2 | 7 2 | 7 2 | 5 2 | 5 2 |
| | 8 2 | 8 2 | 8 2 | 8 2 | 8 2 | 8 2 | 5 2 | 5 2 |
| | 7 2 | 7 2 | 9 2 | 9 2 | 9 2 | 9 2 | 5 2 | 5 2 |
| | 8 2 | 9 2 | 9 2 | 9 2 | 9 2 | 9 2 | 5 2 | 5 2 |
| | 9 2 | 9 2 | 7 2 | 7 2 | 7 2 | 7 2 | 5 2 | 5 2 |
| | 8 2 | 9 2 | 9 2 | 9 2 | 8 2 | 8 2 | 5 2 | 5 2 |
| | 8 2 | 7 2 | 8 2 | 8 2 | 8 2 | 8 2 | 5 2 | 5 2 |
| | 9 2 | 8 2 | 7 2 | 7 2 | 7 2 | 7 2 | 5 2 | 5 2 |
| | 9 2 | 8 2 | 8 2 | 8 2 | 9 2 | 9 2 | 5 2 | 5 2 |
| | 7 2 | 7 2 | 7 2 | 7 2 | 8 2 | 8 2 | 5 2 | 5 2 |
| | 7 2 | 8 2 | 9 2 | 9 2 | 7 2 | 7 2 | 5 2 | 5 2 |
| | 9 2 | 9 2 | 8 2 | 8 2 | 8 2 | 8 2 | 5 2 | 5 2 |
| | 8 2 | 8 2 | 8 2 | 8 2 | 9 2 | 9 2 | 5 2 | 5 2 |
| | 7 2 | 8 2 | 7 2 | 7 2 | 7 2 | 7 2 | 5 2 | 5 2 |
| | 9 2 | 7 2 | 8 2 | 8 2 | 9 2 | 9 2 | 5 2 | 5 2 |
| 189.1.20 | 413276 | 48 | Ga1La1Mg1 | | | | | |
| | -A | -GM | -HA | -H | -KA | -K | -L | -M |
| | 9 2 | 9 2 | 7 2 | 7 2 | 7 2 | 7 2 | 5 2 | 5 2 |
| | 7 2 | 7 2 | 8 2 | 8 2 | 8 2 | 8 2 | 5 2 | 5 2 |
| | 8 2 | 8 2 | 9 2 | 9 2 | 9 2 | 9 2 | 5 2 | 5 2 |
| | 9 2 | 9 2 | 8 2 | 8 2 | 8 2 | 8 2 | 5 2 | 5 2 |
| | 7 2 | 7 2 | 9 2 | 9 2 | 9 2 | 9 2 | 5 2 | 5 2 |
| | 8 2 | 8 2 | 7 2 | 7 2 | 7 2 | 7 2 | 5 2 | 5 2 |
| | 8 2 | 9 2 | 7 2 | 7 2 | 7 2 | 7 2 | 5 2 | 5 2 |
| | 9 2 | 8 2 | 9 2 | 9 2 | 9 2 | 9 2 | 5 2 | 5 2 |
| | 7 2 | 7 2 | 8 2 | 8 2 | 8 2 | 8 2 | 5 2 | 5 2 |
| | 9 2 | 9 2 | 7 2 | 7 2 | 7 2 | 7 2 | 5 2 | 5 2 |
| | 8 2 | 8 2 | 8 2 | 8 2 | 8 2 | 8 2 | 5 2 | 5 2 |
| | 7 2 | 7 2 | 9 2 | 9 2 | 9 2 | 9 2 | 5 2 | 5 2 |
| | 9 2 | 9 2 | 9 2 | 9 2 | 9 2 | 9 2 | 5 2 | 5 2 |
| | 8 2 | 9 2 | 7 2 | 7 2 | 7 2 | 7 2 | 5 2 | 5 2 |
| | 8 2 | 9 2 | 9 2 | 9 2 | 8 2 | 8 2 | 5 2 | 5 2 |
| | 8 2 | 7 2 | 8 2 | 8 2 | 8 2 | 8 2 | 5 2 | 5 2 |
| | 9 2 | 8 2 | 7 2 | 7 2 | 7 2 | 7 2 | 5 2 | 5 2 |
| | 7 2 | 7 2 | 8 2 | 8 2 | 9 2 | 9 2 | 5 2 | 5 2 |
| | 9 2 | 8 2 | 7 2 | 7 2 | 8 2 | 8 2 | 5 2 | 5 2 |
| | 7 2 | 8 2 | 9 2 | 9 2 | 7 2 | 7 2 | 5 2 | 5 2 |
| | 8 2 | 9 2 | 8 2 | 8 2 | 8 2 | 8 2 | 5 2 | 5 2 |
| | 9 2 | 8 2 | 7 2 | 7 2 | 9 2 | 9 2 | 5 2 | 5 2 |
| | 7 2 | 7 2 | 8 2 | 8 2 | 7 2 | 7 2 | 5 2 | 5 2 |
| | 9 2 | 8 2 | 8 2 | 8 2 | 9 2 | 9 2 | 5 2 | 5 2 |
| 189.1.21 | 152689 | 144 | Fe1Ho6Sb2 | | | | | |
| | -A | -GM | -HA | -H | -KA | -K | -L | -M |
| | 7 2 | 9 2 | 7 2 | 7 2 | 7 2 | 7 2 | 5 2 | 5 2 |
| | 8 2 | 7 2 | 8 2 | 8 2 | 8 2 | 8 2 | 5 2 | 5 2 |
| | 9 2 | 8 2 | 9 2 | 9 2 | 9 2 | 9 2 | 5 2 | 5 2 |
| | 9 2 | 9 2 | 9 2 | 9 2 | 9 2 | 9 2 | 5 2 | 5 2 |
| | 7 2 | 7 2 | 7 2 | 7 2 | 7 2 | 7 2 | 5 2 | 5 2 |
| | 8 2 | 8 2 | 8 2 | 8 2 | 8 2 | 8 2 | 5 2 | 5 2 |
| | 7 2 | 8 2 | 9 2 | 9 2 | 8 2 | 8 2 | 5 2 | 5 2 |
| | 8 2 | 9 2 | 7 2 | 7 2 | 7 2 | 7 2 | 5 2 | 5 2 |
| | 9 2 | 7 2 | 8 2 | 8 2 | 9 2 | 9 2 | 5 2 | 5 2 |
| | 7 2 | 9 2 | 7 2 | 7 2 | 7 2 | 7 2 | 5 2 | 5 2 |
| | 9 2 | 7 2 | 9 2 | 9 2 | 9 2 | 9 2 | 5 2 | 5 2 |
| | 8 2 | 8 2 | 8 2 | 8 2 | 8 2 | 8 2 | 5 2 | 5 2 |
| | 8 2 | 8 2 | 7 2 | 7 2 | 8 2 | 8 2 | 5 2 | 5 2 |
| | 9 2 | 9 2 | 8 2 | 8 2 | 7 2 | 7 2 | 5 2 | 5 2 |
| | 9 2 | 7 2 | 9 2 | 9 2 | 9 2 | 9 2 | 5 2 | 5 2 |
| | 7 2 | 9 2 | 8 2 | 8 2 | 9 2 | 9 2 | 5 2 | 5 2 |

Continued on next column

**TABLE LXXVI– continued**

| TQC | ICSD | NoE | SF | | | | | |
|---|---|---|---|---|---|---|---|---|
| | 8 2 | 8 2 | 9 2 | 9 2 | 8 2 | 8 2 | 5 2 | 5 2 |
| | 7 2 | 7 2 | 7 2 | 7 2 | 7 2 | 7 2 | 5 2 | 5 2 |
| | 9 2 | 8 2 | 9 2 | 9 2 | 9 2 | 9 2 | 5 2 | 5 2 |
| | 8 2 | 9 2 | 8 2 | 8 2 | 8 2 | 8 2 | 5 2 | 5 2 |
| | 7 2 | 7 2 | 7 2 | 7 2 | 7 2 | 7 2 | 5 2 | 5 2 |
| | 8 2 | 9 2 | 9 2 | 9 2 | 8 2 | 8 2 | 5 2 | 5 2 |
| | 9 2 | 8 2 | 8 2 | 8 2 | 9 2 | 9 2 | 5 2 | 5 2 |
| | 7 2 | 7 2 | 7 2 | 7 2 | 7 2 | 7 2 | 5 2 | 5 2 |
| | 9 2 | 9 2 | 7 2 | 7 2 | 8 2 | 8 2 | 5 2 | 5 2 |
| | 9 2 | 9 2 | 8 2 | 8 2 | 7 2 | 7 2 | 5 2 | 5 2 |
| | 8 2 | 8 2 | 7 2 | 7 2 | 9 2 | 9 2 | 5 2 | 5 2 |
| | 8 2 | 7 2 | 9 2 | 9 2 | 7 2 | 7 2 | 5 2 | 5 2 |
| | 7 2 | 9 2 | 9 2 | 9 2 | 8 2 | 8 2 | 5 2 | 5 2 |
| | 8 2 | 8 2 | 7 2 | 7 2 | 9 2 | 9 2 | 5 2 | 5 2 |
| | 8 2 | 7 2 | 8 2 | 8 2 | 9 2 | 9 2 | 5 2 | 5 2 |
| | 7 2 | 9 2 | 9 2 | 9 2 | 9 2 | 9 2 | 5 2 | 5 2 |
| | 8 2 | 8 2 | 8 2 | 8 2 | 7 2 | 7 2 | 5 2 | 5 2 |
| | 9 2 | 7 2 | 7 2 | 7 2 | 8 2 | 8 2 | 5 2 | 5 2 |
| | 7 2 | 8 2 | 8 2 | 8 2 | 7 2 | 7 2 | 5 2 | 5 2 |
| | 8 2 | 7 2 | 9 2 | 9 2 | 9 2 | 9 2 | 5 2 | 5 2 |
| | 9 2 | 8 2 | 7 2 | 7 2 | 7 2 | 7 2 | 5 2 | 5 2 |
| | 7 2 | 8 2 | 8 2 | 8 2 | 8 2 | 8 2 | 5 2 | 5 2 |
| | 9 2 | 9 2 | 9 2 | 9 2 | 9 2 | 9 2 | 5 2 | 5 2 |
| | 8 2 | 7 2 | 7 2 | 7 2 | 7 2 | 7 2 | 5 2 | 5 2 |
| | 7 2 | 9 2 | 8 2 | 8 2 | 8 2 | 8 2 | 5 2 | 5 2 |
| | 9 2 | 7 2 | 9 2 | 9 2 | 7 2 | 7 2 | 5 2 | 5 2 |
| | 8 2 | 8 2 | 7 2 | 7 2 | 8 2 | 8 2 | 5 2 | 5 2 |
| | 9 2 | 9 2 | 8 2 | 8 2 | 9 2 | 9 2 | 5 2 | 5 2 |
| | 7 2 | 8 2 | 9 2 | 9 2 | 7 2 | 7 2 | 5 2 | 5 2 |
| | 7 2 | 9 2 | 7 2 | 7 2 | 9 2 | 9 2 | 5 2 | 5 2 |
| | 8 2 | 7 2 | 9 2 | 9 2 | 8 2 | 8 2 | 5 2 | 5 2 |
| | 9 2 | 9 2 | 8 2 | 8 2 | 9 2 | 9 2 | 5 2 | 5 2 |
| | 7 2 | 9 2 | 7 2 | 7 2 | 8 2 | 8 2 | 5 2 | 5 2 |
| | 8 2 | 8 2 | 9 2 | 9 2 | 7 2 | 7 2 | 5 2 | 5 2 |
| | 8 2 | 7 2 | 8 2 | 8 2 | 9 2 | 9 2 | 5 2 | 5 2 |
| | 9 2 | 8 2 | 7 2 | 7 2 | 8 2 | 8 2 | 5 2 | 5 2 |
| | 7 2 | 7 2 | 8 2 | 8 2 | 9 2 | 9 2 | 5 2 | 5 2 |
| | 8 2 | 9 2 | 7 2 | 7 2 | 7 2 | 7 2 | 5 2 | 5 2 |
| | 7 2 | 8 2 | 9 2 | 9 2 | 7 2 | 7 2 | 5 2 | 5 2 |
| | 9 2 | 7 2 | 9 2 | 9 2 | 9 2 | 9 2 | 5 2 | 5 2 |
| | 8 2 | 7 2 | 8 2 | 8 2 | 7 2 | 7 2 | 5 2 | 5 2 |
| | 7 2 | 8 2 | 9 2 | 9 2 | 8 2 | 8 2 | 5 2 | 5 2 |
| | 9 2 | 9 2 | 7 2 | 7 2 | 9 2 | 9 2 | 5 2 | 5 2 |
| | 7 2 | 8 2 | 8 2 | 8 2 | 8 2 | 8 2 | 5 2 | 5 2 |
| | 8 2 | 7 2 | 9 2 | 9 2 | 7 2 | 7 2 | 5 2 | 5 2 |
| | 9 2 | 9 2 | 8 2 | 8 2 | 8 2 | 8 2 | 5 2 | 5 2 |
| | 9 2 | 8 2 | 9 2 | 9 2 | 8 2 | 8 2 | 5 2 | 5 2 |
| | 7 2 | 9 2 | 7 2 | 7 2 | 9 2 | 9 2 | 5 2 | 5 2 |
| | 8 2 | 9 2 | 8 2 | 8 2 | 8 2 | 8 2 | 5 2 | 5 2 |
| | 8 2 | 7 2 | 7 2 | 7 2 | 7 2 | 7 2 | 5 2 | 5 2 |
| | 9 2 | 8 2 | 9 2 | 9 2 | 9 2 | 9 2 | 5 2 | 5 2 |
| | 7 2 | 7 2 | 7 2 | 7 2 | 7 2 | 7 2 | 5 2 | 5 2 |
| | 7 2 | 7 2 | 8 2 | 8 2 | 9 2 | 9 2 | 5 2 | 5 2 |
| | 9 2 | 8 2 | 9 2 | 9 2 | 7 2 | 7 2 | 5 2 | 5 2 |
| | 8 2 | 9 2 | 8 2 | 8 2 | 8 2 | 8 2 | 5 2 | 5 2 |
| | 8 2 | 9 2 | 7 2 | 7 2 | 8 2 | 8 2 | 5 2 | 5 2 |
| 189.1.22 | 418214 | 90 | Cu1La1Mg4 | | | | | |
| | -A | -GM | -HA | -H | -KA | -K | -L | -M |
| | 7 2 | 7 2 | 7 2 | 7 2 | 7 2 | 7 2 | 5 2 | 5 2 |
| | 8 2 | 8 2 | 8 2 | 8 2 | 8 2 | 8 2 | 5 2 | 5 2 |
| | 9 2 | 9 2 | 9 2 | 9 2 | 9 2 | 9 2 | 5 2 | 5 2 |

Continued on next column

**TABLE LXXVI– continued**

| TQC | ICSD | | NoE | SF | |
|---|---|---|---|---|---|
| | 7 2 | 7 2 | 8 2 8 2 | 8 2 | 8 2 5 2 5 2 |
| | 9 2 | 9 2 | 7 2 7 2 | 7 2 | 7 2 5 2 5 2 |
| | 8 2 | 8 2 | 9 2 9 2 | 9 2 | 9 2 5 2 5 2 |
| | 8 2 | 9 2 | 7 2 7 2 | 7 2 | 7 2 5 2 5 2 |
| | 7 2 | 8 2 | 9 2 9 2 | 8 2 | 8 2 5 2 5 2 |
| | 9 2 | 7 2 | 8 2 8 2 | 9 2 | 9 2 5 2 5 2 |
| | 9 2 | 9 2 | 7 2 7 2 | 8 2 | 8 2 5 2 5 2 |
| | 8 2 | 7 2 | 8 2 8 2 | 7 2 | 7 2 5 2 5 2 |
| | 7 2 | 8 2 | 9 2 9 2 | 9 2 | 9 2 5 2 5 2 |
| | 9 2 | 9 2 | 9 2 9 2 | 9 2 | 9 2 5 2 5 2 |
| | 8 2 | 9 2 | 8 2 8 2 | 7 2 | 7 2 5 2 5 2 |
| | 9 2 | 9 2 | 9 2 9 2 | 8 2 | 8 2 5 2 5 2 |
| | 7 2 | 7 2 | 7 2 7 2 | 8 2 | 8 2 5 2 5 2 |
| | 8 2 | 8 2 | 7 2 7 2 | 7 2 | 7 2 5 2 5 2 |
| | 9 2 | 7 2 | 8 2 8 2 | 9 2 | 9 2 5 2 5 2 |
| | 7 2 | 8 2 | 9 2 9 2 | 7 2 | 7 2 5 2 5 2 |
| | 8 2 | 9 2 | 7 2 7 2 | 9 2 | 9 2 5 2 5 2 |
| | 9 2 | 7 2 | 7 2 7 2 | 8 2 | 8 2 5 2 5 2 |
| | 8 2 | 9 2 | 8 2 8 2 | 9 2 | 9 2 5 2 5 2 |
| | 7 2 | 9 2 | 8 2 8 2 | 7 2 | 7 2 5 2 5 2 |
| | 9 2 | 7 2 | 7 2 7 2 | 8 2 | 8 2 5 2 5 2 |
| | 7 2 | 7 2 | 9 2 9 2 | 9 2 | 9 2 5 2 5 2 |
| | 7 2 | 8 2 | 8 2 8 2 | 7 2 | 7 2 5 2 5 2 |
| | 9 2 | 9 2 | 9 2 9 2 | 8 2 | 8 2 5 2 5 2 |
| | 8 2 | 7 2 | 8 2 8 2 | 7 2 | 7 2 5 2 5 2 |
| | 7 2 | 9 2 | 9 2 9 2 | 9 2 | 9 2 5 2 5 2 |
| | 9 2 | 8 2 | 7 2 7 2 | 8 2 | 8 2 5 2 5 2 |
| | 9 2 | 7 2 | 9 2 9 2 | 9 2 | 9 2 5 2 5 2 |
| | 8 2 | 7 2 | 7 2 7 2 | 8 2 | 8 2 5 2 5 2 |
| | 9 2 | 9 2 | 8 2 8 2 | 7 2 | 7 2 5 2 5 2 |
| | 9 2 | 8 2 | 8 2 8 2 | 8 2 | 8 2 5 2 5 2 |
| | 7 2 | 9 2 | 7 2 7 2 | 7 2 | 7 2 5 2 5 2 |
| | 7 2 | 7 2 | 9 2 9 2 | 9 2 | 9 2 5 2 5 2 |
| | 8 2 | 8 2 | 7 2 7 2 | 9 2 | 9 2 5 2 5 2 |
| | 8 2 | 9 2 | 8 2 8 2 | 7 2 | 7 2 5 2 5 2 |
| | 7 2 | 8 2 | 9 2 9 2 | 8 2 | 8 2 5 2 5 2 |
| | 9 2 | 7 2 | 7 2 7 2 | 8 2 | 8 2 5 2 5 2 |
| | 7 2 | 9 2 | 9 2 9 2 | 7 2 | 7 2 5 2 5 2 |
| | 7 2 | 8 2 | 8 2 8 2 | 8 2 | 8 2 5 2 5 2 |
| | 8 2 | 8 2 | 8 2 8 2 | 9 2 | 9 2 5 2 5 2 |
| | 9 2 | 8 2 | 9 2 9 2 | 7 2 | 7 2 5 2 5 2 |
| | 8 2 | 7 2 | 7 2 7 2 | 9 2 | 9 2 5 2 5 2 |
| 189.1.23 | 428473 | | 54 | Cu1Ge1Sc1 | |
| | -A | -GM | -HA -H | -KA | -K -L -M |
| | 8 2 | 9 2 | 9 2 9 2 | 9 2 | 9 2 5 2 5 2 |
| | 9 2 | 9 2 | 7 2 7 2 | 7 2 | 7 2 5 2 5 2 |
| | 8 2 | 9 2 | 9 2 9 2 | 8 2 | 8 2 5 2 5 2 |
| | 8 2 | 8 2 | 8 2 8 2 | 8 2 | 8 2 5 2 5 2 |
| | 9 2 | 7 2 | 8 2 8 2 | 7 2 | 7 2 5 2 5 2 |
| | 9 2 | 8 2 | 7 2 7 2 | 9 2 | 9 2 5 2 5 2 |
| | 7 2 | 7 2 | 8 2 8 2 | 8 2 | 8 2 5 2 5 2 |
| | 7 2 | 8 2 | 9 2 9 2 | 8 2 | 8 2 5 2 5 2 |
| | 8 2 | 9 2 | 7 2 7 2 | 9 2 | 9 2 5 2 5 2 |
| | 7 2 | 8 2 | 7 2 7 2 | 7 2 | 7 2 5 2 5 2 |
| | 9 2 | 7 2 | 8 2 8 2 | 7 2 | 7 2 5 2 5 2 |
| | 9 2 | 7 2 | 8 2 8 2 | 9 2 | 9 2 5 2 5 2 |
| | 9 2 | 8 2 | 9 2 9 2 | 9 2 | 9 2 5 2 5 2 |
| | 8 2 | 9 2 | 8 2 8 2 | 7 2 | 7 2 5 2 5 2 |
| | 8 2 | 8 2 | 7 2 7 2 | 9 2 | 9 2 5 2 5 2 |
| | 7 2 | 7 2 | 9 2 9 2 | 8 2 | 8 2 5 2 5 2 |
| | 7 2 | 9 2 | 9 2 9 2 | 7 2 | 7 2 5 2 5 2 |
| | 8 2 | 9 2 | 7 2 7 2 | 8 2 | 8 2 5 2 5 2 |
| | 9 2 | 7 2 | 9 2 9 2 | 9 2 | 9 2 5 2 5 2 |
| | 7 2 | 8 2 | 8 2 8 2 | 8 2 | 8 2 5 2 5 2 |
| | 9 2 | 7 2 | 7 2 7 2 | 7 2 | 7 2 5 2 5 2 |
| | 8 2 | 9 2 | 7 2 7 2 | 7 2 | 7 2 5 2 5 2 |
| | 8 2 | 8 2 | 9 2 9 2 | 9 2 | 9 2 5 2 5 2 |
| | 7 2 | 8 2 | 8 2 8 2 | 8 2 | 8 2 5 2 5 2 |
| | 9 2 | 9 2 | 8 2 8 2 | 8 2 | 8 2 5 2 5 2 |
| | 7 2 | 8 2 | 7 2 7 2 | 7 2 | 7 2 5 2 5 2 |
| | 9 2 | 7 2 | 8 2 8 2 | 9 2 | 9 2 5 2 5 2 |
| 189.1.24 | 98233 | | 108 | Ce2Ge3Zn6 | |
| | -A | -GM | -HA -H | -KA | -K -L -M |
| | 9 2 | 9 2 | 7 2 7 2 | 7 2 | 7 2 5 2 5 2 |
| | 9 2 | 9 2 | 8 2 8 2 | 8 2 | 8 2 5 2 5 2 |
| | 8 2 | 8 2 | 7 2 7 2 | 7 2 | 7 2 5 2 5 2 |
| | 8 2 | 8 2 | 9 2 9 2 | 9 2 | 9 2 5 2 5 2 |
| | 8 2 | 7 2 | 9 2 9 2 | 9 2 | 9 2 5 2 5 2 |
| | 8 2 | 7 2 | 7 2 7 2 | 8 2 | 8 2 5 2 5 2 |
| | 7 2 | 8 2 | 8 2 8 2 | 7 2 | 7 2 5 2 5 2 |
| | 7 2 | 8 2 | 9 2 9 2 | 9 2 | 9 2 5 2 5 2 |
| | 8 2 | 9 2 | 8 2 8 2 | 9 2 | 9 2 5 2 5 2 |
| | 9 2 | 8 2 | 9 2 9 2 | 7 2 | 7 2 5 2 5 2 |
| | 7 2 | 7 2 | 7 2 7 2 | 8 2 | 8 2 5 2 5 2 |
| | 9 2 | 9 2 | 9 2 9 2 | 9 2 | 9 2 5 2 5 2 |
| | 8 2 | 8 2 | 8 2 8 2 | 7 2 | 7 2 5 2 5 2 |
| | 7 2 | 7 2 | 7 2 7 2 | 8 2 | 8 2 5 2 5 2 |
| | 8 2 | 9 2 | 7 2 7 2 | 8 2 | 8 2 5 2 5 2 |
| | 7 2 | 8 2 | 9 2 9 2 | 7 2 | 7 2 5 2 5 2 |
| | 9 2 | 7 2 | 8 2 8 2 | 9 2 | 9 2 5 2 5 2 |
| | 9 2 | 9 2 | 9 2 9 2 | 9 2 | 9 2 5 2 5 2 |
| | 8 2 | 7 2 | 8 2 8 2 | 9 2 | 9 2 5 2 5 2 |
| | 7 2 | 8 2 | 7 2 7 2 | 8 2 | 8 2 5 2 5 2 |
| | 7 2 | 9 2 | 7 2 7 2 | 8 2 | 8 2 5 2 5 2 |
| | 9 2 | 7 2 | 9 2 9 2 | 7 2 | 7 2 5 2 5 2 |
| | 8 2 | 8 2 | 8 2 8 2 | 7 2 | 7 2 5 2 5 2 |
| | 9 2 | 8 2 | 9 2 9 2 | 8 2 | 8 2 5 2 5 2 |
| | 8 2 | 9 2 | 8 2 8 2 | 9 2 | 9 2 5 2 5 2 |
| | 9 2 | 7 2 | 9 2 9 2 | 7 2 | 7 2 5 2 5 2 |
| | 7 2 | 9 2 | 8 2 8 2 | 7 2 | 7 2 5 2 5 2 |
| | 8 2 | 9 2 | 7 2 7 2 | 9 2 | 9 2 5 2 5 2 |
| | 7 2 | 8 2 | 7 2 7 2 | 8 2 | 8 2 5 2 5 2 |
| | 9 2 | 7 2 | 9 2 9 2 | 8 2 | 8 2 5 2 5 2 |
| | 7 2 | 8 2 | 7 2 7 2 | 9 2 | 9 2 5 2 5 2 |
| | 9 2 | 9 2 | 8 2 8 2 | 7 2 | 7 2 5 2 5 2 |
| | 8 2 | 7 2 | 7 2 7 2 | 9 2 | 9 2 5 2 5 2 |
| | 7 2 | 8 2 | 9 2 9 2 | 8 2 | 8 2 5 2 5 2 |
| | 8 2 | 9 2 | 8 2 8 2 | 7 2 | 7 2 5 2 5 2 |
| | 8 2 | 7 2 | 8 2 8 2 | 9 2 | 9 2 5 2 5 2 |
| | 8 2 | 9 2 | 7 2 7 2 | 8 2 | 8 2 5 2 5 2 |
| | 7 2 | 7 2 | 8 2 8 2 | 7 2 | 7 2 5 2 5 2 |
| | 9 2 | 8 2 | 9 2 9 2 | 7 2 | 7 2 5 2 5 2 |
| | 9 2 | 8 2 | 9 2 9 2 | 9 2 | 9 2 5 2 5 2 |
| | 7 2 | 7 2 | 7 2 7 2 | 8 2 | 8 2 5 2 5 2 |
| | 9 2 | 9 2 | 9 2 9 2 | 9 2 | 9 2 5 2 5 2 |
| | 7 2 | 8 2 | 8 2 8 2 | 7 2 | 7 2 5 2 5 2 |
| | 9 2 | 7 2 | 7 2 7 2 | 8 2 | 8 2 5 2 5 2 |
| | 8 2 | 9 2 | 8 2 8 2 | 7 2 | 7 2 5 2 5 2 |
| | 7 2 | 8 2 | 9 2 9 2 | 8 2 | 8 2 5 2 5 2 |
| | 8 2 | 7 2 | 7 2 7 2 | 9 2 | 9 2 5 2 5 2 |
| | 9 2 | 8 2 | 9 2 9 2 | 8 2 | 8 2 5 2 5 2 |
| | 8 2 | 9 2 | 8 2 8 2 | 9 2 | 9 2 5 2 5 2 |

Continued on next column

**TABLE LXXVI– continued**

| TQC | ICSD | NoE | SF | | |
|---|---|---|---|---|---|
| | 8 2 | 7 2 | 7 2 7 2 | 7 2 | 7 2 5 2 5 2 |
| | 7 2 | 9 2 | 9 2 9 2 | 8 2 | 8 2 5 2 5 2 |
| | 9 2 | 9 2 | 7 2 7 2 | 7 2 | 7 2 5 2 5 2 |
| | 7 2 | 8 2 | 8 2 8 2 | 9 2 | 9 2 5 2 5 2 |
| | 9 2 | 7 2 | 8 2 8 2 | 8 2 | 8 2 5 2 5 2 |
| 189.1.25 | 415492 | 78 | Ce1Pt1Sn1 | | |
| | -A | -GM | -HA -H | -KA | -K -L -M |
| | 9 2 | 9 2 | 7 2 7 2 | 7 2 | 7 2 5 2 5 2 |
| | 7 2 | 7 2 | 8 2 8 2 | 8 2 | 8 2 5 2 5 2 |
| | 8 2 | 8 2 | 9 2 9 2 | 9 2 | 9 2 5 2 5 2 |
| | 7 2 | 9 2 | 8 2 8 2 | 8 2 | 8 2 5 2 5 2 |
| | 9 2 | 7 2 | 9 2 9 2 | 9 2 | 9 2 5 2 5 2 |
| | 8 2 | 8 2 | 7 2 7 2 | 7 2 | 7 2 5 2 5 2 |
| | 8 2 | 9 2 | 9 2 9 2 | 7 2 | 7 2 5 2 5 2 |
| | 9 2 | 8 2 | 7 2 7 2 | 9 2 | 9 2 5 2 5 2 |
| | 7 2 | 7 2 | 8 2 8 2 | 8 2 | 8 2 5 2 5 2 |
| | 8 2 | 9 2 | 7 2 7 2 | 8 2 | 8 2 5 2 5 2 |
| | 9 2 | 8 2 | 9 2 9 2 | 7 2 | 7 2 5 2 5 2 |
| | 7 2 | 7 2 | 8 2 8 2 | 9 2 | 9 2 5 2 5 2 |
| | 8 2 | 9 2 | 8 2 8 2 | 9 2 | 9 2 5 2 5 2 |
| | 9 2 | 7 2 | 9 2 9 2 | 7 2 | 7 2 5 2 5 2 |
| | 7 2 | 8 2 | 7 2 7 2 | 8 2 | 8 2 5 2 5 2 |
| | 9 2 | 9 2 | 9 2 9 2 | 9 2 | 9 2 5 2 5 2 |
| | 8 2 | 9 2 | 9 2 9 2 | 8 2 | 8 2 5 2 5 2 |
| | 7 2 | 7 2 | 7 2 7 2 | 7 2 | 7 2 5 2 5 2 |
| | 8 2 | 9 2 | 7 2 7 2 | 9 2 | 9 2 5 2 5 2 |
| | 7 2 | 7 2 | 8 2 8 2 | 7 2 | 7 2 5 2 5 2 |
| | 8 2 | 9 2 | 9 2 9 2 | 8 2 | 8 2 5 2 5 2 |
| | 9 2 | 7 2 | 7 2 7 2 | 7 2 | 7 2 5 2 5 2 |
| | 7 2 | 8 2 | 8 2 8 2 | 9 2 | 9 2 5 2 5 2 |
| | 8 2 | 9 2 | 9 2 9 2 | 8 2 | 8 2 5 2 5 2 |
| | 8 2 | 7 2 | 8 2 8 2 | 7 2 | 7 2 5 2 5 2 |
| | 7 2 | 8 2 | 9 2 9 2 | 8 2 | 8 2 5 2 5 2 |
| | 9 2 | 9 2 | 7 2 7 2 | 9 2 | 9 2 5 2 5 2 |
| | 9 2 | 7 2 | 9 2 9 2 | 7 2 | 7 2 5 2 5 2 |
| | 7 2 | 9 2 | 8 2 8 2 | 9 2 | 9 2 5 2 5 2 |
| | 8 2 | 9 2 | 7 2 7 2 | 8 2 | 8 2 5 2 5 2 |
| | 7 2 | 7 2 | 9 2 9 2 | 9 2 | 9 2 5 2 5 2 |
| | 8 2 | 7 2 | 7 2 7 2 | 8 2 | 8 2 5 2 5 2 |
| | 9 2 | 8 2 | 9 2 9 2 | 7 2 | 7 2 5 2 5 2 |
| | 9 2 | 8 2 | 8 2 8 2 | 8 2 | 8 2 5 2 5 2 |
| | 7 2 | 8 2 | 7 2 7 2 | 7 2 | 7 2 5 2 5 2 |
| | 8 2 | 8 2 | 8 2 8 2 | 8 2 | 8 2 5 2 5 2 |
| | 9 2 | 9 2 | 8 2 8 2 | 9 2 | 9 2 5 2 5 2 |
| | 7 2 | 7 2 | 7 2 7 2 | 9 2 | 9 2 5 2 5 2 |
| | 9 2 | 9 2 | 8 2 8 2 | 7 2 | 7 2 5 2 5 2 |
| 189.1.26 | 416969 | 78 | Ce1Pd1Sn1 | | |
| | -A | -GM | -HA -H | -KA | -K -L -M |
| | 9 2 | 9 2 | 7 2 7 2 | 7 2 | 7 2 5 2 5 2 |
| | 7 2 | 7 2 | 8 2 8 2 | 8 2 | 8 2 5 2 5 2 |
| | 8 2 | 8 2 | 9 2 9 2 | 9 2 | 9 2 5 2 5 2 |
| | 7 2 | 7 2 | 8 2 8 2 | 8 2 | 8 2 5 2 5 2 |
| | 9 2 | 9 2 | 9 2 9 2 | 9 2 | 9 2 5 2 5 2 |
| | 8 2 | 8 2 | 7 2 7 2 | 7 2 | 7 2 5 2 5 2 |
| | 8 2 | 8 2 | 7 2 7 2 | 7 2 | 7 2 5 2 5 2 |
| | 9 2 | 9 2 | 9 2 9 2 | 9 2 | 9 2 5 2 5 2 |
| | 7 2 | 7 2 | 8 2 8 2 | 8 2 | 8 2 5 2 5 2 |
| | 8 2 | 9 2 | 7 2 7 2 | 8 2 | 8 2 5 2 5 2 |
| | 9 2 | 8 2 | 9 2 9 2 | 7 2 | 7 2 5 2 5 2 |
| | 7 2 | 7 2 | 8 2 8 2 | 9 2 | 9 2 5 2 5 2 |
| | 8 2 | 9 2 | 8 2 8 2 | 9 2 | 9 2 5 2 5 2 |

Continued on next column

**TABLE LXXVI– continued**

| TQC | ICSD | NoE | SF | | |
|---|---|---|---|---|---|
| | 9 2 | 7 2 | 9 2 9 2 | 7 2 | 7 2 5 2 5 2 |
| | 7 2 | 8 2 | 7 2 7 2 | 8 2 | 8 2 5 2 5 2 |
| | 9 2 | 9 2 | 9 2 9 2 | 9 2 | 9 2 5 2 5 2 |
| | 7 2 | 9 2 | 7 2 7 2 | 8 2 | 8 2 5 2 5 2 |
| | 8 2 | 9 2 | 9 2 9 2 | 7 2 | 7 2 5 2 5 2 |
| | 8 2 | 7 2 | 7 2 7 2 | 9 2 | 9 2 5 2 5 2 |
| | 7 2 | 7 2 | 8 2 8 2 | 8 2 | 8 2 5 2 5 2 |
| | 9 2 | 8 2 | 8 2 8 2 | 7 2 | 7 2 5 2 5 2 |
| | 8 2 | 7 2 | 7 2 7 2 | 7 2 | 7 2 5 2 5 2 |
| | 7 2 | 9 2 | 9 2 9 2 | 9 2 | 9 2 5 2 5 2 |
| | 8 2 | 9 2 | 8 2 8 2 | 8 2 | 8 2 5 2 5 2 |
| | 7 2 | 8 2 | 9 2 9 2 | 7 2 | 7 2 5 2 5 2 |
| | 9 2 | 7 2 | 9 2 9 2 | 8 2 | 8 2 5 2 5 2 |
| | 8 2 | 7 2 | 8 2 8 2 | 9 2 | 9 2 5 2 5 2 |
| | 7 2 | 9 2 | 7 2 7 2 | 7 2 | 7 2 5 2 5 2 |
| | 9 2 | 9 2 | 9 2 9 2 | 9 2 | 9 2 5 2 5 2 |
| | 8 2 | 7 2 | 9 2 9 2 | 8 2 | 8 2 5 2 5 2 |
| | 7 2 | 8 2 | 7 2 7 2 | 9 2 | 9 2 5 2 5 2 |
| | 8 2 | 7 2 | 7 2 7 2 | 8 2 | 8 2 5 2 5 2 |
| | 9 2 | 8 2 | 8 2 8 2 | 7 2 | 7 2 5 2 5 2 |
| | 9 2 | 9 2 | 9 2 9 2 | 8 2 | 8 2 5 2 5 2 |
| | 7 2 | 8 2 | 7 2 7 2 | 7 2 | 7 2 5 2 5 2 |
| | 8 2 | 7 2 | 8 2 8 2 | 8 2 | 8 2 5 2 5 2 |
| | 7 2 | 8 2 | 8 2 8 2 | 9 2 | 9 2 5 2 5 2 |
| | 9 2 | 9 2 | 8 2 8 2 | 7 2 | 7 2 5 2 5 2 |
| | 9 2 | 9 2 | 7 2 7 2 | 9 2 | 9 2 5 2 5 2 |
| 189.1.27 | 100002 | 72 | Ce1In1Ir1 | | |
| | -A | -GM | -HA -H | -KA | -K -L -M |
| | 9 2 | 9 2 | 7 2 7 2 | 7 2 | 7 2 5 2 5 2 |
| | 7 2 | 7 2 | 8 2 8 2 | 8 2 | 8 2 5 2 5 2 |
| | 8 2 | 8 2 | 9 2 9 2 | 9 2 | 9 2 5 2 5 2 |
| | 7 2 | 9 2 | 8 2 8 2 | 8 2 | 8 2 5 2 5 2 |
| | 9 2 | 7 2 | 9 2 9 2 | 9 2 | 9 2 5 2 5 2 |
| | 8 2 | 8 2 | 7 2 7 2 | 7 2 | 7 2 5 2 5 2 |
| | 9 2 | 9 2 | 7 2 7 2 | 7 2 | 7 2 5 2 5 2 |
| | 8 2 | 8 2 | 9 2 9 2 | 9 2 | 9 2 5 2 5 2 |
| | 7 2 | 7 2 | 8 2 8 2 | 8 2 | 8 2 5 2 5 2 |
| | 8 2 | 9 2 | 7 2 7 2 | 8 2 | 8 2 5 2 5 2 |
| | 9 2 | 8 2 | 9 2 9 2 | 7 2 | 7 2 5 2 5 2 |
| | 7 2 | 7 2 | 8 2 8 2 | 9 2 | 9 2 5 2 5 2 |
| | 8 2 | 9 2 | 8 2 8 2 | 9 2 | 9 2 5 2 5 2 |
| | 9 2 | 7 2 | 9 2 9 2 | 7 2 | 7 2 5 2 5 2 |
| | 7 2 | 8 2 | 7 2 7 2 | 8 2 | 8 2 5 2 5 2 |
| | 9 2 | 9 2 | 9 2 9 2 | 8 2 | 8 2 5 2 5 2 |
| | 8 2 | 9 2 | 9 2 9 2 | 7 2 | 7 2 5 2 5 2 |
| | 7 2 | 7 2 | 7 2 7 2 | 9 2 | 9 2 5 2 5 2 |
| | 8 2 | 9 2 | 7 2 7 2 | 8 2 | 8 2 5 2 5 2 |
| | 8 2 | 7 2 | 8 2 8 2 | 7 2 | 7 2 5 2 5 2 |
| | 7 2 | 9 2 | 8 2 8 2 | 9 2 | 9 2 5 2 5 2 |
| | 8 2 | 8 2 | 9 2 9 2 | 7 2 | 7 2 5 2 5 2 |
| | 9 2 | 7 2 | 9 2 9 2 | 9 2 | 9 2 5 2 5 2 |
| | 7 2 | 9 2 | 7 2 7 2 | 8 2 | 8 2 5 2 5 2 |
| | 7 2 | 8 2 | 8 2 8 2 | 7 2 | 7 2 5 2 5 2 |
| | 9 2 | 9 2 | 7 2 7 2 | 8 2 | 8 2 5 2 5 2 |
| | 8 2 | 9 2 | 9 2 9 2 | 9 2 | 9 2 5 2 5 2 |
| | 7 2 | 7 2 | 8 2 8 2 | 9 2 | 9 2 5 2 5 2 |
| | 9 2 | 7 2 | 9 2 9 2 | 7 2 | 7 2 5 2 5 2 |
| | 8 2 | 9 2 | 7 2 7 2 | 8 2 | 8 2 5 2 5 2 |
| | 8 2 | 7 2 | 9 2 9 2 | 9 2 | 9 2 5 2 5 2 |
| | 7 2 | 7 2 | 7 2 7 2 | 8 2 | 8 2 5 2 5 2 |
| | 9 2 | 8 2 | 8 2 8 2 | 7 2 | 7 2 5 2 5 2 |

Continued on next column

**TABLE LXXVI– continued**

| TQC | ICSD | NoE | SF | | | | | |
|---|---|---|---|---|---|---|---|---|
| | 9 2 | 8 2 | 9 2 | 9 2 | 7 2 | 7 2 | 5 2 | 5 2 |
| | 7 2 | 8 2 | 7 2 | 7 2 | 8 2 | 8 2 | 5 2 | 5 2 |
| | 8 2 | 9 2 | 8 2 | 8 2 | 9 2 | 9 2 | 5 2 | 5 2 |
| 189.1.28 | 428851 | 102 | Cd1Ce1Pt1 | | | | | |
| | -A | -GM | -HA | -H | -KA | -K | -L | -M |
| | 9 2 | 9 2 | 7 2 | 7 2 | 7 2 | 7 2 | 5 2 | 5 2 |
| | 7 2 | 7 2 | 8 2 | 8 2 | 8 2 | 8 2 | 5 2 | 5 2 |
| | 8 2 | 8 2 | 9 2 | 9 2 | 9 2 | 9 2 | 5 2 | 5 2 |
| | 7 2 | 9 2 | 8 2 | 8 2 | 8 2 | 8 2 | 5 2 | 5 2 |
| | 9 2 | 7 2 | 9 2 | 9 2 | 7 2 | 7 2 | 5 2 | 5 2 |
| | 8 2 | 8 2 | 7 2 | 7 2 | 9 2 | 9 2 | 5 2 | 5 2 |
| | 9 2 | 9 2 | 7 2 | 7 2 | 7 2 | 7 2 | 5 2 | 5 2 |
| | 8 2 | 8 2 | 8 2 | 8 2 | 8 2 | 8 2 | 5 2 | 5 2 |
| | 7 2 | 7 2 | 9 2 | 9 2 | 9 2 | 9 2 | 5 2 | 5 2 |
| | 8 2 | 9 2 | 7 2 | 7 2 | 8 2 | 8 2 | 5 2 | 5 2 |
| | 9 2 | 8 2 | 9 2 | 9 2 | 7 2 | 7 2 | 5 2 | 5 2 |
| | 7 2 | 7 2 | 8 2 | 8 2 | 9 2 | 9 2 | 5 2 | 5 2 |
| | 9 2 | 9 2 | 9 2 | 9 2 | 9 2 | 9 2 | 5 2 | 5 2 |
| | 8 2 | 7 2 | 7 2 | 7 2 | 7 2 | 7 2 | 5 2 | 5 2 |
| | 7 2 | 8 2 | 9 2 | 9 2 | 8 2 | 8 2 | 5 2 | 5 2 |
| | 9 2 | 9 2 | 8 2 | 8 2 | 7 2 | 7 2 | 5 2 | 5 2 |
| | 7 2 | 8 2 | 8 2 | 8 2 | 9 2 | 9 2 | 5 2 | 5 2 |
| | 8 2 | 7 2 | 7 2 | 7 2 | 8 2 | 8 2 | 5 2 | 5 2 |
| | 8 2 | 9 2 | 9 2 | 9 2 | 9 2 | 9 2 | 5 2 | 5 2 |
| | 9 2 | 7 2 | 8 2 | 8 2 | 7 2 | 7 2 | 5 2 | 5 2 |
| | 7 2 | 9 2 | 7 2 | 7 2 | 8 2 | 8 2 | 5 2 | 5 2 |
| | 8 2 | 8 2 | 9 2 | 9 2 | 9 2 | 9 2 | 5 2 | 5 2 |
| | 9 2 | 8 2 | 7 2 | 7 2 | 8 2 | 8 2 | 5 2 | 5 2 |
| | 7 2 | 7 2 | 8 2 | 8 2 | 7 2 | 7 2 | 5 2 | 5 2 |
| | 8 2 | 9 2 | 9 2 | 9 2 | 8 2 | 8 2 | 5 2 | 5 2 |
| | 9 2 | 8 2 | 8 2 | 8 2 | 9 2 | 9 2 | 5 2 | 5 2 |
| | 7 2 | 7 2 | 7 2 | 7 2 | 7 2 | 7 2 | 5 2 | 5 2 |
| | 7 2 | 9 2 | 8 2 | 8 2 | 9 2 | 9 2 | 5 2 | 5 2 |
| | 9 2 | 7 2 | 9 2 | 9 2 | 7 2 | 7 2 | 5 2 | 5 2 |
| | 8 2 | 9 2 | 7 2 | 7 2 | 8 2 | 8 2 | 5 2 | 5 2 |
| | 9 2 | 9 2 | 9 2 | 9 2 | 8 2 | 8 2 | 5 2 | 5 2 |
| | 8 2 | 8 2 | 7 2 | 7 2 | 7 2 | 7 2 | 5 2 | 5 2 |
| | 7 2 | 7 2 | 9 2 | 9 2 | 9 2 | 9 2 | 5 2 | 5 2 |
| | 8 2 | 9 2 | 8 2 | 8 2 | 8 2 | 8 2 | 5 2 | 5 2 |
| | 7 2 | 7 2 | 7 2 | 7 2 | 7 2 | 7 2 | 5 2 | 5 2 |
| | 8 2 | 9 2 | 9 2 | 9 2 | 9 2 | 9 2 | 5 2 | 5 2 |
| | 7 2 | 9 2 | 7 2 | 7 2 | 7 2 | 7 2 | 5 2 | 5 2 |
| | 9 2 | 7 2 | 9 2 | 9 2 | 8 2 | 8 2 | 5 2 | 5 2 |
| | 7 2 | 8 2 | 8 2 | 8 2 | 7 2 | 7 2 | 5 2 | 5 2 |
| | 8 2 | 9 2 | 9 2 | 9 2 | 9 2 | 9 2 | 5 2 | 5 2 |
| | 8 2 | 7 2 | 7 2 | 7 2 | 8 2 | 8 2 | 5 2 | 5 2 |
| | 9 2 | 8 2 | 8 2 | 8 2 | 9 2 | 9 2 | 5 2 | 5 2 |
| | 7 2 | 9 2 | 9 2 | 9 2 | 8 2 | 8 2 | 5 2 | 5 2 |
| | 8 2 | 9 2 | 8 2 | 8 2 | 9 2 | 9 2 | 5 2 | 5 2 |
| | 8 2 | 7 2 | 7 2 | 7 2 | 7 2 | 7 2 | 5 2 | 5 2 |
| | 9 2 | 7 2 | 9 2 | 9 2 | 9 2 | 9 2 | 5 2 | 5 2 |
| | 9 2 | 7 2 | 7 2 | 7 2 | 8 2 | 8 2 | 5 2 | 5 2 |
| | 7 2 | 8 2 | 8 2 | 8 2 | 7 2 | 7 2 | 5 2 | 5 2 |
| | 9 2 | 8 2 | 9 2 | 9 2 | 8 2 | 8 2 | 5 2 | 5 2 |
| | 7 2 | 8 2 | 7 2 | 7 2 | 7 2 | 7 2 | 5 2 | 5 2 |
| | 8 2 | 9 2 | 8 2 | 8 2 | 9 2 | 9 2 | 5 2 | 5 2 |
| 189.1.29 | 51528 | 74 | C2Rb2 | | | | | |
| | -A | -GM | -HA | -H | -KA | -K | -L | -M |
| | 8 2 | 9 2 | 7 2 | 7 2 | 7 2 | 7 2 | 5 2 | 5 2 |
| | 7 2 | 7 2 | 9 2 | 9 2 | 8 2 | 8 2 | 5 2 | 5 2 |
| | 9 2 | 8 2 | 8 2 | 8 2 | 9 2 | 9 2 | 5 2 | 5 2 |

Continued on next column

**TABLE LXXVI– continued**

| TQC | ICSD | NoE | SF | | | | | |
|---|---|---|---|---|---|---|---|---|
| | 9 2 | 9 2 | 9 2 | 9 2 | 9 2 | 9 2 | 5 2 | 5 2 |
| | 7 2 | 7 2 | 7 2 | 7 2 | 7 2 | 7 2 | 5 2 | 5 2 |
| | 8 2 | 8 2 | 8 2 | 8 2 | 8 2 | 8 2 | 5 2 | 5 2 |
| | 8 2 | 8 2 | 9 2 | 9 2 | 7 2 | 7 2 | 5 2 | 5 2 |
| | 7 2 | 7 2 | 7 2 | 7 2 | 9 2 | 9 2 | 5 2 | 5 2 |
| | 8 2 | 8 2 | 8 2 | 8 2 | 8 2 | 8 2 | 5 2 | 5 2 |
| | 7 2 | 7 2 | 9 2 | 9 2 | 9 2 | 9 2 | 5 2 | 5 2 |
| | 9 2 | 9 2 | 7 2 | 7 2 | 7 2 | 7 2 | 5 2 | 5 2 |
| | 9 2 | 9 2 | 8 2 | 8 2 | 8 2 | 8 2 | 5 2 | 5 2 |
| | 7 2 | 8 2 | 9 2 | 9 2 | 9 2 | 9 2 | 5 2 | 5 2 |
| | 8 2 | 7 2 | 8 2 | 8 2 | 7 2 | 7 2 | 5 2 | 5 2 |
| | 9 2 | 9 2 | 7 2 | 7 2 | 8 2 | 8 2 | 5 2 | 5 2 |
| | 7 2 | 7 2 | 9 2 | 9 2 | 9 2 | 9 2 | 5 2 | 5 2 |
| | 8 2 | 9 2 | 7 2 | 7 2 | 7 2 | 7 2 | 5 2 | 5 2 |
| | 9 2 | 8 2 | 8 2 | 8 2 | 8 2 | 8 2 | 5 2 | 5 2 |
| | 8 2 | 8 2 | 7 2 | 7 2 | 9 2 | 9 2 | 5 2 | 5 2 |
| | 9 2 | 7 2 | 9 2 | 9 2 | 8 2 | 8 2 | 5 2 | 5 2 |
| | 7 2 | 9 2 | 7 2 | 7 2 | 7 2 | 7 2 | 5 2 | 5 2 |
| | 8 2 | 8 2 | 8 2 | 8 2 | 7 2 | 7 2 | 5 2 | 5 2 |
| | 9 2 | 9 2 | 9 2 | 9 2 | 8 2 | 8 2 | 5 2 | 5 2 |
| | 7 2 | 7 2 | 8 2 | 8 2 | 9 2 | 9 2 | 5 2 | 5 2 |
| | 8 2 | 8 2 | 9 2 | 9 2 | 9 2 | 9 2 | 5 2 | 5 2 |
| | 9 2 | 9 2 | 8 2 | 8 2 | 8 2 | 8 2 | 5 2 | 5 2 |
| | 7 2 | 7 2 | 7 2 | 7 2 | 7 2 | 7 2 | 5 2 | 5 2 |
| | 8 2 | 9 2 | 9 2 | 9 2 | 9 2 | 9 2 | 5 2 | 5 2 |
| | 7 2 | 8 2 | 8 2 | 8 2 | 8 2 | 8 2 | 5 2 | 5 2 |
| | 9 2 | 7 2 | 7 2 | 7 2 | 7 2 | 7 2 | 5 2 | 5 2 |
| | 8 2 | 9 2 | 8 2 | 8 2 | 9 2 | 9 2 | 5 2 | 5 2 |
| | 8 2 | 8 2 | 7 2 | 7 2 | 7 2 | 7 2 | 5 2 | 5 2 |
| | 8 2 | 8 2 | 9 2 | 9 2 | 9 2 | 9 2 | 5 2 | 5 2 |
| | 9 2 | 9 2 | 9 2 | 9 2 | 7 2 | 7 2 | 5 2 | 5 2 |
| | 7 2 | 7 2 | 7 2 | 7 2 | 8 2 | 8 2 | 5 2 | 5 2 |
| | 9 2 | 8 2 | 9 2 | 9 2 | 9 2 | 9 2 | 5 2 | 5 2 |
| | 9 2 | 8 2 | 8 2 | 8 2 | 8 2 | 8 2 | 5 2 | 5 2 |
| 189.1.30 | 413810 | 48 | Ba1Bi1Na1 | | | | | |
| | -A | -GM | -HA | -H | -KA | -K | -L | -M |
| | 9 2 | 9 2 | 7 2 | 7 2 | 7 2 | 7 2 | 5 2 | 5 2 |
| | 7 2 | 7 2 | 8 2 | 8 2 | 8 2 | 8 2 | 5 2 | 5 2 |
| | 8 2 | 8 2 | 9 2 | 9 2 | 9 2 | 9 2 | 5 2 | 5 2 |
| | 9 2 | 9 2 | 8 2 | 8 2 | 8 2 | 8 2 | 5 2 | 5 2 |
| | 7 2 | 7 2 | 9 2 | 9 2 | 7 2 | 7 2 | 5 2 | 5 2 |
| | 8 2 | 8 2 | 7 2 | 7 2 | 9 2 | 9 2 | 5 2 | 5 2 |
| | 8 2 | 9 2 | 9 2 | 9 2 | 7 2 | 7 2 | 5 2 | 5 2 |
| | 9 2 | 8 2 | 7 2 | 7 2 | 9 2 | 9 2 | 5 2 | 5 2 |
| | 7 2 | 7 2 | 8 2 | 8 2 | 8 2 | 8 2 | 5 2 | 5 2 |
| | 9 2 | 9 2 | 7 2 | 7 2 | 7 2 | 7 2 | 5 2 | 5 2 |
| | 8 2 | 8 2 | 8 2 | 8 2 | 8 2 | 8 2 | 5 2 | 5 2 |
| | 7 2 | 7 2 | 9 2 | 9 2 | 9 2 | 9 2 | 5 2 | 5 2 |
| | 9 2 | 9 2 | 9 2 | 9 2 | 9 2 | 9 2 | 5 2 | 5 2 |
| | 8 2 | 9 2 | 7 2 | 7 2 | 7 2 | 7 2 | 5 2 | 5 2 |
| | 8 2 | 9 2 | 9 2 | 9 2 | 8 2 | 8 2 | 5 2 | 5 2 |
| | 9 2 | 8 2 | 8 2 | 8 2 | 8 2 | 8 2 | 5 2 | 5 2 |
| | 8 2 | 8 2 | 8 2 | 8 2 | 9 2 | 9 2 | 5 2 | 5 2 |
| | 9 2 | 7 2 | 7 2 | 7 2 | 7 2 | 7 2 | 5 2 | 5 2 |
| | 7 2 | 8 2 | 8 2 | 8 2 | 7 2 | 7 2 | 5 2 | 5 2 |
| | 9 2 | 7 2 | 7 2 | 7 2 | 9 2 | 9 2 | 5 2 | 5 2 |
| | 8 2 | 8 2 | 8 2 | 8 2 | 8 2 | 8 2 | 5 2 | 5 2 |
| | 7 2 | 8 2 | 9 2 | 9 2 | 7 2 | 7 2 | 5 2 | 5 2 |
| | 9 2 | 7 2 | 7 2 | 7 2 | 8 2 | 8 2 | 5 2 | 5 2 |
| | 7 2 | 9 2 | 8 2 | 8 2 | 9 2 | 9 2 | 5 2 | 5 2 |
| 189.1.31 | 261317 | 100 | Au1Ge2In3Yb3 | | | | | |

Continued on next column

**TABLE LXXVI– continued**

| TQC | ICSD | NoE | SF | | | | | |
|---|---|---|---|---|---|---|---|---|
| | -A | -GM | -HA | -H | -KA | -K | -L | -M |
| | 8 2 | 9 2 | 7 2 | 7 2 | 7 2 | 7 2 | 5 2 | 5 2 |
| | 7 2 | 7 2 | 9 2 | 9 2 | 8 2 | 8 2 | 5 2 | 5 2 |
| | 9 2 | 8 2 | 8 2 | 8 2 | 9 2 | 9 2 | 5 2 | 5 2 |
| | 7 2 | 7 2 | 9 2 | 9 2 | 8 2 | 8 2 | 5 2 | 5 2 |
| | 8 2 | 9 2 | 7 2 | 7 2 | 7 2 | 7 2 | 5 2 | 5 2 |
| | 9 2 | 8 2 | 8 2 | 8 2 | 9 2 | 9 2 | 5 2 | 5 2 |
| | 8 2 | 9 2 | 7 2 | 7 2 | 7 2 | 7 2 | 5 2 | 5 2 |
| | 9 2 | 8 2 | 8 2 | 8 2 | 9 2 | 9 2 | 5 2 | 5 2 |
| | 7 2 | 7 2 | 9 2 | 9 2 | 8 2 | 8 2 | 5 2 | 5 2 |
| | 8 2 | 9 2 | 7 2 | 7 2 | 7 2 | 7 2 | 5 2 | 5 2 |
| | 9 2 | 8 2 | 9 2 | 9 2 | 8 2 | 8 2 | 5 2 | 5 2 |
| | 7 2 | 7 2 | 8 2 | 8 2 | 9 2 | 9 2 | 5 2 | 5 2 |
| | 9 2 | 9 2 | 8 2 | 8 2 | 7 2 | 7 2 | 5 2 | 5 2 |
| | 9 2 | 9 2 | 7 2 | 7 2 | 8 2 | 8 2 | 5 2 | 5 2 |
| | 8 2 | 9 2 | 9 2 | 9 2 | 9 2 | 9 2 | 5 2 | 5 2 |
| | 7 2 | 7 2 | 8 2 | 8 2 | 9 2 | 9 2 | 5 2 | 5 2 |
| | 8 2 | 8 2 | 7 2 | 7 2 | 7 2 | 7 2 | 5 2 | 5 2 |
| | 9 2 | 9 2 | 7 2 | 7 2 | 8 2 | 8 2 | 5 2 | 5 2 |
| | 7 2 | 7 2 | 8 2 | 8 2 | 7 2 | 7 2 | 5 2 | 5 2 |
| | 8 2 | 9 2 | 8 2 | 8 2 | 9 2 | 9 2 | 5 2 | 5 2 |
| | 9 2 | 7 2 | 9 2 | 9 2 | 7 2 | 7 2 | 5 2 | 5 2 |
| | 8 2 | 8 2 | 7 2 | 7 2 | 8 2 | 8 2 | 5 2 | 5 2 |
| | 7 2 | 7 2 | 8 2 | 8 2 | 9 2 | 9 2 | 5 2 | 5 2 |
| | 8 2 | 8 2 | 9 2 | 9 2 | 9 2 | 9 2 | 5 2 | 5 2 |
| | 8 2 | 8 2 | 7 2 | 7 2 | 7 2 | 7 2 | 5 2 | 5 2 |
| | 7 2 | 8 2 | 9 2 | 9 2 | 8 2 | 8 2 | 5 2 | 5 2 |
| | 8 2 | 7 2 | 9 2 | 9 2 | 9 2 | 9 2 | 5 2 | 5 2 |
| | 7 2 | 9 2 | 9 2 | 9 2 | 7 2 | 7 2 | 5 2 | 5 2 |
| | 9 2 | 7 2 | 7 2 | 7 2 | 8 2 | 8 2 | 5 2 | 5 2 |
| | 7 2 | 9 2 | 8 2 | 8 2 | 7 2 | 7 2 | 5 2 | 5 2 |
| | 9 2 | 8 2 | 7 2 | 7 2 | 9 2 | 9 2 | 5 2 | 5 2 |
| | 8 2 | 9 2 | 8 2 | 8 2 | 8 2 | 8 2 | 5 2 | 5 2 |
| | 8 2 | 7 2 | 9 2 | 9 2 | 9 2 | 9 2 | 5 2 | 5 2 |
| | 9 2 | 8 2 | 7 2 | 7 2 | 7 2 | 7 2 | 5 2 | 5 2 |
| | 7 2 | 8 2 | 8 2 | 8 2 | 8 2 | 8 2 | 5 2 | 5 2 |
| | 8 2 | 9 2 | 9 2 | 9 2 | 9 2 | 9 2 | 5 2 | 5 2 |
| | 9 2 | 8 2 | 9 2 | 9 2 | 7 2 | 7 2 | 5 2 | 5 2 |
| | 7 2 | 7 2 | 7 2 | 7 2 | 8 2 | 8 2 | 5 2 | 5 2 |
| | 8 2 | 9 2 | 7 2 | 7 2 | 7 2 | 7 2 | 5 2 | 5 2 |
| | 9 2 | 7 2 | 8 2 | 8 2 | 8 2 | 8 2 | 5 2 | 5 2 |
| | 7 2 | 8 2 | 8 2 | 8 2 | 9 2 | 9 2 | 5 2 | 5 2 |
| | 7 2 | 9 2 | 9 2 | 9 2 | 8 2 | 8 2 | 5 2 | 5 2 |
| | 8 2 | 7 2 | 7 2 | 7 2 | 7 2 | 7 2 | 5 2 | 5 2 |
| | 9 2 | 9 2 | 8 2 | 8 2 | 9 2 | 9 2 | 5 2 | 5 2 |
| | 7 2 | 7 2 | 7 2 | 7 2 | 8 2 | 8 2 | 5 2 | 5 2 |
| | 8 2 | 8 2 | 9 2 | 9 2 | 9 2 | 9 2 | 5 2 | 5 2 |
| | 9 2 | 9 2 | 9 2 | 9 2 | 7 2 | 7 2 | 5 2 | 5 2 |
| | 9 2 | 8 2 | 8 2 | 8 2 | 7 2 | 7 2 | 5 2 | 5 2 |
| | 8 2 | 8 2 | 7 2 | 7 2 | 8 2 | 8 2 | 5 2 | 5 2 |
| | 7 2 | 7 2 | 9 2 | 9 2 | 9 2 | 9 2 | 5 2 | 5 2 |
| 189.1.32 | 104011 | 96 | Au1Gd1In1 | | | | | |
| | -A | -GM | -HA | -H | -KA | -K | -L | -M |
| | 9 2 | 9 2 | 7 2 | 7 2 | 7 2 | 7 2 | 5 2 | 5 2 |
| | 7 2 | 7 2 | 8 2 | 8 2 | 8 2 | 8 2 | 5 2 | 5 2 |
| | 8 2 | 8 2 | 9 2 | 9 2 | 9 2 | 9 2 | 5 2 | 5 2 |
| | 7 2 | 9 2 | 8 2 | 8 2 | 8 2 | 8 2 | 5 2 | 5 2 |
| | 9 2 | 7 2 | 9 2 | 9 2 | 7 2 | 7 2 | 5 2 | 5 2 |
| | 8 2 | 8 2 | 7 2 | 7 2 | 9 2 | 9 2 | 5 2 | 5 2 |
| | 8 2 | 9 2 | 9 2 | 9 2 | 7 2 | 7 2 | 5 2 | 5 2 |
| | 9 2 | 8 2 | 7 2 | 7 2 | 9 2 | 9 2 | 5 2 | 5 2 |

Continued on next column

**TABLE LXXVI– continued**

| TQC | ICSD | NoE | SF | | | | | |
|---|---|---|---|---|---|---|---|---|
| | 7 2 | 7 2 | 8 2 | 8 2 | 8 2 | 8 2 | 5 2 | 5 2 |
| | 8 2 | 9 2 | 7 2 | 7 2 | 8 2 | 8 2 | 5 2 | 5 2 |
| | 9 2 | 8 2 | 9 2 | 9 2 | 7 2 | 7 2 | 5 2 | 5 2 |
| | 7 2 | 7 2 | 8 2 | 8 2 | 9 2 | 9 2 | 5 2 | 5 2 |
| | 8 2 | 9 2 | 9 2 | 9 2 | 9 2 | 9 2 | 5 2 | 5 2 |
| | 9 2 | 7 2 | 8 2 | 8 2 | 7 2 | 7 2 | 5 2 | 5 2 |
| | 7 2 | 9 2 | 7 2 | 7 2 | 8 2 | 8 2 | 5 2 | 5 2 |
| | 7 2 | 8 2 | 9 2 | 9 2 | 9 2 | 9 2 | 5 2 | 5 2 |
| | 8 2 | 9 2 | 7 2 | 7 2 | 8 2 | 8 2 | 5 2 | 5 2 |
| | 7 2 | 7 2 | 9 2 | 9 2 | 7 2 | 7 2 | 5 2 | 5 2 |
| | 9 2 | 7 2 | 8 2 | 8 2 | 7 2 | 7 2 | 5 2 | 5 2 |
| | 8 2 | 9 2 | 7 2 | 7 2 | 9 2 | 9 2 | 5 2 | 5 2 |
| | 7 2 | 9 2 | 9 2 | 9 2 | 8 2 | 8 2 | 5 2 | 5 2 |
| | 9 2 | 7 2 | 7 2 | 7 2 | 7 2 | 7 2 | 5 2 | 5 2 |
| | 8 2 | 9 2 | 9 2 | 9 2 | 9 2 | 9 2 | 5 2 | 5 2 |
| | 7 2 | 8 2 | 8 2 | 8 2 | 8 2 | 8 2 | 5 2 | 5 2 |
| | 8 2 | 8 2 | 9 2 | 9 2 | 7 2 | 7 2 | 5 2 | 5 2 |
| | 9 2 | 7 2 | 8 2 | 8 2 | 8 2 | 8 2 | 5 2 | 5 2 |
| | 7 2 | 9 2 | 8 2 | 8 2 | 9 2 | 9 2 | 5 2 | 5 2 |
| | 8 2 | 9 2 | 7 2 | 7 2 | 7 2 | 7 2 | 5 2 | 5 2 |
| | 9 2 | 7 2 | 9 2 | 9 2 | 8 2 | 8 2 | 5 2 | 5 2 |
| | 8 2 | 9 2 | 7 2 | 7 2 | 9 2 | 9 2 | 5 2 | 5 2 |
| | 8 2 | 7 2 | 9 2 | 9 2 | 8 2 | 8 2 | 5 2 | 5 2 |
| | 7 2 | 8 2 | 7 2 | 7 2 | 7 2 | 7 2 | 5 2 | 5 2 |
| | 9 2 | 8 2 | 8 2 | 8 2 | 9 2 | 9 2 | 5 2 | 5 2 |
| | 9 2 | 7 2 | 9 2 | 9 2 | 7 2 | 7 2 | 5 2 | 5 2 |
| | 7 2 | 9 2 | 7 2 | 7 2 | 8 2 | 8 2 | 5 2 | 5 2 |
| | 8 2 | 7 2 | 8 2 | 8 2 | 9 2 | 9 2 | 5 2 | 5 2 |
| | 9 2 | 8 2 | 8 2 | 8 2 | 7 2 | 7 2 | 5 2 | 5 2 |
| | 7 2 | 8 2 | 7 2 | 7 2 | 8 2 | 8 2 | 5 2 | 5 2 |
| | 9 2 | 9 2 | 8 2 | 8 2 | 9 2 | 9 2 | 5 2 | 5 2 |
| | 7 2 | 7 2 | 9 2 | 9 2 | 9 2 | 9 2 | 5 2 | 5 2 |
| | 8 2 | 8 2 | 7 2 | 7 2 | 7 2 | 7 2 | 5 2 | 5 2 |
| | 9 2 | 9 2 | 9 2 | 9 2 | 8 2 | 8 2 | 5 2 | 5 2 |
| | 8 2 | 7 2 | 8 2 | 8 2 | 7 2 | 7 2 | 5 2 | 5 2 |
| | 8 2 | 8 2 | 9 2 | 9 2 | 9 2 | 9 2 | 5 2 | 5 2 |
| | 7 2 | 9 2 | 7 2 | 7 2 | 8 2 | 8 2 | 5 2 | 5 2 |
| | 9 2 | 8 2 | 8 2 | 8 2 | 8 2 | 8 2 | 5 2 | 5 2 |
| | 7 2 | 7 2 | 7 2 | 7 2 | 7 2 | 7 2 | 5 2 | 5 2 |
| | 9 2 | 9 2 | 8 2 | 8 2 | 9 2 | 9 2 | 5 2 | 5 2 |
| 189.1.33 | 192086 | 48 | Au1Ga1Mg1 | | | | | |
| | -A | -GM | -HA | -H | -KA | -K | -L | -M |
| | 8 2 | 9 2 | 9 2 | 9 2 | 9 2 | 9 2 | 5 2 | 5 2 |
| | 9 2 | 7 2 | 7 2 | 7 2 | 8 2 | 8 2 | 5 2 | 5 2 |
| | 7 2 | 8 2 | 8 2 | 8 2 | 7 2 | 7 2 | 5 2 | 5 2 |
| | 7 2 | 9 2 | 9 2 | 9 2 | 7 2 | 7 2 | 5 2 | 5 2 |
| | 8 2 | 7 2 | 9 2 | 9 2 | 8 2 | 8 2 | 5 2 | 5 2 |
| | 9 2 | 9 2 | 7 2 | 7 2 | 9 2 | 9 2 | 5 2 | 5 2 |
| | 7 2 | 7 2 | 7 2 | 7 2 | 9 2 | 9 2 | 5 2 | 5 2 |
| | 8 2 | 9 2 | 8 2 | 8 2 | 7 2 | 7 2 | 5 2 | 5 2 |
| | 7 2 | 9 2 | 9 2 | 9 2 | 8 2 | 8 2 | 5 2 | 5 2 |
| | 9 2 | 8 2 | 7 2 | 7 2 | 7 2 | 7 2 | 5 2 | 5 2 |
| | 7 2 | 9 2 | 9 2 | 9 2 | 8 2 | 8 2 | 5 2 | 5 2 |
| | 7 2 | 8 2 | 8 2 | 8 2 | 9 2 | 9 2 | 5 2 | 5 2 |
| | 8 2 | 7 2 | 8 2 | 8 2 | 9 2 | 9 2 | 5 2 | 5 2 |
| | 9 2 | 9 2 | 9 2 | 9 2 | 7 2 | 7 2 | 5 2 | 5 2 |
| | 8 2 | 7 2 | 8 2 | 8 2 | 8 2 | 8 2 | 5 2 | 5 2 |
| | 8 2 | 9 2 | 7 2 | 7 2 | 7 2 | 7 2 | 5 2 | 5 2 |
| | 9 2 | 9 2 | 7 2 | 7 2 | 8 2 | 8 2 | 5 2 | 5 2 |
| | 9 2 | 7 2 | 9 2 | 9 2 | 9 2 | 9 2 | 5 2 | 5 2 |
| | 7 2 | 7 2 | 8 2 | 8 2 | 8 2 | 8 2 | 5 2 | 5 2 |

Continued on next column

**TABLE LXXVI– continued**

| TQC | ICSD | NoE | | SF | | | | |
|---|---|---|---|---|---|---|---|---|
| | 8 2 | 7 2 | 9 2 | 9 2 | 7 2 | 7 2 | 5 2 | 5 2 |
| | 8 2 | 8 2 | 7 2 | 7 2 | 9 2 | 9 2 | 5 2 | 5 2 |
| | 8 2 | 8 2 | 8 2 | 8 2 | 7 2 | 7 2 | 5 2 | 5 2 |
| | 9 2 | 8 2 | 7 2 | 7 2 | 8 2 | 8 2 | 5 2 | 5 2 |
| | 7 2 | 9 2 | 9 2 | 9 2 | 9 2 | 9 2 | 5 2 | 5 2 |
| 189.1.34 | 411543 | 102 | Au1Cd1La1 | | | | | |
| | -A | -GM | -HA | -H | -KA | -K | -L | -M |
| | 9 2 | 9 2 | 7 2 | 7 2 | 7 2 | 7 2 | 5 2 | 5 2 |
| | 7 2 | 7 2 | 8 2 | 8 2 | 8 2 | 8 2 | 5 2 | 5 2 |
| | 8 2 | 8 2 | 9 2 | 9 2 | 9 2 | 9 2 | 5 2 | 5 2 |
| | 7 2 | 9 2 | 8 2 | 8 2 | 8 2 | 8 2 | 5 2 | 5 2 |
| | 9 2 | 7 2 | 9 2 | 9 2 | 7 2 | 7 2 | 5 2 | 5 2 |
| | 8 2 | 8 2 | 7 2 | 7 2 | 9 2 | 9 2 | 5 2 | 5 2 |
| | 9 2 | 9 2 | 7 2 | 7 2 | 7 2 | 7 2 | 5 2 | 5 2 |
| | 8 2 | 8 2 | 8 2 | 8 2 | 8 2 | 8 2 | 5 2 | 5 2 |
| | 7 2 | 7 2 | 9 2 | 9 2 | 9 2 | 9 2 | 5 2 | 5 2 |
| | 8 2 | 9 2 | 7 2 | 7 2 | 7 2 | 7 2 | 5 2 | 5 2 |
| | 9 2 | 8 2 | 9 2 | 9 2 | 8 2 | 8 2 | 5 2 | 5 2 |
| | 7 2 | 7 2 | 8 2 | 8 2 | 9 2 | 9 2 | 5 2 | 5 2 |
| | 9 2 | 9 2 | 9 2 | 9 2 | 9 2 | 9 2 | 5 2 | 5 2 |
| | 7 2 | 7 2 | 7 2 | 7 2 | 7 2 | 7 2 | 5 2 | 5 2 |
| | 8 2 | 8 2 | 9 2 | 9 2 | 8 2 | 8 2 | 5 2 | 5 2 |
| | 7 2 | 9 2 | 8 2 | 8 2 | 7 2 | 7 2 | 5 2 | 5 2 |
| | 9 2 | 8 2 | 8 2 | 8 2 | 9 2 | 9 2 | 5 2 | 5 2 |
| | 8 2 | 7 2 | 7 2 | 7 2 | 8 2 | 8 2 | 5 2 | 5 2 |
| | 9 2 | 9 2 | 9 2 | 9 2 | 9 2 | 9 2 | 5 2 | 5 2 |
| | 8 2 | 7 2 | 8 2 | 8 2 | 7 2 | 7 2 | 5 2 | 5 2 |
| | 7 2 | 9 2 | 7 2 | 7 2 | 8 2 | 8 2 | 5 2 | 5 2 |
| | 8 2 | 8 2 | 9 2 | 9 2 | 9 2 | 9 2 | 5 2 | 5 2 |
| | 9 2 | 8 2 | 7 2 | 7 2 | 8 2 | 8 2 | 5 2 | 5 2 |
| | 7 2 | 7 2 | 8 2 | 8 2 | 7 2 | 7 2 | 5 2 | 5 2 |
| | 8 2 | 9 2 | 9 2 | 9 2 | 9 2 | 9 2 | 5 2 | 5 2 |
| | 9 2 | 8 2 | 8 2 | 8 2 | 8 2 | 8 2 | 5 2 | 5 2 |
| | 7 2 | 7 2 | 7 2 | 7 2 | 7 2 | 7 2 | 5 2 | 5 2 |
| | 7 2 | 9 2 | 9 2 | 9 2 | 9 2 | 9 2 | 5 2 | 5 2 |
| | 9 2 | 9 2 | 8 2 | 8 2 | 8 2 | 8 2 | 5 2 | 5 2 |
| | 8 2 | 7 2 | 7 2 | 7 2 | 7 2 | 7 2 | 5 2 | 5 2 |
| | 8 2 | 9 2 | 9 2 | 9 2 | 8 2 | 8 2 | 5 2 | 5 2 |
| | 7 2 | 7 2 | 7 2 | 7 2 | 9 2 | 9 2 | 5 2 | 5 2 |
| | 7 2 | 9 2 | 9 2 | 9 2 | 7 2 | 7 2 | 5 2 | 5 2 |
| | 8 2 | 7 2 | 7 2 | 7 2 | 8 2 | 8 2 | 5 2 | 5 2 |
| | 9 2 | 8 2 | 8 2 | 8 2 | 7 2 | 7 2 | 5 2 | 5 2 |
| | 7 2 | 9 2 | 9 2 | 9 2 | 9 2 | 9 2 | 5 2 | 5 2 |
| | 9 2 | 9 2 | 7 2 | 7 2 | 7 2 | 7 2 | 5 2 | 5 2 |
| | 8 2 | 7 2 | 8 2 | 8 2 | 8 2 | 8 2 | 5 2 | 5 2 |
| | 7 2 | 9 2 | 9 2 | 9 2 | 9 2 | 9 2 | 5 2 | 5 2 |
| | 8 2 | 8 2 | 9 2 | 9 2 | 7 2 | 7 2 | 5 2 | 5 2 |
| | 7 2 | 7 2 | 7 2 | 7 2 | 8 2 | 8 2 | 5 2 | 5 2 |
| | 9 2 | 8 2 | 8 2 | 8 2 | 9 2 | 9 2 | 5 2 | 5 2 |
| | 8 2 | 9 2 | 9 2 | 9 2 | 8 2 | 8 2 | 5 2 | 5 2 |
| | 8 2 | 7 2 | 7 2 | 7 2 | 9 2 | 9 2 | 5 2 | 5 2 |
| | 9 2 | 9 2 | 8 2 | 8 2 | 7 2 | 7 2 | 5 2 | 5 2 |
| | 8 2 | 7 2 | 9 2 | 9 2 | 8 2 | 8 2 | 5 2 | 5 2 |
| | 9 2 | 8 2 | 7 2 | 7 2 | 7 2 | 7 2 | 5 2 | 5 2 |
| | 7 2 | 8 2 | 8 2 | 8 2 | 9 2 | 9 2 | 5 2 | 5 2 |
| | 9 2 | 7 2 | 9 2 | 9 2 | 9 2 | 9 2 | 5 2 | 5 2 |
| | 8 2 | 8 2 | 7 2 | 7 2 | 7 2 | 7 2 | 5 2 | 5 2 |
| | 7 2 | 8 2 | 8 2 | 8 2 | 8 2 | 8 2 | 5 2 | 5 2 |
| 189.1.35 | 810 | 132 | As2Eu2 | | | | | |
| | -A | -GM | -HA | -H | -KA | -K | -L | -M |
| | 8 2 | 9 2 | 7 2 | 7 2 | 7 2 | 7 2 | 5 2 | 5 2 |

Continued on next column

**TABLE LXXVI– continued**

| TQC | ICSD | NoE | | SF | | | | |
|---|---|---|---|---|---|---|---|---|
| | 7 2 | 7 2 | 8 2 | 8 2 | 8 2 | 8 2 | 5 2 | 5 2 |
| | 9 2 | 8 2 | 9 2 | 9 2 | 9 2 | 9 2 | 5 2 | 5 2 |
| | 9 2 | 7 2 | 7 2 | 7 2 | 7 2 | 7 2 | 5 2 | 5 2 |
| | 7 2 | 8 2 | 8 2 | 8 2 | 8 2 | 8 2 | 5 2 | 5 2 |
| | 8 2 | 9 2 | 9 2 | 9 2 | 9 2 | 9 2 | 5 2 | 5 2 |
| | 8 2 | 9 2 | 8 2 | 8 2 | 8 2 | 8 2 | 5 2 | 5 2 |
| | 7 2 | 8 2 | 9 2 | 9 2 | 9 2 | 9 2 | 5 2 | 5 2 |
| | 9 2 | 7 2 | 7 2 | 7 2 | 7 2 | 7 2 | 5 2 | 5 2 |
| | 9 2 | 9 2 | 7 2 | 7 2 | 7 2 | 7 2 | 5 2 | 5 2 |
| | 7 2 | 7 2 | 9 2 | 9 2 | 9 2 | 9 2 | 5 2 | 5 2 |
| | 8 2 | 8 2 | 8 2 | 8 2 | 8 2 | 8 2 | 5 2 | 5 2 |
| | 8 2 | 9 2 | 8 2 | 8 2 | 7 2 | 7 2 | 5 2 | 5 2 |
| | 9 2 | 8 2 | 7 2 | 7 2 | 9 2 | 9 2 | 5 2 | 5 2 |
| | 8 2 | 9 2 | 8 2 | 8 2 | 8 2 | 8 2 | 5 2 | 5 2 |
| | 9 2 | 8 2 | 9 2 | 9 2 | 8 2 | 8 2 | 5 2 | 5 2 |
| | 7 2 | 7 2 | 7 2 | 7 2 | 9 2 | 9 2 | 5 2 | 5 2 |
| | 9 2 | 8 2 | 9 2 | 9 2 | 7 2 | 7 2 | 5 2 | 5 2 |
| | 7 2 | 7 2 | 9 2 | 9 2 | 8 2 | 8 2 | 5 2 | 5 2 |
| | 8 2 | 9 2 | 7 2 | 7 2 | 9 2 | 9 2 | 5 2 | 5 2 |
| | 7 2 | 7 2 | 8 2 | 8 2 | 7 2 | 7 2 | 5 2 | 5 2 |
| | 7 2 | 9 2 | 9 2 | 9 2 | 8 2 | 8 2 | 5 2 | 5 2 |
| | 8 2 | 7 2 | 7 2 | 7 2 | 9 2 | 9 2 | 5 2 | 5 2 |
| | 9 2 | 8 2 | 8 2 | 8 2 | 7 2 | 7 2 | 5 2 | 5 2 |
| | 8 2 | 9 2 | 8 2 | 8 2 | 9 2 | 9 2 | 5 2 | 5 2 |
| | 9 2 | 9 2 | 7 2 | 7 2 | 7 2 | 7 2 | 5 2 | 5 2 |
| | 9 2 | 9 2 | 8 2 | 8 2 | 8 2 | 8 2 | 5 2 | 5 2 |
| | 8 2 | 8 2 | 7 2 | 7 2 | 7 2 | 7 2 | 5 2 | 5 2 |
| | 9 2 | 8 2 | 9 2 | 9 2 | 9 2 | 9 2 | 5 2 | 5 2 |
| | 8 2 | 8 2 | 9 2 | 9 2 | 8 2 | 8 2 | 5 2 | 5 2 |
| | 8 2 | 9 2 | 9 2 | 9 2 | 9 2 | 9 2 | 5 2 | 5 2 |
| | 9 2 | 9 2 | 9 2 | 9 2 | 7 2 | 7 2 | 5 2 | 5 2 |
| | 7 2 | 9 2 | 7 2 | 7 2 | 8 2 | 8 2 | 5 2 | 5 2 |
| | 7 2 | 7 2 | 7 2 | 7 2 | 8 2 | 8 2 | 5 2 | 5 2 |
| | 8 2 | 8 2 | 8 2 | 8 2 | 7 2 | 7 2 | 5 2 | 5 2 |
| | 8 2 | 7 2 | 8 2 | 8 2 | 9 2 | 9 2 | 5 2 | 5 2 |
| | 7 2 | 8 2 | 8 2 | 8 2 | 8 2 | 8 2 | 5 2 | 5 2 |
| | 9 2 | 8 2 | 7 2 | 7 2 | 7 2 | 7 2 | 5 2 | 5 2 |
| | 9 2 | 7 2 | 9 2 | 9 2 | 9 2 | 9 2 | 5 2 | 5 2 |
| | 9 2 | 9 2 | 8 2 | 8 2 | 7 2 | 7 2 | 5 2 | 5 2 |
| | 7 2 | 9 2 | 7 2 | 7 2 | 9 2 | 9 2 | 5 2 | 5 2 |
| | 7 2 | 7 2 | 9 2 | 9 2 | 8 2 | 8 2 | 5 2 | 5 2 |
| | 8 2 | 7 2 | 8 2 | 8 2 | 8 2 | 8 2 | 5 2 | 5 2 |
| | 9 2 | 9 2 | 7 2 | 7 2 | 7 2 | 7 2 | 5 2 | 5 2 |
| | 7 2 | 7 2 | 8 2 | 8 2 | 9 2 | 9 2 | 5 2 | 5 2 |
| | 8 2 | 9 2 | 9 2 | 9 2 | 7 2 | 7 2 | 5 2 | 5 2 |
| | 9 2 | 8 2 | 7 2 | 7 2 | 8 2 | 8 2 | 5 2 | 5 2 |
| | 8 2 | 8 2 | 9 2 | 9 2 | 9 2 | 9 2 | 5 2 | 5 2 |
| | 8 2 | 7 2 | 7 2 | 7 2 | 7 2 | 7 2 | 5 2 | 5 2 |
| | 9 2 | 7 2 | 8 2 | 8 2 | 8 2 | 8 2 | 5 2 | 5 2 |
| | 7 2 | 8 2 | 9 2 | 9 2 | 9 2 | 9 2 | 5 2 | 5 2 |
| | 7 2 | 9 2 | 7 2 | 7 2 | 7 2 | 7 2 | 5 2 | 5 2 |
| | 8 2 | 9 2 | 9 2 | 9 2 | 9 2 | 9 2 | 5 2 | 5 2 |
| | 9 2 | 7 2 | 8 2 | 8 2 | 8 2 | 8 2 | 5 2 | 5 2 |
| | 8 2 | 8 2 | 7 2 | 7 2 | 8 2 | 8 2 | 5 2 | 5 2 |
| | 9 2 | 7 2 | 8 2 | 8 2 | 7 2 | 7 2 | 5 2 | 5 2 |
| | 7 2 | 8 2 | 9 2 | 9 2 | 9 2 | 9 2 | 5 2 | 5 2 |
| | 9 2 | 7 2 | 8 2 | 8 2 | 7 2 | 7 2 | 5 2 | 5 2 |
| | 7 2 | 8 2 | 7 2 | 7 2 | 8 2 | 8 2 | 5 2 | 5 2 |
| | 8 2 | 9 2 | 9 2 | 9 2 | 9 2 | 9 2 | 5 2 | 5 2 |
| | 7 2 | 9 2 | 7 2 | 7 2 | 8 2 | 8 2 | 5 2 | 5 2 |
| | 7 2 | 9 2 | 8 2 | 8 2 | 7 2 | 7 2 | 5 2 | 5 2 |

Continued on next column

**TABLE LXXVI– continued**

| TQC | ICSD | NoE | SF | | | | | |
|---|---|---|---|---|---|---|---|---|
| | 9 2 | 7 2 | 8 2 | 8 2 | 9 2 | 9 2 | 5 2 | 5 2 |
| | 8 2 | 9 2 | 9 2 | 9 2 | 9 2 | 9 2 | 5 2 | 5 2 |
| | 9 2 | 8 2 | 7 2 | 7 2 | 7 2 | 7 2 | 5 2 | 5 2 |
| | 8 2 | 8 2 | 9 2 | 9 2 | 8 2 | 8 2 | 5 2 | 5 2 |
| 189.1.36 | 193686 | 80 | Al3Co1Gd3Ge2 | | | | | |
| | -A | -GM | -HA | -H | -KA | -K | -L | -M |
| | 9 2 | 9 2 | 7 2 | 7 2 | 7 2 | 7 2 | 5 2 | 5 2 |
| | 7 2 | 7 2 | 8 2 | 8 2 | 8 2 | 8 2 | 5 2 | 5 2 |
| | 8 2 | 8 2 | 9 2 | 9 2 | 9 2 | 9 2 | 5 2 | 5 2 |
| | 9 2 | 9 2 | 8 2 | 8 2 | 8 2 | 8 2 | 5 2 | 5 2 |
| | 7 2 | 7 2 | 7 2 | 7 2 | 7 2 | 7 2 | 5 2 | 5 2 |
| | 8 2 | 8 2 | 9 2 | 9 2 | 9 2 | 9 2 | 5 2 | 5 2 |
| | 9 2 | 9 2 | 7 2 | 7 2 | 7 2 | 7 2 | 5 2 | 5 2 |
| | 8 2 | 8 2 | 8 2 | 8 2 | 8 2 | 8 2 | 5 2 | 5 2 |
| | 7 2 | 7 2 | 9 2 | 9 2 | 9 2 | 9 2 | 5 2 | 5 2 |
| | 8 2 | 8 2 | 7 2 | 7 2 | 7 2 | 7 2 | 5 2 | 5 2 |
| | 9 2 | 9 2 | 8 2 | 8 2 | 8 2 | 8 2 | 5 2 | 5 2 |
| | 7 2 | 7 2 | 9 2 | 9 2 | 9 2 | 9 2 | 5 2 | 5 2 |
| | 8 2 | 9 2 | 7 2 | 7 2 | 7 2 | 7 2 | 5 2 | 5 2 |
| | 8 2 | 9 2 | 9 2 | 9 2 | 8 2 | 8 2 | 5 2 | 5 2 |
| | 8 2 | 9 2 | 8 2 | 8 2 | 9 2 | 9 2 | 5 2 | 5 2 |
| | 9 2 | 7 2 | 7 2 | 7 2 | 8 2 | 8 2 | 5 2 | 5 2 |
| | 7 2 | 8 2 | 9 2 | 9 2 | 7 2 | 7 2 | 5 2 | 5 2 |
| | 9 2 | 8 2 | 9 2 | 9 2 | 9 2 | 9 2 | 5 2 | 5 2 |
| | 9 2 | 7 2 | 7 2 | 7 2 | 9 2 | 9 2 | 5 2 | 5 2 |
| | 7 2 | 9 2 | 8 2 | 8 2 | 7 2 | 7 2 | 5 2 | 5 2 |
| | 9 2 | 9 2 | 8 2 | 8 2 | 8 2 | 8 2 | 5 2 | 5 2 |
| | 7 2 | 7 2 | 9 2 | 9 2 | 7 2 | 7 2 | 5 2 | 5 2 |
| | 8 2 | 8 2 | 7 2 | 7 2 | 9 2 | 9 2 | 5 2 | 5 2 |
| | 9 2 | 7 2 | 8 2 | 8 2 | 9 2 | 9 2 | 5 2 | 5 2 |
| | 7 2 | 8 2 | 7 2 | 7 2 | 7 2 | 7 2 | 5 2 | 5 2 |
| | 9 2 | 8 2 | 8 2 | 8 2 | 9 2 | 9 2 | 5 2 | 5 2 |
| | 9 2 | 8 2 | 9 2 | 9 2 | 8 2 | 8 2 | 5 2 | 5 2 |
| | 7 2 | 9 2 | 7 2 | 7 2 | 7 2 | 7 2 | 5 2 | 5 2 |
| | 8 2 | 7 2 | 8 2 | 8 2 | 8 2 | 8 2 | 5 2 | 5 2 |
| | 8 2 | 9 2 | 9 2 | 9 2 | 8 2 | 8 2 | 5 2 | 5 2 |
| | 7 2 | 7 2 | 7 2 | 7 2 | 7 2 | 7 2 | 5 2 | 5 2 |
| | 9 2 | 9 2 | 9 2 | 9 2 | 9 2 | 9 2 | 5 2 | 5 2 |
| | 8 2 | 7 2 | 7 2 | 7 2 | 9 2 | 9 2 | 5 2 | 5 2 |
| | 9 2 | 8 2 | 8 2 | 8 2 | 7 2 | 7 2 | 5 2 | 5 2 |
| | 9 2 | 8 2 | 8 2 | 8 2 | 8 2 | 8 2 | 5 2 | 5 2 |
| | 7 2 | 9 2 | 9 2 | 9 2 | 9 2 | 9 2 | 5 2 | 5 2 |
| | 8 2 | 8 2 | 7 2 | 7 2 | 7 2 | 7 2 | 5 2 | 5 2 |
| | 7 2 | 7 2 | 8 2 | 8 2 | 8 2 | 8 2 | 5 2 | 5 2 |
| | 7 2 | 8 2 | 9 2 | 9 2 | 7 2 | 7 2 | 5 2 | 5 2 |
| | 9 2 | 7 2 | 7 2 | 7 2 | 8 2 | 8 2 | 5 2 | 5 2 |
| 189.1.37 | 54936 | 96 | Al1Pd1Tb1 | | | | | |
| | -A | -GM | -HA | -H | -KA | -K | -L | -M |
| | 9 2 | 9 2 | 7 2 | 7 2 | 7 2 | 7 2 | 5 2 | 5 2 |
| | 7 2 | 7 2 | 8 2 | 8 2 | 8 2 | 8 2 | 5 2 | 5 2 |
| | 8 2 | 8 2 | 9 2 | 9 2 | 9 2 | 9 2 | 5 2 | 5 2 |
| | 7 2 | 9 2 | 8 2 | 8 2 | 8 2 | 8 2 | 5 2 | 5 2 |
| | 9 2 | 7 2 | 9 2 | 9 2 | 7 2 | 7 2 | 5 2 | 5 2 |
| | 8 2 | 8 2 | 7 2 | 7 2 | 9 2 | 9 2 | 5 2 | 5 2 |
| | 9 2 | 9 2 | 7 2 | 7 2 | 7 2 | 7 2 | 5 2 | 5 2 |
| | 8 2 | 8 2 | 8 2 | 8 2 | 8 2 | 8 2 | 5 2 | 5 2 |
| | 7 2 | 7 2 | 9 2 | 9 2 | 9 2 | 9 2 | 5 2 | 5 2 |
| | 8 2 | 9 2 | 7 2 | 7 2 | 7 2 | 7 2 | 5 2 | 5 2 |
| | 9 2 | 8 2 | 9 2 | 9 2 | 8 2 | 8 2 | 5 2 | 5 2 |
| | 7 2 | 7 2 | 8 2 | 8 2 | 9 2 | 9 2 | 5 2 | 5 2 |
| | 8 2 | 9 2 | 8 2 | 8 2 | 9 2 | 9 2 | 5 2 | 5 2 |
| | 9 2 | 7 2 | 9 2 | 9 2 | 7 2 | 7 2 | 5 2 | 5 2 |
| | 7 2 | 9 2 | 7 2 | 7 2 | 8 2 | 8 2 | 5 2 | 5 2 |
| | 9 2 | 8 2 | 9 2 | 9 2 | 9 2 | 9 2 | 5 2 | 5 2 |
| | 7 2 | 9 2 | 9 2 | 9 2 | 8 2 | 8 2 | 5 2 | 5 2 |
| | 8 2 | 9 2 | 7 2 | 7 2 | 7 2 | 7 2 | 5 2 | 5 2 |
| | 7 2 | 7 2 | 8 2 | 8 2 | 9 2 | 9 2 | 5 2 | 5 2 |
| | 8 2 | 7 2 | 7 2 | 7 2 | 7 2 | 7 2 | 5 2 | 5 2 |
| | 9 2 | 7 2 | 7 2 | 7 2 | 8 2 | 8 2 | 5 2 | 5 2 |
| | 7 2 | 8 2 | 8 2 | 8 2 | 7 2 | 7 2 | 5 2 | 5 2 |
| | 7 2 | 9 2 | 9 2 | 9 2 | 9 2 | 9 2 | 5 2 | 5 2 |
| | 8 2 | 9 2 | 9 2 | 9 2 | 8 2 | 8 2 | 5 2 | 5 2 |
| | 9 2 | 8 2 | 8 2 | 8 2 | 7 2 | 7 2 | 5 2 | 5 2 |
| | 8 2 | 7 2 | 9 2 | 9 2 | 8 2 | 8 2 | 5 2 | 5 2 |
| | 7 2 | 9 2 | 8 2 | 8 2 | 7 2 | 7 2 | 5 2 | 5 2 |
| | 9 2 | 9 2 | 7 2 | 7 2 | 9 2 | 9 2 | 5 2 | 5 2 |
| | 8 2 | 7 2 | 9 2 | 9 2 | 8 2 | 8 2 | 5 2 | 5 2 |
| | 8 2 | 7 2 | 9 2 | 9 2 | 9 2 | 9 2 | 5 2 | 5 2 |
| | 8 2 | 7 2 | 7 2 | 7 2 | 9 2 | 9 2 | 5 2 | 5 2 |
| | 7 2 | 8 2 | 7 2 | 7 2 | 8 2 | 8 2 | 5 2 | 5 2 |
| | 9 2 | 8 2 | 8 2 | 8 2 | 7 2 | 7 2 | 5 2 | 5 2 |
| | 9 2 | 9 2 | 9 2 | 9 2 | 8 2 | 8 2 | 5 2 | 5 2 |
| | 7 2 | 8 2 | 7 2 | 7 2 | 7 2 | 7 2 | 5 2 | 5 2 |
| | 7 2 | 7 2 | 8 2 | 8 2 | 9 2 | 9 2 | 5 2 | 5 2 |
| | 9 2 | 9 2 | 7 2 | 7 2 | 7 2 | 7 2 | 5 2 | 5 2 |
| | 8 2 | 8 2 | 8 2 | 8 2 | 8 2 | 8 2 | 5 2 | 5 2 |
| | 9 2 | 9 2 | 9 2 | 9 2 | 9 2 | 9 2 | 5 2 | 5 2 |
| | 7 2 | 7 2 | 8 2 | 8 2 | 9 2 | 9 2 | 5 2 | 5 2 |
| | 8 2 | 8 2 | 7 2 | 7 2 | 7 2 | 7 2 | 5 2 | 5 2 |
| | 9 2 | 9 2 | 9 2 | 9 2 | 8 2 | 8 2 | 5 2 | 5 2 |
| | 8 2 | 7 2 | 9 2 | 9 2 | 8 2 | 8 2 | 5 2 | 5 2 |
| | 8 2 | 8 2 | 7 2 | 7 2 | 7 2 | 7 2 | 5 2 | 5 2 |
| | 7 2 | 8 2 | 8 2 | 8 2 | 9 2 | 9 2 | 5 2 | 5 2 |
| | 9 2 | 7 2 | 8 2 | 8 2 | 8 2 | 8 2 | 5 2 | 5 2 |
| | 7 2 | 9 2 | 7 2 | 7 2 | 7 2 | 7 2 | 5 2 | 5 2 |
| | 9 2 | 8 2 | 8 2 | 8 2 | 9 2 | 9 2 | 5 2 | 5 2 |
| 189.1.38 | 160908 | 48 | Al1Mg1Y1 | | | | | |
| | -A | -GM | -HA | -H | -KA | -K | -L | -M |
| | 9 2 | 9 2 | 7 2 | 7 2 | 7 2 | 7 2 | 5 2 | 5 2 |
| | 7 2 | 7 2 | 8 2 | 8 2 | 8 2 | 8 2 | 5 2 | 5 2 |
| | 8 2 | 8 2 | 9 2 | 9 2 | 9 2 | 9 2 | 5 2 | 5 2 |
| | 7 2 | 7 2 | 8 2 | 8 2 | 8 2 | 8 2 | 5 2 | 5 2 |
| | 9 2 | 9 2 | 9 2 | 9 2 | 9 2 | 9 2 | 5 2 | 5 2 |
| | 8 2 | 8 2 | 7 2 | 7 2 | 7 2 | 7 2 | 5 2 | 5 2 |
| | 8 2 | 8 2 | 7 2 | 7 2 | 7 2 | 7 2 | 5 2 | 5 2 |
| | 9 2 | 9 2 | 9 2 | 9 2 | 9 2 | 9 2 | 5 2 | 5 2 |
| | 7 2 | 7 2 | 8 2 | 8 2 | 8 2 | 8 2 | 5 2 | 5 2 |
| | 9 2 | 9 2 | 7 2 | 7 2 | 7 2 | 7 2 | 5 2 | 5 2 |
| | 8 2 | 8 2 | 8 2 | 8 2 | 8 2 | 8 2 | 5 2 | 5 2 |
| | 7 2 | 7 2 | 9 2 | 9 2 | 9 2 | 9 2 | 5 2 | 5 2 |
| | 8 2 | 9 2 | 9 2 | 9 2 | 9 2 | 9 2 | 5 2 | 5 2 |
| | 9 2 | 9 2 | 7 2 | 7 2 | 7 2 | 7 2 | 5 2 | 5 2 |
| | 8 2 | 9 2 | 9 2 | 9 2 | 8 2 | 8 2 | 5 2 | 5 2 |
| | 8 2 | 7 2 | 8 2 | 8 2 | 8 2 | 8 2 | 5 2 | 5 2 |
| | 9 2 | 8 2 | 7 2 | 7 2 | 7 2 | 7 2 | 5 2 | 5 2 |
| | 7 2 | 7 2 | 8 2 | 8 2 | 9 2 | 9 2 | 5 2 | 5 2 |
| | 9 2 | 8 2 | 7 2 | 7 2 | 8 2 | 8 2 | 5 2 | 5 2 |
| | 7 2 | 8 2 | 9 2 | 9 2 | 8 2 | 8 2 | 5 2 | 5 2 |
| | 9 2 | 9 2 | 8 2 | 8 2 | 7 2 | 7 2 | 5 2 | 5 2 |
| | 7 2 | 8 2 | 8 2 | 8 2 | 7 2 | 7 2 | 5 2 | 5 2 |
| | 8 2 | 7 2 | 7 2 | 7 2 | 9 2 | 9 2 | 5 2 | 5 2 |
| | 9 2 | 9 2 | 8 2 | 8 2 | 9 2 | 9 2 | 5 2 | 5 2 |

Continued on next column

**TABLE LXXVI– continued**

| TQC | ICSD | NoE | SF | | | | | |
|---|---|---|---|---|---|---|---|---|
| 189.1.39 | 107275 | 78 | Ag1Pb1Y1 | | | | | |
| | -A | -GM | -HA | -H | -KA | -K | -L | -M |
| | 8 2 | 9 2 | 7 2 | 7 2 | 7 2 | 7 2 | 5 2 | 5 2 |
| | 7 2 | 7 2 | 9 2 | 9 2 | 8 2 | 8 2 | 5 2 | 5 2 |
| | 9 2 | 8 2 | 8 2 | 8 2 | 9 2 | 9 2 | 5 2 | 5 2 |
| | 8 2 | 9 2 | 9 2 | 9 2 | 8 2 | 8 2 | 5 2 | 5 2 |
| | 7 2 | 7 2 | 7 2 | 7 2 | 7 2 | 7 2 | 5 2 | 5 2 |
| | 9 2 | 8 2 | 8 2 | 8 2 | 9 2 | 9 2 | 5 2 | 5 2 |
| | 8 2 | 9 2 | 7 2 | 7 2 | 7 2 | 7 2 | 5 2 | 5 2 |
| | 9 2 | 8 2 | 8 2 | 8 2 | 9 2 | 9 2 | 5 2 | 5 2 |
| | 7 2 | 7 2 | 9 2 | 9 2 | 8 2 | 8 2 | 5 2 | 5 2 |
| | 8 2 | 9 2 | 7 2 | 7 2 | 7 2 | 7 2 | 5 2 | 5 2 |
| | 9 2 | 8 2 | 9 2 | 9 2 | 8 2 | 8 2 | 5 2 | 5 2 |
| | 7 2 | 7 2 | 8 2 | 8 2 | 9 2 | 9 2 | 5 2 | 5 2 |
| | 9 2 | 9 2 | 8 2 | 8 2 | 9 2 | 9 2 | 5 2 | 5 2 |
| | 8 2 | 9 2 | 7 2 | 7 2 | 7 2 | 7 2 | 5 2 | 5 2 |
| | 9 2 | 9 2 | 8 2 | 8 2 | 8 2 | 8 2 | 5 2 | 5 2 |
| | 9 2 | 8 2 | 9 2 | 9 2 | 9 2 | 9 2 | 5 2 | 5 2 |
| | 7 2 | 7 2 | 7 2 | 7 2 | 8 2 | 8 2 | 5 2 | 5 2 |
| | 7 2 | 8 2 | 9 2 | 9 2 | 7 2 | 7 2 | 5 2 | 5 2 |
| | 8 2 | 9 2 | 8 2 | 8 2 | 8 2 | 8 2 | 5 2 | 5 2 |
| | 9 2 | 7 2 | 7 2 | 7 2 | 7 2 | 7 2 | 5 2 | 5 2 |
| | 8 2 | 8 2 | 9 2 | 9 2 | 9 2 | 9 2 | 5 2 | 5 2 |
| | 9 2 | 8 2 | 9 2 | 9 2 | 8 2 | 8 2 | 5 2 | 5 2 |
| | 8 2 | 9 2 | 7 2 | 7 2 | 9 2 | 9 2 | 5 2 | 5 2 |
| | 7 2 | 7 2 | 8 2 | 8 2 | 7 2 | 7 2 | 5 2 | 5 2 |
| | 8 2 | 7 2 | 9 2 | 9 2 | 7 2 | 7 2 | 5 2 | 5 2 |
| | 8 2 | 8 2 | 7 2 | 7 2 | 9 2 | 9 2 | 5 2 | 5 2 |
| | 7 2 | 9 2 | 8 2 | 8 2 | 8 2 | 8 2 | 5 2 | 5 2 |
| | 9 2 | 7 2 | 8 2 | 8 2 | 9 2 | 9 2 | 5 2 | 5 2 |
| | 7 2 | 9 2 | 7 2 | 7 2 | 8 2 | 8 2 | 5 2 | 5 2 |
| | 9 2 | 8 2 | 9 2 | 9 2 | 7 2 | 7 2 | 5 2 | 5 2 |
| | 9 2 | 9 2 | 8 2 | 8 2 | 9 2 | 9 2 | 5 2 | 5 2 |
| | 8 2 | 7 2 | 9 2 | 9 2 | 7 2 | 7 2 | 5 2 | 5 2 |
| | 7 2 | 8 2 | 7 2 | 7 2 | 8 2 | 8 2 | 5 2 | 5 2 |
| | 8 2 | 7 2 | 7 2 | 7 2 | 7 2 | 7 2 | 5 2 | 5 2 |
| | 8 2 | 8 2 | 8 2 | 8 2 | 9 2 | 9 2 | 5 2 | 5 2 |
| | 9 2 | 8 2 | 9 2 | 9 2 | 8 2 | 8 2 | 5 2 | 5 2 |
| | 7 2 | 9 2 | 9 2 | 9 2 | 7 2 | 7 2 | 5 2 | 5 2 |
| | 8 2 | 8 2 | 7 2 | 7 2 | 9 2 | 9 2 | 5 2 | 5 2 |
| | 7 2 | 7 2 | 9 2 | 9 2 | 8 2 | 8 2 | 5 2 | 5 2 |
| 189.1.40 | 416377 | 120 | Ag1Lu1Sn1 | | | | | |
| | -A | -GM | -HA | -H | -KA | -K | -L | -M |
| | 9 2 | 9 2 | 7 2 | 7 2 | 7 2 | 7 2 | 5 2 | 5 2 |
| | 7 2 | 7 2 | 8 2 | 8 2 | 8 2 | 8 2 | 5 2 | 5 2 |
| | 8 2 | 8 2 | 9 2 | 9 2 | 9 2 | 9 2 | 5 2 | 5 2 |
| | 9 2 | 7 2 | 8 2 | 8 2 | 8 2 | 8 2 | 5 2 | 5 2 |
| | 7 2 | 9 2 | 7 2 | 7 2 | 7 2 | 7 2 | 5 2 | 5 2 |
| | 8 2 | 8 2 | 9 2 | 9 2 | 9 2 | 9 2 | 5 2 | 5 2 |
| | 9 2 | 9 2 | 7 2 | 7 2 | 7 2 | 7 2 | 5 2 | 5 2 |
| | 8 2 | 8 2 | 9 2 | 9 2 | 9 2 | 9 2 | 5 2 | 5 2 |
| | 7 2 | 7 2 | 8 2 | 8 2 | 8 2 | 8 2 | 5 2 | 5 2 |
| | 9 2 | 9 2 | 7 2 | 7 2 | 7 2 | 7 2 | 5 2 | 5 2 |
| | 8 2 | 8 2 | 8 2 | 8 2 | 8 2 | 8 2 | 5 2 | 5 2 |
| | 7 2 | 7 2 | 9 2 | 9 2 | 9 2 | 9 2 | 5 2 | 5 2 |
| | 8 2 | 9 2 | 9 2 | 9 2 | 9 2 | 9 2 | 5 2 | 5 2 |
| | 9 2 | 9 2 | 7 2 | 7 2 | 7 2 | 7 2 | 5 2 | 5 2 |
| | 8 2 | 9 2 | 9 2 | 9 2 | 8 2 | 8 2 | 5 2 | 5 2 |
| | 7 2 | 7 2 | 8 2 | 8 2 | 8 2 | 8 2 | 5 2 | 5 2 |
| | 8 2 | 9 2 | 7 2 | 7 2 | 9 2 | 9 2 | 5 2 | 5 2 |
| | 9 2 | 8 2 | 8 2 | 8 2 | 7 2 | 7 2 | 5 2 | 5 2 |
| | 7 2 | 7 2 | 9 2 | 9 2 | 8 2 | 8 2 | 5 2 | 5 2 |
| | 9 2 | 8 2 | 7 2 | 7 2 | 7 2 | 7 2 | 5 2 | 5 2 |
| | 8 2 | 8 2 | 9 2 | 9 2 | 9 2 | 9 2 | 5 2 | 5 2 |
| | 7 2 | 9 2 | 8 2 | 8 2 | 7 2 | 7 2 | 5 2 | 5 2 |
| | 8 2 | 7 2 | 7 2 | 7 2 | 9 2 | 9 2 | 5 2 | 5 2 |
| | 9 2 | 9 2 | 9 2 | 9 2 | 8 2 | 8 2 | 5 2 | 5 2 |
| | 8 2 | 8 2 | 8 2 | 8 2 | 9 2 | 9 2 | 5 2 | 5 2 |
| | 7 2 | 7 2 | 7 2 | 7 2 | 8 2 | 8 2 | 5 2 | 5 2 |
| | 9 2 | 9 2 | 8 2 | 8 2 | 7 2 | 7 2 | 5 2 | 5 2 |
| | 7 2 | 8 2 | 9 2 | 9 2 | 8 2 | 8 2 | 5 2 | 5 2 |
| | 9 2 | 7 2 | 7 2 | 7 2 | 7 2 | 7 2 | 5 2 | 5 2 |
| | 8 2 | 8 2 | 8 2 | 8 2 | 9 2 | 9 2 | 5 2 | 5 2 |
| | 8 2 | 8 2 | 8 2 | 8 2 | 8 2 | 8 2 | 5 2 | 5 2 |
| | 9 2 | 9 2 | 7 2 | 7 2 | 7 2 | 7 2 | 5 2 | 5 2 |
| | 9 2 | 7 2 | 9 2 | 9 2 | 9 2 | 9 2 | 5 2 | 5 2 |
| | 7 2 | 7 2 | 8 2 | 8 2 | 7 2 | 7 2 | 5 2 | 5 2 |
| | 7 2 | 9 2 | 7 2 | 7 2 | 9 2 | 9 2 | 5 2 | 5 2 |
| | 9 2 | 8 2 | 9 2 | 9 2 | 8 2 | 8 2 | 5 2 | 5 2 |
| | 7 2 | 7 2 | 8 2 | 8 2 | 9 2 | 9 2 | 5 2 | 5 2 |
| | 8 2 | 8 2 | 9 2 | 9 2 | 8 2 | 8 2 | 5 2 | 5 2 |
| | 8 2 | 8 2 | 7 2 | 7 2 | 7 2 | 7 2 | 5 2 | 5 2 |
| | 9 2 | 9 2 | 7 2 | 7 2 | 9 2 | 9 2 | 5 2 | 5 2 |
| | 7 2 | 7 2 | 8 2 | 8 2 | 7 2 | 7 2 | 5 2 | 5 2 |
| | 8 2 | 8 2 | 9 2 | 9 2 | 8 2 | 8 2 | 5 2 | 5 2 |
| | 9 2 | 7 2 | 9 2 | 9 2 | 9 2 | 9 2 | 5 2 | 5 2 |
| | 7 2 | 9 2 | 7 2 | 7 2 | 7 2 | 7 2 | 5 2 | 5 2 |
| | 8 2 | 8 2 | 8 2 | 8 2 | 8 2 | 8 2 | 5 2 | 5 2 |
| | 9 2 | 9 2 | 7 2 | 7 2 | 9 2 | 9 2 | 5 2 | 5 2 |
| | 7 2 | 7 2 | 8 2 | 8 2 | 7 2 | 7 2 | 5 2 | 5 2 |
| | 9 2 | 9 2 | 9 2 | 9 2 | 9 2 | 9 2 | 5 2 | 5 2 |
| | 8 2 | 7 2 | 8 2 | 8 2 | 8 2 | 8 2 | 5 2 | 5 2 |
| | 7 2 | 9 2 | 9 2 | 9 2 | 7 2 | 7 2 | 5 2 | 5 2 |
| | 8 2 | 7 2 | 7 2 | 7 2 | 8 2 | 8 2 | 5 2 | 5 2 |
| | 8 2 | 8 2 | 9 2 | 9 2 | 9 2 | 9 2 | 5 2 | 5 2 |
| | 9 2 | 9 2 | 8 2 | 8 2 | 8 2 | 8 2 | 5 2 | 5 2 |
| | 7 2 | 8 2 | 7 2 | 7 2 | 7 2 | 7 2 | 5 2 | 5 2 |
| | 9 2 | 7 2 | 7 2 | 7 2 | 7 2 | 7 2 | 5 2 | 5 2 |
| | 7 2 | 8 2 | 9 2 | 9 2 | 9 2 | 9 2 | 5 2 | 5 2 |
| | 9 2 | 8 2 | 8 2 | 8 2 | 8 2 | 8 2 | 5 2 | 5 2 |
| | 8 2 | 9 2 | 8 2 | 8 2 | 7 2 | 7 2 | 5 2 | 5 2 |
| | 9 2 | 8 2 | 7 2 | 7 2 | 8 2 | 8 2 | 5 2 | 5 2 |
| | 7 2 | 7 2 | 8 2 | 8 2 | 9 2 | 9 2 | 5 2 | 5 2 |
| 189.1.41 | 245236 | 120 | Ag1Lu1Sn1 | | | | | |
| | -A | -GM | -HA | -H | -KA | -K | -L | -M |
| | 9 2 | 9 2 | 7 2 | 7 2 | 7 2 | 7 2 | 5 2 | 5 2 |
| | 7 2 | 7 2 | 8 2 | 8 2 | 8 2 | 8 2 | 5 2 | 5 2 |
| | 8 2 | 8 2 | 9 2 | 9 2 | 9 2 | 9 2 | 5 2 | 5 2 |
| | 9 2 | 7 2 | 8 2 | 8 2 | 8 2 | 8 2 | 5 2 | 5 2 |
| | 7 2 | 9 2 | 7 2 | 7 2 | 7 2 | 7 2 | 5 2 | 5 2 |
| | 8 2 | 8 2 | 9 2 | 9 2 | 9 2 | 9 2 | 5 2 | 5 2 |
| | 9 2 | 9 2 | 7 2 | 7 2 | 7 2 | 7 2 | 5 2 | 5 2 |
| | 8 2 | 8 2 | 9 2 | 9 2 | 9 2 | 9 2 | 5 2 | 5 2 |
| | 7 2 | 7 2 | 8 2 | 8 2 | 8 2 | 8 2 | 5 2 | 5 2 |
| | 9 2 | 9 2 | 7 2 | 7 2 | 7 2 | 7 2 | 5 2 | 5 2 |
| | 8 2 | 8 2 | 8 2 | 8 2 | 8 2 | 8 2 | 5 2 | 5 2 |
| | 7 2 | 7 2 | 9 2 | 9 2 | 9 2 | 9 2 | 5 2 | 5 2 |
| | 8 2 | 9 2 | 9 2 | 9 2 | 9 2 | 9 2 | 5 2 | 5 2 |
| | 9 2 | 9 2 | 7 2 | 7 2 | 7 2 | 7 2 | 5 2 | 5 2 |
| | 8 2 | 9 2 | 9 2 | 9 2 | 8 2 | 8 2 | 5 2 | 5 2 |
| | 7 2 | 7 2 | 8 2 | 8 2 | 8 2 | 8 2 | 5 2 | 5 2 |
| | 8 2 | 9 2 | 7 2 | 7 2 | 9 2 | 9 2 | 5 2 | 5 2 |

Continued on next column

**TABLE LXXVI– continued**

| TQC | ICSD | NoE | SF | | |
|---|---|---|---|---|---|
| | 9 2 | 8 2 | 8 2 8 2 | 7 2 | 7 2 5 2 5 2 |
| | 7 2 | 7 2 | 9 2 9 2 | 8 2 | 8 2 5 2 5 2 |
| | 9 2 | 8 2 | 7 2 7 2 | 7 2 | 7 2 5 2 5 2 |
| | 8 2 | 8 2 | 9 2 9 2 | 9 2 | 9 2 5 2 5 2 |
| | 7 2 | 9 2 | 8 2 8 2 | 7 2 | 7 2 5 2 5 2 |
| | 8 2 | 7 2 | 7 2 7 2 | 9 2 | 9 2 5 2 5 2 |
| | 9 2 | 9 2 | 9 2 9 2 | 8 2 | 8 2 5 2 5 2 |
| | 8 2 | 8 2 | 8 2 8 2 | 9 2 | 9 2 5 2 5 2 |
| | 7 2 | 7 2 | 7 2 7 2 | 8 2 | 8 2 5 2 5 2 |
| | 9 2 | 9 2 | 8 2 8 2 | 7 2 | 7 2 5 2 5 2 |
| | 7 2 | 8 2 | 9 2 9 2 | 8 2 | 8 2 5 2 5 2 |
| | 9 2 | 7 2 | 7 2 7 2 | 7 2 | 7 2 5 2 5 2 |
| | 8 2 | 8 2 | 8 2 8 2 | 9 2 | 9 2 5 2 5 2 |
| | 8 2 | 8 2 | 8 2 8 2 | 8 2 | 8 2 5 2 5 2 |
| | 9 2 | 9 2 | 7 2 7 2 | 7 2 | 7 2 5 2 5 2 |
| | 9 2 | 7 2 | 9 2 9 2 | 9 2 | 9 2 5 2 5 2 |
| | 7 2 | 7 2 | 8 2 8 2 | 7 2 | 7 2 5 2 5 2 |
| | 7 2 | 9 2 | 7 2 7 2 | 9 2 | 9 2 5 2 5 2 |
| | 9 2 | 8 2 | 9 2 9 2 | 8 2 | 8 2 5 2 5 2 |
| | 7 2 | 7 2 | 8 2 8 2 | 9 2 | 9 2 5 2 5 2 |
| | 8 2 | 8 2 | 9 2 9 2 | 8 2 | 8 2 5 2 5 2 |
| | 8 2 | 8 2 | 7 2 7 2 | 7 2 | 7 2 5 2 5 2 |
| | 9 2 | 9 2 | 7 2 7 2 | 9 2 | 9 2 5 2 5 2 |
| | 7 2 | 7 2 | 8 2 8 2 | 7 2 | 7 2 5 2 5 2 |
| | 8 2 | 8 2 | 9 2 9 2 | 8 2 | 8 2 5 2 5 2 |
| | 9 2 | 7 2 | 9 2 9 2 | 9 2 | 9 2 5 2 5 2 |
| | 7 2 | 9 2 | 7 2 7 2 | 7 2 | 7 2 5 2 5 2 |
| | 8 2 | 8 2 | 8 2 8 2 | 8 2 | 8 2 5 2 5 2 |
| | 9 2 | 9 2 | 7 2 7 2 | 9 2 | 9 2 5 2 5 2 |
| | 7 2 | 7 2 | 8 2 8 2 | 7 2 | 7 2 5 2 5 2 |
| | 9 2 | 9 2 | 9 2 9 2 | 9 2 | 9 2 5 2 5 2 |
| | 8 2 | 7 2 | 8 2 8 2 | 8 2 | 8 2 5 2 5 2 |
| | 7 2 | 9 2 | 9 2 9 2 | 7 2 | 7 2 5 2 5 2 |
| | 8 2 | 7 2 | 7 2 7 2 | 8 2 | 8 2 5 2 5 2 |
| | 8 2 | 8 2 | 9 2 9 2 | 9 2 | 9 2 5 2 5 2 |
| | 9 2 | 9 2 | 8 2 8 2 | 8 2 | 8 2 5 2 5 2 |
| | 7 2 | 8 2 | 7 2 7 2 | 7 2 | 7 2 5 2 5 2 |
| | 9 2 | 7 2 | 7 2 7 2 | 7 2 | 7 2 5 2 5 2 |
| | 7 2 | 8 2 | 9 2 9 2 | 9 2 | 9 2 5 2 5 2 |
| | 9 2 | 8 2 | 8 2 8 2 | 8 2 | 8 2 5 2 5 2 |
| | 8 2 | 9 2 | 8 2 8 2 | 7 2 | 7 2 5 2 5 2 |
| | 9 2 | 8 2 | 7 2 7 2 | 8 2 | 8 2 5 2 5 2 |
| | 7 2 | 7 2 | 8 2 8 2 | 9 2 | 9 2 5 2 5 2 |
| 189.1.42 | 107278 | 120 | Ag1Lu1Pb1 | | |
| | -A | -GM | -HA -H | -KA | -K -L -M |
| | 8 2 | 9 2 | 7 2 7 2 | 7 2 | 7 2 5 2 5 2 |
| | 7 2 | 7 2 | 8 2 8 2 | 8 2 | 8 2 5 2 5 2 |
| | 9 2 | 8 2 | 9 2 9 2 | 9 2 | 9 2 5 2 5 2 |
| | 8 2 | 7 2 | 9 2 9 2 | 8 2 | 8 2 5 2 5 2 |
| | 7 2 | 9 2 | 7 2 7 2 | 7 2 | 7 2 5 2 5 2 |
| | 9 2 | 8 2 | 8 2 8 2 | 9 2 | 9 2 5 2 5 2 |
| | 8 2 | 9 2 | 7 2 7 2 | 7 2 | 7 2 5 2 5 2 |
| | 9 2 | 8 2 | 8 2 8 2 | 9 2 | 9 2 5 2 5 2 |
| | 7 2 | 7 2 | 9 2 9 2 | 8 2 | 8 2 5 2 5 2 |
| | 8 2 | 9 2 | 7 2 7 2 | 7 2 | 7 2 5 2 5 2 |
| | 9 2 | 8 2 | 9 2 9 2 | 8 2 | 8 2 5 2 5 2 |
| | 7 2 | 7 2 | 8 2 8 2 | 9 2 | 9 2 5 2 5 2 |
| | 9 2 | 9 2 | 8 2 8 2 | 9 2 | 9 2 5 2 5 2 |
| | 8 2 | 9 2 | 7 2 7 2 | 7 2 | 7 2 5 2 5 2 |
| | 9 2 | 9 2 | 8 2 8 2 | 8 2 | 8 2 5 2 5 2 |
| | 7 2 | 7 2 | 7 2 7 2 | 8 2 | 8 2 5 2 5 2 |

Continued on next column

**TABLE LXXVI– continued**

| TQC | ICSD | NoE | SF | | |
|---|---|---|---|---|---|
| | 9 2 | 9 2 | 9 2 9 2 | 9 2 | 9 2 5 2 5 2 |
| | 8 2 | 8 2 | 8 2 8 2 | 7 2 | 7 2 5 2 5 2 |
| | 7 2 | 7 2 | 7 2 7 2 | 8 2 | 8 2 5 2 5 2 |
| | 8 2 | 8 2 | 9 2 9 2 | 9 2 | 9 2 5 2 5 2 |
| | 9 2 | 8 2 | 8 2 8 2 | 7 2 | 7 2 5 2 5 2 |
| | 7 2 | 9 2 | 9 2 9 2 | 7 2 | 7 2 5 2 5 2 |
| | 8 2 | 7 2 | 7 2 7 2 | 9 2 | 9 2 5 2 5 2 |
| | 9 2 | 9 2 | 8 2 8 2 | 8 2 | 8 2 5 2 5 2 |
| | 9 2 | 8 2 | 9 2 9 2 | 9 2 | 9 2 5 2 5 2 |
| | 8 2 | 7 2 | 7 2 7 2 | 8 2 | 8 2 5 2 5 2 |
| | 7 2 | 8 2 | 9 2 9 2 | 7 2 | 7 2 5 2 5 2 |
| | 7 2 | 9 2 | 8 2 8 2 | 8 2 | 8 2 5 2 5 2 |
| | 8 2 | 7 2 | 7 2 7 2 | 7 2 | 7 2 5 2 5 2 |
| | 9 2 | 8 2 | 9 2 9 2 | 9 2 | 9 2 5 2 5 2 |
| | 9 2 | 8 2 | 9 2 9 2 | 8 2 | 8 2 5 2 5 2 |
| | 8 2 | 9 2 | 7 2 7 2 | 7 2 | 7 2 5 2 5 2 |
| | 8 2 | 7 2 | 8 2 8 2 | 9 2 | 9 2 5 2 5 2 |
| | 7 2 | 8 2 | 9 2 9 2 | 7 2 | 7 2 5 2 5 2 |
| | 7 2 | 9 2 | 8 2 8 2 | 9 2 | 9 2 5 2 5 2 |
| | 8 2 | 7 2 | 9 2 9 2 | 8 2 | 8 2 5 2 5 2 |
| | 9 2 | 8 2 | 7 2 7 2 | 7 2 | 7 2 5 2 5 2 |
| | 7 2 | 7 2 | 7 2 7 2 | 8 2 | 8 2 5 2 5 2 |
| | 9 2 | 9 2 | 8 2 8 2 | 9 2 | 9 2 5 2 5 2 |
| | 7 2 | 8 2 | 9 2 9 2 | 7 2 | 7 2 5 2 5 2 |
| | 8 2 | 7 2 | 7 2 7 2 | 8 2 | 8 2 5 2 5 2 |
| | 9 2 | 9 2 | 8 2 8 2 | 9 2 | 9 2 5 2 5 2 |
| | 8 2 | 8 2 | 7 2 7 2 | 8 2 | 8 2 5 2 5 2 |
| | 9 2 | 7 2 | 9 2 9 2 | 9 2 | 9 2 5 2 5 2 |
| | 8 2 | 9 2 | 8 2 8 2 | 7 2 | 7 2 5 2 5 2 |
| | 7 2 | 7 2 | 7 2 7 2 | 7 2 | 7 2 5 2 5 2 |
| | 8 2 | 8 2 | 9 2 9 2 | 9 2 | 9 2 5 2 5 2 |
| | 7 2 | 9 2 | 8 2 8 2 | 9 2 | 9 2 5 2 5 2 |
| | 9 2 | 7 2 | 8 2 8 2 | 8 2 | 8 2 5 2 5 2 |
| | 7 2 | 9 2 | 9 2 9 2 | 8 2 | 8 2 5 2 5 2 |
| | 9 2 | 8 2 | 7 2 7 2 | 7 2 | 7 2 5 2 5 2 |
| | 9 2 | 7 2 | 8 2 8 2 | 9 2 | 9 2 5 2 5 2 |
| | 8 2 | 9 2 | 9 2 9 2 | 7 2 | 7 2 5 2 5 2 |
| | 8 2 | 8 2 | 7 2 7 2 | 8 2 | 8 2 5 2 5 2 |
| | 7 2 | 7 2 | 7 2 7 2 | 7 2 | 7 2 5 2 5 2 |
| | 8 2 | 8 2 | 8 2 8 2 | 9 2 | 9 2 5 2 5 2 |
| | 9 2 | 8 2 | 9 2 9 2 | 8 2 | 8 2 5 2 5 2 |
| | 8 2 | 9 2 | 9 2 9 2 | 7 2 | 7 2 5 2 5 2 |
| | 7 2 | 8 2 | 7 2 7 2 | 9 2 | 9 2 5 2 5 2 |
| | 7 2 | 7 2 | 9 2 9 2 | 8 2 | 8 2 5 2 5 2 |
| 189.1.43 | 10016 | 54 | Ag1Ca1P1 | | |
| | -A | -GM | -HA -H | -KA | -K -L -M |
| | 8 2 | 9 2 | 8 2 8 2 | 9 2 | 9 2 5 2 5 2 |
| | 9 2 | 9 2 | 7 2 7 2 | 7 2 | 7 2 5 2 5 2 |
| | 9 2 | 9 2 | 8 2 8 2 | 8 2 | 8 2 5 2 5 2 |
| | 9 2 | 8 2 | 9 2 9 2 | 8 2 | 8 2 5 2 5 2 |
| | 7 2 | 7 2 | 7 2 7 2 | 9 2 | 9 2 5 2 5 2 |
| | 7 2 | 8 2 | 9 2 9 2 | 7 2 | 7 2 5 2 5 2 |
| | 8 2 | 9 2 | 8 2 8 2 | 8 2 | 8 2 5 2 5 2 |
| | 9 2 | 7 2 | 7 2 7 2 | 7 2 | 7 2 5 2 5 2 |
| | 8 2 | 8 2 | 9 2 9 2 | 8 2 | 8 2 5 2 5 2 |
| | 9 2 | 8 2 | 9 2 9 2 | 9 2 | 9 2 5 2 5 2 |
| | 8 2 | 9 2 | 7 2 7 2 | 7 2 | 7 2 5 2 5 2 |
| | 7 2 | 7 2 | 8 2 8 2 | 9 2 | 9 2 5 2 5 2 |
| | 8 2 | 7 2 | 9 2 9 2 | 7 2 | 7 2 5 2 5 2 |
| | 8 2 | 8 2 | 7 2 7 2 | 9 2 | 9 2 5 2 5 2 |
| | 7 2 | 9 2 | 8 2 8 2 | 9 2 | 9 2 5 2 5 2 |

Continued on next column

**TABLE LXXVI– continued**

| TQC | ICSD | NoE | SF | | | | | |
|---|---|---|---|---|---|---|---|---|
| | 9 2 | 9 2 | 9 2 | 9 2 | 8 2 | 8 2 | 5 2 | 5 2 |
| | 7 2 | 7 2 | 8 2 | 8 2 | 8 2 | 8 2 | 5 2 | 5 2 |
| | 9 2 | 8 2 | 7 2 | 7 2 | 7 2 | 7 2 | 5 2 | 5 2 |
| | 9 2 | 7 2 | 8 2 | 8 2 | 9 2 | 9 2 | 5 2 | 5 2 |
| | 8 2 | 9 2 | 9 2 | 9 2 | 8 2 | 8 2 | 5 2 | 5 2 |
| | 7 2 | 7 2 | 7 2 | 7 2 | 7 2 | 7 2 | 5 2 | 5 2 |
| | 8 2 | 8 2 | 8 2 | 8 2 | 7 2 | 7 2 | 5 2 | 5 2 |
| | 8 2 | 8 2 | 7 2 | 7 2 | 9 2 | 9 2 | 5 2 | 5 2 |
| | 7 2 | 8 2 | 9 2 | 9 2 | 8 2 | 8 2 | 5 2 | 5 2 |
| | 9 2 | 7 2 | 7 2 | 7 2 | 7 2 | 7 2 | 5 2 | 5 2 |
| | 7 2 | 8 2 | 9 2 | 9 2 | 8 2 | 8 2 | 5 2 | 5 2 |
| | 8 2 | 9 2 | 9 2 | 9 2 | 9 2 | 9 2 | 5 2 | 5 2 |
| 189.1.44 | 10017 | 54 | Ag1As1Ca1 | | | | | |
| | -A | -GM | -HA | -H | -KA | -K | -L | -M |
| | 8 2 | 9 2 | 8 2 | 8 2 | 9 2 | 9 2 | 5 2 | 5 2 |
| | 9 2 | 9 2 | 7 2 | 7 2 | 7 2 | 7 2 | 5 2 | 5 2 |
| | 9 2 | 9 2 | 8 2 | 8 2 | 8 2 | 8 2 | 5 2 | 5 2 |
| | 9 2 | 8 2 | 9 2 | 9 2 | 8 2 | 8 2 | 5 2 | 5 2 |
| | 7 2 | 7 2 | 7 2 | 7 2 | 9 2 | 9 2 | 5 2 | 5 2 |
| | 7 2 | 8 2 | 9 2 | 9 2 | 7 2 | 7 2 | 5 2 | 5 2 |
| | 8 2 | 9 2 | 8 2 | 8 2 | 8 2 | 8 2 | 5 2 | 5 2 |
| | 9 2 | 7 2 | 7 2 | 7 2 | 7 2 | 7 2 | 5 2 | 5 2 |
| | 8 2 | 8 2 | 9 2 | 9 2 | 8 2 | 8 2 | 5 2 | 5 2 |
| | 9 2 | 8 2 | 9 2 | 9 2 | 9 2 | 9 2 | 5 2 | 5 2 |
| | 8 2 | 9 2 | 7 2 | 7 2 | 7 2 | 7 2 | 5 2 | 5 2 |
| | 7 2 | 7 2 | 8 2 | 8 2 | 9 2 | 9 2 | 5 2 | 5 2 |
| | 8 2 | 7 2 | 9 2 | 9 2 | 7 2 | 7 2 | 5 2 | 5 2 |
| | 8 2 | 8 2 | 7 2 | 7 2 | 9 2 | 9 2 | 5 2 | 5 2 |
| | 7 2 | 9 2 | 8 2 | 8 2 | 9 2 | 9 2 | 5 2 | 5 2 |
| | 9 2 | 9 2 | 9 2 | 9 2 | 8 2 | 8 2 | 5 2 | 5 2 |
| | 7 2 | 7 2 | 8 2 | 8 2 | 8 2 | 8 2 | 5 2 | 5 2 |
| | 9 2 | 8 2 | 7 2 | 7 2 | 7 2 | 7 2 | 5 2 | 5 2 |
| | 9 2 | 7 2 | 8 2 | 8 2 | 9 2 | 9 2 | 5 2 | 5 2 |
| | 8 2 | 9 2 | 9 2 | 9 2 | 8 2 | 8 2 | 5 2 | 5 2 |
| | 7 2 | 8 2 | 7 2 | 7 2 | 7 2 | 7 2 | 5 2 | 5 2 |
| | 8 2 | 7 2 | 7 2 | 7 2 | 7 2 | 7 2 | 5 2 | 5 2 |
| | 8 2 | 8 2 | 8 2 | 8 2 | 9 2 | 9 2 | 5 2 | 5 2 |
| | 9 2 | 8 2 | 9 2 | 9 2 | 8 2 | 8 2 | 5 2 | 5 2 |
| | 7 2 | 9 2 | 9 2 | 9 2 | 7 2 | 7 2 | 5 2 | 5 2 |
| | 7 2 | 8 2 | 7 2 | 7 2 | 8 2 | 8 2 | 5 2 | 5 2 |
| | 8 2 | 7 2 | 9 2 | 9 2 | 9 2 | 9 2 | 5 2 | 5 2 |

The End

TABLE LXXVII: The computed irreducible representations for the valence bands for the compounds of SEBR and NLC in space group 190. See the list of tables.

| TQC | ICSD | NoE | SF | | | | | |
|---|---|---|---|---|---|---|---|---|
| 190.2.1 | 81072 | 204 | Ba3Cr2O9W1 | | | | | |
| | -A | -GM | -H | -HA | -K | -KA | -L | -M |
| | 8+9 4 | 9 2 | 4 1 | 4 1 | 7 2 | 7 2 | 2+3 2 | 5 2 |
| | 8+9 4 | 9 2 | 5 1 | 5 1 | 8 2 | 8 2 | 4+5 2 | 5 2 |
| | 8+9 4 | 8 2 | 9 2 | 9 2 | 7 2 | 7 2 | 2+3 2 | 5 2 |
| | 8+9 4 | 8 2 | 6 1 | 6 1 | 9 2 | 9 2 | 4+5 2 | 5 2 |
| | 8+9 4 | 9 2 | 7 1 | 7 1 | 9 2 | 9 2 | 2+3 2 | 5 2 |
| | 4+7 2 | 9 2 | 8 2 | 8 2 | 9 2 | 9 2 | 4+5 2 | 5 2 |
| | 5+6 2 | 8 2 | 8 2 | 8 2 | 8 2 | 8 2 | 2+3 2 | 5 2 |
| | 8+9 4 | 9 2 | 9 2 | 9 2 | 9 2 | 9 2 | 4+5 2 | 5 2 |
| | 8+9 4 | 8 2 | 8 2 | 8 2 | 8 2 | 8 2 | 2+3 2 | 5 2 |
| | 8+9 4 | 7 2 | 9 2 | 9 2 | 7 2 | 7 2 | 4+5 2 | 5 2 |

Continued on next column

**TABLE LXXVII– continued**

| TQC | ICSD | NoE | SF | | | | | |
|---|---|---|---|---|---|---|---|---|
| | 4+7 2 | 9 2 | 8 2 | 8 2 | 7 2 | 7 2 | 2+3 2 | 5 2 |
| | 5+6 2 | 7 2 | 6 1 | 6 1 | 9 2 | 9 2 | 4+5 2 | 5 2 |
| | 8+9 4 | 9 2 | 7 1 | 7 1 | 8 2 | 8 2 | 2+3 2 | 5 2 |
| | 4+7 2 | 8 2 | 4 1 | 4 1 | 7 2 | 7 2 | 4+5 2 | 5 2 |
| | 5+6 2 | 9 2 | 5 1 | 5 1 | 9 2 | 9 2 | 4+5 2 | 5 2 |
| | 8+9 4 | 9 2 | 9 2 | 9 2 | 8 2 | 8 2 | 2+3 2 | 5 2 |
| | 8+9 4 | 8 2 | 8 2 | 8 2 | 7 2 | 7 2 | 4+5 2 | 5 2 |
| | 8+9 4 | 7 2 | 6 1 | 6 1 | 8 2 | 8 2 | 2+3 2 | 5 2 |
| | 4+7 2 | 7 2 | 7 1 | 7 1 | 7 2 | 7 2 | 2+3 2 | 5 2 |
| | 5+6 2 | 9 2 | 9 2 | 9 2 | 9 2 | 9 2 | 4+5 2 | 5 2 |
| | 8+9 4 | 8 2 | 8 2 | 8 2 | 8 2 | 8 2 | 2+3 2 | 5 2 |
| | 4+7 2 | 8 2 | 4 1 | 4 1 | 7 2 | 7 2 | 4+5 2 | 5 2 |
| | 5+6 2 | 7 2 | 5 1 | 5 1 | 9 2 | 9 2 | 2+3 2 | 5 2 |
| | 8+9 4 | 7 2 | 9 2 | 9 2 | 9 2 | 9 2 | 4+5 2 | 5 2 |
| | 4+7 2 | 8 2 | 9 2 | 9 2 | 7 2 | 7 2 | 4+5 2 | 5 2 |
| | 5+6 2 | 9 2 | 5 1 | 5 1 | 8 2 | 8 2 | 2+3 2 | 5 2 |
| | 8+9 4 | 8 2 | 4 1 | 4 1 | 7 2 | 7 2 | 2+3 2 | 5 2 |
| | 8+9 4 | 9 2 | 8 2 | 8 2 | 9 2 | 9 2 | 4+5 2 | 5 2 |
| | 8+9 4 | 8 2 | 6 1 | 6 1 | 8 2 | 8 2 | 2+3 2 | 5 2 |
| | 4+7 2 | 8 2 | 7 1 | 7 1 | 8 2 | 8 2 | 4+5 2 | 5 2 |
| | 5+6 2 | 7 2 | 8 2 | 8 2 | 8 2 | 8 2 | 2+3 2 | 5 2 |
| | 8+9 4 | 8 2 | 9 2 | 9 2 | 9 2 | 9 2 | 4+5 2 | 5 2 |
| | 8+9 4 | 7 2 | 5 1 | 5 1 | 7 2 | 7 2 | 4+5 2 | 5 2 |
| | 8+9 4 | 7 2 | 7 1 | 7 1 | 8 2 | 8 2 | 2+3 2 | 5 2 |
| | 4+7 2 | 7 2 | 9 2 | 9 2 | 7 2 | 7 2 | 2+3 2 | 5 2 |
| | 5+6 2 | 9 2 | 4 1 | 4 1 | 8 2 | 8 2 | 4+5 2 | 5 2 |
| | 8+9 4 | 8 2 | 6 1 | 6 1 | 9 2 | 9 2 | 4+5 2 | 5 2 |
| | 8+9 4 | 9 2 | 8 2 | 8 2 | 8 2 | 8 2 | 2+3 2 | 5 2 |
| | 4+7 2 | 7 2 | 8 2 | 8 2 | 9 2 | 9 2 | 4+5 2 | 5 2 |
| | 5+6 2 | 7 2 | 9 2 | 9 2 | 8 2 | 8 2 | 2+3 2 | 5 2 |
| | 8+9 4 | 8 2 | 8 2 | 8 2 | 7 2 | 7 2 | 2+3 2 | 5 2 |
| | 8+9 4 | 8 2 | 9 2 | 9 2 | 7 2 | 7 2 | 4+5 2 | 5 2 |
| | 4+7 2 | 8 2 | 6 1 | 6 1 | 9 2 | 9 2 | 2+3 2 | 5 2 |
| | 5+6 2 | 9 2 | 7 1 | 7 1 | 7 2 | 7 2 | 4+5 2 | 5 2 |
| | 8+9 4 | 8 2 | 4 1 | 4 1 | 8 2 | 8 2 | 2+3 2 | 5 2 |
| | 4+7 2 | 9 2 | 5 1 | 5 1 | 7 2 | 7 2 | 4+5 2 | 5 2 |
| | 5+6 2 | 7 2 | 9 2 | 9 2 | 9 2 | 9 2 | 2+3 2 | 5 2 |
| | 8+9 4 | 7 2 | 8 2 | 8 2 | 8 2 | 8 2 | 4+5 2 | 5 2 |
| | 4+7 2 | 9 2 | 9 2 | 9 2 | 9 2 | 9 2 | 4+5 2 | 5 2 |
| | 5+6 2 | 9 2 | 8 2 | 8 2 | 7 2 | 7 2 | 2+3 2 | 5 2 |
| | 8+9 4 | 9 2 | 9 2 | 9 2 | 9 2 | 9 2 | 2+3 2 | 5 2 |
| | 8+9 4 | 8 2 | 8 2 | 8 2 | 7 2 | 7 2 | 4+5 2 | 5 2 |
| | 4+7 2 | 7 2 | 4 1 | 4 1 | 8 2 | 8 2 | 4+5 2 | 5 2 |
| | 5+6 2 | 7 2 | 5 1 | 5 1 | 8 2 | 8 2 | 2+3 2 | 5 2 |
| | 8+9 4 | 8 2 | 6 1 | 6 1 | 9 2 | 9 2 | 2+3 2 | 5 2 |
| | 8+9 4 | 8 2 | 7 1 | 7 1 | 7 2 | 7 2 | 4+5 2 | 5 2 |
| | 8+9 4 | 9 2 | 5 1 | 5 1 | 8 2 | 8 2 | 2+3 2 | 5 2 |
| | 4+7 2 | 9 2 | 9 2 | 9 2 | 7 2 | 7 2 | 4+5 2 | 5 2 |
| | 5+6 2 | 8 2 | 4 1 | 4 1 | 8 2 | 8 2 | 4+5 2 | 5 2 |
| | 8+9 4 | 7 2 | 8 2 | 8 2 | 9 2 | 9 2 | 2+3 2 | 5 2 |
| | 4+7 2 | 7 2 | 7 1 | 7 1 | 7 2 | 7 2 | 4+5 2 | 5 2 |
| | 5+6 2 | 9 2 | 6 1 | 6 1 | 9 2 | 9 2 | 2+3 2 | 5 2 |
| | 8+9 4 | 8 2 | 8 2 | 8 2 | 8 2 | 8 2 | 2+3 2 | 5 2 |
| | 8+9 4 | 8 2 | 9 2 | 9 2 | 9 2 | 9 2 | 4+5 2 | 5 2 |
| | 4+7 2 | 8 2 | 6 1 | 6 1 | 7 2 | 7 2 | 4+5 2 | 5 2 |
| | 5+6 2 | 9 2 | 7 1 | 7 1 | 8 2 | 8 2 | 2+3 2 | 5 2 |
| | 8+9 4 | 7 2 | 4 1 | 4 1 | 8 2 | 8 2 | 2+3 2 | 5 2 |
| | | 9 2 | 5 1 | 5 1 | 7 2 | 7 2 | 4+5 2 | 5 2 |
| | | 7 2 | 8 2 | 8 2 | 9 2 | 9 2 | 4+5 2 | 5 2 |
| | | 7 2 | 9 2 | 9 2 | 8 2 | 8 2 | 2+3 2 | 5 2 |
| | | 7 2 | 8 2 | 8 2 | 9 2 | 9 2 | 4+5 2 | 5 2 |

Continued on next column

**TABLE LXXVII– continued**

| TQC | ICSD | NoE | SF | | | | | |
|---|---|---|---|---|---|---|---|---|
| | | 9 2 | 9 2 | 9 2 | 7 2 | 7 2 | 2+3 2 | 5 2 |
| | | 9 2 | 9 2 | 9 2 | 8 2 | 8 2 | 2+3 2 | 5 2 |
| | | 7 2 | 5 1 | 5 1 | 9 2 | 9 2 | 4+5 2 | 5 2 |
| | | 8 2 | 4 1 | 4 1 | 7 2 | 7 2 | 4+5 2 | 5 2 |
| | | 8 2 | 8 2 | 8 2 | 9 2 | 9 2 | 2+3 2 | 5 2 |
| | | 7 2 | 6 1 | 6 1 | 7 2 | 7 2 | 2+3 2 | 5 2 |
| | | 9 2 | 7 1 | 7 1 | 9 2 | 9 2 | 4+5 2 | 5 2 |
| | | 8 2 | 9 2 | 9 2 | 7 2 | 7 2 | 4+5 2 | 5 2 |
| | | 8 2 | 4 1 | 4 1 | 8 2 | 8 2 | 2+3 2 | 5 2 |
| | | 7 2 | 5 1 | 5 1 | 8 2 | 8 2 | 2+3 2 | 5 2 |
| | | 9 2 | 8 2 | 8 2 | 9 2 | 9 2 | 4+5 2 | 5 2 |
| | | 7 2 | 9 2 | 9 2 | 8 2 | 8 2 | 4+5 2 | 5 2 |
| | | 8 2 | 6 1 | 6 1 | 9 2 | 9 2 | 2+3 2 | 5 2 |
| | | 7 2 | 7 1 | 7 1 | 7 2 | 7 2 | 2+3 2 | 5 2 |
| | | 9 2 | 8 2 | 8 2 | 7 2 | 7 2 | 4+5 2 | 5 2 |
| | | 8 2 | 9 2 | 9 2 | 8 2 | 8 2 | 2+3 2 | 5 2 |
| | | 8 2 | 8 2 | 8 2 | 9 2 | 9 2 | 4+5 2 | 5 2 |
| | | 9 2 | 9 2 | 9 2 | 8 2 | 8 2 | 4+5 2 | 5 2 |
| | | 7 2 | 5 1 | 5 1 | 7 2 | 7 2 | 2+3 2 | 5 2 |
| | | 7 2 | 4 1 | 4 1 | 9 2 | 9 2 | 2+3 2 | 5 2 |
| | | 8 2 | 9 2 | 9 2 | 9 2 | 9 2 | 4+5 2 | 5 2 |
| | | 9 2 | 8 2 | 8 2 | 7 2 | 7 2 | 4+5 2 | 5 2 |
| | | 8 2 | 8 2 | 8 2 | 8 2 | 8 2 | 2+3 2 | 5 2 |
| | | 7 2 | 7 1 | 7 1 | 8 2 | 8 2 | 2+3 2 | 5 2 |
| | | 9 2 | 6 1 | 6 1 | 7 2 | 7 2 | 4+5 2 | 5 2 |
| | | 9 2 | 6 1 | 6 1 | 7 2 | 7 2 | 2+3 2 | 5 2 |
| | | 9 2 | 8 2 | 8 2 | 8 2 | 8 2 | 4+5 2 | 5 2 |
| | | 7 2 | 7 1 | 7 1 | 8 2 | 8 2 | 4+5 2 | 5 2 |
| | | 9 2 | 9 2 | 9 2 | 9 2 | 9 2 | 2+3 2 | 5 2 |
| | | 7 2 | 8 2 | 8 2 | 7 2 | 7 2 | 2+3 2 | 5 2 |
| | | 8 2 | 5 1 | 5 1 | 8 2 | 8 2 | 4+5 2 | 5 2 |
| | | | 4 1 | 4 1 | | | | |
| | | | 9 2 | 9 2 | | | | |
| | | | 8 2 | 8 2 | | | | |
| | | | 6 1 | 6 1 | | | | |
| | | | 7 1 | 7 1 | | | | |
| | | | 8 2 | 8 2 | | | | |
| | | | 9 2 | 9 2 | | | | |
| | | | 9 2 | 9 2 | | | | |
| | | | 8 2 | 8 2 | | | | |
| | | | 5 1 | 5 1 | | | | |
| | | | 4 1 | 4 1 | | | | |
| | | | 7 1 | 7 1 | | | | |
| | | | 6 1 | 6 1 | | | | |
| | | | 9 2 | 9 2 | | | | |
| | | | 4 1 | 4 1 | | | | |
| | | | 5 1 | 5 1 | | | | |
| | | | 9 2 | 9 2 | | | | |
| | | | 8 2 | 8 2 | | | | |
| | | | 8 2 | 8 2 | | | | |
| | | | 9 2 | 9 2 | | | | |
| | | | 8 2 | 8 2 | | | | |
| | | | 6 1 | 6 1 | | | | |
| | | | 7 1 | 7 1 | | | | |
| | | | 5 1 | 5 1 | | | | |
| | | | 4 1 | 4 1 | | | | |
| | | | 9 2 | 9 2 | | | | |
| | | | 6 1 | 6 1 | | | | |
| | | | 7 1 | 7 1 | | | | |
| | | | 8 2 | 8 2 | | | | |
| | | | 9 2 | 9 2 | | | | |

Continued on next column

**TABLE LXXVII– continued**

| TQC | ICSD | NoE | SF | | | | | |
|---|---|---|---|---|---|---|---|---|
| | | | 8 2 | 8 2 | | | | |
| | | | 8 2 | 8 2 | | | | |
| | | | 7 1 | 7 1 | | | | |
| | | | 6 1 | 6 1 | | | | |
| 190.2.2 | 73183 | 220 | Ba3Ca1O9Ru2 | | | | | |
| | -A | -GM | -H | -HA | -K | -KA | -L | -M |
| | 8+9 4 | 8 2 | 8 2 | 8 2 | 8 2 | 8 2 | 2+3 2 | 5 2 |
| | 8+9 4 | 9 2 | 9 2 | 9 2 | 9 2 | 9 2 | 4+5 2 | 5 2 |
| | 8+9 4 | 9 2 | 4 1 | 4 1 | 7 2 | 7 2 | 2+3 2 | 5 2 |
| | 8+9 4 | 8 2 | 5 1 | 5 1 | 8 2 | 8 2 | 4+5 2 | 5 2 |
| | 8+9 4 | 9 2 | 9 2 | 9 2 | 7 2 | 7 2 | 2+3 2 | 5 2 |
| | 8+9 4 | 8 2 | 6 1 | 6 1 | 9 2 | 9 2 | 4+5 2 | 5 2 |
| | 4+7 2 | 9 2 | 7 1 | 7 1 | 9 2 | 9 2 | 2+3 2 | 5 2 |
| | 5+6 2 | 9 2 | 8 2 | 8 2 | 9 2 | 9 2 | 4+5 2 | 5 2 |
| | 8+9 4 | 8 2 | 8 2 | 8 2 | 8 2 | 8 2 | 2+3 2 | 5 2 |
| | 8+9 4 | 9 2 | 9 2 | 9 2 | 9 2 | 9 2 | 4+5 2 | 5 2 |
| | 4+7 2 | 8 2 | 8 2 | 8 2 | 8 2 | 8 2 | 2+3 2 | 5 2 |
| | 5+6 2 | 9 2 | 9 2 | 9 2 | 7 2 | 7 2 | 4+5 2 | 5 2 |
| | 8+9 4 | 7 2 | 8 2 | 8 2 | 9 2 | 9 2 | 2+3 2 | 5 2 |
| | 8+9 4 | 7 2 | 9 2 | 9 2 | 7 2 | 7 2 | 4+5 2 | 5 2 |
| | 4+7 2 | 9 2 | 6 1 | 6 1 | 8 2 | 8 2 | 2+3 2 | 5 2 |
| | 5+6 2 | 8 2 | 7 1 | 7 1 | 7 2 | 7 2 | 4+5 2 | 5 2 |
| | 8+9 4 | 9 2 | 4 1 | 4 1 | 8 2 | 8 2 | 2+3 2 | 5 2 |
| | 8+9 4 | 8 2 | 5 1 | 5 1 | 9 2 | 9 2 | 4+5 2 | 5 2 |
| | 4+7 2 | 7 2 | 8 2 | 8 2 | 7 2 | 7 2 | 4+5 2 | 5 2 |
| | 5+6 2 | 9 2 | 6 1 | 6 1 | 9 2 | 9 2 | 2+3 2 | 5 2 |
| | 8+9 4 | 7 2 | 7 1 | 7 1 | 8 2 | 8 2 | 2+3 2 | 5 2 |
| | 8+9 4 | 8 2 | 8 2 | 8 2 | 7 2 | 7 2 | 4+5 2 | 5 2 |
| | 8+9 4 | 8 2 | 9 2 | 9 2 | 9 2 | 9 2 | 4+5 2 | 5 2 |
| | 4+7 2 | 9 2 | 4 1 | 4 1 | 7 2 | 7 2 | 2+3 2 | 5 2 |
| | 5+6 2 | 7 2 | 5 1 | 5 1 | 8 2 | 8 2 | 2+3 2 | 5 2 |
| | 8+9 4 | 7 2 | 9 2 | 9 2 | 9 2 | 9 2 | 4+5 2 | 5 2 |
| | 4+7 2 | 9 2 | 8 2 | 8 2 | 8 2 | 8 2 | 2+3 2 | 5 2 |
| | 5+6 2 | 9 2 | 6 1 | 6 1 | 7 2 | 7 2 | 4+5 2 | 5 2 |
| | 8+9 4 | 8 2 | 7 1 | 7 1 | 8 2 | 8 2 | 2+3 2 | 5 2 |
| | 8+9 4 | 8 2 | 9 2 | 9 2 | 7 2 | 7 2 | 4+5 2 | 5 2 |
| | 4+7 2 | 7 2 | 4 1 | 4 1 | 9 2 | 9 2 | 2+3 2 | 5 2 |
| | 5+6 2 | 7 2 | 5 1 | 5 1 | 9 2 | 9 2 | 4+5 2 | 5 2 |
| | 5+6 2 | 8 2 | 8 2 | 8 2 | 8 2 | 8 2 | 2+3 2 | 5 2 |
| | 4+7 2 | 9 2 | 9 2 | 9 2 | 7 2 | 7 2 | 4+5 2 | 5 2 |
| | 8+9 4 | 8 2 | 4 1 | 4 1 | 9 2 | 9 2 | 4+5 2 | 5 2 |
| | 8+9 4 | 9 2 | 5 1 | 5 1 | 7 2 | 7 2 | 2+3 2 | 5 2 |
| | 8+9 4 | 8 2 | 8 2 | 8 2 | 8 2 | 8 2 | 2+3 2 | 5 2 |
| | 4+7 2 | 8 2 | 9 2 | 9 2 | 8 2 | 8 2 | 4+5 2 | 5 2 |
| | 5+6 2 | 7 2 | 6 1 | 6 1 | 8 2 | 8 2 | 4+5 2 | 5 2 |
| | 8+9 4 | 7 2 | 7 1 | 7 1 | 9 2 | 9 2 | 2+3 2 | 5 2 |
| | 8+9 4 | 8 2 | 9 2 | 9 2 | 7 2 | 7 2 | 2+3 2 | 5 2 |
| | 4+7 2 | 7 2 | 8 2 | 8 2 | 8 2 | 8 2 | 4+5 2 | 5 2 |
| | 5+6 2 | 9 2 | 9 2 | 9 2 | 7 2 | 7 2 | 4+5 2 | 5 2 |
| | 8+9 4 | 7 2 | 7 1 | 7 1 | 9 2 | 9 2 | 2+3 2 | 5 2 |
| | 5+6 2 | 8 2 | 5 1 | 5 1 | 8 2 | 8 2 | 2+3 2 | 5 2 |
| | 4+7 2 | 9 2 | 8 2 | 8 2 | 9 2 | 9 2 | 4+5 2 | 5 2 |
| | 8+9 4 | 8 2 | 6 1 | 6 1 | 8 2 | 8 2 | 2+3 2 | 5 2 |
| | 8+9 4 | 7 2 | 4 1 | 4 1 | 8 2 | 8 2 | 4+5 2 | 5 2 |
| | 8+9 4 | 7 2 | 8 2 | 8 2 | 7 2 | 7 2 | 2+3 2 | 5 2 |
| | 8+9 4 | 8 2 | 9 2 | 9 2 | 7 2 | 7 2 | 4+5 2 | 5 2 |
| | 4+7 2 | 7 2 | 8 2 | 8 2 | 8 2 | 8 2 | 2+3 2 | 5 2 |
| | 5+6 2 | 9 2 | 9 2 | 9 2 | 9 2 | 9 2 | 4+5 2 | 5 2 |
| | 8+9 4 | 9 2 | 6 1 | 6 1 | 9 2 | 9 2 | 2+3 2 | 5 2 |
| | 4+7 2 | 7 2 | 4 1 | 4 1 | 7 2 | 7 2 | 4+5 2 | 5 2 |
| | 5+6 2 | 9 2 | 7 1 | 7 1 | 8 2 | 8 2 | 4+5 2 | 5 2 |

Continued on next column

**TABLE LXXVII– continued**

| TQC | ICSD | NoE | SF | | |
|---|---|---|---|---|---|
| | 8+9 4 | 8 2 | 8 2 | 8 2 7 2 | 7 2 2+3 2 5 2 |
| | 8+9 4 | 7 2 | 5 1 | 5 1 9 2 | 9 2 2+3 2 5 2 |
| | 4+7 2 | 9 2 | 9 2 | 9 2 8 2 | 8 2 4+5 2 5 2 |
| | 5+6 2 | 8 2 | 9 2 | 9 2 9 2 | 9 2 2+3 2 5 2 |
| | 8+9 4 | 8 2 | 8 2 | 8 2 8 2 | 8 2 4+5 2 5 2 |
| | 8+9 4 | 7 2 | 9 2 | 9 2 7 2 | 7 2 2+3 2 5 2 |
| | 8+9 4 | 8 2 | 8 2 | 8 2 7 2 | 7 2 4+5 2 5 2 |
| | 4+7 2 | 8 2 | 4 1 | 4 1 8 2 | 8 2 4+5 2 5 2 |
| | 5+6 2 | 7 2 | 5 1 | 5 1 9 2 | 9 2 2+3 2 5 2 |
| | 8+9 4 | 7 2 | 6 1 | 6 1 7 2 | 7 2 2+3 2 5 2 |
| | 4+7 2 | 8 2 | 7 1 | 7 1 9 2 | 9 2 4+5 2 5 2 |
| | 5+6 2 | 9 2 | 9 2 | 9 2 8 2 | 8 2 2+3 2 5 2 |
| | 8+9 4 | 9 2 | 8 2 | 8 2 7 2 | 7 2 4+5 2 5 2 |
| | 4+7 2 | 9 2 | 8 2 | 8 2 9 2 | 9 2 4+5 2 5 2 |
| | 8+9 4 | 7 2 | 6 1 | 6 1 8 2 | 8 2 2+3 2 5 2 |
| | 5+6 2 | 7 2 | 7 1 | 7 1 7 2 | 7 2 2+3 2 5 2 |
| | 8+9 4 | 8 2 | 9 2 | 9 2 8 2 | 8 2 4+5 2 5 2 |
| | | 9 2 | 4 1 | 4 1 7 2 | 7 2 2+3 2 5 2 |
| | | 8 2 | 5 1 | 5 1 9 2 | 9 2 4+5 2 5 2 |
| | | 9 2 | 9 2 | 9 2 8 2 | 8 2 2+3 2 5 2 |
| | | 9 2 | 8 2 | 8 2 7 2 | 7 2 4+5 2 5 2 |
| | | 9 2 | 9 2 | 9 2 9 2 | 9 2 4+5 2 5 2 |
| | | 7 2 | 4 1 | 4 1 8 2 | 8 2 2+3 2 5 2 |
| | | 9 2 | 5 1 | 5 1 7 2 | 7 2 2+3 2 5 2 |
| | | 7 2 | 8 2 | 8 2 7 2 | 7 2 4+5 2 5 2 |
| | | 8 2 | 6 1 | 6 1 8 2 | 8 2 2+3 2 5 2 |
| | | 8 2 | 7 1 | 7 1 9 2 | 9 2 4+5 2 5 2 |
| | | 7 2 | 8 2 | 8 2 8 2 | 8 2 4+5 2 5 2 |
| | | 7 2 | 9 2 | 9 2 9 2 | 9 2 2+3 2 5 2 |
| | | 9 2 | 5 1 | 5 1 7 2 | 7 2 2+3 2 5 2 |
| | | 8 2 | 4 1 | 4 1 8 2 | 8 2 4+5 2 5 2 |
| | | 8 2 | 9 2 | 9 2 9 2 | 9 2 2+3 2 5 2 |
| | | 9 2 | 7 1 | 7 1 7 2 | 7 2 4+5 2 5 2 |
| | | 8 2 | 6 1 | 6 1 9 2 | 9 2 2+3 2 5 2 |
| | | 7 2 | 8 2 | 8 2 8 2 | 8 2 4+5 2 5 2 |
| | | 8 2 | 9 2 | 9 2 9 2 | 9 2 4+5 2 5 2 |
| | | 7 2 | 8 2 | 8 2 7 2 | 7 2 2+3 2 5 2 |
| | | 8 2 | 7 1 | 7 1 7 2 | 7 2 2+3 2 5 2 |
| | | 9 2 | 6 1 | 6 1 9 2 | 9 2 4+5 2 5 2 |
| | | 8 2 | 8 2 | 8 2 8 2 | 8 2 2+3 2 5 2 |
| | | 7 2 | 5 1 | 5 1 7 2 | 7 2 4+5 2 5 2 |
| | | 9 2 | 9 2 | 9 2 9 2 | 9 2 2+3 2 5 2 |
| | | 7 2 | 4 1 | 4 1 8 2 | 8 2 4+5 2 5 2 |
| | | 8 2 | 6 1 | 6 1 9 2 | 9 2 2+3 2 5 2 |
| | | 8 2 | 7 1 | 7 1 9 2 | 9 2 4+5 2 5 2 |
| | | 7 2 | 8 2 | 8 2 7 2 | 7 2 2+3 2 5 2 |
| | | 7 2 | 8 2 | 8 2 8 2 | 8 2 4+5 2 5 2 |
| | | 9 2 | 9 2 | 9 2 8 2 | 8 2 2+3 2 5 2 |
| | | 9 2 | 4 1 | 4 1 7 2 | 7 2 4+5 2 5 2 |
| | | 7 2 | 5 1 | 5 1 8 2 | 8 2 2+3 2 5 2 |
| | | 9 2 | 9 2 | 9 2 9 2 | 9 2 4+5 2 5 2 |
| | | 9 2 | 7 1 | 7 1 7 2 | 7 2 4+5 2 5 2 |
| | | 8 2 | 8 2 | 8 2 8 2 | 8 2 2+3 2 5 2 |
| | | 9 2 | 6 1 | 6 1 7 2 | 7 2 4+5 2 5 2 |
| | | 7 2 | 8 2 | 8 2 8 2 | 8 2 2+3 2 5 2 |
| | | | 9 2 | 9 2 | |
| | | | 6 1 | 6 1 | |
| | | | 7 1 | 7 1 | |
| | | | 8 2 | 8 2 | |
| | | | 5 1 | 5 1 | |
| | | | 4 1 | 4 1 | |

Continued on next column

**TABLE LXXVII– continued**

| TQC | ICSD | NoE | SF |
|---|---|---|---|
| | | 9 2 | 9 2 |
| | | 8 2 | 8 2 |
| | | 9 2 | 9 2 |
| | | 9 2 | 9 2 |
| | | 4 1 | 4 1 |
| | | 5 1 | 5 1 |
| | | 4 1 | 4 1 |
| | | 5 1 | 5 1 |
| | | 9 2 | 9 2 |
| | | 7 1 | 7 1 |
| | | 6 1 | 6 1 |
| | | 8 2 | 8 2 |
| | | 8 2 | 8 2 |
| | | 9 2 | 9 2 |
| | | 9 2 | 9 2 |
| | | 8 2 | 8 2 |
| | | 4 1 | 4 1 |
| | | 5 1 | 5 1 |
| | | 9 2 | 9 2 |
| | | 8 2 | 8 2 |
| | | 6 1 | 6 1 |
| | | 7 1 | 7 1 |
| | | 8 2 | 8 2 |
| | | 9 2 | 9 2 |
| | | 6 1 | 6 1 |
| | | 7 1 | 7 1 |
| | | 8 2 | 8 2 |
| | | 9 2 | 9 2 |
| | | 4 1 | 4 1 |
| | | 5 1 | 5 1 |

The End

TABLE LXXVIII: The computed irreducible representations for the valence bands for the compounds of SEBR and NLC in space group 191. See the list of tables.

| TQC | ICSD | | NoE | SF | | |
|---|---|---|---|---|---|---|
| 191.1.1 | 173575 | | 160 | La10O3Si8 | | |
| | -A | -GM | -H | -K | -L | -M |
| | 9 2 | 9 2 | 7 2 | 7 2 | 6 2 | 6 2 |
| | 10 2 | 10 2 | 8 2 | 8 2 | 5 2 | 5 2 |
| | 12 2 | 12 2 | 9 2 | 9 2 | 5 2 | 6 2 |
| | 11 2 | 11 2 | 7 2 | 7 2 | 6 2 | 5 2 |
| | 7 2 | 7 2 | 8 2 | 8 2 | 5 2 | 5 2 |
| | 8 2 | 8 2 | 9 2 | 9 2 | 6 2 | 6 2 |
| | 9 2 | 9 2 | 9 2 | 9 2 | 5 2 | 5 2 |
| | 10 2 | 7 2 | 7 2 | 9 2 | 5 2 | 6 2 |
| | 11 2 | 8 2 | 8 2 | 7 2 | 5 2 | 5 2 |
| | 12 2 | 9 2 | 9 2 | 8 2 | 6 2 | 6 2 |
| | 7 2 | 7 2 | 9 2 | 9 2 | 6 2 | 6 2 |
| | 8 2 | 8 2 | 8 2 | 8 2 | 5 2 | 5 2 |
| | 9 2 | 9 2 | 7 2 | 7 2 | 6 2 | 6 2 |
| | 10 2 | 11 2 | 8 2 | 8 2 | 5 2 | 5 2 |
| | 11 2 | 10 2 | 7 2 | 7 2 | 6 2 | 6 2 |
| | 9 2 | 12 2 | 9 2 | 9 2 | 5 2 | 5 2 |
| | 8 2 | 9 2 | 8 2 | 7 2 | 6 2 | 6 2 |
| | 7 2 | 8 2 | 9 2 | 8 2 | 6 2 | 5 2 |
| | 12 2 | 12 2 | 7 2 | 8 2 | 5 2 | 6 2 |
| | 12 2 | 7 2 | 8 2 | 9 2 | 6 2 | 6 2 |
| | 9 2 | 11 2 | 7 2 | 8 2 | 6 2 | 5 2 |

Continued on next column

**TABLE LXXVIII– continued**

| TQC | ICSD | | NoE | | SF | |
|---|---|---|---|---|---|---|
| | 8 2 | 10 2 | 8 2 | 9 2 | 5 2 | 6 2 |
| | 7 2 | 12 2 | 9 2 | 7 2 | 6 2 | 5 2 |
| | 9 2 | 12 2 | 7 2 | 7 2 | 6 2 | 5 2 |
| | 12 2 | 11 2 | 8 2 | 9 2 | 5 2 | 6 2 |
| | 11 2 | 10 2 | 7 2 | 8 2 | 5 2 | 5 2 |
| | 8 2 | 9 2 | 9 2 | 8 2 | 6 2 | 5 2 |
| | 10 2 | 10 2 | 8 2 | 7 2 | 6 2 | 6 2 |
| | 11 2 | 12 2 | 8 2 | 9 2 | 5 2 | 6 2 |
| | 12 2 | 11 2 | 7 2 | 8 2 | 6 2 | 6 2 |
| | 7 2 | 8 2 | 9 2 | 7 2 | 5 2 | 5 2 |
| | 9 2 | 7 2 | 9 2 | 7 2 | 5 2 | 5 2 |
| | 10 2 | 7 2 | 7 2 | 9 2 | 6 2 | 6 2 |
| | 7 2 | 9 2 | 8 2 | 8 2 | 6 2 | 5 2 |
| | 12 2 | 10 2 | 9 2 | 9 2 | 5 2 | 6 2 |
| | 8 2 | 12 2 | 8 2 | 8 2 | 6 2 | 6 2 |
| | 10 2 | 8 2 | 7 2 | 7 2 | 6 2 | 6 2 |
| | 7 2 | 11 2 | 9 2 | 8 2 | 6 2 | 5 2 |
| | 8 2 | 10 2 | 8 2 | 9 2 | 5 2 | 6 2 |
| | 9 2 | 12 2 | 7 2 | 7 2 | 6 2 | 5 2 |
| | 9 2 | 12 2 | 7 2 | 7 2 | 6 2 | 5 2 |
| | 8 2 | 11 2 | 8 2 | 9 2 | 5 2 | 6 2 |
| | 7 2 | 10 2 | 9 2 | 8 2 | 6 2 | 5 2 |
| | 12 2 | 9 2 | 8 2 | 9 2 | 6 2 | 5 2 |
| | 7 2 | 12 2 | 7 2 | 8 2 | 5 2 | 6 2 |
| | 9 2 | 10 2 | 9 2 | 7 2 | 5 2 | 6 2 |
| | 12 2 | 9 2 | 7 2 | 8 2 | 6 2 | 5 2 |
| | 10 2 | 8 2 | 9 2 | 7 2 | 5 2 | 6 2 |
| | 11 2 | 7 2 | 9 2 | 8 2 | 6 2 | 5 2 |
| | 8 2 | 11 2 | 7 2 | 7 2 | 6 2 | 5 2 |
| | 10 2 | 10 2 | 8 2 | 9 2 | 5 2 | 5 2 |
| | 12 2 | 12 2 | 7 2 | 8 2 | 6 2 | 6 2 |
| | 10 2 | 11 2 | 9 2 | 7 2 | 5 2 | 5 2 |
| | 11 2 | 12 2 | 9 2 | 9 2 | 5 2 | 6 2 |
| | 12 2 | 10 2 | 7 2 | 9 2 | 6 2 | 5 2 |
| | 11 2 | 11 2 | 8 2 | 7 2 | 5 2 | 6 2 |
| | 10 2 | 11 2 | 9 2 | 8 2 | 5 2 | 5 2 |
| | 12 2 | 9 2 | 8 2 | 9 2 | 6 2 | 6 2 |
| | 11 2 | 11 2 | 7 2 | 9 2 | 5 2 | 5 2 |
| | 8 2 | 10 2 | 8 2 | 8 2 | 5 2 | 6 2 |
| | 9 2 | 12 2 | 8 2 | 7 2 | 5 2 | 5 2 |
| | 8 2 | 11 2 | 9 2 | 9 2 | 5 2 | 5 2 |
| | 12 2 | 8 2 | 8 2 | 8 2 | 6 2 | 6 2 |
| | 10 2 | 7 2 | 7 2 | 7 2 | 6 2 | 5 2 |
| | 11 2 | 12 2 | 9 2 | 7 2 | 5 2 | 6 2 |
| | 12 2 | 7 2 | 7 2 | 9 2 | 5 2 | 5 2 |
| | 10 2 | 9 2 | 8 2 | 8 2 | 6 2 | 6 2 |
| | 11 2 | 10 2 | 8 2 | 9 2 | 5 2 | 5 2 |
| | 8 2 | 12 2 | 7 2 | 8 2 | 6 2 | 5 2 |
| | 10 2 | 7 2 | 8 2 | 7 2 | 5 2 | 6 2 |
| | 9 2 | 8 2 | 7 2 | 9 2 | 5 2 | 6 2 |
| | 11 2 | 8 2 | 9 2 | 7 2 | 6 2 | 5 2 |
| | 7 2 | 8 2 | 9 2 | 7 2 | 6 2 | 6 2 |
| | 7 2 | 9 2 | 8 2 | 8 2 | 5 2 | 6 2 |
| | 9 2 | 11 2 | 7 2 | 9 2 | 6 2 | 5 2 |
| | 7 2 | 10 2 | 7 2 | 7 2 | 5 2 | 5 2 |
| | 9 2 | 12 2 | 9 2 | 9 2 | 6 2 | 5 2 |
| | 11 2 | 7 2 | 8 2 | 8 2 | 6 2 | 5 2 |
| | 12 2 | 8 2 | 9 2 | 7 2 | 5 2 | 6 2 |
| | 11 2 | 12 2 | 7 2 | 8 2 | 5 2 | 5 2 |
| 191.1.2 | 417224 | | 16 | | Ga2Sr1 | |
| | -A | -GM | -H | -K | -L | -M |
| | 9 2 | 9 2 | 9 2 | 9 2 | 5 2 | 5 2 |
| | 12 2 | 12 2 | 8 2 | 8 2 | 6 2 | 6 2 |
| | 12 2 | 12 2 | 7 2 | 7 2 | 6 2 | 6 2 |
| | 10 2 | 10 2 | 8 2 | 8 2 | 6 2 | 6 2 |
| | 12 2 | 9 2 | 7 2 | 8 2 | 5 2 | 6 2 |
| | 9 2 | 12 2 | 9 2 | 7 2 | 6 2 | 5 2 |
| | 11 2 | 8 2 | 8 2 | 9 2 | 5 2 | 6 2 |
| | 10 2 | 7 2 | 8 2 | 9 2 | 6 2 | 5 2 |
| 191.1.3 | 260654 | | 16 | | Ga2Sr1 | |
| | -A | -GM | -H | -K | -L | -M |
| | 9 2 | 9 2 | 9 2 | 9 2 | 5 2 | 5 2 |
| | 12 2 | 12 2 | 8 2 | 8 2 | 6 2 | 6 2 |
| | 12 2 | 12 2 | 7 2 | 7 2 | 6 2 | 6 2 |
| | 10 2 | 10 2 | 8 2 | 8 2 | 6 2 | 6 2 |
| | 12 2 | 9 2 | 7 2 | 8 2 | 5 2 | 6 2 |
| | 9 2 | 12 2 | 9 2 | 7 2 | 6 2 | 5 2 |
| | 11 2 | 8 2 | 8 2 | 9 2 | 5 2 | 6 2 |
| | 10 2 | 7 2 | 8 2 | 9 2 | 6 2 | 5 2 |
| 191.1.4 | 246236 | | 16 | | Ga2Sr1 | |
| | -A | -GM | -H | -K | -L | -M |
| | 9 2 | 9 2 | 9 2 | 9 2 | 5 2 | 5 2 |
| | 12 2 | 12 2 | 8 2 | 8 2 | 6 2 | 6 2 |
| | 12 2 | 12 2 | 7 2 | 7 2 | 6 2 | 6 2 |
| | 10 2 | 10 2 | 8 2 | 8 2 | 6 2 | 6 2 |
| | 12 2 | 9 2 | 7 2 | 8 2 | 5 2 | 6 2 |
| | 9 2 | 12 2 | 9 2 | 7 2 | 6 2 | 5 2 |
| | 11 2 | 8 2 | 8 2 | 9 2 | 5 2 | 6 2 |
| | 10 2 | 7 2 | 8 2 | 9 2 | 6 2 | 5 2 |
| 191.1.5 | 107425 | | 24 | | Ga2Gd1 | |
| | -A | -GM | -H | -K | -L | -M |
| | 9 2 | 9 2 | 9 2 | 9 2 | 5 2 | 5 2 |
| | 12 2 | 12 2 | 8 2 | 8 2 | 6 2 | 6 2 |
| | 12 2 | 12 2 | 8 2 | 7 2 | 6 2 | 6 2 |
| | 10 2 | 10 2 | 7 2 | 8 2 | 6 2 | 6 2 |
| | 12 2 | 9 2 | 7 2 | 8 2 | 5 2 | 6 2 |
| | 9 2 | 12 2 | 9 2 | 7 2 | 6 2 | 5 2 |
| | 10 2 | 11 2 | 8 2 | 9 2 | 5 2 | 6 2 |
| | 11 2 | 8 2 | 8 2 | 7 2 | 5 2 | 6 2 |
| | 12 2 | 10 2 | 7 2 | 8 2 | 6 2 | 6 2 |
| | 11 2 | 7 2 | 9 2 | 9 2 | 6 2 | 5 2 |
| | 10 2 | 12 2 | 8 2 | 9 2 | 6 2 | 6 2 |
| | 11 2 | 11 2 | 7 2 | 7 2 | 6 2 | 5 2 |
| 191.1.6 | 419163 | | 68 | | Ce1Pd2Zn3 | |
| | -A | -GM | -H | -K | -L | -M |
| | 9 2 | 9 2 | 9 2 | 9 2 | 5 2 | 5 2 |
| | 12 2 | 12 2 | 8 2 | 8 2 | 6 2 | 6 2 |
| | 12 2 | 10 2 | 8 2 | 7 2 | 6 2 | 6 2 |
| | 10 2 | 12 2 | 7 2 | 8 2 | 6 2 | 6 2 |
| | 12 2 | 9 2 | 8 2 | 8 2 | 5 2 | 6 2 |
| | 11 2 | 7 2 | 9 2 | 7 2 | 5 2 | 6 2 |
| | 10 2 | 8 2 | 7 2 | 9 2 | 5 2 | 6 2 |
| | 11 2 | 9 2 | 8 2 | 9 2 | 6 2 | 5 2 |
| | 10 2 | 8 2 | 9 2 | 7 2 | 5 2 | 5 2 |
| | 12 2 | 7 2 | 7 2 | 8 2 | 6 2 | 6 2 |
| | 11 2 | 9 2 | 7 2 | 7 2 | 5 2 | 6 2 |
| | 12 2 | 9 2 | 8 2 | 8 2 | 6 2 | 5 2 |
| | 10 2 | 8 2 | 7 2 | 9 2 | 5 2 | 5 2 |
| | 12 2 | 7 2 | 9 2 | 7 2 | 6 2 | 6 2 |
| | 10 2 | 9 2 | 9 2 | 8 2 | 5 2 | 6 2 |
| | 11 2 | 7 2 | 7 2 | 7 2 | 6 2 | 5 2 |
| | 11 2 | 8 2 | 8 2 | 8 2 | 5 2 | 6 2 |

Continued on next column

**TABLE LXXVIII– continued**

| TQC | ICSD | | NoE | SF | | |
|---|---|---|---|---|---|---|
| | 10 2 | 8 2 | 9 2 | 9 2 | 5 2 | 6 2 |
| | 12 2 | 7 2 | 8 2 | 9 2 | 5 2 | 6 2 |
| | 12 2 | 9 2 | 9 2 | 8 2 | 5 2 | 6 2 |
| | 9 2 | 11 2 | 7 2 | 7 2 | 6 2 | 5 2 |
| | 11 2 | 11 2 | 8 2 | 9 2 | 5 2 | 5 2 |
| | 10 2 | 9 2 | 8 2 | 9 2 | 5 2 | 5 2 |
| | 9 2 | 10 2 | 7 2 | 8 2 | 6 2 | 6 2 |
| | 7 2 | 7 2 | 9 2 | 8 2 | 6 2 | 6 2 |
| | 7 2 | 8 2 | 7 2 | 7 2 | 6 2 | 5 2 |
| | 9 2 | 10 2 | 8 2 | 7 2 | 5 2 | 6 2 |
| | 8 2 | 12 2 | 8 2 | 9 2 | 5 2 | 5 2 |
| | 10 2 | 9 2 | 9 2 | 8 2 | 6 2 | 5 2 |
| | 12 2 | 7 2 | 7 2 | 8 2 | 6 2 | 6 2 |
| | 11 2 | 7 2 | 9 2 | 9 2 | 5 2 | 6 2 |
| | 8 2 | 8 2 | 8 2 | 7 2 | 5 2 | 6 2 |
| | 7 2 | 11 2 | 7 2 | 7 2 | 6 2 | 5 2 |
| | 8 2 | 12 2 | 9 2 | 9 2 | 6 2 | 6 2 |
| 191.1.7 | 418614 | | 62 | Ca1Zn5 | | |
| | -A | -GM | -H | -K | -L | -M |
| | 12 2 | 9 2 | 7 2 | 8 2 | 5 2 | 6 2 |
| | 11 2 | 9 2 | 8 2 | 7 2 | 5 2 | 5 2 |
| | 11 2 | 7 2 | 8 2 | 8 2 | 5 2 | 6 2 |
| | 10 2 | 9 2 | 9 2 | 7 2 | 6 2 | 6 2 |
| | 10 2 | 7 2 | 7 2 | 9 2 | 6 2 | 5 2 |
| | 11 2 | 8 2 | 9 2 | 9 2 | 5 2 | 6 2 |
| | 10 2 | 8 2 | 8 2 | 7 2 | 5 2 | 5 2 |
| | 9 2 | 7 2 | 9 2 | 8 2 | 6 2 | 6 2 |
| | 12 2 | 11 2 | 8 2 | 8 2 | 6 2 | 5 2 |
| | 7 2 | 10 2 | 7 2 | 9 2 | 5 2 | 6 2 |
| | 12 2 | 9 2 | 7 2 | 7 2 | 6 2 | 6 2 |
| | 9 2 | 8 2 | 7 2 | 7 2 | 5 2 | 5 2 |
| | 10 2 | 11 2 | 9 2 | 8 2 | 5 2 | 5 2 |
| | 8 2 | 10 2 | 9 2 | 8 2 | 6 2 | 6 2 |
| | 7 2 | 12 2 | 8 2 | 9 2 | 6 2 | 5 2 |
| | 12 2 | 9 2 | 8 2 | 9 2 | 5 2 | 6 2 |
| | 11 2 | 7 2 | 9 2 | 8 2 | 5 2 | 6 2 |
| | 10 2 | 9 2 | 7 2 | 7 2 | 6 2 | 6 2 |
| | 11 2 | 8 2 | 8 2 | 9 2 | 5 2 | 5 2 |
| | 12 2 | 7 2 | 9 2 | 8 2 | 6 2 | 5 2 |
| | 11 2 | 8 2 | 7 2 | 7 2 | 5 2 | 6 2 |
| | 10 2 | 7 2 | 8 2 | 9 2 | 6 2 | 5 2 |
| | 12 2 | 8 2 | 8 2 | 7 2 | 5 2 | 6 2 |
| | 11 2 | 9 2 | 9 2 | 9 2 | 5 2 | 6 2 |
| | 9 2 | 7 2 | 7 2 | 8 2 | 5 2 | 6 2 |
| | 10 2 | 9 2 | 7 2 | 8 2 | 5 2 | 6 2 |
| | 12 2 | 11 2 | 9 2 | 7 2 | 6 2 | 5 2 |
| | 8 2 | 8 2 | 7 2 | 9 2 | 6 2 | 6 2 |
| | 8 2 | 7 2 | 8 2 | 7 2 | 5 2 | 5 2 |
| | 7 2 | 12 2 | 8 2 | 8 2 | 5 2 | 5 2 |
| | 7 2 | 10 2 | 9 2 | 9 2 | 6 2 | 5 2 |
| 191.1.8 | 107305 | | 30 | Ba1Sn5 | | |
| | -A | -GM | -H | -K | -L | -M |
| | 12 2 | 9 2 | 8 2 | 9 2 | 6 2 | 5 2 |
| | 9 2 | 12 2 | 9 2 | 8 2 | 5 2 | 6 2 |
| | 9 2 | 12 2 | 9 2 | 7 2 | 5 2 | 6 2 |
| | 7 2 | 10 2 | 7 2 | 8 2 | 5 2 | 6 2 |
| | 9 2 | 9 2 | 7 2 | 8 2 | 6 2 | 6 2 |
| | 12 2 | 12 2 | 8 2 | 7 2 | 5 2 | 5 2 |
| | 9 2 | 9 2 | 9 2 | 9 2 | 5 2 | 5 2 |
| | 11 2 | 11 2 | 9 2 | 7 2 | 6 2 | 5 2 |
| | 8 2 | 9 2 | 7 2 | 9 2 | 5 2 | 6 2 |

Continued on next column

**TABLE LXXVIII– continued**

| TQC | ICSD | | NoE | SF | | |
|---|---|---|---|---|---|---|
| | 8 2 | 8 2 | 8 2 | 8 2 | 6 2 | 6 2 |
| | 7 2 | 8 2 | 9 2 | 8 2 | 6 2 | 6 2 |
| | 12 2 | 7 2 | 8 2 | 7 2 | 5 2 | 5 2 |
| | 10 2 | 12 2 | 7 2 | 9 2 | 6 2 | 5 2 |
| | 11 2 | 10 2 | 7 2 | 7 2 | 5 2 | 6 2 |
| | 12 2 | 11 2 | 8 2 | 8 2 | 5 2 | 5 2 |
| 191.1.9 | 246238 | | 16 | Ba1Ga2 | | |
| | -A | -GM | -H | -K | -L | -M |
| | 9 2 | 9 2 | 9 2 | 9 2 | 5 2 | 5 2 |
| | 12 2 | 12 2 | 8 2 | 8 2 | 6 2 | 6 2 |
| | 12 2 | 12 2 | 7 2 | 7 2 | 6 2 | 6 2 |
| | 10 2 | 10 2 | 8 2 | 8 2 | 6 2 | 6 2 |
| | 12 2 | 9 2 | 7 2 | 8 2 | 5 2 | 6 2 |
| | 9 2 | 12 2 | 9 2 | 7 2 | 6 2 | 5 2 |
| | 11 2 | 8 2 | 8 2 | 9 2 | 5 2 | 5 2 |
| | 10 2 | 7 2 | 8 2 | 9 2 | 5 2 | 6 2 |
| 191.1.10 | 97723 | | 150 | B4Co11Pr3 | | |
| | -A | -GM | -H | -K | -L | -M |
| | 12 2 | 9 2 | 8 2 | 9 2 | 6 2 | 5 2 |
| | 9 2 | 12 2 | 9 2 | 8 2 | 5 2 | 6 2 |
| | 9 2 | 9 2 | 9 2 | 9 2 | 5 2 | 5 2 |
| | 12 2 | 9 2 | 8 2 | 9 2 | 6 2 | 5 2 |
| | 9 2 | 12 2 | 9 2 | 8 2 | 5 2 | 6 2 |
| | 12 2 | 12 2 | 8 2 | 8 2 | 6 2 | 6 2 |
| | 12 2 | 9 2 | 8 2 | 9 2 | 6 2 | 5 2 |
| | 7 2 | 12 2 | 7 2 | 8 2 | 5 2 | 6 2 |
| | 10 2 | 10 2 | 7 2 | 7 2 | 6 2 | 6 2 |
| | 10 2 | 7 2 | 7 2 | 7 2 | 6 2 | 5 2 |
| | 9 2 | 10 2 | 9 2 | 7 2 | 5 2 | 6 2 |
| | 12 2 | 12 2 | 8 2 | 8 2 | 6 2 | 6 2 |
| | 12 2 | 9 2 | 9 2 | 8 2 | 5 2 | 6 2 |
| | 9 2 | 12 2 | 7 2 | 7 2 | 6 2 | 5 2 |
| | 11 2 | 8 2 | 8 2 | 9 2 | 5 2 | 6 2 |
| | 8 2 | 9 2 | 7 2 | 7 2 | 6 2 | 5 2 |
| | 12 2 | 11 2 | 8 2 | 8 2 | 5 2 | 6 2 |
| | 10 2 | 7 2 | 9 2 | 7 2 | 5 2 | 5 2 |
| | 11 2 | 8 2 | 7 2 | 9 2 | 6 2 | 6 2 |
| | 9 2 | 9 2 | 8 2 | 8 2 | 5 2 | 6 2 |
| | 7 2 | 10 2 | 7 2 | 9 2 | 6 2 | 5 2 |
| | 8 2 | 11 2 | 9 2 | 7 2 | 6 2 | 5 2 |
| | 9 2 | 12 2 | 9 2 | 8 2 | 5 2 | 6 2 |
| | 11 2 | 8 2 | 7 2 | 7 2 | 5 2 | 5 2 |
| | 12 2 | 9 2 | 8 2 | 8 2 | 6 2 | 5 2 |
| | 12 2 | 11 2 | 7 2 | 7 2 | 5 2 | 6 2 |
| | 10 2 | 9 2 | 7 2 | 9 2 | 5 2 | 6 2 |
| | 8 2 | 7 2 | 9 2 | 9 2 | 5 2 | 6 2 |
| | 12 2 | 9 2 | 8 2 | 7 2 | 6 2 | 5 2 |
| | 11 2 | 8 2 | 8 2 | 8 2 | 6 2 | 6 2 |
| | 10 2 | 7 2 | 7 2 | 9 2 | 5 2 | 5 2 |
| | 7 2 | 7 2 | 9 2 | 8 2 | 5 2 | 5 2 |
| | 9 2 | 11 2 | 9 2 | 7 2 | 6 2 | 5 2 |
| | 11 2 | 8 2 | 8 2 | 9 2 | 5 2 | 6 2 |
| | 9 2 | 12 2 | 7 2 | 7 2 | 5 2 | 5 2 |
| | 10 2 | 10 2 | 9 2 | 8 2 | 6 2 | 5 2 |
| | 11 2 | 7 2 | 8 2 | 7 2 | 6 2 | 6 2 |
| | 7 2 | 9 2 | 7 2 | 8 2 | 6 2 | 6 2 |
| | 8 2 | 12 2 | 8 2 | 9 2 | 5 2 | 6 2 |
| | 12 2 | 11 2 | 7 2 | 7 2 | 6 2 | 5 2 |
| | 11 2 | 10 2 | 9 2 | 8 2 | 6 2 | 6 2 |
| | 10 2 | 9 2 | 9 2 | 9 2 | 5 2 | 5 2 |
| | 12 2 | 7 2 | 8 2 | 9 2 | 6 2 | 6 2 |

Continued on next column

**TABLE LXXVIII– continued**

| TQC | ICSD | | NoE | SF | | |
|---|---|---|---|---|---|---|
| | 10 2 | 8 2 | 7 2 | 8 2 | 5 2 | 6 2 |
| | 7 2 | 11 2 | 9 2 | 7 2 | 6 2 | 6 2 |
| | 8 2 | 10 2 | 8 2 | 9 2 | 5 2 | 5 2 |
| | 12 2 | 12 2 | 8 2 | 8 2 | 6 2 | 5 2 |
| | 10 2 | 9 2 | 9 2 | 7 2 | 5 2 | 5 2 |
| | 9 2 | 10 2 | 7 2 | 7 2 | 6 2 | 5 2 |
| | 11 2 | 7 2 | 7 2 | 9 2 | 5 2 | 6 2 |
| | 9 2 | 8 2 | 8 2 | 9 2 | 5 2 | 6 2 |
| | 7 2 | 9 2 | 9 2 | 7 2 | 6 2 | 6 2 |
| | 7 2 | 8 2 | 7 2 | 8 2 | 6 2 | 6 2 |
| | 10 2 | 7 2 | 8 2 | 8 2 | 5 2 | 6 2 |
| | 9 2 | 9 2 | 9 2 | 8 2 | 5 2 | 6 2 |
| | 8 2 | 8 2 | 8 2 | 9 2 | 6 2 | 5 2 |
| | 12 2 | 7 2 | 9 2 | 7 2 | 6 2 | 5 2 |
| | 11 2 | 10 2 | 7 2 | 8 2 | 5 2 | 6 2 |
| | 12 2 | 11 2 | 9 2 | 7 2 | 5 2 | 5 2 |
| | 10 2 | 7 2 | 7 2 | 7 2 | 5 2 | 5 2 |
| | 10 2 | 12 2 | 7 2 | 9 2 | 6 2 | 5 2 |
| | 7 2 | 8 2 | 8 2 | 8 2 | 6 2 | 6 2 |
| | 8 2 | 11 2 | 9 2 | 9 2 | 5 2 | 5 2 |
| | 8 2 | 12 2 | 8 2 | 8 2 | 5 2 | 6 2 |
| | 11 2 | 10 2 | 7 2 | 9 2 | 5 2 | 6 2 |
| | 9 2 | 9 2 | 9 2 | 7 2 | 6 2 | 6 2 |
| | 7 2 | 7 2 | 8 2 | 9 2 | 5 2 | 5 2 |
| | 11 2 | 8 2 | 9 2 | 9 2 | 6 2 | 6 2 |
| | 12 2 | 7 2 | 8 2 | 9 2 | 6 2 | 5 2 |
| | 10 2 | 12 2 | 8 2 | 8 2 | 6 2 | 6 2 |
| | 7 2 | 10 2 | 7 2 | 7 2 | 5 2 | 5 2 |
| | 11 2 | 8 2 | 9 2 | 8 2 | 6 2 | 5 2 |
| | 8 2 | 11 2 | 7 2 | 7 2 | 5 2 | 6 2 |
| | 12 2 | 9 2 | 9 2 | 9 2 | 6 2 | 6 2 |
| | 9 2 | 12 2 | 9 2 | 8 2 | 5 2 | 5 2 |
| 191.1.11 | 97731 | | 224 | B3Co7Dy2 | | |
| | -A | -GM | -H | -K | -L | -M |
| | 9 2 | 9 2 | 9 2 | 9 2 | 5 2 | 5 2 |
| | 12 2 | 12 2 | 8 2 | 8 2 | 6 2 | 6 2 |
| | 9 2 | 9 2 | 9 2 | 9 2 | 5 2 | 5 2 |
| | 12 2 | 9 2 | 8 2 | 9 2 | 6 2 | 5 2 |
| | 12 2 | 12 2 | 8 2 | 8 2 | 6 2 | 6 2 |
| | 9 2 | 9 2 | 9 2 | 9 2 | 5 2 | 5 2 |
| | 12 2 | 12 2 | 8 2 | 8 2 | 6 2 | 6 2 |
| | 9 2 | 12 2 | 9 2 | 8 2 | 5 2 | 6 2 |
| | 9 2 | 12 2 | 9 2 | 8 2 | 5 2 | 6 2 |
| | 12 2 | 9 2 | 8 2 | 7 2 | 6 2 | 5 2 |
| | 7 2 | 10 2 | 7 2 | 9 2 | 5 2 | 6 2 |
| | 10 2 | 12 2 | 7 2 | 8 2 | 6 2 | 6 2 |
| | 7 2 | 7 2 | 7 2 | 7 2 | 5 2 | 5 2 |
| | 10 2 | 10 2 | 7 2 | 7 2 | 6 2 | 6 2 |
| | 9 2 | 10 2 | 9 2 | 7 2 | 5 2 | 6 2 |
| | 12 2 | 12 2 | 8 2 | 8 2 | 6 2 | 6 2 |
| | 12 2 | 9 2 | 9 2 | 8 2 | 5 2 | 6 2 |
| | 9 2 | 12 2 | 7 2 | 7 2 | 6 2 | 5 2 |
| | 12 2 | 9 2 | 8 2 | 9 2 | 6 2 | 5 2 |
| | 8 2 | 11 2 | 7 2 | 7 2 | 5 2 | 6 2 |
| | 9 2 | 12 2 | 9 2 | 8 2 | 5 2 | 6 2 |
| | 11 2 | 8 2 | 7 2 | 7 2 | 6 2 | 5 2 |
| | 8 2 | 11 2 | 8 2 | 9 2 | 6 2 | 5 2 |
| | 10 2 | 9 2 | 8 2 | 9 2 | 6 2 | 6 2 |
| | 11 2 | 7 2 | 7 2 | 7 2 | 5 2 | 5 2 |
| | 9 2 | 8 2 | 9 2 | 8 2 | 5 2 | 6 2 |
| | 12 2 | 9 2 | 9 2 | 7 2 | 5 2 | 5 2 |

Continued on next column

**TABLE LXXVIII– continued**

| TQC | ICSD | | NoE | SF | | |
|---|---|---|---|---|---|---|
| | 7 2 | 10 2 | 7 2 | 8 2 | 5 2 | 6 2 |
| | 8 2 | 11 2 | 8 2 | 9 2 | 6 2 | 6 2 |
| | 9 2 | 12 2 | 8 2 | 8 2 | 6 2 | 5 2 |
| | 10 2 | 7 2 | 7 2 | 7 2 | 6 2 | 5 2 |
| | 11 2 | 8 2 | 9 2 | 9 2 | 6 2 | 5 2 |
| | 11 2 | 8 2 | 7 2 | 7 2 | 6 2 | 6 2 |
| | 9 2 | 11 2 | 8 2 | 9 2 | 5 2 | 5 2 |
| | 7 2 | 12 2 | 8 2 | 8 2 | 5 2 | 6 2 |
| | 12 2 | 9 2 | 7 2 | 9 2 | 5 2 | 6 2 |
| | 8 2 | 10 2 | 9 2 | 7 2 | 6 2 | 5 2 |
| | 12 2 | 9 2 | 9 2 | 9 2 | 5 2 | 6 2 |
| | 11 2 | 9 2 | 7 2 | 8 2 | 5 2 | 5 2 |
| | 9 2 | 7 2 | 8 2 | 7 2 | 5 2 | 6 2 |
| | 7 2 | 10 2 | 7 2 | 7 2 | 5 2 | 5 2 |
| | 9 2 | 11 2 | 8 2 | 8 2 | 5 2 | 5 2 |
| | 7 2 | 12 2 | 9 2 | 8 2 | 6 2 | 5 2 |
| | 8 2 | 8 2 | 7 2 | 7 2 | 5 2 | 5 2 |
| | 12 2 | 11 2 | 8 2 | 9 2 | 6 2 | 6 2 |
| | 10 2 | 9 2 | 8 2 | 8 2 | 6 2 | 6 2 |
| | 7 2 | 7 2 | 7 2 | 7 2 | 6 2 | 6 2 |
| | 8 2 | 8 2 | 7 2 | 9 2 | 5 2 | 5 2 |
| | 12 2 | 12 2 | 9 2 | 7 2 | 6 2 | 5 2 |
| | 10 2 | 10 2 | 9 2 | 9 2 | 5 2 | 6 2 |
| | 11 2 | 7 2 | 9 2 | 9 2 | 6 2 | 6 2 |
| | 11 2 | 8 2 | 8 2 | 8 2 | 6 2 | 5 2 |
| | 9 2 | 10 2 | 7 2 | 8 2 | 5 2 | 6 2 |
| | 10 2 | 11 2 | 8 2 | 7 2 | 6 2 | 6 2 |
| | 11 2 | 12 2 | 8 2 | 9 2 | 6 2 | 6 2 |
| | 11 2 | 11 2 | 9 2 | 7 2 | 6 2 | 5 2 |
| | 10 2 | 10 2 | 7 2 | 9 2 | 6 2 | 5 2 |
| | 12 2 | 9 2 | 9 2 | 8 2 | 5 2 | 5 2 |
| | 10 2 | 7 2 | 8 2 | 7 2 | 5 2 | 6 2 |
| | 7 2 | 7 2 | 7 2 | 8 2 | 6 2 | 5 2 |
| | 8 2 | 8 2 | 8 2 | 9 2 | 5 2 | 5 2 |
| | 10 2 | 12 2 | 9 2 | 8 2 | 6 2 | 6 2 |
| | 9 2 | 10 2 | 7 2 | 7 2 | 5 2 | 6 2 |
| | 12 2 | 12 2 | 9 2 | 9 2 | 6 2 | 5 2 |
| | 7 2 | 10 2 | 8 2 | 7 2 | 5 2 | 5 2 |
| | 8 2 | 11 2 | 7 2 | 8 2 | 5 2 | 6 2 |
| | 9 2 | 12 2 | 8 2 | 8 2 | 6 2 | 6 2 |
| | 12 2 | 9 2 | 7 2 | 7 2 | 5 2 | 6 2 |
| | 7 2 | 8 2 | 9 2 | 9 2 | 6 2 | 6 2 |
| | 7 2 | 10 2 | 9 2 | 9 2 | 6 2 | 5 2 |
| | 9 2 | 9 2 | 7 2 | 7 2 | 5 2 | 5 2 |
| | 11 2 | 11 2 | 9 2 | 9 2 | 6 2 | 6 2 |
| | 8 2 | 7 2 | 8 2 | 8 2 | 5 2 | 5 2 |
| | 10 2 | 10 2 | 7 2 | 7 2 | 5 2 | 6 2 |
| | 7 2 | 7 2 | 8 2 | 9 2 | 6 2 | 6 2 |
| | 11 2 | 12 2 | 9 2 | 7 2 | 5 2 | 6 2 |
| | 12 2 | 11 2 | 7 2 | 8 2 | 6 2 | 5 2 |
| | 10 2 | 10 2 | 7 2 | 7 2 | 5 2 | 6 2 |
| | 9 2 | 12 2 | 9 2 | 8 2 | 5 2 | 5 2 |
| | 8 2 | 8 2 | 8 2 | 9 2 | 5 2 | 5 2 |
| | 9 2 | 9 2 | 7 2 | 8 2 | 6 2 | 6 2 |
| | 12 2 | 12 2 | 9 2 | 7 2 | 6 2 | 6 2 |
| | 10 2 | 7 2 | 8 2 | 8 2 | 6 2 | 5 2 |
| | 9 2 | 11 2 | 9 2 | 7 2 | 5 2 | 5 2 |
| | 7 2 | 9 2 | 8 2 | 8 2 | 5 2 | 6 2 |
| | 8 2 | 8 2 | 7 2 | 9 2 | 6 2 | 6 2 |
| | 11 2 | 7 2 | 9 2 | 9 2 | 6 2 | 6 2 |
| | 11 2 | 8 2 | 8 2 | 9 2 | 5 2 | 5 2 |

Continued on next column

**TABLE LXXVIII– continued**

| TQC | ICSD | | NoE | | SF | |
|---|---|---|---|---|---|---|
| | 8 2 | 11 2 | 9 2 | 8 2 | 6 2 | 6 2 |
| | 12 2 | 7 2 | 8 2 | 9 2 | 5 2 | 5 2 |
| | 7 2 | 9 2 | 9 2 | 8 2 | 5 2 | 6 2 |
| | 9 2 | 11 2 | 7 2 | 9 2 | 6 2 | 6 2 |
| | 8 2 | 10 2 | 8 2 | 7 2 | 6 2 | 6 2 |
| | 10 2 | 11 2 | 7 2 | 8 2 | 5 2 | 5 2 |
| | 11 2 | 8 2 | 9 2 | 7 2 | 5 2 | 6 2 |
| | 8 2 | 9 2 | 9 2 | 9 2 | 5 2 | 5 2 |
| | 10 2 | 7 2 | 7 2 | 7 2 | 6 2 | 6 2 |
| | 7 2 | 10 2 | 7 2 | 8 2 | 6 2 | 5 2 |
| | 12 2 | 12 2 | 8 2 | 7 2 | 6 2 | 6 2 |
| | 10 2 | 10 2 | 8 2 | 9 2 | 5 2 | 5 2 |
| | 11 2 | 8 2 | 9 2 | 9 2 | 6 2 | 6 2 |
| | 7 2 | 11 2 | 9 2 | 8 2 | 5 2 | 6 2 |
| | 8 2 | 7 2 | 7 2 | 8 2 | 6 2 | 6 2 |
| | 11 2 | 12 2 | 8 2 | 7 2 | 5 2 | 5 2 |
| | 12 2 | 10 2 | 8 2 | 9 2 | 5 2 | 6 2 |
| | 10 2 | 12 2 | 7 2 | 7 2 | 6 2 | 5 2 |
| | 9 2 | 8 2 | 9 2 | 8 2 | 5 2 | 6 2 |
| | 12 2 | 7 2 | 8 2 | 9 2 | 6 2 | 5 2 |
| | 7 2 | 11 2 | 9 2 | 7 2 | 5 2 | 6 2 |
| | 10 2 | 10 2 | 7 2 | 8 2 | 5 2 | 6 2 |
| | 11 2 | 9 2 | 8 2 | 9 2 | 6 2 | 5 2 |
| | 9 2 | 12 2 | 8 2 | 9 2 | 6 2 | 6 2 |
| 191.1.12 | 78848 | | 10 | | B2Ti1 | |
| | -A | -GM | -H | -K | -L | -M |
| | 12 2 | 9 2 | 7 2 | 8 2 | 5 2 | 5 2 |
| | 9 2 | 7 2 | 9 2 | 7 2 | 6 2 | 6 2 |
| | 11 2 | 8 2 | 8 2 | 9 2 | 5 2 | 5 2 |
| | 10 2 | 12 2 | 8 2 | 9 2 | 5 2 | 6 2 |
| | 9 2 | 9 2 | 7 2 | 7 2 | 6 2 | 5 2 |
| 191.1.13 | 78847 | | 10 | | B2Ti1 | |
| | -A | -GM | -H | -K | -L | -M |
| | 12 2 | 9 2 | 7 2 | 8 2 | 5 2 | 5 2 |
| | 9 2 | 7 2 | 9 2 | 7 2 | 6 2 | 6 2 |
| | 11 2 | 8 2 | 8 2 | 9 2 | 5 2 | 6 2 |
| | 10 2 | 12 2 | 8 2 | 9 2 | 5 2 | 5 2 |
| | 9 2 | 9 2 | 7 2 | 7 2 | 6 2 | 5 2 |
| 191.1.14 | 56723 | | 10 | | B2Ti1 | |
| | -A | -GM | -H | -K | -L | -M |
| | 12 2 | 9 2 | 7 2 | 8 2 | 5 2 | 5 2 |
| | 9 2 | 7 2 | 9 2 | 7 2 | 6 2 | 6 2 |
| | 11 2 | 8 2 | 8 2 | 9 2 | 5 2 | 6 2 |
| | 10 2 | 12 2 | 8 2 | 9 2 | 5 2 | 5 2 |
| | 9 2 | 9 2 | 7 2 | 7 2 | 6 2 | 5 2 |
| 191.1.15 | 184968 | | 44 | | B2Fe3U1 | |
| | -A | -GM | -H | -K | -L | -M |
| | 9 2 | 9 2 | 9 2 | 9 2 | 5 2 | 5 2 |
| | 12 2 | 12 2 | 8 2 | 8 2 | 6 2 | 6 2 |
| | 12 2 | 10 2 | 8 2 | 7 2 | 6 2 | 6 2 |
| | 10 2 | 12 2 | 7 2 | 8 2 | 6 2 | 6 2 |
| | 9 2 | 9 2 | 8 2 | 8 2 | 5 2 | 6 2 |
| | 11 2 | 11 2 | 7 2 | 7 2 | 6 2 | 5 2 |
| | 12 2 | 7 2 | 9 2 | 9 2 | 5 2 | 6 2 |
| | 12 2 | 8 2 | 7 2 | 8 2 | 5 2 | 5 2 |
| | 10 2 | 8 2 | 9 2 | 9 2 | 6 2 | 6 2 |
| | 7 2 | 9 2 | 9 2 | 7 2 | 5 2 | 6 2 |
| | 10 2 | 9 2 | 7 2 | 9 2 | 5 2 | 5 2 |
| | 11 2 | 8 2 | 8 2 | 7 2 | 5 2 | 6 2 |
| | 8 2 | 7 2 | 7 2 | 9 2 | 6 2 | 5 2 |
| | 8 2 | 7 2 | 9 2 | 8 2 | 6 2 | 5 2 |

Continued on next column

**TABLE LXXVIII– continued**

| TQC | ICSD | | NoE | | SF | |
|---|---|---|---|---|---|---|
| | 12 2 | 9 2 | 8 2 | 7 2 | 5 2 | 6 2 |
| | 9 2 | 7 2 | 7 2 | 8 2 | 6 2 | 5 2 |
| | 7 2 | 8 2 | 9 2 | 7 2 | 6 2 | 6 2 |
| | 11 2 | 12 2 | 7 2 | 8 2 | 5 2 | 6 2 |
| | 10 2 | 9 2 | 8 2 | 7 2 | 6 2 | 6 2 |
| | 12 2 | 7 2 | 8 2 | 9 2 | 5 2 | 5 2 |
| | 11 2 | 10 2 | 9 2 | 8 2 | 5 2 | 6 2 |
| | 10 2 | 8 2 | 9 2 | 9 2 | 6 2 | 6 2 |
| 191.1.16 | 237660 | | 134 | | B1Eu1Pt4 | |
| | -A | -GM | -H | -K | -L | -M |
| | 9 2 | 9 2 | 9 2 | 9 2 | 5 2 | 5 2 |
| | 12 2 | 9 2 | 8 2 | 9 2 | 6 2 | 5 2 |
| | 9 2 | 12 2 | 9 2 | 8 2 | 5 2 | 6 2 |
| | 12 2 | 12 2 | 8 2 | 8 2 | 6 2 | 6 2 |
| | 7 2 | 12 2 | 7 2 | 7 2 | 5 2 | 6 2 |
| | 9 2 | 10 2 | 9 2 | 8 2 | 5 2 | 6 2 |
| | 12 2 | 12 2 | 8 2 | 8 2 | 6 2 | 6 2 |
| | 10 2 | 10 2 | 7 2 | 7 2 | 6 2 | 6 2 |
| | 12 2 | 9 2 | 7 2 | 7 2 | 5 2 | 6 2 |
| | 12 2 | 9 2 | 9 2 | 8 2 | 6 2 | 5 2 |
| | 8 2 | 11 2 | 8 2 | 8 2 | 6 2 | 5 2 |
| | 7 2 | 10 2 | 7 2 | 7 2 | 5 2 | 6 2 |
| | 12 2 | 9 2 | 9 2 | 9 2 | 5 2 | 6 2 |
| | 10 2 | 7 2 | 9 2 | 8 2 | 5 2 | 6 2 |
| | 10 2 | 7 2 | 8 2 | 9 2 | 6 2 | 5 2 |
| | 9 2 | 8 2 | 7 2 | 7 2 | 5 2 | 6 2 |
| | 11 2 | 11 2 | 9 2 | 8 2 | 6 2 | 5 2 |
| | 8 2 | 12 2 | 8 2 | 8 2 | 6 2 | 5 2 |
| | 7 2 | 10 2 | 9 2 | 7 2 | 6 2 | 5 2 |
| | 9 2 | 12 2 | 7 2 | 9 2 | 5 2 | 6 2 |
| | 11 2 | 8 2 | 8 2 | 8 2 | 6 2 | 5 2 |
| | 10 2 | 11 2 | 7 2 | 7 2 | 6 2 | 5 2 |
| | 11 2 | 7 2 | 9 2 | 9 2 | 5 2 | 6 2 |
| | 9 2 | 9 2 | 7 2 | 9 2 | 6 2 | 6 2 |
| | 8 2 | 9 2 | 8 2 | 7 2 | 5 2 | 5 2 |
| | 12 2 | 8 2 | 8 2 | 8 2 | 5 2 | 5 2 |
| | 7 2 | 10 2 | 7 2 | 9 2 | 6 2 | 6 2 |
| | 9 2 | 12 2 | 9 2 | 7 2 | 6 2 | 5 2 |
| | 8 2 | 11 2 | 8 2 | 8 2 | 5 2 | 6 2 |
| | 9 2 | 12 2 | 7 2 | 7 2 | 5 2 | 5 2 |
| | 7 2 | 11 2 | 7 2 | 9 2 | 6 2 | 6 2 |
| | 10 2 | 7 2 | 9 2 | 9 2 | 5 2 | 5 2 |
| | 12 2 | 10 2 | 8 2 | 8 2 | 5 2 | 6 2 |
| | 8 2 | 9 2 | 8 2 | 7 2 | 6 2 | 5 2 |
| | 10 2 | 10 2 | 9 2 | 9 2 | 5 2 | 6 2 |
| | 11 2 | 7 2 | 9 2 | 8 2 | 6 2 | 6 2 |
| | 7 2 | 11 2 | 8 2 | 8 2 | 5 2 | 5 2 |
| | 8 2 | 9 2 | 7 2 | 7 2 | 6 2 | 6 2 |
| | 11 2 | 11 2 | 7 2 | 9 2 | 6 2 | 5 2 |
| | 10 2 | 7 2 | 9 2 | 7 2 | 6 2 | 6 2 |
| | 9 2 | 8 2 | 8 2 | 7 2 | 5 2 | 5 2 |
| | 11 2 | 12 2 | 7 2 | 9 2 | 6 2 | 6 2 |
| | 9 2 | 8 2 | 8 2 | 8 2 | 5 2 | 5 2 |
| | 10 2 | 10 2 | 8 2 | 9 2 | 5 2 | 6 2 |
| | 7 2 | 9 2 | 9 2 | 8 2 | 5 2 | 6 2 |
| | 7 2 | 7 2 | 9 2 | 7 2 | 5 2 | 5 2 |
| | 11 2 | 8 2 | 7 2 | 8 2 | 6 2 | 6 2 |
| | 12 2 | 12 2 | 8 2 | 7 2 | 6 2 | 5 2 |
| | 12 2 | 10 2 | 9 2 | 9 2 | 6 2 | 6 2 |
| | 10 2 | 11 2 | 7 2 | 8 2 | 5 2 | 5 2 |
| | 8 2 | 7 2 | 8 2 | 8 2 | 5 2 | 6 2 |

Continued on next column

**TABLE LXXVIII– continued**

| TQC | ICSD | | NoE | SF | | |
|---|---|---|---|---|---|---|
| | 7 2 | 9 2 | 8 2 | 9 2 | 6 2 | 5 2 |
| | 8 2 | 8 2 | 7 2 | 9 2 | 5 2 | 6 2 |
| | 11 2 | 12 2 | 9 2 | 7 2 | 5 2 | 5 2 |
| | 12 2 | 11 2 | 9 2 | 9 2 | 6 2 | 6 2 |
| | 10 2 | 12 2 | 7 2 | 8 2 | 6 2 | 5 2 |
| | 9 2 | 10 2 | 8 2 | 7 2 | 6 2 | 6 2 |
| | 8 2 | 11 2 | 7 2 | 9 2 | 5 2 | 5 2 |
| | 7 2 | 12 2 | 9 2 | 7 2 | 5 2 | 6 2 |
| | 9 2 | 10 2 | 7 2 | 8 2 | 6 2 | 5 2 |
| | 12 2 | 11 2 | 8 2 | 7 2 | 5 2 | 6 2 |
| | 11 2 | 10 2 | 7 2 | 9 2 | 6 2 | 6 2 |
| | 9 2 | 10 2 | 9 2 | 8 2 | 5 2 | 6 2 |
| | 10 2 | 11 2 | 8 2 | 9 2 | 6 2 | 6 2 |
| | 7 2 | 12 2 | 9 2 | 7 2 | 5 2 | 5 2 |
| | 8 2 | 12 2 | 7 2 | 8 2 | 6 2 | 6 2 |
| | 12 2 | 10 2 | 8 2 | 7 2 | 6 2 | 6 2 |
| 191.1.17 | 97722 | | 104 | B1Co4Pr1 | | |
| | -A | -GM | -H | -K | -L | -M |
| | 12 2 | 9 2 | 8 2 | 9 2 | 6 2 | 5 2 |
| | 9 2 | 9 2 | 9 2 | 9 2 | 5 2 | 5 2 |
| | 9 2 | 12 2 | 9 2 | 8 2 | 5 2 | 6 2 |
| | 12 2 | 12 2 | 8 2 | 8 2 | 6 2 | 6 2 |
| | 9 2 | 12 2 | 7 2 | 8 2 | 5 2 | 6 2 |
| | 7 2 | 10 2 | 9 2 | 7 2 | 5 2 | 6 2 |
| | 12 2 | 10 2 | 8 2 | 7 2 | 6 2 | 6 2 |
| | 10 2 | 12 2 | 7 2 | 8 2 | 6 2 | 6 2 |
| | 12 2 | 9 2 | 9 2 | 8 2 | 5 2 | 6 2 |
| | 9 2 | 11 2 | 7 2 | 7 2 | 6 2 | 5 2 |
| | 8 2 | 12 2 | 8 2 | 9 2 | 6 2 | 5 2 |
| | 10 2 | 9 2 | 7 2 | 7 2 | 5 2 | 5 2 |
| | 11 2 | 7 2 | 8 2 | 9 2 | 6 2 | 6 2 |
| | 9 2 | 8 2 | 9 2 | 8 2 | 5 2 | 6 2 |
| | 11 2 | 9 2 | 7 2 | 7 2 | 5 2 | 6 2 |
| | 12 2 | 8 2 | 8 2 | 8 2 | 5 2 | 5 2 |
| | 9 2 | 9 2 | 7 2 | 7 2 | 6 2 | 5 2 |
| | 7 2 | 11 2 | 8 2 | 9 2 | 6 2 | 5 2 |
| | 8 2 | 11 2 | 8 2 | 9 2 | 5 2 | 6 2 |
| | 7 2 | 10 2 | 9 2 | 7 2 | 5 2 | 5 2 |
| | 12 2 | 12 2 | 7 2 | 9 2 | 6 2 | 5 2 |
| | 11 2 | 10 2 | 9 2 | 8 2 | 5 2 | 5 2 |
| | 9 2 | 7 2 | 8 2 | 8 2 | 5 2 | 6 2 |
| | 10 2 | 9 2 | 7 2 | 7 2 | 6 2 | 6 2 |
| | 11 2 | 7 2 | 7 2 | 9 2 | 5 2 | 6 2 |
| | 7 2 | 8 2 | 9 2 | 7 2 | 6 2 | 5 2 |
| | 8 2 | 12 2 | 7 2 | 8 2 | 6 2 | 6 2 |
| | 12 2 | 11 2 | 8 2 | 9 2 | 6 2 | 6 2 |
| | 10 2 | 10 2 | 8 2 | 8 2 | 6 2 | 5 2 |
| | 11 2 | 12 2 | 9 2 | 7 2 | 6 2 | 5 2 |
| | 10 2 | 7 2 | 9 2 | 7 2 | 5 2 | 6 2 |
| | 9 2 | 10 2 | 7 2 | 9 2 | 5 2 | 5 2 |
| | 8 2 | 8 2 | 9 2 | 8 2 | 6 2 | 5 2 |
| | 7 2 | 9 2 | 9 2 | 7 2 | 6 2 | 6 2 |
| | 9 2 | 12 2 | 7 2 | 9 2 | 5 2 | 5 2 |
| | 10 2 | 7 2 | 8 2 | 9 2 | 5 2 | 6 2 |
| | 12 2 | 9 2 | 8 2 | 8 2 | 5 2 | 6 2 |
| | 7 2 | 11 2 | 7 2 | 8 2 | 5 2 | 6 2 |
| | 9 2 | 7 2 | 8 2 | 9 2 | 6 2 | 6 2 |
| | 8 2 | 10 2 | 7 2 | 8 2 | 6 2 | 5 2 |
| | 7 2 | 8 2 | 9 2 | 7 2 | 6 2 | 6 2 |
| | 10 2 | 11 2 | 9 2 | 7 2 | 5 2 | 5 2 |
| | 11 2 | 8 2 | 8 2 | 7 2 | 5 2 | 5 2 |

Continued on next column

**TABLE LXXVIII– continued**

| TQC | ICSD | | NoE | SF | | |
|---|---|---|---|---|---|---|
| | 8 2 | 9 2 | 7 2 | 9 2 | 5 2 | 6 2 |
| | 11 2 | 7 2 | 9 2 | 8 2 | 6 2 | 5 2 |
| | 12 2 | 8 2 | 9 2 | 9 2 | 6 2 | 6 2 |
| | 12 2 | 12 2 | 9 2 | 8 2 | 5 2 | 6 2 |
| | 10 2 | 10 2 | 8 2 | 9 2 | 6 2 | 5 2 |
| | 7 2 | 11 2 | 7 2 | 8 2 | 5 2 | 6 2 |
| | 8 2 | 7 2 | 8 2 | 7 2 | 5 2 | 6 2 |
| | 11 2 | 12 2 | 7 2 | 7 2 | 6 2 | 5 2 |
| | 9 2 | 8 2 | 8 2 | 9 2 | 6 2 | 5 2 |
| 191.1.18 | 417869 | | 114 | B1Co4Gd1 | | |
| | -A | -GM | -H | -K | -L | -M |
| | 9 2 | 9 2 | 9 2 | 9 2 | 5 2 | 5 2 |
| | 12 2 | 9 2 | 8 2 | 9 2 | 6 2 | 5 2 |
| | 12 2 | 12 2 | 8 2 | 8 2 | 6 2 | 6 2 |
| | 9 2 | 12 2 | 9 2 | 8 2 | 5 2 | 6 2 |
| | 9 2 | 12 2 | 7 2 | 8 2 | 5 2 | 6 2 |
| | 12 2 | 10 2 | 9 2 | 7 2 | 5 2 | 6 2 |
| | 7 2 | 10 2 | 8 2 | 7 2 | 6 2 | 6 2 |
| | 10 2 | 12 2 | 7 2 | 8 2 | 6 2 | 6 2 |
| | 12 2 | 9 2 | 9 2 | 8 2 | 5 2 | 6 2 |
| | 9 2 | 11 2 | 7 2 | 7 2 | 6 2 | 5 2 |
| | 8 2 | 12 2 | 8 2 | 9 2 | 6 2 | 5 2 |
| | 10 2 | 9 2 | 7 2 | 7 2 | 5 2 | 5 2 |
| | 11 2 | 7 2 | 8 2 | 9 2 | 6 2 | 6 2 |
| | 9 2 | 8 2 | 9 2 | 8 2 | 5 2 | 6 2 |
| | 11 2 | 9 2 | 8 2 | 7 2 | 5 2 | 6 2 |
| | 12 2 | 8 2 | 7 2 | 9 2 | 5 2 | 5 2 |
| | 9 2 | 9 2 | 7 2 | 8 2 | 6 2 | 6 2 |
| | 7 2 | 11 2 | 9 2 | 7 2 | 6 2 | 5 2 |
| | 12 2 | 11 2 | 8 2 | 9 2 | 5 2 | 5 2 |
| | 8 2 | 10 2 | 8 2 | 8 2 | 5 2 | 5 2 |
| | 7 2 | 12 2 | 7 2 | 7 2 | 5 2 | 5 2 |
| | 11 2 | 10 2 | 9 2 | 9 2 | 5 2 | 5 2 |
| | 9 2 | 7 2 | 8 2 | 8 2 | 6 2 | 6 2 |
| | 10 2 | 9 2 | 7 2 | 7 2 | 6 2 | 6 2 |
| | 11 2 | 7 2 | 7 2 | 9 2 | 6 2 | 6 2 |
| | 7 2 | 8 2 | 9 2 | 7 2 | 5 2 | 5 2 |
| | 8 2 | 12 2 | 7 2 | 8 2 | 6 2 | 6 2 |
| | 12 2 | 11 2 | 8 2 | 9 2 | 6 2 | 6 2 |
| | 10 2 | 10 2 | 9 2 | 8 2 | 6 2 | 6 2 |
| | 11 2 | 12 2 | 8 2 | 7 2 | 6 2 | 5 2 |
| | 10 2 | 10 2 | 9 2 | 7 2 | 6 2 | 5 2 |
| | 9 2 | 7 2 | 8 2 | 9 2 | 5 2 | 5 2 |
| | 8 2 | 8 2 | 7 2 | 8 2 | 5 2 | 5 2 |
| | 7 2 | 12 2 | 9 2 | 7 2 | 6 2 | 6 2 |
| | 9 2 | 9 2 | 9 2 | 9 2 | 5 2 | 5 2 |
| | 10 2 | 11 2 | 7 2 | 9 2 | 5 2 | 6 2 |
| | 12 2 | 9 2 | 8 2 | 8 2 | 5 2 | 6 2 |
| | 9 2 | 7 2 | 7 2 | 8 2 | 6 2 | 6 2 |
| | 7 2 | 10 2 | 7 2 | 8 2 | 5 2 | 6 2 |
| | 8 2 | 12 2 | 8 2 | 7 2 | 6 2 | 6 2 |
| | 7 2 | 11 2 | 9 2 | 9 2 | 6 2 | 5 2 |
| | 12 2 | 10 2 | 8 2 | 7 2 | 5 2 | 5 2 |
| | 10 2 | 12 2 | 9 2 | 7 2 | 5 2 | 5 2 |
| | 11 2 | 11 2 | 7 2 | 8 2 | 5 2 | 6 2 |
| | 9 2 | 7 2 | 9 2 | 9 2 | 6 2 | 6 2 |
| | 8 2 | 8 2 | 8 2 | 9 2 | 6 2 | 6 2 |
| | 11 2 | 8 2 | 9 2 | 8 2 | 6 2 | 6 2 |
| | 10 2 | 9 2 | 9 2 | 9 2 | 5 2 | 6 2 |
| | 12 2 | 8 2 | 7 2 | 7 2 | 5 2 | 5 2 |
| | 7 2 | 11 2 | 8 2 | 8 2 | 6 2 | 6 2 |

Continued on next column

**TABLE LXXVIII– continued**

| TQC | ICSD | | NoE | | SF | |
|---|---|---|---|---|---|---|
| | 11 2 | 12 2 | 7 2 | 7 2 | 5 2 | 6 2 |
| | 8 2 | 7 2 | 8 2 | 9 2 | 6 2 | 6 2 |
| | 12 2 | 10 2 | 7 2 | 8 2 | 5 2 | 5 2 |
| | 10 2 | 10 2 | 9 2 | 7 2 | 6 2 | 6 2 |
| | 7 2 | 11 2 | 9 2 | 8 2 | 5 2 | 6 2 |
| | 8 2 | 10 2 | 8 2 | 9 2 | 6 2 | 5 2 |
| | 9 2 | 12 2 | 7 2 | 7 2 | 5 2 | 6 2 |
| 191.1.19 | 36504 | | 90 | B1Ca1Ni4 | | |
| | -A | -GM | -H | -K | -L | -M |
| | 9 2 | 9 2 | 7 2 | 7 2 | 6 2 | 6 2 |
| | 12 2 | 11 2 | 8 2 | 8 2 | 5 2 | 5 2 |
| | 11 2 | 12 2 | 9 2 | 9 2 | 5 2 | 5 2 |
| | 7 2 | 9 2 | 7 2 | 9 2 | 6 2 | 5 2 |
| | 8 2 | 8 2 | 9 2 | 7 2 | 6 2 | 6 2 |
| | 12 2 | 7 2 | 8 2 | 8 2 | 5 2 | 6 2 |
| | 8 2 | 8 2 | 7 2 | 7 2 | 6 2 | 6 2 |
| | 9 2 | 9 2 | 9 2 | 7 2 | 6 2 | 5 2 |
| | 10 2 | 10 2 | 7 2 | 8 2 | 5 2 | 6 2 |
| | 12 2 | 12 2 | 9 2 | 9 2 | 5 2 | 5 2 |
| | 9 2 | 9 2 | 8 2 | 9 2 | 6 2 | 6 2 |
| | 11 2 | 11 2 | 9 2 | 8 2 | 6 2 | 5 2 |
| | 10 2 | 11 2 | 8 2 | 7 2 | 5 2 | 5 2 |
| | 12 2 | 10 2 | 7 2 | 7 2 | 5 2 | 5 2 |
| | 7 2 | 12 2 | 9 2 | 8 2 | 5 2 | 5 2 |
| | 8 2 | 7 2 | 7 2 | 9 2 | 6 2 | 6 2 |
| | 8 2 | 8 2 | 7 2 | 8 2 | 6 2 | 6 2 |
| | 7 2 | 7 2 | 8 2 | 7 2 | 6 2 | 5 2 |
| | 9 2 | 9 2 | 9 2 | 9 2 | 5 2 | 6 2 |
| | 10 2 | 11 2 | 7 2 | 9 2 | 5 2 | 6 2 |
| | 11 2 | 10 2 | 8 2 | 7 2 | 5 2 | 6 2 |
| | 8 2 | 12 2 | 9 2 | 8 2 | 5 2 | 6 2 |
| | 7 2 | 10 2 | 8 2 | 9 2 | 6 2 | 5 2 |
| | 12 2 | 12 2 | 7 2 | 7 2 | 5 2 | 5 2 |
| | 11 2 | 7 2 | 9 2 | 9 2 | 6 2 | 5 2 |
| | 10 2 | 9 2 | 8 2 | 8 2 | 6 2 | 5 2 |
| | 12 2 | 8 2 | 8 2 | 7 2 | 5 2 | 6 2 |
| | 7 2 | 7 2 | 7 2 | 8 2 | 6 2 | 5 2 |
| | 9 2 | 11 2 | 9 2 | 8 2 | 5 2 | 6 2 |
| | 10 2 | 9 2 | 7 2 | 9 2 | 6 2 | 6 2 |
| | 11 2 | 7 2 | 8 2 | 8 2 | 6 2 | 5 2 |
| | 12 2 | 8 2 | 8 2 | 7 2 | 5 2 | 6 2 |
| | 7 2 | 10 2 | 9 2 | 9 2 | 5 2 | 5 2 |
| | 8 2 | 11 2 | 7 2 | 8 2 | 6 2 | 6 2 |
| | 10 2 | 8 2 | 9 2 | 7 2 | 6 2 | 6 2 |
| | 11 2 | 10 2 | 7 2 | 9 2 | 5 2 | 5 2 |
| | 8 2 | 12 2 | 8 2 | 7 2 | 6 2 | 5 2 |
| | 9 2 | 8 2 | 9 2 | 8 2 | 5 2 | 6 2 |
| | 7 2 | 9 2 | 8 2 | 9 2 | 6 2 | 6 2 |
| | 9 2 | 7 2 | 7 2 | 8 2 | 6 2 | 5 2 |
| | 10 2 | 11 2 | 9 2 | 7 2 | 5 2 | 6 2 |
| | 11 2 | 10 2 | 9 2 | 9 2 | 6 2 | 5 2 |
| | 9 2 | 9 2 | 8 2 | 7 2 | 5 2 | 6 2 |
| | 12 2 | 12 2 | 9 2 | 8 2 | 5 2 | 5 2 |
| | 8 2 | 11 2 | 7 2 | 9 2 | 5 2 | 6 2 |
| 191.1.20 | 409394 | | 62 | Al9Co2Eu1 | | |
| | -A | -GM | -H | -K | -L | -M |
| | 9 2 | 9 2 | 9 2 | 9 2 | 5 2 | 5 2 |
| | 12 2 | 12 2 | 8 2 | 8 2 | 6 2 | 6 2 |
| | 12 2 | 10 2 | 8 2 | 7 2 | 6 2 | 6 2 |
| | 10 2 | 12 2 | 7 2 | 8 2 | 6 2 | 6 2 |
| | 12 2 | 9 2 | 7 2 | 8 2 | 5 2 | 6 2 |

Continued on next column

**TABLE LXXVIII– continued**

| TQC | ICSD | | NoE | | SF | |
|---|---|---|---|---|---|---|
| | 9 2 | 11 2 | 9 2 | 7 2 | 6 2 | 5 2 |
| | 11 2 | 8 2 | 8 2 | 9 2 | 6 2 | 6 2 |
| | 10 2 | 7 2 | 7 2 | 8 2 | 5 2 | 5 2 |
| | 8 2 | 12 2 | 8 2 | 7 2 | 5 2 | 6 2 |
| | 8 2 | 10 2 | 9 2 | 9 2 | 6 2 | 5 2 |
| | 7 2 | 9 2 | 7 2 | 8 2 | 6 2 | 5 2 |
| | 9 2 | 9 2 | 9 2 | 7 2 | 5 2 | 6 2 |
| | 11 2 | 7 2 | 9 2 | 9 2 | 5 2 | 6 2 |
| | 7 2 | 8 2 | 8 2 | 8 2 | 6 2 | 5 2 |
| | 12 2 | 10 2 | 8 2 | 7 2 | 6 2 | 6 2 |
| | 10 2 | 11 2 | 8 2 | 8 2 | 5 2 | 5 2 |
| | 12 2 | 12 2 | 7 2 | 9 2 | 5 2 | 5 2 |
| | 9 2 | 11 2 | 8 2 | 7 2 | 6 2 | 6 2 |
| | 7 2 | 10 2 | 7 2 | 9 2 | 5 2 | 5 2 |
| | 7 2 | 9 2 | 7 2 | 8 2 | 6 2 | 5 2 |
| | 8 2 | 7 2 | 9 2 | 9 2 | 5 2 | 6 2 |
| | 10 2 | 12 2 | 9 2 | 8 2 | 6 2 | 6 2 |
| | 12 2 | 11 2 | 8 2 | 7 2 | 5 2 | 6 2 |
| | 9 2 | 12 2 | 7 2 | 8 2 | 6 2 | 5 2 |
| | 11 2 | 10 2 | 7 2 | 7 2 | 5 2 | 5 2 |
| | 11 2 | 11 2 | 9 2 | 7 2 | 6 2 | 5 2 |
| | 10 2 | 10 2 | 8 2 | 9 2 | 6 2 | 6 2 |
| | 11 2 | 12 2 | 9 2 | 9 2 | 5 2 | 6 2 |
| | 10 2 | 11 2 | 8 2 | 8 2 | 6 2 | 6 2 |
| | 12 2 | 10 2 | 9 2 | 7 2 | 6 2 | 6 2 |
| | 9 2 | 12 2 | 7 2 | 9 2 | 6 2 | 6 2 |

The End

TABLE LXXIX: The computed irreducible representations for the valence bands for the compounds of SEBR and NLC in space group 193. See the list of tables.

| TQC | ICSD | | NoE | | SF | |
|---|---|---|---|---|---|---|
| 193.1.1 | 150562 | | 64 | | Sn3Zr5 | |
| | -A | -GM | -H | -K | -L | -M |
| | 6 4 | 9 2 | 7+7 4 | 7 2 | 3+4 4 | 6 2 |
| | 6 4 | 8 2 | 8+9 4 | 8 2 | 3+4 4 | 5 2 |
| | 4+5 4 | 7 2 | 8+9 4 | 9 2 | 3+4 4 | 6 2 |
| | 4+5 4 | 10 2 | 8+9 4 | 9 2 | 3+4 4 | 5 2 |
| | 6 4 | 12 2 | 7+7 4 | 7 2 | 3+4 4 | 6 2 |
| | 6 4 | 11 2 | 8+9 4 | 8 2 | 3+4 4 | 5 2 |
| | 6 4 | 11 2 | 7+7 4 | 9 2 | 3+4 4 | 5 2 |
| | 6 4 | 8 2 | 8+9 4 | 7 2 | 3+4 4 | 6 2 |
| | 6 4 | 10 2 | 8+9 4 | 8 2 | 3+4 4 | 6 2 |
| | 4+5 4 | 12 2 | 8+9 4 | 7 2 | 3+4 4 | 5 2 |
| | 4+5 4 | 9 2 | 7+7 4 | 9 2 | 3+4 4 | 5 2 |
| | 6 4 | 7 2 | 8+9 4 | 7 2 | 3+4 4 | 6 2 |
| | 6 4 | 8 2 | 8+9 4 | 8 2 | 3+4 4 | 6 2 |
| | 6 4 | 9 2 | 7+7 4 | 8 2 | 3+4 4 | 6 2 |
| | 4+5 4 | 7 2 | 8+9 4 | 9 2 | 3+4 4 | 5 2 |
| | 6 4 | 8 2 | 8+9 4 | 9 2 | 3+4 4 | 5 2 |
| | | 7 2 | | 7 2 | | 6 2 |
| | | 9 2 | | 8 2 | | 5 2 |
| | | 11 2 | | 8 2 | | 6 2 |
| | | 12 2 | | 7 2 | | 6 2 |
| | | 10 2 | | 9 2 | | 5 2 |
| | | 10 2 | | 7 2 | | 5 2 |
| | | 9 2 | | 9 2 | | 5 2 |
| | | 8 2 | | 8 2 | | 5 2 |
| | | 11 2 | | 7 2 | | 6 2 |

Continued on next column

**TABLE LXXIX– continued**

| TQC | ICSD | NoE | SF | | | |
|---|---|---|---|---|---|---|
| | | 7 2 | | 9 2 | | 6 2 |
| | | 12 2 | | 8 2 | | 5 2 |
| | | 8 2 | | 8 2 | | 5 2 |
| | | 11 2 | | 8 2 | | 6 2 |
| | | 9 2 | | 8 2 | | 6 2 |
| | | 7 2 | | 7 2 | | 6 2 |
| | | 9 2 | | 9 2 | | 5 2 |
| 193.1.2 | 109292 | 64 | Sn3Ti5 | | | |
| | -A | -GM | -H | -K | -L | -M |
| | 6 4 | 9 2 | 7+7 4 | 7 2 | 3+4 4 | 6 2 |
| | 4+5 4 | 7 2 | 8+9 4 | 8 2 | 3+4 4 | 5 2 |
| | 6 4 | 8 2 | 8+9 4 | 9 2 | 3+4 4 | 6 2 |
| | 4+5 4 | 10 2 | 8+9 4 | 9 2 | 3+4 4 | 5 2 |
| | 6 4 | 12 2 | 7+7 4 | 7 2 | 3+4 4 | 6 2 |
| | 6 4 | 11 2 | 8+9 4 | 8 2 | 3+4 4 | 5 2 |
| | 6 4 | 11 2 | 7+7 4 | 8 2 | 3+4 4 | 5 2 |
| | 6 4 | 8 2 | 8+9 4 | 7 2 | 3+4 4 | 6 2 |
| | 6 4 | 9 2 | 8+9 4 | 9 2 | 3+4 4 | 5 2 |
| | 4+5 4 | 10 2 | 8+9 4 | 7 2 | 3+4 4 | 6 2 |
| | 4+5 4 | 12 2 | 7+7 4 | 9 2 | 3+4 4 | 5 2 |
| | 6 4 | 7 2 | 8+9 4 | 7 2 | 3+4 4 | 6 2 |
| | 6 4 | 9 2 | 8+9 4 | 8 2 | 3+4 4 | 6 2 |
| | 4+5 4 | 8 2 | 7+7 4 | 8 2 | 3+4 4 | 6 2 |
| | 6 4 | 7 2 | 8+9 4 | 9 2 | 3+4 4 | 5 2 |
| | 6 4 | 8 2 | 8+9 4 | 9 2 | 3+4 4 | 6 2 |
| | | 7 2 | | 8 2 | | 6 2 |
| | | 11 2 | | 7 2 | | 5 2 |
| | | 10 2 | | 8 2 | | 5 2 |
| | | 12 2 | | 7 2 | | 6 2 |
| | | 11 2 | | 9 2 | | 5 2 |
| | | 9 2 | | 7 2 | | 5 2 |
| | | 10 2 | | 9 2 | | 5 2 |
| | | 9 2 | | 8 2 | | 5 2 |
| | | 7 2 | | 9 2 | | 6 2 |
| | | 8 2 | | 8 2 | | 6 2 |
| | | 8 2 | | 7 2 | | 5 2 |
| | | 12 2 | | 8 2 | | 6 2 |
| | | 9 2 | | 8 2 | | 5 2 |
| | | 11 2 | | 8 2 | | 6 2 |
| | | 9 2 | | 7 2 | | 6 2 |
| | | 7 2 | | 9 2 | | 5 2 |
| 193.1.3 | 62591 | 64 | Si3Ti5 | | | |
| | -A | -GM | -H | -K | -L | -M |
| | 6 4 | 9 2 | 7+7 4 | 7 2 | 3+4 4 | 6 2 |
| | 4+5 4 | 8 2 | 8+9 4 | 8 2 | 3+4 4 | 5 2 |
| | 6 4 | 7 2 | 8+9 4 | 9 2 | 3+4 4 | 6 2 |
| | 6 4 | 10 2 | 8+9 4 | 9 2 | 3+4 4 | 5 2 |
| | 4+5 4 | 12 2 | 7+7 4 | 7 2 | 3+4 4 | 6 2 |
| | 6 4 | 11 2 | 8+9 4 | 8 2 | 3+4 4 | 5 2 |
| | 6 4 | 11 2 | 7+7 4 | 9 2 | 3+4 4 | 5 2 |
| | 6 4 | 8 2 | 8+9 4 | 7 2 | 3+4 4 | 6 2 |
| | 4+5 4 | 10 2 | 8+9 4 | 9 2 | 3+4 4 | 5 2 |
| | 6 4 | 12 2 | 7+7 4 | 7 2 | 3+4 4 | 6 2 |
| | 4+5 4 | 9 2 | 8+9 4 | 8 2 | 3+4 4 | 5 2 |
| | 6 4 | 9 2 | 8+9 4 | 8 2 | 3+4 4 | 6 2 |
| | 6 4 | 7 2 | 8+9 4 | 7 2 | 3+4 4 | 6 2 |
| | 6 4 | 7 2 | 7+7 4 | 8 2 | 3+4 4 | 6 2 |
| | 4+5 4 | 8 2 | 8+9 4 | 9 2 | 3+4 4 | 5 2 |
| | 6 4 | 8 2 | 8+9 4 | 9 2 | 3+4 4 | 6 2 |
| | | 7 2 | | 7 2 | | 5 2 |
| | | 11 2 | | 8 2 | | 5 2 |

Continued on next column

**TABLE LXXIX– continued**

| TQC | ICSD | NoE | SF | | | |
|---|---|---|---|---|---|---|
| | | 10 2 | | 7 2 | | 6 2 |
| | | 12 2 | | 8 2 | | 6 2 |
| | | 9 2 | | 9 2 | | 5 2 |
| | | 10 2 | | 7 2 | | 5 2 |
| | | 11 2 | | 9 2 | | 5 2 |
| | | 12 2 | | 8 2 | | 5 2 |
| | | 8 2 | | 9 2 | | 6 2 |
| | | 7 2 | | 7 2 | | 6 2 |
| | | 9 2 | | 8 2 | | 5 2 |
| | | 9 2 | | 8 2 | | 6 2 |
| | | 8 2 | | 8 2 | | 5 2 |
| | | 7 2 | | 7 2 | | 6 2 |
| | | 9 2 | | 8 2 | | 6 2 |
| | | 11 2 | | 9 2 | | 5 2 |
| 193.1.4 | 105649 | 64 | Pb3Zr5 | | | |
| | -A | -GM | -H | -K | -L | -M |
| | 6 4 | 9 2 | 7+7 4 | 7 2 | 3+4 4 | 6 2 |
| | 4+5 4 | 8 2 | 8+9 4 | 8 2 | 3+4 4 | 5 2 |
| | 6 4 | 7 2 | 8+9 4 | 9 2 | 3+4 4 | 6 2 |
| | 4+5 4 | 10 2 | 8+9 4 | 9 2 | 3+4 4 | 5 2 |
| | 6 4 | 12 2 | 7+7 4 | 7 2 | 3+4 4 | 6 2 |
| | 6 4 | 11 2 | 8+9 4 | 8 2 | 3+4 4 | 5 2 |
| | 6 4 | 11 2 | 8+9 4 | 9 2 | 3+4 4 | 5 2 |
| | 6 4 | 8 2 | 7+7 4 | 7 2 | 3+4 4 | 6 2 |
| | 6 4 | 10 2 | 8+9 4 | 8 2 | 3+4 4 | 5 2 |
| | 4+5 4 | 7 2 | 8+9 4 | 7 2 | 3+4 4 | 5 2 |
| | 4+5 4 | 9 2 | 7+7 4 | 9 2 | 3+4 4 | 6 2 |
| | 6 4 | 12 2 | 8+9 4 | 7 2 | 3+4 4 | 6 2 |
| | 6 4 | 8 2 | 8+9 4 | 8 2 | 3+4 4 | 6 2 |
| | 4+5 4 | 9 2 | 8+9 4 | 9 2 | 3+4 4 | 6 2 |
| | 6 4 | 7 2 | 7+7 4 | 8 2 | 3+4 4 | 5 2 |
| | 6 4 | 8 2 | 8+9 4 | 9 2 | 3+4 4 | 5 2 |
| | | 7 2 | | 7 2 | | 6 2 |
| | | 12 2 | | 8 2 | | 6 2 |
| | | 11 2 | | 8 2 | | 5 2 |
| | | 10 2 | | 9 2 | | 6 2 |
| | | 9 2 | | 7 2 | | 5 2 |
| | | 9 2 | | 7 2 | | 5 2 |
| | | 7 2 | | 9 2 | | 6 2 |
| | | 10 2 | | 8 2 | | 5 2 |
| | | 8 2 | | 7 2 | | 5 2 |
| | | 12 2 | | 8 2 | | 6 2 |
| | | 11 2 | | 9 2 | | 5 2 |
| | | 8 2 | | 8 2 | | 5 2 |
| | | 11 2 | | 8 2 | | 6 2 |
| | | 9 2 | | 8 2 | | 6 2 |
| | | 9 2 | | 7 2 | | 6 2 |
| | | 7 2 | | 9 2 | | 5 2 |
| 193.1.5 | 154390 | 160 | Nb1Sb5U3 | | | |
| | -A | -GM | -H | -K | -L | -M |
| | 6 4 | 9 2 | 8+9 4 | 9 2 | 3+4 4 | 5 2 |
| | 4+5 4 | 8 2 | 7+7 4 | 8 2 | 3+4 4 | 5 2 |
| | 6 4 | 9 2 | 8+9 4 | 7 2 | 3+4 4 | 5 2 |
| | 6 4 | 7 2 | 8+9 4 | 8 2 | 3+4 4 | 6 2 |
| | 6 4 | 8 2 | 8+9 4 | 9 2 | 3+4 4 | 6 2 |
| | 4+5 4 | 10 2 | 7+7 4 | 7 2 | 3+4 4 | 5 2 |
| | 6 4 | 11 2 | 8+9 4 | 8 2 | 3+4 4 | 5 2 |
| | 4+5 4 | 12 2 | 8+9 4 | 9 2 | 3+4 4 | 6 2 |
| | 6 4 | 11 2 | 7+7 4 | 9 2 | 3+4 4 | 6 2 |
| | 6 4 | 12 2 | 8+9 4 | 8 2 | 3+4 4 | 6 2 |
| | 6 4 | 11 2 | 7+7 4 | 9 2 | 3+4 4 | 6 2 |

Continued on next column

**TABLE LXXIX– continued**

| TQC | ICSD | | NoE | | SF | |
|---|---|---|---|---|---|---|
| | 6 4 | 10 2 | 8+9 4 | 7 2 | 3+4 4 | 6 2 |
| | 4+5 4 | 10 2 | 8+9 4 | 7 2 | 3+4 4 | 6 2 |
| | 6 4 | 12 2 | 8+9 4 | 8 2 | 3+4 4 | 6 2 |
| | 4+5 4 | 8 2 | 7+7 4 | 7 2 | 3+4 4 | 5 2 |
| | 6 4 | 7 2 | 8+9 4 | 9 2 | 3+4 4 | 6 2 |
| | 6 4 | 9 2 | 7+7 4 | 8 2 | 3+4 4 | 6 2 |
| | 6 4 | 10 2 | 8+9 4 | 8 2 | 3+4 4 | 5 2 |
| | 4+5 4 | 11 2 | 8+9 4 | 7 2 | 3+4 4 | 6 2 |
| | 6 4 | 12 2 | 7+7 4 | 9 2 | 3+4 4 | 5 2 |
| | 6 4 | 11 2 | 8+9 4 | 7 2 | 3+4 4 | 5 2 |
| | 6 4 | 8 2 | 8+9 4 | 9 2 | 3+4 4 | 6 2 |
| | 4+5 4 | 9 2 | 8+9 4 | 8 2 | 3+4 4 | 5 2 |
| | 6 4 | 7 2 | 7+7 4 | 7 2 | 3+4 4 | 6 2 |
| | 6 4 | 12 2 | 8+9 4 | 8 2 | 3+4 4 | 6 2 |
| | 4+5 4 | 8 2 | 8+9 4 | 8 2 | 3+4 4 | 5 2 |
| | 4+5 4 | 9 2 | 7+7 4 | 9 2 | 3+4 4 | 6 2 |
| | 6 4 | 7 2 | 8+9 4 | 7 2 | 3+4 4 | 6 2 |
| | 6 4 | 10 2 | 8+9 4 | 9 2 | 3+4 4 | 5 2 |
| | 6 4 | 10 2 | 7+7 4 | 9 2 | 3+4 4 | 5 2 |
| | 6 4 | 11 2 | 8+9 4 | 8 2 | 3+4 4 | 6 2 |
| | 4+5 4 | 12 2 | 7+7 4 | 7 2 | 3+4 4 | 5 2 |
| | 6 4 | 9 2 | 8+9 4 | 7 2 | 3+4 4 | 6 2 |
| | 6 4 | 11 2 | 8+9 4 | 8 2 | 3+4 4 | 5 2 |
| | 6 4 | 9 2 | 7+7 4 | 9 2 | 3+4 4 | 5 2 |
| | 4+5 4 | 10 2 | 8+9 4 | 7 2 | 3+4 4 | 6 2 |
| | 6 4 | 7 2 | 7+7 4 | 8 2 | 3+4 4 | 5 2 |
| | 4+5 4 | 11 2 | 8+9 4 | 8 2 | 3+4 4 | 6 2 |
| | 6 4 | 8 2 | 8+9 4 | 7 2 | 3+4 4 | 5 2 |
| | 4+5 4 | 12 2 | 8+9 4 | 9 2 | 3+4 4 | 6 2 |
| | | 12 2 | | 7 2 | | 5 2 |
| | | 8 2 | | 9 2 | | 6 2 |
| | | 7 2 | | 9 2 | | 5 2 |
| | | 11 2 | | 7 2 | | 5 2 |
| | | 8 2 | | 8 2 | | 6 2 |
| | | 9 2 | | 8 2 | | 5 2 |
| | | 8 2 | | 9 2 | | 6 2 |
| | | 12 2 | | 8 2 | | 5 2 |
| | | 9 2 | | 7 2 | | 5 2 |
| | | 10 2 | | 7 2 | | 6 2 |
| | | 10 2 | | 9 2 | | 5 2 |
| | | 11 2 | | 8 2 | | 6 2 |
| | | 12 2 | | 8 2 | | 5 2 |
| | | 10 2 | | 7 2 | | 6 2 |
| | | 8 2 | | 9 2 | | 5 2 |
| | | 7 2 | | 9 2 | | 6 2 |
| | | 9 2 | | 7 2 | | 5 2 |
| | | 7 2 | | 8 2 | | 6 2 |
| | | 11 2 | | 7 2 | | 6 2 |
| | | 8 2 | | 9 2 | | 5 2 |
| | | 7 2 | | 9 2 | | 5 2 |
| | | 11 2 | | 7 2 | | 6 2 |
| | | 11 2 | | 8 2 | | 5 2 |
| | | 10 2 | | 7 2 | | 6 2 |
| | | 9 2 | | 9 2 | | 5 2 |
| | | 7 2 | | 8 2 | | 6 2 |
| | | 9 2 | | 9 2 | | 6 2 |
| | | 9 2 | | 8 2 | | 5 2 |
| | | 12 2 | | 7 2 | | 5 2 |
| | | 8 2 | | 9 2 | | 6 2 |
| | | 7 2 | | 8 2 | | 6 2 |
| | | 7 2 | | 7 2 | | 5 2 |

Continued on next column

**TABLE LXXIX– continued**

| TQC | ICSD | | NoE | | SF | |
|---|---|---|---|---|---|---|
| | | 9 2 | | 8 2 | | 5 2 |
| | | 10 2 | | 9 2 | | 5 2 |
| | | 12 2 | | 7 2 | | 6 2 |
| | | 7 2 | | 8 2 | | 6 2 |
| | | 10 2 | | 9 2 | | 5 2 |
| | | 12 2 | | 9 2 | | 6 2 |
| | | 8 2 | | 7 2 | | 5 2 |
| | | 11 2 | | 8 2 | | 6 2 |
| 193.1.6 | 99999 | | 64 | | K3N1 | |
| | -A | -GM | -H | -K | -L | -M |
| | 6 4 | 9 2 | 8+9 4 | 9 2 | 3+4 4 | 5 2 |
| | 4+5 4 | 11 2 | 7+7 4 | 9 2 | 3+4 4 | 6 2 |
| | 6 4 | 10 2 | 8+9 4 | 7 2 | 3+4 4 | 6 2 |
| | 6 4 | 12 2 | 8+9 4 | 8 2 | 3+4 4 | 6 2 |
| | 4+5 4 | 7 2 | 7+7 4 | 7 2 | 3+4 4 | 5 2 |
| | 6 4 | 8 2 | 8+9 4 | 8 2 | 3+4 4 | 5 2 |
| | 6 4 | 9 2 | 8+9 4 | 9 2 | 3+4 4 | 5 2 |
| | 6 4 | 10 2 | 8+9 4 | 7 2 | 3+4 4 | 6 2 |
| | 4+5 4 | 7 2 | 7+7 4 | 7 2 | 3+4 4 | 5 2 |
| | 6 4 | 11 2 | 8+9 4 | 9 2 | 3+4 4 | 6 2 |
| | 6 4 | 8 2 | 8+9 4 | 8 2 | 3+4 4 | 5 2 |
| | 4+5 4 | 12 2 | 7+7 4 | 8 2 | 3+4 4 | 6 2 |
| | 6 4 | 8 2 | 8+9 4 | 8 2 | 3+4 4 | 5 2 |
| | 6 4 | 9 2 | 8+9 4 | 9 2 | 3+4 4 | 5 2 |
| | 6 4 | 11 2 | 7+7 4 | 9 2 | 3+4 4 | 6 2 |
| | 4+5 4 | 12 2 | 8+9 4 | 8 2 | 3+4 4 | 6 2 |
| | | 10 2 | | 7 2 | | 6 2 |
| | | 11 2 | | 9 2 | | 6 2 |
| | | 8 2 | | 7 2 | | 5 2 |
| | | 7 2 | | 8 2 | | 5 2 |
| | | 9 2 | | 9 2 | | 5 2 |
| | | 10 2 | | 7 2 | | 6 2 |
| | | 12 2 | | 8 2 | | 6 2 |
| | | 7 2 | | 7 2 | | 5 2 |
| | | 9 2 | | 9 2 | | 5 2 |
| | | 8 2 | | 8 2 | | 5 2 |
| | | 9 2 | | 8 2 | | 6 2 |
| | | 11 2 | | 7 2 | | 6 2 |
| | | 12 2 | | 9 2 | | 6 2 |
| | | 10 2 | | 9 2 | | 6 2 |
| | | 11 2 | | 7 2 | | 6 2 |
| | | 10 2 | | 8 2 | | 5 2 |
| 193.1.7 | 44361 | | 64 | | Ge3Hf5 | |
| | -A | -GM | -H | -K | -L | -M |
| | 6 4 | 9 2 | 7+7 4 | 7 2 | 3+4 4 | 6 2 |
| | 6 4 | 8 2 | 8+9 4 | 8 2 | 3+4 4 | 5 2 |
| | 4+5 4 | 7 2 | 8+9 4 | 9 2 | 3+4 4 | 6 2 |
| | 6 4 | 10 2 | 8+9 4 | 9 2 | 3+4 4 | 5 2 |
| | 4+5 4 | 12 2 | 7+7 4 | 7 2 | 3+4 4 | 6 2 |
| | 6 4 | 11 2 | 8+9 4 | 8 2 | 3+4 4 | 5 2 |
| | 6 4 | 11 2 | 7+7 4 | 9 2 | 3+4 4 | 5 2 |
| | 6 4 | 8 2 | 8+9 4 | 7 2 | 3+4 4 | 6 2 |
| | 6 4 | 12 2 | 8+9 4 | 8 2 | 3+4 4 | 5 2 |
| | 4+5 4 | 10 2 | 7+7 4 | 7 2 | 3+4 4 | 6 2 |
| | 6 4 | 9 2 | 8+9 4 | 9 2 | 3+4 4 | 5 2 |
| | 4+5 4 | 8 2 | 8+9 4 | 7 2 | 3+4 4 | 6 2 |
| | 6 4 | 7 2 | 8+9 4 | 8 2 | 3+4 4 | 6 2 |
| | 6 4 | 9 2 | 7+7 4 | 9 2 | 3+4 4 | 6 2 |
| | 4+5 4 | 7 2 | 8+9 4 | 8 2 | 3+4 4 | 5 2 |
| | 6 4 | 8 2 | 8+9 4 | 9 2 | 3+4 4 | 6 2 |
| | | 7 2 | | 7 2 | | 5 2 |

Continued on next column

**TABLE LXXIX– continued**

| TQC | ICSD | | NoE | | SF | |
|---|---|---|---|---|---|---|
| | | 9 2 | | 8 2 | | 5 2 |
| | | 11 2 | | 7 2 | | 6 2 |
| | | 12 2 | | 8 2 | | 6 2 |
| | | 10 2 | | 9 2 | | 5 2 |
| | | 10 2 | | 7 2 | | 5 2 |
| | | 12 2 | | 9 2 | | 5 2 |
| | | 11 2 | | 8 2 | | 5 2 |
| | | 9 2 | | 7 2 | | 6 2 |
| | | 8 2 | | 8 2 | | 6 2 |
| | | 7 2 | | 9 2 | | 5 2 |
| | | 8 2 | | 8 2 | | 6 2 |
| | | 9 2 | | 8 2 | | 5 2 |
| | | 7 2 | | 7 2 | | 6 2 |
| | | 9 2 | | 8 2 | | 6 2 |
| | | 11 2 | | 9 2 | | 5 2 |
| 193.1.8 | 173358 | | 136 | | Bi5Ce3Mn1 | |
| | -A | -GM | -H | -K | -L | -M |
| | 4+5 4 | 9 2 | 7+7 4 | 7 2 | 3+4 4 | 5 2 |
| | 6 4 | 7 2 | 8+9 4 | 8 2 | 3+4 4 | 6 2 |
| | 6 4 | 8 2 | 8+9 4 | 9 2 | 3+4 4 | 6 2 |
| | 4+5 4 | 10 2 | 8+9 4 | 7 2 | 3+4 4 | 5 2 |
| | 6 4 | 11 2 | 7+7 4 | 8 2 | 3+4 4 | 5 2 |
| | 6 4 | 12 2 | 8+9 4 | 9 2 | 3+4 4 | 6 2 |
| | 6 4 | 8 2 | 7+7 4 | 7 2 | 3+4 4 | 5 2 |
| | 6 4 | 9 2 | 8+9 4 | 9 2 | 3+4 4 | 6 2 |
| | 4+5 4 | 7 2 | 8+9 4 | 8 2 | 3+4 4 | 6 2 |
| | 6 4 | 11 2 | 8+9 4 | 8 2 | 3+4 4 | 5 2 |
| | 4+5 4 | 10 2 | 7+7 4 | 7 2 | 3+4 4 | 6 2 |
| | 6 4 | 12 2 | 8+9 4 | 9 2 | 3+4 4 | 5 2 |
| | 6 4 | 8 2 | 7+7 4 | 7 2 | 3+4 4 | 6 2 |
| | 6 4 | 11 2 | 8+9 4 | 9 2 | 3+4 4 | 5 2 |
| | 4+5 4 | 9 2 | 8+9 4 | 8 2 | 3+4 4 | 6 2 |
| | 6 4 | 7 2 | 7+7 4 | 7 2 | 3+4 4 | 5 2 |
| | 6 4 | 12 2 | 8+9 4 | 8 2 | 3+4 4 | 6 2 |
| | 6 4 | 8 2 | 8+9 4 | 8 2 | 3+4 4 | 5 2 |
| | 4+5 4 | 9 2 | 7+7 4 | 9 2 | 3+4 4 | 6 2 |
| | 6 4 | 7 2 | 8+9 4 | 7 2 | 3+4 4 | 6 2 |
| | 6 4 | 10 2 | 8+9 4 | 9 2 | 3+4 4 | 5 2 |
| | 4+5 4 | 10 2 | 7+7 4 | 9 2 | 3+4 4 | 5 2 |
| | 6 4 | 11 2 | 8+9 4 | 8 2 | 3+4 4 | 6 2 |
| | 4+5 4 | 12 2 | 8+9 4 | 7 2 | 3+4 4 | 5 2 |
| | 6 4 | 9 2 | 8+9 4 | 7 2 | 3+4 4 | 6 2 |
| | 6 4 | 11 2 | 7+7 4 | 8 2 | 3+4 4 | 5 2 |
| | 6 4 | 9 2 | 8+9 4 | 9 2 | 3+4 4 | 5 2 |
| | 6 4 | 10 2 | 7+7 4 | 7 2 | 3+4 4 | 6 2 |
| | 4+5 4 | 12 2 | 8+9 4 | 8 2 | 3+4 4 | 6 2 |
| | 6 4 | 11 2 | 8+9 4 | 8 2 | 3+4 4 | 5 2 |
| | 6 4 | 7 2 | 7+7 4 | 7 2 | 3+4 4 | 5 2 |
| | 4+5 4 | 8 2 | 8+9 4 | 9 2 | 3+4 4 | 6 2 |
| | 6 4 | 12 2 | 7+7 4 | 7 2 | 3+4 4 | 5 2 |
| | 4+5 4 | 8 2 | 8+9 4 | 9 2 | 3+4 4 | 6 2 |
| | | 11 2 | | 9 2 | | 5 2 |
| | | 9 2 | | 8 2 | | 5 2 |
| | | 7 2 | | 7 2 | | 6 2 |
| | | 8 2 | | 8 2 | | 5 2 |
| | | 9 2 | | 9 2 | | 6 2 |
| | | 8 2 | | 7 2 | | 6 2 |
| | | 12 2 | | 9 2 | | 5 2 |
| | | 10 2 | | 8 2 | | 6 2 |
| | | 11 2 | | 7 2 | | 6 2 |
| | | 10 2 | | 7 2 | | 5 2 |

Continued on next column

**TABLE LXXIX– continued**

| TQC | ICSD | | NoE | | SF | |
|---|---|---|---|---|---|---|
| | | 10 2 | | 9 2 | | 5 2 |
| | | 12 2 | | 8 2 | | 5 2 |
| | | 8 2 | | 9 2 | | 6 2 |
| | | 9 2 | | 8 2 | | 6 2 |
| | | 11 2 | | 7 2 | | 6 2 |
| | | 7 2 | | 8 2 | | 5 2 |
| | | 7 2 | | 7 2 | | 5 2 |
| | | 11 2 | | 9 2 | | 5 2 |
| | | 8 2 | | 7 2 | | 6 2 |
| | | 11 2 | | 9 2 | | 5 2 |
| | | 7 2 | | 8 2 | | 6 2 |
| | | 9 2 | | 7 2 | | 5 2 |
| | | 10 2 | | 8 2 | | 5 2 |
| | | 7 2 | | 9 2 | | 6 2 |
| | | 9 2 | | 8 2 | | 6 2 |
| | | 12 2 | | 9 2 | | 5 2 |
| | | 9 2 | | 7 2 | | 5 2 |
| | | 7 2 | | 9 2 | | 5 2 |
| | | 8 2 | | 7 2 | | 6 2 |
| | | 9 2 | | 9 2 | | 5 2 |
| | | 7 2 | | 8 2 | | 5 2 |
| | | 12 2 | | 7 2 | | 6 2 |
| | | 7 2 | | 8 2 | | 5 2 |
| | | 8 2 | | 7 2 | | 5 2 |

The End

TABLE LXXX: The computed irreducible representations for the valence bands for the compounds of SEBR and NLC in space group 194. See the list of tables.

| TQC | ICSD | | NoE | | SF | |
|---|---|---|---|---|---|---|
| 194.1.1 | 421015 | | 24 | | Zn1 | |
| | -A | -GM | -H | -K | -L | -M |
| | 6 4 | 9 2 | 4+5 2 | 8 2 | 3+4 4 | 5 2 |
| | 4+5 4 | 11 2 | 9 2 | 8 2 | 3+4 4 | 6 2 |
| | 6 4 | 9 2 | 9 2 | 9 2 | 3+4 4 | 5 2 |
| | 6 4 | 10 2 | 8 2 | 7 2 | 3+4 4 | 6 2 |
| | 6 4 | 7 2 | 8 2 | 7 2 | 3+4 4 | 6 2 |
| | 4+5 4 | 12 2 | 9 2 | 9 2 | 3+4 4 | 6 2 |
| | | 8 2 | 6+7 2 | 9 2 | | 5 2 |
| | | 9 2 | 8 2 | 8 2 | | 5 2 |
| | | 7 2 | 4+5 2 | 8 2 | | 6 2 |
| | | 11 2 | 9 2 | 7 2 | | 5 2 |
| | | 10 2 | 6+7 2 | 7 2 | | 5 2 |
| | | 8 2 | 8 2 | 8 2 | | 6 2 |
| 194.1.2 | 421014 | | 24 | | Zn1 | |
| | -A | -GM | -H | -K | -L | -M |
| | 6 4 | 9 2 | 4+5 2 | 8 2 | 3+4 4 | 5 2 |
| | 4+5 4 | 11 2 | 9 2 | 8 2 | 3+4 4 | 6 2 |
| | 6 4 | 9 2 | 9 2 | 7 2 | 3+4 4 | 5 2 |
| | 6 4 | 10 2 | 8 2 | 9 2 | 3+4 4 | 6 2 |
| | 6 4 | 7 2 | 8 2 | 7 2 | 3+4 4 | 6 2 |
| | 4+5 4 | 12 2 | 9 2 | 9 2 | 3+4 4 | 6 2 |
| | | 8 2 | 6+7 2 | 9 2 | | 5 2 |
| | | 9 2 | 8 2 | 8 2 | | 5 2 |
| | | 7 2 | 4+5 2 | 8 2 | | 6 2 |
| | | 11 2 | 9 2 | 7 2 | | 5 2 |
| | | 10 2 | 6+7 2 | 7 2 | | 5 2 |
| | | 8 2 | 8 2 | 8 2 | | 6 2 |
| 194.1.3 | 421013 | | 24 | | Zn1 | |

Continued on next column

**TABLE LXXX– continued**

| TQC | ICSD | | NoE | | SF | |
|---|---|---|---|---|---|---|
| | -A | -GM | -H | -K | -L | -M |
| | 6 4 | 9 2 | 4+5 2 | 8 2 | 3+4 4 | 5 2 |
| | 4+5 4 | 11 2 | 9 2 | 8 2 | 3+4 4 | 6 2 |
| | 6 4 | 9 2 | 9 2 | 7 2 | 3+4 4 | 5 2 |
| | 6 4 | 10 2 | 8 2 | 9 2 | 3+4 4 | 6 2 |
| | 6 4 | 7 2 | 8 2 | 7 2 | 3+4 4 | 6 2 |
| | 4+5 4 | 12 2 | 9 2 | 9 2 | 3+4 4 | 6 2 |
| | | 8 2 | 6+7 2 | 9 2 | | 5 2 |
| | | 9 2 | 8 2 | 8 2 | | 5 2 |
| | | 7 2 | 4+5 2 | 8 2 | | 6 2 |
| | | 11 2 | 9 2 | 7 2 | | 5 2 |
| | | 10 2 | 6+7 2 | 7 2 | | 5 2 |
| | | 8 2 | 8 2 | 8 2 | | 6 2 |
| 194.1.4 | 247162 | | 24 | | Zn1 | |
| | -A | -GM | -H | -K | -L | -M |
| | 6 4 | 9 2 | 4+5 2 | 8 2 | 3+4 4 | 5 2 |
| | 4+5 4 | 11 2 | 9 2 | 8 2 | 3+4 4 | 6 2 |
| | 6 4 | 9 2 | 9 2 | 7 2 | 3+4 4 | 5 2 |
| | 6 4 | 10 2 | 8 2 | 9 2 | 3+4 4 | 6 2 |
| | 6 4 | 7 2 | 8 2 | 7 2 | 3+4 4 | 6 2 |
| | 4+5 4 | 12 2 | 9 2 | 9 2 | 3+4 4 | 6 2 |
| | | 8 2 | 6+7 2 | 9 2 | | 5 2 |
| | | 9 2 | 8 2 | 8 2 | | 5 2 |
| | | 7 2 | 4+5 2 | 8 2 | | 6 2 |
| | | 11 2 | 9 2 | 7 2 | | 5 2 |
| | | 10 2 | 6+7 2 | 7 2 | | 5 2 |
| | | 8 2 | 8 2 | 8 2 | | 6 2 |
| 194.1.5 | 247161 | | 24 | | Zn1 | |
| | -A | -GM | -H | -K | -L | -M |
| | 6 4 | 9 2 | 4+5 2 | 8 2 | 3+4 4 | 5 2 |
| | 4+5 4 | 11 2 | 9 2 | 8 2 | 3+4 4 | 6 2 |
| | 6 4 | 9 2 | 9 2 | 7 2 | 3+4 4 | 5 2 |
| | 6 4 | 10 2 | 8 2 | 9 2 | 3+4 4 | 6 2 |
| | 6 4 | 7 2 | 8 2 | 7 2 | 3+4 4 | 6 2 |
| | 4+5 4 | 12 2 | 9 2 | 9 2 | 3+4 4 | 6 2 |
| | | 8 2 | 6+7 2 | 9 2 | | 5 2 |
| | | 9 2 | 8 2 | 8 2 | | 5 2 |
| | | 7 2 | 4+5 2 | 8 2 | | 6 2 |
| | | 11 2 | 9 2 | 7 2 | | 5 2 |
| | | 10 2 | 6+7 2 | 7 2 | | 5 2 |
| | | 8 2 | 8 2 | 8 2 | | 6 2 |
| 194.1.6 | 247160 | | 24 | | Zn1 | |
| | -A | -GM | -H | -K | -L | -M |
| | 6 4 | 9 2 | 4+5 2 | 8 2 | 3+4 4 | 5 2 |
| | 4+5 4 | 11 2 | 9 2 | 8 2 | 3+4 4 | 6 2 |
| | 6 4 | 9 2 | 9 2 | 7 2 | 3+4 4 | 5 2 |
| | 6 4 | 10 2 | 8 2 | 9 2 | 3+4 4 | 6 2 |
| | 6 4 | 7 2 | 8 2 | 7 2 | 3+4 4 | 6 2 |
| | 4+5 4 | 12 2 | 9 2 | 9 2 | 3+4 4 | 6 2 |
| | | 8 2 | 6+7 2 | 9 2 | | 5 2 |
| | | 9 2 | 8 2 | 8 2 | | 5 2 |
| | | 7 2 | 4+5 2 | 8 2 | | 6 2 |
| | | 11 2 | 9 2 | 7 2 | | 5 2 |
| | | 10 2 | 6+7 2 | 7 2 | | 5 2 |
| | | 8 2 | 8 2 | 8 2 | | 6 2 |
| 194.1.7 | 247159 | | 24 | | Zn1 | |
| | -A | -GM | -H | -K | -L | -M |
| | 6 4 | 9 2 | 4+5 2 | 8 2 | 3+4 4 | 5 2 |
| | 4+5 4 | 11 2 | 9 2 | 8 2 | 3+4 4 | 6 2 |
| | 6 4 | 9 2 | 9 2 | 7 2 | 3+4 4 | 5 2 |
| | 6 4 | 10 2 | 8 2 | 9 2 | 3+4 4 | 6 2 |

Continued on next column

| TQC | ICSD | | NoE | | SF | |
|---|---|---|---|---|---|---|
| | 6 4 | 7 2 | 8 2 | 7 2 | 3+4 4 | 6 2 |
| | 4+5 4 | 12 2 | 9 2 | 9 2 | 3+4 4 | 6 2 |
| | | 8 2 | 6+7 2 | 9 2 | | 5 2 |
| | | 9 2 | 8 2 | 8 2 | | 5 2 |
| | | 7 2 | 4+5 2 | 8 2 | | 6 2 |
| | | 11 2 | 9 2 | 7 2 | | 5 2 |
| | | 10 2 | 6+7 2 | 7 2 | | 5 2 |
| | | 8 2 | 8 2 | 8 2 | | 6 2 |
| 194.1.8 | 247158 | | 24 | | Zn1 | |
| | -A | -GM | -H | -K | -L | -M |
| | 6 4 | 9 2 | 4+5 2 | 8 2 | 3+4 4 | 5 2 |
| | 4+5 4 | 11 2 | 9 2 | 8 2 | 3+4 4 | 6 2 |
| | 6 4 | 9 2 | 9 2 | 7 2 | 3+4 4 | 5 2 |
| | 6 4 | 10 2 | 8 2 | 9 2 | 3+4 4 | 6 2 |
| | 6 4 | 7 2 | 8 2 | 7 2 | 3+4 4 | 6 2 |
| | 4+5 4 | 12 2 | 9 2 | 9 2 | 3+4 4 | 6 2 |
| | | 8 2 | 6+7 2 | 9 2 | | 5 2 |
| | | 9 2 | 8 2 | 8 2 | | 5 2 |
| | | 7 2 | 4+5 2 | 8 2 | | 6 2 |
| | | 11 2 | 9 2 | 7 2 | | 5 2 |
| | | 10 2 | 6+7 2 | 7 2 | | 5 2 |
| | | 8 2 | 8 2 | 8 2 | | 6 2 |
| 194.1.9 | 247157 | | 24 | | Zn1 | |
| | -A | -GM | -H | -K | -L | -M |
| | 6 4 | 9 2 | 4+5 2 | 8 2 | 3+4 4 | 5 2 |
| | 4+5 4 | 11 2 | 9 2 | 8 2 | 3+4 4 | 6 2 |
| | 6 4 | 9 2 | 9 2 | 7 2 | 3+4 4 | 5 2 |
| | 6 4 | 10 2 | 8 2 | 9 2 | 3+4 4 | 6 2 |
| | 6 4 | 7 2 | 8 2 | 7 2 | 3+4 4 | 6 2 |
| | 4+5 4 | 12 2 | 9 2 | 9 2 | 3+4 4 | 6 2 |
| | | 8 2 | 6+7 2 | 9 2 | | 5 2 |
| | | 9 2 | 8 2 | 8 2 | | 5 2 |
| | | 7 2 | 4+5 2 | 8 2 | | 6 2 |
| | | 11 2 | 9 2 | 7 2 | | 5 2 |
| | | 10 2 | 6+7 2 | 7 2 | | 5 2 |
| | | 8 2 | 8 2 | 8 2 | | 6 2 |
| 194.1.10 | 247156 | | 24 | | Zn1 | |
| | -A | -GM | -H | -K | -L | -M |
| | 6 4 | 9 2 | 4+5 2 | 8 2 | 3+4 4 | 5 2 |
| | 4+5 4 | 11 2 | 9 2 | 8 2 | 3+4 4 | 6 2 |
| | 6 4 | 9 2 | 9 2 | 7 2 | 3+4 4 | 5 2 |
| | 6 4 | 10 2 | 8 2 | 9 2 | 3+4 4 | 6 2 |
| | 6 4 | 7 2 | 8 2 | 7 2 | 3+4 4 | 6 2 |
| | 4+5 4 | 12 2 | 9 2 | 9 2 | 3+4 4 | 6 2 |
| | | 8 2 | 6+7 2 | 9 2 | | 5 2 |
| | | 9 2 | 8 2 | 8 2 | | 5 2 |
| | | 7 2 | 4+5 2 | 8 2 | | 6 2 |
| | | 11 2 | 9 2 | 7 2 | | 5 2 |
| | | 10 2 | 6+7 2 | 7 2 | | 5 2 |
| | | 8 2 | 8 2 | 8 2 | | 6 2 |
| 194.1.11 | 247155 | | 24 | | Zn1 | |
| | -A | -GM | -H | -K | -L | -M |
| | 6 4 | 9 2 | 4+5 2 | 8 2 | 3+4 4 | 5 2 |
| | 4+5 4 | 11 2 | 9 2 | 8 2 | 3+4 4 | 6 2 |
| | 6 4 | 9 2 | 9 2 | 7 2 | 3+4 4 | 5 2 |
| | 6 4 | 10 2 | 8 2 | 9 2 | 3+4 4 | 6 2 |
| | 6 4 | 7 2 | 8 2 | 7 2 | 3+4 4 | 6 2 |
| | 4+5 4 | 12 2 | 9 2 | 9 2 | 3+4 4 | 6 2 |
| | | 8 2 | 6+7 2 | 9 2 | | 5 2 |
| | | 9 2 | 8 2 | 8 2 | | 5 2 |
| | | 7 2 | 4+5 2 | 8 2 | | 6 2 |

Continued on next column

**TABLE LXXX– continued**

| TQC | ICSD | | NoE | | SF | |
|---|---|---|---|---|---|---|
| | | 11 2 | 9 2 | 7 2 | | 5 2 |
| | | 10 2 | 6+7 2 | 7 2 | | 5 2 |
| | | 8 2 | 8 2 | 8 2 | | 6 2 |
| 194.1.12 | 247154 | | 24 | | Zn1 | |
| | -A | -GM | -H | -K | -L | -M |
| | 6 4 | 9 2 | 4+5 2 | 8 2 | 3+4 4 | 5 2 |
| | 4+5 4 | 11 2 | 9 2 | 8 2 | 3+4 4 | 6 2 |
| | 6 4 | 9 2 | 9 2 | 7 2 | 3+4 4 | 5 2 |
| | 6 4 | 10 2 | 8 2 | 9 2 | 3+4 4 | 6 2 |
| | 6 4 | 7 2 | 8 2 | 7 2 | 3+4 4 | 6 2 |
| | 4+5 4 | 12 2 | 9 2 | 9 2 | 3+4 4 | 6 2 |
| | | 8 2 | 6+7 2 | 9 2 | | 5 2 |
| | | 9 2 | 8 2 | 8 2 | | 5 2 |
| | | 7 2 | 4+5 2 | 8 2 | | 6 2 |
| | | 11 2 | 9 2 | 7 2 | | 5 2 |
| | | 10 2 | 6+7 2 | 7 2 | | 5 2 |
| | | 8 2 | 8 2 | 8 2 | | 6 2 |
| 194.1.13 | 247153 | | 24 | | Zn1 | |
| | -A | -GM | -H | -K | -L | -M |
| | 6 4 | 9 2 | 4+5 2 | 8 2 | 3+4 4 | 5 2 |
| | 4+5 4 | 11 2 | 9 2 | 8 2 | 3+4 4 | 6 2 |
| | 6 4 | 9 2 | 9 2 | 7 2 | 3+4 4 | 5 2 |
| | 6 4 | 10 2 | 8 2 | 9 2 | 3+4 4 | 6 2 |
| | 6 4 | 7 2 | 8 2 | 7 2 | 3+4 4 | 6 2 |
| | 4+5 4 | 12 2 | 9 2 | 9 2 | 3+4 4 | 6 2 |
| | | 8 2 | 6+7 2 | 9 2 | | 5 2 |
| | | 9 2 | 8 2 | 8 2 | | 5 2 |
| | | 7 2 | 4+5 2 | 8 2 | | 6 2 |
| | | 11 2 | 9 2 | 7 2 | | 5 2 |
| | | 10 2 | 6+7 2 | 7 2 | | 5 2 |
| | | 8 2 | 8 2 | 8 2 | | 6 2 |
| 194.1.14 | 247152 | | 24 | | Zn1 | |
| | -A | -GM | -H | -K | -L | -M |
| | 6 4 | 9 2 | 4+5 2 | 8 2 | 3+4 4 | 5 2 |
| | 4+5 4 | 11 2 | 9 2 | 8 2 | 3+4 4 | 6 2 |
| | 6 4 | 9 2 | 9 2 | 7 2 | 3+4 4 | 5 2 |
| | 6 4 | 10 2 | 8 2 | 9 2 | 3+4 4 | 6 2 |
| | 6 4 | 7 2 | 8 2 | 7 2 | 3+4 4 | 6 2 |
| | 4+5 4 | 12 2 | 9 2 | 9 2 | 3+4 4 | 6 2 |
| | | 8 2 | 6+7 2 | 9 2 | | 5 2 |
| | | 9 2 | 8 2 | 8 2 | | 5 2 |
| | | 7 2 | 4+5 2 | 8 2 | | 6 2 |
| | | 11 2 | 9 2 | 7 2 | | 5 2 |
| | | 10 2 | 6+7 2 | 7 2 | | 5 2 |
| | | 8 2 | 8 2 | 8 2 | | 6 2 |
| 194.1.15 | 247151 | | 24 | | Zn1 | |
| | -A | -GM | -H | -K | -L | -M |
| | 6 4 | 9 2 | 4+5 2 | 8 2 | 3+4 4 | 5 2 |
| | 4+5 4 | 11 2 | 9 2 | 8 2 | 3+4 4 | 6 2 |
| | 6 4 | 9 2 | 9 2 | 7 2 | 3+4 4 | 5 2 |
| | 6 4 | 10 2 | 8 2 | 9 2 | 3+4 4 | 6 2 |
| | 6 4 | 7 2 | 8 2 | 7 2 | 3+4 4 | 6 2 |
| | 4+5 4 | 12 2 | 9 2 | 9 2 | 3+4 4 | 6 2 |
| | | 8 2 | 6+7 2 | 9 2 | | 5 2 |
| | | 9 2 | 8 2 | 8 2 | | 5 2 |
| | | 7 2 | 4+5 2 | 8 2 | | 6 2 |
| | | 11 2 | 9 2 | 7 2 | | 5 2 |
| | | 10 2 | 6+7 2 | 7 2 | | 5 2 |
| | | 8 2 | 8 2 | 8 2 | | 6 2 |
| 194.1.16 | 247150 | | 24 | | Zn1 | |

Continued on next column

**TABLE LXXX– continued**

| TQC | ICSD | | NoE | | SF | |
|---|---|---|---|---|---|---|
| | -A | -GM | -H | -K | -L | -M |
| | 6 4 | 9 2 | 4+5 2 | 8 2 | 3+4 4 | 5 2 |
| | 4+5 4 | 11 2 | 9 2 | 8 2 | 3+4 4 | 6 2 |
| | 6 4 | 9 2 | 9 2 | 7 2 | 3+4 4 | 5 2 |
| | 6 4 | 10 2 | 8 2 | 9 2 | 3+4 4 | 6 2 |
| | 6 4 | 7 2 | 8 2 | 7 2 | 3+4 4 | 6 2 |
| | 4+5 4 | 12 2 | 9 2 | 9 2 | 3+4 4 | 6 2 |
| | | 8 2 | 6+7 2 | 9 2 | | 5 2 |
| | | 9 2 | 8 2 | 8 2 | | 5 2 |
| | | 7 2 | 4+5 2 | 8 2 | | 6 2 |
| | | 11 2 | 9 2 | 7 2 | | 5 2 |
| | | 10 2 | 6+7 2 | 7 2 | | 5 2 |
| | | 8 2 | 8 2 | 8 2 | | 6 2 |
| 194.1.17 | 247149 | | 24 | | Zn1 | |
| | -A | -GM | -H | -K | -L | -M |
| | 6 4 | 9 2 | 4+5 2 | 8 2 | 3+4 4 | 5 2 |
| | 4+5 4 | 11 2 | 9 2 | 8 2 | 3+4 4 | 6 2 |
| | 6 4 | 9 2 | 9 2 | 7 2 | 3+4 4 | 5 2 |
| | 6 4 | 10 2 | 8 2 | 9 2 | 3+4 4 | 6 2 |
| | 6 4 | 7 2 | 8 2 | 7 2 | 3+4 4 | 6 2 |
| | 4+5 4 | 12 2 | 9 2 | 9 2 | 3+4 4 | 6 2 |
| | | 8 2 | 6+7 2 | 9 2 | | 5 2 |
| | | 9 2 | 8 2 | 8 2 | | 5 2 |
| | | 7 2 | 4+5 2 | 8 2 | | 6 2 |
| | | 11 2 | 9 2 | 7 2 | | 5 2 |
| | | 10 2 | 6+7 2 | 7 2 | | 5 2 |
| | | 8 2 | 8 2 | 8 2 | | 6 2 |
| 194.1.18 | 247148 | | 24 | | Zn1 | |
| | -A | -GM | -H | -K | -L | -M |
| | 6 4 | 9 2 | 4+5 2 | 8 2 | 3+4 4 | 5 2 |
| | 4+5 4 | 11 2 | 9 2 | 8 2 | 3+4 4 | 6 2 |
| | 6 4 | 9 2 | 9 2 | 7 2 | 3+4 4 | 5 2 |
| | 6 4 | 10 2 | 8 2 | 9 2 | 3+4 4 | 6 2 |
| | 6 4 | 7 2 | 8 2 | 7 2 | 3+4 4 | 6 2 |
| | 4+5 4 | 12 2 | 9 2 | 9 2 | 3+4 4 | 6 2 |
| | | 8 2 | 6+7 2 | 9 2 | | 5 2 |
| | | 9 2 | 8 2 | 8 2 | | 5 2 |
| | | 7 2 | 4+5 2 | 8 2 | | 6 2 |
| | | 11 2 | 9 2 | 7 2 | | 5 2 |
| | | 10 2 | 6+7 2 | 7 2 | | 5 2 |
| | | 8 2 | 8 2 | 8 2 | | 6 2 |
| 194.1.19 | 247147 | | 24 | | Zn1 | |
| | -A | -GM | -H | -K | -L | -M |
| | 6 4 | 9 2 | 4+5 2 | 8 2 | 3+4 4 | 5 2 |
| | 4+5 4 | 11 2 | 9 2 | 8 2 | 3+4 4 | 6 2 |
| | 6 4 | 9 2 | 9 2 | 7 2 | 3+4 4 | 5 2 |
| | 6 4 | 10 2 | 8 2 | 9 2 | 3+4 4 | 6 2 |
| | 6 4 | 7 2 | 8 2 | 7 2 | 3+4 4 | 6 2 |
| | 4+5 4 | 12 2 | 9 2 | 9 2 | 3+4 4 | 6 2 |
| | | 8 2 | 6+7 2 | 9 2 | | 5 2 |
| | | 9 2 | 8 2 | 8 2 | | 5 2 |
| | | 7 2 | 4+5 2 | 8 2 | | 6 2 |
| | | 11 2 | 9 2 | 7 2 | | 5 2 |
| | | 10 2 | 6+7 2 | 7 2 | | 5 2 |
| | | 8 2 | 8 2 | 8 2 | | 6 2 |
| 194.1.20 | 422990 | | 72 | | Mn1O3Y1 | |
| | -A | -GM | -H | -K | -L | -M |
| | 6 4 | 8 2 | 8 2 | 8 2 | 3+4 4 | 5 2 |
| | 6 4 | 9 2 | 9 2 | 9 2 | 3+4 4 | 5 2 |
| | 6 4 | 11 2 | 9 2 | 8 2 | 3+4 4 | 6 2 |
| | 4+5 4 | 12 2 | 8 2 | 9 2 | 3+4 4 | 6 2 |

Continued on next column

**TABLE LXXX– continued**

| TQC | ICSD | | NoE | | SF | |
|---|---|---|---|---|---|---|
| | 6 4 | 11 2 | 4+5 2 | 7 2 | 3+4 4 | 6 2 |
| | 6 4 | 12 2 | 6+7 2 | 7 2 | 3+4 4 | 6 2 |
| | 6 4 | 10 2 | 9 2 | 9 2 | 3+4 4 | 6 2 |
| | 6 4 | 10 2 | 8 2 | 8 2 | 3+4 4 | 6 2 |
| | 6 4 | 9 2 | 9 2 | 9 2 | 3+4 4 | 5 2 |
| | 4+5 4 | 8 2 | 8 2 | 9 2 | 3+4 4 | 5 2 |
| | 6 4 | 11 2 | 9 2 | 8 2 | 3+4 4 | 5 2 |
| | 4+5 4 | 9 2 | 4+5 2 | 7 2 | 3+4 4 | 6 2 |
| | 6 4 | 11 2 | 8 2 | 9 2 | 3+4 4 | 6 2 |
| | 6 4 | 12 2 | 6+7 2 | 7 2 | 3+4 4 | 6 2 |
| | 4+5 4 | 9 2 | 9 2 | 8 2 | 3+4 4 | 5 2 |
| | 6 4 | 8 2 | 4+5 2 | 8 2 | 3+4 4 | 5 2 |
| | 6 4 | 9 2 | 9 2 | 7 2 | 3+4 4 | 6 2 |
| | 4+5 4 | 7 2 | 8 2 | 7 2 | 3+4 4 | 6 2 |
| | | 8 2 | 8 2 | 8 2 | | 6 2 |
| | | 7 2 | 6+7 2 | 9 2 | | 5 2 |
| | | 11 2 | 9 2 | 8 2 | | 5 2 |
| | | 12 2 | 4+5 2 | 7 2 | | 5 2 |
| | | 10 2 | 9 2 | 8 2 | | 5 2 |
| | | 7 2 | 4+5 2 | 7 2 | | 6 2 |
| | | 8 2 | 8 2 | 9 2 | | 5 2 |
| | | 12 2 | 6+7 2 | 7 2 | | 6 2 |
| | | 11 2 | 8 2 | 9 2 | | 6 2 |
| | | 10 2 | 9 2 | 8 2 | | 5 2 |
| | | 8 2 | 8 2 | 7 2 | | 6 2 |
| | | 12 2 | 6+7 2 | 9 2 | | 6 2 |
| | | 10 2 | 9 2 | 8 2 | | 5 2 |
| | | 12 2 | 8 2 | 8 2 | | 6 2 |
| | | 9 2 | 8 2 | 8 2 | | 5 2 |
| | | 7 2 | 6+7 2 | 7 2 | | 6 2 |
| | | 11 2 | 9 2 | 9 2 | | 5 2 |
| | | 10 2 | 4+5 2 | 7 2 | | 6 2 |
| 194.1.21 | 422989 | | 72 | | Mn1O3Y1 | |
| | -A | -GM | -H | -K | -L | -M |
| | 6 4 | 8 2 | 8 2 | 8 2 | 3+4 4 | 5 2 |
| | 6 4 | 9 2 | 9 2 | 9 2 | 3+4 4 | 5 2 |
| | 6 4 | 11 2 | 9 2 | 8 2 | 3+4 4 | 6 2 |
| | 4+5 4 | 12 2 | 8 2 | 9 2 | 3+4 4 | 6 2 |
| | 6 4 | 11 2 | 4+5 2 | 7 2 | 3+4 4 | 6 2 |
| | 6 4 | 12 2 | 6+7 2 | 7 2 | 3+4 4 | 6 2 |
| | 6 4 | 10 2 | 9 2 | 9 2 | 3+4 4 | 6 2 |
| | 6 4 | 10 2 | 8 2 | 8 2 | 3+4 4 | 6 2 |
| | 6 4 | 9 2 | 8 2 | 9 2 | 3+4 4 | 5 2 |
| | 4+5 4 | 8 2 | 9 2 | 9 2 | 3+4 4 | 5 2 |
| | 6 4 | 11 2 | 9 2 | 8 2 | 3+4 4 | 5 2 |
| | 4+5 4 | 9 2 | 4+5 2 | 7 2 | 3+4 4 | 6 2 |
| | 6 4 | 11 2 | 8 2 | 9 2 | 3+4 4 | 6 2 |
| | 6 4 | 12 2 | 6+7 2 | 7 2 | 3+4 4 | 6 2 |
| | 4+5 4 | 9 2 | 9 2 | 8 2 | 3+4 4 | 5 2 |
| | 6 4 | 8 2 | 4+5 2 | 8 2 | 3+4 4 | 5 2 |
| | 6 4 | 9 2 | 9 2 | 7 2 | 3+4 4 | 6 2 |
| | 4+5 4 | 7 2 | 8 2 | 7 2 | 3+4 4 | 6 2 |
| | | 8 2 | 8 2 | 8 2 | | 6 2 |
| | | 7 2 | 6+7 2 | 9 2 | | 5 2 |
| | | 11 2 | 9 2 | 8 2 | | 5 2 |
| | | 12 2 | 4+5 2 | 7 2 | | 5 2 |
| | | 10 2 | 9 2 | 8 2 | | 5 2 |
| | | 7 2 | 4+5 2 | 7 2 | | 6 2 |
| | | 8 2 | 8 2 | 9 2 | | 5 2 |
| | | 12 2 | 6+7 2 | 7 2 | | 6 2 |
| | | 11 2 | 8 2 | 9 2 | | 6 2 |

Continued on next column

**TABLE LXXX– continued**

| TQC | ICSD | | NoE | | SF | |
|---|---|---|---|---|---|---|
| | | 10 2 | 9 2 | 8 2 | | 5 2 |
| | | 8 2 | 8 2 | 7 2 | | 6 2 |
| | | 12 2 | 6+7 2 | 9 2 | | 6 2 |
| | | 10 2 | 9 2 | 8 2 | | 5 2 |
| | | 12 2 | 8 2 | 8 2 | | 6 2 |
| | | 9 2 | 8 2 | 8 2 | | 5 2 |
| | | 7 2 | 6+7 2 | 7 2 | | 6 2 |
| | | 11 2 | 9 2 | 9 2 | | 5 2 |
| | | 10 2 | 4+5 2 | 7 2 | | 6 2 |
| 194.1.22 | 422982 | | 72 | | Mn1O3Y1 | |
| | -A | -GM | -H | -K | -L | -M |
| | 6 4 | 8 2 | 8 2 | 8 2 | 3+4 4 | 5 2 |
| | 6 4 | 9 2 | 9 2 | 9 2 | 3+4 4 | 5 2 |
| | 6 4 | 11 2 | 9 2 | 8 2 | 3+4 4 | 6 2 |
| | 4+5 4 | 12 2 | 8 2 | 9 2 | 3+4 4 | 6 2 |
| | 6 4 | 11 2 | 4+5 2 | 7 2 | 3+4 4 | 6 2 |
| | 6 4 | 12 2 | 6+7 2 | 7 2 | 3+4 4 | 6 2 |
| | 6 4 | 10 2 | 9 2 | 9 2 | 3+4 4 | 6 2 |
| | 6 4 | 10 2 | 8 2 | 8 2 | 3+4 4 | 6 2 |
| | 6 4 | 9 2 | 9 2 | 9 2 | 3+4 4 | 5 2 |
| | 4+5 4 | 8 2 | 8 2 | 9 2 | 3+4 4 | 5 2 |
| | 6 4 | 11 2 | 9 2 | 8 2 | 3+4 4 | 5 2 |
| | 4+5 4 | 9 2 | 4+5 2 | 7 2 | 3+4 4 | 6 2 |
| | 6 4 | 11 2 | 8 2 | 9 2 | 3+4 4 | 6 2 |
| | 6 4 | 12 2 | 6+7 2 | 7 2 | 3+4 4 | 6 2 |
| | 4+5 4 | 9 2 | 9 2 | 8 2 | 3+4 4 | 5 2 |
| | 6 4 | 8 2 | 4+5 2 | 8 2 | 3+4 4 | 5 2 |
| | 6 4 | 9 2 | 9 2 | 7 2 | 3+4 4 | 6 2 |
| | 4+5 4 | 7 2 | 8 2 | 7 2 | 3+4 4 | 6 2 |
| | | 8 2 | 8 2 | 8 2 | | 6 2 |
| | | 7 2 | 6+7 2 | 9 2 | | 5 2 |
| | | 11 2 | 9 2 | 8 2 | | 5 2 |
| | | 12 2 | 4+5 2 | 7 2 | | 5 2 |
| | | 10 2 | 9 2 | 8 2 | | 5 2 |
| | | 7 2 | 4+5 2 | 7 2 | | 6 2 |
| | | 8 2 | 8 2 | 9 2 | | 5 2 |
| | | 12 2 | 6+7 2 | 7 2 | | 6 2 |
| | | 11 2 | 8 2 | 9 2 | | 6 2 |
| | | 10 2 | 9 2 | 8 2 | | 5 2 |
| | | 8 2 | 8 2 | 7 2 | | 6 2 |
| | | 12 2 | 6+7 2 | 9 2 | | 6 2 |
| | | 10 2 | 9 2 | 8 2 | | 5 2 |
| | | 12 2 | 8 2 | 8 2 | | 6 2 |
| | | 9 2 | 8 2 | 8 2 | | 5 2 |
| | | 7 2 | 6+7 2 | 7 2 | | 6 2 |
| | | 11 2 | 9 2 | 9 2 | | 5 2 |
| | | 10 2 | 4+5 2 | 7 2 | | 6 2 |
| 194.1.23 | 182284 | | 72 | | Mn1O3Y1 | |
| | -A | -GM | -H | -K | -L | -M |
| | 6 4 | 8 2 | 8 2 | 8 2 | 3+4 4 | 5 2 |
| | 6 4 | 9 2 | 9 2 | 9 2 | 3+4 4 | 5 2 |
| | 6 4 | 11 2 | 9 2 | 8 2 | 3+4 4 | 6 2 |
| | 4+5 4 | 12 2 | 8 2 | 9 2 | 3+4 4 | 6 2 |
| | 6 4 | 11 2 | 4+5 2 | 7 2 | 3+4 4 | 6 2 |
| | 6 4 | 12 2 | 6+7 2 | 7 2 | 3+4 4 | 6 2 |
| | 6 4 | 10 2 | 9 2 | 9 2 | 3+4 4 | 6 2 |
| | 6 4 | 10 2 | 8 2 | 8 2 | 3+4 4 | 6 2 |
| | 6 4 | 9 2 | 8 2 | 9 2 | 3+4 4 | 5 2 |
| | 4+5 4 | 8 2 | 9 2 | 9 2 | 3+4 4 | 5 2 |
| | 6 4 | 11 2 | 9 2 | 8 2 | 3+4 4 | 5 2 |
| | 4+5 4 | 9 2 | 4+5 2 | 7 2 | 3+4 4 | 6 2 |

Continued on next column

**TABLE LXXX– continued**

| TQC | ICSD | | NoE | | SF | |
|---|---|---|---|---|---|---|
| | 6 4 | 11 2 | 8 2 | 9 2 | 3+4 4 | 6 2 |
| | 6 4 | 12 2 | 6+7 2 | 7 2 | 3+4 4 | 6 2 |
| | 4+5 4 | 9 2 | 9 2 | 8 2 | 3+4 4 | 5 2 |
| | 6 4 | 8 2 | 4+5 2 | 8 2 | 3+4 4 | 5 2 |
| | 6 4 | 8 2 | 9 2 | 7 2 | 3+4 4 | 6 2 |
| | 4+5 4 | 7 2 | 8 2 | 7 2 | 3+4 4 | 6 2 |
| | | 9 2 | 8 2 | 8 2 | | 5 2 |
| | | 7 2 | 6+7 2 | 9 2 | | 6 2 |
| | | 12 2 | 9 2 | 8 2 | | 5 2 |
| | | 10 2 | 4+5 2 | 8 2 | | 5 2 |
| | | 11 2 | 9 2 | 7 2 | | 5 2 |
| | | 7 2 | 4+5 2 | 7 2 | | 6 2 |
| | | 8 2 | 8 2 | 9 2 | | 5 2 |
| | | 12 2 | 6+7 2 | 7 2 | | 6 2 |
| | | 11 2 | 8 2 | 9 2 | | 6 2 |
| | | 10 2 | 9 2 | 8 2 | | 5 2 |
| | | 8 2 | 8 2 | 7 2 | | 6 2 |
| | | 12 2 | 6+7 2 | 9 2 | | 6 2 |
| | | 10 2 | 9 2 | 8 2 | | 5 2 |
| | | 12 2 | 8 2 | 8 2 | | 6 2 |
| | | 9 2 | 8 2 | 8 2 | | 5 2 |
| | | 7 2 | 6+7 2 | 7 2 | | 6 2 |
| | | 11 2 | 9 2 | 9 2 | | 5 2 |
| | | 10 2 | 4+5 2 | 7 2 | | 6 2 |
| 194.1.24 | 181182 | | 72 | | Mn1O3Y1 | |
| | -A | -GM | -H | -K | -L | -M |
| | 6 4 | 8 2 | 8 2 | 8 2 | 3+4 4 | 5 2 |
| | 6 4 | 9 2 | 9 2 | 9 2 | 3+4 4 | 5 2 |
| | 6 4 | 11 2 | 9 2 | 8 2 | 3+4 4 | 6 2 |
| | 4+5 4 | 12 2 | 8 2 | 9 2 | 3+4 4 | 6 2 |
| | 6 4 | 11 2 | 4+5 2 | 7 2 | 3+4 4 | 6 2 |
| | 6 4 | 12 2 | 6+7 2 | 7 2 | 3+4 4 | 6 2 |
| | 6 4 | 10 2 | 9 2 | 9 2 | 3+4 4 | 6 2 |
| | 6 4 | 10 2 | 8 2 | 8 2 | 3+4 4 | 6 2 |
| | 6 4 | 9 2 | 8 2 | 9 2 | 3+4 4 | 5 2 |
| | 4+5 4 | 8 2 | 9 2 | 9 2 | 3+4 4 | 5 2 |
| | 6 4 | 11 2 | 9 2 | 8 2 | 3+4 4 | 5 2 |
| | 4+5 4 | 9 2 | 4+5 2 | 7 2 | 3+4 4 | 6 2 |
| | 6 4 | 11 2 | 8 2 | 9 2 | 3+4 4 | 6 2 |
| | 6 4 | 12 2 | 6+7 2 | 7 2 | 3+4 4 | 6 2 |
| | 4+5 4 | 9 2 | 9 2 | 8 2 | 3+4 4 | 5 2 |
| | 6 4 | 8 2 | 4+5 2 | 8 2 | 3+4 4 | 5 2 |
| | 6 4 | 8 2 | 9 2 | 7 2 | 3+4 4 | 6 2 |
| | 4+5 4 | 7 2 | 8 2 | 7 2 | 3+4 4 | 6 2 |
| | | 9 2 | 8 2 | 8 2 | | 5 2 |
| | | 7 2 | 6+7 2 | 9 2 | | 6 2 |
| | | 12 2 | 9 2 | 8 2 | | 5 2 |
| | | 10 2 | 4+5 2 | 8 2 | | 5 2 |
| | | 11 2 | 9 2 | 7 2 | | 5 2 |
| | | 7 2 | 4+5 2 | 7 2 | | 6 2 |
| | | 8 2 | 8 2 | 9 2 | | 5 2 |
| | | 12 2 | 6+7 2 | 7 2 | | 6 2 |
| | | 11 2 | 8 2 | 9 2 | | 6 2 |
| | | 10 2 | 9 2 | 8 2 | | 5 2 |
| | | 8 2 | 8 2 | 7 2 | | 6 2 |
| | | 12 2 | 6+7 2 | 9 2 | | 6 2 |
| | | 10 2 | 9 2 | 8 2 | | 5 2 |
| | | 12 2 | 8 2 | 8 2 | | 6 2 |
| | | 9 2 | 8 2 | 8 2 | | 5 2 |
| | | 7 2 | 6+7 2 | 7 2 | | 6 2 |
| | | 11 2 | 9 2 | 9 2 | | 5 2 |

Continued on next column

**TABLE LXXX– continued**

| TQC | ICSD | | NoE | | SF | |
|---|---|---|---|---|---|---|
| | | 10 2 | 4+5 2 | 7 2 | | 6 2 |
| 194.1.25 | 193568 | | 176 | | Mg1Ni2 | |
| | -A | -GM | -H | -K | -L | -M |
| | 6 4 | 9 2 | 4+5 2 | 7 2 | 3+4 4 | 6 2 |
| | 6 4 | 11 2 | 9 2 | 8 2 | 3+4 4 | 5 2 |
| | 6 4 | 8 2 | 6+7 2 | 9 2 | 3+4 4 | 6 2 |
| | 6 4 | 9 2 | 8 2 | 7 2 | 3+4 4 | 6 2 |
| | 6 4 | 12 2 | 8 2 | 9 2 | 3+4 4 | 5 2 |
| | 6 4 | 8 2 | 9 2 | 9 2 | 3+4 4 | 5 2 |
| | 6 4 | 9 2 | 6+7 2 | 7 2 | 3+4 4 | 6 2 |
| | 4+5 4 | 11 2 | 8 2 | 8 2 | 3+4 4 | 5 2 |
| | 6 4 | 9 2 | 4+5 2 | 8 2 | 3+4 4 | 5 2 |
| | 6 4 | 8 2 | 9 2 | 7 2 | 3+4 4 | 6 2 |
| | 4+5 4 | 12 2 | 8 2 | 9 2 | 3+4 4 | 6 2 |
| | 6 4 | 11 2 | 9 2 | 7 2 | 3+4 4 | 6 2 |
| | 4+5 4 | 8 2 | 9 2 | 8 2 | 3+4 4 | 5 2 |
| | 6 4 | 8 2 | 4+5 2 | 9 2 | 3+4 4 | 5 2 |
| | 4+5 4 | 7 2 | 8 2 | 8 2 | 3+4 4 | 6 2 |
| | 6 4 | 9 2 | 9 2 | 9 2 | 3+4 4 | 6 2 |
| | 4+5 4 | 7 2 | 6+7 2 | 8 2 | 3+4 4 | 6 2 |
| | 6 4 | 9 2 | 8 2 | 7 2 | 3+4 4 | 6 2 |
| | 4+5 4 | 12 2 | 4+5 2 | 9 2 | 3+4 4 | 6 2 |
| | 6 4 | 10 2 | 9 2 | 7 2 | 3+4 4 | 5 2 |
| | 6 4 | 8 2 | 6+7 2 | 7 2 | 3+4 4 | 6 2 |
| | 6 4 | 9 2 | 8 2 | 9 2 | 3+4 4 | 5 2 |
| | 4+5 4 | 7 2 | 9 2 | 7 2 | 3+4 4 | 6 2 |
| | 6 4 | 8 2 | 9 2 | 8 2 | 3+4 4 | 5 2 |
| | 6 4 | 7 2 | 8 2 | 9 2 | 3+4 4 | 5 2 |
| | 4+5 4 | 9 2 | 8 2 | 8 2 | 3+4 4 | 6 2 |
| | 6 4 | 10 2 | 4+5 2 | 8 2 | 3+4 4 | 5 2 |
| | 4+5 4 | 7 2 | 6+7 2 | 9 2 | 3+4 4 | 5 2 |
| | 6 4 | 11 2 | 8 2 | 7 2 | 3+4 4 | 6 2 |
| | 6 4 | 8 2 | 9 2 | 8 2 | 3+4 4 | 6 2 |
| | 4+5 4 | 7 2 | 9 2 | 8 2 | 3+4 4 | 6 2 |
| | 4+5 4 | 8 2 | 8 2 | 8 2 | 3+4 4 | 5 2 |
| | 6 4 | 10 2 | 8 2 | 9 2 | 3+4 4 | 5 2 |
| | 6 4 | 7 2 | 4+5 2 | 7 2 | 3+4 4 | 6 2 |
| | 6 4 | 9 2 | 9 2 | 7 2 | 3+4 4 | 6 2 |
| | 6 4 | 12 2 | 8 2 | 9 2 | 3+4 4 | 6 2 |
| | 4+5 4 | 7 2 | 6+7 2 | 7 2 | 3+4 4 | 5 2 |
| | 6 4 | 9 2 | 6+7 2 | 9 2 | 3+4 4 | 6 2 |
| | 4+5 4 | 10 2 | 9 2 | 8 2 | 3+4 4 | 5 2 |
| | 6 4 | 11 2 | 8 2 | 8 2 | 3+4 4 | 5 2 |
| | 6 4 | 12 2 | 4+5 2 | 7 2 | 3+4 4 | 5 2 |
| | 4+5 4 | 9 2 | 6+7 2 | 9 2 | 3+4 4 | 6 2 |
| | 6 4 | 10 2 | 9 2 | 9 2 | 3+4 4 | 6 2 |
| | 6 4 | 11 2 | 9 2 | 8 2 | 3+4 4 | 6 2 |
| | | 8 2 | 8 2 | 7 2 | | 5 2 |
| | | 7 2 | 8 2 | 8 2 | | 5 2 |
| | | 12 2 | 9 2 | 9 2 | | 5 2 |
| | | 8 2 | 8 2 | 8 2 | | 6 2 |
| | | 10 2 | 4+5 2 | 7 2 | | 5 2 |
| | | 9 2 | 9 2 | 9 2 | | 6 2 |
| | | 11 2 | 6+7 2 | 7 2 | | 5 2 |
| | | 11 2 | 8 2 | 8 2 | | 5 2 |
| | | 7 2 | 9 2 | 8 2 | | 5 2 |
| | | 8 2 | 4+5 2 | 7 2 | | 6 2 |
| | | 7 2 | 8 2 | 9 2 | | 6 2 |
| | | 9 2 | 9 2 | 8 2 | | 5 2 |
| | | 12 2 | 6+7 2 | 7 2 | | 5 2 |
| | | 10 2 | 8 2 | 9 2 | | 6 2 |

Continued on next column

**TABLE LXXX– continued**

| TQC | ICSD | | NoE | | SF | |
|---|---|---|---|---|---|---|
| | | 7 2 | 9 2 | 9 2 | | 6 2 |
| | | 8 2 | 4+5 2 | 7 2 | | 6 2 |
| | | 7 2 | 9 2 | 9 2 | | 6 2 |
| | | 9 2 | 8 2 | 9 2 | | 5 2 |
| | | 7 2 | 8 2 | 7 2 | | 6 2 |
| | | 12 2 | 6+7 2 | 8 2 | | 5 2 |
| | | 8 2 | 4+5 2 | 7 2 | | 5 2 |
| | | 10 2 | 9 2 | 8 2 | | 6 2 |
| | | 8 2 | 8 2 | 9 2 | | 6 2 |
| | | 9 2 | 4+5 2 | 7 2 | | 6 2 |
| | | 8 2 | 6+7 2 | 9 2 | | 5 2 |
| | | 7 2 | 9 2 | 8 2 | | 5 2 |
| | | 9 2 | 9 2 | 7 2 | | 5 2 |
| | | 8 2 | 8 2 | 8 2 | | 6 2 |
| | | 12 2 | 9 2 | 9 2 | | 6 2 |
| | | 8 2 | 8 2 | 9 2 | | 5 2 |
| | | 9 2 | 4+5 2 | 8 2 | | 6 2 |
| | | 7 2 | 9 2 | 7 2 | | 6 2 |
| | | 7 2 | 9 2 | 8 2 | | 6 2 |
| | | 11 2 | 6+7 2 | 7 2 | | 6 2 |
| | | 10 2 | 8 2 | 7 2 | | 5 2 |
| | | 9 2 | 4+5 2 | 9 2 | | 6 2 |
| | | 7 2 | 8 2 | 8 2 | | 6 2 |
| | | 8 2 | 9 2 | 9 2 | | 6 2 |
| | | 7 2 | 6+7 2 | 8 2 | | 5 2 |
| | | 9 2 | 8 2 | 7 2 | | 6 2 |
| | | 11 2 | 9 2 | 8 2 | | 5 2 |
| | | 8 2 | 8 2 | 9 2 | | 5 2 |
| | | 9 2 | 9 2 | 7 2 | | 6 2 |
| | | 12 2 | 4+5 2 | 9 2 | | 6 2 |
| 194.1.26 | 171451 | | 36 | | I1La1 | |
| | -A | -GM | -H | -K | -L | -M |
| | 6 4 | 9 2 | 8 2 | 9 2 | 3+4 4 | 5 2 |
| | 6 4 | 8 2 | 9 2 | 8 2 | 3+4 4 | 5 2 |
| | 6 4 | 11 2 | 9 2 | 8 2 | 3+4 4 | 6 2 |
| | 4+5 4 | 12 2 | 8 2 | 9 2 | 3+4 4 | 6 2 |
| | 6 4 | 11 2 | 4+5 2 | 7 2 | 3+4 4 | 6 2 |
| | 6 4 | 12 2 | 9 2 | 9 2 | 3+4 4 | 6 2 |
| | 6 4 | 10 2 | 6+7 2 | 7 2 | 3+4 4 | 6 2 |
| | 4+5 4 | 10 2 | 8 2 | 8 2 | 3+4 4 | 6 2 |
| | 6 4 | 9 2 | 9 2 | 8 2 | 3+4 4 | 5 2 |
| | | 11 2 | 4+5 2 | 7 2 | | 6 2 |
| | | 8 2 | 6+7 2 | 9 2 | | 5 2 |
| | | 12 2 | 8 2 | 7 2 | | 6 2 |
| | | 8 2 | 8 2 | 8 2 | | 5 2 |
| | | 12 2 | 9 2 | 9 2 | | 6 2 |
| | | 7 2 | 8 2 | 7 2 | | 5 2 |
| | | 10 2 | 6+7 2 | 9 2 | | 6 2 |
| | | 9 2 | 9 2 | 9 2 | | 5 2 |
| | | 8 2 | 4+5 2 | 7 2 | | 5 2 |
| 194.1.27 | 163344 | | 136 | | Gd1Ge1Zn1 | |
| | -A | -GM | -H | -K | -L | -M |
| | 6 4 | 8 2 | 8 2 | 8 2 | 3+4 4 | 5 2 |
| | 6 4 | 9 2 | 9 2 | 9 2 | 3+4 4 | 5 2 |
| | 6 4 | 9 2 | 8 2 | 9 2 | 3+4 4 | 5 2 |
| | 6 4 | 11 2 | 9 2 | 9 2 | 3+4 4 | 6 2 |
| | 6 4 | 12 2 | 8 2 | 8 2 | 3+4 4 | 6 2 |
| | 6 4 | 11 2 | 9 2 | 9 2 | 3+4 4 | 6 2 |
| | 4+5 4 | 8 2 | 8 2 | 8 2 | 3+4 4 | 5 2 |
| | 4+5 4 | 12 2 | 9 2 | 8 2 | 3+4 4 | 6 2 |
| | 6 4 | 12 2 | 8 2 | 8 2 | 3+4 4 | 6 2 |
| | 6 4 | 11 2 | 9 2 | 7 2 | 3+4 4 | 6 2 |
| | 4+5 4 | 8 2 | 4+5 2 | 7 2 | 3+4 4 | 6 2 |
| | 4+5 4 | 10 2 | 6+7 2 | 9 2 | 3+4 4 | 5 2 |
| | 6 4 | 10 2 | 4+5 2 | 8 2 | 3+4 4 | 6 2 |
| | 6 4 | 12 2 | 6+7 2 | 7 2 | 3+4 4 | 6 2 |
| | 6 4 | 7 2 | 8 2 | 7 2 | 3+4 4 | 5 2 |
| | 6 4 | 10 2 | 9 2 | 8 2 | 3+4 4 | 6 2 |
| | 4+5 4 | 9 2 | 8 2 | 8 2 | 3+4 4 | 5 2 |
| | 4+5 4 | 8 2 | 9 2 | 9 2 | 3+4 4 | 5 2 |
| | 6 4 | 11 2 | 6+7 2 | 7 2 | 3+4 4 | 6 2 |
| | 6 4 | 12 2 | 4+5 2 | 7 2 | 3+4 4 | 6 2 |
| | 6 4 | 7 2 | 9 2 | 9 2 | 3+4 4 | 5 2 |
| | 6 4 | 10 2 | 8 2 | 8 2 | 3+4 4 | 5 2 |
| | 4+5 4 | 7 2 | 8 2 | 9 2 | 3+4 4 | 6 2 |
| | 6 4 | 10 2 | 9 2 | 8 2 | 3+4 4 | 6 2 |
| | 4+5 4 | 9 2 | 4+5 2 | 7 2 | 3+4 4 | 5 2 |
| | 6 4 | 11 2 | 6+7 2 | 7 2 | 3+4 4 | 5 2 |
| | 6 4 | 8 2 | 9 2 | 9 2 | 3+4 4 | 6 2 |
| | 6 4 | 12 2 | 8 2 | 8 2 | 3+4 4 | 6 2 |
| | 6 4 | 9 2 | 8 2 | 9 2 | 3+4 4 | 5 2 |
| | 4+5 4 | 11 2 | 9 2 | 8 2 | 3+4 4 | 5 2 |
| | 6 4 | 8 2 | 8 2 | 8 2 | 3+4 4 | 6 2 |
| | 4+5 4 | 12 2 | 9 2 | 9 2 | 3+4 4 | 6 2 |
| | 6 4 | 7 2 | 4+5 2 | 7 2 | 3+4 4 | 5 2 |
| | 6 4 | 10 2 | 9 2 | 8 2 | 3+4 4 | 6 2 |
| | | 9 2 | 6+7 2 | 9 2 | | 5 2 |
| | | 7 2 | 8 2 | 7 2 | | 6 2 |
| | | 10 2 | 9 2 | 9 2 | | 5 2 |
| | | 11 2 | 8 2 | 8 2 | | 6 2 |
| | | 8 2 | 4+5 2 | 7 2 | | 5 2 |
| | | 12 2 | 6+7 2 | 7 2 | | 6 2 |
| | | 9 2 | 9 2 | 8 2 | | 5 2 |
| | | 11 2 | 4+5 2 | 7 2 | | 6 2 |
| | | 8 2 | 8 2 | 9 2 | | 5 2 |
| | | 8 2 | 6+7 2 | 7 2 | | 6 2 |
| | | 7 2 | 9 2 | 9 2 | | 5 2 |
| | | 12 2 | 8 2 | 9 2 | | 5 2 |
| | | 7 2 | 8 2 | 8 2 | | 6 2 |
| | | 10 2 | 9 2 | 8 2 | | 5 2 |
| | | 10 2 | 8 2 | 8 2 | | 6 2 |
| | | 9 2 | 6+7 2 | 7 2 | | 6 2 |
| | | 11 2 | 9 2 | 9 2 | | 5 2 |
| | | 12 2 | 4+5 2 | 7 2 | | 5 2 |
| | | 11 2 | 9 2 | 8 2 | | 6 2 |
| | | 10 2 | 4+5 2 | 7 2 | | 6 2 |
| | | 11 2 | 6+7 2 | 7 2 | | 6 2 |
| | | 8 2 | 8 2 | 8 2 | | 6 2 |
| | | 12 2 | 9 2 | 9 2 | | 5 2 |
| | | 9 2 | 8 2 | 8 2 | | 6 2 |
| | | 12 2 | 8 2 | 9 2 | | 6 2 |
| | | 11 2 | 9 2 | 7 2 | | 5 2 |
| | | 7 2 | 6+7 2 | 9 2 | | 6 2 |
| | | 10 2 | 8 2 | 8 2 | | 5 2 |
| | | 12 2 | 9 2 | 9 2 | | 6 2 |
| | | 9 2 | 4+5 2 | 7 2 | | 6 2 |
| | | 11 2 | 9 2 | 8 2 | | 6 2 |
| | | 10 2 | 4+5 2 | 7 2 | | 6 2 |
| | | 11 2 | 6+7 2 | 7 2 | | 5 2 |
| | | 12 2 | 8 2 | 9 2 | | 5 2 |
| 194.1.28 | 417150 | | 152 | | Fe0.25S2Ta1 | |
| | -A | -GM | -H | -K | -L | -M |

Continued on next column

**TABLE LXXX– continued**

| TQC | ICSD | | NoE | | SF | |
|---|---|---|---|---|---|---|
| | 6 4 | 9 2 | 9 2 | 9 2 | 3+4 4 | 5 2 |
| | 6 4 | 11 2 | 8 2 | 8 2 | 3+4 4 | 6 2 |
| | 4+5 4 | 8 2 | 9 2 | 7 2 | 3+4 4 | 6 2 |
| | 6 4 | 12 2 | 4+5 2 | 9 2 | 3+4 4 | 5 2 |
| | 6 4 | 10 2 | 6+7 2 | 7 2 | 3+4 4 | 5 2 |
| | 4+5 4 | 12 2 | 8 2 | 8 2 | 3+4 4 | 6 2 |
| | 6 4 | 7 2 | 8 2 | 8 2 | 3+4 4 | 6 2 |
| | 6 4 | 8 2 | 9 2 | 9 2 | 3+4 4 | 5 2 |
| | 6 4 | 11 2 | 8 2 | 7 2 | 3+4 4 | 5 2 |
| | 6 4 | 10 2 | 6+7 2 | 8 2 | 3+4 4 | 5 2 |
| | 4+5 4 | 11 2 | 4+5 2 | 7 2 | 3+4 4 | 6 2 |
| | 6 4 | 9 2 | 9 2 | 9 2 | 3+4 4 | 6 2 |
| | 6 4 | 7 2 | 8 2 | 8 2 | 3+4 4 | 6 2 |
| | 4+5 4 | 9 2 | 6+7 2 | 7 2 | 3+4 4 | 5 2 |
| | 6 4 | 12 2 | 9 2 | 9 2 | 3+4 4 | 5 2 |
| | 4+5 4 | 8 2 | 4+5 2 | 7 2 | 3+4 4 | 6 2 |
| | 6 4 | 9 2 | 9 2 | 9 2 | 3+4 4 | 5 2 |
| | 6 4 | 11 2 | 8 2 | 7 2 | 3+4 4 | 6 2 |
| | 4+5 4 | 9 2 | 8 2 | 8 2 | 3+4 4 | 5 2 |
| | 6 4 | 12 2 | 6+7 2 | 9 2 | 3+4 4 | 6 2 |
| | 6 4 | 10 2 | 9 2 | 7 2 | 3+4 4 | 6 2 |
| | 4+5 4 | 8 2 | 4+5 2 | 9 2 | 3+4 4 | 5 2 |
| | 6 4 | 7 2 | 6+7 2 | 9 2 | 3+4 4 | 5 2 |
| | 4+5 4 | 8 2 | 8 2 | 8 2 | 3+4 4 | 6 2 |
| | 6 4 | 11 2 | 9 2 | 7 2 | 3+4 4 | 5 2 |
| | 6 4 | 7 2 | 8 2 | 8 2 | 3+4 4 | 5 2 |
| | 6 4 | 9 2 | 9 2 | 9 2 | 3+4 4 | 6 2 |
| | 4+5 4 | 12 2 | 4+5 2 | 8 2 | 3+4 4 | 6 2 |
| | 6 4 | 7 2 | 8 2 | 7 2 | 3+4 4 | 5 2 |
| | 6 4 | 8 2 | 6+7 2 | 8 2 | 3+4 4 | 6 2 |
| | 6 4 | 10 2 | 8 2 | 7 2 | 3+4 4 | 5 2 |
| | 4+5 4 | 11 2 | 9 2 | 9 2 | 3+4 4 | 5 2 |
| | 6 4 | 12 2 | 9 2 | 8 2 | 3+4 4 | 5 2 |
| | 6 4 | 10 2 | 4+5 2 | 9 2 | 3+4 4 | 6 2 |
| | 4+5 4 | 9 2 | 8 2 | 7 2 | 3+4 4 | 6 2 |
| | 6 4 | 11 2 | 6+7 2 | 7 2 | 3+4 4 | 5 2 |
| | 4+5 4 | 7 2 | 8 2 | 9 2 | 3+4 4 | 5 2 |
| | 6 4 | 9 2 | 9 2 | 9 2 | 3+4 4 | 6 2 |
| | | 8 2 | 4+5 2 | 7 2 | | 6 2 |
| | | 10 2 | 9 2 | 8 2 | | 6 2 |
| | | 11 2 | 8 2 | 8 2 | | 6 2 |
| | | 12 2 | 6+7 2 | 7 2 | | 5 2 |
| | | 8 2 | 9 2 | 9 2 | | 5 2 |
| | | 11 2 | 9 2 | 7 2 | | 5 2 |
| | | 10 2 | 8 2 | 8 2 | | 6 2 |
| | | 7 2 | 4+5 2 | 8 2 | | 5 2 |
| | | 8 2 | 8 2 | 8 2 | | 6 2 |
| | | 12 2 | 6+7 2 | 7 2 | | 6 2 |
| | | 10 2 | 8 2 | 9 2 | | 5 2 |
| | | 12 2 | 9 2 | 8 2 | | 5 2 |
| | | 9 2 | 9 2 | 8 2 | | 6 2 |
| | | 7 2 | 4+5 2 | 9 2 | | 5 2 |
| | | 7 2 | 8 2 | 7 2 | | 6 2 |
| | | 8 2 | 6+7 2 | 7 2 | | 6 2 |
| | | 11 2 | 4+5 2 | 7 2 | | 5 2 |
| | | 10 2 | 8 2 | 8 2 | | 6 2 |
| | | 12 2 | 9 2 | 9 2 | | 5 2 |
| | | 9 2 | 9 2 | 8 2 | | 6 2 |
| | | 9 2 | 8 2 | 9 2 | | 5 2 |
| | | 8 2 | 6+7 2 | 7 2 | | 5 2 |
| | | 11 2 | 9 2 | 9 2 | | 6 2 |
| | | 7 2 | 8 2 | 9 2 | | 5 2 |
| | | 9 2 | 9 2 | 7 2 | | 6 2 |
| | | 10 2 | 4+5 2 | 9 2 | | 5 2 |
| | | 12 2 | 9 2 | 8 2 | | 6 2 |
| | | 8 2 | 8 2 | 8 2 | | 5 2 |
| | | 7 2 | 9 2 | 7 2 | | 5 2 |
| | | 11 2 | 6+7 2 | 8 2 | | 5 2 |
| | | 12 2 | 4+5 2 | 7 2 | | 6 2 |
| | | 10 2 | 8 2 | 9 2 | | 5 2 |
| | | 7 2 | 4+5 2 | 7 2 | | 5 2 |
| | | 9 2 | 9 2 | 9 2 | | 6 2 |
| | | 8 2 | 6+7 2 | 7 2 | | 6 2 |
| | | 8 2 | 8 2 | 8 2 | | 5 2 |
| | | 7 2 | 9 2 | 8 2 | | 5 2 |
| | | 9 2 | 4+5 2 | 7 2 | | 6 2 |
| 194.1.29 | 249600 | | 96 | | Eu1Ga1Sn1 | |
| | -A | -GM | -H | -K | -L | -M |
| | 6 4 | 8 2 | 8 2 | 8 2 | 3+4 4 | 5 2 |
| | 6 4 | 9 2 | 9 2 | 9 2 | 3+4 4 | 5 2 |
| | 6 4 | 9 2 | 8 2 | 9 2 | 3+4 4 | 5 2 |
| | 6 4 | 11 2 | 9 2 | 9 2 | 3+4 4 | 6 2 |
| | 6 4 | 11 2 | 8 2 | 8 2 | 3+4 4 | 6 2 |
| | 4+5 4 | 12 2 | 9 2 | 9 2 | 3+4 4 | 6 2 |
| | 6 4 | 8 2 | 8 2 | 8 2 | 3+4 4 | 5 2 |
| | 4+5 4 | 12 2 | 9 2 | 8 2 | 3+4 4 | 6 2 |
| | 6 4 | 12 2 | 8 2 | 8 2 | 3+4 4 | 6 2 |
| | 6 4 | 11 2 | 9 2 | 9 2 | 3+4 4 | 6 2 |
| | 6 4 | 10 2 | 4+5 2 | 7 2 | 3+4 4 | 6 2 |
| | 6 4 | 10 2 | 6+7 2 | 7 2 | 3+4 4 | 6 2 |
| | 6 4 | 8 2 | 6+7 2 | 7 2 | 3+4 4 | 5 2 |
| | 4+5 4 | 12 2 | 4+5 2 | 7 2 | 3+4 4 | 6 2 |
| | 6 4 | 7 2 | 8 2 | 8 2 | 3+4 4 | 5 2 |
| | 4+5 4 | 10 2 | 9 2 | 8 2 | 3+4 4 | 6 2 |
| | 6 4 | 9 2 | 9 2 | 8 2 | 3+4 4 | 5 2 |
| | 6 4 | 11 2 | 8 2 | 9 2 | 3+4 4 | 5 2 |
| | 6 4 | 8 2 | 6+7 2 | 7 2 | 3+4 4 | 6 2 |
| | 4+5 4 | 12 2 | 4+5 2 | 7 2 | 3+4 4 | 6 2 |
| | 6 4 | 9 2 | 8 2 | 8 2 | 3+4 4 | 5 2 |
| | 6 4 | 11 2 | 6+7 2 | 7 2 | 3+4 4 | 6 2 |
| | 4+5 4 | 8 2 | 4+5 2 | 7 2 | 3+4 4 | 5 2 |
| | 6 4 | 12 2 | 9 2 | 9 2 | 3+4 4 | 6 2 |
| | | 8 2 | 8 2 | 9 2 | | 5 2 |
| | | 7 2 | 9 2 | 9 2 | | 6 2 |
| | | 12 2 | 9 2 | 8 2 | | 5 2 |
| | | 10 2 | 8 2 | 8 2 | | 6 2 |
| | | 9 2 | 8 2 | 8 2 | | 5 2 |
| | | 9 2 | 6+7 2 | 7 2 | | 5 2 |
| | | 7 2 | 9 2 | 8 2 | | 6 2 |
| | | 10 2 | 4+5 2 | 7 2 | | 5 2 |
| | | 11 2 | 8 2 | 9 2 | | 6 2 |
| | | 11 2 | 6+7 2 | 7 2 | | 6 2 |
| | | 11 2 | 4+5 2 | 7 2 | | 5 2 |
| | | 10 2 | 8 2 | 9 2 | | 5 2 |
| | | 12 2 | 9 2 | 8 2 | | 6 2 |
| | | 8 2 | 9 2 | 9 2 | | 6 2 |
| | | 11 2 | 8 2 | 9 2 | | 6 2 |
| | | 12 2 | 9 2 | 8 2 | | 6 2 |
| | | 9 2 | 4+5 2 | 7 2 | | 6 2 |
| | | 10 2 | 6+7 2 | 7 2 | | 6 2 |
| | | 12 2 | 8 2 | 8 2 | | 6 2 |
| | | 7 2 | 9 2 | 9 2 | | 5 2 |

Continued on next column

**TABLE LXXX– continued**

| TQC | -A | -GM | -H | -K | -L | -M |
|---|---|---|---|---|---|---|
| TQC | ICSD | | NoE | | SF | |
| | | 11 2 | 6+7 2 | 7 2 | | 6 2 |
| | | 9 2 | 8 2 | 8 2 | | 6 2 |
| | | 10 2 | 9 2 | 9 2 | | 5 2 |
| | | 11 2 | 4+5 2 | 7 2 | | 5 2 |
| 194.1.30 | 249599 | | 96 | | Eu1Ga1Ge1 | |
| | -A | -GM | -H | -K | -L | -M |
| | 6 4 | 8 2 | 8 2 | 8 2 | 3+4 4 | 5 2 |
| | 6 4 | 9 2 | 9 2 | 9 2 | 3+4 4 | 5 2 |
| | 6 4 | 9 2 | 8 2 | 9 2 | 3+4 4 | 5 2 |
| | 6 4 | 11 2 | 9 2 | 9 2 | 3+4 4 | 6 2 |
| | 6 4 | 11 2 | 8 2 | 8 2 | 3+4 4 | 6 2 |
| | 6 4 | 12 2 | 9 2 | 9 2 | 3+4 4 | 6 2 |
| | 4+5 4 | 8 2 | 8 2 | 8 2 | 3+4 4 | 5 2 |
| | 4+5 4 | 12 2 | 9 2 | 8 2 | 3+4 4 | 6 2 |
| | 6 4 | 12 2 | 8 2 | 8 2 | 3+4 4 | 6 2 |
| | 6 4 | 11 2 | 9 2 | 9 2 | 3+4 4 | 6 2 |
| | 6 4 | 8 2 | 4+5 2 | 7 2 | 3+4 4 | 6 2 |
| | 6 4 | 12 2 | 6+7 2 | 7 2 | 3+4 4 | 5 2 |
| | 6 4 | 10 2 | 6+7 2 | 7 2 | 3+4 4 | 6 2 |
| | 4+5 4 | 10 2 | 8 2 | 8 2 | 3+4 4 | 6 2 |
| | 6 4 | 7 2 | 9 2 | 8 2 | 3+4 4 | 5 2 |
| | 4+5 4 | 10 2 | 4+5 2 | 7 2 | 3+4 4 | 6 2 |
| | 6 4 | 9 2 | 8 2 | 9 2 | 3+4 4 | 5 2 |
| | 6 4 | 11 2 | 6+7 2 | 7 2 | 3+4 4 | 5 2 |
| | 6 4 | 8 2 | 9 2 | 8 2 | 3+4 4 | 6 2 |
| | 6 4 | 12 2 | 4+5 2 | 7 2 | 3+4 4 | 6 2 |
| | 4+5 4 | 9 2 | 8 2 | 8 2 | 3+4 4 | 5 2 |
| | 6 4 | 11 2 | 4+5 2 | 7 2 | 3+4 4 | 6 2 |
| | 4+5 4 | 8 2 | 9 2 | 9 2 | 3+4 4 | 5 2 |
| | 6 4 | 12 2 | 6+7 2 | 7 2 | 3+4 4 | 6 2 |
| | | 8 2 | 8 2 | 9 2 | | 5 2 |
| | | 7 2 | 9 2 | 9 2 | | 6 2 |
| | | 12 2 | 9 2 | 8 2 | | 5 2 |
| | | 9 2 | 8 2 | 8 2 | | 5 2 |
| | | 10 2 | 9 2 | 8 2 | | 6 2 |
| | | 7 2 | 8 2 | 7 2 | | 6 2 |
| | | 9 2 | 4+5 2 | 9 2 | | 5 2 |
| | | 10 2 | 6+7 2 | 7 2 | | 5 2 |
| | | 11 2 | 8 2 | 8 2 | | 6 2 |
| | | 11 2 | 6+7 2 | 7 2 | | 6 2 |
| | | 12 2 | 4+5 2 | 7 2 | | 5 2 |
| | | 11 2 | 9 2 | 9 2 | | 5 2 |
| | | 10 2 | 9 2 | 9 2 | | 6 2 |
| | | 8 2 | 8 2 | 8 2 | | 6 2 |
| | | 12 2 | 8 2 | 9 2 | | 6 2 |
| | | 11 2 | 9 2 | 8 2 | | 6 2 |
| | | 9 2 | 6+7 2 | 7 2 | | 6 2 |
| | | 12 2 | 9 2 | 8 2 | | 6 2 |
| | | 10 2 | 8 2 | 9 2 | | 6 2 |
| | | 7 2 | 4+5 2 | 7 2 | | 5 2 |
| | | 11 2 | 8 2 | 8 2 | | 6 2 |
| | | 9 2 | 6+7 2 | 7 2 | | 6 2 |
| | | 10 2 | 9 2 | 9 2 | | 5 2 |
| | | 8 2 | 4+5 2 | 7 2 | | 5 2 |
| 194.1.31 | 158979 | | 64 | | Cr1Ge3Nd1 | |
| | -A | -GM | -H | -K | -L | -M |
| | 6 4 | 9 2 | 6+7 2 | 7 2 | 3+4 4 | 6 2 |
| | 6 4 | 11 2 | 8 2 | 8 2 | 3+4 4 | 5 2 |
| | 6 4 | 8 2 | 4+5 2 | 7 2 | 3+4 4 | 6 2 |
| | 4+5 4 | 12 2 | 9 2 | 9 2 | 3+4 4 | 5 2 |
| | 6 4 | 7 2 | 9 2 | 9 2 | 3+4 4 | 6 2 |

Continued on next column

**TABLE LXXX– continued**

| TQC | -A | -GM | -H | -K | -L | -M |
|---|---|---|---|---|---|---|
| TQC | ICSD | | NoE | | SF | |
| | 6 4 | 8 2 | 4+5 2 | 8 2 | 3+4 4 | 5 2 |
| | 4+5 4 | 10 2 | 8 2 | 7 2 | 3+4 4 | 6 2 |
| | 6 4 | 12 2 | 9 2 | 9 2 | 3+4 4 | 5 2 |
| | 6 4 | 9 2 | 4+5 2 | 7 2 | 3+4 4 | 6 2 |
| | 4+5 4 | 11 2 | 9 2 | 8 2 | 3+4 4 | 5 2 |
| | 6 4 | 12 2 | 9 2 | 9 2 | 3+4 4 | 5 2 |
| | 4+5 4 | 10 2 | 8 2 | 7 2 | 3+4 4 | 6 2 |
| | 6 4 | 8 2 | 6+7 2 | 8 2 | 3+4 4 | 6 2 |
| | 6 4 | 7 2 | 8 2 | 9 2 | 3+4 4 | 5 2 |
| | 4+5 4 | 9 2 | 8 2 | 8 2 | 3+4 4 | 5 2 |
| | 6 4 | 8 2 | 6+7 2 | 9 2 | 3+4 4 | 5 2 |
| | | 11 2 | 4+5 2 | 7 2 | | 6 2 |
| | | 8 2 | 9 2 | 8 2 | | 5 2 |
| | | 7 2 | 9 2 | 7 2 | | 6 2 |
| | | 7 2 | 8 2 | 8 2 | | 6 2 |
| | | 9 2 | 4+5 2 | 9 2 | | 6 2 |
| | | 12 2 | 9 2 | 7 2 | | 5 2 |
| | | 10 2 | 8 2 | 9 2 | | 5 2 |
| | | 11 2 | 6+7 2 | 9 2 | | 5 2 |
| | | 10 2 | 8 2 | 7 2 | | 5 2 |
| | | 9 2 | 9 2 | 9 2 | | 5 2 |
| | | 7 2 | 8 2 | 7 2 | | 6 2 |
| | | 8 2 | 6+7 2 | 8 2 | | 5 2 |
| | | 12 2 | 4+5 2 | 7 2 | | 5 2 |
| | | 11 2 | 9 2 | 8 2 | | 6 2 |
| | | 7 2 | 9 2 | 8 2 | | 6 2 |
| | | 9 2 | 8 2 | 9 2 | | 6 2 |
| 194.1.32 | 420050 | | 112 | | Ce1Sn1Zn1 | |
| | -A | -GM | -H | -K | -L | -M |
| | 6 4 | 8 2 | 8 2 | 8 2 | 3+4 4 | 5 2 |
| | 6 4 | 9 2 | 9 2 | 9 2 | 3+4 4 | 5 2 |
| | 6 4 | 9 2 | 8 2 | 9 2 | 3+4 4 | 5 2 |
| | 6 4 | 11 2 | 9 2 | 9 2 | 3+4 4 | 6 2 |
| | 6 4 | 12 2 | 8 2 | 8 2 | 3+4 4 | 6 2 |
| | 4+5 4 | 11 2 | 9 2 | 9 2 | 3+4 4 | 6 2 |
| | 6 4 | 8 2 | 8 2 | 8 2 | 3+4 4 | 5 2 |
| | 4+5 4 | 12 2 | 9 2 | 8 2 | 3+4 4 | 6 2 |
| | 6 4 | 12 2 | 4+5 2 | 8 2 | 3+4 4 | 6 2 |
| | 6 4 | 11 2 | 6+7 2 | 7 2 | 3+4 4 | 6 2 |
| | 4+5 4 | 10 2 | 8 2 | 7 2 | 3+4 4 | 6 2 |
| | 4+5 4 | 10 2 | 9 2 | 9 2 | 3+4 4 | 6 2 |
| | 6 4 | 8 2 | 4+5 2 | 7 2 | 3+4 4 | 5 2 |
| | 6 4 | 12 2 | 6+7 2 | 7 2 | 3+4 4 | 6 2 |
| | 6 4 | 7 2 | 8 2 | 8 2 | 3+4 4 | 5 2 |
| | 6 4 | 10 2 | 9 2 | 8 2 | 3+4 4 | 6 2 |
| | 4+5 4 | 9 2 | 8 2 | 8 2 | 3+4 4 | 5 2 |
| | 6 4 | 11 2 | 9 2 | 9 2 | 3+4 4 | 5 2 |
| | 4+5 4 | 8 2 | 6+7 2 | 7 2 | 3+4 4 | 6 2 |
| | 6 4 | 12 2 | 4+5 2 | 7 2 | 3+4 4 | 6 2 |
| | 6 4 | 7 2 | 9 2 | 9 2 | 3+4 4 | 5 2 |
| | 6 4 | 10 2 | 8 2 | 8 2 | 3+4 4 | 5 2 |
| | 6 4 | 7 2 | 8 2 | 9 2 | 3+4 4 | 6 2 |
| | 4+5 4 | 10 2 | 9 2 | 8 2 | 3+4 4 | 6 2 |
| | 6 4 | 9 2 | 4+5 2 | 7 2 | 3+4 4 | 5 2 |
| | 4+5 4 | 8 2 | 6+7 2 | 7 2 | 3+4 4 | 5 2 |
| | 6 4 | 11 2 | 9 2 | 9 2 | 3+4 4 | 6 2 |
| | 6 4 | 12 2 | 8 2 | 8 2 | 3+4 4 | 6 2 |
| | | 9 2 | 8 2 | 9 2 | | 5 2 |
| | | 11 2 | 9 2 | 8 2 | | 6 2 |
| | | 8 2 | 8 2 | 8 2 | | 5 2 |
| | | 12 2 | 9 2 | 7 2 | | 5 2 |

Continued on next column

**TABLE LXXX– continued**

| TQC | ICSD | | NoE | | SF | |
|---|---|---|---|---|---|---|
| | | 7 2 | 4+5 2 | 9 2 | | 6 2 |
| | | 10 2 | 9 2 | 8 2 | | 6 2 |
| | | 9 2 | 8 2 | 9 2 | | 5 2 |
| | | 7 2 | 6+7 2 | 7 2 | | 6 2 |
| | | 10 2 | 9 2 | 9 2 | | 5 2 |
| | | 11 2 | 4+5 2 | 7 2 | | 5 2 |
| | | 8 2 | 8 2 | 8 2 | | 6 2 |
| | | 12 2 | 6+7 2 | 7 2 | | 6 2 |
| | | 9 2 | 9 2 | 8 2 | | 5 2 |
| | | 11 2 | 4+5 2 | 7 2 | | 6 2 |
| | | 8 2 | 6+7 2 | 7 2 | | 5 2 |
| | | 8 2 | 8 2 | 9 2 | | 6 2 |
| | | 7 2 | 9 2 | 9 2 | | 5 2 |
| | | 12 2 | 8 2 | 9 2 | | 5 2 |
| | | 9 2 | 8 2 | 8 2 | | 6 2 |
| | | 10 2 | 9 2 | 8 2 | | 5 2 |
| | | 7 2 | 8 2 | 8 2 | | 5 2 |
| | | 9 2 | 6+7 2 | 7 2 | | 6 2 |
| | | 12 2 | 9 2 | 9 2 | | 5 2 |
| | | 11 2 | 4+5 2 | 7 2 | | 6 2 |
| | | 10 2 | 4+5 2 | 8 2 | | 6 2 |
| | | 11 2 | 9 2 | 7 2 | | 6 2 |
| | | 9 2 | 6+7 2 | 7 2 | | 5 2 |
| | | 11 2 | 8 2 | 8 2 | | 5 2 |
| 194.1.33 | 152624 | | 112 | | Ce1Sn1Zn1 | |
| | -A | -GM | -H | -K | -L | -M |
| | 6 4 | 8 2 | 8 2 | 8 2 | 3+4 4 | 5 2 |
| | 6 4 | 9 2 | 9 2 | 9 2 | 3+4 4 | 5 2 |
| | 6 4 | 9 2 | 8 2 | 9 2 | 3+4 4 | 5 2 |
| | 6 4 | 11 2 | 9 2 | 9 2 | 3+4 4 | 6 2 |
| | 6 4 | 12 2 | 8 2 | 8 2 | 3+4 4 | 6 2 |
| | 4+5 4 | 11 2 | 9 2 | 9 2 | 3+4 4 | 6 2 |
| | 6 4 | 8 2 | 8 2 | 8 2 | 3+4 4 | 5 2 |
| | 4+5 4 | 12 2 | 9 2 | 8 2 | 3+4 4 | 6 2 |
| | 6 4 | 12 2 | 4+5 2 | 8 2 | 3+4 4 | 6 2 |
| | 6 4 | 11 2 | 6+7 2 | 7 2 | 3+4 4 | 6 2 |
| | 4+5 4 | 10 2 | 8 2 | 7 2 | 3+4 4 | 6 2 |
| | 4+5 4 | 10 2 | 9 2 | 9 2 | 3+4 4 | 6 2 |
| | 6 4 | 8 2 | 4+5 2 | 7 2 | 3+4 4 | 5 2 |
| | 6 4 | 12 2 | 6+7 2 | 7 2 | 3+4 4 | 6 2 |
| | 6 4 | 7 2 | 8 2 | 8 2 | 3+4 4 | 5 2 |
| | 6 4 | 10 2 | 9 2 | 8 2 | 3+4 4 | 6 2 |
| | 4+5 4 | 9 2 | 8 2 | 8 2 | 3+4 4 | 5 2 |
| | 4+5 4 | 11 2 | 9 2 | 9 2 | 3+4 4 | 5 2 |
| | 6 4 | 8 2 | 6+7 2 | 7 2 | 3+4 4 | 6 2 |
| | 6 4 | 12 2 | 4+5 2 | 7 2 | 3+4 4 | 6 2 |
| | 6 4 | 7 2 | 9 2 | 9 2 | 3+4 4 | 5 2 |
| | 6 4 | 10 2 | 8 2 | 8 2 | 3+4 4 | 5 2 |
| | 6 4 | 7 2 | 8 2 | 9 2 | 3+4 4 | 6 2 |
| | 4+5 4 | 10 2 | 9 2 | 8 2 | 3+4 4 | 6 2 |
| | 6 4 | 9 2 | 4+5 2 | 7 2 | 3+4 4 | 5 2 |
| | 4+5 4 | 8 2 | 6+7 2 | 7 2 | 3+4 4 | 5 2 |
| | 6 4 | 11 2 | 9 2 | 9 2 | 3+4 4 | 6 2 |
| | 6 4 | 12 2 | 8 2 | 8 2 | 3+4 4 | 6 2 |
| | | 9 2 | 8 2 | 9 2 | | 5 2 |
| | | 11 2 | 9 2 | 8 2 | | 6 2 |
| | | 8 2 | 8 2 | 8 2 | | 5 2 |
| | | 12 2 | 9 2 | 9 2 | | 6 2 |
| | | 7 2 | 4+5 2 | 7 2 | | 5 2 |
| | | 10 2 | 9 2 | 8 2 | | 6 2 |
| | | 9 2 | 8 2 | 9 2 | | 5 2 |

Continued on next column

**TABLE LXXX– continued**

| TQC | ICSD | | NoE | | SF | |
|---|---|---|---|---|---|---|
| | | 7 2 | 6+7 2 | 7 2 | | 6 2 |
| | | 10 2 | 9 2 | 9 2 | | 5 2 |
| | | 11 2 | 4+5 2 | 7 2 | | 5 2 |
| | | 8 2 | 8 2 | 8 2 | | 6 2 |
| | | 12 2 | 6+7 2 | 7 2 | | 6 2 |
| | | 9 2 | 9 2 | 8 2 | | 5 2 |
| | | 11 2 | 4+5 2 | 7 2 | | 6 2 |
| | | 8 2 | 6+7 2 | 7 2 | | 5 2 |
| | | 8 2 | 8 2 | 9 2 | | 6 2 |
| | | 7 2 | 9 2 | 9 2 | | 5 2 |
| | | 9 2 | 8 2 | 9 2 | | 5 2 |
| | | 12 2 | 8 2 | 8 2 | | 6 2 |
| | | 7 2 | 9 2 | 8 2 | | 5 2 |
| | | 10 2 | 8 2 | 8 2 | | 5 2 |
| | | 12 2 | 6+7 2 | 7 2 | | 6 2 |
| | | 9 2 | 9 2 | 9 2 | | 5 2 |
| | | 11 2 | 4+5 2 | 7 2 | | 6 2 |
| | | 10 2 | 4+5 2 | 8 2 | | 6 2 |
| | | 11 2 | 9 2 | 7 2 | | 6 2 |
| | | 9 2 | 6+7 2 | 7 2 | | 5 2 |
| | | 11 2 | 8 2 | 8 2 | | 5 2 |
| 194.1.34 | 420186 | | 112 | | Ce1Ge1Zn1 | |
| | -A | -GM | -H | -K | -L | -M |
| | 6 4 | 8 2 | 8 2 | 8 2 | 3+4 4 | 5 2 |
| | 6 4 | 9 2 | 9 2 | 9 2 | 3+4 4 | 5 2 |
| | 6 4 | 9 2 | 8 2 | 9 2 | 3+4 4 | 5 2 |
| | 6 4 | 11 2 | 9 2 | 9 2 | 3+4 4 | 6 2 |
| | 6 4 | 12 2 | 8 2 | 8 2 | 3+4 4 | 6 2 |
| | 6 4 | 11 2 | 9 2 | 9 2 | 3+4 4 | 6 2 |
| | 4+5 4 | 8 2 | 8 2 | 8 2 | 3+4 4 | 5 2 |
| | 4+5 4 | 12 2 | 9 2 | 8 2 | 3+4 4 | 6 2 |
| | 6 4 | 12 2 | 4+5 2 | 8 2 | 3+4 4 | 6 2 |
| | 6 4 | 11 2 | 6+7 2 | 7 2 | 3+4 4 | 6 2 |
| | 4+5 4 | 8 2 | 8 2 | 7 2 | 3+4 4 | 6 2 |
| | 4+5 4 | 12 2 | 9 2 | 9 2 | 3+4 4 | 5 2 |
| | 6 4 | 10 2 | 6+7 2 | 7 2 | 3+4 4 | 6 2 |
| | 6 4 | 10 2 | 4+5 2 | 8 2 | 3+4 4 | 6 2 |
| | 6 4 | 7 2 | 8 2 | 7 2 | 3+4 4 | 5 2 |
| | 6 4 | 10 2 | 9 2 | 8 2 | 3+4 4 | 6 2 |
| | 4+5 4 | 9 2 | 9 2 | 9 2 | 3+4 4 | 5 2 |
| | 4+5 4 | 8 2 | 8 2 | 8 2 | 3+4 4 | 5 2 |
| | 6 4 | 11 2 | 4+5 2 | 7 2 | 3+4 4 | 6 2 |
| | 6 4 | 12 2 | 6+7 2 | 7 2 | 3+4 4 | 6 2 |
| | 6 4 | 7 2 | 9 2 | 9 2 | 3+4 4 | 5 2 |
| | 6 4 | 10 2 | 8 2 | 8 2 | 3+4 4 | 5 2 |
| | 6 4 | 7 2 | 4+5 2 | 7 2 | 3+4 4 | 6 2 |
| | 4+5 4 | 10 2 | 8 2 | 9 2 | 3+4 4 | 6 2 |
| | 4+5 4 | 9 2 | 9 2 | 8 2 | 3+4 4 | 5 2 |
| | 6 4 | 11 2 | 6+7 2 | 7 2 | 3+4 4 | 5 2 |
| | 6 4 | 8 2 | 9 2 | 9 2 | 3+4 4 | 6 2 |
| | 6 4 | 12 2 | 8 2 | 8 2 | 3+4 4 | 6 2 |
| | | 9 2 | 8 2 | 9 2 | | 5 2 |
| | | 8 2 | 9 2 | 8 2 | | 5 2 |
| | | 11 2 | 9 2 | 9 2 | | 6 2 |
| | | 12 2 | 4+5 2 | 7 2 | | 6 2 |
| | | 7 2 | 8 2 | 8 2 | | 5 2 |
| | | 10 2 | 9 2 | 8 2 | | 6 2 |
| | | 7 2 | 6+7 2 | 9 2 | | 5 2 |
| | | 10 2 | 8 2 | 7 2 | | 6 2 |
| | | 9 2 | 9 2 | 9 2 | | 5 2 |
| | | 11 2 | 8 2 | 8 2 | | 5 2 |

Continued on next column

**TABLE LXXX– continued**

| TQC | ICSD NoE SF |
|---|---|
| | 8 2 4+5 2 7 2 6 2 |
| | 12 2 6+7 2 7 2 6 2 |
| | 9 2 8 2 9 2 5 2 |
| | 11 2 6+7 2 7 2 5 2 |
| | 8 2 9 2 8 2 6 2 |
| | 8 2 4+5 2 7 2 6 2 |
| | 7 2 9 2 9 2 5 2 |
| | 9 2 8 2 9 2 5 2 |
| | 7 2 8 2 8 2 6 2 |
| | 12 2 9 2 8 2 5 2 |
| | 10 2 8 2 8 2 5 2 |
| | 12 2 6+7 2 7 2 6 2 |
| | 10 2 9 2 9 2 6 2 |
| | 11 2 4+5 2 7 2 5 2 |
| | 9 2 9 2 8 2 6 2 |
| | 11 2 4+5 2 7 2 6 2 |
| | 9 2 6+7 2 7 2 5 2 |
| | 11 2 8 2 9 2 5 2 |
| 194.1.35 | 163343 112 Ce1Ge1Zn1 |
| | -A -GM -H -K -L -M |
| | 6 4 8 2 8 2 8 2 3+4 4 5 2 |
| | 6 4 9 2 9 2 9 2 3+4 4 5 2 |
| | 6 4 9 2 8 2 9 2 3+4 4 5 2 |
| | 6 4 11 2 9 2 9 2 3+4 4 6 2 |
| | 6 4 12 2 8 2 8 2 3+4 4 6 2 |
| | 6 4 11 2 9 2 9 2 3+4 4 6 2 |
| | 4+5 4 8 2 8 2 8 2 3+4 4 5 2 |
| | 4+5 4 12 2 9 2 8 2 3+4 4 6 2 |
| | 6 4 12 2 4+5 2 8 2 3+4 4 6 2 |
| | 6 4 11 2 6+7 2 7 2 3+4 4 6 2 |
| | 4+5 4 8 2 8 2 7 2 3+4 4 6 2 |
| | 4+5 4 12 2 9 2 9 2 3+4 4 5 2 |
| | 6 4 10 2 6+7 2 7 2 3+4 4 6 2 |
| | 6 4 10 2 4+5 2 8 2 3+4 4 6 2 |
| | 6 4 7 2 8 2 7 2 3+4 4 5 2 |
| | 6 4 10 2 9 2 8 2 3+4 4 6 2 |
| | 4+5 4 9 2 9 2 9 2 3+4 4 5 2 |
| | 4+5 4 8 2 8 2 8 2 3+4 4 5 2 |
| | 6 4 11 2 4+5 2 7 2 3+4 4 6 2 |
| | 6 4 12 2 6+7 2 7 2 3+4 4 6 2 |
| | 6 4 7 2 9 2 9 2 3+4 4 5 2 |
| | 6 4 10 2 8 2 8 2 3+4 4 5 2 |
| | 6 4 7 2 4+5 2 7 2 3+4 4 6 2 |
| | 4+5 4 10 2 8 2 9 2 3+4 4 6 2 |
| | 4+5 4 9 2 9 2 8 2 3+4 4 5 2 |
| | 6 4 11 2 6+7 2 7 2 3+4 4 5 2 |
| | 6 4 8 2 9 2 9 2 3+4 4 6 2 |
| | 6 4 12 2 8 2 8 2 3+4 4 6 2 |
| | 9 2 8 2 9 2 5 2 |
| | 8 2 9 2 8 2 5 2 |
| | 11 2 9 2 9 2 6 2 |
| | 12 2 4+5 2 7 2 6 2 |
| | 7 2 8 2 8 2 5 2 |
| | 10 2 9 2 8 2 6 2 |
| | 7 2 6+7 2 9 2 5 2 |
| | 10 2 8 2 7 2 6 2 |
| | 9 2 9 2 9 2 5 2 |
| | 11 2 8 2 8 2 5 2 |
| | 8 2 4+5 2 7 2 6 2 |
| | 12 2 6+7 2 7 2 6 2 |
| | 9 2 8 2 9 2 5 2 |
| | 11 2 6+7 2 7 2 5 2 |
| | 8 2 9 2 8 2 6 2 |
| | 8 2 4+5 2 7 2 6 2 |
| | 7 2 9 2 9 2 5 2 |
| | 9 2 8 2 9 2 5 2 |
| | 7 2 8 2 8 2 6 2 |
| | 12 2 9 2 8 2 5 2 |
| | 10 2 8 2 8 2 5 2 |
| | 12 2 6+7 2 7 2 6 2 |
| | 10 2 9 2 9 2 6 2 |
| | 11 2 4+5 2 7 2 5 2 |
| | 9 2 9 2 8 2 6 2 |
| | 11 2 4+5 2 7 2 6 2 |
| | 9 2 6+7 2 7 2 5 2 |
| | 11 2 8 2 9 2 5 2 |
| 194.1.36 | 425673 152 C6Co1N6Pr1 |
| | -A -GM -H -K -L -M |
| | 6 4 9 2 4+5 2 7 2 3+4 4 5 2 |
| | 6 4 11 2 9 2 8 2 3+4 4 6 2 |
| | 6 4 8 2 8 2 9 2 3+4 4 5 2 |
| | 4+5 4 12 2 6+7 2 7 2 3+4 4 6 2 |
| | 6 4 8 2 9 2 9 2 3+4 4 5 2 |
| | 6 4 12 2 8 2 9 2 3+4 4 6 2 |
| | 4+5 4 10 2 6+7 2 7 2 3+4 4 6 2 |
| | 6 4 7 2 8 2 8 2 3+4 4 5 2 |
| | 4+5 4 9 2 8 2 8 2 3+4 4 5 2 |
| | 6 4 11 2 9 2 7 2 3+4 4 6 2 |
| | 6 4 10 2 4+5 2 8 2 3+4 4 5 2 |
| | 6 4 11 2 9 2 8 2 3+4 4 6 2 |
| | 4+5 4 7 2 4+5 2 7 2 3+4 4 6 2 |
| | 6 4 9 2 9 2 9 2 3+4 4 5 2 |
| | 6 4 8 2 8 2 9 2 3+4 4 6 2 |
| | 4+5 4 12 2 6+7 2 7 2 3+4 4 5 2 |
| | 6 4 10 2 9 2 9 2 3+4 4 5 2 |
| | 6 4 7 2 8 2 8 2 3+4 4 6 2 |
| | 4+5 4 8 2 8 2 9 2 3+4 4 6 2 |
| | 6 4 12 2 6+7 2 7 2 3+4 4 5 2 |
| | 6 4 9 2 8 2 8 2 3+4 4 5 2 |
| | 4+5 4 8 2 9 2 9 2 3+4 4 5 2 |
| | 6 4 11 2 9 2 8 2 3+4 4 6 2 |
| | 4+5 4 11 2 4+5 2 7 2 3+4 4 5 2 |
| | 6 4 9 2 9 2 8 2 3+4 4 6 2 |
| | 4+5 4 10 2 4+5 2 9 2 3+4 4 5 2 |
| | 6 4 7 2 8 2 7 2 3+4 4 6 2 |
| | 6 4 8 2 9 2 7 2 3+4 4 5 2 |
| | 6 4 12 2 6+7 2 8 2 3+4 4 6 2 |
| | 4+5 4 7 2 8 2 9 2 3+4 4 5 2 |
| | 6 4 10 2 8 2 9 2 3+4 4 6 2 |
| | 6 4 12 2 6+7 2 7 2 3+4 4 6 2 |
| | 6 4 9 2 9 2 8 2 3+4 4 5 2 |
| | 4+5 4 7 2 4+5 2 7 2 3+4 4 5 2 |
| | 6 4 9 2 8 2 8 2 3+4 4 5 2 |
| | 4+5 4 11 2 9 2 9 2 3+4 4 5 2 |
| | 6 4 9 2 9 2 9 2 3+4 4 6 2 |
| | 6 4 8 2 4+5 2 7 2 3+4 4 5 2 |
| | 7 2 8 2 7 2 5 2 |
| | 8 2 9 2 8 2 6 2 |
| | 7 2 6+7 2 9 2 5 2 |
| | 9 2 8 2 7 2 6 2 |
| | 8 2 4+5 2 9 2 5 2 |
| | 11 2 9 2 8 2 6 2 |

Continued on next column

**TABLE LXXX– continued**

| TQC | ICSD | | NoE | | SF | |
|---|---|---|---|---|---|---|
| | | 10 2 | 6+7 2 | 8 2 | | 5 2 |
| | | 7 2 | 8 2 | 9 2 | | 5 2 |
| | | 12 2 | 8 2 | 7 2 | | 5 2 |
| | | 10 2 | 9 2 | 8 2 | | 6 2 |
| | | 8 2 | 8 2 | 9 2 | | 5 2 |
| | | 12 2 | 6+7 2 | 7 2 | | 6 2 |
| | | 12 2 | 4+5 2 | 7 2 | | 6 2 |
| | | 10 2 | 9 2 | 8 2 | | 6 2 |
| | | 10 2 | 8 2 | 9 2 | | 5 2 |
| | | 11 2 | 9 2 | 8 2 | | 6 2 |
| | | 11 2 | 6+7 2 | 8 2 | | 6 2 |
| | | 12 2 | 8 2 | 7 2 | | 6 2 |
| | | 11 2 | 8 2 | 7 2 | | 5 2 |
| | | 8 2 | 6+7 2 | 9 2 | | 6 2 |
| | | 7 2 | 8 2 | 9 2 | | 5 2 |
| | | 9 2 | 9 2 | 8 2 | | 5 2 |
| | | 8 2 | 4+5 2 | 9 2 | | 5 2 |
| | | 9 2 | 9 2 | 7 2 | | 6 2 |
| | | 7 2 | 9 2 | 8 2 | | 5 2 |
| | | 12 2 | 4+5 2 | 7 2 | | 6 2 |
| | | 10 2 | 8 2 | 9 2 | | 6 2 |
| | | 11 2 | 6+7 2 | 8 2 | | 6 2 |
| | | 12 2 | 8 2 | 9 2 | | 6 2 |
| | | 10 2 | 9 2 | 7 2 | | 6 2 |
| | | 9 2 | 4+5 2 | 9 2 | | 5 2 |
| | | 7 2 | 6+7 2 | 7 2 | | 5 2 |
| | | 8 2 | 9 2 | 8 2 | | 5 2 |
| | | 7 2 | 8 2 | 7 2 | | 5 2 |
| | | 9 2 | 9 2 | 9 2 | | 5 2 |
| | | 8 2 | 8 2 | 8 2 | | 5 2 |
| | | 11 2 | 4+5 2 | 7 2 | | 6 2 |
| | | 12 2 | 9 2 | 8 2 | | 5 2 |
| 194.1.37 | 180423 | | 48 | | C2Si1Ti3 | |
| | -A | -GM | -H | -K | -L | -M |
| | 6 4 | 9 2 | 4+5 2 | 7 2 | 3+4 4 | 5 2 |
| | 6 4 | 8 2 | 6+7 2 | 7 2 | 3+4 4 | 5 2 |
| | 6 4 | 11 2 | 8 2 | 8 2 | 3+4 4 | 6 2 |
| | 6 4 | 12 2 | 9 2 | 9 2 | 3+4 4 | 6 2 |
| | 6 4 | 9 2 | 8 2 | 9 2 | 3+4 4 | 5 2 |
| | 4+5 4 | 11 2 | 9 2 | 9 2 | 3+4 4 | 6 2 |
| | 6 4 | 8 2 | 8 2 | 8 2 | 3+4 4 | 5 2 |
| | 6 4 | 9 2 | 9 2 | 9 2 | 3+4 4 | 5 2 |
| | 4+5 4 | 12 2 | 9 2 | 8 2 | 3+4 4 | 6 2 |
| | 6 4 | 7 2 | 4+5 2 | 7 2 | 3+4 4 | 5 2 |
| | 6 4 | 8 2 | 8 2 | 9 2 | 3+4 4 | 6 2 |
| | 4+5 4 | 7 2 | 6+7 2 | 7 2 | 3+4 4 | 6 2 |
| | | 9 2 | 6+7 2 | 7 2 | | 5 2 |
| | | 11 2 | 8 2 | 8 2 | | 5 2 |
| | | 8 2 | 8 2 | 8 2 | | 6 2 |
| | | 10 2 | 9 2 | 9 2 | | 5 2 |
| | | 12 2 | 9 2 | 8 2 | | 5 2 |
| | | 10 2 | 4+5 2 | 7 2 | | 6 2 |
| | | 11 2 | 4+5 2 | 7 2 | | 5 2 |
| | | 12 2 | 9 2 | 9 2 | | 6 2 |
| | | 9 2 | 8 2 | 8 2 | | 6 2 |
| | | 7 2 | 6+7 2 | 7 2 | | 6 2 |
| | | 8 2 | 9 2 | 8 2 | | 5 2 |
| | | 7 2 | 8 2 | 8 2 | | 5 2 |
| 194.1.38 | 180422 | | 48 | | C2Si1Ti3 | |
| | -A | -GM | -H | -K | -L | -M |
| | 6 4 | 9 2 | 4+5 2 | 7 2 | 3+4 4 | 5 2 |

Continued on next column

| TQC | ICSD | | NoE | | SF | |
|---|---|---|---|---|---|---|
| | 6 4 | 8 2 | 6+7 2 | 7 2 | 3+4 4 | 5 2 |
| | 6 4 | 11 2 | 8 2 | 8 2 | 3+4 4 | 6 2 |
| | 6 4 | 12 2 | 9 2 | 9 2 | 3+4 4 | 6 2 |
| | 6 4 | 9 2 | 8 2 | 9 2 | 3+4 4 | 5 2 |
| | 4+5 4 | 11 2 | 9 2 | 9 2 | 3+4 4 | 6 2 |
| | 6 4 | 8 2 | 8 2 | 8 2 | 3+4 4 | 5 2 |
| | 6 4 | 9 2 | 9 2 | 9 2 | 3+4 4 | 5 2 |
| | 4+5 4 | 12 2 | 9 2 | 8 2 | 3+4 4 | 6 2 |
| | 6 4 | 7 2 | 4+5 2 | 7 2 | 3+4 4 | 5 2 |
| | 6 4 | 8 2 | 8 2 | 9 2 | 3+4 4 | 6 2 |
| | 4+5 4 | 7 2 | 6+7 2 | 7 2 | 3+4 4 | 6 2 |
| | | 9 2 | 6+7 2 | 7 2 | | 5 2 |
| | | 11 2 | 8 2 | 8 2 | | 5 2 |
| | | 8 2 | 8 2 | 8 2 | | 6 2 |
| | | 10 2 | 9 2 | 9 2 | | 5 2 |
| | | 12 2 | 9 2 | 8 2 | | 5 2 |
| | | 10 2 | 4+5 2 | 7 2 | | 6 2 |
| | | 11 2 | 4+5 2 | 7 2 | | 5 2 |
| | | 12 2 | 9 2 | 9 2 | | 6 2 |
| | | 9 2 | 8 2 | 8 2 | | 6 2 |
| | | 7 2 | 6+7 2 | 7 2 | | 6 2 |
| | | 8 2 | 9 2 | 8 2 | | 5 2 |
| | | 7 2 | 8 2 | 8 2 | | 5 2 |
| 194.1.39 | 180421 | | 48 | | C2Si1Ti3 | |
| | -A | -GM | -H | -K | -L | -M |
| | 6 4 | 9 2 | 4+5 2 | 7 2 | 3+4 4 | 5 2 |
| | 6 4 | 8 2 | 6+7 2 | 7 2 | 3+4 4 | 5 2 |
| | 6 4 | 11 2 | 8 2 | 8 2 | 3+4 4 | 6 2 |
| | 6 4 | 12 2 | 9 2 | 9 2 | 3+4 4 | 6 2 |
| | 6 4 | 9 2 | 8 2 | 9 2 | 3+4 4 | 5 2 |
| | 4+5 4 | 11 2 | 9 2 | 9 2 | 3+4 4 | 6 2 |
| | 6 4 | 8 2 | 8 2 | 8 2 | 3+4 4 | 5 2 |
| | 6 4 | 9 2 | 9 2 | 9 2 | 3+4 4 | 5 2 |
| | 4+5 4 | 12 2 | 9 2 | 8 2 | 3+4 4 | 6 2 |
| | 6 4 | 7 2 | 4+5 2 | 7 2 | 3+4 4 | 5 2 |
| | 6 4 | 8 2 | 8 2 | 9 2 | 3+4 4 | 6 2 |
| | 4+5 4 | 7 2 | 6+7 2 | 7 2 | 3+4 4 | 6 2 |
| | | 9 2 | 6+7 2 | 7 2 | | 5 2 |
| | | 11 2 | 8 2 | 8 2 | | 5 2 |
| | | 8 2 | 8 2 | 8 2 | | 6 2 |
| | | 10 2 | 9 2 | 9 2 | | 5 2 |
| | | 12 2 | 9 2 | 8 2 | | 5 2 |
| | | 10 2 | 4+5 2 | 7 2 | | 6 2 |
| | | 11 2 | 4+5 2 | 7 2 | | 5 2 |
| | | 12 2 | 9 2 | 9 2 | | 6 2 |
| | | 9 2 | 8 2 | 8 2 | | 6 2 |
| | | 7 2 | 6+7 2 | 7 2 | | 6 2 |
| | | 8 2 | 9 2 | 8 2 | | 5 2 |
| | | 7 2 | 8 2 | 8 2 | | 5 2 |
| 194.1.40 | 180420 | | 48 | | C2Si1Ti3 | |
| | -A | -GM | -H | -K | -L | -M |
| | 6 4 | 9 2 | 4+5 2 | 7 2 | 3+4 4 | 5 2 |
| | 6 4 | 8 2 | 6+7 2 | 7 2 | 3+4 4 | 5 2 |
| | 6 4 | 11 2 | 8 2 | 8 2 | 3+4 4 | 6 2 |
| | 6 4 | 12 2 | 9 2 | 9 2 | 3+4 4 | 6 2 |
| | 6 4 | 9 2 | 8 2 | 9 2 | 3+4 4 | 5 2 |
| | 4+5 4 | 11 2 | 9 2 | 9 2 | 3+4 4 | 6 2 |
| | 6 4 | 8 2 | 8 2 | 8 2 | 3+4 4 | 5 2 |
| | 6 4 | 9 2 | 9 2 | 9 2 | 3+4 4 | 5 2 |
| | 4+5 4 | 12 2 | 9 2 | 8 2 | 3+4 4 | 6 2 |
| | 6 4 | 7 2 | 4+5 2 | 7 2 | 3+4 4 | 5 2 |

Continued on next column

**TABLE LXXX– continued**

| TQC | ICSD | | NoE | | SF | |
|---|---|---|---|---|---|---|
| | 6 4 | 8 2 | 8 2 | 9 2 | 3+4 4 | 6 2 |
| | 4+5 4 | 7 2 | 6+7 2 | 7 2 | 3+4 4 | 6 2 |
| | | 9 2 | 6+7 2 | 7 2 | | 5 2 |
| | | 11 2 | 8 2 | 8 2 | | 5 2 |
| | | 8 2 | 8 2 | 8 2 | | 6 2 |
| | | 10 2 | 9 2 | 9 2 | | 5 2 |
| | | 12 2 | 9 2 | 8 2 | | 5 2 |
| | | 10 2 | 4+5 2 | 7 2 | | 6 2 |
| | | 11 2 | 4+5 2 | 7 2 | | 5 2 |
| | | 12 2 | 9 2 | 9 2 | | 6 2 |
| | | 9 2 | 8 2 | 8 2 | | 6 2 |
| | | 7 2 | 6+7 2 | 7 2 | | 6 2 |
| | | 8 2 | 9 2 | 8 2 | | 5 2 |
| | | 7 2 | 8 2 | 8 2 | | 5 2 |
| 194.1.41 | 180419 | | 48 | | C2Si1Ti3 | |
| | -A | -GM | -H | -K | -L | -M |
| | 6 4 | 9 2 | 4+5 2 | 7 2 | 3+4 4 | 5 2 |
| | 6 4 | 8 2 | 6+7 2 | 7 2 | 3+4 4 | 5 2 |
| | 6 4 | 11 2 | 8 2 | 8 2 | 3+4 4 | 6 2 |
| | 6 4 | 12 2 | 9 2 | 9 2 | 3+4 4 | 6 2 |
| | 6 4 | 9 2 | 8 2 | 9 2 | 3+4 4 | 5 2 |
| | 4+5 4 | 11 2 | 9 2 | 9 2 | 3+4 4 | 6 2 |
| | 6 4 | 8 2 | 8 2 | 8 2 | 3+4 4 | 5 2 |
| | 6 4 | 9 2 | 9 2 | 9 2 | 3+4 4 | 5 2 |
| | 4+5 4 | 12 2 | 9 2 | 8 2 | 3+4 4 | 6 2 |
| | 6 4 | 7 2 | 4+5 2 | 7 2 | 3+4 4 | 5 2 |
| | 6 4 | 8 2 | 8 2 | 9 2 | 3+4 4 | 6 2 |
| | 4+5 4 | 7 2 | 6+7 2 | 7 2 | 3+4 4 | 6 2 |
| | | 9 2 | 6+7 2 | 7 2 | | 5 2 |
| | | 11 2 | 8 2 | 8 2 | | 5 2 |
| | | 8 2 | 8 2 | 8 2 | | 6 2 |
| | | 10 2 | 9 2 | 9 2 | | 5 2 |
| | | 12 2 | 9 2 | 8 2 | | 5 2 |
| | | 10 2 | 4+5 2 | 7 2 | | 6 2 |
| | | 11 2 | 4+5 2 | 7 2 | | 5 2 |
| | | 12 2 | 9 2 | 9 2 | | 6 2 |
| | | 9 2 | 8 2 | 8 2 | | 6 2 |
| | | 7 2 | 6+7 2 | 7 2 | | 6 2 |
| | | 8 2 | 9 2 | 8 2 | | 5 2 |
| | | 7 2 | 8 2 | 8 2 | | 5 2 |
| 194.1.42 | 153264 | | 48 | | C2Si1Ti3 | |
| | -A | -GM | -H | -K | -L | -M |
| | 6 4 | 9 2 | 4+5 2 | 7 2 | 3+4 4 | 5 2 |
| | 6 4 | 8 2 | 6+7 2 | 7 2 | 3+4 4 | 5 2 |
| | 6 4 | 11 2 | 8 2 | 8 2 | 3+4 4 | 6 2 |
| | 6 4 | 12 2 | 9 2 | 9 2 | 3+4 4 | 6 2 |
| | 6 4 | 9 2 | 9 2 | 9 2 | 3+4 4 | 5 2 |
| | 4+5 4 | 11 2 | 8 2 | 9 2 | 3+4 4 | 6 2 |
| | 6 4 | 8 2 | 8 2 | 8 2 | 3+4 4 | 5 2 |
| | 6 4 | 9 2 | 9 2 | 9 2 | 3+4 4 | 5 2 |
| | 4+5 4 | 12 2 | 8 2 | 8 2 | 3+4 4 | 6 2 |
| | 6 4 | 7 2 | 6+7 2 | 7 2 | 3+4 4 | 5 2 |
| | 6 4 | 8 2 | 9 2 | 9 2 | 3+4 4 | 6 2 |
| | 4+5 4 | 7 2 | 4+5 2 | 7 2 | 3+4 4 | 6 2 |
| | | 9 2 | 4+5 2 | 7 2 | | 5 2 |
| | | 11 2 | 9 2 | 8 2 | | 5 2 |
| | | 8 2 | 9 2 | 8 2 | | 6 2 |
| | | 10 2 | 8 2 | 9 2 | | 5 2 |
| | | 12 2 | 8 2 | 8 2 | | 5 2 |
| | | 10 2 | 6+7 2 | 7 2 | | 6 2 |
| | | 11 2 | 6+7 2 | 7 2 | | 5 2 |

Continued on next column

**TABLE LXXX– continued**

| TQC | ICSD | | NoE | | SF | |
|---|---|---|---|---|---|---|
| | | 12 2 | 8 2 | 9 2 | | 6 2 |
| | | 9 2 | 9 2 | 8 2 | | 6 2 |
| | | 7 2 | 4+5 2 | 7 2 | | 6 2 |
| | | 8 2 | 8 2 | 8 2 | | 5 2 |
| | | 7 2 | 9 2 | 8 2 | | 5 2 |
| 194.1.43 | 180431 | | 48 | | C2Ge1Ti3 | |
| | -A | -GM | -H | -K | -L | -M |
| | 6 4 | 9 2 | 4+5 2 | 7 2 | 3+4 4 | 5 2 |
| | 6 4 | 8 2 | 6+7 2 | 7 2 | 3+4 4 | 5 2 |
| | 6 4 | 11 2 | 8 2 | 8 2 | 3+4 4 | 6 2 |
| | 6 4 | 9 2 | 9 2 | 9 2 | 3+4 4 | 6 2 |
| | 6 4 | 12 2 | 8 2 | 9 2 | 3+4 4 | 5 2 |
| | 4+5 4 | 11 2 | 9 2 | 9 2 | 3+4 4 | 6 2 |
| | 6 4 | 8 2 | 8 2 | 8 2 | 3+4 4 | 5 2 |
| | 6 4 | 9 2 | 9 2 | 9 2 | 3+4 4 | 5 2 |
| | 4+5 4 | 12 2 | 9 2 | 8 2 | 3+4 4 | 6 2 |
| | 6 4 | 7 2 | 4+5 2 | 7 2 | 3+4 4 | 5 2 |
| | 6 4 | 8 2 | 8 2 | 9 2 | 3+4 4 | 6 2 |
| | 4+5 4 | 7 2 | 6+7 2 | 7 2 | 3+4 4 | 6 2 |
| | | 9 2 | 8 2 | 8 2 | | 5 2 |
| | | 11 2 | 6+7 2 | 7 2 | | 5 2 |
| | | 8 2 | 8 2 | 8 2 | | 6 2 |
| | | 10 2 | 9 2 | 9 2 | | 5 2 |
| | | 12 2 | 9 2 | 8 2 | | 5 2 |
| | | 10 2 | 4+5 2 | 7 2 | | 6 2 |
| | | 11 2 | 9 2 | 9 2 | | 5 2 |
| | | 12 2 | 4+5 2 | 7 2 | | 6 2 |
| | | 9 2 | 8 2 | 8 2 | | 6 2 |
| | | 7 2 | 6+7 2 | 7 2 | | 6 2 |
| | | 8 2 | 9 2 | 8 2 | | 5 2 |
| | | 7 2 | 8 2 | 8 2 | | 5 2 |
| 194.1.44 | 180430 | | 48 | | C2Ge1Ti3 | |
| | -A | -GM | -H | -K | -L | -M |
| | 6 4 | 9 2 | 4+5 2 | 7 2 | 3+4 4 | 5 2 |
| | 6 4 | 8 2 | 6+7 2 | 7 2 | 3+4 4 | 5 2 |
| | 6 4 | 11 2 | 8 2 | 8 2 | 3+4 4 | 6 2 |
| | 6 4 | 9 2 | 9 2 | 9 2 | 3+4 4 | 6 2 |
| | 6 4 | 12 2 | 8 2 | 9 2 | 3+4 4 | 5 2 |
| | 4+5 4 | 11 2 | 9 2 | 9 2 | 3+4 4 | 6 2 |
| | 6 4 | 8 2 | 8 2 | 8 2 | 3+4 4 | 5 2 |
| | 6 4 | 9 2 | 9 2 | 9 2 | 3+4 4 | 5 2 |
| | 4+5 4 | 12 2 | 9 2 | 8 2 | 3+4 4 | 6 2 |
| | 6 4 | 7 2 | 4+5 2 | 7 2 | 3+4 4 | 5 2 |
| | 6 4 | 8 2 | 8 2 | 9 2 | 3+4 4 | 6 2 |
| | 4+5 4 | 7 2 | 6+7 2 | 7 2 | 3+4 4 | 6 2 |
| | | 9 2 | 8 2 | 8 2 | | 5 2 |
| | | 11 2 | 6+7 2 | 7 2 | | 5 2 |
| | | 8 2 | 8 2 | 8 2 | | 6 2 |
| | | 10 2 | 9 2 | 9 2 | | 5 2 |
| | | 12 2 | 9 2 | 8 2 | | 5 2 |
| | | 10 2 | 4+5 2 | 7 2 | | 6 2 |
| | | 11 2 | 9 2 | 9 2 | | 5 2 |
| | | 12 2 | 4+5 2 | 7 2 | | 6 2 |
| | | 9 2 | 8 2 | 8 2 | | 6 2 |
| | | 7 2 | 6+7 2 | 7 2 | | 6 2 |
| | | 8 2 | 9 2 | 8 2 | | 5 2 |
| | | 7 2 | 8 2 | 8 2 | | 5 2 |
| 194.1.45 | 180429 | | 48 | | C2Ge1Ti3 | |
| | -A | -GM | -H | -K | -L | -M |
| | 6 4 | 9 2 | 4+5 2 | 7 2 | 3+4 4 | 5 2 |
| | 6 4 | 8 2 | 6+7 2 | 7 2 | 3+4 4 | 5 2 |

Continued on next column

**TABLE LXXX– continued**

| TQC | ICSD | | NoE | | SF | |
|---|---|---|---|---|---|---|
| | 6 4 | 11 2 | 8 2 | 8 2 | 3+4 4 | 6 2 |
| | 6 4 | 9 2 | 9 2 | 9 2 | 3+4 4 | 6 2 |
| | 6 4 | 12 2 | 8 2 | 9 2 | 3+4 4 | 5 2 |
| | 4+5 4 | 11 2 | 9 2 | 9 2 | 3+4 4 | 6 2 |
| | 6 4 | 8 2 | 8 2 | 8 2 | 3+4 4 | 5 2 |
| | 6 4 | 9 2 | 9 2 | 9 2 | 3+4 4 | 5 2 |
| | 4+5 4 | 12 2 | 9 2 | 8 2 | 3+4 4 | 6 2 |
| | 6 4 | 7 2 | 4+5 2 | 7 2 | 3+4 4 | 5 2 |
| | 6 4 | 8 2 | 8 2 | 9 2 | 3+4 4 | 6 2 |
| | 4+5 4 | 7 2 | 6+7 2 | 7 2 | 3+4 4 | 6 2 |
| | | 9 2 | 8 2 | 8 2 | | 5 2 |
| | | 11 2 | 6+7 2 | 7 2 | | 5 2 |
| | | 8 2 | 8 2 | 8 2 | | 6 2 |
| | | 10 2 | 9 2 | 9 2 | | 5 2 |
| | | 12 2 | 9 2 | 8 2 | | 5 2 |
| | | 10 2 | 4+5 2 | 7 2 | | 6 2 |
| | | 11 2 | 9 2 | 9 2 | | 5 2 |
| | | 12 2 | 4+5 2 | 7 2 | | 6 2 |
| | | 9 2 | 8 2 | 8 2 | | 6 2 |
| | | 7 2 | 6+7 2 | 7 2 | | 6 2 |
| | | 8 2 | 9 2 | 8 2 | | 5 2 |
| | | 7 2 | 8 2 | 8 2 | | 5 2 |
| 194.1.46 | 180428 | | 48 | | C2Ge1Ti3 | |
| | -A | -GM | -H | -K | -L | -M |
| | 6 4 | 9 2 | 4+5 2 | 7 2 | 3+4 4 | 5 2 |
| | 6 4 | 8 2 | 6+7 2 | 7 2 | 3+4 4 | 5 2 |
| | 6 4 | 11 2 | 8 2 | 8 2 | 3+4 4 | 6 2 |
| | 6 4 | 9 2 | 9 2 | 9 2 | 3+4 4 | 6 2 |
| | 6 4 | 12 2 | 8 2 | 9 2 | 3+4 4 | 5 2 |
| | 4+5 4 | 11 2 | 9 2 | 9 2 | 3+4 4 | 6 2 |
| | 6 4 | 8 2 | 8 2 | 8 2 | 3+4 4 | 5 2 |
| | 6 4 | 9 2 | 9 2 | 9 2 | 3+4 4 | 5 2 |
| | 4+5 4 | 12 2 | 9 2 | 8 2 | 3+4 4 | 6 2 |
| | 6 4 | 7 2 | 4+5 2 | 7 2 | 3+4 4 | 5 2 |
| | 6 4 | 8 2 | 8 2 | 9 2 | 3+4 4 | 6 2 |
| | 4+5 4 | 7 2 | 6+7 2 | 7 2 | 3+4 4 | 6 2 |
| | | 9 2 | 8 2 | 8 2 | | 5 2 |
| | | 11 2 | 6+7 2 | 7 2 | | 5 2 |
| | | 8 2 | 8 2 | 8 2 | | 6 2 |
| | | 10 2 | 9 2 | 9 2 | | 5 2 |
| | | 12 2 | 9 2 | 8 2 | | 5 2 |
| | | 10 2 | 4+5 2 | 7 2 | | 6 2 |
| | | 11 2 | 9 2 | 9 2 | | 5 2 |
| | | 12 2 | 4+5 2 | 7 2 | | 6 2 |
| | | 9 2 | 8 2 | 8 2 | | 6 2 |
| | | 7 2 | 6+7 2 | 7 2 | | 6 2 |
| | | 8 2 | 9 2 | 8 2 | | 5 2 |
| | | 7 2 | 8 2 | 8 2 | | 5 2 |
| 194.1.47 | 180427 | | 48 | | C2Ge1Ti3 | |
| | -A | -GM | -H | -K | -L | -M |
| | 6 4 | 9 2 | 4+5 2 | 7 2 | 3+4 4 | 5 2 |
| | 6 4 | 8 2 | 6+7 2 | 7 2 | 3+4 4 | 5 2 |
| | 6 4 | 11 2 | 8 2 | 8 2 | 3+4 4 | 6 2 |
| | 6 4 | 9 2 | 9 2 | 9 2 | 3+4 4 | 6 2 |
| | 6 4 | 12 2 | 8 2 | 9 2 | 3+4 4 | 5 2 |
| | 4+5 4 | 11 2 | 9 2 | 9 2 | 3+4 4 | 6 2 |
| | 6 4 | 8 2 | 8 2 | 8 2 | 3+4 4 | 5 2 |
| | 6 4 | 9 2 | 9 2 | 9 2 | 3+4 4 | 5 2 |
| | 4+5 4 | 12 2 | 9 2 | 8 2 | 3+4 4 | 6 2 |
| | 6 4 | 7 2 | 4+5 2 | 7 2 | 3+4 4 | 5 2 |
| | 4+5 4 | 8 2 | 8 2 | 9 2 | 3+4 4 | 6 2 |
| | 6 4 | 7 2 | 6+7 2 | 7 2 | 3+4 4 | 6 2 |
| | | 9 2 | 8 2 | 8 2 | | 5 2 |
| | | 11 2 | 6+7 2 | 7 2 | | 5 2 |
| | | 8 2 | 8 2 | 8 2 | | 6 2 |
| | | 10 2 | 9 2 | 9 2 | | 5 2 |
| | | 12 2 | 9 2 | 8 2 | | 5 2 |
| | | 10 2 | 4+5 2 | 7 2 | | 6 2 |
| | | 11 2 | 9 2 | 9 2 | | 5 2 |
| | | 12 2 | 4+5 2 | 7 2 | | 6 2 |
| | | 9 2 | 8 2 | 8 2 | | 6 2 |
| | | 7 2 | 6+7 2 | 7 2 | | 6 2 |
| | | 8 2 | 9 2 | 8 2 | | 5 2 |
| | | 7 2 | 8 2 | 8 2 | | 5 2 |
| 194.1.48 | 180426 | | 48 | | C2Ge1Ti3 | |
| | -A | -GM | -H | -K | -L | -M |
| | 6 4 | 9 2 | 4+5 2 | 7 2 | 3+4 4 | 5 2 |
| | 6 4 | 8 2 | 6+7 2 | 7 2 | 3+4 4 | 5 2 |
| | 6 4 | 11 2 | 8 2 | 8 2 | 3+4 4 | 6 2 |
| | 6 4 | 9 2 | 9 2 | 9 2 | 3+4 4 | 6 2 |
| | 6 4 | 12 2 | 8 2 | 9 2 | 3+4 4 | 5 2 |
| | 4+5 4 | 11 2 | 9 2 | 9 2 | 3+4 4 | 6 2 |
| | 6 4 | 8 2 | 8 2 | 8 2 | 3+4 4 | 5 2 |
| | 6 4 | 9 2 | 9 2 | 9 2 | 3+4 4 | 5 2 |
| | 4+5 4 | 12 2 | 9 2 | 8 2 | 3+4 4 | 6 2 |
| | 6 4 | 7 2 | 4+5 2 | 7 2 | 3+4 4 | 5 2 |
| | 4+5 4 | 8 2 | 8 2 | 9 2 | 3+4 4 | 6 2 |
| | 6 4 | 7 2 | 6+7 2 | 7 2 | 3+4 4 | 6 2 |
| | | 9 2 | 8 2 | 8 2 | | 5 2 |
| | | 11 2 | 6+7 2 | 7 2 | | 5 2 |
| | | 8 2 | 8 2 | 8 2 | | 6 2 |
| | | 10 2 | 9 2 | 9 2 | | 5 2 |
| | | 12 2 | 9 2 | 8 2 | | 5 2 |
| | | 10 2 | 4+5 2 | 7 2 | | 6 2 |
| | | 11 2 | 9 2 | 9 2 | | 5 2 |
| | | 12 2 | 4+5 2 | 7 2 | | 6 2 |
| | | 9 2 | 8 2 | 8 2 | | 6 2 |
| | | 7 2 | 6+7 2 | 7 2 | | 6 2 |
| | | 8 2 | 9 2 | 8 2 | | 5 2 |
| | | 7 2 | 8 2 | 8 2 | | 5 2 |
| 194.1.49 | 180425 | | 48 | | C2Ge1Ti3 | |
| | -A | -GM | -H | -K | -L | -M |
| | 6 4 | 9 2 | 4+5 2 | 7 2 | 3+4 4 | 5 2 |
| | 6 4 | 8 2 | 6+7 2 | 7 2 | 3+4 4 | 5 2 |
| | 6 4 | 11 2 | 8 2 | 8 2 | 3+4 4 | 6 2 |
| | 6 4 | 9 2 | 9 2 | 9 2 | 3+4 4 | 6 2 |
| | 6 4 | 12 2 | 8 2 | 9 2 | 3+4 4 | 5 2 |
| | 4+5 4 | 11 2 | 9 2 | 9 2 | 3+4 4 | 6 2 |
| | 6 4 | 8 2 | 8 2 | 8 2 | 3+4 4 | 5 2 |
| | 6 4 | 9 2 | 9 2 | 9 2 | 3+4 4 | 5 2 |
| | 4+5 4 | 12 2 | 9 2 | 8 2 | 3+4 4 | 6 2 |
| | 6 4 | 7 2 | 4+5 2 | 7 2 | 3+4 4 | 5 2 |
| | 4+5 4 | 8 2 | 8 2 | 9 2 | 3+4 4 | 6 2 |
| | 6 4 | 7 2 | 6+7 2 | 7 2 | 3+4 4 | 6 2 |
| | | 9 2 | 8 2 | 8 2 | | 5 2 |
| | | 11 2 | 6+7 2 | 7 2 | | 5 2 |
| | | 8 2 | 8 2 | 8 2 | | 6 2 |
| | | 10 2 | 9 2 | 9 2 | | 5 2 |
| | | 12 2 | 9 2 | 8 2 | | 5 2 |
| | | 10 2 | 4+5 2 | 7 2 | | 6 2 |
| | | 11 2 | 9 2 | 9 2 | | 5 2 |
| | | 12 2 | 4+5 2 | 7 2 | | 6 2 |

Continued on next column

**TABLE LXXX– continued**

| TQC | ICSD | | NoE | | SF | |
|---|---|---|---|---|---|---|
| | | 9 2 | 8 2 | 8 2 | | 6 2 |
| | | 7 2 | 6+7 2 | 7 2 | | 6 2 |
| | | 8 2 | 9 2 | 8 2 | | 5 2 |
| | | 7 2 | 8 2 | 8 2 | | 5 2 |
| 194.1.50 | 180424 | | 48 | | C2Ge1Ti3 | |
| | -A | -GM | -H | -K | -L | -M |
| | 6 4 | 9 2 | 4+5 2 | 7 2 | 3+4 4 | 5 2 |
| | 6 4 | 8 2 | 6+7 2 | 7 2 | 3+4 4 | 5 2 |
| | 6 4 | 11 2 | 8 2 | 8 2 | 3+4 4 | 6 2 |
| | 6 4 | 9 2 | 9 2 | 9 2 | 3+4 4 | 6 2 |
| | 6 4 | 12 2 | 8 2 | 9 2 | 3+4 4 | 5 2 |
| | 4+5 4 | 11 2 | 9 2 | 9 2 | 3+4 4 | 6 2 |
| | 6 4 | 8 2 | 8 2 | 8 2 | 3+4 4 | 5 2 |
| | 6 4 | 9 2 | 9 2 | 9 2 | 3+4 4 | 5 2 |
| | 4+5 4 | 12 2 | 9 2 | 8 2 | 3+4 4 | 6 2 |
| | 6 4 | 7 2 | 4+5 2 | 7 2 | 3+4 4 | 5 2 |
| | 4+5 4 | 8 2 | 8 2 | 9 2 | 3+4 4 | 6 2 |
| | 6 4 | 7 2 | 6+7 2 | 7 2 | 3+4 4 | 6 2 |
| | | 9 2 | 8 2 | 8 2 | | 5 2 |
| | | 11 2 | 6+7 2 | 7 2 | | 5 2 |
| | | 8 2 | 8 2 | 8 2 | | 6 2 |
| | | 10 2 | 9 2 | 9 2 | | 5 2 |
| | | 12 2 | 9 2 | 8 2 | | 5 2 |
| | | 10 2 | 4+5 2 | 7 2 | | 6 2 |
| | | 11 2 | 9 2 | 9 2 | | 5 2 |
| | | 12 2 | 4+5 2 | 7 2 | | 6 2 |
| | | 9 2 | 8 2 | 8 2 | | 6 2 |
| | | 7 2 | 6+7 2 | 7 2 | | 6 2 |
| | | 8 2 | 9 2 | 8 2 | | 5 2 |
| | | 7 2 | 8 2 | 8 2 | | 5 2 |
| 194.1.51 | 249399 | | 52 | | C2Al1Ta3 | |
| | -A | -GM | -H | -K | -L | -M |
| | 6 4 | 9 2 | 4+5 2 | 7 2 | 3+4 4 | 5 2 |
| | 6 4 | 8 2 | 6+7 2 | 7 2 | 3+4 4 | 5 2 |
| | 6 4 | 11 2 | 8 2 | 8 2 | 3+4 4 | 6 2 |
| | 6 4 | 12 2 | 9 2 | 9 2 | 3+4 4 | 6 2 |
| | 6 4 | 9 2 | 9 2 | 9 2 | 3+4 4 | 5 2 |
| | 4+5 4 | 11 2 | 8 2 | 8 2 | 3+4 4 | 5 2 |
| | 6 4 | 8 2 | 9 2 | 8 2 | 3+4 4 | 6 2 |
| | 4+5 4 | 9 2 | 8 2 | 9 2 | 3+4 4 | 6 2 |
| | 6 4 | 12 2 | 4+5 2 | 7 2 | 3+4 4 | 5 2 |
| | 6 4 | 7 2 | 6+7 2 | 7 2 | 3+4 4 | 5 2 |
| | 6 4 | 7 2 | 9 2 | 9 2 | 3+4 4 | 5 2 |
| | 4+5 4 | 11 2 | 8 2 | 8 2 | 3+4 4 | 5 2 |
| | 6 4 | 8 2 | 6+7 2 | 7 2 | 3+4 4 | 6 2 |
| | | 9 2 | 4+5 2 | 7 2 | | 6 2 |
| | | 8 2 | 8 2 | 9 2 | | 5 2 |
| | | 10 2 | 9 2 | 8 2 | | 6 2 |
| | | 10 2 | 8 2 | 9 2 | | 5 2 |
| | | 12 2 | 9 2 | 9 2 | | 5 2 |
| | | 11 2 | 6+7 2 | 7 2 | | 5 2 |
| | | 12 2 | 8 2 | 8 2 | | 6 2 |
| | | 9 2 | 9 2 | 8 2 | | 6 2 |
| | | 8 2 | 4+5 2 | 7 2 | | 6 2 |
| | | 7 2 | 8 2 | 8 2 | | 6 2 |
| | | 7 2 | 9 2 | 8 2 | | 5 2 |
| | | 8 2 | 4+5 2 | 7 2 | | 6 2 |
| | | 9 2 | 9 2 | 9 2 | | 6 2 |
| 194.1.52 | 246019 | | 216 | | Ba1O3Ru1 | |
| | -A | -GM | -H | -K | -L | -M |
| | 6 4 | 9 2 | 8 2 | 9 2 | 3+4 4 | 5 2 |

Continued on next column

**TABLE LXXX– continued**

| TQC | ICSD | | NoE | | SF | |
|---|---|---|---|---|---|---|
| | 6 4 | 11 2 | 9 2 | 9 2 | 3+4 4 | 6 2 |
| | 6 4 | 8 2 | 4+5 2 | 7 2 | 3+4 4 | 6 2 |
| | 6 4 | 9 2 | 9 2 | 8 2 | 3+4 4 | 6 2 |
| | 6 4 | 12 2 | 6+7 2 | 7 2 | 3+4 4 | 5 2 |
| | 6 4 | 11 2 | 8 2 | 9 2 | 3+4 4 | 5 2 |
| | 4+5 4 | 9 2 | 8 2 | 8 2 | 3+4 4 | 5 2 |
| | 6 4 | 11 2 | 6+7 2 | 7 2 | 3+4 4 | 6 2 |
| | 4+5 4 | 8 2 | 8 2 | 8 2 | 3+4 4 | 5 2 |
| | 6 4 | 8 2 | 9 2 | 9 2 | 3+4 4 | 5 2 |
| | 6 4 | 12 2 | 4+5 2 | 7 2 | 3+4 4 | 6 2 |
| | 4+5 4 | 7 2 | 9 2 | 8 2 | 3+4 4 | 5 2 |
| | 6 4 | 10 2 | 9 2 | 9 2 | 3+4 4 | 6 2 |
| | 6 4 | 9 2 | 8 2 | 9 2 | 3+4 4 | 5 2 |
| | 6 4 | 12 2 | 4+5 2 | 7 2 | 3+4 4 | 5 2 |
| | 4+5 4 | 9 2 | 9 2 | 9 2 | 3+4 4 | 6 2 |
| | 6 4 | 7 2 | 6+7 2 | 7 2 | 3+4 4 | 5 2 |
| | 4+5 4 | 8 2 | 8 2 | 8 2 | 3+4 4 | 5 2 |
| | 6 4 | 7 2 | 9 2 | 9 2 | 3+4 4 | 5 2 |
| | 4+5 4 | 11 2 | 4+5 2 | 7 2 | 3+4 4 | 6 2 |
| | 6 4 | 12 2 | 8 2 | 8 2 | 3+4 4 | 6 2 |
| | 6 4 | 10 2 | 6+7 2 | 8 2 | 3+4 4 | 6 2 |
| | 4+5 4 | 11 2 | 8 2 | 7 2 | 3+4 4 | 6 2 |
| | 6 4 | 10 2 | 9 2 | 9 2 | 3+4 4 | 6 2 |
| | 6 4 | 8 2 | 9 2 | 8 2 | 3+4 4 | 5 2 |
| | 6 4 | 12 2 | 8 2 | 8 2 | 3+4 4 | 6 2 |
| | 4+5 4 | 9 2 | 9 2 | 8 2 | 3+4 4 | 6 2 |
| | 6 4 | 11 2 | 4+5 2 | 7 2 | 3+4 4 | 6 2 |
| | 4+5 4 | 12 2 | 6+7 2 | 7 2 | 3+4 4 | 5 2 |
| | 6 4 | 8 2 | 8 2 | 9 2 | 3+4 4 | 5 2 |
| | 6 4 | 7 2 | 4+5 2 | 7 2 | 3+4 4 | 5 2 |
| | 6 4 | 8 2 | 8 2 | 8 2 | 3+4 4 | 6 2 |
| | 4+5 4 | 10 2 | 9 2 | 8 2 | 3+4 4 | 6 2 |
| | 6 4 | 7 2 | 6+7 2 | 7 2 | 3+4 4 | 5 2 |
| | 6 4 | 9 2 | 9 2 | 8 2 | 3+4 4 | 6 2 |
| | 4+5 4 | 12 2 | 8 2 | 9 2 | 3+4 4 | 5 2 |
| | 6 4 | 10 2 | 8 2 | 9 2 | 3+4 4 | 5 2 |
| | 4+5 4 | 7 2 | 9 2 | 7 2 | 3+4 4 | 6 2 |
| | 6 4 | 11 2 | 4+5 2 | 8 2 | 3+4 4 | 6 2 |
| | 4+5 4 | 10 2 | 6+7 2 | 9 2 | 3+4 4 | 5 2 |
| | 6 4 | 12 2 | 8 2 | 7 2 | 3+4 4 | 6 2 |
| | 6 4 | 8 2 | 9 2 | 8 2 | 3+4 4 | 5 2 |
| | 6 4 | 7 2 | 4+5 2 | 7 2 | 3+4 4 | 5 2 |
| | 6 4 | 9 2 | 8 2 | 9 2 | 3+4 4 | 5 2 |
| | 4+5 4 | 8 2 | 9 2 | 7 2 | 3+4 4 | 5 2 |
| | 6 4 | 7 2 | 6+7 2 | 8 2 | 3+4 4 | 5 2 |
| | 4+5 4 | 11 2 | 9 2 | 8 2 | 3+4 4 | 5 2 |
| | 6 4 | 8 2 | 8 2 | 9 2 | 3+4 4 | 5 2 |
| | 6 4 | 9 2 | 8 2 | 7 2 | 3+4 4 | 6 2 |
| | 4+5 4 | 11 2 | 6+7 2 | 8 2 | 3+4 4 | 5 2 |
| | 6 4 | 10 2 | 9 2 | 9 2 | 3+4 4 | 6 2 |
| | 6 4 | 12 2 | 4+5 2 | 7 2 | 3+4 4 | 6 2 |
| | 4+5 4 | 9 2 | 9 2 | 9 2 | 3+4 4 | 5 2 |
| | 6 4 | 10 2 | 8 2 | 9 2 | 3+4 4 | 6 2 |
| | | 8 2 | 6+7 2 | 8 2 | | 5 2 |
| | | 12 2 | 8 2 | 7 2 | | 6 2 |
| | | 9 2 | 9 2 | 8 2 | | 6 2 |
| | | 7 2 | 8 2 | 7 2 | | 5 2 |
| | | 9 2 | 9 2 | 9 2 | | 6 2 |
| | | 12 2 | 4+5 2 | 9 2 | | 5 2 |
| | | 7 2 | 6+7 2 | 8 2 | | 6 2 |
| | | 8 2 | 8 2 | 7 2 | | 5 2 |

Continued on next column

**TABLE LXXX– continued**

| TQC | ICSD | | NoE | | SF | |
|---|---|---|---|---|---|---|
| | | 9 2 | 4+5 2 | 7 2 | | 5 2 |
| | | 10 2 | 9 2 | 8 2 | | 5 2 |
| | | 8 2 | 9 2 | 9 2 | | 5 2 |
| | | 7 2 | 8 2 | 8 2 | | 5 2 |
| | | 7 2 | 9 2 | 7 2 | | 6 2 |
| | | 11 2 | 4+5 2 | 9 2 | | 6 2 |
| | | 12 2 | 9 2 | 8 2 | | 5 2 |
| | | 11 2 | 8 2 | 8 2 | | 6 2 |
| | | 10 2 | 6+7 2 | 7 2 | | 5 2 |
| | | 8 2 | 8 2 | 9 2 | | 6 2 |
| | | 11 2 | 6+7 2 | 9 2 | | 6 2 |
| | | 10 2 | 8 2 | 8 2 | | 6 2 |
| | | 11 2 | 9 2 | 7 2 | | 5 2 |
| | | 12 2 | 8 2 | 7 2 | | 6 2 |
| | | 10 2 | 8 2 | 9 2 | | 5 2 |
| | | 7 2 | 9 2 | 9 2 | | 5 2 |
| | | 8 2 | 6+7 2 | 8 2 | | 6 2 |
| | | 11 2 | 8 2 | 8 2 | | 6 2 |
| | | 12 2 | 9 2 | 7 2 | | 6 2 |
| | | 10 2 | 4+5 2 | 8 2 | | 5 2 |
| | | 12 2 | 4+5 2 | 9 2 | | 5 2 |
| | | 9 2 | 9 2 | 8 2 | | 5 2 |
| | | 11 2 | 8 2 | 7 2 | | 6 2 |
| | | 12 2 | 6+7 2 | 7 2 | | 6 2 |
| | | 10 2 | 9 2 | 9 2 | | 6 2 |
| | | 11 2 | 4+5 2 | 8 2 | | 6 2 |
| | | 12 2 | 8 2 | 7 2 | | 6 2 |
| | | 9 2 | 9 2 | 9 2 | | 6 2 |
| | | 10 2 | 8 2 | 7 2 | | 6 2 |
| | | 7 2 | 9 2 | 9 2 | | 5 2 |
| | | 8 2 | 8 2 | 9 2 | | 6 2 |
| | | 9 2 | 6+7 2 | 8 2 | | 6 2 |
| | | 7 2 | 4+5 2 | 8 2 | | 5 2 |
| | | 8 2 | 9 2 | 7 2 | | 5 2 |
| | | 7 2 | 6+7 2 | 8 2 | | 5 2 |
| | | 9 2 | 8 2 | 7 2 | | 5 2 |
| | | 9 2 | 9 2 | 8 2 | | 6 2 |
| | | 8 2 | 8 2 | 9 2 | | 6 2 |
| | | 7 2 | 4+5 2 | 7 2 | | 6 2 |
| | | 11 2 | 9 2 | 8 2 | | 5 2 |
| | | 8 2 | 8 2 | 9 2 | | 5 2 |
| | | 10 2 | 6+7 2 | 8 2 | | 6 2 |
| | | 11 2 | 8 2 | 7 2 | | 5 2 |
| | | 10 2 | 6+7 2 | 7 2 | | 5 2 |
| | | 12 2 | 9 2 | 9 2 | | 5 2 |
| | | 12 2 | 8 2 | 9 2 | | 5 2 |
| 194.1.53 | 173866 | | 216 | | Ba1O3Ru1 | |
| | -A | -GM | -H | -K | -L | -M |
| | 6 4 | 9 2 | 8 2 | 9 2 | 3+4 4 | 5 2 |
| | 6 4 | 11 2 | 9 2 | 9 2 | 3+4 4 | 6 2 |
| | 6 4 | 8 2 | 4+5 2 | 7 2 | 3+4 4 | 6 2 |
| | 6 4 | 9 2 | 9 2 | 8 2 | 3+4 4 | 6 2 |
| | 6 4 | 12 2 | 6+7 2 | 7 2 | 3+4 4 | 5 2 |
| | 6 4 | 11 2 | 8 2 | 9 2 | 3+4 4 | 5 2 |
| | 4+5 4 | 9 2 | 8 2 | 8 2 | 3+4 4 | 5 2 |
| | 6 4 | 11 2 | 6+7 2 | 7 2 | 3+4 4 | 6 2 |
| | 4+5 4 | 8 2 | 8 2 | 8 2 | 3+4 4 | 5 2 |
| | 6 4 | 8 2 | 9 2 | 9 2 | 3+4 4 | 5 2 |
| | 6 4 | 12 2 | 4+5 2 | 7 2 | 3+4 4 | 6 2 |
| | 4+5 4 | 7 2 | 9 2 | 8 2 | 3+4 4 | 5 2 |
| | 6 4 | 10 2 | 9 2 | 9 2 | 3+4 4 | 6 2 |
| | 6 4 | 9 2 | 8 2 | 9 2 | 3+4 4 | 5 2 |
| | 6 4 | 12 2 | 4+5 2 | 7 2 | 3+4 4 | 5 2 |
| | 4+5 4 | 9 2 | 9 2 | 9 2 | 3+4 4 | 6 2 |
| | 6 4 | 7 2 | 6+7 2 | 7 2 | 3+4 4 | 5 2 |
| | 4+5 4 | 8 2 | 8 2 | 8 2 | 3+4 4 | 5 2 |
| | 6 4 | 7 2 | 9 2 | 9 2 | 3+4 4 | 5 2 |
| | 4+5 4 | 11 2 | 4+5 2 | 7 2 | 3+4 4 | 6 2 |
| | 6 4 | 12 2 | 8 2 | 8 2 | 3+4 4 | 6 2 |
| | 6 4 | 10 2 | 6+7 2 | 8 2 | 3+4 4 | 6 2 |
| | 4+5 4 | 11 2 | 8 2 | 7 2 | 3+4 4 | 6 2 |
| | 6 4 | 10 2 | 9 2 | 9 2 | 3+4 4 | 6 2 |
| | 6 4 | 8 2 | 9 2 | 8 2 | 3+4 4 | 5 2 |
| | 6 4 | 12 2 | 8 2 | 8 2 | 3+4 4 | 6 2 |
| | 4+5 4 | 9 2 | 9 2 | 8 2 | 3+4 4 | 6 2 |
| | 6 4 | 11 2 | 4+5 2 | 7 2 | 3+4 4 | 6 2 |
| | 4+5 4 | 12 2 | 6+7 2 | 7 2 | 3+4 4 | 5 2 |
| | 6 4 | 8 2 | 8 2 | 9 2 | 3+4 4 | 5 2 |
| | 6 4 | 7 2 | 4+5 2 | 7 2 | 3+4 4 | 5 2 |
| | 6 4 | 8 2 | 8 2 | 8 2 | 3+4 4 | 6 2 |
| | 4+5 4 | 10 2 | 9 2 | 8 2 | 3+4 4 | 6 2 |
| | 6 4 | 7 2 | 6+7 2 | 7 2 | 3+4 4 | 5 2 |
| | 6 4 | 9 2 | 9 2 | 8 2 | 3+4 4 | 6 2 |
| | 4+5 4 | 12 2 | 8 2 | 9 2 | 3+4 4 | 5 2 |
| | 6 4 | 10 2 | 8 2 | 9 2 | 3+4 4 | 5 2 |
| | 4+5 4 | 7 2 | 9 2 | 7 2 | 3+4 4 | 6 2 |
| | 6 4 | 11 2 | 4+5 2 | 8 2 | 3+4 4 | 6 2 |
| | 4+5 4 | 10 2 | 6+7 2 | 9 2 | 3+4 4 | 5 2 |
| | 6 4 | 12 2 | 8 2 | 7 2 | 3+4 4 | 6 2 |
| | 6 4 | 8 2 | 9 2 | 8 2 | 3+4 4 | 5 2 |
| | 6 4 | 7 2 | 4+5 2 | 7 2 | 3+4 4 | 5 2 |
| | 6 4 | 9 2 | 8 2 | 9 2 | 3+4 4 | 5 2 |
| | 4+5 4 | 8 2 | 9 2 | 7 2 | 3+4 4 | 5 2 |
| | 6 4 | 7 2 | 6+7 2 | 8 2 | 3+4 4 | 5 2 |
| | 4+5 4 | 11 2 | 9 2 | 8 2 | 3+4 4 | 5 2 |
| | 6 4 | 8 2 | 8 2 | 9 2 | 3+4 4 | 5 2 |
| | 6 4 | 9 2 | 8 2 | 7 2 | 3+4 4 | 6 2 |
| | 4+5 4 | 11 2 | 6+7 2 | 8 2 | 3+4 4 | 5 2 |
| | 6 4 | 10 2 | 9 2 | 9 2 | 3+4 4 | 6 2 |
| | 6 4 | 12 2 | 4+5 2 | 7 2 | 3+4 4 | 6 2 |
| | 4+5 4 | 9 2 | 9 2 | 9 2 | 3+4 4 | 5 2 |
| | 6 4 | 10 2 | 8 2 | 9 2 | 3+4 4 | 6 2 |
| | | 8 2 | 6+7 2 | 8 2 | | 5 2 |
| | | 12 2 | 8 2 | 7 2 | | 6 2 |
| | | 9 2 | 9 2 | 8 2 | | 6 2 |
| | | 7 2 | 8 2 | 7 2 | | 5 2 |
| | | 9 2 | 9 2 | 9 2 | | 6 2 |
| | | 12 2 | 4+5 2 | 9 2 | | 5 2 |
| | | 7 2 | 6+7 2 | 8 2 | | 6 2 |
| | | 8 2 | 8 2 | 7 2 | | 5 2 |
| | | 9 2 | 4+5 2 | 7 2 | | 5 2 |
| | | 10 2 | 9 2 | 8 2 | | 5 2 |
| | | 8 2 | 9 2 | 9 2 | | 5 2 |
| | | 7 2 | 8 2 | 8 2 | | 5 2 |
| | | 7 2 | 9 2 | 7 2 | | 6 2 |
| | | 11 2 | 4+5 2 | 9 2 | | 6 2 |
| | | 12 2 | 9 2 | 8 2 | | 5 2 |
| | | 11 2 | 8 2 | 8 2 | | 6 2 |
| | | 10 2 | 6+7 2 | 7 2 | | 5 2 |
| | | 8 2 | 8 2 | 9 2 | | 6 2 |
| | | 11 2 | 6+7 2 | 9 2 | | 6 2 |
| | | 10 2 | 8 2 | 8 2 | | 6 2 |

Continued on next column

Continued on next column

**TABLE LXXX– continued**

| TQC | ICSD | | NoE | | SF | |
|---|---|---|---|---|---|---|
| | | 11 2 | 9 2 | 7 2 | | 5 2 |
| | | 12 2 | 8 2 | 7 2 | | 6 2 |
| | | 10 2 | 8 2 | 9 2 | | 5 2 |
| | | 7 2 | 9 2 | 9 2 | | 5 2 |
| | | 8 2 | 6+7 2 | 8 2 | | 6 2 |
| | | 11 2 | 8 2 | 8 2 | | 6 2 |
| | | 12 2 | 9 2 | 7 2 | | 6 2 |
| | | 10 2 | 4+5 2 | 8 2 | | 5 2 |
| | | 12 2 | 4+5 2 | 9 2 | | 5 2 |
| | | 9 2 | 9 2 | 8 2 | | 5 2 |
| | | 11 2 | 8 2 | 7 2 | | 6 2 |
| | | 12 2 | 6+7 2 | 7 2 | | 6 2 |
| | | 10 2 | 9 2 | 9 2 | | 6 2 |
| | | 11 2 | 4+5 2 | 8 2 | | 6 2 |
| | | 12 2 | 8 2 | 7 2 | | 6 2 |
| | | 9 2 | 9 2 | 9 2 | | 6 2 |
| | | 10 2 | 8 2 | 7 2 | | 6 2 |
| | | 7 2 | 9 2 | 9 2 | | 5 2 |
| | | 8 2 | 8 2 | 9 2 | | 6 2 |
| | | 9 2 | 6+7 2 | 8 2 | | 6 2 |
| | | 7 2 | 4+5 2 | 8 2 | | 5 2 |
| | | 8 2 | 9 2 | 7 2 | | 5 2 |
| | | 7 2 | 6+7 2 | 8 2 | | 5 2 |
| | | 9 2 | 8 2 | 7 2 | | 5 2 |
| | | 9 2 | 9 2 | 8 2 | | 6 2 |
| | | 8 2 | 8 2 | 9 2 | | 6 2 |
| | | 7 2 | 4+5 2 | 7 2 | | 6 2 |
| | | 11 2 | 9 2 | 8 2 | | 5 2 |
| | | 8 2 | 8 2 | 9 2 | | 5 2 |
| | | 10 2 | 6+7 2 | 8 2 | | 6 2 |
| | | 11 2 | 8 2 | 7 2 | | 5 2 |
| | | 10 2 | 6+7 2 | 7 2 | | 5 2 |
| | | 12 2 | 9 2 | 9 2 | | 5 2 |
| | | 12 2 | 8 2 | 9 2 | | 5 2 |
| 194.1.54 | 160287 | | 216 | | Ba1O3Ru1 | |
| | -A | -GM | -H | -K | -L | -M |
| | 6 4 | 9 2 | 8 2 | 9 2 | 3+4 4 | 5 2 |
| | 6 4 | 11 2 | 9 2 | 9 2 | 3+4 4 | 6 2 |
| | 6 4 | 8 2 | 4+5 2 | 7 2 | 3+4 4 | 6 2 |
| | 6 4 | 9 2 | 9 2 | 8 2 | 3+4 4 | 6 2 |
| | 6 4 | 12 2 | 6+7 2 | 7 2 | 3+4 4 | 5 2 |
| | 6 4 | 11 2 | 8 2 | 9 2 | 3+4 4 | 5 2 |
| | 4+5 4 | 9 2 | 8 2 | 8 2 | 3+4 4 | 5 2 |
| | 6 4 | 11 2 | 6+7 2 | 7 2 | 3+4 4 | 6 2 |
| | 4+5 4 | 8 2 | 8 2 | 8 2 | 3+4 4 | 5 2 |
| | 6 4 | 8 2 | 9 2 | 9 2 | 3+4 4 | 5 2 |
| | 6 4 | 12 2 | 4+5 2 | 7 2 | 3+4 4 | 6 2 |
| | 4+5 4 | 7 2 | 9 2 | 8 2 | 3+4 4 | 5 2 |
| | 6 4 | 10 2 | 9 2 | 9 2 | 3+4 4 | 6 2 |
| | 6 4 | 9 2 | 8 2 | 9 2 | 3+4 4 | 5 2 |
| | 6 4 | 12 2 | 4+5 2 | 7 2 | 3+4 4 | 5 2 |
| | 4+5 4 | 9 2 | 9 2 | 9 2 | 3+4 4 | 6 2 |
| | 6 4 | 7 2 | 6+7 2 | 7 2 | 3+4 4 | 5 2 |
| | 4+5 4 | 8 2 | 8 2 | 8 2 | 3+4 4 | 5 2 |
| | 6 4 | 7 2 | 9 2 | 9 2 | 3+4 4 | 5 2 |
| | 4+5 4 | 11 2 | 4+5 2 | 7 2 | 3+4 4 | 6 2 |
| | 6 4 | 12 2 | 8 2 | 8 2 | 3+4 4 | 6 2 |
| | 6 4 | 10 2 | 6+7 2 | 8 2 | 3+4 4 | 6 2 |
| | 4+5 4 | 11 2 | 8 2 | 7 2 | 3+4 4 | 6 2 |
| | 6 4 | 10 2 | 9 2 | 9 2 | 3+4 4 | 6 2 |
| | 6 4 | 8 2 | 9 2 | 8 2 | 3+4 4 | 5 2 |

Continued on next column

**TABLE LXXX– continued**

| TQC | ICSD | | NoE | | SF | |
|---|---|---|---|---|---|---|
| | 6 4 | 9 2 | 8 2 | 8 2 | 3+4 4 | 6 2 |
| | 4+5 4 | 12 2 | 9 2 | 8 2 | 3+4 4 | 6 2 |
| | 6 4 | 11 2 | 4+5 2 | 7 2 | 3+4 4 | 6 2 |
| | 4+5 4 | 12 2 | 6+7 2 | 7 2 | 3+4 4 | 5 2 |
| | 6 4 | 8 2 | 8 2 | 9 2 | 3+4 4 | 5 2 |
| | 6 4 | 7 2 | 4+5 2 | 7 2 | 3+4 4 | 5 2 |
| | 6 4 | 8 2 | 8 2 | 8 2 | 3+4 4 | 6 2 |
| | 4+5 4 | 10 2 | 9 2 | 8 2 | 3+4 4 | 6 2 |
| | 6 4 | 7 2 | 6+7 2 | 7 2 | 3+4 4 | 5 2 |
| | 6 4 | 9 2 | 9 2 | 8 2 | 3+4 4 | 6 2 |
| | 4+5 4 | 12 2 | 8 2 | 9 2 | 3+4 4 | 5 2 |
| | 6 4 | 10 2 | 8 2 | 9 2 | 3+4 4 | 5 2 |
| | 4+5 4 | 7 2 | 9 2 | 7 2 | 3+4 4 | 6 2 |
| | 6 4 | 11 2 | 4+5 2 | 8 2 | 3+4 4 | 6 2 |
| | 4+5 4 | 10 2 | 6+7 2 | 9 2 | 3+4 4 | 5 2 |
| | 6 4 | 12 2 | 8 2 | 7 2 | 3+4 4 | 6 2 |
| | 6 4 | 8 2 | 9 2 | 8 2 | 3+4 4 | 5 2 |
| | 6 4 | 7 2 | 4+5 2 | 7 2 | 3+4 4 | 5 2 |
| | 6 4 | 9 2 | 9 2 | 9 2 | 3+4 4 | 5 2 |
| | 4+5 4 | 8 2 | 8 2 | 7 2 | 3+4 4 | 5 2 |
| | 6 4 | 7 2 | 6+7 2 | 8 2 | 3+4 4 | 5 2 |
| | 4+5 4 | 11 2 | 9 2 | 8 2 | 3+4 4 | 5 2 |
| | 6 4 | 8 2 | 8 2 | 9 2 | 3+4 4 | 5 2 |
| | 6 4 | 9 2 | 8 2 | 7 2 | 3+4 4 | 6 2 |
| | 4+5 4 | 11 2 | 6+7 2 | 8 2 | 3+4 4 | 5 2 |
| | 6 4 | 10 2 | 9 2 | 9 2 | 3+4 4 | 6 2 |
| | 6 4 | 12 2 | 4+5 2 | 7 2 | 3+4 4 | 6 2 |
| | 4+5 4 | 9 2 | 9 2 | 9 2 | 3+4 4 | 5 2 |
| | 6 4 | 10 2 | 8 2 | 9 2 | 3+4 4 | 6 2 |
| | | 8 2 | 6+7 2 | 8 2 | | 5 2 |
| | | 12 2 | 8 2 | 7 2 | | 6 2 |
| | | 9 2 | 9 2 | 8 2 | | 6 2 |
| | | 7 2 | 8 2 | 7 2 | | 5 2 |
| | | 9 2 | 9 2 | 9 2 | | 6 2 |
| | | 12 2 | 4+5 2 | 9 2 | | 5 2 |
| | | 7 2 | 6+7 2 | 8 2 | | 6 2 |
| | | 8 2 | 8 2 | 7 2 | | 5 2 |
| | | 9 2 | 4+5 2 | 7 2 | | 5 2 |
| | | 10 2 | 9 2 | 8 2 | | 5 2 |
| | | 8 2 | 9 2 | 9 2 | | 5 2 |
| | | 7 2 | 8 2 | 8 2 | | 5 2 |
| | | 7 2 | 9 2 | 7 2 | | 6 2 |
| | | 11 2 | 4+5 2 | 9 2 | | 6 2 |
| | | 12 2 | 9 2 | 8 2 | | 5 2 |
| | | 11 2 | 8 2 | 8 2 | | 6 2 |
| | | 10 2 | 6+7 2 | 7 2 | | 5 2 |
| | | 8 2 | 8 2 | 9 2 | | 6 2 |
| | | 11 2 | 6+7 2 | 9 2 | | 6 2 |
| | | 10 2 | 8 2 | 8 2 | | 6 2 |
| | | 11 2 | 9 2 | 7 2 | | 5 2 |
| | | 12 2 | 8 2 | 7 2 | | 6 2 |
| | | 10 2 | 8 2 | 9 2 | | 5 2 |
| | | 7 2 | 9 2 | 9 2 | | 5 2 |
| | | 8 2 | 6+7 2 | 8 2 | | 6 2 |
| | | 11 2 | 8 2 | 8 2 | | 6 2 |
| | | 12 2 | 9 2 | 7 2 | | 6 2 |
| | | 10 2 | 4+5 2 | 8 2 | | 5 2 |
| | | 12 2 | 4+5 2 | 9 2 | | 5 2 |
| | | 9 2 | 9 2 | 8 2 | | 5 2 |
| | | 11 2 | 8 2 | 7 2 | | 6 2 |
| | | 12 2 | 6+7 2 | 7 2 | | 6 2 |

Continued on next column

**TABLE LXXX– continued**

| TQC | ICSD | | NoE | | SF | |
|---|---|---|---|---|---|---|
| | | 10 2 | 9 2 | 9 2 | | 6 2 |
| | | 12 2 | 4+5 2 | 8 2 | | 6 2 |
| | | 9 2 | 8 2 | 7 2 | | 6 2 |
| | | 11 2 | 9 2 | 9 2 | | 6 2 |
| | | 10 2 | 8 2 | 7 2 | | 6 2 |
| | | 7 2 | 9 2 | 9 2 | | 5 2 |
| | | 8 2 | 8 2 | 9 2 | | 6 2 |
| | | 9 2 | 6+7 2 | 8 2 | | 6 2 |
| | | 7 2 | 4+5 2 | 8 2 | | 5 2 |
| | | 8 2 | 9 2 | 7 2 | | 5 2 |
| | | 7 2 | 6+7 2 | 8 2 | | 5 2 |
| | | 9 2 | 8 2 | 7 2 | | 5 2 |
| | | 9 2 | 9 2 | 8 2 | | 6 2 |
| | | 8 2 | 8 2 | 9 2 | | 6 2 |
| | | 7 2 | 4+5 2 | 7 2 | | 6 2 |
| | | 11 2 | 9 2 | 8 2 | | 5 2 |
| | | 8 2 | 8 2 | 9 2 | | 5 2 |
| | | 10 2 | 6+7 2 | 8 2 | | 6 2 |
| | | 11 2 | 8 2 | 7 2 | | 5 2 |
| | | 10 2 | 6+7 2 | 7 2 | | 5 2 |
| | | 12 2 | 9 2 | 9 2 | | 5 2 |
| | | 12 2 | 8 2 | 9 2 | | 5 2 |
| 194.1.55 | 192940 | | 204 | | Ba1Cr1O3 | |
| | -A | -GM | -H | -K | -L | -M |
| | 6 4 | 9 2 | 8 2 | 9 2 | 3+4 4 | 5 2 |
| | 6 4 | 11 2 | 9 2 | 9 2 | 3+4 4 | 6 2 |
| | 6 4 | 8 2 | 4+5 2 | 7 2 | 3+4 4 | 6 2 |
| | 6 4 | 9 2 | 9 2 | 8 2 | 3+4 4 | 5 2 |
| | 6 4 | 12 2 | 6+7 2 | 7 2 | 3+4 4 | 6 2 |
| | 6 4 | 11 2 | 8 2 | 9 2 | 3+4 4 | 5 2 |
| | 4+5 4 | 11 2 | 8 2 | 8 2 | 3+4 4 | 5 2 |
| | 6 4 | 9 2 | 6+7 2 | 7 2 | 3+4 4 | 6 2 |
| | 6 4 | 8 2 | 9 2 | 8 2 | 3+4 4 | 5 2 |
| | 4+5 4 | 9 2 | 8 2 | 9 2 | 3+4 4 | 5 2 |
| | 6 4 | 12 2 | 9 2 | 8 2 | 3+4 4 | 6 2 |
| | 4+5 4 | 8 2 | 4+5 2 | 7 2 | 3+4 4 | 6 2 |
| | 6 4 | 10 2 | 4+5 2 | 7 2 | 3+4 4 | 5 2 |
| | 6 4 | 7 2 | 8 2 | 9 2 | 3+4 4 | 5 2 |
| | 6 4 | 12 2 | 9 2 | 9 2 | 3+4 4 | 6 2 |
| | 4+5 4 | 8 2 | 9 2 | 9 2 | 3+4 4 | 5 2 |
| | 6 4 | 7 2 | 6+7 2 | 8 2 | 3+4 4 | 6 2 |
| | 4+5 4 | 12 2 | 8 2 | 7 2 | 3+4 4 | 6 2 |
| | 6 4 | 10 2 | 8 2 | 8 2 | 3+4 4 | 6 2 |
| | 4+5 4 | 9 2 | 9 2 | 8 2 | 3+4 4 | 6 2 |
| | 6 4 | 7 2 | 9 2 | 7 2 | 3+4 4 | 5 2 |
| | 6 4 | 11 2 | 4+5 2 | 7 2 | 3+4 4 | 5 2 |
| | 6 4 | 10 2 | 6+7 2 | 9 2 | 3+4 4 | 5 2 |
| | 4+5 4 | 11 2 | 8 2 | 9 2 | 3+4 4 | 6 2 |
| | 6 4 | 8 2 | 8 2 | 8 2 | 3+4 4 | 5 2 |
| | 6 4 | 12 2 | 9 2 | 8 2 | 3+4 4 | 6 2 |
| | 4+5 4 | 9 2 | 9 2 | 8 2 | 3+4 4 | 6 2 |
| | 6 4 | 11 2 | 4+5 2 | 7 2 | 3+4 4 | 5 2 |
| | 6 4 | 12 2 | 6+7 2 | 7 2 | 3+4 4 | 6 2 |
| | 6 4 | 8 2 | 8 2 | 9 2 | 3+4 4 | 5 2 |
| | 6 4 | 7 2 | 8 2 | 8 2 | 3+4 4 | 5 2 |
| | 4+5 4 | 8 2 | 4+5 2 | 7 2 | 3+4 4 | 6 2 |
| | 4+5 4 | 10 2 | 9 2 | 8 2 | 3+4 4 | 6 2 |
| | 6 4 | 7 2 | 6+7 2 | 7 2 | 3+4 4 | 5 2 |
| | 6 4 | 9 2 | 9 2 | 8 2 | 3+4 4 | 6 2 |
| | 4+5 4 | 10 2 | 8 2 | 9 2 | 3+4 4 | 5 2 |
| | 6 4 | 12 2 | 4+5 2 | 7 2 | 3+4 4 | 5 2 |

Continued on next column

**TABLE LXXX– continued**

| TQC | ICSD | | NoE | | SF | |
|---|---|---|---|---|---|---|
| | 4+5 4 | 11 2 | 8 2 | 8 2 | 3+4 4 | 5 2 |
| | 6 4 | 7 2 | 9 2 | 9 2 | 3+4 4 | 6 2 |
| | 6 4 | 10 2 | 6+7 2 | 7 2 | 3+4 4 | 6 2 |
| | 4+5 4 | 12 2 | 8 2 | 9 2 | 3+4 4 | 6 2 |
| | 6 4 | 8 2 | 9 2 | 8 2 | 3+4 4 | 5 2 |
| | 6 4 | 8 2 | 4+5 2 | 9 2 | 3+4 4 | 5 2 |
| | 6 4 | 11 2 | 9 2 | 7 2 | 3+4 4 | 5 2 |
| | 4+5 4 | 9 2 | 8 2 | 9 2 | 3+4 4 | 5 2 |
| | 6 4 | 8 2 | 9 2 | 8 2 | 3+4 4 | 6 2 |
| | 4+5 4 | 7 2 | 8 2 | 8 2 | 3+4 4 | 5 2 |
| | 6 4 | 9 2 | 6+7 2 | 7 2 | 3+4 4 | 6 2 |
| | 6 4 | 7 2 | 9 2 | 8 2 | 3+4 4 | 5 2 |
| | 4+5 4 | 9 2 | 8 2 | 7 2 | 3+4 4 | 5 2 |
| | 6 4 | 11 2 | 4+5 2 | 9 2 | 3+4 4 | 5 2 |
| | | 10 2 | 9 2 | 9 2 | | 5 2 |
| | | 12 2 | 6+7 2 | 9 2 | | 6 2 |
| | | 12 2 | 8 2 | 7 2 | | 6 2 |
| | | 9 2 | 6+7 2 | 7 2 | | 5 2 |
| | | 10 2 | 8 2 | 8 2 | | 5 2 |
| | | 9 2 | 9 2 | 9 2 | | 6 2 |
| | | 7 2 | 4+5 2 | 7 2 | | 6 2 |
| | | 8 2 | 9 2 | 8 2 | | 6 2 |
| | | 12 2 | 8 2 | 9 2 | | 6 2 |
| | | 11 2 | 6+7 2 | 8 2 | | 5 2 |
| | | 7 2 | 8 2 | 7 2 | | 6 2 |
| | | 8 2 | 9 2 | 8 2 | | 6 2 |
| | | 10 2 | 4+5 2 | 9 2 | | 5 2 |
| | | 12 2 | 8 2 | 7 2 | | 5 2 |
| | | 9 2 | 9 2 | 7 2 | | 6 2 |
| | | 11 2 | 4+5 2 | 8 2 | | 5 2 |
| | | 8 2 | 9 2 | 9 2 | | 5 2 |
| | | 7 2 | 9 2 | 8 2 | | 5 2 |
| | | 10 2 | 8 2 | 8 2 | | 6 2 |
| | | 7 2 | 6+7 2 | 8 2 | | 6 2 |
| | | 11 2 | 8 2 | 7 2 | | 6 2 |
| | | 10 2 | 8 2 | 9 2 | | 5 2 |
| | | 8 2 | 6+7 2 | 7 2 | | 6 2 |
| | | 12 2 | 9 2 | 9 2 | | 5 2 |
| | | 11 2 | 8 2 | 9 2 | | 5 2 |
| | | 10 2 | 6+7 2 | 7 2 | | 5 2 |
| | | 11 2 | 8 2 | 9 2 | | 6 2 |
| | | 8 2 | 8 2 | 8 2 | | 6 2 |
| | | 9 2 | 9 2 | 8 2 | | 5 2 |
| | | 7 2 | 8 2 | 7 2 | | 5 2 |
| | | 12 2 | 4+5 2 | 8 2 | | 6 2 |
| | | 12 2 | 9 2 | 9 2 | | 5 2 |
| | | 10 2 | 6+7 2 | 7 2 | | 6 2 |
| | | 11 2 | 4+5 2 | 8 2 | | 5 2 |
| | | 11 2 | 9 2 | 9 2 | | 6 2 |
| | | 10 2 | 9 2 | 7 2 | | 6 2 |
| | | 8 2 | 4+5 2 | 8 2 | | 6 2 |
| | | 8 2 | 8 2 | 7 2 | | 6 2 |
| | | 9 2 | 9 2 | 9 2 | | 6 2 |
| | | 7 2 | 8 2 | 9 2 | | 6 2 |
| | | 7 2 | 6+7 2 | 7 2 | | 5 2 |
| | | 12 2 | 4+5 2 | 9 2 | | 6 2 |
| | | 12 2 | 9 2 | 8 2 | | 6 2 |
| | | 9 2 | 8 2 | 7 2 | | 5 2 |
| | | 10 2 | 9 2 | 8 2 | | 5 2 |
| | | 9 2 | 6+7 2 | 8 2 | | 5 2 |
| | | 7 2 | 8 2 | 7 2 | | 5 2 |

Continued on next column

**TABLE LXXX– continued**

| TQC | ICSD | | NoE | | SF | |
|---|---|---|---|---|---|---|
| | | 9 2 | 9 2 | 8 2 | | 6 2 |
| | | 7 2 | 8 2 | 9 2 | | 6 2 |
| | | 8 2 | 4+5 2 | 7 2 | | 6 2 |
| | | 8 2 | 9 2 | 9 2 | | 5 2 |
| 194.1.56 | 192939 | | 204 | | Ba1Cr1O3 | |
| | -A | -GM | -H | -K | -L | -M |
| | 6 4 | 9 2 | 8 2 | 9 2 | 3+4 4 | 5 2 |
| | 6 4 | 11 2 | 9 2 | 9 2 | 3+4 4 | 6 2 |
| | 6 4 | 8 2 | 6+7 2 | 7 2 | 3+4 4 | 6 2 |
| | 6 4 | 9 2 | 8 2 | 8 2 | 3+4 4 | 5 2 |
| | 6 4 | 12 2 | 4+5 2 | 7 2 | 3+4 4 | 6 2 |
| | 4+5 4 | 11 2 | 9 2 | 9 2 | 3+4 4 | 5 2 |
| | 6 4 | 11 2 | 9 2 | 8 2 | 3+4 4 | 5 2 |
| | 6 4 | 9 2 | 4+5 2 | 7 2 | 3+4 4 | 6 2 |
| | 6 4 | 8 2 | 8 2 | 8 2 | 3+4 4 | 6 2 |
| | 4+5 4 | 12 2 | 6+7 2 | 7 2 | 3+4 4 | 5 2 |
| | 6 4 | 8 2 | 8 2 | 8 2 | 3+4 4 | 5 2 |
| | 4+5 4 | 10 2 | 9 2 | 9 2 | 3+4 4 | 6 2 |
| | 6 4 | 7 2 | 8 2 | 9 2 | 3+4 4 | 5 2 |
| | 6 4 | 9 2 | 9 2 | 9 2 | 3+4 4 | 5 2 |
| | 6 4 | 12 2 | 6+7 2 | 7 2 | 3+4 4 | 6 2 |
| | 4+5 4 | 8 2 | 8 2 | 9 2 | 3+4 4 | 5 2 |
| | 6 4 | 11 2 | 4+5 2 | 8 2 | 3+4 4 | 6 2 |
| | 4+5 4 | 7 2 | 9 2 | 7 2 | 3+4 4 | 6 2 |
| | 6 4 | 12 2 | 9 2 | 8 2 | 3+4 4 | 6 2 |
| | 4+5 4 | 10 2 | 8 2 | 8 2 | 3+4 4 | 5 2 |
| | 6 4 | 9 2 | 8 2 | 7 2 | 3+4 4 | 6 2 |
| | 6 4 | 7 2 | 6+7 2 | 9 2 | 3+4 4 | 5 2 |
| | 6 4 | 11 2 | 4+5 2 | 7 2 | 3+4 4 | 5 2 |
| | 6 4 | 10 2 | 9 2 | 9 2 | 3+4 4 | 6 2 |
| | 4+5 4 | 8 2 | 8 2 | 8 2 | 3+4 4 | 5 2 |
| | 6 4 | 12 2 | 9 2 | 8 2 | 3+4 4 | 6 2 |
| | 4+5 4 | 9 2 | 8 2 | 8 2 | 3+4 4 | 6 2 |
| | 6 4 | 11 2 | 6+7 2 | 7 2 | 3+4 4 | 5 2 |
| | 6 4 | 12 2 | 4+5 2 | 7 2 | 3+4 4 | 6 2 |
| | 4+5 4 | 8 2 | 9 2 | 9 2 | 3+4 4 | 5 2 |
| | 6 4 | 7 2 | 9 2 | 8 2 | 3+4 4 | 5 2 |
| | 6 4 | 10 2 | 6+7 2 | 7 2 | 3+4 4 | 6 2 |
| | 6 4 | 8 2 | 8 2 | 8 2 | 3+4 4 | 6 2 |
| | 4+5 4 | 7 2 | 4+5 2 | 7 2 | 3+4 4 | 5 2 |
| | 6 4 | 9 2 | 8 2 | 8 2 | 3+4 4 | 6 2 |
| | 4+5 4 | 10 2 | 6+7 2 | 9 2 | 3+4 4 | 5 2 |
| | 6 4 | 12 2 | 9 2 | 7 2 | 3+4 4 | 5 2 |
| | 4+5 4 | 11 2 | 9 2 | 8 2 | 3+4 4 | 6 2 |
| | 6 4 | 7 2 | 8 2 | 9 2 | 3+4 4 | 5 2 |
| | 6 4 | 10 2 | 4+5 2 | 7 2 | 3+4 4 | 6 2 |
| | 4+5 4 | 12 2 | 9 2 | 8 2 | 3+4 4 | 6 2 |
| | 6 4 | 8 2 | 8 2 | 9 2 | 3+4 4 | 5 2 |
| | 6 4 | 8 2 | 6+7 2 | 9 2 | 3+4 4 | 5 2 |
| | 6 4 | 11 2 | 8 2 | 7 2 | 3+4 4 | 5 2 |
| | 4+5 4 | 9 2 | 8 2 | 9 2 | 3+4 4 | 6 2 |
| | 6 4 | 8 2 | 9 2 | 8 2 | 3+4 4 | 5 2 |
| | 4+5 4 | 7 2 | 4+5 2 | 7 2 | 3+4 4 | 6 2 |
| | 6 4 | 7 2 | 9 2 | 7 2 | 3+4 4 | 5 2 |
| | 6 4 | 9 2 | 8 2 | 8 2 | 3+4 4 | 5 2 |
| | 4+5 4 | 9 2 | 9 2 | 9 2 | 3+4 4 | 5 2 |
| | 6 4 | 12 2 | 6+7 2 | 9 2 | 3+4 4 | 5 2 |
| | | 11 2 | 8 2 | 8 2 | | 5 2 |
| | | 10 2 | 4+5 2 | 9 2 | | 6 2 |
| | | 12 2 | 9 2 | 7 2 | | 6 2 |
| | | 10 2 | 8 2 | 8 2 | | 5 2 |

Continued on next column

**TABLE LXXX– continued**

| TQC | ICSD | | NoE | | SF | |
|---|---|---|---|---|---|---|
| | | 8 2 | 6+7 2 | 7 2 | | 6 2 |
| | | 9 2 | 4+5 2 | 9 2 | | 5 2 |
| | | 7 2 | 9 2 | 8 2 | | 6 2 |
| | | 9 2 | 8 2 | 7 2 | | 6 2 |
| | | 12 2 | 9 2 | 9 2 | | 6 2 |
| | | 7 2 | 4+5 2 | 8 2 | | 5 2 |
| | | 11 2 | 9 2 | 7 2 | | 6 2 |
| | | 10 2 | 8 2 | 8 2 | | 5 2 |
| | | 8 2 | 9 2 | 9 2 | | 6 2 |
| | | 12 2 | 6+7 2 | 7 2 | | 5 2 |
| | | 9 2 | 8 2 | 8 2 | | 5 2 |
| | | 11 2 | 6+7 2 | 7 2 | | 6 2 |
| | | 8 2 | 8 2 | 9 2 | | 5 2 |
| | | 7 2 | 8 2 | 8 2 | | 5 2 |
| | | 10 2 | 9 2 | 8 2 | | 6 2 |
| | | 7 2 | 4+5 2 | 9 2 | | 6 2 |
| | | 8 2 | 9 2 | 7 2 | | 5 2 |
| | | 12 2 | 8 2 | 8 2 | | 6 2 |
| | | 11 2 | 9 2 | 9 2 | | 6 2 |
| | | 10 2 | 4+5 2 | 7 2 | | 5 2 |
| | | 10 2 | 9 2 | 9 2 | | 5 2 |
| | | 11 2 | 4+5 2 | 8 2 | | 5 2 |
| | | 8 2 | 9 2 | 7 2 | | 6 2 |
| | | 11 2 | 9 2 | 7 2 | | 5 2 |
| | | 7 2 | 8 2 | 9 2 | | 6 2 |
| | | 9 2 | 6+7 2 | 8 2 | | 5 2 |
| | | 12 2 | 8 2 | 8 2 | | 6 2 |
| | | 10 2 | 9 2 | 9 2 | | 6 2 |
| | | 12 2 | 4+5 2 | 7 2 | | 5 2 |
| | | 11 2 | 6+7 2 | 8 2 | | 6 2 |
| | | 11 2 | 8 2 | 7 2 | | 5 2 |
| | | 10 2 | 6+7 2 | 9 2 | | 6 2 |
| | | 9 2 | 8 2 | 7 2 | | 6 2 |
| | | 12 2 | 9 2 | 8 2 | | 6 2 |
| | | 7 2 | 8 2 | 9 2 | | 5 2 |
| | | 8 2 | 4+5 2 | 9 2 | | 6 2 |
| | | 8 2 | 9 2 | 7 2 | | 6 2 |
| | | 7 2 | 8 2 | 9 2 | | 6 2 |
| | | 10 2 | 9 2 | 8 2 | | 6 2 |
| | | 12 2 | 6+7 2 | 7 2 | | 5 2 |
| | | 9 2 | 8 2 | 8 2 | | 5 2 |
| | | 9 2 | 4+5 2 | 7 2 | | 5 2 |
| | | 7 2 | 9 2 | 8 2 | | 5 2 |
| | | 9 2 | 8 2 | 8 2 | | 6 2 |
| | | 11 2 | 9 2 | 9 2 | | 6 2 |
| | | 7 2 | 6+7 2 | 7 2 | | 6 2 |
| | | 8 2 | 8 2 | 9 2 | | 5 2 |
| 194.1.57 | 251687 | | 112 | | Au1Ce1Sb1 | |
| | -A | -GM | -H | -K | -L | -M |
| | 6 4 | 8 2 | 8 2 | 8 2 | 3+4 4 | 5 2 |
| | 6 4 | 9 2 | 9 2 | 9 2 | 3+4 4 | 5 2 |
| | 6 4 | 9 2 | 8 2 | 9 2 | 3+4 4 | 5 2 |
| | 6 4 | 11 2 | 9 2 | 9 2 | 3+4 4 | 6 2 |
| | 6 4 | 12 2 | 8 2 | 8 2 | 3+4 4 | 6 2 |
| | 4+5 4 | 11 2 | 9 2 | 9 2 | 3+4 4 | 6 2 |
| | 6 4 | 8 2 | 8 2 | 8 2 | 3+4 4 | 5 2 |
| | 4+5 4 | 12 2 | 9 2 | 8 2 | 3+4 4 | 6 2 |
| | 6 4 | 12 2 | 4+5 2 | 7 2 | 3+4 4 | 6 2 |
| | 6 4 | 11 2 | 6+7 2 | 8 2 | 3+4 4 | 6 2 |
| | 4+5 4 | 10 2 | 8 2 | 7 2 | 3+4 4 | 6 2 |
| | 4+5 4 | 10 2 | 9 2 | 9 2 | 3+4 4 | 6 2 |

Continued on next column

**TABLE LXXX– continued**

| TQC | ICSD | | NoE | | SF | |
|---|---|---|---|---|---|---|
| | 6 4 | 8 2 | 4+5 2 | 7 2 | 3+4 4 | 5 2 |
| | 6 4 | 12 2 | 6+7 2 | 7 2 | 3+4 4 | 6 2 |
| | 6 4 | 7 2 | 8 2 | 8 2 | 3+4 4 | 5 2 |
| | 6 4 | 10 2 | 9 2 | 8 2 | 3+4 4 | 6 2 |
| | 4+5 4 | 9 2 | 9 2 | 8 2 | 3+4 4 | 5 2 |
| | 6 4 | 8 2 | 8 2 | 9 2 | 3+4 4 | 5 2 |
| | 6 4 | 11 2 | 4+5 2 | 7 2 | 3+4 4 | 6 2 |
| | 4+5 4 | 12 2 | 6+7 2 | 7 2 | 3+4 4 | 6 2 |
| | 6 4 | 7 2 | 9 2 | 9 2 | 3+4 4 | 5 2 |
| | 6 4 | 10 2 | 4+5 2 | 7 2 | 3+4 4 | 5 2 |
| | 6 4 | 7 2 | 8 2 | 9 2 | 3+4 4 | 6 2 |
| | 4+5 4 | 10 2 | 8 2 | 8 2 | 3+4 4 | 6 2 |
| | 6 4 | 9 2 | 9 2 | 8 2 | 3+4 4 | 5 2 |
| | 4+5 4 | 11 2 | 6+7 2 | 7 2 | 3+4 4 | 5 2 |
| | 6 4 | 8 2 | 8 2 | 9 2 | 3+4 4 | 6 2 |
| | 6 4 | 12 2 | 9 2 | 9 2 | 3+4 4 | 5 2 |
| | | 9 2 | 9 2 | 8 2 | | 6 2 |
| | | 11 2 | 8 2 | 8 2 | | 6 2 |
| | | 8 2 | 6+7 2 | 7 2 | | 5 2 |
| | | 12 2 | 4+5 2 | 7 2 | | 6 2 |
| | | 7 2 | 8 2 | 8 2 | | 5 2 |
| | | 10 2 | 9 2 | 9 2 | | 6 2 |
| | | 9 2 | 8 2 | 8 2 | | 5 2 |
| | | 8 2 | 9 2 | 9 2 | | 6 2 |
| | | 9 2 | 6+7 2 | 7 2 | | 5 2 |
| | | 11 2 | 8 2 | 8 2 | | 5 2 |
| | | 7 2 | 4+5 2 | 7 2 | | 6 2 |
| | | 10 2 | 9 2 | 8 2 | | 6 2 |
| | | 12 2 | 6+7 2 | 9 2 | | 5 2 |
| | | 11 2 | 8 2 | 7 2 | | 6 2 |
| | | 8 2 | 9 2 | 9 2 | | 5 2 |
| | | 8 2 | 4+5 2 | 7 2 | | 6 2 |
| | | 7 2 | 9 2 | 9 2 | | 5 2 |
| | | 12 2 | 8 2 | 9 2 | | 5 2 |
| | | 9 2 | 8 2 | 8 2 | | 5 2 |
| | | 7 2 | 9 2 | 8 2 | | 6 2 |
| | | 12 2 | 6+7 2 | 7 2 | | 6 2 |
| | | 10 2 | 9 2 | 7 2 | | 5 2 |
| | | 11 2 | 8 2 | 9 2 | | 6 2 |
| | | 9 2 | 4+5 2 | 8 2 | | 6 2 |
| | | 10 2 | 6+7 2 | 7 2 | | 5 2 |
| | | 11 2 | 8 2 | 8 2 | | 6 2 |
| | | 9 2 | 4+5 2 | 7 2 | | 5 2 |
| | | 11 2 | 9 2 | 9 2 | | 5 2 |

The End

TABLE LXXXI: The computed irreducible representations for the valence bands for the compounds of SEBR and NLC in space group 204. See the list of tables.

| TQC | ICSD | | NoE | SF |
|---|---|---|---|---|
| 204.1.1 | 237751 | | 114 | Mn3Na1O12V4 |
| | -GM | -H | -P | -N |
| | 5 2 | 10+9 4 | 5 2 | 5+6 2 |
| | 5 2 | 8 2 | 6+7 4 | 3+4 2 |
| | 6+7 4 | 6+7 4 | 5 2 | 3+4 2 |
| | 6+7 4 | 5 2 | 5 2 | 5+6 2 |
| | 8 2 | 8 2 | 6+7 4 | 5+6 2 |
| | 10+9 4 | 10+9 4 | 6+7 4 | 3+4 2 |
| | 10+9 4 | 5 2 | 5 2 | 5+6 2 |
| | 8 2 | 6+7 4 | 6+7 4 | 3+4 2 |
| | 6+7 4 | 8 2 | 5 2 | 5+6 2 |
| | 5 2 | 10+9 4 | 6+7 4 | 3+4 2 |
| | 5 2 | 10+9 4 | 5 2 | 3+4 2 |
| | 6+7 4 | 8 2 | 6+7 4 | 5+6 2 |
| | 5 2 | 8 2 | 6+7 4 | 5+6 2 |
| | 6+7 4 | 10+9 4 | 5 2 | 5+6 2 |
| | 8 2 | 5 2 | 5 2 | 5+6 2 |
| | 10+9 4 | 6+7 4 | 6+7 4 | 3+4 2 |
| | 5 2 | 5 2 | 5 2 | 3+4 2 |
| | 6+7 4 | 6+7 4 | 6+7 4 | 3+4 2 |
| | 10+9 4 | 6+7 4 | 5 2 | 5+6 2 |
| | 8 2 | 5 2 | 6+7 4 | 5+6 2 |
| | 5 2 | 5 2 | 6+7 4 | 3+4 2 |
| | 8 2 | 8 2 | 6+7 4 | 3+4 2 |
| | 10+9 4 | 10+9 4 | 5 2 | 3+4 2 |
| | 5 2 | 6+7 4 | 5 2 | 3+4 2 |
| | 6+7 4 | 10+9 4 | 5 2 | 5+6 2 |
| | 10+9 4 | 8 2 | 6+7 4 | 5+6 2 |
| | 6+7 4 | 5 2 | 5 2 | 5+6 2 |
| | 8 2 | 6+7 4 | 6+7 4 | 3+4 2 |
| | 10+9 4 | 8 2 | 5 2 | 3+4 2 |
| | 8 2 | 10+9 4 | 6+7 4 | 5+6 2 |
| | 8 2 | 6+7 4 | 5 2 | 5+6 2 |
| | 10+9 4 | 5 2 | 6+7 4 | 5+6 2 |
| | 6+7 4 | 8 2 | 6+7 4 | 3+4 2 |
| | 5 2 | 10+9 4 | 5 2 | 3+4 2 |
| | 6+7 4 | 6+7 4 | 5 2 | 5+6 2 |
| | 5 2 | 5 2 | 6+7 4 | 5+6 2 |
| | 5 2 | 5 2 | 6+7 4 | 3+4 2 |
| | 6+7 4 | 6+7 4 | 5 2 | 3+4 2 |
| | | | | 5+6 2 |
| | | | | 3+4 2 |
| | | | | 5+6 2 |
| | | | | 5+6 2 |
| | | | | 3+4 2 |
| | | | | 5+6 2 |
| | | | | 3+4 2 |
| | | | | 3+4 2 |
| | | | | 3+4 2 |
| | | | | 5+6 2 |
| | | | | 3+4 2 |
| | | | | 5+6 2 |
| | | | | 5+6 2 |
| | | | | 5+6 2 |
| | | | | 3+4 2 |
| | | | | 5+6 2 |
| | | | | 3+4 2 |
| | | | | 5+6 2 |
| | | | | 5+6 2 |
| 204.1.2 | 195439 | | 136 | Cu4O12V4 |
| | -GM | -H | -P | -N |
| | 5 2 | 10+9 4 | 5 2 | 5+6 2 |
| | 6+7 4 | 8 2 | 6+7 4 | 3+4 2 |
| | 5 2 | 6+7 4 | 5 2 | 5+6 2 |
| | 6+7 4 | 5 2 | 5 2 | 5+6 2 |
| | 10+9 4 | 8 2 | 6+7 4 | 3+4 2 |
| | 8 2 | 10+9 4 | 6+7 4 | 3+4 2 |
| | 8 2 | 5 2 | 5 2 | 3+4 2 |
| | 10+9 4 | 6+7 4 | 6+7 4 | 5+6 2 |
| | 6+7 4 | 8 2 | 5 2 | 3+4 2 |

Continued on next column

**TABLE LXXXI– continued**

| TQC | ICSD | NoE | SF | |
|---|---|---|---|---|
| | 5 2 | 10+9 4 | 6+7 4 | 5+6 2 |
| | 5 2 | 10+9 4 | 5 2 | 3+4 2 |
| | 6+7 4 | 8 2 | 6+7 4 | 5+6 2 |
| | 5 2 | 8 2 | 5 2 | 5+6 2 |
| | 6+7 4 | 10+9 4 | 6+7 4 | 5+6 2 |
| | 8 2 | 5 2 | 5 2 | 3+4 2 |
| | 5 2 | 6+7 4 | 6+7 4 | 3+4 2 |
| | 6+7 4 | 5 2 | 5 2 | 3+4 2 |
| | 10+9 4 | 6+7 4 | 6+7 4 | 5+6 2 |
| | 5 2 | 5 2 | 5 2 | 5+6 2 |
| | 10+9 4 | 6+7 4 | 6+7 4 | 3+4 2 |
| | 8 2 | 8 2 | 6+7 4 | 5+6 2 |
| | 8 2 | 10+9 4 | 5 2 | 3+4 2 |
| | 10+9 4 | 6+7 4 | 5 2 | 3+4 2 |
| | 6+7 4 | 5 2 | 6+7 4 | 5+6 2 |
| | 6+7 4 | 10+9 4 | 6+7 4 | 3+4 2 |
| | 5 2 | 8 2 | 5 2 | 5+6 2 |
| | 10+9 4 | 6+7 4 | 5 2 | 5+6 2 |
| | 8 2 | 5 2 | 6+7 4 | 3+4 2 |
| | 10+9 4 | 8 2 | 5 2 | 3+4 2 |
| | 8 2 | 10+9 4 | 6+7 4 | 5+6 2 |
| | 5 2 | 6+7 4 | 5 2 | 5+6 2 |
| | 6+7 4 | 5 2 | 6+7 4 | 5+6 2 |
| | 8 2 | 5 2 | 5 2 | 5+6 2 |
| | 6+7 4 | 6+7 4 | 6+7 4 | 3+4 2 |
| | 5 2 | 5 2 | 5 2 | 3+4 2 |
| | 10+9 4 | 6+7 4 | 6+7 4 | 3+4 2 |
| | 5 2 | 5 2 | 5 2 | 5+6 2 |
| | 6+7 4 | 6+7 4 | 6+7 4 | 5+6 2 |
| | 5 2 | 5 2 | 6+7 4 | 3+4 2 |
| | 6+7 4 | 6+7 4 | 5 2 | 3+4 2 |
| | 5 2 | 6+7 4 | 6+7 4 | 5+6 2 |
| | 6+7 4 | 8 2 | 5 2 | 5+6 2 |
| | 6+7 4 | 5 2 | 6+7 4 | 3+4 2 |
| | 5 2 | 10+9 4 | 6+7 4 | 3+4 2 |
| | 6+7 4 | 6+7 4 | 5 2 | 5+6 2 |
| | | | | 3+4 2 |
| | | | | 3+4 2 |
| | | | | 5+6 2 |
| | | | | 5+6 2 |
| | | | | 3+4 2 |
| | | | | 5+6 2 |
| | | | | 3+4 2 |
| | | | | 5+6 2 |
| | | | | 5+6 2 |
| | | | | 5+6 2 |
| | | | | 3+4 2 |
| | | | | 5+6 2 |
| | | | | 3+4 2 |
| | | | | 5+6 2 |
| | | | | 5+6 2 |
| | | | | 3+4 2 |
| | | | | 3+4 2 |
| | | | | 3+4 2 |
| | | | | 3+4 2 |
| | | | | 5+6 2 |
| | | | | 3+4 2 |
| | | | | 3+4 2 |
| | | | | 5+6 2 |
| 204.1.3 | 195438 | 136 | Cu4O12V4 | |
| | -GM | -H | -P | -N |

Continued on next column

**TABLE LXXXI– continued**

| TQC | ICSD | NoE | SF | |
|---|---|---|---|---|
| | 5 2 | 10+9 4 | 5 2 | 5+6 2 |
| | 6+7 4 | 8 2 | 6+7 4 | 3+4 2 |
| | 5 2 | 6+7 4 | 5 2 | 5+6 2 |
| | 6+7 4 | 5 2 | 5 2 | 5+6 2 |
| | 10+9 4 | 8 2 | 6+7 4 | 3+4 2 |
| | 8 2 | 10+9 4 | 6+7 4 | 3+4 2 |
| | 8 2 | 5 2 | 5 2 | 3+4 2 |
| | 10+9 4 | 6+7 4 | 6+7 4 | 5+6 2 |
| | 6+7 4 | 8 2 | 5 2 | 3+4 2 |
| | 5 2 | 10+9 4 | 6+7 4 | 5+6 2 |
| | 5 2 | 10+9 4 | 5 2 | 3+4 2 |
| | 6+7 4 | 8 2 | 6+7 4 | 5+6 2 |
| | 5 2 | 8 2 | 5 2 | 5+6 2 |
| | 6+7 4 | 10+9 4 | 6+7 4 | 5+6 2 |
| | 8 2 | 5 2 | 5 2 | 3+4 2 |
| | 5 2 | 6+7 4 | 6+7 4 | 3+4 2 |
| | 6+7 4 | 5 2 | 5 2 | 3+4 2 |
| | 10+9 4 | 6+7 4 | 6+7 4 | 5+6 2 |
| | 5 2 | 5 2 | 5 2 | 5+6 2 |
| | 10+9 4 | 6+7 4 | 6+7 4 | 3+4 2 |
| | 8 2 | 8 2 | 6+7 4 | 5+6 2 |
| | 8 2 | 10+9 4 | 5 2 | 3+4 2 |
| | 10+9 4 | 6+7 4 | 5 2 | 3+4 2 |
| | 6+7 4 | 5 2 | 6+7 4 | 5+6 2 |
| | 6+7 4 | 10+9 4 | 6+7 4 | 3+4 2 |
| | 5 2 | 8 2 | 5 2 | 5+6 2 |
| | 10+9 4 | 6+7 4 | 5 2 | 5+6 2 |
| | 8 2 | 5 2 | 6+7 4 | 3+4 2 |
| | 10+9 4 | 8 2 | 5 2 | 3+4 2 |
| | 8 2 | 10+9 4 | 6+7 4 | 5+6 2 |
| | 5 2 | 6+7 4 | 5 2 | 5+6 2 |
| | 6+7 4 | 5 2 | 6+7 4 | 5+6 2 |
| | 8 2 | 5 2 | 5 2 | 5+6 2 |
| | 6+7 4 | 6+7 4 | 6+7 4 | 3+4 2 |
| | 10+9 4 | 5 2 | 5 2 | 3+4 2 |
| | 5 2 | 6+7 4 | 6+7 4 | 3+4 2 |
| | 5 2 | 5 2 | 5 2 | 5+6 2 |
| | 6+7 4 | 6+7 4 | 6+7 4 | 5+6 2 |
| | 5 2 | 5 2 | 6+7 4 | 3+4 2 |
| | 6+7 4 | 6+7 4 | 5 2 | 3+4 2 |
| | 5 2 | 6+7 4 | 6+7 4 | 5+6 2 |
| | 6+7 4 | 8 2 | 5 2 | 5+6 2 |
| | 6+7 4 | 5 2 | 6+7 4 | 3+4 2 |
| | 5 2 | 10+9 4 | 6+7 4 | 3+4 2 |
| | 6+7 4 | 6+7 4 | 5 2 | 5+6 2 |
| | | | | 3+4 2 |
| | | | | 3+4 2 |
| | | | | 5+6 2 |
| | | | | 5+6 2 |
| | | | | 3+4 2 |
| | | | | 5+6 2 |
| | | | | 3+4 2 |
| | | | | 5+6 2 |
| | | | | 5+6 2 |
| | | | | 5+6 2 |
| | | | | 3+4 2 |
| | | | | 5+6 2 |
| | | | | 3+4 2 |
| | | | | 5+6 2 |
| | | | | 5+6 2 |
| | | | | 3+4 2 |

Continued on next column

**TABLE LXXXI– continued**

| TQC | ICSD | | NoE | SF |
|---|---|---|---|---|
| | | | | 3+4 2 |
| | | | | 3+4 2 |
| | | | | 3+4 2 |
| | | | | 5+6 2 |
| | | | | 3+4 2 |
| | | | | 3+4 2 |
| | | | | 5+6 2 |
| 204.1.4 | 51893 | | 138 | Cu3Na1O12Ru4 |
| | -GM | -H | -P | -N |
| | 5 2 | 10+9 4 | 6+7 4 | 5+6 2 |
| | 6+7 4 | 8 2 | 5 2 | 3+4 2 |
| | 5 2 | 8 2 | 5 2 | 5+6 2 |
| | 6+7 4 | 10+9 4 | 5 2 | 5+6 2 |
| | 10+9 4 | 6+7 4 | 6+7 4 | 3+4 2 |
| | 8 2 | 5 2 | 6+7 4 | 3+4 2 |
| | 8 2 | 5 2 | 5 2 | 3+4 2 |
| | 10+9 4 | 6+7 4 | 6+7 4 | 5+6 2 |
| | 6+7 4 | 10+9 4 | 5 2 | 5+6 2 |
| | 5 2 | 8 2 | 6+7 4 | 3+4 2 |
| | 6+7 4 | 10+9 4 | 5 2 | 3+4 2 |
| | 5 2 | 10+9 4 | 6+7 4 | 5+6 2 |
| | 5 2 | 8 2 | 6+7 4 | 5+6 2 |
| | 6+7 4 | 8 2 | 5 2 | 3+4 2 |
| | 5 2 | 6+7 4 | 6+7 4 | 5+6 2 |
| | 6+7 4 | 5 2 | 5 2 | 3+4 2 |
| | 8 2 | 5 2 | 6+7 4 | 5+6 2 |
| | 10+9 4 | 6+7 4 | 5 2 | 3+4 2 |
| | 5 2 | 10+9 4 | 6+7 4 | 3+4 2 |
| | 6+7 4 | 8 2 | 5 2 | 5+6 2 |
| | 5 2 | 8 2 | 6+7 4 | 5+6 2 |
| | 10+9 4 | 5 2 | 6+7 4 | 3+4 2 |
| | 8 2 | 10+9 4 | 5 2 | 3+4 2 |
| | 6+7 4 | 6+7 4 | 5 2 | 5+6 2 |
| | 8 2 | 6+7 4 | 6+7 4 | 5+6 2 |
| | 10+9 4 | 5 2 | 5 2 | 3+4 2 |
| | 10+9 4 | 5 2 | 5 2 | 5+6 2 |
| | 8 2 | 6+7 4 | 6+7 4 | 3+4 2 |
| | 10+9 4 | 8 2 | 5 2 | 3+4 2 |
| | 8 2 | 10+9 4 | 5 2 | 5+6 2 |
| | 6+7 4 | 5 2 | 6+7 4 | 5+6 2 |
| | 5 2 | 6+7 4 | 6+7 4 | 5+6 2 |
| | 5 2 | 6+7 4 | 6+7 4 | 3+4 2 |
| | 6+7 4 | 5 2 | 5 2 | 5+6 2 |
| | 8 2 | 8 2 | 5 2 | 3+4 2 |
| | 10+9 4 | 5 2 | 6+7 4 | 5+6 2 |
| | 5 2 | 6+7 4 | 5 2 | 3+4 2 |
| | 6+7 4 | 10+9 4 | 6+7 4 | 5+6 2 |
| | 6+7 4 | 6+7 4 | 5 2 | 3+4 2 |
| | 5 2 | 5 2 | 6+7 4 | 5+6 2 |
| | 6+7 4 | 5 2 | 6+7 4 | 3+4 2 |
| | 5 2 | 6+7 4 | 5 2 | 3+4 2 |
| | 5 2 | 10+9 4 | 5 2 | 5+6 2 |
| | 6+7 4 | 8 2 | 6+7 4 | 5+6 2 |
| | 6+7 4 | 5 2 | 5 2 | 3+4 2 |
| | 5 2 | 6+7 4 | 6+7 4 | 3+4 2 |
| | | | | 3+4 2 |
| | | | | 5+6 2 |
| | | | | 5+6 2 |
| | | | | 3+4 2 |
| | | | | 5+6 2 |
| | | | | 3+4 2 |
| | | | | 5+6 2 |
| | | | | 5+6 2 |
| | | | | 5+6 2 |
| | | | | 3+4 2 |
| | | | | 5+6 2 |
| | | | | 3+4 2 |
| | | | | 5+6 2 |
| | | | | 3+4 2 |
| | | | | 5+6 2 |
| | | | | 5+6 2 |
| | | | | 5+6 2 |
| | | | | 5+6 2 |
| | | | | 3+4 2 |
| | | | | 3+4 2 |
| | | | | 5+6 2 |
| | | | | 5+6 2 |
| | | | | 3+4 2 |
| 204.1.5 | 237422 | | 160 | Cu3Fe4O12Tm1 |
| | -GM | -H | -P | -N |
| | 5 2 | 5 2 | 5 2 | 3+4 2 |
| | 8 2 | 8 2 | 5 2 | 5+6 2 |
| | 10+9 4 | 10+9 4 | 6+7 4 | 5+6 2 |
| | 5 2 | 8 2 | 5 2 | 5+6 2 |
| | 6+7 4 | 5 2 | 5 2 | 3+4 2 |
| | 5 2 | 6+7 4 | 6+7 4 | 5+6 2 |
| | 6+7 4 | 10+9 4 | 6+7 4 | 5+6 2 |
| | 8 2 | 5 2 | 5 2 | 3+4 2 |
| | 10+9 4 | 6+7 4 | 6+7 4 | 3+4 2 |
| | 8 2 | 8 2 | 5 2 | 5+6 2 |
| | 10+9 4 | 10+9 4 | 6+7 4 | 3+4 2 |
| | 6+7 4 | 8 2 | 5 2 | 5+6 2 |
| | 5 2 | 8 2 | 6+7 4 | 3+4 2 |
| | 5 2 | 10+9 4 | 5 2 | 3+4 2 |
| | 6+7 4 | 8 2 | 5 2 | 5+6 2 |
| | 5 2 | 10+9 4 | 6+7 4 | 5+6 2 |
| | 6+7 4 | 5 2 | 6+7 4 | 5+6 2 |
| | 8 2 | 6+7 4 | 5 2 | 3+4 2 |
| | 10+9 4 | 5 2 | 6+7 4 | 3+4 2 |
| | 6+7 4 | 6+7 4 | 5 2 | 5+6 2 |
| | 5 2 | 10+9 4 | 6+7 4 | 5+6 2 |
| | 5 2 | 6+7 4 | 5 2 | 3+4 2 |
| | 6+7 4 | 5 2 | 6+7 4 | 3+4 2 |
| | 8 2 | 6+7 4 | 5 2 | 5+6 2 |
| | 10+9 4 | 5 2 | 6+7 4 | 5+6 2 |
| | 8 2 | 8 2 | 5 2 | 3+4 2 |
| | 6+7 4 | 10+9 4 | 6+7 4 | 3+4 2 |
| | 5 2 | 6+7 4 | 6+7 4 | 5+6 2 |
| | 8 2 | 5 2 | 5 2 | 3+4 2 |
| | 10+9 4 | 8 2 | 5 2 | 5+6 2 |
| | 10+9 4 | 10+9 4 | 6+7 4 | 3+4 2 |
| | 8 2 | 8 2 | 5 2 | 5+6 2 |
| | 10+9 4 | 10+9 4 | 6+7 4 | 5+6 2 |
| | 5 2 | 6+7 4 | 5 2 | 3+4 2 |
| | 6+7 4 | 5 2 | 6+7 4 | 5+6 2 |
| | 6+7 4 | 5 2 | 5 2 | 5+6 2 |
| | 5 2 | 6+7 4 | 6+7 4 | 5+6 2 |
| | 5 2 | 6+7 4 | 5 2 | 3+4 2 |
| | 6+7 4 | 5 2 | 6+7 4 | 3+4 2 |
| | 8 2 | 5 2 | 5 2 | 3+4 2 |
| | 10+9 4 | 6+7 4 | 6+7 4 | 5+6 2 |
| | 5 2 | 5 2 | 5 2 | 5+6 2 |

Continued on next column

**TABLE LXXXI– continued**

| TQC | ICSD | | NoE | SF |
|---|---|---|---|---|
| | 6+7 4 | 8 2 | 6+7 4 | 3+4 2 |
| | 5 2 | 6+7 4 | 5 2 | 3+4 2 |
| | 10+9 4 | 10+9 4 | 6+7 4 | 5+6 2 |
| | 8 2 | 8 2 | 6+7 4 | 5+6 2 |
| | 6+7 4 | 10+9 4 | 5 2 | 3+4 2 |
| | 6+7 4 | 8 2 | 5 2 | 5+6 2 |
| | 5 2 | 10+9 4 | 6+7 4 | 3+4 2 |
| | 6+7 4 | 10+9 4 | 5 2 | 3+4 2 |
| | 8 2 | 8 2 | 6+7 4 | 5+6 2 |
| | 8 2 | 8 2 | 5 2 | 3+4 2 |
| | 10+9 4 | 10+9 4 | 6+7 4 | 5+6 2 |
| | 5 2 | 8 2 | 5 2 | 3+4 2 |
| | | | | 5+6 2 |
| | | | | 3+4 2 |
| | | | | 5+6 2 |
| | | | | 5+6 2 |
| | | | | 5+6 2 |
| | | | | 3+4 2 |
| | | | | 5+6 2 |
| | | | | 5+6 2 |
| | | | | 5+6 2 |
| | | | | 3+4 2 |
| | | | | 5+6 2 |
| | | | | 3+4 2 |
| | | | | 5+6 2 |
| | | | | 5+6 2 |
| | | | | 5+6 2 |
| | | | | 5+6 2 |
| | | | | 5+6 2 |
| | | | | 3+4 2 |
| | | | | 3+4 2 |
| | | | | 5+6 2 |
| | | | | 3+4 2 |
| | | | | 5+6 2 |
| | | | | 5+6 2 |
| | | | | 3+4 2 |
| | | | | 5+6 2 |
| | | | | 5+6 2 |
| 204.1.6 | 248615 | | 140 | Cr4Cu3O12Y1 |
| | -GM | -H | -P | -N |
| | 5 2 | 5 2 | 5 2 | 3+4 2 |
| | 8 2 | 8 2 | 5 2 | 5+6 2 |
| | 10+9 4 | 10+9 4 | 6+7 4 | 5+6 2 |
| | 5 2 | 8 2 | 5 2 | 5+6 2 |
| | 6+7 4 | 10+9 4 | 5 2 | 3+4 2 |
| | 5 2 | 5 2 | 6+7 4 | 5+6 2 |
| | 6+7 4 | 6+7 4 | 6+7 4 | 3+4 2 |
| | 8 2 | 8 2 | 5 2 | 3+4 2 |
| | 10+9 4 | 10+9 4 | 6+7 4 | 5+6 2 |
| | 8 2 | 5 2 | 5 2 | 5+6 2 |
| | 10+9 4 | 6+7 4 | 6+7 4 | 3+4 2 |
| | 6+7 4 | 8 2 | 5 2 | 5+6 2 |
| | 5 2 | 10+9 4 | 6+7 4 | 3+4 2 |
| | 5 2 | 8 2 | 5 2 | 5+6 2 |
| | 6+7 4 | 10+9 4 | 6+7 4 | 3+4 2 |
| | 5 2 | 8 2 | 6+7 4 | 5+6 2 |
| | 6+7 4 | 5 2 | 5 2 | 5+6 2 |
| | 8 2 | 6+7 4 | 5 2 | 5+6 2 |
| | 10+9 4 | 5 2 | 6+7 4 | 3+4 2 |
| | 5 2 | 6+7 4 | 6+7 4 | 3+4 2 |
| | 6+7 4 | 10+9 4 | 5 2 | 3+4 2 |

Continued on next column

**TABLE LXXXI– continued**

| TQC | ICSD | | NoE | SF |
|---|---|---|---|---|
| | 5 2 | 6+7 4 | 5 2 | 5+6 2 |
| | 6+7 4 | 5 2 | 6+7 4 | 3+4 2 |
| | 8 2 | 6+7 4 | 6+7 4 | 3+4 2 |
| | 10+9 4 | 5 2 | 5 2 | 5+6 2 |
| | 6+7 4 | 8 2 | 5 2 | 5+6 2 |
| | 5 2 | 10+9 4 | 6+7 4 | 3+4 2 |
| | 8 2 | 6+7 4 | 6+7 4 | 5+6 2 |
| | 10+9 4 | 5 2 | 5 2 | 3+4 2 |
| | 8 2 | 10+9 4 | 5 2 | 5+6 2 |
| | 10+9 4 | 8 2 | 6+7 4 | 3+4 2 |
| | 10+9 4 | 8 2 | 5 2 | 5+6 2 |
| | 8 2 | 10+9 4 | 6+7 4 | 3+4 2 |
| | 5 2 | 5 2 | 5 2 | 5+6 2 |
| | 6+7 4 | 6+7 4 | 6+7 4 | 5+6 2 |
| | 6+7 4 | 6+7 4 | 5 2 | 5+6 2 |
| | 5 2 | 5 2 | 6+7 4 | 3+4 2 |
| | 8 2 | 5 2 | 5 2 | 5+6 2 |
| | 10+9 4 | 6+7 4 | 6+7 4 | 3+4 2 |
| | 5 2 | 6+7 4 | 5 2 | 3+4 2 |
| | 6+7 4 | 5 2 | 6+7 4 | 5+6 2 |
| | 5 2 | 5 2 | 6+7 4 | 5+6 2 |
| | 6+7 4 | 6+7 4 | 5 2 | 3+4 2 |
| | 5 2 | 8 2 | 6+7 4 | 3+4 2 |
| | 6+7 4 | 10+9 4 | 5 2 | 5+6 2 |
| | 6+7 4 | 8 2 | 5 2 | 3+4 2 |
| | 5 2 | 10+9 4 | 6+7 4 | 5+6 2 |
| | | | | 3+4 2 |
| | | | | 3+4 2 |
| | | | | 5+6 2 |
| | | | | 3+4 2 |
| | | | | 5+6 2 |
| | | | | 5+6 2 |
| | | | | 3+4 2 |
| | | | | 5+6 2 |
| | | | | 3+4 2 |
| | | | | 5+6 2 |
| | | | | 5+6 2 |
| | | | | 3+4 2 |
| | | | | 5+6 2 |
| | | | | 5+6 2 |
| | | | | 3+4 2 |
| | | | | 5+6 2 |
| | | | | 5+6 2 |
| | | | | 3+4 2 |
| | | | | 5+6 2 |
| | | | | 5+6 2 |
| | | | | 5+6 2 |
| | | | | 3+4 2 |
| | | | | 3+4 2 |
| 204.1.7 | 248614 | | 140 | Cr4Cu3O12Y1 |
| | -GM | -H | -P | -N |
| | 5 2 | 5 2 | 5 2 | 3+4 2 |
| | 8 2 | 8 2 | 5 2 | 5+6 2 |
| | 10+9 4 | 10+9 4 | 6+7 4 | 5+6 2 |
| | 5 2 | 8 2 | 5 2 | 5+6 2 |
| | 6+7 4 | 10+9 4 | 5 2 | 3+4 2 |
| | 5 2 | 5 2 | 6+7 4 | 5+6 2 |
| | 6+7 4 | 6+7 4 | 6+7 4 | 3+4 2 |
| | 8 2 | 8 2 | 5 2 | 3+4 2 |
| | 10+9 4 | 10+9 4 | 6+7 4 | 5+6 2 |
| | 8 2 | 5 2 | 5 2 | 5+6 2 |

Continued on next column

**TABLE LXXXI– continued**

| TQC | ICSD | NoE | SF | |
|---|---|---|---|---|
| | 10+9 4 | 6+7 4 | 6+7 4 | 3+4 2 |
| | 6+7 4 | 8 2 | 5 2 | 5+6 2 |
| | 5 2 | 10+9 4 | 6+7 4 | 3+4 2 |
| | 5 2 | 8 2 | 5 2 | 5+6 2 |
| | 6+7 4 | 10+9 4 | 6+7 4 | 3+4 2 |
| | 5 2 | 8 2 | 6+7 4 | 5+6 2 |
| | 6+7 4 | 5 2 | 5 2 | 5+6 2 |
| | 8 2 | 6+7 4 | 5 2 | 5+6 2 |
| | 10+9 4 | 5 2 | 6+7 4 | 3+4 2 |
| | 5 2 | 6+7 4 | 5 2 | 3+4 2 |
| | 6+7 4 | 10+9 4 | 6+7 4 | 3+4 2 |
| | 5 2 | 6+7 4 | 5 2 | 5+6 2 |
| | 6+7 4 | 5 2 | 6+7 4 | 3+4 2 |
| | 8 2 | 6+7 4 | 6+7 4 | 3+4 2 |
| | 10+9 4 | 5 2 | 5 2 | 5+6 2 |
| | 6+7 4 | 8 2 | 5 2 | 5+6 2 |
| | 5 2 | 10+9 4 | 6+7 4 | 3+4 2 |
| | 8 2 | 6+7 4 | 6+7 4 | 5+6 2 |
| | 10+9 4 | 5 2 | 5 2 | 3+4 2 |
| | 8 2 | 10+9 4 | 5 2 | 5+6 2 |
| | 10+9 4 | 8 2 | 6+7 4 | 3+4 2 |
| | 10+9 4 | 8 2 | 5 2 | 5+6 2 |
| | 8 2 | 10+9 4 | 6+7 4 | 3+4 2 |
| | 5 2 | 5 2 | 5 2 | 5+6 2 |
| | 6+7 4 | 6+7 4 | 6+7 4 | 5+6 2 |
| | 6+7 4 | 6+7 4 | 5 2 | 5+6 2 |
| | 5 2 | 5 2 | 6+7 4 | 3+4 2 |
| | 8 2 | 5 2 | 5 2 | 5+6 2 |
| | 10+9 4 | 6+7 4 | 6+7 4 | 3+4 2 |
| | 5 2 | 6+7 4 | 5 2 | 3+4 2 |
| | 6+7 4 | 5 2 | 6+7 4 | 5+6 2 |
| | 5 2 | 5 2 | 6+7 4 | 5+6 2 |
| | 6+7 4 | 6+7 4 | 5 2 | 3+4 2 |
| | 5 2 | 8 2 | 6+7 4 | 3+4 2 |
| | 6+7 4 | 10+9 4 | 5 2 | 5+6 2 |
| | 6+7 4 | 8 2 | 5 2 | 3+4 2 |
| | 5 2 | 10+9 4 | 6+7 4 | 5+6 2 |
| | | | | 3+4 2 |
| | | | | 3+4 2 |
| | | | | 5+6 2 |
| | | | | 3+4 2 |
| | | | | 5+6 2 |
| | | | | 5+6 2 |
| | | | | 3+4 2 |
| | | | | 5+6 2 |
| | | | | 3+4 2 |
| | | | | 5+6 2 |
| | | | | 5+6 2 |
| | | | | 3+4 2 |
| | | | | 5+6 2 |
| | | | | 5+6 2 |
| | | | | 3+4 2 |
| | | | | 5+6 2 |
| | | | | 5+6 2 |
| | | | | 3+4 2 |
| | | | | 5+6 2 |
| | | | | 5+6 2 |
| | | | | 5+6 2 |
| | | | | 3+4 2 |
| | | | | 3+4 2 |
| 204.1.8 | 248613 | 140 | Cr4Cu3O12Y1 | |

Continued on next column

**TABLE LXXXI– continued**

| TQC | ICSD | NoE | SF | |
|---|---|---|---|---|
| | -GM | -H | -P | -N |
| | 5 2 | 5 2 | 5 2 | 3+4 2 |
| | 8 2 | 8 2 | 5 2 | 5+6 2 |
| | 10+9 4 | 10+9 4 | 6+7 4 | 5+6 2 |
| | 5 2 | 8 2 | 5 2 | 5+6 2 |
| | 6+7 4 | 10+9 4 | 5 2 | 3+4 2 |
| | 5 2 | 5 2 | 6+7 4 | 5+6 2 |
| | 6+7 4 | 6+7 4 | 6+7 4 | 3+4 2 |
| | 8 2 | 8 2 | 5 2 | 3+4 2 |
| | 10+9 4 | 10+9 4 | 6+7 4 | 5+6 2 |
| | 8 2 | 5 2 | 5 2 | 5+6 2 |
| | 10+9 4 | 6+7 4 | 6+7 4 | 3+4 2 |
| | 6+7 4 | 8 2 | 5 2 | 5+6 2 |
| | 5 2 | 10+9 4 | 6+7 4 | 3+4 2 |
| | 5 2 | 8 2 | 5 2 | 5+6 2 |
| | 6+7 4 | 8 2 | 6+7 4 | 3+4 2 |
| | 5 2 | 10+9 4 | 6+7 4 | 5+6 2 |
| | 6+7 4 | 5 2 | 5 2 | 5+6 2 |
| | 8 2 | 6+7 4 | 5 2 | 5+6 2 |
| | 10+9 4 | 5 2 | 6+7 4 | 3+4 2 |
| | 5 2 | 6+7 4 | 5 2 | 3+4 2 |
| | 6+7 4 | 10+9 4 | 6+7 4 | 3+4 2 |
| | 5 2 | 6+7 4 | 5 2 | 5+6 2 |
| | 8 2 | 6+7 4 | 6+7 4 | 3+4 2 |
| | 6+7 4 | 5 2 | 6+7 4 | 3+4 2 |
| | 10+9 4 | 5 2 | 5 2 | 5+6 2 |
| | 6+7 4 | 8 2 | 5 2 | 5+6 2 |
| | 5 2 | 10+9 4 | 6+7 4 | 3+4 2 |
| | 8 2 | 6+7 4 | 6+7 4 | 5+6 2 |
| | 10+9 4 | 5 2 | 5 2 | 3+4 2 |
| | 8 2 | 10+9 4 | 5 2 | 5+6 2 |
| | 10+9 4 | 8 2 | 6+7 4 | 3+4 2 |
| | 10+9 4 | 8 2 | 5 2 | 5+6 2 |
| | 8 2 | 10+9 4 | 6+7 4 | 3+4 2 |
| | 5 2 | 5 2 | 5 2 | 5+6 2 |
| | 6+7 4 | 6+7 4 | 6+7 4 | 5+6 2 |
| | 6+7 4 | 6+7 4 | 5 2 | 5+6 2 |
| | 5 2 | 5 2 | 6+7 4 | 3+4 2 |
| | 8 2 | 5 2 | 5 2 | 5+6 2 |
| | 10+9 4 | 6+7 4 | 6+7 4 | 3+4 2 |
| | 5 2 | 6+7 4 | 5 2 | 3+4 2 |
| | 6+7 4 | 5 2 | 6+7 4 | 5+6 2 |
| | 5 2 | 5 2 | 6+7 4 | 5+6 2 |
| | 6+7 4 | 6+7 4 | 5 2 | 3+4 2 |
| | 5 2 | 8 2 | 6+7 4 | 3+4 2 |
| | 6+7 4 | 10+9 4 | 5 2 | 5+6 2 |
| | 6+7 4 | 8 2 | 5 2 | 3+4 2 |
| | 5 2 | 10+9 4 | 6+7 4 | 5+6 2 |
| | | | | 3+4 2 |
| | | | | 3+4 2 |
| | | | | 5+6 2 |
| | | | | 3+4 2 |
| | | | | 5+6 2 |
| | | | | 5+6 2 |
| | | | | 3+4 2 |
| | | | | 5+6 2 |
| | | | | 3+4 2 |
| | | | | 5+6 2 |
| | | | | 5+6 2 |
| | | | | 3+4 2 |
| | | | | 5+6 2 |

Continued on next column

**TABLE LXXXI– continued**

| TQC | ICSD | | NoE | SF |
|---|---|---|---|---|
| | | | | 5+6 2 |
| | | | | 3+4 2 |
| | | | | 5+6 2 |
| | | | | 5+6 2 |
| | | | | 3+4 2 |
| | | | | 5+6 2 |
| | | | | 5+6 2 |
| | | | | 5+6 2 |
| | | | | 3+4 2 |
| | | | | 3+4 2 |
| 204.1.9 | 248612 | | 140 | Cr4Cu3O12Y1 |
| | -GM | -H | -P | -N |
| | 5 2 | 5 2 | 5 2 | 3+4 2 |
| | 8 2 | 8 2 | 5 2 | 5+6 2 |
| | 10+9 4 | 10+9 4 | 6+7 4 | 5+6 2 |
| | 5 2 | 8 2 | 5 2 | 5+6 2 |
| | 6+7 4 | 10+9 4 | 5 2 | 3+4 2 |
| | 5 2 | 5 2 | 6+7 4 | 5+6 2 |
| | 6+7 4 | 6+7 4 | 6+7 4 | 3+4 2 |
| | 8 2 | 8 2 | 5 2 | 3+4 2 |
| | 10+9 4 | 10+9 4 | 6+7 4 | 5+6 2 |
| | 8 2 | 5 2 | 5 2 | 5+6 2 |
| | 10+9 4 | 6+7 4 | 6+7 4 | 3+4 2 |
| | 6+7 4 | 8 2 | 5 2 | 5+6 2 |
| | 5 2 | 10+9 4 | 6+7 4 | 3+4 2 |
| | 5 2 | 8 2 | 5 2 | 5+6 2 |
| | 6+7 4 | 8 2 | 6+7 4 | 3+4 2 |
| | 5 2 | 10+9 4 | 6+7 4 | 5+6 2 |
| | 6+7 4 | 5 2 | 5 2 | 5+6 2 |
| | 8 2 | 6+7 4 | 5 2 | 5+6 2 |
| | 10+9 4 | 5 2 | 6+7 4 | 3+4 2 |
| | 5 2 | 6+7 4 | 5 2 | 3+4 2 |
| | 6+7 4 | 10+9 4 | 6+7 4 | 3+4 2 |
| | 5 2 | 6+7 4 | 5 2 | 5+6 2 |
| | 8 2 | 6+7 4 | 6+7 4 | 3+4 2 |
| | 6+7 4 | 5 2 | 6+7 4 | 3+4 2 |
| | 10+9 4 | 5 2 | 5 2 | 5+6 2 |
| | 6+7 4 | 8 2 | 5 2 | 5+6 2 |
| | 5 2 | 10+9 4 | 6+7 4 | 3+4 2 |
| | 8 2 | 6+7 4 | 6+7 4 | 5+6 2 |
| | 10+9 4 | 5 2 | 5 2 | 3+4 2 |
| | 8 2 | 10+9 4 | 5 2 | 5+6 2 |
| | 10+9 4 | 8 2 | 6+7 4 | 3+4 2 |
| | 10+9 4 | 8 2 | 5 2 | 5+6 2 |
| | 8 2 | 10+9 4 | 6+7 4 | 3+4 2 |
| | 5 2 | 5 2 | 5 2 | 5+6 2 |
| | 6+7 4 | 6+7 4 | 6+7 4 | 5+6 2 |
| | 6+7 4 | 6+7 4 | 5 2 | 5+6 2 |
| | 5 2 | 5 2 | 6+7 4 | 3+4 2 |
| | 8 2 | 5 2 | 5 2 | 5+6 2 |
| | 10+9 4 | 6+7 4 | 6+7 4 | 3+4 2 |
| | 5 2 | 6+7 4 | 5 2 | 3+4 2 |
| | 6+7 4 | 5 2 | 6+7 4 | 5+6 2 |
| | 5 2 | 5 2 | 6+7 4 | 5+6 2 |
| | 6+7 4 | 6+7 4 | 5 2 | 3+4 2 |
| | 5 2 | 8 2 | 6+7 4 | 3+4 2 |
| | 6+7 4 | 10+9 4 | 5 2 | 5+6 2 |
| | 6+7 4 | 8 2 | 5 2 | 3+4 2 |
| | 5 2 | 10+9 4 | 6+7 4 | 5+6 2 |
| | | | | 3+4 2 |
| | | | | 3+4 2 |

Continued on next column

**TABLE LXXXI– continued**

| TQC | ICSD | | NoE | SF |
|---|---|---|---|---|
| | | | | 5+6 2 |
| | | | | 3+4 2 |
| | | | | 5+6 2 |
| | | | | 5+6 2 |
| | | | | 3+4 2 |
| | | | | 5+6 2 |
| | | | | 3+4 2 |
| | | | | 5+6 2 |
| | | | | 5+6 2 |
| | | | | 3+4 2 |
| | | | | 5+6 2 |
| | | | | 5+6 2 |
| | | | | 3+4 2 |
| | | | | 5+6 2 |
| | | | | 5+6 2 |
| | | | | 3+4 2 |
| | | | | 5+6 2 |
| | | | | 5+6 2 |
| | | | | 5+6 2 |
| | | | | 3+4 2 |
| | | | | 3+4 2 |
| 204.1.10 | 248611 | | 140 | Cr4Cu3O12Y1 |
| | -GM | -H | -P | -N |
| | 5 2 | 5 2 | 5 2 | 3+4 2 |
| | 8 2 | 8 2 | 5 2 | 5+6 2 |
| | 10+9 4 | 10+9 4 | 6+7 4 | 5+6 2 |
| | 5 2 | 8 2 | 5 2 | 5+6 2 |
| | 6+7 4 | 10+9 4 | 5 2 | 3+4 2 |
| | 5 2 | 5 2 | 6+7 4 | 5+6 2 |
| | 6+7 4 | 6+7 4 | 6+7 4 | 3+4 2 |
| | 8 2 | 8 2 | 5 2 | 3+4 2 |
| | 10+9 4 | 10+9 4 | 6+7 4 | 5+6 2 |
| | 8 2 | 5 2 | 5 2 | 5+6 2 |
| | 10+9 4 | 6+7 4 | 6+7 4 | 3+4 2 |
| | 6+7 4 | 8 2 | 5 2 | 5+6 2 |
| | 5 2 | 10+9 4 | 6+7 4 | 3+4 2 |
| | 5 2 | 8 2 | 5 2 | 5+6 2 |
| | 6+7 4 | 8 2 | 6+7 4 | 3+4 2 |
| | 5 2 | 10+9 4 | 6+7 4 | 5+6 2 |
| | 6+7 4 | 5 2 | 5 2 | 5+6 2 |
| | 8 2 | 6+7 4 | 5 2 | 5+6 2 |
| | 10+9 4 | 5 2 | 6+7 4 | 3+4 2 |
| | 5 2 | 6+7 4 | 5 2 | 3+4 2 |
| | 6+7 4 | 10+9 4 | 6+7 4 | 3+4 2 |
| | 5 2 | 6+7 4 | 5 2 | 5+6 2 |
| | 8 2 | 6+7 4 | 6+7 4 | 3+4 2 |
| | 6+7 4 | 5 2 | 6+7 4 | 3+4 2 |
| | 10+9 4 | 5 2 | 5 2 | 5+6 2 |
| | 6+7 4 | 8 2 | 5 2 | 5+6 2 |
| | 5 2 | 10+9 4 | 6+7 4 | 3+4 2 |
| | 8 2 | 6+7 4 | 6+7 4 | 5+6 2 |
| | 10+9 4 | 5 2 | 5 2 | 3+4 2 |
| | 8 2 | 10+9 4 | 5 2 | 5+6 2 |
| | 10+9 4 | 8 2 | 6+7 4 | 3+4 2 |
| | 10+9 4 | 8 2 | 5 2 | 5+6 2 |
| | 8 2 | 10+9 4 | 6+7 4 | 3+4 2 |
| | 5 2 | 5 2 | 5 2 | 5+6 2 |
| | 6+7 4 | 6+7 4 | 6+7 4 | 5+6 2 |
| | 6+7 4 | 6+7 4 | 5 2 | 5+6 2 |
| | 5 2 | 5 2 | 6+7 4 | 3+4 2 |
| | 8 2 | 5 2 | 5 2 | 5+6 2 |

Continued on next column

**TABLE LXXXI– continued**

| TQC | ICSD | NoE | SF | |
|---|---|---|---|---|
| | 10+9 4 | 6+7 4 | 6+7 4 | 3+4 2 |
| | 5 2 | 6+7 4 | 5 2 | 3+4 2 |
| | 6+7 4 | 5 2 | 6+7 4 | 5+6 2 |
| | 5 2 | 5 2 | 6+7 4 | 5+6 2 |
| | 6+7 4 | 6+7 4 | 5 2 | 3+4 2 |
| | 5 2 | 8 2 | 6+7 4 | 3+4 2 |
| | 6+7 4 | 10+9 4 | 5 2 | 5+6 2 |
| | 6+7 4 | 8 2 | 5 2 | 3+4 2 |
| | 5 2 | 10+9 4 | 6+7 4 | 5+6 2 |
| | | | | 3+4 2 |
| | | | | 3+4 2 |
| | | | | 5+6 2 |
| | | | | 3+4 2 |
| | | | | 5+6 2 |
| | | | | 5+6 2 |
| | | | | 3+4 2 |
| | | | | 5+6 2 |
| | | | | 3+4 2 |
| | | | | 5+6 2 |
| | | | | 5+6 2 |
| | | | | 3+4 2 |
| | | | | 5+6 2 |
| | | | | 5+6 2 |
| | | | | 3+4 2 |
| | | | | 5+6 2 |
| | | | | 5+6 2 |
| | | | | 3+4 2 |
| | | | | 5+6 2 |
| | | | | 5+6 2 |
| | | | | 5+6 2 |
| | | | | 3+4 2 |
| | | | | 3+4 2 |
| 204.1.11 | 248610 | 140 | Cr4Cu3O12Y1 | |
| | -GM | -H | -P | -N |
| | 5 2 | 5 2 | 5 2 | 3+4 2 |
| | 8 2 | 8 2 | 5 2 | 5+6 2 |
| | 10+9 4 | 10+9 4 | 6+7 4 | 5+6 2 |
| | 5 2 | 8 2 | 5 2 | 5+6 2 |
| | 6+7 4 | 10+9 4 | 5 2 | 3+4 2 |
| | 5 2 | 5 2 | 6+7 4 | 5+6 2 |
| | 6+7 4 | 6+7 4 | 6+7 4 | 3+4 2 |
| | 8 2 | 8 2 | 5 2 | 3+4 2 |
| | 10+9 4 | 10+9 4 | 6+7 4 | 5+6 2 |
| | 8 2 | 5 2 | 5 2 | 5+6 2 |
| | 10+9 4 | 6+7 4 | 6+7 4 | 3+4 2 |
| | 6+7 4 | 8 2 | 5 2 | 5+6 2 |
| | 5 2 | 10+9 4 | 6+7 4 | 3+4 2 |
| | 5 2 | 8 2 | 5 2 | 5+6 2 |
| | 6+7 4 | 8 2 | 6+7 4 | 3+4 2 |
| | 5 2 | 10+9 4 | 5 2 | 5+6 2 |
| | 6+7 4 | 5 2 | 6+7 4 | 5+6 2 |
| | 8 2 | 6+7 4 | 5 2 | 5+6 2 |
| | 10+9 4 | 10+9 4 | 6+7 4 | 3+4 2 |
| | 5 2 | 5 2 | 5 2 | 3+4 2 |
| | 6+7 4 | 6+7 4 | 6+7 4 | 3+4 2 |
| | 5 2 | 6+7 4 | 5 2 | 5+6 2 |
| | 8 2 | 6+7 4 | 6+7 4 | 3+4 2 |
| | 10+9 4 | 5 2 | 6+7 4 | 3+4 2 |
| | 6+7 4 | 5 2 | 5 2 | 5+6 2 |
| | 6+7 4 | 8 2 | 5 2 | 5+6 2 |
| | 5 2 | 10+9 4 | 6+7 4 | 3+4 2 |

Continued on next column

**TABLE LXXXI– continued**

| TQC | ICSD | NoE | SF | |
|---|---|---|---|---|
| | 8 2 | 6+7 4 | 6+7 4 | 5+6 2 |
| | 10+9 4 | 5 2 | 5 2 | 3+4 2 |
| | 8 2 | 10+9 4 | 5 2 | 5+6 2 |
| | 10+9 4 | 8 2 | 6+7 4 | 3+4 2 |
| | 10+9 4 | 8 2 | 5 2 | 5+6 2 |
| | 8 2 | 10+9 4 | 6+7 4 | 3+4 2 |
| | 6+7 4 | 5 2 | 5 2 | 5+6 2 |
| | 5 2 | 6+7 4 | 6+7 4 | 5+6 2 |
| | 6+7 4 | 6+7 4 | 5 2 | 5+6 2 |
| | 5 2 | 5 2 | 6+7 4 | 3+4 2 |
| | 8 2 | 5 2 | 5 2 | 5+6 2 |
| | 10+9 4 | 6+7 4 | 6+7 4 | 3+4 2 |
| | 5 2 | 6+7 4 | 5 2 | 3+4 2 |
| | 6+7 4 | 5 2 | 6+7 4 | 5+6 2 |
| | 5 2 | 5 2 | 6+7 4 | 5+6 2 |
| | 6+7 4 | 6+7 4 | 5 2 | 3+4 2 |
| | 5 2 | 8 2 | 6+7 4 | 3+4 2 |
| | 6+7 4 | 10+9 4 | 5 2 | 5+6 2 |
| | 6+7 4 | 8 2 | 5 2 | 3+4 2 |
| | 5 2 | 10+9 4 | 6+7 4 | 5+6 2 |
| | | | | 3+4 2 |
| | | | | 3+4 2 |
| | | | | 5+6 2 |
| | | | | 3+4 2 |
| | | | | 5+6 2 |
| | | | | 5+6 2 |
| | | | | 3+4 2 |
| | | | | 5+6 2 |
| | | | | 3+4 2 |
| | | | | 5+6 2 |
| | | | | 5+6 2 |
| | | | | 3+4 2 |
| | | | | 5+6 2 |
| | | | | 5+6 2 |
| | | | | 3+4 2 |
| | | | | 5+6 2 |
| | | | | 5+6 2 |
| | | | | 3+4 2 |
| | | | | 5+6 2 |
| | | | | 5+6 2 |
| | | | | 5+6 2 |
| | | | | 3+4 2 |
| | | | | 3+4 2 |
| 204.1.12 | 248609 | 140 | Cr4Cu3O12Y1 | |
| | -GM | -H | -P | -N |
| | 5 2 | 5 2 | 5 2 | 3+4 2 |
| | 8 2 | 8 2 | 5 2 | 5+6 2 |
| | 10+9 4 | 10+9 4 | 6+7 4 | 5+6 2 |
| | 5 2 | 8 2 | 5 2 | 5+6 2 |
| | 6+7 4 | 10+9 4 | 5 2 | 3+4 2 |
| | 5 2 | 5 2 | 6+7 4 | 5+6 2 |
| | 6+7 4 | 6+7 4 | 6+7 4 | 3+4 2 |
| | 8 2 | 8 2 | 5 2 | 3+4 2 |
| | 10+9 4 | 10+9 4 | 6+7 4 | 5+6 2 |
| | 8 2 | 5 2 | 5 2 | 5+6 2 |
| | 10+9 4 | 6+7 4 | 6+7 4 | 3+4 2 |
| | 6+7 4 | 8 2 | 5 2 | 5+6 2 |
| | 5 2 | 10+9 4 | 6+7 4 | 3+4 2 |
| | 5 2 | 8 2 | 5 2 | 3+4 2 |
| | 6+7 4 | 8 2 | 6+7 4 | 5+6 2 |
| | 5 2 | 10+9 4 | 5 2 | 5+6 2 |

Continued on next column

**TABLE LXXXI– continued**

| TQC | ICSD | | NoE | SF |
|---|---|---|---|---|
| | 6+7 4 | 5 2 | 6+7 4 | 5+6 2 |
| | 8 2 | 6+7 4 | 5 2 | 5+6 2 |
| | 10+9 4 | 10+9 4 | 6+7 4 | 3+4 2 |
| | 5 2 | 5 2 | 5 2 | 3+4 2 |
| | 6+7 4 | 6+7 4 | 6+7 4 | 3+4 2 |
| | 5 2 | 6+7 4 | 5 2 | 5+6 2 |
| | 8 2 | 6+7 4 | 6+7 4 | 3+4 2 |
| | 10+9 4 | 5 2 | 6+7 4 | 3+4 2 |
| | 6+7 4 | 5 2 | 5 2 | 5+6 2 |
| | 6+7 4 | 8 2 | 5 2 | 5+6 2 |
| | 5 2 | 10+9 4 | 6+7 4 | 3+4 2 |
| | 8 2 | 6+7 4 | 6+7 4 | 5+6 2 |
| | 10+9 4 | 5 2 | 5 2 | 3+4 2 |
| | 8 2 | 10+9 4 | 5 2 | 5+6 2 |
| | 10+9 4 | 8 2 | 6+7 4 | 3+4 2 |
| | 10+9 4 | 8 2 | 5 2 | 5+6 2 |
| | 8 2 | 10+9 4 | 6+7 4 | 3+4 2 |
| | 6+7 4 | 5 2 | 5 2 | 5+6 2 |
| | 5 2 | 6+7 4 | 6+7 4 | 5+6 2 |
| | 8 2 | 6+7 4 | 5 2 | 5+6 2 |
| | 5 2 | 5 2 | 6+7 4 | 3+4 2 |
| | 6+7 4 | 5 2 | 5 2 | 5+6 2 |
| | 10+9 4 | 6+7 4 | 6+7 4 | 3+4 2 |
| | 5 2 | 6+7 4 | 5 2 | 3+4 2 |
| | 6+7 4 | 5 2 | 6+7 4 | 5+6 2 |
| | 5 2 | 5 2 | 6+7 4 | 5+6 2 |
| | 6+7 4 | 6+7 4 | 5 2 | 3+4 2 |
| | 5 2 | 8 2 | 6+7 4 | 3+4 2 |
| | 6+7 4 | 10+9 4 | 5 2 | 5+6 2 |
| | 6+7 4 | 8 2 | 5 2 | 3+4 2 |
| | 5 2 | 10+9 4 | 6+7 4 | 5+6 2 |
| | | | | 3+4 2 |
| | | | | 3+4 2 |
| | | | | 5+6 2 |
| | | | | 3+4 2 |
| | | | | 5+6 2 |
| | | | | 5+6 2 |
| | | | | 3+4 2 |
| | | | | 5+6 2 |
| | | | | 3+4 2 |
| | | | | 5+6 2 |
| | | | | 5+6 2 |
| | | | | 3+4 2 |
| | | | | 5+6 2 |
| | | | | 5+6 2 |
| | | | | 3+4 2 |
| | | | | 5+6 2 |
| | | | | 5+6 2 |
| | | | | 3+4 2 |
| | | | | 5+6 2 |
| | | | | 5+6 2 |
| | | | | 5+6 2 |
| | | | | 3+4 2 |
| | | | | 3+4 2 |
| 204.1.13 | 248608 | | 140 | Cr4Cu3O12Y1 |
| | -GM | -H | -P | -N |
| | 5 2 | 5 2 | 5 2 | 3+4 2 |
| | 8 2 | 8 2 | 5 2 | 5+6 2 |
| | 10+9 4 | 10+9 4 | 6+7 4 | 5+6 2 |
| | 5 2 | 8 2 | 5 2 | 5+6 2 |
| | 6+7 4 | 10+9 4 | 5 2 | 3+4 2 |

Continued on next column

**TABLE LXXXI– continued**

| TQC | ICSD | | NoE | SF |
|---|---|---|---|---|
| | 5 2 | 5 2 | 6+7 4 | 5+6 2 |
| | 6+7 4 | 6+7 4 | 6+7 4 | 3+4 2 |
| | 8 2 | 8 2 | 5 2 | 3+4 2 |
| | 10+9 4 | 10+9 4 | 6+7 4 | 5+6 2 |
| | 8 2 | 5 2 | 5 2 | 5+6 2 |
| | 10+9 4 | 6+7 4 | 6+7 4 | 3+4 2 |
| | 6+7 4 | 8 2 | 5 2 | 5+6 2 |
| | 5 2 | 10+9 4 | 6+7 4 | 3+4 2 |
| | 5 2 | 8 2 | 5 2 | 3+4 2 |
| | 6+7 4 | 8 2 | 6+7 4 | 5+6 2 |
| | 5 2 | 10+9 4 | 5 2 | 5+6 2 |
| | 6+7 4 | 5 2 | 6+7 4 | 5+6 2 |
| | 8 2 | 6+7 4 | 5 2 | 5+6 2 |
| | 10+9 4 | 10+9 4 | 6+7 4 | 3+4 2 |
| | 5 2 | 5 2 | 5 2 | 3+4 2 |
| | 6+7 4 | 6+7 4 | 6+7 4 | 3+4 2 |
| | 5 2 | 6+7 4 | 5 2 | 5+6 2 |
| | 8 2 | 6+7 4 | 6+7 4 | 3+4 2 |
| | 10+9 4 | 5 2 | 6+7 4 | 3+4 2 |
| | 6+7 4 | 5 2 | 5 2 | 5+6 2 |
| | 6+7 4 | 8 2 | 5 2 | 5+6 2 |
| | 5 2 | 10+9 4 | 6+7 4 | 3+4 2 |
| | 8 2 | 6+7 4 | 6+7 4 | 5+6 2 |
| | 10+9 4 | 5 2 | 5 2 | 3+4 2 |
| | 8 2 | 10+9 4 | 5 2 | 5+6 2 |
| | 10+9 4 | 8 2 | 6+7 4 | 3+4 2 |
| | 10+9 4 | 8 2 | 5 2 | 5+6 2 |
| | 8 2 | 10+9 4 | 6+7 4 | 3+4 2 |
| | 6+7 4 | 5 2 | 5 2 | 5+6 2 |
| | 5 2 | 6+7 4 | 6+7 4 | 5+6 2 |
| | 8 2 | 6+7 4 | 5 2 | 5+6 2 |
| | 5 2 | 5 2 | 6+7 4 | 3+4 2 |
| | 6+7 4 | 5 2 | 5 2 | 5+6 2 |
| | 10+9 4 | 6+7 4 | 6+7 4 | 3+4 2 |
| | 5 2 | 6+7 4 | 5 2 | 3+4 2 |
| | 6+7 4 | 5 2 | 6+7 4 | 5+6 2 |
| | 5 2 | 5 2 | 6+7 4 | 5+6 2 |
| | 6+7 4 | 6+7 4 | 5 2 | 3+4 2 |
| | 5 2 | 8 2 | 6+7 4 | 3+4 2 |
| | 6+7 4 | 10+9 4 | 5 2 | 5+6 2 |
| | 6+7 4 | 8 2 | 5 2 | 3+4 2 |
| | 5 2 | 10+9 4 | 6+7 4 | 5+6 2 |
| | | | | 3+4 2 |
| | | | | 3+4 2 |
| | | | | 5+6 2 |
| | | | | 3+4 2 |
| | | | | 5+6 2 |
| | | | | 5+6 2 |
| | | | | 3+4 2 |
| | | | | 5+6 2 |
| | | | | 3+4 2 |
| | | | | 5+6 2 |
| | | | | 5+6 2 |
| | | | | 3+4 2 |
| | | | | 5+6 2 |
| | | | | 5+6 2 |
| | | | | 3+4 2 |
| | | | | 5+6 2 |
| | | | | 5+6 2 |
| | | | | 3+4 2 |
| | | | | 5+6 2 |

Continued on next column

**TABLE LXXXI– continued**

| TQC | ICSD | | NoE | SF |
|---|---|---|---|---|
| | | | | 5+6 2 |
| | | | | 5+6 2 |
| | | | | 3+4 2 |
| | | | | 3+4 2 |
| 204.1.14 | 248607 | | 140 | Cr4Cu3La1O12 |
| | -GM | -H | -P | -N |
| | 5 2 | 5 2 | 5 2 | 3+4 2 |
| | 5 2 | 8 2 | 5 2 | 5+6 2 |
| | 8 2 | 10+9 4 | 6+7 4 | 3+4 2 |
| | 10+9 4 | 6+7 4 | 5 2 | 5+6 2 |
| | 6+7 4 | 5 2 | 5 2 | 5+6 2 |
| | 5 2 | 8 2 | 6+7 4 | 3+4 2 |
| | 6+7 4 | 10+9 4 | 6+7 4 | 3+4 2 |
| | 8 2 | 5 2 | 5 2 | 3+4 2 |
| | 10+9 4 | 6+7 4 | 6+7 4 | 5+6 2 |
| | 8 2 | 8 2 | 5 2 | 3+4 2 |
| | 10+9 4 | 10+9 4 | 6+7 4 | 5+6 2 |
| | 6+7 4 | 8 2 | 5 2 | 5+6 2 |
| | 5 2 | 10+9 4 | 6+7 4 | 3+4 2 |
| | 6+7 4 | 8 2 | 6+7 4 | 5+6 2 |
| | 5 2 | 8 2 | 5 2 | 5+6 2 |
| | 5 2 | 10+9 4 | 6+7 4 | 5+6 2 |
| | 6+7 4 | 6+7 4 | 5 2 | 5+6 2 |
| | 8 2 | 5 2 | 5 2 | 3+4 2 |
| | 10+9 4 | 6+7 4 | 6+7 4 | 5+6 2 |
| | 5 2 | 5 2 | 5 2 | 3+4 2 |
| | 6+7 4 | 10+9 4 | 6+7 4 | 5+6 2 |
| | 5 2 | 6+7 4 | 5 2 | 3+4 2 |
| | 6+7 4 | 6+7 4 | 6+7 4 | 3+4 2 |
| | 8 2 | 5 2 | 6+7 4 | 5+6 2 |
| | 10+9 4 | 5 2 | 5 2 | 3+4 2 |
| | 6+7 4 | 8 2 | 6+7 4 | 5+6 2 |
| | 5 2 | 10+9 4 | 5 2 | 3+4 2 |
| | 8 2 | 6+7 4 | 6+7 4 | 3+4 2 |
| | 10+9 4 | 5 2 | 5 2 | 5+6 2 |
| | 10+9 4 | 8 2 | 5 2 | 3+4 2 |
| | 8 2 | 10+9 4 | 6+7 4 | 5+6 2 |
| | 8 2 | 8 2 | 5 2 | 5+6 2 |
| | 10+9 4 | 10+9 4 | 6+7 4 | 3+4 2 |
| | 6+7 4 | 5 2 | 5 2 | 5+6 2 |
| | 5 2 | 6+7 4 | 6+7 4 | 5+6 2 |
| | 6+7 4 | 6+7 4 | 5 2 | 5+6 2 |
| | 5 2 | 5 2 | 6+7 4 | 3+4 2 |
| | 8 2 | 5 2 | 5 2 | 5+6 2 |
| | 10+9 4 | 6+7 4 | 6+7 4 | 3+4 2 |
| | 5 2 | 6+7 4 | 5 2 | 3+4 2 |
| | 6+7 4 | 5 2 | 6+7 4 | 5+6 2 |
| | 5 2 | 5 2 | 5 2 | 5+6 2 |
| | 6+7 4 | 6+7 4 | 6+7 4 | 3+4 2 |
| | 5 2 | 8 2 | 6+7 4 | 3+4 2 |
| | 6+7 4 | 10+9 4 | 5 2 | 3+4 2 |
| | 6+7 4 | 8 2 | 5 2 | 5+6 2 |
| | 5 2 | 10+9 4 | 6+7 4 | 3+4 2 |
| | | | | 5+6 2 |
| | | | | 3+4 2 |
| | | | | 5+6 2 |
| | | | | 3+4 2 |
| | | | | 5+6 2 |
| | | | | 5+6 2 |
| | | | | 3+4 2 |
| | | | | 5+6 2 |

Continued on next column

**TABLE LXXXI– continued**

| TQC | ICSD | | NoE | SF |
|---|---|---|---|---|
| | | | | 3+4 2 |
| | | | | 5+6 2 |
| | | | | 5+6 2 |
| | | | | 3+4 2 |
| | | | | 5+6 2 |
| | | | | 5+6 2 |
| | | | | 3+4 2 |
| | | | | 5+6 2 |
| | | | | 5+6 2 |
| | | | | 3+4 2 |
| | | | | 5+6 2 |
| | | | | 5+6 2 |
| | | | | 5+6 2 |
| | | | | 3+4 2 |
| | | | | 5+6 2 |
| 204.1.15 | 248606 | | 140 | Cr4Cu3La1O12 |
| | -GM | -H | -P | -N |
| | 5 2 | 5 2 | 5 2 | 3+4 2 |
| | 5 2 | 8 2 | 5 2 | 5+6 2 |
| | 8 2 | 10+9 4 | 6+7 4 | 3+4 2 |
| | 10+9 4 | 6+7 4 | 5 2 | 5+6 2 |
| | 6+7 4 | 5 2 | 5 2 | 5+6 2 |
| | 5 2 | 8 2 | 6+7 4 | 3+4 2 |
| | 6+7 4 | 10+9 4 | 6+7 4 | 3+4 2 |
| | 8 2 | 5 2 | 5 2 | 3+4 2 |
| | 10+9 4 | 6+7 4 | 6+7 4 | 5+6 2 |
| | 8 2 | 8 2 | 5 2 | 3+4 2 |
| | 10+9 4 | 10+9 4 | 6+7 4 | 5+6 2 |
| | 6+7 4 | 8 2 | 5 2 | 5+6 2 |
| | 5 2 | 10+9 4 | 6+7 4 | 3+4 2 |
| | 6+7 4 | 8 2 | 6+7 4 | 5+6 2 |
| | 5 2 | 8 2 | 5 2 | 5+6 2 |
| | 5 2 | 10+9 4 | 6+7 4 | 5+6 2 |
| | 6+7 4 | 6+7 4 | 5 2 | 5+6 2 |
| | 8 2 | 5 2 | 5 2 | 3+4 2 |
| | 10+9 4 | 6+7 4 | 6+7 4 | 5+6 2 |
| | 5 2 | 5 2 | 5 2 | 3+4 2 |
| | 6+7 4 | 10+9 4 | 6+7 4 | 5+6 2 |
| | 5 2 | 6+7 4 | 5 2 | 3+4 2 |
| | 6+7 4 | 6+7 4 | 6+7 4 | 3+4 2 |
| | 8 2 | 5 2 | 6+7 4 | 5+6 2 |
| | 10+9 4 | 5 2 | 5 2 | 3+4 2 |
| | 6+7 4 | 8 2 | 6+7 4 | 5+6 2 |
| | 5 2 | 10+9 4 | 5 2 | 3+4 2 |
| | 8 2 | 6+7 4 | 6+7 4 | 3+4 2 |
| | 10+9 4 | 5 2 | 5 2 | 5+6 2 |
| | 10+9 4 | 8 2 | 5 2 | 3+4 2 |
| | 8 2 | 10+9 4 | 6+7 4 | 5+6 2 |
| | 8 2 | 8 2 | 5 2 | 5+6 2 |
| | 10+9 4 | 10+9 4 | 6+7 4 | 3+4 2 |
| | 6+7 4 | 5 2 | 5 2 | 5+6 2 |
| | 5 2 | 6+7 4 | 6+7 4 | 5+6 2 |
| | 6+7 4 | 6+7 4 | 5 2 | 5+6 2 |
| | 5 2 | 5 2 | 6+7 4 | 3+4 2 |
| | 8 2 | 5 2 | 5 2 | 5+6 2 |
| | 10+9 4 | 6+7 4 | 6+7 4 | 3+4 2 |
| | 5 2 | 6+7 4 | 5 2 | 3+4 2 |
| | 6+7 4 | 5 2 | 6+7 4 | 5+6 2 |
| | 5 2 | 5 2 | 5 2 | 5+6 2 |
| | 6+7 4 | 6+7 4 | 6+7 4 | 3+4 2 |
| | 5 2 | 8 2 | 6+7 4 | 3+4 2 |

Continued on next column

**TABLE LXXXI– continued**

| TQC | ICSD | NoE | SF | |
|---|---|---|---|---|
| | 6+7 4 | 10+9 4 | 5 2 | 3+4 2 |
| | 6+7 4 | 8 2 | 5 2 | 5+6 2 |
| | 5 2 | 10+9 4 | 6+7 4 | 3+4 2 |
| | | | | 5+6 2 |
| | | | | 3+4 2 |
| | | | | 5+6 2 |
| | | | | 3+4 2 |
| | | | | 5+6 2 |
| | | | | 5+6 2 |
| | | | | 3+4 2 |
| | | | | 5+6 2 |
| | | | | 3+4 2 |
| | | | | 5+6 2 |
| | | | | 5+6 2 |
| | | | | 3+4 2 |
| | | | | 5+6 2 |
| | | | | 5+6 2 |
| | | | | 3+4 2 |
| | | | | 5+6 2 |
| | | | | 5+6 2 |
| | | | | 3+4 2 |
| | | | | 5+6 2 |
| | | | | 5+6 2 |
| | | | | 5+6 2 |
| | | | | 3+4 2 |
| | | | | 5+6 2 |
| 204.1.16 | 248605 | | 140 | Cr4Cu3La1O12 |
| | -GM | -H | -P | -N |
| | 5 2 | 5 2 | 5 2 | 3+4 2 |
| | 5 2 | 8 2 | 5 2 | 5+6 2 |
| | 8 2 | 10+9 4 | 6+7 4 | 3+4 2 |
| | 10+9 4 | 6+7 4 | 5 2 | 5+6 2 |
| | 6+7 4 | 5 2 | 5 2 | 5+6 2 |
| | 5 2 | 8 2 | 6+7 4 | 3+4 2 |
| | 6+7 4 | 10+9 4 | 6+7 4 | 3+4 2 |
| | 8 2 | 5 2 | 5 2 | 3+4 2 |
| | 10+9 4 | 6+7 4 | 6+7 4 | 5+6 2 |
| | 8 2 | 8 2 | 5 2 | 3+4 2 |
| | 10+9 4 | 10+9 4 | 6+7 4 | 5+6 2 |
| | 6+7 4 | 8 2 | 5 2 | 5+6 2 |
| | 5 2 | 10+9 4 | 6+7 4 | 3+4 2 |
| | 6+7 4 | 8 2 | 6+7 4 | 5+6 2 |
| | 5 2 | 8 2 | 5 2 | 5+6 2 |
| | 5 2 | 10+9 4 | 6+7 4 | 5+6 2 |
| | 6+7 4 | 6+7 4 | 5 2 | 5+6 2 |
| | 8 2 | 5 2 | 5 2 | 3+4 2 |
| | 10+9 4 | 6+7 4 | 6+7 4 | 5+6 2 |
| | 5 2 | 5 2 | 5 2 | 3+4 2 |
| | 6+7 4 | 10+9 4 | 6+7 4 | 5+6 2 |
| | 5 2 | 6+7 4 | 5 2 | 3+4 2 |
| | 6+7 4 | 6+7 4 | 6+7 4 | 3+4 2 |
| | 8 2 | 5 2 | 6+7 4 | 5+6 2 |
| | 10+9 4 | 5 2 | 5 2 | 3+4 2 |
| | 6+7 4 | 8 2 | 6+7 4 | 5+6 2 |
| | 5 2 | 10+9 4 | 5 2 | 3+4 2 |
| | 8 2 | 6+7 4 | 6+7 4 | 3+4 2 |
| | 10+9 4 | 5 2 | 5 2 | 5+6 2 |
| | 10+9 4 | 8 2 | 5 2 | 3+4 2 |
| | 8 2 | 10+9 4 | 6+7 4 | 5+6 2 |
| | 8 2 | 8 2 | 5 2 | 5+6 2 |
| | 10+9 4 | 10+9 4 | 6+7 4 | 3+4 2 |

Continued on next column

**TABLE LXXXI– continued**

| TQC | ICSD | NoE | SF | |
|---|---|---|---|---|
| | 6+7 4 | 5 2 | 5 2 | 5+6 2 |
| | 5 2 | 6+7 4 | 6+7 4 | 5+6 2 |
| | 6+7 4 | 6+7 4 | 5 2 | 5+6 2 |
| | 5 2 | 5 2 | 6+7 4 | 3+4 2 |
| | 8 2 | 5 2 | 5 2 | 5+6 2 |
| | 10+9 4 | 6+7 4 | 6+7 4 | 3+4 2 |
| | 5 2 | 6+7 4 | 5 2 | 3+4 2 |
| | 6+7 4 | 5 2 | 6+7 4 | 5+6 2 |
| | 5 2 | 5 2 | 5 2 | 5+6 2 |
| | 6+7 4 | 6+7 4 | 6+7 4 | 3+4 2 |
| | 5 2 | 8 2 | 6+7 4 | 3+4 2 |
| | 6+7 4 | 10+9 4 | 5 2 | 3+4 2 |
| | 6+7 4 | 8 2 | 5 2 | 5+6 2 |
| | 5 2 | 10+9 4 | 6+7 4 | 3+4 2 |
| | | | | 5+6 2 |
| | | | | 3+4 2 |
| | | | | 5+6 2 |
| | | | | 3+4 2 |
| | | | | 5+6 2 |
| | | | | 5+6 2 |
| | | | | 3+4 2 |
| | | | | 5+6 2 |
| | | | | 3+4 2 |
| | | | | 5+6 2 |
| | | | | 5+6 2 |
| | | | | 3+4 2 |
| | | | | 5+6 2 |
| | | | | 5+6 2 |
| | | | | 3+4 2 |
| | | | | 5+6 2 |
| | | | | 5+6 2 |
| | | | | 3+4 2 |
| | | | | 5+6 2 |
| | | | | 5+6 2 |
| | | | | 5+6 2 |
| | | | | 3+4 2 |
| | | | | 5+6 2 |
| 204.1.17 | 248604 | | 140 | Cr4Cu3La1O12 |
| | -GM | -H | -P | -N |
| | 5 2 | 5 2 | 5 2 | 3+4 2 |
| | 5 2 | 8 2 | 5 2 | 5+6 2 |
| | 8 2 | 10+9 4 | 6+7 4 | 3+4 2 |
| | 10+9 4 | 6+7 4 | 5 2 | 5+6 2 |
| | 6+7 4 | 5 2 | 5 2 | 5+6 2 |
| | 5 2 | 8 2 | 6+7 4 | 3+4 2 |
| | 6+7 4 | 10+9 4 | 6+7 4 | 3+4 2 |
| | 8 2 | 5 2 | 5 2 | 3+4 2 |
| | 10+9 4 | 6+7 4 | 6+7 4 | 5+6 2 |
| | 8 2 | 8 2 | 5 2 | 3+4 2 |
| | 10+9 4 | 10+9 4 | 6+7 4 | 5+6 2 |
| | 6+7 4 | 8 2 | 5 2 | 5+6 2 |
| | 5 2 | 10+9 4 | 6+7 4 | 3+4 2 |
| | 6+7 4 | 8 2 | 6+7 4 | 5+6 2 |
| | 5 2 | 8 2 | 5 2 | 5+6 2 |
| | 5 2 | 10+9 4 | 6+7 4 | 5+6 2 |
| | 6+7 4 | 6+7 4 | 5 2 | 5+6 2 |
| | 8 2 | 5 2 | 5 2 | 3+4 2 |
| | 10+9 4 | 6+7 4 | 6+7 4 | 5+6 2 |
| | 5 2 | 5 2 | 5 2 | 3+4 2 |
| | 6+7 4 | 10+9 4 | 6+7 4 | 5+6 2 |
| | 5 2 | 6+7 4 | 5 2 | 3+4 2 |

Continued on next column

**TABLE LXXXI– continued**

| TQC | ICSD | | NoE | SF |
|---|---|---|---|---|
| | 6+7 4 | 6+7 4 | 6+7 4 | 3+4 2 |
| | 8 2 | 5 2 | 6+7 4 | 5+6 2 |
| | 10+9 4 | 5 2 | 5 2 | 3+4 2 |
| | 6+7 4 | 8 2 | 6+7 4 | 5+6 2 |
| | 5 2 | 10+9 4 | 5 2 | 3+4 2 |
| | 8 2 | 6+7 4 | 6+7 4 | 3+4 2 |
| | 10+9 4 | 5 2 | 5 2 | 5+6 2 |
| | 10+9 4 | 8 2 | 5 2 | 3+4 2 |
| | 8 2 | 10+9 4 | 6+7 4 | 5+6 2 |
| | 8 2 | 8 2 | 5 2 | 5+6 2 |
| | 10+9 4 | 10+9 4 | 6+7 4 | 3+4 2 |
| | 6+7 4 | 5 2 | 5 2 | 5+6 2 |
| | 5 2 | 6+7 4 | 6+7 4 | 5+6 2 |
| | 6+7 4 | 6+7 4 | 5 2 | 5+6 2 |
| | 5 2 | 5 2 | 6+7 4 | 3+4 2 |
| | 8 2 | 5 2 | 5 2 | 5+6 2 |
| | 10+9 4 | 6+7 4 | 6+7 4 | 3+4 2 |
| | 5 2 | 6+7 4 | 5 2 | 3+4 2 |
| | 6+7 4 | 5 2 | 6+7 4 | 5+6 2 |
| | 5 2 | 5 2 | 5 2 | 5+6 2 |
| | 6+7 4 | 6+7 4 | 6+7 4 | 3+4 2 |
| | 5 2 | 8 2 | 6+7 4 | 3+4 2 |
| | 6+7 4 | 10+9 4 | 5 2 | 3+4 2 |
| | 6+7 4 | 8 2 | 5 2 | 5+6 2 |
| | 5 2 | 10+9 4 | 6+7 4 | 3+4 2 |
| | | | | 5+6 2 |
| | | | | 3+4 2 |
| | | | | 5+6 2 |
| | | | | 3+4 2 |
| | | | | 5+6 2 |
| | | | | 5+6 2 |
| | | | | 3+4 2 |
| | | | | 5+6 2 |
| | | | | 3+4 2 |
| | | | | 5+6 2 |
| | | | | 5+6 2 |
| | | | | 3+4 2 |
| | | | | 5+6 2 |
| | | | | 5+6 2 |
| | | | | 3+4 2 |
| | | | | 5+6 2 |
| | | | | 5+6 2 |
| | | | | 3+4 2 |
| | | | | 5+6 2 |
| | | | | 5+6 2 |
| | | | | 5+6 2 |
| | | | | 3+4 2 |
| | | | | 5+6 2 |
| 204.1.18 | 248603 | | 140 | Cr4Cu3La1O12 |
| | -GM | -H | -P | -N |
| | 5 2 | 5 2 | 5 2 | 3+4 2 |
| | 5 2 | 8 2 | 5 2 | 5+6 2 |
| | 8 2 | 10+9 4 | 6+7 4 | 3+4 2 |
| | 10+9 4 | 6+7 4 | 5 2 | 5+6 2 |
| | 6+7 4 | 5 2 | 5 2 | 5+6 2 |
| | 5 2 | 8 2 | 6+7 4 | 3+4 2 |
| | 6+7 4 | 10+9 4 | 6+7 4 | 3+4 2 |
| | 8 2 | 5 2 | 5 2 | 3+4 2 |
| | 10+9 4 | 6+7 4 | 6+7 4 | 5+6 2 |
| | 8 2 | 8 2 | 5 2 | 3+4 2 |
| | 10+9 4 | 10+9 4 | 6+7 4 | 5+6 2 |

Continued on next column

**TABLE LXXXI– continued**

| TQC | ICSD | | NoE | SF |
|---|---|---|---|---|
| | 6+7 4 | 8 2 | 5 2 | 5+6 2 |
| | 5 2 | 10+9 4 | 6+7 4 | 3+4 2 |
| | 6+7 4 | 8 2 | 6+7 4 | 5+6 2 |
| | 5 2 | 8 2 | 5 2 | 5+6 2 |
| | 5 2 | 10+9 4 | 6+7 4 | 5+6 2 |
| | 6+7 4 | 6+7 4 | 5 2 | 5+6 2 |
| | 8 2 | 5 2 | 5 2 | 3+4 2 |
| | 10+9 4 | 6+7 4 | 6+7 4 | 5+6 2 |
| | 5 2 | 5 2 | 5 2 | 3+4 2 |
| | 6+7 4 | 10+9 4 | 6+7 4 | 5+6 2 |
| | 5 2 | 6+7 4 | 5 2 | 3+4 2 |
| | 6+7 4 | 6+7 4 | 6+7 4 | 3+4 2 |
| | 8 2 | 5 2 | 6+7 4 | 3+4 2 |
| | 10+9 4 | 5 2 | 5 2 | 5+6 2 |
| | 6+7 4 | 8 2 | 6+7 4 | 5+6 2 |
| | 5 2 | 10+9 4 | 5 2 | 3+4 2 |
| | 8 2 | 6+7 4 | 6+7 4 | 3+4 2 |
| | 10+9 4 | 5 2 | 5 2 | 5+6 2 |
| | 10+9 4 | 8 2 | 5 2 | 3+4 2 |
| | 8 2 | 10+9 4 | 6+7 4 | 5+6 2 |
| | 8 2 | 8 2 | 5 2 | 5+6 2 |
| | 10+9 4 | 10+9 4 | 6+7 4 | 3+4 2 |
| | 6+7 4 | 5 2 | 5 2 | 5+6 2 |
| | 5 2 | 6+7 4 | 6+7 4 | 5+6 2 |
| | 6+7 4 | 6+7 4 | 5 2 | 5+6 2 |
| | 5 2 | 5 2 | 6+7 4 | 3+4 2 |
| | 8 2 | 5 2 | 5 2 | 5+6 2 |
| | 10+9 4 | 6+7 4 | 6+7 4 | 3+4 2 |
| | 5 2 | 6+7 4 | 5 2 | 3+4 2 |
| | 6+7 4 | 5 2 | 6+7 4 | 5+6 2 |
| | 5 2 | 5 2 | 5 2 | 5+6 2 |
| | 6+7 4 | 6+7 4 | 6+7 4 | 3+4 2 |
| | 5 2 | 8 2 | 6+7 4 | 3+4 2 |
| | 6+7 4 | 10+9 4 | 5 2 | 3+4 2 |
| | 6+7 4 | 8 2 | 5 2 | 5+6 2 |
| | 5 2 | 10+9 4 | 6+7 4 | 3+4 2 |
| | | | | 5+6 2 |
| | | | | 3+4 2 |
| | | | | 5+6 2 |
| | | | | 3+4 2 |
| | | | | 5+6 2 |
| | | | | 5+6 2 |
| | | | | 3+4 2 |
| | | | | 5+6 2 |
| | | | | 3+4 2 |
| | | | | 5+6 2 |
| | | | | 5+6 2 |
| | | | | 3+4 2 |
| | | | | 5+6 2 |
| | | | | 5+6 2 |
| | | | | 3+4 2 |
| | | | | 5+6 2 |
| | | | | 5+6 2 |
| | | | | 3+4 2 |
| | | | | 5+6 2 |
| | | | | 5+6 2 |
| | | | | 5+6 2 |
| | | | | 3+4 2 |
| | | | | 5+6 2 |
| 204.1.19 | 248601 | | 140 | Cr4Cu3La1O12 |
| | -GM | -H | -P | -N |

Continued on next column

**TABLE LXXXI– continued**

| TQC | ICSD | NoE | SF | |
|---|---|---|---|---|
| | 5 2 | 5 2 | 5 2 | 3+4 2 |
| | 5 2 | 8 2 | 5 2 | 5+6 2 |
| | 8 2 | 10+9 4 | 6+7 4 | 3+4 2 |
| | 10+9 4 | 6+7 4 | 5 2 | 5+6 2 |
| | 6+7 4 | 5 2 | 5 2 | 5+6 2 |
| | 5 2 | 8 2 | 6+7 4 | 3+4 2 |
| | 6+7 4 | 10+9 4 | 6+7 4 | 3+4 2 |
| | 8 2 | 5 2 | 5 2 | 3+4 2 |
| | 10+9 4 | 6+7 4 | 6+7 4 | 5+6 2 |
| | 8 2 | 8 2 | 5 2 | 3+4 2 |
| | 10+9 4 | 10+9 4 | 6+7 4 | 5+6 2 |
| | 6+7 4 | 8 2 | 5 2 | 5+6 2 |
| | 5 2 | 10+9 4 | 6+7 4 | 3+4 2 |
| | 6+7 4 | 8 2 | 6+7 4 | 5+6 2 |
| | 5 2 | 8 2 | 5 2 | 5+6 2 |
| | 5 2 | 10+9 4 | 6+7 4 | 5+6 2 |
| | 6+7 4 | 6+7 4 | 5 2 | 5+6 2 |
| | 8 2 | 10+9 4 | 5 2 | 3+4 2 |
| | 10+9 4 | 5 2 | 6+7 4 | 5+6 2 |
| | 5 2 | 6+7 4 | 5 2 | 3+4 2 |
| | 6+7 4 | 5 2 | 6+7 4 | 5+6 2 |
| | 5 2 | 6+7 4 | 5 2 | 3+4 2 |
| | 8 2 | 6+7 4 | 6+7 4 | 3+4 2 |
| | 10+9 4 | 5 2 | 6+7 4 | 3+4 2 |
| | 6+7 4 | 5 2 | 5 2 | 5+6 2 |
| | 6+7 4 | 8 2 | 6+7 4 | 5+6 2 |
| | 5 2 | 10+9 4 | 5 2 | 3+4 2 |
| | 8 2 | 6+7 4 | 6+7 4 | 3+4 2 |
| | 10+9 4 | 5 2 | 5 2 | 5+6 2 |
| | 8 2 | 8 2 | 5 2 | 3+4 2 |
| | 10+9 4 | 10+9 4 | 5 2 | 5+6 2 |
| | 8 2 | 8 2 | 6+7 4 | 5+6 2 |
| | 10+9 4 | 10+9 4 | 6+7 4 | 3+4 2 |
| | 6+7 4 | 5 2 | 5 2 | 5+6 2 |
| | 5 2 | 6+7 4 | 6+7 4 | 5+6 2 |
| | 6+7 4 | 6+7 4 | 5 2 | 5+6 2 |
| | 5 2 | 5 2 | 6+7 4 | 3+4 2 |
| | 8 2 | 5 2 | 5 2 | 5+6 2 |
| | 10+9 4 | 6+7 4 | 6+7 4 | 3+4 2 |
| | 5 2 | 6+7 4 | 5 2 | 3+4 2 |
| | 6+7 4 | 5 2 | 6+7 4 | 5+6 2 |
| | 5 2 | 5 2 | 5 2 | 5+6 2 |
| | 6+7 4 | 6+7 4 | 6+7 4 | 3+4 2 |
| | 5 2 | 8 2 | 6+7 4 | 3+4 2 |
| | 6+7 4 | 10+9 4 | 5 2 | 3+4 2 |
| | 6+7 4 | 8 2 | 5 2 | 5+6 2 |
| | 5 2 | 10+9 4 | 6+7 4 | 3+4 2 |
| | | | | 5+6 2 |
| | | | | 3+4 2 |
| | | | | 5+6 2 |
| | | | | 3+4 2 |
| | | | | 5+6 2 |
| | | | | 5+6 2 |
| | | | | 3+4 2 |
| | | | | 5+6 2 |
| | | | | 3+4 2 |
| | | | | 5+6 2 |
| | | | | 5+6 2 |
| | | | | 3+4 2 |
| | | | | 5+6 2 |
| | | | | 5+6 2 |

Continued on next column

**TABLE LXXXI– continued**

| TQC | ICSD | NoE | SF | |
|---|---|---|---|---|
| | | | | 3+4 2 |
| | | | | 5+6 2 |
| | | | | 5+6 2 |
| | | | | 3+4 2 |
| | | | | 5+6 2 |
| | | | | 5+6 2 |
| | | | | 5+6 2 |
| | | | | 3+4 2 |
| | | | | 5+6 2 |
| 204.1.20 | 248600 | | 140 | Cr4Cu3La1O12 |
| | -GM | -H | -P | -N |
| | 5 2 | 5 2 | 5 2 | 3+4 2 |
| | 5 2 | 8 2 | 5 2 | 5+6 2 |
| | 8 2 | 10+9 4 | 6+7 4 | 3+4 2 |
| | 10+9 4 | 6+7 4 | 5 2 | 5+6 2 |
| | 6+7 4 | 5 2 | 5 2 | 5+6 2 |
| | 5 2 | 8 2 | 6+7 4 | 3+4 2 |
| | 6+7 4 | 10+9 4 | 6+7 4 | 3+4 2 |
| | 8 2 | 5 2 | 5 2 | 3+4 2 |
| | 10+9 4 | 6+7 4 | 6+7 4 | 5+6 2 |
| | 8 2 | 8 2 | 5 2 | 3+4 2 |
| | 10+9 4 | 10+9 4 | 6+7 4 | 5+6 2 |
| | 6+7 4 | 8 2 | 5 2 | 5+6 2 |
| | 5 2 | 10+9 4 | 6+7 4 | 3+4 2 |
| | 6+7 4 | 8 2 | 6+7 4 | 5+6 2 |
| | 5 2 | 8 2 | 5 2 | 5+6 2 |
| | 5 2 | 10+9 4 | 6+7 4 | 5+6 2 |
| | 6+7 4 | 6+7 4 | 5 2 | 5+6 2 |
| | 8 2 | 10+9 4 | 5 2 | 3+4 2 |
| | 10+9 4 | 5 2 | 6+7 4 | 5+6 2 |
| | 5 2 | 6+7 4 | 5 2 | 3+4 2 |
| | 6+7 4 | 5 2 | 6+7 4 | 5+6 2 |
| | 5 2 | 6+7 4 | 5 2 | 3+4 2 |
| | 8 2 | 6+7 4 | 6+7 4 | 3+4 2 |
| | 10+9 4 | 5 2 | 6+7 4 | 3+4 2 |
| | 6+7 4 | 5 2 | 5 2 | 5+6 2 |
| | 6+7 4 | 8 2 | 6+7 4 | 5+6 2 |
| | 5 2 | 10+9 4 | 5 2 | 3+4 2 |
| | 8 2 | 6+7 4 | 6+7 4 | 3+4 2 |
| | 10+9 4 | 5 2 | 5 2 | 5+6 2 |
| | 8 2 | 8 2 | 5 2 | 3+4 2 |
| | 10+9 4 | 10+9 4 | 5 2 | 5+6 2 |
| | 8 2 | 8 2 | 6+7 4 | 5+6 2 |
| | 10+9 4 | 10+9 4 | 6+7 4 | 3+4 2 |
| | 6+7 4 | 5 2 | 5 2 | 5+6 2 |
| | 5 2 | 6+7 4 | 6+7 4 | 5+6 2 |
| | 6+7 4 | 6+7 4 | 5 2 | 5+6 2 |
| | 5 2 | 5 2 | 6+7 4 | 3+4 2 |
| | 8 2 | 5 2 | 5 2 | 5+6 2 |
| | 10+9 4 | 6+7 4 | 6+7 4 | 3+4 2 |
| | 5 2 | 6+7 4 | 5 2 | 3+4 2 |
| | 6+7 4 | 5 2 | 6+7 4 | 5+6 2 |
| | 5 2 | 5 2 | 5 2 | 5+6 2 |
| | 6+7 4 | 6+7 4 | 6+7 4 | 3+4 2 |
| | 5 2 | 8 2 | 6+7 4 | 3+4 2 |
| | 6+7 4 | 10+9 4 | 5 2 | 3+4 2 |
| | 6+7 4 | 8 2 | 5 2 | 5+6 2 |
| | 5 2 | 10+9 4 | 6+7 4 | 3+4 2 |
| | | | | 5+6 2 |
| | | | | 3+4 2 |
| | | | | 5+6 2 |

Continued on next column

**TABLE LXXXI– continued**

| TQC | ICSD | NoE | SF | |
|---|---|---|---|---|
| | | | | 3+4 2 |
| | | | | 5+6 2 |
| | | | | 5+6 2 |
| | | | | 3+4 2 |
| | | | | 5+6 2 |
| | | | | 3+4 2 |
| | | | | 5+6 2 |
| | | | | 5+6 2 |
| | | | | 3+4 2 |
| | | | | 5+6 2 |
| | | | | 5+6 2 |
| | | | | 3+4 2 |
| | | | | 5+6 2 |
| | | | | 5+6 2 |
| | | | | 3+4 2 |
| | | | | 5+6 2 |
| | | | | 5+6 2 |
| | | | | 5+6 2 |
| | | | | 3+4 2 |
| | | | | 5+6 2 |
| 204.1.21 | 248599 | | 140 | Cr4Cu3La1O12 |
| | -GM | -H | -P | -N |
| | 5 2 | 5 2 | 5 2 | 3+4 2 |
| | 5 2 | 8 2 | 5 2 | 5+6 2 |
| | 8 2 | 10+9 4 | 6+7 4 | 3+4 2 |
| | 10+9 4 | 6+7 4 | 5 2 | 5+6 2 |
| | 6+7 4 | 5 2 | 5 2 | 5+6 2 |
| | 5 2 | 8 2 | 6+7 4 | 3+4 2 |
| | 6+7 4 | 10+9 4 | 6+7 4 | 3+4 2 |
| | 8 2 | 5 2 | 5 2 | 3+4 2 |
| | 10+9 4 | 6+7 4 | 6+7 4 | 5+6 2 |
| | 8 2 | 8 2 | 5 2 | 3+4 2 |
| | 10+9 4 | 10+9 4 | 6+7 4 | 5+6 2 |
| | 6+7 4 | 8 2 | 5 2 | 5+6 2 |
| | 5 2 | 10+9 4 | 6+7 4 | 3+4 2 |
| | 6+7 4 | 8 2 | 6+7 4 | 5+6 2 |
| | 5 2 | 8 2 | 5 2 | 5+6 2 |
| | 5 2 | 10+9 4 | 6+7 4 | 5+6 2 |
| | 6+7 4 | 6+7 4 | 5 2 | 5+6 2 |
| | 8 2 | 10+9 4 | 5 2 | 3+4 2 |
| | 10+9 4 | 5 2 | 6+7 4 | 5+6 2 |
| | 5 2 | 6+7 4 | 5 2 | 3+4 2 |
| | 6+7 4 | 5 2 | 6+7 4 | 5+6 2 |
| | 5 2 | 6+7 4 | 5 2 | 3+4 2 |
| | 8 2 | 6+7 4 | 6+7 4 | 3+4 2 |
| | 10+9 4 | 5 2 | 6+7 4 | 3+4 2 |
| | 6+7 4 | 5 2 | 5 2 | 5+6 2 |
| | 6+7 4 | 8 2 | 6+7 4 | 5+6 2 |
| | 5 2 | 10+9 4 | 5 2 | 3+4 2 |
| | 8 2 | 6+7 4 | 6+7 4 | 3+4 2 |
| | 10+9 4 | 5 2 | 5 2 | 5+6 2 |
| | 8 2 | 8 2 | 5 2 | 3+4 2 |
| | 10+9 4 | 10+9 4 | 5 2 | 5+6 2 |
| | 8 2 | 8 2 | 6+7 4 | 5+6 2 |
| | 10+9 4 | 10+9 4 | 6+7 4 | 3+4 2 |
| | 6+7 4 | 5 2 | 5 2 | 5+6 2 |
| | 5 2 | 6+7 4 | 6+7 4 | 5+6 2 |
| | 6+7 4 | 6+7 4 | 5 2 | 5+6 2 |
| | 5 2 | 5 2 | 6+7 4 | 3+4 2 |
| | 8 2 | 5 2 | 5 2 | 5+6 2 |
| | 10+9 4 | 6+7 4 | 6+7 4 | 3+4 2 |

Continued on next column

**TABLE LXXXI– continued**

| TQC | ICSD | NoE | SF | |
|---|---|---|---|---|
| | 5 2 | 6+7 4 | 5 2 | 3+4 2 |
| | 6+7 4 | 5 2 | 6+7 4 | 5+6 2 |
| | 5 2 | 5 2 | 5 2 | 5+6 2 |
| | 6+7 4 | 6+7 4 | 6+7 4 | 3+4 2 |
| | 5 2 | 8 2 | 6+7 4 | 3+4 2 |
| | 6+7 4 | 10+9 4 | 5 2 | 3+4 2 |
| | 6+7 4 | 8 2 | 5 2 | 5+6 2 |
| | 5 2 | 10+9 4 | 6+7 4 | 3+4 2 |
| | | | | 5+6 2 |
| | | | | 3+4 2 |
| | | | | 5+6 2 |
| | | | | 3+4 2 |
| | | | | 5+6 2 |
| | | | | 5+6 2 |
| | | | | 3+4 2 |
| | | | | 5+6 2 |
| | | | | 3+4 2 |
| | | | | 5+6 2 |
| | | | | 5+6 2 |
| | | | | 3+4 2 |
| | | | | 5+6 2 |
| | | | | 5+6 2 |
| | | | | 3+4 2 |
| | | | | 5+6 2 |
| | | | | 5+6 2 |
| | | | | 3+4 2 |
| | | | | 5+6 2 |
| | | | | 5+6 2 |
| | | | | 5+6 2 |
| | | | | 3+4 2 |
| | | | | 5+6 2 |
| 204.1.22 | 248598 | | 140 | Cr4Cu3La1O12 |
| | -GM | -H | -P | -N |
| | 5 2 | 5 2 | 5 2 | 3+4 2 |
| | 5 2 | 8 2 | 5 2 | 5+6 2 |
| | 8 2 | 10+9 4 | 6+7 4 | 3+4 2 |
| | 10+9 4 | 6+7 4 | 5 2 | 5+6 2 |
| | 6+7 4 | 5 2 | 5 2 | 5+6 2 |
| | 5 2 | 8 2 | 6+7 4 | 3+4 2 |
| | 6+7 4 | 10+9 4 | 6+7 4 | 3+4 2 |
| | 8 2 | 5 2 | 5 2 | 3+4 2 |
| | 10+9 4 | 6+7 4 | 6+7 4 | 5+6 2 |
| | 8 2 | 8 2 | 5 2 | 3+4 2 |
| | 10+9 4 | 10+9 4 | 6+7 4 | 5+6 2 |
| | 6+7 4 | 8 2 | 5 2 | 5+6 2 |
| | 5 2 | 10+9 4 | 6+7 4 | 3+4 2 |
| | 6+7 4 | 8 2 | 6+7 4 | 5+6 2 |
| | 5 2 | 8 2 | 5 2 | 5+6 2 |
| | 5 2 | 10+9 4 | 6+7 4 | 5+6 2 |
| | 6+7 4 | 6+7 4 | 5 2 | 5+6 2 |
| | 8 2 | 10+9 4 | 5 2 | 3+4 2 |
| | 10+9 4 | 5 2 | 6+7 4 | 5+6 2 |
| | 5 2 | 6+7 4 | 5 2 | 3+4 2 |
| | 6+7 4 | 5 2 | 6+7 4 | 5+6 2 |
| | 5 2 | 6+7 4 | 5 2 | 3+4 2 |
| | 8 2 | 6+7 4 | 6+7 4 | 3+4 2 |
| | 10+9 4 | 5 2 | 6+7 4 | 3+4 2 |
| | 6+7 4 | 5 2 | 5 2 | 5+6 2 |
| | 6+7 4 | 8 2 | 6+7 4 | 5+6 2 |
| | 5 2 | 10+9 4 | 5 2 | 3+4 2 |
| | 8 2 | 6+7 4 | 6+7 4 | 3+4 2 |

Continued on next column

**TABLE LXXXI– continued**

| TQC | ICSD | | NoE | SF |
|---|---|---|---|---|
| | 10+9 4 | 5 2 | 5 2 | 5+6 2 |
| | 8 2 | 8 2 | 5 2 | 3+4 2 |
| | 10+9 4 | 10+9 4 | 5 2 | 5+6 2 |
| | 8 2 | 8 2 | 6+7 4 | 3+4 2 |
| | 10+9 4 | 10+9 4 | 6+7 4 | 5+6 2 |
| | 6+7 4 | 5 2 | 5 2 | 5+6 2 |
| | 5 2 | 6+7 4 | 6+7 4 | 5+6 2 |
| | 6+7 4 | 6+7 4 | 5 2 | 5+6 2 |
| | 5 2 | 5 2 | 6+7 4 | 3+4 2 |
| | 8 2 | 5 2 | 5 2 | 5+6 2 |
| | 10+9 4 | 6+7 4 | 6+7 4 | 3+4 2 |
| | 5 2 | 6+7 4 | 5 2 | 3+4 2 |
| | 6+7 4 | 5 2 | 6+7 4 | 5+6 2 |
| | 5 2 | 5 2 | 5 2 | 5+6 2 |
| | 6+7 4 | 6+7 4 | 6+7 4 | 3+4 2 |
| | 5 2 | 8 2 | 6+7 4 | 3+4 2 |
| | 6+7 4 | 10+9 4 | 5 2 | 3+4 2 |
| | 6+7 4 | 8 2 | 5 2 | 5+6 2 |
| | 5 2 | 10+9 4 | 6+7 4 | 3+4 2 |
| | | | | 5+6 2 |
| | | | | 3+4 2 |
| | | | | 5+6 2 |
| | | | | 3+4 2 |
| | | | | 5+6 2 |
| | | | | 5+6 2 |
| | | | | 3+4 2 |
| | | | | 5+6 2 |
| | | | | 3+4 2 |
| | | | | 5+6 2 |
| | | | | 5+6 2 |
| | | | | 3+4 2 |
| | | | | 5+6 2 |
| | | | | 5+6 2 |
| | | | | 3+4 2 |
| | | | | 5+6 2 |
| | | | | 5+6 2 |
| | | | | 3+4 2 |
| | | | | 5+6 2 |
| | | | | 5+6 2 |
| | | | | 5+6 2 |
| | | | | 3+4 2 |
| | | | | 5+6 2 |
| 204.1.23 | 174551 | | 98 | Ba1Ge12Pt4 |
| | -GM | -H | -P | -N |
| | 5 2 | 5 2 | 5 2 | 3+4 2 |
| | 8 2 | 8 2 | 5 2 | 5+6 2 |
| | 10+9 4 | 10+9 4 | 6+7 4 | 5+6 2 |
| | 5 2 | 6+7 4 | 5 2 | 5+6 2 |
| | 6+7 4 | 8 2 | 6+7 4 | 5+6 2 |
| | 8 2 | 10+9 4 | 5 2 | 3+4 2 |
| | 10+9 4 | 5 2 | 5 2 | 5+6 2 |
| | 6+7 4 | 8 2 | 6+7 4 | 3+4 2 |
| | 5 2 | 10+9 4 | 6+7 4 | 3+4 2 |
| | 8 2 | 5 2 | 5 2 | 5+6 2 |
| | 10+9 4 | 6+7 4 | 6+7 4 | 5+6 2 |
| | 6+7 4 | 10+9 4 | 6+7 4 | 3+4 2 |
| | 5 2 | 8 2 | 5 2 | 3+4 2 |
| | 6+7 4 | 10+9 4 | 6+7 4 | 5+6 2 |
| | 5 2 | 8 2 | 5 2 | 3+4 2 |
| | 6+7 4 | 8 2 | 6+7 4 | 5+6 2 |
| | 5 2 | 10+9 4 | 5 2 | 5+6 2 |
| | 6+7 4 | 8 2 | 6+7 4 | 3+4 2 |
| | 5 2 | 5 2 | 5 2 | 3+4 2 |
| | 5 2 | 10+9 4 | 6+7 4 | 5+6 2 |
| | 6+7 4 | 10+9 4 | 5 2 | 3+4 2 |
| | 6+7 4 | 8 2 | 6+7 4 | 3+4 2 |
| | 5 2 | 10+9 4 | 5 2 | 5+6 2 |
| | 6+7 4 | 8 2 | 6+7 4 | 5+6 2 |
| | 5 2 | 6+7 4 | 5 2 | 3+4 2 |
| | 8 2 | 10+9 4 | 6+7 4 | 5+6 2 |
| | 10+9 4 | 8 2 | 5 2 | 5+6 2 |
| | 6+7 4 | 6+7 4 | 6+7 4 | 3+4 2 |
| | 8 2 | 5 2 | 5 2 | 5+6 2 |
| | 10+9 4 | 5 2 | 6+7 4 | 3+4 2 |
| | 5 2 | 8 2 | 5 2 | 5+6 2 |
| | 8 2 | 10+9 4 | 6+7 4 | 3+4 2 |
| | 10+9 4 | 6+7 4 | 5 2 | 3+4 2 |
| | | | | 5+6 2 |
| | | | | 3+4 2 |
| | | | | 5+6 2 |
| | | | | 5+6 2 |
| | | | | 3+4 2 |
| | | | | 5+6 2 |
| | | | | 3+4 2 |
| | | | | 3+4 2 |
| | | | | 5+6 2 |
| | | | | 3+4 2 |
| | | | | 3+4 2 |
| | | | | 5+6 2 |
| | | | | 5+6 2 |
| | | | | 5+6 2 |
| | | | | 3+4 2 |
| | | | | 3+4 2 |
| 204.1.24 | 160797 | | 98 | Ba1Ge12Pt4 |
| | -GM | -H | -P | -N |
| | 5 2 | 5 2 | 5 2 | 3+4 2 |
| | 8 2 | 8 2 | 5 2 | 5+6 2 |
| | 10+9 4 | 10+9 4 | 6+7 4 | 5+6 2 |
| | 5 2 | 6+7 4 | 5 2 | 5+6 2 |
| | 6+7 4 | 8 2 | 6+7 4 | 5+6 2 |
| | 8 2 | 10+9 4 | 5 2 | 3+4 2 |
| | 10+9 4 | 5 2 | 5 2 | 5+6 2 |
| | 6+7 4 | 8 2 | 6+7 4 | 3+4 2 |
| | 5 2 | 10+9 4 | 6+7 4 | 3+4 2 |
| | 8 2 | 5 2 | 5 2 | 5+6 2 |
| | 10+9 4 | 6+7 4 | 6+7 4 | 5+6 2 |
| | 6+7 4 | 10+9 4 | 6+7 4 | 3+4 2 |
| | 5 2 | 8 2 | 5 2 | 3+4 2 |
| | 6+7 4 | 10+9 4 | 6+7 4 | 5+6 2 |
| | 5 2 | 8 2 | 5 2 | 3+4 2 |
| | 6+7 4 | 8 2 | 6+7 4 | 5+6 2 |
| | 5 2 | 10+9 4 | 5 2 | 5+6 2 |
| | 6+7 4 | 8 2 | 6+7 4 | 3+4 2 |
| | 5 2 | 5 2 | 5 2 | 3+4 2 |
| | 5 2 | 10+9 4 | 6+7 4 | 5+6 2 |
| | 6+7 4 | 10+9 4 | 5 2 | 3+4 2 |
| | 6+7 4 | 8 2 | 6+7 4 | 3+4 2 |
| | 5 2 | 10+9 4 | 5 2 | 5+6 2 |
| | 6+7 4 | 8 2 | 6+7 4 | 5+6 2 |
| | 5 2 | 6+7 4 | 5 2 | 3+4 2 |
| | 8 2 | 10+9 4 | 6+7 4 | 5+6 2 |
| | 10+9 4 | 8 2 | 5 2 | 5+6 2 |

Continued on next column

**TABLE LXXXI– continued**

| TQC | ICSD | | NoE | SF |
|---|---|---|---|---|
| | 6+7 4 | 6+7 4 | 6+7 4 | 3+4 2 |
| | 8 2 | 5 2 | 5 2 | 5+6 2 |
| | 10+9 4 | 5 2 | 6+7 4 | 3+4 2 |
| | 5 2 | 8 2 | 5 2 | 5+6 2 |
| | 8 2 | 10+9 4 | 6+7 4 | 3+4 2 |
| | 10+9 4 | 6+7 4 | 5 2 | 3+4 2 |
| | | | | 5+6 2 |
| | | | | 3+4 2 |
| | | | | 5+6 2 |
| | | | | 5+6 2 |
| | | | | 3+4 2 |
| | | | | 5+6 2 |
| | | | | 3+4 2 |
| | | | | 3+4 2 |
| | | | | 5+6 2 |
| | | | | 3+4 2 |
| | | | | 3+4 2 |
| | | | | 5+6 2 |
| | | | | 5+6 2 |
| | | | | 5+6 2 |
| | | | | 3+4 2 |
| | | | | 3+4 2 |

The End

TABLE LXXXII: The computed irreducible representations for the valence bands for the compounds of SEBR and NLC in space group 205. See the list of tables.

| TQC | ICSD | | NoE | SF |
|---|---|---|---|---|
| 205.2.1 | 77109 | | 80 | Pd1Sb2 |
| | -GM | -R | -M | -X |
| | 5 2 | 8+9 2 | 3+3 4 | 3+4 4 |
| | 6+7 4 | 11+11 6 | 4+4 4 | 3+4 4 |
| | 5 2 | 10+10 6 | 3+3 4 | 3+4 4 |
| | 8 2 | 5+6 2 | 4+4 4 | 3+4 4 |
| | 10+9 4 | 5+6 2 | 4+4 4 | 3+4 4 |
| | 8 2 | 10+10 6 | 3+3 4 | 3+4 4 |
| | 6+7 4 | 4+4 2 | 4+4 4 | 3+4 4 |
| | 5 2 | 10+10 6 | 3+3 4 | 3+4 4 |
| | 6+7 4 | 10+10 6 | 3+3 4 | 3+4 4 |
| | 5 2 | 5+6 2 | 4+4 4 | 3+4 4 |
| | 6+7 4 | 4+4 2 | 4+4 4 | 3+4 4 |
| | 5 2 | 10+10 6 | 3+3 4 | 3+4 4 |
| | 6+7 4 | 5+6 2 | 3+3 4 | 3+4 4 |
| | 8 2 | 10+10 6 | 4+4 4 | 3+4 4 |
| | 10+9 4 | 8+9 2 | 4+4 4 | 3+4 4 |
| | 5 2 | 11+11 6 | 3+3 4 | 3+4 4 |
| | 6+7 4 | 11+11 6 | 3+3 4 | 3+4 4 |
| | 6+7 4 | 7+7 2 | 4+4 4 | 3+4 4 |
| | 5 2 | 8+9 2 | 4+4 4 | 3+4 4 |
| | 6+7 4 | 11+11 6 | 3+3 4 | 3+4 4 |
| | 5 2 | | | |
| | 10+9 4 | | | |
| | 5 2 | | | |
| | 6+7 4 | | | |
| | 5 2 | | | |
| | 6+7 4 | | | |
| | 5 2 | | | |
| 205.2.2 | 300225 | | 80 | Os1Te2 |
| | -GM | -R | -M | -X |

Continued on next column

**TABLE LXXXII– continued**

| TQC | ICSD | | NoE | SF |
|---|---|---|---|---|
| | 5 2 | 8+9 2 | 3+3 4 | 3+4 4 |
| | 6+7 4 | 11+11 6 | 4+4 4 | 3+4 4 |
| | 5 2 | 10+10 6 | 3+3 4 | 3+4 4 |
| | 8 2 | 5+6 2 | 4+4 4 | 3+4 4 |
| | 10+9 4 | 5+6 2 | 4+4 4 | 3+4 4 |
| | 8 2 | 10+10 6 | 3+3 4 | 3+4 4 |
| | 6+7 4 | 10+10 6 | 3+3 4 | 3+4 4 |
| | 5 2 | 4+4 2 | 4+4 4 | 3+4 4 |
| | 5 2 | 8+9 2 | 4+4 4 | 3+4 4 |
| | 6+7 4 | 11+11 6 | 3+3 4 | 3+4 4 |
| | 8 2 | 11+11 6 | 4+4 4 | 3+4 4 |
| | 10+9 4 | 10+10 6 | 3+3 4 | 3+4 4 |
| | 6+7 4 | 5+6 2 | 3+3 4 | 3+4 4 |
| | 10+9 4 | 10+10 6 | 4+4 4 | 3+4 4 |
| | 5 2 | 8+9 2 | 3+3 4 | 3+4 4 |
| | 6+7 4 | 7+7 2 | 4+4 4 | 3+4 4 |
| | 5 2 | 4+4 2 | 4+4 4 | 3+4 4 |
| | 8 2 | 11+11 6 | 3+3 4 | 3+4 4 |
| | 5 2 | 5+6 2 | 4+4 4 | 3+4 4 |
| | 6+7 4 | 10+10 6 | 3+3 4 | 3+4 4 |
| | 6+7 4 | | | |
| | 10+9 4 | | | |
| | 5 2 | | | |
| | 6+7 4 | | | |
| | 5 2 | | | |
| | 8 2 | | | |
| | 6+7 4 | | | |
| 205.2.3 | 86354 | | 88 | Ni1S2 |
| | -GM | -R | -M | -X |
| | 5 2 | 8+9 2 | 3+3 4 | 3+4 4 |
| | 5 2 | 11+11 6 | 4+4 4 | 3+4 4 |
| | 6+7 4 | 5+6 2 | 3+3 4 | 3+4 4 |
| | 8 2 | 10+10 6 | 4+4 4 | 3+4 4 |
| | 10+9 4 | 5+6 2 | 4+4 4 | 3+4 4 |
| | 8 2 | 10+10 6 | 3+3 4 | 3+4 4 |
| | 5 2 | 8+9 2 | 3+3 4 | 3+4 4 |
| | 6+7 4 | 11+11 6 | 4+4 4 | 3+4 4 |
| | 8 2 | 10+10 6 | 3+3 4 | 3+4 4 |
| | 10+9 4 | 4+4 2 | 4+4 4 | 3+4 4 |
| | 5 2 | 11+11 6 | 3+3 4 | 3+4 4 |
| | 10+9 4 | 7+7 2 | 4+4 4 | 3+4 4 |
| | 5 2 | 8+9 2 | 4+4 4 | 3+4 4 |
| | 6+7 4 | 11+11 6 | 3+3 4 | 3+4 4 |
| | 6+7 4 | 10+10 6 | 3+3 4 | 3+4 4 |
| | 5 2 | 4+4 2 | 4+4 4 | 3+4 4 |
| | 6+7 4 | 5+6 2 | 4+4 4 | 3+4 4 |
| | 8 2 | 10+10 6 | 3+3 4 | 3+4 4 |
| | 10+9 4 | 5+6 2 | 4+4 4 | 3+4 4 |
| | 5 2 | 10+10 6 | 3+3 4 | 3+4 4 |
| | 6+7 4 | 10+10 6 | 4+4 4 | 3+4 4 |
| | 6+7 4 | 5+6 2 | 3+3 4 | 3+4 4 |
| | 5 2 | | | |
| | 6+7 4 | | | |
| | 6+7 4 | | | |
| | 5 2 | | | |
| | 5 2 | | | |
| | 6+7 4 | | | |
| | 5 2 | | | |
| | 8 2 | | | |
| 205.2.4 | 68169 | | 88 | Ni1S2 |
| | -GM | -R | -M | -X |

Continued on next column

**TABLE LXXXII– continued**

| TQC | ICSD | | NoE | SF |
|---|---|---|---|---|
| | 5 2 | 8+9 2 | 3+3 4 | 3+4 4 |
| | 5 2 | 11+11 6 | 4+4 4 | 3+4 4 |
| | 6+7 4 | 5+6 2 | 3+3 4 | 3+4 4 |
| | 8 2 | 10+10 6 | 4+4 4 | 3+4 4 |
| | 10+9 4 | 5+6 2 | 4+4 4 | 3+4 4 |
| | 8 2 | 10+10 6 | 3+3 4 | 3+4 4 |
| | 5 2 | 8+9 2 | 3+3 4 | 3+4 4 |
| | 6+7 4 | 11+11 6 | 4+4 4 | 3+4 4 |
| | 8 2 | 10+10 6 | 4+4 4 | 3+4 4 |
| | 10+9 4 | 4+4 2 | 3+3 4 | 3+4 4 |
| | 5 2 | 11+11 6 | 3+3 4 | 3+4 4 |
| | 10+9 4 | 7+7 2 | 4+4 4 | 3+4 4 |
| | 5 2 | 8+9 2 | 4+4 4 | 3+4 4 |
| | 6+7 4 | 11+11 6 | 3+3 4 | 3+4 4 |
| | 6+7 4 | 10+10 6 | 3+3 4 | 3+4 4 |
| | 5 2 | 4+4 2 | 4+4 4 | 3+4 4 |
| | 6+7 4 | 5+6 2 | 4+4 4 | 3+4 4 |
| | 8 2 | 10+10 6 | 3+3 4 | 3+4 4 |
| | 10+9 4 | 5+6 2 | 4+4 4 | 3+4 4 |
| | 5 2 | 10+10 6 | 3+3 4 | 3+4 4 |
| | 6+7 4 | 10+10 6 | 4+4 4 | 3+4 4 |
| | 6+7 4 | 5+6 2 | 3+3 4 | 3+4 4 |
| | 5 2 | | | |
| | 6+7 4 | | | |
| | 6+7 4 | | | |
| | 5 2 | | | |
| | 5 2 | | | |
| | 6+7 4 | | | |
| | 5 2 | | | |
| | 8 2 | | | |
| 205.2.5 | 68168 | | 88 | Ni1S2 |
| | -GM | -R | -M | -X |
| | 5 2 | 8+9 2 | 3+3 4 | 3+4 4 |
| | 5 2 | 11+11 6 | 4+4 4 | 3+4 4 |
| | 6+7 4 | 5+6 2 | 3+3 4 | 3+4 4 |
| | 8 2 | 10+10 6 | 4+4 4 | 3+4 4 |
| | 10+9 4 | 5+6 2 | 4+4 4 | 3+4 4 |
| | 8 2 | 10+10 6 | 3+3 4 | 3+4 4 |
| | 5 2 | 8+9 2 | 3+3 4 | 3+4 4 |
| | 6+7 4 | 11+11 6 | 4+4 4 | 3+4 4 |
| | 8 2 | 10+10 6 | 4+4 4 | 3+4 4 |
| | 10+9 4 | 4+4 2 | 3+3 4 | 3+4 4 |
| | 5 2 | 11+11 6 | 3+3 4 | 3+4 4 |
| | 10+9 4 | 7+7 2 | 4+4 4 | 3+4 4 |
| | 5 2 | 8+9 2 | 4+4 4 | 3+4 4 |
| | 6+7 4 | 11+11 6 | 3+3 4 | 3+4 4 |
| | 6+7 4 | 10+10 6 | 3+3 4 | 3+4 4 |
| | 5 2 | 4+4 2 | 4+4 4 | 3+4 4 |
| | 6+7 4 | 5+6 2 | 4+4 4 | 3+4 4 |
| | 8 2 | 10+10 6 | 3+3 4 | 3+4 4 |
| | 10+9 4 | 5+6 2 | 4+4 4 | 3+4 4 |
| | 5 2 | 10+10 6 | 3+3 4 | 3+4 4 |
| | 6+7 4 | 10+10 6 | 4+4 4 | 3+4 4 |
| | 6+7 4 | 5+6 2 | 3+3 4 | 3+4 4 |
| | 5 2 | | | |
| | 6+7 4 | | | |
| | 6+7 4 | | | |
| | 5 2 | | | |
| | 5 2 | | | |
| | 6+7 4 | | | |
| | 5 2 | | | |

Continued on next column

**TABLE LXXXII– continued**

| TQC | ICSD | | NoE | SF |
|---|---|---|---|---|
| | 8 2 | | | |
| 205.2.6 | 68167 | | 88 | Ni1S2 |
| | -GM | -R | -M | -X |
| | 5 2 | 8+9 2 | 3+3 4 | 3+4 4 |
| | 5 2 | 11+11 6 | 4+4 4 | 3+4 4 |
| | 6+7 4 | 5+6 2 | 3+3 4 | 3+4 4 |
| | 8 2 | 10+10 6 | 4+4 4 | 3+4 4 |
| | 10+9 4 | 5+6 2 | 4+4 4 | 3+4 4 |
| | 8 2 | 10+10 6 | 3+3 4 | 3+4 4 |
| | 5 2 | 8+9 2 | 3+3 4 | 3+4 4 |
| | 6+7 4 | 11+11 6 | 4+4 4 | 3+4 4 |
| | 8 2 | 10+10 6 | 4+4 4 | 3+4 4 |
| | 10+9 4 | 4+4 2 | 3+3 4 | 3+4 4 |
| | 5 2 | 11+11 6 | 3+3 4 | 3+4 4 |
| | 10+9 4 | 7+7 2 | 4+4 4 | 3+4 4 |
| | 5 2 | 8+9 2 | 4+4 4 | 3+4 4 |
| | 6+7 4 | 11+11 6 | 3+3 4 | 3+4 4 |
| | 6+7 4 | 10+10 6 | 3+3 4 | 3+4 4 |
| | 5 2 | 4+4 2 | 4+4 4 | 3+4 4 |
| | 6+7 4 | 5+6 2 | 4+4 4 | 3+4 4 |
| | 8 2 | 10+10 6 | 3+3 4 | 3+4 4 |
| | 10+9 4 | 5+6 2 | 4+4 4 | 3+4 4 |
| | 5 2 | 10+10 6 | 3+3 4 | 3+4 4 |
| | 6+7 4 | 10+10 6 | 4+4 4 | 3+4 4 |
| | 6+7 4 | 5+6 2 | 3+3 4 | 3+4 4 |
| | 5 2 | | | |
| | 6+7 4 | | | |
| | 6+7 4 | | | |
| | 5 2 | | | |
| | 5 2 | | | |
| | 6+7 4 | | | |
| | 5 2 | | | |
| | 8 2 | | | |

The End

TABLE LXXXIII: The computed irreducible representations for the valence bands for the compounds of SEBR and NLC in space group 216. See the list of tables.

| TQC | ICSD | | NoE | SF |
|---|---|---|---|---|
| 216.2.1 | 260984 | | 64 | Ni5U1 |
| | -GM | -X | -L | -W |
| | 6 2 | 7 2 | 6 2 | 6 1 |
| | 7 2 | 6 2 | 6 2 | 8 1 |
| | 8 4 | 6 2 | 6 2 | 7 1 |
| | 6 2 | 7 2 | 4+5 2 | 5 1 |
| | 6 2 | 7 2 | 6 2 | 6 1 |
| | 7 2 | 6 2 | 6 2 | 8 1 |
| | 8 4 | 6 2 | 4+5 2 | 7 1 |
| | 7 2 | 7 2 | 6 2 | 5 1 |
| | 8 4 | 6 2 | 6 2 | 7 1 |
| | 8 4 | 7 2 | 6 2 | 5 1 |
| | 8 4 | 7 2 | 4+5 2 | 7 1 |
| | 8 4 | 6 2 | 6 2 | 5 1 |
| | 7 2 | 6 2 | 4+5 2 | 6 1 |
| | 8 4 | 7 2 | 6 2 | 8 1 |
| | 6 2 | 7 2 | 6 2 | 8 1 |
| | 8 4 | 7 2 | 6 2 | 6 1 |
| | 7 2 | 6 2 | 4+5 2 | 8 1 |
| | 8 4 | 6 2 | 6 2 | 6 1 |

Continued on next column

**TABLE LXXXIII– continued**

| TQC | ICSD | NoE | SF |
|---|---|---|---|
| | 6 2 6 2 4+5 2 7 1 | | |
| | 8 4 7 2 6 2 5 1 | | |
| | 7 2 6 2 4+5 2 5 1 | | |
| | 6 2 7 2 6 2 8 1 | | |
| | 7 2 6 2 7 1 | | |
| | 7 2 4+5 2 6 1 | | |
| | 6 2 6 2 6 1 | | |
| | 7 2 6 2 8 1 | | |
| | 6 2 4+5 2 5 1 | | |
| | 6 2 6 2 7 1 | | |
| | 7 2 4+5 2 6 1 | | |
| | 7 2 6 2 5 1 | | |
| | 6 2 6 2 7 1 | | |
| | 6 2 6 2 8 1 | | |
| | 8 1 | | |
| | 7 1 | | |
| | 6 1 | | |
| | 5 1 | | |
| | 5 1 | | |
| | 8 1 | | |
| | 7 1 | | |
| | 6 1 | | |
| | 7 1 | | |
| | 6 1 | | |
| | 8 1 | | |
| | 6 1 | | |
| | 5 1 | | |
| | 7 1 | | |
| | 5 1 | | |
| | 6 1 | | |
| | 8 1 | | |
| | 8 1 | | |
| | 6 1 | | |
| | 7 1 | | |
| | 5 1 | | |
| | 7 1 | | |
| | 6 1 | | |
| | 7 1 | | |
| | 8 1 | | |
| | 5 1 | | |
| | 7 1 | | |
| | 5 1 | | |
| | 6 1 | | |
| | 7 1 | | |
| | 5 1 | | |
| | 8 1 | | |
| 216.2.2 | 54390 | 58 | Ni4Sn1U1 |
| | -GM -X -L -W | | |
| | 6 2 7 2 6 2 6 1 | | |
| | 7 2 6 2 6 2 8 1 | | |
| | 8 4 6 2 6 2 5 1 | | |
| | 6 2 7 2 4+5 2 7 1 | | |
| | 6 2 6 2 6 2 8 1 | | |
| | 7 2 7 2 6 2 6 1 | | |
| | 8 4 6 2 6 2 5 1 | | |
| | 7 2 6 2 4+5 2 7 1 | | |
| | 8 4 7 2 6 2 5 1 | | |
| | 8 4 7 2 6 2 8 1 | | |
| | 8 4 7 2 4+5 2 7 1 | | |
| | 6 2 6 2 6 2 5 1 | | |
| | 7 2 6 2 6 2 6 1 | | |

Continued on next column

**TABLE LXXXIII– continued**

| TQC | ICSD | NoE | SF |
|---|---|---|---|
| | 8 4 7 2 4+5 2 8 1 | | |
| | 8 4 6 2 6 2 7 1 | | |
| | 6 2 7 2 6 2 6 1 | | |
| | 7 2 6 2 4+5 2 7 1 | | |
| | 8 4 7 2 6 2 5 1 | | |
| | 6 2 6 2 6 2 7 1 | | |
| | 8 4 7 2 4+5 2 6 1 | | |
| | 7 2 6 2 6 1 | | |
| | 6 2 4+5 2 8 1 | | |
| | 6 2 6 2 5 1 | | |
| | 6 2 6 2 8 1 | | |
| | 7 2 6 2 6 1 | | |
| | 6 2 4+5 2 8 1 | | |
| | 7 2 6 2 5 1 | | |
| | 7 2 4+5 2 7 1 | | |
| | 7 2 6 2 5 1 | | |
| | 6 1 | | |
| | 7 1 | | |
| | 8 1 | | |
| | 6 1 | | |
| | 8 1 | | |
| | 7 1 | | |
| | 5 1 | | |
| | 7 1 | | |
| | 8 1 | | |
| | 5 1 | | |
| | 6 1 | | |
| | 6 1 | | |
| | 7 1 | | |
| | 5 1 | | |
| | 6 1 | | |
| | 8 1 | | |
| | 8 1 | | |
| | 5 1 | | |
| | 7 1 | | |
| | 7 1 | | |
| | 6 1 | | |
| | 8 1 | | |
| | 5 1 | | |
| | 7 1 | | |
| | 5 1 | | |
| | 8 1 | | |
| | 6 1 | | |
| | 7 1 | | |
| | 6 1 | | |
| 216.2.3 | 54387 | 58 | Ni4Sn1U1 |
| | -GM -X -L -W | | |
| | 6 2 7 2 6 2 6 1 | | |
| | 7 2 6 2 6 2 8 1 | | |
| | 8 4 6 2 6 2 5 1 | | |
| | 6 2 7 2 4+5 2 7 1 | | |
| | 6 2 6 2 6 2 8 1 | | |
| | 7 2 7 2 6 2 6 1 | | |
| | 8 4 6 2 6 2 5 1 | | |
| | 7 2 6 2 4+5 2 7 1 | | |
| | 8 4 7 2 6 2 5 1 | | |
| | 8 4 7 2 6 2 8 1 | | |
| | 6 2 7 2 6 2 7 1 | | |
| | 7 2 6 2 4+5 2 5 1 | | |
| | 8 4 6 2 6 2 6 1 | | |
| | 8 4 7 2 4+5 2 7 1 | | |

Continued on next column

**TABLE LXXXIII– continued**

| TQC | ICSD | NoE | SF |
|---|---|---|---|
| | 6 2 7 2 | 6 2 | 8 1 |
| | 7 2 6 2 | 6 2 | 6 1 |
| | 8 4 6 2 | 6 2 | 7 1 |
| | 8 4 7 2 | 4+5 2 | 6 1 |
| | 6 2 6 2 | 6 2 | 7 1 |
| | 8 4 7 2 | 4+5 2 | 6 1 |
| | 6 2 | 6 2 | 8 1 |
| | 7 2 | 4+5 2 | 5 1 |
| | 6 2 | 6 2 | 6 1 |
| | 7 2 | 6 2 | 5 1 |
| | 6 2 | 6 2 | 8 1 |
| | 7 2 | 4+5 2 | 8 1 |
| | 6 2 | 6 2 | 5 1 |
| | 7 2 | 4+5 2 | 5 1 |
| | 7 2 | 6 2 | 7 1 |
| | | | 6 1 |
| | | | 7 1 |
| | | | 8 1 |
| | | | 8 1 |
| | | | 6 1 |
| | | | 8 1 |
| | | | 5 1 |
| | | | 7 1 |
| | | | 6 1 |
| | | | 5 1 |
| | | | 7 1 |
| | | | 7 1 |
| | | | 6 1 |
| | | | 6 1 |
| | | | 5 1 |
| | | | 8 1 |
| | | | 5 1 |
| | | | 7 1 |
| | | | 8 1 |
| | | | 7 1 |
| | | | 6 1 |
| | | | 8 1 |
| | | | 5 1 |
| | | | 7 1 |
| | | | 5 1 |
| | | | 8 1 |
| | | | 6 1 |
| | | | 7 1 |
| | | | 6 1 |
| 216.2.4 | 418013 | 216 | La4Mg1Ru1 |
| | -GM -X | -L | -W |
| | 6 2 6 2 | 6 2 | 5 1 |
| | 7 2 6 2 | 6 2 | 8 1 |
| | 8 4 7 2 | 4+5 2 | 5 1 |
| | 8 4 7 2 | 6 2 | 6 1 |
| | 6 2 6 2 | 4+5 2 | 6 1 |
| | 7 2 7 2 | 6 2 | 7 1 |
| | 8 4 7 2 | 6 2 | 7 1 |
| | 8 4 6 2 | 6 2 | 8 1 |
| | 6 2 6 2 | 4+5 2 | 6 1 |
| | 7 2 7 2 | 6 2 | 7 1 |
| | 8 4 7 2 | 4+5 2 | 5 1 |
| | 8 4 6 2 | 6 2 | 8 1 |
| | 8 4 6 2 | 6 2 | 5 1 |
| | 6 2 6 2 | 4+5 2 | 7 1 |
| | 7 2 7 2 | 6 2 | 5 1 |

Continued on next column

**TABLE LXXXIII– continued**

| TQC | ICSD | NoE | SF |
|---|---|---|---|
| | 8 4 7 2 | 6 2 | 6 1 |
| | 6 2 7 2 | 4+5 2 | 6 1 |
| | 8 4 6 2 | 6 2 | 7 1 |
| | 7 2 6 2 | 6 2 | 8 1 |
| | 6 2 6 2 | 4+5 2 | 8 1 |
| | 8 4 7 2 | 6 2 | 6 1 |
| | 7 2 7 2 | 6 2 | 8 1 |
| | 7 2 6 2 | 4+5 2 | 5 1 |
| | 6 2 7 2 | 6 2 | 7 1 |
| | 7 2 7 2 | 6 2 | 6 1 |
| | 6 2 6 2 | 4+5 2 | 7 1 |
| | 8 4 7 2 | 6 2 | 5 1 |
| | 8 4 6 2 | 6 2 | 5 1 |
| | 8 4 7 2 | 6 2 | 7 1 |
| | 8 4 6 2 | 4+5 2 | 8 1 |
| | 7 2 6 2 | 6 2 | 6 1 |
| | 7 2 7 2 | 6 2 | 8 1 |
| | 6 2 7 2 | 6 2 | 5 1 |
| | 8 4 7 2 | 6 2 | 8 1 |
| | 6 2 6 2 | 6 2 | 6 1 |
| | 8 4 6 2 | 6 2 | 7 1 |
| | 8 4 6 2 | 4+5 2 | 5 1 |
| | 7 2 7 2 | 6 2 | 8 1 |
| | 8 4 7 2 | 4+5 2 | 6 1 |
| | 6 2 6 2 | 6 2 | 7 1 |
| | 8 4 7 2 | 4+5 2 | 8 1 |
| | 8 4 6 2 | 6 2 | 5 1 |
| | 8 4 6 2 | 6 2 | 7 1 |
| | 6 2 7 2 | 4+5 2 | 6 1 |
| | 6 2 6 2 | 6 2 | 5 1 |
| | 8 4 6 2 | 4+5 2 | 7 1 |
| | 7 2 7 2 | 6 2 | 8 1 |
| | 8 4 7 2 | 6 2 | 6 1 |
| | 7 2 6 2 | 6 2 | 5 1 |
| | 8 4 7 2 | 6 2 | 6 1 |
| | 8 4 6 2 | 4+5 2 | 8 1 |
| | 8 4 6 2 | 6 2 | 7 1 |
| | 6 2 7 2 | 6 2 | 5 1 |
| | 7 2 7 2 | 4+5 2 | 7 1 |
| | 8 4 7 2 | 6 2 | 8 1 |
| | 6 2 6 2 | 6 2 | 6 1 |
| | 8 4 6 2 | 4+5 2 | 6 1 |
| | 7 2 7 2 | 6 2 | 7 1 |
| | 6 2 7 2 | 4+5 2 | 7 1 |
| | 8 4 6 2 | 6 2 | 5 1 |
| | 7 2 6 2 | 6 2 | 6 1 |
| | 6 2 7 2 | 6 2 | 8 1 |
| | 8 4 7 2 | 4+5 2 | 5 1 |
| | 7 2 6 2 | 4+5 2 | 8 1 |
| | 8 4 7 2 | 6 2 | 7 1 |
| | 6 2 6 2 | 6 2 | 8 1 |
| | 8 4 7 2 | 6 2 | 7 1 |
| | 6 2 6 2 | 6 2 | 6 1 |
| | 6 2 7 2 | 6 2 | 6 1 |
| | 8 4 6 2 | 4+5 2 | 5 1 |
| | 7 2 6 2 | 4+5 2 | 6 1 |
| | 8 4 6 2 | 6 2 | 8 1 |
| | 8 4 7 2 | 6 2 | 5 1 |
| | 7 2 | 4+5 2 | 7 1 |
| | 7 2 | 4+5 2 | 5 1 |
| | 7 2 | 6 2 | 6 1 |

Continued on next column

**TABLE LXXXIII– continued**

| TQC | ICSD | NoE | SF |
|---|---|---|---|
| | 6 2 | 6 2 | 8 1 |
| | 6 2 | 6 2 | 6 1 |
| | 7 2 | 6 2 | 7 1 |
| | 6 2 | 4+5 2 | 5 1 |
| | 6 2 | 6 2 | 8 1 |
| | 7 2 | 6 2 | 5 1 |
| | 7 2 | 4+5 2 | 8 1 |
| | 6 2 | 6 2 | 7 1 |
| | 7 2 | 6 2 | 7 1 |
| | 6 2 | 4+5 2 | 6 1 |
| | 6 2 | 6 2 | 8 1 |
| | 7 2 | 4+5 2 | 5 1 |
| | 6 2 | 6 2 | 8 1 |
| | 7 2 | 4+5 2 | 6 1 |
| | 7 2 | 6 2 | 5 1 |
| | 7 2 | 6 2 | 7 1 |
| | 7 2 | 6 2 | 8 1 |
| | 6 2 | 4+5 2 | 7 1 |
| | 6 2 | 6 2 | 6 1 |
| | 6 2 | 6 2 | 5 1 |
| | 7 2 | 4+5 2 | 6 1 |
| | 6 2 | 6 2 | 5 1 |
| | 7 2 | 6 2 | 7 1 |
| | 6 2 | 6 2 | 6 1 |
| | 7 2 | 6 2 | 8 1 |
| | 6 2 | 4+5 2 | 7 1 |
| | 6 2 | 6 2 | 6 1 |
| | 7 2 | 4+5 2 | 8 1 |
| | 7 2 | 6 2 | 5 1 |
| | 6 2 | 6 2 | 8 1 |
| | 6 2 | 4+5 2 | 7 1 |
| | 7 2 | 6 2 | 5 1 |
| | | | 7 1 |
| | | | 8 1 |
| | | | 5 1 |
| | | | 6 1 |
| | | | 6 1 |
| | | | 5 1 |
| | | | 7 1 |
| | | | 5 1 |
| | | | 8 1 |
| | | | 6 1 |
| | | | 8 1 |
| | | | 7 1 |
| | | | 5 1 |
| | | | 8 1 |
| | | | 6 1 |
| | | | 7 1 |
| | | | 7 1 |
| | | | 8 1 |
| | | | 5 1 |
| | | | 6 1 |
| | | | 8 1 |
| | | | 6 1 |
| | | | 8 1 |
| | | | 5 1 |
| | | | 6 1 |
| | | | 7 1 |
| | | | 5 1 |
| | | | 7 1 |
| | | | 5 1 |

Continued on next column

**TABLE LXXXIII– continued**

| TQC | ICSD | NoE | SF |
|---|---|---|---|
| | | | 6 1 |
| | | | 7 1 |
| | | | 7 1 |
| | | | 8 1 |
| | | | 5 1 |
| | | | 6 1 |
| | | | 8 1 |
| | | | 7 1 |
| | | | 6 1 |
| | | | 5 1 |
| | | | 8 1 |
| | | | 6 1 |
| | | | 8 1 |
| | | | 8 1 |
| | | | 5 1 |
| | | | 7 1 |
| | | | 8 1 |
| | | | 5 1 |
| | | | 6 1 |
| | | | 7 1 |
| | | | 5 1 |
| | | | 7 1 |
| | | | 5 1 |
| | | | 6 1 |
| | | | 8 1 |
| | | | 7 1 |
| | | | 6 1 |
| | | | 5 1 |
| | | | 5 1 |
| | | | 7 1 |
| | | | 8 1 |
| | | | 8 1 |
| | | | 5 1 |
| | | | 7 1 |
| | | | 5 1 |
| | | | 6 1 |
| | | | 7 1 |
| | | | 6 1 |
| | | | 8 1 |
| | | | 7 1 |
| | | | 6 1 |
| | | | 5 1 |
| | | | 8 1 |
| | | | 6 1 |
| | | | 7 1 |
| | | | 8 1 |
| | | | 6 1 |
| | | | 5 1 |
| | | | 7 1 |
| | | | 5 1 |
| | | | 8 1 |
| | | | 6 1 |
| | | | 8 1 |
| | | | 7 1 |
| | | | 6 1 |
| | | | 5 1 |
| | | | 7 1 |
| | | | 7 1 |
| | | | 6 1 |
| | | | 8 1 |
| | | | 6 1 |

Continued on next column

**TABLE LXXXIII– continued**

| TQC | ICSD | NoE | SF |
|---|---|---|---|
| | | | 5 1 |
| | | | 8 1 |
| | | | 6 1 |
| | | | 8 1 |
| | | | 8 1 |
| | | | 5 1 |
| | | | 7 1 |
| | | | 5 1 |
| | | | 6 1 |
| | | | 7 1 |
| | | | 7 1 |
| | | | 5 1 |
| | | | 5 1 |
| | | | 8 1 |
| | | | 6 1 |
| | | | 8 1 |
| | | | 7 1 |
| | | | 6 1 |
| 216.2.5 | 417519 | 416 | In1Rh1Tm4 |
| | -GM -X | -L | -W |
| | 6 2 6 2 | 6 2 | 5 1 |
| | 7 2 6 2 | 4+5 2 | 8 1 |
| | 8 4 7 2 | 6 2 | 5 1 |
| | 8 4 7 2 | 6 2 | 6 1 |
| | 6 2 6 2 | 4+5 2 | 6 1 |
| | 7 2 7 2 | 6 2 | 7 1 |
| | 8 4 7 2 | 6 2 | 7 1 |
| | 8 4 6 2 | 4+5 2 | 8 1 |
| | 6 2 6 2 | 6 2 | 5 1 |
| | 7 2 7 2 | 6 2 | 6 1 |
| | 8 4 6 2 | 4+5 2 | 7 1 |
| | 8 4 7 2 | 6 2 | 8 1 |
| | 8 4 6 2 | 6 2 | 6 1 |
| | 6 2 6 2 | 4+5 2 | 8 1 |
| | 7 2 7 2 | 6 2 | 5 1 |
| | 8 4 7 2 | 6 2 | 5 1 |
| | 6 2 7 2 | 4+5 2 | 6 1 |
| | 8 4 6 2 | 6 2 | 7 1 |
| | 7 2 6 2 | 6 2 | 7 1 |
| | 6 2 6 2 | 4+5 2 | 8 1 |
| | 8 4 7 2 | 6 2 | 5 1 |
| | 7 2 7 2 | 6 2 | 6 1 |
| | 7 2 6 2 | 4+5 2 | 7 1 |
| | 6 2 7 2 | 6 2 | 8 1 |
| | 7 2 6 2 | 6 2 | 6 1 |
| | 8 4 7 2 | 4+5 2 | 7 1 |
| | 6 2 7 2 | 6 2 | 5 1 |
| | 8 4 6 2 | 6 2 | 5 1 |
| | 8 4 6 2 | 4+5 2 | 6 1 |
| | 8 4 7 2 | 6 2 | 7 1 |
| | 7 2 6 2 | 6 2 | 8 1 |
| | 6 2 7 2 | 6 2 | 8 1 |
| | 7 2 7 2 | 6 2 | 5 1 |
| | 8 4 6 2 | 6 2 | 8 1 |
| | 8 4 7 2 | 6 2 | 6 1 |
| | 8 4 6 2 | 4+5 2 | 7 1 |
| | 6 2 7 2 | 6 2 | 5 1 |
| | 8 4 6 2 | 6 2 | 6 1 |
| | 7 2 6 2 | 4+5 2 | 7 1 |
| | 6 2 7 2 | 6 2 | 8 1 |
| | 8 4 7 2 | 4+5 2 | 5 1 |

Continued on next column

**TABLE LXXXIII– continued**

| TQC | ICSD | NoE | SF |
|---|---|---|---|
| | 8 4 6 2 | 6 2 | 8 1 |
| | 8 4 6 2 | 4+5 2 | 6 1 |
| | 6 2 7 2 | 6 2 | 7 1 |
| | 8 4 6 2 | 6 2 | 5 1 |
| | 7 2 6 2 | 6 2 | 6 1 |
| | 6 2 7 2 | 4+5 2 | 8 1 |
| | 7 2 7 2 | 6 2 | 7 1 |
| | 6 2 6 2 | 6 2 | 5 1 |
| | 8 4 6 2 | 4+5 2 | 6 1 |
| | 8 4 7 2 | 6 2 | 8 1 |
| | 8 4 7 2 | 6 2 | 8 1 |
| | 8 4 6 2 | 4+5 2 | 6 1 |
| | 7 2 7 2 | 6 2 | 7 1 |
| | 6 2 6 2 | 6 2 | 5 1 |
| | 7 2 7 2 | 4+5 2 | 7 1 |
| | 8 4 6 2 | 6 2 | 6 1 |
| | 8 4 7 2 | 6 2 | 7 1 |
| | 8 4 6 2 | 6 2 | 7 1 |
| | 6 2 7 2 | 4+5 2 | 5 1 |
| | 7 2 6 2 | 6 2 | 6 1 |
| | 6 2 7 2 | 6 2 | 8 1 |
| | 8 4 7 2 | 4+5 2 | 5 1 |
| | 8 4 6 2 | 4+5 2 | 8 1 |
| | 6 2 7 2 | 6 2 | 6 1 |
| | 7 2 6 2 | 6 2 | 7 1 |
| | 8 4 7 2 | 4+5 2 | 5 1 |
| | 8 4 6 2 | 6 2 | 6 1 |
| | 6 2 6 2 | 6 2 | 8 1 |
| | 8 4 7 2 | 6 2 | 7 1 |
| | 7 2 6 2 | 4+5 2 | 5 1 |
| | 6 2 6 2 | 6 2 | 7 1 |
| | 8 4 7 2 | 6 2 | 8 1 |
| | 7 2 6 2 | 6 2 | 6 1 |
| | 8 4 7 2 | 4+5 2 | 5 1 |
| | 8 4 7 2 | 6 2 | 7 1 |
| | 6 2 6 2 | 4+5 2 | 8 1 |
| | 8 4 7 2 | 6 2 | 6 1 |
| | 7 2 6 2 | 4+5 2 | 5 1 |
| | 8 4 6 2 | 6 2 | 8 1 |
| | 6 2 7 2 | 6 2 | 7 1 |
| | 7 2 6 2 | 6 2 | 5 1 |
| | 8 4 7 2 | 4+5 2 | 8 1 |
| | 6 2 7 2 | 6 2 | 6 1 |
| | 6 2 7 2 | 4+5 2 | 7 1 |
| | 7 2 6 2 | 6 2 | 5 1 |
| | 8 4 7 2 | 6 2 | 6 1 |
| | 8 4 6 2 | 4+5 2 | 8 1 |
| | 7 2 7 2 | 6 2 | 5 1 |
| | 8 4 6 2 | 6 2 | 7 1 |
| | 8 4 7 2 | 4+5 2 | 6 1 |
| | 8 4 6 2 | 6 2 | 5 1 |
| | 6 2 7 2 | 6 2 | 8 1 |
| | 8 4 6 2 | 4+5 2 | 8 1 |
| | 7 2 7 2 | 6 2 | 6 1 |
| | 8 4 6 2 | 6 2 | 7 1 |
| | 7 2 6 2 | 4+5 2 | 8 1 |
| | 6 2 7 2 | 6 2 | 7 1 |
| | 8 4 6 2 | 6 2 | 5 1 |
| | 6 2 7 2 | 6 2 | 6 1 |
| | 7 2 6 2 | 6 2 | 7 1 |
| | 8 4 7 2 | 4+5 2 | 5 1 |

Continued on next column

**TABLE LXXXIII– continued**

| TQC | ICSD | NoE | SF |
|---|---|---|---|
| | 6 2 7 2 4+5 2 6 1 | | |
| | 7 2 6 2 6 2 7 1 | | |
| | 8 4 6 2 6 2 8 1 | | |
| | 6 2 7 2 6 2 6 1 | | |
| | 8 4 6 2 4+5 2 8 1 | | |
| | 6 2 7 2 6 2 5 1 | | |
| | 8 4 6 2 6 2 7 1 | | |
| | 7 2 7 2 6 2 6 1 | | |
| | 8 4 6 2 4+5 2 8 1 | | |
| | 6 2 6 2 6 2 7 1 | | |
| | 8 4 7 2 6 2 5 1 | | |
| | 7 2 7 2 4+5 2 5 1 | | |
| | 8 4 7 2 6 2 6 1 | | |
| | 7 2 6 2 4+5 2 8 1 | | |
| | 7 2 6 2 6 2 6 1 | | |
| | 8 4 7 2 6 2 7 1 | | |
| | 8 4 7 2 6 2 5 1 | | |
| | 6 2 6 2 4+5 2 8 1 | | |
| | 8 4 7 2 6 2 5 1 | | |
| | 7 2 6 2 4+5 2 6 1 | | |
| | 8 4 6 2 6 2 8 1 | | |
| | 6 2 7 2 6 2 7 1 | | |
| | 8 4 7 2 6 2 6 1 | | |
| | 8 4 7 2 4+5 2 5 1 | | |
| | 7 2 6 2 6 2 8 1 | | |
| | 6 2 6 2 6 2 7 1 | | |
| | 8 4 7 2 4+5 2 5 1 | | |
| | 8 4 6 2 4+5 2 7 1 | | |
| | 8 4 6 2 6 2 5 1 | | |
| | 6 2 7 2 4+5 2 8 1 | | |
| | 7 2 6 2 6 2 7 1 | | |
| | 6 2 7 2 6 2 6 1 | | |
| | 8 4 6 2 6 2 8 1 | | |
| | 8 4 7 2 4+5 2 6 1 | | |
| | 7 2 6 2 6 2 6 1 | | |
| | 8 4 7 2 6 2 7 1 | | |
| | 7 2 6 2 6 2 8 1 | | |
| | 8 4 7 2 4+5 2 6 1 | | |
| | 7 2 6 2 8 1 | | |
| | 7 2 4+5 2 5 1 | | |
| | 6 2 6 2 7 1 | | |
| | 6 2 6 2 5 1 | | |
| | 6 2 6 2 6 1 | | |
| | 6 2 4+5 2 7 1 | | |
| | 7 2 6 2 7 1 | | |
| | 7 2 6 2 6 1 | | |
| | 6 2 4+5 2 5 1 | | |
| | 7 2 6 2 8 1 | | |
| | 7 2 6 2 6 1 | | |
| | 6 2 4+5 2 7 1 | | |
| | 7 2 6 2 5 1 | | |
| | 7 2 6 2 8 1 | | |
| | 6 2 6 2 8 1 | | |
| | 7 2 4+5 2 5 1 | | |
| | 6 2 6 2 7 1 | | |
| | 6 2 6 2 5 1 | | |
| | 7 2 4+5 2 8 1 | | |
| | 6 2 6 2 6 1 | | |
| | 6 2 6 2 5 1 | | |
| | 7 2 6 2 6 1 | | |
| | 6 2 6 2 8 1 | | |

Continued on next column

**TABLE LXXXIII– continued**

| TQC | ICSD | NoE | SF |
|---|---|---|---|
| | 7 2 6 2 5 1 | | |
| | 6 2 4+5 2 7 1 | | |
| | 7 2 6 2 8 1 | | |
| | 7 2 4+5 2 6 1 | | |
| | 7 2 6 2 5 1 | | |
| | 6 2 4+5 2 8 1 | | |
| | 6 2 6 2 7 1 | | |
| | 6 2 4+5 2 8 1 | | |
| | 7 2 6 2 6 1 | | |
| | 6 2 6 2 7 1 | | |
| | 7 2 4+5 2 5 1 | | |
| | 7 2 6 2 6 1 | | |
| | 7 2 6 2 8 1 | | |
| | 6 2 4+5 2 7 1 | | |
| | 7 2 6 2 5 1 | | |
| | 6 2 6 2 5 1 | | |
| | 6 2 4+5 2 8 1 | | |
| | 7 2 6 2 6 1 | | |
| | 6 2 6 2 7 1 | | |
| | 7 2 4+5 2 7 1 | | |
| | 6 2 6 2 6 1 | | |
| | 6 2 4+5 2 8 1 | | |
| | 7 2 6 2 5 1 | | |
| | 6 2 6 2 6 1 | | |
| | 7 2 6 2 5 1 | | |
| | 6 2 4+5 2 7 1 | | |
| | 6 2 6 2 8 1 | | |
| | 7 2 6 2 7 1 | | |
| | 6 2 4+5 2 6 1 | | |
| | 7 2 6 2 5 1 | | |
| | 6 2 4+5 2 7 1 | | |
| | 7 2 6 2 8 1 | | |
| | 7 2 6 2 6 1 | | |
| | 6 2 6 2 7 1 | | |
| | 7 2 6 2 5 1 | | |
| | 6 2 4+5 2 8 1 | | |
| | 7 2 6 2 6 1 | | |
| | 7 2 6 2 6 1 | | |
| | 6 2 6 2 5 1 | | |
| | 6 2 4+5 2 7 1 | | |
| | 7 2 4+5 2 8 1 | | |
| | 6 2 6 2 6 1 | | |
| | 7 2 6 2 7 1 | | |
| | 6 2 4+5 2 6 1 | | |
| | 7 2 6 2 8 1 | | |
| | 5 1 | | |
| | 7 1 | | |
| | 5 1 | | |
| | 8 1 | | |
| | 5 1 | | |
| | 6 1 | | |
| | 8 1 | | |
| | 7 1 | | |
| | 8 1 | | |
| | 5 1 | | |
| | 6 1 | | |
| | 8 1 | | |
| | 5 1 | | |
| | 5 1 | | |
| | 6 1 | | |
| | 7 1 | | |

Continued on next column

**TABLE LXXXIII– continued**

| TQC | ICSD | NoE | SF |
|---|---|---|---|
| | | 8 | 1 |
| | | 7 | 1 |
| | | 8 | 1 |
| | | 5 | 1 |
| | | 6 | 1 |
| | | 7 | 1 |
| | | 5 | 1 |
| | | 6 | 1 |
| | | 8 | 1 |
| | | 5 | 1 |
| | | 6 | 1 |
| | | 7 | 1 |
| | | 8 | 1 |
| | | 8 | 1 |
| | | 6 | 1 |
| | | 7 | 1 |
| | | 5 | 1 |
| | | 6 | 1 |
| | | 7 | 1 |
| | | 7 | 1 |
| | | 5 | 1 |
| | | 6 | 1 |
| | | 7 | 1 |
| | | 8 | 1 |
| | | 8 | 1 |
| | | 5 | 1 |
| | | 6 | 1 |
| | | 7 | 1 |
| | | 8 | 1 |
| | | 5 | 1 |
| | | 6 | 1 |
| | | 7 | 1 |
| | | 8 | 1 |
| | | 5 | 1 |
| | | 6 | 1 |
| | | 8 | 1 |
| | | 5 | 1 |
| | | 7 | 1 |
| | | 5 | 1 |
| | | 6 | 1 |
| | | 7 | 1 |
| | | 6 | 1 |
| | | 8 | 1 |
| | | 5 | 1 |
| | | 7 | 1 |
| | | 7 | 1 |
| | | 8 | 1 |
| | | 6 | 1 |
| | | 8 | 1 |
| | | 6 | 1 |
| | | 7 | 1 |
| | | 6 | 1 |
| | | 7 | 1 |
| | | 5 | 1 |
| | | 8 | 1 |
| | | 5 | 1 |
| | | 6 | 1 |
| | | 5 | 1 |
| | | 6 | 1 |
| | | 7 | 1 |
| | | 8 | 1 |

Continued on next column

**TABLE LXXXIII– continued**

| TQC | ICSD | NoE | SF |
|---|---|---|---|
| | | 7 | 1 |
| | | 5 | 1 |
| | | 6 | 1 |
| | | 7 | 1 |
| | | 5 | 1 |
| | | 8 | 1 |
| | | 6 | 1 |
| | | 8 | 1 |
| | | 7 | 1 |
| | | 8 | 1 |
| | | 5 | 1 |
| | | 8 | 1 |
| | | 6 | 1 |
| | | 5 | 1 |
| | | 7 | 1 |
| | | 5 | 1 |
| | | 8 | 1 |
| | | 6 | 1 |
| | | 8 | 1 |
| | | 7 | 1 |
| | | 5 | 1 |
| | | 6 | 1 |
| | | 8 | 1 |
| | | 7 | 1 |
| | | 5 | 1 |
| | | 7 | 1 |
| | | 5 | 1 |
| | | 6 | 1 |
| | | 7 | 1 |
| | | 8 | 1 |
| | | 6 | 1 |
| | | 5 | 1 |
| | | 8 | 1 |
| | | 7 | 1 |
| | | 5 | 1 |
| | | 6 | 1 |
| | | 7 | 1 |
| | | 8 | 1 |
| | | 5 | 1 |
| | | 8 | 1 |
| | | 6 | 1 |
| | | 7 | 1 |
| | | 6 | 1 |
| | | 8 | 1 |
| | | 6 | 1 |
| | | 7 | 1 |
| | | 5 | 1 |
| | | 6 | 1 |
| | | 5 | 1 |
| | | 8 | 1 |
| | | 5 | 1 |
| | | 6 | 1 |
| | | 7 | 1 |
| | | 8 | 1 |
| | | 5 | 1 |
| | | 6 | 1 |
| | | 7 | 1 |
| | | 8 | 1 |
| | | 7 | 1 |
| | | 8 | 1 |
| | | 8 | 1 |

Continued on next column

**TABLE LXXXIII– continued**

| TQC | ICSD | NoE | SF |
|---|---|---|---|
| | | 6 1 | |
| | | 6 1 | |
| | | 7 1 | |
| | | 5 1 | |
| | | 5 1 | |
| | | 6 1 | |
| | | 7 1 | |
| | | 8 1 | |
| | | 7 1 | |
| | | 8 1 | |
| | | 5 1 | |
| | | 6 1 | |
| | | 8 1 | |
| | | 5 1 | |
| | | 6 1 | |
| | | 7 1 | |
| | | 5 1 | |
| | | 7 1 | |
| | | 8 1 | |
| | | 6 1 | |
| | | 7 1 | |
| | | 8 1 | |
| | | 5 1 | |
| | | 5 1 | |
| | | 6 1 | |
| | | 6 1 | |
| | | 7 1 | |
| | | 5 1 | |
| | | 7 1 | |
| | | 6 1 | |
| | | 7 1 | |
| | | 8 1 | |
| | | 5 1 | |
| | | 6 1 | |
| | | 8 1 | |
| | | 5 1 | |
| | | 6 1 | |
| | | 6 1 | |
| | | 7 1 | |
| | | 8 1 | |
| | | 5 1 | |
| | | 7 1 | |
| | | 5 1 | |
| | | 6 1 | |
| | | 7 1 | |
| | | 8 1 | |
| | | 7 1 | |
| | | 5 1 | |
| | | 6 1 | |
| | | 8 1 | |
| | | 5 1 | |
| | | 8 1 | |
| | | 5 1 | |
| | | 6 1 | |
| | | 7 1 | |
| | | 8 1 | |
| | | 5 1 | |
| | | 7 1 | |
| | | 8 1 | |
| | | 6 1 | |
| | | 7 1 | |

Continued on next column

**TABLE LXXXIII– continued**

| TQC | ICSD | NoE | SF |
|---|---|---|---|
| | | 8 1 | |
| | | 8 1 | |
| | | 5 1 | |
| | | 6 1 | |
| | | 7 1 | |
| | | 5 1 | |
| | | 6 1 | |
| | | 7 1 | |
| | | 7 1 | |
| 216.2.6 | 81917 | 18 | Hg1S1 |
| | -GM -X | -L -W | |
| | 6 2 6 2 | 6 2 5 1 | |
| | 8 4 7 2 | 4+5 2 8 1 | |
| | 7 2 6 2 | 6 2 5 1 | |
| | 8 4 7 2 | 6 2 6 1 | |
| | 6 2 6 2 | 4+5 2 7 1 | |
| | 8 4 6 2 | 6 2 8 1 | |
| | 7 2 | 6 2 7 1 | |
| | 6 2 | 4+5 2 7 1 | |
| | 7 2 | 6 2 6 1 | |
| | | 5 1 | |
| | | 8 1 | |
| | | 6 1 | |
| | | 5 1 | |
| | | 8 1 | |
| | | 7 1 | |
| | | 6 1 | |
| | | 5 1 | |
| | | 7 1 | |
| 216.2.7 | 419068 | 368 | Cd1Gd4Ru1 |
| | -GM -X | -L -W | |
| | 6 2 7 2 | 6 2 5 1 | |
| | 7 2 6 2 | 6 2 7 1 | |
| | 8 4 6 2 | 4+5 2 5 1 | |
| | 8 4 7 2 | 6 2 6 1 | |
| | 6 2 7 2 | 4+5 2 6 1 | |
| | 7 2 6 2 | 6 2 7 1 | |
| | 8 4 6 2 | 6 2 8 1 | |
| | 8 4 6 2 | 4+5 2 8 1 | |
| | 6 2 7 2 | 6 2 5 1 | |
| | 7 2 7 2 | 6 2 6 1 | |
| | 8 4 6 2 | 4+5 2 7 1 | |
| | 8 4 7 2 | 6 2 8 1 | |
| | 8 4 7 2 | 6 2 5 1 | |
| | 6 2 6 2 | 4+5 2 8 1 | |
| | 7 2 6 2 | 6 2 5 1 | |
| | 8 4 7 2 | 6 2 6 1 | |
| | 6 2 6 2 | 4+5 2 6 1 | |
| | 8 4 7 2 | 6 2 7 1 | |
| | 7 2 7 2 | 6 2 7 1 | |
| | 6 2 6 2 | 4+5 2 8 1 | |
| | 8 4 7 2 | 6 2 5 1 | |
| | 7 2 6 2 | 6 2 6 1 | |
| | 7 2 7 2 | 4+5 2 7 1 | |
| | 6 2 6 2 | 6 2 8 1 | |
| | 7 2 6 2 | 6 2 6 1 | |
| | 8 4 7 2 | 4+5 2 8 1 | |
| | 6 2 6 2 | 6 2 5 1 | |
| | 8 4 7 2 | 6 2 5 1 | |
| | 8 4 6 2 | 6 2 6 1 | |
| | 8 4 7 2 | 4+5 2 7 1 | |

Continued on next column

**TABLE LXXXIII– continued**

| TQC | ICSD | NoE | SF |
|---|---|---|---|
| | 7 2 7 2 | 6 2 7 1 | |
| | 6 2 6 2 | 6 2 8 1 | |
| | 7 2 6 2 | 6 2 5 1 | |
| | 8 4 7 2 | 6 2 7 1 | |
| | 8 4 6 2 | 6 2 6 1 | |
| | 6 2 7 2 | 6 2 8 1 | |
| | 8 4 6 2 | 4+5 2 5 1 | |
| | 7 2 7 2 | 6 2 6 1 | |
| | 8 4 6 2 | 4+5 2 7 1 | |
| | 6 2 7 2 | 6 2 8 1 | |
| | 8 4 6 2 | 4+5 2 5 1 | |
| | 8 4 7 2 | 6 2 7 1 | |
| | 8 4 7 2 | 4+5 2 6 1 | |
| | 6 2 6 2 | 6 2 8 1 | |
| | 8 4 7 2 | 6 2 5 1 | |
| | 8 4 7 2 | 4+5 2 8 1 | |
| | 7 2 6 2 | 6 2 7 1 | |
| | 8 4 6 2 | 6 2 6 1 | |
| | 7 2 7 2 | 6 2 5 1 | |
| | 8 4 6 2 | 4+5 2 7 1 | |
| | 7 2 7 2 | 6 2 6 1 | |
| | 6 2 7 2 | 6 2 8 1 | |
| | 8 4 6 2 | 6 2 5 1 | |
| | 8 4 6 2 | 4+5 2 8 1 | |
| | 6 2 6 2 | 6 2 6 1 | |
| | 8 4 7 2 | 6 2 7 1 | |
| | 6 2 7 2 | 4+5 2 6 1 | |
| | 6 2 6 2 | 6 2 8 1 | |
| | 8 4 6 2 | 6 2 8 1 | |
| | 7 2 7 2 | 4+5 2 5 1 | |
| | 7 2 7 2 | 6 2 6 1 | |
| | 8 4 6 2 | 6 2 7 1 | |
| | 8 4 6 2 | 4+5 2 5 1 | |
| | 6 2 7 2 | 4+5 2 7 1 | |
| | 8 4 6 2 | 6 2 8 1 | |
| | 8 4 7 2 | 4+5 2 7 1 | |
| | 7 2 6 2 | 6 2 6 1 | |
| | 6 2 6 2 | 6 2 8 1 | |
| | 7 2 7 2 | 4+5 2 6 1 | |
| | 8 4 7 2 | 6 2 5 1 | |
| | 6 2 7 2 | 4+5 2 6 1 | |
| | 8 4 6 2 | 6 2 7 1 | |
| | 8 4 7 2 | 6 2 5 1 | |
| | 7 2 6 2 | 6 2 8 1 | |
| | 6 2 7 2 | 4+5 2 7 1 | |
| | 8 4 7 2 | 6 2 5 1 | |
| | 8 4 6 2 | 6 2 6 1 | |
| | 7 2 7 2 | 4+5 2 8 1 | |
| | 6 2 6 2 | 6 2 6 1 | |
| | 8 4 7 2 | 4+5 2 5 1 | |
| | 6 2 6 2 | 6 2 7 1 | |
| | 8 4 6 2 | 6 2 7 1 | |
| | 8 4 6 2 | 4+5 2 5 1 | |
| | 6 2 7 2 | 6 2 8 1 | |
| | 7 2 6 2 | 6 2 8 1 | |
| | 8 4 7 2 | 6 2 6 1 | |
| | 6 2 6 2 | 4+5 2 7 1 | |
| | 7 2 7 2 | 6 2 5 1 | |
| | 8 4 6 2 | 6 2 7 1 | |
| | 6 2 7 2 | 6 2 5 1 | |
| | 8 4 7 2 | 6 2 6 1 | |

Continued on next column

**TABLE LXXXIII– continued**

| TQC | ICSD | NoE | SF |
|---|---|---|---|
| | 7 2 6 2 | 4+5 2 8 1 | |
| | 8 4 6 2 | 6 2 7 1 | |
| | 7 2 7 2 | 6 2 6 1 | |
| | 8 4 7 2 | 4+5 2 8 1 | |
| | 8 4 6 2 | 6 2 5 1 | |
| | 7 2 7 2 | 4+5 2 5 1 | |
| | 6 2 6 2 | 6 2 6 1 | |
| | 8 4 6 2 | 6 2 8 1 | |
| | 7 2 7 2 | 4+5 2 7 1 | |
| | 8 4 6 2 | 6 2 6 1 | |
| | 8 4 7 2 | 6 2 8 1 | |
| | 6 2 6 2 | 4+5 2 6 1 | |
| | 8 4 6 2 | 6 2 7 1 | |
| | 8 4 7 2 | 4+5 2 5 1 | |
| | 6 2 7 2 | 6 2 7 1 | |
| | 8 4 7 2 | 6 2 8 1 | |
| | 8 4 6 2 | 4+5 2 5 1 | |
| | 7 2 6 2 | 6 2 6 1 | |
| | 6 2 7 2 | 6 2 8 1 | |
| | 7 2 6 2 | 6 2 5 1 | |
| | 8 4 7 2 | 6 2 7 1 | |
| | 6 2 6 2 | 4+5 2 6 1 | |
| | 7 2 7 2 | 4+5 2 5 1 | |
| | 6 2 6 2 | 6 2 8 1 | |
| | 7 2 7 2 | 4+5 2 7 1 | |
| | 8 4 7 2 | 6 2 7 1 | |
| | 6 2 6 2 | 6 2 8 1 | |
| | 8 4 7 2 | 6 2 5 1 | |
| | 6 2 6 2 | 6 2 6 1 | |
| | 7 2 6 2 | 6 2 5 1 | |
| | 8 4 7 2 | 4+5 2 7 1 | |
| | 8 4 7 2 | 6 2 8 1 | |
| | 8 4 6 2 | 4+5 2 6 1 | |
| | 7 2 | 6 2 8 1 | |
| | 6 2 | 4+5 2 7 1 | |
| | 7 2 | 6 2 5 1 | |
| | 6 2 | 6 2 6 1 | |
| | 7 2 | 6 2 7 1 | |
| | 6 2 | 4+5 2 6 1 | |
| | 6 2 | 6 2 6 1 | |
| | 7 2 | 6 2 5 1 | |
| | 6 2 | 6 2 7 1 | |
| | 7 2 | 6 2 8 1 | |
| | 7 2 | 4+5 2 8 1 | |
| | 6 2 | 6 2 5 1 | |
| | 6 2 | 6 2 5 1 | |
| | 7 2 | 4+5 2 6 1 | |
| | 6 2 | 6 2 7 1 | |
| | 7 2 | 6 2 8 1 | |
| | 7 2 | 6 2 6 1 | |
| | 6 2 | 4+5 2 7 1 | |
| | 7 2 | 4+5 2 8 1 | |
| | 6 2 | 6 2 5 1 | |
| | 7 2 | 6 2 8 1 | |
| | 6 2 | 4+5 2 5 1 | |
| | 6 2 | 6 2 5 1 | |
| | 7 2 | 4+5 2 6 1 | |
| | 6 2 | 6 2 8 1 | |
| | 7 2 | 6 2 7 1 | |
| | 6 2 | 6 2 5 1 | |
| | 7 2 | 4+5 2 8 1 | |

Continued on next column

**TABLE LXXXIII– continued**

| TQC | ICSD | NoE | SF |
|---|---|---|---|
| | 7 2 | 4+5 2 | 6 1 |
| | 6 2 | 6 2 | 7 1 |
| | 6 2 | 6 2 | 6 1 |
| | 7 2 | 6 2 | 8 1 |
| | 7 2 | 6 2 | 7 1 |
| | 7 2 | 4+5 2 | 5 1 |
| | 6 2 | 6 2 | 5 1 |
| | 6 2 | 6 2 | 6 1 |
| | 6 2 | 4+5 2 | 7 1 |
| | 7 2 | 6 2 | 8 1 |
| | 7 2 | 6 2 | 6 1 |
| | 6 2 | 4+5 2 | 7 1 |
| | 7 2 | 4+5 2 | 5 1 |
| | 6 2 | 6 2 | 8 1 |
| | 7 2 | 6 2 | 6 1 |
| | 6 2 | 6 2 | 7 1 |
| | 7 2 | 4+5 2 | 7 1 |
| | 6 2 | 6 2 | 6 1 |
| | 7 2 | 6 2 | 7 1 |
| | 6 2 | 6 2 | 5 1 |
| | 6 2 | 4+5 2 | 6 1 |
| | 6 2 | 6 2 | 8 1 |
| | 7 2 | 6 2 | 5 1 |
| | 7 2 | 6 2 | 8 1 |
| | 7 2 | 4+5 2 | 5 1 |
| | 6 2 | 6 2 | 8 1 |
| | 7 2 | 4+5 2 | 8 1 |
| | 6 2 | 6 2 | 6 1 |
| | 6 2 | 6 2 | 5 1 |
| | 6 2 | 6 2 | 7 1 |
| | 7 2 | 4+5 2 | 6 1 |
| | 7 2 | 6 2 | 8 1 |
| | | | 7 1 |
| | | | 6 1 |
| | | | 6 1 |
| | | | 7 1 |
| | | | 8 1 |
| | | | 6 1 |
| | | | 5 1 |
| | | | 8 1 |
| | | | 7 1 |
| | | | 5 1 |
| | | | 7 1 |
| | | | 6 1 |
| | | | 5 1 |
| | | | 7 1 |
| | | | 5 1 |
| | | | 8 1 |
| | | | 7 1 |
| | | | 5 1 |
| | | | 7 1 |
| | | | 8 1 |
| | | | 5 1 |
| | | | 6 1 |
| | | | 5 1 |
| | | | 8 1 |
| | | | 7 1 |
| | | | 5 1 |
| | | | 6 1 |
| | | | 8 1 |
| | | | 6 1 |

Continued on next column

**TABLE LXXXIII– continued**

| TQC | ICSD | NoE | SF |
|---|---|---|---|
| | | | 5 1 |
| | | | 7 1 |
| | | | 8 1 |
| | | | 8 1 |
| | | | 6 1 |
| | | | 5 1 |
| | | | 7 1 |
| | | | 7 1 |
| | | | 6 1 |
| | | | 6 1 |
| | | | 8 1 |
| | | | 5 1 |
| | | | 8 1 |
| | | | 5 1 |
| | | | 7 1 |
| | | | 7 1 |
| | | | 6 1 |
| | | | 5 1 |
| | | | 8 1 |
| | | | 6 1 |
| | | | 8 1 |
| | | | 7 1 |
| | | | 5 1 |
| | | | 6 1 |
| | | | 7 1 |
| | | | 8 1 |
| | | | 6 1 |
| | | | 8 1 |
| | | | 6 1 |
| | | | 7 1 |
| | | | 5 1 |
| | | | 7 1 |
| | | | 5 1 |
| | | | 8 1 |
| | | | 6 1 |
| | | | 5 1 |
| | | | 7 1 |
| | | | 8 1 |
| | | | 5 1 |
| | | | 6 1 |
| | | | 8 1 |
| | | | 6 1 |
| | | | 7 1 |
| | | | 7 1 |
| | | | 8 1 |
| | | | 5 1 |
| | | | 7 1 |
| | | | 8 1 |
| | | | 5 1 |
| | | | 6 1 |
| | | | 5 1 |
| | | | 6 1 |
| | | | 5 1 |
| | | | 6 1 |
| | | | 8 1 |
| | | | 8 1 |
| | | | 6 1 |
| | | | 7 1 |
| | | | 5 1 |
| | | | 7 1 |
| | | | 6 1 |

Continued on next column

**TABLE LXXXIII– continued**

| TQC | ICSD | NoE | SF |
|---|---|---|---|
| | | 7 1 | |
| | | 5 1 | |
| | | 8 1 | |
| | | 6 1 | |
| | | 6 1 | |
| | | 7 1 | |
| | | 8 1 | |
| | | 5 1 | |
| | | 5 1 | |
| | | 7 1 | |
| | | 7 1 | |
| | | 8 1 | |
| | | 8 1 | |
| | | 8 1 | |
| | | 5 1 | |
| | | 6 1 | |
| | | 7 1 | |
| | | 6 1 | |
| | | 7 1 | |
| | | 8 1 | |
| | | 5 1 | |
| | | 7 1 | |
| | | 5 1 | |
| | | 6 1 | |
| | | 6 1 | |
| | | 7 1 | |
| | | 8 1 | |
| | | 5 1 | |
| | | 6 1 | |
| | | 8 1 | |
| | | 7 1 | |
| | | 8 1 | |
| | | 5 1 | |
| | | 5 1 | |
| | | 6 1 | |
| | | 7 1 | |
| | | 8 1 | |
| | | 6 1 | |
| | | 8 1 | |
| | | 5 1 | |
| | | 8 1 | |
| | | 6 1 | |
| | | 6 1 | |
| | | 5 1 | |
| | | 7 1 | |
| | | 8 1 | |
| | | 7 1 | |
| | | 7 1 | |
| | | 5 1 | |
| | | 8 1 | |
| | | 6 1 | |
| | | 5 1 | |
| | | 5 1 | |
| | | 6 1 | |
| | | 7 1 | |
| | | 8 1 | |
| | | 6 1 | |
| | | 7 1 | |
| | | 8 1 | |
| | | 6 1 | |
| | | 5 1 | |

Continued on next column

**TABLE LXXXIII– continued**

| TQC | ICSD | NoE | SF |
|---|---|---|---|
| | | 7 1 | |
| | | 5 1 | |
| | | 7 1 | |
| | | 6 1 | |
| | | 8 1 | |
| | | 7 1 | |
| | | 8 1 | |
| | | 5 1 | |
| | | 6 1 | |
| | | 5 1 | |
| | | 7 1 | |
| | | 8 1 | |
| | | 5 1 | |
| | | 6 1 | |
| | | 6 1 | |
| | | 7 1 | |
| | | 8 1 | |
| | | 7 1 | |
| | | 5 1 | |
| | | 8 1 | |
| | | 6 1 | |
| | | 5 1 | |
| | | 6 1 | |
| | | 7 1 | |
| | | 8 1 | |
| | | 7 1 | |
| | | 5 1 | |
| | | 6 1 | |
| | | 8 1 | |
| | | 6 1 | |
| | | 7 1 | |
| | | 5 1 | |
| | | 8 1 | |
| 216.2.8 | 415194 | 64 | Cd1Ce1Ni4 |
| | -GM -X | -L -W | |
| | 6 2 6 2 | 6 2 5 1 | |
| | 7 2 7 2 | 6 2 8 1 | |
| | 8 4 7 2 | 6 2 7 1 | |
| | 8 4 6 2 | 4+5 2 6 1 | |
| | 8 4 6 2 | 6 2 5 1 | |
| | 7 2 7 2 | 4+5 2 8 1 | |
| | 6 2 6 2 | 6 2 7 1 | |
| | 6 2 7 2 | 4+5 2 6 1 | |
| | 7 2 6 2 | 6 2 7 1 | |
| | 8 4 6 2 | 6 2 6 1 | |
| | 7 2 7 2 | 6 2 8 1 | |
| | 8 4 7 2 | 6 2 5 1 | |
| | 8 4 7 2 | 4+5 2 7 1 | |
| | 8 4 6 2 | 6 2 5 1 | |
| | 8 4 6 2 | 6 2 6 1 | |
| | 7 2 7 2 | 4+5 2 8 1 | |
| | 8 4 6 2 | 6 2 5 1 | |
| | 6 2 6 2 | 6 2 7 1 | |
| | 7 2 7 2 | 4+5 2 6 1 | |
| | 8 4 7 2 | 6 2 8 1 | |
| | 6 2 6 2 | 6 2 7 1 | |
| | 6 2 7 2 | 6 2 6 1 | |
| | 6 2 | 4+5 2 7 1 | |
| | 7 2 | 4+5 2 5 1 | |
| | 6 2 | 6 2 5 1 | |
| | 6 2 | 6 2 8 1 | |

Continued on next column

**TABLE LXXXIII– continued**

| TQC | ICSD | NoE | SF |
|---|---|---|---|
| | 7 2 | 4+5 2 | 7 1 |
| | 7 2 | 6 2 | 6 1 |
| | 6 2 | 6 2 | 8 1 |
| | 7 2 | 4+5 2 | 5 1 |
| | 6 2 | 6 2 | 8 1 |
| | 7 2 | 6 2 | 6 1 |
| | | | 5 1 |
| | | | 6 1 |
| | | | 7 1 |
| | | | 5 1 |
| | | | 8 1 |
| | | | 7 1 |
| | | | 6 1 |
| | | | 8 1 |
| | | | 7 1 |
| | | | 5 1 |
| | | | 8 1 |
| | | | 5 1 |
| | | | 7 1 |
| | | | 6 1 |
| | | | 7 1 |
| | | | 5 1 |
| | | | 8 1 |
| | | | 6 1 |
| | | | 5 1 |
| | | | 5 1 |
| | | | 7 1 |
| | | | 6 1 |
| | | | 8 1 |
| | | | 6 1 |
| | | | 7 1 |
| | | | 8 1 |
| | | | 7 1 |
| | | | 5 1 |
| | | | 6 1 |
| | | | 7 1 |
| | | | 8 1 |
| | | | 6 1 |

The End

TABLE LXXXIV: The computed irreducible representations for the valence bands for the compounds of SEBR and NLC in space group 217. See the list of tables.

| TQC | ICSD | NoE | SF |
|---|---|---|---|
| 217.1.1 | 427048 | 128 | F6Ni3O6Sb4 |
| | -GM -H | -PA | -P -N |
| | 6 2 8 4 | 7 2 | 7 2 5 2 |
| | 6 2 7 2 | 8 4 | 8 4 5 2 |
| | 8 4 8 4 | 6 2 | 6 2 5 2 |
| | 8 4 7 2 | 8 4 | 8 4 5 2 |
| | 6 2 6 2 | 6 2 | 6 2 5 2 |
| | 8 4 8 4 | 8 4 | 8 4 5 2 |
| | 7 2 7 2 | 7 2 | 7 2 5 2 |
| | 8 4 8 4 | 8 4 | 8 4 5 2 |
| | 7 2 7 2 | 7 2 | 7 2 5 2 |
| | 8 4 8 4 | 8 4 | 8 4 5 2 |
| | 6 2 6 2 | 6 2 | 6 2 5 2 |
| | 6 2 6 2 | 6 2 | 6 2 5 2 |
| | 8 4 6 2 | 8 4 | 8 4 5 2 |

Continued on next column

**TABLE LXXXIV– continued**

| TQC | ICSD | NoE | SF |
|---|---|---|---|
| | 8 4 8 4 | 6 2 | 6 2 5 2 |
| | 8 4 8 4 | 8 4 | 8 4 5 2 |
| | 7 2 7 2 | 7 2 | 7 2 5 2 |
| | 7 2 8 4 | 6 2 | 6 2 5 2 |
| | 6 2 8 4 | 8 4 | 8 4 5 2 |
| | 8 4 7 2 | 8 4 | 8 4 5 2 |
| | 7 2 6 2 | 7 2 | 7 2 5 2 |
| | 8 4 8 4 | 8 4 | 8 4 5 2 |
| | 7 2 6 2 | 7 2 | 7 2 5 2 |
| | 8 4 8 4 | 8 4 | 8 4 5 2 |
| | 7 2 7 2 | 8 4 | 8 4 5 2 |
| | 8 4 8 4 | 6 2 | 6 2 5 2 |
| | 8 4 6 2 | 8 4 | 8 4 5 2 |
| | 6 2 8 4 | 7 2 | 7 2 5 2 |
| | 8 4 8 4 | 8 4 | 8 4 5 2 |
| | 6 2 6 2 | 7 2 | 7 2 5 2 |
| | 7 2 7 2 | 6 2 | 6 2 5 2 |
| | 8 4 8 4 | 8 4 | 8 4 5 2 |
| | 8 4 6 2 | 6 2 | 6 2 5 2 |
| | 6 2 8 4 | 8 4 | 8 4 5 2 |
| | 7 2 7 2 | 7 2 | 7 2 5 2 |
| | 8 4 8 4 | 8 4 | 8 4 5 2 |
| | 8 4 7 2 | 7 2 | 7 2 5 2 |
| | 7 2 8 4 | 8 4 | 8 4 5 2 |
| | 6 2 6 2 | 6 2 | 6 2 5 2 |
| | 8 4 7 2 | 8 4 | 8 4 5 2 |
| | 8 4 8 4 | 8 4 | 8 4 5 2 |
| | 7 2 8 4 | 6 2 | 6 2 5 2 |
| | 6 2 7 2 | 8 4 | 8 4 5 2 |
| | 8 4 8 4 | 6 2 | 6 2 5 2 |
| | | | 5 2 |
| | | | 5 2 |
| | | | 5 2 |
| | | | 5 2 |
| | | | 5 2 |
| | | | 5 2 |
| | | | 5 2 |
| | | | 5 2 |
| | | | 5 2 |
| | | | 5 2 |
| | | | 5 2 |
| | | | 5 2 |
| | | | 5 2 |
| | | | 5 2 |
| | | | 5 2 |
| | | | 5 2 |
| | | | 5 2 |
| | | | 5 2 |
| | | | 5 2 |
| | | | 5 2 |
| | | | 5 2 |
| 217.1.2 | 427047 | 128 | F6Ni3O6Sb4 |
| | -GM -H | -PA | -P -N |
| | 6 2 8 4 | 7 2 | 7 2 5 2 |
| | 6 2 7 2 | 8 4 | 8 4 5 2 |
| | 8 4 8 4 | 6 2 | 6 2 5 2 |
| | 8 4 7 2 | 8 4 | 8 4 5 2 |
| | 6 2 6 2 | 6 2 | 6 2 5 2 |
| | 8 4 8 4 | 8 4 | 8 4 5 2 |
| | 7 2 7 2 | 7 2 | 7 2 5 2 |
| | 8 4 8 4 | 8 4 | 8 4 5 2 |

Continued on next column

**TABLE LXXXIV– continued**

| TQC | ICSD | NoE | SF |
|---|---|---|---|
| | 7 2 7 2 | 7 2 | 7 2 5 2 |
| | 8 4 8 4 | 8 4 | 8 4 5 2 |
| | 6 2 6 2 | 6 2 | 6 2 5 2 |
| | 6 2 6 2 | 6 2 | 6 2 5 2 |
| | 8 4 6 2 | 8 4 | 8 4 5 2 |
| | 8 4 8 4 | 6 2 | 6 2 5 2 |
| | 8 4 8 4 | 8 4 | 8 4 5 2 |
| | 7 2 7 2 | 7 2 | 7 2 5 2 |
| | 7 2 8 4 | 6 2 | 6 2 5 2 |
| | 6 2 8 4 | 8 4 | 8 4 5 2 |
| | 8 4 7 2 | 8 4 | 8 4 5 2 |
| | 7 2 6 2 | 7 2 | 7 2 5 2 |
| | 8 4 8 4 | 8 4 | 8 4 5 2 |
| | 7 2 6 2 | 7 2 | 7 2 5 2 |
| | 8 4 8 4 | 8 4 | 8 4 5 2 |
| | 7 2 7 2 | 8 4 | 8 4 5 2 |
| | 8 4 8 4 | 6 2 | 6 2 5 2 |
| | 8 4 6 2 | 8 4 | 8 4 5 2 |
| | 6 2 8 4 | 7 2 | 7 2 5 2 |
| | 8 4 8 4 | 8 4 | 8 4 5 2 |
| | 6 2 6 2 | 6 2 | 6 2 5 2 |
| | 7 2 7 2 | 7 2 | 7 2 5 2 |
| | 8 4 8 4 | 8 4 | 8 4 5 2 |
| | 8 4 6 2 | 6 2 | 6 2 5 2 |
| | 6 2 8 4 | 8 4 | 8 4 5 2 |
| | 7 2 7 2 | 7 2 | 7 2 5 2 |
| | 8 4 8 4 | 8 4 | 8 4 5 2 |
| | 8 4 7 2 | 7 2 | 7 2 5 2 |
| | 7 2 8 4 | 8 4 | 8 4 5 2 |
| | 6 2 6 2 | 6 2 | 6 2 5 2 |
| | 8 4 7 2 | 8 4 | 8 4 5 2 |
| | 8 4 8 4 | 8 4 | 8 4 5 2 |
| | 7 2 8 4 | 6 2 | 6 2 5 2 |
| | 6 2 7 2 | 8 4 | 8 4 5 2 |
| | 8 4 8 4 | 6 2 | 6 2 5 2 |
| | | | 5 2 |
| | | | 5 2 |
| | | | 5 2 |
| | | | 5 2 |
| | | | 5 2 |
| | | | 5 2 |
| | | | 5 2 |
| | | | 5 2 |
| | | | 5 2 |
| | | | 5 2 |
| | | | 5 2 |
| | | | 5 2 |
| | | | 5 2 |
| | | | 5 2 |
| | | | 5 2 |
| | | | 5 2 |
| | | | 5 2 |
| | | | 5 2 |
| | | | 5 2 |
| | | | 5 2 |
| | | | 5 2 |
| 217.1.3 | 84572 | 230 | Cu12S13Sb4 |
| | -GM -H | -P | -PA -N |
| | 6 2 6 2 | 6 2 | 6 2 5 2 |
| | 6 2 7 2 | 7 2 | 7 2 5 2 |
| | 7 2 8 4 | 8 4 | 8 4 5 2 |

Continued on next column

**TABLE LXXXIV– continued**

| TQC | ICSD | NoE | SF |
|---|---|---|---|
| | 8 4 6 2 | 6 2 | 6 2 5 2 |
| | 8 4 8 4 | 7 2 | 7 2 5 2 |
| | 6 2 7 2 | 8 4 | 8 4 5 2 |
| | 8 4 8 4 | 8 4 | 8 4 5 2 |
| | 7 2 6 2 | 6 2 | 6 2 5 2 |
| | 8 4 8 4 | 8 4 | 8 4 5 2 |
| | 6 2 6 2 | 6 2 | 6 2 5 2 |
| | 7 2 7 2 | 7 2 | 7 2 5 2 |
| | 8 4 8 4 | 8 4 | 8 4 5 2 |
| | 7 2 7 2 | 7 2 | 7 2 5 2 |
| | 8 4 8 4 | 8 4 | 8 4 5 2 |
| | 8 4 7 2 | 8 4 | 8 4 5 2 |
| | 8 4 8 4 | 7 2 | 7 2 5 2 |
| | 6 2 6 2 | 6 2 | 6 2 5 2 |
| | 6 2 8 4 | 8 4 | 8 4 5 2 |
| | 8 4 8 4 | 8 4 | 8 4 5 2 |
| | 8 4 8 4 | 6 2 | 6 2 5 2 |
| | 7 2 7 2 | 7 2 | 7 2 5 2 |
| | 7 2 6 2 | 8 4 | 8 4 5 2 |
| | 8 4 7 2 | 7 2 | 7 2 5 2 |
| | 6 2 8 4 | 6 2 | 6 2 5 2 |
| | 8 4 6 2 | 8 4 | 8 4 5 2 |
| | 7 2 8 4 | 8 4 | 8 4 5 2 |
| | 8 4 8 4 | 8 4 | 8 4 5 2 |
| | 6 2 7 2 | 7 2 | 7 2 5 2 |
| | 6 2 8 4 | 8 4 | 8 4 5 2 |
| | 8 4 6 2 | 8 4 | 8 4 5 2 |
| | 8 4 8 4 | 6 2 | 6 2 5 2 |
| | 7 2 7 2 | 6 2 | 6 2 5 2 |
| | 8 4 8 4 | 7 2 | 7 2 5 2 |
| | 7 2 8 4 | 8 4 | 8 4 5 2 |
| | 6 2 6 2 | 6 2 | 6 2 5 2 |
| | 8 4 6 2 | 8 4 | 8 4 5 2 |
| | 7 2 7 2 | 7 2 | 7 2 5 2 |
| | 8 4 8 4 | 8 4 | 8 4 5 2 |
| | 8 4 7 2 | 7 2 | 7 2 5 2 |
| | 7 2 8 4 | 8 4 | 8 4 5 2 |
| | 6 2 6 2 | 8 4 | 8 4 5 2 |
| | 8 4 8 4 | 6 2 | 6 2 5 2 |
| | 8 4 8 4 | 8 4 | 8 4 5 2 |
| | 8 4 8 4 | 8 4 | 8 4 5 2 |
| | 7 2 7 2 | 7 2 | 7 2 5 2 |
| | 6 2 6 2 | 6 2 | 6 2 5 2 |
| | 8 4 8 4 | 8 4 | 8 4 5 2 |
| | 7 2 6 2 | 8 4 | 8 4 5 2 |
| | 6 2 6 2 | 6 2 | 6 2 5 2 |
| | 8 4 8 4 | 7 2 | 7 2 5 2 |
| | 8 4 7 2 | 6 2 | 6 2 5 2 |
| | 6 2 8 4 | 8 4 | 8 4 5 2 |
| | 8 4 8 4 | 8 4 | 8 4 5 2 |
| | 7 2 6 2 | 7 2 | 7 2 5 2 |
| | 8 4 8 4 | 6 2 | 6 2 5 2 |
| | 7 2 7 2 | 8 4 | 8 4 5 2 |
| | 6 2 7 2 | 7 2 | 7 2 5 2 |
| | 8 4 8 4 | 8 4 | 8 4 5 2 |
| | 8 4 8 4 | 8 4 | 8 4 5 2 |
| | 7 2 7 2 | 6 2 | 6 2 5 2 |
| | 8 4 6 2 | 8 4 | 8 4 5 2 |
| | 6 2 8 4 | 7 2 | 7 2 5 2 |
| | 6 2 8 4 | 8 4 | 8 4 5 2 |
| | 8 4 6 2 | 8 4 | 8 4 5 2 |

Continued on next column

**TABLE LXXXIV– continued**

| TQC | ICSD | NoE | SF |
|---|---|---|---|
| | 7 2 6 2 | 7 2 | 7 2 5 2 |
| | 8 4 8 4 | 6 2 | 6 2 5 2 |
| | 7 2 7 2 | 8 4 | 8 4 5 2 |
| | 8 4 8 4 | 7 2 | 7 2 5 2 |
| | 8 4 6 2 | 8 4 | 8 4 5 2 |
| | 8 4 8 4 | 6 2 | 6 2 5 2 |
| | 6 2 8 4 | 6 2 | 6 2 5 2 |
| | 7 2 7 2 | 8 4 | 8 4 5 2 |
| | 8 4 6 2 | 7 2 | 7 2 5 2 |
| | 6 2 8 4 | 7 2 | 7 2 5 2 |
| | 8 4 7 2 | 8 4 | 8 4 5 2 |
| | 8 4 8 4 | 8 4 | 8 4 5 2 |
| | 7 2 8 4 | 8 4 | 8 4 5 2 |
| | | | 5 2 |
| | | | 5 2 |
| | | | 5 2 |
| | | | 5 2 |
| | | | 5 2 |
| | | | 5 2 |
| | | | 5 2 |
| | | | 5 2 |
| | | | 5 2 |
| | | | 5 2 |
| | | | 5 2 |
| | | | 5 2 |
| | | | 5 2 |
| | | | 5 2 |
| | | | 5 2 |
| | | | 5 2 |
| | | | 5 2 |
| | | | 5 2 |
| | | | 5 2 |
| | | | 5 2 |
| | | | 5 2 |
| | | | 5 2 |
| | | | 5 2 |
| | | | 5 2 |
| | | | 5 2 |
| | | | 5 2 |
| | | | 5 2 |
| | | | 5 2 |
| | | | 5 2 |
| | | | 5 2 |
| | | | 5 2 |
| | | | 5 2 |
| | | | 5 2 |
| | | | 5 2 |
| | | | 5 2 |
| | | | 5 2 |
| | | | 5 2 |
| | | | 5 2 |
| 217.1.4 | 84571 | 230 | Cu12S13Sb4 |
| | -GM -H | -P | -PA -N |
| | 6 2 6 2 | 6 2 | 6 2 5 2 |
| | 6 2 7 2 | 7 2 | 7 2 5 2 |
| | 7 2 8 4 | 8 4 | 8 4 5 2 |
| | 8 4 6 2 | 6 2 | 6 2 5 2 |
| | 8 4 8 4 | 7 2 | 7 2 5 2 |
| | 6 2 7 2 | 8 4 | 8 4 5 2 |
| | 8 4 8 4 | 8 4 | 8 4 5 2 |
| | 7 2 6 2 | 6 2 | 6 2 5 2 |

Continued on next column

**TABLE LXXXIV– continued**

| TQC | ICSD | NoE | SF |
|---|---|---|---|
| | 8 4 8 4 | 8 4 | 8 4 5 2 |
| | 6 2 6 2 | 6 2 | 6 2 5 2 |
| | 7 2 7 2 | 7 2 | 7 2 5 2 |
| | 8 4 8 4 | 8 4 | 8 4 5 2 |
| | 7 2 7 2 | 7 2 | 7 2 5 2 |
| | 8 4 8 4 | 8 4 | 8 4 5 2 |
| | 8 4 7 2 | 8 4 | 8 4 5 2 |
| | 8 4 8 4 | 6 2 | 6 2 5 2 |
| | 6 2 6 2 | 7 2 | 7 2 5 2 |
| | 6 2 8 4 | 8 4 | 8 4 5 2 |
| | 8 4 8 4 | 8 4 | 8 4 5 2 |
| | 8 4 8 4 | 6 2 | 6 2 5 2 |
| | 7 2 7 2 | 7 2 | 7 2 5 2 |
| | 7 2 6 2 | 8 4 | 8 4 5 2 |
| | 8 4 7 2 | 7 2 | 7 2 5 2 |
| | 6 2 8 4 | 6 2 | 6 2 5 2 |
| | 8 4 6 2 | 8 4 | 8 4 5 2 |
| | 7 2 8 4 | 8 4 | 8 4 5 2 |
| | 8 4 8 4 | 8 4 | 8 4 5 2 |
| | 6 2 7 2 | 7 2 | 7 2 5 2 |
| | 6 2 8 4 | 8 4 | 8 4 5 2 |
| | 8 4 6 2 | 8 4 | 8 4 5 2 |
| | 8 4 8 4 | 6 2 | 6 2 5 2 |
| | 7 2 7 2 | 6 2 | 6 2 5 2 |
| | 8 4 8 4 | 7 2 | 7 2 5 2 |
| | 7 2 8 4 | 8 4 | 8 4 5 2 |
| | 6 2 6 2 | 8 4 | 8 4 5 2 |
| | 8 4 6 2 | 6 2 | 6 2 5 2 |
| | 7 2 7 2 | 7 2 | 7 2 5 2 |
| | 8 4 8 4 | 8 4 | 8 4 5 2 |
| | 8 4 7 2 | 7 2 | 7 2 5 2 |
| | 7 2 8 4 | 8 4 | 8 4 5 2 |
| | 6 2 6 2 | 8 4 | 8 4 5 2 |
| | 8 4 8 4 | 6 2 | 6 2 5 2 |
| | 8 4 8 4 | 8 4 | 8 4 5 2 |
| | 8 4 8 4 | 8 4 | 8 4 5 2 |
| | 7 2 7 2 | 7 2 | 7 2 5 2 |
| | 6 2 6 2 | 6 2 | 6 2 5 2 |
| | 8 4 8 4 | 8 4 | 8 4 5 2 |
| | 7 2 6 2 | 8 4 | 8 4 5 2 |
| | 6 2 6 2 | 7 2 | 7 2 5 2 |
| | 8 4 8 4 | 6 2 | 6 2 5 2 |
| | 8 4 7 2 | 6 2 | 6 2 5 2 |
| | 6 2 8 4 | 8 4 | 8 4 5 2 |
| | 8 4 8 4 | 8 4 | 8 4 5 2 |
| | 7 2 6 2 | 7 2 | 7 2 5 2 |
| | 8 4 8 4 | 6 2 | 6 2 5 2 |
| | 6 2 7 2 | 8 4 | 8 4 5 2 |
| | 7 2 7 2 | 8 4 | 8 4 5 2 |
| | 8 4 8 4 | 7 2 | 7 2 5 2 |
| | 8 4 8 4 | 8 4 | 8 4 5 2 |
| | 7 2 7 2 | 8 4 | 8 4 5 2 |
| | 8 4 6 2 | 6 2 | 6 2 5 2 |
| | 6 2 8 4 | 7 2 | 7 2 5 2 |
| | 6 2 8 4 | 8 4 | 8 4 5 2 |
| | 8 4 6 2 | 8 4 | 8 4 5 2 |
| | 7 2 6 2 | 7 2 | 7 2 5 2 |
| | 8 4 8 4 | 6 2 | 6 2 5 2 |
| | 7 2 7 2 | 8 4 | 8 4 5 2 |
| | 8 4 8 4 | 7 2 | 7 2 5 2 |
| | 8 4 6 2 | 8 4 | 8 4 5 2 |

Continued on next column

**TABLE LXXXIV– continued**

| TQC | ICSD | NoE | SF |
|---|---|---|---|
| | 6 2 8 4 | 6 2 | 6 2 5 2 |
| | 8 4 8 4 | 6 2 | 6 2 5 2 |
| | 8 4 7 2 | 8 4 | 8 4 5 2 |
| | 7 2 6 2 | 7 2 | 7 2 5 2 |
| | 6 2 8 4 | 7 2 | 7 2 5 2 |
| | 8 4 7 2 | 8 4 | 8 4 5 2 |
| | 8 4 8 4 | 8 4 | 8 4 5 2 |
| | 7 2 8 4 | 8 4 | 8 4 5 2 |
| | | | 5 2 |
| | | | 5 2 |
| | | | 5 2 |
| | | | 5 2 |
| | | | 5 2 |
| | | | 5 2 |
| | | | 5 2 |
| | | | 5 2 |
| | | | 5 2 |
| | | | 5 2 |
| | | | 5 2 |
| | | | 5 2 |
| | | | 5 2 |
| | | | 5 2 |
| | | | 5 2 |
| | | | 5 2 |
| | | | 5 2 |
| | | | 5 2 |
| | | | 5 2 |
| | | | 5 2 |
| | | | 5 2 |
| | | | 5 2 |
| | | | 5 2 |
| | | | 5 2 |
| | | | 5 2 |
| | | | 5 2 |
| | | | 5 2 |
| | | | 5 2 |
| | | | 5 2 |
| | | | 5 2 |
| | | | 5 2 |
| | | | 5 2 |
| | | | 5 2 |
| | | | 5 2 |
| | | | 5 2 |
| | | | 5 2 |
| | | | 5 2 |
| | | | 5 2 |
| 217.1.5 | 62116 | 230 | Cu12S13Sb4 |
| | -GM -H | -P | -PA -N |
| | 6 2 6 2 | 6 2 | 6 2 5 2 |
| | 6 2 7 2 | 7 2 | 7 2 5 2 |
| | 7 2 8 4 | 8 4 | 8 4 5 2 |
| | 8 4 6 2 | 6 2 | 6 2 5 2 |
| | 8 4 8 4 | 7 2 | 7 2 5 2 |
| | 6 2 7 2 | 8 4 | 8 4 5 2 |
| | 8 4 8 4 | 8 4 | 8 4 5 2 |
| | 7 2 6 2 | 6 2 | 6 2 5 2 |
| | 8 4 8 4 | 8 4 | 8 4 5 2 |
| | 6 2 6 2 | 6 2 | 6 2 5 2 |
| | 7 2 7 2 | 7 2 | 7 2 5 2 |
| | 8 4 8 4 | 8 4 | 8 4 5 2 |
| | 7 2 7 2 | 7 2 | 7 2 5 2 |

Continued on next column

**TABLE LXXXIV– continued**

| TQC | ICSD | NoE | SF |
|---|---|---|---|
| | 8 4 8 4 | 8 4 | 8 4 5 2 |
| | 8 4 7 2 | 8 4 | 8 4 5 2 |
| | 8 4 8 4 | 7 2 | 7 2 5 2 |
| | 6 2 6 2 | 8 4 | 8 4 5 2 |
| | 6 2 8 4 | 6 2 | 6 2 5 2 |
| | 8 4 8 4 | 8 4 | 8 4 5 2 |
| | 8 4 8 4 | 6 2 | 6 2 5 2 |
| | 7 2 7 2 | 7 2 | 7 2 5 2 |
| | 7 2 6 2 | 8 4 | 8 4 5 2 |
| | 8 4 8 4 | 6 2 | 6 2 5 2 |
| | 6 2 7 2 | 7 2 | 7 2 5 2 |
| | 8 4 6 2 | 8 4 | 8 4 5 2 |
| | 7 2 8 4 | 8 4 | 8 4 5 2 |
| | 8 4 8 4 | 8 4 | 8 4 5 2 |
| | 6 2 7 2 | 7 2 | 7 2 5 2 |
| | 6 2 8 4 | 8 4 | 8 4 5 2 |
| | 8 4 6 2 | 8 4 | 8 4 5 2 |
| | 8 4 8 4 | 6 2 | 6 2 5 2 |
| | 7 2 7 2 | 6 2 | 6 2 5 2 |
| | 8 4 8 4 | 8 4 | 8 4 5 2 |
| | 7 2 8 4 | 7 2 | 7 2 5 2 |
| | 6 2 6 2 | 6 2 | 6 2 5 2 |
| | 8 4 6 2 | 8 4 | 8 4 5 2 |
| | 7 2 7 2 | 7 2 | 7 2 5 2 |
| | 8 4 8 4 | 8 4 | 8 4 5 2 |
| | 8 4 7 2 | 7 2 | 7 2 5 2 |
| | 7 2 8 4 | 8 4 | 8 4 5 2 |
| | 6 2 6 2 | 8 4 | 8 4 5 2 |
| | 8 4 8 4 | 6 2 | 6 2 5 2 |
| | 8 4 8 4 | 8 4 | 8 4 5 2 |
| | 8 4 8 4 | 8 4 | 8 4 5 2 |
| | 7 2 7 2 | 6 2 | 6 2 5 2 |
| | 6 2 6 2 | 7 2 | 7 2 5 2 |
| | 8 4 8 4 | 8 4 | 8 4 5 2 |
| | 7 2 6 2 | 8 4 | 8 4 5 2 |
| | 6 2 6 2 | 6 2 | 6 2 5 2 |
| | 8 4 8 4 | 7 2 | 7 2 5 2 |
| | 8 4 7 2 | 8 4 | 8 4 5 2 |
| | 6 2 8 4 | 6 2 | 6 2 5 2 |
| | 8 4 8 4 | 8 4 | 8 4 5 2 |
| | 7 2 6 2 | 7 2 | 7 2 5 2 |
| | 8 4 7 2 | 6 2 | 6 2 5 2 |
| | 7 2 8 4 | 8 4 | 8 4 5 2 |
| | 6 2 7 2 | 7 2 | 7 2 5 2 |
| | 8 4 8 4 | 8 4 | 8 4 5 2 |
| | 8 4 8 4 | 8 4 | 8 4 5 2 |
| | 7 2 7 2 | 8 4 | 8 4 5 2 |
| | 8 4 8 4 | 6 2 | 6 2 5 2 |
| | 6 2 6 2 | 7 2 | 7 2 5 2 |
| | 6 2 8 4 | 8 4 | 8 4 5 2 |
| | 8 4 6 2 | 8 4 | 8 4 5 2 |
| | 7 2 6 2 | 7 2 | 7 2 5 2 |
| | 8 4 8 4 | 6 2 | 6 2 5 2 |
| | 7 2 7 2 | 8 4 | 8 4 5 2 |
| | 8 4 8 4 | 7 2 | 7 2 5 2 |
| | 8 4 6 2 | 8 4 | 8 4 5 2 |
| | 8 4 8 4 | 6 2 | 6 2 5 2 |
| | 6 2 7 2 | 6 2 | 6 2 5 2 |
| | 7 2 8 4 | 8 4 | 8 4 5 2 |
| | 8 4 6 2 | 7 2 | 7 2 5 2 |
| | 6 2 8 4 | 7 2 | 7 2 5 2 |

Continued on next column

**TABLE LXXXIV– continued**

| TQC | ICSD | NoE | SF |
|---|---|---|---|
| | 8 4 7 2 | 8 4 | 8 4 5 2 |
| | 8 4 8 4 | 8 4 | 8 4 5 2 |
| | 7 2 8 4 | 8 4 | 8 4 5 2 |
| | | | 5 2 |
| | | | 5 2 |
| | | | 5 2 |
| | | | 5 2 |
| | | | 5 2 |
| | | | 5 2 |
| | | | 5 2 |
| | | | 5 2 |
| | | | 5 2 |
| | | | 5 2 |
| | | | 5 2 |
| | | | 5 2 |
| | | | 5 2 |
| | | | 5 2 |
| | | | 5 2 |
| | | | 5 2 |
| | | | 5 2 |
| | | | 5 2 |
| | | | 5 2 |
| | | | 5 2 |
| | | | 5 2 |
| | | | 5 2 |
| | | | 5 2 |
| | | | 5 2 |
| | | | 5 2 |
| | | | 5 2 |
| | | | 5 2 |
| | | | 5 2 |
| | | | 5 2 |
| | | | 5 2 |
| | | | 5 2 |
| | | | 5 2 |
| | | | 5 2 |
| | | | 5 2 |
| | | | 5 2 |
| | | | 5 2 |
| | | | 5 2 |
| | | | 5 2 |
| 217.1.6 | 25707 | 230 | Cu12S13Sb4 |
| | -GM -H | -P | -PA -N |
| | 6 2 6 2 | 6 2 | 6 2 5 2 |
| | 6 2 7 2 | 7 2 | 7 2 5 2 |
| | 7 2 8 4 | 8 4 | 8 4 5 2 |
| | 8 4 6 2 | 6 2 | 6 2 5 2 |
| | 8 4 8 4 | 7 2 | 7 2 5 2 |
| | 6 2 7 2 | 8 4 | 8 4 5 2 |
| | 8 4 8 4 | 8 4 | 8 4 5 2 |
| | 7 2 6 2 | 6 2 | 6 2 5 2 |
| | 8 4 8 4 | 8 4 | 8 4 5 2 |
| | 6 2 6 2 | 6 2 | 6 2 5 2 |
| | 7 2 7 2 | 7 2 | 7 2 5 2 |
| | 8 4 8 4 | 8 4 | 8 4 5 2 |
| | 7 2 7 2 | 7 2 | 7 2 5 2 |
| | 8 4 8 4 | 8 4 | 8 4 5 2 |
| | 8 4 7 2 | 8 4 | 8 4 5 2 |
| | 6 2 8 4 | 7 2 | 7 2 5 2 |
| | 8 4 6 2 | 8 4 | 8 4 5 2 |
| | 6 2 8 4 | 6 2 | 6 2 5 2 |

Continued on next column

**TABLE LXXXIV– continued**

| TQC | ICSD | NoE | SF |
|---|---|---|---|
| | 8 4 8 4 | 8 4 | 8 4 5 2 |
| | 8 4 7 2 | 6 2 | 6 2 5 2 |
| | 7 2 8 4 | 7 2 | 7 2 5 2 |
| | 7 2 6 2 | 8 4 | 8 4 5 2 |
| | 8 4 8 4 | 6 2 | 6 2 5 2 |
| | 6 2 7 2 | 7 2 | 7 2 5 2 |
| | 8 4 6 2 | 8 4 | 8 4 5 2 |
| | 6 2 8 4 | 8 4 | 8 4 5 2 |
| | 8 4 8 4 | 8 4 | 8 4 5 2 |
| | 7 2 7 2 | 7 2 | 7 2 5 2 |
| | 6 2 8 4 | 8 4 | 8 4 5 2 |
| | 8 4 6 2 | 8 4 | 8 4 5 2 |
| | 8 4 8 4 | 6 2 | 6 2 5 2 |
| | 7 2 7 2 | 6 2 | 6 2 5 2 |
| | 8 4 8 4 | 8 4 | 8 4 5 2 |
| | 7 2 8 4 | 7 2 | 7 2 5 2 |
| | 6 2 6 2 | 6 2 | 6 2 5 2 |
| | 8 4 6 2 | 8 4 | 8 4 5 2 |
| | 7 2 7 2 | 7 2 | 7 2 5 2 |
| | 8 4 8 4 | 8 4 | 8 4 5 2 |
| | 8 4 7 2 | 7 2 | 7 2 5 2 |
| | 7 2 8 4 | 8 4 | 8 4 5 2 |
| | 6 2 6 2 | 8 4 | 8 4 5 2 |
| | 8 4 8 4 | 6 2 | 6 2 5 2 |
| | 8 4 8 4 | 8 4 | 8 4 5 2 |
| | 8 4 8 4 | 8 4 | 8 4 5 2 |
| | 7 2 7 2 | 6 2 | 6 2 5 2 |
| | 6 2 8 4 | 7 2 | 7 2 5 2 |
| | 8 4 6 2 | 8 4 | 8 4 5 2 |
| | 6 2 6 2 | 6 2 | 6 2 5 2 |
| | 7 2 6 2 | 8 4 | 8 4 5 2 |
| | 8 4 8 4 | 7 2 | 7 2 5 2 |
| | 6 2 7 2 | 8 4 | 8 4 5 2 |
| | 8 4 8 4 | 6 2 | 6 2 5 2 |
| | 8 4 8 4 | 7 2 | 7 2 5 2 |
| | 7 2 6 2 | 8 4 | 8 4 5 2 |
| | 8 4 7 2 | 8 4 | 8 4 5 2 |
| | 7 2 8 4 | 6 2 | 6 2 5 2 |
| | 6 2 7 2 | 7 2 | 7 2 5 2 |
| | 8 4 8 4 | 8 4 | 8 4 5 2 |
| | 8 4 8 4 | 8 4 | 8 4 5 2 |
| | 7 2 7 2 | 6 2 | 6 2 5 2 |
| | 8 4 8 4 | 8 4 | 8 4 5 2 |
| | 6 2 6 2 | 7 2 | 7 2 5 2 |
| | 6 2 8 4 | 8 4 | 8 4 5 2 |
| | 8 4 6 2 | 8 4 | 8 4 5 2 |
| | 7 2 6 2 | 7 2 | 7 2 5 2 |
| | 8 4 8 4 | 6 2 | 6 2 5 2 |
| | 7 2 7 2 | 8 4 | 8 4 5 2 |
| | 8 4 8 4 | 7 2 | 7 2 5 2 |
| | 8 4 6 2 | 8 4 | 8 4 5 2 |
| | 8 4 8 4 | 6 2 | 6 2 5 2 |
| | 6 2 7 2 | 6 2 | 6 2 5 2 |
| | 7 2 8 4 | 8 4 | 8 4 5 2 |
| | 8 4 6 2 | 7 2 | 7 2 5 2 |
| | 6 2 8 4 | 8 4 | 8 4 5 2 |
| | 8 4 7 2 | 7 2 | 7 2 5 2 |
| | 8 4 8 4 | 8 4 | 8 4 5 2 |
| | 7 2 8 4 | 8 4 | 8 4 5 2 |
| | | | 5 2 |
| | | | 5 2 |

Continued on next column

**TABLE LXXXIV– continued**

| TQC | ICSD | NoE | SF |
|---|---|---|---|
| | | | 5 2 |
| | | | 5 2 |
| | | | 5 2 |
| | | | 5 2 |
| | | | 5 2 |
| | | | 5 2 |
| | | | 5 2 |
| | | | 5 2 |
| | | | 5 2 |
| | | | 5 2 |
| | | | 5 2 |
| | | | 5 2 |
| | | | 5 2 |
| | | | 5 2 |
| | | | 5 2 |
| | | | 5 2 |
| | | | 5 2 |
| | | | 5 2 |
| | | | 5 2 |
| | | | 5 2 |
| | | | 5 2 |
| | | | 5 2 |
| | | | 5 2 |
| | | | 5 2 |
| | | | 5 2 |
| | | | 5 2 |
| | | | 5 2 |
| | | | 5 2 |
| | | | 5 2 |
| | | | 5 2 |
| | | | 5 2 |
| | | | 5 2 |
| | | | 5 2 |
| | | | 5 2 |
| | | | 5 2 |
| | | | 5 2 |
| 217.1.7 | 33588 | 224 | As4Cu12S12 |
| | -GM -H | -P | -PA -N |
| | 6 2 7 2 | 7 2 | 7 2 5 2 |
| | 7 2 8 4 | 8 4 | 8 4 5 2 |
| | 8 4 6 2 | 6 2 | 6 2 5 2 |
| | 8 4 8 4 | 7 2 | 7 2 5 2 |
| | 6 2 7 2 | 8 4 | 8 4 5 2 |
| | 8 4 8 4 | 8 4 | 8 4 5 2 |
| | 7 2 6 2 | 6 2 | 6 2 5 2 |
| | 8 4 8 4 | 8 4 | 8 4 5 2 |
| | 6 2 6 2 | 6 2 | 6 2 5 2 |
| | 7 2 7 2 | 7 2 | 7 2 5 2 |
| | 8 4 8 4 | 8 4 | 8 4 5 2 |
| | 8 4 7 2 | 8 4 | 8 4 5 2 |
| | 6 2 8 4 | 7 2 | 7 2 5 2 |
| | 6 2 6 2 | 8 4 | 8 4 5 2 |
| | 8 4 8 4 | 6 2 | 6 2 5 2 |
| | 8 4 8 4 | 7 2 | 7 2 5 2 |
| | 8 4 7 2 | 8 4 | 8 4 5 2 |
| | 7 2 6 2 | 8 4 | 8 4 5 2 |
| | 7 2 8 4 | 6 2 | 6 2 5 2 |
| | 8 4 8 4 | 6 2 | 6 2 5 2 |
| | 6 2 8 4 | 8 4 | 8 4 5 2 |
| | 8 4 7 2 | 7 2 | 7 2 5 2 |
| | 7 2 8 4 | 8 4 | 8 4 5 2 |
| | 8 4 6 2 | 8 4 | 8 4 5 2 |
| | 6 2 7 2 | 7 2 | 7 2 5 2 |
| | 8 4 8 4 | 8 4 | 8 4 5 2 |
| | 6 2 6 2 | 6 2 | 6 2 5 2 |
| | 7 2 8 4 | 8 4 | 8 4 5 2 |
| | 8 4 7 2 | 6 2 | 6 2 5 2 |
| | 8 4 8 4 | 6 2 | 6 2 5 2 |
| | 7 2 8 4 | 8 4 | 8 4 5 2 |
| | 6 2 6 2 | 8 4 | 8 4 5 2 |
| | 8 4 6 2 | 7 2 | 7 2 5 2 |
| | 7 2 7 2 | 7 2 | 7 2 5 2 |
| | 8 4 8 4 | 8 4 | 8 4 5 2 |
| | 6 2 6 2 | 6 2 | 6 2 5 2 |
| | 8 4 6 2 | 8 4 | 8 4 5 2 |
| | 8 4 8 4 | 7 2 | 7 2 5 2 |
| | 7 2 7 2 | 8 4 | 8 4 5 2 |
| | 6 2 6 2 | 7 2 | 7 2 5 2 |
| | 7 2 8 4 | 8 4 | 8 4 5 2 |
| | 8 4 8 4 | 6 2 | 6 2 5 2 |
| | 6 2 7 2 | 8 4 | 8 4 5 2 |
| | 8 4 8 4 | 6 2 | 6 2 5 2 |
| | 7 2 6 2 | 8 4 | 8 4 5 2 |
| | 8 4 8 4 | 7 2 | 7 2 5 2 |
| | 8 4 7 2 | 8 4 | 8 4 5 2 |
| | 7 2 6 2 | 6 2 | 6 2 5 2 |
| | 8 4 8 4 | 7 2 | 7 2 5 2 |
| | 6 2 8 4 | 8 4 | 8 4 5 2 |
| | 8 4 7 2 | 8 4 | 8 4 5 2 |
| | 8 4 8 4 | 6 2 | 6 2 5 2 |
| | 7 2 8 4 | 8 4 | 8 4 5 2 |
| | 8 4 6 2 | 8 4 | 8 4 5 2 |
| | 8 4 8 4 | 7 2 | 7 2 5 2 |
| | 6 2 7 2 | 6 2 | 6 2 5 2 |
| | 7 2 8 4 | 7 2 | 7 2 5 2 |
| | 8 4 8 4 | 8 4 | 8 4 5 2 |
| | 7 2 6 2 | 8 4 | 8 4 5 2 |
| | 6 2 7 2 | 8 4 | 8 4 5 2 |
| | 8 4 8 4 | 7 2 | 7 2 5 2 |
| | 6 2 7 2 | 8 4 | 8 4 5 2 |
| | 8 4 8 4 | 6 2 | 6 2 5 2 |
| | 7 2 6 2 | 7 2 | 7 2 5 2 |
| | 8 4 8 4 | 8 4 | 8 4 5 2 |
| | 7 2 7 2 | 8 4 | 8 4 5 2 |
| | 8 4 6 2 | 6 2 | 6 2 5 2 |
| | 8 4 8 4 | 7 2 | 7 2 5 2 |
| | 6 2 7 2 | 8 4 | 8 4 5 2 |
| | 8 4 8 4 | 6 2 | 6 2 5 2 |
| | 6 2 8 4 | 6 2 | 6 2 5 2 |
| | 6 2 6 2 | 8 4 | 8 4 5 2 |
| | 8 4 8 4 | 7 2 | 7 2 5 2 |
| | 8 4 8 4 | 8 4 | 8 4 5 2 |
| | 7 2 6 2 | 8 4 | 8 4 5 2 |
| | | | 5 2 |
| | | | 5 2 |
| | | | 5 2 |
| | | | 5 2 |
| | | | 5 2 |
| | | | 5 2 |
| | | | 5 2 |
| | | | 5 2 |
| | | | 5 2 |

Continued on next column

Continued on next column

**TABLE LXXXIV– continued**

| TQC | ICSD | NoE | SF |
|---|---|---|---|
| | | | 5 2 |
| | | | 5 2 |
| | | | 5 2 |
| | | | 5 2 |
| | | | 5 2 |
| | | | 5 2 |
| | | | 5 2 |
| | | | 5 2 |
| | | | 5 2 |
| | | | 5 2 |
| | | | 5 2 |
| | | | 5 2 |
| | | | 5 2 |
| | | | 5 2 |
| | | | 5 2 |
| | | | 5 2 |
| | | | 5 2 |
| | | | 5 2 |
| | | | 5 2 |
| | | | 5 2 |
| | | | 5 2 |
| | | | 5 2 |
| | | | 5 2 |
| | | | 5 2 |
| | | | 5 2 |
| | | | 5 2 |
| | | | 5 2 |
| | | | 5 2 |

The End

TABLE LXXXV: The computed irreducible representations for the valence bands for the compounds of SEBR and NLC in space group 220. See the list of tables.

| TQC | ICSD | NoE | SF |
|---|---|---|---|
| 220.2.1 | 155383 | 136 | C3Y2 |
| | -GM -H | -PA -P | -N |
| | 6 2 6+7 4 | 6 2 6 2 | 2+5 2 |
| | 7 2 8+8 8 | 7 3 7 3 | 3+4 2 |
| | 7 2 6+7 4 | 8 3 8 3 | 2+5 2 |
| | 8 4 6+7 4 | 6 2 6 2 | 3+4 2 |
| | 6 2 8+8 8 | 7 3 7 3 | 2+5 2 |
| | 8 4 6+7 4 | 8 3 8 3 | 3+4 2 |
| | 8 4 8+8 8 | 7 3 7 3 | 2+5 2 |
| | 6 2 6+7 4 | 6 2 6 2 | 3+4 2 |
| | 8 4 8+8 8 | 8 3 8 3 | 3+4 2 |
| | 7 2 6+7 4 | 7 3 7 3 | 2+5 2 |
| | 6 2 8+8 8 | 6 2 6 2 | 2+5 2 |
| | 7 2 6+7 4 | 8 3 8 3 | 3+4 2 |
| | 8 4 8+8 8 | 5 1 5 1 | 3+4 2 |
| | 8 4 6+7 4 | 8 3 8 3 | 2+5 2 |
| | 7 2 8+8 8 | 4 1 4 1 | 2+5 2 |
| | 6 2 6+7 4 | 7 3 7 3 | 3+4 2 |
| | 8 4 8+8 8 | 7 3 7 3 | 3+4 2 |
| | 8 4 8+8 8 | 8 3 8 3 | 2+5 2 |
| | 8 4 6+7 4 | 6 2 6 2 | 3+4 2 |
| | 6 2 8+8 8 | 8 3 8 3 | 2+5 2 |
| | 7 2 6+7 4 | 6 2 6 2 | 3+4 2 |
| | 8 4 6+7 4 | 7 3 7 3 | 2+5 2 |
| | 6 2 8+8 8 | 4 1 4 1 | 2+5 2 |

Continued on next column

**TABLE LXXXV– continued**

| TQC | ICSD | NoE | SF |
|---|---|---|---|
| | 8 4 | 5 1 5 1 | 3+4 2 |
| | 6 2 | 7 3 7 3 | 2+5 2 |
| | 8 4 | 8 3 8 3 | 3+4 2 |
| | 7 2 | 7 3 7 3 | 2+5 2 |
| | 8 4 | 6 2 6 2 | 3+4 2 |
| | 7 2 | 7 3 7 3 | 2+5 2 |
| | 8 4 | 4 1 4 1 | 3+4 2 |
| | 6 2 | 8 3 8 3 | 3+4 2 |
| | 6 2 | 6 2 6 2 | 2+5 2 |
| | 8 4 | 7 3 7 3 | 2+5 2 |
| | 6 2 | 8 3 8 3 | 3+4 2 |
| | 8 4 | 8 3 8 3 | 3+4 2 |
| | 7 2 | 5 1 5 1 | 2+5 2 |
| | 8 4 | 7 3 7 3 | 3+4 2 |
| | 7 2 | 6 2 6 2 | 2+5 2 |
| | 8 4 | 8 3 8 3 | 3+4 2 |
| | 8 4 | 8 3 8 3 | 2+5 2 |
| | 6 2 | 5 1 5 1 | 2+5 2 |
| | 8 4 | 6 2 6 2 | 3+4 2 |
| | 7 2 | 7 3 7 3 | 2+5 2 |
| | 8 4 | 8 3 8 3 | 3+4 2 |
| | 8 4 | 7 3 7 3 | 2+5 2 |
| | 6 2 | 4 1 4 1 | 3+4 2 |
| | | 6 2 6 2 | 3+4 2 |
| | | 8 3 8 3 | 2+5 2 |
| | | 7 3 7 3 | 2+5 2 |
| | | 7 3 7 3 | 3+4 2 |
| | | 4 1 4 1 | 2+5 2 |
| | | 8 3 8 3 | 3+4 2 |
| | | 6 2 6 2 | 3+4 2 |
| | | 7 3 7 3 | 2+5 2 |
| | | 8 3 8 3 | 2+5 2 |
| | | 5 1 5 1 | 3+4 2 |
| | | | 2+5 2 |
| | | | 3+4 2 |
| | | | 2+5 2 |
| | | | 3+4 2 |
| | | | 2+5 2 |
| | | | 3+4 2 |
| | | | 2+5 2 |
| | | | 3+4 2 |
| | | | 2+5 2 |
| | | | 3+4 2 |
| | | | 2+5 2 |
| | | | 3+4 2 |
| 220.2.2 | 155382 | 136 | C3Y2 |
| | -GM -H | -PA -P | -N |
| | 6 2 6+7 4 | 6 2 6 2 | 2+5 2 |
| | 7 2 8+8 8 | 7 3 7 3 | 3+4 2 |
| | 7 2 6+7 4 | 8 3 8 3 | 2+5 2 |
| | 8 4 6+7 4 | 6 2 6 2 | 3+4 2 |
| | 6 2 8+8 8 | 7 3 7 3 | 2+5 2 |
| | 8 4 6+7 4 | 8 3 8 3 | 3+4 2 |
| | 8 4 8+8 8 | 7 3 7 3 | 2+5 2 |
| | 6 2 6+7 4 | 6 2 6 2 | 3+4 2 |
| | 8 4 8+8 8 | 8 3 8 3 | 3+4 2 |
| | 7 2 6+7 4 | 7 3 7 3 | 2+5 2 |
| | 6 2 8+8 8 | 6 2 6 2 | 2+5 2 |
| | 7 2 6+7 4 | 8 3 8 3 | 3+4 2 |
| | 8 4 8+8 8 | 5 1 5 1 | 3+4 2 |
| | 8 4 6+7 4 | 8 3 8 3 | 2+5 2 |

Continued on next column

**TABLE LXXXV– continued**

| TQC | ICSD | | NoE | SF | |
|---|---|---|---|---|---|
| | 7 2 | 8+8 8 | 4 1 | 4 1 | 2+5 2 |
| | 6 2 | 6+7 4 | 7 3 | 7 3 | 3+4 2 |
| | 8 4 | 8+8 8 | 7 3 | 7 3 | 3+4 2 |
| | 8 4 | 8+8 8 | 8 3 | 8 3 | 2+5 2 |
| | 8 4 | 6+7 4 | 6 2 | 6 2 | 3+4 2 |
| | 6 2 | 8+8 8 | 8 3 | 8 3 | 2+5 2 |
| | 7 2 | 6+7 4 | 6 2 | 6 2 | 3+4 2 |
| | 8 4 | 6+7 4 | 7 3 | 7 3 | 2+5 2 |
| | 6 2 | 8+8 8 | 4 1 | 4 1 | 2+5 2 |
| | 8 4 | | 5 1 | 5 1 | 3+4 2 |
| | 6 2 | | 7 3 | 7 3 | 2+5 2 |
| | 8 4 | | 8 3 | 8 3 | 3+4 2 |
| | 7 2 | | 7 3 | 7 3 | 2+5 2 |
| | 8 4 | | 6 2 | 6 2 | 3+4 2 |
| | 7 2 | | 7 3 | 7 3 | 2+5 2 |
| | 8 4 | | 4 1 | 4 1 | 3+4 2 |
| | 6 2 | | 8 3 | 8 3 | 3+4 2 |
| | 6 2 | | 6 2 | 6 2 | 2+5 2 |
| | 8 4 | | 7 3 | 7 3 | 2+5 2 |
| | 6 2 | | 8 3 | 8 3 | 3+4 2 |
| | 8 4 | | 8 3 | 8 3 | 3+4 2 |
| | 7 2 | | 5 1 | 5 1 | 2+5 2 |
| | 8 4 | | 7 3 | 7 3 | 3+4 2 |
| | 7 2 | | 6 2 | 6 2 | 2+5 2 |
| | 8 4 | | 8 3 | 8 3 | 3+4 2 |
| | 8 4 | | 8 3 | 8 3 | 2+5 2 |
| | 6 2 | | 5 1 | 5 1 | 2+5 2 |
| | 8 4 | | 6 2 | 6 2 | 3+4 2 |
| | 7 2 | | 7 3 | 7 3 | 2+5 2 |
| | 8 4 | | 8 3 | 8 3 | 3+4 2 |
| | 8 4 | | 7 3 | 7 3 | 2+5 2 |
| | 6 2 | | 4 1 | 4 1 | 3+4 2 |
| | | | 6 2 | 6 2 | 3+4 2 |
| | | | 8 3 | 8 3 | 2+5 2 |
| | | | 7 3 | 7 3 | 2+5 2 |
| | | | 7 3 | 7 3 | 3+4 2 |
| | | | 4 1 | 4 1 | 2+5 2 |
| | | | 8 3 | 8 3 | 3+4 2 |
| | | | 6 2 | 6 2 | 3+4 2 |
| | | | 7 3 | 7 3 | 2+5 2 |
| | | | 8 3 | 8 3 | 2+5 2 |
| | | | 5 1 | 5 1 | 3+4 2 |
| | | | | | 2+5 2 |
| | | | | | 3+4 2 |
| | | | | | 2+5 2 |
| | | | | | 3+4 2 |
| | | | | | 2+5 2 |
| | | | | | 3+4 2 |
| | | | | | 2+5 2 |
| | | | | | 3+4 2 |
| | | | | | 2+5 2 |
| | | | | | 3+4 2 |
| | | | | | 2+5 2 |
| | | | | | 3+4 2 |
| 220.2.3 | 26477 | | 160 | C3U2 | |
| | -GM | -H | -PA | -P | -N |
| | 6 2 | 6+7 4 | 6 2 | 6 2 | 2+5 2 |
| | 7 2 | 8+8 8 | 7 3 | 7 3 | 3+4 2 |
| | 7 2 | 6+7 4 | 8 3 | 8 3 | 2+5 2 |
| | 8 4 | 6+7 4 | 6 2 | 6 2 | 3+4 2 |
| | 6 2 | 8+8 8 | 7 3 | 7 3 | 2+5 2 |

Continued on next column

**TABLE LXXXV– continued**

| TQC | ICSD | | NoE | SF | |
|---|---|---|---|---|---|
| | 8 4 | 6+7 4 | 8 3 | 8 3 | 3+4 2 |
| | 8 4 | 6+7 4 | 7 3 | 7 3 | 2+5 2 |
| | 7 2 | 8+8 8 | 6 2 | 6 2 | 3+4 2 |
| | 6 2 | 8+8 8 | 8 3 | 8 3 | 3+4 2 |
| | 8 4 | 6+7 4 | 7 3 | 7 3 | 2+5 2 |
| | 6 2 | 8+8 8 | 6 2 | 6 2 | 2+5 2 |
| | 7 2 | 6+7 4 | 8 3 | 8 3 | 3+4 2 |
| | 8 4 | 8+8 8 | 8 3 | 8 3 | 3+4 2 |
| | 6 2 | 6+7 4 | 5 1 | 5 1 | 2+5 2 |
| | 8 4 | 8+8 8 | 7 3 | 7 3 | 3+4 2 |
| | 7 2 | 8+8 8 | 4 1 | 4 1 | 2+5 2 |
| | 8 4 | 6+7 4 | 7 3 | 7 3 | 2+5 2 |
| | 8 4 | 8+8 8 | 6 2 | 6 2 | 3+4 2 |
| | 6 2 | 6+7 4 | 8 3 | 8 3 | 3+4 2 |
| | 8 4 | 8+8 8 | 8 3 | 8 3 | 3+4 2 |
| | 7 2 | 6+7 4 | 6 2 | 6 2 | 2+5 2 |
| | 8 4 | 6+7 4 | 7 3 | 7 3 | 2+5 2 |
| | 6 2 | 6+7 4 | 4 1 | 4 1 | 3+4 2 |
| | 8 4 | 8+8 8 | 8 3 | 8 3 | 2+5 2 |
| | 8 4 | 8+8 8 | 5 1 | 5 1 | 2+5 2 |
| | 6 2 | 6+7 4 | 7 3 | 7 3 | 3+4 2 |
| | 7 2 | 8+8 8 | 7 3 | 7 3 | 2+5 2 |
| | 8 4 | | 4 1 | 4 1 | 3+4 2 |
| | 8 4 | | 7 3 | 7 3 | 2+5 2 |
| | 7 2 | | 6 2 | 6 2 | 3+4 2 |
| | 6 2 | | 8 3 | 8 3 | 2+5 2 |
| | 8 4 | | 6 2 | 6 2 | 3+4 2 |
| | 6 2 | | 7 3 | 7 3 | 3+4 2 |
| | 8 4 | | 8 3 | 8 3 | 2+5 2 |
| | 7 2 | | 8 3 | 8 3 | 3+4 2 |
| | 6 2 | | 5 1 | 5 1 | 2+5 2 |
| | 8 4 | | 7 3 | 7 3 | 3+4 2 |
| | 8 4 | | 6 2 | 6 2 | 2+5 2 |
| | 7 2 | | 8 3 | 8 3 | 3+4 2 |
| | 8 4 | | 7 3 | 7 3 | 2+5 2 |
| | 6 2 | | 6 2 | 6 2 | 3+4 2 |
| | 7 2 | | 8 3 | 8 3 | 2+5 2 |
| | 8 4 | | 7 3 | 7 3 | 2+5 2 |
| | 8 4 | | 4 1 | 4 1 | 3+4 2 |
| | 8 4 | | 8 3 | 8 3 | 3+4 2 |
| | 7 2 | | 5 1 | 5 1 | 2+5 2 |
| | 6 2 | | 7 3 | 7 3 | 3+4 2 |
| | 8 4 | | 4 1 | 4 1 | 2+5 2 |
| | 6 2 | | 6 2 | 6 2 | 2+5 2 |
| | 8 4 | | 7 3 | 7 3 | 3+4 2 |
| | 8 4 | | 8 3 | 8 3 | 3+4 2 |
| | 6 2 | | 6 2 | 6 2 | 2+5 2 |
| | 7 2 | | 8 3 | 8 3 | 2+5 2 |
| | 8 4 | | 7 3 | 7 3 | 3+4 2 |
| | | | 7 3 | 7 3 | 3+4 2 |
| | | | 4 1 | 4 1 | 2+5 2 |
| | | | 8 3 | 8 3 | 3+4 2 |
| | | | 5 1 | 5 1 | 2+5 2 |
| | | | 6 2 | 6 2 | 3+4 2 |
| | | | 7 3 | 7 3 | 2+5 2 |
| | | | 8 3 | 8 3 | 3+4 2 |
| | | | 5 1 | 5 1 | 2+5 2 |
| | | | 8 3 | 8 3 | 2+5 2 |
| | | | 7 3 | 7 3 | 3+4 2 |
| | | | 6 2 | 6 2 | 3+4 2 |
| | | | 8 3 | 8 3 | 2+5 2 |

Continued on next column

**TABLE LXXXV– continued**

| TQC | ICSD / -GM | -H | NoE / -PA | -P | SF / -N |
|---|---|---|---|---|---|
| | | | | | 2+5 2 |
| | | | | | 3+4 2 |
| | | | | | 2+5 2 |
| | | | | | 3+4 2 |
| | | | | | 2+5 2 |
| | | | | | 3+4 2 |
| | | | | | 2+5 2 |
| | | | | | 3+4 2 |
| | | | | | 2+5 2 |
| | | | | | 3+4 2 |
| | | | | | 3+4 2 |
| | | | | | 2+5 2 |
| | | | | | 3+4 2 |
| | | | | | 2+5 2 |
| 220.2.4 | 2447 | | 160 | | C3Nd2 |
| | -GM | -H | -PA | -P | -N |
| | 6 2 | 8+8 8 | 6 2 | 7 3 | 2+5 2 |
| | 7 2 | 6+7 4 | 8 3 | 8 3 | 3+4 2 |
| | 8 4 | 6+7 4 | 7 3 | 6 2 | 2+5 2 |
| | 7 2 | 6+7 4 | 7 3 | 6 2 | 3+4 2 |
| | 6 2 | 8+8 8 | 8 3 | 8 3 | 2+5 2 |
| | 8 4 | 6+7 4 | 6 2 | 7 3 | 3+4 2 |
| | 8 4 | 8+8 8 | 8 3 | 7 3 | 3+4 2 |
| | 6 2 | 6+7 4 | 7 3 | 6 2 | 2+5 2 |
| | 7 2 | 8+8 8 | 6 2 | 8 3 | 3+4 2 |
| | 8 4 | 6+7 4 | 7 3 | 8 3 | 2+5 2 |
| | 6 2 | 8+8 8 | 6 2 | 7 3 | 3+4 2 |
| | 7 2 | 6+7 4 | 8 3 | 6 2 | 2+5 2 |
| | 8 4 | 8+8 8 | 8 3 | 5 1 | 3+4 2 |
| | 6 2 | 6+7 4 | 4 1 | 8 3 | 2+5 2 |
| | 8 4 | 8+8 8 | 7 3 | 4 1 | 3+4 2 |
| | 7 2 | 8+8 8 | 5 1 | 7 3 | 2+5 2 |
| | 8 4 | 6+7 4 | 7 3 | 7 3 | 2+5 2 |
| | 8 4 | 8+8 8 | 6 2 | 8 3 | 3+4 2 |
| | 6 2 | 6+7 4 | 8 3 | 6 2 | 3+4 2 |
| | 8 4 | 8+8 8 | 8 3 | 8 3 | 2+5 2 |
| | 7 2 | 6+7 4 | 6 2 | 6 2 | 2+5 2 |
| | 8 4 | 6+7 4 | 7 3 | 7 3 | 3+4 2 |
| | 6 2 | 6+7 4 | 4 1 | 4 1 | 2+5 2 |
| | 8 4 | 8+8 8 | 5 1 | 7 3 | 3+4 2 |
| | 8 4 | 8+8 8 | 8 3 | 8 3 | 2+5 2 |
| | 6 2 | 6+7 4 | 7 3 | 5 1 | 3+4 2 |
| | 7 2 | 8+8 8 | 7 3 | 7 3 | 2+5 2 |
| | 8 4 | | 4 1 | 4 1 | 3+4 2 |
| | 8 4 | | 6 2 | 6 2 | 2+5 2 |
| | 7 2 | | 7 3 | 7 3 | 3+4 2 |
| | 6 2 | | 8 3 | 8 3 | 3+4 2 |
| | 6 2 | | 6 2 | 6 2 | 2+5 2 |
| | 6 2 | | 7 3 | 7 3 | 3+4 2 |
| | 8 4 | | 8 3 | 8 3 | 2+5 2 |
| | 7 2 | | 8 3 | 8 3 | 3+4 2 |
| | 8 4 | | 5 1 | 5 1 | 2+5 2 |
| | 8 4 | | 7 3 | 7 3 | 3+4 2 |
| | 7 2 | | 6 2 | 8 3 | 2+5 2 |
| | 8 4 | | 8 3 | 6 2 | 3+4 2 |
| | 8 4 | | 6 2 | 6 2 | 2+5 2 |
| | 6 2 | | 8 3 | 7 3 | 3+4 2 |
| | 8 4 | | 7 3 | 8 3 | 2+5 2 |
| | 8 4 | | 7 3 | 8 3 | 2+5 2 |
| | 7 2 | | 8 3 | 7 3 | 3+4 2 |
| | 6 2 | | 4 1 | 4 1 | 2+5 2 |

Continued on next column

**TABLE LXXXV– continued**

| TQC | ICSD / -GM | -H | NoE / -PA | -P | SF / -N |
|---|---|---|---|---|---|
| | 7 2 | | 5 1 | 5 1 | 3+4 2 |
| | 8 4 | | 7 3 | 7 3 | 3+4 2 |
| | 8 4 | | 4 1 | 4 1 | 2+5 2 |
| | 6 2 | | 6 2 | 6 2 | 2+5 2 |
| | 8 4 | | 7 3 | 8 3 | 3+4 2 |
| | 7 2 | | 8 3 | 7 3 | 2+5 2 |
| | 8 4 | | 6 2 | 6 2 | 3+4 2 |
| | 6 2 | | 7 3 | 7 3 | 2+5 2 |
| | 8 4 | | 8 3 | 8 3 | 3+4 2 |
| | | | 4 1 | 8 3 | 3+4 2 |
| | | | 7 3 | 7 3 | 2+5 2 |
| | | | 8 3 | 5 1 | 3+4 2 |
| | | | 5 1 | 8 3 | 2+5 2 |
| | | | 6 2 | 4 1 | 3+4 2 |
| | | | 8 3 | 7 3 | 2+5 2 |
| | | | 7 3 | 6 2 | 2+5 2 |
| | | | 8 3 | 8 3 | 3+4 2 |
| | | | 6 2 | 6 2 | 2+5 2 |
| | | | 5 1 | 7 3 | 3+4 2 |
| | | | 7 3 | 5 1 | 2+5 2 |
| | | | 8 3 | 8 3 | 3+4 2 |
| | | | | | 3+4 2 |
| | | | | | 2+5 2 |
| | | | | | 3+4 2 |
| | | | | | 2+5 2 |
| | | | | | 2+5 2 |
| | | | | | 3+4 2 |
| | | | | | 2+5 2 |
| | | | | | 3+4 2 |
| | | | | | 2+5 2 |
| | | | | | 3+4 2 |
| | | | | | 2+5 2 |
| | | | | | 3+4 2 |
| | | | | | 2+5 2 |
| | | | | | 3+4 2 |

The End

TABLE LXXXVI: The computed irreducible representations for the valence bands for the compounds of SEBR and NLC in space group 221. See the list of tables.

| TQC | ICSD / -GM | -R | NoE / -M | SF / -X |
|---|---|---|---|---|
| 221.1.1 | 201036 | | 40 | O1Sn1Sr3 |
| | -GM | -R | -M | -X |
| | 6 2 | 7 2 | 6 2 | 8 2 |
| | 10 4 | 10 4 | 8 2 | 6 2 |
| | 11 4 | 9 2 | 9 2 | 9 2 |
| | 8 2 | 11 4 | 6 2 | 8 2 |
| | 6 2 | 9 2 | 7 2 | 7 2 |
| | 8 2 | 8 2 | 8 2 | 6 2 |
| | 11 4 | 11 4 | 6 2 | 8 2 |
| | 9 2 | 11 4 | 7 2 | 6 2 |
| | 11 4 | 9 2 | 8 2 | 7 2 |
| | 6 2 | 6 2 | 9 2 | 9 2 |
| | 11 4 | 7 2 | 6 2 | 6 2 |
| | 8 2 | 10 4 | 7 2 | 7 2 |
| | 10 4 | 8 2 | 6 2 | 8 2 |
| | 8 2 | 11 4 | 7 2 | 6 2 |
| | | | 8 2 | 6 2 |
| | | | 9 2 | 7 2 |

Continued on next column

**TABLE LXXXVI– continued**

| TQC | ICSD | | NoE | SF |
|---|---|---|---|---|
| | | | 8 2 | 6 2 |
| | | | 9 2 | 8 2 |
| | | | 8 2 | 8 2 |
| | | | 9 2 | 9 2 |
| 221.1.2 | 181127 | | 34 | Ni3Sn1 |
| | -GM | -R | -M | -X |
| | 6 2 | 6 2 | 7 2 | 6 2 |
| | 6 2 | 8 2 | 9 2 | 8 2 |
| | 10 4 | 7 2 | 8 2 | 8 2 |
| | 7 2 | 10 4 | 9 2 | 8 2 |
| | 10 4 | 11 4 | 6 2 | 9 2 |
| | 10 4 | 10 4 | 8 2 | 9 2 |
| | 7 2 | 6 2 | 8 2 | 7 2 |
| | 8 2 | 10 4 | 9 2 | 6 2 |
| | 10 4 | 7 2 | 6 2 | 7 2 |
| | 11 4 | 10 4 | 7 2 | 8 2 |
| | 7 2 | 9 2 | 7 2 | 7 2 |
| | 6 2 | 6 2 | 9 2 | 8 2 |
| | | | 8 2 | 9 2 |
| | | | 8 2 | 9 2 |
| | | | 6 2 | 6 2 |
| | | | 9 2 | 9 2 |
| | | | 7 2 | 8 2 |
| 221.1.3 | 163195 | | 12 | Mg1Pd1 |
| | -GM | -R | -M | -X |
| | 6 2 | 11 4 | 7 2 | 8 2 |
| | 10 4 | 8 2 | 6 2 | 9 2 |
| | 10 4 | 11 4 | 6 2 | 6 2 |
| | 7 2 | 7 2 | 7 2 | 8 2 |
| | | | 7 2 | 9 2 |
| | | | 8 2 | 9 2 |
| 221.1.4 | 247065 | | 38 | In1N1Ni3 |
| | -GM | -R | -M | -X |
| | 6 2 | 9 2 | 6 2 | 8 2 |
| | 6 2 | 10 4 | 9 2 | 6 2 |
| | 11 4 | 7 2 | 8 2 | 6 2 |
| | 8 2 | 6 2 | 7 2 | 7 2 |
| | 10 4 | 8 2 | 8 2 | 6 2 |
| | 7 2 | 11 4 | 9 2 | 8 2 |
| | 10 4 | 7 2 | 8 2 | 9 2 |
| | 10 4 | 10 4 | 9 2 | 9 2 |
| | 10 4 | 10 4 | 6 2 | 8 2 |
| | 7 2 | 6 2 | 8 2 | 7 2 |
| | 6 2 | 10 4 | 7 2 | 8 2 |
| | 10 4 | 6 2 | 9 2 | 9 2 |
| | 7 2 | 10 4 | 6 2 | 7 2 |
| | | | 7 2 | 8 2 |
| | | | 9 2 | 9 2 |
| | | | 8 2 | 9 2 |
| | | | 7 2 | 8 2 |
| | | | 9 2 | 6 2 |
| | | | 8 2 | 7 2 |
| 221.1.5 | 247273 | | 34 | H1Pd3Tl1 |
| | -GM | -R | -M | -X |
| | 6 2 | 6 2 | 7 2 | 8 2 |
| | 6 2 | 9 2 | 6 2 | 6 2 |
| | 10 4 | 10 4 | 9 2 | 8 2 |
| | 7 2 | 7 2 | 8 2 | 9 2 |
| | 10 4 | 10 4 | 9 2 | 8 2 |
| | 10 4 | 6 2 | 8 2 | 9 2 |
| | 7 2 | 7 2 | 9 2 | 6 2 |

Continued on next column

**TABLE LXXXVI– continued**

| TQC | ICSD | | NoE | SF |
|---|---|---|---|---|
| | 10 4 | 10 4 | 8 2 | 7 2 |
| | 7 2 | 8 2 | 6 2 | 7 2 |
| | 6 2 | 10 4 | 7 2 | 8 2 |
| | 10 4 | 11 4 | 8 2 | 6 2 |
| | 8 2 | 6 2 | 9 2 | 7 2 |
| | | | 7 2 | 8 2 |
| | | | 8 2 | 9 2 |
| | | | 9 2 | 9 2 |
| | | | 6 2 | 9 2 |
| | | | 7 2 | 8 2 |
| 221.1.6 | 100533 | | 12 | Fe1Ti1 |
| | -GM | -R | -M | -X |
| | 6 2 | 10 4 | 7 2 | 6 2 |
| | 10 4 | 7 2 | 6 2 | 8 2 |
| | 10 4 | 8 2 | 7 2 | 7 2 |
| | 7 2 | 11 4 | 6 2 | 6 2 |
| | | | 6 2 | 7 2 |
| | | | 7 2 | 9 2 |
| 221.1.7 | 262924 | | 34 | Cs1I3Sn1 |
| | -GM | -R | -M | -X |
| | 6 2 | 6 2 | 7 2 | 6 2 |
| | 6 2 | 7 2 | 6 2 | 8 2 |
| | 10 4 | 10 4 | 8 2 | 6 2 |
| | 8 2 | 8 2 | 9 2 | 9 2 |
| | 11 4 | 11 4 | 9 2 | 8 2 |
| | 6 2 | 9 2 | 8 2 | 9 2 |
| | 11 4 | 11 4 | 9 2 | 8 2 |
| | 8 2 | 8 2 | 6 2 | 8 2 |
| | 8 2 | 11 4 | 8 2 | 7 2 |
| | 11 4 | 9 2 | 7 2 | 6 2 |
| | 9 2 | 11 4 | 8 2 | 6 2 |
| | 11 4 | 7 2 | 6 2 | 8 2 |
| | | | 7 2 | 7 2 |
| | | | 9 2 | 6 2 |
| | | | 7 2 | 9 2 |
| | | | 6 2 | 7 2 |
| | | | 6 2 | 8 2 |
| 221.1.8 | 100787 | | 16 | Ca3O1Sn1 |
| | -GM | -R | -M | -X |
| | 6 2 | 9 2 | 6 2 | 8 2 |
| | 6 2 | 6 2 | 7 2 | 6 2 |
| | 8 2 | 7 2 | 8 2 | 6 2 |
| | 11 4 | 10 4 | 9 2 | 6 2 |
| | 8 2 | 8 2 | 8 2 | 7 2 |
| | 10 4 | 11 4 | 9 2 | 8 2 |
| | | | 8 2 | 8 2 |
| | | | 9 2 | 9 2 |
| 221.1.9 | 100789 | | 16 | Ca3O1Pb1 |
| | -GM | -R | -M | -X |
| | 6 2 | 9 2 | 6 2 | 8 2 |
| | 6 2 | 6 2 | 7 2 | 6 2 |
| | 8 2 | 7 2 | 9 2 | 6 2 |
| | 11 4 | 10 4 | 8 2 | 6 2 |
| | 8 2 | 8 2 | 8 2 | 7 2 |
| | 10 4 | 11 4 | 9 2 | 8 2 |
| | | | 8 2 | 8 2 |
| | | | 9 2 | 9 2 |
| 221.1.10 | 402087 | | 16 | Ca3Ge1O1 |
| | -GM | -R | -M | -X |
| | 6 2 | 9 2 | 6 2 | 8 2 |

Continued on next column

**TABLE LXXXVI– continued**

| TQC | ICSD | NoE | SF |
|---|---|---|---|
| | 6 2 | 6 2 | 7 2 6 2 |
| | 8 2 | 7 2 | 8 2 6 2 |
| | 11 4 | 10 4 | 9 2 6 2 |
| | 8 2 | 8 2 | 8 2 7 2 |
| | 10 4 | 11 4 | 9 2 8 2 |
| | | | 8 2 8 2 |
| | | | 9 2 9 2 |
| 221.1.11 | 80320 | 34 | Br3Cs1Ge1 |
| | -GM | -R | -M -X |
| | 6 2 | 6 2 | 7 2 6 2 |
| | 6 2 | 7 2 | 6 2 8 2 |
| | 10 4 | 10 4 | 8 2 6 2 |
| | 8 2 | 9 2 | 9 2 9 2 |
| | 11 4 | 8 2 | 9 2 8 2 |
| | 6 2 | 11 4 | 6 2 8 2 |
| | 11 4 | 11 4 | 8 2 9 2 |
| | 8 2 | 8 2 | 9 2 8 2 |
| | 11 4 | 11 4 | 8 2 7 2 |
| | 8 2 | 9 2 | 7 2 6 2 |
| | 9 2 | 11 4 | 7 2 6 2 |
| | 11 4 | 7 2 | 6 2 8 2 |
| | | | 8 2 7 2 |
| | | | 9 2 9 2 |
| | | | 7 2 6 2 |
| | | | 6 2 7 2 |
| | | | 6 2 8 2 |
| 221.1.12 | 409846 | 60 | Br3Ce3Ga1 |
| | -GM | -R | -M -X |
| | 6 2 | 7 2 | 6 2 8 2 |
| | 10 4 | 10 4 | 8 2 6 2 |
| | 11 4 | 11 4 | 9 2 9 2 |
| | 8 2 | 9 2 | 7 2 7 2 |
| | 11 4 | 8 2 | 8 2 6 2 |
| | 9 2 | 11 4 | 6 2 8 2 |
| | 6 2 | 9 2 | 9 2 7 2 |
| | 8 2 | 11 4 | 7 2 6 2 |
| | 10 4 | 11 4 | 8 2 9 2 |
| | 11 4 | 8 2 | 6 2 6 2 |
| | 6 2 | 10 4 | 6 2 8 2 |
| | 11 4 | 6 2 | 7 2 8 2 |
| | 8 2 | 7 2 | 8 2 7 2 |
| | 11 4 | 6 2 | 9 2 6 2 |
| | 8 2 | 10 4 | 7 2 7 2 |
| | 9 2 | 10 4 | 9 2 8 2 |
| | 11 4 | 9 2 | 6 2 9 2 |
| | 7 2 | 7 2 | 8 2 8 2 |
| | 10 4 | 10 4 | 7 2 9 2 |
| | 11 4 | 11 4 | 7 2 6 2 |
| | | | 6 2 8 2 |
| | | | 9 2 8 2 |
| | | | 7 2 7 2 |
| | | | 6 2 6 2 |
| | | | 6 2 9 2 |
| | | | 8 2 6 2 |
| | | | 9 2 9 2 |
| | | | 8 2 7 2 |
| | | | 9 2 6 2 |
| | | | 6 2 7 2 |
| 221.1.13 | 50313 | 28 | B6Sr1 |
| | -GM | -R | -M -X |
| | 6 2 | 6 2 | 7 2 6 2 |

Continued on next column

**TABLE LXXXVI– continued**

| TQC | ICSD | NoE | SF |
|---|---|---|---|
| | 8 2 | 8 2 | 9 2 8 2 |
| | 11 4 | 11 4 | 8 2 8 2 |
| | 6 2 | 9 2 | 9 2 9 2 |
| | 6 2 | 7 2 | 6 2 8 2 |
| | 8 2 | 10 4 | 9 2 6 2 |
| | 11 4 | 8 2 | 8 2 6 2 |
| | 10 4 | 11 4 | 6 2 7 2 |
| | 10 4 | 10 4 | 8 2 8 2 |
| | 7 2 | 7 2 | 7 2 9 2 |
| | | | 6 2 8 2 |
| | | | 7 2 9 2 |
| | | | 8 2 6 2 |
| | | | 9 2 7 2 |
| 221.1.14 | 51495 | 28 | B6Ba1 |
| | -GM | -R | -M -X |
| | 6 2 | 6 2 | 7 2 6 2 |
| | 6 2 | 8 2 | 9 2 8 2 |
| | 8 2 | 9 2 | 6 2 8 2 |
| | 11 4 | 11 4 | 8 2 9 2 |
| | 6 2 | 7 2 | 9 2 8 2 |
| | 8 2 | 10 4 | 9 2 6 2 |
| | 11 4 | 8 2 | 8 2 6 2 |
| | 10 4 | 11 4 | 6 2 7 2 |
| | 10 4 | 10 4 | 8 2 8 2 |
| | 7 2 | 7 2 | 7 2 9 2 |
| | | | 7 2 8 2 |
| | | | 6 2 9 2 |
| | | | 8 2 6 2 |
| | | | 9 2 7 2 |

The End

TABLE LXXXVII: The computed irreducible representations for the valence bands for the compounds of SEBR and NLC in space group 225. See the list of tables.

| TQC | ICSD | NoE | SF |
|---|---|---|---|
| 225.1.1 | 63098 | 10 | Pb1Te1 |
| | -GM -X | -L -W | |
| | 6 2 6 2 | 8 2 7 2 | |
| | 6 2 6 2 | 9 2 6 2 | |
| | 8 2 8 2 | 9 2 7 2 | |
| | 11 4 8 2 | 6+7 2 6 2 | |
| | 9 2 | 8 2 6 2 | |
| 225.1.2 | 153711 | 10 | Pb1Te1 |
| | -GM -X | -L -W | |
| | 6 2 6 2 | 8 2 7 2 | |
| | 6 2 6 2 | 9 2 6 2 | |
| | 8 2 8 2 | 9 2 7 2 | |
| | 11 4 8 2 | 6+7 2 6 2 | |
| | 9 2 | 8 2 6 2 | |
| 225.1.3 | 63097 | 10 | Pb1Se1 |
| | -GM -X | -L -W | |
| | 6 2 6 2 | 8 2 7 2 | |
| | 6 2 6 2 | 9 2 6 2 | |
| | 8 2 8 2 | 9 2 7 2 | |
| | 11 4 8 2 | 6+7 2 6 2 | |
| | 9 2 | 8 2 6 2 | |
| 225.1.4 | 63096 | 10 | Pb1Se1 |
| | -GM -X | -L -W | |
| | 6 2 6 2 | 8 2 7 2 | |

Continued on next column

**TABLE LXXXVII– continued**

| TQC | ICSD | NoE | SF |
|---|---|---|---|
| | 6 2 6 2 | 9 2 6 2 | |
| | 8 2 8 2 | 9 2 7 2 | |
| | 11 4 8 2 | 6+7 2 6 2 | |
| | 9 2 | 8 2 6 2 | |
| 225.1.5 | 62196 | 10 | Pb1Se1 |
| | -GM -X | -L -W | |
| | 6 2 6 2 | 8 2 7 2 | |
| | 6 2 6 2 | 9 2 6 2 | |
| | 8 2 8 2 | 9 2 7 2 | |
| | 11 4 8 2 | 6+7 2 6 2 | |
| | 9 2 | 8 2 6 2 | |
| 225.1.6 | 62195 | 10 | Pb1Se1 |
| | -GM -X | -L -W | |
| | 6 2 6 2 | 8 2 7 2 | |
| | 6 2 6 2 | 9 2 6 2 | |
| | 8 2 8 2 | 9 2 7 2 | |
| | 11 4 8 2 | 6+7 2 6 2 | |
| | 9 2 | 8 2 6 2 | |
| 225.1.7 | 38294 | 10 | Pb1Se1 |
| | -GM -X | -L -W | |
| | 6 2 6 2 | 8 2 7 2 | |
| | 6 2 6 2 | 9 2 6 2 | |
| | 8 2 8 2 | 9 2 7 2 | |
| | 11 4 8 2 | 6+7 2 6 2 | |
| | 9 2 | 8 2 6 2 | |
| 225.1.8 | 62192 | 10 | Pb1S1 |
| | -GM -X | -L -W | |
| | 6 2 6 2 | 8 2 7 2 | |
| | 6 2 6 2 | 9 2 6 2 | |
| | 8 2 8 2 | 9 2 7 2 | |
| | 11 4 8 2 | 6+7 2 6 2 | |
| | 9 2 | 8 2 6 2 | |
| 225.1.9 | 62191 | 10 | Pb1S1 |
| | -GM -X | -L -W | |
| | 6 2 6 2 | 8 2 7 2 | |
| | 6 2 6 2 | 9 2 6 2 | |
| | 8 2 8 2 | 9 2 7 2 | |
| | 11 4 8 2 | 6+7 2 6 2 | |
| | 9 2 | 8 2 6 2 | |
| 225.1.10 | 62190 | 10 | Pb1S1 |
| | -GM -X | -L -W | |
| | 6 2 6 2 | 8 2 7 2 | |
| | 6 2 6 2 | 9 2 6 2 | |
| | 8 2 8 2 | 9 2 7 2 | |
| | 11 4 8 2 | 6+7 2 6 2 | |
| | 9 2 | 8 2 6 2 | |
| 225.1.11 | 35204 | 26 | O2U1 |
| | -GM -X | -L -W | |
| | 6 2 6 2 | 8 2 7 2 | |
| | 8 2 8 2 | 9 2 6 2 | |
| | 6 2 8 2 | 9 2 7 2 | |
| | 9 2 7 2 | 8 2 6 2 | |
| | 11 4 9 2 | 6+7 2 6 2 | |
| | 10 4 8 2 | 9 2 7 2 | |
| | 7 2 6 2 | 4+5 2 7 2 | |
| | 8 2 8 2 | 8 2 6 2 | |
| | 11 4 6 2 | 8 2 6 2 | |
| | 9 2 9 2 | 9 2 7 2 | |
| | 7 2 | 9 2 7 2 | |
| | 9 2 | 6+7 2 6 2 | |

Continued on next column

**TABLE LXXXVII– continued**

| TQC | ICSD | NoE | SF |
|---|---|---|---|
| | 9 2 | 9 2 6 2 | |
| 225.1.12 | 246851 | 26 | O2U1 |
| | -GM -X | -L -W | |
| | 6 2 6 2 | 8 2 7 2 | |
| | 8 2 8 2 | 9 2 6 2 | |
| | 6 2 8 2 | 9 2 7 2 | |
| | 9 2 7 2 | 8 2 6 2 | |
| | 11 4 9 2 | 6+7 2 6 2 | |
| | 10 4 8 2 | 9 2 7 2 | |
| | 7 2 6 2 | 4+5 2 7 2 | |
| | 8 2 8 2 | 8 2 6 2 | |
| | 11 4 6 2 | 8 2 6 2 | |
| | 9 2 9 2 | 9 2 7 2 | |
| | 7 2 | 9 2 7 2 | |
| | 9 2 | 6+7 2 6 2 | |
| | 9 2 | 9 2 6 2 | |
| 225.1.13 | 419774 | 248 | Ni21P6Sn2 |
| | -GM -X | -L -W | |
| | 6 2 6 2 | 9 2 6 2 | |
| | 11 4 8 2 | 8 2 7 2 | |
| | 8 2 9 2 | 4+5 2 6 2 | |
| | 10 4 8 2 | 8 2 7 2 | |
| | 6 2 6 2 | 6+7 2 7 2 | |
| | 9 2 7 2 | 9 2 6 2 | |
| | 7 2 7 2 | 8 2 6 2 | |
| | 10 4 8 2 | 9 2 7 2 | |
| | 6 2 9 2 | 8 2 7 2 | |
| | 8 2 8 2 | 4+5 2 6 2 | |
| | 11 4 6 2 | 9 2 7 2 | |
| | 9 2 7 2 | 6+7 2 6 2 | |
| | 11 4 6 2 | 8 2 6 2 | |
| | 10 4 7 2 | 9 2 7 2 | |
| | 11 4 9 2 | 8 2 7 2 | |
| | 8 2 8 2 | 4+5 2 6 2 | |
| | 10 4 6 2 | 8 2 7 2 | |
| | 6 2 6 2 | 9 2 6 2 | |
| | 7 2 7 2 | 6+7 2 7 2 | |
| | 6 2 9 2 | 9 2 6 2 | |
| | 10 4 8 2 | 9 2 7 2 | |
| | 8 2 8 2 | 6+7 2 6 2 | |
| | 11 4 7 2 | 9 2 7 2 | |
| | 10 4 9 2 | 4+5 2 6 2 | |
| | 11 4 8 2 | 8 2 7 2 | |
| | 9 2 6 2 | 8 2 6 2 | |
| | 11 4 6 2 | 9 2 7 2 | |
| | 8 2 9 2 | 8 2 6 2 | |
| | 11 4 8 2 | 8 2 6 2 | |
| | 10 4 9 2 | 4+5 2 7 2 | |
| | 7 2 6 2 | 9 2 6 2 | |
| | 10 4 6 2 | 6+7 2 7 2 | |
| | 9 2 7 2 | 9 2 6 2 | |
| | 10 4 6 2 | 6+7 2 7 2 | |
| | 11 4 9 2 | 9 2 7 2 | |
| | 6 2 7 2 | 8 2 6 2 | |
| | 7 2 8 2 | 8 2 6 2 | |
| | 8 2 9 2 | 4+5 2 7 2 | |
| | 11 4 7 2 | 8 2 6 2 | |
| | 10 4 6 2 | 9 2 6 2 | |
| | 9 2 7 2 | 6+7 2 7 2 | |
| | 6 2 7 2 | 4+5 2 7 2 | |
| | 10 4 6 2 | 8 2 6 2 | |

Continued on next column

**TABLE LXXXVII– continued**

| TQC | ICSD | NoE | SF |
|---|---|---|---|
| | 7 2 9 2 | 6+7 2 7 2 | |
| | 10 4 8 2 | 9 2 7 2 | |
| | 11 4 9 2 | 8 2 6 2 | |
| | 7 2 7 2 | 8 2 7 2 | |
| | 10 4 8 2 | 4+5 2 6 2 | |
| | 8 2 9 2 | 8 2 7 2 | |
| | 11 4 8 2 | 9 2 6 2 | |
| | 11 4 7 2 | 4+5 2 6 2 | |
| | 6 2 6 2 | 8 2 7 2 | |
| | 10 4 8 2 | 6+7 2 6 2 | |
| | 9 2 6 2 | 9 2 6 2 | |
| | 8 2 7 2 | 8 2 7 2 | |
| | 6 2 9 2 | 6+7 2 6 2 | |
| | 11 4 8 2 | 9 2 6 2 | |
| | 10 4 6 2 | 8 2 7 2 | |
| | 9 2 9 2 | 9 2 6 2 | |
| | 11 4 8 2 | 4+5 2 7 2 | |
| | 7 2 6 2 | 9 2 7 2 | |
| | 9 2 9 2 | 9 2 7 2 | |
| | 10 4 7 2 | 6+7 2 6 2 | |
| | 7 2 9 2 | 8 2 6 2 | |
| | 11 4 6 2 | 9 2 7 2 | |
| | 10 4 8 2 | 4+5 2 7 2 | |
| | 8 2 7 2 | 8 2 6 2 | |
| | 11 4 8 2 | 9 2 7 2 | |
| | 7 2 7 2 | 4+5 2 6 2 | |
| | 6 2 9 2 | 6+7 2 7 2 | |
| | 8 2 6 2 | 8 2 7 2 | |
| | 10 4 8 2 | 8 2 6 2 | |
| | 11 4 9 2 | 9 2 7 2 | |
| | 10 4 8 2 | 8 2 6 2 | |
| | 7 2 7 2 | 8 2 6 2 | |
| | 10 4 6 2 | 4+5 2 7 2 | |
| | 9 2 6 2 | 9 2 6 2 | |
| | 11 4 9 2 | 6+7 2 7 2 | |
| | 6 2 7 2 | 6+7 2 6 2 | |
| | 10 4 7 2 | 9 2 7 2 | |
| | 9 2 8 2 | 8 2 6 2 | |
| | 11 4 6 2 | 9 2 7 2 | |
| | 8 2 8 2 | 4+5 2 7 2 | |
| | 9 2 | 9 2 6 2 | |
| | 9 2 | 8 2 7 2 | |
| | 8 2 | 8 2 6 2 | |
| | 6 2 | 4+5 2 7 2 | |
| | 7 2 | 8 2 6 2 | |
| | 7 2 | 9 2 6 2 | |
| | 7 2 | 9 2 7 2 | |
| | 8 2 | 6+7 2 7 2 | |
| | 9 2 | 4+5 2 6 2 | |
| | 6 2 | 9 2 7 2 | |
| | 9 2 | 8 2 6 2 | |
| | 6 2 | 6+7 2 6 2 | |
| | 7 2 | 9 2 6 2 | |
| | 9 2 | 8 2 7 2 | |
| | 8 2 | 8 2 7 2 | |
| | 9 2 | 4+5 2 6 2 | |
| | 6 2 | 9 2 7 2 | |
| | 7 2 | 9 2 6 2 | |
| | 6 2 | 6+7 2 6 2 | |
| | 8 2 | 8 2 7 2 | |
| | 9 2 | 4+5 2 7 2 | |
| | 7 2 | 8 2 6 2 | |
| | 6 2 | 9 2 6 2 | |
| | 7 2 | 4+5 2 7 2 | |
| | 7 2 | 8 2 6 2 | |
| | 8 2 | 9 2 6 2 | |
| | 6 2 | 6+7 2 7 2 | |
| | 6 2 | 9 2 6 2 | |
| | 7 2 | 9 2 7 2 | |
| | 9 2 | 6+7 2 7 2 | |
| | 8 2 | 8 2 7 2 | |
| | 9 2 | 8 2 6 2 | |
| | 6 2 | 4+5 2 7 2 | |
| | 6 2 | 9 2 6 2 | |
| | 7 2 | 8 2 6 2 | |
| | 8 2 | 8 2 7 2 | |
| | 7 2 | 8 2 7 2 | |
| | 8 2 | 4+5 2 6 2 | |
| | 8 2 | 9 2 6 2 | |
| | 9 2 | 6+7 2 7 2 | |
| | 7 2 | 9 2 6 2 | |
| 225.1.14 | 159261 | 200 | Mg6Ni16Si7 |
| | -GM -X | -L -W | |
| | 6 2 8 2 | 9 2 6 2 | |
| | 7 2 6 2 | 8 2 7 2 | |
| | 10 4 7 2 | 9 2 6 2 | |
| | 6 2 9 2 | 8 2 7 2 | |
| | 10 4 8 2 | 4+5 2 6 2 | |
| | 9 2 6 2 | 9 2 6 2 | |
| | 8 2 9 2 | 6+7 2 7 2 | |
| | 11 4 6 2 | 8 2 7 2 | |
| | 8 2 7 2 | 9 2 6 2 | |
| | 11 4 8 2 | 4+5 2 7 2 | |
| | 11 4 9 2 | 8 2 7 2 | |
| | 8 2 6 2 | 9 2 6 2 | |
| | 6 2 7 2 | 6+7 2 6 2 | |
| | 6 2 6 2 | 9 2 7 2 | |
| | 10 4 8 2 | 8 2 6 2 | |
| | 11 4 9 2 | 8 2 7 2 | |
| | 10 4 7 2 | 4+5 2 6 2 | |
| | 7 2 6 2 | 8 2 6 2 | |
| | 11 4 8 2 | 6+7 2 7 2 | |
| | 9 2 9 2 | 9 2 7 2 | |
| | 9 2 8 2 | 4+5 2 6 2 | |
| | 11 4 7 2 | 8 2 7 2 | |
| | 11 4 9 2 | 6+7 2 7 2 | |
| | 8 2 8 2 | 9 2 7 2 | |
| | 10 4 6 2 | 4+5 2 7 2 | |
| | 7 2 7 2 | 8 2 6 2 | |
| | 9 2 6 2 | 9 2 6 2 | |
| | 10 4 7 2 | 8 2 6 2 | |
| | 8 2 8 2 | 8 2 7 2 | |
| | 11 4 6 2 | 9 2 6 2 | |
| | 11 4 8 2 | 6+7 2 7 2 | |
| | 10 4 9 2 | 9 2 6 2 | |
| | 9 2 9 2 | 8 2 6 2 | |
| | 6 2 8 2 | 9 2 6 2 | |
| | 10 4 7 2 | 6+7 2 7 2 | |
| | 7 2 6 2 | 9 2 6 2 | |
| | 10 4 7 2 | 8 2 7 2 | |
| | 6 2 6 2 | 8 2 7 2 | |
| | 10 4 7 2 | 4+5 2 6 2 | |

Continued on next column

Continued on next column

**TABLE LXXXVII– continued**

| TQC | ICSD | NoE | SF |
|---|---|---|---|
| | 11 4 9 2 | 6+7 2 7 2 | |
| | 11 4 6 2 | 9 2 6 2 | |
| | 6 2 8 2 | 4+5 2 7 2 | |
| | 8 2 9 2 | 8 2 7 2 | |
| | 11 4 7 2 | 8 2 7 2 | |
| | 7 2 6 2 | 4+5 2 6 2 | |
| | 11 4 9 2 | 8 2 7 2 | |
| | 8 2 7 2 | 9 2 6 2 | |
| | 10 4 8 2 | 6+7 2 7 2 | |
| | 10 4 7 2 | 9 2 6 2 | |
| | 11 4 7 2 | 9 2 7 2 | |
| | 9 2 9 2 | 8 2 6 2 | |
| | 10 4 8 2 | 4+5 2 6 2 | |
| | 7 2 8 2 | 6+7 2 7 2 | |
| | 9 2 6 2 | 8 2 6 2 | |
| | 6 2 9 2 | 9 2 7 2 | |
| | 10 4 8 2 | 6+7 2 6 2 | |
| | 10 4 9 2 | 9 2 7 2 | |
| | 7 2 6 2 | 9 2 6 2 | |
| | 8 2 7 2 | 4+5 2 7 2 | |
| | 11 4 8 2 | 8 2 6 2 | |
| | 11 4 8 2 | 8 2 6 2 | |
| | 10 4 6 2 | 4+5 2 7 2 | |
| | 6 2 7 2 | 6+7 2 7 2 | |
| | 9 2 9 2 | 9 2 6 2 | |
| | 11 4 7 2 | 8 2 6 2 | |
| | 11 4 6 2 | 9 2 7 2 | |
| | 6 2 | 4+5 2 7 2 | |
| | 9 2 | 8 2 6 2 | |
| | 8 2 | 9 2 7 2 | |
| | 9 2 | 8 2 6 2 | |
| | 8 2 | 6+7 2 6 2 | |
| | 6 2 | 9 2 7 2 | |
| | 8 2 | 9 2 6 2 | |
| | 9 2 | 8 2 7 2 | |
| | 6 2 | 9 2 6 2 | |
| | 7 2 | 6+7 2 7 2 | |
| | 6 2 | 4+5 2 6 2 | |
| | 8 2 | 8 2 6 2 | |
| | 7 2 | 9 2 7 2 | |
| | 9 2 | 8 2 7 2 | |
| | 6 2 | 6+7 2 6 2 | |
| | 7 2 | 9 2 7 2 | |
| | 8 2 | 8 2 7 2 | |
| | 9 2 | 4+5 2 6 2 | |
| | 6 2 | 6+7 2 7 2 | |
| | 9 2 | 9 2 7 2 | |
| | 7 2 | 8 2 6 2 | |
| | 6 2 | 8 2 6 2 | |
| | 8 2 | 4+5 2 6 2 | |
| | 9 2 | 9 2 6 2 | |
| | 7 2 | 9 2 7 2 | |
| | 9 2 | 6+7 2 7 2 | |
| | 8 2 | 8 2 6 2 | |
| | 7 2 | 4+5 2 7 2 | |
| | 9 2 | 9 2 6 2 | |
| | 7 2 | 8 2 7 2 | |
| | 9 2 | 8 2 6 2 | |
| | 8 2 | 4+5 2 6 2 | |
| | 8 2 | 6+7 2 7 2 | |
| | 6 2 | 9 2 6 2 | |

Continued on next column

**TABLE LXXXVII– continued**

| TQC | ICSD | NoE | SF |
|---|---|---|---|
| 225.1.15 | 106715 | 16 | Ga2Li1Rh1 |
| | -GM -X | -L -W | |
| | 6 2 8 2 | 8 2 6 2 | |
| | 10 4 7 2 | 9 2 7 2 | |
| | 7 2 6 2 | 4+5 2 7 2 | |
| | 10 4 8 2 | 8 2 6 2 | |
| | 9 2 9 2 | 8 2 6 2 | |
| | 8 2 6 2 | 4+5 2 7 2 | |
| | 7 2 | 8 2 6 2 | |
| | 7 2 | 9 2 7 2 | |
| 225.1.16 | 160154 | 240 | Fe16Si7U6 |
| | -GM -X | -L -W | |
| | 6 2 6 2 | 8 2 7 2 | |
| | 8 2 8 2 | 9 2 6 2 | |
| | 11 4 8 2 | 6+7 2 6 2 | |
| | 10 4 9 2 | 9 2 7 2 | |
| | 11 4 6 2 | 4+5 2 6 2 | |
| | 10 4 7 2 | 8 2 7 2 | |
| | 6 2 9 2 | 6+7 2 6 2 | |
| | 8 2 8 2 | 9 2 7 2 | |
| | 7 2 7 2 | 4+5 2 6 2 | |
| | 8 2 6 2 | 8 2 7 2 | |
| | 6 2 6 2 | 8 2 7 2 | |
| | 11 4 8 2 | 9 2 6 2 | |
| | 10 4 7 2 | 8 2 6 2 | |
| | 9 2 8 2 | 9 2 6 2 | |
| | 11 4 6 2 | 8 2 7 2 | |
| | 10 4 8 2 | 9 2 6 2 | |
| | 6 2 9 2 | 6+7 2 7 2 | |
| | 10 4 7 2 | 4+5 2 7 2 | |
| | 6 2 6 2 | 8 2 6 2 | |
| | 10 4 9 2 | 9 2 7 2 | |
| | 7 2 8 2 | 9 2 7 2 | |
| | 9 2 6 2 | 8 2 6 2 | |
| | 11 4 9 2 | 6+7 2 7 2 | |
| | 8 2 7 2 | 4+5 2 6 2 | |
| | 8 2 6 2 | 8 2 6 2 | |
| | 11 4 8 2 | 9 2 7 2 | |
| | 11 4 8 2 | 4+5 2 6 2 | |
| | 9 2 9 2 | 6+7 2 7 2 | |
| | 7 2 6 2 | 9 2 6 2 | |
| | 10 4 7 2 | 8 2 6 2 | |
| | 8 2 9 2 | 9 2 7 2 | |
| | 6 2 7 2 | 9 2 6 2 | |
| | 11 4 6 2 | 8 2 7 2 | |
| | 11 4 9 2 | 4+5 2 7 2 | |
| | 7 2 8 2 | 8 2 7 2 | |
| | 10 4 6 2 | 9 2 6 2 | |
| | 8 2 9 2 | 9 2 6 2 | |
| | 11 4 7 2 | 6+7 2 7 2 | |
| | 9 2 6 2 | 8 2 6 2 | |
| | 10 4 8 2 | 8 2 7 2 | |
| | 6 2 7 2 | 4+5 2 6 2 | |
| | 10 4 6 2 | 6+7 2 7 2 | |
| | 11 4 9 2 | 9 2 7 2 | |
| | 8 2 7 2 | 8 2 6 2 | |
| | 11 4 8 2 | 8 2 7 2 | |
| | 9 2 8 2 | 4+5 2 6 2 | |
| | 11 4 7 2 | 9 2 7 2 | |
| | 9 2 6 2 | 6+7 2 6 2 | |
| | 11 4 7 2 | 8 2 7 2 | |

Continued on next column

**TABLE LXXXVII– continued**

| TQC | ICSD | NoE | SF |
|---|---|---|---|
| | 6 2 8 2 | 4+5 2 6 2 | |
| | 10 4 9 2 | 9 2 7 2 | |
| | 7 2 6 2 | 8 2 6 2 | |
| | 10 4 8 2 | 9 2 7 2 | |
| | 10 4 9 2 | 8 2 6 2 | |
| | 6 2 8 2 | 8 2 7 2 | |
| | 10 4 8 2 | 9 2 6 2 | |
| | 7 2 9 2 | 6+7 2 6 2 | |
| | 10 4 7 2 | 9 2 7 2 | |
| | 9 2 6 2 | 9 2 6 2 | |
| | 11 4 7 2 | 6+7 2 7 2 | |
| | 11 4 6 2 | 6+7 2 7 2 | |
| | 8 2 7 2 | 4+5 2 6 2 | |
| | 6 2 7 2 | 9 2 6 2 | |
| | 11 4 6 2 | 8 2 7 2 | |
| | 7 2 9 2 | 8 2 7 2 | |
| | 10 4 8 2 | 8 2 6 2 | |
| | 10 4 9 2 | 9 2 6 2 | |
| | 6 2 6 2 | 8 2 7 2 | |
| | 8 2 7 2 | 4+5 2 6 2 | |
| | 10 4 8 2 | 4+5 2 6 2 | |
| | 11 4 9 2 | 8 2 7 2 | |
| | 8 2 8 2 | 4+5 2 6 2 | |
| | 10 4 9 2 | 8 2 7 2 | |
| | 11 4 9 2 | 9 2 7 2 | |
| | 9 2 7 2 | 6+7 2 6 2 | |
| | 7 2 9 2 | 8 2 6 2 | |
| | 11 4 6 2 | 6+7 2 7 2 | |
| | 9 2 7 2 | 9 2 6 2 | |
| | 11 4 8 2 | 9 2 7 2 | |
| | 9 2 9 2 | 8 2 6 2 | |
| | 8 2 | 6+7 2 7 2 | |
| | 6 2 | 9 2 6 2 | |
| | 7 2 | 4+5 2 7 2 | |
| | 7 2 | 9 2 7 2 | |
| | 8 2 | 8 2 7 2 | |
| | 6 2 | 9 2 6 2 | |
| | 9 2 | 4+5 2 7 2 | |
| | 9 2 | 6+7 2 6 2 | |
| | 6 2 | 9 2 7 2 | |
| | 7 2 | 8 2 7 2 | |
| | 6 2 | 9 2 6 2 | |
| | 8 2 | 8 2 7 2 | |
| | 8 2 | 4+5 2 6 2 | |
| | 6 2 | 8 2 6 2 | |
| | 7 2 | 9 2 7 2 | |
| | 8 2 | 6+7 2 7 2 | |
| | 6 2 | 9 2 6 2 | |
| | 7 2 | 8 2 6 2 | |
| | 9 2 | 9 2 7 2 | |
| | 6 2 | 4+5 2 7 2 | |
| | 8 2 | 8 2 6 2 | |
| | 9 2 | 8 2 7 2 | |
| | 6 2 | 6+7 2 7 2 | |
| | 9 2 | 9 2 6 2 | |
| | 6 2 | 8 2 6 2 | |
| | 8 2 | 4+5 2 7 2 | |
| | 7 2 | 6+7 2 6 2 | |
| | 9 2 | 9 2 7 2 | |
| | 6 2 | 8 2 6 2 | |
| | 8 2 | 6+7 2 7 2 | |
| | 7 2 | 8 2 6 2 | |
| | 9 2 | 4+5 2 7 2 | |
| | 7 2 | 9 2 6 2 | |
| | 6 2 | 9 2 7 2 | |
| | 6 2 | 8 2 6 2 | |
| | 7 2 | 6+7 2 7 2 | |
| | 9 2 | 9 2 6 2 | |
| | 8 2 | 8 2 6 2 | |
| | 7 2 | 4+5 2 7 2 | |
| | 9 2 | 9 2 6 2 | |
| 225.1.17 | 188351 | 132 | Cl1Cu6O8Y1 |
| | -GM -X | -L -W | |
| | 6 2 6 2 | 8 2 7 2 | |
| | 8 2 8 2 | 9 2 6 2 | |
| | 11 4 8 2 | 9 2 7 2 | |
| | 6 2 9 2 | 6+7 2 6 2 | |
| | 7 2 6 2 | 8 2 7 2 | |
| | 10 4 7 2 | 8 2 6 2 | |
| | 9 2 8 2 | 9 2 7 2 | |
| | 8 2 7 2 | 4+5 2 6 2 | |
| | 11 4 6 2 | 8 2 6 2 | |
| | 6 2 8 2 | 9 2 7 2 | |
| | 10 4 9 2 | 9 2 7 2 | |
| | 6 2 9 2 | 6+7 2 6 2 | |
| | 10 4 6 2 | 9 2 6 2 | |
| | 7 2 6 2 | 4+5 2 6 2 | |
| | 10 4 8 2 | 8 2 7 2 | |
| | 6 2 9 2 | 6+7 2 7 2 | |
| | 8 2 9 2 | 9 2 6 2 | |
| | 11 4 7 2 | 8 2 6 2 | |
| | 10 4 8 2 | 9 2 7 2 | |
| | 7 2 8 2 | 8 2 6 2 | |
| | 9 2 9 2 | 8 2 7 2 | |
| | 11 4 8 2 | 9 2 7 2 | |
| | 9 2 6 2 | 6+7 2 6 2 | |
| | 11 4 7 2 | 4+5 2 6 2 | |
| | 7 2 7 2 | 8 2 7 2 | |
| | 10 4 9 2 | 9 2 7 2 | |
| | 10 4 8 2 | 9 2 6 2 | |
| | 6 2 7 2 | 6+7 2 6 2 | |
| | 10 4 6 2 | 9 2 7 2 | |
| | 6 2 7 2 | 8 2 7 2 | |
| | 10 4 6 2 | 4+5 2 6 2 | |
| | 7 2 6 2 | 8 2 7 2 | |
| | 10 4 8 2 | 4+5 2 6 2 | |
| | 6 2 9 2 | 9 2 7 2 | |
| | 8 2 8 2 | 8 2 6 2 | |
| | 7 2 9 2 | 6+7 2 6 2 | |
| | 10 4 8 2 | 8 2 7 2 | |
| | 11 4 9 2 | 4+5 2 7 2 | |
| | 10 4 9 2 | 9 2 7 2 | |
| | 8 2 9 2 | 8 2 6 2 | |
| | 10 4 6 2 | 6+7 2 6 2 | |
| | 11 4 7 2 | 9 2 7 2 | |
| | 7 2 8 2 | 8 2 6 2 | |
| | 6 2 9 2 | 9 2 6 2 | |
| | 10 4 8 2 | 8 2 6 2 | |
| | 6 2 | 4+5 2 7 2 | |
| | 8 2 | 9 2 6 2 | |
| | 7 2 | 8 2 7 2 | |
| | 9 2 | 8 2 7 2 | |

Continued on next column

**TABLE LXXXVII– continued**

| TQC | ICSD | NoE | SF |
|---|---|---|---|
| | 8 2 4+5 2 6 2 | | |
| | 7 2 6+7 2 7 2 | | |
| | 7 2 9 2 7 2 | | |
| | 9 2 9 2 6 2 | | |
| | 6 2 6+7 2 7 2 | | |
| | 9 2 9 2 7 2 | | |
| | 6 2 8 2 6 2 | | |
| | 6 2 4+5 2 7 2 | | |
| | 7 2 9 2 6 2 | | |
| | 8 2 8 2 7 2 | | |
| | 8 2 8 2 6 2 | | |
| | 9 2 4+5 2 7 2 | | |
| | 8 2 9 2 6 2 | | |
| | 9 2 6+7 2 7 2 | | |
| | 8 2 8 2 6 2 | | |
| | 8 2 9 2 6 2 | | |
| | 9 2 4+5 2 7 2 | | |
| 225.1.18 | 69612 | 124 | Cl1Cu6In1O8 |
| | -GM -X -L -W | | |
| | 6 2 6 2 9 2 6 2 | | |
| | 8 2 8 2 8 2 7 2 | | |
| | 11 4 8 2 4+5 2 6 2 | | |
| | 7 2 9 2 8 2 7 2 | | |
| | 10 4 7 2 9 2 6 2 | | |
| | 9 2 7 2 6+7 2 7 2 | | |
| | 6 2 6 2 9 2 7 2 | | |
| | 6 2 9 2 8 2 6 2 | | |
| | 10 4 6 2 8 2 7 2 | | |
| | 10 4 6 2 9 2 6 2 | | |
| | 7 2 8 2 6+7 2 7 2 | | |
| | 10 4 9 2 9 2 7 2 | | |
| | 6 2 8 2 8 2 6 2 | | |
| | 8 2 9 2 4+5 2 7 2 | | |
| | 11 4 6 2 8 2 6 2 | | |
| | 9 2 8 2 8 2 7 2 | | |
| | 11 4 9 2 4+5 2 6 2 | | |
| | 10 4 8 2 9 2 6 2 | | |
| | 7 2 7 2 9 2 7 2 | | |
| | 9 2 6 2 6+7 2 6 2 | | |
| | 8 2 7 2 9 2 7 2 | | |
| | 7 2 7 2 8 2 6 2 | | |
| | 11 4 8 2 8 2 6 2 | | |
| | 10 4 9 2 4+5 2 7 2 | | |
| | 11 4 7 2 8 2 6 2 | | |
| | 10 4 6 2 9 2 7 2 | | |
| | 6 2 7 2 6+7 2 7 2 | | |
| | 10 4 9 2 8 2 6 2 | | |
| | 6 2 8 2 9 2 7 2 | | |
| | 10 4 6 2 4+5 2 7 2 | | |
| | 7 2 8 2 6+7 2 6 2 | | |
| | 6 2 9 2 9 2 6 2 | | |
| | 10 4 8 2 9 2 7 2 | | |
| | 7 2 9 2 6+7 2 6 2 | | |
| | 10 4 9 2 8 2 6 2 | | |
| | 6 2 9 2 9 2 7 2 | | |
| | 8 2 8 2 8 2 6 2 | | |
| | 11 4 6 2 4+5 2 6 2 | | |
| | 10 4 9 2 9 2 7 2 | | |
| | 10 4 6 2 8 2 7 2 | | |
| | 7 2 8 2 9 2 7 2 | | |
| | 10 4 7 2 6+7 2 6 2 | | |
| | 8 2 8 2 7 2 | | |
| | 6 2 4+5 2 7 2 | | |
| | 9 2 9 2 6 2 | | |
| | 7 2 8 2 6 2 | | |
| | 8 2 9 2 6 2 | | |
| | 7 2 8 2 7 2 | | |
| | 9 2 6+7 2 6 2 | | |
| | 7 2 4+5 2 7 2 | | |
| | 9 2 9 2 6 2 | | |
| | 8 2 8 2 7 2 | | |
| | 6 2 8 2 6 2 | | |
| | 8 2 8 2 7 2 | | |
| | 9 2 4+5 2 7 2 | | |
| | 6 2 9 2 7 2 | | |
| | 7 2 6+7 2 6 2 | | |
| | 8 2 9 2 6 2 | | |
| | 9 2 6+7 2 6 2 | | |
| | 8 2 9 2 7 2 | | |
| | 9 2 8 2 7 2 | | |
| | 8 2 6+7 2 6 2 | | |
| 225.1.19 | 75013 | 260 | Ce3Pd20Si6 |
| | -GM -X -L -W | | |
| | 6 2 6 2 8 2 7 2 | | |
| | 9 2 7 2 8 2 6 2 | | |
| | 6 2 8 2 9 2 7 2 | | |
| | 8 2 8 2 9 2 6 2 | | |
| | 7 2 9 2 8 2 6 2 | | |
| | 8 2 6 2 9 2 7 2 | | |
| | 11 4 8 2 6+7 2 7 2 | | |
| | 10 4 9 2 9 2 6 2 | | |
| | 11 4 6 2 4+5 2 7 2 | | |
| | 6 2 8 2 8 2 6 2 | | |
| | 8 2 7 2 9 2 6 2 | | |
| | 11 4 9 2 6+7 2 7 2 | | |
| | 10 4 6 2 9 2 6 2 | | |
| | 7 2 8 2 8 2 7 2 | | |
| | 10 4 6 2 4+5 2 6 2 | | |
| | 10 4 8 2 8 2 7 2 | | |
| | 8 2 9 2 9 2 7 2 | | |
| | 11 4 7 2 6+7 2 6 2 | | |
| | 9 2 7 2 9 2 7 2 | | |
| | 7 2 6 2 8 2 6 2 | | |
| | 11 4 9 2 4+5 2 6 2 | | |
| | 6 2 8 2 9 2 6 2 | | |
| | 10 4 7 2 6+7 2 7 2 | | |
| | 11 4 6 2 8 2 7 2 | | |
| | 8 2 7 2 8 2 7 2 | | |
| | 9 2 6 2 4+5 2 6 2 | | |
| | 10 4 8 2 8 2 7 2 | | |
| | 6 2 7 2 9 2 6 2 | | |
| | 11 4 9 2 8 2 7 2 | | |
| | 9 2 9 2 6+7 2 6 2 | | |
| | 10 4 8 2 8 2 6 2 | | |
| | 8 2 7 2 9 2 6 2 | | |
| | 11 4 6 2 4+5 2 7 2 | | |
| | 11 4 8 2 9 2 6 2 | | |
| | 6 2 9 2 4+5 2 7 2 | | |
| | 7 2 8 2 6+7 2 7 2 | | |
| | 10 4 9 2 8 2 6 2 | | |
| | 11 4 6 2 9 2 7 2 | | |
| | 11 4 7 2 8 2 6 2 | | |

Continued on next column

Continued on next column

**TABLE LXXXVII– continued**

| TQC | ICSD | NoE | SF |
|---|---|---|---|
| | 10 4 6 2 | 9 2 7 2 | |
| | 8 2 9 2 | 6+7 2 6 2 | |
| | 9 2 8 2 | 4+5 2 7 2 | |
| | 10 4 6 2 | 8 2 6 2 | |
| | 11 4 7 2 | 9 2 7 2 | |
| | 6 2 7 2 | 8 2 6 2 | |
| | 10 4 6 2 | 4+5 2 7 2 | |
| | 7 2 7 2 | 8 2 6 2 | |
| | 9 2 9 2 | 6+7 2 7 2 | |
| | 10 4 7 2 | 9 2 6 2 | |
| | 7 2 9 2 | 9 2 6 2 | |
| | 6 2 8 2 | 9 2 7 2 | |
| | 6 2 8 2 | 4+5 2 7 2 | |
| | 10 4 7 2 | 8 2 6 2 | |
| | 11 4 6 2 | 9 2 7 2 | |
| | 8 2 6 2 | 8 2 6 2 | |
| | 10 4 8 2 | 6+7 2 7 2 | |
| | 6 2 9 2 | 9 2 6 2 | |
| | 10 4 8 2 | 4+5 2 7 2 | |
| | 11 4 7 2 | 8 2 6 2 | |
| | 7 2 9 2 | 6+7 2 7 2 | |
| | 10 4 6 2 | 8 2 6 2 | |
| | 7 2 8 2 | 8 2 7 2 | |
| | 8 2 6 2 | 9 2 7 2 | |
| | 9 2 8 2 | 6+7 2 6 2 | |
| | 11 4 9 2 | 9 2 7 2 | |
| | 11 4 9 2 | 8 2 6 2 | |
| | 9 2 9 2 | 8 2 6 2 | |
| | 10 4 7 2 | 4+5 2 7 2 | |
| | 11 4 8 2 | 9 2 7 2 | |
| | 8 2 6 2 | 9 2 7 2 | |
| | 7 2 6 2 | 9 2 6 2 | |
| | 7 2 7 2 | 6+7 2 6 2 | |
| | 11 4 6 2 | 8 2 6 2 | |
| | 9 2 7 2 | 4+5 2 7 2 | |
| | 6 2 8 2 | 9 2 7 2 | |
| | 10 4 9 2 | 9 2 7 2 | |
| | 8 2 9 2 | 4+5 2 6 2 | |
| | 10 4 7 2 | 8 2 7 2 | |
| | 11 4 6 2 | 6+7 2 6 2 | |
| | 8 2 8 2 | 8 2 6 2 | |
| | 11 4 6 2 | 4+5 2 7 2 | |
| | 9 2 8 2 | 8 2 7 2 | |
| | 6 2 9 2 | 9 2 6 2 | |
| | 11 4 7 2 | 9 2 7 2 | |
| | 10 4 8 2 | 8 2 6 2 | |
| | 6 2 8 2 | 6+7 2 6 2 | |
| | 9 2 9 2 | 9 2 7 2 | |
| | 11 4 8 2 | 9 2 7 2 | |
| | 6 2 | 4+5 2 6 2 | |
| | 6 2 | 9 2 7 2 | |
| | 7 2 | 6+7 2 6 2 | |
| | 9 2 | 8 2 6 2 | |
| | 7 2 | 8 2 6 2 | |
| | 9 2 | 8 2 7 2 | |
| | 7 2 | 9 2 6 2 | |
| | 8 2 | 8 2 7 2 | |
| | 9 2 | 6+7 2 6 2 | |
| | 6 2 | 4+5 2 7 2 | |
| | 7 2 | 9 2 7 2 | |
| | 9 2 | 8 2 6 2 | |

Continued on next column

**TABLE LXXXVII– continued**

| TQC | ICSD | NoE | SF |
|---|---|---|---|
| | 6 2 | 6+7 2 6 2 | |
| | 9 2 | 9 2 6 2 | |
| | 7 2 | 4+5 2 7 2 | |
| | 6 2 | 8 2 6 2 | |
| | 8 2 | 9 2 7 2 | |
| | 9 2 | 6+7 2 7 2 | |
| | 8 2 | 8 2 6 2 | |
| | 9 2 | 4+5 2 7 2 | |
| | 9 2 | 9 2 6 2 | |
| | 7 2 | 9 2 6 2 | |
| | 8 2 | 6+7 2 7 2 | |
| | 6 2 | 9 2 7 2 | |
| | 7 2 | 8 2 6 2 | |
| | 6 2 | 9 2 7 2 | |
| | 8 2 | 8 2 7 2 | |
| | 7 2 | 6+7 2 6 2 | |
| | 6 2 | 8 2 7 2 | |
| | 7 2 | 4+5 2 7 2 | |
| | 8 2 | 8 2 6 2 | |
| | 8 2 | 8 2 6 2 | |
| | 9 2 | 9 2 7 2 | |
| | 8 2 | 8 2 7 2 | |
| | 9 2 | 4+5 2 6 2 | |
| | 6 2 | 9 2 6 2 | |
| | 7 2 | 6+7 2 6 2 | |
| | 6 2 | 8 2 7 2 | |
| | 6 2 | 4+5 2 6 2 | |
| | 8 2 | 9 2 7 2 | |
| | 9 2 | 9 2 7 2 | |
| | 7 2 | 8 2 6 2 | |
| 225.1.20 | 237543 | 332 | Ca3Cd8Pt4 |
| | -GM -X | -L -W | |
| | 6 2 6 2 | 8 2 7 2 | |
| | 8 2 6 2 | 4+5 2 6 2 | |
| | 10 4 7 2 | 6+7 2 6 2 | |
| | 11 4 8 2 | 8 2 6 2 | |
| | 6 2 8 2 | 9 2 7 2 | |
| | 11 4 9 2 | 9 2 7 2 | |
| | 10 4 6 2 | 6+7 2 6 2 | |
| | 8 2 8 2 | 8 2 7 2 | |
| | 6 2 9 2 | 9 2 7 2 | |
| | 8 2 6 2 | 4+5 2 6 2 | |
| | 7 2 7 2 | 8 2 7 2 | |
| | 11 4 8 2 | 9 2 6 2 | |
| | 10 4 6 2 | 8 2 6 2 | |
| | 9 2 8 2 | 9 2 7 2 | |
| | 10 4 7 2 | 8 2 6 2 | |
| | 11 4 8 2 | 6+7 2 6 2 | |
| | 10 4 9 2 | 9 2 7 2 | |
| | 10 4 6 2 | 4+5 2 6 2 | |
| | 7 2 7 2 | 8 2 7 2 | |
| | 6 2 9 2 | 9 2 7 2 | |
| | 7 2 6 2 | 4+5 2 6 2 | |
| | 8 2 7 2 | 6+7 2 6 2 | |
| | 10 4 8 2 | 8 2 7 2 | |
| | 11 4 9 2 | 9 2 7 2 | |
| | 9 2 6 2 | 6+7 2 6 2 | |
| | 6 2 6 2 | 8 2 7 2 | |
| | 11 4 8 2 | 9 2 6 2 | |
| | 10 4 9 2 | 8 2 7 2 | |
| | 11 4 8 2 | 4+5 2 6 2 | |

Continued on next column

**TABLE LXXXVII– continued**

| TQC | ICSD | NoE | SF |
|---|---|---|---|
| | 10 4 9 2 | 9 2 7 2 | |
| | 9 2 7 2 | 9 2 6 2 | |
| | 11 4 9 2 | 6+7 2 7 2 | |
| | 8 2 8 2 | 8 2 7 2 | |
| | 10 4 7 2 | 4+5 2 6 2 | |
| | 11 4 6 2 | 9 2 6 2 | |
| | 11 4 7 2 | 8 2 7 2 | |
| | 6 2 8 2 | 6+7 2 6 2 | |
| | 7 2 9 2 | 8 2 7 2 | |
| | 10 4 6 2 | 9 2 6 2 | |
| | 8 2 7 2 | 9 2 7 2 | |
| | 10 4 9 2 | 4+5 2 7 2 | |
| | 11 4 7 2 | 8 2 6 2 | |
| | 9 2 9 2 | 9 2 6 2 | |
| | 10 4 6 2 | 6+7 2 7 2 | |
| | 7 2 7 2 | 4+5 2 6 2 | |
| | 6 2 8 2 | 9 2 7 2 | |
| | 9 2 6 2 | 8 2 6 2 | |
| | 10 4 9 2 | 4+5 2 7 2 | |
| | 11 4 7 2 | 8 2 6 2 | |
| | 8 2 8 2 | 9 2 6 2 | |
| | 7 2 8 2 | 6+7 2 7 2 | |
| | 11 4 6 2 | 8 2 7 2 | |
| | 11 4 8 2 | 9 2 6 2 | |
| | 6 2 9 2 | 4+5 2 7 2 | |
| | 8 2 6 2 | 6+7 2 7 2 | |
| | 9 2 7 2 | 8 2 6 2 | |
| | 10 4 9 2 | 9 2 6 2 | |
| | 10 4 7 2 | 8 2 7 2 | |
| | 6 2 9 2 | 9 2 6 2 | |
| | 8 2 6 2 | 4+5 2 7 2 | |
| | 11 4 8 2 | 8 2 6 2 | |
| | 10 4 8 2 | 6+7 2 7 2 | |
| | 11 4 6 2 | 9 2 7 2 | |
| | 7 2 7 2 | 8 2 6 2 | |
| | 9 2 8 2 | 6+7 2 6 2 | |
| | 11 4 6 2 | 8 2 7 2 | |
| | 10 4 9 2 | 9 2 6 2 | |
| | 6 2 6 2 | 4+5 2 7 2 | |
| | 11 4 8 2 | 9 2 6 2 | |
| | 9 2 6 2 | 8 2 7 2 | |
| | 7 2 7 2 | 9 2 7 2 | |
| | 10 4 9 2 | 6+7 2 6 2 | |
| | 6 2 9 2 | 8 2 6 2 | |
| | 11 4 8 2 | 8 2 7 2 | |
| | 8 2 7 2 | 4+5 2 6 2 | |
| | 10 4 7 2 | 9 2 7 2 | |
| | 6 2 8 2 | 6+7 2 7 2 | |
| | 10 4 7 2 | 9 2 6 2 | |
| | 7 2 9 2 | 8 2 7 2 | |
| | 10 4 6 2 | 4+5 2 6 2 | |
| | 11 4 6 2 | 9 2 6 2 | |
| | 8 2 8 2 | 8 2 7 2 | |
| | 10 4 9 2 | 9 2 6 2 | |
| | 10 4 7 2 | 8 2 7 2 | |
| | 7 2 7 2 | 6+7 2 6 2 | |
| | 6 2 9 2 | 9 2 7 2 | |
| | 10 4 7 2 | 8 2 6 2 | |
| | 6 2 6 2 | 6+7 2 7 2 | |
| | 10 4 8 2 | 9 2 7 2 | |
| | 7 2 9 2 | 4+5 2 6 2 | |

Continued on next column

**TABLE LXXXVII– continued**

| TQC | ICSD | NoE | SF |
|---|---|---|---|
| | 7 2 6 2 | 8 2 6 2 | |
| | 6 2 8 2 | 8 2 7 2 | |
| | 10 4 6 2 | 4+5 2 7 2 | |
| | 10 4 7 2 | 6+7 2 6 2 | |
| | 10 4 8 2 | 9 2 7 2 | |
| | 7 2 9 2 | 8 2 7 2 | |
| | 6 2 8 2 | 4+5 2 6 2 | |
| | 10 4 9 2 | 9 2 6 2 | |
| | 10 4 7 2 | 8 2 7 2 | |
| | 7 2 6 2 | 9 2 6 2 | |
| | 7 2 8 2 | 4+5 2 7 2 | |
| | 10 4 6 2 | 6+7 2 7 2 | |
| | 6 2 9 2 | 8 2 6 2 | |
| | 10 4 7 2 | 9 2 6 2 | |
| | 10 4 8 2 | 9 2 6 2 | |
| | 8 2 6 2 | 8 2 7 2 | |
| | 11 4 7 2 | 9 2 6 2 | |
| | 11 4 9 2 | 8 2 7 2 | |
| | 9 2 8 2 | 4+5 2 7 2 | |
| | 9 2 6 2 | 6+7 2 6 2 | |
| | 6 2 | 9 2 6 2 | |
| | 9 2 | 8 2 7 2 | |
| | 7 2 | 9 2 7 2 | |
| | 8 2 | 8 2 6 2 | |
| | 9 2 | 4+5 2 7 2 | |
| | 9 2 | 8 2 6 2 | |
| | 6 2 | 9 2 7 2 | |
| | 7 2 | 6+7 2 6 2 | |
| | 8 2 | 8 2 7 2 | |
| | 7 2 | 9 2 6 2 | |
| | 9 2 | 6+7 2 7 2 | |
| | 6 2 | 9 2 7 2 | |
| | 8 2 | 4+5 2 6 2 | |
| | 7 2 | 8 2 6 2 | |
| | 8 2 | 6+7 2 6 2 | |
| | 6 2 | 4+5 2 7 2 | |
| | 6 2 | 8 2 7 2 | |
| | 9 2 | 9 2 6 2 | |
| | 7 2 | 9 2 7 2 | |
| | 8 2 | 8 2 6 2 | |
| | 9 2 | 9 2 7 2 | |
| | 7 2 | 4+5 2 6 2 | |
| | 6 2 | 8 2 7 2 | |
| | 9 2 | 9 2 6 2 | |
| | 7 2 | 8 2 7 2 | |
| | 8 2 | 6+7 2 6 2 | |
| | 6 2 | 8 2 7 2 | |
| | 8 2 | 9 2 6 2 | |
| | 9 2 | 9 2 7 2 | |
| | 6 2 | 8 2 6 2 | |
| | 7 2 | 4+5 2 7 2 | |
| | 8 2 | 6+7 2 7 2 | |
| | 9 2 | 9 2 6 2 | |
| | 7 2 | 8 2 6 2 | |
| | 8 2 | 8 2 7 2 | |
| | 6 2 | 8 2 7 2 | |
| | 9 2 | 4+5 2 6 2 | |
| | 7 2 | 9 2 6 2 | |
| | 8 2 | 4+5 2 7 2 | |
| | 6 2 | 6+7 2 6 2 | |
| | 7 2 | 8 2 7 2 | |

Continued on next column

**TABLE LXXXVII– continued**

| TQC | ICSD | NoE | SF |
|---|---|---|---|
| | 7 2 | 8 2 6 2 | |
| | 9 2 | 9 2 6 2 | |
| | 6 2 | 4+5 2 7 2 | |
| | 7 2 | 6+7 2 7 2 | |
| | 6 2 | 9 2 6 2 | |
| | 8 2 | 6+7 2 7 2 | |
| | 9 2 | 9 2 6 2 | |
| | 9 2 | 8 2 7 2 | |
| | 8 2 | 6+7 2 6 2 | |
| | 9 2 | 4+5 2 7 2 | |
| | 8 2 | 9 2 7 2 | |
| | 7 2 | 8 2 6 2 | |
| | 6 2 | 8 2 6 2 | |
| | 7 2 | 9 2 7 2 | |
| | 8 2 | 9 2 6 2 | |
| 225.1.21 | 62667 | 162 | C6Cr23 |
| | -GM -X | -L -W | |
| | 6 2 6 2 | 9 2 6 2 | |
| | 11 4 8 2 | 8 2 7 2 | |
| | 8 2 9 2 | 4+5 2 6 2 | |
| | 10 4 8 2 | 8 2 7 2 | |
| | 7 2 6 2 | 6+7 2 7 2 | |
| | 10 4 7 2 | 9 2 6 2 | |
| | 8 2 7 2 | 8 2 6 2 | |
| | 11 4 8 2 | 9 2 7 2 | |
| | 6 2 8 2 | 8 2 7 2 | |
| | 6 2 9 2 | 4+5 2 6 2 | |
| | 10 4 6 2 | 8 2 7 2 | |
| | 9 2 6 2 | 9 2 7 2 | |
| | 11 4 6 2 | 6+7 2 6 2 | |
| | 11 4 7 2 | 9 2 6 2 | |
| | 8 2 8 2 | 4+5 2 6 2 | |
| | 9 2 9 2 | 8 2 7 2 | |
| | 10 4 8 2 | 6+7 2 6 2 | |
| | 6 2 7 2 | 9 2 7 2 | |
| | 6 2 6 2 | 9 2 7 2 | |
| | 7 2 7 2 | 8 2 7 2 | |
| | 10 4 9 2 | 8 2 6 2 | |
| | 8 2 8 2 | 4+5 2 6 2 | |
| | 11 4 6 2 | 6+7 2 7 2 | |
| | 11 4 9 2 | 9 2 6 2 | |
| | 10 4 8 2 | 9 2 7 2 | |
| | 8 2 7 2 | 8 2 6 2 | |
| | 11 4 7 2 | 8 2 7 2 | |
| | 9 2 6 2 | 4+5 2 6 2 | |
| | 11 4 6 2 | 8 2 6 2 | |
| | 10 4 8 2 | 9 2 7 2 | |
| | 7 2 9 2 | 8 2 6 2 | |
| | 10 4 9 2 | 6+7 2 7 2 | |
| | 10 4 6 2 | 9 2 6 2 | |
| | 7 2 6 2 | 9 2 7 2 | |
| | 11 4 7 2 | 9 2 6 2 | |
| | 6 2 8 2 | 6+7 2 6 2 | |
| | 11 4 9 2 | 4+5 2 7 2 | |
| | 8 2 8 2 | 8 2 6 2 | |
| | 9 2 7 2 | 8 2 7 2 | |
| | 11 4 9 2 | 4+5 2 6 2 | |
| | 9 2 6 2 | 8 2 7 2 | |
| | 10 4 7 2 | 8 2 6 2 | |
| | 7 2 9 2 | 9 2 7 2 | |
| | 10 4 6 2 | 6+7 2 6 2 | |

Continued on next column

**TABLE LXXXVII– continued**

| TQC | ICSD | NoE | SF |
|---|---|---|---|
| | 8 2 7 2 | 6+7 2 6 2 | |
| | 11 4 9 2 | 9 2 7 2 | |
| | 10 4 6 2 | 8 2 7 2 | |
| | 6 2 8 2 | 4+5 2 6 2 | |
| | 10 4 7 2 | 8 2 7 2 | |
| | 7 2 6 2 | 9 2 7 2 | |
| | 10 4 7 2 | 6+7 2 7 2 | |
| | 7 2 8 2 | 9 2 6 2 | |
| | 6 2 9 2 | 4+5 2 6 2 | |
| | 10 4 8 2 | 8 2 7 2 | |
| | 6 2 | 4+5 2 6 2 | |
| | 7 2 | 8 2 7 2 | |
| | 8 2 | 9 2 6 2 | |
| | 9 2 | 6+7 2 6 2 | |
| | 7 2 | 9 2 7 2 | |
| | 6 2 | 8 2 7 2 | |
| | 6 2 | 6+7 2 6 2 | |
| | 7 2 | 9 2 6 2 | |
| | 8 2 | 9 2 7 2 | |
| | 9 2 | 9 2 7 2 | |
| | 9 2 | 4+5 2 6 2 | |
| | 7 2 | 8 2 6 2 | |
| | 8 2 | 8 2 7 2 | |
| | 9 2 | 4+5 2 7 2 | |
| | 6 2 | 8 2 6 2 | |
| | 8 2 | 9 2 6 2 | |
| | 9 2 | 6+7 2 6 2 | |
| | 9 2 | 8 2 7 2 | |
| | 8 2 | 4+5 2 7 2 | |
| | 6 2 | 8 2 6 2 | |
| | 7 2 | 9 2 7 2 | |
| | 8 2 | 9 2 6 2 | |
| | 6 2 | 9 2 7 2 | |
| | 9 2 | 6+7 2 7 2 | |
| | 7 2 | 4+5 2 6 2 | |
| | 6 2 | 8 2 6 2 | |
| | 7 2 | 8 2 7 2 | |
| 225.1.22 | 159874 | 8 | C1Zr1 |
| | -GM -X | -L -W | |
| | 6 2 6 2 | 9 2 6 2 | |
| | 8 2 8 2 | 8 2 7 2 | |
| | 10 4 7 2 | 4+5 2 6 2 | |
| | 8 2 | 8 2 7 2 | |
| 225.1.23 | 424342 | 10 | C1W1 |
| | -GM -X | -L -W | |
| | 6 2 6 2 | 9 2 6 2 | |
| | 10 4 7 2 | 8 2 7 2 | |
| | 7 2 8 2 | 4+5 2 6 2 | |
| | 8 2 8 2 | 8 2 7 2 | |
| | 9 2 | 8 2 6 2 | |
| 225.1.24 | 159873 | 8 | C1Hf1 |
| | -GM -X | -L -W | |
| | 6 2 6 2 | 9 2 6 2 | |
| | 8 2 8 2 | 8 2 7 2 | |
| | 11 4 7 2 | 4+5 2 6 2 | |
| | 8 2 | 8 2 7 2 | |
| 225.1.25 | 155551 | 78 | Ba2Nd1O6Ru1 |
| | -GM -X | -L -W | |
| | 6 2 6 2 | 8 2 7 2 | |
| | 6 2 8 2 | 9 2 6 2 | |
| | 9 2 7 2 | 8 2 7 2 | |

Continued on next column

**TABLE LXXXVII– continued**

| TQC | ICSD | NoE | SF |
|---|---|---|---|
| | 8 2 8 2 | 9 2 6 2 | |
| | 11 4 9 2 | 9 2 6 2 | |
| | 6 2 8 2 | 6+7 2 7 2 | |
| | 10 4 6 2 | 8 2 6 2 | |
| | 8 2 6 2 | 8 2 7 2 | |
| | 11 4 7 2 | 4+5 2 7 2 | |
| | 7 2 8 2 | 9 2 6 2 | |
| | 8 2 9 2 | 9 2 6 2 | |
| | 10 4 8 2 | 6+7 2 7 2 | |
| | 11 4 9 2 | 9 2 7 2 | |
| | 10 4 6 2 | 8 2 6 2 | |
| | 6 2 8 2 | 6+7 2 7 2 | |
| | 7 2 7 2 | 4+5 2 6 2 | |
| | 10 4 6 2 | 8 2 7 2 | |
| | 8 2 9 2 | 9 2 6 2 | |
| | 11 4 7 2 | 9 2 7 2 | |
| | 11 4 6 2 | 9 2 6 2 | |
| | 9 2 7 2 | 6+7 2 6 2 | |
| | 8 2 6 2 | 9 2 7 2 | |
| | 11 4 6 2 | 6+7 2 7 2 | |
| | 6 2 7 2 | 8 2 6 2 | |
| | 10 4 8 2 | 8 2 7 2 | |
| | 10 4 8 2 | 9 2 7 2 | |
| | 7 2 9 2 | 4+5 2 6 2 | |
| | 9 2 | 8 2 7 2 | |
| | 6 2 | 4+5 2 6 2 | |
| | 7 2 | 8 2 6 2 | |
| | 9 2 | 8 2 6 2 | |
| | 8 2 | 4+5 2 7 2 | |
| | 8 2 | 8 2 6 2 | |
| | 8 2 | 9 2 7 2 | |
| | 9 2 | 6+7 2 6 2 | |
| | 6 2 | 9 2 7 2 | |
| | 7 2 | 6+7 2 6 2 | |
| | 9 2 | 9 2 7 2 | |
| | 8 2 | 9 2 7 2 | |
| 225.1.26 | 171491 | 84 | Ba2Ir1O6Tb1 |
| | -GM -X | -L -W | |
| | 6 2 6 2 | 9 2 6 2 | |
| | 6 2 8 2 | 8 2 6 2 | |
| | 9 2 7 2 | 9 2 7 2 | |
| | 8 2 8 2 | 8 2 7 2 | |
| | 11 4 9 2 | 8 2 7 2 | |
| | 6 2 8 2 | 4+5 2 6 2 | |
| | 10 4 6 2 | 8 2 7 2 | |
| | 8 2 6 2 | 9 2 6 2 | |
| | 11 4 7 2 | 9 2 6 2 | |
| | 7 2 8 2 | 6+7 2 7 2 | |
| | 8 2 8 2 | 8 2 7 2 | |
| | 10 4 9 2 | 4+5 2 6 2 | |
| | 11 4 9 2 | 8 2 6 2 | |
| | 10 4 6 2 | 9 2 7 2 | |
| | 6 2 8 2 | 4+5 2 6 2 | |
| | 7 2 7 2 | 6+7 2 7 2 | |
| | 10 4 6 2 | 8 2 6 2 | |
| | 8 2 9 2 | 9 2 7 2 | |
| | 11 4 6 2 | 8 2 6 2 | |
| | 11 4 7 2 | 4+5 2 7 2 | |
| | 9 2 6 2 | 8 2 7 2 | |
| | 8 2 7 2 | 8 2 6 2 | |
| | 11 4 6 2 | 4+5 2 7 2 | |

Continued on next column

**TABLE LXXXVII– continued**

| TQC | ICSD | NoE | SF |
|---|---|---|---|
| | 6 2 7 2 | 8 2 6 2 | |
| | 10 4 8 2 | 9 2 6 2 | |
| | 10 4 8 2 | 9 2 6 2 | |
| | 9 2 9 2 | 6+7 2 7 2 | |
| | 11 4 9 2 | 9 2 6 2 | |
| | 7 2 6 2 | 9 2 7 2 | |
| | 7 2 | 6+7 2 7 2 | |
| | 8 2 | 9 2 7 2 | |
| | 9 2 | 6+7 2 6 2 | |
| | 8 2 | 8 2 7 2 | |
| | 8 2 | 9 2 6 2 | |
| | 9 2 | 4+5 2 7 2 | |
| | 6 2 | 8 2 6 2 | |
| | 7 2 | 4+5 2 7 2 | |
| | 9 2 | 8 2 6 2 | |
| | 8 2 | 8 2 6 2 | |
| | 9 2 | 8 2 7 2 | |
| | 6 2 | 4+5 2 6 2 | |
| | 9 2 | 8 2 6 2 | |
| 225.1.27 | 153057 | 78 | Ba2Ir1O6Pr1 |
| | -GM -X | -L -W | |
| | 6 2 6 2 | 8 2 7 2 | |
| | 6 2 8 2 | 9 2 6 2 | |
| | 9 2 7 2 | 8 2 7 2 | |
| | 8 2 8 2 | 9 2 6 2 | |
| | 11 4 9 2 | 9 2 6 2 | |
| | 6 2 8 2 | 6+7 2 7 2 | |
| | 10 4 6 2 | 8 2 6 2 | |
| | 8 2 7 2 | 4+5 2 7 2 | |
| | 11 4 6 2 | 8 2 7 2 | |
| | 7 2 8 2 | 9 2 6 2 | |
| | 8 2 8 2 | 9 2 6 2 | |
| | 10 4 9 2 | 6+7 2 7 2 | |
| | 11 4 9 2 | 8 2 6 2 | |
| | 10 4 6 2 | 9 2 7 2 | |
| | 6 2 8 2 | 4+5 2 6 2 | |
| | 7 2 7 2 | 6+7 2 7 2 | |
| | 10 4 9 2 | 8 2 7 2 | |
| | 8 2 6 2 | 9 2 6 2 | |
| | 11 4 6 2 | 9 2 6 2 | |
| | 9 2 7 2 | 6+7 2 7 2 | |
| | 11 4 6 2 | 9 2 6 2 | |
| | 8 2 7 2 | 8 2 7 2 | |
| | 11 4 6 2 | 9 2 6 2 | |
| | 6 2 7 2 | 8 2 7 2 | |
| | 10 4 8 2 | 4+5 2 7 2 | |
| | 10 4 8 2 | 6+7 2 7 2 | |
| | 7 2 9 2 | 9 2 6 2 | |
| | 6 2 | 8 2 6 2 | |
| | 9 2 | 4+5 2 7 2 | |
| | 7 2 | 8 2 6 2 | |
| | 8 2 | 9 2 7 2 | |
| | 9 2 | 8 2 6 2 | |
| | 8 2 | 9 2 6 2 | |
| | 8 2 | 4+5 2 7 2 | |
| | 9 2 | 6+7 2 6 2 | |
| | 6 2 | 8 2 7 2 | |
| | 7 2 | 6+7 2 6 2 | |
| | 6 2 | 9 2 7 2 | |
| | 9 2 | 9 2 7 2 | |
| 225.1.28 | 59744 | 86 | Ba2Er1O6Ru1 |

Continued on next column

**TABLE LXXXVII– continued**

| TQC | ICSD | NoE | SF |
|---|---|---|---|
| | -GM -X | -L -W | |
| | 6 2 6 2 | 9 2 6 2 | |
| | 6 2 8 2 | 8 2 7 2 | |
| | 9 2 7 2 | 9 2 6 2 | |
| | 8 2 8 2 | 8 2 7 2 | |
| | 11 4 9 2 | 8 2 7 2 | |
| | 6 2 8 2 | 4+5 2 6 2 | |
| | 10 4 6 2 | 8 2 7 2 | |
| | 8 2 6 2 | 9 2 6 2 | |
| | 11 4 7 2 | 9 2 6 2 | |
| | 7 2 8 2 | 6+7 2 7 2 | |
| | 8 2 9 2 | 8 2 7 2 | |
| | 10 4 8 2 | 4+5 2 6 2 | |
| | 11 4 9 2 | 8 2 6 2 | |
| | 10 4 6 2 | 9 2 7 2 | |
| | 6 2 8 2 | 4+5 2 6 2 | |
| | 7 2 7 2 | 6+7 2 7 2 | |
| | 10 4 9 2 | 8 2 6 2 | |
| | 8 2 6 2 | 9 2 7 2 | |
| | 11 4 6 2 | 8 2 6 2 | |
| | 11 4 7 2 | 4+5 2 7 2 | |
| | 9 2 7 2 | 8 2 7 2 | |
| | 11 4 6 2 | 8 2 6 2 | |
| | 8 2 6 2 | 4+5 2 6 2 | |
| | 6 2 7 2 | 9 2 7 2 | |
| | 10 4 8 2 | 8 2 6 2 | |
| | 9 2 9 2 | 9 2 6 2 | |
| | 11 4 8 2 | 6+7 2 7 2 | |
| | 10 4 9 2 | 9 2 6 2 | |
| | 7 2 6 2 | 9 2 7 2 | |
| | 9 2 7 2 | 6+7 2 7 2 | |
| | 9 2 | 9 2 7 2 | |
| | 8 2 | 8 2 6 2 | |
| | 8 2 | 6+7 2 7 2 | |
| | 9 2 | 4+5 2 6 2 | |
| | 8 2 | 9 2 7 2 | |
| | 6 2 | 8 2 6 2 | |
| | 9 2 | 4+5 2 6 2 | |
| | 8 2 | 8 2 7 2 | |
| | 9 2 | 8 2 6 2 | |
| | 7 2 | 8 2 6 2 | |
| | 9 2 | 8 2 7 2 | |
| | 8 2 | 4+5 2 6 2 | |
| | 9 2 | 8 2 7 2 | |
| 225.1.29 | 59743 | 86 | Ba2Er1O6Ru1 |
| | -GM -X | -L -W | |
| | 6 2 6 2 | 9 2 6 2 | |
| | 6 2 8 2 | 8 2 7 2 | |
| | 9 2 7 2 | 9 2 6 2 | |
| | 8 2 8 2 | 8 2 7 2 | |
| | 11 4 9 2 | 8 2 7 2 | |
| | 6 2 8 2 | 4+5 2 6 2 | |
| | 10 4 6 2 | 8 2 7 2 | |
| | 8 2 6 2 | 9 2 6 2 | |
| | 11 4 7 2 | 9 2 6 2 | |
| | 7 2 8 2 | 6+7 2 7 2 | |
| | 8 2 9 2 | 8 2 7 2 | |
| | 10 4 8 2 | 4+5 2 6 2 | |
| | 11 4 9 2 | 8 2 6 2 | |
| | 10 4 6 2 | 9 2 7 2 | |
| | 6 2 8 2 | 4+5 2 6 2 | |

Continued on next column

**TABLE LXXXVII– continued**

| TQC | ICSD | NoE | SF |
|---|---|---|---|
| | 7 2 7 2 | 6+7 2 7 2 | |
| | 10 4 9 2 | 8 2 6 2 | |
| | 8 2 6 2 | 9 2 7 2 | |
| | 11 4 6 2 | 8 2 6 2 | |
| | 11 4 7 2 | 4+5 2 7 2 | |
| | 9 2 7 2 | 8 2 7 2 | |
| | 11 4 6 2 | 8 2 6 2 | |
| | 8 2 6 2 | 4+5 2 6 2 | |
| | 6 2 7 2 | 9 2 7 2 | |
| | 10 4 8 2 | 8 2 6 2 | |
| | 9 2 9 2 | 9 2 6 2 | |
| | 11 4 8 2 | 6+7 2 7 2 | |
| | 10 4 9 2 | 9 2 6 2 | |
| | 7 2 6 2 | 9 2 7 2 | |
| | 9 2 7 2 | 6+7 2 7 2 | |
| | 9 2 | 9 2 7 2 | |
| | 8 2 | 6+7 2 6 2 | |
| | 8 2 | 8 2 7 2 | |
| | 9 2 | 4+5 2 6 2 | |
| | 8 2 | 9 2 7 2 | |
| | 6 2 | 8 2 6 2 | |
| | 9 2 | 4+5 2 6 2 | |
| | 8 2 | 8 2 7 2 | |
| | 9 2 | 8 2 6 2 | |
| | 7 2 | 8 2 6 2 | |
| | 9 2 | 8 2 7 2 | |
| | 8 2 | 4+5 2 6 2 | |
| | 9 2 | 8 2 7 2 | |
| 225.1.30 | 409959 | 234 | B6Ni21Sc2 |
| | -GM -X | -L -W | |
| | 6 2 6 2 | 9 2 6 2 | |
| | 8 2 8 2 | 8 2 7 2 | |
| | 11 4 9 2 | 4+5 2 6 2 | |
| | 10 4 8 2 | 8 2 7 2 | |
| | 6 2 6 2 | 6+7 2 7 2 | |
| | 7 2 7 2 | 9 2 6 2 | |
| | 10 4 7 2 | 8 2 6 2 | |
| | 8 2 6 2 | 9 2 7 2 | |
| | 11 4 8 2 | 8 2 7 2 | |
| | 10 4 9 2 | 9 2 6 2 | |
| | 6 2 8 2 | 8 2 7 2 | |
| | 9 2 6 2 | 4+5 2 7 2 | |
| | 11 4 7 2 | 6+7 2 6 2 | |
| | 11 4 6 2 | 4+5 2 6 2 | |
| | 8 2 9 2 | 8 2 6 2 | |
| | 9 2 8 2 | 9 2 7 2 | |
| | 6 2 7 2 | 9 2 7 2 | |
| | 7 2 8 2 | 6+7 2 6 2 | |
| | 10 4 6 2 | 9 2 7 2 | |
| | 10 4 9 2 | 8 2 7 2 | |
| | 6 2 7 2 | 4+5 2 6 2 | |
| | 8 2 8 2 | 8 2 7 2 | |
| | 11 4 6 2 | 6+7 2 6 2 | |
| | 10 4 7 2 | 9 2 7 2 | |
| | 11 4 6 2 | 9 2 6 2 | |
| | 8 2 9 2 | 8 2 7 2 | |
| | 11 4 6 2 | 9 2 6 2 | |
| | 9 2 8 2 | 4+5 2 6 2 | |
| | 11 4 7 2 | 8 2 6 2 | |
| | 10 4 9 2 | 9 2 7 2 | |
| | 7 2 9 2 | 6+7 2 6 2 | |

Continued on next column

**TABLE LXXXVII– continued**

| TQC | ICSD | | NoE | SF |
|---|---|---|---|---|
| | 10 4 | 8 2 | 8 2 | 6 2 |
| | 9 2 | 6 2 | 8 2 | 7 2 |
| | 10 4 | 7 2 | 9 2 | 6 2 |
| | 11 4 | 7 2 | 9 2 | 7 2 |
| | 7 2 | 9 2 | 6+7 2 | 7 2 |
| | 6 2 | 7 2 | 4+5 2 | 6 2 |
| | 11 4 | 6 2 | 8 2 | 6 2 |
| | 9 2 | 9 2 | 8 2 | 7 2 |
| | 10 4 | 9 2 | 4+5 2 | 6 2 |
| | 11 4 | 8 2 | 6+7 2 | 7 2 |
| | 10 4 | 6 2 | 8 2 | 6 2 |
| | 7 2 | 7 2 | 9 2 | 6 2 |
| | 8 2 | 8 2 | 9 2 | 7 2 |
| | 10 4 | 8 2 | 6+7 2 | 7 2 |
| | 6 2 | 9 2 | 8 2 | 7 2 |
| | 8 2 | 7 2 | 4+5 2 | 7 2 |
| | 11 4 | 8 2 | 6+7 2 | 6 2 |
| | 10 4 | 9 2 | 9 2 | 7 2 |
| | 7 2 | 6 2 | 8 2 | 6 2 |
| | 11 4 | 8 2 | 9 2 | 6 2 |
| | 10 4 | 6 2 | 8 2 | 6 2 |
| | 6 2 | 7 2 | 4+5 2 | 7 2 |
| | 9 2 | 6 2 | 8 2 | 6 2 |
| | 10 4 | 7 2 | 8 2 | 7 2 |
| | 7 2 | 6 2 | 9 2 | 6 2 |
| | 8 2 | 9 2 | 6+7 2 | 6 2 |
| | 10 4 | 8 2 | 9 2 | 7 2 |
| | 11 4 | 6 2 | 8 2 | 7 2 |
| | 9 2 | 7 2 | 4+5 2 | 6 2 |
| | 11 4 | 8 2 | 6+7 2 | 7 2 |
| | 6 2 | 9 2 | 9 2 | 6 2 |
| | 7 2 | 7 2 | 9 2 | 7 2 |
| | 10 4 | 6 2 | 8 2 | 6 2 |
| | 9 2 | 8 2 | 9 2 | 7 2 |
| | 11 4 | 9 2 | 4+5 2 | 6 2 |
| | 7 2 | 7 2 | 8 2 | 7 2 |
| | 10 4 | 9 2 | 9 2 | 7 2 |
| | 11 4 | 8 2 | 8 2 | 6 2 |
| | 10 4 | 9 2 | 4+5 2 | 7 2 |
| | 8 2 | 8 2 | 6+7 2 | 6 2 |
| | 7 2 | 6 2 | 9 2 | 7 2 |
| | 8 2 | 7 2 | 8 2 | 7 2 |
| | 6 2 | 6 2 | 8 2 | 6 2 |
| | 9 2 | 9 2 | 4+5 2 | 7 2 |
| | 10 4 | 9 2 | 4+5 2 | 6 2 |
| | 11 4 | 8 2 | 8 2 | 6 2 |
| | 11 4 | 6 2 | 9 2 | 7 2 |
| | | 7 2 | 6+7 2 | 6 2 |
| | | 8 2 | 8 2 | 7 2 |
| | | 7 2 | 8 2 | 6 2 |
| | | 9 2 | 6+7 2 | 6 2 |
| | | 8 2 | 9 2 | 7 2 |
| | | 9 2 | 9 2 | 7 2 |
| | | 7 2 | 9 2 | 6 2 |
| | | 9 2 | 8 2 | 7 2 |
| | | 6 2 | 4+5 2 | 6 2 |
| | | 7 2 | 8 2 | 7 2 |
| | | 6 2 | 6+7 2 | 6 2 |
| | | 7 2 | 9 2 | 7 2 |
| | | 6 2 | 4+5 2 | 6 2 |
| | | 8 2 | 8 2 | 6 2 |

Continued on next column

**TABLE LXXXVII– continued**

| TQC | ICSD | | NoE | SF |
|---|---|---|---|---|
| | | 7 2 | 9 2 | 7 2 |
| | | 9 2 | 8 2 | 7 2 |
| | | 6 2 | 9 2 | 6 2 |
| | | 9 2 | 4+5 2 | 7 2 |
| | | 7 2 | 6+7 2 | 6 2 |
| | | 6 2 | 9 2 | 7 2 |
| | | 8 2 | 8 2 | 6 2 |
| | | 6 2 | 6+7 2 | 6 2 |
| | | 8 2 | 9 2 | 6 2 |
| | | 9 2 | 4+5 2 | 7 2 |
| | | 7 2 | 8 2 | 7 2 |
| | | 8 2 | 8 2 | 7 2 |
| | | 9 2 | 9 2 | 6 2 |
| | | 7 2 | 4+5 2 | 7 2 |
| | | 6 2 | 8 2 | 6 2 |
| | | 9 2 | 6+7 2 | 6 2 |
| | | 8 2 | 9 2 | 7 2 |
| | | 7 2 | 9 2 | 6 2 |
| | | 8 2 | 8 2 | 7 2 |
| | | 6 2 | 9 2 | 7 2 |
| | | 7 2 | 8 2 | 6 2 |
| | | 6 2 | 4+5 2 | 6 2 |
| | | 8 2 | 9 2 | 7 2 |
| | | 9 2 | 6+7 2 | 6 2 |
| | | 7 2 | 8 2 | 6 2 |

The End

TABLE LXXXVIII: The computed irreducible representations for the valence bands for the compounds of SEBR and NLC in space group 227. See the list of tables.

| TQC | ICSD | | NoE | SF |
|---|---|---|---|---|
| 227.1.1 | 72042 | | 64 | S2Ti1 |
| | -GM | -X | -L | -W |
| | 6 2 | 5 4 | 9 2 | 5+6 2 |
| | 9 2 | 5 4 | 8 2 | 7 2 |
| | 8 2 | 5 4 | 9 2 | 3+4 2 |
| | 11 4 | 5 4 | 9 2 | 7 2 |
| | 7 2 | 5 4 | 6+7 2 | 3+4 2 |
| | 10 4 | 5 4 | 4+5 2 | 7 2 |
| | 6 2 | 5 4 | 8 2 | 5+6 2 |
| | 7 2 | 5 4 | 8 2 | 7 2 |
| | 10 4 | 5 4 | 9 2 | 3+4 2 |
| | 11 4 | 5 4 | 6+7 2 | 7 2 |
| | 9 2 | 5 4 | 9 2 | 7 2 |
| | 10 4 | 5 4 | 4+5 2 | 5+6 2 |
| | 6 2 | 5 4 | 8 2 | 7 2 |
| | 10 4 | 5 4 | 8 2 | 3+4 2 |
| | 7 2 | 5 4 | 9 2 | 5+6 2 |
| | 10 4 | 5 4 | 9 2 | 7 2 |
| | 9 2 | | 6+7 2 | 5+6 2 |
| | 11 4 | | 9 2 | 7 2 |
| | 11 4 | | 8 2 | 7 2 |
| | 8 2 | | 4+5 2 | 3+4 2 |
| | 10 4 | | 8 2 | 5+6 2 |
| | 6 2 | | 9 2 | 7 2 |
| | | | 9 2 | 3+4 2 |
| | | | 6+7 2 | 7 2 |
| | | | 8 2 | 3+4 2 |
| | | | 4+5 2 | 7 2 |

Continued on next column

**TABLE LXXXVIII– continued**

| TQC | ICSD | | NoE | SF |
|---|---|---|---|---|
| | | | 8 2 | 5+6 2 |
| | | | 9 2 | 7 2 |
| | | | 6+7 2 | 7 2 |
| | | | 8 2 | 5+6 2 |
| | | | 8 2 | 3+4 2 |
| | | | 4+5 2 | 7 2 |
| 227.1.2 | 160865 | | 148 | O7V2Y2 |
| | -GM | -X | -L | -W |
| | 6 2 | 5 4 | 8 2 | 3+4 2 |
| | 7 2 | 5 4 | 9 2 | 7 2 |
| | 10 4 | 5 4 | 4+5 2 | 5+6 2 |
| | 9 2 | 5 4 | 8 2 | 7 2 |
| | 11 4 | 5 4 | 9 2 | 3+4 2 |
| | 8 2 | 5 4 | 8 2 | 7 2 |
| | 9 2 | 5 4 | 6+7 2 | 7 2 |
| | 11 4 | 5 4 | 9 2 | 5+6 2 |
| | 8 2 | 5 4 | 9 2 | 7 2 |
| | 11 4 | 5 4 | 8 2 | 3+4 2 |
| | 11 4 | 5 4 | 9 2 | 7 2 |
| | 6 2 | 5 4 | 6+7 2 | 5+6 2 |
| | 6 2 | 5 4 | 9 2 | 3+4 2 |
| | 7 2 | 5 4 | 4+5 2 | 7 2 |
| | 10 4 | 5 4 | 9 2 | 5+6 2 |
| | 10 4 | 5 4 | 6+7 2 | 7 2 |
| | 11 4 | 5 4 | 8 2 | 5+6 2 |
| | 8 2 | 5 4 | 9 2 | 7 2 |
| | 11 4 | 5 4 | 8 2 | 7 2 |
| | 9 2 | 5 4 | 6+7 2 | 3+4 2 |
| | 9 2 | 5 4 | 9 2 | 7 2 |
| | 7 2 | 5 4 | 8 2 | 7 2 |
| | 10 4 | 5 4 | 4+5 2 | 3+4 2 |
| | 7 2 | 5 4 | 8 2 | 5+6 2 |
| | 10 4 | 5 4 | 6+7 2 | 7 2 |
| | 6 2 | 5 4 | 9 2 | 3+4 2 |
| | 6 2 | 5 4 | 4+5 2 | 7 2 |
| | 10 4 | 5 4 | 9 2 | 5+6 2 |
| | 7 2 | 5 4 | 8 2 | 3+4 2 |
| | 10 4 | 5 4 | 9 2 | 7 2 |
| | 10 4 | 5 4 | 9 2 | 3+4 2 |
| | 9 2 | 5 4 | 6+7 2 | 7 2 |
| | 9 2 | 5 4 | 9 2 | 3+4 2 |
| | 11 4 | 5 4 | 8 2 | 7 2 |
| | 11 4 | 5 4 | 4+5 2 | 5+6 2 |
| | 8 2 | 5 4 | 8 2 | 7 2 |
| | 11 4 | 5 4 | 9 2 | 5+6 2 |
| | 8 2 | | 6+7 2 | 7 2 |
| | 10 4 | | 9 2 | 3+4 2 |
| | 6 2 | | 8 2 | 7 2 |
| | 7 2 | | 4+5 2 | 5+6 2 |
| | 10 4 | | 8 2 | 7 2 |
| | 8 2 | | 9 2 | 7 2 |
| | 11 4 | | 8 2 | 5+6 2 |
| | 11 4 | | 4+5 2 | 7 2 |
| | 9 2 | | 9 2 | 5+6 2 |
| | 11 4 | | 6+7 2 | 7 2 |
| | 8 2 | | 8 2 | 5+6 2 |
| | 11 4 | | 9 2 | 3+4 2 |
| | 10 4 | | 9 2 | 7 2 |
| | | | 8 2 | 3+4 2 |
| | | | 6+7 2 | 7 2 |
| | | | 9 2 | 5+6 2 |

Continued on next column

**TABLE LXXXVIII– continued**

| TQC | ICSD | | NoE | SF |
|---|---|---|---|---|
| | | | 4+5 2 | 7 2 |
| | | | 8 2 | 7 2 |
| | | | 8 2 | 3+4 2 |
| | | | 8 2 | 5+6 2 |
| | | | 4+5 2 | 7 2 |
| | | | 9 2 | 3+4 2 |
| | | | 6+7 2 | 7 2 |
| | | | 9 2 | 3+4 2 |
| | | | 9 2 | 7 2 |
| | | | 8 2 | 5+6 2 |
| | | | 6+7 2 | 7 2 |
| | | | 9 2 | 7 2 |
| | | | 8 2 | 3+4 2 |
| | | | 4+5 2 | 5+6 2 |
| | | | 8 2 | 7 2 |
| | | | 6+7 2 | 3+4 2 |
| | | | 9 2 | 7 2 |
| | | | 8 2 | 7 2 |
| | | | 4+5 2 | 5+6 2 |
| | | | 9 2 | 7 2 |
| | | | 6+7 2 | 5+6 2 |
| 227.1.3 | 151747 | | 176 | O7Tb2Ti2 |
| | -GM | -X | -L | -W |
| | 6 2 | 5 4 | 9 2 | 5+6 2 |
| | 7 2 | 5 4 | 8 2 | 7 2 |
| | 10 4 | 5 4 | 6+7 2 | 3+4 2 |
| | 9 2 | 5 4 | 9 2 | 7 2 |
| | 11 4 | 5 4 | 8 2 | 5+6 2 |
| | 8 2 | 5 4 | 9 2 | 7 2 |
| | 9 2 | 5 4 | 4+5 2 | 7 2 |
| | 11 4 | 5 4 | 8 2 | 3+4 2 |
| | 8 2 | 5 4 | 8 2 | 7 2 |
| | 11 4 | 5 4 | 9 2 | 5+6 2 |
| | 11 4 | 5 4 | 8 2 | 7 2 |
| | 6 2 | 5 4 | 4+5 2 | 3+4 2 |
| | 6 2 | 5 4 | 8 2 | 5+6 2 |
| | 7 2 | 5 4 | 6+7 2 | 7 2 |
| | 10 4 | 5 4 | 8 2 | 3+4 2 |
| | 10 4 | 5 4 | 4+5 2 | 7 2 |
| | 9 2 | 5 4 | 9 2 | 7 2 |
| | 8 2 | 5 4 | 8 2 | 5+6 2 |
| | 11 4 | 5 4 | 9 2 | 3+4 2 |
| | 11 4 | 5 4 | 6+7 2 | 7 2 |
| | 9 2 | 5 4 | 4+5 2 | 7 2 |
| | 7 2 | 5 4 | 9 2 | 5+6 2 |
| | 10 4 | 5 4 | 8 2 | 7 2 |
| | 7 2 | 5 4 | 8 2 | 3+4 2 |
| | 10 4 | 5 4 | 9 2 | 7 2 |
| | 6 2 | 5 4 | 4+5 2 | 5+6 2 |
| | 6 2 | 5 4 | 9 2 | 7 2 |
| | 10 4 | 5 4 | 8 2 | 3+4 2 |
| | 10 4 | 5 4 | 6+7 2 | 7 2 |
| | 7 2 | 5 4 | 8 2 | 5+6 2 |
| | 9 2 | 5 4 | 8 2 | 5+6 2 |
| | 10 4 | 5 4 | 4+5 2 | 7 2 |
| | 9 2 | 5 4 | 8 2 | 7 2 |
| | 11 4 | 5 4 | 9 2 | 3+4 2 |
| | 11 4 | 5 4 | 6+7 2 | 5+6 2 |
| | 8 2 | 5 4 | 9 2 | 7 2 |
| | 11 4 | 5 4 | 8 2 | 3+4 2 |
| | 8 2 | 5 4 | 4+5 2 | 7 2 |

Continued on next column

**TABLE LXXXVIII– continued**

| TQC | ICSD | NoE | SF |
|---|---|---|---|
| | 11 4 5 4 | 8 2 | 3+4 2 |
| | 8 2 5 4 | 9 2 | 5+6 2 |
| | 10 4 5 4 | 6+7 2 | 7 2 |
| | 6 2 5 4 | 9 2 | 7 2 |
| | 7 2 5 4 | 8 2 | 7 2 |
| | 10 4 5 4 | 9 2 | 3+4 2 |
| | 11 4 | 6+7 2 | 7 2 |
| | 9 2 | 8 2 | 3+4 2 |
| | 11 4 | 4+5 2 | 5+6 2 |
| | 8 2 | 8 2 | 7 2 |
| | 11 4 | 9 2 | 7 2 |
| | 8 2 | 8 2 | 3+4 2 |
| | 11 4 | 9 2 | 5+6 2 |
| | 9 2 | 4+5 2 | 7 2 |
| | 8 2 | 8 2 | 3+4 2 |
| | 11 4 | 9 2 | 7 2 |
| | 9 2 | 6+7 2 | 7 2 |
| | 11 4 | 9 2 | 3+4 2 |
| | 11 4 | 4+5 2 | 7 2 |
| | 11 4 | 8 2 | 5+6 2 |
| | 9 2 | 9 2 | 7 2 |
| | 8 2 | 6+7 2 | 5+6 2 |
| | | 8 2 | 5+6 2 |
| | | 9 2 | 7 2 |
| | | 4+5 2 | 3+4 2 |
| | | 8 2 | 7 2 |
| | | 8 2 | 7 2 |
| | | 9 2 | 3+4 2 |
| | | 8 2 | 5+6 2 |
| | | 4+5 2 | 7 2 |
| | | 6+7 2 | 5+6 2 |
| | | 9 2 | 7 2 |
| | | 9 2 | 7 2 |
| | | 6+7 2 | 3+4 2 |
| | | 8 2 | 7 2 |
| | | 4+5 2 | 3+4 2 |
| | | 8 2 | 5+6 2 |
| | | 8 2 | 7 2 |
| | | 9 2 | 7 2 |
| | | 6+7 2 | 5+6 2 |
| | | 8 2 | 3+4 2 |
| | | 4+5 2 | 7 2 |
| | | 8 2 | 7 2 |
| | | 9 2 | 3+4 2 |
| | | 8 2 | 5+6 2 |
| | | 4+5 2 | 7 2 |
| | | 4+5 2 | 7 2 |
| | | 8 2 | 5+6 2 |
| | | 8 2 | 3+4 2 |
| | | 8 2 | 7 2 |
| 227.1.4 | 429192 | 152 | O7Pr2Zr2 |
| | -GM -X | -L | -W |
| | 6 2 5 4 | 9 2 | 5+6 2 |
| | 7 2 5 4 | 8 2 | 7 2 |
| | 10 4 5 4 | 6+7 2 | 3+4 2 |
| | 9 2 5 4 | 9 2 | 7 2 |
| | 11 4 5 4 | 8 2 | 5+6 2 |
| | 8 2 5 4 | 9 2 | 7 2 |
| | 6 2 5 4 | 4+5 2 | 7 2 |
| | 6 2 5 4 | 8 2 | 3+4 2 |
| | 11 4 5 4 | 8 2 | 7 2 |

Continued on next column

**TABLE LXXXVIII– continued**

| TQC | ICSD | NoE | SF |
|---|---|---|---|
| | 7 2 5 4 | 9 2 | 5+6 2 |
| | 10 4 5 4 | 8 2 | 7 2 |
| | 9 2 5 4 | 4+5 2 | 3+4 2 |
| | 10 4 5 4 | 6+7 2 | 5+6 2 |
| | 8 2 5 4 | 9 2 | 7 2 |
| | 11 4 5 4 | 8 2 | 3+4 2 |
| | 9 2 5 4 | 9 2 | 5+6 2 |
| | 11 4 5 4 | 6+7 2 | 7 2 |
| | 8 2 5 4 | 8 2 | 7 2 |
| | 11 4 5 4 | 8 2 | 7 2 |
| | 11 4 5 4 | 4+5 2 | 3+4 2 |
| | 9 2 5 4 | 9 2 | 7 2 |
| | 7 2 5 4 | 9 2 | 5+6 2 |
| | 10 4 5 4 | 8 2 | 3+4 2 |
| | 10 4 5 4 | 4+5 2 | 7 2 |
| | 7 2 5 4 | 8 2 | 7 2 |
| | 10 4 5 4 | 4+5 2 | 5+6 2 |
| | 6 2 5 4 | 6+7 2 | 3+4 2 |
| | 6 2 5 4 | 8 2 | 7 2 |
| | 10 4 5 4 | 9 2 | 7 2 |
| | 7 2 5 4 | 8 2 | 5+6 2 |
| | 10 4 5 4 | 4+5 2 | 7 2 |
| | 9 2 5 4 | 8 2 | 5+6 2 |
| | 9 2 5 4 | 8 2 | 7 2 |
| | 11 4 5 4 | 8 2 | 3+4 2 |
| | 8 2 5 4 | 9 2 | 5+6 2 |
| | 11 4 5 4 | 6+7 2 | 7 2 |
| | 11 4 5 4 | 9 2 | 5+6 2 |
| | 10 4 5 4 | 6+7 2 | 7 2 |
| | 8 2 | 8 2 | 3+4 2 |
| | 6 2 | 4+5 2 | 7 2 |
| | 7 2 | 9 2 | 3+4 2 |
| | 10 4 | 9 2 | 7 2 |
| | 11 4 | 8 2 | 7 2 |
| | 8 2 | 9 2 | 3+4 2 |
| | 11 4 | 6+7 2 | 7 2 |
| | 9 2 | 8 2 | 3+4 2 |
| | 11 4 | 4+5 2 | 5+6 2 |
| | 8 2 | 8 2 | 7 2 |
| | 11 4 | 8 2 | 7 2 |
| | 8 2 | 9 2 | 3+4 2 |
| | 11 4 | 9 2 | 7 2 |
| | 9 2 | 4+5 2 | 5+6 2 |
| | | 8 2 | 3+4 2 |
| | | 6+7 2 | 7 2 |
| | | 9 2 | 5+6 2 |
| | | 9 2 | 7 2 |
| | | 4+5 2 | 7 2 |
| | | 8 2 | 5+6 2 |
| | | 9 2 | 7 2 |
| | | 8 2 | 3+4 2 |
| | | 6+7 2 | 5+6 2 |
| | | 9 2 | 7 2 |
| | | 4+5 2 | 3+4 2 |
| | | 8 2 | 7 2 |
| | | 9 2 | 7 2 |
| | | 8 2 | 5+6 2 |
| | | 6+7 2 | 3+4 2 |
| | | 4+5 2 | 7 2 |
| | | 8 2 | 5+6 2 |
| | | 9 2 | 7 2 |

Continued on next column

**TABLE LXXXVIII– continued**

| TQC | ICSD | | NoE | SF |
|---|---|---|---|---|
| | | | 9 2 | 3+4 2 |
| | | | 6+7 2 | 7 2 |
| | | | 8 2 | 3+4 2 |
| | | | 4+5 2 | 7 2 |
| | | | 8 2 | 7 2 |
| | | | 8 2 | 5+6 2 |
| 227.1.5 | 261854 | | 156 | Nd2O7Zr2 |
| | -GM | -X | -L | -W |
| | 6 2 | 5 4 | 9 2 | 5+6 2 |
| | 7 2 | 5 4 | 8 2 | 7 2 |
| | 10 4 | 5 4 | 6+7 2 | 3+4 2 |
| | 9 2 | 5 4 | 9 2 | 7 2 |
| | 11 4 | 5 4 | 8 2 | 5+6 2 |
| | 8 2 | 5 4 | 9 2 | 7 2 |
| | 6 2 | 5 4 | 4+5 2 | 7 2 |
| | 9 2 | 5 4 | 8 2 | 3+4 2 |
| | 11 4 | 5 4 | 8 2 | 7 2 |
| | 6 2 | 5 4 | 9 2 | 5+6 2 |
| | 7 2 | 5 4 | 8 2 | 7 2 |
| | 10 4 | 5 4 | 4+5 2 | 3+4 2 |
| | 10 4 | 5 4 | 8 2 | 5+6 2 |
| | 8 2 | 5 4 | 6+7 2 | 7 2 |
| | 11 4 | 5 4 | 9 2 | 3+4 2 |
| | 9 2 | 5 4 | 8 2 | 7 2 |
| | 11 4 | 5 4 | 9 2 | 5+6 2 |
| | 8 2 | 5 4 | 4+5 2 | 7 2 |
| | 11 4 | 5 4 | 6+7 2 | 7 2 |
| | 11 4 | 5 4 | 8 2 | 3+4 2 |
| | 9 2 | 5 4 | 9 2 | 7 2 |
| | 7 2 | 5 4 | 9 2 | 5+6 2 |
| | 10 4 | 5 4 | 8 2 | 3+4 2 |
| | 10 4 | 5 4 | 4+5 2 | 7 2 |
| | 7 2 | 5 4 | 8 2 | 7 2 |
| | 10 4 | 5 4 | 4+5 2 | 5+6 2 |
| | 6 2 | 5 4 | 6+7 2 | 3+4 2 |
| | 6 2 | 5 4 | 8 2 | 7 2 |
| | 10 4 | 5 4 | 9 2 | 7 2 |
| | 7 2 | 5 4 | 8 2 | 5+6 2 |
| | 10 4 | 5 4 | 4+5 2 | 7 2 |
| | 9 2 | 5 4 | 8 2 | 5+6 2 |
| | 9 2 | 5 4 | 8 2 | 7 2 |
| | 11 4 | 5 4 | 8 2 | 3+4 2 |
| | 8 2 | 5 4 | 9 2 | 5+6 2 |
| | 11 4 | 5 4 | 6+7 2 | 7 2 |
| | 11 4 | 5 4 | 9 2 | 5+6 2 |
| | 10 4 | 5 4 | 6+7 2 | 7 2 |
| | 7 2 | 5 4 | 8 2 | 3+4 2 |
| | 10 4 | | 4+5 2 | 7 2 |
| | 6 2 | | 9 2 | 3+4 2 |
| | 8 2 | | 9 2 | 7 2 |
| | 11 4 | | 8 2 | 7 2 |
| | 8 2 | | 9 2 | 3+4 2 |
| | 11 4 | | 6+7 2 | 3+4 2 |
| | 9 2 | | 8 2 | 7 2 |
| | 11 4 | | 4+5 2 | 5+6 2 |
| | 8 2 | | 8 2 | 7 2 |
| | 11 4 | | 8 2 | 7 2 |
| | 8 2 | | 9 2 | 3+4 2 |
| | 11 4 | | 4+5 2 | 7 2 |
| | 9 2 | | 9 2 | 5+6 2 |
| | 11 4 | | 8 2 | 3+4 2 |

Continued on next column

**TABLE LXXXVIII– continued**

| TQC | ICSD | | NoE | SF |
|---|---|---|---|---|
| | | | 6+7 2 | 7 2 |
| | | | 9 2 | 5+6 2 |
| | | | 9 2 | 7 2 |
| | | | 4+5 2 | 7 2 |
| | | | 8 2 | 5+6 2 |
| | | | 9 2 | 7 2 |
| | | | 8 2 | 3+4 2 |
| | | | 6+7 2 | 5+6 2 |
| | | | 9 2 | 7 2 |
| | | | 4+5 2 | 3+4 2 |
| | | | 8 2 | 7 2 |
| | | | 8 2 | 3+4 2 |
| | | | 9 2 | 7 2 |
| | | | 6+7 2 | 7 2 |
| | | | 4+5 2 | 5+6 2 |
| | | | 8 2 | 5+6 2 |
| | | | 9 2 | 7 2 |
| | | | 9 2 | 3+4 2 |
| | | | 6+7 2 | 7 2 |
| | | | 8 2 | 3+4 2 |
| | | | 4+5 2 | 7 2 |
| | | | 8 2 | 7 2 |
| | | | 8 2 | 5+6 2 |
| | | | 6+7 2 | 3+4 2 |
| | | | 9 2 | 7 2 |
| 227.1.6 | 249974 | | 156 | Nd2O7Zr2 |
| | -GM | -X | -L | -W |
| | 6 2 | 5 4 | 9 2 | 5+6 2 |
| | 7 2 | 5 4 | 8 2 | 7 2 |
| | 10 4 | 5 4 | 6+7 2 | 3+4 2 |
| | 9 2 | 5 4 | 9 2 | 7 2 |
| | 11 4 | 5 4 | 4+5 2 | 7 2 |
| | 8 2 | 5 4 | 8 2 | 5+6 2 |
| | 11 4 | 5 4 | 9 2 | 3+4 2 |
| | 9 2 | 5 4 | 8 2 | 7 2 |
| | 8 2 | 5 4 | 8 2 | 7 2 |
| | 6 2 | 5 4 | 8 2 | 5+6 2 |
| | 10 4 | 5 4 | 4+5 2 | 7 2 |
| | 7 2 | 5 4 | 6+7 2 | 3+4 2 |
| | 10 4 | 5 4 | 9 2 | 7 2 |
| | 11 4 | 5 4 | 9 2 | 5+6 2 |
| | 6 2 | 5 4 | 6+7 2 | 3+4 2 |
| | 9 2 | 5 4 | 9 2 | 7 2 |
| | 11 4 | 5 4 | 8 2 | 7 2 |
| | 11 4 | 5 4 | 8 2 | 3+4 2 |
| | 8 2 | 5 4 | 9 2 | 5+6 2 |
| | 11 4 | 5 4 | 4+5 2 | 7 2 |
| | 9 2 | 5 4 | 8 2 | 5+6 2 |
| | 10 4 | 5 4 | 9 2 | 7 2 |
| | 7 2 | 5 4 | 4+5 2 | 7 2 |
| | 10 4 | 5 4 | 8 2 | 3+4 2 |
| | 7 2 | 5 4 | 8 2 | 7 2 |
| | 10 4 | 5 4 | 6+7 2 | 5+6 2 |
| | 6 2 | 5 4 | 9 2 | 7 2 |
| | 10 4 | 5 4 | 4+5 2 | 5+6 2 |
| | 6 2 | 5 4 | 8 2 | 3+4 2 |
| | 8 2 | 5 4 | 8 2 | 7 2 |
| | 11 4 | 5 4 | 9 2 | 7 2 |
| | 9 2 | 5 4 | 6+7 2 | 3+4 2 |
| | 11 4 | 5 4 | 9 2 | 3+4 2 |
| | 11 4 | 5 4 | 9 2 | 7 2 |

Continued on next column

**TABLE LXXXVIII– continued**

| TQC | ICSD | NoE | SF |
|---|---|---|---|
| | 10 4 5 4 | 4+5 2 | 5+6 2 |
| | 7 2 5 4 | 6+7 2 | 7 2 |
| | 11 4 5 4 | 8 2 | 5+6 2 |
| | 10 4 5 4 | 8 2 | 7 2 |
| | 6 2 5 4 | 9 2 | 3+4 2 |
| | 8 2 | 8 2 | 7 2 |
| | 9 2 | 4+5 2 | 7 2 |
| | 8 2 | 9 2 | 5+6 2 |
| | 11 4 | 6+7 2 | 7 2 |
| | 6 2 | 9 2 | 3+4 2 |
| | 9 2 | 8 2 | 5+6 2 |
| | 7 2 | 4+5 2 | 7 2 |
| | 10 4 | 8 2 | 3+4 2 |
| | 11 4 | 4+5 2 | 7 2 |
| | 8 2 | 8 2 | 7 2 |
| | 11 4 | 9 2 | 3+4 2 |
| | 10 4 | 8 2 | 7 2 |
| | 7 2 | 9 2 | 5+6 2 |
| | 8 2 | 8 2 | 7 2 |
| | 9 2 | 6+7 2 | 5+6 2 |
| | | 9 2 | 3+4 2 |
| | | 6+7 2 | 7 2 |
| | | 9 2 | 5+6 2 |
| | | 8 2 | 7 2 |
| | | 8 2 | 3+4 2 |
| | | 4+5 2 | 7 2 |
| | | 9 2 | 7 2 |
| | | 9 2 | 5+6 2 |
| | | 6+7 2 | 7 2 |
| | | 8 2 | 3+4 2 |
| | | 4+5 2 | 5+6 2 |
| | | 8 2 | 7 2 |
| | | 8 2 | 3+4 2 |
| | | 9 2 | 7 2 |
| | | 8 2 | 3+4 2 |
| | | 9 2 | 7 2 |
| | | 6+7 2 | 7 2 |
| | | 4+5 2 | 5+6 2 |
| | | 8 2 | 3+4 2 |
| | | 9 2 | 7 2 |
| | | 8 2 | 7 2 |
| | | 4+5 2 | 5+6 2 |
| | | 8 2 | 3+4 2 |
| | | 8 2 | 7 2 |
| 227.1.7 | 174187 | 156 | Nd2O7Zr2 |
| | -GM -X | -L | -W |
| | 6 2 5 4 | 9 2 | 5+6 2 |
| | 7 2 5 4 | 8 2 | 7 2 |
| | 10 4 5 4 | 6+7 2 | 3+4 2 |
| | 9 2 5 4 | 9 2 | 7 2 |
| | 11 4 5 4 | 8 2 | 5+6 2 |
| | 8 2 5 4 | 9 2 | 7 2 |
| | 9 2 5 4 | 4+5 2 | 7 2 |
| | 6 2 5 4 | 8 2 | 3+4 2 |
| | 6 2 5 4 | 8 2 | 7 2 |
| | 11 4 5 4 | 9 2 | 5+6 2 |
| | 7 2 5 4 | 8 2 | 7 2 |
| | 10 4 5 4 | 4+5 2 | 3+4 2 |
| | 10 4 5 4 | 8 2 | 5+6 2 |
| | 8 2 5 4 | 6+7 2 | 7 2 |
| | 11 4 5 4 | 8 2 | 3+4 2 |

Continued on next column

**TABLE LXXXVIII– continued**

| TQC | ICSD | NoE | SF |
|---|---|---|---|
| | 9 2 5 4 | 9 2 | 7 2 |
| | 11 4 5 4 | 4+5 2 | 5+6 2 |
| | 8 2 5 4 | 8 2 | 7 2 |
| | 11 4 5 4 | 9 2 | 7 2 |
| | 11 4 5 4 | 6+7 2 | 3+4 2 |
| | 9 2 5 4 | 9 2 | 5+6 2 |
| | 10 4 5 4 | 9 2 | 7 2 |
| | 7 2 5 4 | 8 2 | 3+4 2 |
| | 10 4 5 4 | 4+5 2 | 7 2 |
| | 7 2 5 4 | 8 2 | 7 2 |
| | 6 2 5 4 | 4+5 2 | 5+6 2 |
| | 10 4 5 4 | 6+7 2 | 3+4 2 |
| | 6 2 5 4 | 8 2 | 7 2 |
| | 10 4 5 4 | 9 2 | 7 2 |
| | 7 2 5 4 | 8 2 | 5+6 2 |
| | 10 4 5 4 | 4+5 2 | 7 2 |
| | 9 2 5 4 | 8 2 | 5+6 2 |
| | 11 4 5 4 | 8 2 | 3+4 2 |
| | 9 2 5 4 | 8 2 | 7 2 |
| | 8 2 5 4 | 9 2 | 5+6 2 |
| | 11 4 5 4 | 6+7 2 | 7 2 |
| | 8 2 5 4 | 8 2 | 3+4 2 |
| | 11 4 5 4 | 4+5 2 | 7 2 |
| | 10 4 5 4 | 9 2 | 7 2 |
| | 6 2 | 6+7 2 | 5+6 2 |
| | 10 4 | 9 2 | 3+4 2 |
| | 7 2 | 8 2 | 7 2 |
| | 11 4 | 9 2 | 7 2 |
| | 8 2 | 9 2 | 3+4 2 |
| | 11 4 | 6+7 2 | 7 2 |
| | 9 2 | 8 2 | 3+4 2 |
| | 11 4 | 8 2 | 5+6 2 |
| | 8 2 | 4+5 2 | 7 2 |
| | 11 4 | 8 2 | 7 2 |
| | 8 2 | 9 2 | 3+4 2 |
| | 11 4 | 4+5 2 | 5+6 2 |
| | 11 4 | 8 2 | 7 2 |
| | 9 2 | 9 2 | 3+4 2 |
| | | 6+7 2 | 7 2 |
| | | 9 2 | 7 2 |
| | | 9 2 | 5+6 2 |
| | | 4+5 2 | 7 2 |
| | | 8 2 | 3+4 2 |
| | | 9 2 | 7 2 |
| | | 8 2 | 5+6 2 |
| | | 6+7 2 | 5+6 2 |
| | | 9 2 | 7 2 |
| | | 4+5 2 | 3+4 2 |
| | | 8 2 | 7 2 |
| | | 9 2 | 7 2 |
| | | 8 2 | 5+6 2 |
| | | 6+7 2 | 3+4 2 |
| | | 4+5 2 | 7 2 |
| | | 8 2 | 5+6 2 |
| | | 9 2 | 7 2 |
| | | 9 2 | 3+4 2 |
| | | 6+7 2 | 7 2 |
| | | 8 2 | 3+4 2 |
| | | 4+5 2 | 7 2 |
| | | 8 2 | 5+6 2 |
| | | 8 2 | 7 2 |

Continued on next column

**TABLE LXXXVIII– continued**

| TQC | ICSD | NoE | SF |
|---|---|---|---|
| | | 6+7 2 | 3+4 2 |
| | | 9 2 | 7 2 |
| 227.1.8 | 153227 | 156 | Nd2O7Zr2 |
| | -GM -X | -L | -W |
| | 6 2 5 4 | 9 2 | 5+6 2 |
| | 7 2 5 4 | 8 2 | 7 2 |
| | 10 4 5 4 | 6+7 2 | 3+4 2 |
| | 9 2 5 4 | 9 2 | 7 2 |
| | 11 4 5 4 | 8 2 | 5+6 2 |
| | 8 2 5 4 | 9 2 | 7 2 |
| | 9 2 5 4 | 4+5 2 | 7 2 |
| | 6 2 5 4 | 8 2 | 3+4 2 |
| | 11 4 5 4 | 8 2 | 7 2 |
| | 6 2 5 4 | 9 2 | 5+6 2 |
| | 7 2 5 4 | 8 2 | 7 2 |
| | 10 4 5 4 | 4+5 2 | 3+4 2 |
| | 10 4 5 4 | 8 2 | 5+6 2 |
| | 8 2 5 4 | 6+7 2 | 7 2 |
| | 11 4 5 4 | 9 2 | 3+4 2 |
| | 9 2 5 4 | 8 2 | 7 2 |
| | 11 4 5 4 | 4+5 2 | 5+6 2 |
| | 8 2 5 4 | 9 2 | 7 2 |
| | 11 4 5 4 | 6+7 2 | 7 2 |
| | 11 4 5 4 | 8 2 | 3+4 2 |
| | 9 2 5 4 | 9 2 | 7 2 |
| | 7 2 5 4 | 9 2 | 5+6 2 |
| | 10 4 5 4 | 8 2 | 3+4 2 |
| | 10 4 5 4 | 4+5 2 | 7 2 |
| | 7 2 5 4 | 8 2 | 7 2 |
| | 10 4 5 4 | 4+5 2 | 5+6 2 |
| | 6 2 5 4 | 6+7 2 | 3+4 2 |
| | 6 2 5 4 | 8 2 | 7 2 |
| | 10 4 5 4 | 9 2 | 7 2 |
| | 7 2 5 4 | 8 2 | 5+6 2 |
| | 10 4 5 4 | 4+5 2 | 7 2 |
| | 9 2 5 4 | 8 2 | 5+6 2 |
| | 9 2 5 4 | 8 2 | 7 2 |
| | 11 4 5 4 | 8 2 | 3+4 2 |
| | 8 2 5 4 | 9 2 | 5+6 2 |
| | 11 4 5 4 | 6+7 2 | 7 2 |
| | 11 4 5 4 | 9 2 | 5+6 2 |
| | 8 2 5 4 | 6+7 2 | 7 2 |
| | 10 4 5 4 | 8 2 | 3+4 2 |
| | 7 2 | 4+5 2 | 7 2 |
| | 10 4 | 9 2 | 3+4 2 |
| | 6 2 | 9 2 | 7 2 |
| | 11 4 | 8 2 | 7 2 |
| | 8 2 | 9 2 | 3+4 2 |
| | 11 4 | 6+7 2 | 7 2 |
| | 9 2 | 8 2 | 3+4 2 |
| | 11 4 | 4+5 2 | 5+6 2 |
| | 8 2 | 8 2 | 7 2 |
| | 11 4 | 8 2 | 7 2 |
| | 8 2 | 4+5 2 | 3+4 2 |
| | 11 4 | 9 2 | 7 2 |
| | 9 2 | 9 2 | 5+6 2 |
| | 11 4 | 8 2 | 3+4 2 |
| | | 6+7 2 | 7 2 |
| | | 9 2 | 5+6 2 |
| | | 9 2 | 7 2 |
| | | 4+5 2 | 7 2 |

Continued on next column

**TABLE LXXXVIII– continued**

| TQC | ICSD | NoE | SF |
|---|---|---|---|
| | | 8 2 | 5+6 2 |
| | | 9 2 | 7 2 |
| | | 8 2 | 3+4 2 |
| | | 6+7 2 | 5+6 2 |
| | | 9 2 | 7 2 |
| | | 4+5 2 | 3+4 2 |
| | | 8 2 | 7 2 |
| | | 8 2 | 7 2 |
| | | 9 2 | 3+4 2 |
| | | 6+7 2 | 5+6 2 |
| | | 4+5 2 | 7 2 |
| | | 8 2 | 5+6 2 |
| | | 9 2 | 7 2 |
| | | 9 2 | 3+4 2 |
| | | 6+7 2 | 7 2 |
| | | 8 2 | 3+4 2 |
| | | 4+5 2 | 7 2 |
| | | 8 2 | 5+6 2 |
| | | 8 2 | 7 2 |
| | | 6+7 2 | 3+4 2 |
| | | 9 2 | 7 2 |
| 227.1.9 | 82958 | 156 | Nd2O7Sn2 |
| | -GM -X | -L | -W |
| | 6 2 5 4 | 9 2 | 5+6 2 |
| | 7 2 5 4 | 8 2 | 7 2 |
| | 10 4 5 4 | 6+7 2 | 3+4 2 |
| | 9 2 5 4 | 9 2 | 7 2 |
| | 11 4 5 4 | 8 2 | 5+6 2 |
| | 8 2 5 4 | 9 2 | 7 2 |
| | 6 2 5 4 | 8 2 | 3+4 2 |
| | 11 4 5 4 | 4+5 2 | 7 2 |
| | 6 2 5 4 | 8 2 | 7 2 |
| | 9 2 5 4 | 9 2 | 5+6 2 |
| | 8 2 5 4 | 9 2 | 7 2 |
| | 7 2 5 4 | 6+7 2 | 3+4 2 |
| | 10 4 5 4 | 8 2 | 5+6 2 |
| | 11 4 5 4 | 4+5 2 | 7 2 |
| | 9 2 5 4 | 8 2 | 7 2 |
| | 10 4 5 4 | 9 2 | 3+4 2 |
| | 11 4 5 4 | 9 2 | 7 2 |
| | 8 2 5 4 | 6+7 2 | 3+4 2 |
| | 11 4 5 4 | 9 2 | 5+6 2 |
| | 11 4 5 4 | 4+5 2 | 7 2 |
| | 9 2 5 4 | 8 2 | 7 2 |
| | 10 4 5 4 | 8 2 | 5+6 2 |
| | 7 2 5 4 | 8 2 | 3+4 2 |
| | 9 2 5 4 | 4+5 2 | 7 2 |
| | 11 4 5 4 | 8 2 | 7 2 |
| | 6 2 5 4 | 6+7 2 | 5+6 2 |
| | 9 2 5 4 | 8 2 | 3+4 2 |
| | 7 2 5 4 | 4+5 2 | 7 2 |
| | 10 4 5 4 | 9 2 | 7 2 |
| | 8 2 5 4 | 8 2 | 5+6 2 |
| | 11 4 5 4 | 8 2 | 5+6 2 |
| | 10 4 5 4 | 4+5 2 | 7 2 |
| | 7 2 5 4 | 9 2 | 7 2 |
| | 11 4 5 4 | 9 2 | 3+4 2 |
| | 8 2 5 4 | 8 2 | 7 2 |
| | 10 4 5 4 | 9 2 | 3+4 2 |
| | 10 4 5 4 | 6+7 2 | 3+4 2 |
| | 6 2 5 4 | 9 2 | 7 2 |

Continued on next column

**TABLE LXXXVIII– continued**

| TQC | ICSD | NoE | SF |
|---|---|---|---|
| | 7 2 5 4 | 9 2 | 5+6 2 |
| | 10 4 | 6+7 2 | 7 2 |
| | 11 4 | 8 2 | 3+4 2 |
| | 8 2 | 8 2 | 7 2 |
| | 11 4 | 4+5 2 | 5+6 2 |
| | 9 2 | 9 2 | 7 2 |
| | 11 4 | 6+7 2 | 7 2 |
| | 8 2 | 8 2 | 3+4 2 |
| | 11 4 | 9 2 | 5+6 2 |
| | 10 4 | 6+7 2 | 7 2 |
| | 6 2 | 9 2 | 5+6 2 |
| | 8 2 | 9 2 | 7 2 |
| | 11 4 | 8 2 | 7 2 |
| | 11 4 | 4+5 2 | 3+4 2 |
| | 9 2 | 8 2 | 7 2 |
| | | 9 2 | 5+6 2 |
| | | 4+5 2 | 7 2 |
| | | 9 2 | 3+4 2 |
| | | 6+7 2 | 5+6 2 |
| | | 8 2 | 7 2 |
| | | 8 2 | 3+4 2 |
| | | 8 2 | 7 2 |
| | | 9 2 | 3+4 2 |
| | | 6+7 2 | 7 2 |
| | | 4+5 2 | 5+6 2 |
| | | 8 2 | 7 2 |
| | | 4+5 2 | 7 2 |
| | | 8 2 | 5+6 2 |
| | | 9 2 | 7 2 |
| | | 6+7 2 | 3+4 2 |
| | | 9 2 | 5+6 2 |
| | | 4+5 2 | 7 2 |
| | | 8 2 | 7 2 |
| | | 8 2 | 3+4 2 |
| | | 8 2 | 3+4 2 |
| | | 4+5 2 | 7 2 |
| | | 8 2 | 5+6 2 |
| | | 8 2 | 7 2 |
| | | 6+7 2 | 7 2 |
| | | 9 2 | 3+4 2 |
| 227.1.10 | 202522 | 152 | Mo2O7Y2 |
| | -GM -X | -L | -W |
| | 6 2 5 4 | 9 2 | 5+6 2 |
| | 7 2 5 4 | 8 2 | 7 2 |
| | 10 4 5 4 | 6+7 2 | 3+4 2 |
| | 9 2 5 4 | 9 2 | 7 2 |
| | 11 4 5 4 | 8 2 | 5+6 2 |
| | 8 2 5 4 | 4+5 2 | 7 2 |
| | 9 2 5 4 | 9 2 | 7 2 |
| | 11 4 5 4 | 8 2 | 3+4 2 |
| | 8 2 5 4 | 8 2 | 7 2 |
| | 11 4 5 4 | 8 2 | 5+6 2 |
| | 11 4 5 4 | 9 2 | 7 2 |
| | 6 2 5 4 | 4+5 2 | 3+4 2 |
| | 7 2 5 4 | 6+7 2 | 5+6 2 |
| | 10 4 5 4 | 8 2 | 7 2 |
| | 10 4 5 4 | 8 2 | 3+4 2 |
| | 6 2 5 4 | 4+5 2 | 7 2 |
| | 11 4 5 4 | 8 2 | 3+4 2 |
| | 8 2 5 4 | 9 2 | 7 2 |
| | 11 4 5 4 | 4+5 2 | 5+6 2 |

Continued on next column

**TABLE LXXXVIII– continued**

| TQC | ICSD | NoE | SF |
|---|---|---|---|
| | 9 2 5 4 | 8 2 | 7 2 |
| | 9 2 5 4 | 9 2 | 7 2 |
| | 10 4 5 4 | 6+7 2 | 5+6 2 |
| | 7 2 5 4 | 4+5 2 | 7 2 |
| | 10 4 5 4 | 8 2 | 3+4 2 |
| | 7 2 5 4 | 8 2 | 5+6 2 |
| | 10 4 5 4 | 9 2 | 7 2 |
| | 6 2 5 4 | 6+7 2 | 3+4 2 |
| | 6 2 5 4 | 9 2 | 7 2 |
| | 10 4 5 4 | 9 2 | 5+6 2 |
| | 7 2 5 4 | 8 2 | 7 2 |
| | 10 4 5 4 | 4+5 2 | 7 2 |
| | 9 2 5 4 | 8 2 | 5+6 2 |
| | 11 4 5 4 | 8 2 | 7 2 |
| | 9 2 5 4 | 8 2 | 3+4 2 |
| | 8 2 5 4 | 8 2 | 5+6 2 |
| | 11 4 5 4 | 6+7 2 | 7 2 |
| | 6 2 5 4 | 9 2 | 5+6 2 |
| | 10 4 5 4 | 4+5 2 | 7 2 |
| | 7 2 | 9 2 | 3+4 2 |
| | 10 4 | 9 2 | 7 2 |
| | 11 4 | 6+7 2 | 3+4 2 |
| | 8 2 | 8 2 | 7 2 |
| | 11 4 | 9 2 | 7 2 |
| | 9 2 | 8 2 | 3+4 2 |
| | 11 4 | 4+5 2 | 7 2 |
| | 8 2 | 8 2 | 3+4 2 |
| | 11 4 | 9 2 | 5+6 2 |
| | 8 2 | 6+7 2 | 7 2 |
| | 11 4 | 9 2 | 7 2 |
| | 6 2 | 4+5 2 | 3+4 2 |
| | 10 4 | 8 2 | 5+6 2 |
| | 7 2 | 9 2 | 7 2 |
| | | 8 2 | 3+4 2 |
| | | 6+7 2 | 7 2 |
| | | 9 2 | 7 2 |
| | | 8 2 | 5+6 2 |
| | | 9 2 | 7 2 |
| | | 6+7 2 | 5+6 2 |
| | | 9 2 | 3+4 2 |
| | | 8 2 | 7 2 |
| | | 4+5 2 | 5+6 2 |
| | | 8 2 | 7 2 |
| | | 4+5 2 | 3+4 2 |
| | | 9 2 | 7 2 |
| | | 9 2 | 7 2 |
| | | 8 2 | 5+6 2 |
| | | 6+7 2 | 3+4 2 |
| | | 9 2 | 7 2 |
| | | 8 2 | 5+6 2 |
| | | 4+5 2 | 7 2 |
| | | 9 2 | 7 2 |
| | | 6+7 2 | 3+4 2 |
| | | 8 2 | 7 2 |
| | | 8 2 | 3+4 2 |
| | | 4+5 2 | 7 2 |
| | | 8 2 | 3+4 2 |
| 227.1.11 | 83277 | 124 | Mn2O7Sc2 |
| | -GM -X | -L | -W |
| | 6 2 5 4 | 8 2 | 3+4 2 |
| | 6 2 5 4 | 9 2 | 7 2 |

Continued on next column

**TABLE LXXXVIII– continued**

| TQC | ICSD | | NoE | SF |
|---|---|---|---|---|
| | 7 2 | 5 4 | 8 2 | 5+6 2 |
| | 10 4 | 5 4 | 4+5 2 | 7 2 |
| | 11 4 | 5 4 | 9 2 | 5+6 2 |
| | 11 4 | 5 4 | 6+7 2 | 7 2 |
| | 8 2 | 5 4 | 9 2 | 3+4 2 |
| | 9 2 | 5 4 | 8 2 | 7 2 |
| | 10 4 | 5 4 | 8 2 | 7 2 |
| | 9 2 | 5 4 | 4+5 2 | 5+6 2 |
| | 10 4 | 5 4 | 6+7 2 | 3+4 2 |
| | 7 2 | 5 4 | 9 2 | 7 2 |
| | 10 4 | 5 4 | 9 2 | 5+6 2 |
| | 7 2 | 5 4 | 8 2 | 7 2 |
| | 6 2 | 5 4 | 4+5 2 | 5+6 2 |
| | 10 4 | 5 4 | 8 2 | 7 2 |
| | 6 2 | 5 4 | 8 2 | 7 2 |
| | 9 2 | 5 4 | 8 2 | 5+6 2 |
| | 11 4 | 5 4 | 4+5 2 | 3+4 2 |
| | 7 2 | 5 4 | 9 2 | 7 2 |
| | 10 4 | 5 4 | 6+7 2 | 7 2 |
| | 10 4 | 5 4 | 9 2 | 3+4 2 |
| | 9 2 | 5 4 | 8 2 | 5+6 2 |
| | 8 2 | 5 4 | 9 2 | 7 2 |
| | 11 4 | 5 4 | 8 2 | 3+4 2 |
| | 8 2 | 5 4 | 9 2 | 7 2 |
| | 11 4 | 5 4 | 6+7 2 | 3+4 2 |
| | 6 2 | 5 4 | 9 2 | 7 2 |
| | 10 4 | 5 4 | 4+5 2 | 7 2 |
| | 7 2 | 5 4 | 8 2 | 3+4 2 |
| | 10 4 | 5 4 | 9 2 | 3+4 2 |
| | 11 4 | | 6+7 2 | 7 2 |
| | 9 2 | | 8 2 | 5+6 2 |
| | 11 4 | | 4+5 2 | 7 2 |
| | 8 2 | | 8 2 | 7 2 |
| | 11 4 | | 8 2 | 5+6 2 |
| | 8 2 | | 9 2 | 3+4 2 |
| | 11 4 | | 6+7 2 | 7 2 |
| | 10 4 | | 9 2 | 7 2 |
| | 7 2 | | 6+7 2 | 5+6 2 |
| | 10 4 | | 9 2 | 3+4 2 |
| | 6 2 | | 8 2 | 7 2 |
| | | | 9 2 | 7 2 |
| | | | 4+5 2 | 5+6 2 |
| | | | 8 2 | 5+6 2 |
| | | | 8 2 | 7 2 |
| | | | 4+5 2 | 3+4 2 |
| | | | 9 2 | 7 2 |
| | | | 8 2 | 5+6 2 |
| | | | 9 2 | 7 2 |
| | | | 6+7 2 | 3+4 2 |
| | | | 9 2 | 7 2 |
| | | | 9 2 | 5+6 2 |
| | | | 8 2 | 7 2 |
| | | | 4+5 2 | 7 2 |
| | | | 6+7 2 | 3+4 2 |
| | | | 8 2 | 7 2 |
| | | | 4+5 2 | 3+4 2 |
| | | | 8 2 | 5+6 2 |
| | | | 8 2 | 7 2 |
| | | | 6+7 2 | 7 2 |
| | | | 9 2 | 3+4 2 |
| 227.1.12 | 187996 | | 120 | Mg3Ni2Ti1 |

Continued on next column

**TABLE LXXXVIII– continued**

| TQC | ICSD | | NoE | SF |
|---|---|---|---|---|
| | -GM | -X | -L | -W |
| | 6 2 | 5 4 | 8 2 | 7 2 |
| | 9 2 | 5 4 | 9 2 | 3+4 2 |
| | 8 2 | 5 4 | 9 2 | 5+6 2 |
| | 11 4 | 5 4 | 8 2 | 7 2 |
| | 6 2 | 5 4 | 4+5 2 | 3+4 2 |
| | 7 2 | 5 4 | 6+7 2 | 7 2 |
| | 10 4 | 5 4 | 9 2 | 7 2 |
| | 7 2 | 5 4 | 8 2 | 5+6 2 |
| | 10 4 | 5 4 | 8 2 | 3+4 2 |
| | 11 4 | 5 4 | 6+7 2 | 7 2 |
| | 10 4 | 5 4 | 9 2 | 5+6 2 |
| | 6 2 | 5 4 | 4+5 2 | 7 2 |
| | 9 2 | 5 4 | 8 2 | 3+4 2 |
| | 10 4 | 5 4 | 8 2 | 7 2 |
| | 7 2 | 5 4 | 9 2 | 5+6 2 |
| | 10 4 | 5 4 | 4+5 2 | 7 2 |
| | 7 2 | 5 4 | 8 2 | 3+4 2 |
| | 11 4 | 5 4 | 9 2 | 7 2 |
| | 8 2 | 5 4 | 4+5 2 | 7 2 |
| | 11 4 | 5 4 | 8 2 | 5+6 2 |
| | 9 2 | 5 4 | 9 2 | 7 2 |
| | 10 4 | 5 4 | 8 2 | 5+6 2 |
| | 6 2 | 5 4 | 6+7 2 | 5+6 2 |
| | 10 4 | 5 4 | 9 2 | 7 2 |
| | 11 4 | 5 4 | 6+7 2 | 3+4 2 |
| | 11 4 | 5 4 | 9 2 | 7 2 |
| | 8 2 | 5 4 | 8 2 | 7 2 |
| | 9 2 | 5 4 | 4+5 2 | 3+4 2 |
| | 9 2 | 5 4 | 8 2 | 5+6 2 |
| | 7 2 | 5 4 | 6+7 2 | 7 2 |
| | 11 4 | | 8 2 | 3+4 2 |
| | 10 4 | | 9 2 | 7 2 |
| | 11 4 | | 4+5 2 | 3+4 2 |
| | 6 2 | | 8 2 | 5+6 2 |
| | 8 2 | | 8 2 | 7 2 |
| | 10 4 | | 4+5 2 | 7 2 |
| | 11 4 | | 9 2 | 5+6 2 |
| | 10 4 | | 9 2 | 7 2 |
| | 6 2 | | 8 2 | 7 2 |
| | 10 4 | | 9 2 | 3+4 2 |
| | | | 6+7 2 | 7 2 |
| | | | 8 2 | 5+6 2 |
| | | | 4+5 2 | 3+4 2 |
| | | | 6+7 2 | 7 2 |
| | | | 9 2 | 7 2 |
| | | | 8 2 | 7 2 |
| | | | 6+7 2 | 3+4 2 |
| | | | 9 2 | 5+6 2 |
| | | | 9 2 | 3+4 2 |
| | | | 8 2 | 7 2 |
| | | | 4+5 2 | 5+6 2 |
| | | | 9 2 | 7 2 |
| | | | 6+7 2 | 7 2 |
| | | | 9 2 | 5+6 2 |
| | | | 8 2 | 3+4 2 |
| | | | 8 2 | 7 2 |
| | | | 4+5 2 | 7 2 |
| | | | 6+7 2 | 3+4 2 |
| | | | 9 2 | 7 2 |
| | | | 9 2 | 5+6 2 |

Continued on next column

**TABLE LXXXVIII– continued**

| TQC | ICSD | | NoE | SF |
|---|---|---|---|---|
| 227.1.13 | 184696 | | 68 | Mg1O4Ti2 |
| | -GM | -X | -L | -W |
| | 6 2 | 5 4 | 9 2 | 5+6 2 |
| | 7 2 | 5 4 | 8 2 | 7 2 |
| | 10 4 | 5 4 | 9 2 | 3+4 2 |
| | 9 2 | 5 4 | 9 2 | 7 2 |
| | 8 2 | 5 4 | 6+7 2 | 3+4 2 |
| | 11 4 | 5 4 | 4+5 2 | 7 2 |
| | 6 2 | 5 4 | 8 2 | 5+6 2 |
| | 7 2 | 5 4 | 8 2 | 7 2 |
| | 10 4 | 5 4 | 9 2 | 5+6 2 |
| | 9 2 | 5 4 | 8 2 | 7 2 |
| | 11 4 | 5 4 | 9 2 | 3+4 2 |
| | 10 4 | 5 4 | 6+7 2 | 7 2 |
| | 7 2 | 5 4 | 9 2 | 3+4 2 |
| | 10 4 | 5 4 | 4+5 2 | 7 2 |
| | 6 2 | 5 4 | 8 2 | 5+6 2 |
| | 10 4 | 5 4 | 9 2 | 7 2 |
| | 11 4 | 5 4 | 6+7 2 | 5+6 2 |
| | 8 2 | | 8 2 | 7 2 |
| | 9 2 | | 9 2 | 3+4 2 |
| | 11 4 | | 9 2 | 7 2 |
| | 8 2 | | 6+7 2 | 5+6 2 |
| | 11 4 | | 9 2 | 7 2 |
| | 10 4 | | 8 2 | 7 2 |
| | | | 4+5 2 | 3+4 2 |
| | | | 8 2 | 7 2 |
| | | | 4+5 2 | 5+6 2 |
| | | | 8 2 | 3+4 2 |
| | | | 9 2 | 7 2 |
| | | | 6+7 2 | 7 2 |
| | | | 8 2 | 5+6 2 |
| | | | 8 2 | 3+4 2 |
| | | | 4+5 2 | 7 2 |
| | | | 9 2 | 5+6 2 |
| | | | 6+7 2 | 7 2 |
| 227.1.14 | 32688 | | 204 | Lu2O7V2 |
| | -GM | -X | -L | -W |
| | 6 2 | 5 4 | 8 2 | 3+4 2 |
| | 7 2 | 5 4 | 9 2 | 7 2 |
| | 10 4 | 5 4 | 4+5 2 | 5+6 2 |
| | 9 2 | 5 4 | 8 2 | 7 2 |
| | 11 4 | 5 4 | 9 2 | 3+4 2 |
| | 8 2 | 5 4 | 6+7 2 | 7 2 |
| | 9 2 | 5 4 | 8 2 | 7 2 |
| | 11 4 | 5 4 | 9 2 | 5+6 2 |
| | 8 2 | 5 4 | 9 2 | 7 2 |
| | 11 4 | 5 4 | 8 2 | 3+4 2 |
| | 11 4 | 5 4 | 9 2 | 7 2 |
| | 6 2 | 5 4 | 6+7 2 | 5+6 2 |
| | 6 2 | 5 4 | 9 2 | 3+4 2 |
| | 7 2 | 5 4 | 4+5 2 | 7 2 |
| | 10 4 | 5 4 | 9 2 | 5+6 2 |
| | 10 4 | 5 4 | 6+7 2 | 7 2 |
| | 11 4 | 5 4 | 9 2 | 5+6 2 |
| | 8 2 | 5 4 | 8 2 | 7 2 |
| | 11 4 | 5 4 | 8 2 | 7 2 |
| | 9 2 | 5 4 | 6+7 2 | 3+4 2 |
| | 9 2 | 5 4 | 9 2 | 3+4 2 |
| | 7 2 | 5 4 | 8 2 | 7 2 |
| | 10 4 | 5 4 | 4+5 2 | 7 2 |

Continued on next column

**TABLE LXXXVIII– continued**

| TQC | ICSD | | NoE | SF |
|---|---|---|---|---|
| | 7 2 | 5 4 | 6+7 2 | 5+6 2 |
| | 10 4 | 5 4 | 9 2 | 3+4 2 |
| | 6 2 | 5 4 | 8 2 | 7 2 |
| | 10 4 | 5 4 | 4+5 2 | 3+4 2 |
| | 6 2 | 5 4 | 9 2 | 7 2 |
| | 10 4 | 5 4 | 8 2 | 5+6 2 |
| | 7 2 | 5 4 | 9 2 | 7 2 |
| | 11 4 | 5 4 | 6+7 2 | 3+4 2 |
| | 9 2 | 5 4 | 9 2 | 7 2 |
| | 11 4 | 5 4 | 9 2 | 3+4 2 |
| | 9 2 | 5 4 | 8 2 | 7 2 |
| | 8 2 | 5 4 | 4+5 2 | 7 2 |
| | 11 4 | 5 4 | 6+7 2 | 5+6 2 |
| | 11 4 | 5 4 | 9 2 | 5+6 2 |
| | 8 2 | 5 4 | 8 2 | 7 2 |
| | 10 4 | 5 4 | 8 2 | 3+4 2 |
| | 11 4 | 5 4 | 9 2 | 5+6 2 |
| | 9 2 | 5 4 | 8 2 | 7 2 |
| | 9 2 | 5 4 | 4+5 2 | 7 2 |
| | 8 2 | 5 4 | 9 2 | 3+4 2 |
| | 11 4 | 5 4 | 9 2 | 7 2 |
| | 8 2 | 5 4 | 6+7 2 | 5+6 2 |
| | 11 4 | 5 4 | 8 2 | 7 2 |
| | 11 4 | 5 4 | 9 2 | 5+6 2 |
| | 8 2 | 5 4 | 6+7 2 | 7 2 |
| | 11 4 | 5 4 | 4+5 2 | 7 2 |
| | 8 2 | 5 4 | 8 2 | 5+6 2 |
| | 9 2 | 5 4 | 9 2 | 7 2 |
| | 11 4 | | 9 2 | 7 2 |
| | 11 4 | | 6+7 2 | 3+4 2 |
| | 9 2 | | 8 2 | 3+4 2 |
| | 10 4 | | 9 2 | 5+6 2 |
| | 7 2 | | 9 2 | 7 2 |
| | 10 4 | | 4+5 2 | 7 2 |
| | 6 2 | | 9 2 | 5+6 2 |
| | 9 2 | | 6+7 2 | 7 2 |
| | 8 2 | | 8 2 | 3+4 2 |
| | 11 4 | | 9 2 | 7 2 |
| | 11 4 | | 9 2 | 5+6 2 |
| | 9 2 | | 8 2 | 7 2 |
| | 11 4 | | 6+7 2 | 3+4 2 |
| | 11 4 | | 9 2 | 5+6 2 |
| | 8 2 | | 8 2 | 7 2 |
| | 8 2 | | 4+5 2 | 5+6 2 |
| | 11 4 | | 9 2 | 7 2 |
| | 10 4 | | 8 2 | 3+4 2 |
| | | | 6+7 2 | 7 2 |
| | | | 4+5 2 | 7 2 |
| | | | 8 2 | 3+4 2 |
| | | | 9 2 | 7 2 |
| | | | 9 2 | 5+6 2 |
| | | | 8 2 | 7 2 |
| | | | 6+7 2 | 5+6 2 |
| | | | 9 2 | 3+4 2 |
| | | | 6+7 2 | 7 2 |
| | | | 9 2 | 3+4 2 |
| | | | 6+7 2 | 7 2 |
| | | | 8 2 | 5+6 2 |
| | | | 9 2 | 7 2 |
| | | | 9 2 | 7 2 |
| | | | 4+5 2 | 3+4 2 |

Continued on next column

**TABLE LXXXVIII– continued**

| TQC | ICSD | NoE | SF |
|---|---|---|---|
| | | 9 2 | 7 2 |
| | | 9 2 | 3+4 2 |
| | | 8 2 | 3+4 2 |
| | | 6+7 2 | 5+6 2 |
| | | 9 2 | 7 2 |
| | | 6+7 2 | 7 2 |
| | | 8 2 | 5+6 2 |
| | | 8 2 | 7 2 |
| | | 9 2 | 7 2 |
| | | 4+5 2 | 3+4 2 |
| | | 9 2 | 7 2 |
| | | 8 2 | 5+6 2 |
| | | 8 2 | 3+4 2 |
| | | 4+5 2 | 7 2 |
| | | 9 2 | 7 2 |
| | | 6+7 2 | 5+6 2 |
| | | 9 2 | 7 2 |
| | | 6+7 2 | 5+6 2 |
| 227.1.15 | 159290 | 204 | Lu2O7V2 |
| | -GM -X | -L | -W |
| | 6 2 5 4 | 8 2 | 3+4 2 |
| | 7 2 5 4 | 9 2 | 7 2 |
| | 10 4 5 4 | 4+5 2 | 5+6 2 |
| | 9 2 5 4 | 8 2 | 7 2 |
| | 11 4 5 4 | 9 2 | 3+4 2 |
| | 8 2 5 4 | 6+7 2 | 7 2 |
| | 9 2 5 4 | 8 2 | 7 2 |
| | 8 2 5 4 | 9 2 | 5+6 2 |
| | 11 4 5 4 | 9 2 | 7 2 |
| | 11 4 5 4 | 8 2 | 3+4 2 |
| | 11 4 5 4 | 9 2 | 5+6 2 |
| | 6 2 5 4 | 6+7 2 | 7 2 |
| | 6 2 5 4 | 9 2 | 3+4 2 |
| | 7 2 5 4 | 4+5 2 | 7 2 |
| | 10 4 5 4 | 9 2 | 5+6 2 |
| | 10 4 5 4 | 6+7 2 | 7 2 |
| | 11 4 5 4 | 8 2 | 5+6 2 |
| | 8 2 5 4 | 9 2 | 7 2 |
| | 11 4 5 4 | 8 2 | 7 2 |
| | 9 2 5 4 | 6+7 2 | 3+4 2 |
| | 9 2 5 4 | 9 2 | 3+4 2 |
| | 7 2 5 4 | 9 2 | 7 2 |
| | 10 4 5 4 | 8 2 | 7 2 |
| | 7 2 5 4 | 4+5 2 | 3+4 2 |
| | 10 4 5 4 | 6+7 2 | 7 2 |
| | 6 2 5 4 | 9 2 | 5+6 2 |
| | 6 2 5 4 | 8 2 | 3+4 2 |
| | 10 4 5 4 | 9 2 | 7 2 |
| | 10 4 5 4 | 4+5 2 | 5+6 2 |
| | 7 2 5 4 | 8 2 | 7 2 |
| | 11 4 5 4 | 9 2 | 3+4 2 |
| | 9 2 5 4 | 6+7 2 | 7 2 |
| | 11 4 5 4 | 9 2 | 5+6 2 |
| | 8 2 5 4 | 8 2 | 7 2 |
| | 9 2 5 4 | 4+5 2 | 3+4 2 |
| | 11 4 5 4 | 6+7 2 | 7 2 |
| | 11 4 5 4 | 9 2 | 7 2 |
| | 8 2 5 4 | 8 2 | 5+6 2 |
| | 8 2 5 4 | 9 2 | 3+4 2 |
| | 9 2 5 4 | 8 2 | 7 2 |
| | 11 4 5 4 | 9 2 | 5+6 2 |

Continued on next column

**TABLE LXXXVIII– continued**

| TQC | ICSD | NoE | SF |
|---|---|---|---|
| | 9 2 5 4 | 8 2 | 7 2 |
| | 10 4 5 4 | 9 2 | 3+4 2 |
| | 11 4 5 4 | 4+5 2 | 7 2 |
| | 8 2 5 4 | 6+7 2 | 7 2 |
| | 11 4 5 4 | 8 2 | 5+6 2 |
| | 11 4 5 4 | 6+7 2 | 5+6 2 |
| | 8 2 5 4 | 9 2 | 7 2 |
| | 11 4 5 4 | 9 2 | 7 2 |
| | 8 2 5 4 | 6+7 2 | 3+4 2 |
| | 11 4 5 4 | 9 2 | 5+6 2 |
| | 9 2 | 4+5 2 | 7 2 |
| | 11 4 | 9 2 | 3+4 2 |
| | 9 2 | 8 2 | 7 2 |
| | 9 2 | 9 2 | 7 2 |
| | 10 4 | 4+5 2 | 5+6 2 |
| | 6 2 | 8 2 | 7 2 |
| | 10 4 | 9 2 | 7 2 |
| | 7 2 | 6+7 2 | 5+6 2 |
| | 11 4 | 9 2 | 3+4 2 |
| | 8 2 | 8 2 | 5+6 2 |
| | 11 4 | 8 2 | 7 2 |
| | 11 4 | 9 2 | 7 2 |
| | 8 2 | 9 2 | 3+4 2 |
| | 9 2 | 6+7 2 | 5+6 2 |
| | 11 4 | 8 2 | 7 2 |
| | 8 2 | 9 2 | 5+6 2 |
| | 11 4 | 6+7 2 | 7 2 |
| | 10 4 | 8 2 | 7 2 |
| | | 4+5 2 | 3+4 2 |
| | | 9 2 | 7 2 |
| | | 9 2 | 3+4 2 |
| | | 6+7 2 | 7 2 |
| | | 9 2 | 5+6 2 |
| | | 8 2 | 7 2 |
| | | 6+7 2 | 7 2 |
| | | 9 2 | 5+6 2 |
| | | 9 2 | 3+4 2 |
| | | 4+5 2 | 3+4 2 |
| | | 6+7 2 | 7 2 |
| | | 8 2 | 5+6 2 |
| | | 9 2 | 7 2 |
| | | 8 2 | 3+4 2 |
| | | 9 2 | 7 2 |
| | | 4+5 2 | 7 2 |
| | | 9 2 | 3+4 2 |
| | | 8 2 | 5+6 2 |
| | | 6+7 2 | 7 2 |
| | | 9 2 | 3+4 2 |
| | | 6+7 2 | 7 2 |
| | | 8 2 | 7 2 |
| | | 9 2 | 5+6 2 |
| | | 8 2 | 7 2 |
| | | 4+5 2 | 3+4 2 |
| | | 9 2 | 5+6 2 |
| | | 8 2 | 7 2 |
| | | 8 2 | 3+4 2 |
| | | 4+5 2 | 7 2 |
| | | 9 2 | 7 2 |
| | | 6+7 2 | 5+6 2 |
| | | 9 2 | 7 2 |
| | | 6+7 2 | 5+6 2 |

Continued on next column

**TABLE LXXXVIII– continued**

| TQC | ICSD | NoE | SF |
|---|---|---|---|
| 227.1.16 | 159286 | 204 | Lu2O7V2 |
| | -GM -X | -L | -W |
| | 6 2 5 4 | 8 2 | 3+4 2 |
| | 7 2 5 4 | 9 2 | 7 2 |
| | 10 4 5 4 | 4+5 2 | 5+6 2 |
| | 9 2 5 4 | 8 2 | 7 2 |
| | 11 4 5 4 | 9 2 | 3+4 2 |
| | 8 2 5 4 | 6+7 2 | 7 2 |
| | 9 2 5 4 | 8 2 | 7 2 |
| | 8 2 5 4 | 9 2 | 5+6 2 |
| | 11 4 5 4 | 9 2 | 7 2 |
| | 11 4 5 4 | 8 2 | 3+4 2 |
| | 11 4 5 4 | 9 2 | 5+6 2 |
| | 6 2 5 4 | 6+7 2 | 7 2 |
| | 6 2 5 4 | 9 2 | 3+4 2 |
| | 7 2 5 4 | 4+5 2 | 7 2 |
| | 10 4 5 4 | 9 2 | 5+6 2 |
| | 10 4 5 4 | 6+7 2 | 7 2 |
| | 11 4 5 4 | 8 2 | 5+6 2 |
| | 8 2 5 4 | 9 2 | 7 2 |
| | 11 4 5 4 | 8 2 | 7 2 |
| | 9 2 5 4 | 6+7 2 | 3+4 2 |
| | 9 2 5 4 | 9 2 | 3+4 2 |
| | 7 2 5 4 | 8 2 | 7 2 |
| | 10 4 5 4 | 4+5 2 | 7 2 |
| | 7 2 5 4 | 9 2 | 3+4 2 |
| | 10 4 5 4 | 6+7 2 | 7 2 |
| | 6 2 5 4 | 9 2 | 5+6 2 |
| | 6 2 5 4 | 8 2 | 3+4 2 |
| | 10 4 5 4 | 4+5 2 | 7 2 |
| | 10 4 5 4 | 9 2 | 5+6 2 |
| | 7 2 5 4 | 8 2 | 7 2 |
| | 11 4 5 4 | 9 2 | 3+4 2 |
| | 9 2 5 4 | 6+7 2 | 7 2 |
| | 11 4 5 4 | 9 2 | 5+6 2 |
| | 8 2 5 4 | 8 2 | 7 2 |
| | 9 2 5 4 | 4+5 2 | 3+4 2 |
| | 11 4 5 4 | 6+7 2 | 7 2 |
| | 11 4 5 4 | 9 2 | 7 2 |
| | 8 2 5 4 | 8 2 | 5+6 2 |
| | 8 2 5 4 | 9 2 | 3+4 2 |
| | 9 2 5 4 | 8 2 | 7 2 |
| | 11 4 5 4 | 9 2 | 5+6 2 |
| | 9 2 5 4 | 8 2 | 7 2 |
| | 10 4 5 4 | 4+5 2 | 3+4 2 |
| | 11 4 5 4 | 9 2 | 7 2 |
| | 8 2 5 4 | 6+7 2 | 7 2 |
| | 11 4 5 4 | 8 2 | 5+6 2 |
| | 11 4 5 4 | 6+7 2 | 5+6 2 |
| | 8 2 5 4 | 9 2 | 7 2 |
| | 11 4 5 4 | 9 2 | 7 2 |
| | 8 2 5 4 | 4+5 2 | 3+4 2 |
| | 11 4 5 4 | 6+7 2 | 5+6 2 |
| | 9 2 | 9 2 | 7 2 |
| | 11 4 | 9 2 | 7 2 |
| | 9 2 | 8 2 | 3+4 2 |
| | 9 2 | 9 2 | 7 2 |
| | 10 4 | 4+5 2 | 5+6 2 |
| | 6 2 | 8 2 | 7 2 |
| | 10 4 | 9 2 | 7 2 |
| | 7 2 | 6+7 2 | 5+6 2 |

Continued on next column

**TABLE LXXXVIII– continued**

| TQC | ICSD | NoE | SF |
|---|---|---|---|
| | 11 4 | 9 2 | 3+4 2 |
| | 8 2 | 8 2 | 5+6 2 |
| | 11 4 | 8 2 | 7 2 |
| | 11 4 | 9 2 | 7 2 |
| | 8 2 | 9 2 | 3+4 2 |
| | 9 2 | 6+7 2 | 5+6 2 |
| | 11 4 | 8 2 | 7 2 |
| | 8 2 | 9 2 | 5+6 2 |
| | 11 4 | 6+7 2 | 7 2 |
| | 10 4 | 8 2 | 7 2 |
| | | 4+5 2 | 3+4 2 |
| | | 9 2 | 7 2 |
| | | 9 2 | 3+4 2 |
| | | 6+7 2 | 7 2 |
| | | 8 2 | 5+6 2 |
| | | 9 2 | 7 2 |
| | | 4+5 2 | 5+6 2 |
| | | 6+7 2 | 7 2 |
| | | 9 2 | 3+4 2 |
| | | 9 2 | 3+4 2 |
| | | 6+7 2 | 7 2 |
| | | 8 2 | 5+6 2 |
| | | 9 2 | 7 2 |
| | | 8 2 | 3+4 2 |
| | | 9 2 | 7 2 |
| | | 4+5 2 | 7 2 |
| | | 9 2 | 3+4 2 |
| | | 8 2 | 5+6 2 |
| | | 6+7 2 | 7 2 |
| | | 9 2 | 3+4 2 |
| | | 6+7 2 | 7 2 |
| | | 8 2 | 7 2 |
| | | 9 2 | 5+6 2 |
| | | 8 2 | 7 2 |
| | | 4+5 2 | 3+4 2 |
| | | 9 2 | 5+6 2 |
| | | 8 2 | 7 2 |
| | | 8 2 | 3+4 2 |
| | | 4+5 2 | 7 2 |
| | | 9 2 | 7 2 |
| | | 6+7 2 | 5+6 2 |
| | | 9 2 | 7 2 |
| | | 6+7 2 | 5+6 2 |
| 227.1.17 | 48128 | 68 | Li1O2Ti1 |
| | -GM -X | -L | -W |
| | 6 2 5 4 | 8 2 | 7 2 |
| | 7 2 5 4 | 9 2 | 3+4 2 |
| | 10 4 5 4 | 8 2 | 5+6 2 |
| | 9 2 5 4 | 8 2 | 7 2 |
| | 8 2 5 4 | 4+5 2 | 5+6 2 |
| | 11 4 5 4 | 6+7 2 | 7 2 |
| | 6 2 5 4 | 9 2 | 3+4 2 |
| | 7 2 5 4 | 9 2 | 7 2 |
| | 10 4 5 4 | 8 2 | 5+6 2 |
| | 9 2 5 4 | 8 2 | 7 2 |
| | 11 4 5 4 | 4+5 2 | 7 2 |
| | 10 4 5 4 | 9 2 | 3+4 2 |
| | 6 2 5 4 | 6+7 2 | 5+6 2 |
| | 10 4 5 4 | 9 2 | 7 2 |
| | 7 2 5 4 | 8 2 | 3+4 2 |
| | 10 4 5 4 | 8 2 | 7 2 |

Continued on next column

**TABLE LXXXVIII– continued**

| TQC | ICSD | NoE | SF |
|---|---|---|---|
| | 11 4 5 4 | 4+5 2 | 3+4 2 |
| | 8 2 | 8 2 | 7 2 |
| | 9 2 | 8 2 | 7 2 |
| | 11 4 | 9 2 | 5+6 2 |
| | 8 2 | 6+7 2 | 3+4 2 |
| | 11 4 | 9 2 | 7 2 |
| | 10 4 | 4+5 2 | 5+6 2 |
| | | 8 2 | 7 2 |
| | | 9 2 | 5+6 2 |
| | | 6+7 2 | 7 2 |
| | | 9 2 | 3+4 2 |
| | | 8 2 | 7 2 |
| | | 4+5 2 | 7 2 |
| | | 9 2 | 3+4 2 |
| | | 9 2 | 5+6 2 |
| | | 6+7 2 | 7 2 |
| | | 8 2 | 3+4 2 |
| | | 4+5 2 | 7 2 |
| 227.1.18 | 423608 | 16 | In2Mg1 |
| | -GM -X | -L | -W |
| | 6 2 5 4 | 8 2 | 7 2 |
| | 7 2 5 4 | 9 2 | 3+4 2 |
| | 10 4 5 4 | 8 2 | 5+6 2 |
| | 9 2 5 4 | 4+5 2 | 7 2 |
| | 6 2 | 9 2 | 7 2 |
| | 11 4 | 6+7 2 | 5+6 2 |
| | | 9 2 | 3+4 2 |
| | | 8 2 | 7 2 |
| 227.1.19 | 420818 | 164 | Hg2O7Ru2 |
| | -GM -X | -L | -W |
| | 6 2 5 4 | 8 2 | 7 2 |
| | 7 2 5 4 | 9 2 | 3+4 2 |
| | 10 4 5 4 | 8 2 | 5+6 2 |
| | 6 2 5 4 | 4+5 2 | 7 2 |
| | 10 4 5 4 | 9 2 | 5+6 2 |
| | 9 2 5 4 | 8 2 | 7 2 |
| | 11 4 5 4 | 8 2 | 5+6 2 |
| | 11 4 5 4 | 4+5 2 | 7 2 |
| | 8 2 5 4 | 6+7 2 | 3+4 2 |
| | 9 2 5 4 | 9 2 | 7 2 |
| | 10 4 5 4 | 8 2 | 7 2 |
| | 7 2 5 4 | 6+7 2 | 5+6 2 |
| | 10 4 5 4 | 9 2 | 3+4 2 |
| | 10 4 5 4 | 9 2 | 7 2 |
| | 6 2 5 4 | 6+7 2 | 7 2 |
| | 7 2 5 4 | 4+5 2 | 3+4 2 |
| | 10 4 5 4 | 9 2 | 5+6 2 |
| | 6 2 5 4 | 8 2 | 3+4 2 |
| | 7 2 5 4 | 8 2 | 7 2 |
| | 10 4 5 4 | 8 2 | 5+6 2 |
| | 10 4 5 4 | 9 2 | 7 2 |
| | 6 2 5 4 | 6+7 2 | 7 2 |
| | 7 2 5 4 | 9 2 | 7 2 |
| | 10 4 5 4 | 8 2 | 3+4 2 |
| | 6 2 5 4 | 9 2 | 5+6 2 |
| | 10 4 5 4 | 4+5 2 | 7 2 |
| | 7 2 5 4 | 9 2 | 3+4 2 |
| | 10 4 5 4 | 8 2 | 7 2 |
| | 10 4 5 4 | 9 2 | 7 2 |
| | 11 4 5 4 | 6+7 2 | 3+4 2 |
| | 8 2 5 4 | 4+5 2 | 7 2 |

Continued on next column

**TABLE LXXXVIII– continued**

| TQC | ICSD | NoE | SF |
|---|---|---|---|
| | 9 2 5 4 | 8 2 | 5+6 2 |
| | 11 4 5 4 | 6+7 2 | 5+6 2 |
| | 9 2 5 4 | 9 2 | 7 2 |
| | 6 2 5 4 | 8 2 | 3+4 2 |
| | 8 2 5 4 | 4+5 2 | 7 2 |
| | 10 4 5 4 | 9 2 | 3+4 2 |
| | 11 4 5 4 | 8 2 | 7 2 |
| | 10 4 5 4 | 9 2 | 7 2 |
| | 6 2 5 4 | 6+7 2 | 5+6 2 |
| | 10 4 5 4 | 9 2 | 7 2 |
| | 11 4 | 8 2 | 3+4 2 |
| | 7 2 | 4+5 2 | 5+6 2 |
| | 8 2 | 9 2 | 7 2 |
| | 11 4 | 8 2 | 7 2 |
| | 9 2 | 6+7 2 | 5+6 2 |
| | 11 4 | 9 2 | 7 2 |
| | 10 4 | 6+7 2 | 3+4 2 |
| | 8 2 | 9 2 | 5+6 2 |
| | 7 2 | 9 2 | 7 2 |
| | 11 4 | 8 2 | 7 2 |
| | 10 4 | 8 2 | 3+4 2 |
| | 6 2 | 4+5 2 | 7 2 |
| | 7 2 | 9 2 | 5+6 2 |
| | 10 4 | 9 2 | 7 2 |
| | | 6+7 2 | 5+6 2 |
| | | 9 2 | 7 2 |
| | | 4+5 2 | 3+4 2 |
| | | 8 2 | 3+4 2 |
| | | 6+7 2 | 7 2 |
| | | 9 2 | 7 2 |
| | | 8 2 | 3+4 2 |
| | | 4+5 2 | 5+6 2 |
| | | 9 2 | 7 2 |
| | | 8 2 | 3+4 2 |
| | | 8 2 | 7 2 |
| | | 9 2 | 5+6 2 |
| | | 6+7 2 | 7 2 |
| | | 8 2 | 5+6 2 |
| | | 4+5 2 | 7 2 |
| | | 9 2 | 3+4 2 |
| | | 6+7 2 | 7 2 |
| | | 9 2 | 5+6 2 |
| | | 9 2 | 7 2 |
| | | 6+7 2 | 7 2 |
| | | 9 2 | 3+4 2 |
| | | 8 2 | 5+6 2 |
| | | 4+5 2 | 7 2 |
| | | 8 2 | 7 2 |
| | | 8 2 | 3+4 2 |
| | | 4+5 2 | 7 2 |
| | | 9 2 | 5+6 2 |
| 227.1.20 | 159809 | 164 | Hg2O7Os2 |
| | -GM -X | -L | -W |
| | 6 2 5 4 | 8 2 | 7 2 |
| | 7 2 5 4 | 9 2 | 3+4 2 |
| | 10 4 5 4 | 8 2 | 5+6 2 |
| | 10 4 5 4 | 4+5 2 | 7 2 |
| | 6 2 5 4 | 8 2 | 5+6 2 |
| | 11 4 5 4 | 4+5 2 | 7 2 |
| | 9 2 5 4 | 6+7 2 | 3+4 2 |
| | 8 2 5 4 | 9 2 | 7 2 |

Continued on next column

**TABLE LXXXVIII– continued**

| TQC | ICSD | NoE | SF |
|---|---|---|---|
| | 11 4 5 4 | 8 2 | 7 2 |
| | 9 2 5 4 | 9 2 | 5+6 2 |
| | 10 4 5 4 | 6+7 2 | 3+4 2 |
| | 7 2 5 4 | 8 2 | 7 2 |
| | 10 4 5 4 | 9 2 | 5+6 2 |
| | 6 2 5 4 | 9 2 | 7 2 |
| | 10 4 5 4 | 4+5 2 | 7 2 |
| | 7 2 5 4 | 8 2 | 3+4 2 |
| | 6 2 5 4 | 8 2 | 5+6 2 |
| | 10 4 5 4 | 6+7 2 | 7 2 |
| | 7 2 5 4 | 8 2 | 7 2 |
| | 10 4 5 4 | 9 2 | 3+4 2 |
| | 10 4 5 4 | 9 2 | 5+6 2 |
| | 7 2 5 4 | 6+7 2 | 7 2 |
| | 6 2 5 4 | 9 2 | 3+4 2 |
| | 10 4 5 4 | 8 2 | 5+6 2 |
| | 6 2 5 4 | 8 2 | 7 2 |
| | 10 4 5 4 | 4+5 2 | 7 2 |
| | 7 2 5 4 | 9 2 | 3+4 2 |
| | 9 2 5 4 | 9 2 | 7 2 |
| | 8 2 5 4 | 4+5 2 | 7 2 |
| | 11 4 5 4 | 8 2 | 3+4 2 |
| | 10 4 5 4 | 6+7 2 | 7 2 |
| | 10 4 5 4 | 9 2 | 5+6 2 |
| | 11 4 5 4 | 6+7 2 | 7 2 |
| | 9 2 5 4 | 9 2 | 5+6 2 |
| | 6 2 5 4 | 4+5 2 | 3+4 2 |
| | 10 4 5 4 | 8 2 | 7 2 |
| | 8 2 5 4 | 9 2 | 3+4 2 |
| | 11 4 5 4 | 8 2 | 7 2 |
| | 10 4 5 4 | 8 2 | 5+6 2 |
| | 11 4 5 4 | 9 2 | 7 2 |
| | 6 2 5 4 | 6+7 2 | 7 2 |
| | 8 2 | 9 2 | 3+4 2 |
| | 11 4 | 4+5 2 | 5+6 2 |
| | 10 4 | 8 2 | 7 2 |
| | 7 2 | 9 2 | 7 2 |
| | 9 2 | 6+7 2 | 3+4 2 |
| | 11 4 | 9 2 | 7 2 |
| | 8 2 | 9 2 | 5+6 2 |
| | 10 4 | 6+7 2 | 5+6 2 |
| | 11 4 | 8 2 | 7 2 |
| | 7 2 | 8 2 | 7 2 |
| | 6 2 | 4+5 2 | 7 2 |
| | 10 4 | 9 2 | 3+4 2 |
| | 7 2 | 9 2 | 7 2 |
| | 10 4 | 9 2 | 5+6 2 |
| | | 6+7 2 | 5+6 2 |
| | | 9 2 | 3+4 2 |
| | | 4+5 2 | 7 2 |
| | | 6+7 2 | 3+4 2 |
| | | 8 2 | 7 2 |
| | | 9 2 | 7 2 |
| | | 9 2 | 3+4 2 |
| | | 8 2 | 5+6 2 |
| | | 4+5 2 | 7 2 |
| | | 9 2 | 7 2 |
| | | 8 2 | 3+4 2 |
| | | 8 2 | 5+6 2 |
| | | 6+7 2 | 7 2 |
| | | 8 2 | 5+6 2 |

Continued on next column

**TABLE LXXXVIII– continued**

| TQC | ICSD | NoE | SF |
|---|---|---|---|
| | | 4+5 2 | 7 2 |
| | | 9 2 | 3+4 2 |
| | | 6+7 2 | 7 2 |
| | | 9 2 | 5+6 2 |
| | | 9 2 | 7 2 |
| | | 6+7 2 | 7 2 |
| | | 9 2 | 3+4 2 |
| | | 8 2 | 5+6 2 |
| | | 8 2 | 7 2 |
| | | 8 2 | 7 2 |
| | | 4+5 2 | 3+4 2 |
| | | 9 2 | 7 2 |
| | | 4+5 2 | 5+6 2 |
| 227.1.21 | 159808 | 164 | Hg2O7Os2 |
| | -GM -X | -L | -W |
| | 6 2 5 4 | 8 2 | 7 2 |
| | 7 2 5 4 | 9 2 | 3+4 2 |
| | 10 4 5 4 | 8 2 | 5+6 2 |
| | 10 4 5 4 | 4+5 2 | 7 2 |
| | 6 2 5 4 | 8 2 | 5+6 2 |
| | 11 4 5 4 | 4+5 2 | 7 2 |
| | 9 2 5 4 | 6+7 2 | 3+4 2 |
| | 8 2 5 4 | 9 2 | 7 2 |
| | 11 4 5 4 | 8 2 | 7 2 |
| | 9 2 5 4 | 9 2 | 5+6 2 |
| | 10 4 5 4 | 6+7 2 | 3+4 2 |
| | 7 2 5 4 | 8 2 | 7 2 |
| | 10 4 5 4 | 9 2 | 5+6 2 |
| | 6 2 5 4 | 9 2 | 7 2 |
| | 10 4 5 4 | 4+5 2 | 7 2 |
| | 7 2 5 4 | 8 2 | 3+4 2 |
| | 6 2 5 4 | 8 2 | 5+6 2 |
| | 10 4 5 4 | 6+7 2 | 7 2 |
| | 7 2 5 4 | 8 2 | 7 2 |
| | 10 4 5 4 | 9 2 | 3+4 2 |
| | 10 4 5 4 | 9 2 | 5+6 2 |
| | 7 2 5 4 | 6+7 2 | 7 2 |
| | 6 2 5 4 | 9 2 | 3+4 2 |
| | 10 4 5 4 | 8 2 | 5+6 2 |
| | 6 2 5 4 | 8 2 | 7 2 |
| | 10 4 5 4 | 4+5 2 | 7 2 |
| | 7 2 5 4 | 9 2 | 3+4 2 |
| | 9 2 5 4 | 9 2 | 7 2 |
| | 8 2 5 4 | 4+5 2 | 7 2 |
| | 11 4 5 4 | 8 2 | 3+4 2 |
| | 10 4 5 4 | 6+7 2 | 7 2 |
| | 10 4 5 4 | 9 2 | 5+6 2 |
| | 11 4 5 4 | 6+7 2 | 7 2 |
| | 9 2 5 4 | 9 2 | 5+6 2 |
| | 6 2 5 4 | 4+5 2 | 3+4 2 |
| | 10 4 5 4 | 8 2 | 7 2 |
| | 8 2 5 4 | 9 2 | 3+4 2 |
| | 11 4 5 4 | 8 2 | 7 2 |
| | 10 4 5 4 | 8 2 | 5+6 2 |
| | 11 4 5 4 | 9 2 | 7 2 |
| | 6 2 5 4 | 6+7 2 | 7 2 |
| | 11 4 | 9 2 | 3+4 2 |
| | 8 2 | 4+5 2 | 5+6 2 |
| | 10 4 | 8 2 | 7 2 |
| | 7 2 | 9 2 | 7 2 |
| | 9 2 | 6+7 2 | 3+4 2 |

Continued on next column

**TABLE LXXXVIII– continued**

| TQC | ICSD | NoE | SF |
|---|---|---|---|
| | 11 4 | 9 2 | 7 2 |
| | 8 2 | 9 2 | 5+6 2 |
| | 10 4 | 6+7 2 | 5+6 2 |
| | 11 4 | 8 2 | 7 2 |
| | 7 2 | 8 2 | 7 2 |
| | 6 2 | 4+5 2 | 7 2 |
| | 10 4 | 9 2 | 3+4 2 |
| | 7 2 | 9 2 | 7 2 |
| | 10 4 | 9 2 | 5+6 2 |
| | | 6+7 2 | 5+6 2 |
| | | 9 2 | 3+4 2 |
| | | 4+5 2 | 7 2 |
| | | 6+7 2 | 3+4 2 |
| | | 8 2 | 7 2 |
| | | 9 2 | 7 2 |
| | | 9 2 | 3+4 2 |
| | | 8 2 | 5+6 2 |
| | | 4+5 2 | 7 2 |
| | | 9 2 | 7 2 |
| | | 8 2 | 3+4 2 |
| | | 8 2 | 5+6 2 |
| | | 6+7 2 | 7 2 |
| | | 8 2 | 5+6 2 |
| | | 4+5 2 | 7 2 |
| | | 9 2 | 3+4 2 |
| | | 6+7 2 | 7 2 |
| | | 9 2 | 5+6 2 |
| | | 9 2 | 7 2 |
| | | 6+7 2 | 7 2 |
| | | 9 2 | 3+4 2 |
| | | 8 2 | 5+6 2 |
| | | 8 2 | 7 2 |
| | | 8 2 | 7 2 |
| | | 4+5 2 | 3+4 2 |
| | | 9 2 | 7 2 |
| | | 4+5 2 | 5+6 2 |
| 227.1.22 | 159807 | 164 | Hg2O7Os2 |
| | -GM -X | -L | -W |
| | 6 2 5 4 | 8 2 | 7 2 |
| | 7 2 5 4 | 9 2 | 3+4 2 |
| | 10 4 5 4 | 8 2 | 5+6 2 |
| | 10 4 5 4 | 4+5 2 | 7 2 |
| | 6 2 5 4 | 8 2 | 5+6 2 |
| | 11 4 5 4 | 4+5 2 | 7 2 |
| | 9 2 5 4 | 6+7 2 | 3+4 2 |
| | 8 2 5 4 | 9 2 | 7 2 |
| | 11 4 5 4 | 8 2 | 7 2 |
| | 9 2 5 4 | 9 2 | 5+6 2 |
| | 10 4 5 4 | 6+7 2 | 3+4 2 |
| | 7 2 5 4 | 8 2 | 7 2 |
| | 10 4 5 4 | 9 2 | 5+6 2 |
| | 6 2 5 4 | 9 2 | 7 2 |
| | 10 4 5 4 | 4+5 2 | 7 2 |
| | 7 2 5 4 | 8 2 | 3+4 2 |
| | 6 2 5 4 | 8 2 | 5+6 2 |
| | 10 4 5 4 | 6+7 2 | 7 2 |
| | 7 2 5 4 | 8 2 | 7 2 |
| | 10 4 5 4 | 9 2 | 3+4 2 |
| | 10 4 5 4 | 9 2 | 5+6 2 |
| | 7 2 5 4 | 6+7 2 | 7 2 |
| | 6 2 5 4 | 9 2 | 3+4 2 |

Continued on next column

**TABLE LXXXVIII– continued**

| TQC | ICSD | NoE | SF |
|---|---|---|---|
| | 10 4 5 4 | 8 2 | 5+6 2 |
| | 6 2 5 4 | 8 2 | 7 2 |
| | 10 4 5 4 | 4+5 2 | 7 2 |
| | 7 2 5 4 | 9 2 | 3+4 2 |
| | 9 2 5 4 | 9 2 | 7 2 |
| | 8 2 5 4 | 4+5 2 | 7 2 |
| | 11 4 5 4 | 8 2 | 3+4 2 |
| | 10 4 5 4 | 6+7 2 | 7 2 |
| | 10 4 5 4 | 9 2 | 5+6 2 |
| | 11 4 5 4 | 6+7 2 | 7 2 |
| | 9 2 5 4 | 9 2 | 3+4 2 |
| | 6 2 5 4 | 4+5 2 | 5+6 2 |
| | 10 4 5 4 | 8 2 | 7 2 |
| | 8 2 5 4 | 9 2 | 3+4 2 |
| | 11 4 5 4 | 8 2 | 7 2 |
| | 10 4 5 4 | 8 2 | 5+6 2 |
| | 11 4 5 4 | 9 2 | 7 2 |
| | 6 2 5 4 | 6+7 2 | 7 2 |
| | 11 4 | 9 2 | 3+4 2 |
| | 8 2 | 4+5 2 | 5+6 2 |
| | 10 4 | 8 2 | 7 2 |
| | 7 2 | 9 2 | 7 2 |
| | 9 2 | 6+7 2 | 3+4 2 |
| | 11 4 | 9 2 | 7 2 |
| | 8 2 | 9 2 | 5+6 2 |
| | 10 4 | 6+7 2 | 5+6 2 |
| | 11 4 | 8 2 | 7 2 |
| | 7 2 | 8 2 | 7 2 |
| | 6 2 | 4+5 2 | 7 2 |
| | 10 4 | 9 2 | 3+4 2 |
| | 7 2 | 9 2 | 7 2 |
| | 10 4 | 9 2 | 5+6 2 |
| | | 6+7 2 | 5+6 2 |
| | | 9 2 | 3+4 2 |
| | | 4+5 2 | 7 2 |
| | | 6+7 2 | 3+4 2 |
| | | 8 2 | 7 2 |
| | | 9 2 | 7 2 |
| | | 9 2 | 3+4 2 |
| | | 8 2 | 5+6 2 |
| | | 4+5 2 | 7 2 |
| | | 9 2 | 7 2 |
| | | 8 2 | 3+4 2 |
| | | 8 2 | 5+6 2 |
| | | 6+7 2 | 7 2 |
| | | 8 2 | 5+6 2 |
| | | 4+5 2 | 7 2 |
| | | 9 2 | 3+4 2 |
| | | 6+7 2 | 7 2 |
| | | 9 2 | 5+6 2 |
| | | 9 2 | 7 2 |
| | | 6+7 2 | 7 2 |
| | | 9 2 | 3+4 2 |
| | | 8 2 | 5+6 2 |
| | | 8 2 | 7 2 |
| | | 8 2 | 7 2 |
| | | 4+5 2 | 3+4 2 |
| | | 9 2 | 7 2 |
| | | 4+5 2 | 5+6 2 |
| 227.1.23 | 159806 | 164 | Hg2O7Os2 |
| | -GM -X | -L | -W |

Continued on next column

**TABLE LXXXVIII– continued**

| TQC | ICSD | NoE | SF |
|---|---|---|---|
| | 6 2 5 4 | 8 2 | 7 2 |
| | 7 2 5 4 | 9 2 | 3+4 2 |
| | 10 4 5 4 | 8 2 | 5+6 2 |
| | 10 4 5 4 | 4+5 2 | 7 2 |
| | 6 2 5 4 | 8 2 | 5+6 2 |
| | 11 4 5 4 | 4+5 2 | 7 2 |
| | 9 2 5 4 | 6+7 2 | 3+4 2 |
| | 8 2 5 4 | 9 2 | 7 2 |
| | 11 4 5 4 | 8 2 | 7 2 |
| | 9 2 5 4 | 9 2 | 5+6 2 |
| | 10 4 5 4 | 6+7 2 | 3+4 2 |
| | 7 2 5 4 | 8 2 | 7 2 |
| | 10 4 5 4 | 9 2 | 5+6 2 |
| | 6 2 5 4 | 9 2 | 7 2 |
| | 10 4 5 4 | 4+5 2 | 7 2 |
| | 7 2 5 4 | 8 2 | 3+4 2 |
| | 6 2 5 4 | 8 2 | 5+6 2 |
| | 10 4 5 4 | 6+7 2 | 7 2 |
| | 7 2 5 4 | 8 2 | 7 2 |
| | 10 4 5 4 | 9 2 | 3+4 2 |
| | 10 4 5 4 | 9 2 | 5+6 2 |
| | 7 2 5 4 | 6+7 2 | 7 2 |
| | 6 2 5 4 | 9 2 | 3+4 2 |
| | 10 4 5 4 | 8 2 | 5+6 2 |
| | 6 2 5 4 | 8 2 | 7 2 |
| | 10 4 5 4 | 4+5 2 | 7 2 |
| | 7 2 5 4 | 9 2 | 3+4 2 |
| | 9 2 5 4 | 9 2 | 7 2 |
| | 8 2 5 4 | 4+5 2 | 7 2 |
| | 11 4 5 4 | 8 2 | 3+4 2 |
| | 10 4 5 4 | 6+7 2 | 7 2 |
| | 10 4 5 4 | 9 2 | 5+6 2 |
| | 11 4 5 4 | 6+7 2 | 7 2 |
| | 9 2 5 4 | 9 2 | 3+4 2 |
| | 6 2 5 4 | 4+5 2 | 5+6 2 |
| | 10 4 5 4 | 8 2 | 7 2 |
| | 8 2 5 4 | 9 2 | 3+4 2 |
| | 11 4 5 4 | 8 2 | 7 2 |
| | 10 4 5 4 | 8 2 | 5+6 2 |
| | 11 4 5 4 | 9 2 | 7 2 |
| | 6 2 5 4 | 6+7 2 | 7 2 |
| | 11 4 | 9 2 | 3+4 2 |
| | 8 2 | 4+5 2 | 5+6 2 |
| | 10 4 | 8 2 | 7 2 |
| | 7 2 | 9 2 | 7 2 |
| | 9 2 | 6+7 2 | 3+4 2 |
| | 11 4 | 9 2 | 7 2 |
| | 8 2 | 9 2 | 5+6 2 |
| | 10 4 | 6+7 2 | 5+6 2 |
| | 11 4 | 8 2 | 7 2 |
| | 7 2 | 8 2 | 7 2 |
| | 6 2 | 4+5 2 | 7 2 |
| | 10 4 | 9 2 | 3+4 2 |
| | 7 2 | 9 2 | 7 2 |
| | 10 4 | 9 2 | 5+6 2 |
| | | 6+7 2 | 5+6 2 |
| | | 9 2 | 3+4 2 |
| | | 4+5 2 | 7 2 |
| | | 6+7 2 | 3+4 2 |
| | | 8 2 | 7 2 |
| | | 9 2 | 7 2 |
| | | 9 2 | 3+4 2 |
| | | 8 2 | 5+6 2 |
| | | 4+5 2 | 7 2 |
| | | 9 2 | 7 2 |
| | | 8 2 | 3+4 2 |
| | | 8 2 | 5+6 2 |
| | | 6+7 2 | 7 2 |
| | | 8 2 | 5+6 2 |
| | | 4+5 2 | 7 2 |
| | | 9 2 | 3+4 2 |
| | | 6+7 2 | 7 2 |
| | | 9 2 | 5+6 2 |
| | | 9 2 | 7 2 |
| | | 6+7 2 | 7 2 |
| | | 9 2 | 3+4 2 |
| | | 8 2 | 5+6 2 |
| | | 8 2 | 7 2 |
| | | 8 2 | 7 2 |
| | | 4+5 2 | 3+4 2 |
| | | 9 2 | 7 2 |
| | | 4+5 2 | 5+6 2 |
| 227.1.24 | 159805 | 164 | Hg2O7Os2 |
| | -GM -X | -L | -W |
| | 6 2 5 4 | 8 2 | 7 2 |
| | 7 2 5 4 | 9 2 | 3+4 2 |
| | 10 4 5 4 | 8 2 | 5+6 2 |
| | 10 4 5 4 | 4+5 2 | 7 2 |
| | 6 2 5 4 | 8 2 | 5+6 2 |
| | 11 4 5 4 | 4+5 2 | 7 2 |
| | 9 2 5 4 | 6+7 2 | 3+4 2 |
| | 8 2 5 4 | 9 2 | 7 2 |
| | 11 4 5 4 | 8 2 | 7 2 |
| | 9 2 5 4 | 9 2 | 5+6 2 |
| | 10 4 5 4 | 6+7 2 | 3+4 2 |
| | 7 2 5 4 | 8 2 | 7 2 |
| | 10 4 5 4 | 9 2 | 5+6 2 |
| | 6 2 5 4 | 9 2 | 7 2 |
| | 10 4 5 4 | 4+5 2 | 7 2 |
| | 7 2 5 4 | 8 2 | 3+4 2 |
| | 6 2 5 4 | 8 2 | 5+6 2 |
| | 10 4 5 4 | 6+7 2 | 7 2 |
| | 7 2 5 4 | 8 2 | 7 2 |
| | 10 4 5 4 | 9 2 | 3+4 2 |
| | 10 4 5 4 | 9 2 | 5+6 2 |
| | 7 2 5 4 | 6+7 2 | 7 2 |
| | 6 2 5 4 | 9 2 | 3+4 2 |
| | 10 4 5 4 | 8 2 | 5+6 2 |
| | 6 2 5 4 | 8 2 | 7 2 |
| | 10 4 5 4 | 4+5 2 | 7 2 |
| | 7 2 5 4 | 9 2 | 3+4 2 |
| | 9 2 5 4 | 9 2 | 7 2 |
| | 8 2 5 4 | 4+5 2 | 7 2 |
| | 11 4 5 4 | 8 2 | 3+4 2 |
| | 10 4 5 4 | 6+7 2 | 7 2 |
| | 10 4 5 4 | 9 2 | 5+6 2 |
| | 11 4 5 4 | 6+7 2 | 7 2 |
| | 9 2 5 4 | 9 2 | 3+4 2 |
| | 6 2 5 4 | 4+5 2 | 5+6 2 |
| | 10 4 5 4 | 8 2 | 7 2 |
| | 8 2 5 4 | 9 2 | 3+4 2 |
| | 11 4 5 4 | 8 2 | 7 2 |

Continued on next column

**TABLE LXXXVIII– continued**

| TQC | ICSD | NoE | SF |
|---|---|---|---|
| | 10 4 5 4 | 8 2 | 5+6 2 |
| | 11 4 5 4 | 9 2 | 7 2 |
| | 6 2 5 4 | 6+7 2 | 7 2 |
| | 11 4 | 9 2 | 3+4 2 |
| | 8 2 | 4+5 2 | 5+6 2 |
| | 10 4 | 8 2 | 7 2 |
| | 7 2 | 9 2 | 7 2 |
| | 9 2 | 6+7 2 | 3+4 2 |
| | 11 4 | 9 2 | 7 2 |
| | 8 2 | 9 2 | 5+6 2 |
| | 10 4 | 6+7 2 | 5+6 2 |
| | 11 4 | 8 2 | 7 2 |
| | 7 2 | 8 2 | 7 2 |
| | 6 2 | 4+5 2 | 7 2 |
| | 10 4 | 9 2 | 3+4 2 |
| | 7 2 | 9 2 | 7 2 |
| | 10 4 | 9 2 | 5+6 2 |
| | | 6+7 2 | 5+6 2 |
| | | 9 2 | 3+4 2 |
| | | 4+5 2 | 7 2 |
| | | 6+7 2 | 3+4 2 |
| | | 8 2 | 7 2 |
| | | 9 2 | 7 2 |
| | | 9 2 | 5+6 2 |
| | | 8 2 | 3+4 2 |
| | | 4+5 2 | 7 2 |
| | | 9 2 | 7 2 |
| | | 8 2 | 3+4 2 |
| | | 8 2 | 5+6 2 |
| | | 6+7 2 | 7 2 |
| | | 8 2 | 5+6 2 |
| | | 4+5 2 | 7 2 |
| | | 9 2 | 3+4 2 |
| | | 6+7 2 | 7 2 |
| | | 9 2 | 5+6 2 |
| | | 9 2 | 7 2 |
| | | 6+7 2 | 7 2 |
| | | 9 2 | 3+4 2 |
| | | 8 2 | 5+6 2 |
| | | 8 2 | 7 2 |
| | | 8 2 | 7 2 |
| | | 4+5 2 | 3+4 2 |
| | | 9 2 | 7 2 |
| | | 4+5 2 | 5+6 2 |
| 227.1.25 | 159804 | 164 | Hg2O7Os2 |
| | -GM -X | -L | -W |
| | 6 2 5 4 | 8 2 | 7 2 |
| | 7 2 5 4 | 9 2 | 3+4 2 |
| | 10 4 5 4 | 8 2 | 5+6 2 |
| | 10 4 5 4 | 4+5 2 | 7 2 |
| | 6 2 5 4 | 8 2 | 5+6 2 |
| | 11 4 5 4 | 4+5 2 | 7 2 |
| | 9 2 5 4 | 6+7 2 | 3+4 2 |
| | 8 2 5 4 | 9 2 | 7 2 |
| | 11 4 5 4 | 8 2 | 7 2 |
| | 9 2 5 4 | 9 2 | 5+6 2 |
| | 10 4 5 4 | 6+7 2 | 3+4 2 |
| | 7 2 5 4 | 8 2 | 7 2 |
| | 10 4 5 4 | 9 2 | 5+6 2 |
| | 6 2 5 4 | 9 2 | 7 2 |
| | 10 4 5 4 | 4+5 2 | 7 2 |

Continued on next column

**TABLE LXXXVIII– continued**

| TQC | ICSD | NoE | SF |
|---|---|---|---|
| | 7 2 5 4 | 8 2 | 3+4 2 |
| | 6 2 5 4 | 8 2 | 5+6 2 |
| | 10 4 5 4 | 6+7 2 | 7 2 |
| | 7 2 5 4 | 8 2 | 7 2 |
| | 10 4 5 4 | 9 2 | 3+4 2 |
| | 10 4 5 4 | 9 2 | 5+6 2 |
| | 7 2 5 4 | 6+7 2 | 7 2 |
| | 6 2 5 4 | 9 2 | 3+4 2 |
| | 10 4 5 4 | 8 2 | 5+6 2 |
| | 6 2 5 4 | 8 2 | 7 2 |
| | 10 4 5 4 | 4+5 2 | 7 2 |
| | 7 2 5 4 | 9 2 | 3+4 2 |
| | 9 2 5 4 | 9 2 | 7 2 |
| | 8 2 5 4 | 4+5 2 | 7 2 |
| | 11 4 5 4 | 8 2 | 3+4 2 |
| | 10 4 5 4 | 6+7 2 | 7 2 |
| | 10 4 5 4 | 9 2 | 5+6 2 |
| | 11 4 5 4 | 6+7 2 | 7 2 |
| | 9 2 5 4 | 9 2 | 3+4 2 |
| | 6 2 5 4 | 4+5 2 | 5+6 2 |
| | 10 4 5 4 | 8 2 | 7 2 |
| | 8 2 5 4 | 9 2 | 3+4 2 |
| | 11 4 5 4 | 8 2 | 7 2 |
| | 10 4 5 4 | 8 2 | 5+6 2 |
| | 11 4 5 4 | 9 2 | 7 2 |
| | 6 2 5 4 | 6+7 2 | 7 2 |
| | 11 4 | 9 2 | 3+4 2 |
| | 8 2 | 4+5 2 | 5+6 2 |
| | 10 4 | 8 2 | 7 2 |
| | 7 2 | 9 2 | 7 2 |
| | 9 2 | 6+7 2 | 3+4 2 |
| | 11 4 | 9 2 | 7 2 |
| | 8 2 | 9 2 | 5+6 2 |
| | 10 4 | 6+7 2 | 5+6 2 |
| | 11 4 | 8 2 | 7 2 |
| | 7 2 | 8 2 | 7 2 |
| | 6 2 | 4+5 2 | 7 2 |
| | 10 4 | 9 2 | 3+4 2 |
| | 7 2 | 9 2 | 7 2 |
| | 10 4 | 9 2 | 5+6 2 |
| | | 6+7 2 | 5+6 2 |
| | | 9 2 | 3+4 2 |
| | | 4+5 2 | 7 2 |
| | | 6+7 2 | 3+4 2 |
| | | 8 2 | 7 2 |
| | | 9 2 | 7 2 |
| | | 9 2 | 5+6 2 |
| | | 8 2 | 3+4 2 |
| | | 4+5 2 | 7 2 |
| | | 9 2 | 7 2 |
| | | 8 2 | 3+4 2 |
| | | 8 2 | 5+6 2 |
| | | 6+7 2 | 7 2 |
| | | 8 2 | 5+6 2 |
| | | 4+5 2 | 7 2 |
| | | 9 2 | 3+4 2 |
| | | 6+7 2 | 7 2 |
| | | 9 2 | 5+6 2 |
| | | 9 2 | 7 2 |
| | | 6+7 2 | 7 2 |
| | | 9 2 | 3+4 2 |

Continued on next column

**TABLE LXXXVIII– continued**

| TQC | ICSD | NoE | SF |
|---|---|---|---|
| | | 8 2 | 5+6 2 |
| | | 8 2 | 7 2 |
| | | 8 2 | 7 2 |
| | | 4+5 2 | 3+4 2 |
| | | 9 2 | 7 2 |
| | | 4+5 2 | 5+6 2 |
| 227.1.26 | 159803 | 164 | Hg2O7Os2 |
| | -GM -X | -L | -W |
| | 6 2 5 4 | 8 2 | 7 2 |
| | 7 2 5 4 | 9 2 | 3+4 2 |
| | 10 4 5 4 | 8 2 | 5+6 2 |
| | 10 4 5 4 | 4+5 2 | 7 2 |
| | 6 2 5 4 | 8 2 | 5+6 2 |
| | 11 4 5 4 | 4+5 2 | 7 2 |
| | 9 2 5 4 | 6+7 2 | 3+4 2 |
| | 8 2 5 4 | 9 2 | 7 2 |
| | 11 4 5 4 | 8 2 | 7 2 |
| | 9 2 5 4 | 9 2 | 5+6 2 |
| | 10 4 5 4 | 6+7 2 | 3+4 2 |
| | 7 2 5 4 | 8 2 | 7 2 |
| | 10 4 5 4 | 9 2 | 5+6 2 |
| | 6 2 5 4 | 9 2 | 7 2 |
| | 10 4 5 4 | 4+5 2 | 7 2 |
| | 7 2 5 4 | 8 2 | 3+4 2 |
| | 6 2 5 4 | 8 2 | 5+6 2 |
| | 10 4 5 4 | 6+7 2 | 7 2 |
| | 7 2 5 4 | 8 2 | 7 2 |
| | 10 4 5 4 | 9 2 | 3+4 2 |
| | 10 4 5 4 | 9 2 | 5+6 2 |
| | 7 2 5 4 | 6+7 2 | 7 2 |
| | 6 2 5 4 | 9 2 | 3+4 2 |
| | 10 4 5 4 | 8 2 | 5+6 2 |
| | 6 2 5 4 | 8 2 | 7 2 |
| | 10 4 5 4 | 4+5 2 | 7 2 |
| | 7 2 5 4 | 9 2 | 3+4 2 |
| | 9 2 5 4 | 9 2 | 7 2 |
| | 8 2 5 4 | 4+5 2 | 7 2 |
| | 11 4 5 4 | 8 2 | 3+4 2 |
| | 10 4 5 4 | 6+7 2 | 7 2 |
| | 10 4 5 4 | 9 2 | 5+6 2 |
| | 11 4 5 4 | 6+7 2 | 7 2 |
| | 9 2 5 4 | 9 2 | 3+4 2 |
| | 6 2 5 4 | 4+5 2 | 5+6 2 |
| | 10 4 5 4 | 8 2 | 7 2 |
| | 8 2 5 4 | 9 2 | 3+4 2 |
| | 11 4 5 4 | 8 2 | 7 2 |
| | 10 4 5 4 | 8 2 | 5+6 2 |
| | 11 4 5 4 | 9 2 | 7 2 |
| | 6 2 5 4 | 6+7 2 | 7 2 |
| | 11 4 | 9 2 | 3+4 2 |
| | 8 2 | 4+5 2 | 5+6 2 |
| | 10 4 | 8 2 | 7 2 |
| | 7 2 | 9 2 | 7 2 |
| | 9 2 | 6+7 2 | 3+4 2 |
| | 11 4 | 9 2 | 7 2 |
| | 8 2 | 9 2 | 5+6 2 |
| | 10 4 | 6+7 2 | 5+6 2 |
| | 11 4 | 8 2 | 7 2 |
| | 7 2 | 8 2 | 7 2 |
| | 6 2 | 4+5 2 | 7 2 |
| | 10 4 | 9 2 | 3+4 2 |

Continued on next column

**TABLE LXXXVIII– continued**

| TQC | ICSD | NoE | SF |
|---|---|---|---|
| | 7 2 | 9 2 | 7 2 |
| | 10 4 | 9 2 | 5+6 2 |
| | | 6+7 2 | 5+6 2 |
| | | 9 2 | 3+4 2 |
| | | 4+5 2 | 7 2 |
| | | 6+7 2 | 3+4 2 |
| | | 8 2 | 7 2 |
| | | 9 2 | 7 2 |
| | | 9 2 | 5+6 2 |
| | | 8 2 | 3+4 2 |
| | | 4+5 2 | 7 2 |
| | | 9 2 | 7 2 |
| | | 8 2 | 3+4 2 |
| | | 8 2 | 5+6 2 |
| | | 6+7 2 | 7 2 |
| | | 8 2 | 5+6 2 |
| | | 4+5 2 | 7 2 |
| | | 9 2 | 3+4 2 |
| | | 6+7 2 | 7 2 |
| | | 9 2 | 5+6 2 |
| | | 9 2 | 7 2 |
| | | 6+7 2 | 7 2 |
| | | 9 2 | 3+4 2 |
| | | 8 2 | 5+6 2 |
| | | 8 2 | 7 2 |
| | | 8 2 | 7 2 |
| | | 4+5 2 | 3+4 2 |
| | | 9 2 | 7 2 |
| | | 4+5 2 | 5+6 2 |
| 227.1.27 | 159802 | 164 | Hg2O7Os2 |
| | -GM -X | -L | -W |
| | 6 2 5 4 | 8 2 | 7 2 |
| | 7 2 5 4 | 9 2 | 3+4 2 |
| | 10 4 5 4 | 8 2 | 5+6 2 |
| | 10 4 5 4 | 4+5 2 | 7 2 |
| | 6 2 5 4 | 8 2 | 5+6 2 |
| | 11 4 5 4 | 4+5 2 | 7 2 |
| | 9 2 5 4 | 6+7 2 | 3+4 2 |
| | 8 2 5 4 | 9 2 | 7 2 |
| | 11 4 5 4 | 8 2 | 7 2 |
| | 9 2 5 4 | 9 2 | 5+6 2 |
| | 10 4 5 4 | 6+7 2 | 3+4 2 |
| | 7 2 5 4 | 8 2 | 7 2 |
| | 10 4 5 4 | 9 2 | 5+6 2 |
| | 6 2 5 4 | 9 2 | 7 2 |
| | 10 4 5 4 | 4+5 2 | 7 2 |
| | 7 2 5 4 | 8 2 | 3+4 2 |
| | 6 2 5 4 | 8 2 | 5+6 2 |
| | 10 4 5 4 | 6+7 2 | 7 2 |
| | 7 2 5 4 | 8 2 | 7 2 |
| | 10 4 5 4 | 9 2 | 3+4 2 |
| | 10 4 5 4 | 9 2 | 5+6 2 |
| | 7 2 5 4 | 6+7 2 | 7 2 |
| | 6 2 5 4 | 9 2 | 3+4 2 |
| | 10 4 5 4 | 8 2 | 5+6 2 |
| | 6 2 5 4 | 8 2 | 7 2 |
| | 10 4 5 4 | 4+5 2 | 7 2 |
| | 7 2 5 4 | 9 2 | 3+4 2 |
| | 9 2 5 4 | 9 2 | 7 2 |
| | 8 2 5 4 | 4+5 2 | 7 2 |
| | 11 4 5 4 | 8 2 | 3+4 2 |

Continued on next column

**TABLE LXXXVIII– continued**

| TQC | ICSD | NoE | SF |
|---|---|---|---|
| | 10 4 5 4 | 6+7 2 | 7 2 |
| | 10 4 5 4 | 9 2 | 5+6 2 |
| | 11 4 5 4 | 6+7 2 | 7 2 |
| | 9 2 5 4 | 9 2 | 3+4 2 |
| | 6 2 5 4 | 4+5 2 | 5+6 2 |
| | 10 4 5 4 | 8 2 | 7 2 |
| | 8 2 5 4 | 9 2 | 3+4 2 |
| | 11 4 5 4 | 8 2 | 7 2 |
| | 10 4 5 4 | 8 2 | 5+6 2 |
| | 11 4 5 4 | 9 2 | 7 2 |
| | 6 2 5 4 | 6+7 2 | 7 2 |
| | 11 4 | 9 2 | 3+4 2 |
| | 8 2 | 4+5 2 | 5+6 2 |
| | 10 4 | 8 2 | 7 2 |
| | 7 2 | 9 2 | 7 2 |
| | 9 2 | 6+7 2 | 3+4 2 |
| | 11 4 | 9 2 | 7 2 |
| | 8 2 | 9 2 | 5+6 2 |
| | 10 4 | 6+7 2 | 5+6 2 |
| | 11 4 | 8 2 | 7 2 |
| | 7 2 | 8 2 | 7 2 |
| | 6 2 | 4+5 2 | 7 2 |
| | 10 4 | 9 2 | 3+4 2 |
| | 7 2 | 9 2 | 7 2 |
| | 10 4 | 9 2 | 5+6 2 |
| | | 6+7 2 | 5+6 2 |
| | | 9 2 | 3+4 2 |
| | | 4+5 2 | 7 2 |
| | | 6+7 2 | 3+4 2 |
| | | 8 2 | 7 2 |
| | | 9 2 | 7 2 |
| | | 9 2 | 5+6 2 |
| | | 8 2 | 3+4 2 |
| | | 4+5 2 | 7 2 |
| | | 9 2 | 7 2 |
| | | 8 2 | 3+4 2 |
| | | 8 2 | 5+6 2 |
| | | 6+7 2 | 7 2 |
| | | 8 2 | 5+6 2 |
| | | 4+5 2 | 7 2 |
| | | 9 2 | 3+4 2 |
| | | 6+7 2 | 7 2 |
| | | 9 2 | 5+6 2 |
| | | 9 2 | 7 2 |
| | | 6+7 2 | 7 2 |
| | | 9 2 | 3+4 2 |
| | | 8 2 | 5+6 2 |
| | | 8 2 | 7 2 |
| | | 8 2 | 7 2 |
| | | 4+5 2 | 3+4 2 |
| | | 9 2 | 7 2 |
| | | 4+5 2 | 5+6 2 |
| 227.1.28 | 159801 | 164 | Hg2O7Os2 |
| | -GM -X | -L | -W |
| | 6 2 5 4 | 8 2 | 7 2 |
| | 7 2 5 4 | 9 2 | 3+4 2 |
| | 10 4 5 4 | 8 2 | 5+6 2 |
| | 10 4 5 4 | 4+5 2 | 7 2 |
| | 6 2 5 4 | 8 2 | 5+6 2 |
| | 11 4 5 4 | 4+5 2 | 7 2 |
| | 9 2 5 4 | 6+7 2 | 3+4 2 |

Continued on next column

**TABLE LXXXVIII– continued**

| TQC | ICSD | NoE | SF |
|---|---|---|---|
| | 8 2 5 4 | 9 2 | 7 2 |
| | 11 4 5 4 | 8 2 | 7 2 |
| | 9 2 5 4 | 9 2 | 5+6 2 |
| | 10 4 5 4 | 6+7 2 | 3+4 2 |
| | 7 2 5 4 | 8 2 | 7 2 |
| | 10 4 5 4 | 9 2 | 5+6 2 |
| | 6 2 5 4 | 9 2 | 7 2 |
| | 10 4 5 4 | 4+5 2 | 7 2 |
| | 7 2 5 4 | 8 2 | 3+4 2 |
| | 6 2 5 4 | 8 2 | 5+6 2 |
| | 10 4 5 4 | 6+7 2 | 7 2 |
| | 7 2 5 4 | 8 2 | 7 2 |
| | 10 4 5 4 | 9 2 | 3+4 2 |
| | 10 4 5 4 | 6+7 2 | 5+6 2 |
| | 7 2 5 4 | 9 2 | 7 2 |
| | 6 2 5 4 | 9 2 | 3+4 2 |
| | 10 4 5 4 | 8 2 | 5+6 2 |
| | 6 2 5 4 | 8 2 | 7 2 |
| | 10 4 5 4 | 4+5 2 | 7 2 |
| | 7 2 5 4 | 9 2 | 3+4 2 |
| | 9 2 5 4 | 9 2 | 7 2 |
| | 8 2 5 4 | 4+5 2 | 7 2 |
| | 11 4 5 4 | 8 2 | 3+4 2 |
| | 10 4 5 4 | 6+7 2 | 7 2 |
| | 10 4 5 4 | 9 2 | 5+6 2 |
| | 11 4 5 4 | 6+7 2 | 7 2 |
| | 9 2 5 4 | 9 2 | 3+4 2 |
| | 6 2 5 4 | 4+5 2 | 5+6 2 |
| | 10 4 5 4 | 8 2 | 7 2 |
| | 8 2 5 4 | 9 2 | 3+4 2 |
| | 11 4 5 4 | 8 2 | 7 2 |
| | 10 4 5 4 | 8 2 | 5+6 2 |
| | 11 4 5 4 | 9 2 | 7 2 |
| | 6 2 5 4 | 6+7 2 | 7 2 |
| | 11 4 | 9 2 | 3+4 2 |
| | 8 2 | 4+5 2 | 5+6 2 |
| | 10 4 | 8 2 | 7 2 |
| | 7 2 | 9 2 | 7 2 |
| | 9 2 | 6+7 2 | 3+4 2 |
| | 11 4 | 9 2 | 7 2 |
| | 8 2 | 9 2 | 5+6 2 |
| | 10 4 | 6+7 2 | 5+6 2 |
| | 11 4 | 8 2 | 7 2 |
| | 7 2 | 8 2 | 7 2 |
| | 6 2 | 4+5 2 | 7 2 |
| | 10 4 | 9 2 | 3+4 2 |
| | 7 2 | 9 2 | 7 2 |
| | 10 4 | 9 2 | 5+6 2 |
| | | 6+7 2 | 5+6 2 |
| | | 9 2 | 3+4 2 |
| | | 4+5 2 | 7 2 |
| | | 6+7 2 | 3+4 2 |
| | | 8 2 | 7 2 |
| | | 9 2 | 7 2 |
| | | 9 2 | 5+6 2 |
| | | 8 2 | 3+4 2 |
| | | 4+5 2 | 7 2 |
| | | 9 2 | 7 2 |
| | | 8 2 | 3+4 2 |
| | | 8 2 | 5+6 2 |
| | | 6+7 2 | 7 2 |

Continued on next column

**TABLE LXXXVIII– continued**

| TQC | ICSD | | NoE | SF |
|---|---|---|---|---|
| | | | 8 2 | 5+6 2 |
| | | | 4+5 2 | 7 2 |
| | | | 9 2 | 3+4 2 |
| | | | 6+7 2 | 7 2 |
| | | | 9 2 | 5+6 2 |
| | | | 9 2 | 7 2 |
| | | | 6+7 2 | 7 2 |
| | | | 9 2 | 3+4 2 |
| | | | 8 2 | 5+6 2 |
| | | | 8 2 | 7 2 |
| | | | 8 2 | 7 2 |
| | | | 4+5 2 | 3+4 2 |
| | | | 9 2 | 7 2 |
| | | | 4+5 2 | 5+6 2 |
| 227.1.29 | 159800 | | 164 | Hg2O7Os2 |
| | -GM | -X | -L | -W |
| | 6 2 | 5 4 | 8 2 | 7 2 |
| | 7 2 | 5 4 | 9 2 | 3+4 2 |
| | 10 4 | 5 4 | 8 2 | 5+6 2 |
| | 10 4 | 5 4 | 4+5 2 | 7 2 |
| | 6 2 | 5 4 | 8 2 | 5+6 2 |
| | 11 4 | 5 4 | 4+5 2 | 7 2 |
| | 9 2 | 5 4 | 6+7 2 | 3+4 2 |
| | 8 2 | 5 4 | 9 2 | 7 2 |
| | 11 4 | 5 4 | 8 2 | 7 2 |
| | 9 2 | 5 4 | 9 2 | 5+6 2 |
| | 10 4 | 5 4 | 6+7 2 | 3+4 2 |
| | 7 2 | 5 4 | 8 2 | 7 2 |
| | 10 4 | 5 4 | 9 2 | 5+6 2 |
| | 6 2 | 5 4 | 9 2 | 7 2 |
| | 10 4 | 5 4 | 4+5 2 | 7 2 |
| | 7 2 | 5 4 | 8 2 | 3+4 2 |
| | 6 2 | 5 4 | 8 2 | 5+6 2 |
| | 10 4 | 5 4 | 6+7 2 | 7 2 |
| | 7 2 | 5 4 | 8 2 | 7 2 |
| | 10 4 | 5 4 | 9 2 | 3+4 2 |
| | 10 4 | 5 4 | 6+7 2 | 5+6 2 |
| | 7 2 | 5 4 | 9 2 | 7 2 |
| | 6 2 | 5 4 | 9 2 | 3+4 2 |
| | 10 4 | 5 4 | 8 2 | 5+6 2 |
| | 6 2 | 5 4 | 8 2 | 7 2 |
| | 10 4 | 5 4 | 4+5 2 | 7 2 |
| | 7 2 | 5 4 | 9 2 | 3+4 2 |
| | 9 2 | 5 4 | 9 2 | 7 2 |
| | 8 2 | 5 4 | 4+5 2 | 7 2 |
| | 11 4 | 5 4 | 8 2 | 3+4 2 |
| | 10 4 | 5 4 | 6+7 2 | 7 2 |
| | 10 4 | 5 4 | 9 2 | 5+6 2 |
| | 11 4 | 5 4 | 6+7 2 | 7 2 |
| | 9 2 | 5 4 | 9 2 | 3+4 2 |
| | 6 2 | 5 4 | 4+5 2 | 5+6 2 |
| | 10 4 | 5 4 | 8 2 | 7 2 |
| | 8 2 | 5 4 | 9 2 | 3+4 2 |
| | 11 4 | 5 4 | 8 2 | 7 2 |
| | 10 4 | 5 4 | 8 2 | 5+6 2 |
| | 11 4 | 5 4 | 9 2 | 7 2 |
| | 6 2 | 5 4 | 6+7 2 | 7 2 |
| | 11 4 | | 9 2 | 3+4 2 |
| | 8 2 | | 4+5 2 | 5+6 2 |
| | 10 4 | | 8 2 | 7 2 |
| | 7 2 | | 9 2 | 7 2 |

Continued on next column

**TABLE LXXXVIII– continued**

| TQC | ICSD | | NoE | SF |
|---|---|---|---|---|
| | 9 2 | | 6+7 2 | 3+4 2 |
| | 11 4 | | 9 2 | 7 2 |
| | 8 2 | | 9 2 | 5+6 2 |
| | 10 4 | | 6+7 2 | 5+6 2 |
| | 11 4 | | 8 2 | 7 2 |
| | 7 2 | | 8 2 | 7 2 |
| | 6 2 | | 4+5 2 | 7 2 |
| | 10 4 | | 9 2 | 3+4 2 |
| | 7 2 | | 9 2 | 7 2 |
| | 10 4 | | 9 2 | 5+6 2 |
| | | | 6+7 2 | 5+6 2 |
| | | | 9 2 | 3+4 2 |
| | | | 4+5 2 | 7 2 |
| | | | 6+7 2 | 3+4 2 |
| | | | 8 2 | 7 2 |
| | | | 9 2 | 7 2 |
| | | | 9 2 | 5+6 2 |
| | | | 8 2 | 3+4 2 |
| | | | 4+5 2 | 7 2 |
| | | | 9 2 | 7 2 |
| | | | 8 2 | 3+4 2 |
| | | | 8 2 | 5+6 2 |
| | | | 6+7 2 | 7 2 |
| | | | 8 2 | 5+6 2 |
| | | | 4+5 2 | 7 2 |
| | | | 9 2 | 3+4 2 |
| | | | 6+7 2 | 7 2 |
| | | | 9 2 | 5+6 2 |
| | | | 9 2 | 7 2 |
| | | | 6+7 2 | 7 2 |
| | | | 9 2 | 3+4 2 |
| | | | 8 2 | 5+6 2 |
| | | | 8 2 | 7 2 |
| | | | 8 2 | 7 2 |
| | | | 4+5 2 | 3+4 2 |
| | | | 9 2 | 7 2 |
| | | | 4+5 2 | 5+6 2 |
| 227.1.30 | 159799 | | 164 | Hg2O7Os2 |
| | -GM | -X | -L | -W |
| | 6 2 | 5 4 | 8 2 | 7 2 |
| | 7 2 | 5 4 | 9 2 | 3+4 2 |
| | 10 4 | 5 4 | 8 2 | 5+6 2 |
| | 10 4 | 5 4 | 4+5 2 | 7 2 |
| | 6 2 | 5 4 | 8 2 | 5+6 2 |
| | 11 4 | 5 4 | 4+5 2 | 7 2 |
| | 9 2 | 5 4 | 6+7 2 | 3+4 2 |
| | 8 2 | 5 4 | 9 2 | 7 2 |
| | 11 4 | 5 4 | 8 2 | 7 2 |
| | 9 2 | 5 4 | 9 2 | 5+6 2 |
| | 10 4 | 5 4 | 6+7 2 | 3+4 2 |
| | 7 2 | 5 4 | 8 2 | 7 2 |
| | 10 4 | 5 4 | 9 2 | 5+6 2 |
| | 6 2 | 5 4 | 9 2 | 7 2 |
| | 10 4 | 5 4 | 4+5 2 | 7 2 |
| | 7 2 | 5 4 | 8 2 | 3+4 2 |
| | 6 2 | 5 4 | 8 2 | 5+6 2 |
| | 10 4 | 5 4 | 6+7 2 | 7 2 |
| | 7 2 | 5 4 | 8 2 | 7 2 |
| | 10 4 | 5 4 | 9 2 | 3+4 2 |
| | 10 4 | 5 4 | 9 2 | 5+6 2 |
| | 7 2 | 5 4 | 6+7 2 | 7 2 |

Continued on next column

**TABLE LXXXVIII– continued**

| TQC | ICSD | NoE | SF |
|---|---|---|---|
| | 6 2 5 4 | 9 2 | 3+4 2 |
| | 10 4 5 4 | 8 2 | 5+6 2 |
| | 6 2 5 4 | 8 2 | 7 2 |
| | 10 4 5 4 | 4+5 2 | 7 2 |
| | 7 2 5 4 | 9 2 | 3+4 2 |
| | 9 2 5 4 | 9 2 | 7 2 |
| | 8 2 5 4 | 4+5 2 | 7 2 |
| | 11 4 5 4 | 8 2 | 3+4 2 |
| | 10 4 5 4 | 6+7 2 | 7 2 |
| | 10 4 5 4 | 9 2 | 5+6 2 |
| | 11 4 5 4 | 6+7 2 | 7 2 |
| | 9 2 5 4 | 9 2 | 3+4 2 |
| | 6 2 5 4 | 4+5 2 | 5+6 2 |
| | 10 4 5 4 | 8 2 | 7 2 |
| | 8 2 5 4 | 9 2 | 3+4 2 |
| | 11 4 5 4 | 8 2 | 7 2 |
| | 10 4 5 4 | 8 2 | 5+6 2 |
| | 11 4 5 4 | 9 2 | 7 2 |
| | 6 2 5 4 | 6+7 2 | 7 2 |
| | 11 4 | 9 2 | 3+4 2 |
| | 8 2 | 4+5 2 | 5+6 2 |
| | 10 4 | 8 2 | 7 2 |
| | 7 2 | 9 2 | 7 2 |
| | 9 2 | 6+7 2 | 3+4 2 |
| | 11 4 | 9 2 | 7 2 |
| | 8 2 | 9 2 | 5+6 2 |
| | 10 4 | 6+7 2 | 5+6 2 |
| | 11 4 | 8 2 | 7 2 |
| | 7 2 | 8 2 | 7 2 |
| | 6 2 | 4+5 2 | 7 2 |
| | 10 4 | 9 2 | 3+4 2 |
| | 7 2 | 9 2 | 7 2 |
| | 10 4 | 9 2 | 5+6 2 |
| | | 6+7 2 | 5+6 2 |
| | | 9 2 | 3+4 2 |
| | | 4+5 2 | 7 2 |
| | | 6+7 2 | 3+4 2 |
| | | 8 2 | 7 2 |
| | | 9 2 | 7 2 |
| | | 9 2 | 5+6 2 |
| | | 8 2 | 3+4 2 |
| | | 4+5 2 | 7 2 |
| | | 9 2 | 7 2 |
| | | 8 2 | 3+4 2 |
| | | 8 2 | 5+6 2 |
| | | 6+7 2 | 7 2 |
| | | 8 2 | 5+6 2 |
| | | 4+5 2 | 7 2 |
| | | 9 2 | 3+4 2 |
| | | 6+7 2 | 7 2 |
| | | 9 2 | 5+6 2 |
| | | 9 2 | 7 2 |
| | | 6+7 2 | 7 2 |
| | | 9 2 | 3+4 2 |
| | | 8 2 | 5+6 2 |
| | | 8 2 | 7 2 |
| | | 8 2 | 7 2 |
| | | 4+5 2 | 3+4 2 |
| | | 9 2 | 7 2 |
| | | 4+5 2 | 5+6 2 |
| 227.1.31 | 159798 | 164 | Hg2O7Os2 |

Continued on next column

**TABLE LXXXVIII– continued**

| TQC | ICSD | NoE | SF |
|---|---|---|---|
| | -GM -X | -L | -W |
| | 6 2 5 4 | 8 2 | 7 2 |
| | 7 2 5 4 | 9 2 | 3+4 2 |
| | 10 4 5 4 | 8 2 | 5+6 2 |
| | 10 4 5 4 | 4+5 2 | 7 2 |
| | 6 2 5 4 | 8 2 | 5+6 2 |
| | 11 4 5 4 | 4+5 2 | 7 2 |
| | 9 2 5 4 | 6+7 2 | 3+4 2 |
| | 8 2 5 4 | 9 2 | 7 2 |
| | 11 4 5 4 | 8 2 | 7 2 |
| | 9 2 5 4 | 9 2 | 5+6 2 |
| | 10 4 5 4 | 6+7 2 | 3+4 2 |
| | 7 2 5 4 | 8 2 | 7 2 |
| | 10 4 5 4 | 9 2 | 5+6 2 |
| | 6 2 5 4 | 9 2 | 7 2 |
| | 10 4 5 4 | 4+5 2 | 7 2 |
| | 7 2 5 4 | 8 2 | 3+4 2 |
| | 6 2 5 4 | 8 2 | 5+6 2 |
| | 10 4 5 4 | 6+7 2 | 7 2 |
| | 7 2 5 4 | 8 2 | 7 2 |
| | 10 4 5 4 | 9 2 | 3+4 2 |
| | 10 4 5 4 | 9 2 | 5+6 2 |
| | 7 2 5 4 | 6+7 2 | 7 2 |
| | 6 2 5 4 | 9 2 | 3+4 2 |
| | 10 4 5 4 | 8 2 | 5+6 2 |
| | 6 2 5 4 | 8 2 | 7 2 |
| | 10 4 5 4 | 4+5 2 | 7 2 |
| | 7 2 5 4 | 9 2 | 3+4 2 |
| | 9 2 5 4 | 9 2 | 7 2 |
| | 8 2 5 4 | 4+5 2 | 7 2 |
| | 11 4 5 4 | 8 2 | 3+4 2 |
| | 10 4 5 4 | 6+7 2 | 7 2 |
| | 10 4 5 4 | 9 2 | 5+6 2 |
| | 11 4 5 4 | 6+7 2 | 7 2 |
| | 9 2 5 4 | 9 2 | 3+4 2 |
| | 6 2 5 4 | 4+5 2 | 5+6 2 |
| | 10 4 5 4 | 8 2 | 7 2 |
| | 8 2 5 4 | 9 2 | 3+4 2 |
| | 11 4 5 4 | 8 2 | 7 2 |
| | 10 4 5 4 | 8 2 | 5+6 2 |
| | 11 4 5 4 | 9 2 | 7 2 |
| | 6 2 5 4 | 6+7 2 | 7 2 |
| | 11 4 | 9 2 | 3+4 2 |
| | 8 2 | 4+5 2 | 5+6 2 |
| | 10 4 | 8 2 | 7 2 |
| | 7 2 | 9 2 | 7 2 |
| | 9 2 | 6+7 2 | 3+4 2 |
| | 11 4 | 9 2 | 7 2 |
| | 8 2 | 9 2 | 5+6 2 |
| | 10 4 | 6+7 2 | 5+6 2 |
| | 11 4 | 8 2 | 7 2 |
| | 7 2 | 8 2 | 7 2 |
| | 6 2 | 4+5 2 | 7 2 |
| | 10 4 | 9 2 | 3+4 2 |
| | 7 2 | 9 2 | 7 2 |
| | 10 4 | 9 2 | 5+6 2 |
| | | 6+7 2 | 5+6 2 |
| | | 9 2 | 3+4 2 |
| | | 4+5 2 | 7 2 |
| | | 6+7 2 | 3+4 2 |
| | | 8 2 | 7 2 |

Continued on next column

**TABLE LXXXVIII– continued**

| TQC | ICSD | NoE | SF |
|---|---|---|---|
| | | 9 2 | 7 2 |
| | | 9 2 | 5+6 2 |
| | | 8 2 | 3+4 2 |
| | | 4+5 2 | 7 2 |
| | | 9 2 | 7 2 |
| | | 8 2 | 3+4 2 |
| | | 8 2 | 5+6 2 |
| | | 6+7 2 | 7 2 |
| | | 8 2 | 5+6 2 |
| | | 4+5 2 | 7 2 |
| | | 9 2 | 3+4 2 |
| | | 6+7 2 | 7 2 |
| | | 9 2 | 5+6 2 |
| | | 9 2 | 7 2 |
| | | 6+7 2 | 7 2 |
| | | 9 2 | 3+4 2 |
| | | 8 2 | 5+6 2 |
| | | 8 2 | 7 2 |
| | | 8 2 | 7 2 |
| | | 4+5 2 | 3+4 2 |
| | | 9 2 | 7 2 |
| | | 4+5 2 | 5+6 2 |
| 227.1.32 | 159797 | 164 | Hg2O7Os2 |
| | -GM -X | -L | -W |
| | 6 2 5 4 | 8 2 | 7 2 |
| | 7 2 5 4 | 9 2 | 3+4 2 |
| | 10 4 5 4 | 8 2 | 5+6 2 |
| | 10 4 5 4 | 4+5 2 | 7 2 |
| | 6 2 5 4 | 8 2 | 5+6 2 |
| | 11 4 5 4 | 4+5 2 | 7 2 |
| | 9 2 5 4 | 6+7 2 | 3+4 2 |
| | 8 2 5 4 | 9 2 | 7 2 |
| | 11 4 5 4 | 8 2 | 7 2 |
| | 9 2 5 4 | 9 2 | 5+6 2 |
| | 10 4 5 4 | 6+7 2 | 3+4 2 |
| | 7 2 5 4 | 8 2 | 7 2 |
| | 10 4 5 4 | 9 2 | 5+6 2 |
| | 6 2 5 4 | 9 2 | 7 2 |
| | 10 4 5 4 | 4+5 2 | 7 2 |
| | 7 2 5 4 | 8 2 | 3+4 2 |
| | 6 2 5 4 | 8 2 | 5+6 2 |
| | 10 4 5 4 | 6+7 2 | 7 2 |
| | 7 2 5 4 | 8 2 | 7 2 |
| | 10 4 5 4 | 9 2 | 3+4 2 |
| | 10 4 5 4 | 6+7 2 | 5+6 2 |
| | 7 2 5 4 | 9 2 | 7 2 |
| | 6 2 5 4 | 9 2 | 3+4 2 |
| | 10 4 5 4 | 8 2 | 5+6 2 |
| | 6 2 5 4 | 8 2 | 7 2 |
| | 10 4 5 4 | 4+5 2 | 7 2 |
| | 7 2 5 4 | 9 2 | 3+4 2 |
| | 9 2 5 4 | 9 2 | 7 2 |
| | 8 2 5 4 | 4+5 2 | 7 2 |
| | 11 4 5 4 | 8 2 | 3+4 2 |
| | 10 4 5 4 | 6+7 2 | 7 2 |
| | 10 4 5 4 | 9 2 | 5+6 2 |
| | 11 4 5 4 | 6+7 2 | 7 2 |
| | 9 2 5 4 | 9 2 | 3+4 2 |
| | 6 2 5 4 | 4+5 2 | 5+6 2 |
| | 10 4 5 4 | 8 2 | 7 2 |
| | 8 2 5 4 | 9 2 | 3+4 2 |

Continued on next column

**TABLE LXXXVIII– continued**

| TQC | ICSD | NoE | SF |
|---|---|---|---|
| | 11 4 5 4 | 8 2 | 7 2 |
| | 10 4 5 4 | 8 2 | 5+6 2 |
| | 11 4 5 4 | 9 2 | 7 2 |
| | 6 2 5 4 | 6+7 2 | 7 2 |
| | 11 4 | 9 2 | 3+4 2 |
| | 8 2 | 4+5 2 | 5+6 2 |
| | 10 4 | 8 2 | 7 2 |
| | 7 2 | 9 2 | 7 2 |
| | 9 2 | 6+7 2 | 3+4 2 |
| | 11 4 | 9 2 | 7 2 |
| | 8 2 | 9 2 | 5+6 2 |
| | 10 4 | 6+7 2 | 5+6 2 |
| | 11 4 | 8 2 | 7 2 |
| | 7 2 | 8 2 | 7 2 |
| | 6 2 | 4+5 2 | 7 2 |
| | 10 4 | 9 2 | 3+4 2 |
| | 7 2 | 9 2 | 7 2 |
| | 10 4 | 9 2 | 5+6 2 |
| | | 6+7 2 | 5+6 2 |
| | | 9 2 | 3+4 2 |
| | | 4+5 2 | 7 2 |
| | | 6+7 2 | 3+4 2 |
| | | 8 2 | 7 2 |
| | | 9 2 | 7 2 |
| | | 9 2 | 5+6 2 |
| | | 8 2 | 3+4 2 |
| | | 4+5 2 | 7 2 |
| | | 9 2 | 7 2 |
| | | 8 2 | 3+4 2 |
| | | 8 2 | 5+6 2 |
| | | 6+7 2 | 7 2 |
| | | 8 2 | 5+6 2 |
| | | 4+5 2 | 7 2 |
| | | 9 2 | 3+4 2 |
| | | 6+7 2 | 7 2 |
| | | 9 2 | 5+6 2 |
| | | 9 2 | 7 2 |
| | | 6+7 2 | 7 2 |
| | | 9 2 | 3+4 2 |
| | | 8 2 | 5+6 2 |
| | | 8 2 | 7 2 |
| | | 8 2 | 7 2 |
| | | 4+5 2 | 3+4 2 |
| | | 9 2 | 7 2 |
| | | 4+5 2 | 5+6 2 |
| 227.1.33 | 159796 | 164 | Hg2O7Os2 |
| | -GM -X | -L | -W |
| | 6 2 5 4 | 8 2 | 7 2 |
| | 7 2 5 4 | 9 2 | 3+4 2 |
| | 10 4 5 4 | 8 2 | 5+6 2 |
| | 10 4 5 4 | 4+5 2 | 7 2 |
| | 6 2 5 4 | 8 2 | 5+6 2 |
| | 11 4 5 4 | 4+5 2 | 7 2 |
| | 9 2 5 4 | 6+7 2 | 3+4 2 |
| | 8 2 5 4 | 9 2 | 7 2 |
| | 11 4 5 4 | 8 2 | 7 2 |
| | 9 2 5 4 | 9 2 | 5+6 2 |
| | 10 4 5 4 | 6+7 2 | 3+4 2 |
| | 7 2 5 4 | 8 2 | 7 2 |
| | 10 4 5 4 | 9 2 | 5+6 2 |
| | 6 2 5 4 | 9 2 | 7 2 |

Continued on next column

**TABLE LXXXVIII– continued**

| TQC | ICSD | NoE | SF |
|---|---|---|---|
| | 10 4 5 4 | 4+5 2 | 7 2 |
| | 7 2 5 4 | 8 2 | 3+4 2 |
| | 6 2 5 4 | 8 2 | 5+6 2 |
| | 10 4 5 4 | 6+7 2 | 7 2 |
| | 7 2 5 4 | 8 2 | 7 2 |
| | 10 4 5 4 | 9 2 | 3+4 2 |
| | 10 4 5 4 | 9 2 | 5+6 2 |
| | 7 2 5 4 | 6+7 2 | 7 2 |
| | 6 2 5 4 | 9 2 | 3+4 2 |
| | 10 4 5 4 | 8 2 | 5+6 2 |
| | 6 2 5 4 | 8 2 | 7 2 |
| | 10 4 5 4 | 4+5 2 | 7 2 |
| | 7 2 5 4 | 9 2 | 3+4 2 |
| | 9 2 5 4 | 9 2 | 7 2 |
| | 8 2 5 4 | 4+5 2 | 7 2 |
| | 11 4 5 4 | 8 2 | 3+4 2 |
| | 10 4 5 4 | 6+7 2 | 7 2 |
| | 10 4 5 4 | 9 2 | 5+6 2 |
| | 11 4 5 4 | 6+7 2 | 7 2 |
| | 9 2 5 4 | 9 2 | 3+4 2 |
| | 6 2 5 4 | 4+5 2 | 5+6 2 |
| | 10 4 5 4 | 8 2 | 7 2 |
| | 8 2 5 4 | 9 2 | 3+4 2 |
| | 11 4 5 4 | 8 2 | 7 2 |
| | 10 4 5 4 | 8 2 | 5+6 2 |
| | 11 4 5 4 | 9 2 | 7 2 |
| | 6 2 5 4 | 6+7 2 | 7 2 |
| | 11 4 | 9 2 | 3+4 2 |
| | 8 2 | 4+5 2 | 5+6 2 |
| | 10 4 | 8 2 | 7 2 |
| | 7 2 | 9 2 | 7 2 |
| | 9 2 | 6+7 2 | 3+4 2 |
| | 11 4 | 9 2 | 7 2 |
| | 8 2 | 9 2 | 5+6 2 |
| | 10 4 | 6+7 2 | 5+6 2 |
| | 11 4 | 8 2 | 7 2 |
| | 7 2 | 8 2 | 7 2 |
| | 6 2 | 4+5 2 | 7 2 |
| | 10 4 | 9 2 | 3+4 2 |
| | 7 2 | 9 2 | 7 2 |
| | 10 4 | 9 2 | 5+6 2 |
| | | 6+7 2 | 5+6 2 |
| | | 9 2 | 3+4 2 |
| | | 4+5 2 | 7 2 |
| | | 6+7 2 | 3+4 2 |
| | | 8 2 | 7 2 |
| | | 9 2 | 7 2 |
| | | 9 2 | 5+6 2 |
| | | 8 2 | 3+4 2 |
| | | 4+5 2 | 7 2 |
| | | 9 2 | 7 2 |
| | | 8 2 | 3+4 2 |
| | | 8 2 | 5+6 2 |
| | | 6+7 2 | 7 2 |
| | | 8 2 | 5+6 2 |
| | | 4+5 2 | 7 2 |
| | | 9 2 | 3+4 2 |
| | | 6+7 2 | 7 2 |
| | | 9 2 | 5+6 2 |
| | | 9 2 | 7 2 |
| | | 6+7 2 | 7 2 |

Continued on next column

**TABLE LXXXVIII– continued**

| TQC | ICSD | NoE | SF |
|---|---|---|---|
| | | 9 2 | 3+4 2 |
| | | 8 2 | 5+6 2 |
| | | 8 2 | 7 2 |
| | | 8 2 | 7 2 |
| | | 4+5 2 | 3+4 2 |
| | | 9 2 | 7 2 |
| | | 4+5 2 | 5+6 2 |
| 227.1.34 | 159795 | 164 | Hg2O7Os2 |
| | -GM -X | -L | -W |
| | 6 2 5 4 | 8 2 | 7 2 |
| | 7 2 5 4 | 9 2 | 3+4 2 |
| | 10 4 5 4 | 8 2 | 5+6 2 |
| | 10 4 5 4 | 4+5 2 | 7 2 |
| | 6 2 5 4 | 8 2 | 5+6 2 |
| | 11 4 5 4 | 4+5 2 | 7 2 |
| | 9 2 5 4 | 6+7 2 | 3+4 2 |
| | 8 2 5 4 | 9 2 | 7 2 |
| | 11 4 5 4 | 8 2 | 7 2 |
| | 9 2 5 4 | 9 2 | 5+6 2 |
| | 10 4 5 4 | 6+7 2 | 3+4 2 |
| | 7 2 5 4 | 8 2 | 7 2 |
| | 10 4 5 4 | 9 2 | 5+6 2 |
| | 6 2 5 4 | 9 2 | 7 2 |
| | 10 4 5 4 | 4+5 2 | 7 2 |
| | 7 2 5 4 | 8 2 | 3+4 2 |
| | 6 2 5 4 | 8 2 | 5+6 2 |
| | 10 4 5 4 | 6+7 2 | 7 2 |
| | 7 2 5 4 | 8 2 | 7 2 |
| | 10 4 5 4 | 9 2 | 3+4 2 |
| | 10 4 5 4 | 9 2 | 5+6 2 |
| | 7 2 5 4 | 6+7 2 | 7 2 |
| | 6 2 5 4 | 9 2 | 3+4 2 |
| | 10 4 5 4 | 8 2 | 5+6 2 |
| | 6 2 5 4 | 8 2 | 7 2 |
| | 10 4 5 4 | 4+5 2 | 7 2 |
| | 7 2 5 4 | 9 2 | 3+4 2 |
| | 9 2 5 4 | 9 2 | 7 2 |
| | 8 2 5 4 | 4+5 2 | 7 2 |
| | 11 4 5 4 | 8 2 | 3+4 2 |
| | 10 4 5 4 | 6+7 2 | 7 2 |
| | 10 4 5 4 | 9 2 | 5+6 2 |
| | 11 4 5 4 | 6+7 2 | 7 2 |
| | 9 2 5 4 | 9 2 | 3+4 2 |
| | 6 2 5 4 | 4+5 2 | 5+6 2 |
| | 10 4 5 4 | 8 2 | 7 2 |
| | 8 2 5 4 | 9 2 | 3+4 2 |
| | 11 4 5 4 | 8 2 | 7 2 |
| | 10 4 5 4 | 8 2 | 5+6 2 |
| | 11 4 5 4 | 9 2 | 7 2 |
| | 6 2 5 4 | 6+7 2 | 7 2 |
| | 11 4 | 9 2 | 3+4 2 |
| | 8 2 | 4+5 2 | 5+6 2 |
| | 10 4 | 8 2 | 7 2 |
| | 7 2 | 9 2 | 7 2 |
| | 9 2 | 6+7 2 | 3+4 2 |
| | 11 4 | 9 2 | 7 2 |
| | 8 2 | 9 2 | 5+6 2 |
| | 10 4 | 6+7 2 | 5+6 2 |
| | 11 4 | 8 2 | 7 2 |
| | 7 2 | 8 2 | 7 2 |
| | 6 2 | 4+5 2 | 7 2 |

Continued on next column

**TABLE LXXXVIII– continued**

| TQC | ICSD | | NoE | SF |
|---|---|---|---|---|
| | 10 4 | | 9 2 | 3+4 2 |
| | 7 2 | | 9 2 | 7 2 |
| | 10 4 | | 9 2 | 5+6 2 |
| | | | 6+7 2 | 5+6 2 |
| | | | 9 2 | 3+4 2 |
| | | | 4+5 2 | 7 2 |
| | | | 6+7 2 | 3+4 2 |
| | | | 8 2 | 7 2 |
| | | | 9 2 | 7 2 |
| | | | 9 2 | 5+6 2 |
| | | | 8 2 | 3+4 2 |
| | | | 4+5 2 | 7 2 |
| | | | 9 2 | 7 2 |
| | | | 8 2 | 3+4 2 |
| | | | 8 2 | 5+6 2 |
| | | | 6+7 2 | 7 2 |
| | | | 8 2 | 5+6 2 |
| | | | 4+5 2 | 7 2 |
| | | | 9 2 | 3+4 2 |
| | | | 6+7 2 | 7 2 |
| | | | 9 2 | 5+6 2 |
| | | | 9 2 | 7 2 |
| | | | 6+7 2 | 7 2 |
| | | | 9 2 | 3+4 2 |
| | | | 8 2 | 5+6 2 |
| | | | 8 2 | 7 2 |
| | | | 8 2 | 7 2 |
| | | | 4+5 2 | 3+4 2 |
| | | | 9 2 | 7 2 |
| | | | 4+5 2 | 5+6 2 |
| 227.1.35 | 159794 | | 164 | Hg2O7Os2 |
| | -GM | -X | -L | -W |
| | 6 2 | 5 4 | 8 2 | 7 2 |
| | 7 2 | 5 4 | 9 2 | 3+4 2 |
| | 10 4 | 5 4 | 8 2 | 5+6 2 |
| | 10 4 | 5 4 | 4+5 2 | 7 2 |
| | 6 2 | 5 4 | 8 2 | 5+6 2 |
| | 11 4 | 5 4 | 4+5 2 | 7 2 |
| | 9 2 | 5 4 | 6+7 2 | 3+4 2 |
| | 8 2 | 5 4 | 9 2 | 7 2 |
| | 11 4 | 5 4 | 8 2 | 7 2 |
| | 9 2 | 5 4 | 9 2 | 5+6 2 |
| | 10 4 | 5 4 | 6+7 2 | 3+4 2 |
| | 7 2 | 5 4 | 8 2 | 7 2 |
| | 10 4 | 5 4 | 9 2 | 5+6 2 |
| | 6 2 | 5 4 | 9 2 | 7 2 |
| | 10 4 | 5 4 | 4+5 2 | 7 2 |
| | 7 2 | 5 4 | 8 2 | 3+4 2 |
| | 6 2 | 5 4 | 8 2 | 5+6 2 |
| | 10 4 | 5 4 | 6+7 2 | 7 2 |
| | 7 2 | 5 4 | 8 2 | 7 2 |
| | 10 4 | 5 4 | 9 2 | 3+4 2 |
| | 10 4 | 5 4 | 6+7 2 | 5+6 2 |
| | 7 2 | 5 4 | 9 2 | 7 2 |
| | 6 2 | 5 4 | 9 2 | 3+4 2 |
| | 10 4 | 5 4 | 8 2 | 5+6 2 |
| | 6 2 | 5 4 | 8 2 | 7 2 |
| | 10 4 | 5 4 | 4+5 2 | 7 2 |
| | 7 2 | 5 4 | 9 2 | 3+4 2 |
| | 9 2 | 5 4 | 9 2 | 7 2 |
| | 8 2 | 5 4 | 4+5 2 | 7 2 |

Continued on next column

**TABLE LXXXVIII– continued**

| TQC | ICSD | | NoE | SF |
|---|---|---|---|---|
| | 11 4 | 5 4 | 8 2 | 3+4 2 |
| | 10 4 | 5 4 | 6+7 2 | 7 2 |
| | 10 4 | 5 4 | 9 2 | 5+6 2 |
| | 11 4 | 5 4 | 6+7 2 | 7 2 |
| | 9 2 | 5 4 | 9 2 | 3+4 2 |
| | 6 2 | 5 4 | 4+5 2 | 5+6 2 |
| | 10 4 | 5 4 | 8 2 | 7 2 |
| | 8 2 | 5 4 | 9 2 | 3+4 2 |
| | 11 4 | 5 4 | 8 2 | 7 2 |
| | 10 4 | 5 4 | 8 2 | 5+6 2 |
| | 11 4 | 5 4 | 9 2 | 7 2 |
| | 6 2 | 5 4 | 6+7 2 | 7 2 |
| | 11 4 | | 9 2 | 3+4 2 |
| | 8 2 | | 4+5 2 | 5+6 2 |
| | 10 4 | | 8 2 | 7 2 |
| | 7 2 | | 9 2 | 7 2 |
| | 9 2 | | 6+7 2 | 3+4 2 |
| | 11 4 | | 9 2 | 7 2 |
| | 8 2 | | 9 2 | 5+6 2 |
| | 10 4 | | 6+7 2 | 5+6 2 |
| | 11 4 | | 8 2 | 7 2 |
| | 7 2 | | 8 2 | 7 2 |
| | 6 2 | | 4+5 2 | 7 2 |
| | 10 4 | | 9 2 | 3+4 2 |
| | 7 2 | | 9 2 | 7 2 |
| | 10 4 | | 9 2 | 5+6 2 |
| | | | 6+7 2 | 5+6 2 |
| | | | 9 2 | 3+4 2 |
| | | | 4+5 2 | 7 2 |
| | | | 6+7 2 | 3+4 2 |
| | | | 8 2 | 7 2 |
| | | | 9 2 | 7 2 |
| | | | 9 2 | 5+6 2 |
| | | | 8 2 | 3+4 2 |
| | | | 4+5 2 | 7 2 |
| | | | 9 2 | 7 2 |
| | | | 8 2 | 3+4 2 |
| | | | 8 2 | 5+6 2 |
| | | | 6+7 2 | 7 2 |
| | | | 8 2 | 5+6 2 |
| | | | 4+5 2 | 7 2 |
| | | | 9 2 | 3+4 2 |
| | | | 6+7 2 | 7 2 |
| | | | 9 2 | 5+6 2 |
| | | | 9 2 | 7 2 |
| | | | 6+7 2 | 7 2 |
| | | | 9 2 | 3+4 2 |
| | | | 8 2 | 5+6 2 |
| | | | 8 2 | 7 2 |
| | | | 8 2 | 7 2 |
| | | | 4+5 2 | 3+4 2 |
| | | | 9 2 | 7 2 |
| | | | 4+5 2 | 5+6 2 |
| 227.1.36 | 84753 | | 172 | Gd2O7Sn2 |
| | -GM | -X | -L | -W |
| | 6 2 | 5 4 | 9 2 | 5+6 2 |
| | 7 2 | 5 4 | 8 2 | 7 2 |
| | 10 4 | 5 4 | 6+7 2 | 3+4 2 |
| | 9 2 | 5 4 | 9 2 | 7 2 |
| | 11 4 | 5 4 | 8 2 | 5+6 2 |
| | 8 2 | 5 4 | 9 2 | 7 2 |

Continued on next column

**TABLE LXXXVIII– continued**

| TQC | ICSD | NoE | SF |
|---|---|---|---|
| | 9 2 5 4 | 4+5 2 | 7 2 |
| | 11 4 5 4 | 8 2 | 3+4 2 |
| | 6 2 5 4 | 8 2 | 7 2 |
| | 8 2 5 4 | 8 2 | 5+6 2 |
| | 11 4 5 4 | 9 2 | 7 2 |
| | 11 4 5 4 | 8 2 | 3+4 2 |
| | 6 2 5 4 | 4+5 2 | 5+6 2 |
| | 7 2 5 4 | 6+7 2 | 7 2 |
| | 10 4 5 4 | 9 2 | 3+4 2 |
| | 9 2 5 4 | 8 2 | 7 2 |
| | 8 2 5 4 | 4+5 2 | 7 2 |
| | 11 4 5 4 | 8 2 | 5+6 2 |
| | 11 4 5 4 | 9 2 | 3+4 2 |
| | 10 4 5 4 | 9 2 | 7 2 |
| | 9 2 5 4 | 6+7 2 | 7 2 |
| | 10 4 5 4 | 4+5 2 | 3+4 2 |
| | 7 2 5 4 | 9 2 | 5+6 2 |
| | 9 2 5 4 | 8 2 | 7 2 |
| | 11 4 5 4 | 6+7 2 | 7 2 |
| | 6 2 5 4 | 8 2 | 7 2 |
| | 9 2 5 4 | 8 2 | 3+4 2 |
| | 8 2 5 4 | 9 2 | 5+6 2 |
| | 7 2 5 4 | 4+5 2 | 5+6 2 |
| | 10 4 5 4 | 8 2 | 7 2 |
| | 11 4 5 4 | 8 2 | 5+6 2 |
| | 10 4 5 4 | 4+5 2 | 7 2 |
| | 7 2 5 4 | 9 2 | 7 2 |
| | 11 4 5 4 | 9 2 | 3+4 2 |
| | 8 2 5 4 | 8 2 | 7 2 |
| | 10 4 5 4 | 9 2 | 3+4 2 |
| | 11 4 5 4 | 6+7 2 | 3+4 2 |
| | 10 4 5 4 | 9 2 | 7 2 |
| | 6 2 5 4 | 8 2 | 7 2 |
| | 7 2 5 4 | 9 2 | 5+6 2 |
| | 10 4 5 4 | 6+7 2 | 5+6 2 |
| | 8 2 5 4 | 8 2 | 3+4 2 |
| | 11 4 5 4 | 4+5 2 | 7 2 |
| | 9 2 | 9 2 | 7 2 |
| | 11 4 | 6+7 2 | 7 2 |
| | 8 2 | 8 2 | 5+6 2 |
| | 11 4 | 9 2 | 3+4 2 |
| | 10 4 | 6+7 2 | 7 2 |
| | 6 2 | 9 2 | 7 2 |
| | 8 2 | 9 2 | 5+6 2 |
| | 11 4 | 8 2 | 3+4 2 |
| | 9 2 | 4+5 2 | 7 2 |
| | 11 4 | 8 2 | 7 2 |
| | 8 2 | 9 2 | 7 2 |
| | 11 4 | 4+5 2 | 3+4 2 |
| | 9 2 | 8 2 | 5+6 2 |
| | 11 4 | 8 2 | 5+6 2 |
| | 11 4 | 9 2 | 7 2 |
| | | 6+7 2 | 3+4 2 |
| | | 8 2 | 7 2 |
| | | 4+5 2 | 3+4 2 |
| | | 9 2 | 7 2 |
| | | 6+7 2 | 5+6 2 |
| | | 8 2 | 7 2 |
| | | 9 2 | 7 2 |
| | | 4+5 2 | 5+6 2 |
| | | 8 2 | 7 2 |
| | | 6+7 2 | 3+4 2 |
| | | 9 2 | 5+6 2 |
| | | 8 2 | 7 2 |
| | | 4+5 2 | 3+4 2 |
| | | 8 2 | 7 2 |
| | | 8 2 | 7 2 |
| | | 4+5 2 | 3+4 2 |
| | | 8 2 | 5+6 2 |
| | | 8 2 | 7 2 |
| | | 9 2 | 7 2 |
| | | 6+7 2 | 5+6 2 |
| | | 4+5 2 | 3+4 2 |
| | | 8 2 | 7 2 |
| | | 9 2 | 7 2 |
| | | 8 2 | 3+4 2 |
| | | 8 2 | 5+6 2 |
| | | 4+5 2 | 7 2 |
| | | 4+5 2 | 7 2 |
| | | 8 2 | 5+6 2 |
| 227.1.37 | 160866 | 192 | Er2O7V2 |
| | -GM -X | -L | -W |
| | 6 2 5 4 | 8 2 | 3+4 2 |
| | 7 2 5 4 | 9 2 | 7 2 |
| | 10 4 5 4 | 4+5 2 | 5+6 2 |
| | 9 2 5 4 | 8 2 | 7 2 |
| | 11 4 5 4 | 9 2 | 3+4 2 |
| | 8 2 5 4 | 6+7 2 | 7 2 |
| | 9 2 5 4 | 8 2 | 7 2 |
| | 11 4 5 4 | 9 2 | 5+6 2 |
| | 8 2 5 4 | 9 2 | 7 2 |
| | 11 4 5 4 | 8 2 | 3+4 2 |
| | 11 4 5 4 | 9 2 | 7 2 |
| | 6 2 5 4 | 6+7 2 | 5+6 2 |
| | 7 2 5 4 | 9 2 | 3+4 2 |
| | 10 4 5 4 | 4+5 2 | 7 2 |
| | 6 2 5 4 | 9 2 | 5+6 2 |
| | 10 4 5 4 | 6+7 2 | 7 2 |
| | 11 4 5 4 | 8 2 | 7 2 |
| | 8 2 5 4 | 9 2 | 5+6 2 |
| | 9 2 5 4 | 8 2 | 3+4 2 |
| | 11 4 5 4 | 6+7 2 | 7 2 |
| | 9 2 5 4 | 4+5 2 | 7 2 |
| | 7 2 5 4 | 9 2 | 3+4 2 |
| | 10 4 5 4 | 8 2 | 7 2 |
| | 10 4 5 4 | 8 2 | 5+6 2 |
| | 7 2 5 4 | 9 2 | 7 2 |
| | 6 2 5 4 | 6+7 2 | 3+4 2 |
| | 10 4 5 4 | 4+5 2 | 7 2 |
| | 6 2 5 4 | 9 2 | 5+6 2 |
| | 10 4 5 4 | 8 2 | 3+4 2 |
| | 7 2 5 4 | 9 2 | 7 2 |
| | 9 2 5 4 | 9 2 | 3+4 2 |
| | 11 4 5 4 | 6+7 2 | 7 2 |
| | 10 4 5 4 | 9 2 | 7 2 |
| | 9 2 5 4 | 8 2 | 3+4 2 |
| | 11 4 5 4 | 8 2 | 5+6 2 |
| | 8 2 5 4 | 4+5 2 | 7 2 |
| | 11 4 5 4 | 9 2 | 3+4 2 |
| | 8 2 5 4 | 6+7 2 | 5+6 2 |
| | 10 4 5 4 | 9 2 | 7 2 |
| | 7 2 5 4 | 4+5 2 | 7 2 |

Continued on next column

**TABLE LXXXVIII– continued**

| TQC | ICSD | NoE | SF |
|---|---|---|---|
| | 10 4 5 4 | 8 2 | 7 2 |
| | 6 2 5 4 | 8 2 | 5+6 2 |
| | 11 4 5 4 | 9 2 | 7 2 |
| | 8 2 5 4 | 8 2 | 5+6 2 |
| | 9 2 5 4 | 6+7 2 | 7 2 |
| | 11 4 5 4 | 9 2 | 5+6 2 |
| | 11 4 5 4 | 8 2 | 7 2 |
| | 8 2 5 4 | 4+5 2 | 5+6 2 |
| | 11 4 | 9 2 | 3+4 2 |
| | 8 2 | 9 2 | 7 2 |
| | 11 4 | 6+7 2 | 3+4 2 |
| | 9 2 | 9 2 | 7 2 |
| | 11 4 | 8 2 | 5+6 2 |
| | 9 2 | 4+5 2 | 7 2 |
| | 8 2 | 8 2 | 7 2 |
| | 11 4 | 8 2 | 3+4 2 |
| | 11 4 | 4+5 2 | 7 2 |
| | 10 4 | 8 2 | 3+4 2 |
| | 11 4 | 9 2 | 7 2 |
| | 7 2 | 6+7 2 | 5+6 2 |
| | 10 4 | 9 2 | 3+4 2 |
| | 9 2 | 8 2 | 7 2 |
| | 11 4 | 6+7 2 | 5+6 2 |
| | 8 2 | 9 2 | 7 2 |
| | 6 2 | 9 2 | 7 2 |
| | | 8 2 | 3+4 2 |
| | | 4+5 2 | 7 2 |
| | | 9 2 | 5+6 2 |
| | | 6+7 2 | 3+4 2 |
| | | 8 2 | 7 2 |
| | | 8 2 | 7 2 |
| | | 4+5 2 | 5+6 2 |
| | | 9 2 | 7 2 |
| | | 6+7 2 | 5+6 2 |
| | | 9 2 | 3+4 2 |
| | | 9 2 | 7 2 |
| | | 4+5 2 | 7 2 |
| | | 9 2 | 3+4 2 |
| | | 8 2 | 7 2 |
| | | 6+7 2 | 5+6 2 |
| | | 8 2 | 7 2 |
| | | 9 2 | 5+6 2 |
| | | 9 2 | 3+4 2 |
| | | 6+7 2 | 7 2 |
| | | 9 2 | 7 2 |
| | | 6+7 2 | 5+6 2 |
| | | 9 2 | 3+4 2 |
| | | 6+7 2 | 7 2 |
| | | 9 2 | 7 2 |
| | | 9 2 | 5+6 2 |
| | | 9 2 | 7 2 |
| | | 8 2 | 3+4 2 |
| | | 6+7 2 | 7 2 |
| | | 4+5 2 | 3+4 2 |
| | | 8 2 | 7 2 |
| | | 9 2 | 5+6 2 |
| 227.1.38 | 50047 | 96 | Cr2O4Zn1 |
| | -GM -X | -L | -W |
| | 6 2 5 4 | 9 2 | 5+6 2 |
| | 9 2 5 4 | 8 2 | 7 2 |
| | 7 2 5 4 | 9 2 | 3+4 2 |

Continued on next column

**TABLE LXXXVIII– continued**

| TQC | ICSD | NoE | SF |
|---|---|---|---|
| | 10 4 5 4 | 9 2 | 7 2 |
| | 8 2 5 4 | 6+7 2 | 7 2 |
| | 11 4 5 4 | 8 2 | 3+4 2 |
| | 6 2 5 4 | 4+5 2 | 5+6 2 |
| | 7 2 5 4 | 8 2 | 7 2 |
| | 10 4 5 4 | 9 2 | 7 2 |
| | 9 2 5 4 | 8 2 | 5+6 2 |
| | 11 4 5 4 | 9 2 | 3+4 2 |
| | 10 4 5 4 | 6+7 2 | 7 2 |
| | 7 2 5 4 | 9 2 | 7 2 |
| | 11 4 5 4 | 6+7 2 | 5+6 2 |
| | 8 2 5 4 | 4+5 2 | 3+4 2 |
| | 10 4 5 4 | 8 2 | 7 2 |
| | 10 4 5 4 | 9 2 | 5+6 2 |
| | 11 4 5 4 | 4+5 2 | 3+4 2 |
| | 6 2 5 4 | 8 2 | 7 2 |
| | 10 4 5 4 | 8 2 | 7 2 |
| | 11 4 5 4 | 4+5 2 | 7 2 |
| | 8 2 5 4 | 8 2 | 5+6 2 |
| | 10 4 5 4 | 9 2 | 7 2 |
| | 7 2 5 4 | 6+7 2 | 3+4 2 |
| | 9 2 | 9 2 | 7 2 |
| | 11 4 | 6+7 2 | 5+6 2 |
| | 8 2 | 9 2 | 7 2 |
| | 11 4 | 8 2 | 3+4 2 |
| | 10 4 | 9 2 | 7 2 |
| | 6 2 | 6+7 2 | 5+6 2 |
| | 10 4 | 9 2 | 7 2 |
| | 7 2 | 4+5 2 | 5+6 2 |
| | | 8 2 | 3+4 2 |
| | | 9 2 | 7 2 |
| | | 8 2 | 3+4 2 |
| | | 8 2 | 7 2 |
| | | 4+5 2 | 7 2 |
| | | 8 2 | 5+6 2 |
| | | 6+7 2 | 5+6 2 |
| | | 9 2 | 7 2 |
| | | 4+5 2 | 3+4 2 |
| | | 8 2 | 7 2 |
| | | 9 2 | 7 2 |
| | | 6+7 2 | 3+4 2 |
| | | 8 2 | 5+6 2 |
| | | 9 2 | 7 2 |
| | | 9 2 | 5+6 2 |
| | | 6+7 2 | 7 2 |
| 227.1.39 | 196119 | 96 | Cr2O4Zn1 |
| | -GM -X | -L | -W |
| | 6 2 5 4 | 9 2 | 5+6 2 |
| | 7 2 5 4 | 8 2 | 7 2 |
| | 9 2 5 4 | 9 2 | 3+4 2 |
| | 10 4 5 4 | 9 2 | 7 2 |
| | 8 2 5 4 | 6+7 2 | 7 2 |
| | 11 4 5 4 | 8 2 | 3+4 2 |
| | 6 2 5 4 | 4+5 2 | 5+6 2 |
| | 7 2 5 4 | 8 2 | 7 2 |
| | 10 4 5 4 | 9 2 | 7 2 |
| | 9 2 5 4 | 8 2 | 5+6 2 |
| | 11 4 5 4 | 9 2 | 3+4 2 |
| | 10 4 5 4 | 6+7 2 | 7 2 |
| | 7 2 5 4 | 9 2 | 7 2 |
| | 11 4 5 4 | 6+7 2 | 5+6 2 |

Continued on next column

**TABLE LXXXVIII– continued**

| TQC | ICSD | NoE | SF |
|---|---|---|---|
| | 8 2 5 4 | 4+5 2 | 3+4 2 |
| | 10 4 5 4 | 8 2 | 7 2 |
| | 10 4 5 4 | 9 2 | 5+6 2 |
| | 11 4 5 4 | 4+5 2 | 3+4 2 |
| | 6 2 5 4 | 8 2 | 7 2 |
| | 10 4 5 4 | 8 2 | 7 2 |
| | 11 4 5 4 | 4+5 2 | 7 2 |
| | 8 2 5 4 | 8 2 | 5+6 2 |
| | 10 4 5 4 | 9 2 | 7 2 |
| | 7 2 5 4 | 6+7 2 | 3+4 2 |
| | 9 2 | 9 2 | 7 2 |
| | 11 4 | 6+7 2 | 5+6 2 |
| | 8 2 | 9 2 | 7 2 |
| | 11 4 | 8 2 | 3+4 2 |
| | 10 4 | 9 2 | 7 2 |
| | 6 2 | 6+7 2 | 5+6 2 |
| | 10 4 | 9 2 | 7 2 |
| | 7 2 | 4+5 2 | 5+6 2 |
| | | 8 2 | 3+4 2 |
| | | 9 2 | 7 2 |
| | | 8 2 | 3+4 2 |
| | | 8 2 | 7 2 |
| | | 4+5 2 | 7 2 |
| | | 8 2 | 5+6 2 |
| | | 6+7 2 | 5+6 2 |
| | | 9 2 | 7 2 |
| | | 4+5 2 | 3+4 2 |
| | | 8 2 | 7 2 |
| | | 9 2 | 7 2 |
| | | 6+7 2 | 3+4 2 |
| | | 8 2 | 5+6 2 |
| | | 9 2 | 7 2 |
| | | 9 2 | 5+6 2 |
| | | 6+7 2 | 7 2 |
| 227.1.40 | 171899 | 96 | Cr2O4Zn1 |
| | -GM -X | -L | -W |
| | 6 2 5 4 | 9 2 | 5+6 2 |
| | 9 2 5 4 | 8 2 | 7 2 |
| | 7 2 5 4 | 9 2 | 3+4 2 |
| | 10 4 5 4 | 9 2 | 7 2 |
| | 8 2 5 4 | 6+7 2 | 7 2 |
| | 11 4 5 4 | 8 2 | 3+4 2 |
| | 6 2 5 4 | 4+5 2 | 5+6 2 |
| | 7 2 5 4 | 8 2 | 7 2 |
| | 10 4 5 4 | 9 2 | 7 2 |
| | 11 4 5 4 | 8 2 | 5+6 2 |
| | 9 2 5 4 | 9 2 | 3+4 2 |
| | 10 4 5 4 | 6+7 2 | 7 2 |
| | 7 2 5 4 | 9 2 | 7 2 |
| | 11 4 5 4 | 6+7 2 | 5+6 2 |
| | 8 2 5 4 | 4+5 2 | 3+4 2 |
| | 10 4 5 4 | 8 2 | 7 2 |
| | 10 4 5 4 | 9 2 | 5+6 2 |
| | 11 4 5 4 | 4+5 2 | 7 2 |
| | 6 2 5 4 | 8 2 | 3+4 2 |
| | 10 4 5 4 | 8 2 | 7 2 |
| | 11 4 5 4 | 4+5 2 | 7 2 |
| | 8 2 5 4 | 8 2 | 5+6 2 |
| | 10 4 5 4 | 9 2 | 7 2 |
| | 7 2 5 4 | 6+7 2 | 3+4 2 |
| | 9 2 | 6+7 2 | 7 2 |

Continued on next column

**TABLE LXXXVIII– continued**

| TQC | ICSD | NoE | SF |
|---|---|---|---|
| | 11 4 | 9 2 | 5+6 2 |
| | 8 2 | 9 2 | 7 2 |
| | 11 4 | 8 2 | 3+4 2 |
| | 10 4 | 9 2 | 7 2 |
| | 6 2 | 6+7 2 | 5+6 2 |
| | 10 4 | 9 2 | 7 2 |
| | 7 2 | 9 2 | 5+6 2 |
| | | 4+5 2 | 3+4 2 |
| | | 8 2 | 7 2 |
| | | 8 2 | 3+4 2 |
| | | 8 2 | 7 2 |
| | | 4+5 2 | 5+6 2 |
| | | 8 2 | 7 2 |
| | | 9 2 | 5+6 2 |
| | | 6+7 2 | 7 2 |
| | | 8 2 | 3+4 2 |
| | | 4+5 2 | 7 2 |
| | | 9 2 | 7 2 |
| | | 6+7 2 | 3+4 2 |
| | | 8 2 | 5+6 2 |
| | | 9 2 | 7 2 |
| | | 6+7 2 | 3+4 2 |
| | | 9 2 | 7 2 |
| 227.1.41 | 171896 | 96 | Cr2O4Zn1 |
| | -GM -X | -L | -W |
| | 6 2 5 4 | 9 2 | 5+6 2 |
| | 9 2 5 4 | 8 2 | 7 2 |
| | 7 2 5 4 | 9 2 | 3+4 2 |
| | 10 4 5 4 | 9 2 | 7 2 |
| | 8 2 5 4 | 6+7 2 | 7 2 |
| | 11 4 5 4 | 8 2 | 3+4 2 |
| | 6 2 5 4 | 4+5 2 | 5+6 2 |
| | 7 2 5 4 | 8 2 | 7 2 |
| | 10 4 5 4 | 9 2 | 7 2 |
| | 11 4 5 4 | 8 2 | 5+6 2 |
| | 9 2 5 4 | 9 2 | 3+4 2 |
| | 10 4 5 4 | 6+7 2 | 7 2 |
| | 7 2 5 4 | 9 2 | 7 2 |
| | 11 4 5 4 | 6+7 2 | 5+6 2 |
| | 8 2 5 4 | 4+5 2 | 3+4 2 |
| | 10 4 5 4 | 8 2 | 7 2 |
| | 10 4 5 4 | 9 2 | 5+6 2 |
| | 11 4 5 4 | 4+5 2 | 3+4 2 |
| | 6 2 5 4 | 8 2 | 7 2 |
| | 10 4 5 4 | 8 2 | 7 2 |
| | 11 4 5 4 | 4+5 2 | 7 2 |
| | 8 2 5 4 | 8 2 | 5+6 2 |
| | 10 4 5 4 | 9 2 | 7 2 |
| | 7 2 5 4 | 6+7 2 | 3+4 2 |
| | 9 2 | 6+7 2 | 7 2 |
| | 11 4 | 9 2 | 5+6 2 |
| | 8 2 | 9 2 | 7 2 |
| | 11 4 | 8 2 | 3+4 2 |
| | 10 4 | 9 2 | 7 2 |
| | 6 2 | 6+7 2 | 5+6 2 |
| | 10 4 | 9 2 | 7 2 |
| | 7 2 | 9 2 | 5+6 2 |
| | | 4+5 2 | 3+4 2 |
| | | 8 2 | 7 2 |
| | | 8 2 | 3+4 2 |
| | | 8 2 | 7 2 |

Continued on next column

**TABLE LXXXVIII– continued**

| TQC | ICSD | | NoE | SF |
|---|---|---|---|---|
| | | | 4+5 2 | 5+6 2 |
| | | | 8 2 | 7 2 |
| | | | 6+7 2 | 5+6 2 |
| | | | 9 2 | 7 2 |
| | | | 8 2 | 3+4 2 |
| | | | 4+5 2 | 7 2 |
| | | | 9 2 | 7 2 |
| | | | 6+7 2 | 3+4 2 |
| | | | 8 2 | 5+6 2 |
| | | | 9 2 | 7 2 |
| | | | 6+7 2 | 3+4 2 |
| | | | 9 2 | 7 2 |
| 227.1.42 | 171894 | | 96 | Cr2O4Zn1 |
| | -GM | -X | -L | -W |
| | 6 2 | 5 4 | 9 2 | 5+6 2 |
| | 9 2 | 5 4 | 8 2 | 7 2 |
| | 7 2 | 5 4 | 9 2 | 3+4 2 |
| | 10 4 | 5 4 | 9 2 | 7 2 |
| | 8 2 | 5 4 | 6+7 2 | 7 2 |
| | 11 4 | 5 4 | 8 2 | 3+4 2 |
| | 6 2 | 5 4 | 4+5 2 | 5+6 2 |
| | 7 2 | 5 4 | 8 2 | 7 2 |
| | 10 4 | 5 4 | 9 2 | 7 2 |
| | 9 2 | 5 4 | 8 2 | 5+6 2 |
| | 11 4 | 5 4 | 9 2 | 3+4 2 |
| | 10 4 | 5 4 | 6+7 2 | 7 2 |
| | 7 2 | 5 4 | 9 2 | 7 2 |
| | 11 4 | 5 4 | 6+7 2 | 5+6 2 |
| | 8 2 | 5 4 | 4+5 2 | 3+4 2 |
| | 10 4 | 5 4 | 8 2 | 7 2 |
| | 10 4 | 5 4 | 9 2 | 5+6 2 |
| | 11 4 | 5 4 | 4+5 2 | 3+4 2 |
| | 6 2 | 5 4 | 8 2 | 7 2 |
| | 10 4 | 5 4 | 8 2 | 7 2 |
| | 11 4 | 5 4 | 4+5 2 | 7 2 |
| | 8 2 | 5 4 | 8 2 | 5+6 2 |
| | 10 4 | 5 4 | 9 2 | 7 2 |
| | 7 2 | 5 4 | 6+7 2 | 3+4 2 |
| | 9 2 | | 6+7 2 | 7 2 |
| | 11 4 | | 9 2 | 5+6 2 |
| | 8 2 | | 9 2 | 7 2 |
| | 11 4 | | 8 2 | 3+4 2 |
| | 10 4 | | 9 2 | 7 2 |
| | 6 2 | | 6+7 2 | 5+6 2 |
| | 10 4 | | 9 2 | 7 2 |
| | 7 2 | | 4+5 2 | 5+6 2 |
| | | | 8 2 | 3+4 2 |
| | | | 9 2 | 7 2 |
| | | | 8 2 | 3+4 2 |
| | | | 8 2 | 7 2 |
| | | | 4+5 2 | 5+6 2 |
| | | | 8 2 | 7 2 |
| | | | 6+7 2 | 5+6 2 |
| | | | 9 2 | 7 2 |
| | | | 8 2 | 3+4 2 |
| | | | 4+5 2 | 7 2 |
| | | | 9 2 | 7 2 |
| | | | 6+7 2 | 3+4 2 |
| | | | 8 2 | 5+6 2 |
| | | | 9 2 | 7 2 |
| | | | 9 2 | 3+4 2 |

Continued on next column

**TABLE LXXXVIII– continued**

| TQC | ICSD | | NoE | SF |
|---|---|---|---|---|
| | | | 6+7 2 | 7 2 |
| 227.1.43 | 171893 | | 96 | Cr2O4Zn1 |
| | -GM | -X | -L | -W |
| | 6 2 | 5 4 | 9 2 | 5+6 2 |
| | 9 2 | 5 4 | 8 2 | 7 2 |
| | 7 2 | 5 4 | 9 2 | 3+4 2 |
| | 10 4 | 5 4 | 9 2 | 7 2 |
| | 8 2 | 5 4 | 6+7 2 | 7 2 |
| | 11 4 | 5 4 | 8 2 | 3+4 2 |
| | 6 2 | 5 4 | 4+5 2 | 5+6 2 |
| | 7 2 | 5 4 | 8 2 | 7 2 |
| | 10 4 | 5 4 | 9 2 | 7 2 |
| | 9 2 | 5 4 | 8 2 | 5+6 2 |
| | 11 4 | 5 4 | 9 2 | 3+4 2 |
| | 10 4 | 5 4 | 6+7 2 | 7 2 |
| | 7 2 | 5 4 | 9 2 | 7 2 |
| | 11 4 | 5 4 | 6+7 2 | 5+6 2 |
| | 8 2 | 5 4 | 4+5 2 | 3+4 2 |
| | 10 4 | 5 4 | 8 2 | 7 2 |
| | 10 4 | 5 4 | 9 2 | 5+6 2 |
| | 11 4 | 5 4 | 4+5 2 | 3+4 2 |
| | 6 2 | 5 4 | 8 2 | 7 2 |
| | 10 4 | 5 4 | 8 2 | 7 2 |
| | 11 4 | 5 4 | 4+5 2 | 7 2 |
| | 8 2 | 5 4 | 8 2 | 5+6 2 |
| | 10 4 | 5 4 | 9 2 | 7 2 |
| | 7 2 | 5 4 | 6+7 2 | 3+4 2 |
| | 9 2 | | 6+7 2 | 7 2 |
| | 11 4 | | 9 2 | 5+6 2 |
| | 8 2 | | 9 2 | 7 2 |
| | 11 4 | | 8 2 | 3+4 2 |
| | 10 4 | | 9 2 | 7 2 |
| | 6 2 | | 6+7 2 | 5+6 2 |
| | 10 4 | | 9 2 | 7 2 |
| | 7 2 | | 4+5 2 | 5+6 2 |
| | | | 8 2 | 3+4 2 |
| | | | 9 2 | 7 2 |
| | | | 8 2 | 3+4 2 |
| | | | 8 2 | 7 2 |
| | | | 4+5 2 | 5+6 2 |
| | | | 8 2 | 7 2 |
| | | | 6+7 2 | 5+6 2 |
| | | | 9 2 | 7 2 |
| | | | 8 2 | 3+4 2 |
| | | | 4+5 2 | 7 2 |
| | | | 9 2 | 7 2 |
| | | | 6+7 2 | 3+4 2 |
| | | | 8 2 | 5+6 2 |
| | | | 9 2 | 7 2 |
| | | | 6+7 2 | 3+4 2 |
| | | | 9 2 | 7 2 |
| 227.1.44 | 97201 | | 76 | Cr2Mg1O4 |
| | -GM | -X | -L | -W |
| | 6 2 | 5 4 | 9 2 | 7 2 |
| | 7 2 | 5 4 | 8 2 | 5+6 2 |
| | 10 4 | 5 4 | 9 2 | 3+4 2 |
| | 9 2 | 5 4 | 9 2 | 7 2 |
| | 8 2 | 5 4 | 6+7 2 | 7 2 |
| | 11 4 | 5 4 | 8 2 | 3+4 2 |
| | 6 2 | 5 4 | 4+5 2 | 5+6 2 |
| | 7 2 | 5 4 | 8 2 | 7 2 |

Continued on next column

**TABLE LXXXVIII– continued**

| TQC | ICSD | NoE | SF |
|---|---|---|---|
| | 10 4 5 4 | 9 2 | 5+6 2 |
| | 9 2 5 4 | 8 2 | 7 2 |
| | 11 4 5 4 | 9 2 | 3+4 2 |
| | 10 4 5 4 | 6+7 2 | 7 2 |
| | 6 2 5 4 | 4+5 2 | 3+4 2 |
| | 10 4 5 4 | 8 2 | 7 2 |
| | 7 2 5 4 | 9 2 | 5+6 2 |
| | 10 4 5 4 | 9 2 | 7 2 |
| | 11 4 5 4 | 6+7 2 | 5+6 2 |
| | 8 2 5 4 | 8 2 | 7 2 |
| | 9 2 5 4 | 9 2 | 7 2 |
| | 11 4 | 4+5 2 | 3+4 2 |
| | 8 2 | 8 2 | 5+6 2 |
| | 11 4 | 9 2 | 7 2 |
| | 10 4 | 6+7 2 | 7 2 |
| | 6 2 | 9 2 | 3+4 2 |
| | 10 4 | 8 2 | 5+6 2 |
| | 7 2 | 4+5 2 | 7 2 |
| | | 8 2 | 3+4 2 |
| | | 9 2 | 7 2 |
| | | 6+7 2 | 7 2 |
| | | 8 2 | 5+6 2 |
| | | 8 2 | 3+4 2 |
| | | 4+5 2 | 7 2 |
| | | 9 2 | 7 2 |
| | | 6+7 2 | 3+4 2 |
| | | 9 2 | 5+6 2 |
| | | 8 2 | 7 2 |
| | | 6+7 2 | 5+6 2 |
| | | 9 2 | 7 2 |
| 227.1.45 | 75623 | 76 | Cr2Mg1O4 |
| | -GM -X | -L | -W |
| | 6 2 5 4 | 9 2 | 7 2 |
| | 7 2 5 4 | 8 2 | 5+6 2 |
| | 10 4 5 4 | 9 2 | 3+4 2 |
| | 9 2 5 4 | 9 2 | 7 2 |
| | 8 2 5 4 | 6+7 2 | 7 2 |
| | 11 4 5 4 | 8 2 | 3+4 2 |
| | 6 2 5 4 | 4+5 2 | 5+6 2 |
| | 7 2 5 4 | 8 2 | 7 2 |
| | 10 4 5 4 | 9 2 | 5+6 2 |
| | 9 2 5 4 | 8 2 | 7 2 |
| | 11 4 5 4 | 9 2 | 3+4 2 |
| | 10 4 5 4 | 6+7 2 | 7 2 |
| | 6 2 5 4 | 4+5 2 | 3+4 2 |
| | 10 4 5 4 | 8 2 | 7 2 |
| | 7 2 5 4 | 9 2 | 5+6 2 |
| | 10 4 5 4 | 9 2 | 7 2 |
| | 11 4 5 4 | 6+7 2 | 5+6 2 |
| | 8 2 5 4 | 9 2 | 7 2 |
| | 9 2 5 4 | 8 2 | 7 2 |
| | 11 4 | 6+7 2 | 3+4 2 |
| | 8 2 | 9 2 | 5+6 2 |
| | 11 4 | 4+5 2 | 7 2 |
| | 10 4 | 8 2 | 7 2 |
| | 6 2 | 9 2 | 3+4 2 |
| | 10 4 | 8 2 | 5+6 2 |
| | 7 2 | 4+5 2 | 7 2 |
| | | 8 2 | 3+4 2 |
| | | 9 2 | 7 2 |
| | | 6+7 2 | 7 2 |

Continued on next column

**TABLE LXXXVIII– continued**

| TQC | ICSD | NoE | SF |
|---|---|---|---|
| | | 8 2 | 5+6 2 |
| | | 8 2 | 3+4 2 |
| | | 4+5 2 | 7 2 |
| | | 9 2 | 7 2 |
| | | 6+7 2 | 3+4 2 |
| | | 9 2 | 5+6 2 |
| | | 8 2 | 7 2 |
| | | 6+7 2 | 5+6 2 |
| | | 9 2 | 7 2 |
| 227.1.46 | 290599 | 76 | Cr2Mg1O4 |
| | -GM -X | -L | -W |
| | 6 2 5 4 | 9 2 | 7 2 |
| | 7 2 5 4 | 8 2 | 5+6 2 |
| | 10 4 5 4 | 9 2 | 3+4 2 |
| | 9 2 5 4 | 9 2 | 7 2 |
| | 8 2 5 4 | 6+7 2 | 7 2 |
| | 11 4 5 4 | 8 2 | 3+4 2 |
| | 6 2 5 4 | 4+5 2 | 5+6 2 |
| | 7 2 5 4 | 8 2 | 7 2 |
| | 10 4 5 4 | 9 2 | 5+6 2 |
| | 9 2 5 4 | 8 2 | 7 2 |
| | 11 4 5 4 | 9 2 | 3+4 2 |
| | 10 4 5 4 | 6+7 2 | 7 2 |
| | 6 2 5 4 | 4+5 2 | 3+4 2 |
| | 10 4 5 4 | 8 2 | 7 2 |
| | 7 2 5 4 | 9 2 | 5+6 2 |
| | 10 4 5 4 | 9 2 | 7 2 |
| | 11 4 5 4 | 6+7 2 | 5+6 2 |
| | 8 2 5 4 | 9 2 | 7 2 |
| | 9 2 5 4 | 8 2 | 7 2 |
| | 11 4 | 6+7 2 | 3+4 2 |
| | 8 2 | 9 2 | 5+6 2 |
| | 11 4 | 4+5 2 | 7 2 |
| | 10 4 | 8 2 | 7 2 |
| | 6 2 | 9 2 | 3+4 2 |
| | 10 4 | 8 2 | 5+6 2 |
| | 7 2 | 4+5 2 | 7 2 |
| | | 8 2 | 3+4 2 |
| | | 9 2 | 7 2 |
| | | 6+7 2 | 7 2 |
| | | 8 2 | 5+6 2 |
| | | 8 2 | 3+4 2 |
| | | 4+5 2 | 7 2 |
| | | 9 2 | 7 2 |
| | | 6+7 2 | 3+4 2 |
| | | 9 2 | 5+6 2 |
| | | 8 2 | 7 2 |
| | | 6+7 2 | 5+6 2 |
| | | 9 2 | 7 2 |
| 227.1.47 | 171107 | 76 | Cr2Mg1O4 |
| | -GM -X | -L | -W |
| | 6 2 5 4 | 9 2 | 7 2 |
| | 7 2 5 4 | 8 2 | 5+6 2 |
| | 10 4 5 4 | 9 2 | 3+4 2 |
| | 9 2 5 4 | 9 2 | 7 2 |
| | 8 2 5 4 | 6+7 2 | 7 2 |
| | 11 4 5 4 | 8 2 | 3+4 2 |
| | 6 2 5 4 | 4+5 2 | 5+6 2 |
| | 7 2 5 4 | 8 2 | 7 2 |
| | 10 4 5 4 | 9 2 | 5+6 2 |
| | 9 2 5 4 | 8 2 | 7 2 |

Continued on next column

**TABLE LXXXVIII– continued**

| TQC | ICSD | NoE | SF |
|---|---|---|---|
| | 11 4 5 4 | 9 2 | 3+4 2 |
| | 10 4 5 4 | 6+7 2 | 7 2 |
| | 6 2 5 4 | 4+5 2 | 3+4 2 |
| | 10 4 5 4 | 8 2 | 7 2 |
| | 7 2 5 4 | 9 2 | 5+6 2 |
| | 10 4 5 4 | 9 2 | 7 2 |
| | 11 4 5 4 | 6+7 2 | 5+6 2 |
| | 8 2 5 4 | 9 2 | 7 2 |
| | 9 2 5 4 | 8 2 | 7 2 |
| | 11 4 | 6+7 2 | 3+4 2 |
| | 8 2 | 9 2 | 5+6 2 |
| | 11 4 | 4+5 2 | 7 2 |
| | 10 4 | 8 2 | 7 2 |
| | 6 2 | 9 2 | 3+4 2 |
| | 10 4 | 8 2 | 5+6 2 |
| | 7 2 | 4+5 2 | 7 2 |
| | | 8 2 | 3+4 2 |
| | | 9 2 | 7 2 |
| | | 6+7 2 | 7 2 |
| | | 8 2 | 5+6 2 |
| | | 8 2 | 3+4 2 |
| | | 4+5 2 | 7 2 |
| | | 9 2 | 7 2 |
| | | 6+7 2 | 3+4 2 |
| | | 9 2 | 5+6 2 |
| | | 8 2 | 7 2 |
| | | 6+7 2 | 5+6 2 |
| | | 9 2 | 7 2 |
| 227.1.48 | 171106 | 76 | Cr2Mg1O4 |
| | -GM -X | -L | -W |
| | 6 2 5 4 | 9 2 | 7 2 |
| | 7 2 5 4 | 8 2 | 5+6 2 |
| | 10 4 5 4 | 9 2 | 3+4 2 |
| | 9 2 5 4 | 9 2 | 7 2 |
| | 8 2 5 4 | 6+7 2 | 7 2 |
| | 11 4 5 4 | 8 2 | 3+4 2 |
| | 6 2 5 4 | 4+5 2 | 5+6 2 |
| | 7 2 5 4 | 8 2 | 7 2 |
| | 10 4 5 4 | 9 2 | 5+6 2 |
| | 9 2 5 4 | 8 2 | 7 2 |
| | 11 4 5 4 | 9 2 | 3+4 2 |
| | 10 4 5 4 | 6+7 2 | 7 2 |
| | 6 2 5 4 | 4+5 2 | 3+4 2 |
| | 10 4 5 4 | 8 2 | 7 2 |
| | 7 2 5 4 | 9 2 | 5+6 2 |
| | 10 4 5 4 | 9 2 | 7 2 |
| | 11 4 5 4 | 6+7 2 | 5+6 2 |
| | 8 2 5 4 | 9 2 | 7 2 |
| | 9 2 5 4 | 8 2 | 7 2 |
| | 11 4 | 6+7 2 | 3+4 2 |
| | 8 2 | 9 2 | 5+6 2 |
| | 11 4 | 4+5 2 | 7 2 |
| | 10 4 | 8 2 | 7 2 |
| | 6 2 | 9 2 | 3+4 2 |
| | 10 4 | 8 2 | 5+6 2 |
| | 7 2 | 4+5 2 | 7 2 |
| | | 8 2 | 3+4 2 |
| | | 9 2 | 7 2 |
| | | 6+7 2 | 7 2 |
| | | 8 2 | 5+6 2 |
| | | 8 2 | 3+4 2 |

Continued on next column

**TABLE LXXXVIII– continued**

| TQC | ICSD | NoE | SF |
|---|---|---|---|
| | | 4+5 2 | 7 2 |
| | | 9 2 | 7 2 |
| | | 6+7 2 | 3+4 2 |
| | | 9 2 | 5+6 2 |
| | | 8 2 | 7 2 |
| | | 6+7 2 | 5+6 2 |
| | | 9 2 | 7 2 |
| 227.1.49 | 160953 | 76 | Cr2Mg1O4 |
| | -GM -X | -L | -W |
| | 6 2 5 4 | 9 2 | 7 2 |
| | 7 2 5 4 | 8 2 | 5+6 2 |
| | 10 4 5 4 | 9 2 | 3+4 2 |
| | 9 2 5 4 | 9 2 | 7 2 |
| | 8 2 5 4 | 6+7 2 | 7 2 |
| | 11 4 5 4 | 8 2 | 3+4 2 |
| | 6 2 5 4 | 4+5 2 | 5+6 2 |
| | 7 2 5 4 | 8 2 | 7 2 |
| | 10 4 5 4 | 9 2 | 5+6 2 |
| | 9 2 5 4 | 8 2 | 7 2 |
| | 11 4 5 4 | 9 2 | 3+4 2 |
| | 10 4 5 4 | 6+7 2 | 7 2 |
| | 6 2 5 4 | 4+5 2 | 3+4 2 |
| | 10 4 5 4 | 8 2 | 7 2 |
| | 7 2 5 4 | 9 2 | 5+6 2 |
| | 10 4 5 4 | 9 2 | 7 2 |
| | 11 4 5 4 | 6+7 2 | 5+6 2 |
| | 8 2 5 4 | 9 2 | 7 2 |
| | 9 2 5 4 | 8 2 | 7 2 |
| | 11 4 | 6+7 2 | 3+4 2 |
| | 8 2 | 9 2 | 5+6 2 |
| | 11 4 | 4+5 2 | 7 2 |
| | 10 4 | 8 2 | 7 2 |
| | 6 2 | 9 2 | 3+4 2 |
| | 10 4 | 8 2 | 5+6 2 |
| | 7 2 | 4+5 2 | 7 2 |
| | | 8 2 | 3+4 2 |
| | | 9 2 | 7 2 |
| | | 6+7 2 | 7 2 |
| | | 8 2 | 5+6 2 |
| | | 8 2 | 3+4 2 |
| | | 4+5 2 | 7 2 |
| | | 9 2 | 7 2 |
| | | 6+7 2 | 3+4 2 |
| | | 9 2 | 5+6 2 |
| | | 8 2 | 7 2 |
| | | 6+7 2 | 5+6 2 |
| | | 9 2 | 7 2 |
| 227.1.50 | 402408 | 96 | Cr2Hg1Se4 |
| | -GM -X | -L | -W |
| | 6 2 5 4 | 9 2 | 5+6 2 |
| | 9 2 5 4 | 8 2 | 7 2 |
| | 8 2 5 4 | 8 2 | 3+4 2 |
| | 11 4 5 4 | 9 2 | 7 2 |
| | 7 2 5 4 | 6+7 2 | 5+6 2 |
| | 10 4 5 4 | 9 2 | 7 2 |
| | 10 4 5 4 | 8 2 | 7 2 |
| | 11 4 5 4 | 4+5 2 | 3+4 2 |
| | 6 2 5 4 | 9 2 | 7 2 |
| | 7 2 5 4 | 8 2 | 5+6 2 |
| | 11 4 5 4 | 4+5 2 | 3+4 2 |
| | 8 2 5 4 | 6+7 2 | 7 2 |

Continued on next column

**TABLE LXXXVIII– continued**

| TQC | ICSD | NoE | SF |
|---|---|---|---|
| | 10 4 5 4 | 9 2 | 5+6 2 |
| | 9 2 5 4 | 8 2 | 7 2 |
| | 7 2 5 4 | 6+7 2 | 7 2 |
| | 10 4 5 4 | 8 2 | 5+6 2 |
| | 11 4 5 4 | 9 2 | 3+4 2 |
| | 10 4 5 4 | 4+5 2 | 7 2 |
| | 6 2 5 4 | 9 2 | 7 2 |
| | 10 4 5 4 | 8 2 | 3+4 2 |
| | 11 4 5 4 | 9 2 | 7 2 |
| | 8 2 5 4 | 6+7 2 | 5+6 2 |
| | 7 2 5 4 | 4+5 2 | 3+4 2 |
| | 10 4 5 4 | 8 2 | 7 2 |
| | 9 2 | 9 2 | 7 2 |
| | 11 4 | 8 2 | 3+4 2 |
| | 6 2 | 6+7 2 | 5+6 2 |
| | 10 4 | 9 2 | 7 2 |
| | 8 2 | 8 2 | 7 2 |
| | 7 2 | 9 2 | 5+6 2 |
| | 10 4 | 8 2 | 5+6 2 |
| | 11 4 | 4+5 2 | 7 2 |
| | | 9 2 | 3+4 2 |
| | | 6+7 2 | 7 2 |
| | | 9 2 | 7 2 |
| | | 8 2 | 3+4 2 |
| | | 4+5 2 | 5+6 2 |
| | | 9 2 | 7 2 |
| | | 8 2 | 7 2 |
| | | 6+7 2 | 5+6 2 |
| | | 8 2 | 7 2 |
| | | 4+5 2 | 3+4 2 |
| | | 8 2 | 7 2 |
| | | 9 2 | 3+4 2 |
| | | 6+7 2 | 5+6 2 |
| | | 9 2 | 7 2 |
| | | 9 2 | 3+4 2 |
| | | 6+7 2 | 7 2 |
| 227.1.51 | 183965 | 88 | Cr2Fe1O4 |
| | -GM -X | -L | -W |
| | 6 2 5 4 | 9 2 | 5+6 2 |
| | 7 2 5 4 | 8 2 | 7 2 |
| | 10 4 5 4 | 9 2 | 3+4 2 |
| | 9 2 5 4 | 9 2 | 7 2 |
| | 8 2 5 4 | 6+7 2 | 7 2 |
| | 11 4 5 4 | 8 2 | 3+4 2 |
| | 6 2 5 4 | 4+5 2 | 5+6 2 |
| | 7 2 5 4 | 8 2 | 7 2 |
| | 10 4 5 4 | 9 2 | 5+6 2 |
| | 9 2 5 4 | 8 2 | 7 2 |
| | 11 4 5 4 | 9 2 | 3+4 2 |
| | 10 4 5 4 | 6+7 2 | 7 2 |
| | 7 2 5 4 | 9 2 | 3+4 2 |
| | 10 4 5 4 | 4+5 2 | 7 2 |
| | 6 2 5 4 | 8 2 | 5+6 2 |
| | 10 4 5 4 | 9 2 | 7 2 |
| | 11 4 5 4 | 6+7 2 | 5+6 2 |
| | 8 2 5 4 | 8 2 | 7 2 |
| | 9 2 5 4 | 4+5 2 | 7 2 |
| | 11 4 5 4 | 8 2 | 3+4 2 |
| | 8 2 5 4 | 9 2 | 5+6 2 |
| | 11 4 5 4 | 9 2 | 7 2 |
| | 10 4 | 6+7 2 | 7 2 |

Continued on next column

**TABLE LXXXVIII– continued**

| TQC | ICSD | NoE | SF |
|---|---|---|---|
| | 10 4 | 8 2 | 3+4 2 |
| | 6 2 | 9 2 | 5+6 2 |
| | 7 2 | 4+5 2 | 7 2 |
| | 10 4 | 8 2 | 3+4 2 |
| | 11 4 | 9 2 | 7 2 |
| | 10 4 | 6+7 2 | 7 2 |
| | | 8 2 | 3+4 2 |
| | | 4+5 2 | 5+6 2 |
| | | 8 2 | 7 2 |
| | | 9 2 | 7 2 |
| | | 6+7 2 | 5+6 2 |
| | | 9 2 | 7 2 |
| | | 8 2 | 3+4 2 |
| | | 4+5 2 | 7 2 |
| | | 9 2 | 5+6 2 |
| | | 6+7 2 | 3+4 2 |
| | | 8 2 | 7 2 |
| | | 6+7 2 | 7 2 |
| | | 9 2 | 3+4 2 |
| | | 4+5 2 | 7 2 |
| | | 8 2 | 5+6 2 |
| 227.1.52 | 183964 | 88 | Cr2Fe1O4 |
| | -GM -X | -L | -W |
| | 6 2 5 4 | 9 2 | 5+6 2 |
| | 7 2 5 4 | 8 2 | 7 2 |
| | 10 4 5 4 | 9 2 | 3+4 2 |
| | 9 2 5 4 | 9 2 | 7 2 |
| | 8 2 5 4 | 6+7 2 | 7 2 |
| | 11 4 5 4 | 8 2 | 3+4 2 |
| | 6 2 5 4 | 4+5 2 | 5+6 2 |
| | 7 2 5 4 | 8 2 | 7 2 |
| | 10 4 5 4 | 9 2 | 5+6 2 |
| | 9 2 5 4 | 8 2 | 7 2 |
| | 11 4 5 4 | 9 2 | 3+4 2 |
| | 10 4 5 4 | 6+7 2 | 7 2 |
| | 7 2 5 4 | 9 2 | 3+4 2 |
| | 10 4 5 4 | 4+5 2 | 7 2 |
| | 6 2 5 4 | 8 2 | 5+6 2 |
| | 10 4 5 4 | 9 2 | 7 2 |
| | 11 4 5 4 | 6+7 2 | 5+6 2 |
| | 8 2 5 4 | 8 2 | 7 2 |
| | 9 2 5 4 | 8 2 | 7 2 |
| | 11 4 5 4 | 4+5 2 | 3+4 2 |
| | 8 2 5 4 | 9 2 | 5+6 2 |
| | 11 4 5 4 | 9 2 | 7 2 |
| | 10 4 | 6+7 2 | 7 2 |
| | 10 4 | 8 2 | 3+4 2 |
| | 7 2 | 9 2 | 5+6 2 |
| | 6 2 | 4+5 2 | 7 2 |
| | 10 4 | 8 2 | 3+4 2 |
| | 11 4 | 9 2 | 7 2 |
| | 10 4 | 6+7 2 | 7 2 |
| | | 8 2 | 3+4 2 |
| | | 4+5 2 | 5+6 2 |
| | | 8 2 | 7 2 |
| | | 9 2 | 7 2 |
| | | 6+7 2 | 5+6 2 |
| | | 9 2 | 7 2 |
| | | 8 2 | 3+4 2 |
| | | 4+5 2 | 7 2 |
| | | 9 2 | 5+6 2 |

Continued on next column

**TABLE LXXXVIII– continued**

| TQC | ICSD | | NoE | SF |
|---|---|---|---|---|
| | | | 6+7 2 | 3+4 2 |
| | | | 8 2 | 7 2 |
| | | | 6+7 2 | 7 2 |
| | | | 9 2 | 3+4 2 |
| | | | 4+5 2 | 7 2 |
| | | | 8 2 | 5+6 2 |
| 227.1.53 | 183963 | | 88 | Cr2Fe1O4 |
| | -GM | -X | -L | -W |
| | 6 2 | 5 4 | 9 2 | 5+6 2 |
| | 7 2 | 5 4 | 8 2 | 7 2 |
| | 10 4 | 5 4 | 9 2 | 3+4 2 |
| | 9 2 | 5 4 | 9 2 | 7 2 |
| | 8 2 | 5 4 | 6+7 2 | 7 2 |
| | 11 4 | 5 4 | 8 2 | 3+4 2 |
| | 6 2 | 5 4 | 4+5 2 | 5+6 2 |
| | 7 2 | 5 4 | 8 2 | 7 2 |
| | 10 4 | 5 4 | 9 2 | 5+6 2 |
| | 9 2 | 5 4 | 8 2 | 7 2 |
| | 11 4 | 5 4 | 9 2 | 3+4 2 |
| | 10 4 | 5 4 | 6+7 2 | 7 2 |
| | 7 2 | 5 4 | 9 2 | 3+4 2 |
| | 10 4 | 5 4 | 4+5 2 | 7 2 |
| | 6 2 | 5 4 | 8 2 | 5+6 2 |
| | 10 4 | 5 4 | 9 2 | 7 2 |
| | 11 4 | 5 4 | 6+7 2 | 5+6 2 |
| | 8 2 | 5 4 | 8 2 | 7 2 |
| | 9 2 | 5 4 | 8 2 | 7 2 |
| | 11 4 | 5 4 | 4+5 2 | 3+4 2 |
| | 8 2 | 5 4 | 9 2 | 5+6 2 |
| | 11 4 | 5 4 | 9 2 | 7 2 |
| | 10 4 | | 6+7 2 | 7 2 |
| | 10 4 | | 8 2 | 3+4 2 |
| | 7 2 | | 9 2 | 5+6 2 |
| | 6 2 | | 4+5 2 | 7 2 |
| | 10 4 | | 8 2 | 3+4 2 |
| | 11 4 | | 9 2 | 7 2 |
| | 10 4 | | 6+7 2 | 7 2 |
| | | | 8 2 | 3+4 2 |
| | | | 4+5 2 | 5+6 2 |
| | | | 8 2 | 7 2 |
| | | | 9 2 | 7 2 |
| | | | 6+7 2 | 5+6 2 |
| | | | 9 2 | 7 2 |
| | | | 8 2 | 3+4 2 |
| | | | 4+5 2 | 7 2 |
| | | | 9 2 | 5+6 2 |
| | | | 6+7 2 | 3+4 2 |
| | | | 8 2 | 7 2 |
| | | | 6+7 2 | 7 2 |
| | | | 9 2 | 3+4 2 |
| | | | 4+5 2 | 7 2 |
| | | | 8 2 | 5+6 2 |
| 227.1.54 | 183962 | | 88 | Cr2Fe1O4 |
| | -GM | -X | -L | -W |
| | 6 2 | 5 4 | 9 2 | 5+6 2 |
| | 7 2 | 5 4 | 8 2 | 7 2 |
| | 10 4 | 5 4 | 9 2 | 3+4 2 |
| | 9 2 | 5 4 | 9 2 | 7 2 |
| | 8 2 | 5 4 | 6+7 2 | 7 2 |
| | 11 4 | 5 4 | 8 2 | 3+4 2 |
| | 6 2 | 5 4 | 4+5 2 | 5+6 2 |

Continued on next column

**TABLE LXXXVIII– continued**

| TQC | ICSD | | NoE | SF |
|---|---|---|---|---|
| | 7 2 | 5 4 | 8 2 | 7 2 |
| | 10 4 | 5 4 | 9 2 | 5+6 2 |
| | 9 2 | 5 4 | 8 2 | 7 2 |
| | 11 4 | 5 4 | 9 2 | 3+4 2 |
| | 10 4 | 5 4 | 6+7 2 | 7 2 |
| | 7 2 | 5 4 | 9 2 | 3+4 2 |
| | 10 4 | 5 4 | 4+5 2 | 7 2 |
| | 6 2 | 5 4 | 8 2 | 5+6 2 |
| | 10 4 | 5 4 | 9 2 | 7 2 |
| | 11 4 | 5 4 | 6+7 2 | 5+6 2 |
| | 8 2 | 5 4 | 8 2 | 7 2 |
| | 9 2 | 5 4 | 8 2 | 7 2 |
| | 11 4 | 5 4 | 4+5 2 | 3+4 2 |
| | 8 2 | 5 4 | 9 2 | 5+6 2 |
| | 11 4 | 5 4 | 9 2 | 7 2 |
| | 10 4 | | 6+7 2 | 7 2 |
| | 10 4 | | 8 2 | 3+4 2 |
| | 7 2 | | 9 2 | 5+6 2 |
| | 6 2 | | 4+5 2 | 7 2 |
| | 10 4 | | 8 2 | 3+4 2 |
| | 11 4 | | 9 2 | 7 2 |
| | 10 4 | | 6+7 2 | 7 2 |
| | | | 8 2 | 3+4 2 |
| | | | 4+5 2 | 5+6 2 |
| | | | 8 2 | 7 2 |
| | | | 9 2 | 7 2 |
| | | | 6+7 2 | 5+6 2 |
| | | | 9 2 | 7 2 |
| | | | 8 2 | 3+4 2 |
| | | | 4+5 2 | 7 2 |
| | | | 9 2 | 5+6 2 |
| | | | 6+7 2 | 3+4 2 |
| | | | 8 2 | 7 2 |
| | | | 6+7 2 | 7 2 |
| | | | 9 2 | 3+4 2 |
| | | | 4+5 2 | 7 2 |
| | | | 8 2 | 5+6 2 |
| 227.1.55 | 183961 | | 88 | Cr2Fe1O4 |
| | -GM | -X | -L | -W |
| | 6 2 | 5 4 | 9 2 | 5+6 2 |
| | 7 2 | 5 4 | 8 2 | 7 2 |
| | 10 4 | 5 4 | 9 2 | 3+4 2 |
| | 9 2 | 5 4 | 9 2 | 7 2 |
| | 8 2 | 5 4 | 6+7 2 | 7 2 |
| | 11 4 | 5 4 | 8 2 | 3+4 2 |
| | 6 2 | 5 4 | 4+5 2 | 5+6 2 |
| | 7 2 | 5 4 | 8 2 | 7 2 |
| | 10 4 | 5 4 | 9 2 | 5+6 2 |
| | 9 2 | 5 4 | 8 2 | 7 2 |
| | 11 4 | 5 4 | 9 2 | 3+4 2 |
| | 10 4 | 5 4 | 6+7 2 | 7 2 |
| | 7 2 | 5 4 | 9 2 | 3+4 2 |
| | 10 4 | 5 4 | 4+5 2 | 7 2 |
| | 6 2 | 5 4 | 8 2 | 5+6 2 |
| | 10 4 | 5 4 | 9 2 | 7 2 |
| | 11 4 | 5 4 | 6+7 2 | 5+6 2 |
| | 8 2 | 5 4 | 8 2 | 7 2 |
| | 9 2 | 5 4 | 8 2 | 7 2 |
| | 11 4 | 5 4 | 4+5 2 | 3+4 2 |
| | 8 2 | 5 4 | 9 2 | 5+6 2 |
| | 11 4 | 5 4 | 9 2 | 7 2 |

Continued on next column

**TABLE LXXXVIII– continued**

| TQC | ICSD | NoE | SF |
|---|---|---|---|
| | 10 4 | 6+7 2 | 7 2 |
| | 10 4 | 8 2 | 3+4 2 |
| | 7 2 | 9 2 | 5+6 2 |
| | 6 2 | 4+5 2 | 7 2 |
| | 10 4 | 8 2 | 3+4 2 |
| | 11 4 | 9 2 | 7 2 |
| | 10 4 | 6+7 2 | 7 2 |
| | | 8 2 | 3+4 2 |
| | | 8 2 | 5+6 2 |
| | | 4+5 2 | 7 2 |
| | | 9 2 | 7 2 |
| | | 6+7 2 | 5+6 2 |
| | | 9 2 | 7 2 |
| | | 8 2 | 3+4 2 |
| | | 4+5 2 | 7 2 |
| | | 9 2 | 5+6 2 |
| | | 6+7 2 | 3+4 2 |
| | | 8 2 | 7 2 |
| | | 6+7 2 | 7 2 |
| | | 9 2 | 3+4 2 |
| | | 4+5 2 | 7 2 |
| | | 8 2 | 5+6 2 |
| 227.1.56 | 183960 | 88 | Cr2Fe1O4 |
| | -GM -X | -L | -W |
| | 6 2 5 4 | 9 2 | 5+6 2 |
| | 9 2 5 4 | 8 2 | 7 2 |
| | 7 2 5 4 | 9 2 | 3+4 2 |
| | 10 4 5 4 | 9 2 | 7 2 |
| | 8 2 5 4 | 6+7 2 | 7 2 |
| | 11 4 5 4 | 8 2 | 3+4 2 |
| | 6 2 5 4 | 4+5 2 | 5+6 2 |
| | 7 2 5 4 | 8 2 | 7 2 |
| | 10 4 5 4 | 9 2 | 5+6 2 |
| | 9 2 5 4 | 8 2 | 7 2 |
| | 11 4 5 4 | 9 2 | 3+4 2 |
| | 10 4 5 4 | 6+7 2 | 7 2 |
| | 7 2 5 4 | 9 2 | 3+4 2 |
| | 10 4 5 4 | 4+5 2 | 7 2 |
| | 6 2 5 4 | 8 2 | 5+6 2 |
| | 10 4 5 4 | 9 2 | 7 2 |
| | 11 4 5 4 | 6+7 2 | 5+6 2 |
| | 8 2 5 4 | 8 2 | 7 2 |
| | 9 2 5 4 | 8 2 | 7 2 |
| | 11 4 5 4 | 4+5 2 | 3+4 2 |
| | 8 2 5 4 | 9 2 | 5+6 2 |
| | 11 4 5 4 | 9 2 | 7 2 |
| | 10 4 | 6+7 2 | 7 2 |
| | 10 4 | 8 2 | 3+4 2 |
| | 7 2 | 9 2 | 5+6 2 |
| | 6 2 | 4+5 2 | 7 2 |
| | 10 4 | 8 2 | 3+4 2 |
| | 11 4 | 9 2 | 7 2 |
| | 10 4 | 6+7 2 | 7 2 |
| | | 8 2 | 3+4 2 |
| | | 8 2 | 5+6 2 |
| | | 4+5 2 | 7 2 |
| | | 9 2 | 7 2 |
| | | 6+7 2 | 5+6 2 |
| | | 9 2 | 7 2 |
| | | 8 2 | 3+4 2 |
| | | 4+5 2 | 7 2 |

Continued on next column

| TQC | ICSD | NoE | SF |
|---|---|---|---|
| | | 9 2 | 5+6 2 |
| | | 6+7 2 | 3+4 2 |
| | | 8 2 | 7 2 |
| | | 6+7 2 | 7 2 |
| | | 9 2 | 3+4 2 |
| | | 4+5 2 | 7 2 |
| | | 8 2 | 5+6 2 |
| 227.1.57 | 171121 | 88 | Cr2Fe1O4 |
| | -GM -X | -L | -W |
| | 6 2 5 4 | 9 2 | 5+6 2 |
| | 9 2 5 4 | 8 2 | 7 2 |
| | 7 2 5 4 | 9 2 | 3+4 2 |
| | 10 4 5 4 | 9 2 | 7 2 |
| | 8 2 5 4 | 6+7 2 | 7 2 |
| | 11 4 5 4 | 8 2 | 3+4 2 |
| | 6 2 5 4 | 4+5 2 | 5+6 2 |
| | 7 2 5 4 | 8 2 | 7 2 |
| | 10 4 5 4 | 9 2 | 5+6 2 |
| | 9 2 5 4 | 8 2 | 7 2 |
| | 11 4 5 4 | 9 2 | 3+4 2 |
| | 10 4 5 4 | 6+7 2 | 7 2 |
| | 7 2 5 4 | 9 2 | 3+4 2 |
| | 10 4 5 4 | 4+5 2 | 7 2 |
| | 6 2 5 4 | 8 2 | 5+6 2 |
| | 10 4 5 4 | 9 2 | 7 2 |
| | 11 4 5 4 | 6+7 2 | 5+6 2 |
| | 8 2 5 4 | 8 2 | 7 2 |
| | 9 2 5 4 | 8 2 | 7 2 |
| | 11 4 5 4 | 4+5 2 | 3+4 2 |
| | 8 2 5 4 | 9 2 | 5+6 2 |
| | 11 4 5 4 | 9 2 | 7 2 |
| | 10 4 | 6+7 2 | 7 2 |
| | 10 4 | 8 2 | 3+4 2 |
| | 7 2 | 9 2 | 5+6 2 |
| | 6 2 | 4+5 2 | 7 2 |
| | 10 4 | 8 2 | 3+4 2 |
| | 11 4 | 9 2 | 7 2 |
| | 10 4 | 6+7 2 | 7 2 |
| | | 8 2 | 3+4 2 |
| | | 8 2 | 5+6 2 |
| | | 4+5 2 | 7 2 |
| | | 9 2 | 7 2 |
| | | 6+7 2 | 5+6 2 |
| | | 9 2 | 7 2 |
| | | 8 2 | 3+4 2 |
| | | 4+5 2 | 7 2 |
| | | 9 2 | 5+6 2 |
| | | 6+7 2 | 3+4 2 |
| | | 8 2 | 7 2 |
| | | 6+7 2 | 7 2 |
| | | 9 2 | 3+4 2 |
| | | 4+5 2 | 7 2 |
| | | 8 2 | 5+6 2 |
| 227.1.58 | 859 | 92 | Co2O4Si1 |
| | -GM -X | -L | -W |
| | 6 2 5 4 | 8 2 | 3+4 2 |
| | 9 2 5 4 | 9 2 | 7 2 |
| | 8 2 5 4 | 9 2 | 5+6 2 |
| | 11 4 5 4 | 9 2 | 7 2 |
| | 7 2 5 4 | 6+7 2 | 7 2 |
| | 10 4 5 4 | 4+5 2 | 3+4 2 |

Continued on next column

**TABLE LXXXVIII– continued**

| TQC | ICSD | NoE | SF |
|---|---|---|---|
| | 6 2 5 4 | 8 2 | 5+6 2 |
| | 9 2 5 4 | 8 2 | 7 2 |
| | 7 2 5 4 | 9 2 | 3+4 2 |
| | 10 4 5 4 | 8 2 | 7 2 |
| | 11 4 5 4 | 9 2 | 5+6 2 |
| | 11 4 5 4 | 4+5 2 | 7 2 |
| | 8 2 5 4 | 8 2 | 5+6 2 |
| | 7 2 5 4 | 8 2 | 7 2 |
| | 10 4 5 4 | 6+7 2 | 7 2 |
| | 10 4 5 4 | 9 2 | 3+4 2 |
| | 6 2 5 4 | 9 2 | 5+6 2 |
| | 10 4 5 4 | 6+7 2 | 7 2 |
| | 9 2 5 4 | 9 2 | 3+4 2 |
| | 11 4 5 4 | 8 2 | 7 2 |
| | 8 2 5 4 | 4+5 2 | 7 2 |
| | 11 4 5 4 | 8 2 | 5+6 2 |
| | 6 2 5 4 | 4+5 2 | 7 2 |
| | 10 4 | 8 2 | 3+4 2 |
| | 10 4 | 9 2 | 5+6 2 |
| | 7 2 | 9 2 | 7 2 |
| | 10 4 | 6+7 2 | 3+4 2 |
| | 6 2 | 9 2 | 7 2 |
| | 7 2 | 6+7 2 | 7 2 |
| | 10 4 | 8 2 | 5+6 2 |
| | 10 4 | 8 2 | 7 2 |
| | | 4+5 2 | 3+4 2 |
| | | 9 2 | 5+6 2 |
| | | 6+7 2 | 7 2 |
| | | 9 2 | 7 2 |
| | | 9 2 | 3+4 2 |
| | | 8 2 | 5+6 2 |
| | | 4+5 2 | 7 2 |
| | | 8 2 | 3+4 2 |
| | | 6+7 2 | 7 2 |
| | | 9 2 | 7 2 |
| | | 9 2 | 5+6 2 |
| | | 9 2 | 3+4 2 |
| | | 6+7 2 | 7 2 |
| | | 6+7 2 | 3+4 2 |
| | | 9 2 | 7 2 |
| 227.1.59 | 154730 | 64 | Co2Nd1 |
| | -GM -X | -L | -W |
| | 6 2 5 4 | 8 2 | 3+4 2 |
| | 9 2 5 4 | 9 2 | 7 2 |
| | 7 2 5 4 | 8 2 | 5+6 2 |
| | 8 2 5 4 | 9 2 | 7 2 |
| | 10 4 5 4 | 4+5 2 | 7 2 |
| | 11 4 5 4 | 9 2 | 3+4 2 |
| | 6 2 5 4 | 8 2 | 5+6 2 |
| | 7 2 5 4 | 6+7 2 | 7 2 |
| | 10 4 5 4 | 9 2 | 5+6 2 |
| | 6 2 5 4 | 8 2 | 7 2 |
| | 10 4 5 4 | 9 2 | 7 2 |
| | 7 2 5 4 | 6+7 2 | 3+4 2 |
| | 10 4 5 4 | 9 2 | 7 2 |
| | 11 4 5 4 | 9 2 | 3+4 2 |
| | 10 4 5 4 | 6+7 2 | 5+6 2 |
| | 7 2 5 4 | 9 2 | 7 2 |
| | 10 4 | 9 2 | 7 2 |
| | 6 2 | 8 2 | 3+4 2 |
| | 10 4 | 9 2 | 5+6 2 |

Continued on next column

**TABLE LXXXVIII– continued**

| TQC | ICSD | NoE | SF |
|---|---|---|---|
| | 9 2 | 4+5 2 | 7 2 |
| | 11 4 | 6+7 2 | 5+6 2 |
| | 6 2 | 8 2 | 7 2 |
| | | 4+5 2 | 3+4 2 |
| | | 8 2 | 7 2 |
| | | 9 2 | 7 2 |
| | | 6+7 2 | 7 2 |
| | | 9 2 | 5+6 2 |
| | | 9 2 | 3+4 2 |
| | | 8 2 | 5+6 2 |
| | | 6+7 2 | 7 2 |
| | | 4+5 2 | 3+4 2 |
| | | 8 2 | 7 2 |
| 227.1.60 | 154729 | 64 | Co2Nd1 |
| | -GM -X | -L | -W |
| | 6 2 5 4 | 8 2 | 3+4 2 |
| | 9 2 5 4 | 9 2 | 7 2 |
| | 7 2 5 4 | 8 2 | 5+6 2 |
| | 8 2 5 4 | 9 2 | 7 2 |
| | 10 4 5 4 | 4+5 2 | 7 2 |
| | 11 4 5 4 | 9 2 | 3+4 2 |
| | 6 2 5 4 | 8 2 | 5+6 2 |
| | 7 2 5 4 | 6+7 2 | 7 2 |
| | 10 4 5 4 | 9 2 | 5+6 2 |
| | 6 2 5 4 | 8 2 | 7 2 |
| | 10 4 5 4 | 9 2 | 7 2 |
| | 7 2 5 4 | 6+7 2 | 3+4 2 |
| | 10 4 5 4 | 9 2 | 7 2 |
| | 11 4 5 4 | 9 2 | 3+4 2 |
| | 10 4 5 4 | 6+7 2 | 5+6 2 |
| | 7 2 5 4 | 9 2 | 7 2 |
| | 10 4 | 9 2 | 7 2 |
| | 6 2 | 8 2 | 3+4 2 |
| | 10 4 | 9 2 | 5+6 2 |
| | 9 2 | 4+5 2 | 7 2 |
| | 11 4 | 6+7 2 | 5+6 2 |
| | 6 2 | 8 2 | 7 2 |
| | | 4+5 2 | 3+4 2 |
| | | 8 2 | 7 2 |
| | | 9 2 | 7 2 |
| | | 6+7 2 | 7 2 |
| | | 9 2 | 5+6 2 |
| | | 9 2 | 3+4 2 |
| | | 8 2 | 5+6 2 |
| | | 6+7 2 | 7 2 |
| | | 4+5 2 | 3+4 2 |
| | | 8 2 | 7 2 |
| 227.1.61 | 154728 | 64 | Co2Nd1 |
| | -GM -X | -L | -W |
| | 6 2 5 4 | 9 2 | 5+6 2 |
| | 9 2 5 4 | 8 2 | 7 2 |
| | 7 2 5 4 | 9 2 | 3+4 2 |
| | 8 2 5 4 | 8 2 | 7 2 |
| | 10 4 5 4 | 6+7 2 | 7 2 |
| | 11 4 5 4 | 8 2 | 5+6 2 |
| | 6 2 5 4 | 9 2 | 3+4 2 |
| | 7 2 5 4 | 4+5 2 | 7 2 |
| | 10 4 5 4 | 8 2 | 3+4 2 |
| | 6 2 5 4 | 9 2 | 7 2 |
| | 10 4 5 4 | 8 2 | 7 2 |
| | 7 2 5 4 | 4+5 2 | 5+6 2 |

Continued on next column

**TABLE LXXXVIII– continued**

| TQC | ICSD | NoE | SF |
|---|---|---|---|
| | 10 4 5 4 | 8 2 | 7 2 |
| | 11 4 5 4 | 8 2 | 5+6 2 |
| | 10 4 5 4 | 4+5 2 | 3+4 2 |
| | 7 2 5 4 | 8 2 | 7 2 |
| | 10 4 | 8 2 | 7 2 |
| | 6 2 | 9 2 | 5+6 2 |
| | 10 4 | 8 2 | 3+4 2 |
| | 9 2 | 6+7 2 | 7 2 |
| | 11 4 | 4+5 2 | 3+4 2 |
| | 6 2 | 9 2 | 7 2 |
| | | 6+7 2 | 5+6 2 |
| | | 9 2 | 7 2 |
| | | 8 2 | 7 2 |
| | | 4+5 2 | 7 2 |
| | | 8 2 | 3+4 2 |
| | | 8 2 | 5+6 2 |
| | | 9 2 | 3+4 2 |
| | | 4+5 2 | 7 2 |
| | | 6+7 2 | 5+6 2 |
| | | 9 2 | 7 2 |
| 227.1.62 | 154727 | 64 | Co2Nd1 |
| | -GM -X | -L | -W |
| | 6 2 5 4 | 9 2 | 5+6 2 |
| | 9 2 5 4 | 8 2 | 7 2 |
| | 7 2 5 4 | 9 2 | 3+4 2 |
| | 8 2 5 4 | 8 2 | 7 2 |
| | 10 4 5 4 | 6+7 2 | 7 2 |
| | 11 4 5 4 | 8 2 | 5+6 2 |
| | 6 2 5 4 | 9 2 | 3+4 2 |
| | 7 2 5 4 | 4+5 2 | 7 2 |
| | 10 4 5 4 | 8 2 | 3+4 2 |
| | 6 2 5 4 | 9 2 | 7 2 |
| | 10 4 5 4 | 8 2 | 7 2 |
| | 7 2 5 4 | 4+5 2 | 5+6 2 |
| | 10 4 5 4 | 8 2 | 7 2 |
| | 11 4 5 4 | 8 2 | 5+6 2 |
| | 10 4 5 4 | 4+5 2 | 3+4 2 |
| | 7 2 5 4 | 8 2 | 7 2 |
| | 10 4 | 8 2 | 7 2 |
| | 6 2 | 9 2 | 5+6 2 |
| | 10 4 | 8 2 | 3+4 2 |
| | 9 2 | 6+7 2 | 7 2 |
| | 11 4 | 4+5 2 | 3+4 2 |
| | 6 2 | 9 2 | 7 2 |
| | | 6+7 2 | 5+6 2 |
| | | 9 2 | 7 2 |
| | | 8 2 | 7 2 |
| | | 4+5 2 | 7 2 |
| | | 8 2 | 3+4 2 |
| | | 8 2 | 5+6 2 |
| | | 9 2 | 3+4 2 |
| | | 4+5 2 | 7 2 |
| | | 6+7 2 | 5+6 2 |
| | | 9 2 | 7 2 |
| 227.1.63 | 154726 | 64 | Co2Nd1 |
| | -GM -X | -L | -W |
| | 6 2 5 4 | 8 2 | 3+4 2 |
| | 9 2 5 4 | 9 2 | 7 2 |
| | 7 2 5 4 | 8 2 | 5+6 2 |
| | 8 2 5 4 | 9 2 | 7 2 |
| | 10 4 5 4 | 4+5 2 | 7 2 |

Continued on next column

**TABLE LXXXVIII– continued**

| TQC | ICSD | NoE | SF |
|---|---|---|---|
| | 11 4 5 4 | 9 2 | 3+4 2 |
| | 6 2 5 4 | 8 2 | 5+6 2 |
| | 7 2 5 4 | 6+7 2 | 7 2 |
| | 10 4 5 4 | 9 2 | 5+6 2 |
| | 6 2 5 4 | 8 2 | 7 2 |
| | 10 4 5 4 | 9 2 | 7 2 |
| | 7 2 5 4 | 6+7 2 | 3+4 2 |
| | 10 4 5 4 | 9 2 | 7 2 |
| | 11 4 5 4 | 9 2 | 3+4 2 |
| | 10 4 5 4 | 6+7 2 | 5+6 2 |
| | 7 2 5 4 | 9 2 | 7 2 |
| | 10 4 | 9 2 | 7 2 |
| | 6 2 | 8 2 | 3+4 2 |
| | 10 4 | 9 2 | 5+6 2 |
| | 9 2 | 4+5 2 | 7 2 |
| | 11 4 | 6+7 2 | 5+6 2 |
| | 6 2 | 8 2 | 7 2 |
| | | 4+5 2 | 3+4 2 |
| | | 8 2 | 7 2 |
| | | 9 2 | 7 2 |
| | | 6+7 2 | 7 2 |
| | | 9 2 | 5+6 2 |
| | | 9 2 | 3+4 2 |
| | | 8 2 | 5+6 2 |
| | | 6+7 2 | 7 2 |
| | | 4+5 2 | 3+4 2 |
| | | 8 2 | 7 2 |
| 227.1.64 | 154725 | 64 | Co2Nd1 |
| | -GM -X | -L | -W |
| | 6 2 5 4 | 9 2 | 5+6 2 |
| | 9 2 5 4 | 8 2 | 7 2 |
| | 7 2 5 4 | 9 2 | 3+4 2 |
| | 8 2 5 4 | 8 2 | 7 2 |
| | 10 4 5 4 | 6+7 2 | 7 2 |
| | 11 4 5 4 | 8 2 | 5+6 2 |
| | 6 2 5 4 | 9 2 | 3+4 2 |
| | 7 2 5 4 | 4+5 2 | 7 2 |
| | 10 4 5 4 | 8 2 | 3+4 2 |
| | 6 2 5 4 | 9 2 | 7 2 |
| | 10 4 5 4 | 8 2 | 7 2 |
| | 7 2 5 4 | 4+5 2 | 5+6 2 |
| | 10 4 5 4 | 8 2 | 7 2 |
| | 11 4 5 4 | 8 2 | 5+6 2 |
| | 10 4 5 4 | 4+5 2 | 3+4 2 |
| | 7 2 5 4 | 8 2 | 7 2 |
| | 10 4 | 8 2 | 7 2 |
| | 6 2 | 9 2 | 5+6 2 |
| | 10 4 | 8 2 | 3+4 2 |
| | 9 2 | 6+7 2 | 7 2 |
| | 11 4 | 4+5 2 | 3+4 2 |
| | 6 2 | 9 2 | 7 2 |
| | | 6+7 2 | 5+6 2 |
| | | 9 2 | 7 2 |
| | | 8 2 | 7 2 |
| | | 4+5 2 | 7 2 |
| | | 8 2 | 3+4 2 |
| | | 8 2 | 5+6 2 |
| | | 9 2 | 3+4 2 |
| | | 4+5 2 | 7 2 |
| | | 6+7 2 | 5+6 2 |
| | | 9 2 | 7 2 |

Continued on next column

**TABLE LXXXVIII– continued**

| TQC | ICSD | NoE | SF |
|---|---|---|---|
| 227.1.65 | 154724 | 64 | Co2Nd1 |
| | -GM -X | -L | -W |
| | 6 2 5 4 | 8 2 | 3+4 2 |
| | 9 2 5 4 | 9 2 | 7 2 |
| | 7 2 5 4 | 8 2 | 5+6 2 |
| | 8 2 5 4 | 9 2 | 7 2 |
| | 10 4 5 4 | 4+5 2 | 7 2 |
| | 11 4 5 4 | 9 2 | 3+4 2 |
| | 6 2 5 4 | 8 2 | 5+6 2 |
| | 7 2 5 4 | 6+7 2 | 7 2 |
| | 10 4 5 4 | 9 2 | 5+6 2 |
| | 6 2 5 4 | 8 2 | 7 2 |
| | 10 4 5 4 | 9 2 | 7 2 |
| | 7 2 5 4 | 6+7 2 | 3+4 2 |
| | 10 4 5 4 | 9 2 | 7 2 |
| | 11 4 5 4 | 9 2 | 3+4 2 |
| | 10 4 5 4 | 6+7 2 | 5+6 2 |
| | 7 2 5 4 | 9 2 | 7 2 |
| | 10 4 | 9 2 | 7 2 |
| | 6 2 | 8 2 | 3+4 2 |
| | 10 4 | 9 2 | 5+6 2 |
| | 9 2 | 4+5 2 | 7 2 |
| | 11 4 | 6+7 2 | 5+6 2 |
| | 6 2 | 8 2 | 7 2 |
| | | 4+5 2 | 3+4 2 |
| | | 8 2 | 7 2 |
| | | 9 2 | 7 2 |
| | | 6+7 2 | 7 2 |
| | | 9 2 | 5+6 2 |
| | | 9 2 | 3+4 2 |
| | | 8 2 | 5+6 2 |
| | | 6+7 2 | 7 2 |
| | | 4+5 2 | 3+4 2 |
| | | 8 2 | 7 2 |
| 227.1.66 | 194667 | 80 | Co2Er1 |
| | -GM -X | -L | -W |
| | 6 2 5 4 | 8 2 | 3+4 2 |
| | 9 2 5 4 | 9 2 | 7 2 |
| | 7 2 5 4 | 8 2 | 5+6 2 |
| | 8 2 5 4 | 9 2 | 7 2 |
| | 10 4 5 4 | 4+5 2 | 7 2 |
| | 11 4 5 4 | 9 2 | 3+4 2 |
| | 6 2 5 4 | 8 2 | 5+6 2 |
| | 6 2 5 4 | 6+7 2 | 7 2 |
| | 7 2 5 4 | 9 2 | 5+6 2 |
| | 10 4 5 4 | 8 2 | 7 2 |
| | 10 4 5 4 | 9 2 | 7 2 |
| | 7 2 5 4 | 6+7 2 | 3+4 2 |
| | 10 4 5 4 | 9 2 | 7 2 |
| | 9 2 5 4 | 9 2 | 3+4 2 |
| | 6 2 5 4 | 9 2 | 5+6 2 |
| | 11 4 5 4 | 6+7 2 | 7 2 |
| | 10 4 5 4 | 9 2 | 7 2 |
| | 7 2 5 4 | 8 2 | 3+4 2 |
| | 10 4 5 4 | 6+7 2 | 5+6 2 |
| | 10 4 5 4 | 9 2 | 7 2 |
| | 6 2 | 4+5 2 | 7 2 |
| | 10 4 | 8 2 | 3+4 2 |
| | 9 2 | 8 2 | 5+6 2 |
| | 11 4 | 9 2 | 7 2 |
| | 11 4 | 8 2 | 3+4 2 |

Continued on next column

**TABLE LXXXVIII– continued**

| TQC | ICSD | NoE | SF |
|---|---|---|---|
| | 8 2 | 4+5 2 | 7 2 |
| | 6 2 | 6+7 2 | 5+6 2 |
| | 7 2 | 9 2 | 7 2 |
| | | 4+5 2 | 3+4 2 |
| | | 8 2 | 7 2 |
| | | 9 2 | 7 2 |
| | | 6+7 2 | 5+6 2 |
| | | 9 2 | 7 2 |
| | | 8 2 | 3+4 2 |
| | | 9 2 | 7 2 |
| | | 4+5 2 | 5+6 2 |
| | | 6+7 2 | 7 2 |
| | | 8 2 | 3+4 2 |
| | | 9 2 | 7 2 |
| | | 8 2 | 5+6 2 |
| 227.1.67 | 168600 | 148 | Ce2O7Zr2 |
| | -GM -X | -L | -W |
| | 6 2 5 4 | 9 2 | 5+6 2 |
| | 7 2 5 4 | 8 2 | 7 2 |
| | 10 4 5 4 | 6+7 2 | 3+4 2 |
| | 9 2 5 4 | 9 2 | 7 2 |
| | 11 4 5 4 | 8 2 | 5+6 2 |
| | 8 2 5 4 | 9 2 | 7 2 |
| | 6 2 5 4 | 4+5 2 | 7 2 |
| | 6 2 5 4 | 8 2 | 3+4 2 |
| | 11 4 5 4 | 8 2 | 7 2 |
| | 7 2 5 4 | 9 2 | 5+6 2 |
| | 10 4 5 4 | 8 2 | 7 2 |
| | 10 4 5 4 | 4+5 2 | 3+4 2 |
| | 9 2 5 4 | 9 2 | 5+6 2 |
| | 8 2 5 4 | 6+7 2 | 7 2 |
| | 9 2 5 4 | 8 2 | 5+6 2 |
| | 11 4 5 4 | 9 2 | 7 2 |
| | 11 4 5 4 | 6+7 2 | 3+4 2 |
| | 8 2 5 4 | 8 2 | 7 2 |
| | 11 4 5 4 | 4+5 2 | 3+4 2 |
| | 11 4 5 4 | 9 2 | 7 2 |
| | 9 2 5 4 | 8 2 | 5+6 2 |
| | 7 2 5 4 | 9 2 | 7 2 |
| | 10 4 5 4 | 8 2 | 3+4 2 |
| | 10 4 5 4 | 4+5 2 | 7 2 |
| | 7 2 5 4 | 8 2 | 7 2 |
| | 10 4 5 4 | 4+5 2 | 5+6 2 |
| | 6 2 5 4 | 6+7 2 | 3+4 2 |
| | 6 2 5 4 | 8 2 | 7 2 |
| | 10 4 5 4 | 9 2 | 7 2 |
| | 7 2 5 4 | 8 2 | 5+6 2 |
| | 10 4 5 4 | 4+5 2 | 7 2 |
| | 9 2 5 4 | 8 2 | 5+6 2 |
| | 9 2 5 4 | 8 2 | 7 2 |
| | 11 4 5 4 | 8 2 | 3+4 2 |
| | 8 2 5 4 | 9 2 | 5+6 2 |
| | 11 4 5 4 | 6+7 2 | 7 2 |
| | 11 4 5 4 | 9 2 | 5+6 2 |
| | 8 2 | 6+7 2 | 3+4 2 |
| | 10 4 | 8 2 | 7 2 |
| | 6 2 | 4+5 2 | 7 2 |
| | 7 2 | 9 2 | 3+4 2 |
| | 10 4 | 9 2 | 7 2 |
| | 11 4 | 8 2 | 7 2 |
| | 8 2 | 9 2 | 3+4 2 |

Continued on next column

**TABLE LXXXVIII– continued**

| TQC | ICSD | | NoE | SF |
|---|---|---|---|---|
| | 11 4 | | 6+7 2 | 7 2 |
| | 9 2 | | 8 2 | 3+4 2 |
| | 11 4 | | 4+5 2 | 5+6 2 |
| | 8 2 | | 8 2 | 7 2 |
| | 11 4 | | 8 2 | 7 2 |
| | 11 4 | | 9 2 | 3+4 2 |
| | | | 9 2 | 7 2 |
| | | | 4+5 2 | 5+6 2 |
| | | | 8 2 | 3+4 2 |
| | | | 6+7 2 | 7 2 |
| | | | 9 2 | 7 2 |
| | | | 9 2 | 5+6 2 |
| | | | 4+5 2 | 7 2 |
| | | | 8 2 | 7 2 |
| | | | 9 2 | 5+6 2 |
| | | | 8 2 | 3+4 2 |
| | | | 6+7 2 | 5+6 2 |
| | | | 9 2 | 7 2 |
| | | | 4+5 2 | 3+4 2 |
| | | | 8 2 | 7 2 |
| | | | 9 2 | 7 2 |
| | | | 8 2 | 5+6 2 |
| | | | 6+7 2 | 3+4 2 |
| | | | 4+5 2 | 7 2 |
| | | | 8 2 | 5+6 2 |
| | | | 9 2 | 7 2 |
| | | | 9 2 | 3+4 2 |
| | | | 6+7 2 | 7 2 |
| | | | 4+5 2 | 7 2 |
| | | | 8 2 | 3+4 2 |
| 227.1.68 | 168595 | | 148 | Ce2O7Zr2 |
| | -GM | -X | -L | -W |
| | 6 2 | 5 4 | 9 2 | 5+6 2 |
| | 7 2 | 5 4 | 8 2 | 7 2 |
| | 10 4 | 5 4 | 6+7 2 | 3+4 2 |
| | 9 2 | 5 4 | 9 2 | 7 2 |
| | 11 4 | 5 4 | 8 2 | 5+6 2 |
| | 8 2 | 5 4 | 9 2 | 7 2 |
| | 6 2 | 5 4 | 4+5 2 | 7 2 |
| | 6 2 | 5 4 | 8 2 | 3+4 2 |
| | 11 4 | 5 4 | 8 2 | 7 2 |
| | 7 2 | 5 4 | 9 2 | 5+6 2 |
| | 10 4 | 5 4 | 8 2 | 7 2 |
| | 10 4 | 5 4 | 4+5 2 | 3+4 2 |
| | 9 2 | 5 4 | 9 2 | 5+6 2 |
| | 8 2 | 5 4 | 6+7 2 | 7 2 |
| | 9 2 | 5 4 | 8 2 | 5+6 2 |
| | 11 4 | 5 4 | 9 2 | 7 2 |
| | 11 4 | 5 4 | 6+7 2 | 3+4 2 |
| | 8 2 | 5 4 | 8 2 | 7 2 |
| | 11 4 | 5 4 | 9 2 | 3+4 2 |
| | 11 4 | 5 4 | 4+5 2 | 7 2 |
| | 9 2 | 5 4 | 8 2 | 5+6 2 |
| | 7 2 | 5 4 | 9 2 | 7 2 |
| | 10 4 | 5 4 | 8 2 | 3+4 2 |
| | 10 4 | 5 4 | 4+5 2 | 7 2 |
| | 7 2 | 5 4 | 8 2 | 7 2 |
| | 10 4 | 5 4 | 4+5 2 | 5+6 2 |
| | 6 2 | 5 4 | 6+7 2 | 3+4 2 |
| | 6 2 | 5 4 | 8 2 | 7 2 |
| | 10 4 | 5 4 | 9 2 | 7 2 |

Continued on next column

**TABLE LXXXVIII– continued**

| TQC | ICSD | | NoE | SF |
|---|---|---|---|---|
| | 7 2 | 5 4 | 8 2 | 5+6 2 |
| | 10 4 | 5 4 | 4+5 2 | 7 2 |
| | 9 2 | 5 4 | 8 2 | 5+6 2 |
| | 9 2 | 5 4 | 8 2 | 7 2 |
| | 11 4 | 5 4 | 8 2 | 3+4 2 |
| | 8 2 | 5 4 | 9 2 | 5+6 2 |
| | 11 4 | 5 4 | 6+7 2 | 7 2 |
| | 11 4 | 5 4 | 9 2 | 5+6 2 |
| | 10 4 | | 6+7 2 | 7 2 |
| | 6 2 | | 8 2 | 3+4 2 |
| | 8 2 | | 4+5 2 | 7 2 |
| | 7 2 | | 9 2 | 3+4 2 |
| | 10 4 | | 9 2 | 7 2 |
| | 11 4 | | 9 2 | 3+4 2 |
| | 8 2 | | 6+7 2 | 7 2 |
| | 11 4 | | 8 2 | 7 2 |
| | 9 2 | | 8 2 | 3+4 2 |
| | 11 4 | | 4+5 2 | 5+6 2 |
| | 8 2 | | 8 2 | 7 2 |
| | 11 4 | | 8 2 | 7 2 |
| | 11 4 | | 9 2 | 3+4 2 |
| | | | 9 2 | 7 2 |
| | | | 4+5 2 | 5+6 2 |
| | | | 8 2 | 3+4 2 |
| | | | 6+7 2 | 7 2 |
| | | | 9 2 | 7 2 |
| | | | 9 2 | 5+6 2 |
| | | | 4+5 2 | 7 2 |
| | | | 8 2 | 5+6 2 |
| | | | 9 2 | 7 2 |
| | | | 8 2 | 3+4 2 |
| | | | 6+7 2 | 5+6 2 |
| | | | 9 2 | 7 2 |
| | | | 4+5 2 | 3+4 2 |
| | | | 8 2 | 7 2 |
| | | | 9 2 | 7 2 |
| | | | 8 2 | 5+6 2 |
| | | | 6+7 2 | 3+4 2 |
| | | | 4+5 2 | 7 2 |
| | | | 8 2 | 5+6 2 |
| | | | 9 2 | 7 2 |
| | | | 9 2 | 3+4 2 |
| | | | 6+7 2 | 7 2 |
| | | | 8 2 | 7 2 |
| | | | 4+5 2 | 3+4 2 |
| 227.1.69 | 155773 | | 164 | Cd2O7Os2 |
| | -GM | -X | -L | -W |
| | 6 2 | 5 4 | 8 2 | 7 2 |
| | 7 2 | 5 4 | 9 2 | 3+4 2 |
| | 10 4 | 5 4 | 8 2 | 5+6 2 |
| | 10 4 | 5 4 | 4+5 2 | 7 2 |
| | 11 4 | 5 4 | 8 2 | 5+6 2 |
| | 8 2 | 5 4 | 4+5 2 | 7 2 |
| | 11 4 | 5 4 | 6+7 2 | 3+4 2 |
| | 9 2 | 5 4 | 9 2 | 7 2 |
| | 6 2 | 5 4 | 8 2 | 7 2 |
| | 9 2 | 5 4 | 6+7 2 | 5+6 2 |
| | 10 4 | 5 4 | 9 2 | 3+4 2 |
| | 7 2 | 5 4 | 9 2 | 7 2 |
| | 10 4 | 5 4 | 9 2 | 5+6 2 |
| | 10 4 | 5 4 | 8 2 | 7 2 |

Continued on next column

**TABLE LXXXVIII– continued**

| TQC | ICSD | | NoE | SF |
|---|---|---|---|---|
| | 6 2 | 5 4 | 4+5 2 | 7 2 |
| | 7 2 | 5 4 | 8 2 | 7 2 |
| | 10 4 | 5 4 | 8 2 | 5+6 2 |
| | 7 2 | 5 4 | 6+7 2 | 3+4 2 |
| | 6 2 | 5 4 | 9 2 | 3+4 2 |
| | 10 4 | 5 4 | 8 2 | 7 2 |
| | 10 4 | 5 4 | 6+7 2 | 7 2 |
| | 10 4 | 5 4 | 9 2 | 5+6 2 |
| | 6 2 | 5 4 | 9 2 | 7 2 |
| | 6 2 | 5 4 | 4+5 2 | 3+4 2 |
| | 10 4 | 5 4 | 9 2 | 5+6 2 |
| | 7 2 | 5 4 | 8 2 | 7 2 |
| | 7 2 | 5 4 | 9 2 | 3+4 2 |
| | 10 4 | 5 4 | 6+7 2 | 7 2 |
| | 10 4 | 5 4 | 9 2 | 7 2 |
| | 6 2 | 5 4 | 9 2 | 3+4 2 |
| | 9 2 | 5 4 | 8 2 | 5+6 2 |
| | 8 2 | 5 4 | 4+5 2 | 7 2 |
| | 11 4 | 5 4 | 6+7 2 | 7 2 |
| | 11 4 | 5 4 | 9 2 | 5+6 2 |
| | 9 2 | 5 4 | 6+7 2 | 3+4 2 |
| | 10 4 | 5 4 | 8 2 | 7 2 |
| | 10 4 | 5 4 | 9 2 | 7 2 |
| | 6 2 | 5 4 | 8 2 | 5+6 2 |
| | 7 2 | 5 4 | 4+5 2 | 3+4 2 |
| | 10 4 | 5 4 | 9 2 | 7 2 |
| | 8 2 | 5 4 | 8 2 | 5+6 2 |
| | 11 4 | | 6+7 2 | 7 2 |
| | 11 4 | | 9 2 | 3+4 2 |
| | 9 2 | | 4+5 2 | 7 2 |
| | 11 4 | | 8 2 | 5+6 2 |
| | 8 2 | | 9 2 | 7 2 |
| | 11 4 | | 8 2 | 7 2 |
| | 8 2 | | 9 2 | 3+4 2 |
| | 10 4 | | 6+7 2 | 5+6 2 |
| | 11 4 | | 9 2 | 7 2 |
| | 7 2 | | 8 2 | 7 2 |
| | 10 4 | | 9 2 | 3+4 2 |
| | 7 2 | | 4+5 2 | 7 2 |
| | 10 4 | | 8 2 | 7 2 |
| | 6 2 | | 9 2 | 3+4 2 |
| | | | 6+7 2 | 5+6 2 |
| | | | 9 2 | 5+6 2 |
| | | | 4+5 2 | 7 2 |
| | | | 8 2 | 3+4 2 |
| | | | 6+7 2 | 7 2 |
| | | | 9 2 | 5+6 2 |
| | | | 9 2 | 7 2 |
| | | | 9 2 | 7 2 |
| | | | 8 2 | 3+4 2 |
| | | | 8 2 | 5+6 2 |
| | | | 6+7 2 | 7 2 |
| | | | 8 2 | 3+4 2 |
| | | | 4+5 2 | 7 2 |
| | | | 8 2 | 5+6 2 |
| | | | 4+5 2 | 7 2 |
| | | | 9 2 | 3+4 2 |
| | | | 6+7 2 | 7 2 |
| | | | 9 2 | 5+6 2 |
| | | | 9 2 | 7 2 |
| | | | 6+7 2 | 3+4 2 |
| | | | 9 2 | 7 2 |
| | | | 8 2 | 7 2 |
| | | | 8 2 | 5+6 2 |
| | | | 4+5 2 | 3+4 2 |
| | | | 8 2 | 7 2 |
| | | | 4+5 2 | 7 2 |
| | | | 9 2 | 5+6 2 |
| 227.1.70 | 155772 | | 164 | Cd2O7Os2 |
| | -GM | -X | -L | -W |
| | 6 2 | 5 4 | 8 2 | 7 2 |
| | 7 2 | 5 4 | 9 2 | 3+4 2 |
| | 10 4 | 5 4 | 8 2 | 5+6 2 |
| | 10 4 | 5 4 | 4+5 2 | 7 2 |
| | 11 4 | 5 4 | 8 2 | 5+6 2 |
| | 8 2 | 5 4 | 4+5 2 | 7 2 |
| | 11 4 | 5 4 | 6+7 2 | 3+4 2 |
| | 9 2 | 5 4 | 9 2 | 7 2 |
| | 6 2 | 5 4 | 8 2 | 7 2 |
| | 9 2 | 5 4 | 6+7 2 | 5+6 2 |
| | 10 4 | 5 4 | 9 2 | 3+4 2 |
| | 7 2 | 5 4 | 9 2 | 7 2 |
| | 10 4 | 5 4 | 9 2 | 5+6 2 |
| | 10 4 | 5 4 | 8 2 | 7 2 |
| | 6 2 | 5 4 | 4+5 2 | 7 2 |
| | 7 2 | 5 4 | 8 2 | 7 2 |
| | 10 4 | 5 4 | 8 2 | 5+6 2 |
| | 7 2 | 5 4 | 6+7 2 | 3+4 2 |
| | 6 2 | 5 4 | 9 2 | 3+4 2 |
| | 10 4 | 5 4 | 8 2 | 7 2 |
| | 10 4 | 5 4 | 6+7 2 | 7 2 |
| | 10 4 | 5 4 | 9 2 | 5+6 2 |
| | 6 2 | 5 4 | 9 2 | 7 2 |
| | 6 2 | 5 4 | 4+5 2 | 3+4 2 |
| | 10 4 | 5 4 | 9 2 | 5+6 2 |
| | 7 2 | 5 4 | 8 2 | 7 2 |
| | 7 2 | 5 4 | 9 2 | 3+4 2 |
| | 10 4 | 5 4 | 6+7 2 | 7 2 |
| | 10 4 | 5 4 | 9 2 | 7 2 |
| | 6 2 | 5 4 | 9 2 | 3+4 2 |
| | 9 2 | 5 4 | 8 2 | 5+6 2 |
| | 8 2 | 5 4 | 4+5 2 | 7 2 |
| | 11 4 | 5 4 | 6+7 2 | 7 2 |
| | 11 4 | 5 4 | 9 2 | 5+6 2 |
| | 9 2 | 5 4 | 6+7 2 | 3+4 2 |
| | 10 4 | 5 4 | 8 2 | 7 2 |
| | 10 4 | 5 4 | 9 2 | 7 2 |
| | 6 2 | 5 4 | 8 2 | 5+6 2 |
| | 7 2 | 5 4 | 4+5 2 | 3+4 2 |
| | 10 4 | 5 4 | 9 2 | 7 2 |
| | 8 2 | 5 4 | 8 2 | 5+6 2 |
| | 11 4 | | 6+7 2 | 7 2 |
| | 11 4 | | 9 2 | 3+4 2 |
| | 9 2 | | 4+5 2 | 7 2 |
| | 11 4 | | 8 2 | 5+6 2 |
| | 8 2 | | 9 2 | 7 2 |
| | 11 4 | | 8 2 | 7 2 |
| | 8 2 | | 9 2 | 3+4 2 |
| | 10 4 | | 6+7 2 | 5+6 2 |
| | 11 4 | | 9 2 | 7 2 |
| | 7 2 | | 8 2 | 7 2 |
| | 10 4 | | 9 2 | 3+4 2 |

Continued on next column

**TABLE LXXXVIII– continued**

| TQC | ICSD | NoE | SF |
|---|---|---|---|
| | 7 2 | 4+5 2 | 7 2 |
| | 10 4 | 8 2 | 7 2 |
| | 6 2 | 9 2 | 3+4 2 |
| | | 6+7 2 | 5+6 2 |
| | | 9 2 | 5+6 2 |
| | | 4+5 2 | 7 2 |
| | | 8 2 | 3+4 2 |
| | | 6+7 2 | 7 2 |
| | | 9 2 | 5+6 2 |
| | | 9 2 | 7 2 |
| | | 9 2 | 7 2 |
| | | 8 2 | 3+4 2 |
| | | 8 2 | 5+6 2 |
| | | 6+7 2 | 7 2 |
| | | 8 2 | 3+4 2 |
| | | 4+5 2 | 7 2 |
| | | 8 2 | 5+6 2 |
| | | 4+5 2 | 7 2 |
| | | 9 2 | 3+4 2 |
| | | 6+7 2 | 7 2 |
| | | 9 2 | 5+6 2 |
| | | 9 2 | 7 2 |
| | | 6+7 2 | 3+4 2 |
| | | 9 2 | 7 2 |
| | | 8 2 | 7 2 |
| | | 8 2 | 5+6 2 |
| | | 4+5 2 | 3+4 2 |
| | | 8 2 | 7 2 |
| | | 4+5 2 | 7 2 |
| | | 9 2 | 5+6 2 |
| 227.1.71 | 155771 | 164 | Cd2O7Os2 |
| | -GM -X | -L | -W |
| | 6 2 5 4 | 8 2 | 7 2 |
| | 7 2 5 4 | 9 2 | 3+4 2 |
| | 10 4 5 4 | 8 2 | 5+6 2 |
| | 10 4 5 4 | 4+5 2 | 7 2 |
| | 11 4 5 4 | 8 2 | 5+6 2 |
| | 8 2 5 4 | 4+5 2 | 7 2 |
| | 11 4 5 4 | 6+7 2 | 3+4 2 |
| | 9 2 5 4 | 9 2 | 7 2 |
| | 6 2 5 4 | 8 2 | 7 2 |
| | 9 2 5 4 | 6+7 2 | 5+6 2 |
| | 10 4 5 4 | 9 2 | 3+4 2 |
| | 7 2 5 4 | 9 2 | 7 2 |
| | 10 4 5 4 | 9 2 | 5+6 2 |
| | 10 4 5 4 | 8 2 | 7 2 |
| | 6 2 5 4 | 4+5 2 | 7 2 |
| | 7 2 5 4 | 8 2 | 7 2 |
| | 10 4 5 4 | 8 2 | 5+6 2 |
| | 7 2 5 4 | 6+7 2 | 3+4 2 |
| | 6 2 5 4 | 9 2 | 3+4 2 |
| | 10 4 5 4 | 8 2 | 7 2 |
| | 10 4 5 4 | 6+7 2 | 7 2 |
| | 10 4 5 4 | 9 2 | 5+6 2 |
| | 6 2 5 4 | 9 2 | 7 2 |
| | 6 2 5 4 | 4+5 2 | 3+4 2 |
| | 10 4 5 4 | 9 2 | 5+6 2 |
| | 7 2 5 4 | 8 2 | 7 2 |
| | 7 2 5 4 | 9 2 | 3+4 2 |
| | 10 4 5 4 | 6+7 2 | 7 2 |
| | 10 4 5 4 | 9 2 | 7 2 |
| | 6 2 5 4 | 9 2 | 3+4 2 |
| | 9 2 5 4 | 8 2 | 5+6 2 |
| | 8 2 5 4 | 4+5 2 | 7 2 |
| | 11 4 5 4 | 6+7 2 | 7 2 |
| | 11 4 5 4 | 9 2 | 5+6 2 |
| | 9 2 5 4 | 6+7 2 | 3+4 2 |
| | 10 4 5 4 | 8 2 | 7 2 |
| | 10 4 5 4 | 9 2 | 7 2 |
| | 6 2 5 4 | 8 2 | 5+6 2 |
| | 7 2 5 4 | 4+5 2 | 3+4 2 |
| | 10 4 5 4 | 9 2 | 7 2 |
| | 8 2 5 4 | 8 2 | 5+6 2 |
| | 11 4 | 6+7 2 | 7 2 |
| | 11 4 | 9 2 | 3+4 2 |
| | 9 2 | 4+5 2 | 7 2 |
| | 11 4 | 8 2 | 5+6 2 |
| | 8 2 | 9 2 | 7 2 |
| | 11 4 | 8 2 | 7 2 |
| | 8 2 | 9 2 | 3+4 2 |
| | 10 4 | 6+7 2 | 5+6 2 |
| | 11 4 | 9 2 | 7 2 |
| | 7 2 | 8 2 | 7 2 |
| | 10 4 | 4+5 2 | 3+4 2 |
| | 7 2 | 9 2 | 7 2 |
| | 10 4 | 8 2 | 7 2 |
| | 6 2 | 9 2 | 3+4 2 |
| | | 6+7 2 | 5+6 2 |
| | | 9 2 | 5+6 2 |
| | | 4+5 2 | 7 2 |
| | | 8 2 | 3+4 2 |
| | | 6+7 2 | 7 2 |
| | | 9 2 | 5+6 2 |
| | | 9 2 | 7 2 |
| | | 9 2 | 7 2 |
| | | 8 2 | 3+4 2 |
| | | 8 2 | 5+6 2 |
| | | 6+7 2 | 7 2 |
| | | 8 2 | 3+4 2 |
| | | 4+5 2 | 7 2 |
| | | 8 2 | 5+6 2 |
| | | 4+5 2 | 7 2 |
| | | 9 2 | 3+4 2 |
| | | 6+7 2 | 7 2 |
| | | 9 2 | 5+6 2 |
| | | 9 2 | 7 2 |
| | | 6+7 2 | 3+4 2 |
| | | 9 2 | 7 2 |
| | | 8 2 | 7 2 |
| | | 8 2 | 5+6 2 |
| | | 4+5 2 | 3+4 2 |
| | | 8 2 | 7 2 |
| | | 4+5 2 | 7 2 |
| | | 9 2 | 5+6 2 |
| 227.1.72 | 155770 | 164 | Cd2O7Os2 |
| | -GM -X | -L | -W |
| | 6 2 5 4 | 8 2 | 7 2 |
| | 7 2 5 4 | 9 2 | 3+4 2 |
| | 10 4 5 4 | 8 2 | 5+6 2 |
| | 10 4 5 4 | 4+5 2 | 7 2 |
| | 11 4 5 4 | 8 2 | 5+6 2 |
| | 8 2 5 4 | 4+5 2 | 7 2 |

Continued on next column

**TABLE LXXXVIII– continued**

| TQC | ICSD | NoE | SF |
|---|---|---|---|
| | 11 4 5 4 | 6+7 2 | 3+4 2 |
| | 9 2 5 4 | 9 2 | 7 2 |
| | 6 2 5 4 | 8 2 | 7 2 |
| | 9 2 5 4 | 6+7 2 | 5+6 2 |
| | 10 4 5 4 | 9 2 | 3+4 2 |
| | 7 2 5 4 | 9 2 | 7 2 |
| | 10 4 5 4 | 9 2 | 5+6 2 |
| | 10 4 5 4 | 8 2 | 7 2 |
| | 6 2 5 4 | 4+5 2 | 7 2 |
| | 7 2 5 4 | 8 2 | 7 2 |
| | 10 4 5 4 | 8 2 | 5+6 2 |
| | 7 2 5 4 | 6+7 2 | 3+4 2 |
| | 6 2 5 4 | 9 2 | 3+4 2 |
| | 10 4 5 4 | 8 2 | 7 2 |
| | 10 4 5 4 | 6+7 2 | 7 2 |
| | 10 4 5 4 | 9 2 | 5+6 2 |
| | 6 2 5 4 | 9 2 | 7 2 |
| | 6 2 5 4 | 4+5 2 | 3+4 2 |
| | 10 4 5 4 | 9 2 | 5+6 2 |
| | 7 2 5 4 | 8 2 | 7 2 |
| | 7 2 5 4 | 9 2 | 3+4 2 |
| | 10 4 5 4 | 6+7 2 | 7 2 |
| | 10 4 5 4 | 9 2 | 7 2 |
| | 6 2 5 4 | 9 2 | 3+4 2 |
| | 9 2 5 4 | 8 2 | 5+6 2 |
| | 8 2 5 4 | 4+5 2 | 7 2 |
| | 11 4 5 4 | 6+7 2 | 7 2 |
| | 11 4 5 4 | 9 2 | 5+6 2 |
| | 9 2 5 4 | 6+7 2 | 3+4 2 |
| | 10 4 5 4 | 8 2 | 7 2 |
| | 10 4 5 4 | 9 2 | 7 2 |
| | 6 2 5 4 | 8 2 | 5+6 2 |
| | 7 2 5 4 | 4+5 2 | 3+4 2 |
| | 10 4 5 4 | 9 2 | 7 2 |
| | 8 2 5 4 | 8 2 | 5+6 2 |
| | 11 4 | 6+7 2 | 7 2 |
| | 11 4 | 9 2 | 3+4 2 |
| | 9 2 | 4+5 2 | 7 2 |
| | 11 4 | 8 2 | 5+6 2 |
| | 8 2 | 9 2 | 7 2 |
| | 11 4 | 8 2 | 7 2 |
| | 8 2 | 9 2 | 3+4 2 |
| | 10 4 | 6+7 2 | 5+6 2 |
| | 11 4 | 9 2 | 7 2 |
| | 7 2 | 8 2 | 7 2 |
| | 10 4 | 9 2 | 3+4 2 |
| | 7 2 | 4+5 2 | 7 2 |
| | 10 4 | 8 2 | 7 2 |
| | 6 2 | 9 2 | 3+4 2 |
| | | 6+7 2 | 5+6 2 |
| | | 9 2 | 5+6 2 |
| | | 4+5 2 | 7 2 |
| | | 8 2 | 3+4 2 |
| | | 6+7 2 | 7 2 |
| | | 9 2 | 5+6 2 |
| | | 9 2 | 7 2 |
| | | 9 2 | 7 2 |
| | | 8 2 | 3+4 2 |
| | | 8 2 | 5+6 2 |
| | | 6+7 2 | 7 2 |
| | | 8 2 | 3+4 2 |

Continued on next column

**TABLE LXXXVIII– continued**

| TQC | ICSD | NoE | SF |
|---|---|---|---|
| | | 4+5 2 | 7 2 |
| | | 8 2 | 5+6 2 |
| | | 4+5 2 | 7 2 |
| | | 9 2 | 3+4 2 |
| | | 6+7 2 | 7 2 |
| | | 9 2 | 5+6 2 |
| | | 9 2 | 7 2 |
| | | 6+7 2 | 3+4 2 |
| | | 9 2 | 7 2 |
| | | 8 2 | 7 2 |
| | | 8 2 | 5+6 2 |
| | | 4+5 2 | 3+4 2 |
| | | 8 2 | 7 2 |
| | | 4+5 2 | 7 2 |
| | | 9 2 | 5+6 2 |
| 227.1.73 | 155769 | 164 | Cd2O7Os2 |
| | -GM -X | -L | -W |
| | 6 2 5 4 | 8 2 | 7 2 |
| | 7 2 5 4 | 9 2 | 3+4 2 |
| | 10 4 5 4 | 8 2 | 5+6 2 |
| | 10 4 5 4 | 4+5 2 | 7 2 |
| | 11 4 5 4 | 8 2 | 5+6 2 |
| | 8 2 5 4 | 4+5 2 | 7 2 |
| | 11 4 5 4 | 6+7 2 | 3+4 2 |
| | 9 2 5 4 | 9 2 | 7 2 |
| | 6 2 5 4 | 8 2 | 7 2 |
| | 9 2 5 4 | 6+7 2 | 5+6 2 |
| | 10 4 5 4 | 9 2 | 3+4 2 |
| | 7 2 5 4 | 9 2 | 7 2 |
| | 10 4 5 4 | 9 2 | 5+6 2 |
| | 10 4 5 4 | 8 2 | 7 2 |
| | 6 2 5 4 | 4+5 2 | 7 2 |
| | 7 2 5 4 | 8 2 | 7 2 |
| | 10 4 5 4 | 8 2 | 5+6 2 |
| | 7 2 5 4 | 6+7 2 | 3+4 2 |
| | 6 2 5 4 | 9 2 | 3+4 2 |
| | 10 4 5 4 | 8 2 | 7 2 |
| | 10 4 5 4 | 6+7 2 | 7 2 |
| | 10 4 5 4 | 9 2 | 5+6 2 |
| | 6 2 5 4 | 9 2 | 7 2 |
| | 6 2 5 4 | 4+5 2 | 3+4 2 |
| | 10 4 5 4 | 9 2 | 5+6 2 |
| | 7 2 5 4 | 8 2 | 7 2 |
| | 7 2 5 4 | 9 2 | 3+4 2 |
| | 10 4 5 4 | 6+7 2 | 7 2 |
| | 10 4 5 4 | 9 2 | 7 2 |
| | 6 2 5 4 | 9 2 | 3+4 2 |
| | 9 2 5 4 | 8 2 | 5+6 2 |
| | 8 2 5 4 | 4+5 2 | 7 2 |
| | 11 4 5 4 | 6+7 2 | 7 2 |
| | 11 4 5 4 | 9 2 | 5+6 2 |
| | 9 2 5 4 | 6+7 2 | 3+4 2 |
| | 10 4 5 4 | 8 2 | 7 2 |
| | 10 4 5 4 | 9 2 | 7 2 |
| | 6 2 5 4 | 8 2 | 5+6 2 |
| | 7 2 5 4 | 4+5 2 | 3+4 2 |
| | 10 4 5 4 | 9 2 | 7 2 |
| | 8 2 5 4 | 8 2 | 5+6 2 |
| | 11 4 | 6+7 2 | 7 2 |
| | 11 4 | 9 2 | 3+4 2 |
| | 9 2 | 4+5 2 | 7 2 |

Continued on next column

**TABLE LXXXVIII– continued**

| TQC | ICSD | NoE | SF |
|---|---|---|---|
| | 11 4 | 8 2 | 5+6 2 |
| | 8 2 | 9 2 | 7 2 |
| | 11 4 | 8 2 | 3+4 2 |
| | 8 2 | 9 2 | 7 2 |
| | 10 4 | 6+7 2 | 5+6 2 |
| | 11 4 | 9 2 | 7 2 |
| | 7 2 | 8 2 | 7 2 |
| | 10 4 | 9 2 | 3+4 2 |
| | 7 2 | 4+5 2 | 7 2 |
| | 10 4 | 8 2 | 7 2 |
| | 6 2 | 9 2 | 3+4 2 |
| | | 6+7 2 | 5+6 2 |
| | | 9 2 | 5+6 2 |
| | | 4+5 2 | 7 2 |
| | | 8 2 | 3+4 2 |
| | | 6+7 2 | 7 2 |
| | | 9 2 | 5+6 2 |
| | | 9 2 | 7 2 |
| | | 9 2 | 7 2 |
| | | 8 2 | 3+4 2 |
| | | 8 2 | 5+6 2 |
| | | 6+7 2 | 7 2 |
| | | 8 2 | 3+4 2 |
| | | 4+5 2 | 7 2 |
| | | 8 2 | 5+6 2 |
| | | 4+5 2 | 7 2 |
| | | 9 2 | 3+4 2 |
| | | 6+7 2 | 7 2 |
| | | 9 2 | 5+6 2 |
| | | 9 2 | 7 2 |
| | | 6+7 2 | 3+4 2 |
| | | 9 2 | 7 2 |
| | | 8 2 | 7 2 |
| | | 8 2 | 5+6 2 |
| | | 4+5 2 | 3+4 2 |
| | | 8 2 | 7 2 |
| | | 4+5 2 | 7 2 |
| | | 9 2 | 5+6 2 |
| 227.1.74 | 155768 | 164 | Cd2O7Os2 |
| | -GM -X | -L | -W |
| | 6 2 5 4 | 8 2 | 7 2 |
| | 7 2 5 4 | 9 2 | 3+4 2 |
| | 10 4 5 4 | 8 2 | 5+6 2 |
| | 10 4 5 4 | 4+5 2 | 7 2 |
| | 11 4 5 4 | 8 2 | 5+6 2 |
| | 8 2 5 4 | 4+5 2 | 7 2 |
| | 11 4 5 4 | 6+7 2 | 3+4 2 |
| | 9 2 5 4 | 9 2 | 7 2 |
| | 6 2 5 4 | 8 2 | 7 2 |
| | 9 2 5 4 | 6+7 2 | 5+6 2 |
| | 10 4 5 4 | 9 2 | 3+4 2 |
| | 7 2 5 4 | 9 2 | 7 2 |
| | 10 4 5 4 | 9 2 | 5+6 2 |
| | 10 4 5 4 | 8 2 | 7 2 |
| | 6 2 5 4 | 4+5 2 | 7 2 |
| | 7 2 5 4 | 8 2 | 7 2 |
| | 10 4 5 4 | 8 2 | 5+6 2 |
| | 7 2 5 4 | 6+7 2 | 3+4 2 |
| | 6 2 5 4 | 9 2 | 3+4 2 |
| | 10 4 5 4 | 8 2 | 7 2 |
| | 10 4 5 4 | 6+7 2 | 7 2 |
| | 10 4 5 4 | 9 2 | 5+6 2 |
| | 6 2 5 4 | 9 2 | 7 2 |
| | 6 2 5 4 | 4+5 2 | 3+4 2 |
| | 10 4 5 4 | 9 2 | 5+6 2 |
| | 7 2 5 4 | 8 2 | 7 2 |
| | 7 2 5 4 | 9 2 | 3+4 2 |
| | 10 4 5 4 | 6+7 2 | 7 2 |
| | 10 4 5 4 | 9 2 | 7 2 |
| | 6 2 5 4 | 9 2 | 3+4 2 |
| | 9 2 5 4 | 8 2 | 5+6 2 |
| | 8 2 5 4 | 4+5 2 | 7 2 |
| | 11 4 5 4 | 6+7 2 | 7 2 |
| | 11 4 5 4 | 9 2 | 5+6 2 |
| | 9 2 5 4 | 6+7 2 | 3+4 2 |
| | 10 4 5 4 | 8 2 | 7 2 |
| | 10 4 5 4 | 9 2 | 7 2 |
| | 6 2 5 4 | 8 2 | 5+6 2 |
| | 7 2 5 4 | 4+5 2 | 3+4 2 |
| | 10 4 5 4 | 9 2 | 7 2 |
| | 8 2 5 4 | 8 2 | 5+6 2 |
| | 11 4 | 6+7 2 | 7 2 |
| | 11 4 | 9 2 | 3+4 2 |
| | 9 2 | 4+5 2 | 7 2 |
| | 11 4 | 8 2 | 5+6 2 |
| | 8 2 | 9 2 | 7 2 |
| | 11 4 | 8 2 | 7 2 |
| | 8 2 | 9 2 | 3+4 2 |
| | 10 4 | 6+7 2 | 5+6 2 |
| | 11 4 | 9 2 | 7 2 |
| | 7 2 | 8 2 | 7 2 |
| | 10 4 | 4+5 2 | 3+4 2 |
| | 7 2 | 9 2 | 7 2 |
| | 10 4 | 8 2 | 7 2 |
| | 6 2 | 9 2 | 3+4 2 |
| | | 6+7 2 | 5+6 2 |
| | | 9 2 | 5+6 2 |
| | | 4+5 2 | 7 2 |
| | | 8 2 | 3+4 2 |
| | | 6+7 2 | 7 2 |
| | | 9 2 | 5+6 2 |
| | | 9 2 | 7 2 |
| | | 9 2 | 7 2 |
| | | 8 2 | 3+4 2 |
| | | 8 2 | 5+6 2 |
| | | 6+7 2 | 7 2 |
| | | 8 2 | 3+4 2 |
| | | 4+5 2 | 7 2 |
| | | 8 2 | 5+6 2 |
| | | 4+5 2 | 7 2 |
| | | 9 2 | 3+4 2 |
| | | 6+7 2 | 7 2 |
| | | 9 2 | 5+6 2 |
| | | 9 2 | 7 2 |
| | | 6+7 2 | 3+4 2 |
| | | 9 2 | 7 2 |
| | | 8 2 | 7 2 |
| | | 8 2 | 5+6 2 |
| | | 4+5 2 | 3+4 2 |
| | | 8 2 | 7 2 |
| | | 4+5 2 | 7 2 |
| | | 9 2 | 5+6 2 |

Continued on next column

**TABLE LXXXVIII– continued**

| TQC | ICSD | NoE | SF |
|---|---|---|---|
| 227.1.75 | 155767 | 164 | Cd2O7Os2 |
| | -GM -X | -L | -W |
| | 6 2 5 4 | 8 2 | 7 2 |
| | 7 2 5 4 | 9 2 | 3+4 2 |
| | 10 4 5 4 | 8 2 | 5+6 2 |
| | 10 4 5 4 | 4+5 2 | 7 2 |
| | 11 4 5 4 | 8 2 | 5+6 2 |
| | 8 2 5 4 | 4+5 2 | 7 2 |
| | 11 4 5 4 | 6+7 2 | 3+4 2 |
| | 9 2 5 4 | 9 2 | 7 2 |
| | 6 2 5 4 | 8 2 | 7 2 |
| | 9 2 5 4 | 6+7 2 | 5+6 2 |
| | 10 4 5 4 | 9 2 | 3+4 2 |
| | 7 2 5 4 | 9 2 | 7 2 |
| | 10 4 5 4 | 9 2 | 5+6 2 |
| | 10 4 5 4 | 8 2 | 7 2 |
| | 6 2 5 4 | 4+5 2 | 7 2 |
| | 7 2 5 4 | 8 2 | 7 2 |
| | 10 4 5 4 | 8 2 | 5+6 2 |
| | 7 2 5 4 | 6+7 2 | 3+4 2 |
| | 6 2 5 4 | 9 2 | 3+4 2 |
| | 10 4 5 4 | 8 2 | 7 2 |
| | 10 4 5 4 | 6+7 2 | 7 2 |
| | 10 4 5 4 | 9 2 | 5+6 2 |
| | 6 2 5 4 | 9 2 | 7 2 |
| | 6 2 5 4 | 4+5 2 | 3+4 2 |
| | 10 4 5 4 | 9 2 | 5+6 2 |
| | 7 2 5 4 | 8 2 | 7 2 |
| | 7 2 5 4 | 9 2 | 3+4 2 |
| | 10 4 5 4 | 6+7 2 | 7 2 |
| | 10 4 5 4 | 9 2 | 7 2 |
| | 6 2 5 4 | 9 2 | 3+4 2 |
| | 9 2 5 4 | 8 2 | 5+6 2 |
| | 8 2 5 4 | 4+5 2 | 7 2 |
| | 11 4 5 4 | 6+7 2 | 7 2 |
| | 11 4 5 4 | 9 2 | 5+6 2 |
| | 9 2 5 4 | 6+7 2 | 3+4 2 |
| | 10 4 5 4 | 8 2 | 7 2 |
| | 10 4 5 4 | 9 2 | 7 2 |
| | 6 2 5 4 | 8 2 | 5+6 2 |
| | 7 2 5 4 | 4+5 2 | 3+4 2 |
| | 10 4 5 4 | 9 2 | 7 2 |
| | 8 2 5 4 | 8 2 | 5+6 2 |
| | 11 4 | 6+7 2 | 7 2 |
| | 11 4 | 9 2 | 3+4 2 |
| | 9 2 | 4+5 2 | 7 2 |
| | 11 4 | 8 2 | 5+6 2 |
| | 8 2 | 9 2 | 7 2 |
| | 11 4 | 8 2 | 3+4 2 |
| | 8 2 | 9 2 | 7 2 |
| | 10 4 | 6+7 2 | 5+6 2 |
| | 11 4 | 9 2 | 7 2 |
| | 7 2 | 8 2 | 7 2 |
| | 10 4 | 9 2 | 3+4 2 |
| | 7 2 | 4+5 2 | 7 2 |
| | 10 4 | 8 2 | 7 2 |
| | 6 2 | 9 2 | 3+4 2 |
| | | 6+7 2 | 5+6 2 |
| | | 9 2 | 5+6 2 |
| | | 4+5 2 | 7 2 |
| | | 8 2 | 3+4 2 |

Continued on next column

**TABLE LXXXVIII– continued**

| TQC | ICSD | NoE | SF |
|---|---|---|---|
| | | 6+7 2 | 7 2 |
| | | 9 2 | 5+6 2 |
| | | 9 2 | 7 2 |
| | | 9 2 | 7 2 |
| | | 8 2 | 3+4 2 |
| | | 8 2 | 5+6 2 |
| | | 6+7 2 | 7 2 |
| | | 8 2 | 3+4 2 |
| | | 4+5 2 | 7 2 |
| | | 8 2 | 5+6 2 |
| | | 4+5 2 | 7 2 |
| | | 9 2 | 3+4 2 |
| | | 6+7 2 | 7 2 |
| | | 9 2 | 5+6 2 |
| | | 9 2 | 7 2 |
| | | 6+7 2 | 3+4 2 |
| | | 9 2 | 7 2 |
| | | 8 2 | 7 2 |
| | | 8 2 | 5+6 2 |
| | | 4+5 2 | 3+4 2 |
| | | 8 2 | 7 2 |
| | | 4+5 2 | 7 2 |
| | | 9 2 | 5+6 2 |
| 227.1.76 | 155766 | 164 | Cd2O7Os2 |
| | -GM -X | -L | -W |
| | 6 2 5 4 | 8 2 | 7 2 |
| | 7 2 5 4 | 9 2 | 3+4 2 |
| | 10 4 5 4 | 8 2 | 5+6 2 |
| | 10 4 5 4 | 4+5 2 | 7 2 |
| | 11 4 5 4 | 8 2 | 5+6 2 |
| | 8 2 5 4 | 4+5 2 | 7 2 |
| | 11 4 5 4 | 6+7 2 | 3+4 2 |
| | 9 2 5 4 | 9 2 | 7 2 |
| | 6 2 5 4 | 8 2 | 7 2 |
| | 9 2 5 4 | 6+7 2 | 5+6 2 |
| | 10 4 5 4 | 9 2 | 3+4 2 |
| | 7 2 5 4 | 9 2 | 7 2 |
| | 10 4 5 4 | 9 2 | 5+6 2 |
| | 10 4 5 4 | 8 2 | 7 2 |
| | 6 2 5 4 | 4+5 2 | 7 2 |
| | 7 2 5 4 | 8 2 | 7 2 |
| | 10 4 5 4 | 8 2 | 5+6 2 |
| | 7 2 5 4 | 6+7 2 | 3+4 2 |
| | 6 2 5 4 | 9 2 | 3+4 2 |
| | 10 4 5 4 | 8 2 | 7 2 |
| | 10 4 5 4 | 6+7 2 | 7 2 |
| | 10 4 5 4 | 9 2 | 5+6 2 |
| | 6 2 5 4 | 9 2 | 7 2 |
| | 6 2 5 4 | 4+5 2 | 3+4 2 |
| | 10 4 5 4 | 9 2 | 5+6 2 |
| | 7 2 5 4 | 8 2 | 7 2 |
| | 7 2 5 4 | 9 2 | 3+4 2 |
| | 10 4 5 4 | 6+7 2 | 7 2 |
| | 10 4 5 4 | 9 2 | 7 2 |
| | 6 2 5 4 | 9 2 | 3+4 2 |
| | 9 2 5 4 | 8 2 | 5+6 2 |
| | 8 2 5 4 | 4+5 2 | 7 2 |
| | 11 4 5 4 | 6+7 2 | 7 2 |
| | 11 4 5 4 | 9 2 | 5+6 2 |
| | 9 2 5 4 | 6+7 2 | 3+4 2 |
| | 10 4 5 4 | 8 2 | 7 2 |

Continued on next column

**TABLE LXXXVIII– continued**

| TQC | ICSD | NoE | SF |
|---|---|---|---|
| | 10 4 5 4 | 9 2 | 7 2 |
| | 6 2 5 4 | 8 2 | 5+6 2 |
| | 7 2 5 4 | 4+5 2 | 3+4 2 |
| | 10 4 5 4 | 9 2 | 7 2 |
| | 8 2 5 4 | 8 2 | 5+6 2 |
| | 11 4 | 6+7 2 | 7 2 |
| | 11 4 | 9 2 | 3+4 2 |
| | 9 2 | 4+5 2 | 7 2 |
| | 11 4 | 8 2 | 5+6 2 |
| | 8 2 | 9 2 | 7 2 |
| | 11 4 | 8 2 | 3+4 2 |
| | 8 2 | 9 2 | 7 2 |
| | 10 4 | 6+7 2 | 5+6 2 |
| | 11 4 | 9 2 | 7 2 |
| | 7 2 | 8 2 | 7 2 |
| | 10 4 | 9 2 | 3+4 2 |
| | 7 2 | 4+5 2 | 7 2 |
| | 10 4 | 8 2 | 7 2 |
| | 6 2 | 9 2 | 3+4 2 |
| | | 6+7 2 | 5+6 2 |
| | | 9 2 | 5+6 2 |
| | | 4+5 2 | 7 2 |
| | | 8 2 | 3+4 2 |
| | | 6+7 2 | 7 2 |
| | | 9 2 | 5+6 2 |
| | | 9 2 | 7 2 |
| | | 9 2 | 7 2 |
| | | 8 2 | 3+4 2 |
| | | 8 2 | 5+6 2 |
| | | 6+7 2 | 7 2 |
| | | 8 2 | 3+4 2 |
| | | 4+5 2 | 7 2 |
| | | 8 2 | 5+6 2 |
| | | 4+5 2 | 7 2 |
| | | 9 2 | 3+4 2 |
| | | 6+7 2 | 7 2 |
| | | 9 2 | 5+6 2 |
| | | 9 2 | 7 2 |
| | | 6+7 2 | 3+4 2 |
| | | 9 2 | 7 2 |
| | | 8 2 | 7 2 |
| | | 8 2 | 5+6 2 |
| | | 4+5 2 | 3+4 2 |
| | | 8 2 | 7 2 |
| | | 4+5 2 | 7 2 |
| | | 9 2 | 5+6 2 |
| 227.1.77 | 155765 | 164 | Cd2O7Os2 |
| | -GM -X | -L | -W |
| | 6 2 5 4 | 8 2 | 7 2 |
| | 7 2 5 4 | 9 2 | 3+4 2 |
| | 10 4 5 4 | 8 2 | 5+6 2 |
| | 10 4 5 4 | 4+5 2 | 7 2 |
| | 11 4 5 4 | 8 2 | 5+6 2 |
| | 8 2 5 4 | 4+5 2 | 7 2 |
| | 11 4 5 4 | 6+7 2 | 3+4 2 |
| | 9 2 5 4 | 9 2 | 7 2 |
| | 6 2 5 4 | 8 2 | 7 2 |
| | 9 2 5 4 | 6+7 2 | 5+6 2 |
| | 10 4 5 4 | 9 2 | 3+4 2 |
| | 7 2 5 4 | 9 2 | 7 2 |
| | 10 4 5 4 | 9 2 | 5+6 2 |

Continued on next column

**TABLE LXXXVIII– continued**

| TQC | ICSD | NoE | SF |
|---|---|---|---|
| | 10 4 5 4 | 8 2 | 7 2 |
| | 6 2 5 4 | 4+5 2 | 7 2 |
| | 7 2 5 4 | 8 2 | 7 2 |
| | 10 4 5 4 | 8 2 | 5+6 2 |
| | 7 2 5 4 | 6+7 2 | 3+4 2 |
| | 6 2 5 4 | 9 2 | 3+4 2 |
| | 10 4 5 4 | 8 2 | 7 2 |
| | 10 4 5 4 | 6+7 2 | 7 2 |
| | 10 4 5 4 | 9 2 | 5+6 2 |
| | 6 2 5 4 | 9 2 | 7 2 |
| | 6 2 5 4 | 4+5 2 | 3+4 2 |
| | 10 4 5 4 | 9 2 | 5+6 2 |
| | 7 2 5 4 | 8 2 | 7 2 |
| | 7 2 5 4 | 9 2 | 3+4 2 |
| | 10 4 5 4 | 6+7 2 | 7 2 |
| | 10 4 5 4 | 9 2 | 7 2 |
| | 6 2 5 4 | 9 2 | 3+4 2 |
| | 9 2 5 4 | 8 2 | 5+6 2 |
| | 8 2 5 4 | 4+5 2 | 7 2 |
| | 11 4 5 4 | 6+7 2 | 7 2 |
| | 11 4 5 4 | 9 2 | 5+6 2 |
| | 9 2 5 4 | 6+7 2 | 3+4 2 |
| | 10 4 5 4 | 8 2 | 7 2 |
| | 10 4 5 4 | 9 2 | 7 2 |
| | 6 2 5 4 | 8 2 | 5+6 2 |
| | 7 2 5 4 | 4+5 2 | 3+4 2 |
| | 10 4 5 4 | 9 2 | 7 2 |
| | 8 2 5 4 | 8 2 | 5+6 2 |
| | 11 4 | 6+7 2 | 7 2 |
| | 11 4 | 9 2 | 3+4 2 |
| | 9 2 | 4+5 2 | 7 2 |
| | 11 4 | 8 2 | 5+6 2 |
| | 8 2 | 9 2 | 7 2 |
| | 11 4 | 8 2 | 3+4 2 |
| | 8 2 | 9 2 | 7 2 |
| | 10 4 | 6+7 2 | 5+6 2 |
| | 11 4 | 9 2 | 7 2 |
| | 7 2 | 8 2 | 7 2 |
| | 10 4 | 9 2 | 3+4 2 |
| | 7 2 | 4+5 2 | 7 2 |
| | 10 4 | 8 2 | 7 2 |
| | 6 2 | 9 2 | 3+4 2 |
| | | 6+7 2 | 5+6 2 |
| | | 9 2 | 5+6 2 |
| | | 4+5 2 | 7 2 |
| | | 8 2 | 3+4 2 |
| | | 6+7 2 | 7 2 |
| | | 9 2 | 5+6 2 |
| | | 9 2 | 7 2 |
| | | 9 2 | 7 2 |
| | | 8 2 | 3+4 2 |
| | | 8 2 | 5+6 2 |
| | | 6+7 2 | 7 2 |
| | | 8 2 | 3+4 2 |
| | | 4+5 2 | 7 2 |
| | | 8 2 | 5+6 2 |
| | | 4+5 2 | 7 2 |
| | | 9 2 | 3+4 2 |
| | | 6+7 2 | 7 2 |
| | | 9 2 | 5+6 2 |
| | | 9 2 | 7 2 |

Continued on next column

**TABLE LXXXVIII– continued**

| TQC | ICSD | | NoE | SF |
|---|---|---|---|---|
| | | | 6+7 2 | 3+4 2 |
| | | | 9 2 | 7 2 |
| | | | 8 2 | 7 2 |
| | | | 8 2 | 5+6 2 |
| | | | 4+5 2 | 3+4 2 |
| | | | 8 2 | 7 2 |
| | | | 4+5 2 | 7 2 |
| | | | 9 2 | 5+6 2 |
| 227.1.78 | 155764 | | 164 | Cd2O7Os2 |
| | -GM | -X | -L | -W |
| | 6 2 | 5 4 | 8 2 | 7 2 |
| | 7 2 | 5 4 | 9 2 | 3+4 2 |
| | 10 4 | 5 4 | 8 2 | 5+6 2 |
| | 10 4 | 5 4 | 4+5 2 | 7 2 |
| | 11 4 | 5 4 | 8 2 | 5+6 2 |
| | 8 2 | 5 4 | 4+5 2 | 7 2 |
| | 11 4 | 5 4 | 6+7 2 | 3+4 2 |
| | 9 2 | 5 4 | 9 2 | 7 2 |
| | 6 2 | 5 4 | 8 2 | 7 2 |
| | 9 2 | 5 4 | 6+7 2 | 5+6 2 |
| | 10 4 | 5 4 | 9 2 | 3+4 2 |
| | 7 2 | 5 4 | 9 2 | 7 2 |
| | 10 4 | 5 4 | 9 2 | 5+6 2 |
| | 10 4 | 5 4 | 8 2 | 7 2 |
| | 6 2 | 5 4 | 4+5 2 | 7 2 |
| | 7 2 | 5 4 | 8 2 | 7 2 |
| | 10 4 | 5 4 | 8 2 | 5+6 2 |
| | 7 2 | 5 4 | 6+7 2 | 3+4 2 |
| | 6 2 | 5 4 | 9 2 | 3+4 2 |
| | 10 4 | 5 4 | 8 2 | 7 2 |
| | 10 4 | 5 4 | 6+7 2 | 7 2 |
| | 10 4 | 5 4 | 9 2 | 5+6 2 |
| | 6 2 | 5 4 | 9 2 | 7 2 |
| | 6 2 | 5 4 | 9 2 | 3+4 2 |
| | 10 4 | 5 4 | 4+5 2 | 5+6 2 |
| | 7 2 | 5 4 | 8 2 | 7 2 |
| | 7 2 | 5 4 | 9 2 | 3+4 2 |
| | 10 4 | 5 4 | 6+7 2 | 7 2 |
| | 10 4 | 5 4 | 9 2 | 7 2 |
| | 6 2 | 5 4 | 9 2 | 3+4 2 |
| | 9 2 | 5 4 | 8 2 | 5+6 2 |
| | 8 2 | 5 4 | 4+5 2 | 7 2 |
| | 11 4 | 5 4 | 6+7 2 | 7 2 |
| | 11 4 | 5 4 | 9 2 | 5+6 2 |
| | 9 2 | 5 4 | 6+7 2 | 3+4 2 |
| | 10 4 | 5 4 | 8 2 | 7 2 |
| | 10 4 | 5 4 | 9 2 | 7 2 |
| | 6 2 | 5 4 | 8 2 | 5+6 2 |
| | 7 2 | 5 4 | 4+5 2 | 3+4 2 |
| | 10 4 | 5 4 | 9 2 | 7 2 |
| | 8 2 | 5 4 | 8 2 | 5+6 2 |
| | 11 4 | | 6+7 2 | 7 2 |
| | 11 4 | | 9 2 | 3+4 2 |
| | 9 2 | | 4+5 2 | 7 2 |
| | 11 4 | | 8 2 | 5+6 2 |
| | 8 2 | | 9 2 | 7 2 |
| | 11 4 | | 8 2 | 3+4 2 |
| | 8 2 | | 9 2 | 7 2 |
| | 10 4 | | 6+7 2 | 5+6 2 |
| | 11 4 | | 9 2 | 7 2 |
| | 7 2 | | 8 2 | 7 2 |

Continued on next column

**TABLE LXXXVIII– continued**

| TQC | ICSD | | NoE | SF |
|---|---|---|---|---|
| | 10 4 | | 9 2 | 3+4 2 |
| | 7 2 | | 4+5 2 | 7 2 |
| | 10 4 | | 8 2 | 7 2 |
| | 6 2 | | 9 2 | 3+4 2 |
| | | | 6+7 2 | 5+6 2 |
| | | | 9 2 | 5+6 2 |
| | | | 4+5 2 | 7 2 |
| | | | 8 2 | 3+4 2 |
| | | | 6+7 2 | 7 2 |
| | | | 9 2 | 5+6 2 |
| | | | 9 2 | 7 2 |
| | | | 9 2 | 7 2 |
| | | | 8 2 | 3+4 2 |
| | | | 8 2 | 5+6 2 |
| | | | 6+7 2 | 7 2 |
| | | | 8 2 | 3+4 2 |
| | | | 4+5 2 | 7 2 |
| | | | 8 2 | 5+6 2 |
| | | | 4+5 2 | 7 2 |
| | | | 9 2 | 3+4 2 |
| | | | 6+7 2 | 7 2 |
| | | | 9 2 | 5+6 2 |
| | | | 9 2 | 7 2 |
| | | | 6+7 2 | 3+4 2 |
| | | | 9 2 | 7 2 |
| | | | 8 2 | 7 2 |
| | | | 8 2 | 5+6 2 |
| | | | 4+5 2 | 3+4 2 |
| | | | 8 2 | 7 2 |
| | | | 4+5 2 | 7 2 |
| | | | 9 2 | 5+6 2 |
| 227.1.79 | 155763 | | 164 | Cd2O7Os2 |
| | -GM | -X | -L | -W |
| | 6 2 | 5 4 | 8 2 | 7 2 |
| | 7 2 | 5 4 | 9 2 | 3+4 2 |
| | 10 4 | 5 4 | 8 2 | 5+6 2 |
| | 10 4 | 5 4 | 4+5 2 | 7 2 |
| | 11 4 | 5 4 | 8 2 | 5+6 2 |
| | 8 2 | 5 4 | 4+5 2 | 7 2 |
| | 11 4 | 5 4 | 6+7 2 | 3+4 2 |
| | 9 2 | 5 4 | 9 2 | 7 2 |
| | 6 2 | 5 4 | 8 2 | 7 2 |
| | 9 2 | 5 4 | 6+7 2 | 5+6 2 |
| | 10 4 | 5 4 | 9 2 | 3+4 2 |
| | 7 2 | 5 4 | 9 2 | 7 2 |
| | 10 4 | 5 4 | 9 2 | 5+6 2 |
| | 10 4 | 5 4 | 8 2 | 7 2 |
| | 6 2 | 5 4 | 4+5 2 | 7 2 |
| | 7 2 | 5 4 | 8 2 | 7 2 |
| | 10 4 | 5 4 | 8 2 | 5+6 2 |
| | 7 2 | 5 4 | 6+7 2 | 3+4 2 |
| | 6 2 | 5 4 | 9 2 | 3+4 2 |
| | 10 4 | 5 4 | 8 2 | 7 2 |
| | 10 4 | 5 4 | 6+7 2 | 7 2 |
| | 10 4 | 5 4 | 9 2 | 5+6 2 |
| | 6 2 | 5 4 | 9 2 | 7 2 |
| | 6 2 | 5 4 | 9 2 | 3+4 2 |
| | 10 4 | 5 4 | 4+5 2 | 5+6 2 |
| | 7 2 | 5 4 | 8 2 | 7 2 |
| | 7 2 | 5 4 | 9 2 | 3+4 2 |
| | 10 4 | 5 4 | 6+7 2 | 7 2 |

Continued on next column

**TABLE LXXXVIII– continued**

| TQC | ICSD | NoE | SF |
|---|---|---|---|
| | 10 4 5 4 | 9 2 | 7 2 |
| | 6 2 5 4 | 9 2 | 3+4 2 |
| | 9 2 5 4 | 8 2 | 5+6 2 |
| | 8 2 5 4 | 4+5 2 | 7 2 |
| | 11 4 5 4 | 6+7 2 | 7 2 |
| | 11 4 5 4 | 9 2 | 5+6 2 |
| | 9 2 5 4 | 6+7 2 | 3+4 2 |
| | 10 4 5 4 | 8 2 | 7 2 |
| | 10 4 5 4 | 9 2 | 7 2 |
| | 6 2 5 4 | 8 2 | 5+6 2 |
| | 7 2 5 4 | 4+5 2 | 3+4 2 |
| | 10 4 5 4 | 9 2 | 7 2 |
| | 8 2 5 4 | 8 2 | 5+6 2 |
| | 11 4 | 6+7 2 | 7 2 |
| | 11 4 | 9 2 | 3+4 2 |
| | 9 2 | 4+5 2 | 7 2 |
| | 11 4 | 8 2 | 5+6 2 |
| | 8 2 | 9 2 | 7 2 |
| | 11 4 | 8 2 | 3+4 2 |
| | 8 2 | 9 2 | 7 2 |
| | 10 4 | 6+7 2 | 5+6 2 |
| | 11 4 | 9 2 | 7 2 |
| | 7 2 | 8 2 | 7 2 |
| | 10 4 | 9 2 | 3+4 2 |
| | 7 2 | 4+5 2 | 7 2 |
| | 10 4 | 8 2 | 7 2 |
| | 6 2 | 9 2 | 3+4 2 |
| | | 6+7 2 | 5+6 2 |
| | | 9 2 | 5+6 2 |
| | | 4+5 2 | 7 2 |
| | | 8 2 | 3+4 2 |
| | | 6+7 2 | 7 2 |
| | | 9 2 | 5+6 2 |
| | | 9 2 | 7 2 |
| | | 9 2 | 7 2 |
| | | 8 2 | 3+4 2 |
| | | 8 2 | 5+6 2 |
| | | 6+7 2 | 7 2 |
| | | 8 2 | 3+4 2 |
| | | 4+5 2 | 7 2 |
| | | 8 2 | 5+6 2 |
| | | 4+5 2 | 7 2 |
| | | 9 2 | 3+4 2 |
| | | 6+7 2 | 7 2 |
| | | 9 2 | 5+6 2 |
| | | 9 2 | 7 2 |
| | | 6+7 2 | 3+4 2 |
| | | 9 2 | 7 2 |
| | | 8 2 | 7 2 |
| | | 8 2 | 5+6 2 |
| | | 4+5 2 | 3+4 2 |
| | | 8 2 | 7 2 |
| | | 4+5 2 | 7 2 |
| | | 9 2 | 5+6 2 |
| 227.1.80 | 155762 | 164 | Cd2O7Os2 |
| | -GM -X | -L | -W |
| | 6 2 5 4 | 8 2 | 7 2 |
| | 7 2 5 4 | 9 2 | 3+4 2 |
| | 10 4 5 4 | 8 2 | 5+6 2 |
| | 10 4 5 4 | 4+5 2 | 7 2 |
| | 11 4 5 4 | 8 2 | 5+6 2 |

Continued on next column

**TABLE LXXXVIII– continued**

| TQC | ICSD | NoE | SF |
|---|---|---|---|
| | 8 2 5 4 | 4+5 2 | 7 2 |
| | 11 4 5 4 | 6+7 2 | 3+4 2 |
| | 9 2 5 4 | 9 2 | 7 2 |
| | 6 2 5 4 | 8 2 | 7 2 |
| | 9 2 5 4 | 6+7 2 | 5+6 2 |
| | 10 4 5 4 | 9 2 | 3+4 2 |
| | 7 2 5 4 | 9 2 | 7 2 |
| | 10 4 5 4 | 9 2 | 5+6 2 |
| | 10 4 5 4 | 8 2 | 7 2 |
| | 6 2 5 4 | 4+5 2 | 7 2 |
| | 7 2 5 4 | 8 2 | 7 2 |
| | 10 4 5 4 | 8 2 | 5+6 2 |
| | 7 2 5 4 | 6+7 2 | 3+4 2 |
| | 6 2 5 4 | 9 2 | 3+4 2 |
| | 10 4 5 4 | 8 2 | 7 2 |
| | 10 4 5 4 | 6+7 2 | 7 2 |
| | 10 4 5 4 | 9 2 | 5+6 2 |
| | 6 2 5 4 | 9 2 | 7 2 |
| | 6 2 5 4 | 9 2 | 3+4 2 |
| | 10 4 5 4 | 4+5 2 | 5+6 2 |
| | 7 2 5 4 | 8 2 | 7 2 |
| | 7 2 5 4 | 9 2 | 3+4 2 |
| | 10 4 5 4 | 6+7 2 | 7 2 |
| | 10 4 5 4 | 9 2 | 7 2 |
| | 6 2 5 4 | 9 2 | 3+4 2 |
| | 9 2 5 4 | 8 2 | 5+6 2 |
| | 8 2 5 4 | 4+5 2 | 7 2 |
| | 11 4 5 4 | 6+7 2 | 7 2 |
| | 11 4 5 4 | 9 2 | 5+6 2 |
| | 9 2 5 4 | 6+7 2 | 3+4 2 |
| | 10 4 5 4 | 8 2 | 7 2 |
| | 10 4 5 4 | 9 2 | 7 2 |
| | 6 2 5 4 | 8 2 | 5+6 2 |
| | 7 2 5 4 | 4+5 2 | 3+4 2 |
| | 10 4 5 4 | 9 2 | 7 2 |
| | 8 2 5 4 | 8 2 | 5+6 2 |
| | 11 4 | 6+7 2 | 7 2 |
| | 11 4 | 9 2 | 3+4 2 |
| | 9 2 | 4+5 2 | 7 2 |
| | 11 4 | 8 2 | 5+6 2 |
| | 8 2 | 9 2 | 7 2 |
| | 11 4 | 8 2 | 3+4 2 |
| | 8 2 | 9 2 | 7 2 |
| | 10 4 | 6+7 2 | 5+6 2 |
| | 11 4 | 9 2 | 7 2 |
| | 7 2 | 8 2 | 7 2 |
| | 10 4 | 9 2 | 3+4 2 |
| | 7 2 | 4+5 2 | 7 2 |
| | 10 4 | 8 2 | 7 2 |
| | 6 2 | 9 2 | 3+4 2 |
| | | 6+7 2 | 5+6 2 |
| | | 9 2 | 5+6 2 |
| | | 4+5 2 | 7 2 |
| | | 8 2 | 3+4 2 |
| | | 6+7 2 | 7 2 |
| | | 9 2 | 5+6 2 |
| | | 9 2 | 7 2 |
| | | 9 2 | 7 2 |
| | | 8 2 | 3+4 2 |
| | | 8 2 | 5+6 2 |
| | | 6+7 2 | 7 2 |

Continued on next column

**TABLE LXXXVIII– continued**

| TQC | ICSD | NoE | SF |
|---|---|---|---|
| | | 8 2 | 3+4 2 |
| | | 4+5 2 | 7 2 |
| | | 8 2 | 5+6 2 |
| | | 4+5 2 | 7 2 |
| | | 9 2 | 3+4 2 |
| | | 6+7 2 | 7 2 |
| | | 9 2 | 5+6 2 |
| | | 9 2 | 7 2 |
| | | 6+7 2 | 3+4 2 |
| | | 9 2 | 7 2 |
| | | 8 2 | 7 2 |
| | | 8 2 | 5+6 2 |
| | | 4+5 2 | 3+4 2 |
| | | 8 2 | 7 2 |
| | | 4+5 2 | 7 2 |
| | | 9 2 | 5+6 2 |
| 227.1.81 | 155761 | 164 | Cd2O7Os2 |
| | -GM -X | -L | -W |
| | 6 2 5 4 | 8 2 | 7 2 |
| | 7 2 5 4 | 9 2 | 3+4 2 |
| | 10 4 5 4 | 8 2 | 5+6 2 |
| | 10 4 5 4 | 4+5 2 | 7 2 |
| | 11 4 5 4 | 8 2 | 5+6 2 |
| | 8 2 5 4 | 4+5 2 | 7 2 |
| | 11 4 5 4 | 6+7 2 | 3+4 2 |
| | 9 2 5 4 | 9 2 | 7 2 |
| | 6 2 5 4 | 8 2 | 7 2 |
| | 9 2 5 4 | 6+7 2 | 5+6 2 |
| | 10 4 5 4 | 9 2 | 3+4 2 |
| | 7 2 5 4 | 9 2 | 7 2 |
| | 10 4 5 4 | 9 2 | 5+6 2 |
| | 10 4 5 4 | 8 2 | 7 2 |
| | 6 2 5 4 | 4+5 2 | 7 2 |
| | 7 2 5 4 | 8 2 | 7 2 |
| | 10 4 5 4 | 8 2 | 5+6 2 |
| | 7 2 5 4 | 6+7 2 | 3+4 2 |
| | 6 2 5 4 | 9 2 | 3+4 2 |
| | 10 4 5 4 | 8 2 | 7 2 |
| | 10 4 5 4 | 6+7 2 | 7 2 |
| | 10 4 5 4 | 9 2 | 5+6 2 |
| | 6 2 5 4 | 9 2 | 7 2 |
| | 6 2 5 4 | 9 2 | 3+4 2 |
| | 10 4 5 4 | 4+5 2 | 5+6 2 |
| | 7 2 5 4 | 8 2 | 7 2 |
| | 7 2 5 4 | 9 2 | 3+4 2 |
| | 10 4 5 4 | 6+7 2 | 7 2 |
| | 10 4 5 4 | 9 2 | 7 2 |
| | 6 2 5 4 | 9 2 | 3+4 2 |
| | 9 2 5 4 | 8 2 | 5+6 2 |
| | 8 2 5 4 | 4+5 2 | 7 2 |
| | 11 4 5 4 | 6+7 2 | 7 2 |
| | 11 4 5 4 | 9 2 | 5+6 2 |
| | 9 2 5 4 | 6+7 2 | 3+4 2 |
| | 10 4 5 4 | 8 2 | 7 2 |
| | 10 4 5 4 | 9 2 | 7 2 |
| | 6 2 5 4 | 8 2 | 5+6 2 |
| | 7 2 5 4 | 4+5 2 | 3+4 2 |
| | 10 4 5 4 | 9 2 | 7 2 |
| | 8 2 5 4 | 8 2 | 5+6 2 |
| | 11 4 | 6+7 2 | 7 2 |
| | 11 4 | 9 2 | 3+4 2 |

Continued on next column

**TABLE LXXXVIII– continued**

| TQC | ICSD | NoE | SF |
|---|---|---|---|
| | 9 2 | 4+5 2 | 7 2 |
| | 11 4 | 8 2 | 5+6 2 |
| | 8 2 | 9 2 | 7 2 |
| | 11 4 | 8 2 | 3+4 2 |
| | 8 2 | 9 2 | 7 2 |
| | 10 4 | 6+7 2 | 5+6 2 |
| | 11 4 | 9 2 | 7 2 |
| | 7 2 | 8 2 | 7 2 |
| | 10 4 | 9 2 | 3+4 2 |
| | 7 2 | 4+5 2 | 7 2 |
| | 10 4 | 8 2 | 3+4 2 |
| | 6 2 | 9 2 | 7 2 |
| | | 6+7 2 | 5+6 2 |
| | | 9 2 | 5+6 2 |
| | | 4+5 2 | 7 2 |
| | | 8 2 | 3+4 2 |
| | | 6+7 2 | 7 2 |
| | | 9 2 | 5+6 2 |
| | | 9 2 | 7 2 |
| | | 9 2 | 7 2 |
| | | 8 2 | 3+4 2 |
| | | 8 2 | 5+6 2 |
| | | 6+7 2 | 7 2 |
| | | 8 2 | 3+4 2 |
| | | 4+5 2 | 7 2 |
| | | 8 2 | 5+6 2 |
| | | 4+5 2 | 7 2 |
| | | 9 2 | 3+4 2 |
| | | 6+7 2 | 7 2 |
| | | 9 2 | 5+6 2 |
| | | 9 2 | 7 2 |
| | | 6+7 2 | 3+4 2 |
| | | 9 2 | 7 2 |
| | | 8 2 | 7 2 |
| | | 8 2 | 5+6 2 |
| | | 4+5 2 | 3+4 2 |
| | | 8 2 | 7 2 |
| | | 4+5 2 | 7 2 |
| | | 9 2 | 5+6 2 |
| 227.1.82 | 37408 | 168 | Cd1Se4Yb2 |
| | -GM -X | -L | -W |
| | 6 2 5 4 | 8 2 | 3+4 2 |
| | 7 2 5 4 | 9 2 | 5+6 2 |
| | 10 4 5 4 | 6+7 2 | 7 2 |
| | 9 2 5 4 | 9 2 | 7 2 |
| | 11 4 5 4 | 4+5 2 | 5+6 2 |
| | 8 2 5 4 | 8 2 | 7 2 |
| | 9 2 5 4 | 9 2 | 3+4 2 |
| | 8 2 5 4 | 8 2 | 7 2 |
| | 11 4 5 4 | 8 2 | 7 2 |
| | 11 4 5 4 | 8 2 | 5+6 2 |
| | 11 4 5 4 | 9 2 | 3+4 2 |
| | 6 2 5 4 | 4+5 2 | 7 2 |
| | 10 4 5 4 | 8 2 | 5+6 2 |
| | 7 2 5 4 | 6+7 2 | 3+4 2 |
| | 9 2 5 4 | 8 2 | 7 2 |
| | 11 4 5 4 | 4+5 2 | 7 2 |
| | 8 2 5 4 | 9 2 | 7 2 |
| | 10 4 5 4 | 8 2 | 5+6 2 |
| | 11 4 5 4 | 9 2 | 3+4 2 |
| | 7 2 5 4 | 9 2 | 7 2 |

Continued on next column

**TABLE LXXXVIII– continued**

| TQC | ICSD | NoE | SF |
|---|---|---|---|
| | 10 4 5 4 | 6+7 2 | 7 2 |
| | 11 4 5 4 | 4+5 2 | 3+4 2 |
| | 8 2 5 4 | 8 2 | 5+6 2 |
| | 6 2 5 4 | 8 2 | 7 2 |
| | 9 2 5 4 | 9 2 | 7 2 |
| | 10 4 5 4 | 4+5 2 | 5+6 2 |
| | 7 2 5 4 | 6+7 2 | 3+4 2 |
| | 11 4 5 4 | 8 2 | 7 2 |
| | 8 2 5 4 | 9 2 | 5+6 2 |
| | 11 4 5 4 | 8 2 | 7 2 |
| | 11 4 5 4 | 6+7 2 | 7 2 |
| | 11 4 5 4 | 4+5 2 | 3+4 2 |
| | 9 2 5 4 | 8 2 | 5+6 2 |
| | 11 4 5 4 | 9 2 | 7 2 |
| | 8 2 5 4 | 9 2 | 7 2 |
| | 9 2 5 4 | 8 2 | 3+4 2 |
| | 11 4 5 4 | 9 2 | 7 2 |
| | 7 2 5 4 | 6+7 2 | 5+6 2 |
| | 10 4 5 4 | 9 2 | 3+4 2 |
| | 6 2 5 4 | 4+5 2 | 7 2 |
| | 9 2 5 4 | 8 2 | 7 2 |
| | 8 2 5 4 | 8 2 | 5+6 2 |
| | 10 4 | 9 2 | 3+4 2 |
| | 10 4 | 9 2 | 5+6 2 |
| | 8 2 | 6+7 2 | 7 2 |
| | 11 4 | 8 2 | 7 2 |
| | 11 4 | 4+5 2 | 7 2 |
| | 11 4 | 8 2 | 3+4 2 |
| | 9 2 | 4+5 2 | 5+6 2 |
| | 8 2 | 6+7 2 | 3+4 2 |
| | 11 4 | 8 2 | 7 2 |
| | 9 2 | 8 2 | 7 2 |
| | 8 2 | 9 2 | 5+6 2 |
| | 11 4 | 4+5 2 | 7 2 |
| | 11 4 | 8 2 | 7 2 |
| | 9 2 | 8 2 | 7 2 |
| | 8 2 | 9 2 | 3+4 2 |
| | | 8 2 | 5+6 2 |
| | | 4+5 2 | 7 2 |
| | | 8 2 | 5+6 2 |
| | | 9 2 | 3+4 2 |
| | | 6+7 2 | 7 2 |
| | | 9 2 | 7 2 |
| | | 8 2 | 3+4 2 |
| | | 4+5 2 | 7 2 |
| | | 9 2 | 5+6 2 |
| | | 8 2 | 7 2 |
| | | 8 2 | 5+6 2 |
| | | 6+7 2 | 3+4 2 |
| | | 4+5 2 | 7 2 |
| | | 4+5 2 | 7 2 |
| | | 8 2 | 5+6 2 |
| | | 8 2 | 3+4 2 |
| | | 9 2 | 7 2 |
| | | 8 2 | 5+6 2 |
| | | 4+5 2 | 7 2 |
| | | 8 2 | 3+4 2 |
| | | 8 2 | 7 2 |
| | | 9 2 | 7 2 |
| | | 4+5 2 | 5+6 2 |
| | | 8 2 | 3+4 2 |

Continued on next column

**TABLE LXXXVIII– continued**

| TQC | ICSD | NoE | SF |
|---|---|---|---|
| | | 9 2 | 7 2 |
| | | 6+7 2 | 7 2 |
| | | 8 2 | 5+6 2 |
| 227.1.83 | 188292 | 264 | Cd1Ni1 |
| | -GM -X | -L | -W |
| | 6 2 5 4 | 9 2 | 5+6 2 |
| | 7 2 5 4 | 6+7 2 | 7 2 |
| | 11 4 5 4 | 8 2 | 7 2 |
| | 10 4 5 4 | 9 2 | 3+4 2 |
| | 9 2 5 4 | 4+5 2 | 5+6 2 |
| | 10 4 5 4 | 8 2 | 7 2 |
| | 11 4 5 4 | 9 2 | 5+6 2 |
| | 10 4 5 4 | 6+7 2 | 3+4 2 |
| | 10 4 5 4 | 9 2 | 7 2 |
| | 8 2 5 4 | 8 2 | 7 2 |
| | 6 2 5 4 | 8 2 | 3+4 2 |
| | 11 4 5 4 | 4+5 2 | 7 2 |
| | 9 2 5 4 | 9 2 | 5+6 2 |
| | 7 2 5 4 | 9 2 | 7 2 |
| | 8 2 5 4 | 6+7 2 | 7 2 |
| | 11 4 5 4 | 4+5 2 | 3+4 2 |
| | 6 2 5 4 | 8 2 | 7 2 |
| | 8 2 5 4 | 8 2 | 3+4 2 |
| | 10 4 5 4 | 9 2 | 5+6 2 |
| | 6 2 5 4 | 8 2 | 7 2 |
| | 7 2 5 4 | 6+7 2 | 7 2 |
| | 9 2 5 4 | 9 2 | 3+4 2 |
| | 11 4 5 4 | 4+5 2 | 7 2 |
| | 8 2 5 4 | 8 2 | 5+6 2 |
| | 10 4 5 4 | 4+5 2 | 7 2 |
| | 11 4 5 4 | 8 2 | 3+4 2 |
| | 8 2 5 4 | 9 2 | 7 2 |
| | 9 2 5 4 | 9 2 | 5+6 2 |
| | 10 4 5 4 | 8 2 | 3+4 2 |
| | 11 4 5 4 | 8 2 | 7 2 |
| | 11 4 5 4 | 6+7 2 | 7 2 |
| | 10 4 5 4 | 4+5 2 | 5+6 2 |
| | 7 2 5 4 | 9 2 | 3+4 2 |
| | 6 2 5 4 | 8 2 | 7 2 |
| | 11 4 5 4 | 9 2 | 5+6 2 |
| | 9 2 5 4 | 6+7 2 | 7 2 |
| | 6 2 5 4 | 9 2 | 7 2 |
| | 10 4 5 4 | 8 2 | 3+4 2 |
| | 11 4 5 4 | 8 2 | 5+6 2 |
| | 10 4 5 4 | 8 2 | 7 2 |
| | 7 2 5 4 | 9 2 | 7 2 |
| | 11 4 5 4 | 4+5 2 | 3+4 2 |
| | 8 2 5 4 | 6+7 2 | 7 2 |
| | 7 2 5 4 | 4+5 2 | 5+6 2 |
| | 10 4 5 4 | 9 2 | 7 2 |
| | 6 2 5 4 | 8 2 | 5+6 2 |
| | 9 2 5 4 | 8 2 | 3+4 2 |
| | 10 4 5 4 | 9 2 | 7 2 |
| | 7 2 5 4 | 6+7 2 | 5+6 2 |
| | 11 4 5 4 | 9 2 | 7 2 |
| | 6 2 5 4 | 8 2 | 3+4 2 |
| | 10 4 5 4 | 4+5 2 | 7 2 |
| | 9 2 5 4 | 9 2 | 7 2 |
| | 7 2 5 4 | 8 2 | 3+4 2 |
| | 10 4 5 4 | 6+7 2 | 5+6 2 |
| | 10 4 5 4 | 4+5 2 | 7 2 |

Continued on next column

**TABLE LXXXVIII– continued**

| TQC | ICSD | | NoE | SF |
|---|---|---|---|---|
| | 7 2 | 5 4 | 6+7 2 | 5+6 2 |
| | 8 2 | 5 4 | 9 2 | 7 2 |
| | 11 4 | 5 4 | 8 2 | 3+4 2 |
| | 10 4 | 5 4 | 8 2 | 7 2 |
| | 6 2 | 5 4 | 9 2 | 5+6 2 |
| | 9 2 | 5 4 | 9 2 | 7 2 |
| | 10 4 | 5 4 | 9 2 | 7 2 |
| | 11 4 | 5 4 | 6+7 2 | 3+4 2 |
| | 10 4 | 5 4 | 4+5 2 | 3+4 2 |
| | 11 4 | 5 4 | 8 2 | 7 2 |
| | 10 4 | | 8 2 | 5+6 2 |
| | 8 2 | | 9 2 | 7 2 |
| | 11 4 | | 6+7 2 | 3+4 2 |
| | 6 2 | | 9 2 | 7 2 |
| | 7 2 | | 8 2 | 5+6 2 |
| | 9 2 | | 4+5 2 | 7 2 |
| | 11 4 | | 8 2 | 7 2 |
| | 11 4 | | 9 2 | 3+4 2 |
| | 9 2 | | 4+5 2 | 5+6 2 |
| | 8 2 | | 8 2 | 7 2 |
| | 6 2 | | 9 2 | 3+4 2 |
| | 10 4 | | 8 2 | 7 2 |
| | 11 4 | | 4+5 2 | 7 2 |
| | 10 4 | | 8 2 | 5+6 2 |
| | 6 2 | | 9 2 | 7 2 |
| | 10 4 | | 8 2 | 5+6 2 |
| | 7 2 | | 6+7 2 | 5+6 2 |
| | 7 2 | | 9 2 | 7 2 |
| | 10 4 | | 8 2 | 3+4 2 |
| | 10 4 | | 8 2 | 7 2 |
| | 10 4 | | 4+5 2 | 7 2 |
| | 6 2 | | 9 2 | 3+4 2 |
| | | | 6+7 2 | 5+6 2 |
| | | | 8 2 | 7 2 |
| | | | 4+5 2 | 7 2 |
| | | | 8 2 | 3+4 2 |
| | | | 6+7 2 | 7 2 |
| | | | 9 2 | 5+6 2 |
| | | | 4+5 2 | 3+4 2 |
| | | | 8 2 | 7 2 |
| | | | 8 2 | 5+6 2 |
| | | | 9 2 | 7 2 |
| | | | 8 2 | 7 2 |
| | | | 6+7 2 | 3+4 2 |
| | | | 9 2 | 5+6 2 |
| | | | 9 2 | 7 2 |
| | | | 6+7 2 | 7 2 |
| | | | 4+5 2 | 3+4 2 |
| | | | 9 2 | 7 2 |
| | | | 8 2 | 3+4 2 |
| | | | 9 2 | 5+6 2 |
| | | | 6+7 2 | 7 2 |
| | | | 4+5 2 | 7 2 |
| | | | 8 2 | 3+4 2 |
| | | | 6+7 2 | 7 2 |
| | | | 8 2 | 5+6 2 |
| | | | 9 2 | 7 2 |
| | | | 9 2 | 3+4 2 |
| | | | 9 2 | 5+6 2 |
| | | | 8 2 | 7 2 |
| | | | 6+7 2 | 3+4 2 |

Continued on next column

**TABLE LXXXVIII– continued**

| TQC | ICSD | | NoE | SF |
|---|---|---|---|---|
| | | | 4+5 2 | 7 2 |
| | | | 8 2 | 3+4 2 |
| | | | 9 2 | 7 2 |
| | | | 8 2 | 5+6 2 |
| | | | 4+5 2 | 7 2 |
| | | | 8 2 | 7 2 |
| | | | 9 2 | 5+6 2 |
| | | | 8 2 | 7 2 |
| | | | 4+5 2 | 3+4 2 |
| | | | 8 2 | 5+6 2 |
| | | | 4+5 2 | 7 2 |
| | | | 8 2 | 3+4 2 |
| | | | 9 2 | 7 2 |
| | | | 6+7 2 | 5+6 2 |
| | | | 8 2 | 7 2 |
| 227.1.84 | 246498 | | 160 | Cd1Er2S4 |
| | -GM | -X | -L | -W |
| | 6 2 | 5 4 | 9 2 | 5+6 2 |
| | 7 2 | 5 4 | 6+7 2 | 7 2 |
| | 10 4 | 5 4 | 8 2 | 3+4 2 |
| | 9 2 | 5 4 | 9 2 | 7 2 |
| | 11 4 | 5 4 | 8 2 | 5+6 2 |
| | 8 2 | 5 4 | 4+5 2 | 7 2 |
| | 9 2 | 5 4 | 9 2 | 3+4 2 |
| | 8 2 | 5 4 | 8 2 | 7 2 |
| | 11 4 | 5 4 | 8 2 | 7 2 |
| | 11 4 | 5 4 | 8 2 | 5+6 2 |
| | 11 4 | 5 4 | 9 2 | 3+4 2 |
| | 6 2 | 5 4 | 4+5 2 | 7 2 |
| | 10 4 | 5 4 | 8 2 | 5+6 2 |
| | 7 2 | 5 4 | 6+7 2 | 3+4 2 |
| | 11 4 | 5 4 | 8 2 | 7 2 |
| | 9 2 | 5 4 | 4+5 2 | 7 2 |
| | 8 2 | 5 4 | 9 2 | 7 2 |
| | 10 4 | 5 4 | 8 2 | 5+6 2 |
| | 11 4 | 5 4 | 9 2 | 3+4 2 |
| | 7 2 | 5 4 | 9 2 | 7 2 |
| | 10 4 | 5 4 | 6+7 2 | 7 2 |
| | 11 4 | 5 4 | 4+5 2 | 3+4 2 |
| | 8 2 | 5 4 | 8 2 | 5+6 2 |
| | 6 2 | 5 4 | 8 2 | 7 2 |
| | 9 2 | 5 4 | 9 2 | 7 2 |
| | 7 2 | 5 4 | 4+5 2 | 5+6 2 |
| | 10 4 | 5 4 | 6+7 2 | 3+4 2 |
| | 11 4 | 5 4 | 8 2 | 7 2 |
| | 8 2 | 5 4 | 9 2 | 5+6 2 |
| | 11 4 | 5 4 | 8 2 | 7 2 |
| | 10 4 | 5 4 | 6+7 2 | 7 2 |
| | 7 2 | 5 4 | 9 2 | 3+4 2 |
| | 6 2 | 5 4 | 4+5 2 | 7 2 |
| | 10 4 | 5 4 | 8 2 | 5+6 2 |
| | 10 4 | 5 4 | 9 2 | 7 2 |
| | 11 4 | 5 4 | 8 2 | 3+4 2 |
| | 11 4 | 5 4 | 9 2 | 5+6 2 |
| | 9 2 | 5 4 | 6+7 2 | 7 2 |
| | 11 4 | 5 4 | 9 2 | 3+4 2 |
| | 9 2 | 5 4 | 8 2 | 7 2 |
| | 8 2 | | 4+5 2 | 5+6 2 |
| | 11 4 | | 8 2 | 7 2 |
| | 9 2 | | 9 2 | 7 2 |
| | 8 2 | | 6+7 2 | 3+4 2 |

Continued on next column

**TABLE LXXXVIII– continued**

| TQC | ICSD | NoE | SF |
|---|---|---|---|
| | 8 2 | 9 2 | 5+6 2 |
| | 11 4 | 4+5 2 | 7 2 |
| | 11 4 | 8 2 | 7 2 |
| | 9 2 | 9 2 | 3+4 2 |
| | 11 4 | 9 2 | 5+6 2 |
| | 11 4 | 8 2 | 7 2 |
| | 8 2 | 6+7 2 | 3+4 2 |
| | 9 2 | 4+5 2 | 7 2 |
| | 8 2 | 4+5 2 | 5+6 2 |
| | 11 4 | 8 2 | 7 2 |
| | | 6+7 2 | 7 2 |
| | | 8 2 | 3+4 2 |
| | | 9 2 | 5+6 2 |
| | | 8 2 | 7 2 |
| | | 8 2 | 7 2 |
| | | 8 2 | 3+4 2 |
| | | 4+5 2 | 5+6 2 |
| | | 8 2 | 7 2 |
| | | 9 2 | 7 2 |
| | | 4+5 2 | 5+6 2 |
| | | 8 2 | 3+4 2 |
| | | 9 2 | 7 2 |
| | | 8 2 | 7 2 |
| | | 6+7 2 | 5+6 2 |
| | | 8 2 | 3+4 2 |
| | | 4+5 2 | 7 2 |
| | | 4+5 2 | 7 2 |
| | | 8 2 | 3+4 2 |
| | | 8 2 | 5+6 2 |
| | | 8 2 | 7 2 |
| | | 9 2 | 5+6 2 |
| | | 4+5 2 | 7 2 |
| | | 8 2 | 7 2 |
| | | 9 2 | 3+4 2 |
| | | 8 2 | 5+6 2 |
| | | 6+7 2 | 7 2 |
| 227.1.85 | 78554 | 96 | Cd1Cr2Se4 |
| | -GM -X | -L | -W |
| | 6 2 5 4 | 9 2 | 5+6 2 |
| | 9 2 5 4 | 8 2 | 7 2 |
| | 11 4 5 4 | 8 2 | 3+4 2 |
| | 8 2 5 4 | 6+7 2 | 7 2 |
| | 10 4 5 4 | 9 2 | 5+6 2 |
| | 7 2 5 4 | 4+5 2 | 7 2 |
| | 10 4 5 4 | 8 2 | 3+4 2 |
| | 11 4 5 4 | 9 2 | 7 2 |
| | 7 2 5 4 | 9 2 | 7 2 |
| | 11 4 5 4 | 8 2 | 5+6 2 |
| | 10 4 5 4 | 4+5 2 | 3+4 2 |
| | 8 2 5 4 | 6+7 2 | 7 2 |
| | 6 2 5 4 | 9 2 | 5+6 2 |
| | 9 2 5 4 | 8 2 | 7 2 |
| | 10 4 5 4 | 6+7 2 | 7 2 |
| | 7 2 5 4 | 8 2 | 5+6 2 |
| | 11 4 5 4 | 4+5 2 | 3+4 2 |
| | 10 4 5 4 | 9 2 | 7 2 |
| | 6 2 5 4 | 9 2 | 7 2 |
| | 11 4 5 4 | 8 2 | 3+4 2 |
| | 10 4 5 4 | 9 2 | 7 2 |
| | 8 2 5 4 | 6+7 2 | 5+6 2 |
| | 7 2 5 4 | 4+5 2 | 3+4 2 |
| | 10 4 5 4 | 8 2 | 7 2 |
| | 9 2 | 9 2 | 5+6 2 |
| | 11 4 | 8 2 | 7 2 |
| | 6 2 | 9 2 | 3+4 2 |
| | 8 2 | 6+7 2 | 7 2 |
| | 10 4 | 8 2 | 5+6 2 |
| | 11 4 | 9 2 | 7 2 |
| | 7 2 | 8 2 | 5+6 2 |
| | 10 4 | 4+5 2 | 7 2 |
| | | 9 2 | 7 2 |
| | | 6+7 2 | 3+4 2 |
| | | 9 2 | 3+4 2 |
| | | 8 2 | 7 2 |
| | | 4+5 2 | 5+6 2 |
| | | 9 2 | 7 2 |
| | | 6+7 2 | 7 2 |
| | | 8 2 | 5+6 2 |
| | | 8 2 | 3+4 2 |
| | | 4+5 2 | 7 2 |
| | | 9 2 | 7 2 |
| | | 6+7 2 | 3+4 2 |
| | | 8 2 | 5+6 2 |
| | | 9 2 | 7 2 |
| | | 9 2 | 3+4 2 |
| | | 6+7 2 | 7 2 |
| 227.1.86 | 39416 | 96 | Cd1Cr2Se4 |
| | -GM -X | -L | -W |
| | 6 2 5 4 | 9 2 | 5+6 2 |
| | 9 2 5 4 | 8 2 | 7 2 |
| | 11 4 5 4 | 8 2 | 3+4 2 |
| | 8 2 5 4 | 6+7 2 | 7 2 |
| | 10 4 5 4 | 9 2 | 5+6 2 |
| | 7 2 5 4 | 4+5 2 | 7 2 |
| | 10 4 5 4 | 8 2 | 3+4 2 |
| | 11 4 5 4 | 9 2 | 7 2 |
| | 7 2 5 4 | 9 2 | 7 2 |
| | 11 4 5 4 | 8 2 | 5+6 2 |
| | 10 4 5 4 | 4+5 2 | 3+4 2 |
| | 8 2 5 4 | 6+7 2 | 7 2 |
| | 6 2 5 4 | 9 2 | 5+6 2 |
| | 9 2 5 4 | 8 2 | 7 2 |
| | 10 4 5 4 | 6+7 2 | 7 2 |
| | 7 2 5 4 | 8 2 | 5+6 2 |
| | 11 4 5 4 | 4+5 2 | 3+4 2 |
| | 10 4 5 4 | 9 2 | 7 2 |
| | 6 2 5 4 | 9 2 | 7 2 |
| | 11 4 5 4 | 8 2 | 3+4 2 |
| | 10 4 5 4 | 9 2 | 7 2 |
| | 8 2 5 4 | 6+7 2 | 5+6 2 |
| | 7 2 5 4 | 4+5 2 | 3+4 2 |
| | 10 4 5 4 | 8 2 | 7 2 |
| | 9 2 | 9 2 | 5+6 2 |
| | 11 4 | 8 2 | 7 2 |
| | 6 2 | 9 2 | 3+4 2 |
| | 8 2 | 6+7 2 | 7 2 |
| | 10 4 | 8 2 | 5+6 2 |
| | 11 4 | 9 2 | 7 2 |
| | 7 2 | 8 2 | 5+6 2 |
| | 10 4 | 4+5 2 | 7 2 |
| | | 9 2 | 7 2 |
| | | 6+7 2 | 3+4 2 |

Continued on next column

**TABLE LXXXVIII– continued**

| TQC | ICSD | NoE | SF |
|---|---|---|---|
| | | 9 2 | 3+4 2 |
| | | 8 2 | 7 2 |
| | | 4+5 2 | 5+6 2 |
| | | 9 2 | 7 2 |
| | | 6+7 2 | 7 2 |
| | | 8 2 | 5+6 2 |
| | | 8 2 | 3+4 2 |
| | | 4+5 2 | 7 2 |
| | | 9 2 | 7 2 |
| | | 6+7 2 | 3+4 2 |
| | | 8 2 | 5+6 2 |
| | | 9 2 | 7 2 |
| | | 9 2 | 3+4 2 |
| | | 6+7 2 | 7 2 |
| 227.1.87 | 39415 | 96 | Cd1Cr2S4 |
| | -GM -X | -L | -W |
| | 6 2 5 4 | 9 2 | 5+6 2 |
| | 9 2 5 4 | 8 2 | 7 2 |
| | 11 4 5 4 | 8 2 | 3+4 2 |
| | 8 2 5 4 | 6+7 2 | 7 2 |
| | 10 4 5 4 | 9 2 | 5+6 2 |
| | 7 2 5 4 | 4+5 2 | 7 2 |
| | 10 4 5 4 | 8 2 | 3+4 2 |
| | 11 4 5 4 | 9 2 | 7 2 |
| | 7 2 5 4 | 9 2 | 7 2 |
| | 11 4 5 4 | 8 2 | 5+6 2 |
| | 10 4 5 4 | 4+5 2 | 3+4 2 |
| | 8 2 5 4 | 6+7 2 | 7 2 |
| | 6 2 5 4 | 9 2 | 5+6 2 |
| | 9 2 5 4 | 8 2 | 7 2 |
| | 7 2 5 4 | 6+7 2 | 7 2 |
| | 10 4 5 4 | 8 2 | 5+6 2 |
| | 11 4 5 4 | 4+5 2 | 3+4 2 |
| | 10 4 5 4 | 9 2 | 7 2 |
| | 6 2 5 4 | 9 2 | 7 2 |
| | 10 4 5 4 | 8 2 | 3+4 2 |
| | 8 2 5 4 | 9 2 | 5+6 2 |
| | 11 4 5 4 | 6+7 2 | 7 2 |
| | 7 2 5 4 | 4+5 2 | 3+4 2 |
| | 10 4 5 4 | 8 2 | 7 2 |
| | 9 2 | 9 2 | 5+6 2 |
| | 11 4 | 6+7 2 | 7 2 |
| | 6 2 | 9 2 | 3+4 2 |
| | 10 4 | 8 2 | 7 2 |
| | 8 2 | 9 2 | 7 2 |
| | 11 4 | 8 2 | 5+6 2 |
| | 7 2 | 4+5 2 | 5+6 2 |
| | 10 4 | 8 2 | 7 2 |
| | | 9 2 | 7 2 |
| | | 6+7 2 | 3+4 2 |
| | | 9 2 | 3+4 2 |
| | | 8 2 | 7 2 |
| | | 4+5 2 | 5+6 2 |
| | | 9 2 | 7 2 |
| | | 6+7 2 | 7 2 |
| | | 8 2 | 5+6 2 |
| | | 8 2 | 7 2 |
| | | 4+5 2 | 3+4 2 |
| | | 9 2 | 7 2 |
| | | 6+7 2 | 3+4 2 |
| | | 8 2 | 5+6 2 |

Continued on next column

**TABLE LXXXVIII– continued**

| TQC | ICSD | NoE | SF |
|---|---|---|---|
| | | 9 2 | 7 2 |
| | | 9 2 | 3+4 2 |
| | | 6+7 2 | 7 2 |
| 227.1.88 | 57512 | 32 | Al2Ba1 |
| | -GM -X | -L | -W |
| | 6 2 5 4 | 8 2 | 3+4 2 |
| | 9 2 5 4 | 9 2 | 7 2 |
| | 7 2 5 4 | 8 2 | 5+6 2 |
| | 8 2 5 4 | 9 2 | 7 2 |
| | 10 4 5 4 | 4+5 2 | 7 2 |
| | 11 4 5 4 | 9 2 | 3+4 2 |
| | 6 2 5 4 | 8 2 | 5+6 2 |
| | 9 2 5 4 | 6+7 2 | 7 2 |
| | 7 2 | 9 2 | 5+6 2 |
| | 10 4 | 8 2 | 7 2 |
| | 8 2 | 9 2 | 7 2 |
| | 11 4 | 6+7 2 | 3+4 2 |
| | | 8 2 | 7 2 |
| | | 8 2 | 3+4 2 |
| | | 4+5 2 | 5+6 2 |
| | | 8 2 | 7 2 |

The End

TABLE LXXXIX: The computed irreducible representations for the valence bands for the compounds of SEBR and NLC in space group 229. See the list of tables.

| TQC | ICSD | NoE | SF |
|---|---|---|---|
| 229.1.1 | 26270 | 112 | As7Re3 |
| | -GM -H | -P | -N |
| | 6 2 11 4 | 6 2 | 5 2 |
| | 10 4 8 2 | 8 4 | 6 2 |
| | 7 2 10 4 | 7 2 | 6 2 |
| | 11 4 6 2 | 8 4 | 5 2 |
| | 6 2 9 2 | 7 2 | 6 2 |
| | 9 2 8 2 | 6 2 | 5 2 |
| | 11 4 11 4 | 8 4 | 6 2 |
| | 8 2 10 4 | 8 4 | 5 2 |
| | 10 4 7 2 | 7 2 | 6 2 |
| | 9 2 7 2 | 6 2 | 5 2 |
| | 10 4 6 2 | 8 4 | 5 2 |
| | 6 2 11 4 | 8 4 | 6 2 |
| | 8 2 10 4 | 7 2 | 5 2 |
| | 11 4 7 2 | 6 2 | 6 2 |
| | 11 4 9 2 | 8 4 | 5 2 |
| | 9 2 11 4 | 6 2 | 6 2 |
| | 8 2 8 2 | 8 4 | 5 2 |
| | 7 2 10 4 | 7 2 | 6 2 |
| | 10 4 11 4 | 6 2 | 6 2 |
| | 11 4 8 2 | 8 4 | 5 2 |
| | 11 4 10 4 | 8 4 | 5 2 |
| | 10 4 11 4 | 8 4 | 6 2 |
| | 7 2 6 2 | 7 2 | 6 2 |
| | 10 4 9 2 | 6 2 | 5 2 |
| | 7 2 6 2 | 8 4 | 5 2 |
| | 9 2 10 4 | 7 2 | 6 2 |
| | 11 4 9 2 | 8 4 | 5 2 |
| | 10 4 11 4 | 6 2 | 6 2 |
| | 7 2 10 4 | 7 2 | 5 2 |
| | 6 2 11 4 | 7 2 | 6 2 |

Continued on next column

**TABLE LXXXIX– continued**

| TQC | ICSD | NoE | SF |
|---|---|---|---|
| | 10 4 10 4 | 8 4 | 5 2 |
| | 8 2 7 2 | 8 4 | 6 2 |
| | 6 2 6 2 | 6 2 | 6 2 |
| | 11 4 8 2 | 8 4 | 5 2 |
| | 10 4 10 4 | 7 2 | 5 2 |
| | 6 2 9 2 | 8 4 | 6 2 |
| | 10 4 11 4 | 8 4 | 6 2 |
| | 9 2 7 2 | 6 2 | 6 2 |
| | | | 5 2 |
| | | | 5 2 |
| | | | 5 2 |
| | | | 6 2 |
| | | | 5 2 |
| | | | 6 2 |
| | | | 5 2 |
| | | | 6 2 |
| | | | 6 2 |
| | | | 5 2 |
| | | | 6 2 |
| | | | 5 2 |
| | | | 6 2 |
| | | | 6 2 |
| | | | 6 2 |
| | | | 5 2 |
| | | | 5 2 |
| | | | 5 2 |

The End

TABLE XC: A list of "Low-Quality" topological compounds. See the list of tables.

| SG | Type | TQC | ICSD | NoE | DG | Stoich. Formula |
|---|---|---|---|---|---|---|
| 70 | SEBR | | | | | |
| 70 | NLC | | | | | |
| 70 | ES | | | | | |
| | | 70.low1 | 402423 | 132 | ∅ | Ag1Nb1Rb2Se4 |
| | | 70.low2 | 174182 | 60 | ∅ | Cu1Mg2 |
| | | 70.low3 | 414272 | 116 | ∅ | Cu1Rb2Se4Ta1 |
| | | 70.low4 | 24128 | 144 | ∅ | Ba1Mn2O8 |
| | | 70.low5 | 23443 | 144 | ∅ | Ba1Mn2O8 |
| | | 70.low6 | 188588 | 190 | ∅ | Al5Cu6U1 |
| | | 70.low7 | 26997 | 226 | ∅ | Nb3O10U1 |
| | | 70.low8 | 44866 | 32 | ∅ | Pu1 |
| | | 70.low9 | 40621 | 70 | ∅ | B2Ir2Y1 |
| | | 70.low10 | 40619 | 82 | ∅ | B2Eu1Ir2 |
| | | 70.low11 | 249644 | 154 | ∅ | Au4K1Sn6 |
| | | 70.low12 | 1819 | 60 | ∅ | Co1In2 |
| | | 70.low13 | 410521 | 228 | ∅ | Ag2Eu4Si6 |
| | | 70.low14 | 26499 | 192 | ∅ | H8O12S2Zr1 |
| | | 70.low15 | 200611 | 184 | ∅ | H4O10Si1U2 |
| | | 70.low16 | 106521 | 56 | ∅ | Cr1Sn2 |
| 70 | ESFD | | | | | |
| | | 70.low17 | 402423 | 132 | ∅ | Ag1Nb1Rb2Se4 |
| | | 70.low18 | 174182 | 60 | ∅ | Cu1Mg2 |
| | | 70.low19 | 414272 | 116 | ∅ | Cu1Rb2Se4Ta1 |
| | | 70.low20 | 24128 | 144 | ∅ | Ba1Mn2O8 |
| | | 70.low21 | 23443 | 144 | ∅ | Ba1Mn2O8 |
| | | 70.low22 | 188588 | 190 | ∅ | Al5Cu6U1 |
| | | 70.low23 | 26997 | 226 | ∅ | Nb3O10U1 |
| | | 70.low24 | 44866 | 32 | ∅ | Pu1 |

Continued on next Column

**TABLE XC– continued**

| SG | Type | TQC | ICSD | NoE | DG | Stoich. Formula |
|---|---|---|---|---|---|---|
| | | 70.low25 | 40621 | 70 | ∅ | B2Ir2Y1 |
| | | 70.low26 | 40619 | 82 | ∅ | B2Eu1Ir2 |
| | | 70.low27 | 249644 | 154 | ∅ | Au4K1Sn6 |
| | | 70.low28 | 1819 | 60 | ∅ | Co1In2 |
| | | 70.low29 | 410521 | 228 | ∅ | Ag2Eu4Si6 |
| | | 70.low30 | 26499 | 192 | ∅ | H8O12S2Zr1 |
| | | 70.low31 | 200611 | 184 | ∅ | H4O10Si1U2 |
| | | 70.low32 | 106521 | 56 | ∅ | Cr1Sn2 |
| 166 | SEBR | | | | | |
| 166 | NLC | | | | | |
| 166 | ES | | | | | |
| | | 166.low1 | 402423 | 132 | ∅ | Ag1Nb1Rb2Se4 |
| | | 166.low2 | 174182 | 60 | ∅ | Cu1Mg2 |
| | | 166.low3 | 414272 | 116 | ∅ | Cu1Rb2Se4Ta1 |
| | | 166.low4 | 24128 | 144 | ∅ | Ba1Mn2O8 |
| | | 166.low5 | 23443 | 144 | ∅ | Ba1Mn2O8 |
| | | 166.low6 | 188588 | 190 | ∅ | Al5Cu6U1 |
| | | 166.low7 | 26997 | 226 | ∅ | Nb3O10U1 |
| | | 166.low8 | 44866 | 32 | ∅ | Pu1 |
| | | 166.low9 | 40621 | 70 | ∅ | B2Ir2Y1 |
| | | 166.low10 | 40619 | 82 | ∅ | B2Eu1Ir2 |
| | | 166.low11 | 249644 | 154 | ∅ | Au4K1Sn6 |
| | | 166.low12 | 1819 | 60 | ∅ | Co1In2 |
| | | 166.low13 | 410521 | 228 | ∅ | Ag2Eu4Si6 |
| | | 166.low14 | 26499 | 192 | ∅ | H8O12S2Zr1 |
| | | 166.low15 | 200611 | 184 | ∅ | H4O10Si1U2 |
| | | 166.low16 | 106521 | 56 | ∅ | Cr1Sn2 |
| 166 | ESFD | | | | | |
| | | 166.low17 | 402423 | 132 | ∅ | Ag1Nb1Rb2Se4 |
| | | 166.low18 | 174182 | 60 | ∅ | Cu1Mg2 |
| | | 166.low19 | 414272 | 116 | ∅ | Cu1Rb2Se4Ta1 |
| | | 166.low20 | 24128 | 144 | ∅ | Ba1Mn2O8 |
| | | 166.low21 | 23443 | 144 | ∅ | Ba1Mn2O8 |
| | | 166.low22 | 188588 | 190 | ∅ | Al5Cu6U1 |
| | | 166.low23 | 26997 | 226 | ∅ | Nb3O10U1 |
| | | 166.low24 | 44866 | 32 | ∅ | Pu1 |
| | | 166.low25 | 40621 | 70 | ∅ | B2Ir2Y1 |
| | | 166.low26 | 40619 | 82 | ∅ | B2Eu1Ir2 |
| | | 166.low27 | 249644 | 154 | ∅ | Au4K1Sn6 |
| | | 166.low28 | 1819 | 60 | ∅ | Co1In2 |
| | | 166.low29 | 410521 | 228 | ∅ | Ag2Eu4Si6 |
| | | 166.low30 | 26499 | 192 | ∅ | H8O12S2Zr1 |
| | | 166.low31 | 200611 | 184 | ∅ | H4O10Si1U2 |
| | | 166.low32 | 106521 | 56 | ∅ | Cr1Sn2 |
| 167 | SEBR | | | | | |
| 167 | NLC | | | | | |
| 167 | ES | | | | | |
| | | 167.low1 | 402423 | 132 | ∅ | Ag1Nb1Rb2Se4 |
| | | 167.low2 | 174182 | 60 | ∅ | Cu1Mg2 |
| | | 167.low3 | 414272 | 116 | ∅ | Cu1Rb2Se4Ta1 |
| | | 167.low4 | 24128 | 144 | ∅ | Ba1Mn2O8 |
| | | 167.low5 | 23443 | 144 | ∅ | Ba1Mn2O8 |
| | | 167.low6 | 188588 | 190 | ∅ | Al5Cu6U1 |
| | | 167.low7 | 26997 | 226 | ∅ | Nb3O10U1 |
| | | 167.low8 | 44866 | 32 | ∅ | Pu1 |
| | | 167.low9 | 40621 | 70 | ∅ | B2Ir2Y1 |
| | | 167.low10 | 40619 | 82 | ∅ | B2Eu1Ir2 |
| | | 167.low11 | 249644 | 154 | ∅ | Au4K1Sn6 |
| | | 167.low12 | 1819 | 60 | ∅ | Co1In2 |
| | | 167.low13 | 410521 | 228 | ∅ | Ag2Eu4Si6 |
| | | 167.low14 | 26499 | 192 | ∅ | H8O12S2Zr1 |

Continued on next Column

**TABLE XC– continued**

| SG | Type | TQC | ICSD | NoE | DG | Stoich. Formula |
|---|---|---|---|---|---|---|
| | | 167.low15 | 200611 | 184 | ∅ | H4O10Si1U2 |
| | | 167.low16 | 106521 | 56 | ∅ | Cr1Sn2 |
| 167 | ESFD | | | | | |
| | | 167.low17 | 402423 | 132 | ∅ | Ag1Nb1Rb2Se4 |
| | | 167.low18 | 174182 | 60 | ∅ | Cu1Mg2 |
| | | 167.low19 | 414272 | 116 | ∅ | Cu1Rb2Se4Ta1 |
| | | 167.low20 | 24128 | 144 | ∅ | Ba1Mn2O8 |
| | | 167.low21 | 23443 | 144 | ∅ | Ba1Mn2O8 |
| | | 167.low22 | 188588 | 190 | ∅ | Al5Cu6U1 |
| | | 167.low23 | 26997 | 226 | ∅ | Nb3O10U1 |
| | | 167.low24 | 44866 | 32 | ∅ | Pu1 |
| | | 167.low25 | 40621 | 70 | ∅ | B2Ir2Y1 |
| | | 167.low26 | 40619 | 82 | ∅ | B2Eu1Ir2 |
| | | 167.low27 | 249644 | 154 | ∅ | Au4K1Sn6 |
| | | 167.low28 | 1819 | 60 | ∅ | Co1In2 |
| | | 167.low29 | 410521 | 228 | ∅ | Ag2Eu4Si6 |
| | | 167.low30 | 26499 | 192 | ∅ | H8O12S2Zr1 |
| | | 167.low31 | 200611 | 184 | ∅ | H4O10Si1U2 |
| | | 167.low32 | 106521 | 56 | ∅ | Cr1Sn2 |

The End

## Appendix U: Figures of all band structures

In this section one can find the link to all figures of band structures of all space groups. Return to page 1.

# List of Figures

### 0.0.1 SG 2

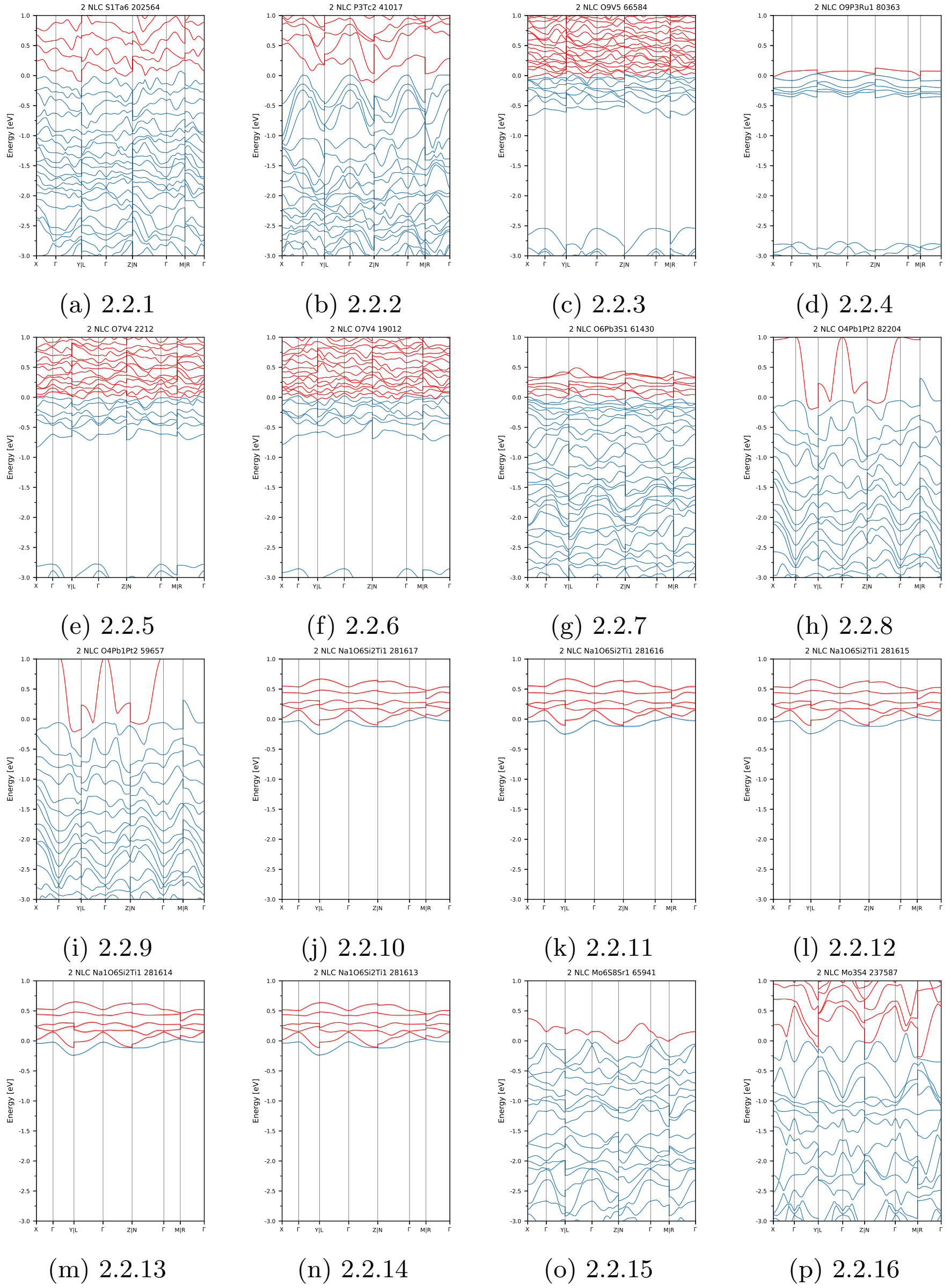

(a) 2.2.1 (b) 2.2.2 (c) 2.2.3 (d) 2.2.4

(e) 2.2.5 (f) 2.2.6 (g) 2.2.7 (h) 2.2.8

(i) 2.2.9 (j) 2.2.10 (k) 2.2.11 (l) 2.2.12

(m) 2.2.13 (n) 2.2.14 (o) 2.2.15 (p) 2.2.16

Figure 1: Band structures for topological materials in SG 2, part 1/7, from 2.2.1 to 2.2.16. See the list of figures.

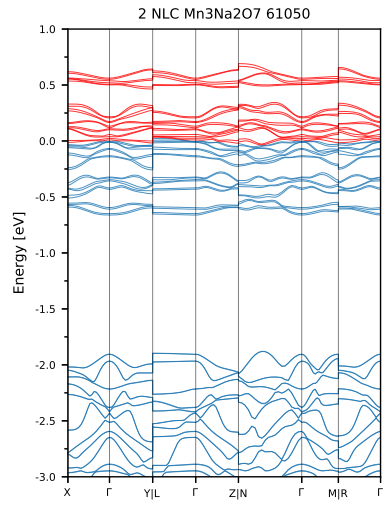

(q) 2.2.17

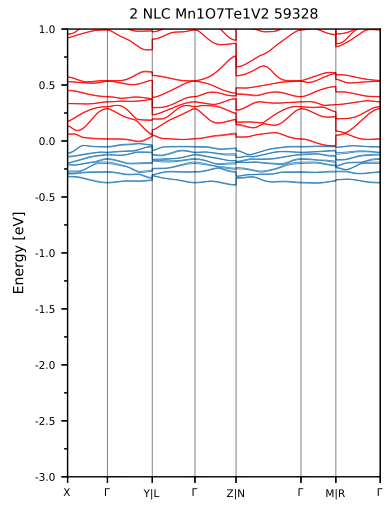

(r) 2.2.18

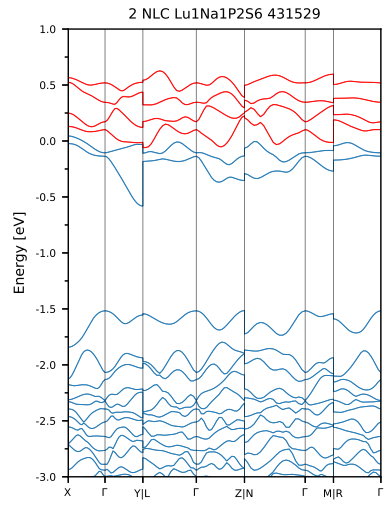

(s) 2.2.19

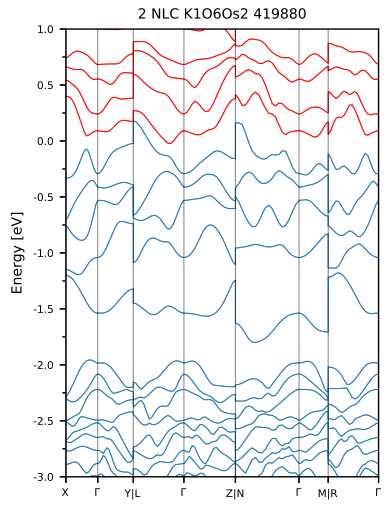

(t) 2.2.20

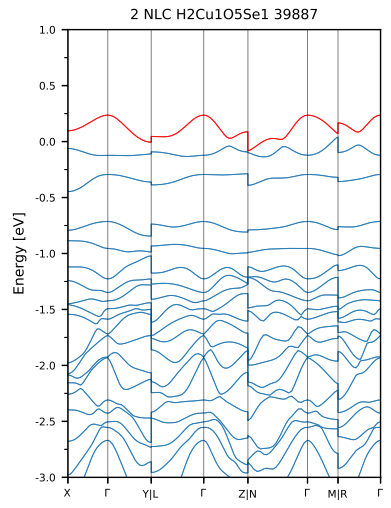

(u) 2.2.21

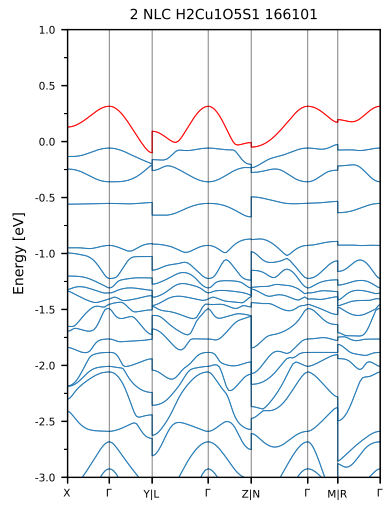

(v) 2.2.22

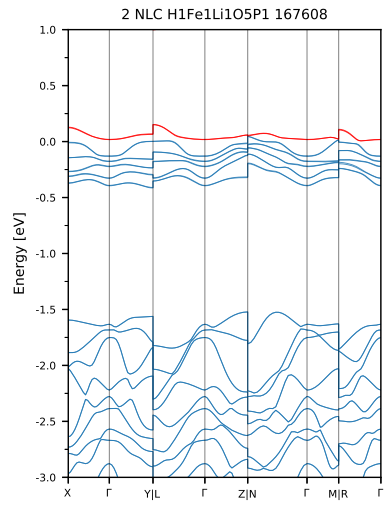

(w) 2.2.23

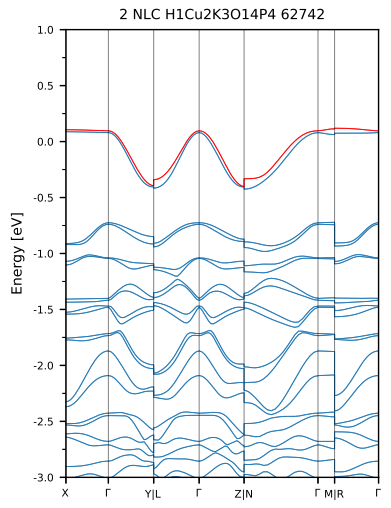

(x) 2.2.24

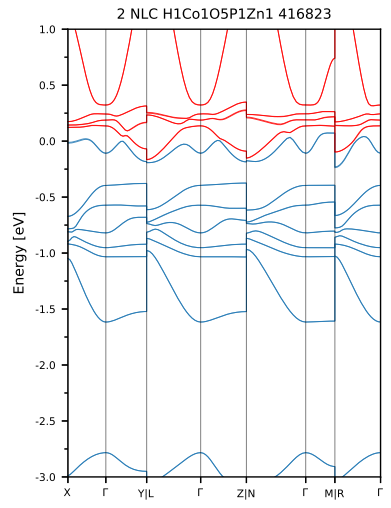

(y) 2.2.25

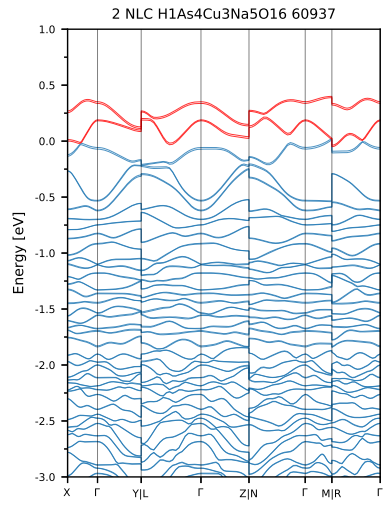

(z) 2.2.26

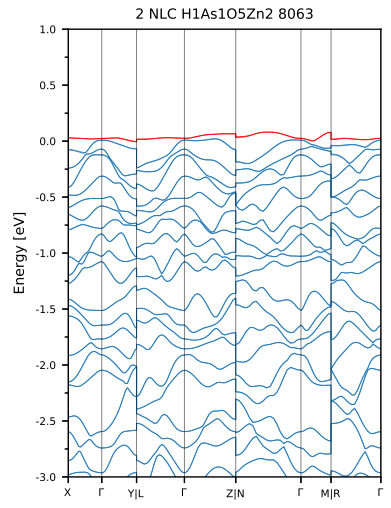

(aa) 2.2.27

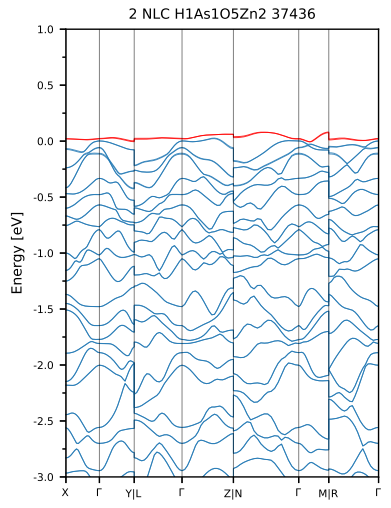

(ab) 2.2.28

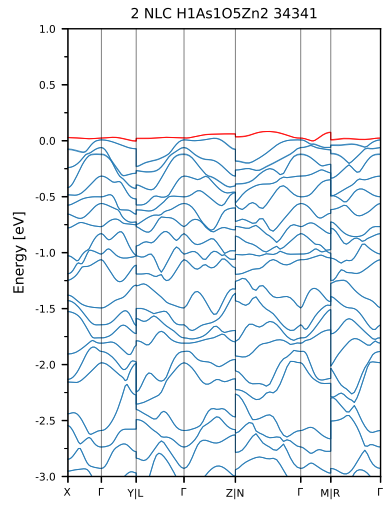

(ac) 2.2.29

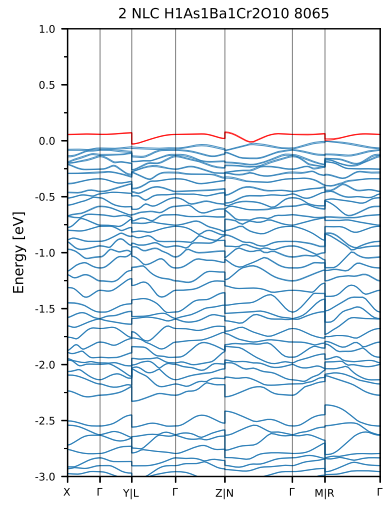

(ad) 2.2.30

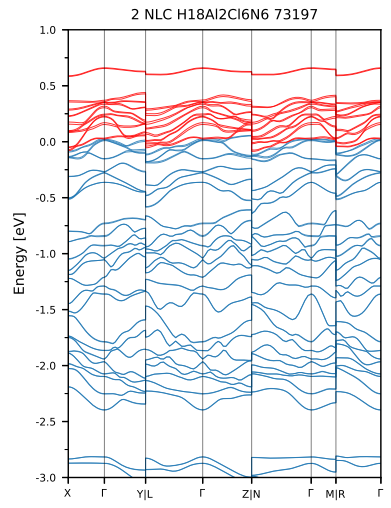

(ae) 2.2.31

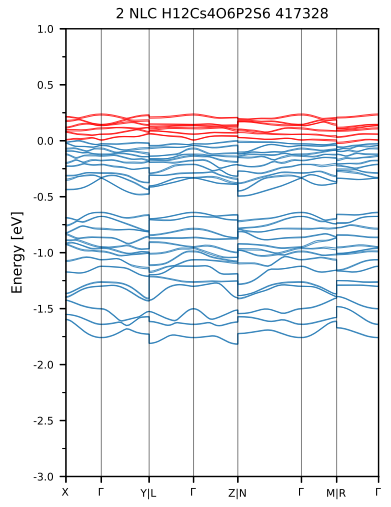

(af) 2.2.32

Figure 2: Band structures for topological materials in SG 2, part 2/7, from 2.2.17 to 2.2.32. See the list of figures.

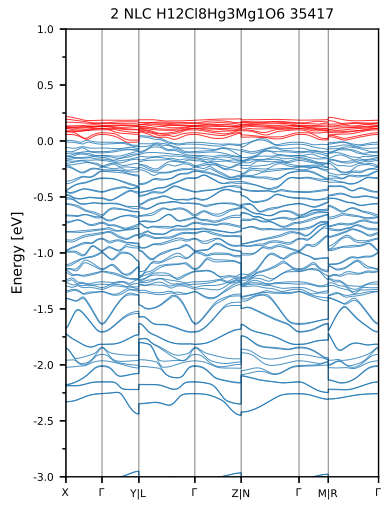

(ag) 2.2.33

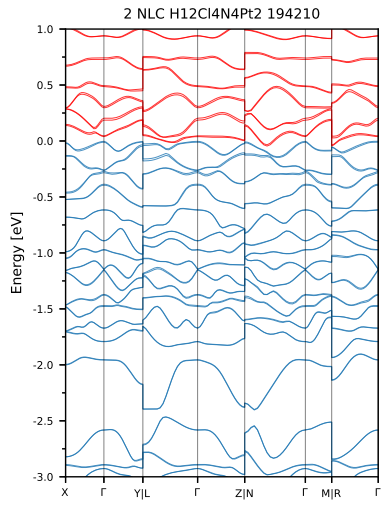

(ah) 2.2.34

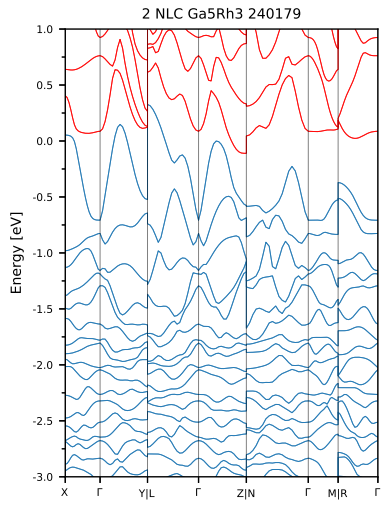

(ai) 2.2.35

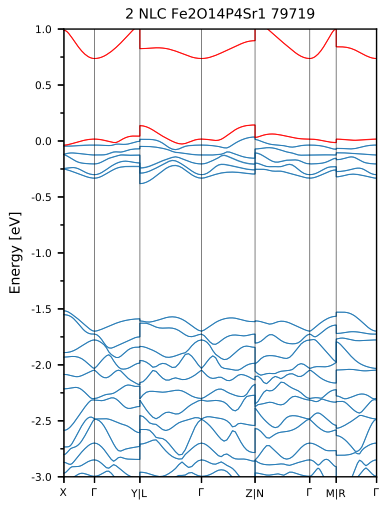

(aj) 2.2.36

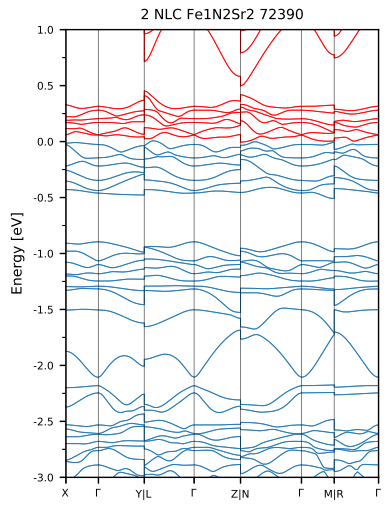

(ak) 2.2.37

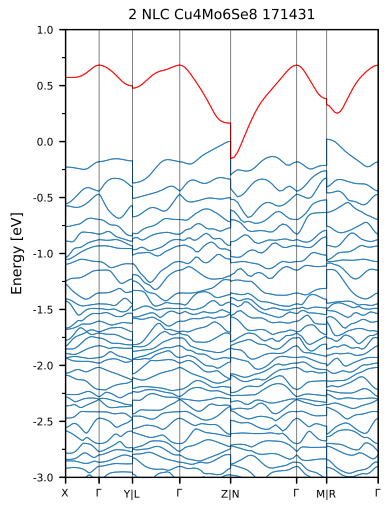

(al) 2.2.38

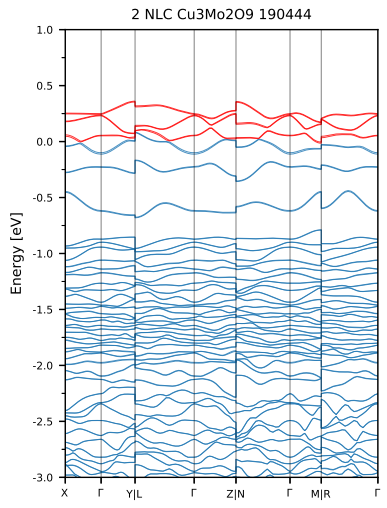

(am) 2.2.39

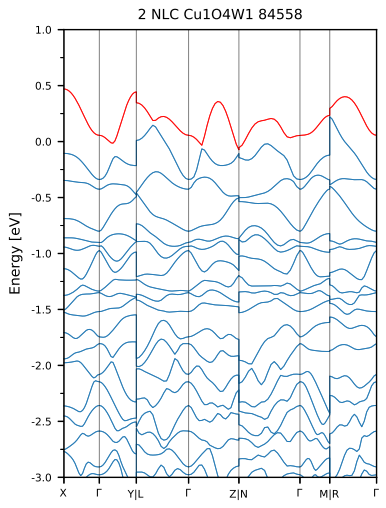

(an) 2.2.40

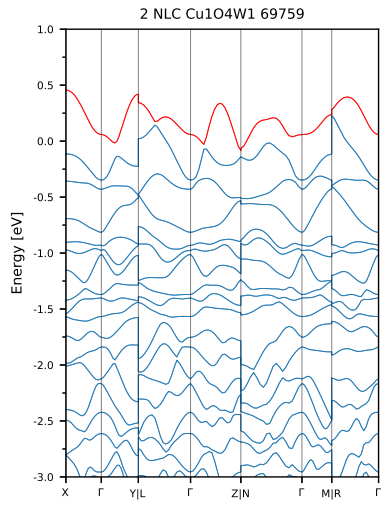

(ao) 2.2.41

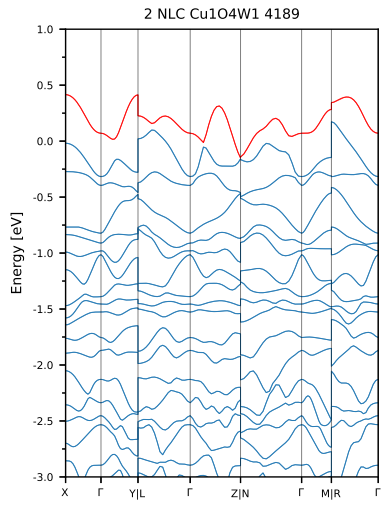

(ap) 2.2.42

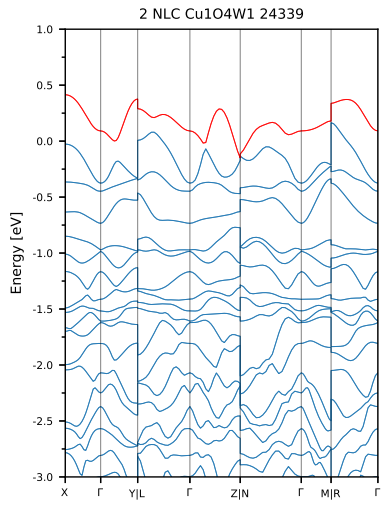

(aq) 2.2.43

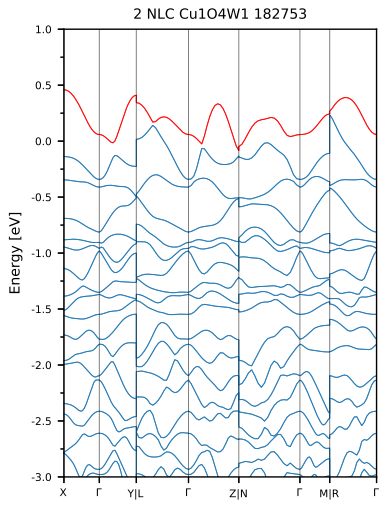

(ar) 2.2.44

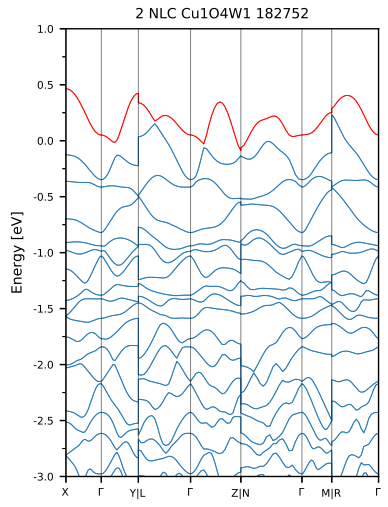

(as) 2.2.45

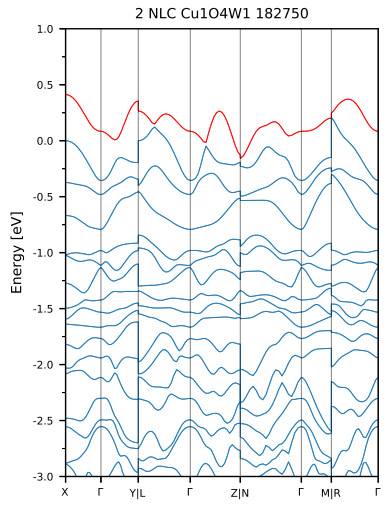

(at) 2.2.46

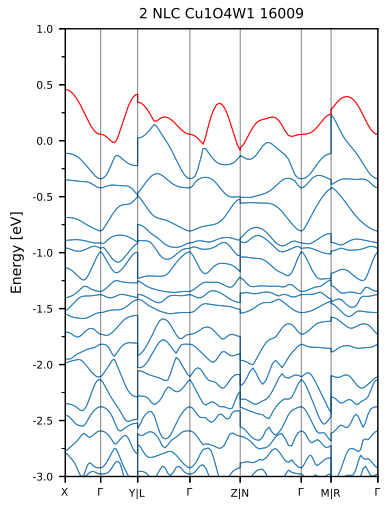

(au) 2.2.47

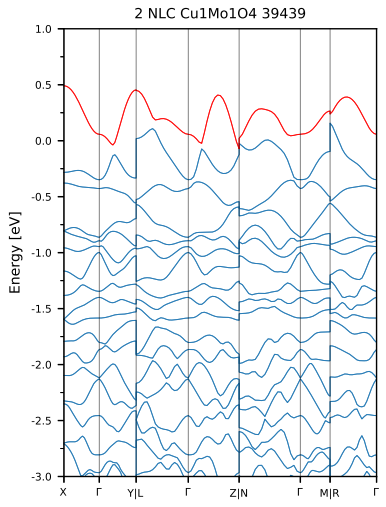

(av) 2.2.48

Figure 3: Band structures for topological materials in SG 2, part 3/7, from 2.2.33 to 2.2.48. See the list of figures.

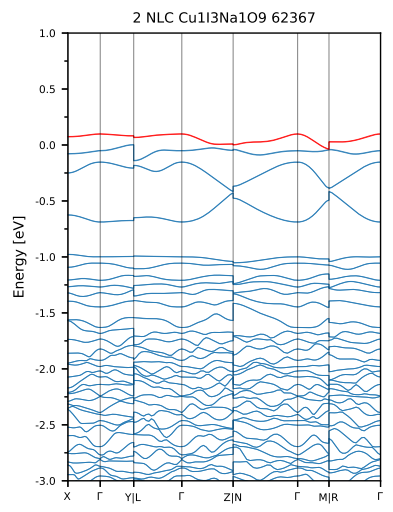

(aw) 2.2.49

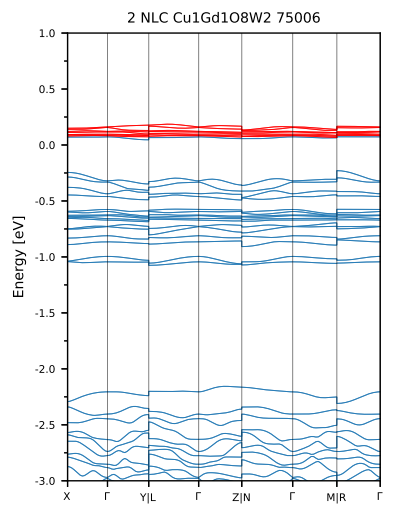

(ax) 2.2.50

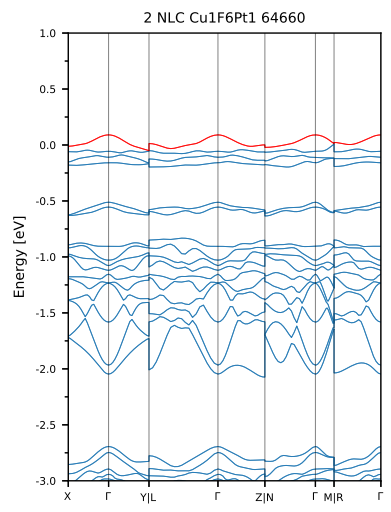

(ay) 2.2.51

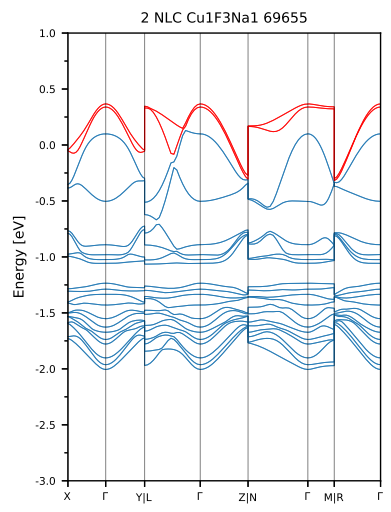

(az) 2.2.52

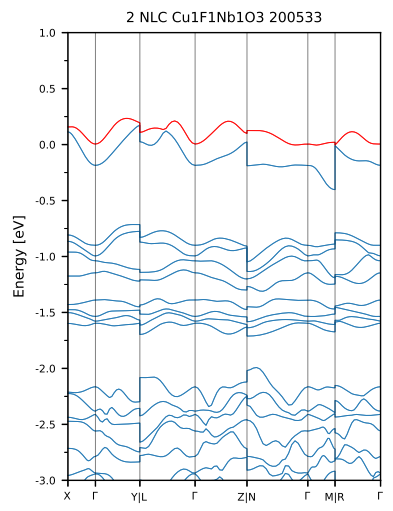

(ba) 2.2.53

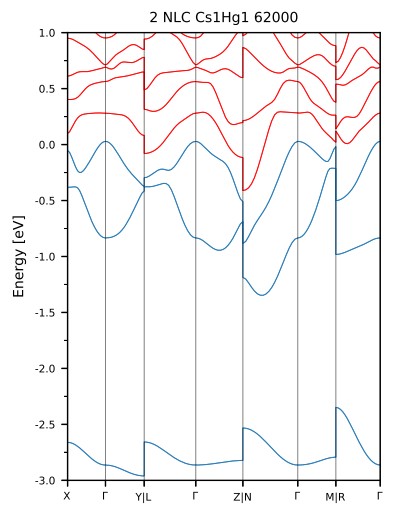

(bb) 2.2.54

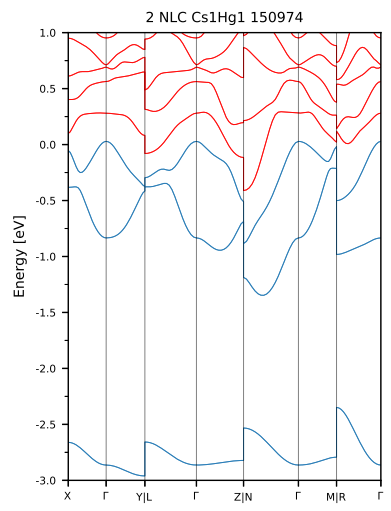

(bc) 2.2.55

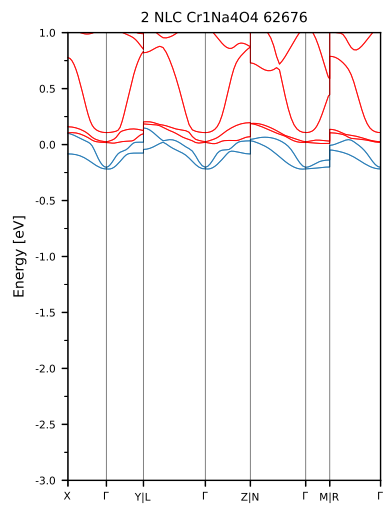

(bd) 2.2.56

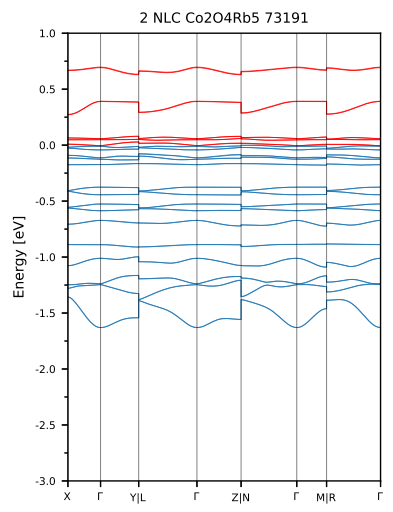

(be) 2.2.57

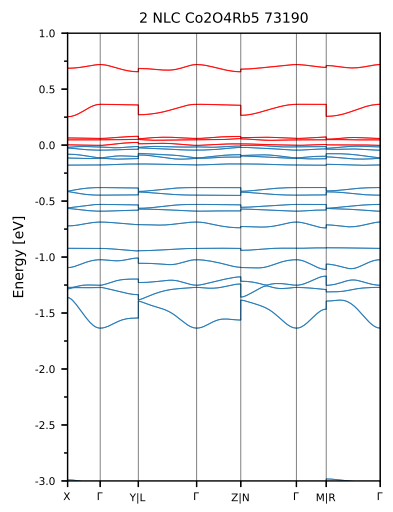

(bf) 2.2.58

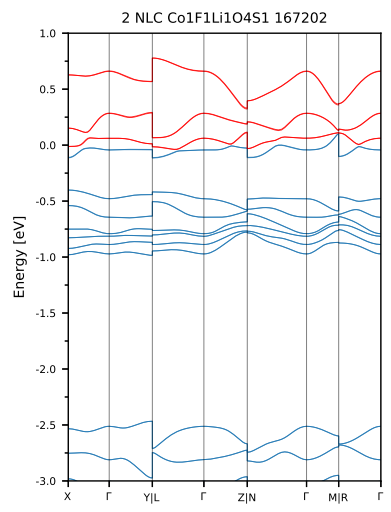

(bg) 2.2.59

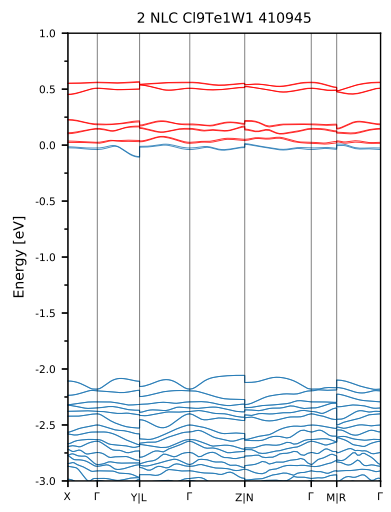

(bh) 2.2.60

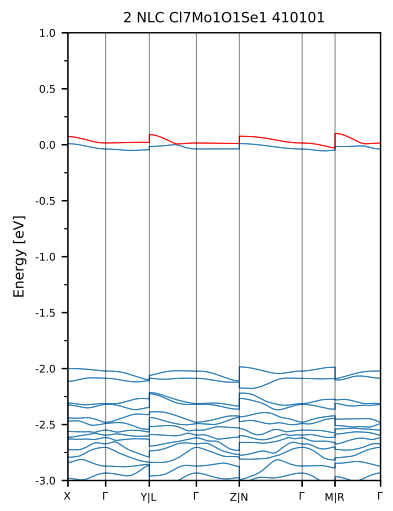

(bi) 2.2.61

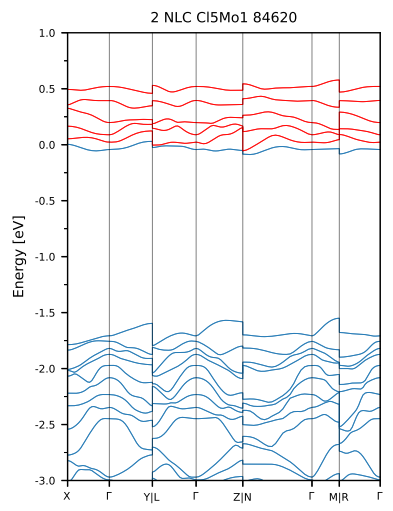

(bj) 2.2.62

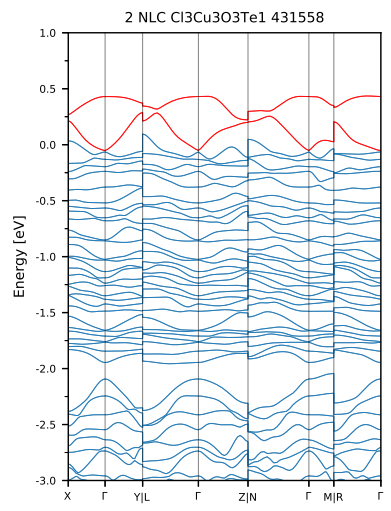

(bk) 2.2.63

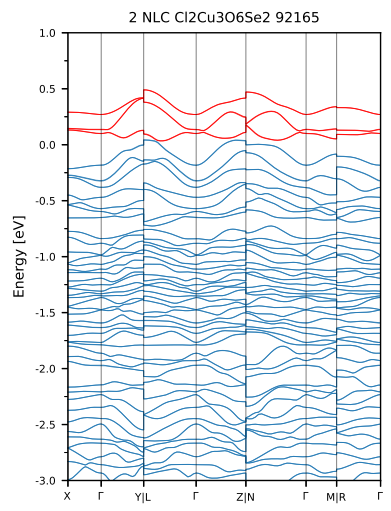

(bl) 2.2.64

Figure 4: Band structures for topological materials in SG 2, part 4/7, from 2.2.49 to 2.2.64. See the list of figures.

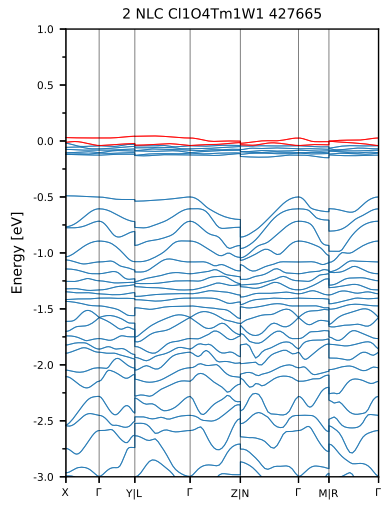

(bm) 2.2.65

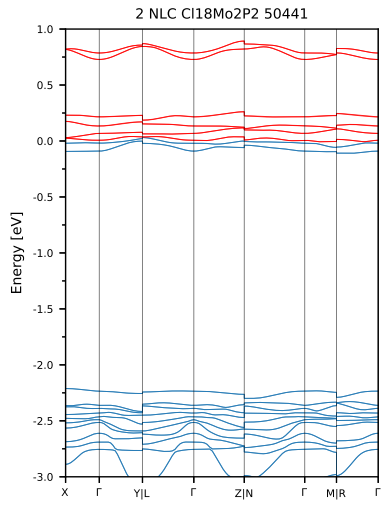

(bn) 2.2.66

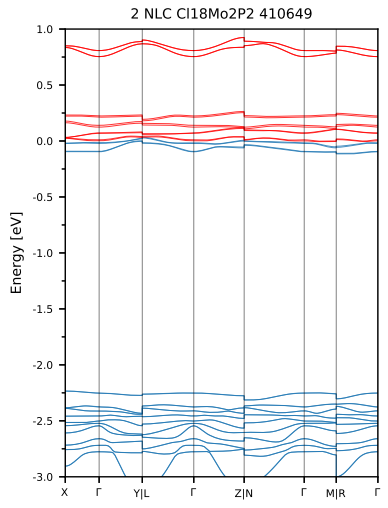

(bo) 2.2.67

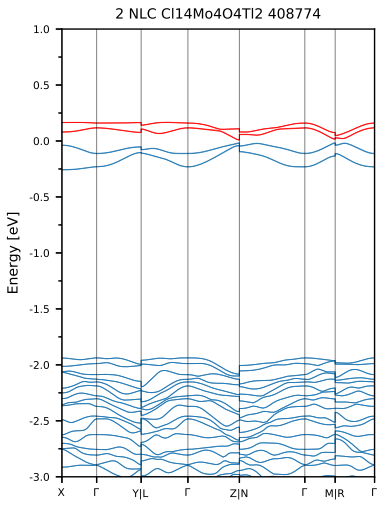

(bp) 2.2.68

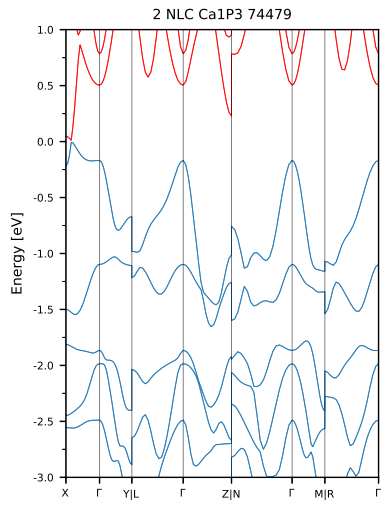

(bq) 2.2.69

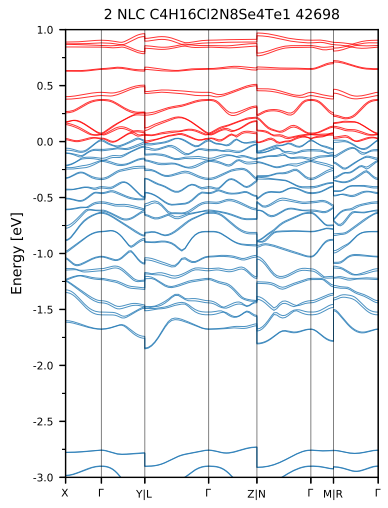

(br) 2.2.70

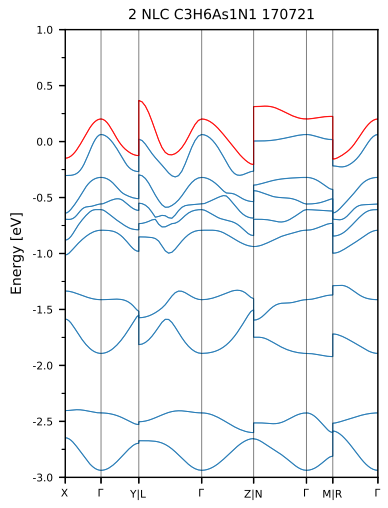

(bs) 2.2.71

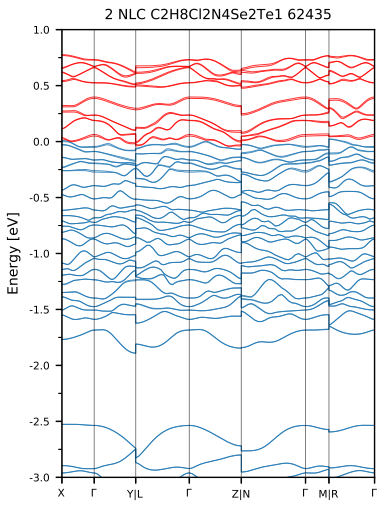

(bt) 2.2.72

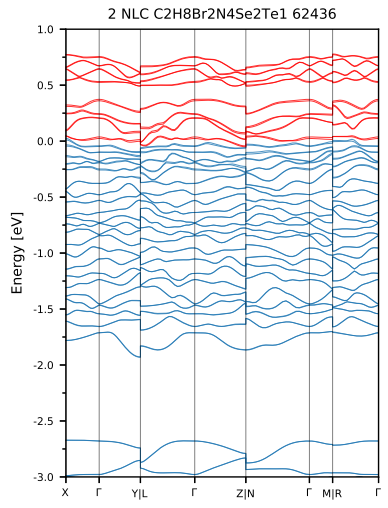

(bu) 2.2.73

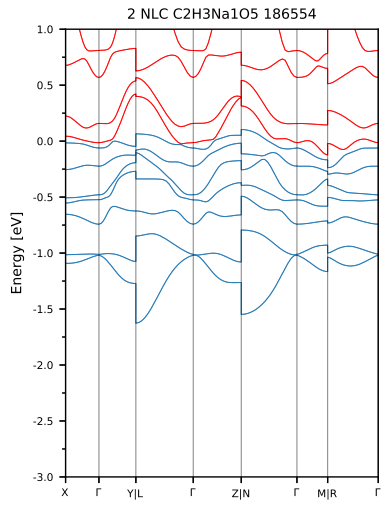

(bv) 2.2.74

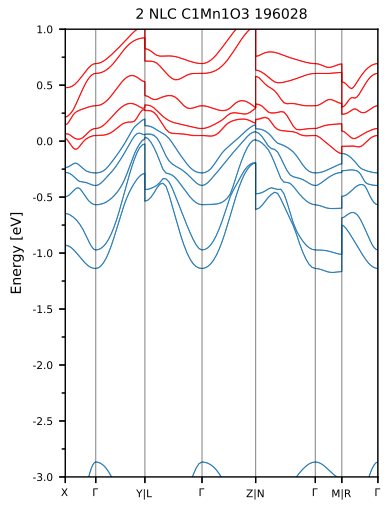

(bw) 2.2.75

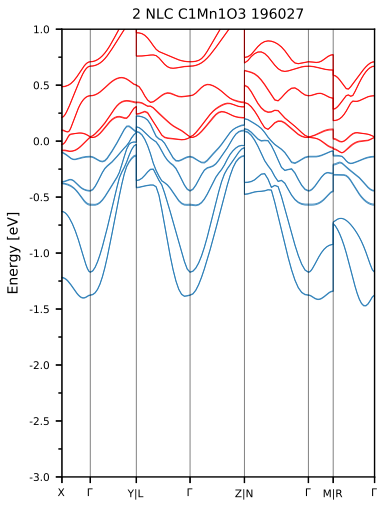

(bx) 2.2.76

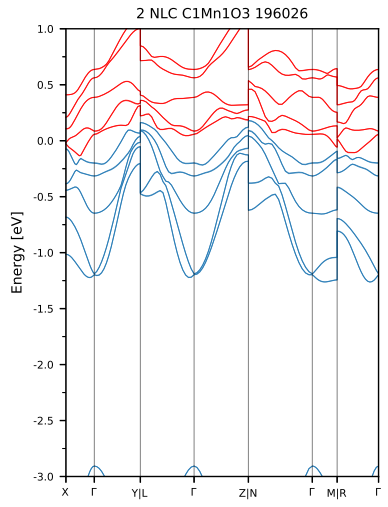

(by) 2.2.77

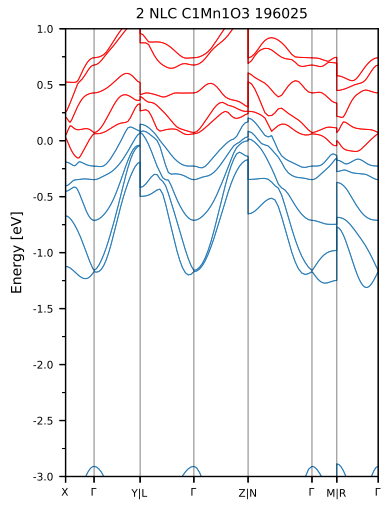

(bz) 2.2.78

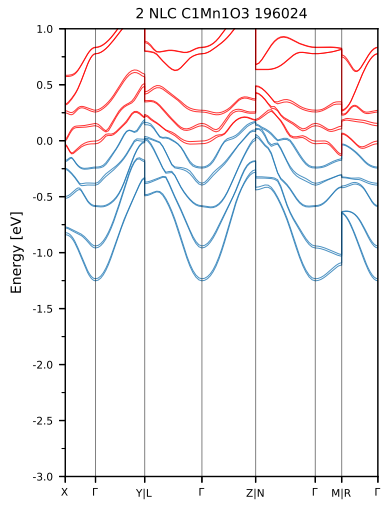

(ca) 2.2.79

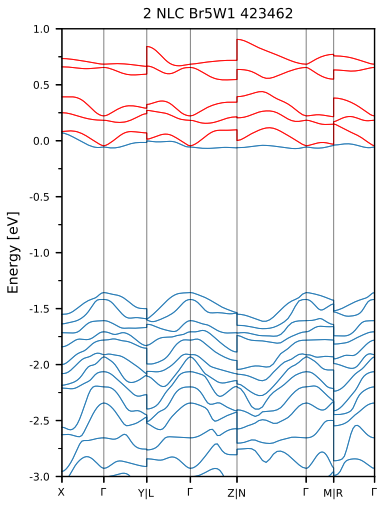

(cb) 2.2.80

Figure 5: Band structures for topological materials in SG 2, part 5/7, from 2.2.65 to 2.2.80. See the list of figures.

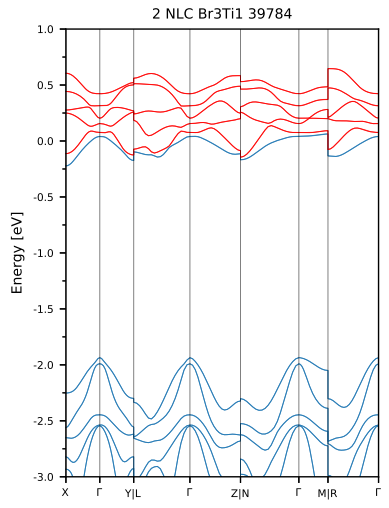

(cc) 2.2.81

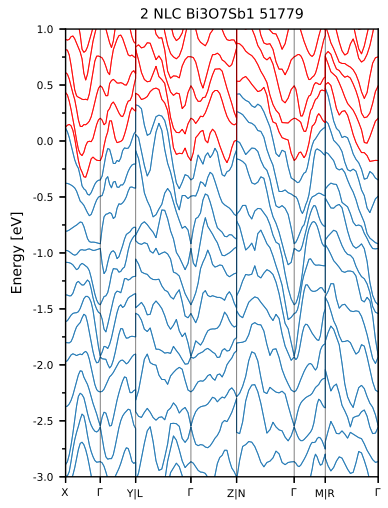

(cd) 2.2.82

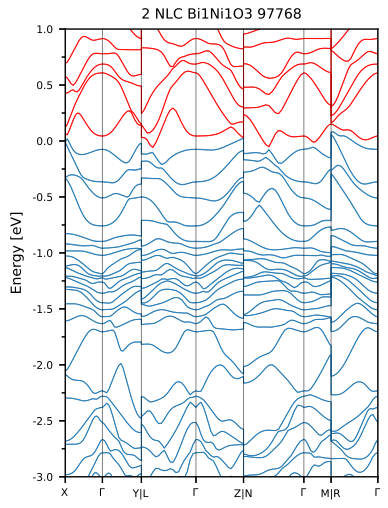

(ce) 2.2.83

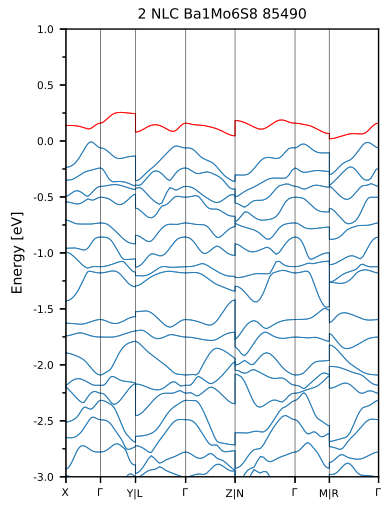

(cf) 2.2.84

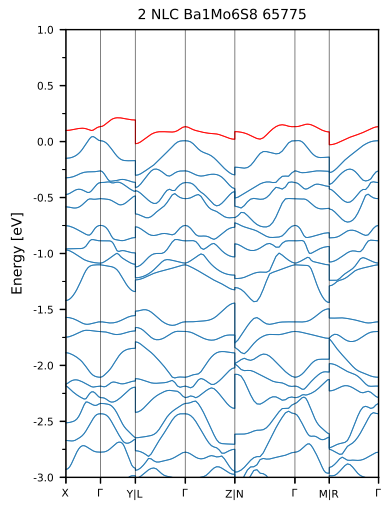

(cg) 2.2.85

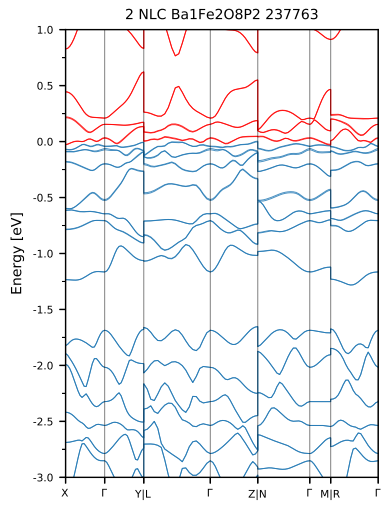

(ch) 2.2.86

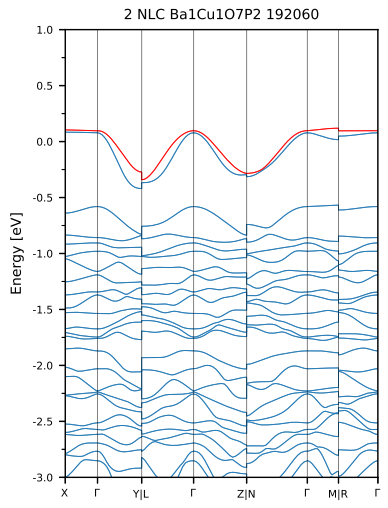

(ci) 2.2.87

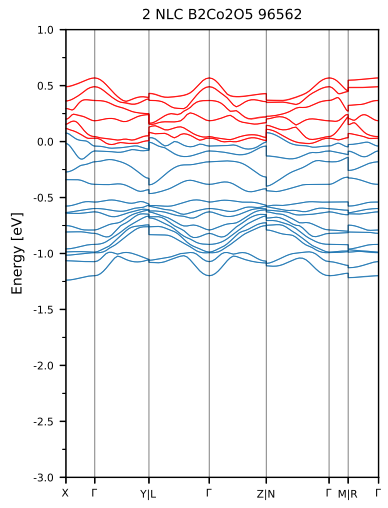

(cj) 2.2.88

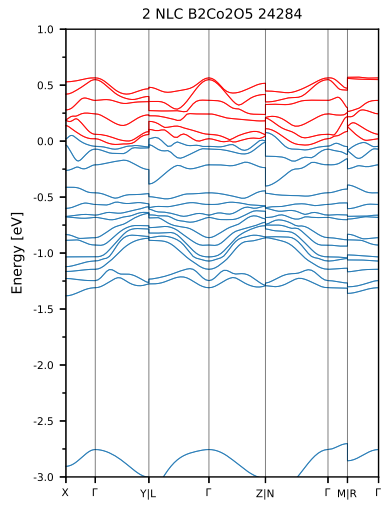

(ck) 2.2.89

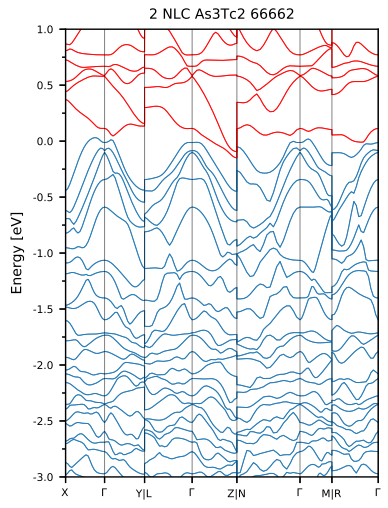

(cl) 2.2.90

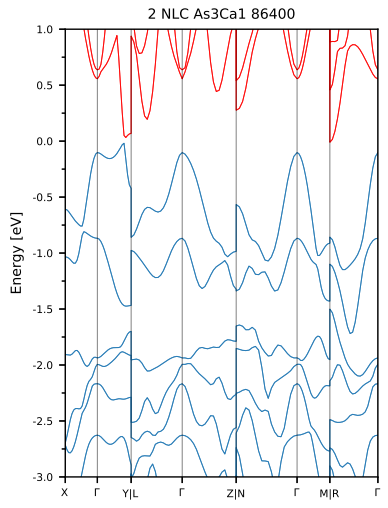

(cm) 2.2.91

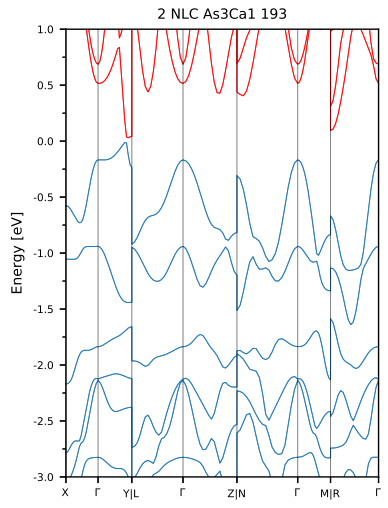

(cn) 2.2.92

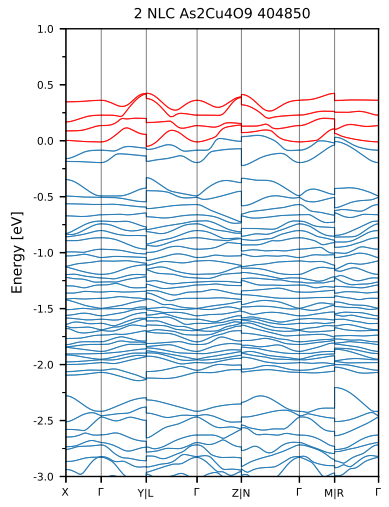

(co) 2.2.93

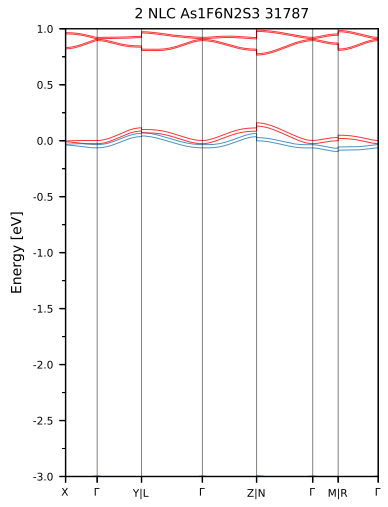

(cp) 2.2.94

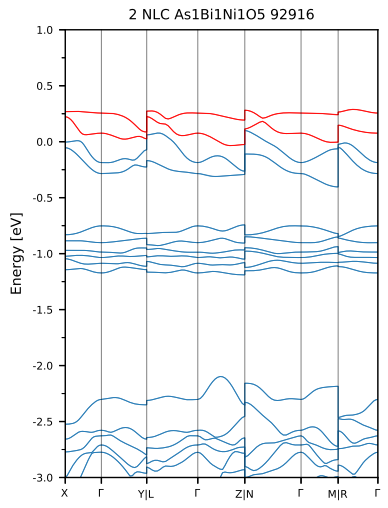

(cq) 2.2.95

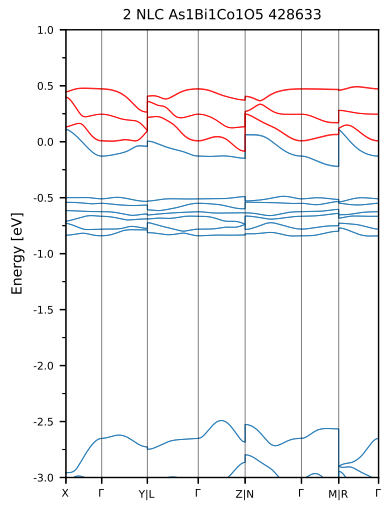

(cr) 2.2.96

Figure 6: Band structures for topological materials in SG 2, part 6/7, from 2.2.81 to 2.2.96. See the list of figures.

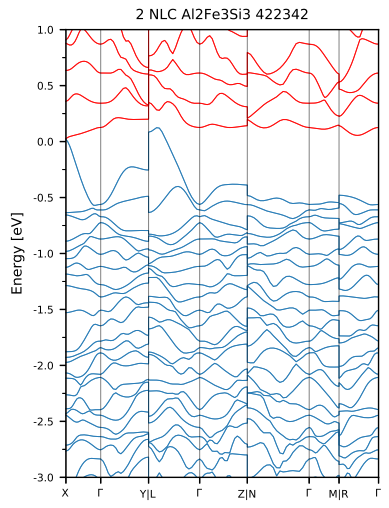

(cs) 2.2.97

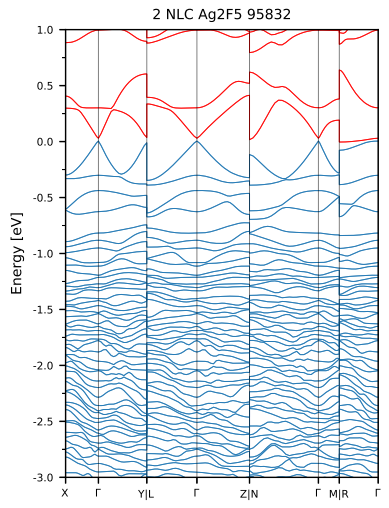

(ct) 2.2.98

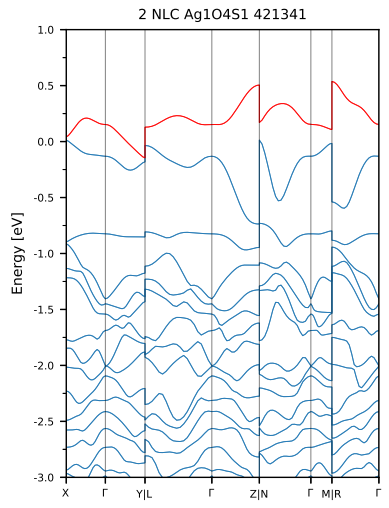

(cu) 2.2.99

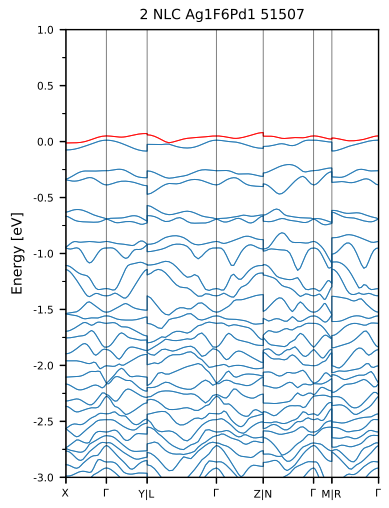

(cv) 2.2.100

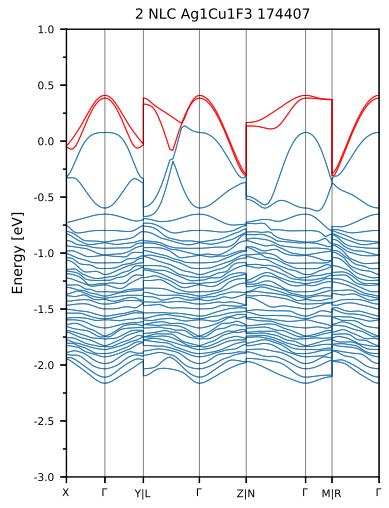

(cw) 2.2.101

Figure 7: Band structures for topological materials in SG 2, part 7/7, from 2.2.97 to 2.2.101. See the list of figures.

### 0.0.2 SG 4

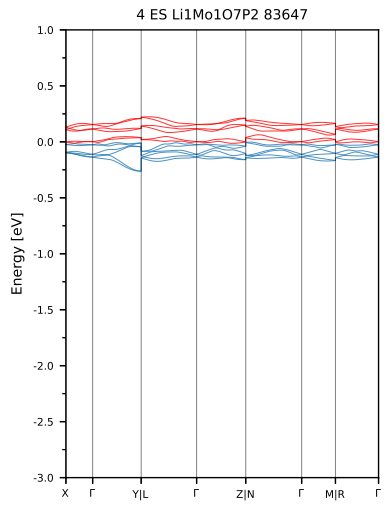

(a) 4.3.1

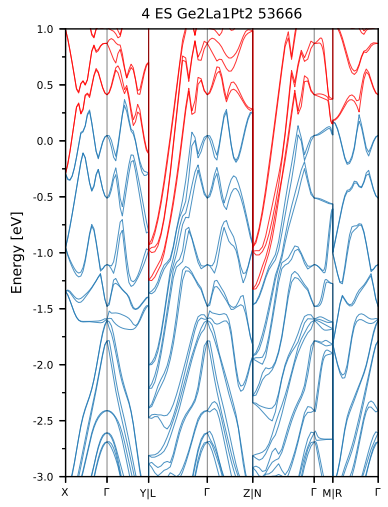

(b) 4.3.2

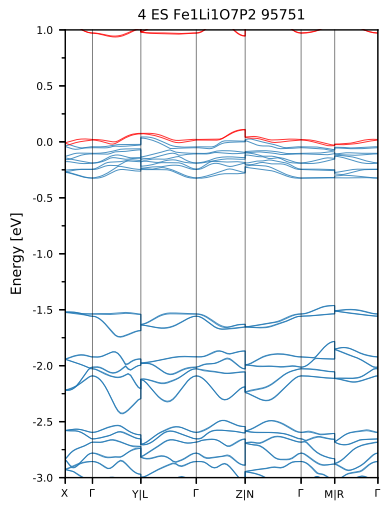

(c) 4.3.3

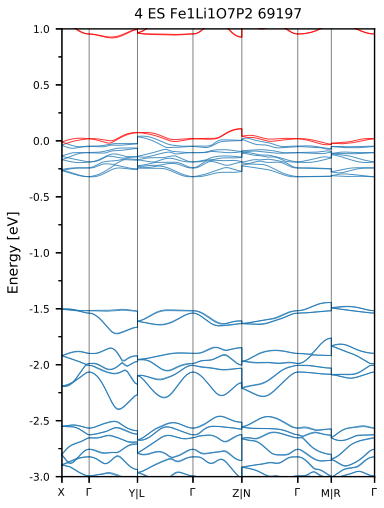

(d) 4.3.4

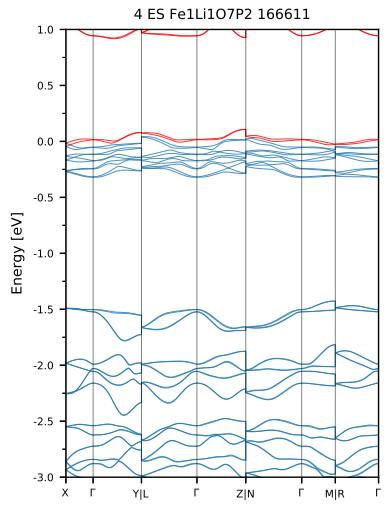

(e) 4.3.5

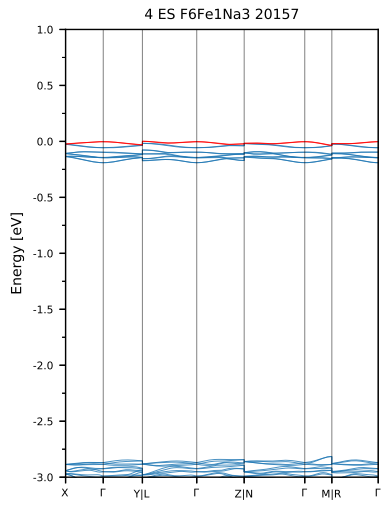

(f) 4.3.6

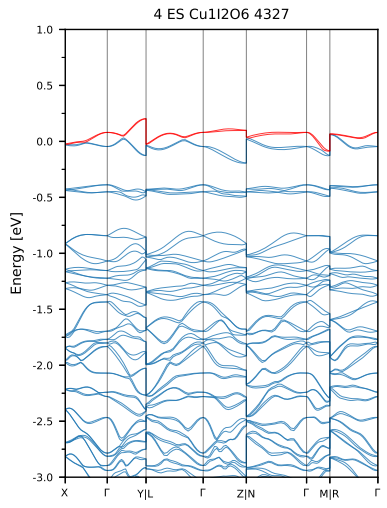

(g) 4.3.7

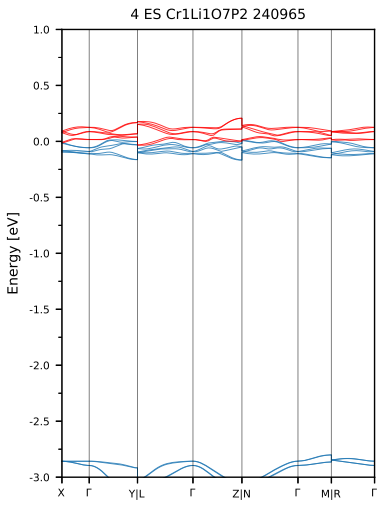

(h) 4.3.8

Figure 8: Band structures for topological materials in SG 4, part 1/1, from 4.3.1 to 4.3.8. See the list of figures.

### 0.0.3 SG 7

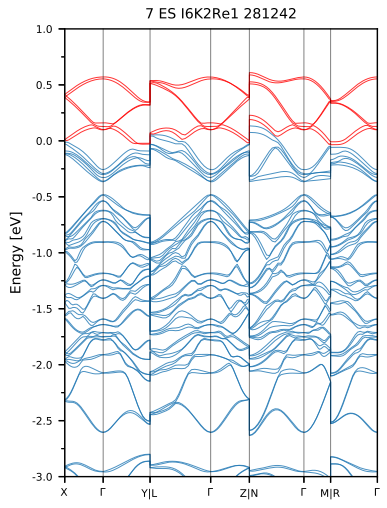

(a) 7.3.1

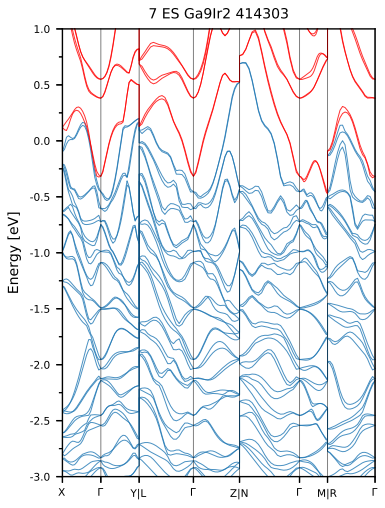

(b) 7.3.2

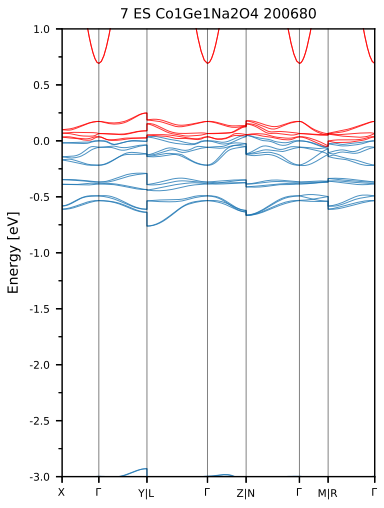

(c) 7.3.3

Figure 9: Band structures for topological materials in SG 7, part 1/1, from 7.3.1 to 7.3.3. See the list of figures.

### 0.0.4 SG 9

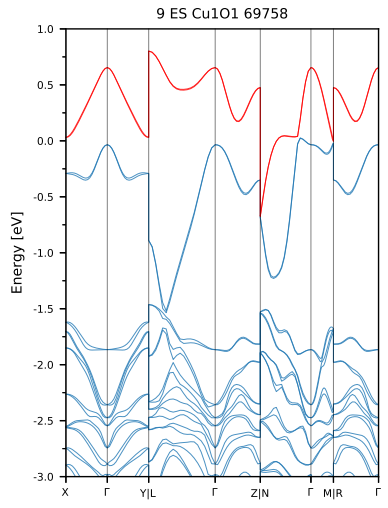

(a) 9.3.1

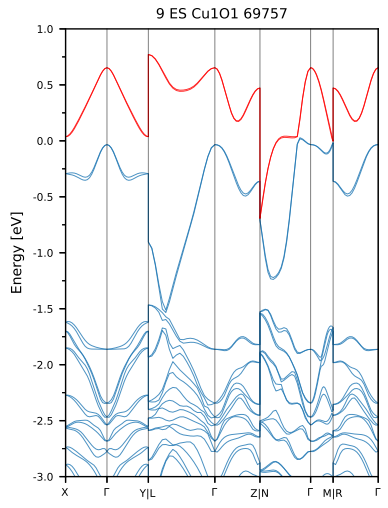

(b) 9.3.2

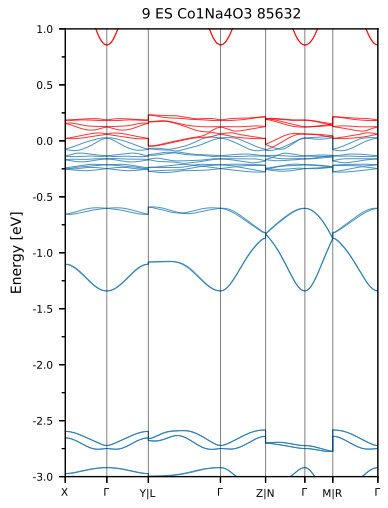

(c) 9.3.3

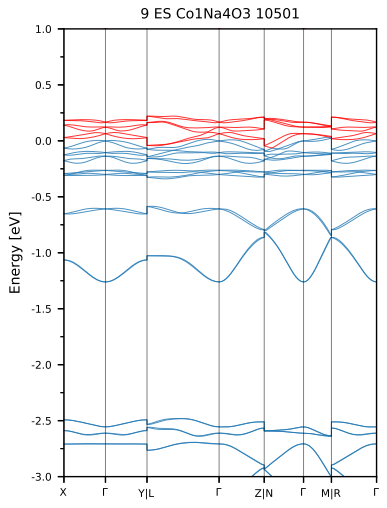

(d) 9.3.4

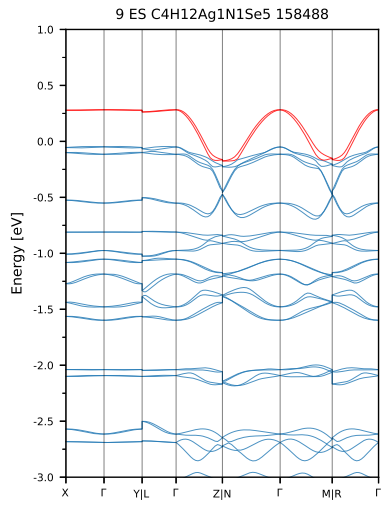

(e) 9.3.5

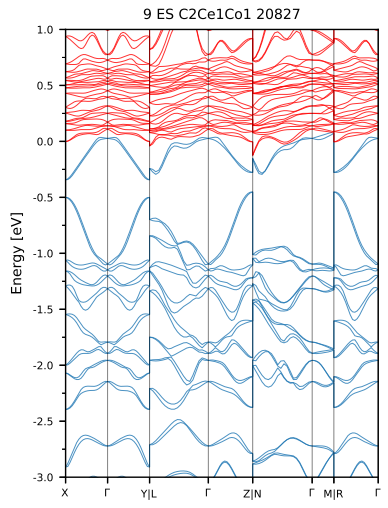

(f) 9.3.6

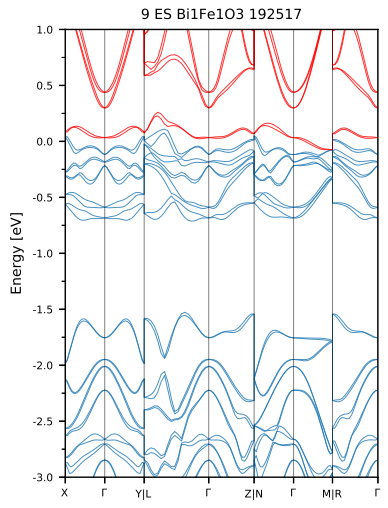

(g) 9.3.7

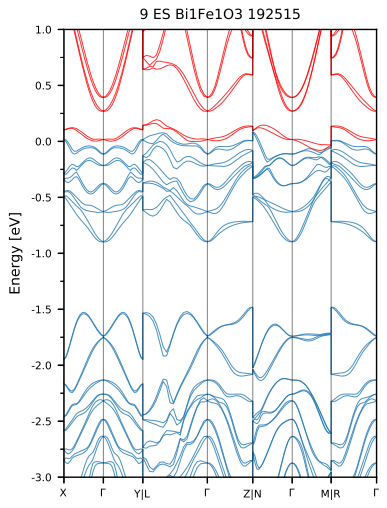

(h) 9.3.8

Figure 10: Band structures for topological materials in SG 9, part 1/1, from 9.3.1 to 9.3.8. See the list of figures.

### 0.0.5 SG 11

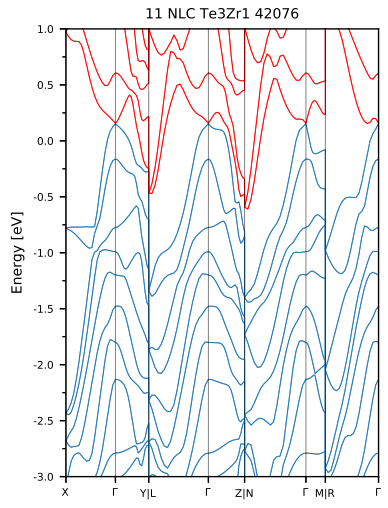

(a) 11.2.1

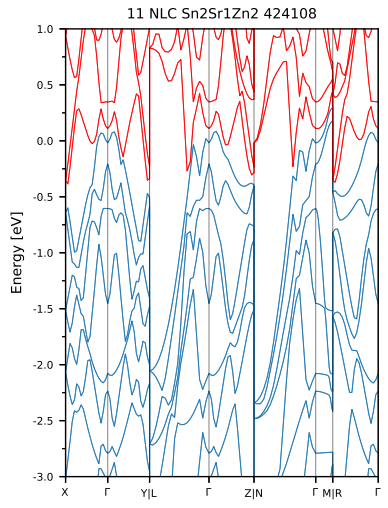

(b) 11.2.2

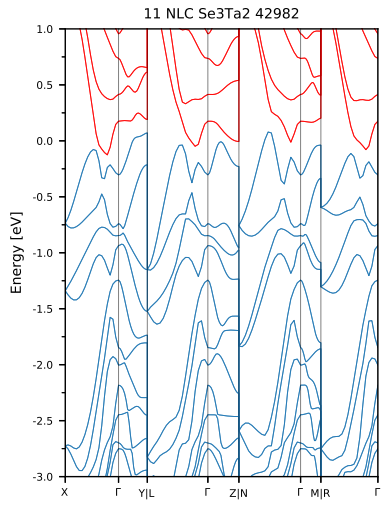

(c) 11.2.3

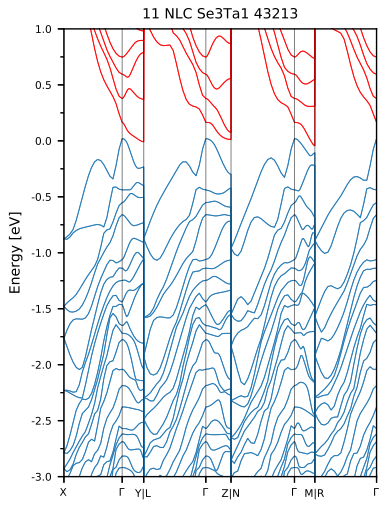

(d) 11.2.4

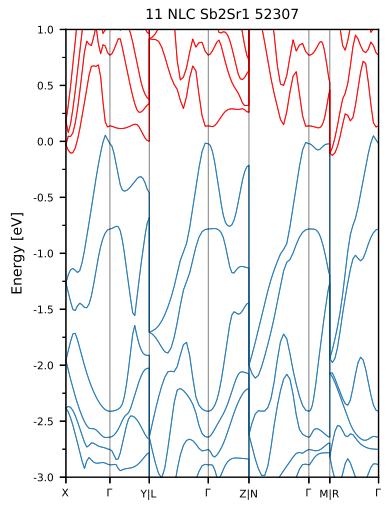

(e) 11.2.5

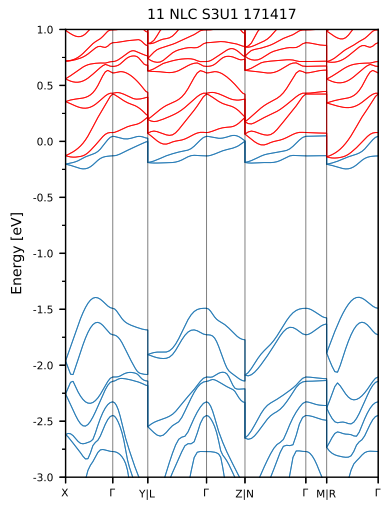

(f) 11.2.6

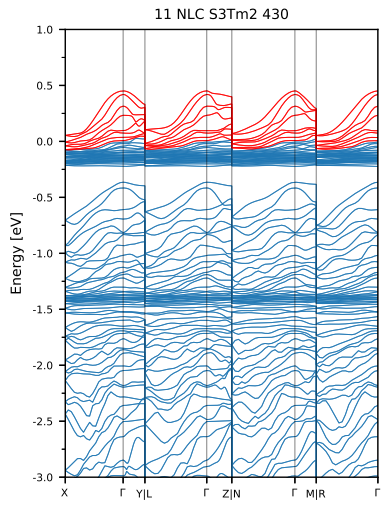

(g) 11.2.7

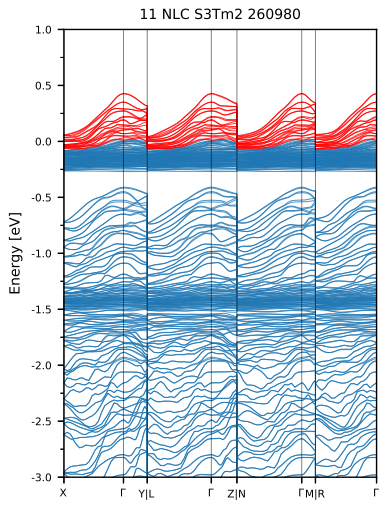

(h) 11.2.8

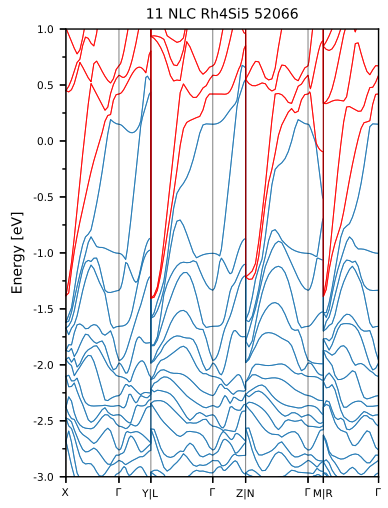

(i) 11.2.9

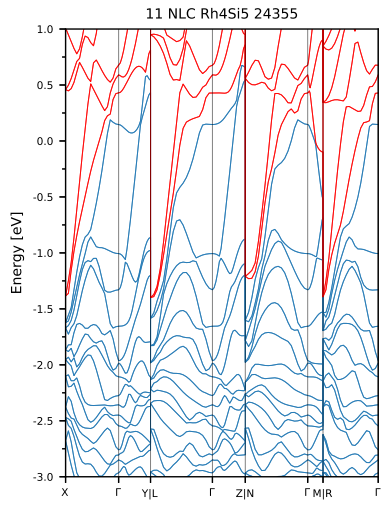

(j) 11.2.10

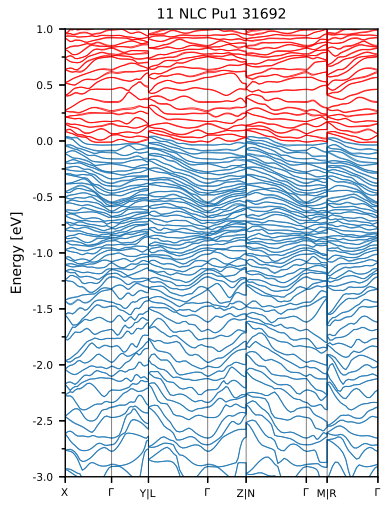

(k) 11.2.11

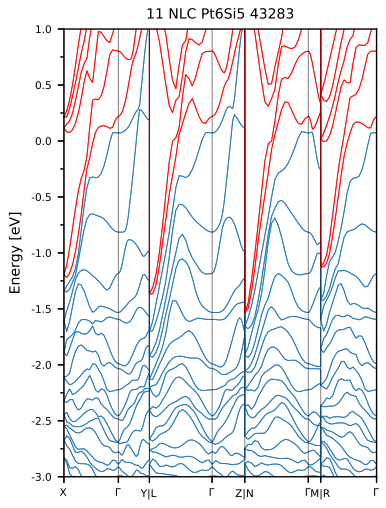

(l) 11.2.12

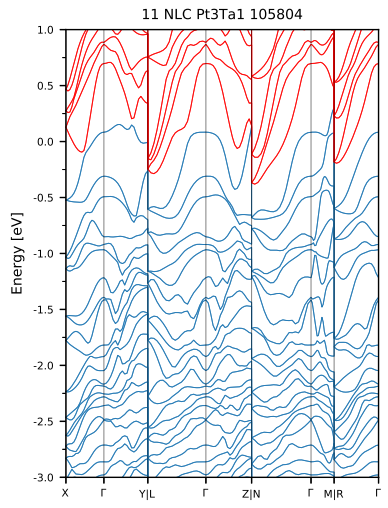

(m) 11.2.13

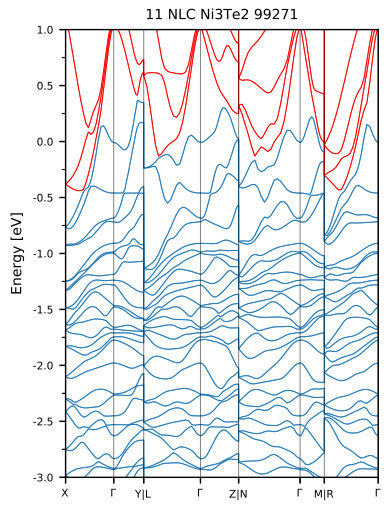

(n) 11.2.14

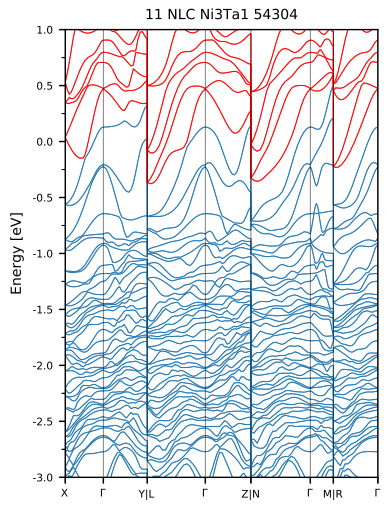

(o) 11.2.15

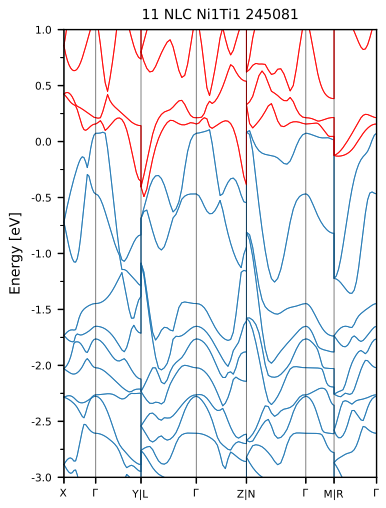

(p) 11.2.16

Figure 11: Band structures for topological materials in SG 11, part 1/13, from 11.2.1 to 11.2.16. See the list of figures.

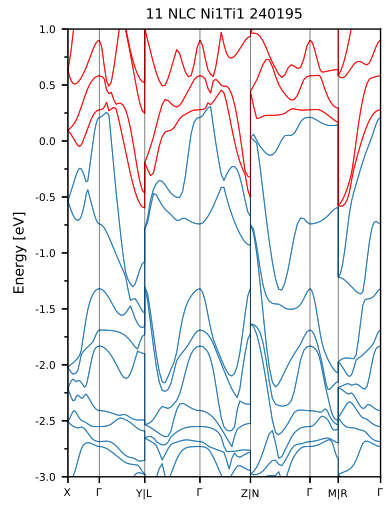

(q) 11.2.17

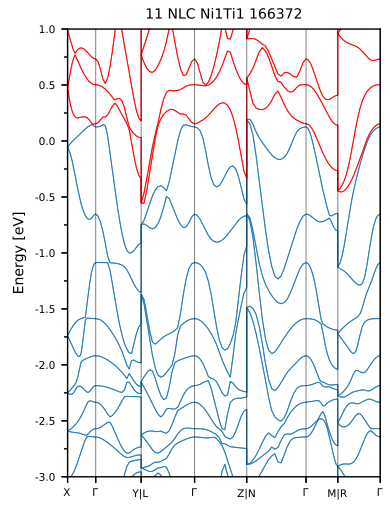

(r) 11.2.18

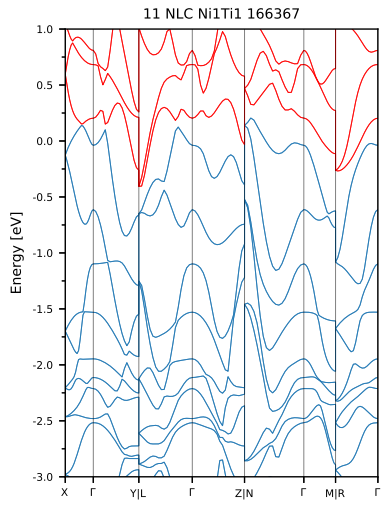

(s) 11.2.19

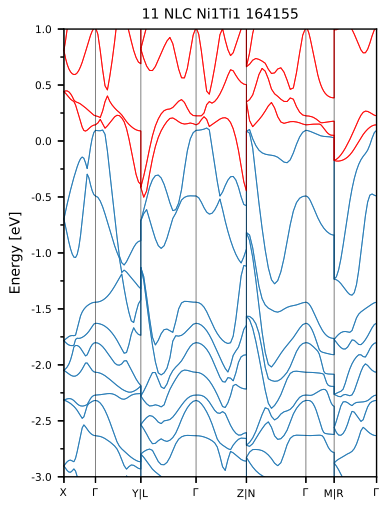

(t) 11.2.20

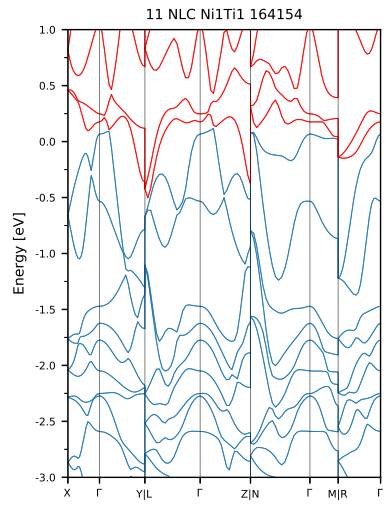

(u) 11.2.21

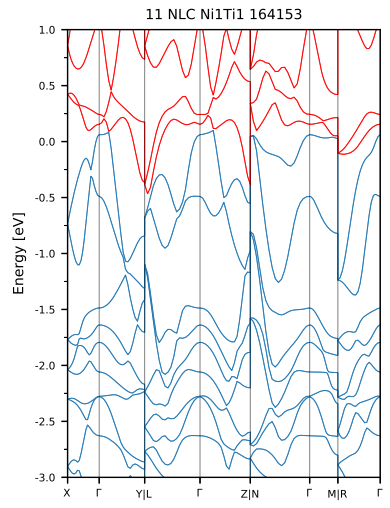

(v) 11.2.22

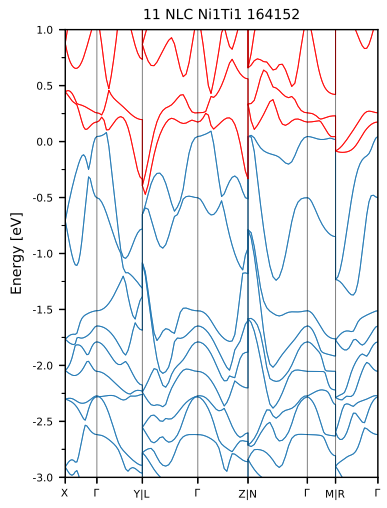

(w) 11.2.23

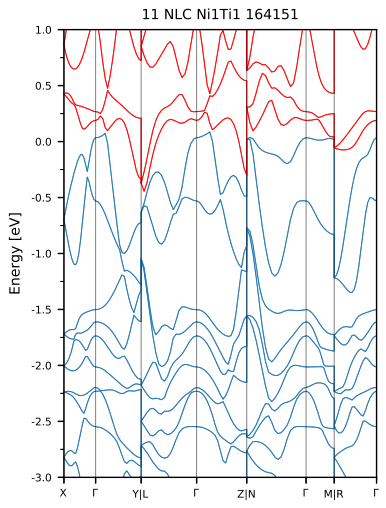

(x) 11.2.24

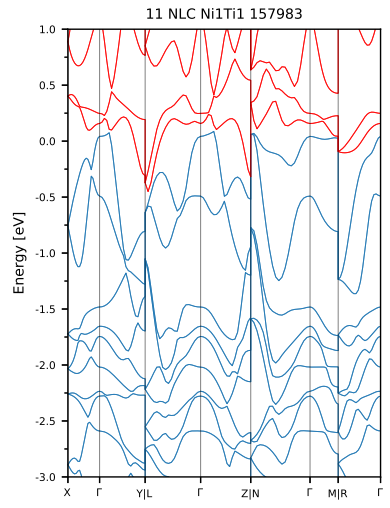

(y) 11.2.25

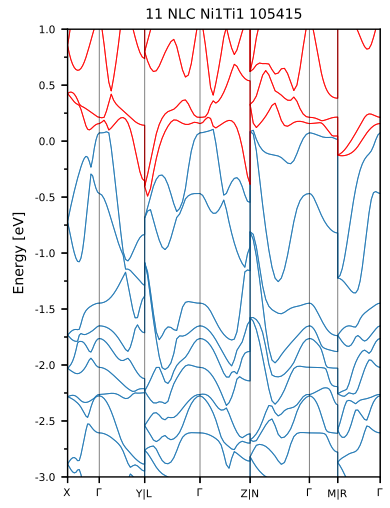

(z) 11.2.26

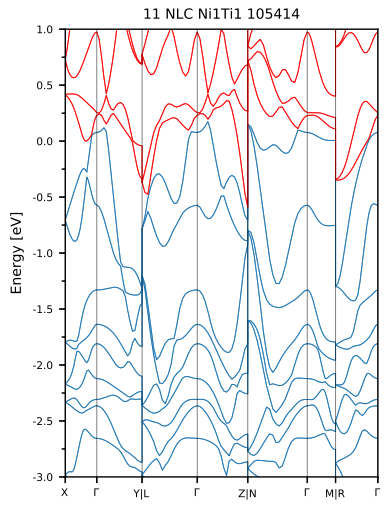

(aa) 11.2.27

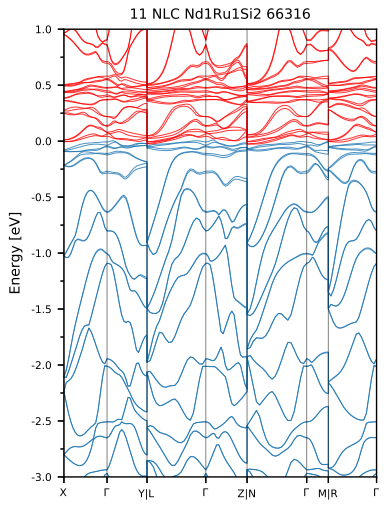

(ab) 11.2.28

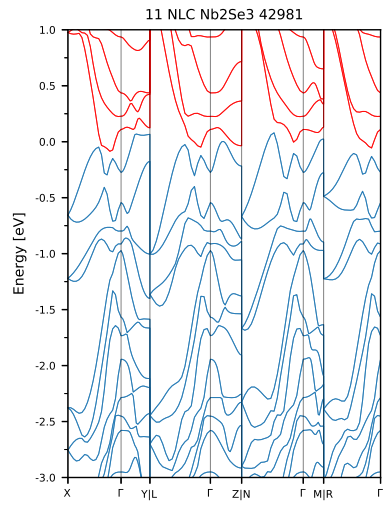

(ac) 11.2.29

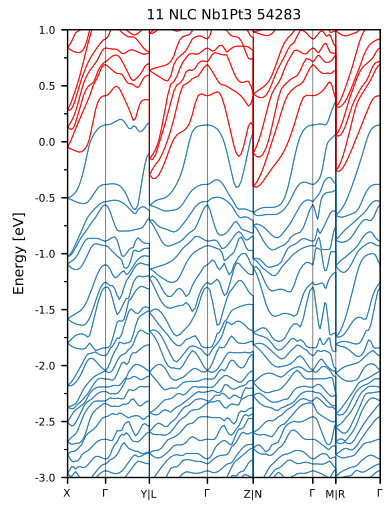

(ad) 11.2.30

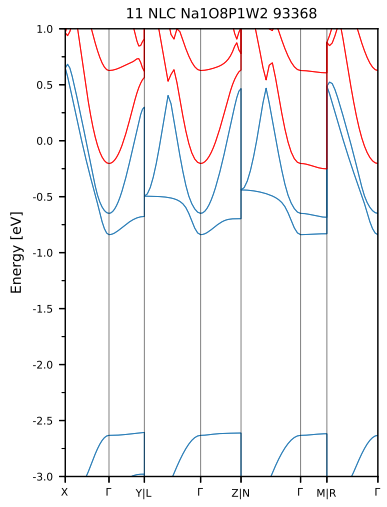

(ae) 11.2.31

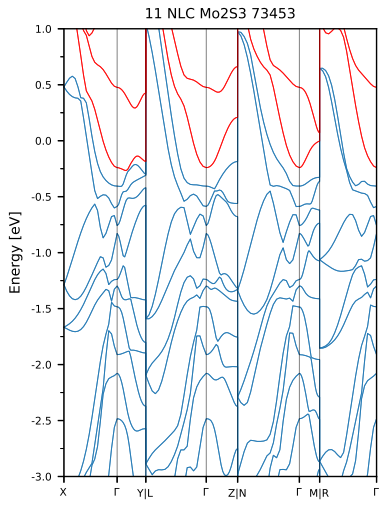

(af) 11.2.32

Figure 12: Band structures for topological materials in SG 11, part 2/13, from 11.2.17 to 11.2.32. See the list of figures.

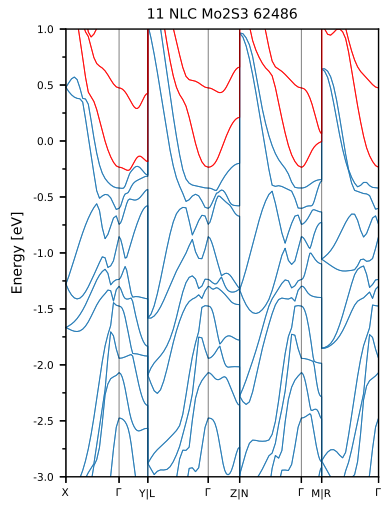

(ag) 11.2.33

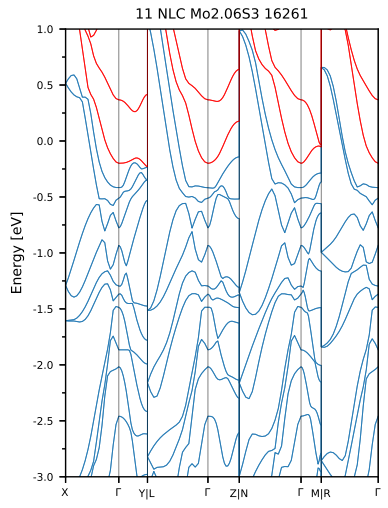

(ah) 11.2.34

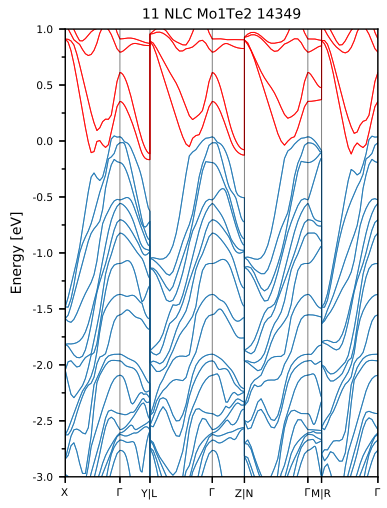

(ai) 11.2.35

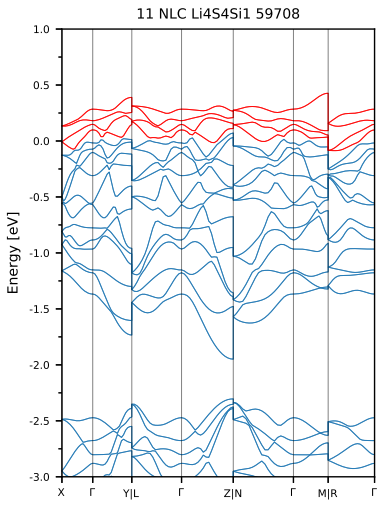

(aj) 11.2.36

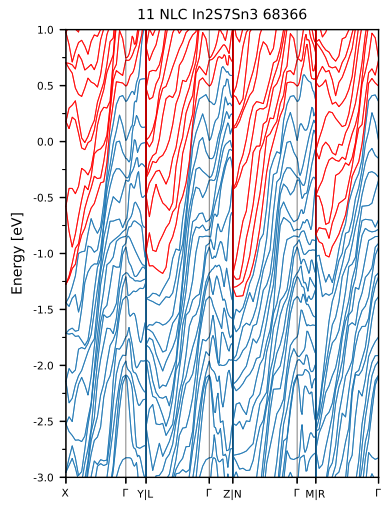

(ak) 11.2.37

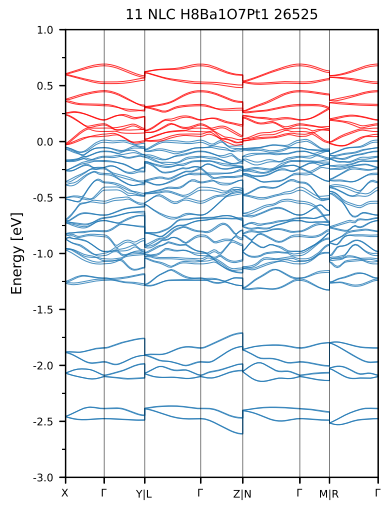

(al) 11.2.38

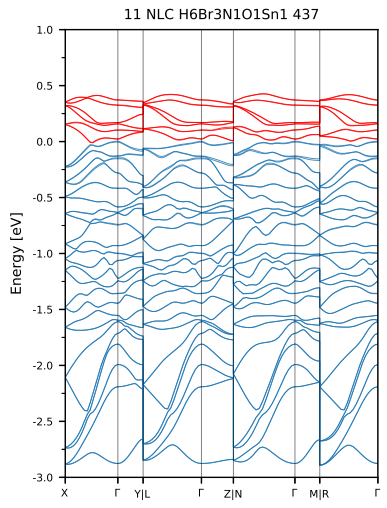

(am) 11.2.39

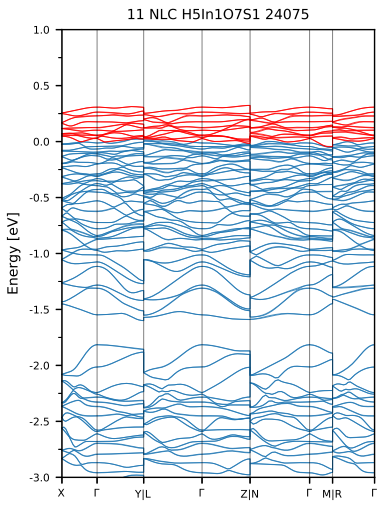

(an) 11.2.40

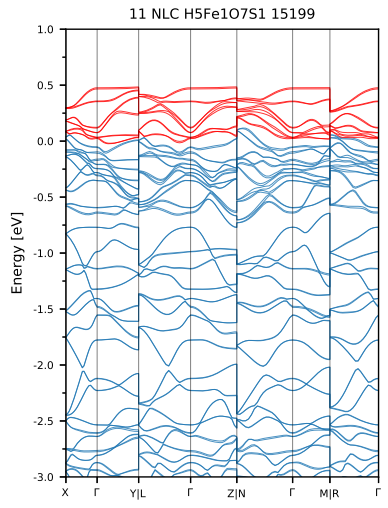

(ao) 11.2.41

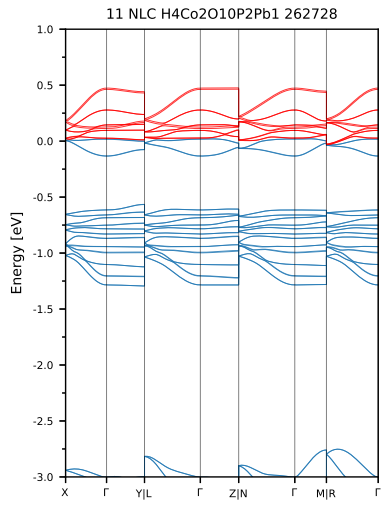

(ap) 11.2.42

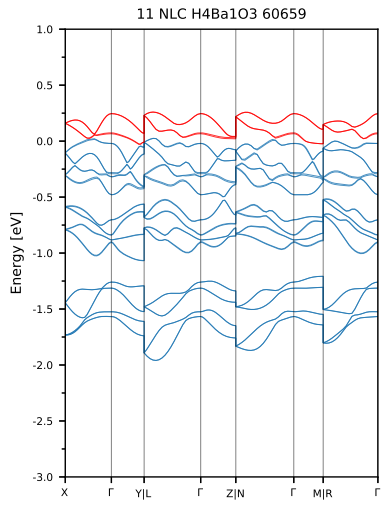

(aq) 11.2.43

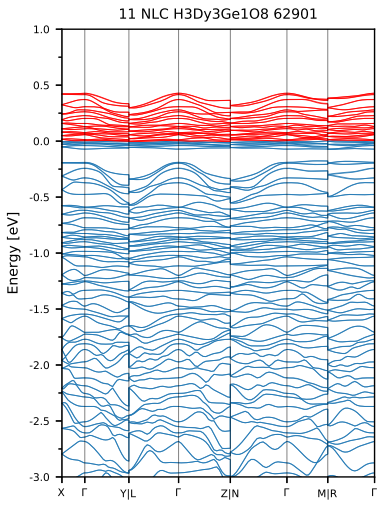

(ar) 11.2.45

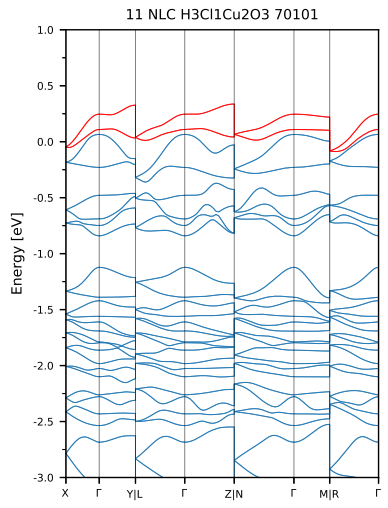

(as) 11.2.46

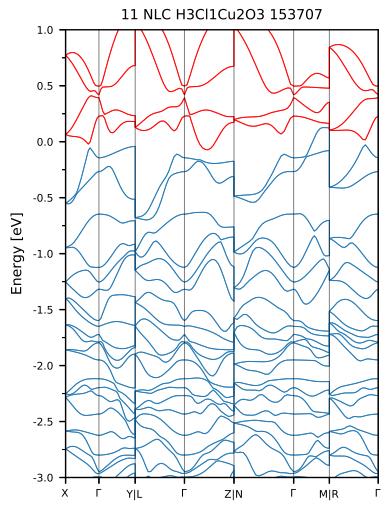

(at) 11.2.47

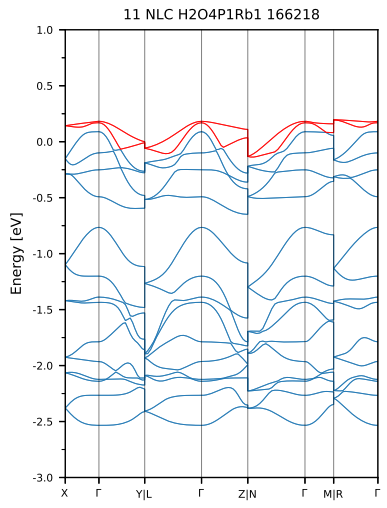

(au) 11.2.48

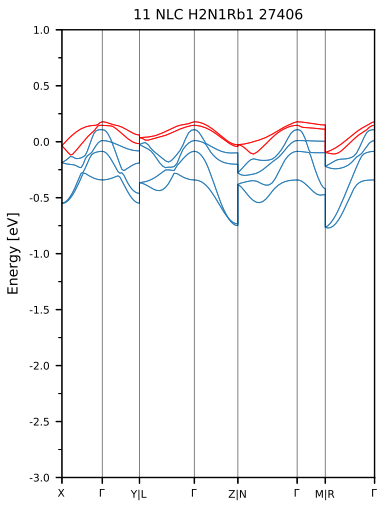

(av) 11.2.49

Figure 13: Band structures for topological materials in SG 11, part 3/13, from 11.2.33 to 11.2.49. See the list of figures.

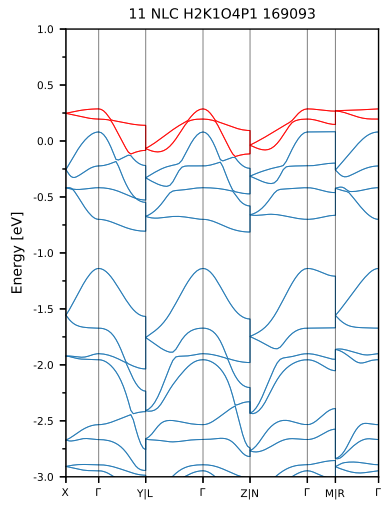

(aw) 11.2.50

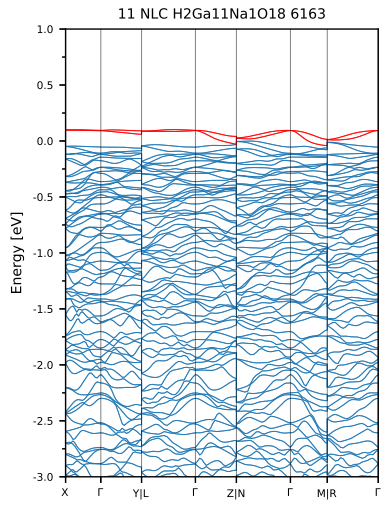

(ax) 11.2.51

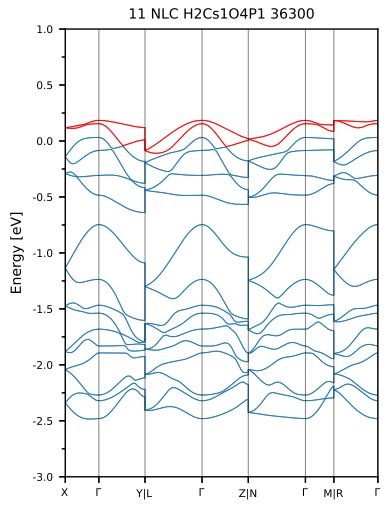

(ay) 11.2.52

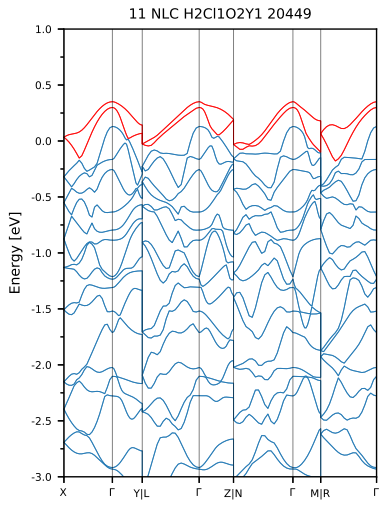

(az) 11.2.53

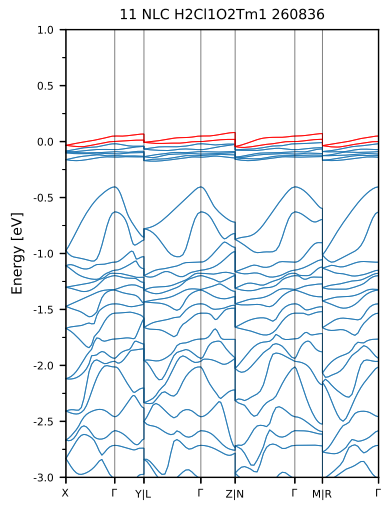

(ba) 11.2.54

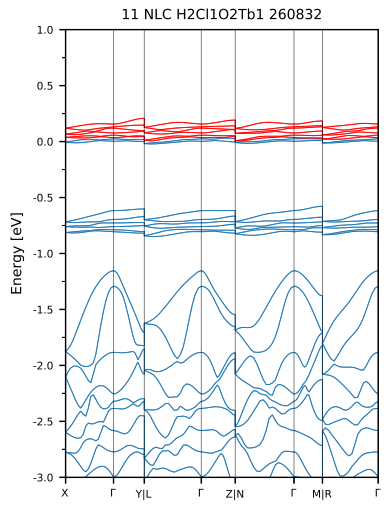

(bb) 11.2.55

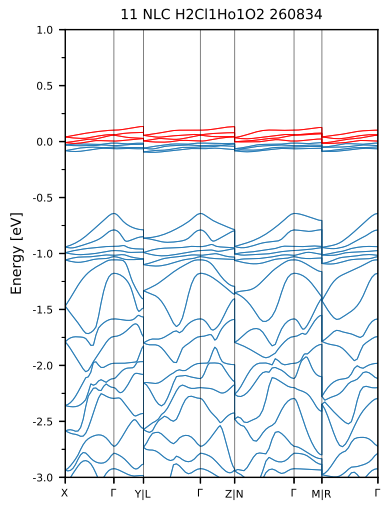

(bc) 11.2.56

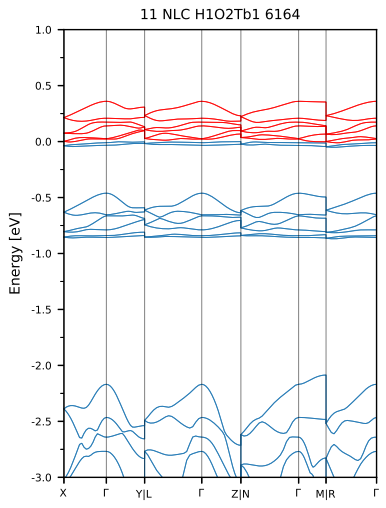

(bd) 11.2.57

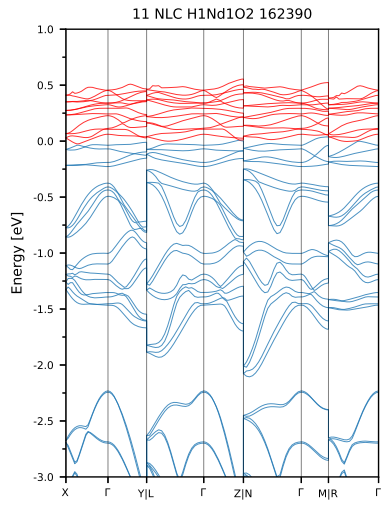

(be) 11.2.58

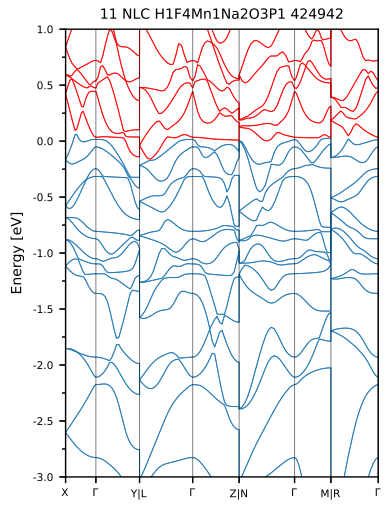

(bf) 11.2.59

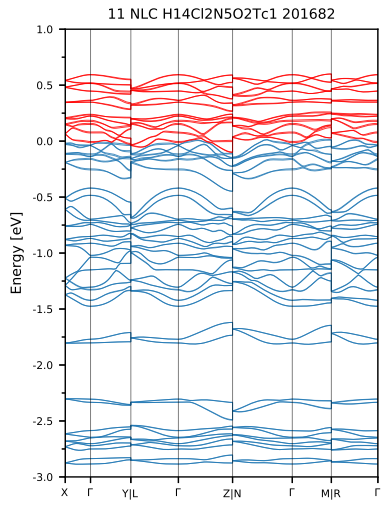

(bg) 11.2.60

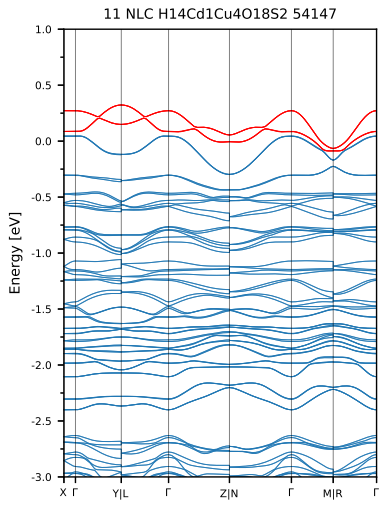

(bh) 11.2.61

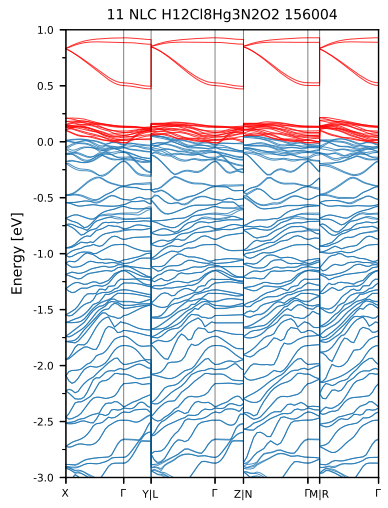

(bi) 11.2.62

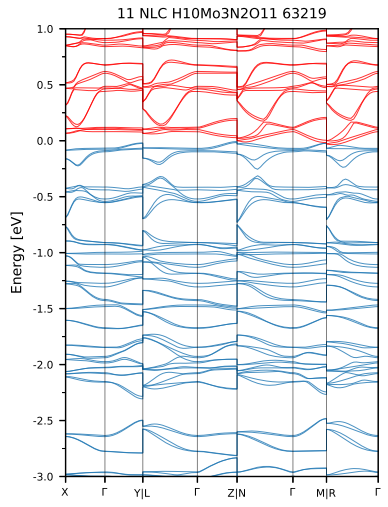

(bj) 11.2.63

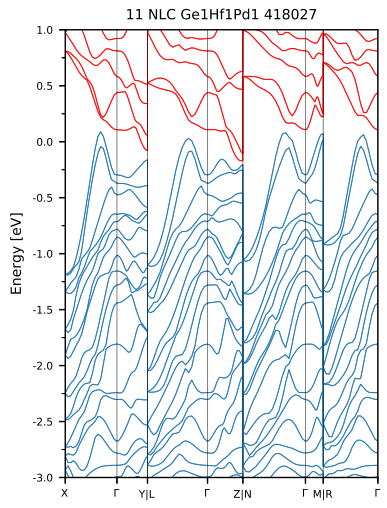

(bk) 11.2.64

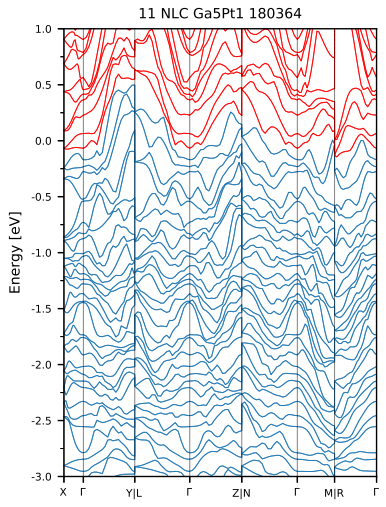

(bl) 11.2.65

Figure 14: Band structures for topological materials in SG 11, part 4/13, from 11.2.50 to 11.2.65. See the list of figures.

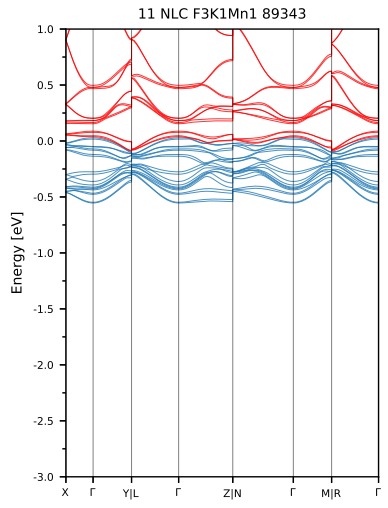

(bm) 11.2.66

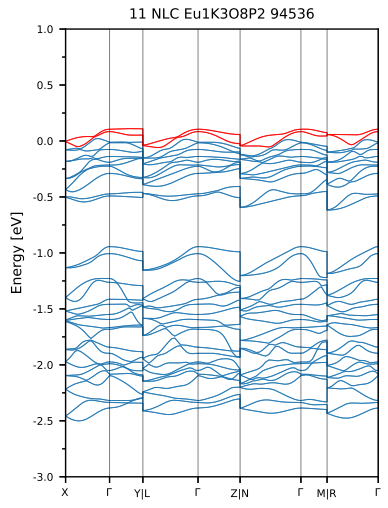

(bn) 11.2.67

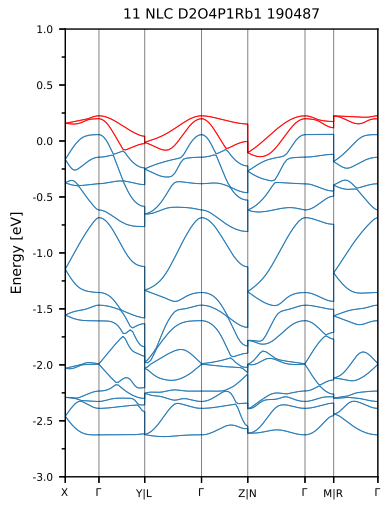

(bo) 11.2.68

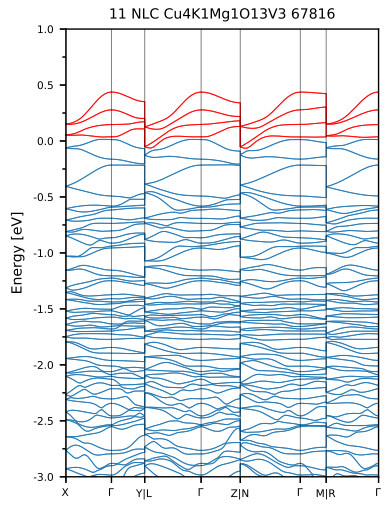

(bp) 11.2.69

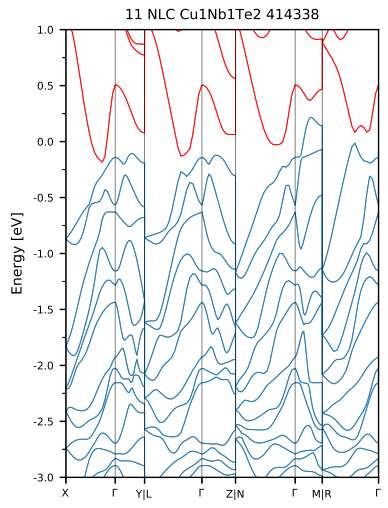

(bq) 11.2.70

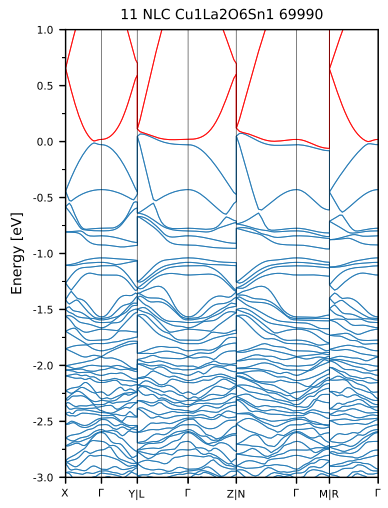

(br) 11.2.71

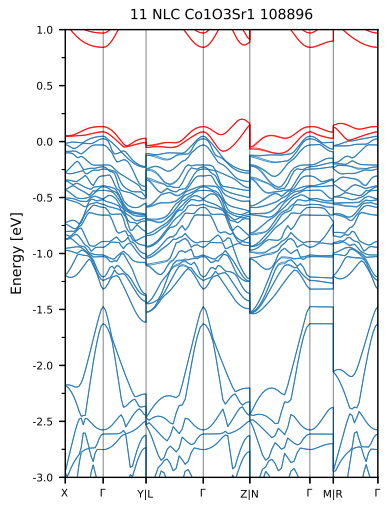

(bs) 11.2.72

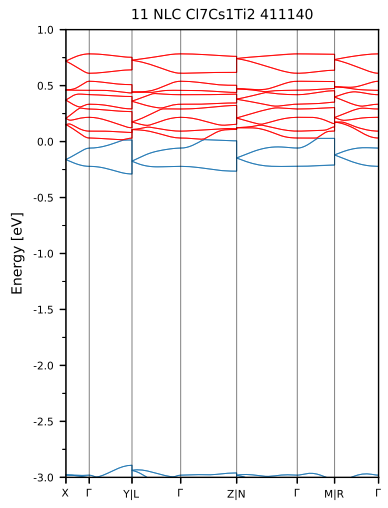

(bt) 11.2.73

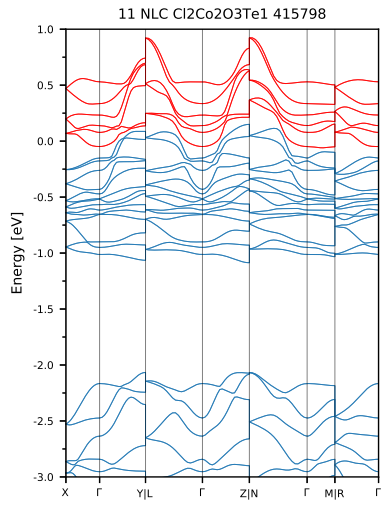

(bu) 11.2.74

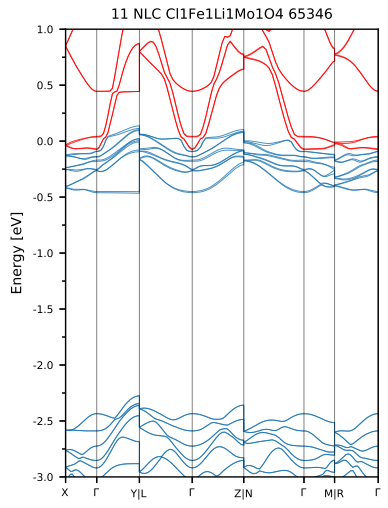

(bv) 11.2.75

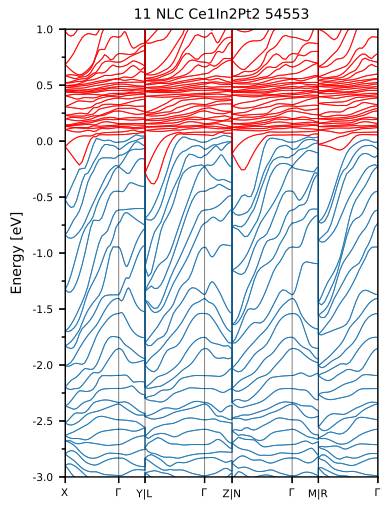

(bw) 11.2.76

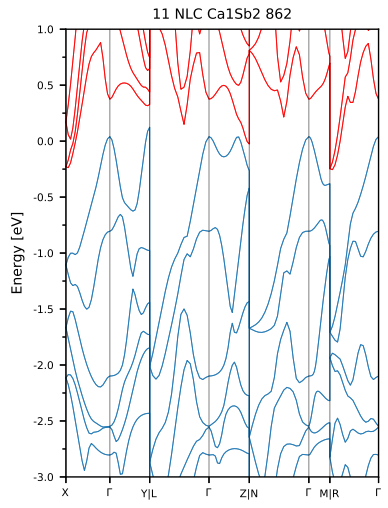

(bx) 11.2.77

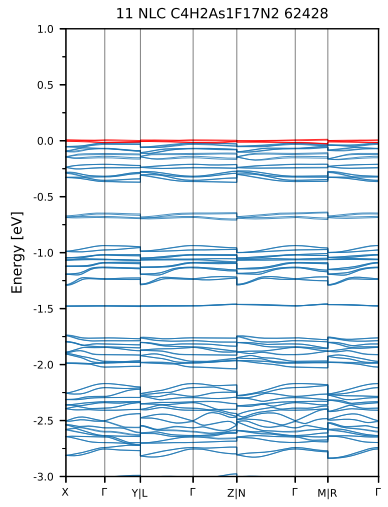

(by) 11.2.78

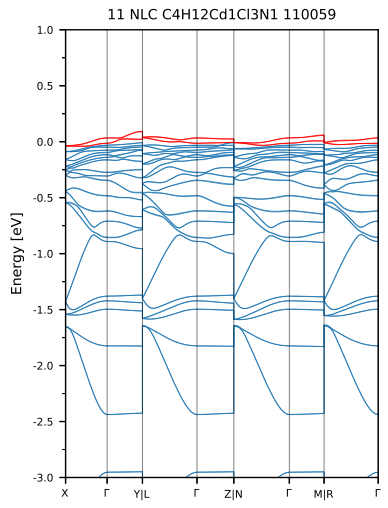

(bz) 11.2.79

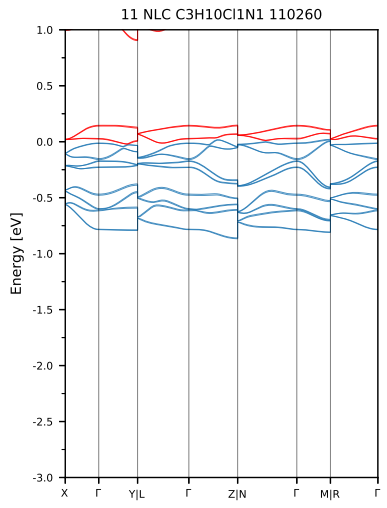

(ca) 11.2.80

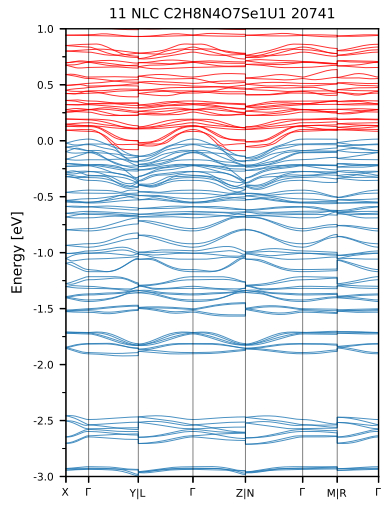

(cb) 11.2.81

Figure 15: Band structures for topological materials in SG 11, part 5/13, from 11.2.66 to 11.2.81. See the list of figures.

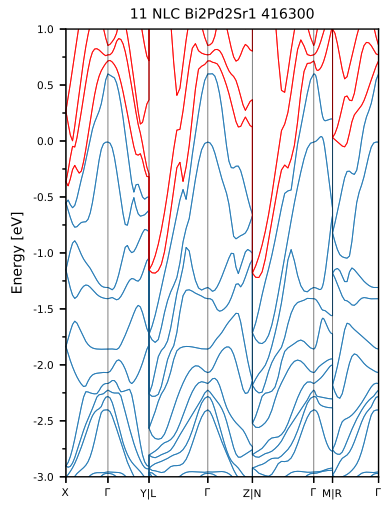

(cc) 11.2.82

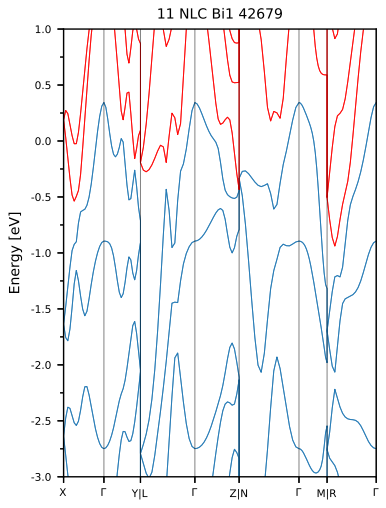

(cd) 11.2.83

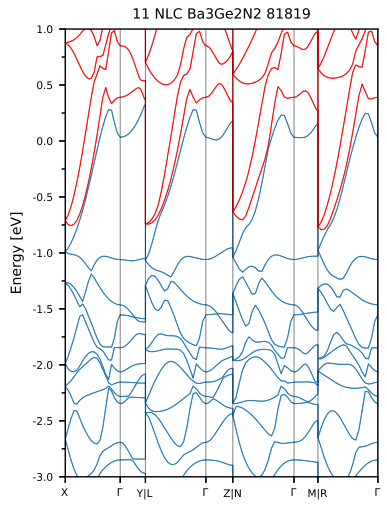

(ce) 11.2.84

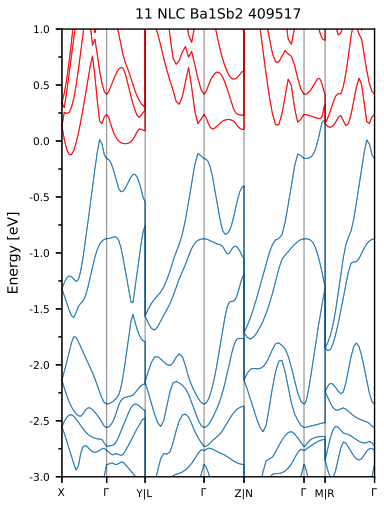

(cf) 11.2.85

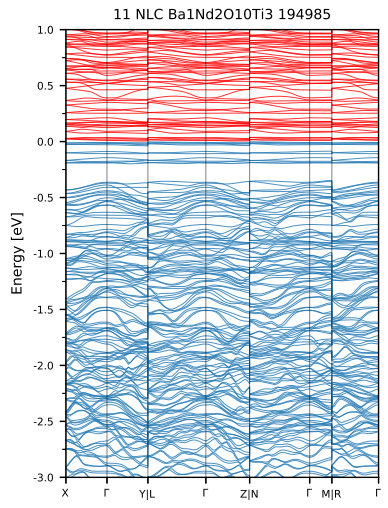

(cg) 11.2.86

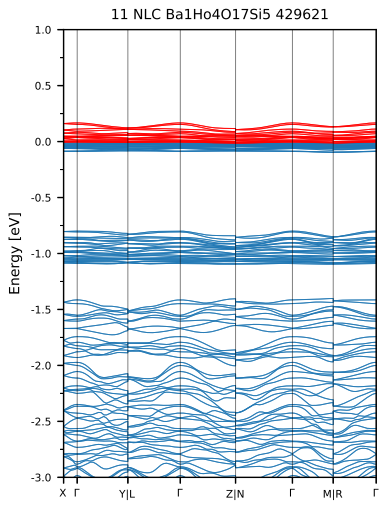

(ch) 11.2.87

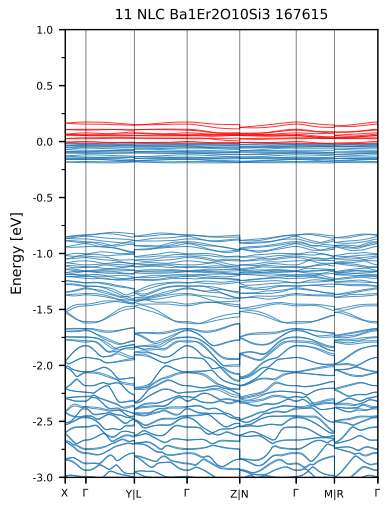

(ci) 11.2.88

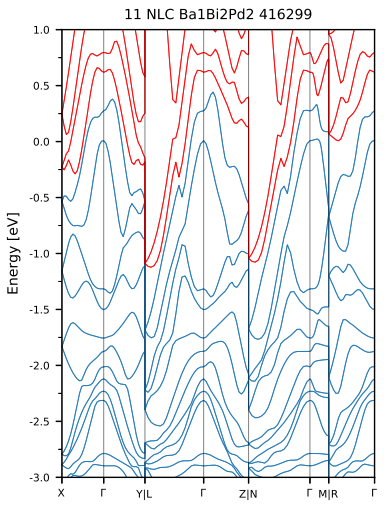

(cj) 11.2.89

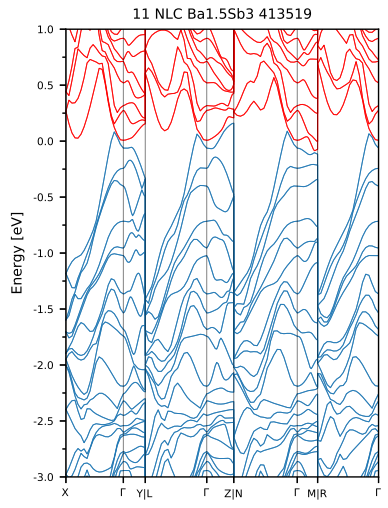

(ck) 11.2.90

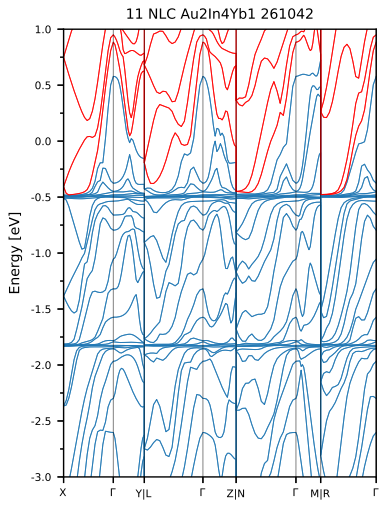

(cl) 11.2.91

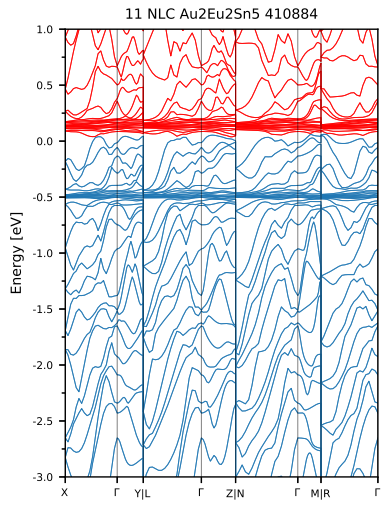

(cm) 11.2.92

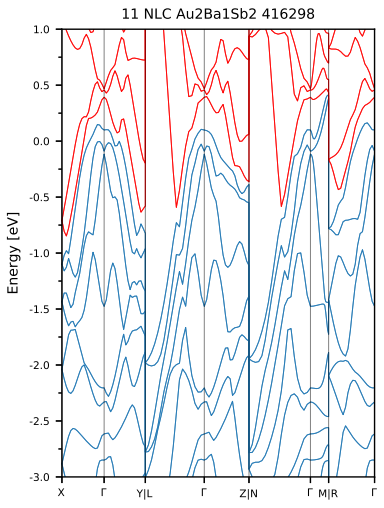

(cn) 11.2.93

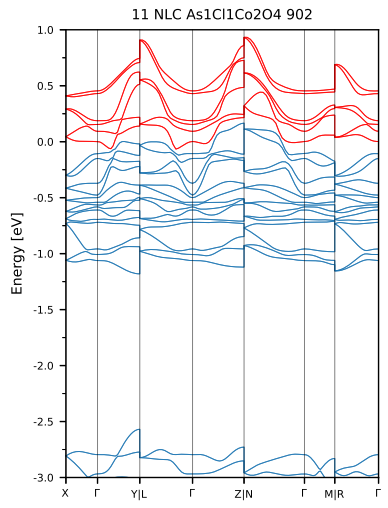

(co) 11.2.94

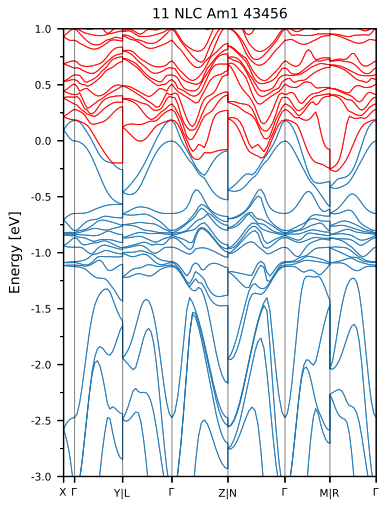

(cp) 11.2.95

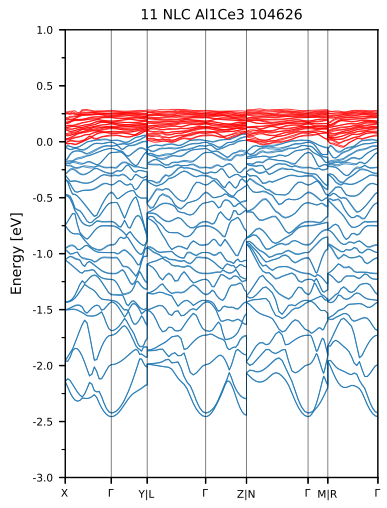

(cq) 11.2.96

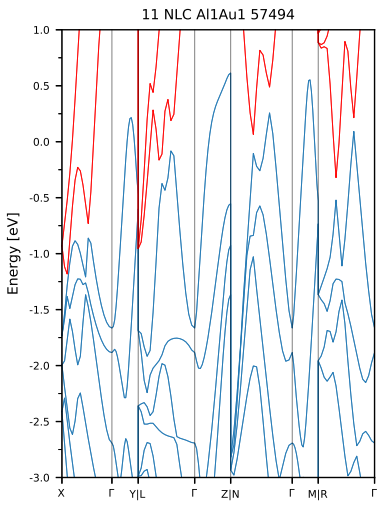

(cr) 11.2.97

Figure 16: Band structures for topological materials in SG 11, part 6/13, from 11.2.82 to 11.2.97. See the list of figures.

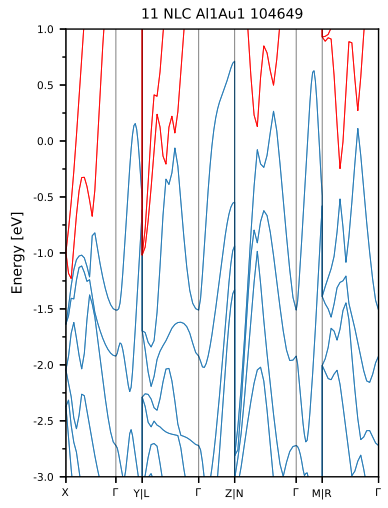

(cs) 11.2.98

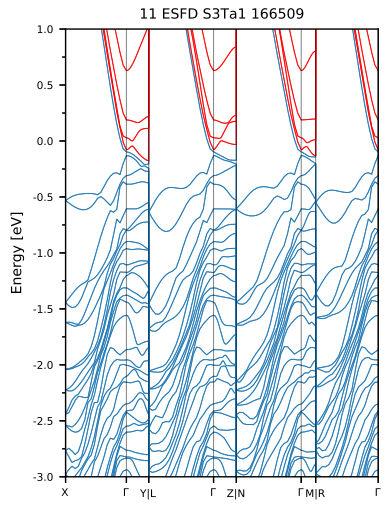

(ct) 11.4.1

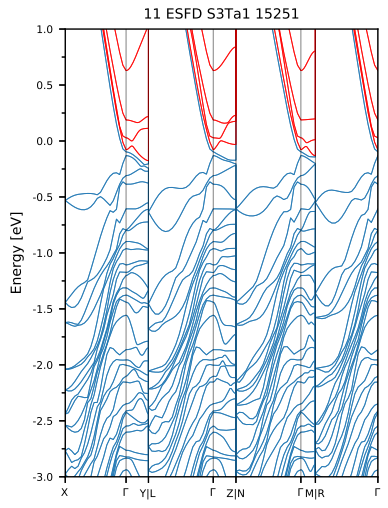

(cu) 11.4.2

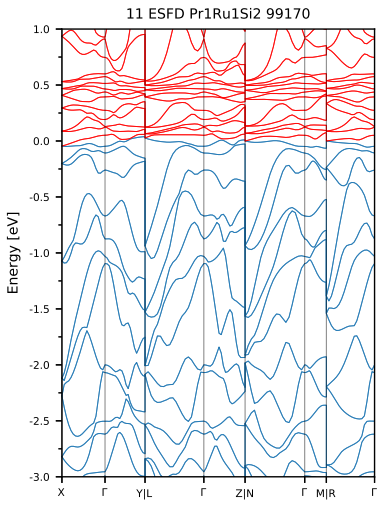

(cv) 11.4.3

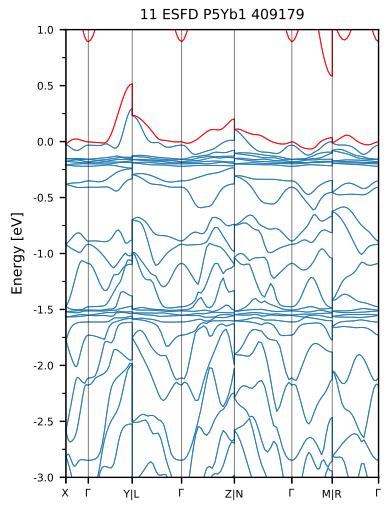

(cw) 11.4.4

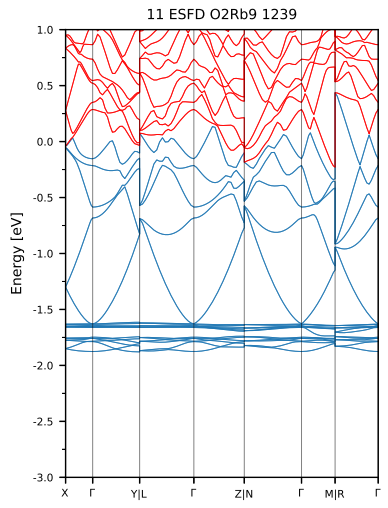

(cx) 11.4.5

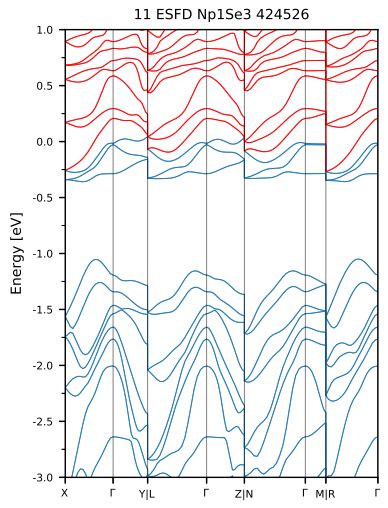

(cy) 11.4.6

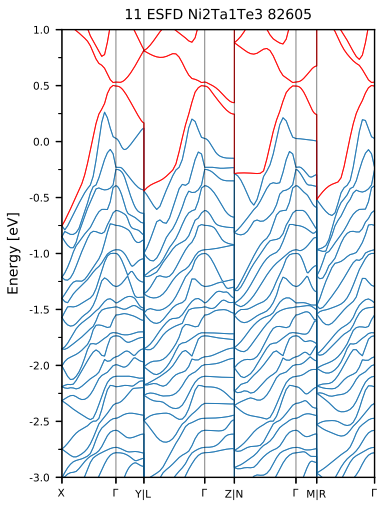

(cz) 11.4.7

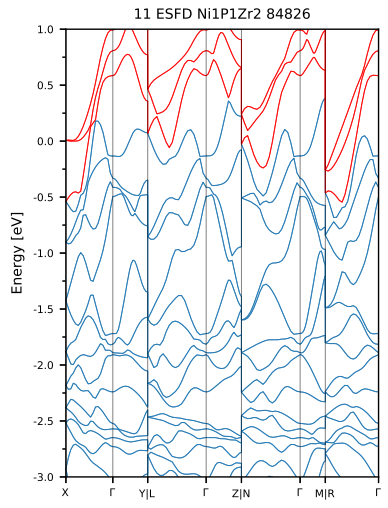

(da) 11.4.8

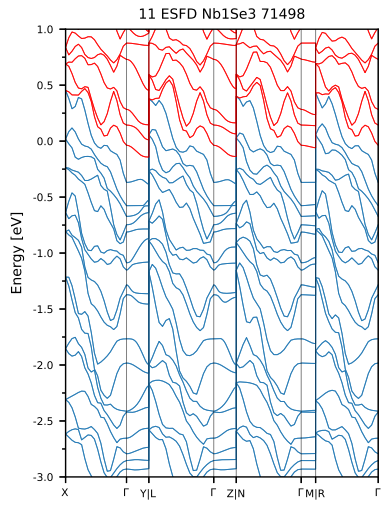

(db) 11.4.9

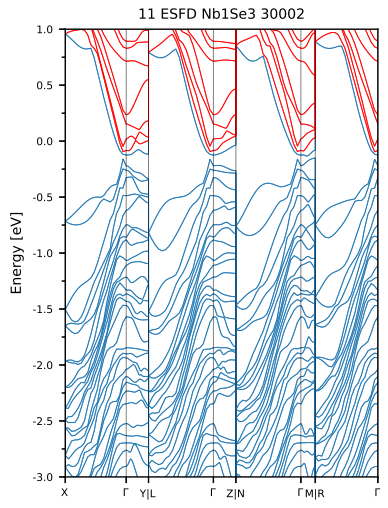

(dc) 11.4.10

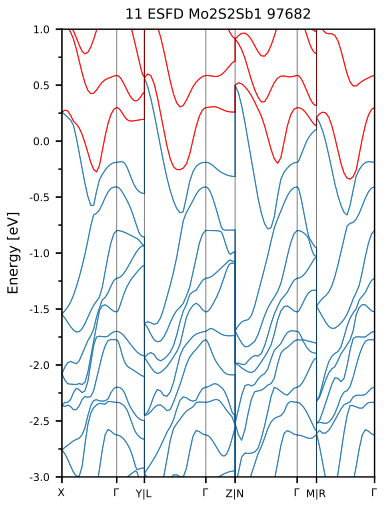

(dd) 11.4.11

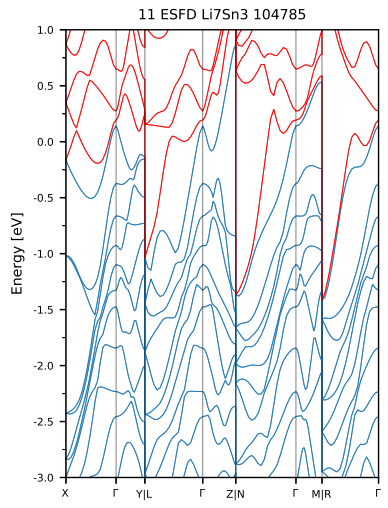

(de) 11.4.12

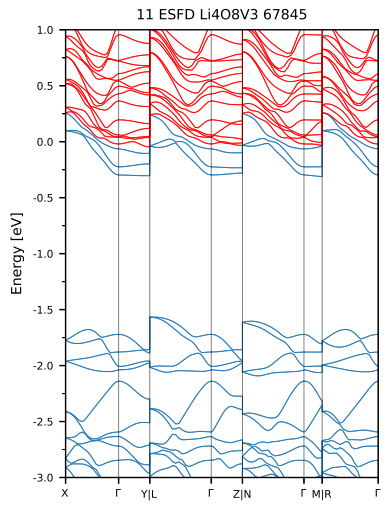

(df) 11.4.13

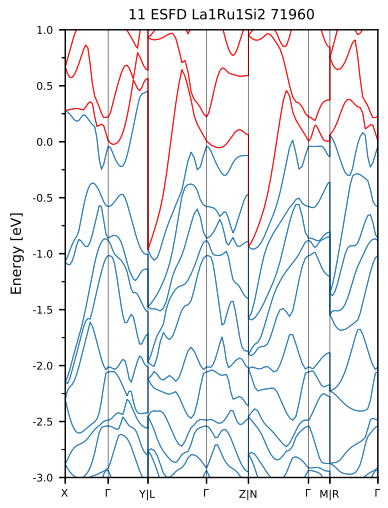

(dg) 11.4.14

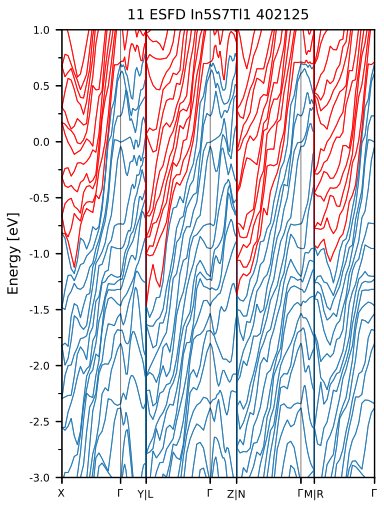

(dh) 11.4.15

Figure 17: Band structures for topological materials in SG 11, part 7/13, from 11.2.98 to 11.4.15. See the list of figures.

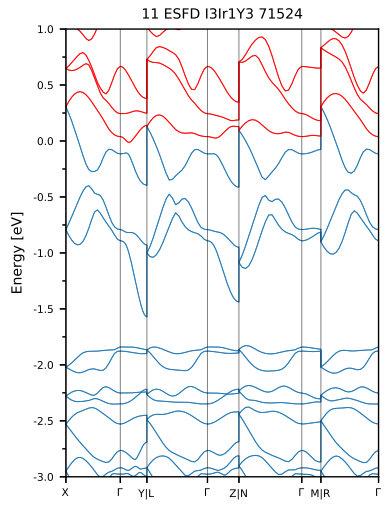

(di) 11.4.16

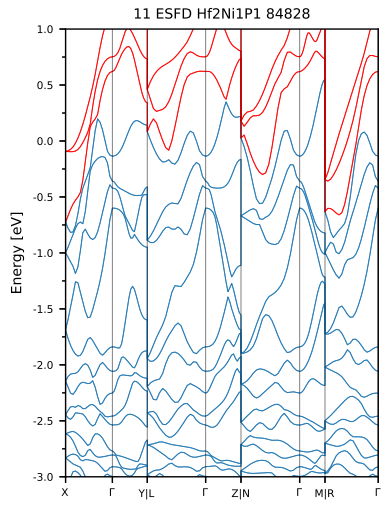

(dj) 11.4.17

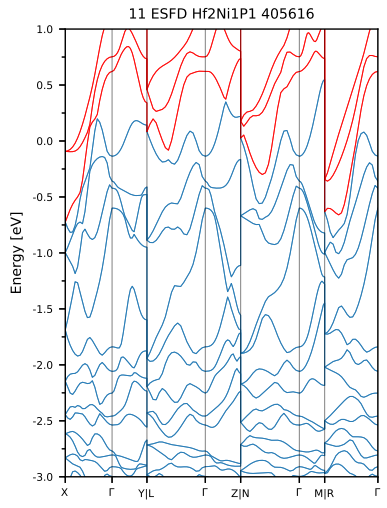

(dk) 11.4.18

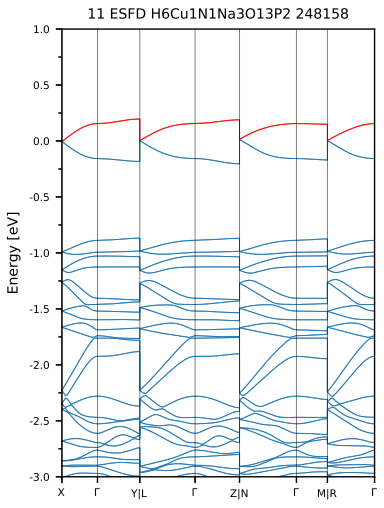

(dl) 11.4.19

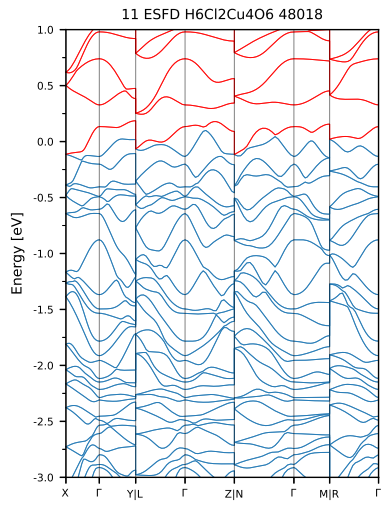

(dm) 11.4.20

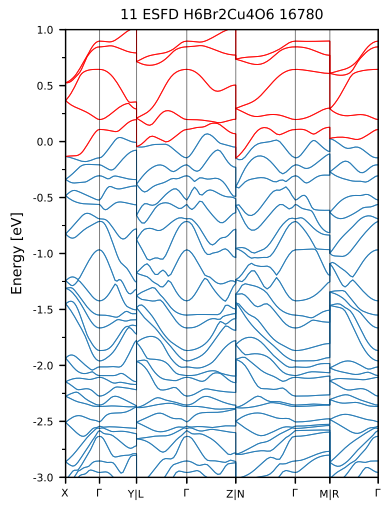

(dn) 11.4.21

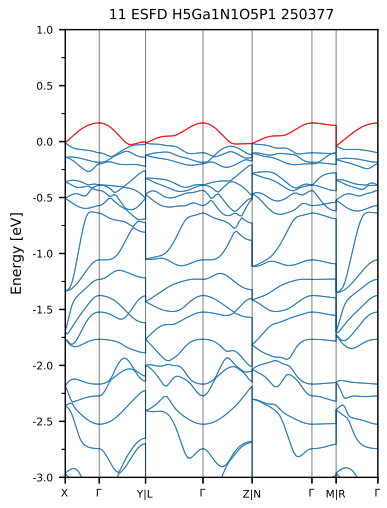

(do) 11.4.22

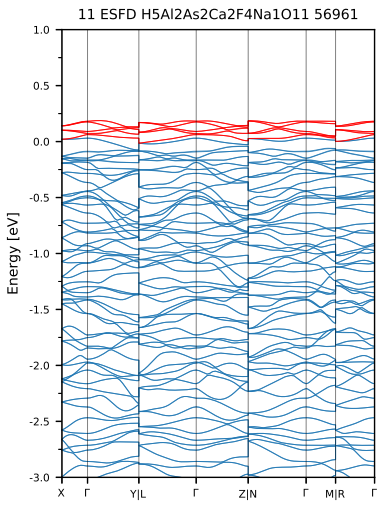

(dp) 11.4.23

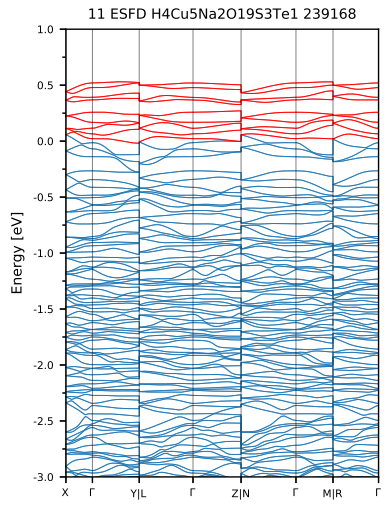

(dq) 11.4.24

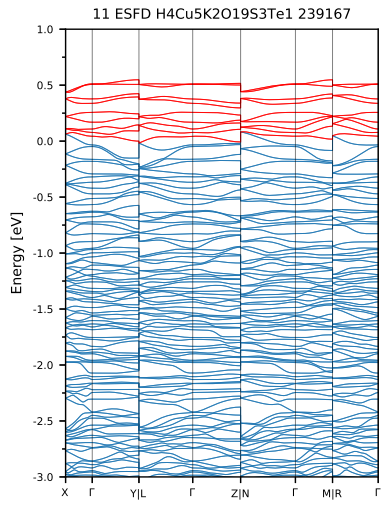

(dr) 11.4.25

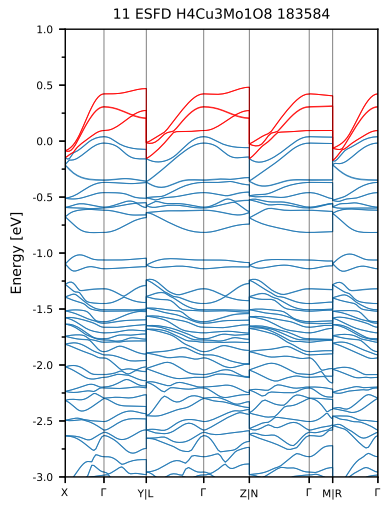

(ds) 11.4.26

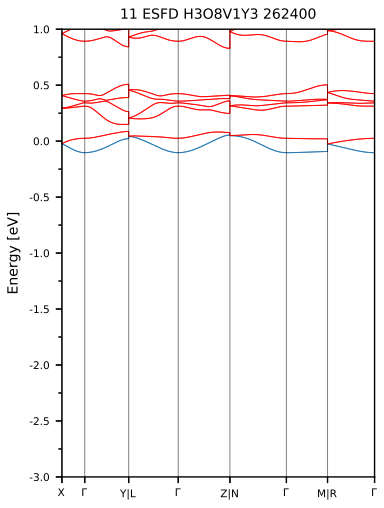

(dt) 11.4.27

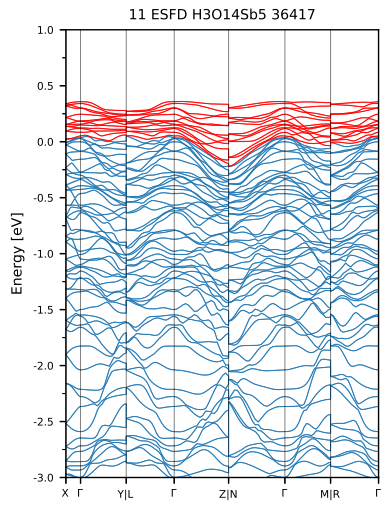

(du) 11.4.28

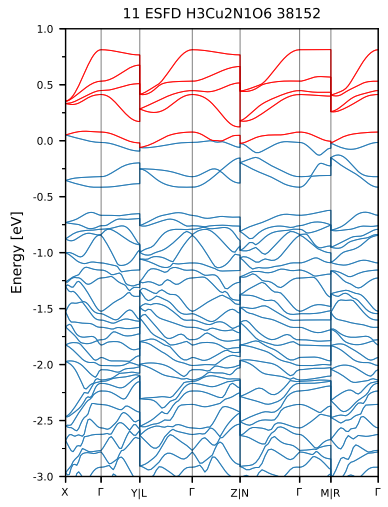

(dv) 11.4.29

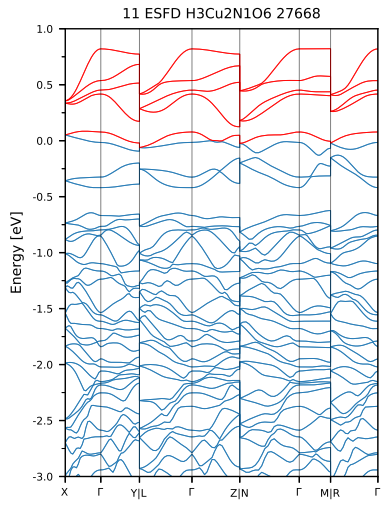

(dw) 11.4.30

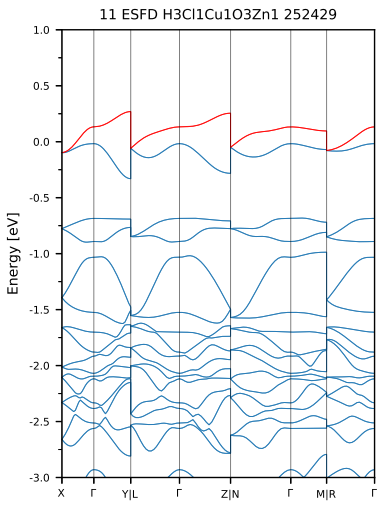

(dx) 11.4.31

Figure 18: Band structures for topological materials in SG 11, part 8/13, from 11.4.16 to 11.4.31. See the list of figures.

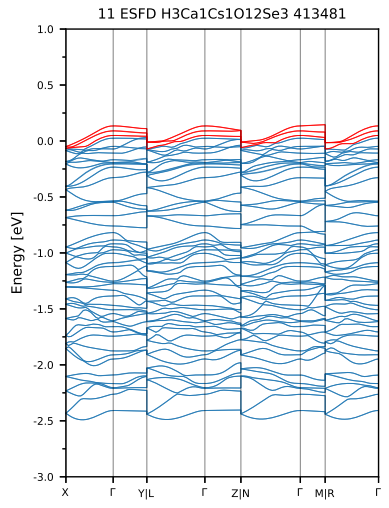

(dy) 11.4.32

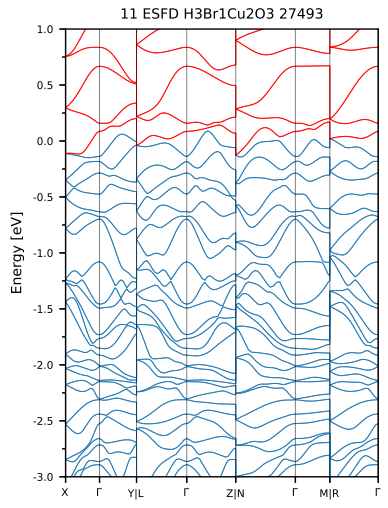

(dz) 11.4.33

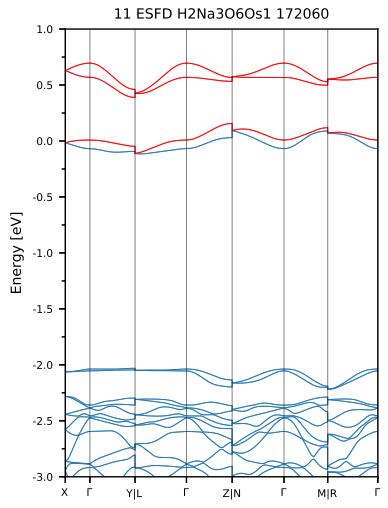

(ea) 11.4.34

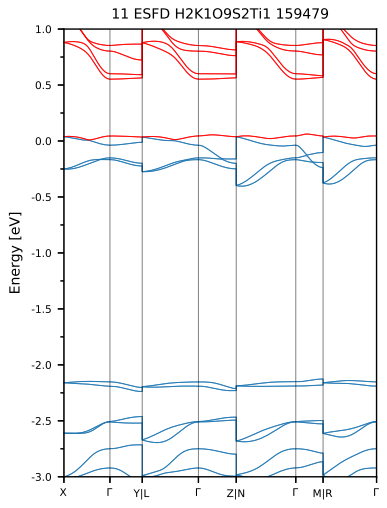

(eb) 11.4.35

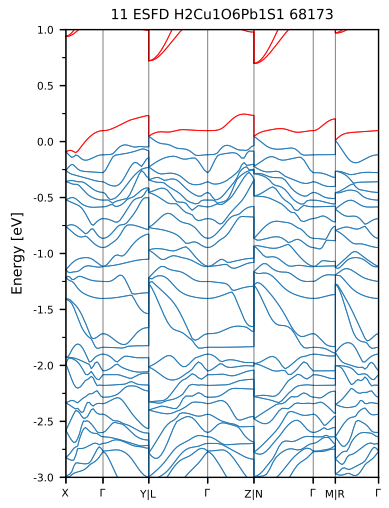

(ec) 11.4.36

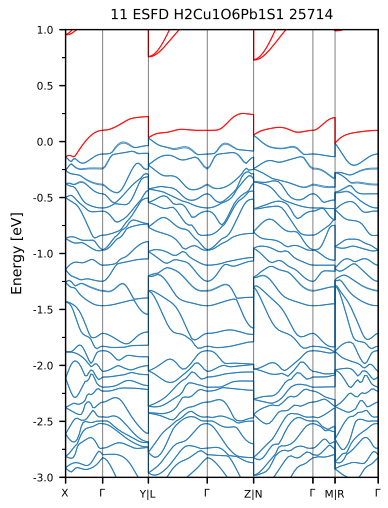

(ed) 11.4.37

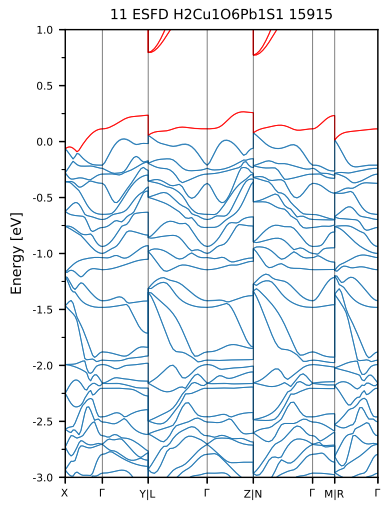

(ee) 11.4.38

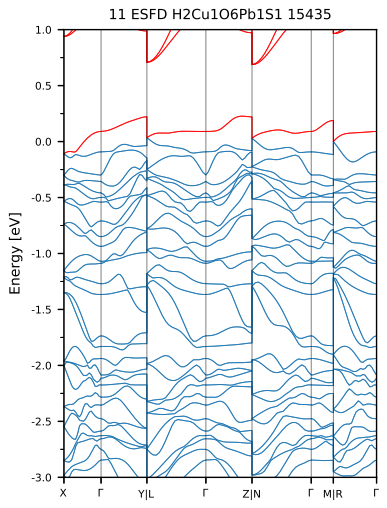

(ef) 11.4.39

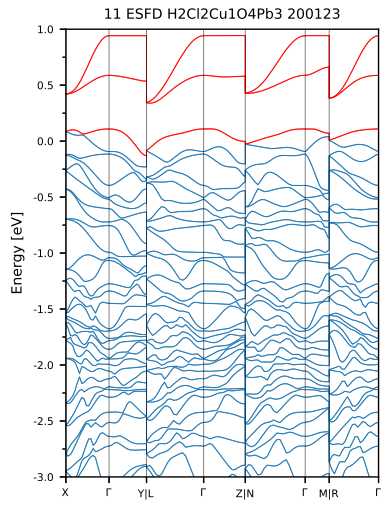

(eg) 11.4.40

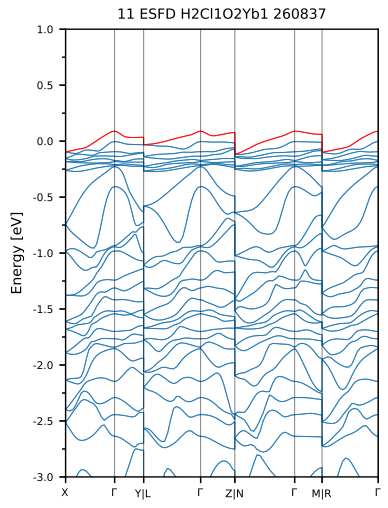

(eh) 11.4.41

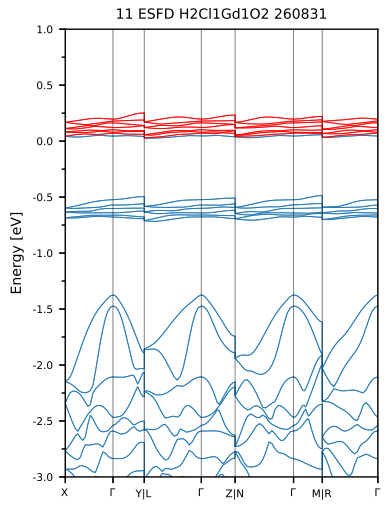

(ei) 11.4.42

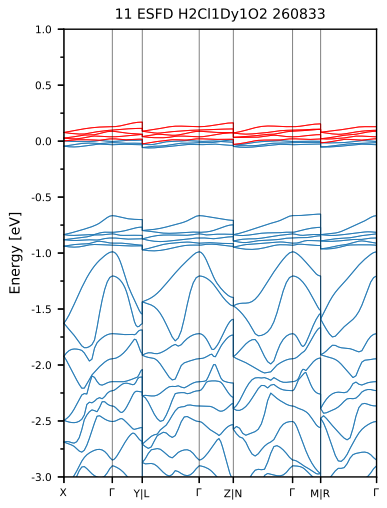

(ej) 11.4.43

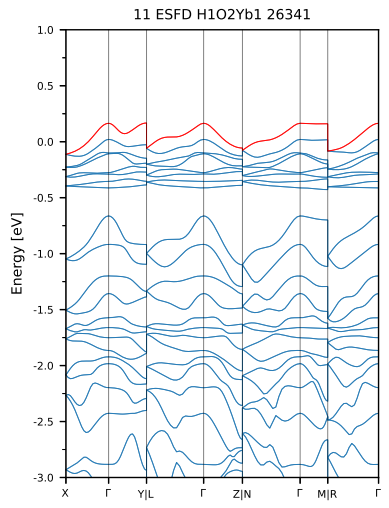

(ek) 11.4.44

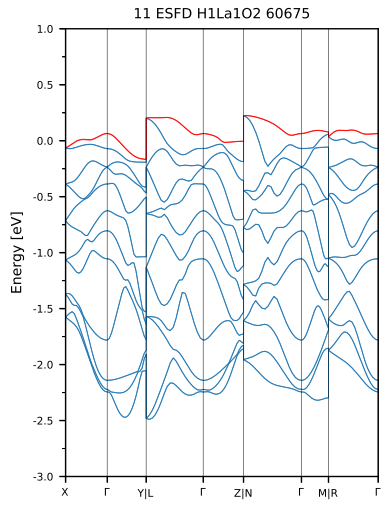

(el) 11.4.45

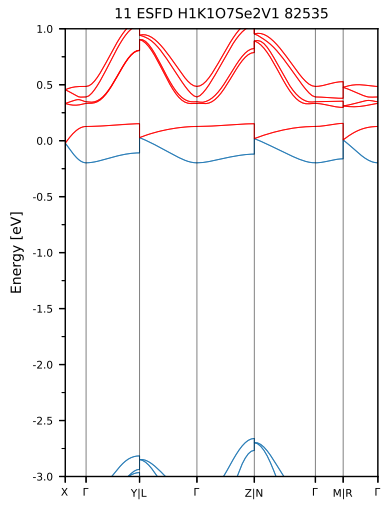

(em) 11.4.46

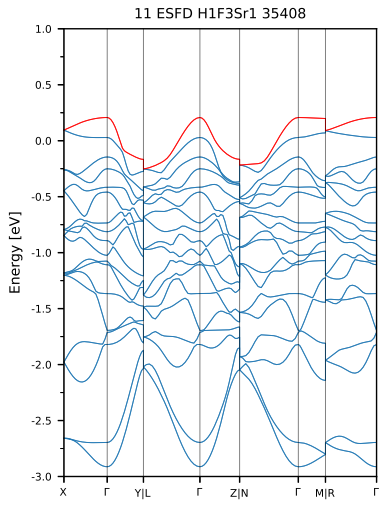

(en) 11.4.47

Figure 19: Band structures for topological materials in SG 11, part 9/13, from 11.4.32 to 11.4.47. See the list of figures.

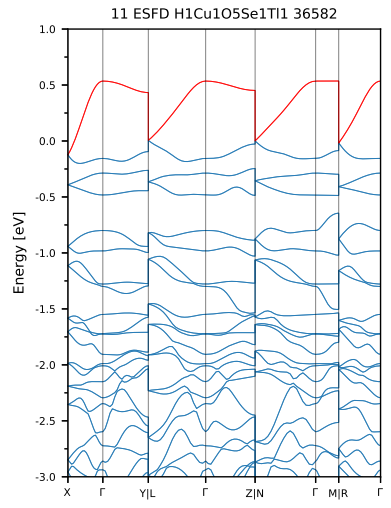

(eo) 11.4.48

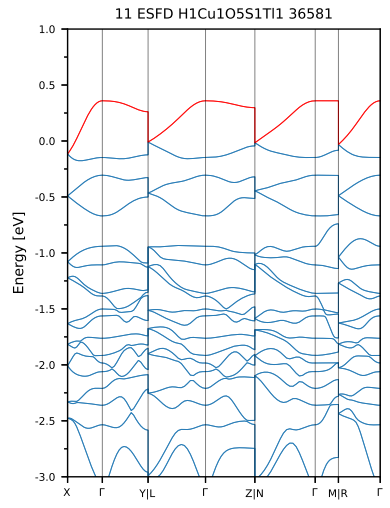

(ep) 11.4.49

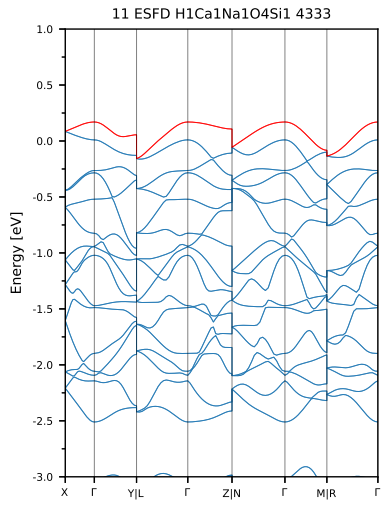

(eq) 11.4.50

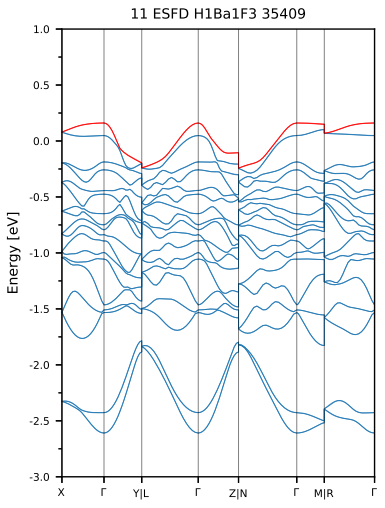

(er) 11.4.51

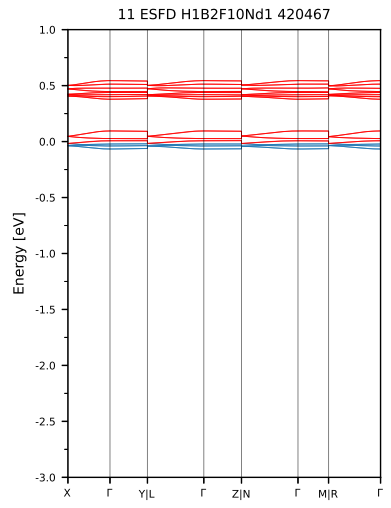

(es) 11.4.52

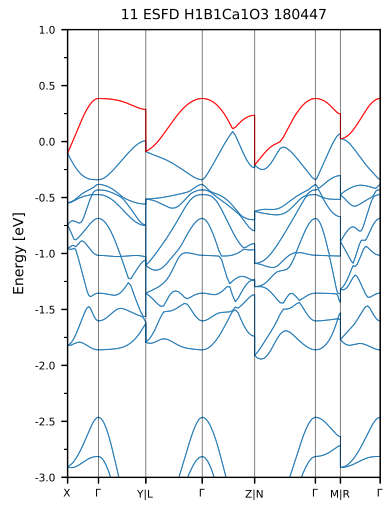

(et) 11.4.53

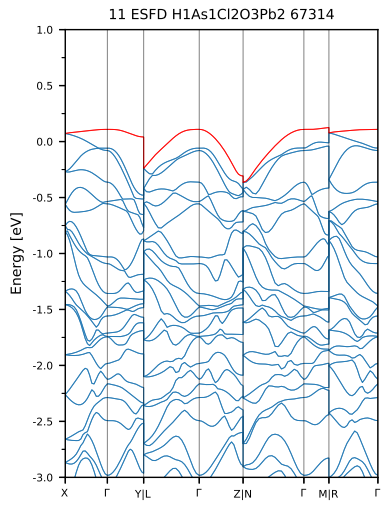

(eu) 11.4.54

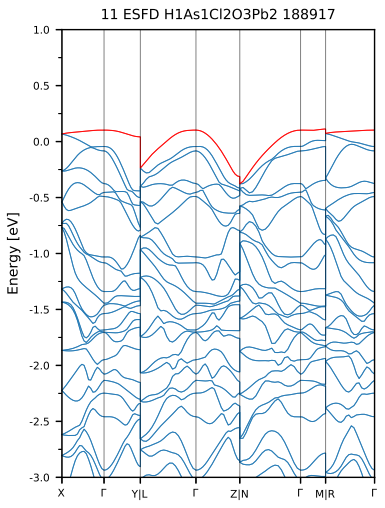

(ev) 11.4.55

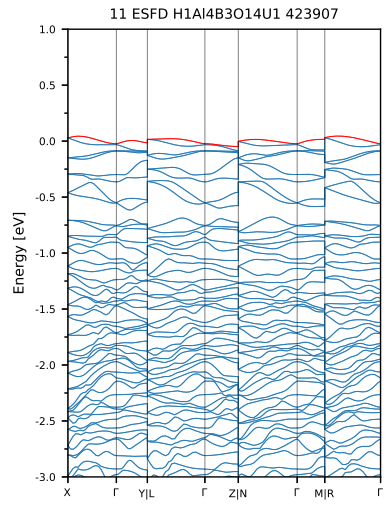

(ew) 11.4.56

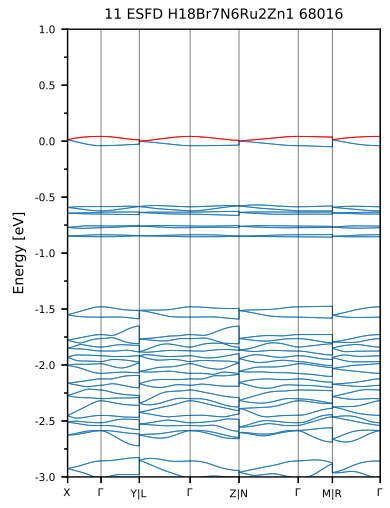

(ex) 11.4.57

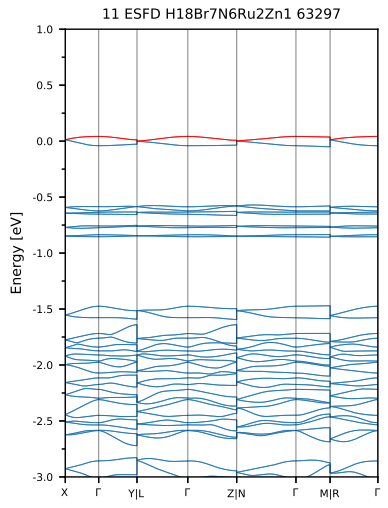

(ey) 11.4.58

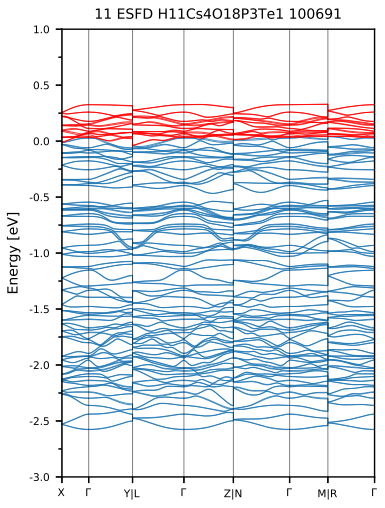

(ez) 11.4.59

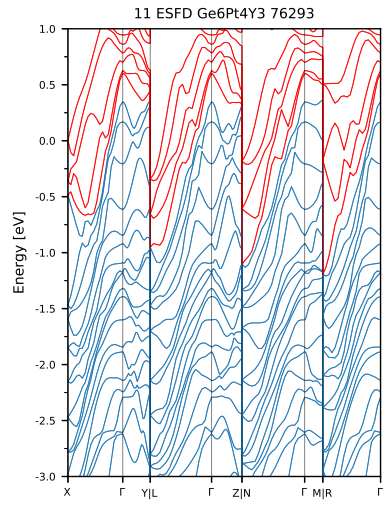

(fa) 11.4.60

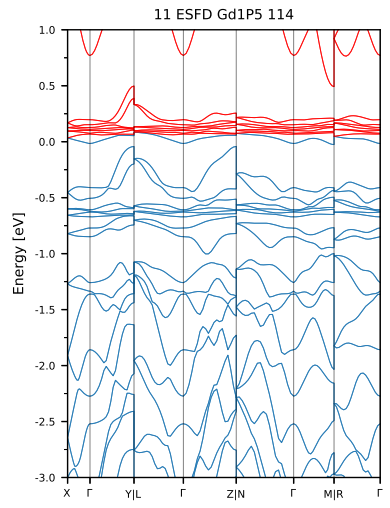

(fb) 11.4.61

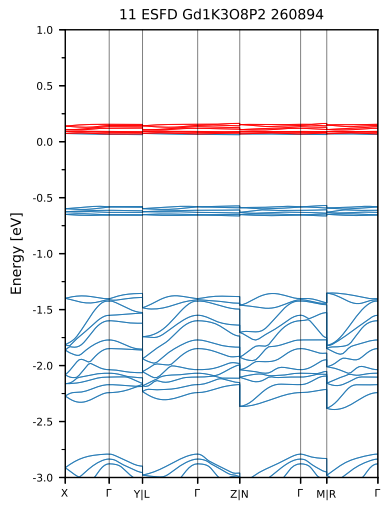

(fc) 11.4.62

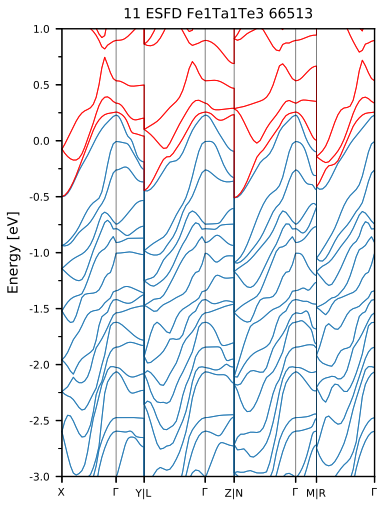

(fd) 11.4.63

Figure 20: Band structures for topological materials in SG 11, part 10/13, from 11.4.48 to 11.4.63. See the list of figures.

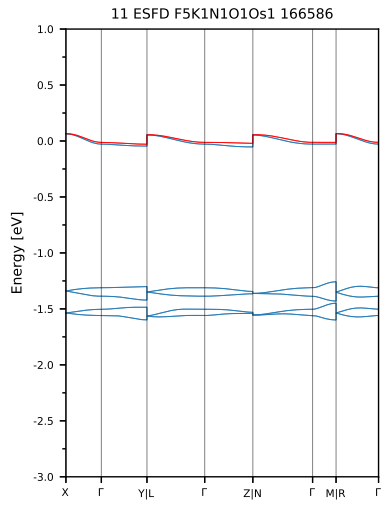

(fe) 11.4.64

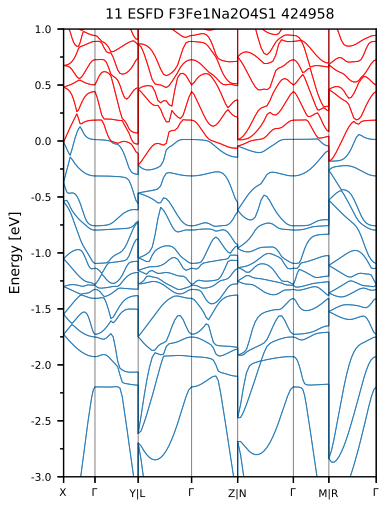

(ff) 11.4.65

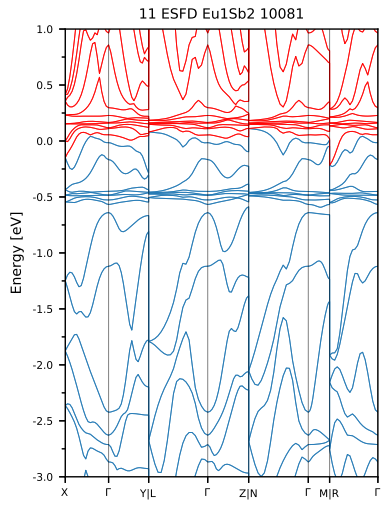

(fg) 11.4.66

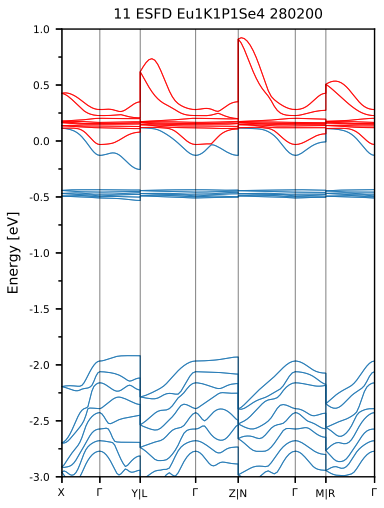

(fh) 11.4.67

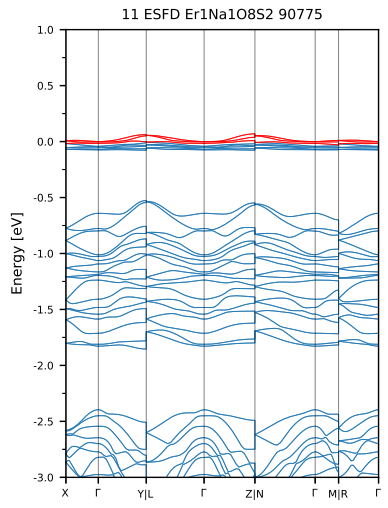

(fi) 11.4.68

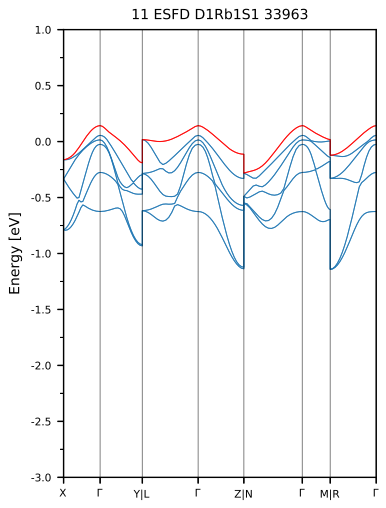

(fj) 11.4.69

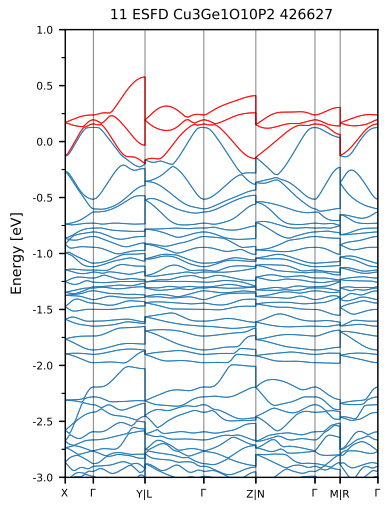

(fk) 11.4.70

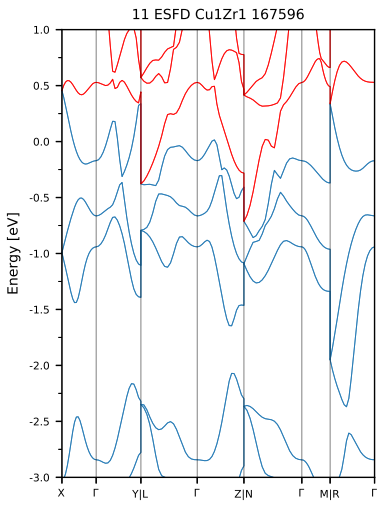

(fl) 11.4.71

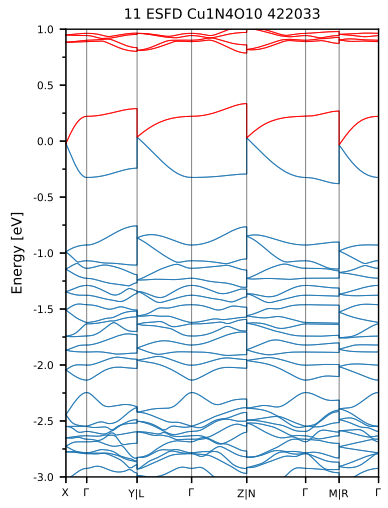

(fm) 11.4.72

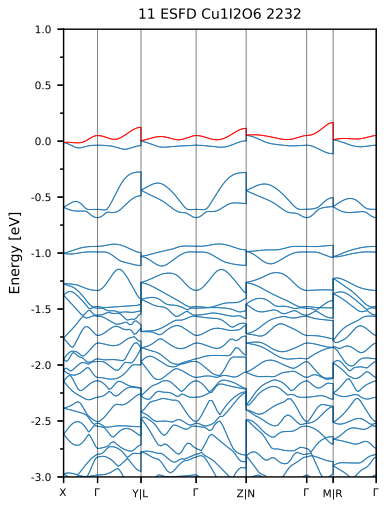

(fn) 11.4.73

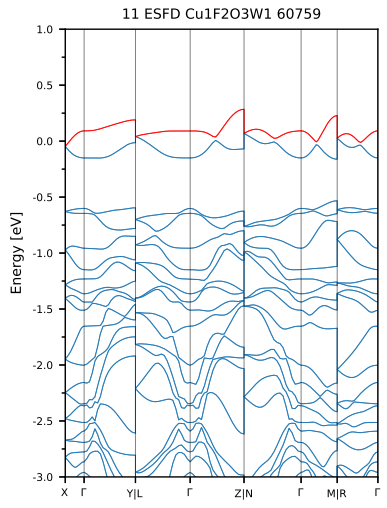

(fo) 11.4.74

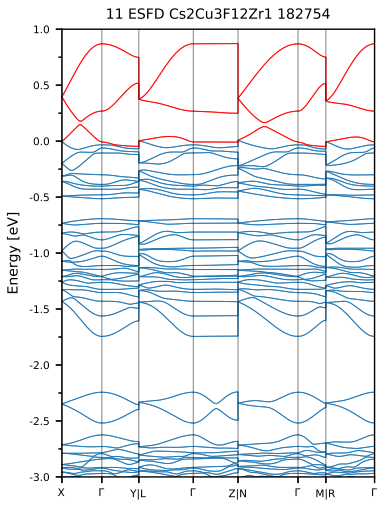

(fp) 11.4.75

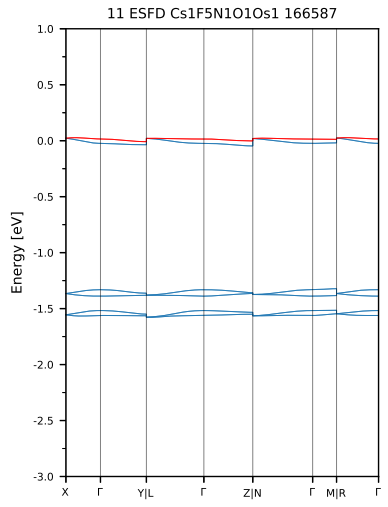

(fq) 11.4.76

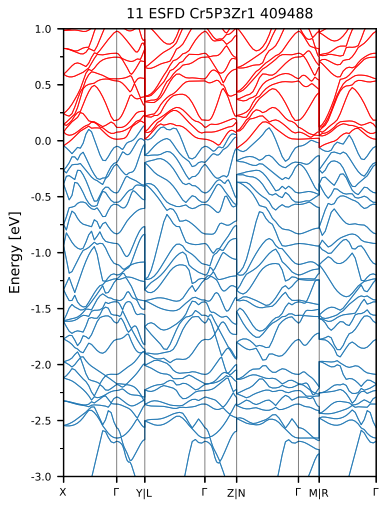

(fr) 11.4.77

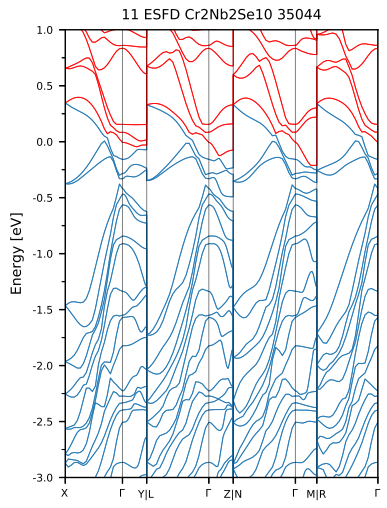

(fs) 11.4.78

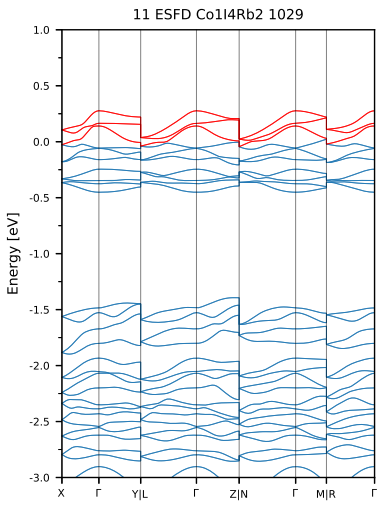

(ft) 11.4.79

Figure 21: Band structures for topological materials in SG 11, part 11/13, from 11.4.64 to 11.4.79. See the list of figures.

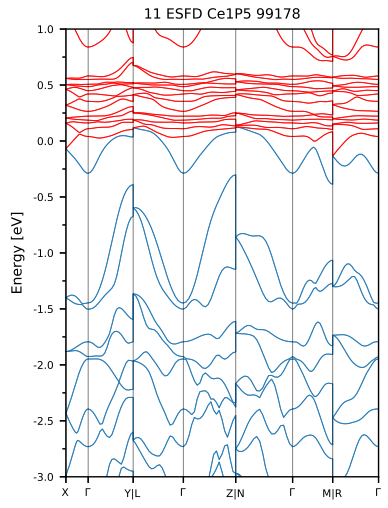

(fu) 11.4.80

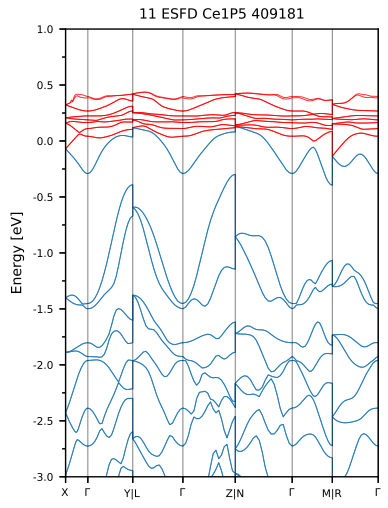

(fv) 11.4.81

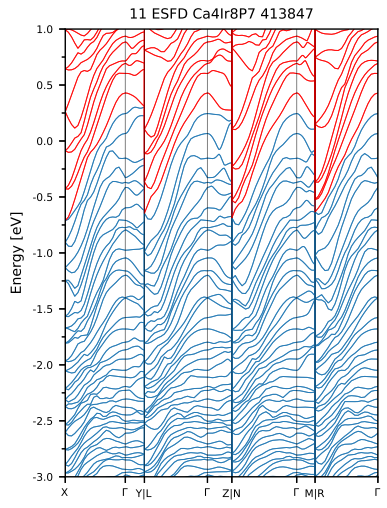

(fw) 11.4.82

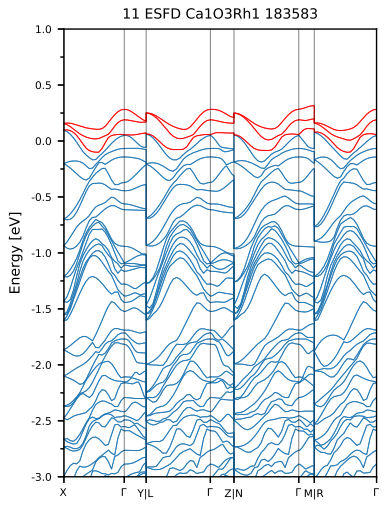

(fx) 11.4.83

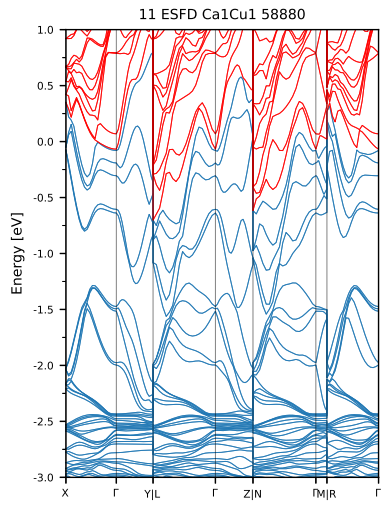

(fy) 11.4.84

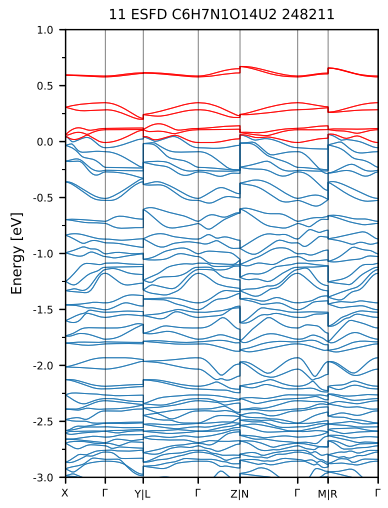

(fz) 11.4.85

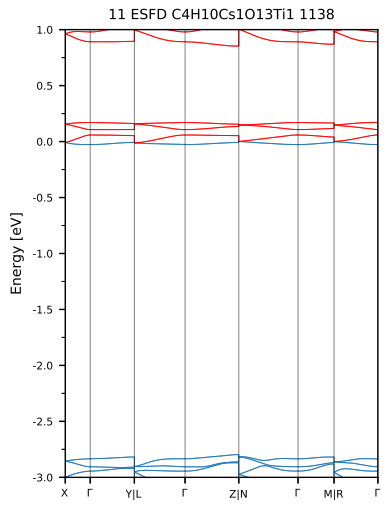

(ga) 11.4.86

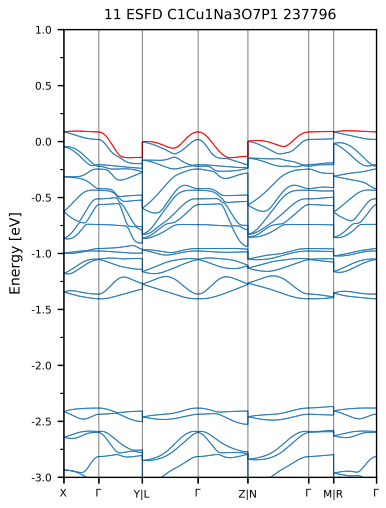

(gb) 11.4.87

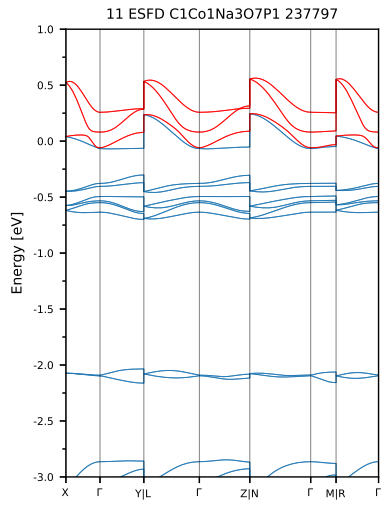

(gc) 11.4.88

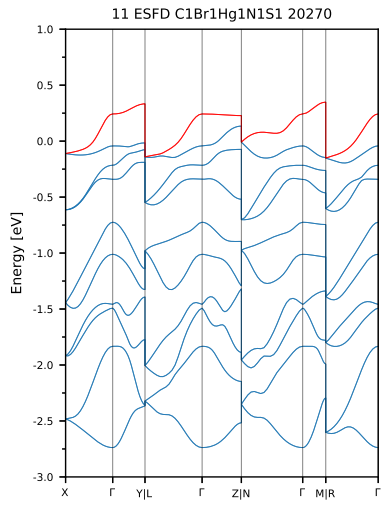

(gd) 11.4.89

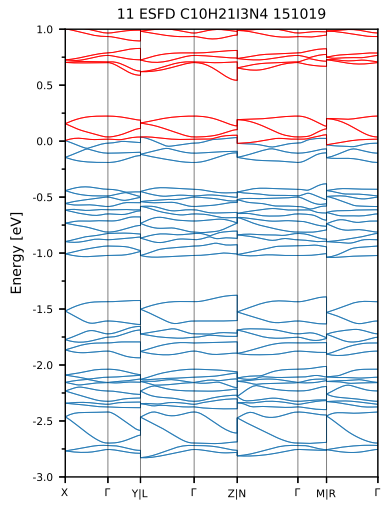

(ge) 11.4.90

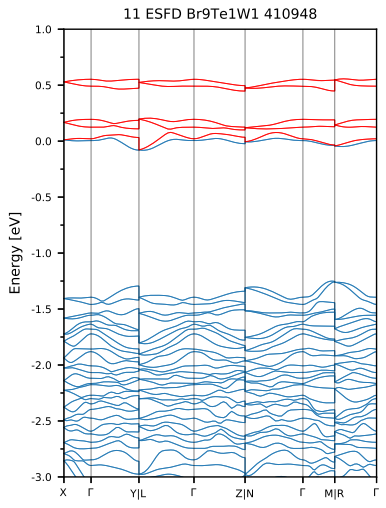

(gf) 11.4.91

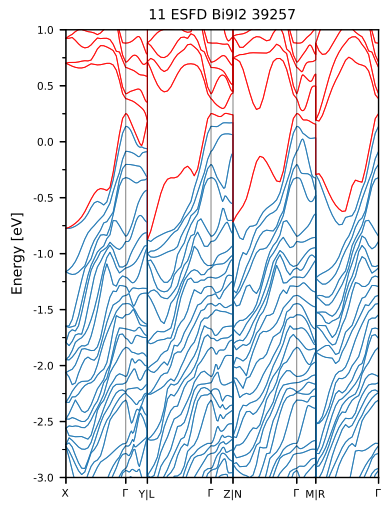

(gg) 11.4.92

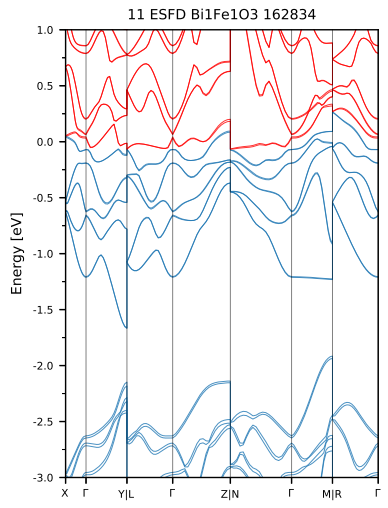

(gh) 11.4.93

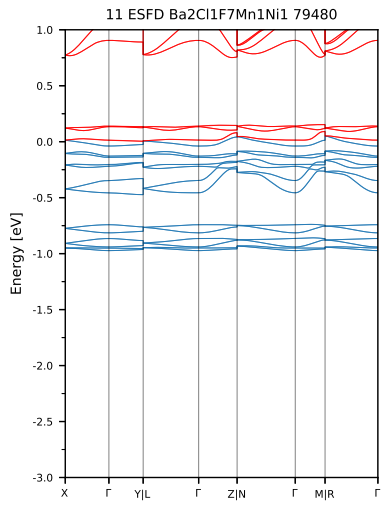

(gi) 11.4.94

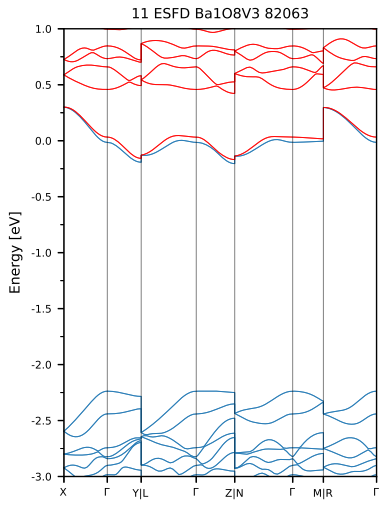

(gj) 11.4.95

Figure 22: Band structures for topological materials in SG 11, part 12/13, from 11.4.80 to 11.4.95. See the list of figures.

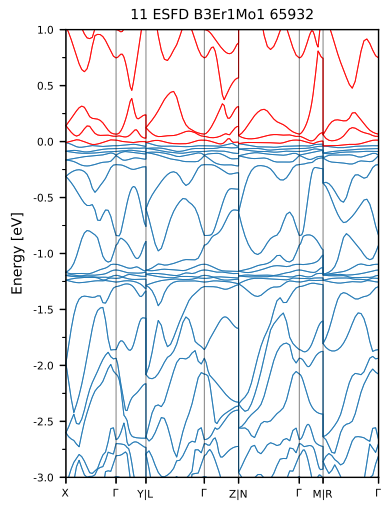

(gk) 11.4.96

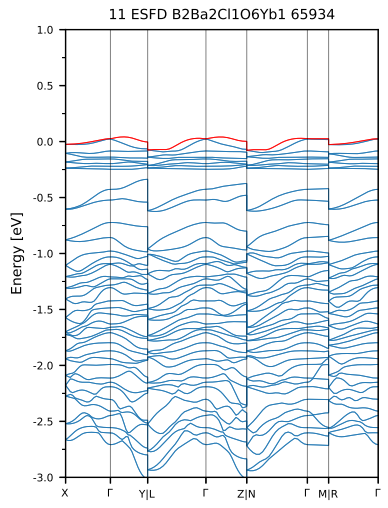

(gl) 11.4.97

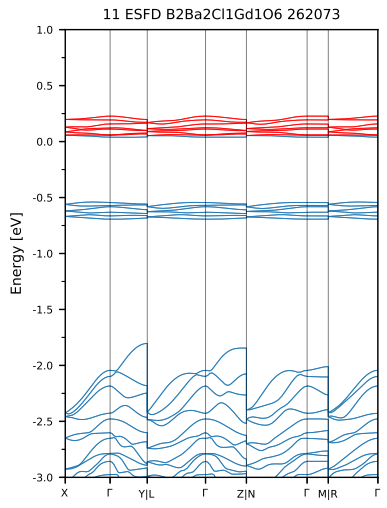

(gm) 11.4.98

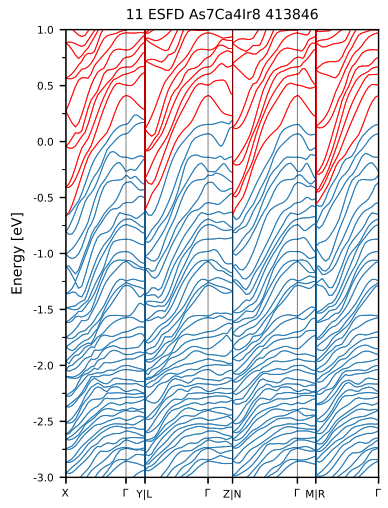

(gn) 11.4.99

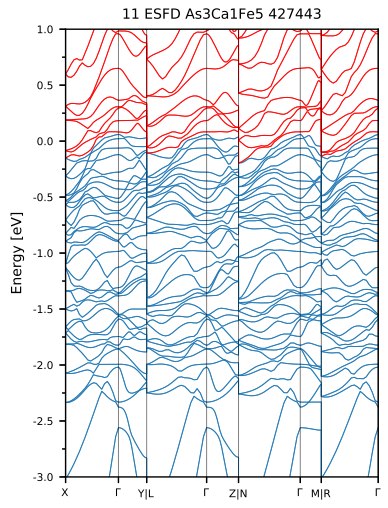

(go) 11.4.100

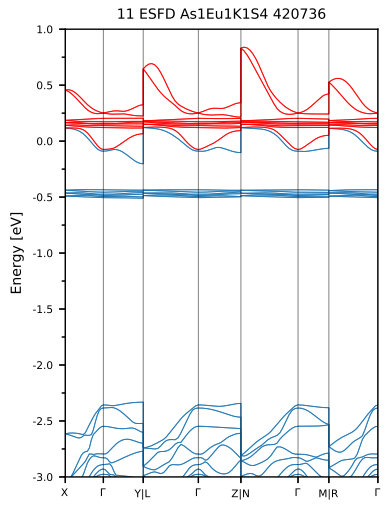

(gp) 11.4.101

Figure 23: Band structures for topological materials in SG 11, part 13/13, from 11.4.96 to 11.4.101. See the list of figures.

### 0.0.6 SG 12

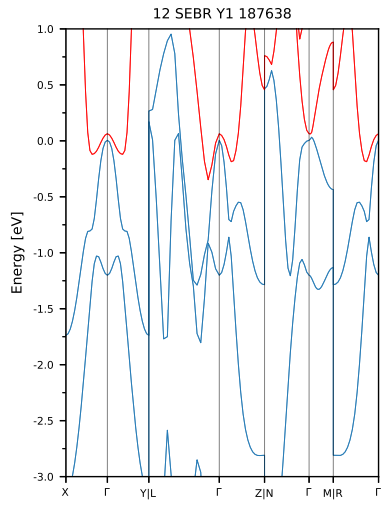

(a) 12.1.1

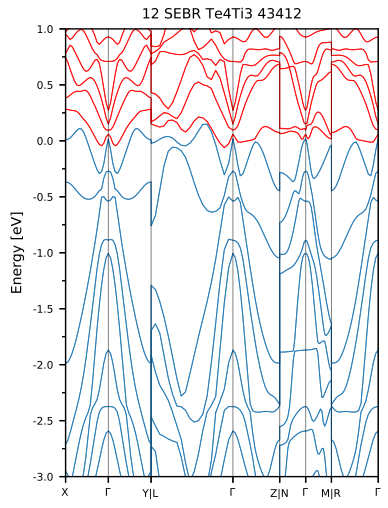

(b) 12.1.2

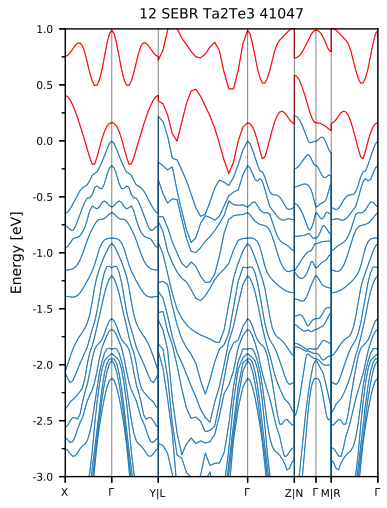

(c) 12.1.3

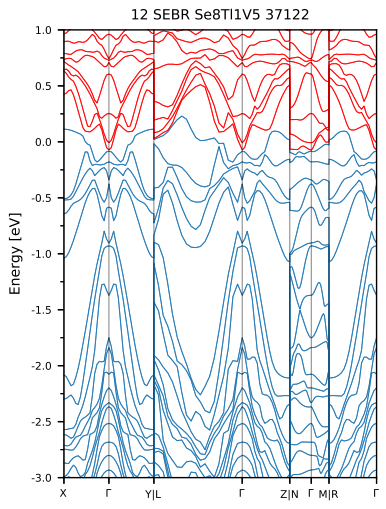

(d) 12.1.4

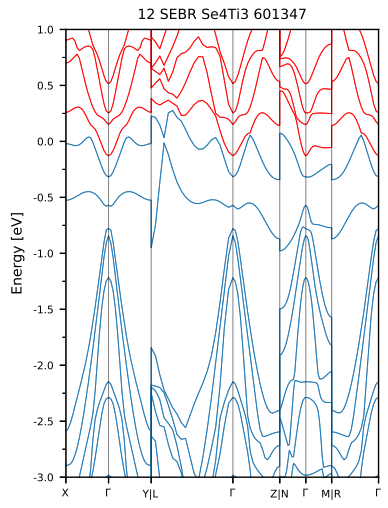

(e) 12.1.5

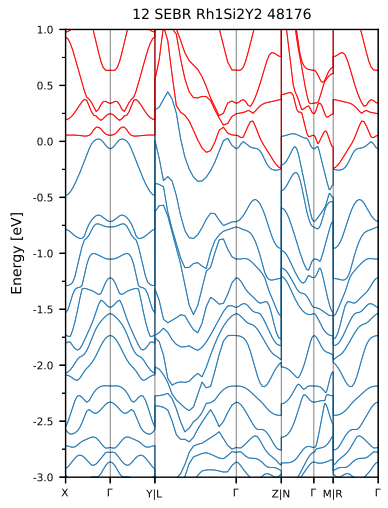

(f) 12.1.6

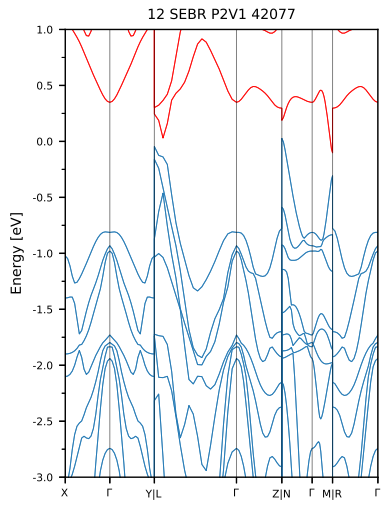

(g) 12.1.7

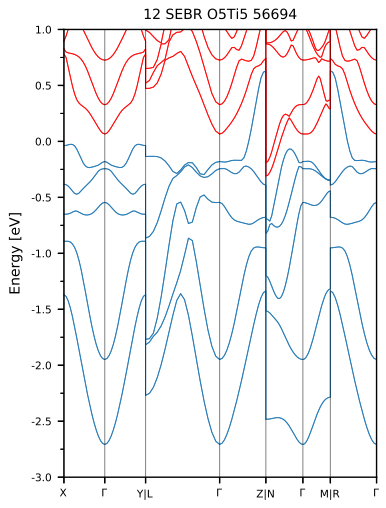

(h) 12.1.8

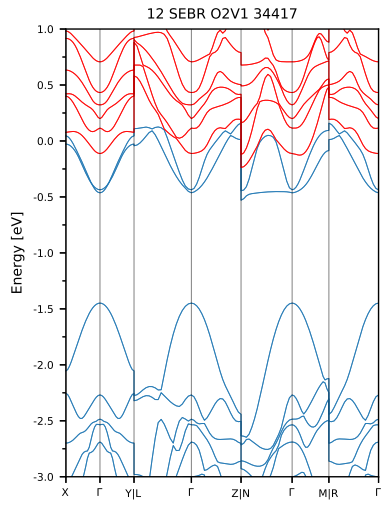

(i) 12.1.9

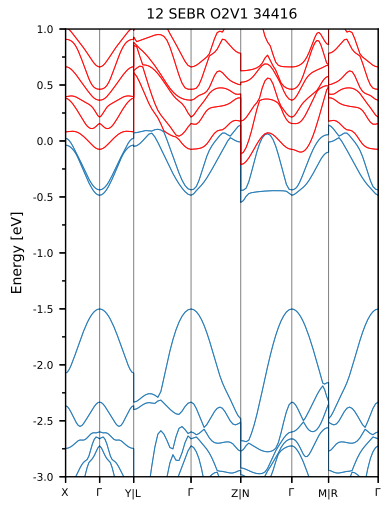

(j) 12.1.10

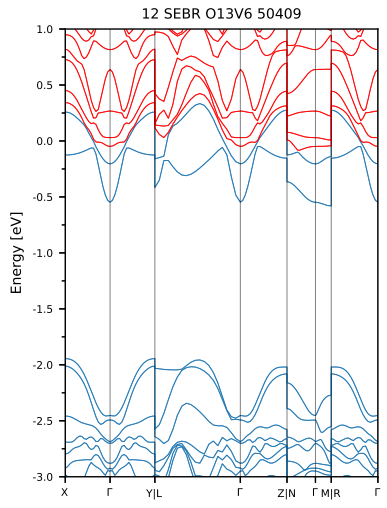

(k) 12.1.11

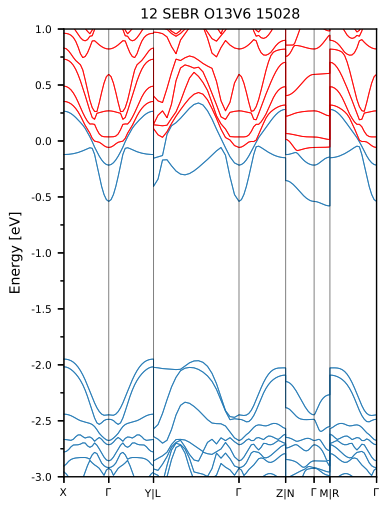

(l) 12.1.12

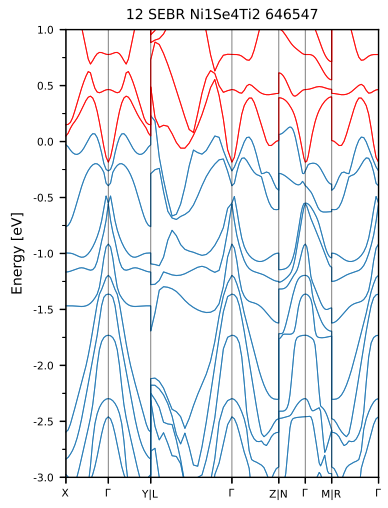

(m) 12.1.13

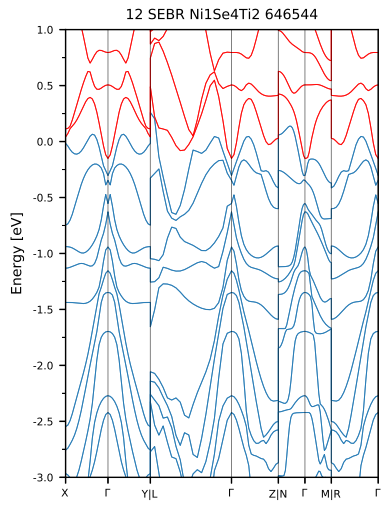

(n) 12.1.14

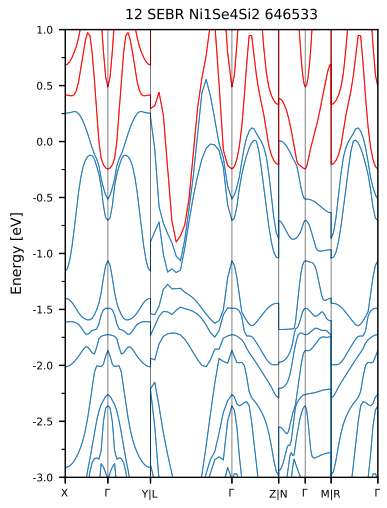

(o) 12.1.15

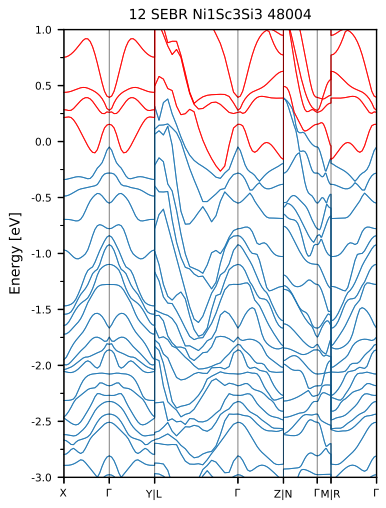

(p) 12.1.16

Figure 24: Band structures for topological materials in SG 12, part 1/25, from 12.1.1 to 12.1.16. See the list of figures.

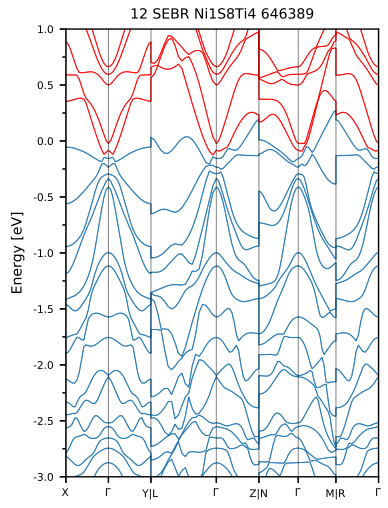

(q) 12.1.17

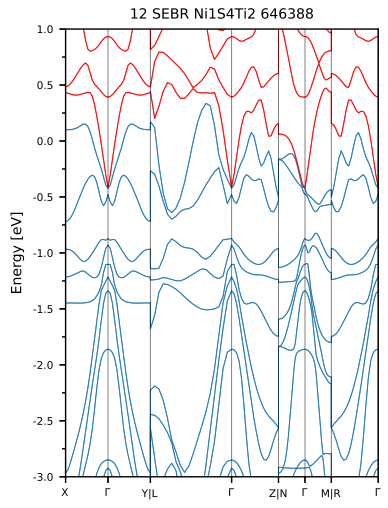

(r) 12.1.18

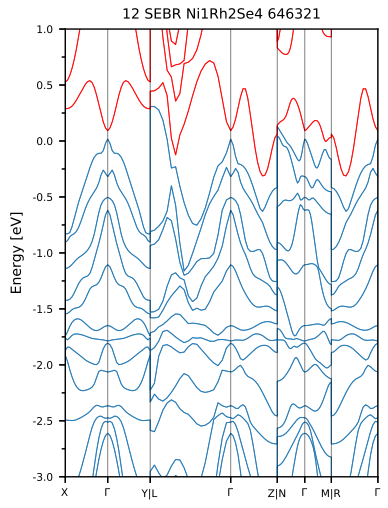

(s) 12.1.19

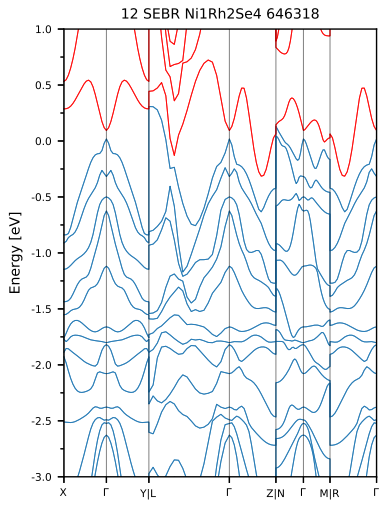

(t) 12.1.20

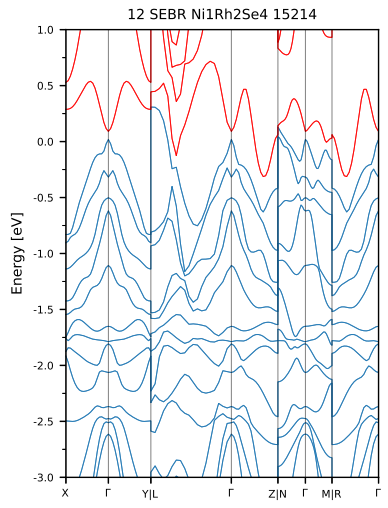

(u) 12.1.21

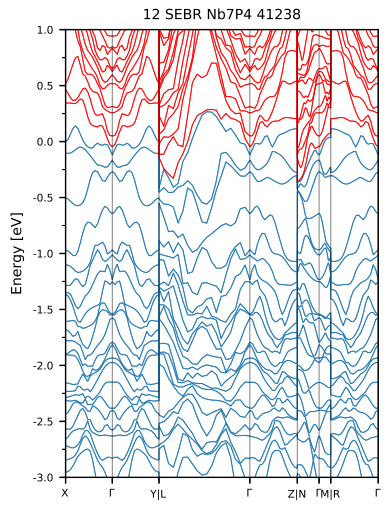

(v) 12.1.22

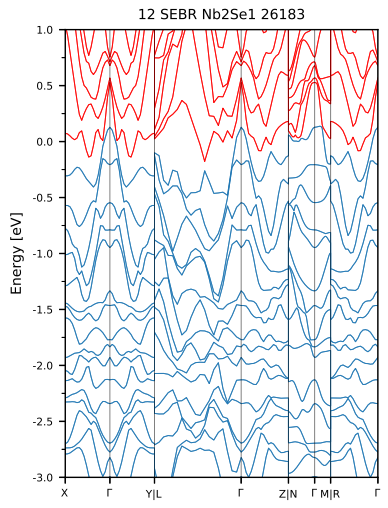

(w) 12.1.23

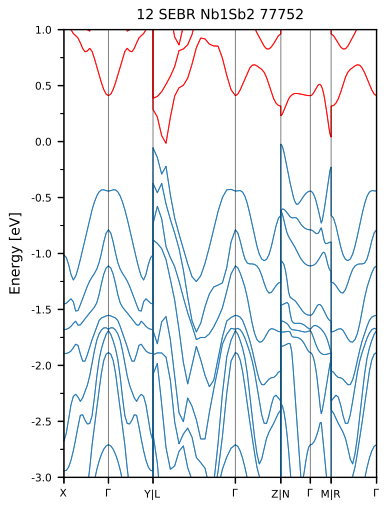

(x) 12.1.24

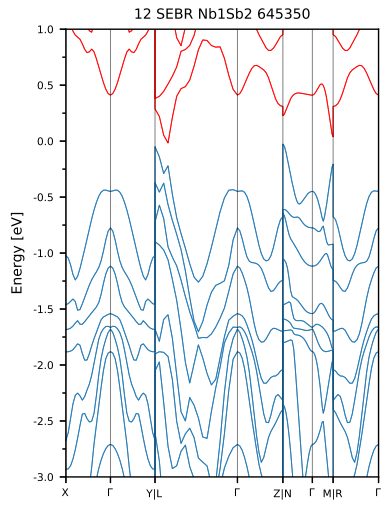

(y) 12.1.25

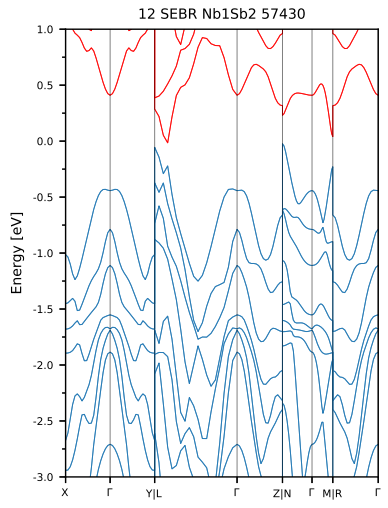

(z) 12.1.26

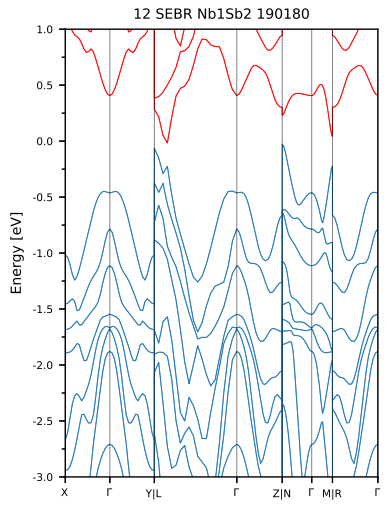

(aa) 12.1.27

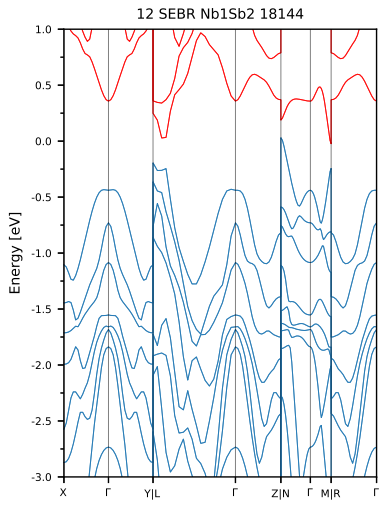

(ab) 12.1.28

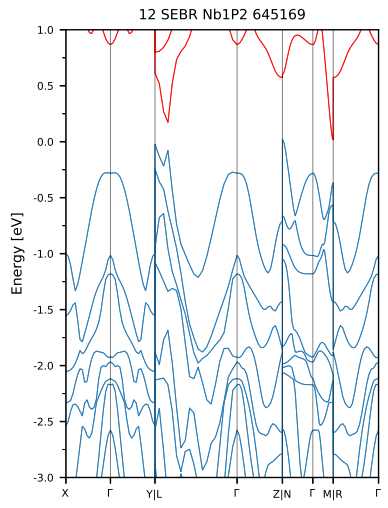

(ac) 12.1.29

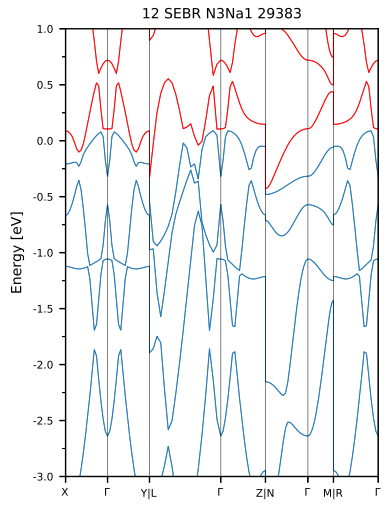

(ad) 12.1.30

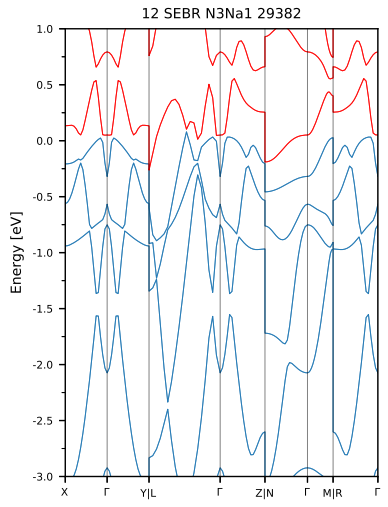

(ae) 12.1.31

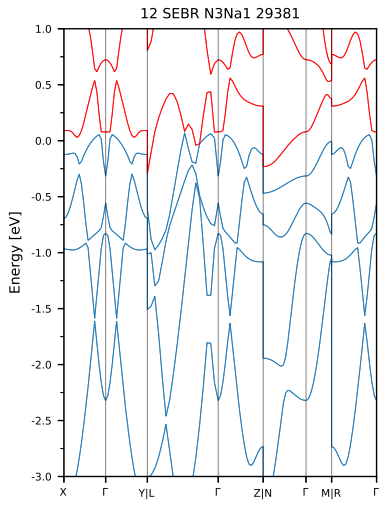

(af) 12.1.32

Figure 25: Band structures for topological materials in SG 12, part 2/25, from 12.1.17 to 12.1.32. See the list of figures.

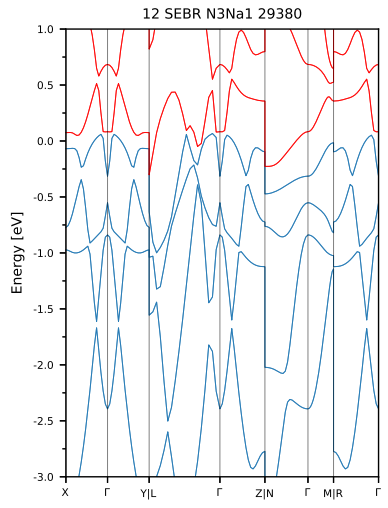

(ag) 12.1.33

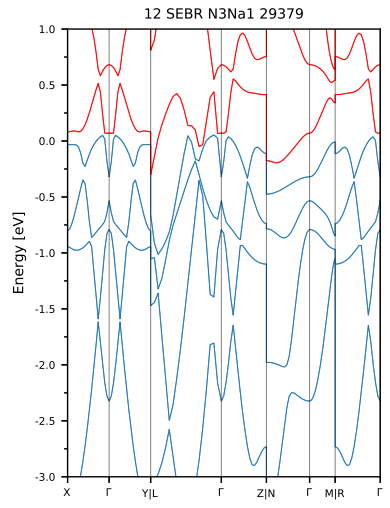

(ah) 12.1.34

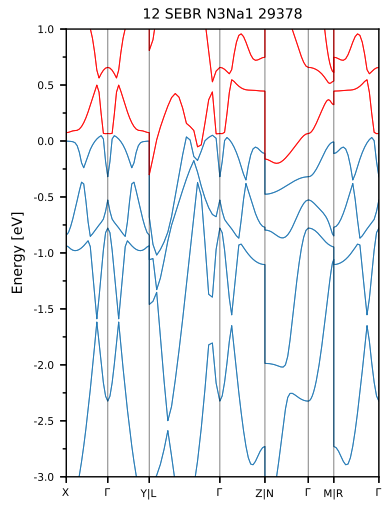

(ai) 12.1.35

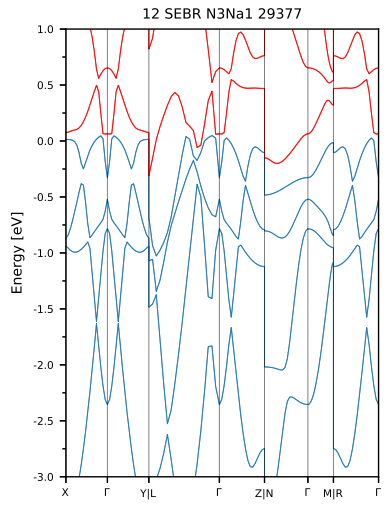

(aj) 12.1.36

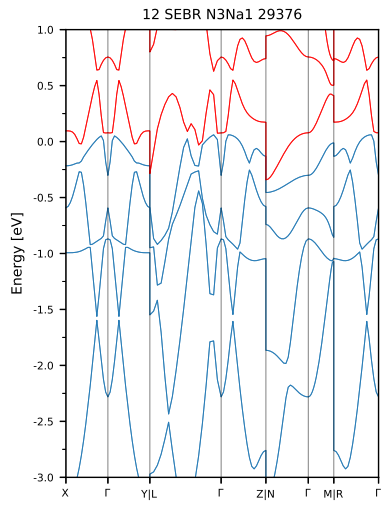

(ak) 12.1.37

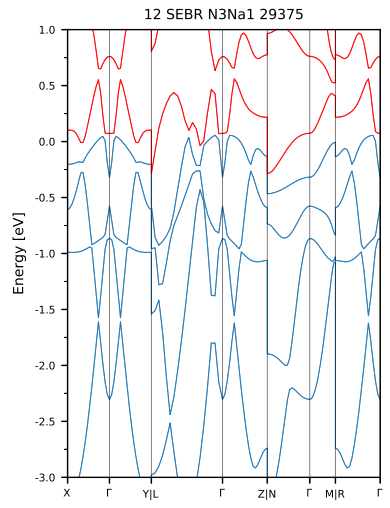

(al) 12.1.38

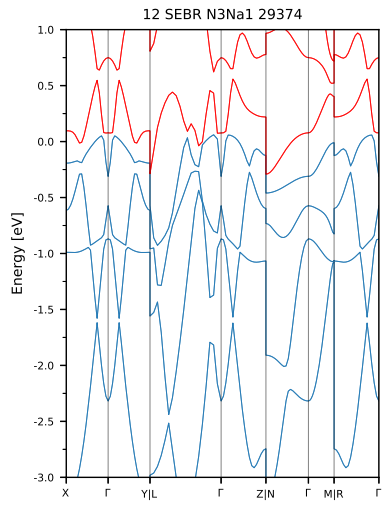

(am) 12.1.39

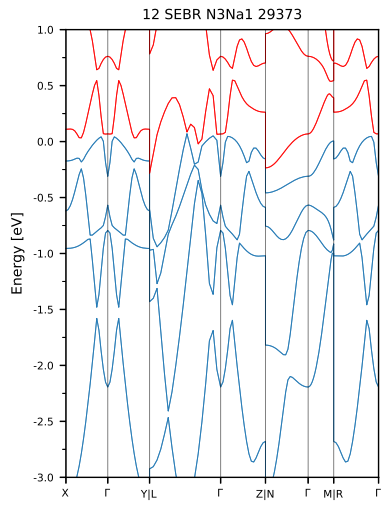

(an) 12.1.40

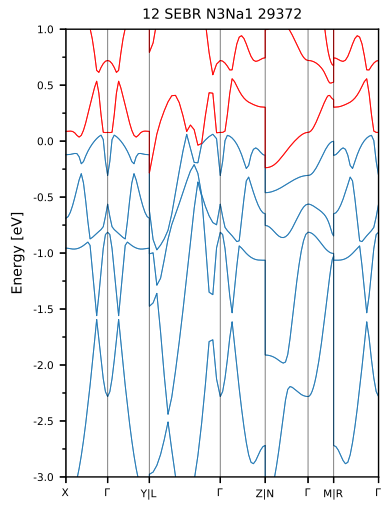

(ao) 12.1.41

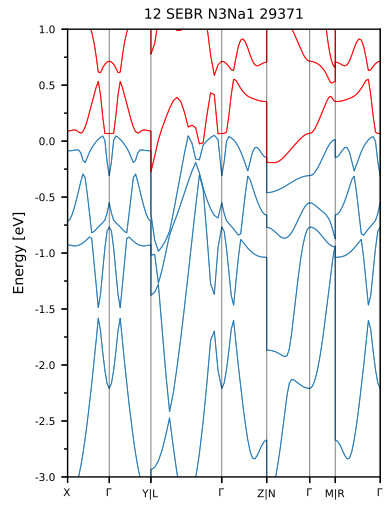

(ap) 12.1.42

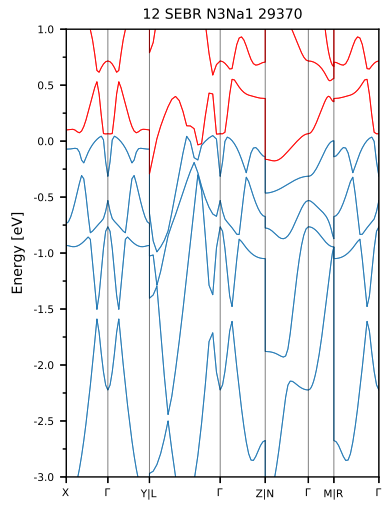

(aq) 12.1.43

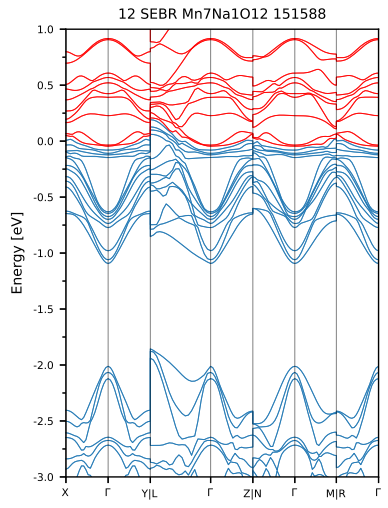

(ar) 12.1.44

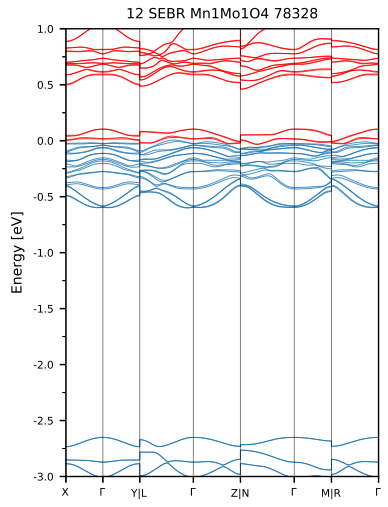

(as) 12.1.45

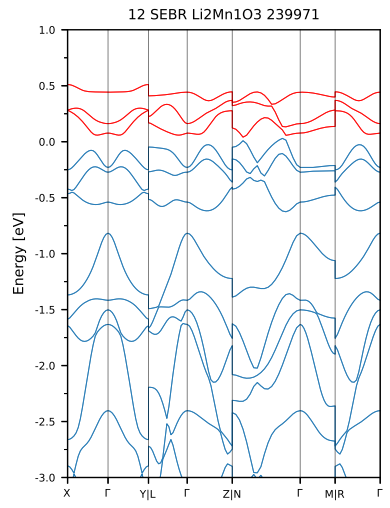

(at) 12.1.46

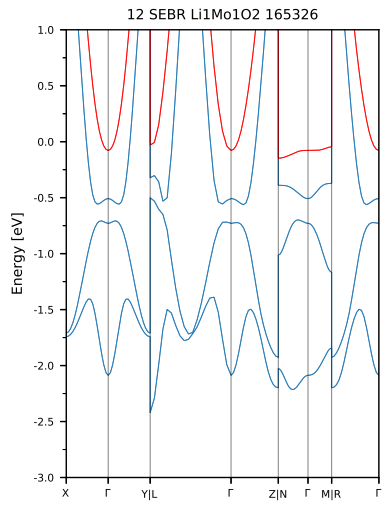

(au) 12.1.47

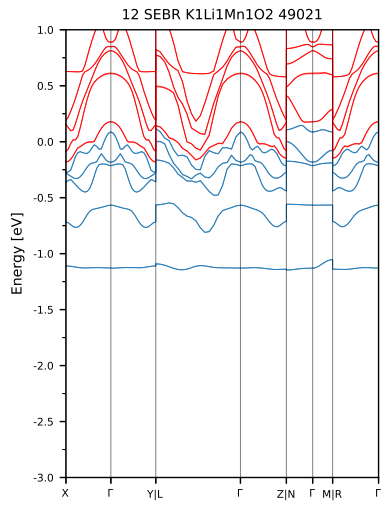

(av) 12.1.48

Figure 26: Band structures for topological materials in SG 12, part 3/25, from 12.1.33 to 12.1.48. See the list of figures.

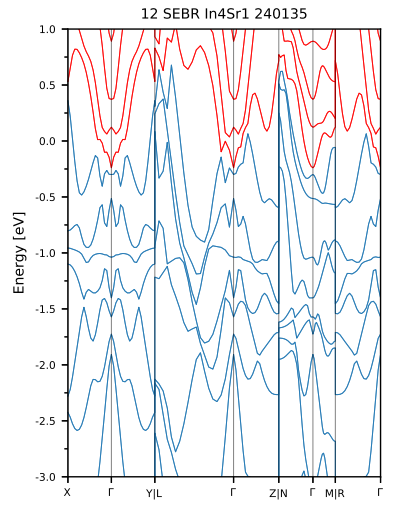

(aw) 12.1.49

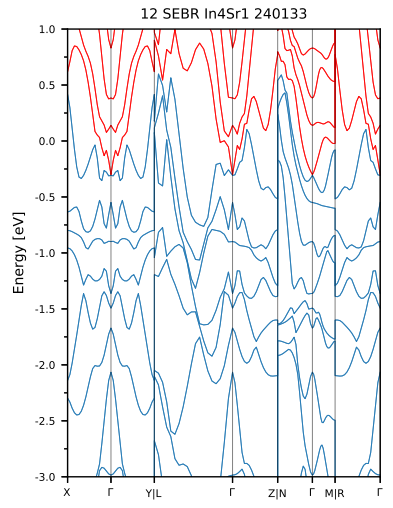

(ax) 12.1.50

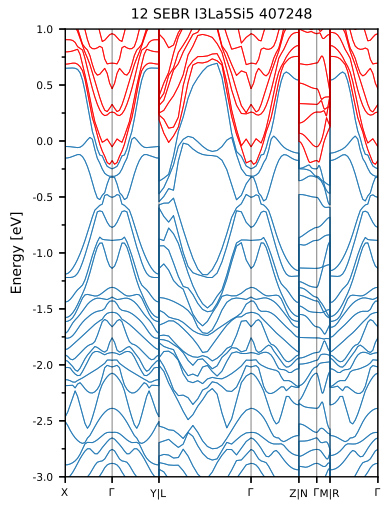

(ay) 12.1.51

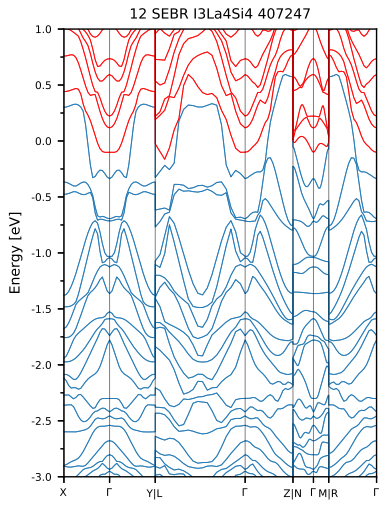

(az) 12.1.52

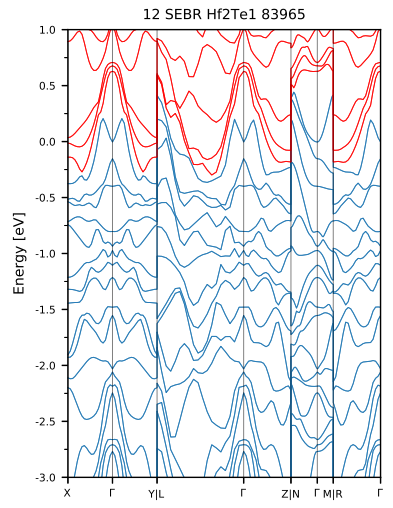

(ba) 12.1.53

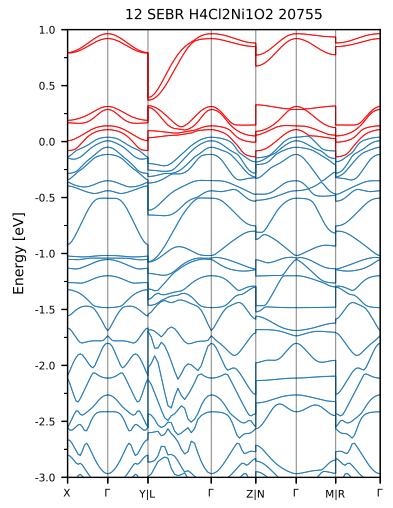

(bb) 12.1.54

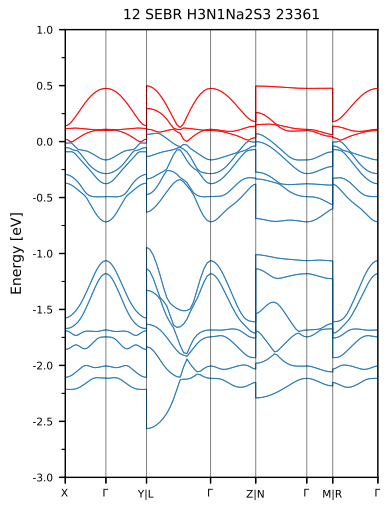

(bc) 12.1.55

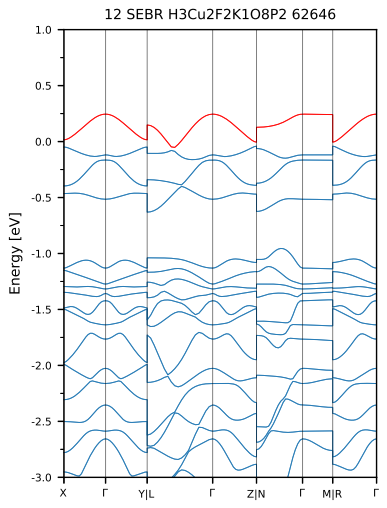

(bd) 12.1.56

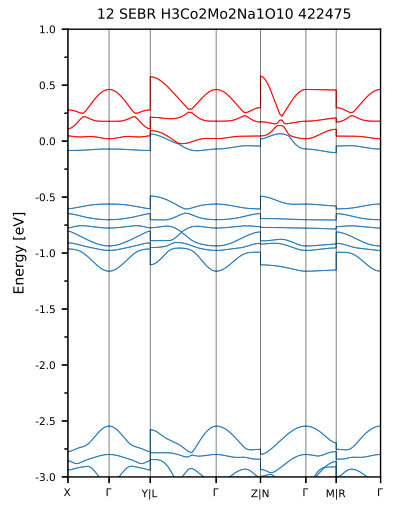

(be) 12.1.57

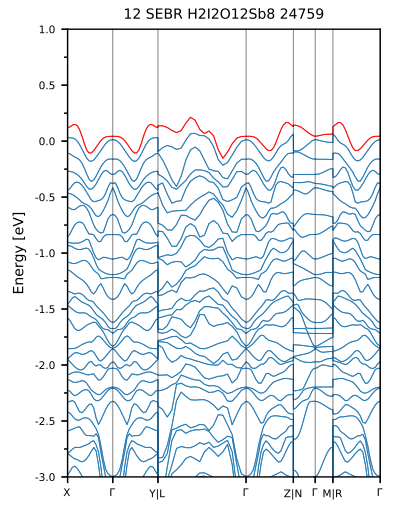

(bf) 12.1.58

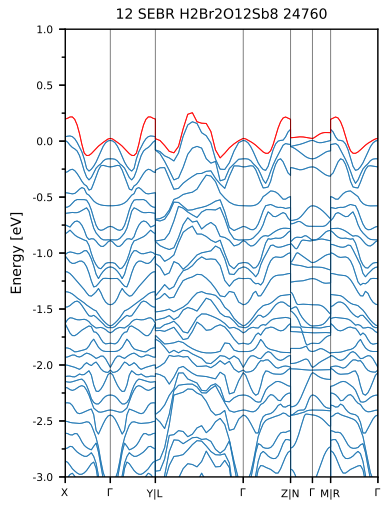

(bg) 12.1.59

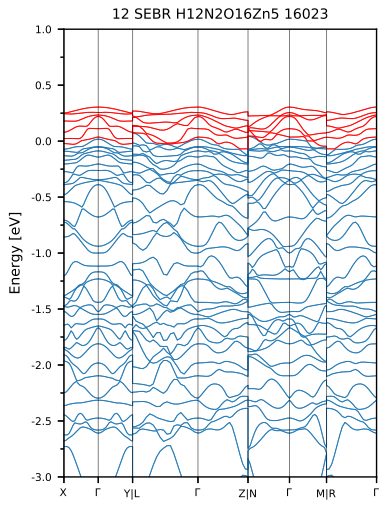

(bh) 12.1.60

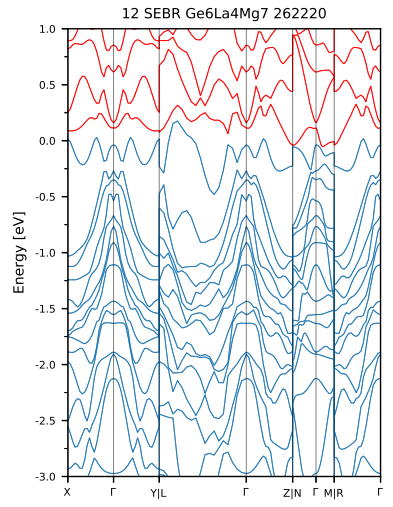

(bi) 12.1.61

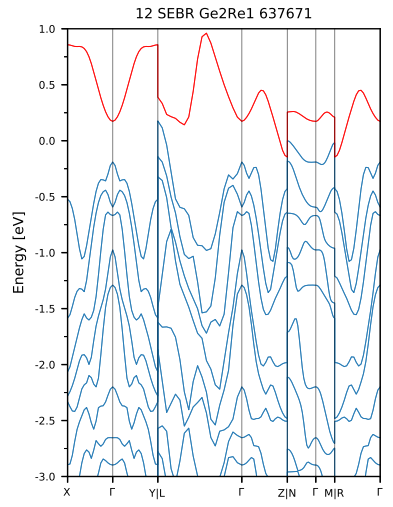

(bj) 12.1.62

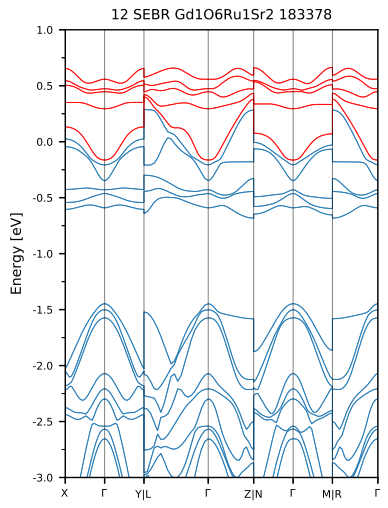

(bk) 12.1.63

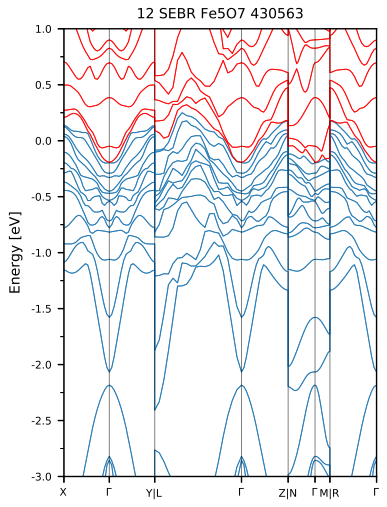

(bl) 12.1.64

Figure 27: Band structures for topological materials in SG 12, part 4/25, from 12.1.49 to 12.1.64. See the list of figures.

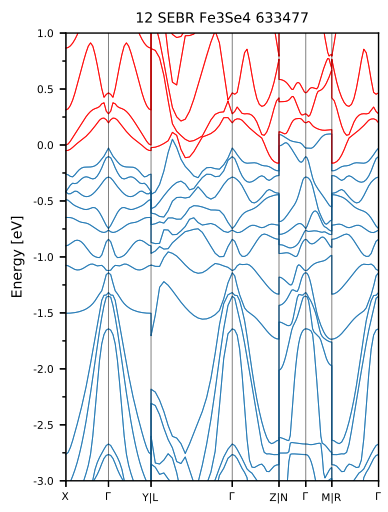

(bm) 12.1.65

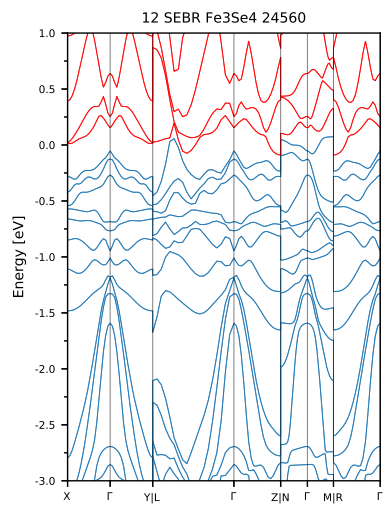

(bn) 12.1.66

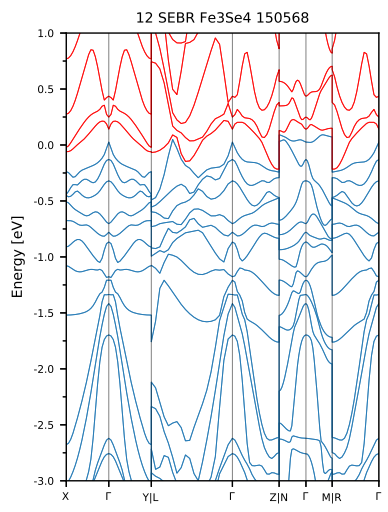

(bo) 12.1.67

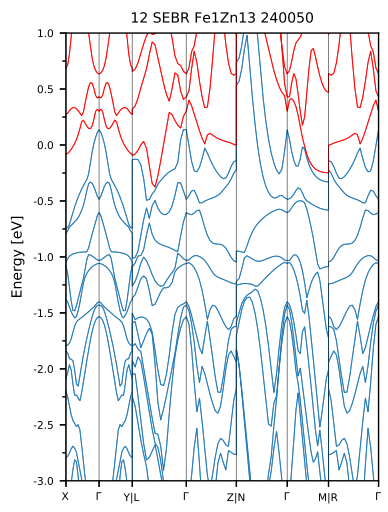

(bp) 12.1.68

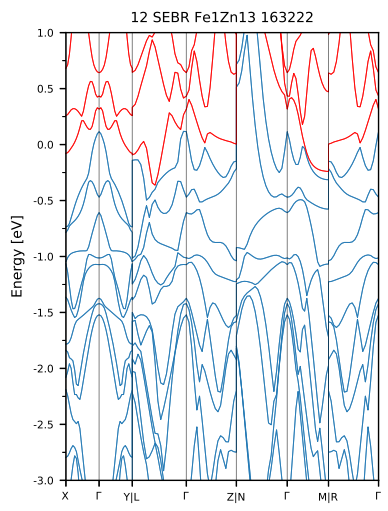

(bq) 12.1.69

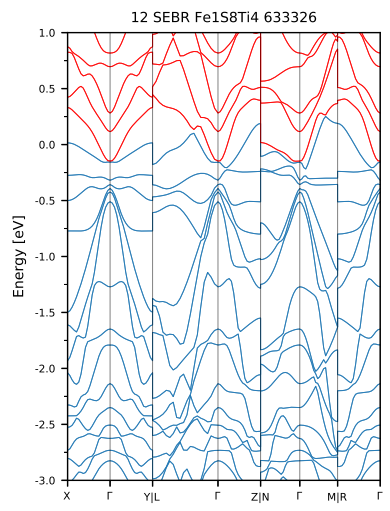

(br) 12.1.70

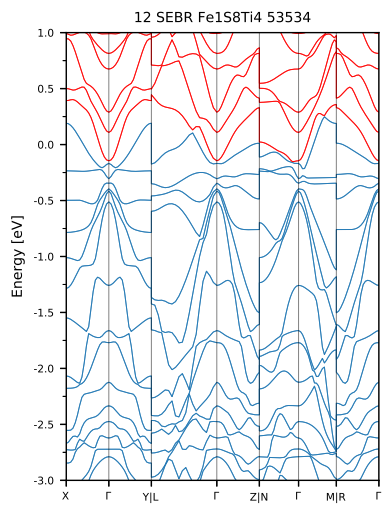

(bs) 12.1.71

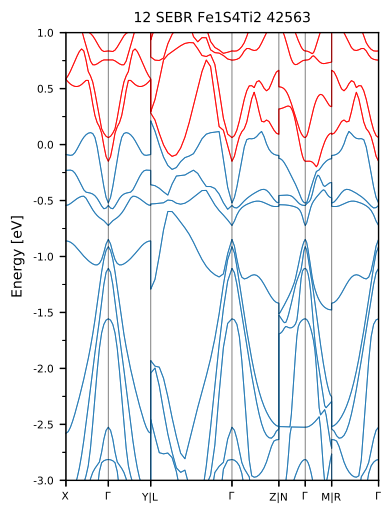

(bt) 12.1.72

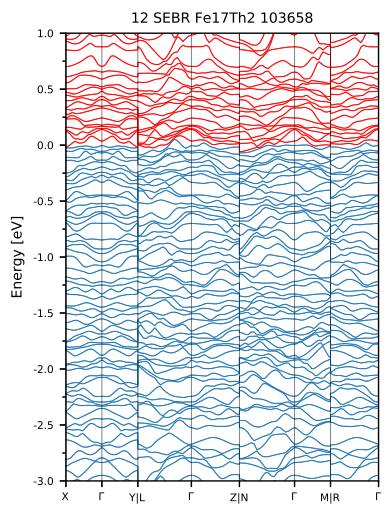

(bu) 12.1.73

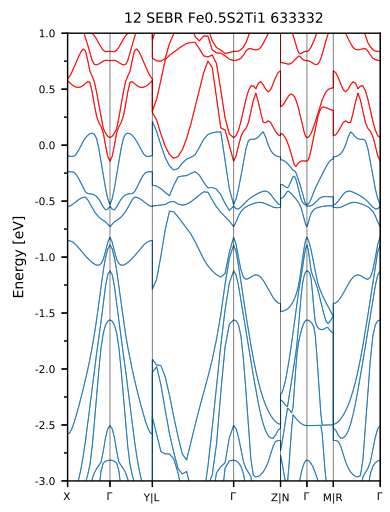

(bv) 12.1.74

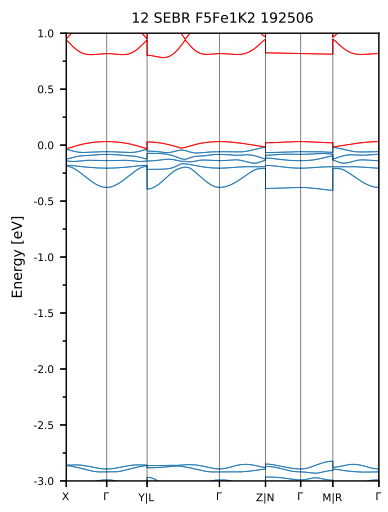

(bw) 12.1.75

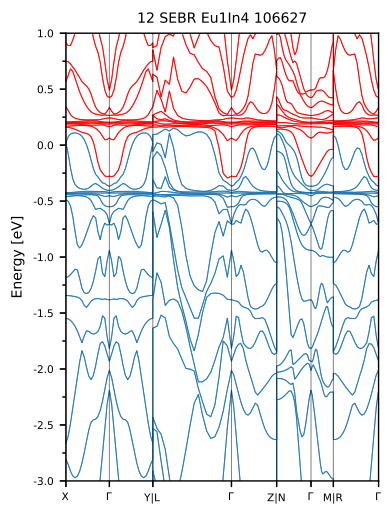

(bx) 12.1.76

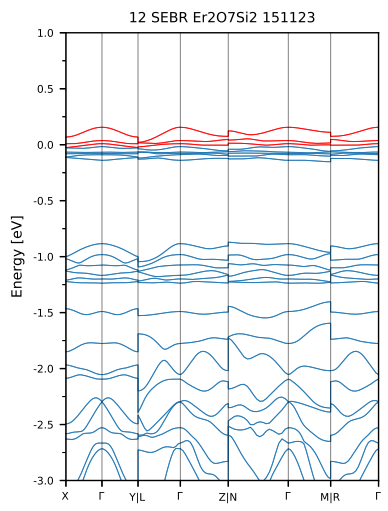

(by) 12.1.77

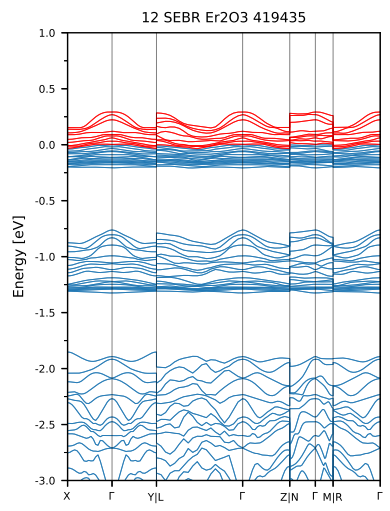

(bz) 12.1.78

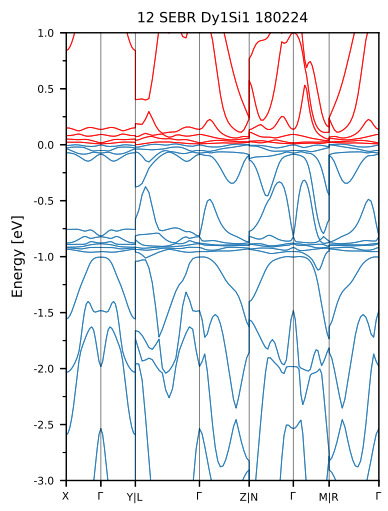

(ca) 12.1.79

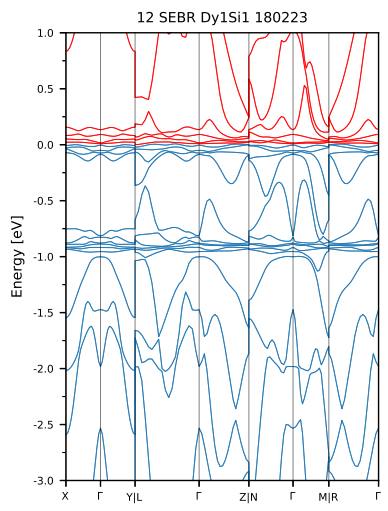

(cb) 12.1.80

Figure 28: Band structures for topological materials in SG 12, part 5/25, from 12.1.65 to 12.1.80. See the list of figures.

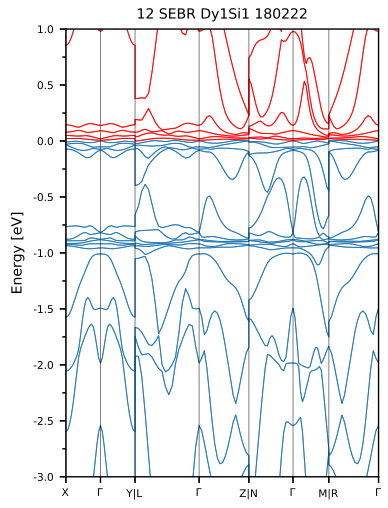

(cc) 12.1.81

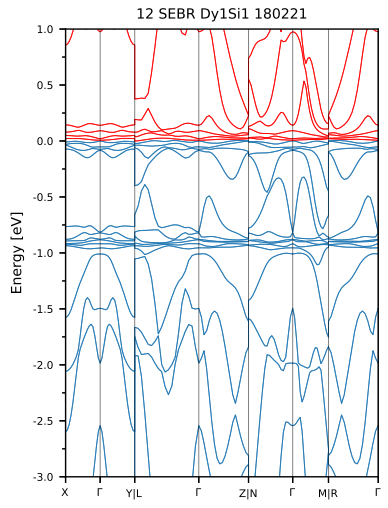

(cd) 12.1.82

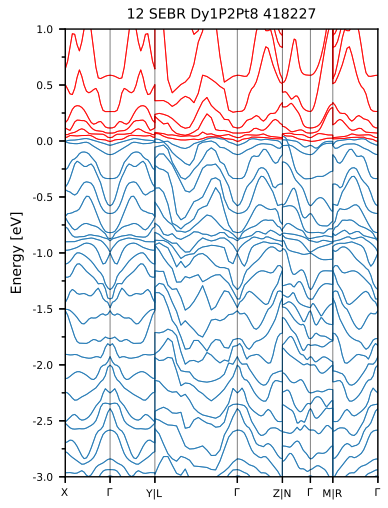

(ce) 12.1.83

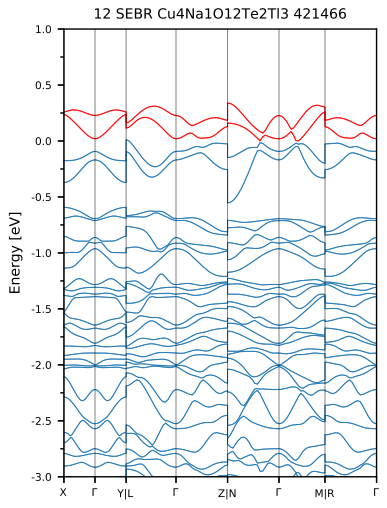

(cf) 12.1.84

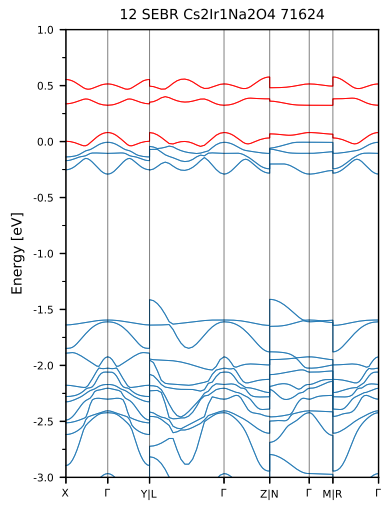

(cg) 12.1.85

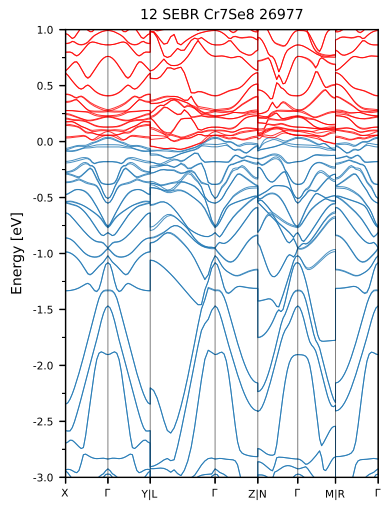

(ch) 12.1.86

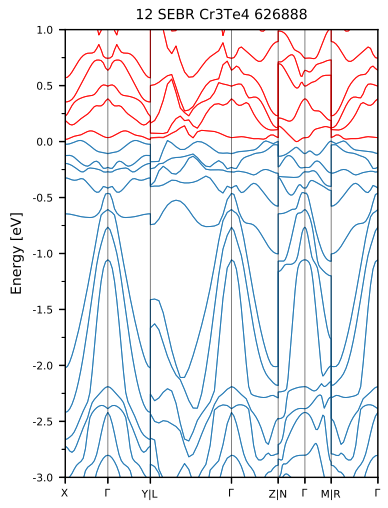

(ci) 12.1.87

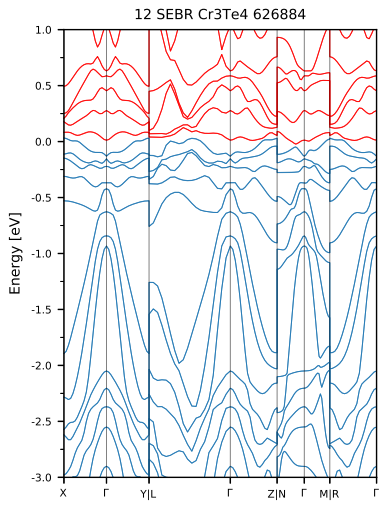

(cj) 12.1.88

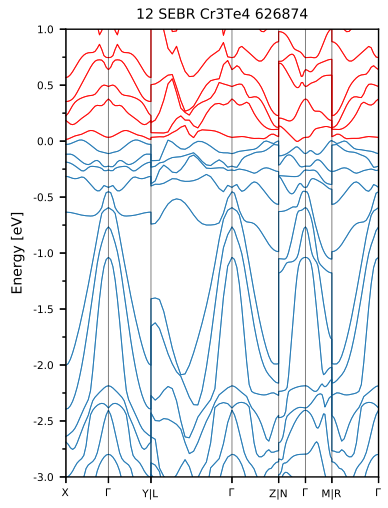

(ck) 12.1.89

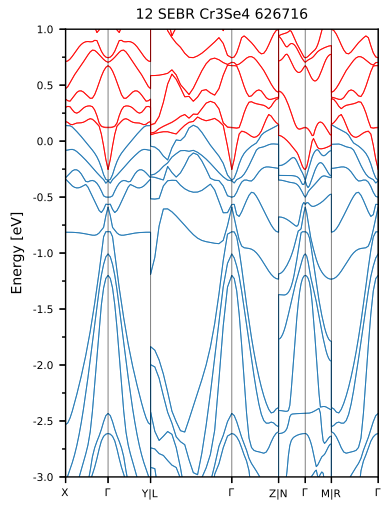

(cl) 12.1.90

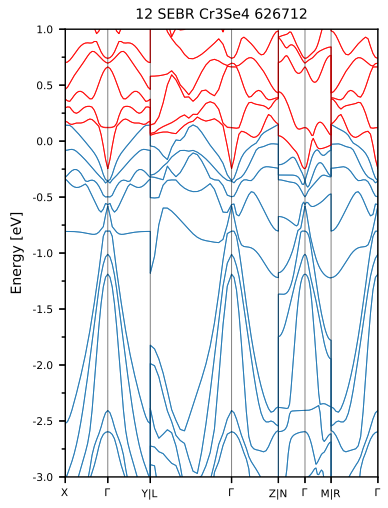

(cm) 12.1.91

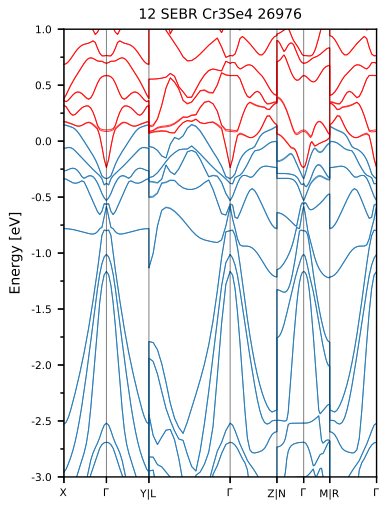

(cn) 12.1.92

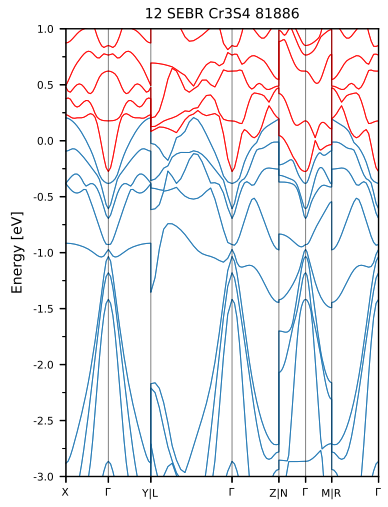

(co) 12.1.93

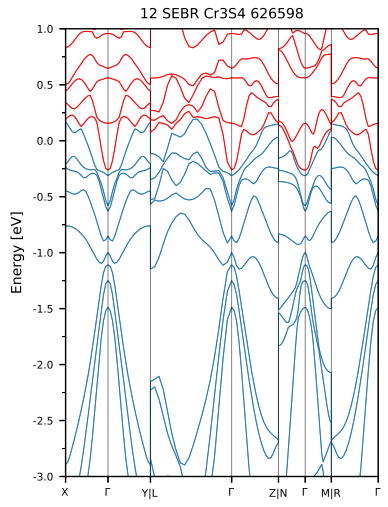

(cp) 12.1.94

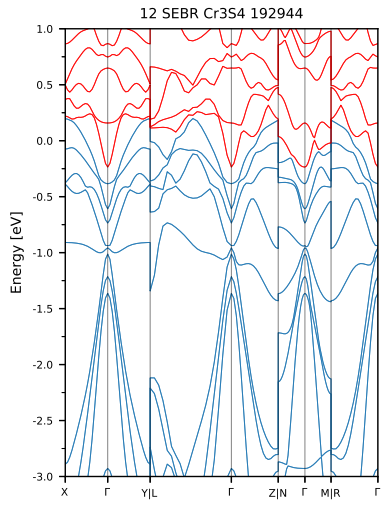

(cq) 12.1.95

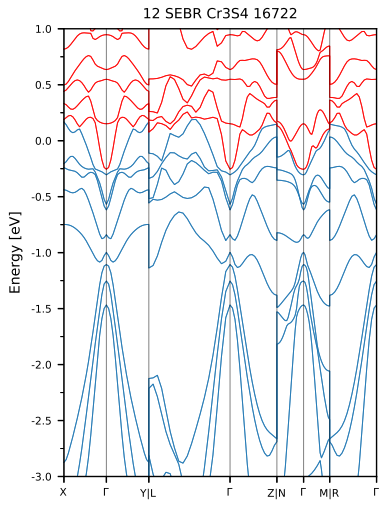

(cr) 12.1.96

Figure 29: Band structures for topological materials in SG 12, part 6/25, from 12.1.81 to 12.1.96. See the list of figures.

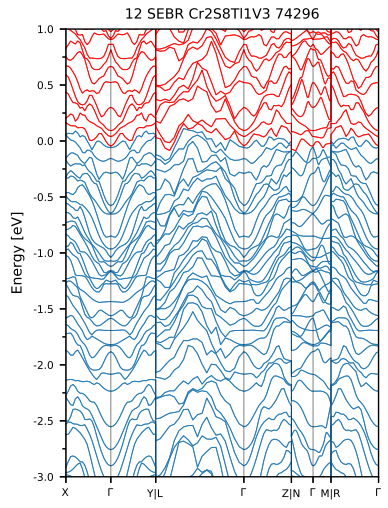

(cs) 12.1.97

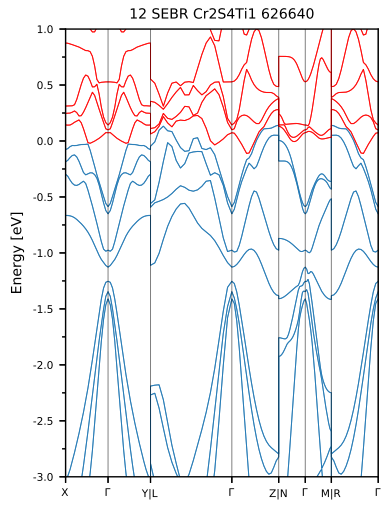

(ct) 12.1.98

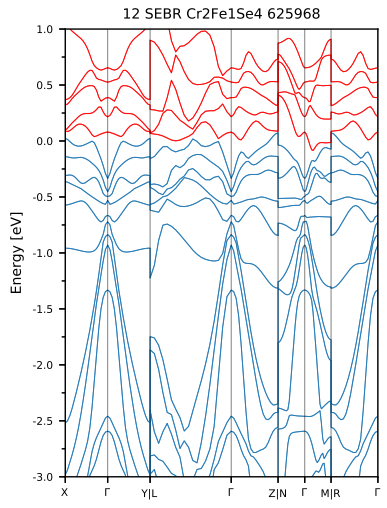

(cu) 12.1.99

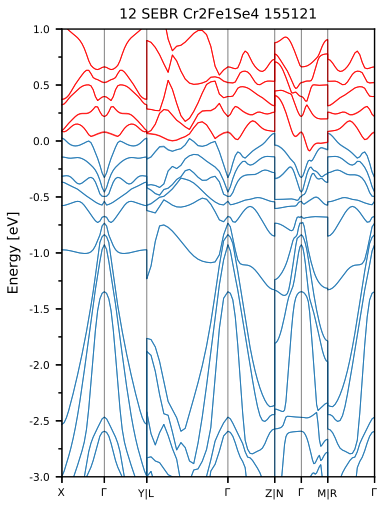

(cv) 12.1.100

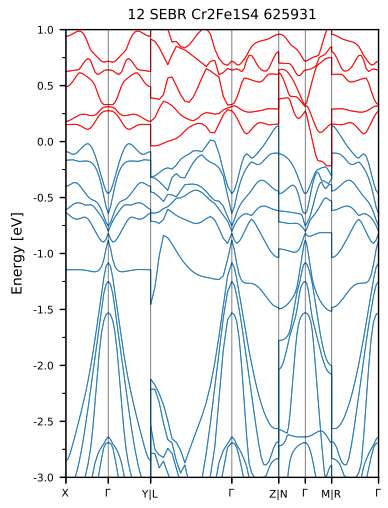

(cw) 12.1.101

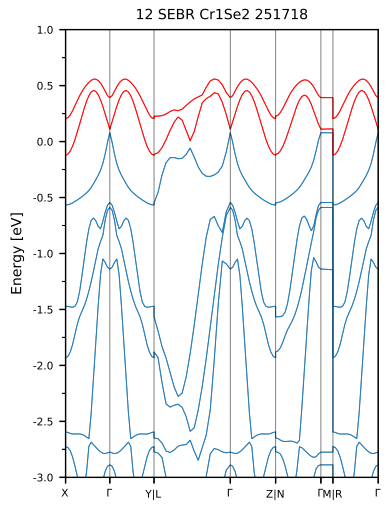

(cx) 12.1.102

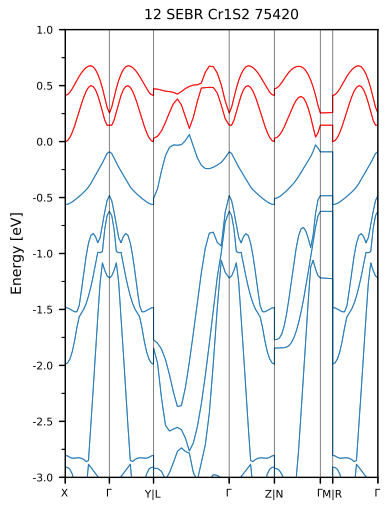

(cy) 12.1.103

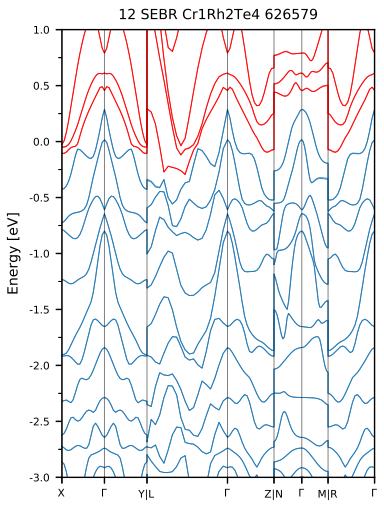

(cz) 12.1.104

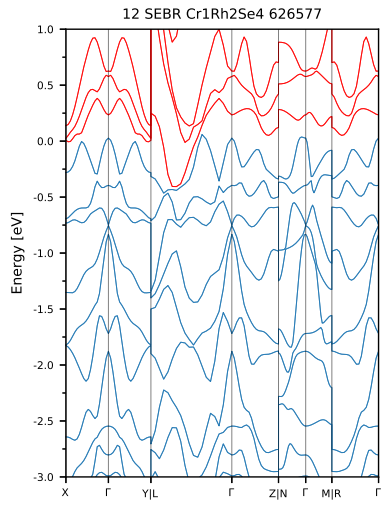

(da) 12.1.105

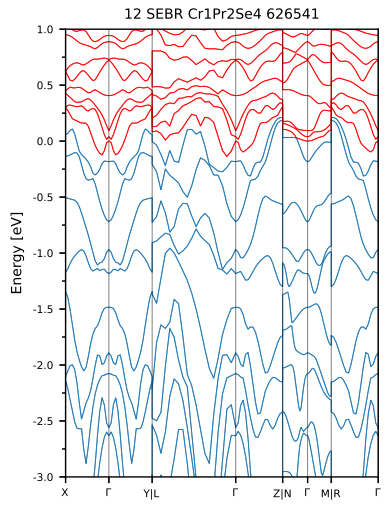

(db) 12.1.106

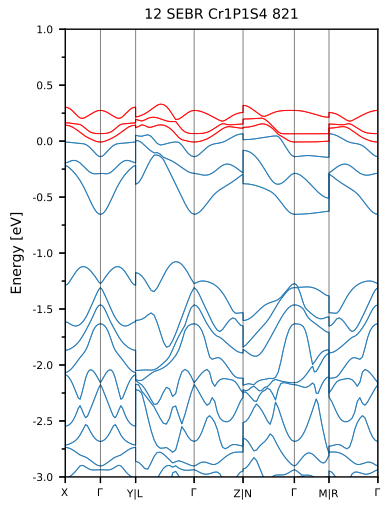

(dc) 12.1.107

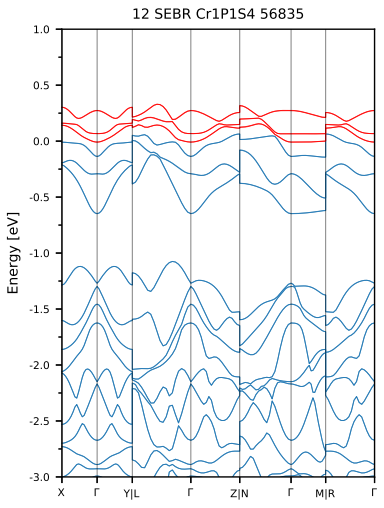

(dd) 12.1.108

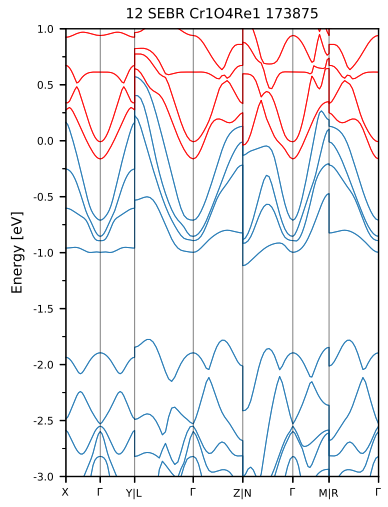

(de) 12.1.109

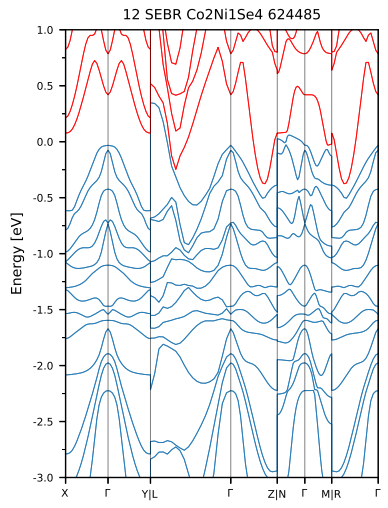

(df) 12.1.110

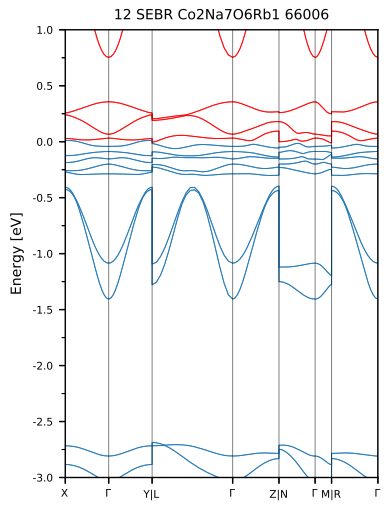

(dg) 12.1.111

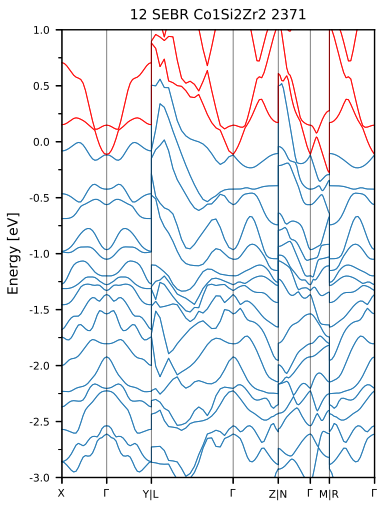

(dh) 12.1.112

Figure 30: Band structures for topological materials in SG 12, part 7/25, from 12.1.97 to 12.1.112. See the list of figures.

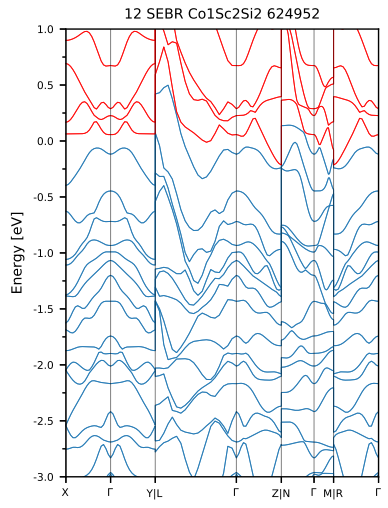

(di) 12.1.113

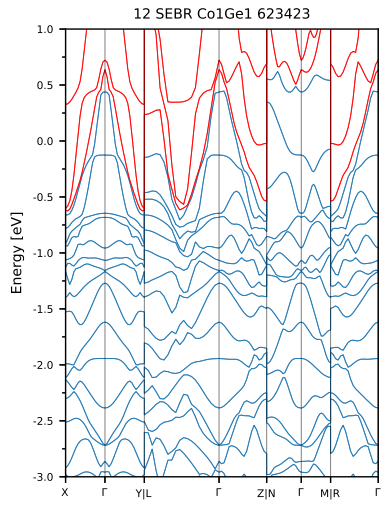

(dj) 12.1.114

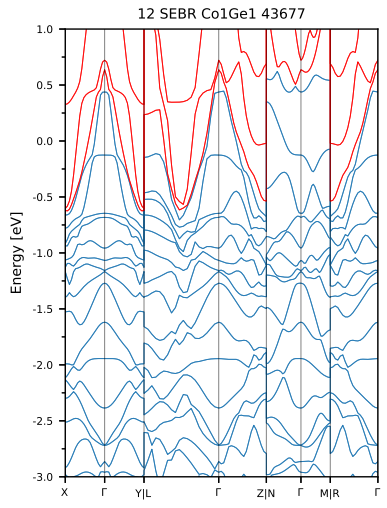

(dk) 12.1.115

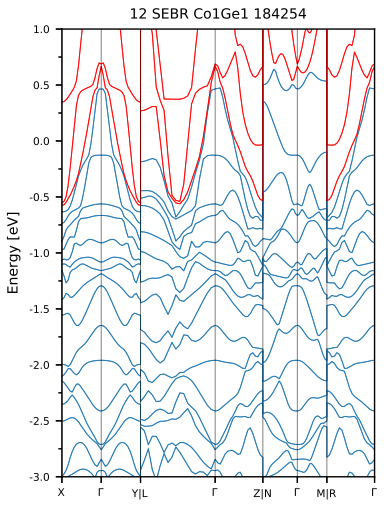

(dl) 12.1.116

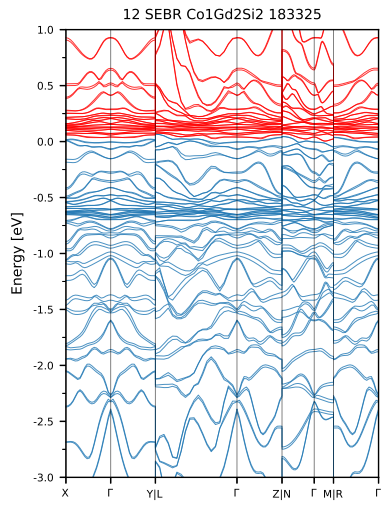

(dm) 12.1.117

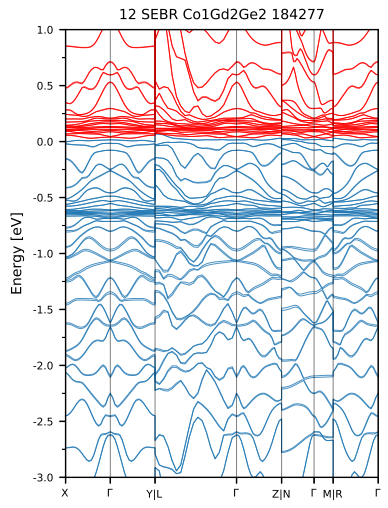

(dn) 12.1.118

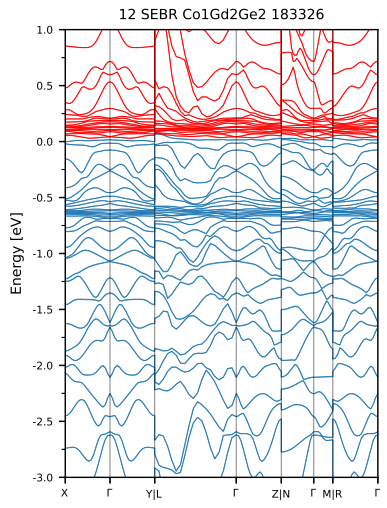

(do) 12.1.119

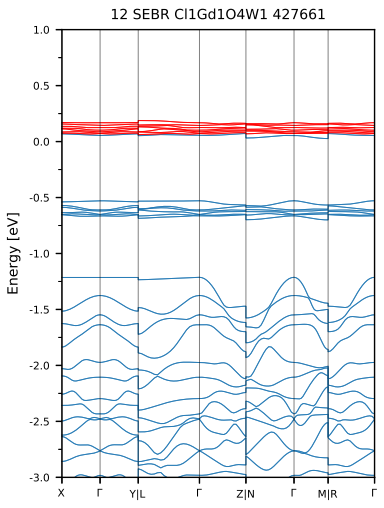

(dp) 12.1.120

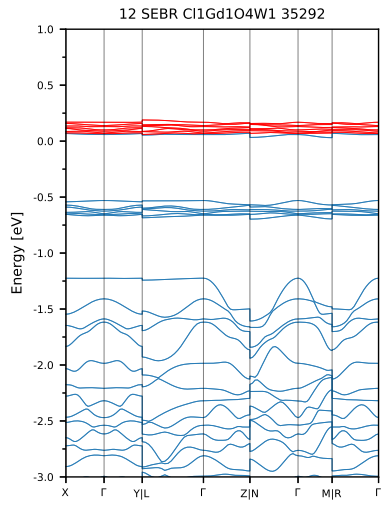

(dq) 12.1.121

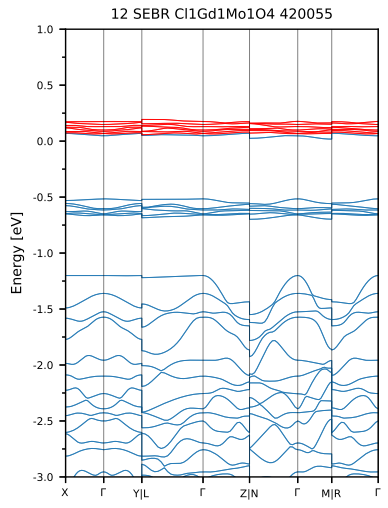

(dr) 12.1.122

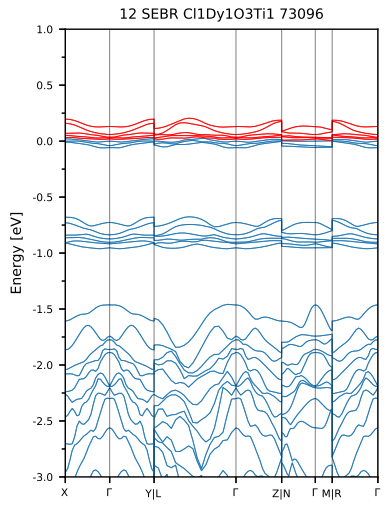

(ds) 12.1.123

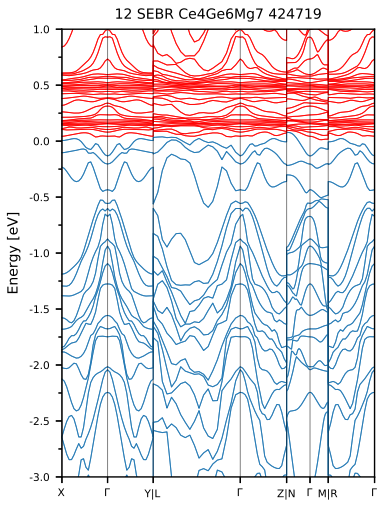

(dt) 12.1.124

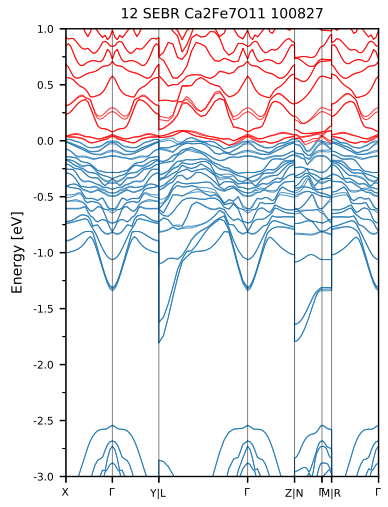

(du) 12.1.125

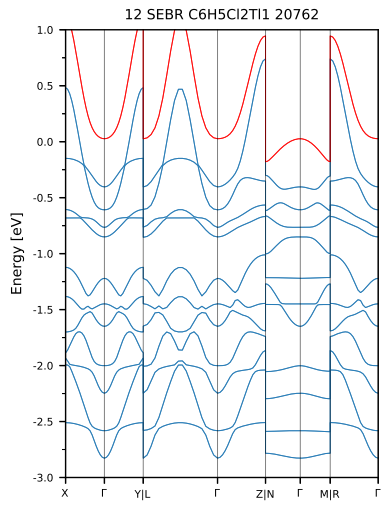

(dv) 12.1.126

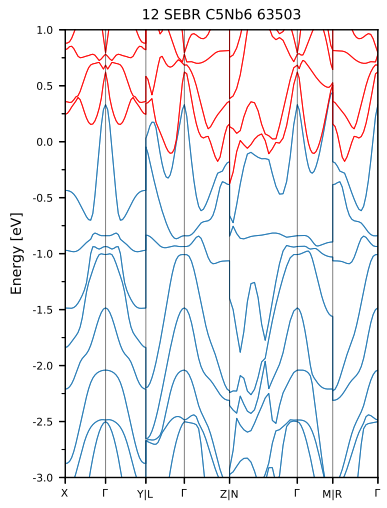

(dw) 12.1.127

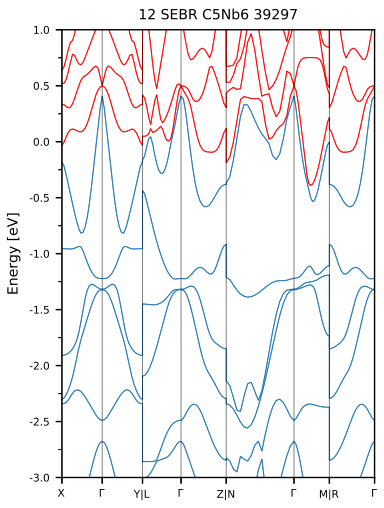

(dx) 12.1.128

Figure 31: Band structures for topological materials in SG 12, part 8/25, from 12.1.113 to 12.1.128. See the list of figures.

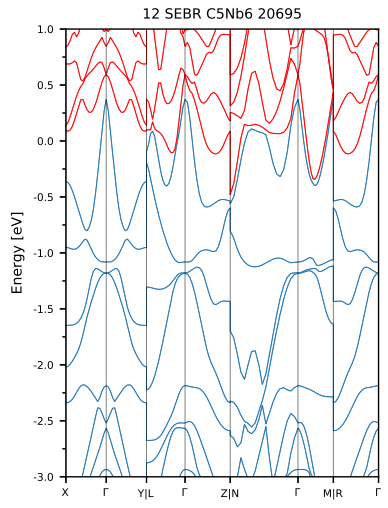

(dy) 12.1.129

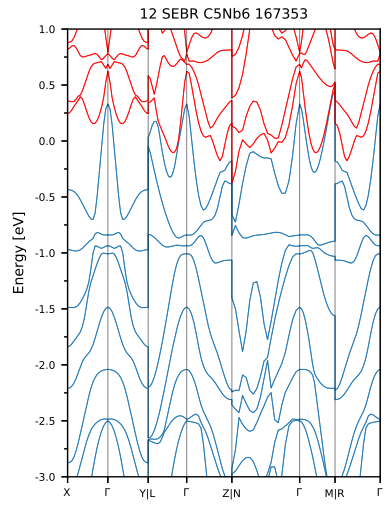

(dz) 12.1.130

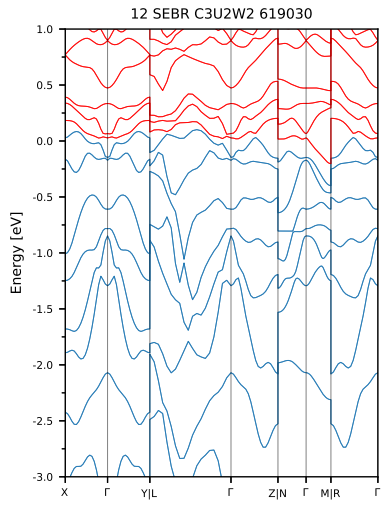

(ea) 12.1.131

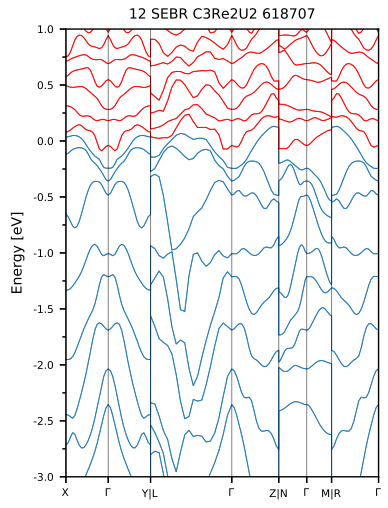

(eb) 12.1.132

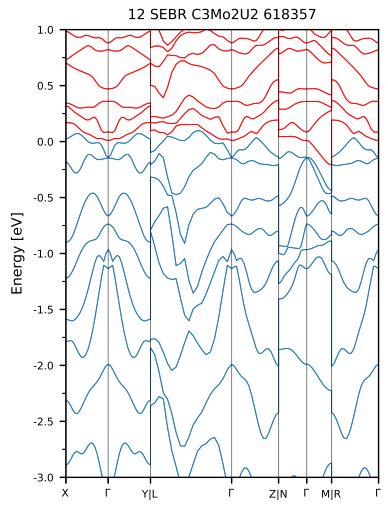

(ec) 12.1.133

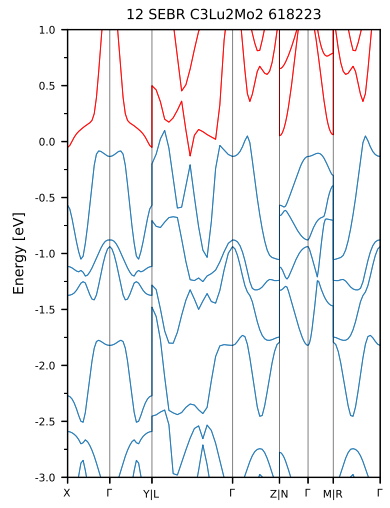

(ed) 12.1.134

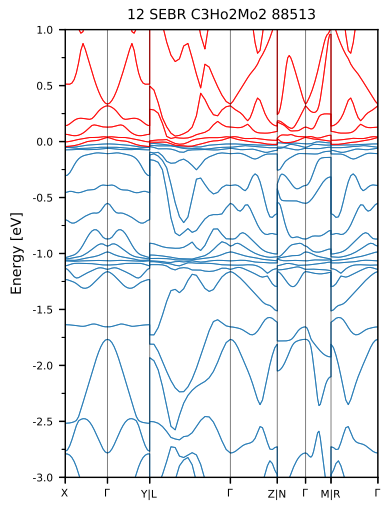

(ee) 12.1.135

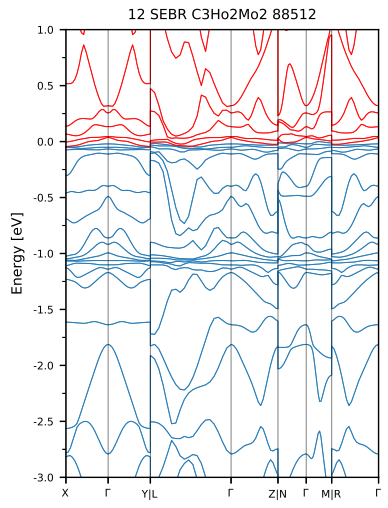

(ef) 12.1.136

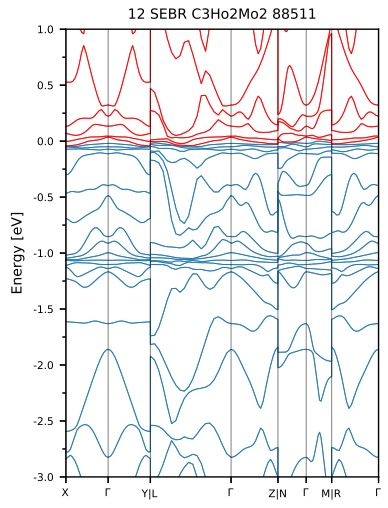

(eg) 12.1.137

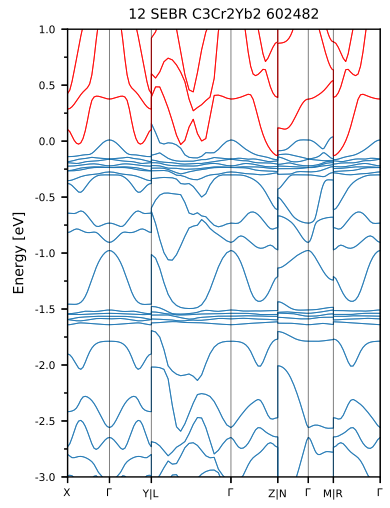

(eh) 12.1.138

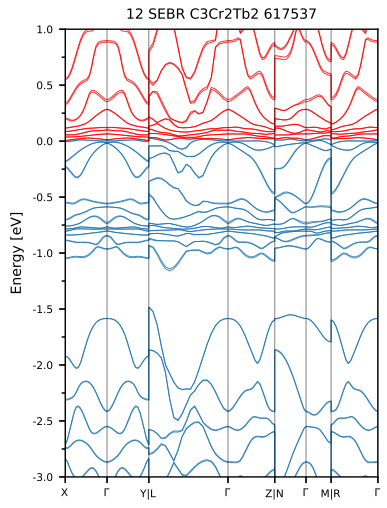

(ei) 12.1.139

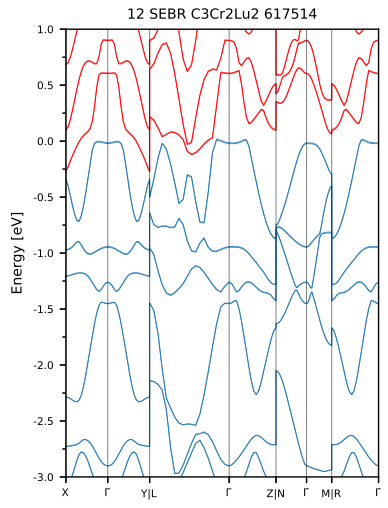

(ej) 12.1.140

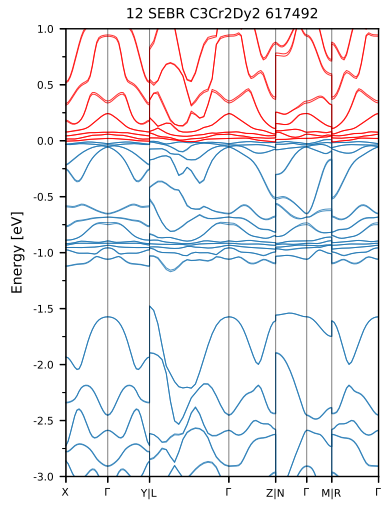

(ek) 12.1.141

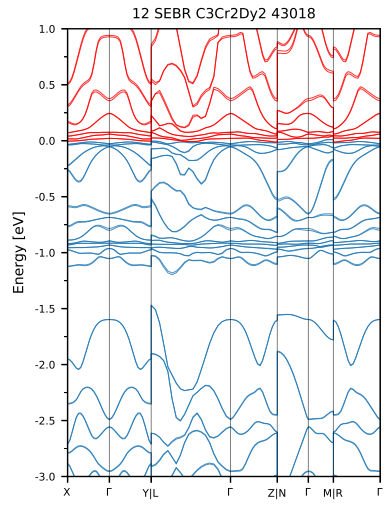

(el) 12.1.142

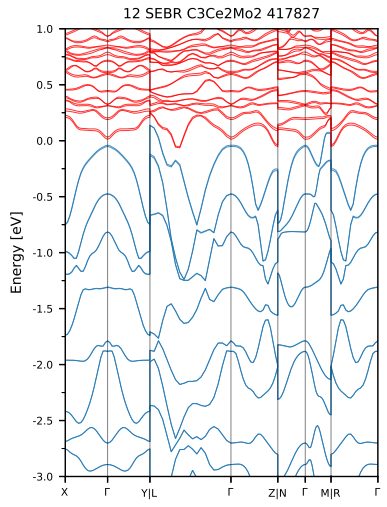

(em) 12.1.143

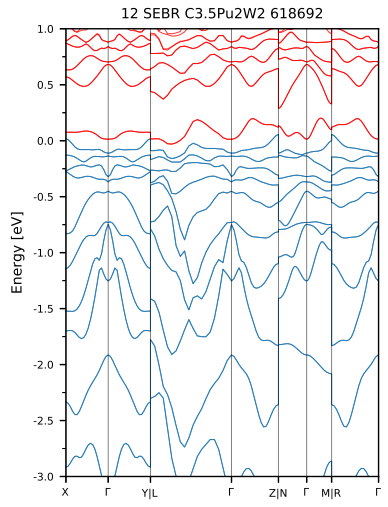

(en) 12.1.144

Figure 32: Band structures for topological materials in SG 12, part 9/25, from 12.1.129 to 12.1.144. See the list of figures.

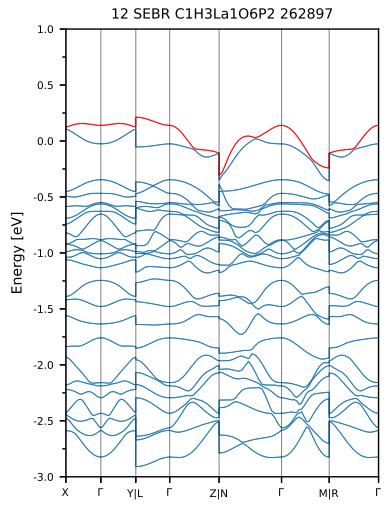

(eo) 12.1.145

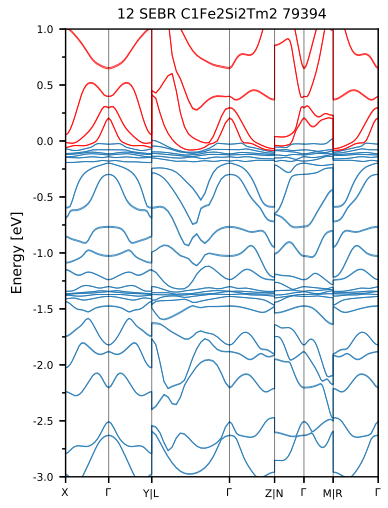

(ep) 12.1.146

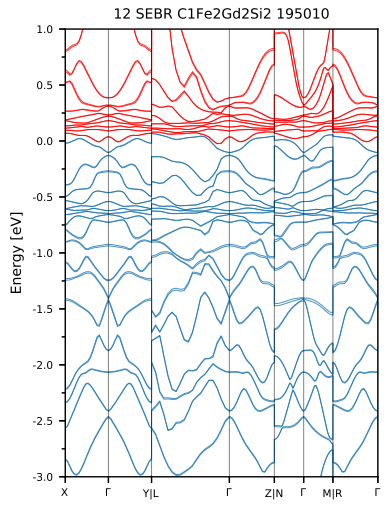

(eq) 12.1.147

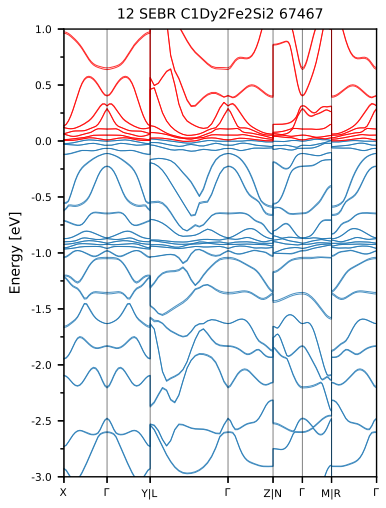

(er) 12.1.148

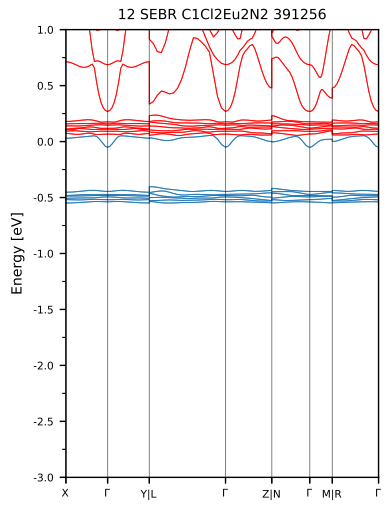

(es) 12.1.149

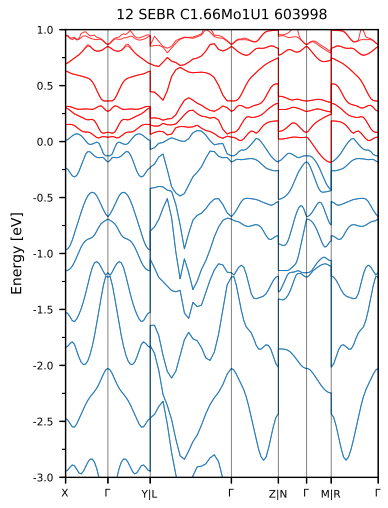

(et) 12.1.150

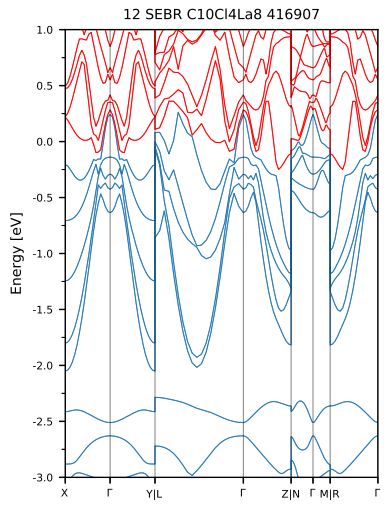

(eu) 12.1.151

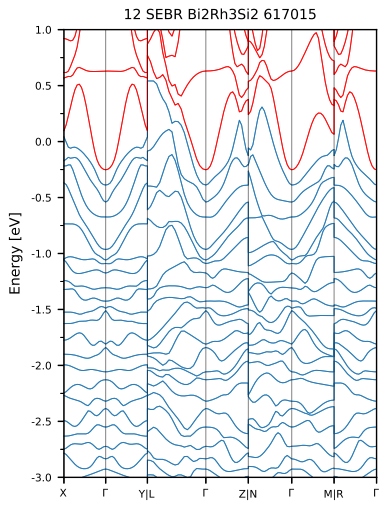

(ev) 12.1.152

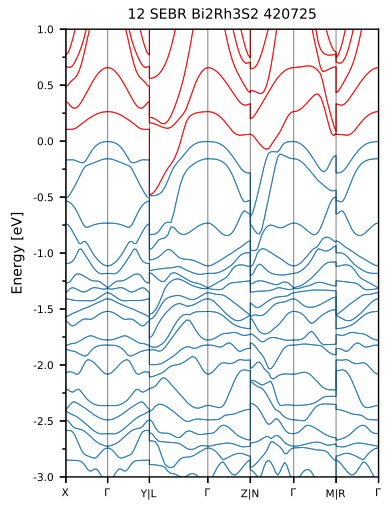

(ew) 12.1.153

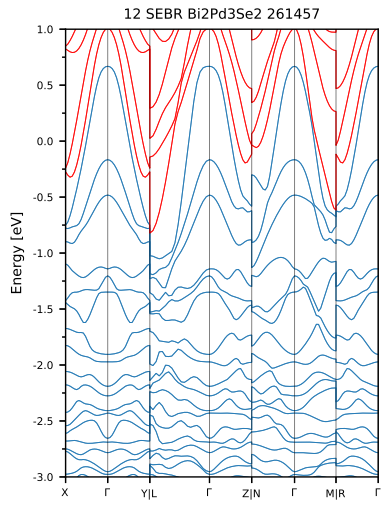

(ex) 12.1.154

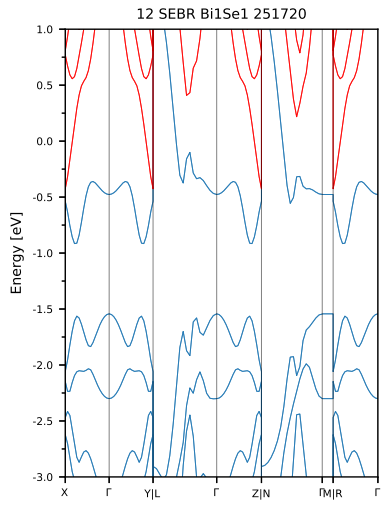

(ey) 12.1.155

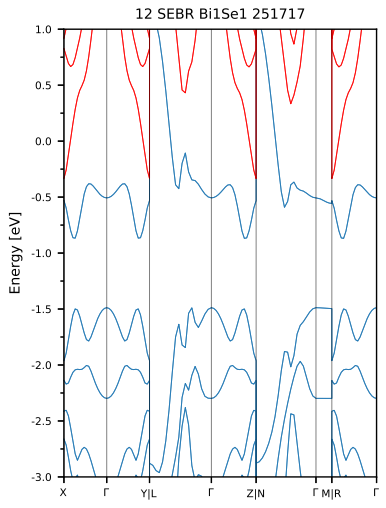

(ez) 12.1.156

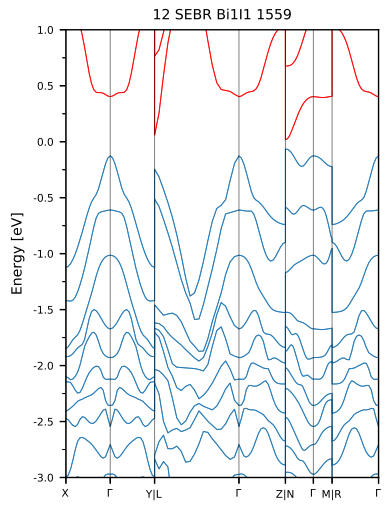

(fa) 12.1.157

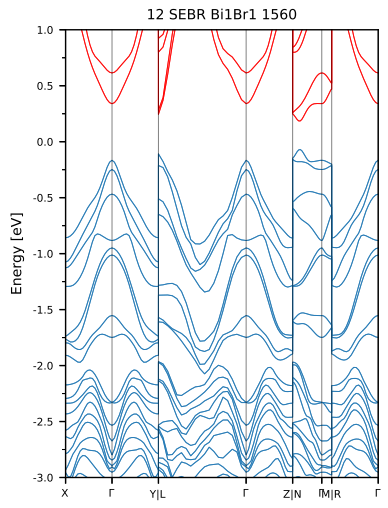

(fb) 12.1.158

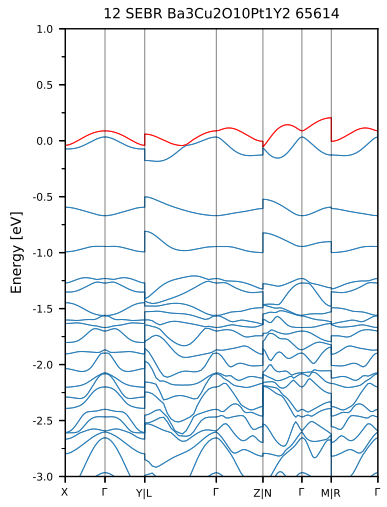

(fc) 12.1.159

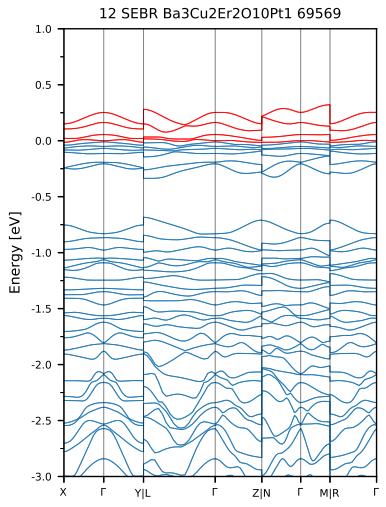

(fd) 12.1.160

Figure 33: Band structures for topological materials in SG 12, part 10/25, from 12.1.145 to 12.1.160. See the list of figures.

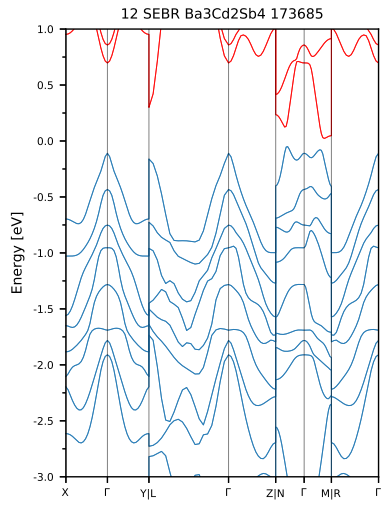

(fe) 12.1.161

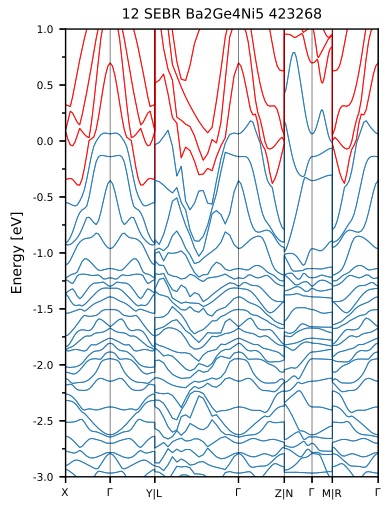

(ff) 12.1.162

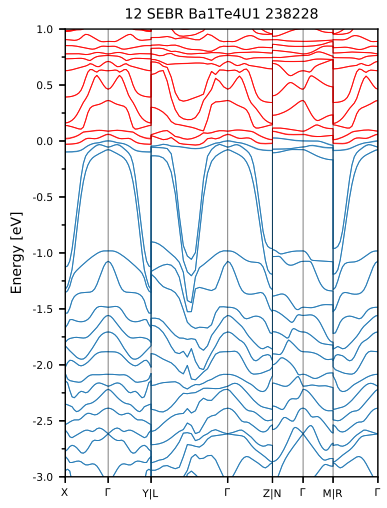

(fg) 12.1.163

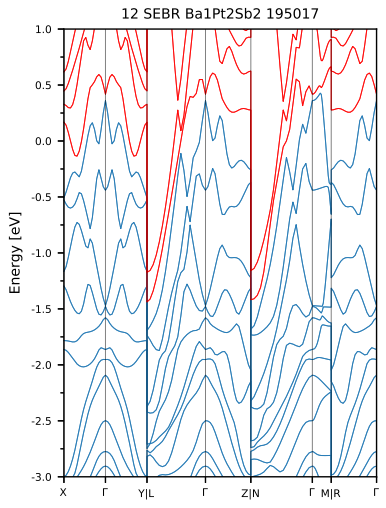

(fh) 12.1.164

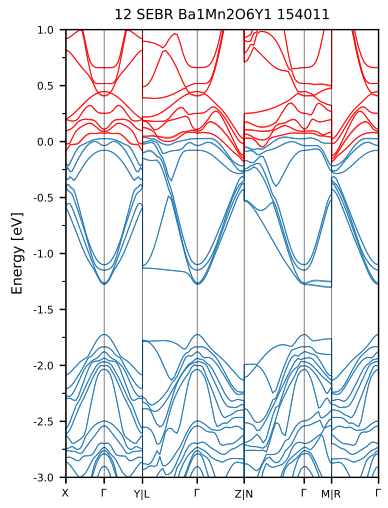

(fi) 12.1.165

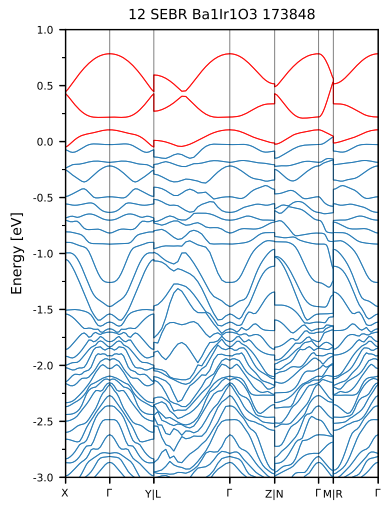

(fj) 12.1.166

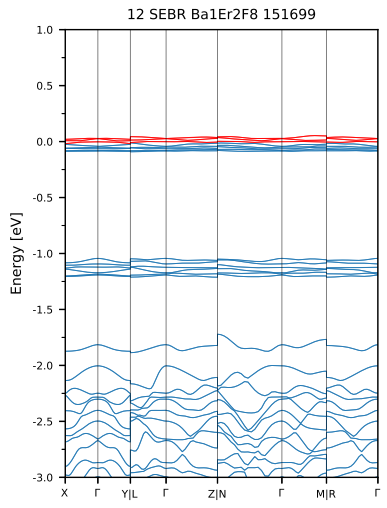

(fk) 12.1.167

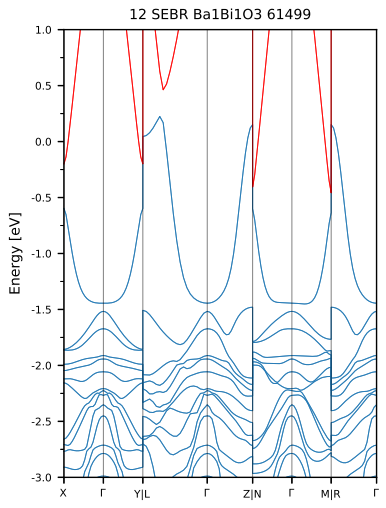

(fl) 12.1.168

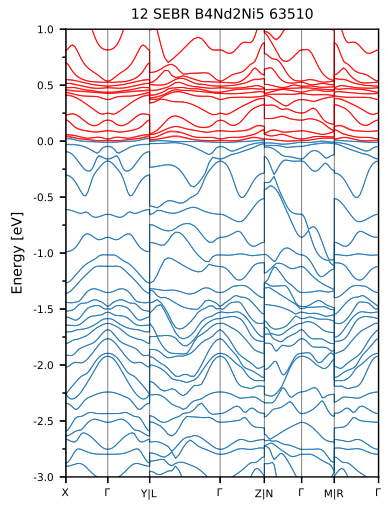

(fm) 12.1.169

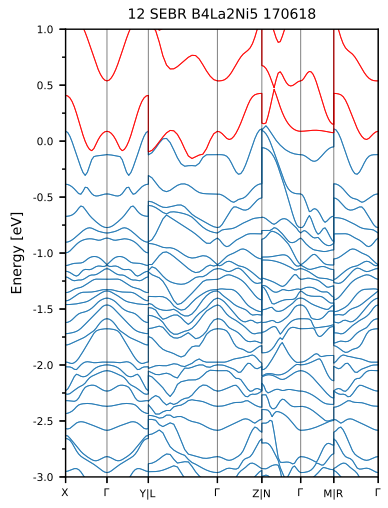

(fn) 12.1.170

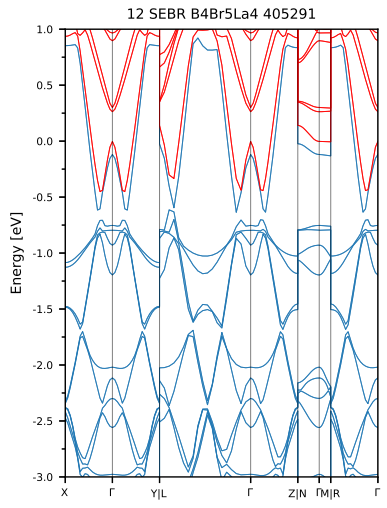

(fo) 12.1.171

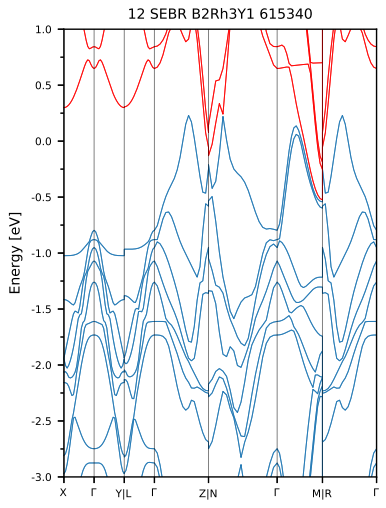

(fp) 12.1.172

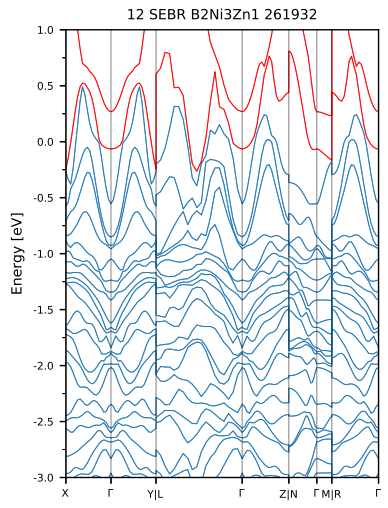

(fq) 12.1.173

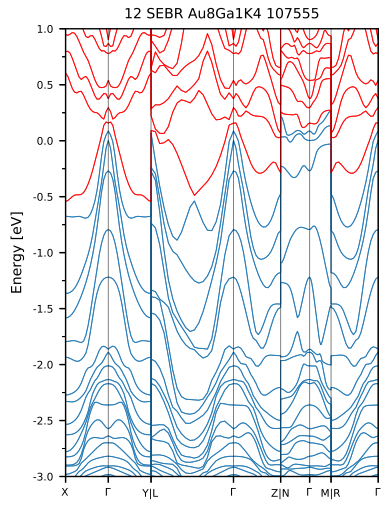

(fr) 12.1.174

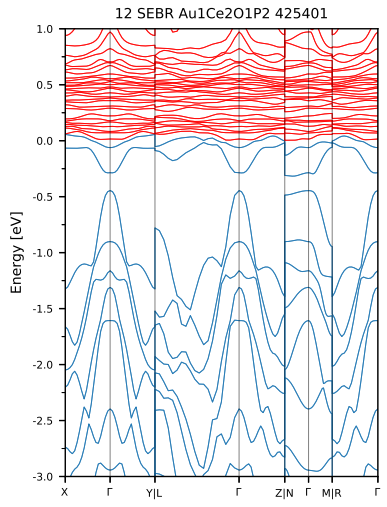

(fs) 12.1.175

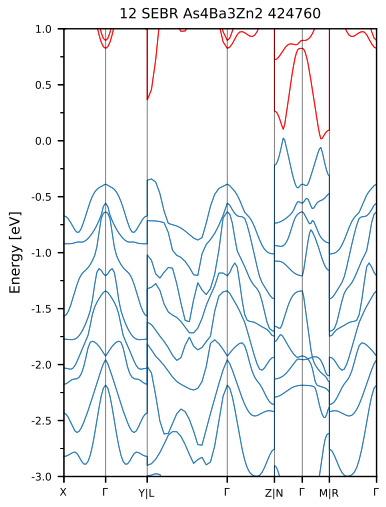

(ft) 12.1.176

Figure 34: Band structures for topological materials in SG 12, part 11/25, from 12.1.161 to 12.1.176. See the list of figures.

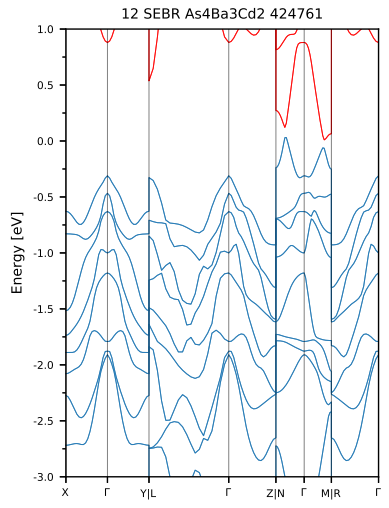

(fu) 12.1.177

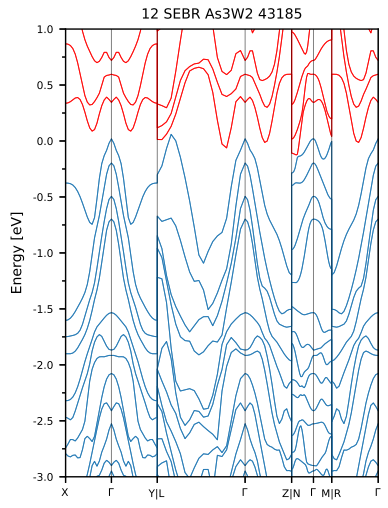

(fv) 12.1.178

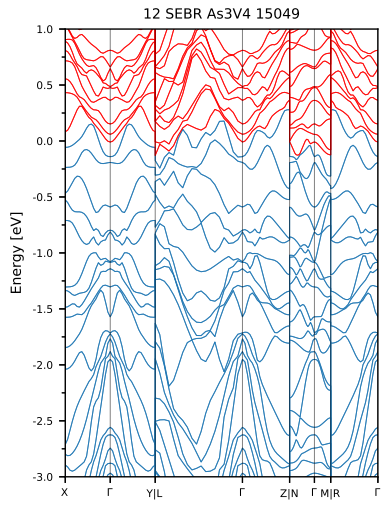

(fw) 12.1.179

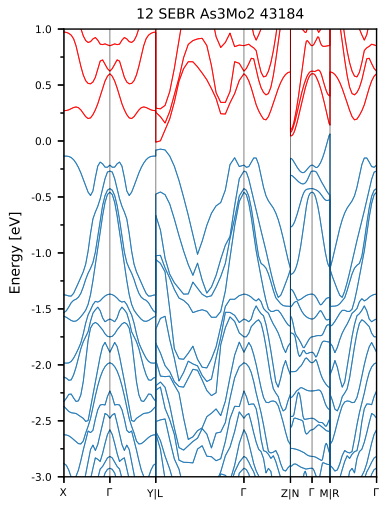

(fx) 12.1.180

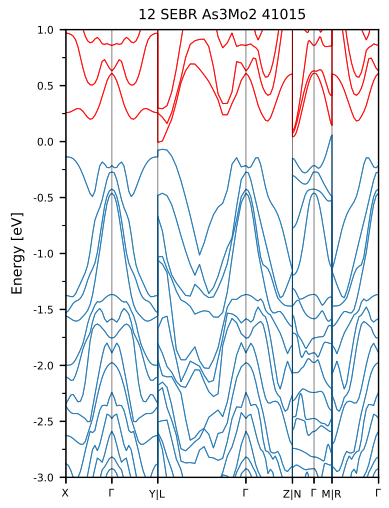

(fy) 12.1.181

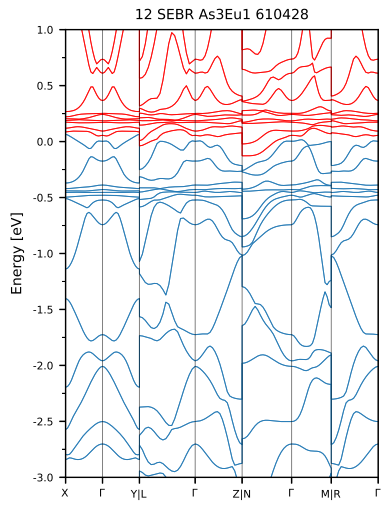

(fz) 12.1.182

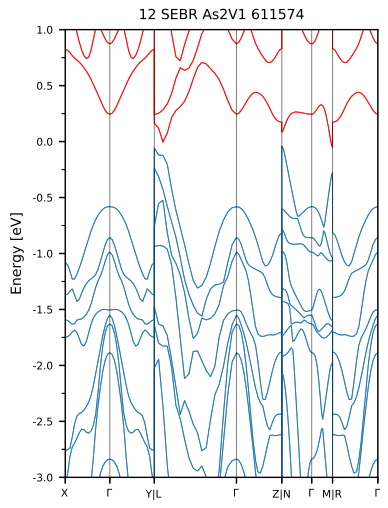

(ga) 12.1.183

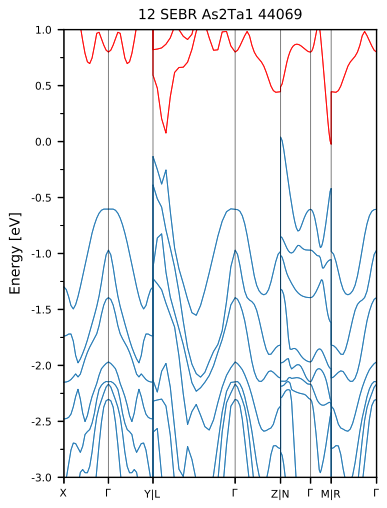

(gb) 12.1.184

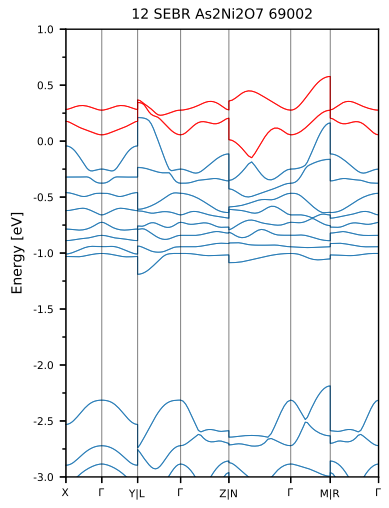

(gc) 12.1.185

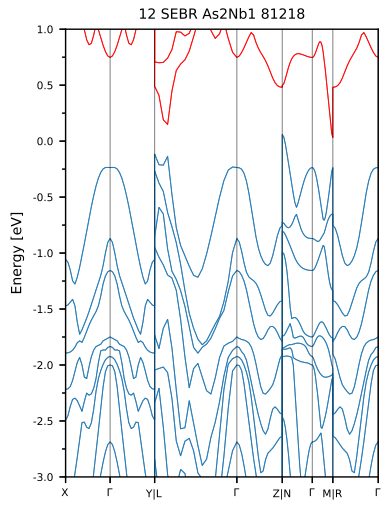

(gd) 12.1.186

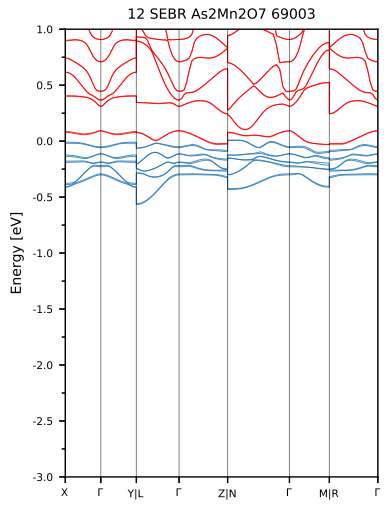

(ge) 12.1.187

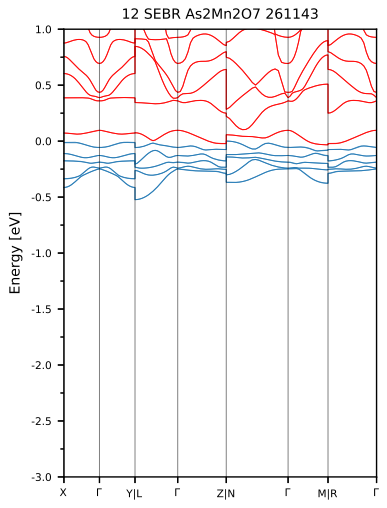

(gf) 12.1.188

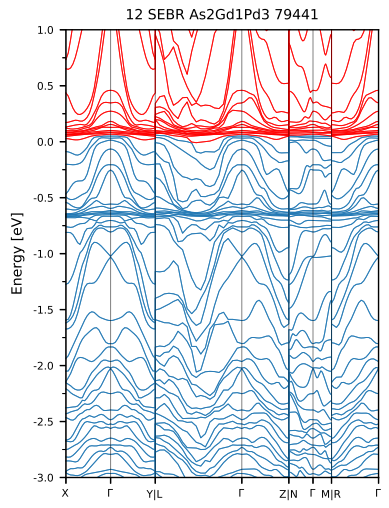

(gg) 12.1.189

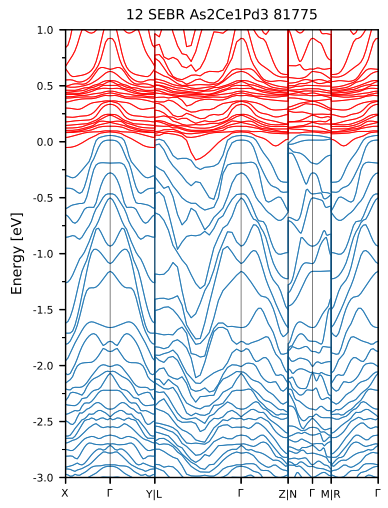

(gh) 12.1.190

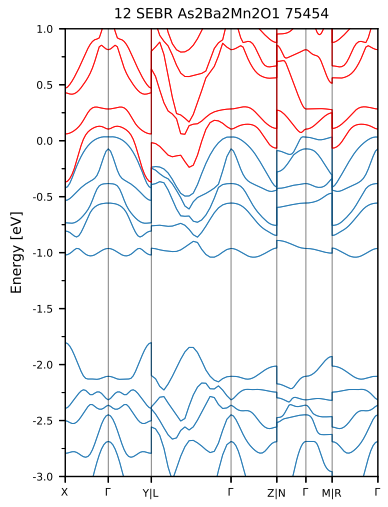

(gi) 12.1.191

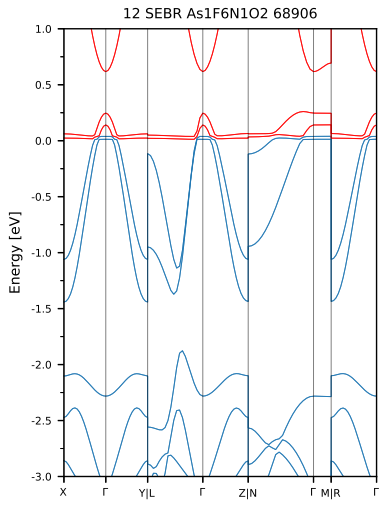

(gj) 12.1.192

Figure 35: Band structures for topological materials in SG 12, part 12/25, from 12.1.177 to 12.1.192. See the list of figures.

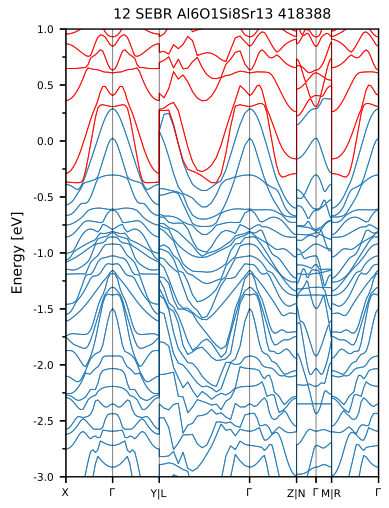

(gk) 12.1.193

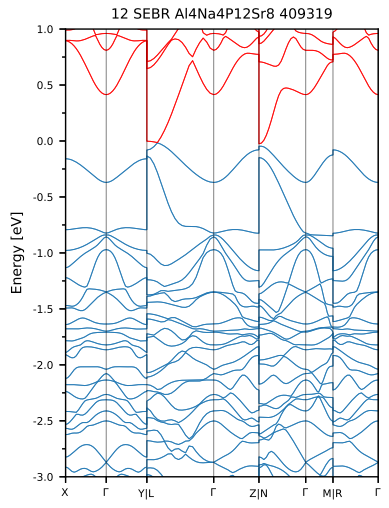

(gl) 12.1.194

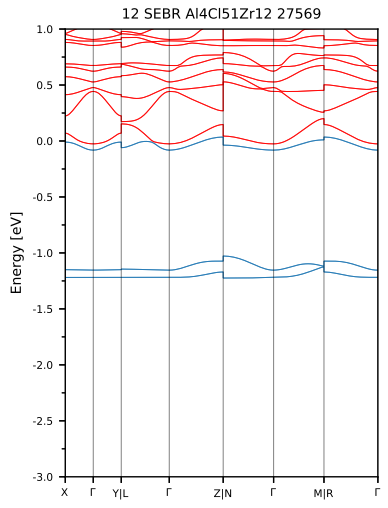

(gm) 12.1.195

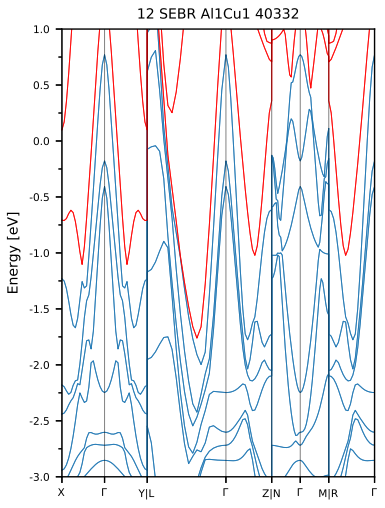

(gn) 12.1.196

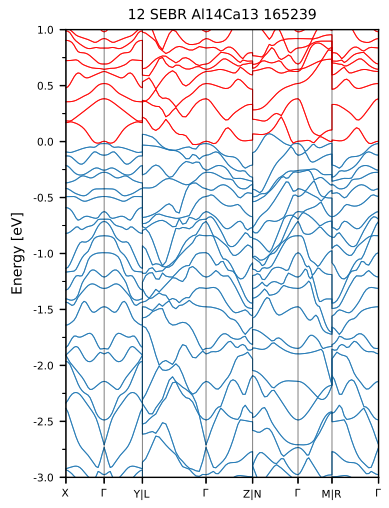

(go) 12.1.197

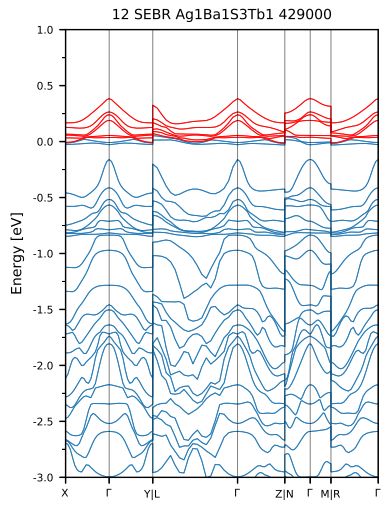

(gp) 12.1.198

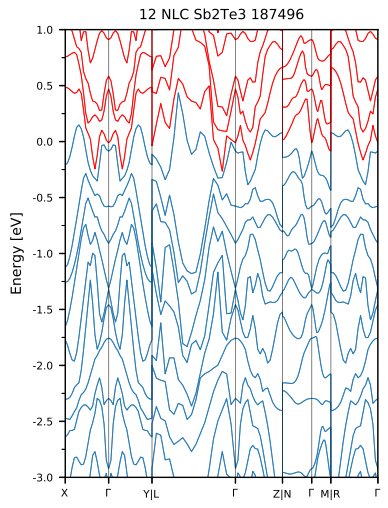

(gq) 12.2.1

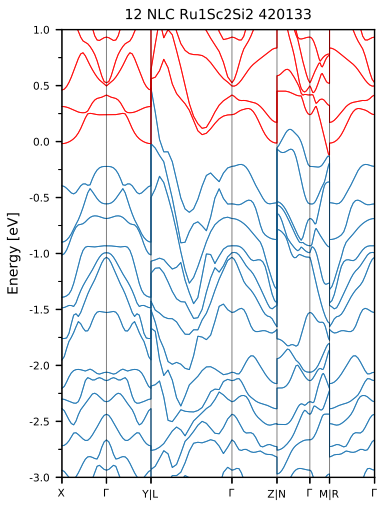

(gr) 12.2.2

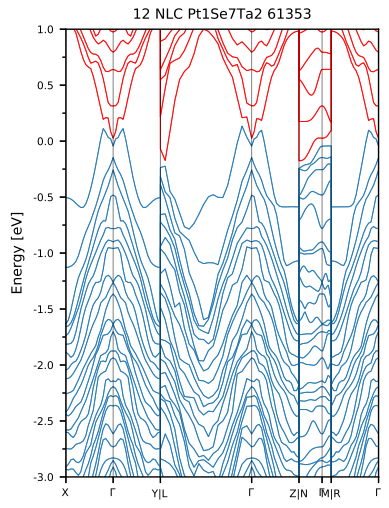

(gs) 12.2.3

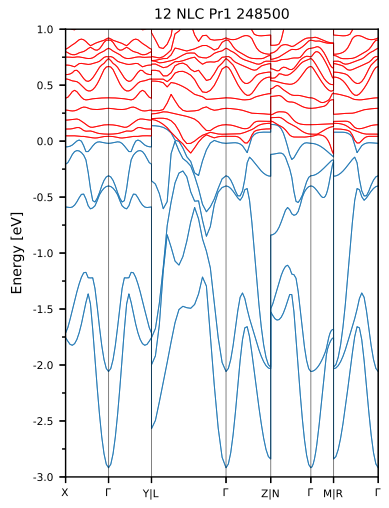

(gt) 12.2.4

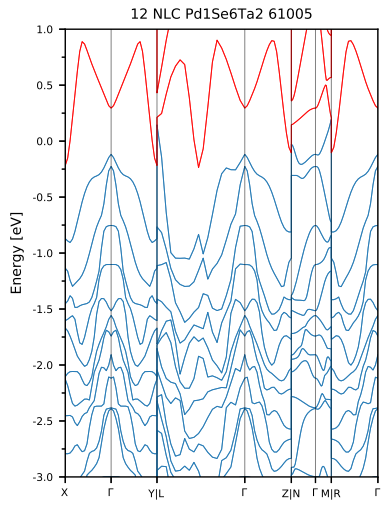

(gu) 12.2.5

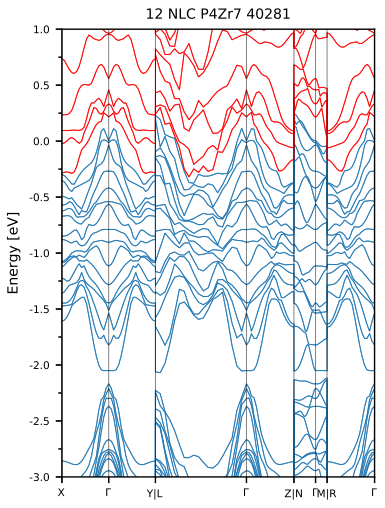

(gv) 12.2.6

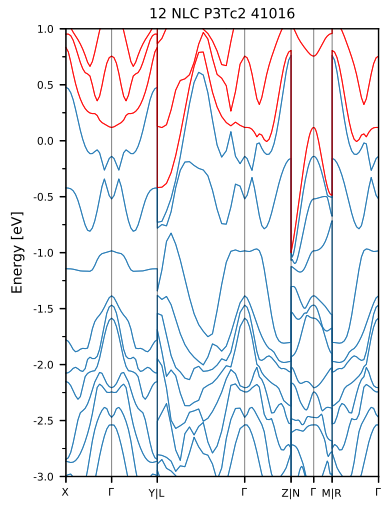

(gw) 12.2.7

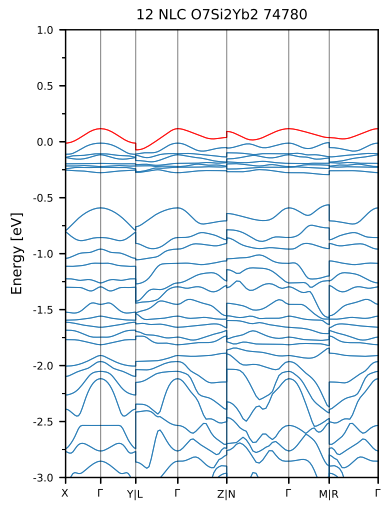

(gx) 12.2.8

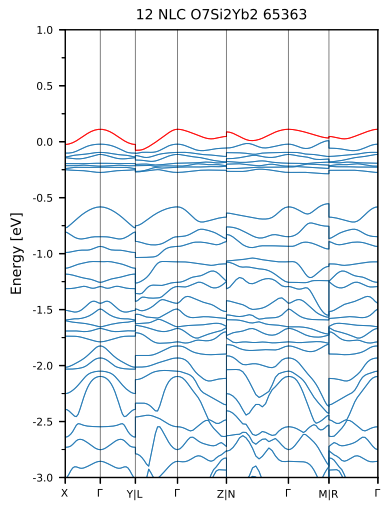

(gy) 12.2.9

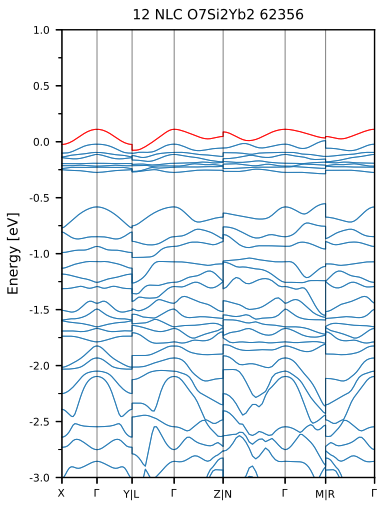

(gz) 12.2.10

Figure 36: Band structures for topological materials in SG 12, part 13/25, from 12.1.193 to 12.2.10. See the list of figures.

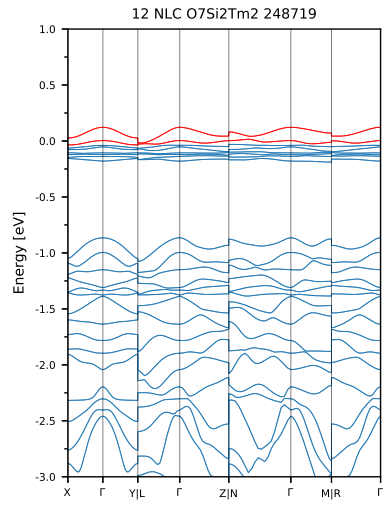

(ha) 12.2.11

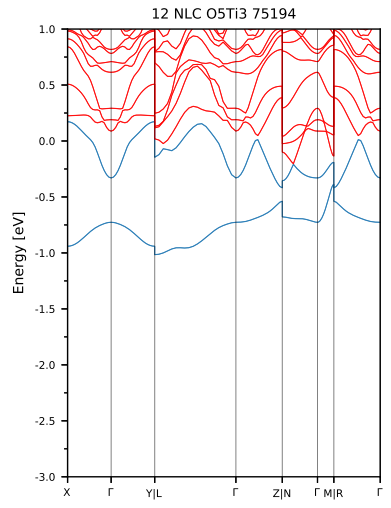

(hb) 12.2.12

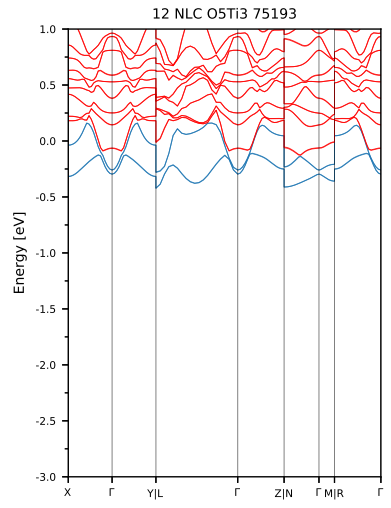

(hc) 12.2.13

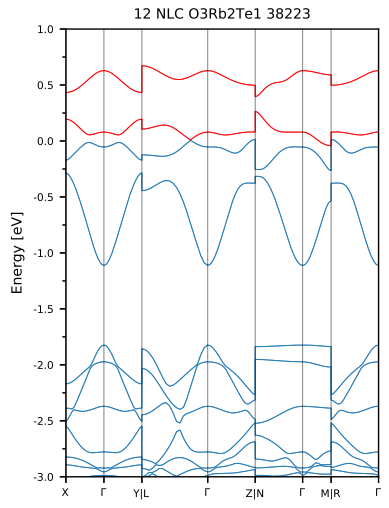

(hd) 12.2.14

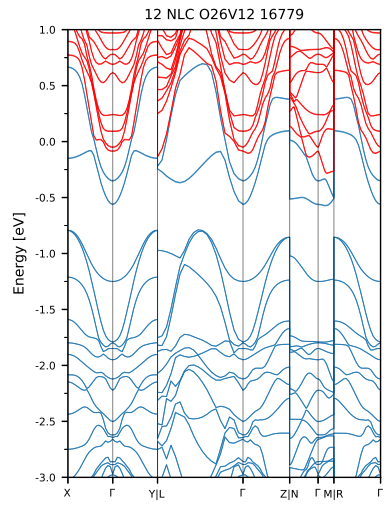

(he) 12.2.15

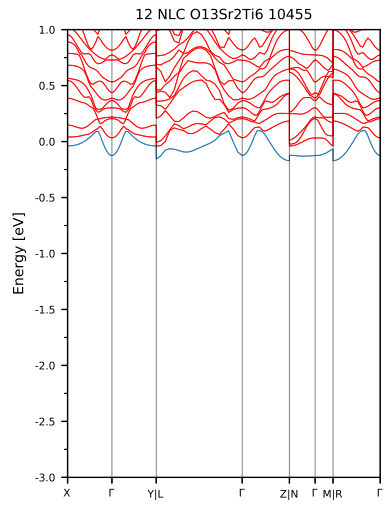

(hf) 12.2.16

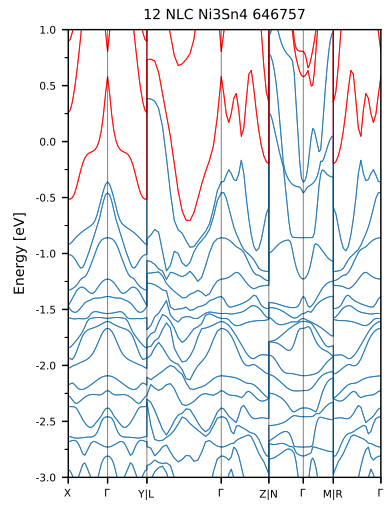

(hg) 12.2.17

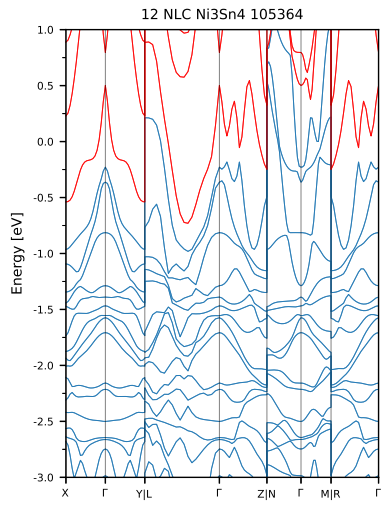

(hh) 12.2.18

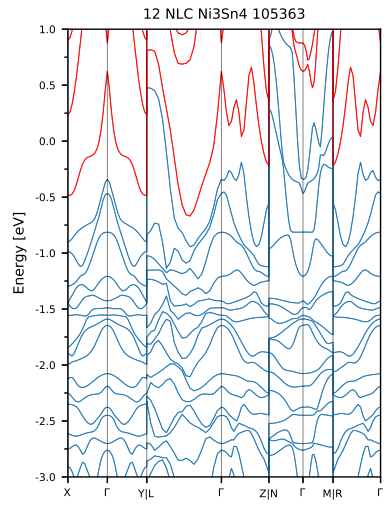

(hi) 12.2.19

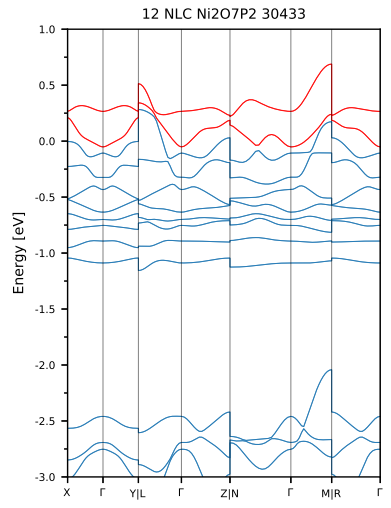

(hj) 12.2.20

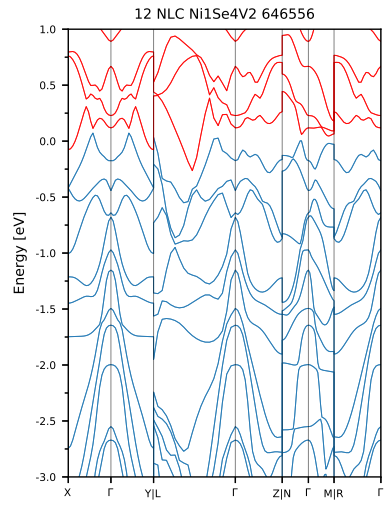

(hk) 12.2.21

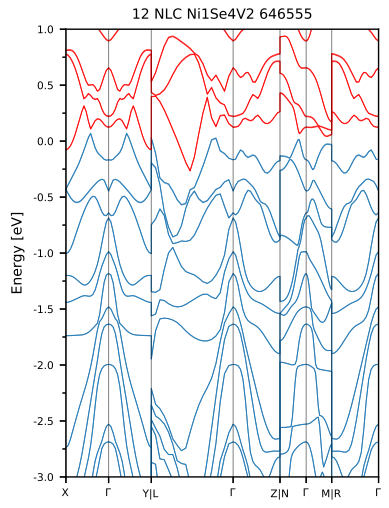

(hl) 12.2.22

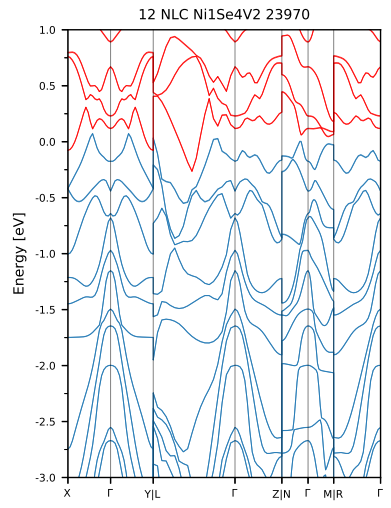

(hm) 12.2.23

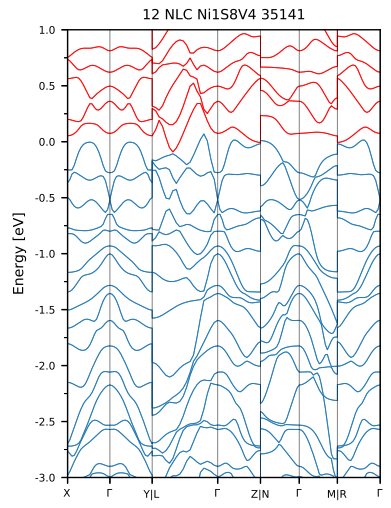

(hn) 12.2.24

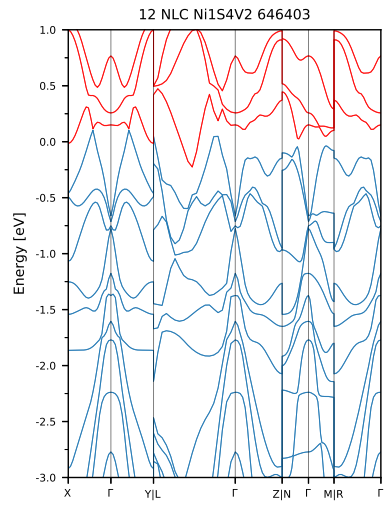

(ho) 12.2.25

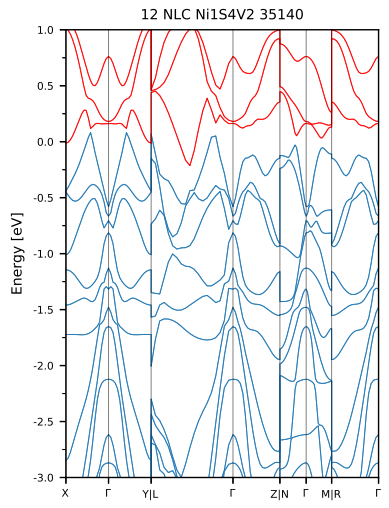

(hp) 12.2.26

Figure 37: Band structures for topological materials in SG 12, part 14/25, from 12.2.11 to 12.2.26. See the list of figures.

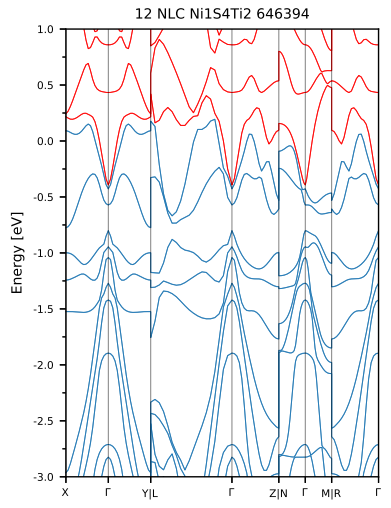

(hq) 12.2.27

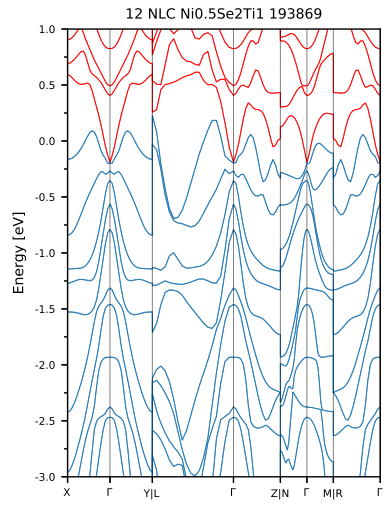

(hr) 12.2.28

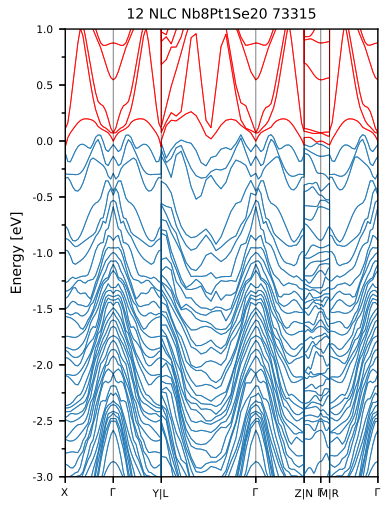

(hs) 12.2.29

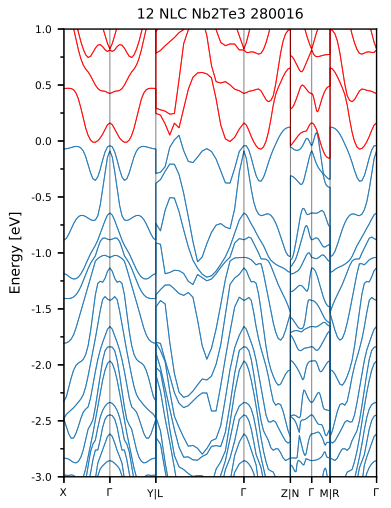

(ht) 12.2.30

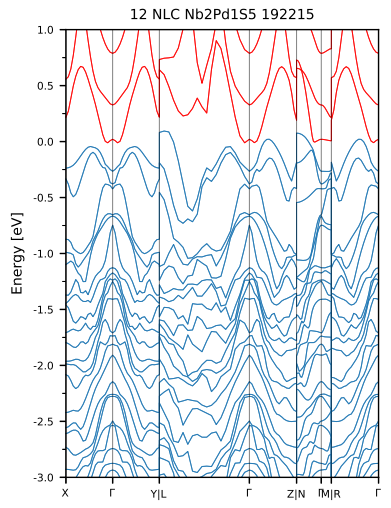

(hu) 12.2.31

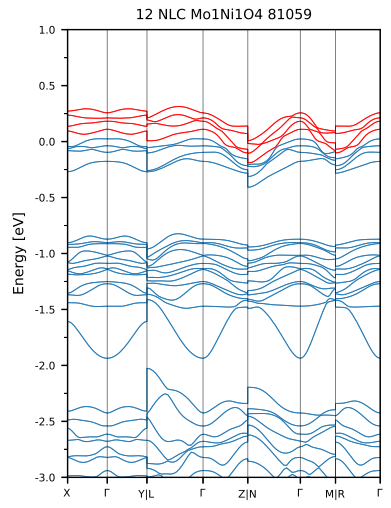

(hv) 12.2.32

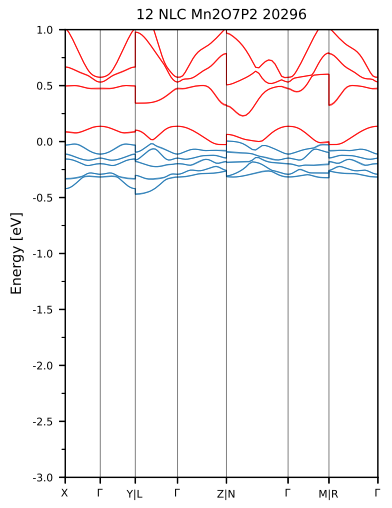

(hw) 12.2.33

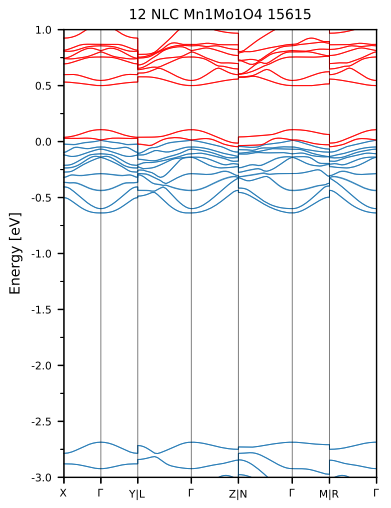

(hx) 12.2.34

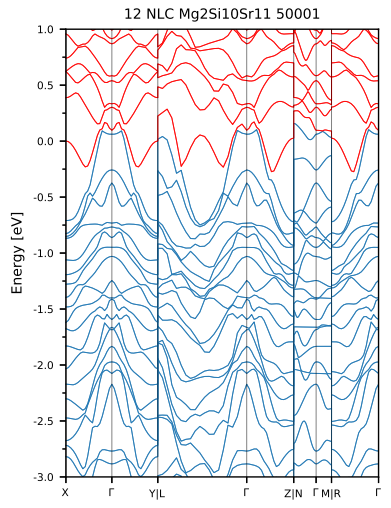

(hy) 12.2.35

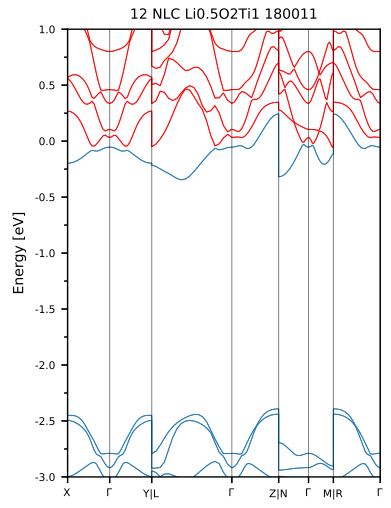

(hz) 12.2.36

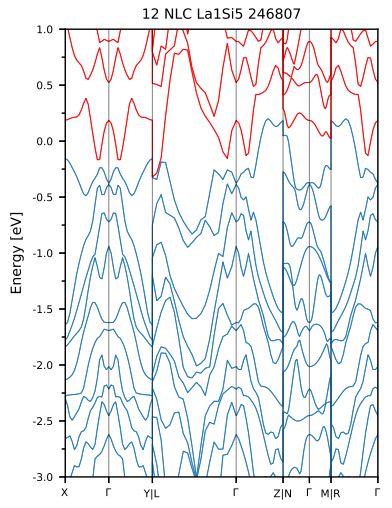

(ia) 12.2.37

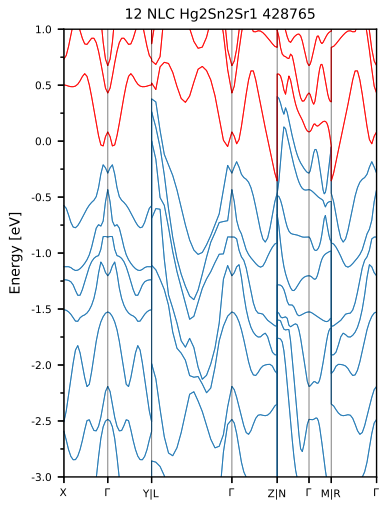

(ib) 12.2.38

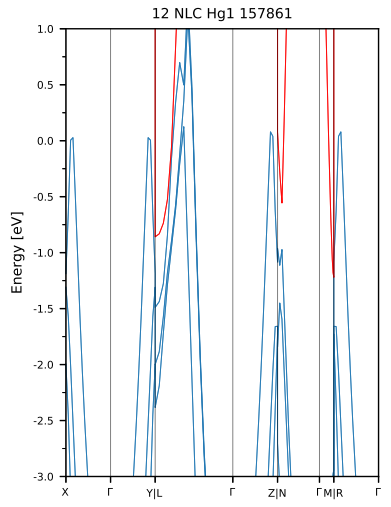

(ic) 12.2.39

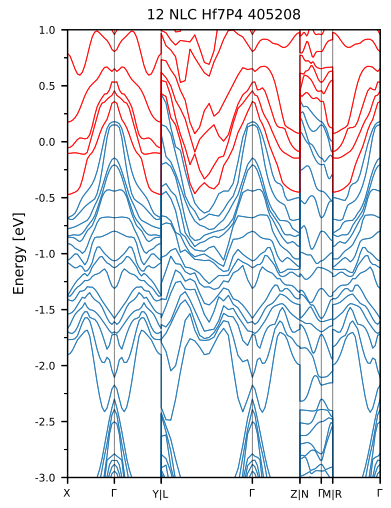

(id) 12.2.40

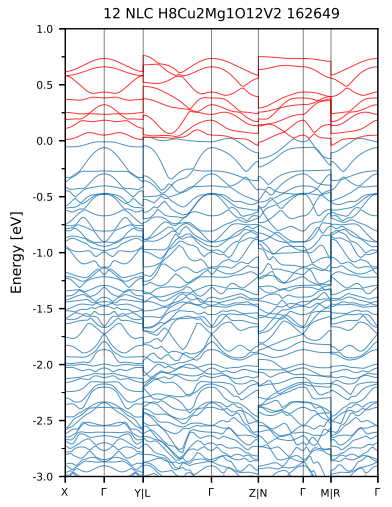

(ie) 12.2.41

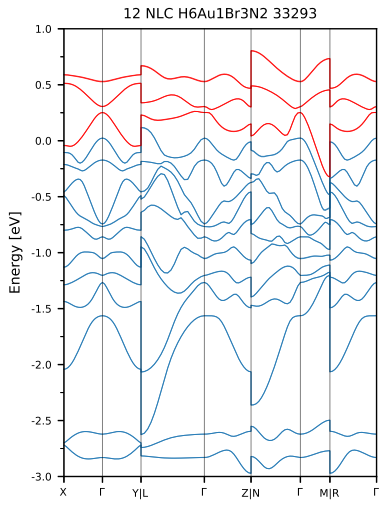

(if) 12.2.42

Figure 38: Band structures for topological materials in SG 12, part 15/25, from 12.2.27 to 12.2.42. See the list of figures.

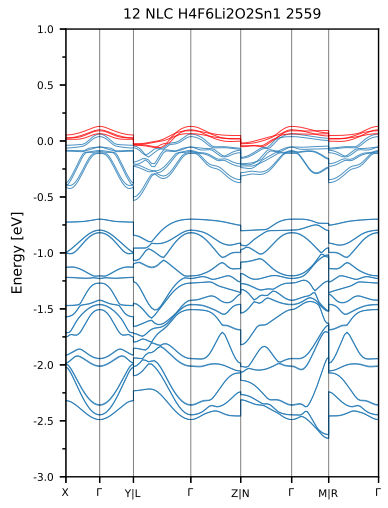

(ig) 12.2.43

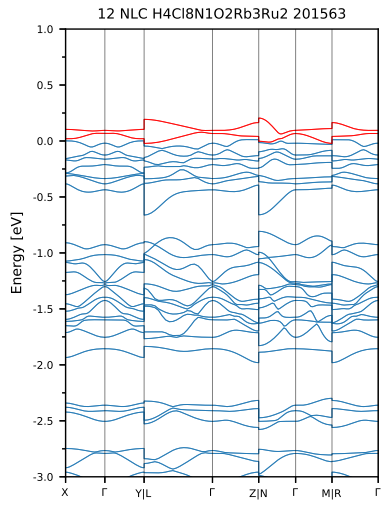

(ih) 12.2.44

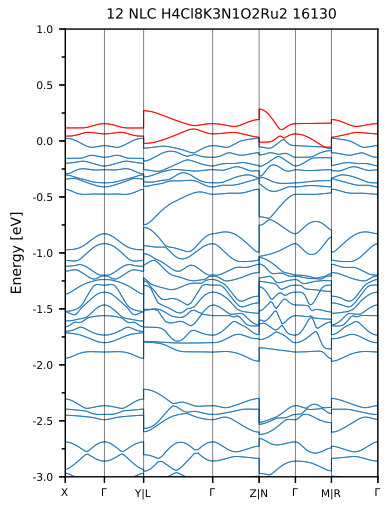

(ii) 12.2.45

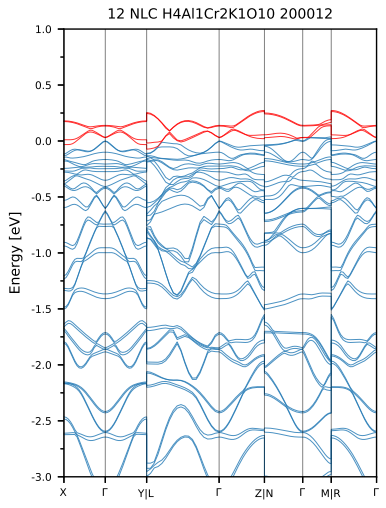

(ij) 12.2.46

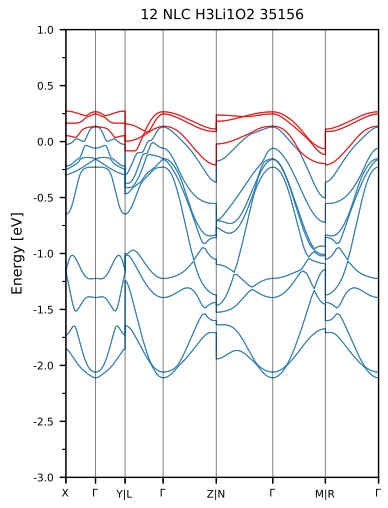

(ik) 12.2.47

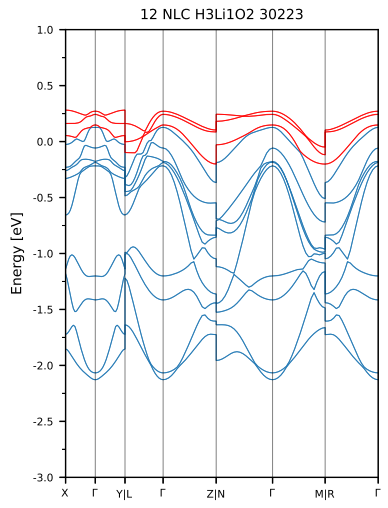

(il) 12.2.48

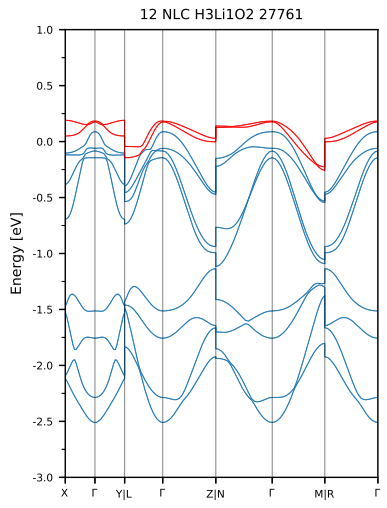

(im) 12.2.49

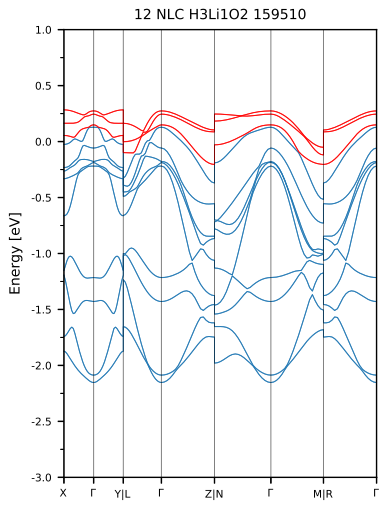

(in) 12.2.50

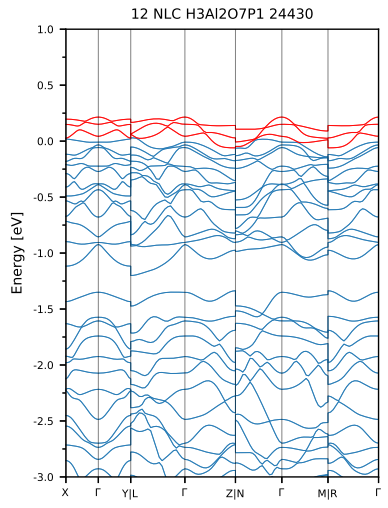

(io) 12.2.51

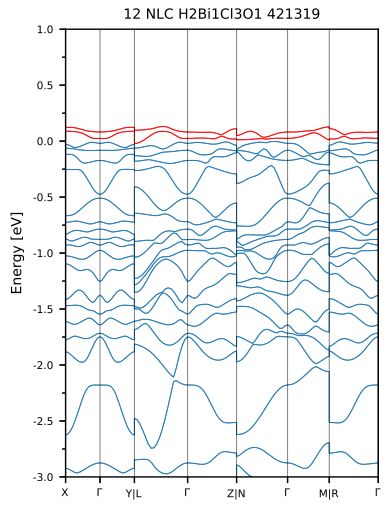

(ip) 12.2.52

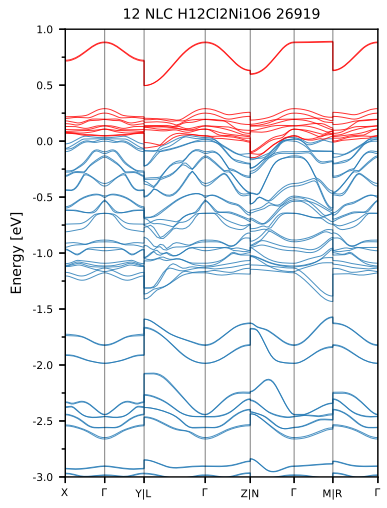

(iq) 12.2.53

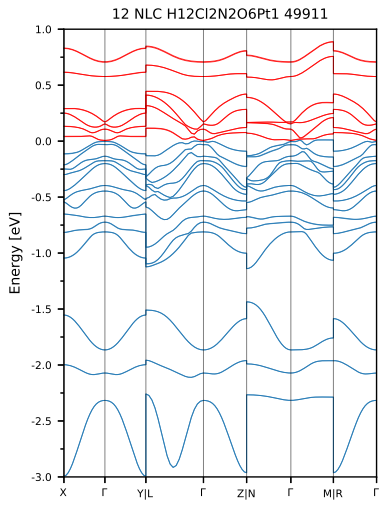

(ir) 12.2.54

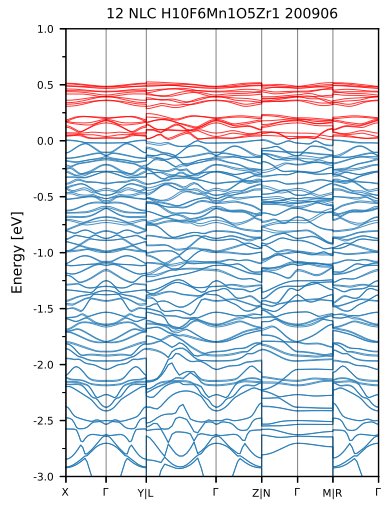

(is) 12.2.55

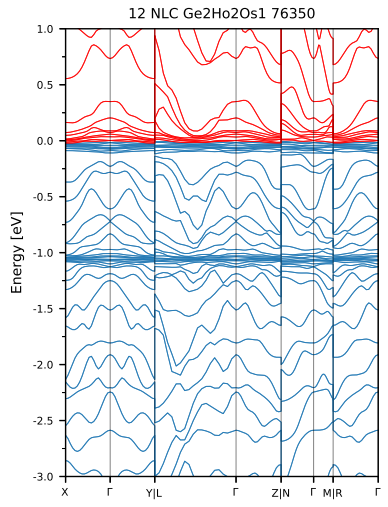

(it) 12.2.56

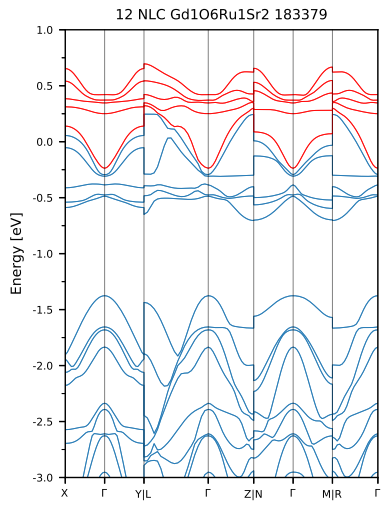

(iu) 12.2.57

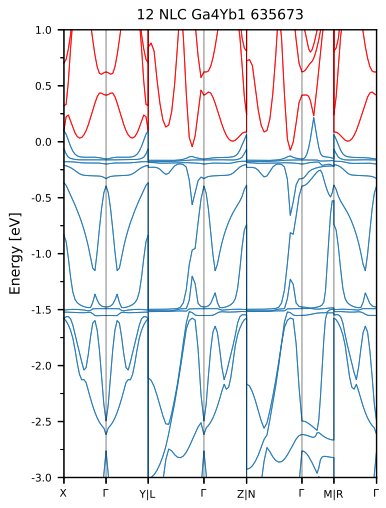

(iv) 12.2.58

Figure 39: Band structures for topological materials in SG 12, part 16/25, from 12.2.43 to 12.2.58. See the list of figures.

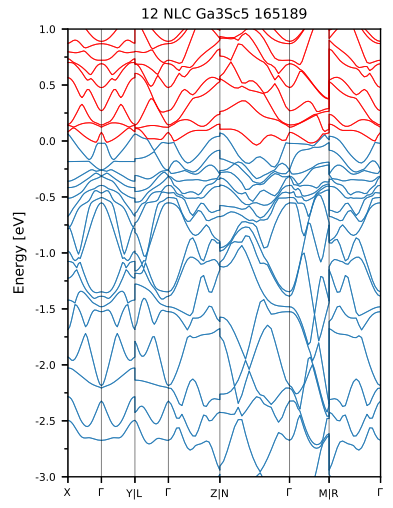

(iw) 12.2.59

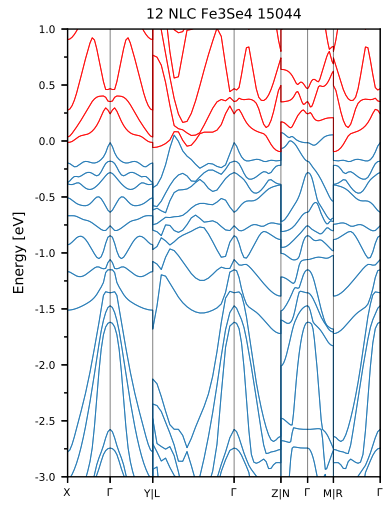

(ix) 12.2.60

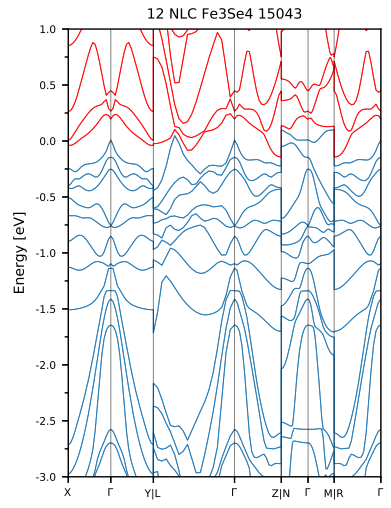

(iy) 12.2.61

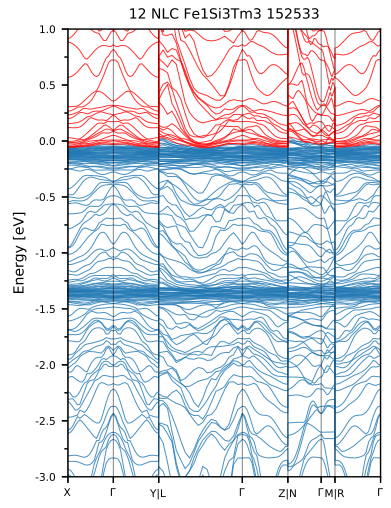

(iz) 12.2.62

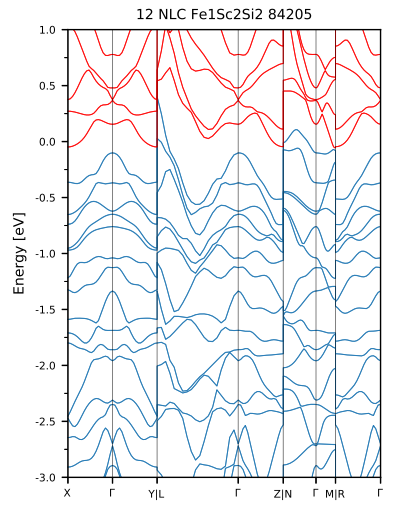

(ja) 12.2.63

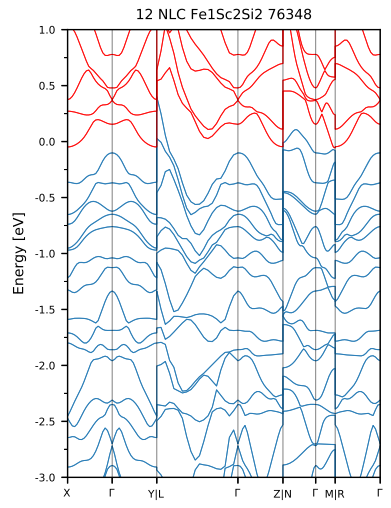

(jb) 12.2.64

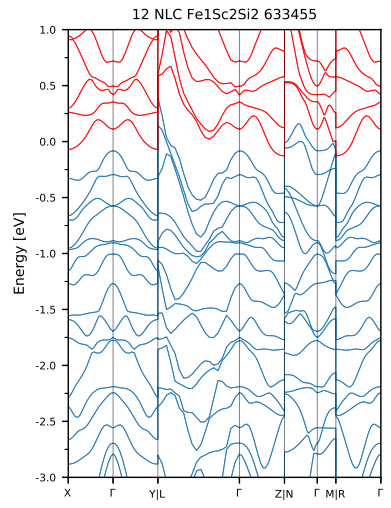

(jc) 12.2.65

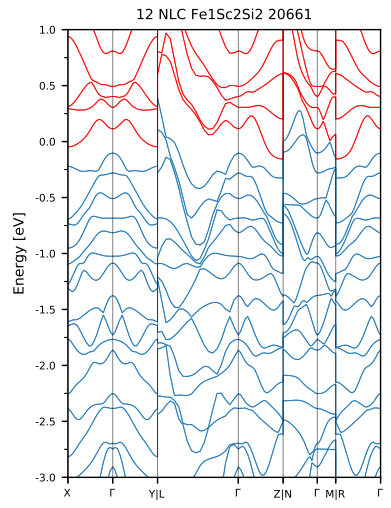

(jd) 12.2.66

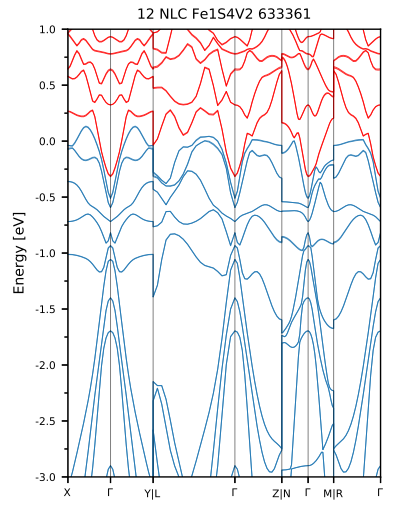

(je) 12.2.67

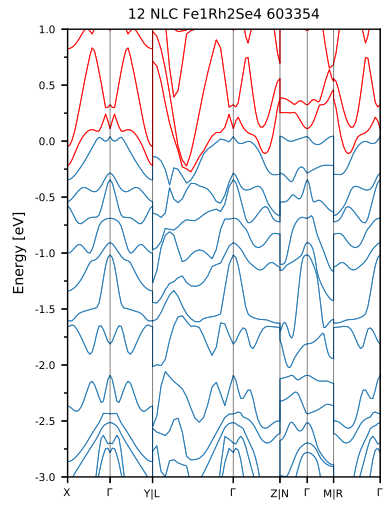

(jf) 12.2.68

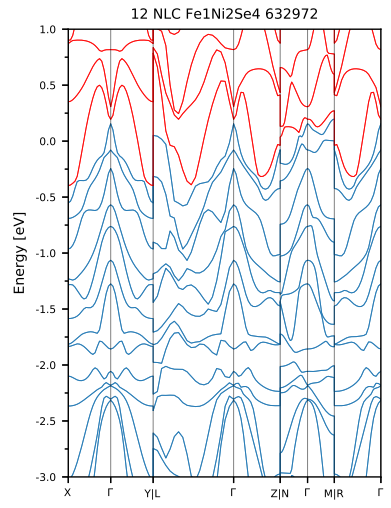

(jg) 12.2.69

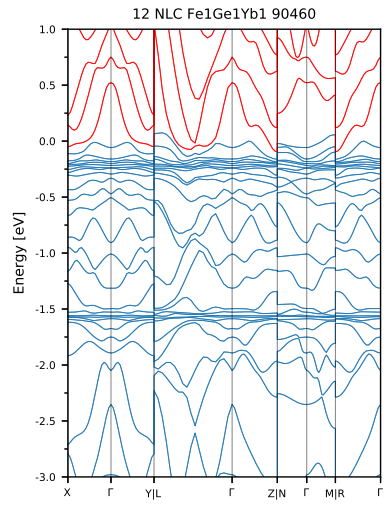

(jh) 12.2.70

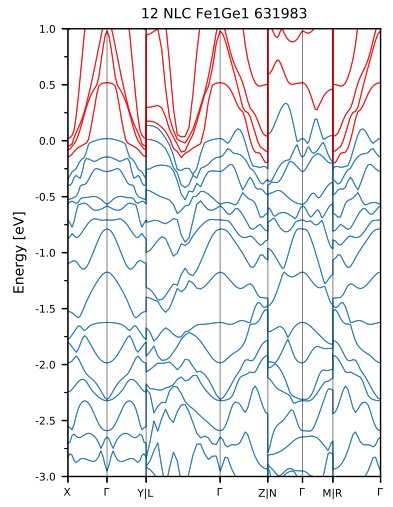

(ji) 12.2.71

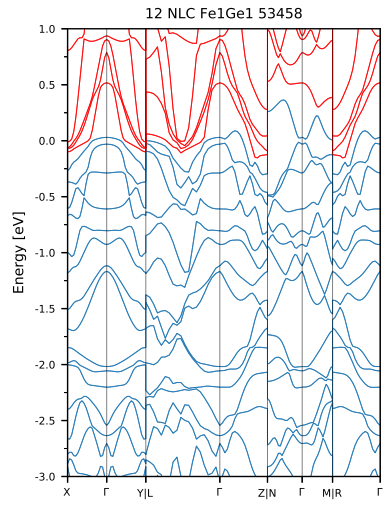

(jj) 12.2.72

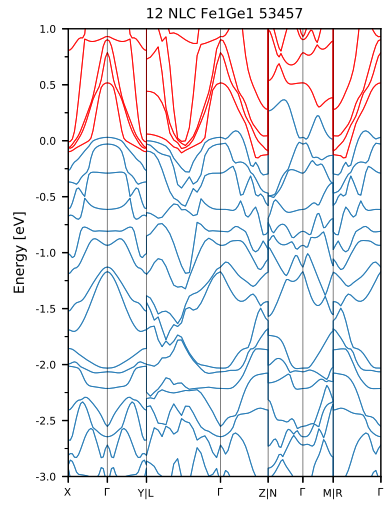

(jk) 12.2.73

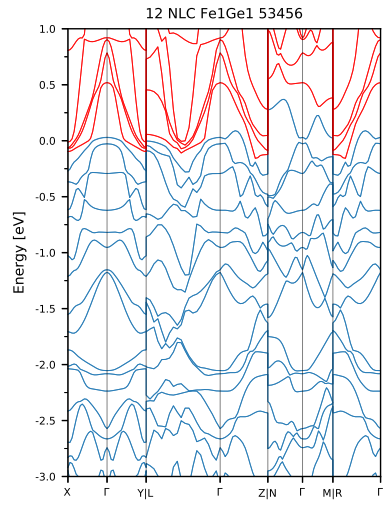

(jl) 12.2.74

Figure 40: Band structures for topological materials in SG 12, part 17/25, from 12.2.59 to 12.2.74. See the list of figures.

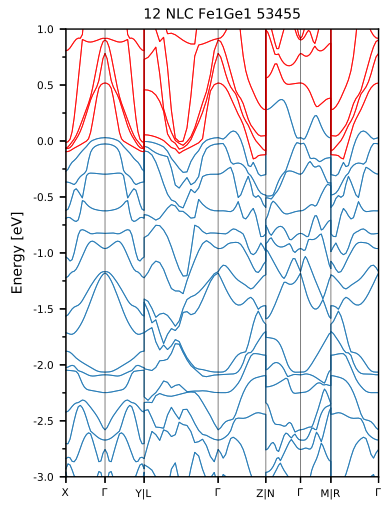

(jm) 12.2.75

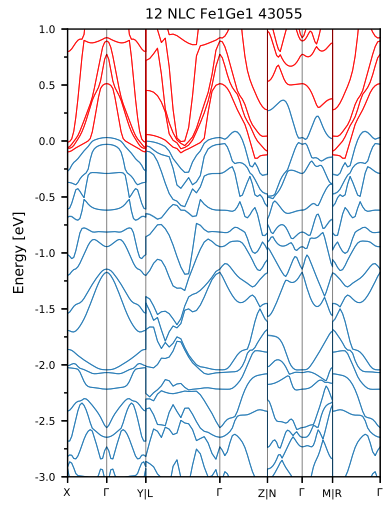

(jn) 12.2.76

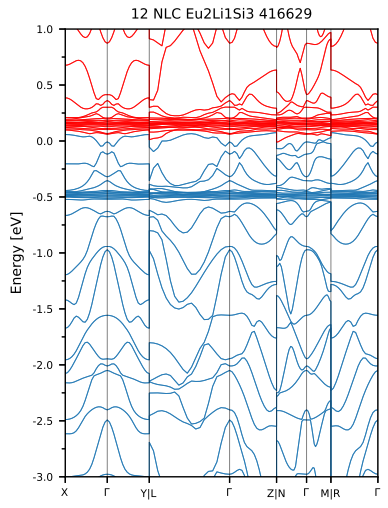

(jo) 12.2.77

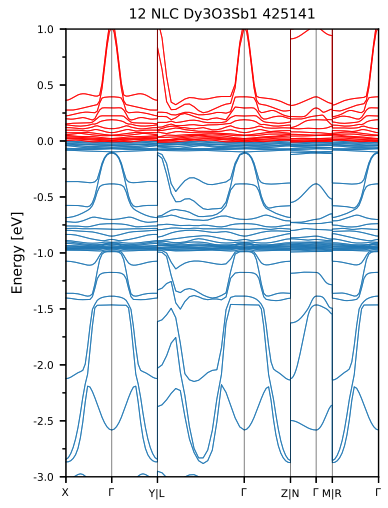

(jp) 12.2.78

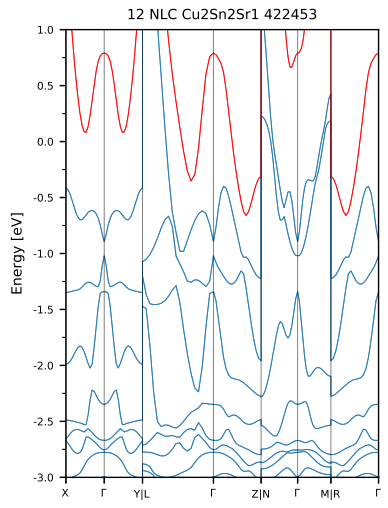

(jq) 12.2.79

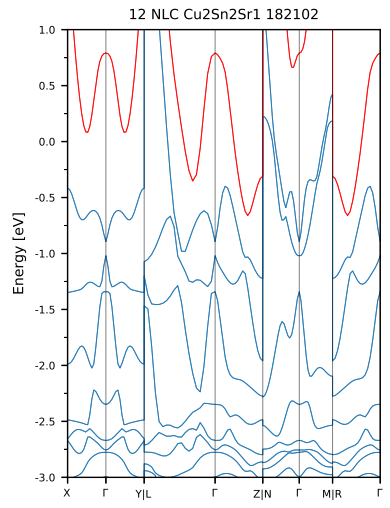

(jr) 12.2.80

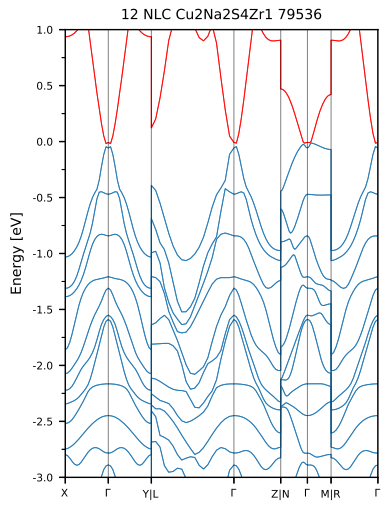

(js) 12.2.81

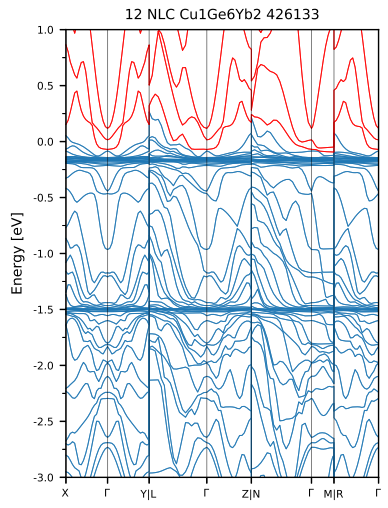

(jt) 12.2.82

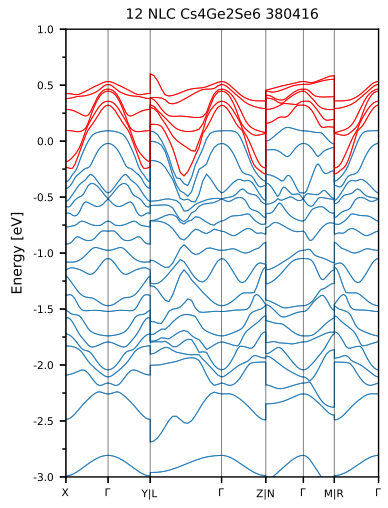

(ju) 12.2.83

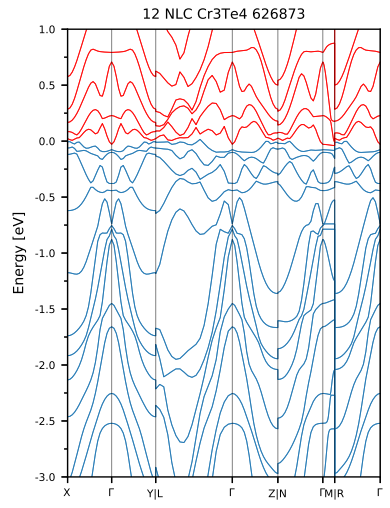

(jv) 12.2.84

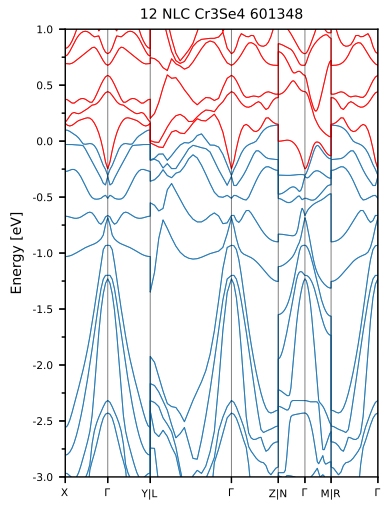

(jw) 12.2.85

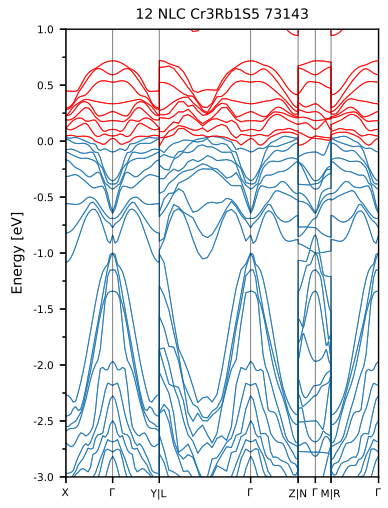

(jx) 12.2.86

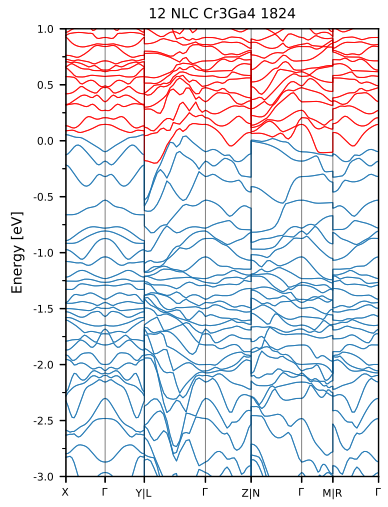

(jy) 12.2.87

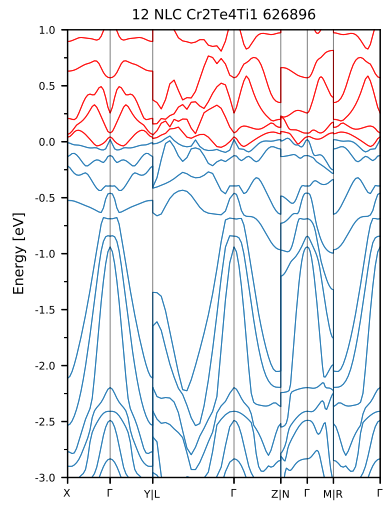

(jz) 12.2.88

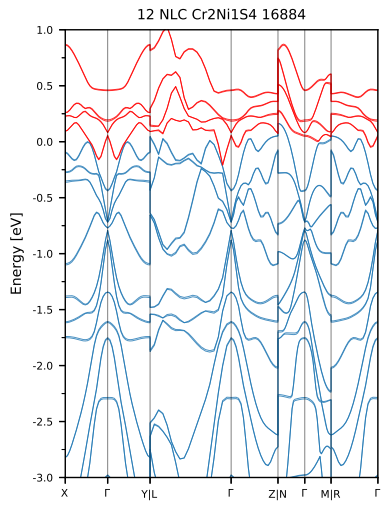

(ka) 12.2.89

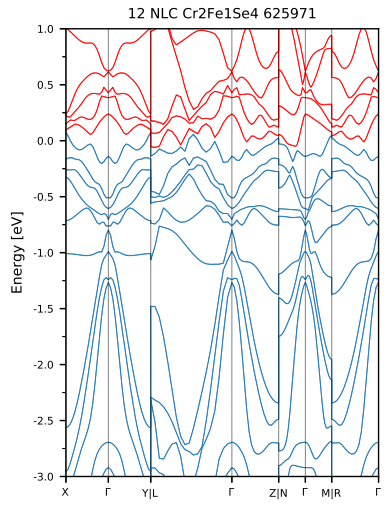

(kb) 12.2.90

Figure 41: Band structures for topological materials in SG 12, part 18/25, from 12.2.75 to 12.2.90. See the list of figures.

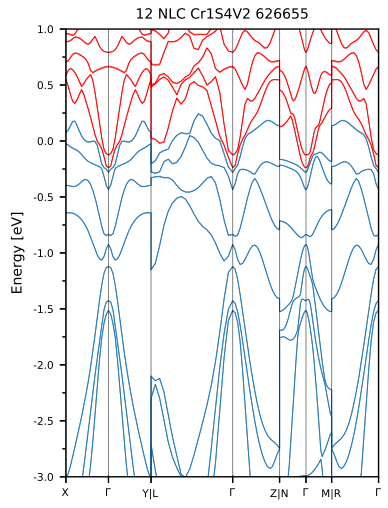

(kc) 12.2.91

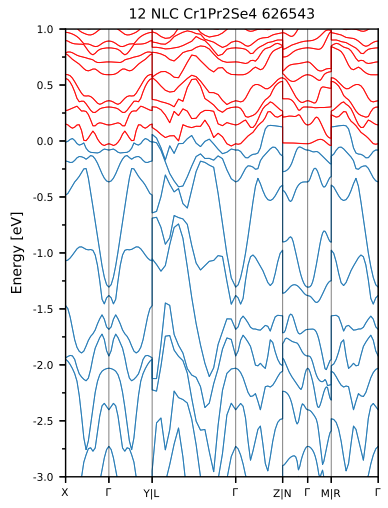

(kd) 12.2.92

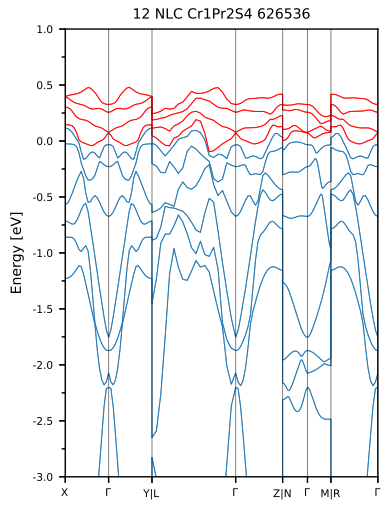

(ke) 12.2.93

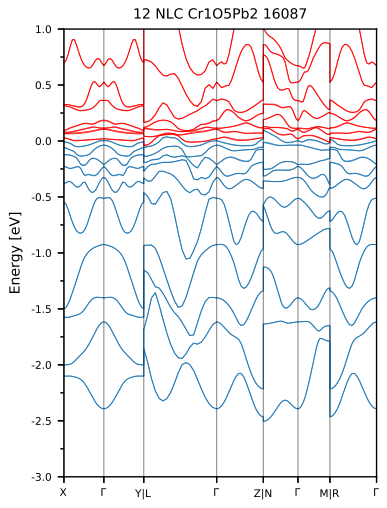

(kf) 12.2.94

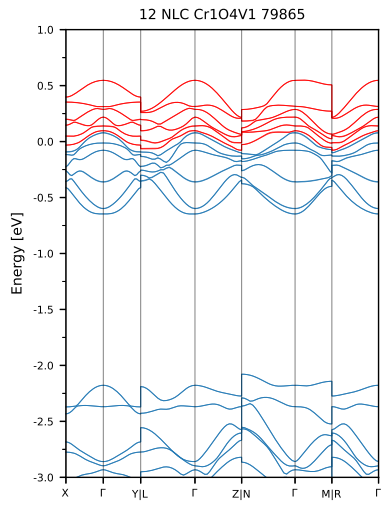

(kg) 12.2.95

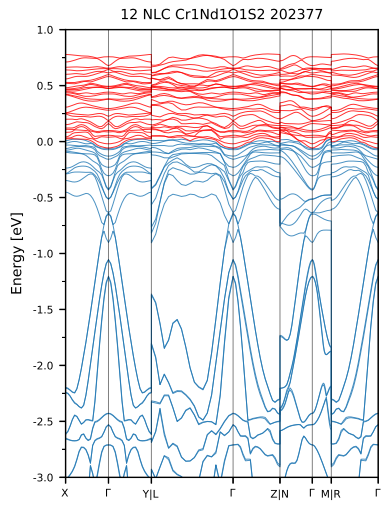

(kh) 12.2.96

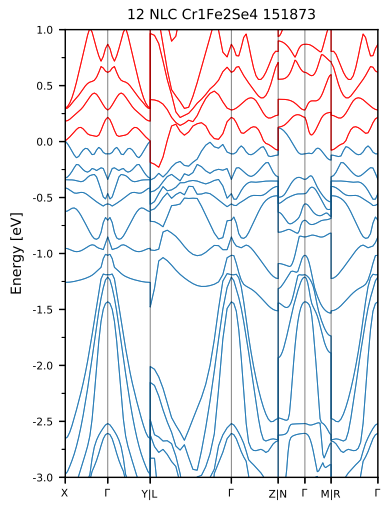

(ki) 12.2.97

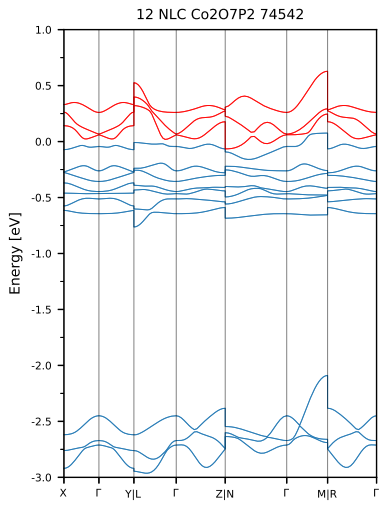

(kj) 12.2.98

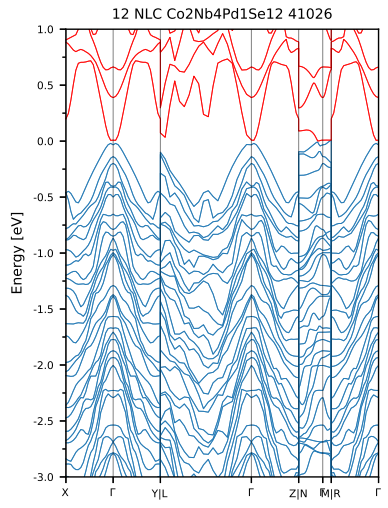

(kk) 12.2.99

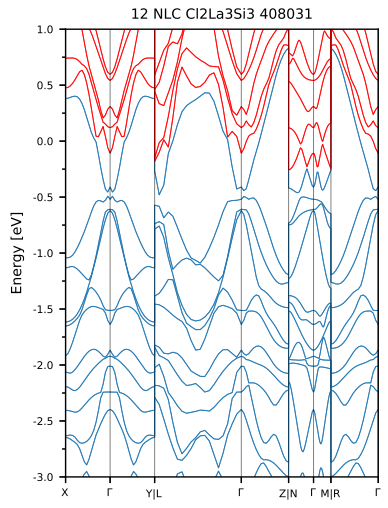

(kl) 12.2.100

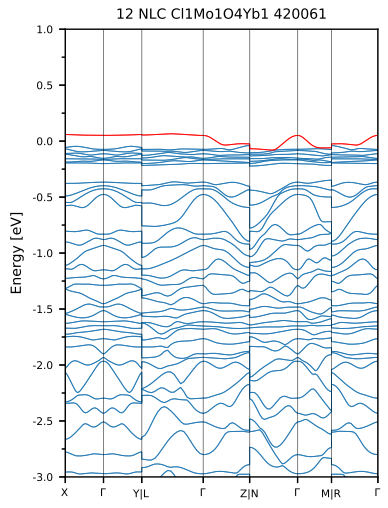

(km) 12.2.101

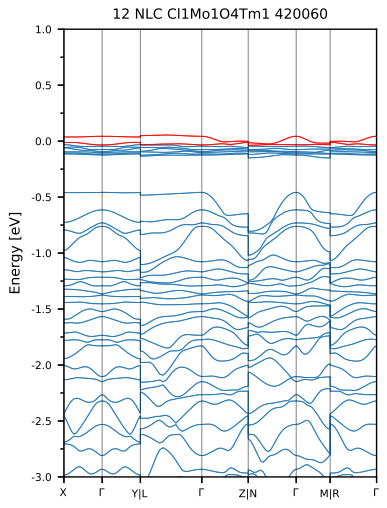

(kn) 12.2.102

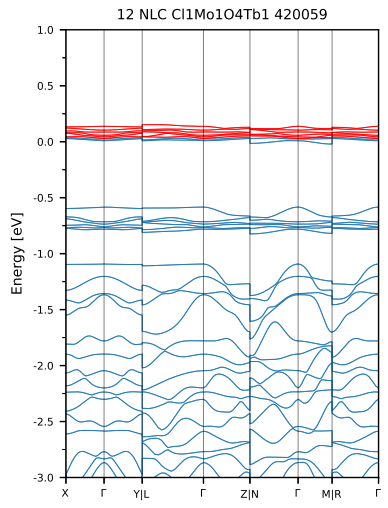

(ko) 12.2.103

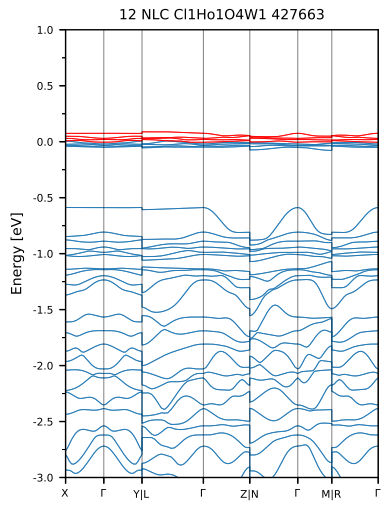

(kp) 12.2.104

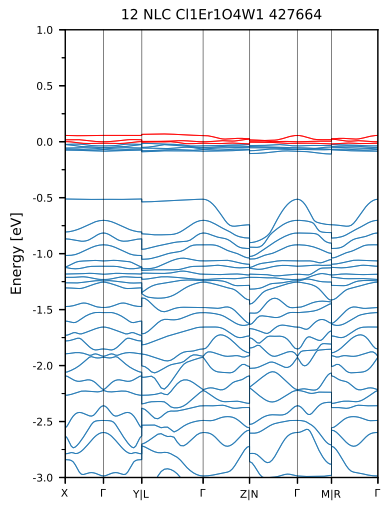

(kq) 12.2.105

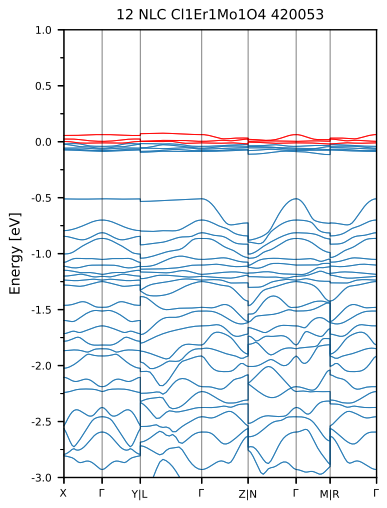

(kr) 12.2.106

Figure 42: Band structures for topological materials in SG 12, part 19/25, from 12.2.91 to 12.2.106. See the list of figures.

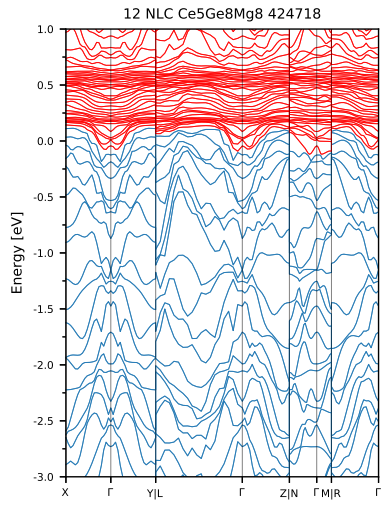

(ks) 12.2.107

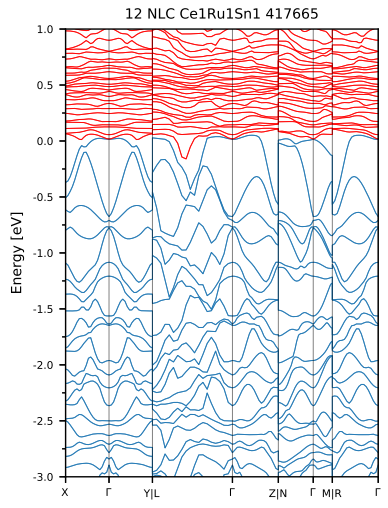

(kt) 12.2.108

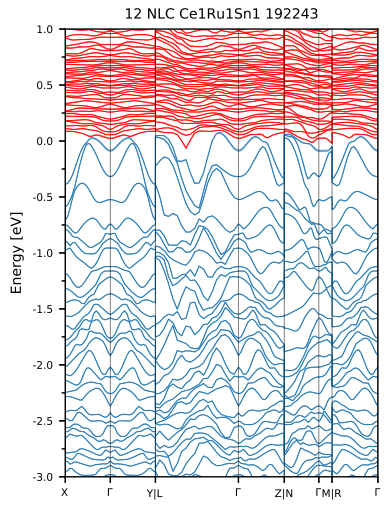

(ku) 12.2.109

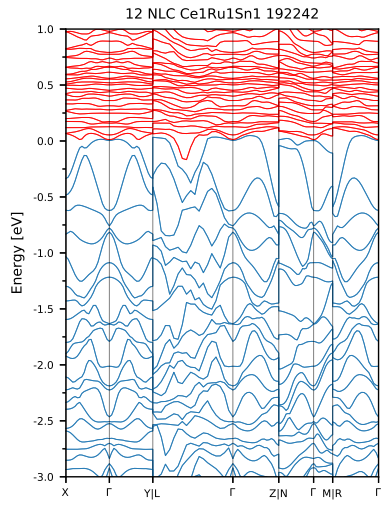

(kv) 12.2.110

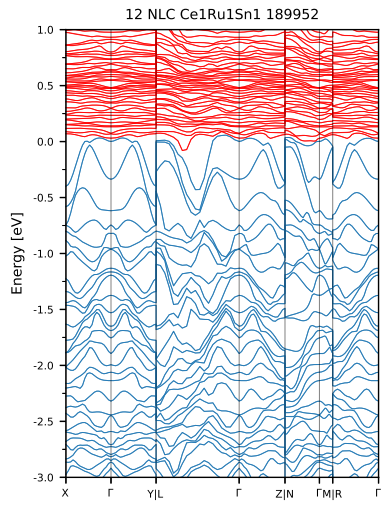

(kw) 12.2.111

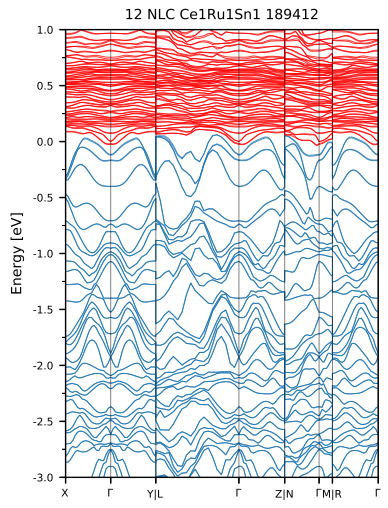

(kx) 12.2.112

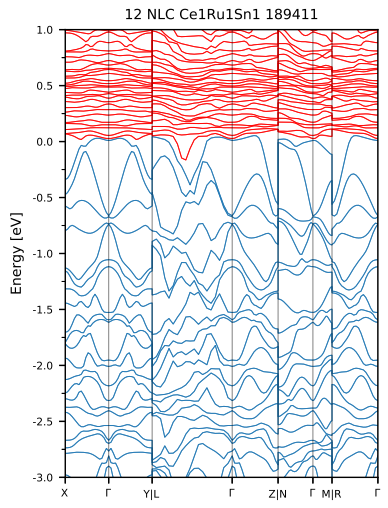

(ky) 12.2.113

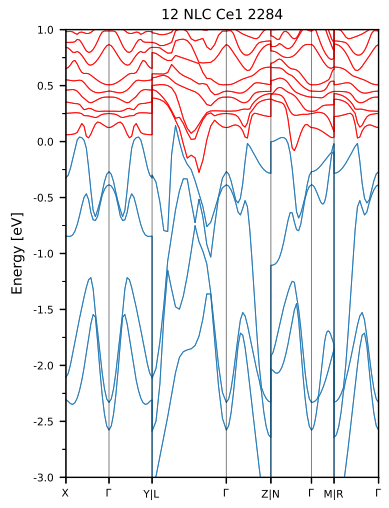

(kz) 12.2.114

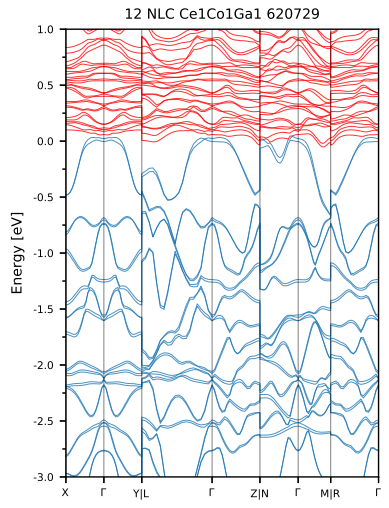

(la) 12.2.115

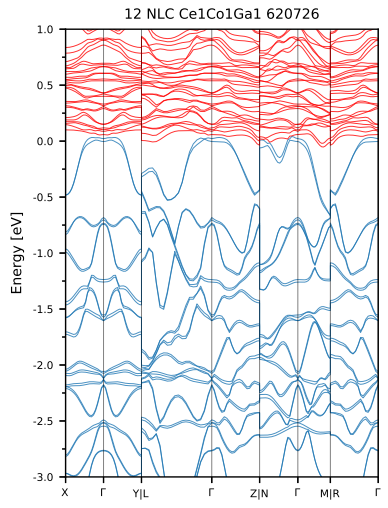

(lb) 12.2.116

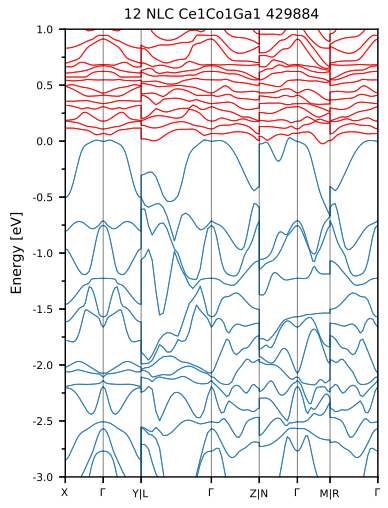

(lc) 12.2.117

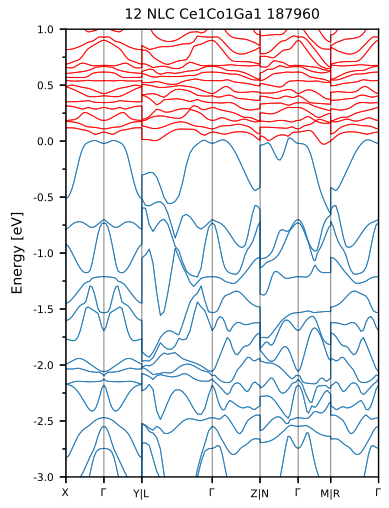

(ld) 12.2.118

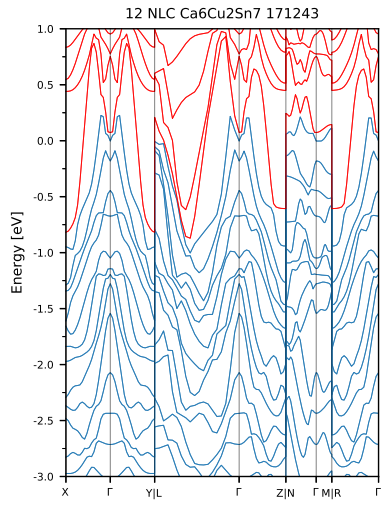

(le) 12.2.119

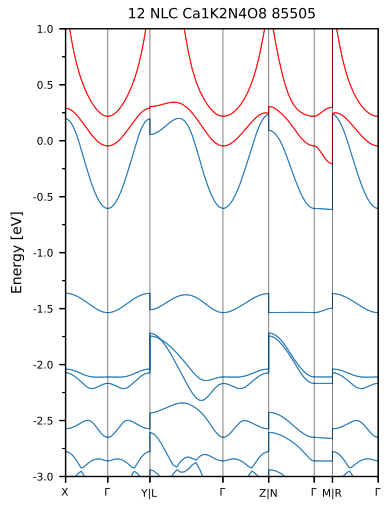

(lf) 12.2.120

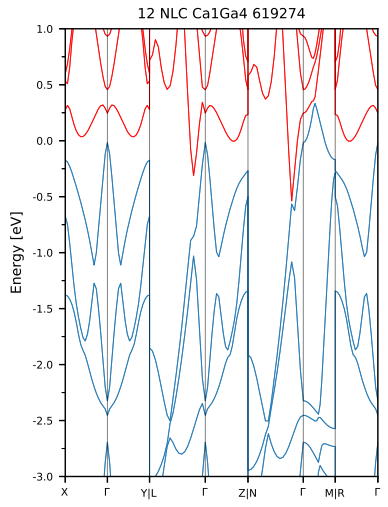

(lg) 12.2.121

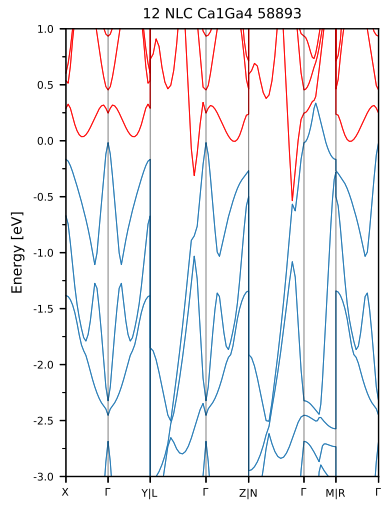

(lh) 12.2.122

Figure 43: Band structures for topological materials in SG 12, part 20/25, from 12.2.107 to 12.2.122. See the list of figures.

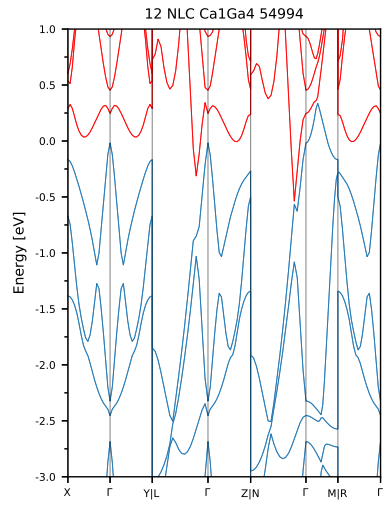

(li) 12.2.123

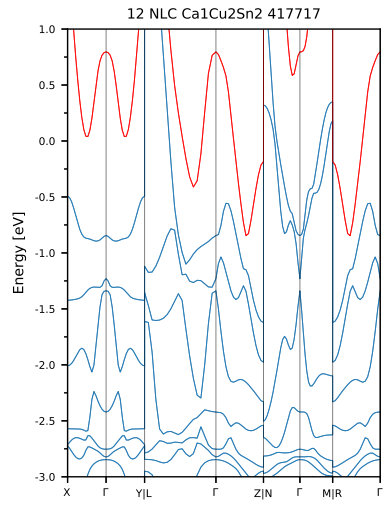

(lj) 12.2.124

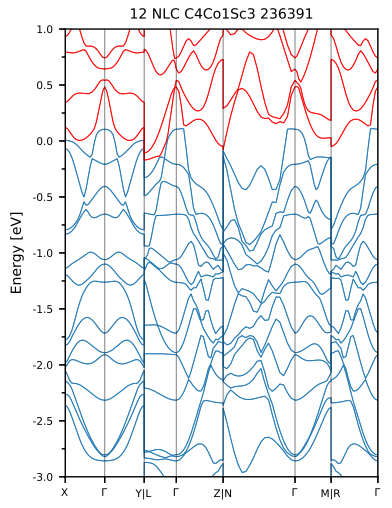

(lk) 12.2.125

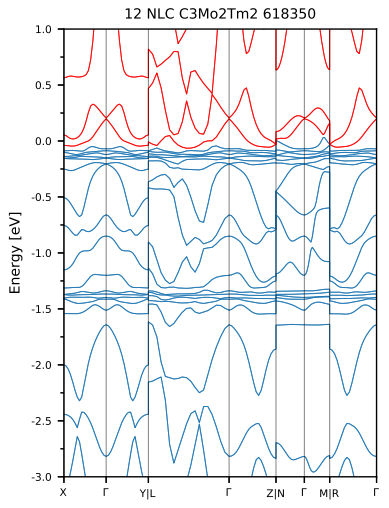

(ll) 12.2.126

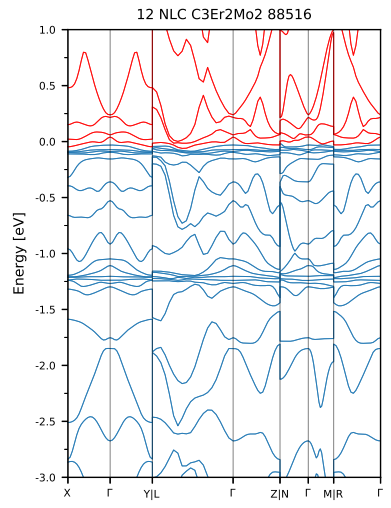

(lm) 12.2.127

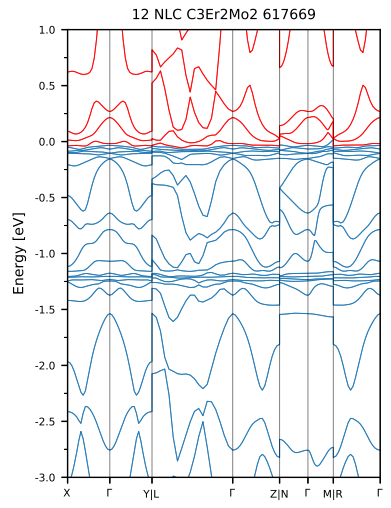

(ln) 12.2.128

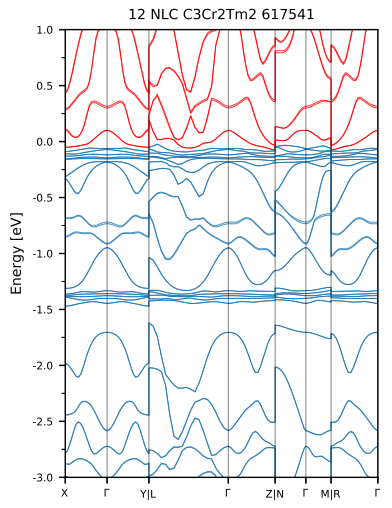

(lo) 12.2.129

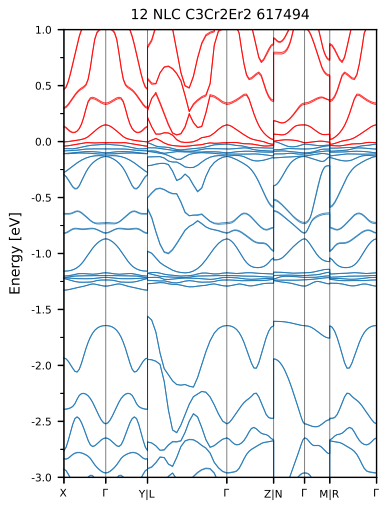

(lp) 12.2.130

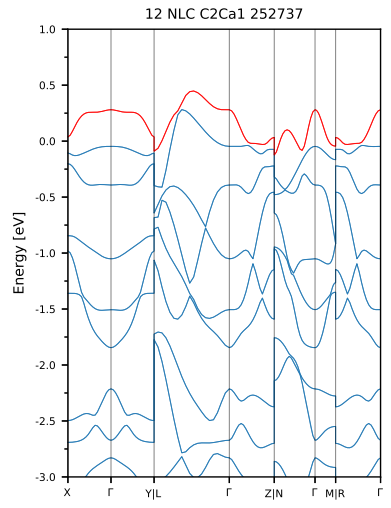

(lq) 12.2.131

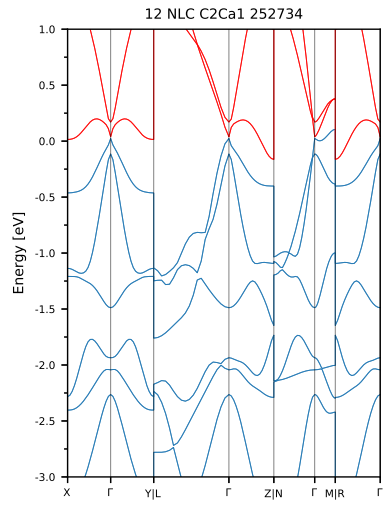

(lr) 12.2.132

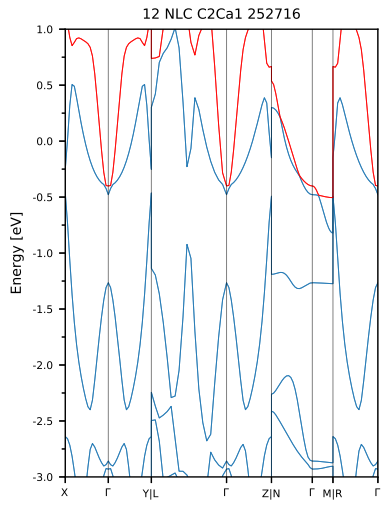

(ls) 12.2.133

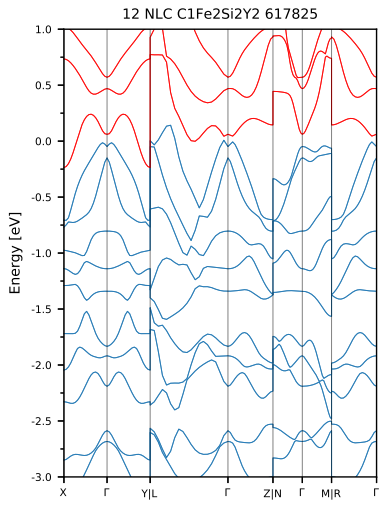

(lt) 12.2.134

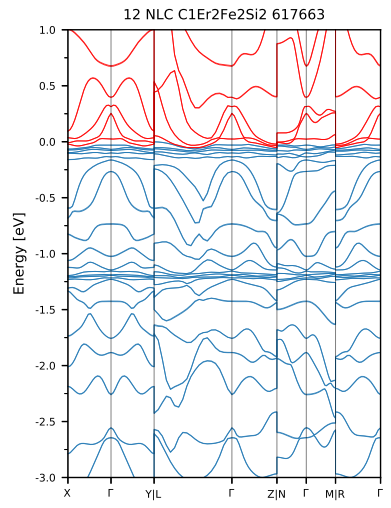

(lu) 12.2.135

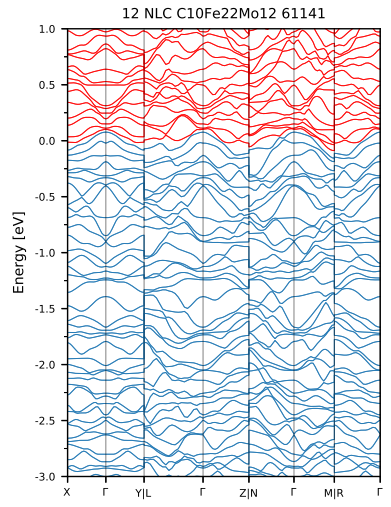

(lv) 12.2.136

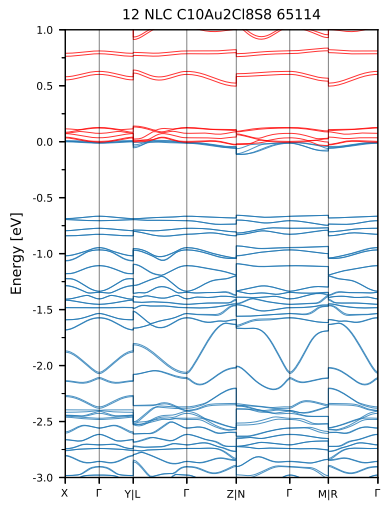

(lw) 12.2.137

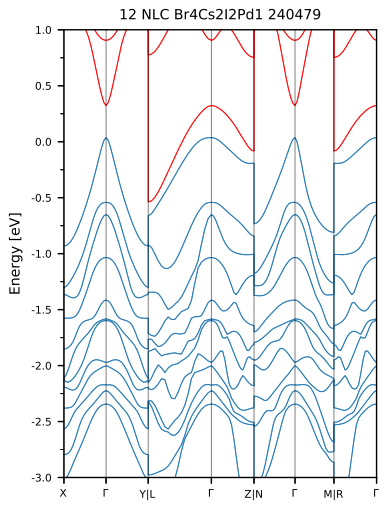

(lx) 12.2.138

Figure 44: Band structures for topological materials in SG 12, part 21/25, from 12.2.123 to 12.2.138. See the list of figures.

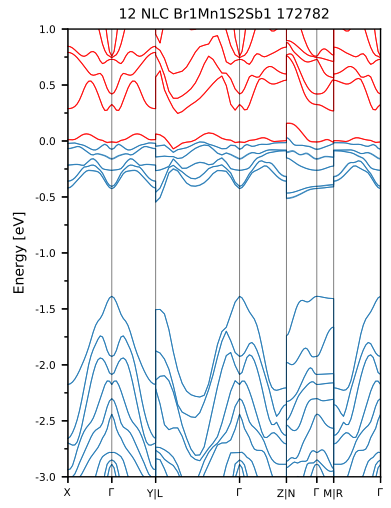

(ly) 12.2.139

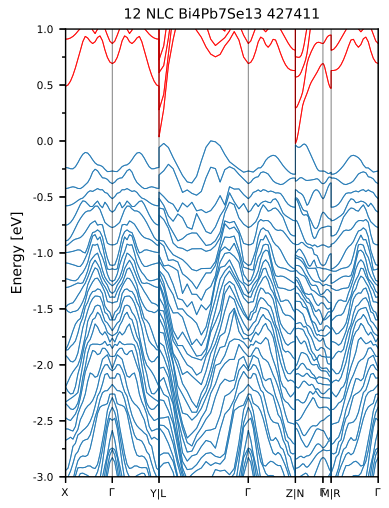

(lz) 12.2.140

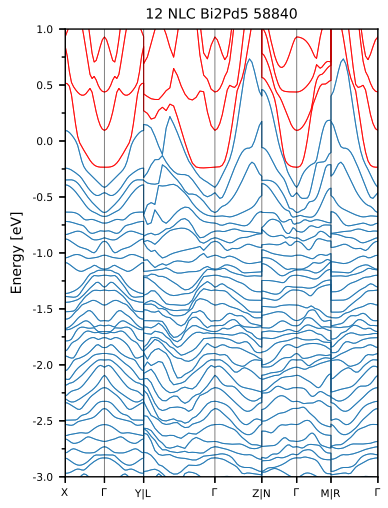

(ma) 12.2.141

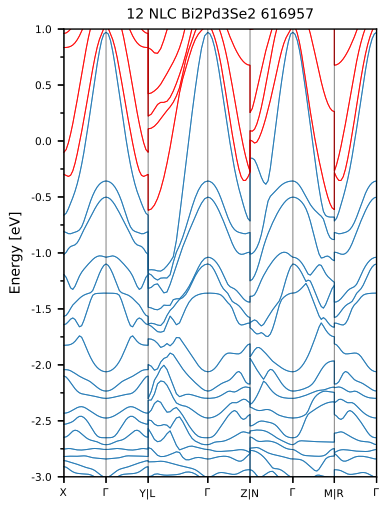

(mb) 12.2.142

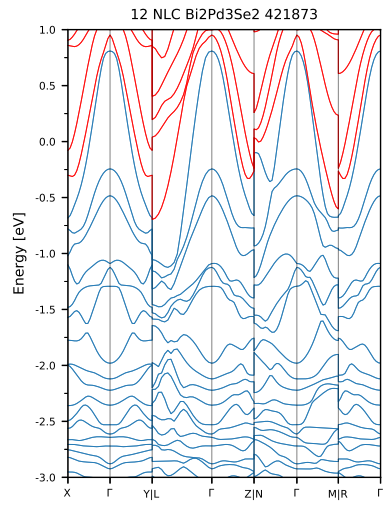

(mc) 12.2.143

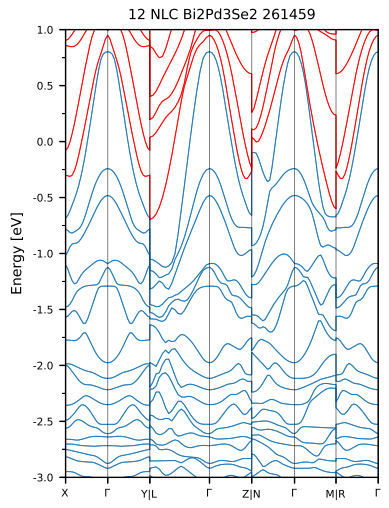

(md) 12.2.144

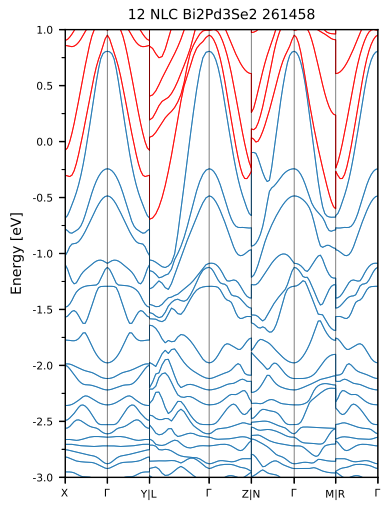

(me) 12.2.145

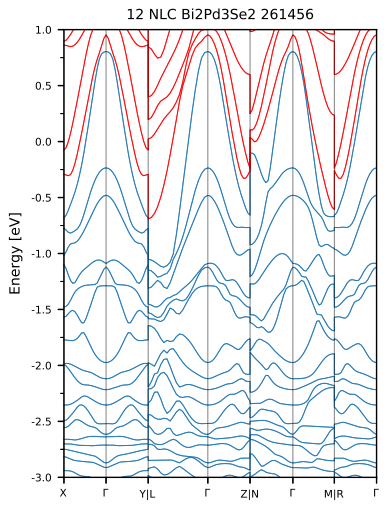

(mf) 12.2.146

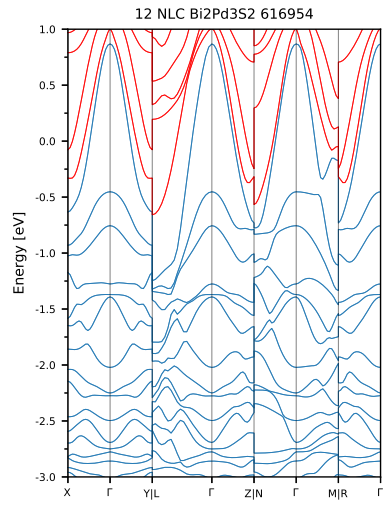

(mg) 12.2.147

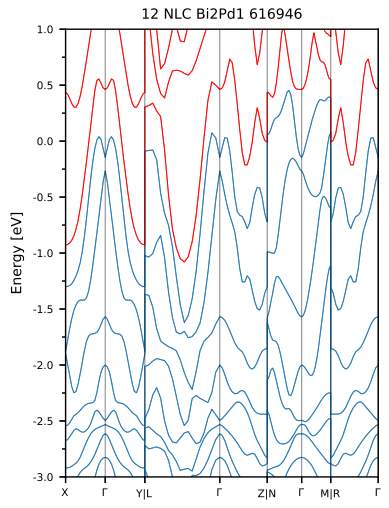

(mh) 12.2.148

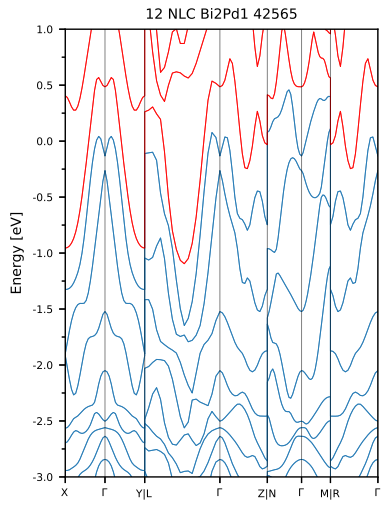

(mi) 12.2.149

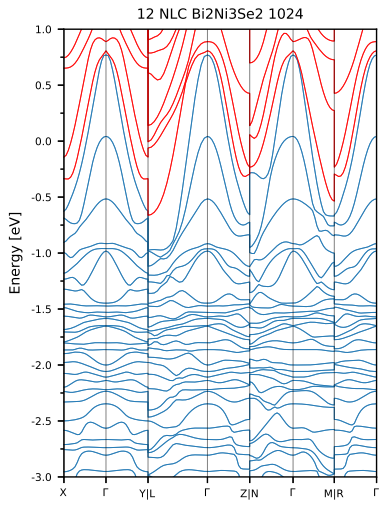

(mj) 12.2.150

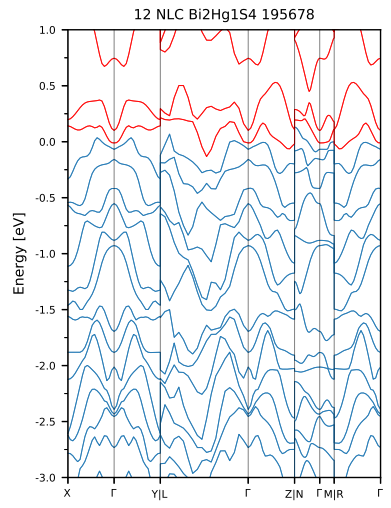

(mk) 12.2.151

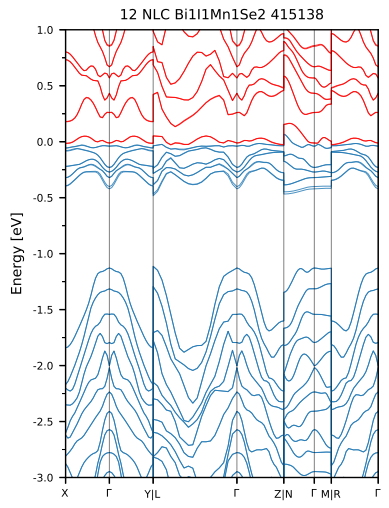

(ml) 12.2.152

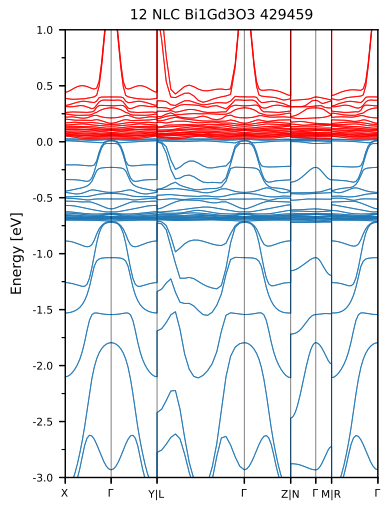

(mm) 12.2.153

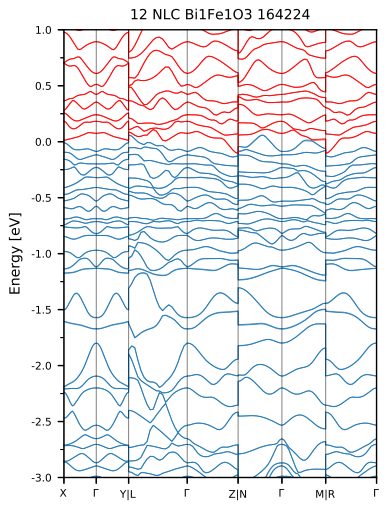

(mn) 12.2.154

Figure 45: Band structures for topological materials in SG 12, part 22/25, from 12.2.139 to 12.2.154. See the list of figures.

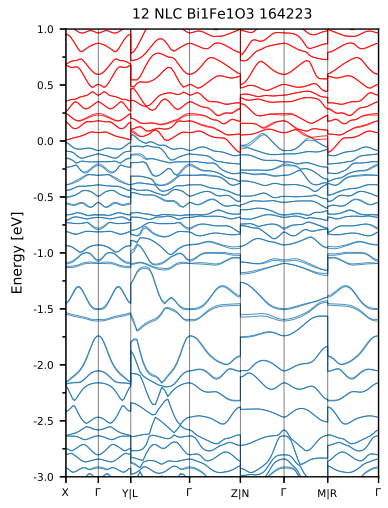

(mo) 12.2.155

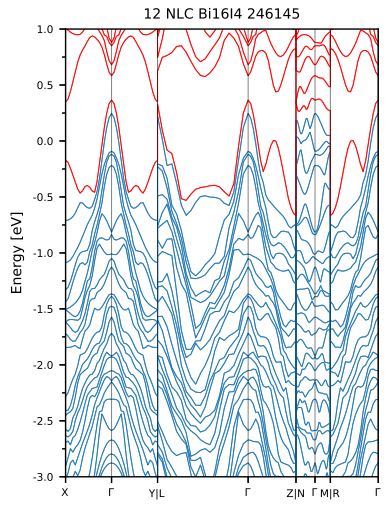

(mp) 12.2.156

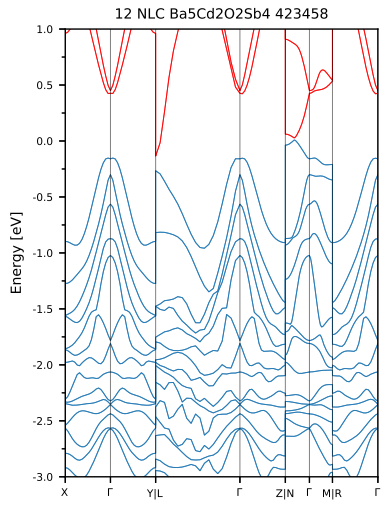

(mq) 12.2.157

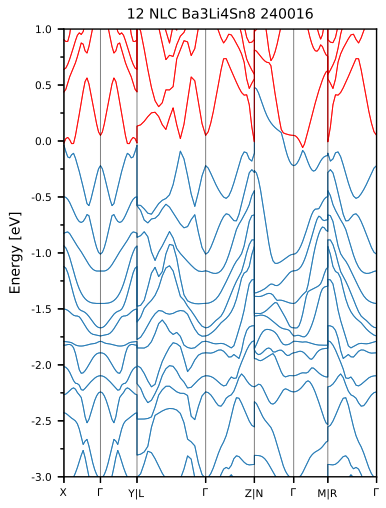

(mr) 12.2.158

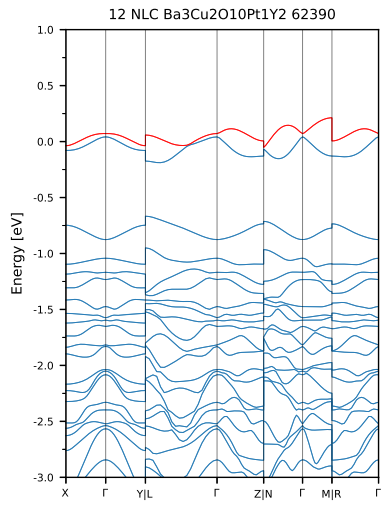

(ms) 12.2.159

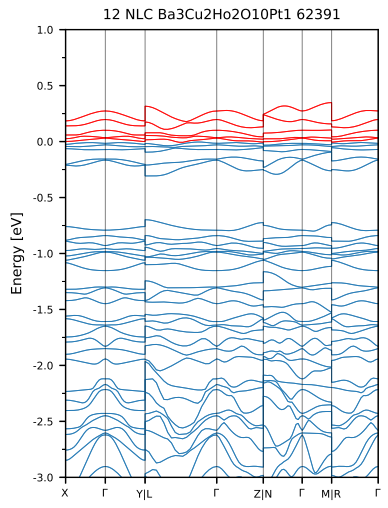

(mt) 12.2.160

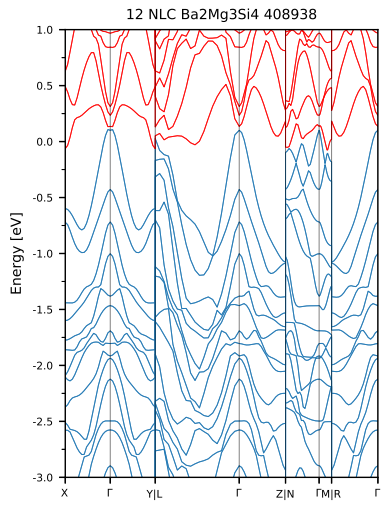

(mu) 12.2.161

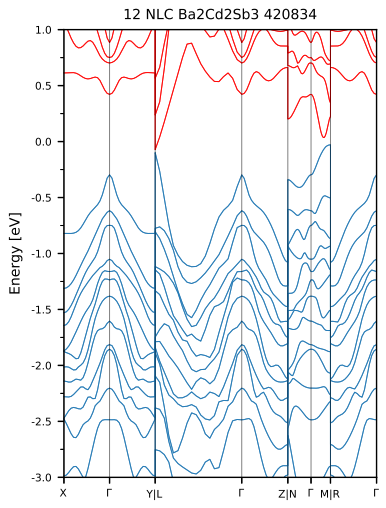

(mv) 12.2.162

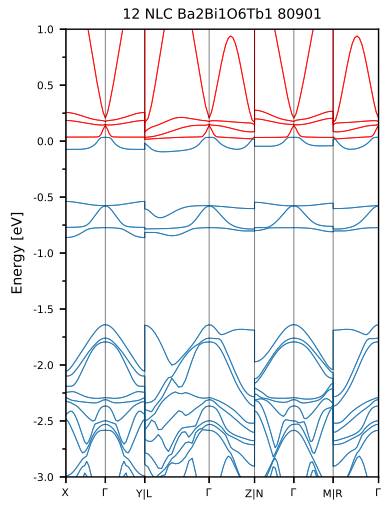

(mw) 12.2.163

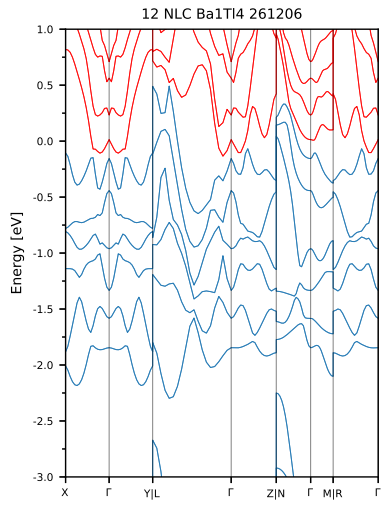

(mx) 12.2.164

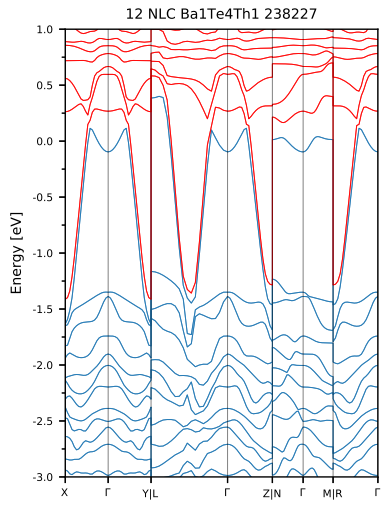

(my) 12.2.165

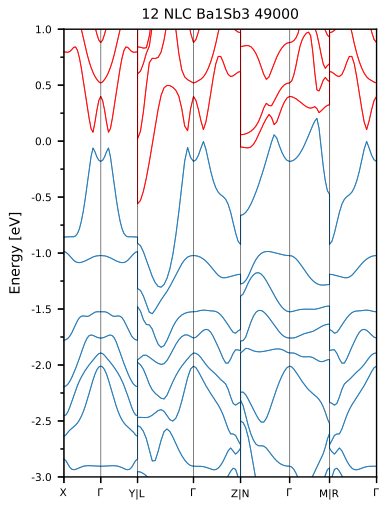

(mz) 12.2.166

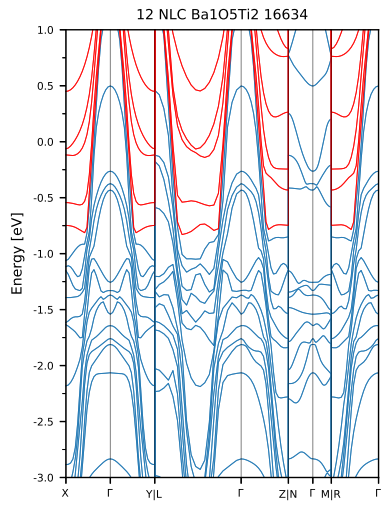

(na) 12.2.167

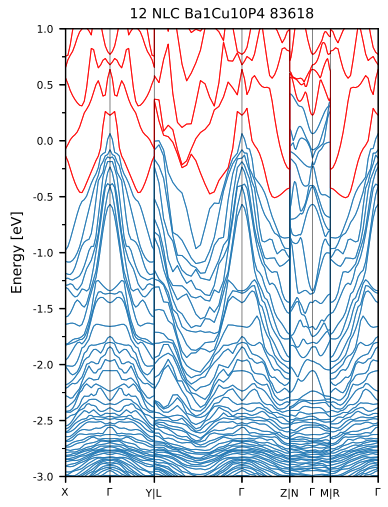

(nb) 12.2.168

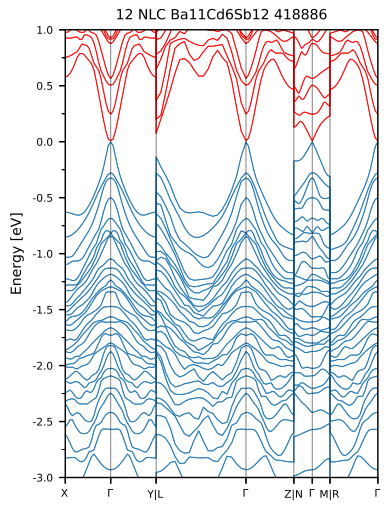

(nc) 12.2.169

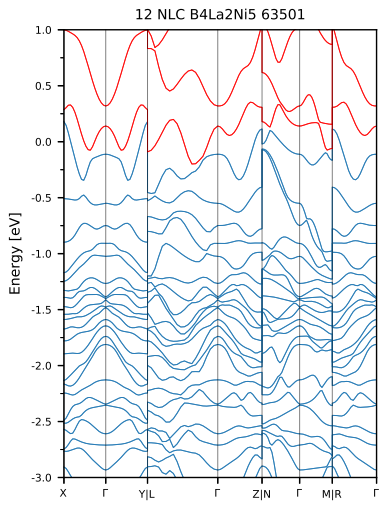

(nd) 12.2.170

Figure 46: Band structures for topological materials in SG 12, part 23/25, from 12.2.155 to 12.2.170. See the list of figures.

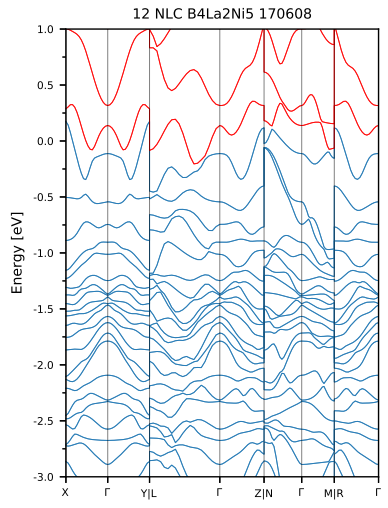

(ne) 12.2.171

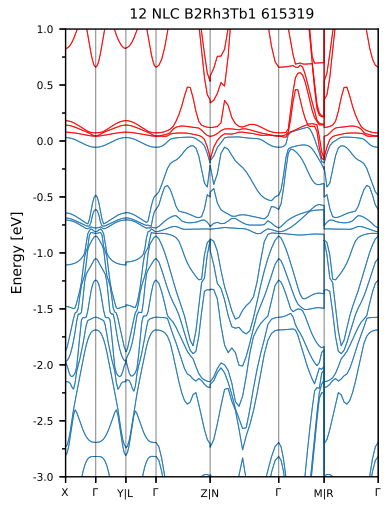

(nf) 12.2.172

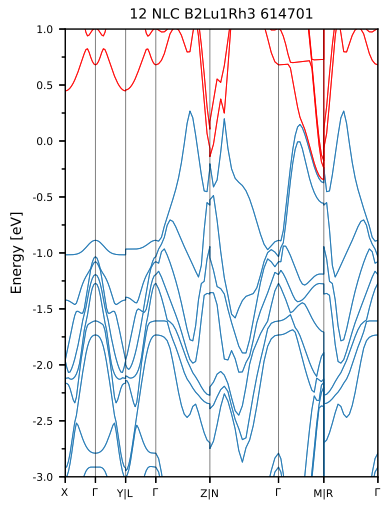

(ng) 12.2.173

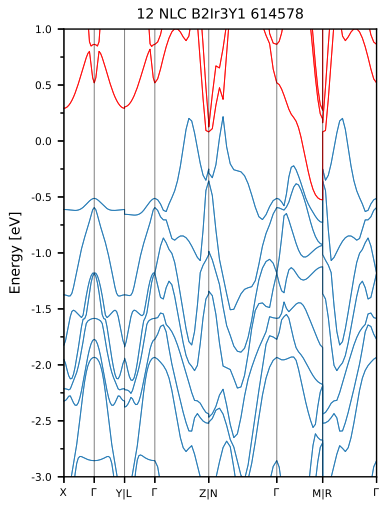

(nh) 12.2.174

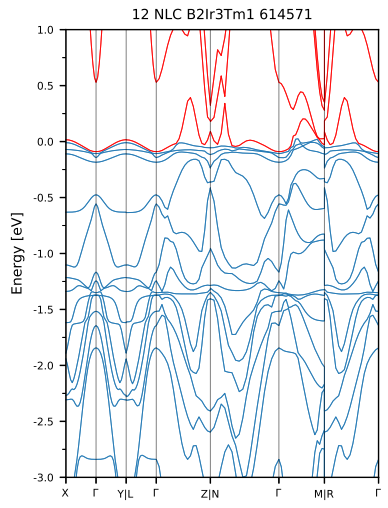

(ni) 12.2.175

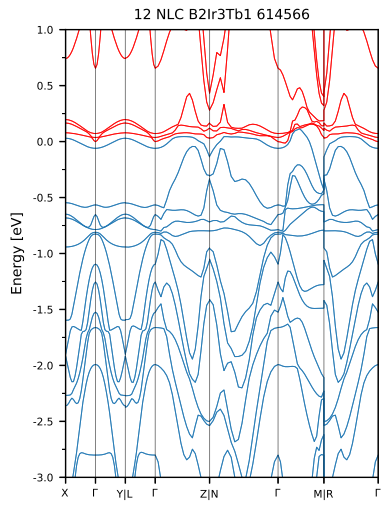

(nj) 12.2.176

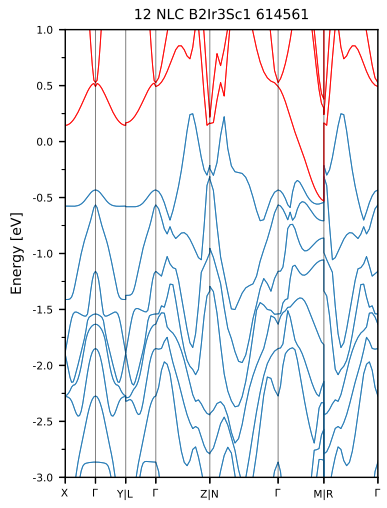

(nk) 12.2.177

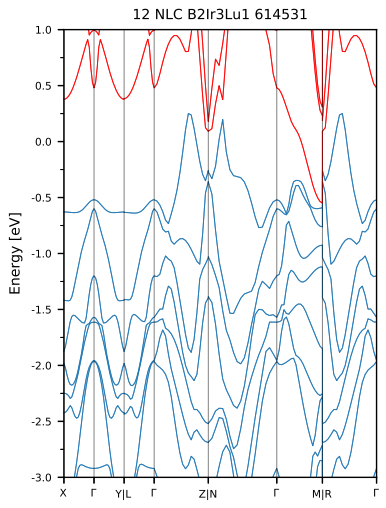

(nl) 12.2.178

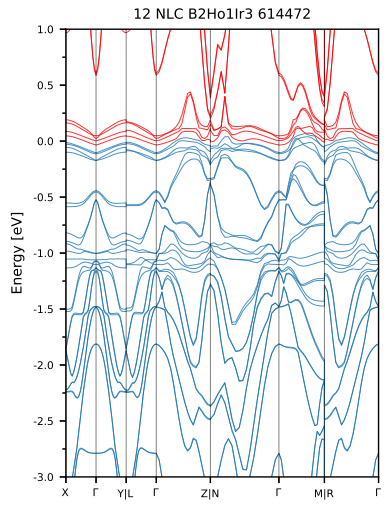

(nm) 12.2.179

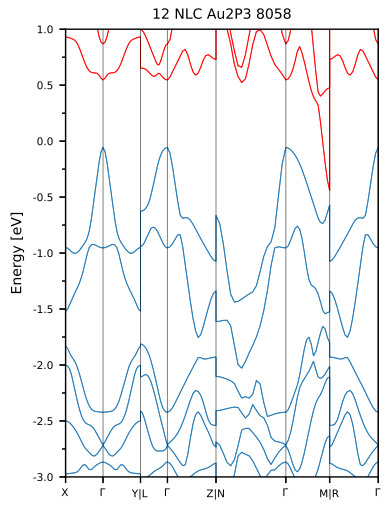

(nn) 12.2.180

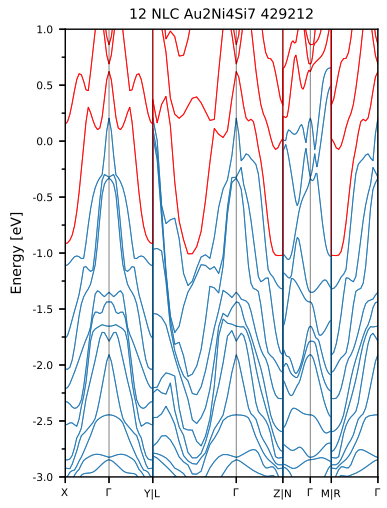

(no) 12.2.181

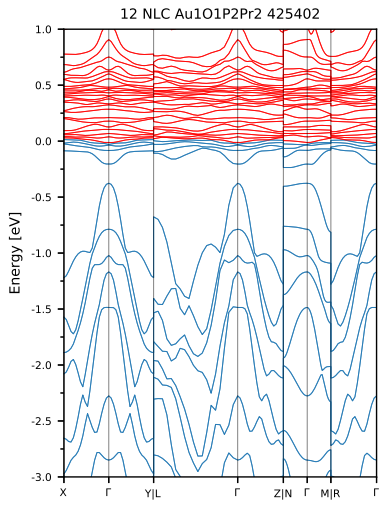

(np) 12.2.182

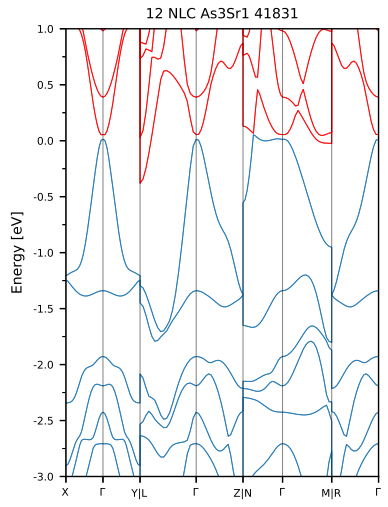

(nq) 12.2.183

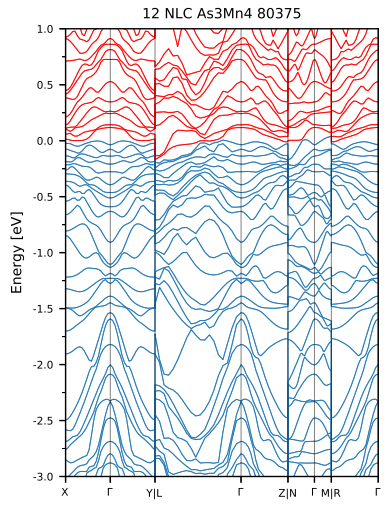

(nr) 12.2.184

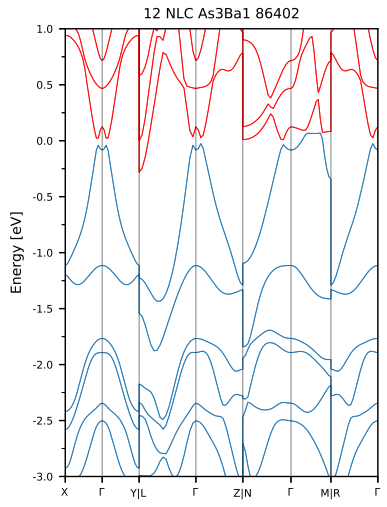

(ns) 12.2.185

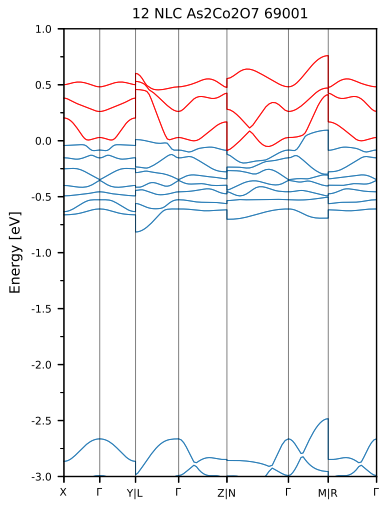

(nt) 12.2.186

Figure 47: Band structures for topological materials in SG 12, part 24/25, from 12.2.171 to 12.2.186. See the list of figures.

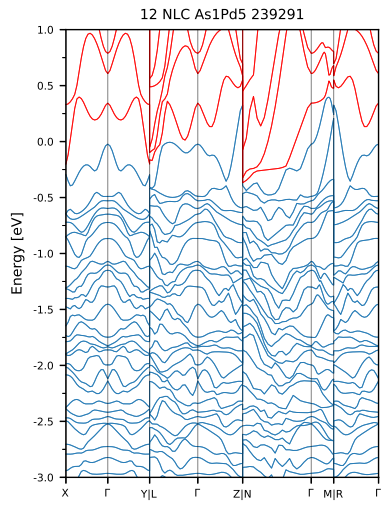

(nu) 12.2.187

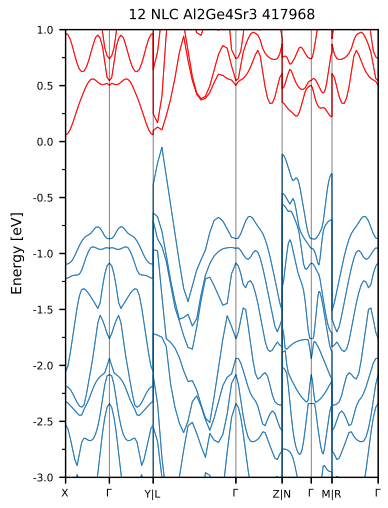

(nv) 12.2.188

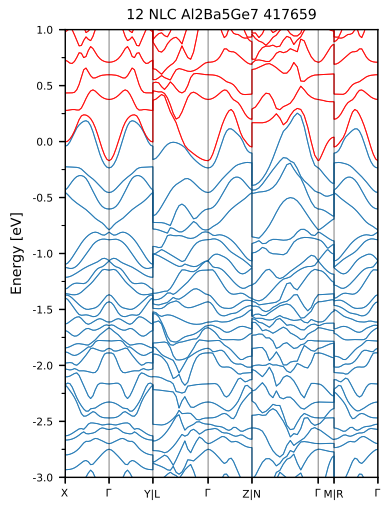

(nw) 12.2.189

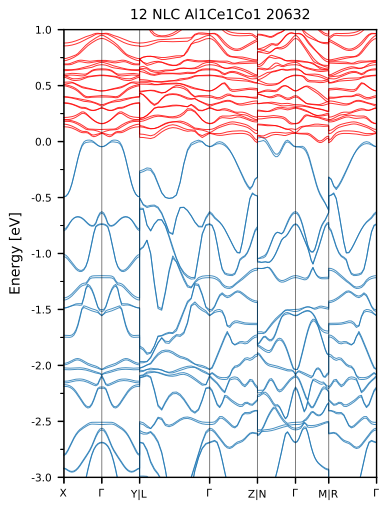

(nx) 12.2.190

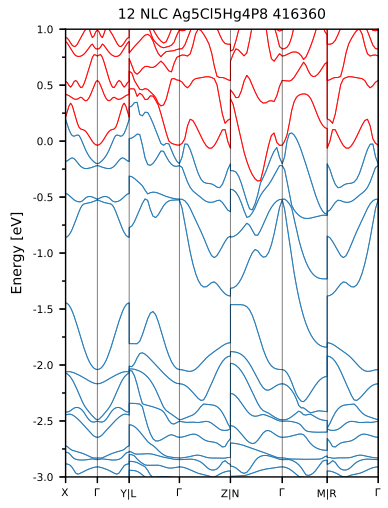

(ny) 12.2.191

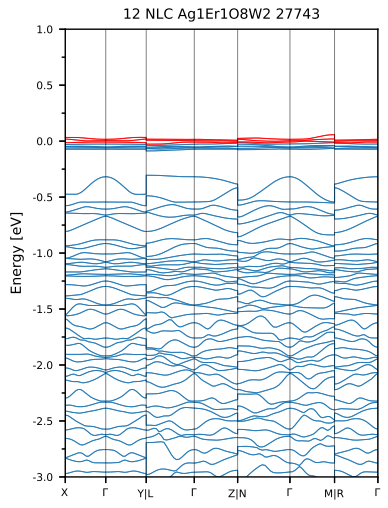

(nz) 12.2.192

Figure 48: Band structures for topological materials in SG 12, part 25/25, from 12.2.187 to 12.2.192. See the list of figures.

### 0.0.7 SG 13

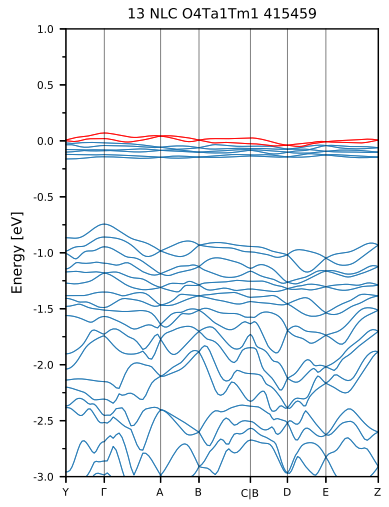

(a) 13.2.1

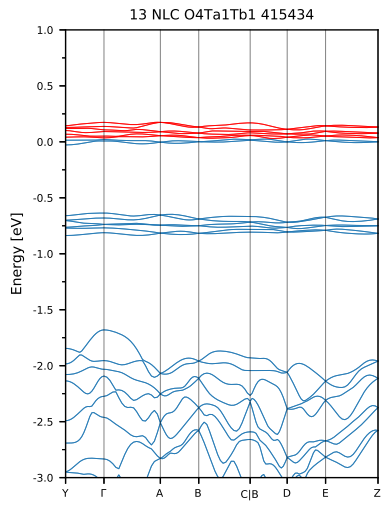

(b) 13.2.2

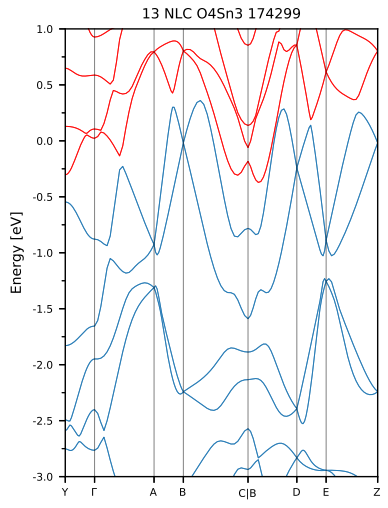

(c) 13.2.3

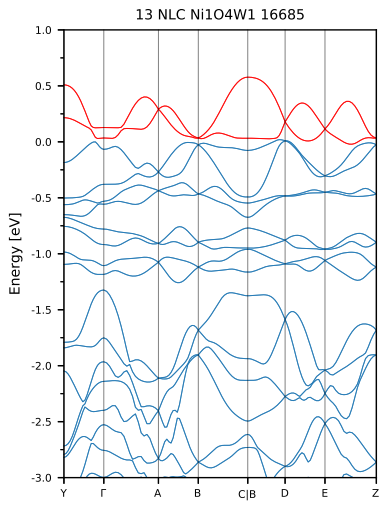

(d) 13.2.4

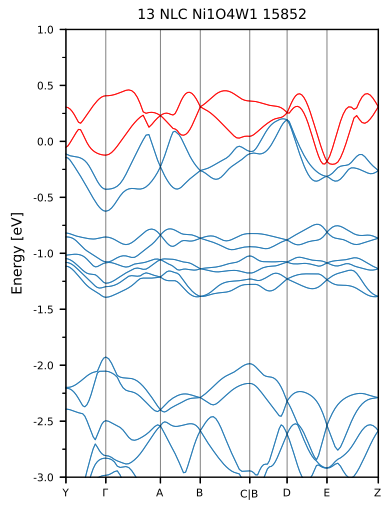

(e) 13.2.5

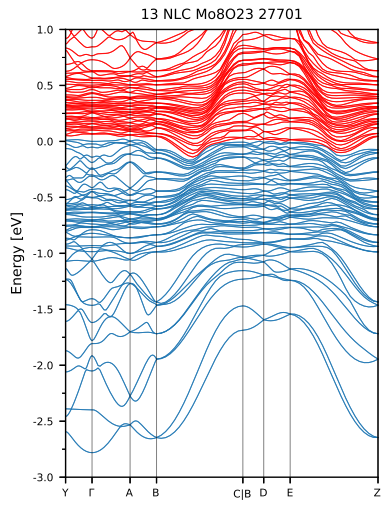

(f) 13.2.6

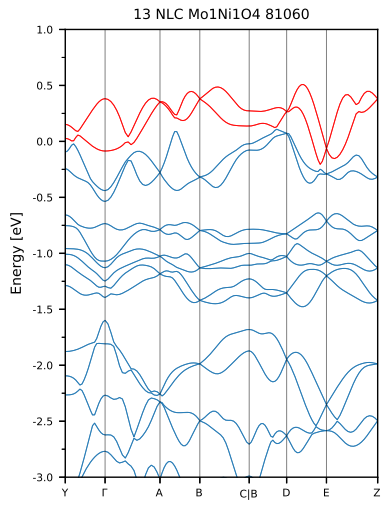

(g) 13.2.7

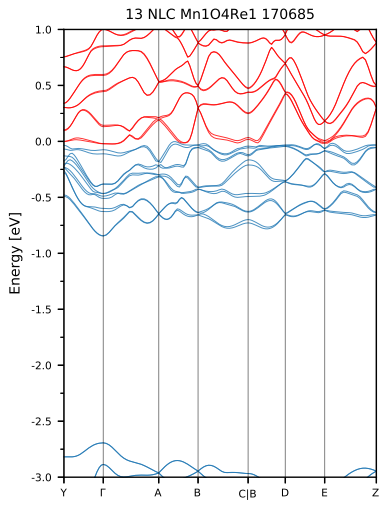

(h) 13.2.8

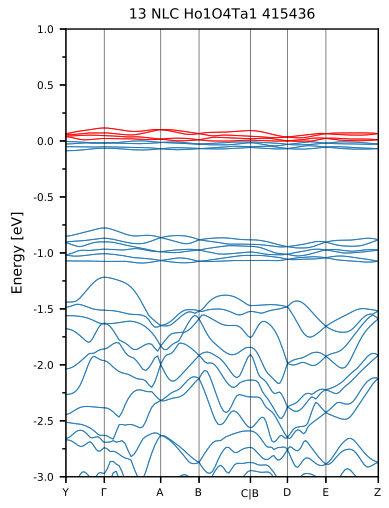

(i) 13.2.9

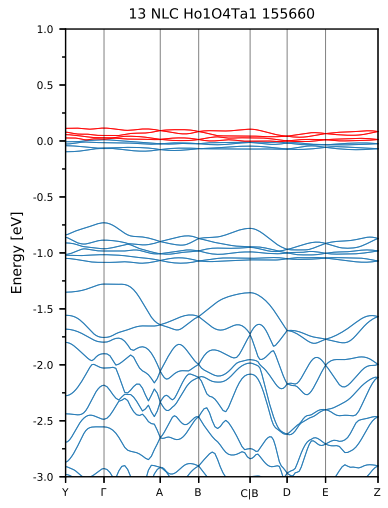

(j) 13.2.10

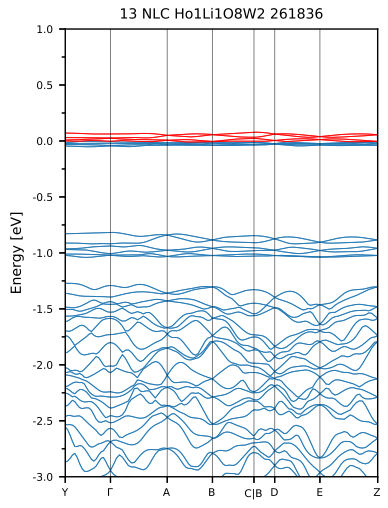

(k) 13.2.11

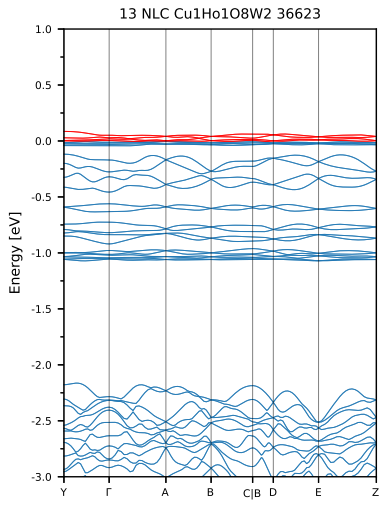

(l) 13.2.12

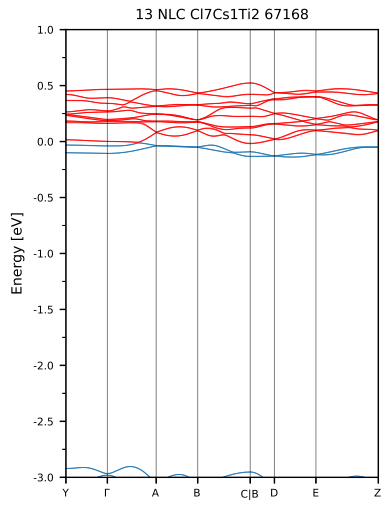

(m) 13.2.13

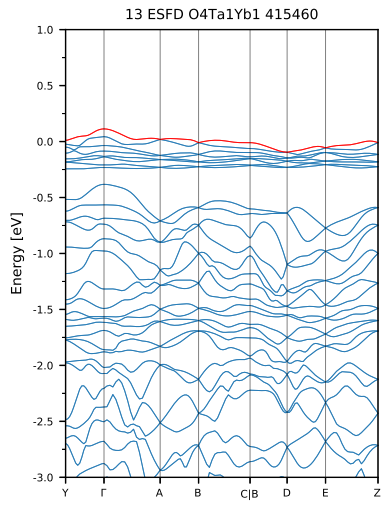

(n) 13.4.1

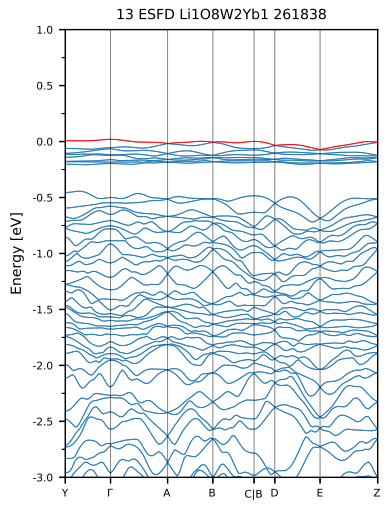

(o) 13.4.2

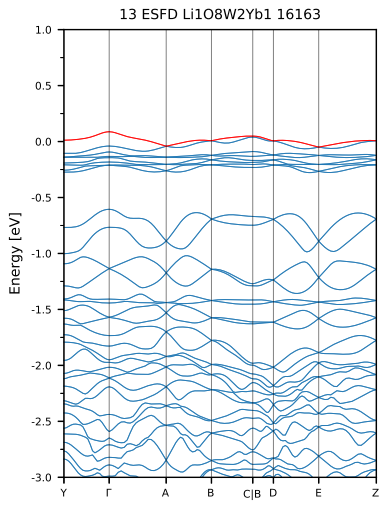

(p) 13.4.3

Figure 49: Band structures for topological materials in SG 13, part 1/3, from 13.2.1 to 13.4.3. See the list of figures.

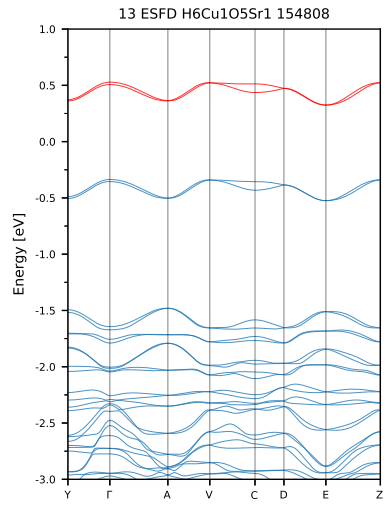

(q) 13.4.4

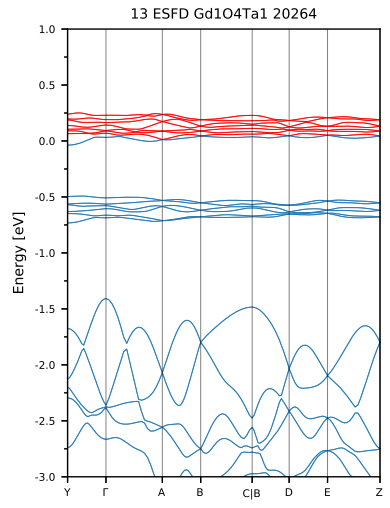

(r) 13.4.5

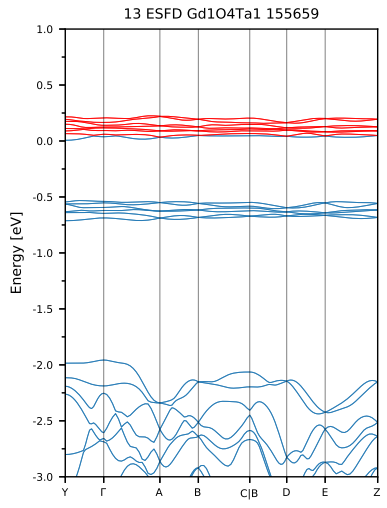

(s) 13.4.6

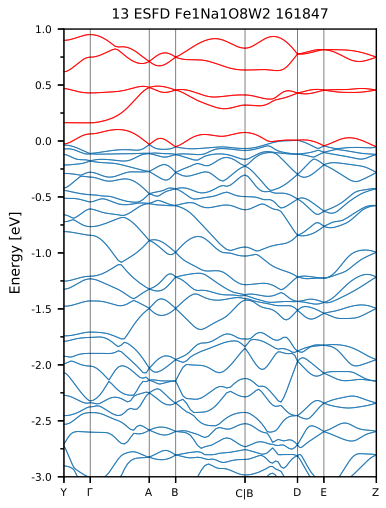

(t) 13.4.7

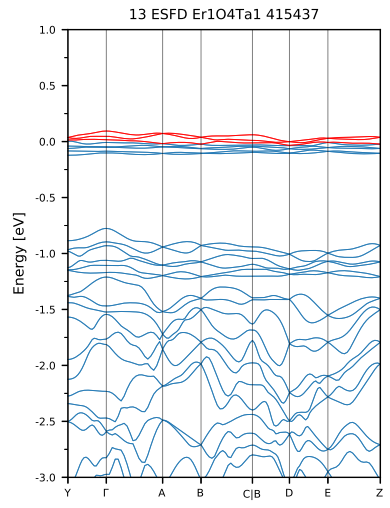

(u) 13.4.8

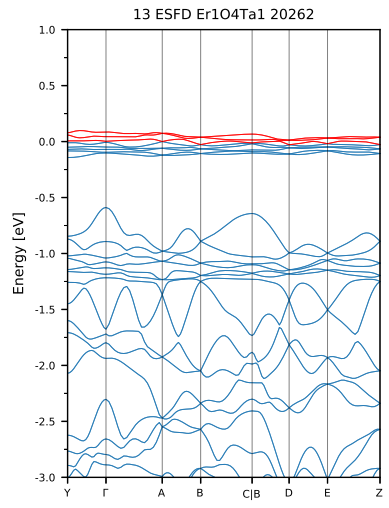

(v) 13.4.9

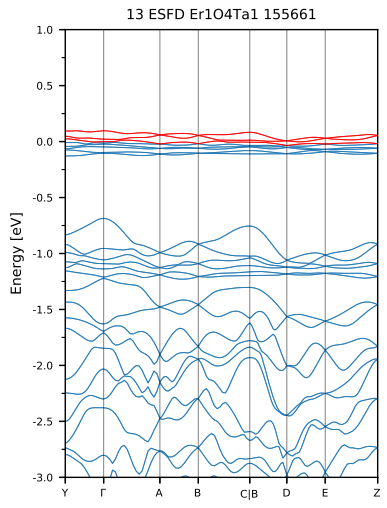

(w) 13.4.10

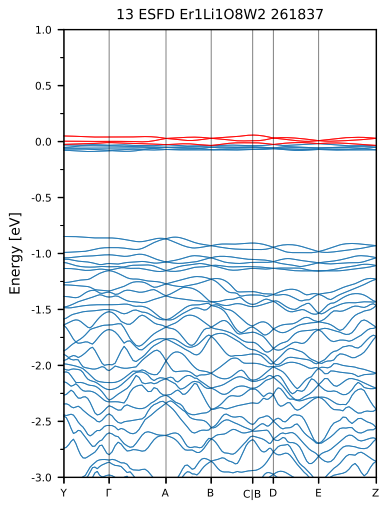

(x) 13.4.11

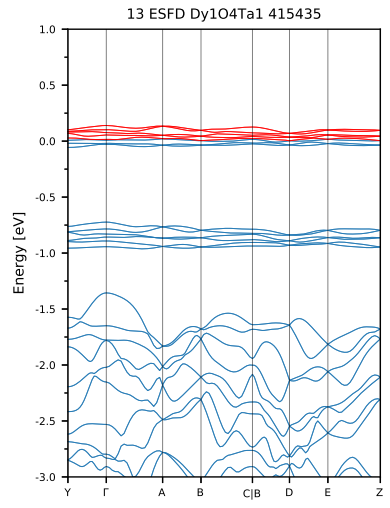

(y) 13.4.12

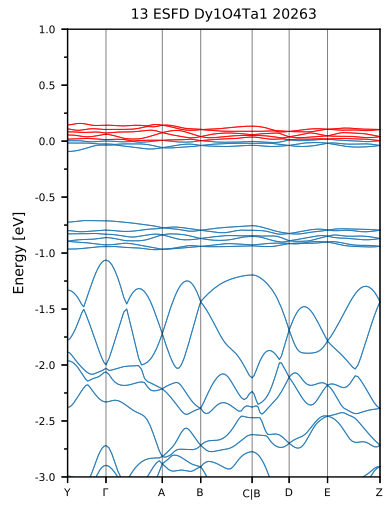

(z) 13.4.13

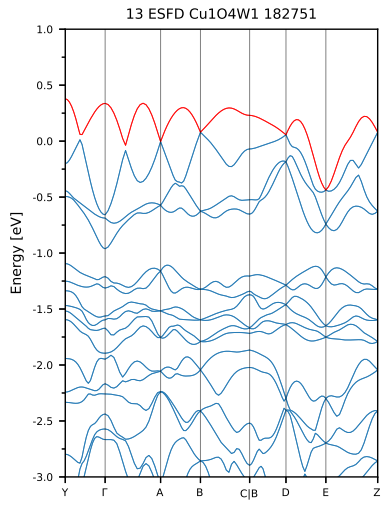

(aa) 13.4.14

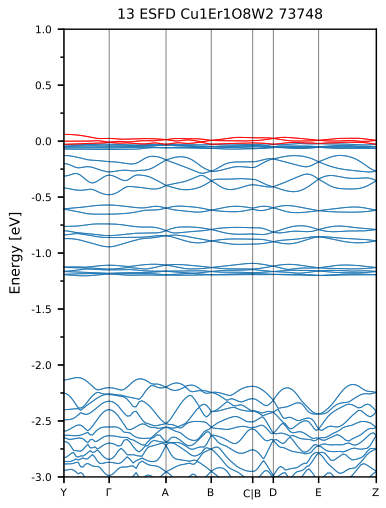

(ab) 13.4.15

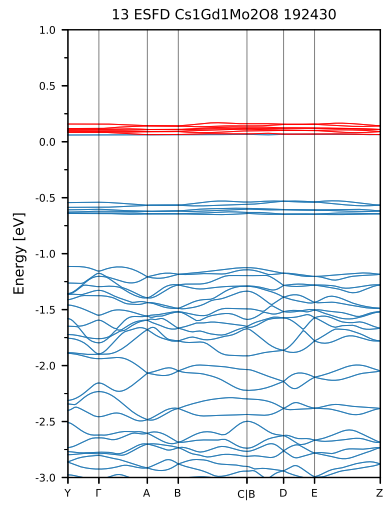

(ac) 13.4.16

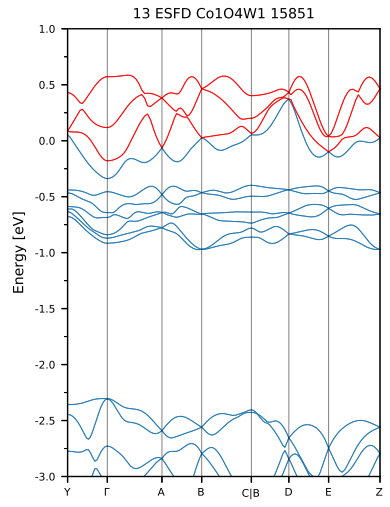

(ad) 13.4.17

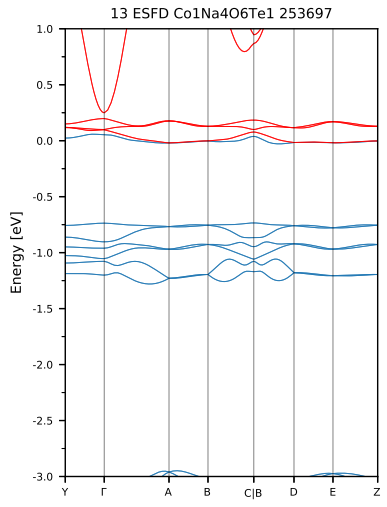

(ae) 13.4.18

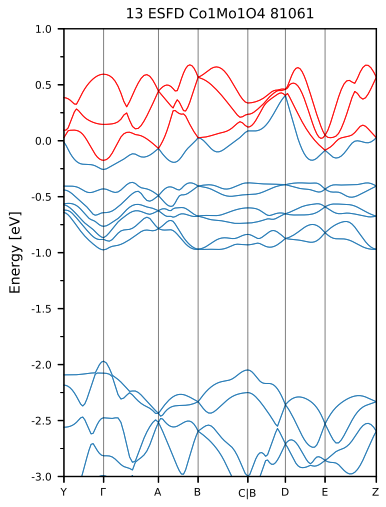

(af) 13.4.19

Figure 50: Band structures for topological materials in SG 13, part 2/3, from 13.4.4 to 13.4.19. See the list of figures.

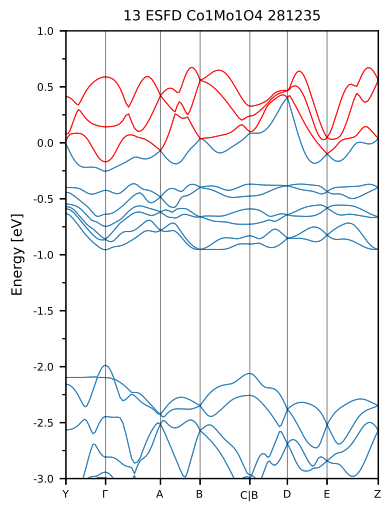

(ag) 13.4.20

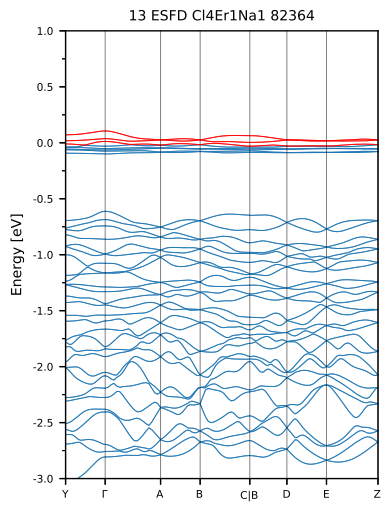

(ah) 13.4.21

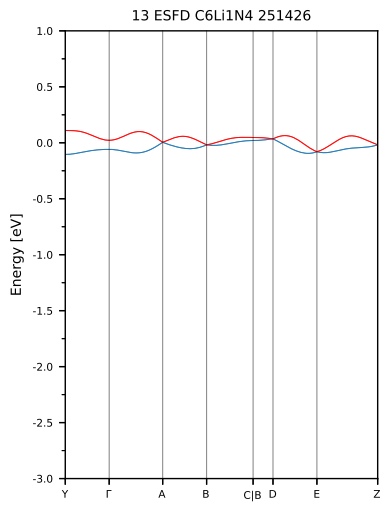

(ai) 13.4.22

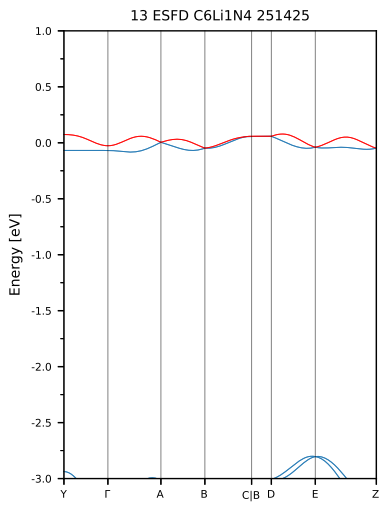

(aj) 13.4.23

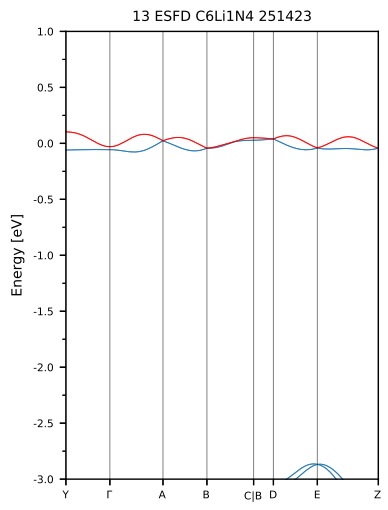

(ak) 13.4.24

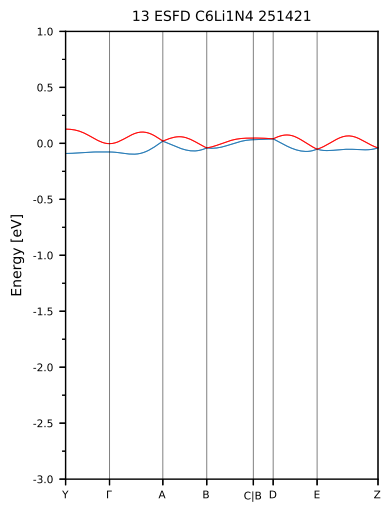

(al) 13.4.25

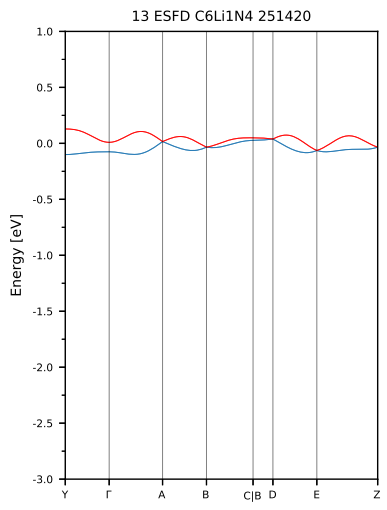

(am) 13.4.26

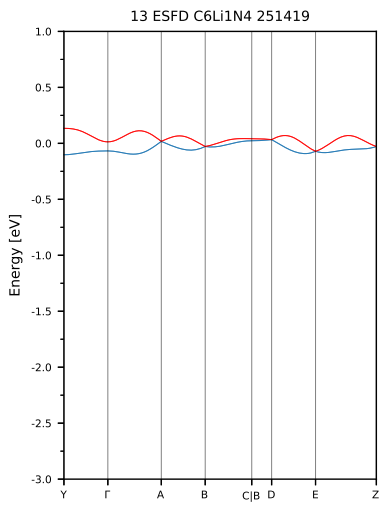

(an) 13.4.27

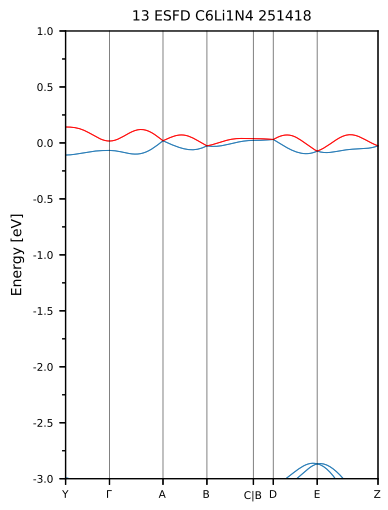

(ao) 13.4.28

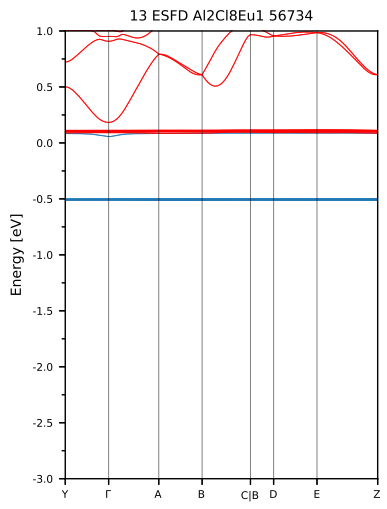

(ap) 13.4.29

Figure 51: Band structures for topological materials in SG 13, part 3/3, from 13.4.20 to 13.4.29. See the list of figures.

### 0.0.8 SG 14

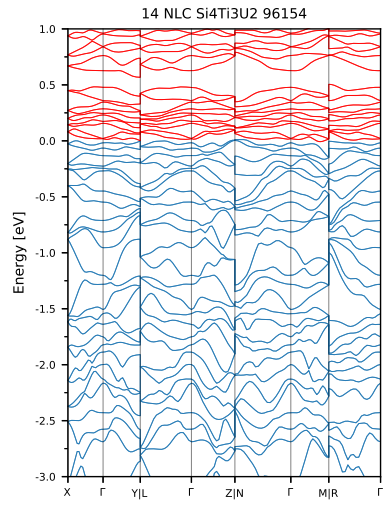

(a) 14.2.1

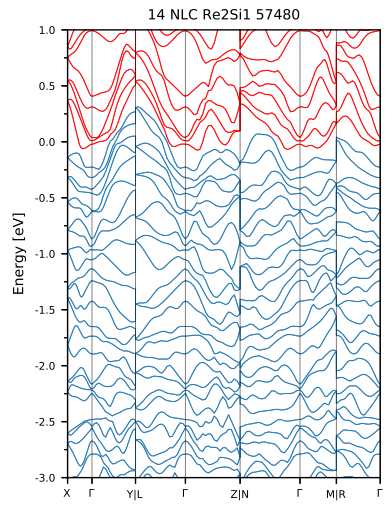

(b) 14.2.2

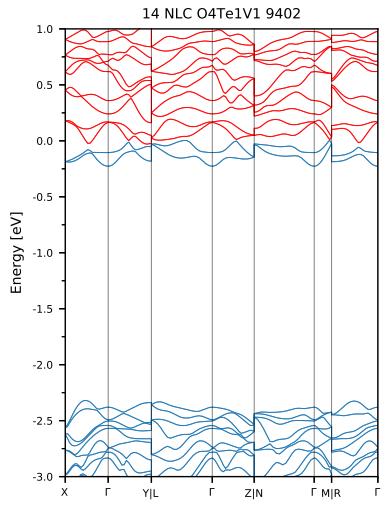

(c) 14.2.3

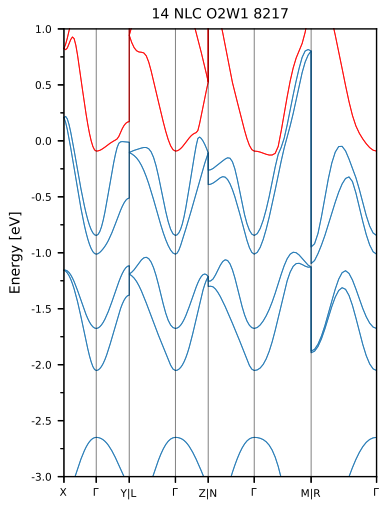

(d) 14.2.4

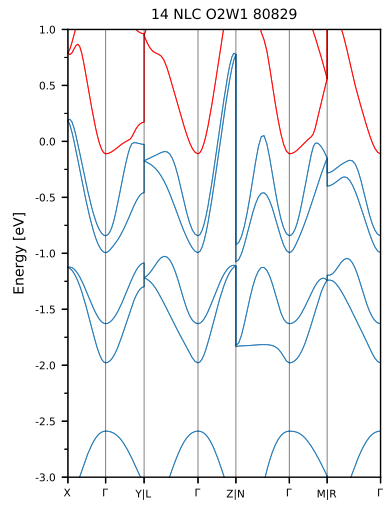

(e) 14.2.5

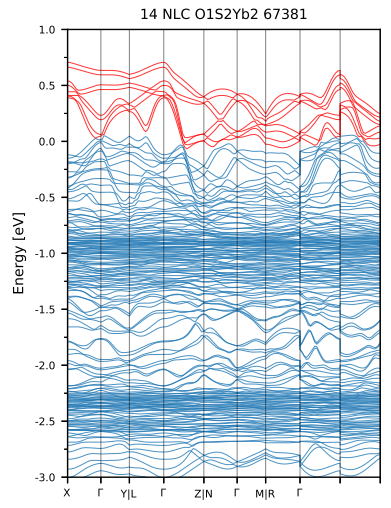

(f) 14.2.6

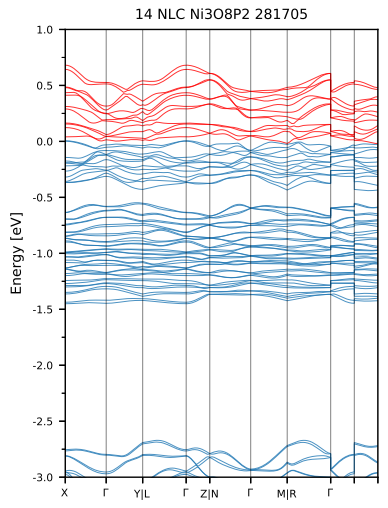

(g) 14.2.7

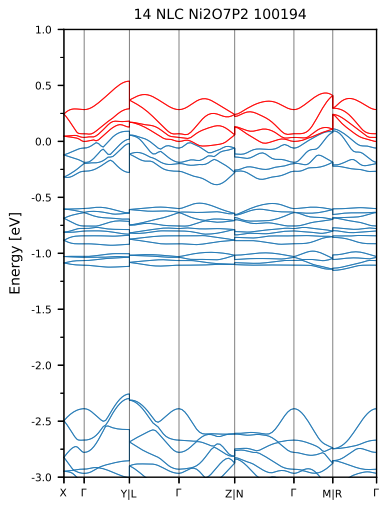

(h) 14.2.8

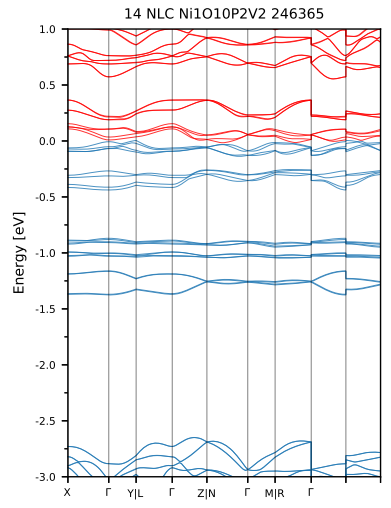

(i) 14.2.9

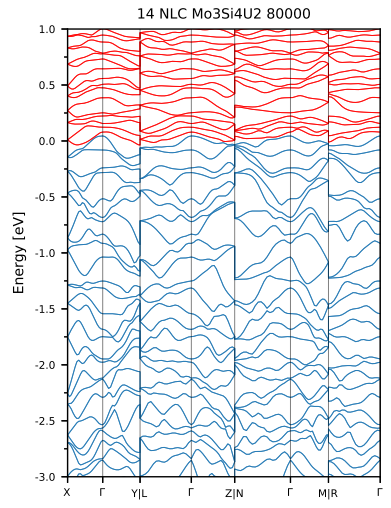

(j) 14.2.10

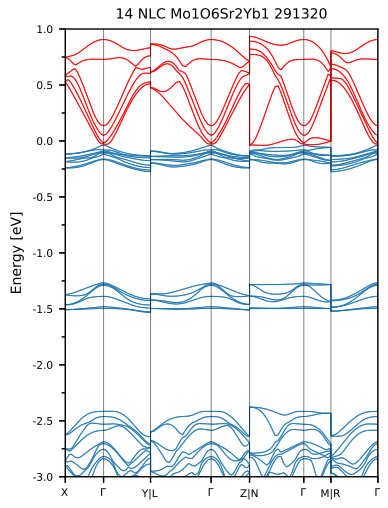

(k) 14.2.11

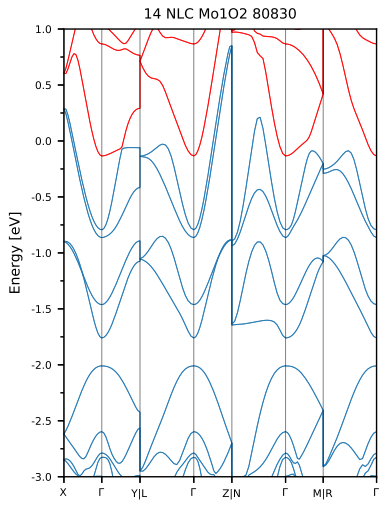

(l) 14.2.12

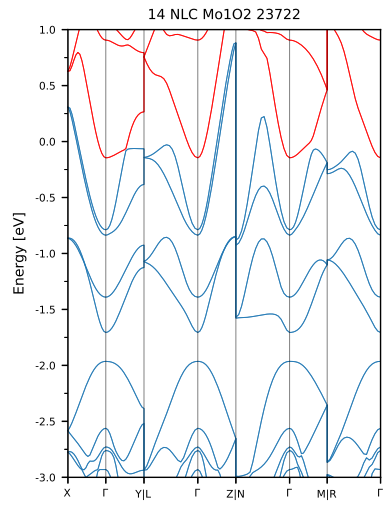

(m) 14.2.13

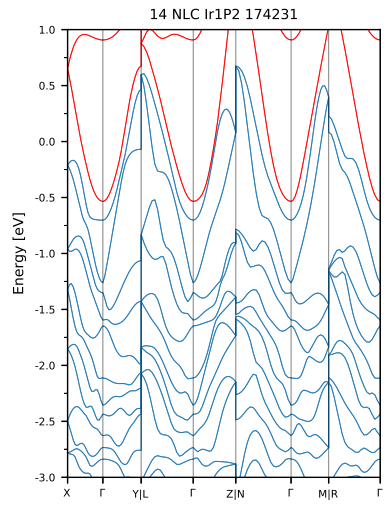

(n) 14.2.14

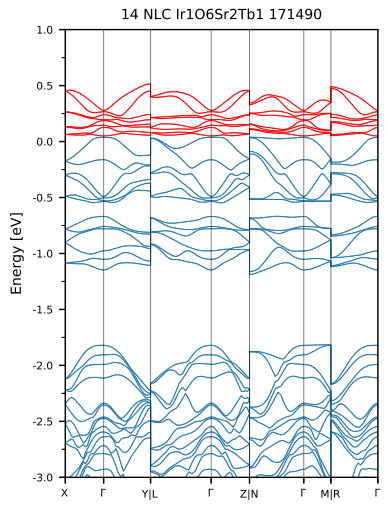

(o) 14.2.15

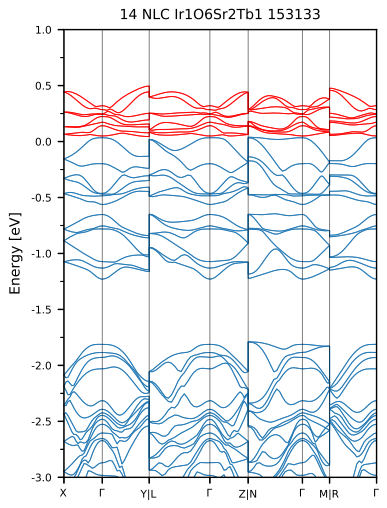

(p) 14.2.16

Figure 52: Band structures for topological materials in SG 14, part 1/10, from 14.2.1 to 14.2.16. See the list of figures.

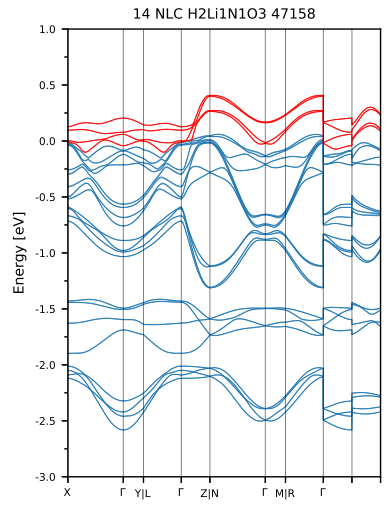

(q) 14.2.17

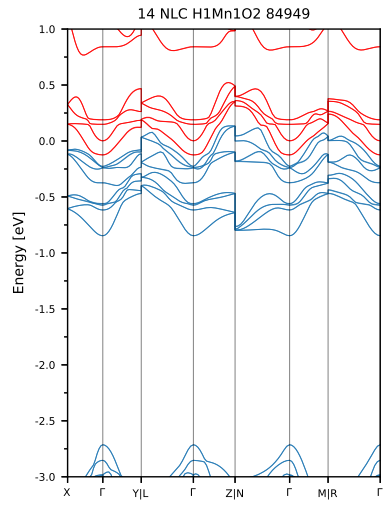

(r) 14.2.18

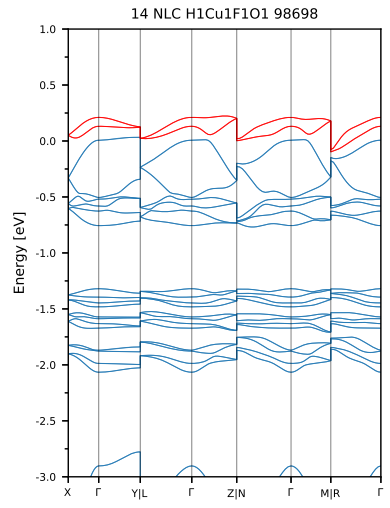

(s) 14.2.19

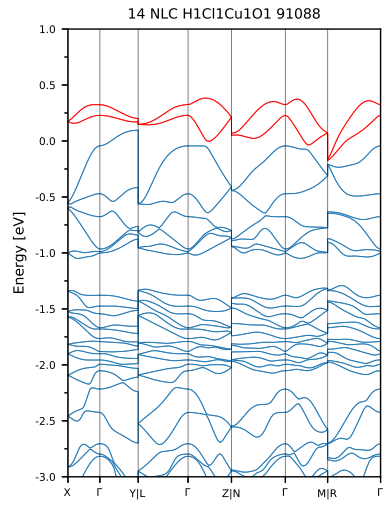

(t) 14.2.20

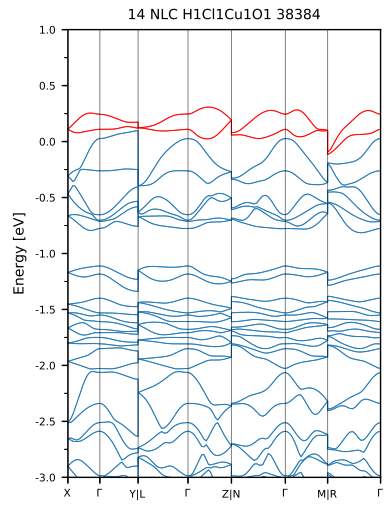

(u) 14.2.21

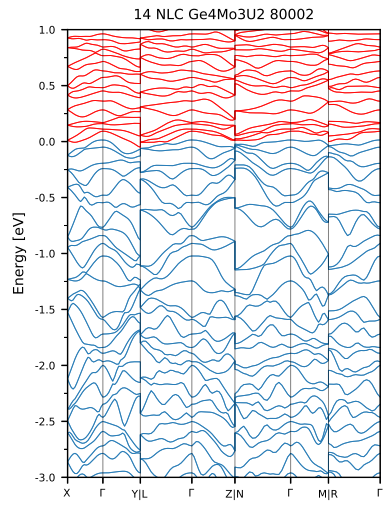

(v) 14.2.22

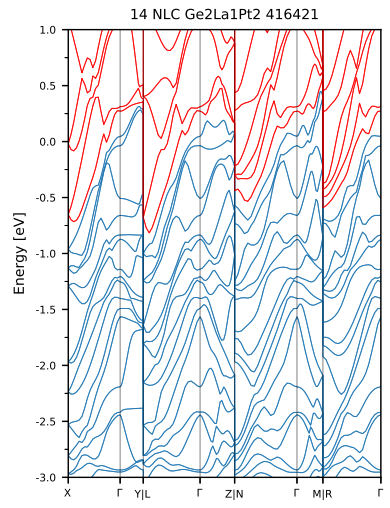

(w) 14.2.23

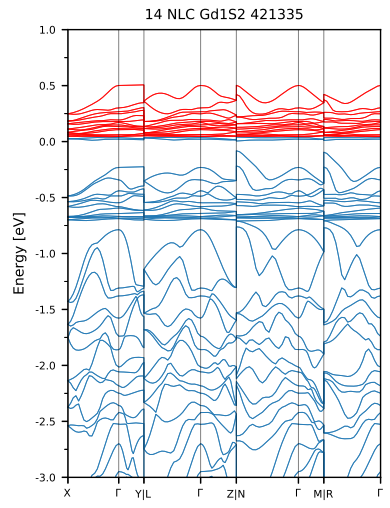

(x) 14.2.24

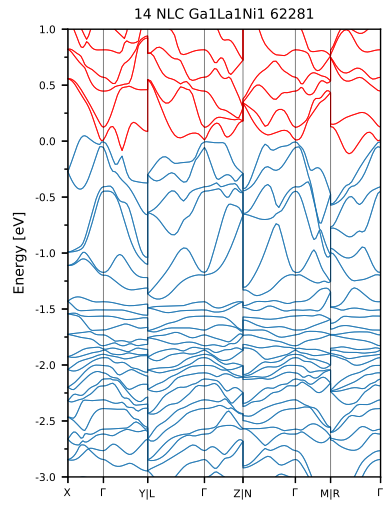

(y) 14.2.25

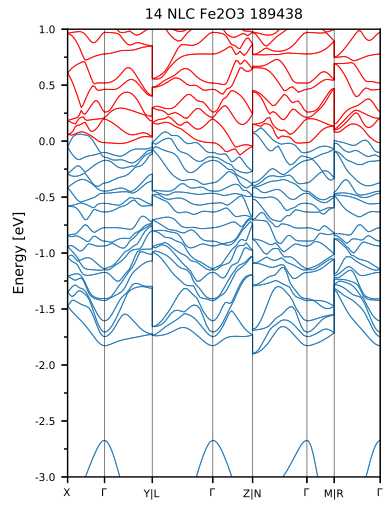

(z) 14.2.26

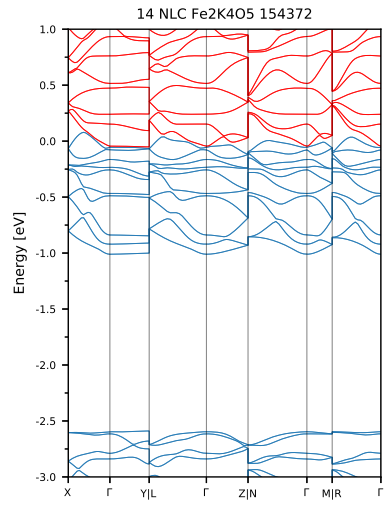

(aa) 14.2.27

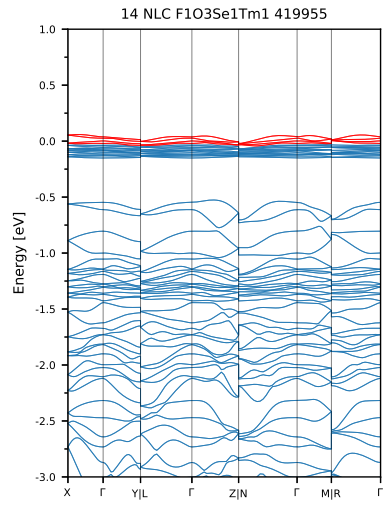

(ab) 14.2.28

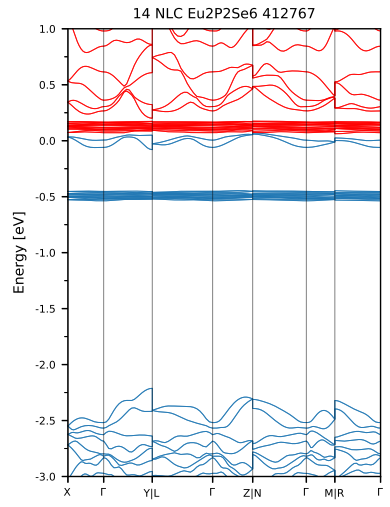

(ac) 14.2.29

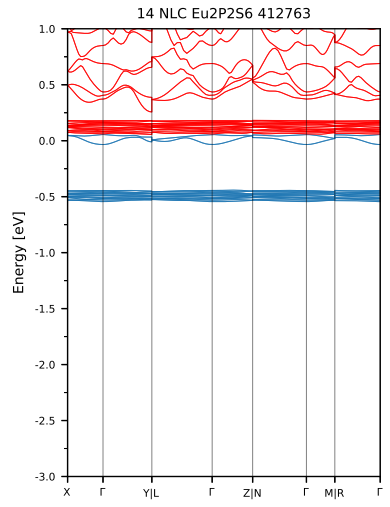

(ad) 14.2.30

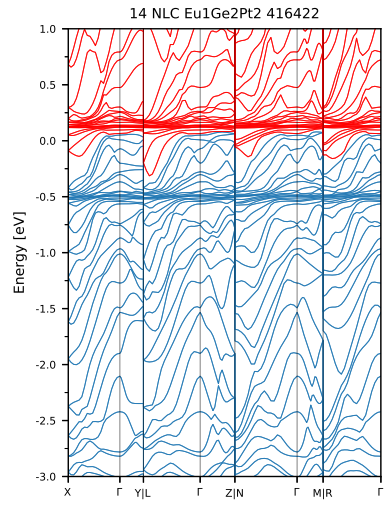

(ae) 14.2.31

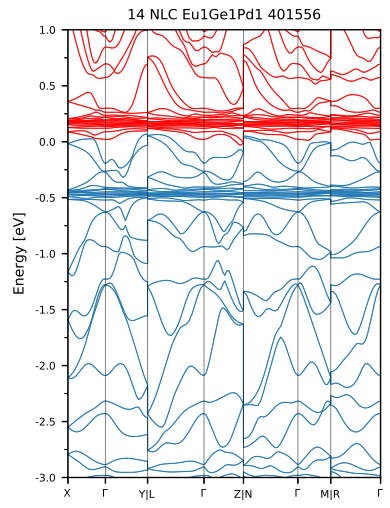

(af) 14.2.32

Figure 53: Band structures for topological materials in SG 14, part 2/10, from 14.2.17 to 14.2.32. See the list of figures.

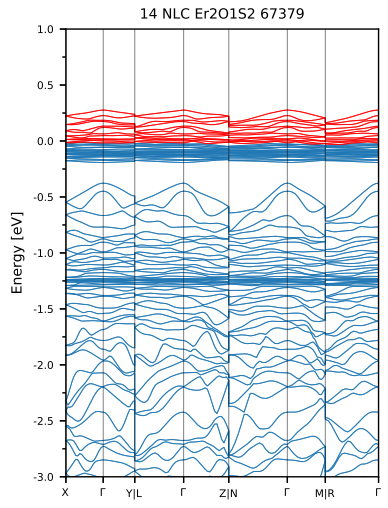

(ag) 14.2.33

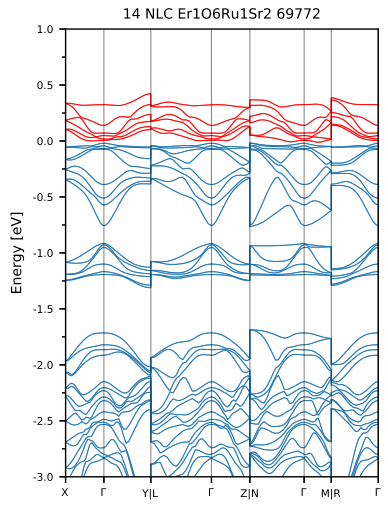

(ah) 14.2.34

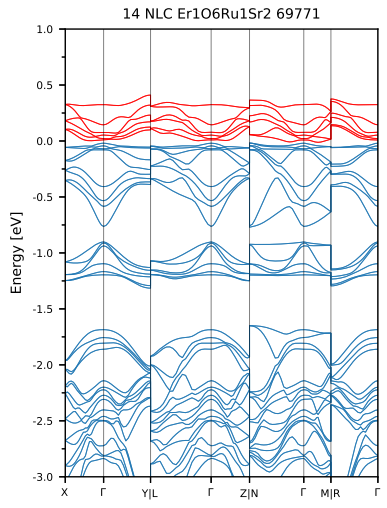

(ai) 14.2.35

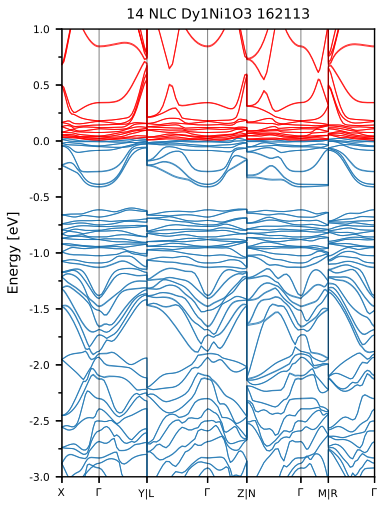

(aj) 14.2.36

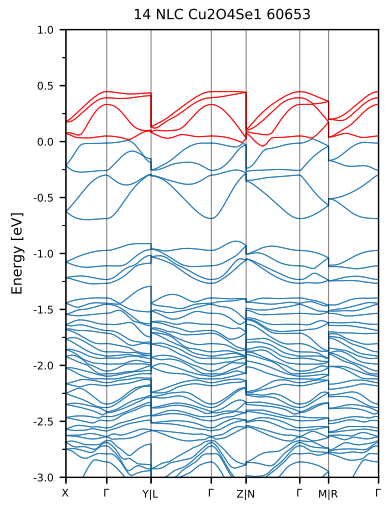

(ak) 14.2.37

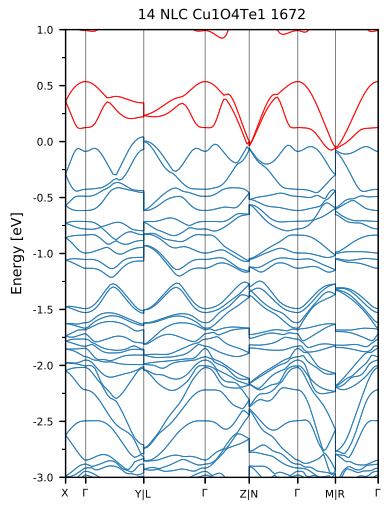

(al) 14.2.38

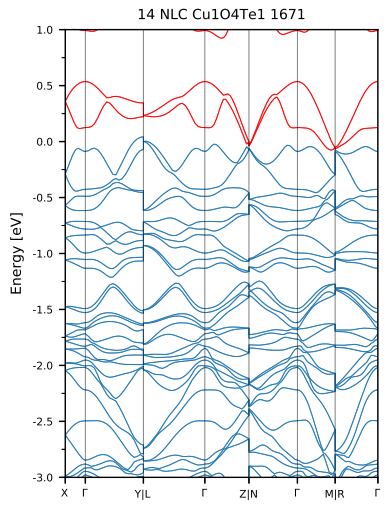

(am) 14.2.39

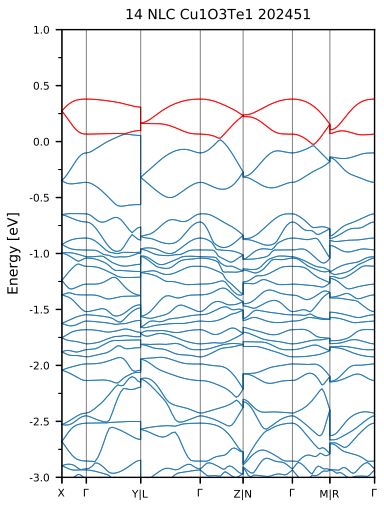

(an) 14.2.40

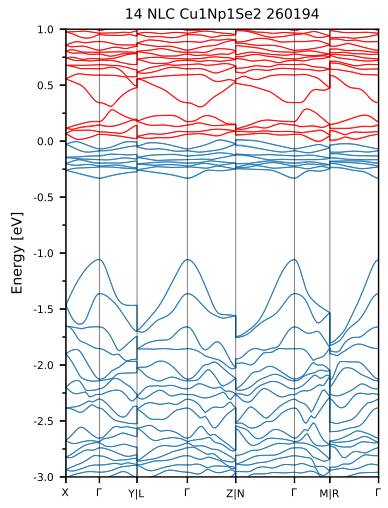

(ao) 14.2.41

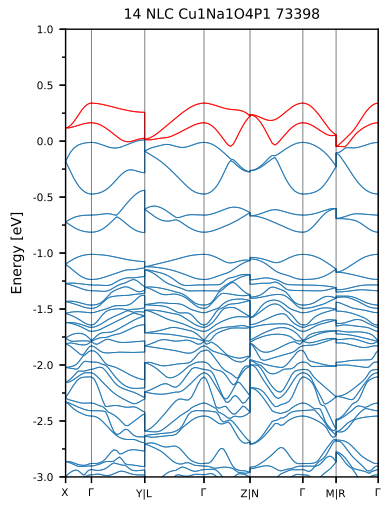

(ap) 14.2.42

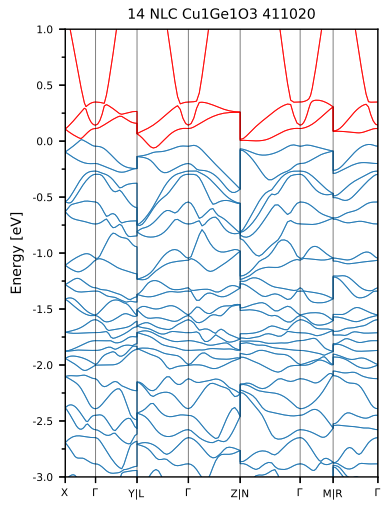

(aq) 14.2.43

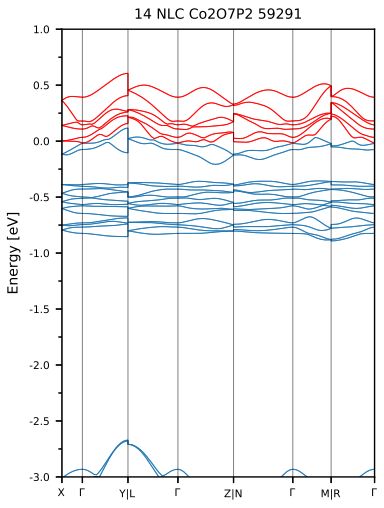

(ar) 14.2.44

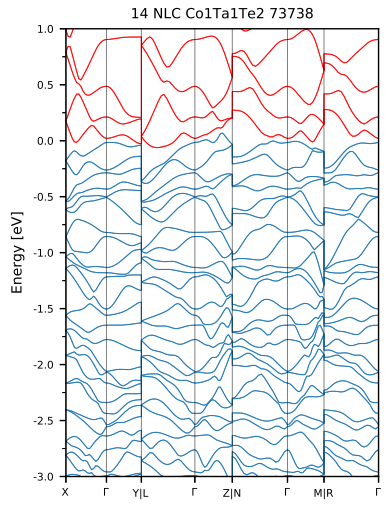

(as) 14.2.45

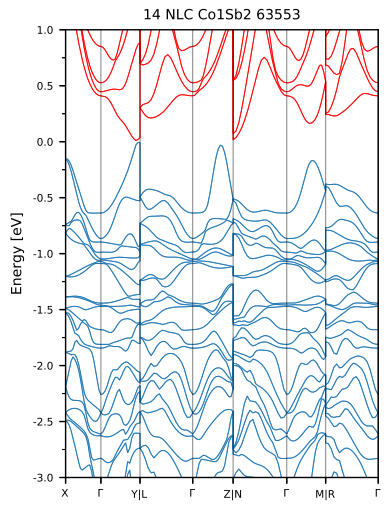

(at) 14.2.46

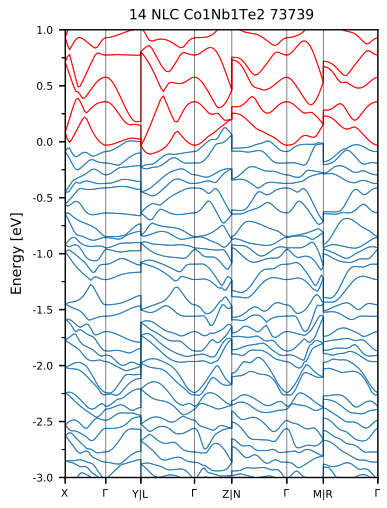

(au) 14.2.47

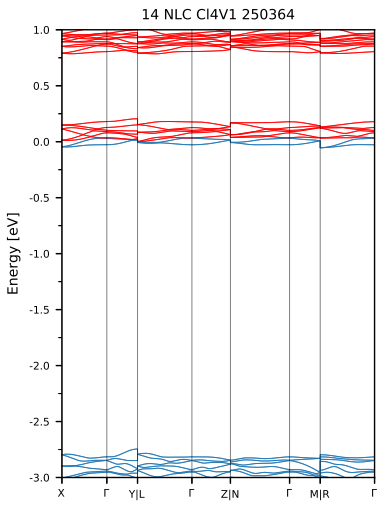

(av) 14.2.48

Figure 54: Band structures for topological materials in SG 14, part 3/10, from 14.2.33 to 14.2.48. See the list of figures.

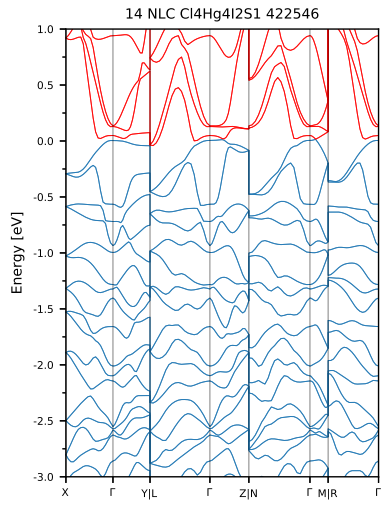

(aw) 14.2.49

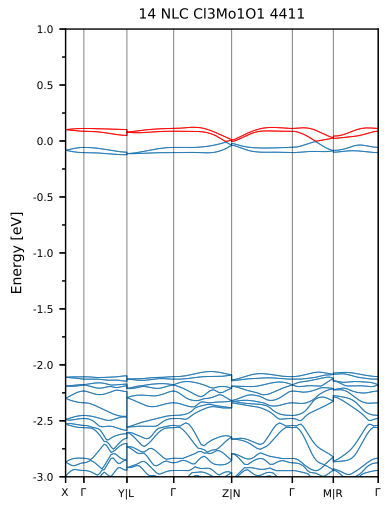

(ax) 14.2.50

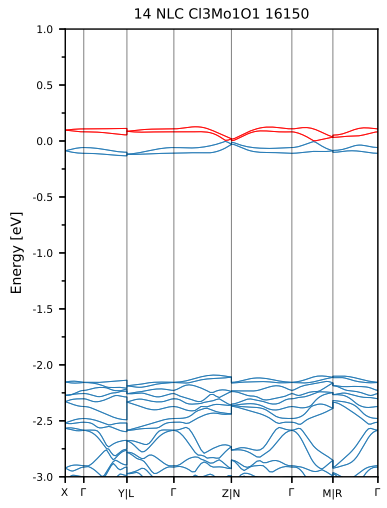

(ay) 14.2.51

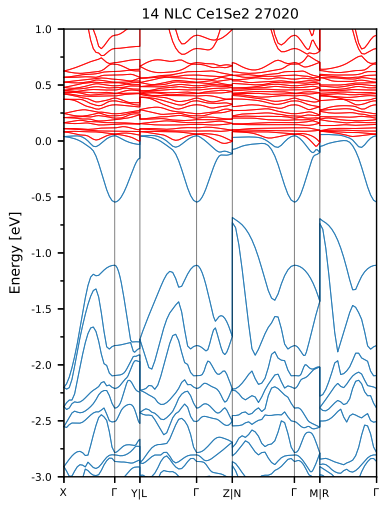

(az) 14.2.52

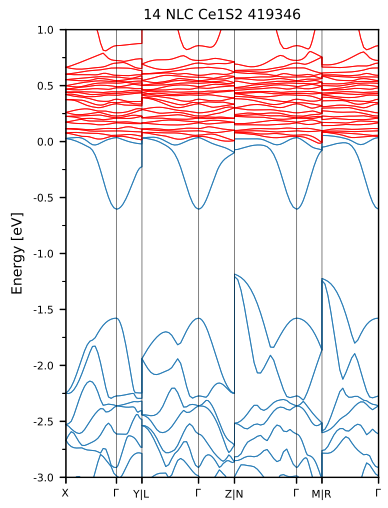

(ba) 14.2.53

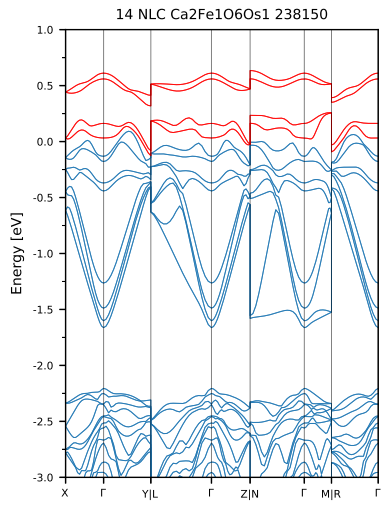

(bb) 14.2.54

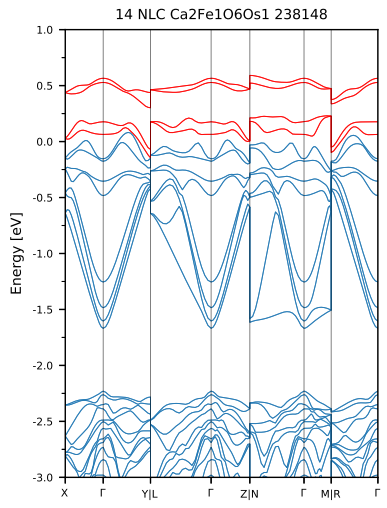

(bc) 14.2.55

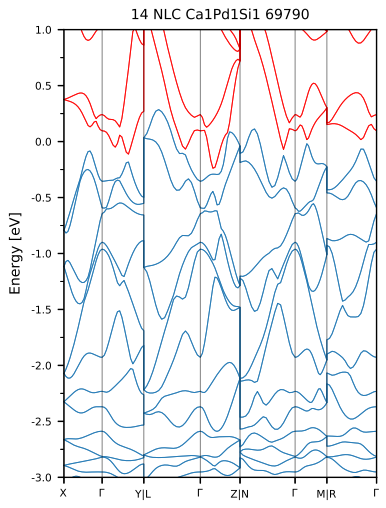

(bd) 14.2.56

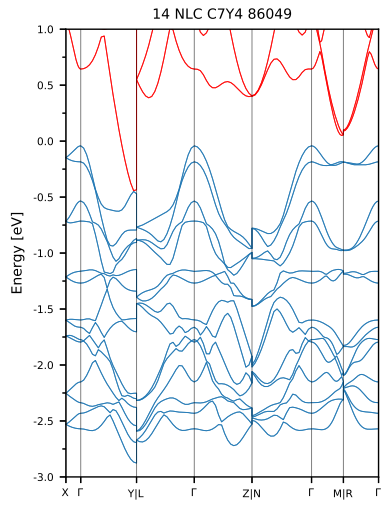

(be) 14.2.57

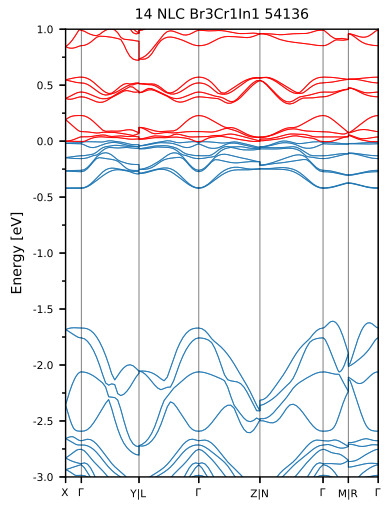

(bf) 14.2.58

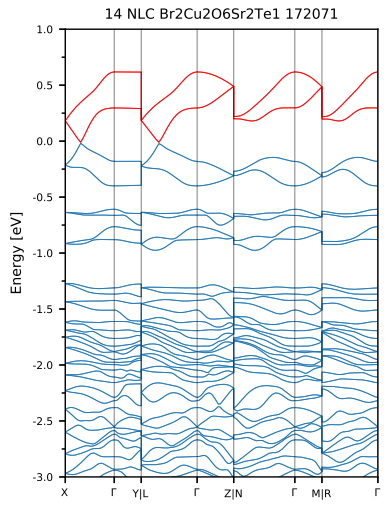

(bg) 14.2.59

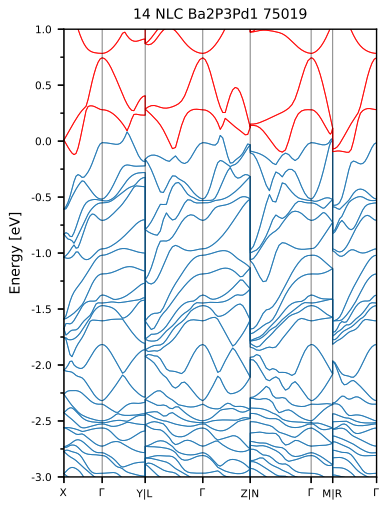

(bh) 14.2.60

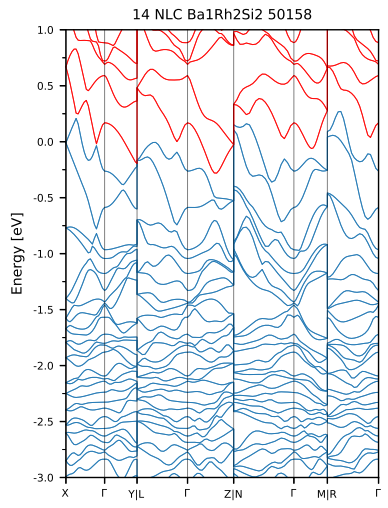

(bi) 14.2.61

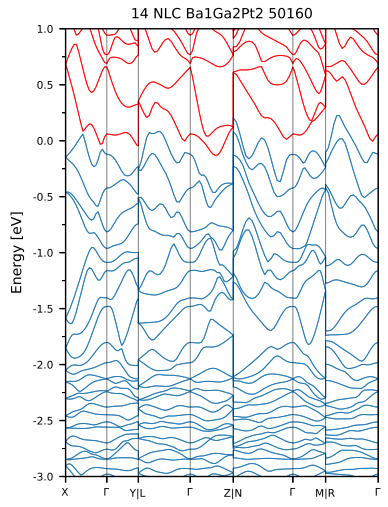

(bj) 14.2.62

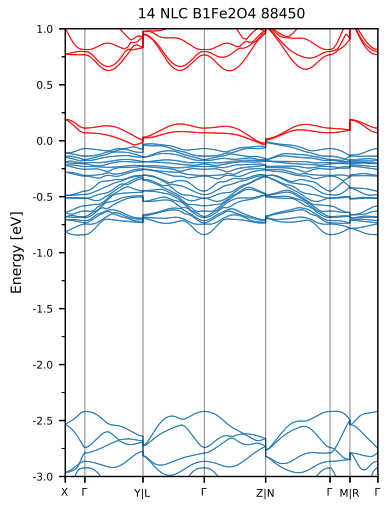

(bk) 14.2.63

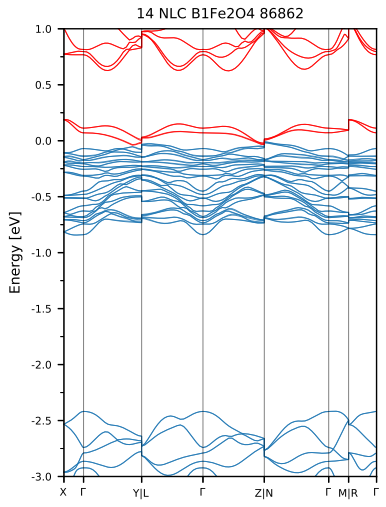

(bl) 14.2.64

Figure 55: Band structures for topological materials in SG 14, part 4/10, from 14.2.49 to 14.2.64. See the list of figures.

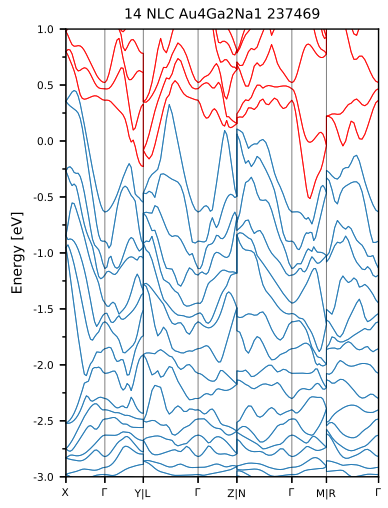

(bm) 14.2.65

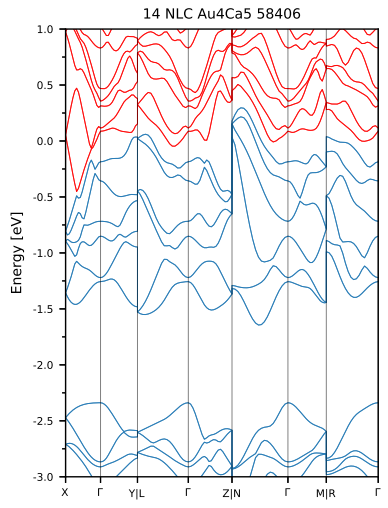

(bn) 14.2.66

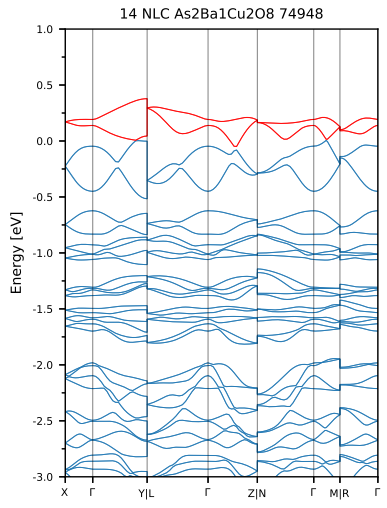

(bo) 14.2.67

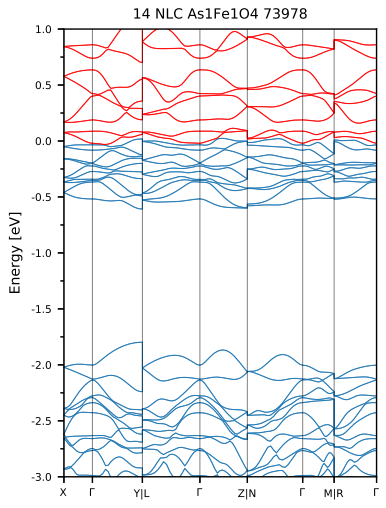

(bp) 14.2.68

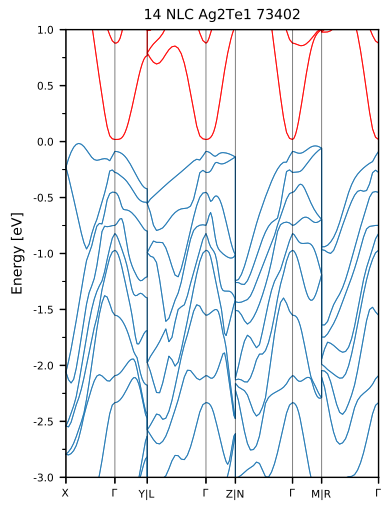

(bq) 14.2.69

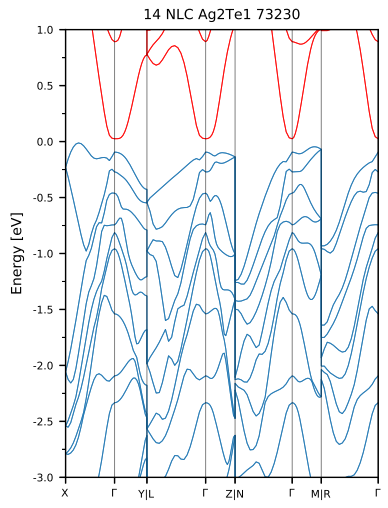

(br) 14.2.70

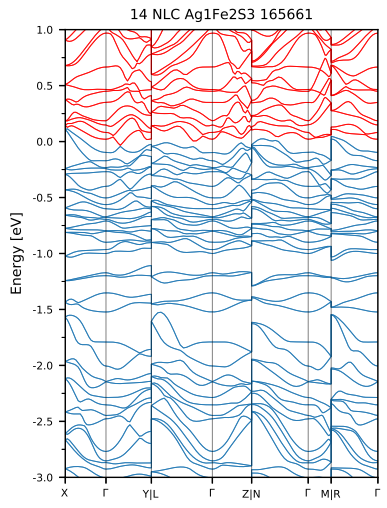

(bs) 14.2.71

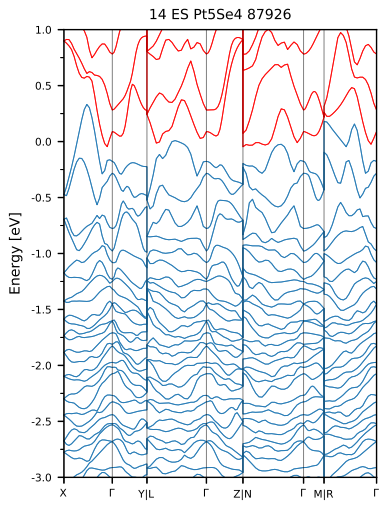

(bt) 14.3.1

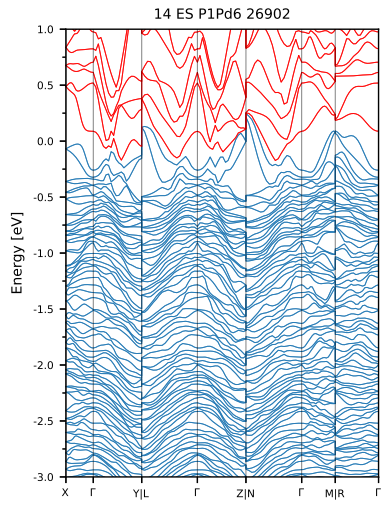

(bu) 14.3.2

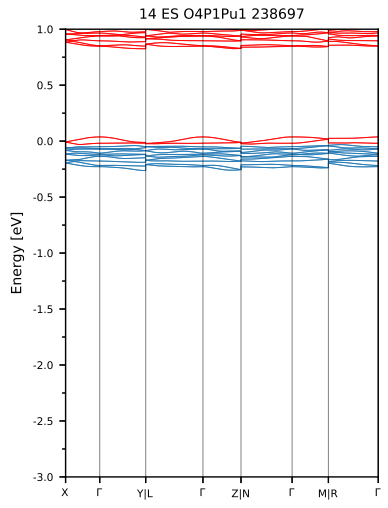

(bv) 14.3.3

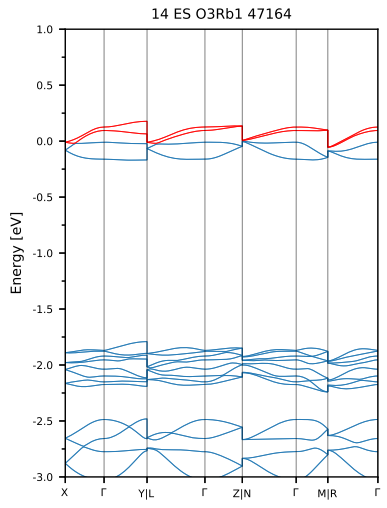

(bw) 14.3.4

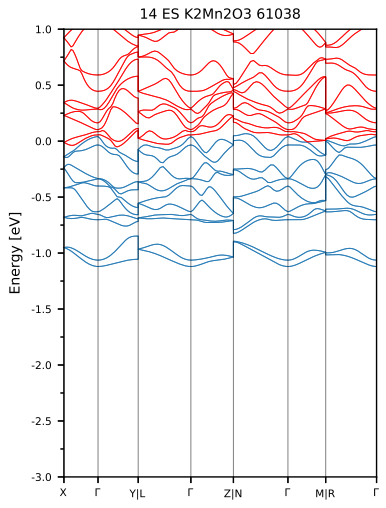

(bx) 14.3.5

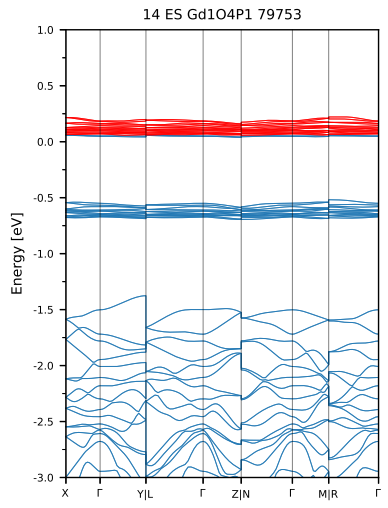

(by) 14.3.6

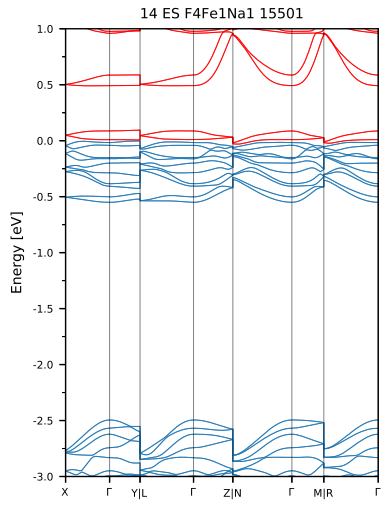

(bz) 14.3.7

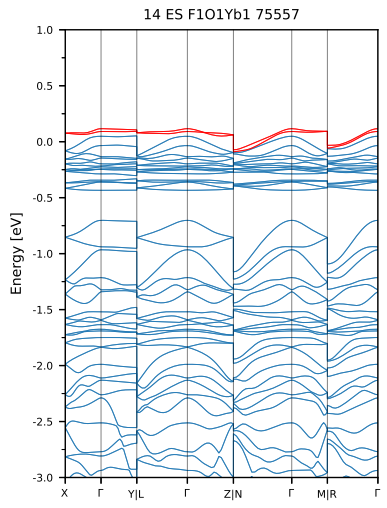

(ca) 14.3.8

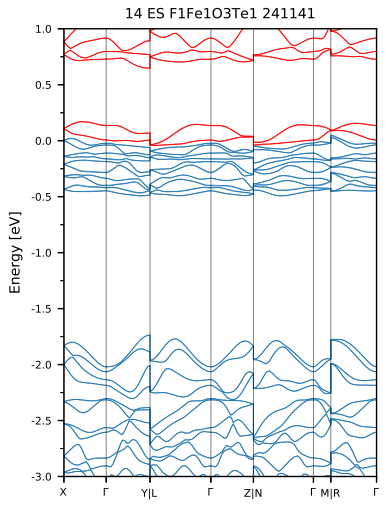

(cb) 14.3.9

Figure 56: Band structures for topological materials in SG 14, part 5/10, from 14.2.65 to 14.3.9. See the list of figures.

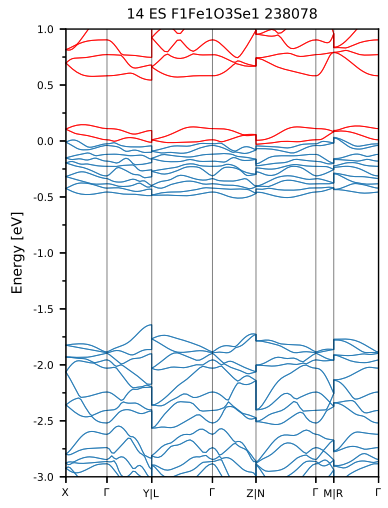

(cc) 14.3.10

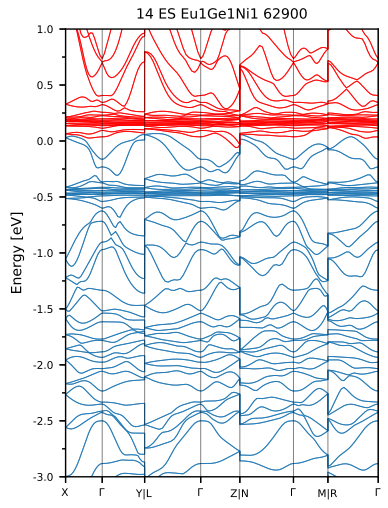

(cd) 14.3.11

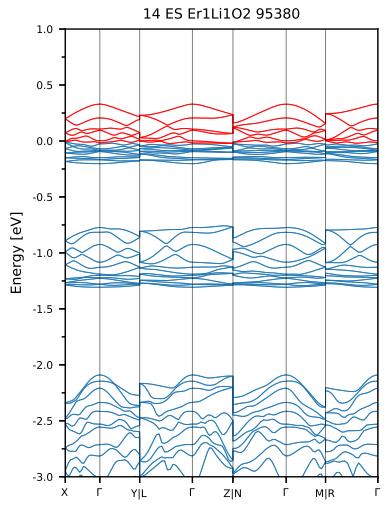

(ce) 14.3.12

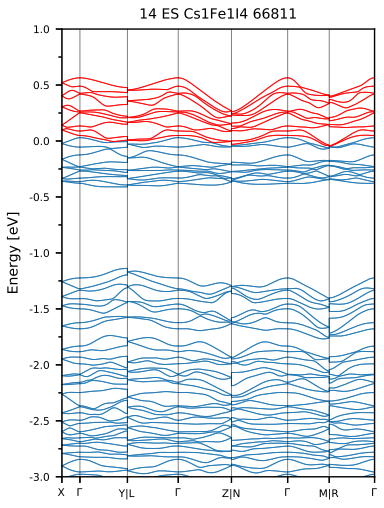

(cf) 14.3.13

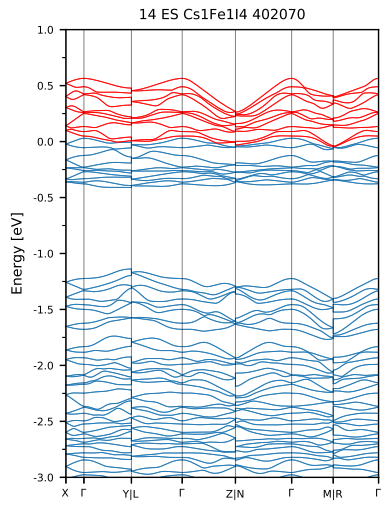

(cg) 14.3.14

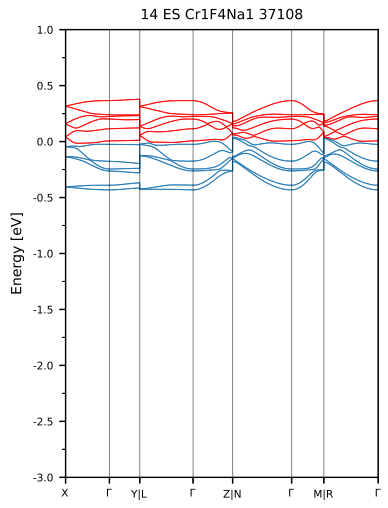

(ch) 14.3.15

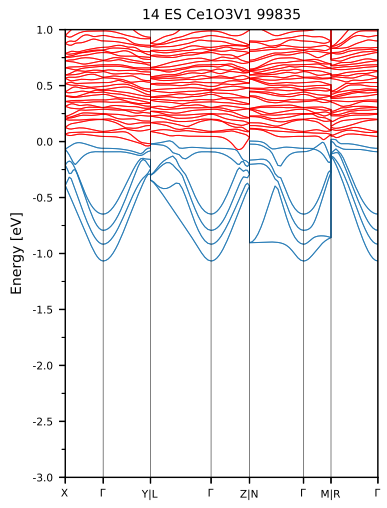

(ci) 14.3.16

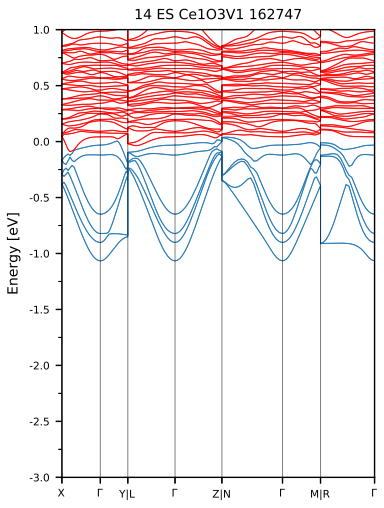

(cj) 14.3.17

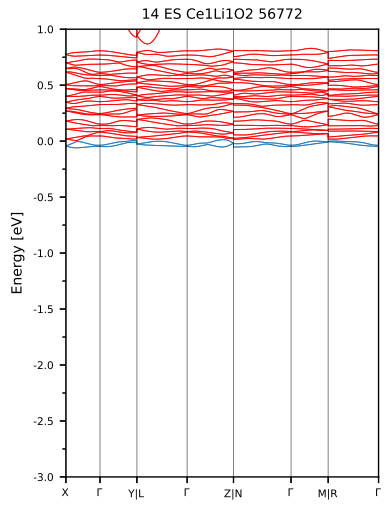

(ck) 14.3.18

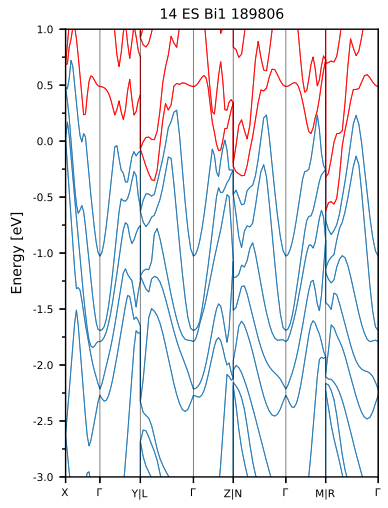

(cl) 14.3.19

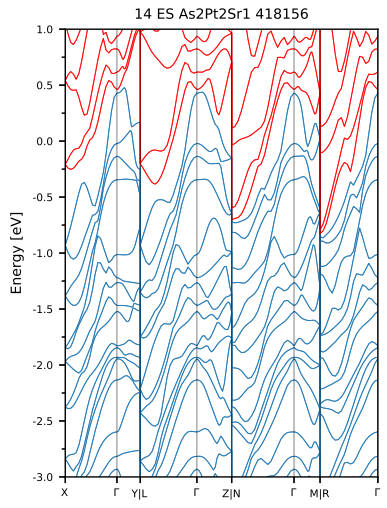

(cm) 14.3.20

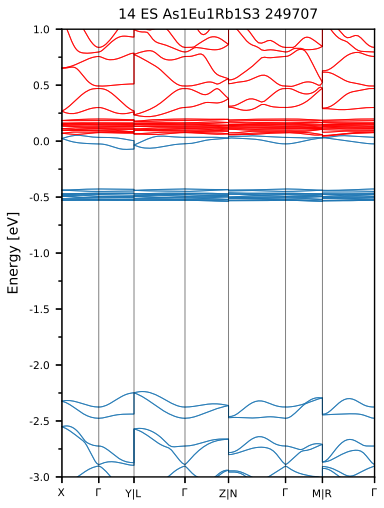

(cn) 14.3.21

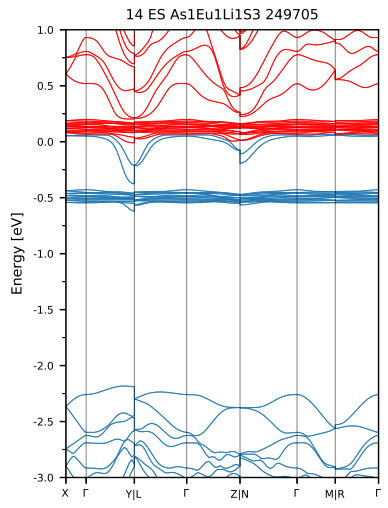

(co) 14.3.22

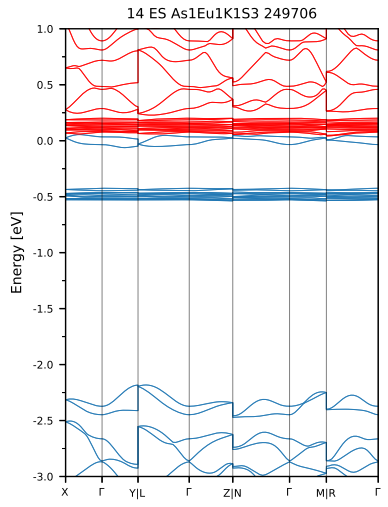

(cp) 14.3.23

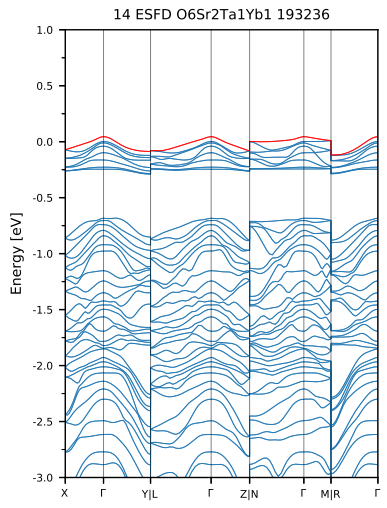

(cq) 14.4.1

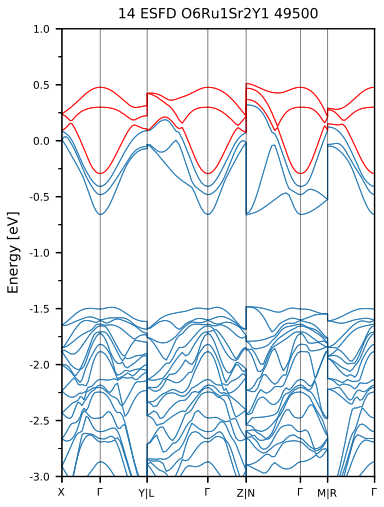

(cr) 14.4.2

Figure 57: Band structures for topological materials in SG 14, part 6/10, from 14.3.10 to 14.4.2. See the list of figures.

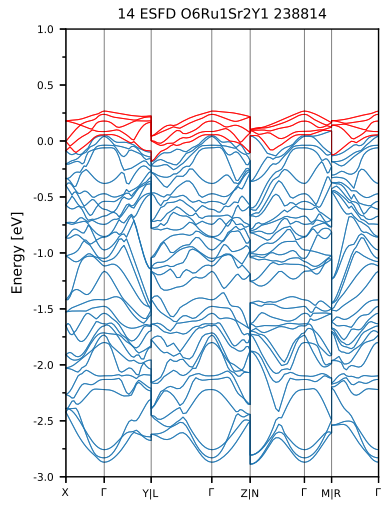

(cs) 14.4.3

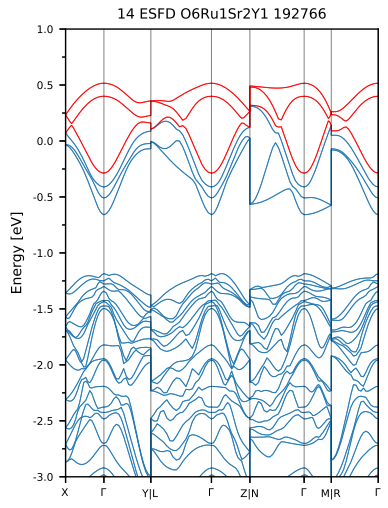

(ct) 14.4.4

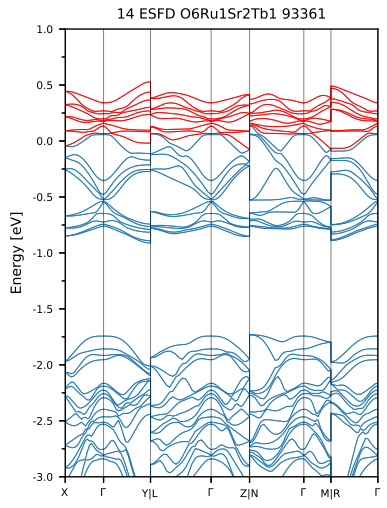

(cu) 14.4.5

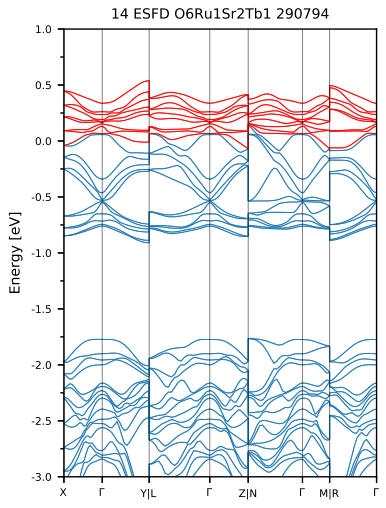

(cv) 14.4.6

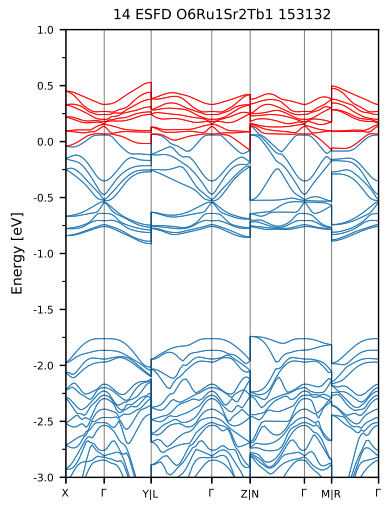

(cw) 14.4.7

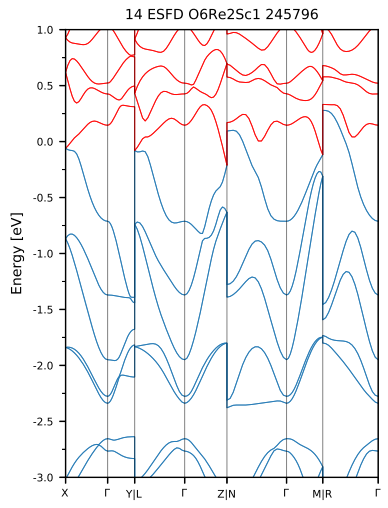

(cx) 14.4.8

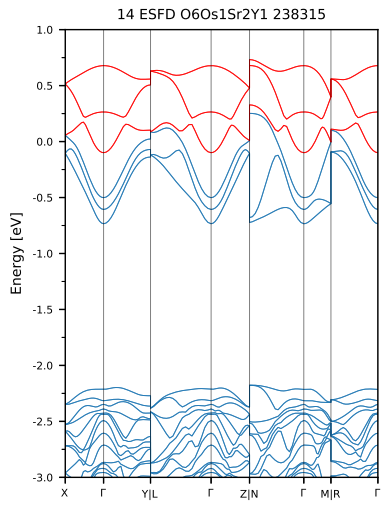

(cy) 14.4.9

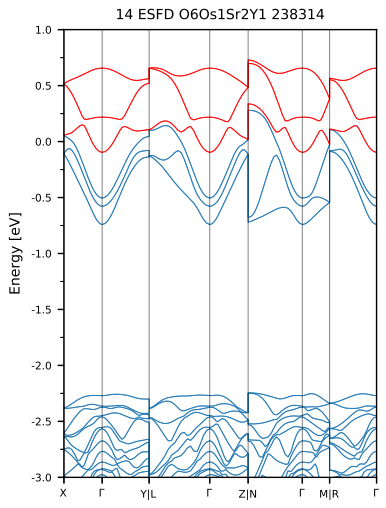

(cz) 14.4.10

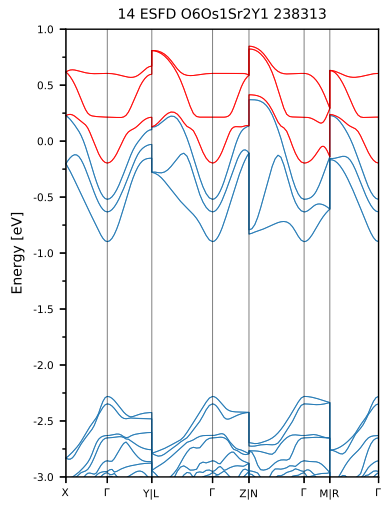

(da) 14.4.11

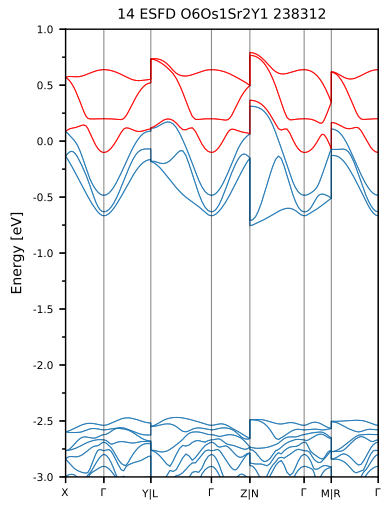

(db) 14.4.12

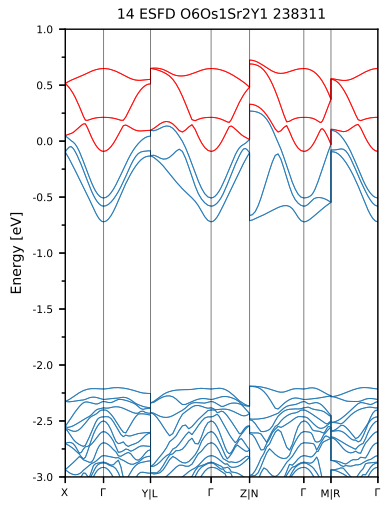

(dc) 14.4.13

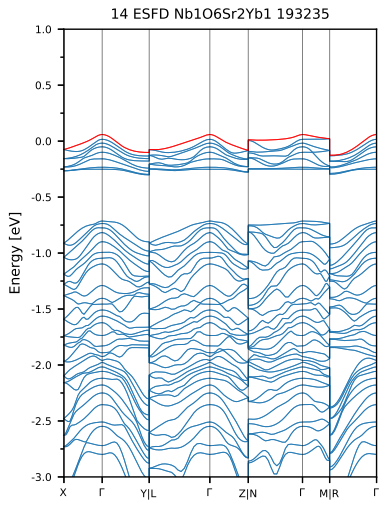

(dd) 14.4.14

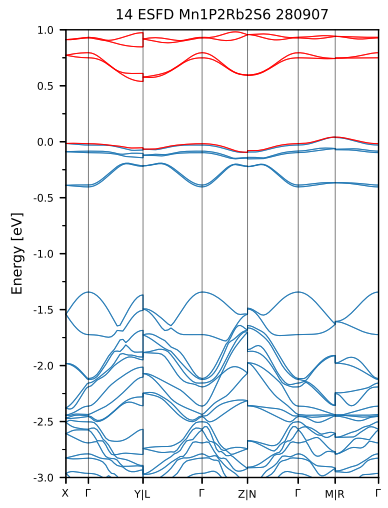

(de) 14.4.15

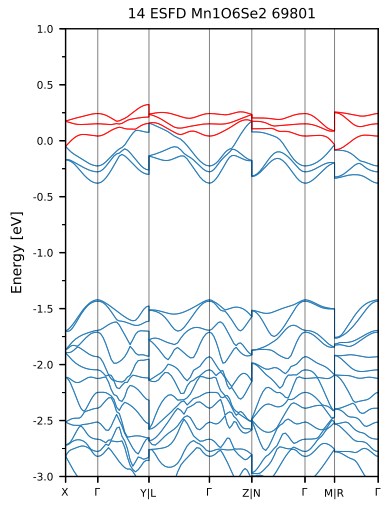

(df) 14.4.16

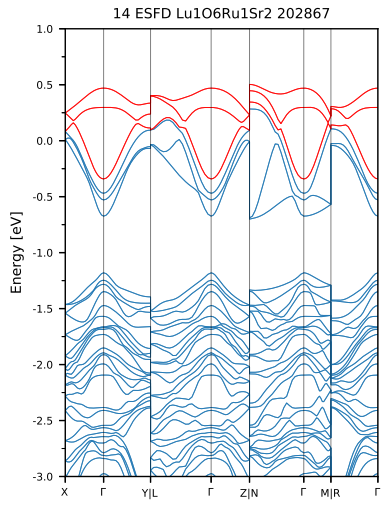

(dg) 14.4.17

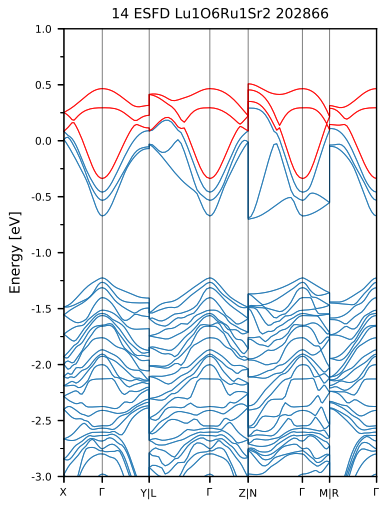

(dh) 14.4.18

Figure 58: Band structures for topological materials in SG 14, part 7/10, from 14.4.3 to 14.4.18. See the list of figures.

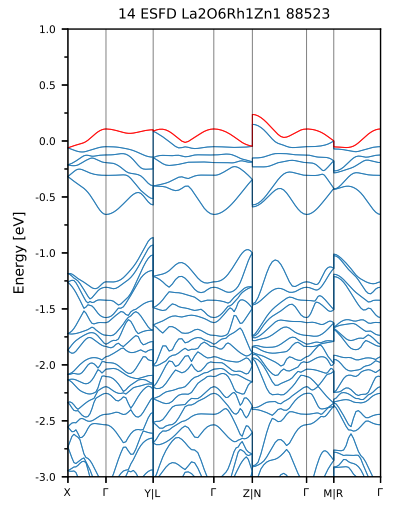

(di) 14.4.19

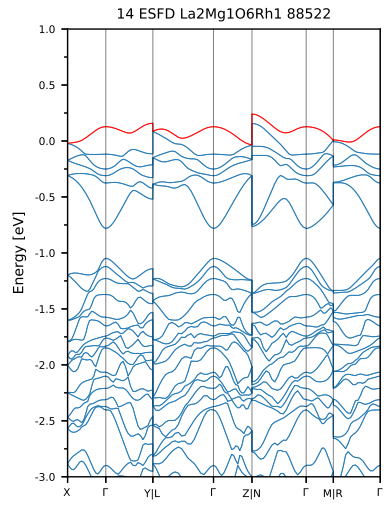

(dj) 14.4.20

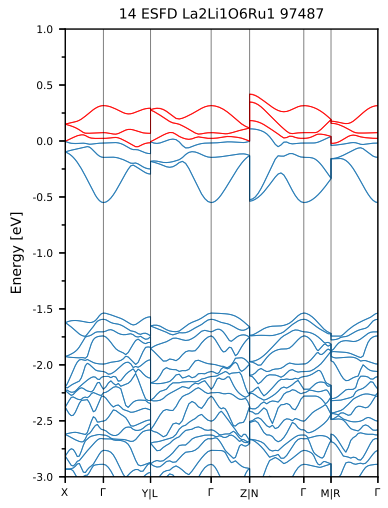

(dk) 14.4.21

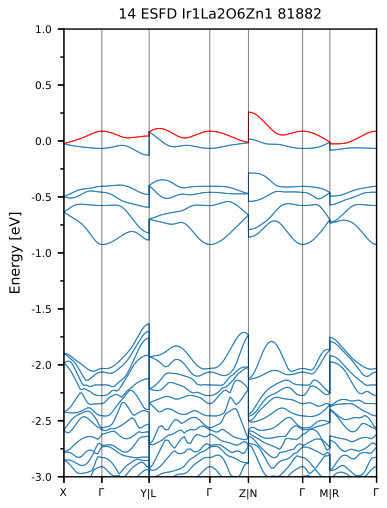

(dl) 14.4.22

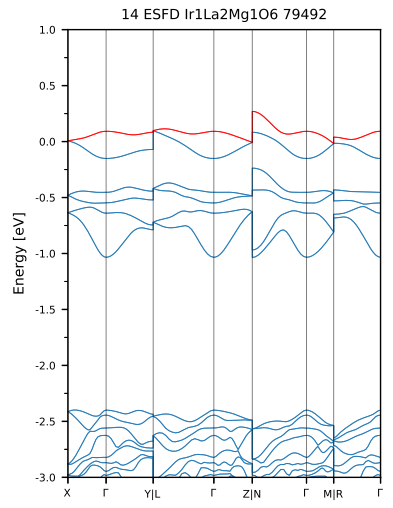

(dm) 14.4.23

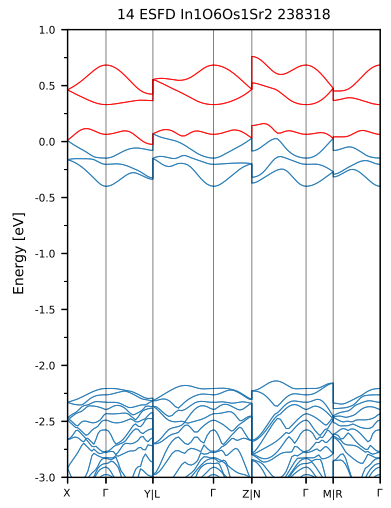

(dn) 14.4.24

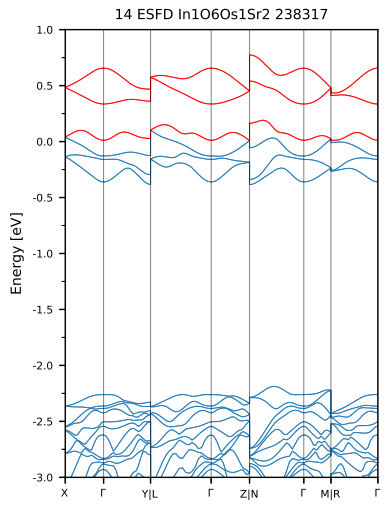

(do) 14.4.25

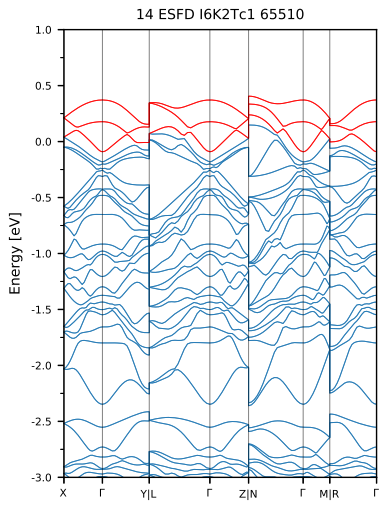

(dp) 14.4.26

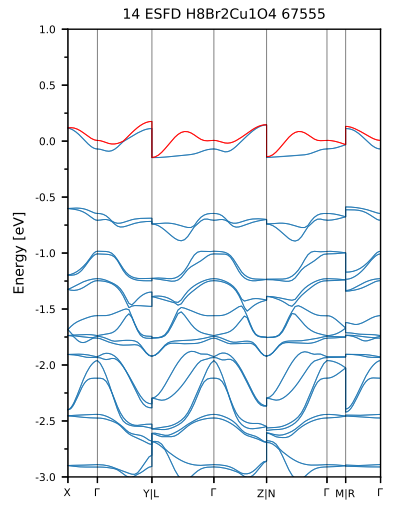

(dq) 14.4.27

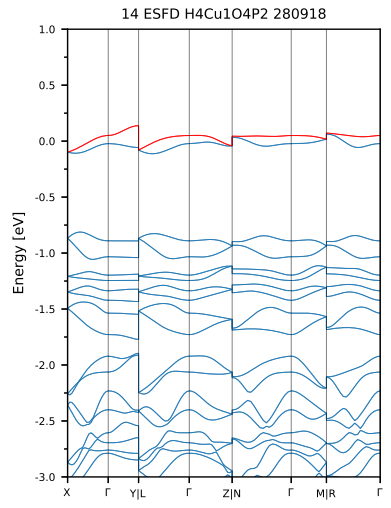

(dr) 14.4.28

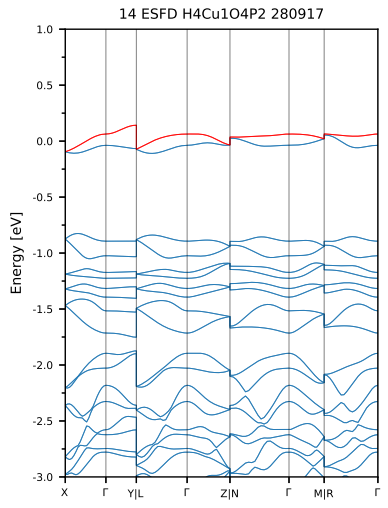

(ds) 14.4.29

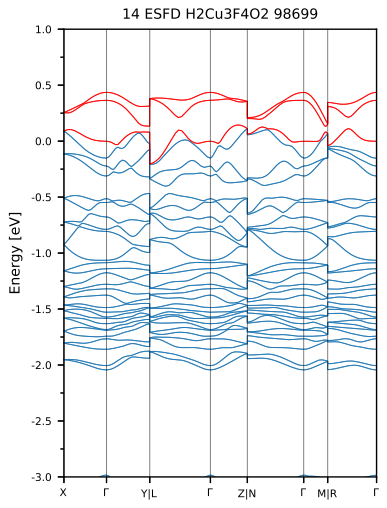

(dt) 14.4.30

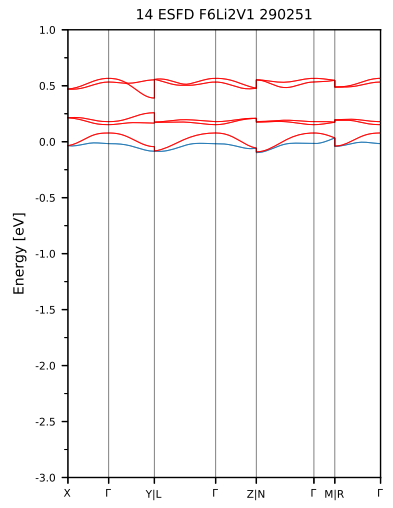

(du) 14.4.31

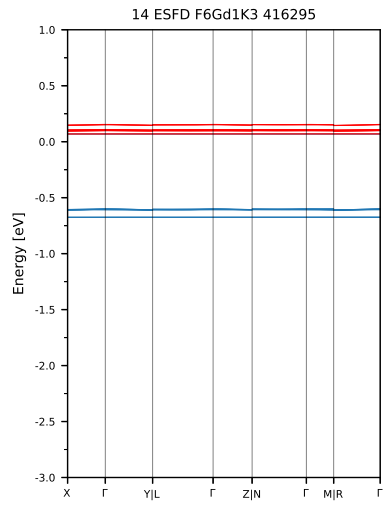

(dv) 14.4.32

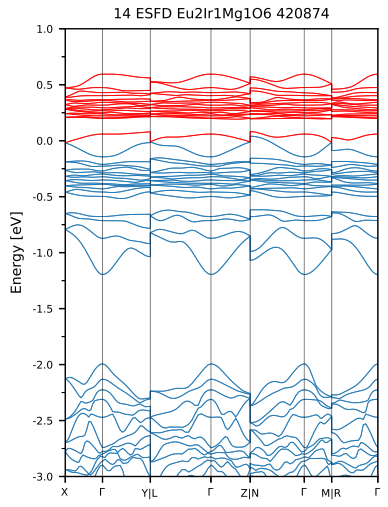

(dw) 14.4.33

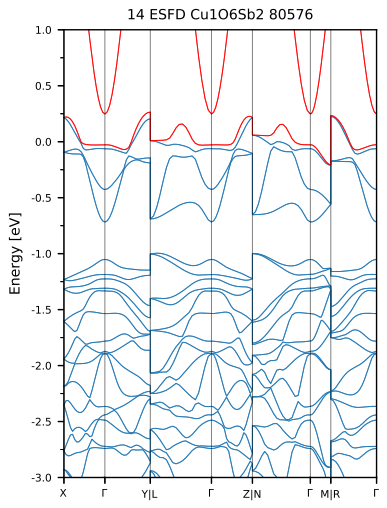

(dx) 14.4.34

Figure 59: Band structures for topological materials in SG 14, part 8/10, from 14.4.19 to 14.4.34. See the list of figures.

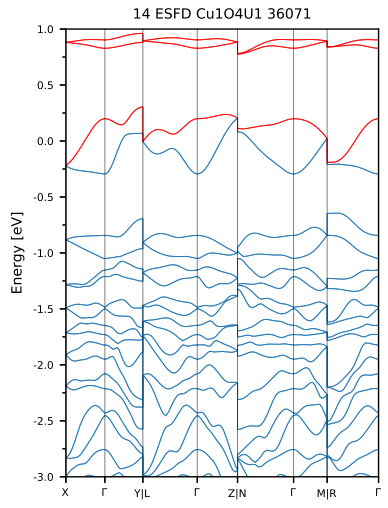

(dy) 14.4.35

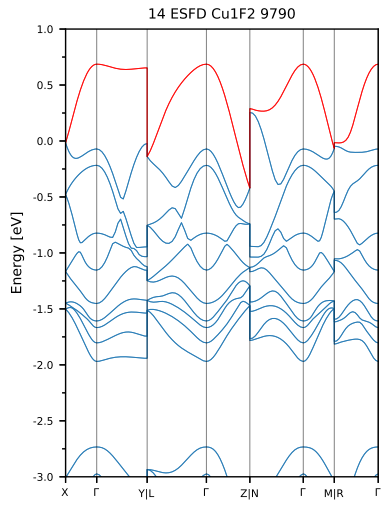

(dz) 14.4.36

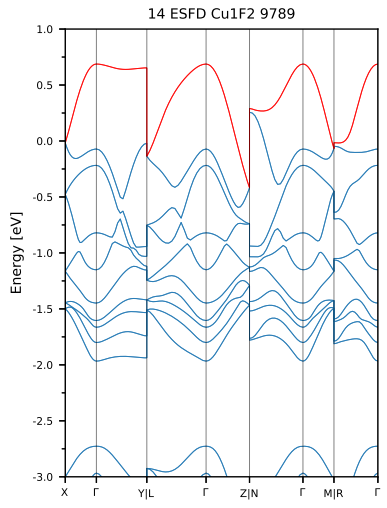

(ea) 14.4.37

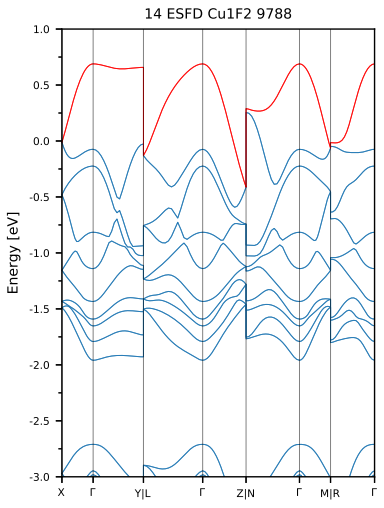

(eb) 14.4.38

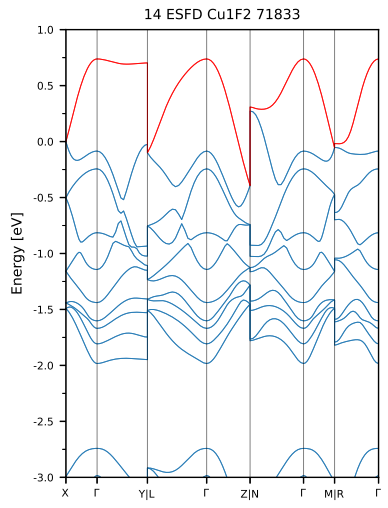

(ec) 14.4.39

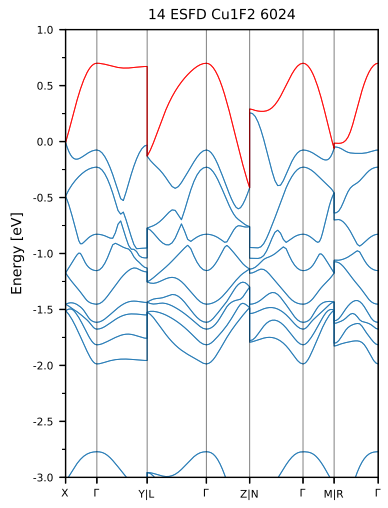

(ed) 14.4.40

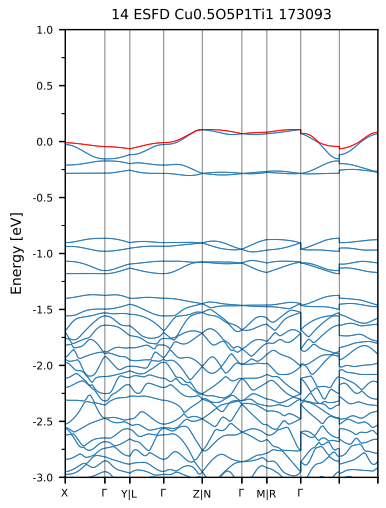

(ee) 14.4.41

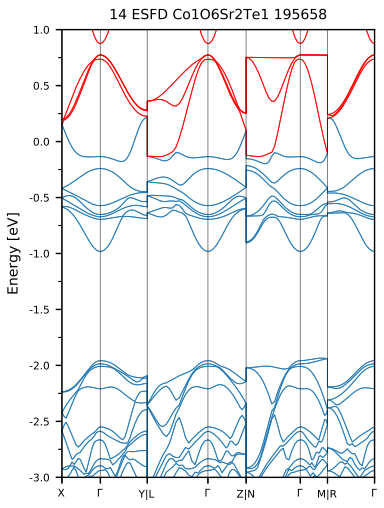

(ef) 14.4.42

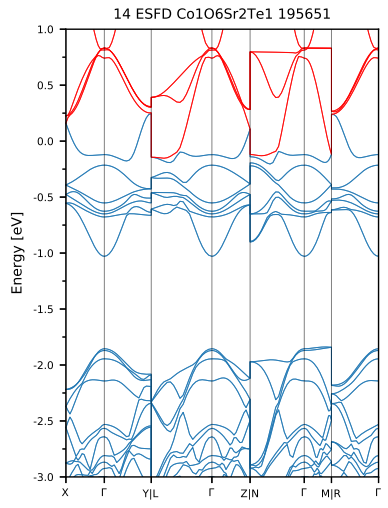

(eg) 14.4.43

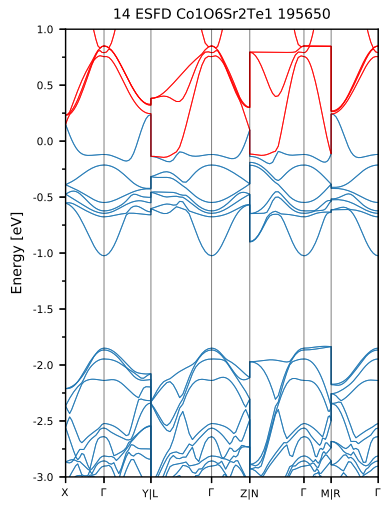

(eh) 14.4.44

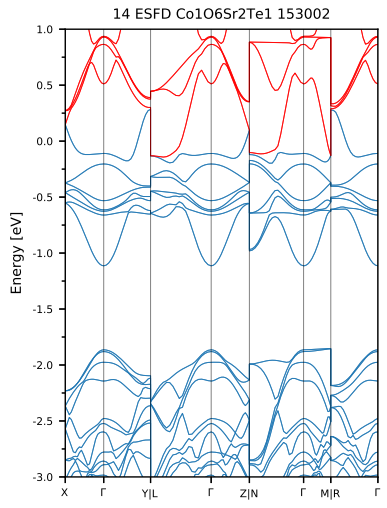

(ei) 14.4.45

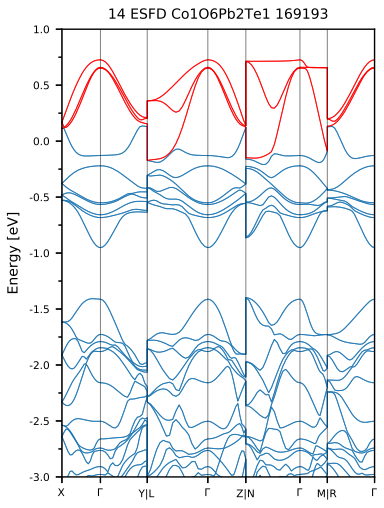

(ej) 14.4.46

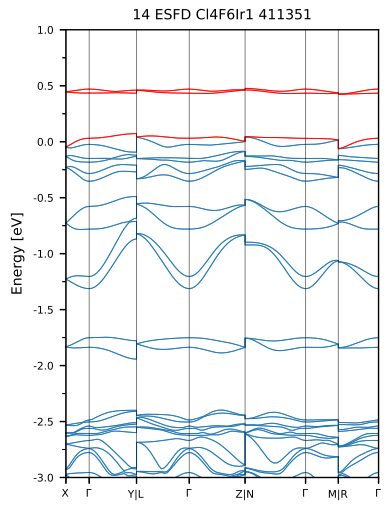

(ek) 14.4.47

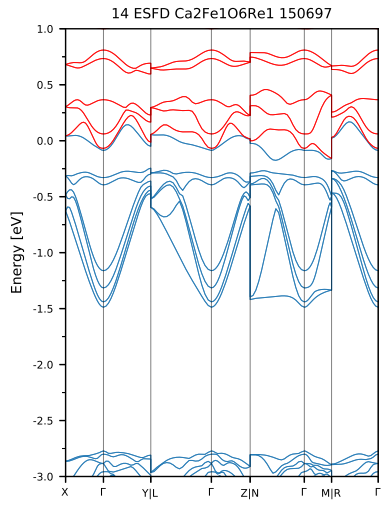

(el) 14.4.48

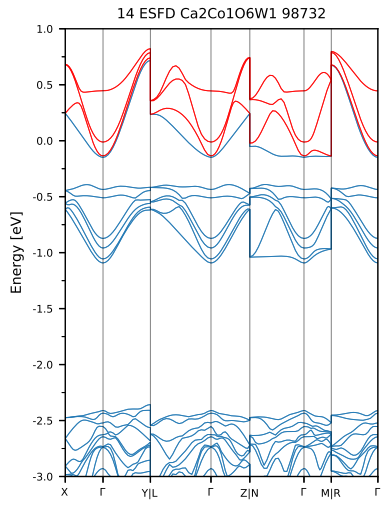

(em) 14.4.49

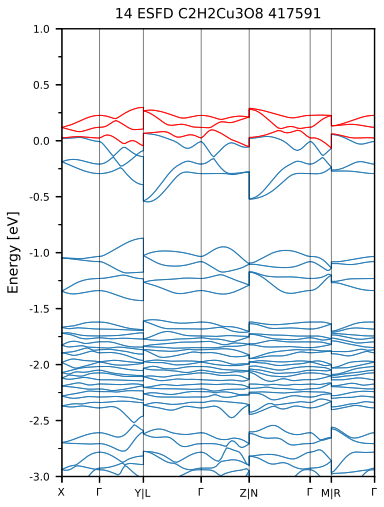

(en) 14.4.50

Figure 60: Band structures for topological materials in SG 14, part 9/10, from 14.4.35 to 14.4.50. See the list of figures.

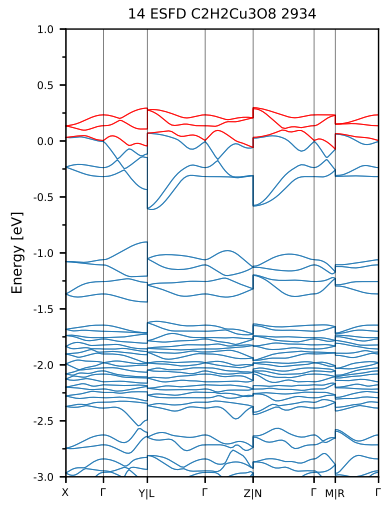

(eo) 14.4.51

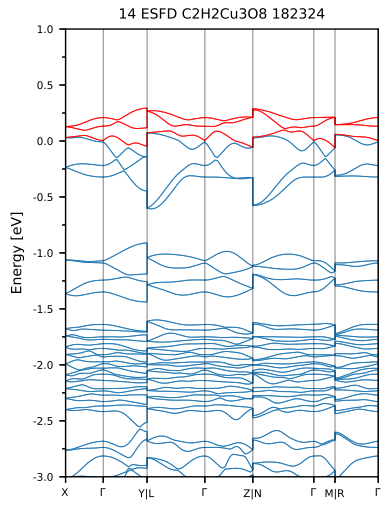

(ep) 14.4.52

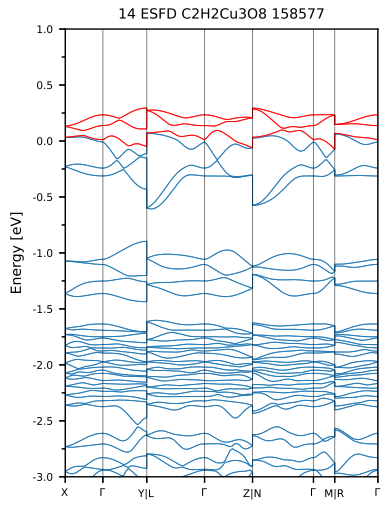

(eq) 14.4.53

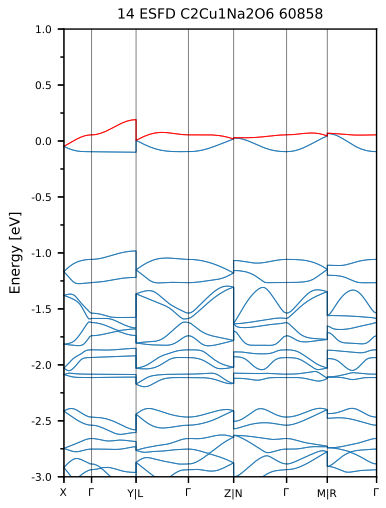

(er) 14.4.54

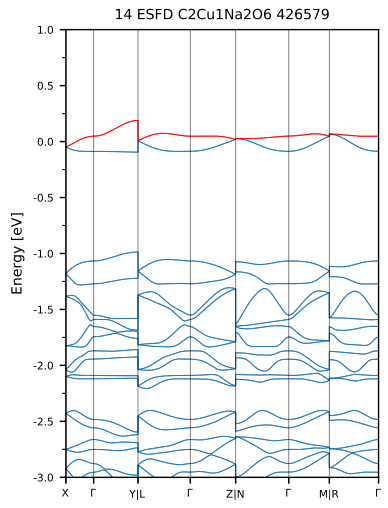

(es) 14.4.55

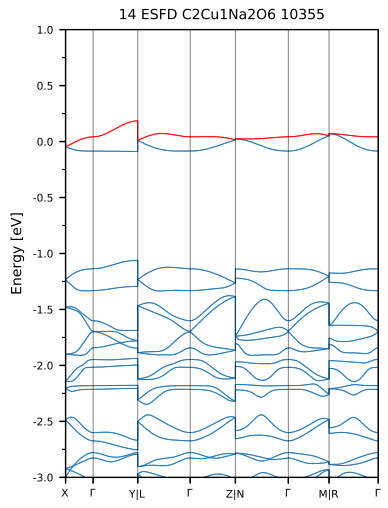

(et) 14.4.56

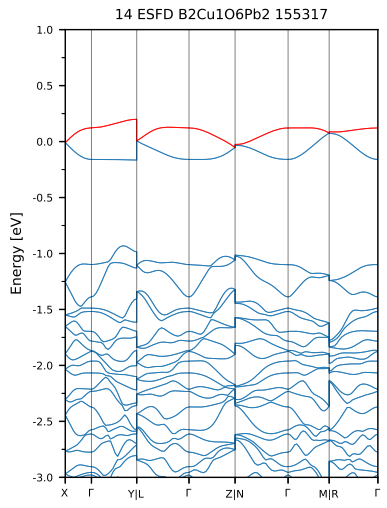

(eu) 14.4.57

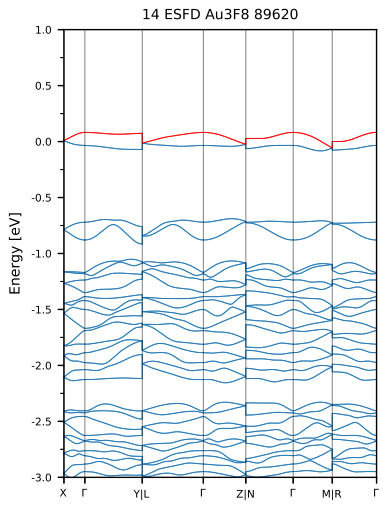

(ev) 14.4.58

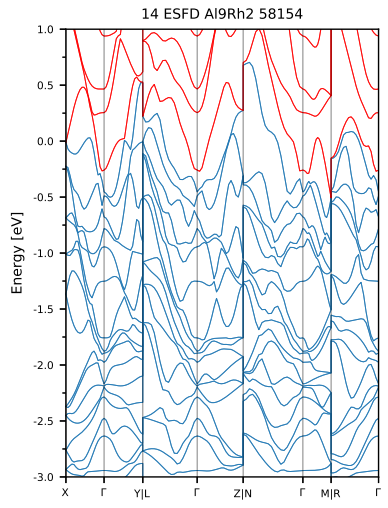

(ew) 14.4.59

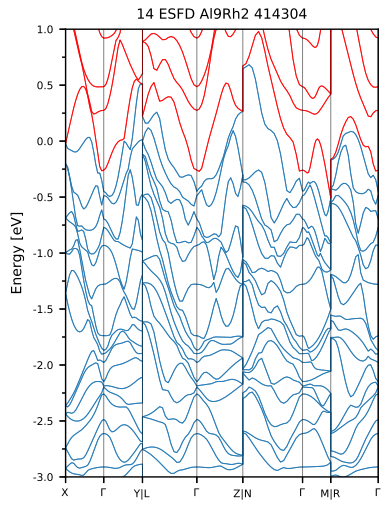

(ex) 14.4.60

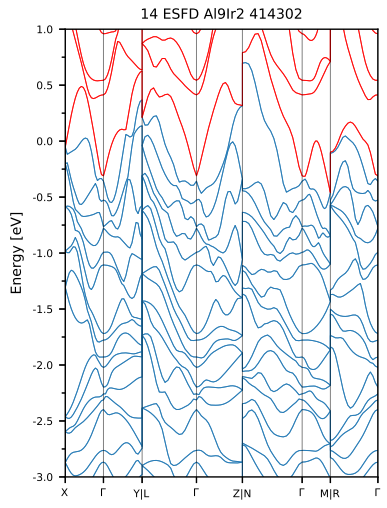

(ey) 14.4.61

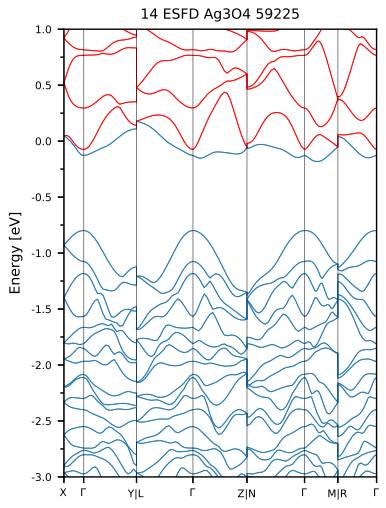

(ez) 14.4.62

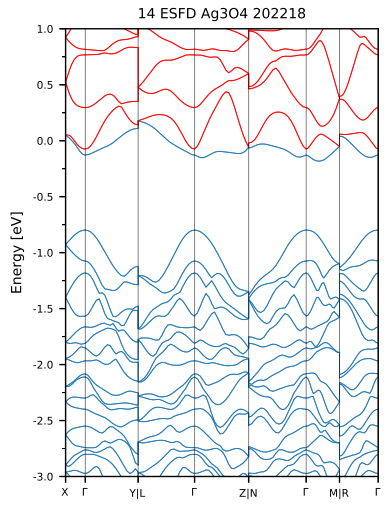

(fa) 14.4.63

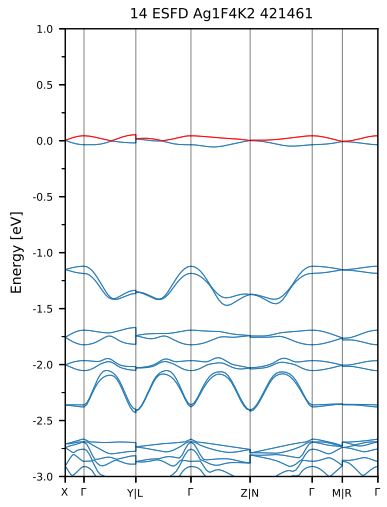

(fb) 14.4.64

Figure 61: Band structures for topological materials in SG 14, part 10/10, from 14.4.51 to 14.4.64. See the list of figures.

### 0.0.9 SG 15

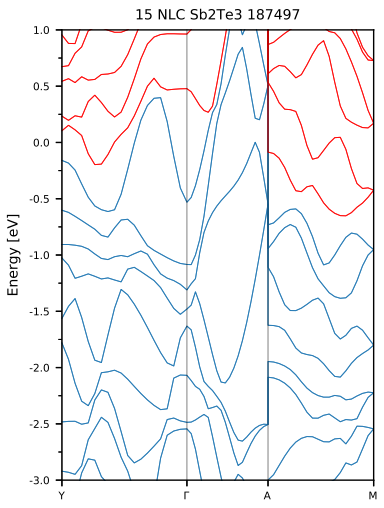

(a) 15.2.1

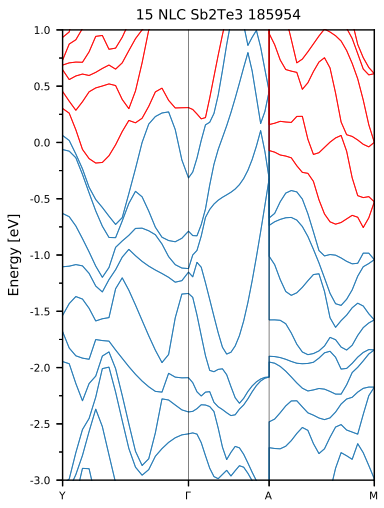

(b) 15.2.2

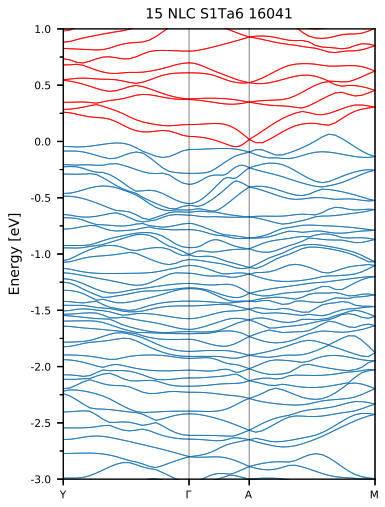

(c) 15.2.3

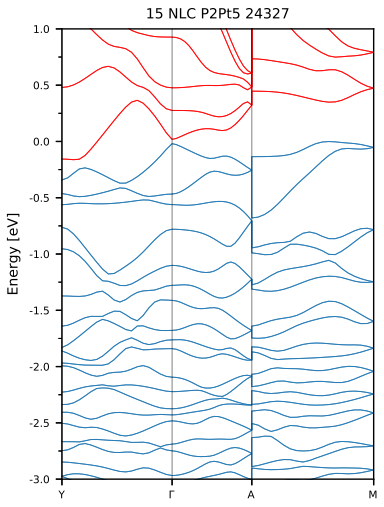

(d) 15.2.4

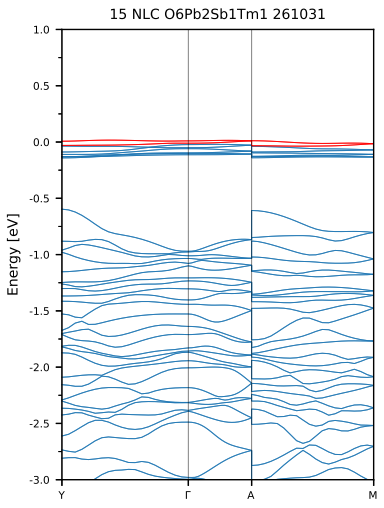

(e) 15.2.5

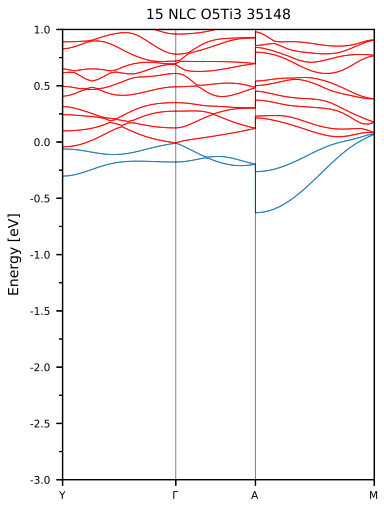

(f) 15.2.6

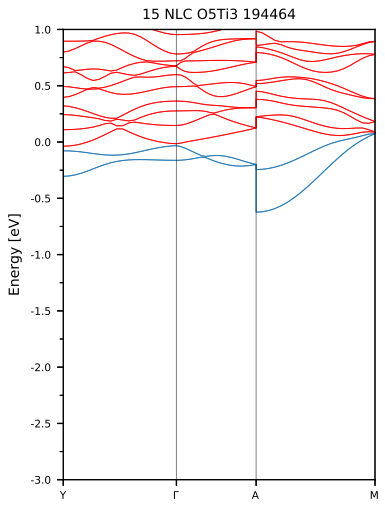

(g) 15.2.7

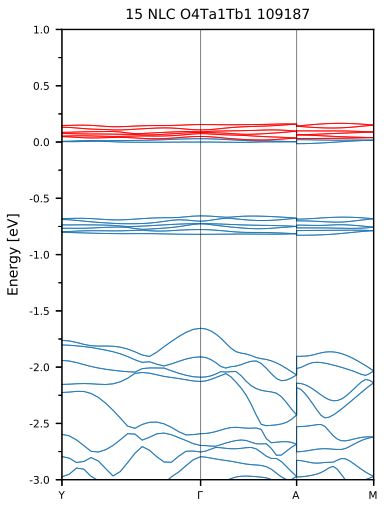

(h) 15.2.8

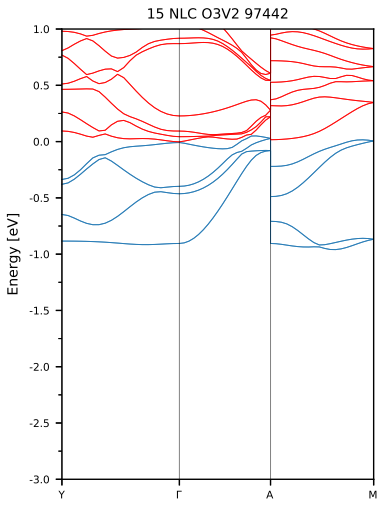

(i) 15.2.9

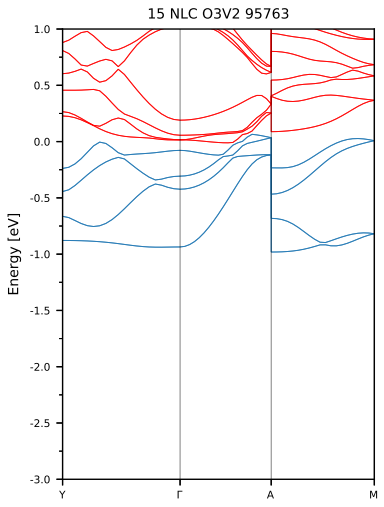

(j) 15.2.10

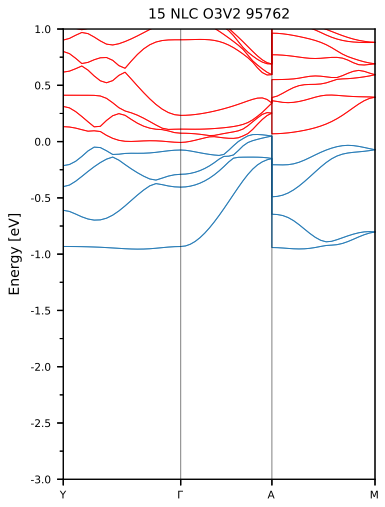

(k) 15.2.11

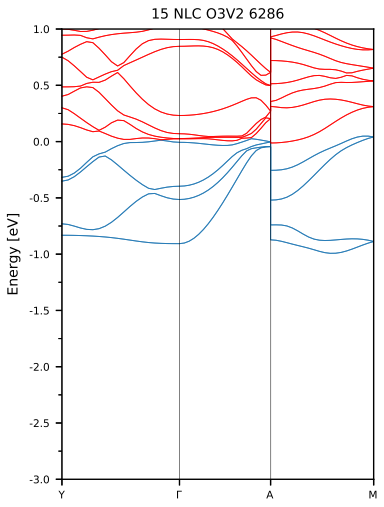

(l) 15.2.12

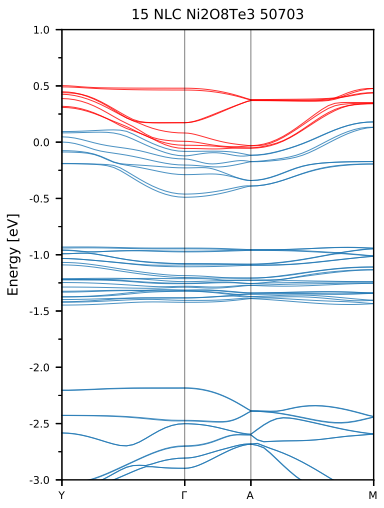

(m) 15.2.13

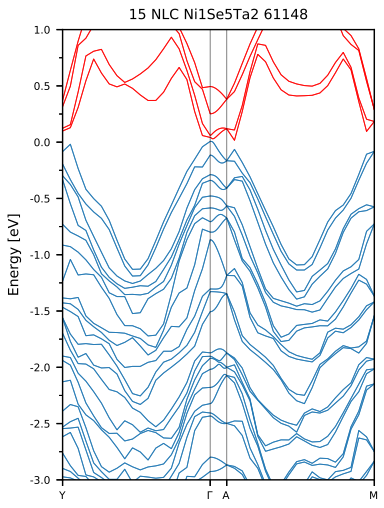

(n) 15.2.14

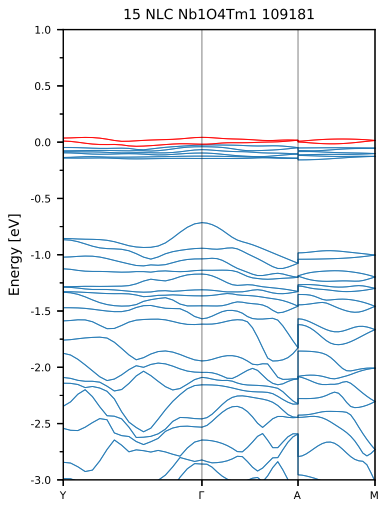

(o) 15.2.15

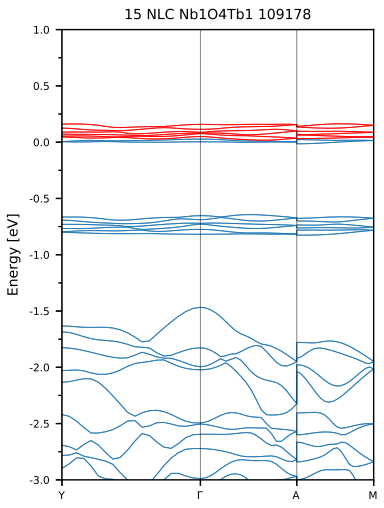

(p) 15.2.16

Figure 62: Band structures for topological materials in SG 15, part 1/17, from 15.2.1 to 15.2.16. See the list of figures.

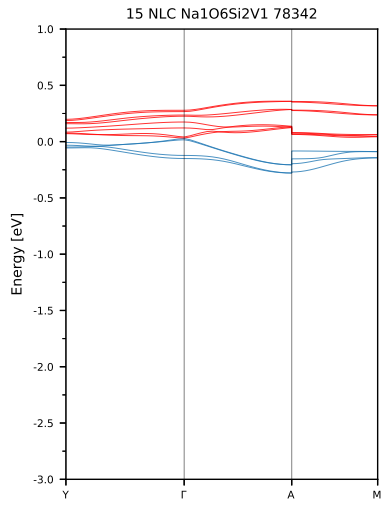

(q) 15.2.17

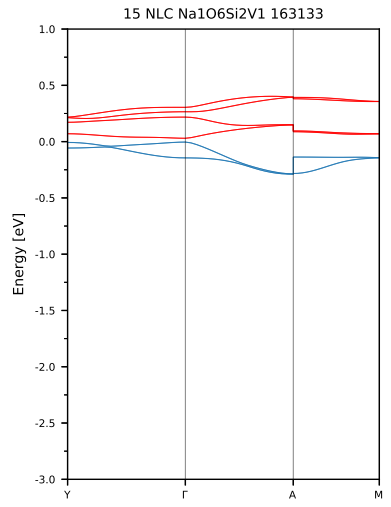

(r) 15.2.18

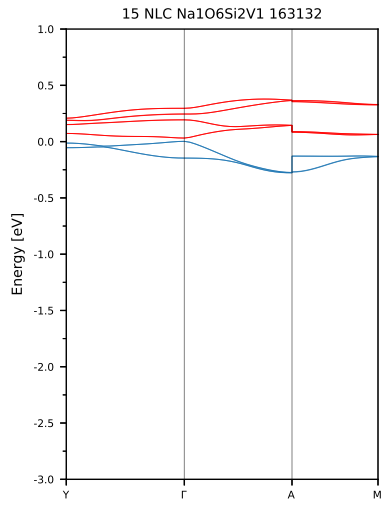

(s) 15.2.19

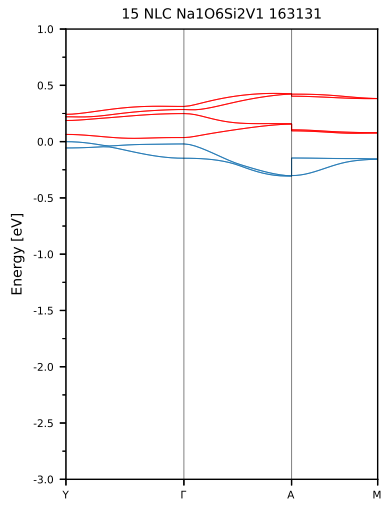

(t) 15.2.20

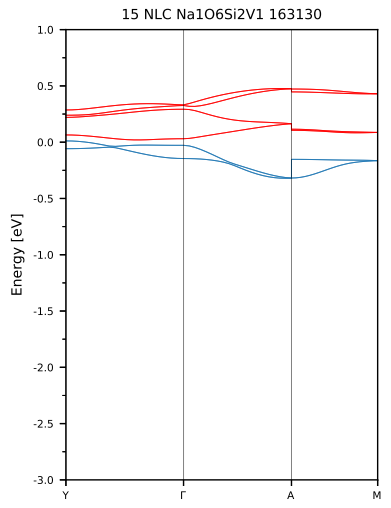

(u) 15.2.21

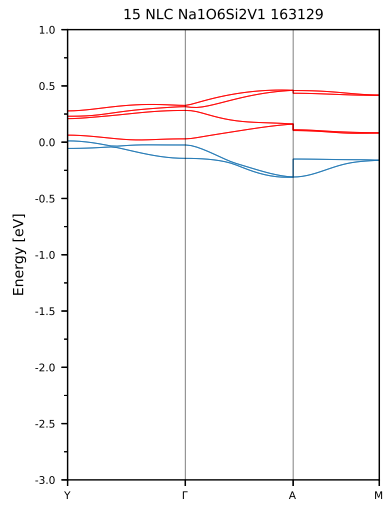

(v) 15.2.22

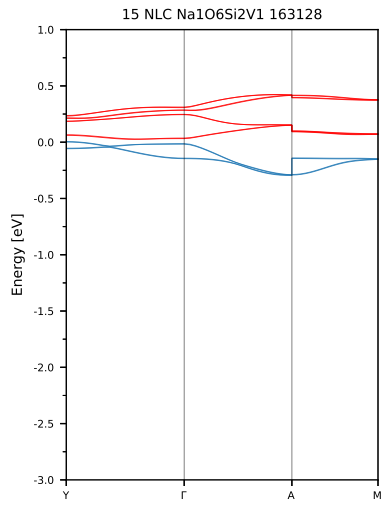

(w) 15.2.23

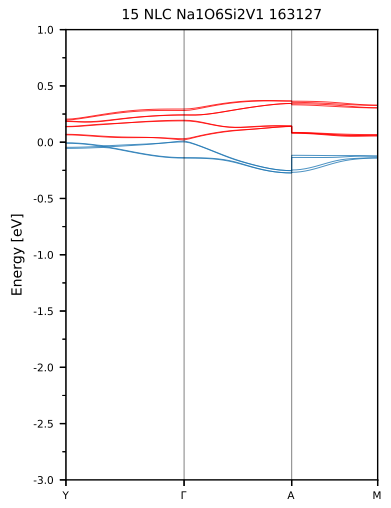

(x) 15.2.24

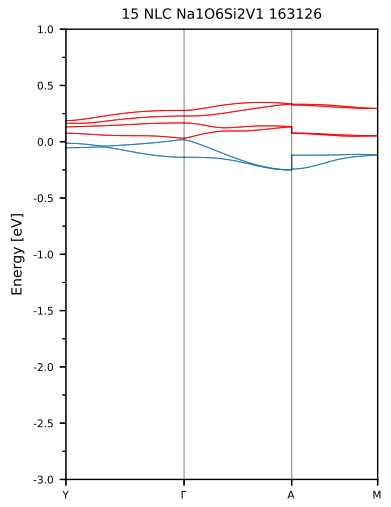

(y) 15.2.25

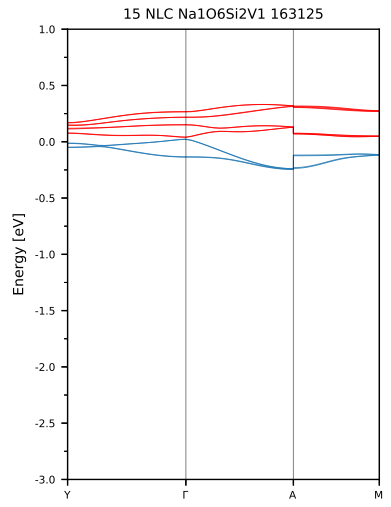

(z) 15.2.26

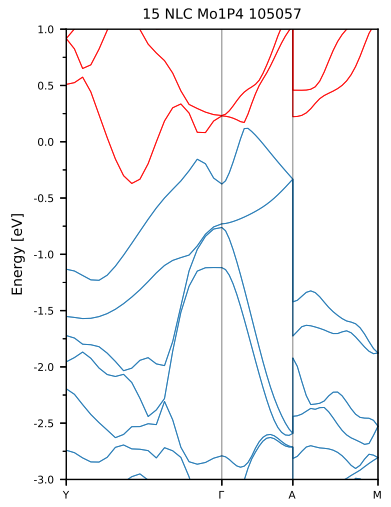

(aa) 15.2.27

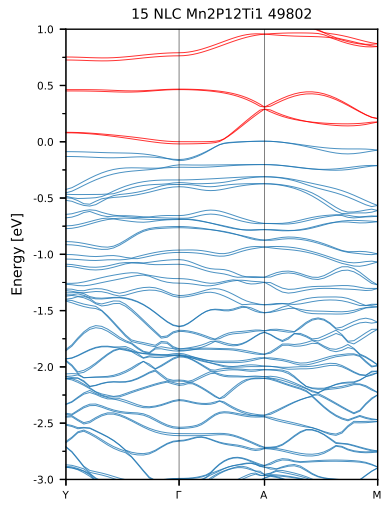

(ab) 15.2.28

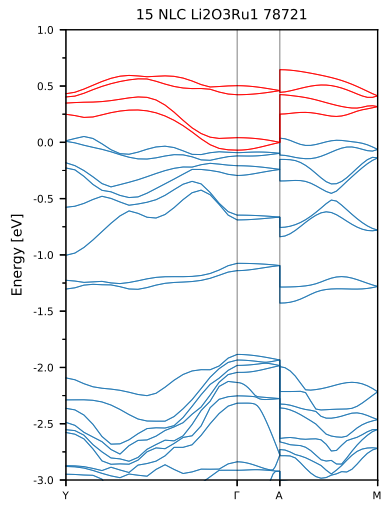

(ac) 15.2.29

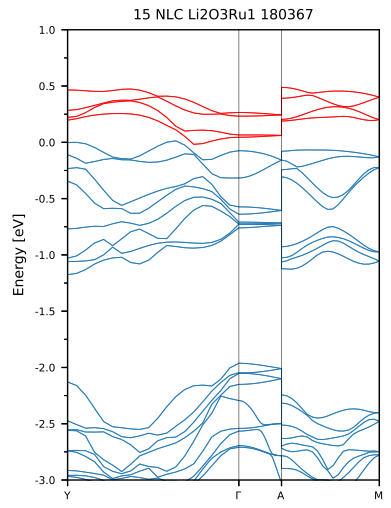

(ad) 15.2.30

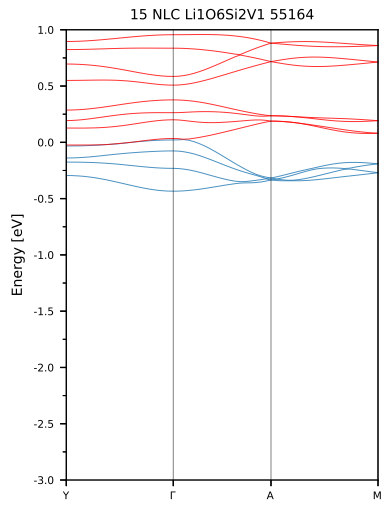

(ae) 15.2.31

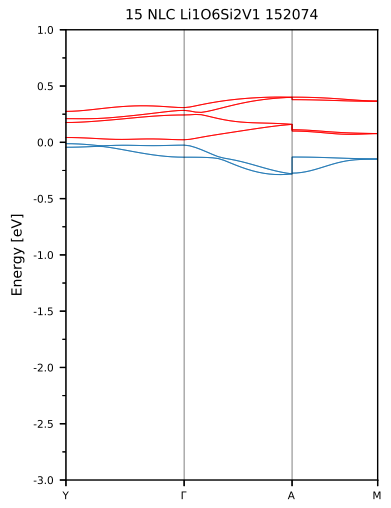

(af) 15.2.32

Figure 63: Band structures for topological materials in SG 15, part 2/17, from 15.2.17 to 15.2.32. See the list of figures.

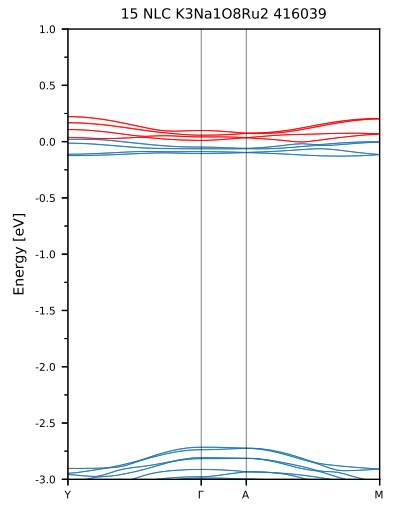

(ag) 15.2.33

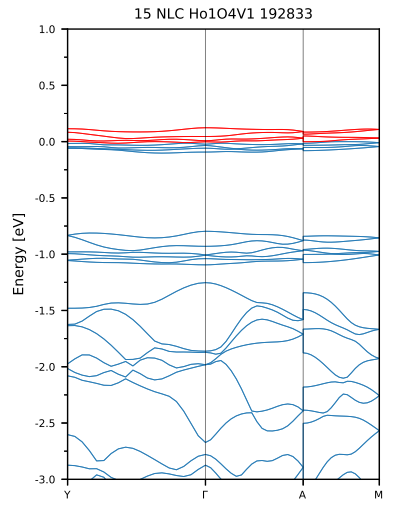

(ah) 15.2.34

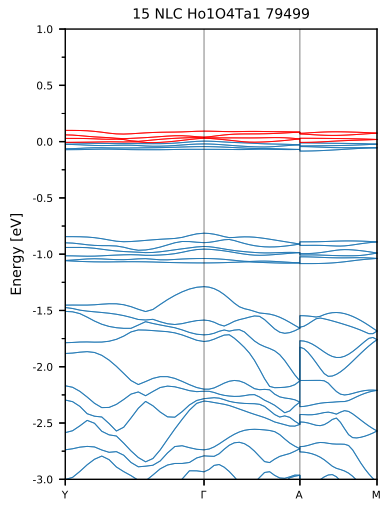

(ai) 15.2.35

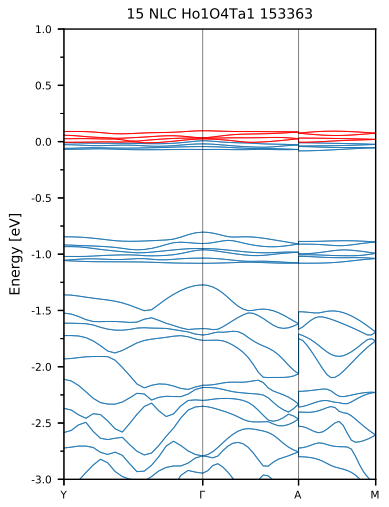

(aj) 15.2.36

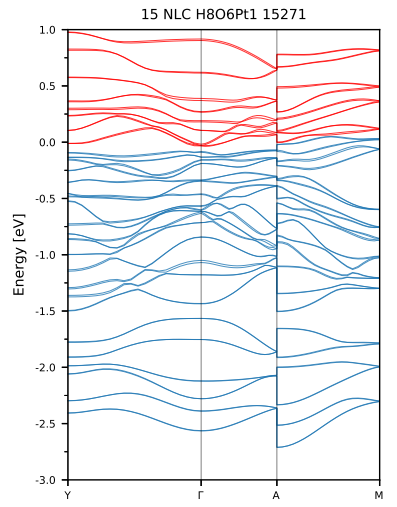

(ak) 15.2.37

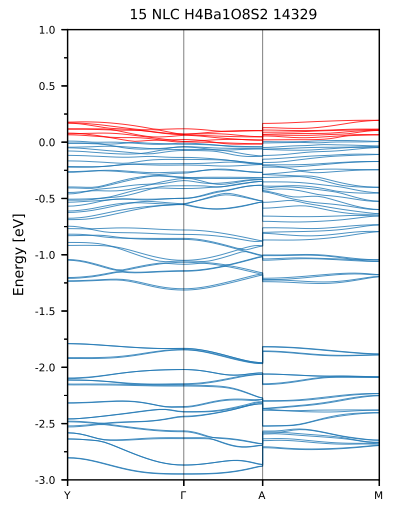

(al) 15.2.38

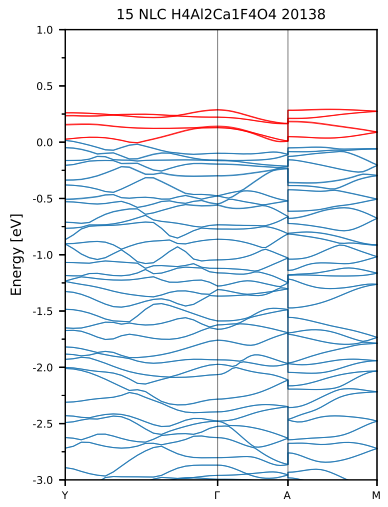

(am) 15.2.39

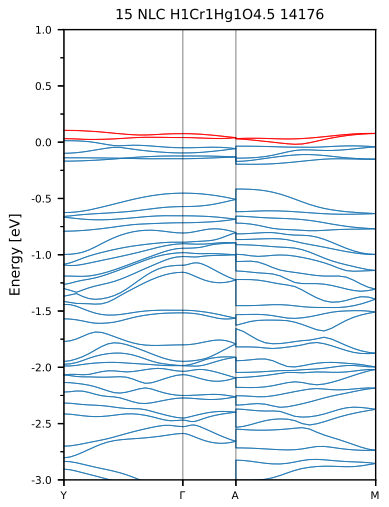

(an) 15.2.40

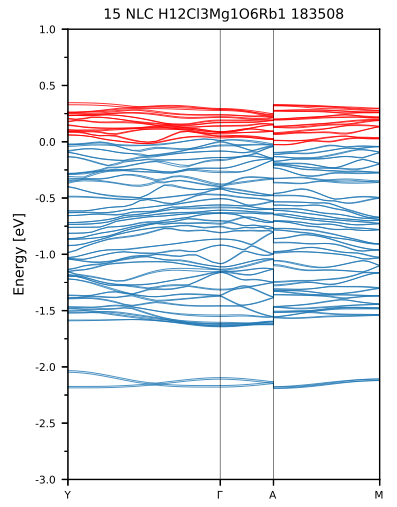

(ao) 15.2.41

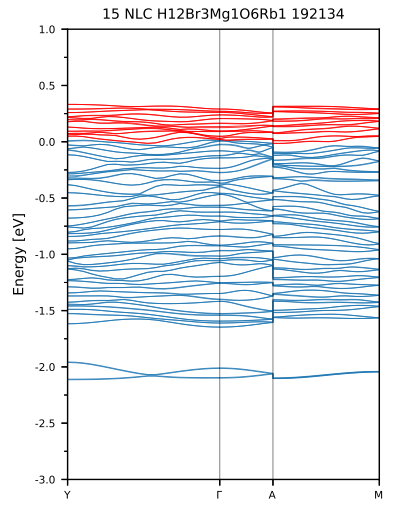

(ap) 15.2.42

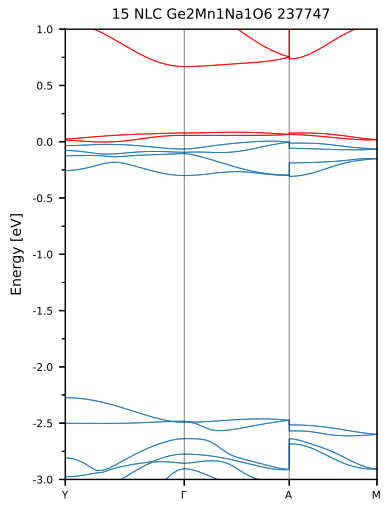

(aq) 15.2.43

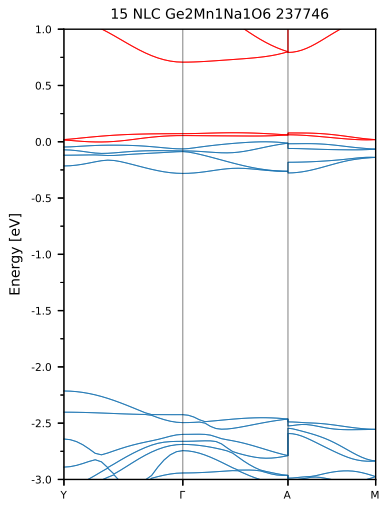

(ar) 15.2.44

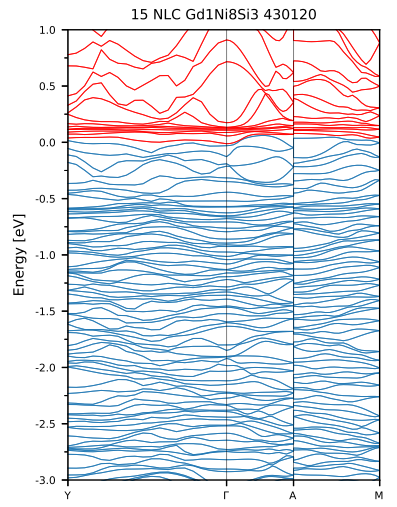

(as) 15.2.45

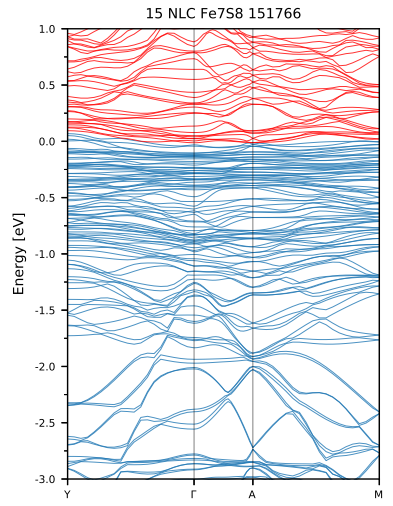

(at) 15.2.46

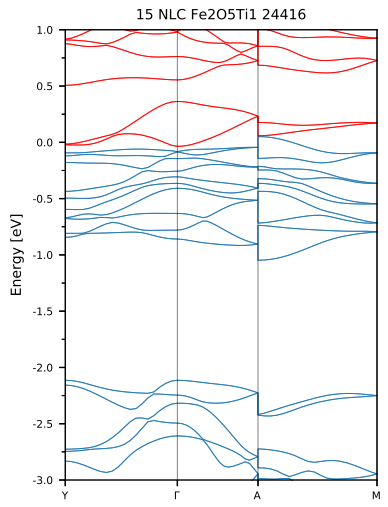

(au) 15.2.47

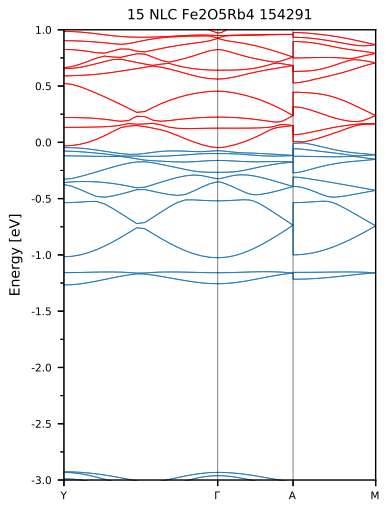

(av) 15.2.48

Figure 64: Band structures for topological materials in SG 15, part 3/17, from 15.2.33 to 15.2.48. See the list of figures.

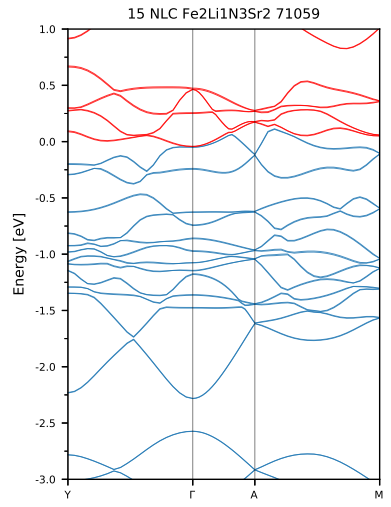

(aw) 15.2.49

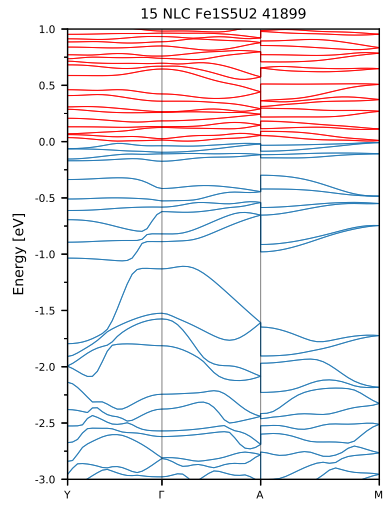

(ax) 15.2.50

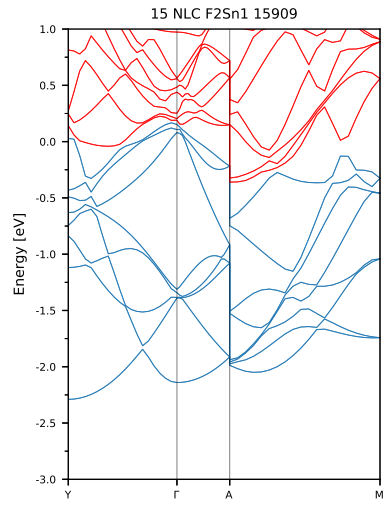

(ay) 15.2.51

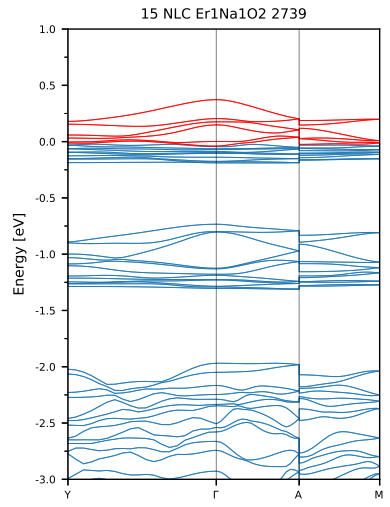

(az) 15.2.52

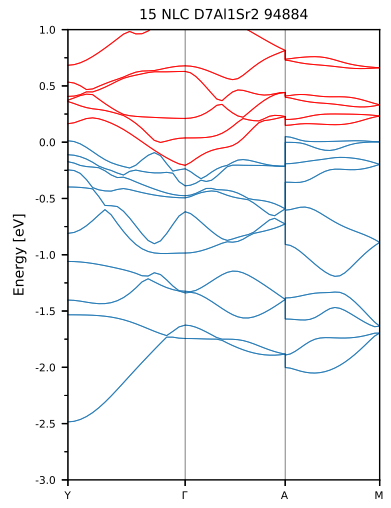

(ba) 15.2.53

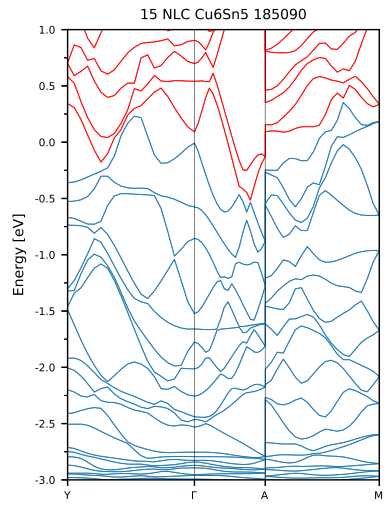

(bb) 15.2.54

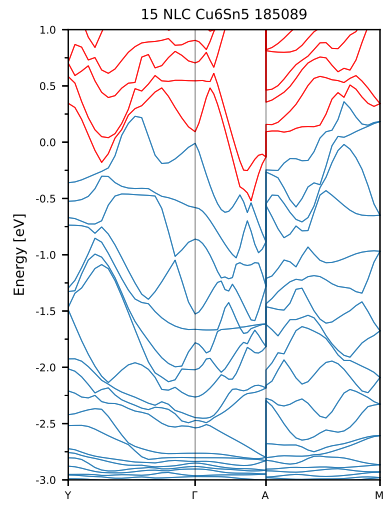

(bc) 15.2.55

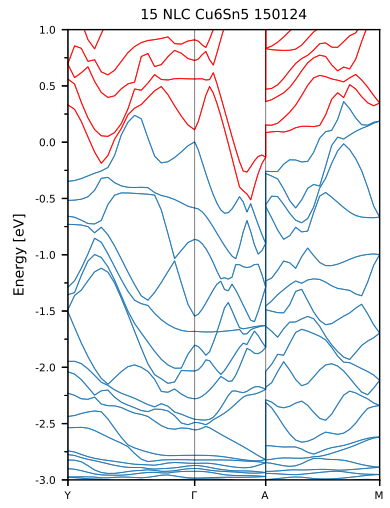

(bd) 15.2.56

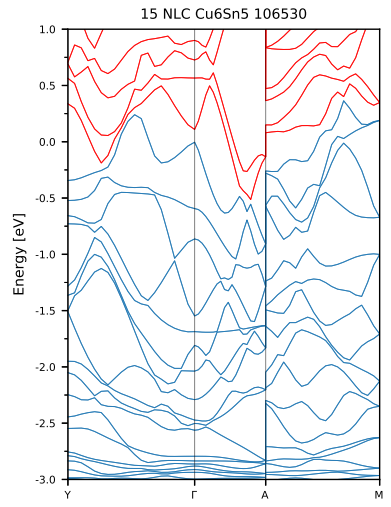

(be) 15.2.57

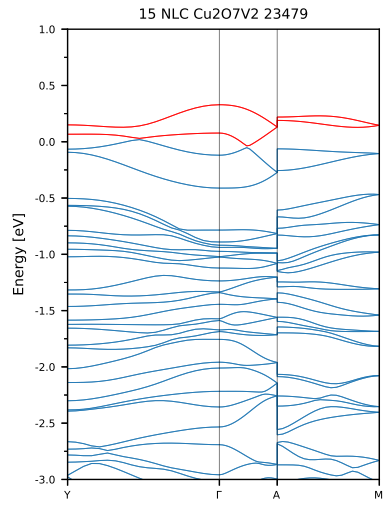

(bf) 15.2.58

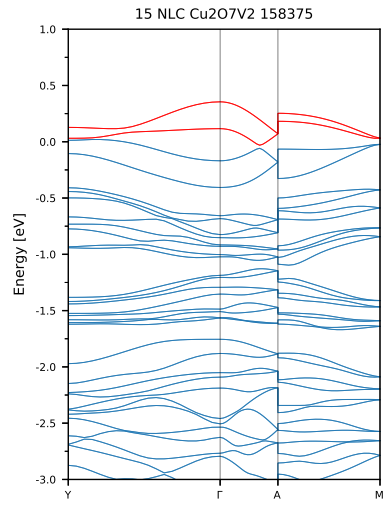

(bg) 15.2.59

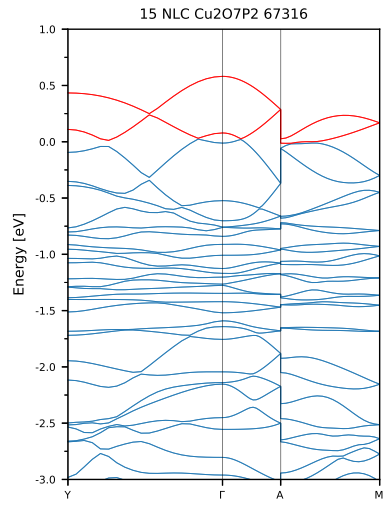

(bh) 15.2.60

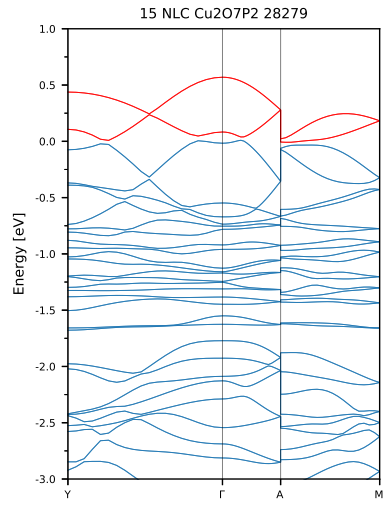

(bi) 15.2.61

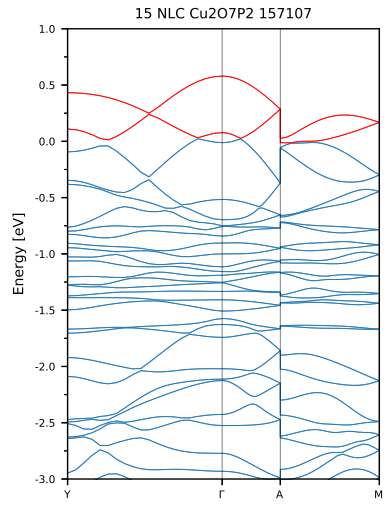

(bj) 15.2.62

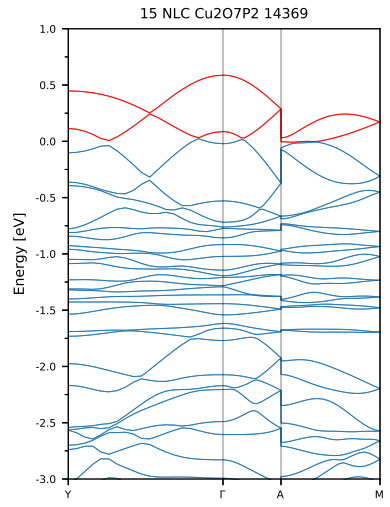

(bk) 15.2.63

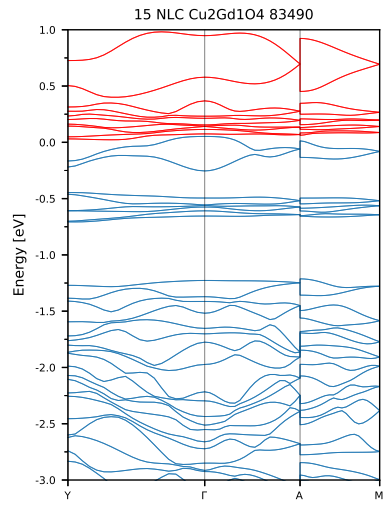

(bl) 15.2.64

Figure 65: Band structures for topological materials in SG 15, part 4/17, from 15.2.49 to 15.2.64. See the list of figures.

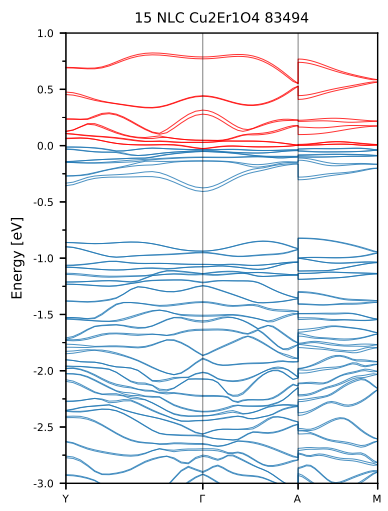

(bm) 15.2.65

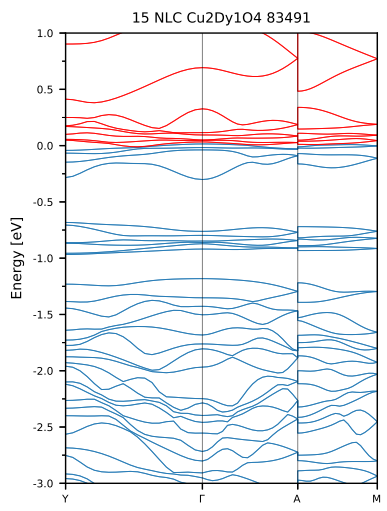

(bn) 15.2.66

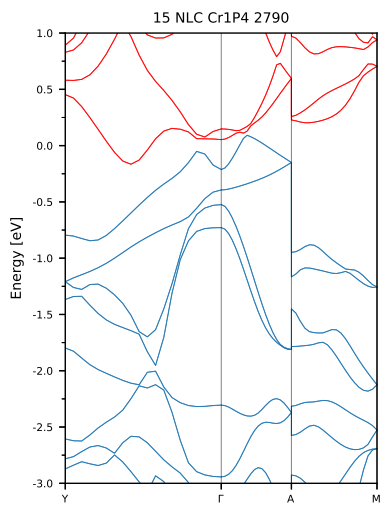

(bo) 15.2.67

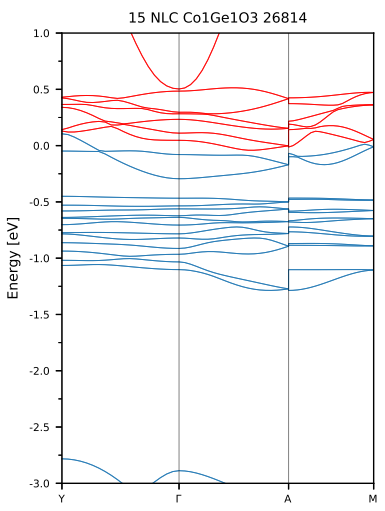

(bp) 15.2.68

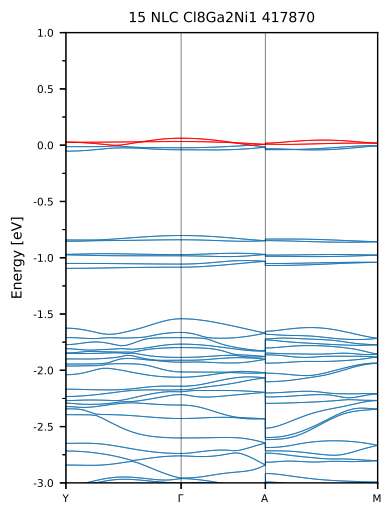

(bq) 15.2.69

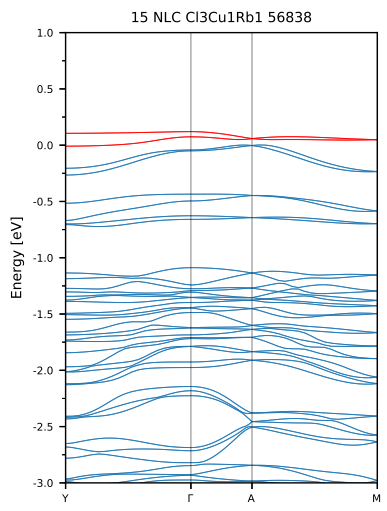

(br) 15.2.70

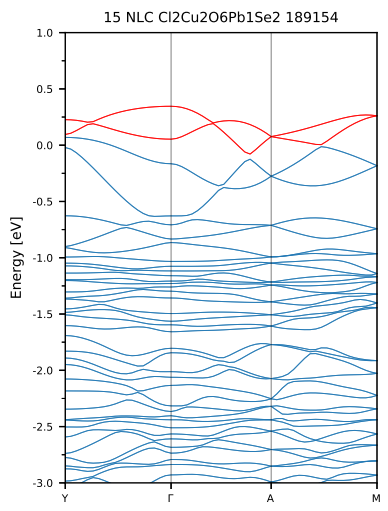

(bs) 15.2.71

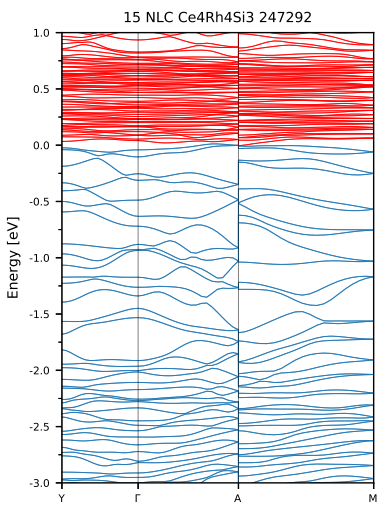

(bt) 15.2.72

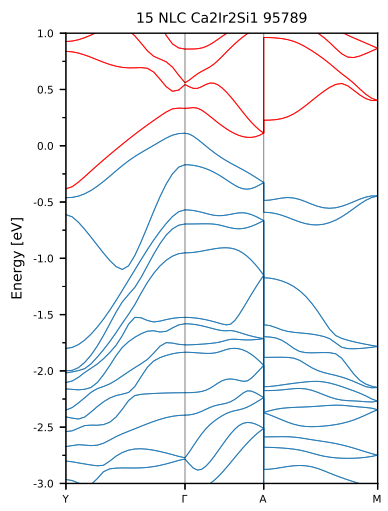

(bu) 15.2.73

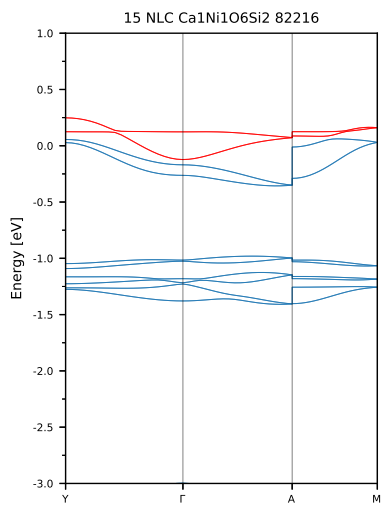

(bv) 15.2.74

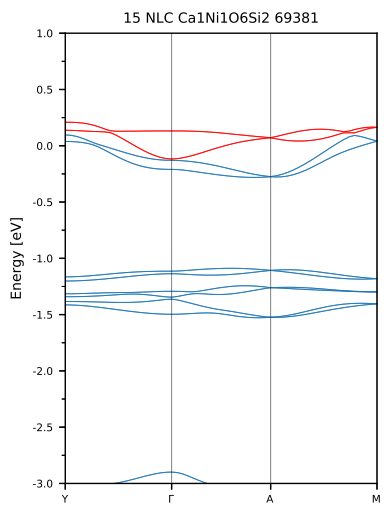

(bw) 15.2.75

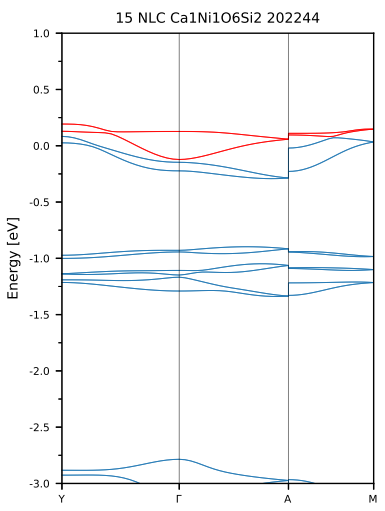

(bx) 15.2.76

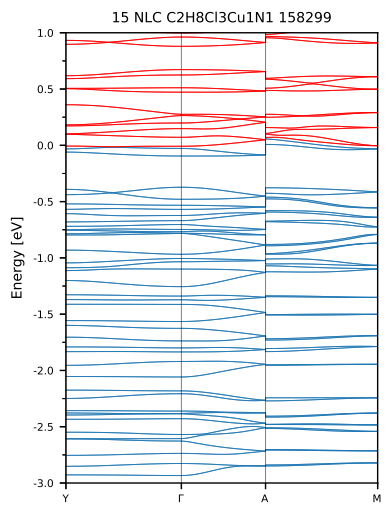

(by) 15.2.77

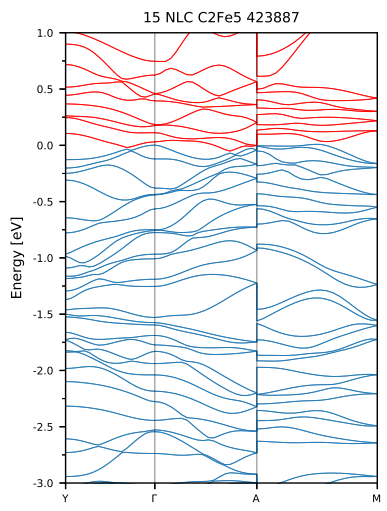

(bz) 15.2.78

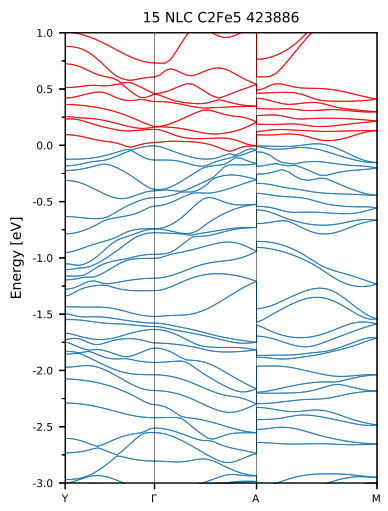

(ca) 15.2.79

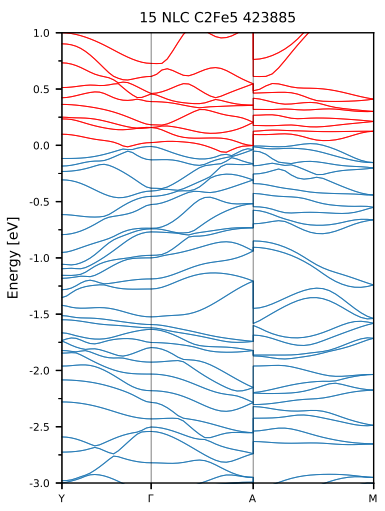

(cb) 15.2.80

Figure 66: Band structures for topological materials in SG 15, part 5/17, from 15.2.65 to 15.2.80. See the list of figures.

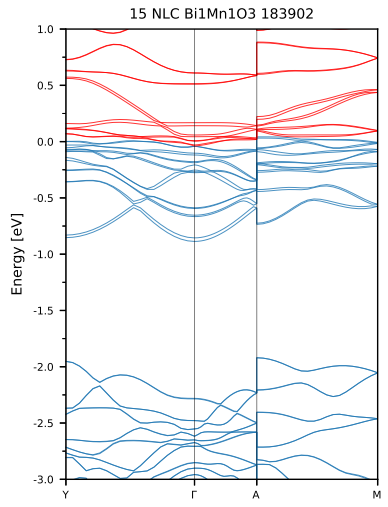

(cc) 15.2.81

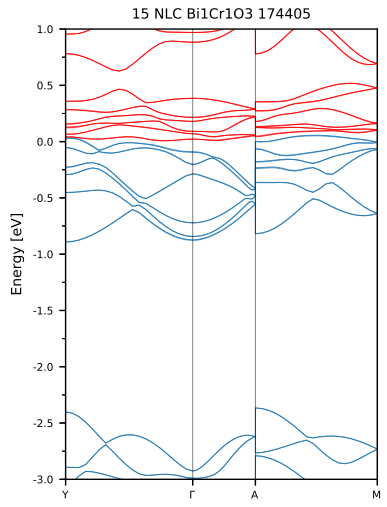

(cd) 15.2.82

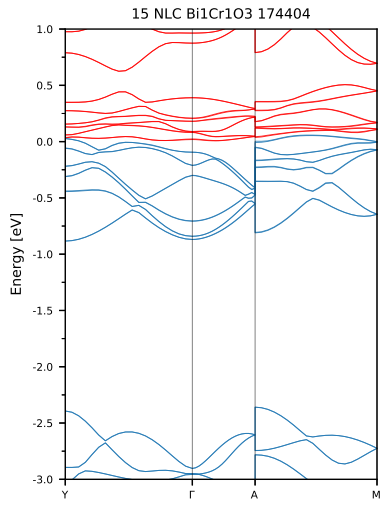

(ce) 15.2.83

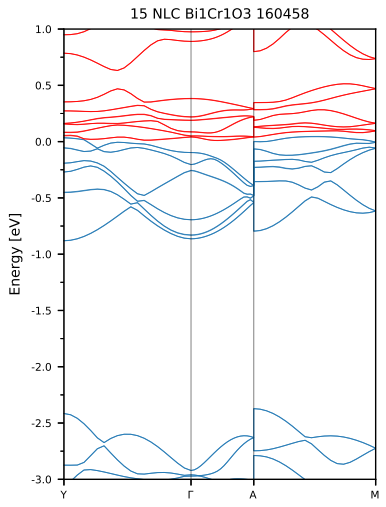

(cf) 15.2.84

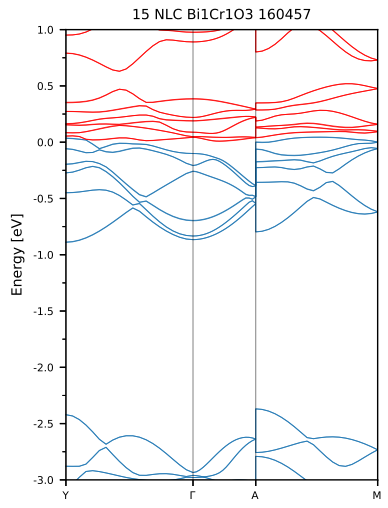

(cg) 15.2.85

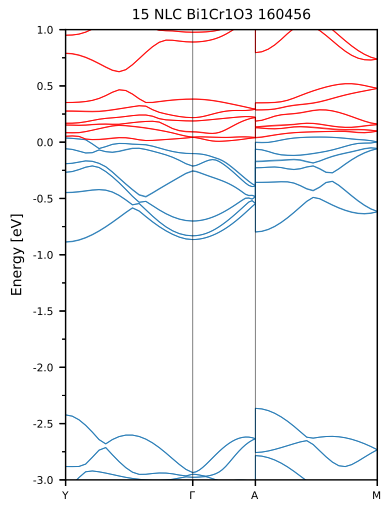

(ch) 15.2.86

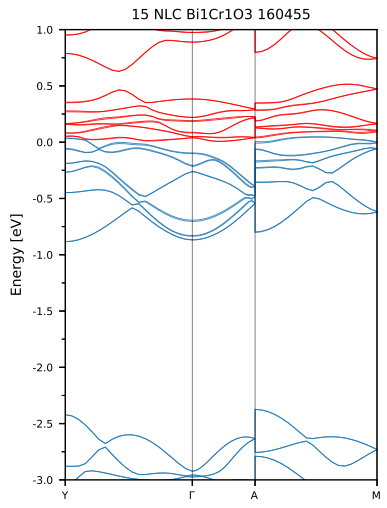

(ci) 15.2.87

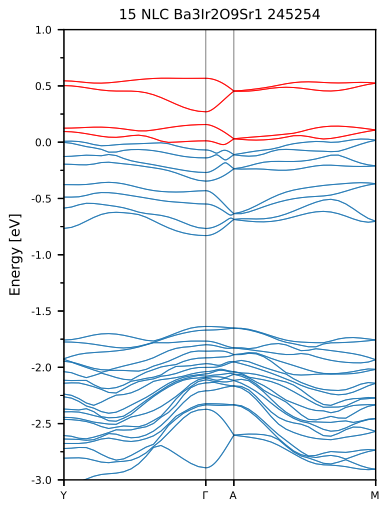

(cj) 15.2.88

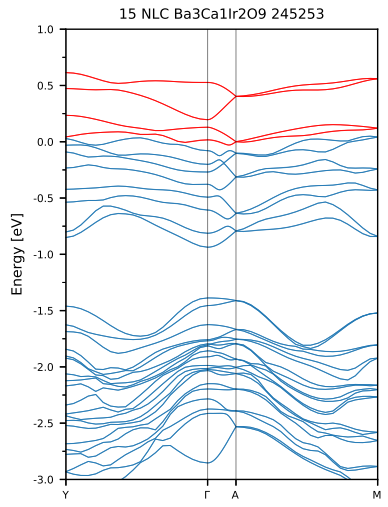

(ck) 15.2.89

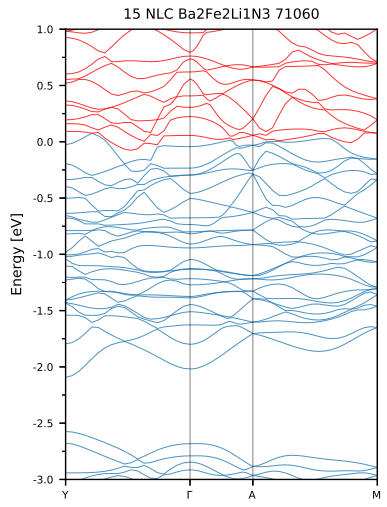

(cl) 15.2.90

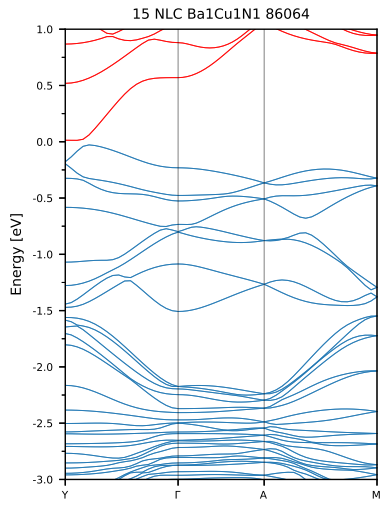

(cm) 15.2.91

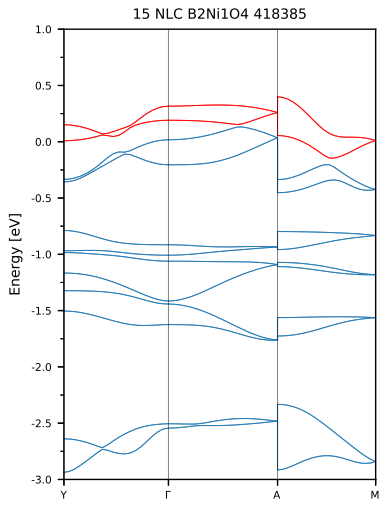

(cn) 15.2.92

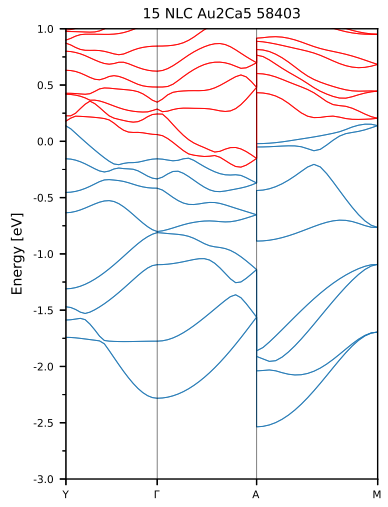

(co) 15.2.93

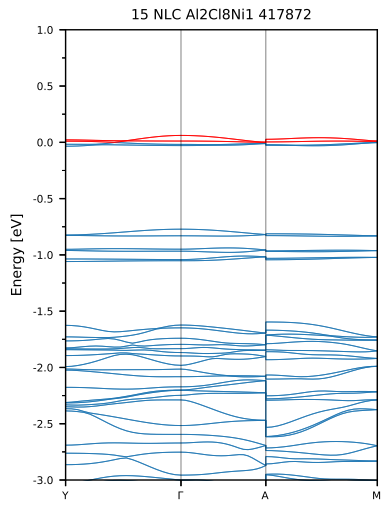

(cp) 15.2.94

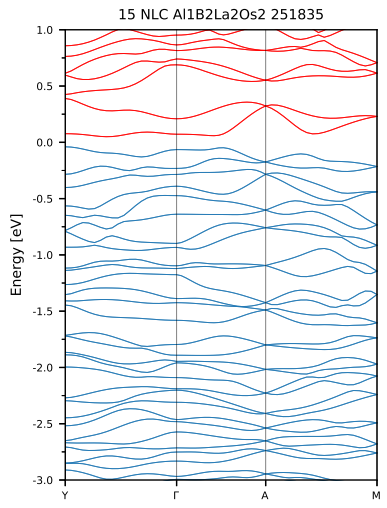

(cq) 15.2.95

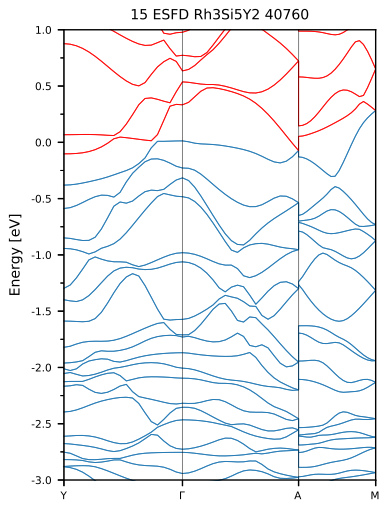

(cr) 15.4.1

Figure 67: Band structures for topological materials in SG 15, part 6/17, from 15.2.81 to 15.4.1. See the list of figures.

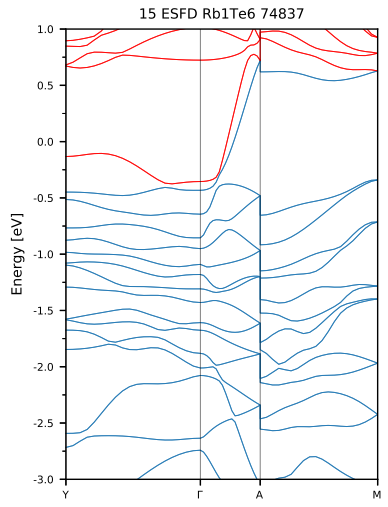

(cs) 15.4.2

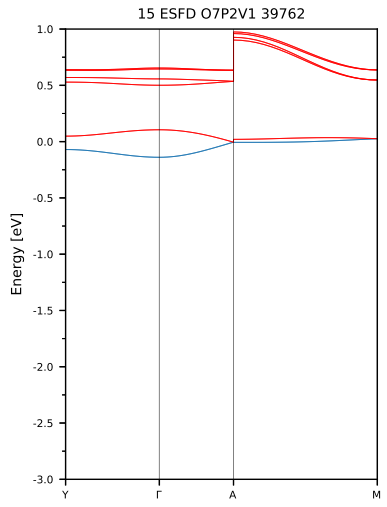

(ct) 15.4.3

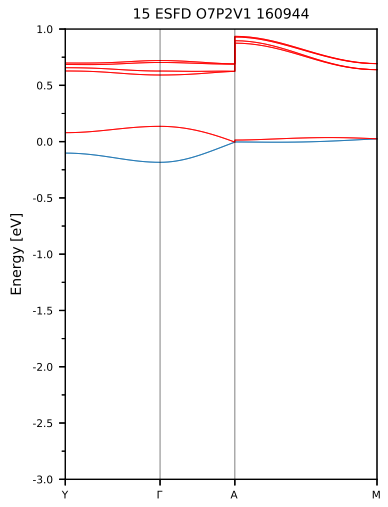

(cu) 15.4.4

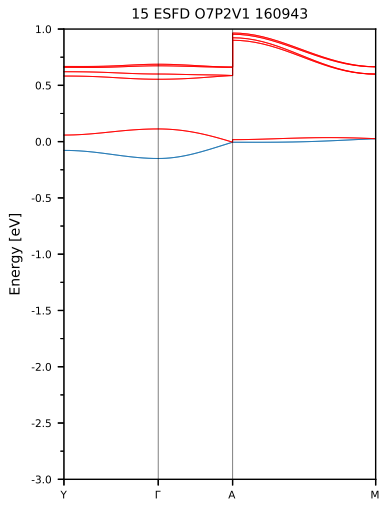

(cv) 15.4.5

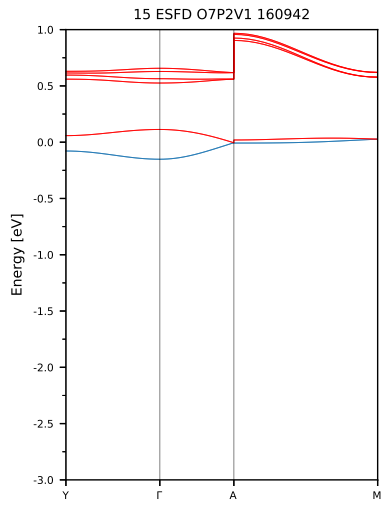

(cw) 15.4.6

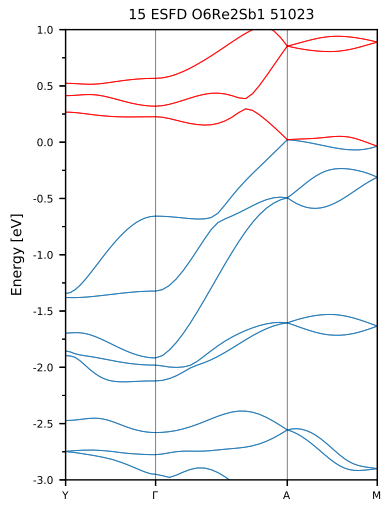

(cx) 15.4.7

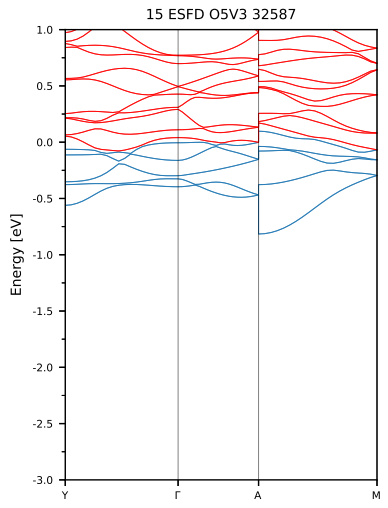

(cy) 15.4.8

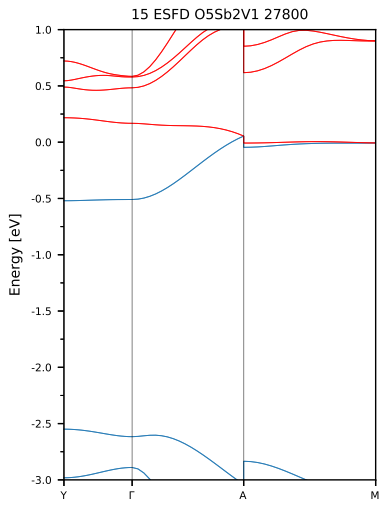

(cz) 15.4.9

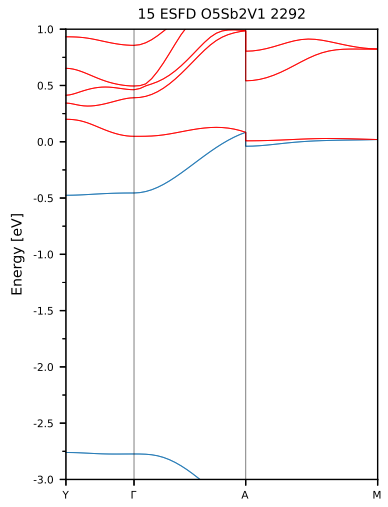

(da) 15.4.10

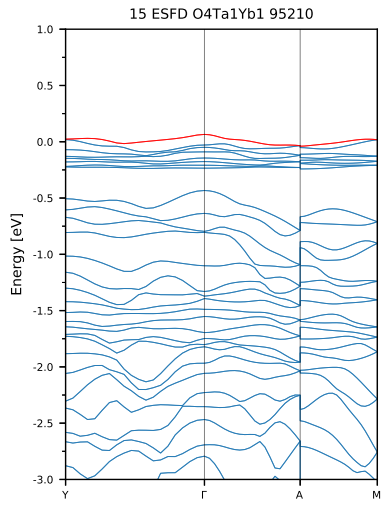

(db) 15.4.11

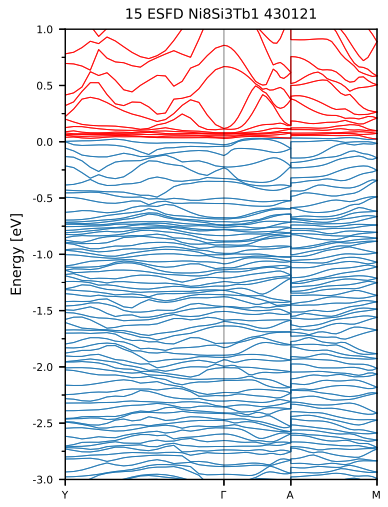

(dc) 15.4.12

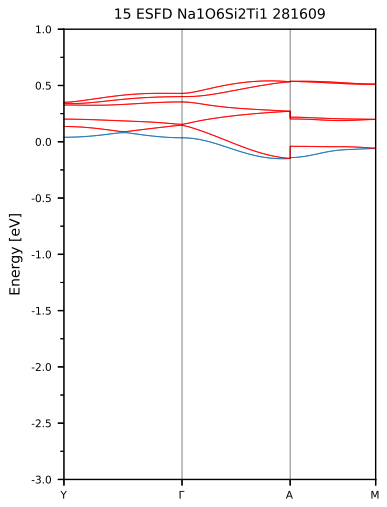

(dd) 15.4.13

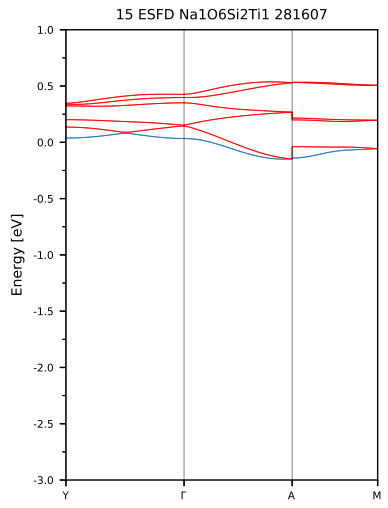

(de) 15.4.14

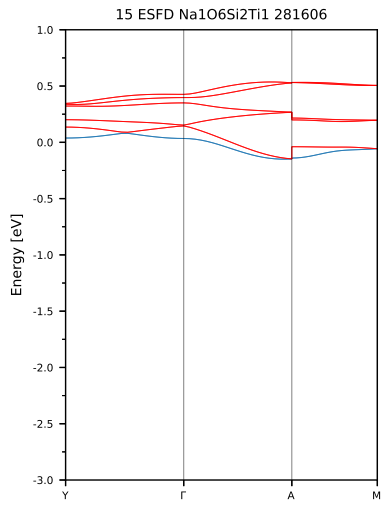

(df) 15.4.15

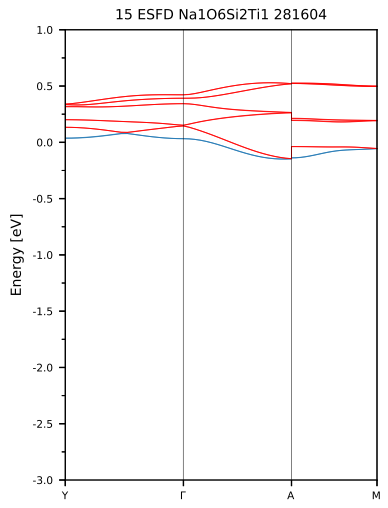

(dg) 15.4.16

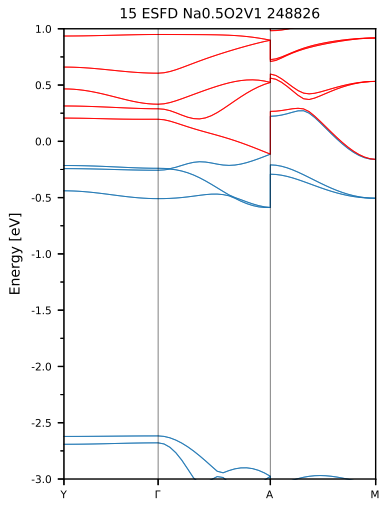

(dh) 15.4.17

Figure 68: Band structures for topological materials in SG 15, part 7/17, from 15.4.2 to 15.4.17. See the list of figures.

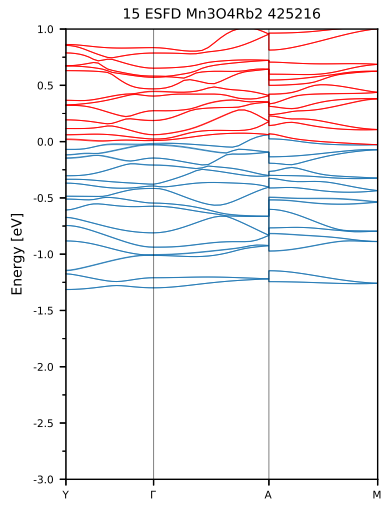

(di) 15.4.18

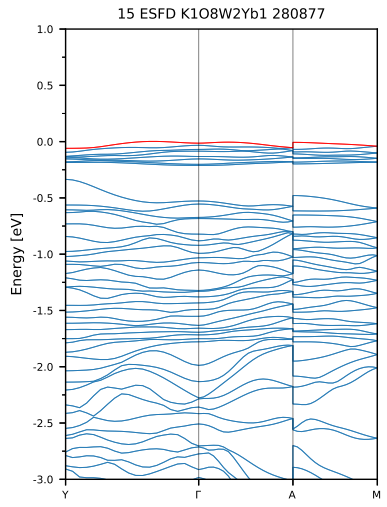

(dj) 15.4.19

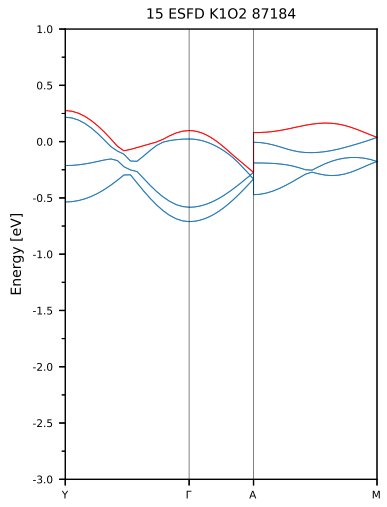

(dk) 15.4.20

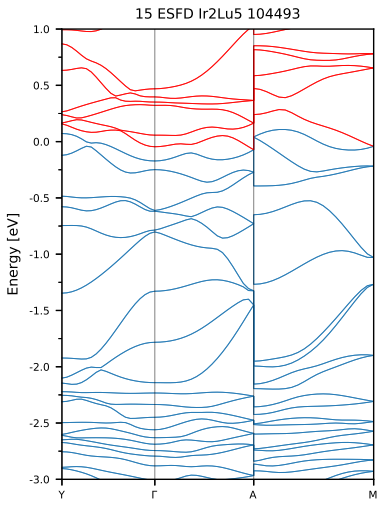

(dl) 15.4.21

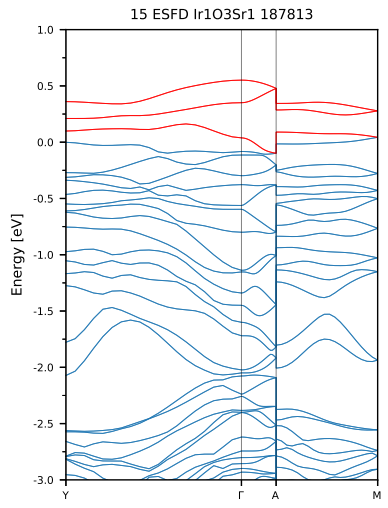

(dm) 15.4.22

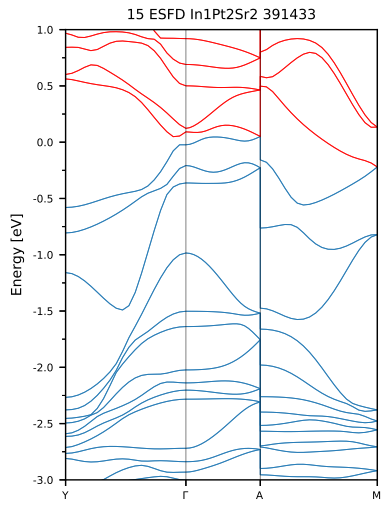

(dn) 15.4.23

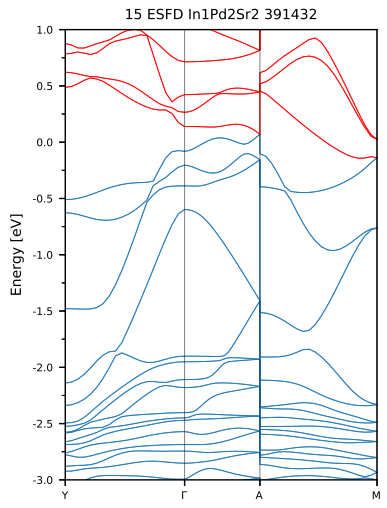

(do) 15.4.24

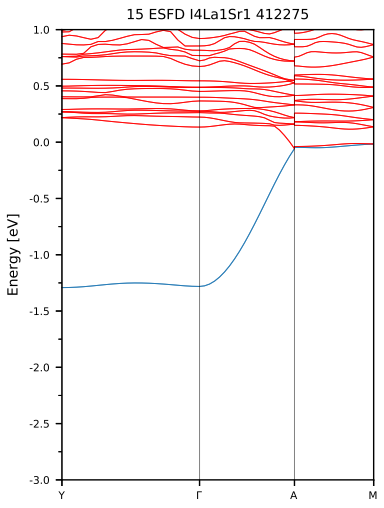

(dp) 15.4.25

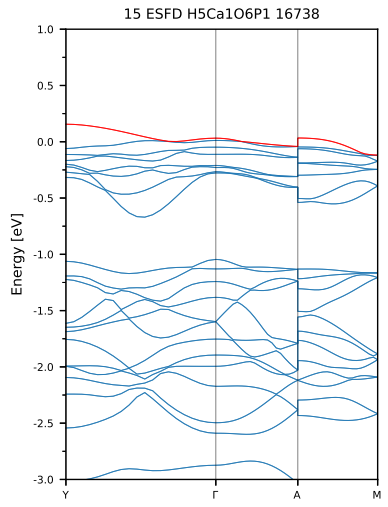

(dq) 15.4.26

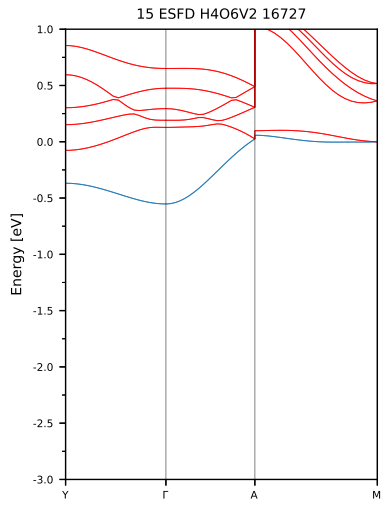

(dr) 15.4.27

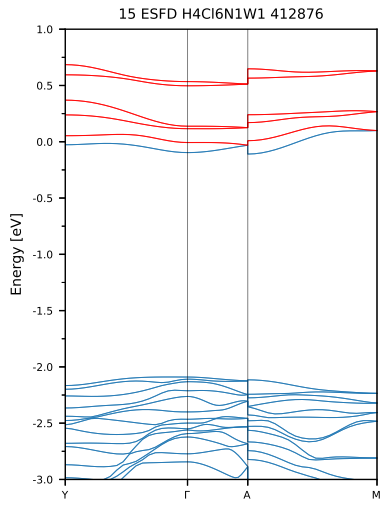

(ds) 15.4.28

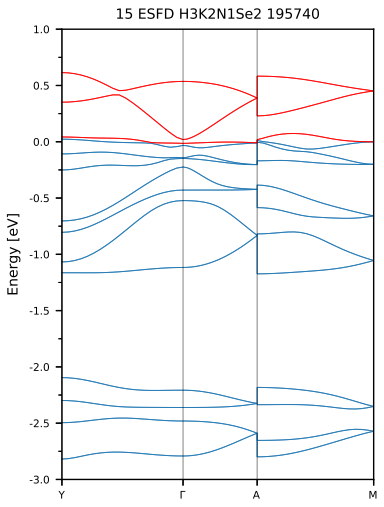

(dt) 15.4.29

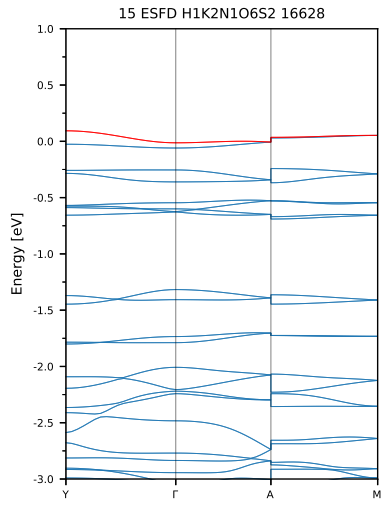

(du) 15.4.30

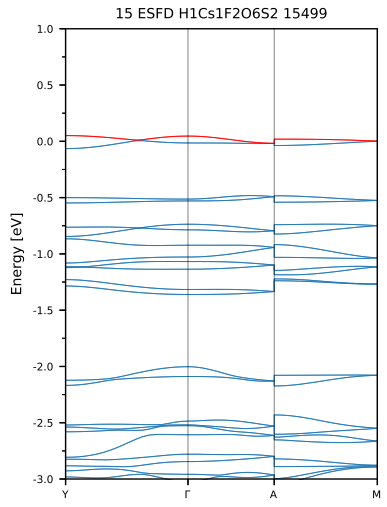

(dv) 15.4.31

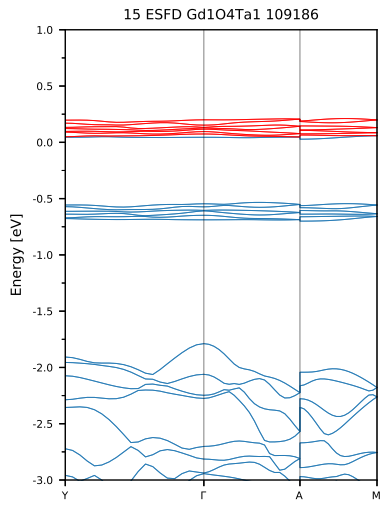

(dw) 15.4.32

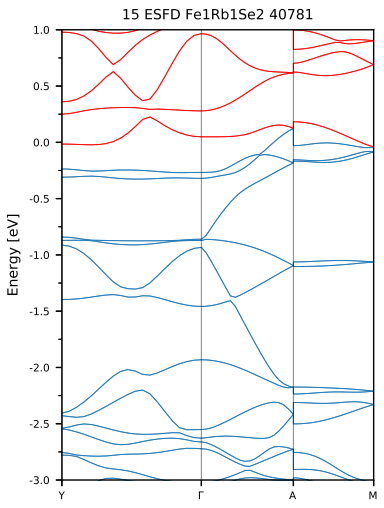

(dx) 15.4.33

Figure 69: Band structures for topological materials in SG 15, part 8/17, from 15.4.18 to 15.4.33. See the list of figures.

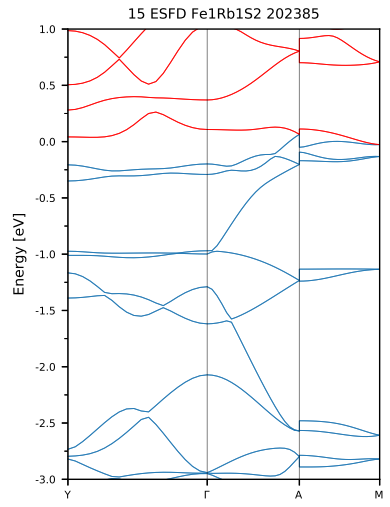

(dy) 15.4.34

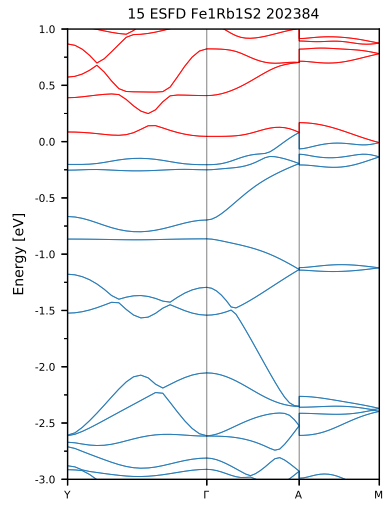

(dz) 15.4.35

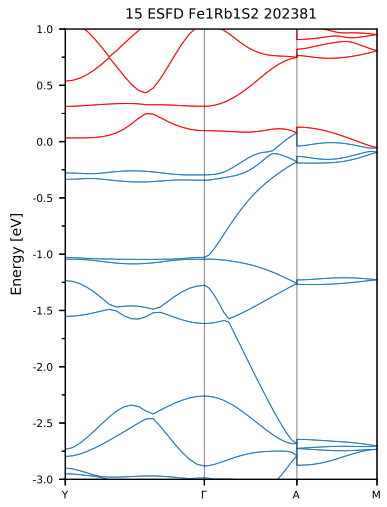

(ea) 15.4.36

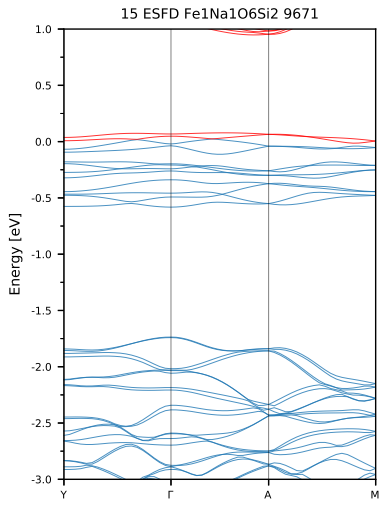

(eb) 15.4.37

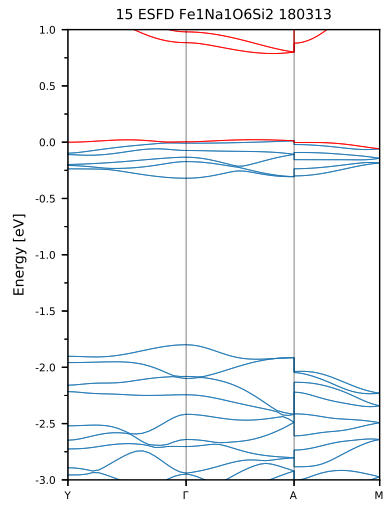

(ec) 15.4.38

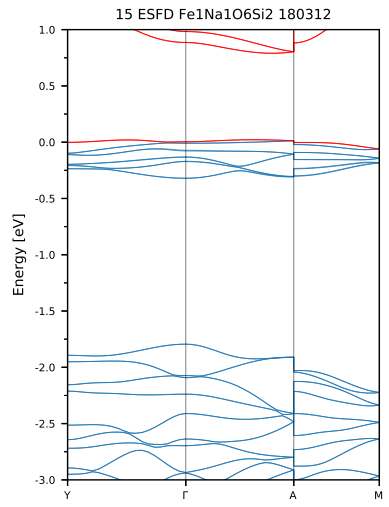

(ed) 15.4.39

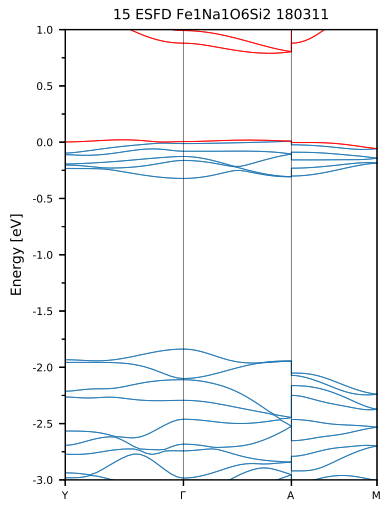

(ee) 15.4.40

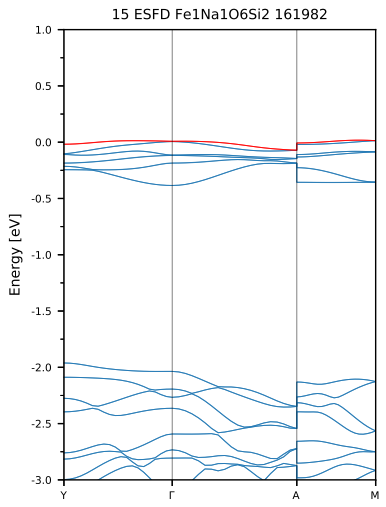

(ef) 15.4.41

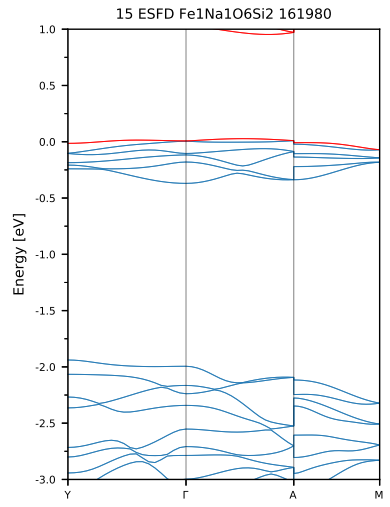

(eg) 15.4.42

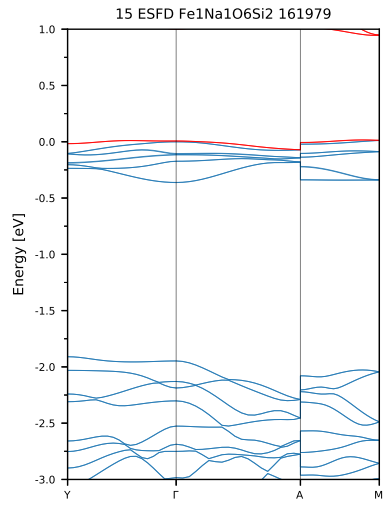

(eh) 15.4.43

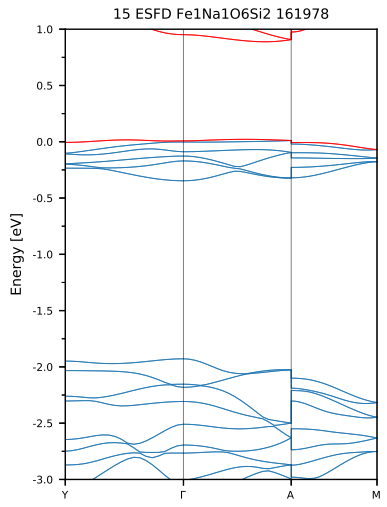

(ei) 15.4.44

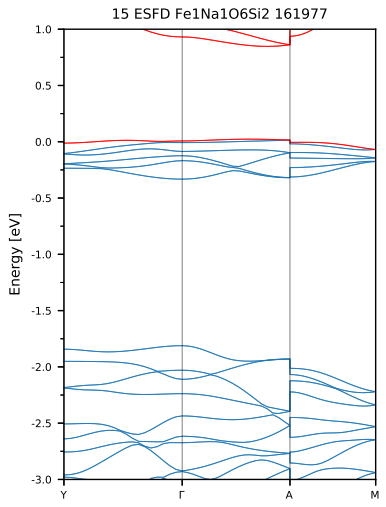

(ej) 15.4.45

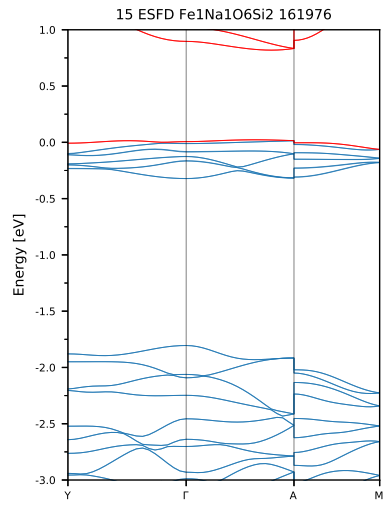

(ek) 15.4.46

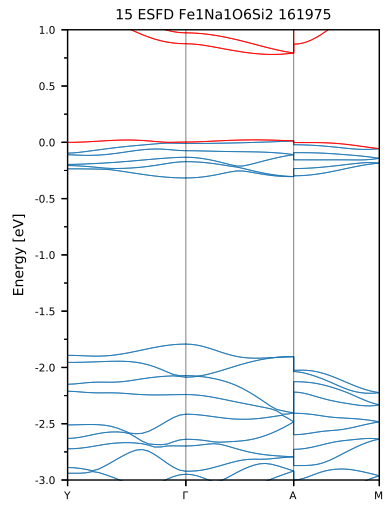

(el) 15.4.47

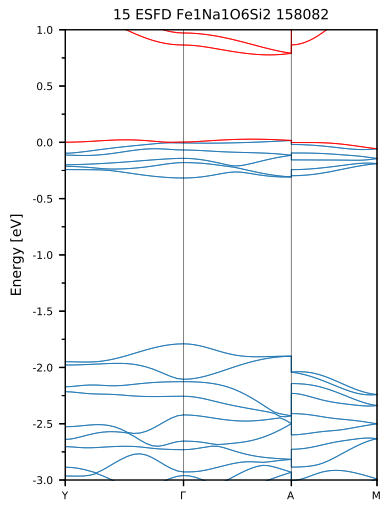

(em) 15.4.48

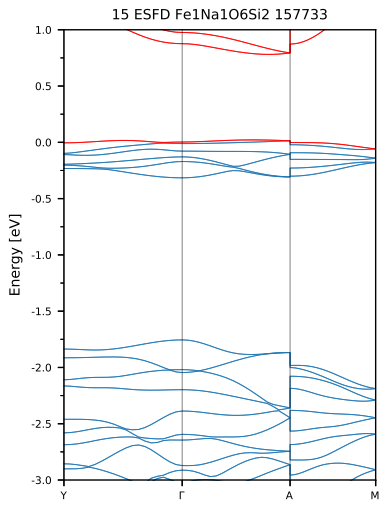

(en) 15.4.49

Figure 70: Band structures for topological materials in SG 15, part 9/17, from 15.4.34 to 15.4.49. See the list of figures.

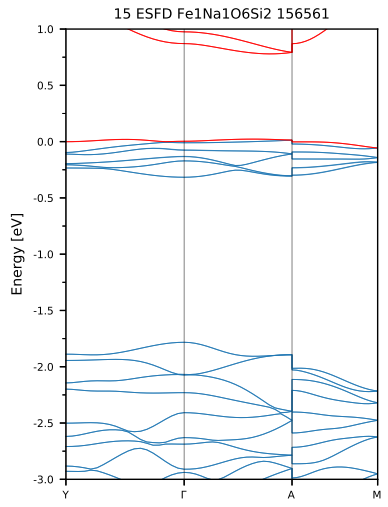

(eo) 15.4.50

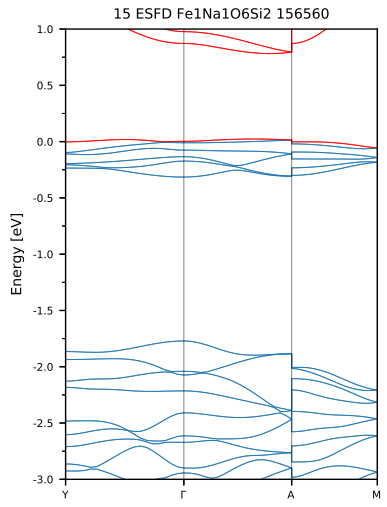

(ep) 15.4.51

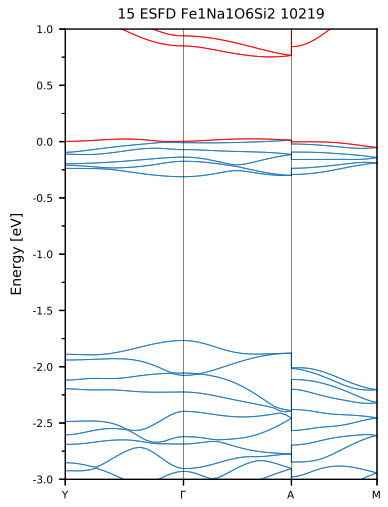

(eq) 15.4.52

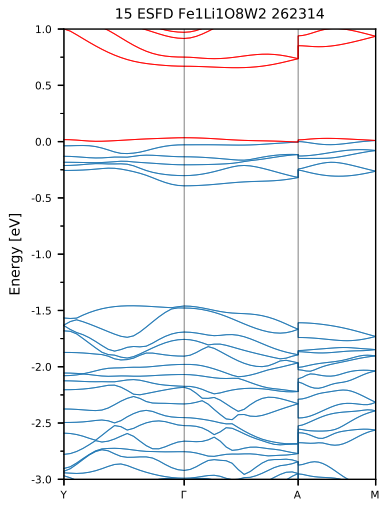

(er) 15.4.53

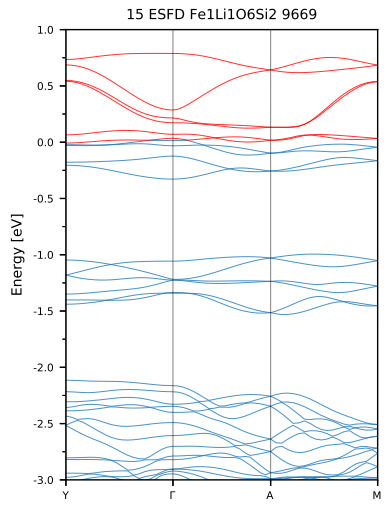

(es) 15.4.54

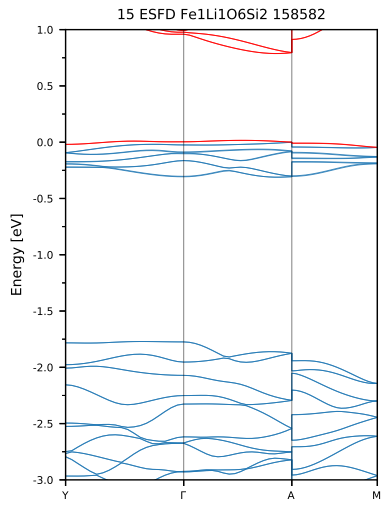

(et) 15.4.55

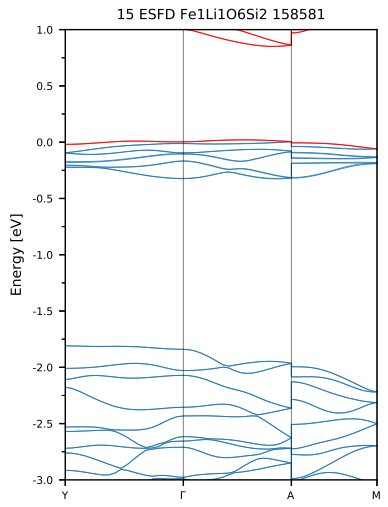

(eu) 15.4.56

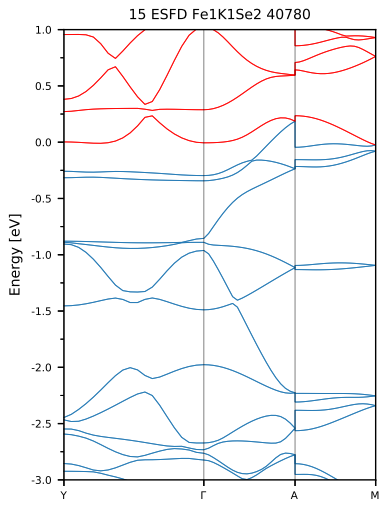

(ev) 15.4.57

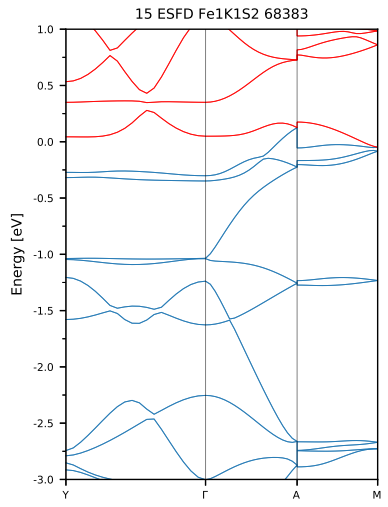

(ew) 15.4.58

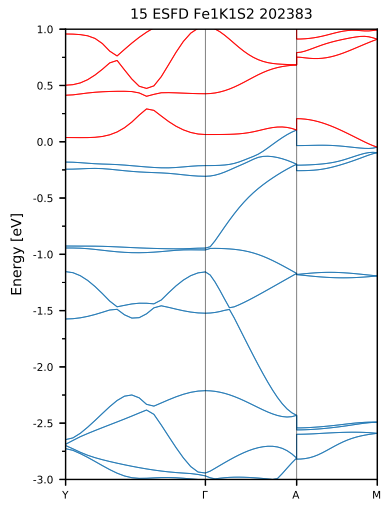

(ex) 15.4.59

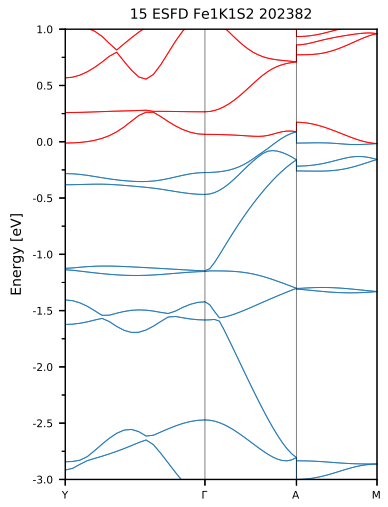

(ey) 15.4.60

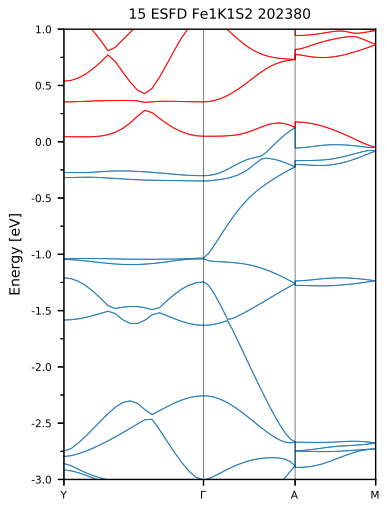

(ez) 15.4.61

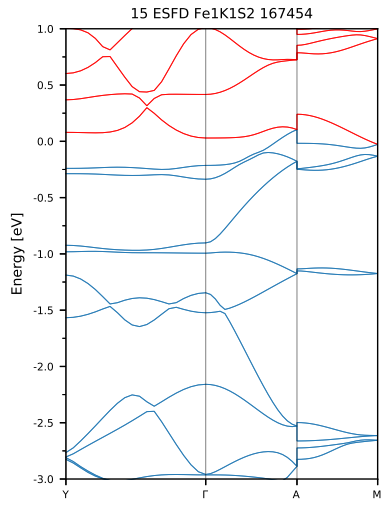

(fa) 15.4.62

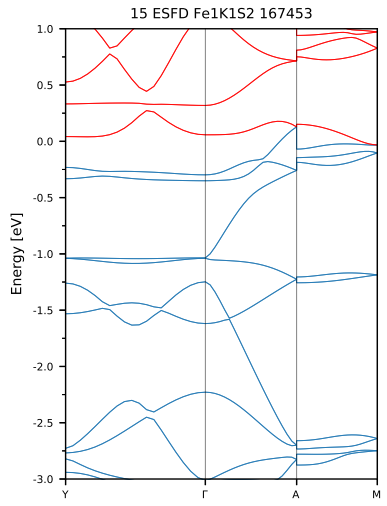

(fb) 15.4.63

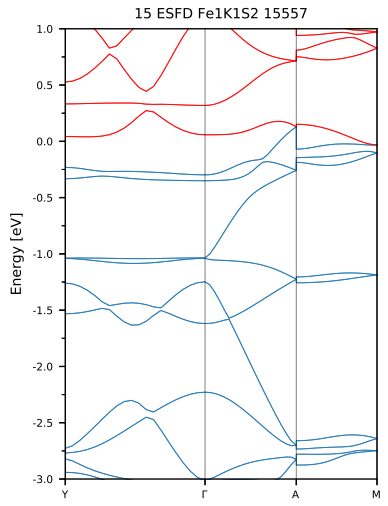

(fc) 15.4.64

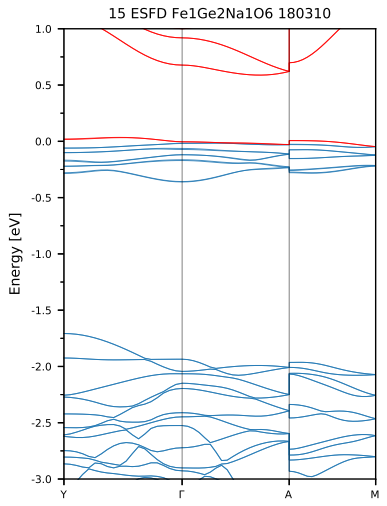

(fd) 15.4.65

Figure 71: Band structures for topological materials in SG 15, part 10/17, from 15.4.50 to 15.4.65. See the list of figures.

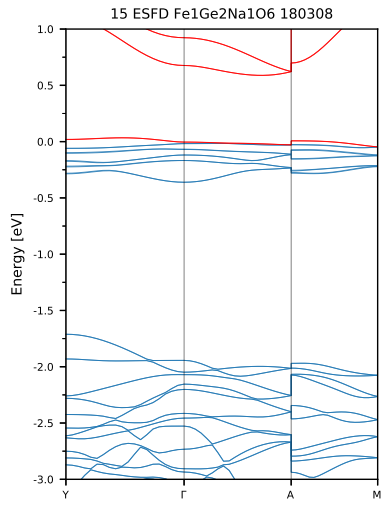

(fe) 15.4.66

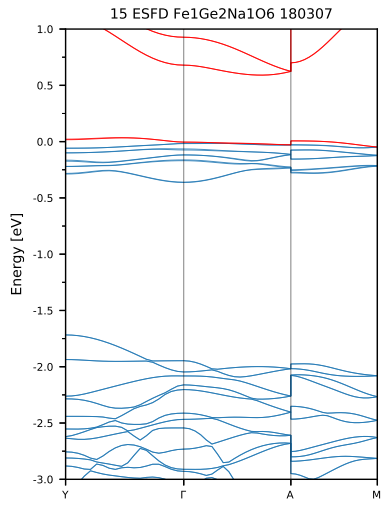

(ff) 15.4.67

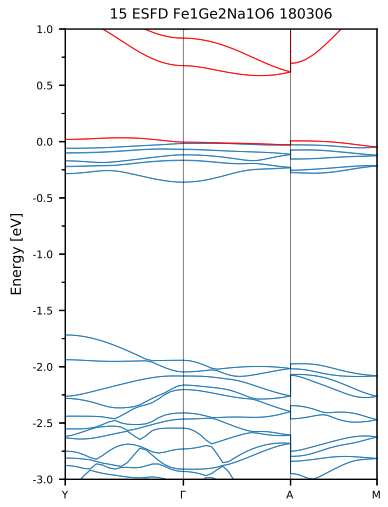

(fg) 15.4.68

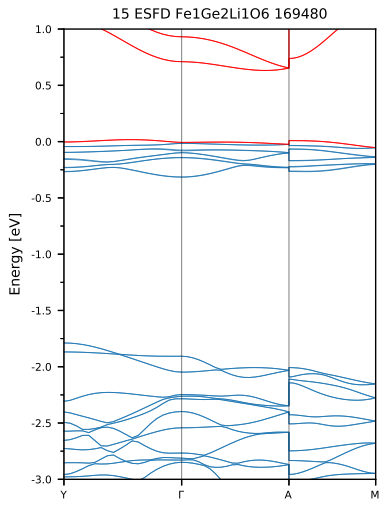

(fh) 15.4.69

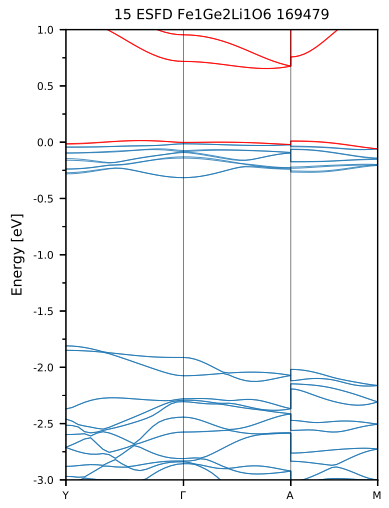

(fi) 15.4.70

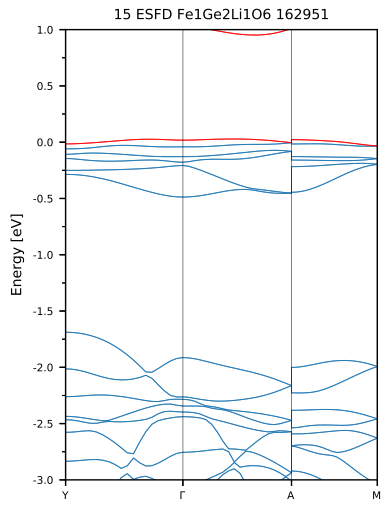

(fj) 15.4.71

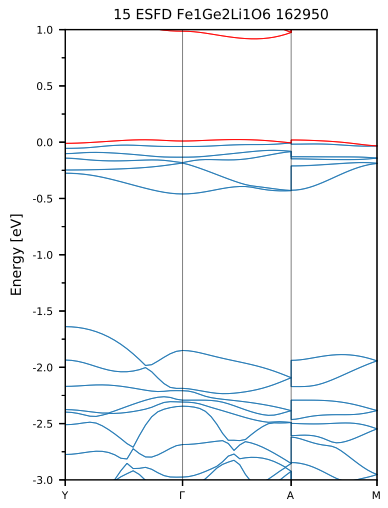

(fk) 15.4.72

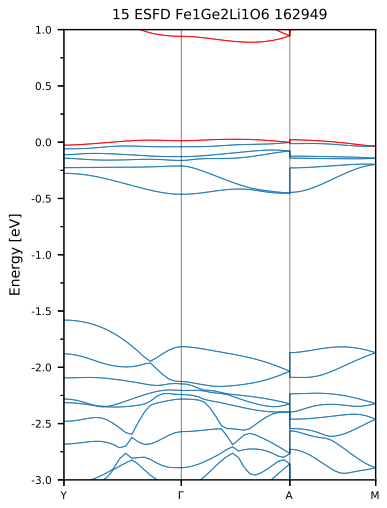

(fl) 15.4.73

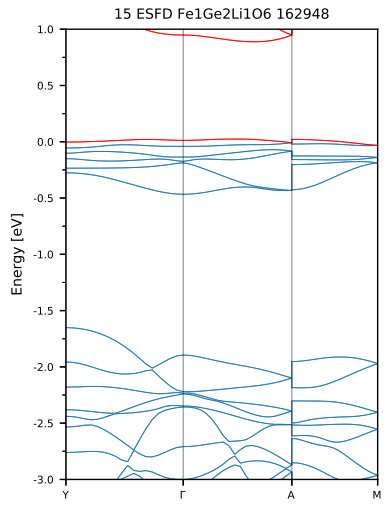

(fm) 15.4.74

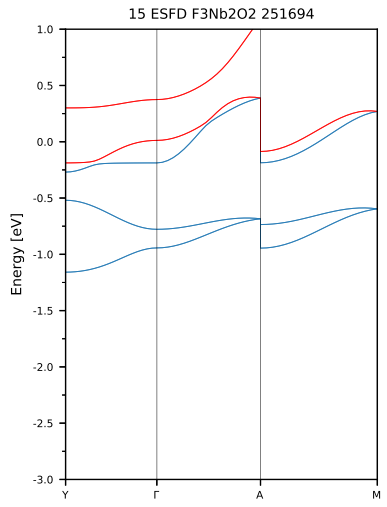

(fn) 15.4.75

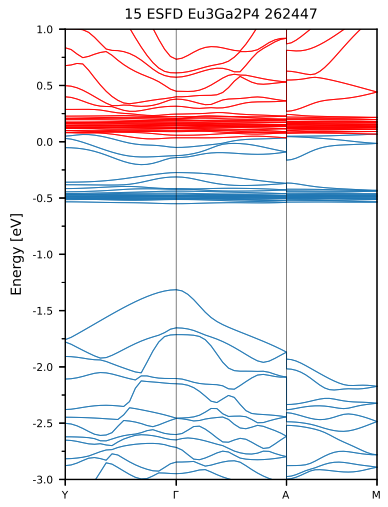

(fo) 15.4.76

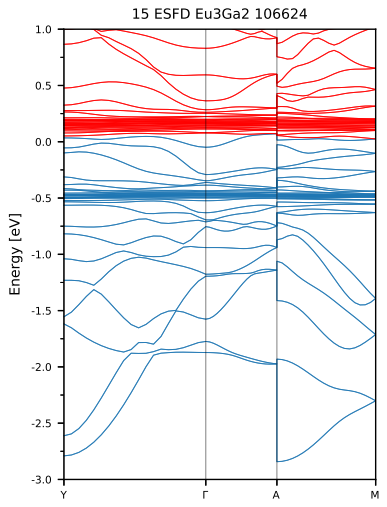

(fp) 15.4.77

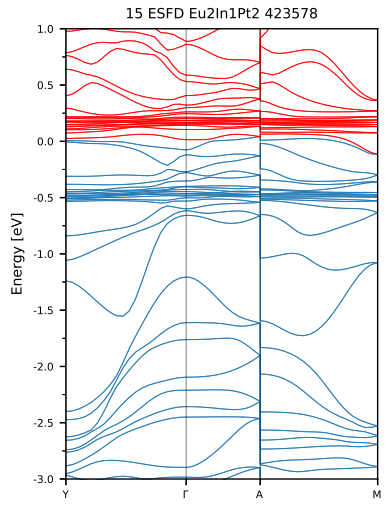

(fq) 15.4.78

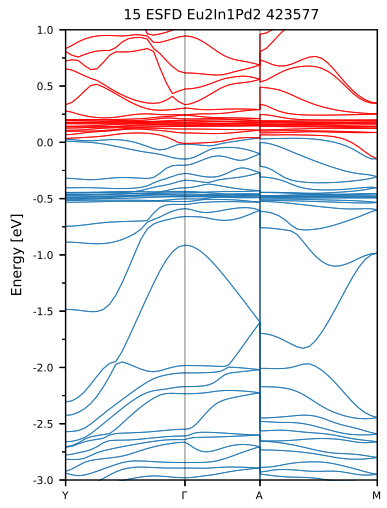

(fr) 15.4.79

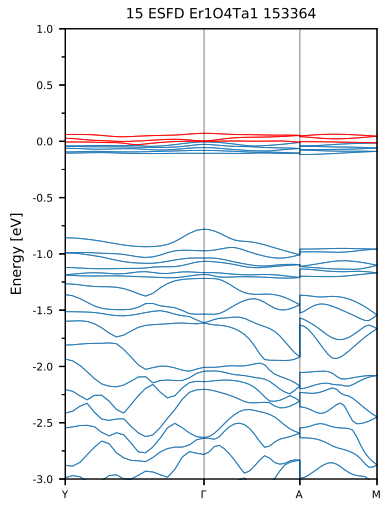

(fs) 15.4.80

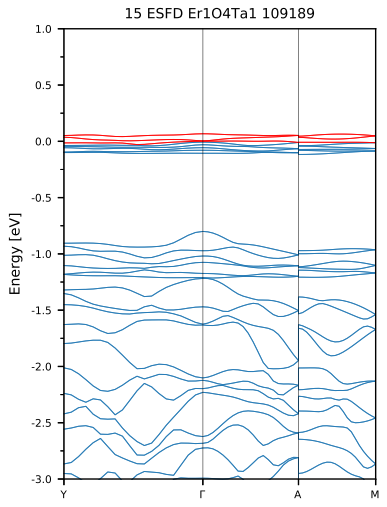

(ft) 15.4.81

Figure 72: Band structures for topological materials in SG 15, part 11/17, from 15.4.66 to 15.4.81. See the list of figures.

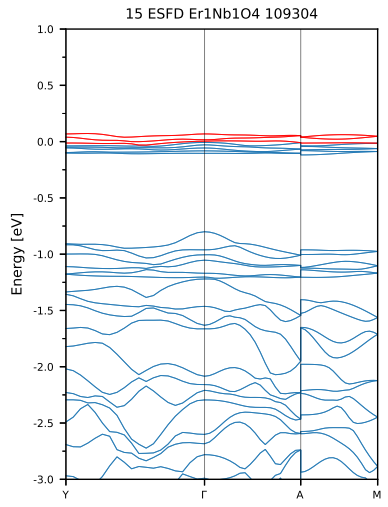

(fu) 15.4.82

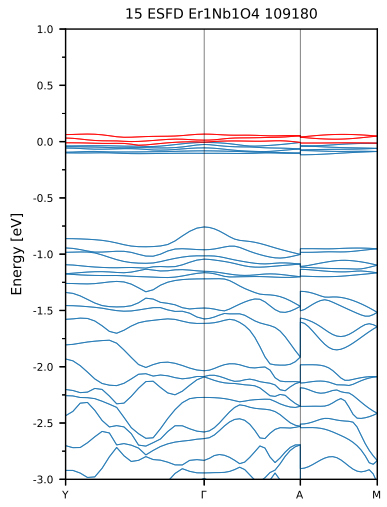

(fv) 15.4.83

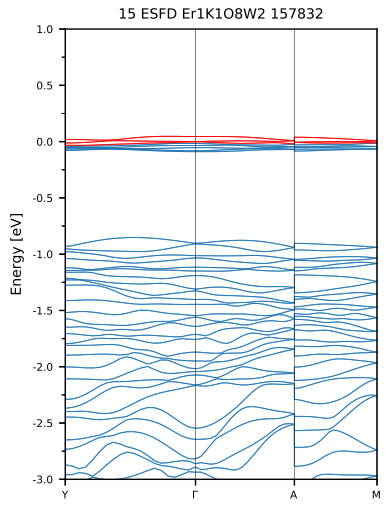

(fw) 15.4.84

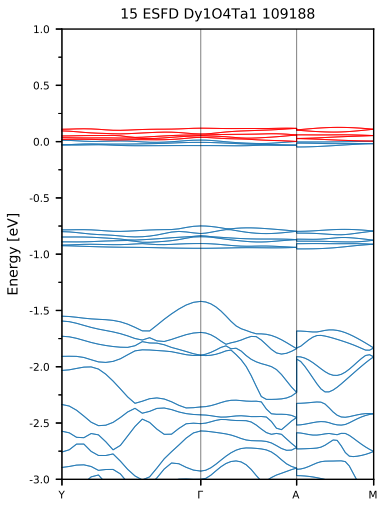

(fx) 15.4.85

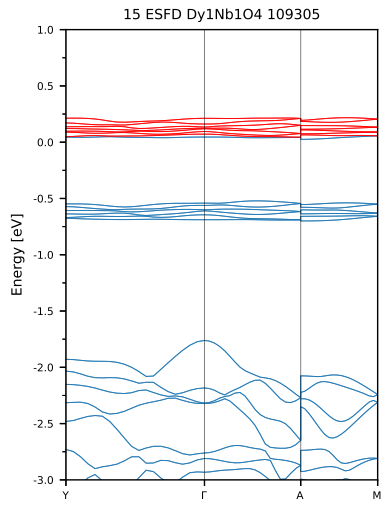

(fy) 15.4.86

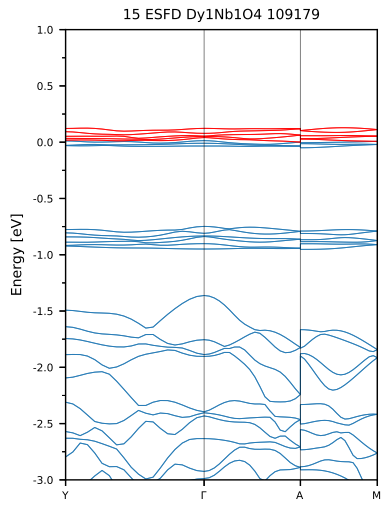

(fz) 15.4.87

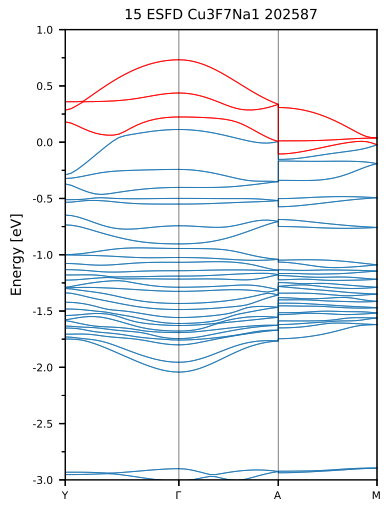

(ga) 15.4.88

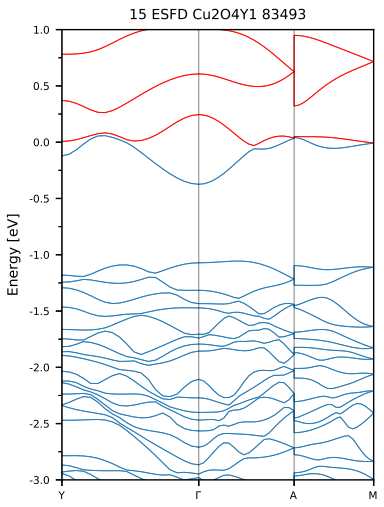

(gb) 15.4.89

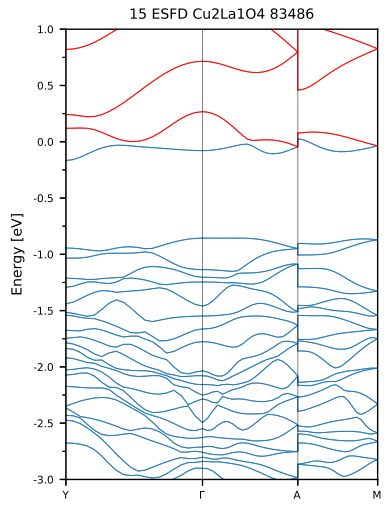

(gc) 15.4.90

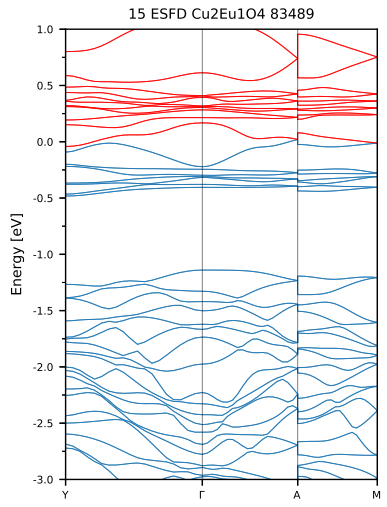

(gd) 15.4.91

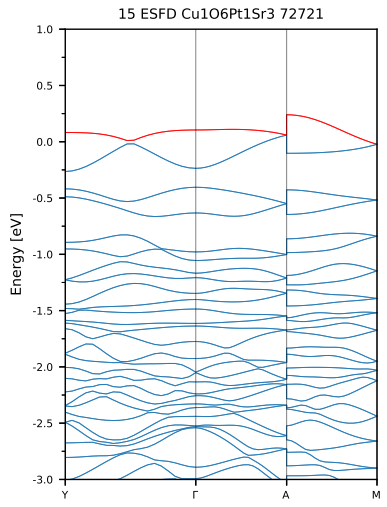

(ge) 15.4.92

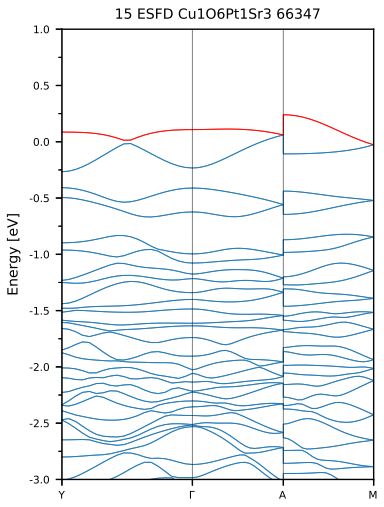

(gf) 15.4.93

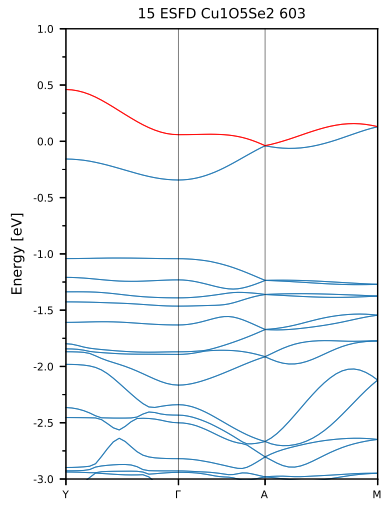

(gg) 15.4.94

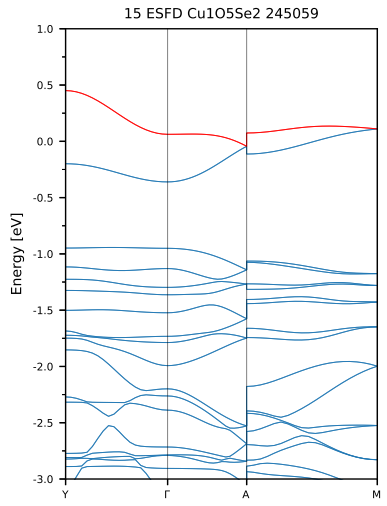

(gh) 15.4.95

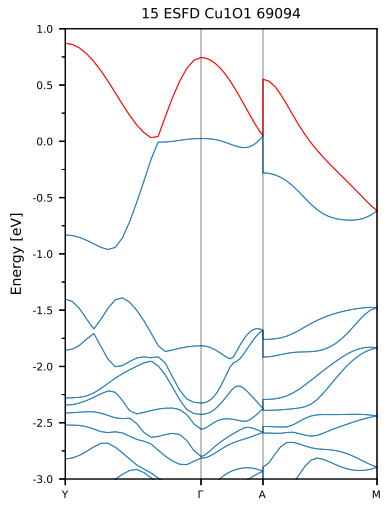

(gi) 15.4.96

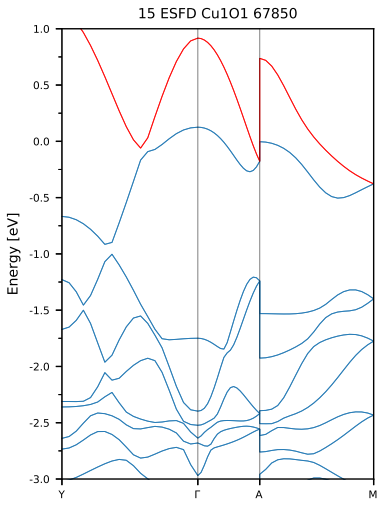

(gj) 15.4.97

Figure 73: Band structures for topological materials in SG 15, part 12/17, from 15.4.82 to 15.4.97. See the list of figures.

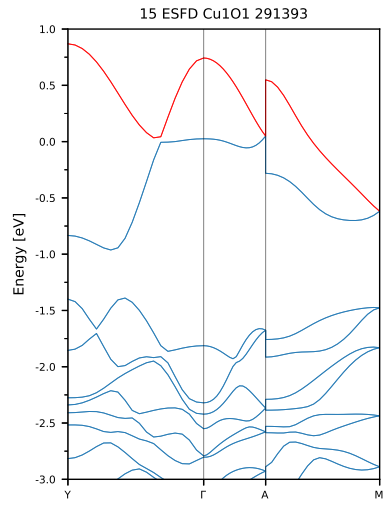

(gk) 15.4.98

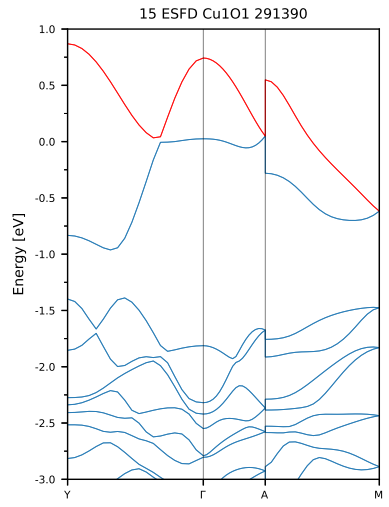

(gl) 15.4.99

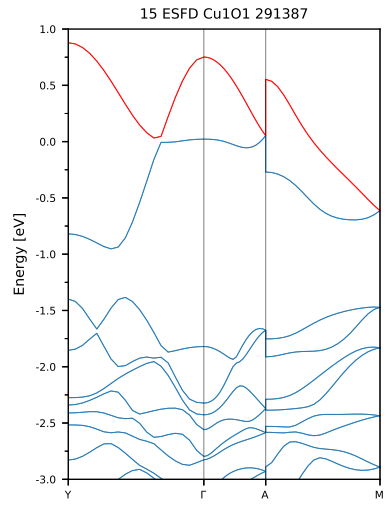

(gm) 15.4.100

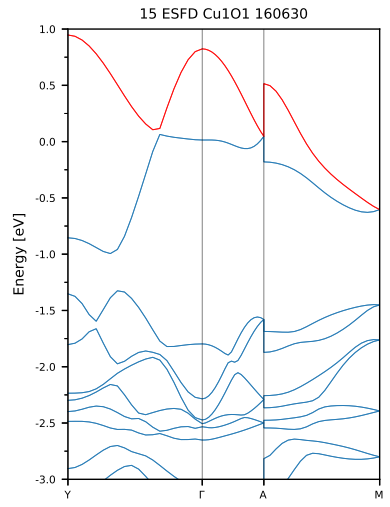

(gn) 15.4.101

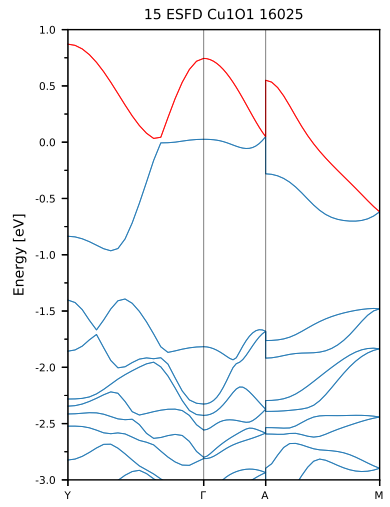

(go) 15.4.102

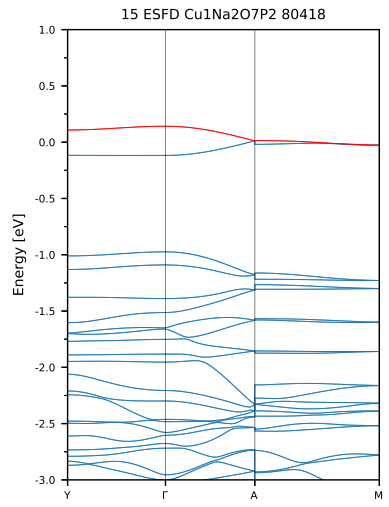

(gp) 15.4.103

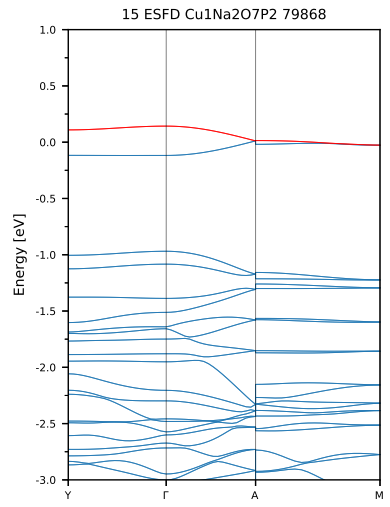

(gq) 15.4.104

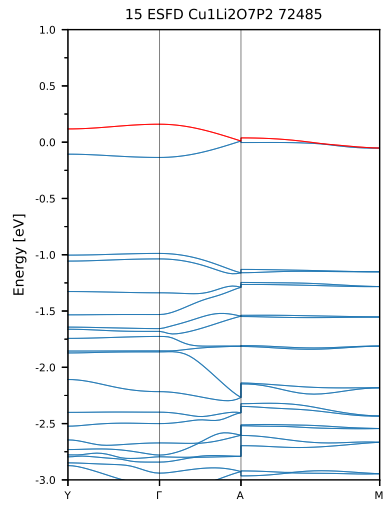

(gr) 15.4.105

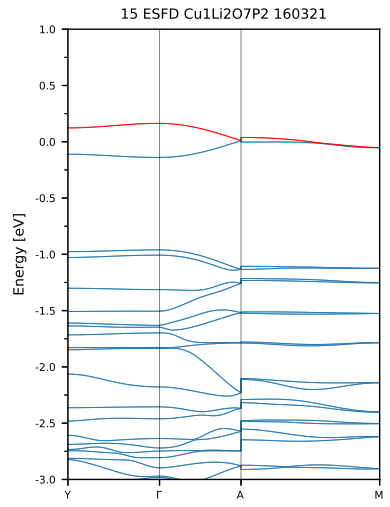

(gs) 15.4.106

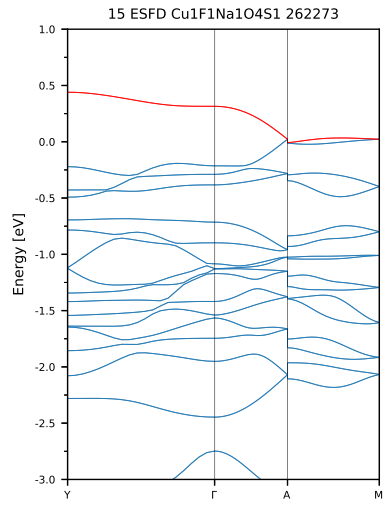

(gt) 15.4.107

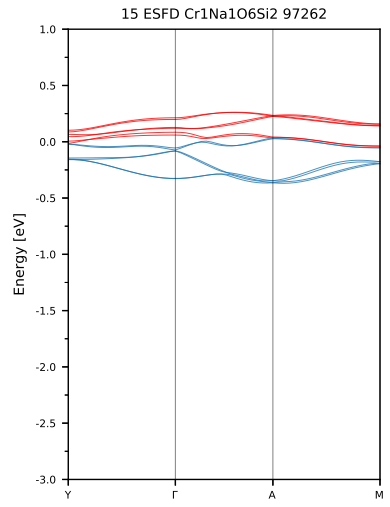

(gu) 15.4.108

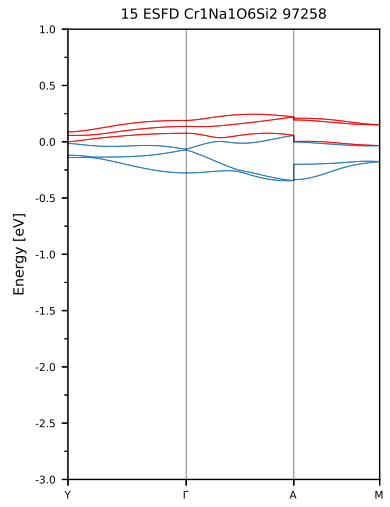

(gv) 15.4.109

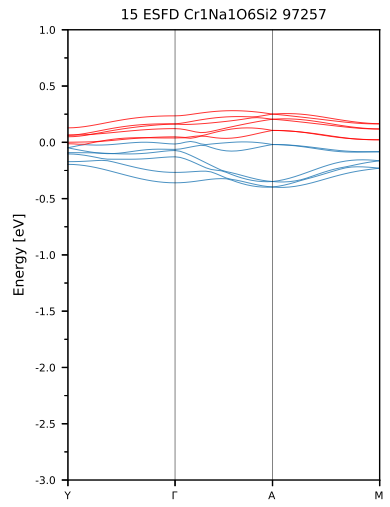

(gw) 15.4.110

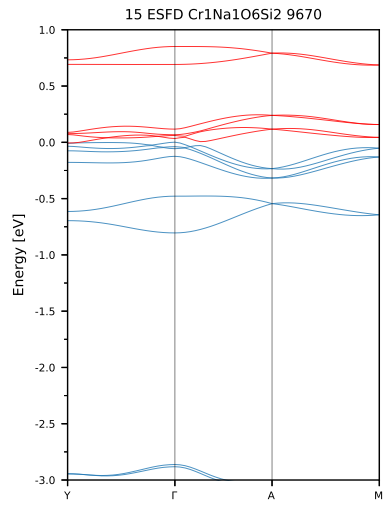

(gx) 15.4.111

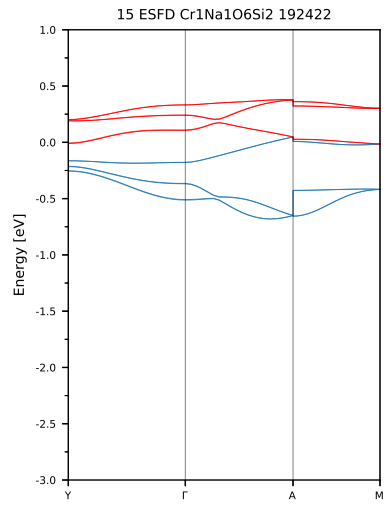

(gy) 15.4.112

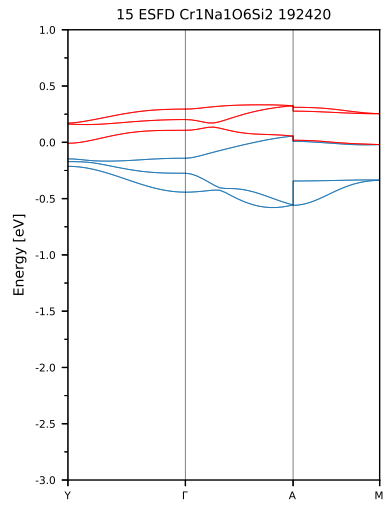

(gz) 15.4.113

Figure 74: Band structures for topological materials in SG 15, part 13/17, from 15.4.98 to 15.4.113. See the list of figures.

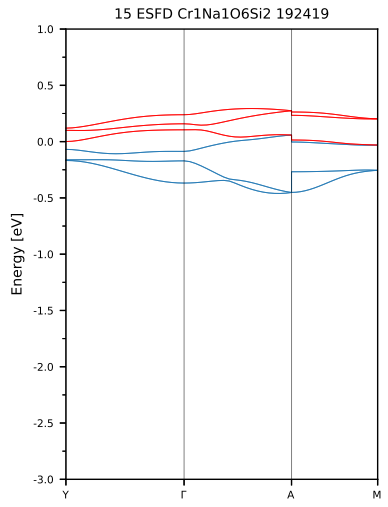

(ha) 15.4.114

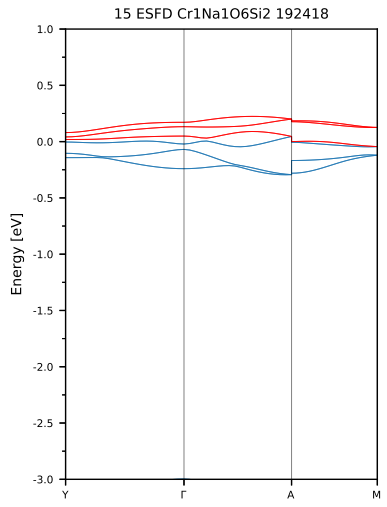

(hb) 15.4.115

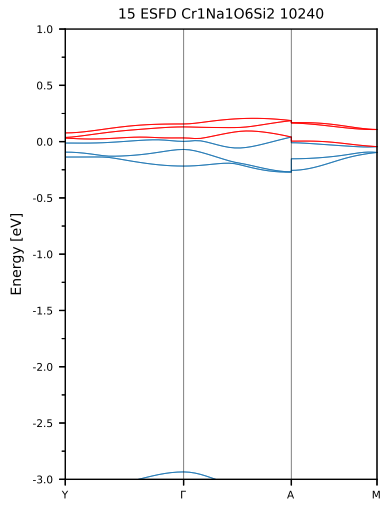

(hc) 15.4.116

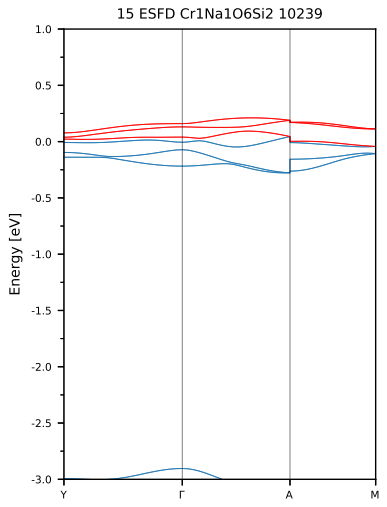

(hd) 15.4.117

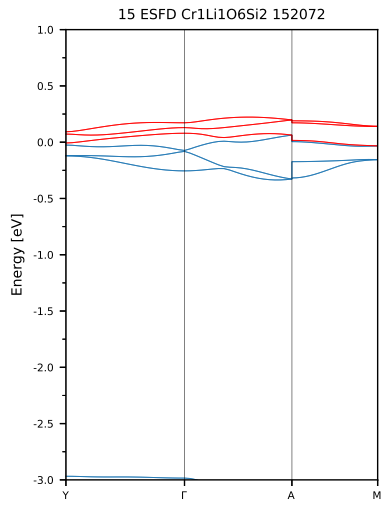

(he) 15.4.118

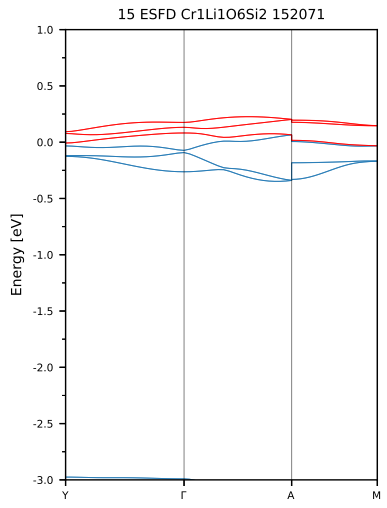

(hf) 15.4.119

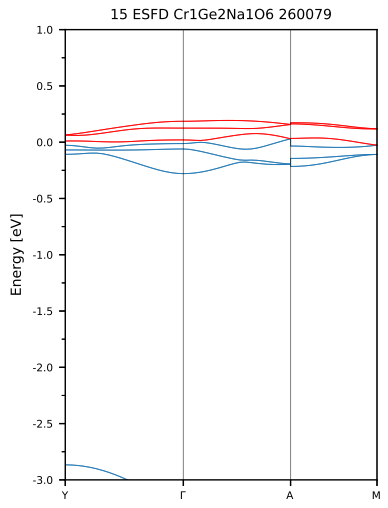

(hg) 15.4.120

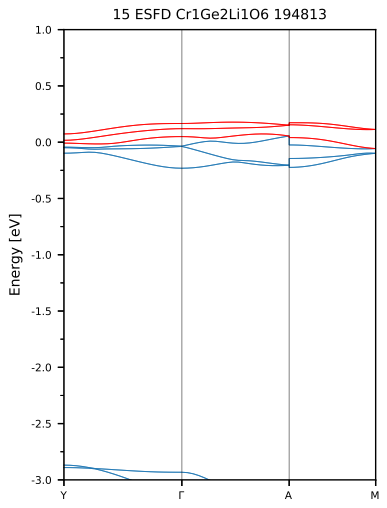

(hh) 15.4.121

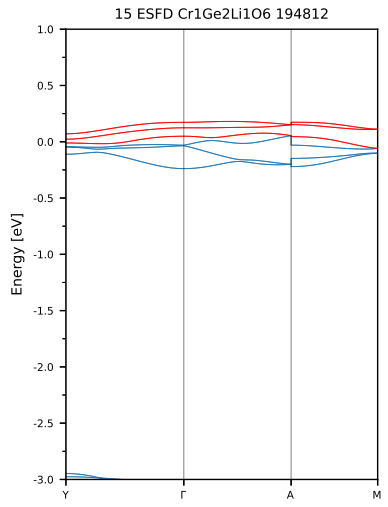

(hi) 15.4.122

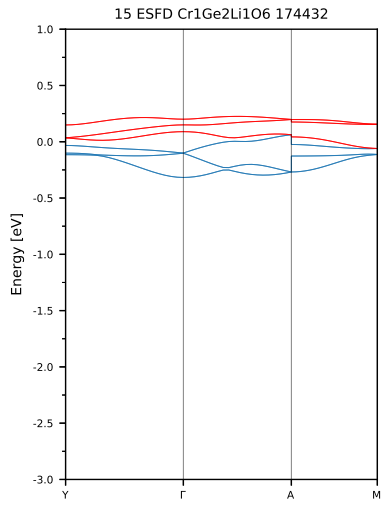

(hj) 15.4.123

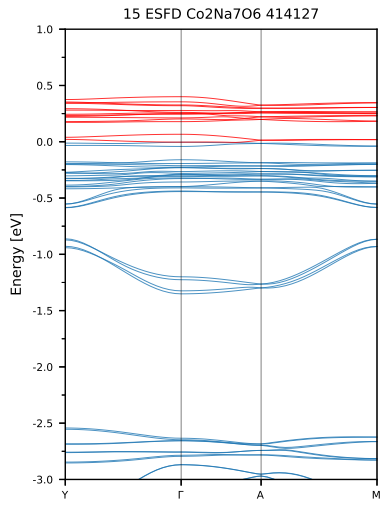

(hk) 15.4.124

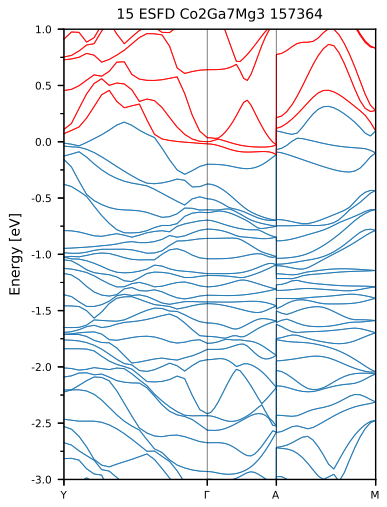

(hl) 15.4.125

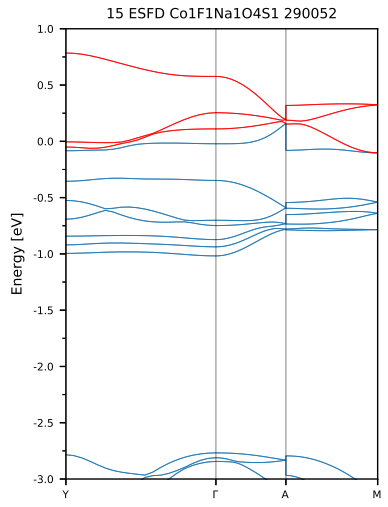

(hm) 15.4.126

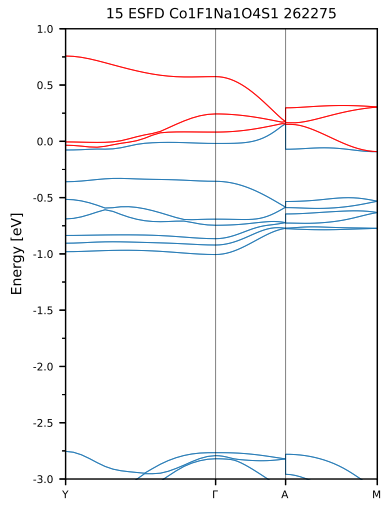

(hn) 15.4.127

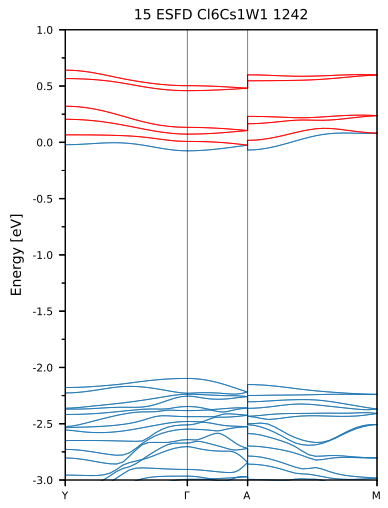

(ho) 15.4.128

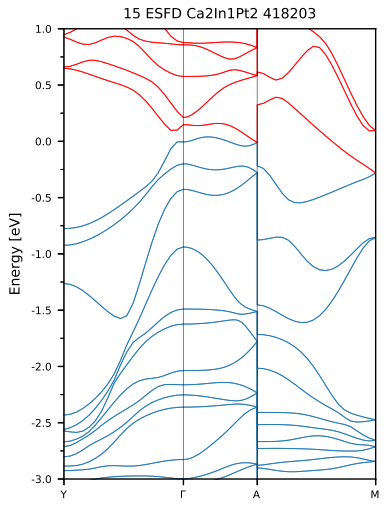

(hp) 15.4.129

Figure 75: Band structures for topological materials in SG 15, part 14/17, from 15.4.114 to 15.4.129. See the list of figures.

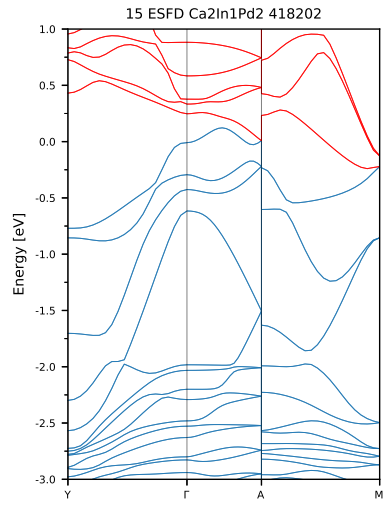

(hq) 15.4.130

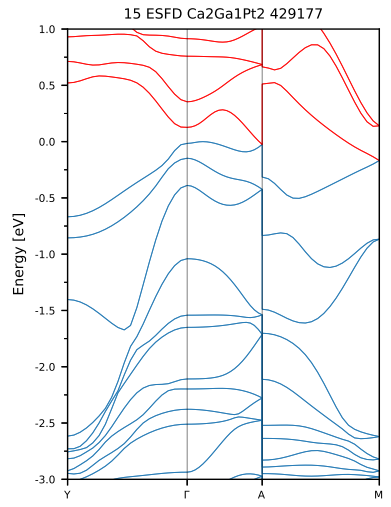

(hr) 15.4.131

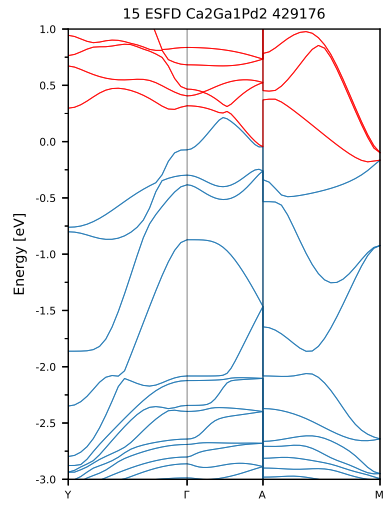

(hs) 15.4.132

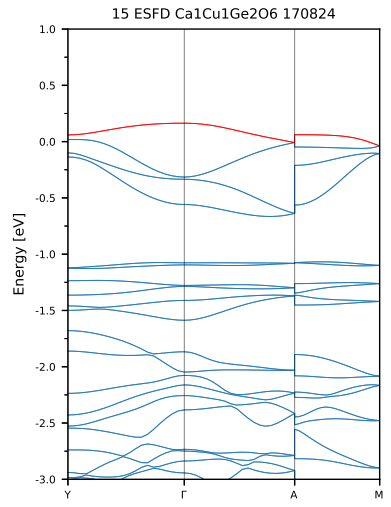

(ht) 15.4.133

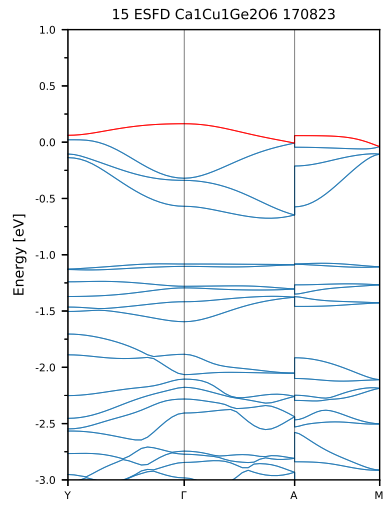

(hu) 15.4.134

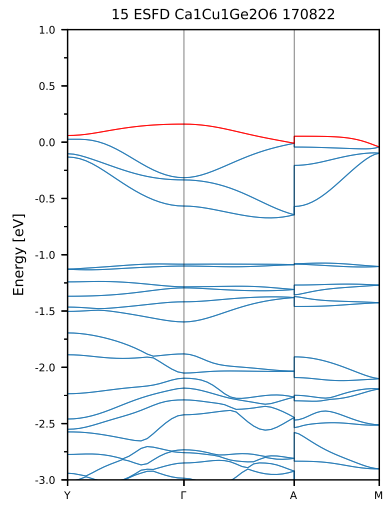

(hv) 15.4.135

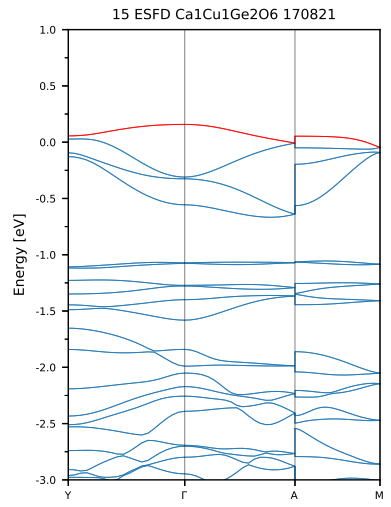

(hw) 15.4.136

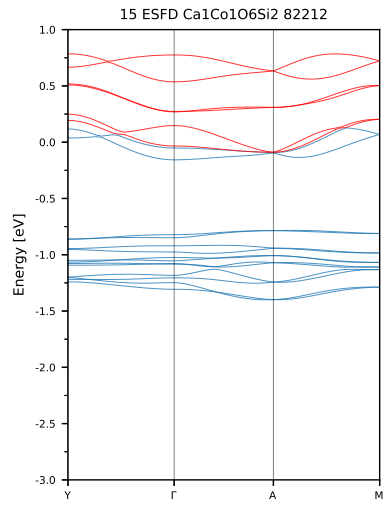

(hx) 15.4.137

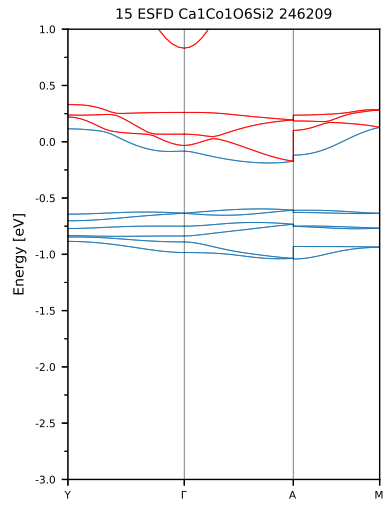

(hy) 15.4.138

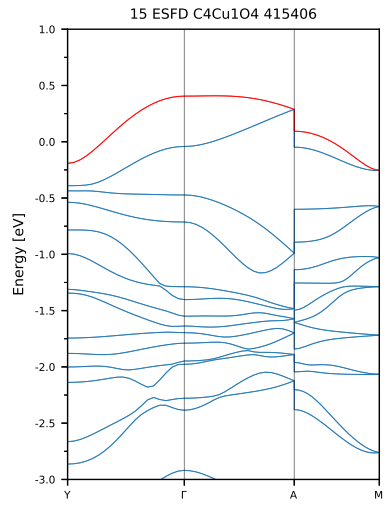

(hz) 15.4.139

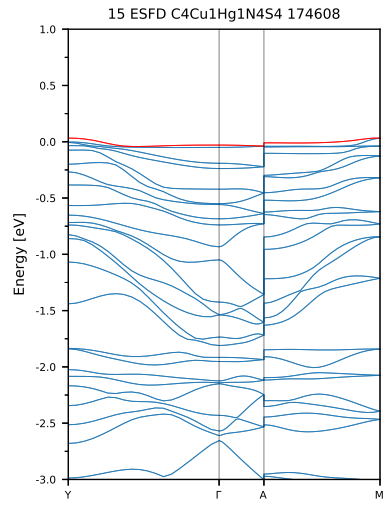

(ia) 15.4.140

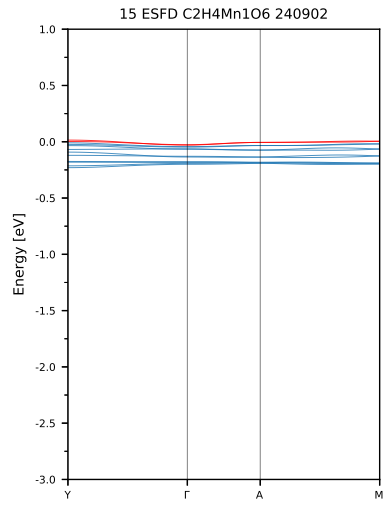

(ib) 15.4.141

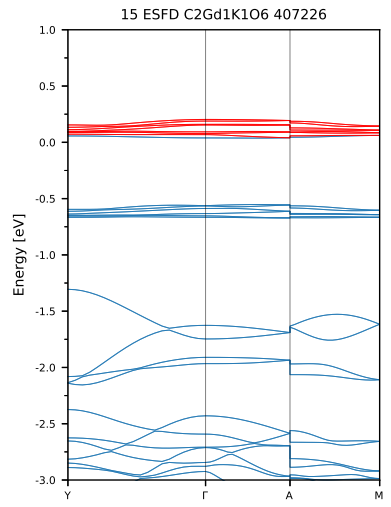

(ic) 15.4.142

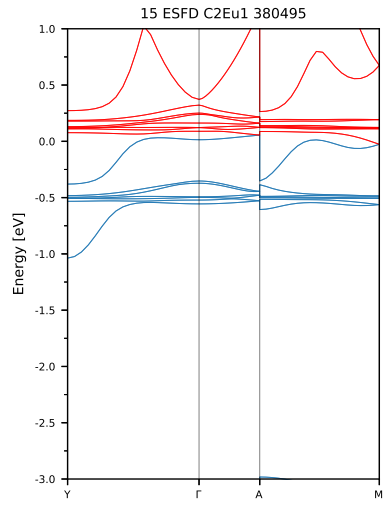

(id) 15.4.143

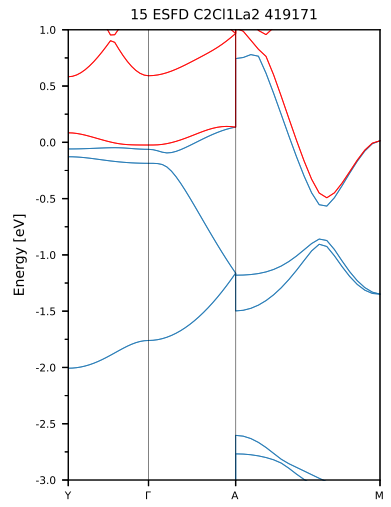

(ie) 15.4.144

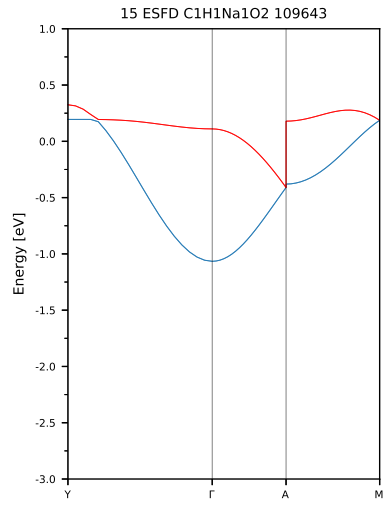

(if) 15.4.145

Figure 76: Band structures for topological materials in SG 15, part 15/17, from 15.4.130 to 15.4.145. See the list of figures.

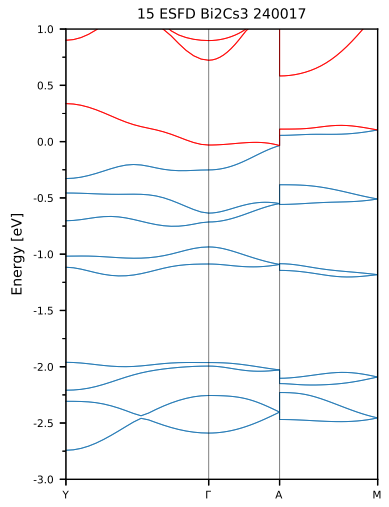

(ig) 15.4.146

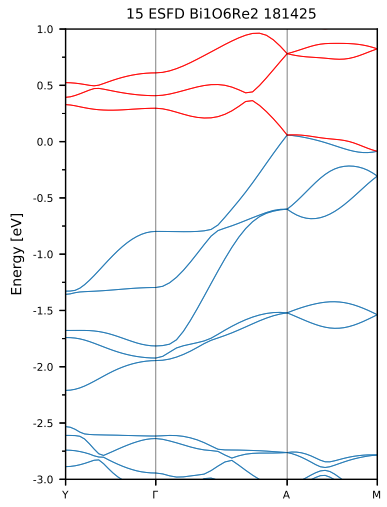

(ih) 15.4.147

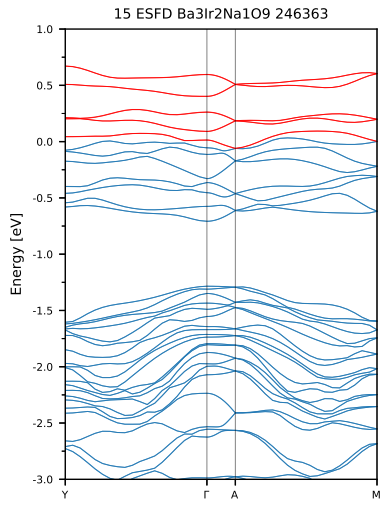

(ii) 15.4.148

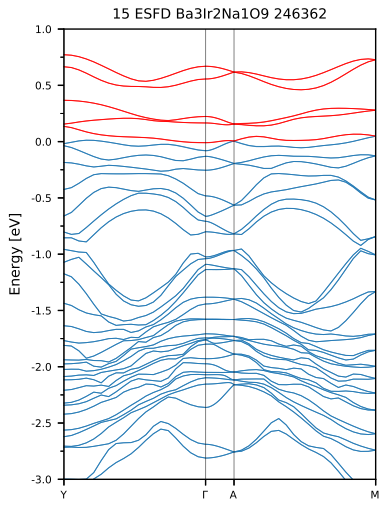

(ij) 15.4.149

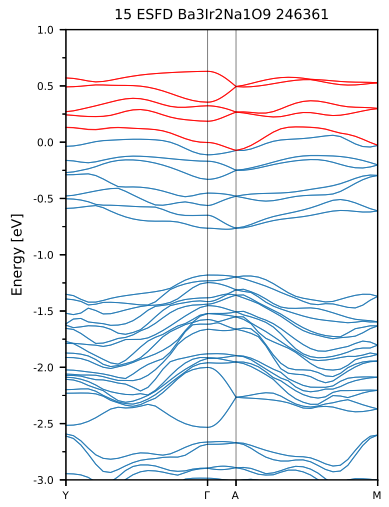

(ik) 15.4.150

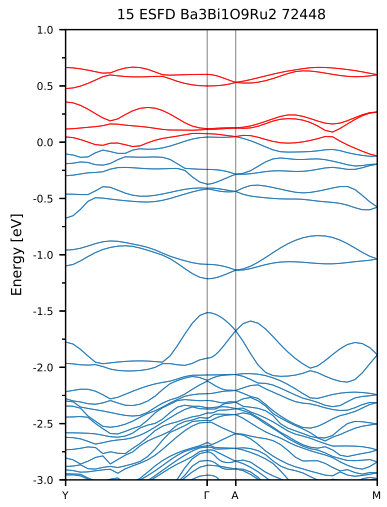

(il) 15.4.151

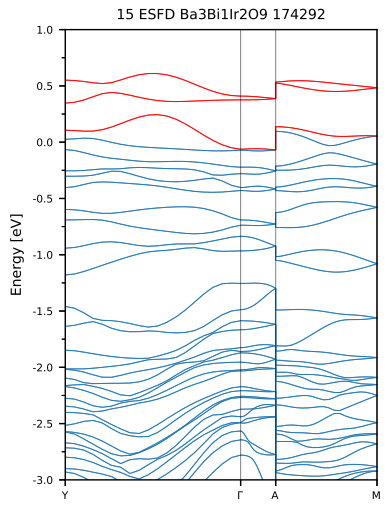

(im) 15.4.152

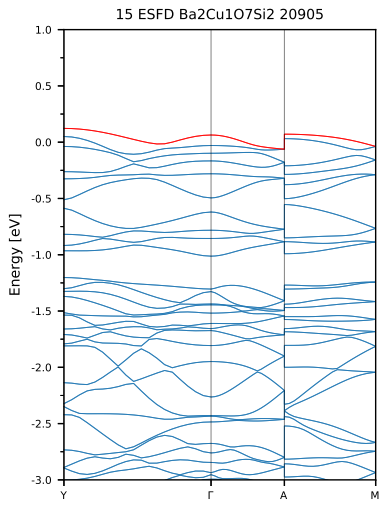

(in) 15.4.153

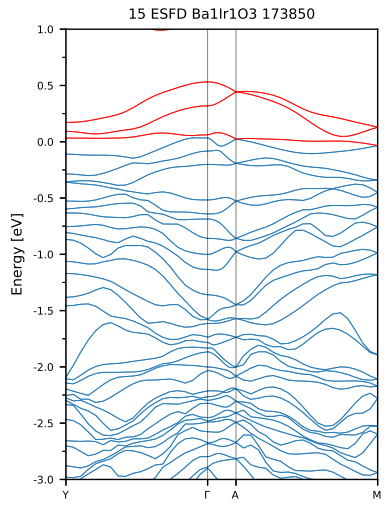

(io) 15.4.154

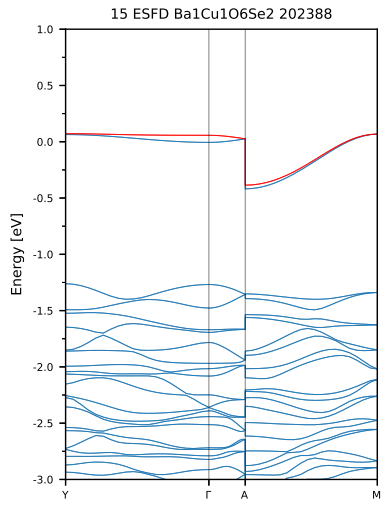

(ip) 15.4.155

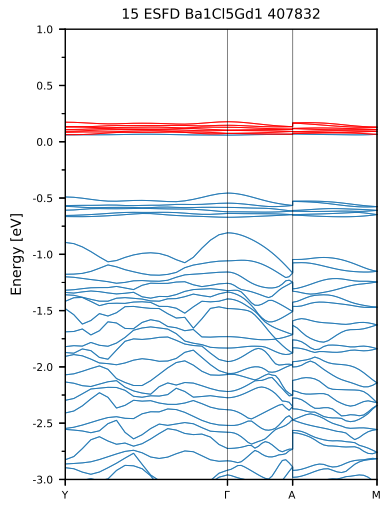

(iq) 15.4.156

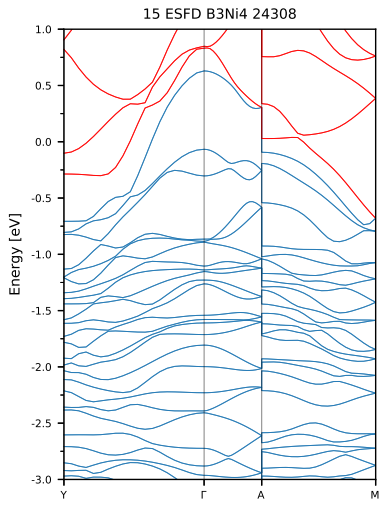

(ir) 15.4.157

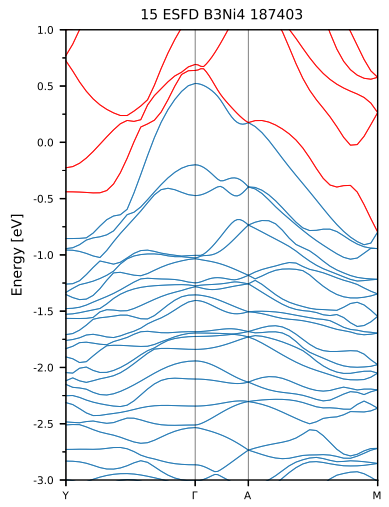

(is) 15.4.158

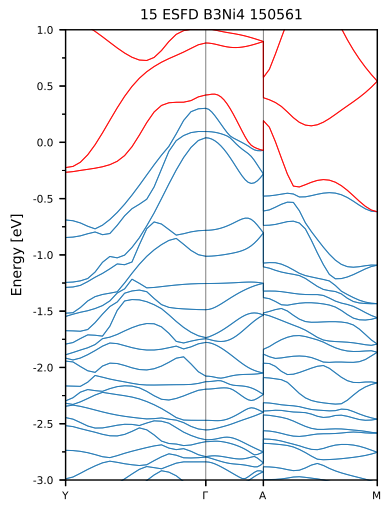

(it) 15.4.159

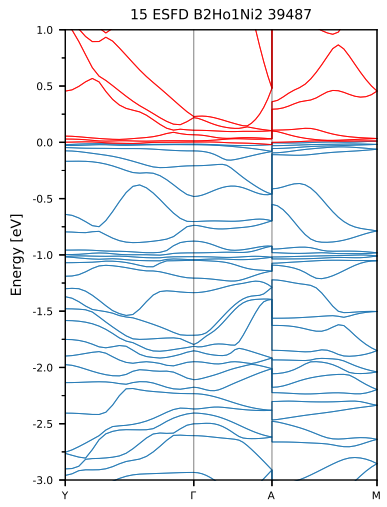

(iu) 15.4.160

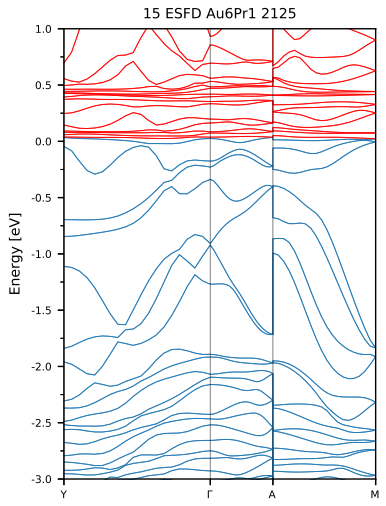

(iv) 15.4.161

Figure 77: Band structures for topological materials in SG 15, part 16/17, from 15.4.146 to 15.4.161. See the list of figures.

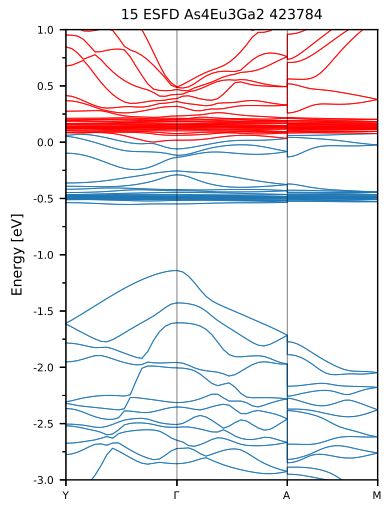

(iw) 15.4.162

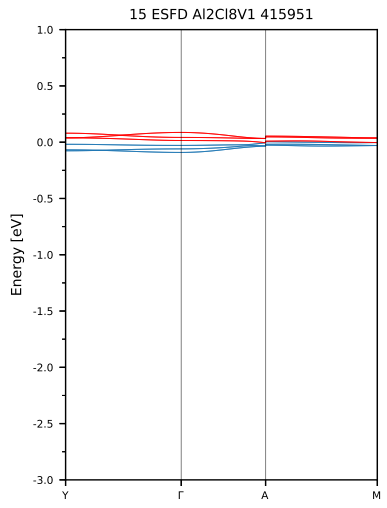

(ix) 15.4.163

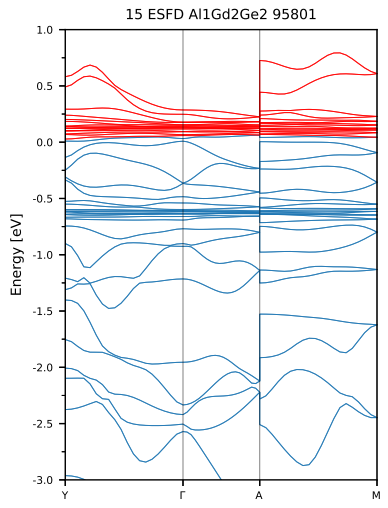

(iy) 15.4.164

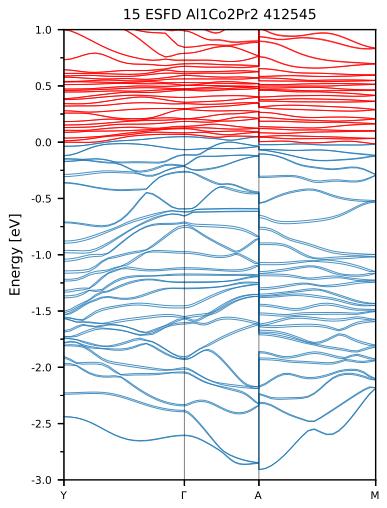

(iz) 15.4.165

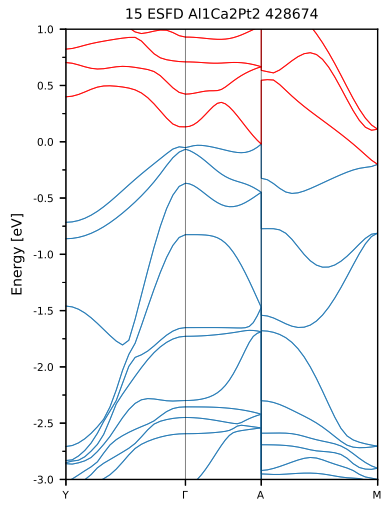

(ja) 15.4.166

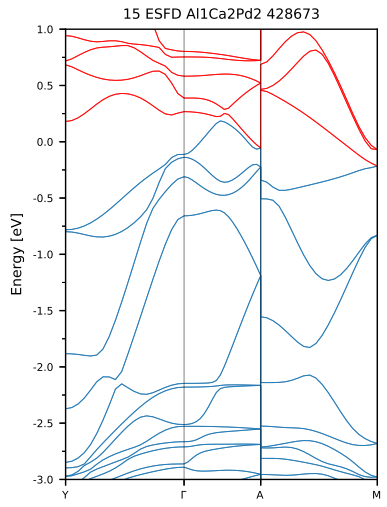

(jb) 15.4.167

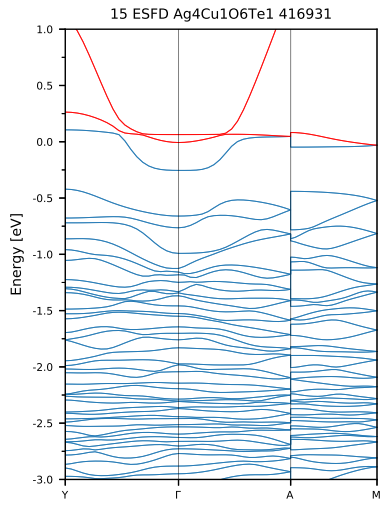

(jc) 15.4.168

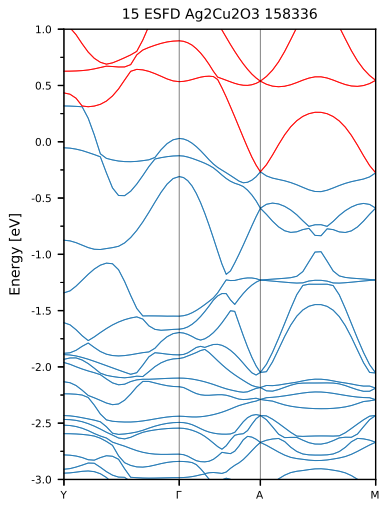

(jd) 15.4.169

Figure 78: Band structures for topological materials in SG 15, part 17/17, from 15.4.162 to 15.4.169. See the list of figures.

## 0.0.10 SG 19

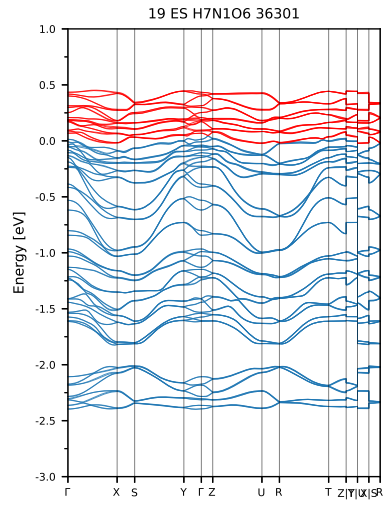

(a) 19.3.1

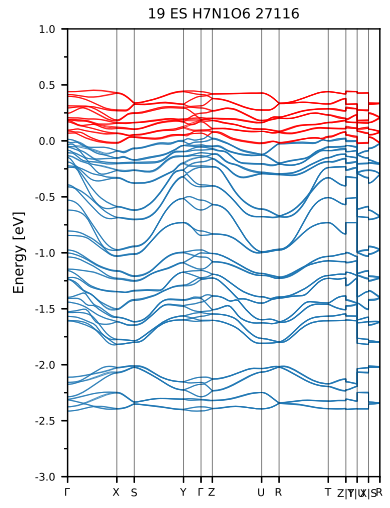

(b) 19.3.2

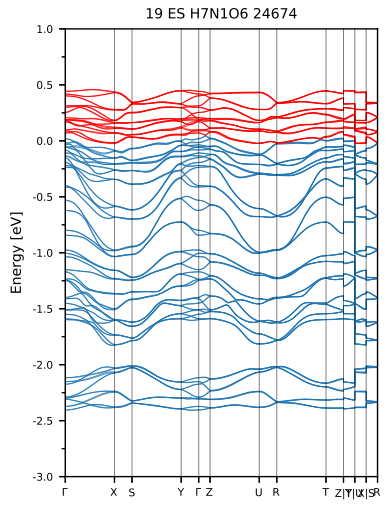

(c) 19.3.3

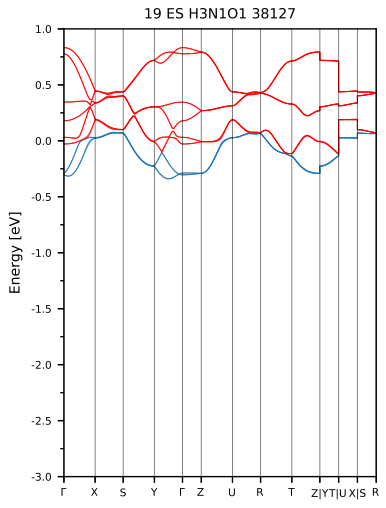

(d) 19.3.4

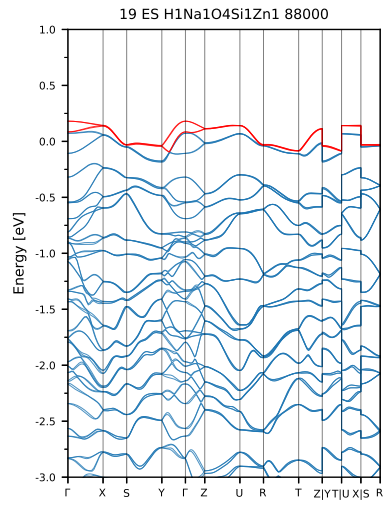

(e) 19.3.5

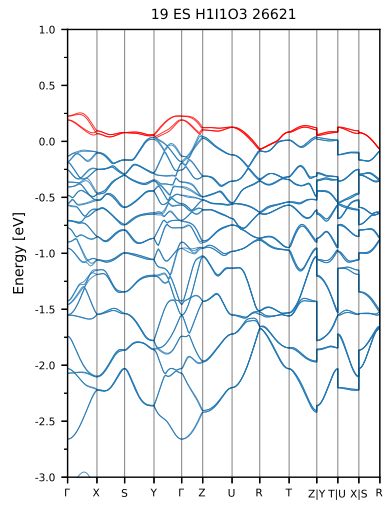

(f) 19.3.6

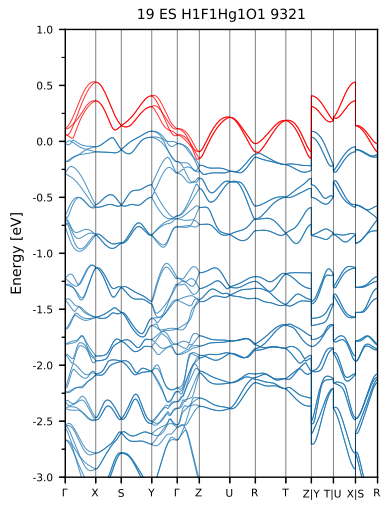

(g) 19.3.7

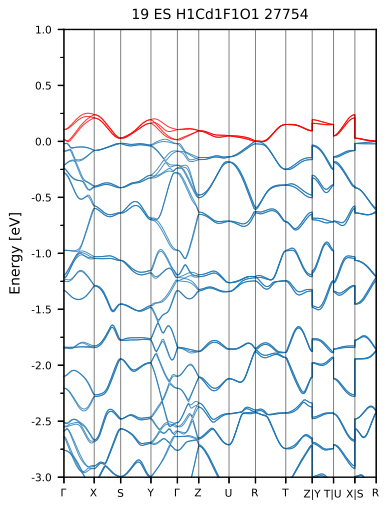

(h) 19.3.8

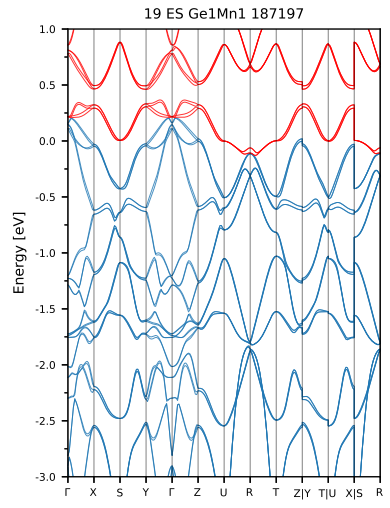

(i) 19.3.9

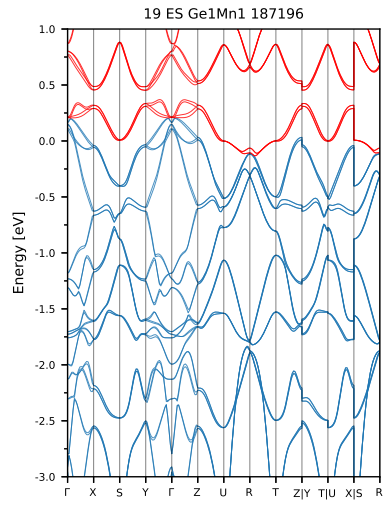

(j) 19.3.10

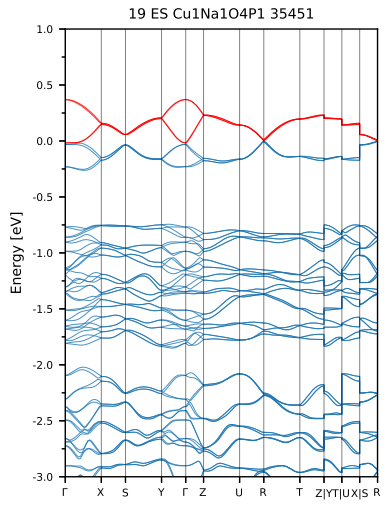

(k) 19.3.11

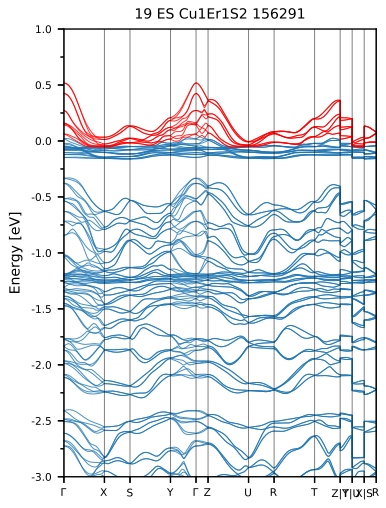

(l) 19.3.12

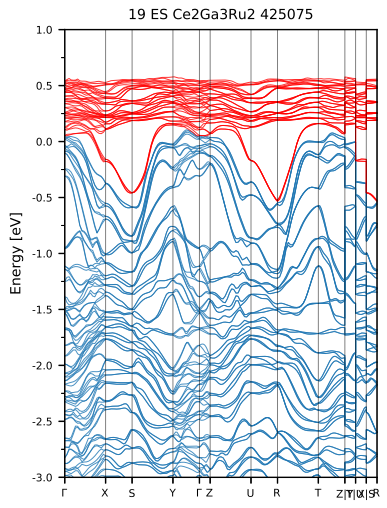

(m) 19.3.13

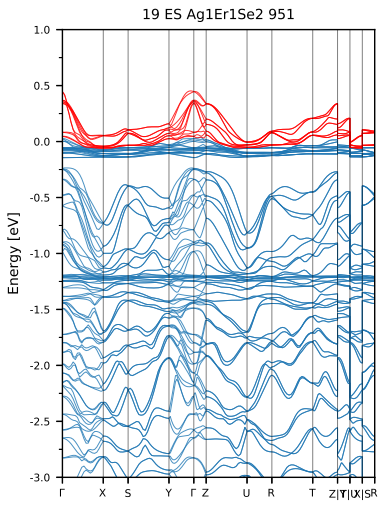

(n) 19.3.14

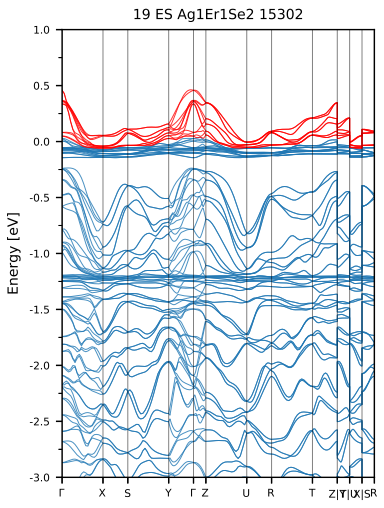

(o) 19.3.15

Figure 79: Band structures for topological materials in SG 19, part 1/1, from 19.3.1 to 19.3.15. See the list of figures.

### 0.0.11 SG 20

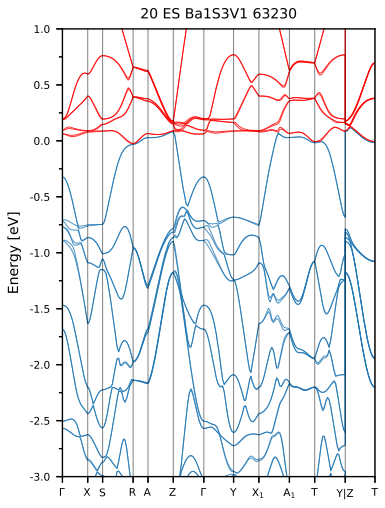

(a) 20.3.1

Figure 80: Band structures for topological materials in SG 20, part 1/1, from 20.3.1 to 20.3.1. See the list of figures.

### 0.0.12 SG 26

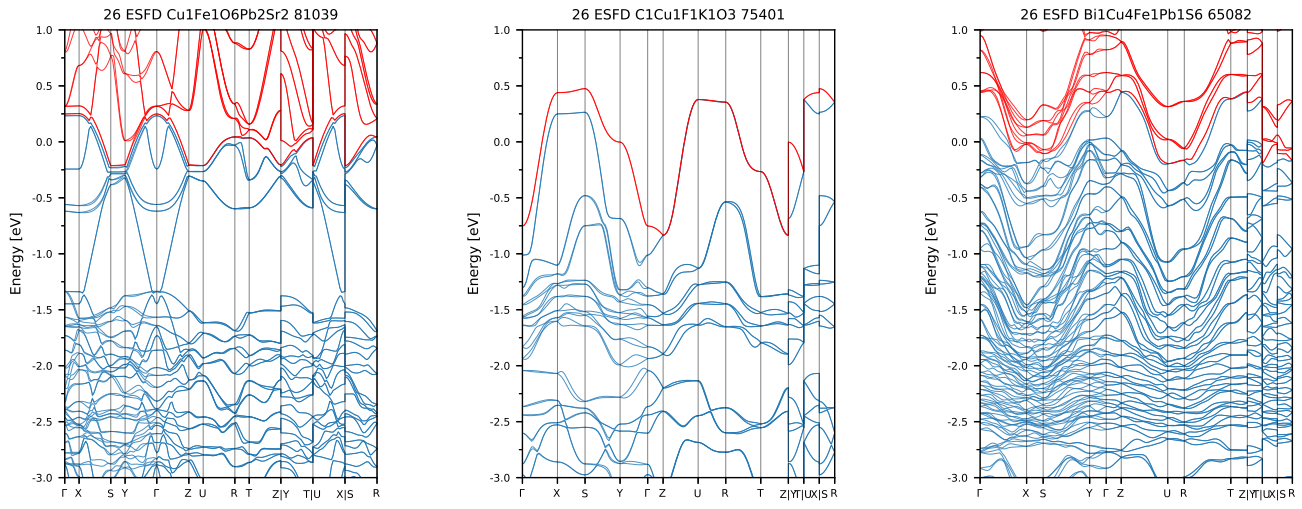

(a) 26.4.1 (b) 26.4.2 (c) 26.4.3

Figure 81: Band structures for topological materials in SG 26, part 1/1, from 26.4.1 to 26.4.3. See the list of figures.

### 0.0.13 SG 33

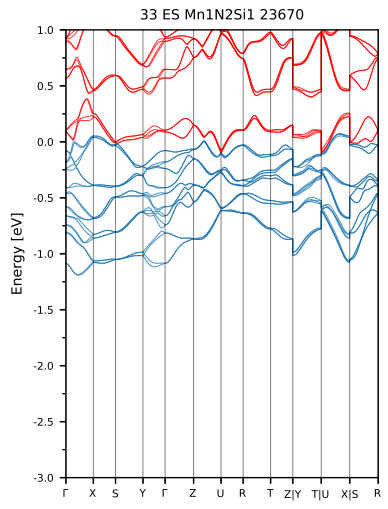

(a) 33.3.1

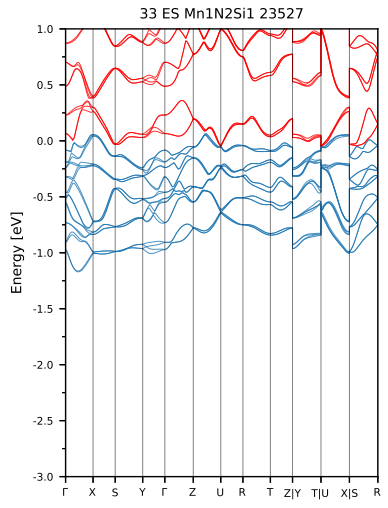

(b) 33.3.2

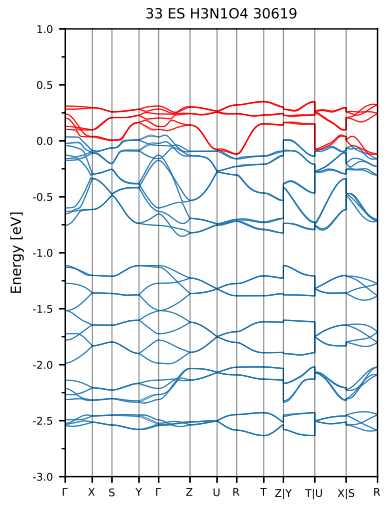

(c) 33.3.3

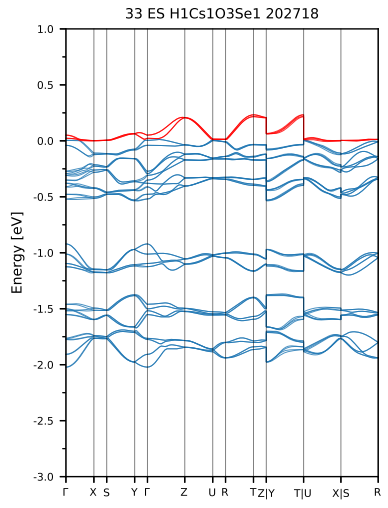

(d) 33.3.4

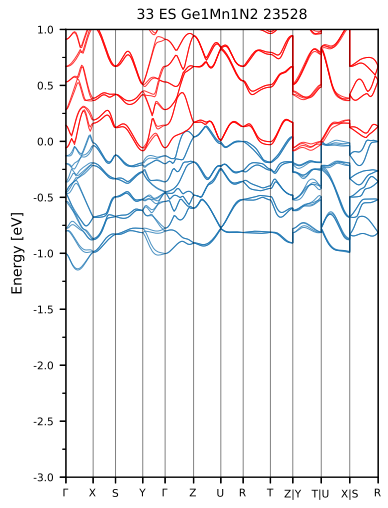

(e) 33.3.5

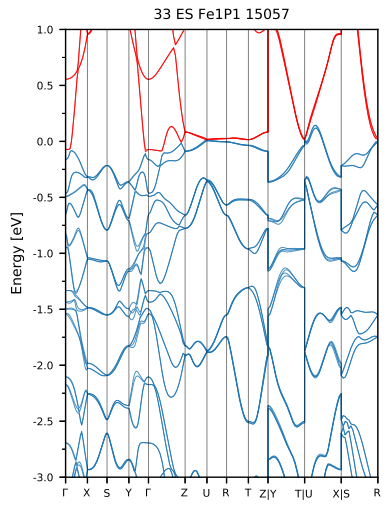

(f) 33.3.6

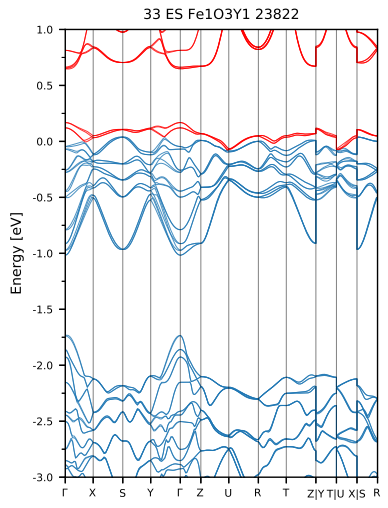

(g) 33.3.7

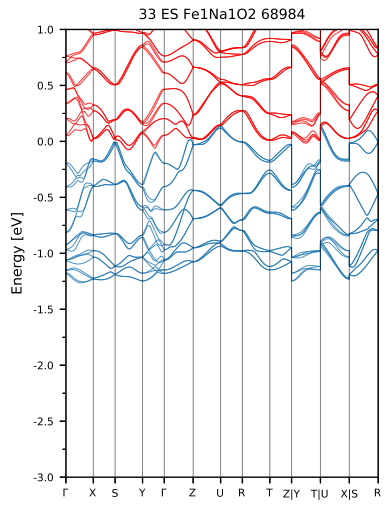

(h) 33.3.8

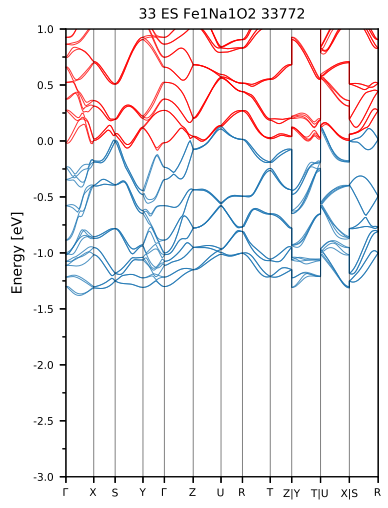

(i) 33.3.9

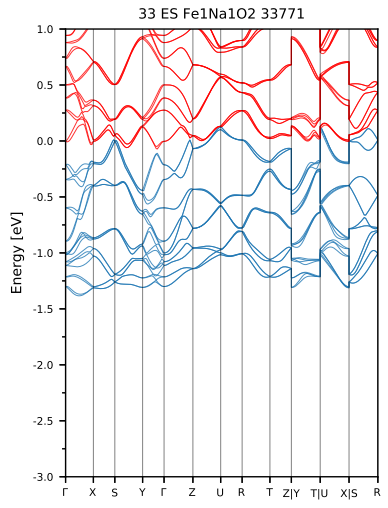

(j) 33.3.10

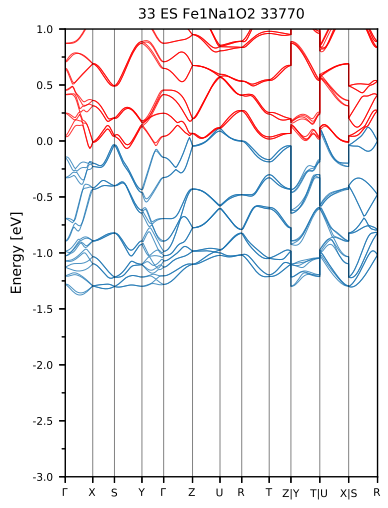

(k) 33.3.11

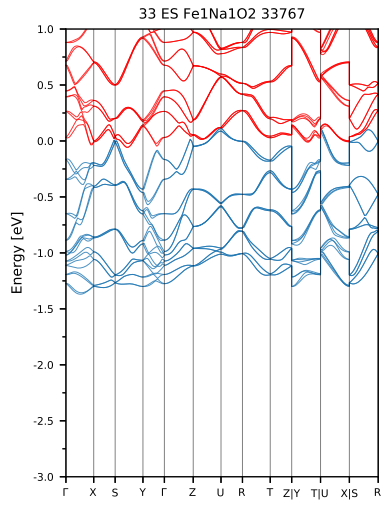

(l) 33.3.12

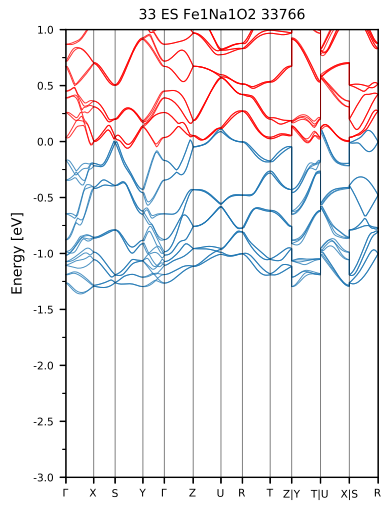

(m) 33.3.13

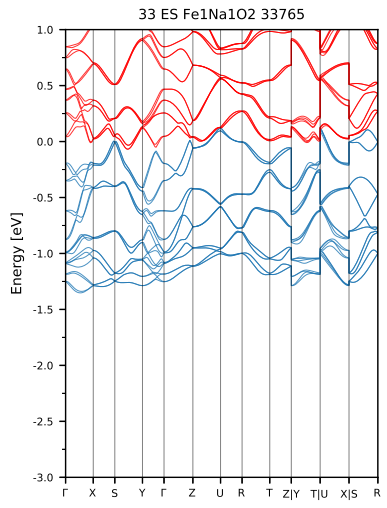

(n) 33.3.14

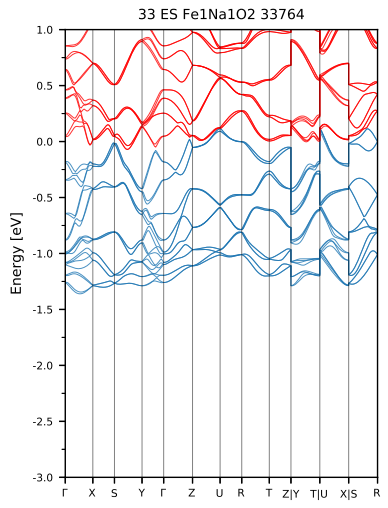

(o) 33.3.15

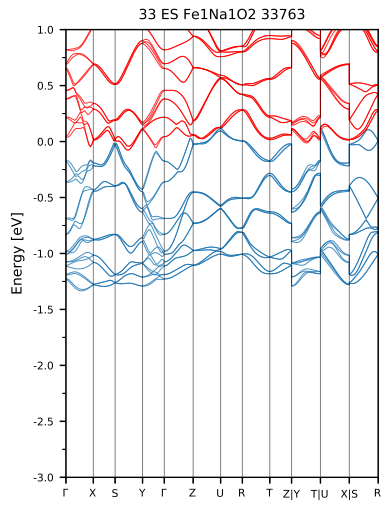

(p) 33.3.16

Figure 82: Band structures for topological materials in SG 33, part 1/2, from 33.3.1 to 33.3.16. See the list of figures.

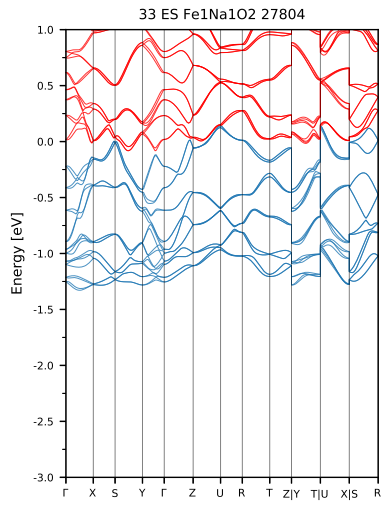

(q) 33.3.17

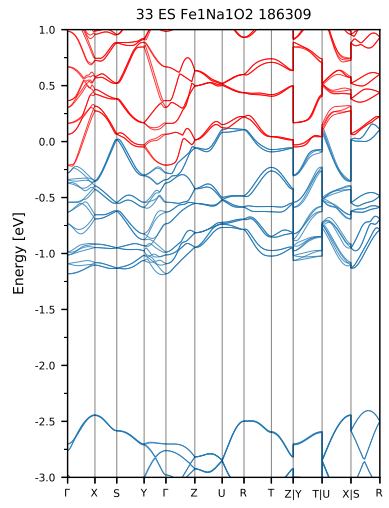

(r) 33.3.18

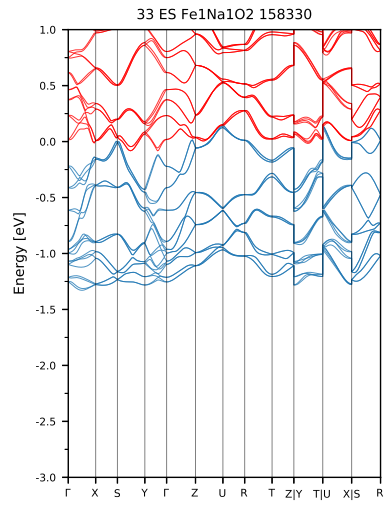

(s) 33.3.19

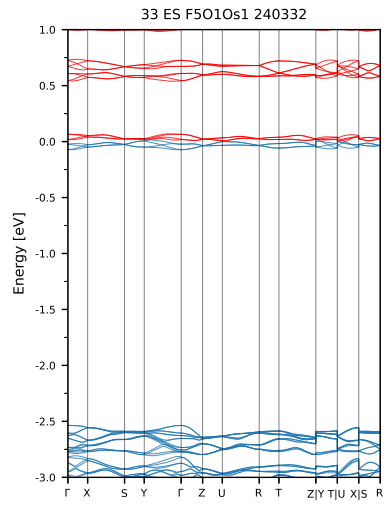

(t) 33.3.20

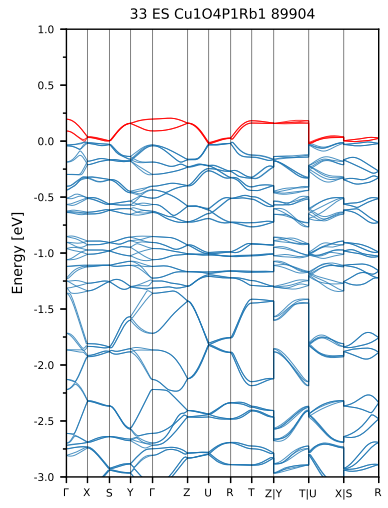

(u) 33.3.21

Figure 83: Band structures for topological materials in SG 33, part 2/2, from 33.3.17 to 33.3.21. See the list of figures.

## 0.0.14 SG 36

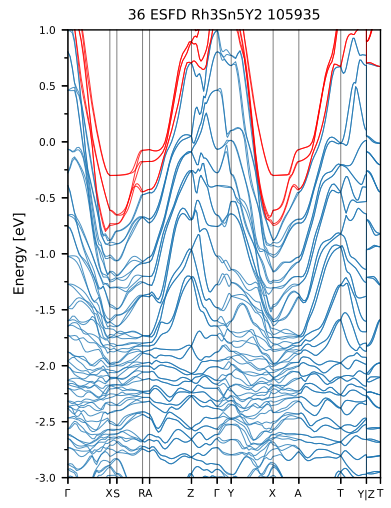

(a) 36.4.1

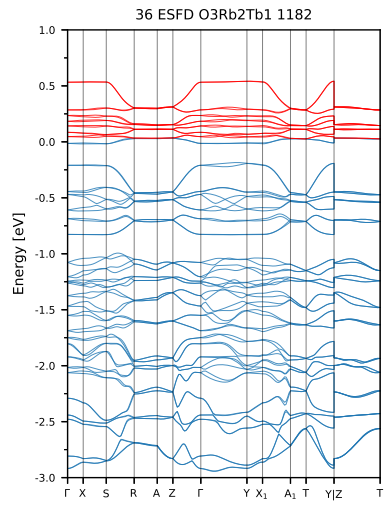

(b) 36.4.2

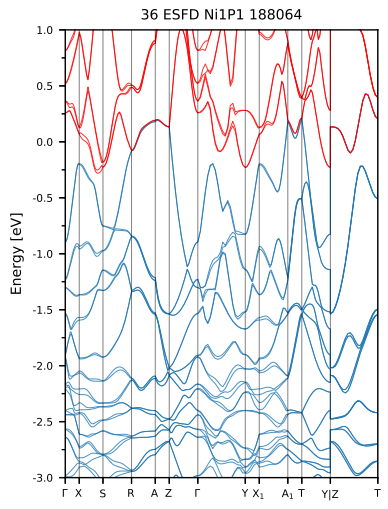

(c) 36.4.3

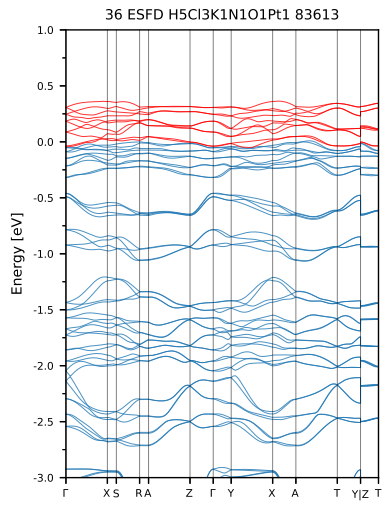

(d) 36.4.4

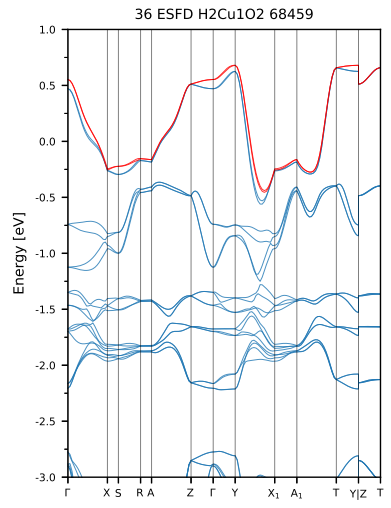

(e) 36.4.5

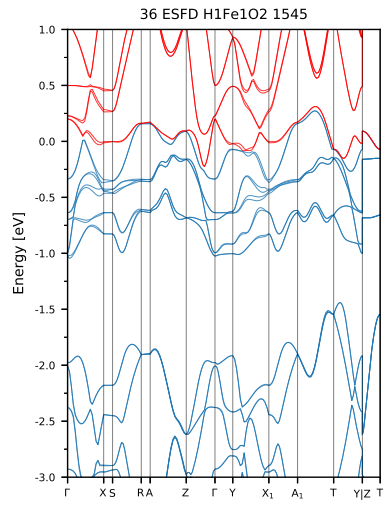

(f) 36.4.6

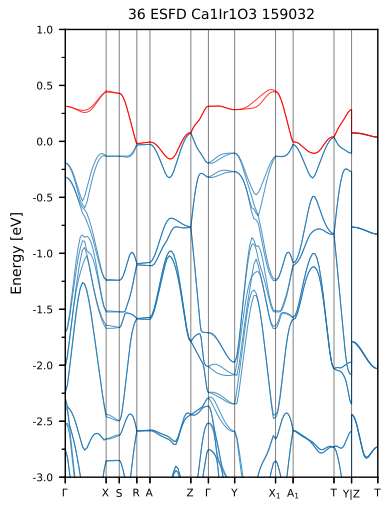

(g) 36.4.7

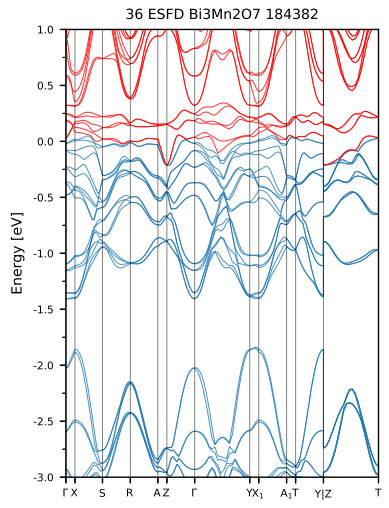

(h) 36.4.8

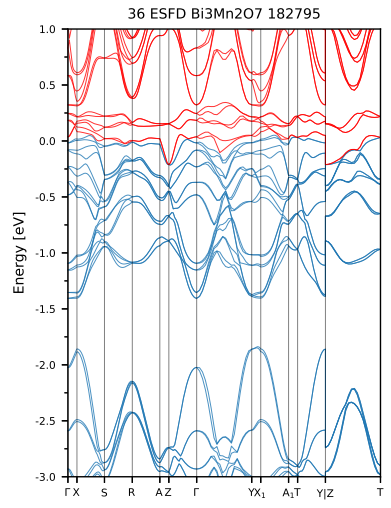

(i) 36.4.9

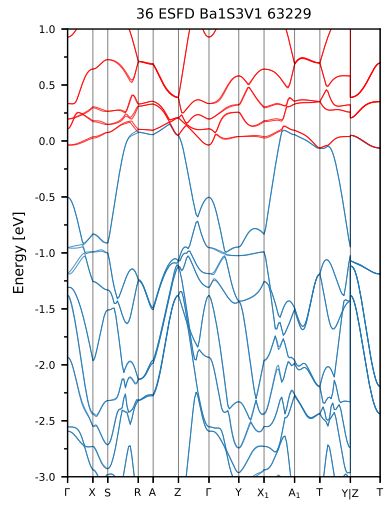

(j) 36.4.10

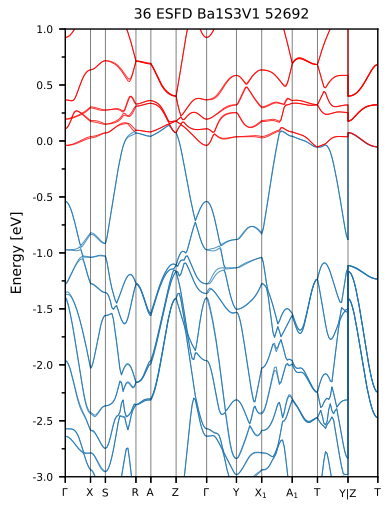

(k) 36.4.11

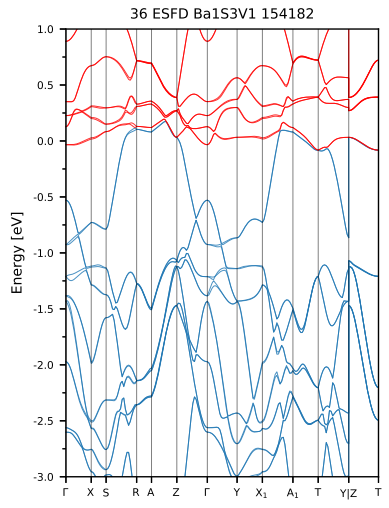

(l) 36.4.12

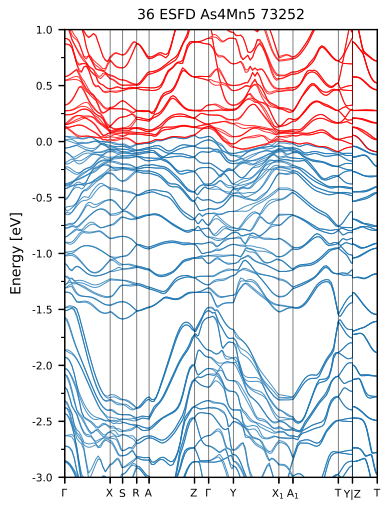

(m) 36.4.13

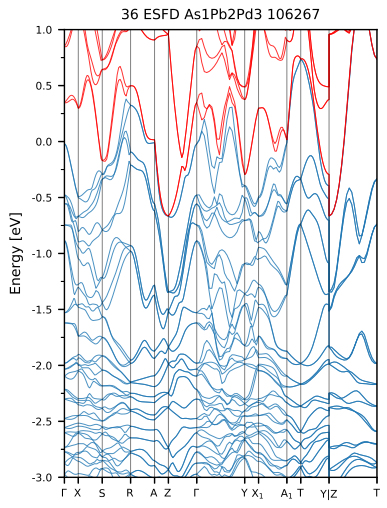

(n) 36.4.14

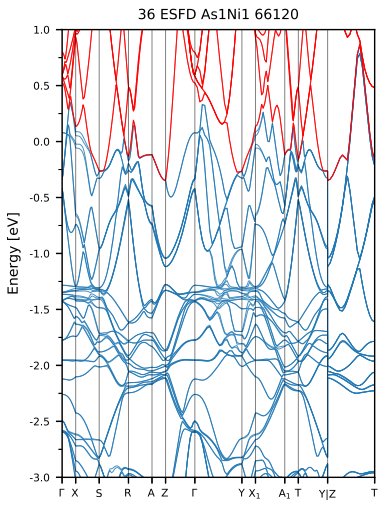

(o) 36.4.15

Figure 84: Band structures for topological materials in SG 36, part 1/1, from 36.4.1 to 36.4.15. See the list of figures.

### 0.0.15 SG 43

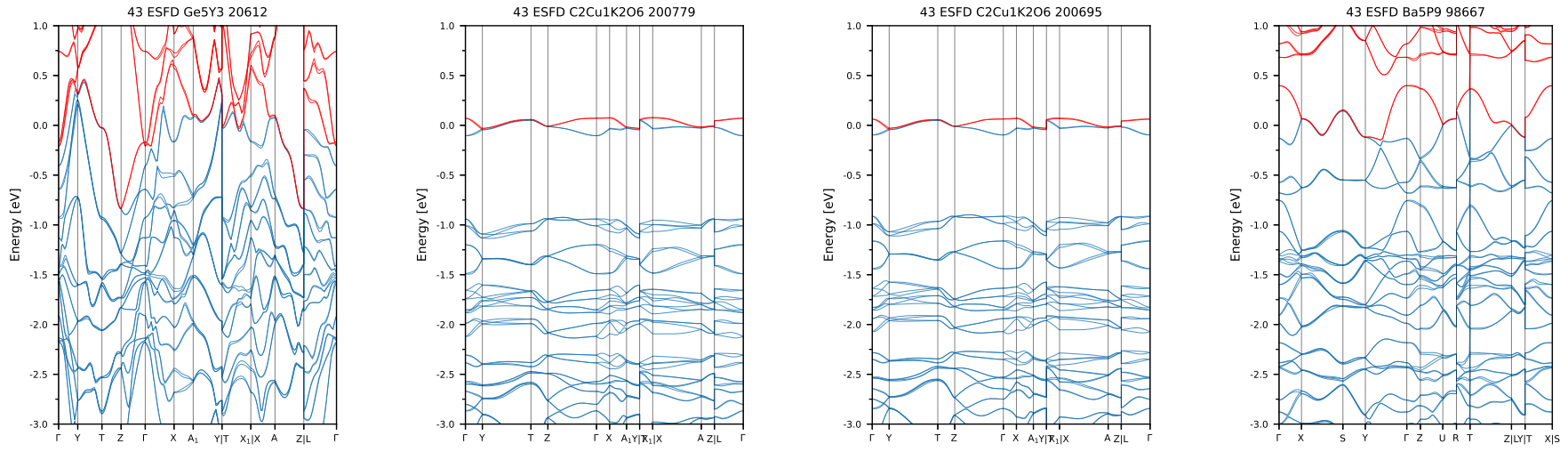

(a) 43.4.1 (b) 43.4.2 (c) 43.4.3 (d) 43.4.4

Figure 85: Band structures for topological materials in SG 43, part 1/1, from 43.4.1 to 43.4.4. See the list of figures.

### 0.0.16 SG 47

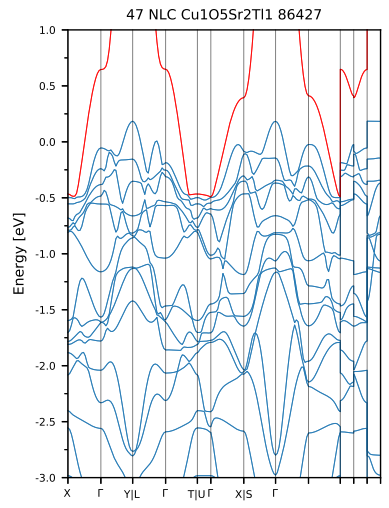

(a) 47.2.1

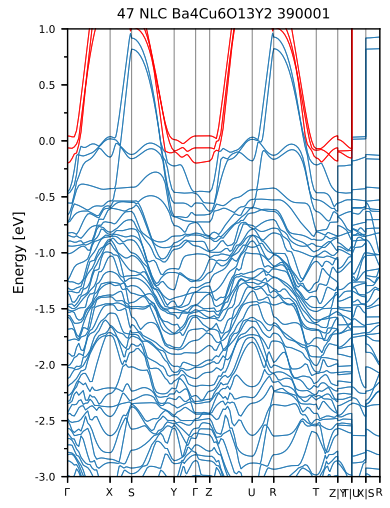

(b) 47.2.2

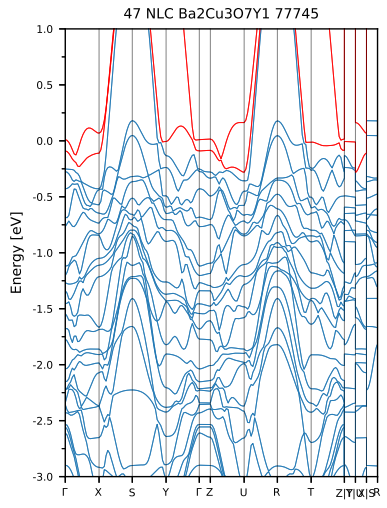

(c) 47.2.3

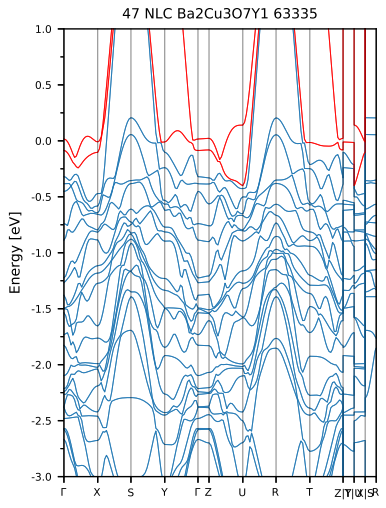

(d) 47.2.4

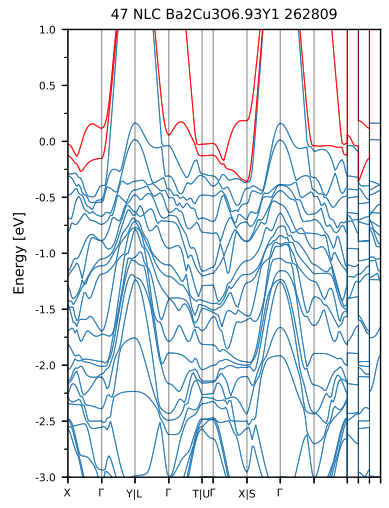

(e) 47.2.5

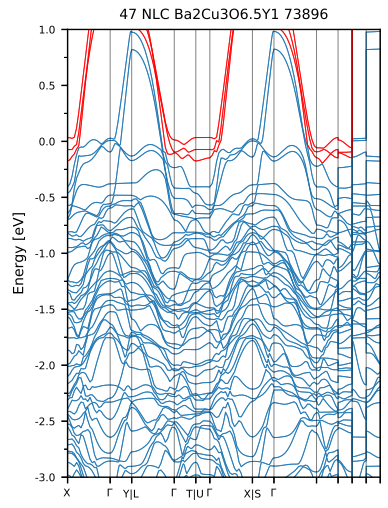

(f) 47.2.6

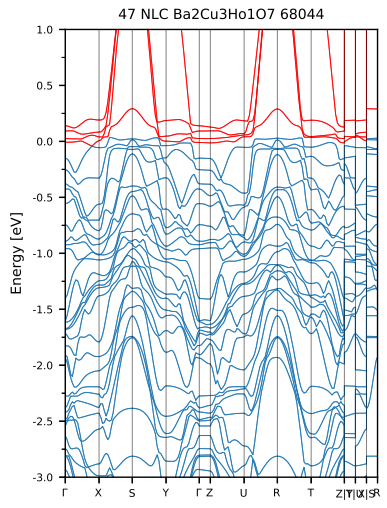

(g) 47.2.7

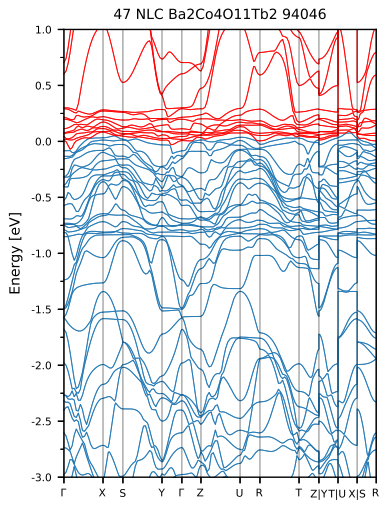

(h) 47.2.8

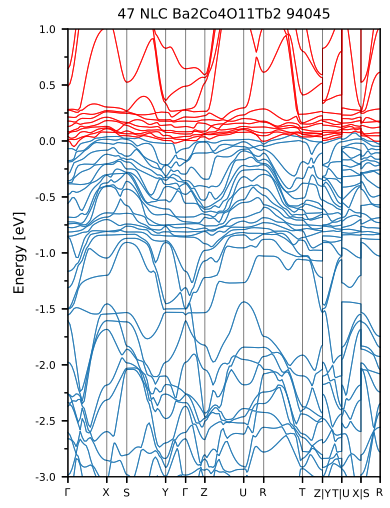

(i) 47.2.9

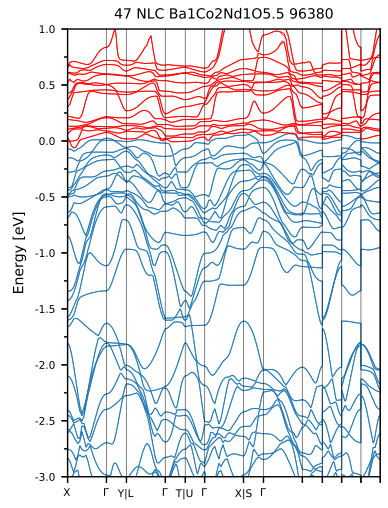

(j) 47.2.10

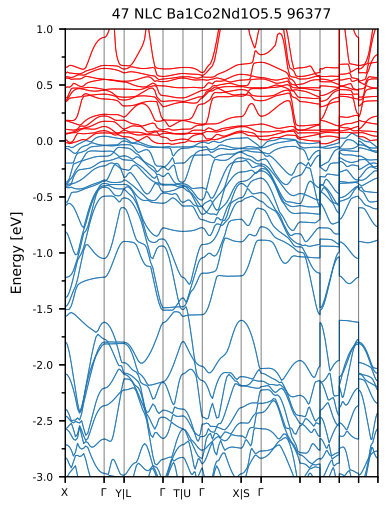

(k) 47.2.11

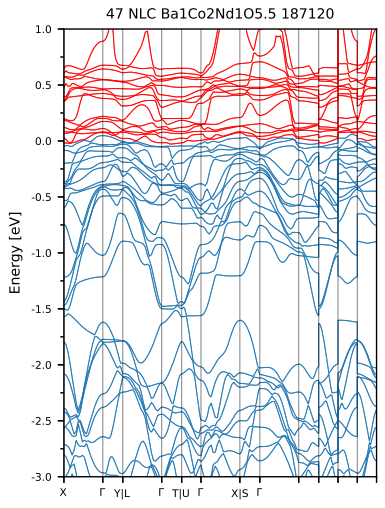

(l) 47.2.12

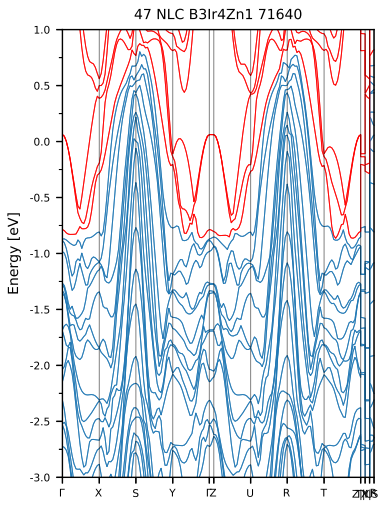

(m) 47.2.13

Figure 86: Band structures for topological materials in SG 47, part 1/1, from 47.2.1 to 47.2.13. See the list of figures.

## 0.0.17 SG 51

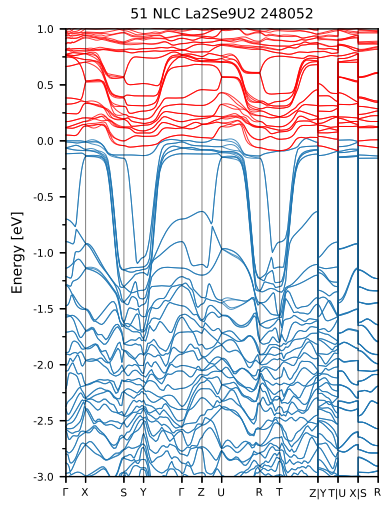

(a) 51.2.1

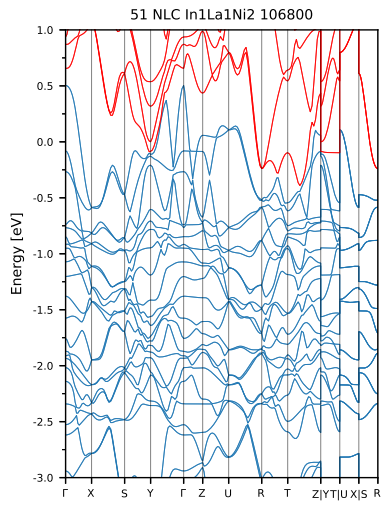

(b) 51.2.2

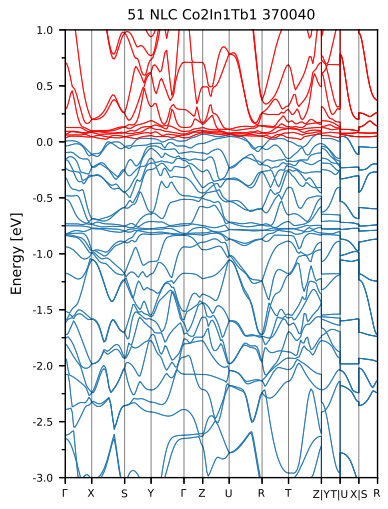

(c) 51.2.3

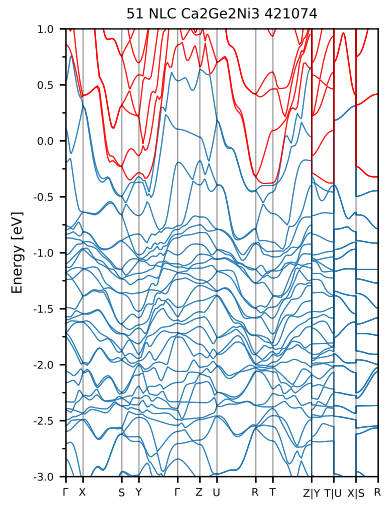

(d) 51.2.4

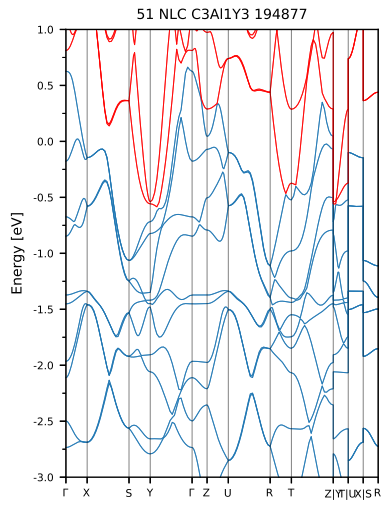

(e) 51.2.5

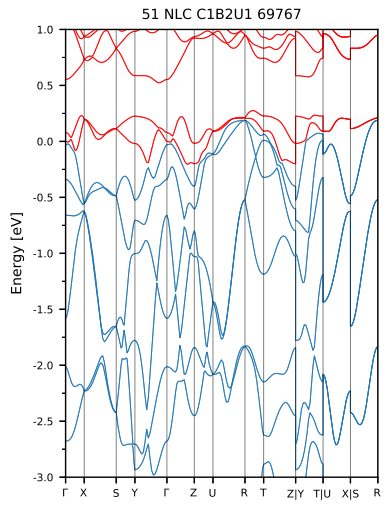

(f) 51.2.6

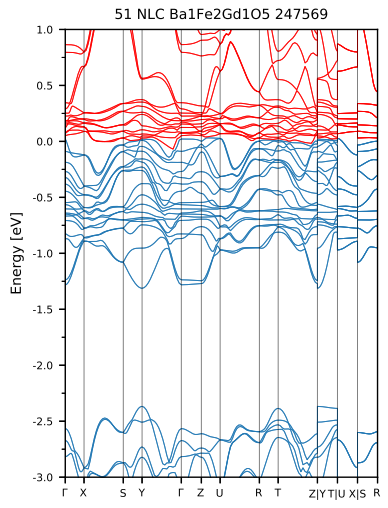

(g) 51.2.7

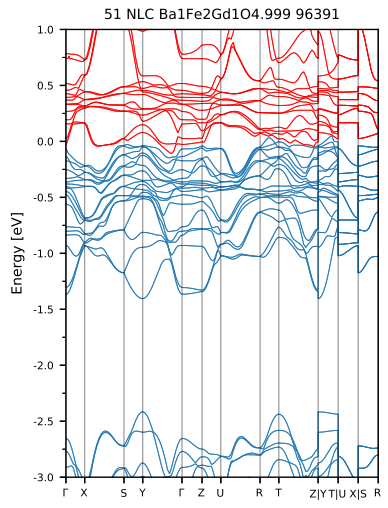

(h) 51.2.8

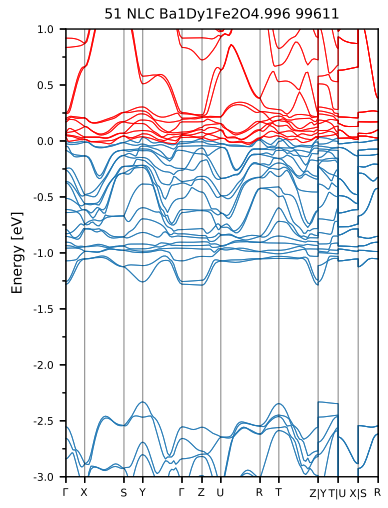

(i) 51.2.9

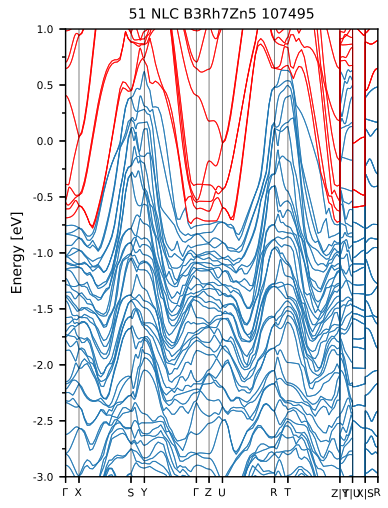

(j) 51.2.10

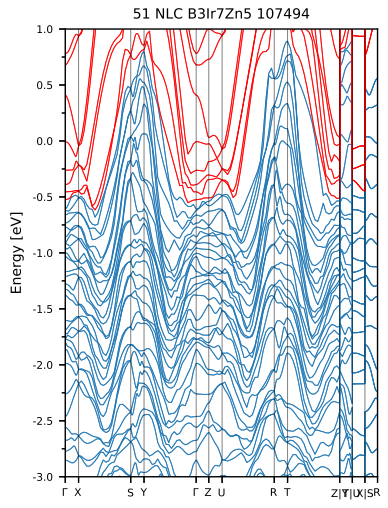

(k) 51.2.11

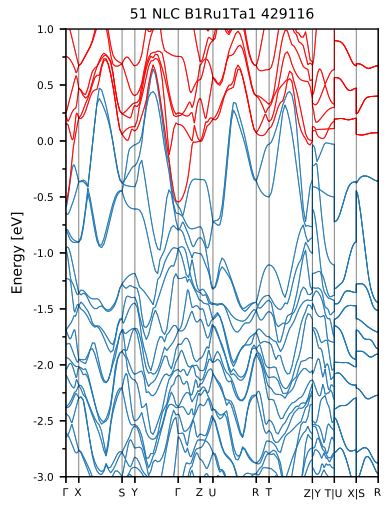

(l) 51.2.12

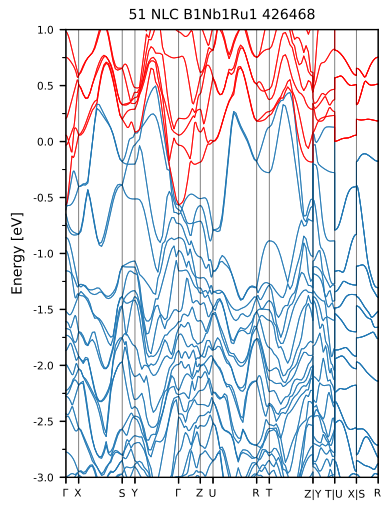

(m) 51.2.13

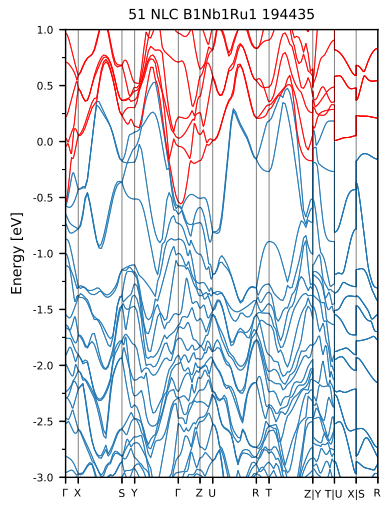

(n) 51.2.14

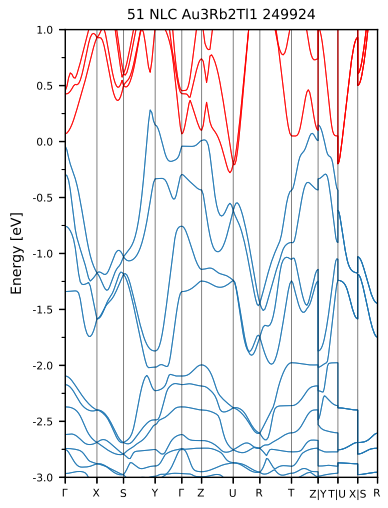

(o) 51.2.15

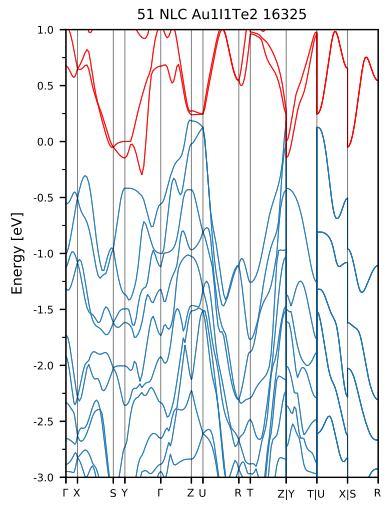

(p) 51.2.16

Figure 87: Band structures for topological materials in SG 51, part 1/3, from 51.2.1 to 51.2.16. See the list of figures.

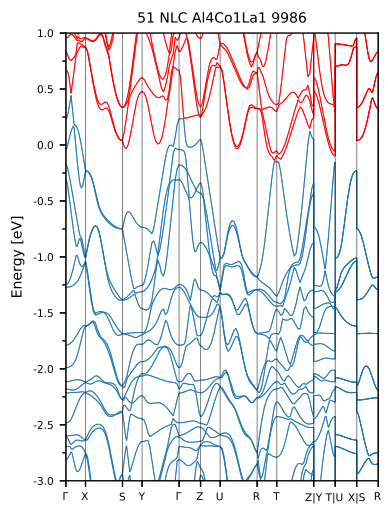

(q) 51.2.17

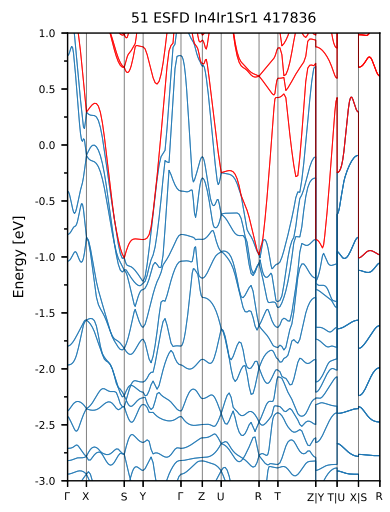

(r) 51.4.1

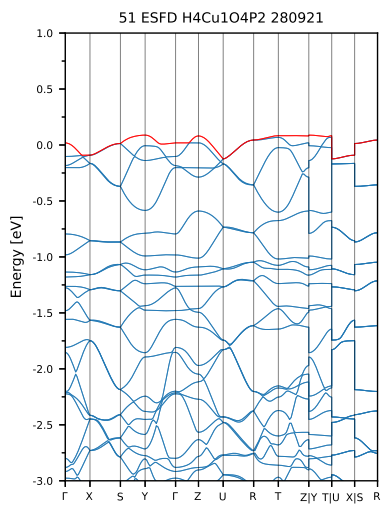

(s) 51.4.2

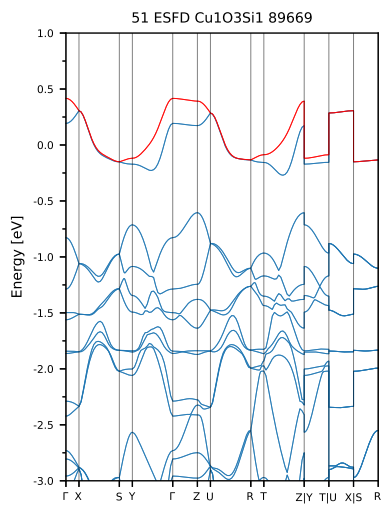

(t) 51.4.3

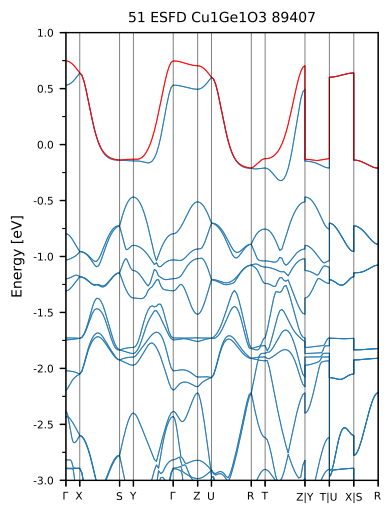

(u) 51.4.4

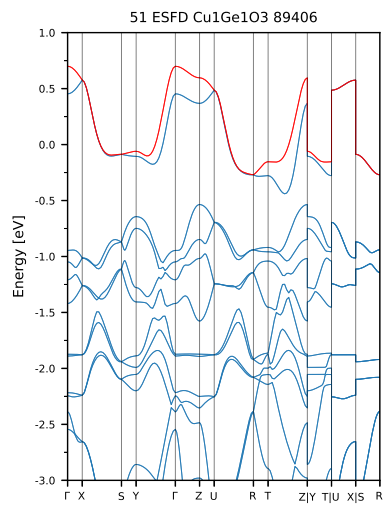

(v) 51.4.5

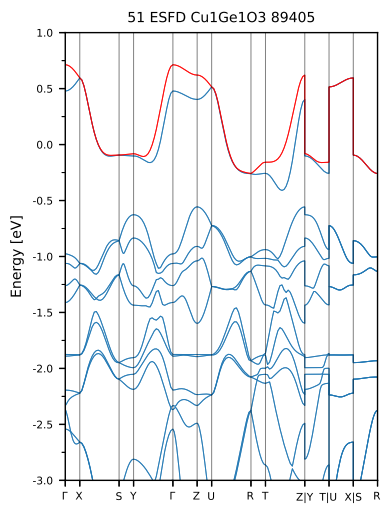

(w) 51.4.6

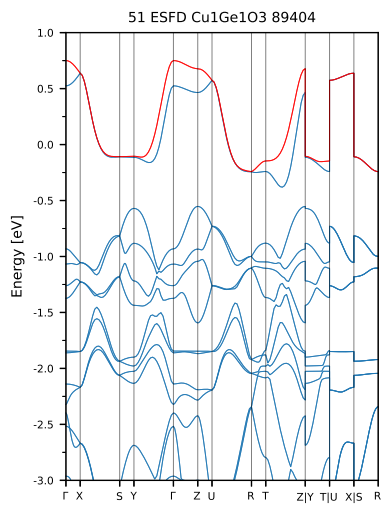

(x) 51.4.7

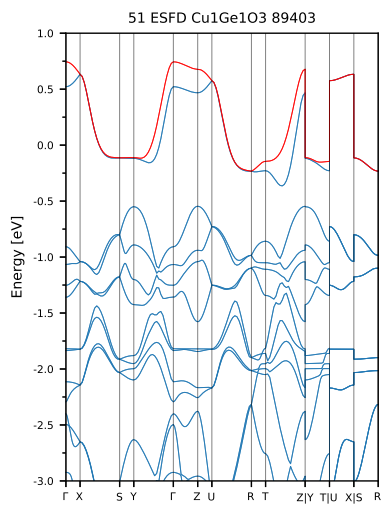

(y) 51.4.8

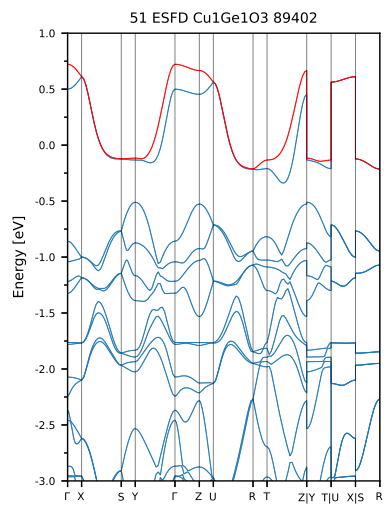

(z) 51.4.9

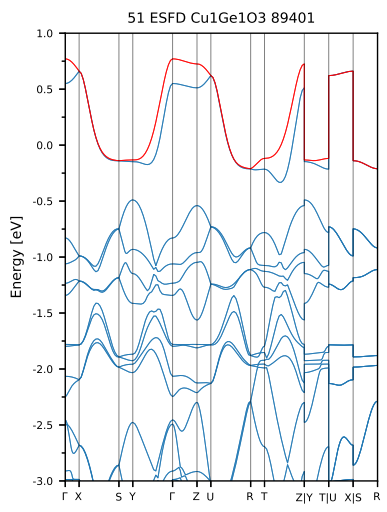

(aa) 51.4.10

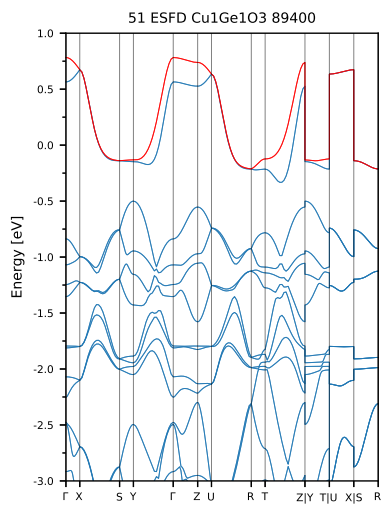

(ab) 51.4.11

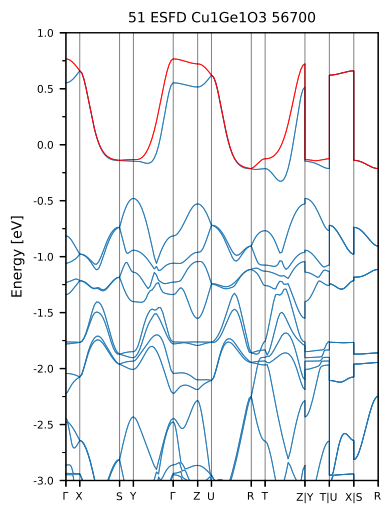

(ac) 51.4.12

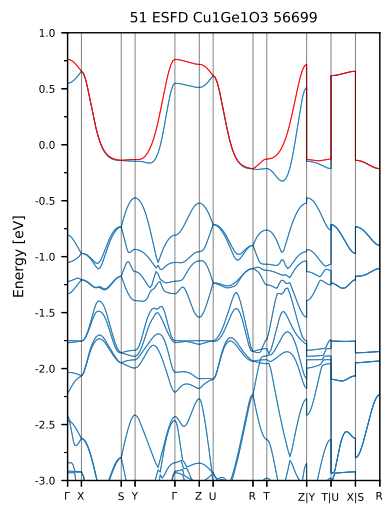

(ad) 51.4.13

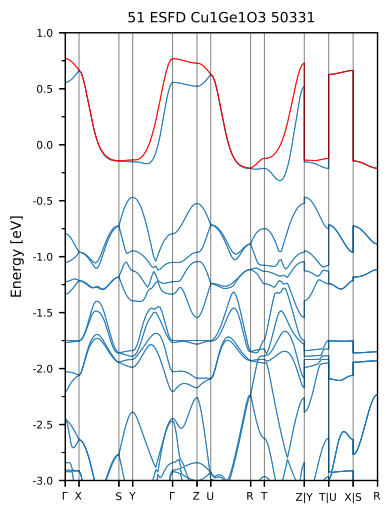

(ae) 51.4.14

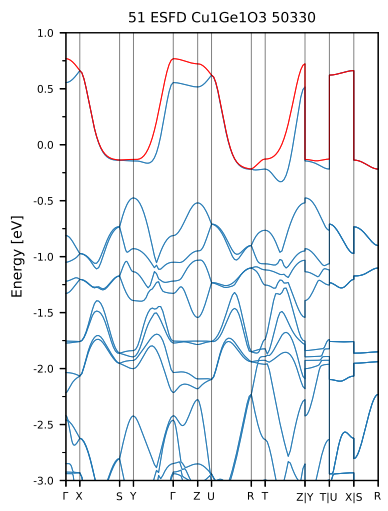

(af) 51.4.15

Figure 88: Band structures for topological materials in SG 51, part 2/3, from 51.2.17 to 51.4.15. See the list of figures.

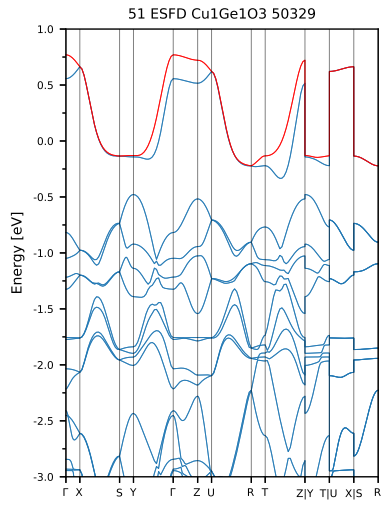

(ag) 51.4.16

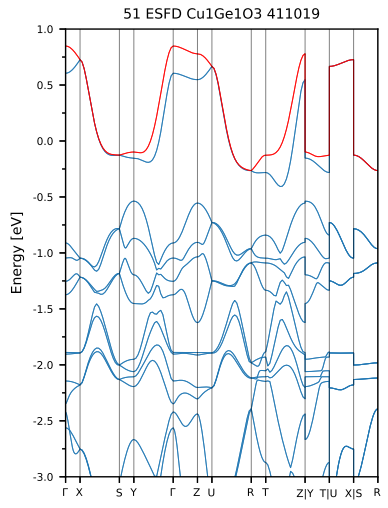

(ah) 51.4.17

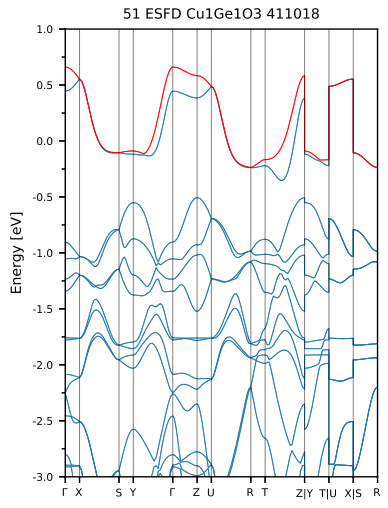

(ai) 51.4.18

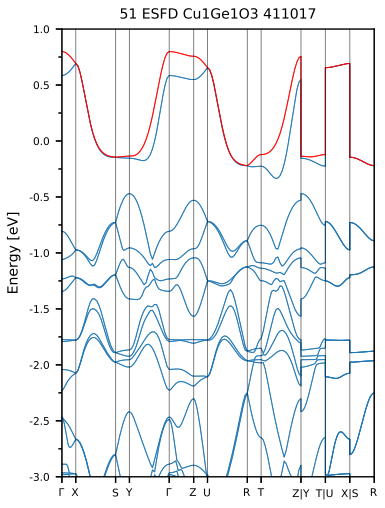

(aj) 51.4.19

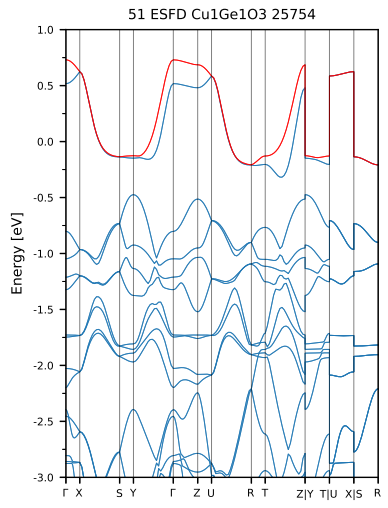

(ak) 51.4.20

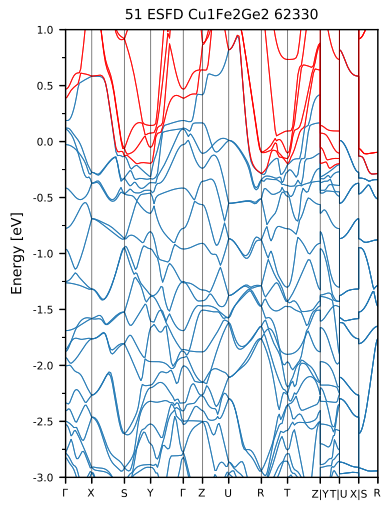

(al) 51.4.21

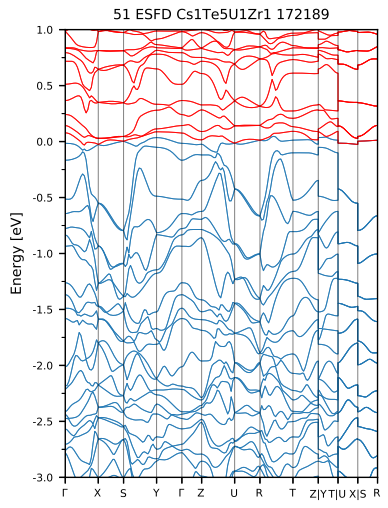

(am) 51.4.22

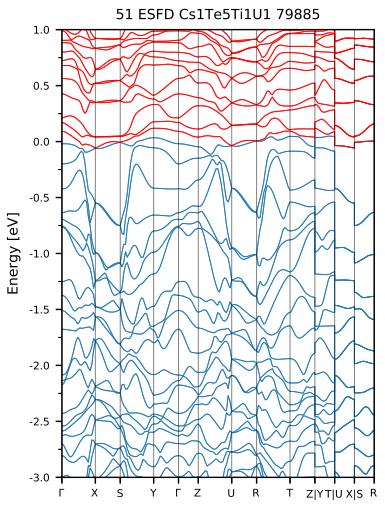

(an) 51.4.23

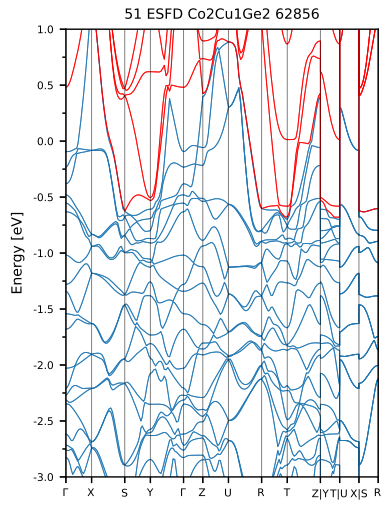

(ao) 51.4.24

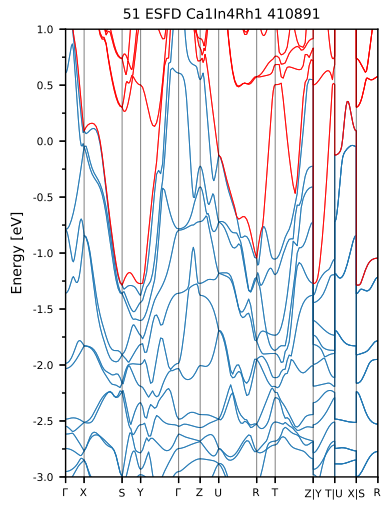

(ap) 51.4.25

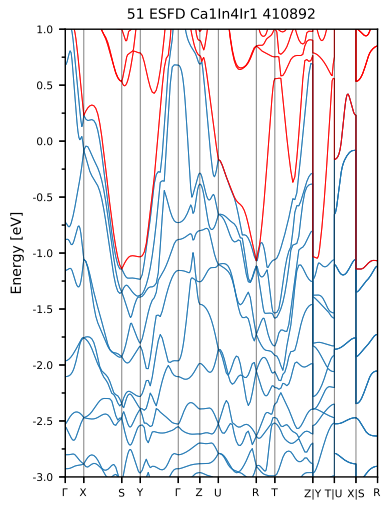

(aq) 51.4.26

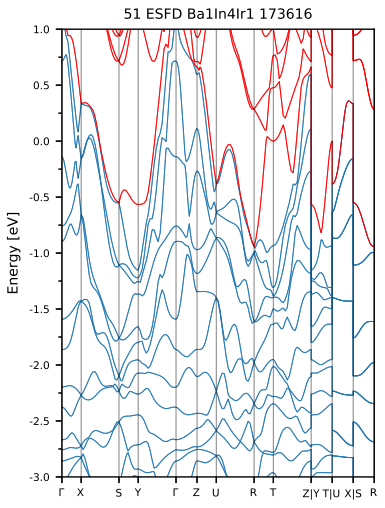

(ar) 51.4.27

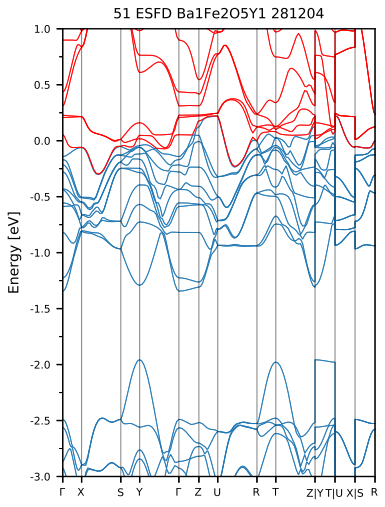

(as) 51.4.28

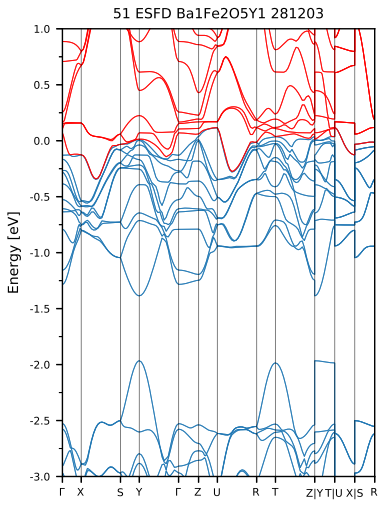

(at) 51.4.29

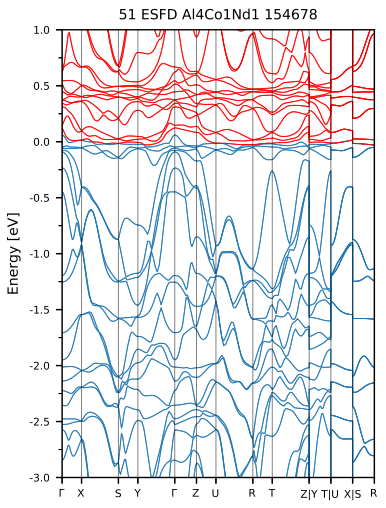

(au) 51.4.30

Figure 89: Band structures for topological materials in SG 51, part 3/3, from 51.4.16 to 51.4.30. See the list of figures.

### 0.0.18 SG 52

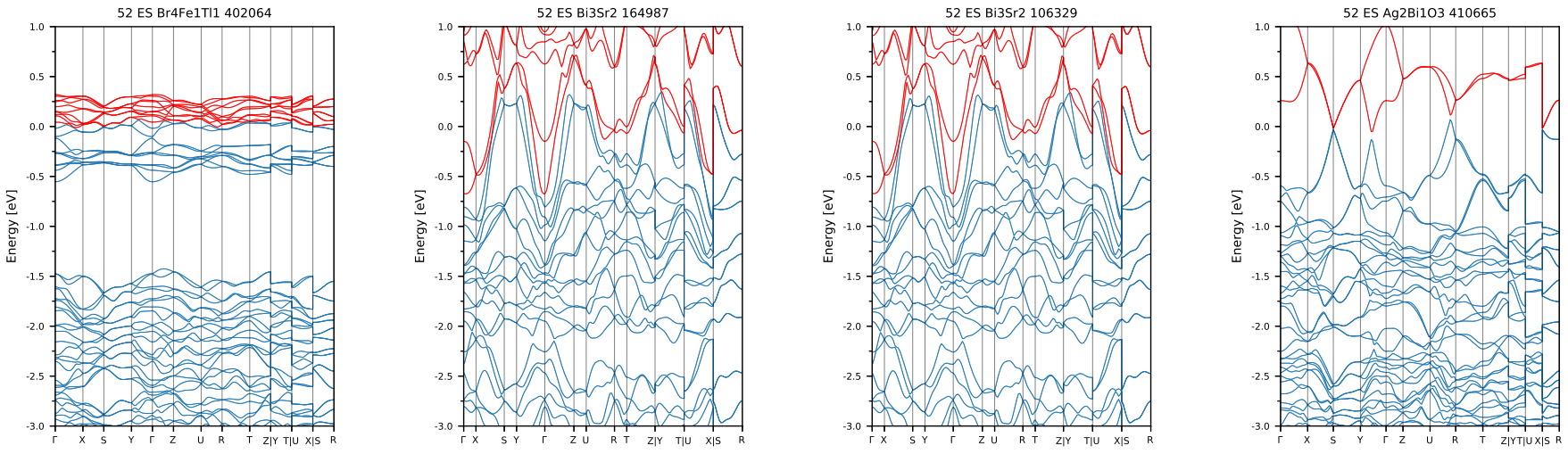

(a) 52.3.1 (b) 52.3.2 (c) 52.3.3 (d) 52.3.4

Figure 90: Band structures for topological materials in SG 52, part 1/1, from 52.3.1 to 52.3.4. See the list of figures.

### 0.0.19 SG 53

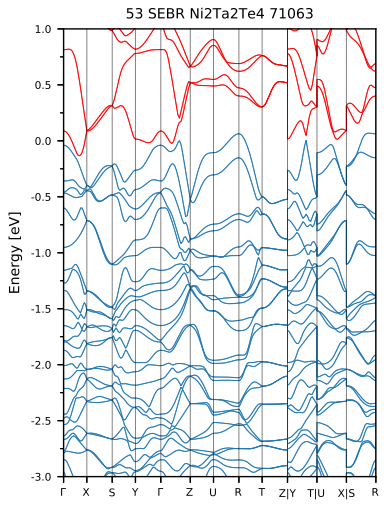

(a) 53.1.1

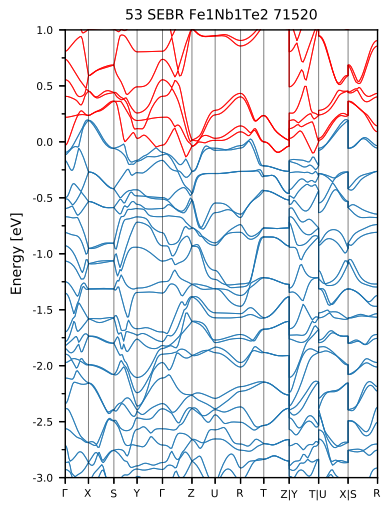

(b) 53.1.2

Figure 91: Band structures for topological materials in SG 53, part 1/1, from 53.1.1 to 53.1.2. See the list of figures.

### 0.0.20 SG 55

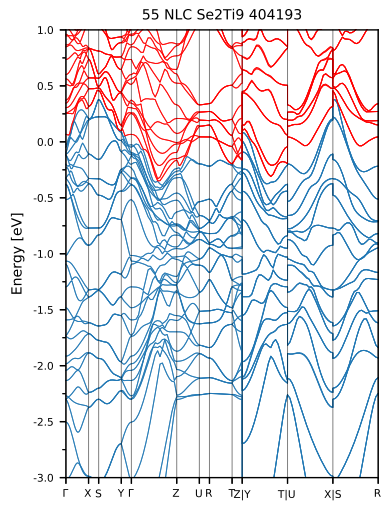

(a) 55.2.1

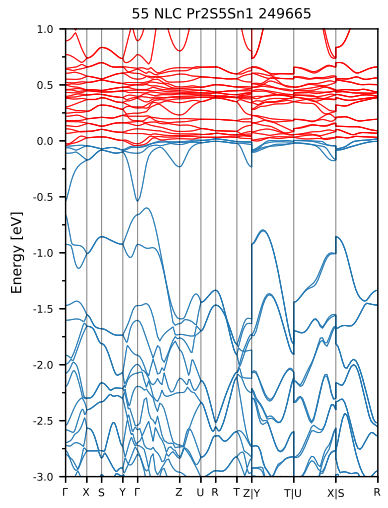

(b) 55.2.2

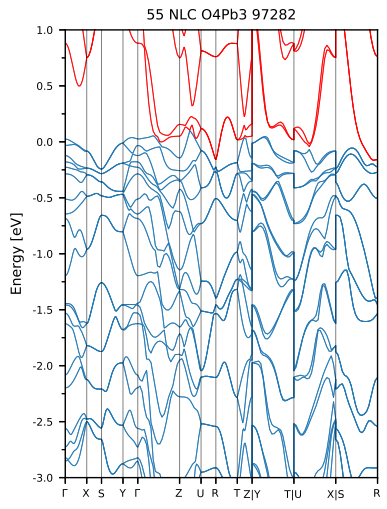

(c) 55.2.3

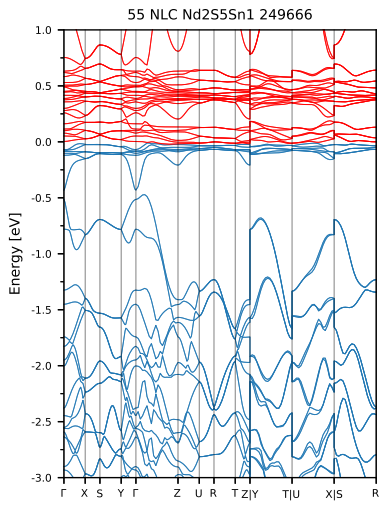

(d) 55.2.4

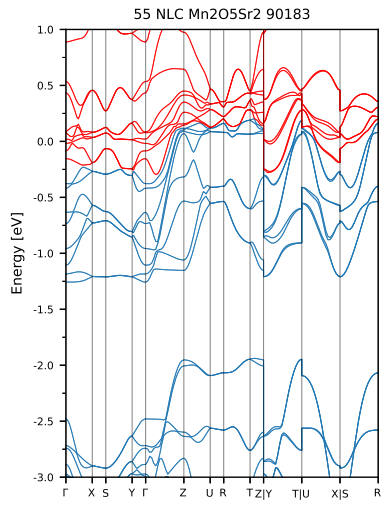

(e) 55.2.5

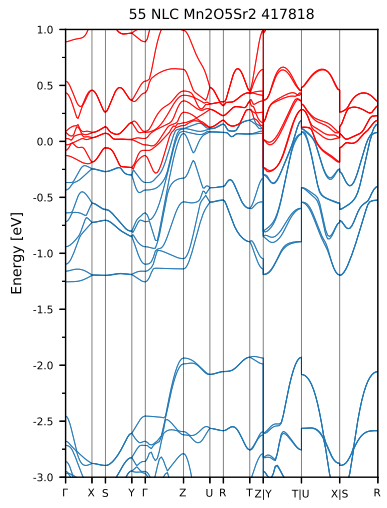

(f) 55.2.6

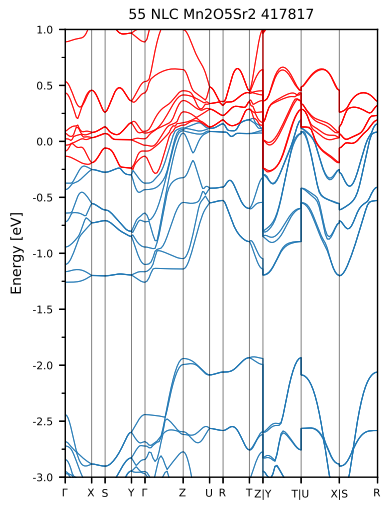

(g) 55.2.7

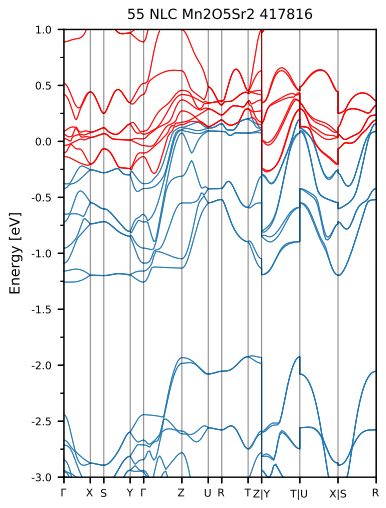

(h) 55.2.8

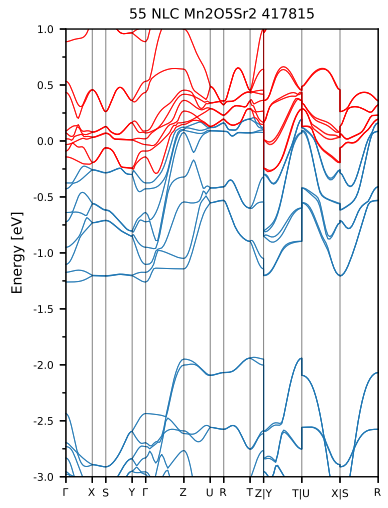

(i) 55.2.9

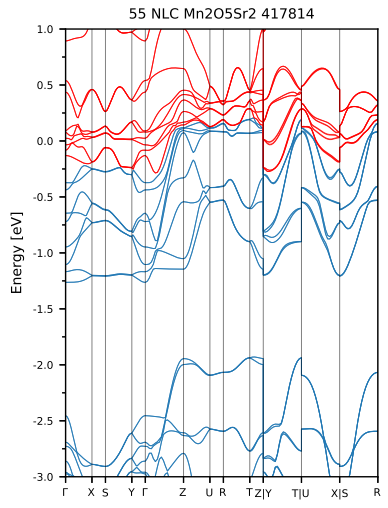

(j) 55.2.10

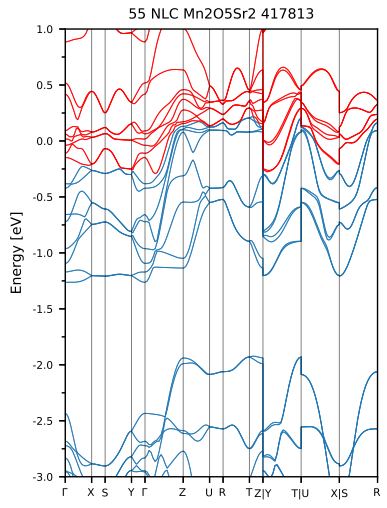

(k) 55.2.11

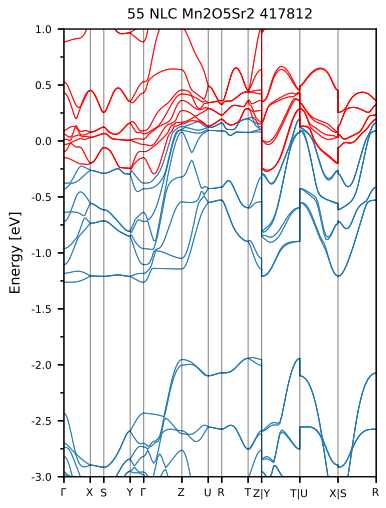

(l) 55.2.12

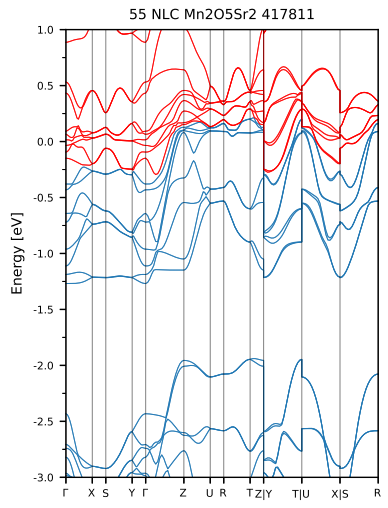

(m) 55.2.13

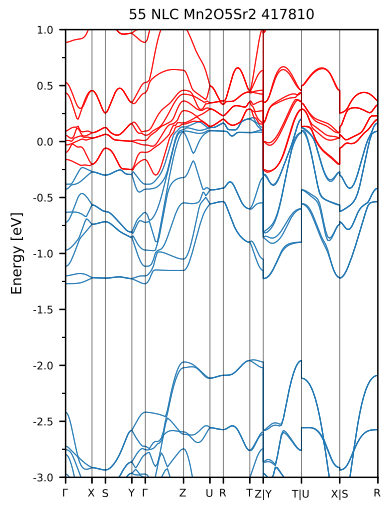

(n) 55.2.14

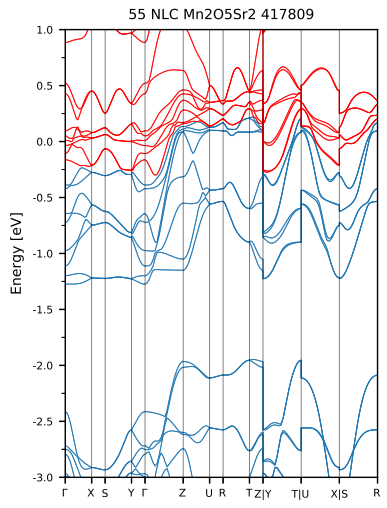

(o) 55.2.15

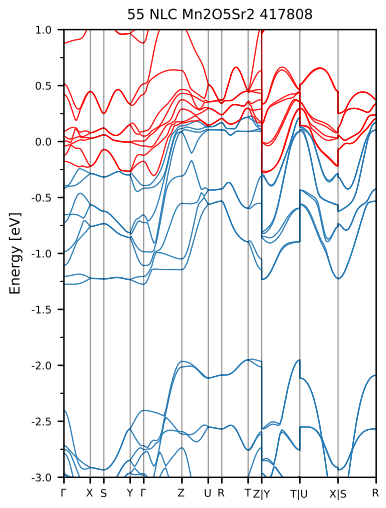

(p) 55.2.16

Figure 92: Band structures for topological materials in SG 55, part 1/6, from 55.2.1 to 55.2.16. See the list of figures.

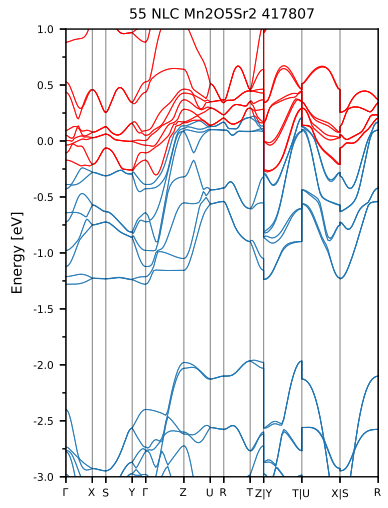

(q) 55.2.17

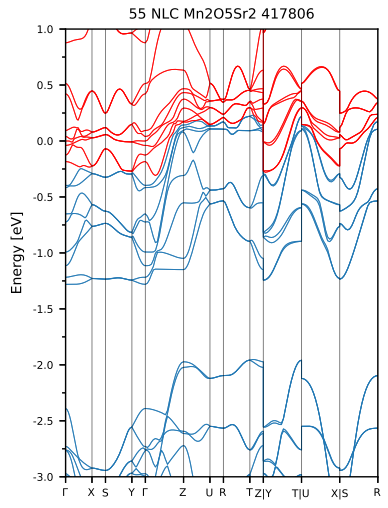

(r) 55.2.18

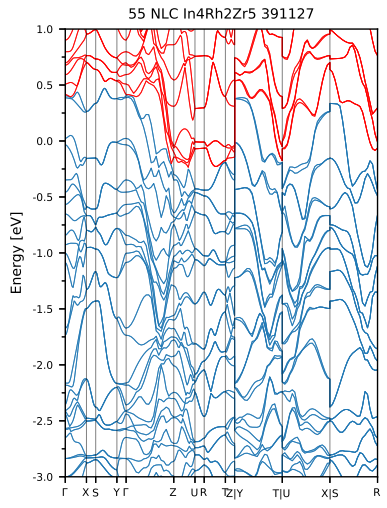

(s) 55.2.19

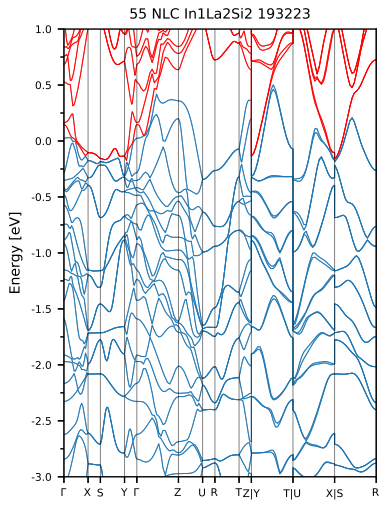

(t) 55.2.20

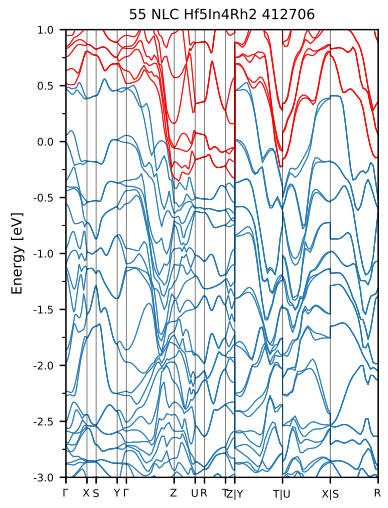

(u) 55.2.21

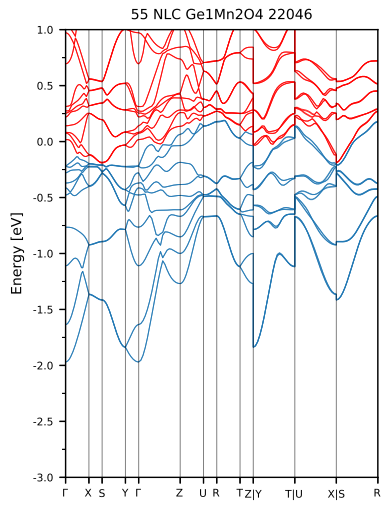

(v) 55.2.22

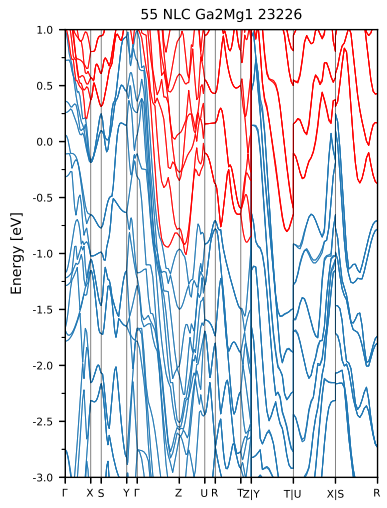

(w) 55.2.23

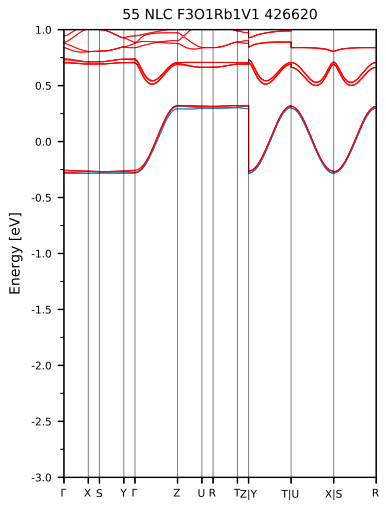

(x) 55.2.24

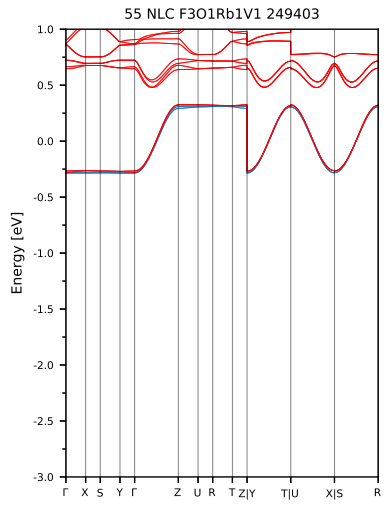

(y) 55.2.25

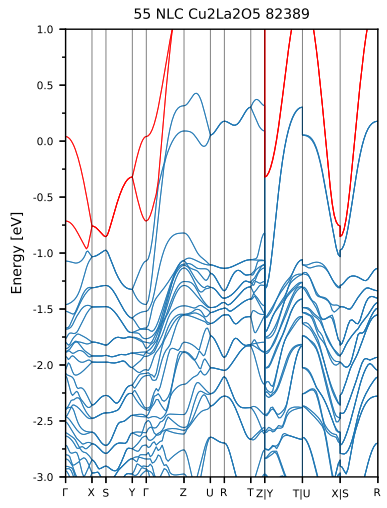

(z) 55.2.26

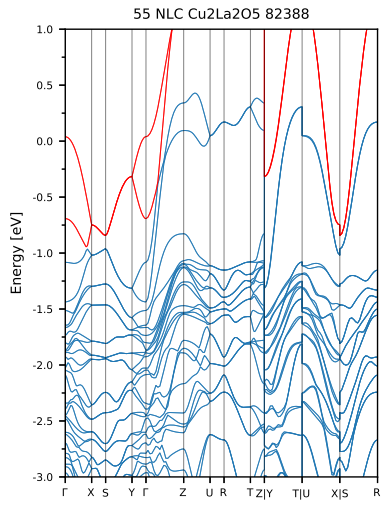

(aa) 55.2.27

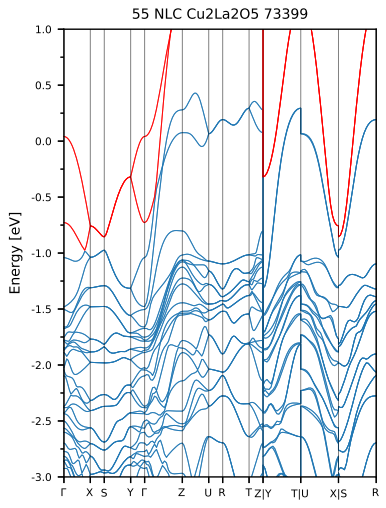

(ab) 55.2.28

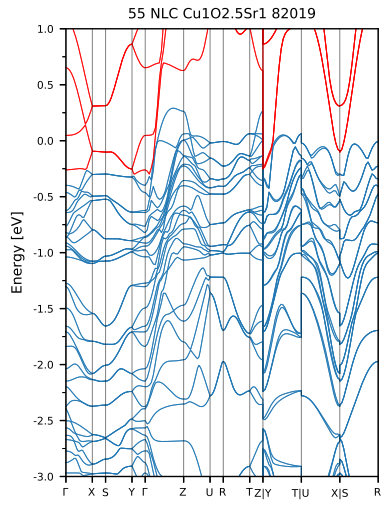

(ac) 55.2.29

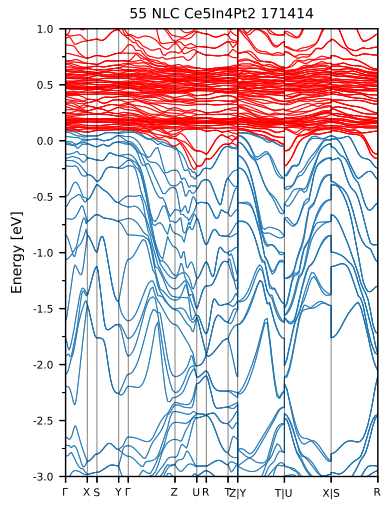

(ad) 55.2.30

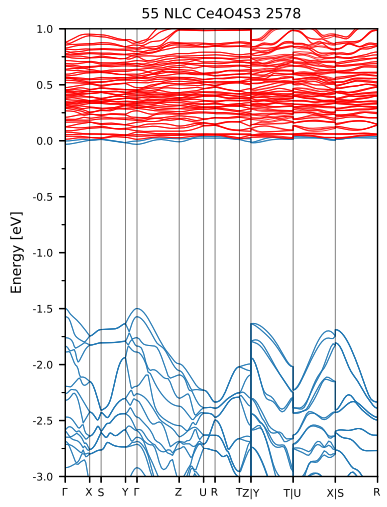

(ae) 55.2.31

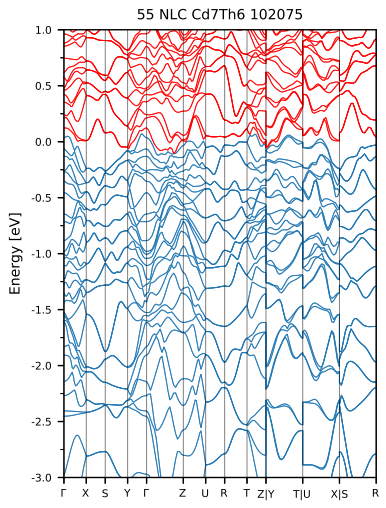

(af) 55.2.32

Figure 93: Band structures for topological materials in SG 55, part 2/6, from 55.2.17 to 55.2.32. See the list of figures.

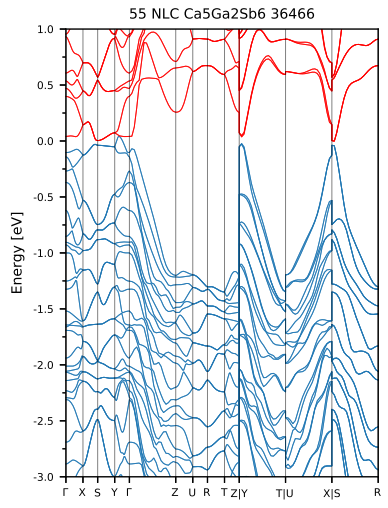

(ag) 55.2.33

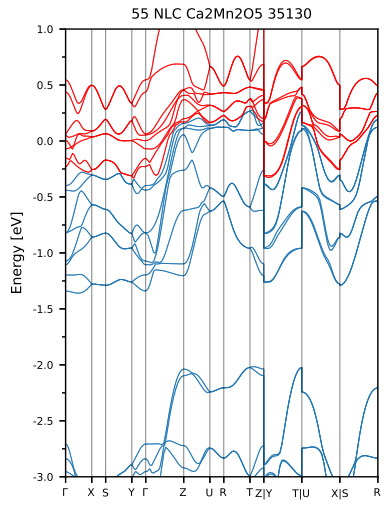

(ah) 55.2.34

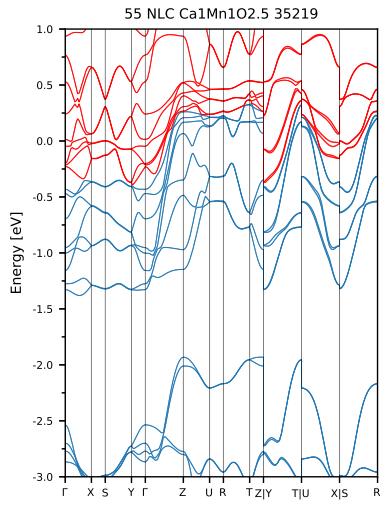

(ai) 55.2.35

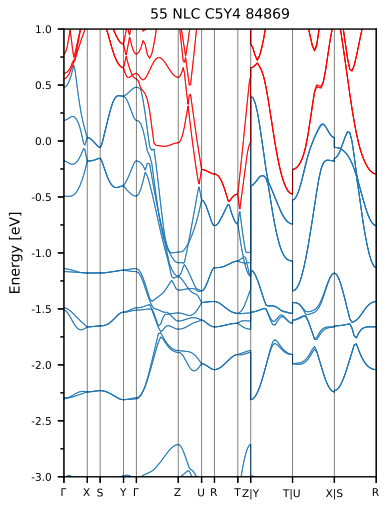

(aj) 55.2.36

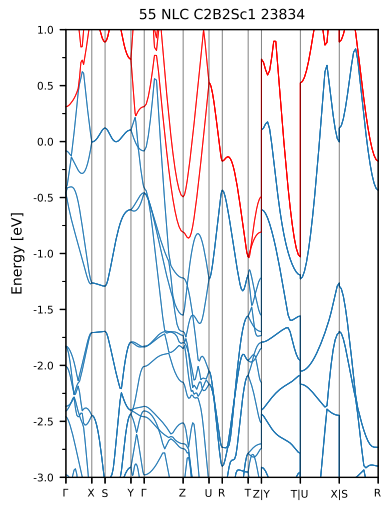

(ak) 55.2.37

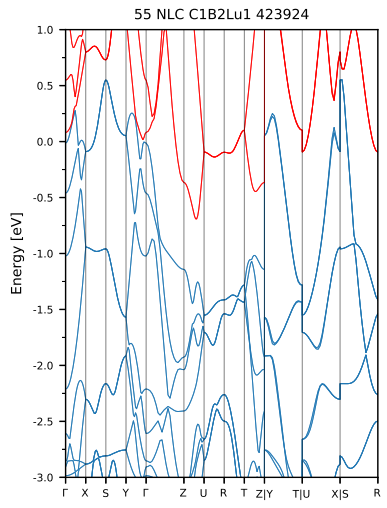

(al) 55.2.38

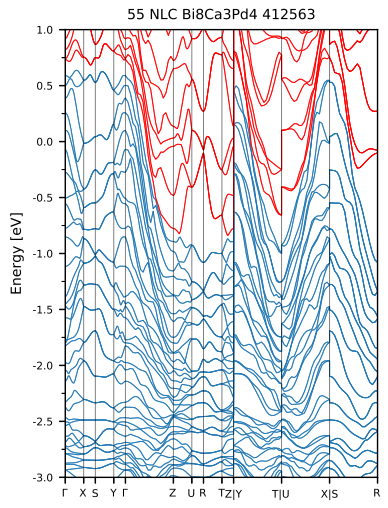

(am) 55.2.39

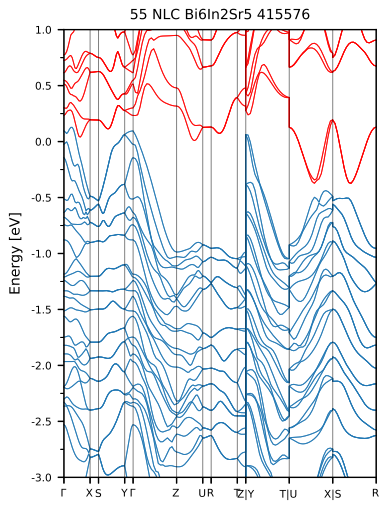

(an) 55.2.40

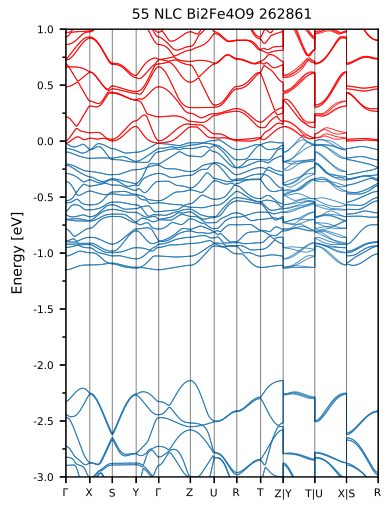

(ao) 55.2.41

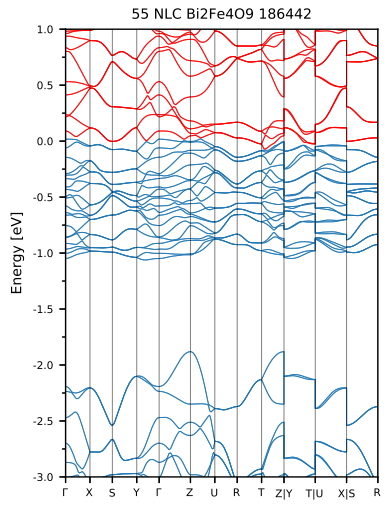

(ap) 55.2.42

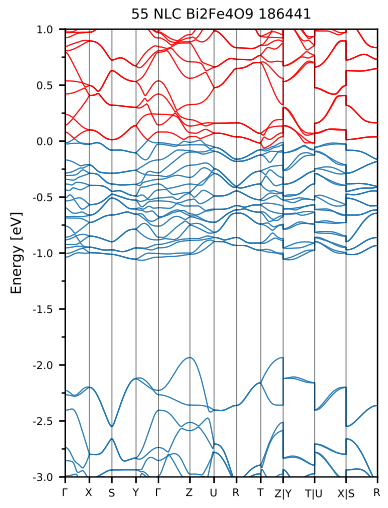

(aq) 55.2.43

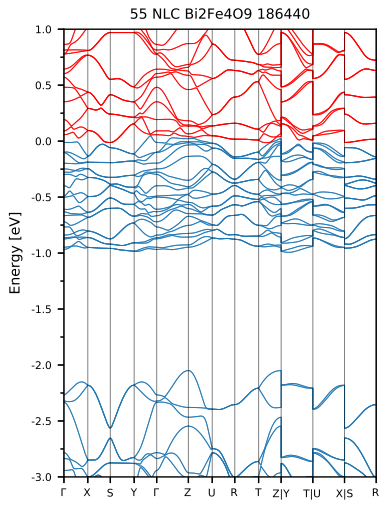

(ar) 55.2.44

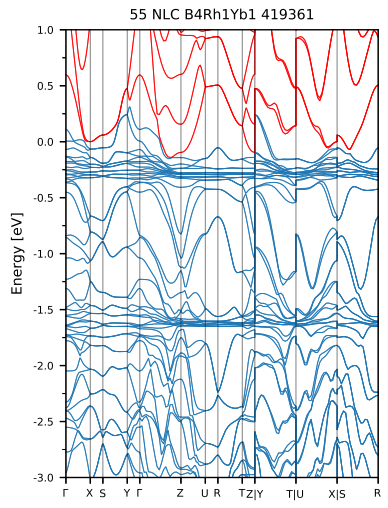

(as) 55.2.45

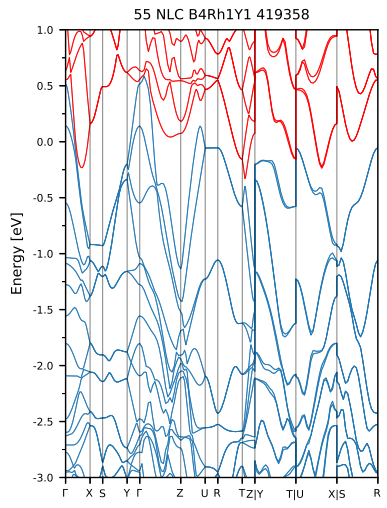

(at) 55.2.46

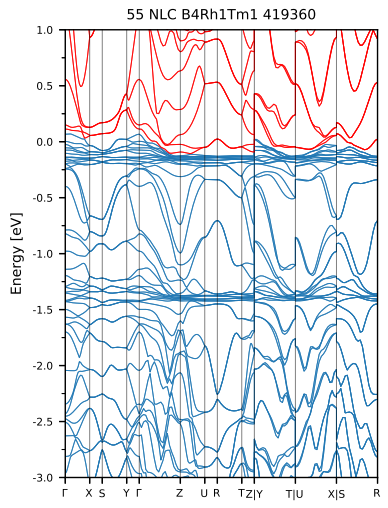

(au) 55.2.47

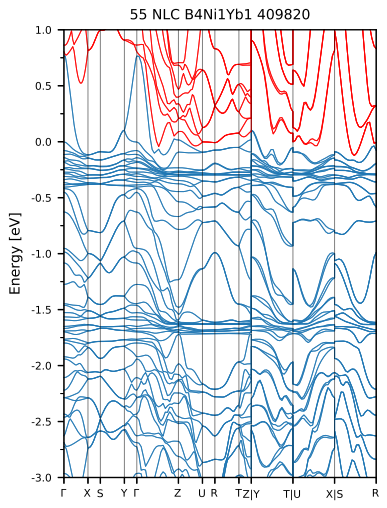

(av) 55.2.48

Figure 94: Band structures for topological materials in SG 55, part 3/6, from 55.2.33 to 55.2.48. See the list of figures.

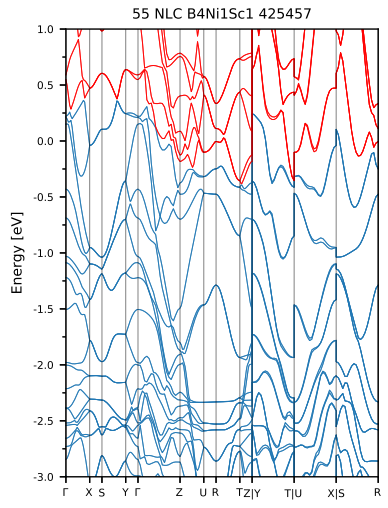

(aw) 55.2.49

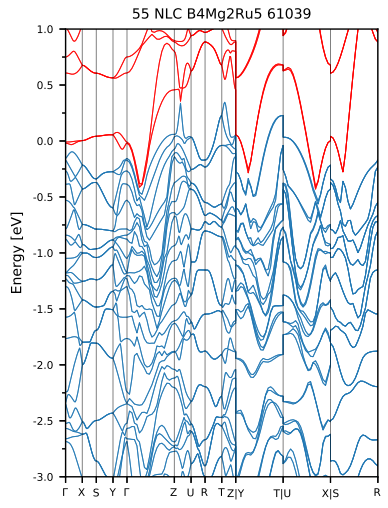

(ax) 55.2.50

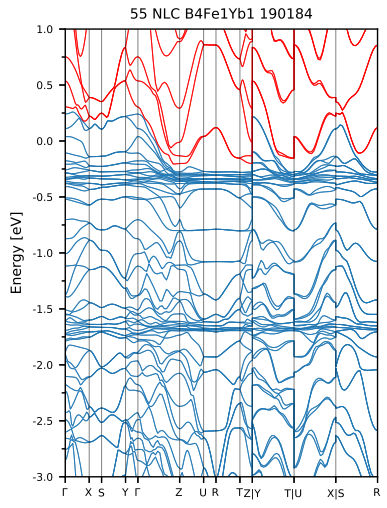

(ay) 55.2.51

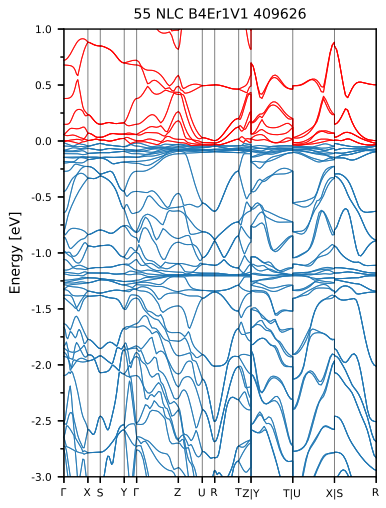

(az) 55.2.52

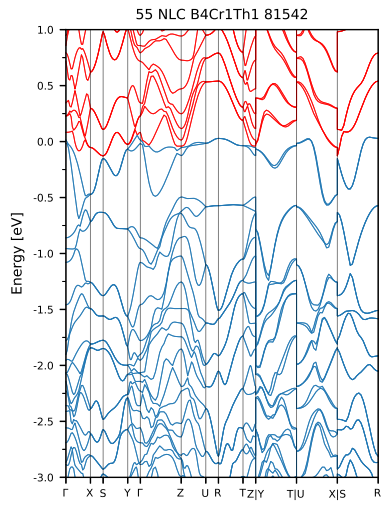

(ba) 55.2.53

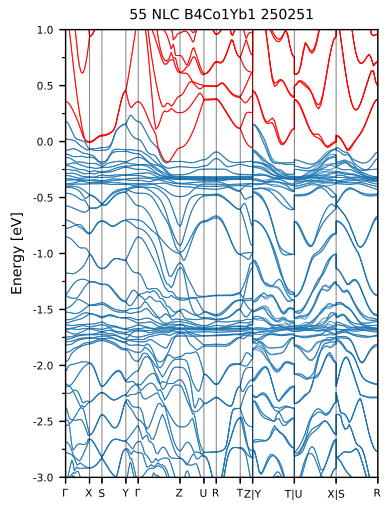

(bb) 55.2.54

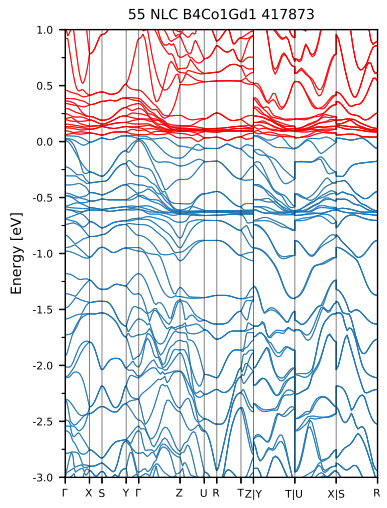

(bc) 55.2.55

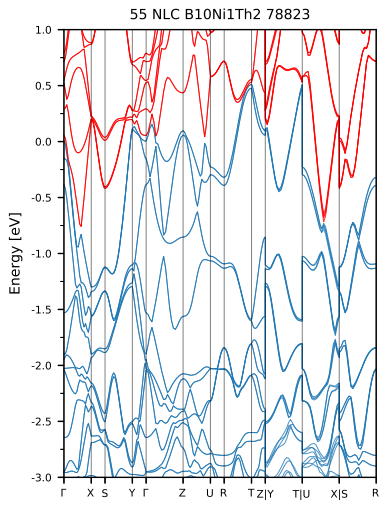

(bd) 55.2.56

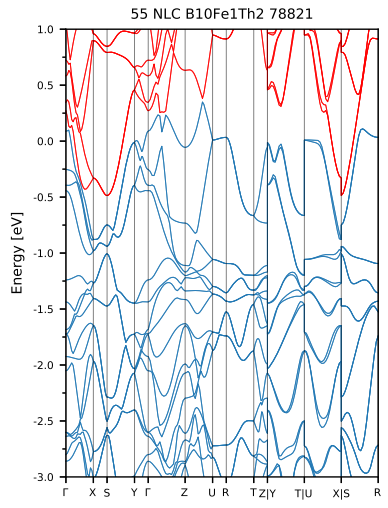

(be) 55.2.57

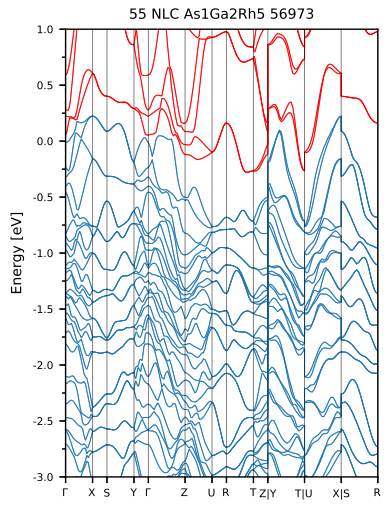

(bf) 55.2.58

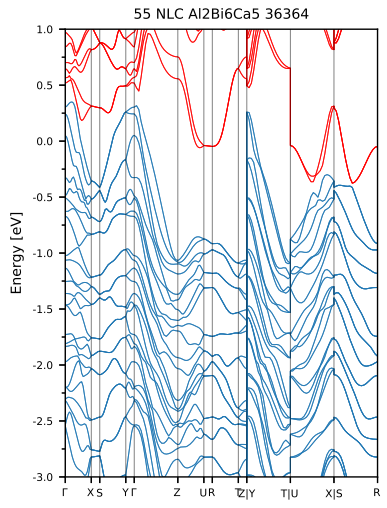

(bg) 55.2.59

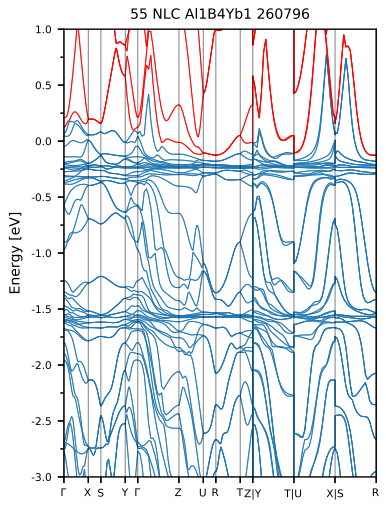

(bh) 55.2.60

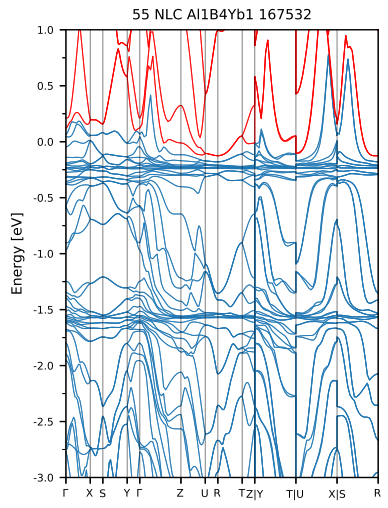

(bi) 55.2.61

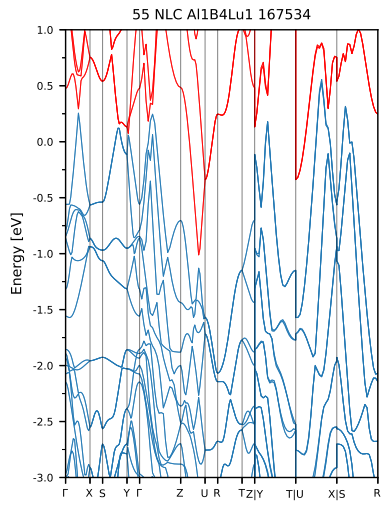

(bj) 55.2.62

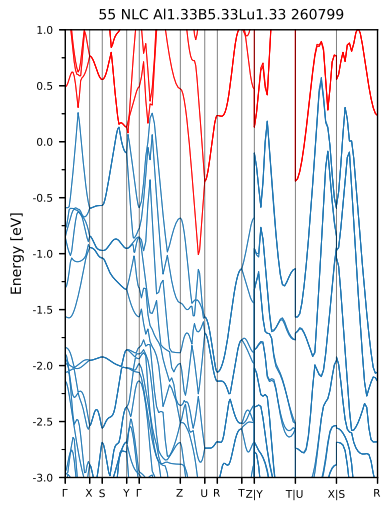

(bk) 55.2.63

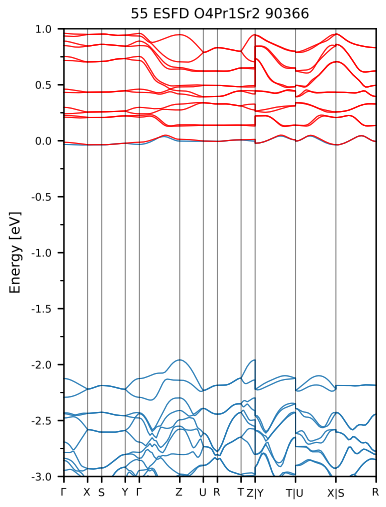

(bl) 55.4.1

Figure 95: Band structures for topological materials in SG 55, part 4/6, from 55.2.49 to 55.4.1. See the list of figures.

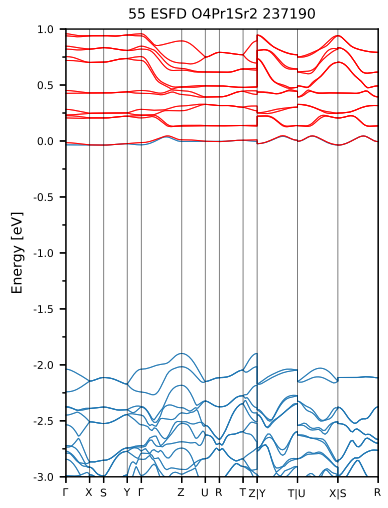

(bm) 55.4.2

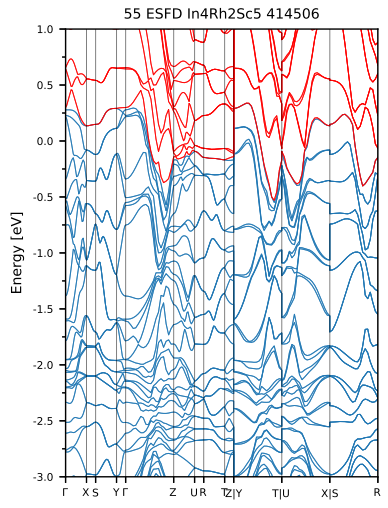

(bn) 55.4.3

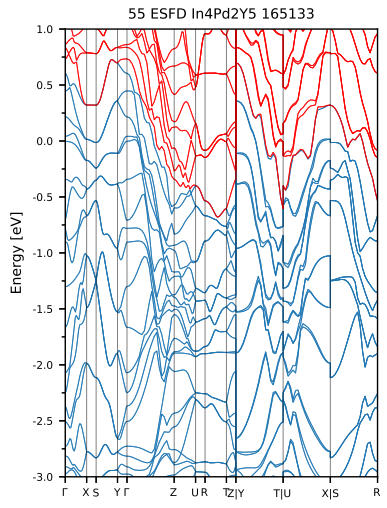

(bo) 55.4.4

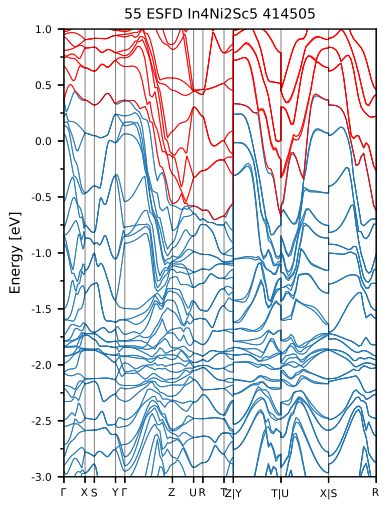

(bp) 55.4.5

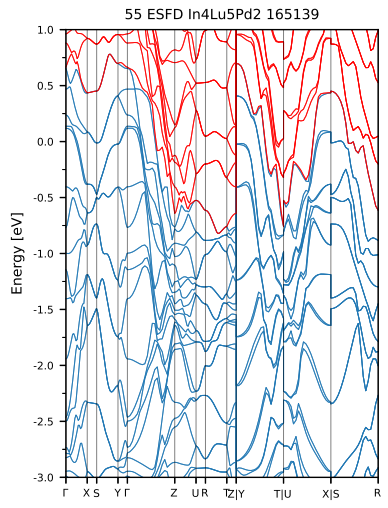

(bq) 55.4.6

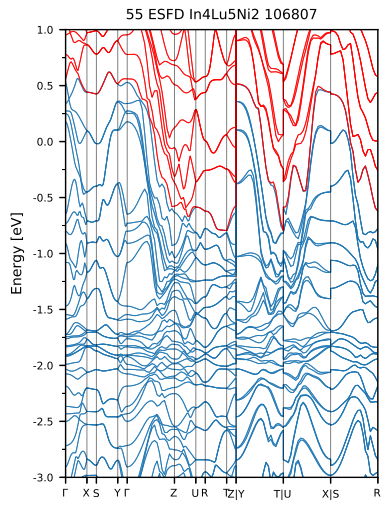

(br) 55.4.7

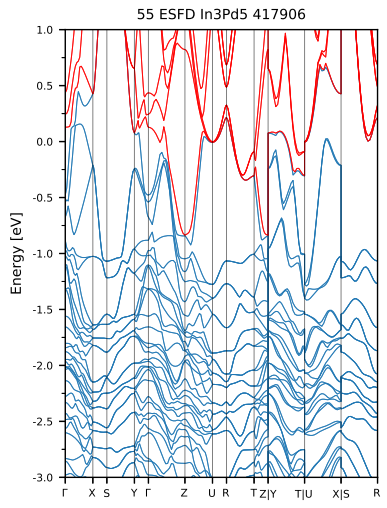

(bs) 55.4.8

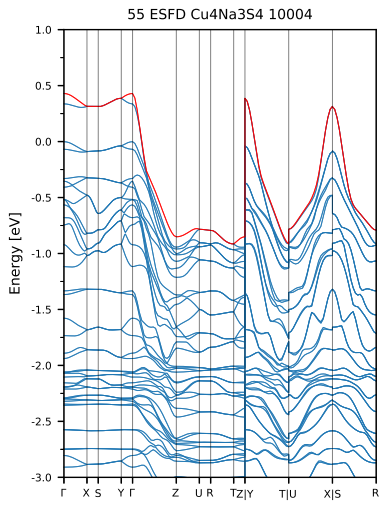

(bt) 55.4.9

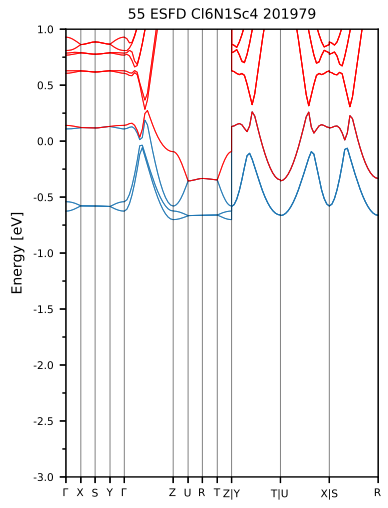

(bu) 55.4.10

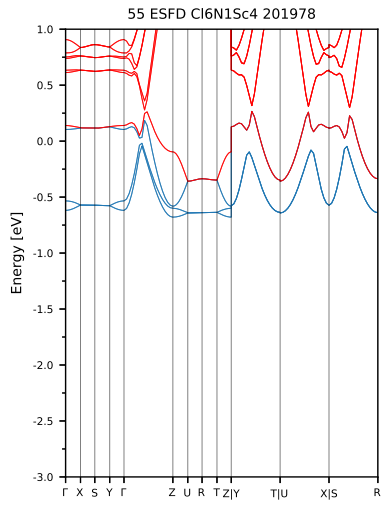

(bv) 55.4.11

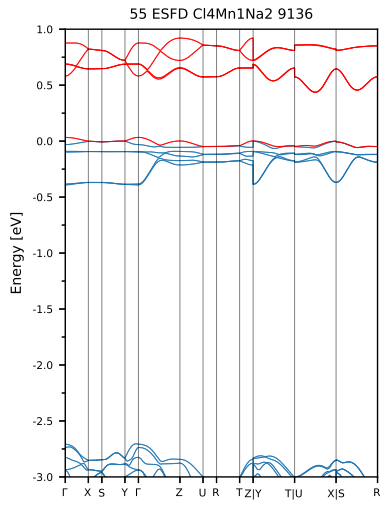

(bw) 55.4.12

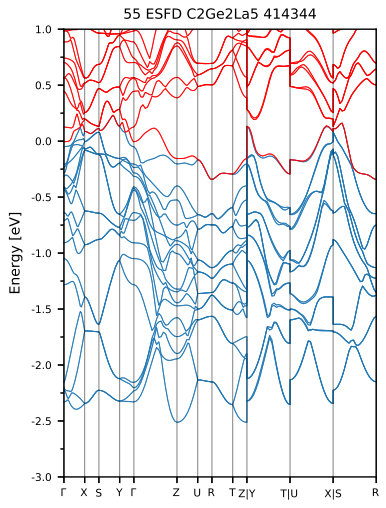

(bx) 55.4.13

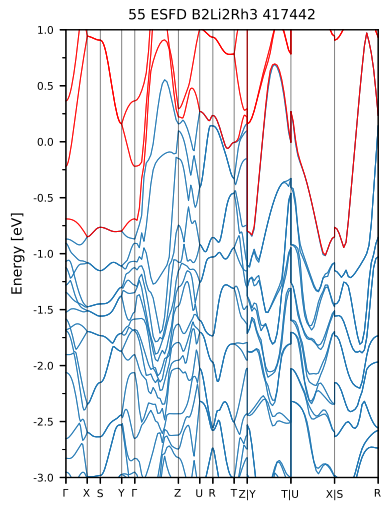

(by) 55.4.14

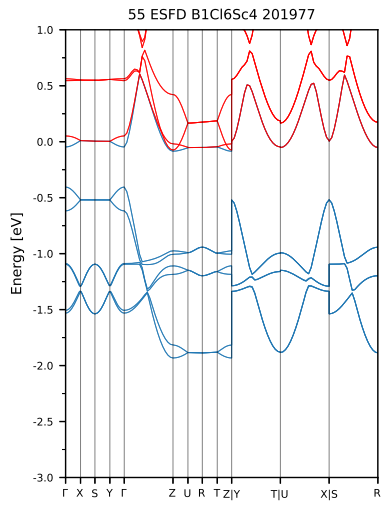

(bz) 55.4.15

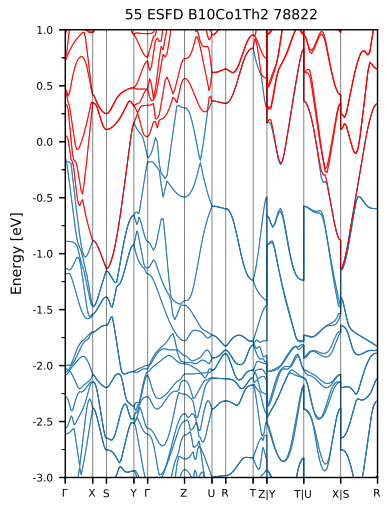

(ca) 55.4.16

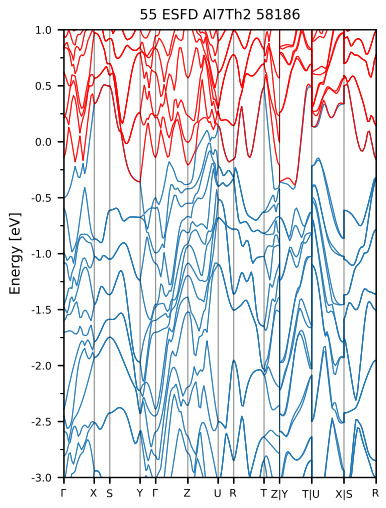

(cb) 55.4.17

Figure 96: Band structures for topological materials in SG 55, part 5/6, from 55.4.2 to 55.4.17. See the list of figures.

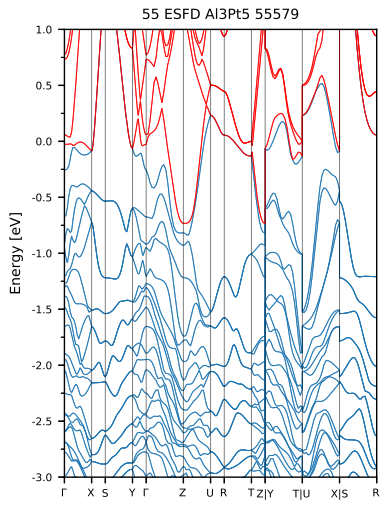

(cc) 55.4.18

Figure 97: Band structures for topological materials in SG 55, part 6/6, from 55.4.18 to 55.4.18. See the list of figures.

### 0.0.21 SG 57

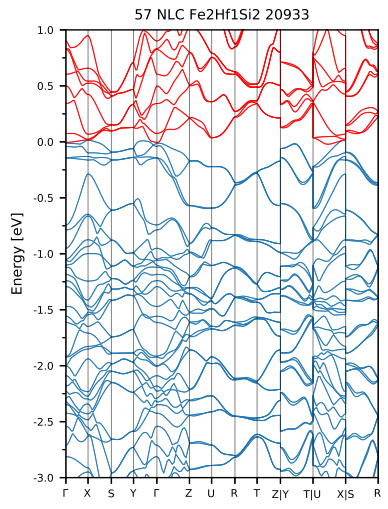

(a) 57.2.1

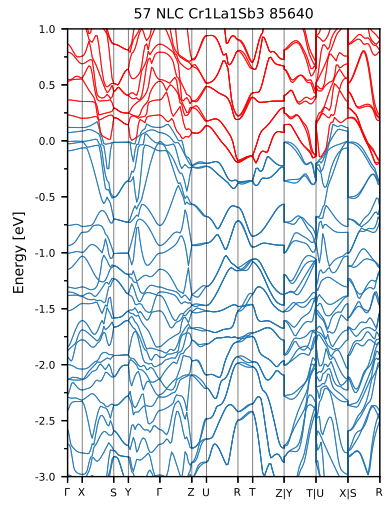

(b) 57.2.2

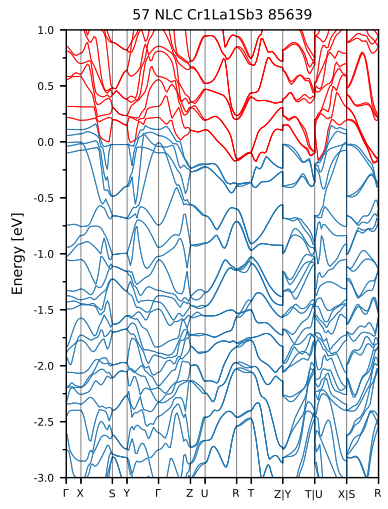

(c) 57.2.3

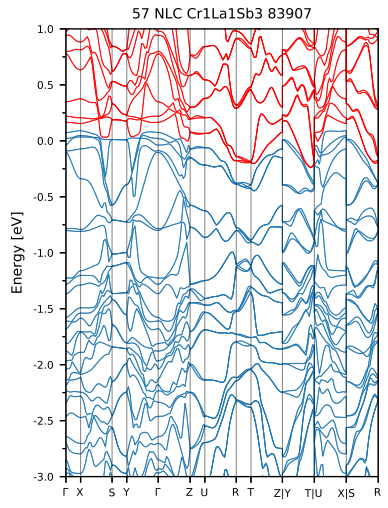

(d) 57.2.4

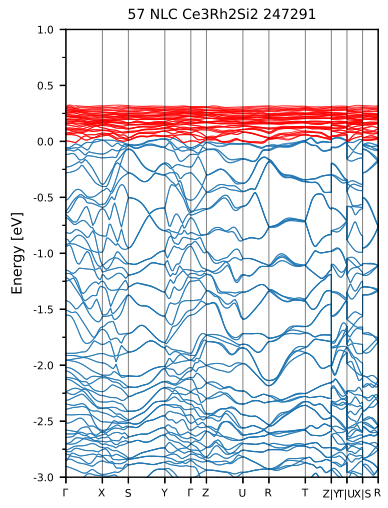

(e) 57.2.5

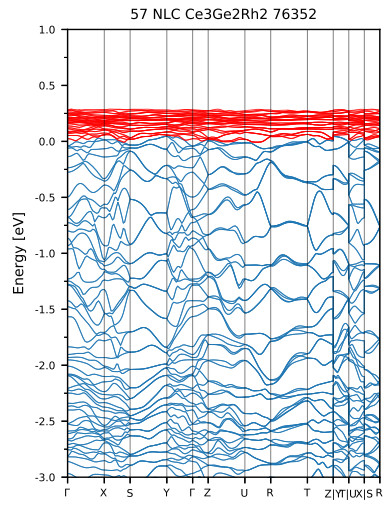

(f) 57.2.6

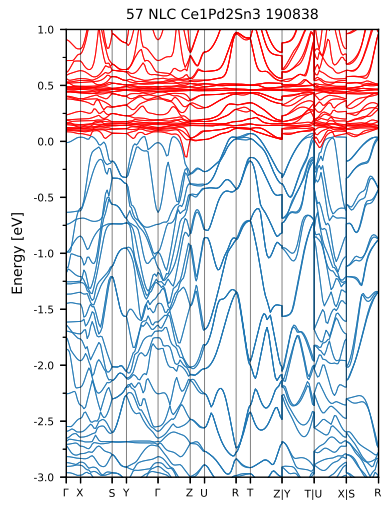

(g) 57.2.7

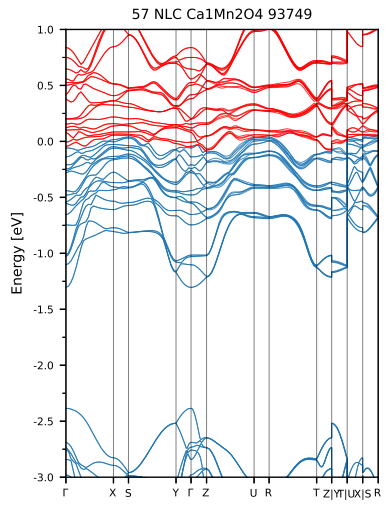

(h) 57.2.8

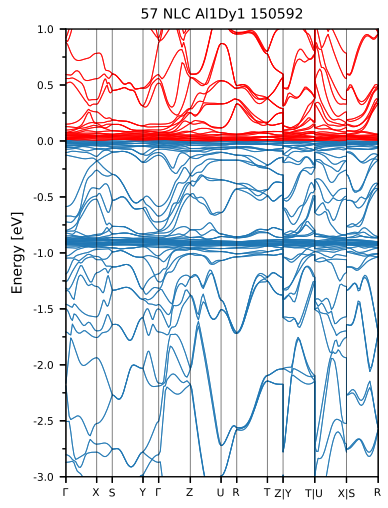

(i) 57.2.9

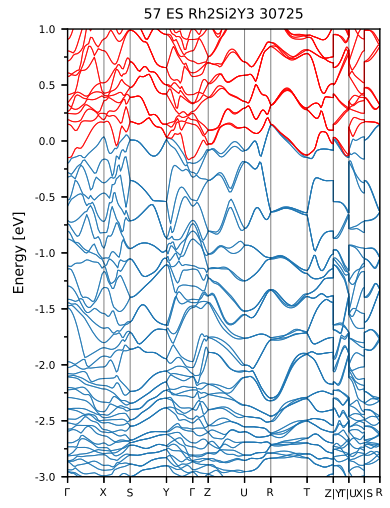

(j) 57.3.1

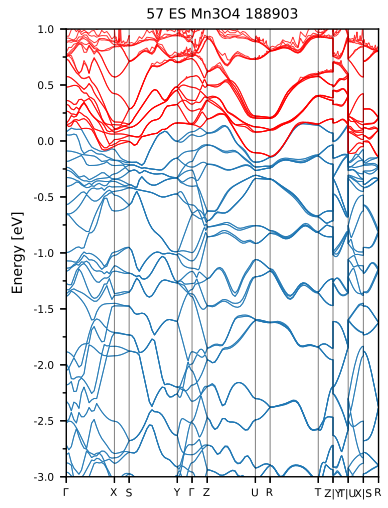

(k) 57.3.2

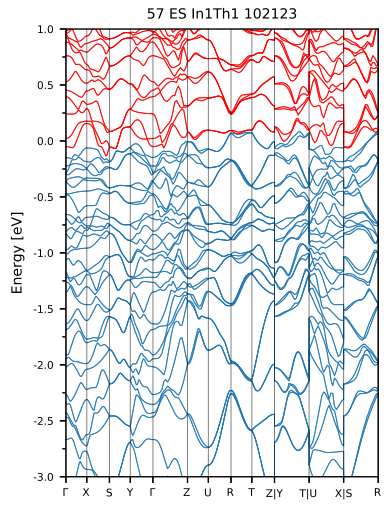

(l) 57.3.3

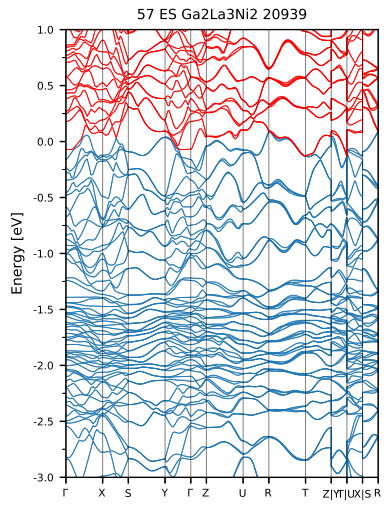

(m) 57.3.4

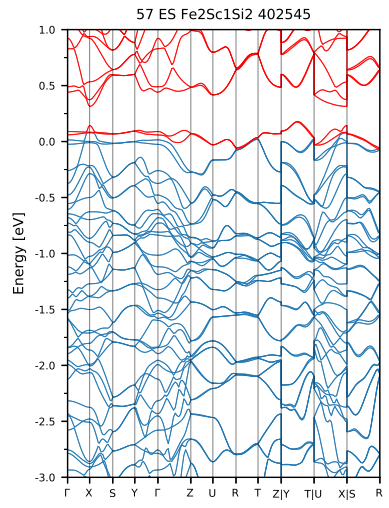

(n) 57.3.5

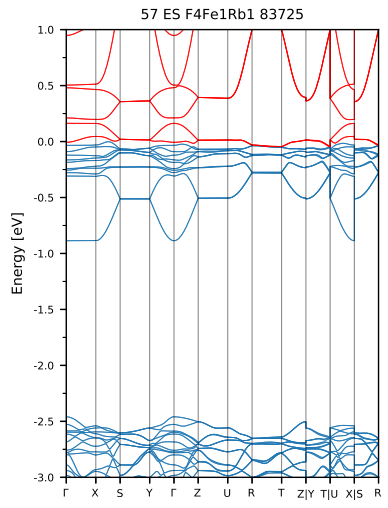

(o) 57.3.6

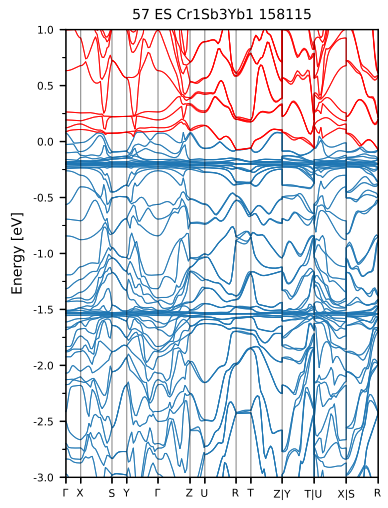

(p) 57.3.7

Figure 98: Band structures for topological materials in SG 57, part 1/2, from 57.2.1 to 57.3.7. See the list of figures.

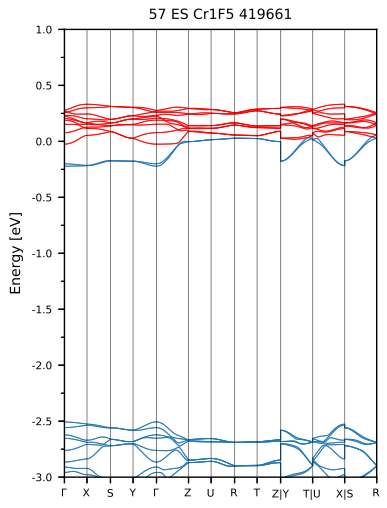

(q) 57.3.8

Figure 99: Band structures for topological materials in SG 57, part 2/2, from 57.3.8 to 57.3.8. See the list of figures.

## 0.0.22 SG 58

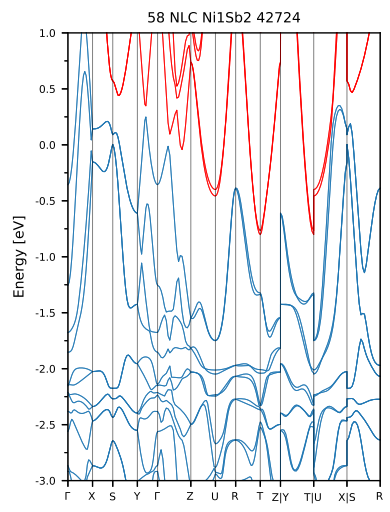

(a) 58.2.1

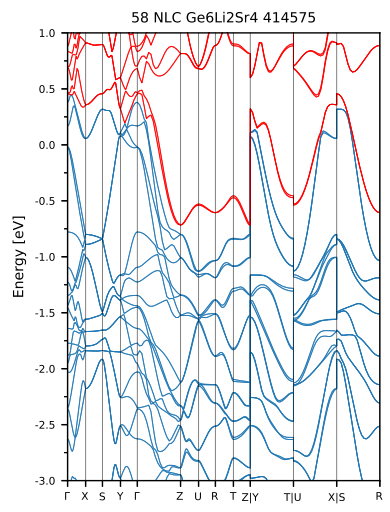

(b) 58.2.2

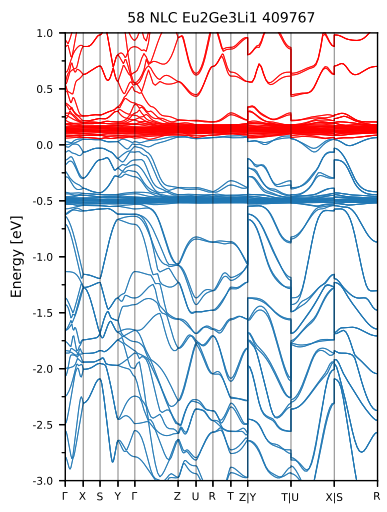

(c) 58.2.3

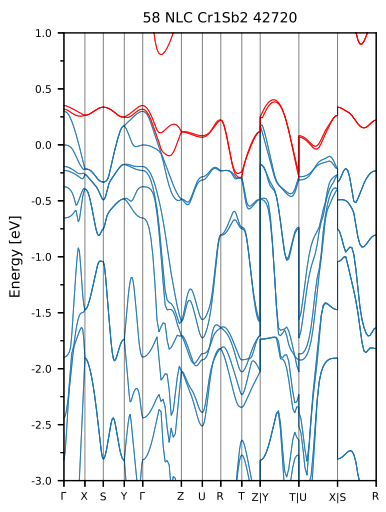

(d) 58.2.4

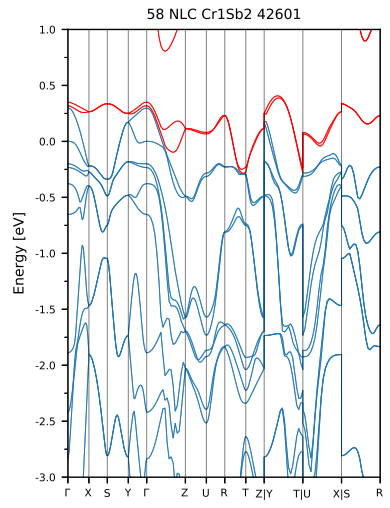

(e) 58.2.5

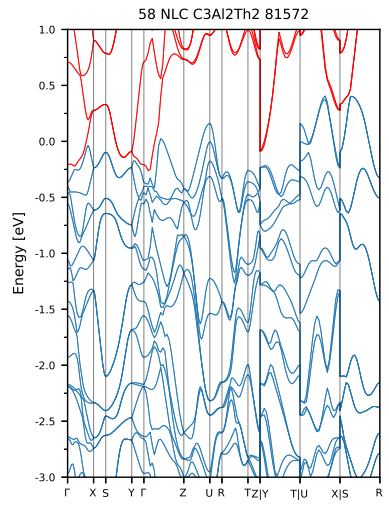

(f) 58.2.6

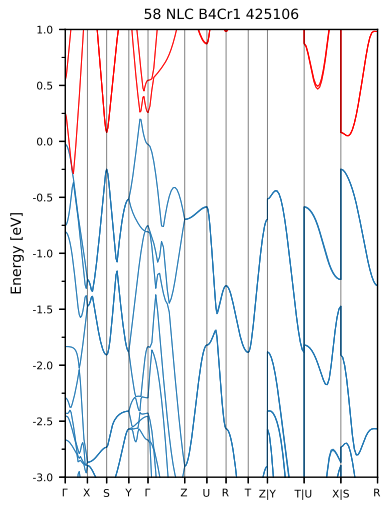

(g) 58.2.7

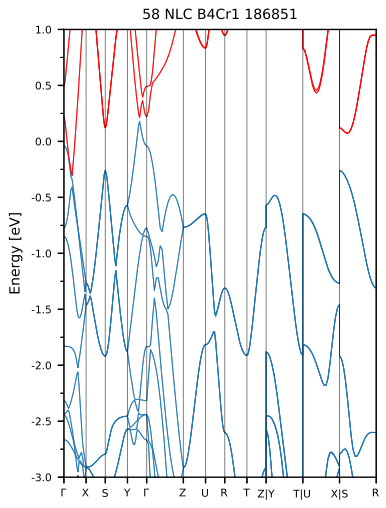

(h) 58.2.8

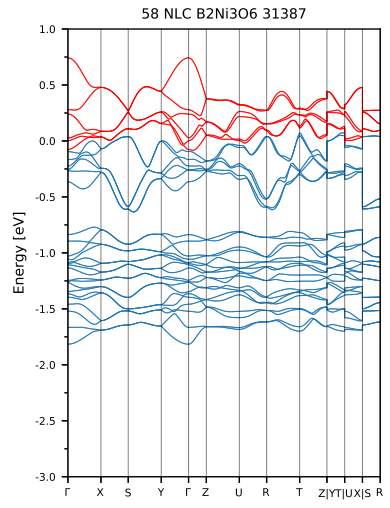

(i) 58.2.9

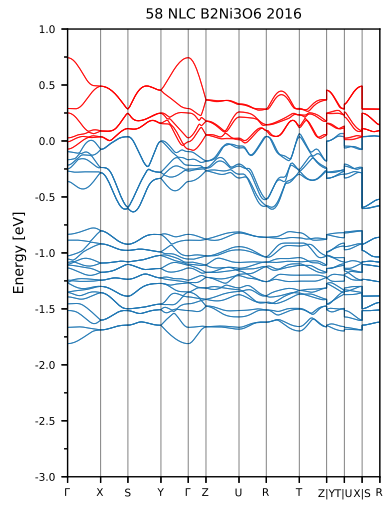

(j) 58.2.10

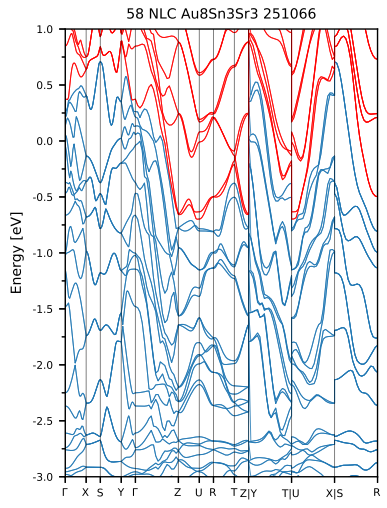

(k) 58.2.11

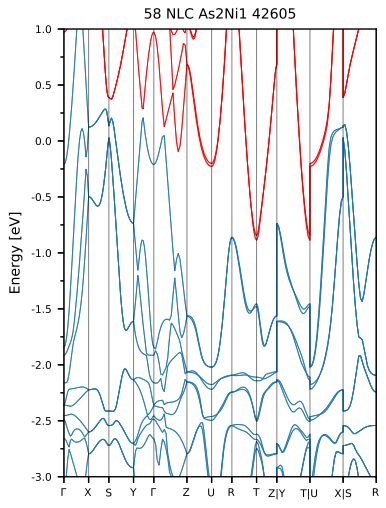

(l) 58.2.12

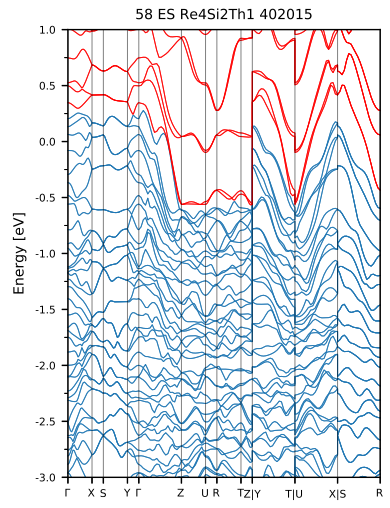

(m) 58.3.1

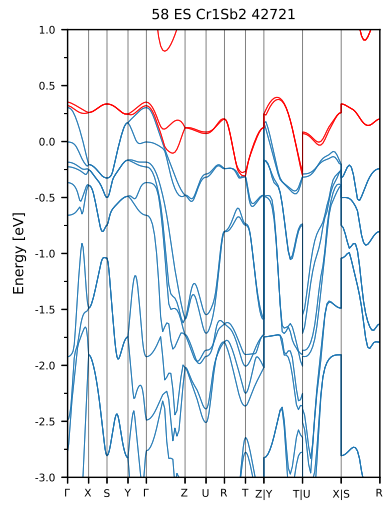

(n) 58.3.2

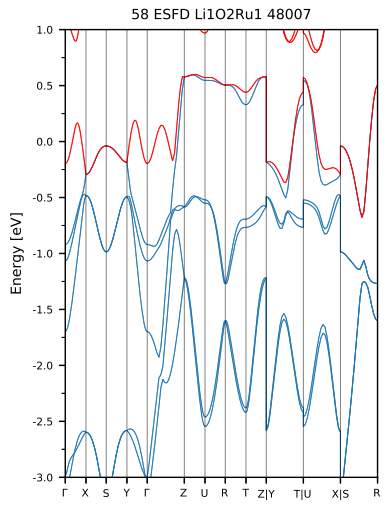

(o) 58.4.1

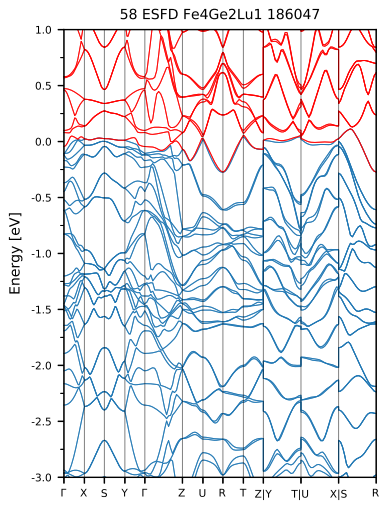

(p) 58.4.2

Figure 100: Band structures for topological materials in SG 58, part 1/2, from 58.2.1 to 58.4.2. See the list of figures.

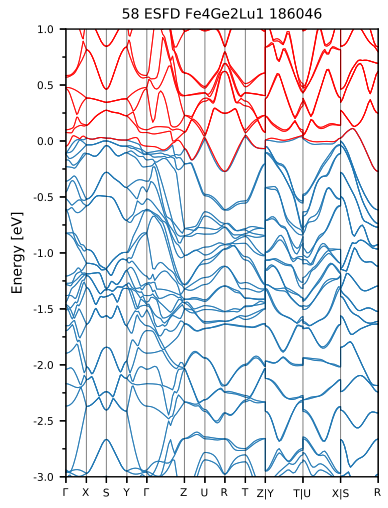

(q) 58.4.3

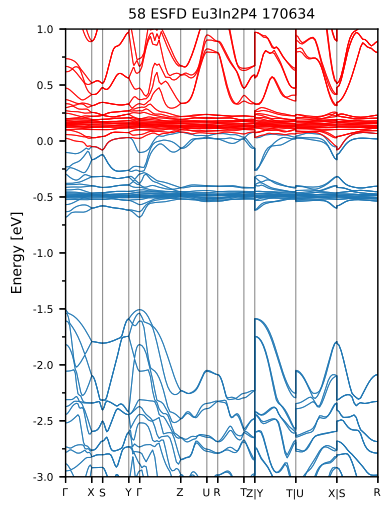

(r) 58.4.4

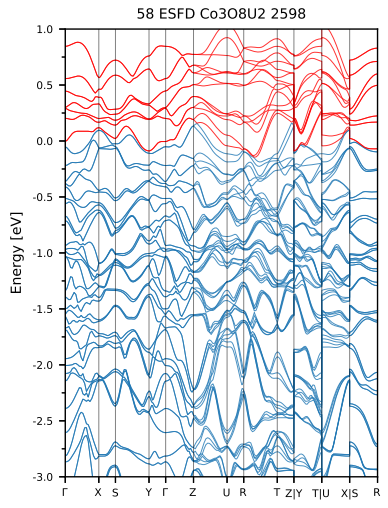

(s) 58.4.5

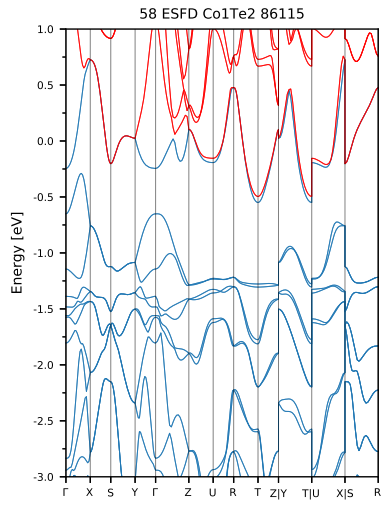

(t) 58.4.6

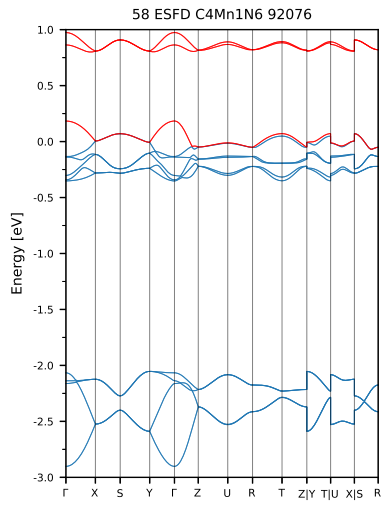

(u) 58.4.7

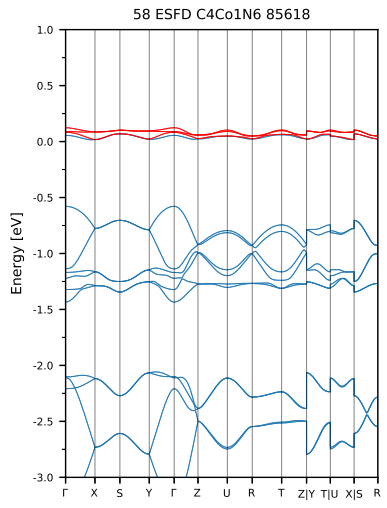

(v) 58.4.8

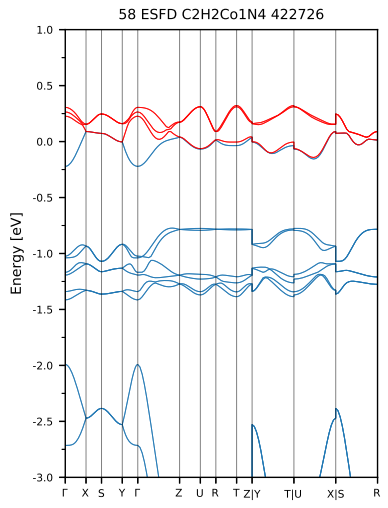

(w) 58.4.9

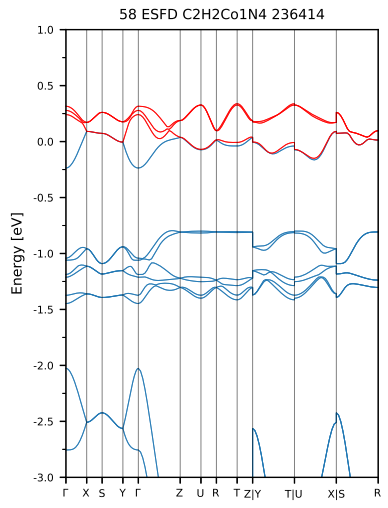

(x) 58.4.10

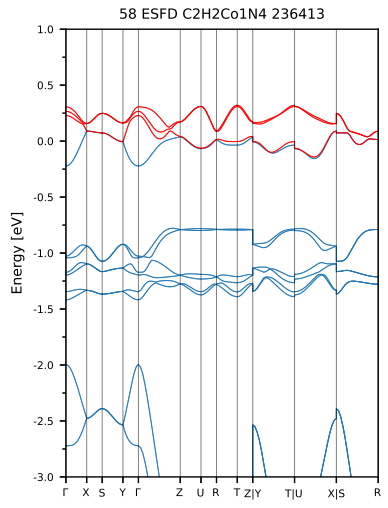

(y) 58.4.11

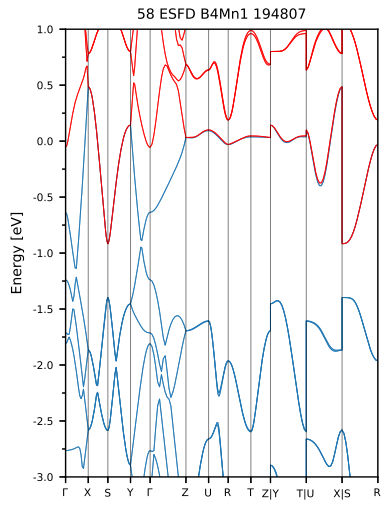

(z) 58.4.12

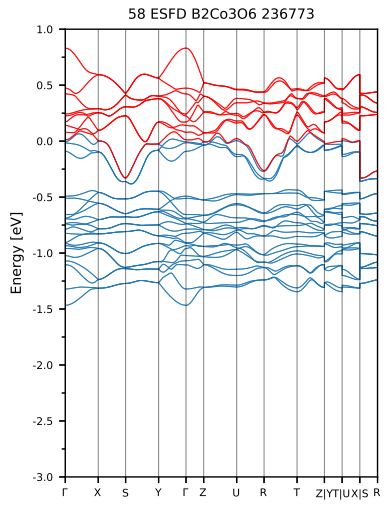

(aa) 58.4.13

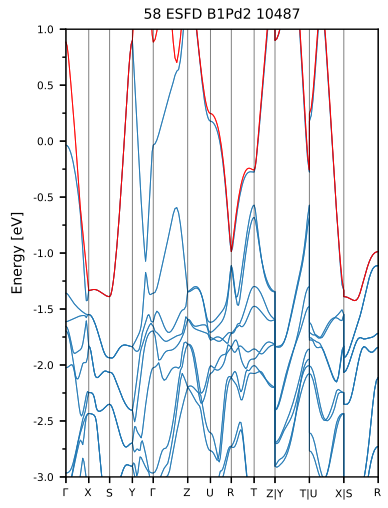

(ab) 58.4.14

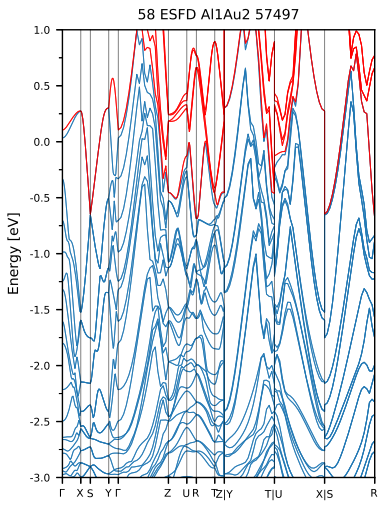

(ac) 58.4.15

Figure 101: Band structures for topological materials in SG 58, part 2/2, from 58.4.3 to 58.4.15. See the list of figures.

### 0.0.23 SG 59

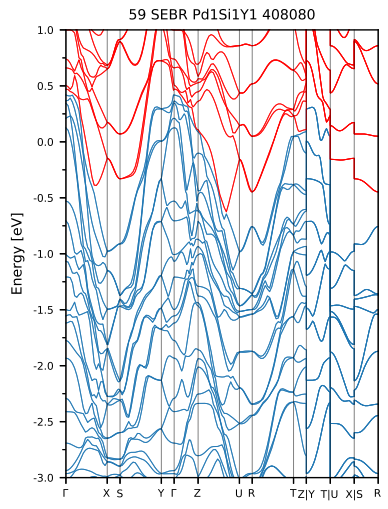

(a) 59.1.1

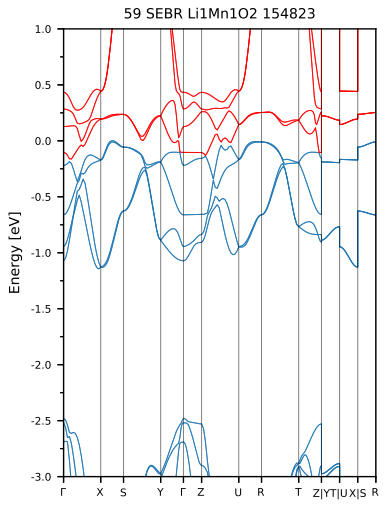

(b) 59.1.2

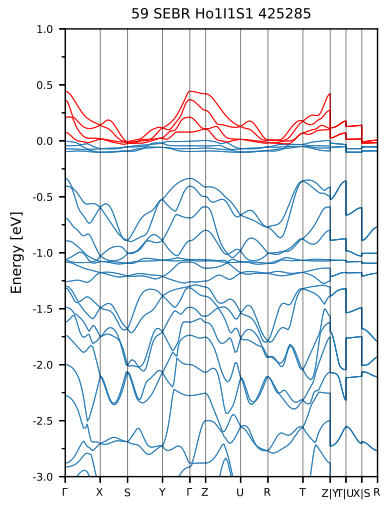

(c) 59.1.3

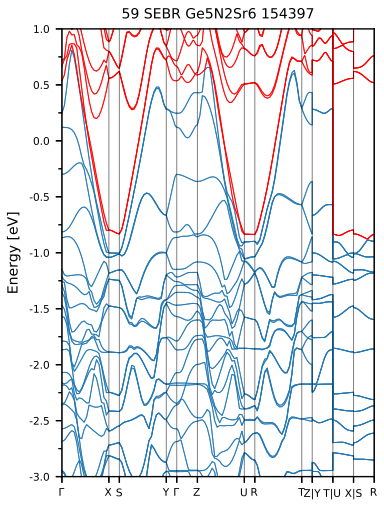

(d) 59.1.4

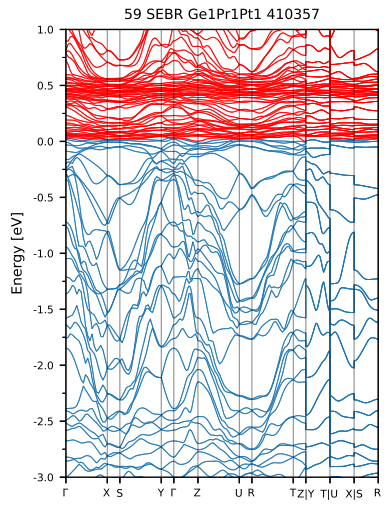

(e) 59.1.5

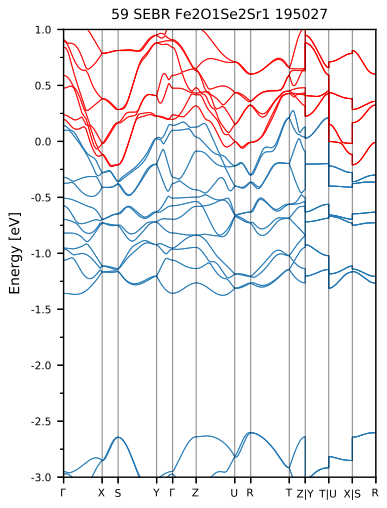

(f) 59.1.6

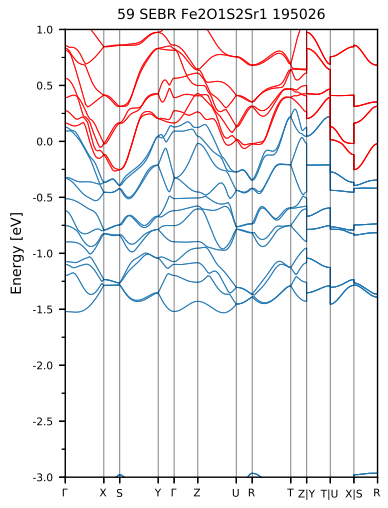

(g) 59.1.7

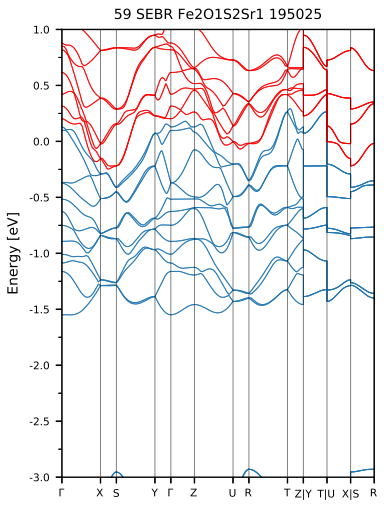

(h) 59.1.8

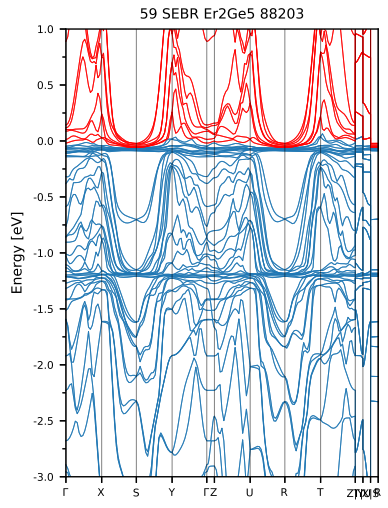

(i) 59.1.9

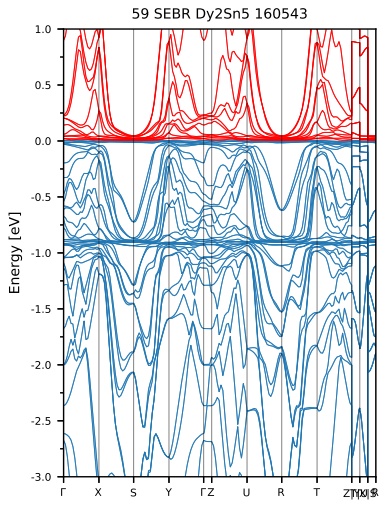

(j) 59.1.10

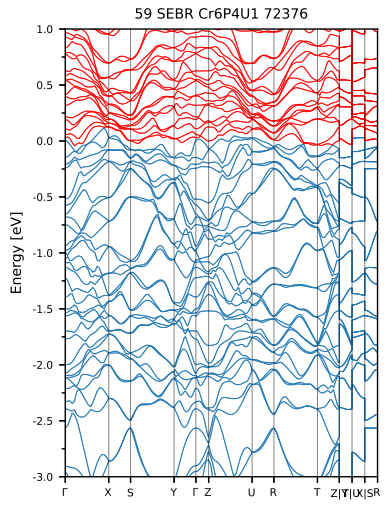

(k) 59.1.11

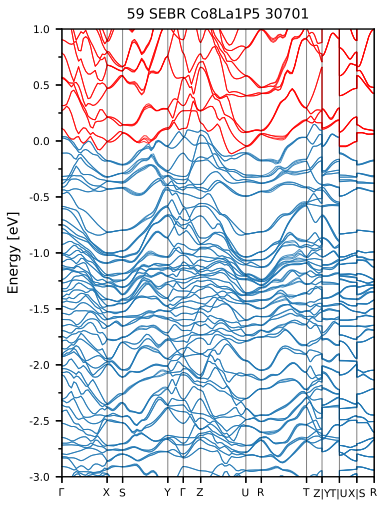

(l) 59.1.12

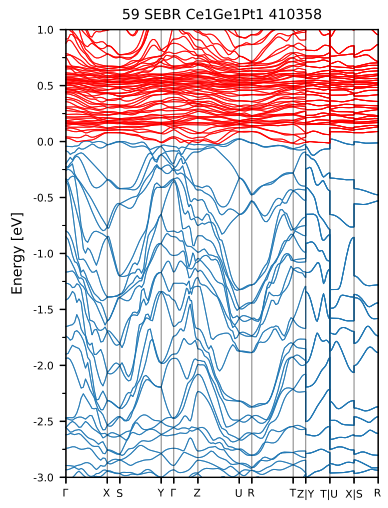

(m) 59.1.13

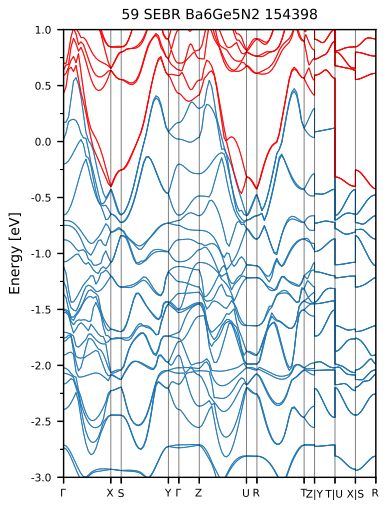

(n) 59.1.14

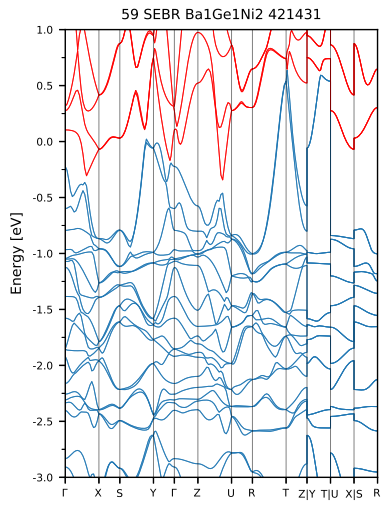

(o) 59.1.15

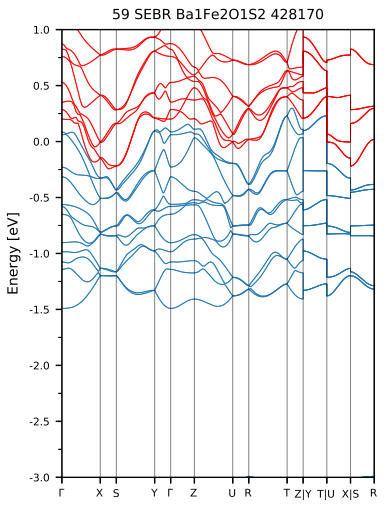

(p) 59.1.16

Figure 102: Band structures for topological materials in SG 59, part 1/5, from 59.1.1 to 59.1.16. See the list of figures.

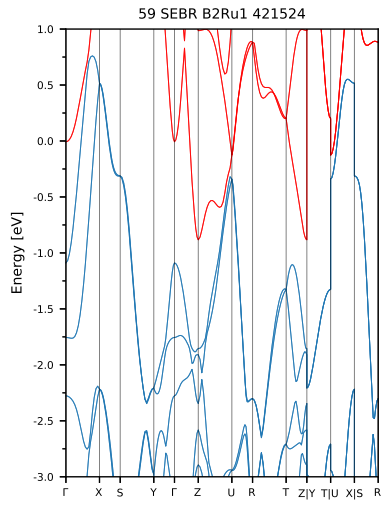

(q) 59.1.17

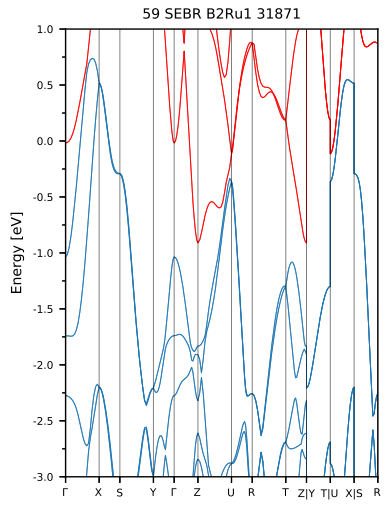

(r) 59.1.18

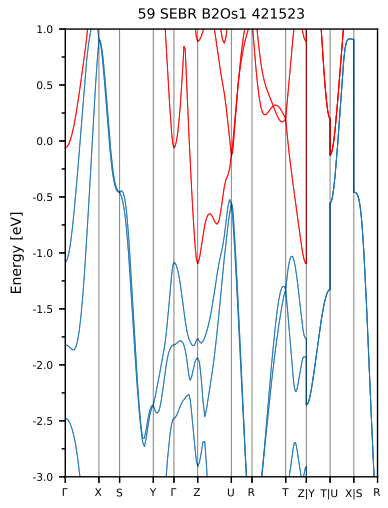

(s) 59.1.19

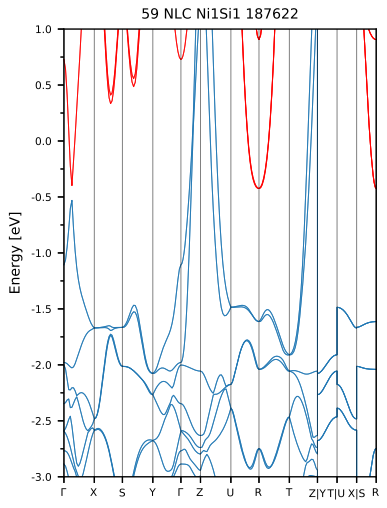

(t) 59.2.1

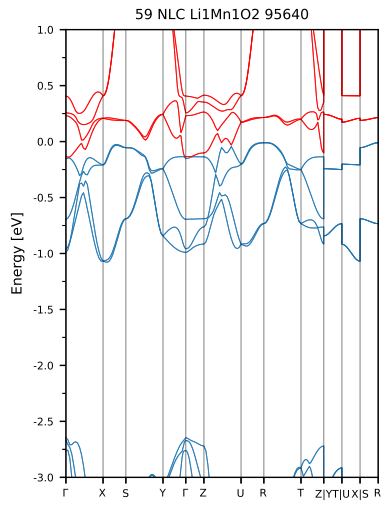

(u) 59.2.2

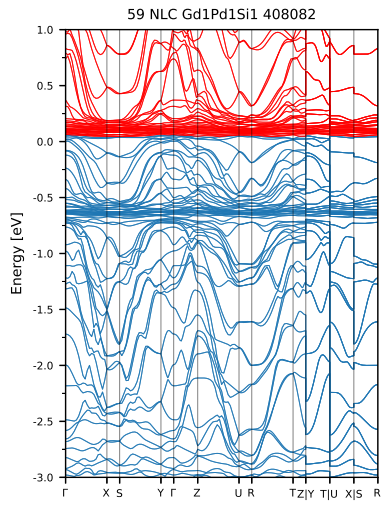

(v) 59.2.3

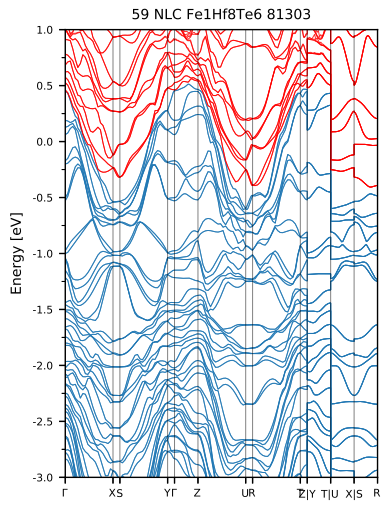

(w) 59.2.4

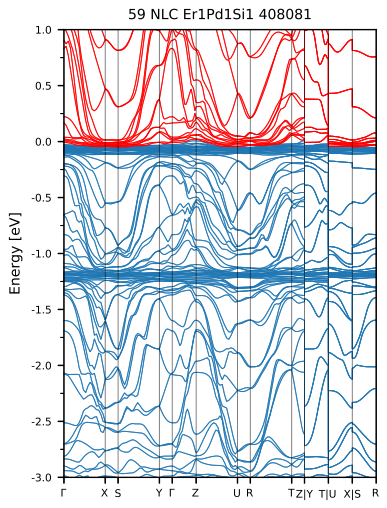

(x) 59.2.5

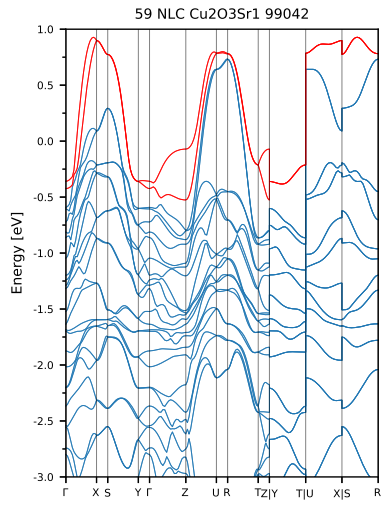

(y) 59.2.6

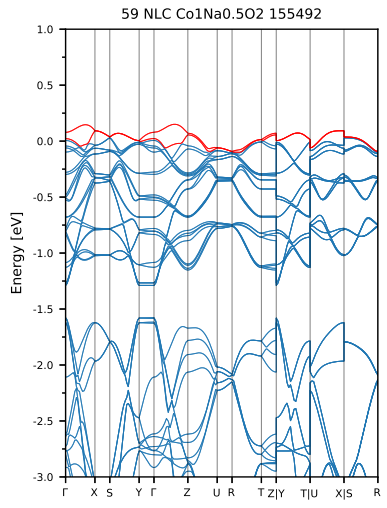

(z) 59.2.7

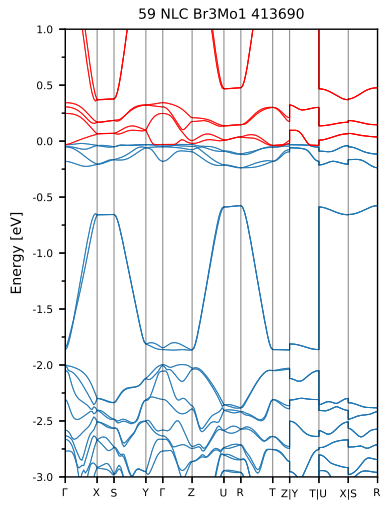

(aa) 59.2.8

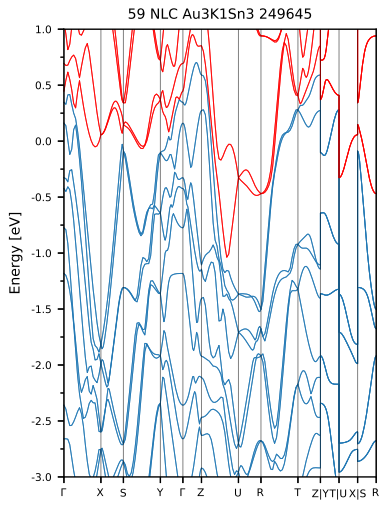

(ab) 59.2.9

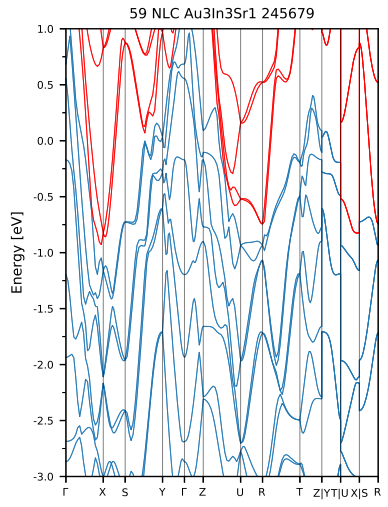

(ac) 59.2.10

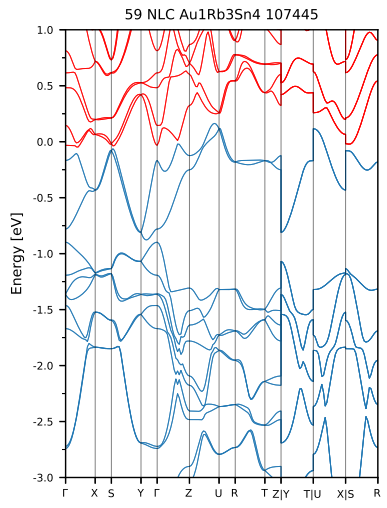

(ad) 59.2.11

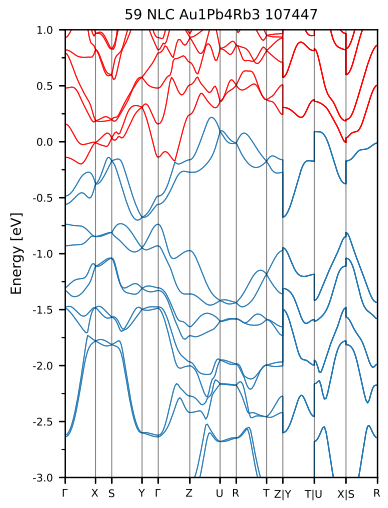

(ae) 59.2.12

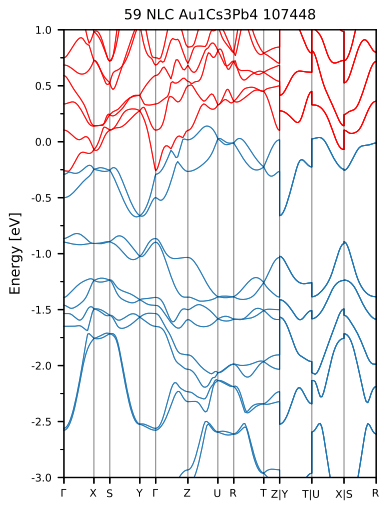

(af) 59.2.13

Figure 103: Band structures for topological materials in SG 59, part 2/5, from 59.1.17 to 59.2.13. See the list of figures.

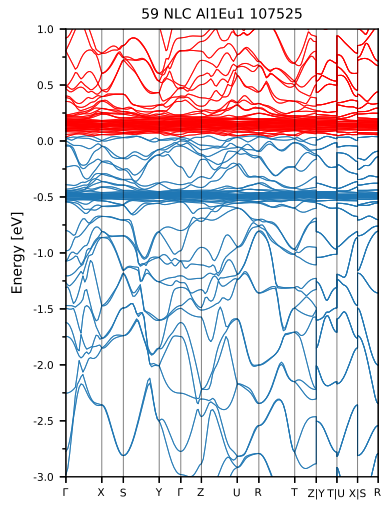

(ag) 59.2.14

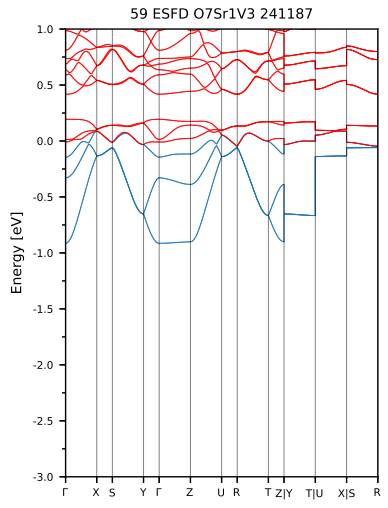

(ah) 59.4.1

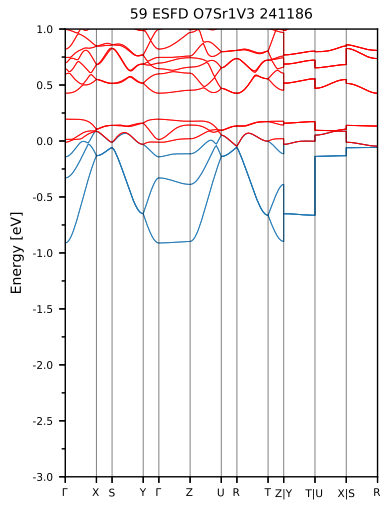

(ai) 59.4.2

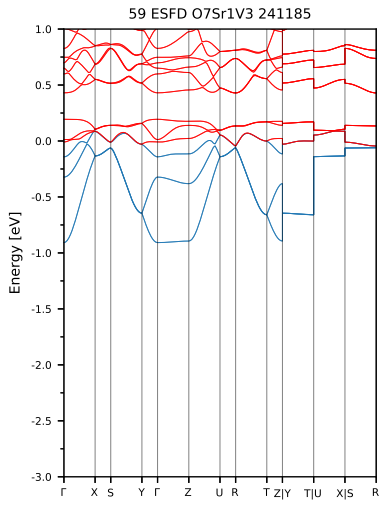

(aj) 59.4.3

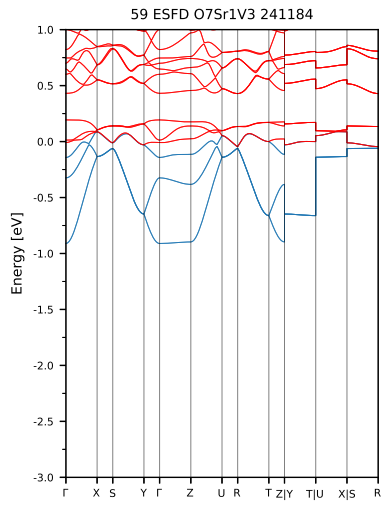

(ak) 59.4.4

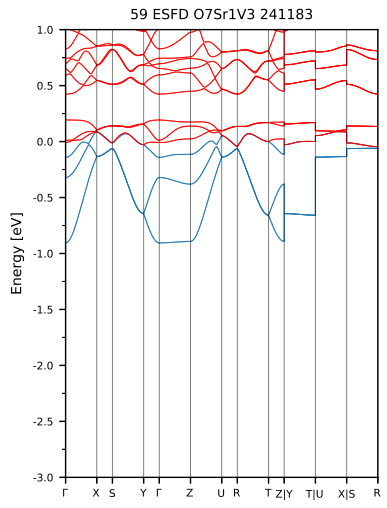

(al) 59.4.5

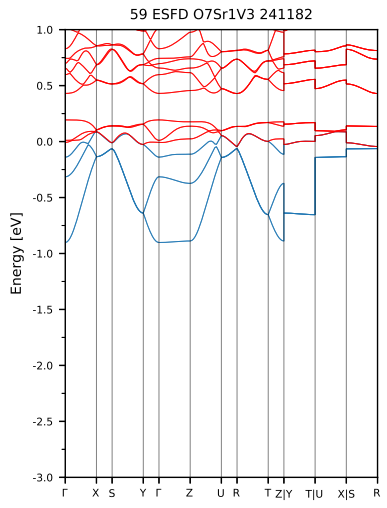

(am) 59.4.6

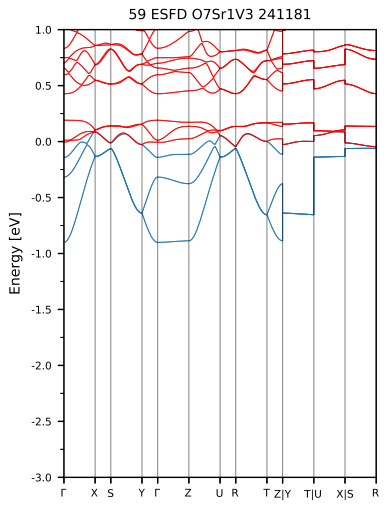

(an) 59.4.7

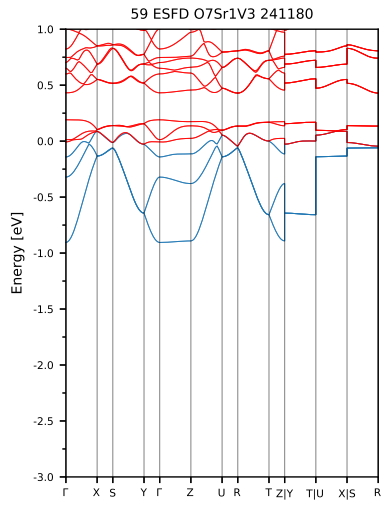

(ao) 59.4.8

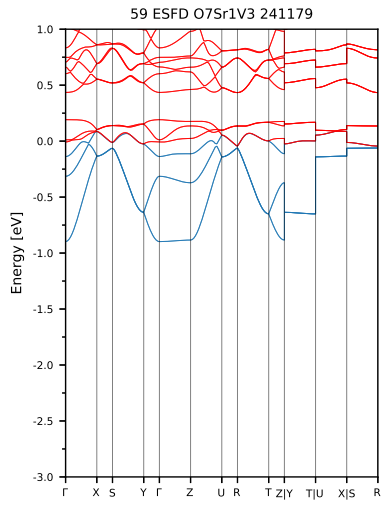

(ap) 59.4.9

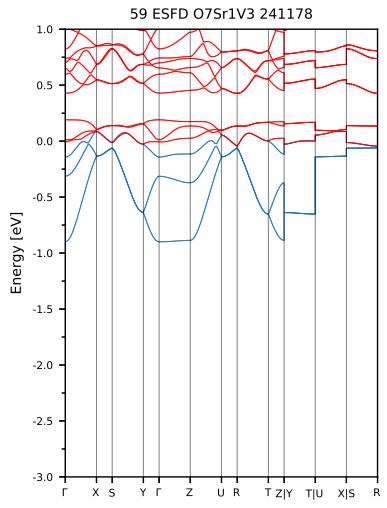

(aq) 59.4.10

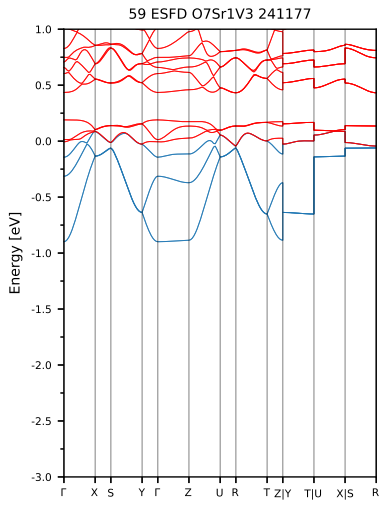

(ar) 59.4.11

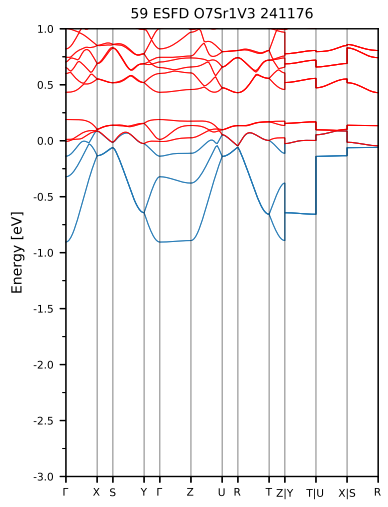

(as) 59.4.12

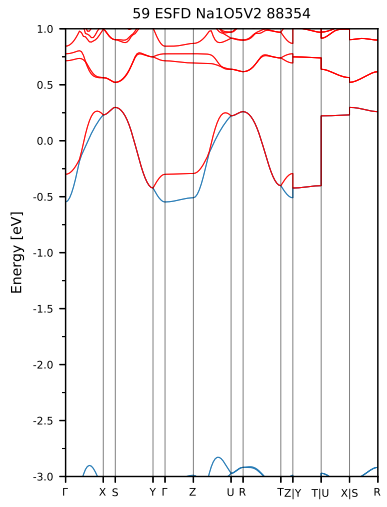

(at) 59.4.13

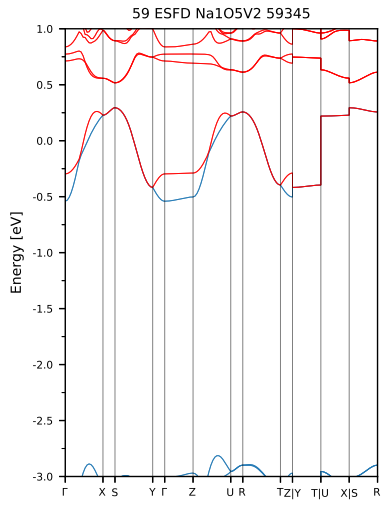

(au) 59.4.14

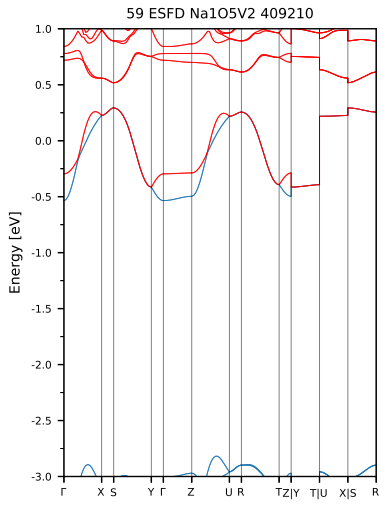

(av) 59.4.15

Figure 104: Band structures for topological materials in SG 59, part 3/5, from 59.2.14 to 59.4.15. See the list of figures.

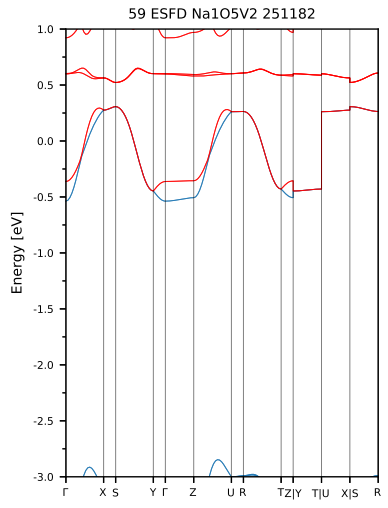

(aw) 59.4.16

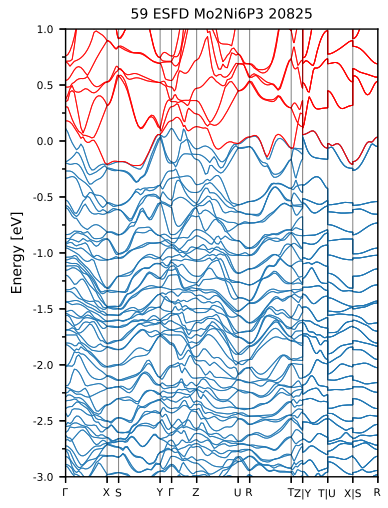

(ax) 59.4.17

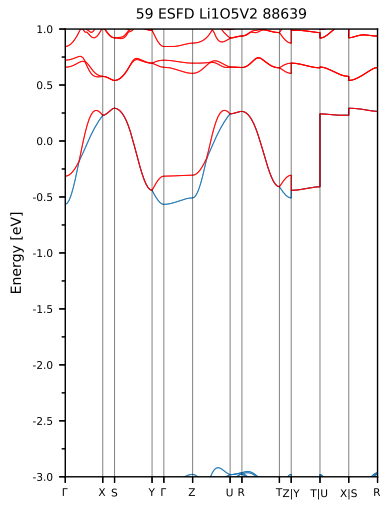

(ay) 59.4.18

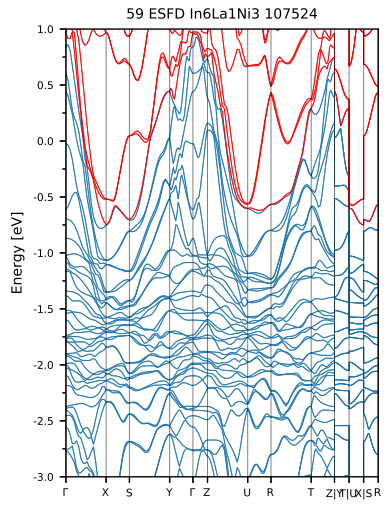

(az) 59.4.19

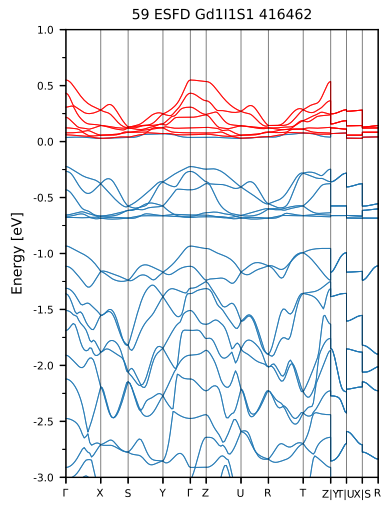

(ba) 59.4.20

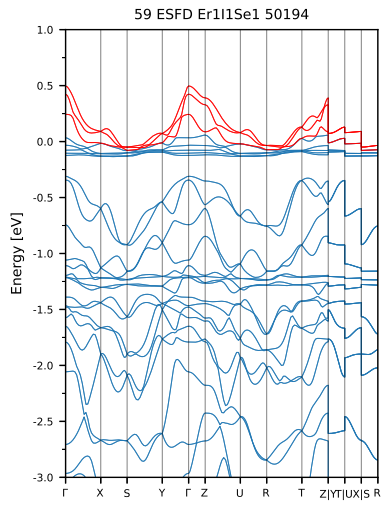

(bb) 59.4.21

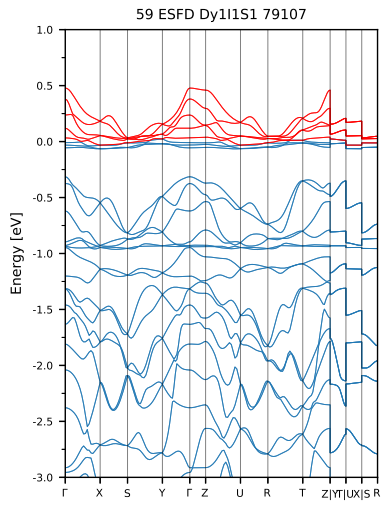

(bc) 59.4.22

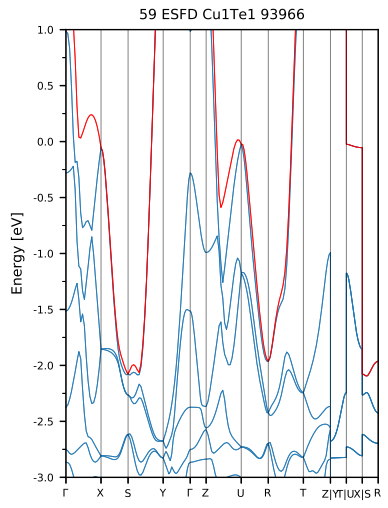

(bd) 59.4.23

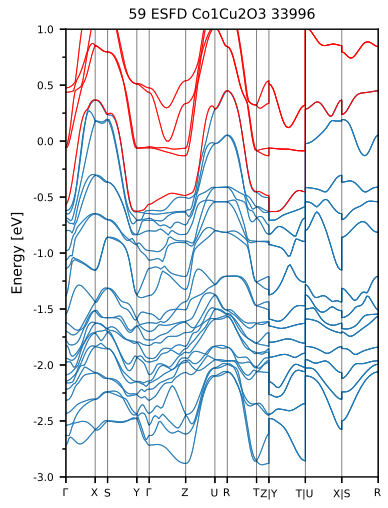

(be) 59.4.24

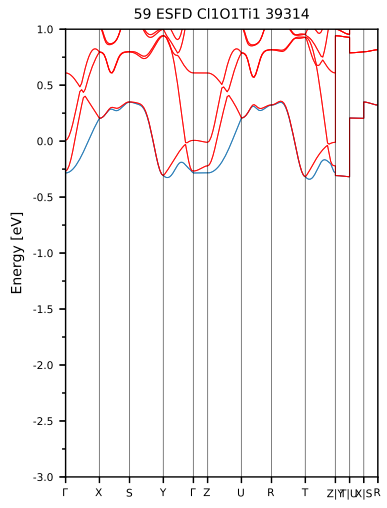

(bf) 59.4.25

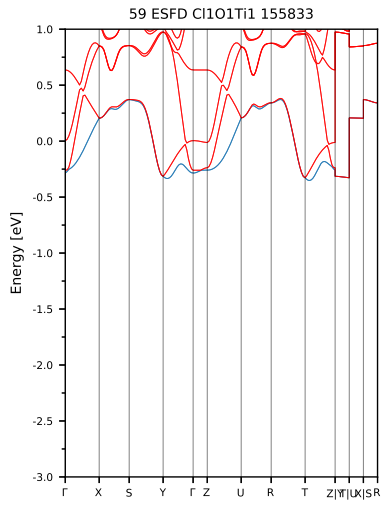

(bg) 59.4.26

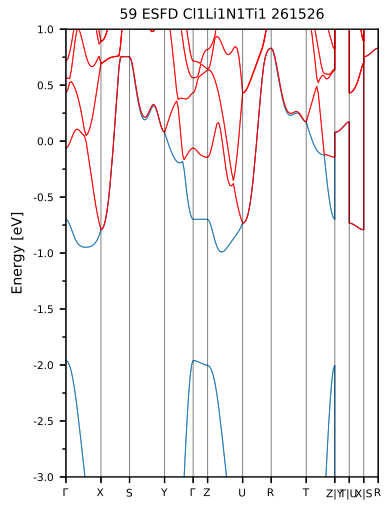

(bh) 59.4.27

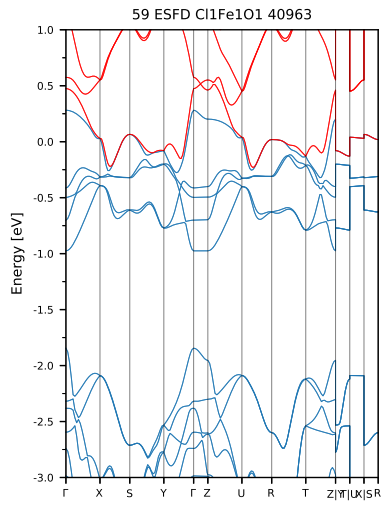

(bi) 59.4.28

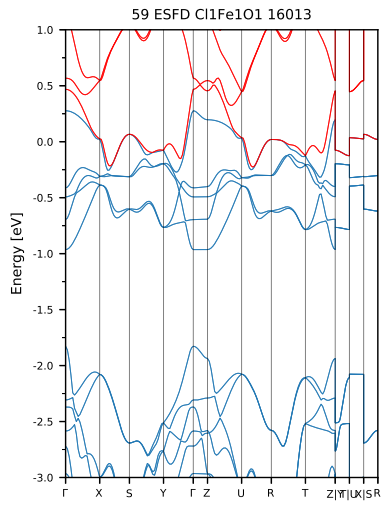

(bj) 59.4.29

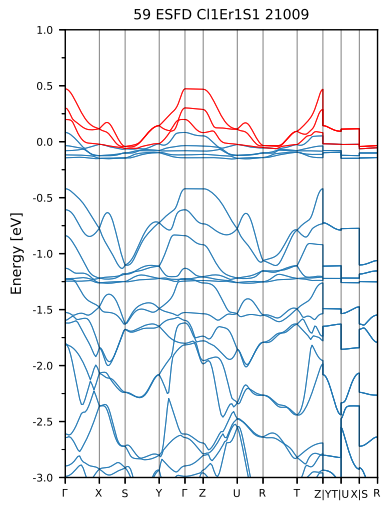

(bk) 59.4.30

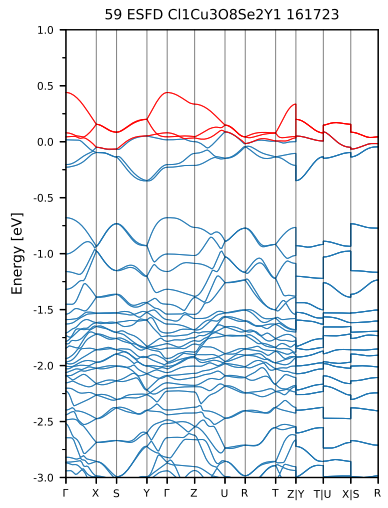

(bl) 59.4.31

Figure 105: Band structures for topological materials in SG 59, part 4/5, from 59.4.16 to 59.4.31. See the list of figures.

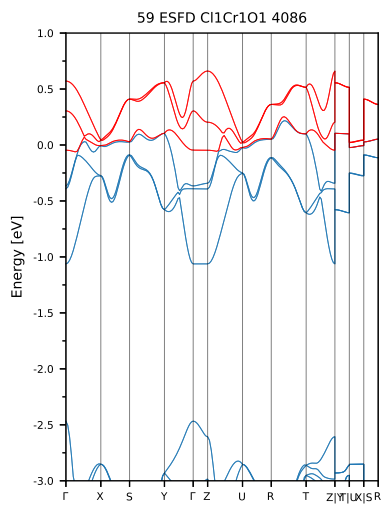

(bm) 59.4.32

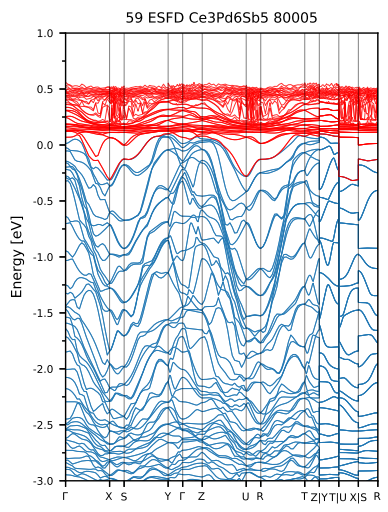

(bn) 59.4.33

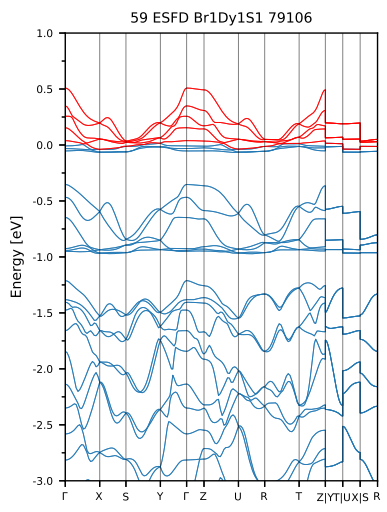

(bo) 59.4.34

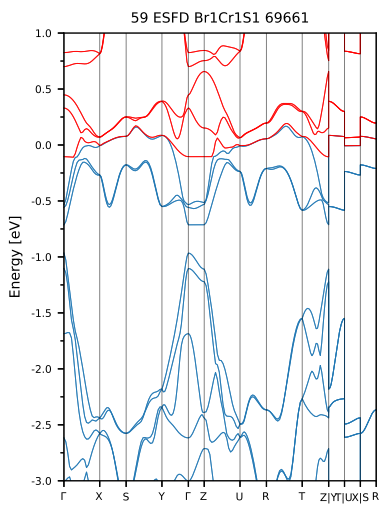

(bp) 59.4.35

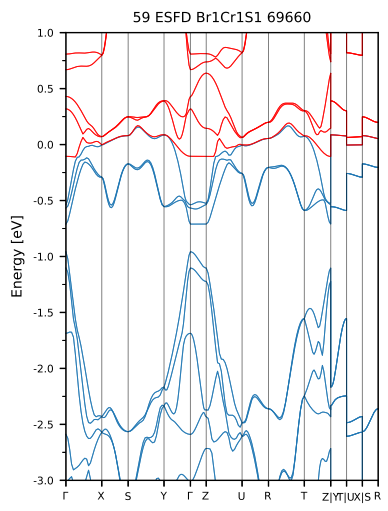

(bq) 59.4.36

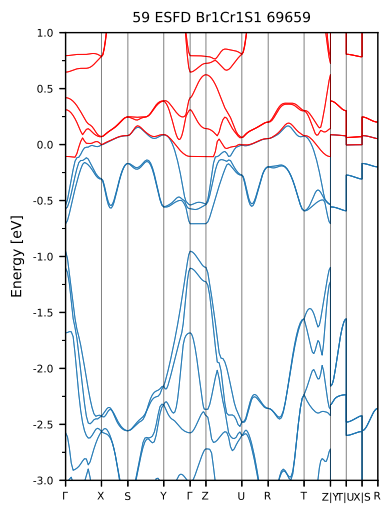

(br) 59.4.37

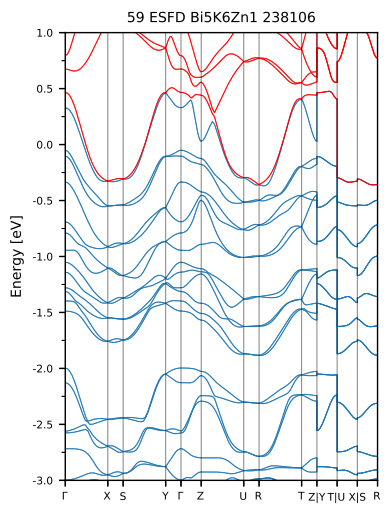

(bs) 59.4.38

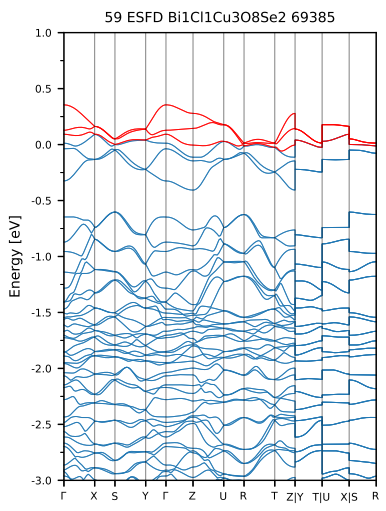

(bt) 59.4.39

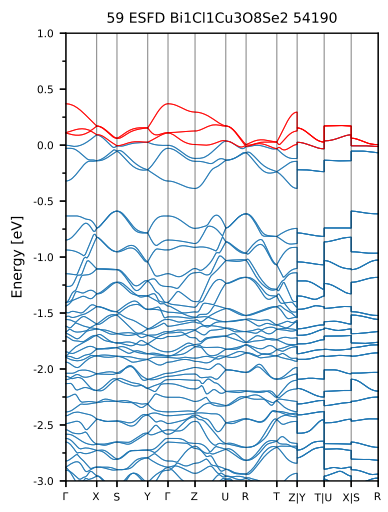

(bu) 59.4.40

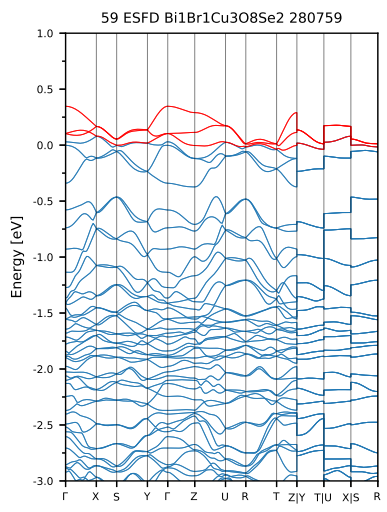

(bv) 59.4.41

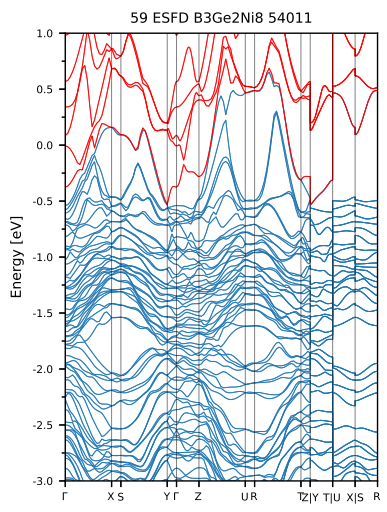

(bw) 59.4.42

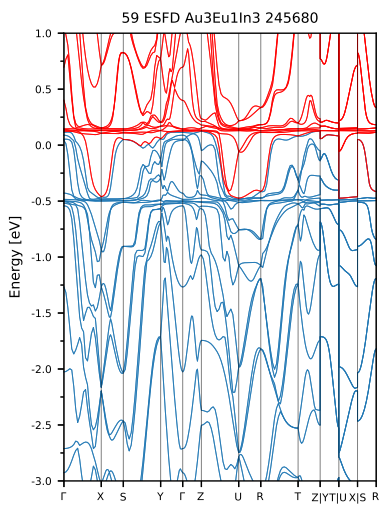

(bx) 59.4.43

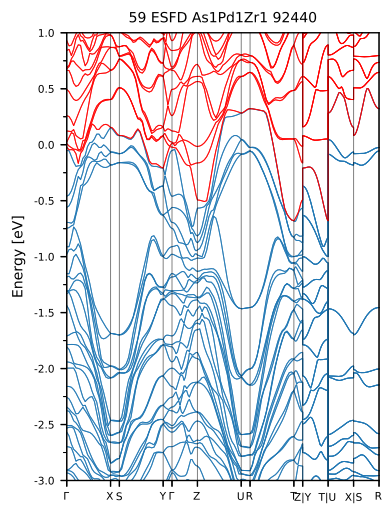

(by) 59.4.44

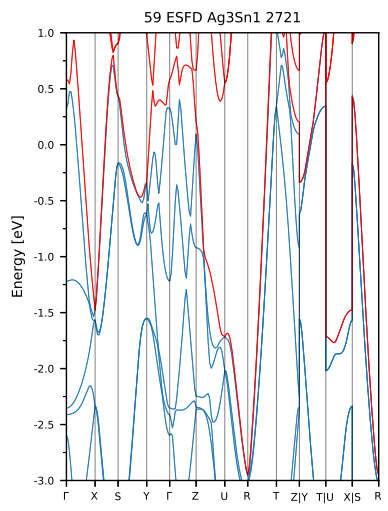

(bz) 59.4.45

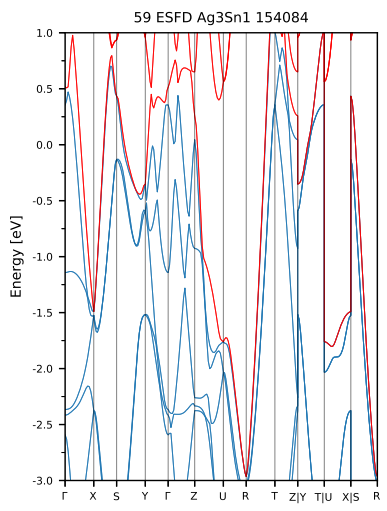

(ca) 59.4.46

Figure 106: Band structures for topological materials in SG 59, part 5/5, from 59.4.32 to 59.4.46. See the list of figures.

### 0.0.24 SG 60

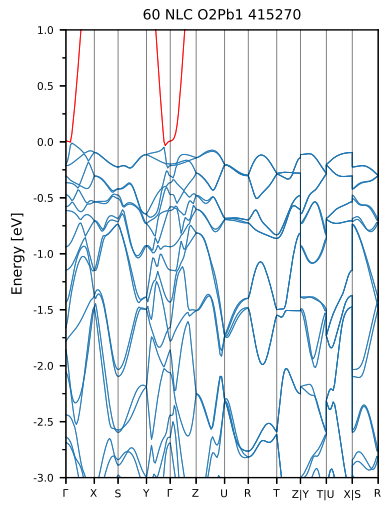

(a) 60.2.1

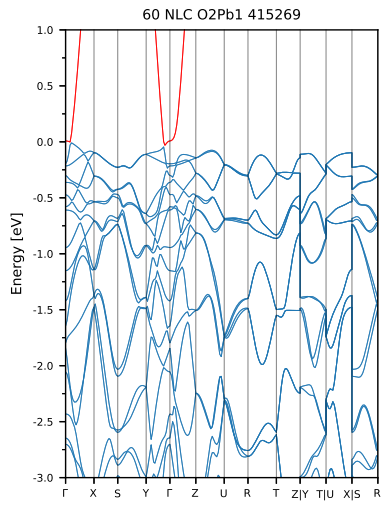

(b) 60.2.2

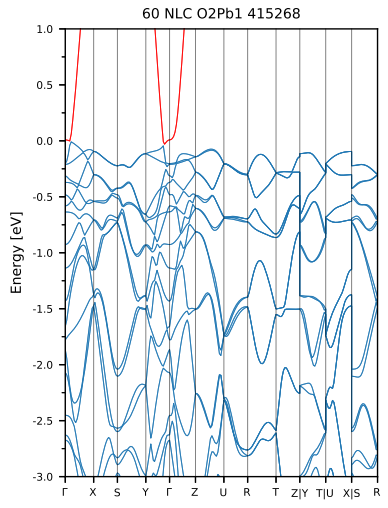

(c) 60.2.3

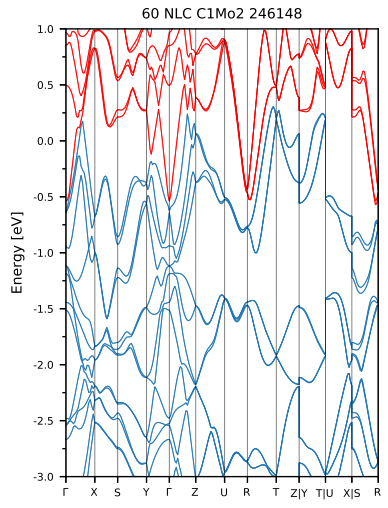

(d) 60.2.4

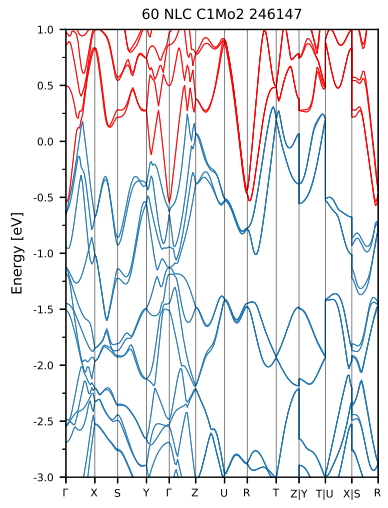

(e) 60.2.5

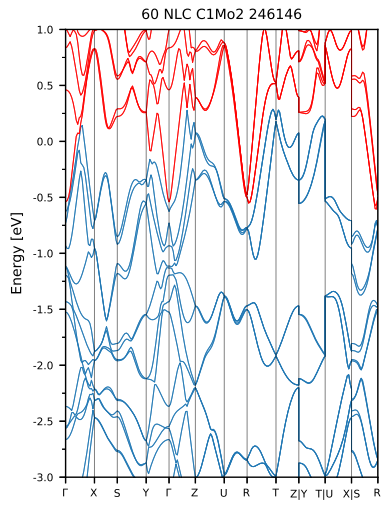

(f) 60.2.6

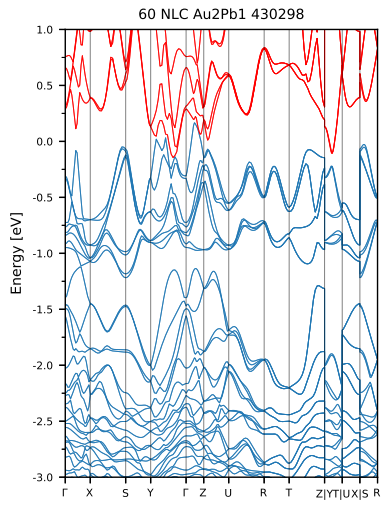

(g) 60.2.7

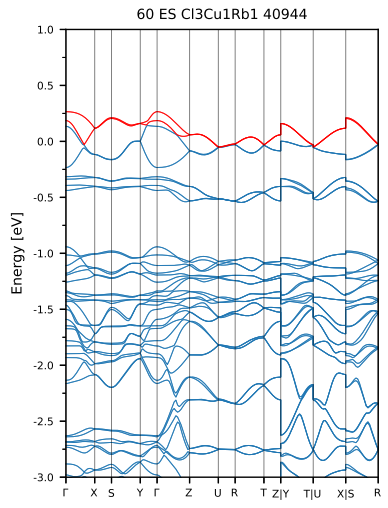

(h) 60.3.1

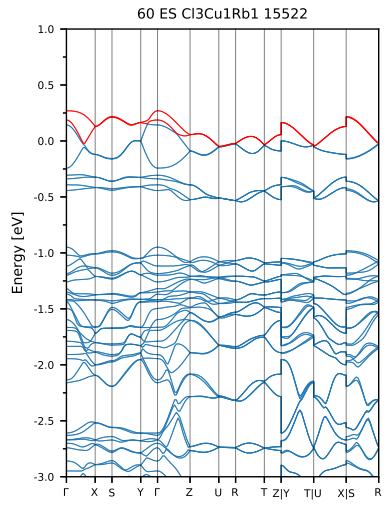

(i) 60.3.2

Figure 107: Band structures for topological materials in SG 60, part 1/1, from 60.2.1 to 60.3.2. See the list of figures.

## 0.0.25 SG 61

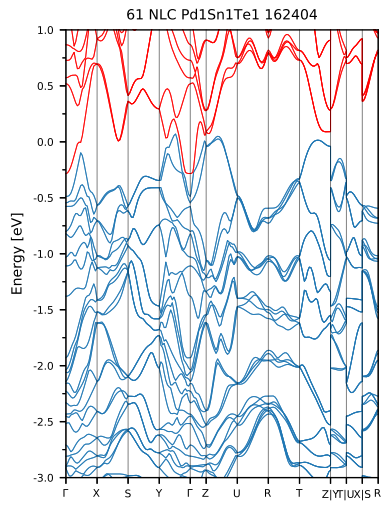

(a) 61.2.1

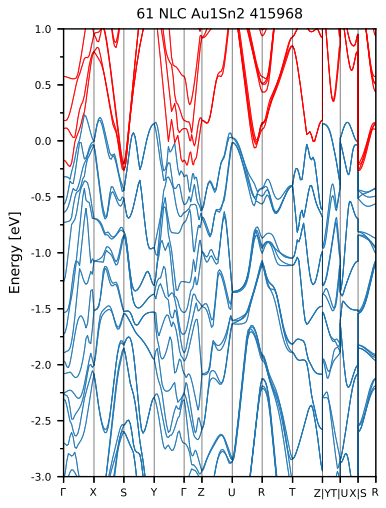

(b) 61.2.2

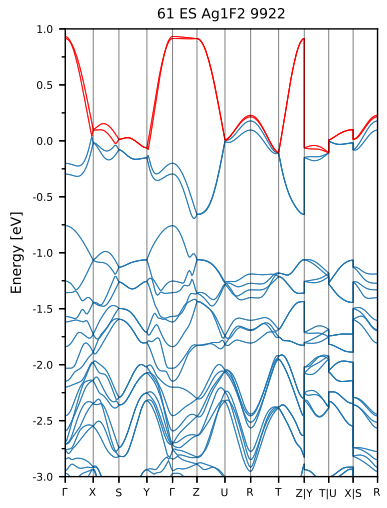

(c) 61.3.1

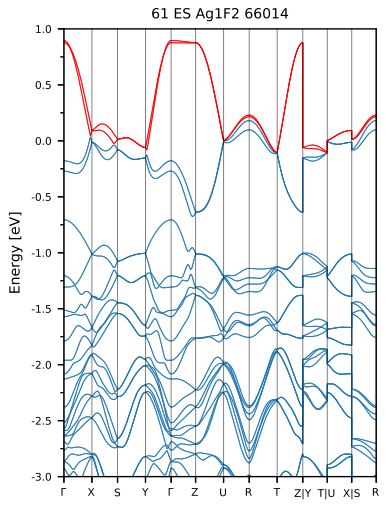

(d) 61.3.2

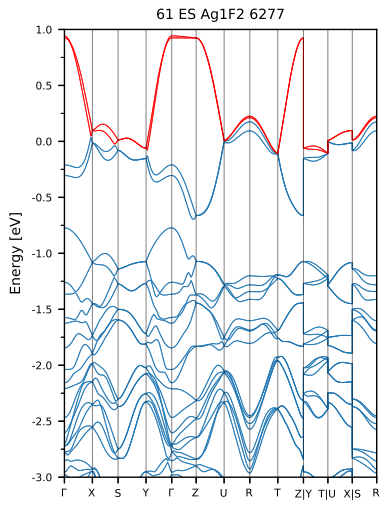

(e) 61.3.3

Figure 108: Band structures for topological materials in SG 61, part 1/1, from 61.2.1 to 61.3.3. See the list of figures.

### 0.0.26 SG 62

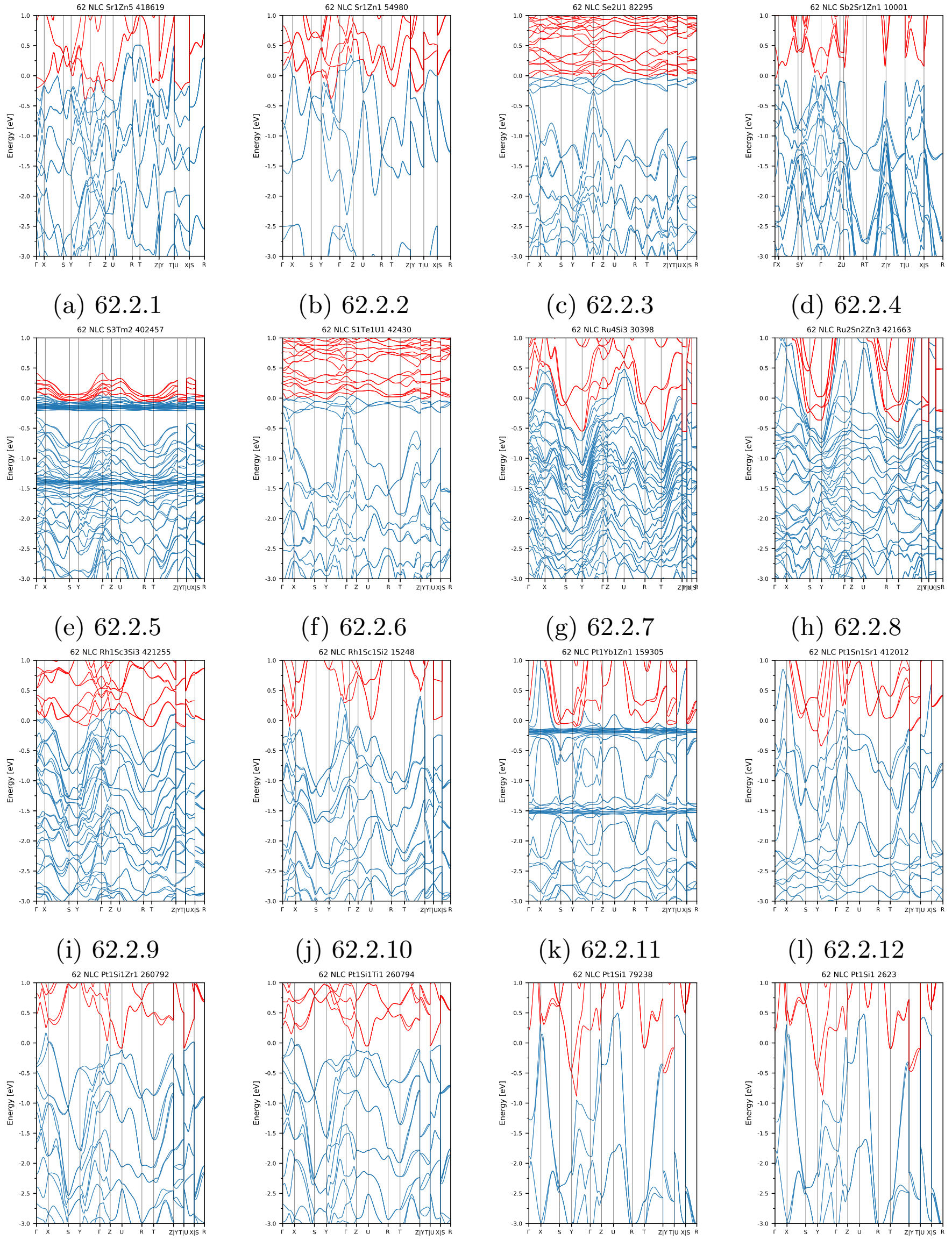

(a) 62.2.1 (b) 62.2.2 (c) 62.2.3 (d) 62.2.4

(e) 62.2.5 (f) 62.2.6 (g) 62.2.7 (h) 62.2.8

(i) 62.2.9 (j) 62.2.10 (k) 62.2.11 (l) 62.2.12

(m) 62.2.13 (n) 62.2.14 (o) 62.2.15 (p) 62.2.16

Figure 109: Band structures for topological materials in SG 62, part 1/36, from 62.2.1 to 62.2.16. See the list of figures.

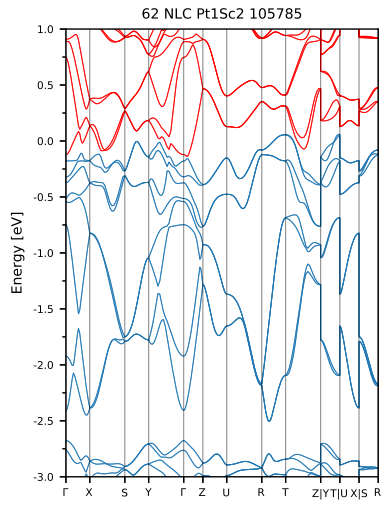

(q) 62.2.17

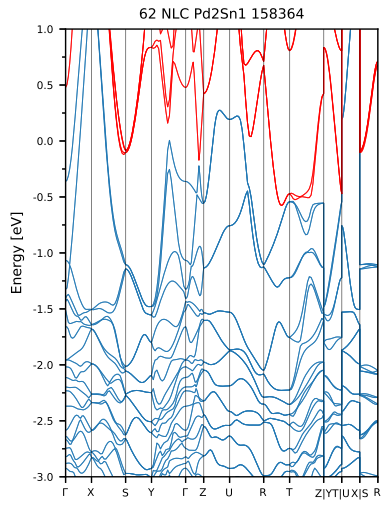

(r) 62.2.18

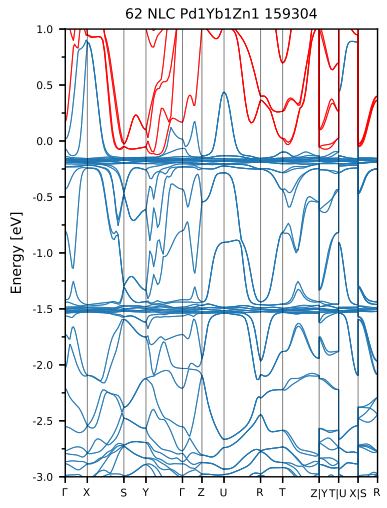

(s) 62.2.19

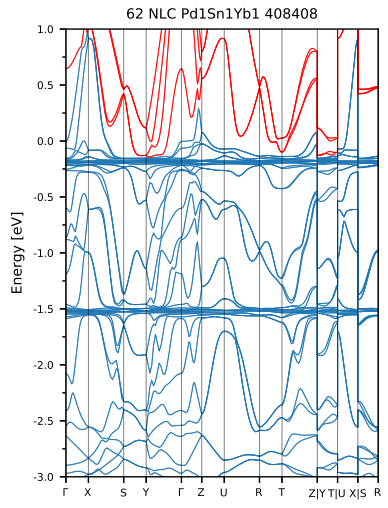

(t) 62.2.20

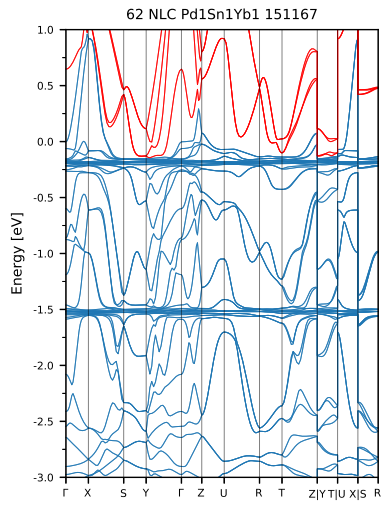

(u) 62.2.21

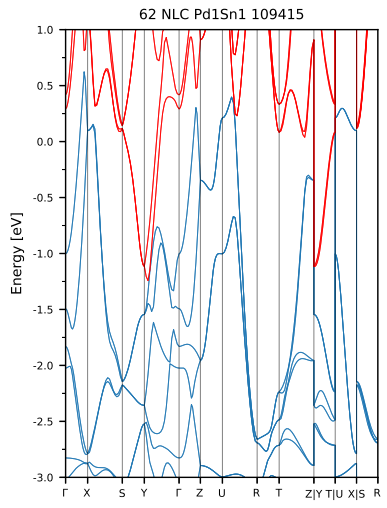

(v) 62.2.22

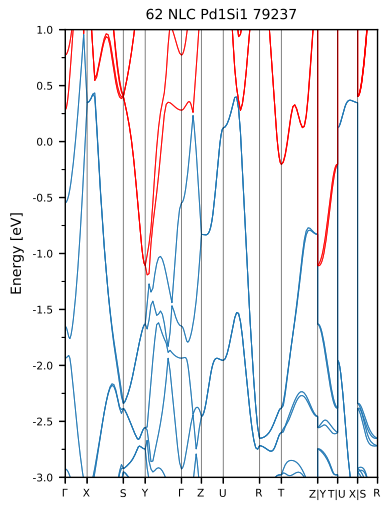

(w) 62.2.23

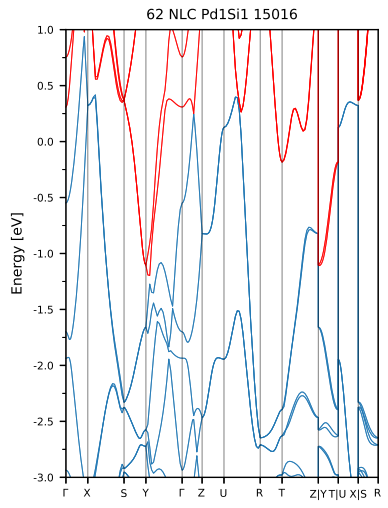

(x) 62.2.24

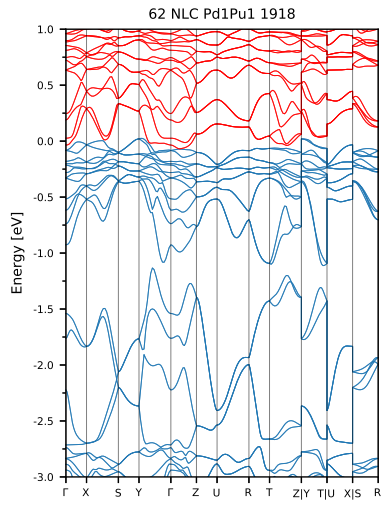

(y) 62.2.25

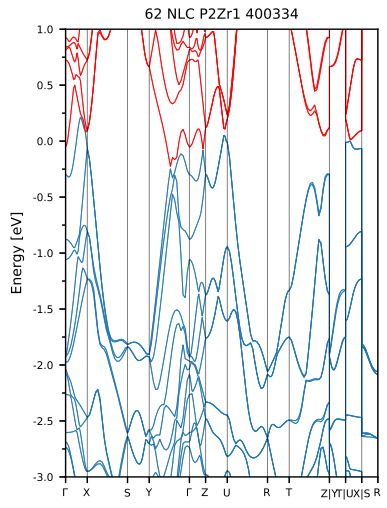

(z) 62.2.26

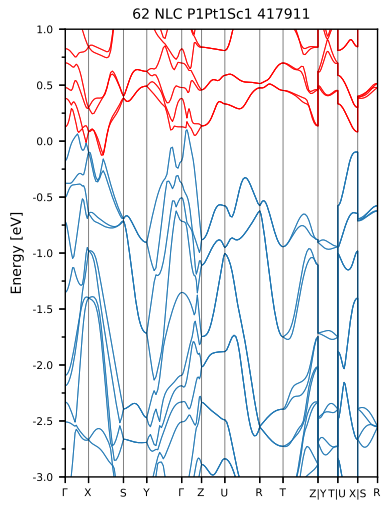

(aa) 62.2.27

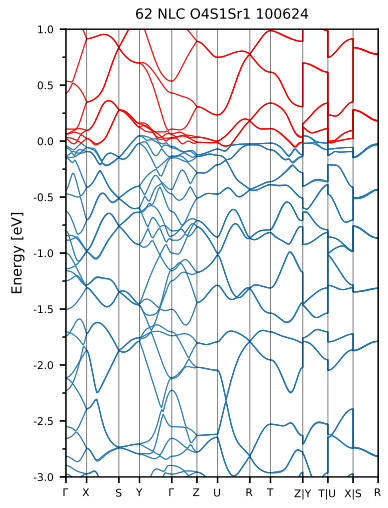

(ab) 62.2.28

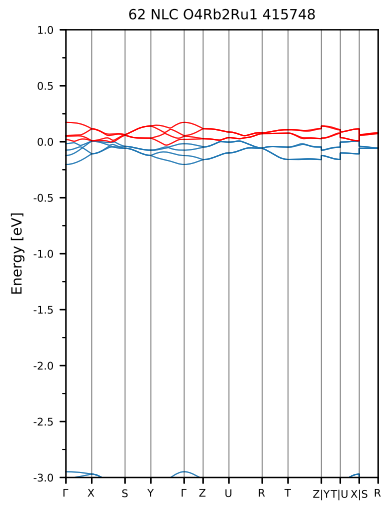

(ac) 62.2.29

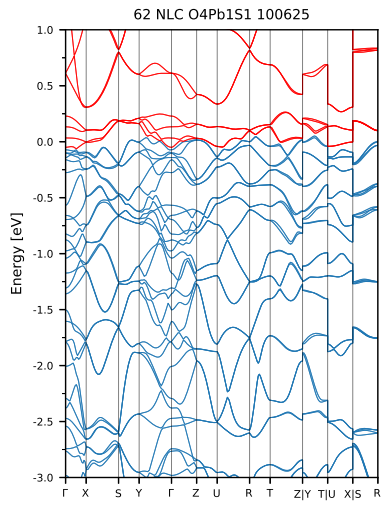

(ad) 62.2.30

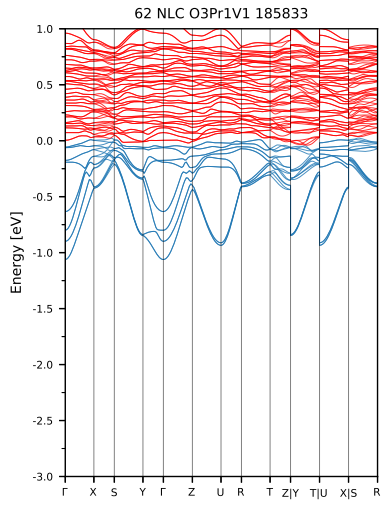

(ae) 62.2.31

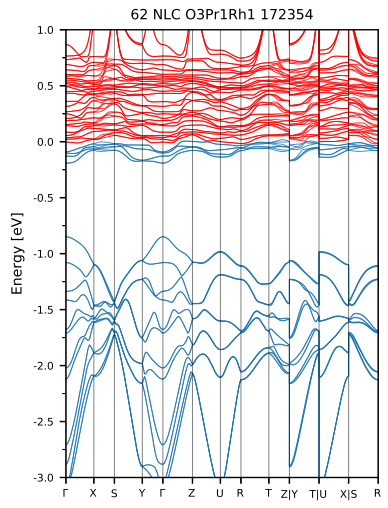

(af) 62.2.32

Figure 110: Band structures for topological materials in SG 62, part 2/36, from 62.2.17 to 62.2.32. See the list of figures.

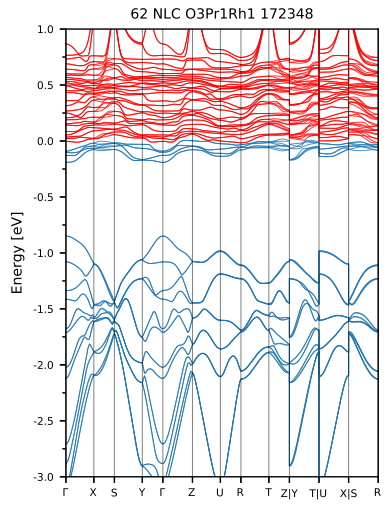

(ag) 62.2.33

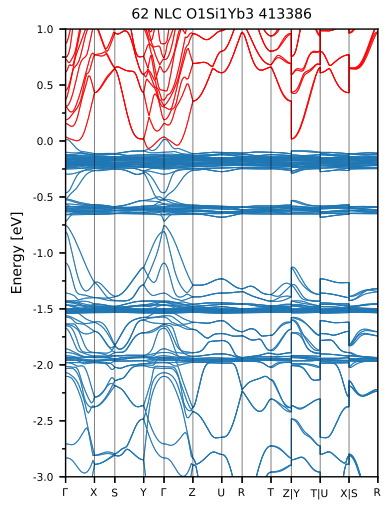

(ah) 62.2.34

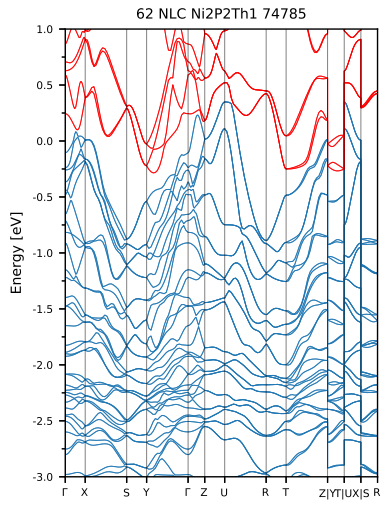

(ai) 62.2.35

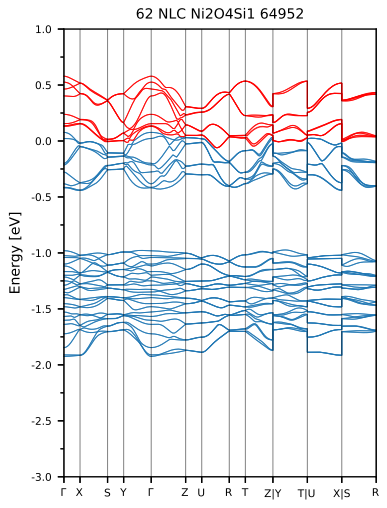

(aj) 62.2.36

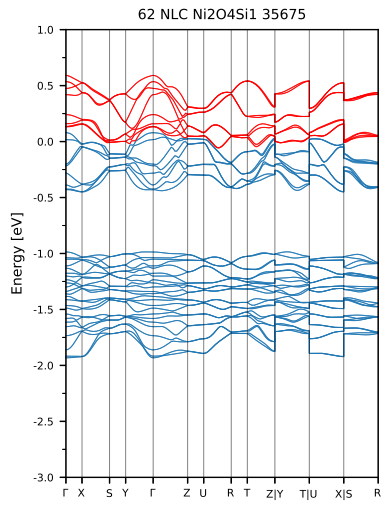

(ak) 62.2.37

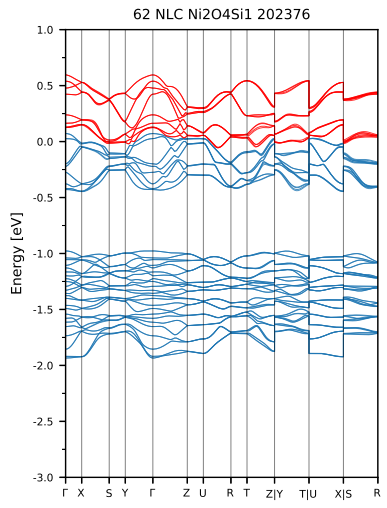

(al) 62.2.38

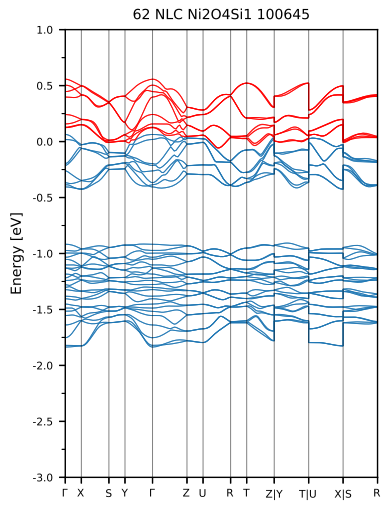

(am) 62.2.39

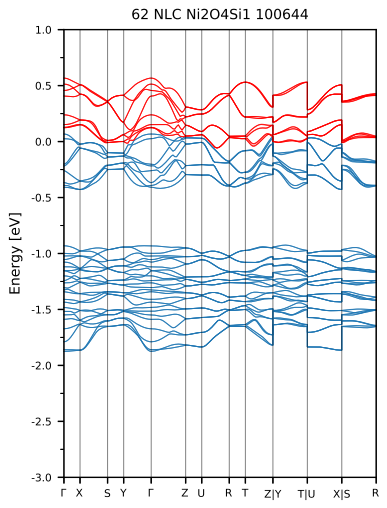

(an) 62.2.40

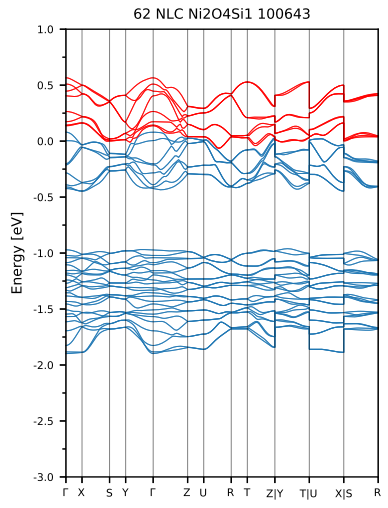

(ao) 62.2.41

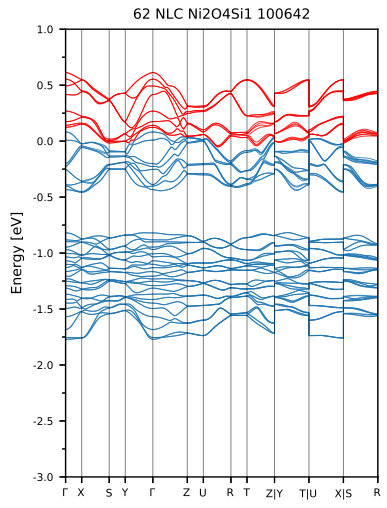

(ap) 62.2.42

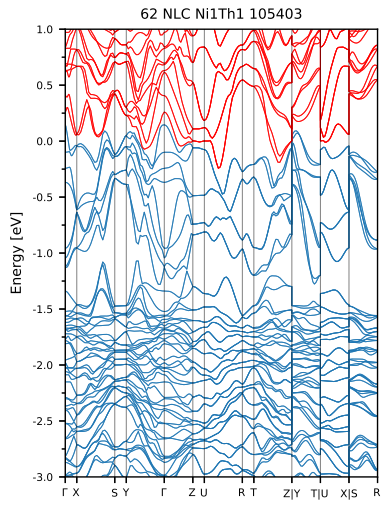

(aq) 62.2.43

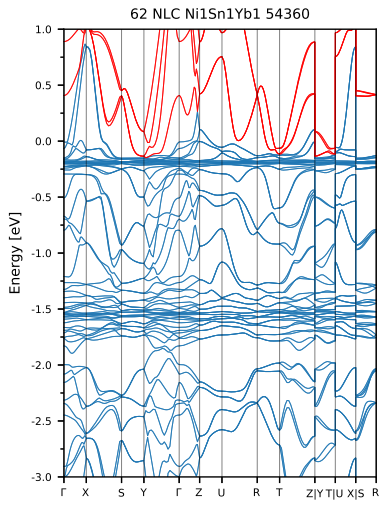

(ar) 62.2.44

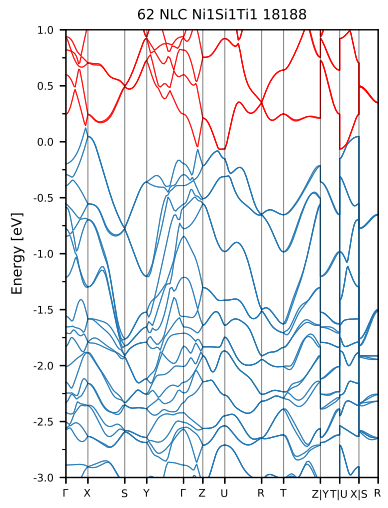

(as) 62.2.45

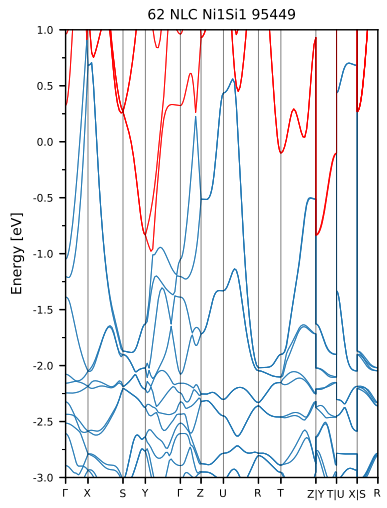

(at) 62.2.46

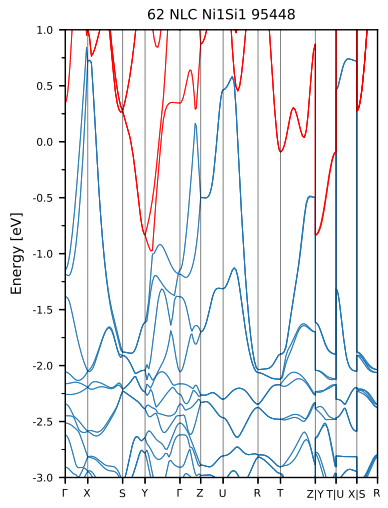

(au) 62.2.47

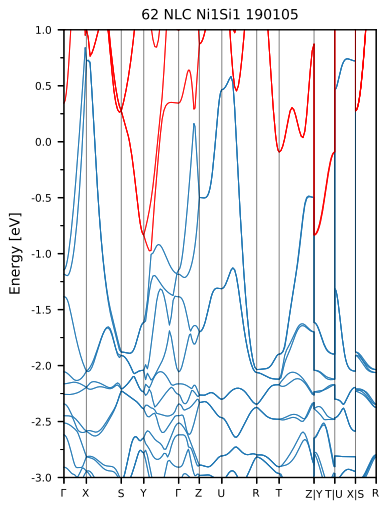

(av) 62.2.48

Figure 111: Band structures for topological materials in SG 62, part 3/36, from 62.2.33 to 62.2.48. See the list of figures.

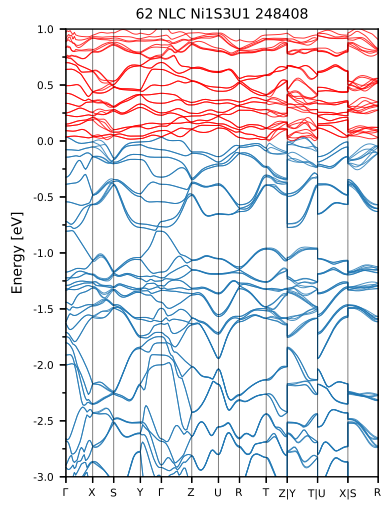

(aw) 62.2.49

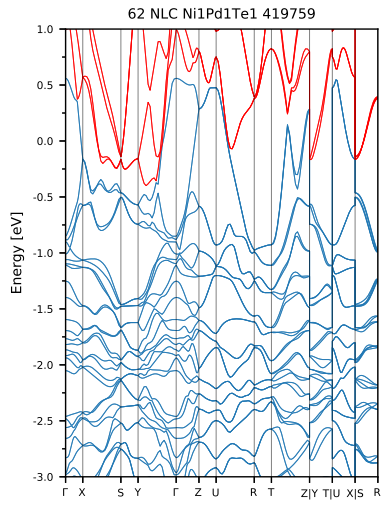

(ax) 62.2.50

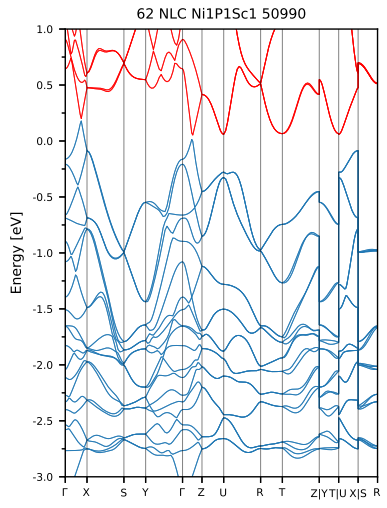

(ay) 62.2.51

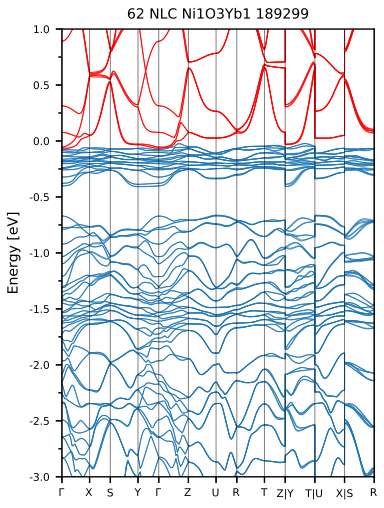

(az) 62.2.52

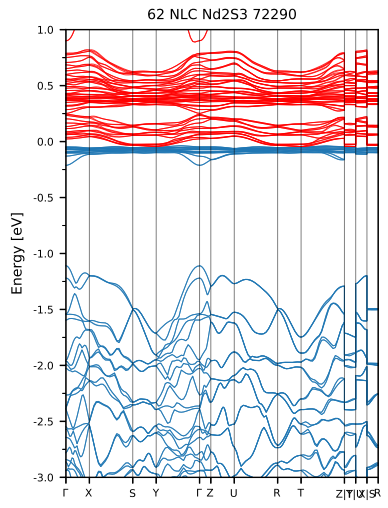

(ba) 62.2.53

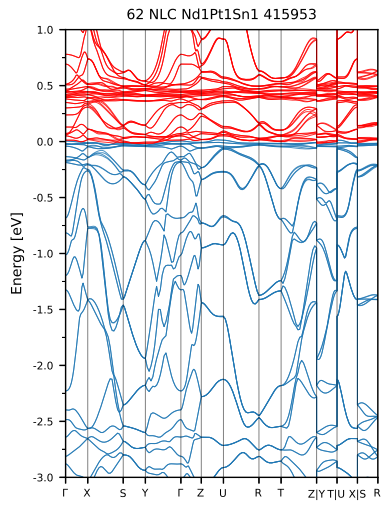

(bb) 62.2.54

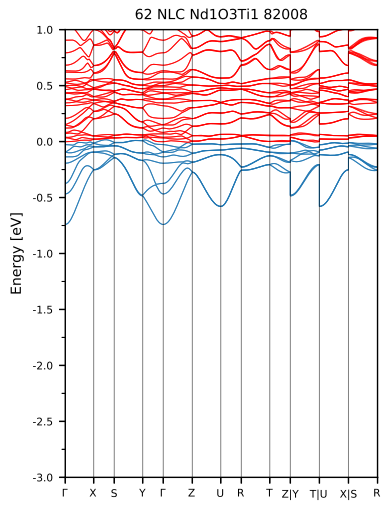

(bc) 62.2.55

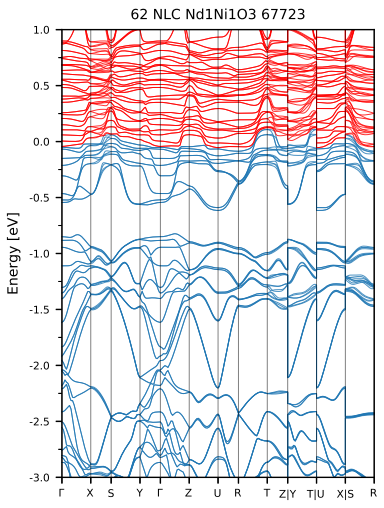

(bd) 62.2.56

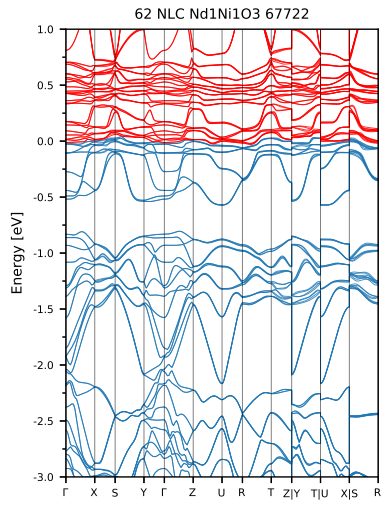

(be) 62.2.57

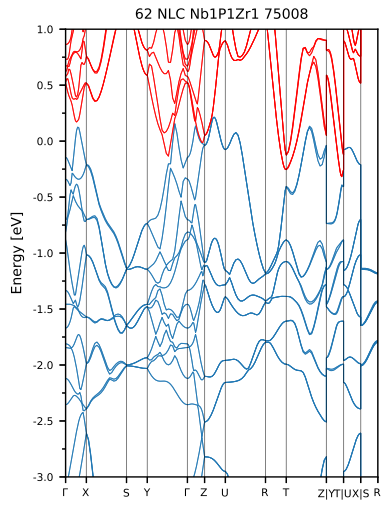

(bf) 62.2.58

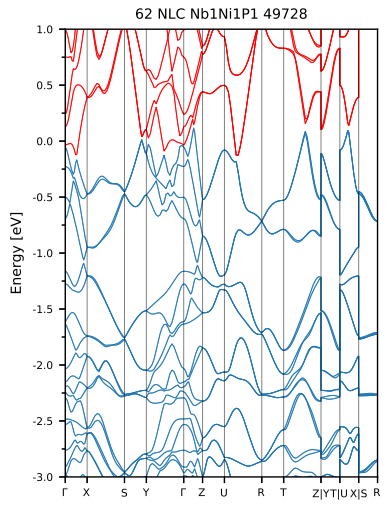

(bg) 62.2.59

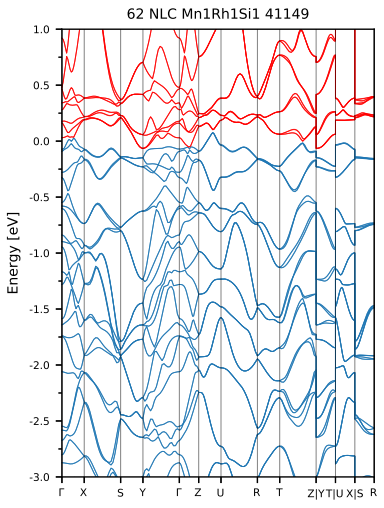

(bh) 62.2.60

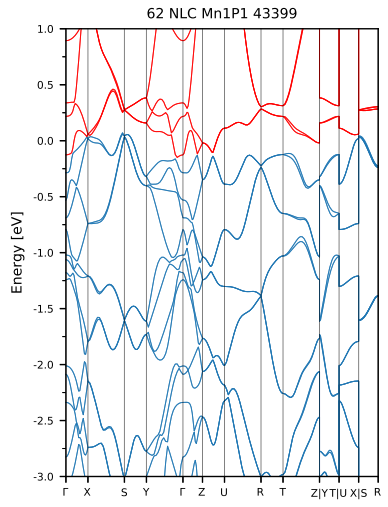

(bi) 62.2.61

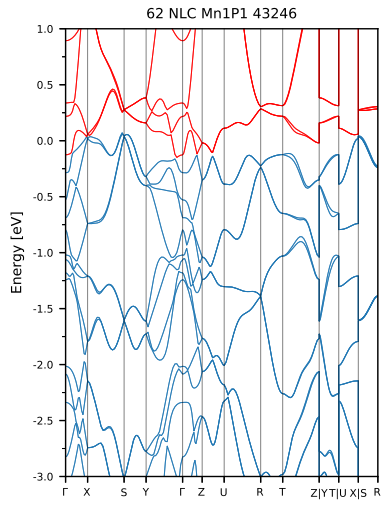

(bj) 62.2.62

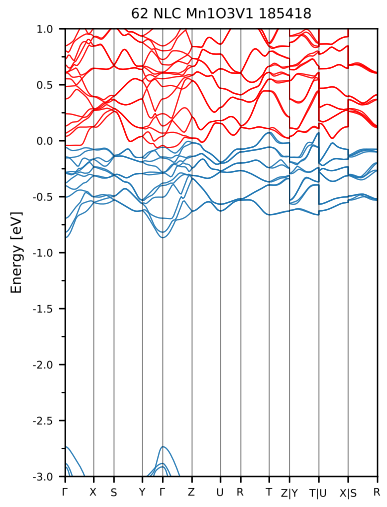

(bk) 62.2.63

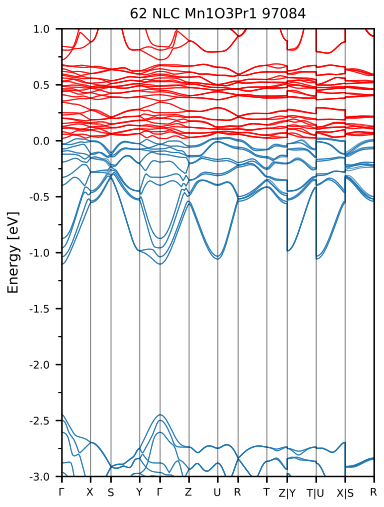

(bl) 62.2.64

Figure 112: Band structures for topological materials in SG 62, part 4/36, from 62.2.49 to 62.2.64. See the list of figures.

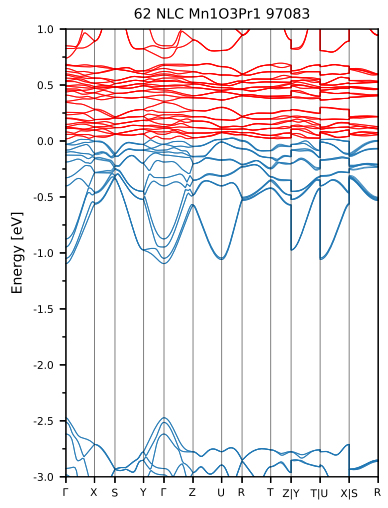

(bm) 62.2.65

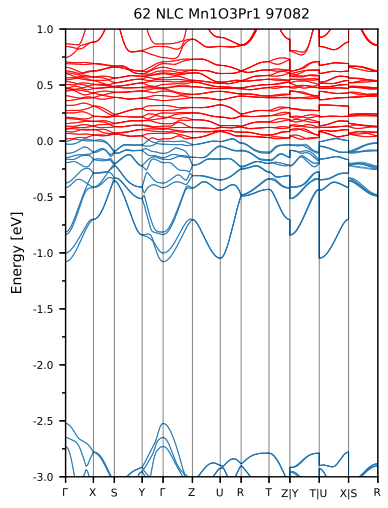

(bn) 62.2.66

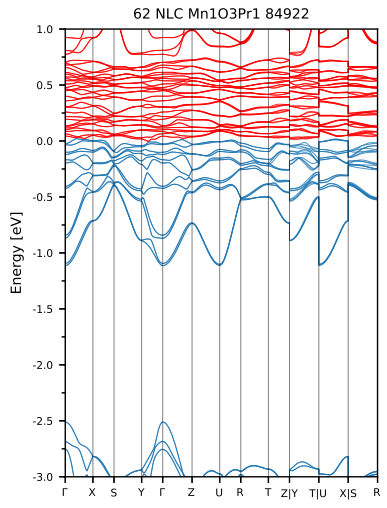

(bo) 62.2.67

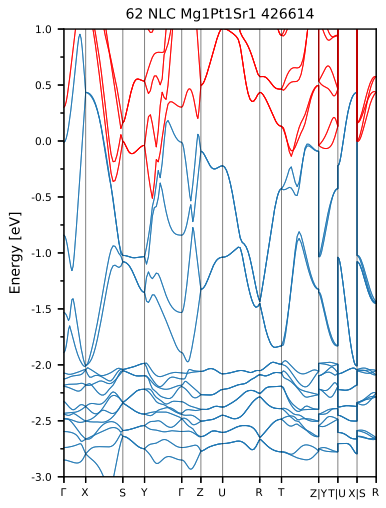

(bp) 62.2.68

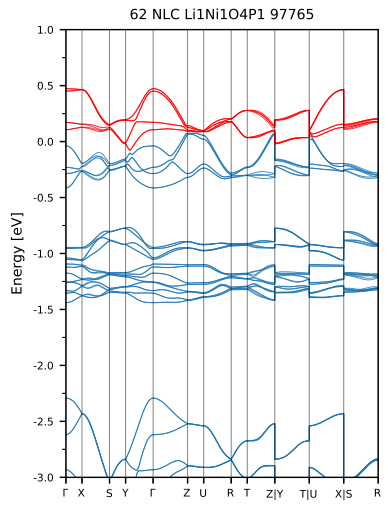

(bq) 62.2.69

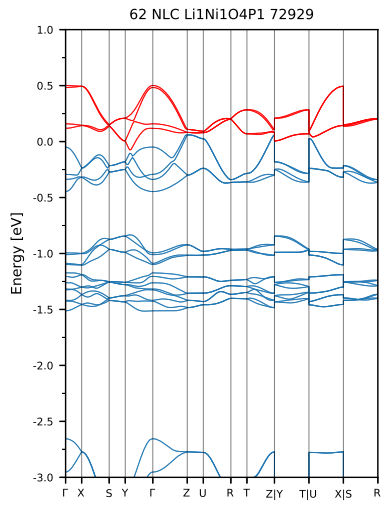

(br) 62.2.70

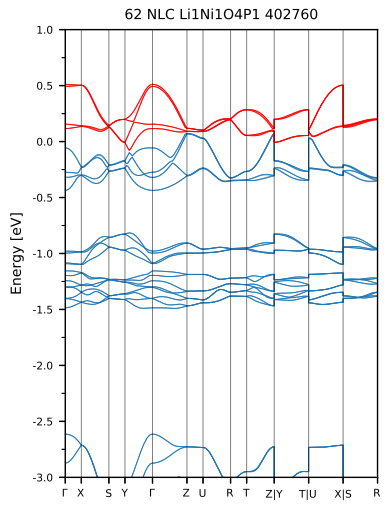

(bs) 62.2.71

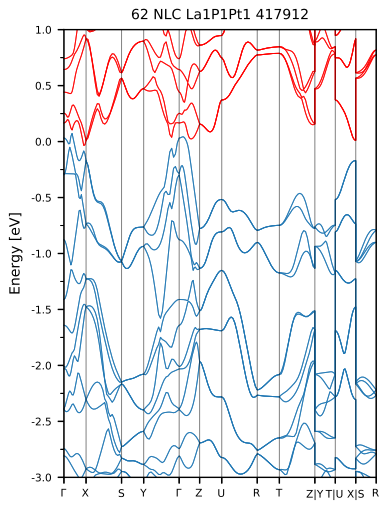

(bt) 62.2.72

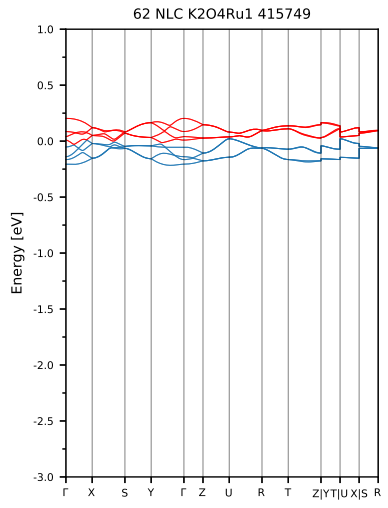

(bu) 62.2.73

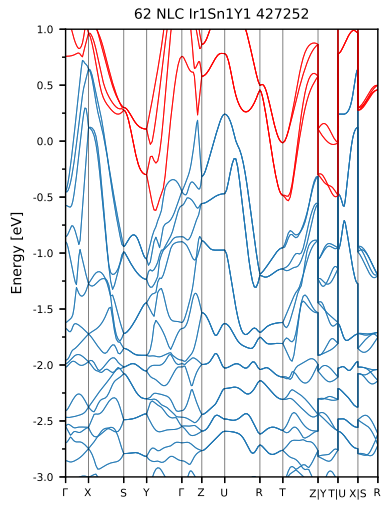

(bv) 62.2.74

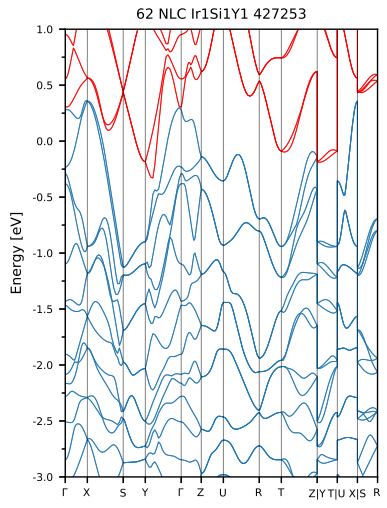

(bw) 62.2.75

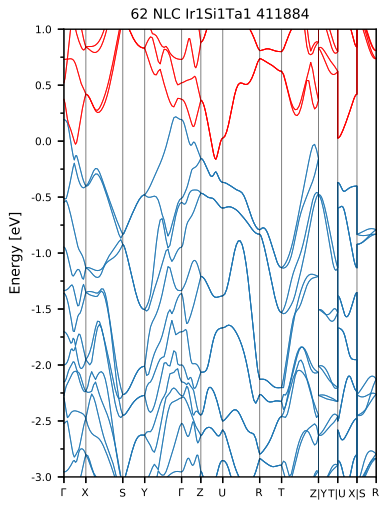

(bx) 62.2.76

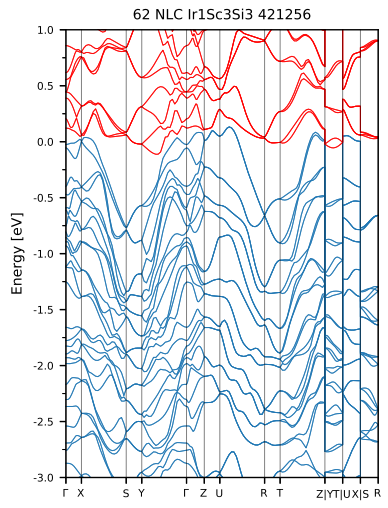

(by) 62.2.77

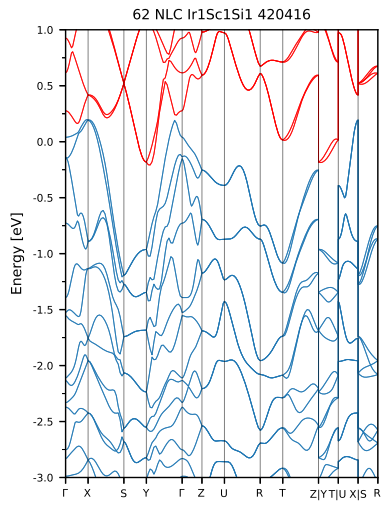

(bz) 62.2.78

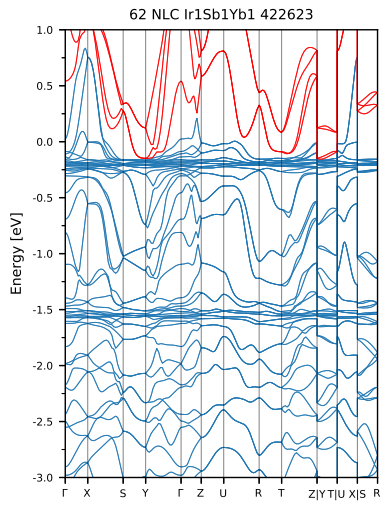

(ca) 62.2.79

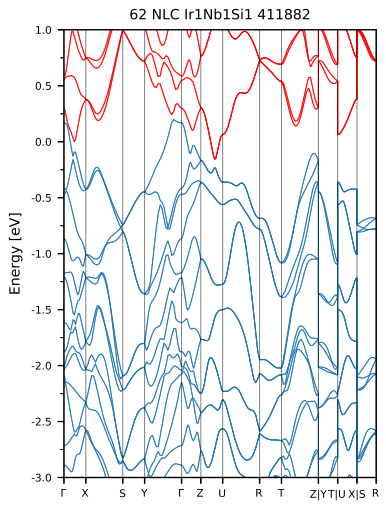

(cb) 62.2.80

Figure 113: Band structures for topological materials in SG 62, part 5/36, from 62.2.65 to 62.2.80. See the list of figures.

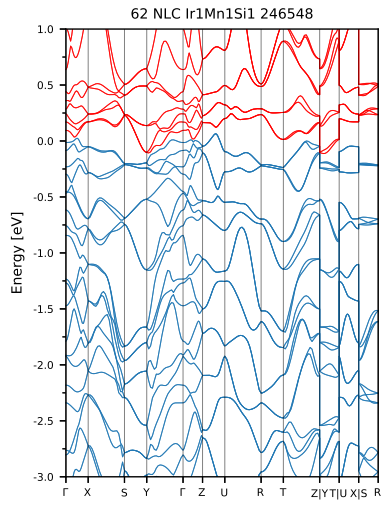

(cc) 62.2.81

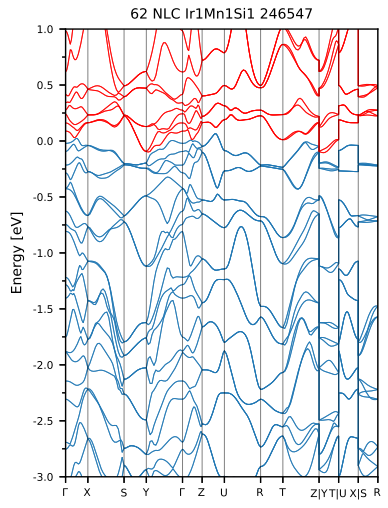

(cd) 62.2.82

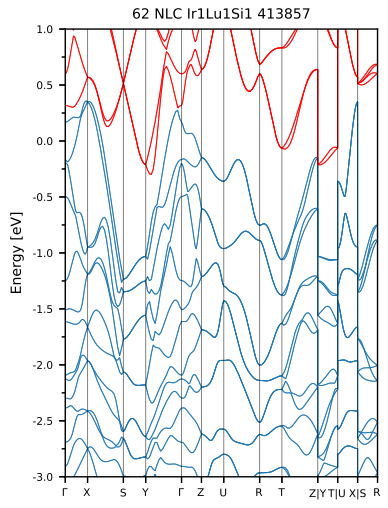

(ce) 62.2.83

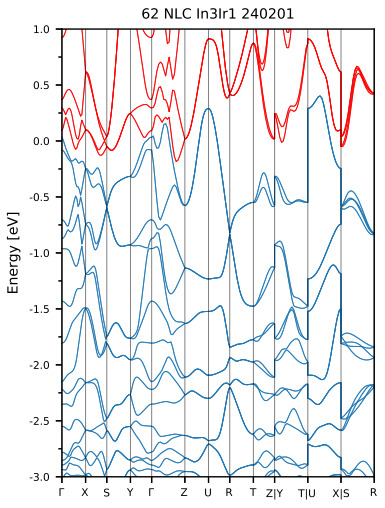

(cf) 62.2.84

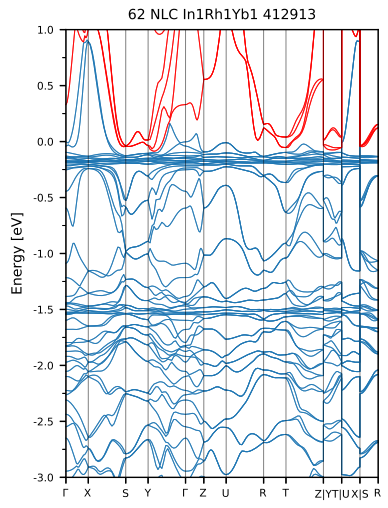

(cg) 62.2.85

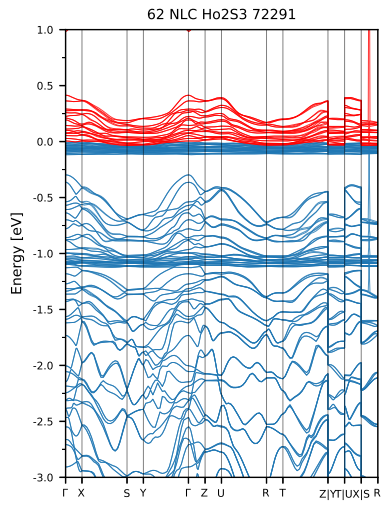

(ch) 62.2.86

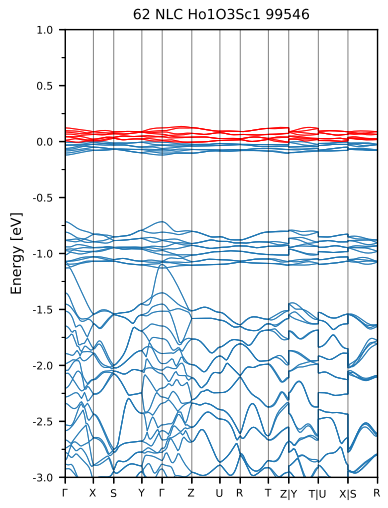

(ci) 62.2.87

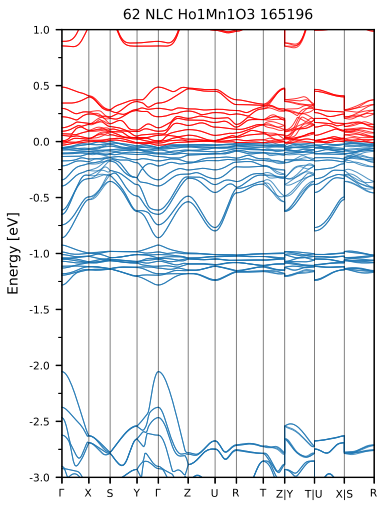

(cj) 62.2.88

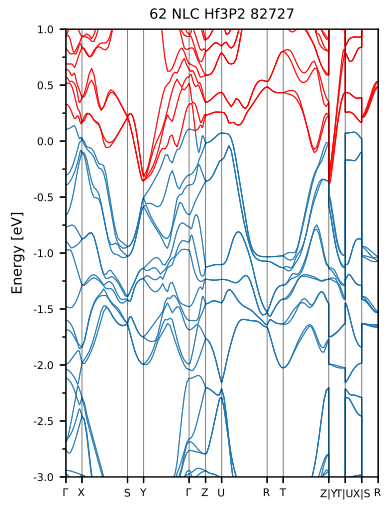

(ck) 62.2.89

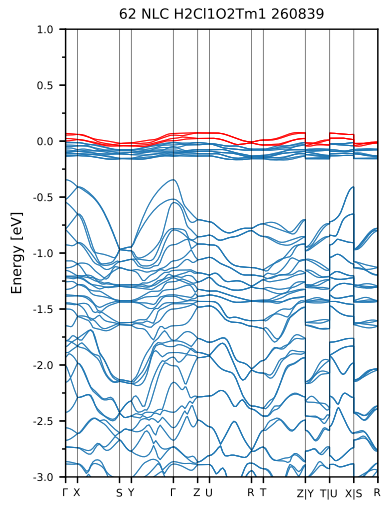

(cl) 62.2.90

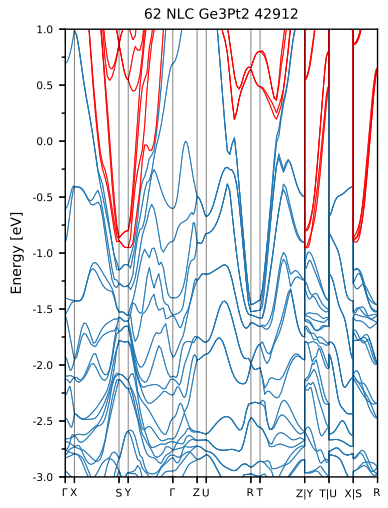

(cm) 62.2.91

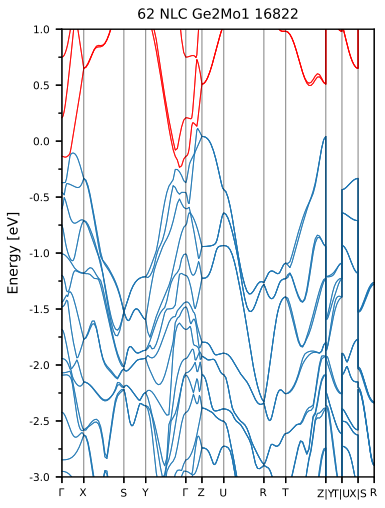

(cn) 62.2.92

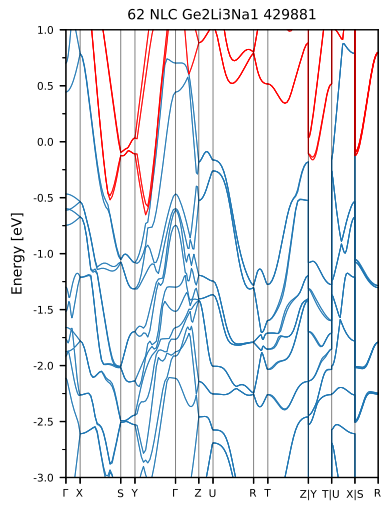

(co) 62.2.93

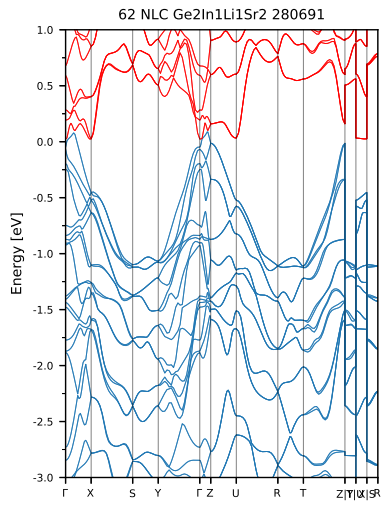

(cp) 62.2.94

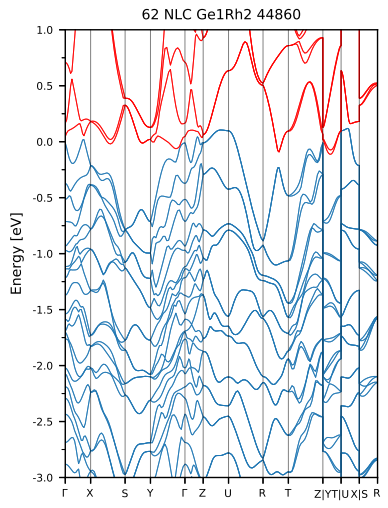

(cq) 62.2.95

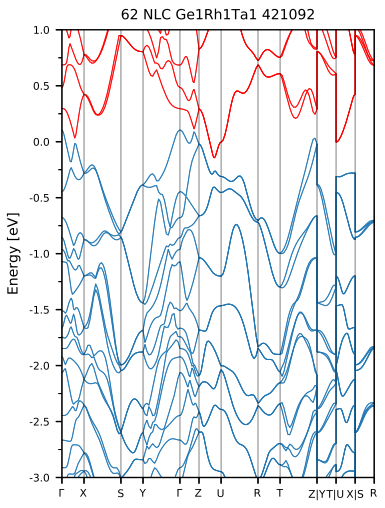

(cr) 62.2.96

Figure 114: Band structures for topological materials in SG 62, part 6/36, from 62.2.81 to 62.2.96. See the list of figures.

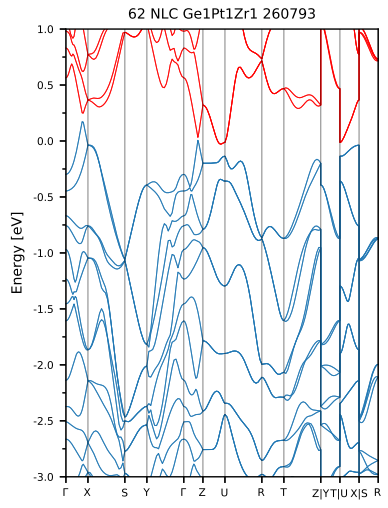

(cs) 62.2.97

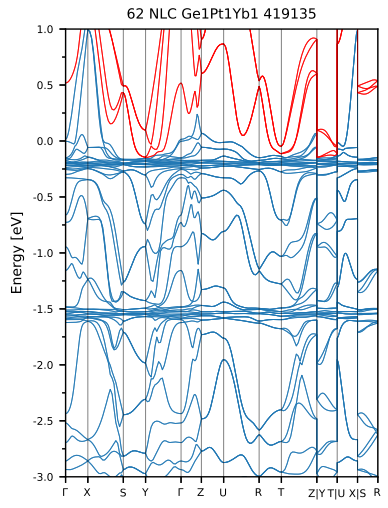

(ct) 62.2.98

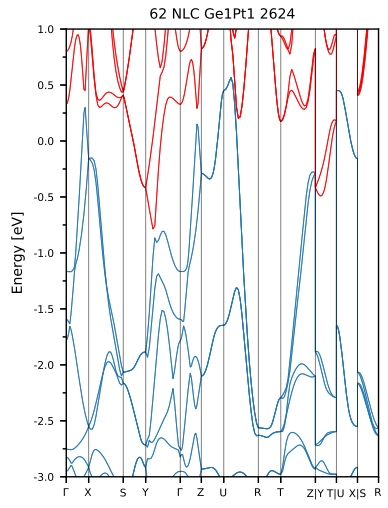

(cu) 62.2.99

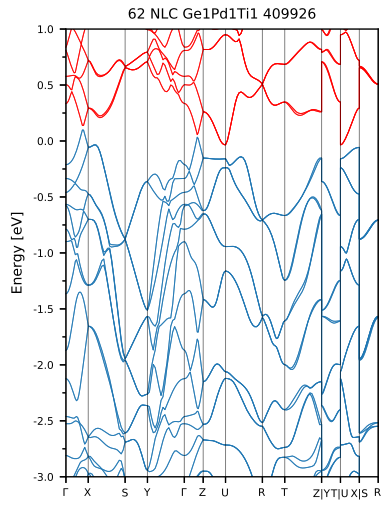

(cv) 62.2.100

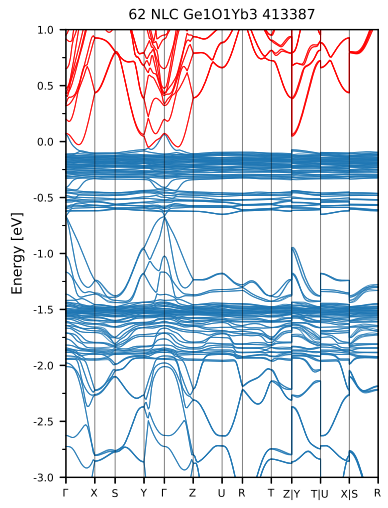

(cw) 62.2.101

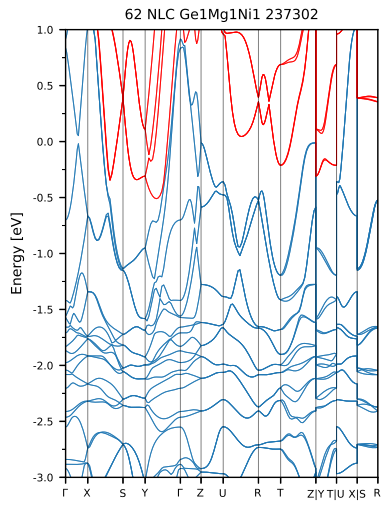

(cx) 62.2.102

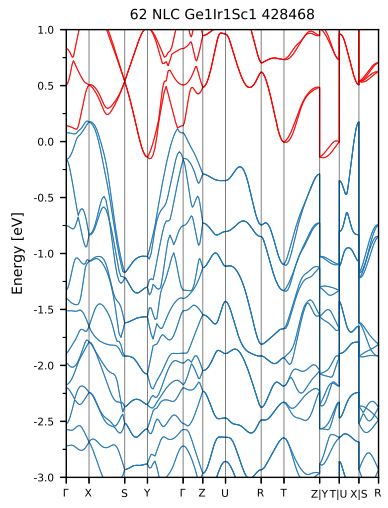

(cy) 62.2.103

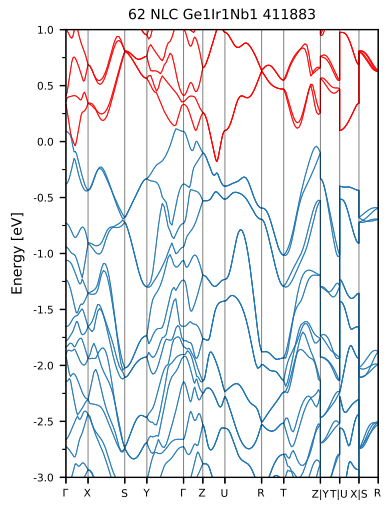

(cz) 62.2.104

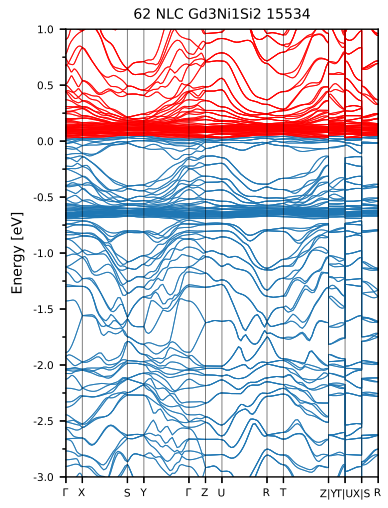

(da) 62.2.105

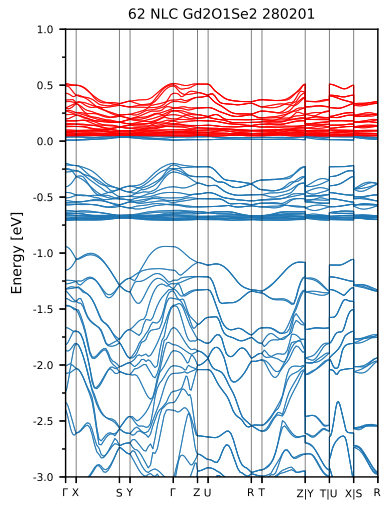

(db) 62.2.106

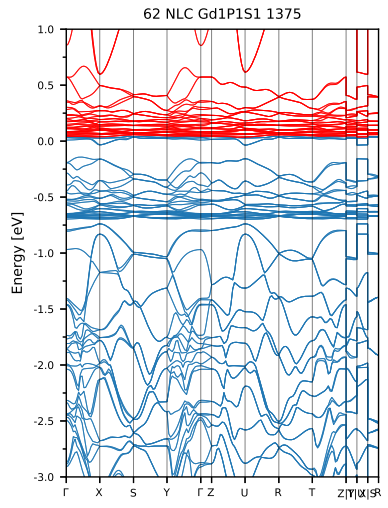

(dc) 62.2.107

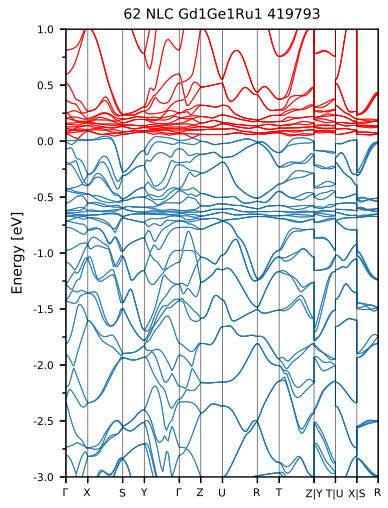

(dd) 62.2.108

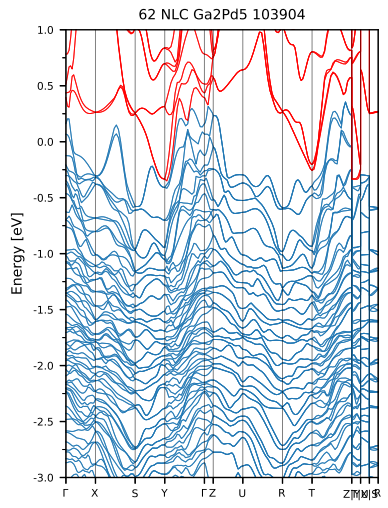

(de) 62.2.109

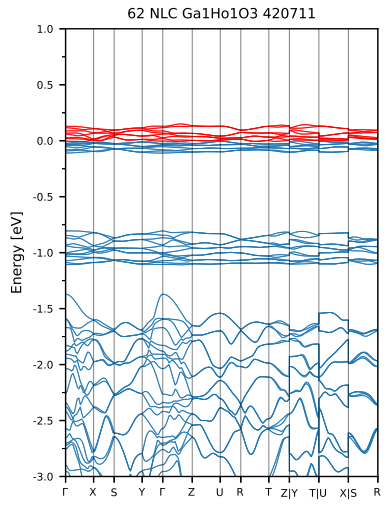

(df) 62.2.110

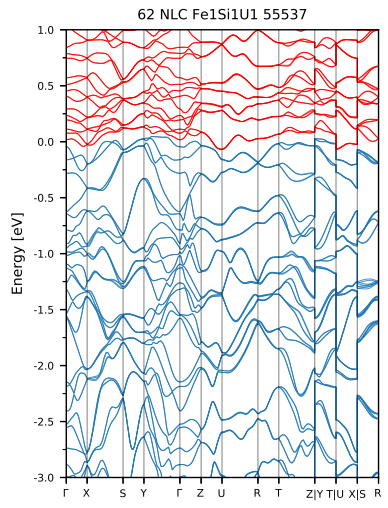

(dg) 62.2.111

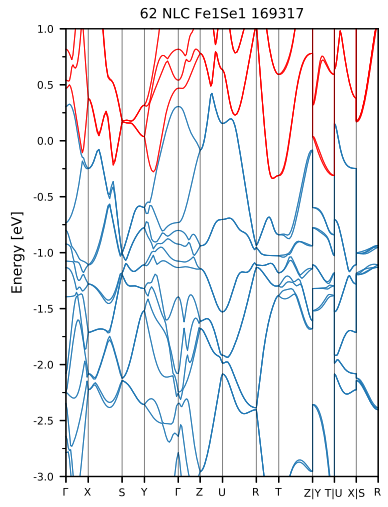

(dh) 62.2.112

Figure 115: Band structures for topological materials in SG 62, part 7/36, from 62.2.97 to 62.2.112. See the list of figures.

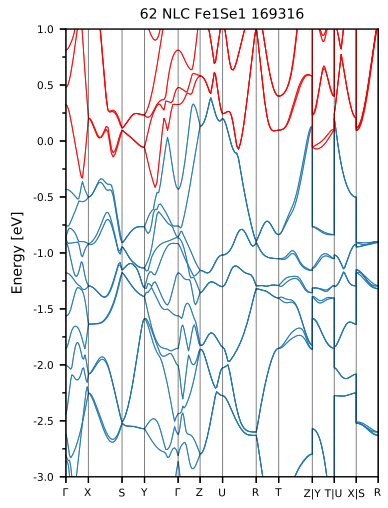

(di) 62.2.113

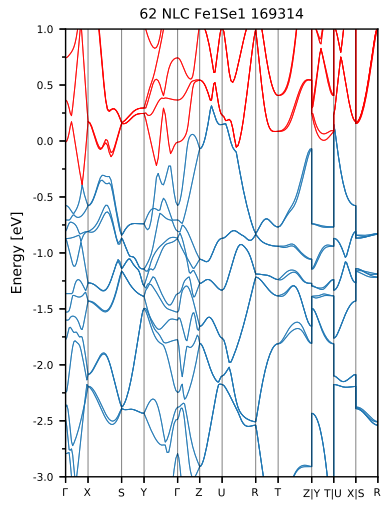

(dj) 62.2.114

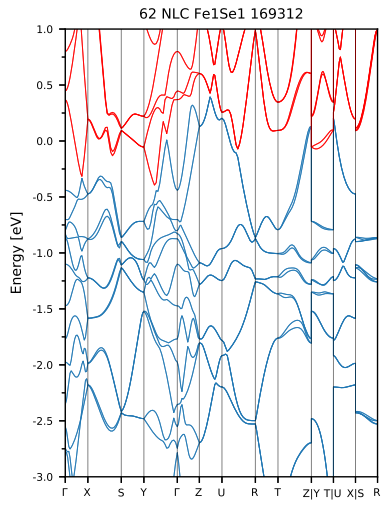

(dk) 62.2.115

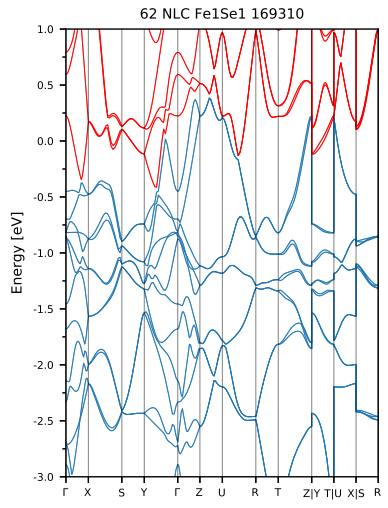

(dl) 62.2.116

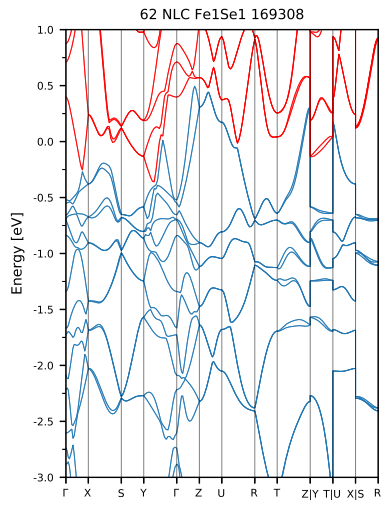

(dm) 62.2.117

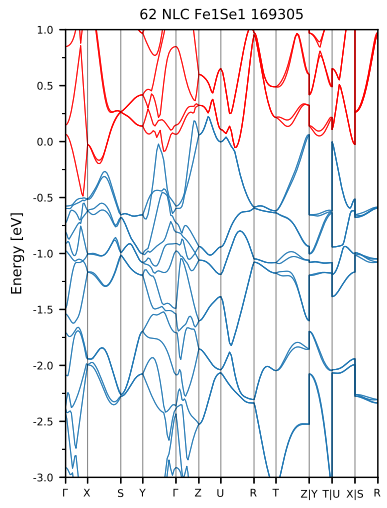

(dn) 62.2.118

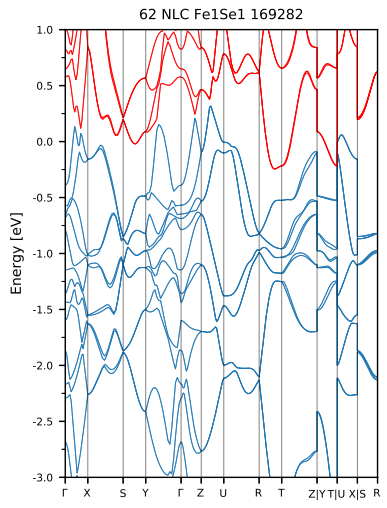

(do) 62.2.119

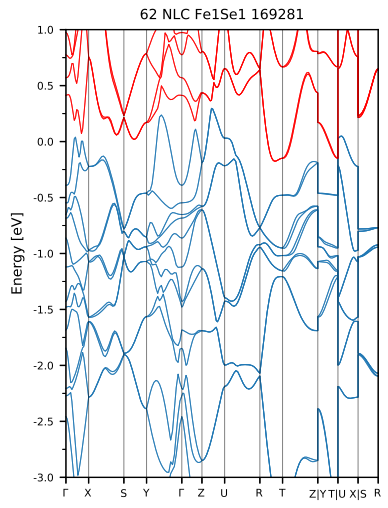

(dp) 62.2.120

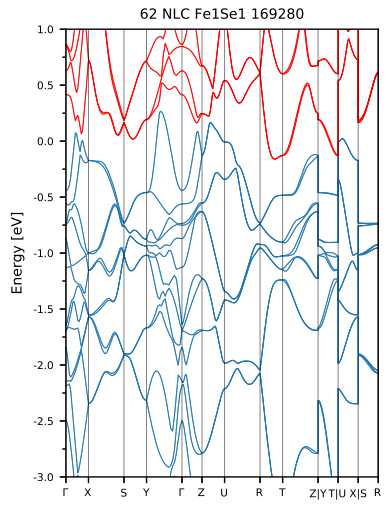

(dq) 62.2.121

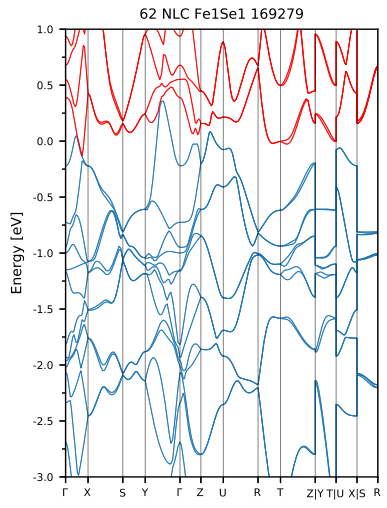

(dr) 62.2.122

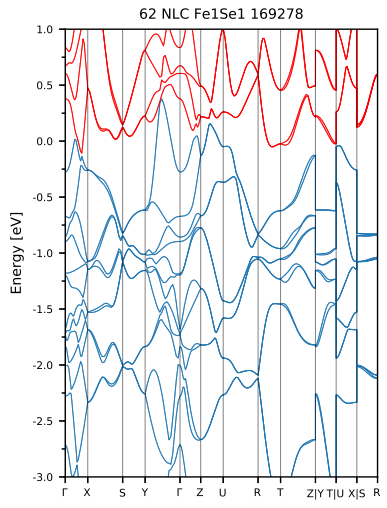

(ds) 62.2.123

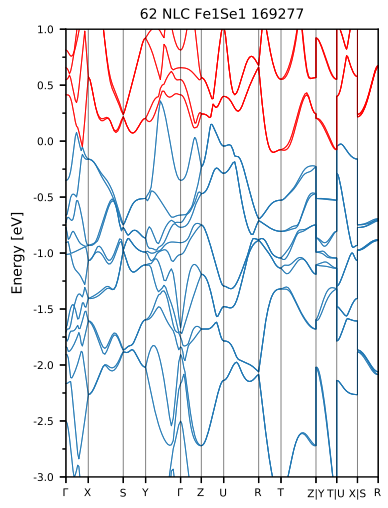

(dt) 62.2.124

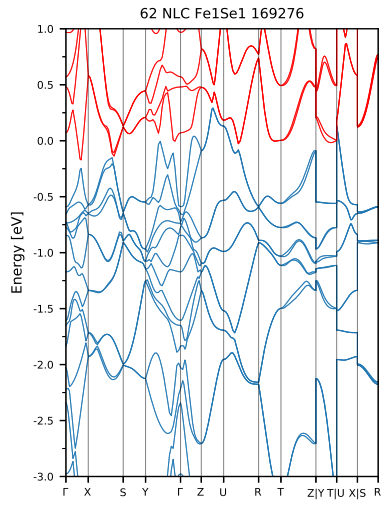

(du) 62.2.125

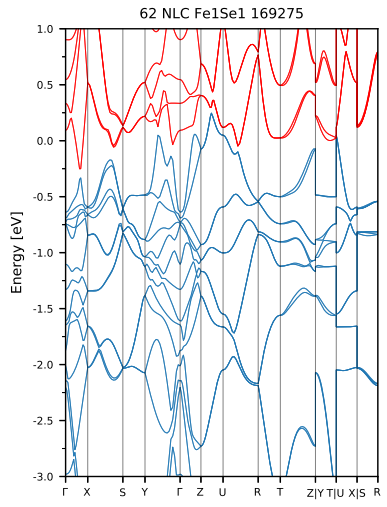

(dv) 62.2.126

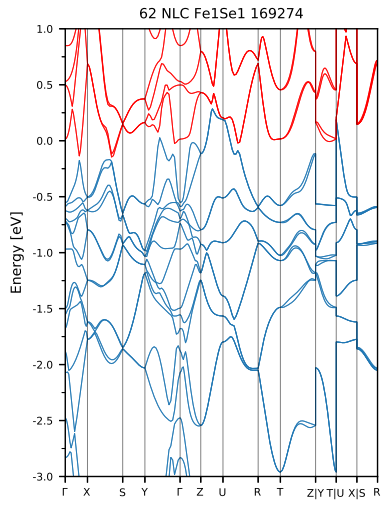

(dw) 62.2.127

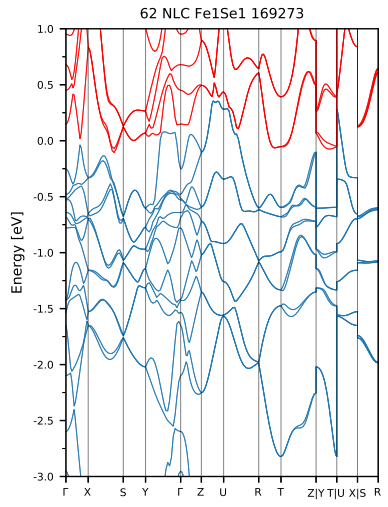

(dx) 62.2.128

Figure 116: Band structures for topological materials in SG 62, part 8/36, from 62.2.113 to 62.2.128. See the list of figures.

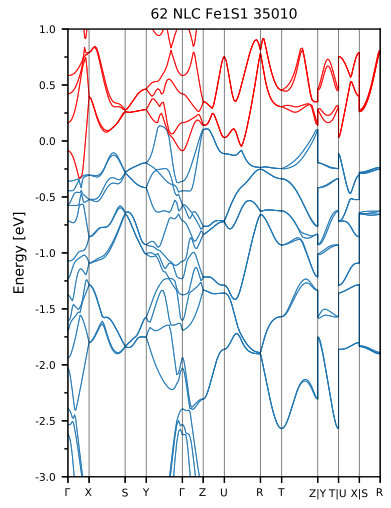

(dy) 62.2.129

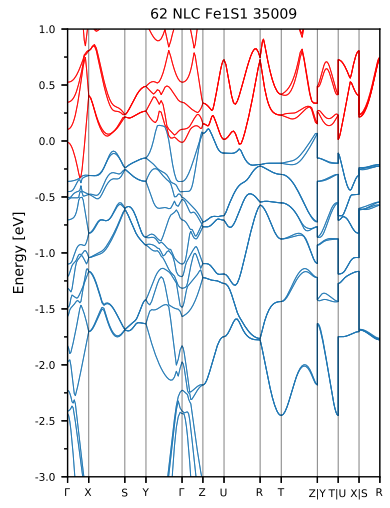

(dz) 62.2.130

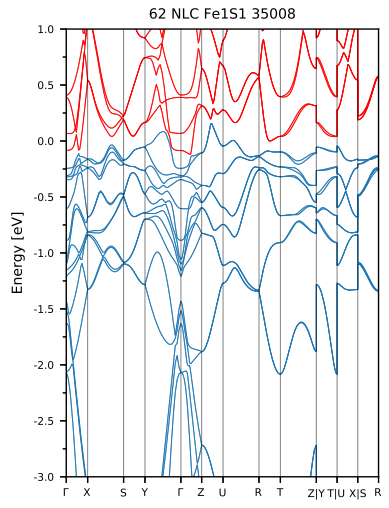

(ea) 62.2.131

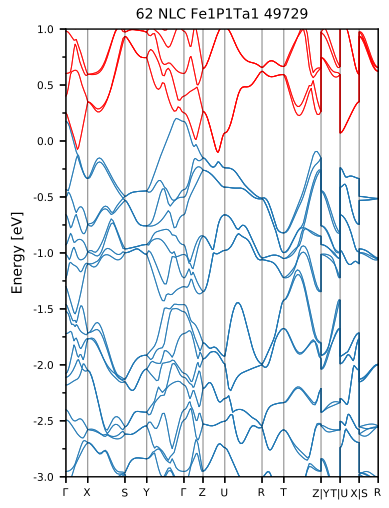

(eb) 62.2.132

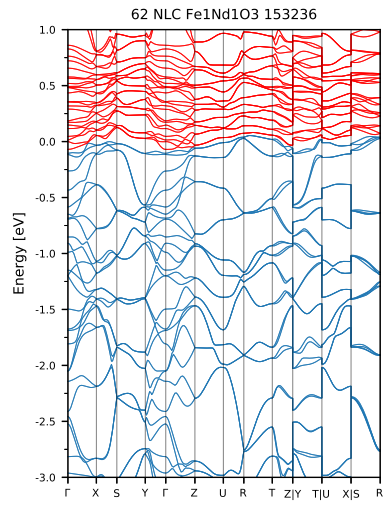

(ec) 62.2.133

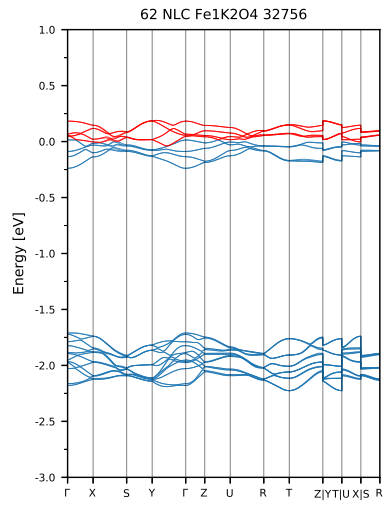

(ed) 62.2.134

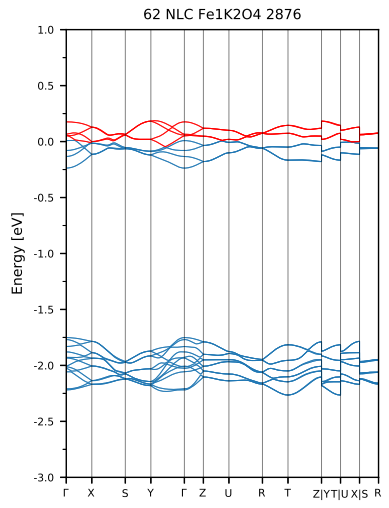

(ee) 62.2.135

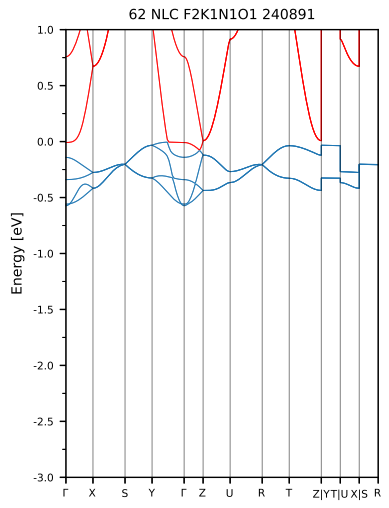

(ef) 62.2.136

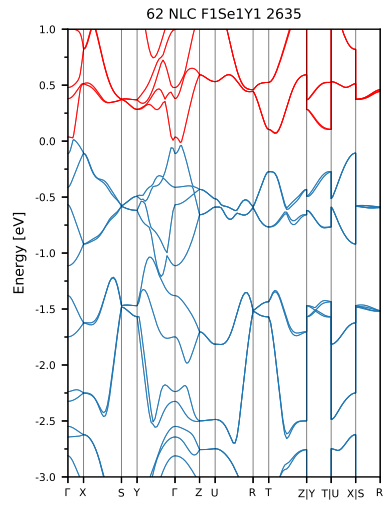

(eg) 62.2.137

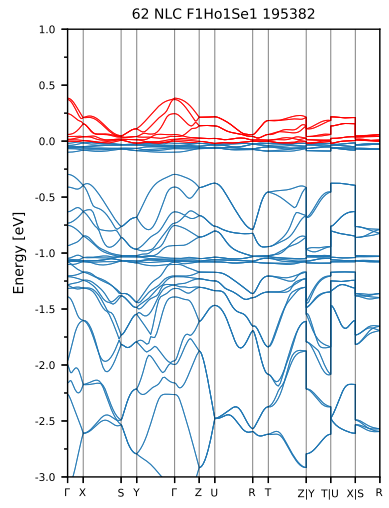

(eh) 62.2.138

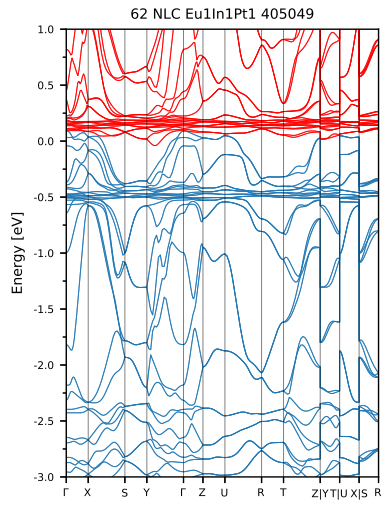

(ei) 62.2.139

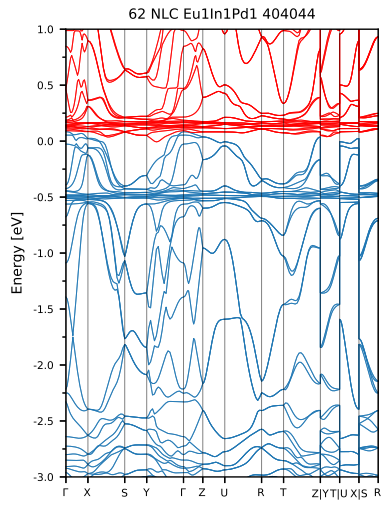

(ej) 62.2.140

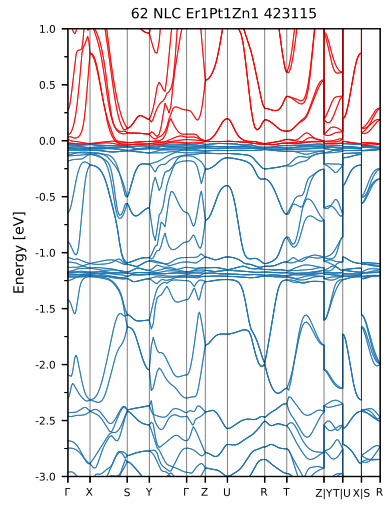

(ek) 62.2.141

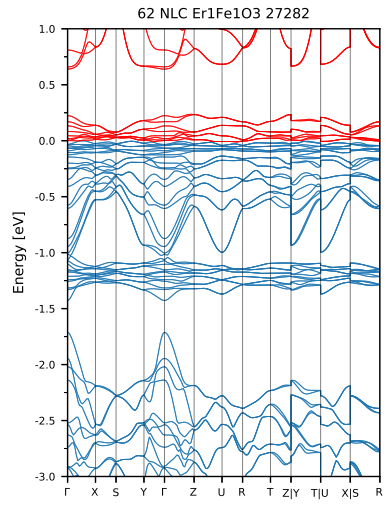

(el) 62.2.142

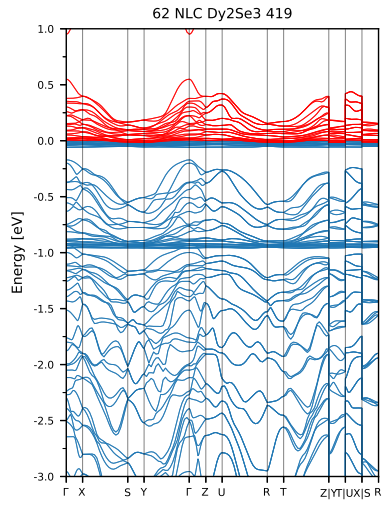

(em) 62.2.143

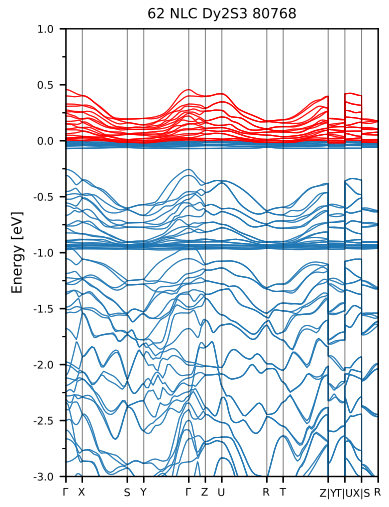

(en) 62.2.144

Figure 117: Band structures for topological materials in SG 62, part 9/36, from 62.2.129 to 62.2.144. See the list of figures.

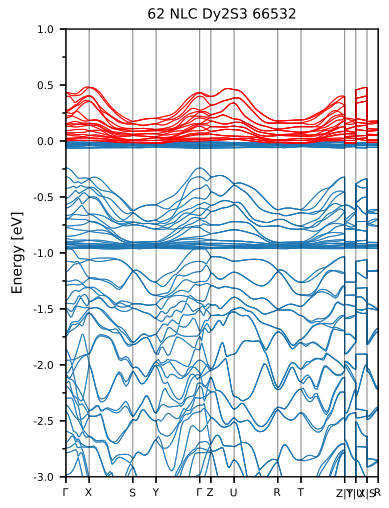

(eo) 62.2.145

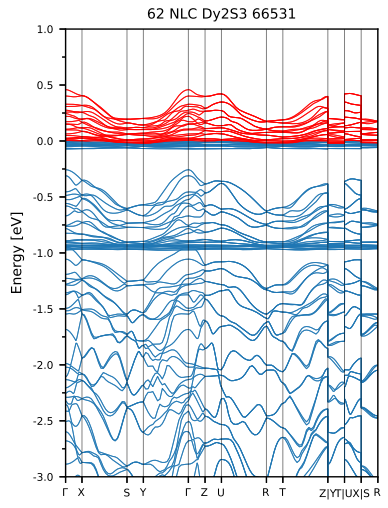

(ep) 62.2.146

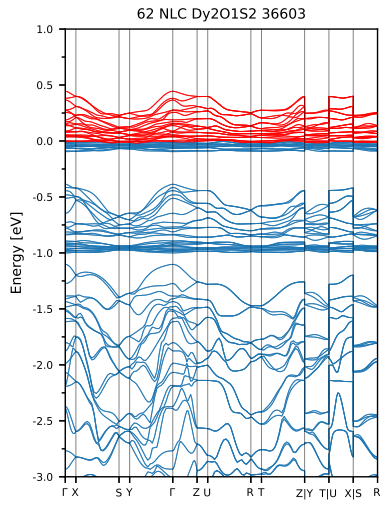

(eq) 62.2.147

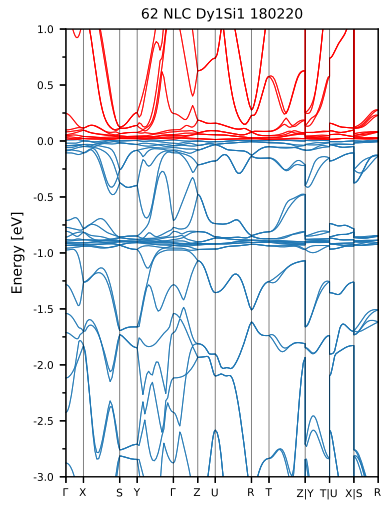

(er) 62.2.148

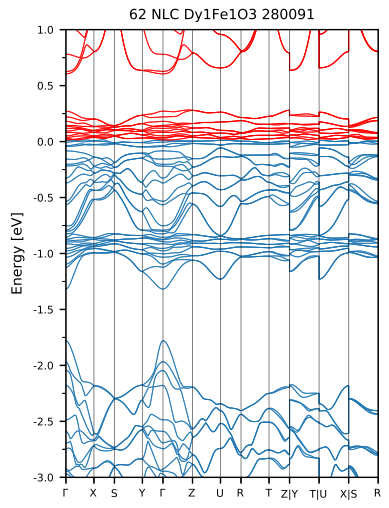

(es) 62.2.149

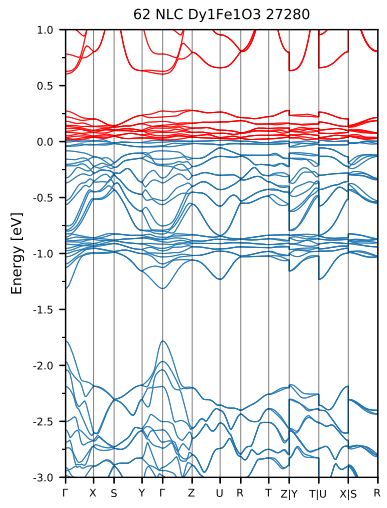

(et) 62.2.150

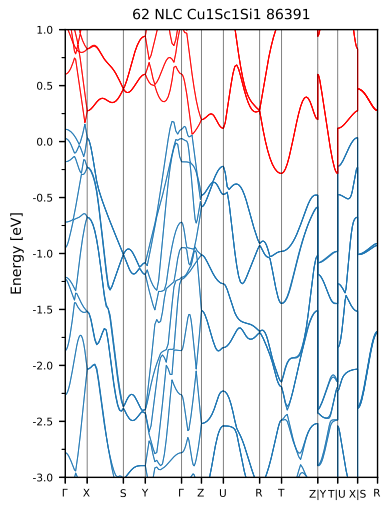

(eu) 62.2.151

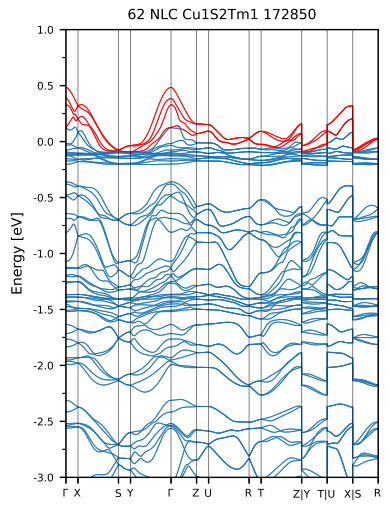

(ev) 62.2.152

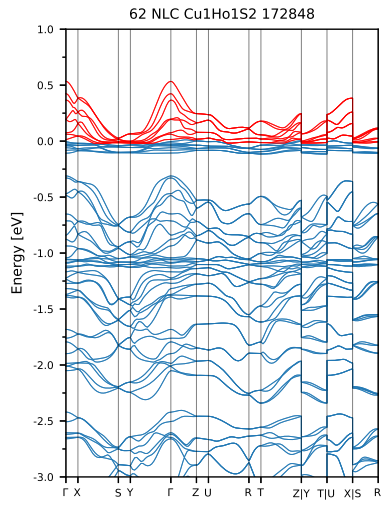

(ew) 62.2.153

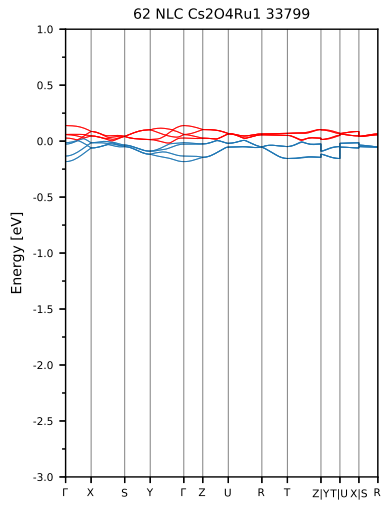

(ex) 62.2.154

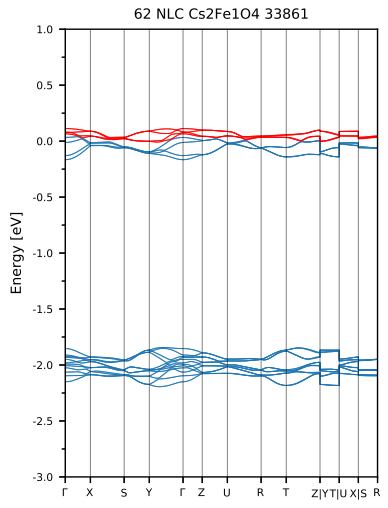

(ey) 62.2.155

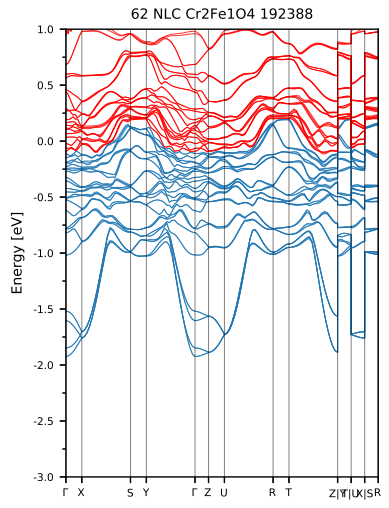

(ez) 62.2.156

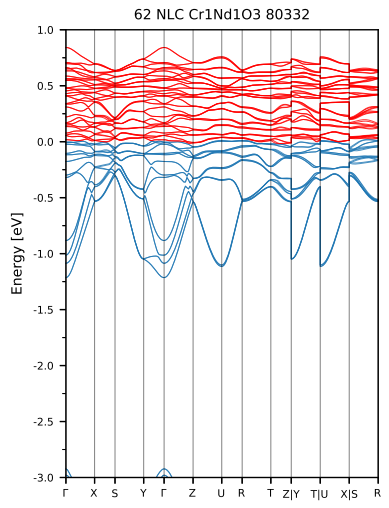

(fa) 62.2.157

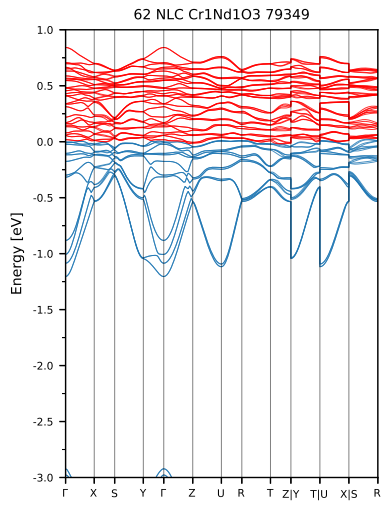

(fb) 62.2.158

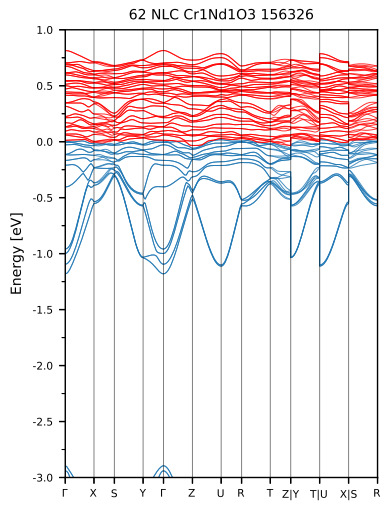

(fc) 62.2.159

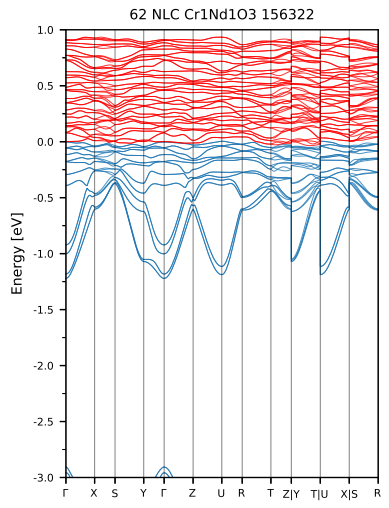

(fd) 62.2.160

Figure 118: Band structures for topological materials in SG 62, part 10/36, from 62.2.145 to 62.2.160. See the list of figures.

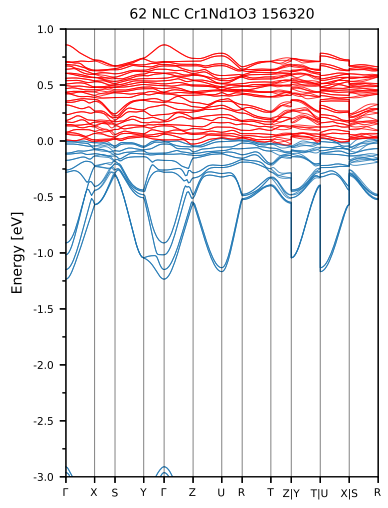

(fe) 62.2.161

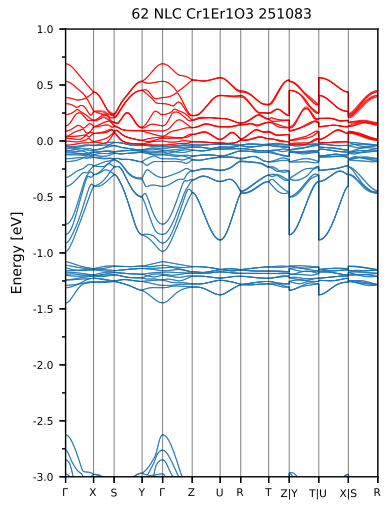

(ff) 62.2.162

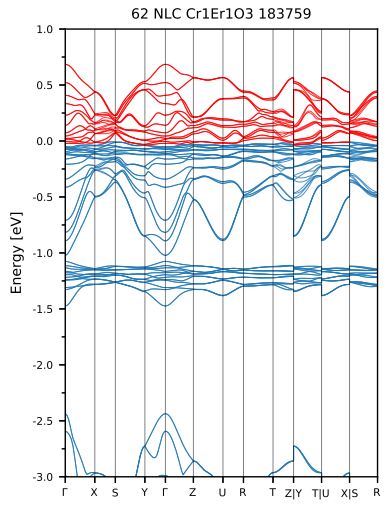

(fg) 62.2.163

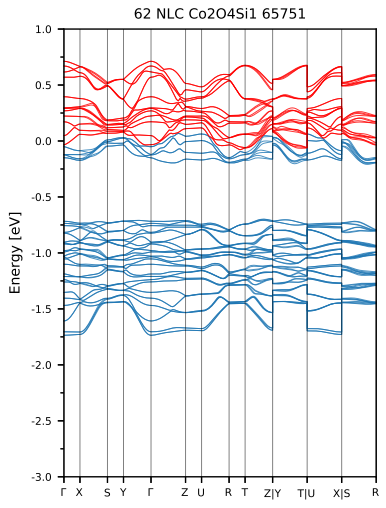

(fh) 62.2.164

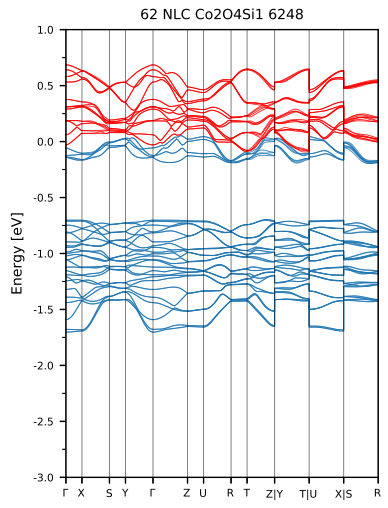

(fi) 62.2.165

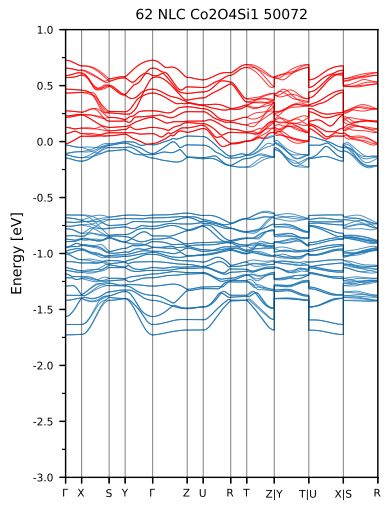

(fj) 62.2.166

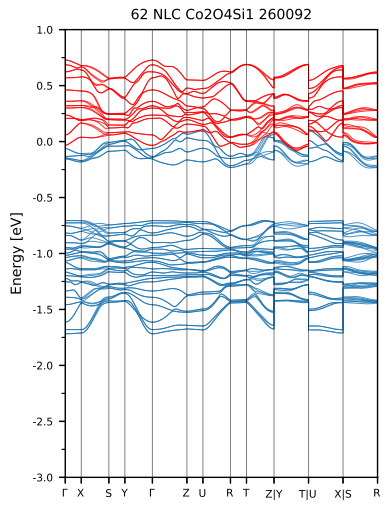

(fk) 62.2.167

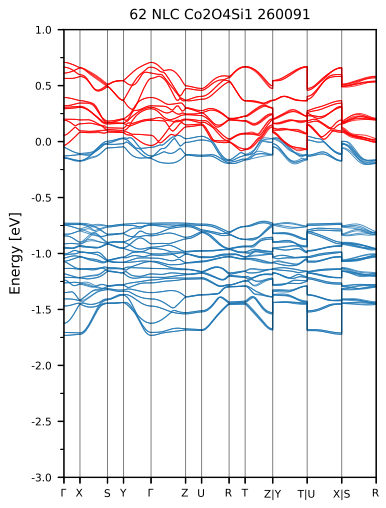

(fl) 62.2.168

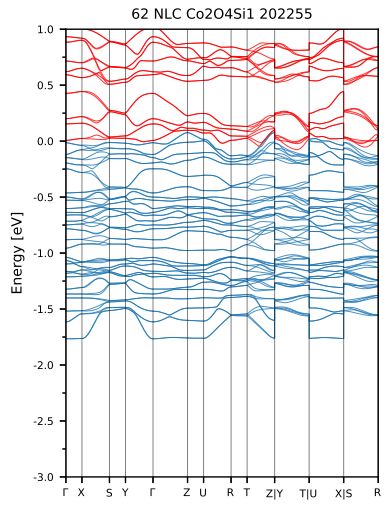

(fm) 62.2.169

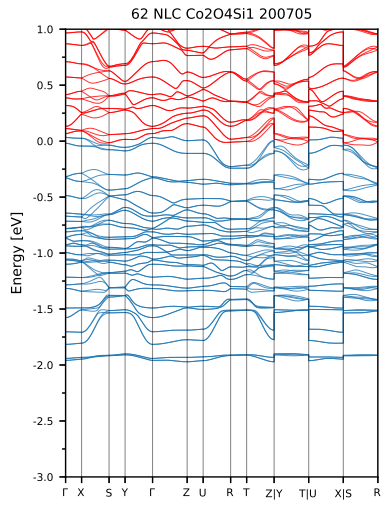

(fn) 62.2.170

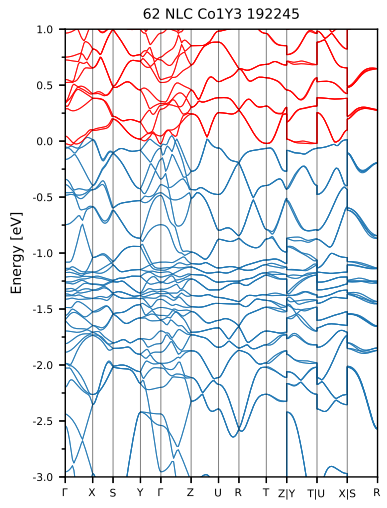

(fo) 62.2.171

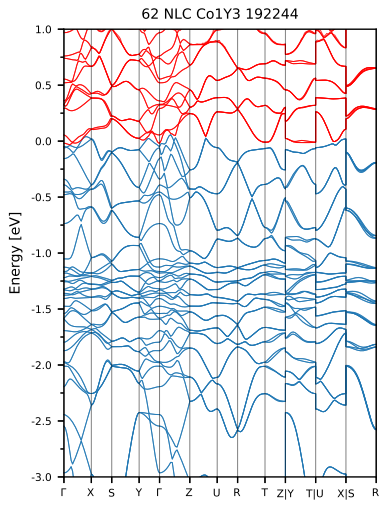

(fp) 62.2.172

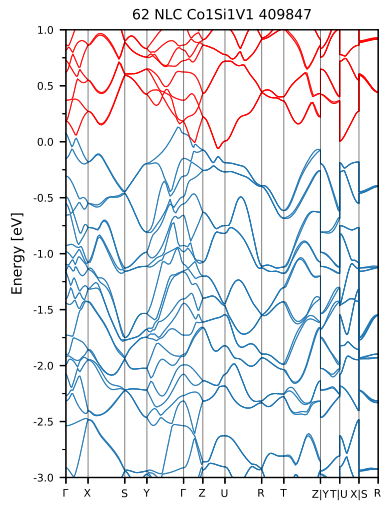

(fq) 62.2.173

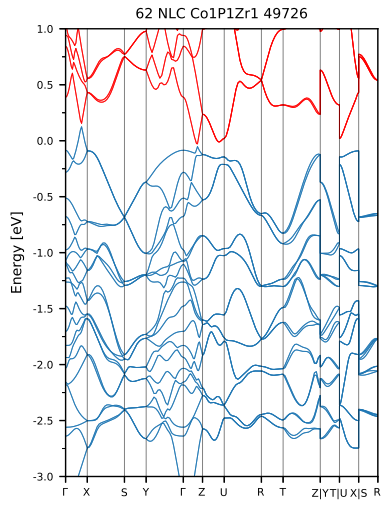

(fr) 62.2.174

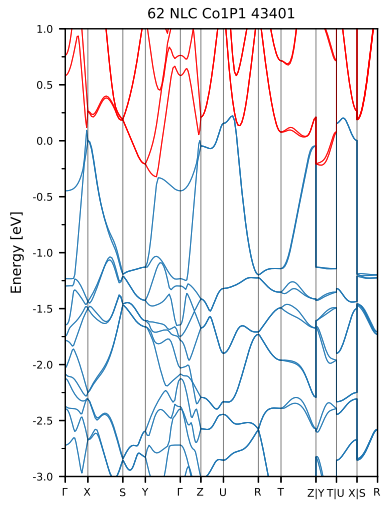

(fs) 62.2.175

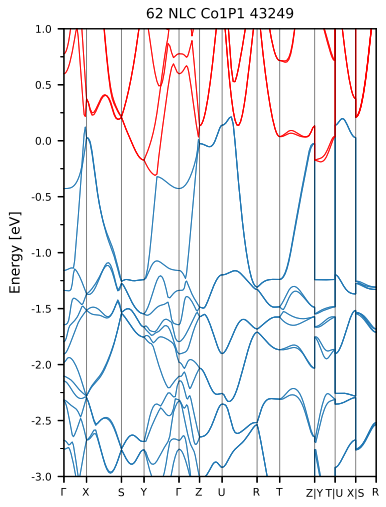

(ft) 62.2.176

Figure 119: Band structures for topological materials in SG 62, part 11/36, from 62.2.161 to 62.2.176. See the list of figures.

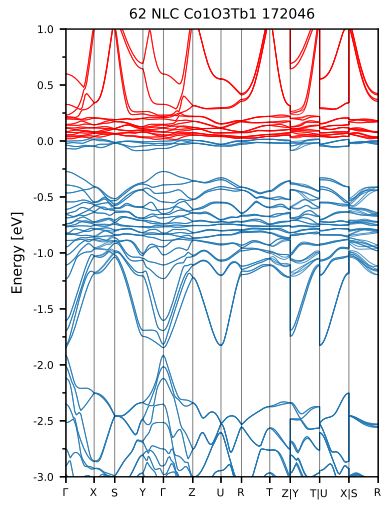

(fu) 62.2.177

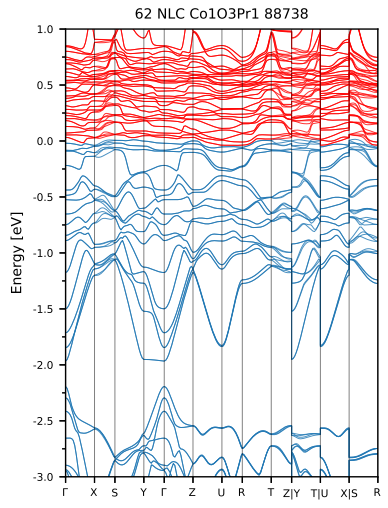

(fv) 62.2.178

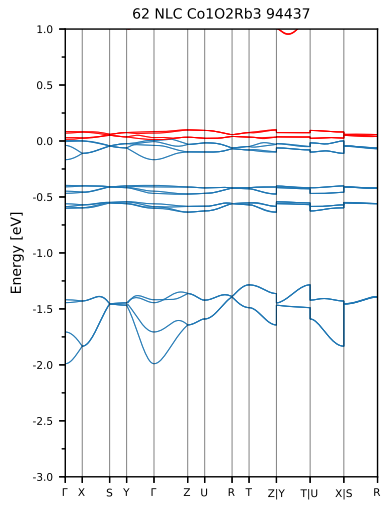

(fw) 62.2.179

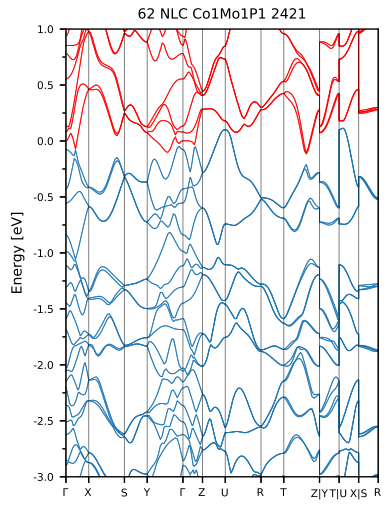

(fx) 62.2.180

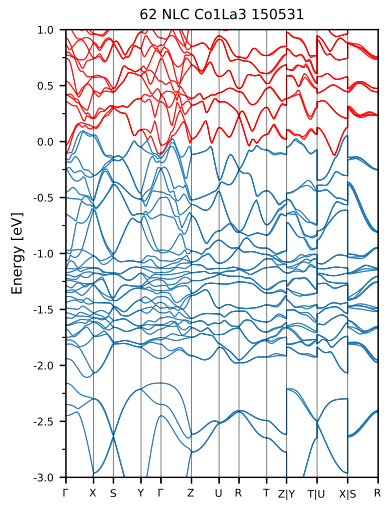

(fy) 62.2.181

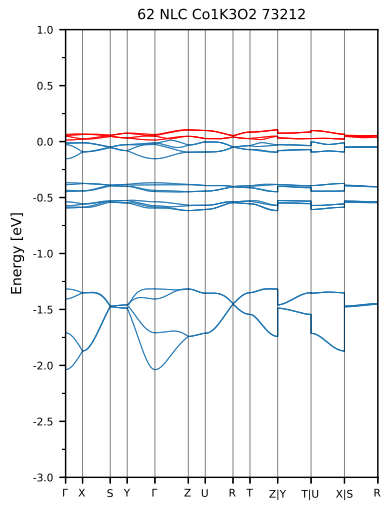

(fz) 62.2.182

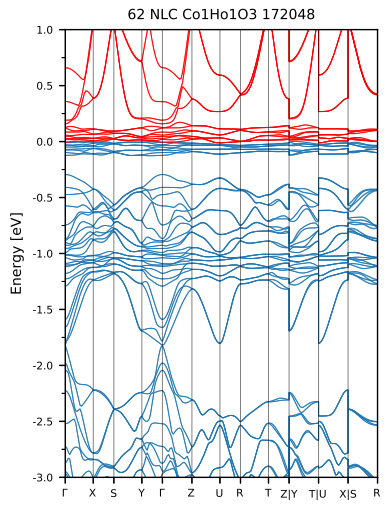

(ga) 62.2.183

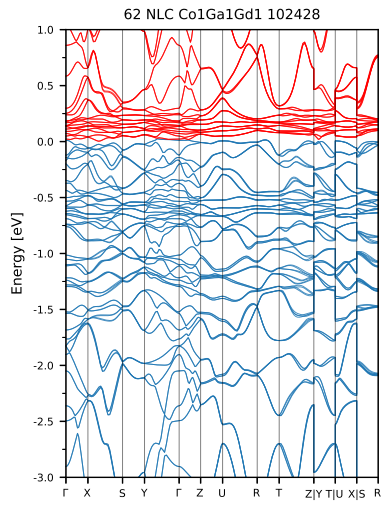

(gb) 62.2.184

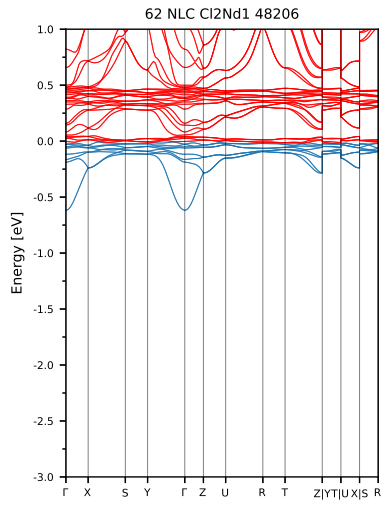

(gc) 62.2.185

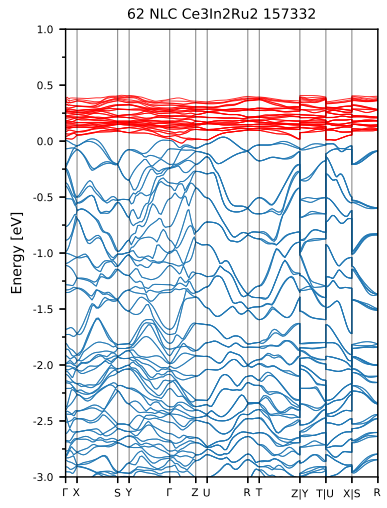

(gd) 62.2.186

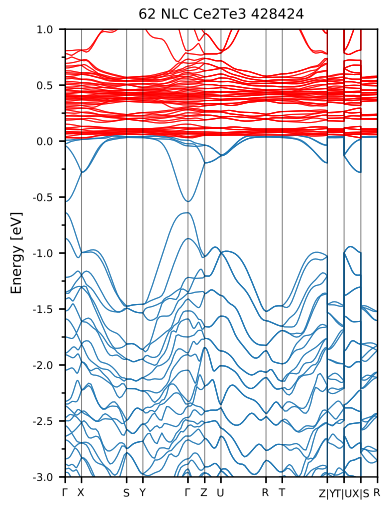

(ge) 62.2.187

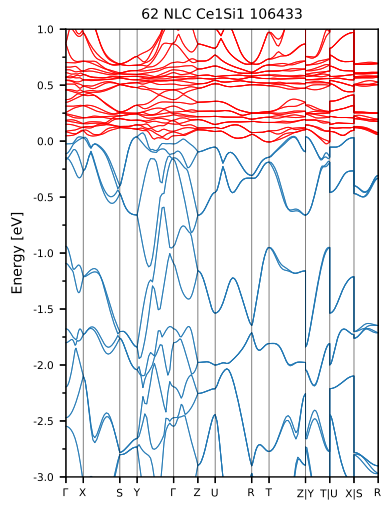

(gf) 62.2.188

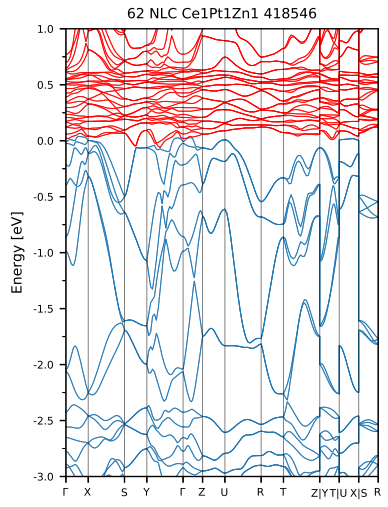

(gg) 62.2.189

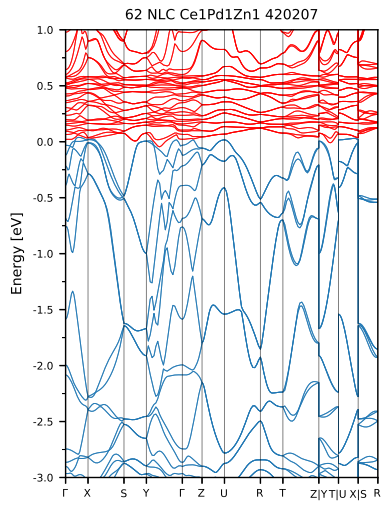

(gh) 62.2.190

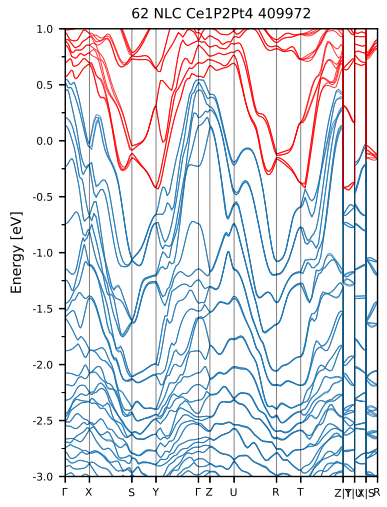

(gi) 62.2.191

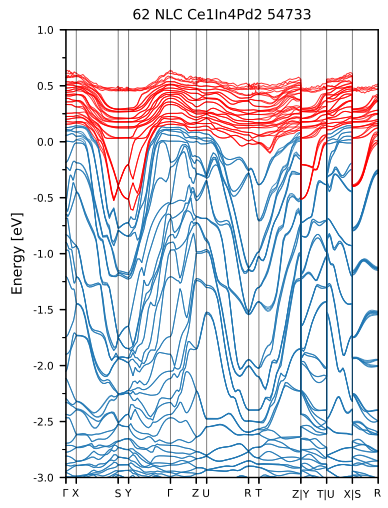

(gj) 62.2.192

Figure 120: Band structures for topological materials in SG 62, part 12/36, from 62.2.177 to 62.2.192. See the list of figures.

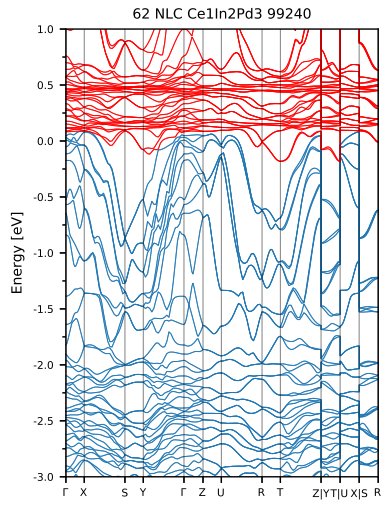

(gk) 62.2.193

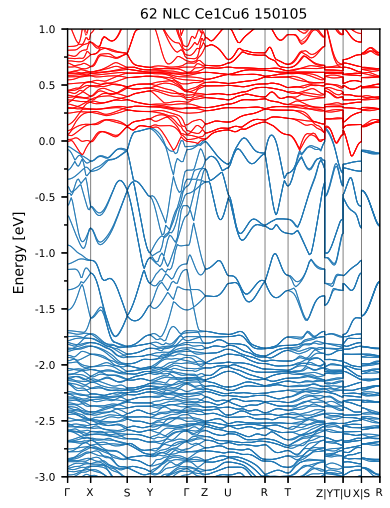

(gl) 62.2.194

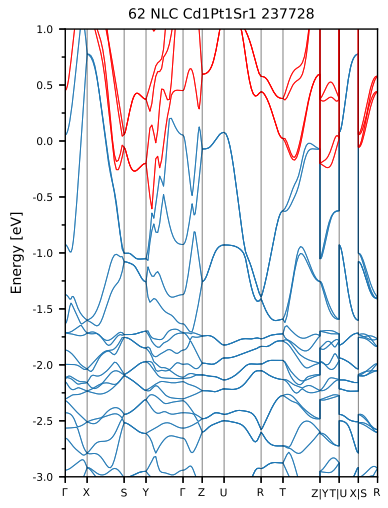

(gm) 62.2.195

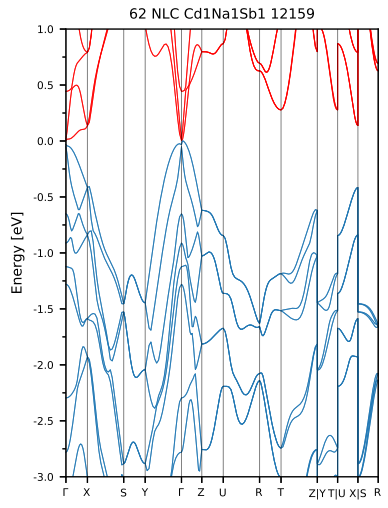

(gn) 62.2.196

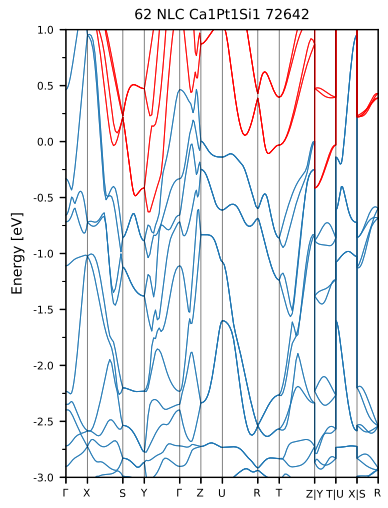

(go) 62.2.197

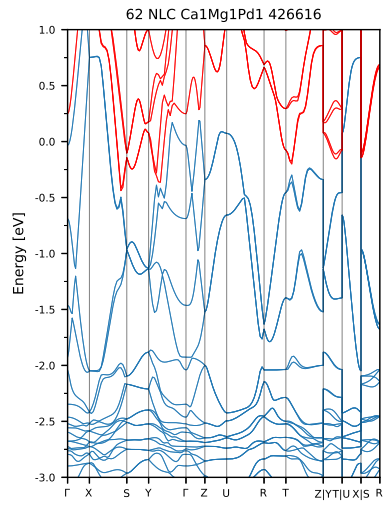

(gp) 62.2.198

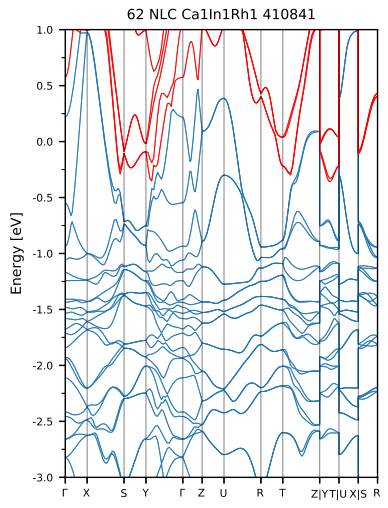

(gq) 62.2.199

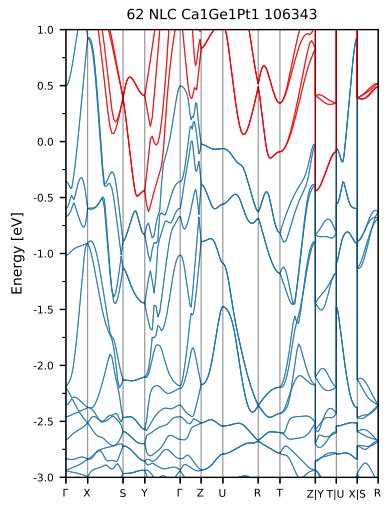

(gr) 62.2.200

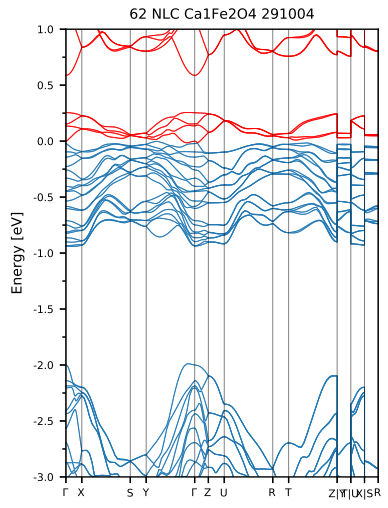

(gs) 62.2.201

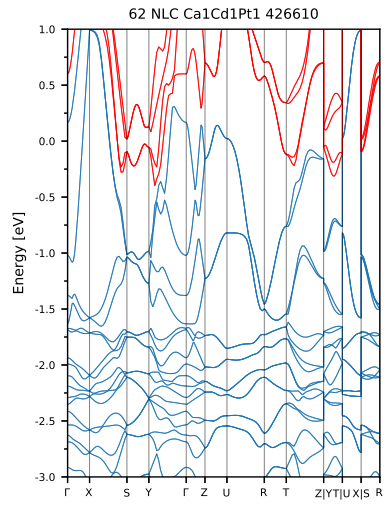

(gt) 62.2.202

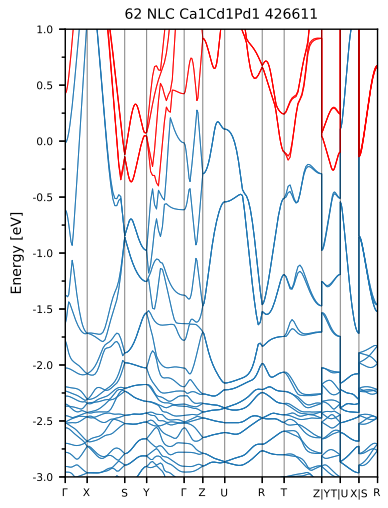

(gu) 62.2.203

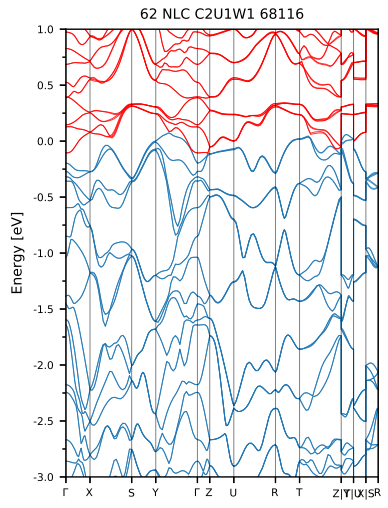

(gv) 62.2.204

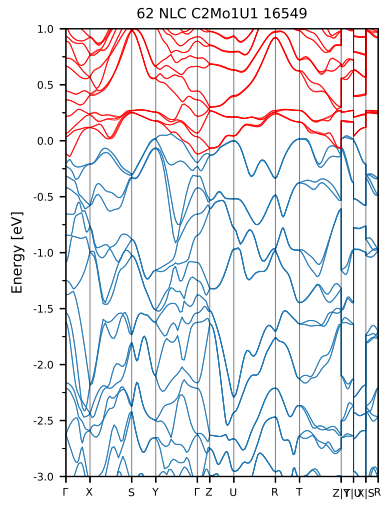

(gw) 62.2.205

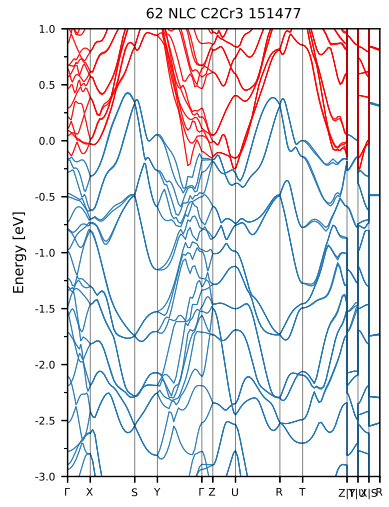

(gx) 62.2.206

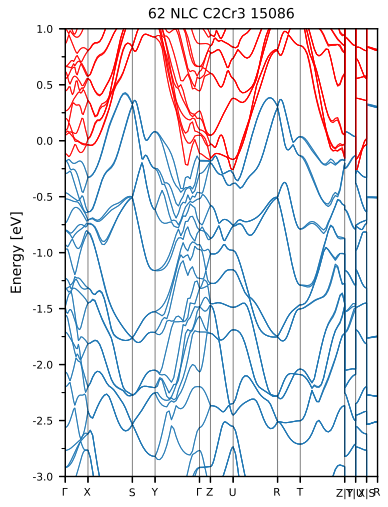

(gy) 62.2.207

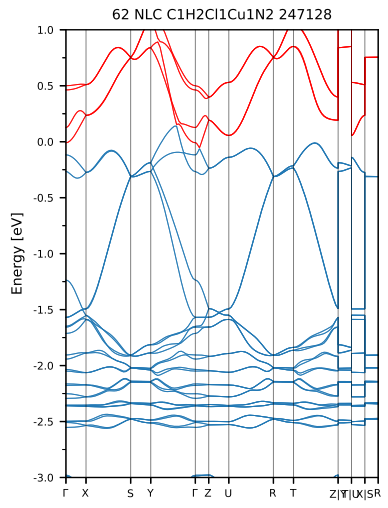

(gz) 62.2.208

Figure 121: Band structures for topological materials in SG 62, part 13/36, from 62.2.193 to 62.2.208. See the list of figures.

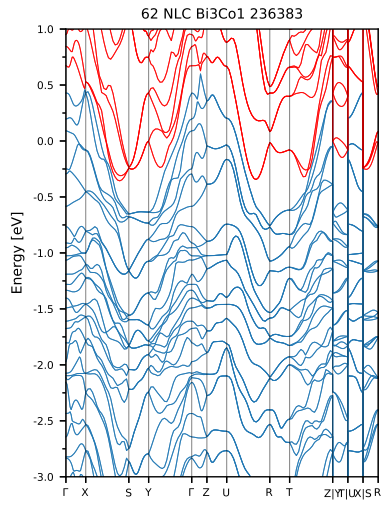

(ha) 62.2.209

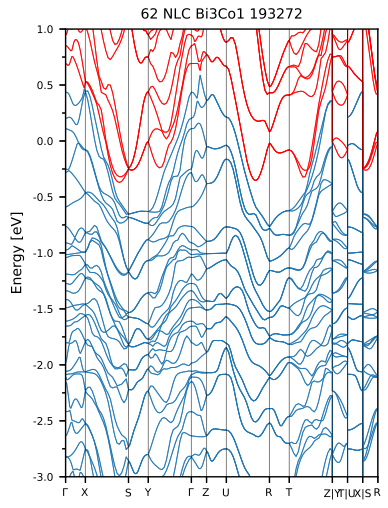

(hb) 62.2.210

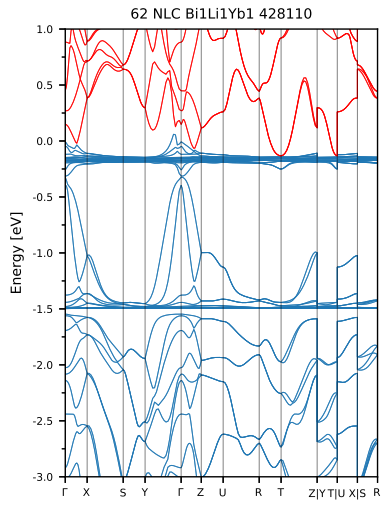

(hc) 62.2.211

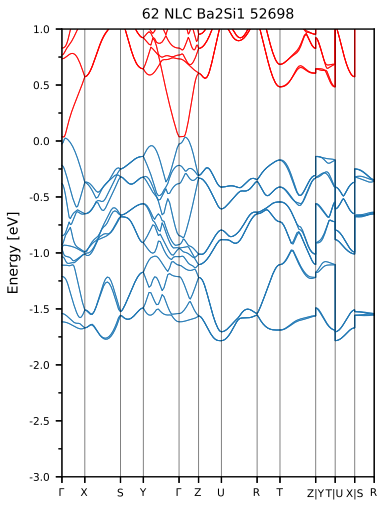

(hd) 62.2.212

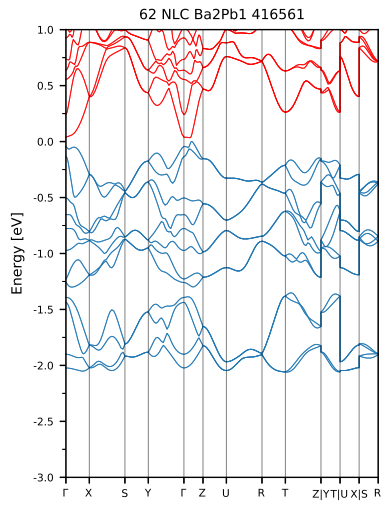

(he) 62.2.213

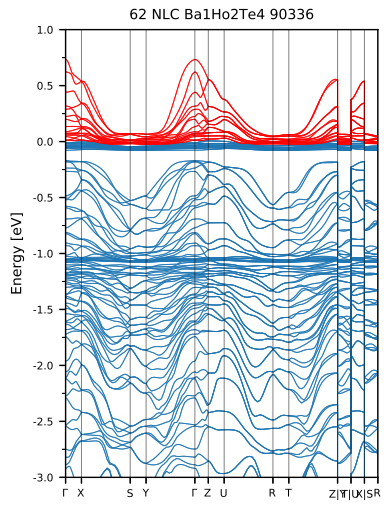

(hf) 62.2.214

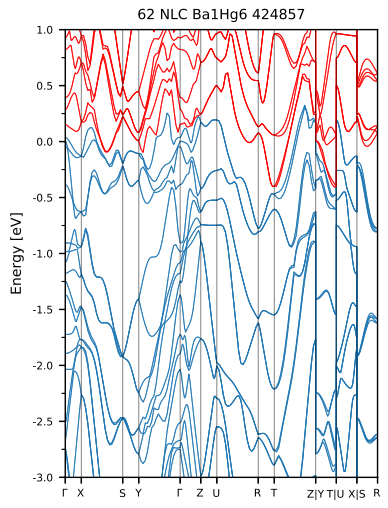

(hg) 62.2.215

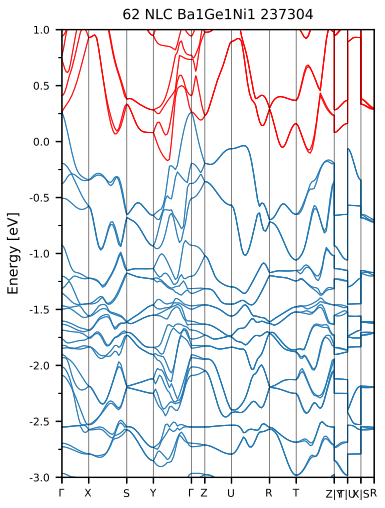

(hh) 62.2.216

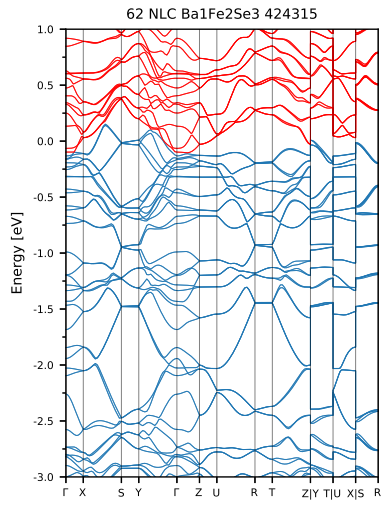

(hi) 62.2.217

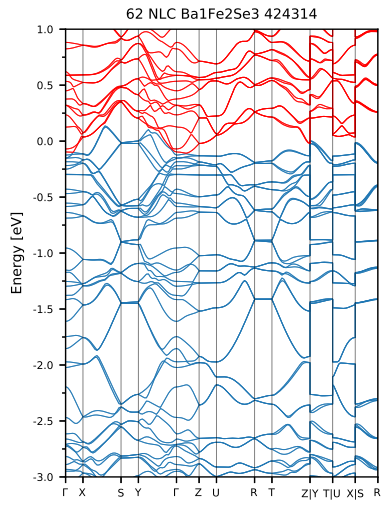

(hj) 62.2.218

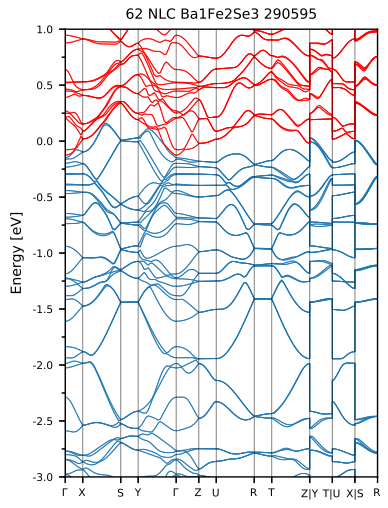

(hk) 62.2.219

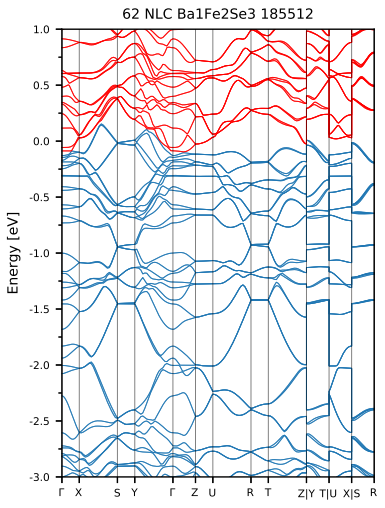

(hl) 62.2.220

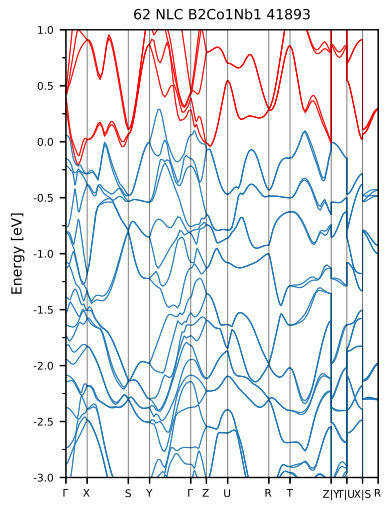

(hm) 62.2.221

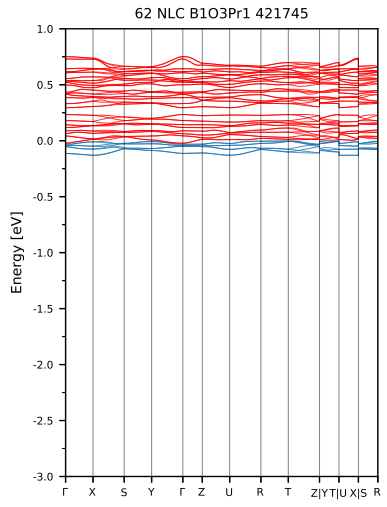

(hn) 62.2.222

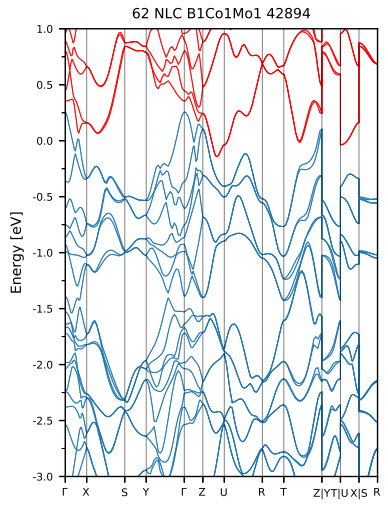

(ho) 62.2.223

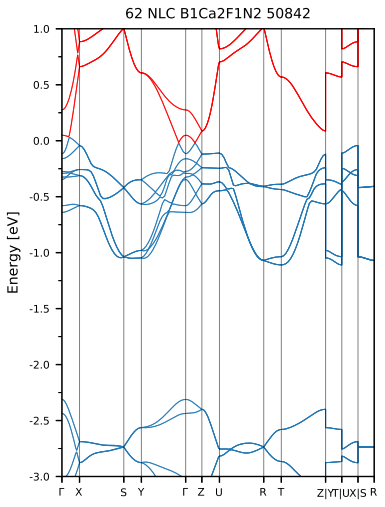

(hp) 62.2.224

Figure 122: Band structures for topological materials in SG 62, part 14/36, from 62.2.209 to 62.2.224. See the list of figures.

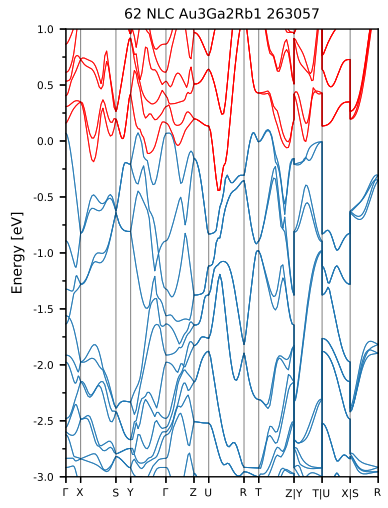

(hq) 62.2.225

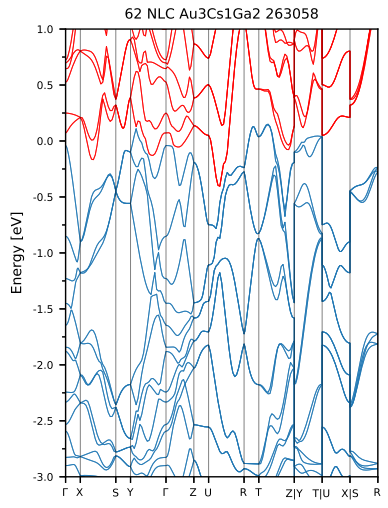

(hr) 62.2.226

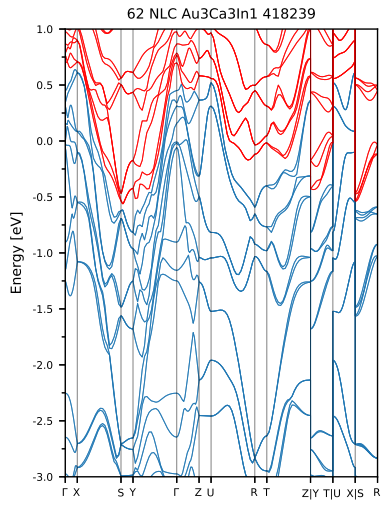

(hs) 62.2.227

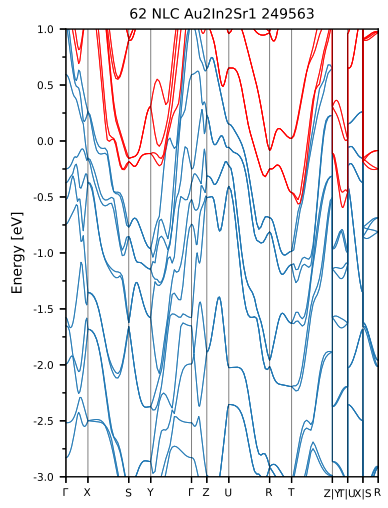

(ht) 62.2.228

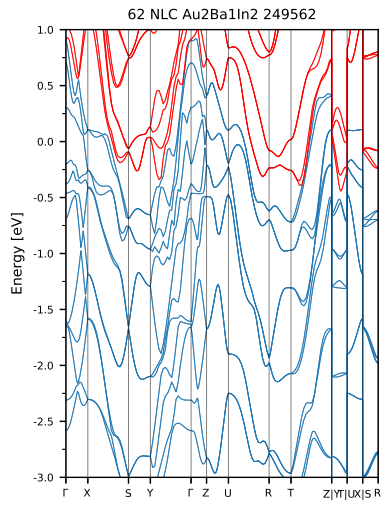

(hu) 62.2.229

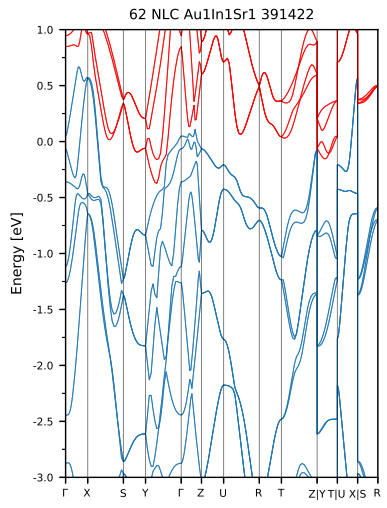

(hv) 62.2.230

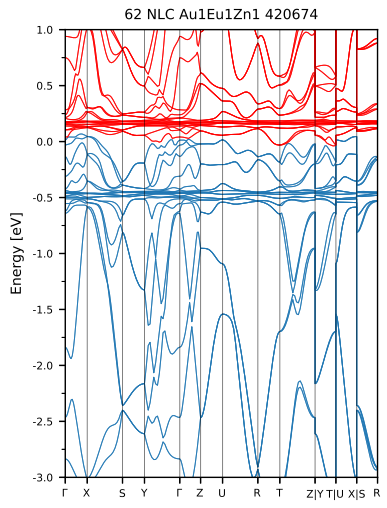

(hw) 62.2.231

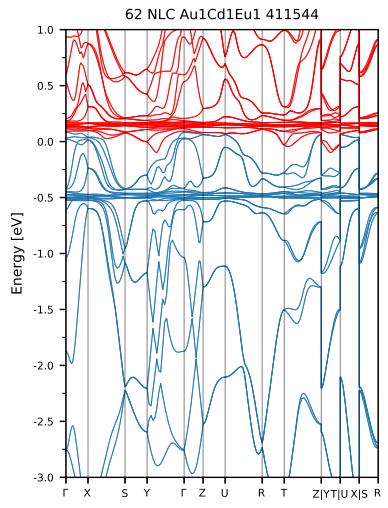

(hx) 62.2.232

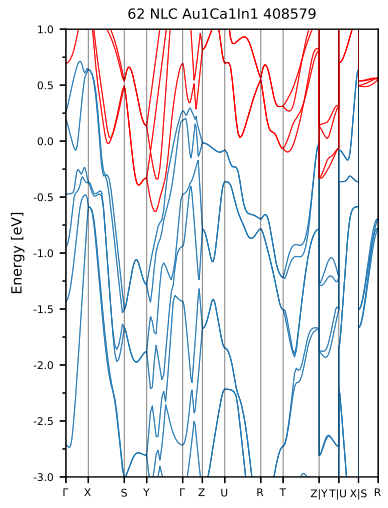

(hy) 62.2.233

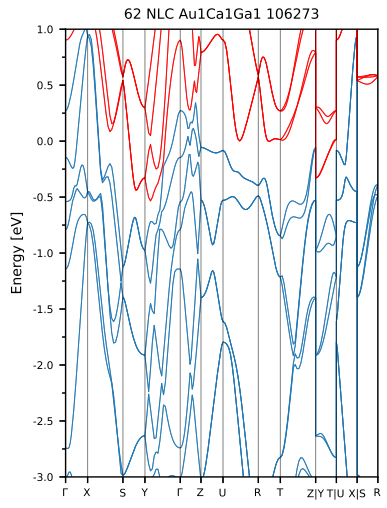

(hz) 62.2.234

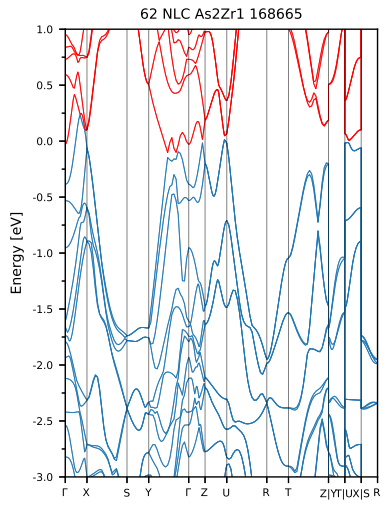

(ia) 62.2.235

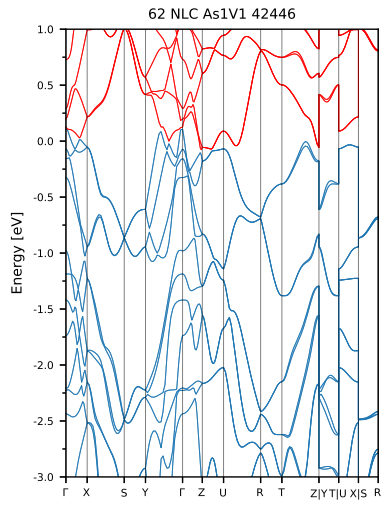

(ib) 62.2.236

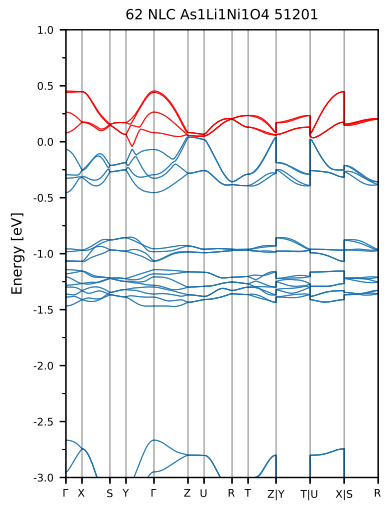

(ic) 62.2.237

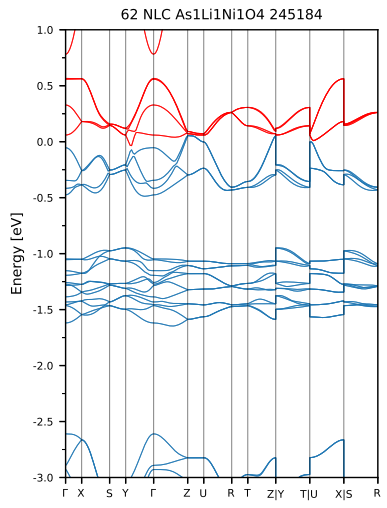

(id) 62.2.238

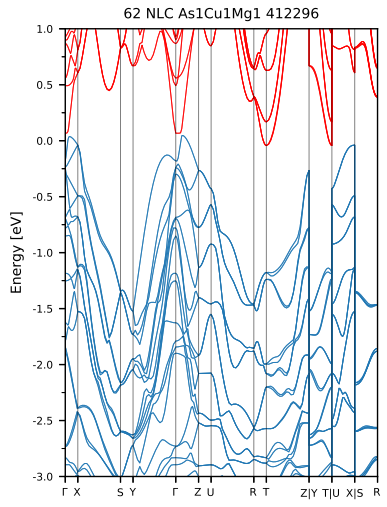

(ie) 62.2.239

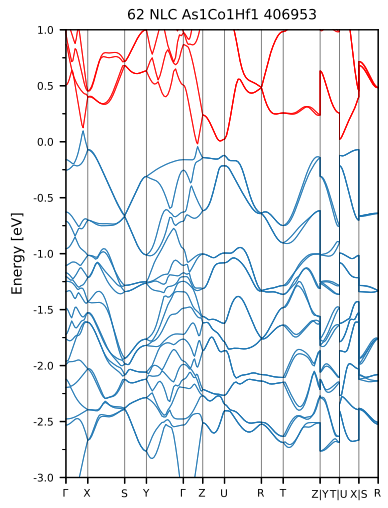

(if) 62.2.240

Figure 123: Band structures for topological materials in SG 62, part 15/36, from 62.2.225 to 62.2.240. See the list of figures.

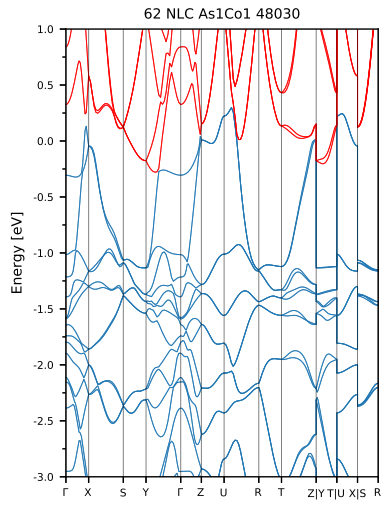

(ig) 62.2.241

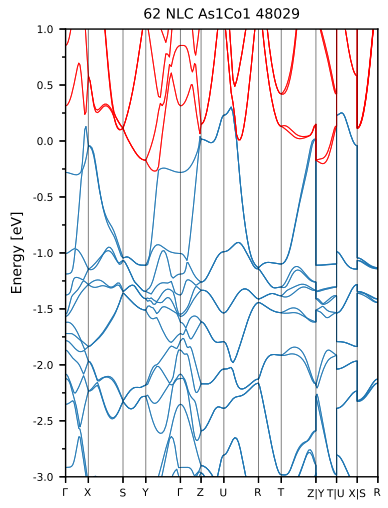

(ih) 62.2.242

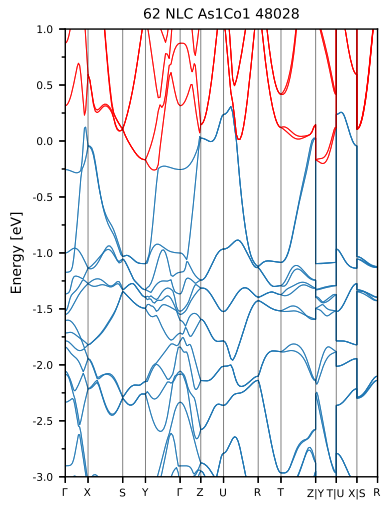

(ii) 62.2.243

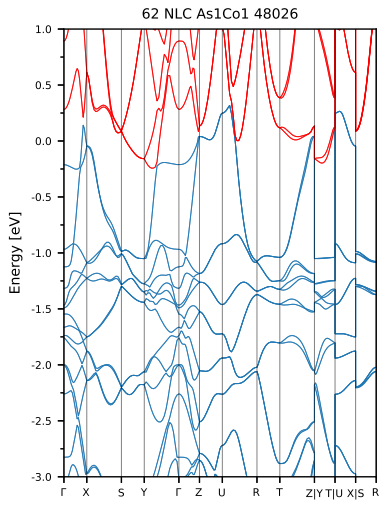

(ij) 62.2.244

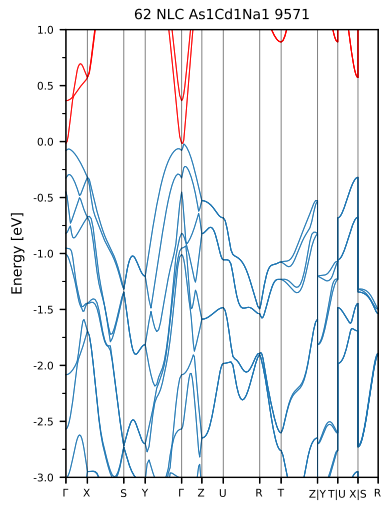

(ik) 62.2.245

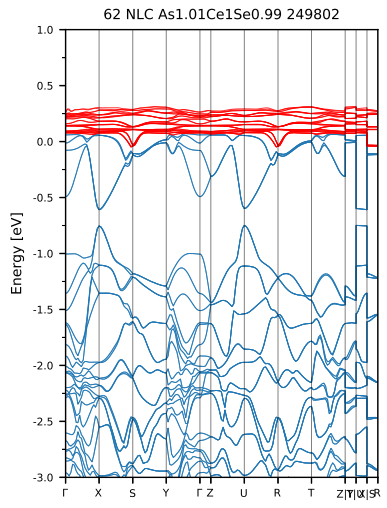

(il) 62.2.246

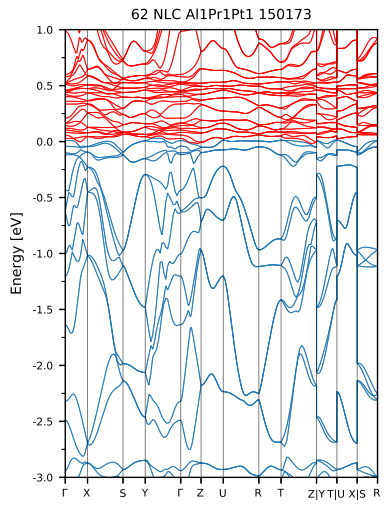

(im) 62.2.247

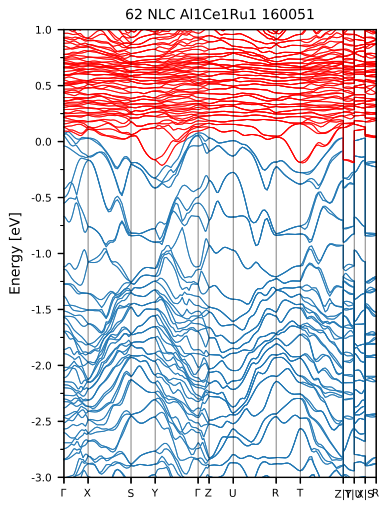

(in) 62.2.248

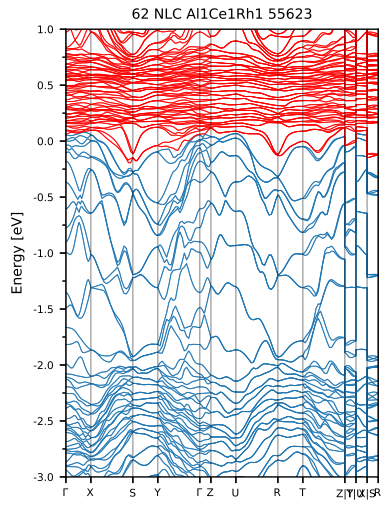

(io) 62.2.249

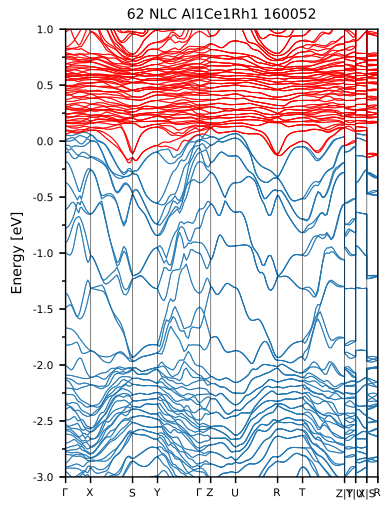

(ip) 62.2.250

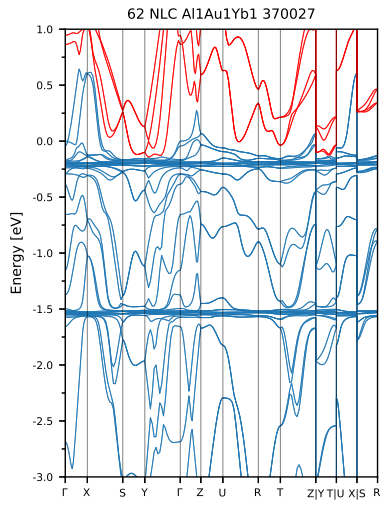

(iq) 62.2.251

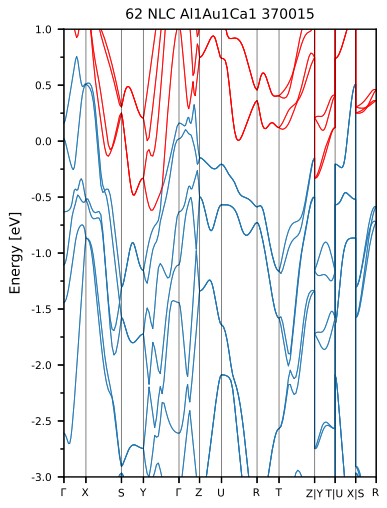

(ir) 62.2.252

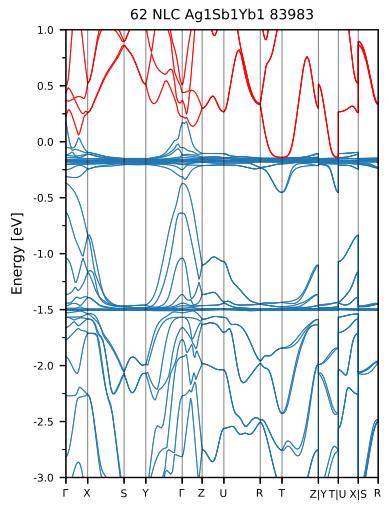

(is) 62.2.253

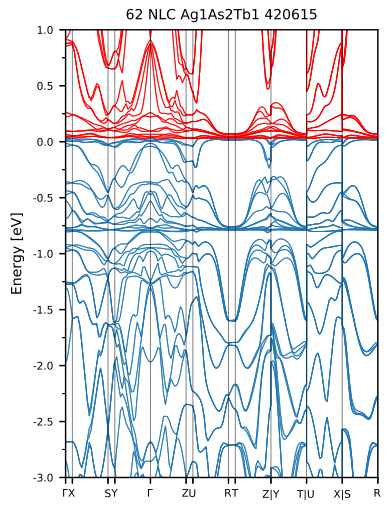

(it) 62.2.254

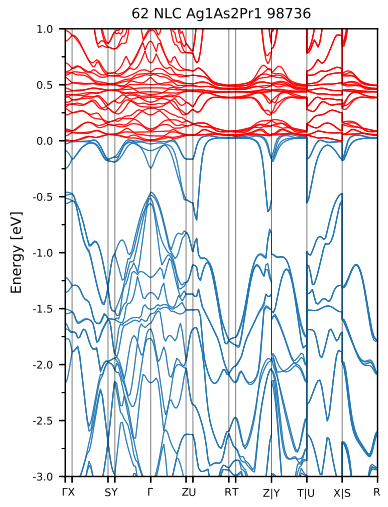

(iu) 62.2.255

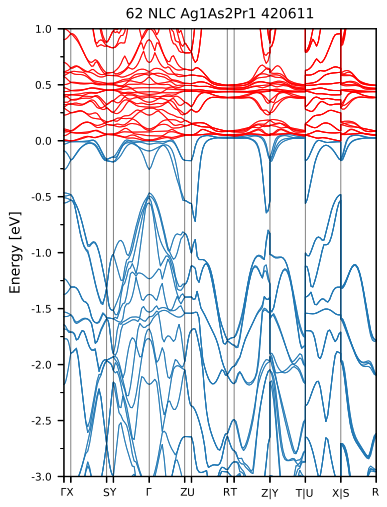

(iv) 62.2.256

Figure 124: Band structures for topological materials in SG 62, part 16/36, from 62.2.241 to 62.2.256. See the list of figures.

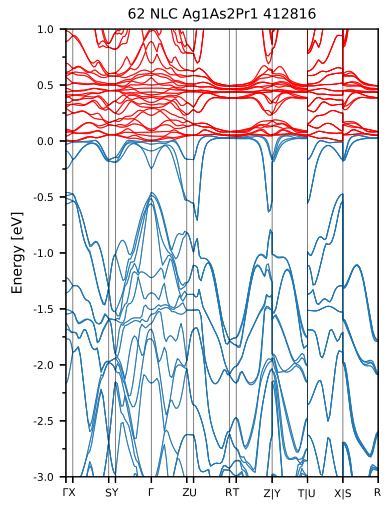

(iw) 62.2.257

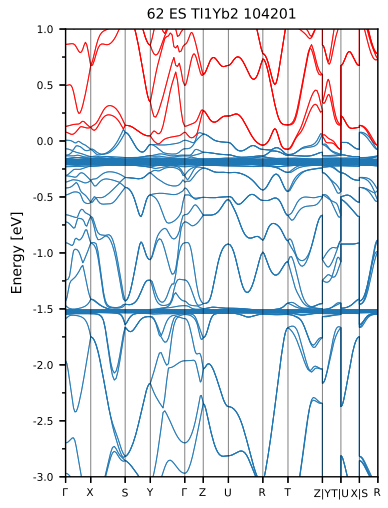

(ix) 62.3.1

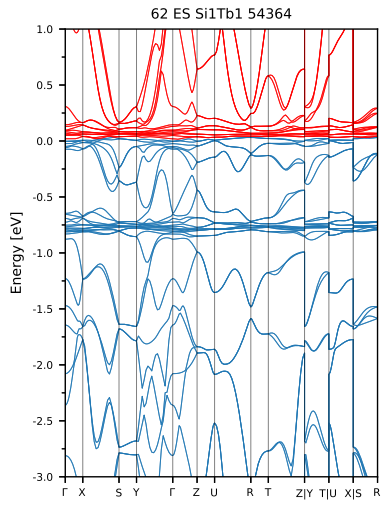

(iy) 62.3.2

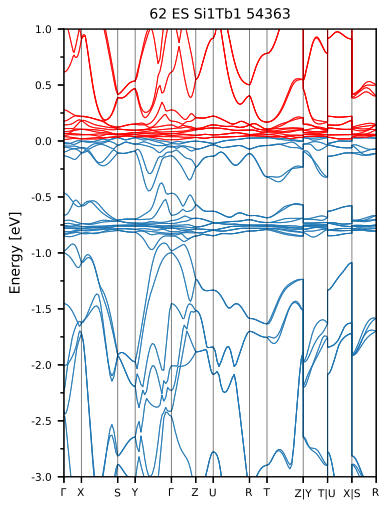

(iz) 62.3.3

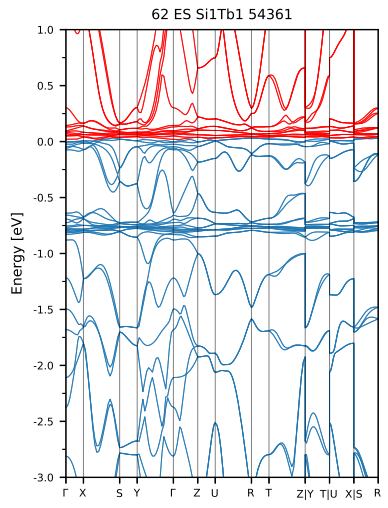

(ja) 62.3.4

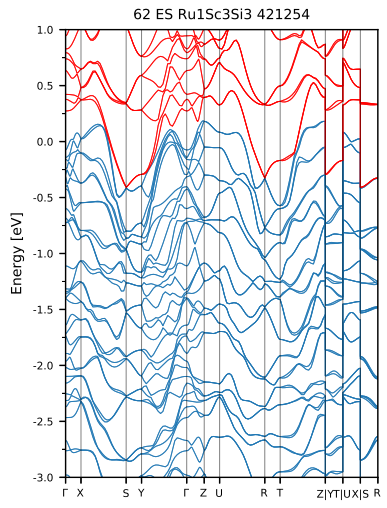

(jb) 62.3.5

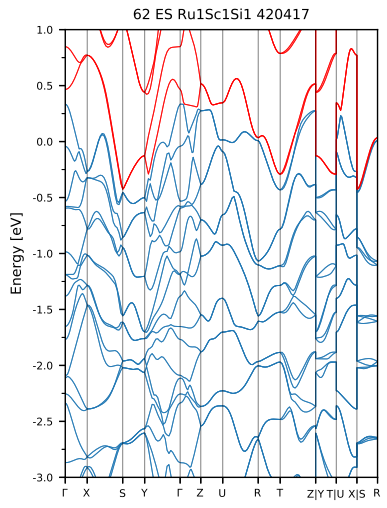

(jc) 62.3.6

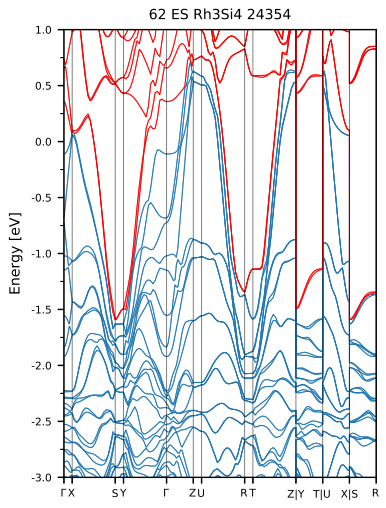

(jd) 62.3.7

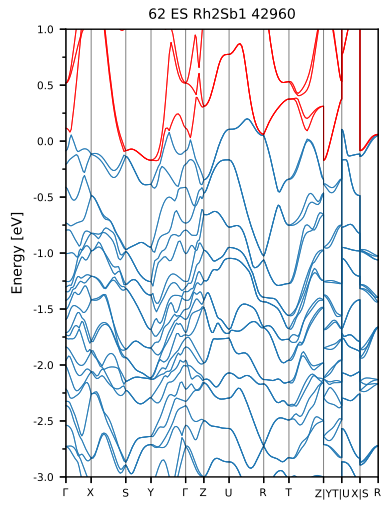

(je) 62.3.8

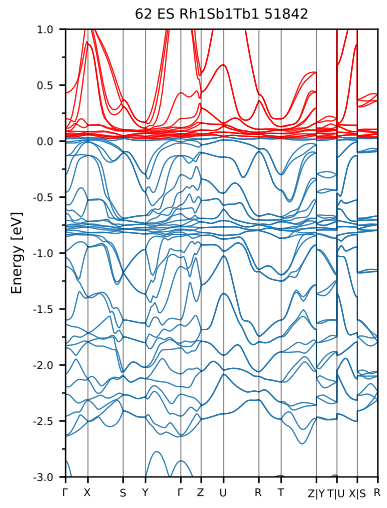

(jf) 62.3.9

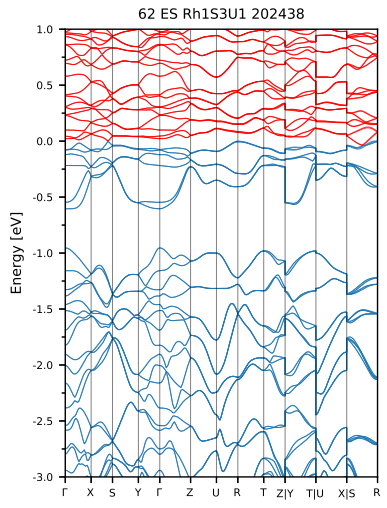

(jg) 62.3.10

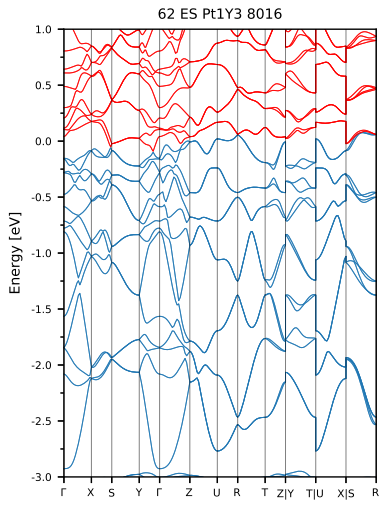

(jh) 62.3.11

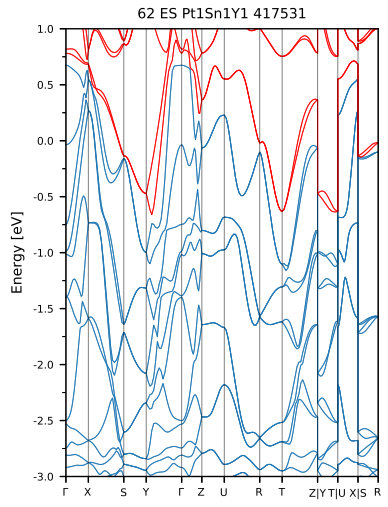

(ji) 62.3.12

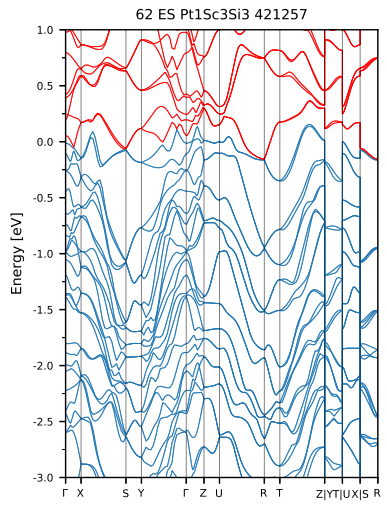

(jj) 62.3.13

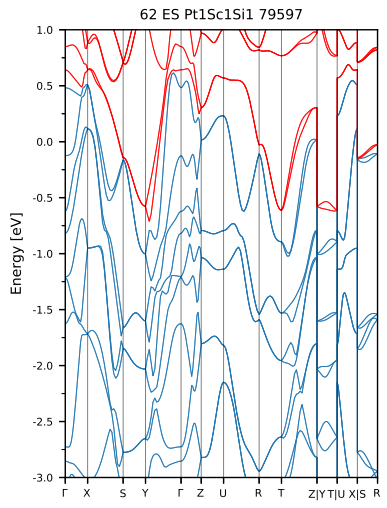

(jk) 62.3.14

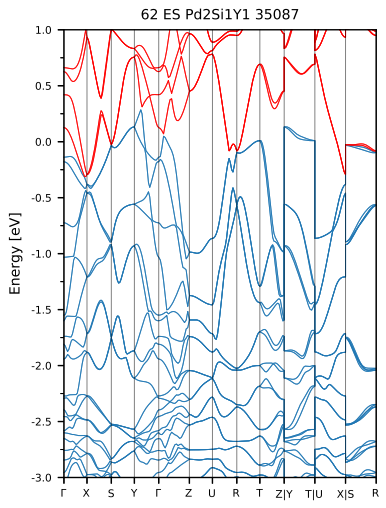

(jl) 62.3.15

Figure 125: Band structures for topological materials in SG 62, part 17/36, from 62.2.257 to 62.3.15. See the list of figures.

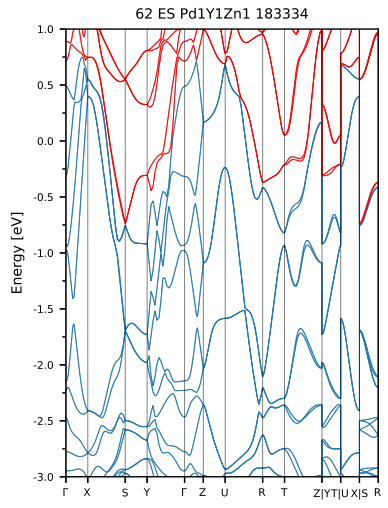

(jm) 62.3.16

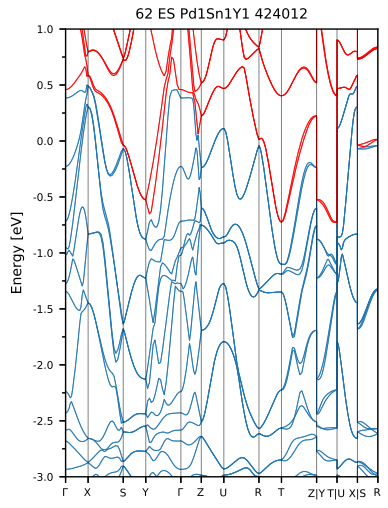

(jn) 62.3.17

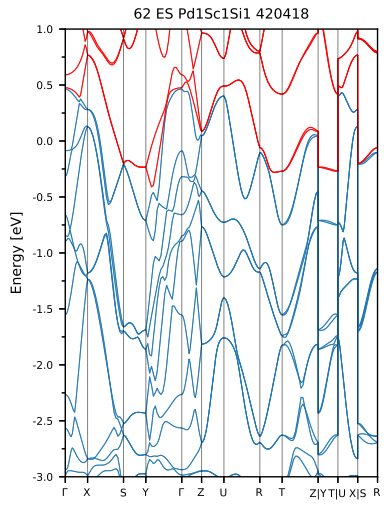

(jo) 62.3.18

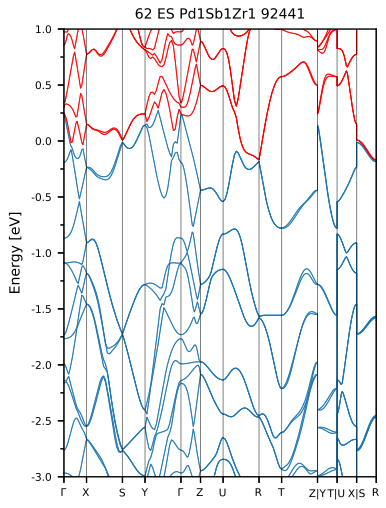

(jp) 62.3.19

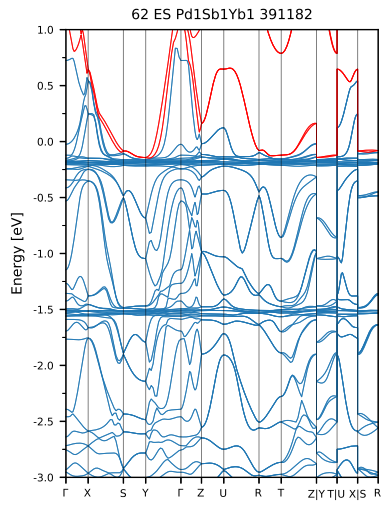

(jq) 62.3.20

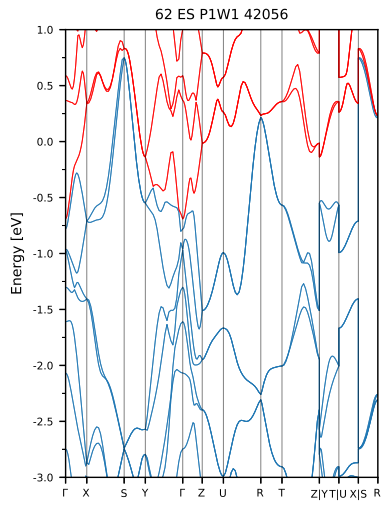

(jr) 62.3.21

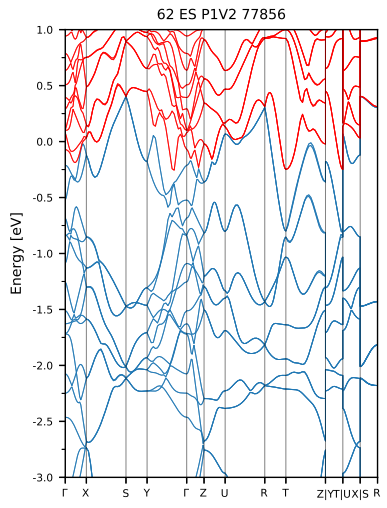

(js) 62.3.22

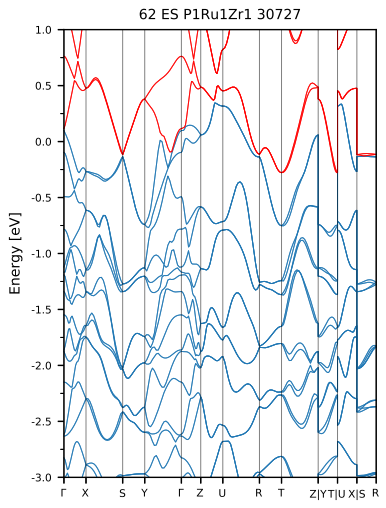

(jt) 62.3.23

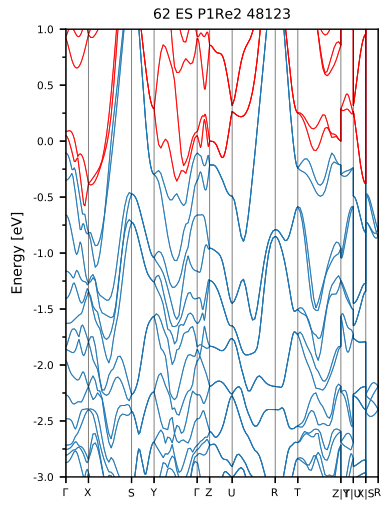

(ju) 62.3.24

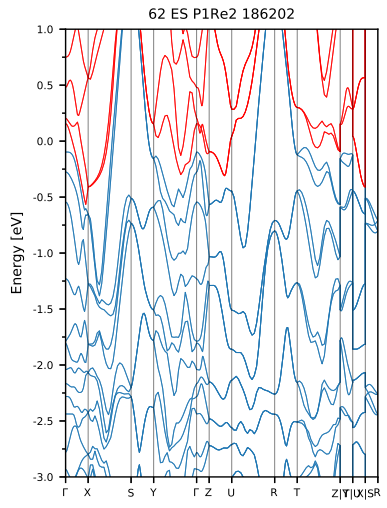

(jv) 62.3.25

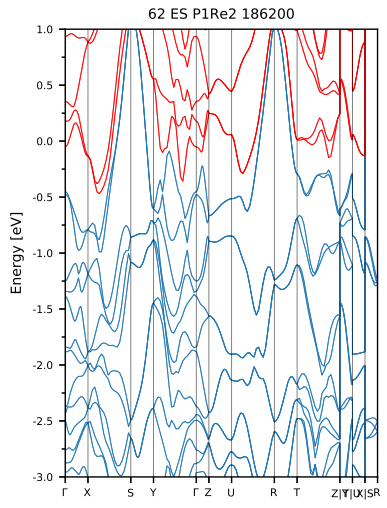

(jw) 62.3.26

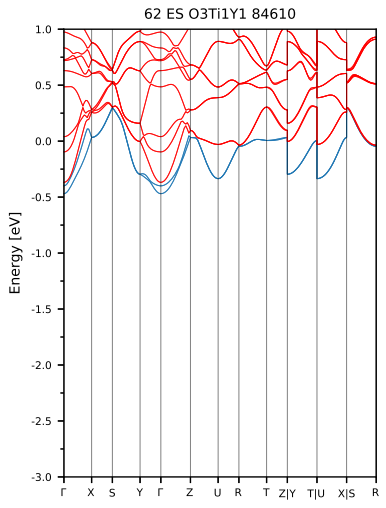

(jx) 62.3.27

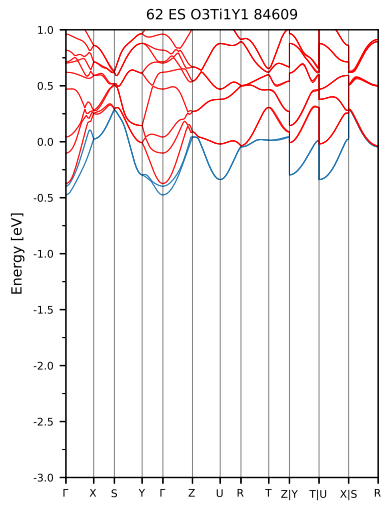

(jy) 62.3.28

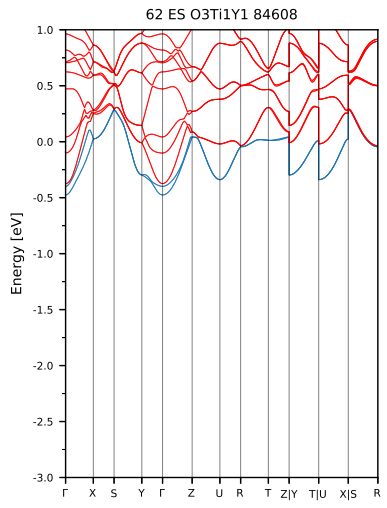

(jz) 62.3.29

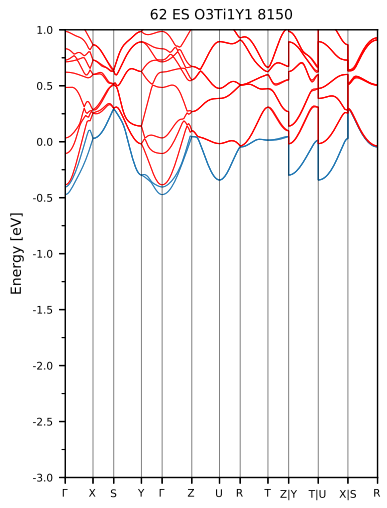

(ka) 62.3.30

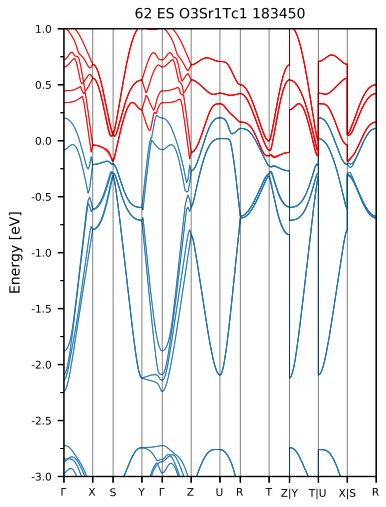

(kb) 62.3.31

Figure 126: Band structures for topological materials in SG 62, part 18/36, from 62.3.16 to 62.3.31. See the list of figures.

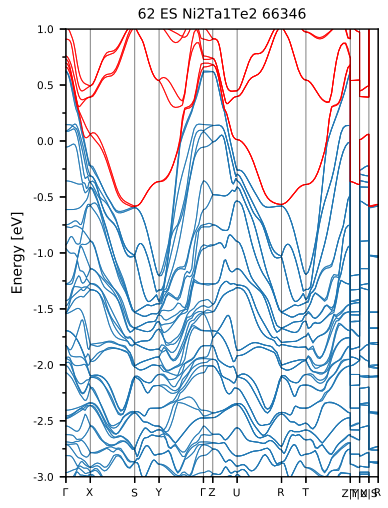

(kc) 62.3.32

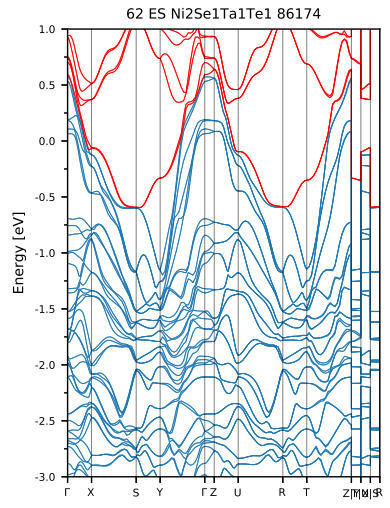

(kd) 62.3.33

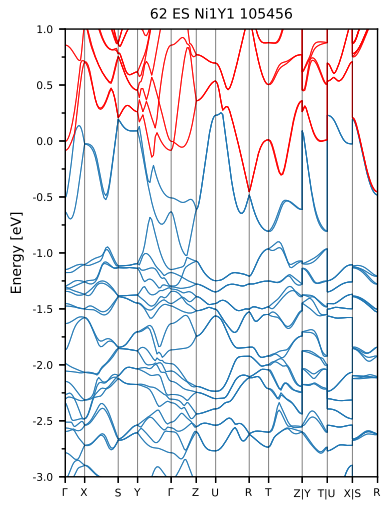

(ke) 62.3.34

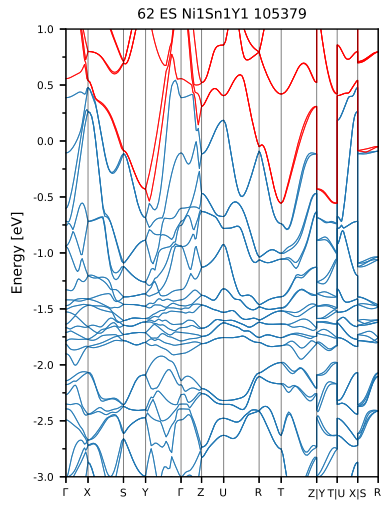

(kf) 62.3.35

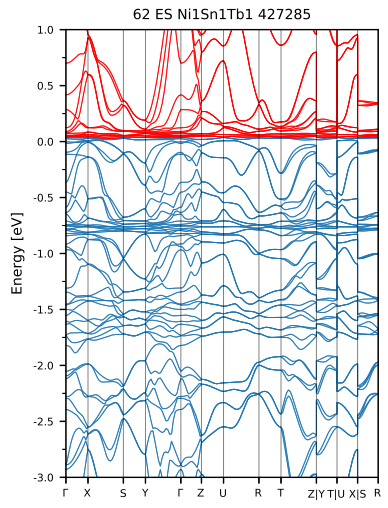

(kg) 62.3.36

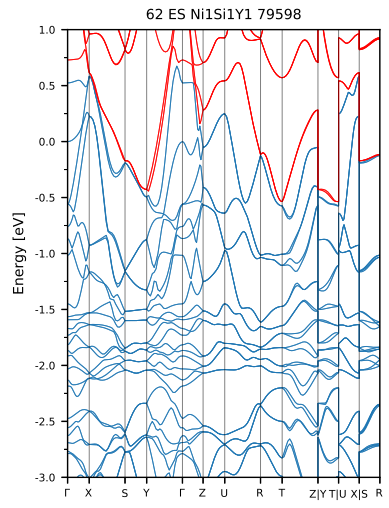

(kh) 62.3.37

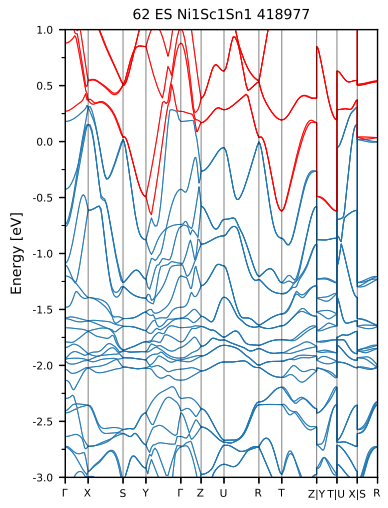

(ki) 62.3.38

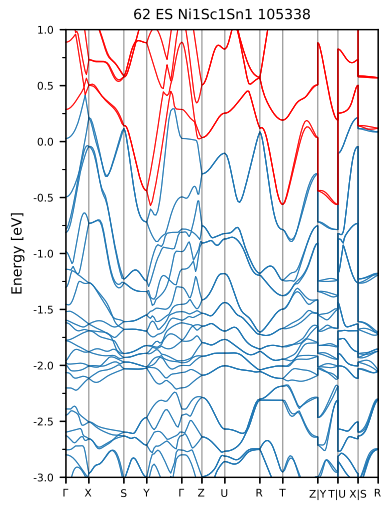

(kj) 62.3.39

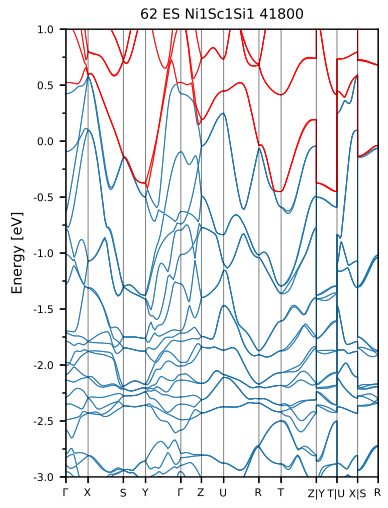

(kk) 62.3.40

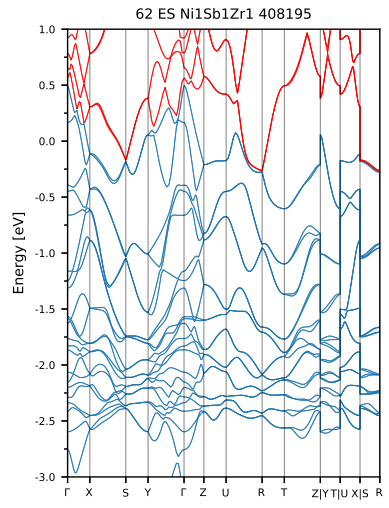

(kl) 62.3.41

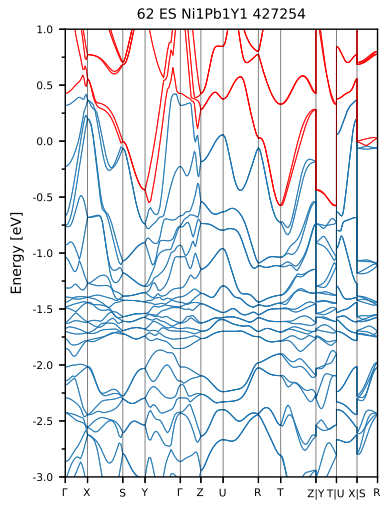

(km) 62.3.42

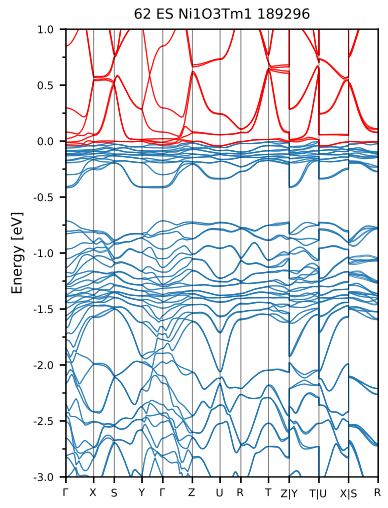

(kn) 62.3.43

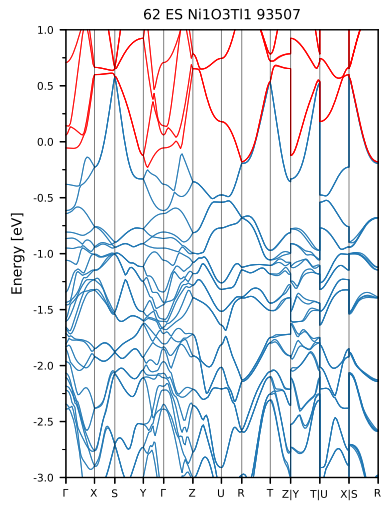

(ko) 62.3.44

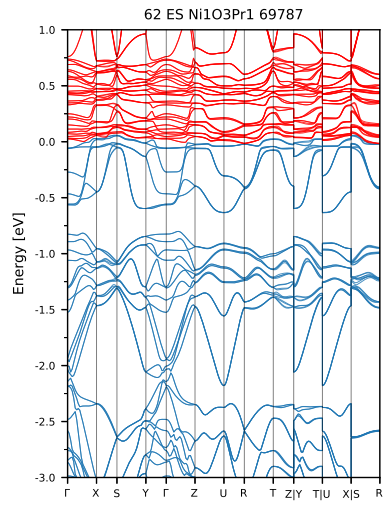

(kp) 62.3.45

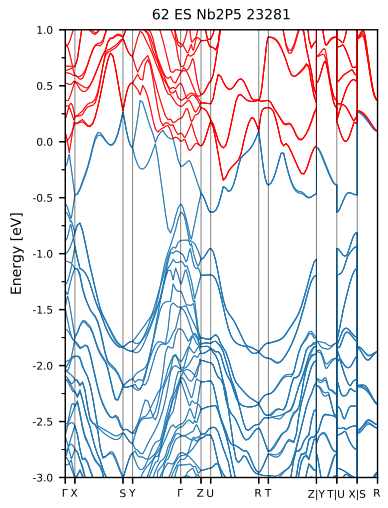

(kq) 62.3.46

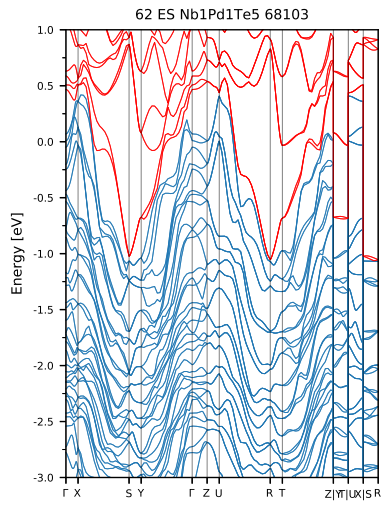

(kr) 62.3.47

Figure 127: Band structures for topological materials in SG 62, part 19/36, from 62.3.32 to 62.3.47. See the list of figures.

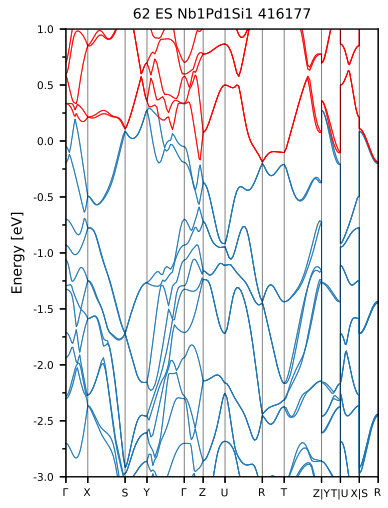

(ks) 62.3.48

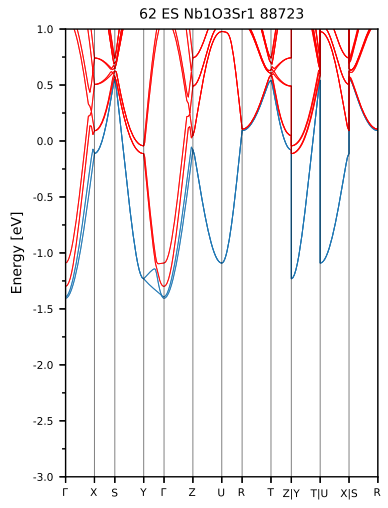

(kt) 62.3.49

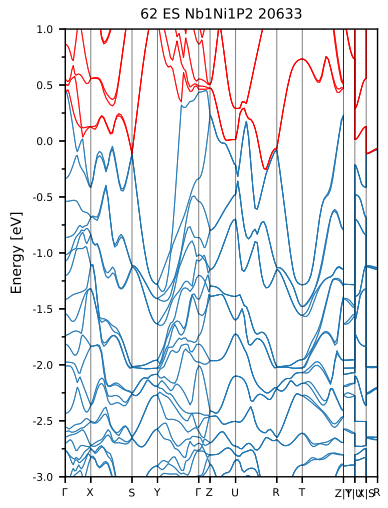

(ku) 62.3.50

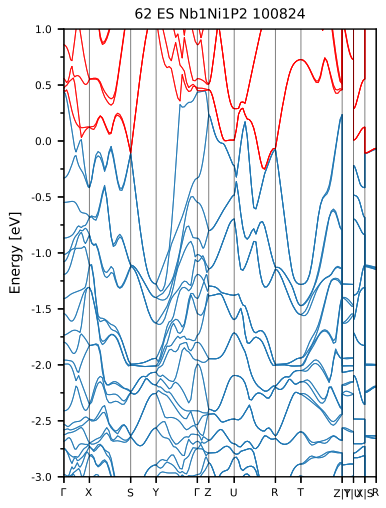

(kv) 62.3.51

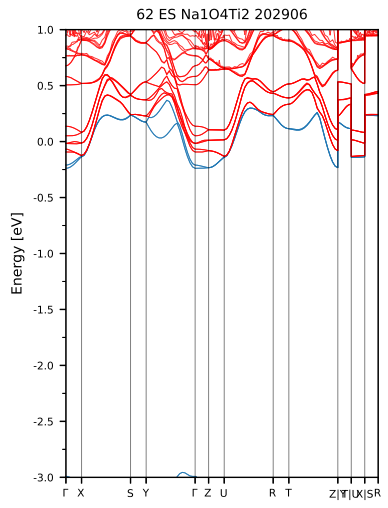

(kw) 62.3.52

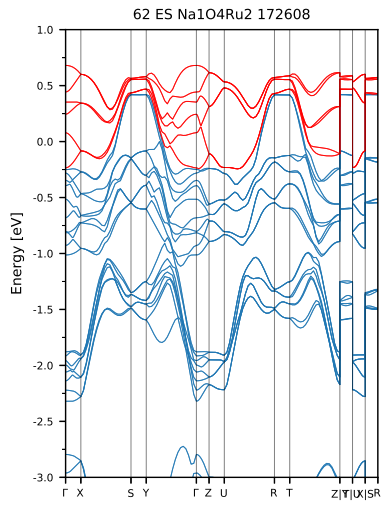

(kx) 62.3.53

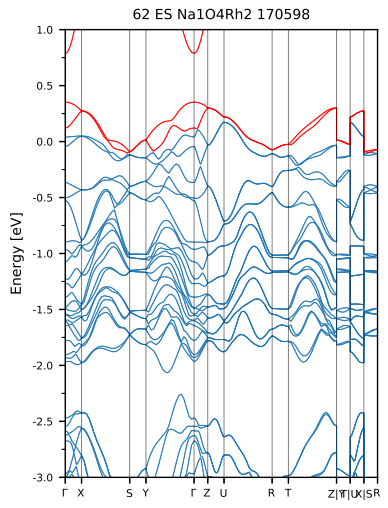

(ky) 62.3.54

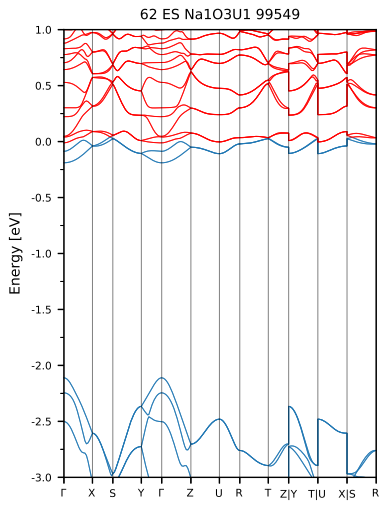

(kz) 62.3.55

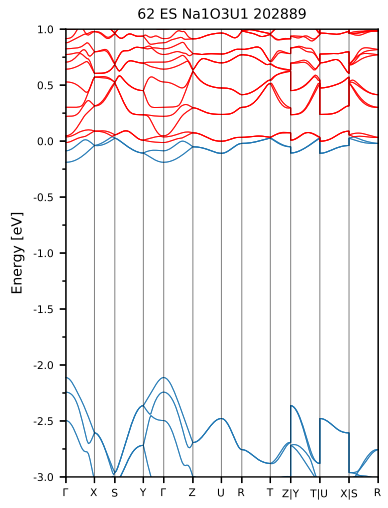

(la) 62.3.56

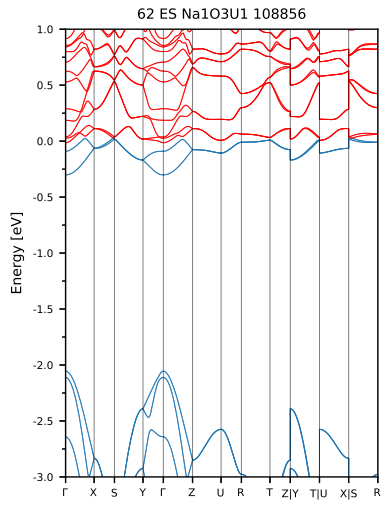

(lb) 62.3.57

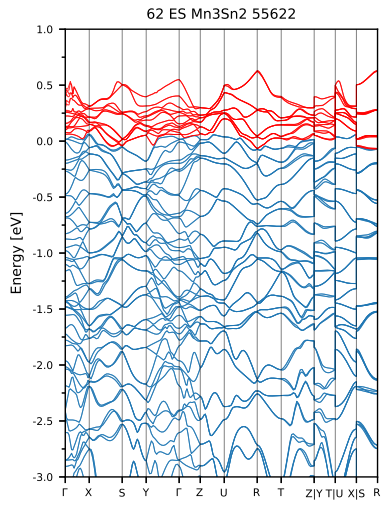

(lc) 62.3.58

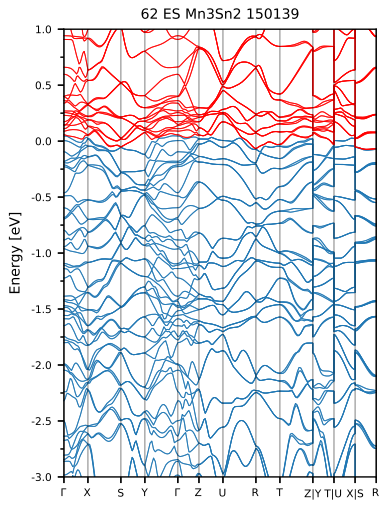

(ld) 62.3.59

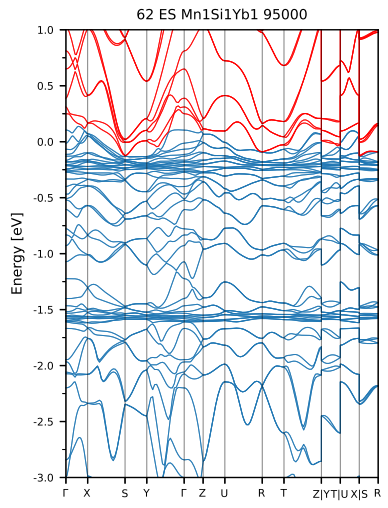

(le) 62.3.60

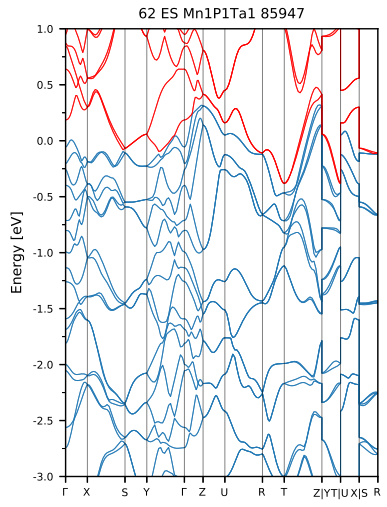

(lf) 62.3.61

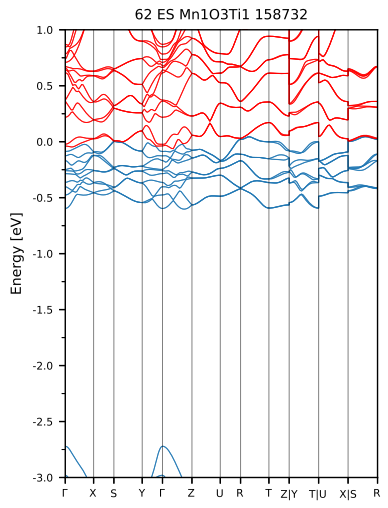

(lg) 62.3.62

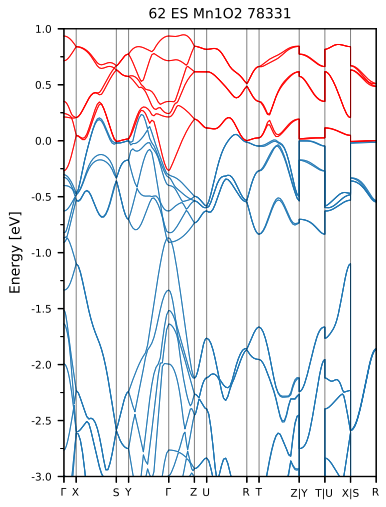

(lh) 62.3.63

Figure 128: Band structures for topological materials in SG 62, part 20/36, from 62.3.48 to 62.3.63. See the list of figures.

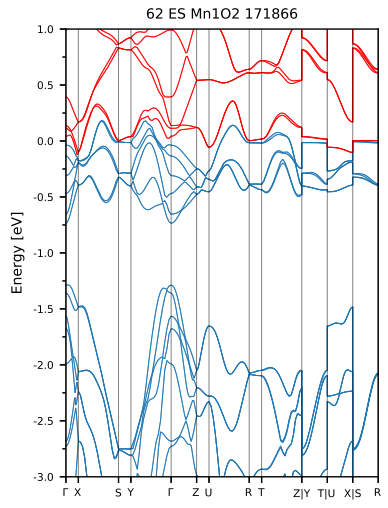

(li) 62.3.64

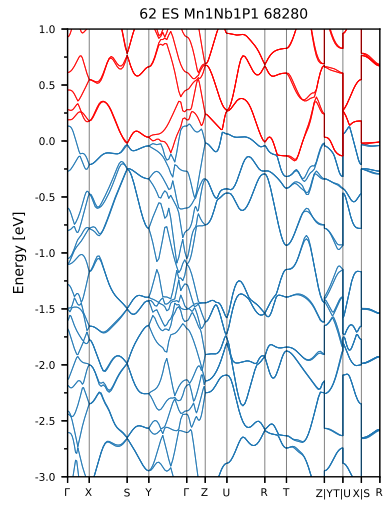

(lj) 62.3.65

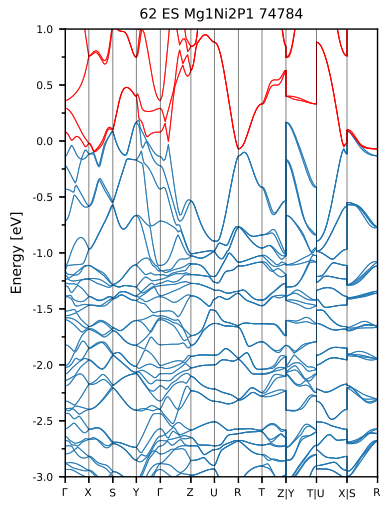

(lk) 62.3.66

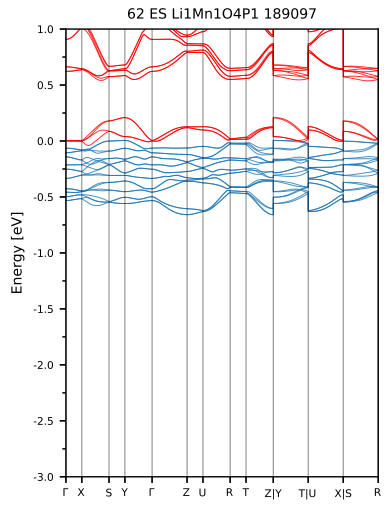

(ll) 62.3.67

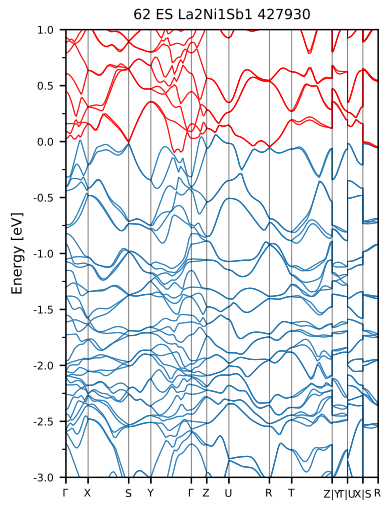

(lm) 62.3.68

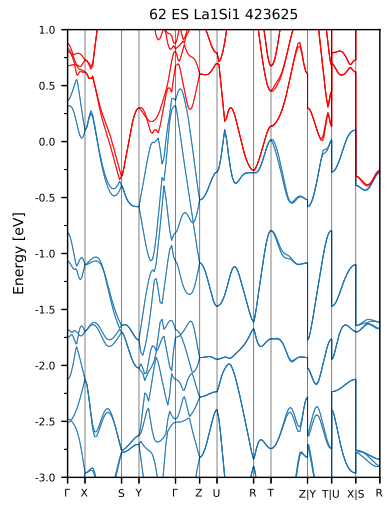

(ln) 62.3.69

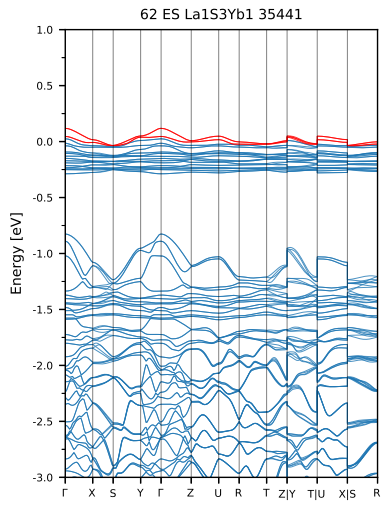

(lo) 62.3.70

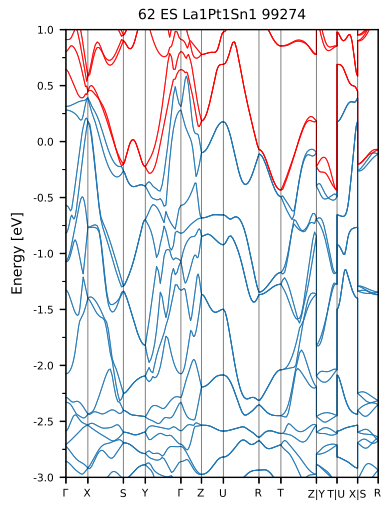

(lp) 62.3.71

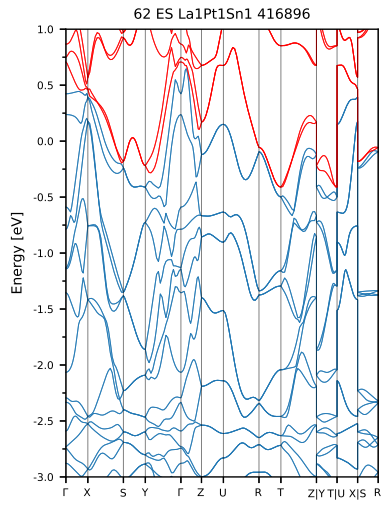

(lq) 62.3.72

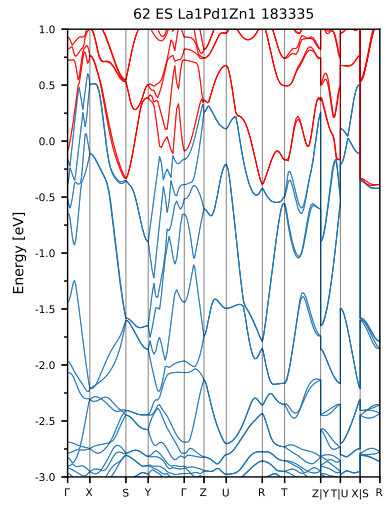

(lr) 62.3.73

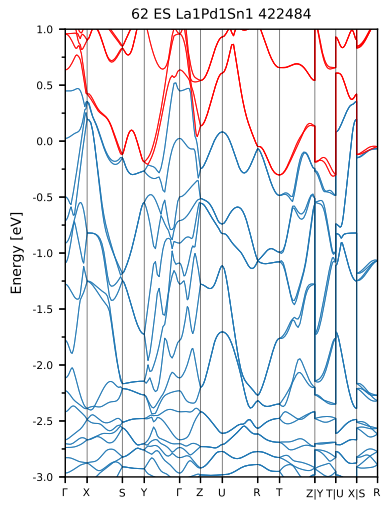

(ls) 62.3.74

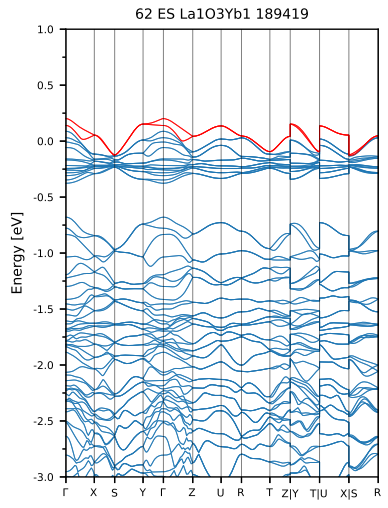

(lt) 62.3.75

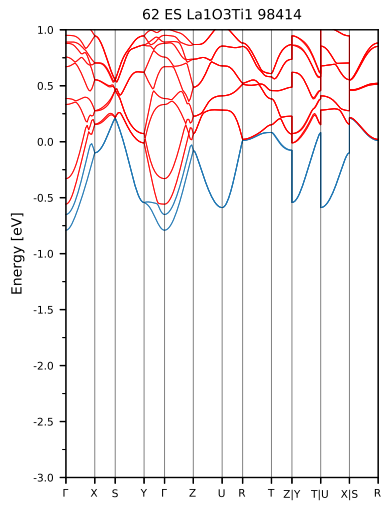

(lu) 62.3.76

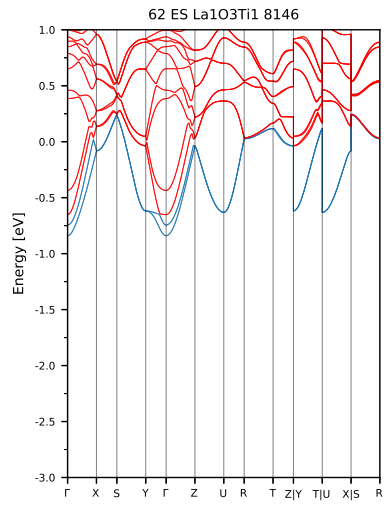

(lv) 62.3.77

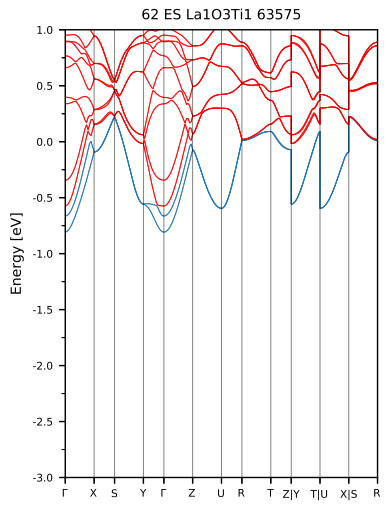

(lw) 62.3.78

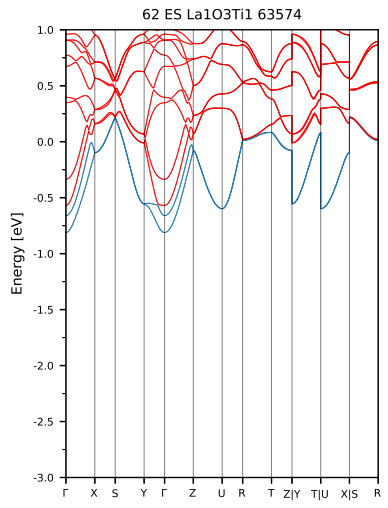

(lx) 62.3.79

Figure 129: Band structures for topological materials in SG 62, part 21/36, from 62.3.64 to 62.3.79. See the list of figures.

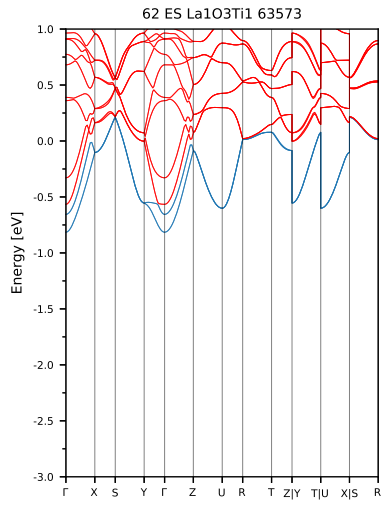

(ly) 62.3.80

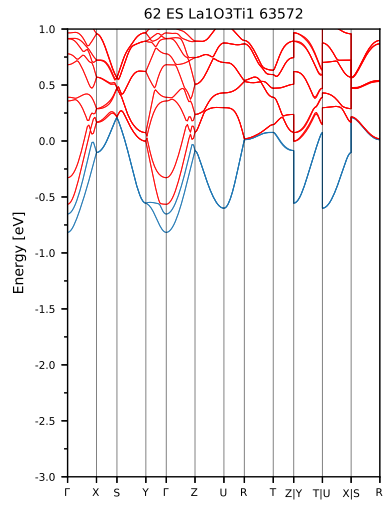

(lz) 62.3.81

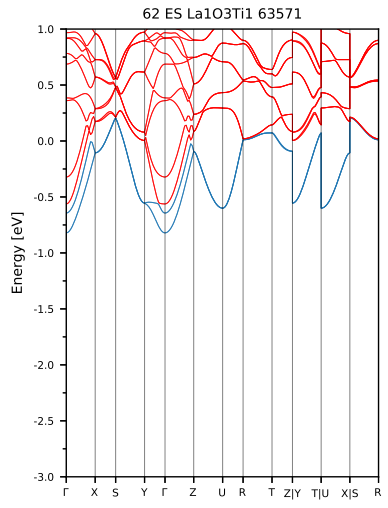

(ma) 62.3.82

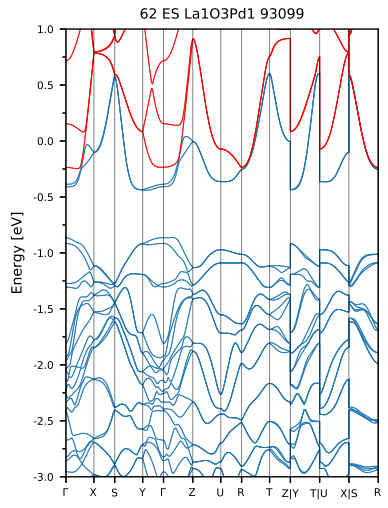

(mb) 62.3.83

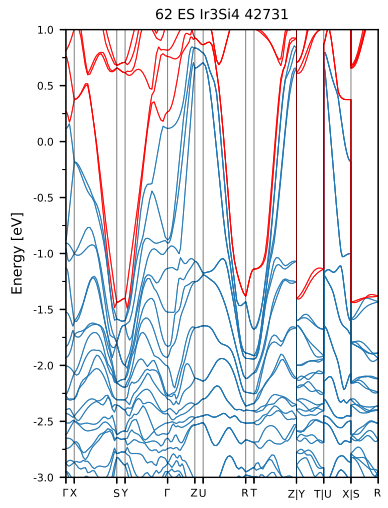

(mc) 62.3.84

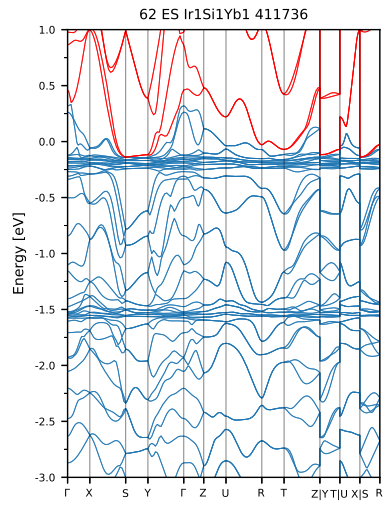

(md) 62.3.85

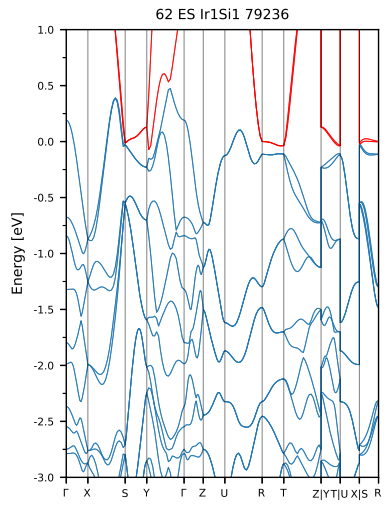

(me) 62.3.86

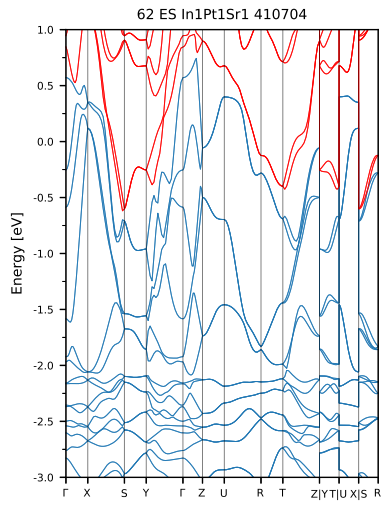

(mf) 62.3.87

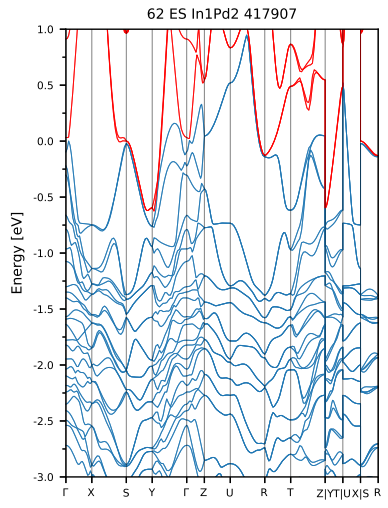

(mg) 62.3.88

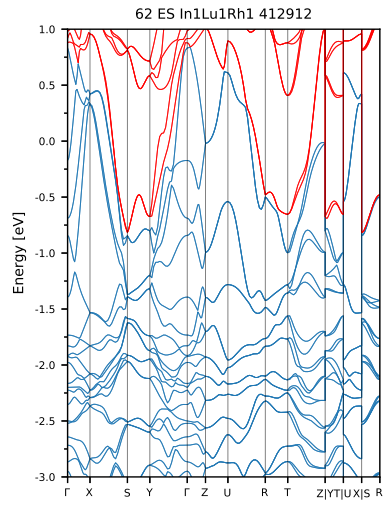

(mh) 62.3.89

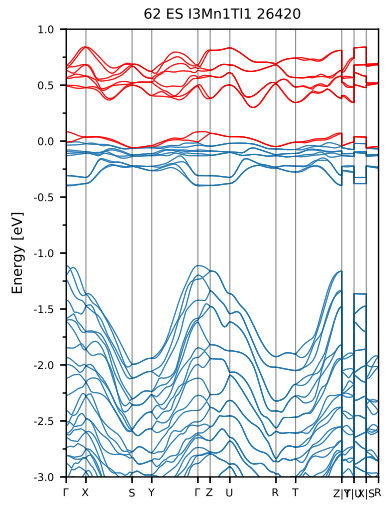

(mi) 62.3.90

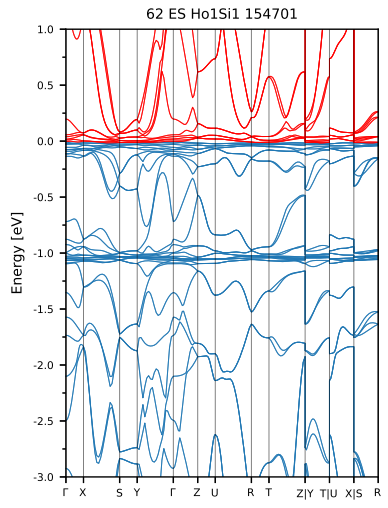

(mj) 62.3.91

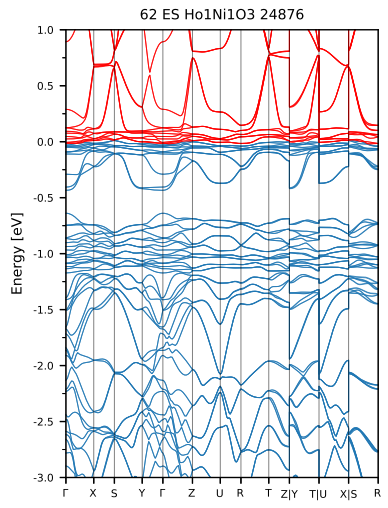

(mk) 62.3.92

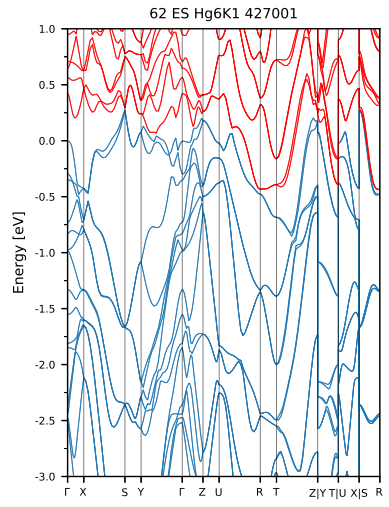

(ml) 62.3.93

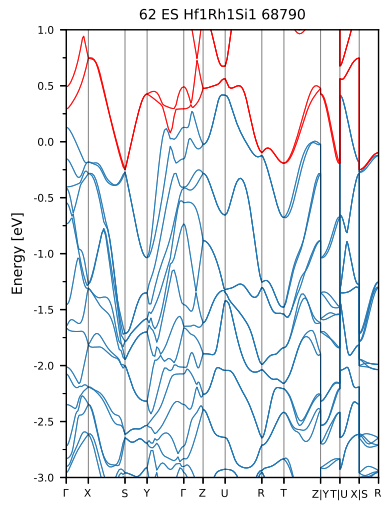

(mm) 62.3.94

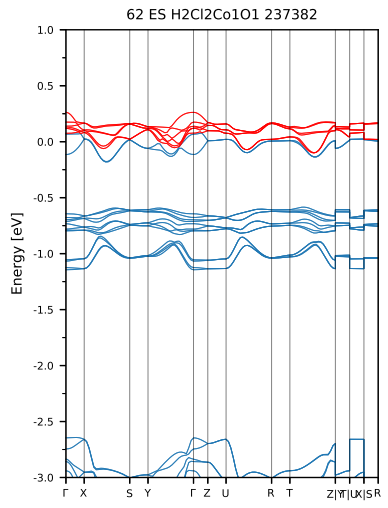

(mn) 62.3.95

Figure 130: Band structures for topological materials in SG 62, part 22/36, from 62.3.80 to 62.3.95. See the list of figures.

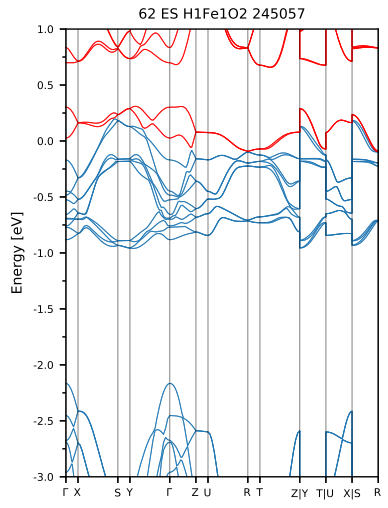

(mo) 62.3.96

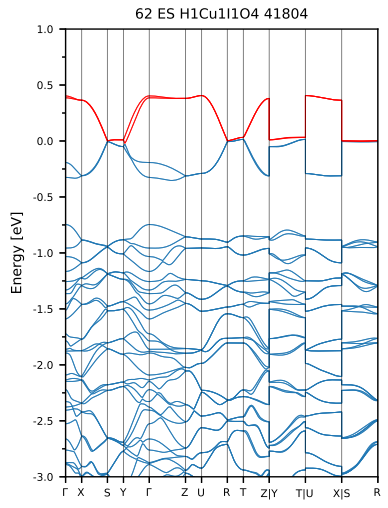

(mp) 62.3.97

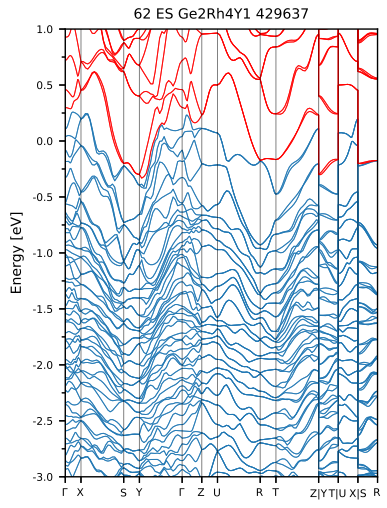

(mq) 62.3.98

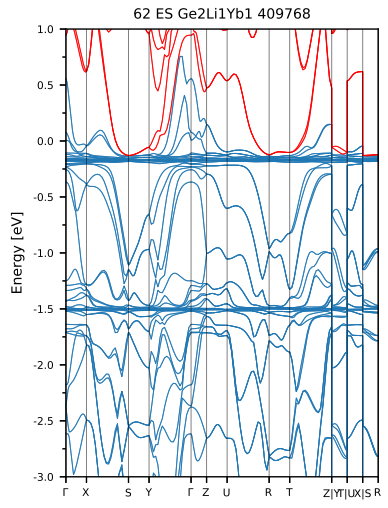

(mr) 62.3.99

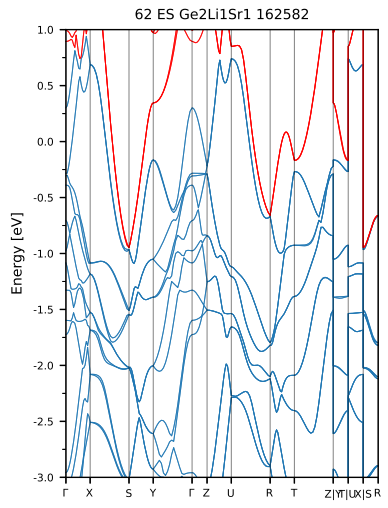

(ms) 62.3.100

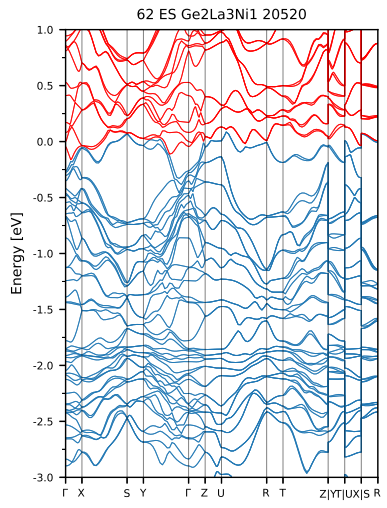

(mt) 62.3.101

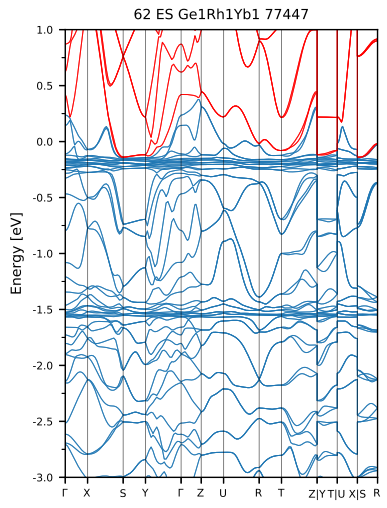

(mu) 62.3.102

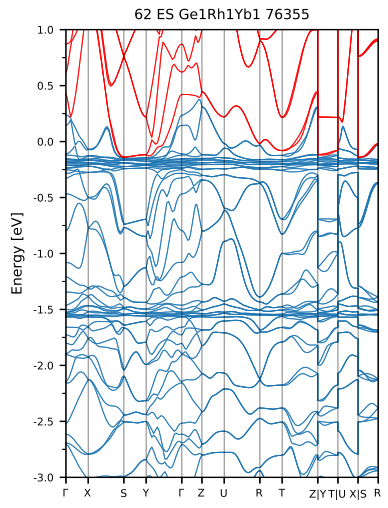

(mv) 62.3.103

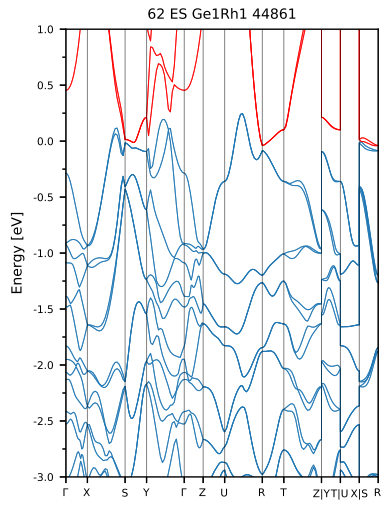

(mw) 62.3.104

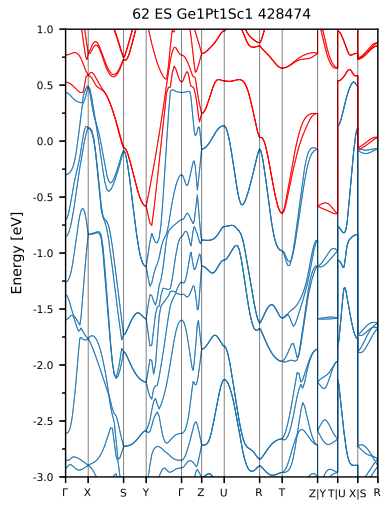

(mx) 62.3.105

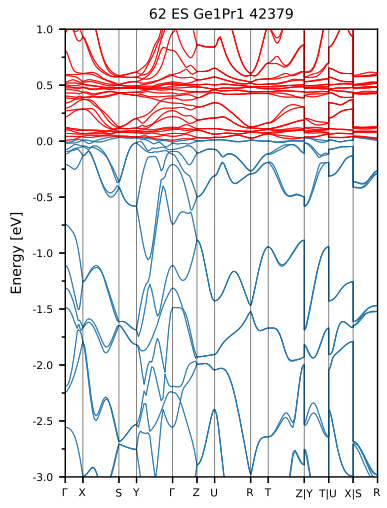

(my) 62.3.106

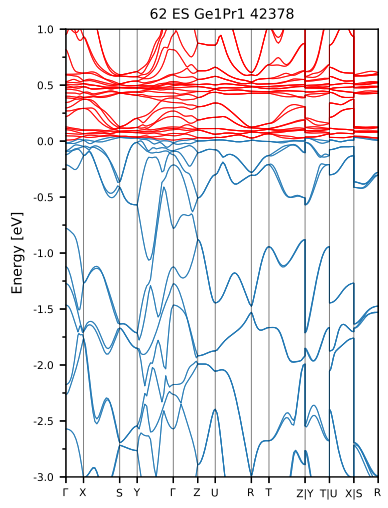

(mz) 62.3.107

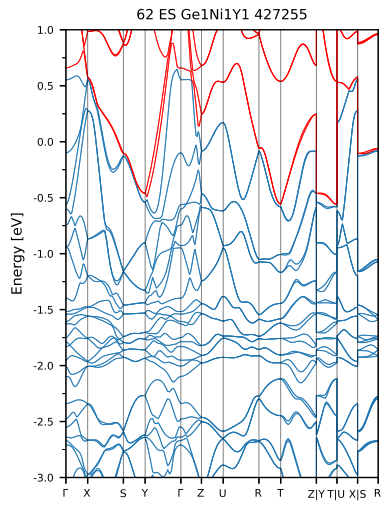

(na) 62.3.108

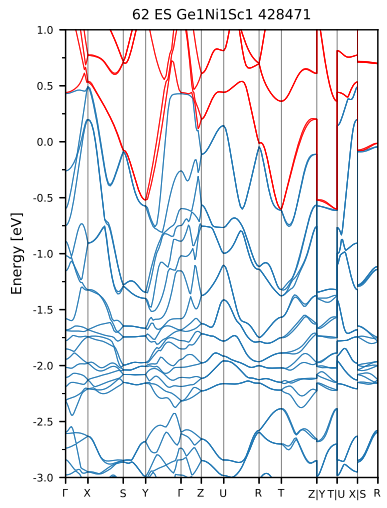

(nb) 62.3.109

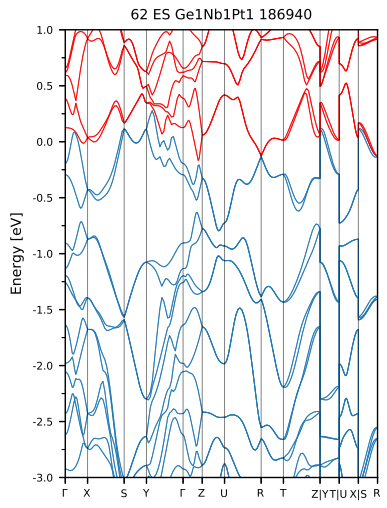

(nc) 62.3.110

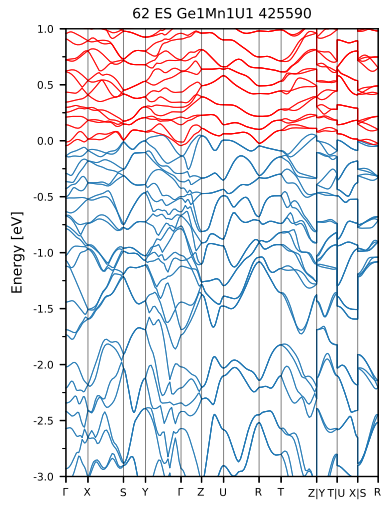

(nd) 62.3.111

Figure 131: Band structures for topological materials in SG 62, part 23/36, from 62.3.96 to 62.3.111. See the list of figures.

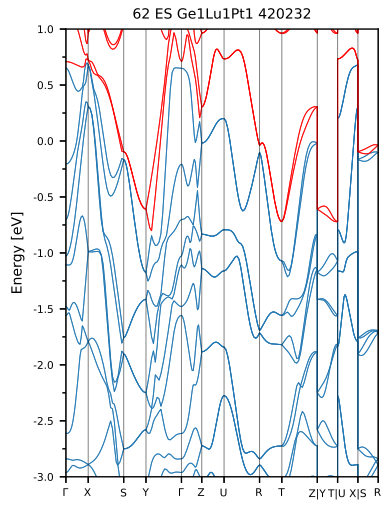

(ne) 62.3.112

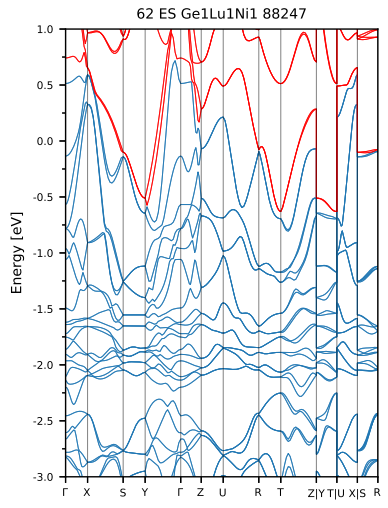

(nf) 62.3.113

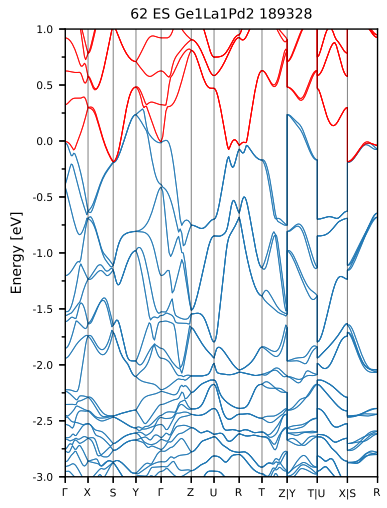

(ng) 62.3.114

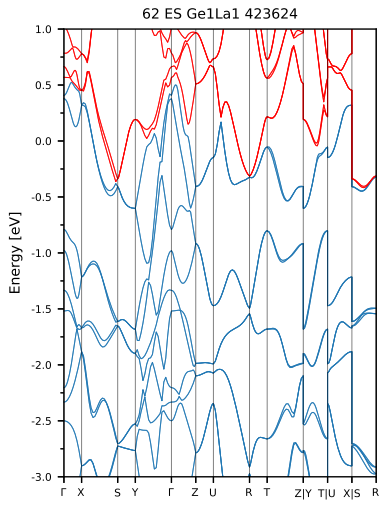

(nh) 62.3.115

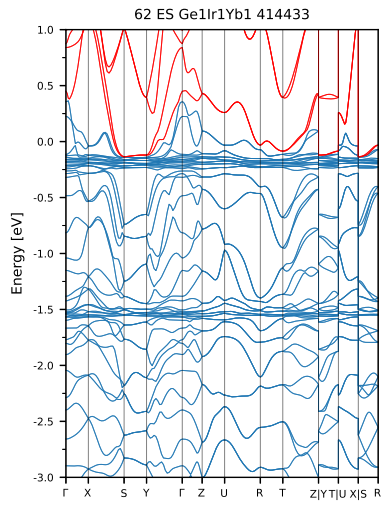

(ni) 62.3.116

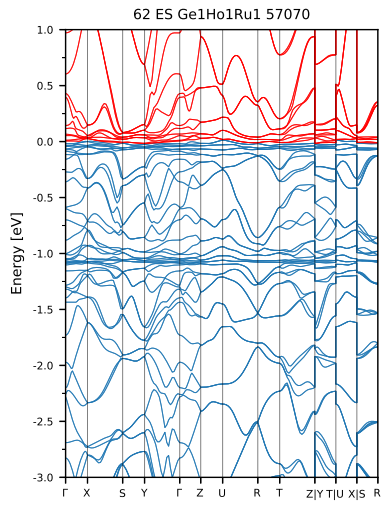

(nj) 62.3.117

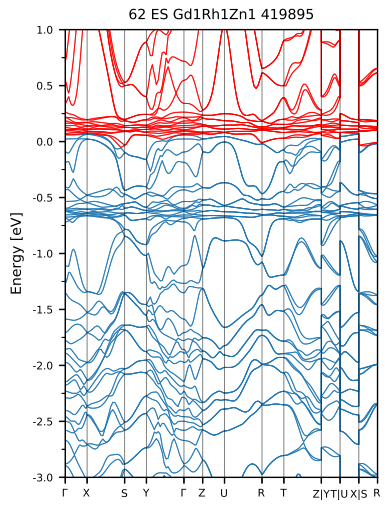

(nk) 62.3.118

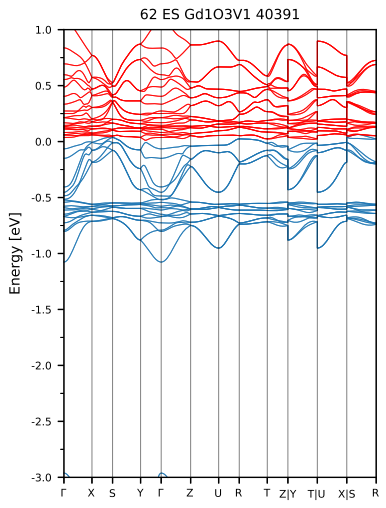

(nl) 62.3.119

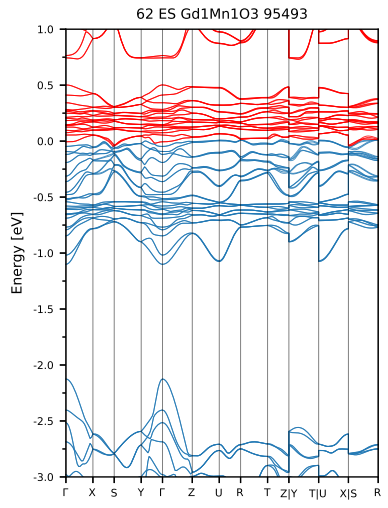

(nm) 62.3.120

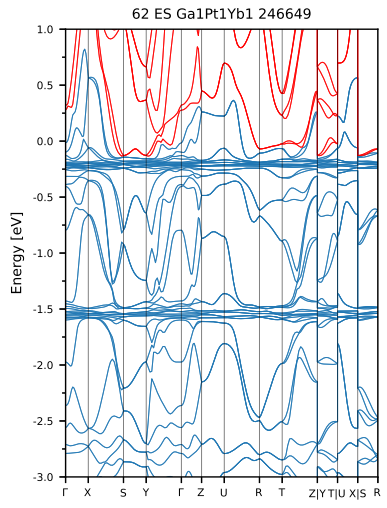

(nn) 62.3.121

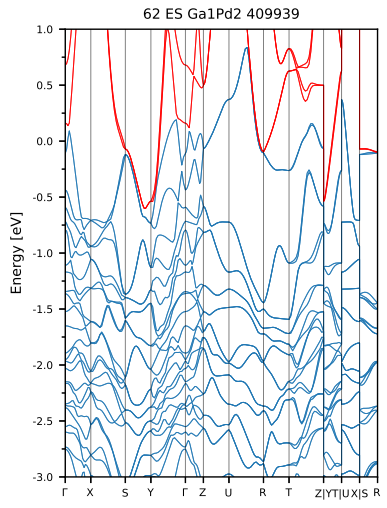

(no) 62.3.122

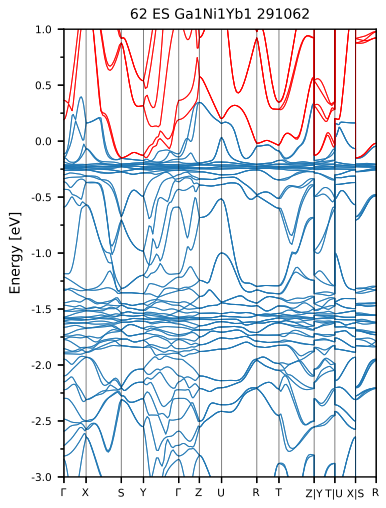

(np) 62.3.123

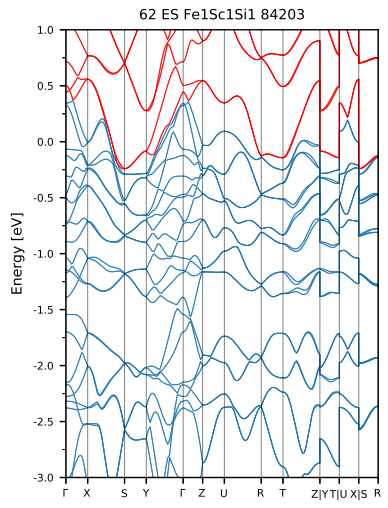

(nq) 62.3.124

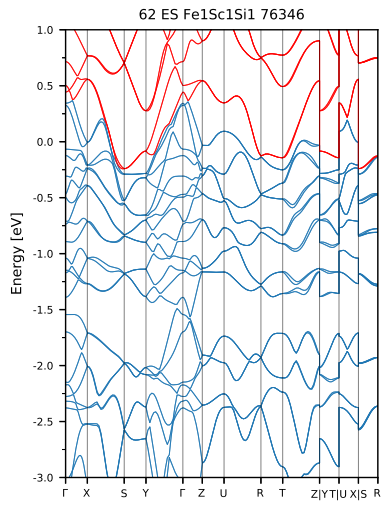

(nr) 62.3.125

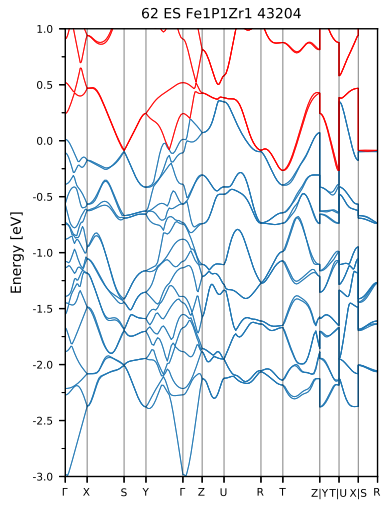

(ns) 62.3.126

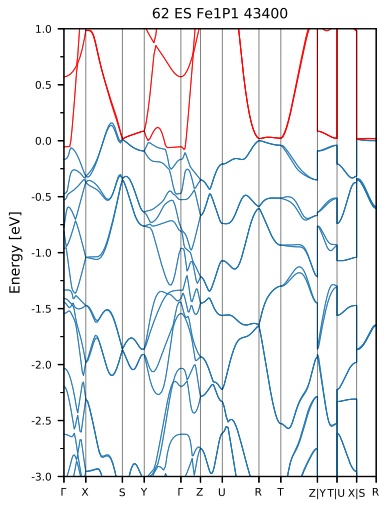

(nt) 62.3.127

Figure 132: Band structures for topological materials in SG 62, part 24/36, from 62.3.112 to 62.3.127. See the list of figures.

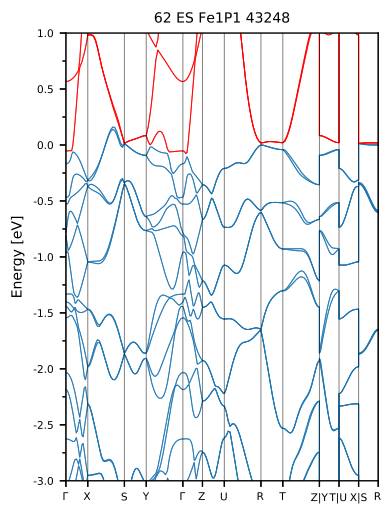

(nu) 62.3.128

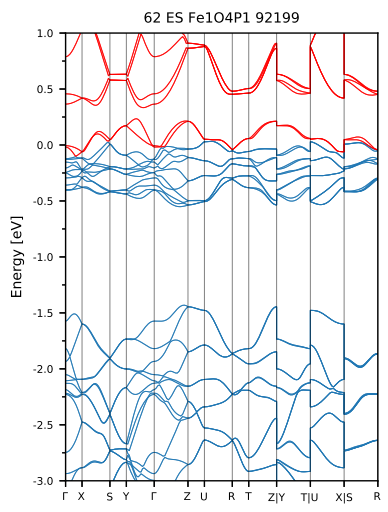

(nv) 62.3.129

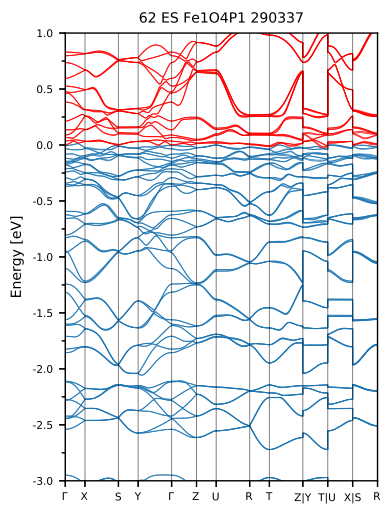

(nw) 62.3.130

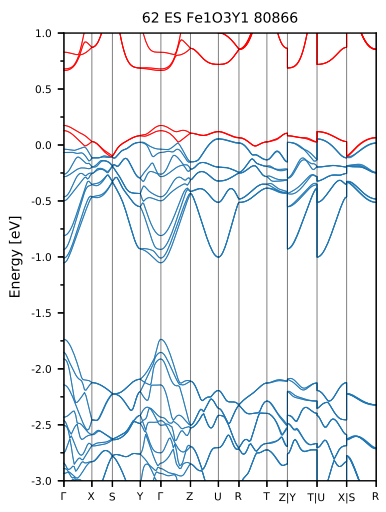

(nx) 62.3.131

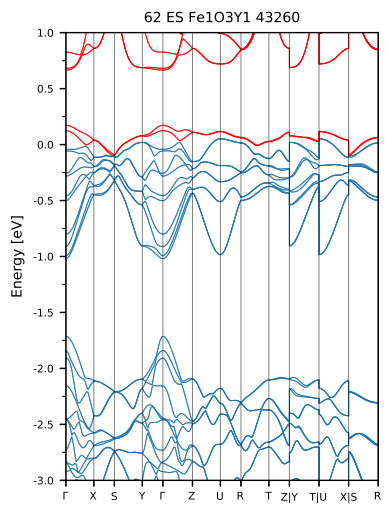

(ny) 62.3.132

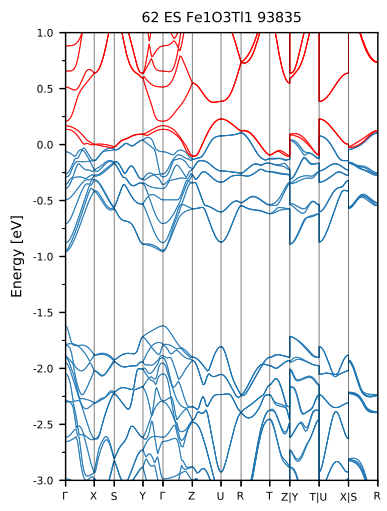

(nz) 62.3.133

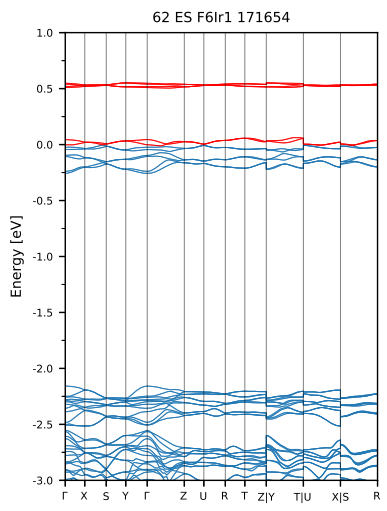

(oa) 62.3.134

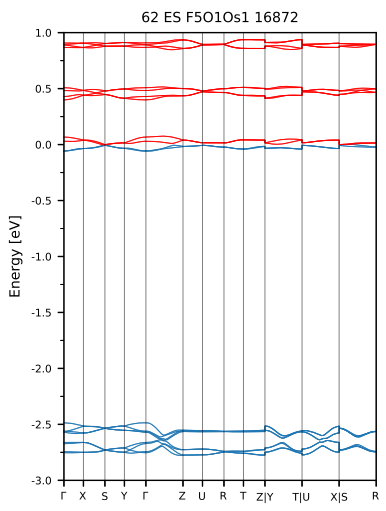

(ob) 62.3.135

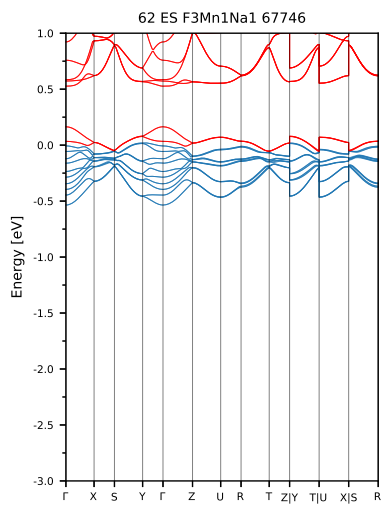

(oc) 62.3.136

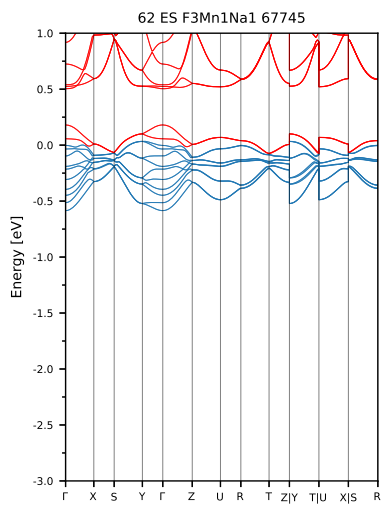

(od) 62.3.137

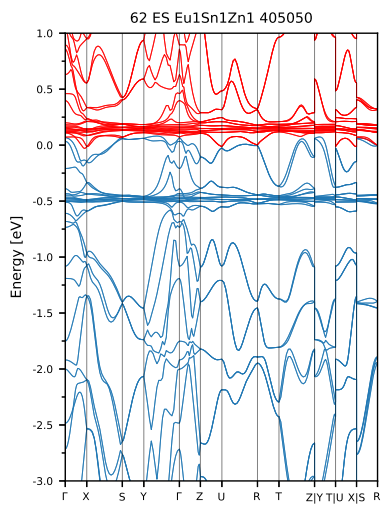

(oe) 62.3.138

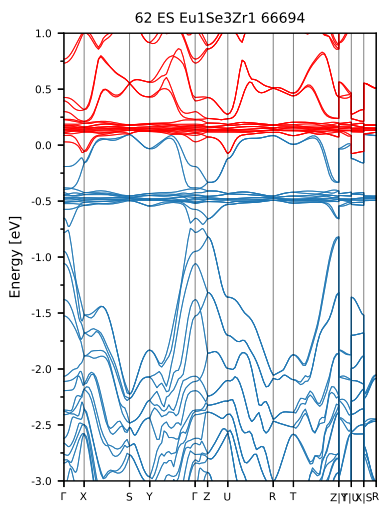

(of) 62.3.139

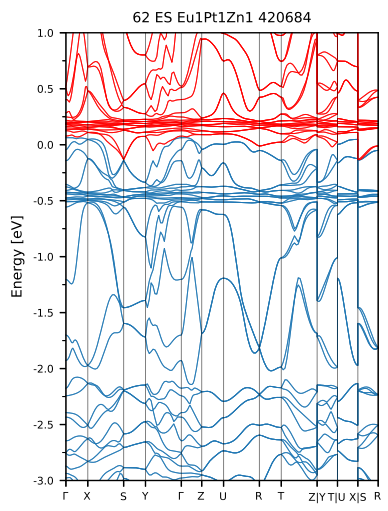

(og) 62.3.140

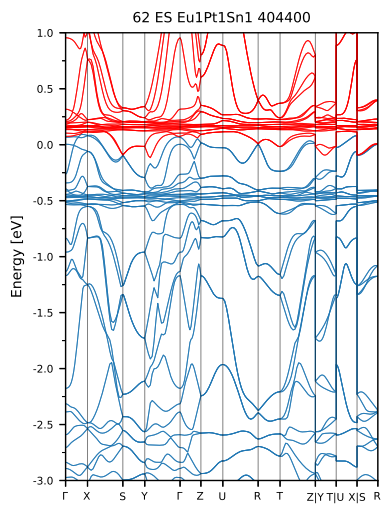

(oh) 62.3.141

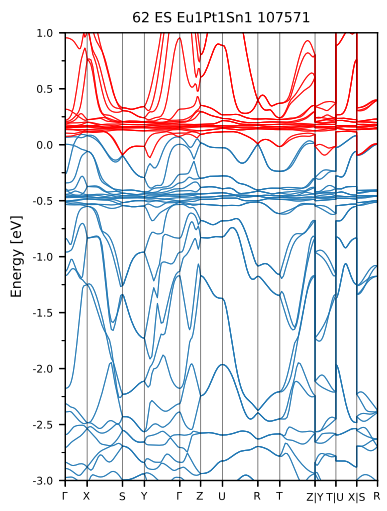

(oi) 62.3.142

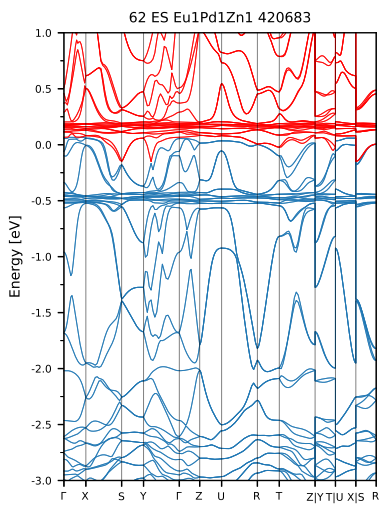

(oj) 62.3.143

Figure 133: Band structures for topological materials in SG 62, part 25/36, from 62.3.128 to 62.3.143. See the list of figures.

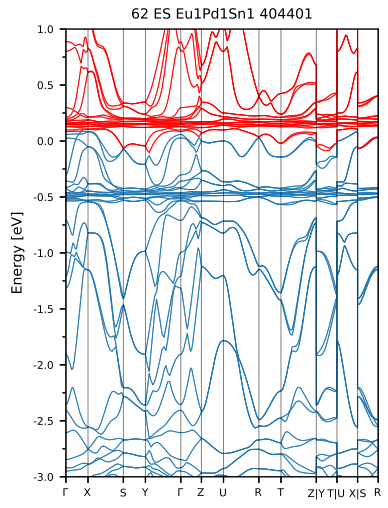

(ok) 62.3.144

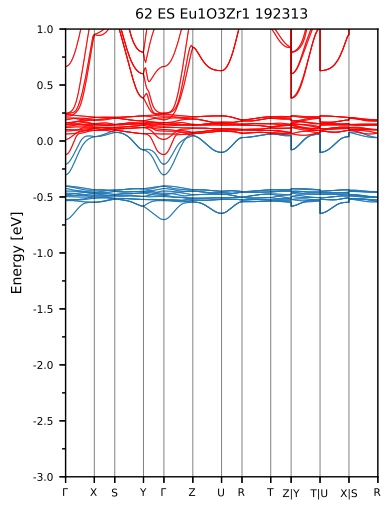

(ol) 62.3.145

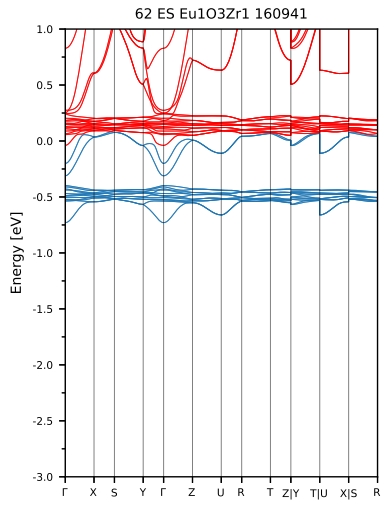

(om) 62.3.146

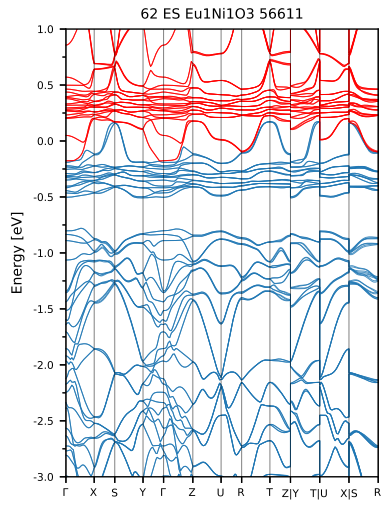

(on) 62.3.147

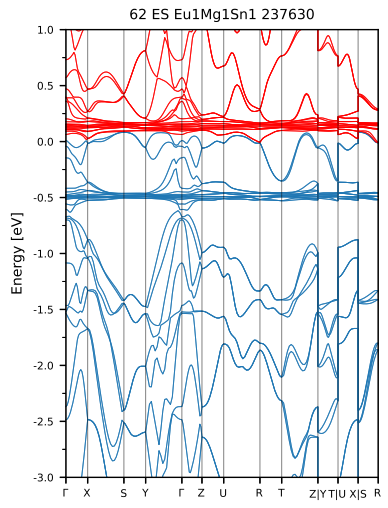

(oo) 62.3.148

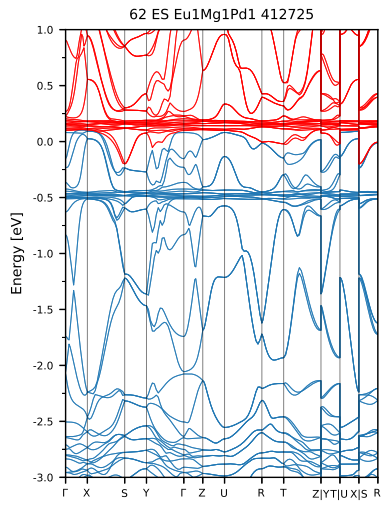

(op) 62.3.149

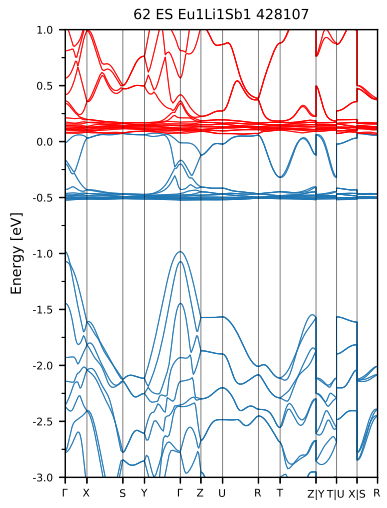

(oq) 62.3.150

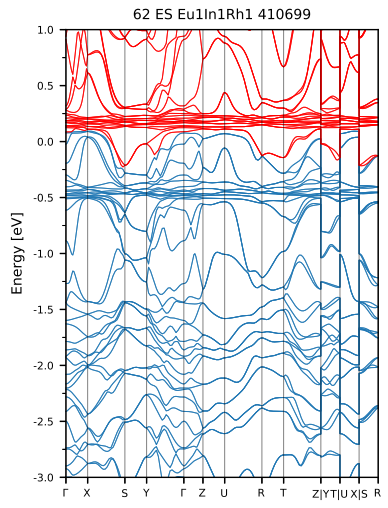

(or) 62.3.151

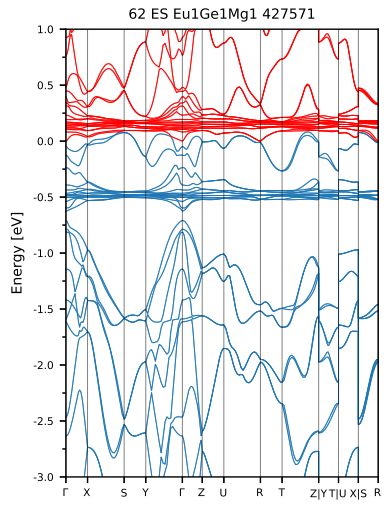

(os) 62.3.152

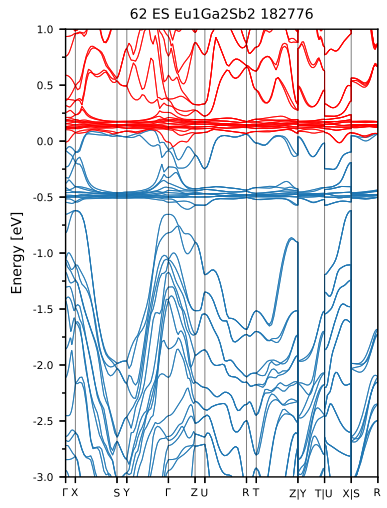

(ot) 62.3.153

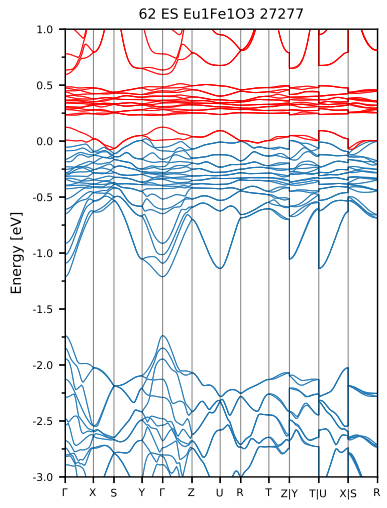

(ou) 62.3.154

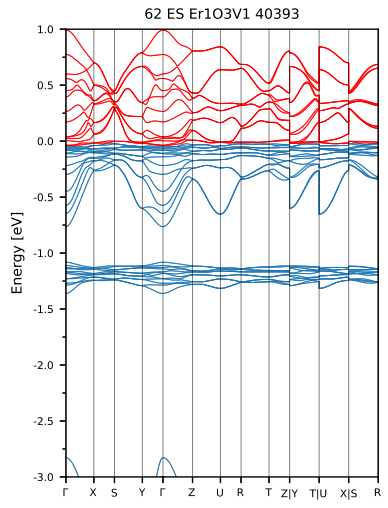

(ov) 62.3.155

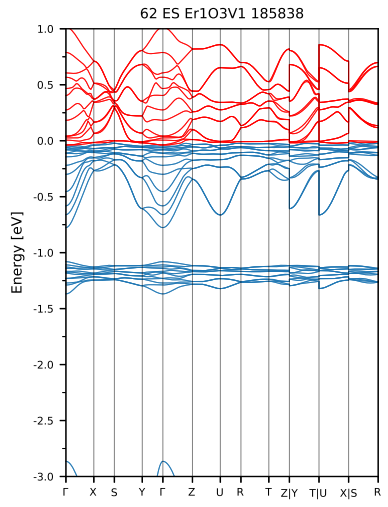

(ow) 62.3.156

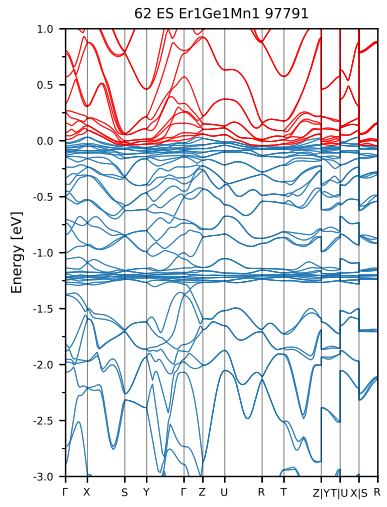

(ox) 62.3.157

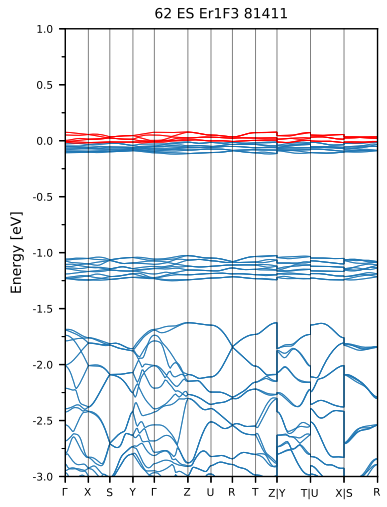

(oy) 62.3.158

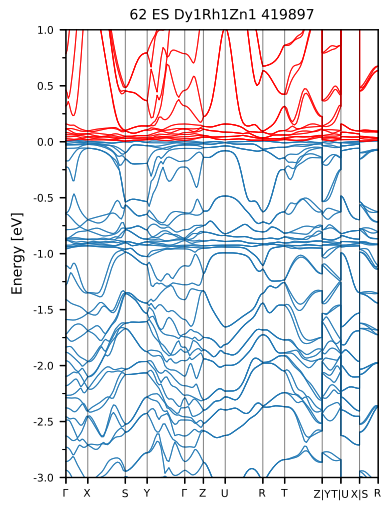

(oz) 62.3.159

Figure 134: Band structures for topological materials in SG 62, part 26/36, from 62.3.144 to 62.3.159. See the list of figures.

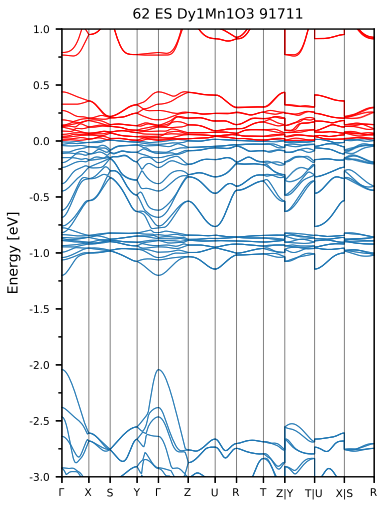

(pa) 62.3.160

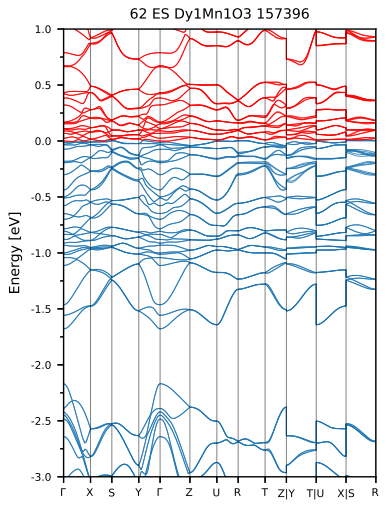

(pb) 62.3.161

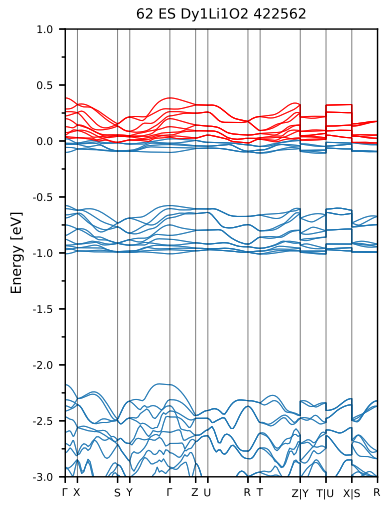

(pc) 62.3.162

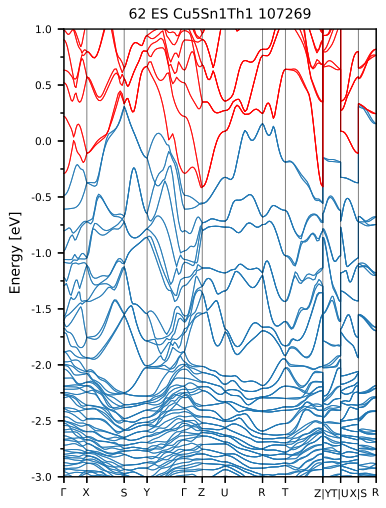

(pd) 62.3.163

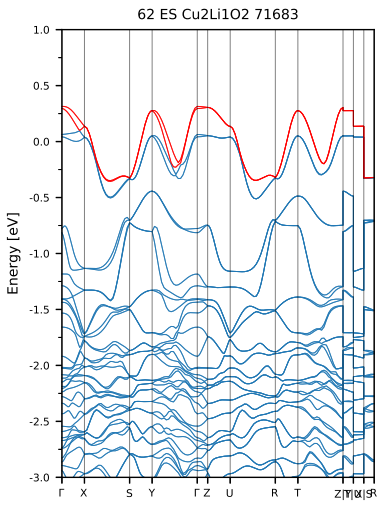

(pe) 62.3.164

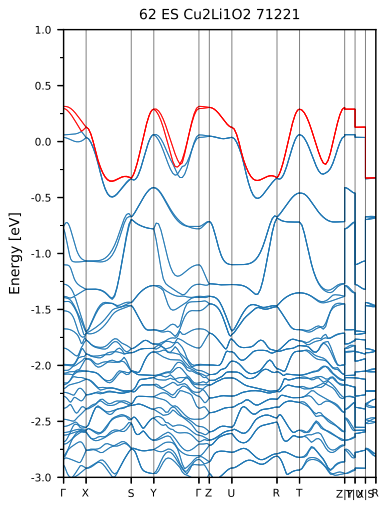

(pf) 62.3.165

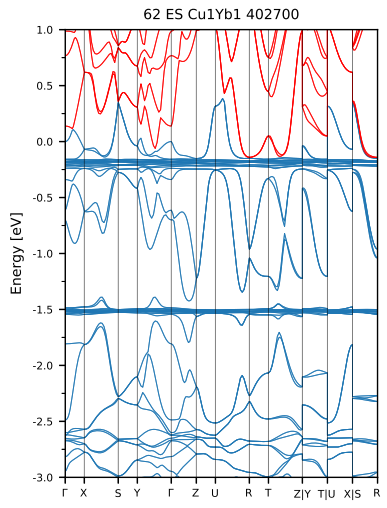

(pg) 62.3.166

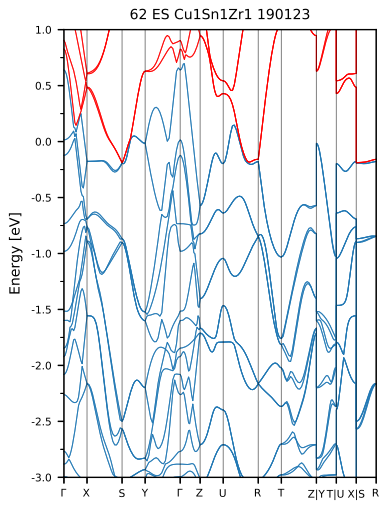

(ph) 62.3.167

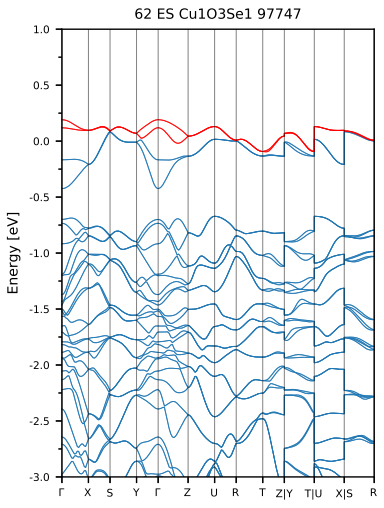

(pi) 62.3.168

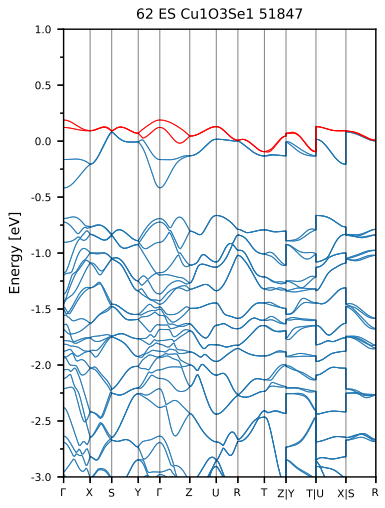

(pj) 62.3.169

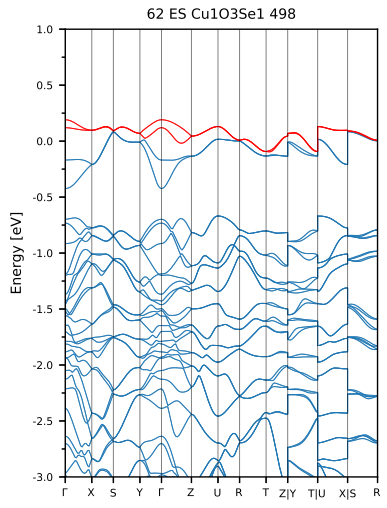

(pk) 62.3.170

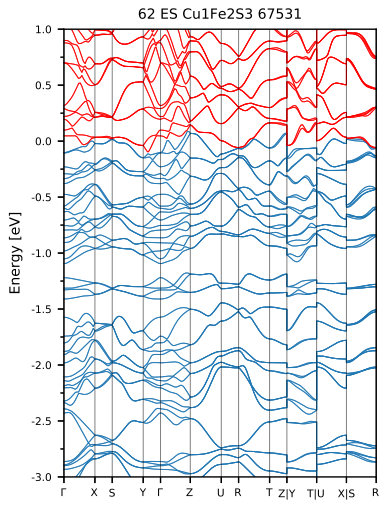

(pl) 62.3.171

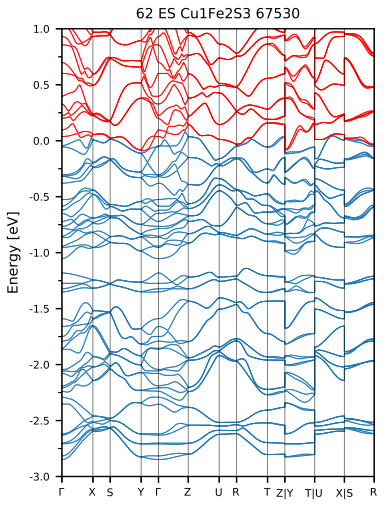

(pm) 62.3.172

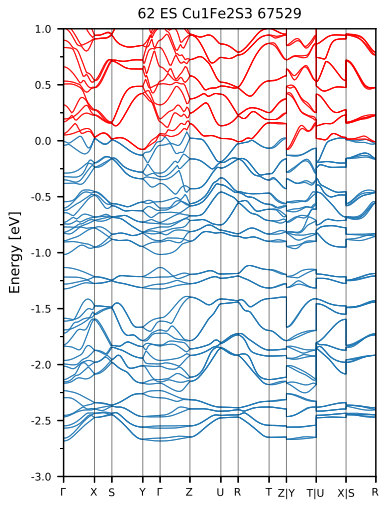

(pn) 62.3.173

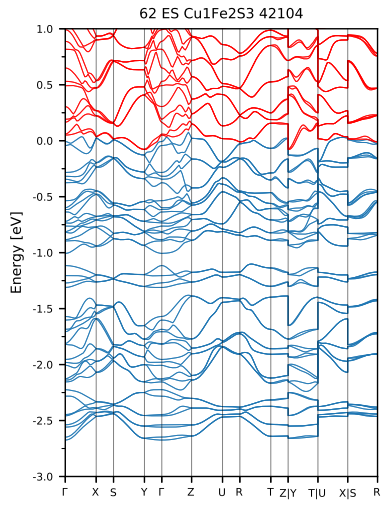

(po) 62.3.174

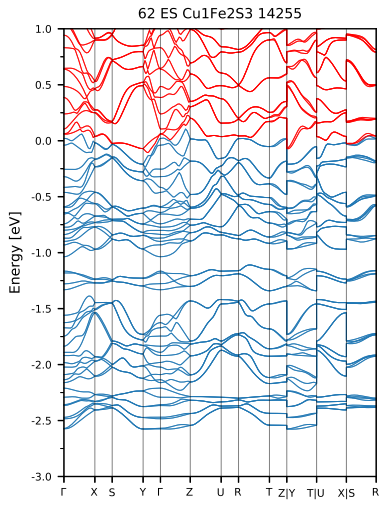

(pp) 62.3.175

Figure 135: Band structures for topological materials in SG 62, part 27/36, from 62.3.160 to 62.3.175. See the list of figures.

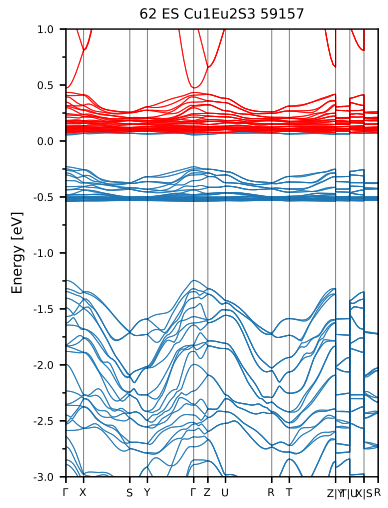

(pq) 62.3.176

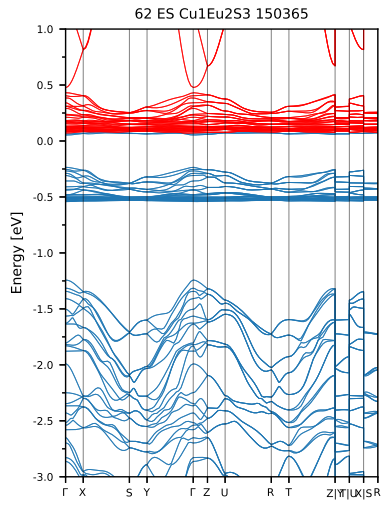

(pr) 62.3.177

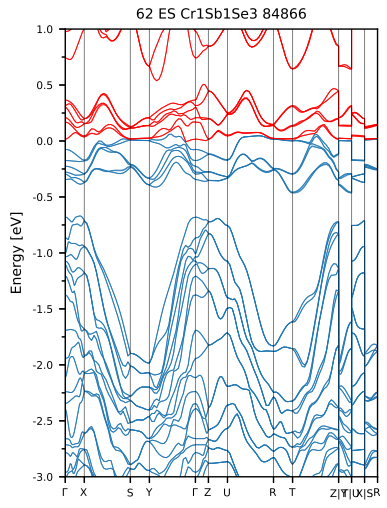

(ps) 62.3.178

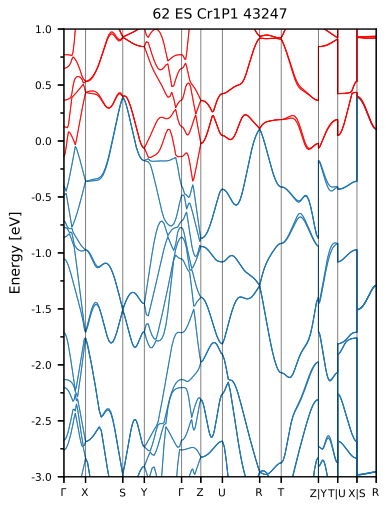

(pt) 62.3.179

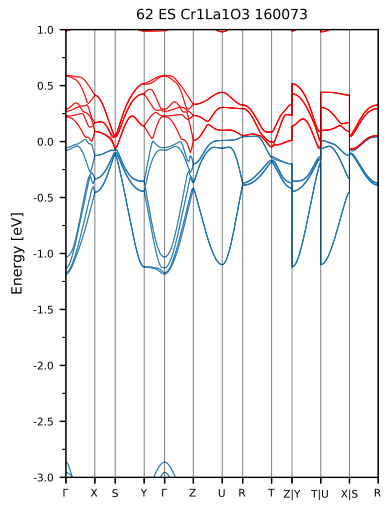

(pu) 62.3.180

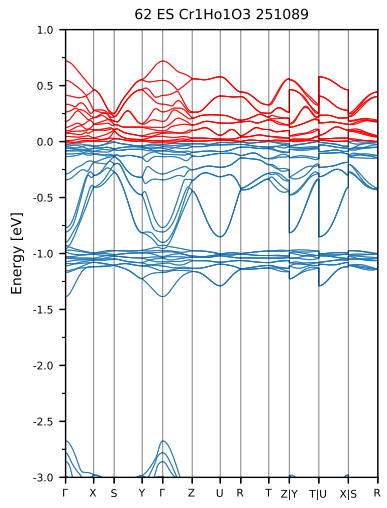

(pv) 62.3.181

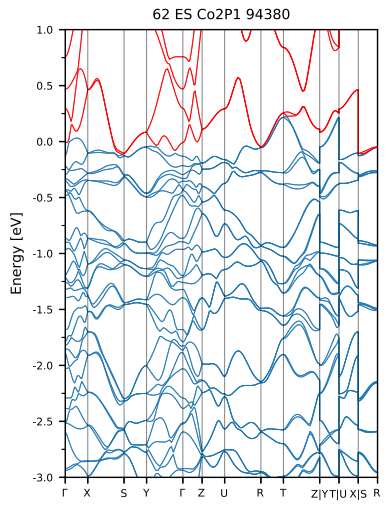

(pw) 62.3.182

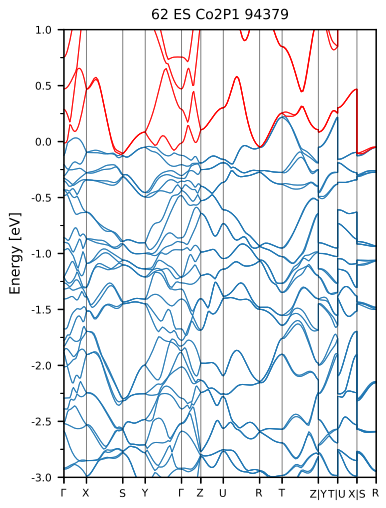

(px) 62.3.183

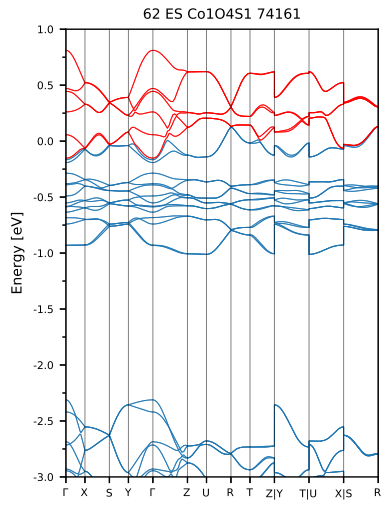

(py) 62.3.184

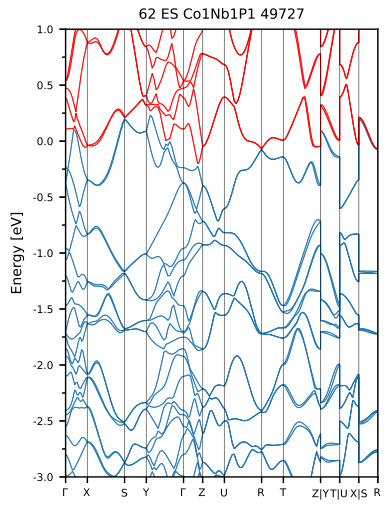

(pz) 62.3.185

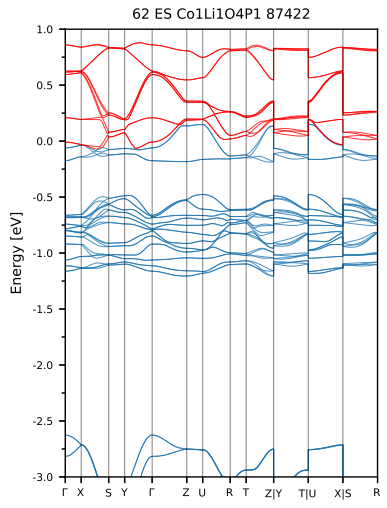

(qa) 62.3.186

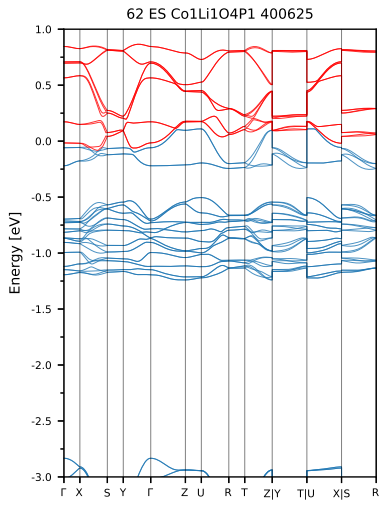

(qb) 62.3.187

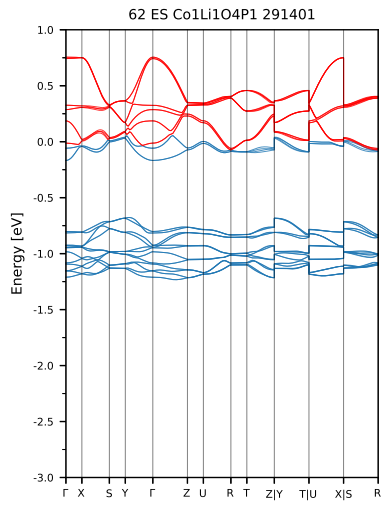

(qc) 62.3.188

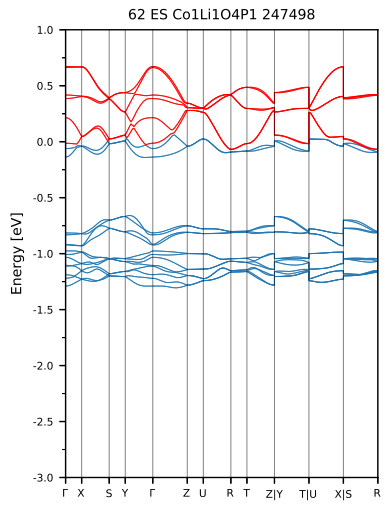

(qd) 62.3.189

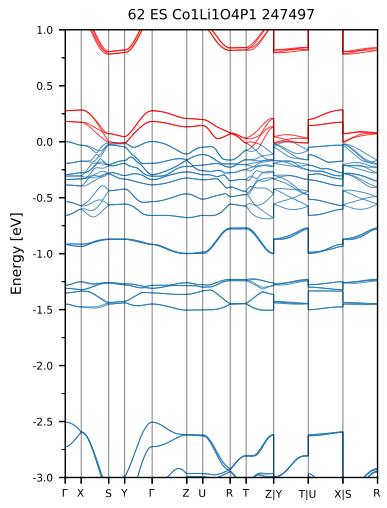

(qe) 62.3.190

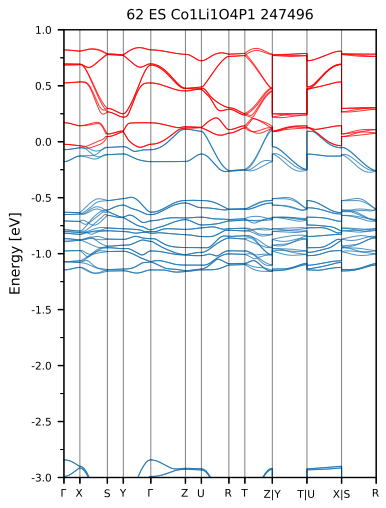

(qf) 62.3.191

Figure 136: Band structures for topological materials in SG 62, part 28/36, from 62.3.176 to 62.3.191. See the list of figures.

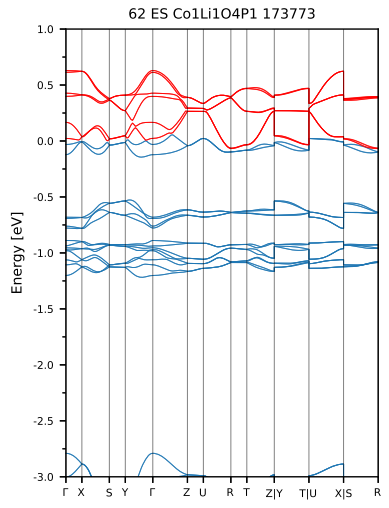

(qg) 62.3.192

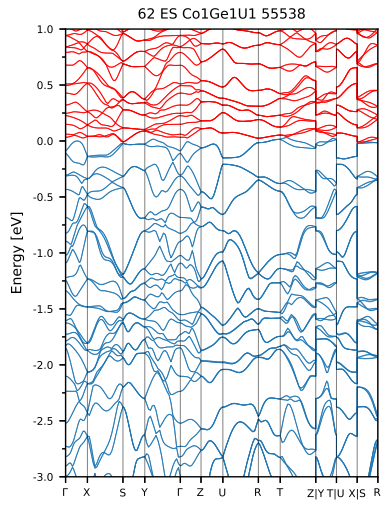

(qh) 62.3.193

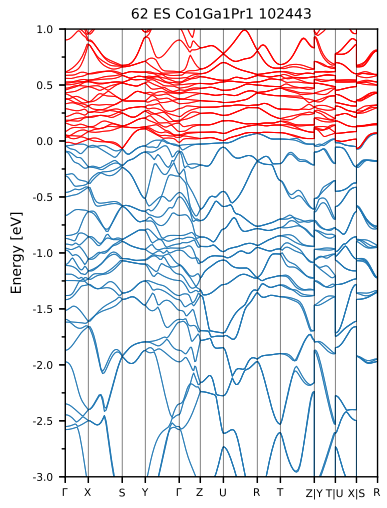

(qi) 62.3.194

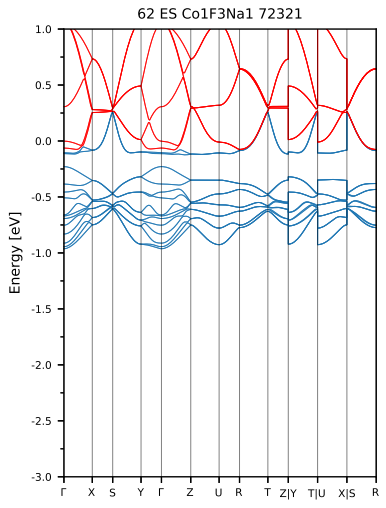

(qj) 62.3.195

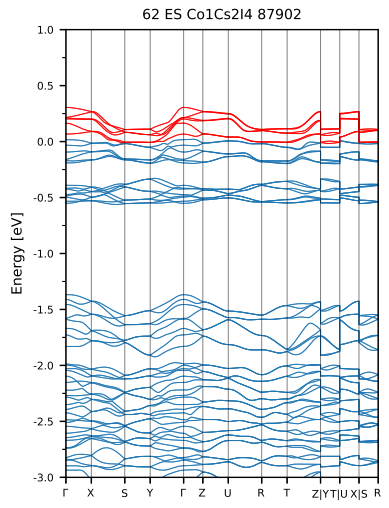

(qk) 62.3.196

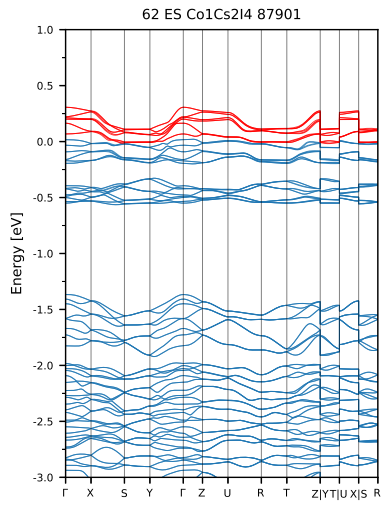

(ql) 62.3.197

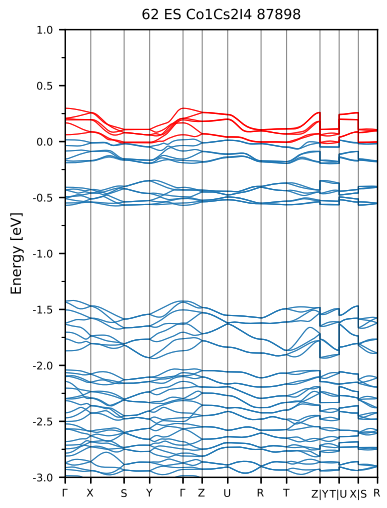

(qm) 62.3.198

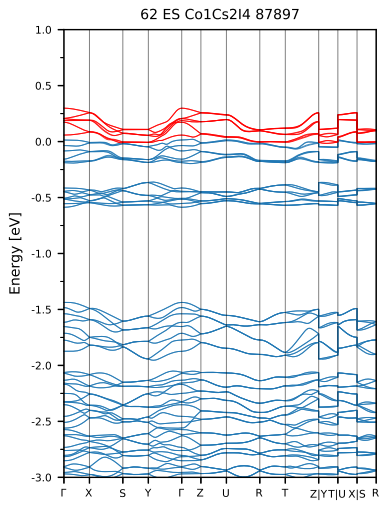

(qn) 62.3.199

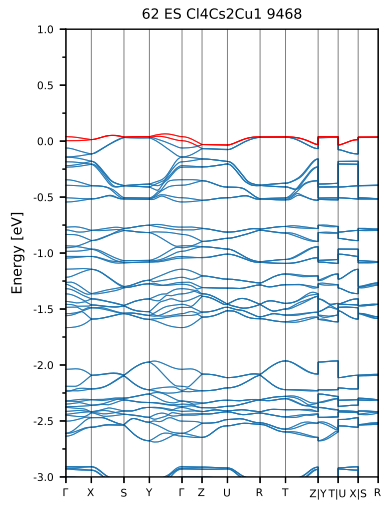

(qo) 62.3.200

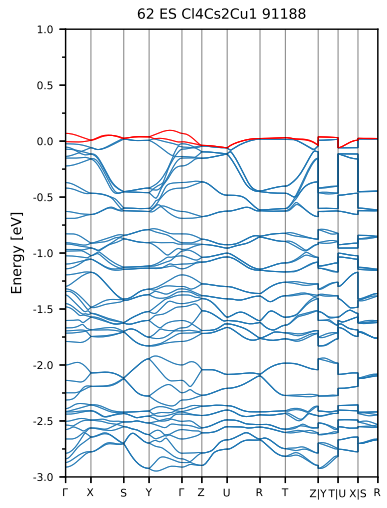

(qp) 62.3.201

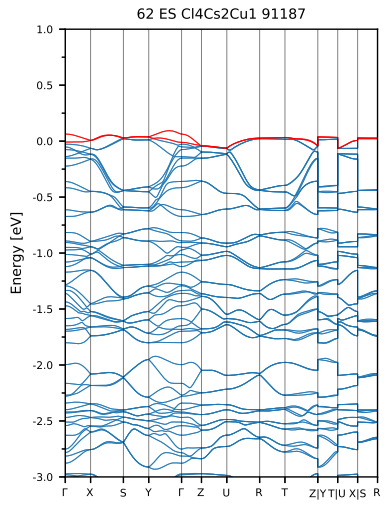

(qq) 62.3.202

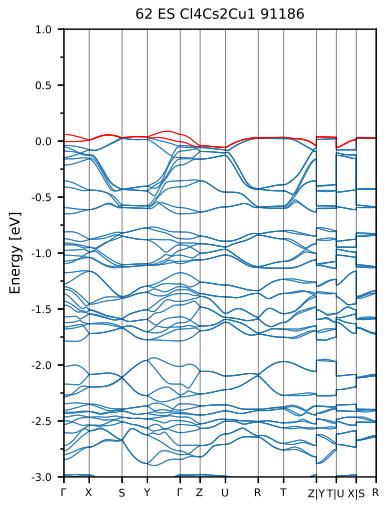

(qr) 62.3.203

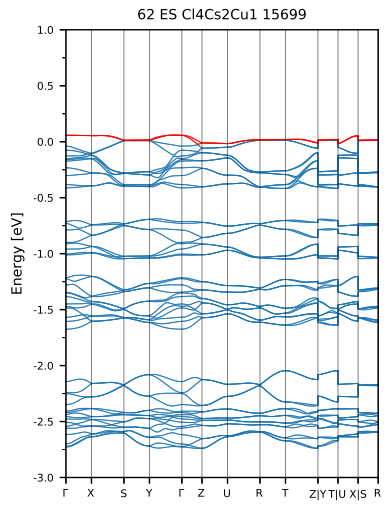

(qs) 62.3.204

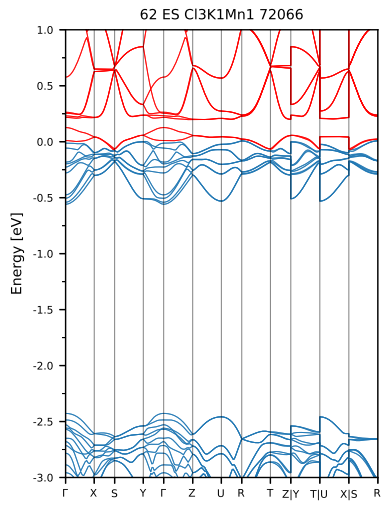

(qt) 62.3.205

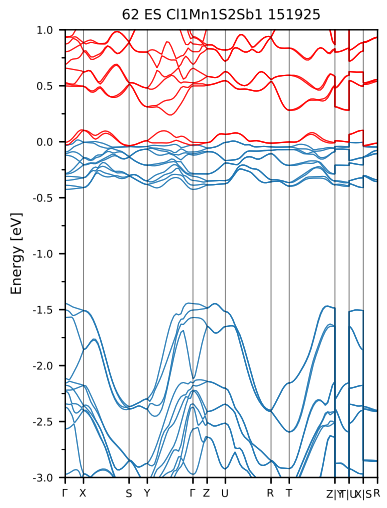

(qu) 62.3.206

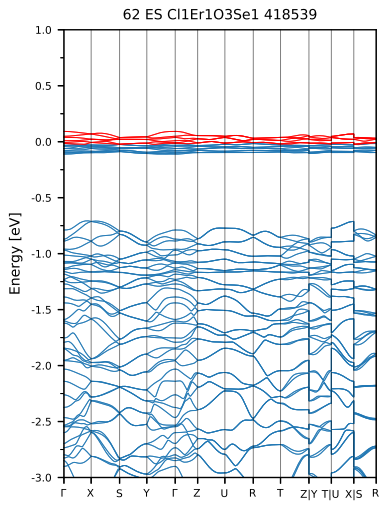

(qv) 62.3.207

Figure 137: Band structures for topological materials in SG 62, part 29/36, from 62.3.192 to 62.3.207. See the list of figures.

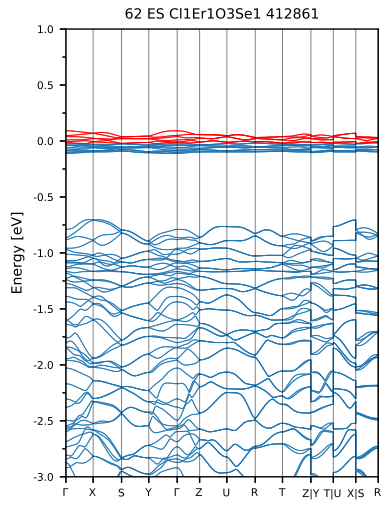

(qw) 62.3.208

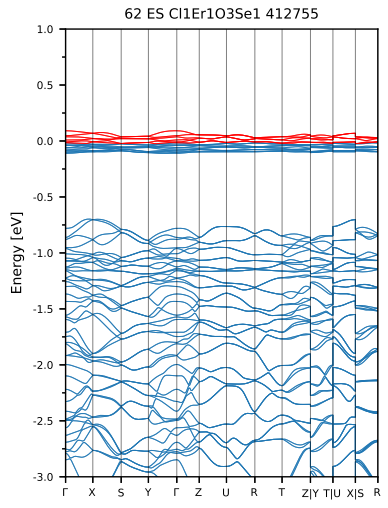

(qx) 62.3.209

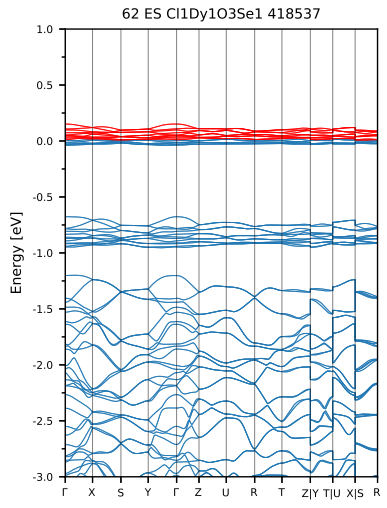

(qy) 62.3.210

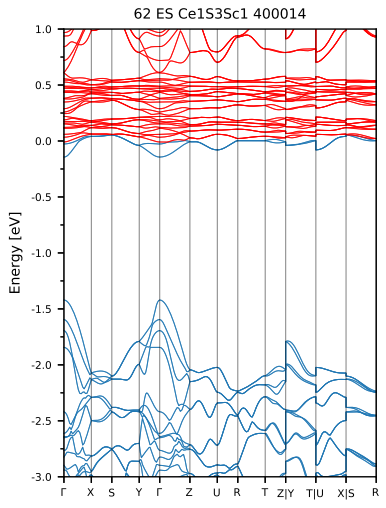

(qz) 62.3.211

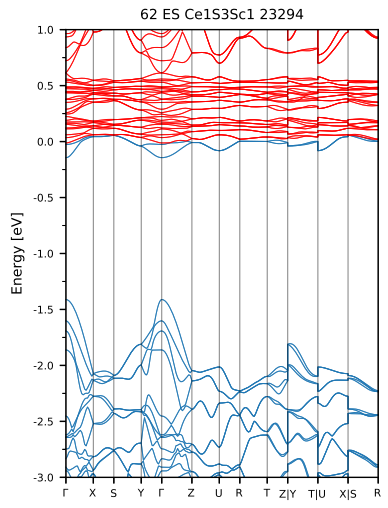

(ra) 62.3.212

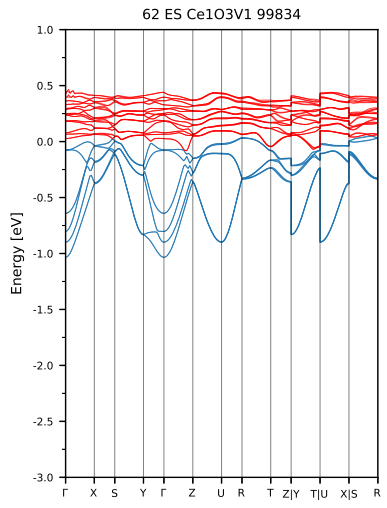

(rb) 62.3.213

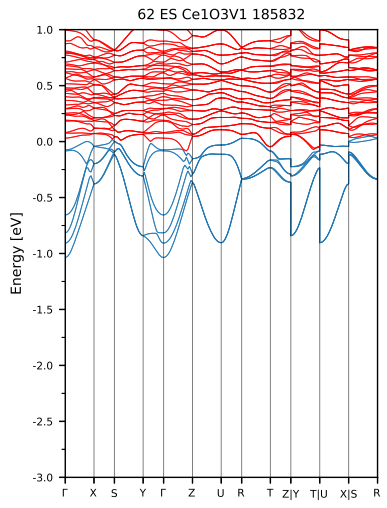

(rc) 62.3.214

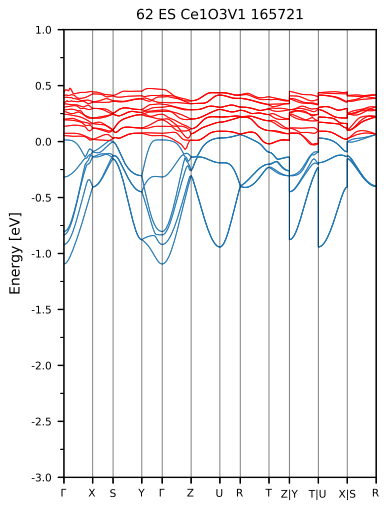

(rd) 62.3.215

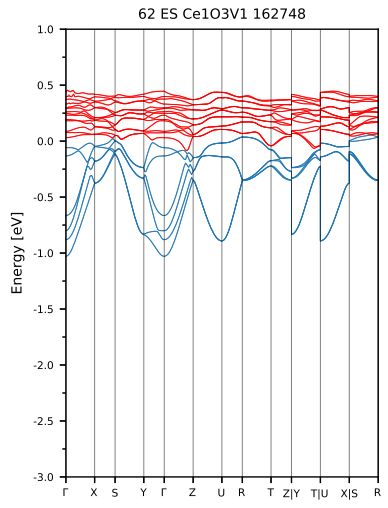

(re) 62.3.216

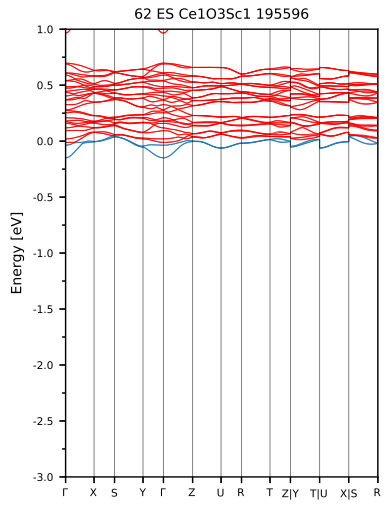

(rf) 62.3.217

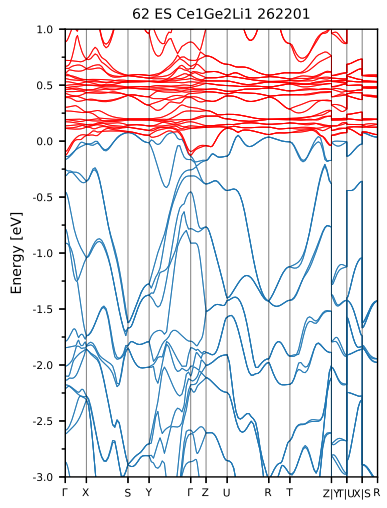

(rg) 62.3.218

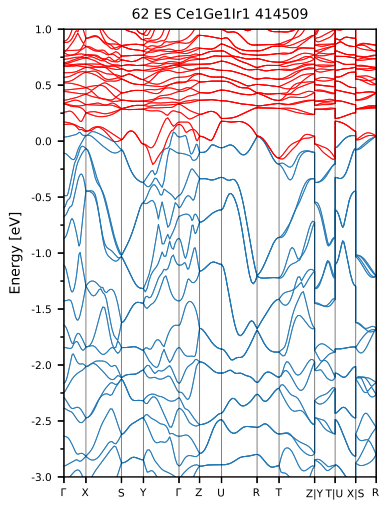

(rh) 62.3.219

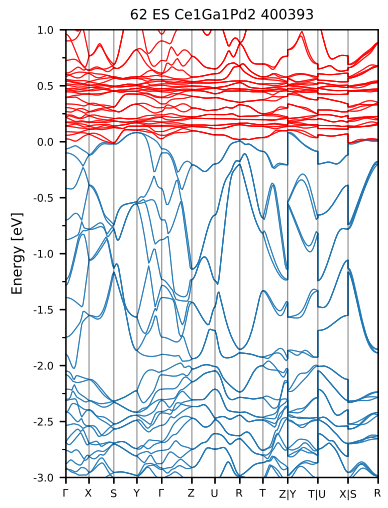

(ri) 62.3.220

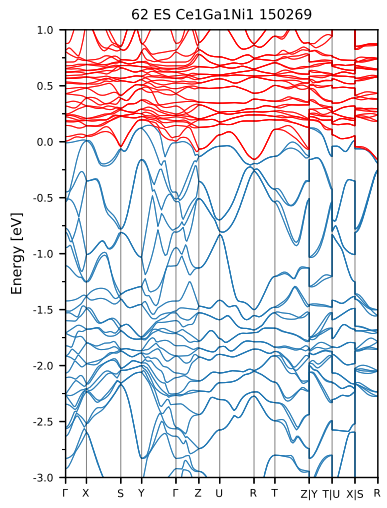

(rj) 62.3.221

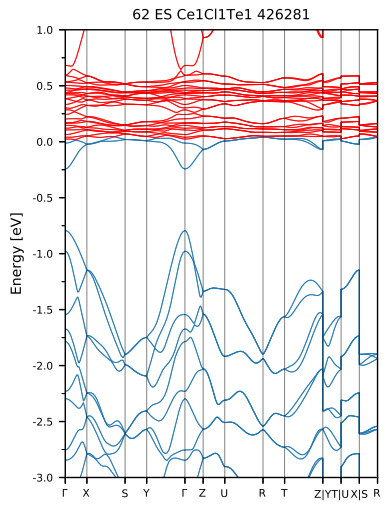

(rk) 62.3.222

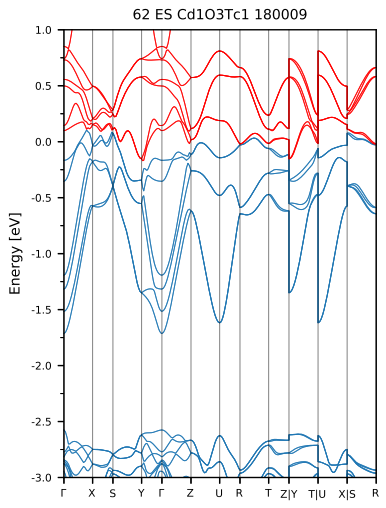

(rl) 62.3.223

Figure 138: Band structures for topological materials in SG 62, part 30/36, from 62.3.208 to 62.3.223. See the list of figures.

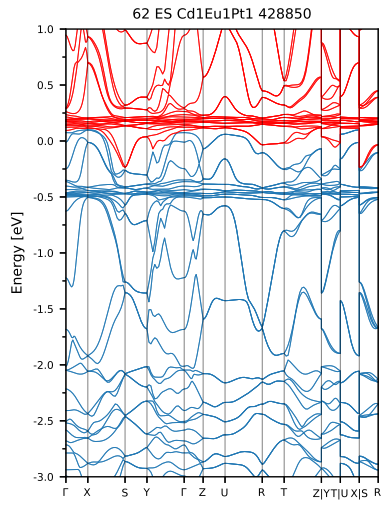

(rm) 62.3.224

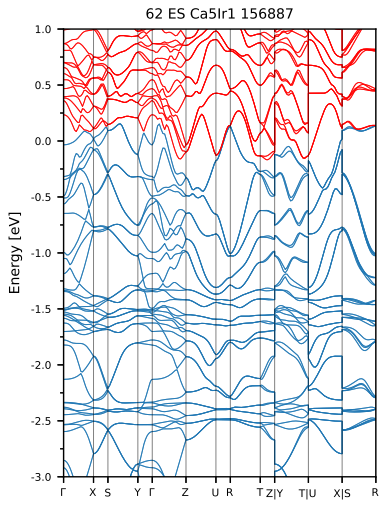

(rn) 62.3.225

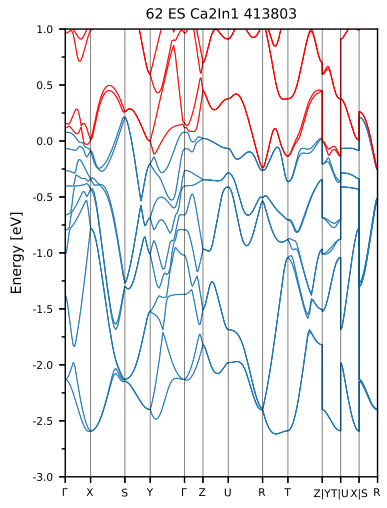

(ro) 62.3.226

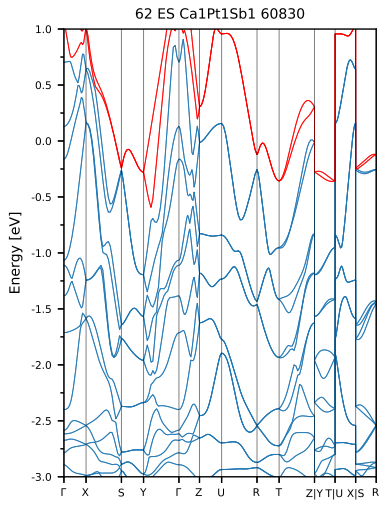

(rp) 62.3.227

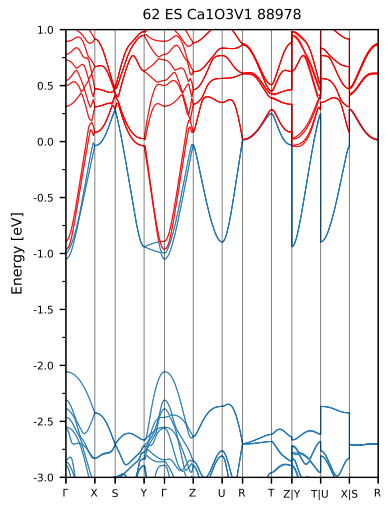

(rq) 62.3.228

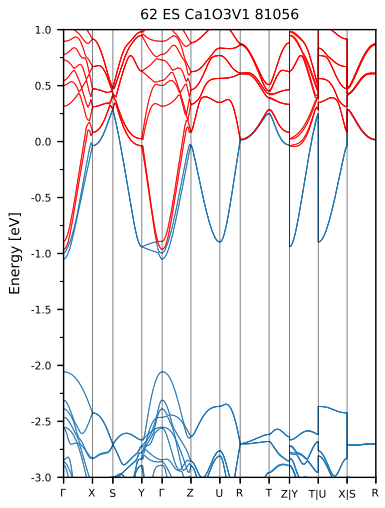

(rr) 62.3.229

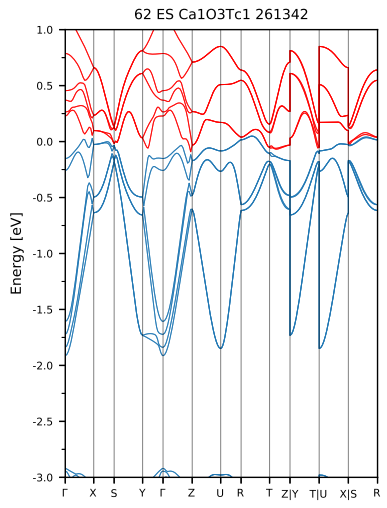

(rs) 62.3.230

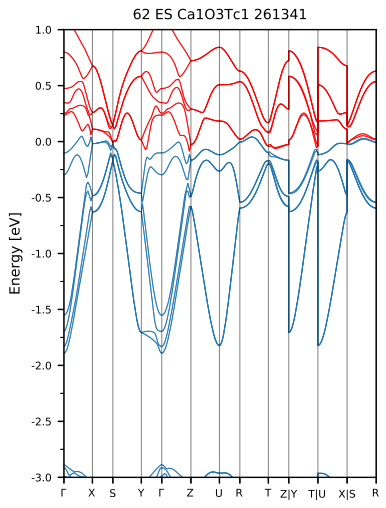

(rt) 62.3.231

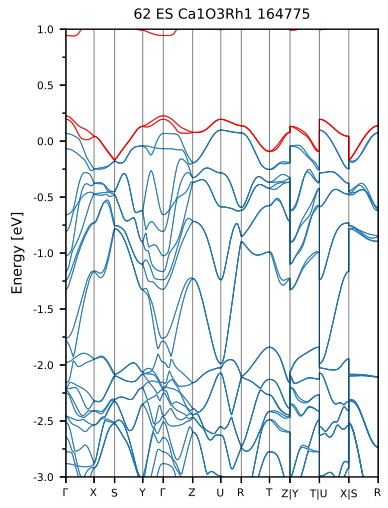

(ru) 62.3.232

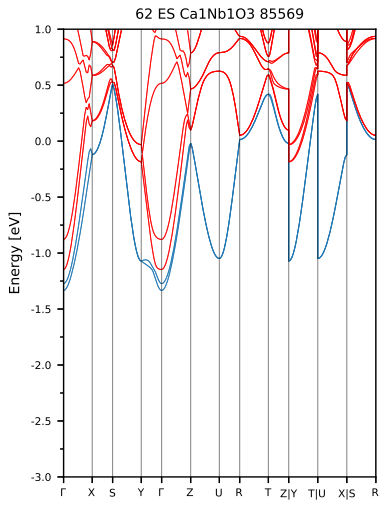

(rv) 62.3.233

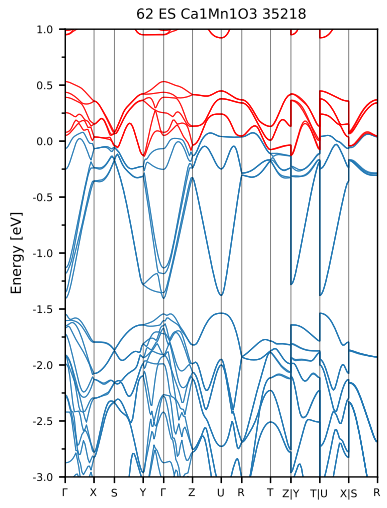

(rw) 62.3.234

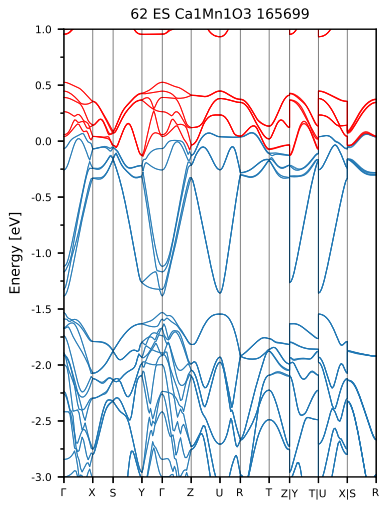

(rx) 62.3.235

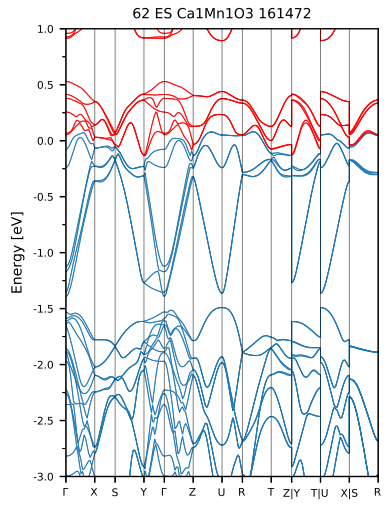

(ry) 62.3.236

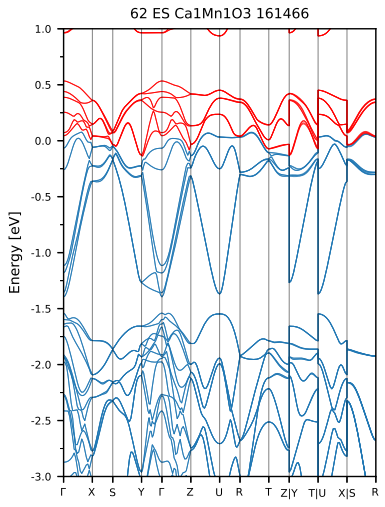

(rz) 62.3.237

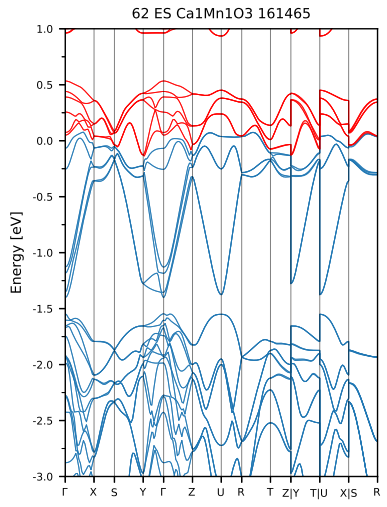

(sa) 62.3.238

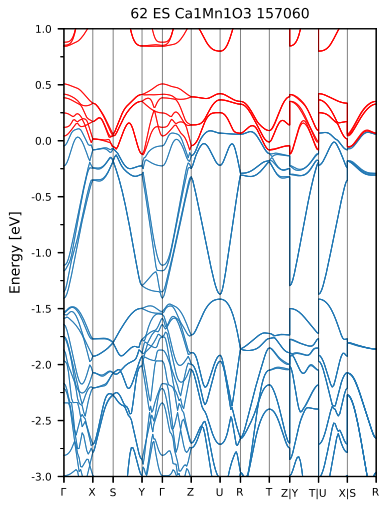

(sb) 62.3.239

Figure 139: Band structures for topological materials in SG 62, part 31/36, from 62.3.224 to 62.3.239. See the list of figures.

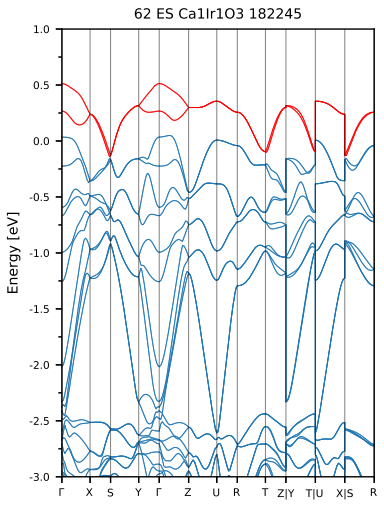

(sc) 62.3.240

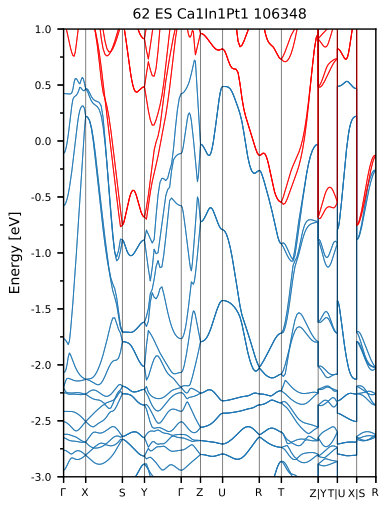

(sd) 62.3.241

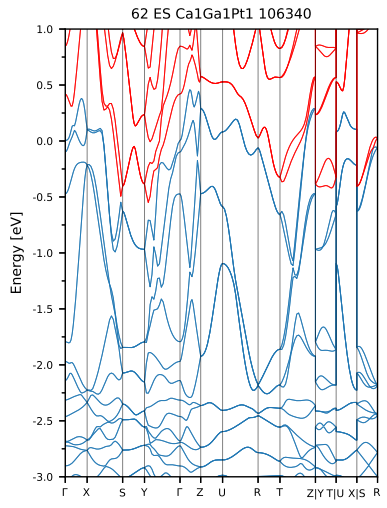

(se) 62.3.242

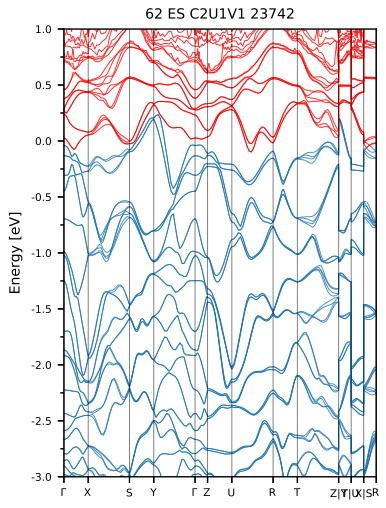

(sf) 62.3.243

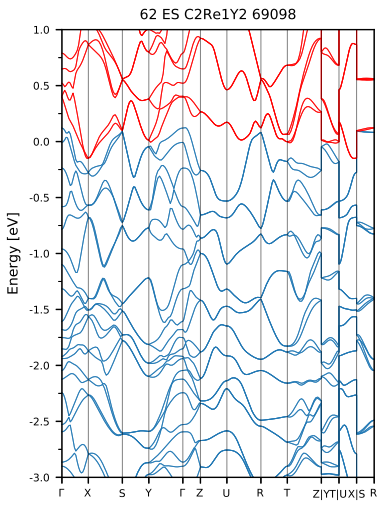

(sg) 62.3.244

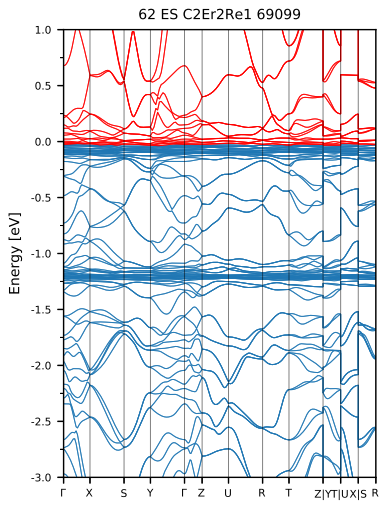

(sh) 62.3.245

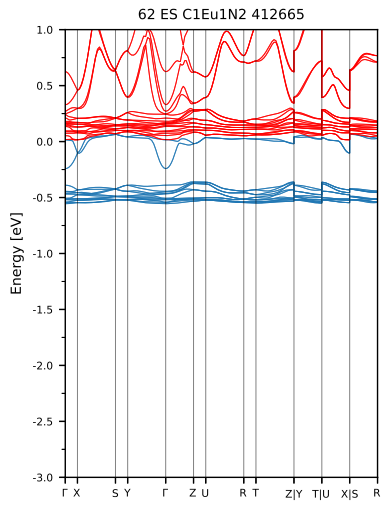

(si) 62.3.246

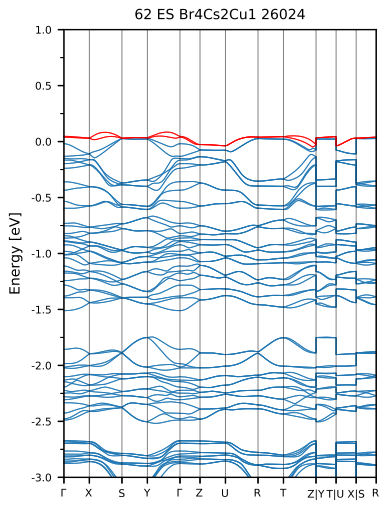

(sj) 62.3.247

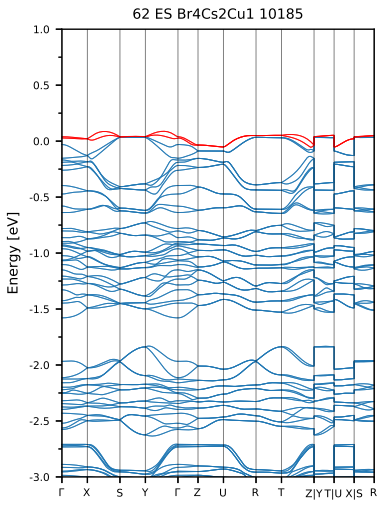

(sk) 62.3.248

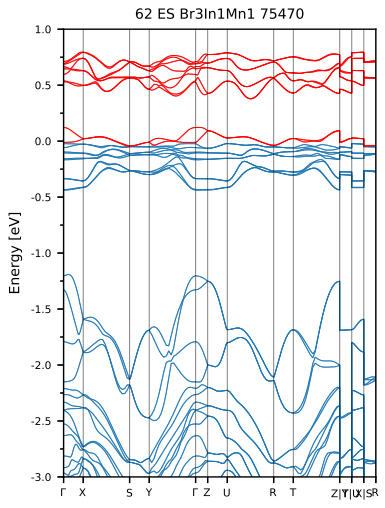

(sl) 62.3.249

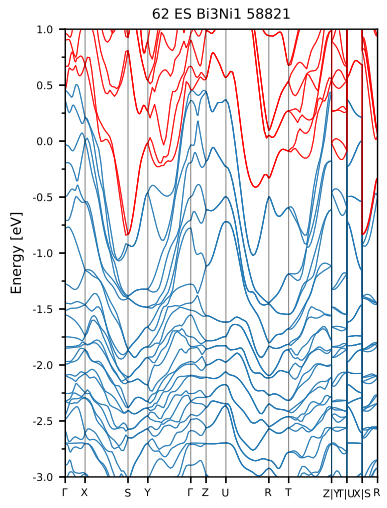

(sm) 62.3.250

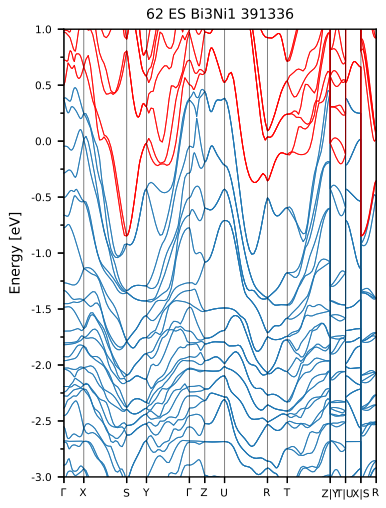

(sn) 62.3.251

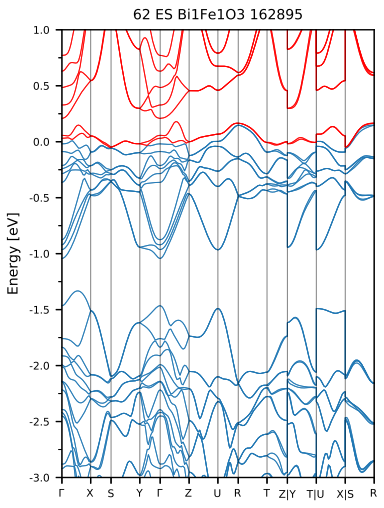

(so) 62.3.252

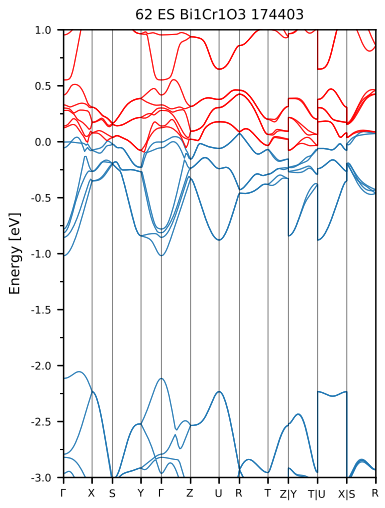

(sp) 62.3.253

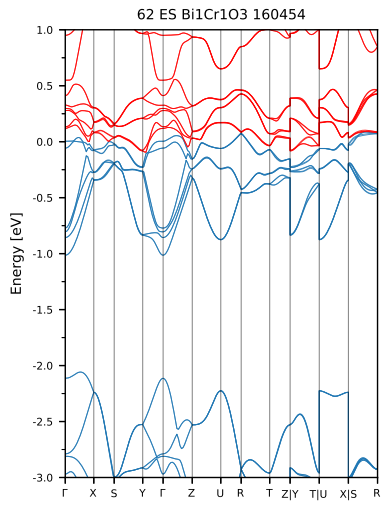

(sq) 62.3.254

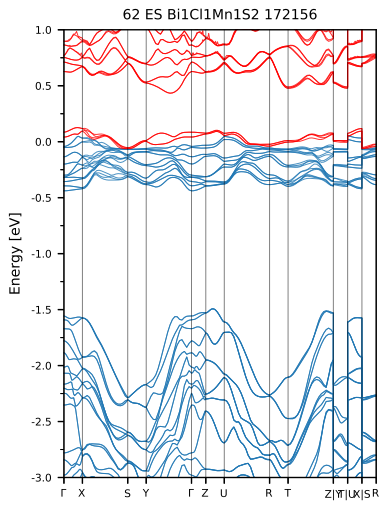

(sr) 62.3.255

Figure 140: Band structures for topological materials in SG 62, part 32/36, from 62.3.240 to 62.3.255. See the list of figures.

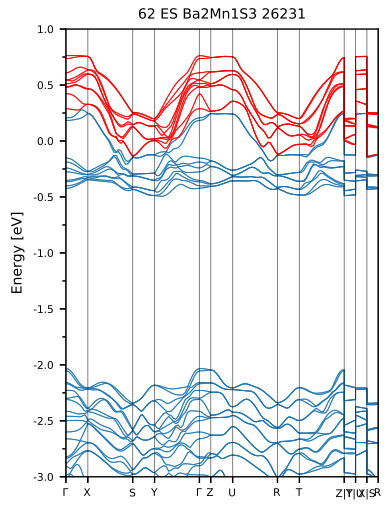

(ss) 62.3.256

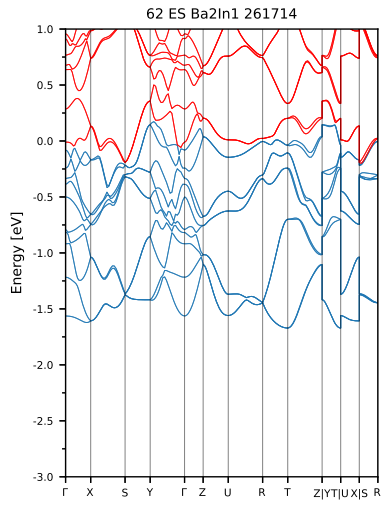

(st) 62.3.257

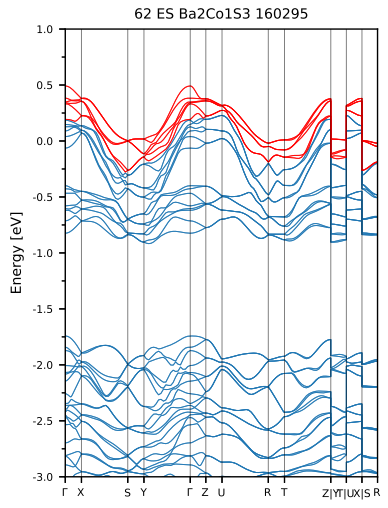

(su) 62.3.258

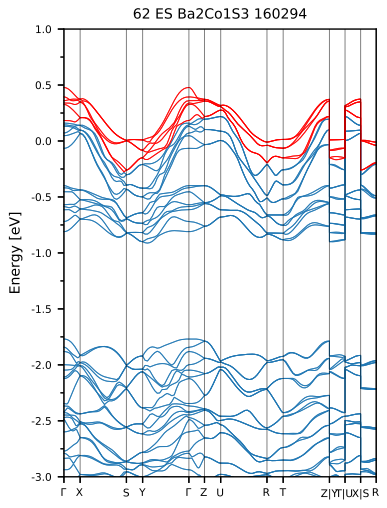

(sv) 62.3.259

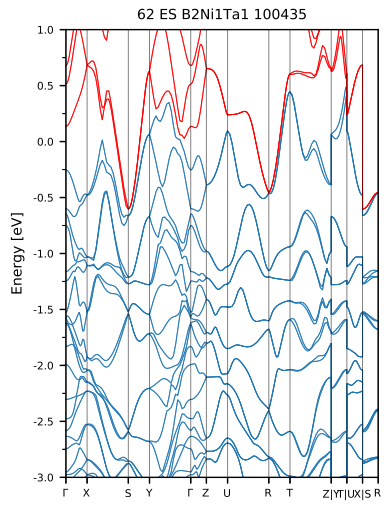

(sw) 62.3.260

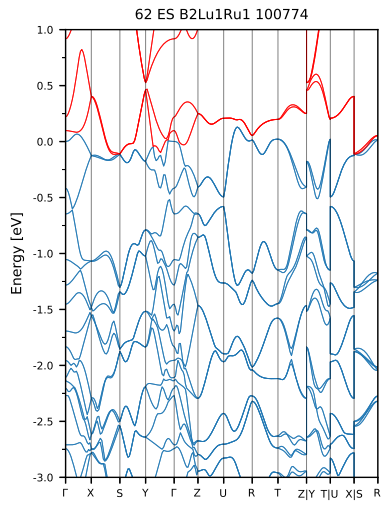

(sx) 62.3.261

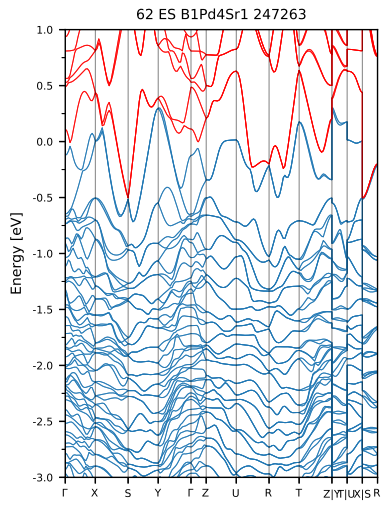

(sy) 62.3.262

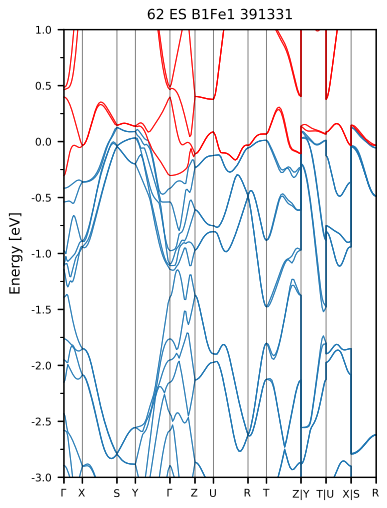

(sz) 62.3.263

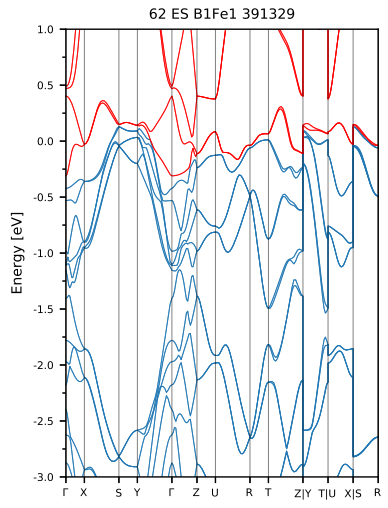

(ta) 62.3.264

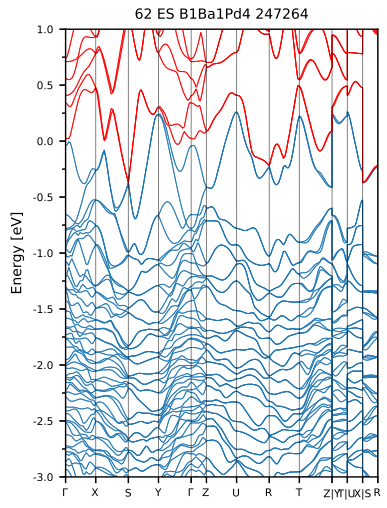

(tb) 62.3.265

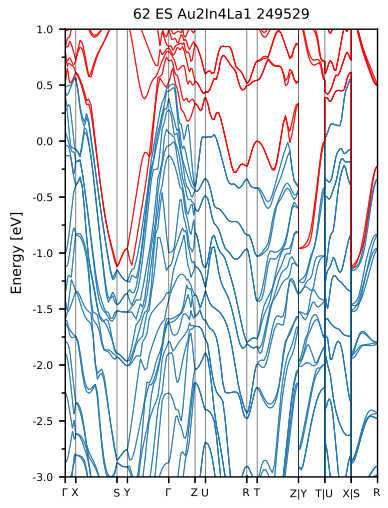

(tc) 62.3.266

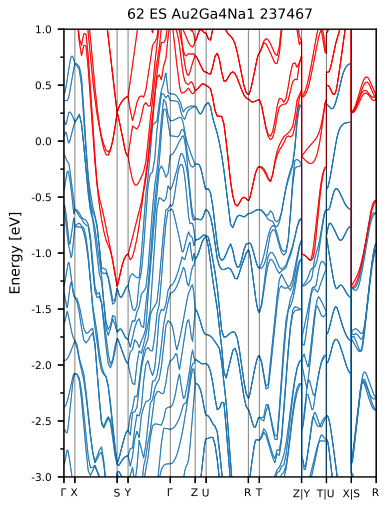

(td) 62.3.267

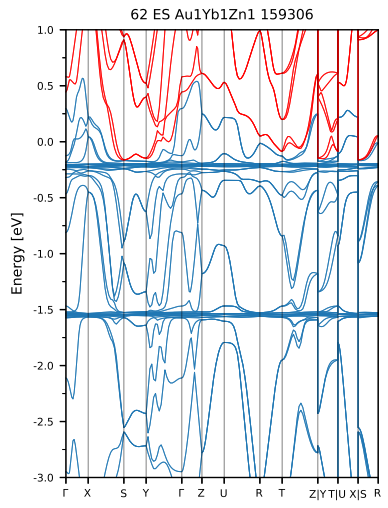

(te) 62.3.268

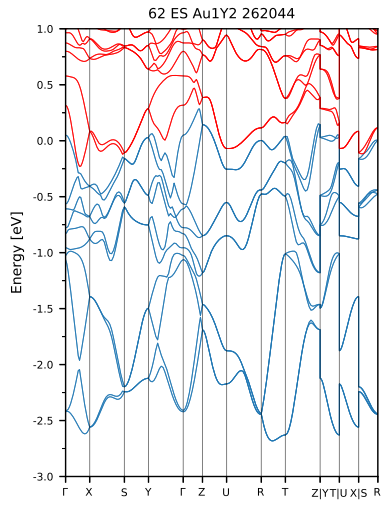

(tf) 62.3.269

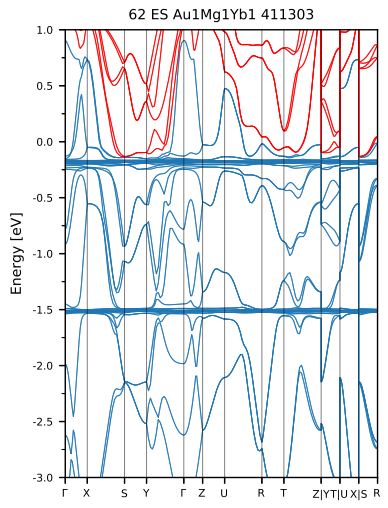

(tg) 62.3.270

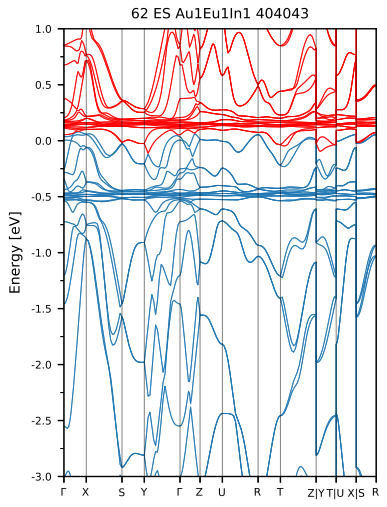

(th) 62.3.271

Figure 141: Band structures for topological materials in SG 62, part 33/36, from 62.3.256 to 62.3.271. See the list of figures.

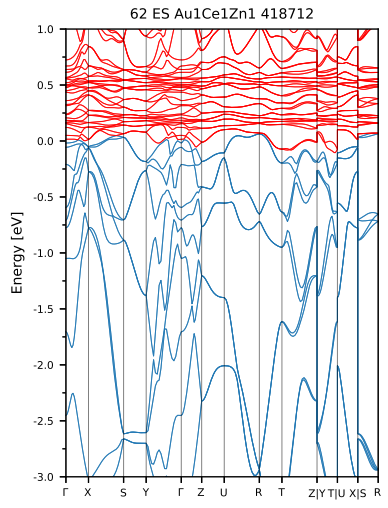

(ti) 62.3.272

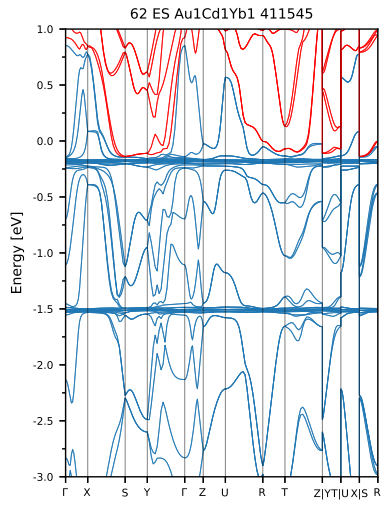

(tj) 62.3.273

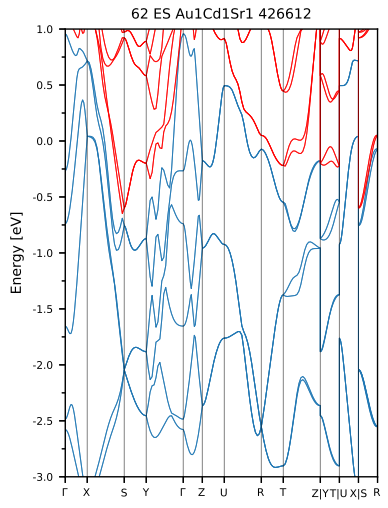

(tk) 62.3.274

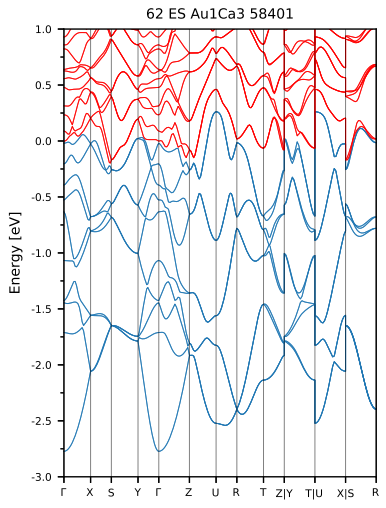

(tl) 62.3.275

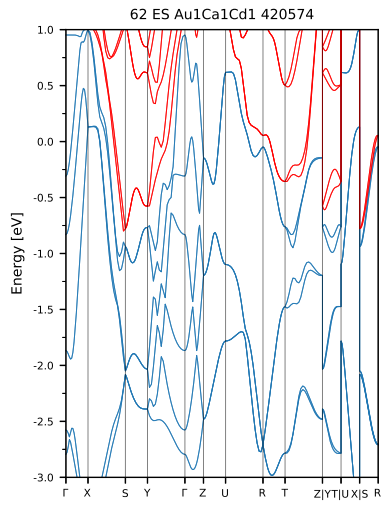

(tm) 62.3.276

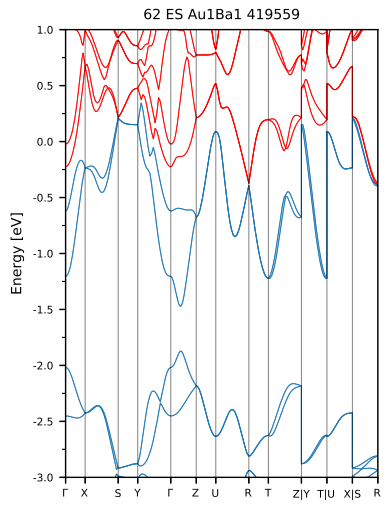

(tn) 62.3.277

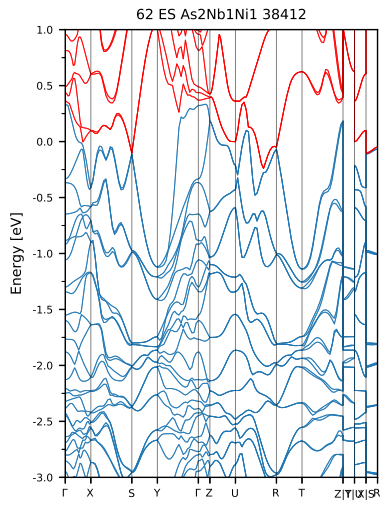

(to) 62.3.278

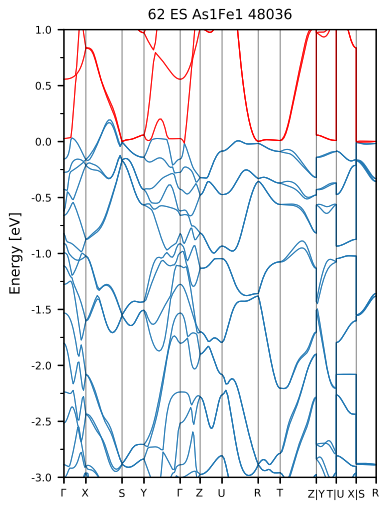

(tp) 62.3.279

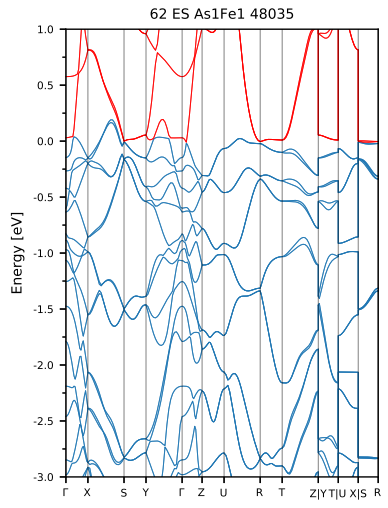

(tq) 62.3.280

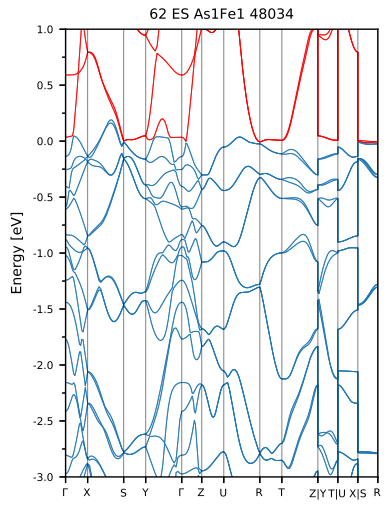

(tr) 62.3.281

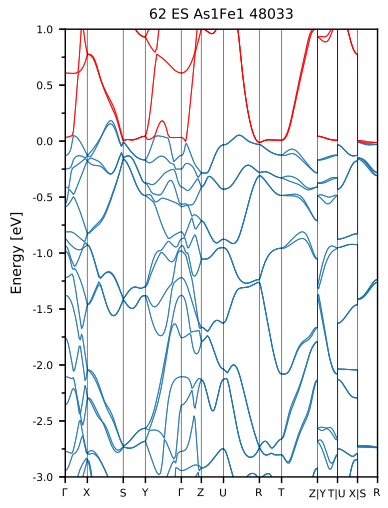

(ts) 62.3.282

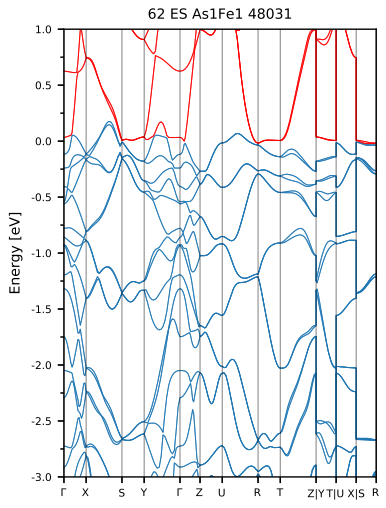

(tt) 62.3.283

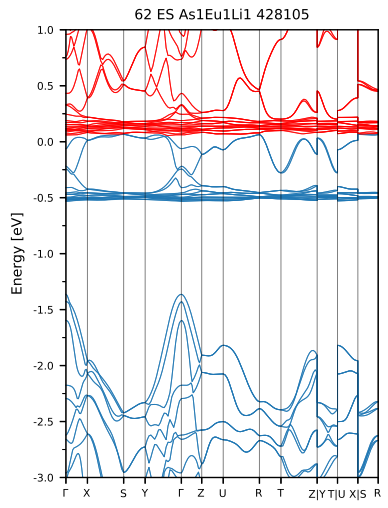

(tu) 62.3.284

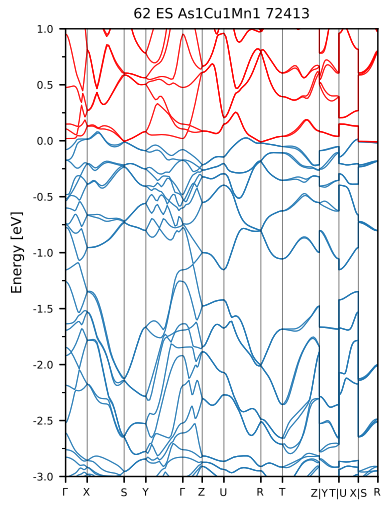

(tv) 62.3.285

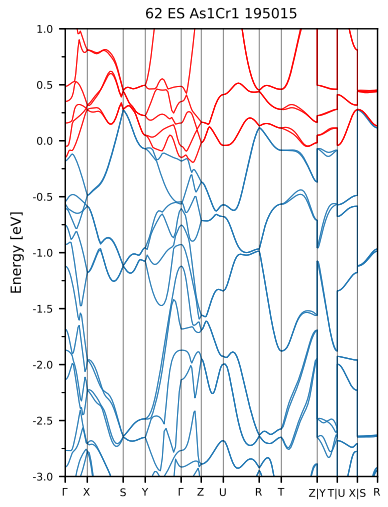

(tw) 62.3.286

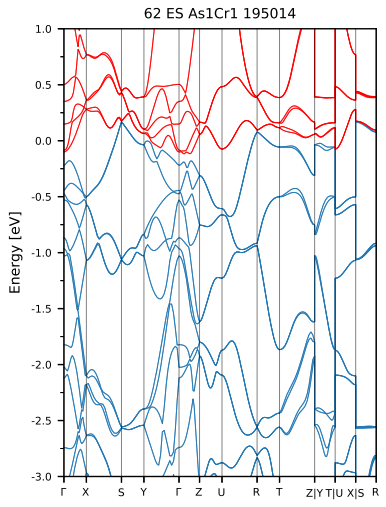

(tx) 62.3.287

Figure 142: Band structures for topological materials in SG 62, part 34/36, from 62.3.272 to 62.3.287. See the list of figures.

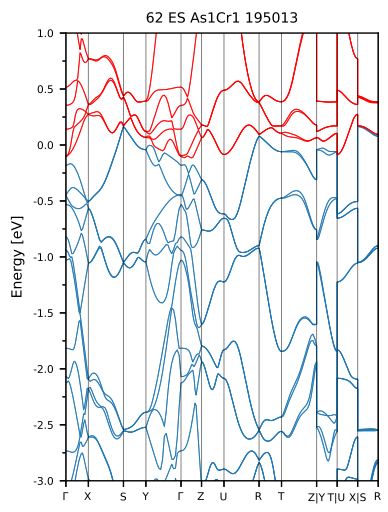

(ty) 62.3.288

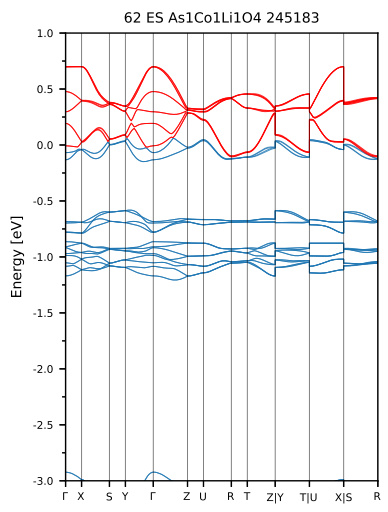

(tz) 62.3.289

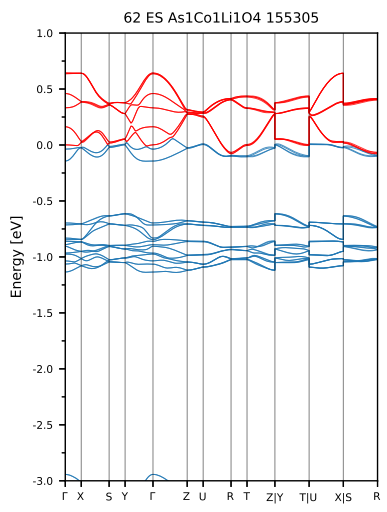

(ua) 62.3.290

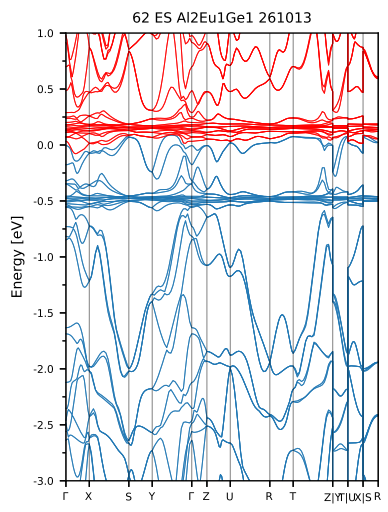

(ub) 62.3.291

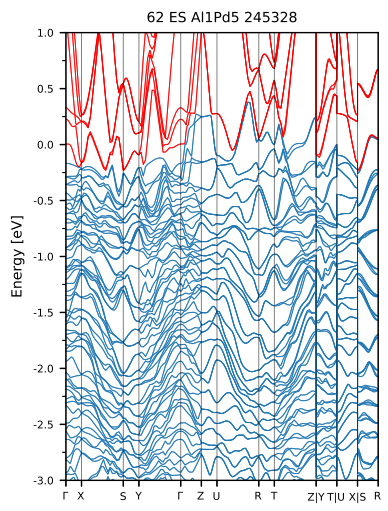

(uc) 62.3.292

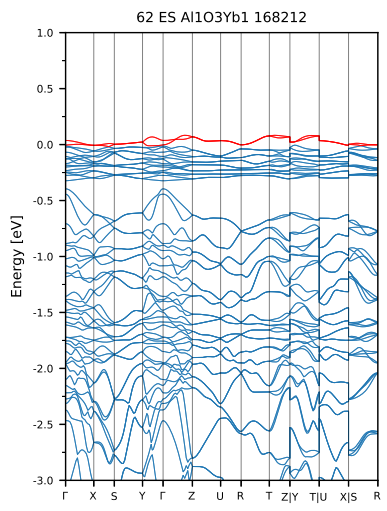

(ud) 62.3.293

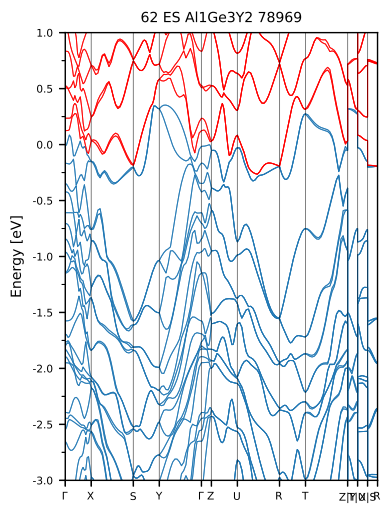

(ue) 62.3.294

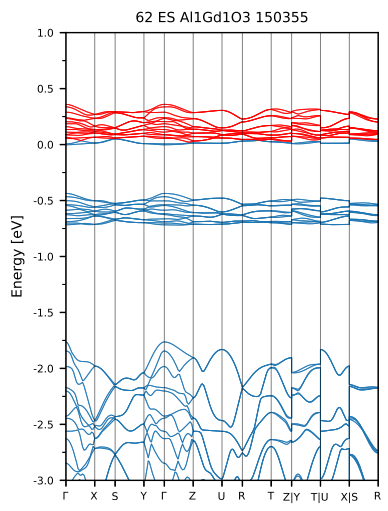

(uf) 62.3.295

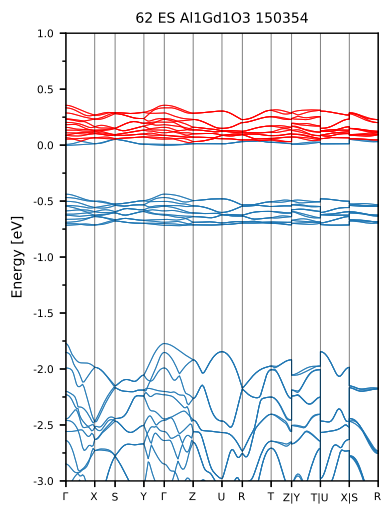

(ug) 62.3.296

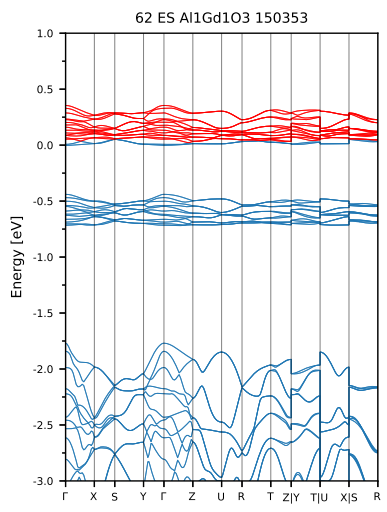

(uh) 62.3.297

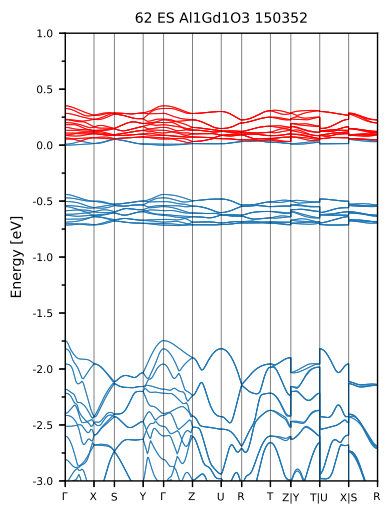

(ui) 62.3.298

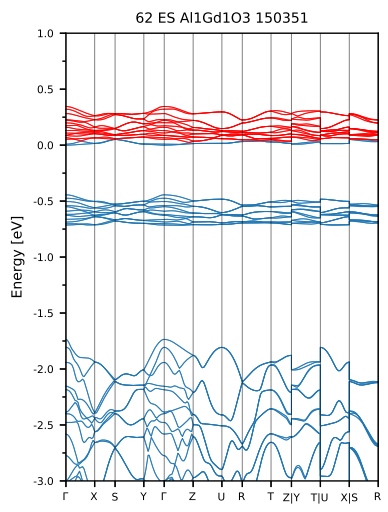

(uj) 62.3.299

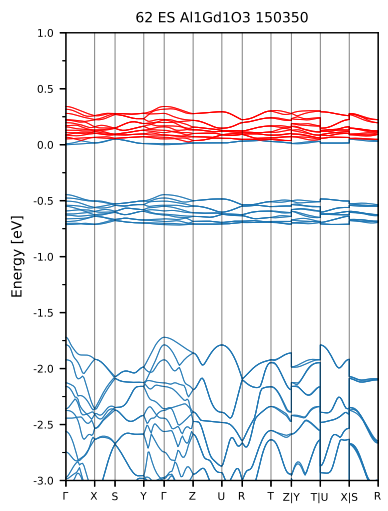

(uk) 62.3.300

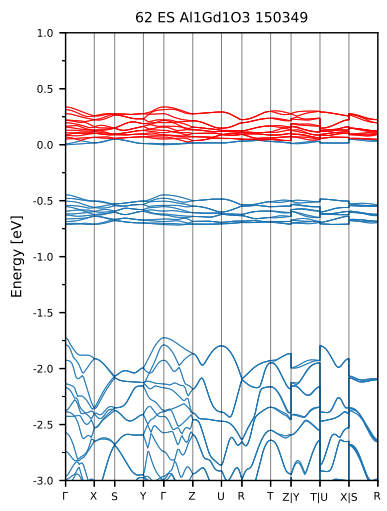

(ul) 62.3.301

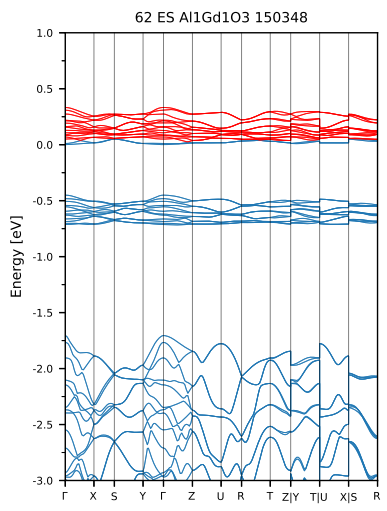

(um) 62.3.302

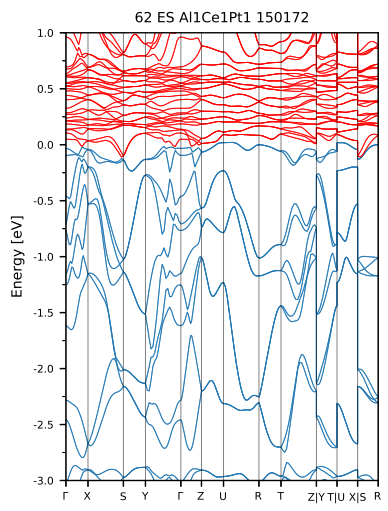

(un) 62.3.303

Figure 143: Band structures for topological materials in SG 62, part 35/36, from 62.3.288 to 62.3.303. See the list of figures.

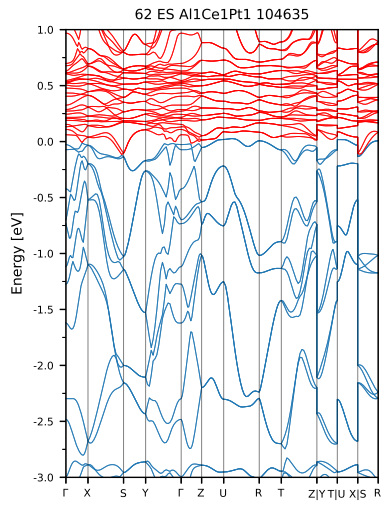

(uo) 62.3.304

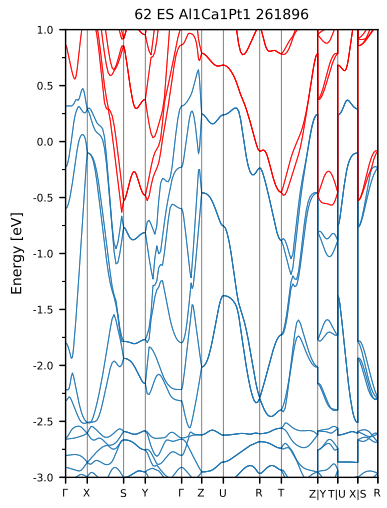

(up) 62.3.305

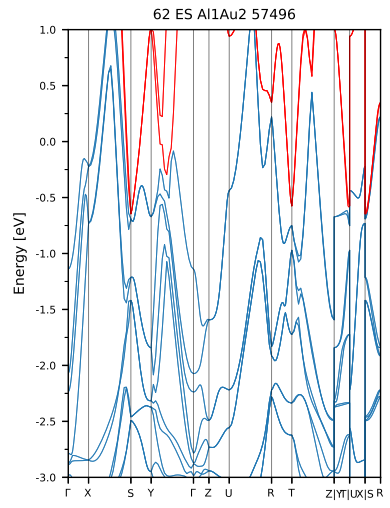

(uq) 62.3.306

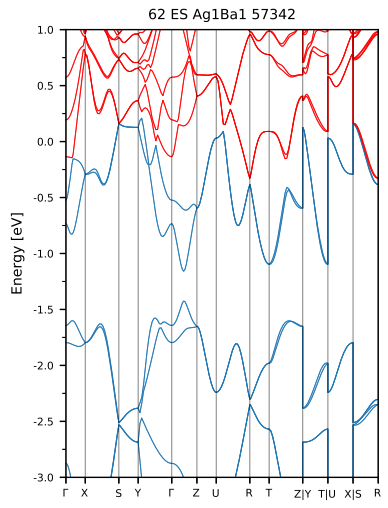

(ur) 62.3.307

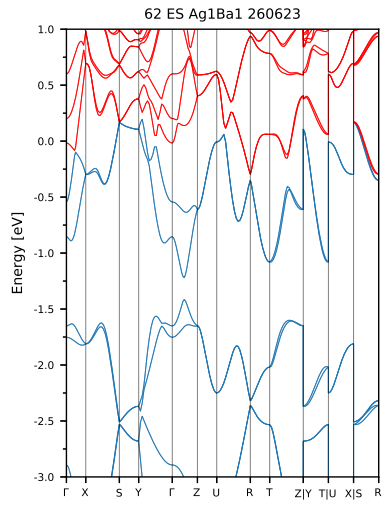

(us) 62.3.308

Figure 144: Band structures for topological materials in SG 62, part 36/36, from 62.3.304 to 62.3.308. See the list of figures.

### 0.0.27 SG 63

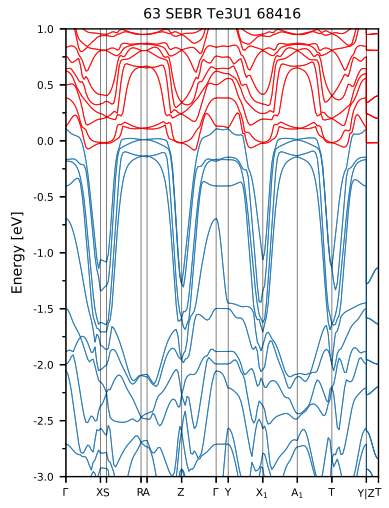

(a) 63.1.1

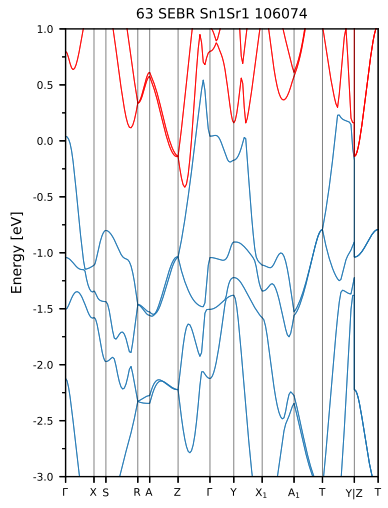

(b) 63.1.2

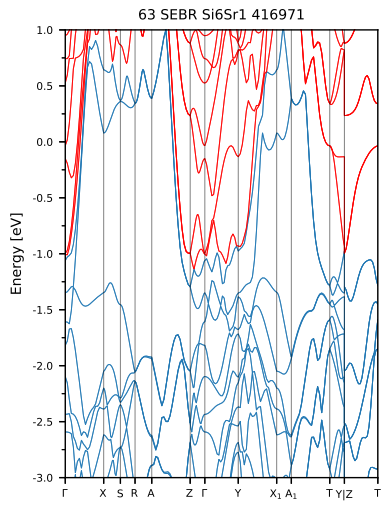

(c) 63.1.3

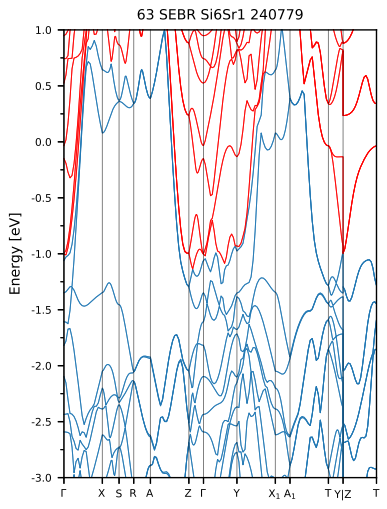

(d) 63.1.4

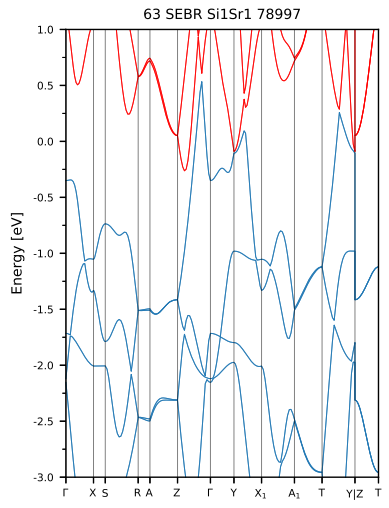

(e) 63.1.5

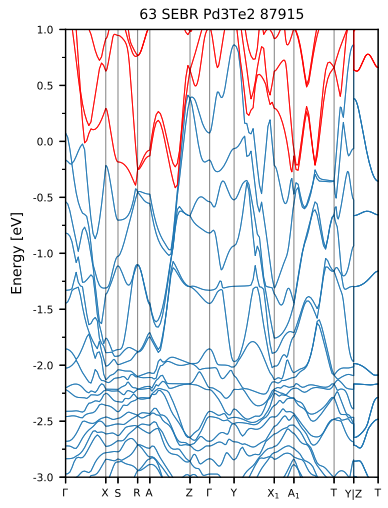

(f) 63.1.6

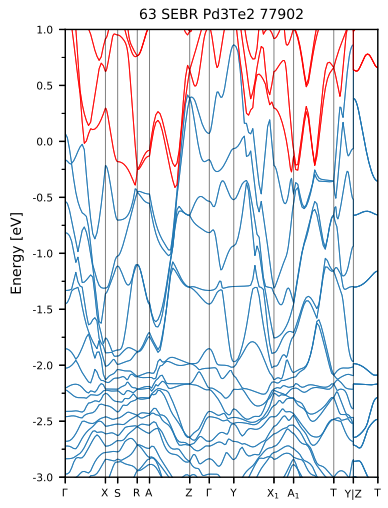

(g) 63.1.7

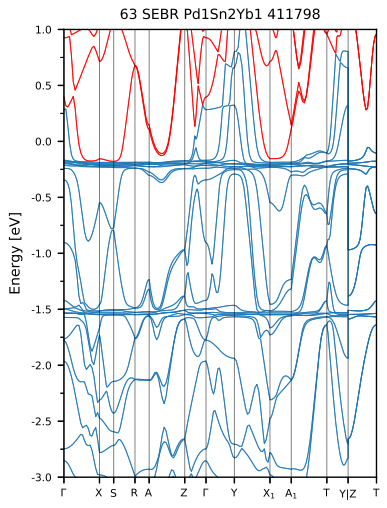

(h) 63.1.8

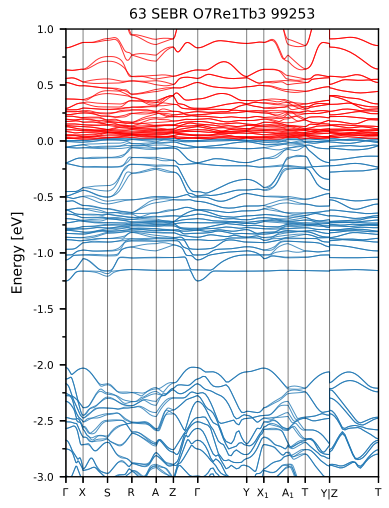

(i) 63.1.9

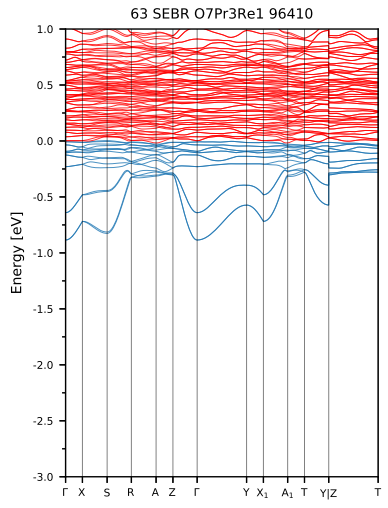

(j) 63.1.10

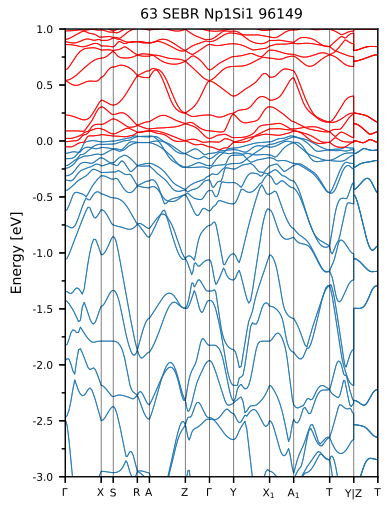

(k) 63.1.11

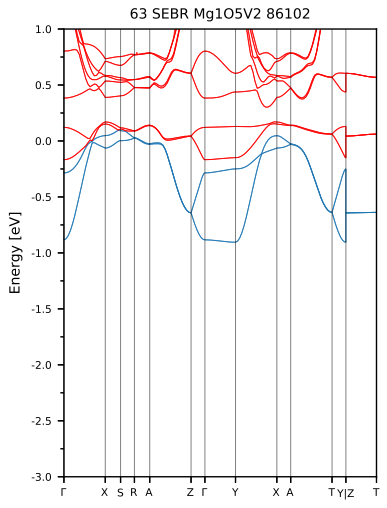

(l) 63.1.12

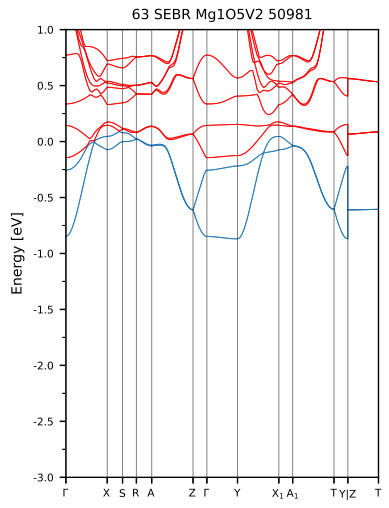

(m) 63.1.13

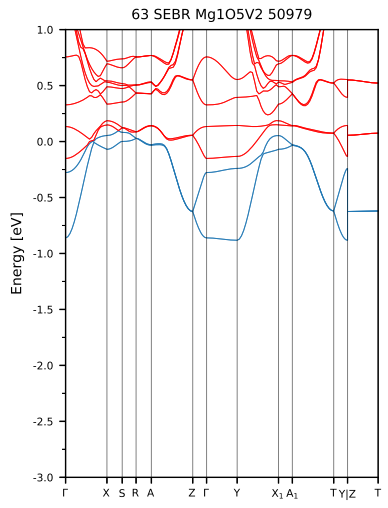

(n) 63.1.14

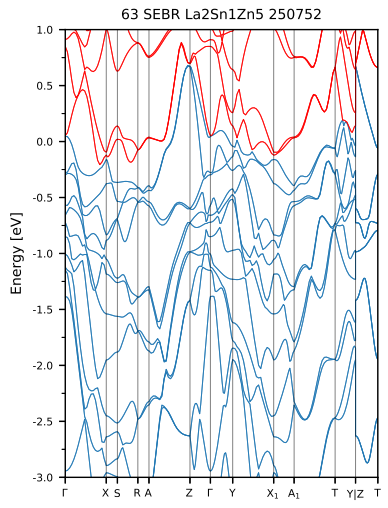

(o) 63.1.15

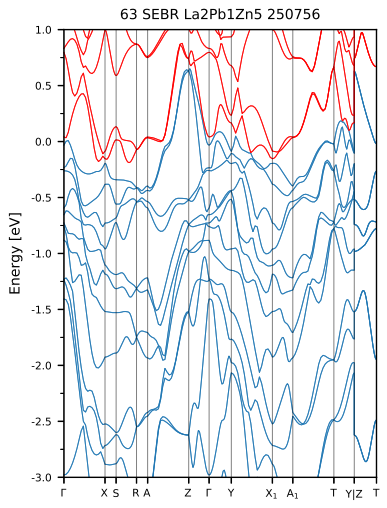

(p) 63.1.16

Figure 145: Band structures for topological materials in SG 63, part 1/29, from 63.1.1 to 63.1.16. See the list of figures.

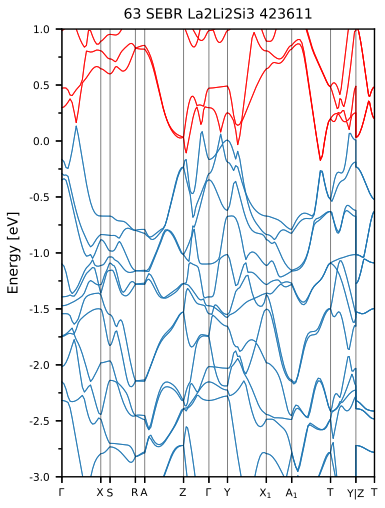

(q) 63.1.17

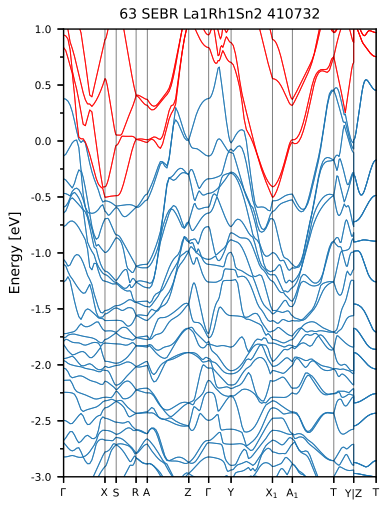

(r) 63.1.18

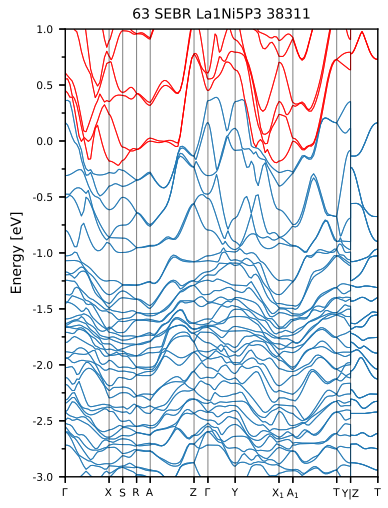

(s) 63.1.19

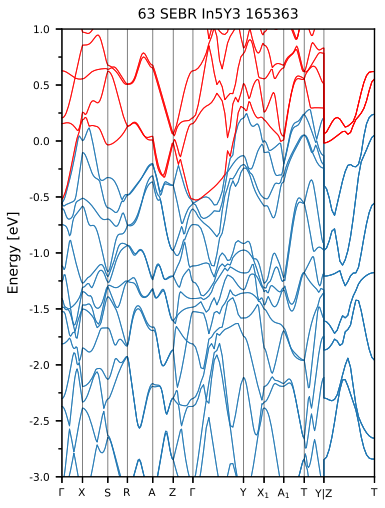

(t) 63.1.20

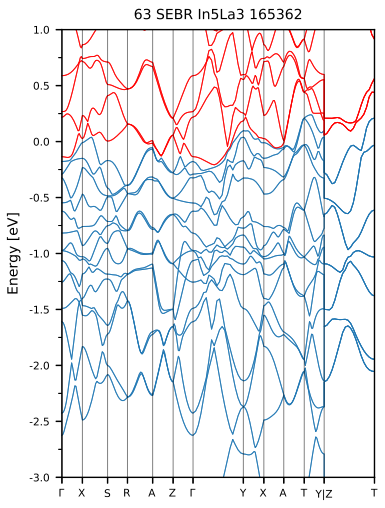

(u) 63.1.21

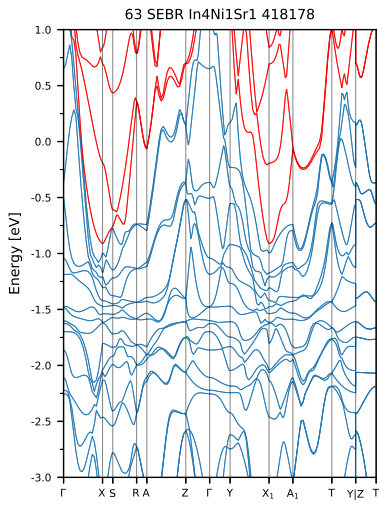

(v) 63.1.22

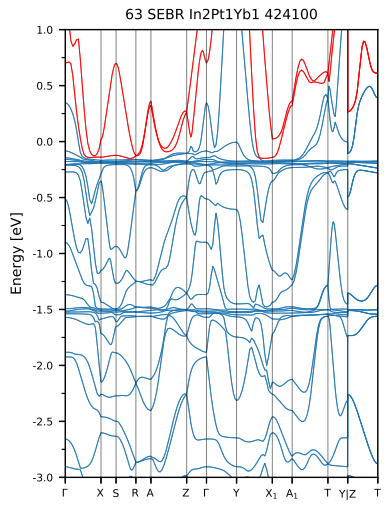

(w) 63.1.23

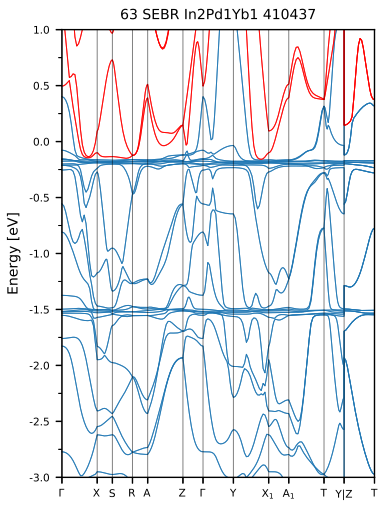

(x) 63.1.24

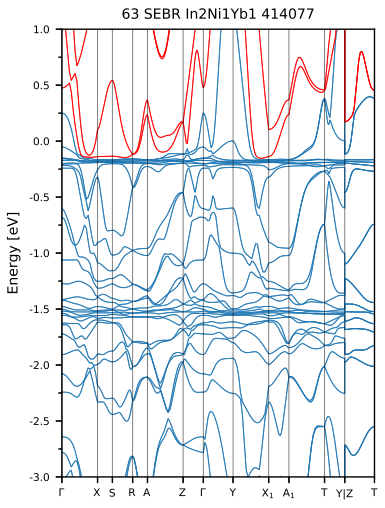

(y) 63.1.25

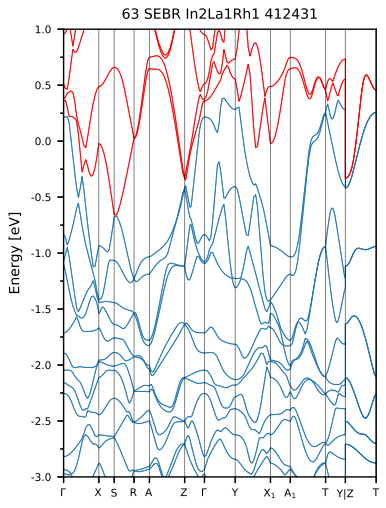

(z) 63.1.26

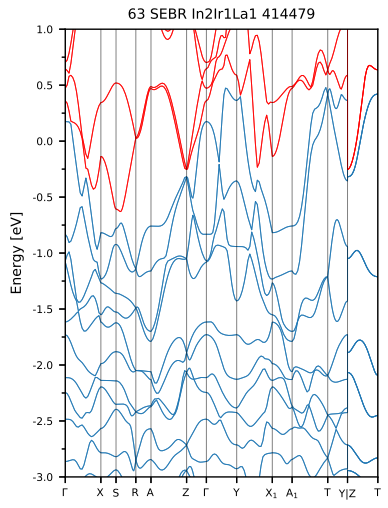

(aa) 63.1.27

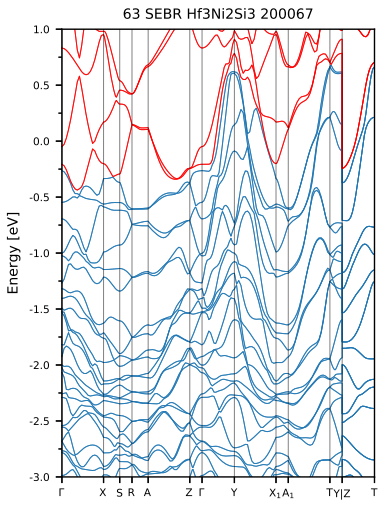

(ab) 63.1.28

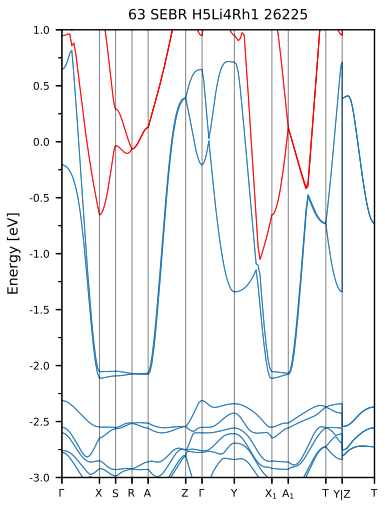

(ac) 63.1.29

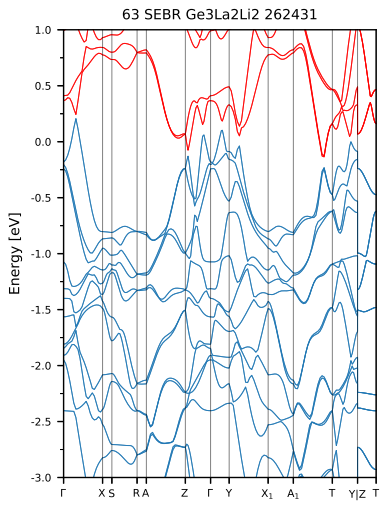

(ad) 63.1.30

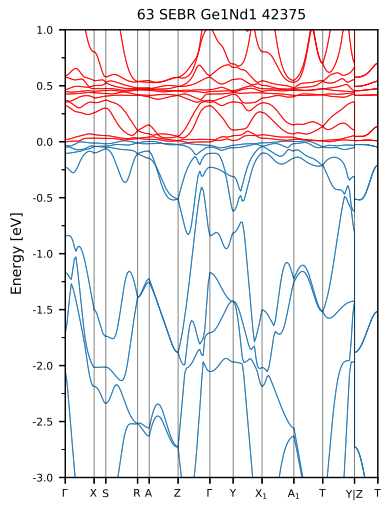

(ae) 63.1.31

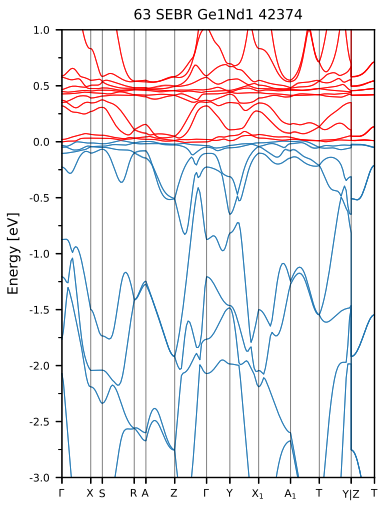

(af) 63.1.32

Figure 146: Band structures for topological materials in SG 63, part 2/29, from 63.1.17 to 63.1.32. See the list of figures.

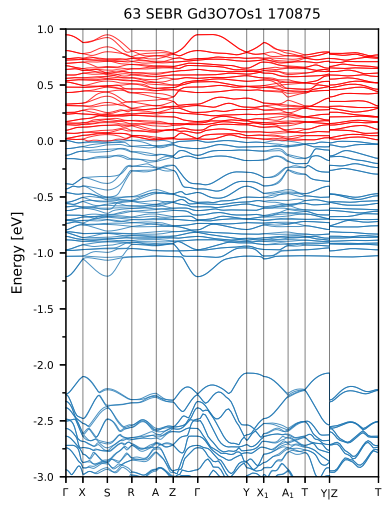

(ag) 63.1.33

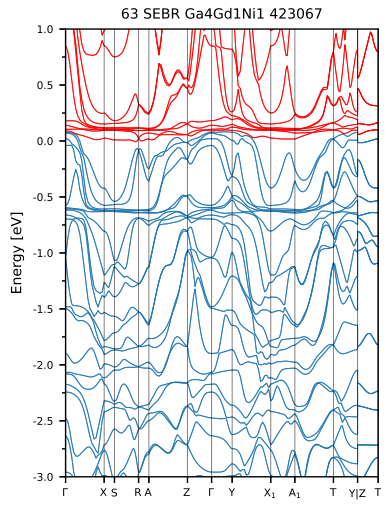

(ah) 63.1.34

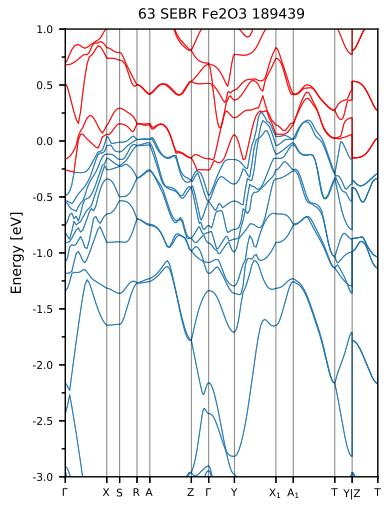

(ai) 63.1.35

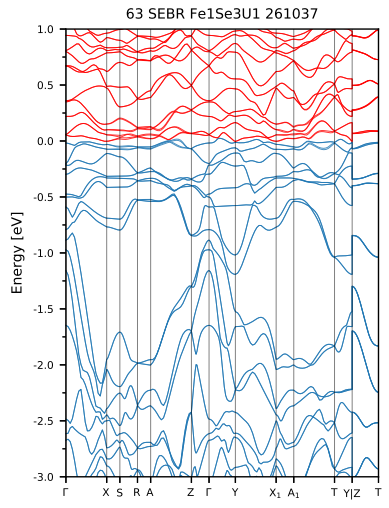

(aj) 63.1.36

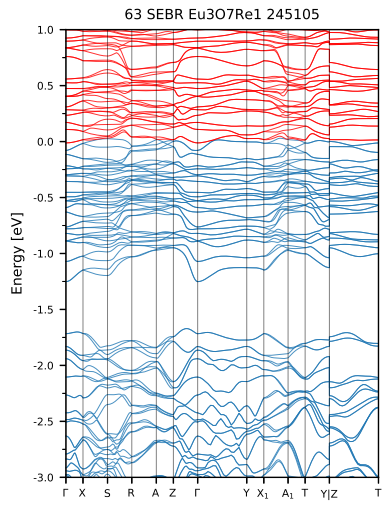

(ak) 63.1.37

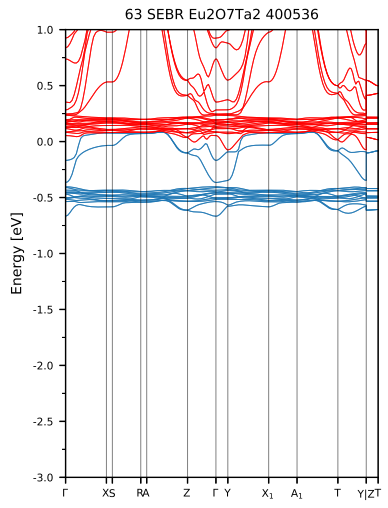

(al) 63.1.38

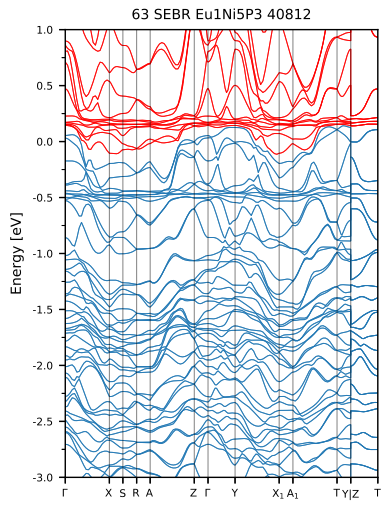

(am) 63.1.39

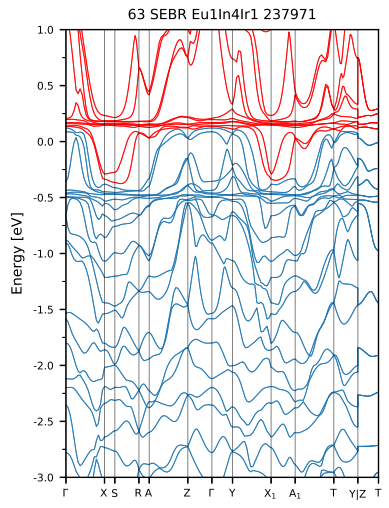

(an) 63.1.40

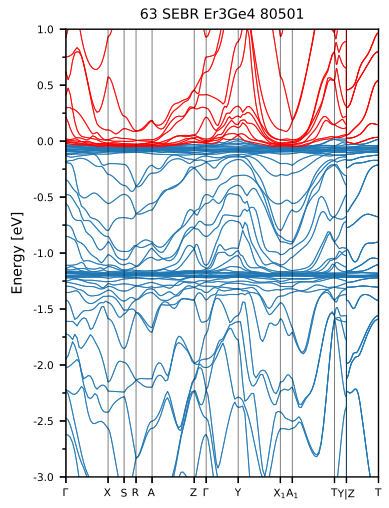

(ao) 63.1.41

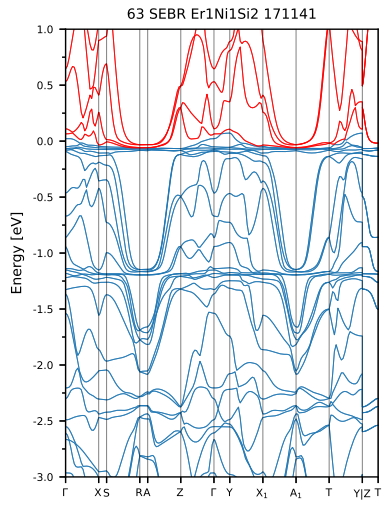

(ap) 63.1.42

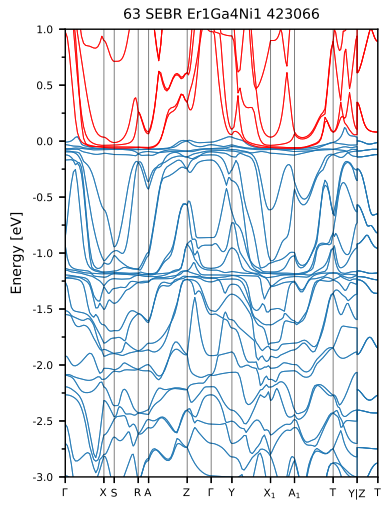

(aq) 63.1.43

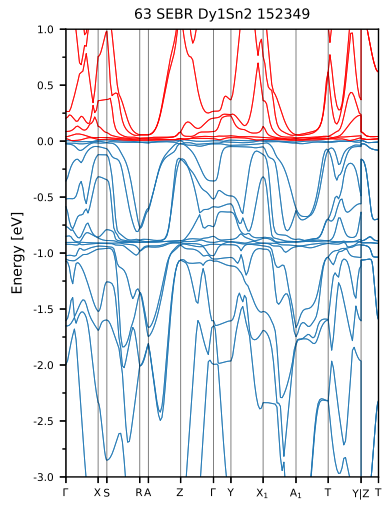

(ar) 63.1.44

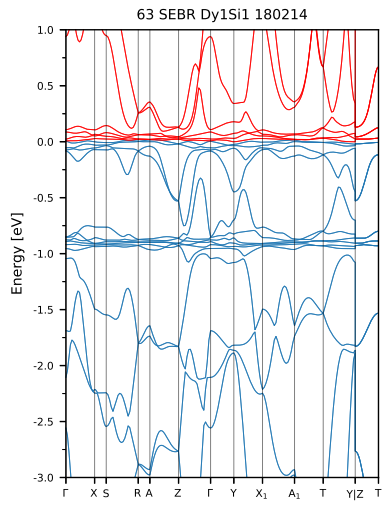

(as) 63.1.45

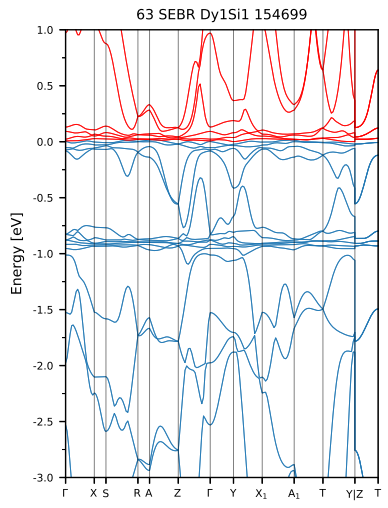

(at) 63.1.46

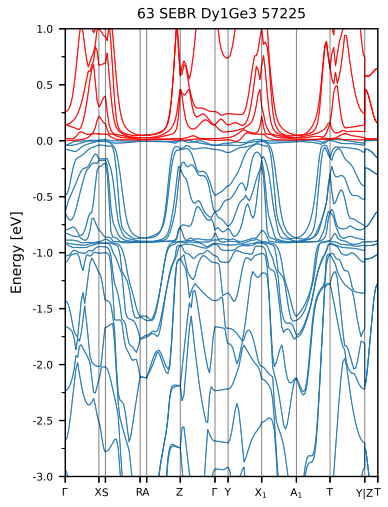

(au) 63.1.47

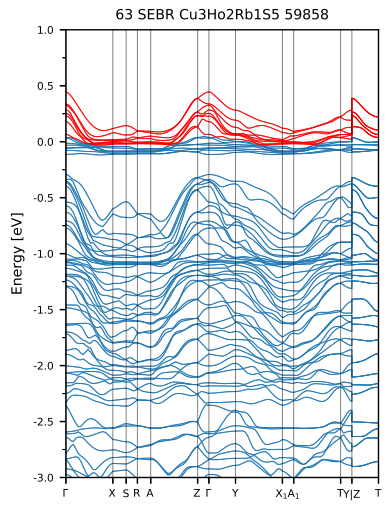

(av) 63.1.48

Figure 147: Band structures for topological materials in SG 63, part 3/29, from 63.1.33 to 63.1.48. See the list of figures.

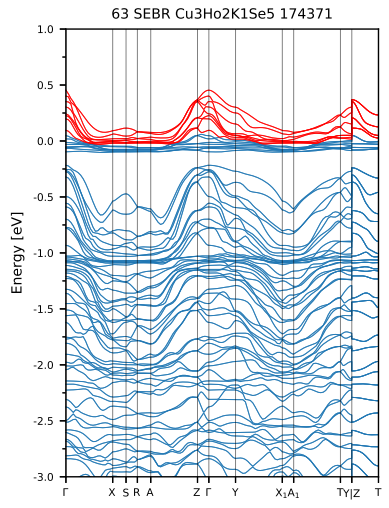

(aw) 63.1.49

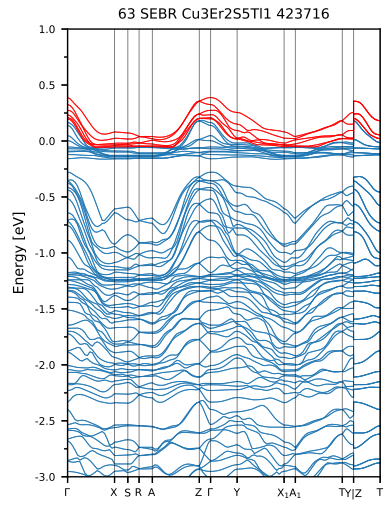

(ax) 63.1.50

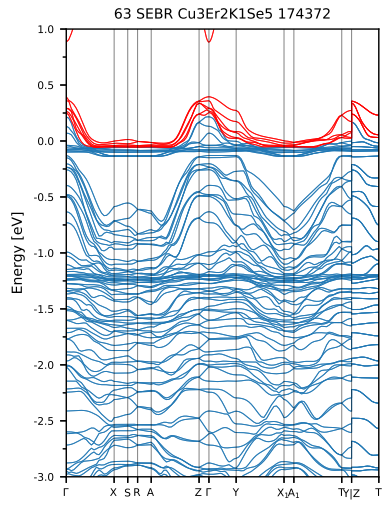

(ay) 63.1.51

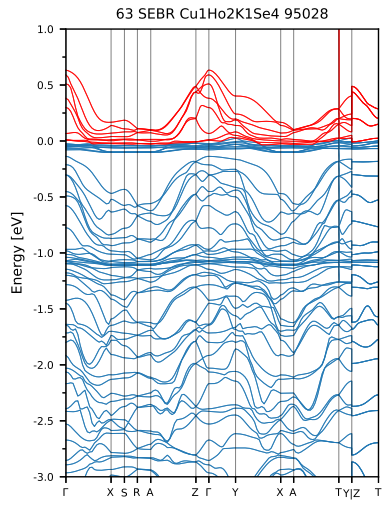

(az) 63.1.52

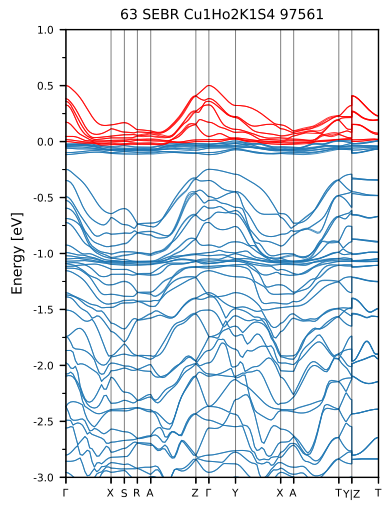

(ba) 63.1.53

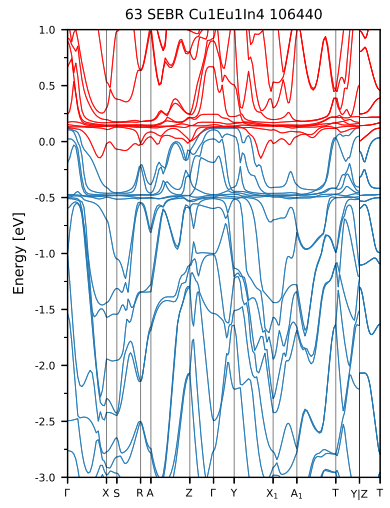

(bb) 63.1.54

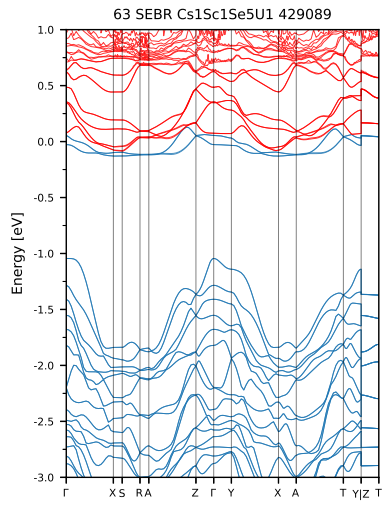

(bc) 63.1.55

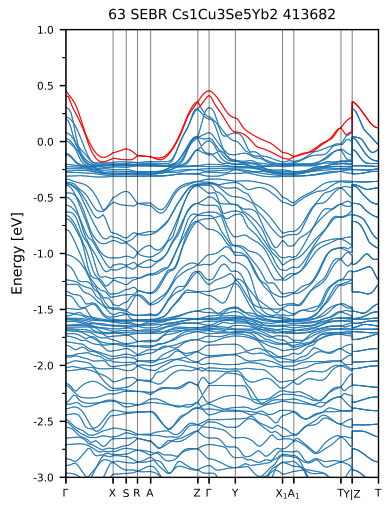

(bd) 63.1.56

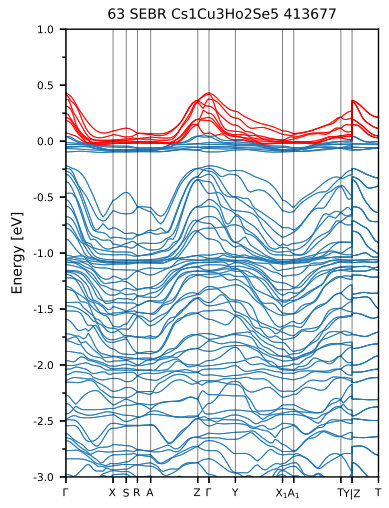

(be) 63.1.57

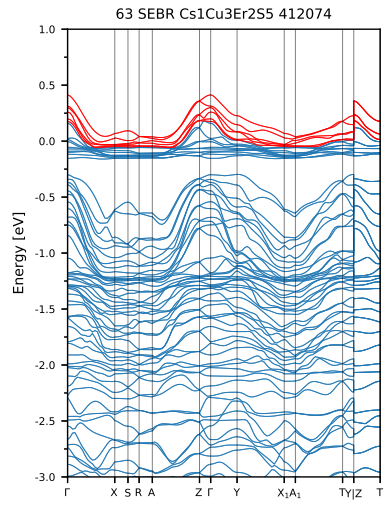

(bf) 63.1.58

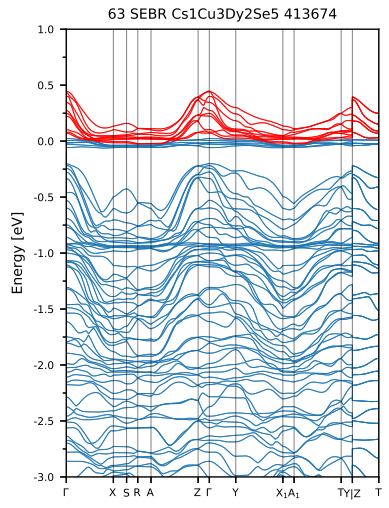

(bg) 63.1.59

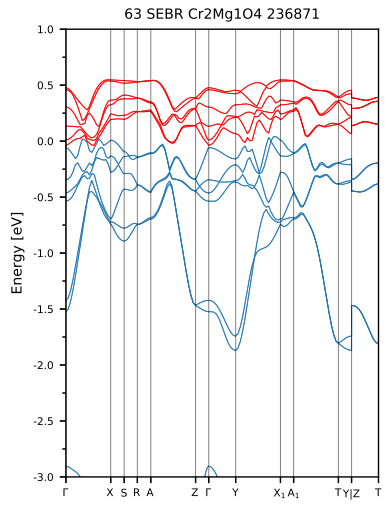

(bh) 63.1.60

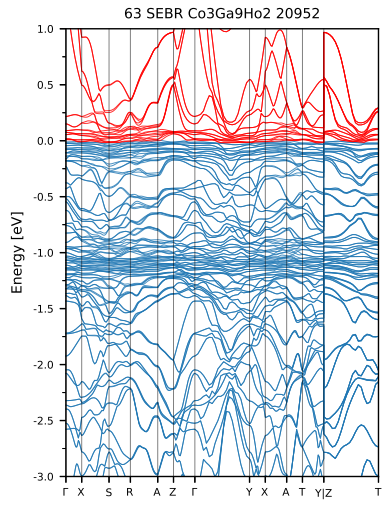

(bi) 63.1.61

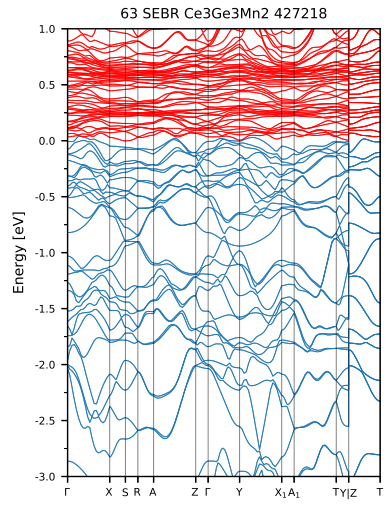

(bj) 63.1.62

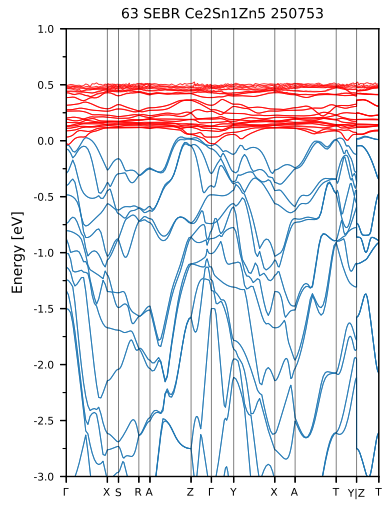

(bk) 63.1.63

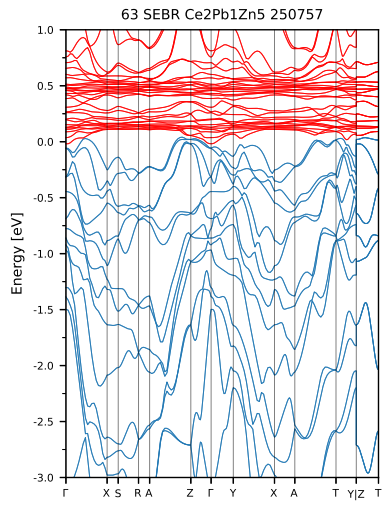

(bl) 63.1.64

Figure 148: Band structures for topological materials in SG 63, part 4/29, from 63.1.49 to 63.1.64. See the list of figures.

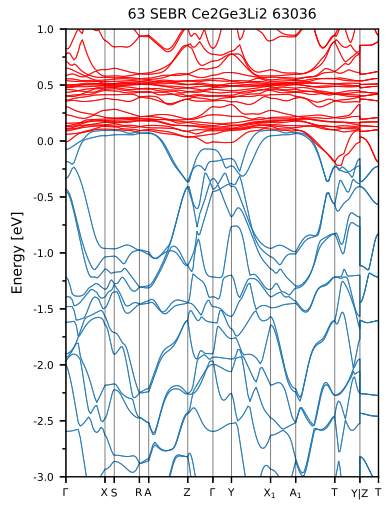

(bm) 63.1.65

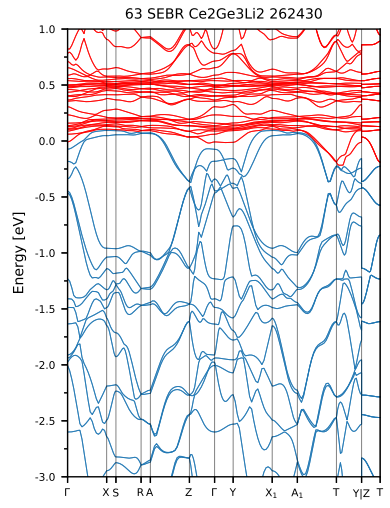

(bn) 63.1.66

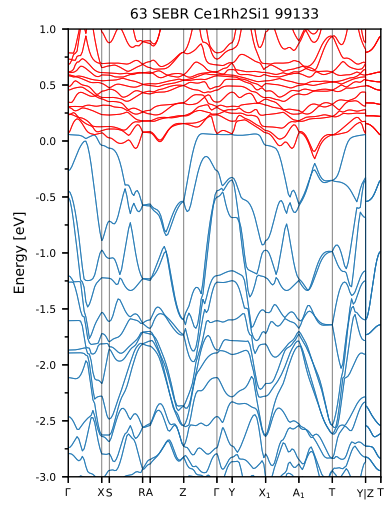

(bo) 63.1.67

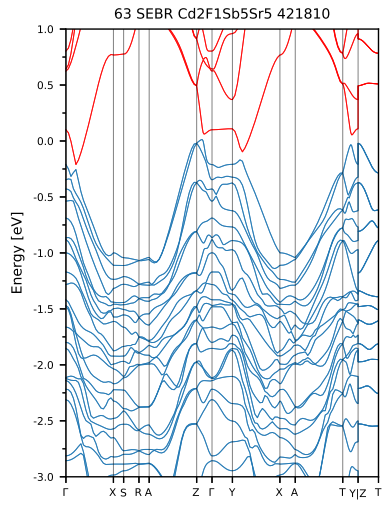

(bp) 63.1.68

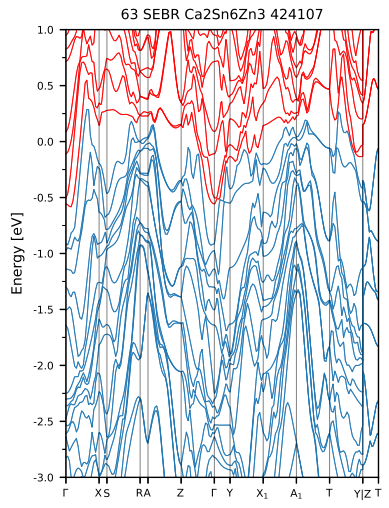

(bq) 63.1.69

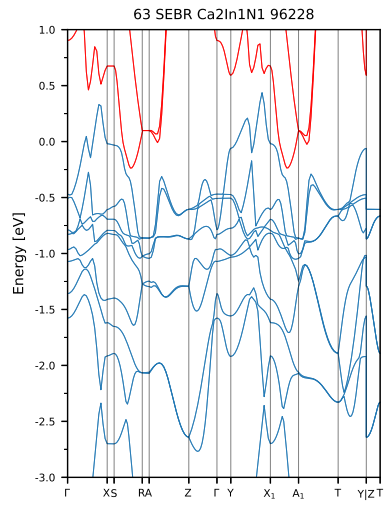

(br) 63.1.70

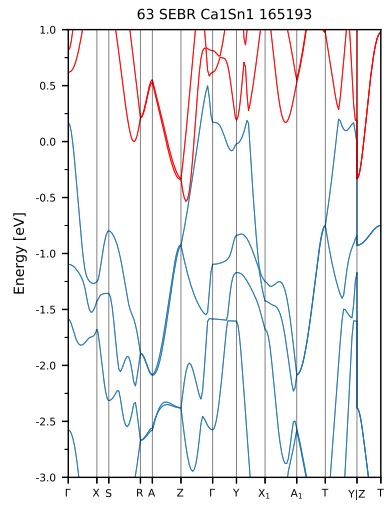

(bs) 63.1.71

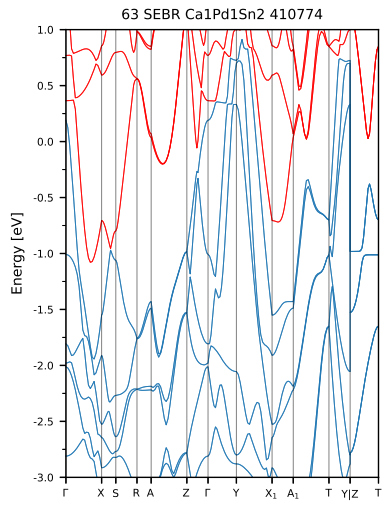

(bt) 63.1.72

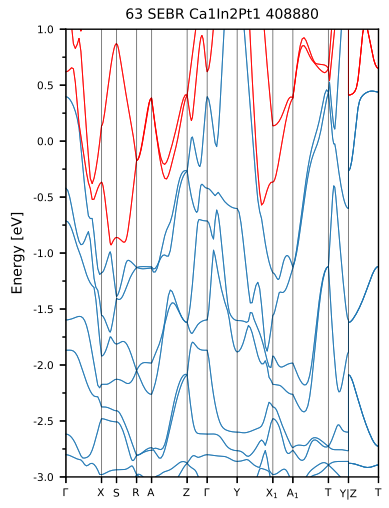

(bu) 63.1.73

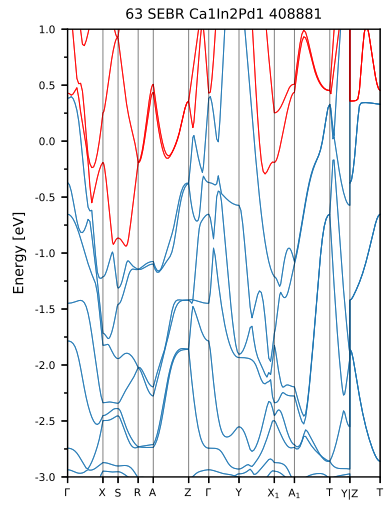

(bv) 63.1.74

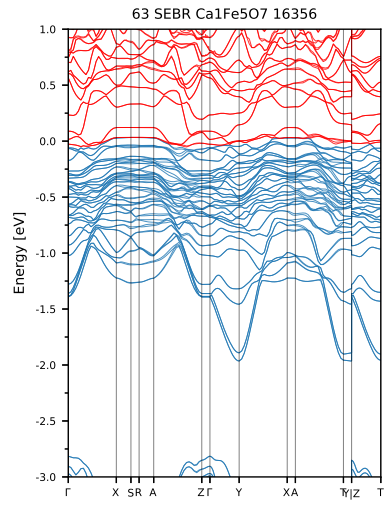

(bw) 63.1.75

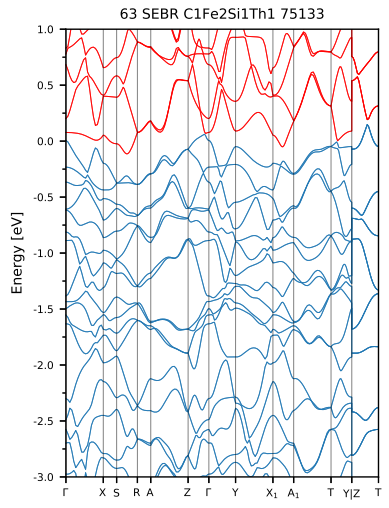

(bx) 63.1.76

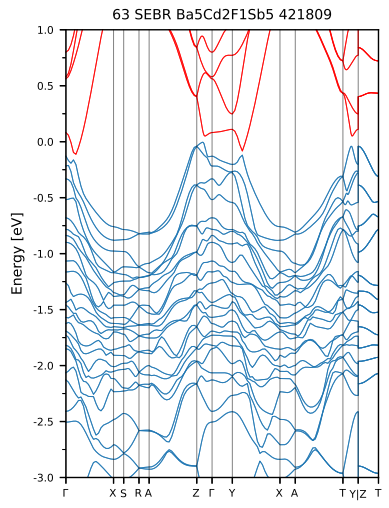

(by) 63.1.77

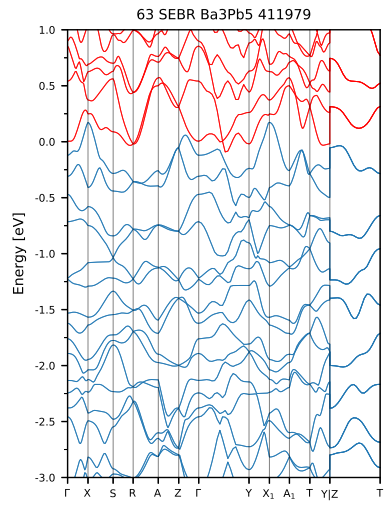

(bz) 63.1.78

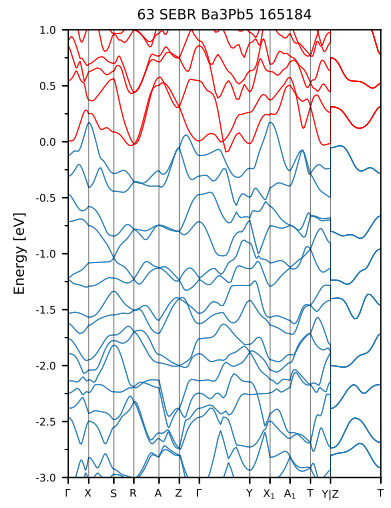

(ca) 63.1.79

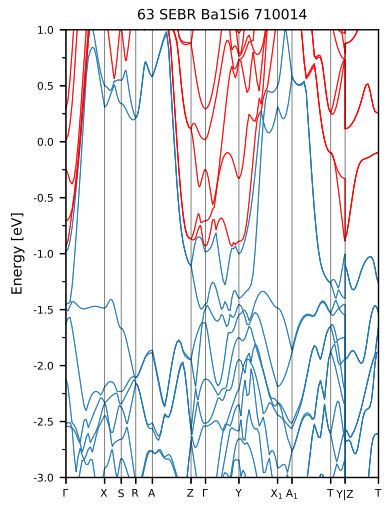

(cb) 63.1.80

Figure 149: Band structures for topological materials in SG 63, part 5/29, from 63.1.65 to 63.1.80. See the list of figures.

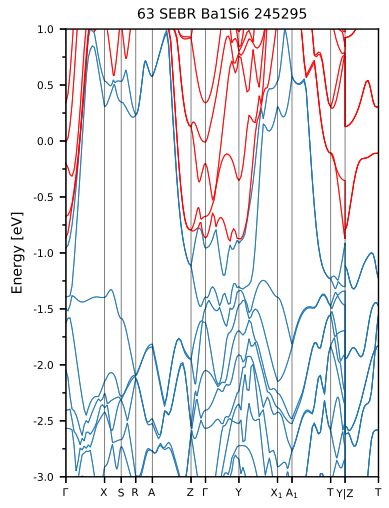

(cc) 63.1.81

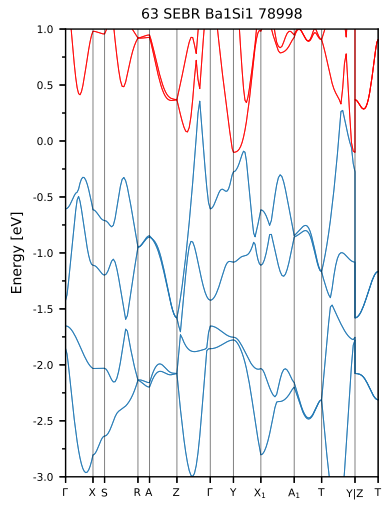

(cd) 63.1.82

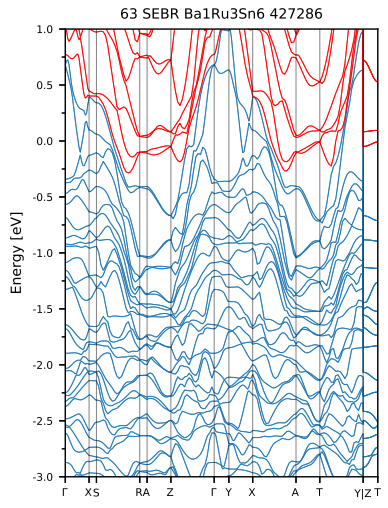

(ce) 63.1.83

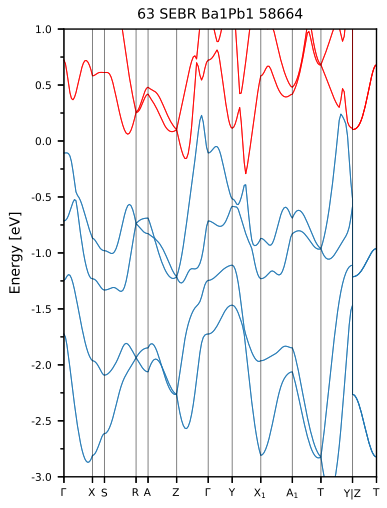

(cf) 63.1.84

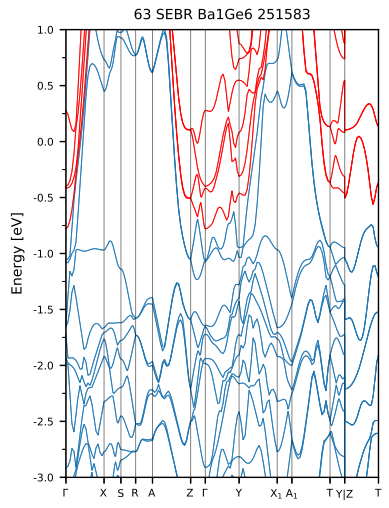

(cg) 63.1.85

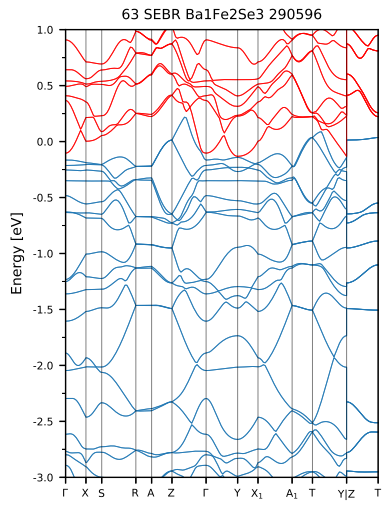

(ch) 63.1.86

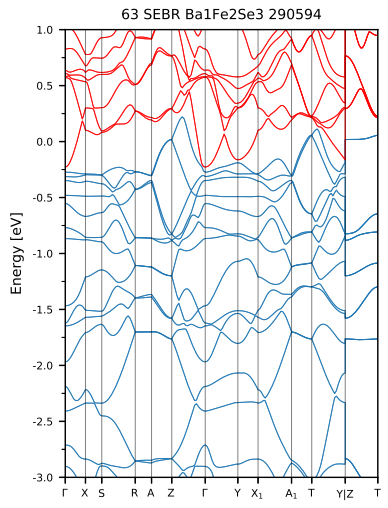

(ci) 63.1.87

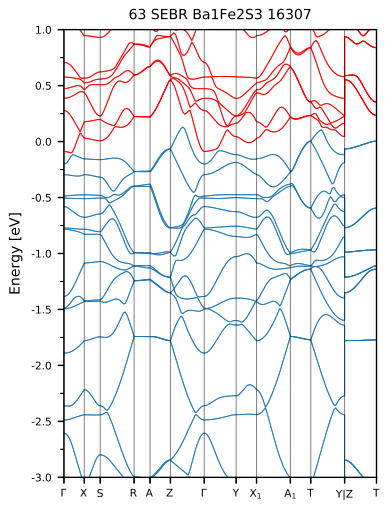

(cj) 63.1.88

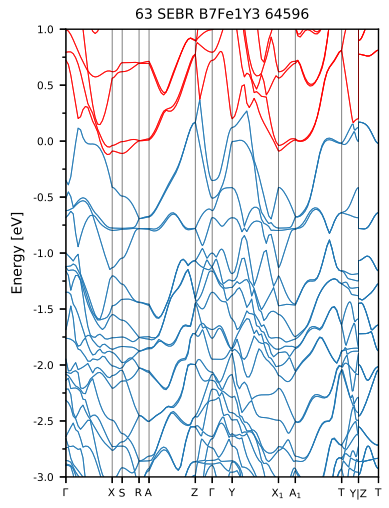

(ck) 63.1.89

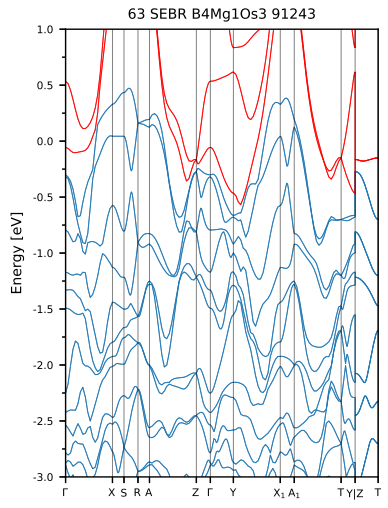

(cl) 63.1.90

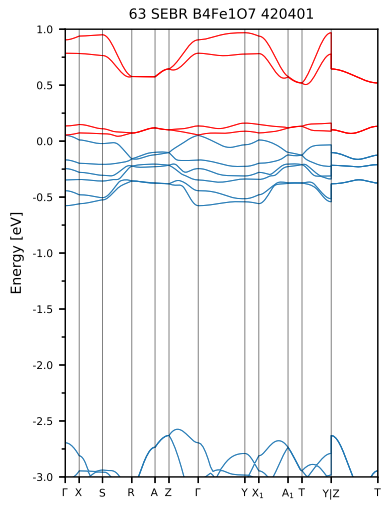

(cm) 63.1.91

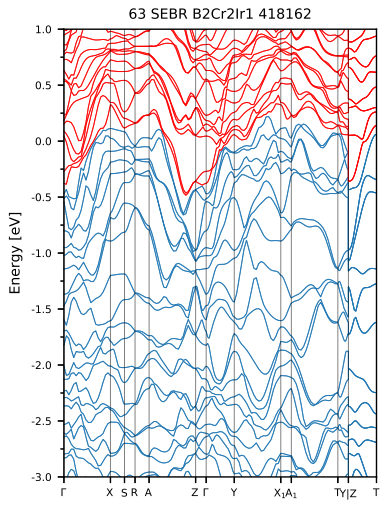

(cn) 63.1.92

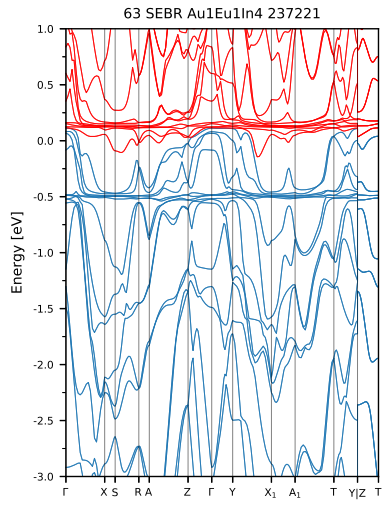

(co) 63.1.93

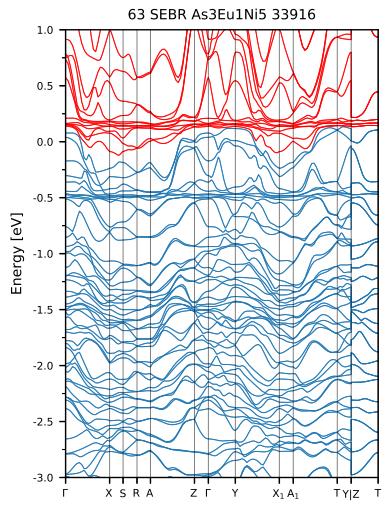

(cp) 63.1.94

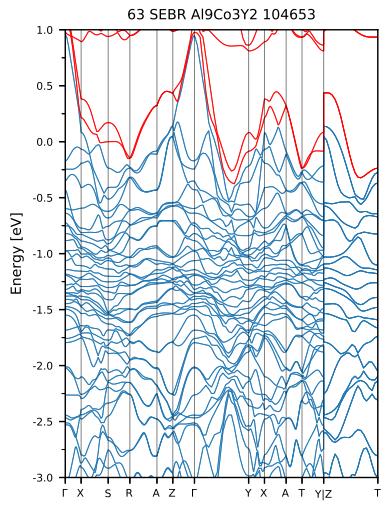

(cq) 63.1.95

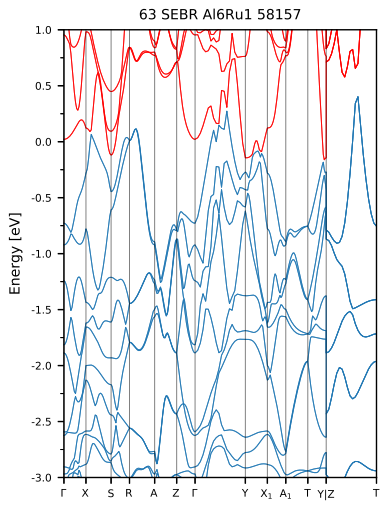

(cr) 63.1.96

Figure 150: Band structures for topological materials in SG 63, part 6/29, from 63.1.81 to 63.1.96. See the list of figures.

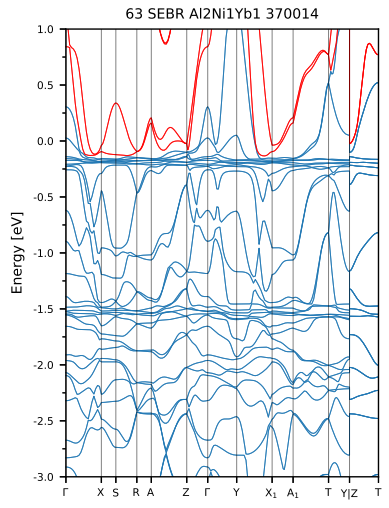

(cs) 63.1.97

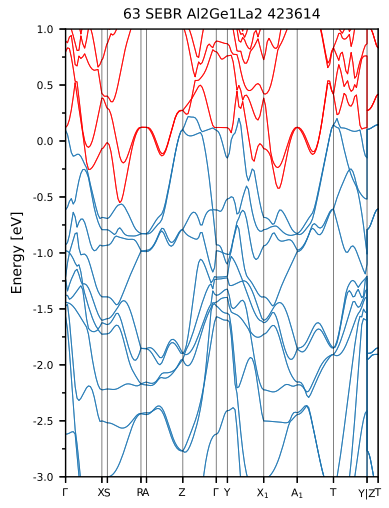

(ct) 63.1.98

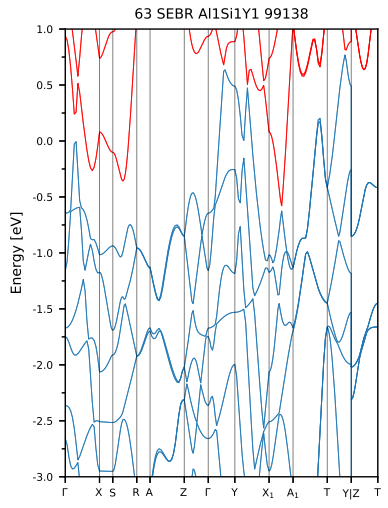

(cu) 63.1.99

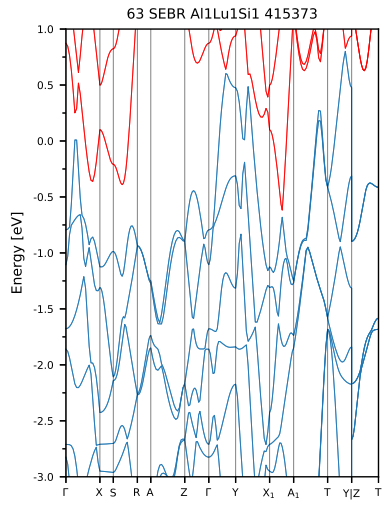

(cv) 63.1.100

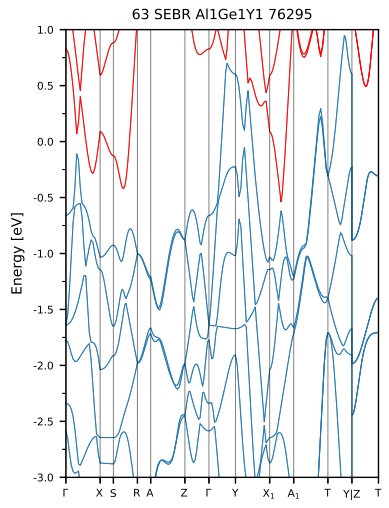

(cw) 63.1.101

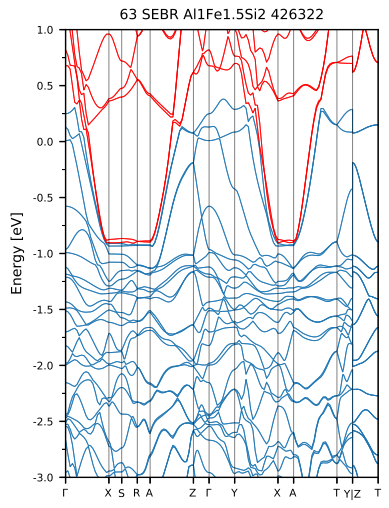

(cx) 63.1.102

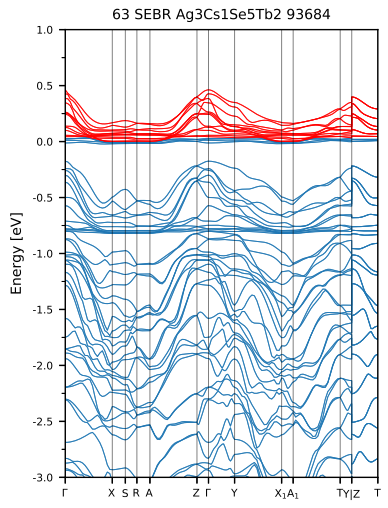

(cy) 63.1.103

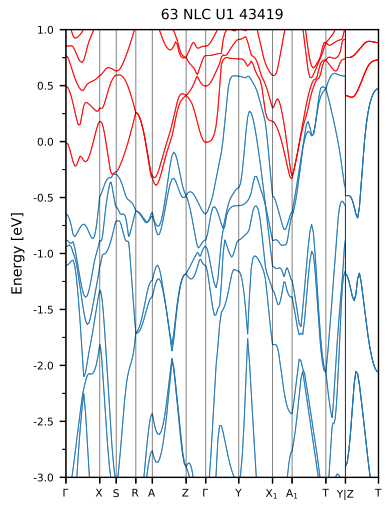

(cz) 63.2.1

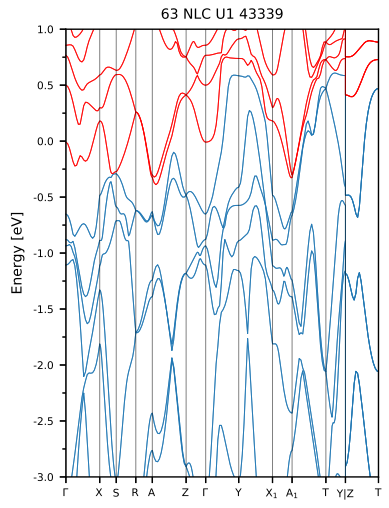

(da) 63.2.2

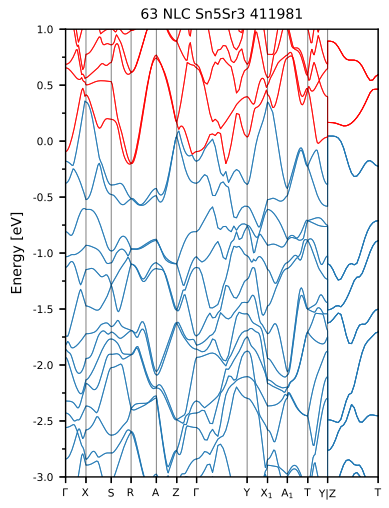

(db) 63.2.3

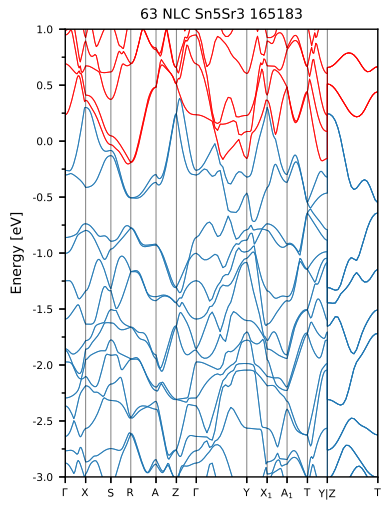

(dc) 63.2.4

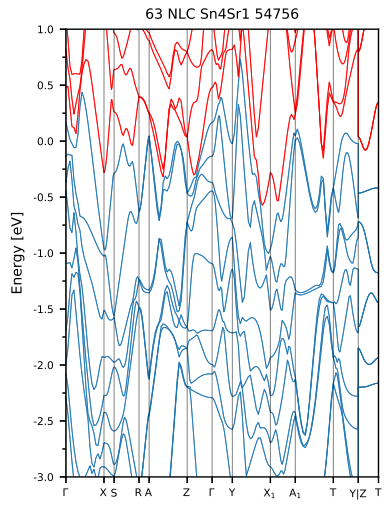

(dd) 63.2.5

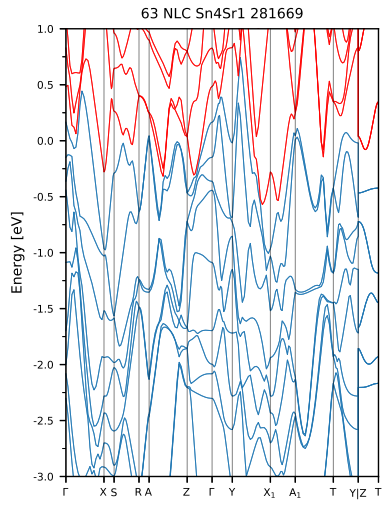

(de) 63.2.6

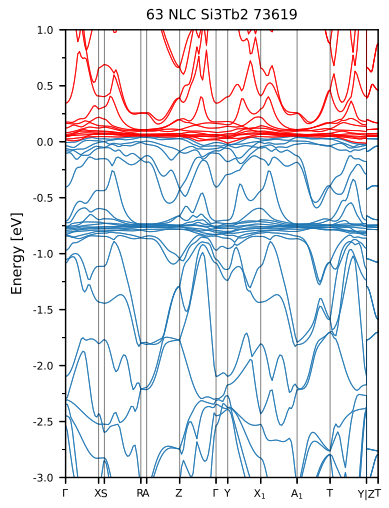

(df) 63.2.7

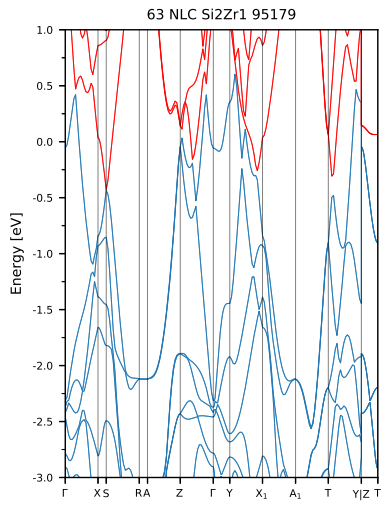

(dg) 63.2.8

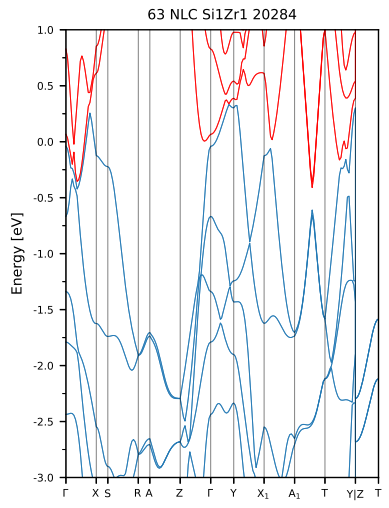

(dh) 63.2.9

Figure 151: Band structures for topological materials in SG 63, part 7/29, from 63.1.97 to 63.2.9. See the list of figures.

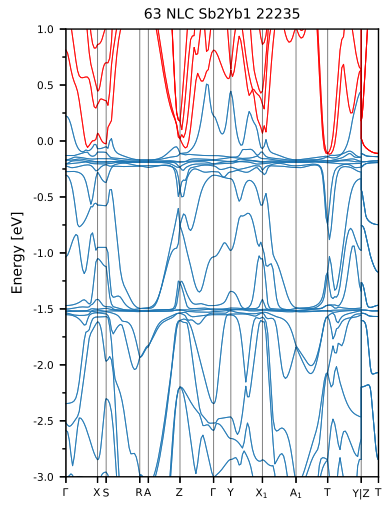

(di) 63.2.10

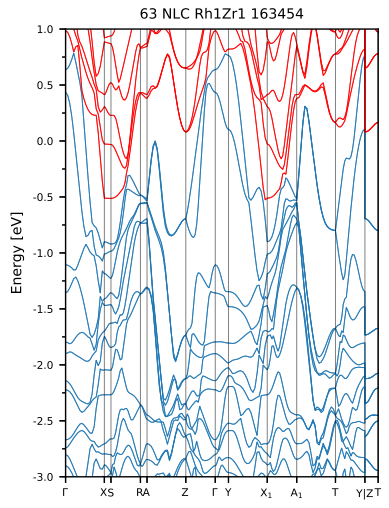

(dj) 63.2.11

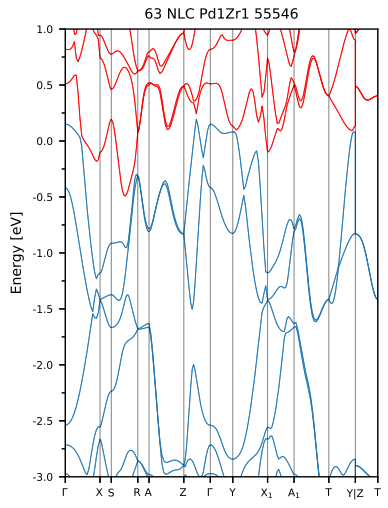

(dk) 63.2.12

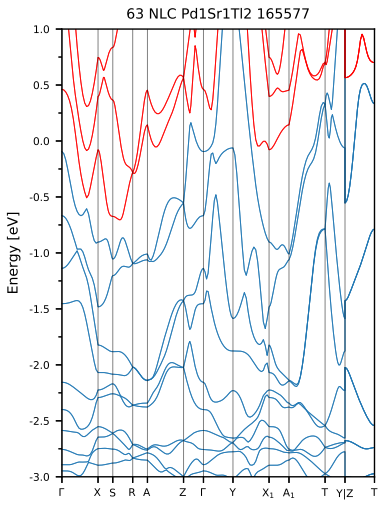

(dl) 63.2.13

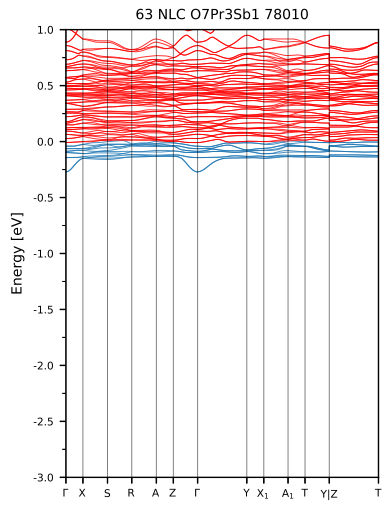

(dm) 63.2.14

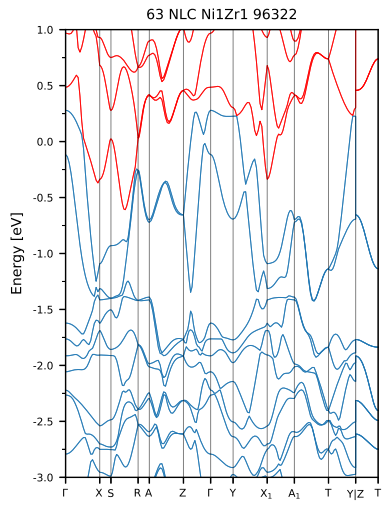

(dn) 63.2.15

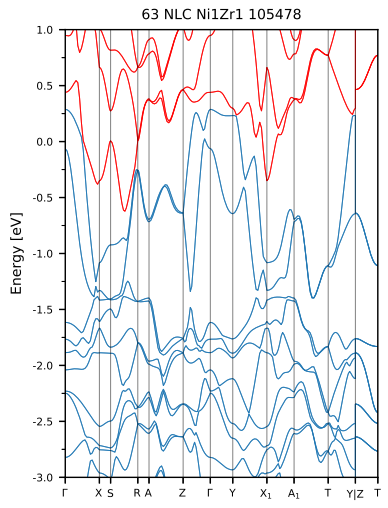

(do) 63.2.16

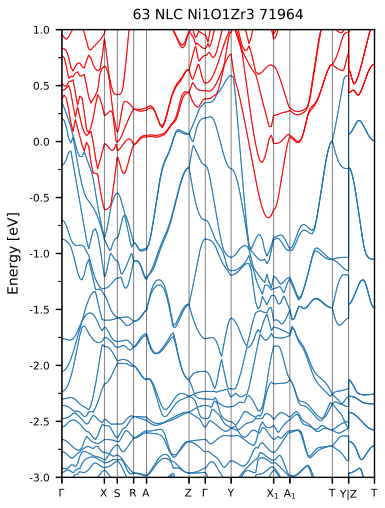

(dp) 63.2.17

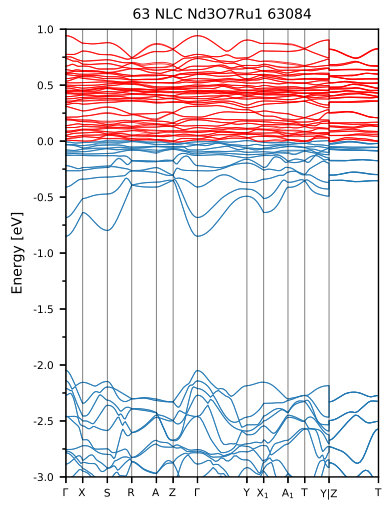

(dq) 63.2.18

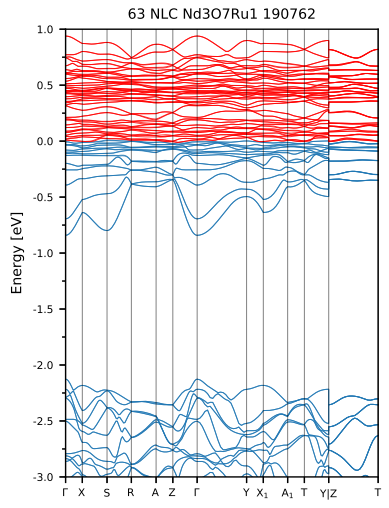

(dr) 63.2.19

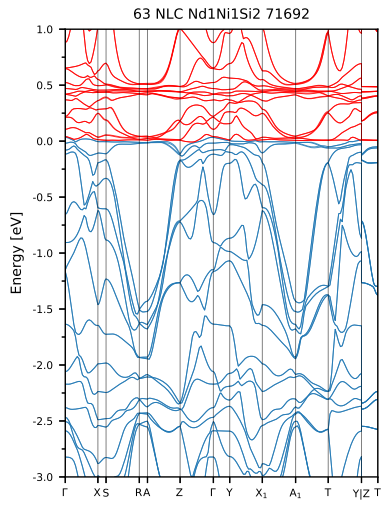

(ds) 63.2.20

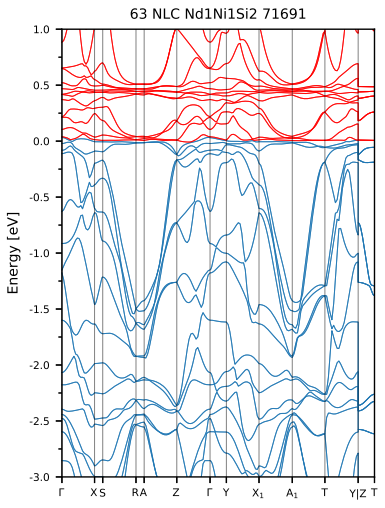

(dt) 63.2.21

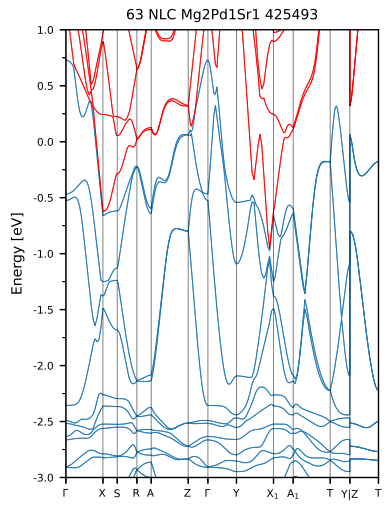

(du) 63.2.22

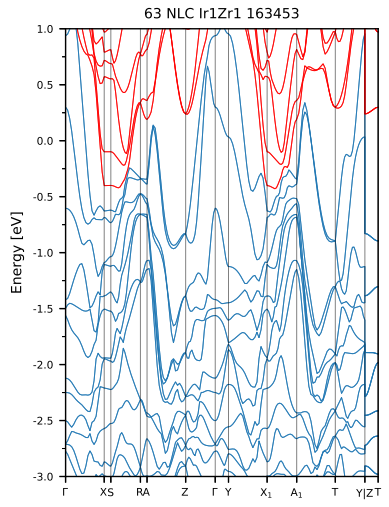

(dv) 63.2.23

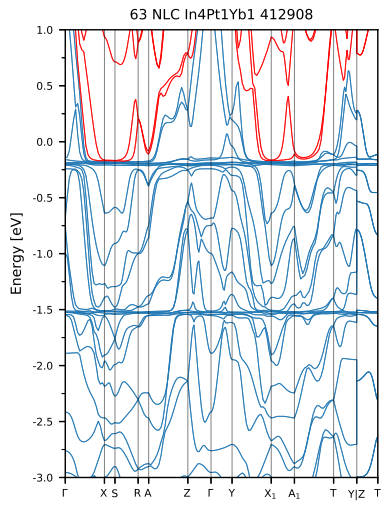

(dw) 63.2.24

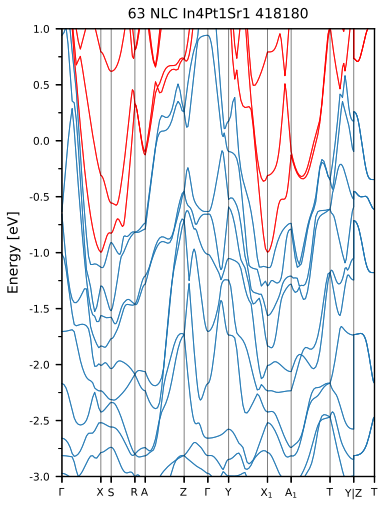

(dx) 63.2.25

Figure 152: Band structures for topological materials in SG 63, part 8/29, from 63.2.10 to 63.2.25. See the list of figures.

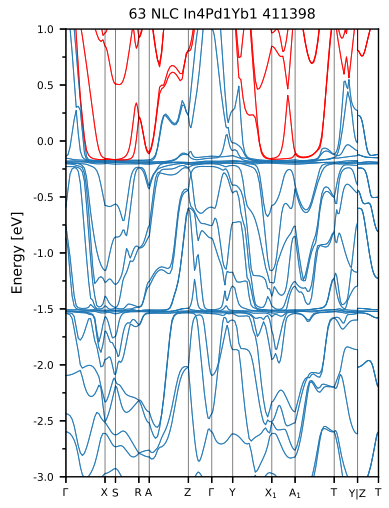

(dy) 63.2.26

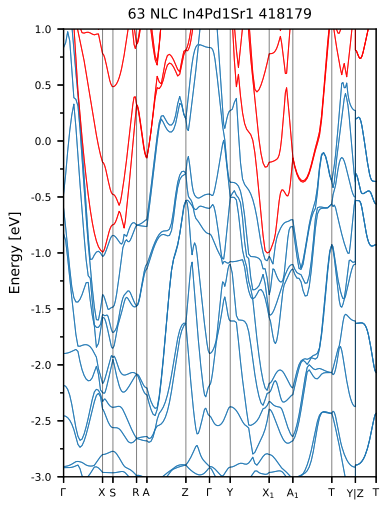

(dz) 63.2.27

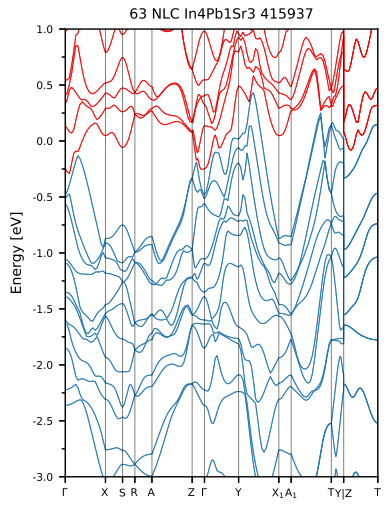

(ea) 63.2.28

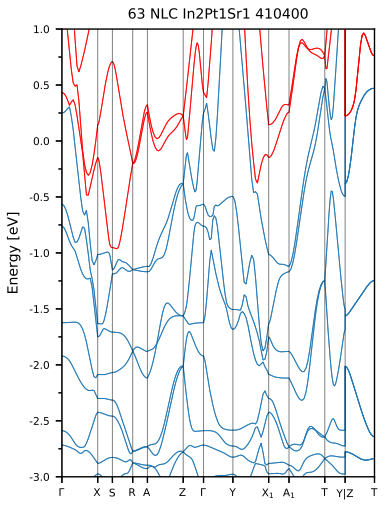

(eb) 63.2.29

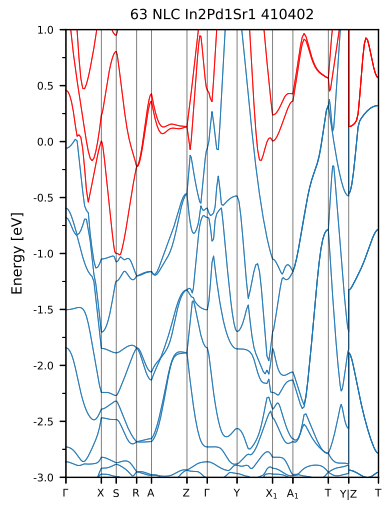

(ec) 63.2.30

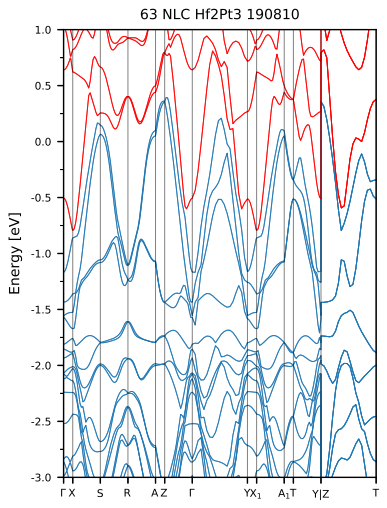

(ed) 63.2.31

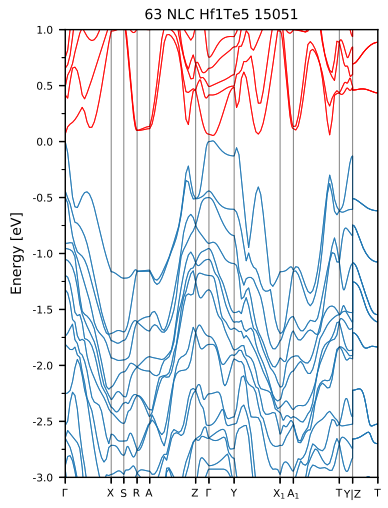

(ee) 63.2.32

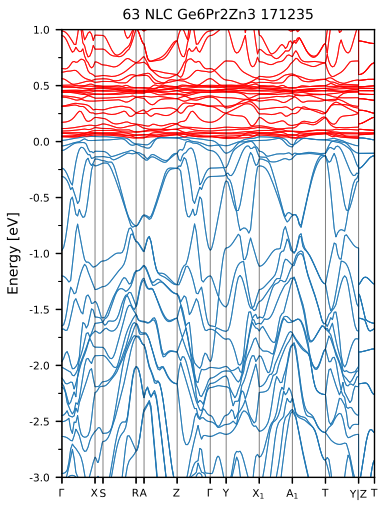

(ef) 63.2.33

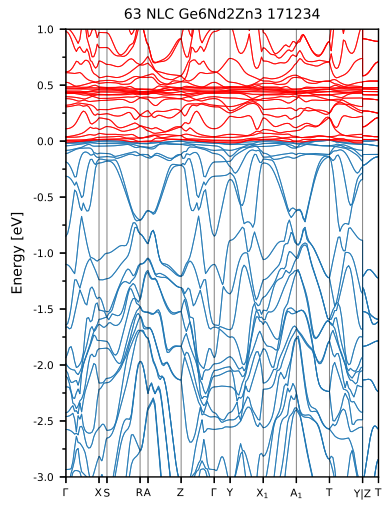

(eg) 63.2.34

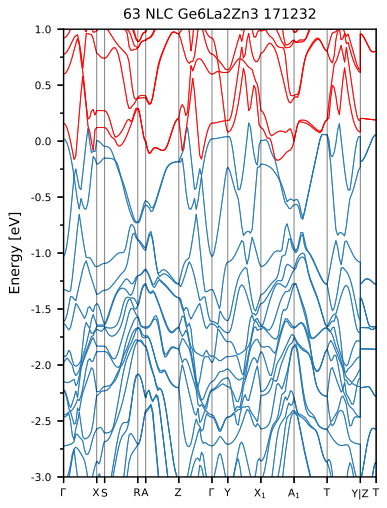

(eh) 63.2.35

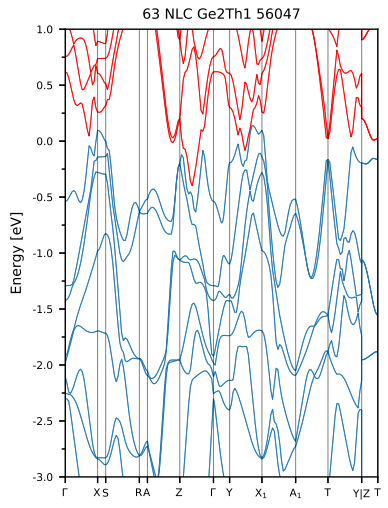

(ei) 63.2.36

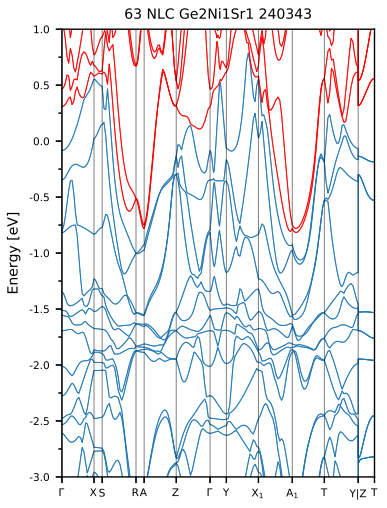

(ej) 63.2.37

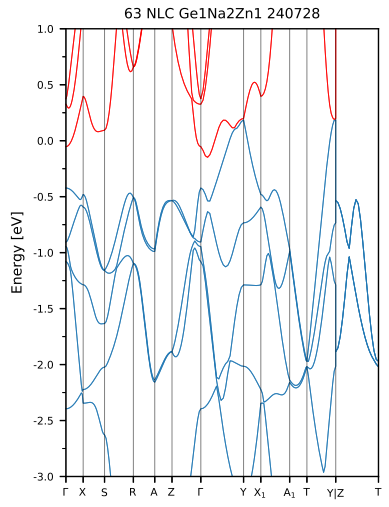

(ek) 63.2.38

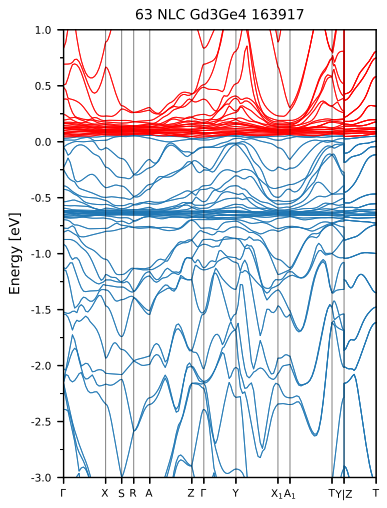

(el) 63.2.39

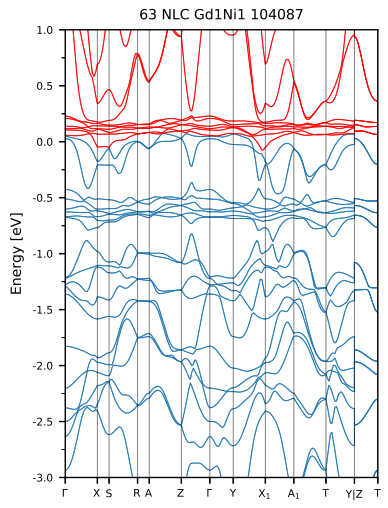

(em) 63.2.40

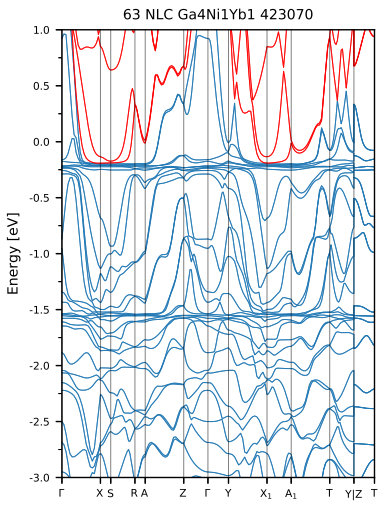

(en) 63.2.41

Figure 153: Band structures for topological materials in SG 63, part 9/29, from 63.2.26 to 63.2.41. See the list of figures.

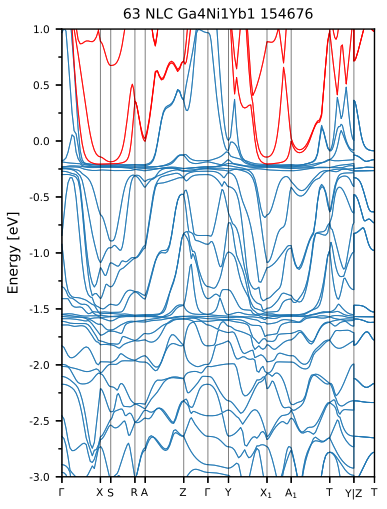

(eo) 63.2.42

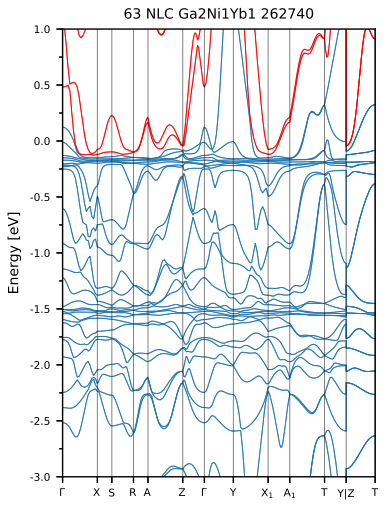

(ep) 63.2.43

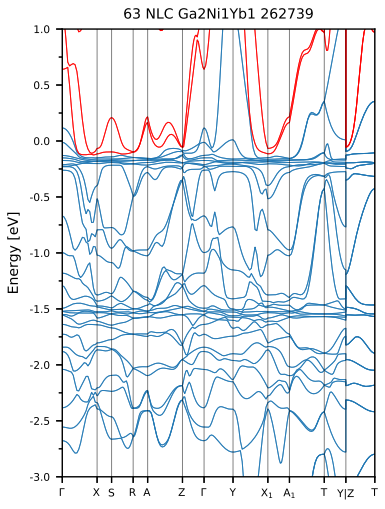

(eq) 63.2.44

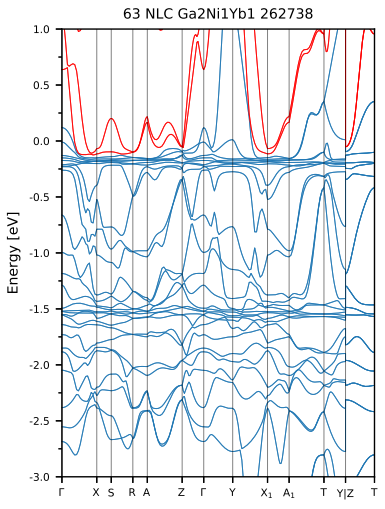

(er) 63.2.45

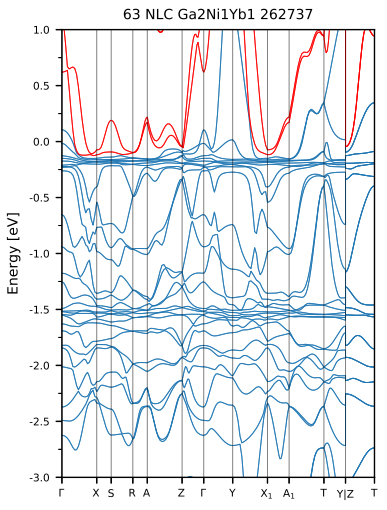

(es) 63.2.46

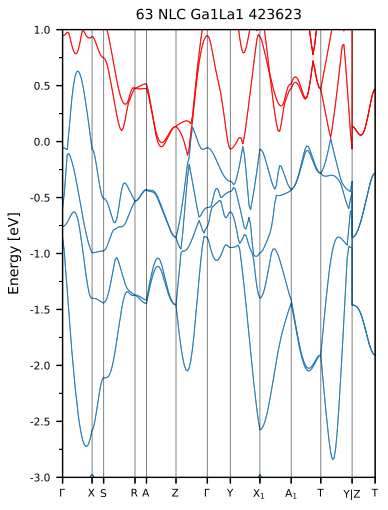

(et) 63.2.47

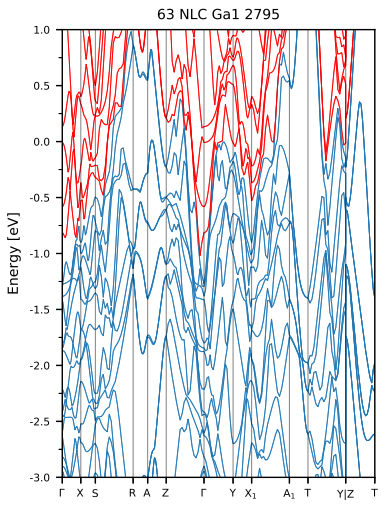

(eu) 63.2.48

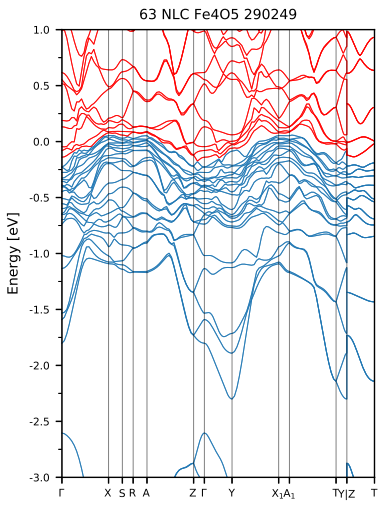

(ev) 63.2.49

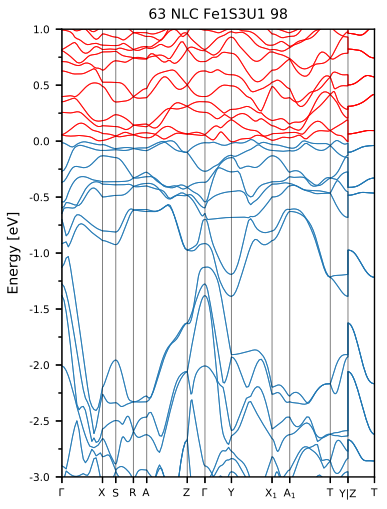

(ew) 63.2.50

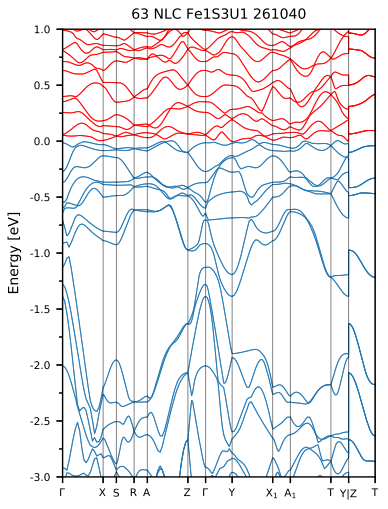

(ex) 63.2.51

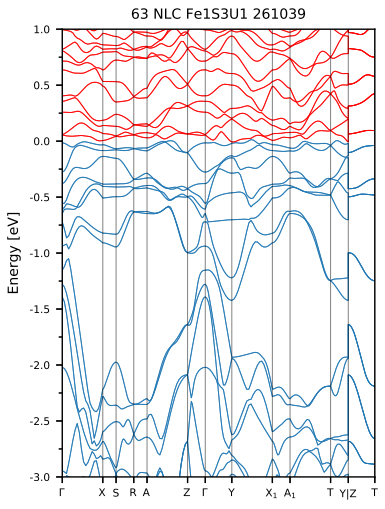

(ey) 63.2.52

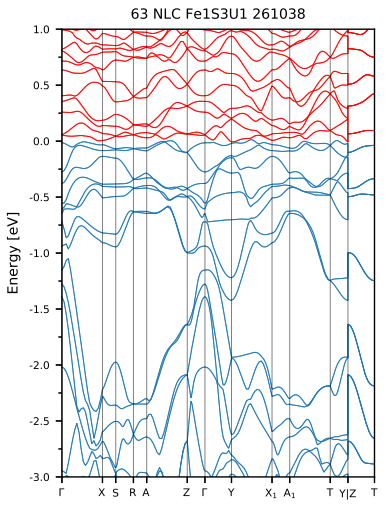

(ez) 63.2.53

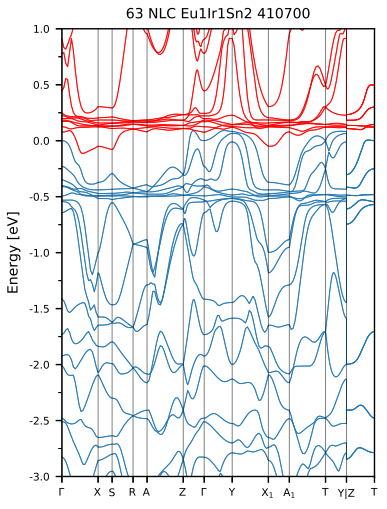

(fa) 63.2.54

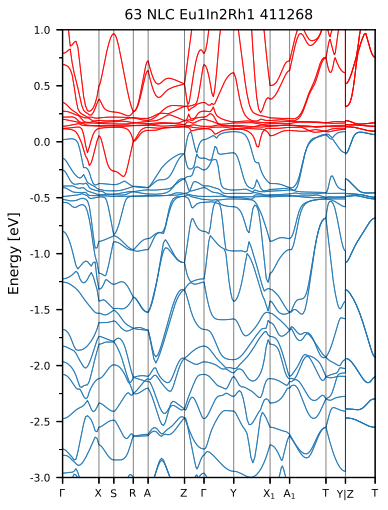

(fb) 63.2.55

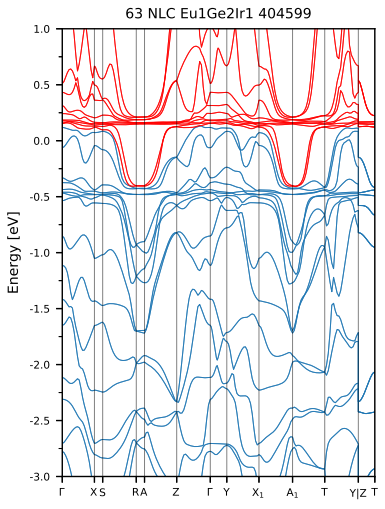

(fc) 63.2.56

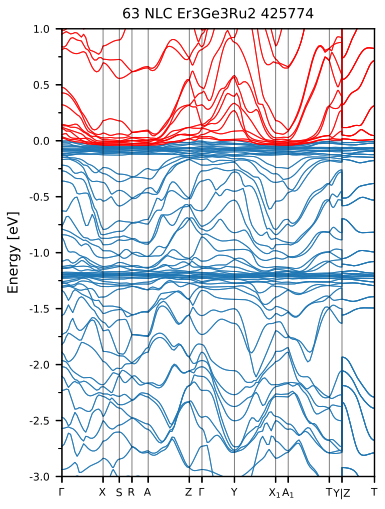

(fd) 63.2.57

Figure 154: Band structures for topological materials in SG 63, part 10/29, from 63.2.42 to 63.2.57. See the list of figures.

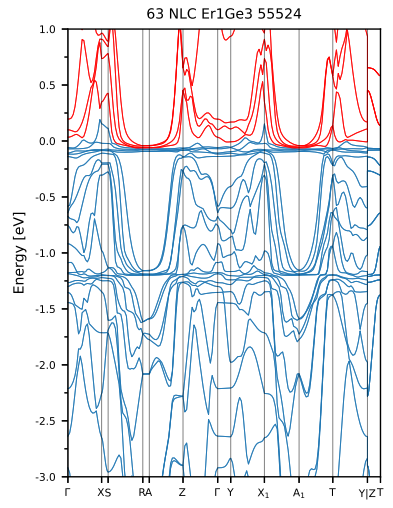

(fe) 63.2.58

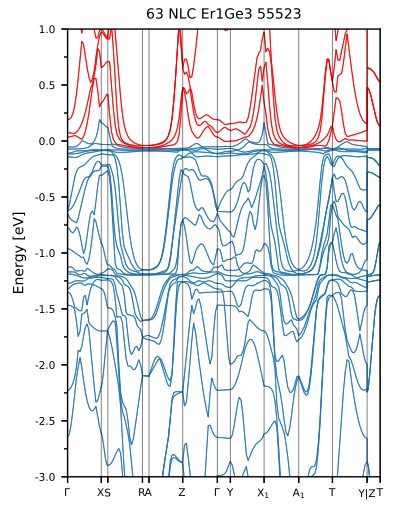

(ff) 63.2.59

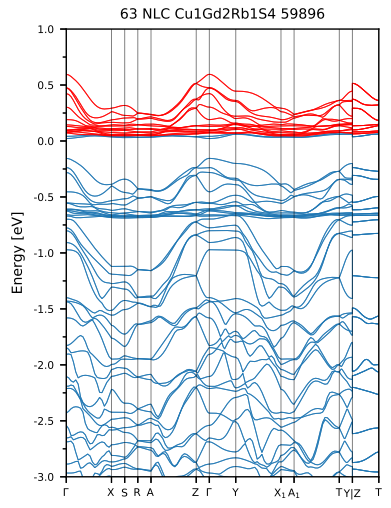

(fg) 63.2.60

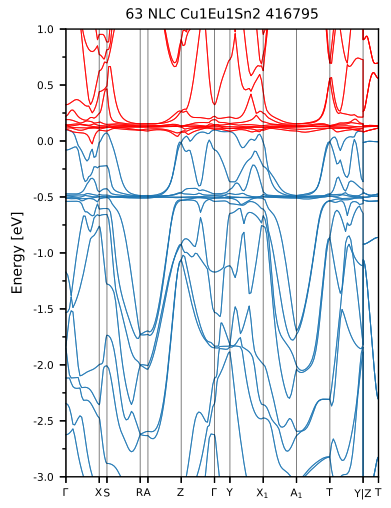

(fh) 63.2.61

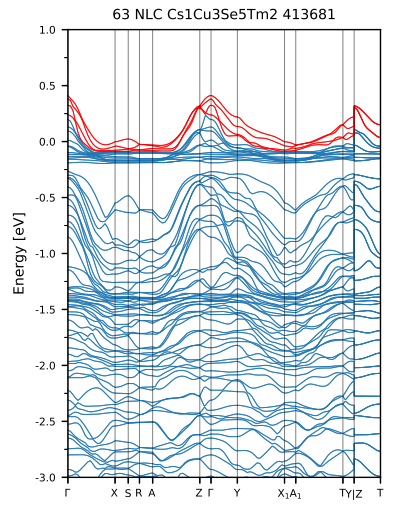

(fi) 63.2.62

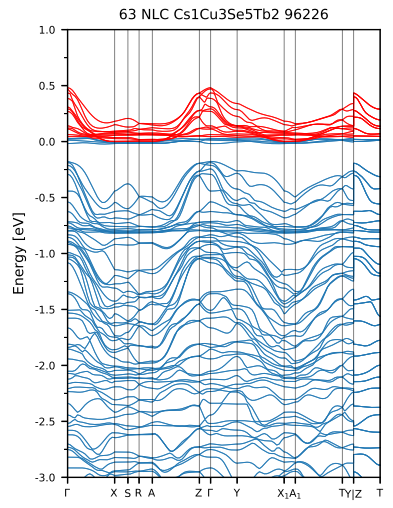

(fj) 63.2.63

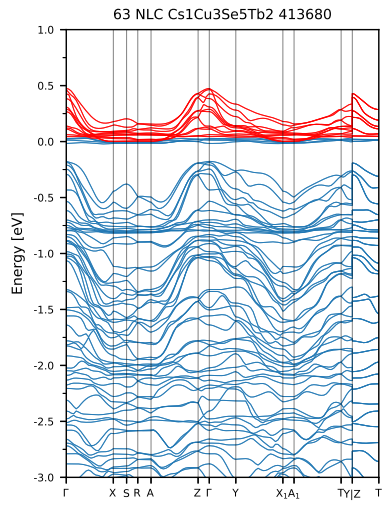

(fk) 63.2.64

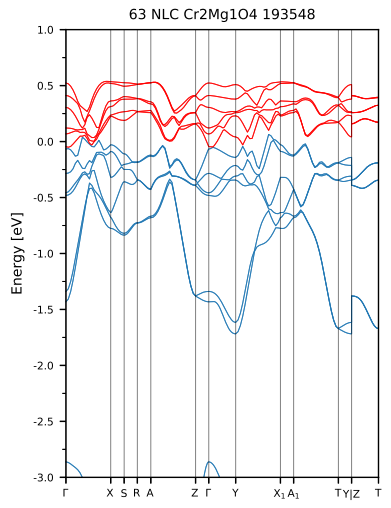

(fl) 63.2.65

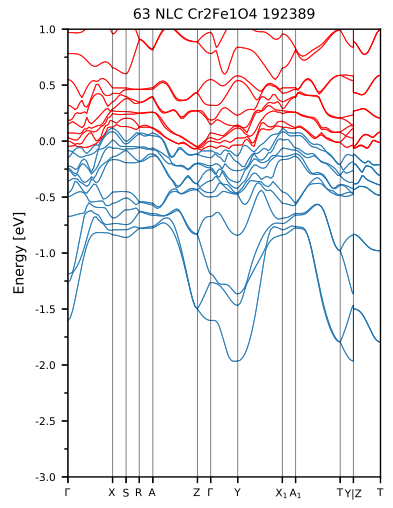

(fm) 63.2.66

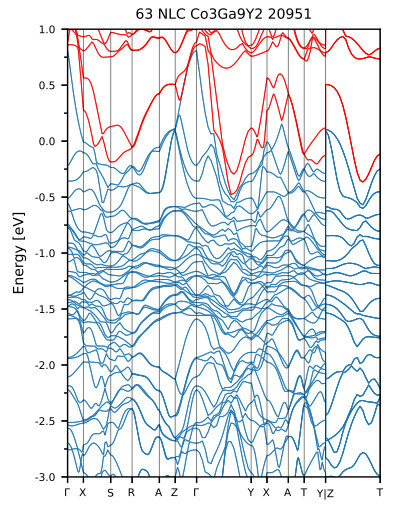

(fn) 63.2.67

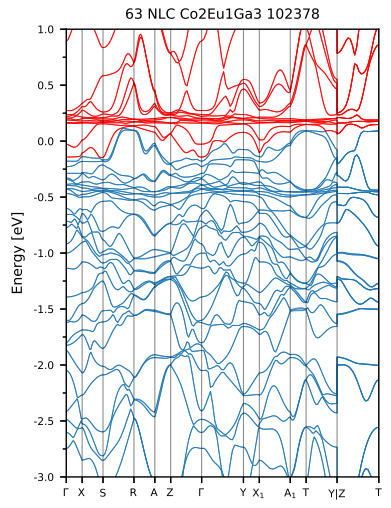

(fo) 63.2.68

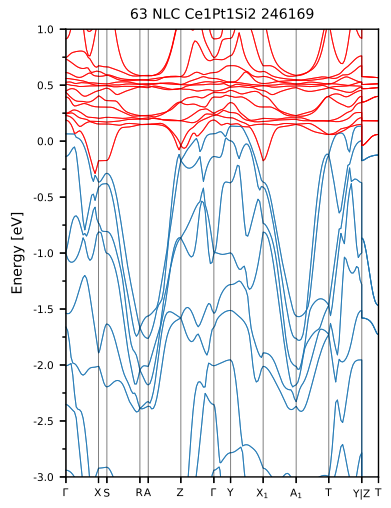

(fp) 63.2.69

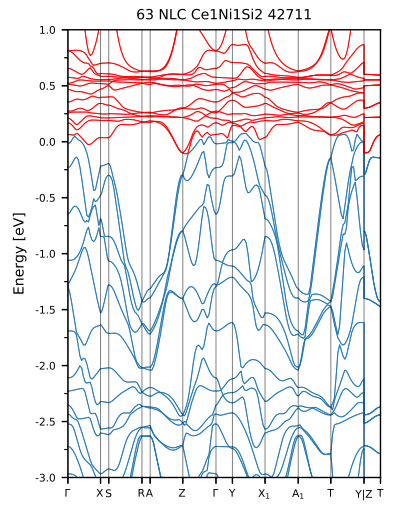

(fq) 63.2.70

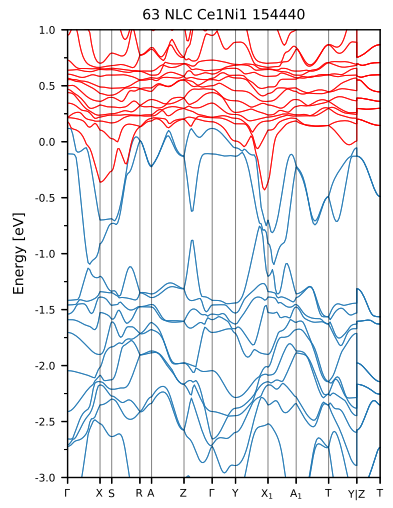

(fr) 63.2.71

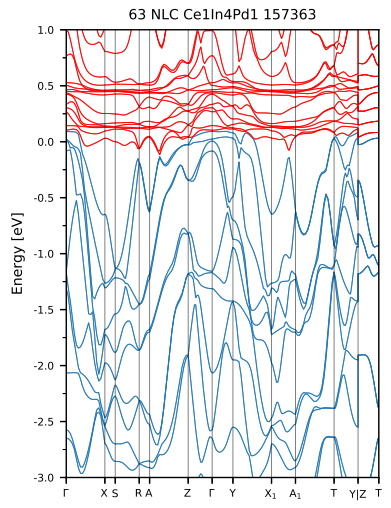

(fs) 63.2.72

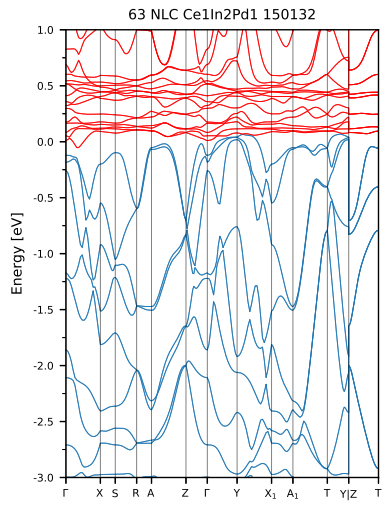

(ft) 63.2.73

Figure 155: Band structures for topological materials in SG 63, part 11/29, from 63.2.58 to 63.2.73. See the list of figures.

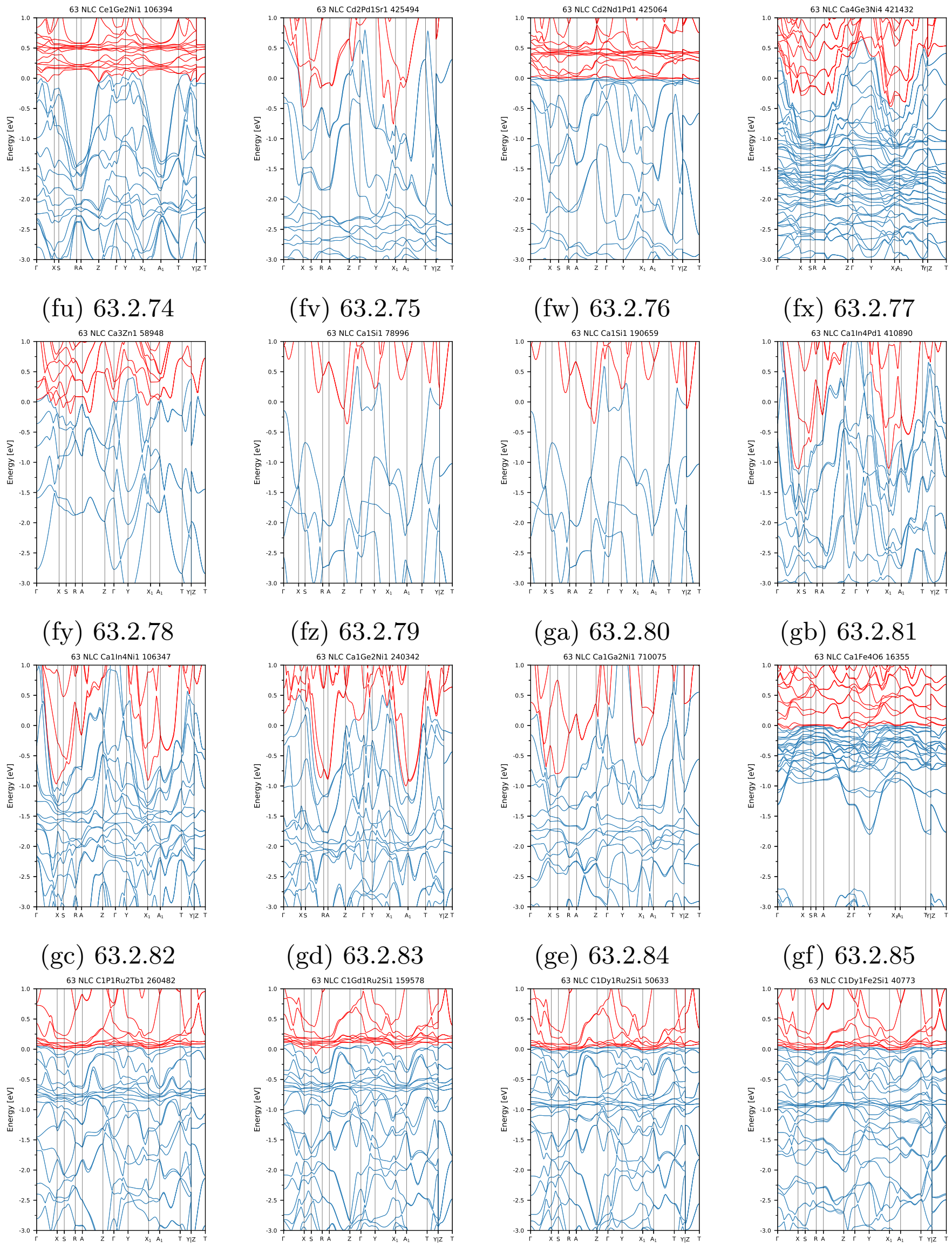

(fu) 63.2.74 (fv) 63.2.75 (fw) 63.2.76 (fx) 63.2.77

(fy) 63.2.78 (fz) 63.2.79 (ga) 63.2.80 (gb) 63.2.81

(gc) 63.2.82 (gd) 63.2.83 (ge) 63.2.84 (gf) 63.2.85

(gg) 63.2.86 (gh) 63.2.87 (gi) 63.2.88 (gj) 63.2.89

Figure 156: Band structures for topological materials in SG 63, part 12/29, from 63.2.74 to 63.2.89. See the list of figures.

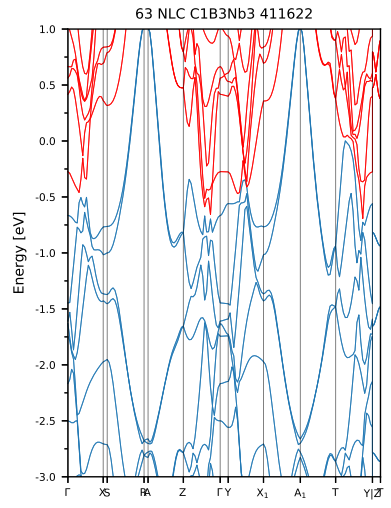

(gk) 63.2.90

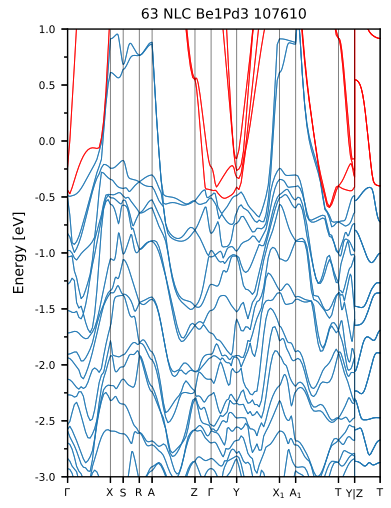

(gl) 63.2.91

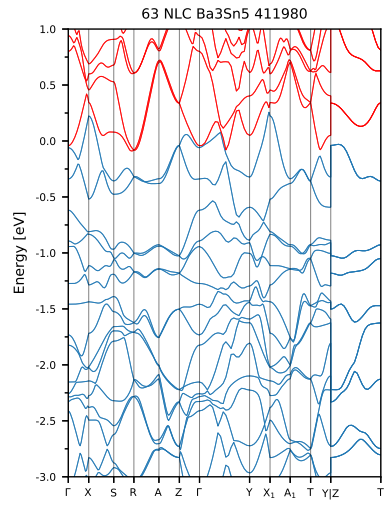

(gm) 63.2.92

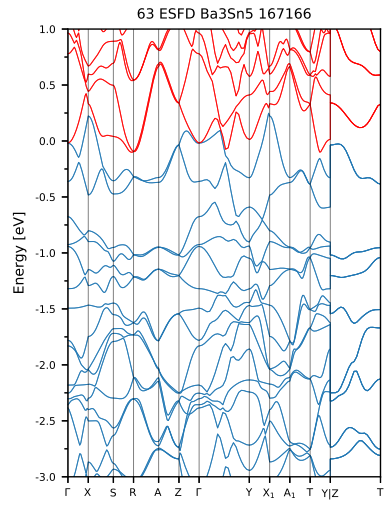

(gn) 63.2.93

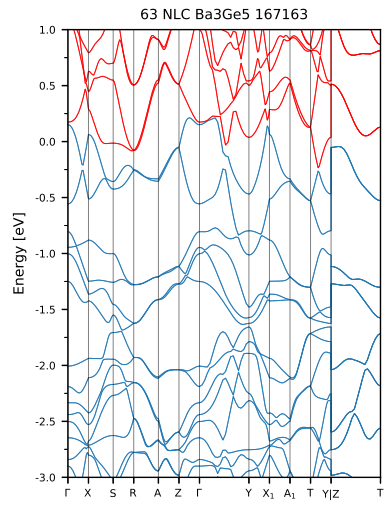

(go) 63.2.94

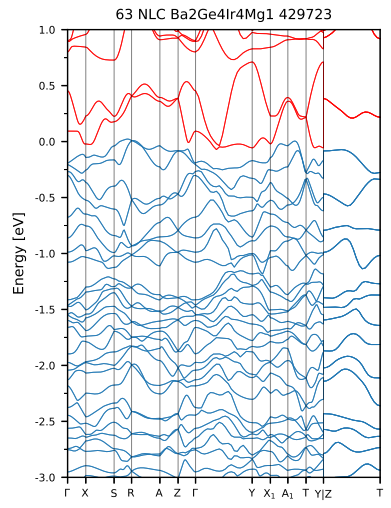

(gp) 63.2.95

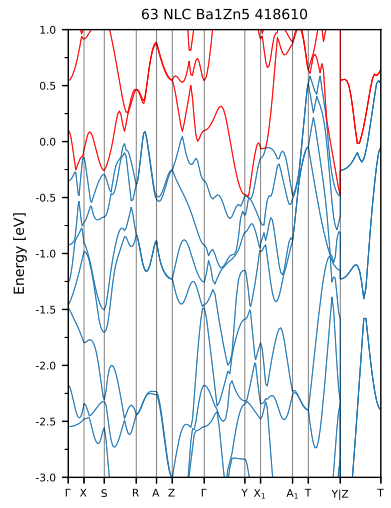

(gq) 63.2.96

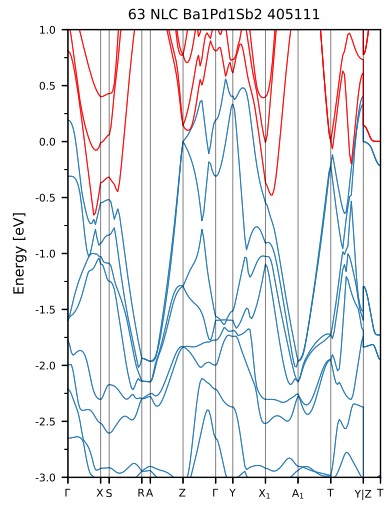

(gr) 63.2.97

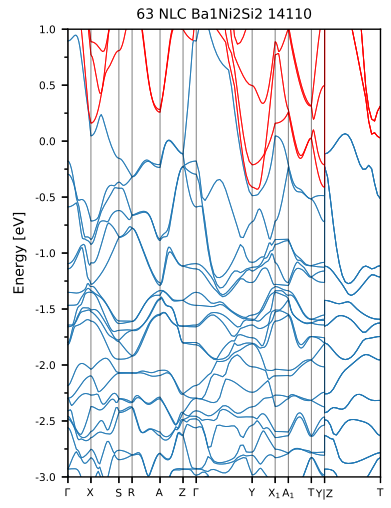

(gs) 63.2.98

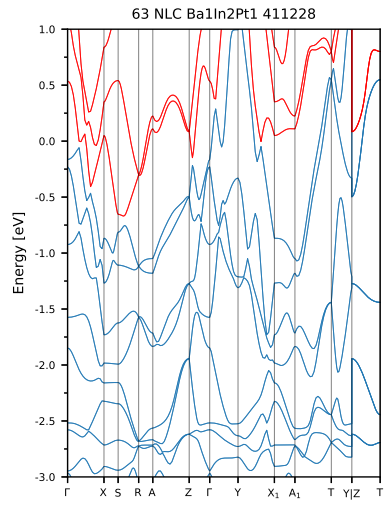

(gt) 63.2.99

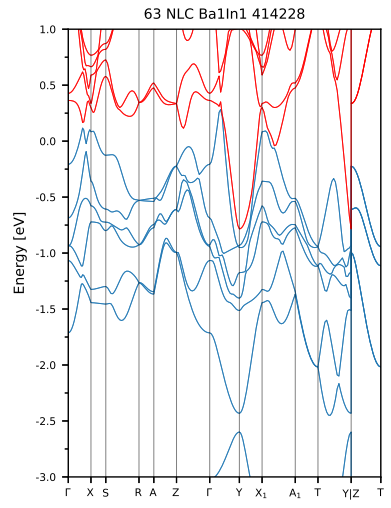

(gu) 63.2.100

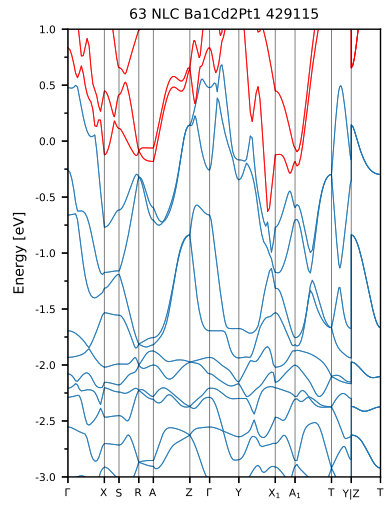

(gv) 63.2.101

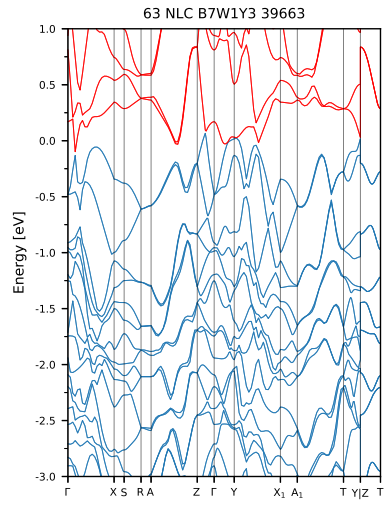

(gw) 63.2.102

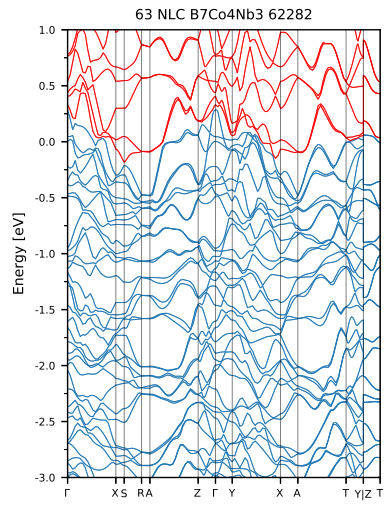

(gx) 63.2.103

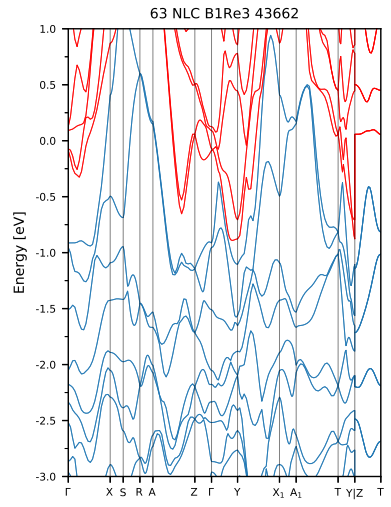

(gy) 63.2.104

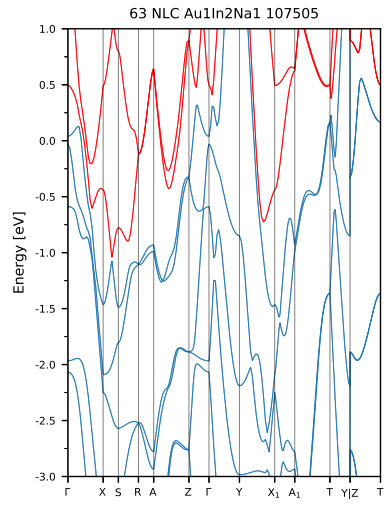

(gz) 63.2.105

Figure 157: Band structures for topological materials in SG 63, part 13/29, from 63.2.90 to 63.2.105. See the list of figures.

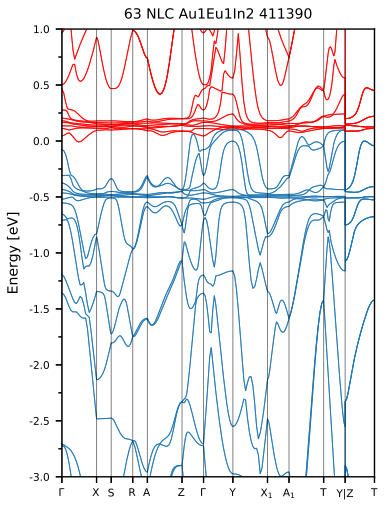

(ha) 63.2.106

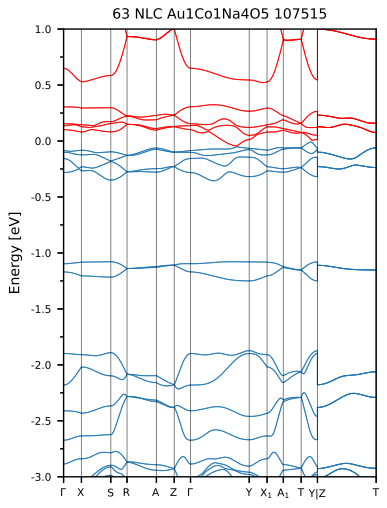

(hb) 63.2.107

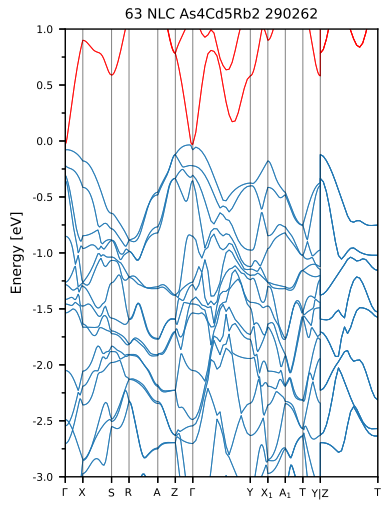

(hc) 63.2.108

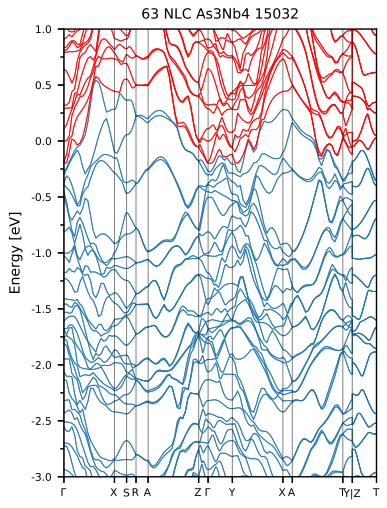

(hd) 63.2.109

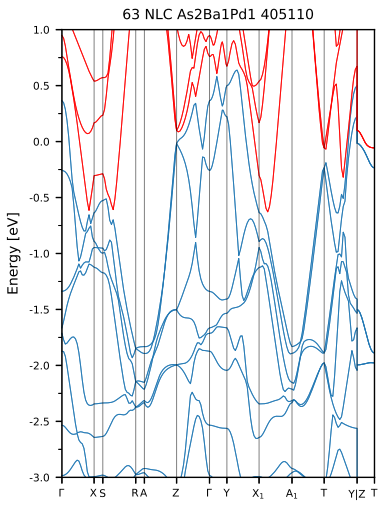

(he) 63.2.110

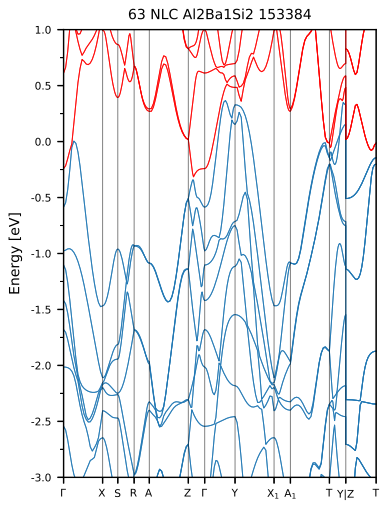

(hf) 63.2.111

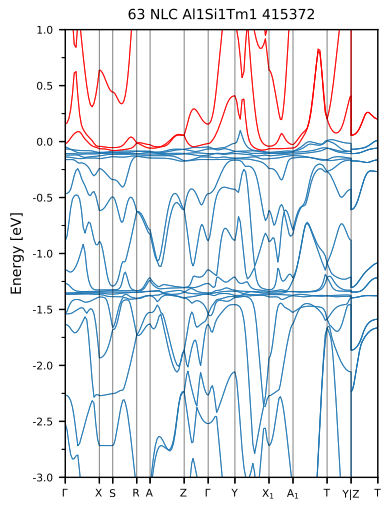

(hg) 63.2.112

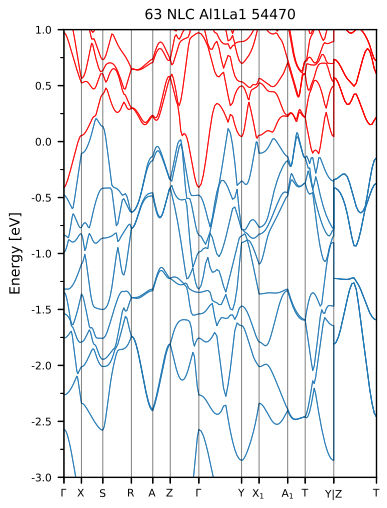

(hh) 63.2.113

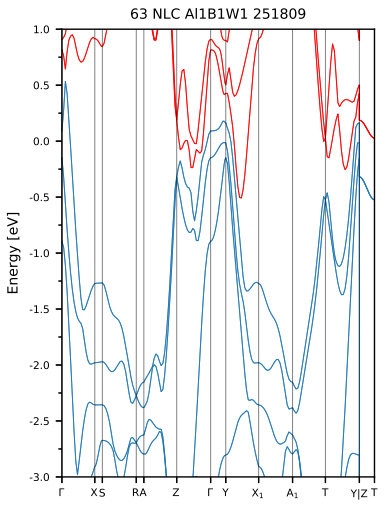

(hi) 63.2.114

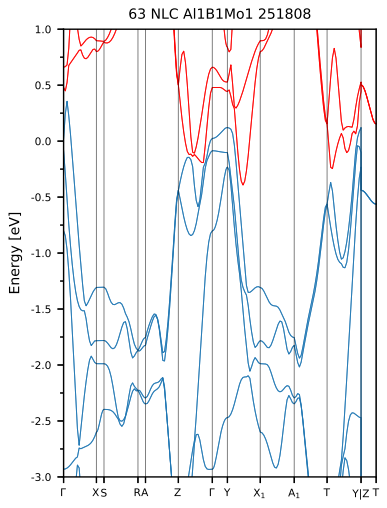

(hj) 63.2.115

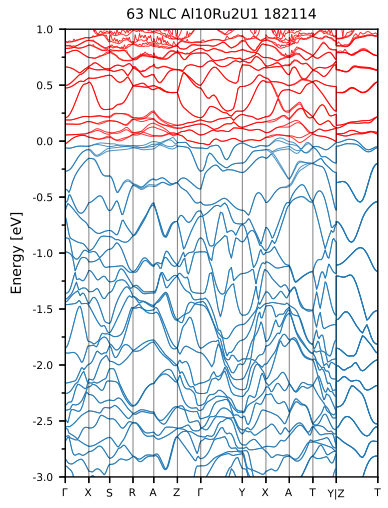

(hk) 63.2.116

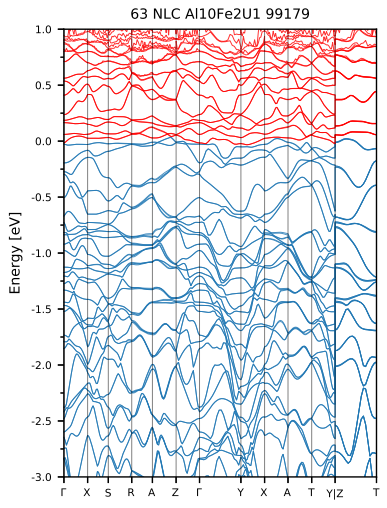

(hl) 63.2.117

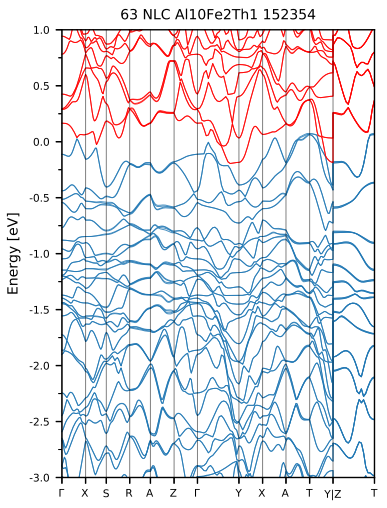

(hm) 63.2.118

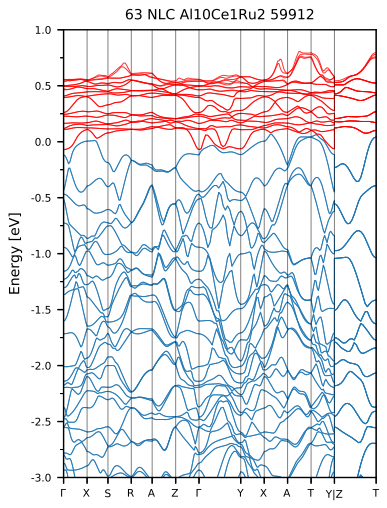

(hn) 63.2.119

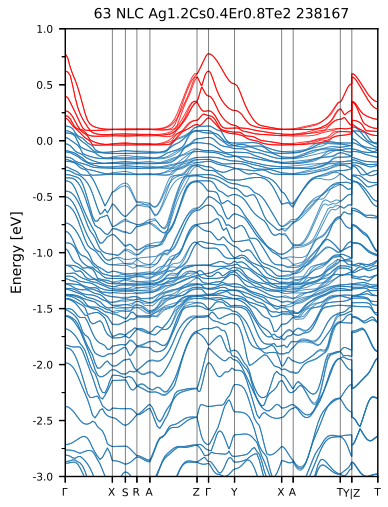

(ho) 63.2.120

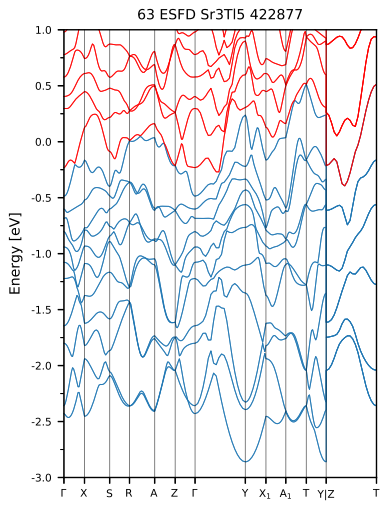

(hp) 63.4.1

Figure 158: Band structures for topological materials in SG 63, part 14/29, from 63.2.106 to 63.4.1. See the list of figures.

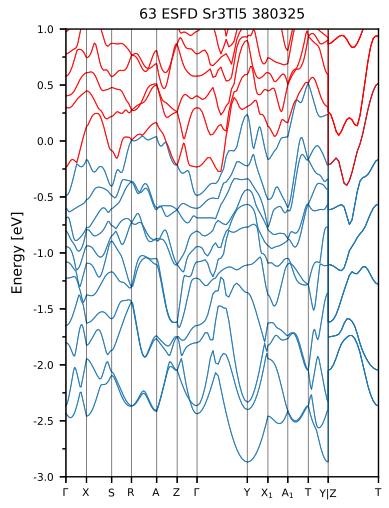

(hq) 63.4.2

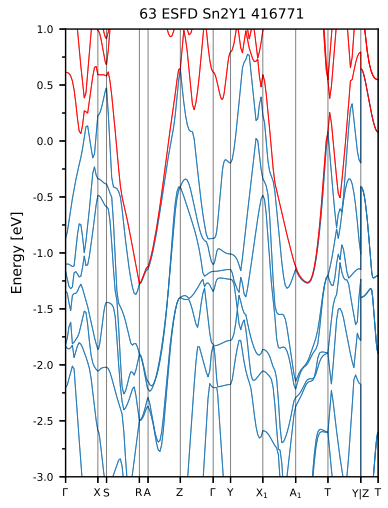

(hr) 63.4.3

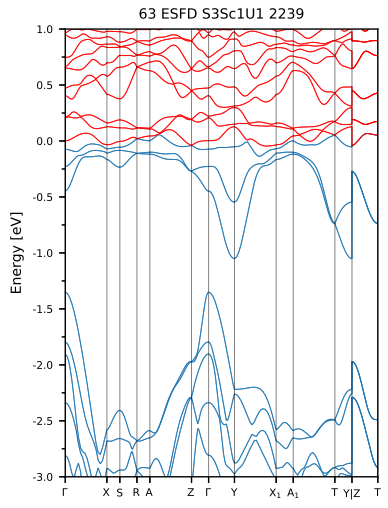

(hs) 63.4.4

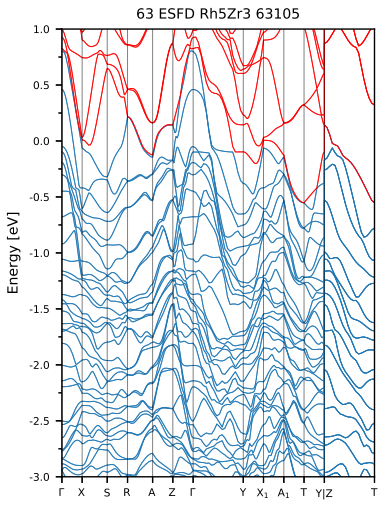

(ht) 63.4.5

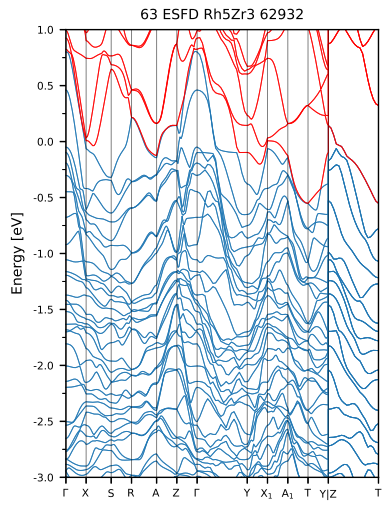

(hu) 63.4.6

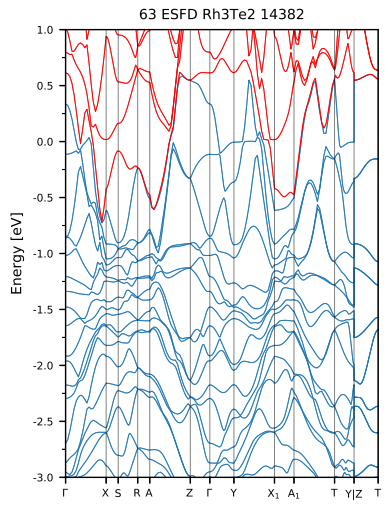

(hv) 63.4.7

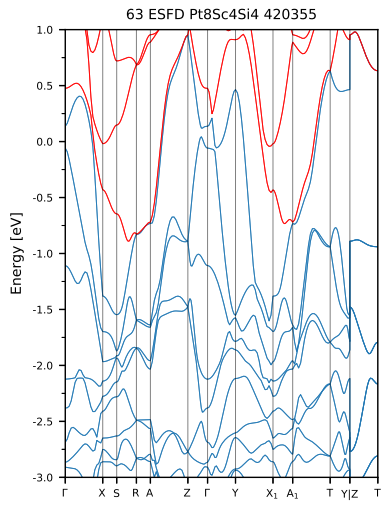

(hw) 63.4.8

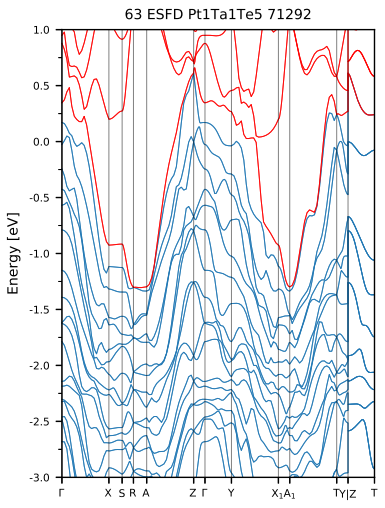

(hx) 63.4.9

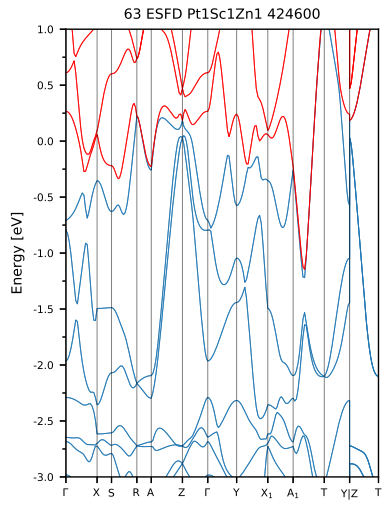

(hy) 63.4.10

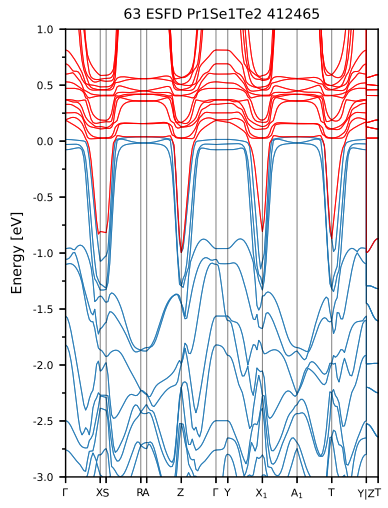

(hz) 63.4.11

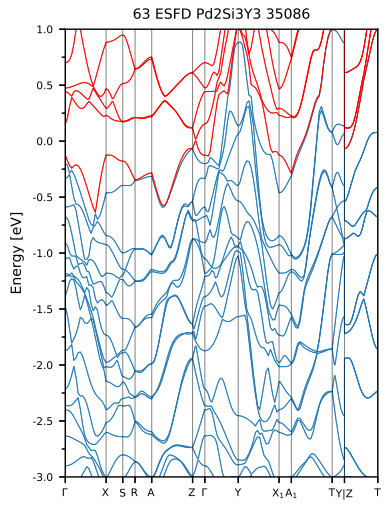

(ia) 63.4.12

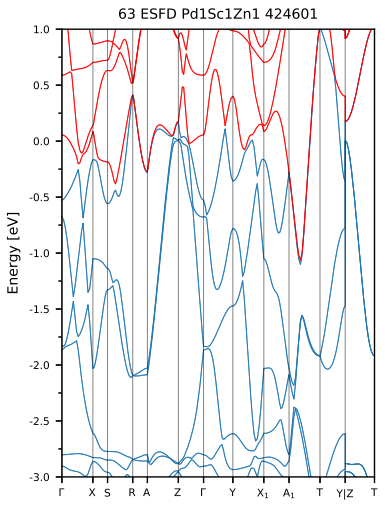

(ib) 63.4.13

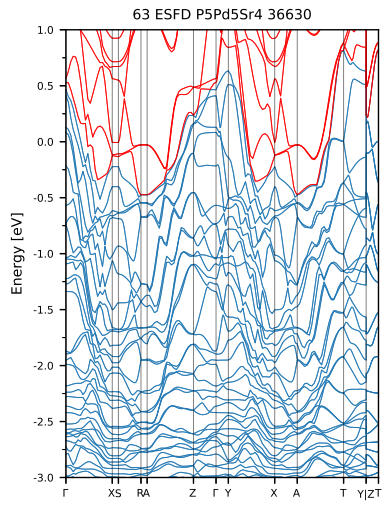

(ic) 63.4.14

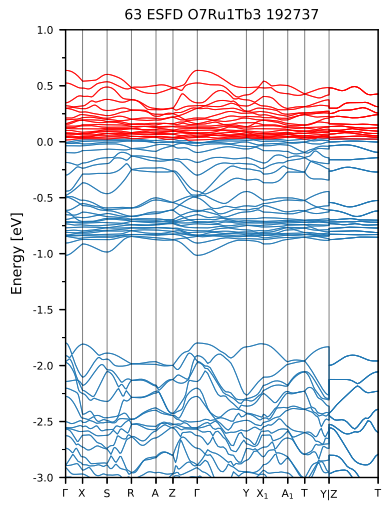

(id) 63.4.15

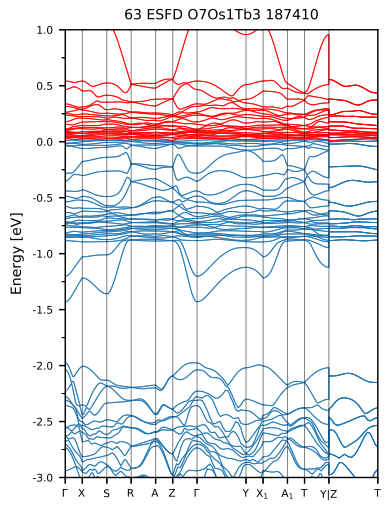

(ie) 63.4.16

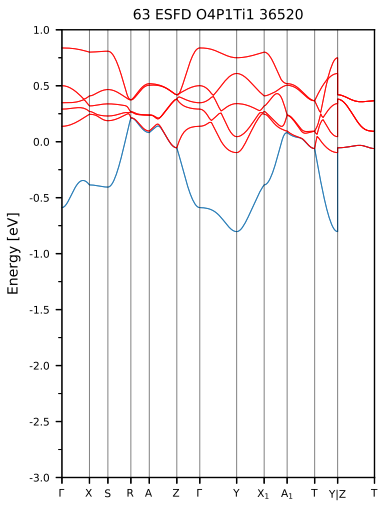

(if) 63.4.17

Figure 159: Band structures for topological materials in SG 63, part 15/29, from 63.4.2 to 63.4.17. See the list of figures.

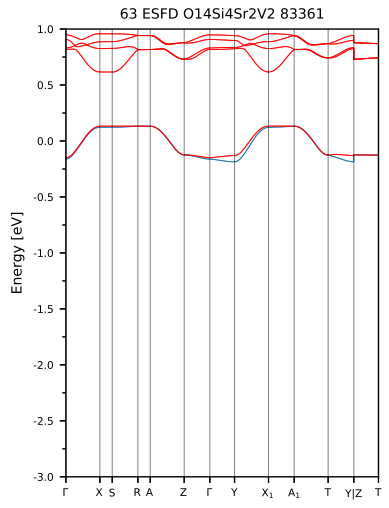

(ig) 63.4.18

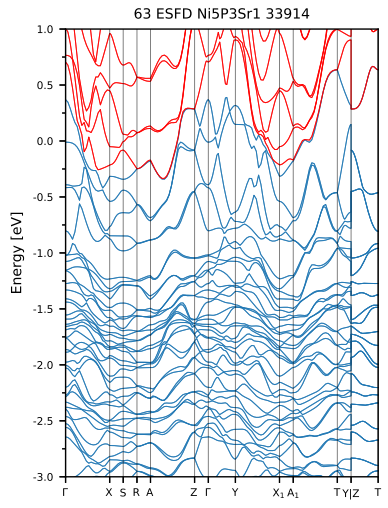

(ih) 63.4.19

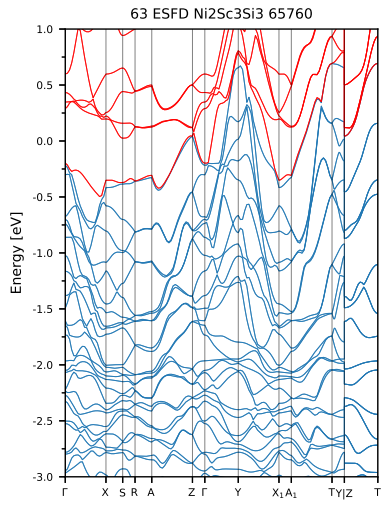

(ii) 63.4.20

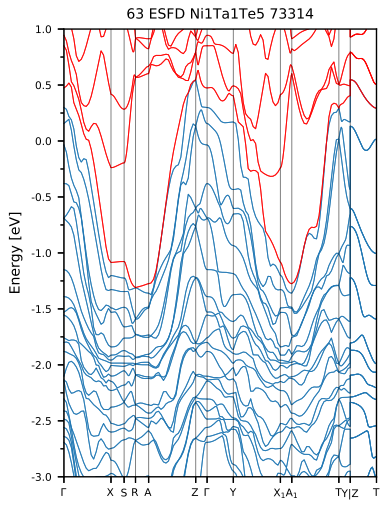

(ij) 63.4.21

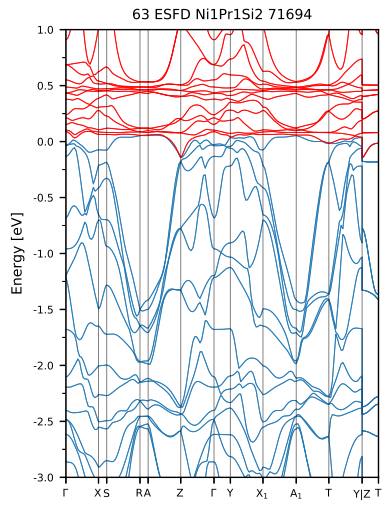

(ik) 63.4.22

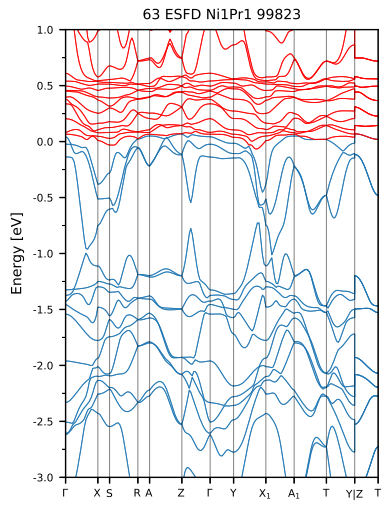

(il) 63.4.23

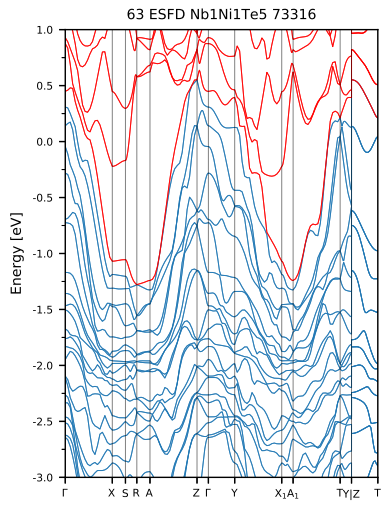

(im) 63.4.24

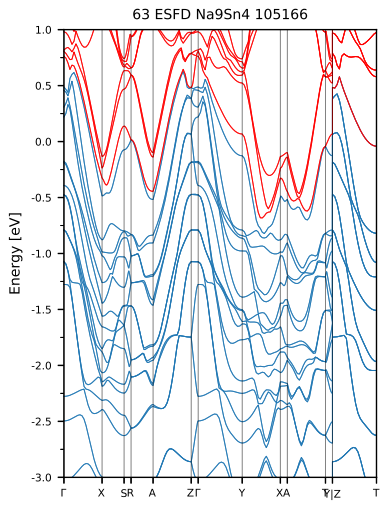

(in) 63.4.25

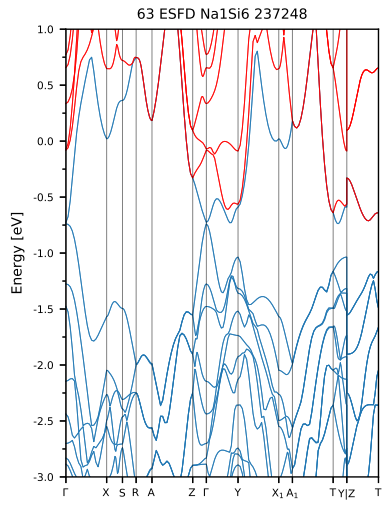

(io) 63.4.26

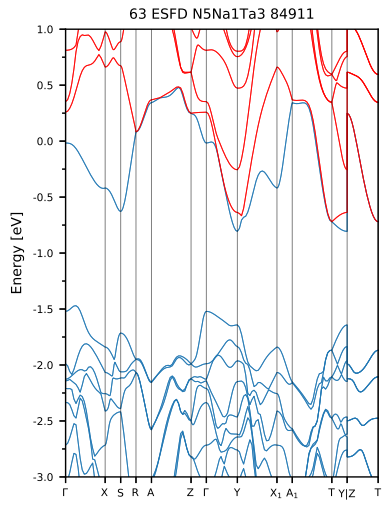

(ip) 63.4.27

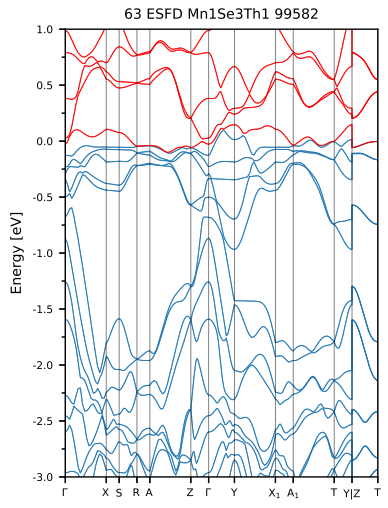

(iq) 63.4.28

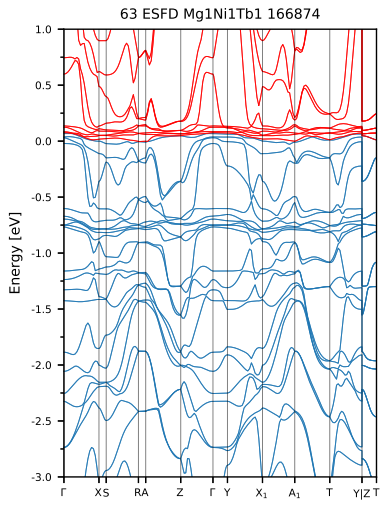

(ir) 63.4.29

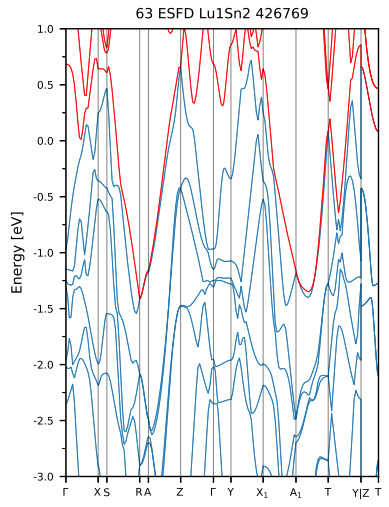

(is) 63.4.30

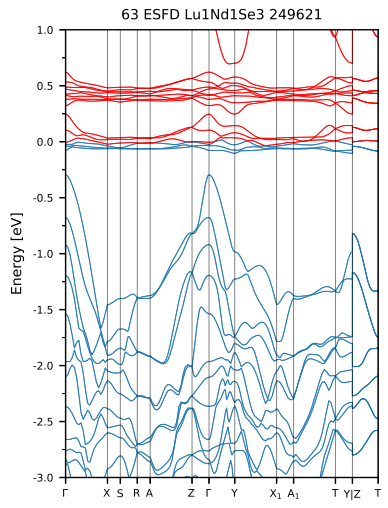

(it) 63.4.31

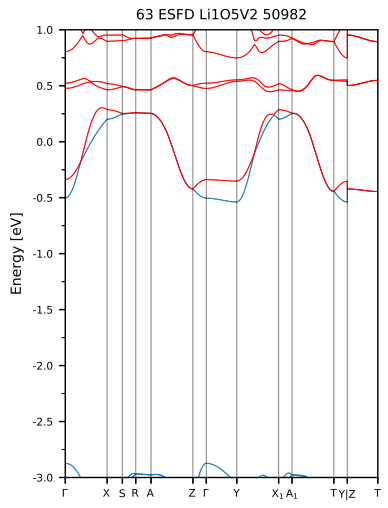

(iu) 63.4.32

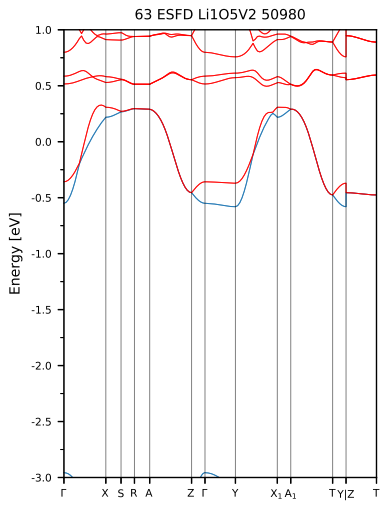

(iv) 63.4.33

Figure 160: Band structures for topological materials in SG 63, part 16/29, from 63.4.18 to 63.4.33. See the list of figures.

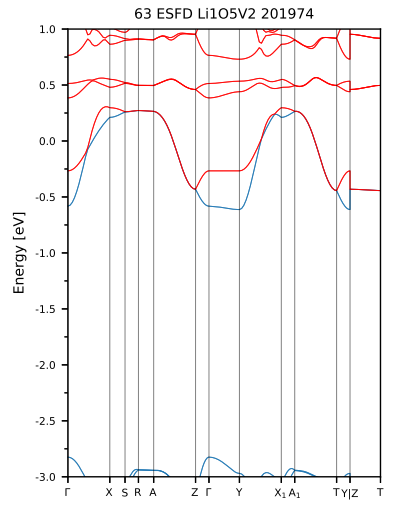

(iw) 63.4.34

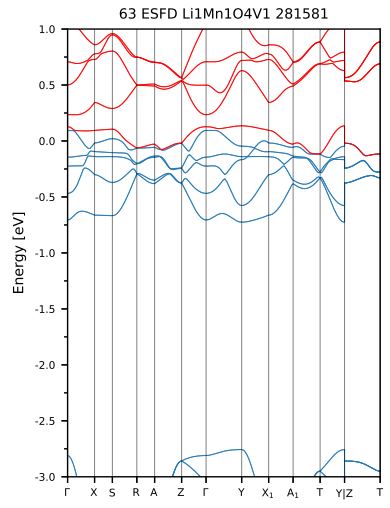

(ix) 63.4.35

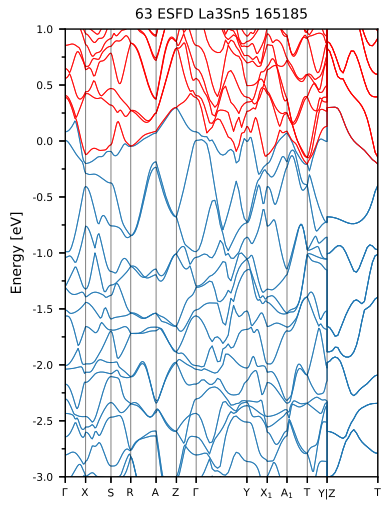

(iy) 63.4.36

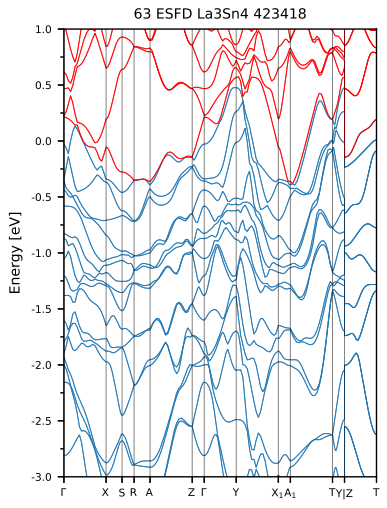

(iz) 63.4.37

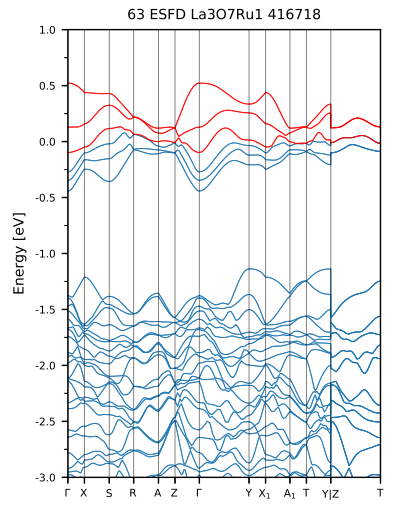

(ja) 63.4.38

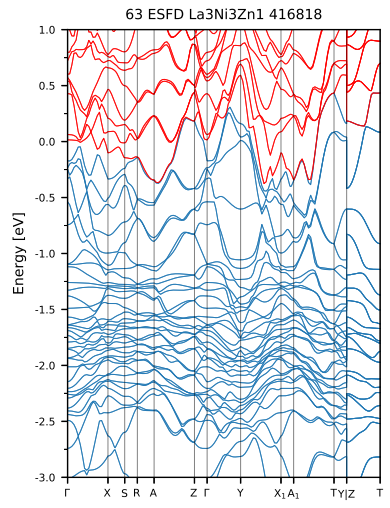

(jb) 63.4.39

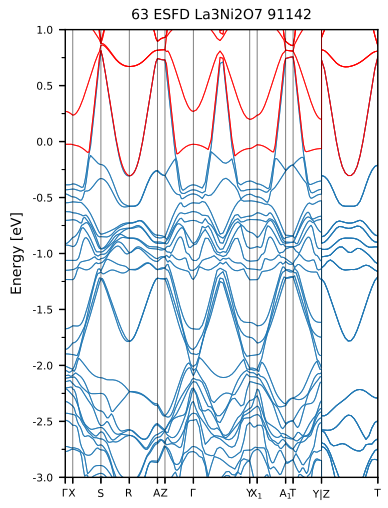

(jc) 63.4.40

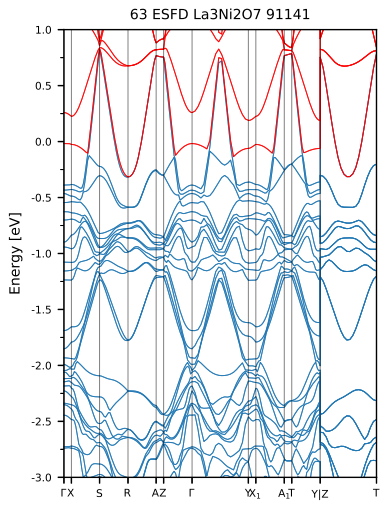

(jd) 63.4.41

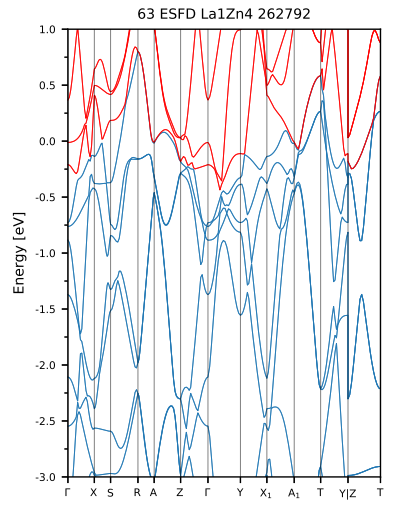

(je) 63.4.42

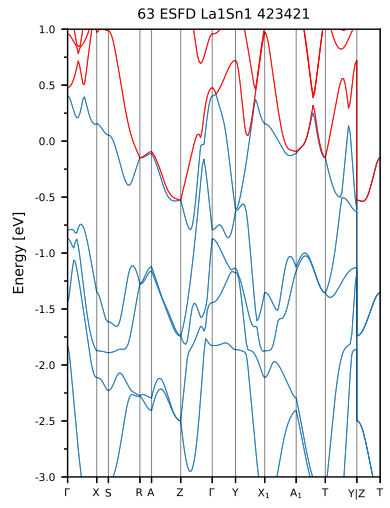

(jf) 63.4.43

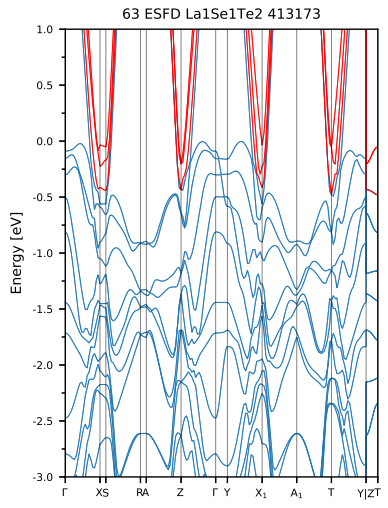

(jg) 63.4.44

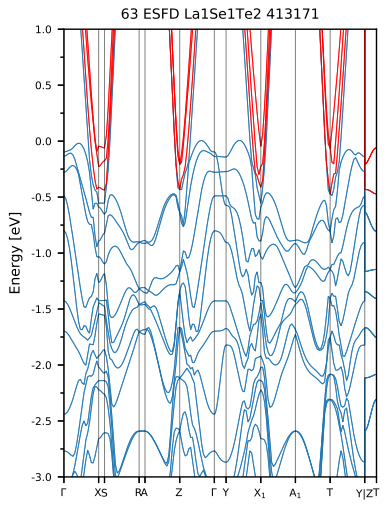

(jh) 63.4.45

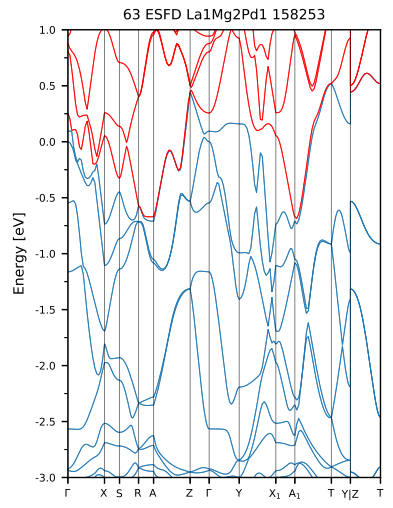

(ji) 63.4.46

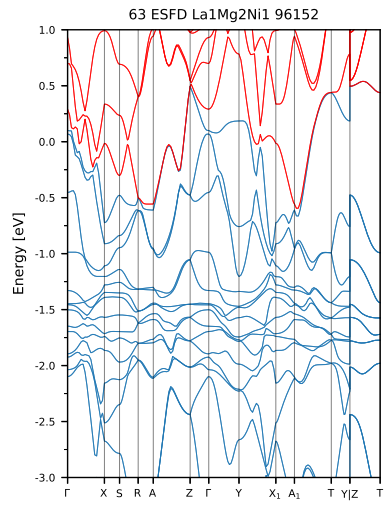

(jj) 63.4.47

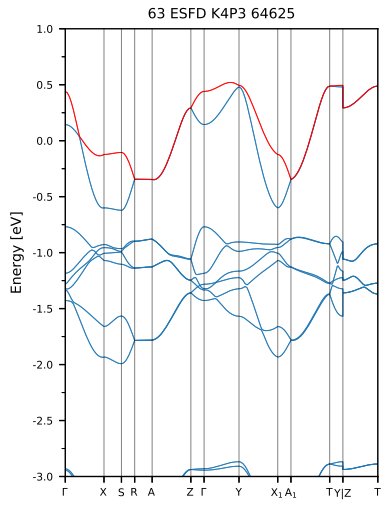

(jk) 63.4.48

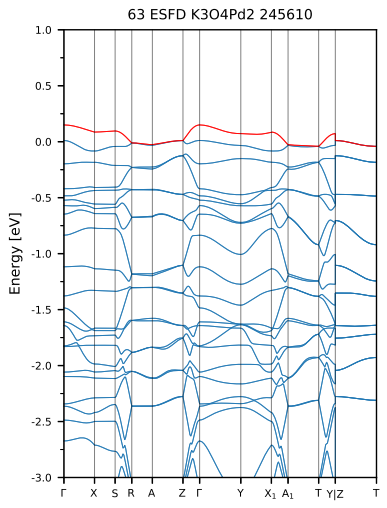

(jl) 63.4.49

Figure 161: Band structures for topological materials in SG 63, part 17/29, from 63.4.34 to 63.4.49. See the list of figures.

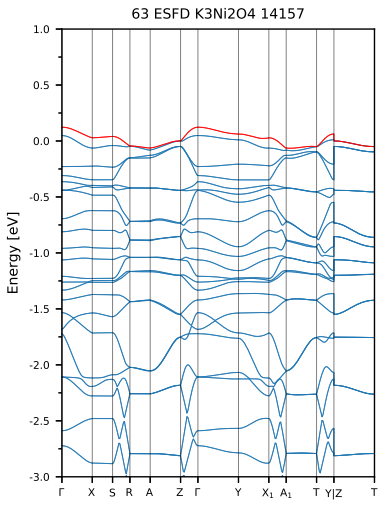

(jm) 63.4.50

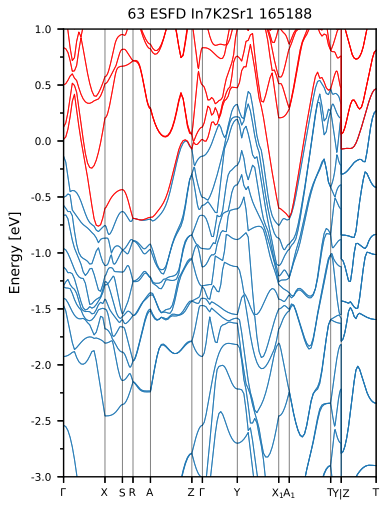

(jn) 63.4.51

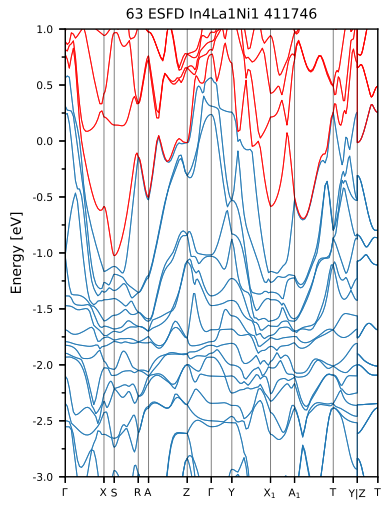

(jo) 63.4.52

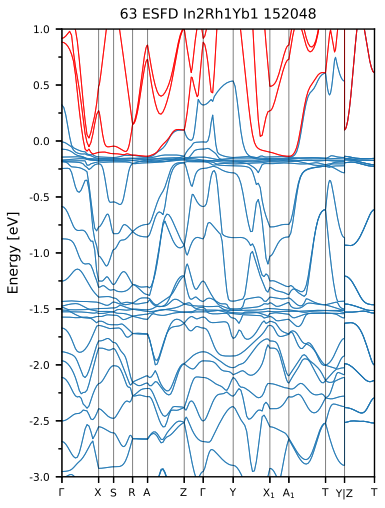

(jp) 63.4.53

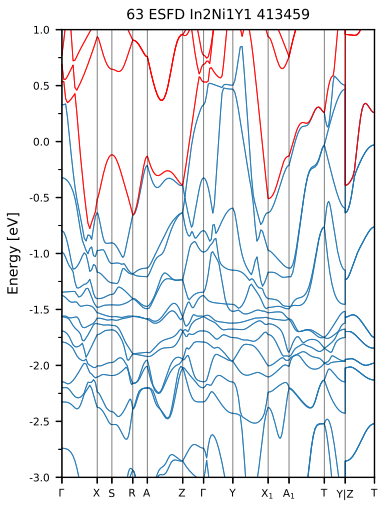

(jq) 63.4.54

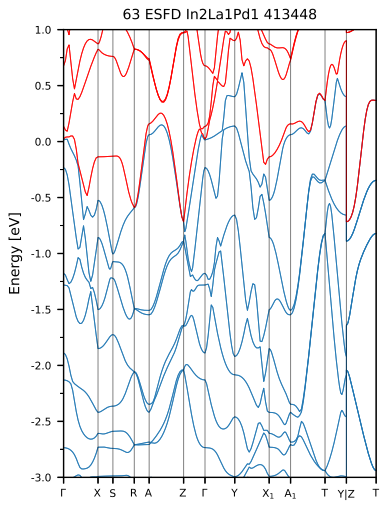

(jr) 63.4.55

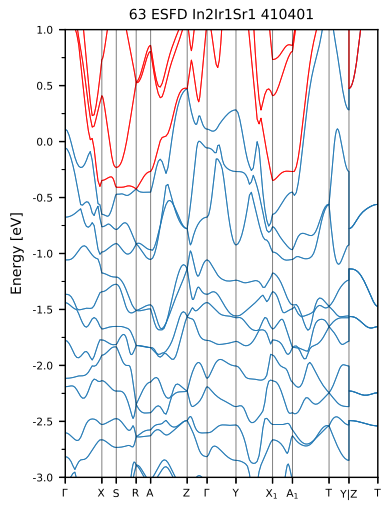

(js) 63.4.56

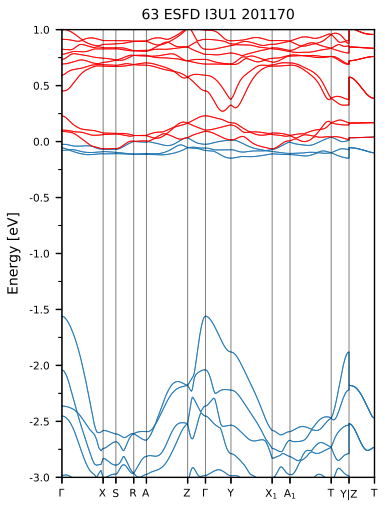

(jt) 63.4.57

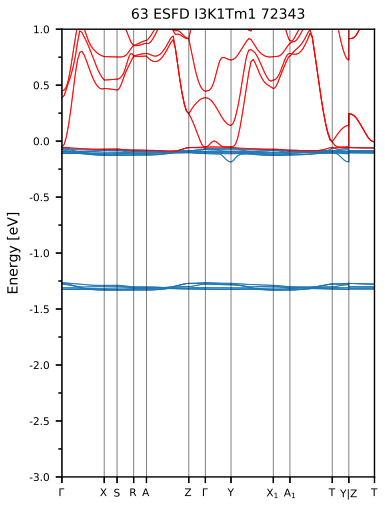

(ju) 63.4.58

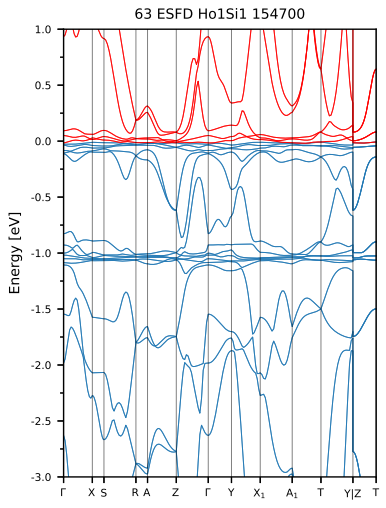

(jv) 63.4.59

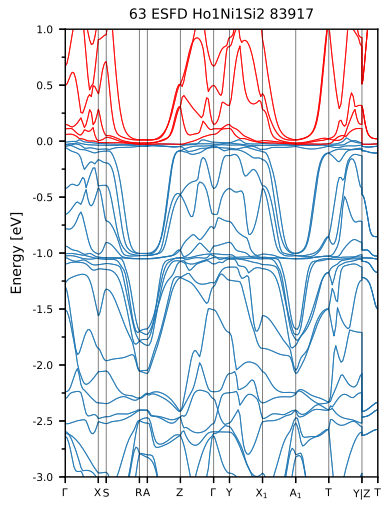

(jw) 63.4.60

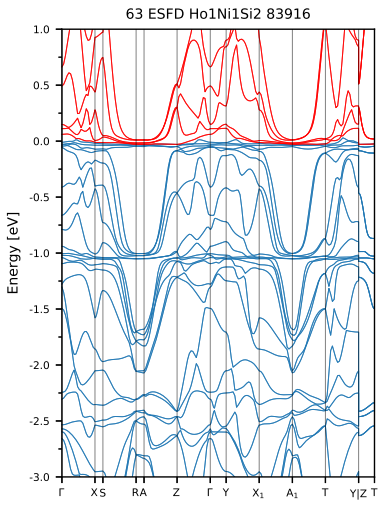

(jx) 63.4.61

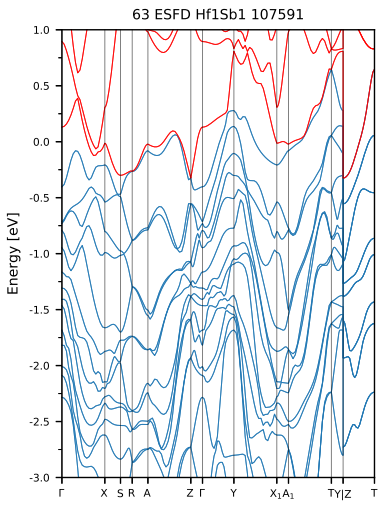

(jy) 63.4.62

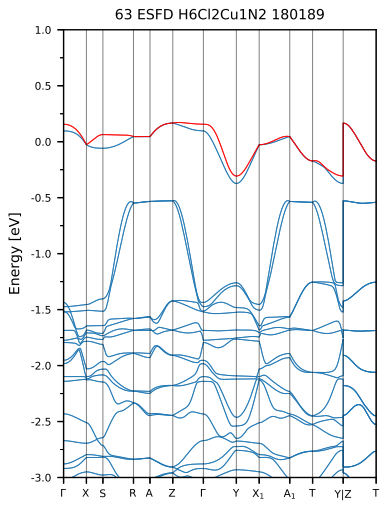

(jz) 63.4.63

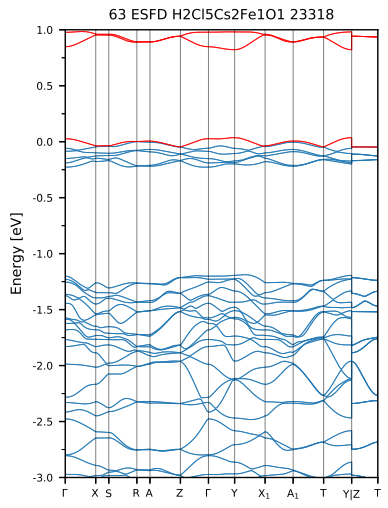

(ka) 63.4.64

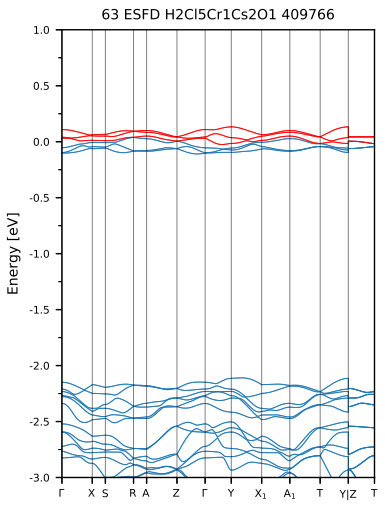

(kb) 63.4.65

Figure 162: Band structures for topological materials in SG 63, part 18/29, from 63.4.50 to 63.4.65. See the list of figures.

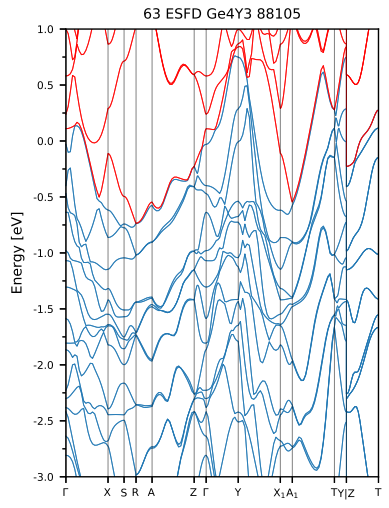

(kc) 63.4.66

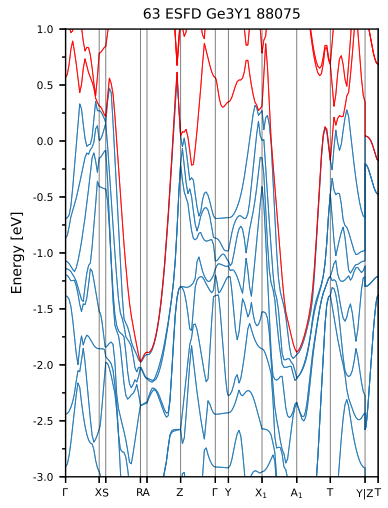

(kd) 63.4.67

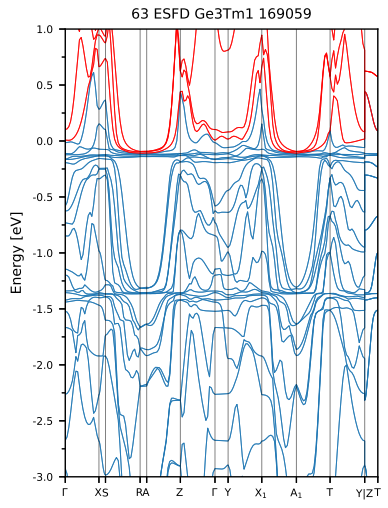

(ke) 63.4.68

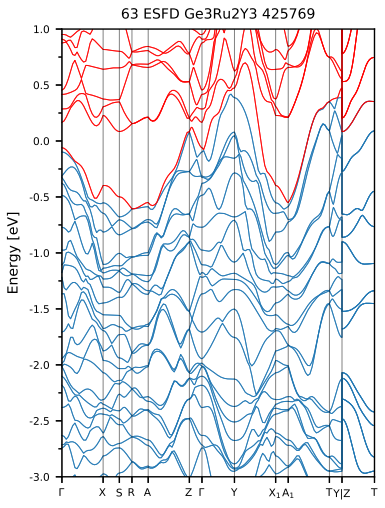

(kf) 63.4.69

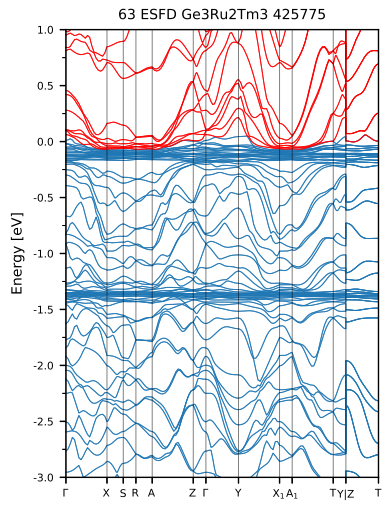

(kg) 63.4.70

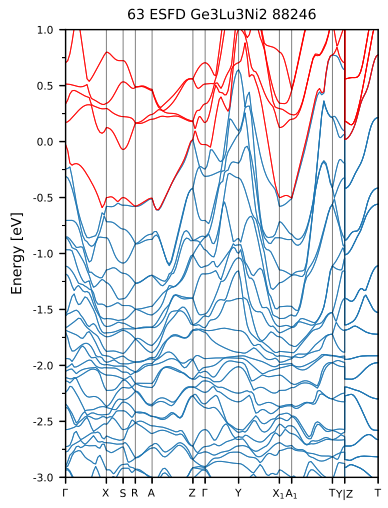

(kh) 63.4.71

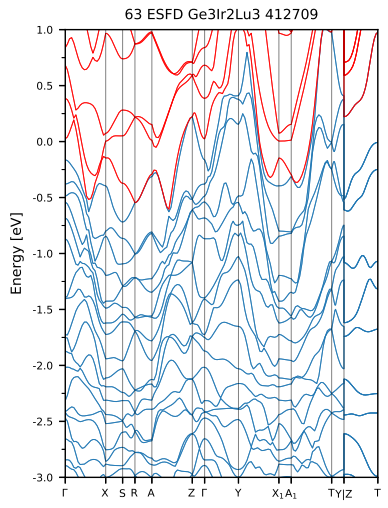

(ki) 63.4.72

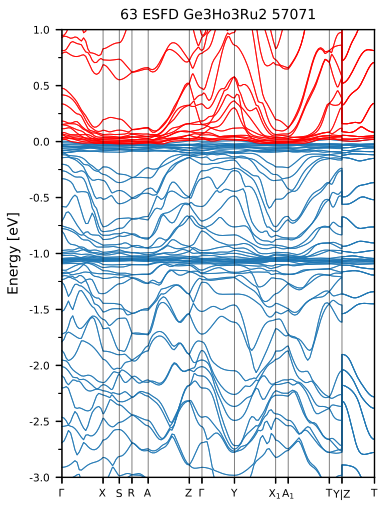

(kj) 63.4.73

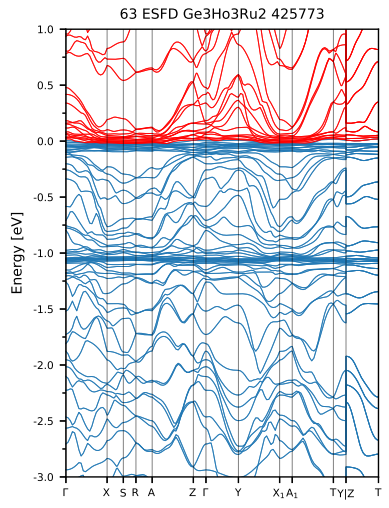

(kk) 63.4.74

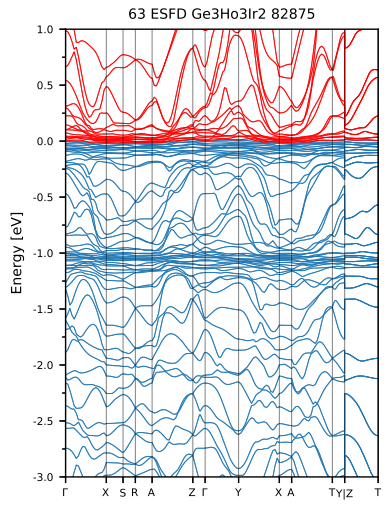

(kl) 63.4.75

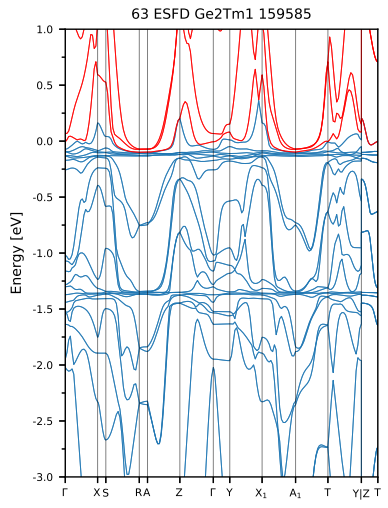

(km) 63.4.76

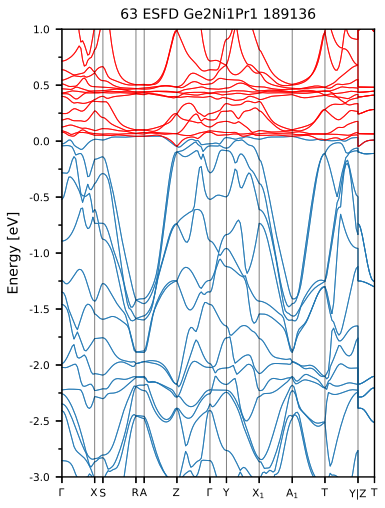

(kn) 63.4.77

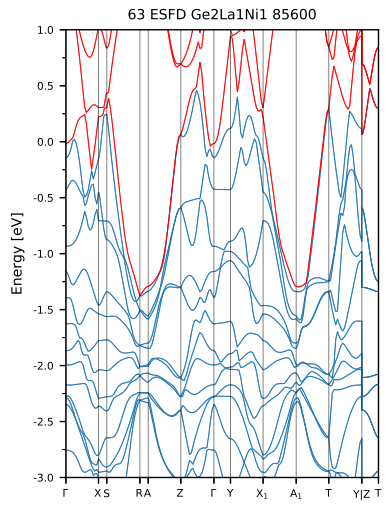

(ko) 63.4.78

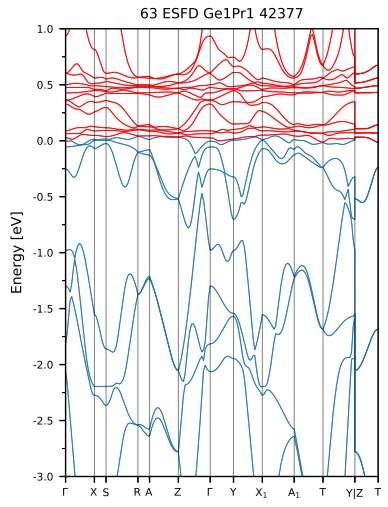

(kp) 63.4.79

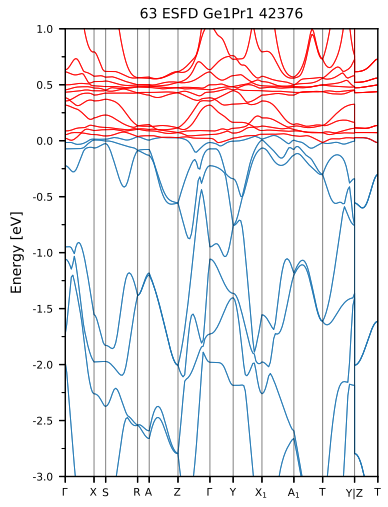

(kq) 63.4.80

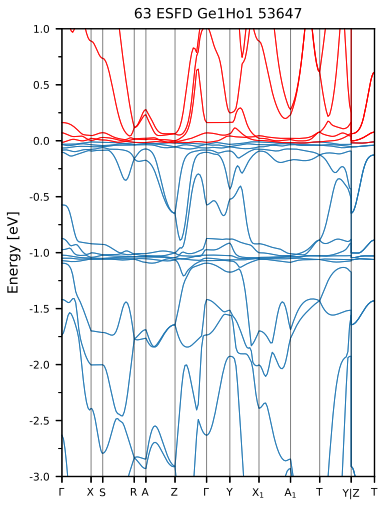

(kr) 63.4.81

Figure 163: Band structures for topological materials in SG 63, part 19/29, from 63.4.66 to 63.4.81. See the list of figures.

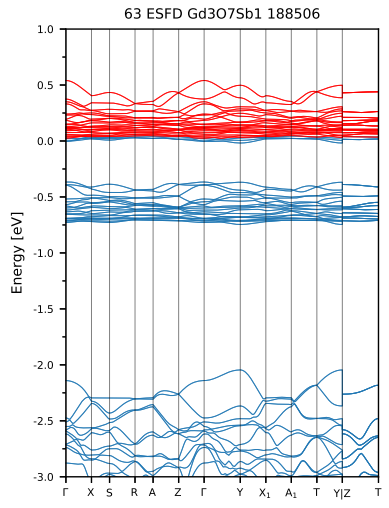

(ks) 63.4.82

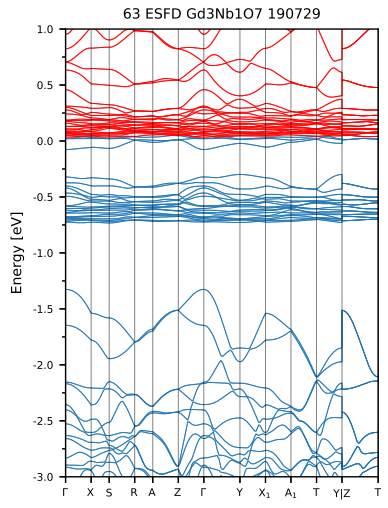

(kt) 63.4.83

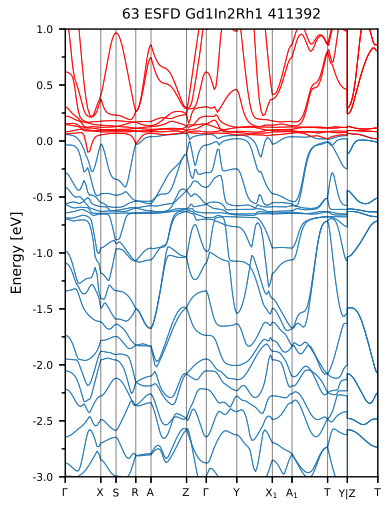

(ku) 63.4.84

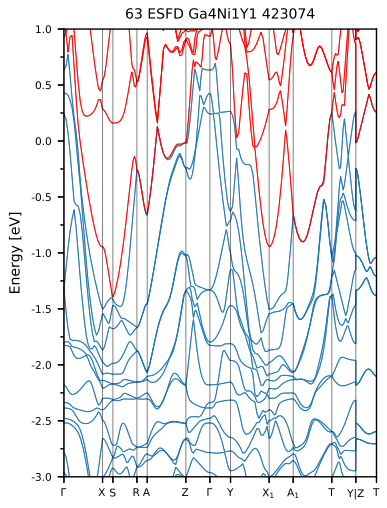

(kv) 63.4.85

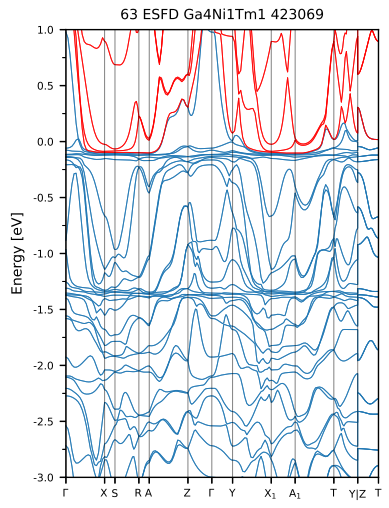

(kw) 63.4.86

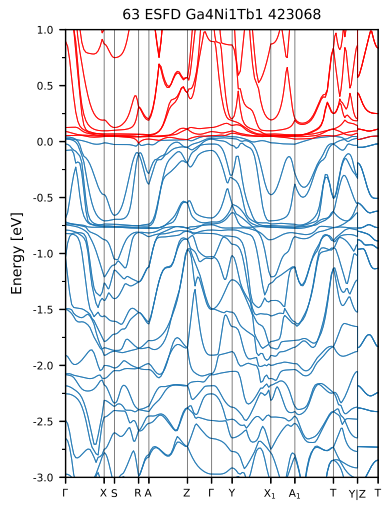

(kx) 63.4.87

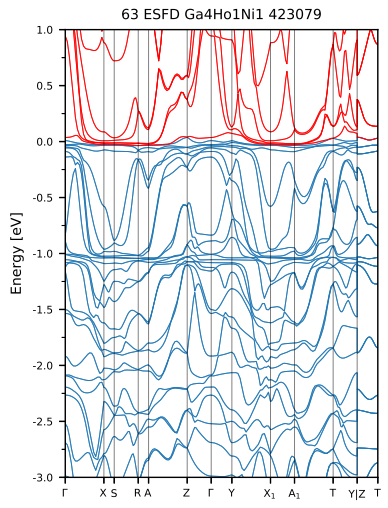

(ky) 63.4.88

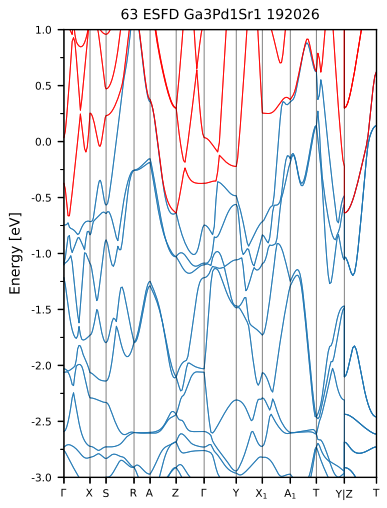

(kz) 63.4.89

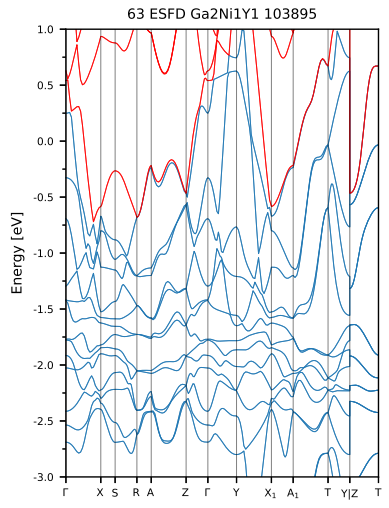

(la) 63.4.90

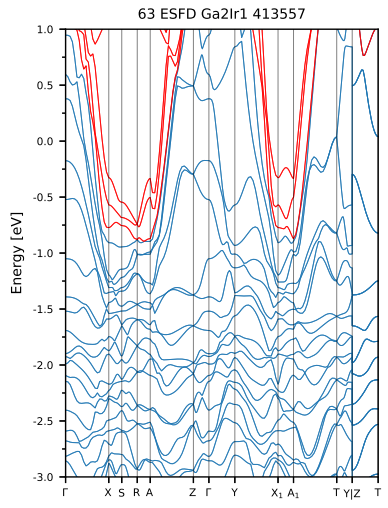

(lb) 63.4.91

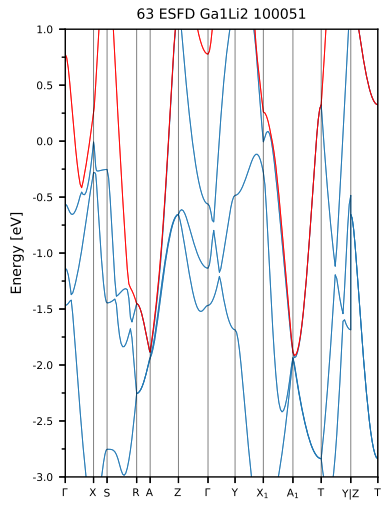

(lc) 63.4.92

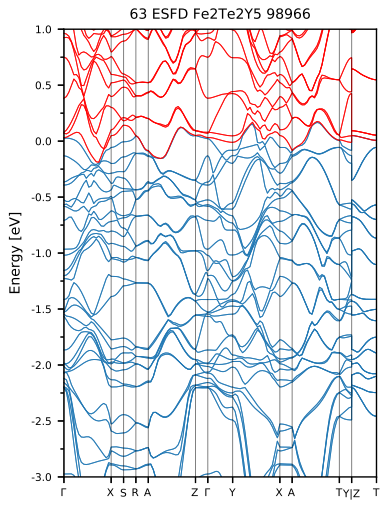

(ld) 63.4.93

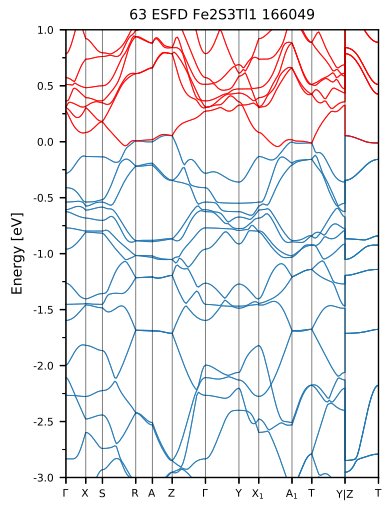

(le) 63.4.94

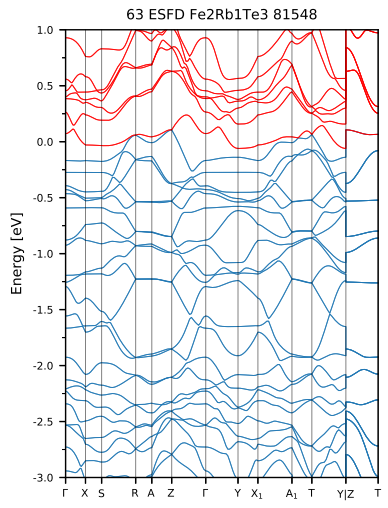

(lf) 63.4.95

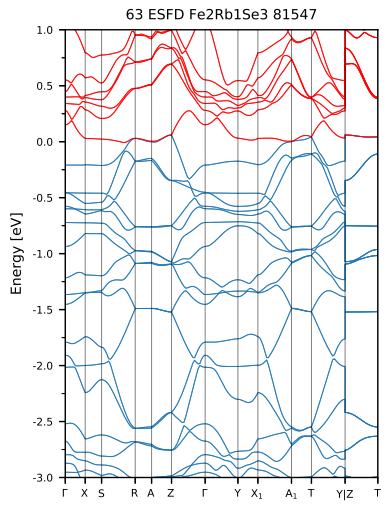

(lg) 63.4.96

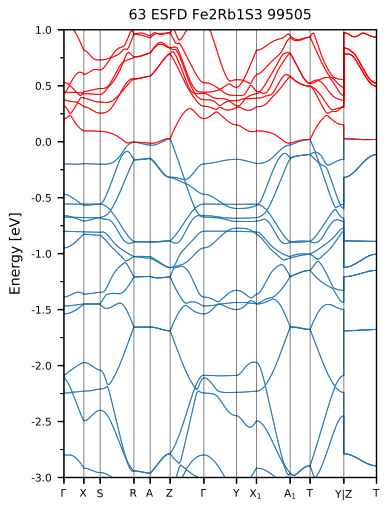

(lh) 63.4.97

Figure 164: Band structures for topological materials in SG 63, part 20/29, from 63.4.82 to 63.4.97. See the list of figures.

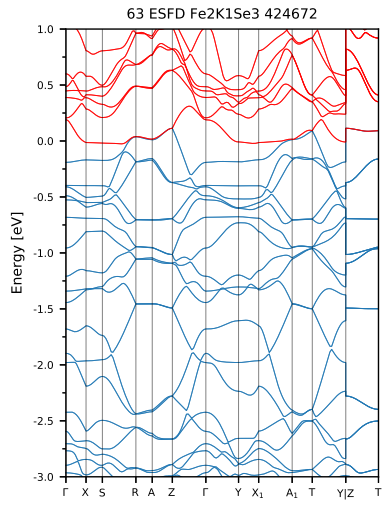

(li) 63.4.98

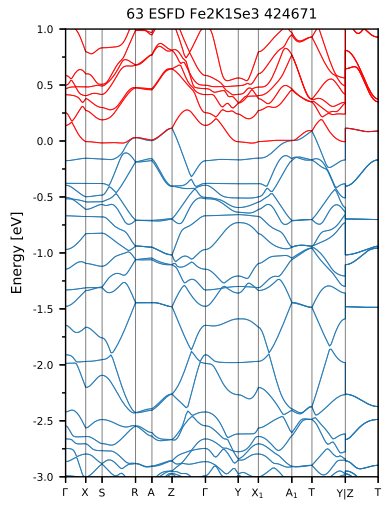

(lj) 63.4.99

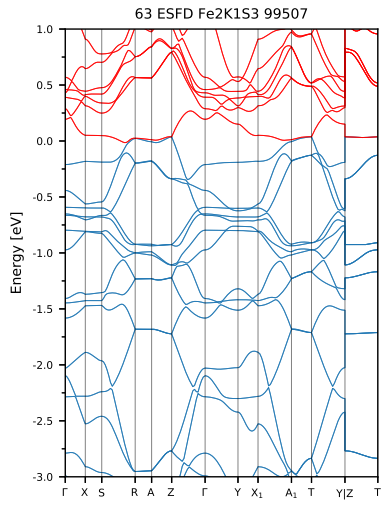

(lk) 63.4.100

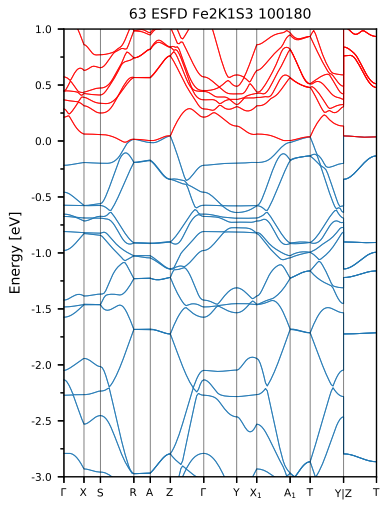

(ll) 63.4.101

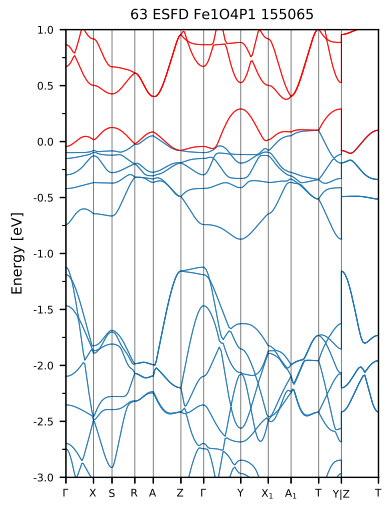

(lm) 63.4.102

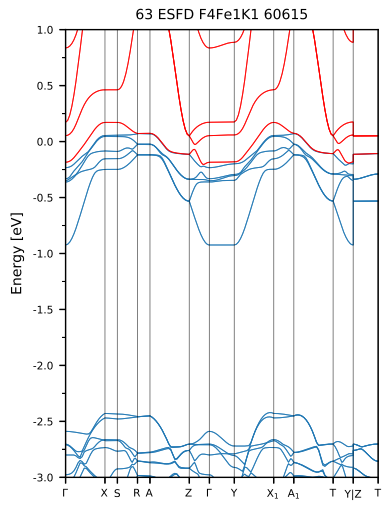

(ln) 63.4.103

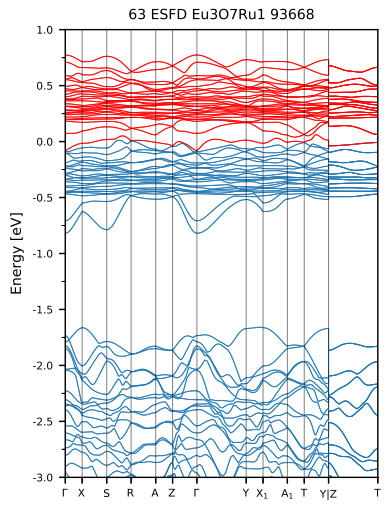

(lo) 63.4.104

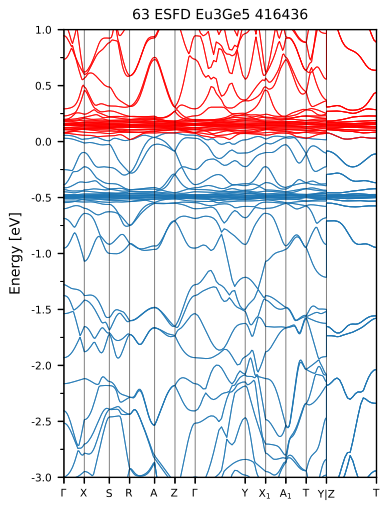

(lp) 63.4.105

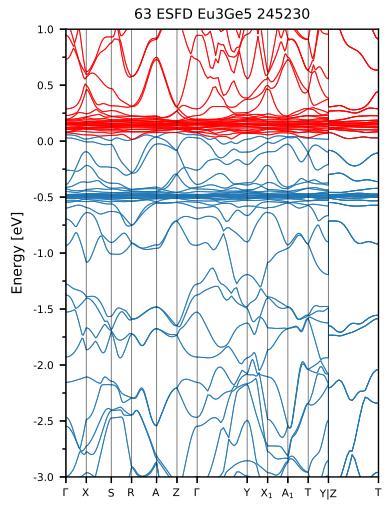

(lq) 63.4.106

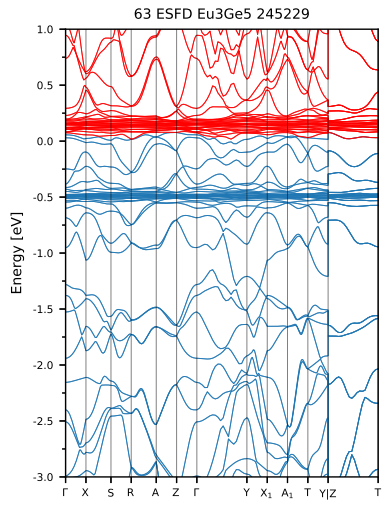

(lr) 63.4.107

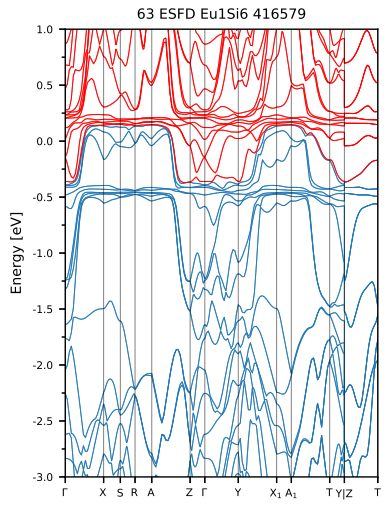

(ls) 63.4.108

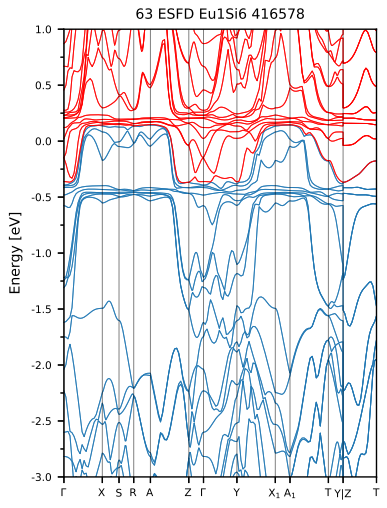

(lt) 63.4.109

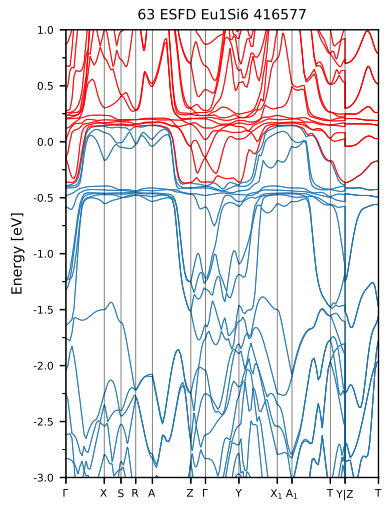

(lu) 63.4.110

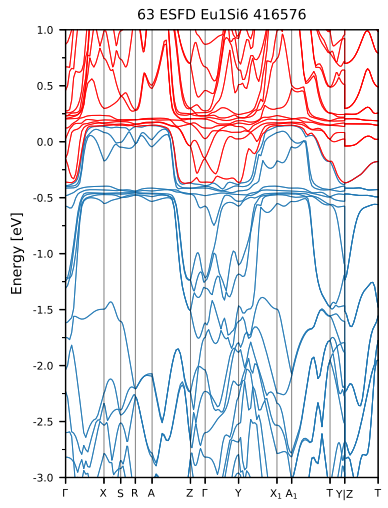

(lv) 63.4.111

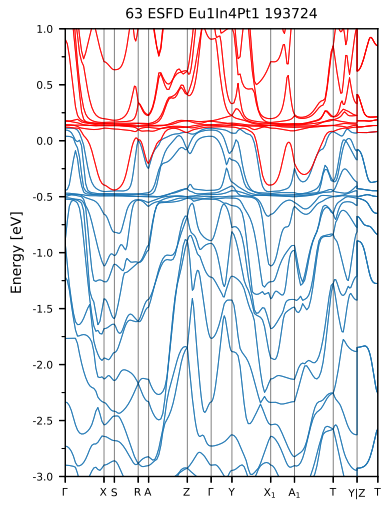

(lw) 63.4.112

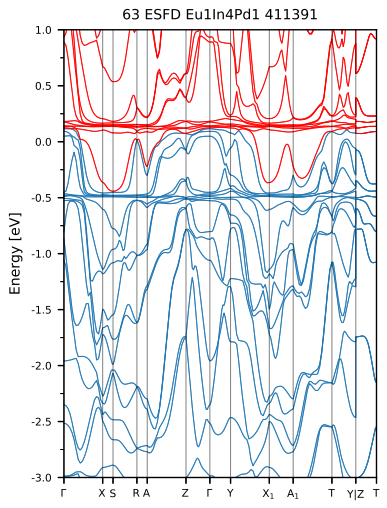

(lx) 63.4.113

Figure 165: Band structures for topological materials in SG 63, part 21/29, from 63.4.98 to 63.4.113. See the list of figures.

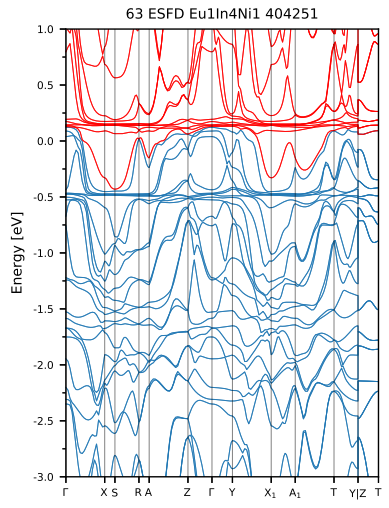

(ly) 63.4.114

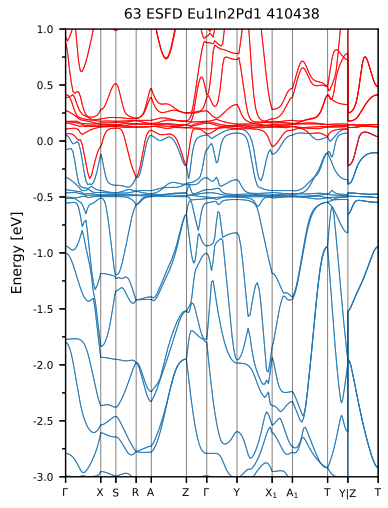

(lz) 63.4.115

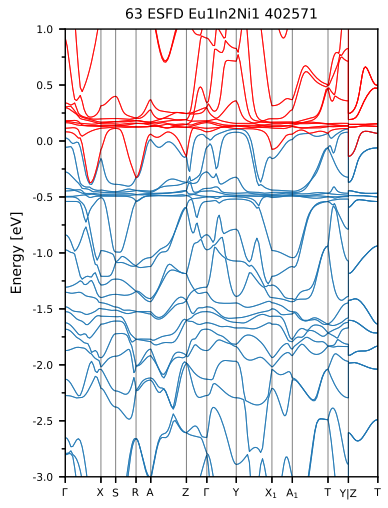

(ma) 63.4.116

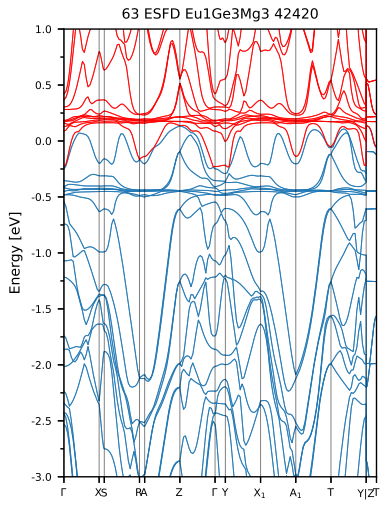

(mb) 63.4.117

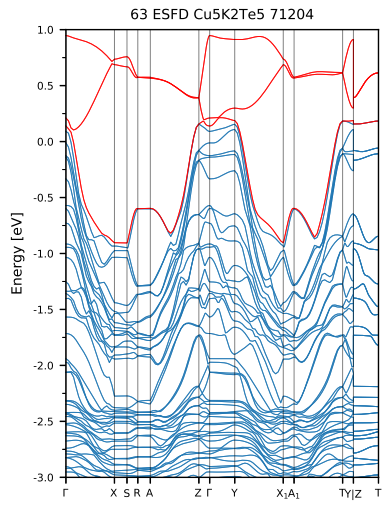

(mc) 63.4.118

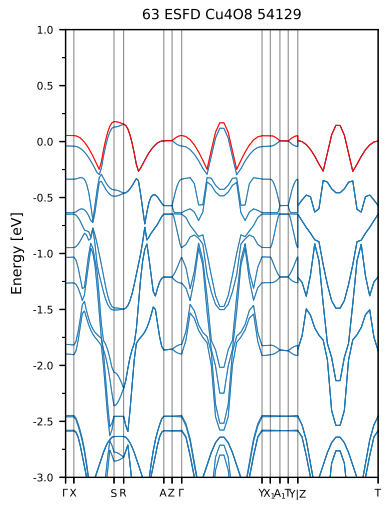

(md) 63.4.119

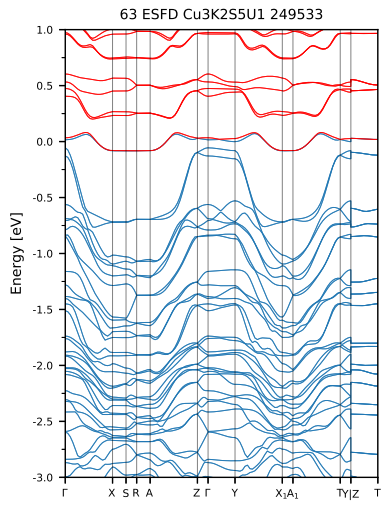

(me) 63.4.120

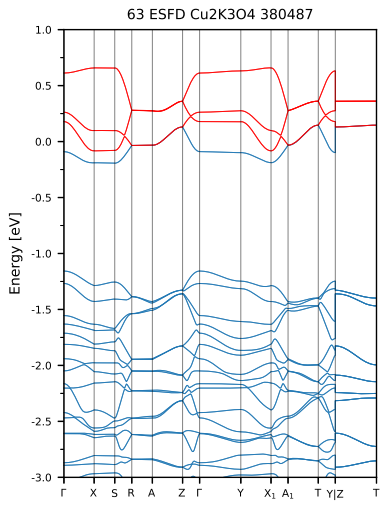

(mf) 63.4.121

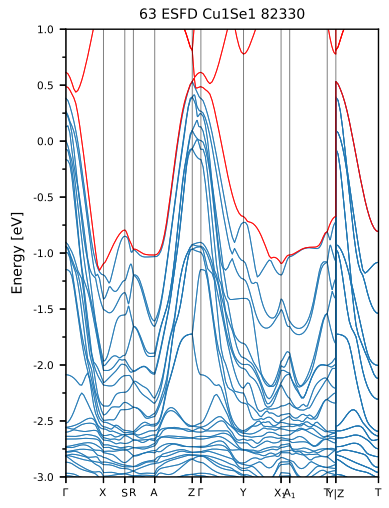

(mg) 63.4.122

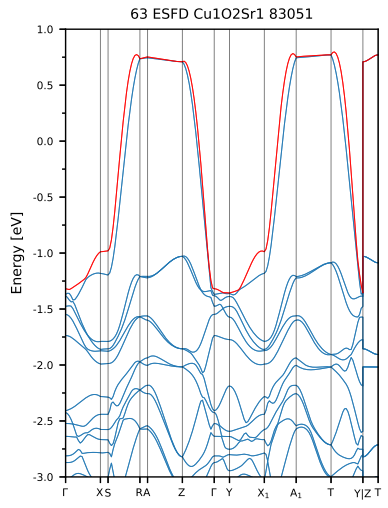

(mh) 63.4.123

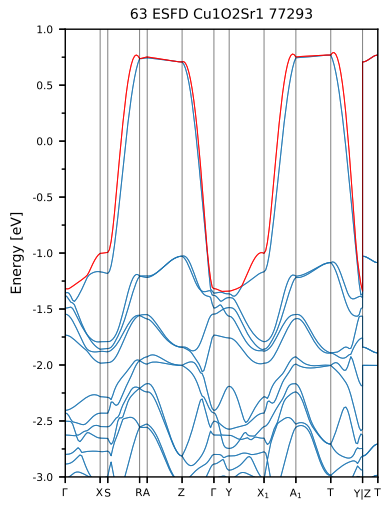

(mi) 63.4.124

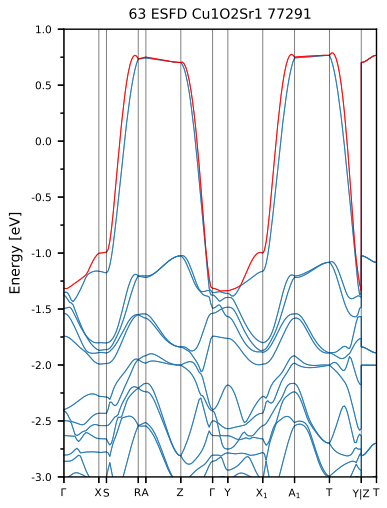

(mj) 63.4.125

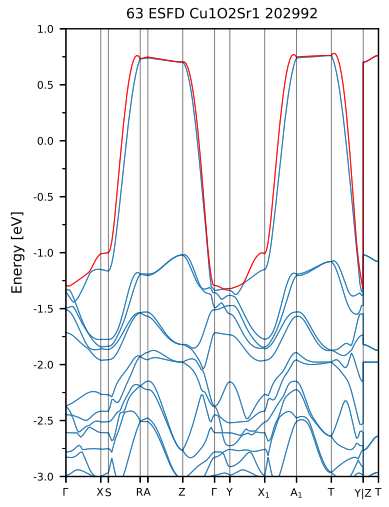

(mk) 63.4.126

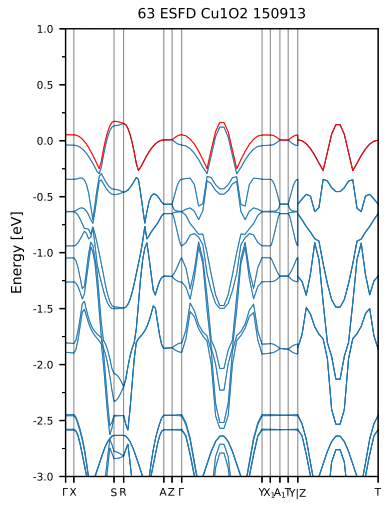

(ml) 63.4.127

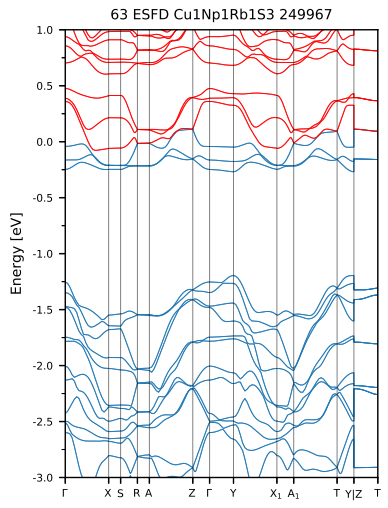

(mm) 63.4.128

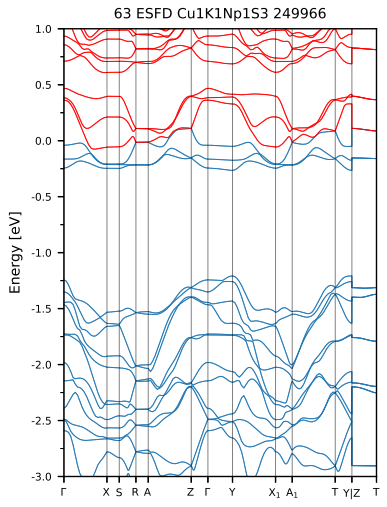

(mn) 63.4.129

Figure 166: Band structures for topological materials in SG 63, part 22/29, from 63.4.114 to 63.4.129. See the list of figures.

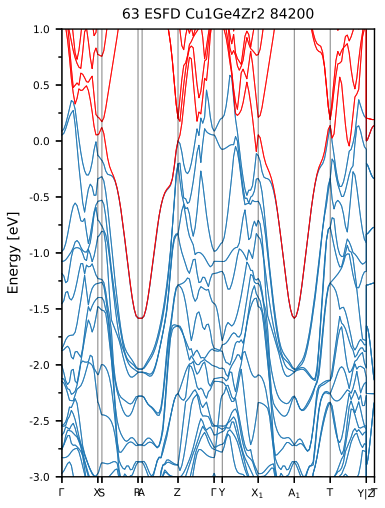

(mo) 63.4.130

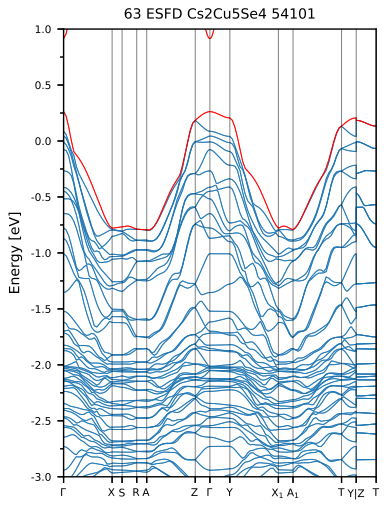

(mp) 63.4.131

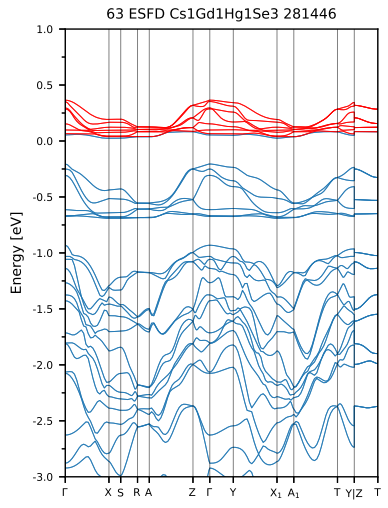

(mq) 63.4.132

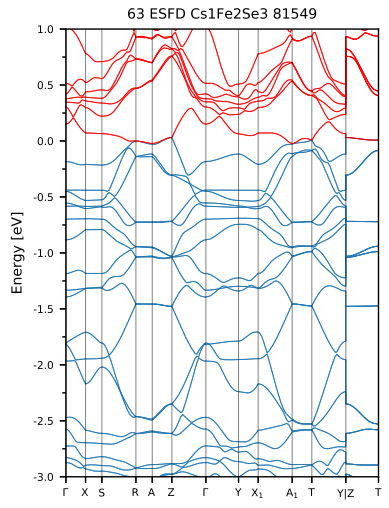

(mr) 63.4.133

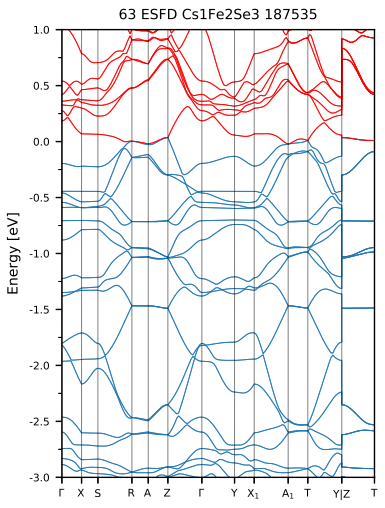

(ms) 63.4.134

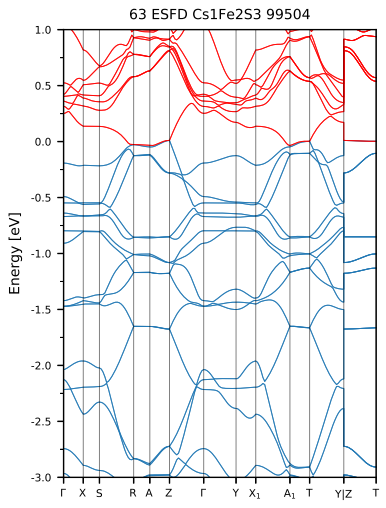

(mt) 63.4.135

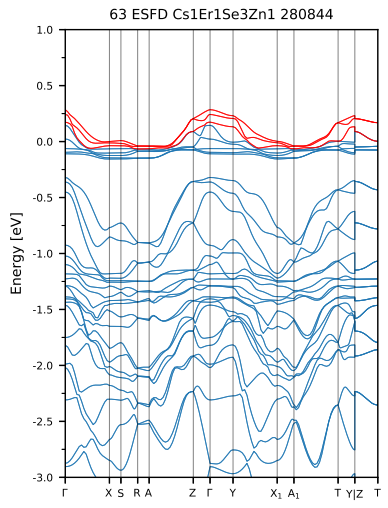

(mu) 63.4.136

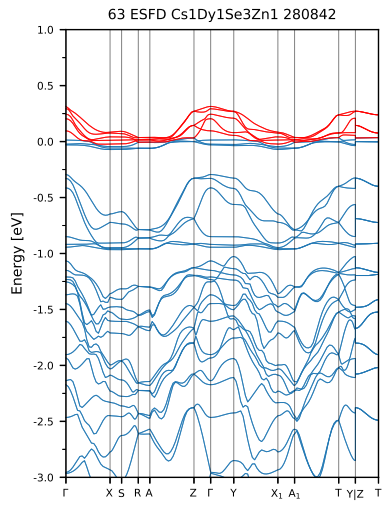

(mv) 63.4.137

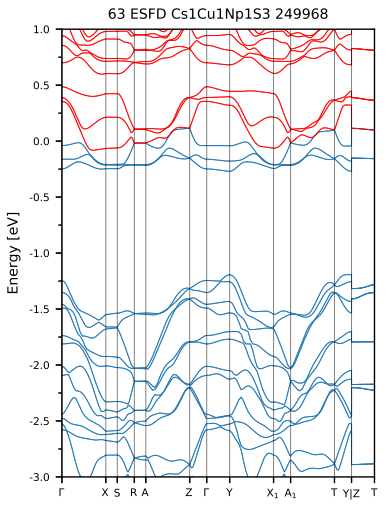

(mw) 63.4.138

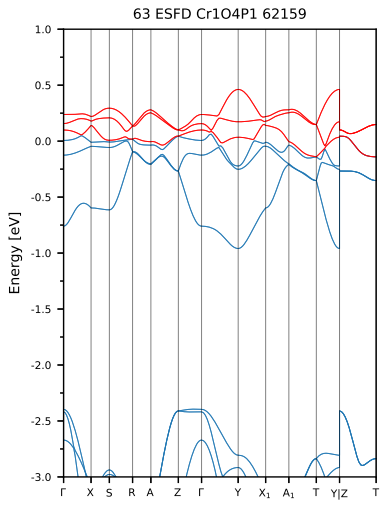

(mx) 63.4.139

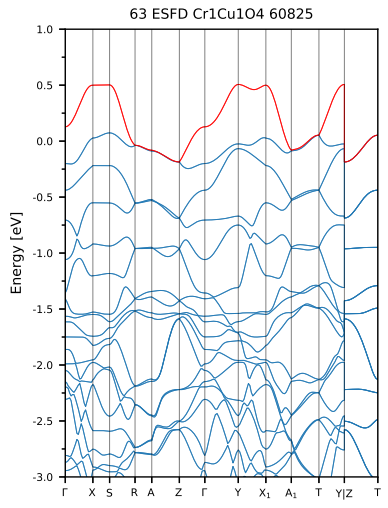

(my) 63.4.140

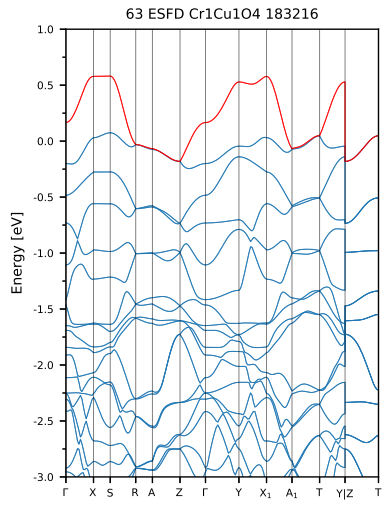

(mz) 63.4.141

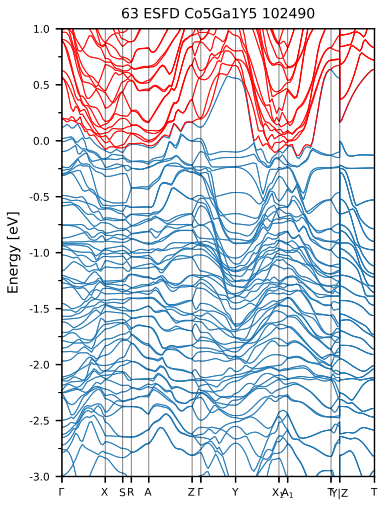

(na) 63.4.142

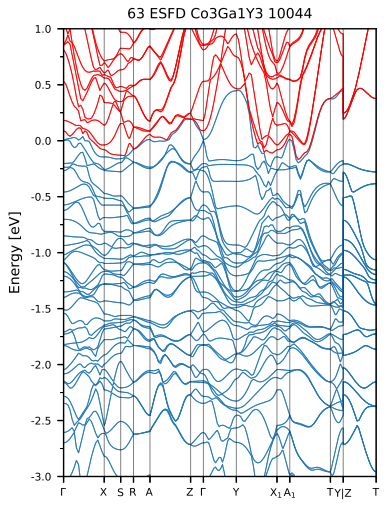

(nb) 63.4.143

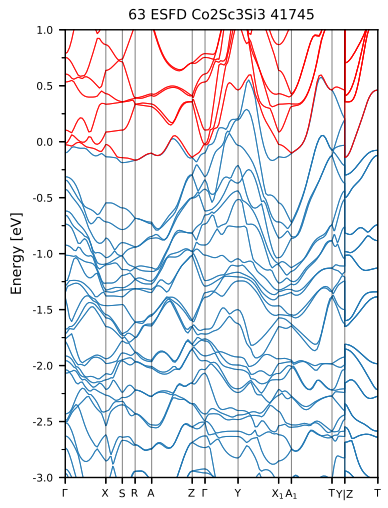

(nc) 63.4.144

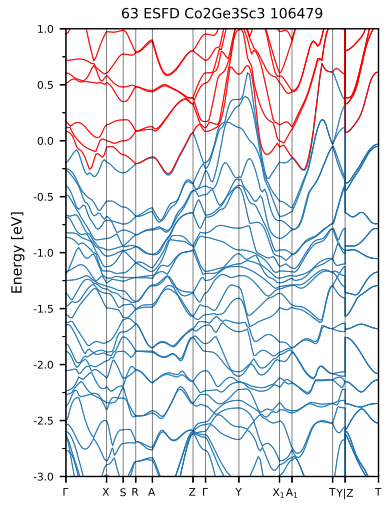

(nd) 63.4.145

Figure 167: Band structures for topological materials in SG 63, part 23/29, from 63.4.130 to 63.4.145. See the list of figures.

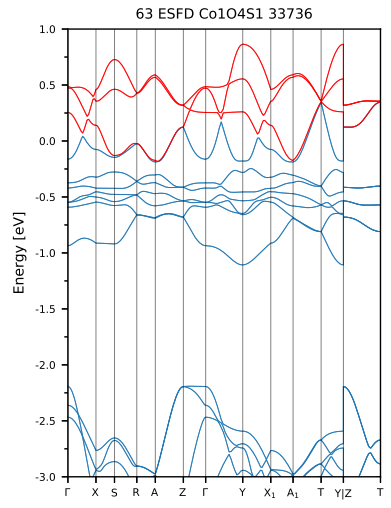

(ne) 63.4.146

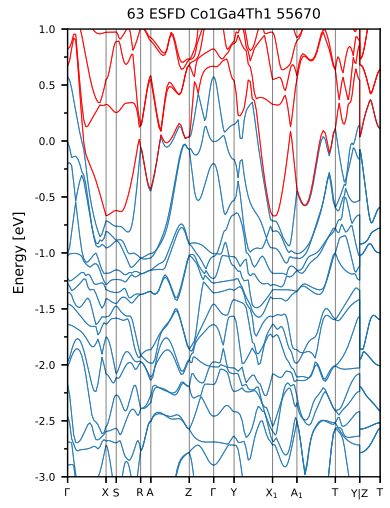

(nf) 63.4.147

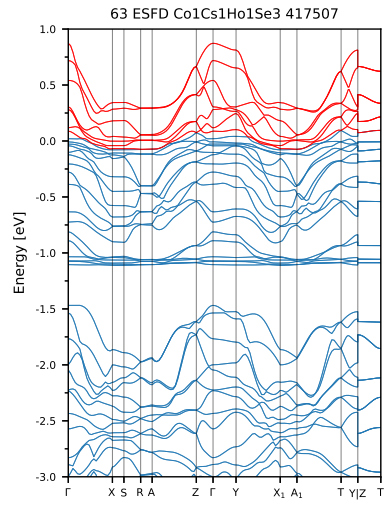

(ng) 63.4.148

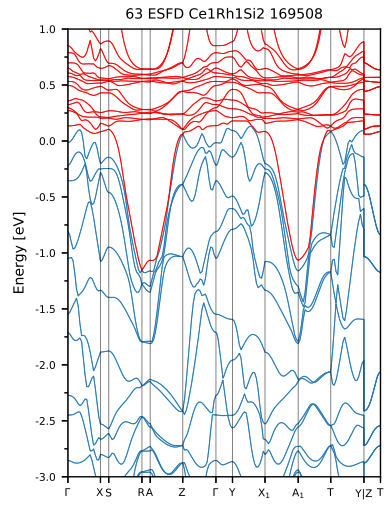

(nh) 63.4.149

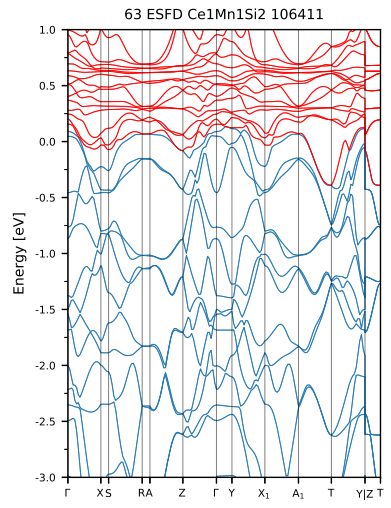

(ni) 63.4.150

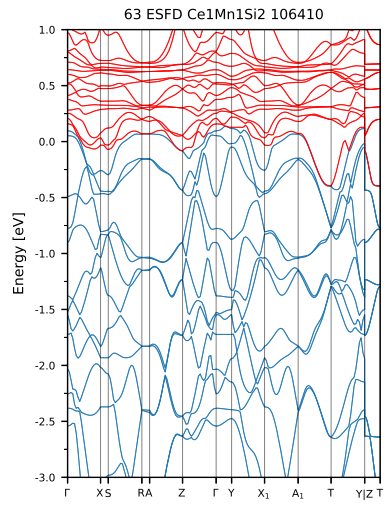

(nj) 63.4.151

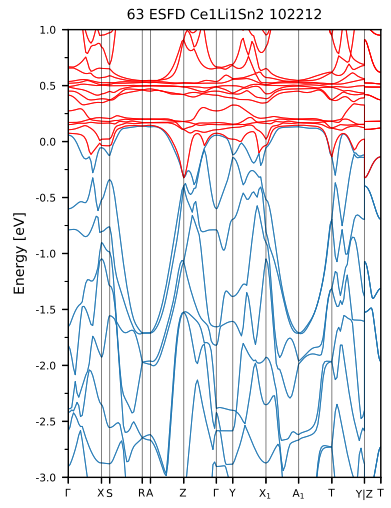

(nk) 63.4.152

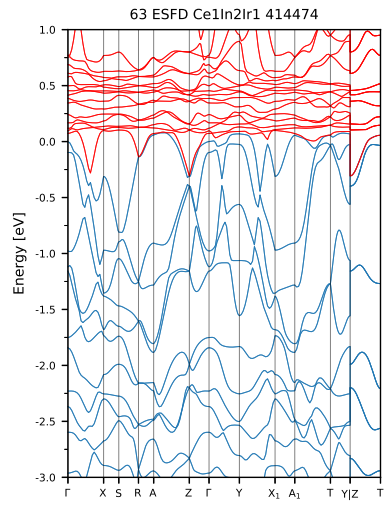

(nl) 63.4.153

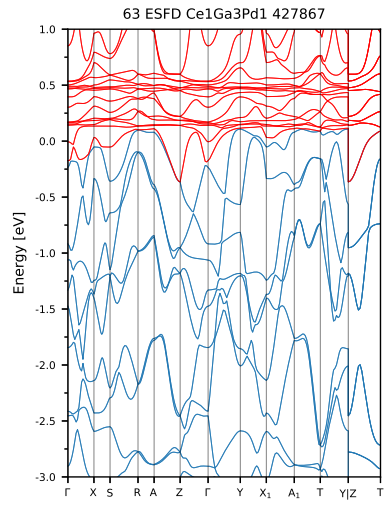

(nm) 63.4.154

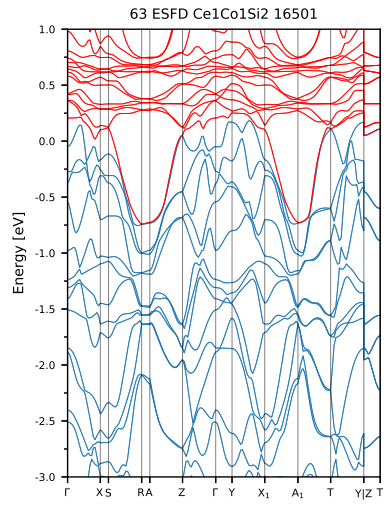

(nn) 63.4.155

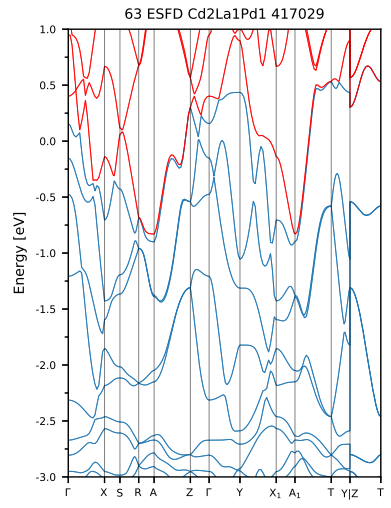

(no) 63.4.156

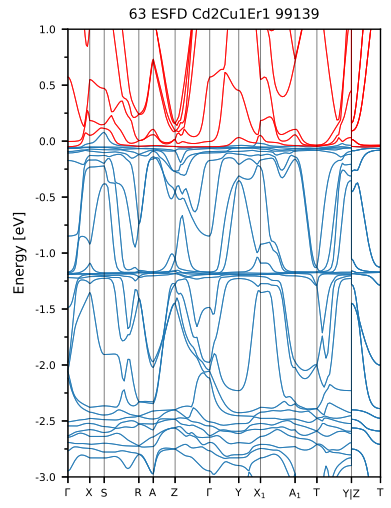

(np) 63.4.157

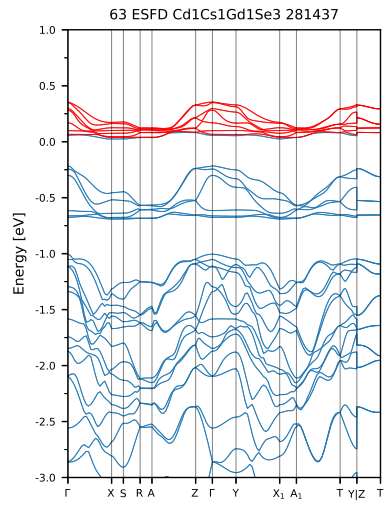

(nq) 63.4.158

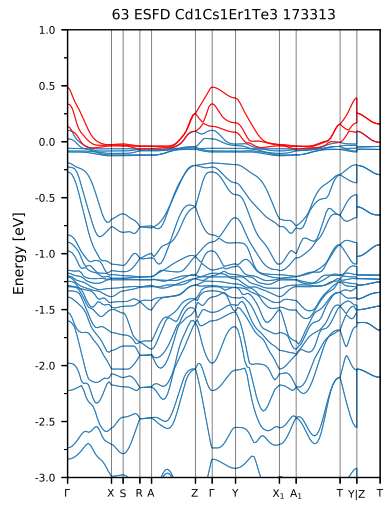

(nr) 63.4.159

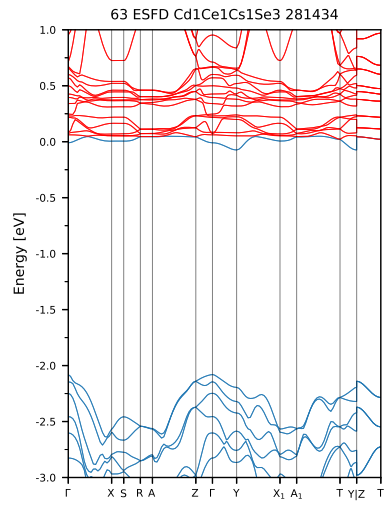

(ns) 63.4.160

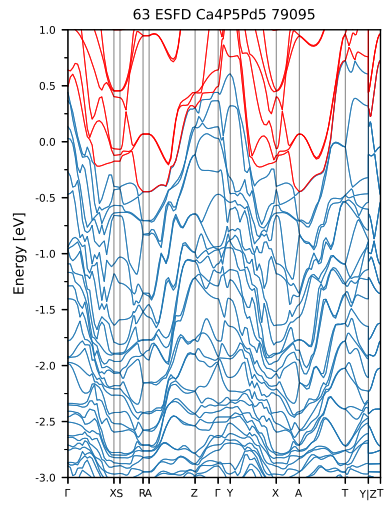

(nt) 63.4.161

Figure 168: Band structures for topological materials in SG 63, part 24/29, from 63.4.146 to 63.4.161. See the list of figures.

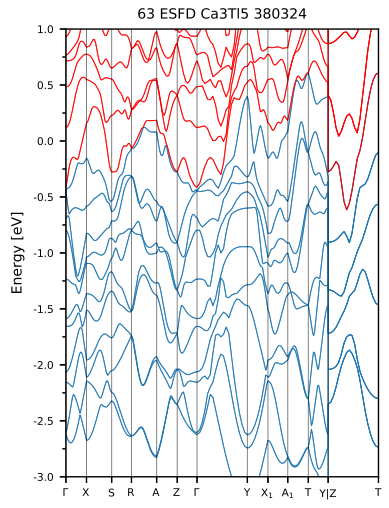

(nu) 63.4.162

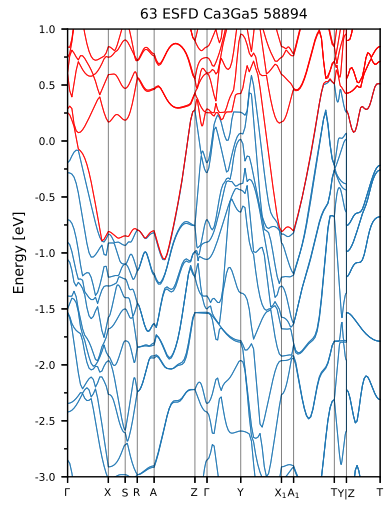

(nv) 63.4.163

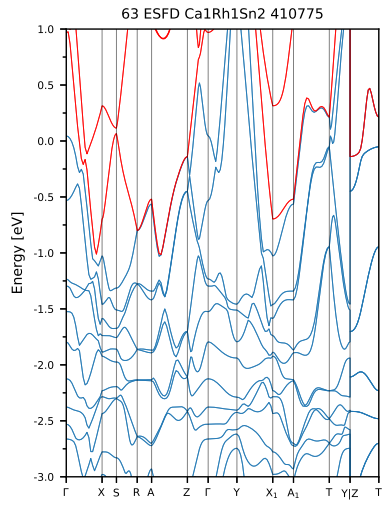

(nw) 63.4.164

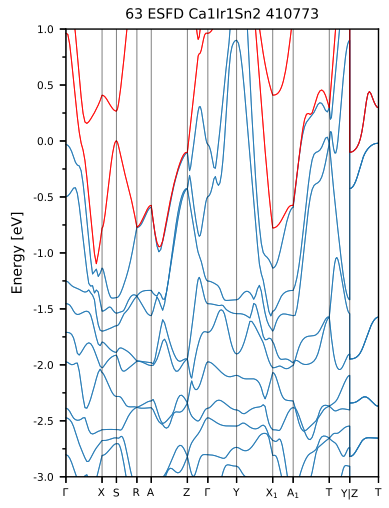

(nx) 63.4.165

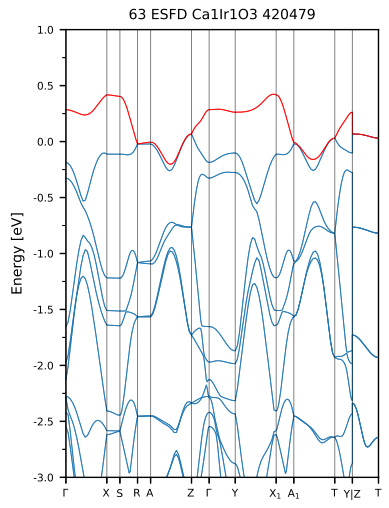

(ny) 63.4.166

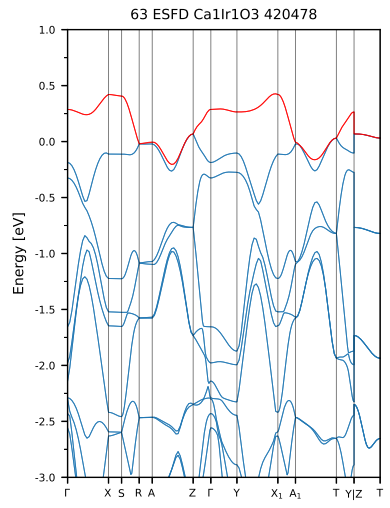

(nz) 63.4.167

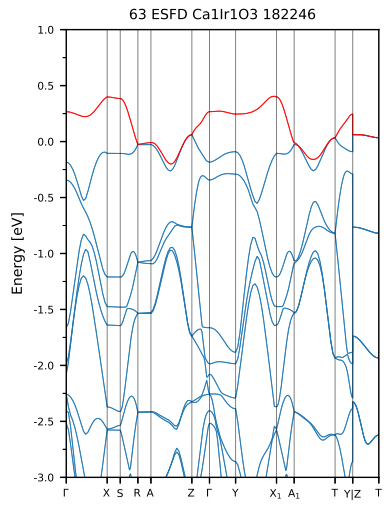

(oa) 63.4.168

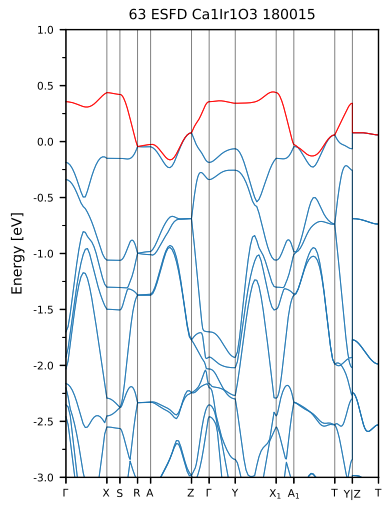

(ob) 63.4.169

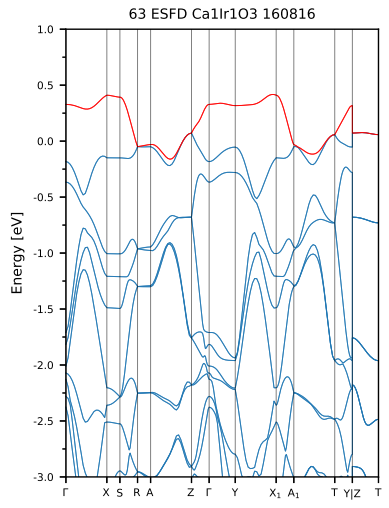

(oc) 63.4.170

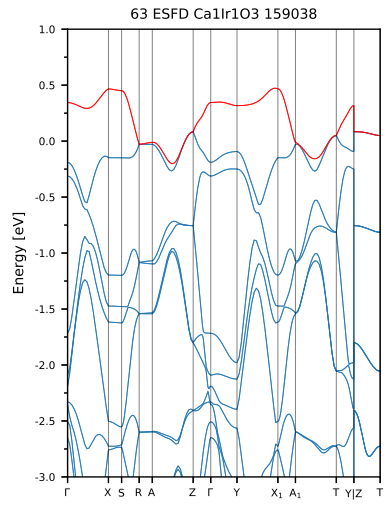

(od) 63.4.171

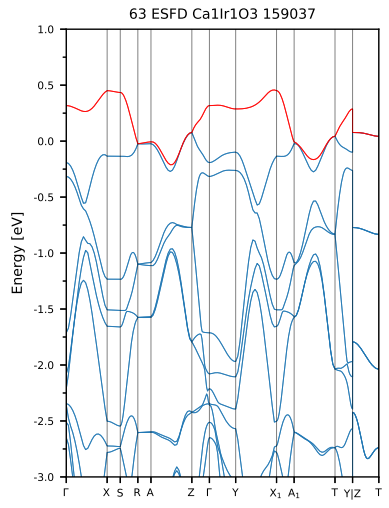

(oe) 63.4.172

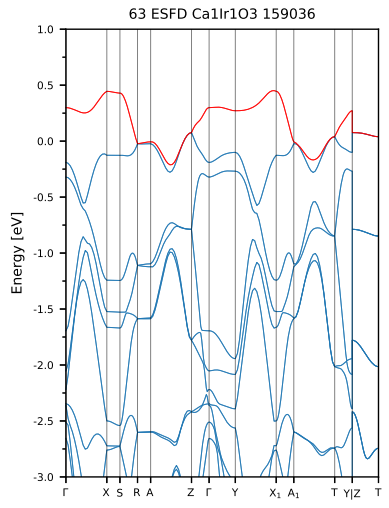

(of) 63.4.173

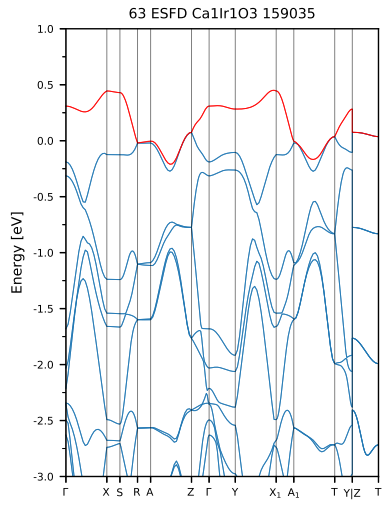

(og) 63.4.174

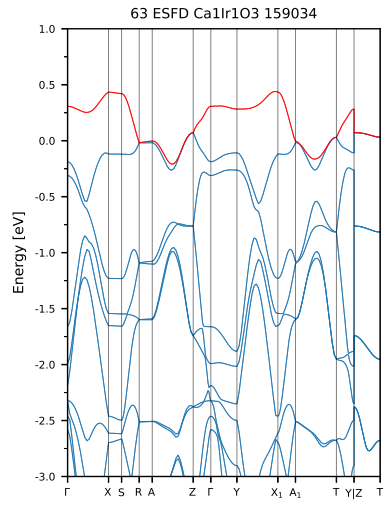

(oh) 63.4.175

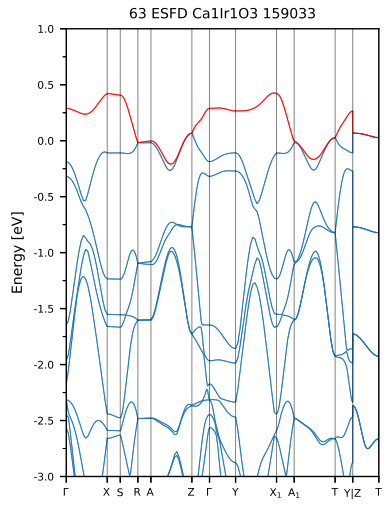

(oi) 63.4.176

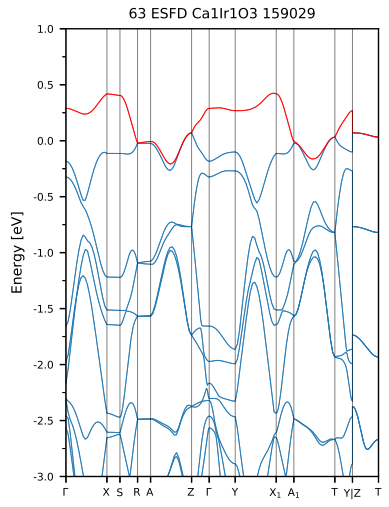

(oj) 63.4.177

Figure 169: Band structures for topological materials in SG 63, part 25/29, from 63.4.162 to 63.4.177. See the list of figures.

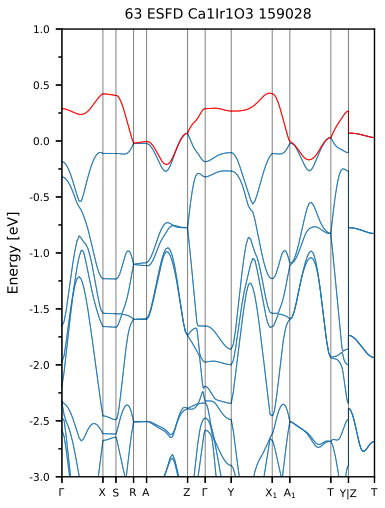

(ok) 63.4.178

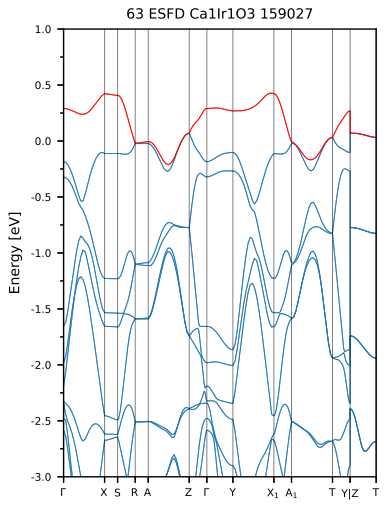

(ol) 63.4.179

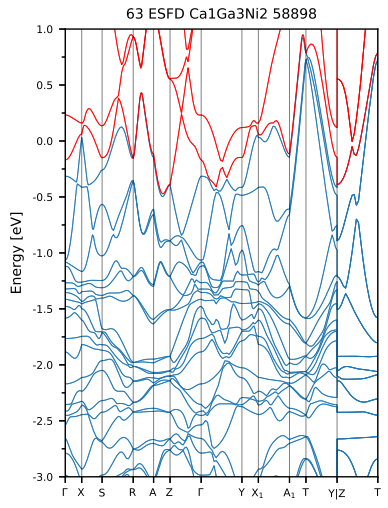

(om) 63.4.180

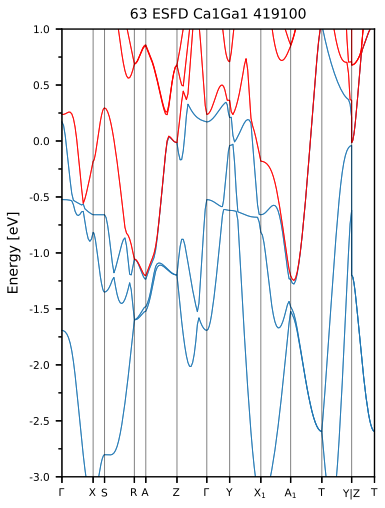

(on) 63.4.181

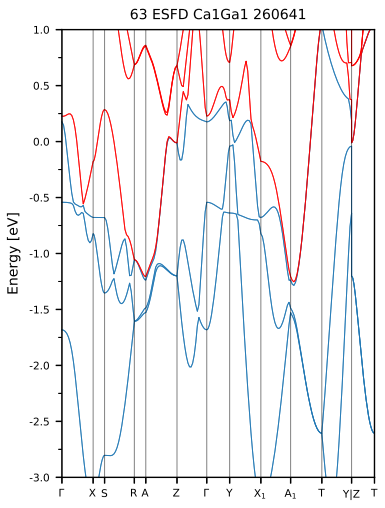

(oo) 63.4.182

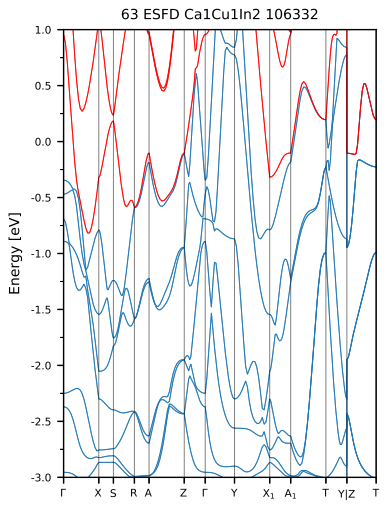

(op) 63.4.183

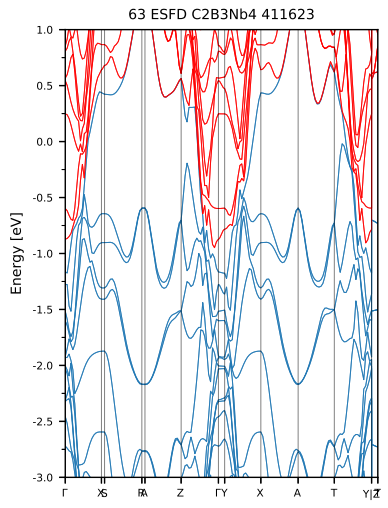

(oq) 63.4.184

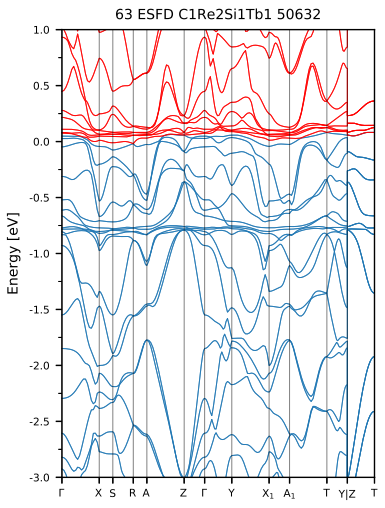

(or) 63.4.185

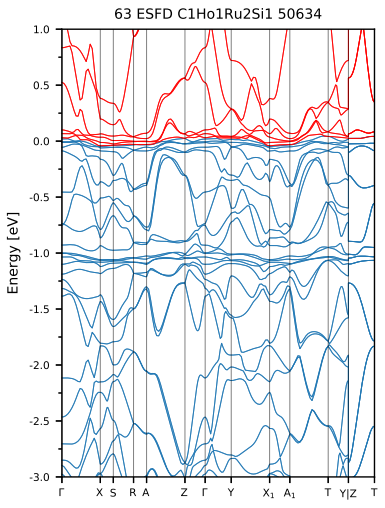

(os) 63.4.186

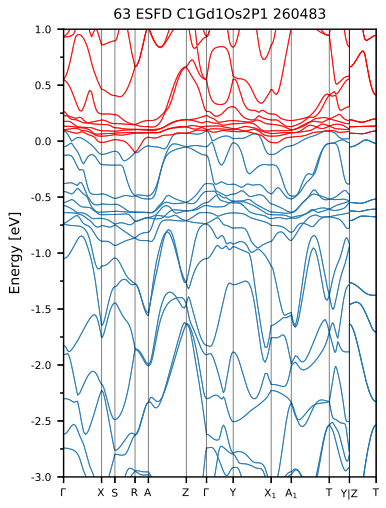

(ot) 63.4.187

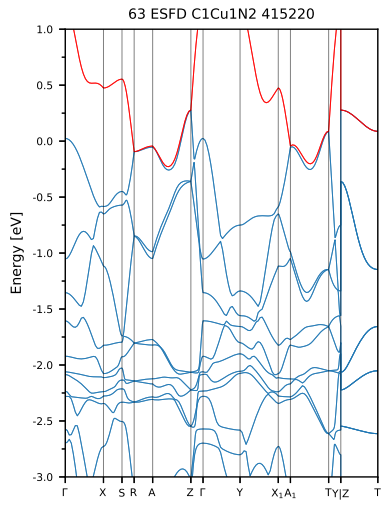

(ou) 63.4.188

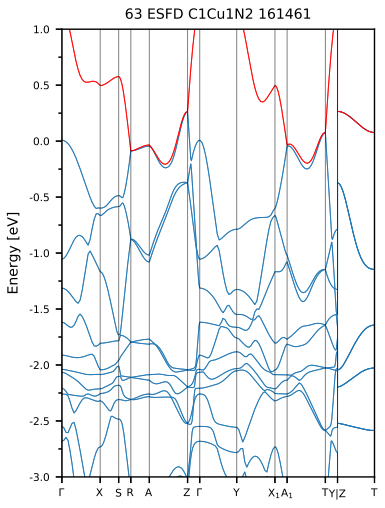

(ov) 63.4.189

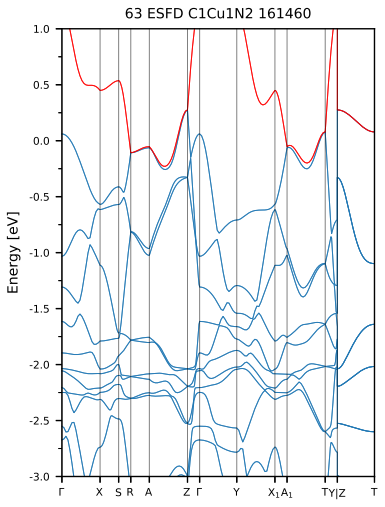

(ow) 63.4.190

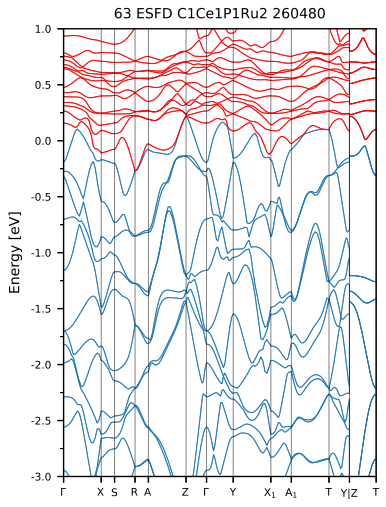

(ox) 63.4.191

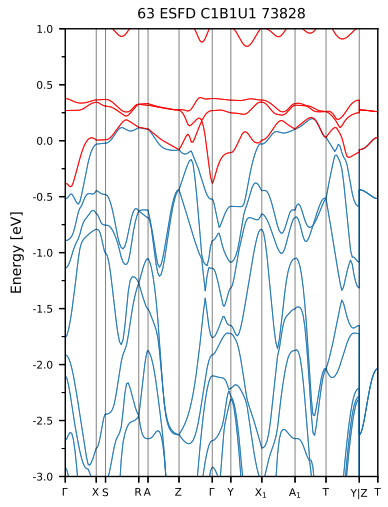

(oy) 63.4.192

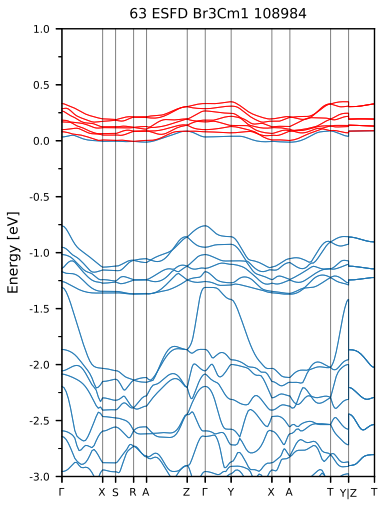

(oz) 63.4.193

Figure 170: Band structures for topological materials in SG 63, part 26/29, from 63.4.178 to 63.4.193. See the list of figures.

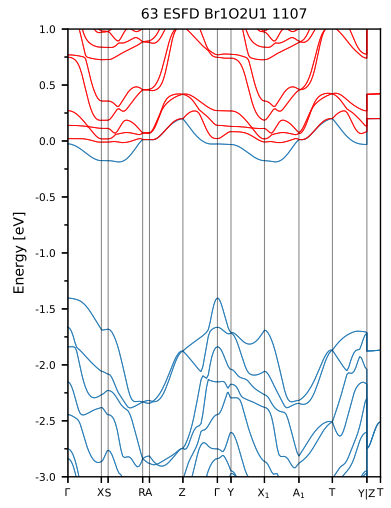

(pa) 63.4.194

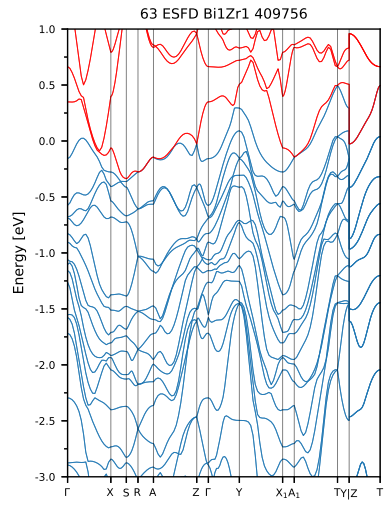

(pb) 63.4.195

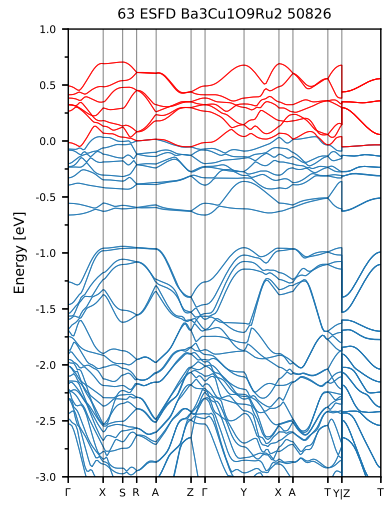

(pc) 63.4.196

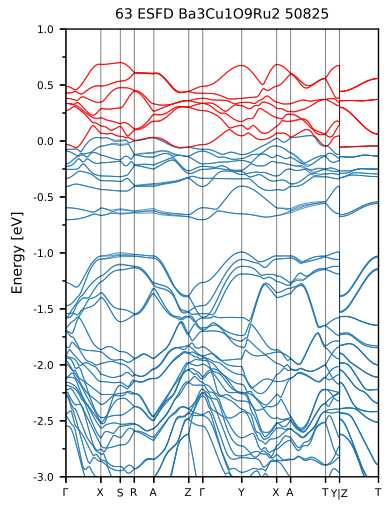

(pd) 63.4.197

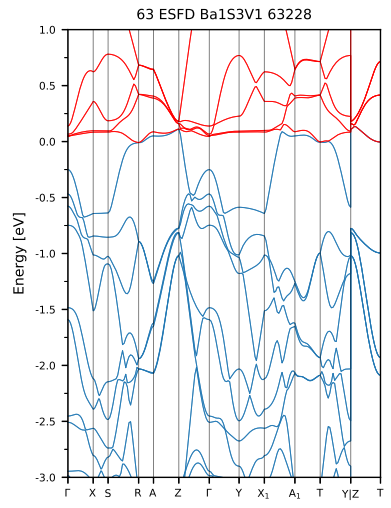

(pe) 63.4.198

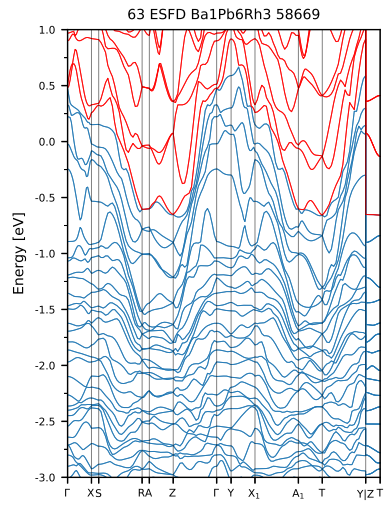

(pf) 63.4.199

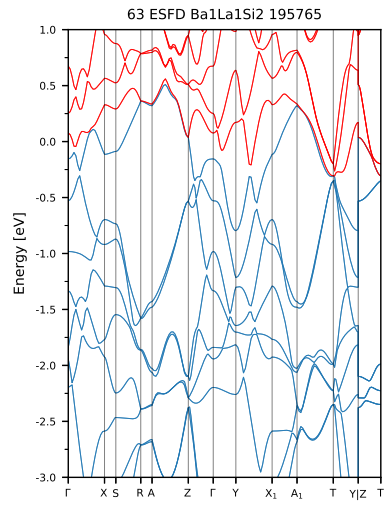

(pg) 63.4.200

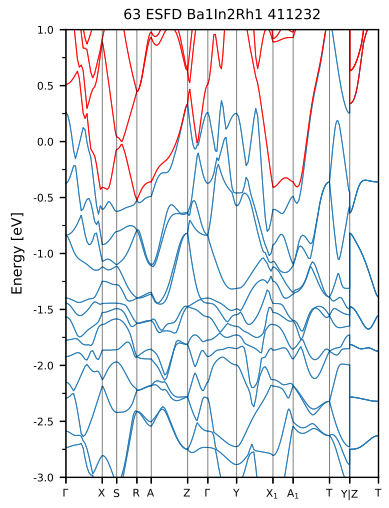

(ph) 63.4.201

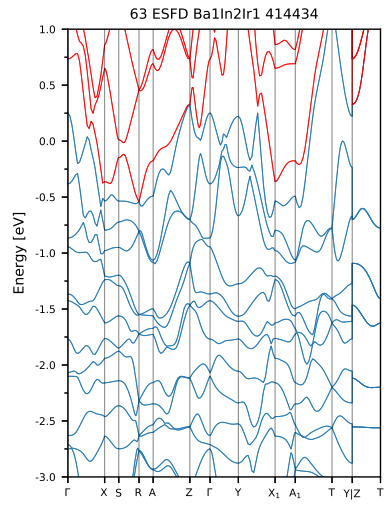

(pi) 63.4.202

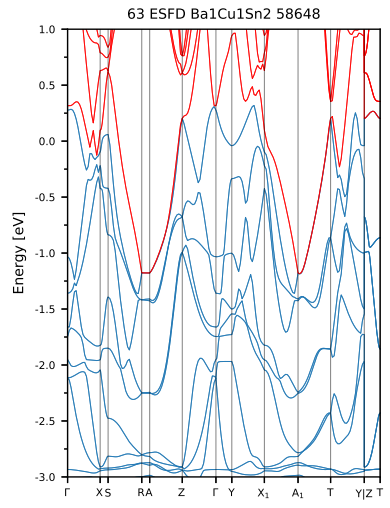

(pj) 63.4.203

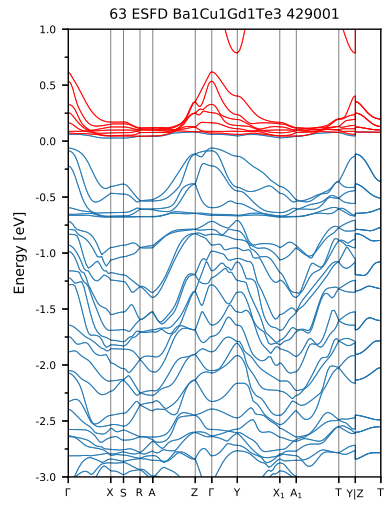

(pk) 63.4.204

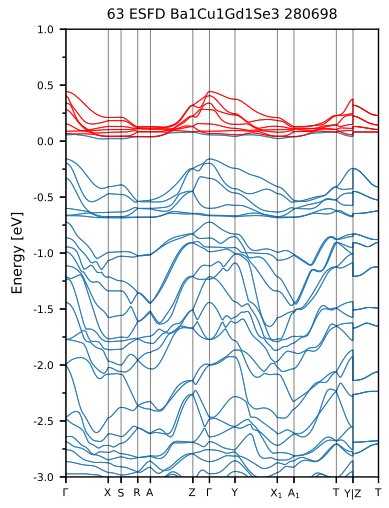

(pl) 63.4.205

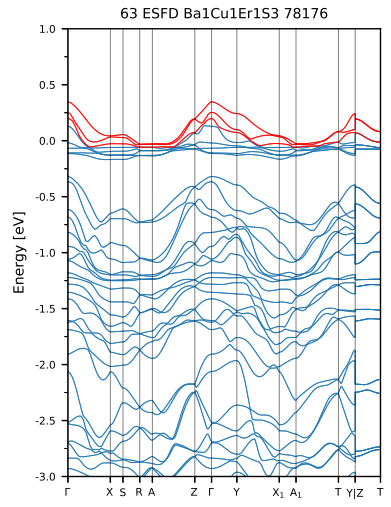

(pm) 63.4.206

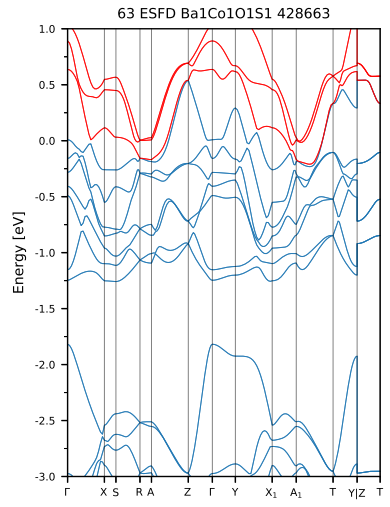

(pn) 63.4.207

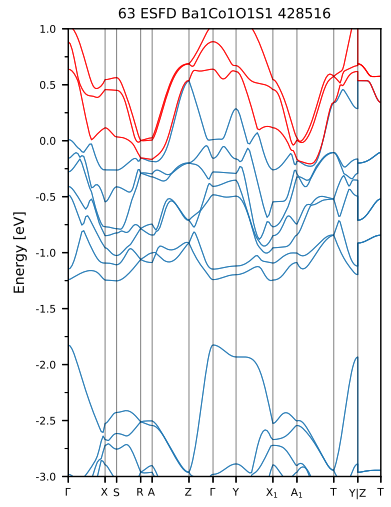

(po) 63.4.208

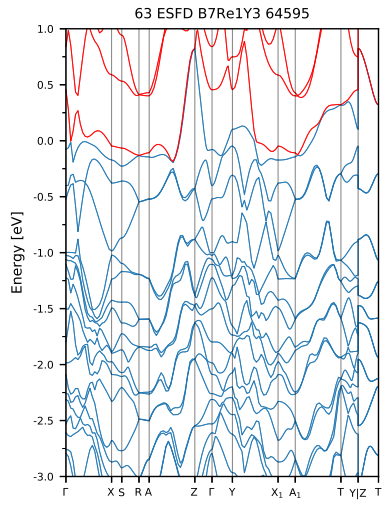

(pp) 63.4.209

Figure 171: Band structures for topological materials in SG 63, part 27/29, from 63.4.194 to 63.4.209. See the list of figures.

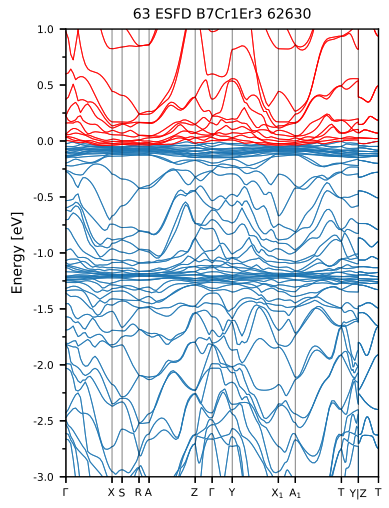

(pq) 63.4.210

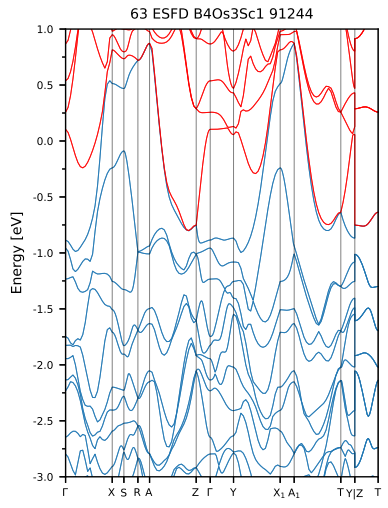

(pr) 63.4.211

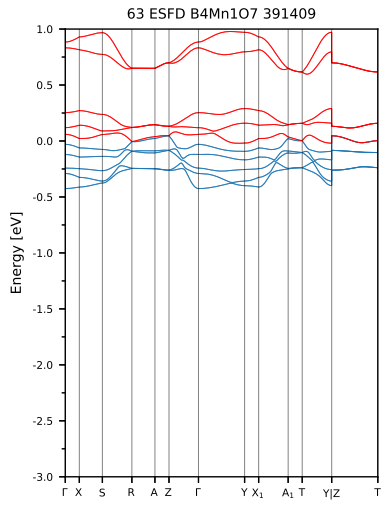

(ps) 63.4.212

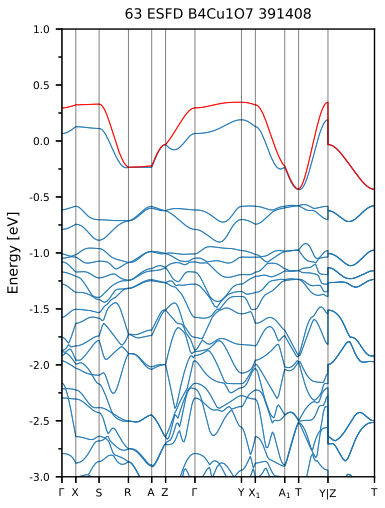

(pt) 63.4.213

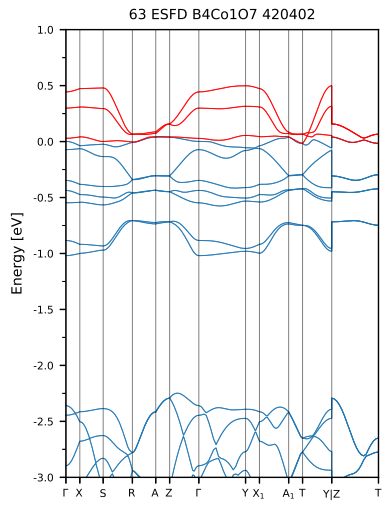

(pu) 63.4.214

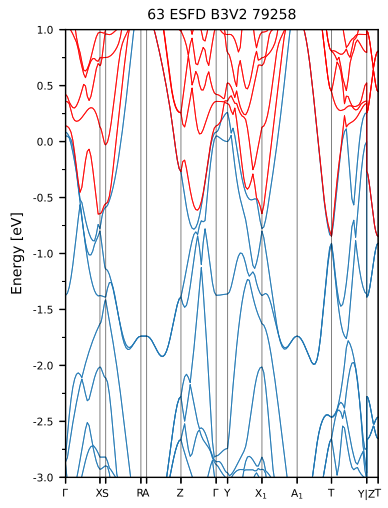

(pv) 63.4.215

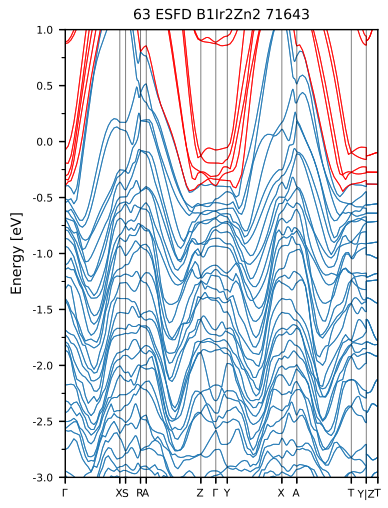

(pw) 63.4.216

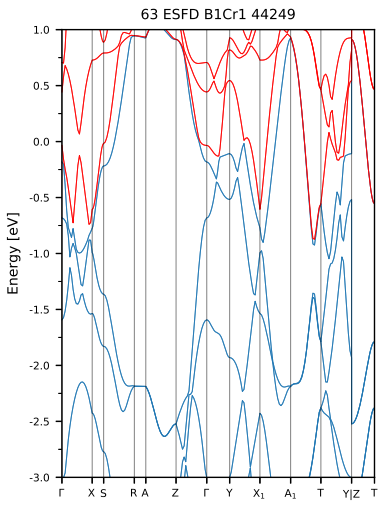

(px) 63.4.217

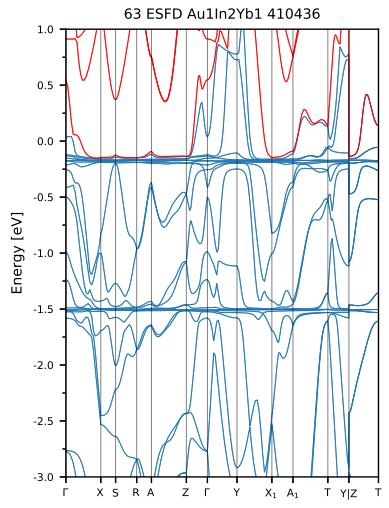

(py) 63.4.218

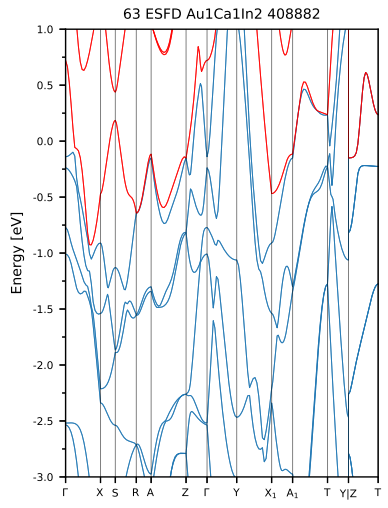

(pz) 63.4.219

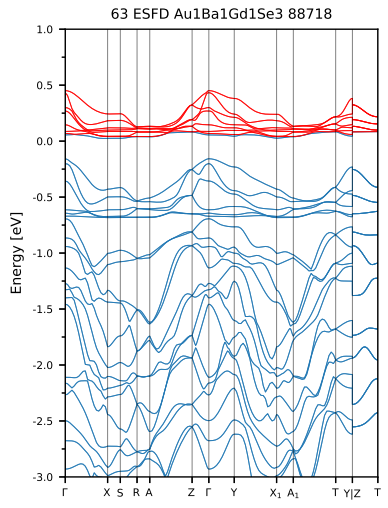

(qa) 63.4.220

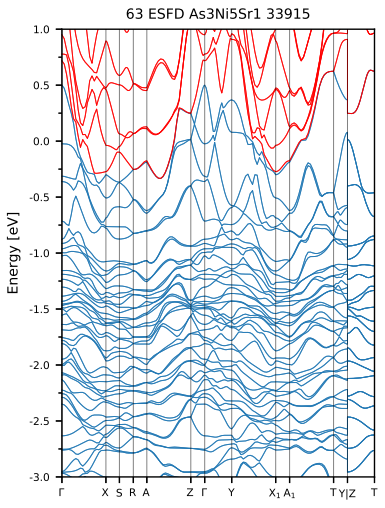

(qb) 63.4.221

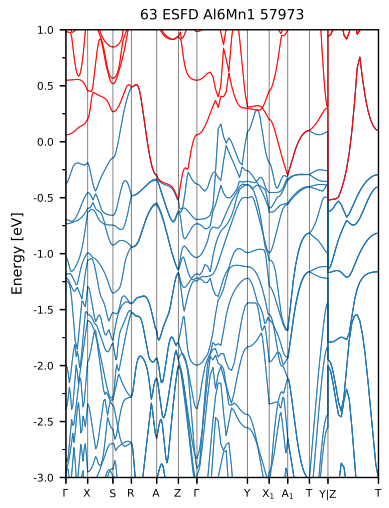

(qc) 63.4.222

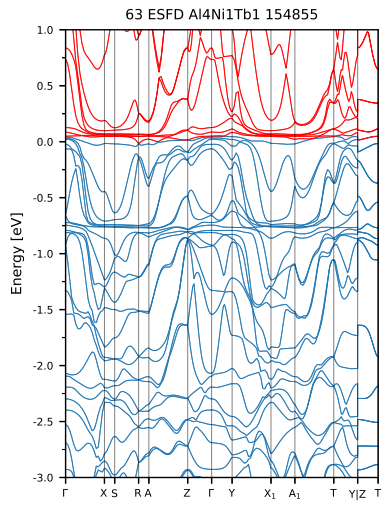

(qd) 63.4.223

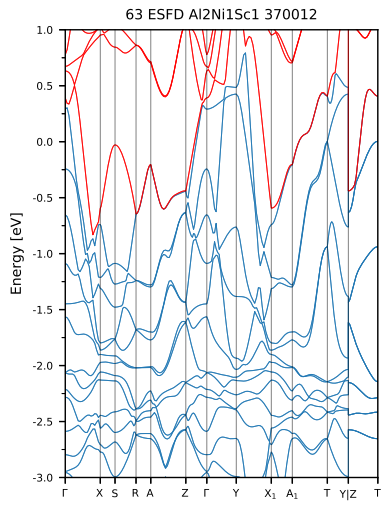

(qe) 63.4.224

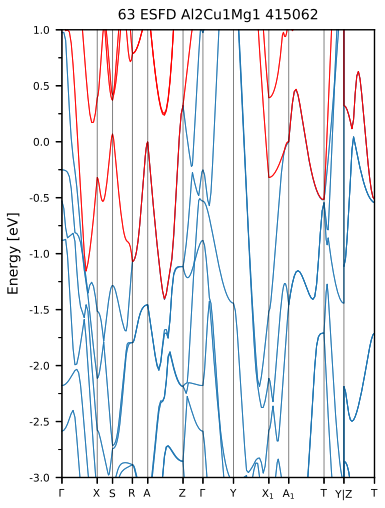

(qf) 63.4.225

Figure 172: Band structures for topological materials in SG 63, part 28/29, from 63.4.210 to 63.4.225. See the list of figures.

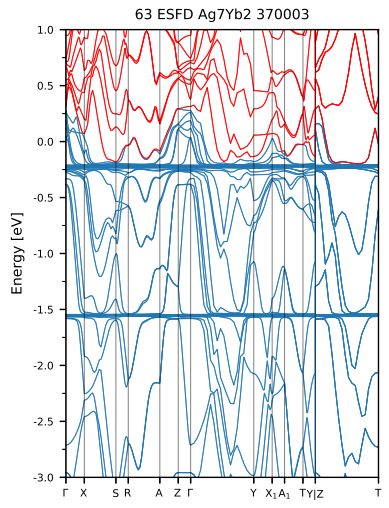

(qg) 63.4.226

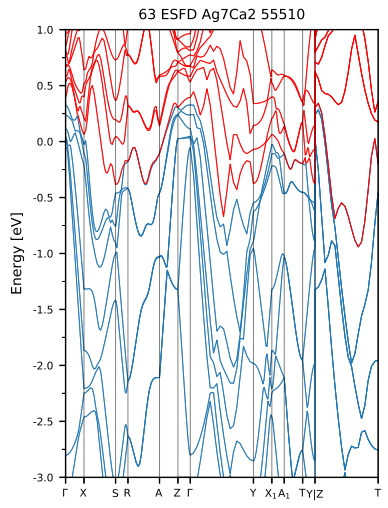

(qh) 63.4.227

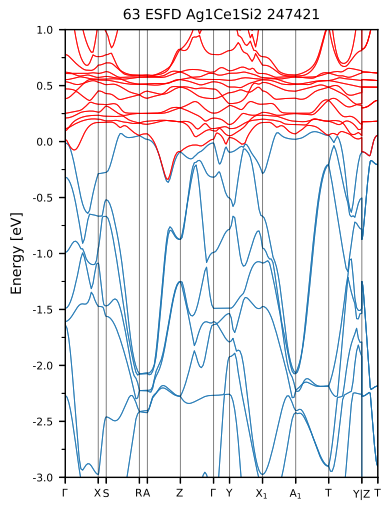

(qi) 63.4.228

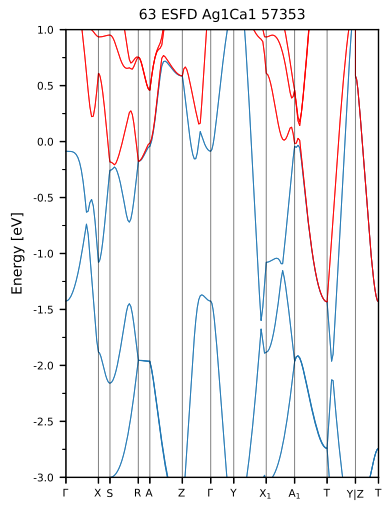

(qj) 63.4.229

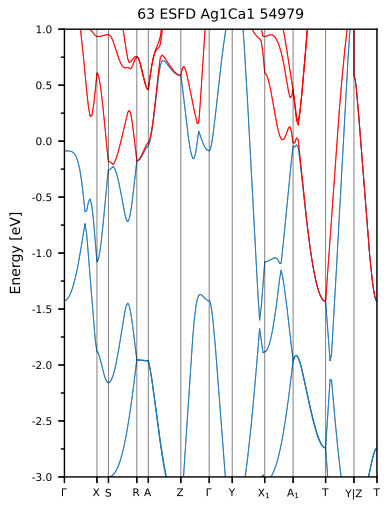

(qk) 63.4.230

Figure 173: Band structures for topological materials in SG 63, part 29/29, from 63.4.226 to 63.4.230. See the list of figures.

## 0.0.28 SG 64

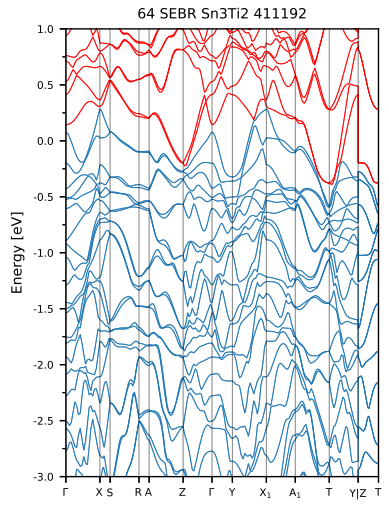

(a) 64.1.1

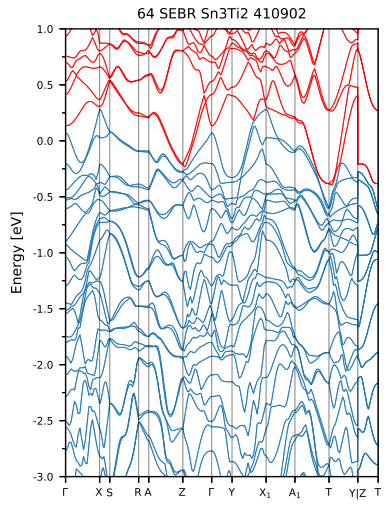

(b) 64.1.2

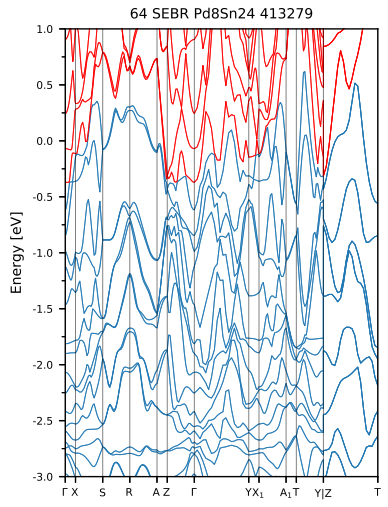

(c) 64.1.3

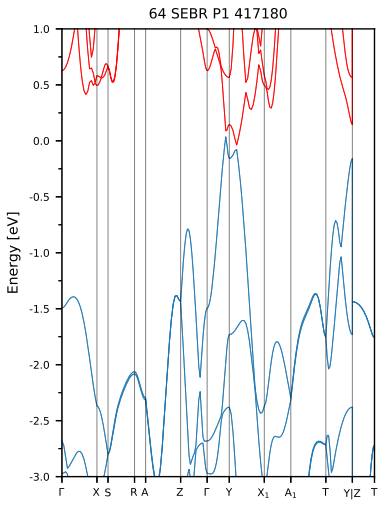

(d) 64.1.4

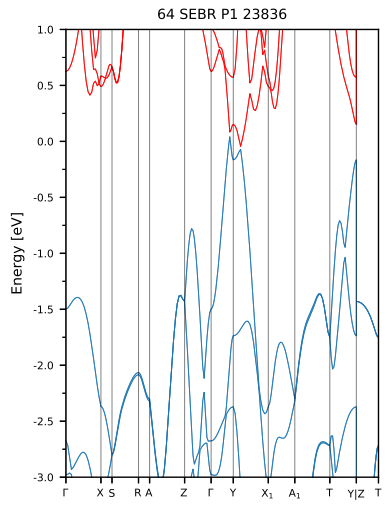

(e) 64.1.5

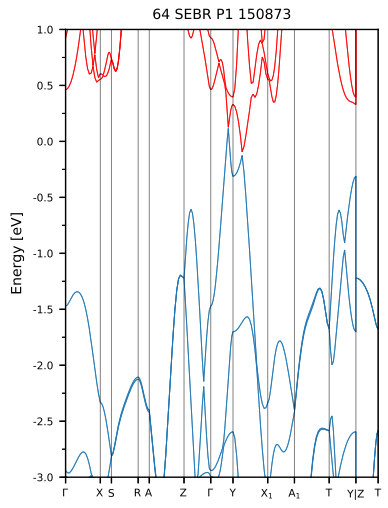

(f) 64.1.6

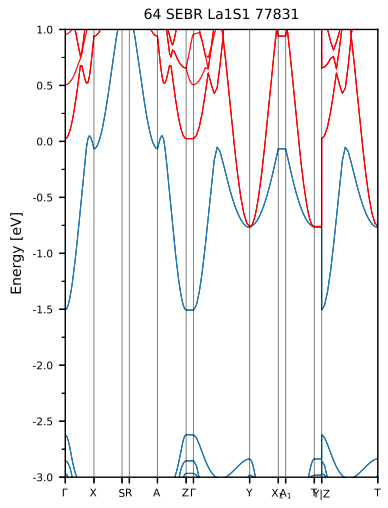

(g) 64.1.7

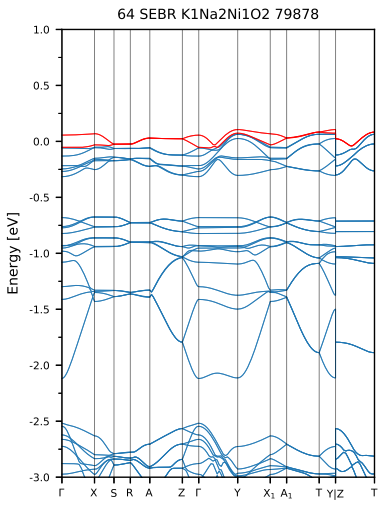

(h) 64.1.8

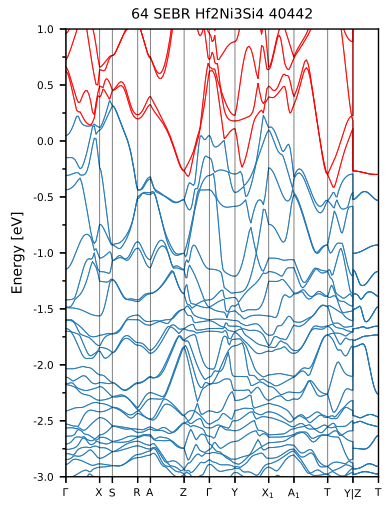

(i) 64.1.9

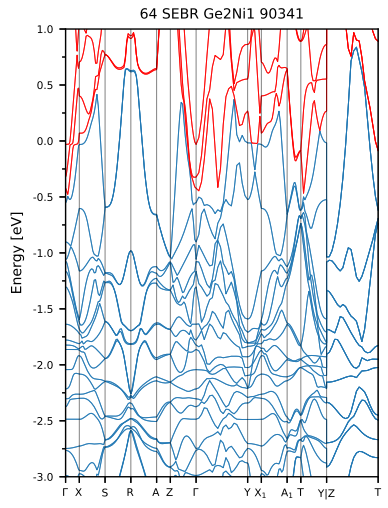

(j) 64.1.10

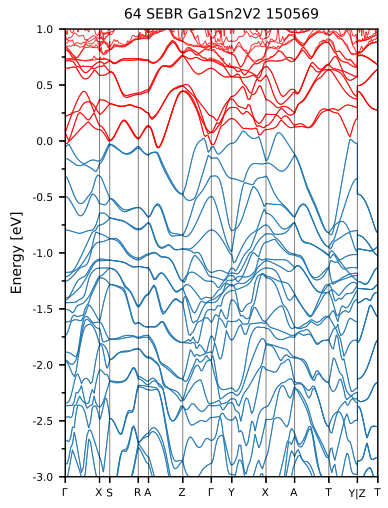

(k) 64.1.11

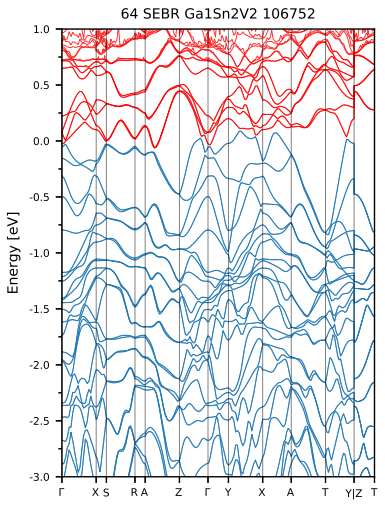

(l) 64.1.12

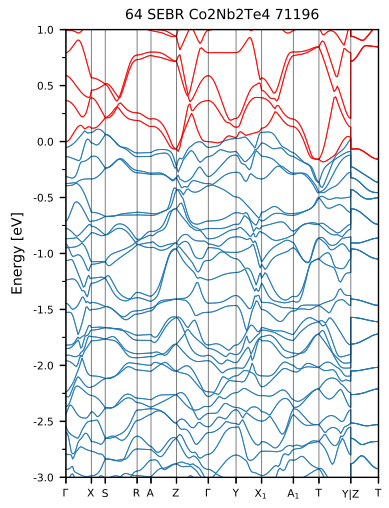

(m) 64.1.13

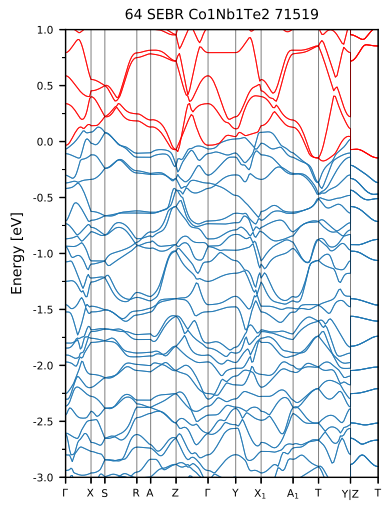

(n) 64.1.14

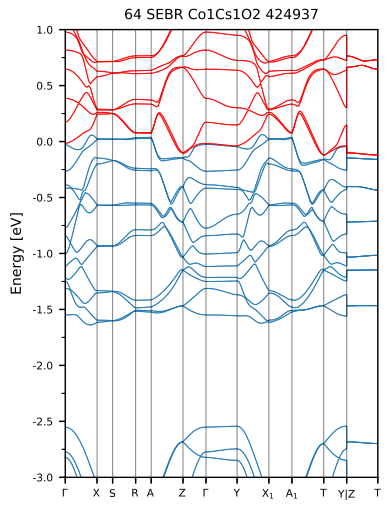

(o) 64.1.15

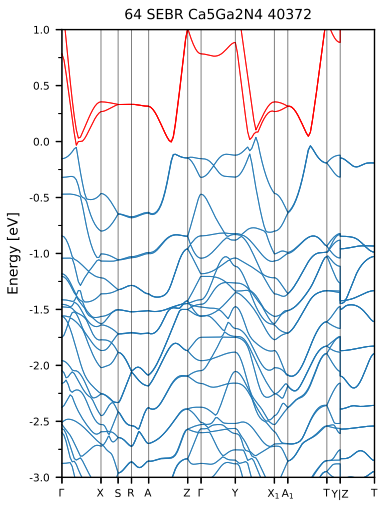

(p) 64.1.16

Figure 174: Band structures for topological materials in SG 64, part 1/4, from 64.1.1 to 64.1.16. See the list of figures.

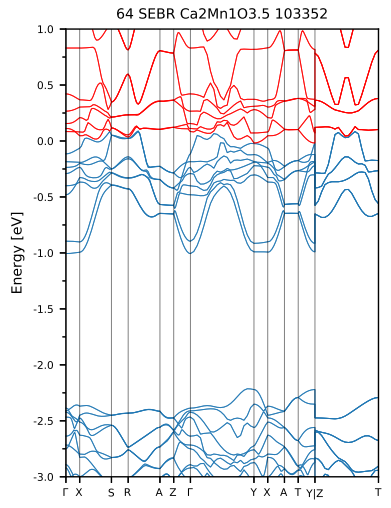

(q) 64.1.17

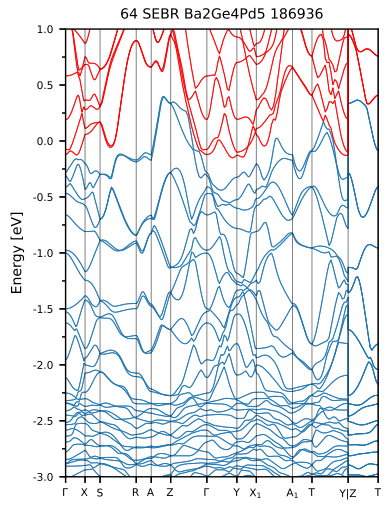

(r) 64.1.18

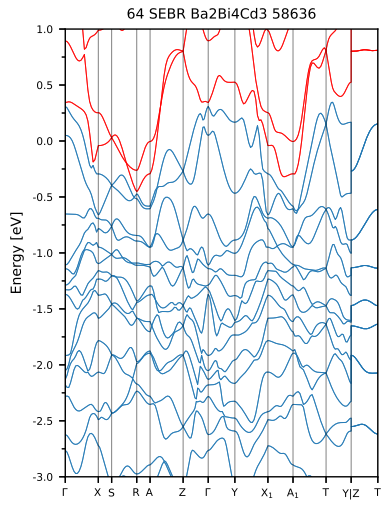

(s) 64.1.19

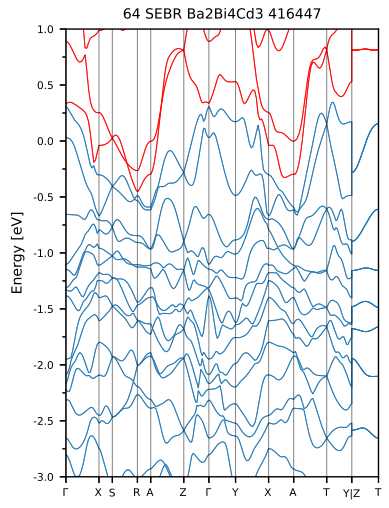

(t) 64.1.20

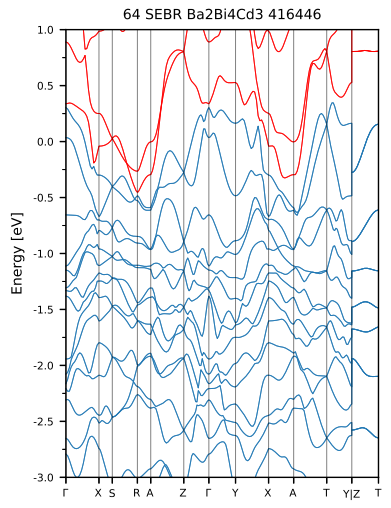

(u) 64.1.21

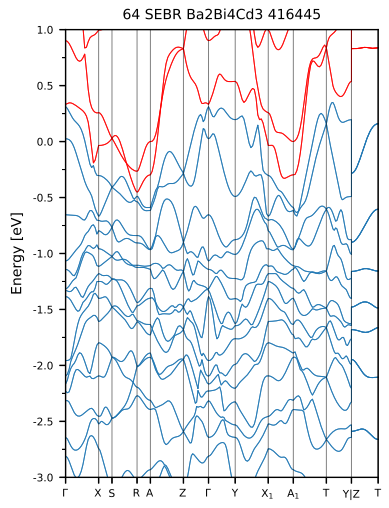

(v) 64.1.22

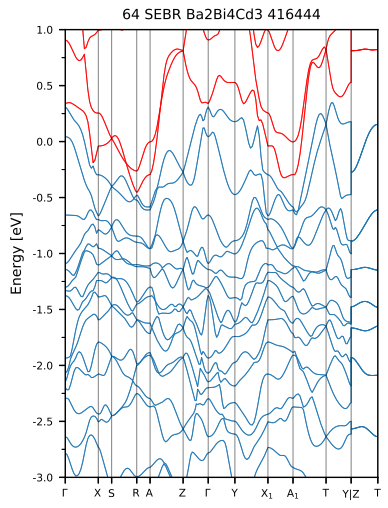

(w) 64.1.23

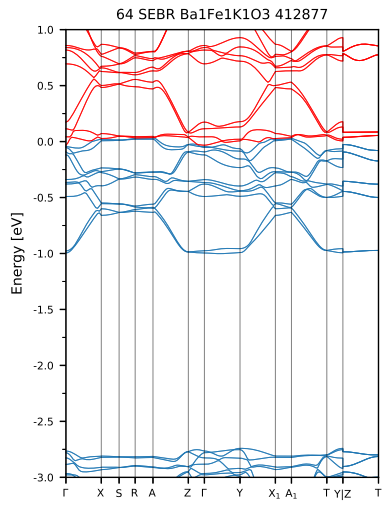

(x) 64.1.24

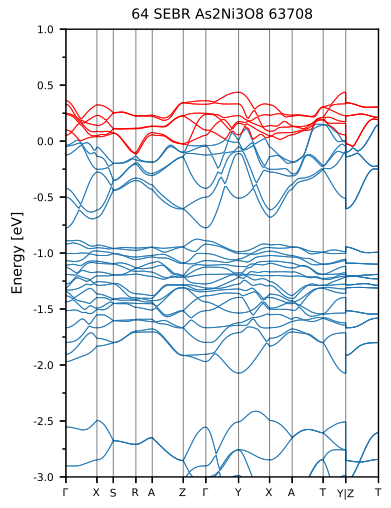

(y) 64.1.25

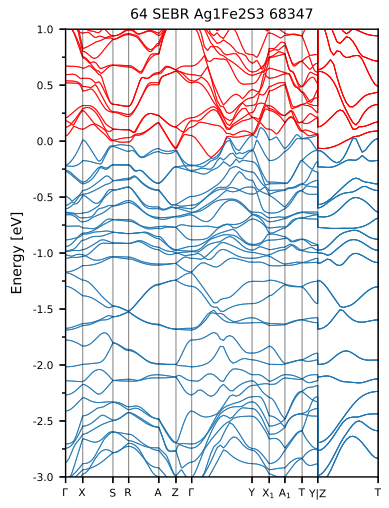

(z) 64.1.26

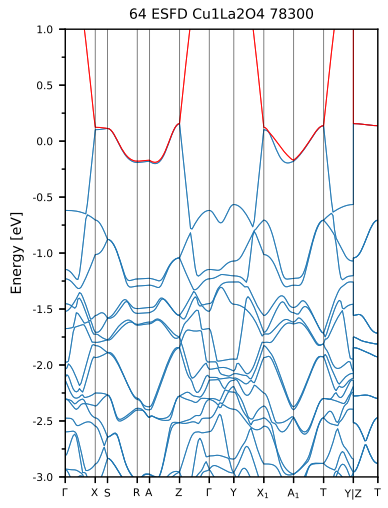

(aa) 64.4.1

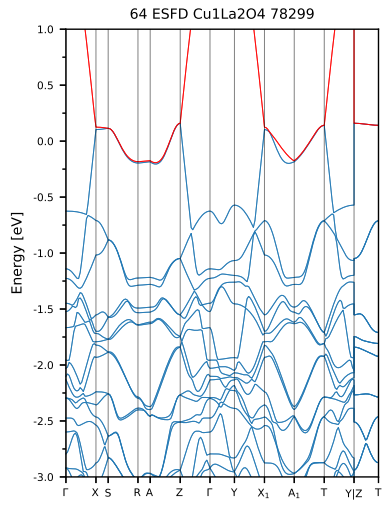

(ab) 64.4.2

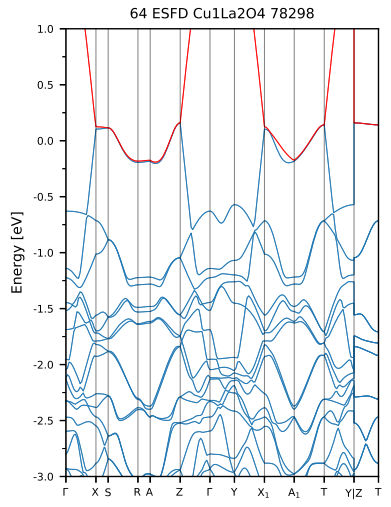

(ac) 64.4.3

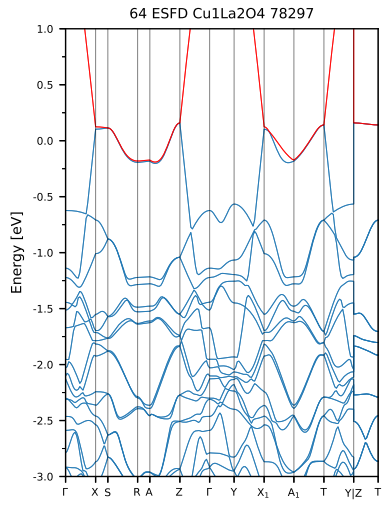

(ad) 64.4.4

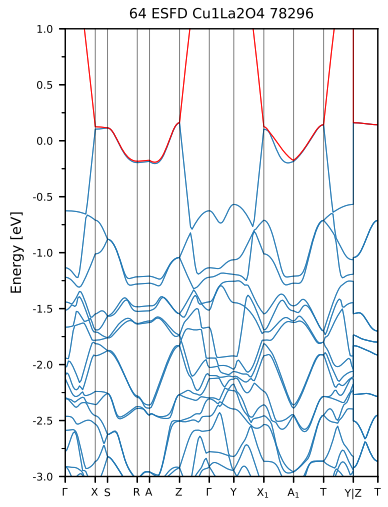

(ae) 64.4.5

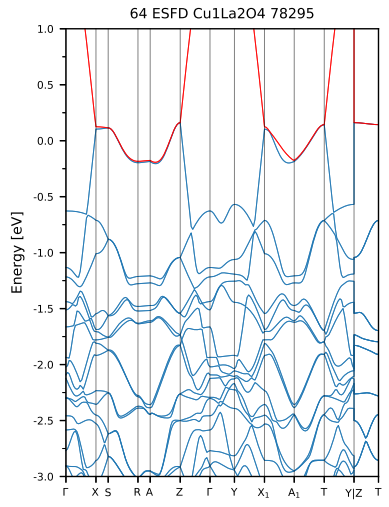

(af) 64.4.6

Figure 175: Band structures for topological materials in SG 64, part 2/4, from 64.1.17 to 64.4.6. See the list of figures.

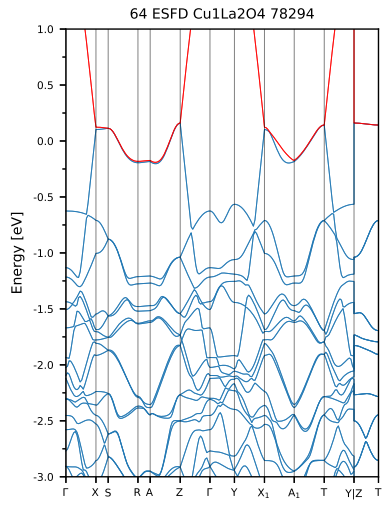

(ag) 64.4.7

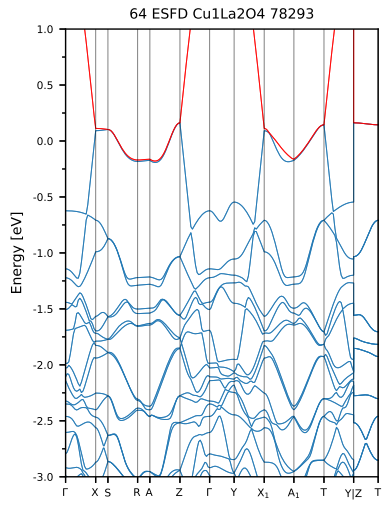

(ah) 64.4.8

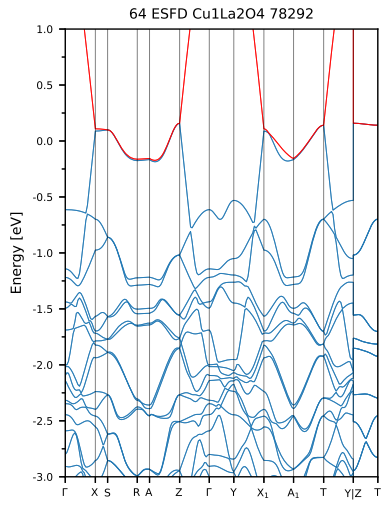

(ai) 64.4.9

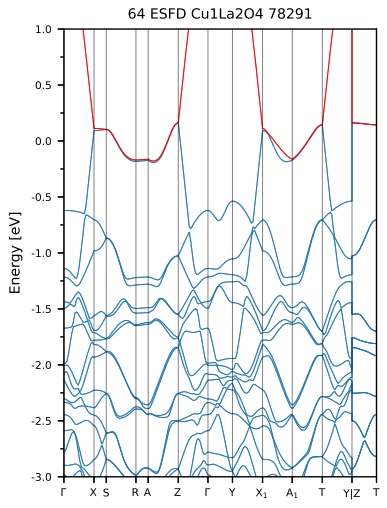

(aj) 64.4.10

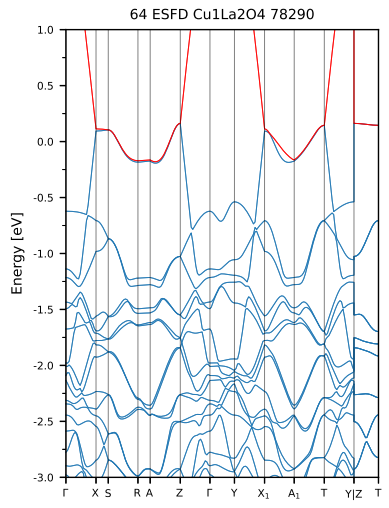

(ak) 64.4.11

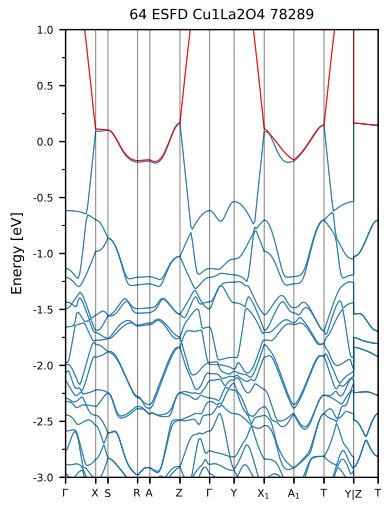

(al) 64.4.12

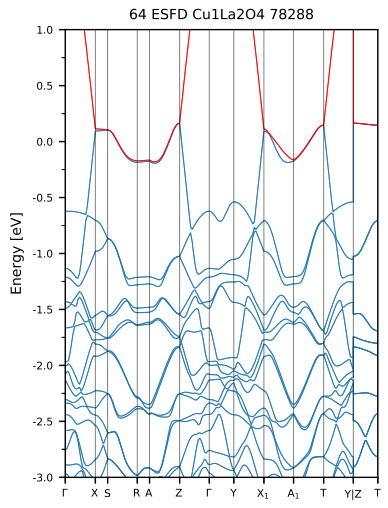

(am) 64.4.13

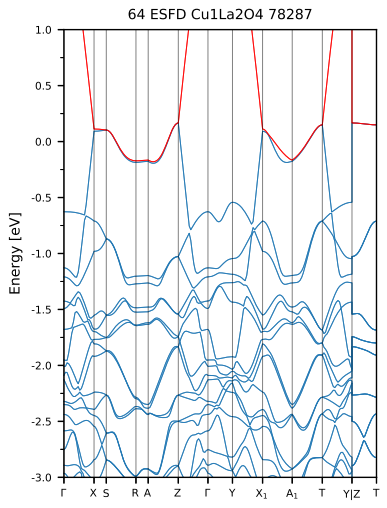

(an) 64.4.14

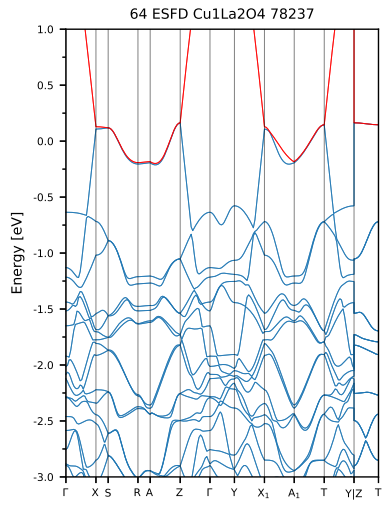

(ao) 64.4.15

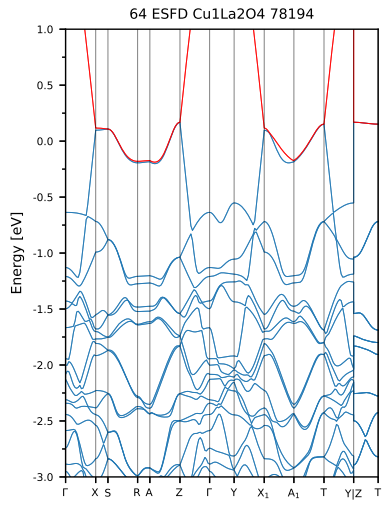

(ap) 64.4.16

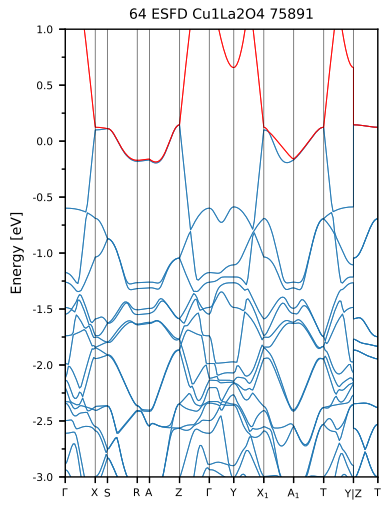

(aq) 64.4.17

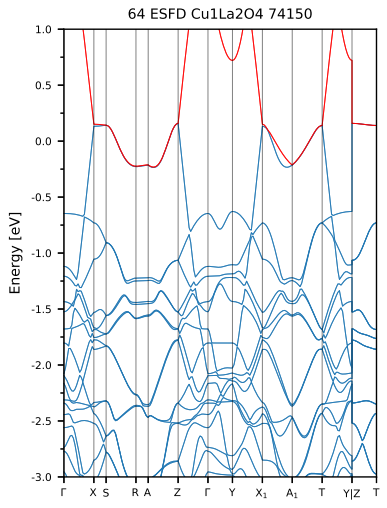

(ar) 64.4.18

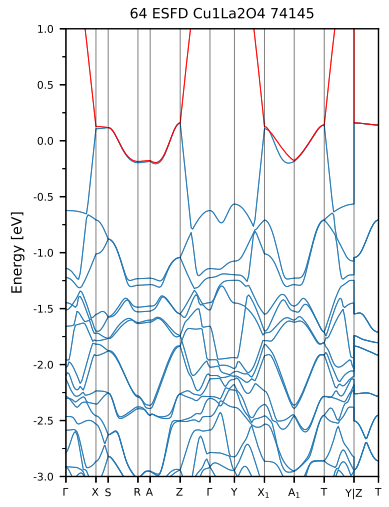

(as) 64.4.19

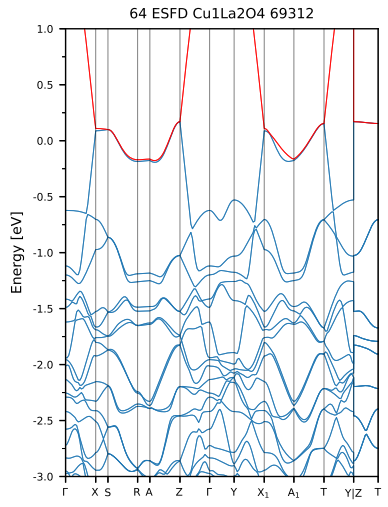

(at) 64.4.20

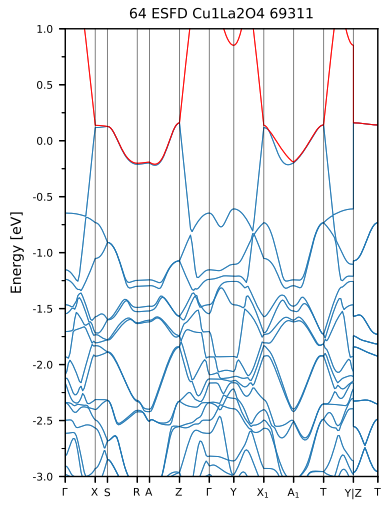

(au) 64.4.21

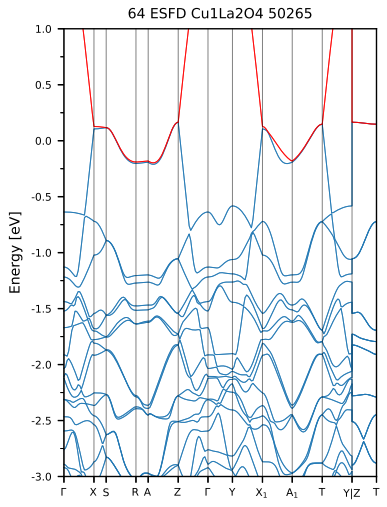

(av) 64.4.22

Figure 176: Band structures for topological materials in SG 64, part 3/4, from 64.4.7 to 64.4.22. See the list of figures.

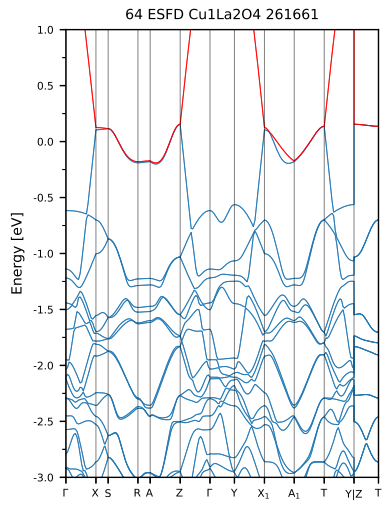

(aw) 64.4.23

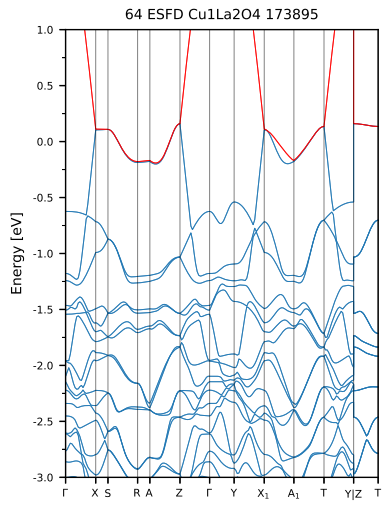

(ax) 64.4.24

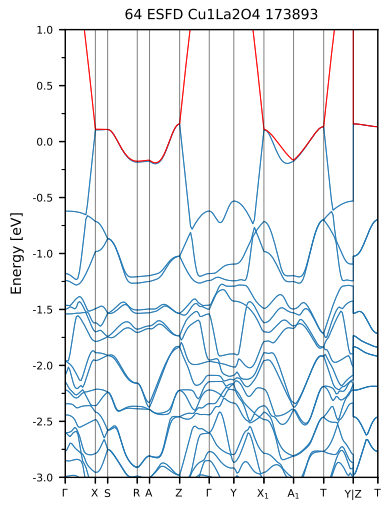

(ay) 64.4.25

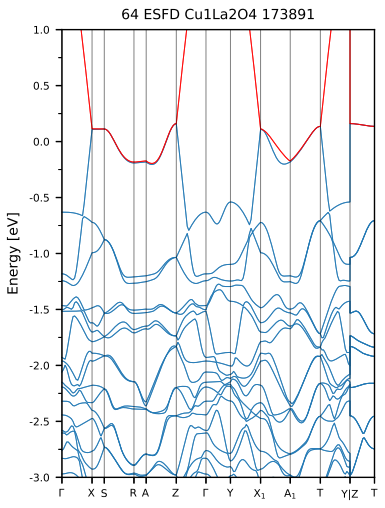

(az) 64.4.26

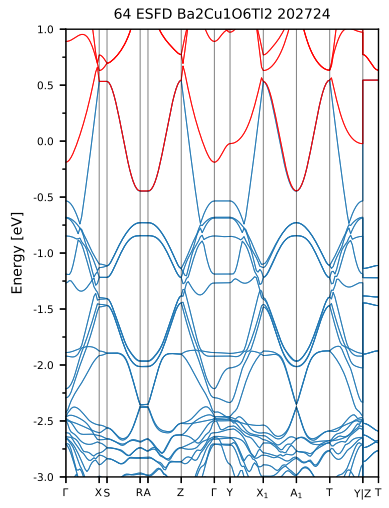

(ba) 64.4.27

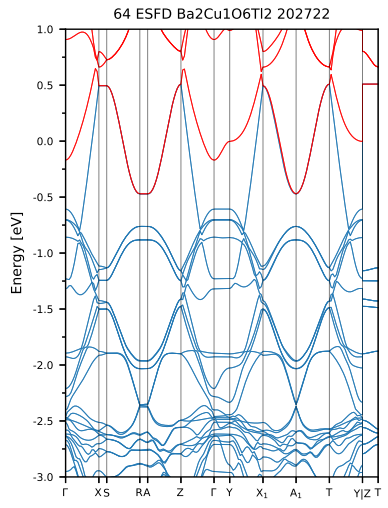

(bb) 64.4.28

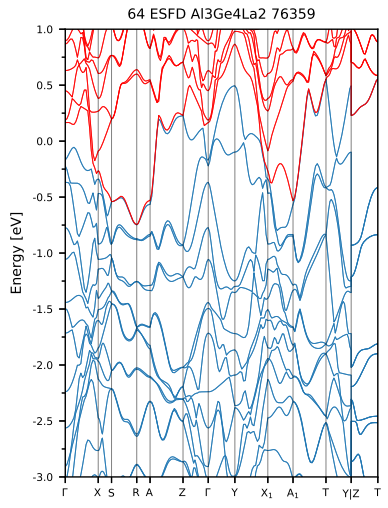

(bc) 64.4.29

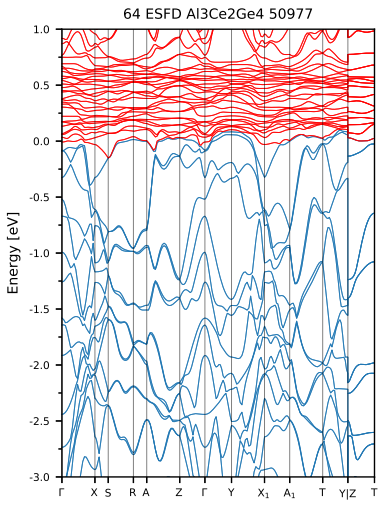

(bd) 64.4.30

Figure 177: Band structures for topological materials in SG 64, part 4/4, from 64.4.23 to 64.4.30. See the list of figures.

### 0.0.29 SG 65

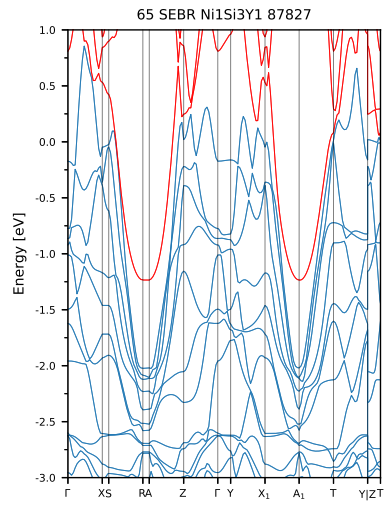

(a) 65.1.1

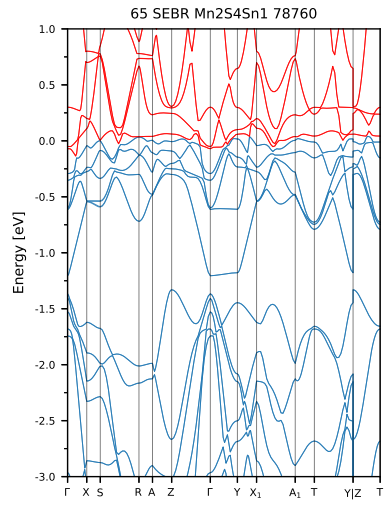

(b) 65.1.2

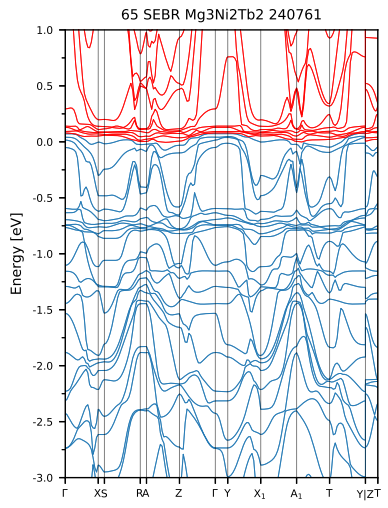

(c) 65.1.3

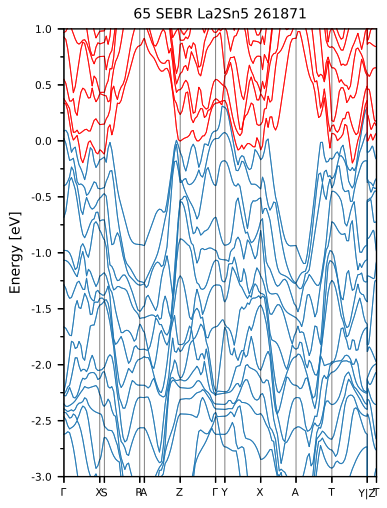

(d) 65.1.4

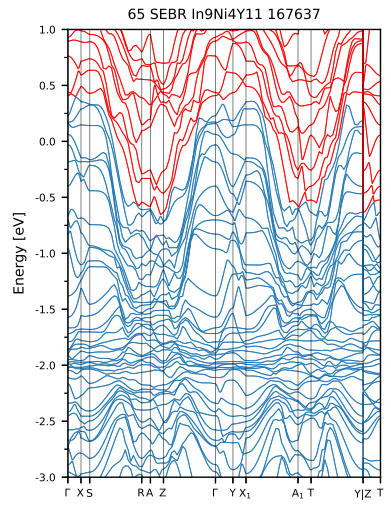

(e) 65.1.5

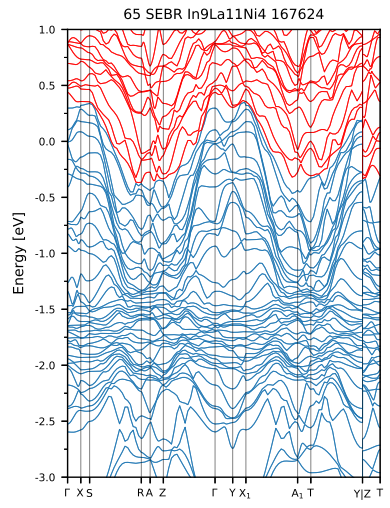

(f) 65.1.6

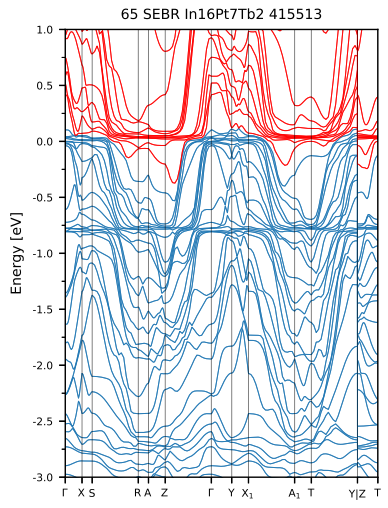

(g) 65.1.7

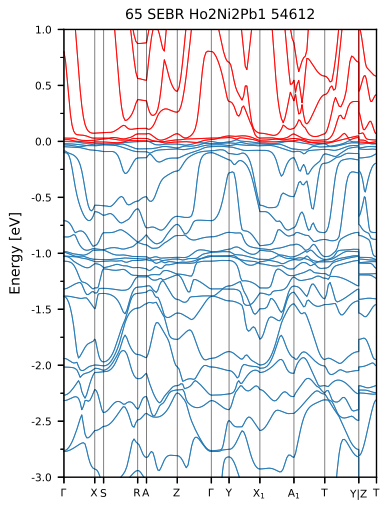

(h) 65.1.8

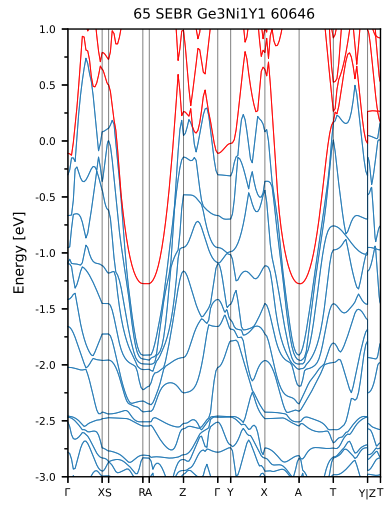

(i) 65.1.9

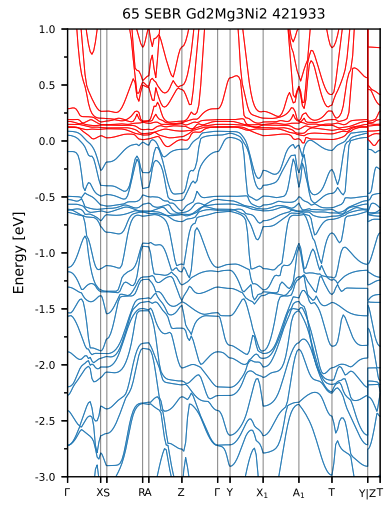

(j) 65.1.10

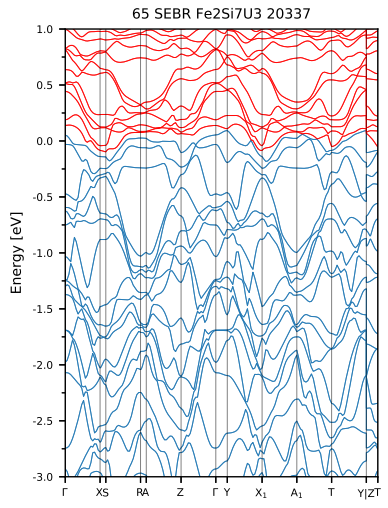

(k) 65.1.11

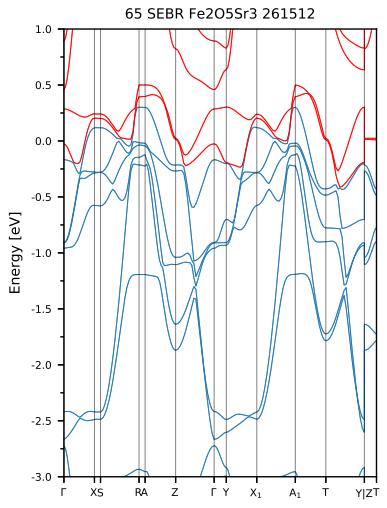

(l) 65.1.12

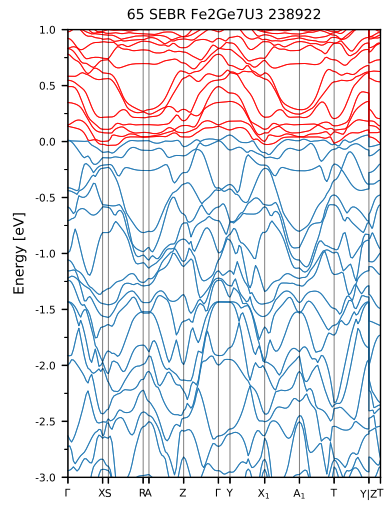

(m) 65.1.13

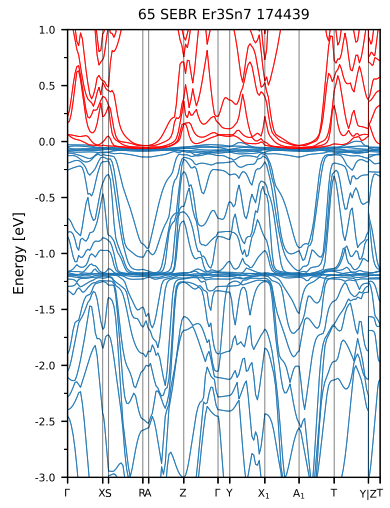

(n) 65.1.14

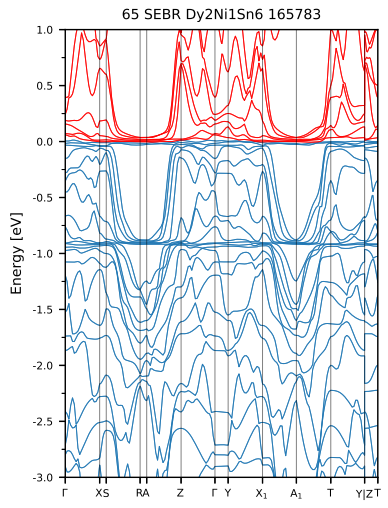

(o) 65.1.15

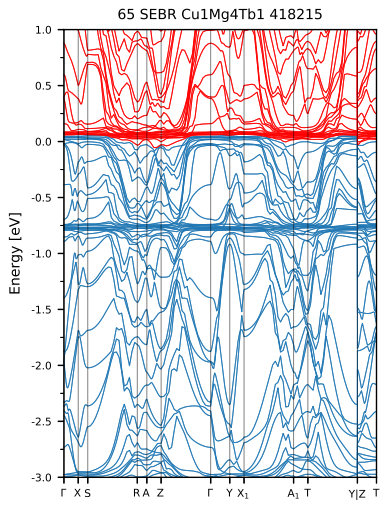

(p) 65.1.16

Figure 178: Band structures for topological materials in SG 65, part 1/4, from 65.1.1 to 65.1.16. See the list of figures.

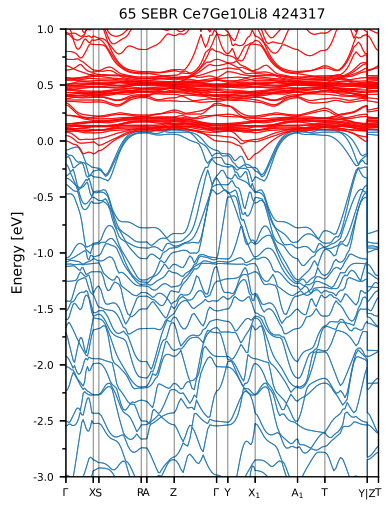

(q) 65.1.17

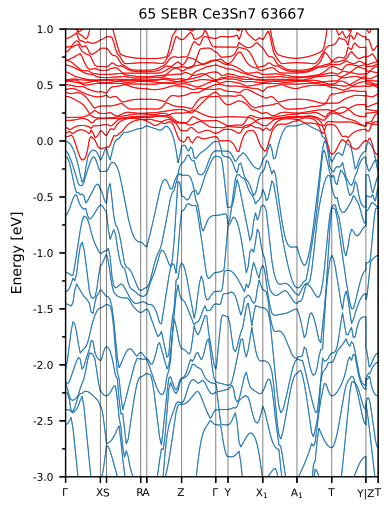

(r) 65.1.18

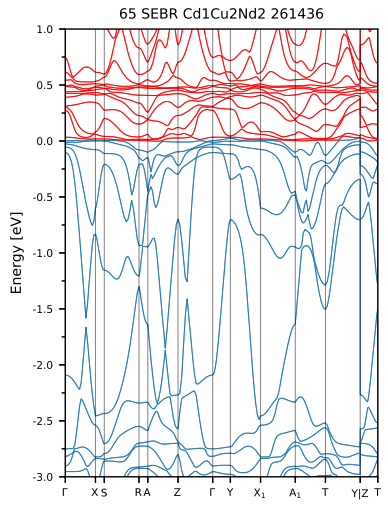

(s) 65.1.19

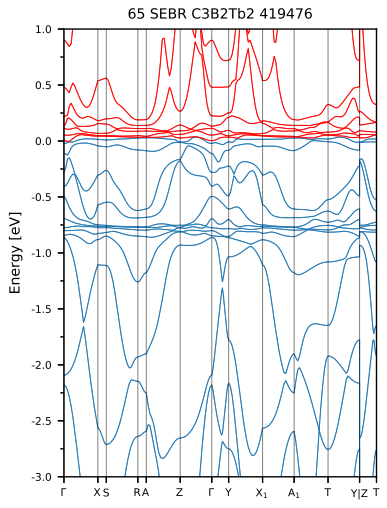

(t) 65.1.20

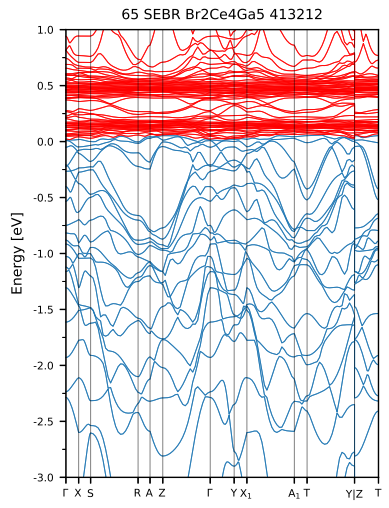

(u) 65.1.21

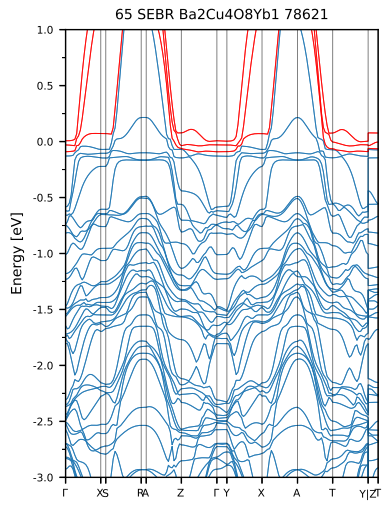

(v) 65.1.22

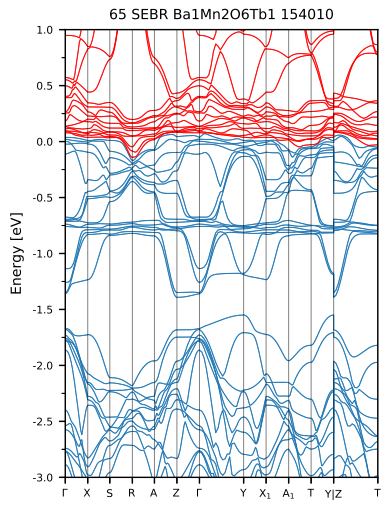

(w) 65.1.23

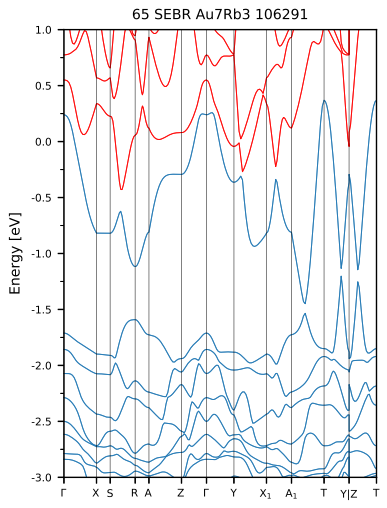

(x) 65.1.24

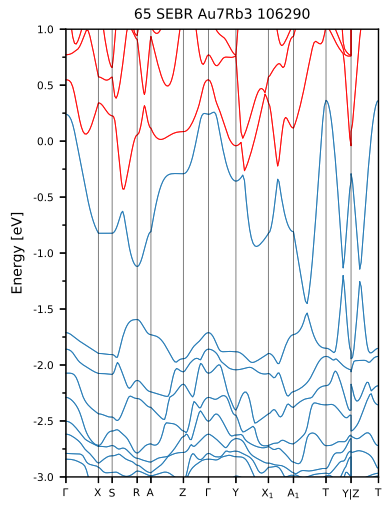

(y) 65.1.25

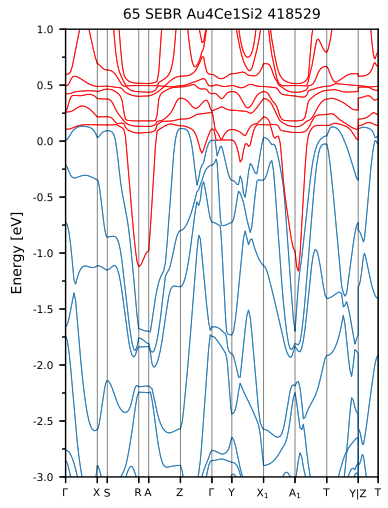

(z) 65.1.26

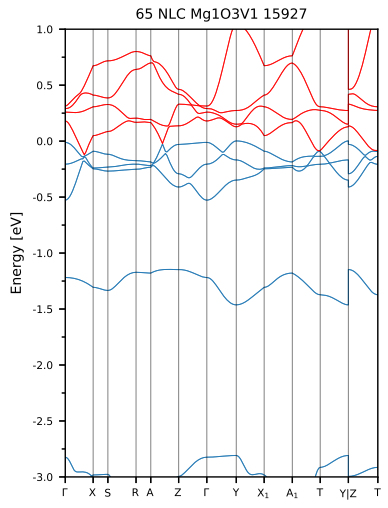

(aa) 65.2.1

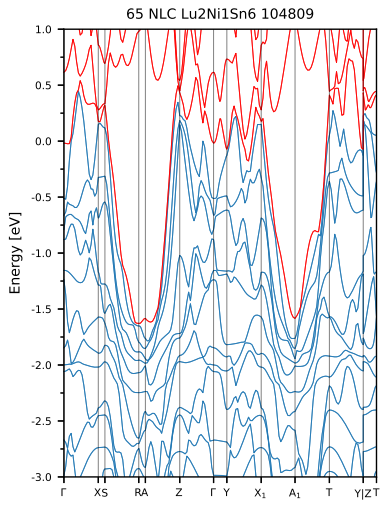

(ab) 65.2.2

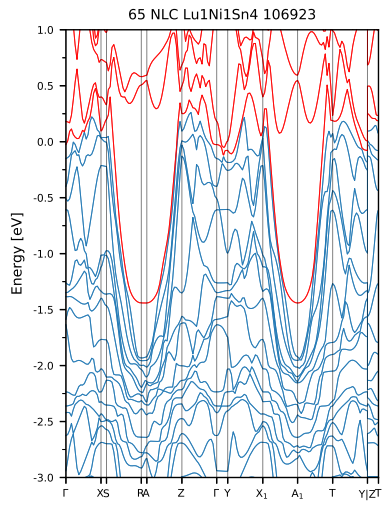

(ac) 65.2.3

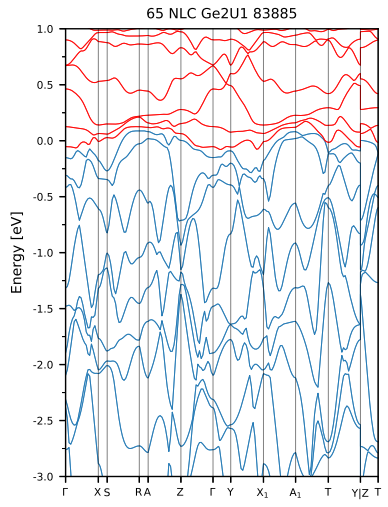

(ad) 65.2.4

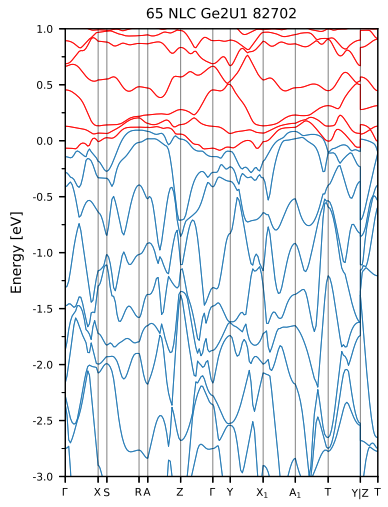

(ae) 65.2.5

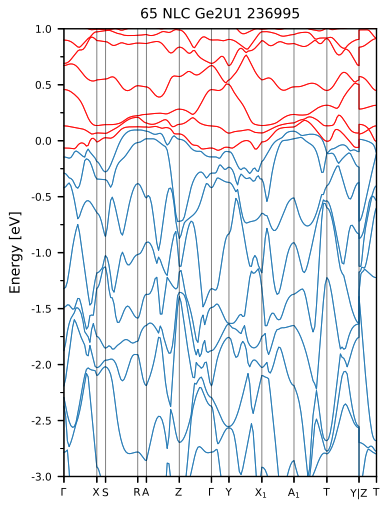

(af) 65.2.6

Figure 179: Band structures for topological materials in SG 65, part 2/4, from 65.1.17 to 65.2.6. See the list of figures.

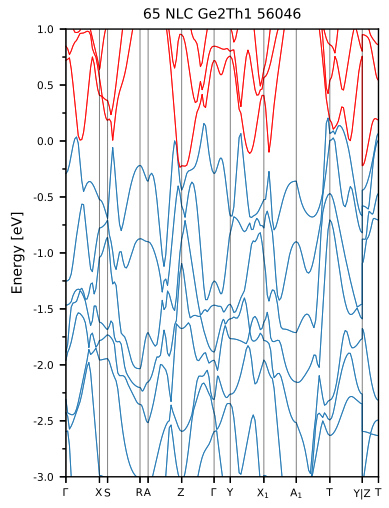

(ag) 65.2.7

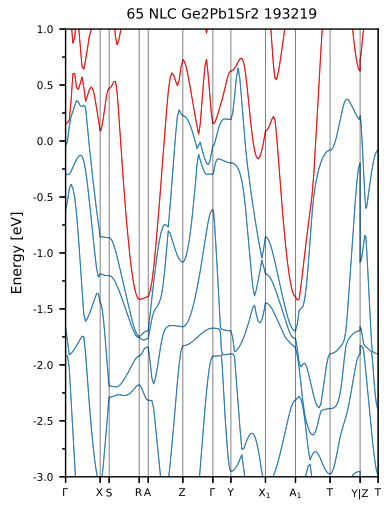

(ah) 65.2.8

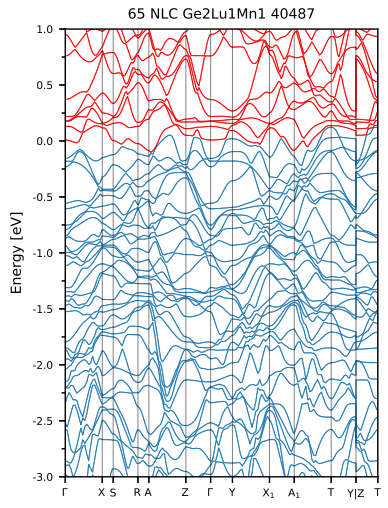

(ai) 65.2.9

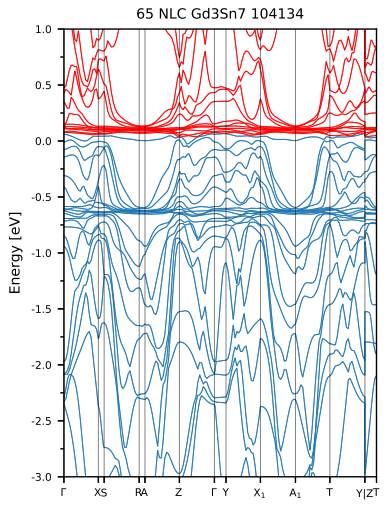

(aj) 65.2.10

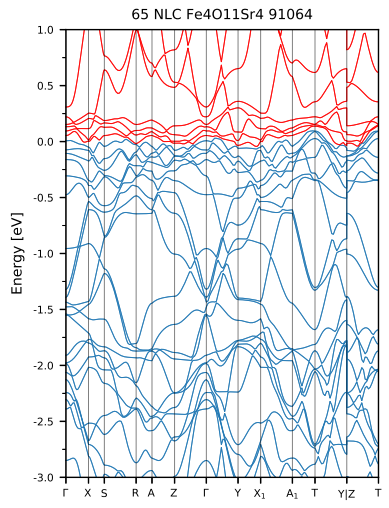

(ak) 65.2.11

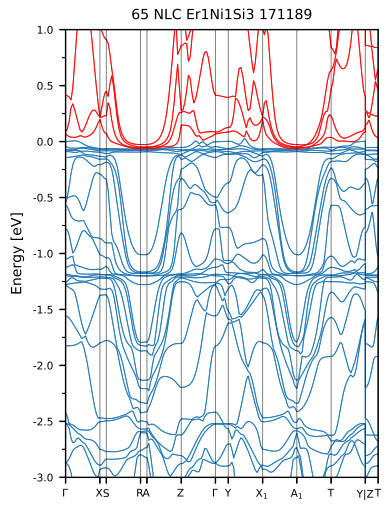

(al) 65.2.12

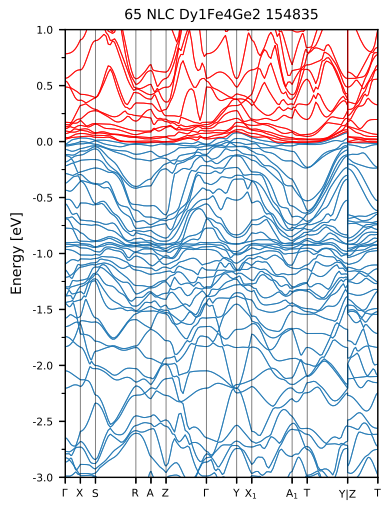

(am) 65.2.13

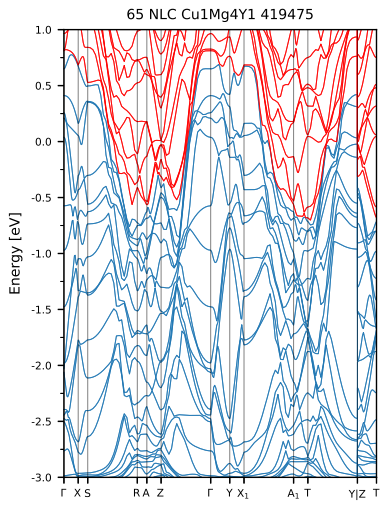

(an) 65.2.14

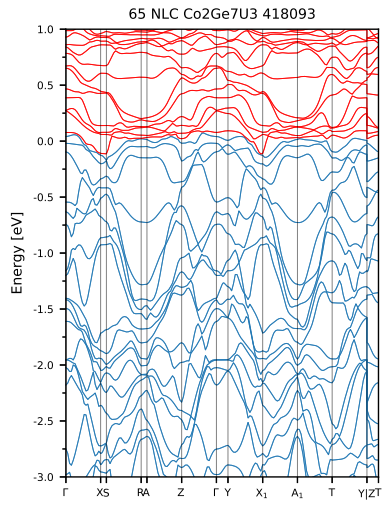

(ao) 65.2.15

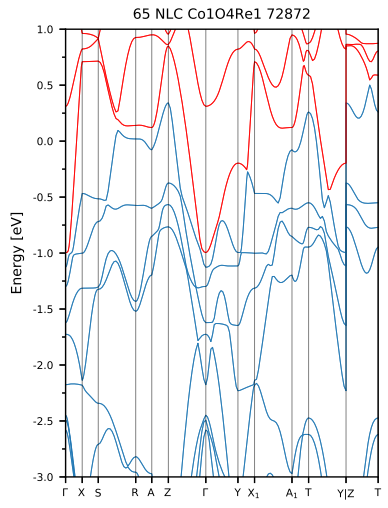

(ap) 65.2.16

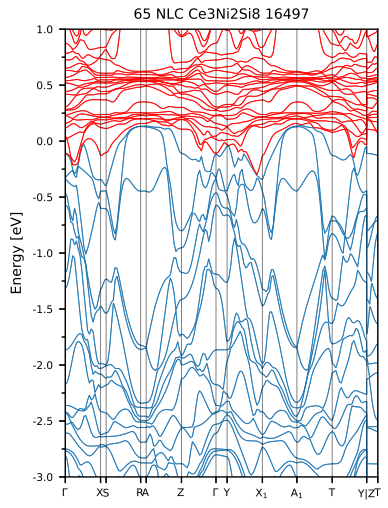

(aq) 65.2.17

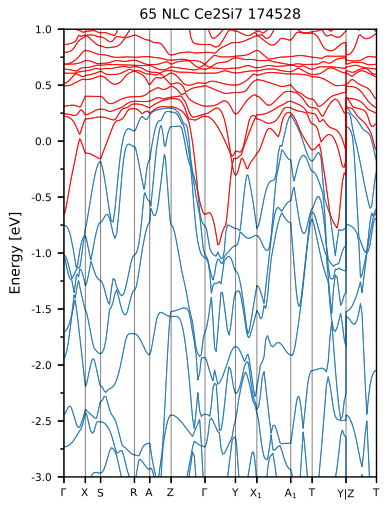

(ar) 65.2.18

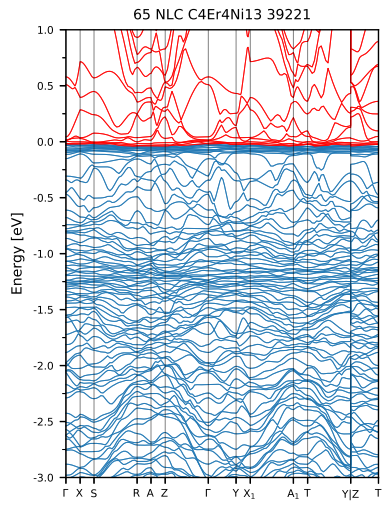

(as) 65.2.19

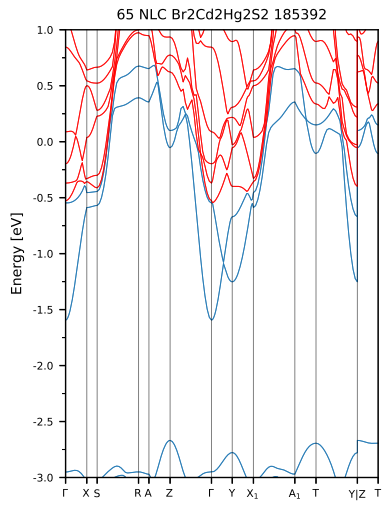

(at) 65.2.20

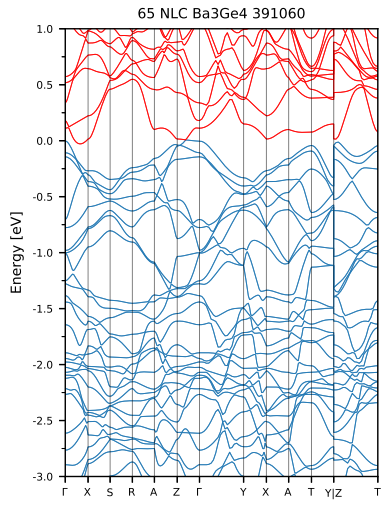

(au) 65.2.21

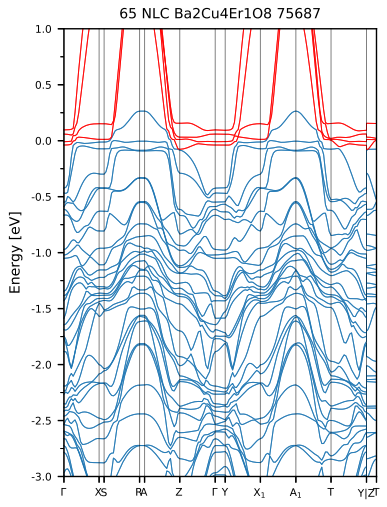

(av) 65.2.22

Figure 180: Band structures for topological materials in SG 65, part 3/4, from 65.2.7 to 65.2.22. See the list of figures.

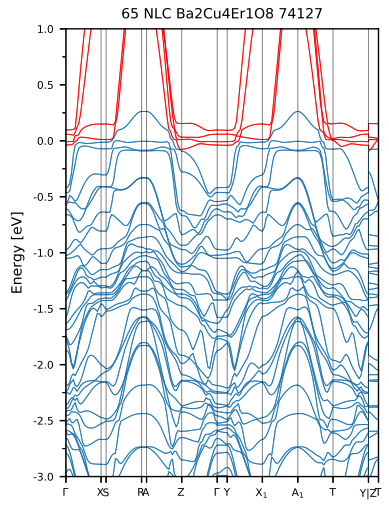

(aw) 65.2.23

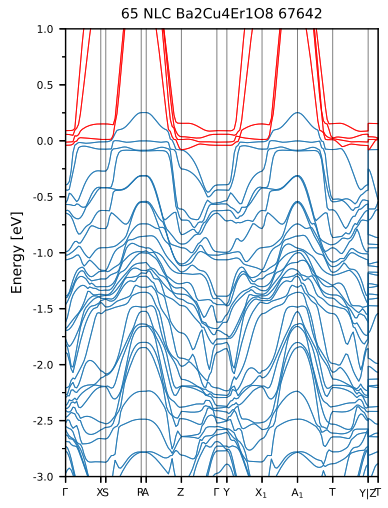

(ax) 65.2.24

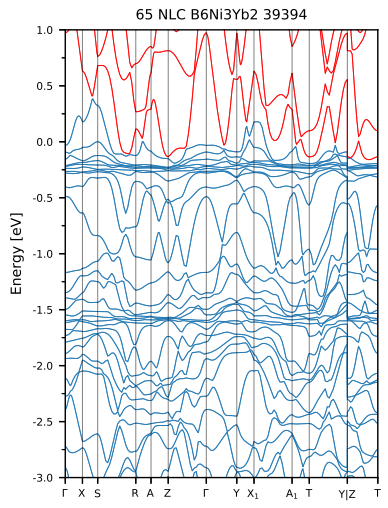

(ay) 65.2.25

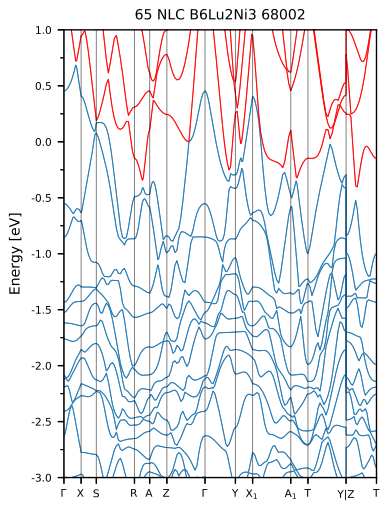

(az) 65.2.26

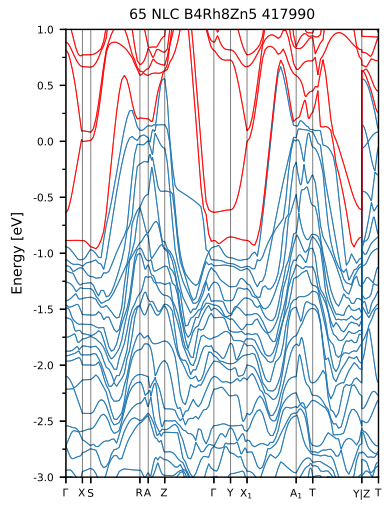

(ba) 65.2.27

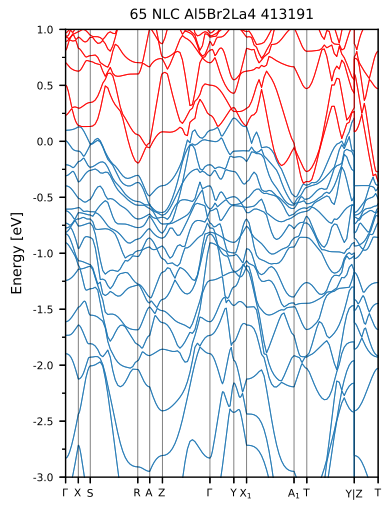

(bb) 65.2.28

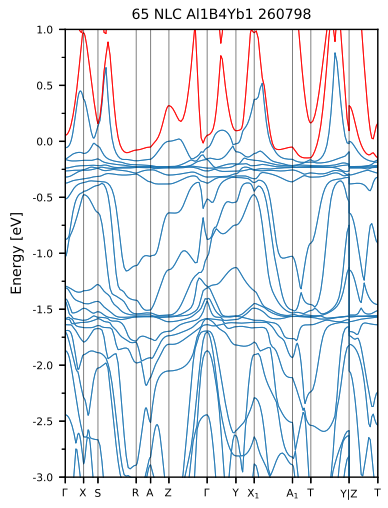

(bc) 65.2.29

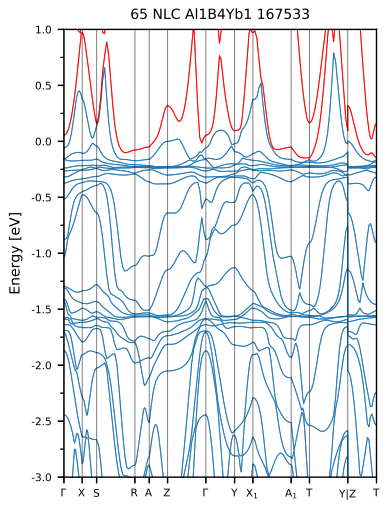

(bd) 65.2.30

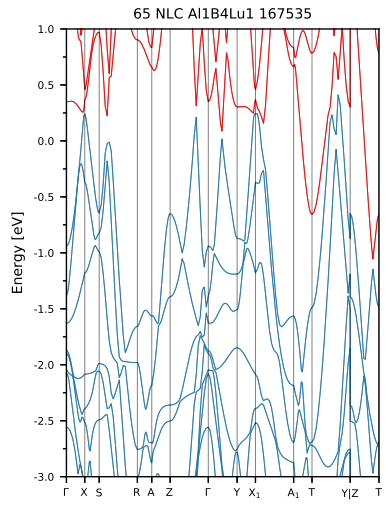

(be) 65.2.31

Figure 181: Band structures for topological materials in SG 65, part 4/4, from 65.2.23 to 65.2.31. See the list of figures.

### 0.0.30 SG 67

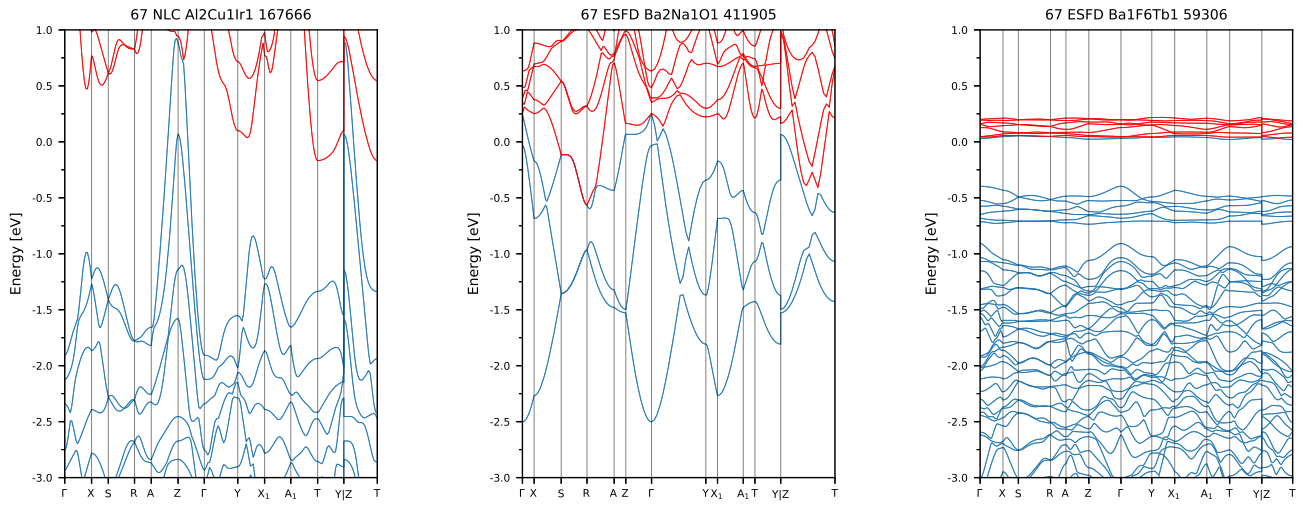

(a) 67.2.1 (b) 67.4.1 (c) 67.4.2

Figure 182: Band structures for topological materials in SG 67, part 1/1, from 67.2.1 to 67.4.2. See the list of figures.

## 0.0.31 SG 69

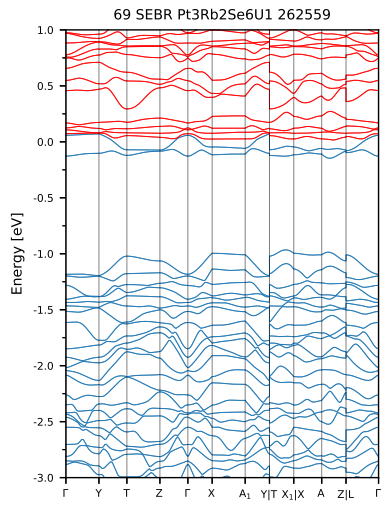

(a) 69.1.1

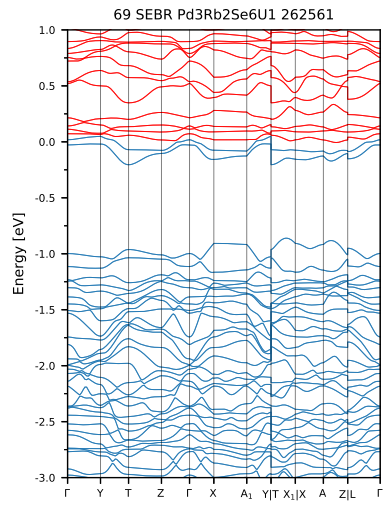

(b) 69.1.2

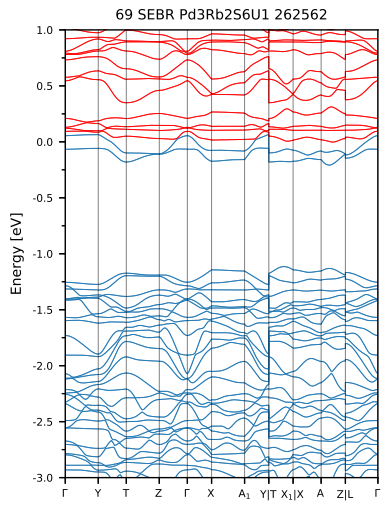

(c) 69.1.3

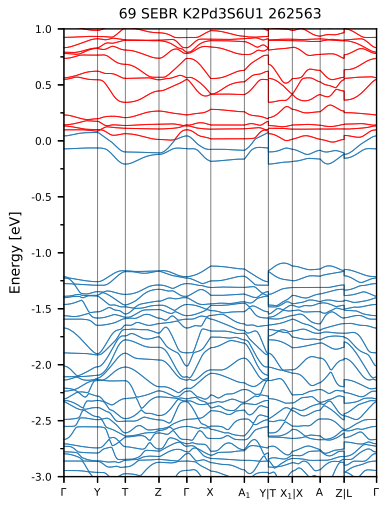

(d) 69.1.4

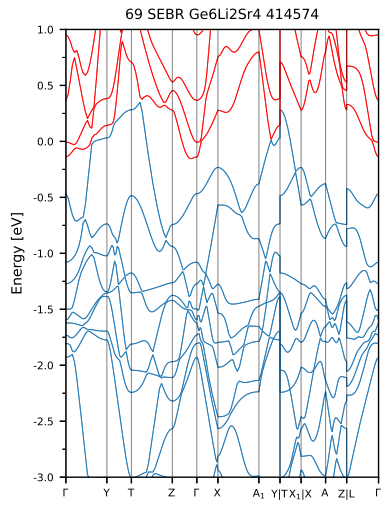

(e) 69.1.5

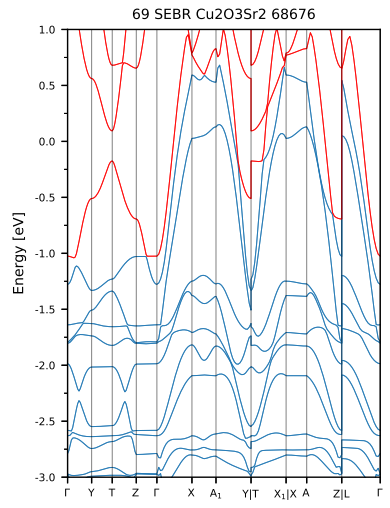

(f) 69.1.6

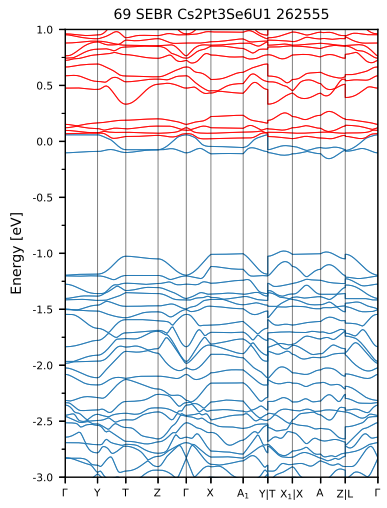

(g) 69.1.7

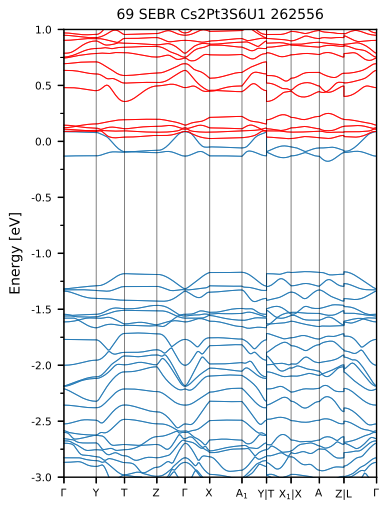

(h) 69.1.8

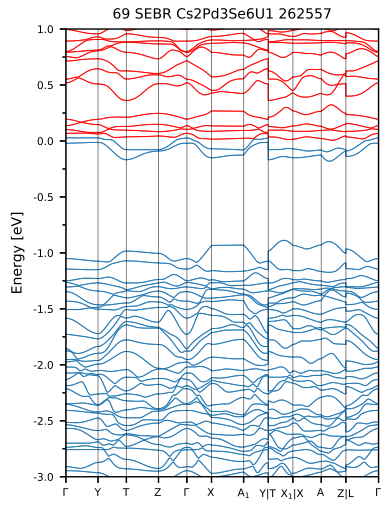

(i) 69.1.9

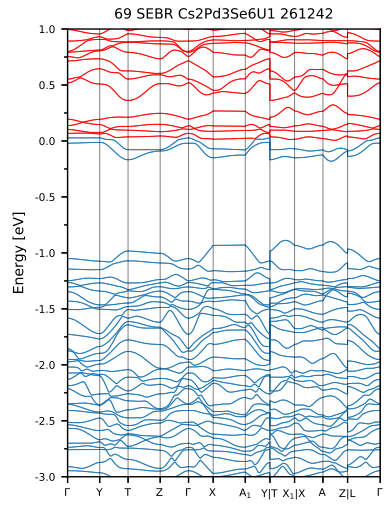

(j) 69.1.10

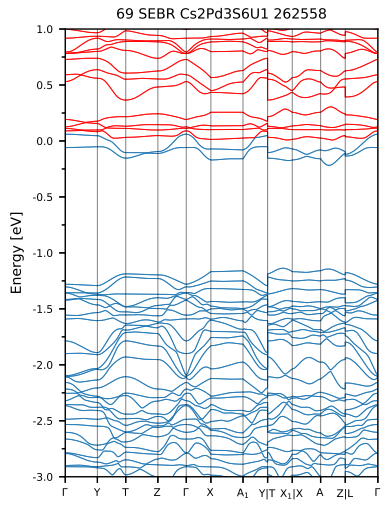

(k) 69.1.11

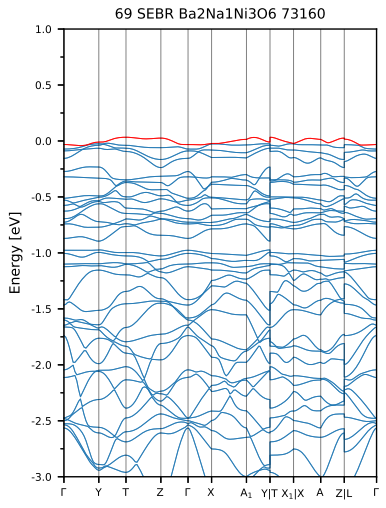

(l) 69.1.12

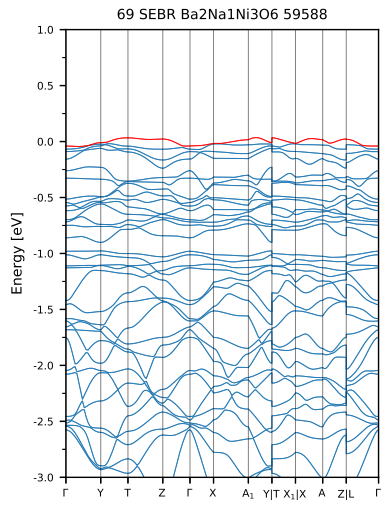

(m) 69.1.13

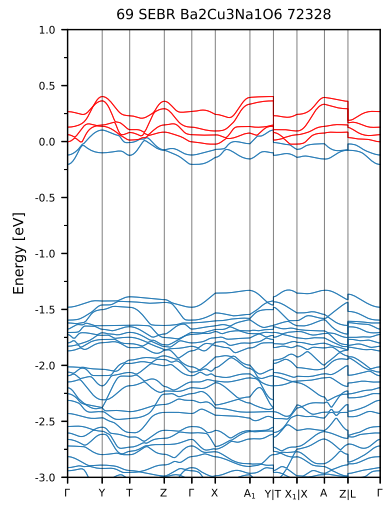

(n) 69.1.14

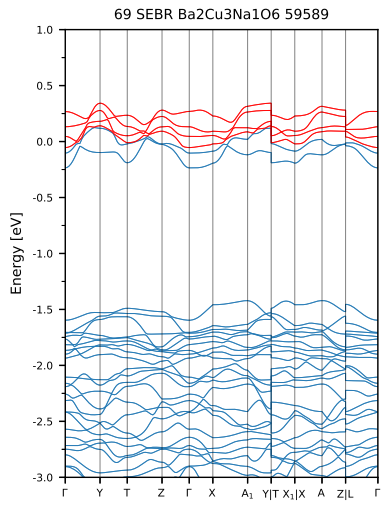

(o) 69.1.15

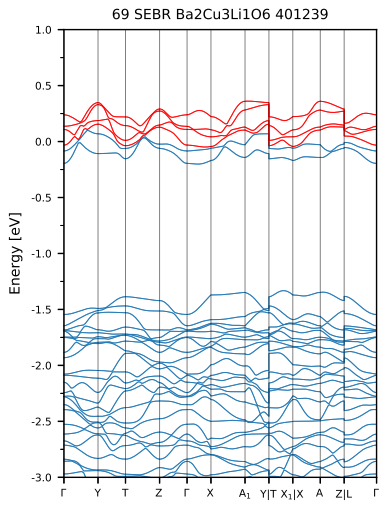

(p) 69.1.16

Figure 183: Band structures for topological materials in SG 69, part 1/2, from 69.1.1 to 69.1.16. See the list of figures.

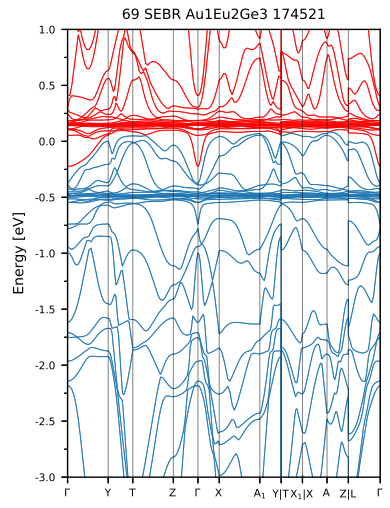

(q) 69.1.17

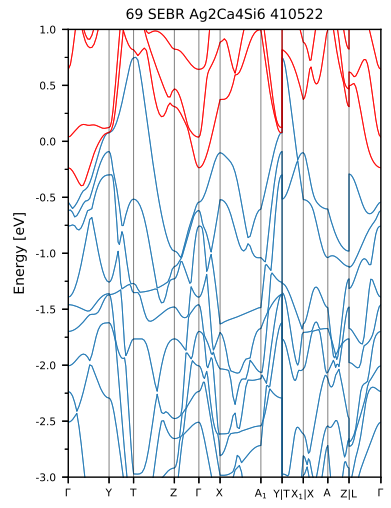

(r) 69.1.18

Figure 184: Band structures for topological materials in SG 69, part 2/2, from 69.1.17 to 69.1.18. See the list of figures.

### 0.0.32 SG 70

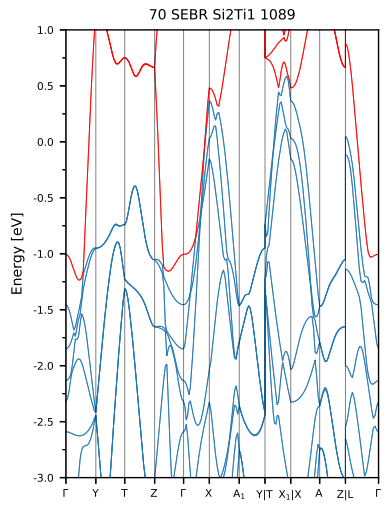

(a) 70.1.1

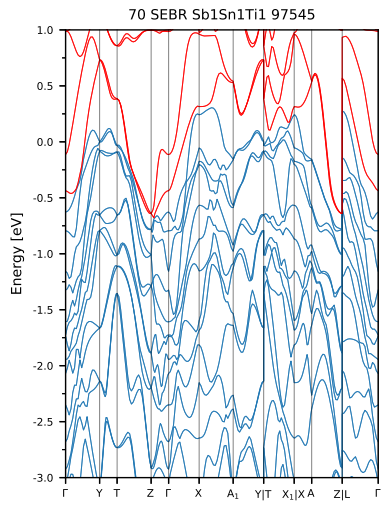

(b) 70.1.2

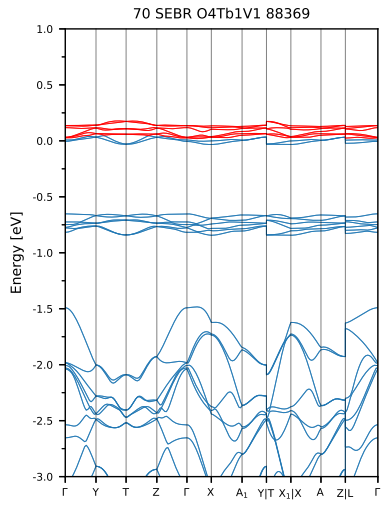

(c) 70.1.3

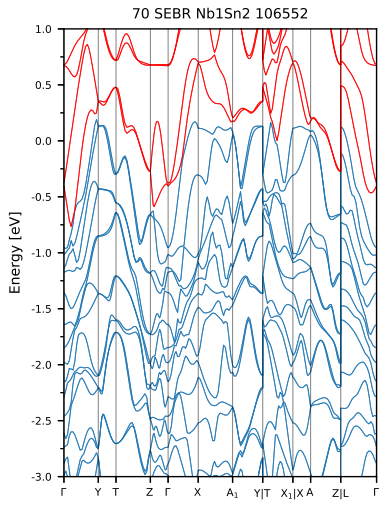

(d) 70.1.4

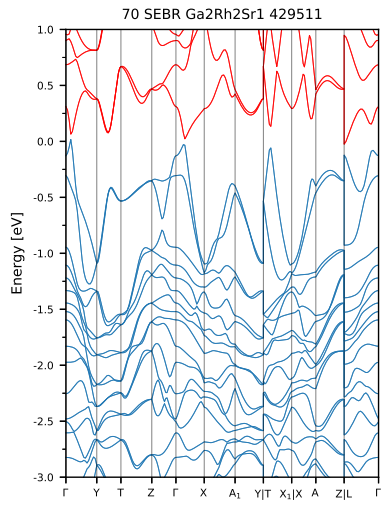

(e) 70.1.5

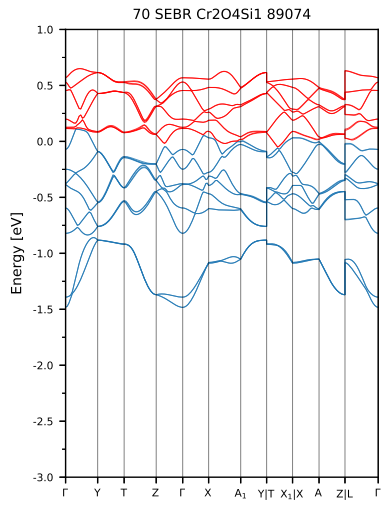

(f) 70.1.6

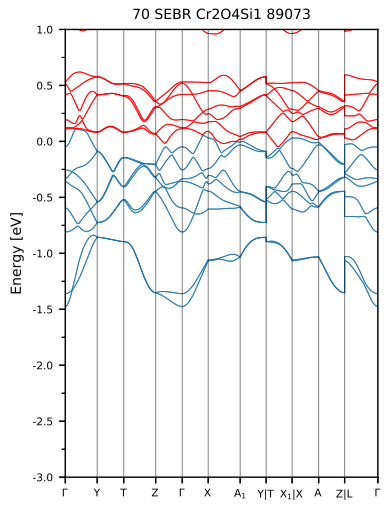

(g) 70.1.7

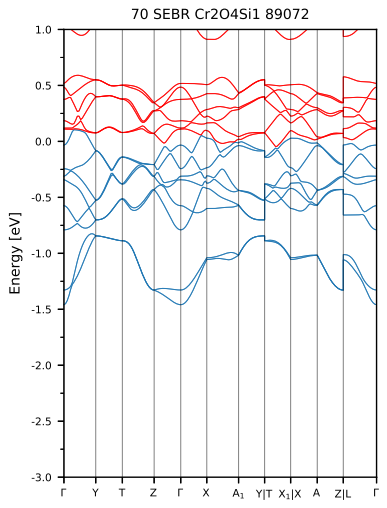

(h) 70.1.8

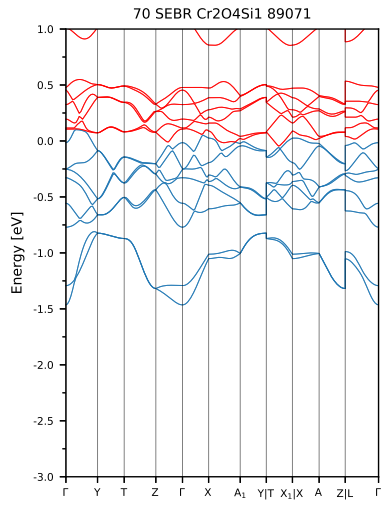

(i) 70.1.9

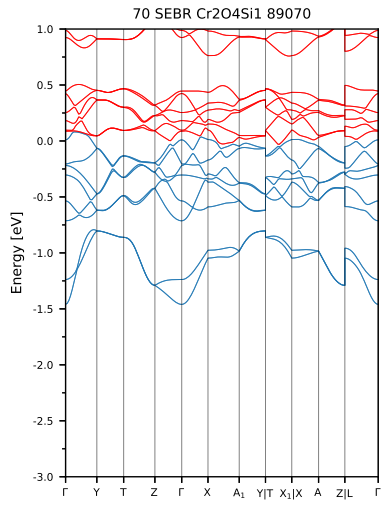

(j) 70.1.10

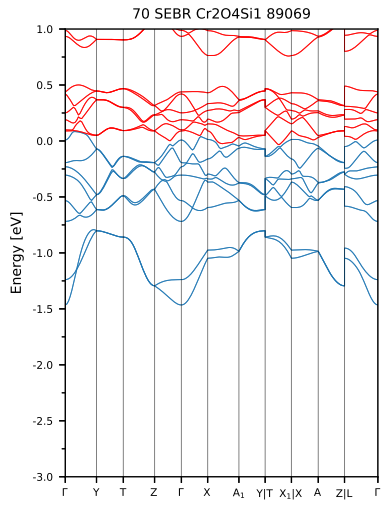

(k) 70.1.11

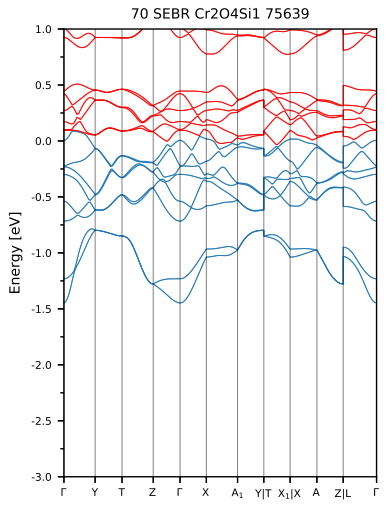

(l) 70.1.12

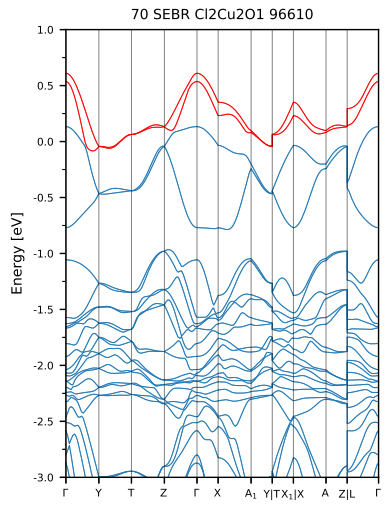

(m) 70.1.13

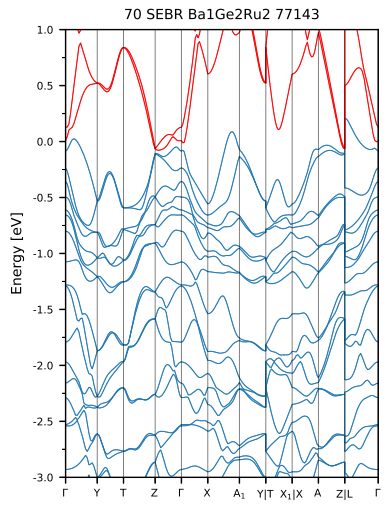

(n) 70.1.14

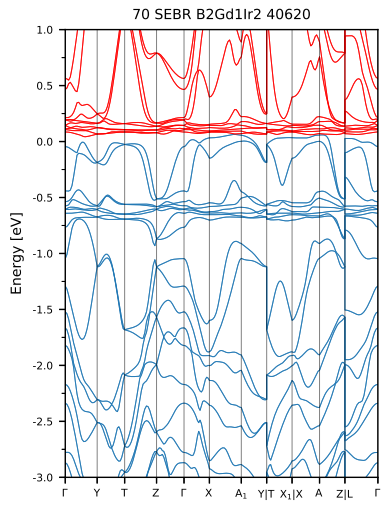

(o) 70.1.15

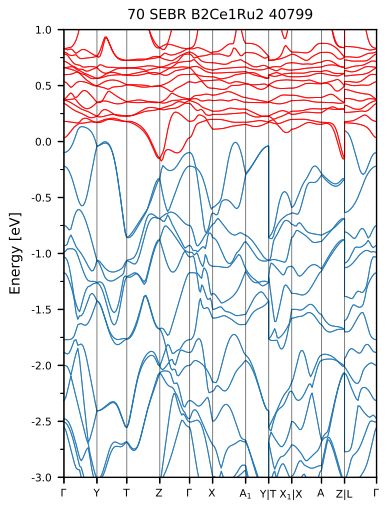

(p) 70.1.16

Figure 185: Band structures for topological materials in SG 70, part 1/2, from 70.1.1 to 70.1.16. See the list of figures.

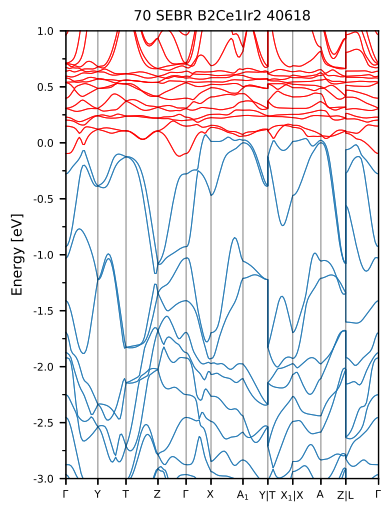

(q) 70.1.17

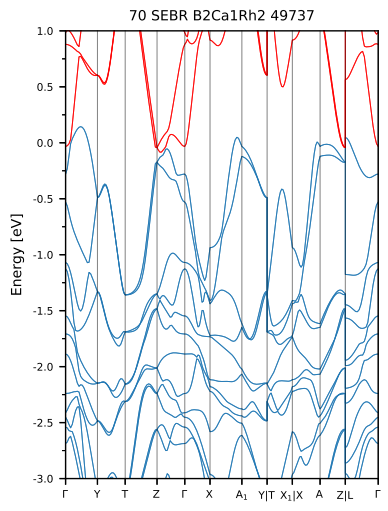

(r) 70.1.18

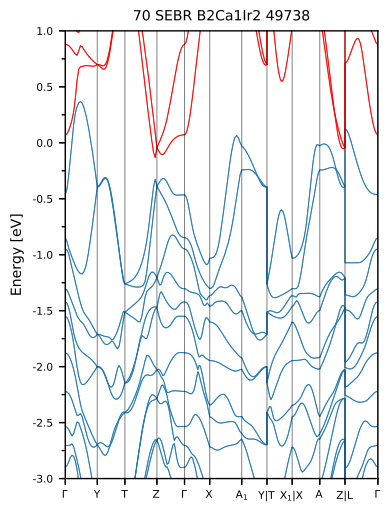

(s) 70.1.19

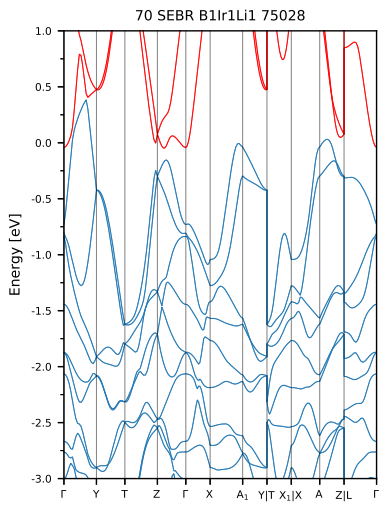

(t) 70.1.20

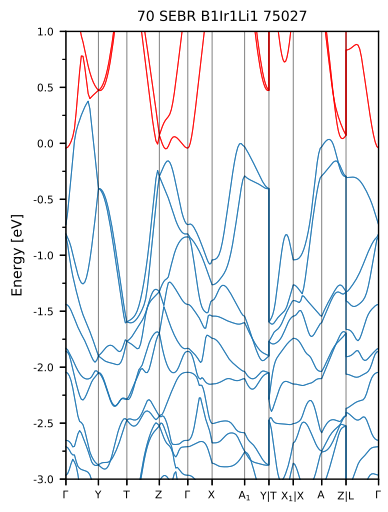

(u) 70.1.21

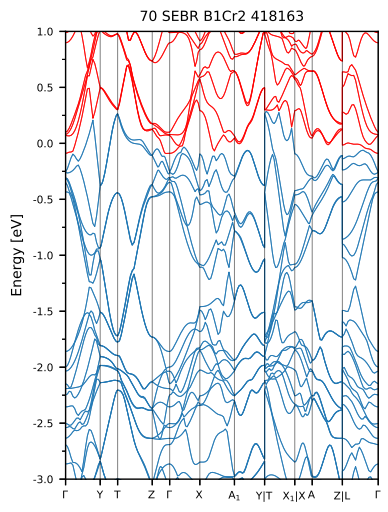

(v) 70.1.22

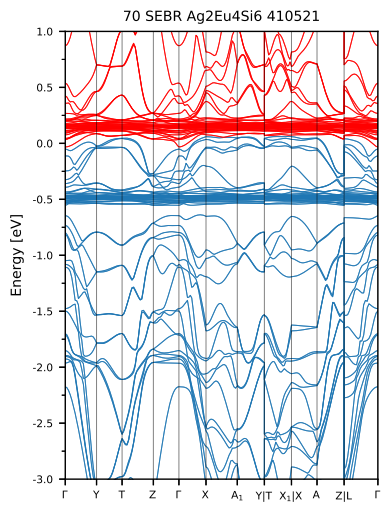

(w) 70.1.23

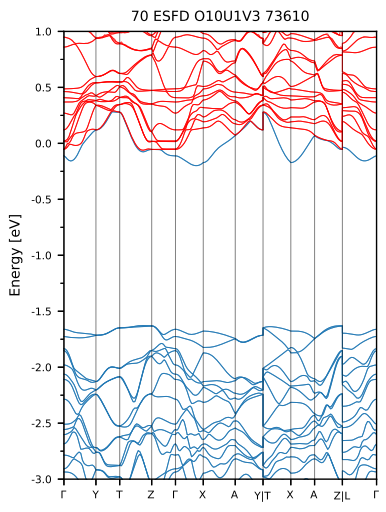

(x) 70.4.1

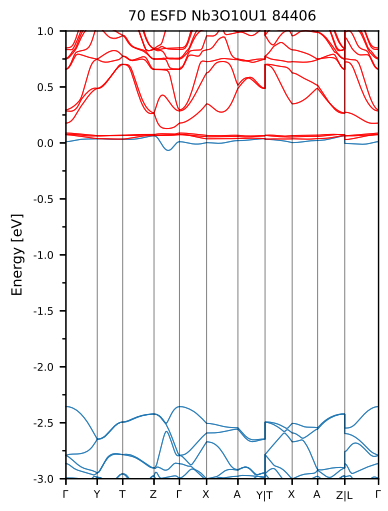

(y) 70.4.2

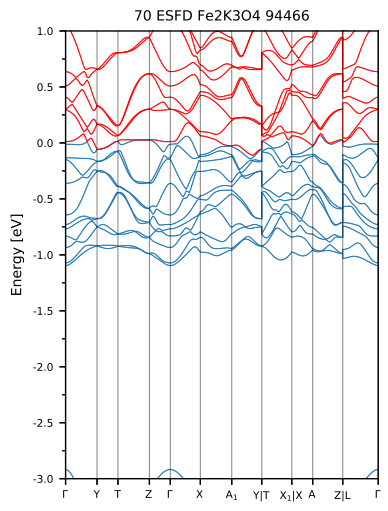

(z) 70.4.3

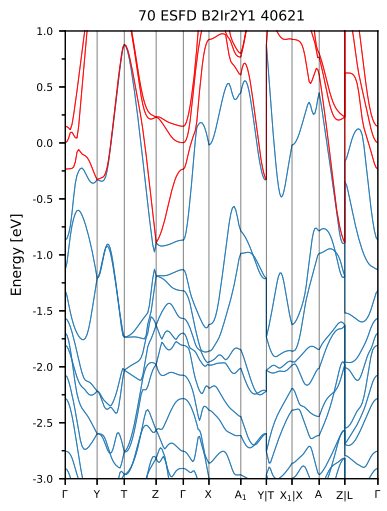

(aa) 70.4.4

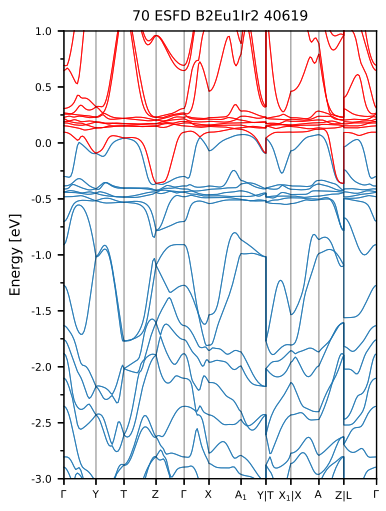

(ab) 70.4.5

Figure 186: Band structures for topological materials in SG 70, part 2/2, from 70.1.17 to 70.4.5. See the list of figures.

### 0.0.33 SG 71

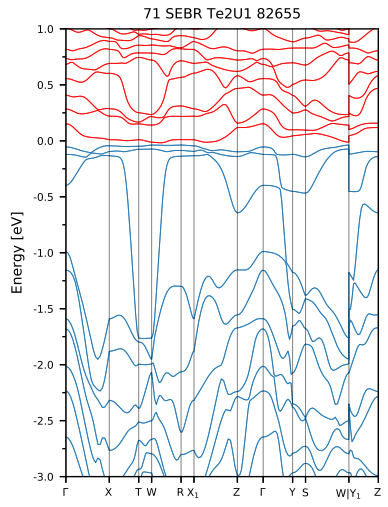

(a) 71.1.1

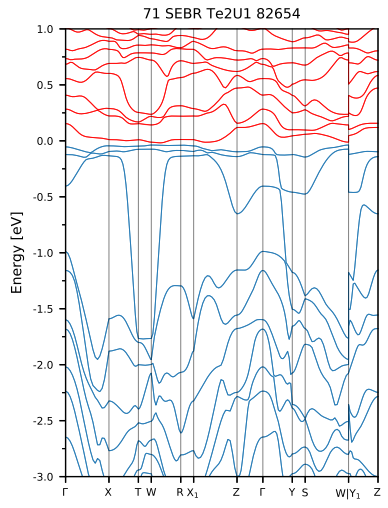

(b) 71.1.2

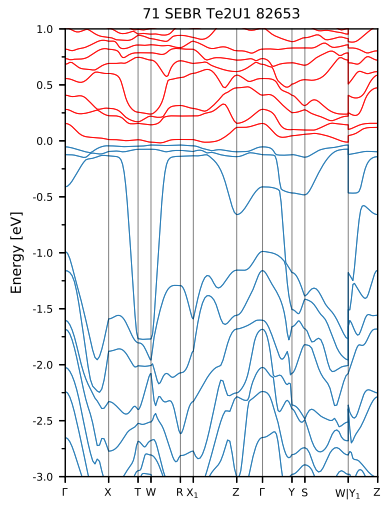

(c) 71.1.3

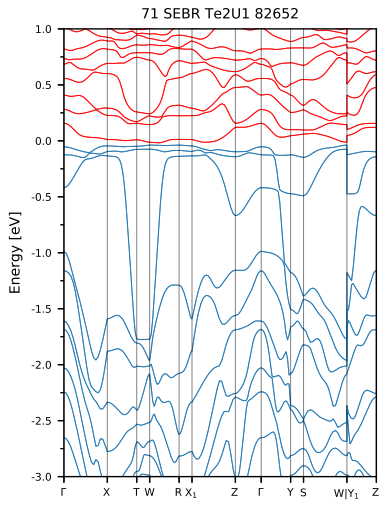

(d) 71.1.4

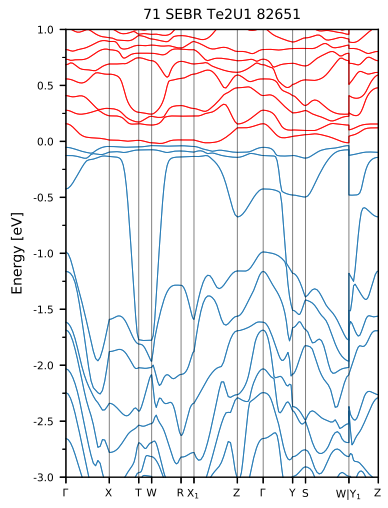

(e) 71.1.5

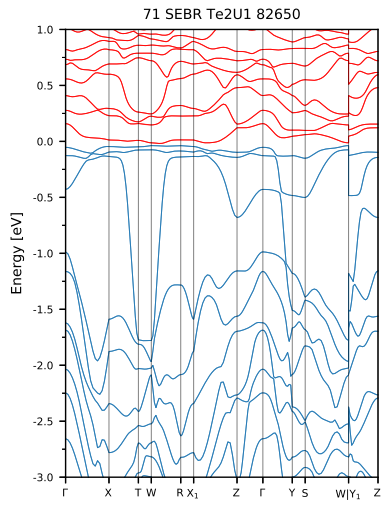

(f) 71.1.6

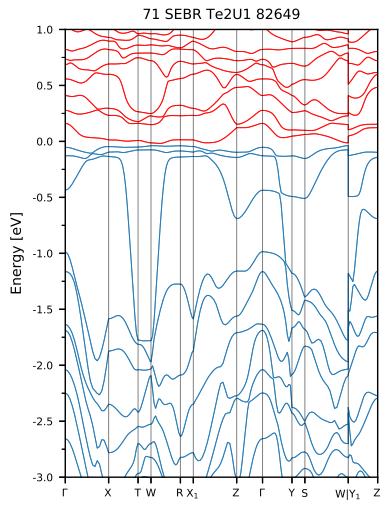

(g) 71.1.7

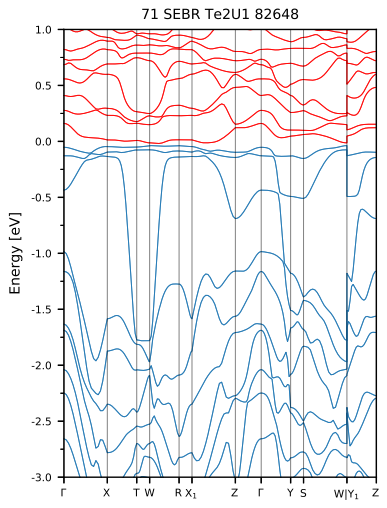

(h) 71.1.8

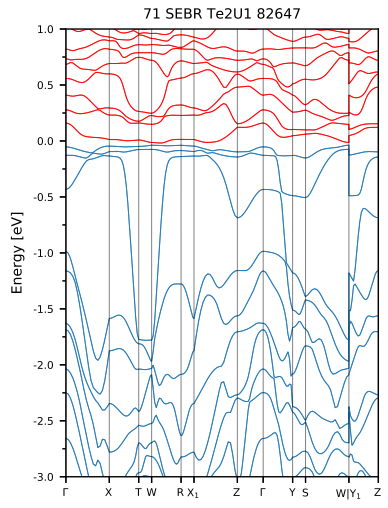

(i) 71.1.9

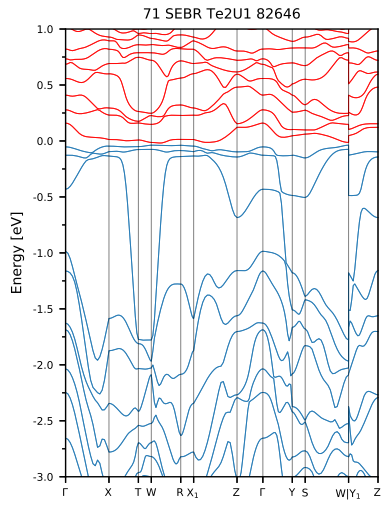

(j) 71.1.10

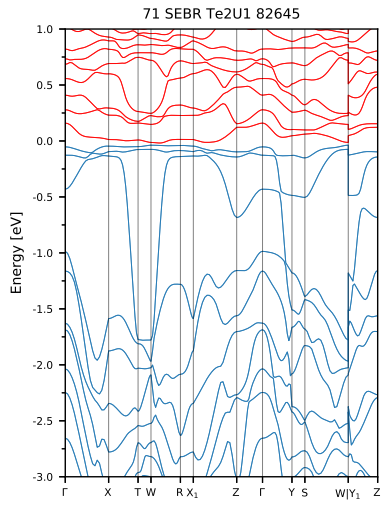

(k) 71.1.11

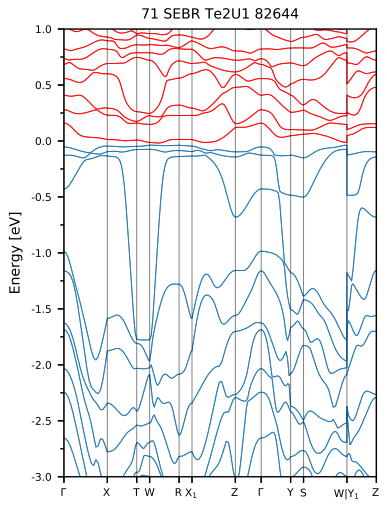

(l) 71.1.12

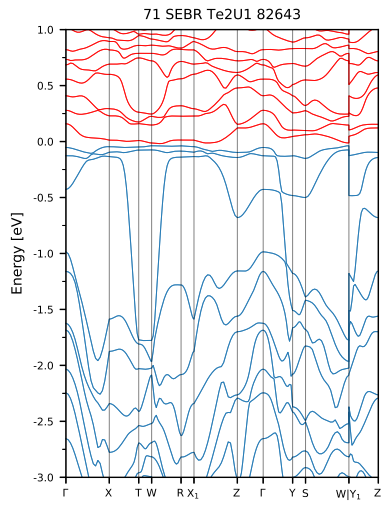

(m) 71.1.13

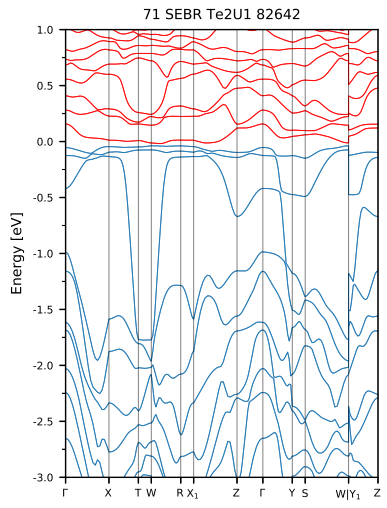

(n) 71.1.14

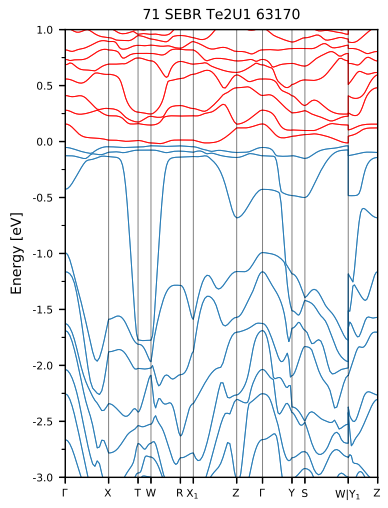

(o) 71.1.15

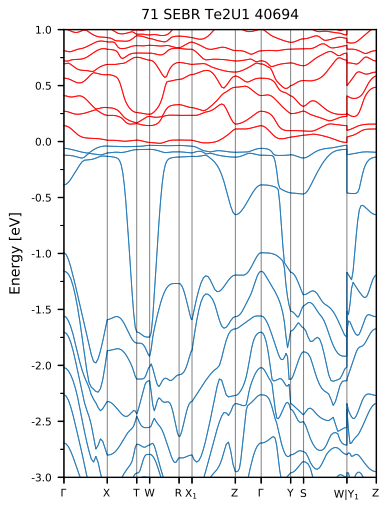

(p) 71.1.16

Figure 187: Band structures for topological materials in SG 71, part 1/7, from 71.1.1 to 71.1.16. See the list of figures.

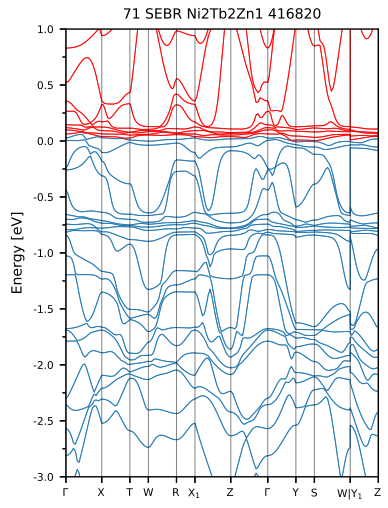

(q) 71.1.17

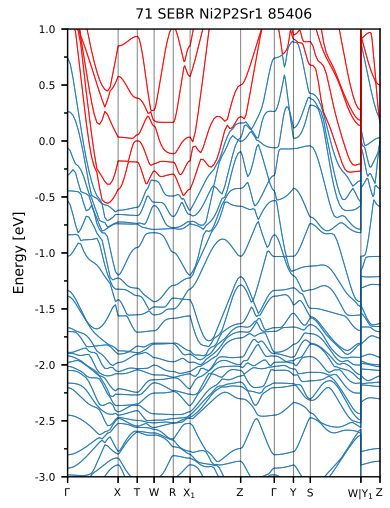

(r) 71.1.18

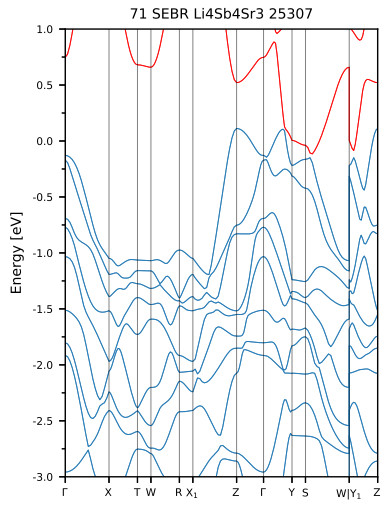

(s) 71.1.19

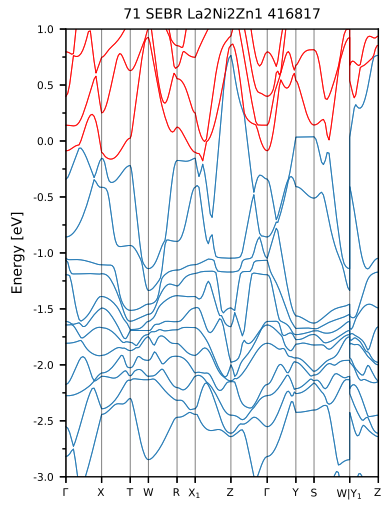

(t) 71.1.20

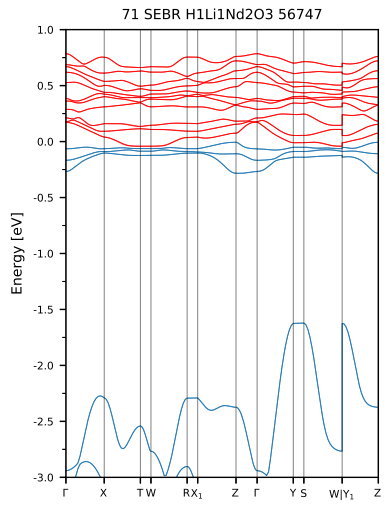

(u) 71.1.21

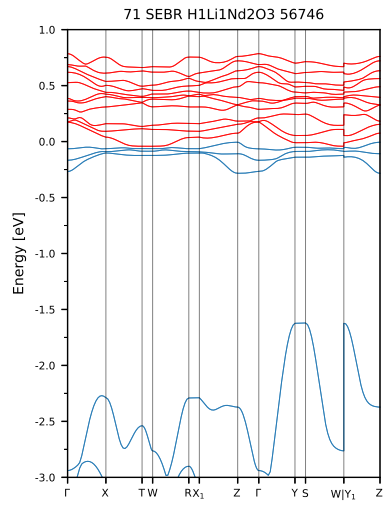

(v) 71.1.22

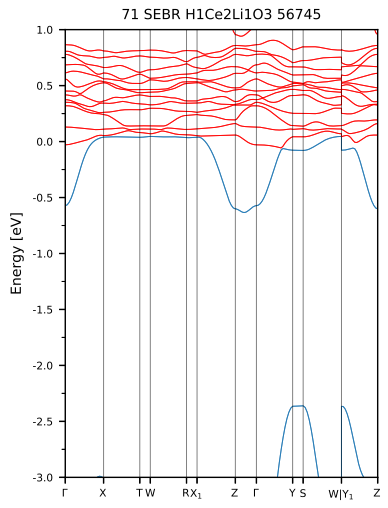

(w) 71.1.23

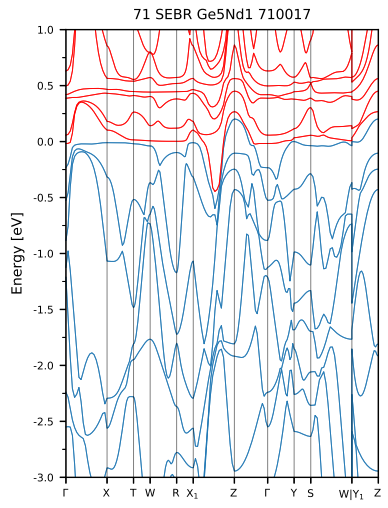

(x) 71.1.24

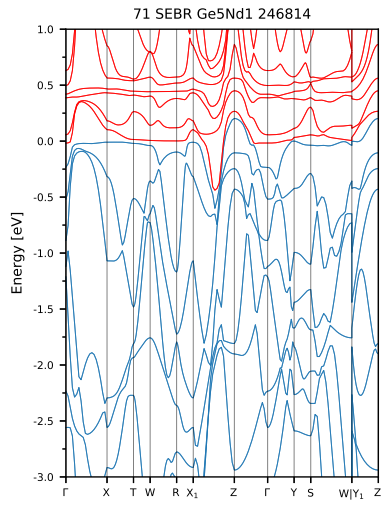

(y) 71.1.25

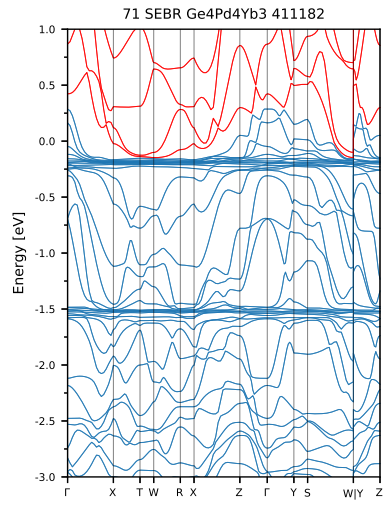

(z) 71.1.26

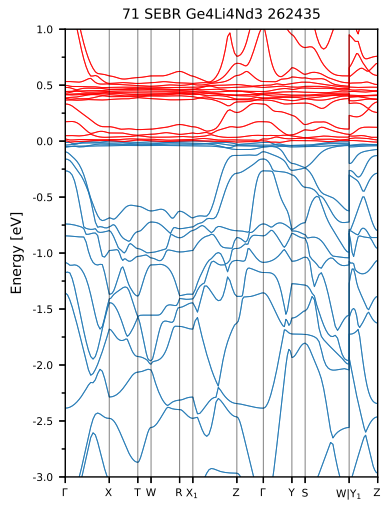

(aa) 71.1.27

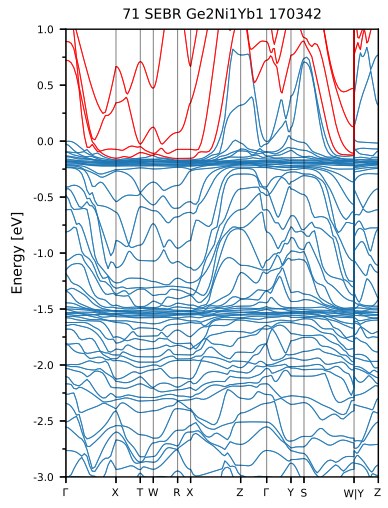

(ab) 71.1.28

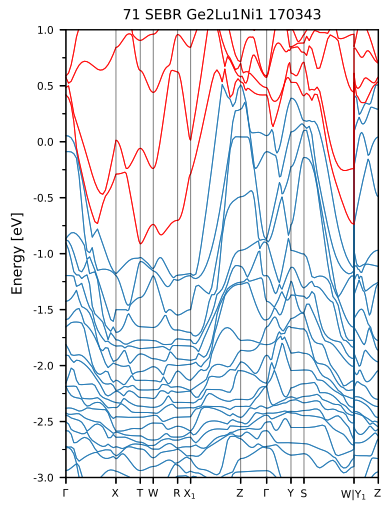

(ac) 71.1.29

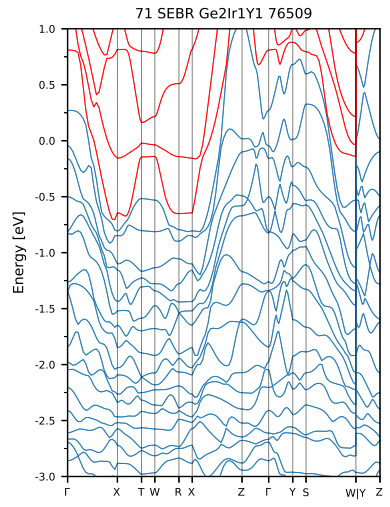

(ad) 71.1.30

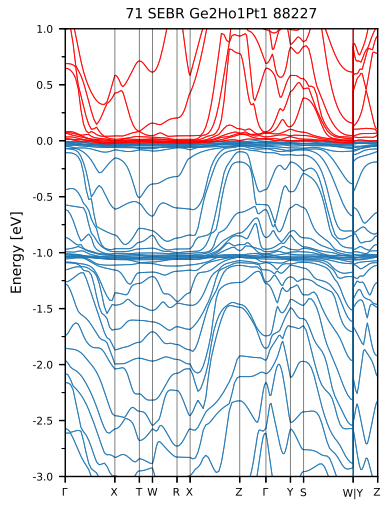

(ae) 71.1.31

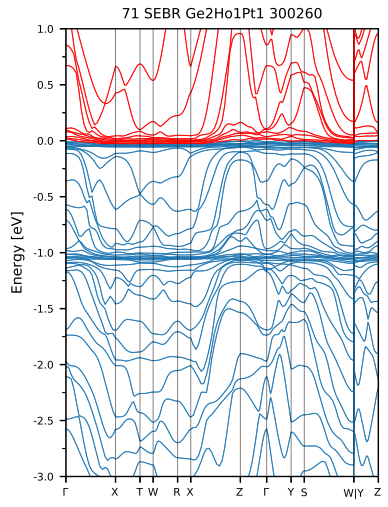

(af) 71.1.32

Figure 188: Band structures for topological materials in SG 71, part 2/7, from 71.1.17 to 71.1.32. See the list of figures.

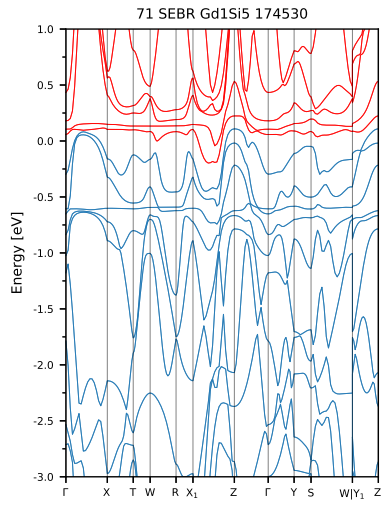

(ag) 71.1.33

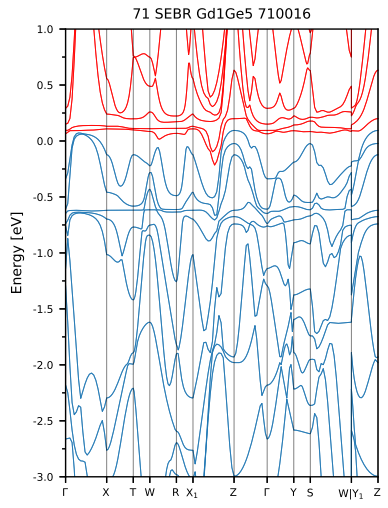

(ah) 71.1.34

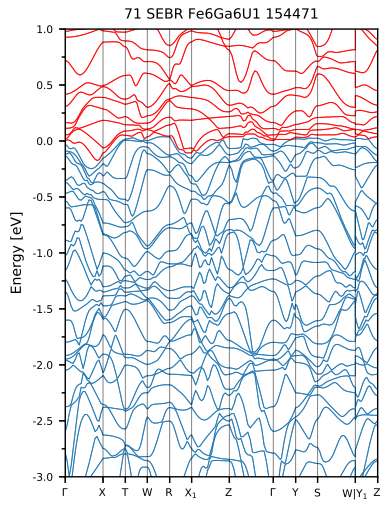

(ai) 71.1.35

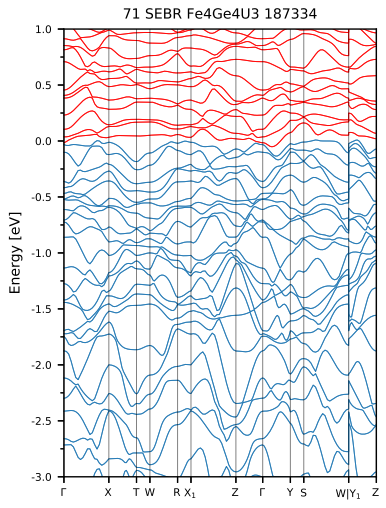

(aj) 71.1.36

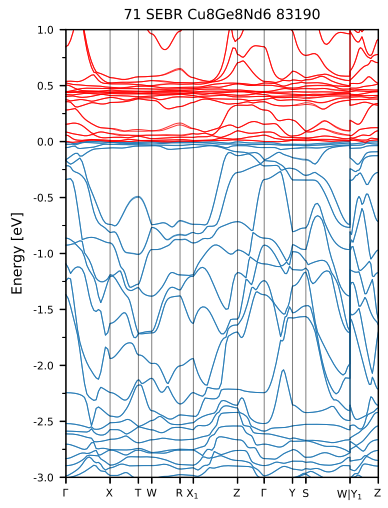

(ak) 71.1.37

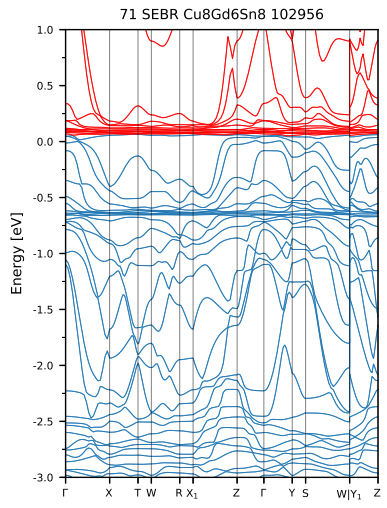

(al) 71.1.38

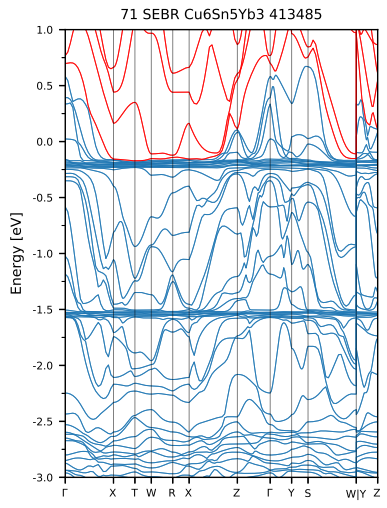

(am) 71.1.39

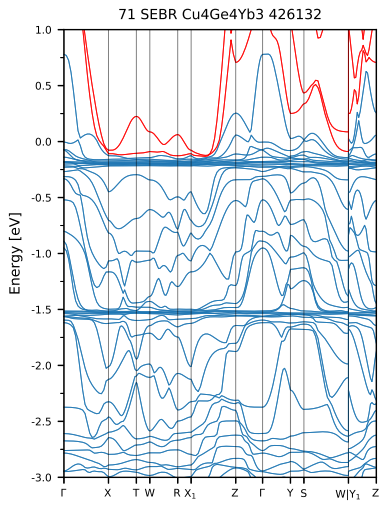

(an) 71.1.40

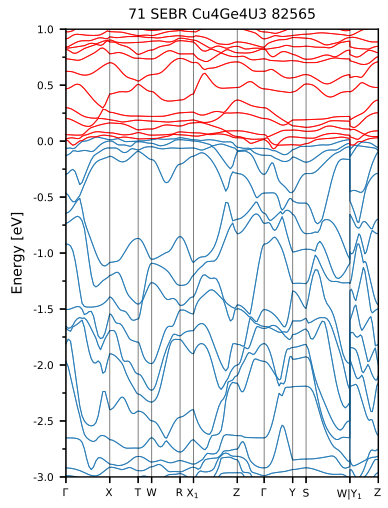

(ao) 71.1.41

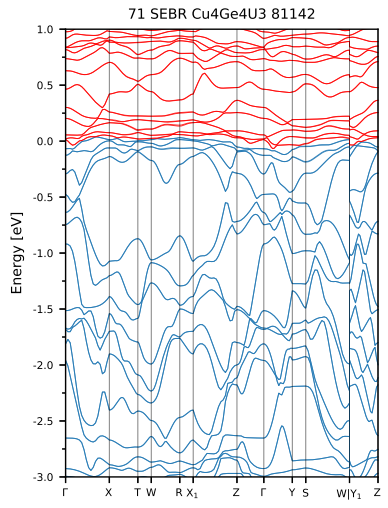

(ap) 71.1.42

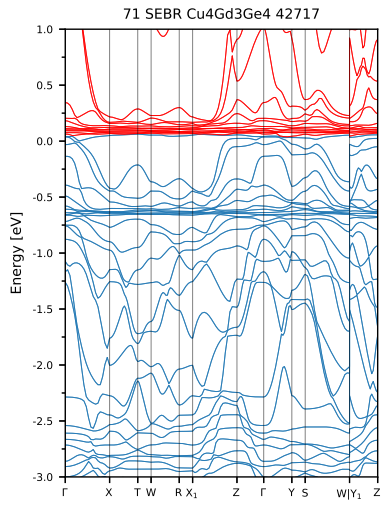

(aq) 71.1.43

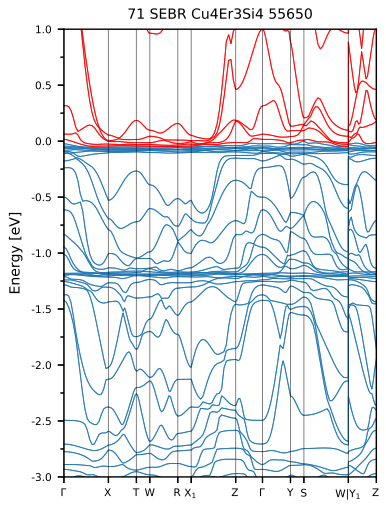

(ar) 71.1.44

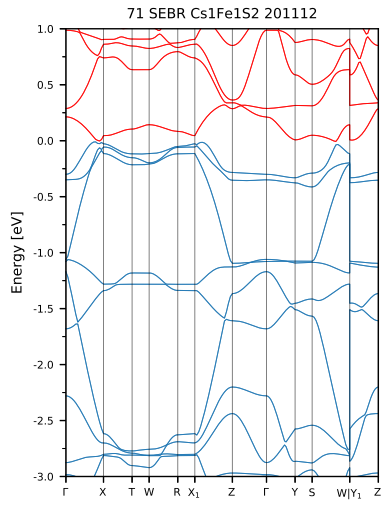

(as) 71.1.45

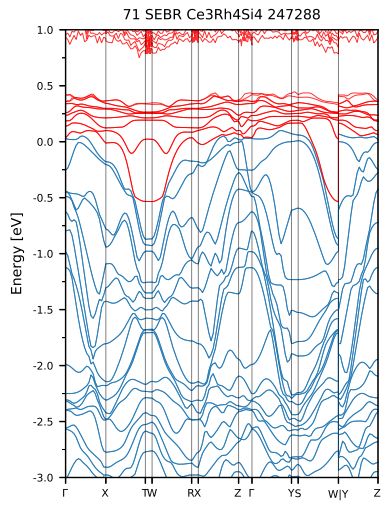

(at) 71.1.46

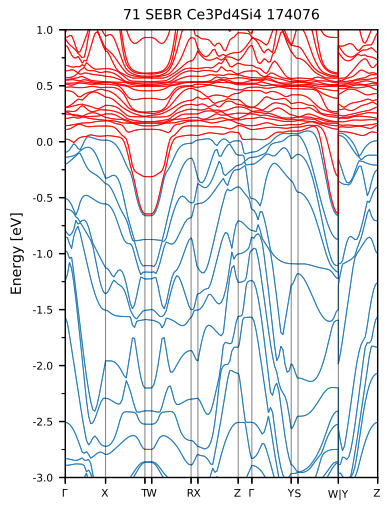

(au) 71.1.47

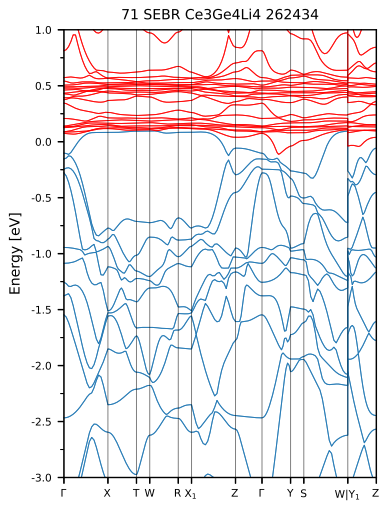

(av) 71.1.48

Figure 189: Band structures for topological materials in SG 71, part 3/7, from 71.1.33 to 71.1.48. See the list of figures.

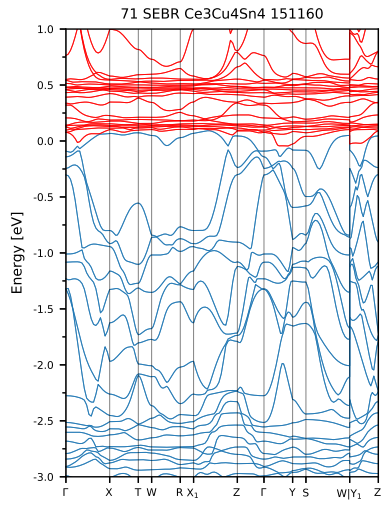

(aw) 71.1.49

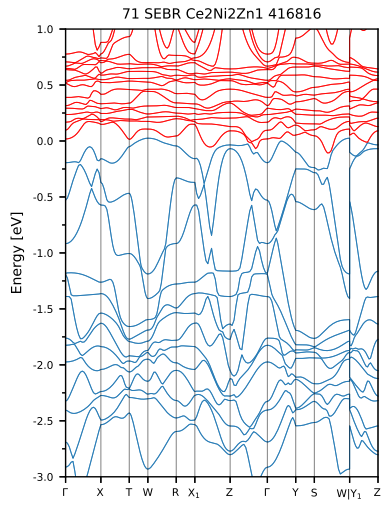

(ax) 71.1.50

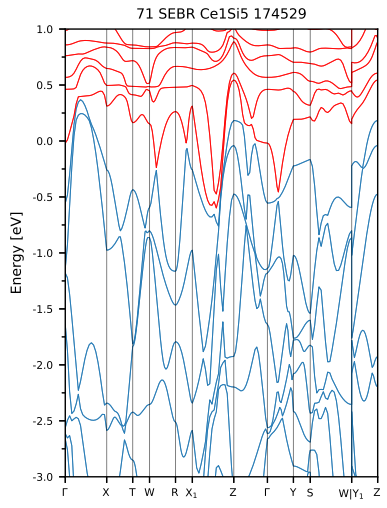

(ay) 71.1.51

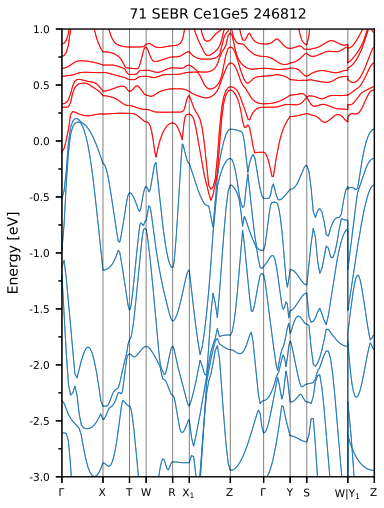

(az) 71.1.52

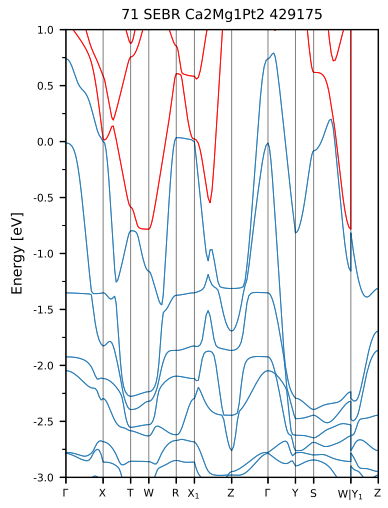

(ba) 71.1.53

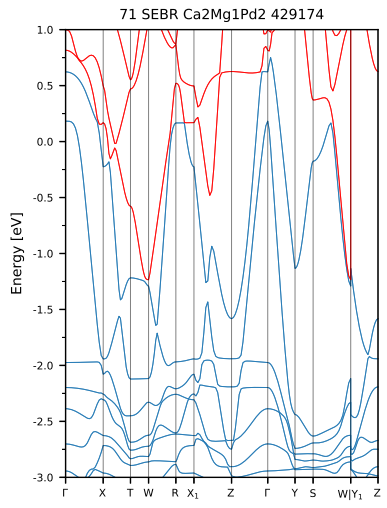

(bb) 71.1.54

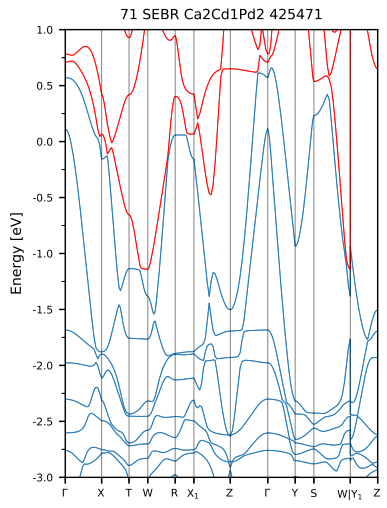

(bc) 71.1.55

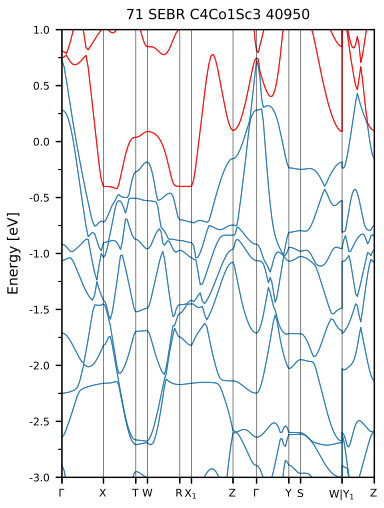

(bd) 71.1.56

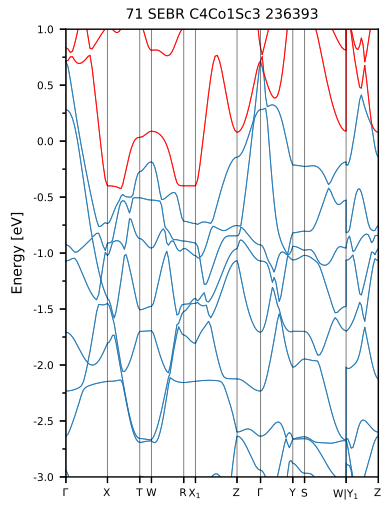

(be) 71.1.57

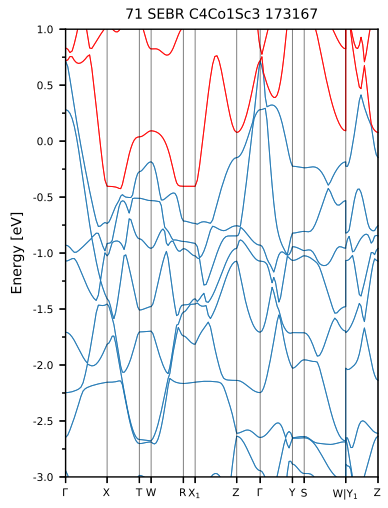

(bf) 71.1.58

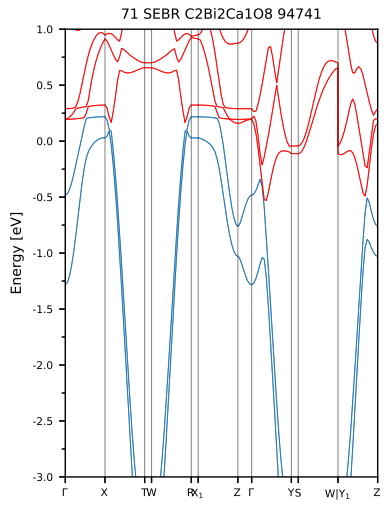

(bg) 71.1.59

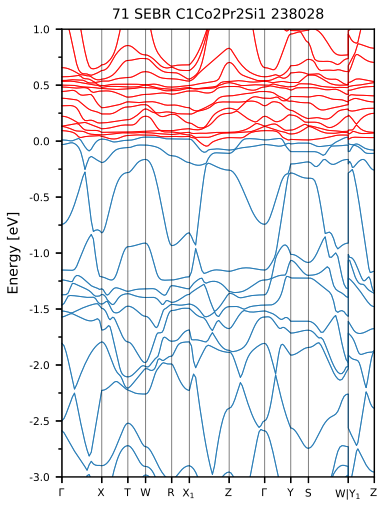

(bh) 71.1.60

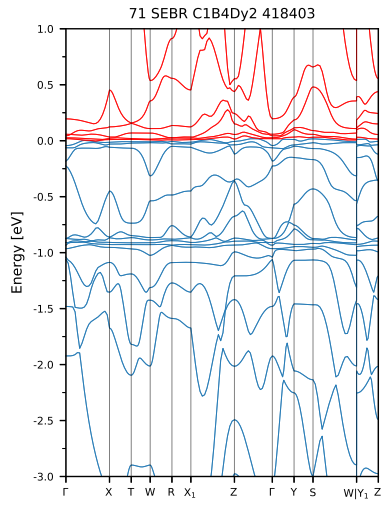

(bi) 71.1.61

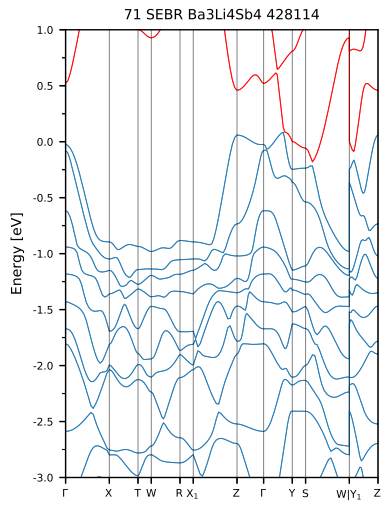

(bj) 71.1.62

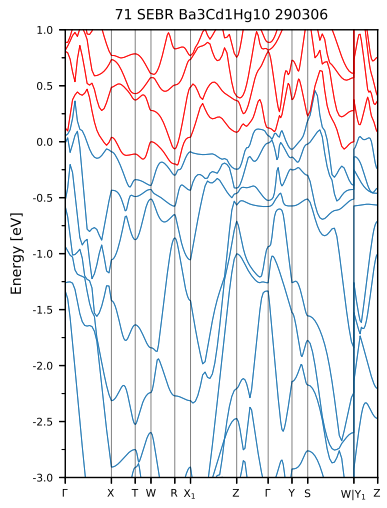

(bk) 71.1.63

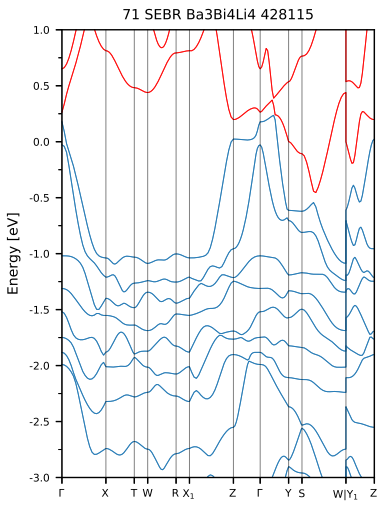

(bl) 71.1.64

Figure 190: Band structures for topological materials in SG 71, part 4/7, from 71.1.49 to 71.1.64. See the list of figures.

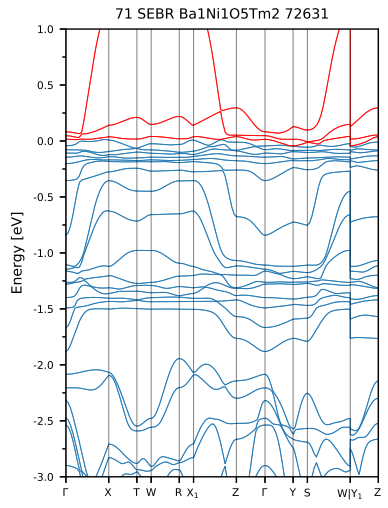

(bm) 71.1.65

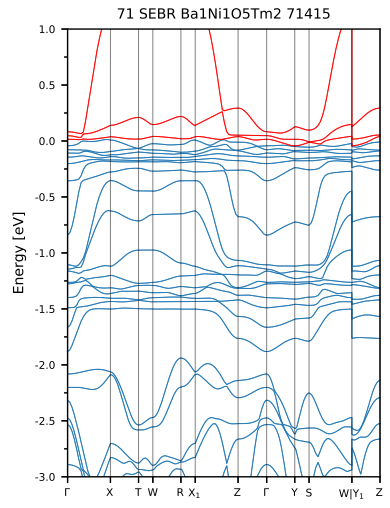

(bn) 71.1.66

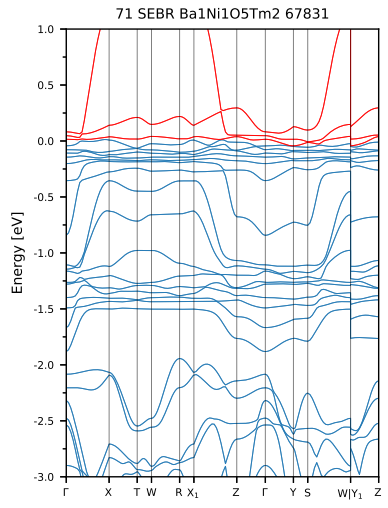

(bo) 71.1.67

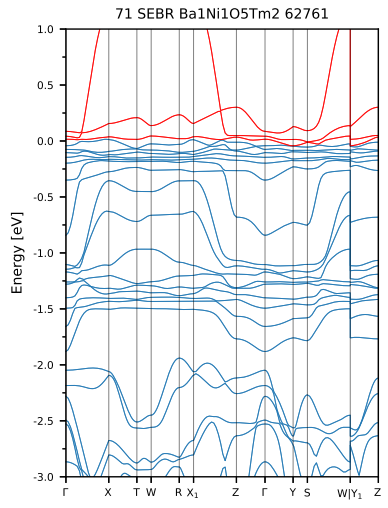

(bp) 71.1.68

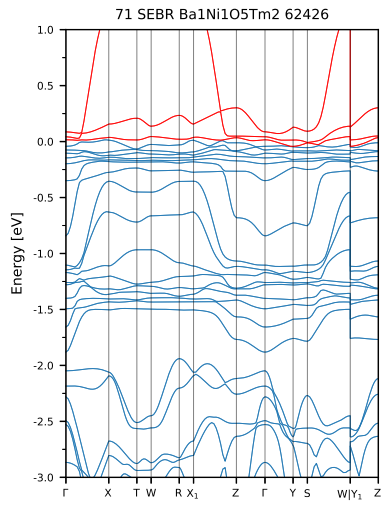

(bq) 71.1.69

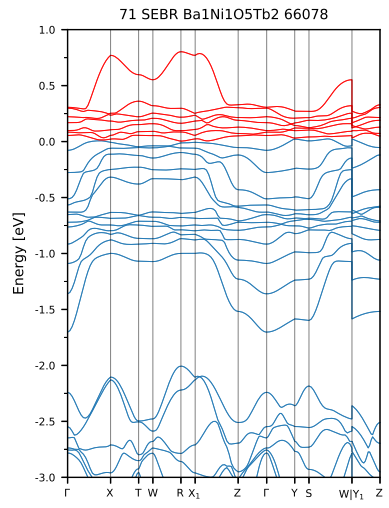

(br) 71.1.70

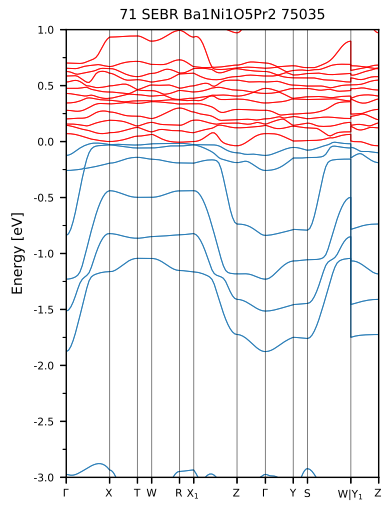

(bs) 71.1.71

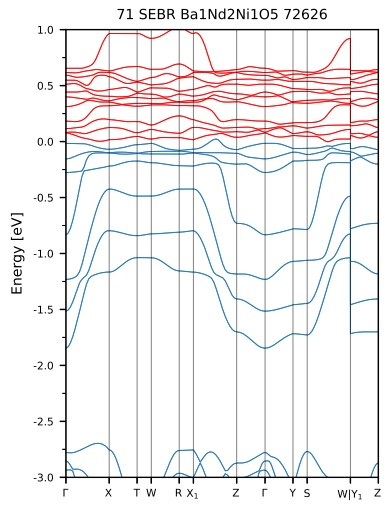

(bt) 71.1.72

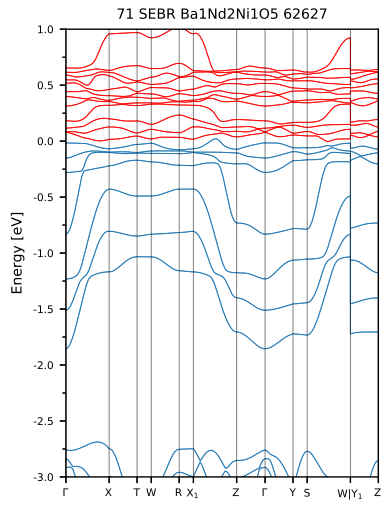

(bu) 71.1.73

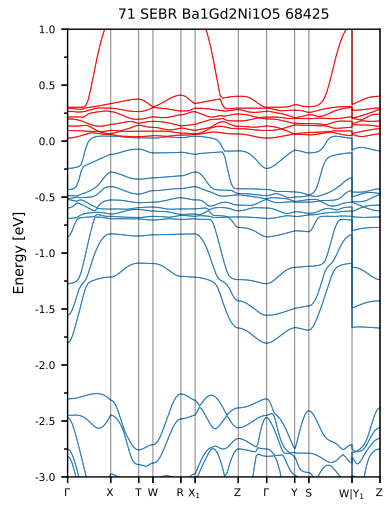

(bv) 71.1.74

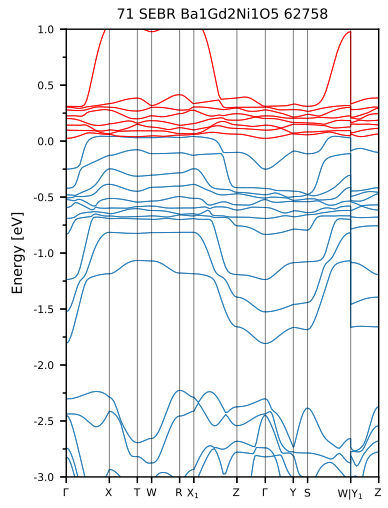

(bw) 71.1.75

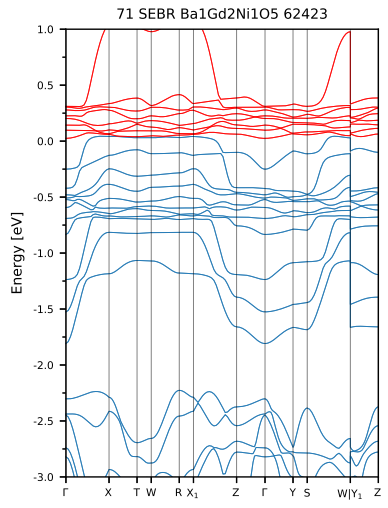

(bx) 71.1.76

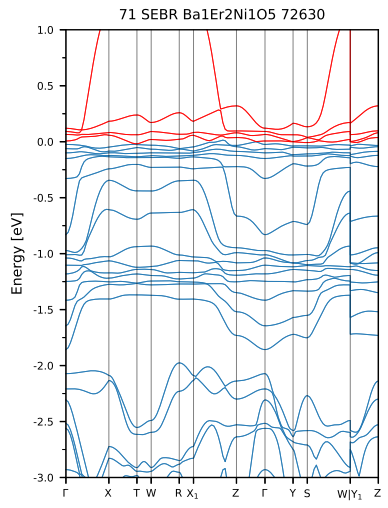

(by) 71.1.77

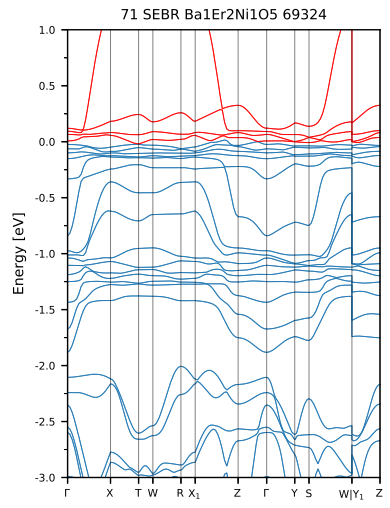

(bz) 71.1.78

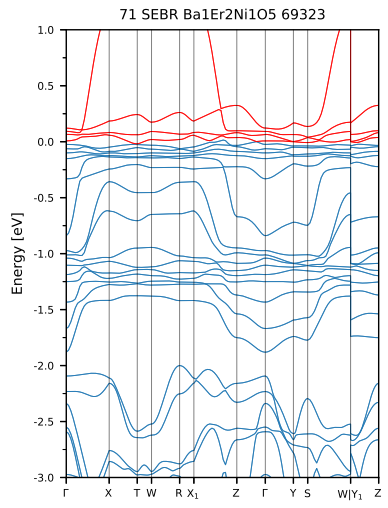

(ca) 71.1.79

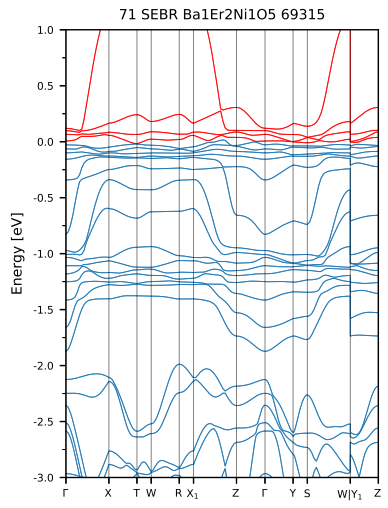

(cb) 71.1.80

Figure 191: Band structures for topological materials in SG 71, part 5/7, from 71.1.65 to 71.1.80. See the list of figures.

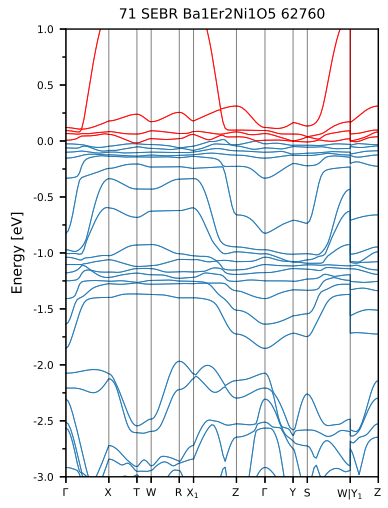

(cc) 71.1.81

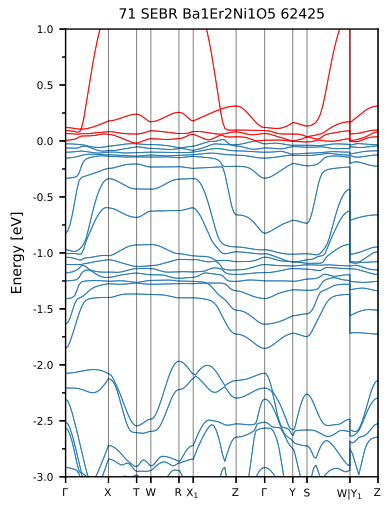

(cd) 71.1.82

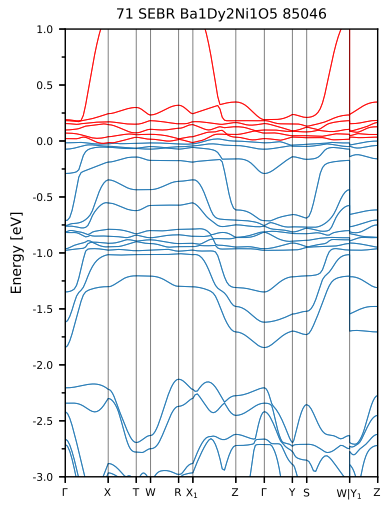

(ce) 71.1.83

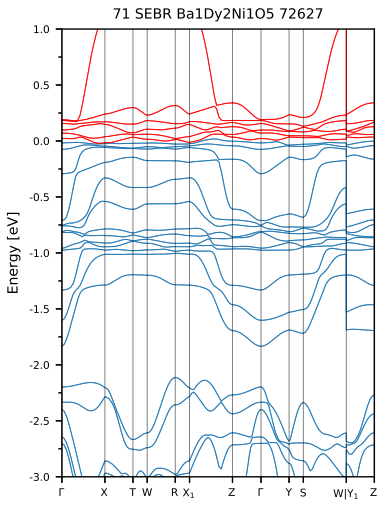

(cf) 71.1.84

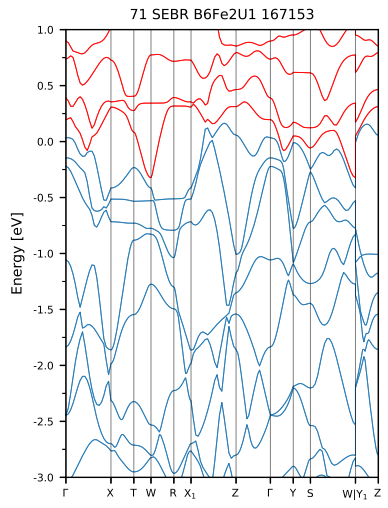

(cg) 71.1.85

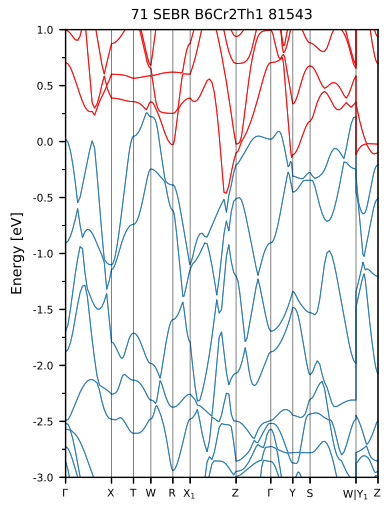

(ch) 71.1.86

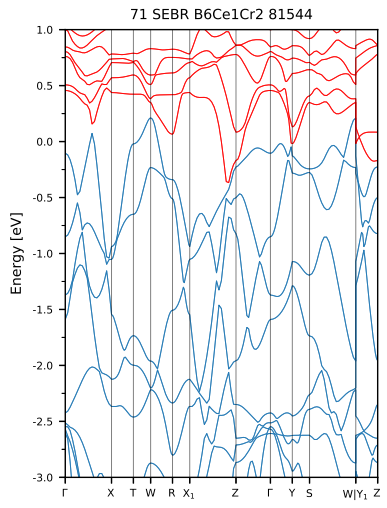

(ci) 71.1.87

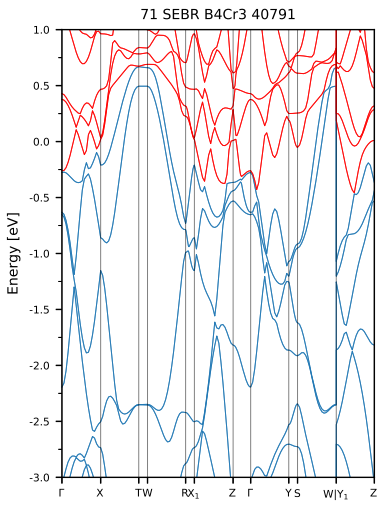

(cj) 71.1.88

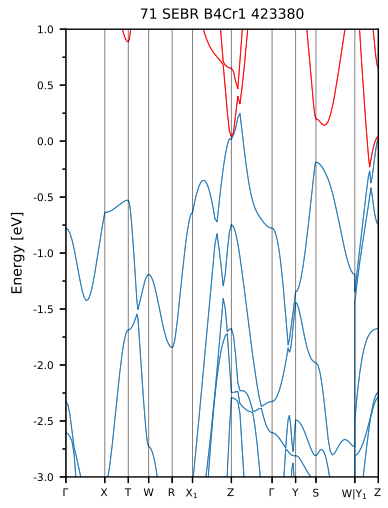

(ck) 71.1.89

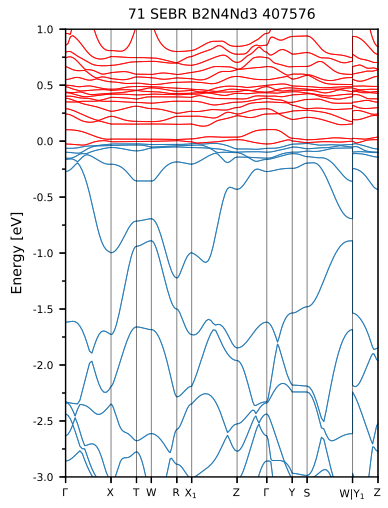

(cl) 71.1.90

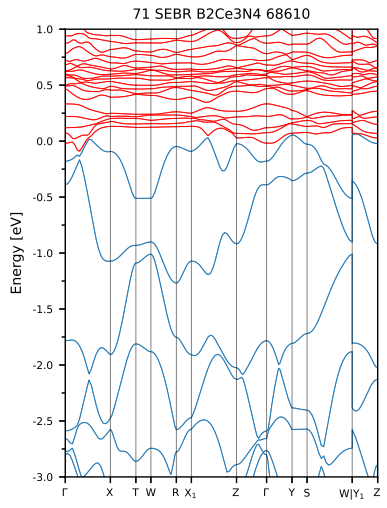

(cm) 71.1.91

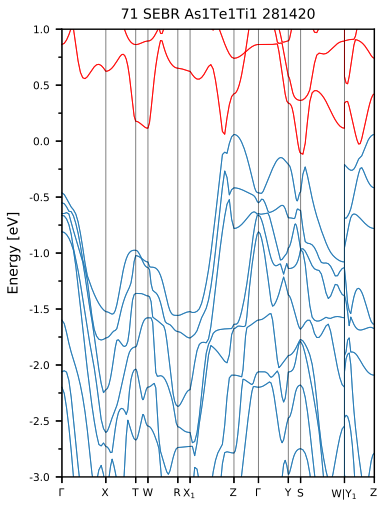

(cn) 71.1.92

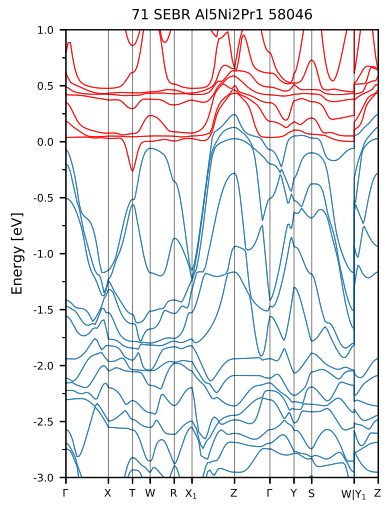

(co) 71.1.93

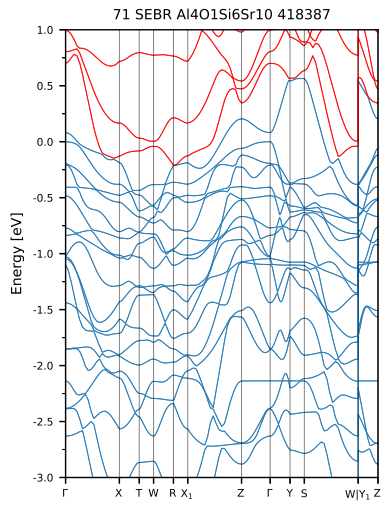

(cp) 71.1.94

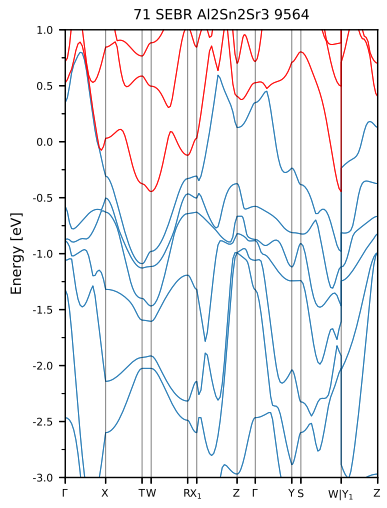

(cq) 71.1.95

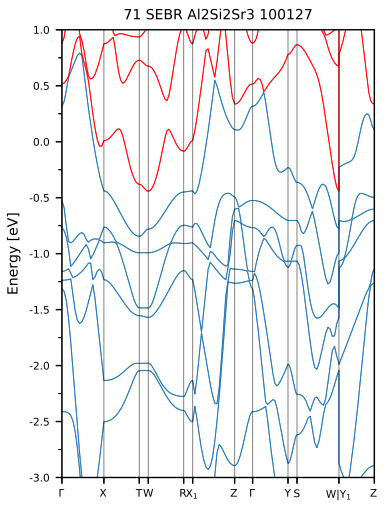

(cr) 71.1.96

Figure 192: Band structures for topological materials in SG 71, part 6/7, from 71.1.81 to 71.1.96. See the list of figures.

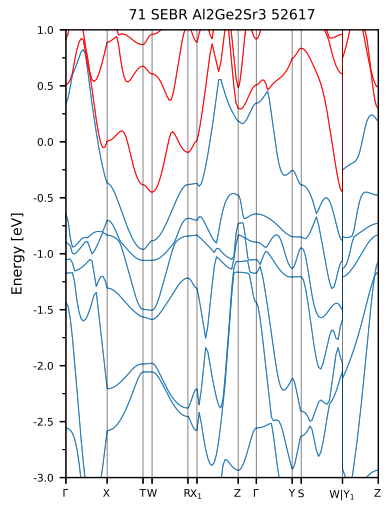

(cs) 71.1.97

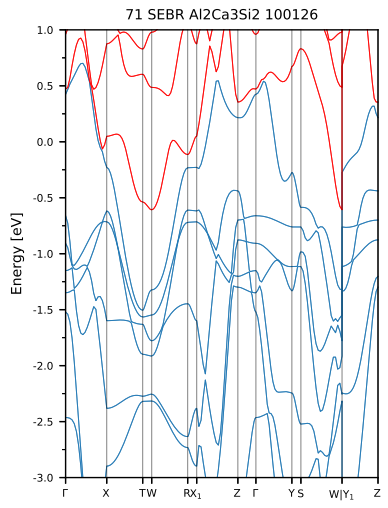

(ct) 71.1.98

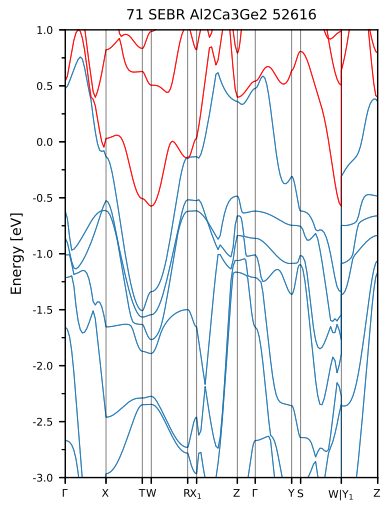

(cu) 71.1.99

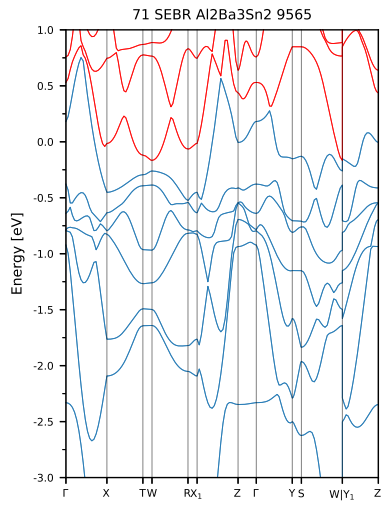

(cv) 71.1.100

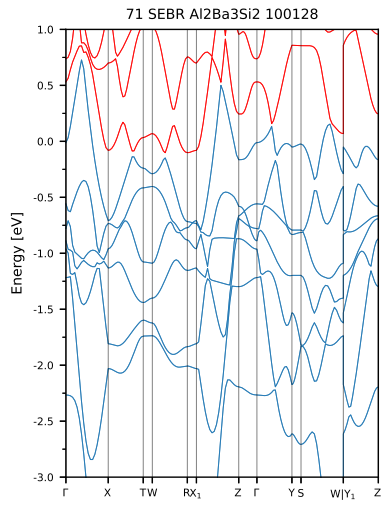

(cw) 71.1.101

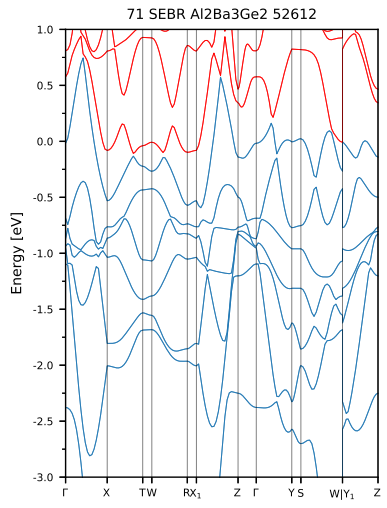

(cx) 71.1.102

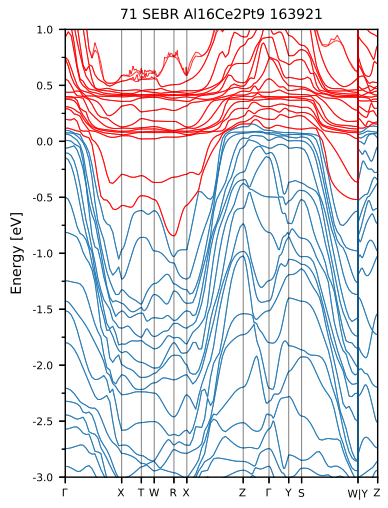

(cy) 71.1.103

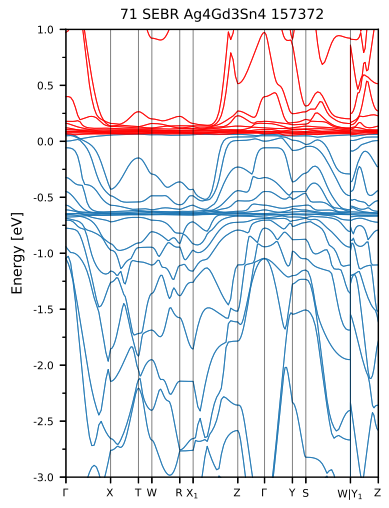

(cz) 71.1.104

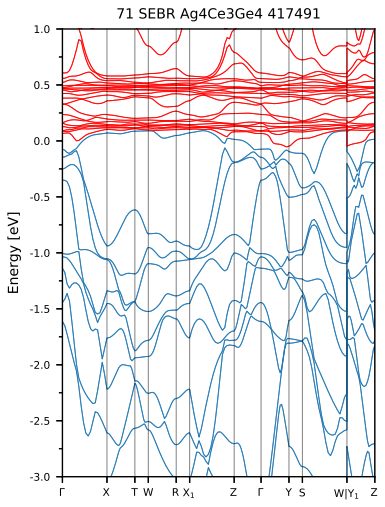

(da) 71.1.105

Figure 193: Band structures for topological materials in SG 71, part 7/7, from 71.1.97 to 71.1.105. See the list of figures.

## 0.0.34 SG 72

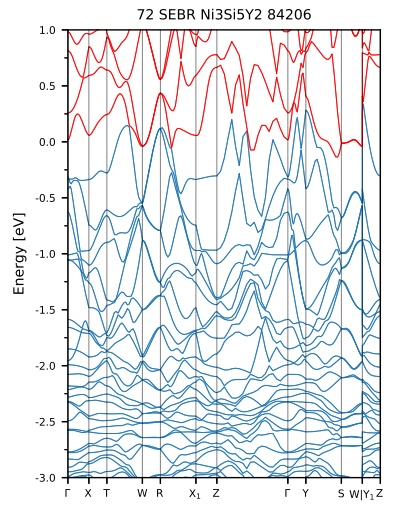

(a) 72.1.1

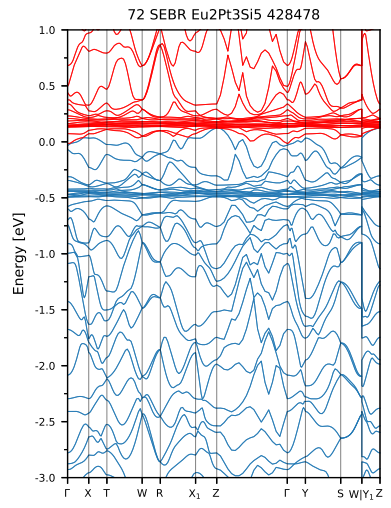

(b) 72.1.2

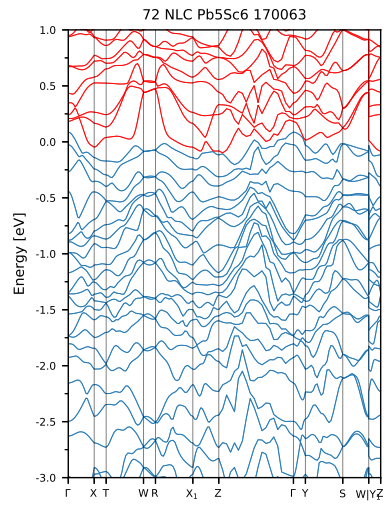

(c) 72.2.1

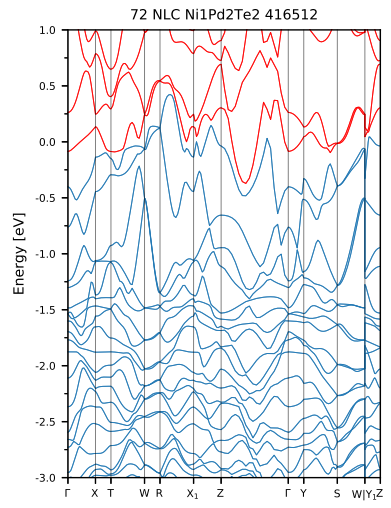

(d) 72.2.2

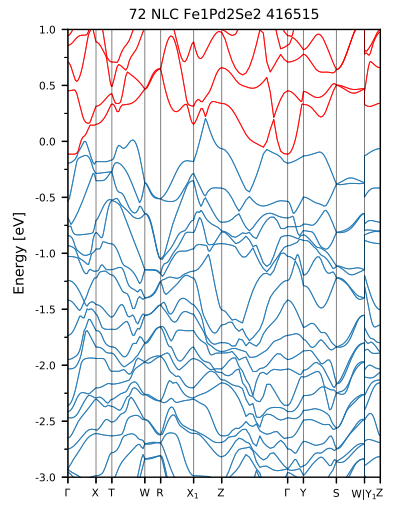

(e) 72.2.3

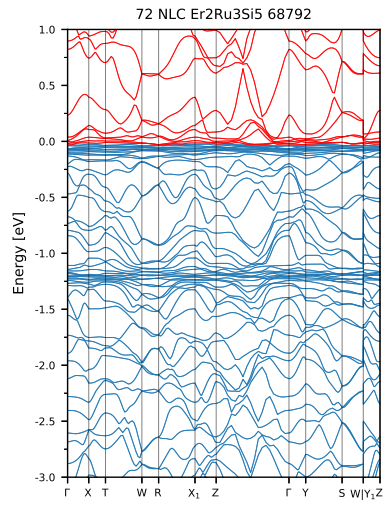

(f) 72.2.4

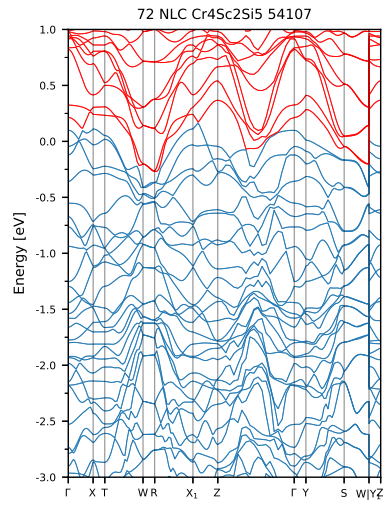

(g) 72.2.5

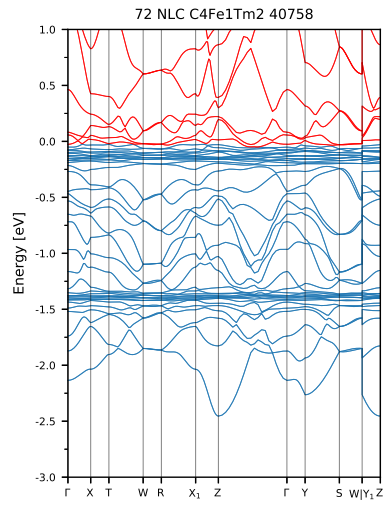

(h) 72.2.6

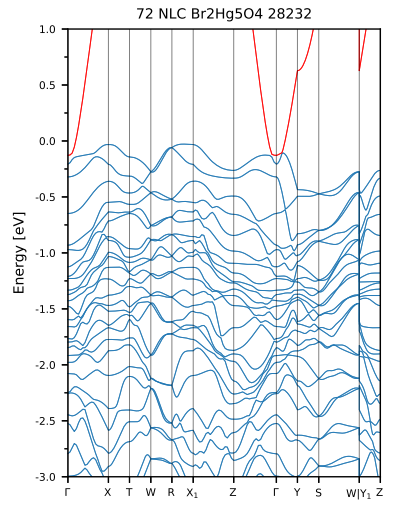

(i) 72.2.7

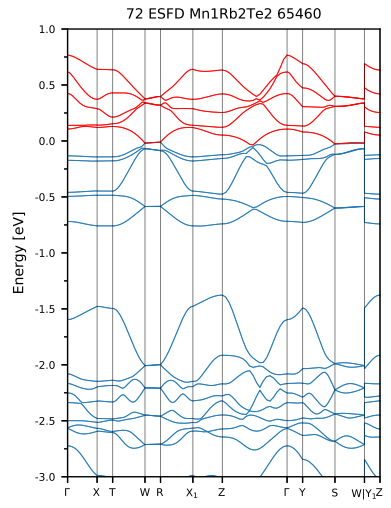

(j) 72.4.1

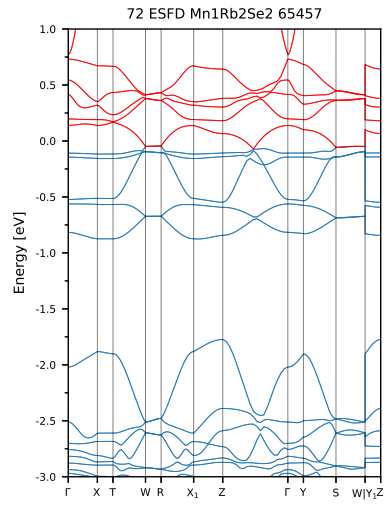

(k) 72.4.2

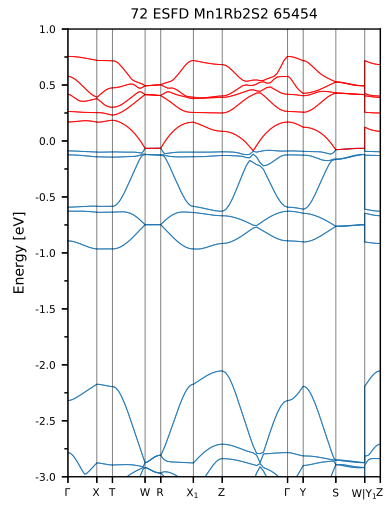

(l) 72.4.3

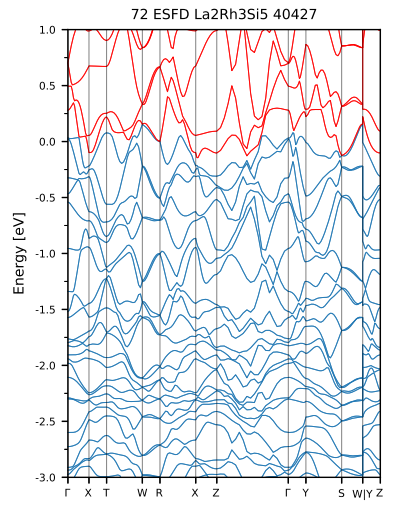

(m) 72.4.4

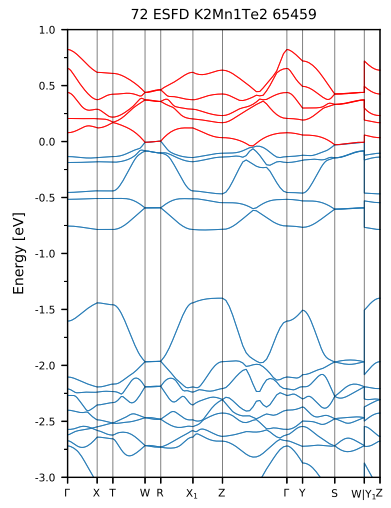

(n) 72.4.5

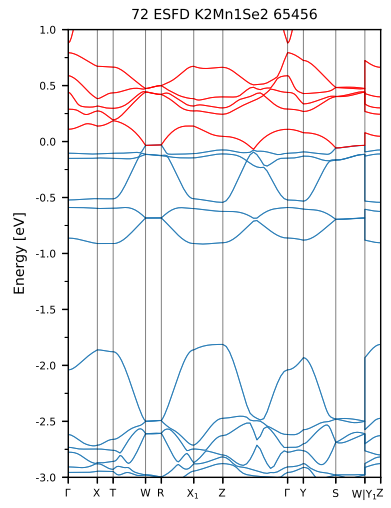

(o) 72.4.6

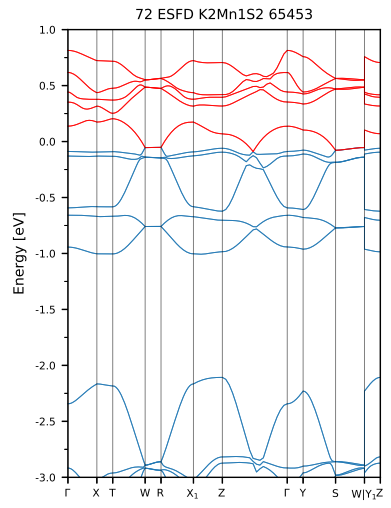

(p) 72.4.7

Figure 194: Band structures for topological materials in SG 72, part 1/3, from 72.1.1 to 72.4.7. See the list of figures.

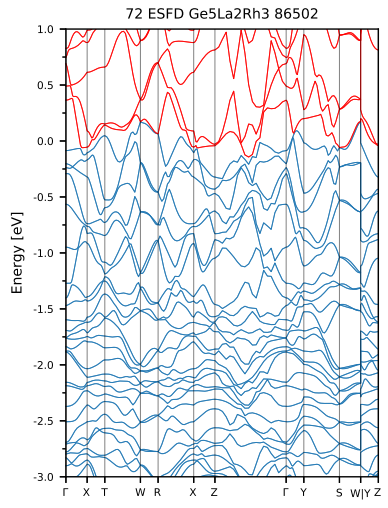

(q) 72.4.8

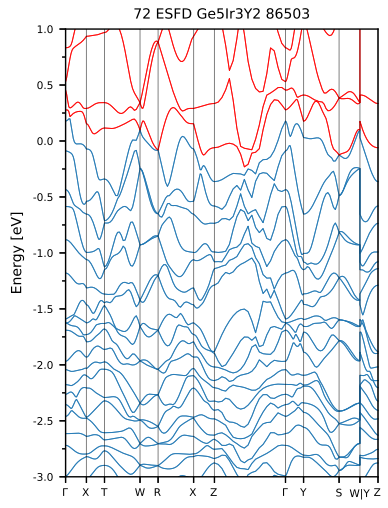

(r) 72.4.9

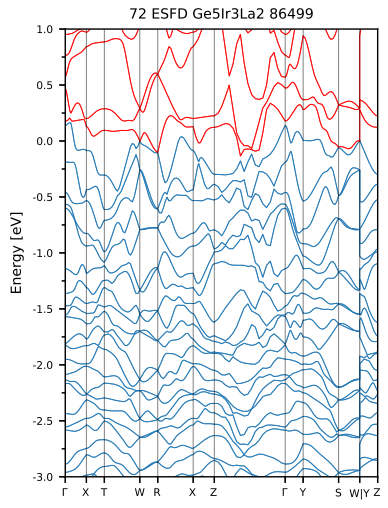

(s) 72.4.10

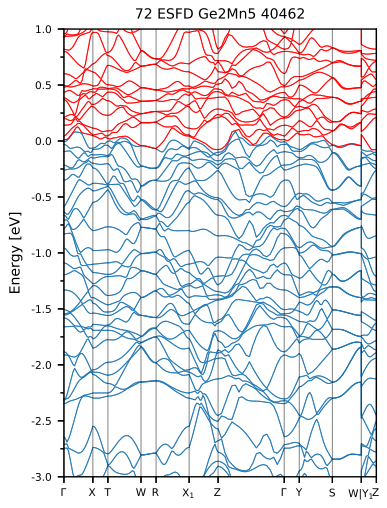

(t) 72.4.11

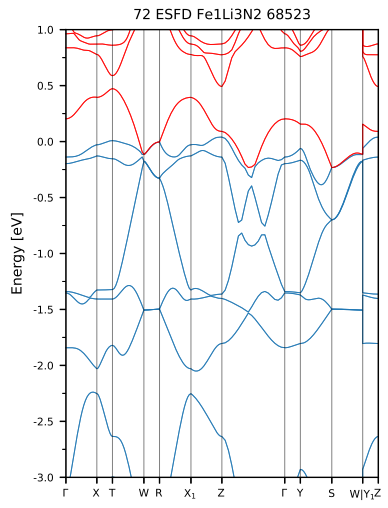

(u) 72.4.12

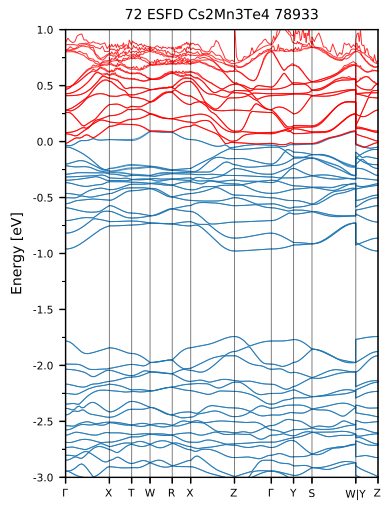

(v) 72.4.13

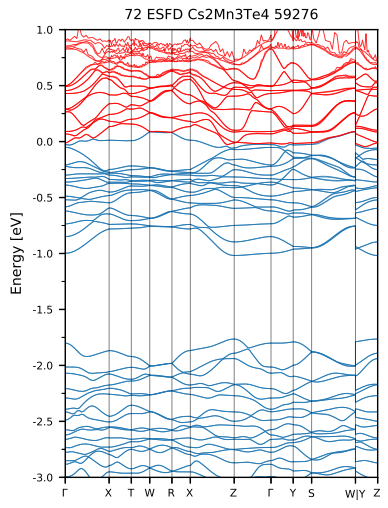

(w) 72.4.14

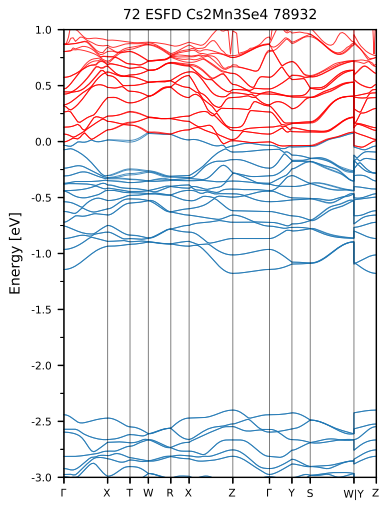

(x) 72.4.15

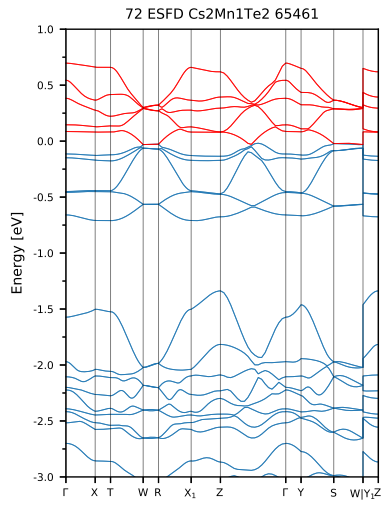

(y) 72.4.16

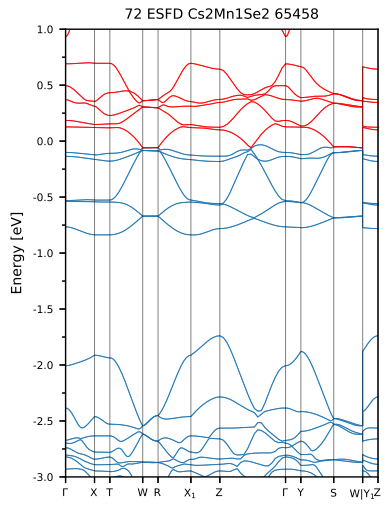

(z) 72.4.17

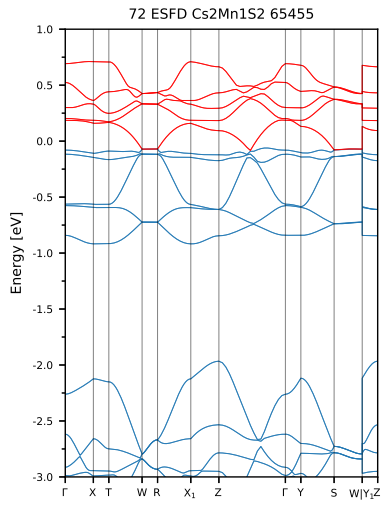

(aa) 72.4.18

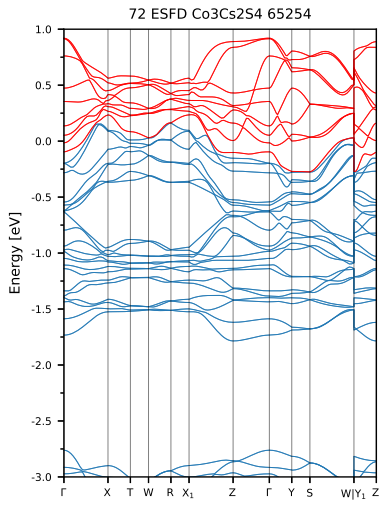

(ab) 72.4.19

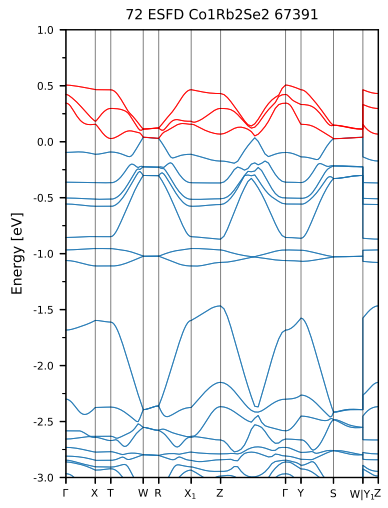

(ac) 72.4.20

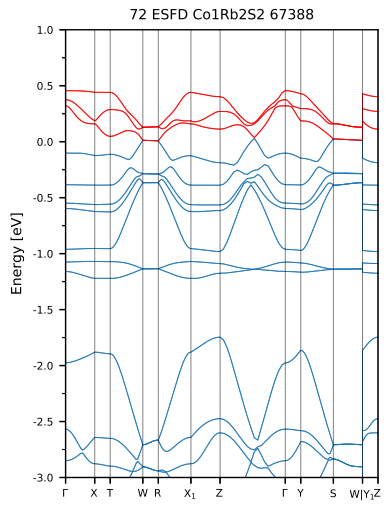

(ad) 72.4.21

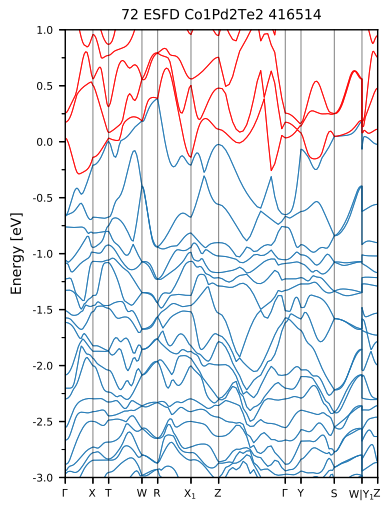

(ae) 72.4.22

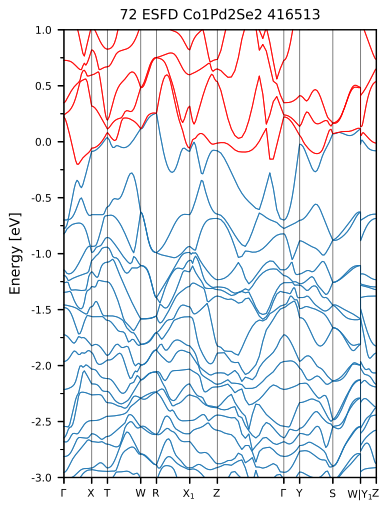

(af) 72.4.23

Figure 195: Band structures for topological materials in SG 72, part 2/3, from 72.4.8 to 72.4.23. See the list of figures.

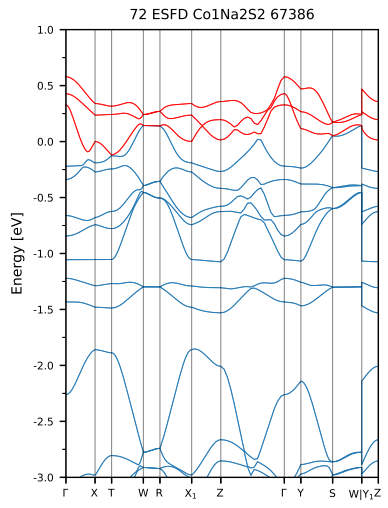

(ag) 72.4.24

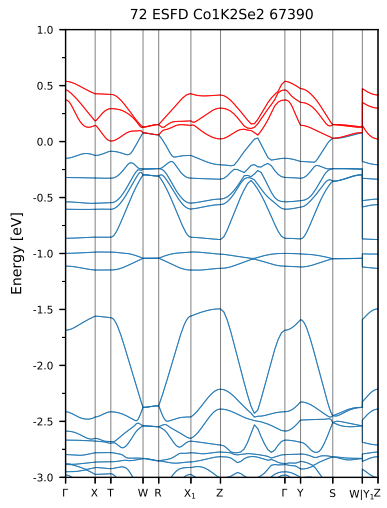

(ah) 72.4.25

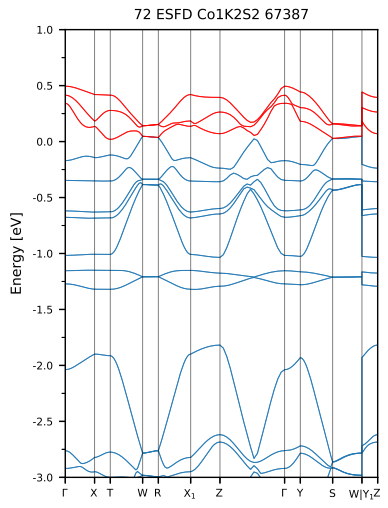

(ai) 72.4.26

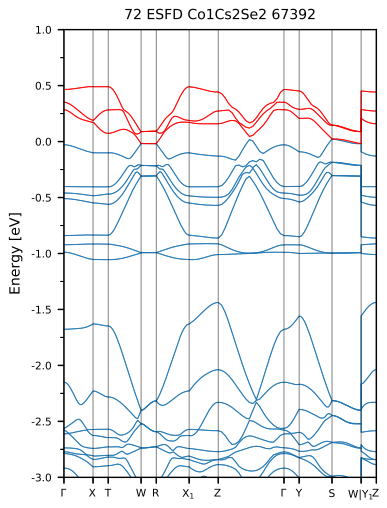

(aj) 72.4.27

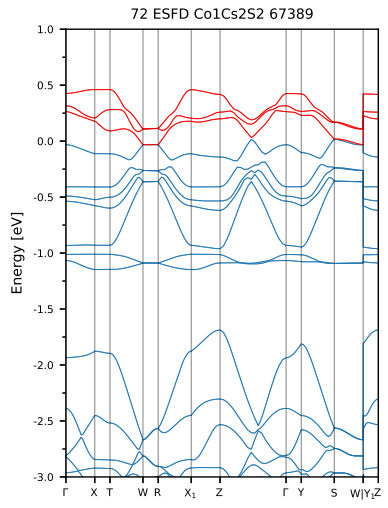

(ak) 72.4.28

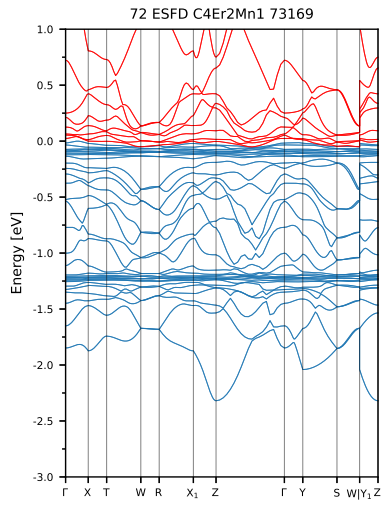

(al) 72.4.29

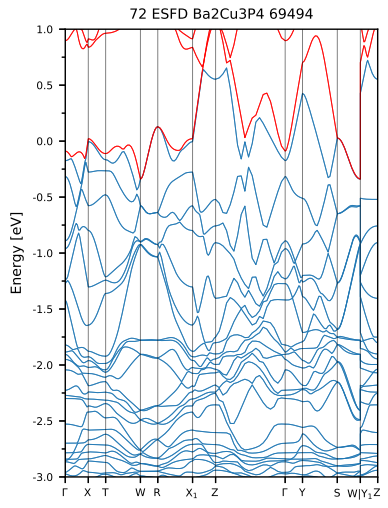

(am) 72.4.30

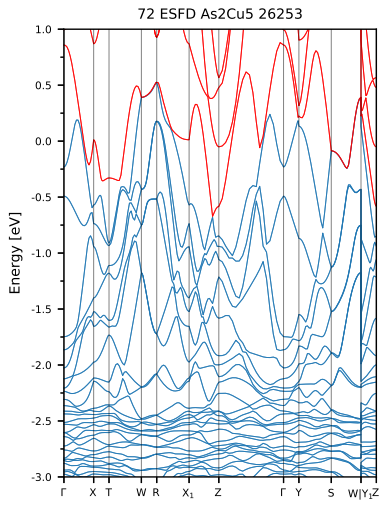

(an) 72.4.31

Figure 196: Band structures for topological materials in SG 72, part 3/3, from 72.4.24 to 72.4.31. See the list of figures.

## 0.0.35 SG 74

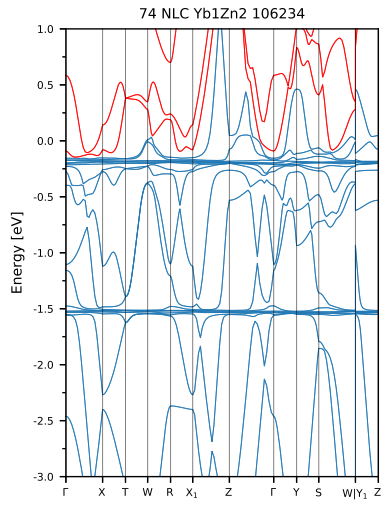

(a) 74.2.1

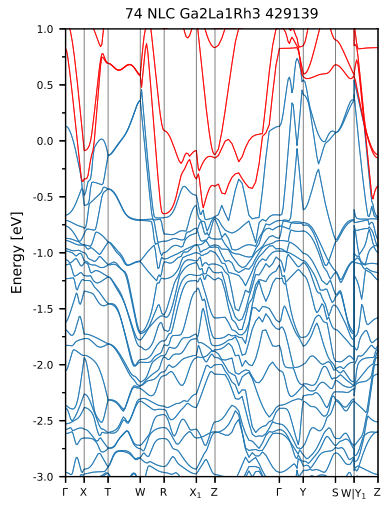

(b) 74.2.2

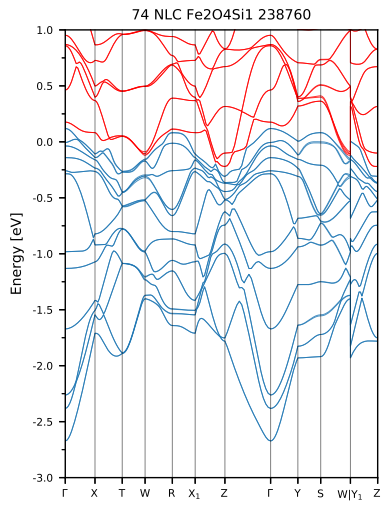

(c) 74.2.3

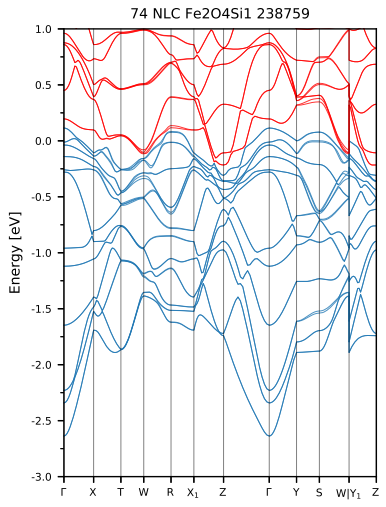

(d) 74.2.4

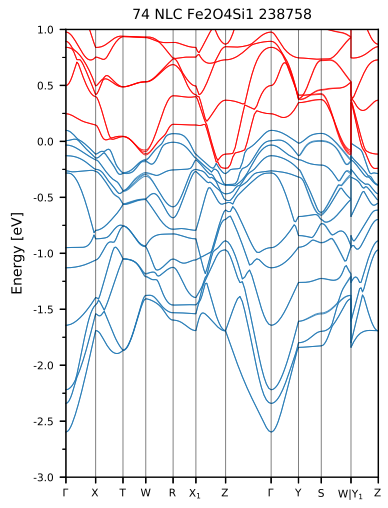

(e) 74.2.5

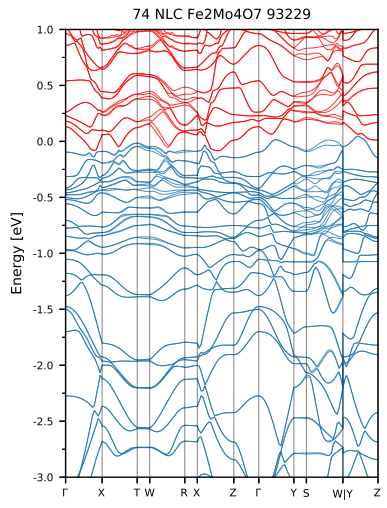

(f) 74.2.6

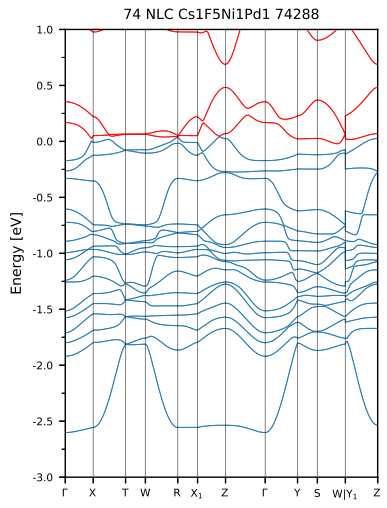

(g) 74.2.7

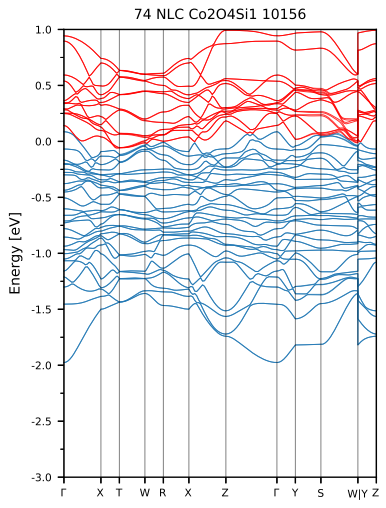

(h) 74.2.8

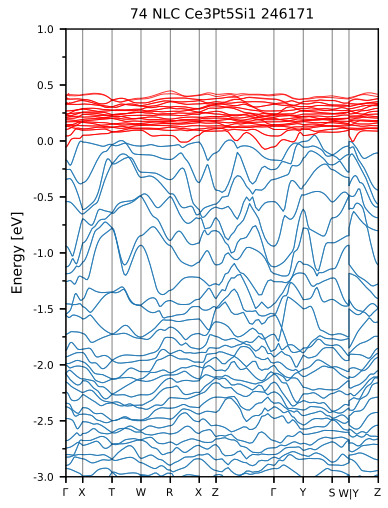

(i) 74.2.9

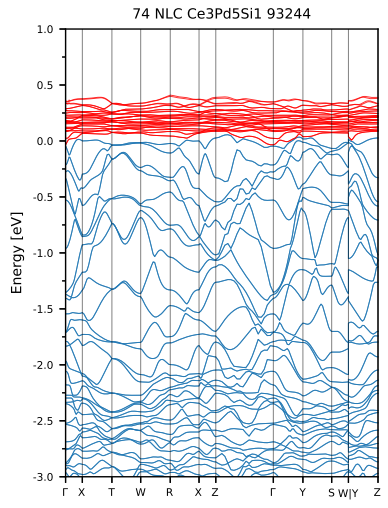

(j) 74.2.10

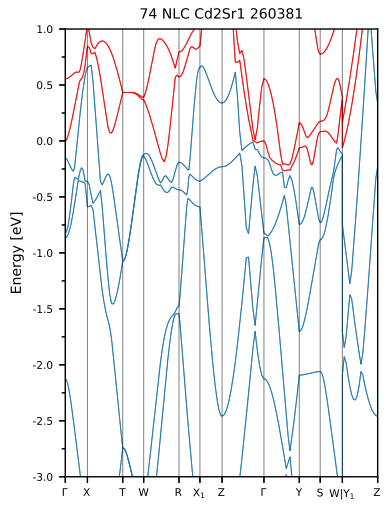

(k) 74.2.11

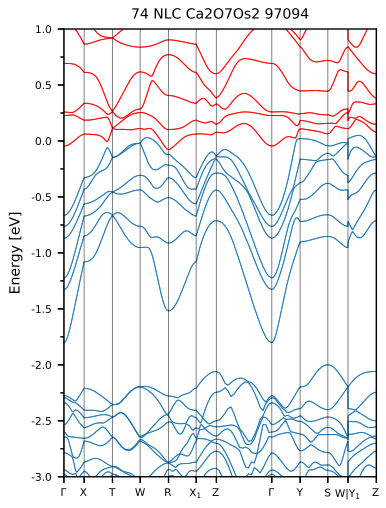

(l) 74.2.12

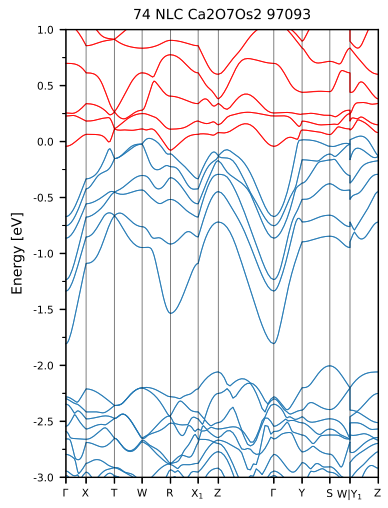

(m) 74.2.13

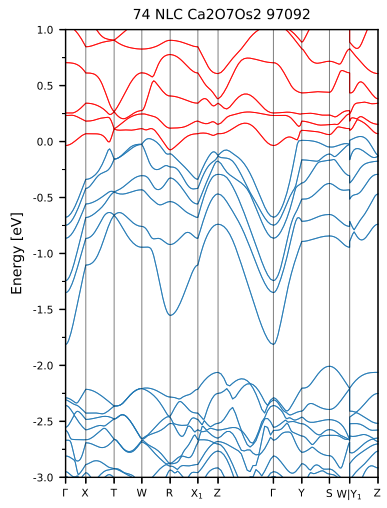

(n) 74.2.14

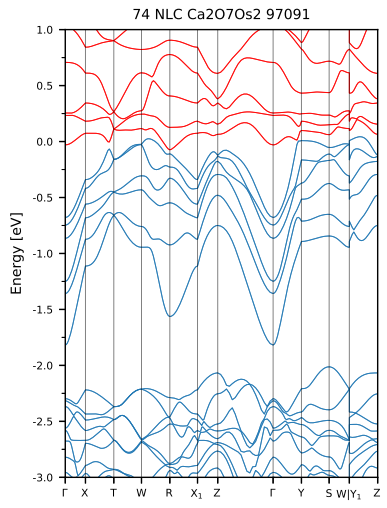

(o) 74.2.15

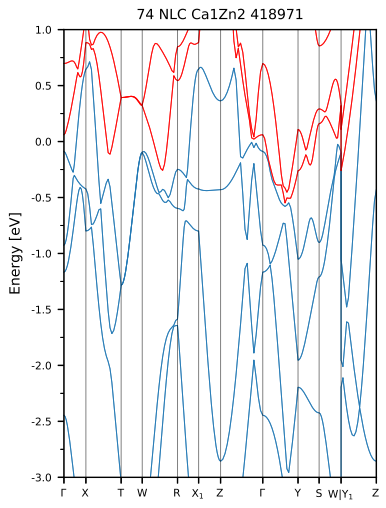

(p) 74.2.16

Figure 197: Band structures for topological materials in SG 74, part 1/3, from 74.2.1 to 74.2.16. See the list of figures.

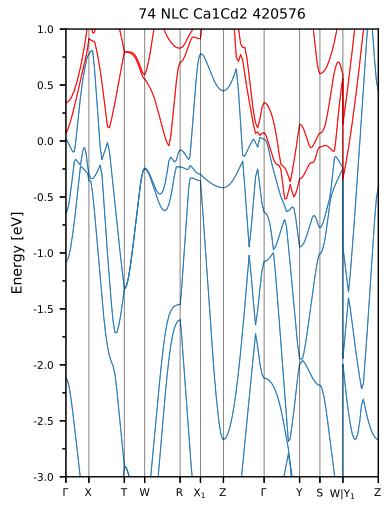

(q) 74.2.17

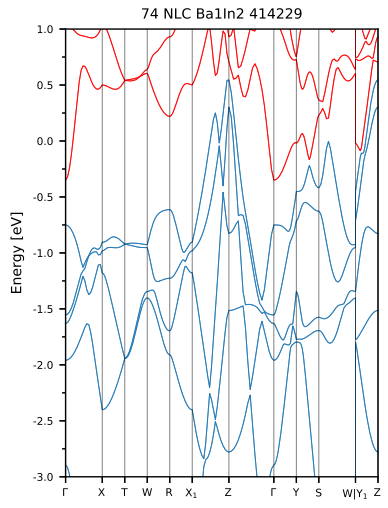

(r) 74.2.18

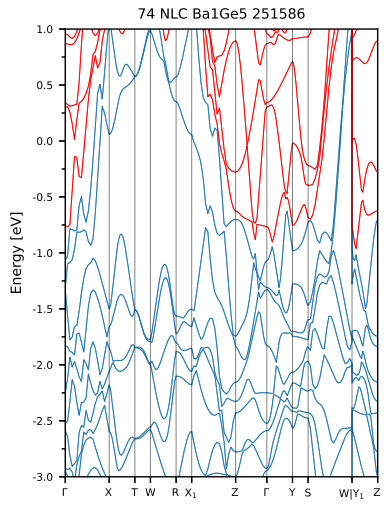

(s) 74.2.19

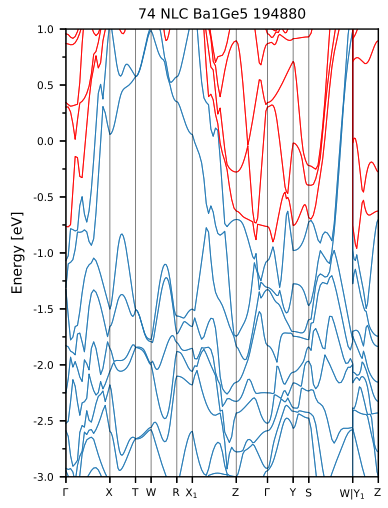

(t) 74.2.20

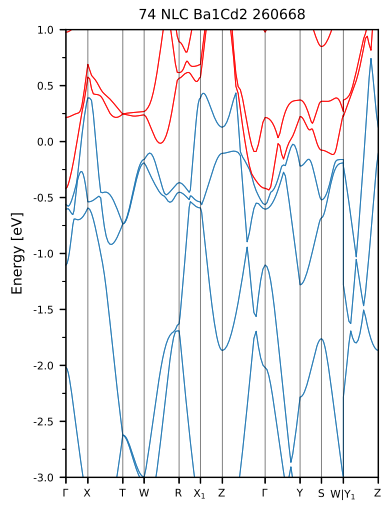

(u) 74.2.21

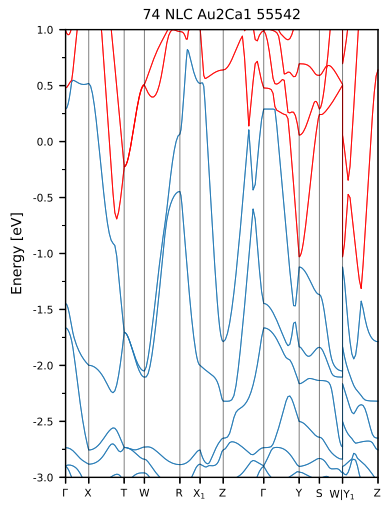

(v) 74.2.22

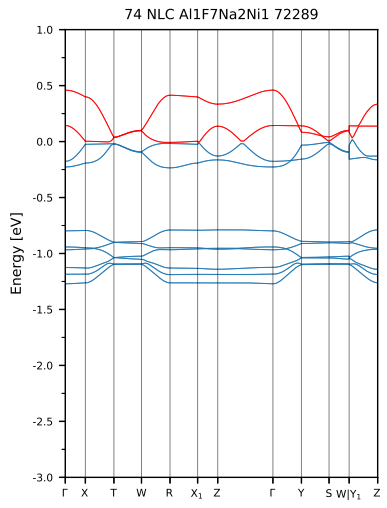

(w) 74.2.23

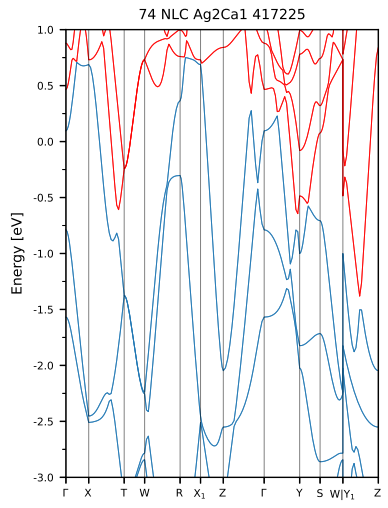

(x) 74.2.24

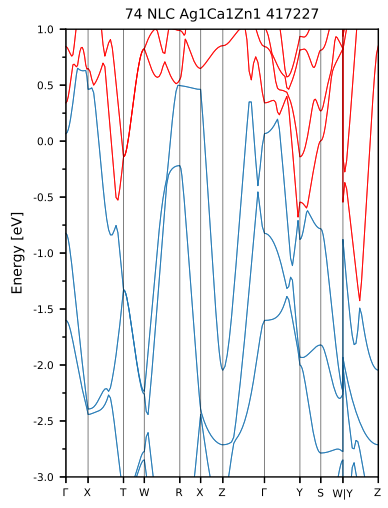

(y) 74.2.25

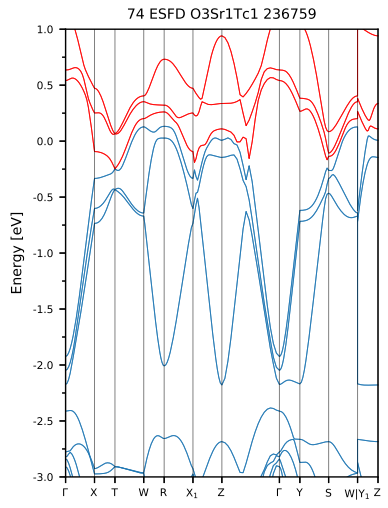

(z) 74.4.1

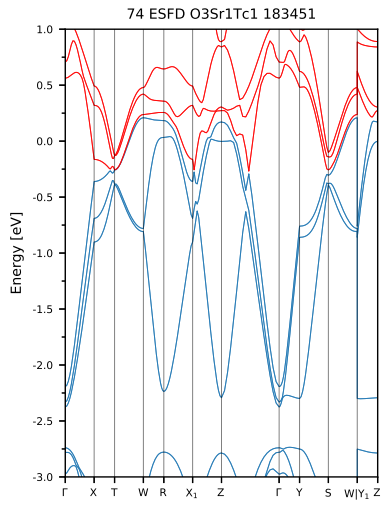

(aa) 74.4.2

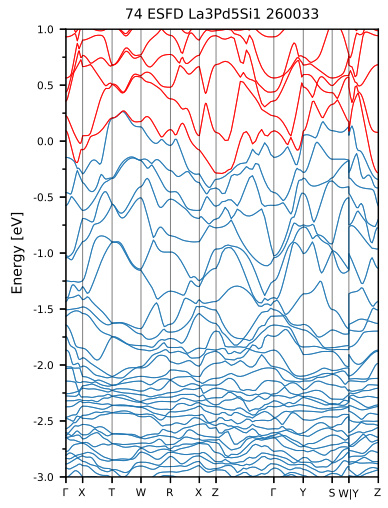

(ab) 74.4.3

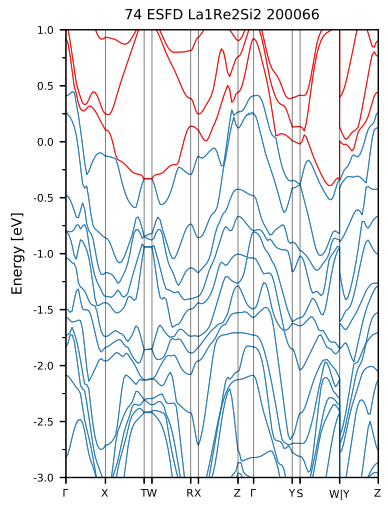

(ac) 74.4.4

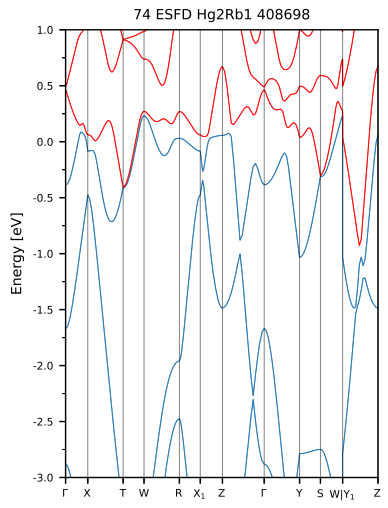

(ad) 74.4.5

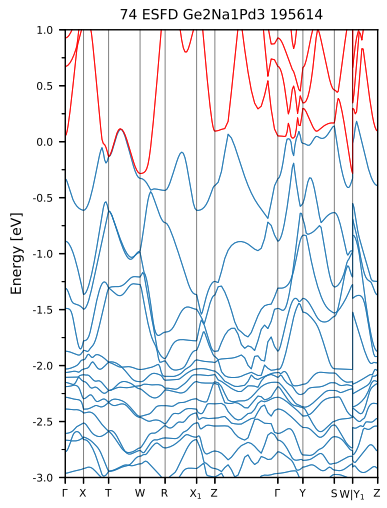

(ae) 74.4.6

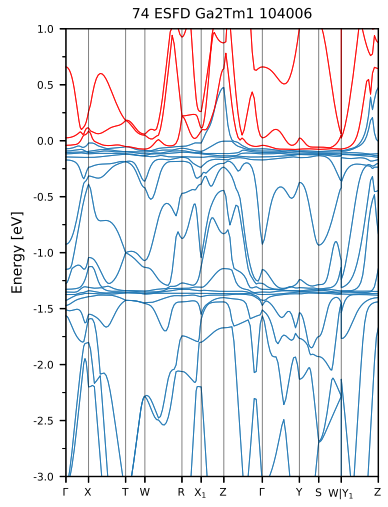

(af) 74.4.7

Figure 198: Band structures for topological materials in SG 74, part 2/3, from 74.2.17 to 74.4.7. See the list of figures.

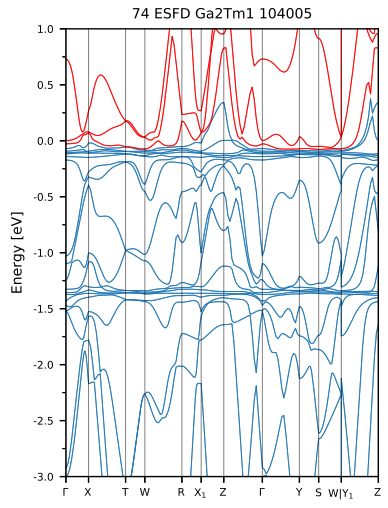

(ag) 74.4.8

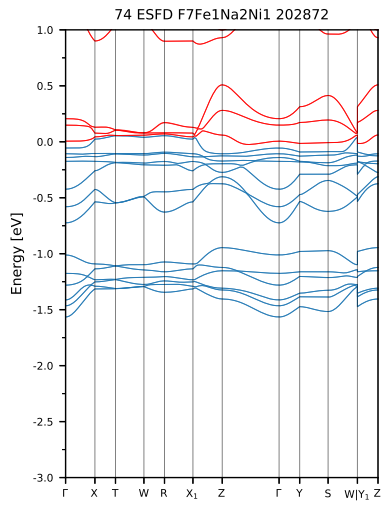

(ah) 74.4.9

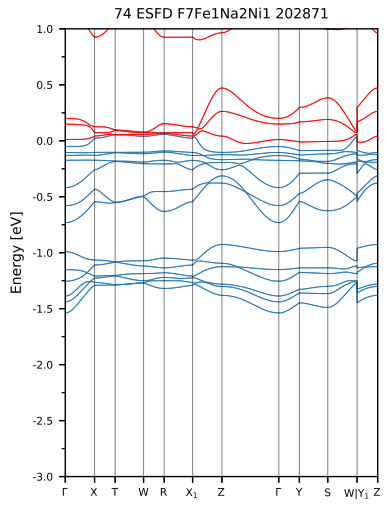

(ai) 74.4.10

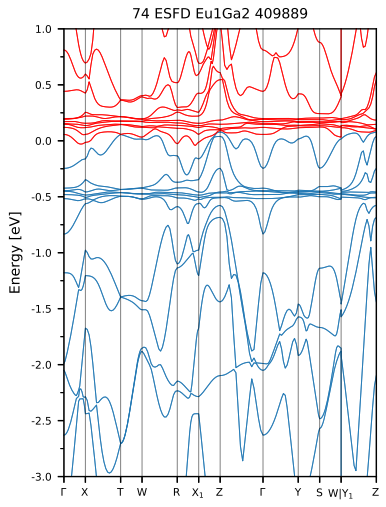

(aj) 74.4.11

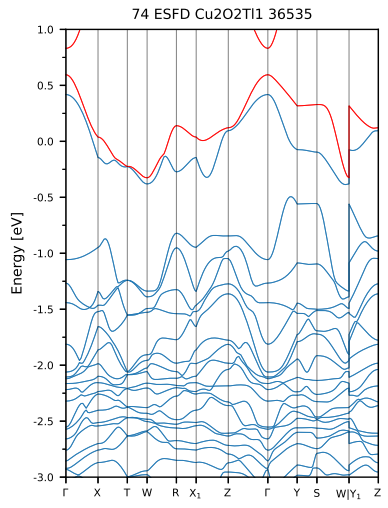

(ak) 74.4.12

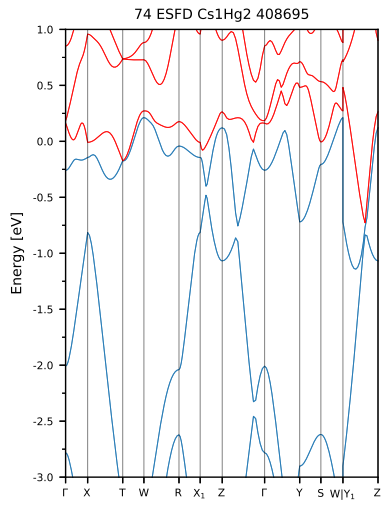

(al) 74.4.13

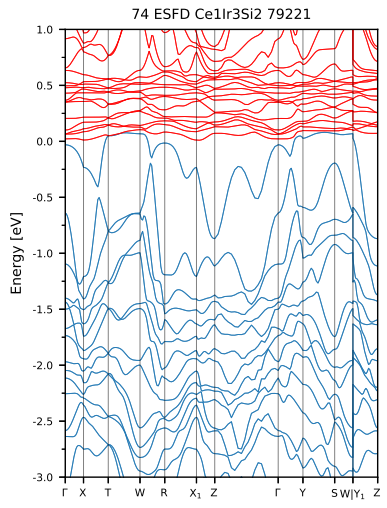

(am) 74.4.14

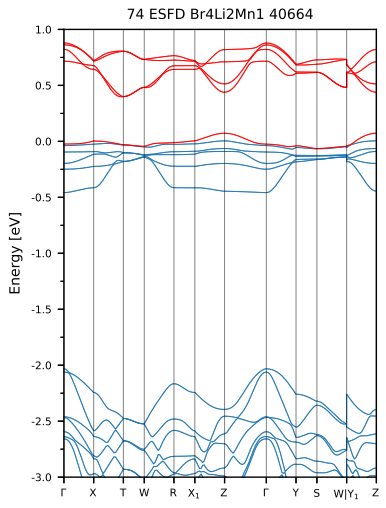

(an) 74.4.15

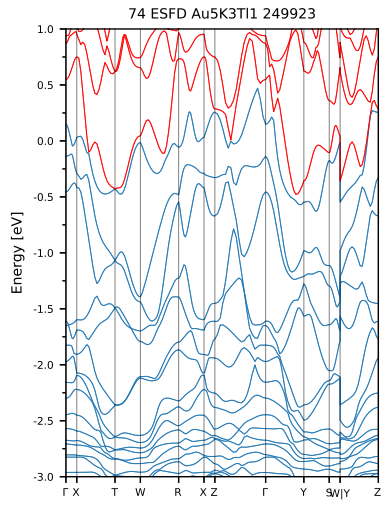

(ao) 74.4.16

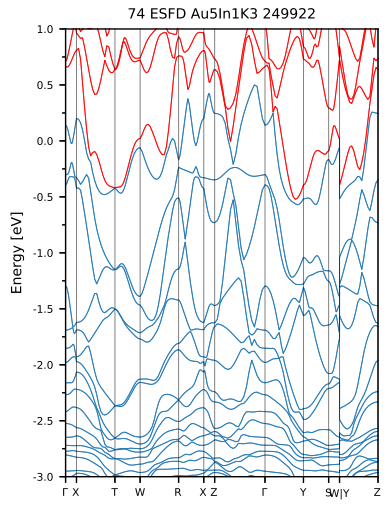

(ap) 74.4.17

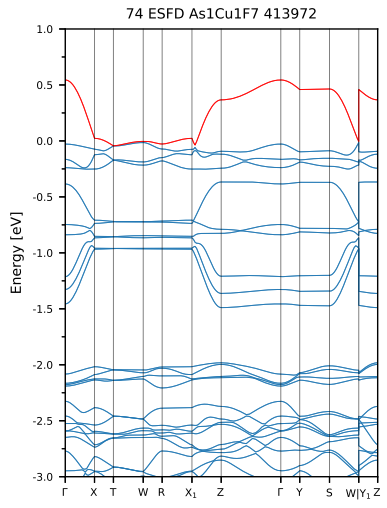

(aq) 74.4.18

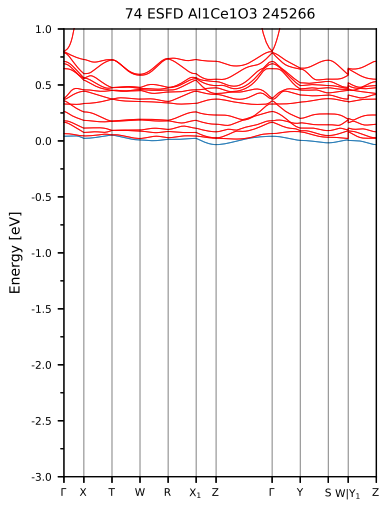

(ar) 74.4.19

Figure 199: Band structures for topological materials in SG 74, part 3/3, from 74.4.8 to 74.4.19. See the list of figures.

### 0.0.36 SG 78

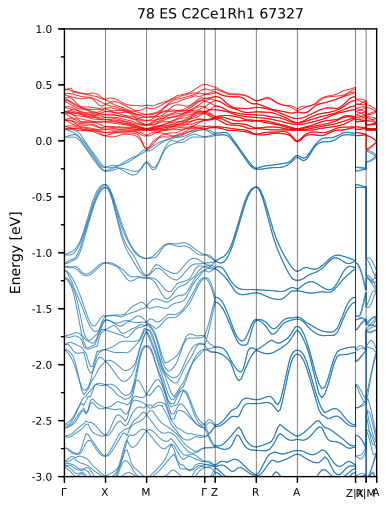

(a) 78.3.1

Figure 200: Band structures for topological materials in SG 78, part 1/1, from 78.3.1 to 78.3.1. See the list of figures.

### 0.0.37 SG 79

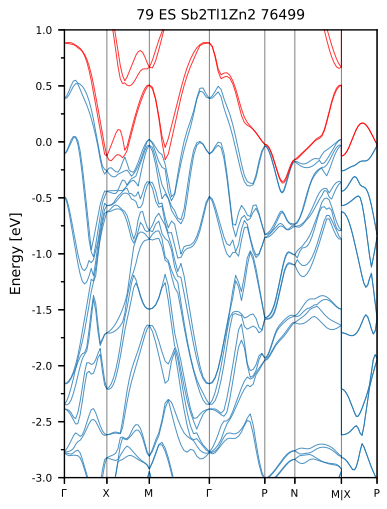

(a) 79.3.1

Figure 201: Band structures for topological materials in SG 79, part 1/1, from 79.3.1 to 79.3.1. See the list of figures.

## 0.0.38 SG 82

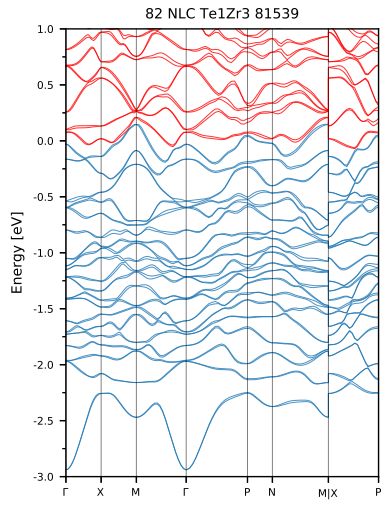

(a) 82.2.1

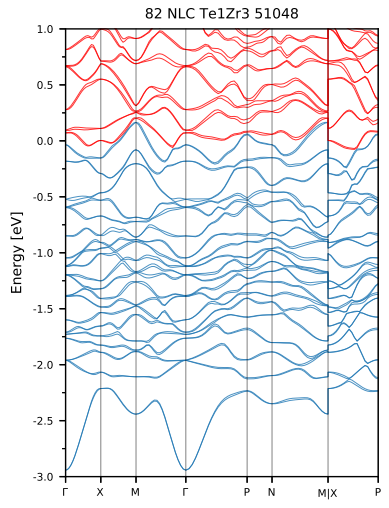

(b) 82.2.2

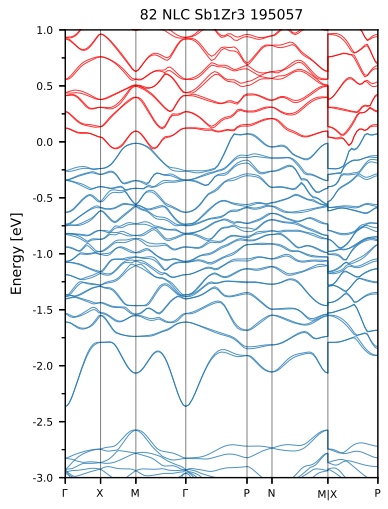

(c) 82.2.3

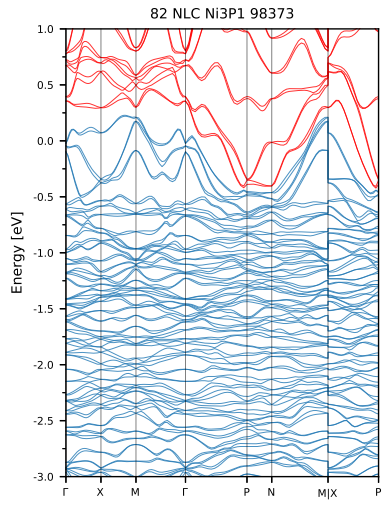

(d) 82.2.4

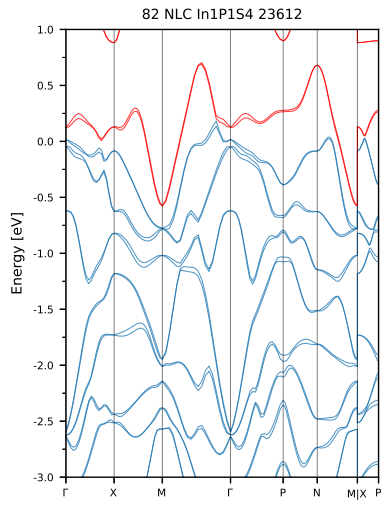

(e) 82.2.5

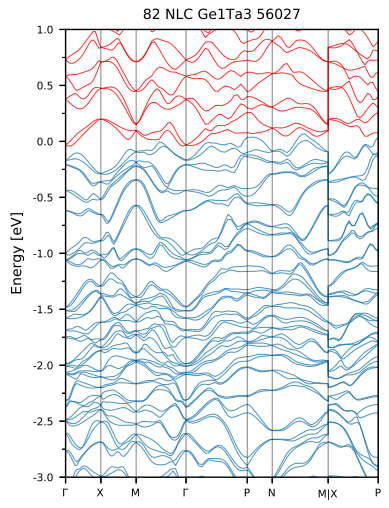

(f) 82.2.6

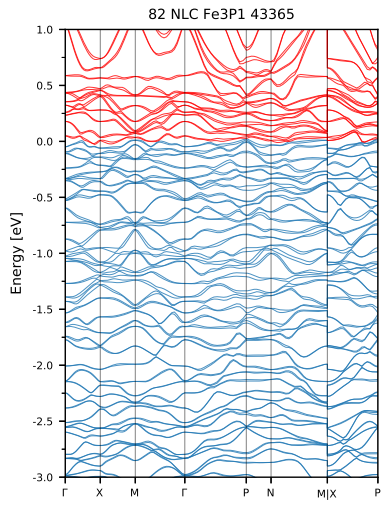

(g) 82.2.7

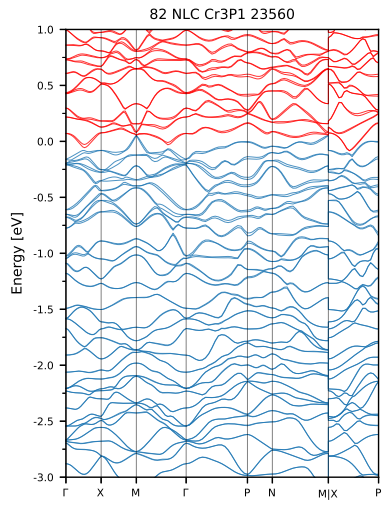

(h) 82.2.8

Figure 202: Band structures for topological materials in SG 82, part 1/1, from 82.2.1 to 82.2.8. See the list of figures.

### 0.0.39 SG 83

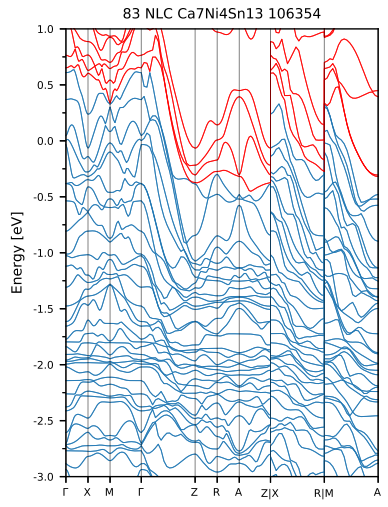

(a) 83.2.1

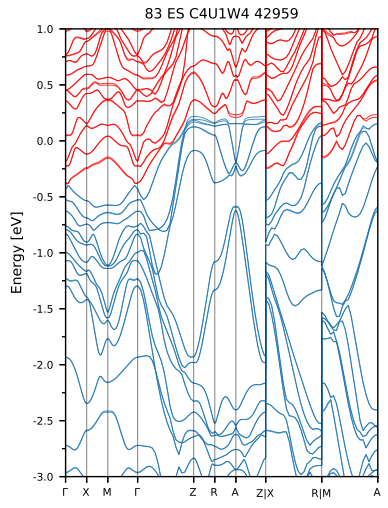

(b) 83.3.1

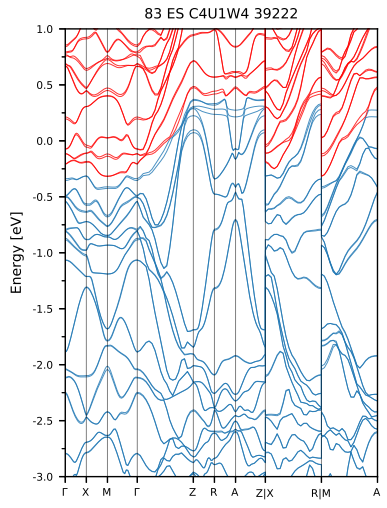

(c) 83.3.2

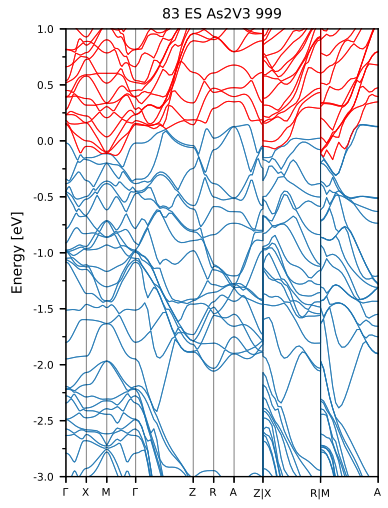

(d) 83.3.3

Figure 203: Band structures for topological materials in SG 83, part 1/1, from 83.2.1 to 83.3.3. See the list of figures.

### 0.0.40 SG 85

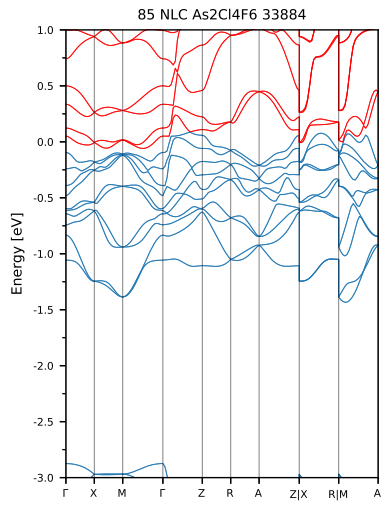

(a) 85.2.1

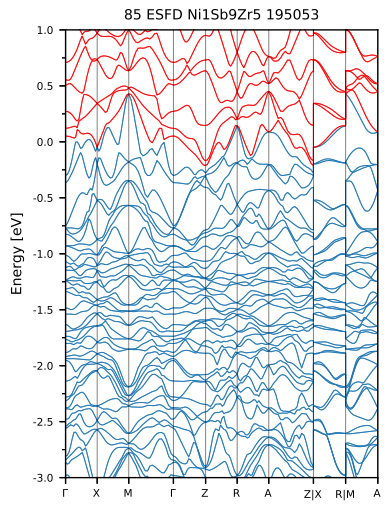

(b) 85.4.1

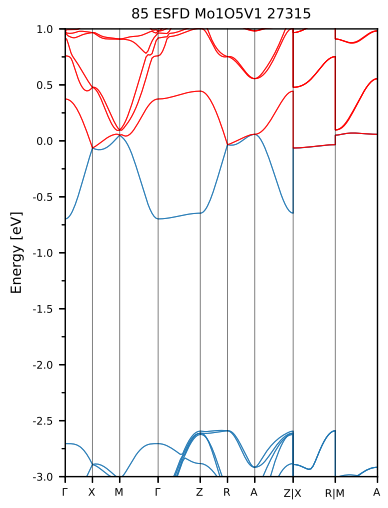

(c) 85.4.2

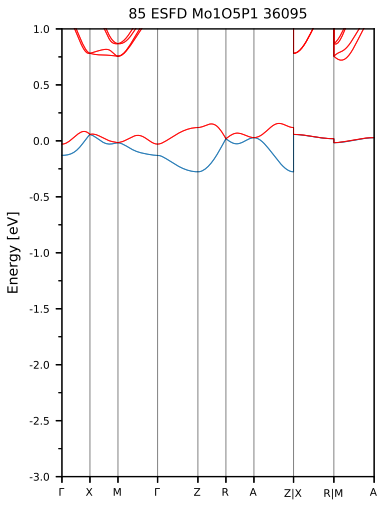

(d) 85.4.3

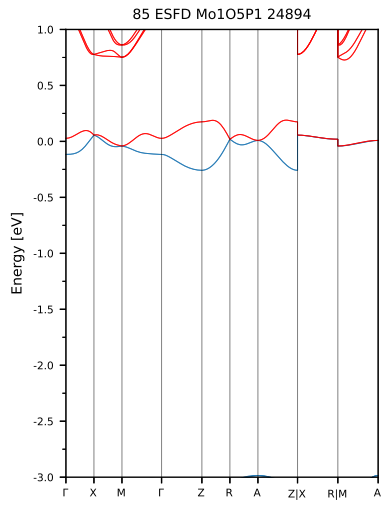

(e) 85.4.4

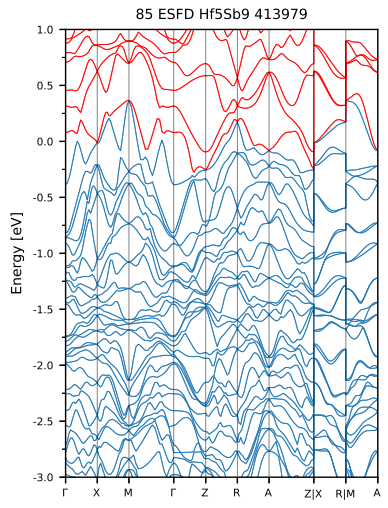

(f) 85.4.5

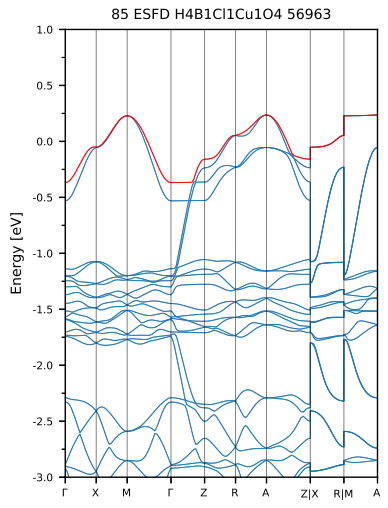

(g) 85.4.6

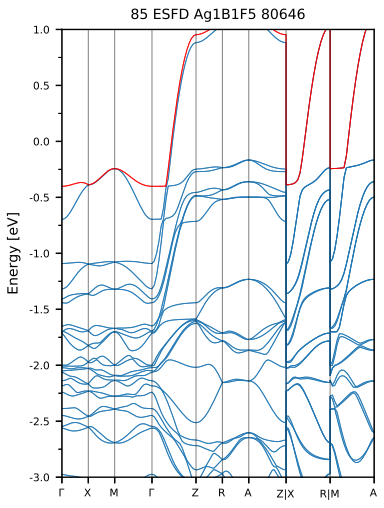

(h) 85.4.7

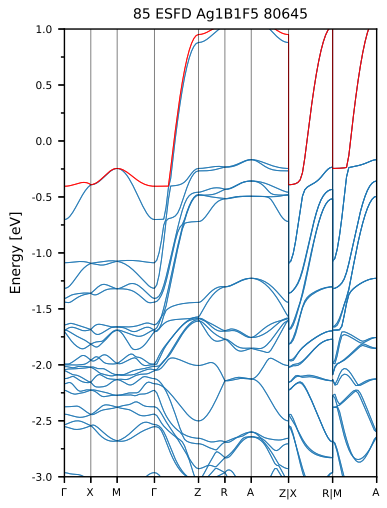

(i) 85.4.8

Figure 204: Band structures for topological materials in SG 85, part 1/1, from 85.2.1 to 85.4.8. See the list of figures.

### 0.0.41 SG 86

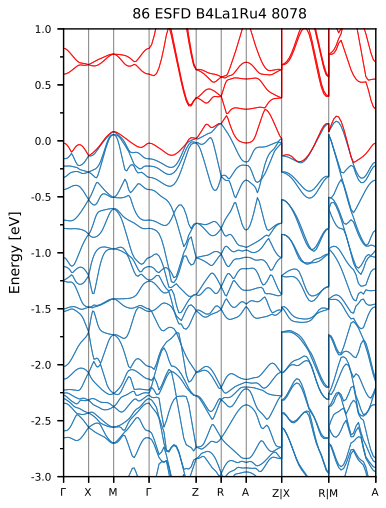

(a) 86.4.1

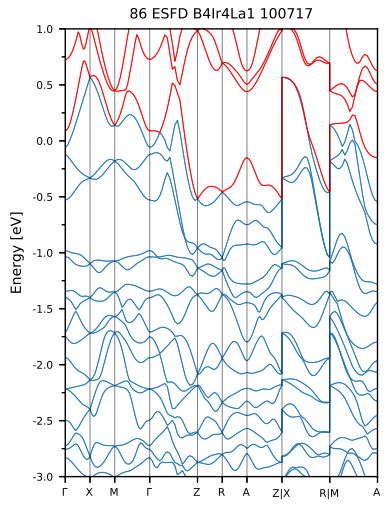

(b) 86.4.2

Figure 205: Band structures for topological materials in SG 86, part 1/1, from 86.4.1 to 86.4.2. See the list of figures.

### 0.0.42 SG 87

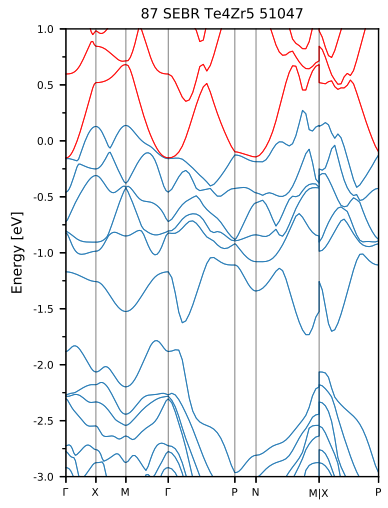

(a) 87.1.1

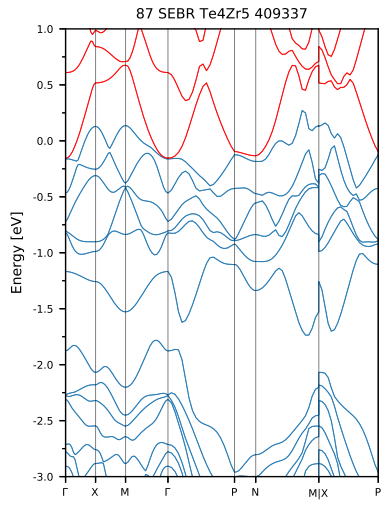

(b) 87.1.2

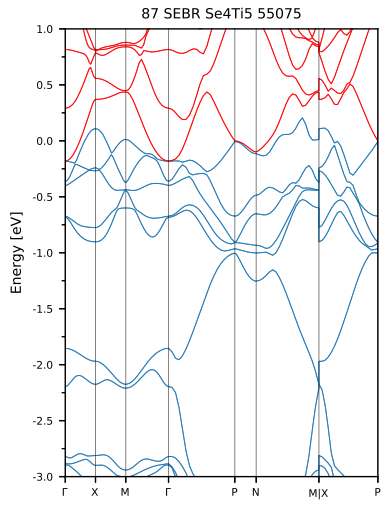

(c) 87.1.3

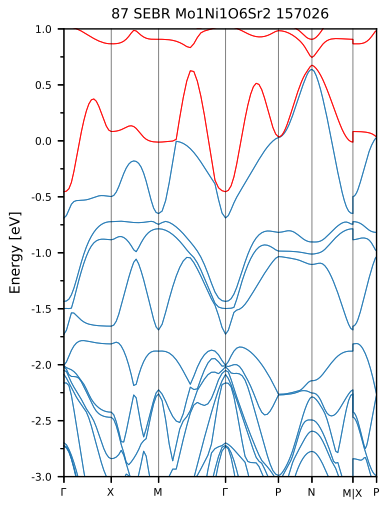

(d) 87.1.4

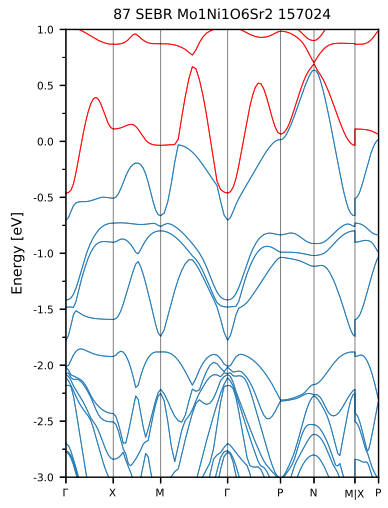

(e) 87.1.5

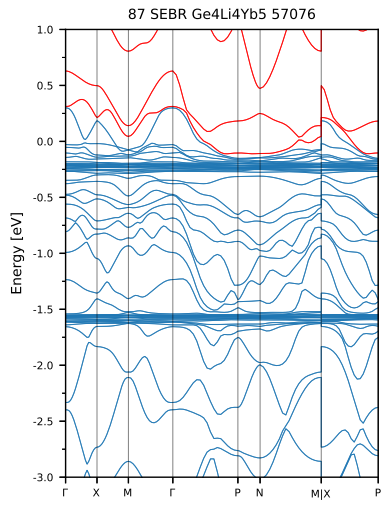

(f) 87.1.6

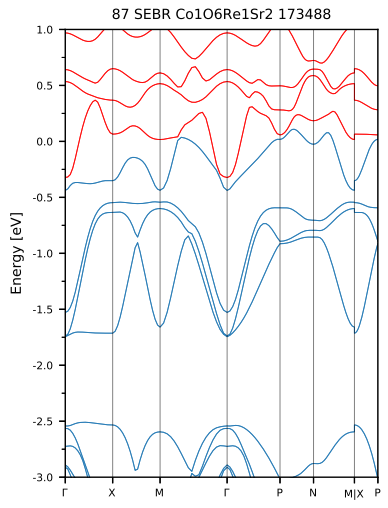

(g) 87.1.7

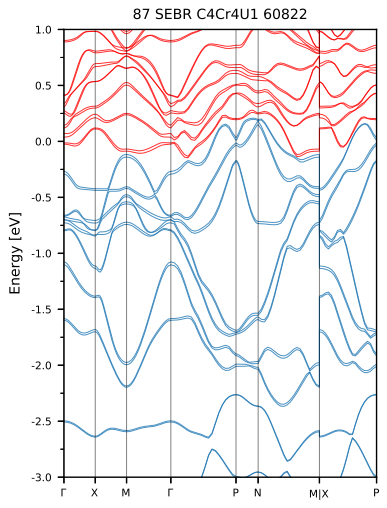

(h) 87.1.8

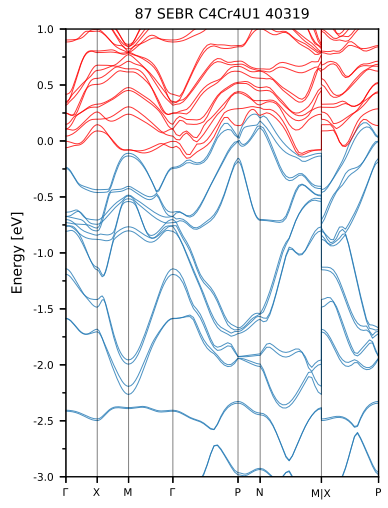

(i) 87.1.9

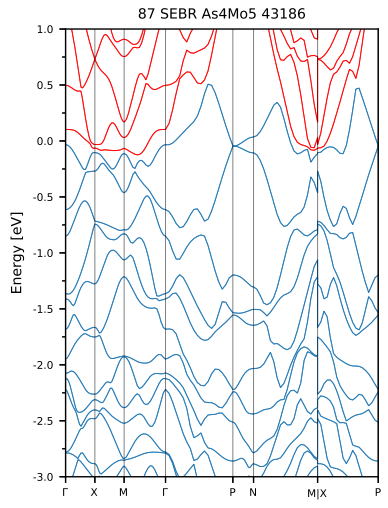

(j) 87.1.10

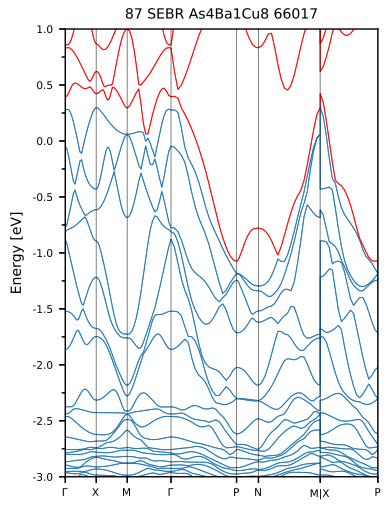

(k) 87.1.11

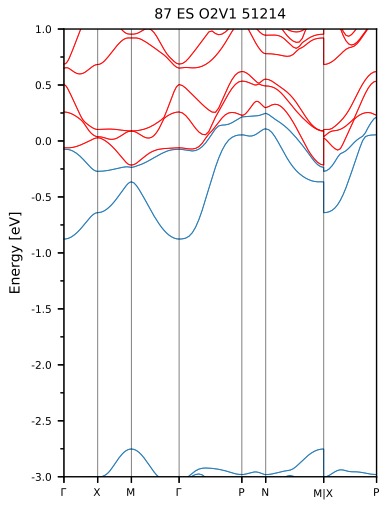

(l) 87.3.1

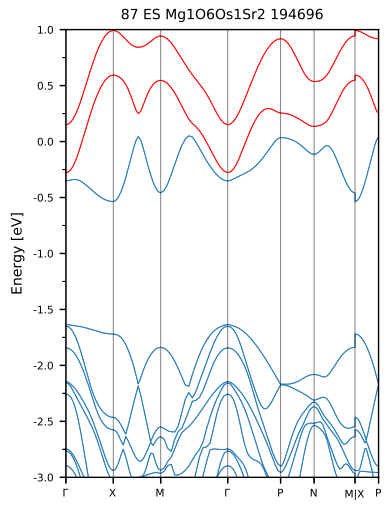

(m) 87.3.2

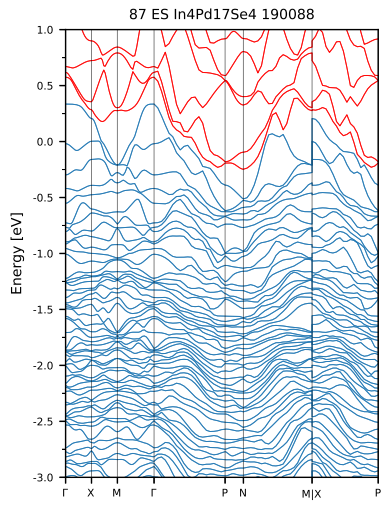

(n) 87.3.3

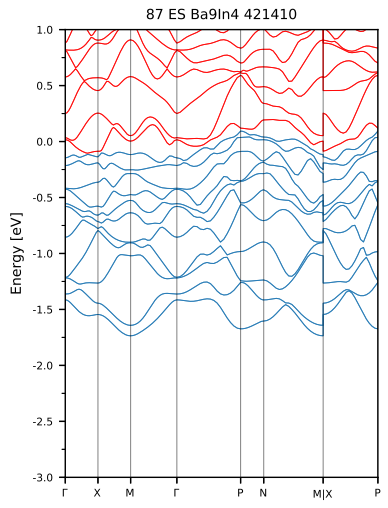

(o) 87.3.4

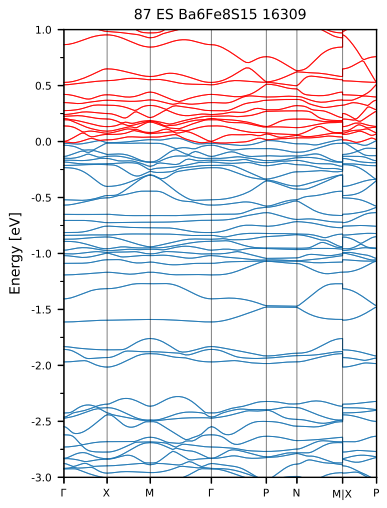

(p) 87.3.5

Figure 206: Band structures for topological materials in SG 87, part 1/2, from 87.1.1 to 87.3.5. See the list of figures.

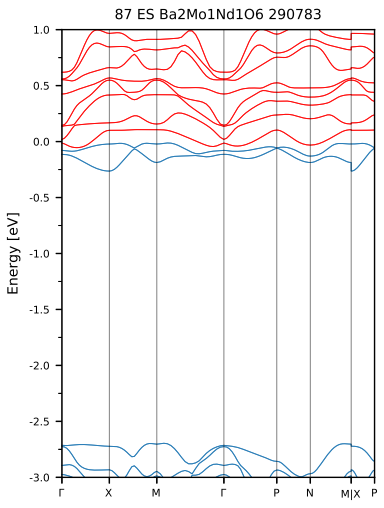

(q) 87.3.6

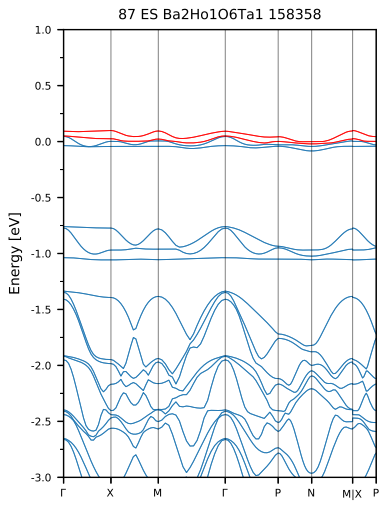

(r) 87.3.7

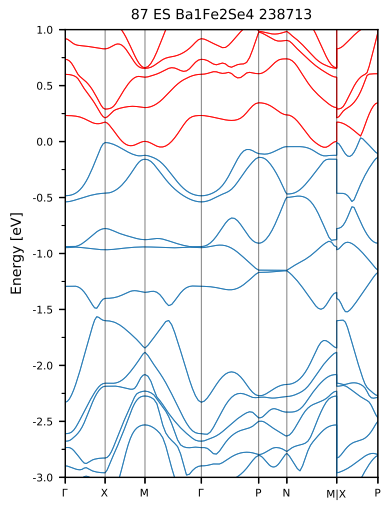

(s) 87.3.8

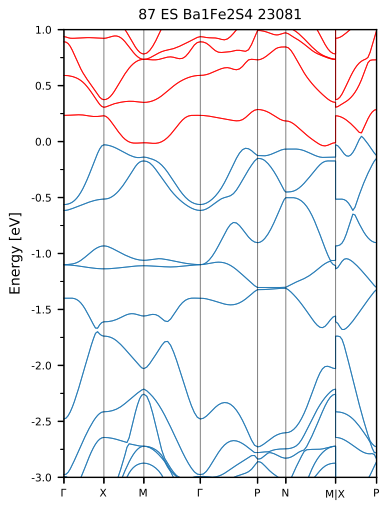

(t) 87.3.9

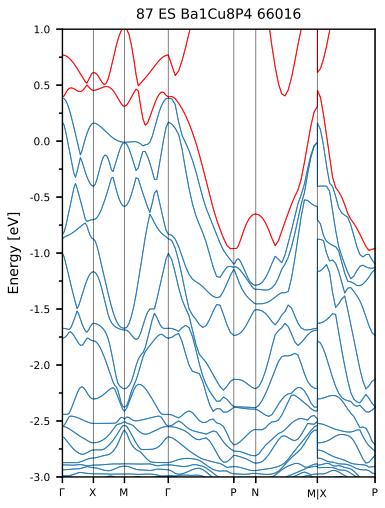

(u) 87.3.10

Figure 207: Band structures for topological materials in SG 87, part 2/2, from 87.3.6 to 87.3.10. See the list of figures.

### 0.0.43 SG 92

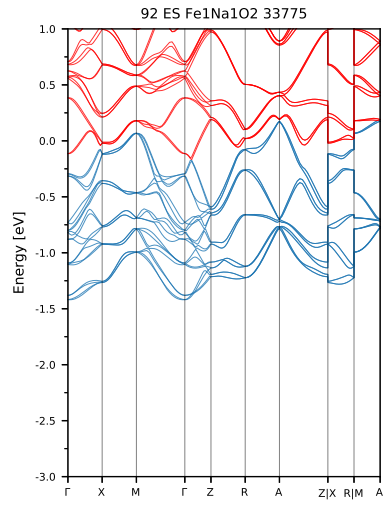

(a) 92.3.1

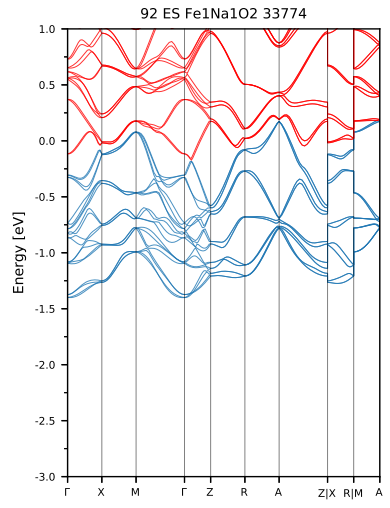

(b) 92.3.2

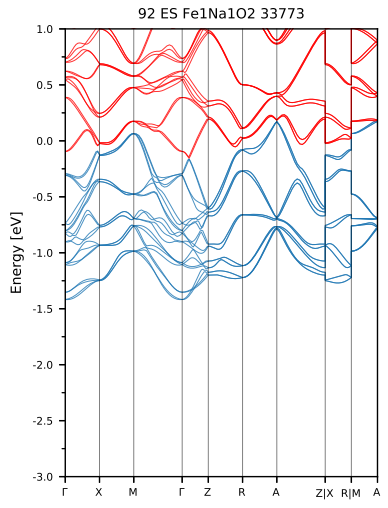

(c) 92.3.3

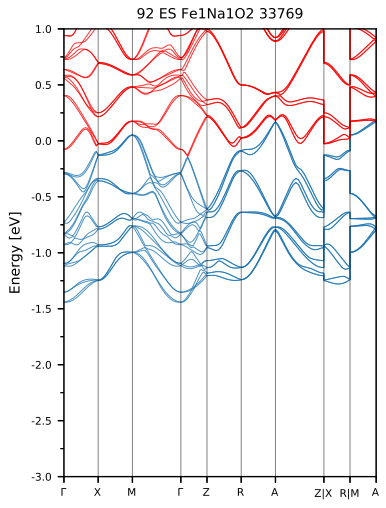

(d) 92.3.4

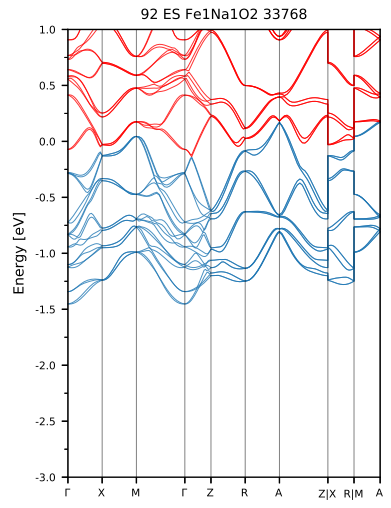

(e) 92.3.5

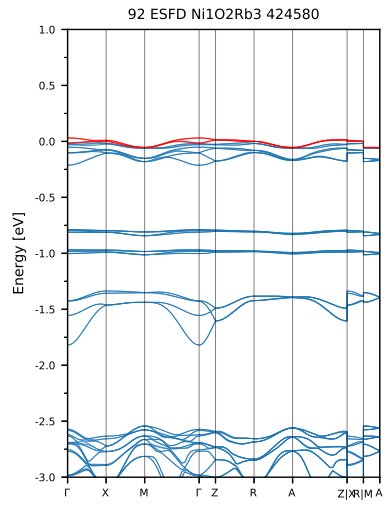

(f) 92.4.1

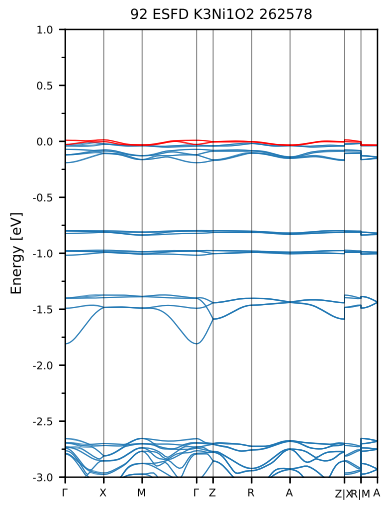

(g) 92.4.2

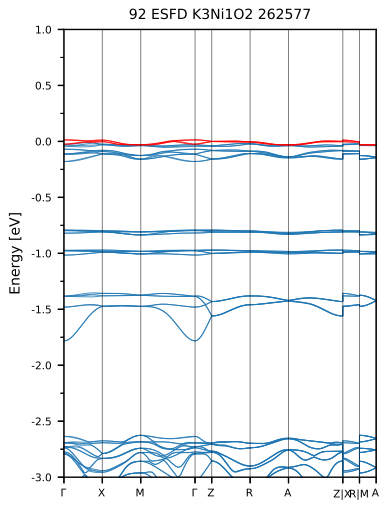

(h) 92.4.3

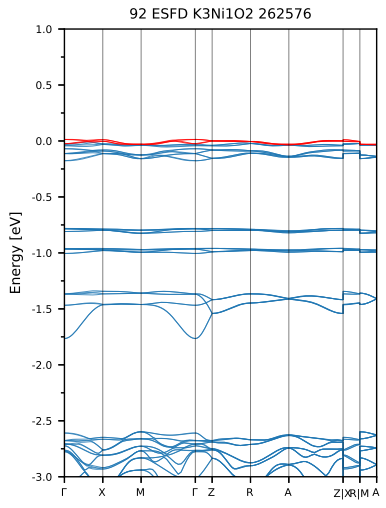

(i) 92.4.4

Figure 208: Band structures for topological materials in SG 92, part 1/1, from 92.3.1 to 92.4.4. See the list of figures.

### 0.0.44 SG 97

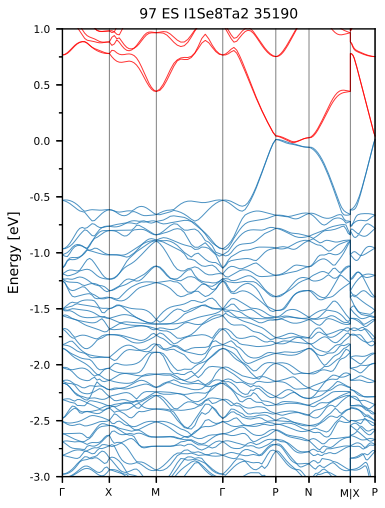

(a) 97.3.1

Figure 209: Band structures for topological materials in SG 97, part 1/1, from 97.3.1 to 97.3.1. See the list of figures.

### 0.0.45 SG 99

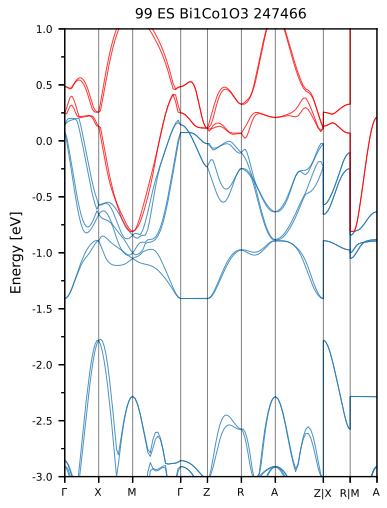

(a) 99.3.1

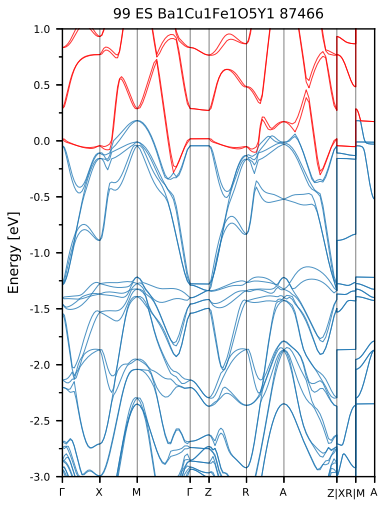

(b) 99.3.2

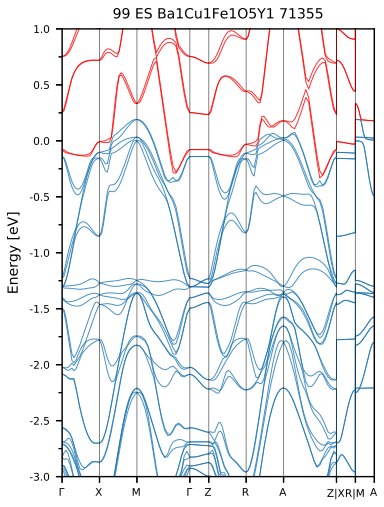

(c) 99.3.3

Figure 210: Band structures for topological materials in SG 99, part 1/1, from 99.3.1 to 99.3.3. See the list of figures.

### 0.0.46 SG 102

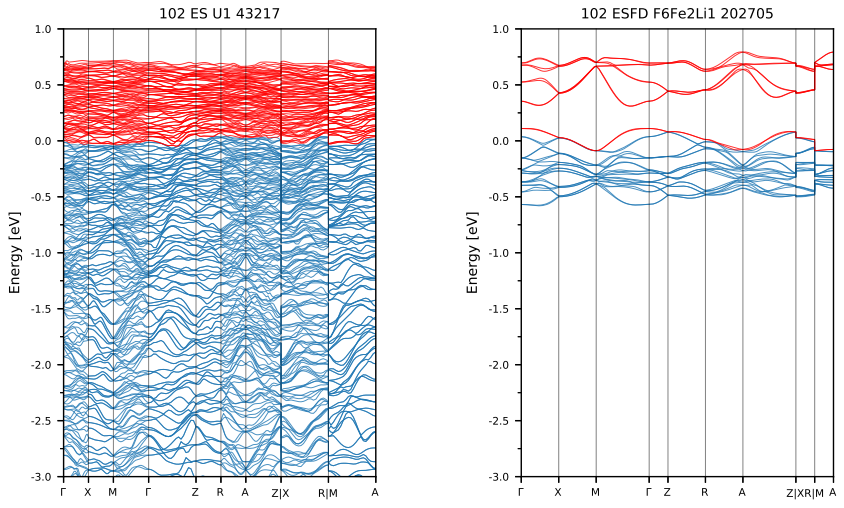

(a) 102.3.1 (b) 102.4.1

Figure 211: Band structures for topological materials in SG 102, part 1/1, from 102.3.1 to 102.4.1. See the list of figures.

### 0.0.47 SG 103

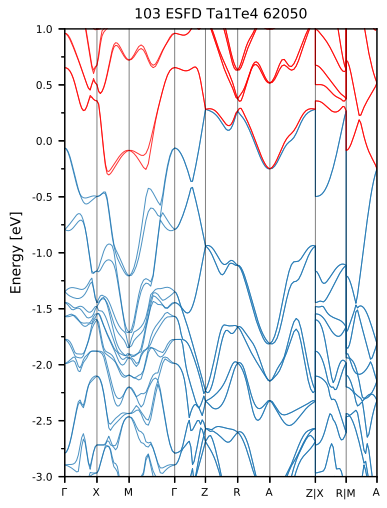

(a) 103.4.1

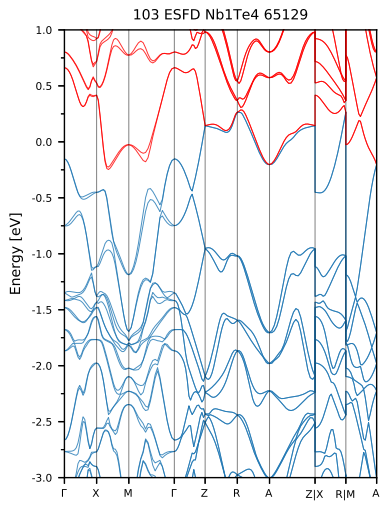

(b) 103.4.2

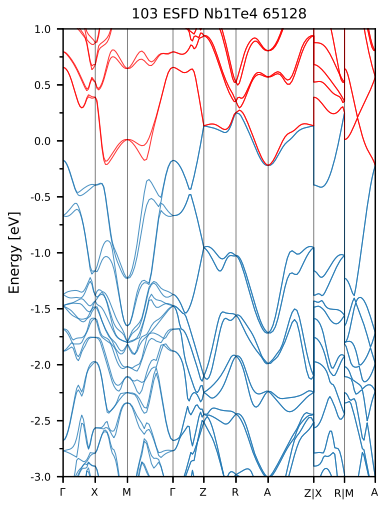

(c) 103.4.3

Figure 212: Band structures for topological materials in SG 103, part 1/1, from 103.4.1 to 103.4.3. See the list of figures.

### 0.0.48 SG 106

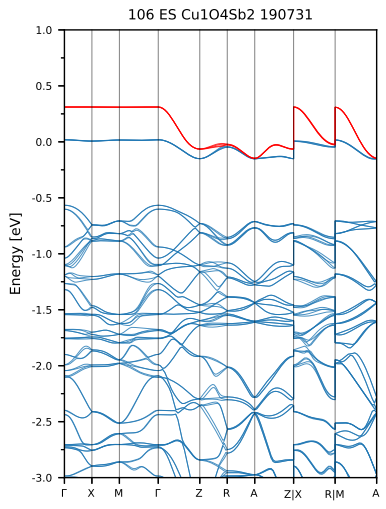

(a) 106.3.1

Figure 213: Band structures for topological materials in SG 106, part 1/1, from 106.3.1 to 106.3.1. See the list of figures.

## 0.0.49 SG 107

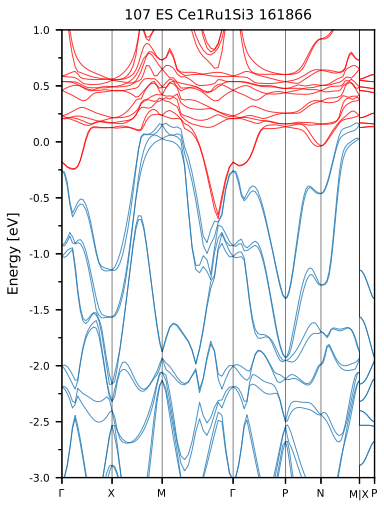

(a) 107.3.1

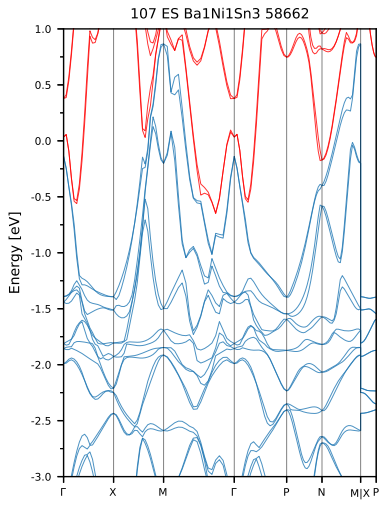

(b) 107.3.2

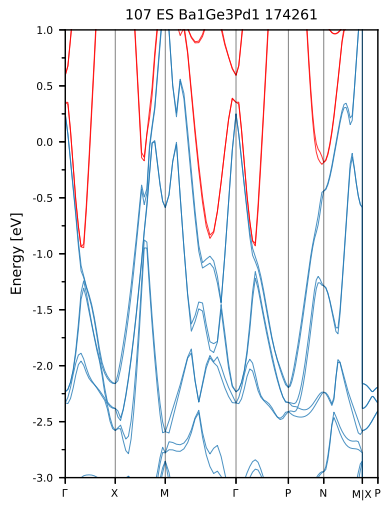

(c) 107.3.3

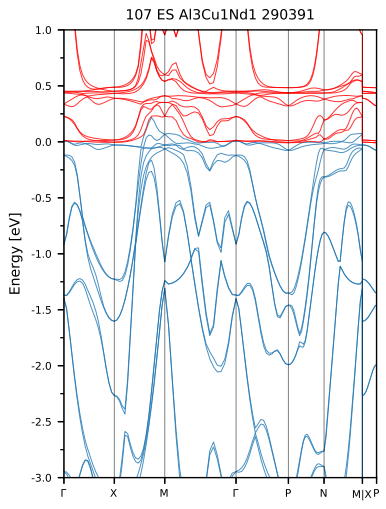

(d) 107.3.4

Figure 214: Band structures for topological materials in SG 107, part 1/1, from 107.3.1 to 107.3.4. See the list of figures.

## 0.0.50 SG 113

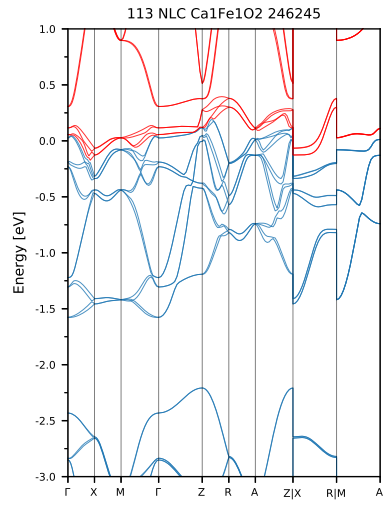

(a) 113.2.1

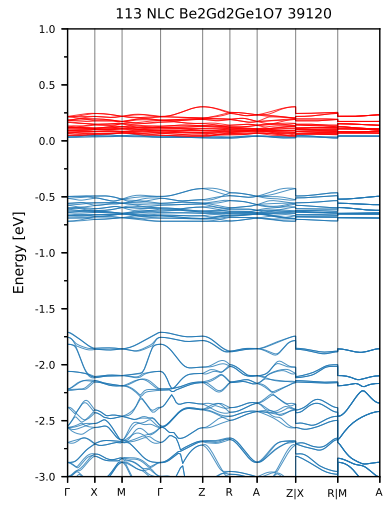

(b) 113.2.2

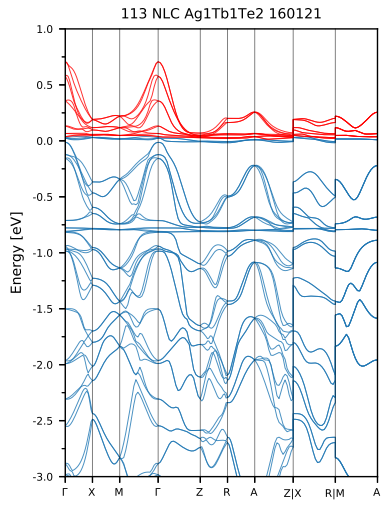

(c) 113.2.3

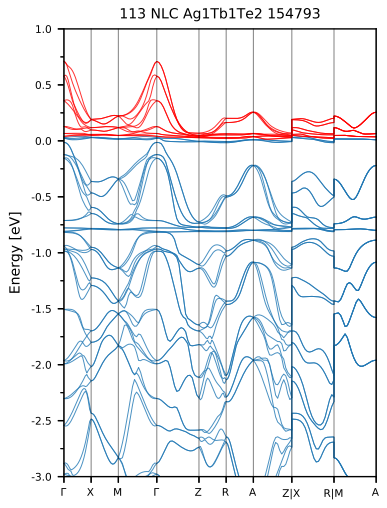

(d) 113.2.4

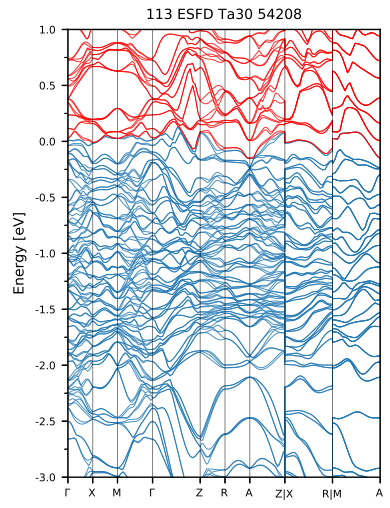

(e) 113.4.1

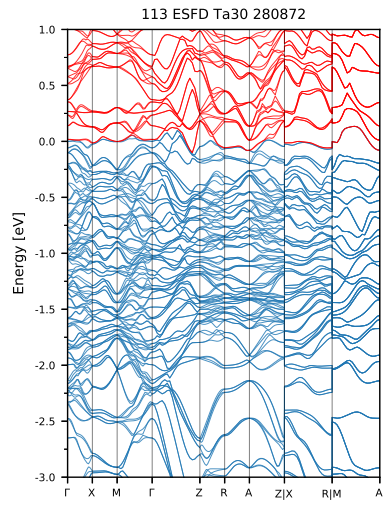

(f) 113.4.2

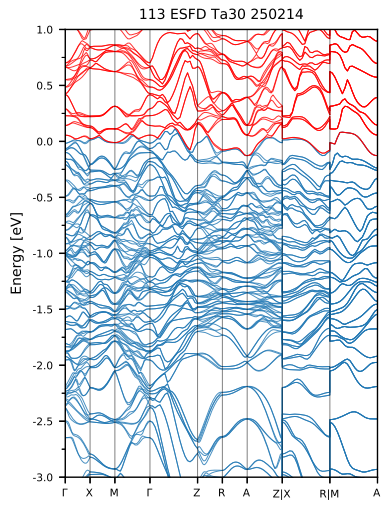

(g) 113.4.3

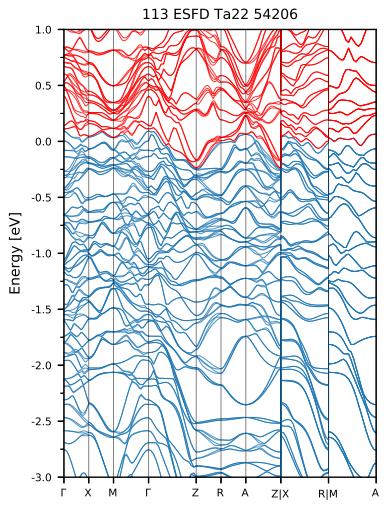

(h) 113.4.4

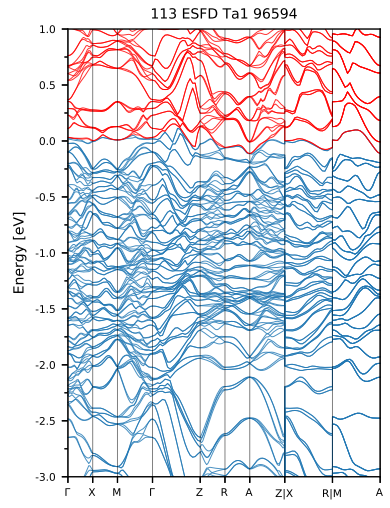

(i) 113.4.5

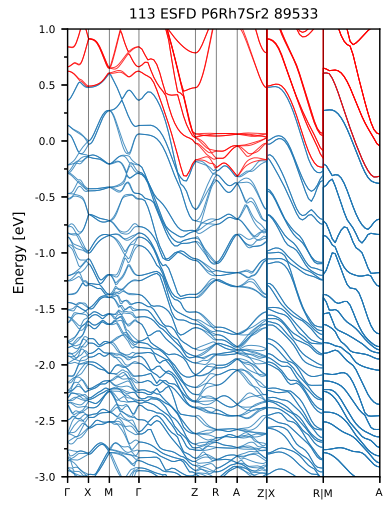

(j) 113.4.6

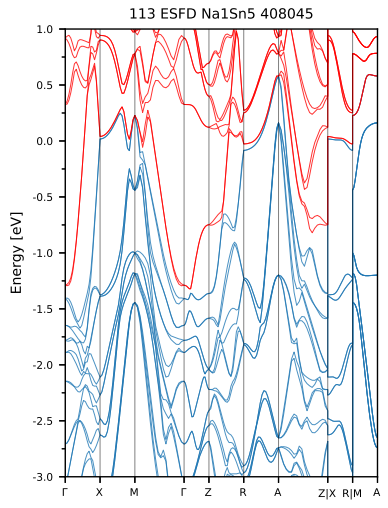

(k) 113.4.7

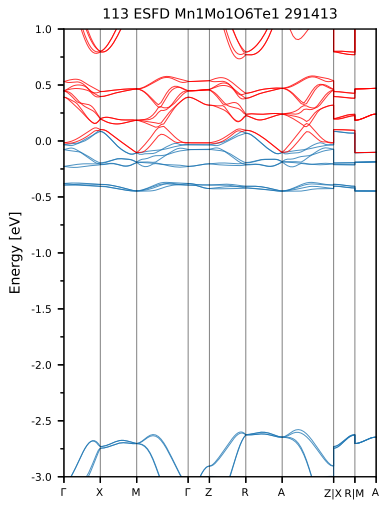

(l) 113.4.8

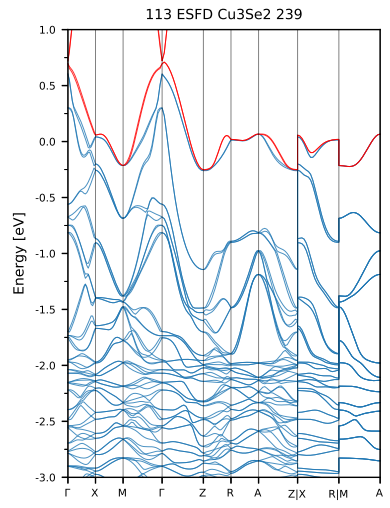

(m) 113.4.9

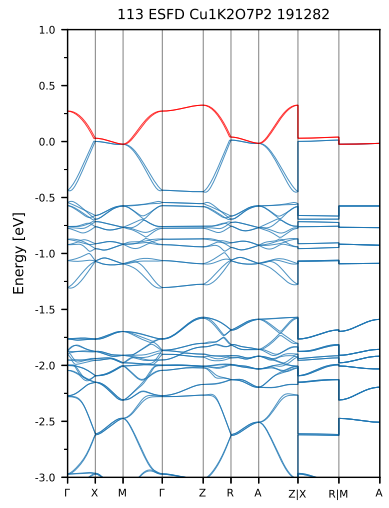

(n) 113.4.10

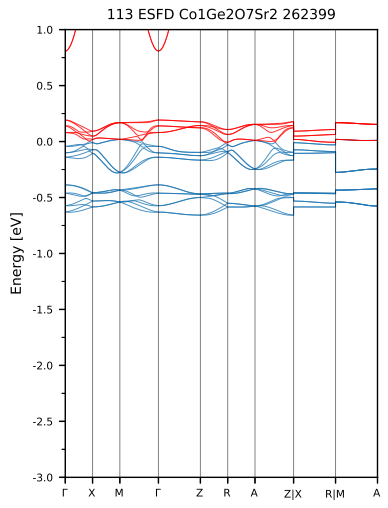

(o) 113.4.11

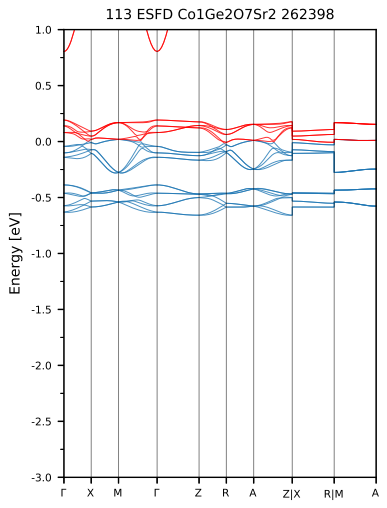

(p) 113.4.12

Figure 215: Band structures for topological materials in SG 113, part 1/2, from 113.2.1 to 113.4.12. See the list of figures.

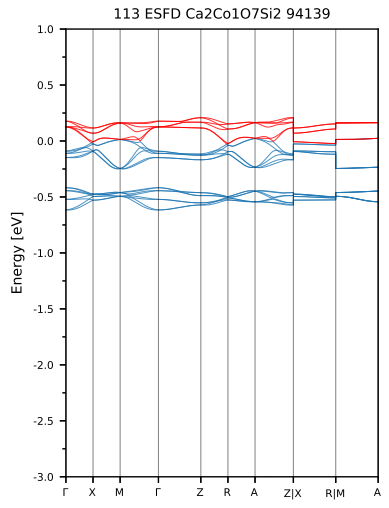

(q) 113.4.13

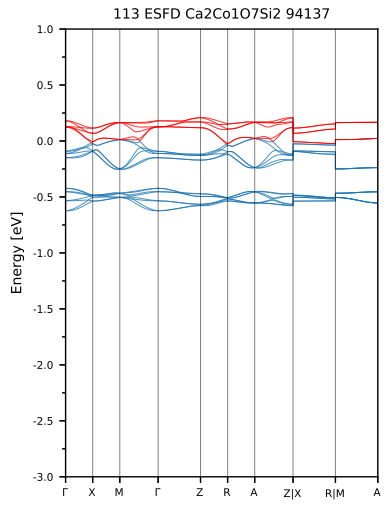

(r) 113.4.14

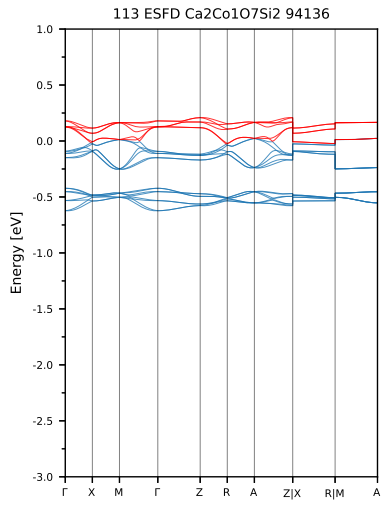

(s) 113.4.15

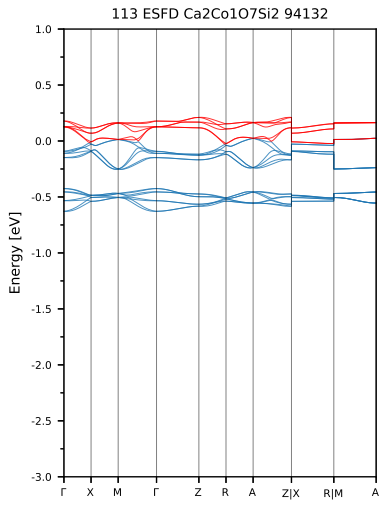

(t) 113.4.16

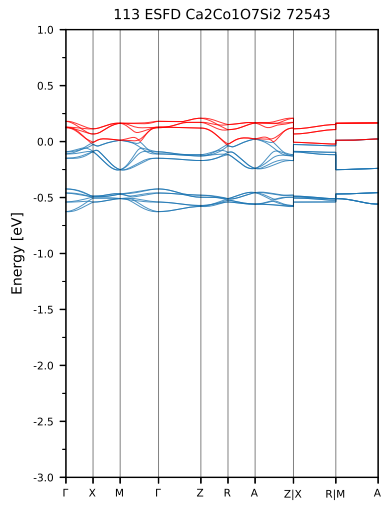

(u) 113.4.17

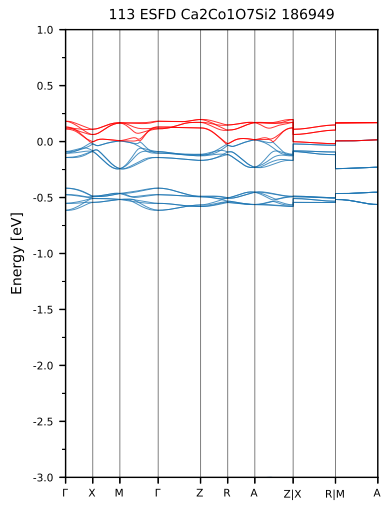

(v) 113.4.18

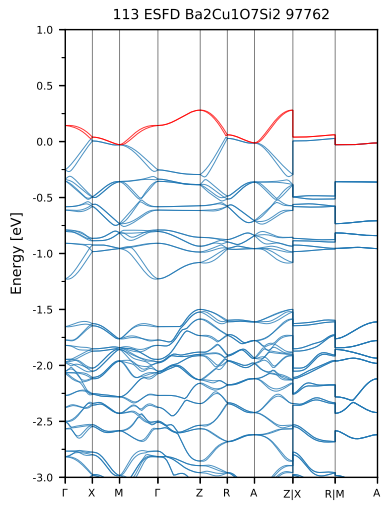

(w) 113.4.19

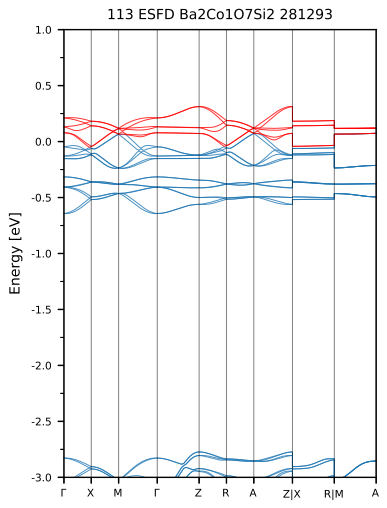

(x) 113.4.20

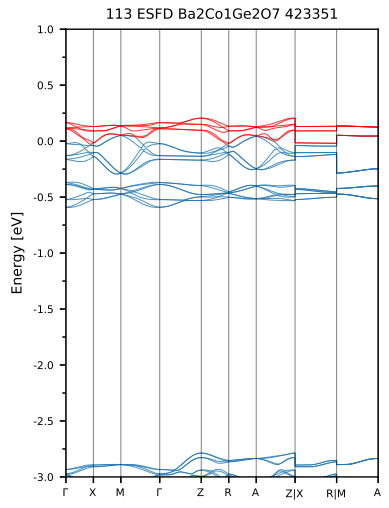

(y) 113.4.21

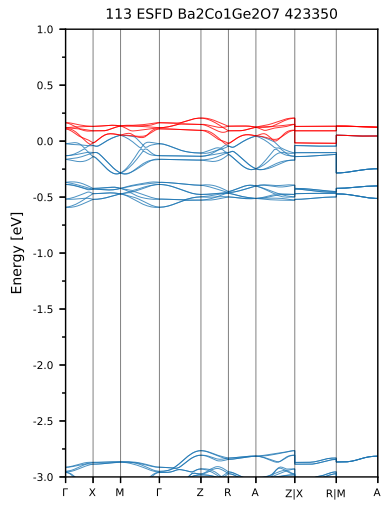

(z) 113.4.22

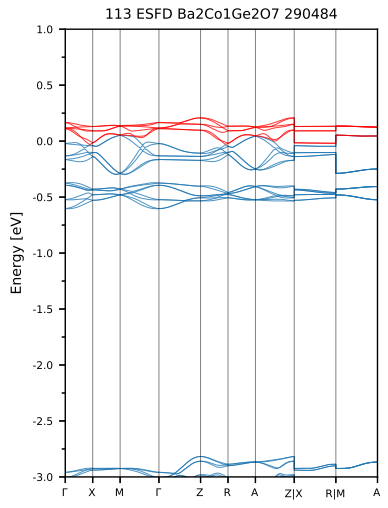

(aa) 113.4.23

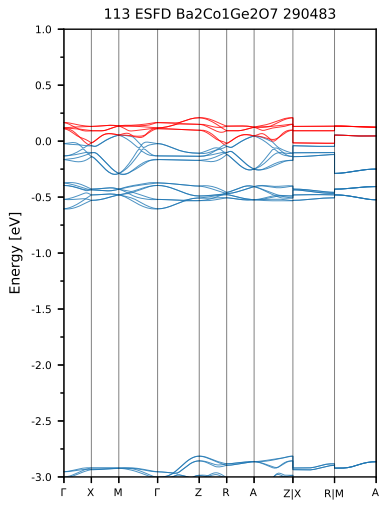

(ab) 113.4.24

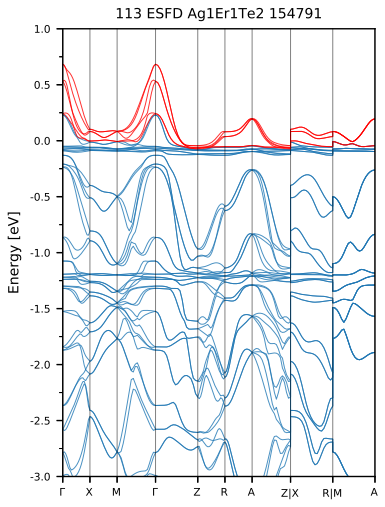

(ac) 113.4.25

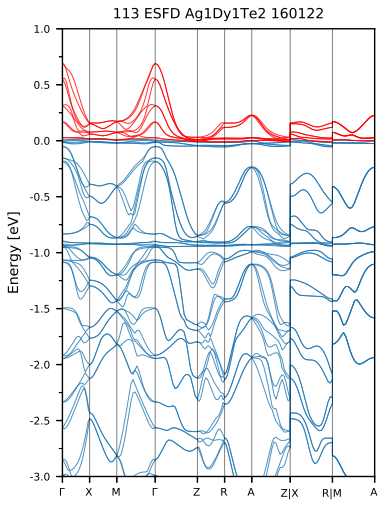

(ad) 113.4.26

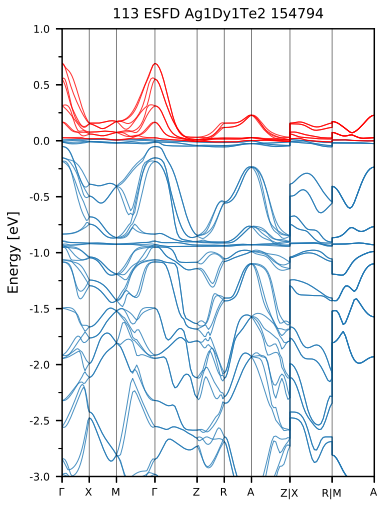

(ae) 113.4.27

Figure 216: Band structures for topological materials in SG 113, part 2/2, from 113.4.13 to 113.4.27. See the list of figures.

### 0.0.51 SG 115

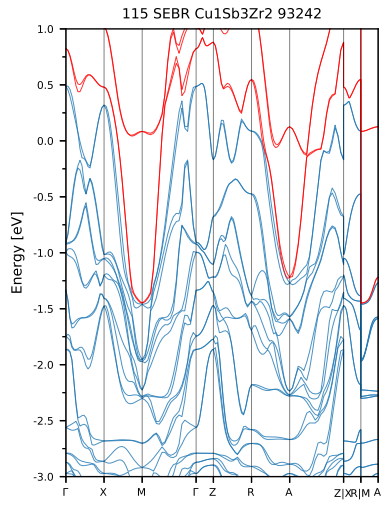

(a) 115.1.1

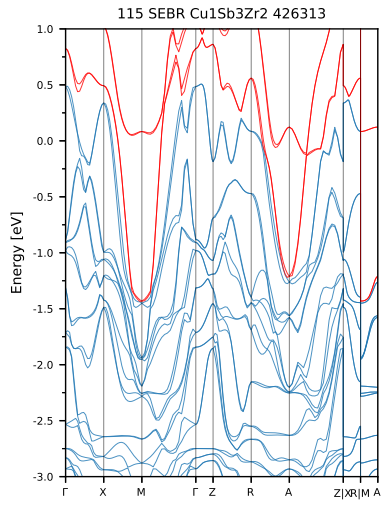

(b) 115.1.2

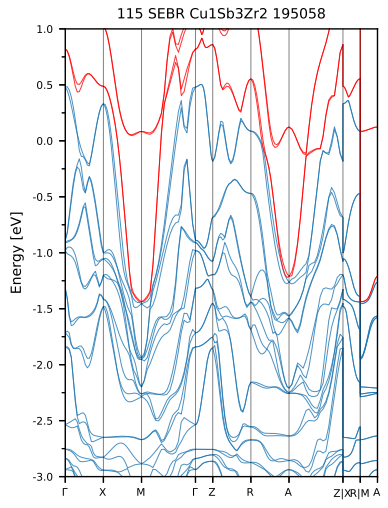

(c) 115.1.3

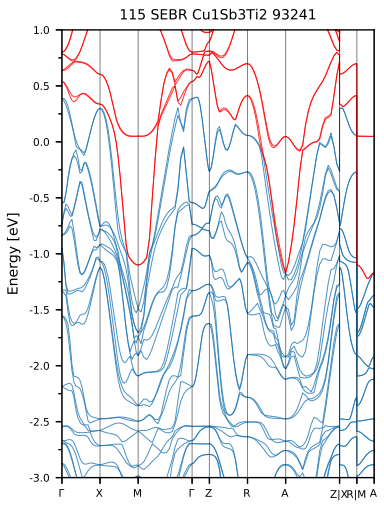

(d) 115.1.4

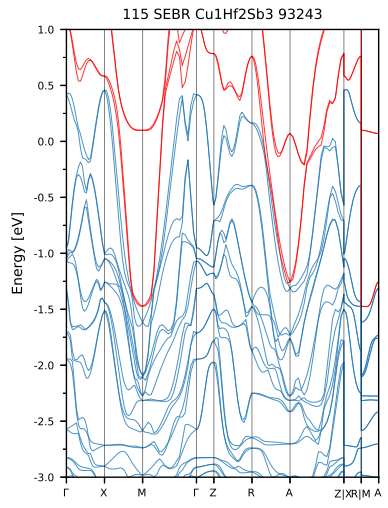

(e) 115.1.5

Figure 217: Band structures for topological materials in SG 115, part 1/1, from 115.1.1 to 115.1.5. See the list of figures.

### 0.0.52 SG 120

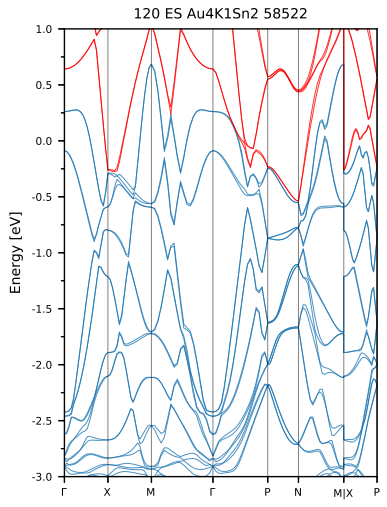

(a) 120.3.1

Figure 218: Band structures for topological materials in SG 120, part 1/1, from 120.3.1 to 120.3.1. See the list of figures.

### 0.0.53 SG 121

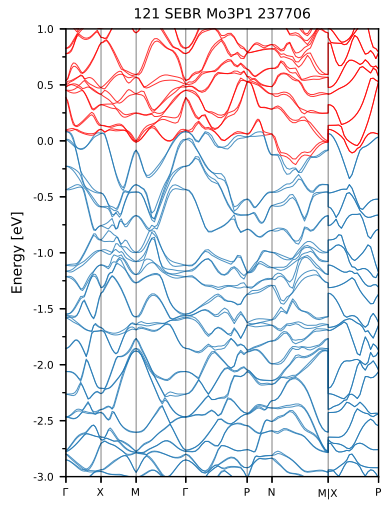

(a) 121.1.1

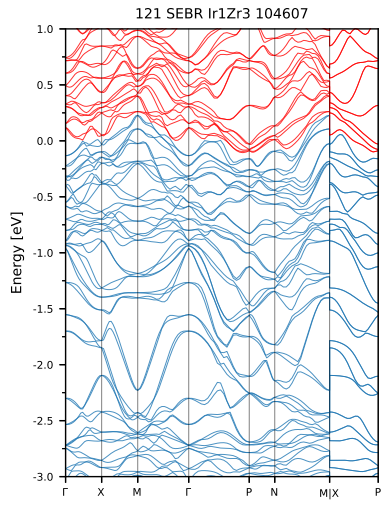

(b) 121.1.2

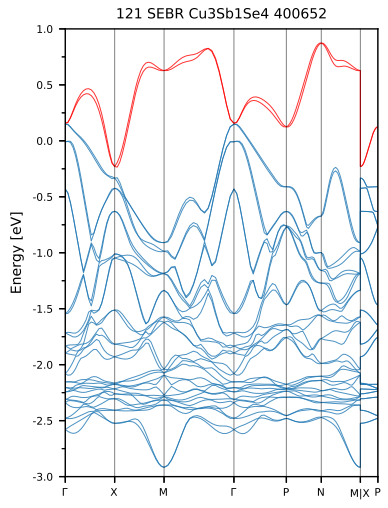

(c) 121.1.3

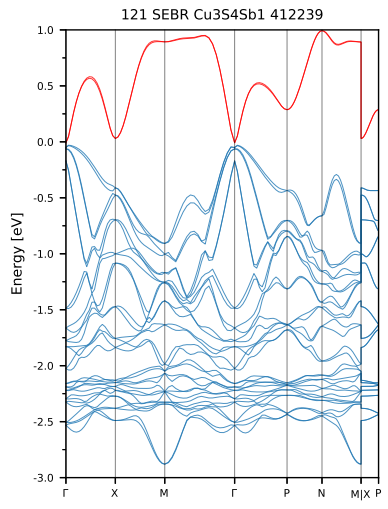

(d) 121.1.4

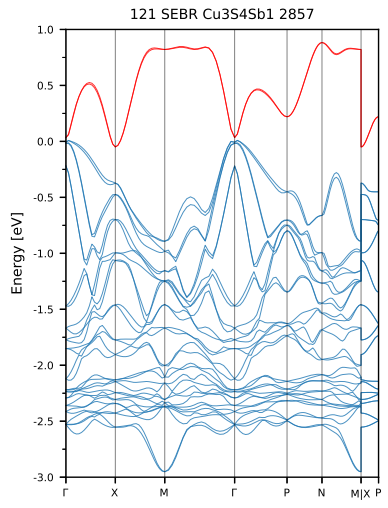

(e) 121.1.5

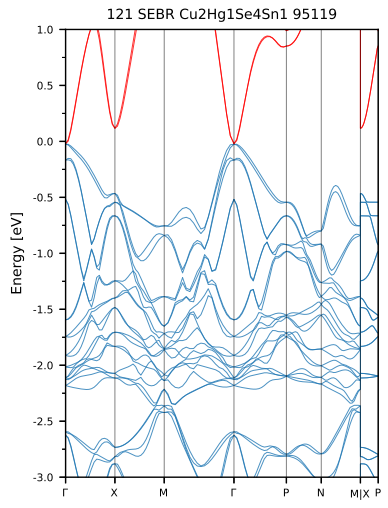

(f) 121.1.6

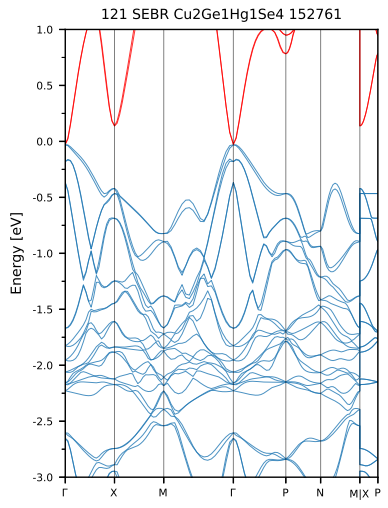

(g) 121.1.7

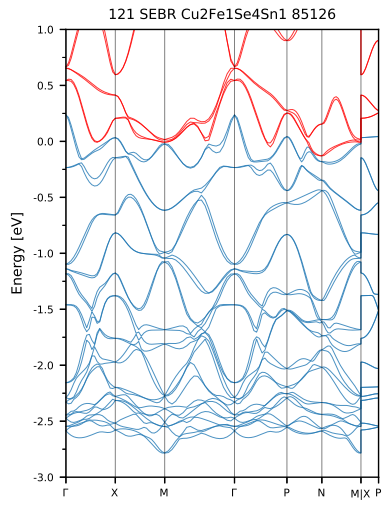

(h) 121.1.8

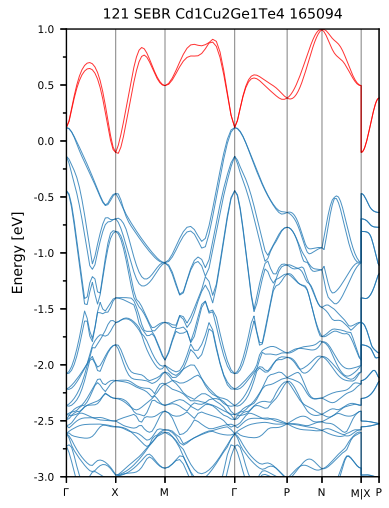

(i) 121.1.9

Figure 219: Band structures for topological materials in SG 121, part 1/1, from 121.1.1 to 121.1.9. See the list of figures.

## 0.0.54 SG 123

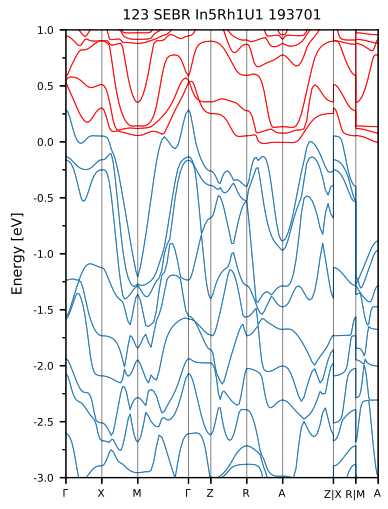

(a) 123.1.1

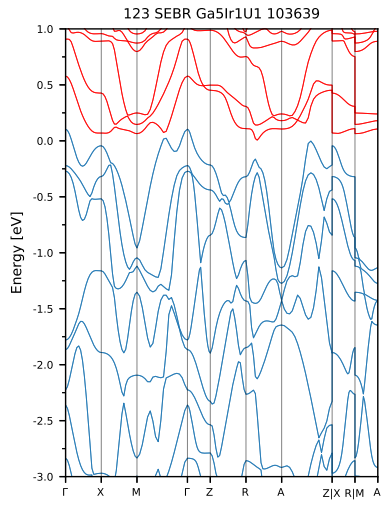

(b) 123.1.2

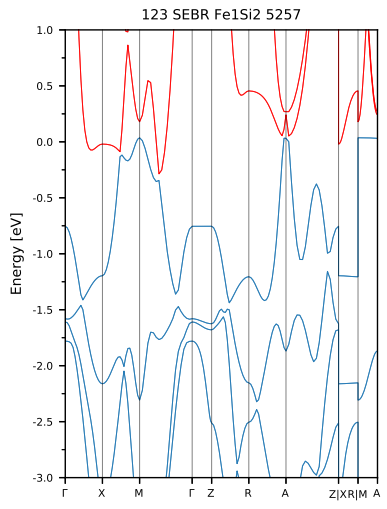

(c) 123.1.3

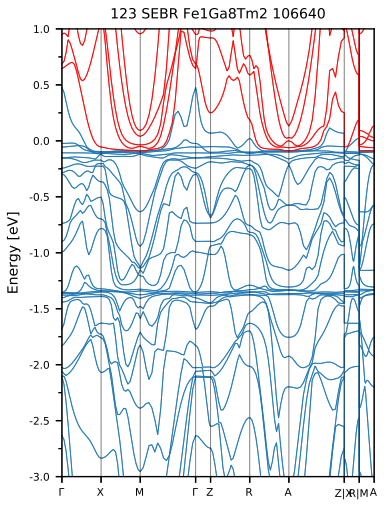

(d) 123.1.4

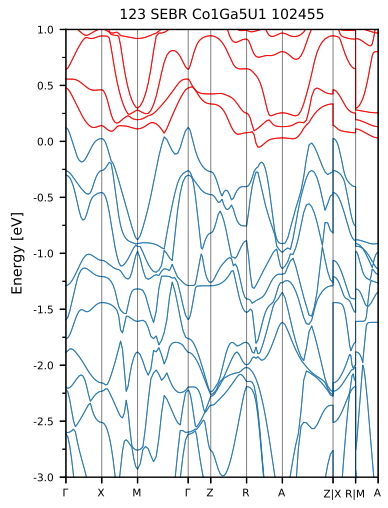

(e) 123.1.5

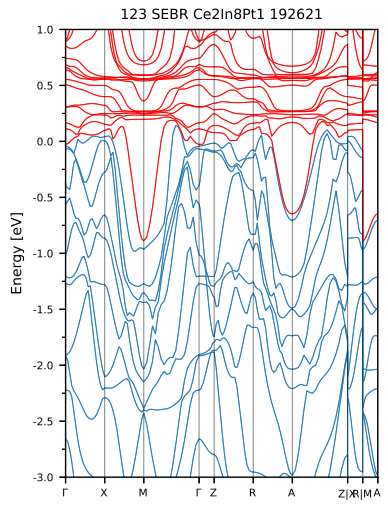

(f) 123.1.6

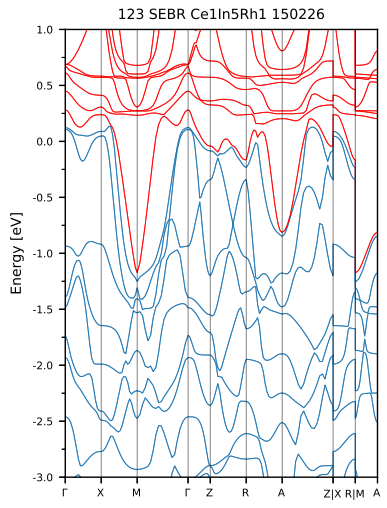

(g) 123.1.7

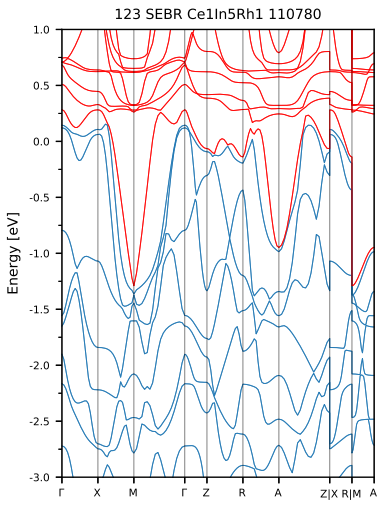

(h) 123.1.8

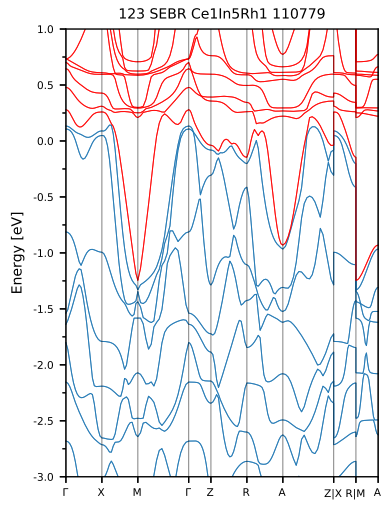

(i) 123.1.9

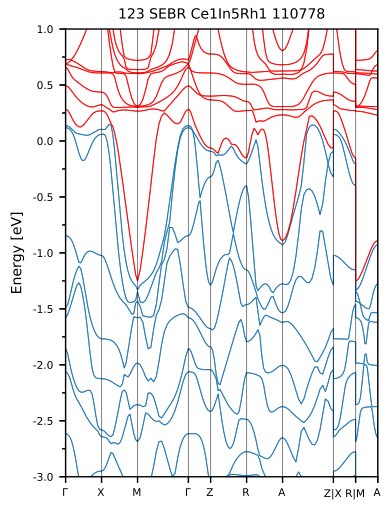

(j) 123.1.10

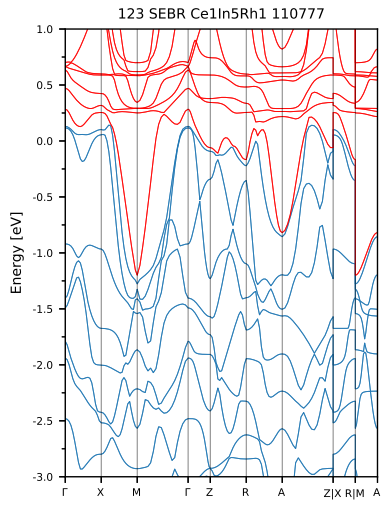

(k) 123.1.11

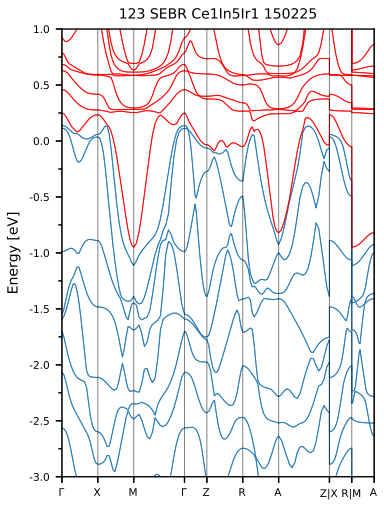

(l) 123.1.12

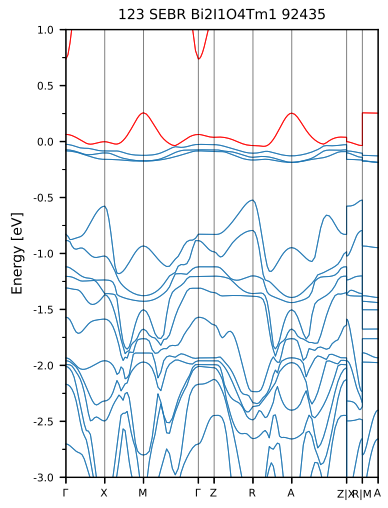

(m) 123.1.13

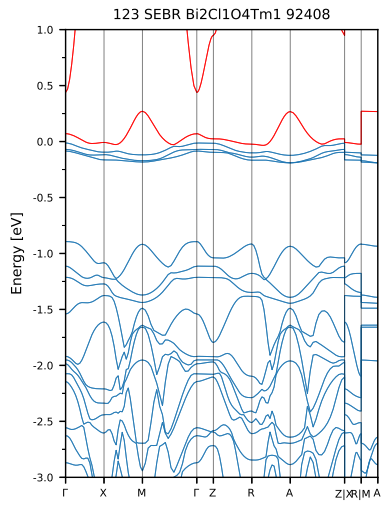

(n) 123.1.14

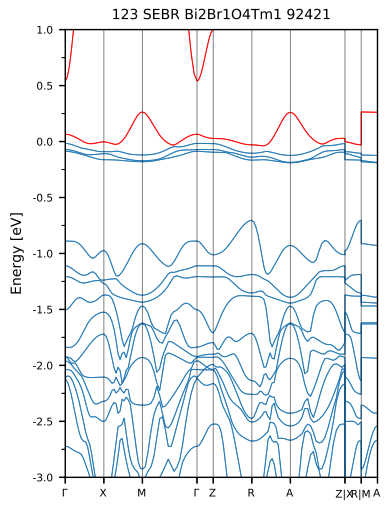

(o) 123.1.15

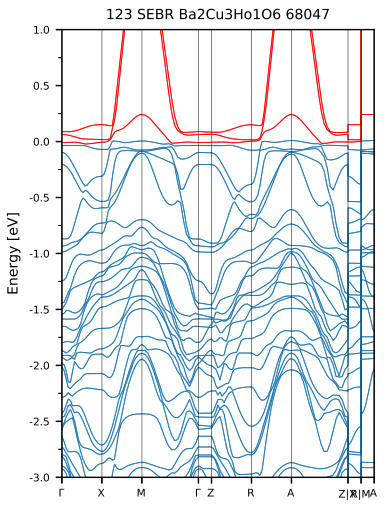

(p) 123.1.16

Figure 220: Band structures for topological materials in SG 123, part 1/5, from 123.1.1 to 123.1.16. See the list of figures.

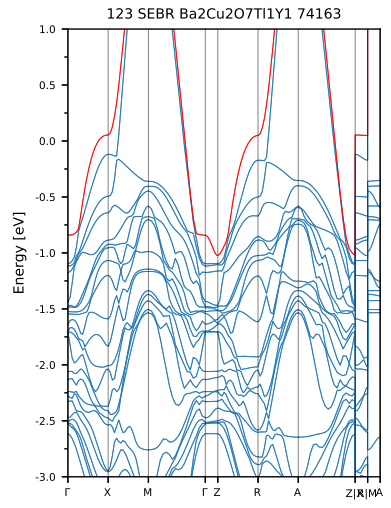

(q) 123.1.17

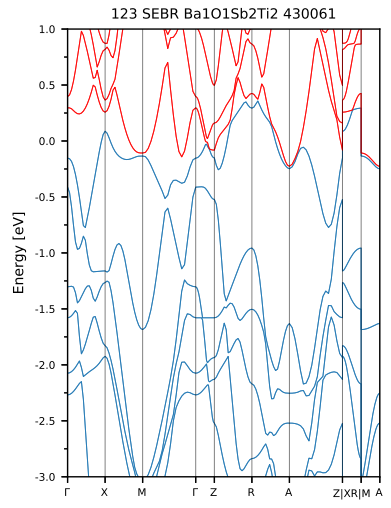

(r) 123.1.18

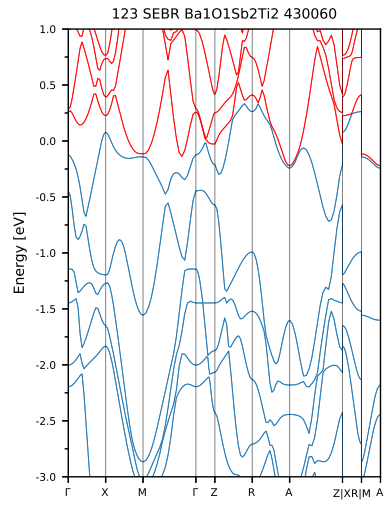

(s) 123.1.19

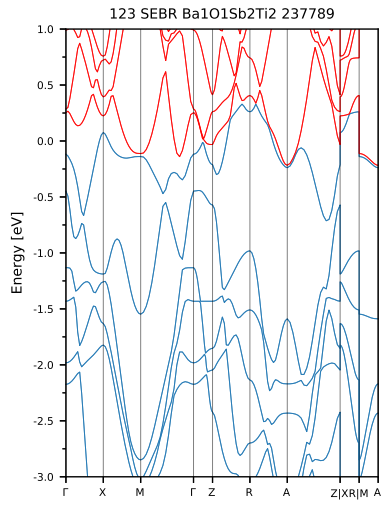

(t) 123.1.20

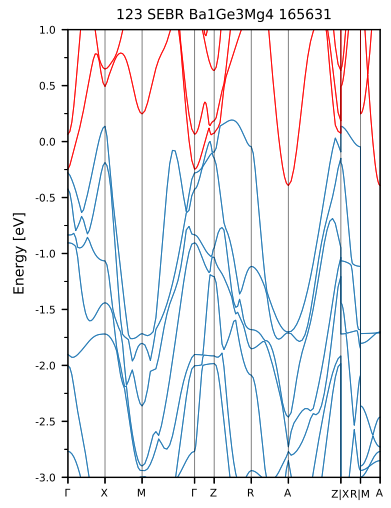

(u) 123.1.21

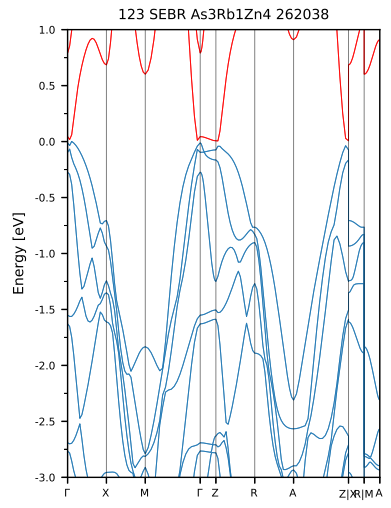

(v) 123.1.22

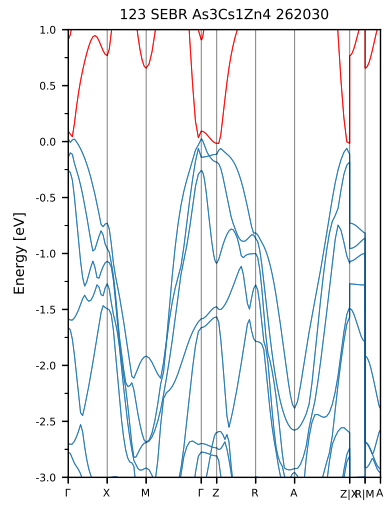

(w) 123.1.23

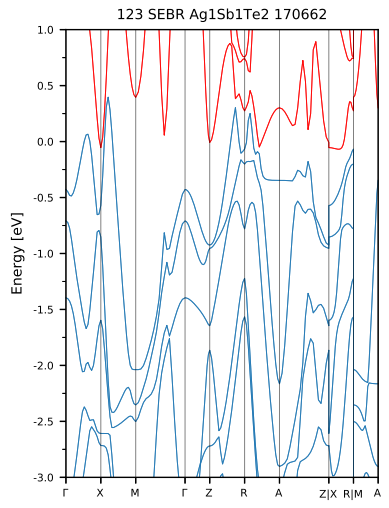

(x) 123.1.24

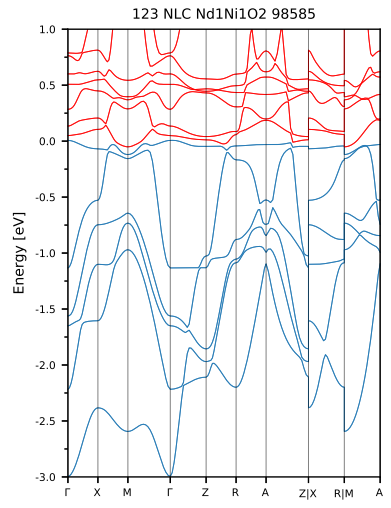

(y) 123.2.1

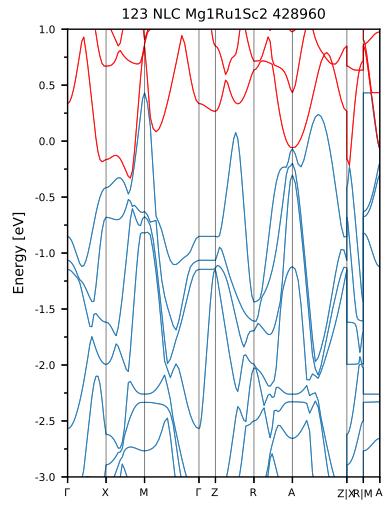

(z) 123.2.2

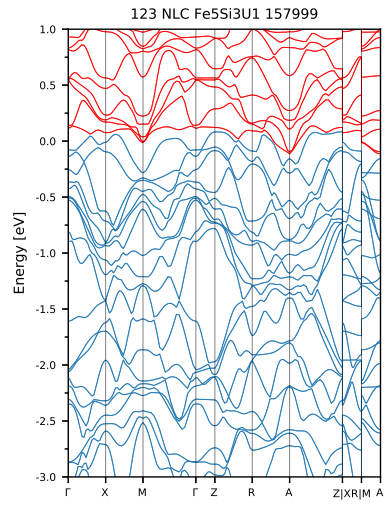

(aa) 123.2.3

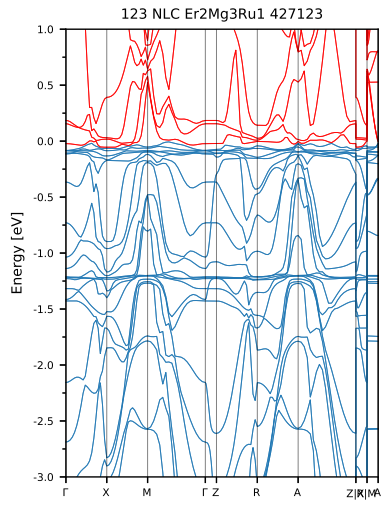

(ab) 123.2.4

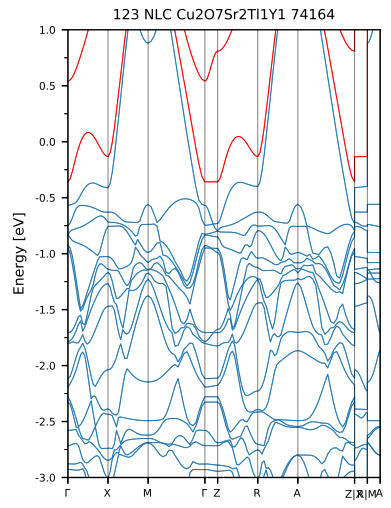

(ac) 123.2.5

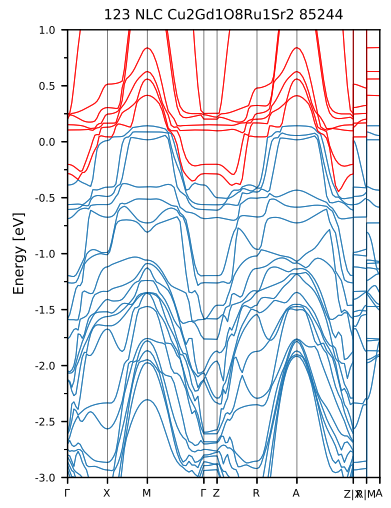

(ad) 123.2.6

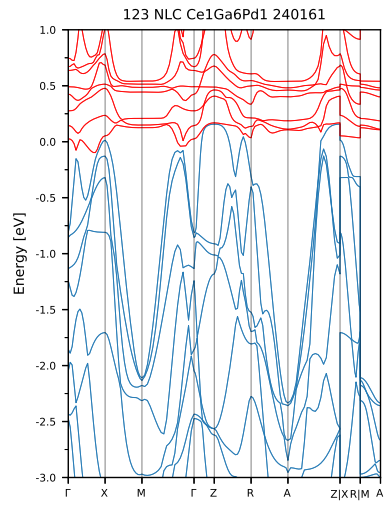

(ae) 123.2.7

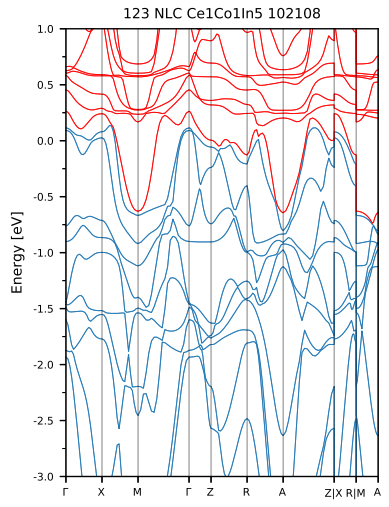

(af) 123.2.8

Figure 221: Band structures for topological materials in SG 123, part 2/5, from 123.1.17 to 123.2.8. See the list of figures.

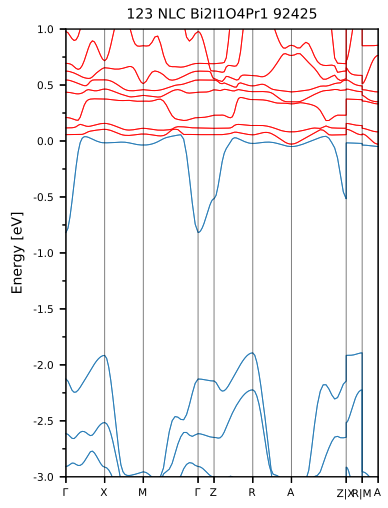

(ag) 123.2.9

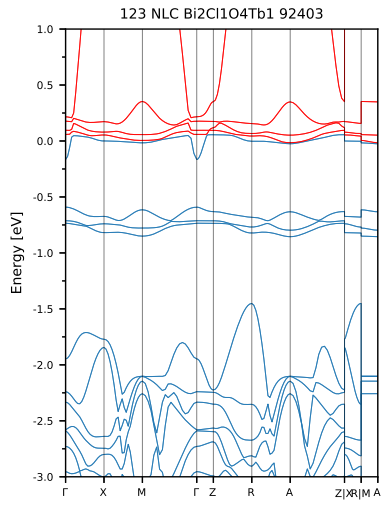

(ah) 123.2.10

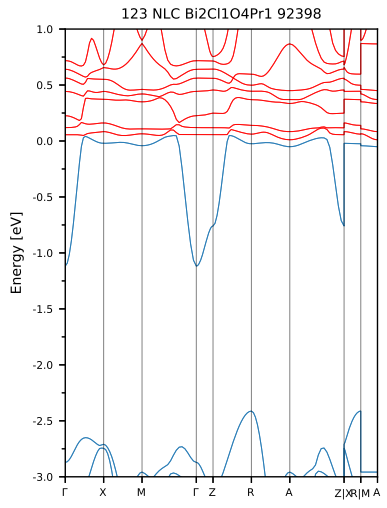

(ai) 123.2.11

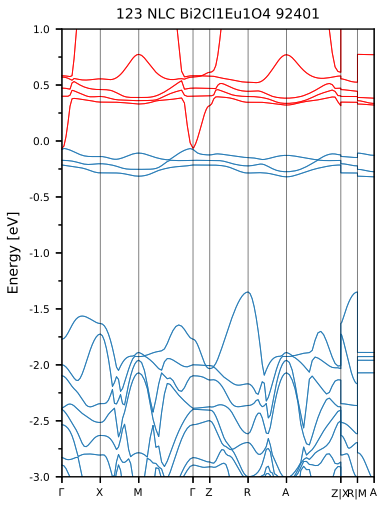

(aj) 123.2.12

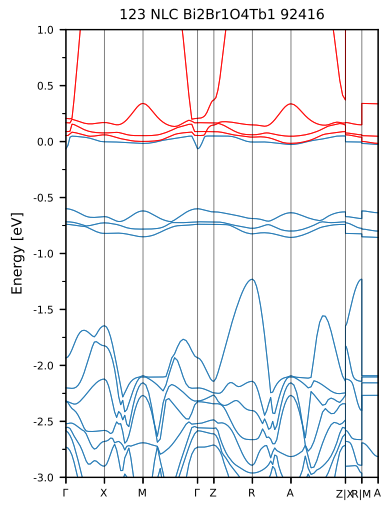

(ak) 123.2.13

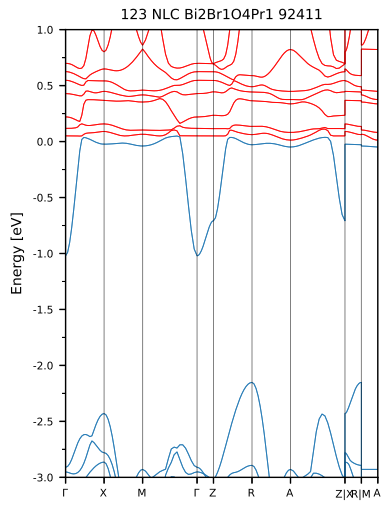

(al) 123.2.14

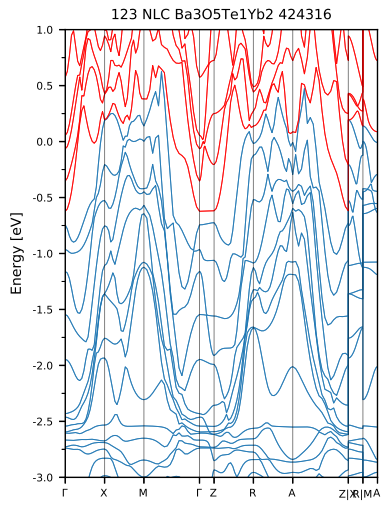

(am) 123.2.15

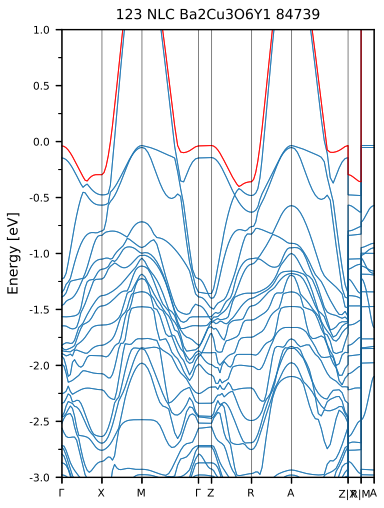

(an) 123.2.16

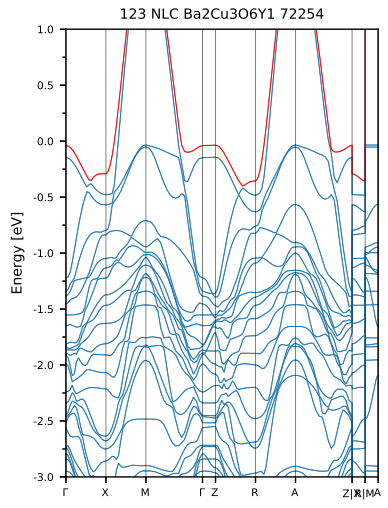

(ao) 123.2.17

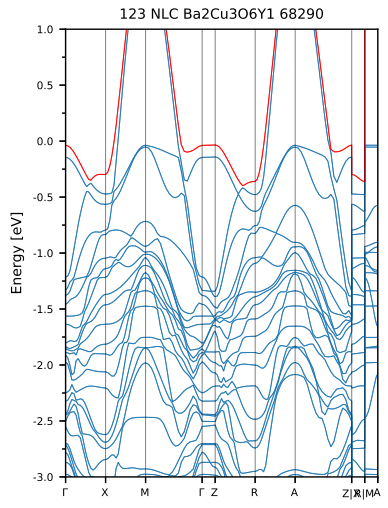

(ap) 123.2.18

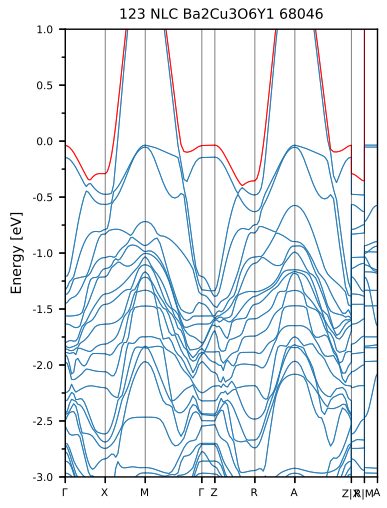

(aq) 123.2.19

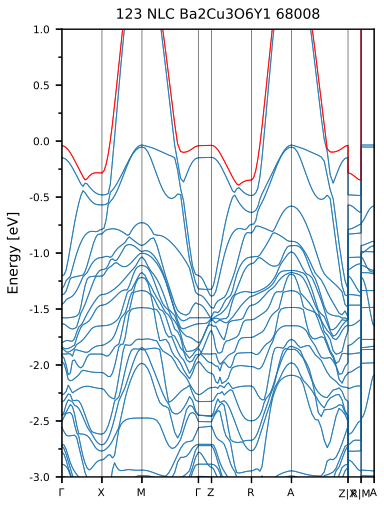

(ar) 123.2.20

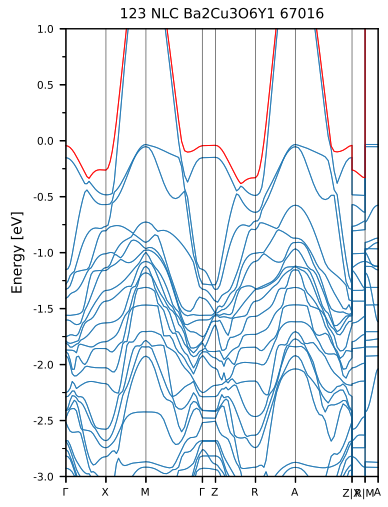

(as) 123.2.21

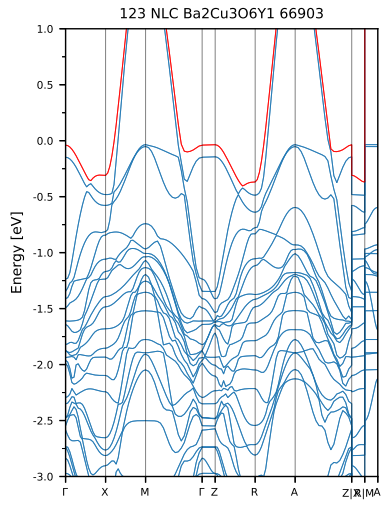

(at) 123.2.22

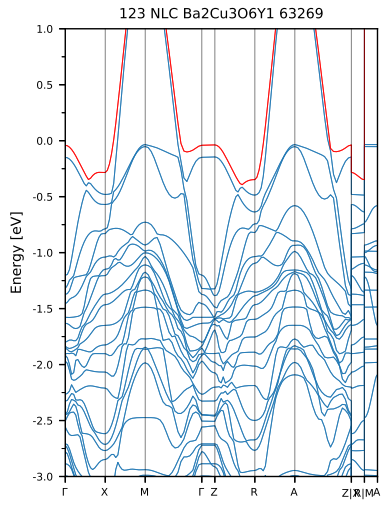

(au) 123.2.23

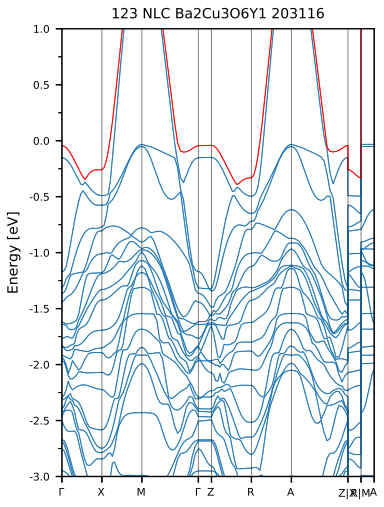

(av) 123.2.24

Figure 222: Band structures for topological materials in SG 123, part 3/5, from 123.2.9 to 123.2.24. See the list of figures.

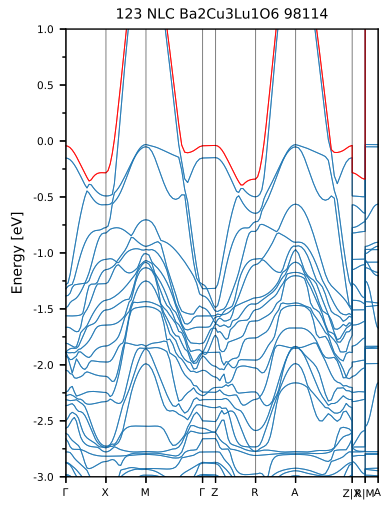

(aw) 123.2.25

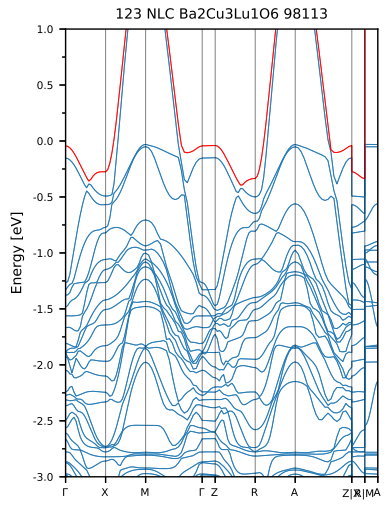

(ax) 123.2.26

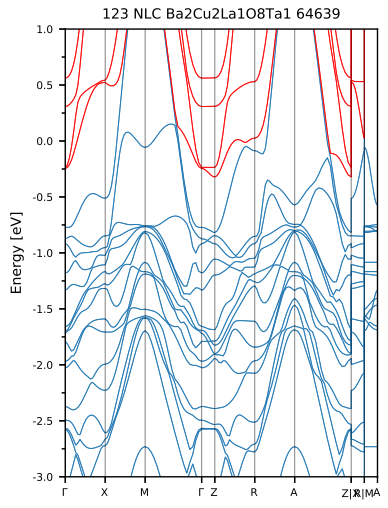

(ay) 123.2.27

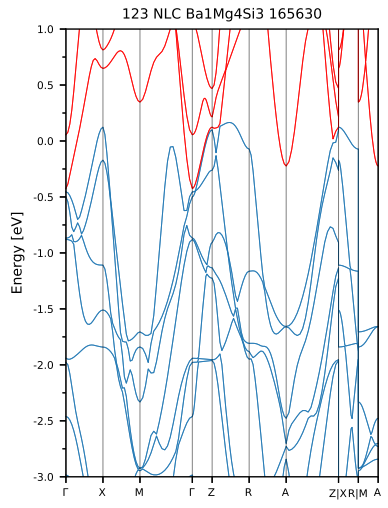

(az) 123.2.28

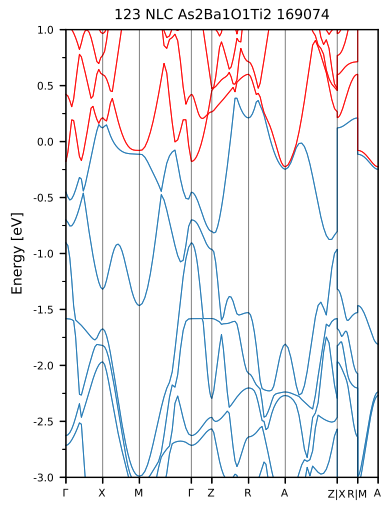

(ba) 123.2.29

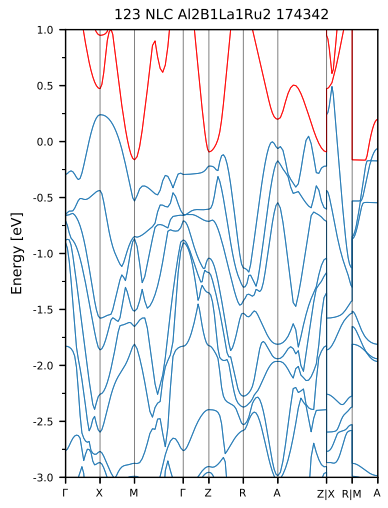

(bb) 123.2.30

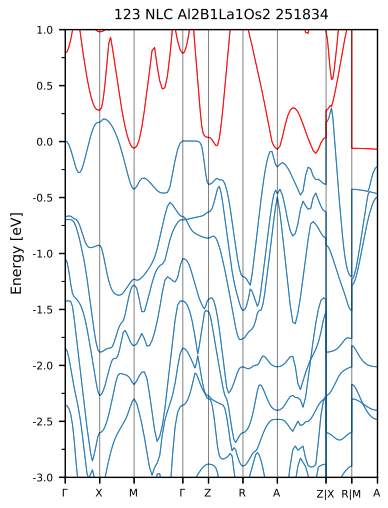

(bc) 123.2.31

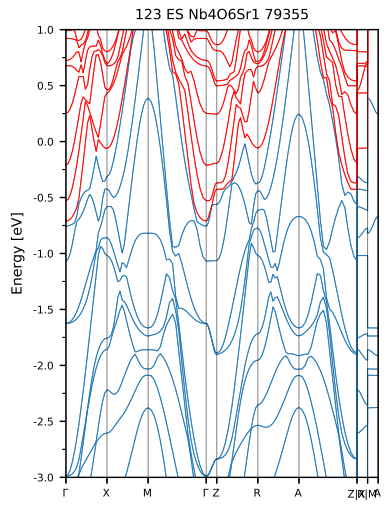

(bd) 123.3.1

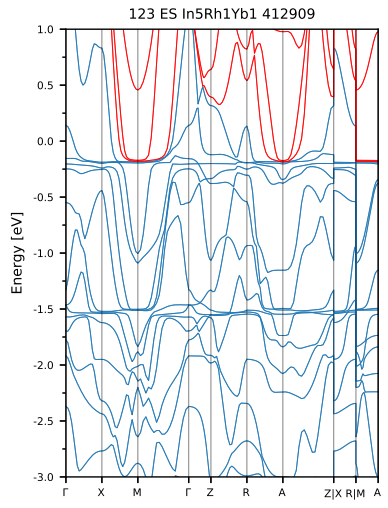

(be) 123.3.2

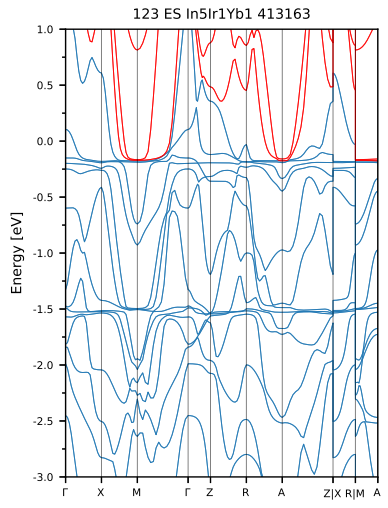

(bf) 123.3.3

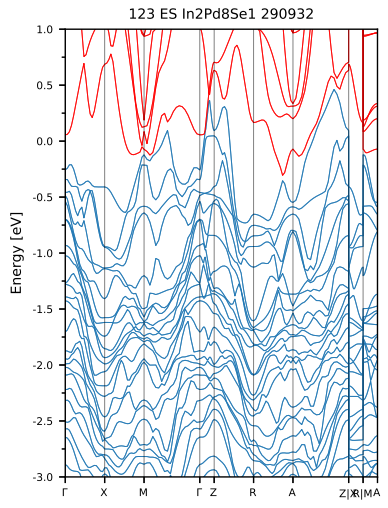

(bg) 123.3.4

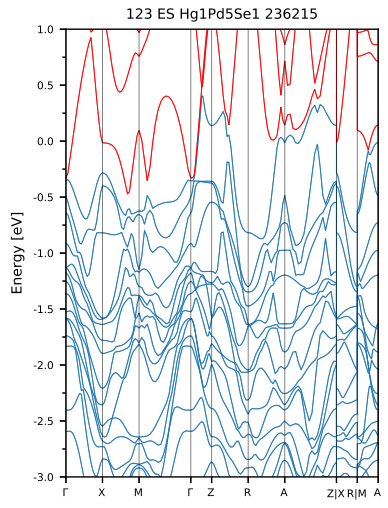

(bh) 123.3.5

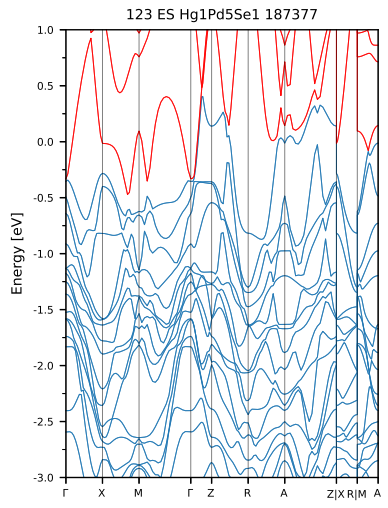

(bi) 123.3.6

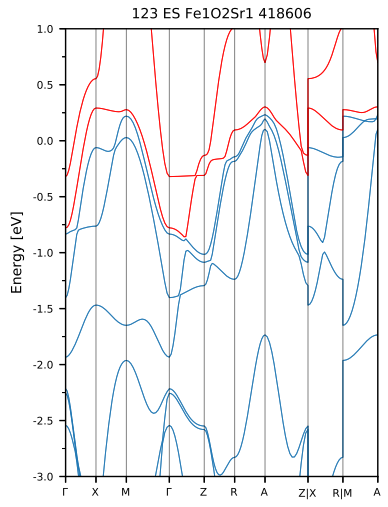

(bj) 123.3.7

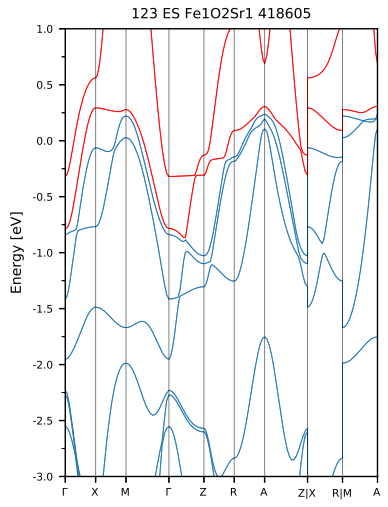

(bk) 123.3.8

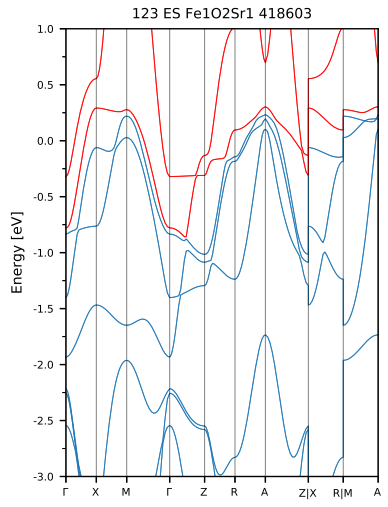

(bl) 123.3.9

Figure 223: Band structures for topological materials in SG 123, part 4/5, from 123.2.25 to 123.3.9. See the list of figures.

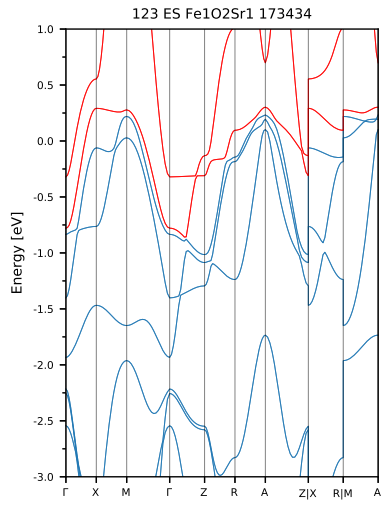

(bm) 123.3.10

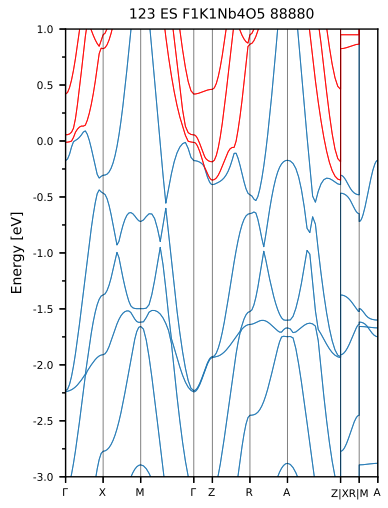

(bn) 123.3.11

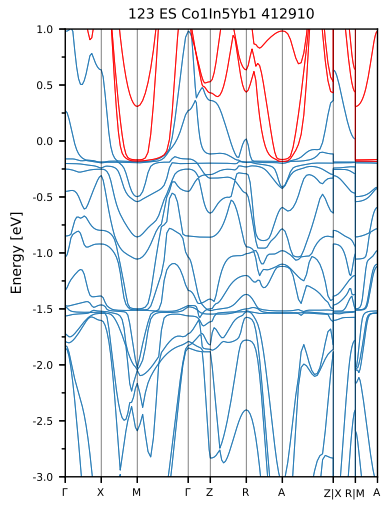

(bo) 123.3.12

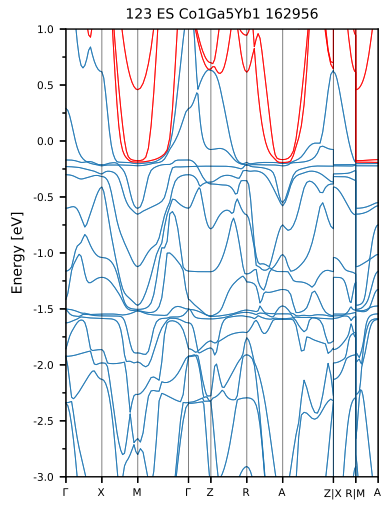

(bp) 123.3.13

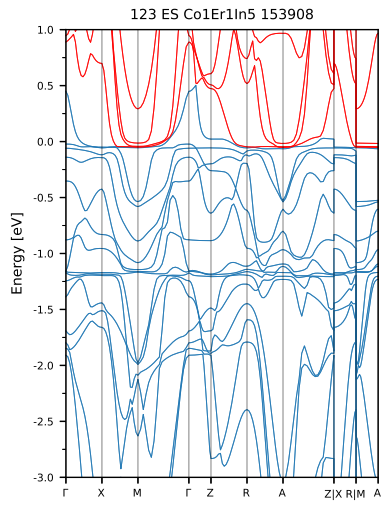

(bq) 123.3.14

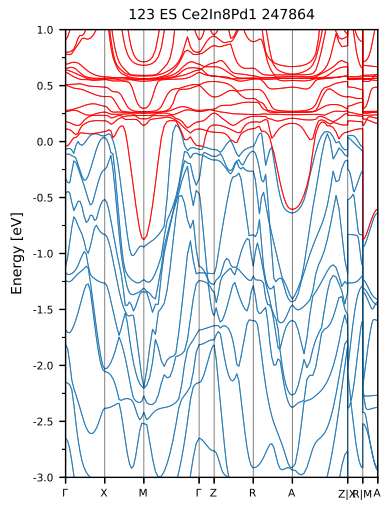

(br) 123.3.15

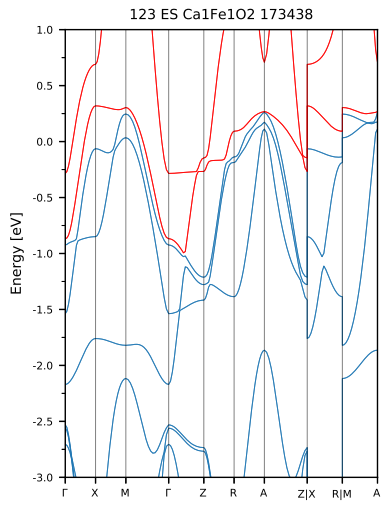

(bs) 123.3.16

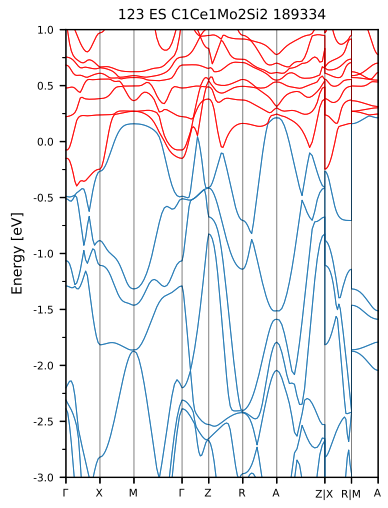

(bt) 123.3.17

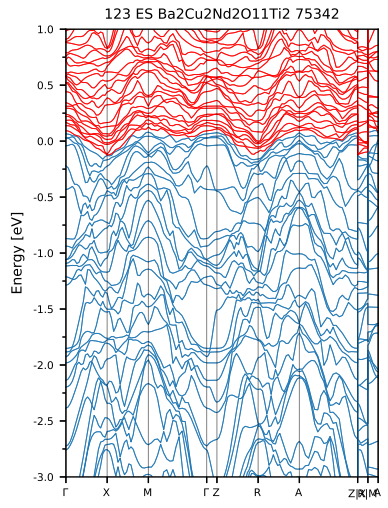

(bu) 123.3.18

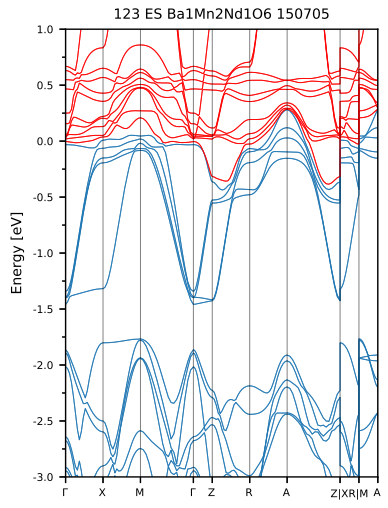

(bv) 123.3.19

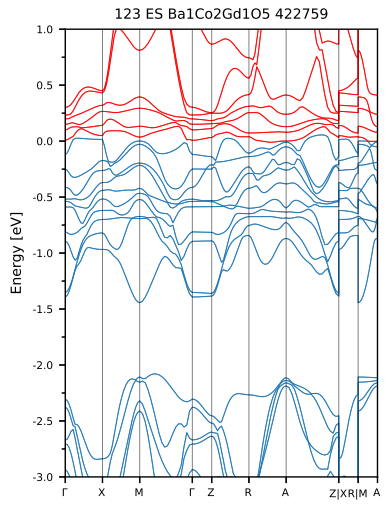

(bw) 123.3.20

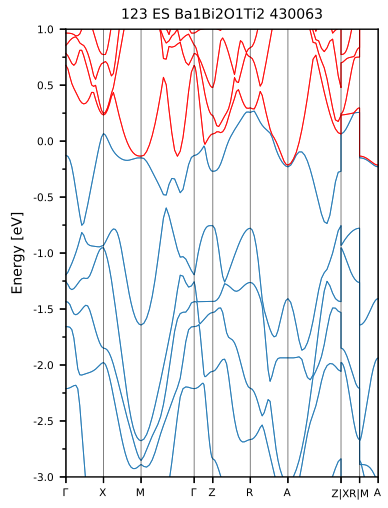

(bx) 123.3.21

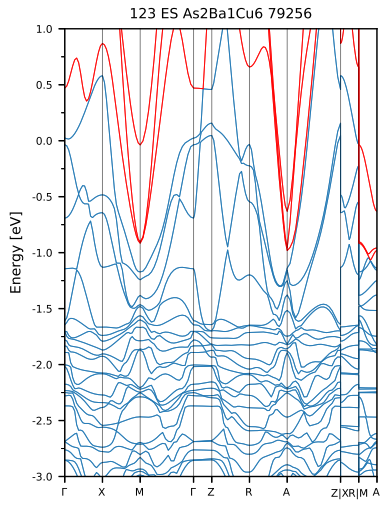

(by) 123.3.22

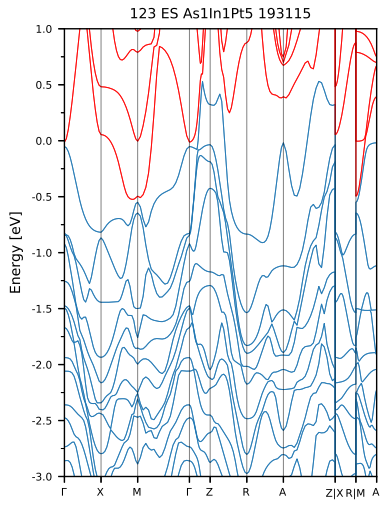

(bz) 123.3.23

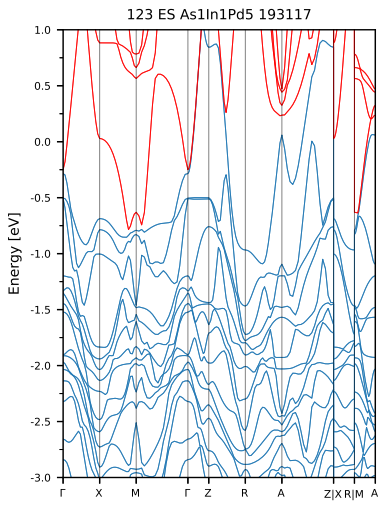

(ca) 123.3.24

Figure 224: Band structures for topological materials in SG 123, part 5/5, from 123.3.10 to 123.3.24. See the list of figures.

### 0.0.55 SG 124

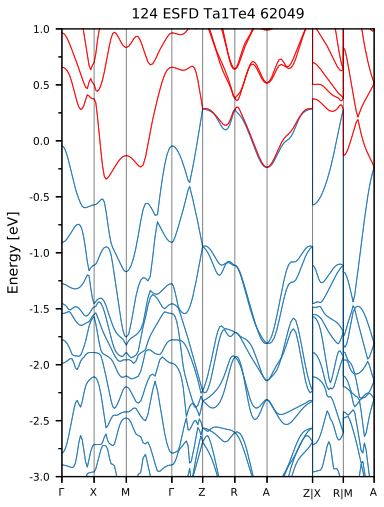

(a) 124.4.1

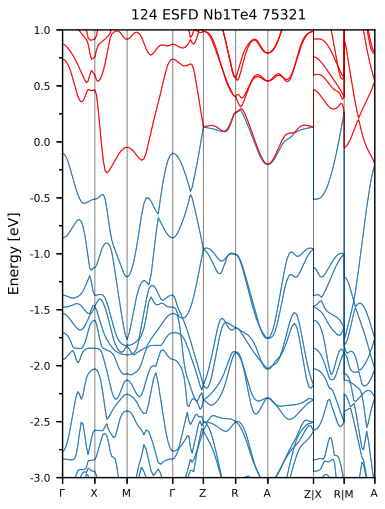

(b) 124.4.2

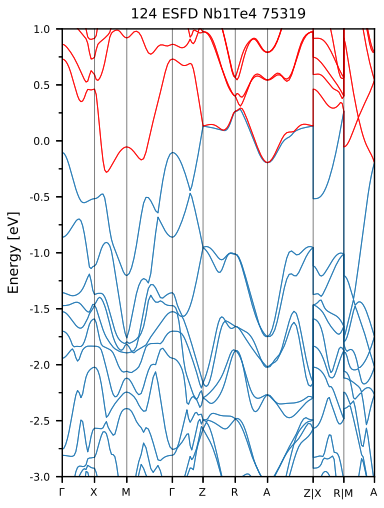

(c) 124.4.3

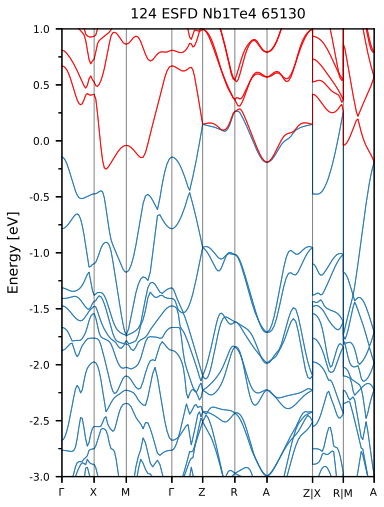

(d) 124.4.4

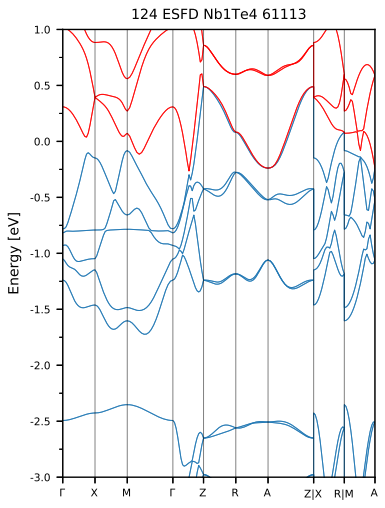

(e) 124.4.5

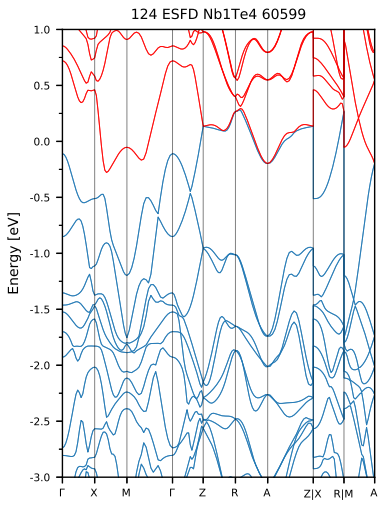

(f) 124.4.6

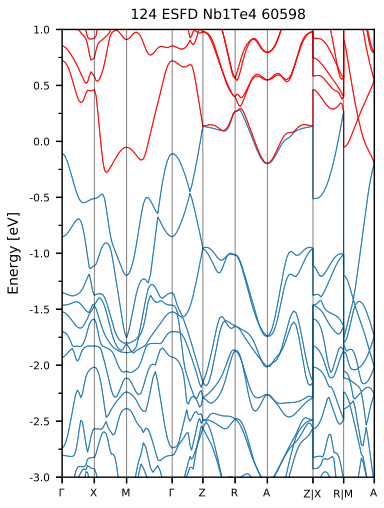

(g) 124.4.7

Figure 225: Band structures for topological materials in SG 124, part 1/1, from 124.4.1 to 124.4.7. See the list of figures.

### 0.0.56 SG 125

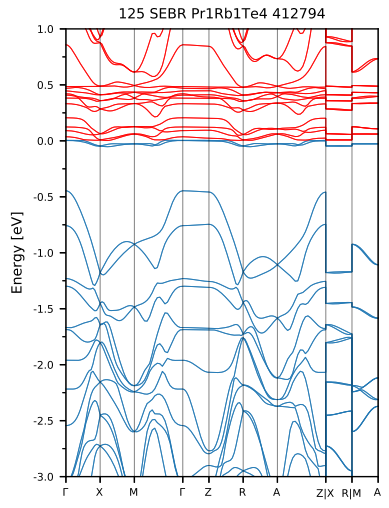

(a) 125.1.1

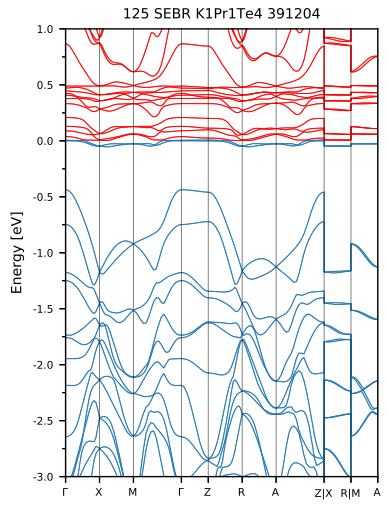

(b) 125.1.2

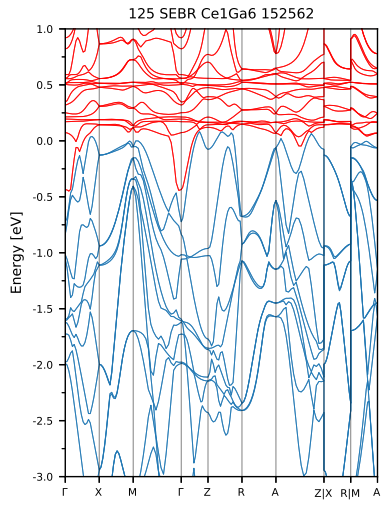

(c) 125.1.3

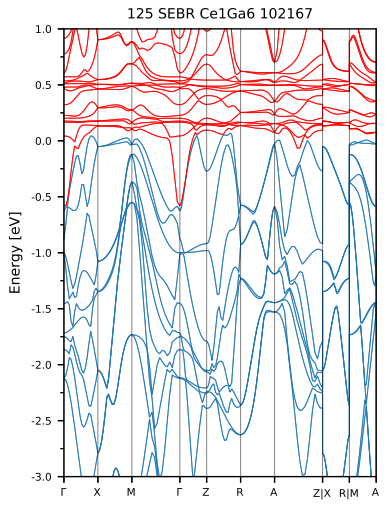

(d) 125.1.4

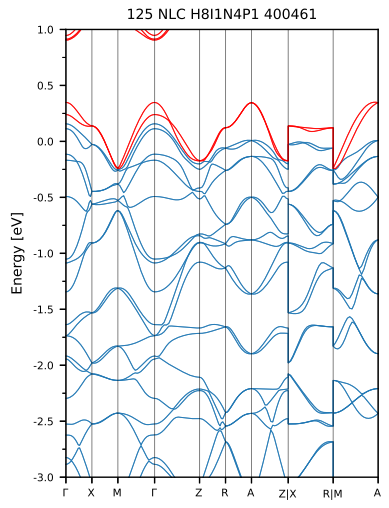

(e) 125.2.1

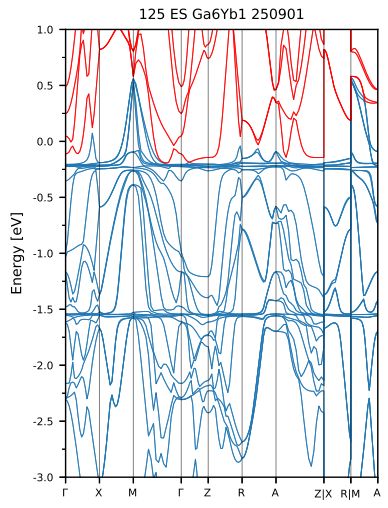

(f) 125.3.1

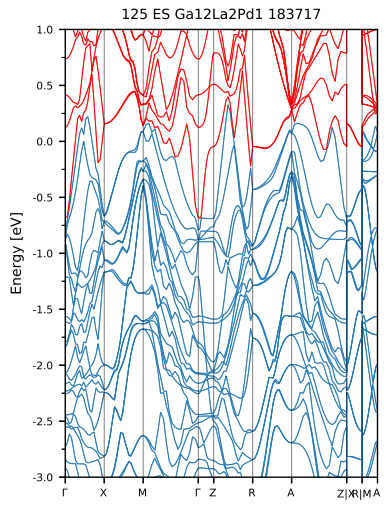

(g) 125.3.2

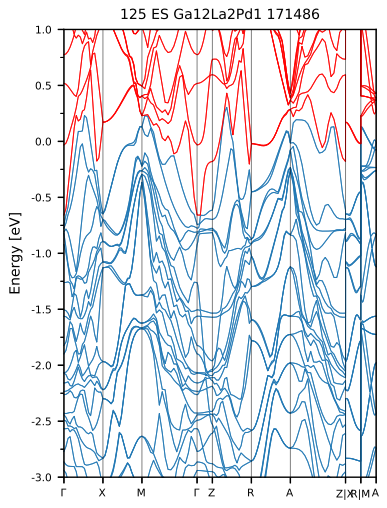

(h) 125.3.3

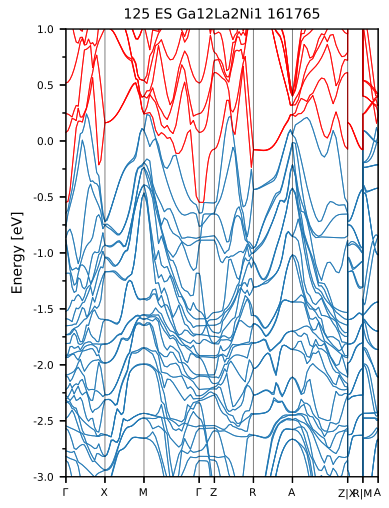

(i) 125.3.4

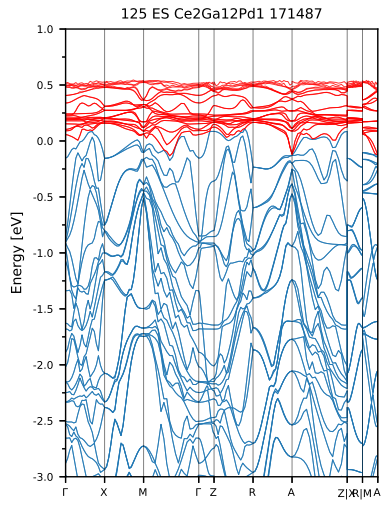

(j) 125.3.5

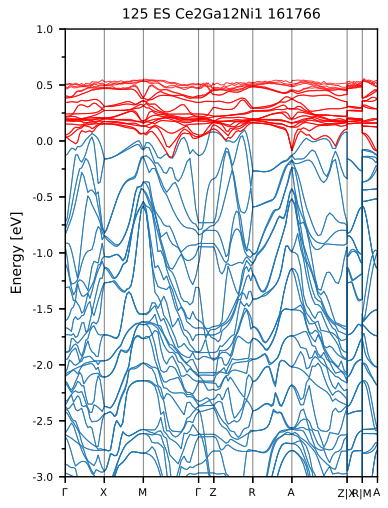

(k) 125.3.6

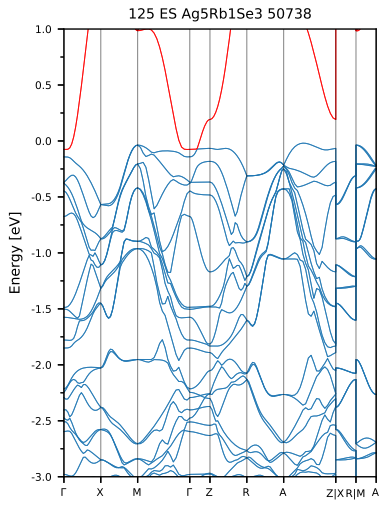

(l) 125.3.7

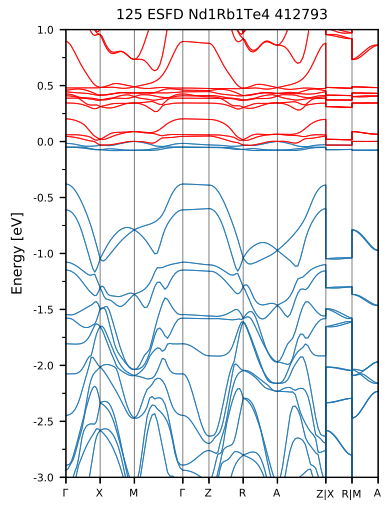

(m) 125.4.1

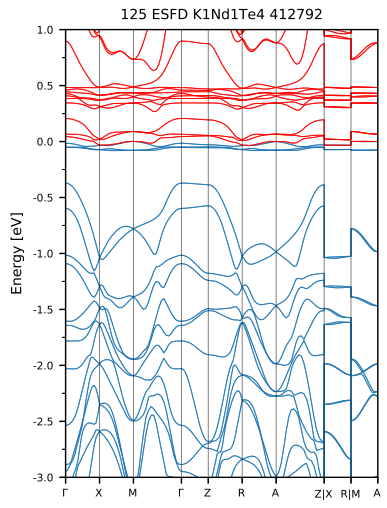

(n) 125.4.2

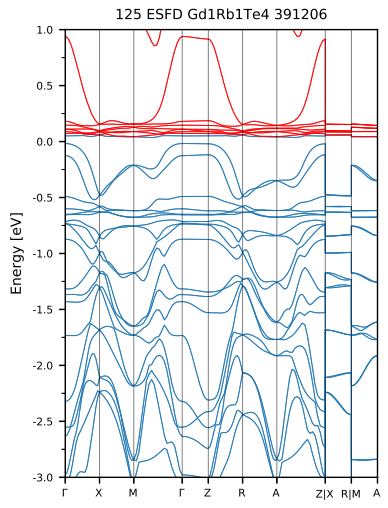

(o) 125.4.3

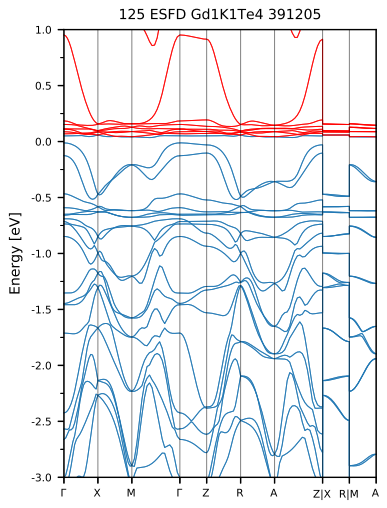

(p) 125.4.4

Figure 226: Band structures for topological materials in SG 125, part 1/2, from 125.1.1 to 125.4.4. See the list of figures.

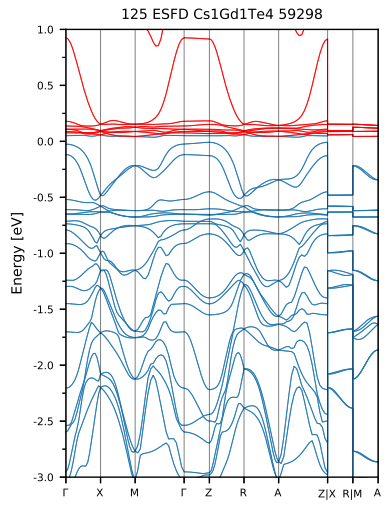

(q) 125.4.5

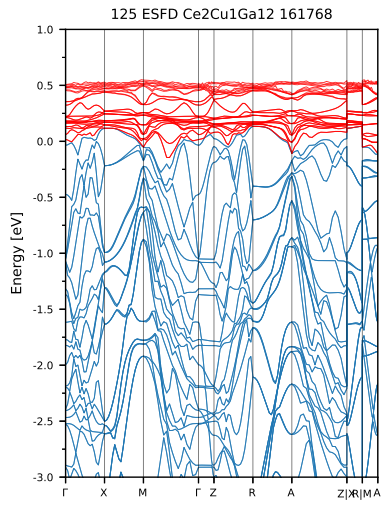

(r) 125.4.6

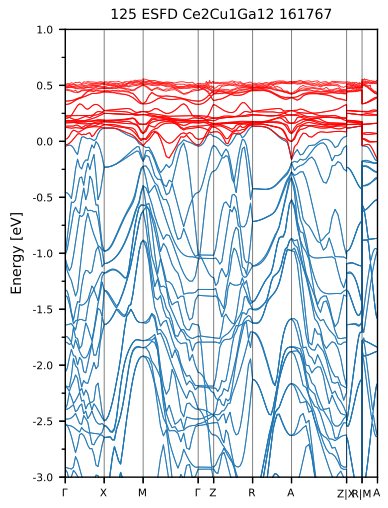

(s) 125.4.7

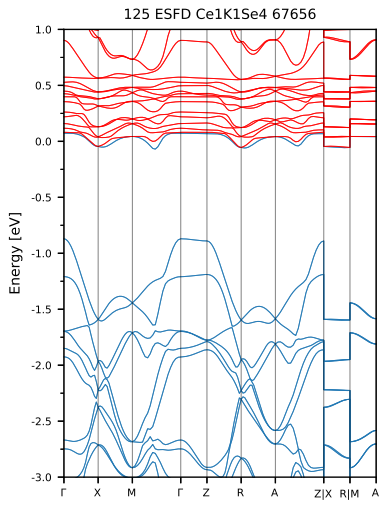

(t) 125.4.8

Figure 227: Band structures for topological materials in SG 125, part 2/2, from 125.4.5 to 125.4.8. See the list of figures.

## 0.0.57 SG 127

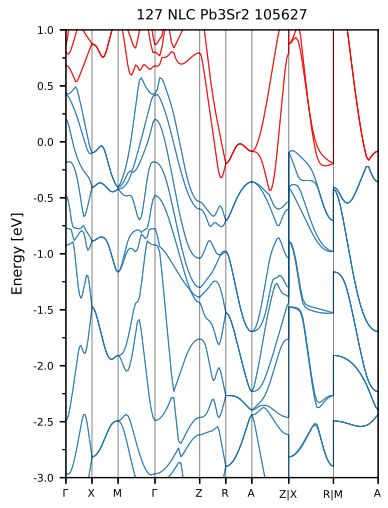

(a) 127.2.1

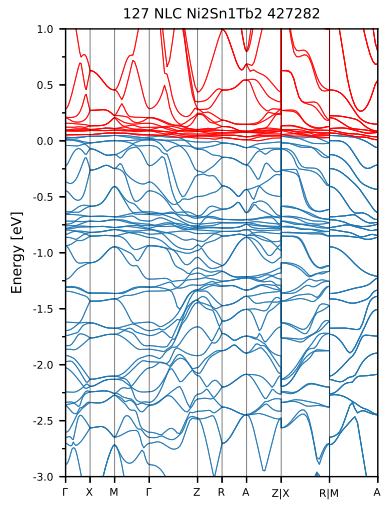

(b) 127.2.2

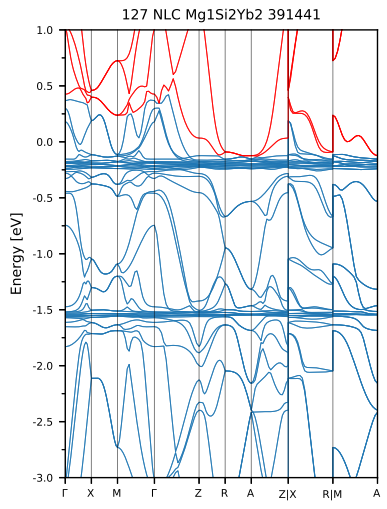

(c) 127.2.3

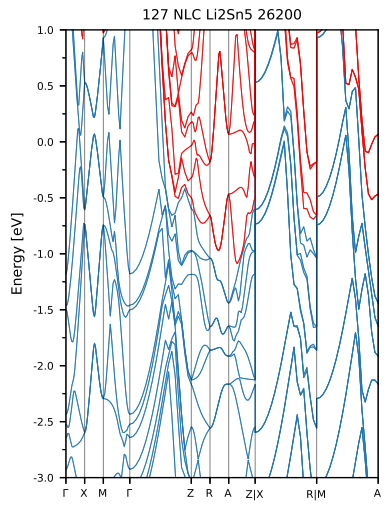

(d) 127.2.4

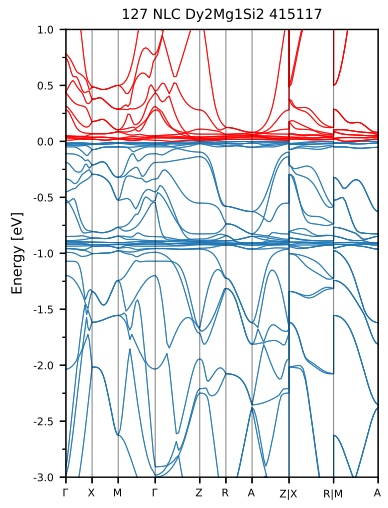

(e) 127.2.5

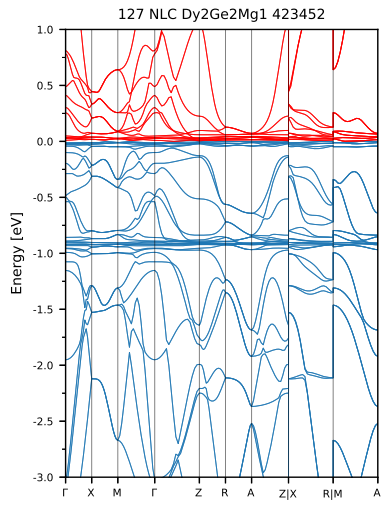

(f) 127.2.6

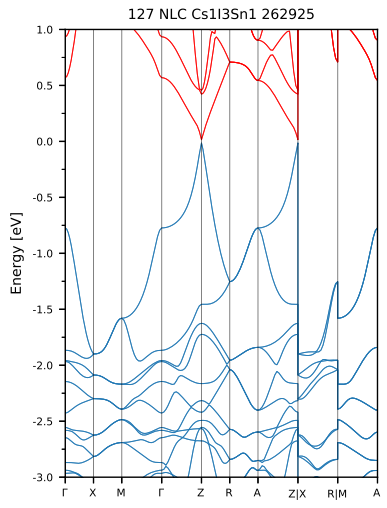

(g) 127.2.7

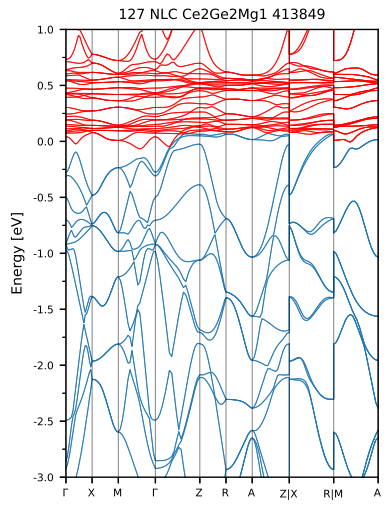

(h) 127.2.8

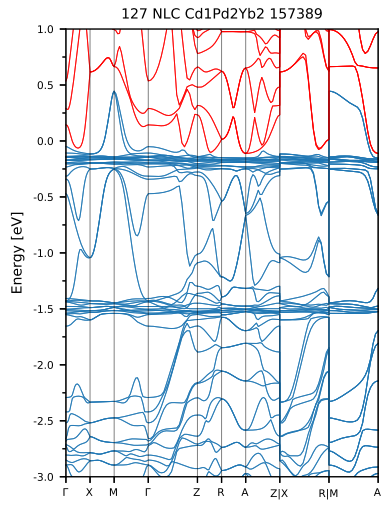

(i) 127.2.9

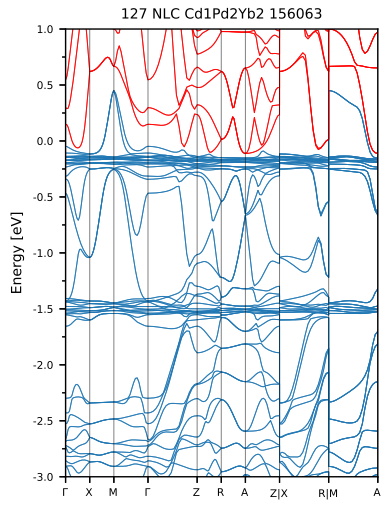

(j) 127.2.10

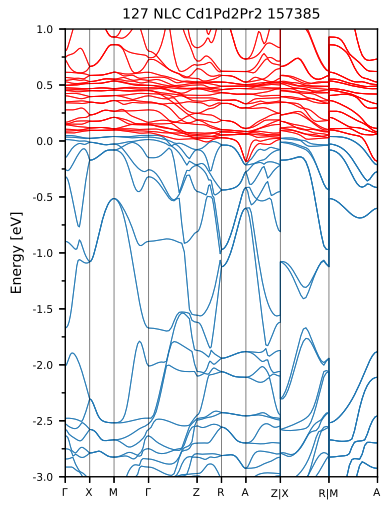

(k) 127.2.11

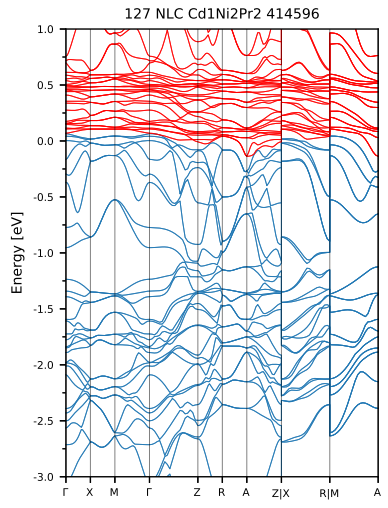

(l) 127.2.12

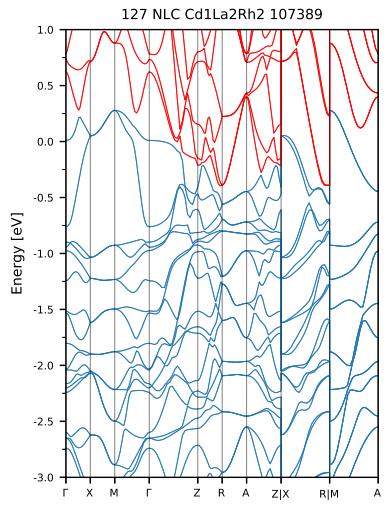

(m) 127.2.13

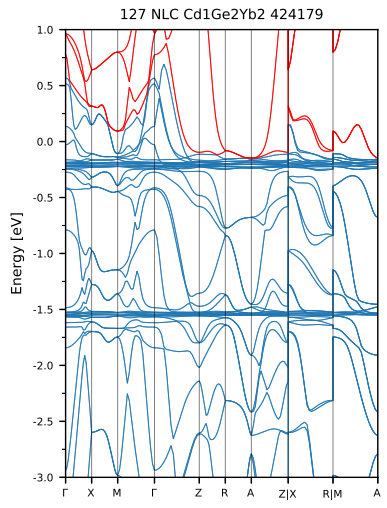

(n) 127.2.14

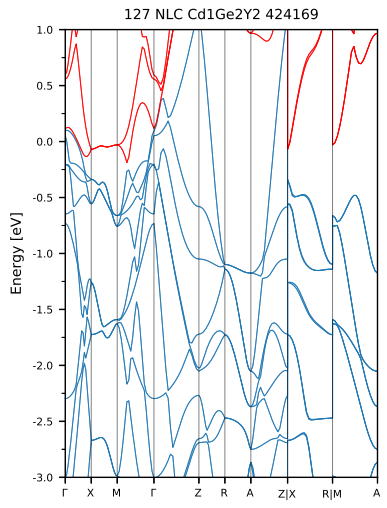

(o) 127.2.15

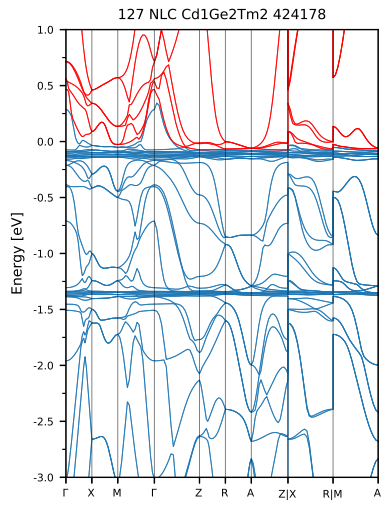

(p) 127.2.16

Figure 228: Band structures for topological materials in SG 127, part 1/10, from 127.2.1 to 127.2.16. See the list of figures.

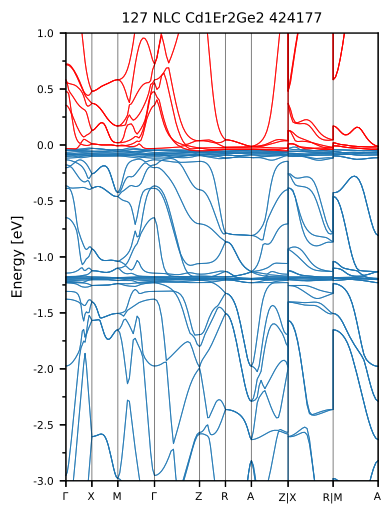

(q) 127.2.17

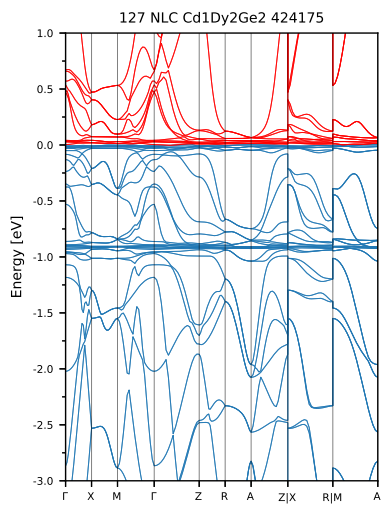

(r) 127.2.18

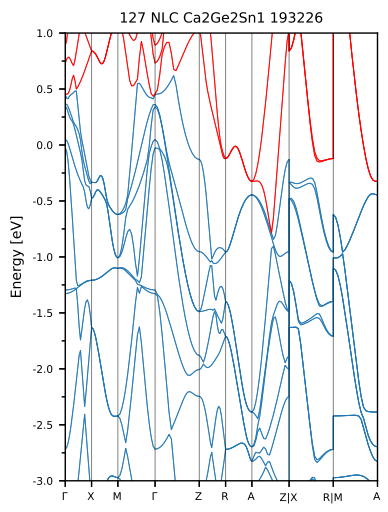

(s) 127.2.19

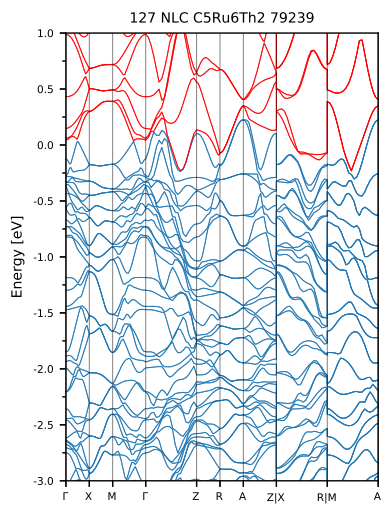

(t) 127.2.20

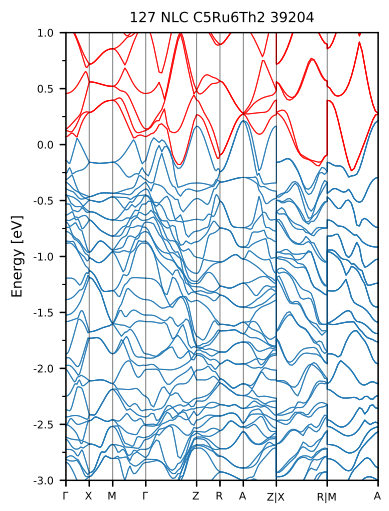

(u) 127.2.21

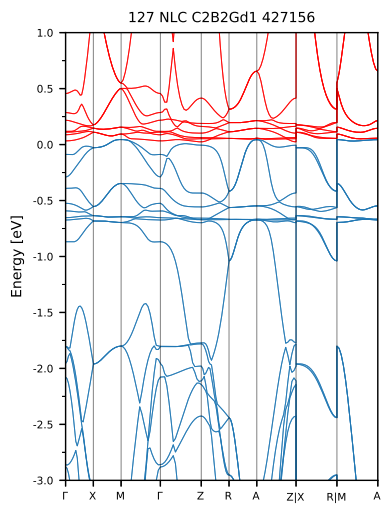

(v) 127.2.22

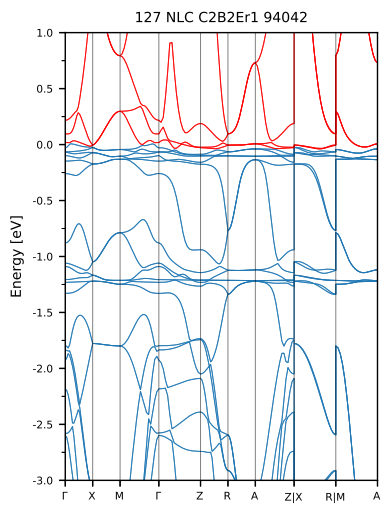

(w) 127.2.23

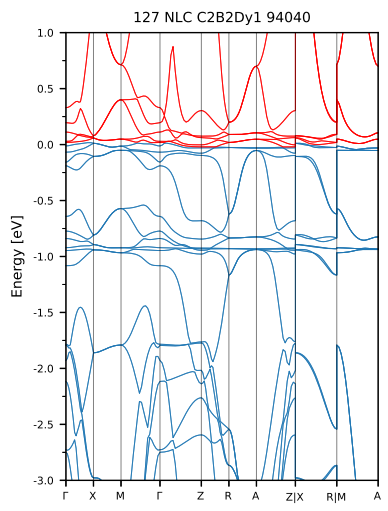

(x) 127.2.24

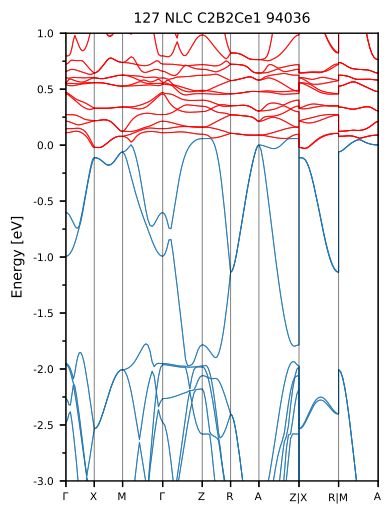

(y) 127.2.25

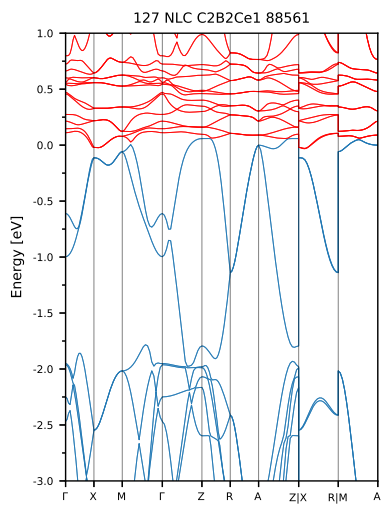

(z) 127.2.26

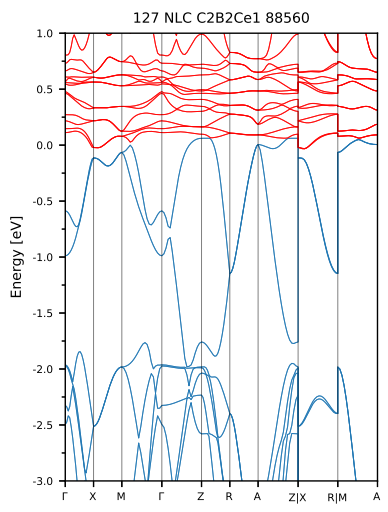

(aa) 127.2.27

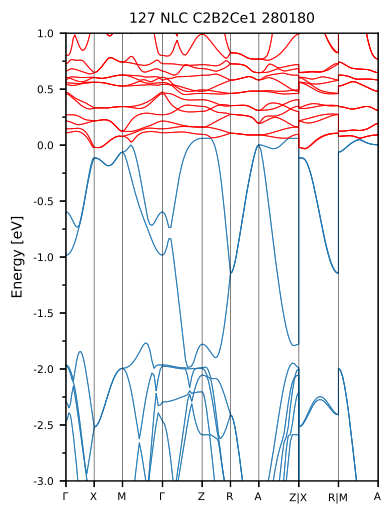

(ab) 127.2.28

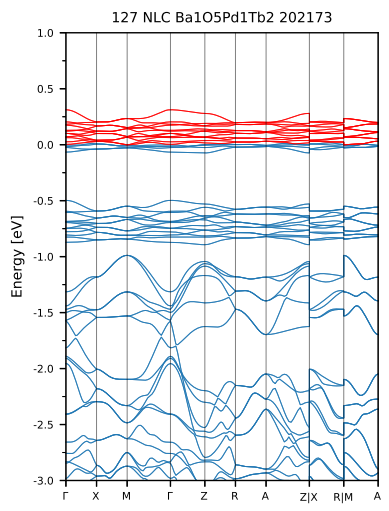

(ac) 127.2.29

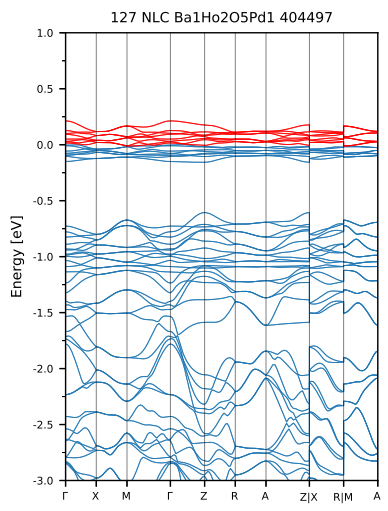

(ad) 127.2.30

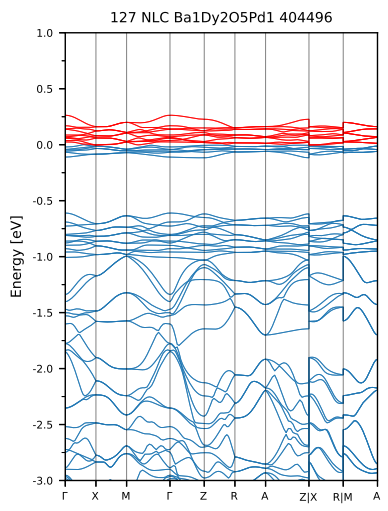

(ae) 127.2.31

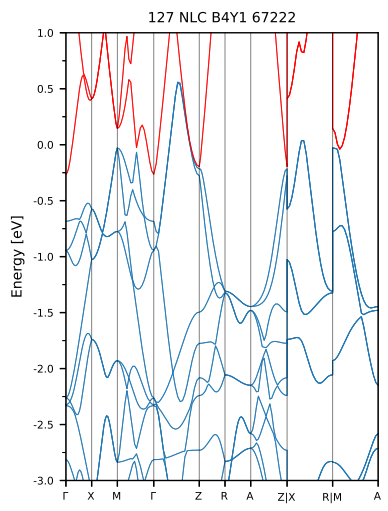

(af) 127.2.32

Figure 229: Band structures for topological materials in SG 127, part 2/10, from 127.2.17 to 127.2.32. See the list of figures.

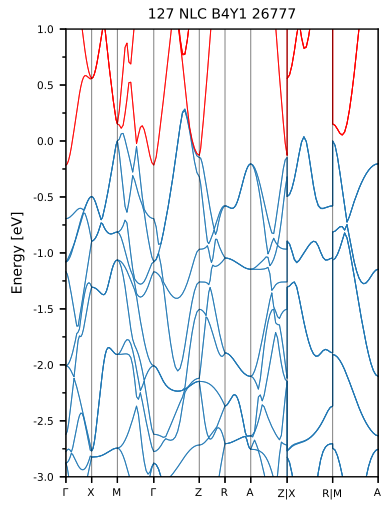

(ag) 127.2.33

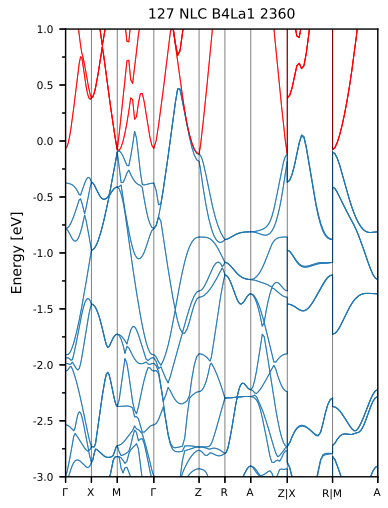

(ah) 127.2.34

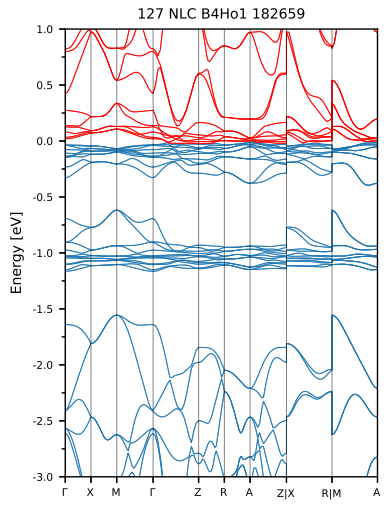

(ai) 127.2.35

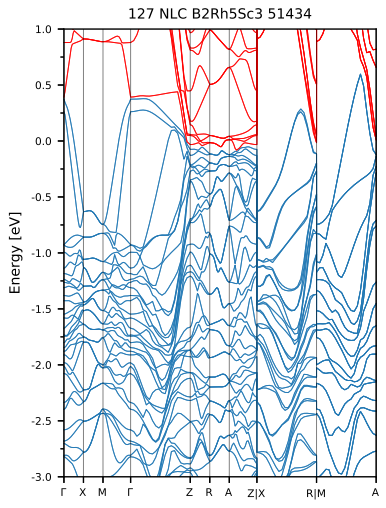

(aj) 127.2.36

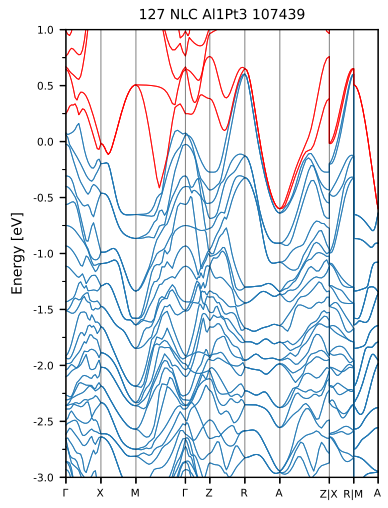

(ak) 127.2.37

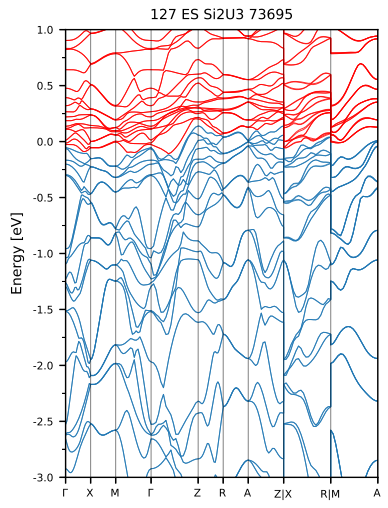

(al) 127.3.1

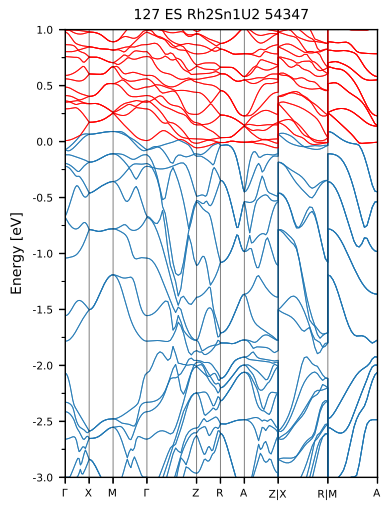

(am) 127.3.2

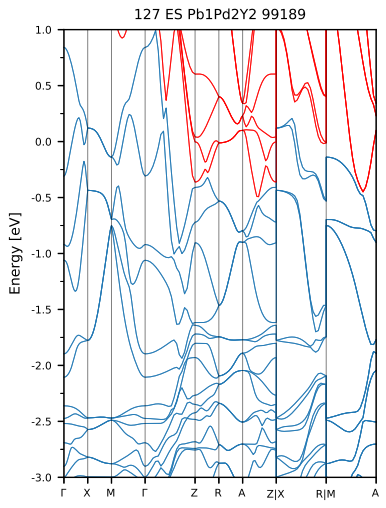

(an) 127.3.3

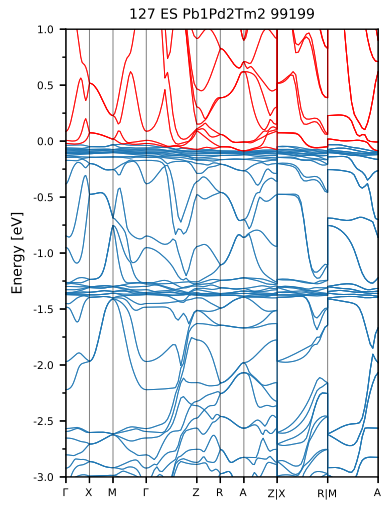

(ao) 127.3.4

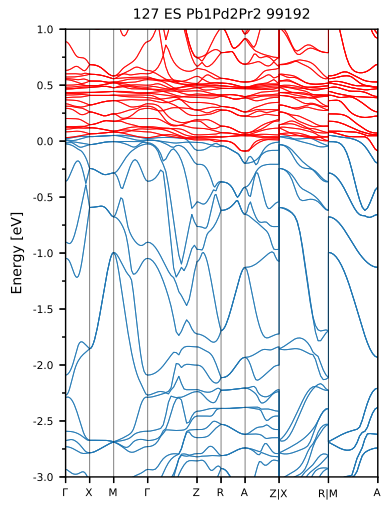

(ap) 127.3.5

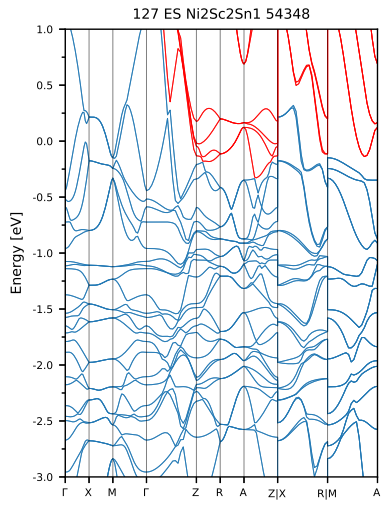

(aq) 127.3.6

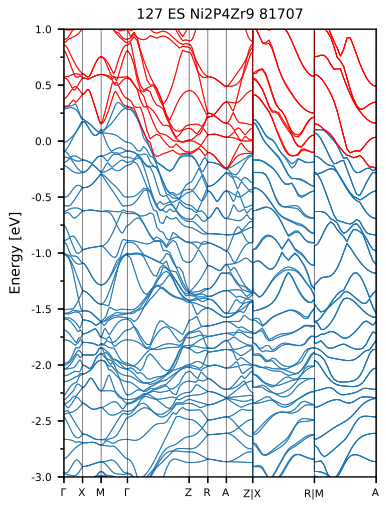

(ar) 127.3.7

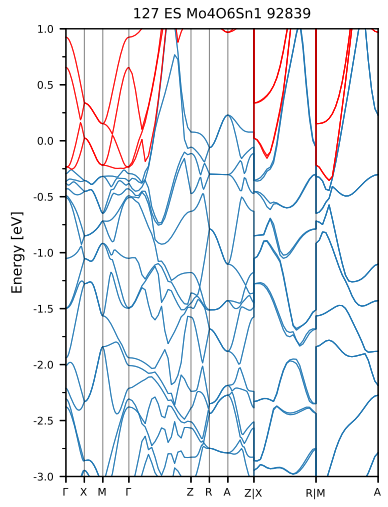

(as) 127.3.8

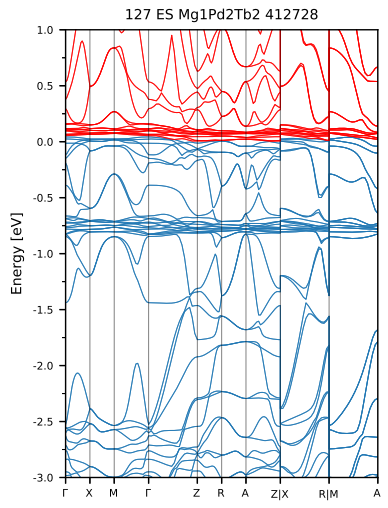

(at) 127.3.9

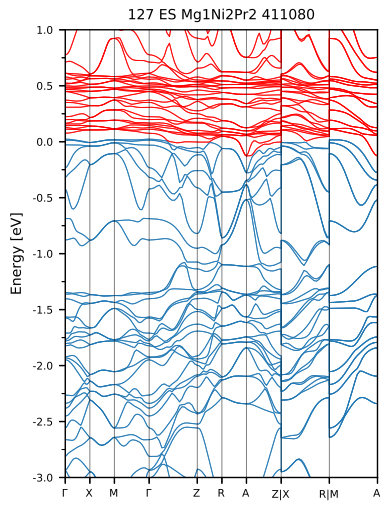

(au) 127.3.10

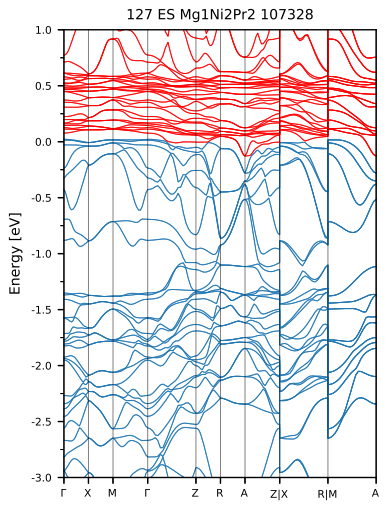

(av) 127.3.11

Figure 230: Band structures for topological materials in SG 127, part 3/10, from 127.2.33 to 127.3.11. See the list of figures.

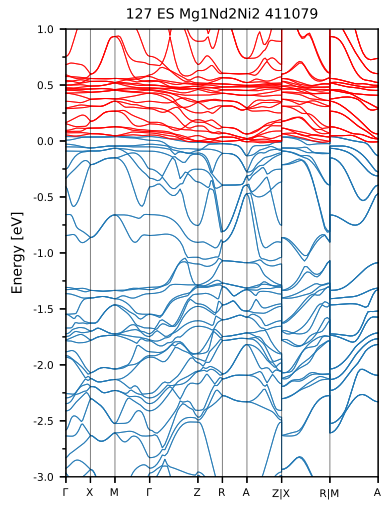

(aw) 127.3.12

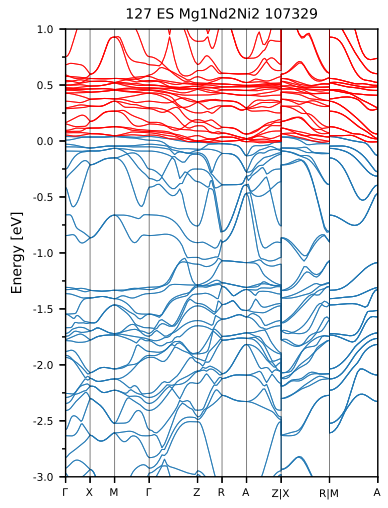

(ax) 127.3.13

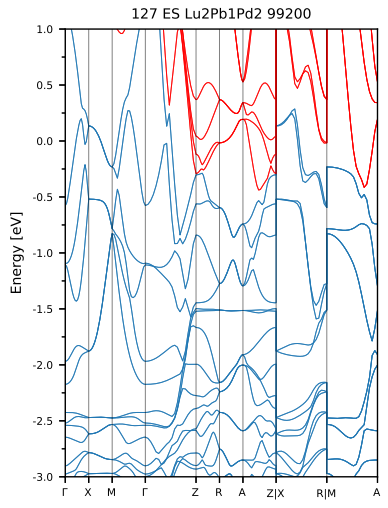

(ay) 127.3.14

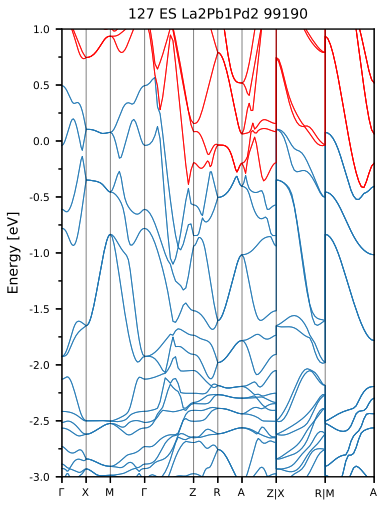

(az) 127.3.15

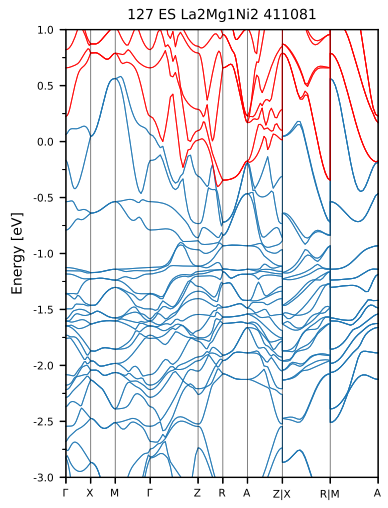

(ba) 127.3.16

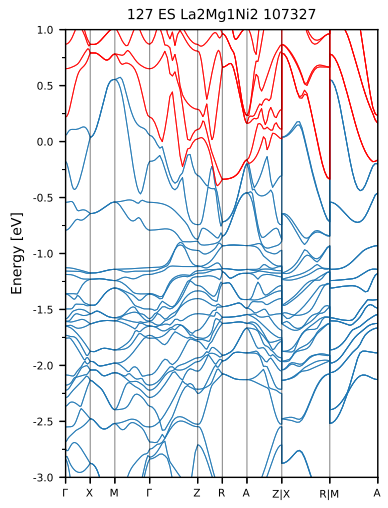

(bb) 127.3.17

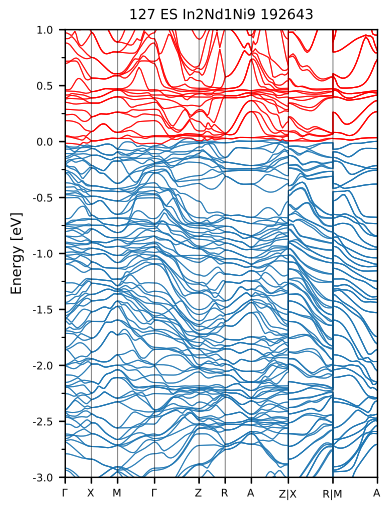

(bc) 127.3.18

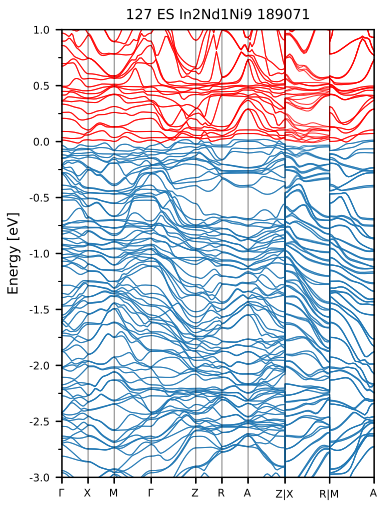

(bd) 127.3.19

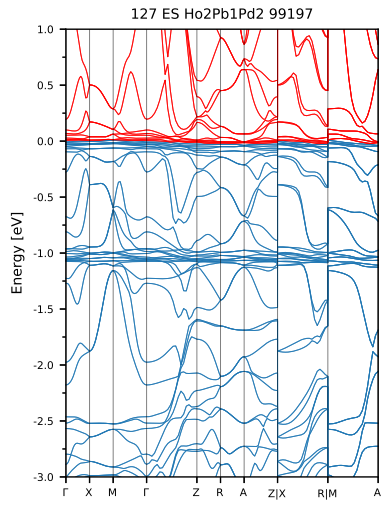

(be) 127.3.20

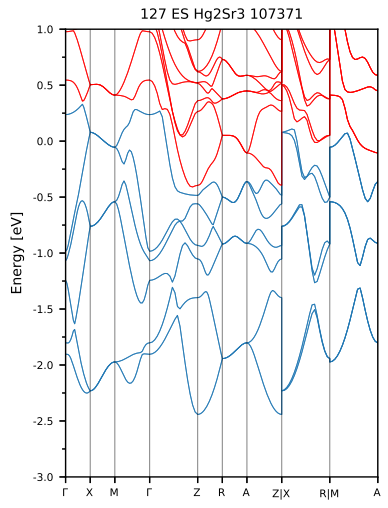

(bf) 127.3.21

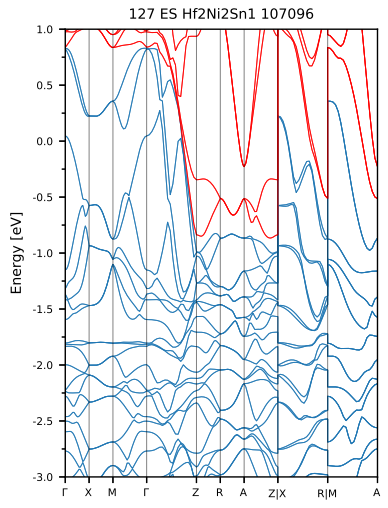

(bg) 127.3.22

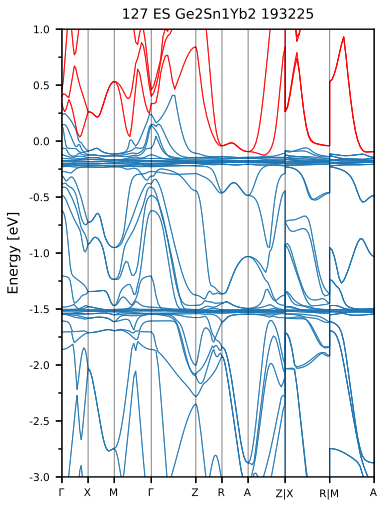

(bh) 127.3.23

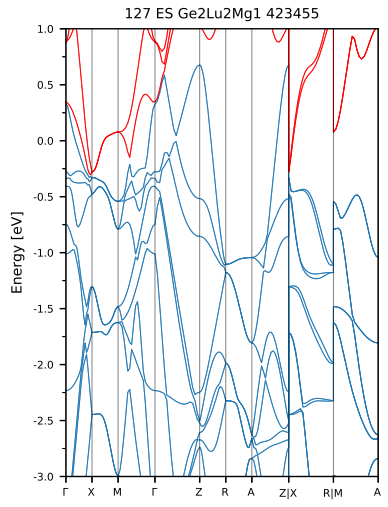

(bi) 127.3.24

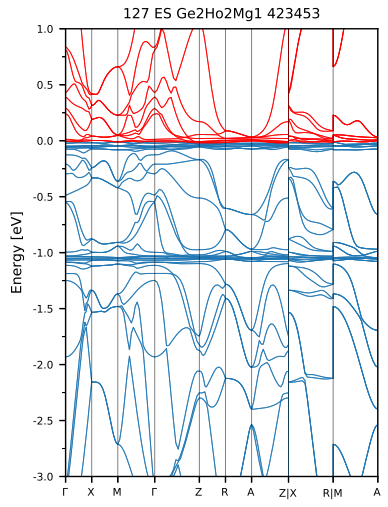

(bj) 127.3.25

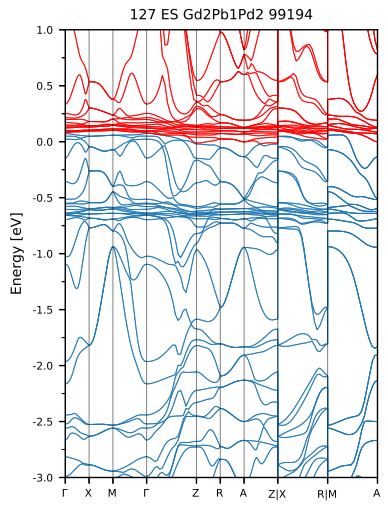

(bk) 127.3.26

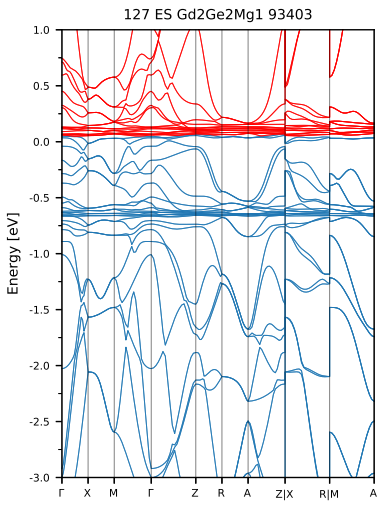

(bl) 127.3.27

Figure 231: Band structures for topological materials in SG 127, part 4/10, from 127.3.12 to 127.3.27. See the list of figures.

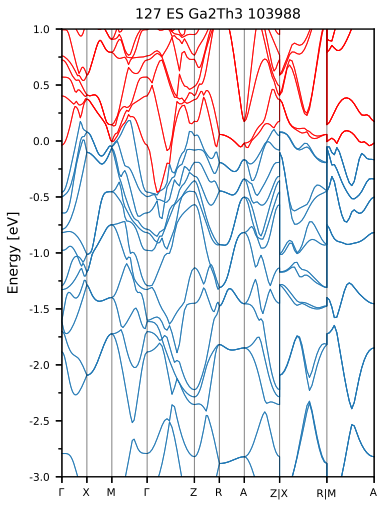

(bm) 127.3.28

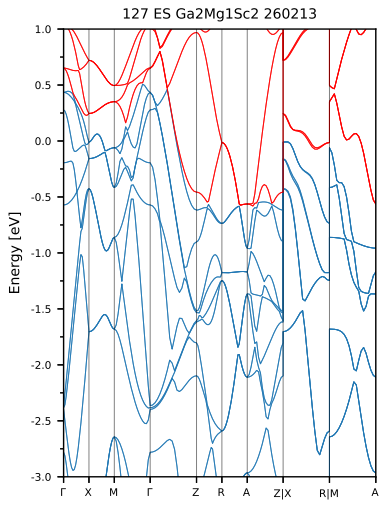

(bn) 127.3.29

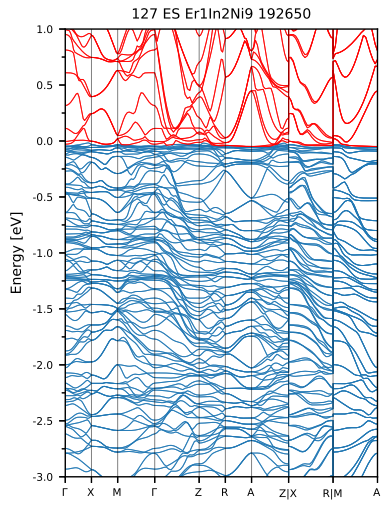

(bo) 127.3.30

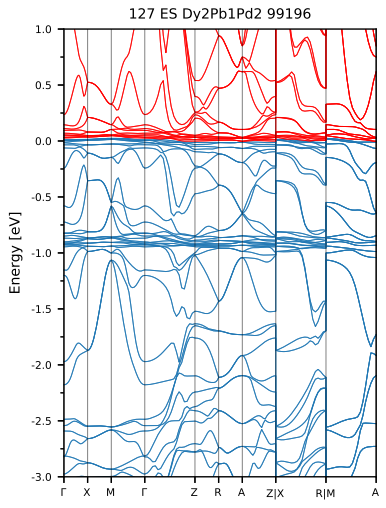

(bp) 127.3.31

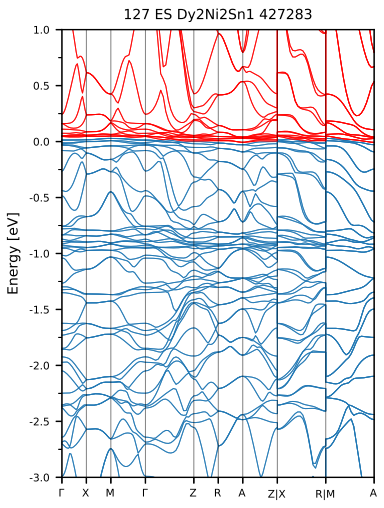

(bq) 127.3.32

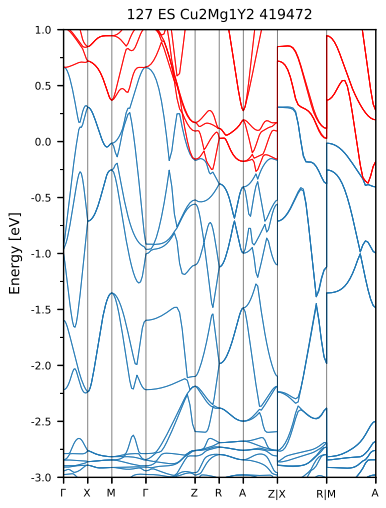

(br) 127.3.33

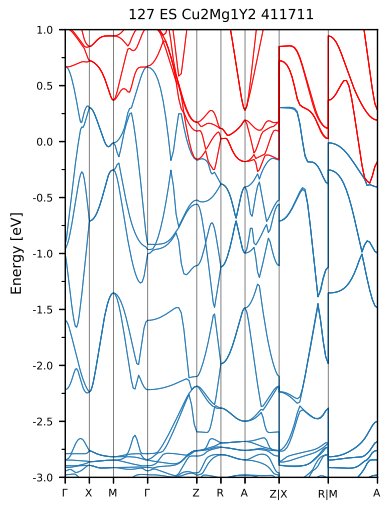

(bs) 127.3.34

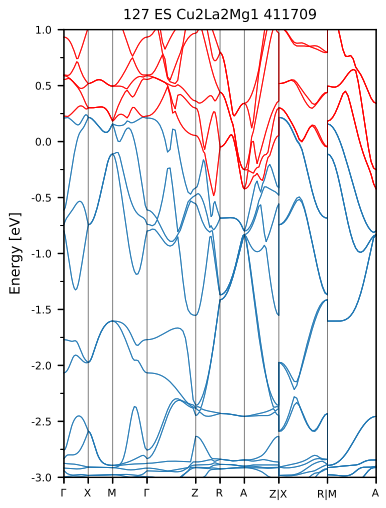

(bt) 127.3.35

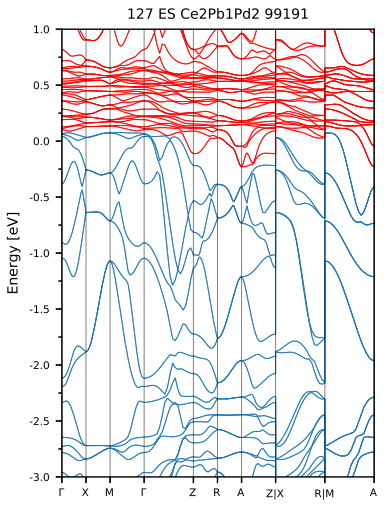

(bu) 127.3.36

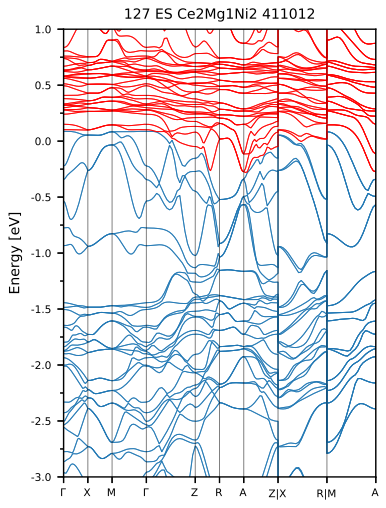

(bv) 127.3.37

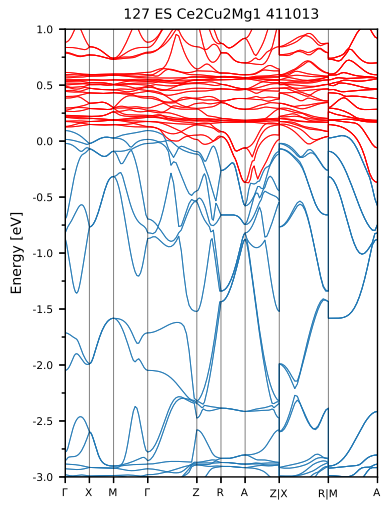

(bw) 127.3.38

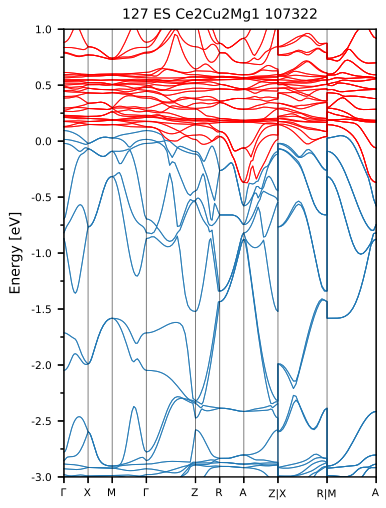

(bx) 127.3.39

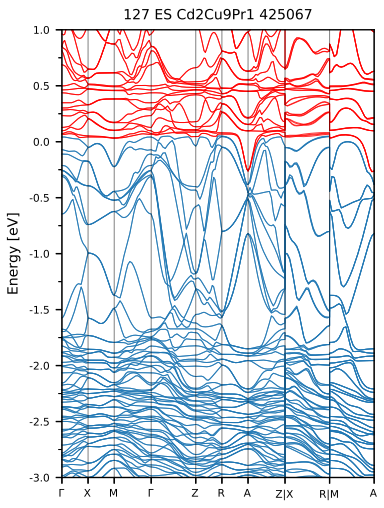

(by) 127.3.40

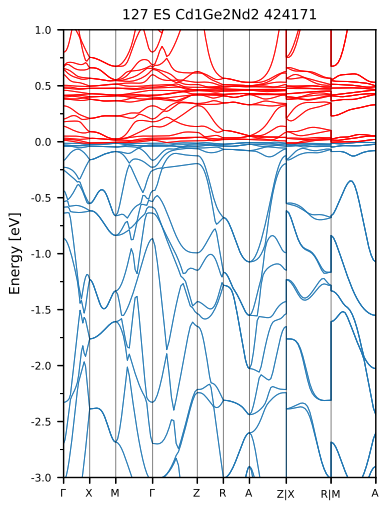

(bz) 127.3.41

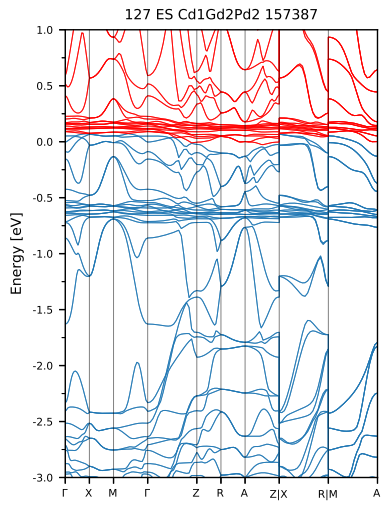

(ca) 127.3.42

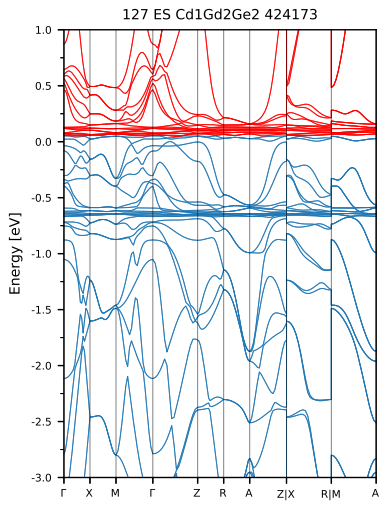

(cb) 127.3.43

Figure 232: Band structures for topological materials in SG 127, part 5/10, from 127.3.28 to 127.3.43. See the list of figures.

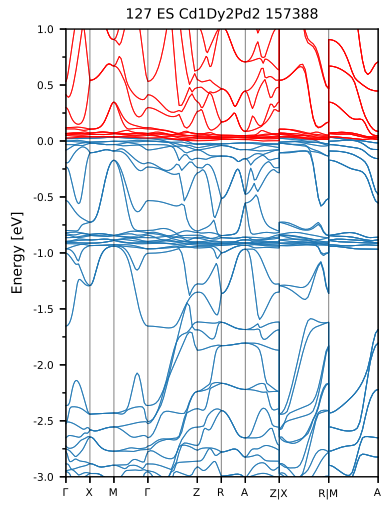

(cc) 127.3.44

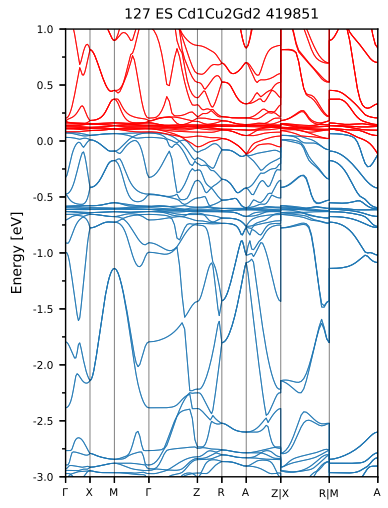

(cd) 127.3.45

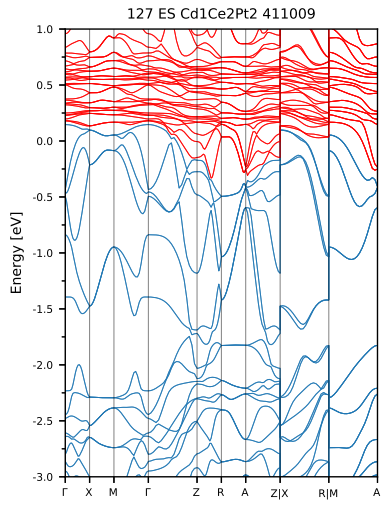

(ce) 127.3.46

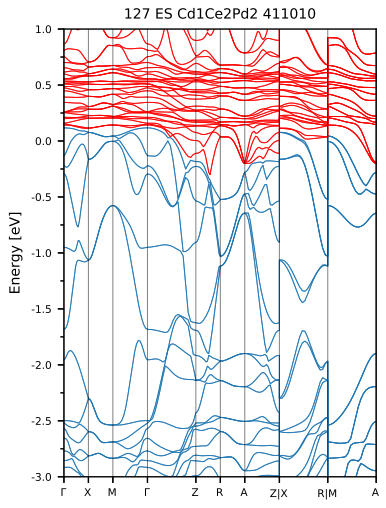

(cf) 127.3.47

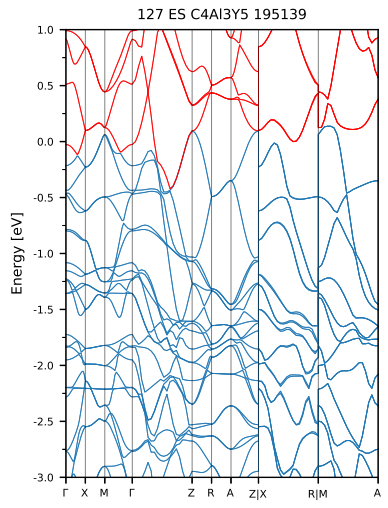

(cg) 127.3.48

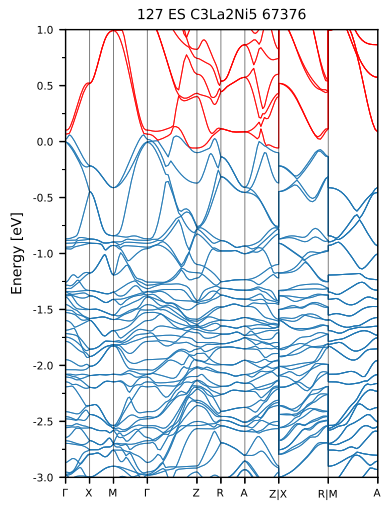

(ch) 127.3.49

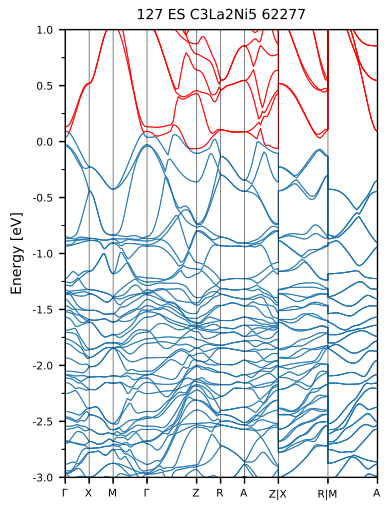

(ci) 127.3.50

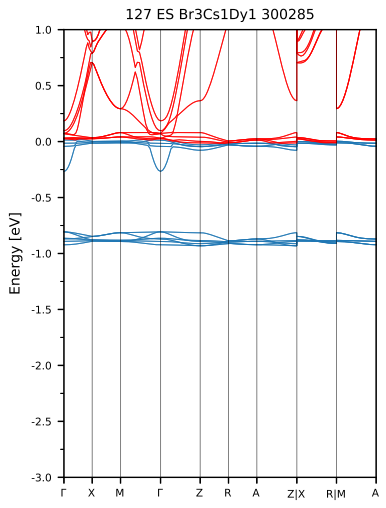

(cj) 127.3.51

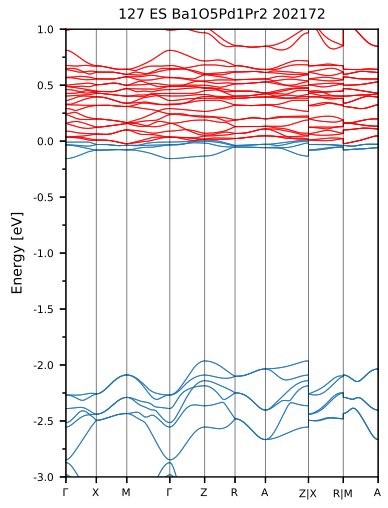

(ck) 127.3.52

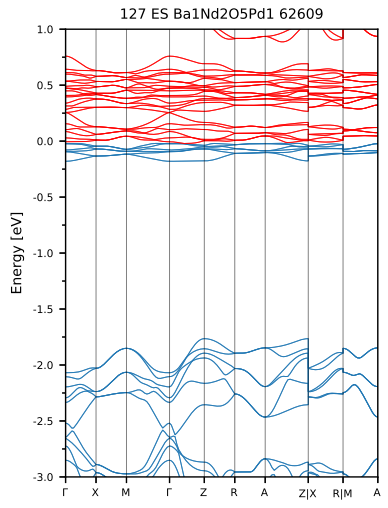

(cl) 127.3.53

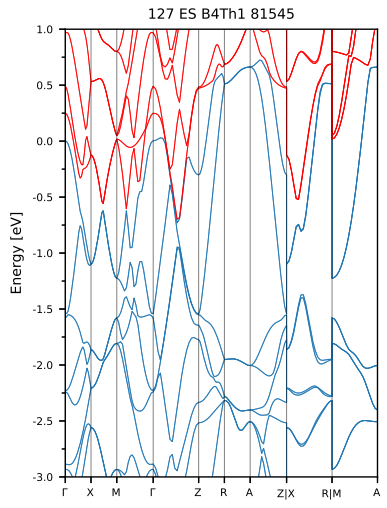

(cm) 127.3.54

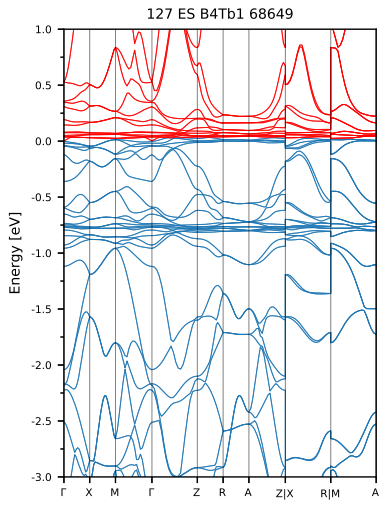

(cn) 127.3.55

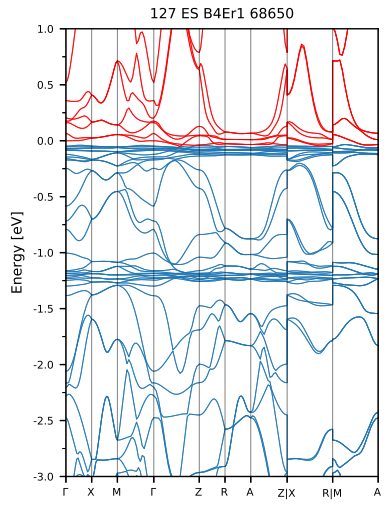

(co) 127.3.56

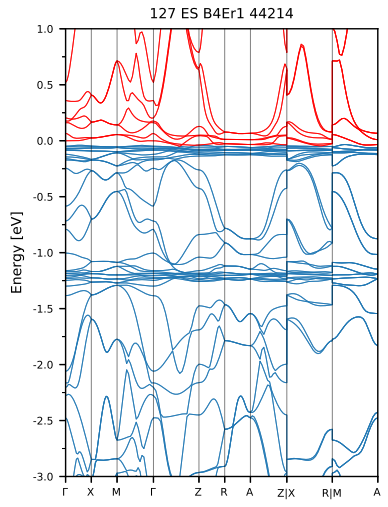

(cp) 127.3.57

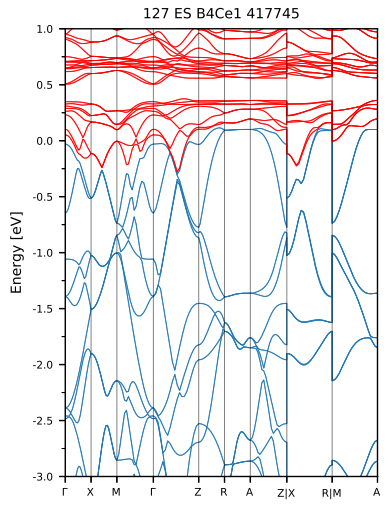

(cq) 127.3.58

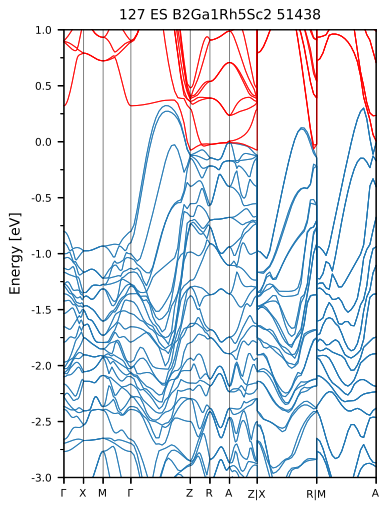

(cr) 127.3.59

Figure 233: Band structures for topological materials in SG 127, part 6/10, from 127.3.44 to 127.3.59. See the list of figures.

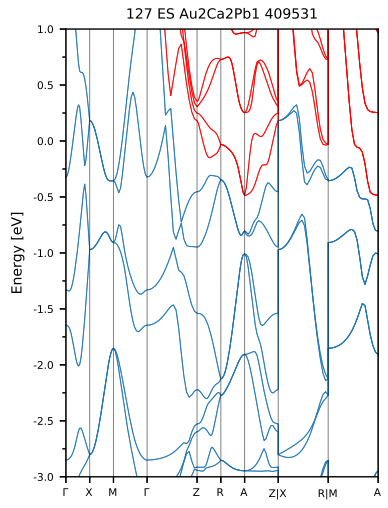

(cs) 127.3.60

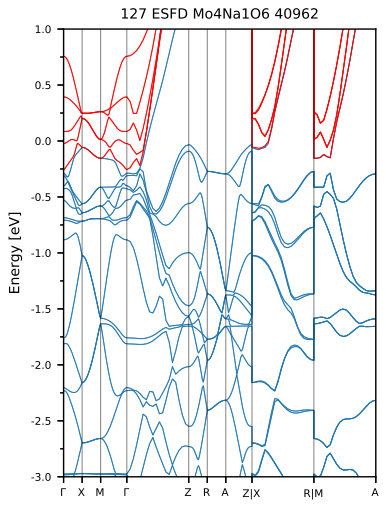

(ct) 127.4.1

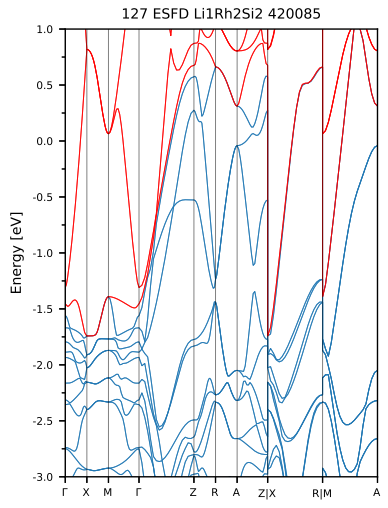

(cu) 127.4.2

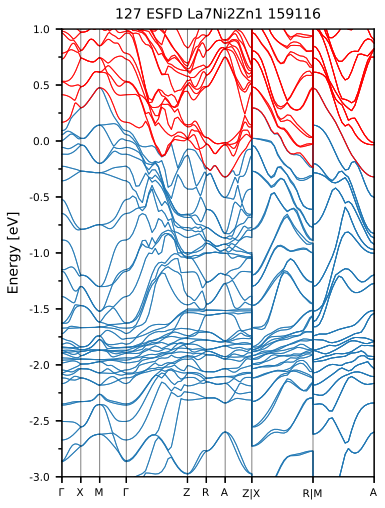

(cv) 127.4.3

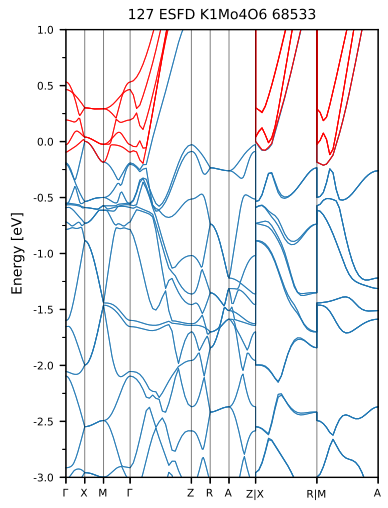

(cw) 127.4.4

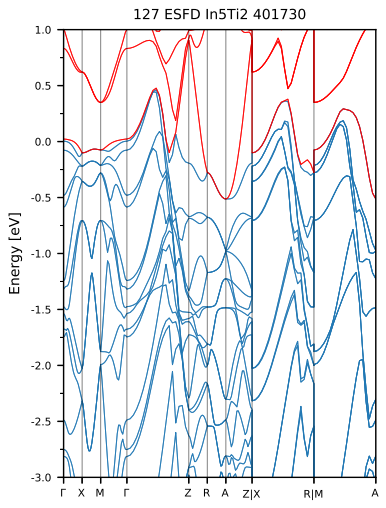

(cx) 127.4.5

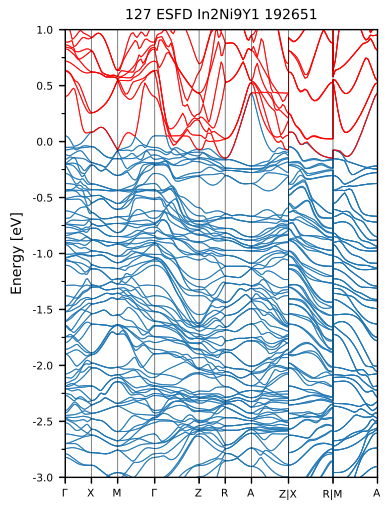

(cy) 127.4.6

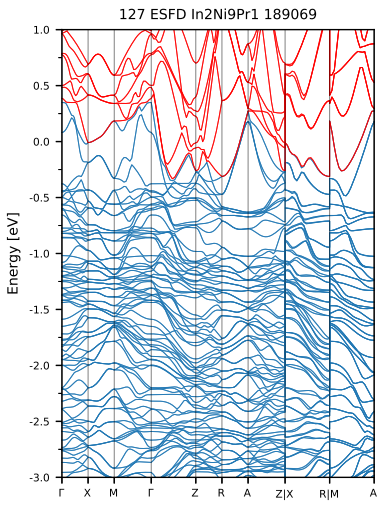

(cz) 127.4.7

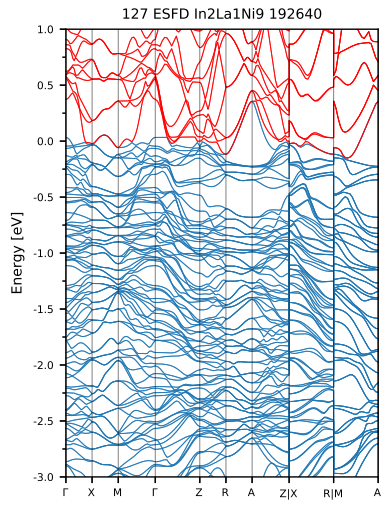

(da) 127.4.8

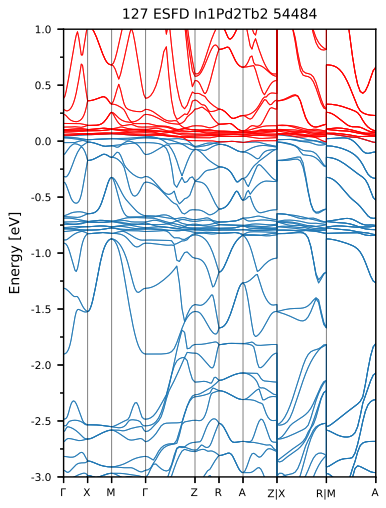

(db) 127.4.9

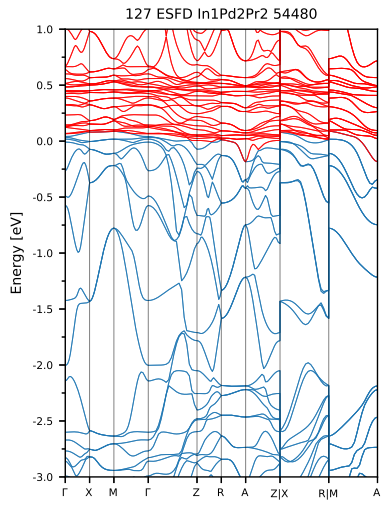

(dc) 127.4.10

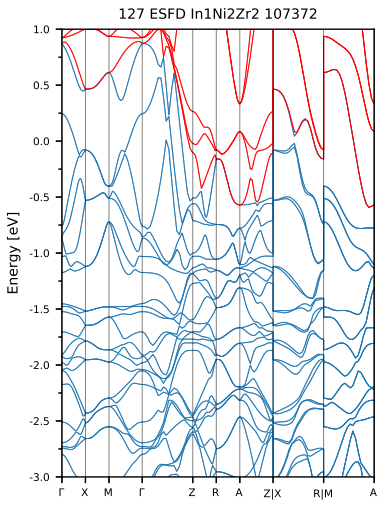

(dd) 127.4.11

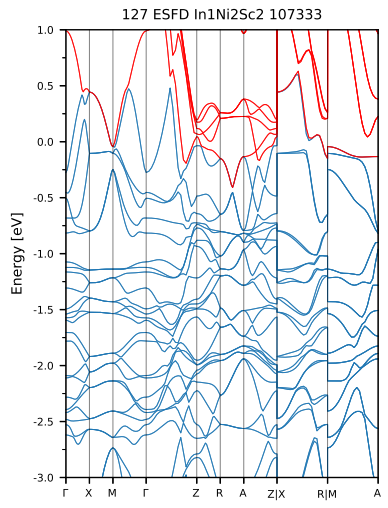

(de) 127.4.12

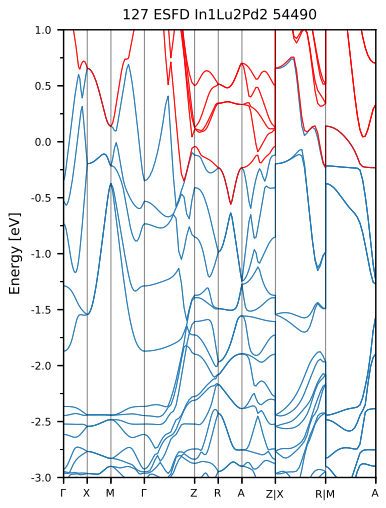

(df) 127.4.13

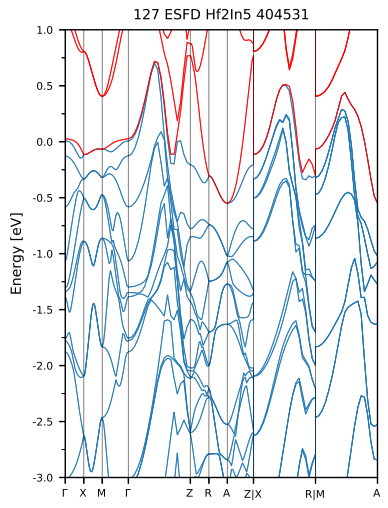

(dg) 127.4.14

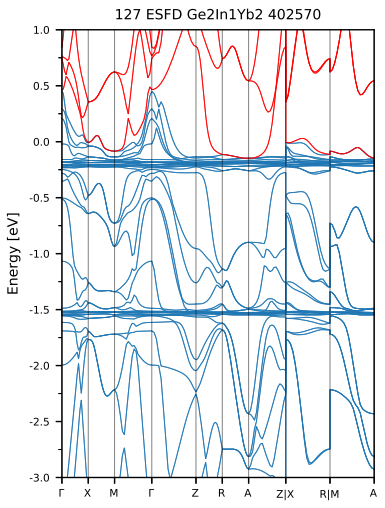

(dh) 127.4.15

Figure 234: Band structures for topological materials in SG 127, part 7/10, from 127.3.60 to 127.4.15. See the list of figures.

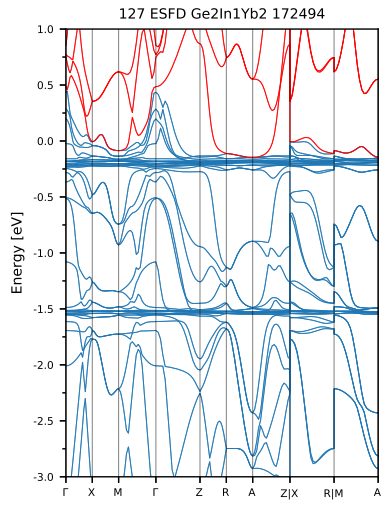

(di) 127.4.16

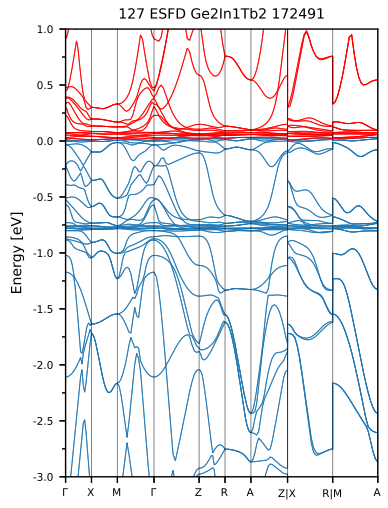

(dj) 127.4.17

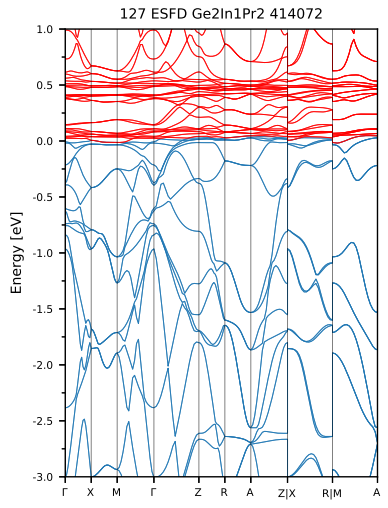

(dk) 127.4.18

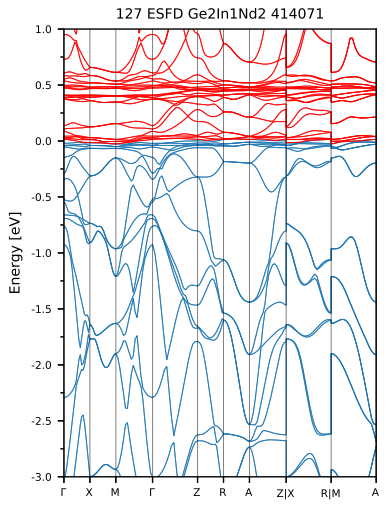

(dl) 127.4.19

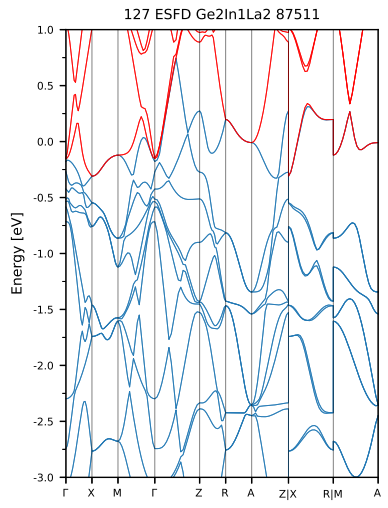

(dm) 127.4.20

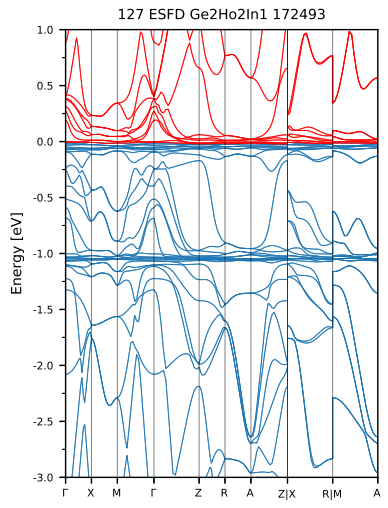

(dn) 127.4.21

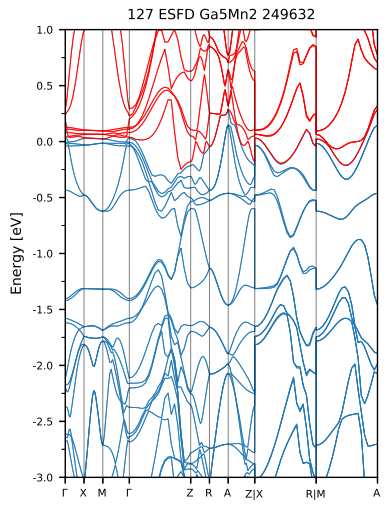

(do) 127.4.22

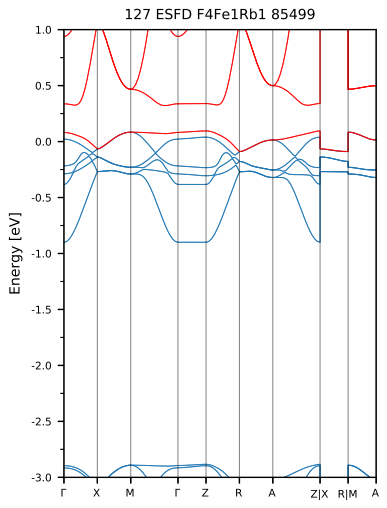

(dp) 127.4.23

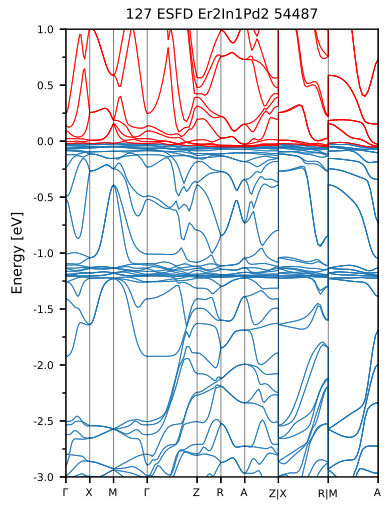

(dq) 127.4.24

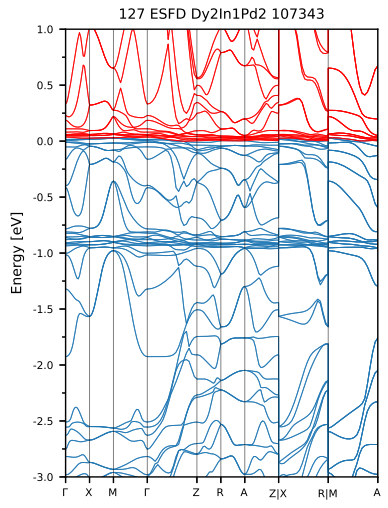

(dr) 127.4.25

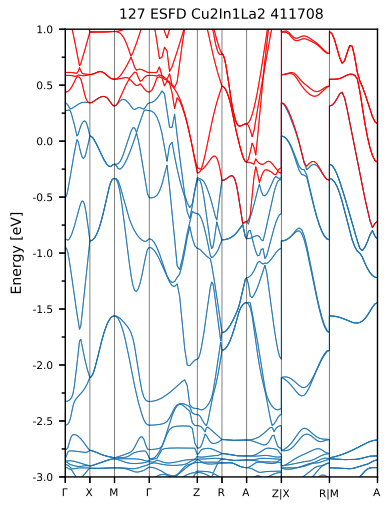

(ds) 127.4.26

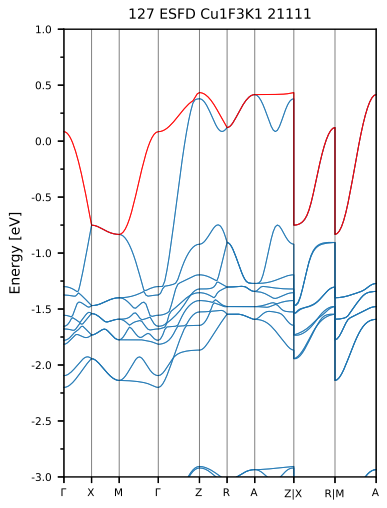

(dt) 127.4.27

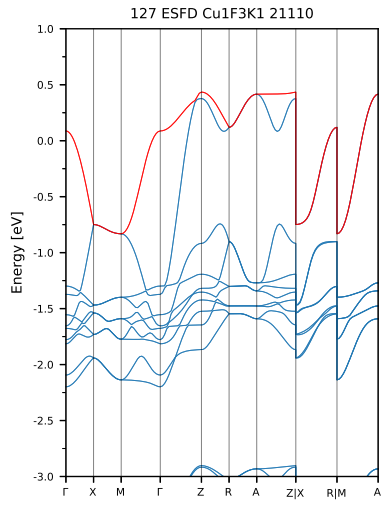

(du) 127.4.28

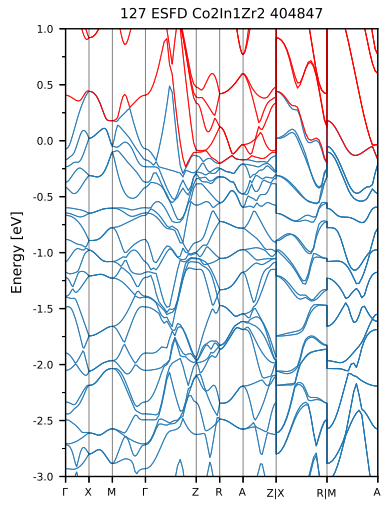

(dv) 127.4.29

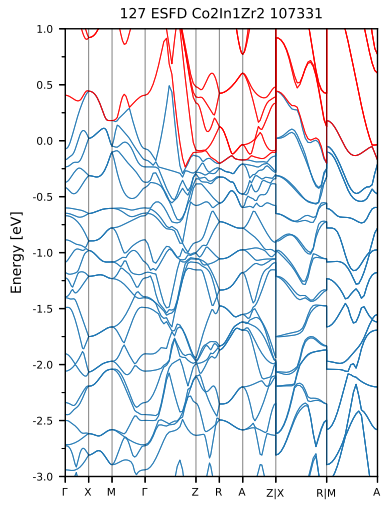

(dw) 127.4.30

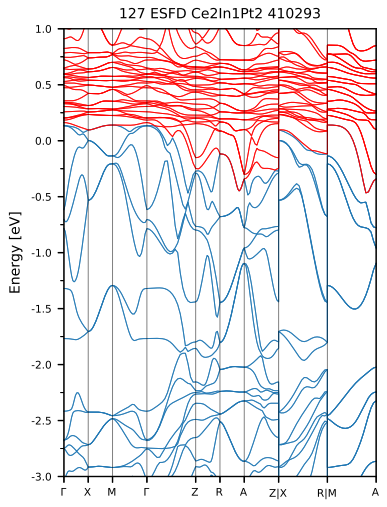

(dx) 127.4.31

Figure 235: Band structures for topological materials in SG 127, part 8/10, from 127.4.16 to 127.4.31. See the list of figures.

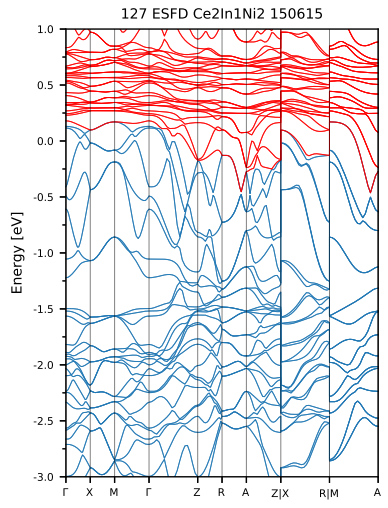

(dy) 127.4.32

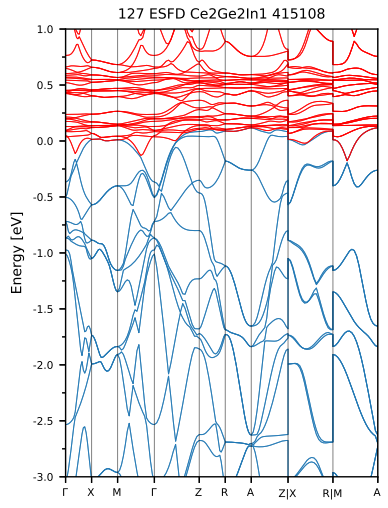

(dz) 127.4.33

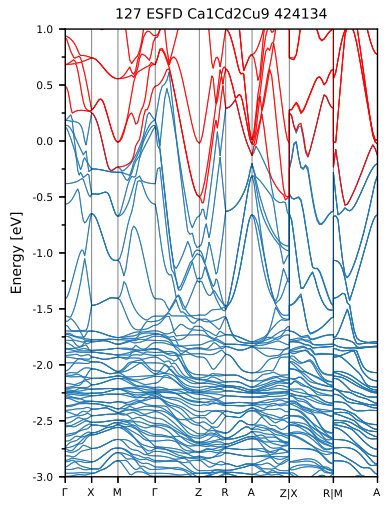

(ea) 127.4.34

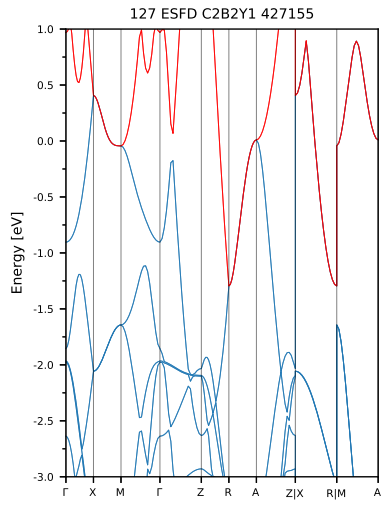

(eb) 127.4.35

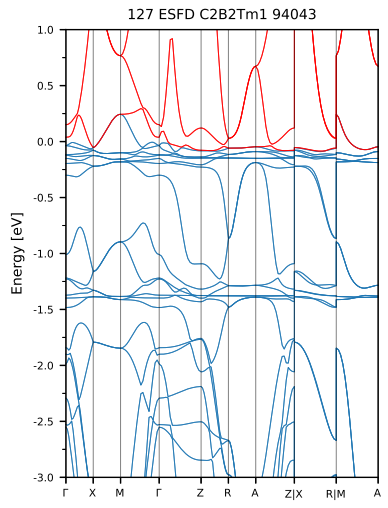

(ec) 127.4.36

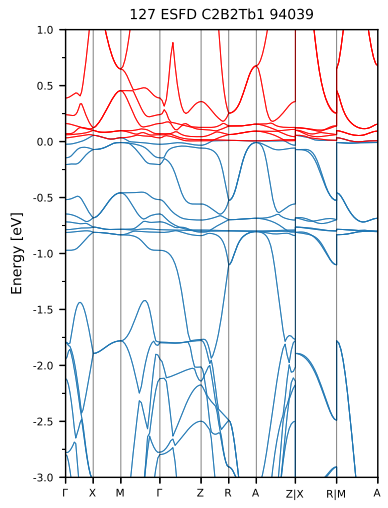

(ed) 127.4.37

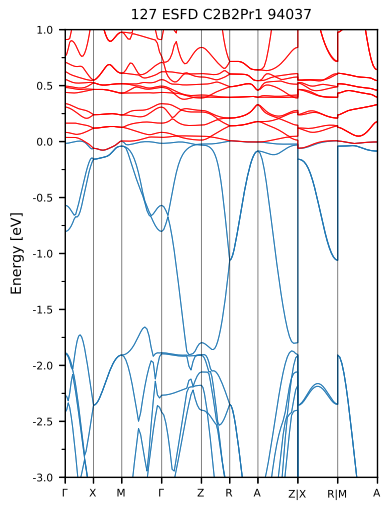

(ee) 127.4.38

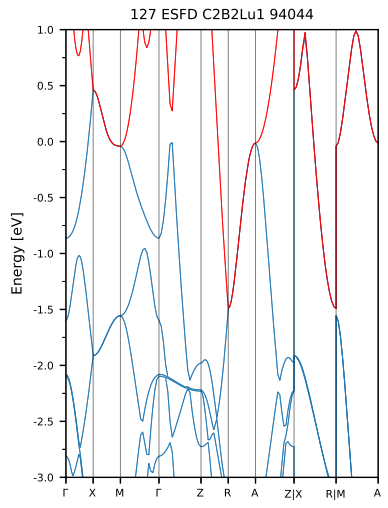

(ef) 127.4.39

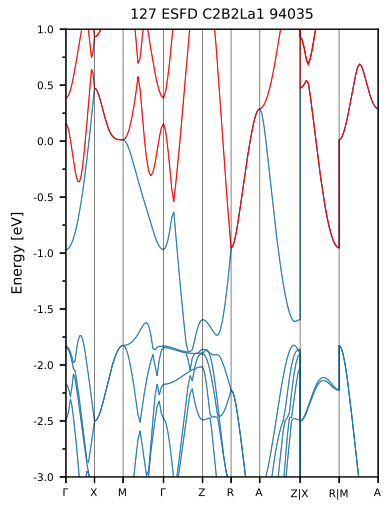

(eg) 127.4.40

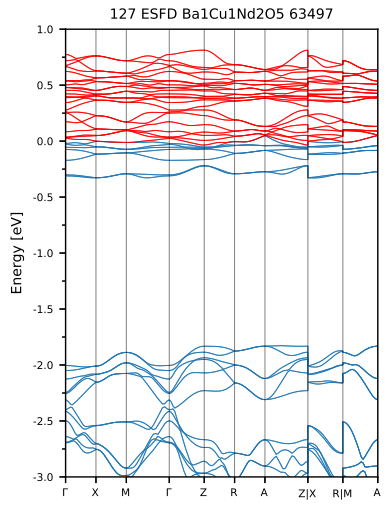

(eh) 127.4.41

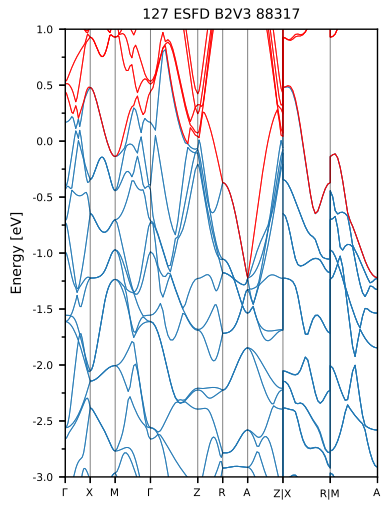

(ei) 127.4.42

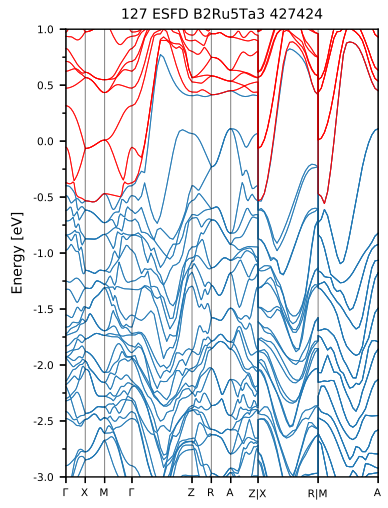

(ej) 127.4.43

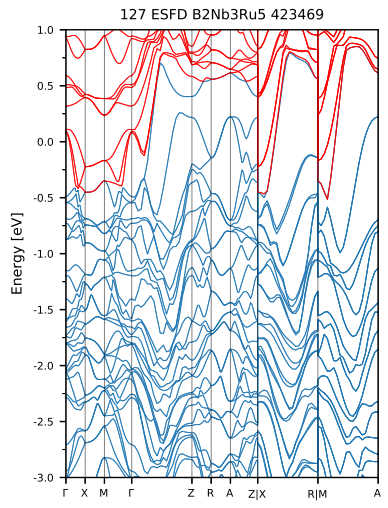

(ek) 127.4.44

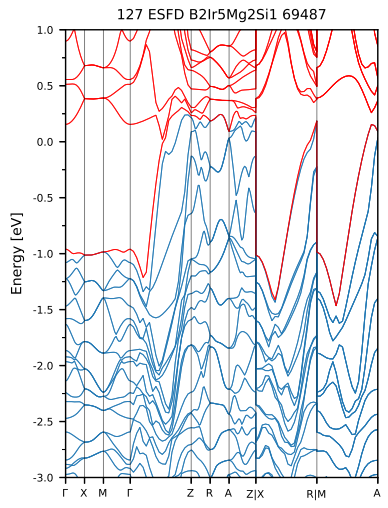

(el) 127.4.45

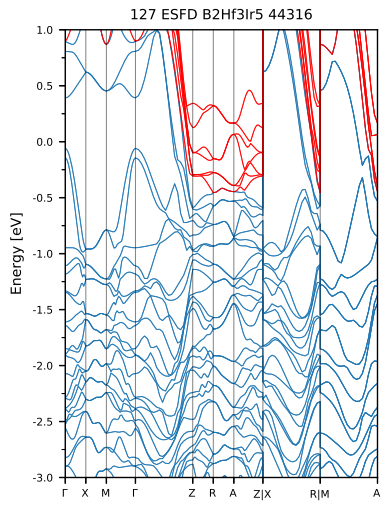

(em) 127.4.46

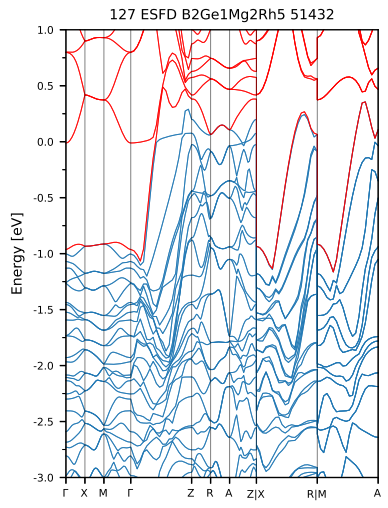

(en) 127.4.47

Figure 236: Band structures for topological materials in SG 127, part 9/10, from 127.4.32 to 127.4.47. See the list of figures.

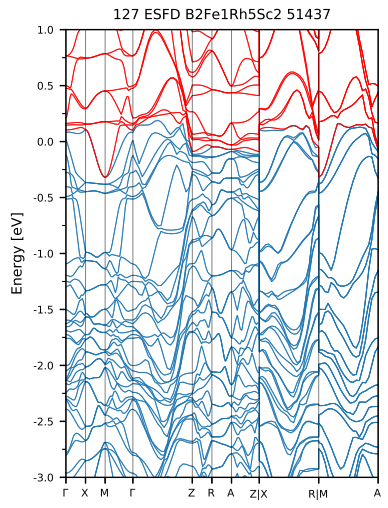

(eo) 127.4.48

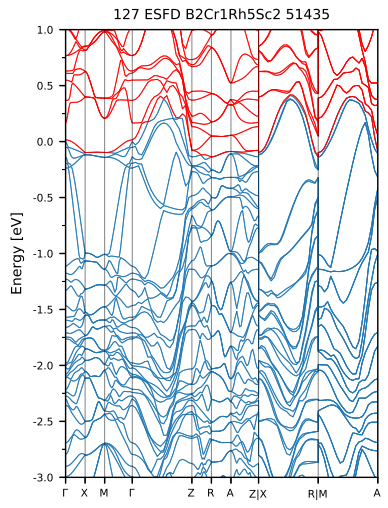

(ep) 127.4.49

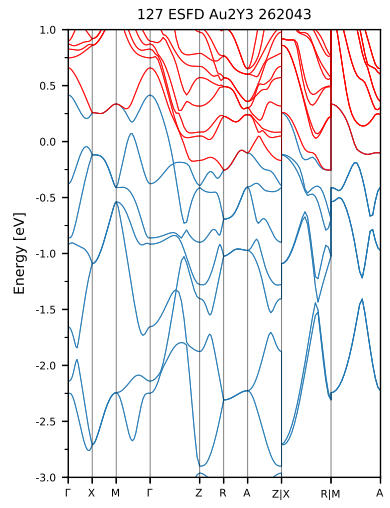

(eq) 127.4.50

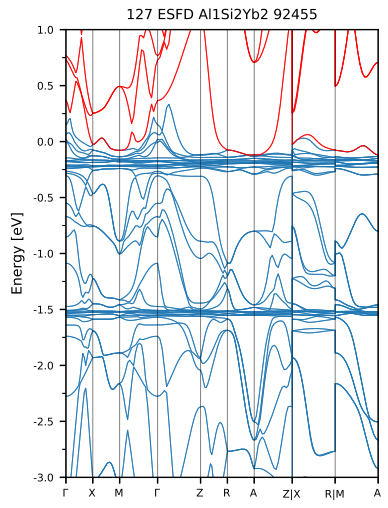

(er) 127.4.51

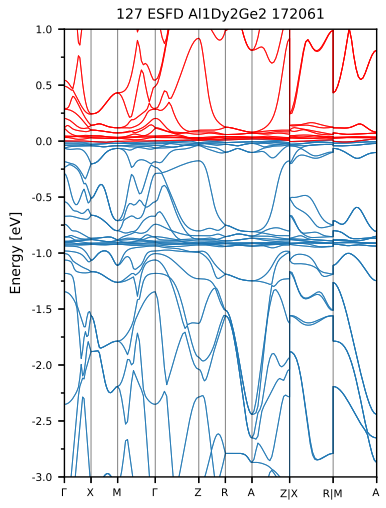

(es) 127.4.52

Figure 237: Band structures for topological materials in SG 127, part 10/10, from 127.4.48 to 127.4.52. See the list of figures.

### 0.0.58 SG 128

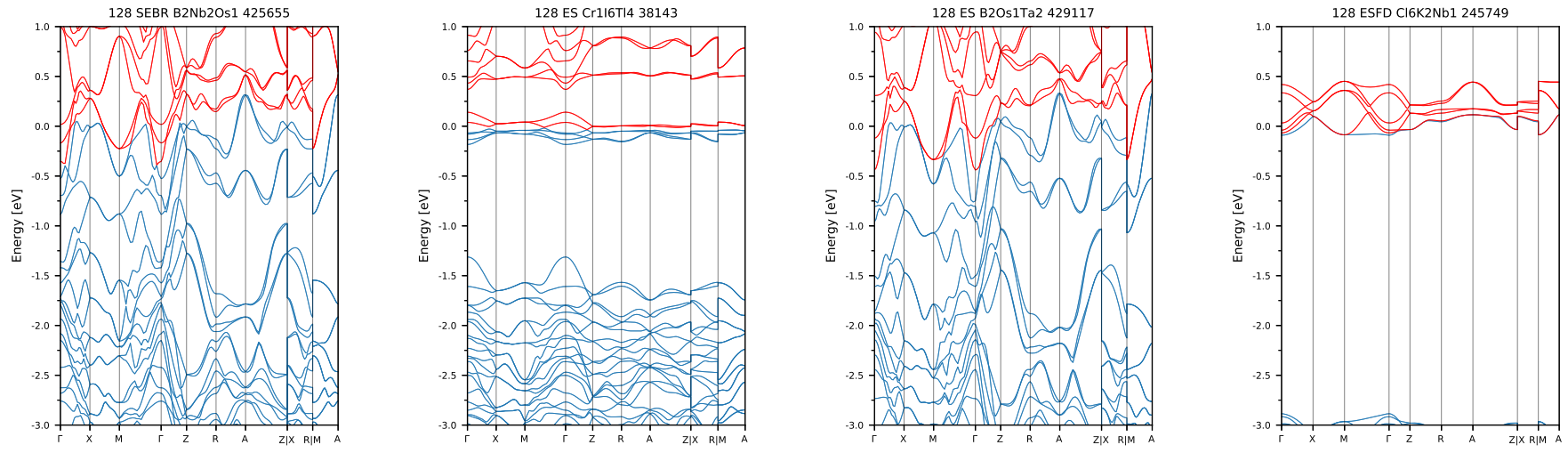

(a) 128.1.1 (b) 128.3.1 (c) 128.3.2 (d) 128.4.1

Figure 238: Band structures for topological materials in SG 128, part 1/1, from 128.1.1 to 128.4.1. See the list of figures.

## 0.0.59 SG 129

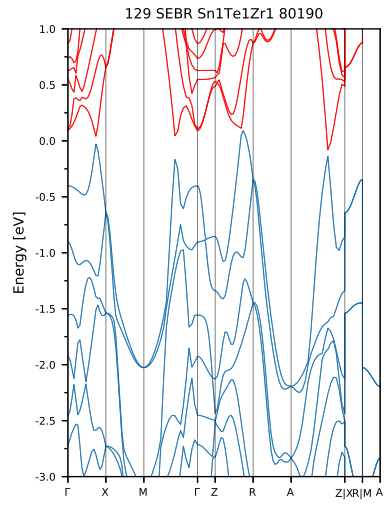

(a) 129.1.1

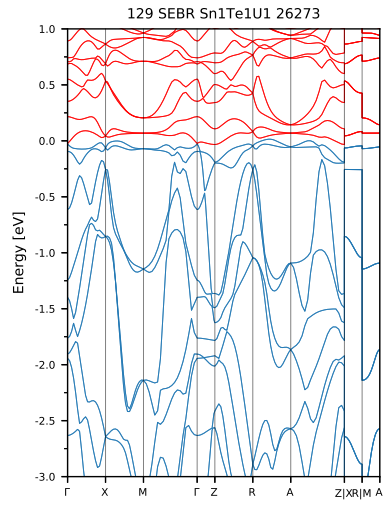

(b) 129.1.2

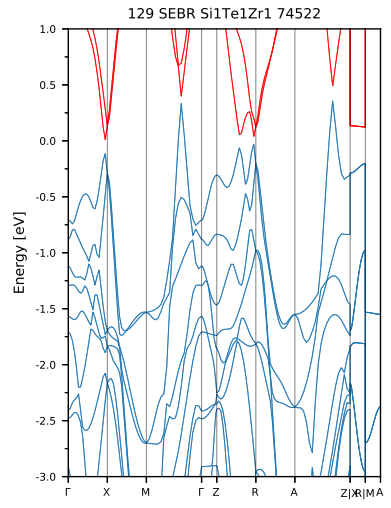

(c) 129.1.3

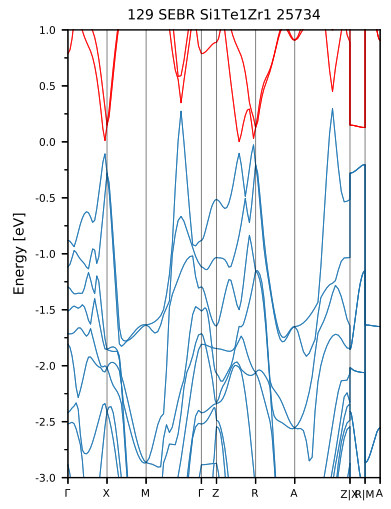

(d) 129.1.4

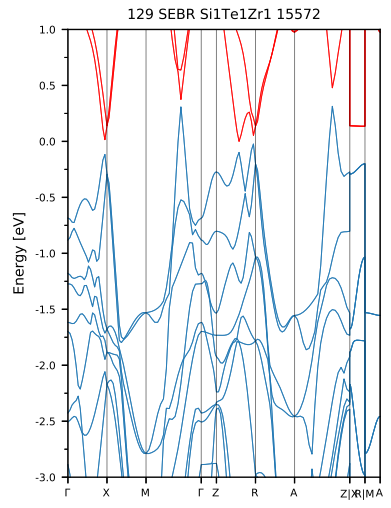

(e) 129.1.5

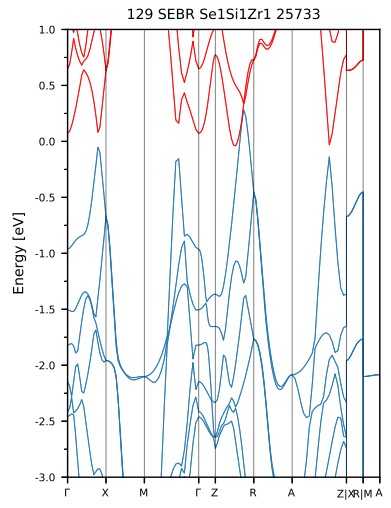

(f) 129.1.6

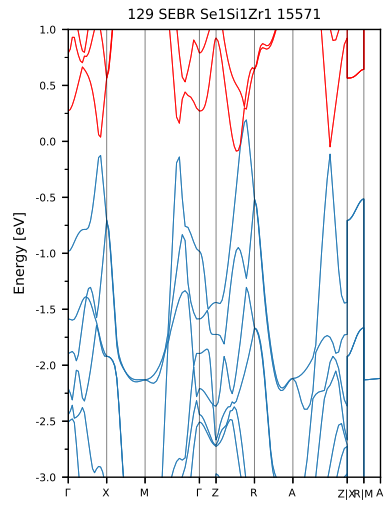

(g) 129.1.7

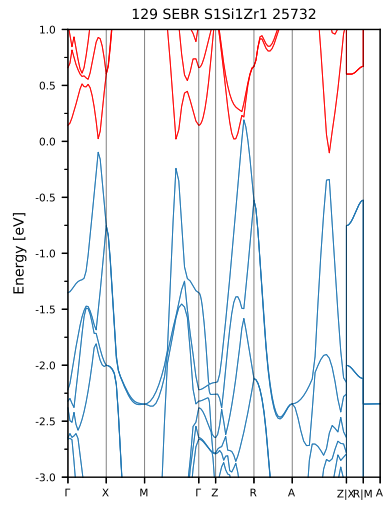

(h) 129.1.8

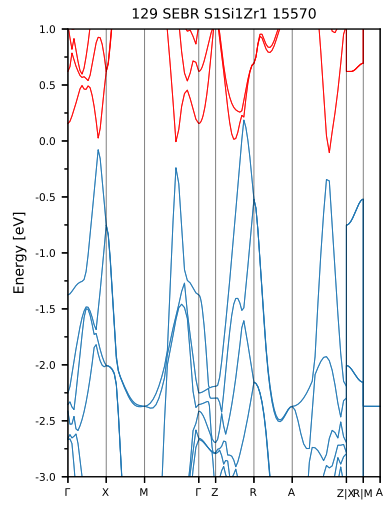

(i) 129.1.9

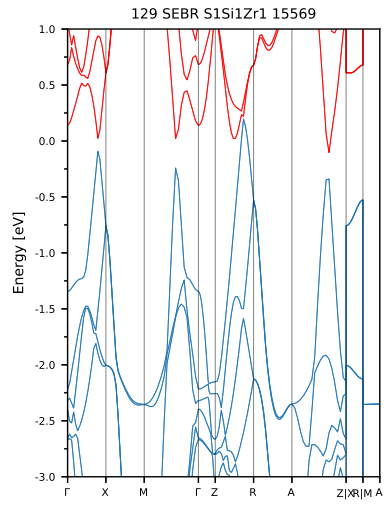

(j) 129.1.10

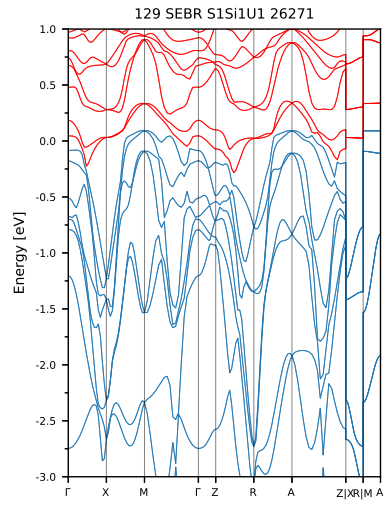

(k) 129.1.11

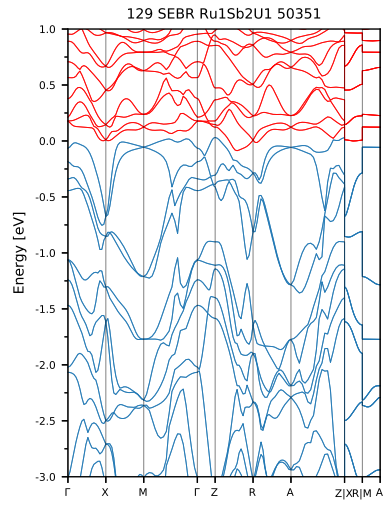

(l) 129.1.12

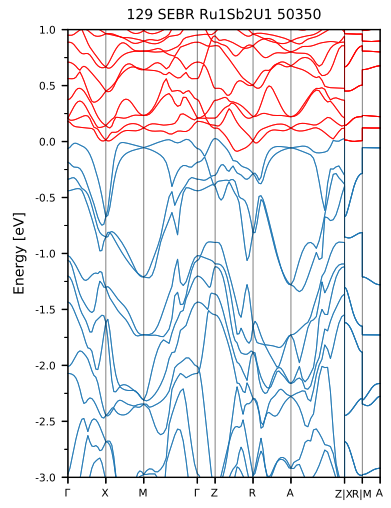

(m) 129.1.13

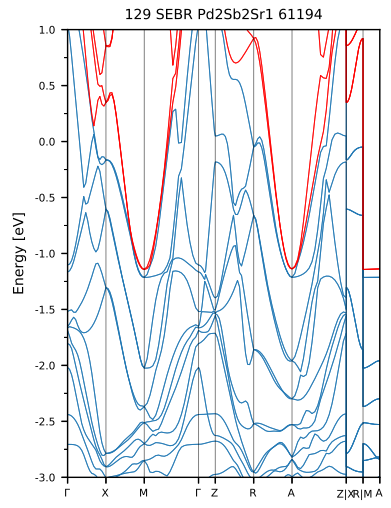

(n) 129.1.14

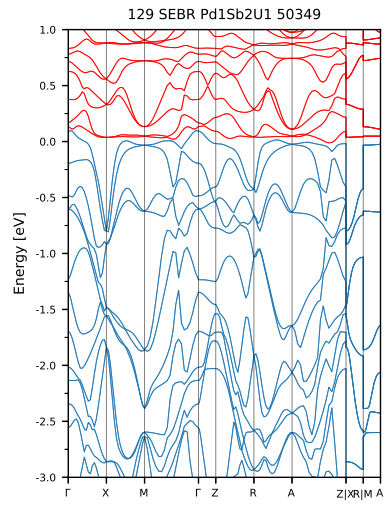

(o) 129.1.15

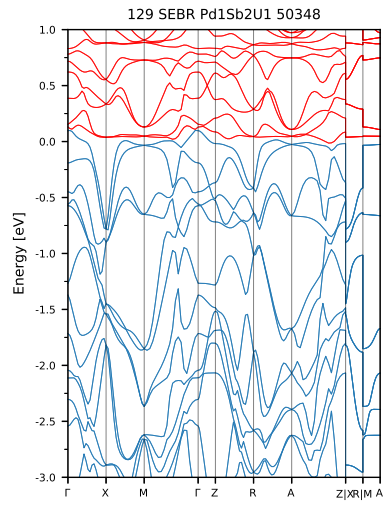

(p) 129.1.16

Figure 239: Band structures for topological materials in SG 129, part 1/39, from 129.1.1 to 129.1.16. See the list of figures.

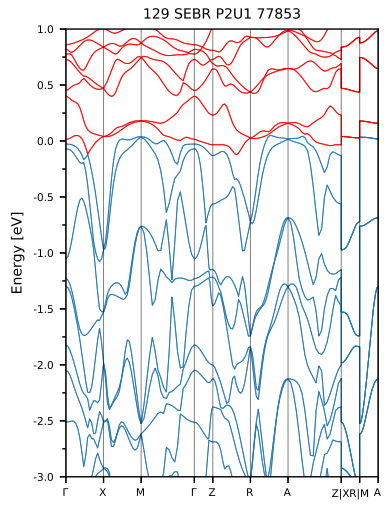

(q) 129.1.17

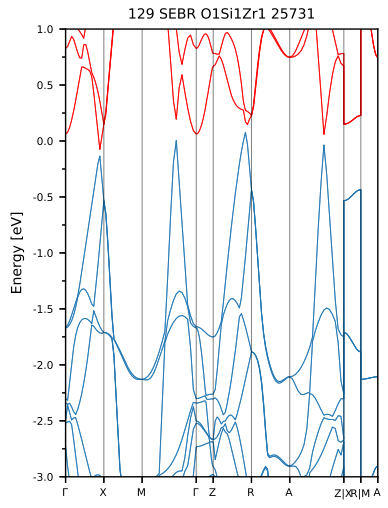

(r) 129.1.18

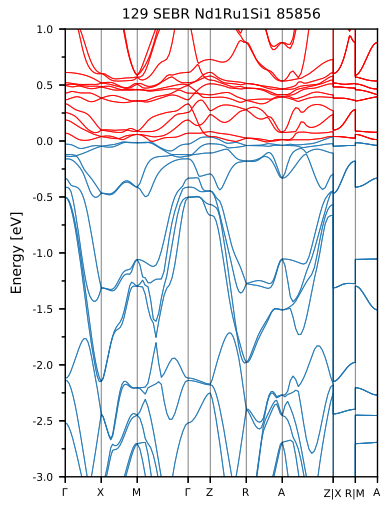

(s) 129.1.19

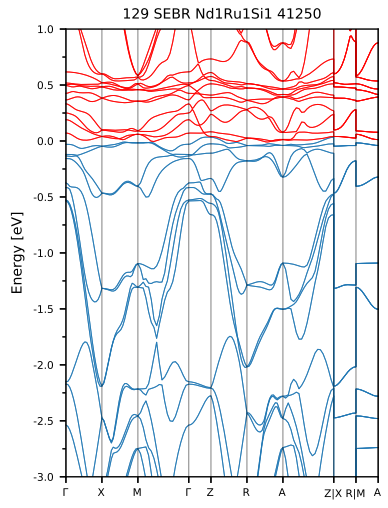

(t) 129.1.20

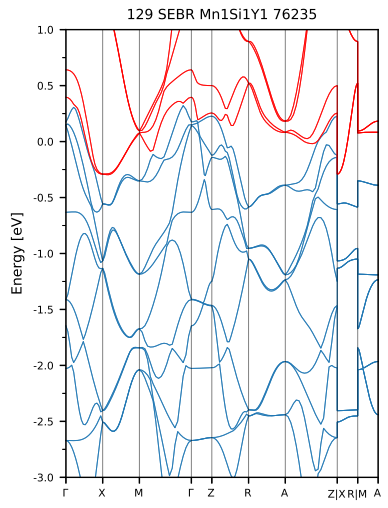

(u) 129.1.21

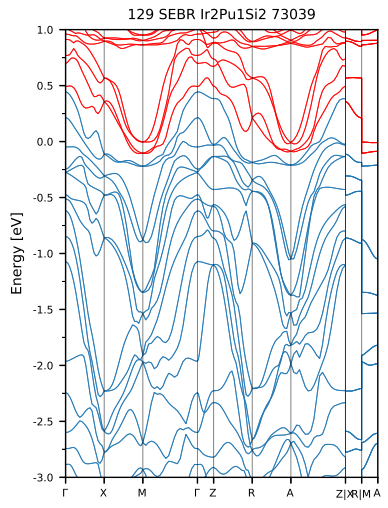

(v) 129.1.22

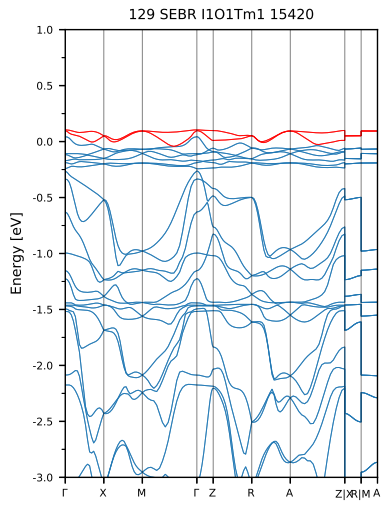

(w) 129.1.23

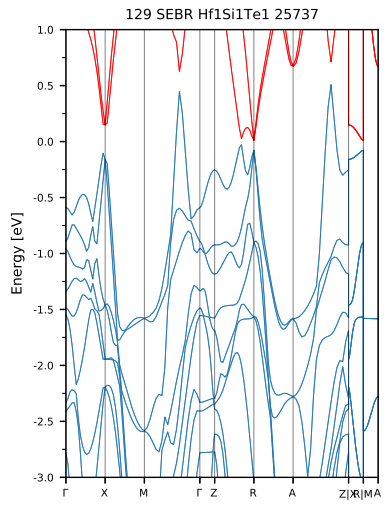

(x) 129.1.24

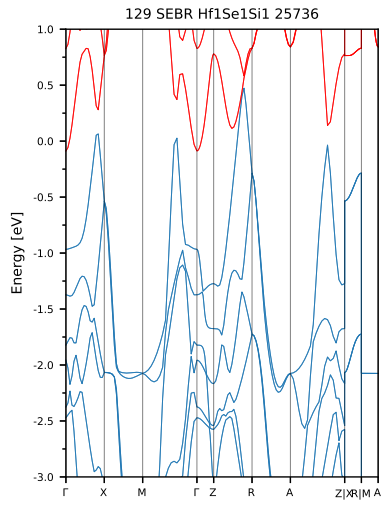

(y) 129.1.25

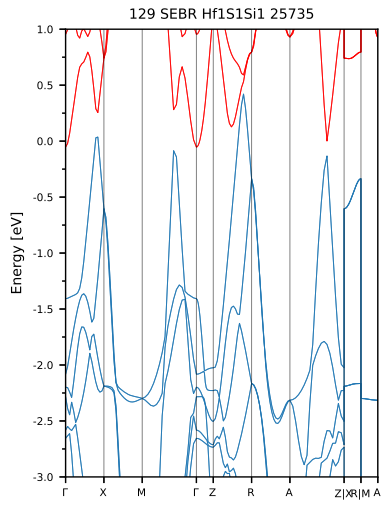

(z) 129.1.26

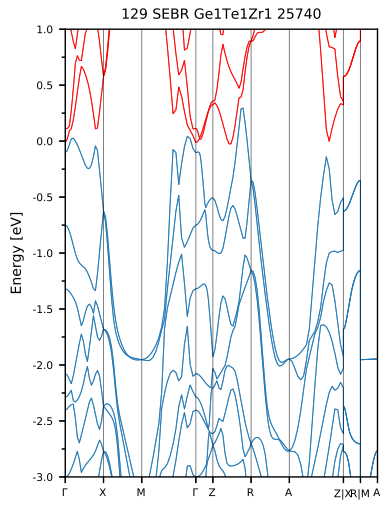

(aa) 129.1.27

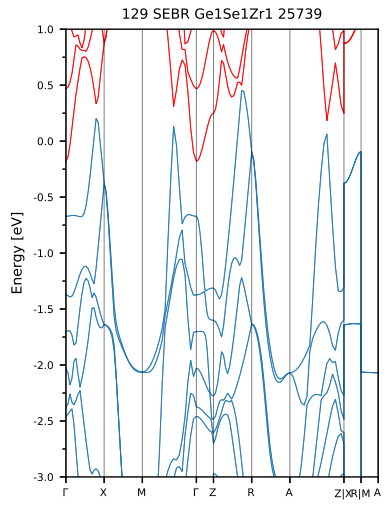

(ab) 129.1.28

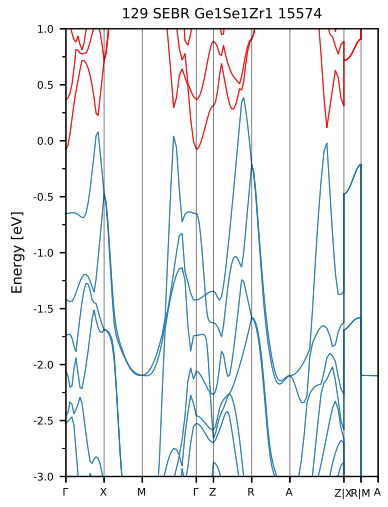

(ac) 129.1.29

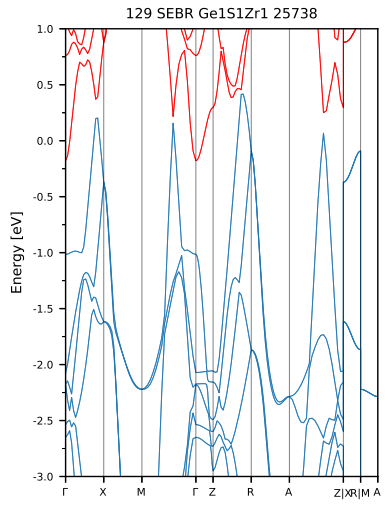

(ad) 129.1.30

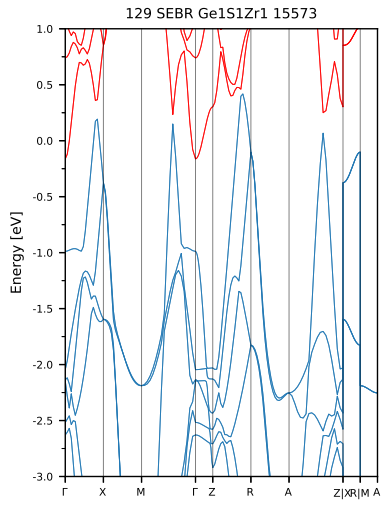

(ae) 129.1.31

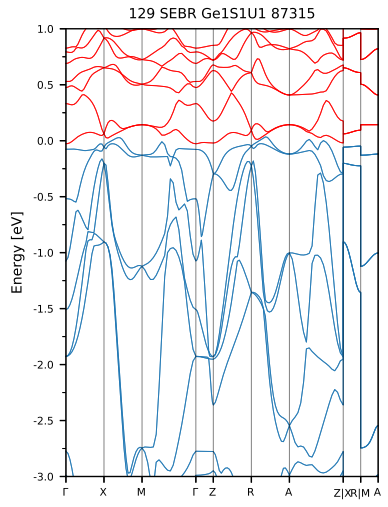

(af) 129.1.32

Figure 240: Band structures for topological materials in SG 129, part 2/39, from 129.1.17 to 129.1.32. See the list of figures.

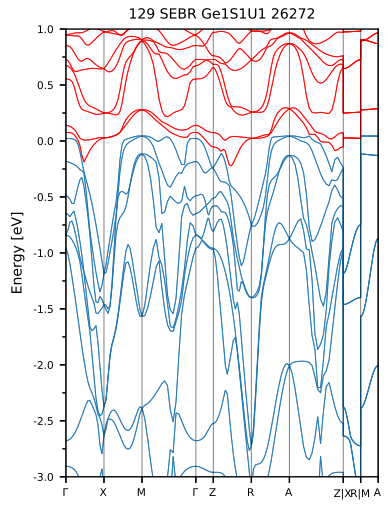

(ag) 129.1.33

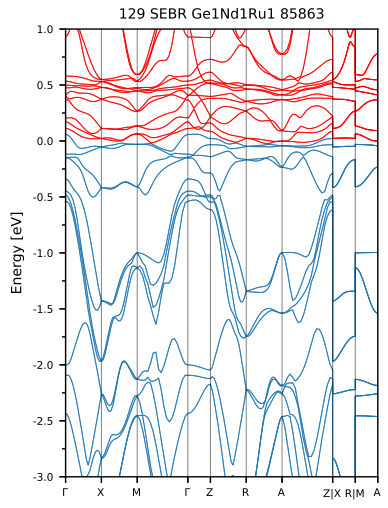

(ah) 129.1.34

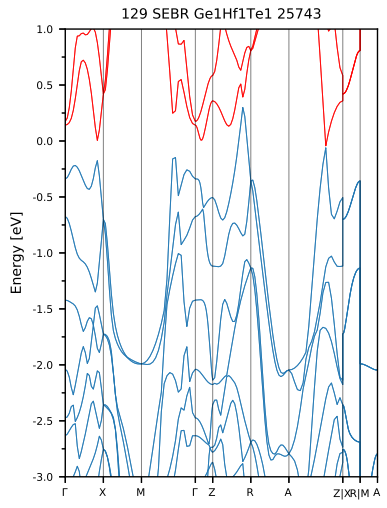

(ai) 129.1.35

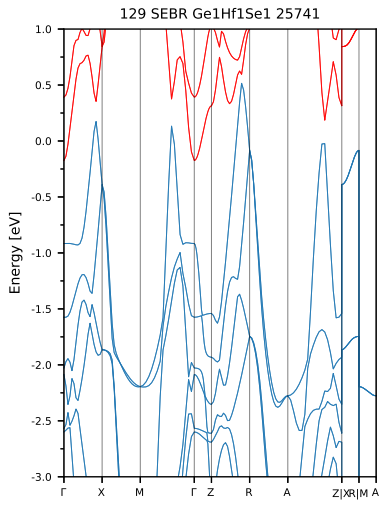

(aj) 129.1.36

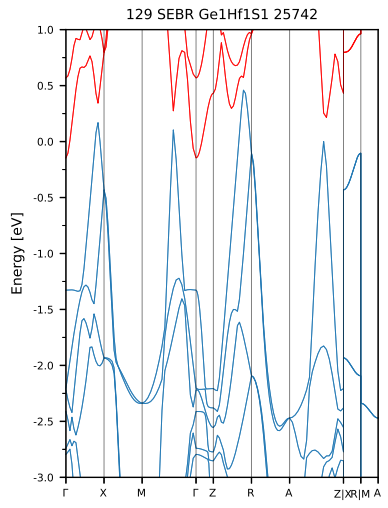

(ak) 129.1.37

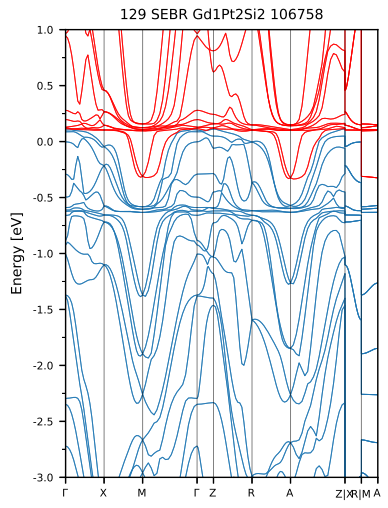

(al) 129.1.38

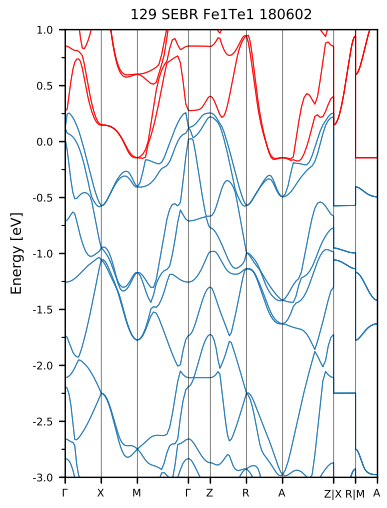

(am) 129.1.39

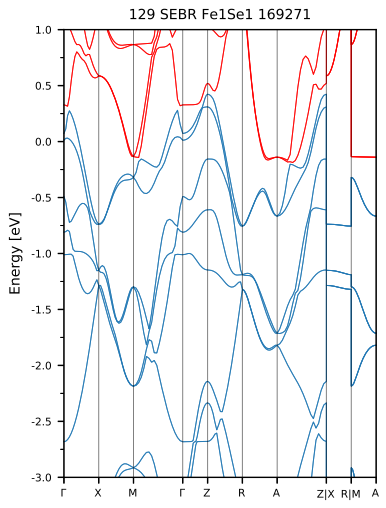

(an) 129.1.40

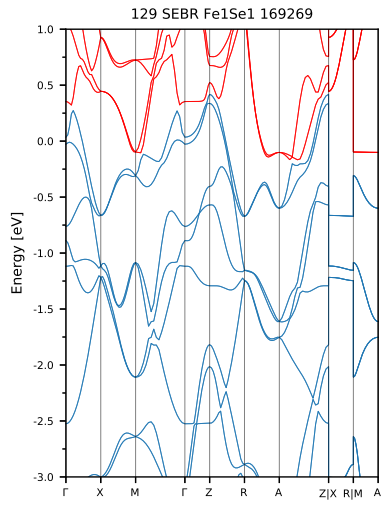

(ao) 129.1.41

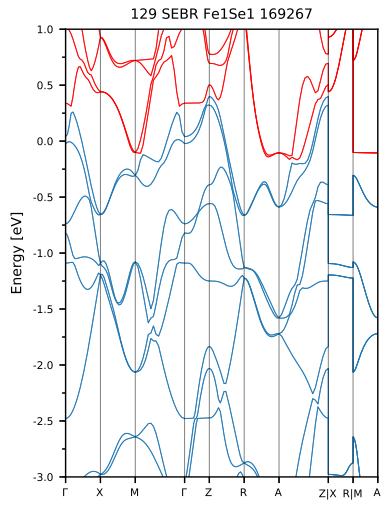

(ap) 129.1.42

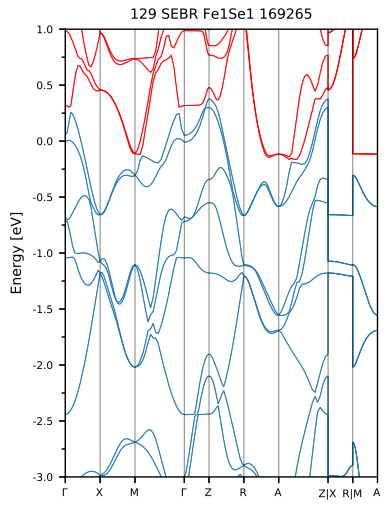

(aq) 129.1.43

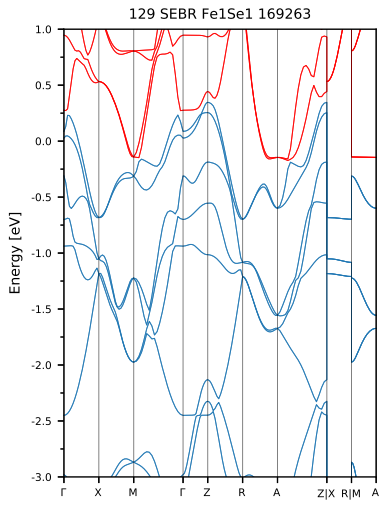

(ar) 129.1.44

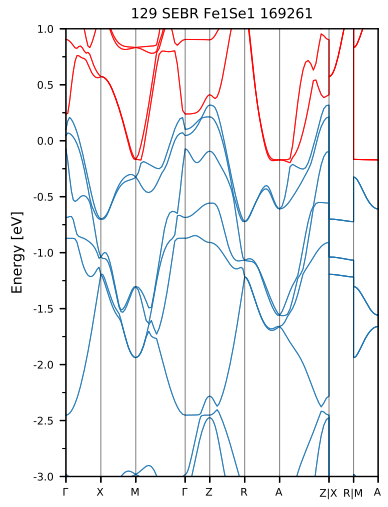

(as) 129.1.45

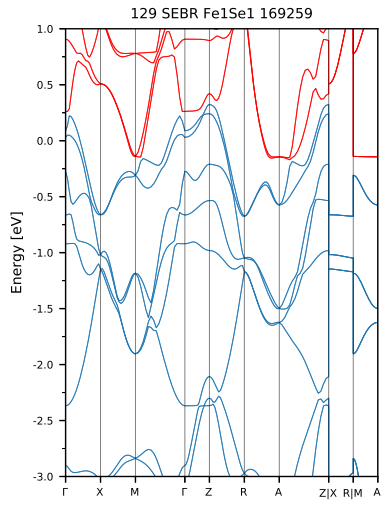

(at) 129.1.46

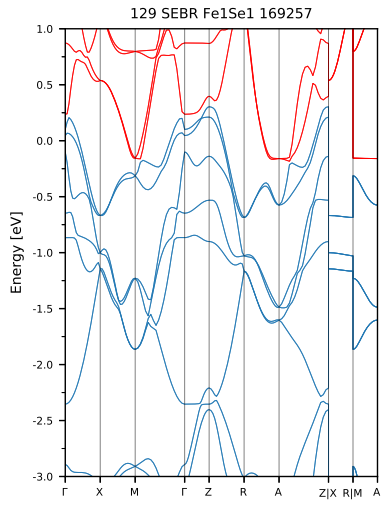

(au) 129.1.47

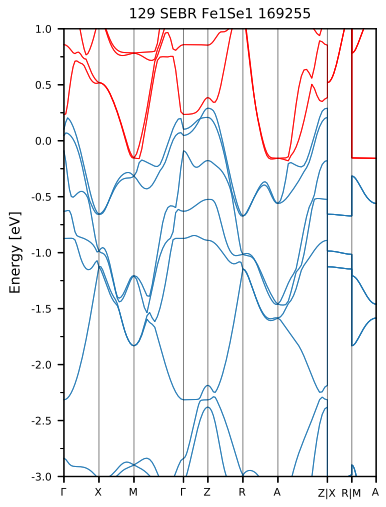

(av) 129.1.48

Figure 241: Band structures for topological materials in SG 129, part 3/39, from 129.1.33 to 129.1.48. See the list of figures.

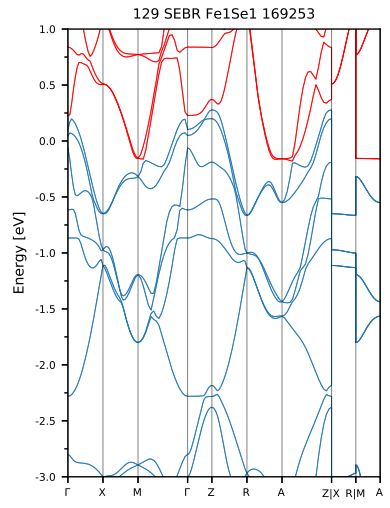

(aw) 129.1.49

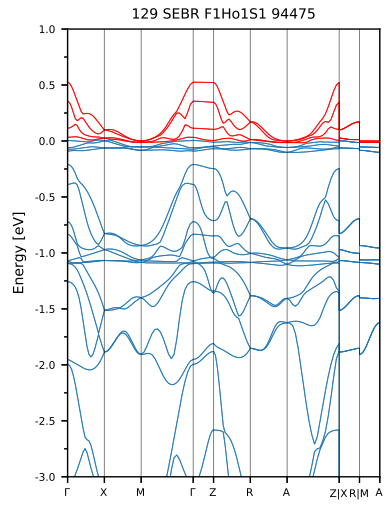

(ax) 129.1.50

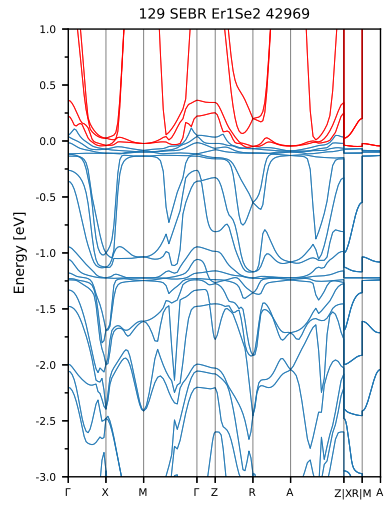

(ay) 129.1.51

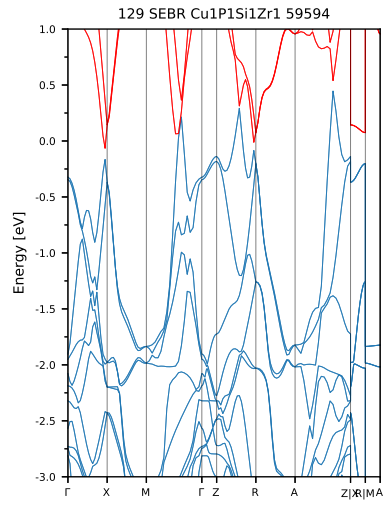

(az) 129.1.52

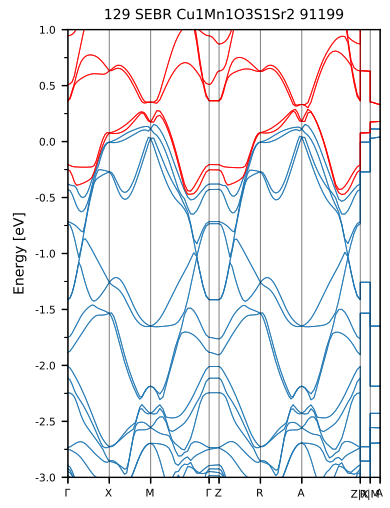

(ba) 129.1.53

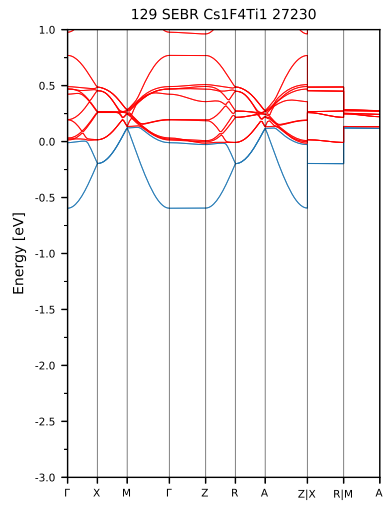

(bb) 129.1.54

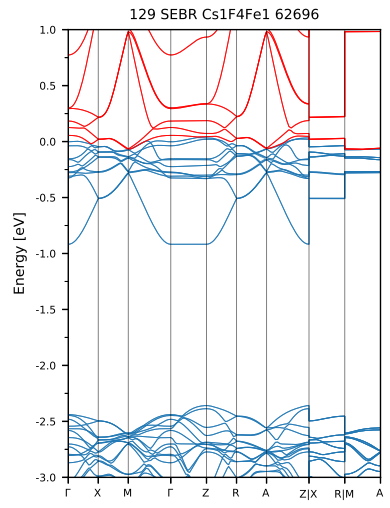

(bc) 129.1.55

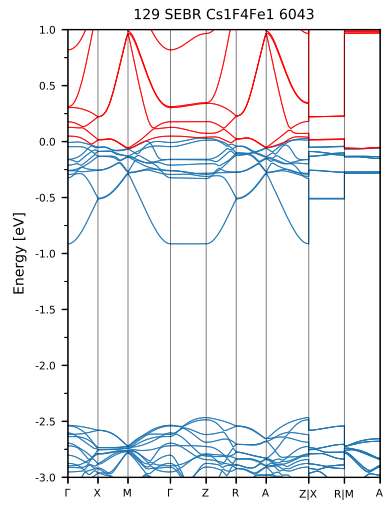

(bd) 129.1.56

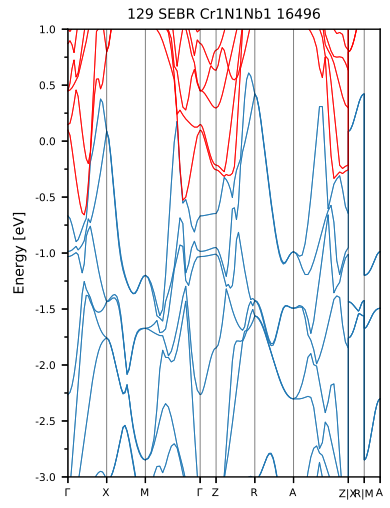

(be) 129.1.57

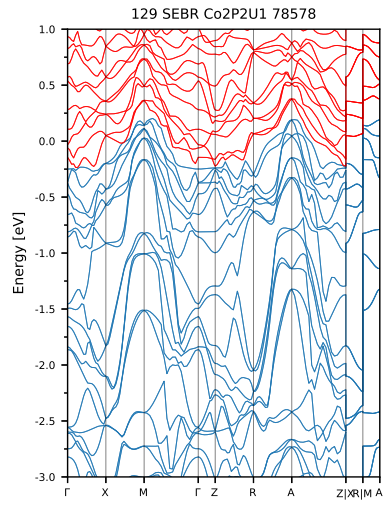

(bf) 129.1.58

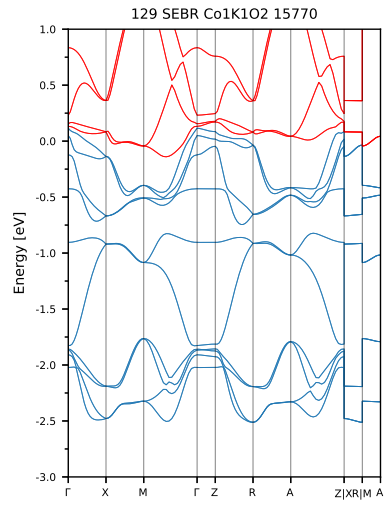

(bg) 129.1.59

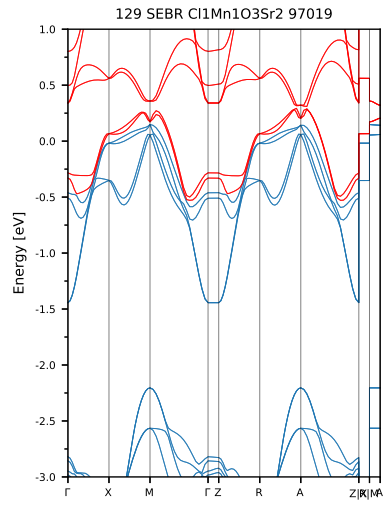

(bh) 129.1.60

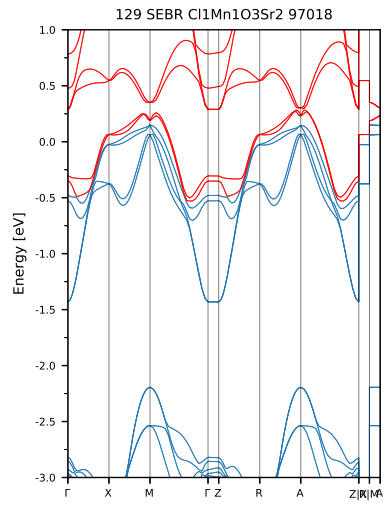

(bi) 129.1.61

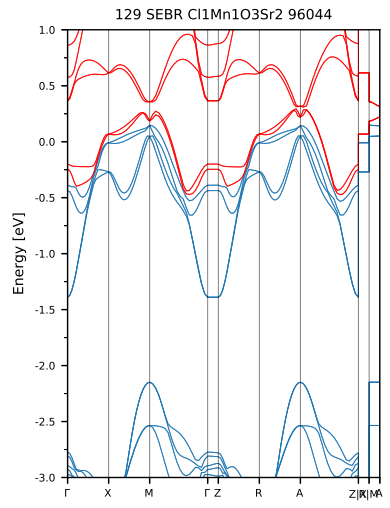

(bj) 129.1.62

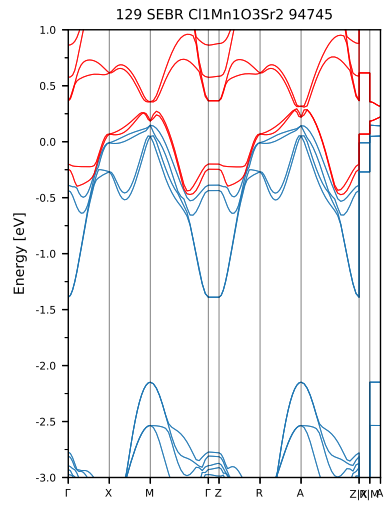

(bk) 129.1.63

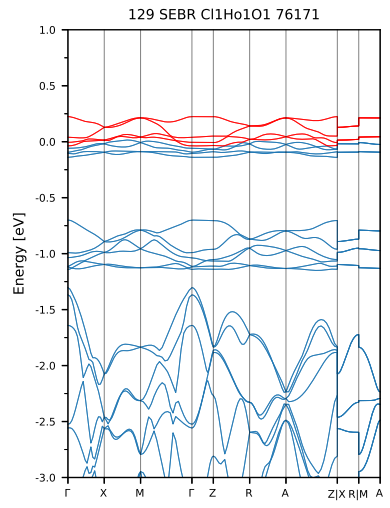

(bl) 129.1.64

Figure 242: Band structures for topological materials in SG 129, part 4/39, from 129.1.49 to 129.1.64. See the list of figures.

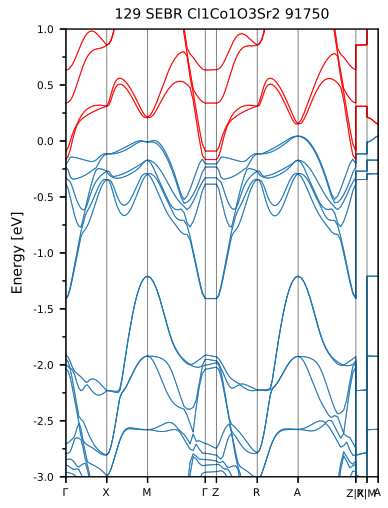

(bm) 129.1.65

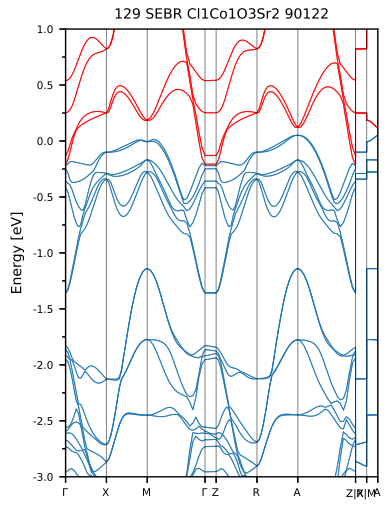

(bn) 129.1.66

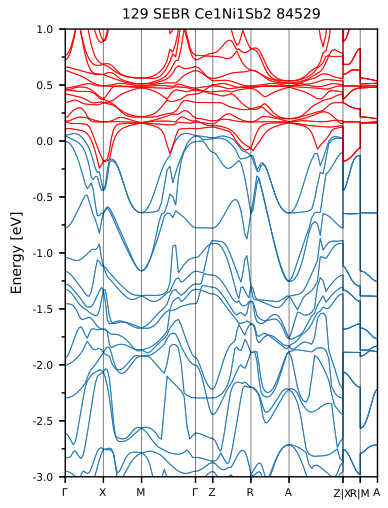

(bo) 129.1.67

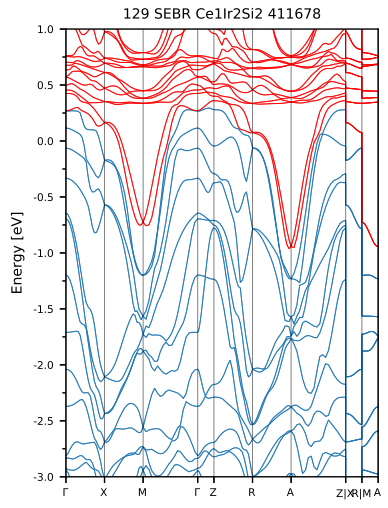

(bp) 129.1.68

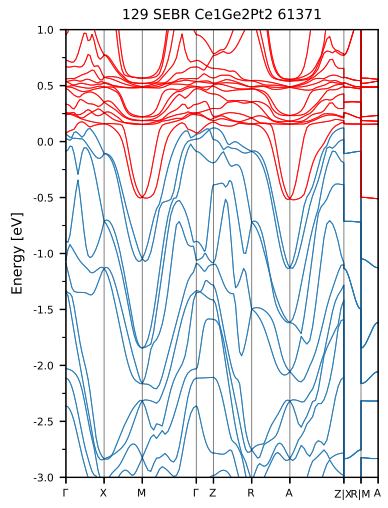

(bq) 129.1.69

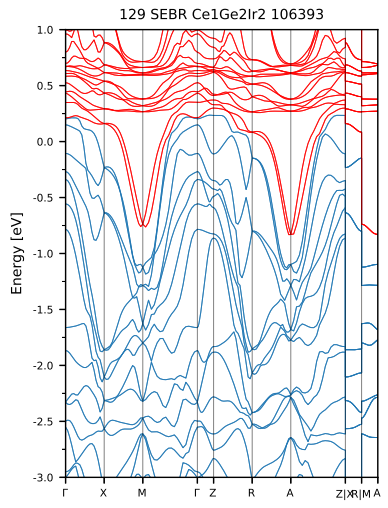

(br) 129.1.70

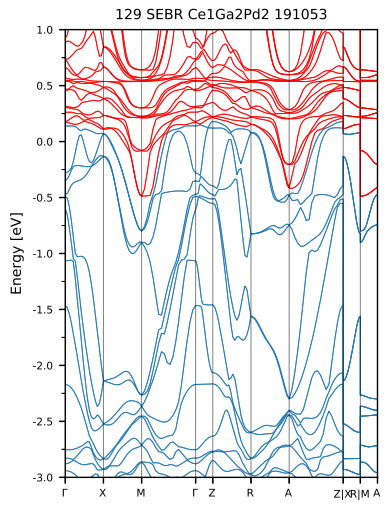

(bs) 129.1.71

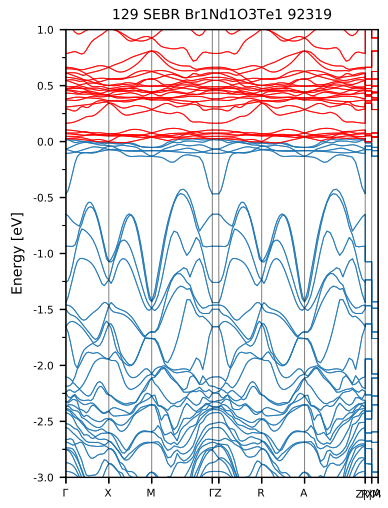

(bt) 129.1.72

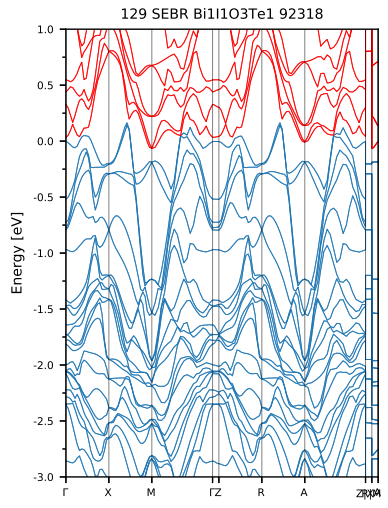

(bu) 129.1.73

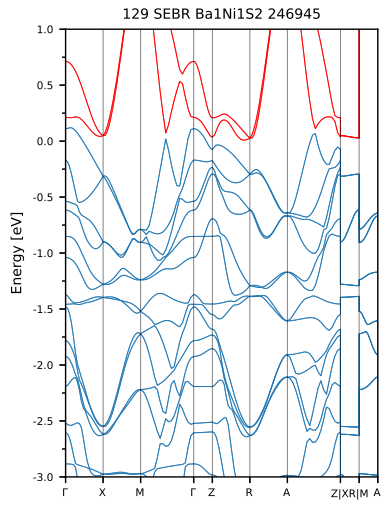

(bv) 129.1.74

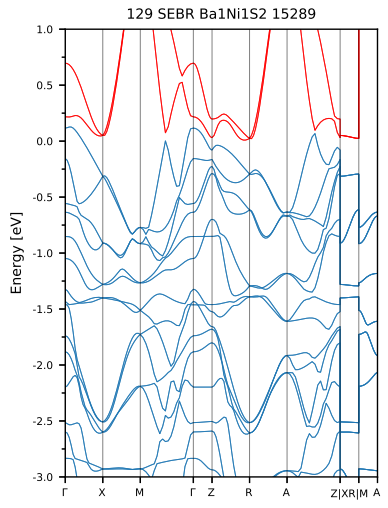

(bw) 129.1.75

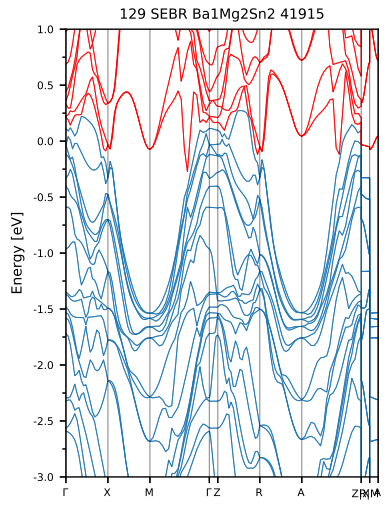

(bx) 129.1.76

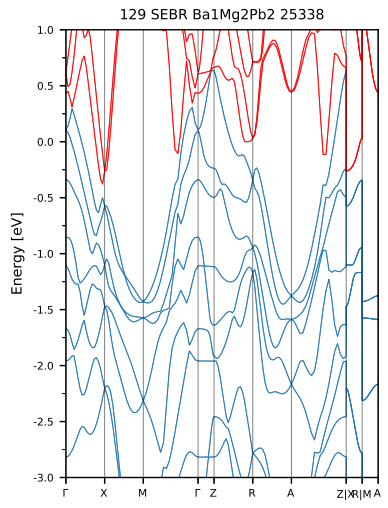

(by) 129.1.77

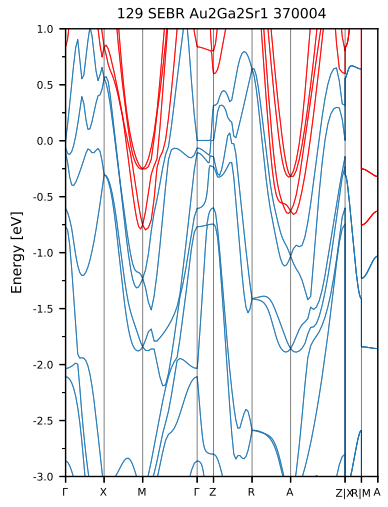

(bz) 129.1.78

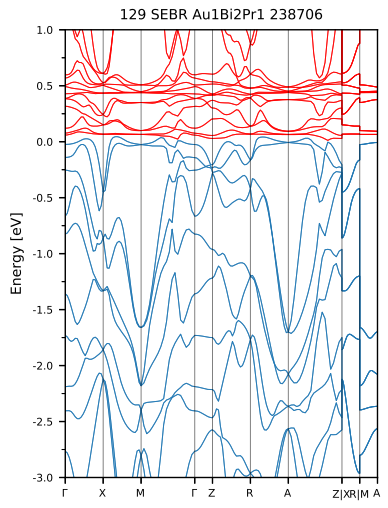

(ca) 129.1.79

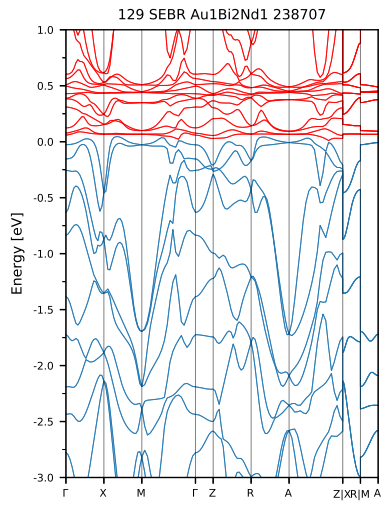

(cb) 129.1.80

Figure 243: Band structures for topological materials in SG 129, part 5/39, from 129.1.65 to 129.1.80. See the list of figures.

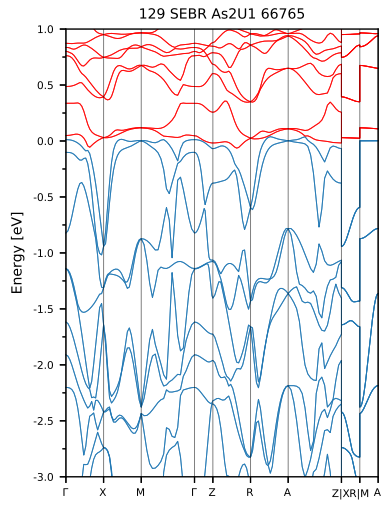

(cc) 129.1.81

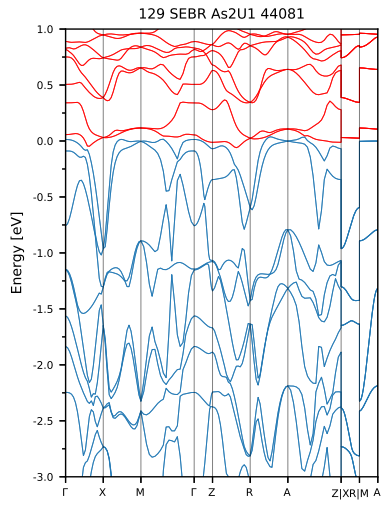

(cd) 129.1.82

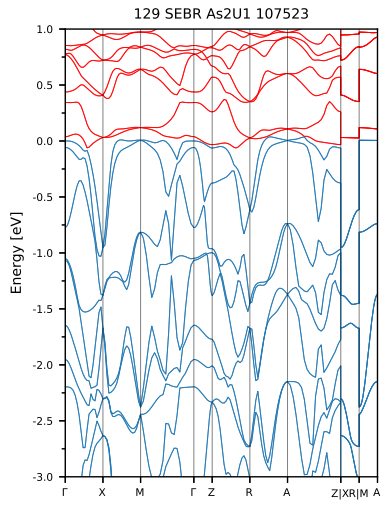

(ce) 129.1.83

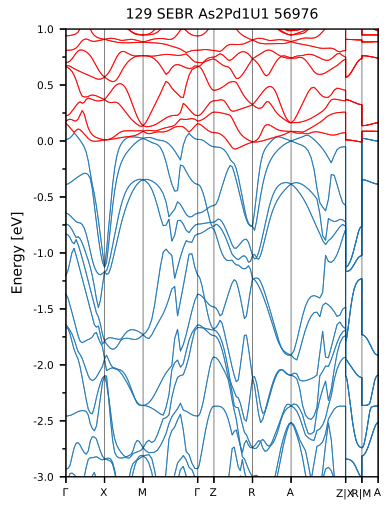

(cf) 129.1.84

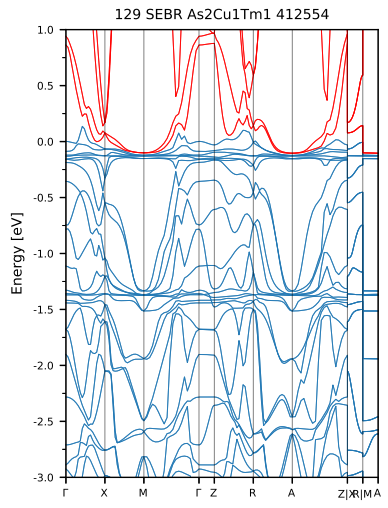

(cg) 129.1.85

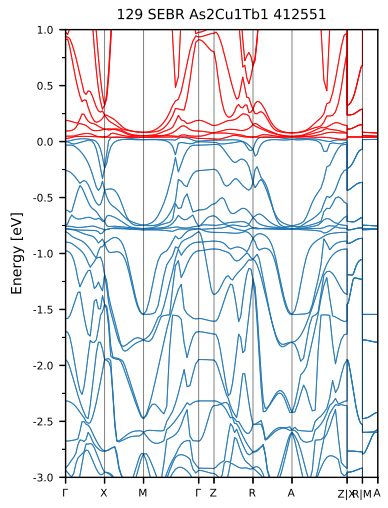

(ch) 129.1.86

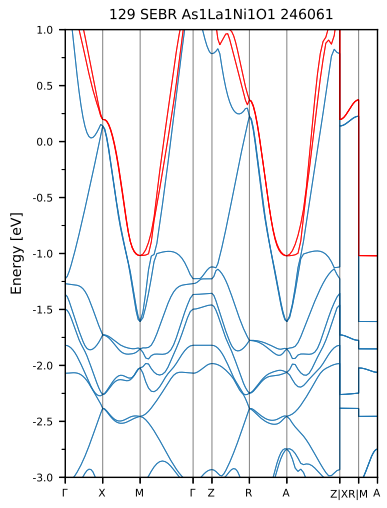

(ci) 129.1.87

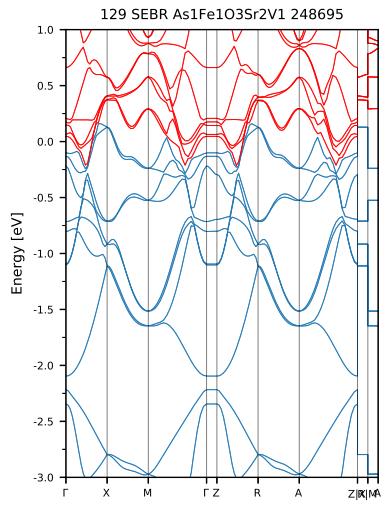

(cj) 129.1.88

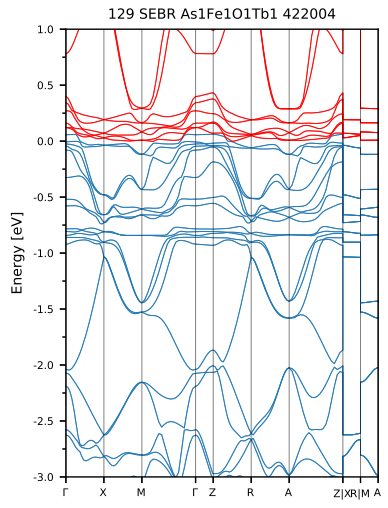

(ck) 129.1.89

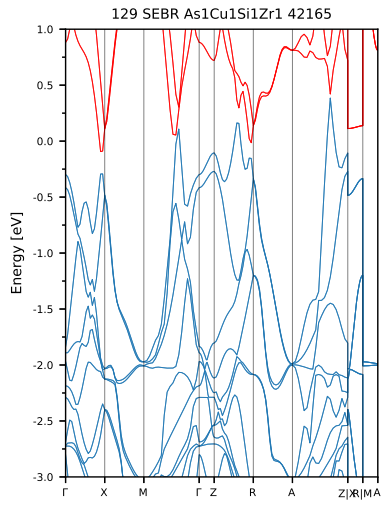

(cl) 129.1.90

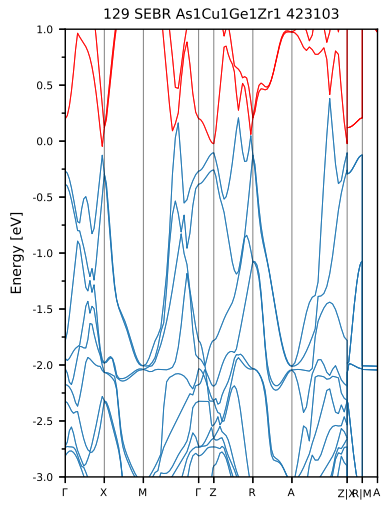

(cm) 129.1.91

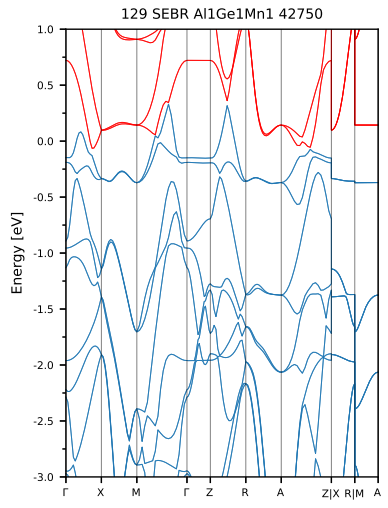

(cn) 129.1.92

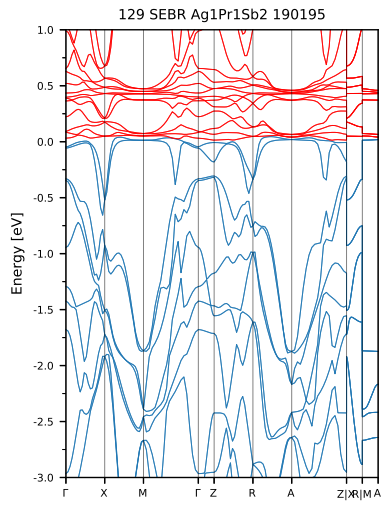

(co) 129.1.93

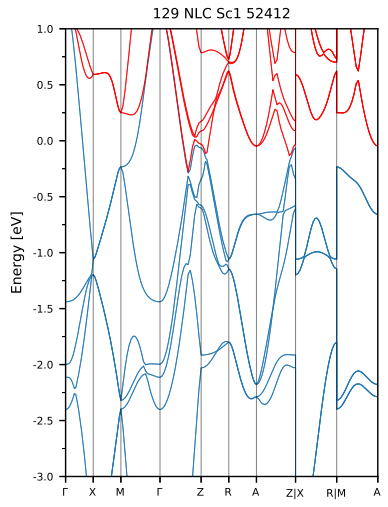

(cp) 129.2.1

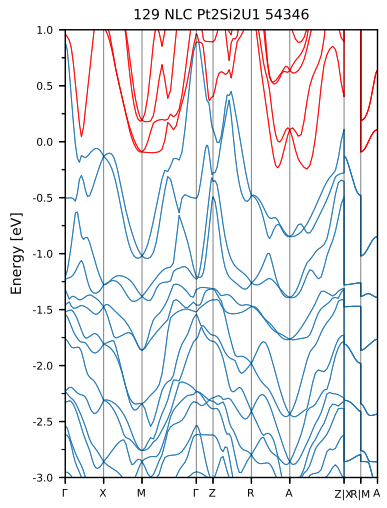

(cq) 129.2.2

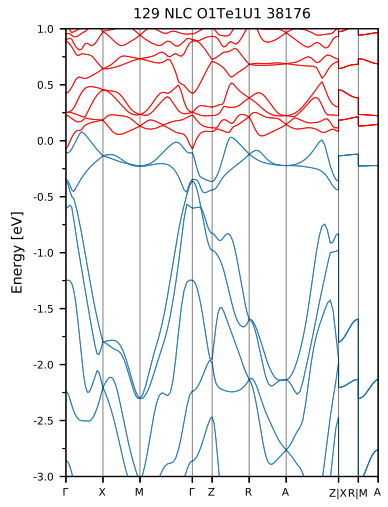

(cr) 129.2.3

Figure 244: Band structures for topological materials in SG 129, part 6/39, from 129.1.81 to 129.2.3. See the list of figures.

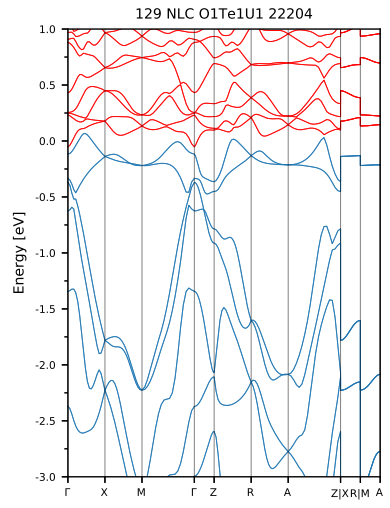

(cs) 129.2.4

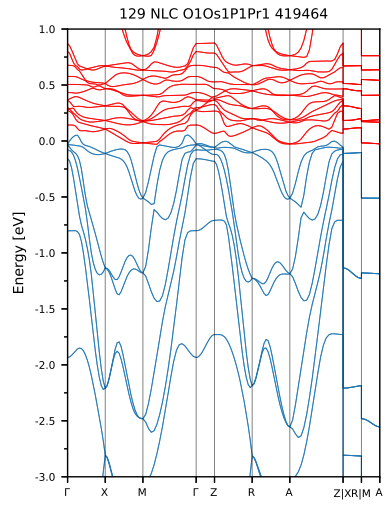

(ct) 129.2.5

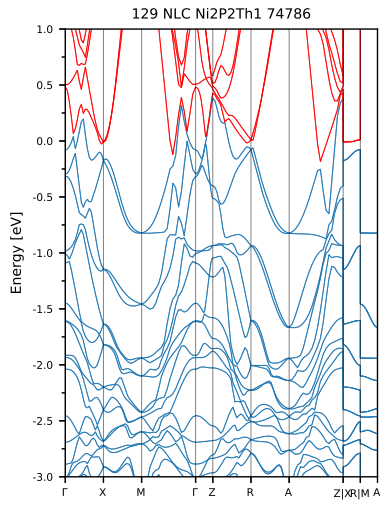

(cu) 129.2.6

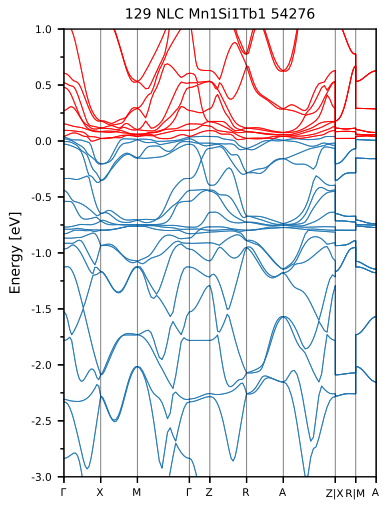

(cv) 129.2.7

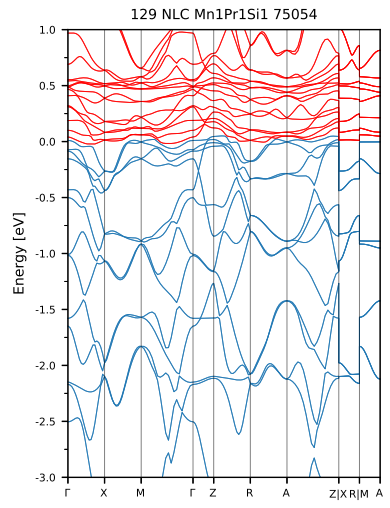

(cw) 129.2.8

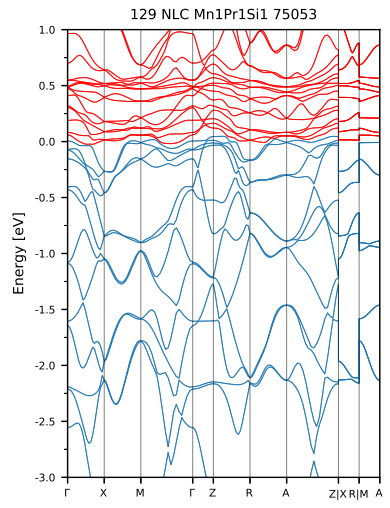

(cx) 129.2.9

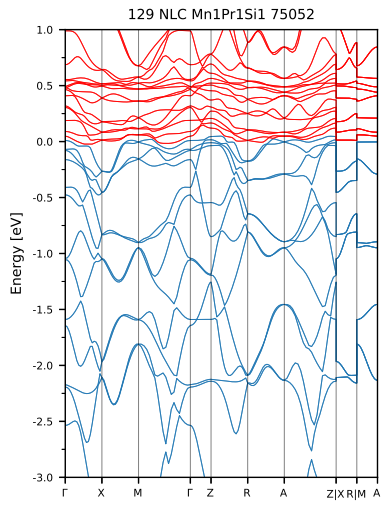

(cy) 129.2.10

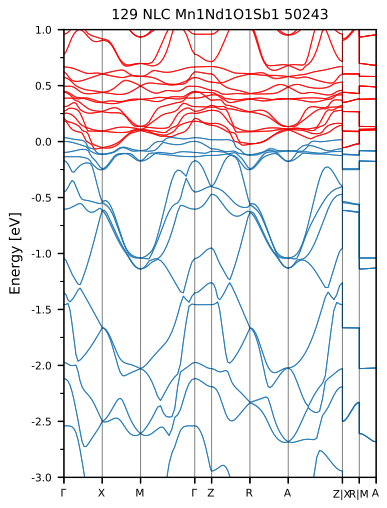

(cz) 129.2.11

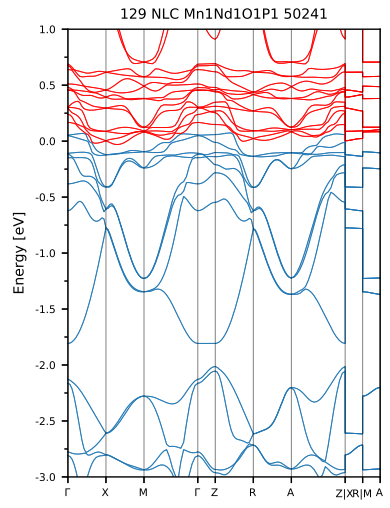

(da) 129.2.12

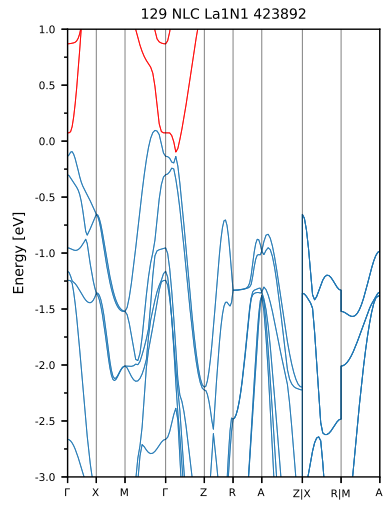

(db) 129.2.13

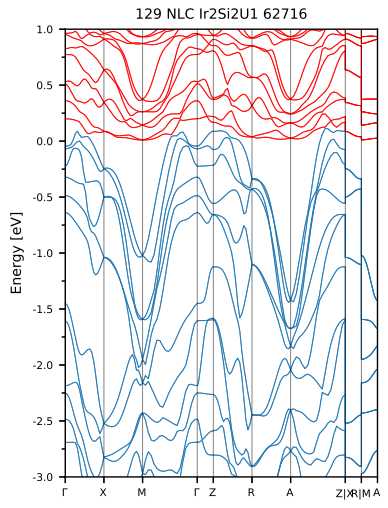

(dc) 129.2.14

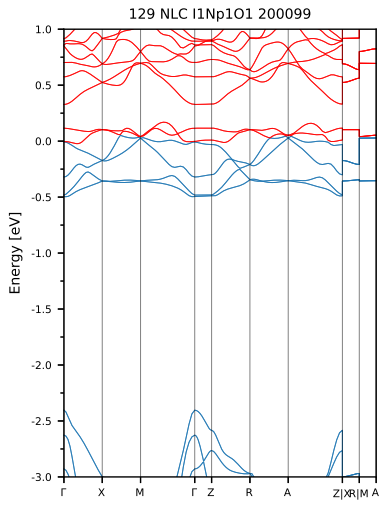

(dd) 129.2.15

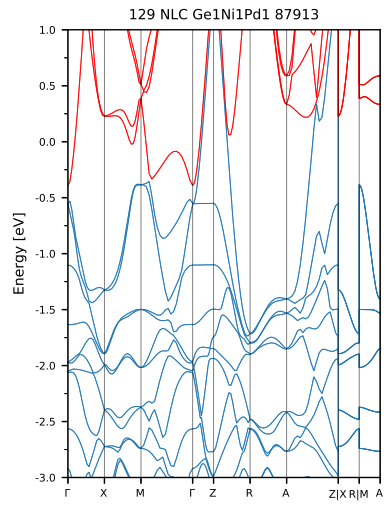

(de) 129.2.16

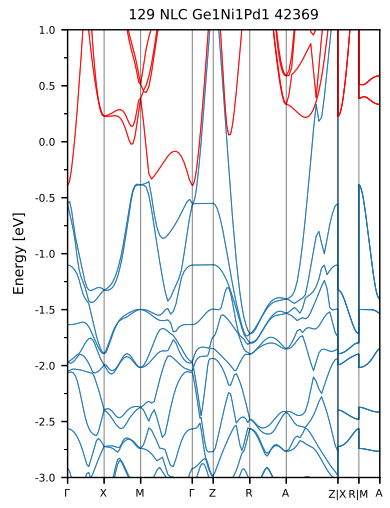

(df) 129.2.17

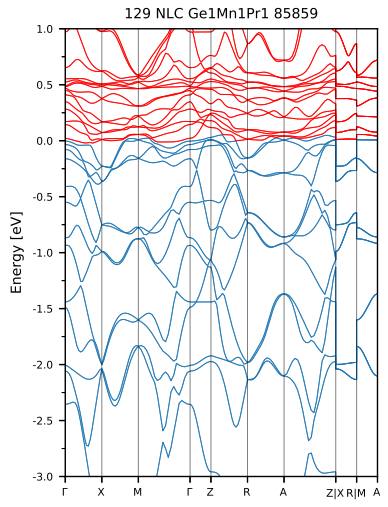

(dg) 129.2.18

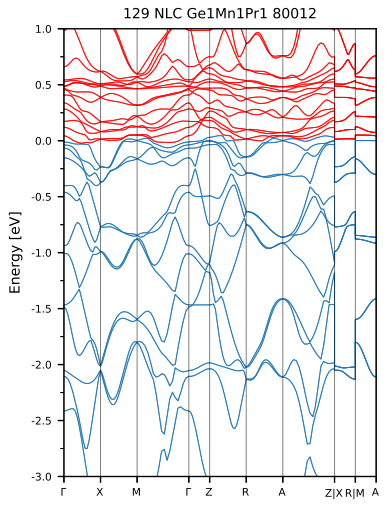

(dh) 129.2.19

Figure 245: Band structures for topological materials in SG 129, part 7/39, from 129.2.4 to 129.2.19. See the list of figures.

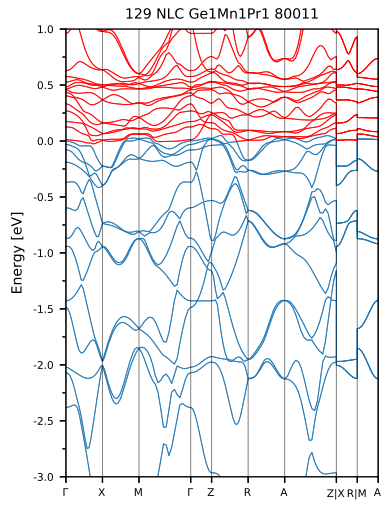

(di) 129.2.20

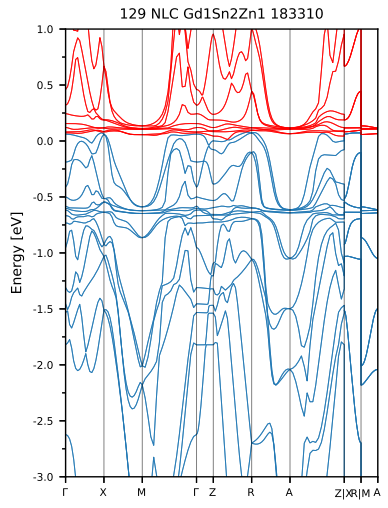

(dj) 129.2.21

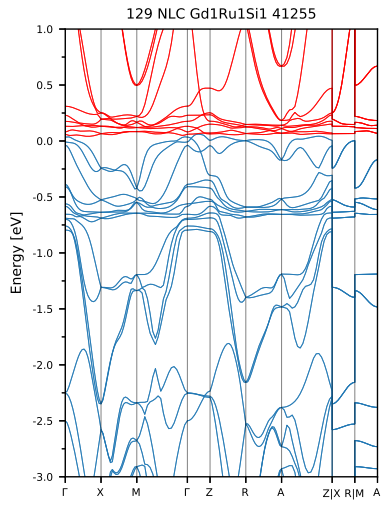

(dk) 129.2.22

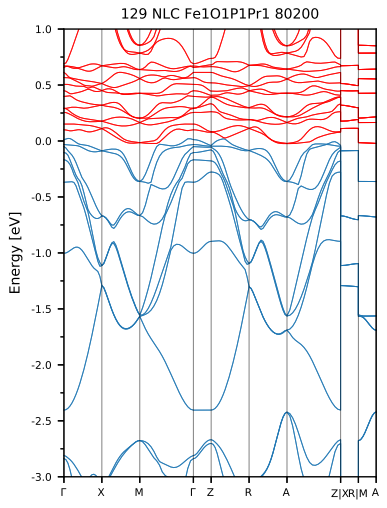

(dl) 129.2.23

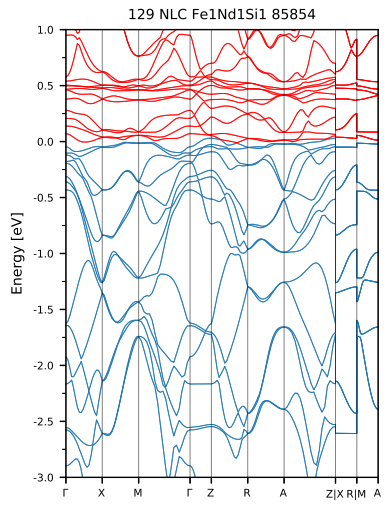

(dm) 129.2.24

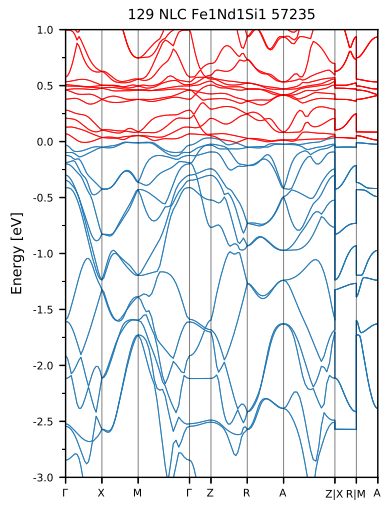

(dn) 129.2.25

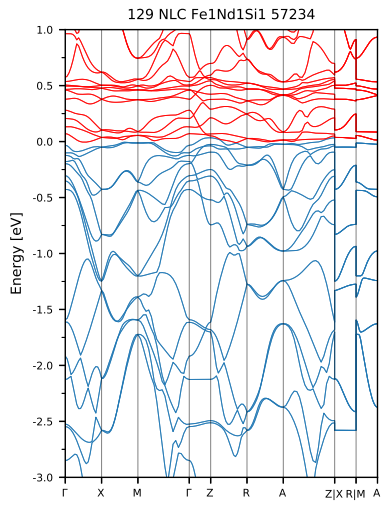

(do) 129.2.26

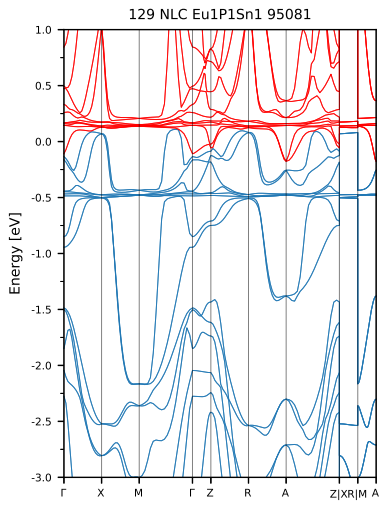

(dp) 129.2.27

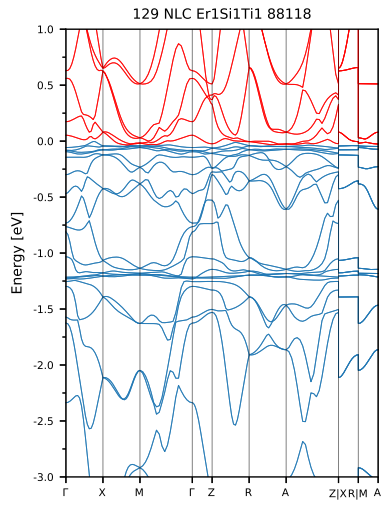

(dq) 129.2.28

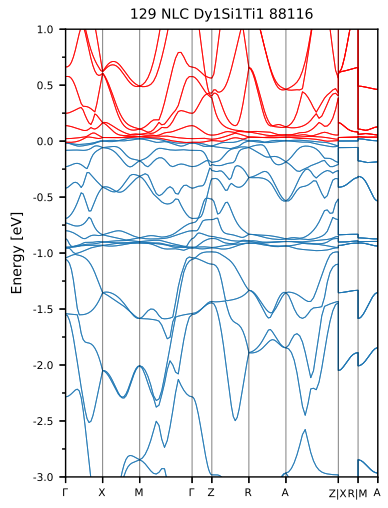

(dr) 129.2.29

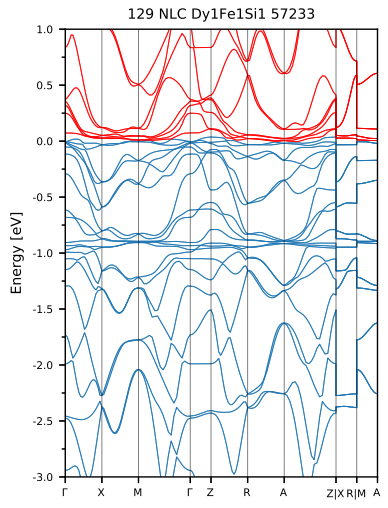

(ds) 129.2.30

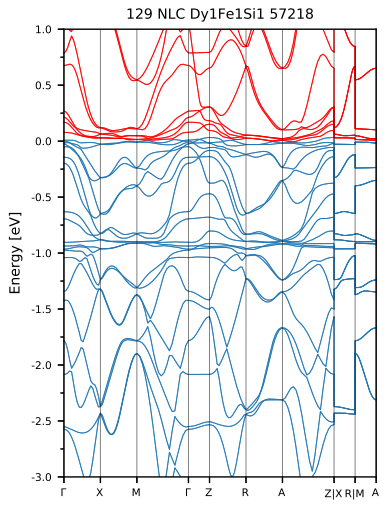

(dt) 129.2.31

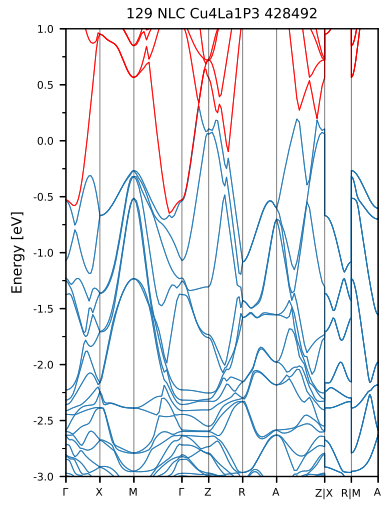

(du) 129.2.32

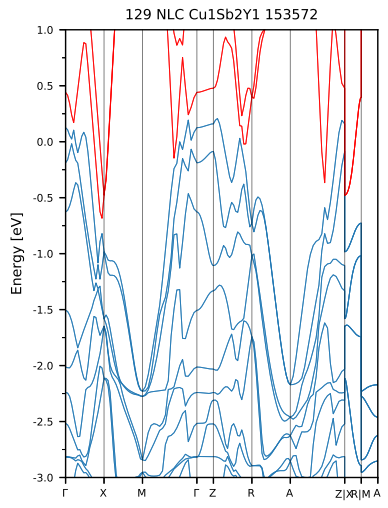

(dv) 129.2.33

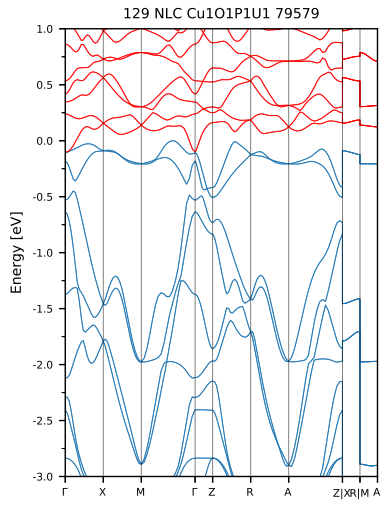

(dw) 129.2.34

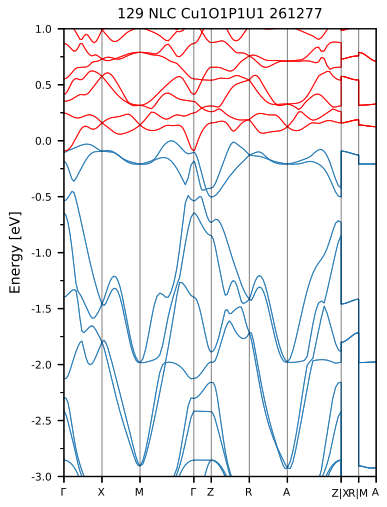

(dx) 129.2.35

Figure 246: Band structures for topological materials in SG 129, part 8/39, from 129.2.20 to 129.2.35. See the list of figures.

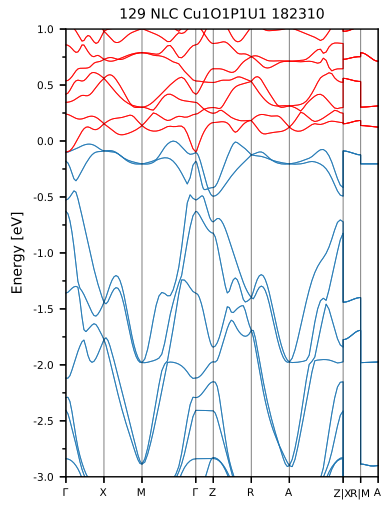

(dy) 129.2.36

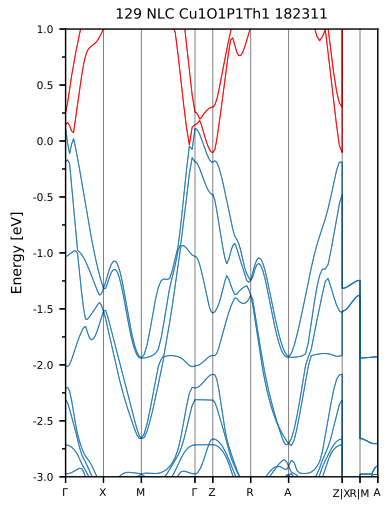

(dz) 129.2.37

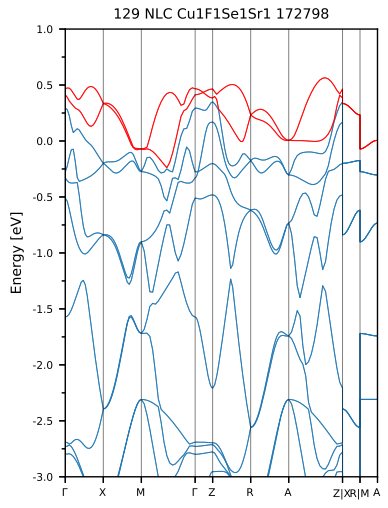

(ea) 129.2.38

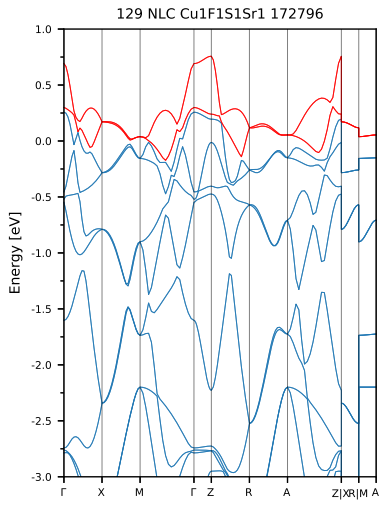

(eb) 129.2.39

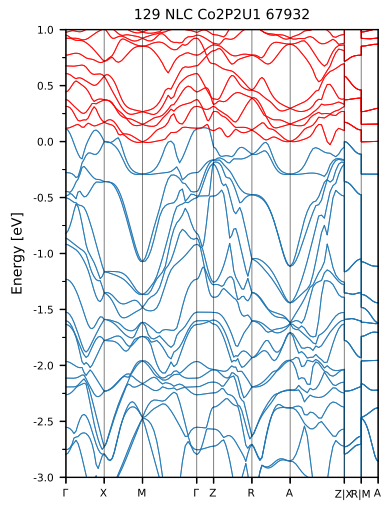

(ec) 129.2.40

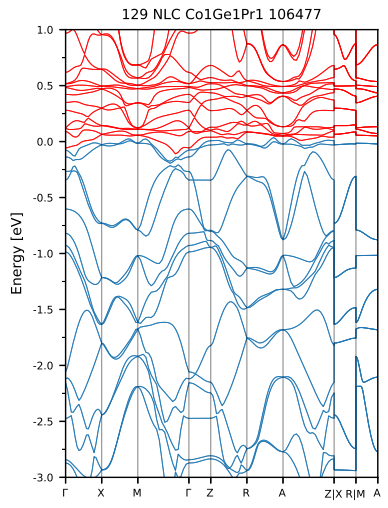

(ed) 129.2.41

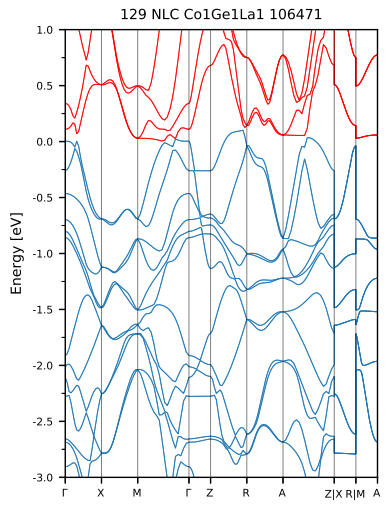

(ee) 129.2.42

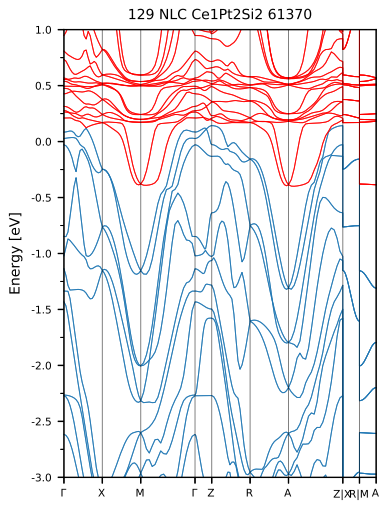

(ef) 129.2.43

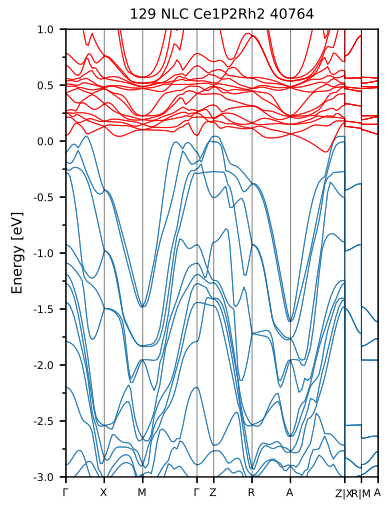

(eg) 129.2.44

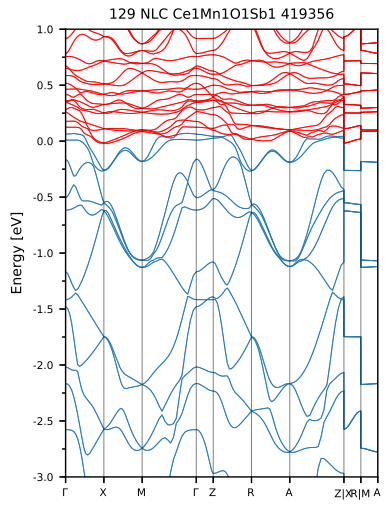

(eh) 129.2.45

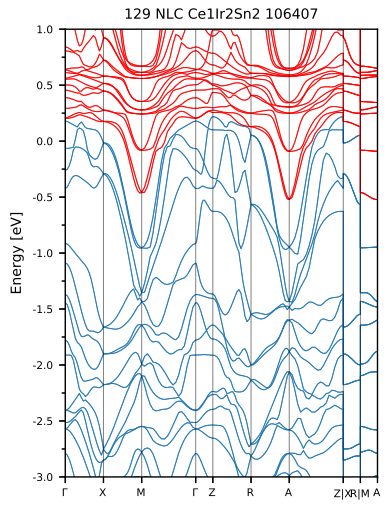

(ei) 129.2.46

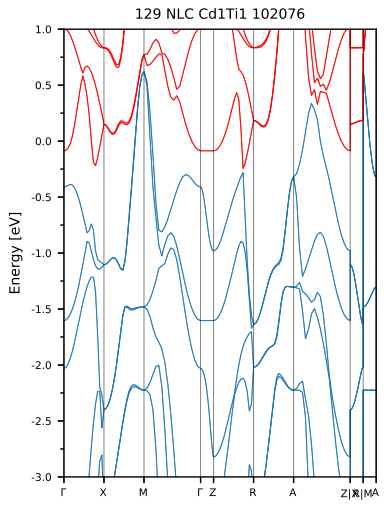

(ej) 129.2.47

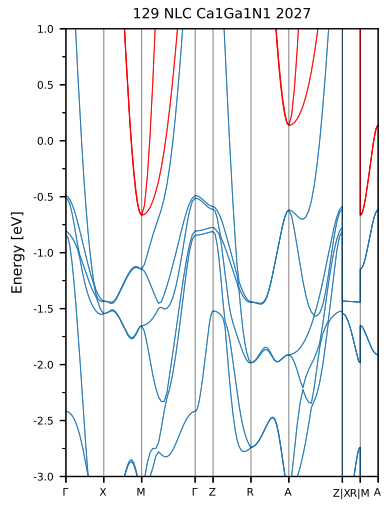

(ek) 129.2.48

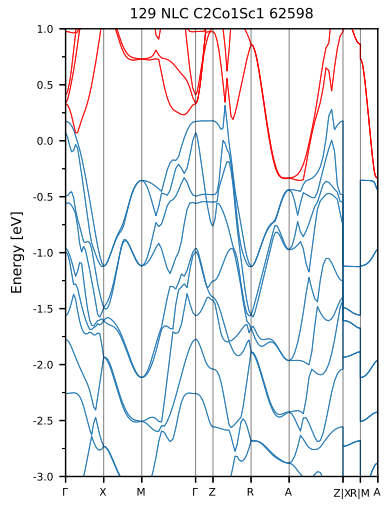

(el) 129.2.49

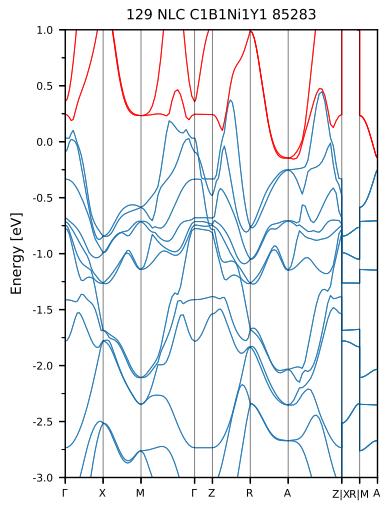

(em) 129.2.50

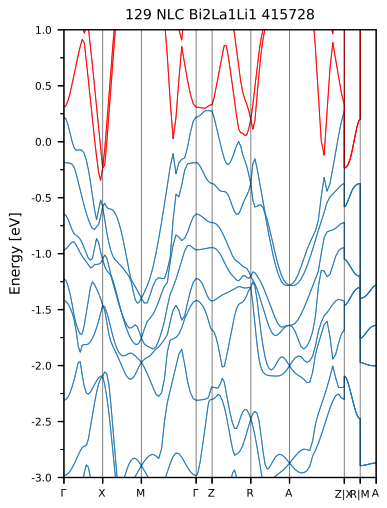

(en) 129.2.51

Figure 247: Band structures for topological materials in SG 129, part 9/39, from 129.2.36 to 129.2.51. See the list of figures.

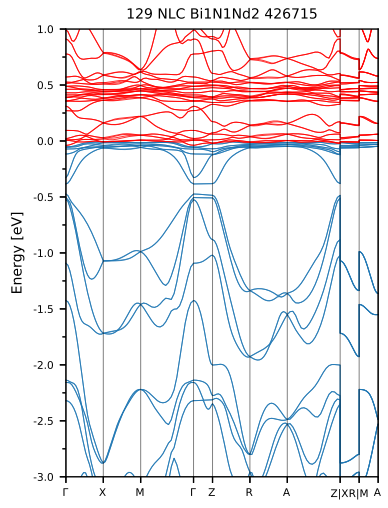

(eo) 129.2.52

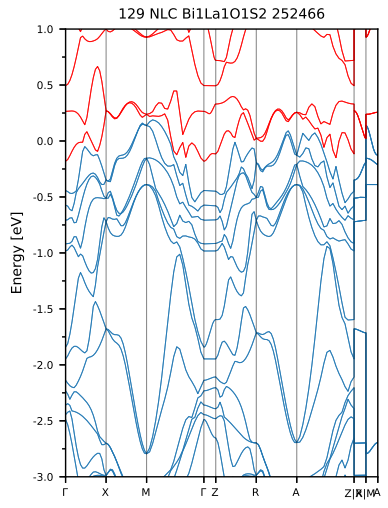

(ep) 129.2.53

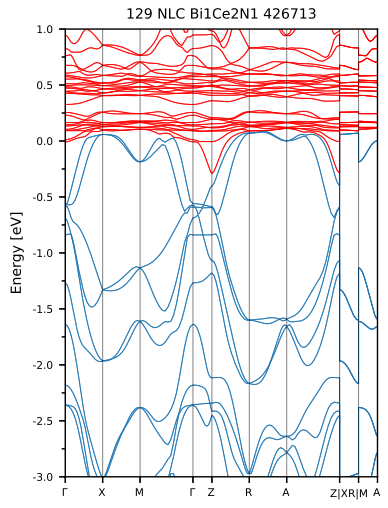

(eq) 129.2.54

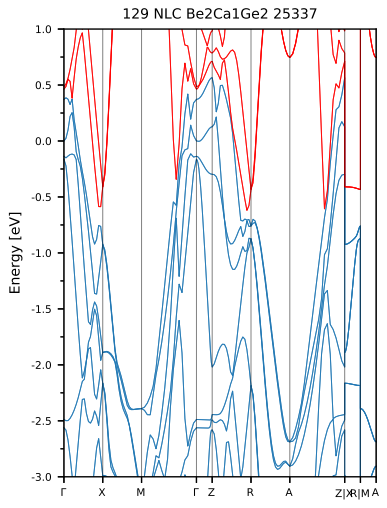

(er) 129.2.55

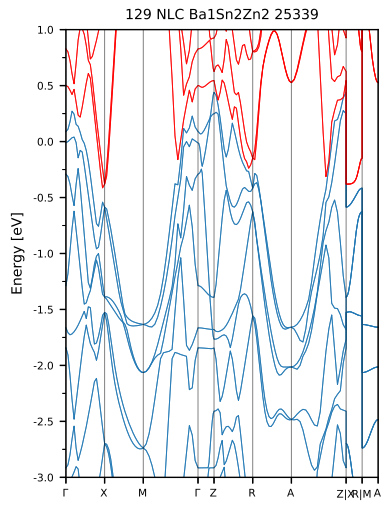

(es) 129.2.56

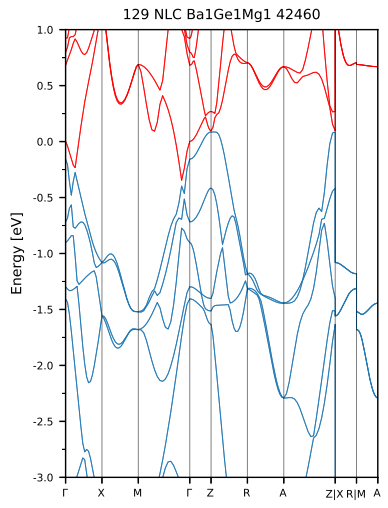

(et) 129.2.57

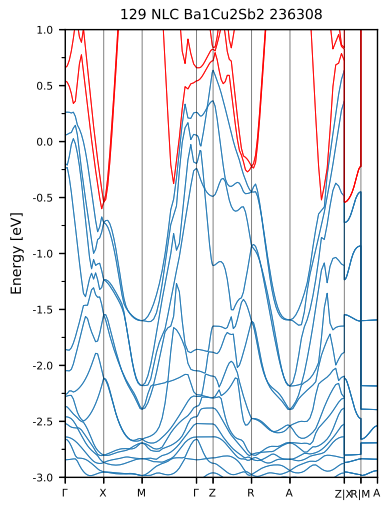

(eu) 129.2.58

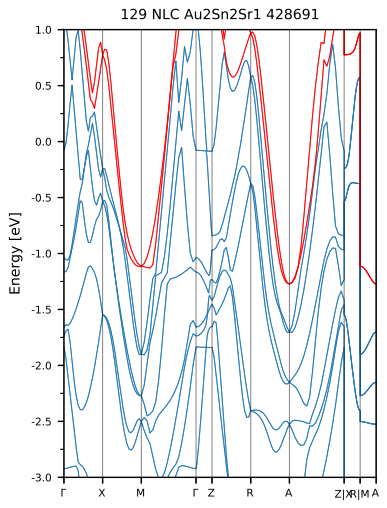

(ev) 129.2.59

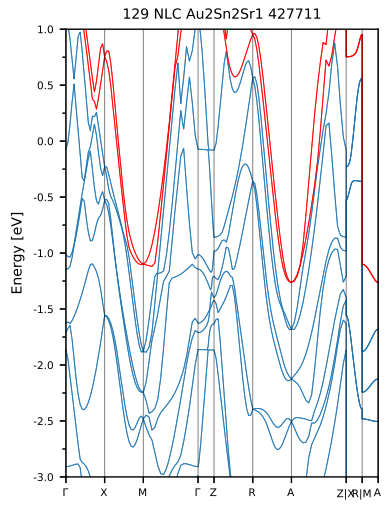

(ew) 129.2.60

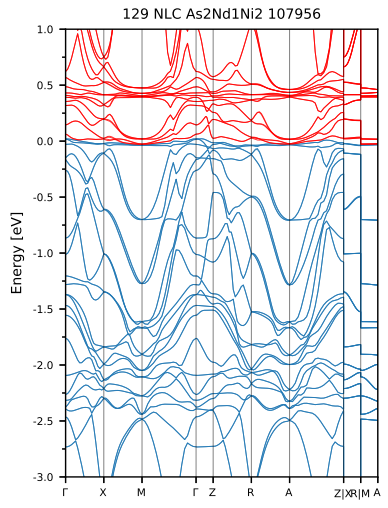

(ex) 129.2.61

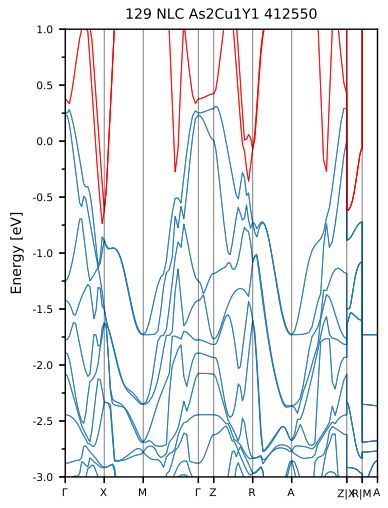

(ey) 129.2.62

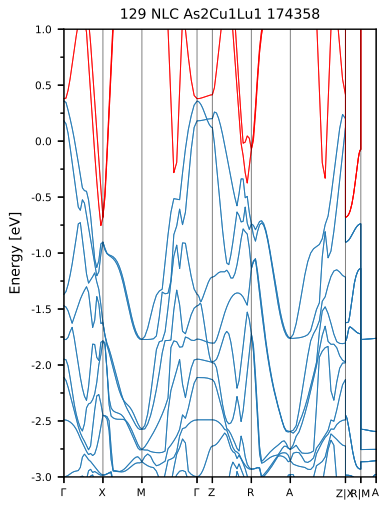

(ez) 129.2.63

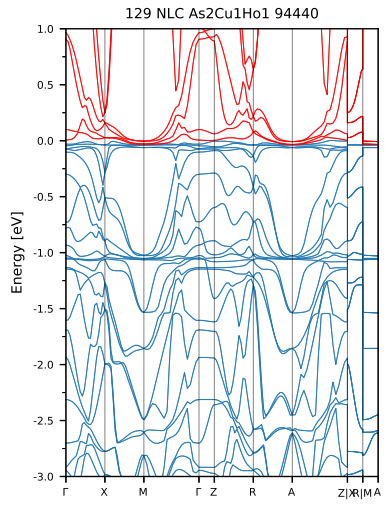

(fa) 129.2.64

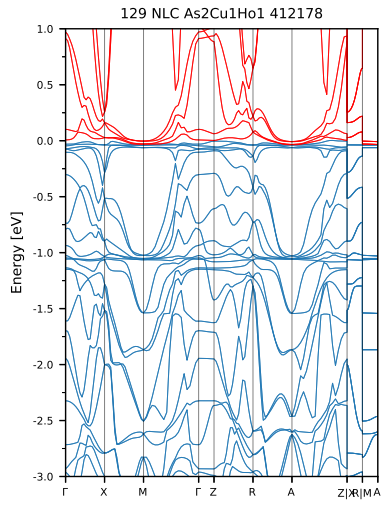

(fb) 129.2.65

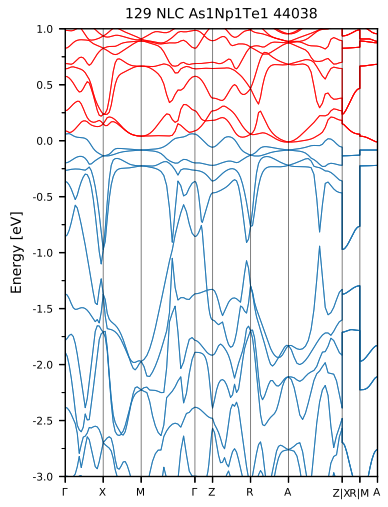

(fc) 129.2.66

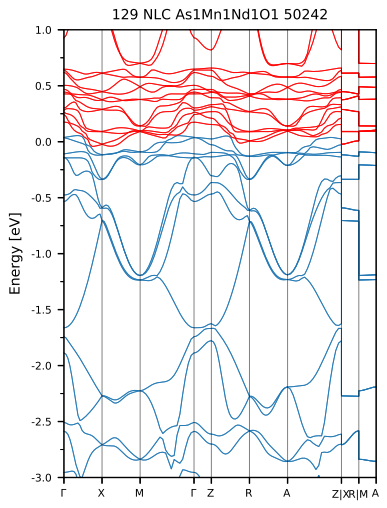

(fd) 129.2.67

Figure 248: Band structures for topological materials in SG 129, part 10/39, from 129.2.52 to 129.2.67. See the list of figures.

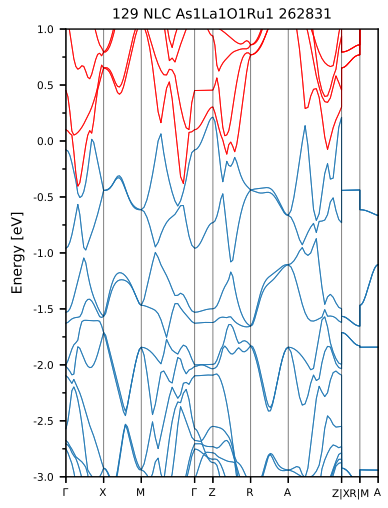

(fe) 129.2.68

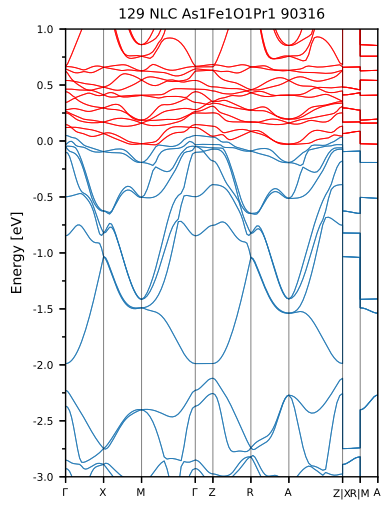

(ff) 129.2.69

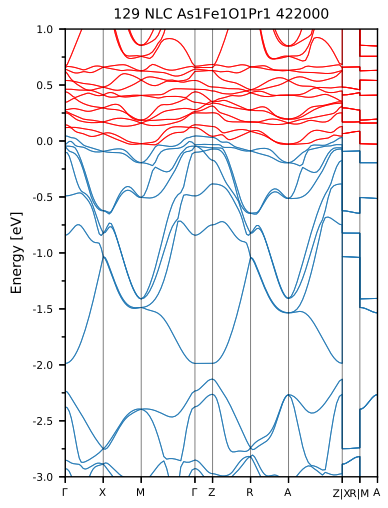

(fg) 129.2.70

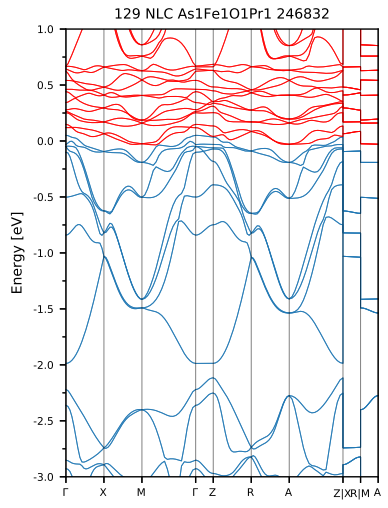

(fh) 129.2.71

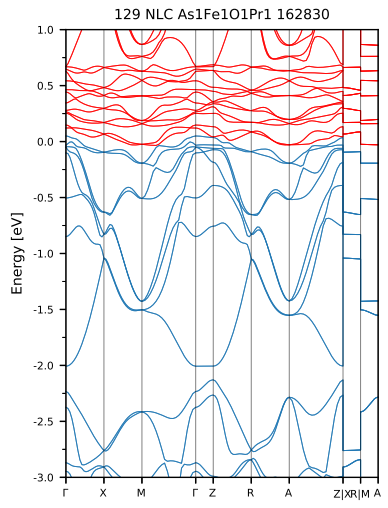

(fi) 129.2.72

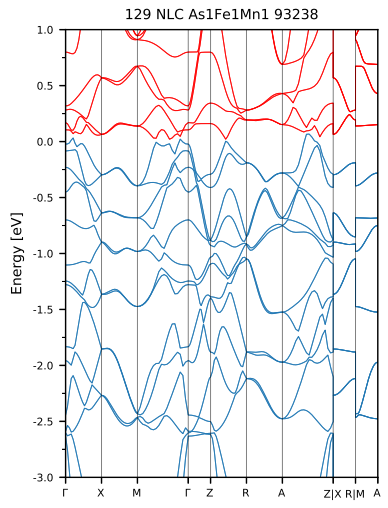

(fj) 129.2.73

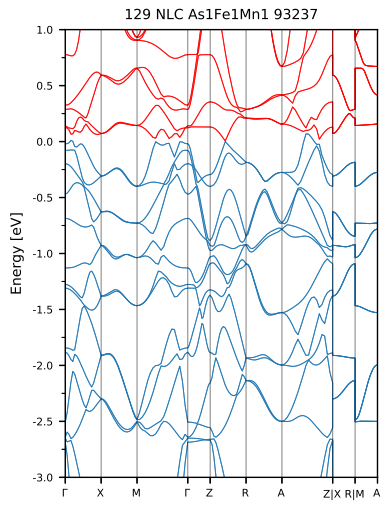

(fk) 129.2.74

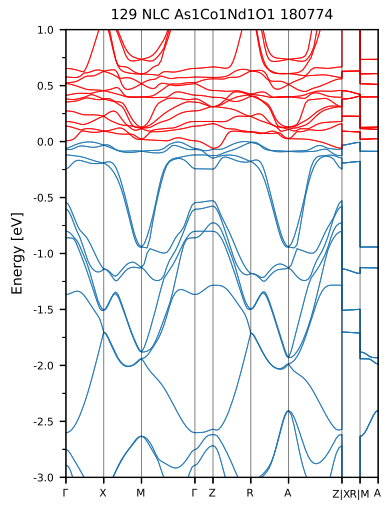

(fl) 129.2.75

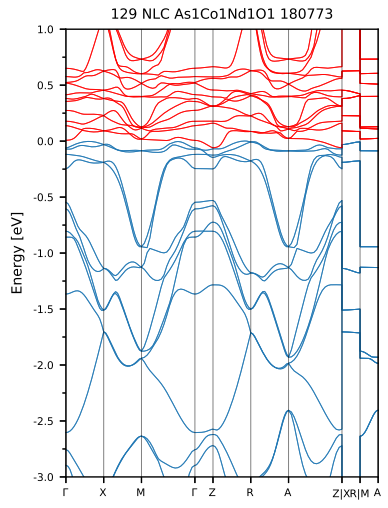

(fm) 129.2.76

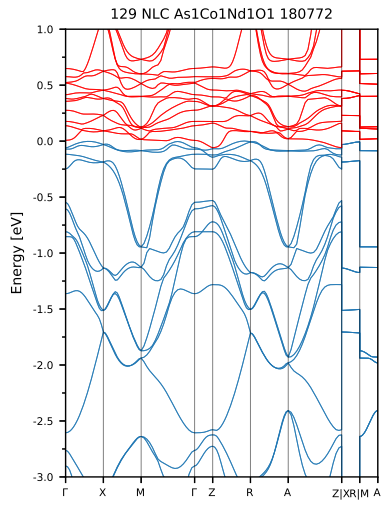

(fn) 129.2.77

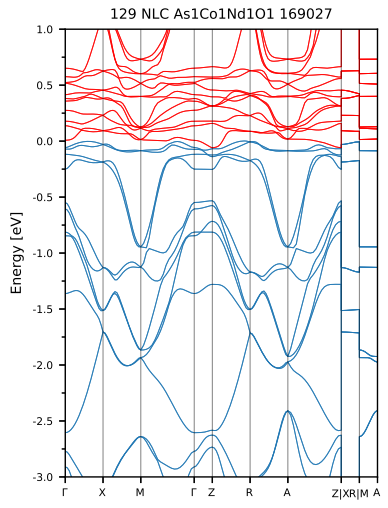

(fo) 129.2.78

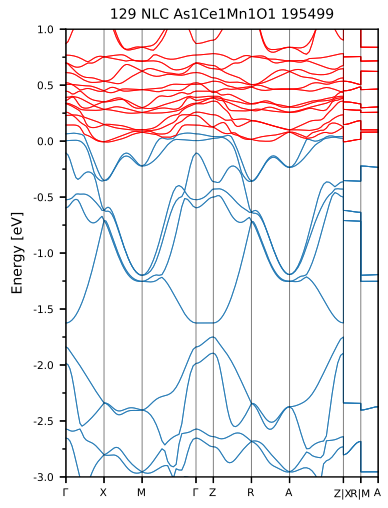

(fp) 129.2.79

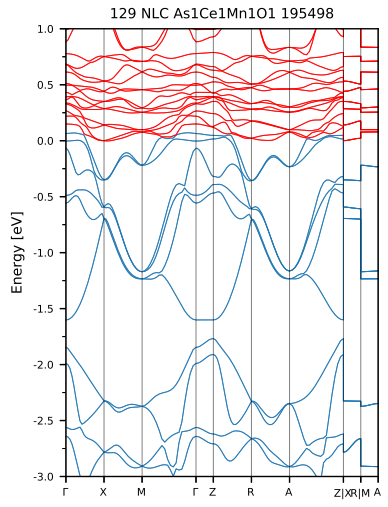

(fq) 129.2.80

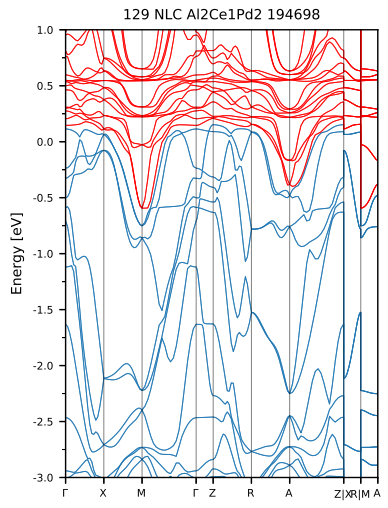

(fr) 129.2.81

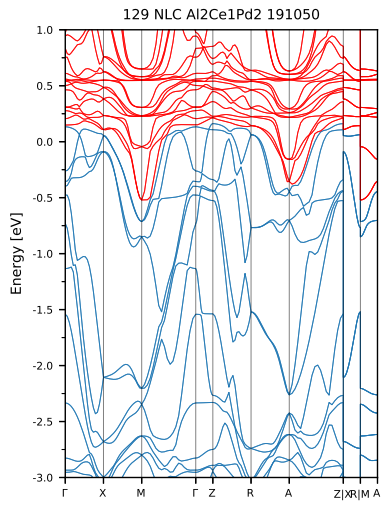

(fs) 129.2.82

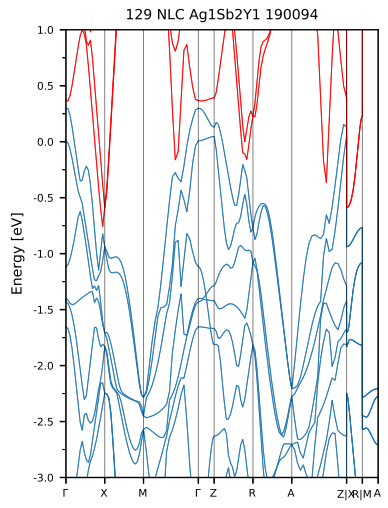

(ft) 129.2.83

Figure 249: Band structures for topological materials in SG 129, part 11/39, from 129.2.68 to 129.2.83. See the list of figures.

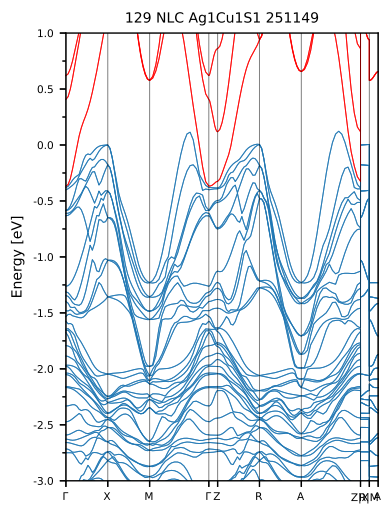

(fu) 129.2.84

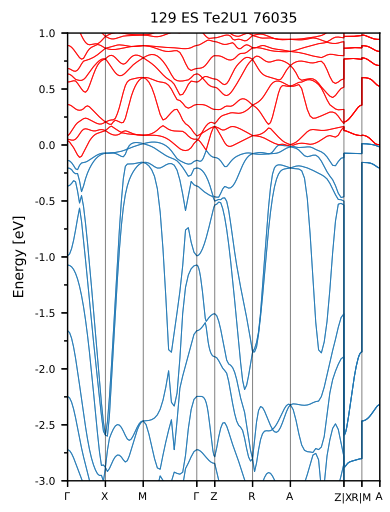

(fv) 129.3.1

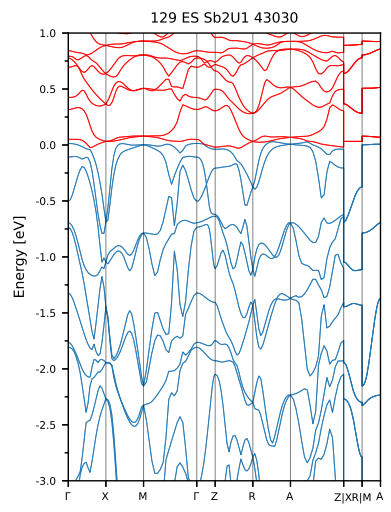

(fw) 129.3.2

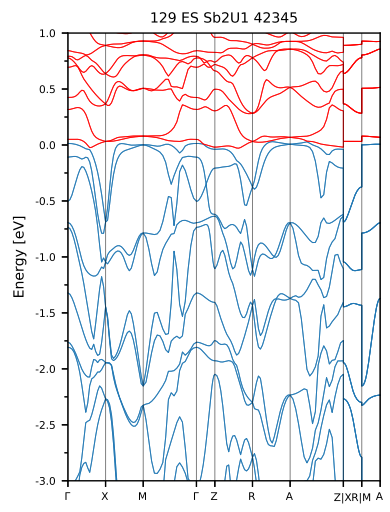

(fx) 129.3.3

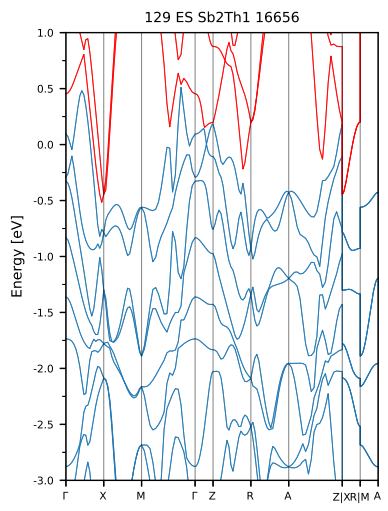

(fy) 129.3.4

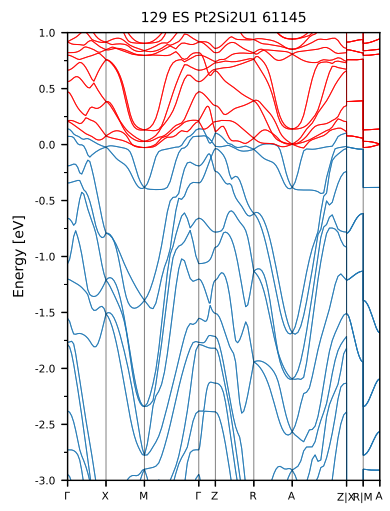

(fz) 129.3.5

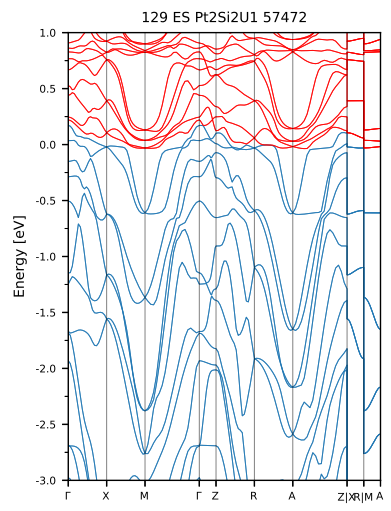

(ga) 129.3.6

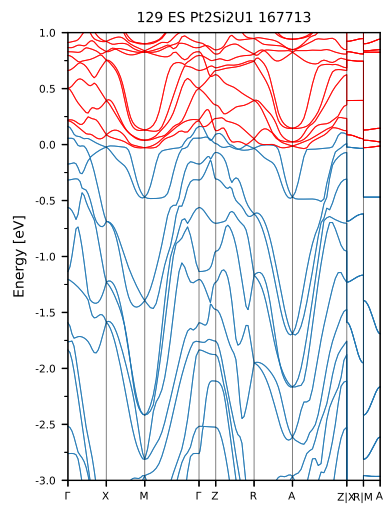

(gb) 129.3.7

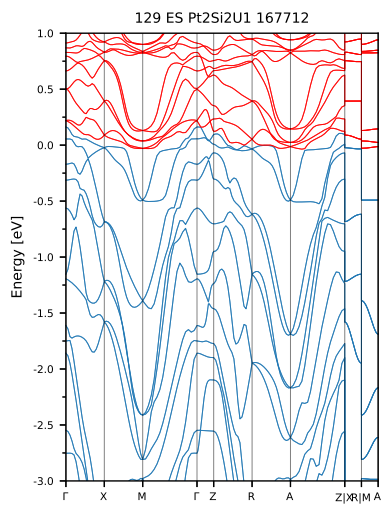

(gc) 129.3.8

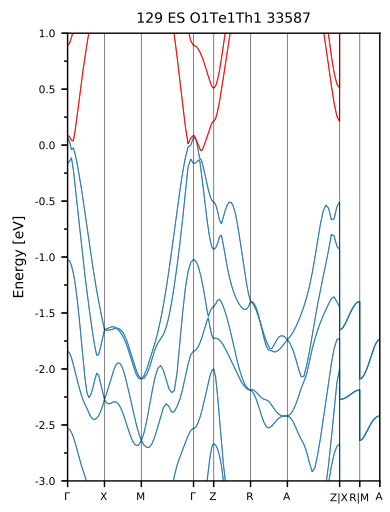

(gd) 129.3.9

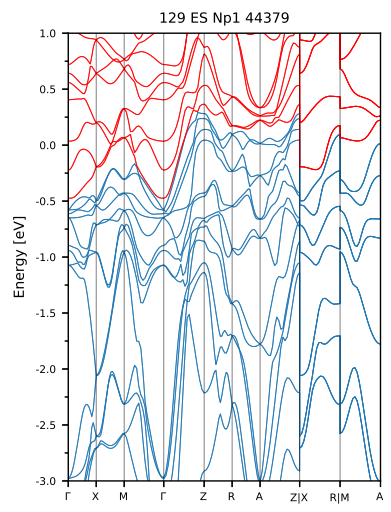

(ge) 129.3.10

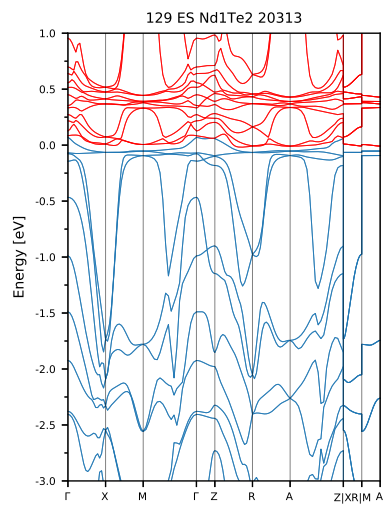

(gf) 129.3.11

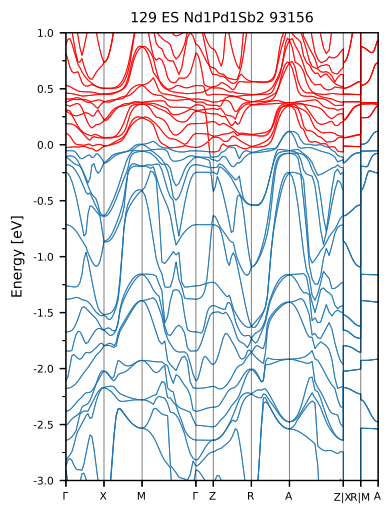

(gg) 129.3.12

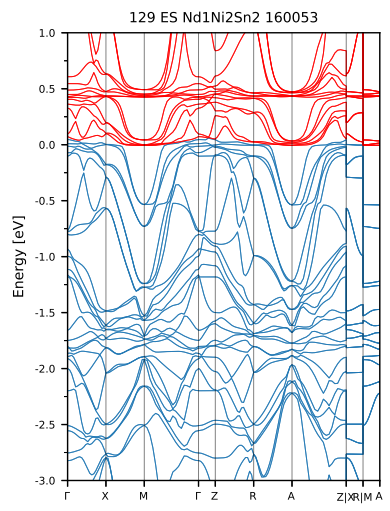

(gh) 129.3.13

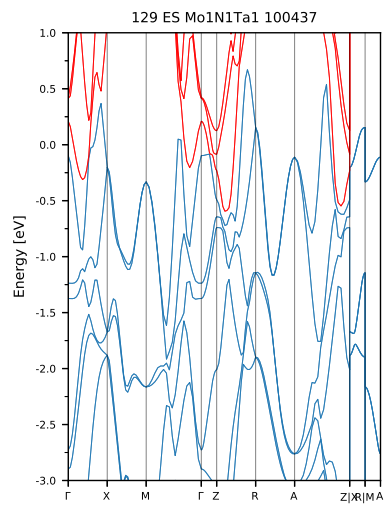

(gi) 129.3.14

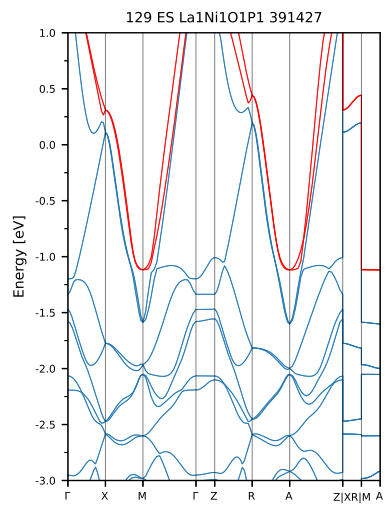

(gj) 129.3.15

Figure 250: Band structures for topological materials in SG 129, part 12/39, from 129.2.84 to 129.3.15. See the list of figures.

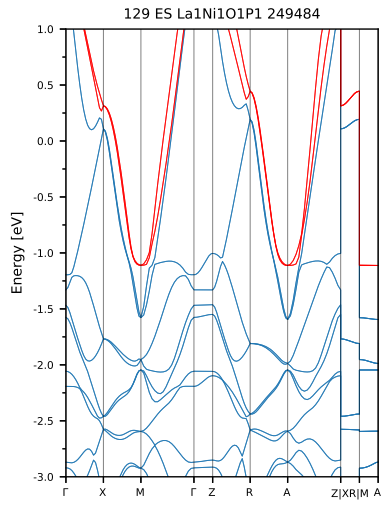

(gk) 129.3.16

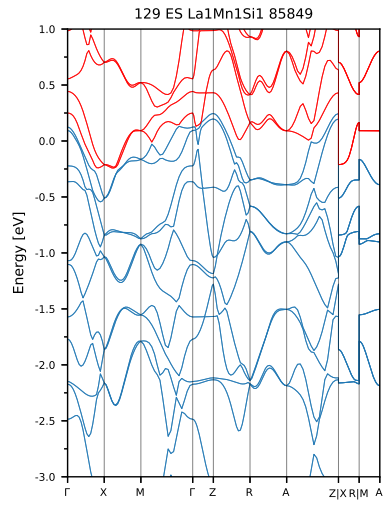

(gl) 129.3.17

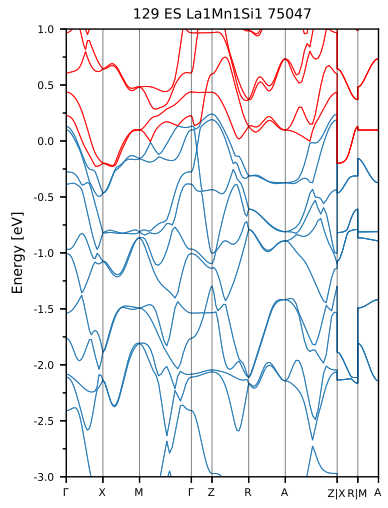

(gm) 129.3.18

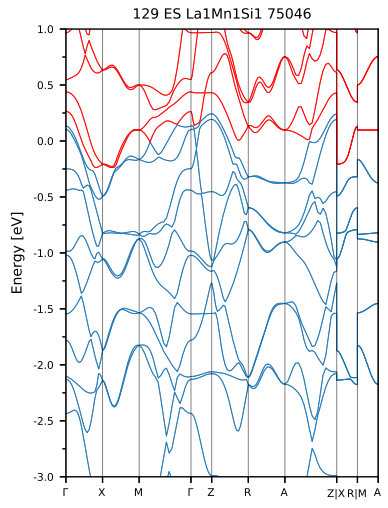

(gn) 129.3.19

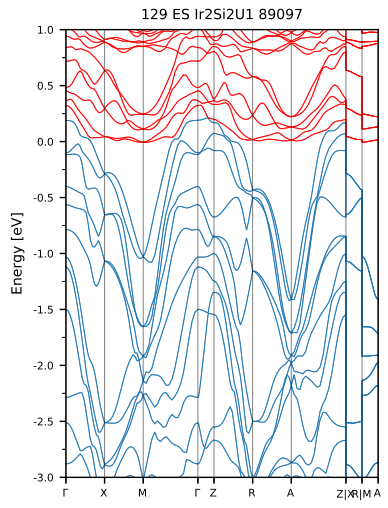

(go) 129.3.20

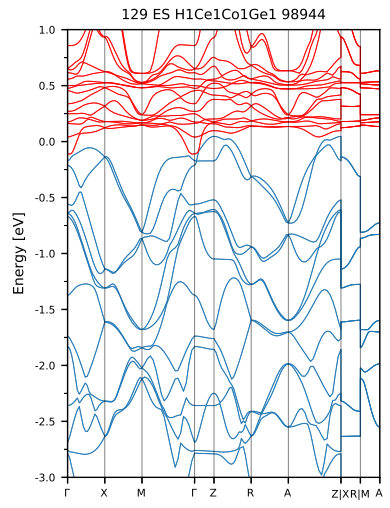

(gp) 129.3.21

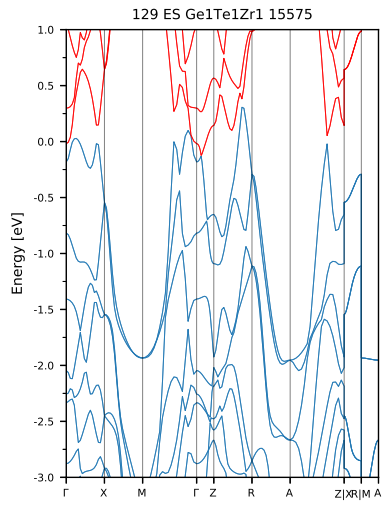

(gq) 129.3.22

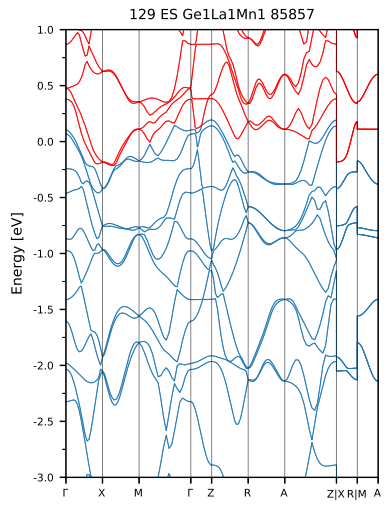

(gr) 129.3.23

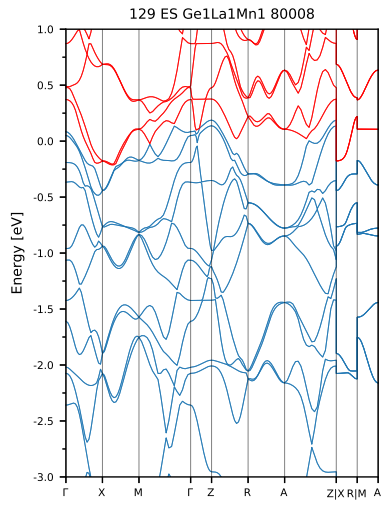

(gs) 129.3.24

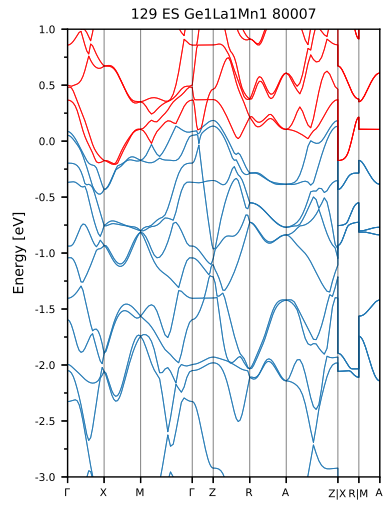

(gt) 129.3.25

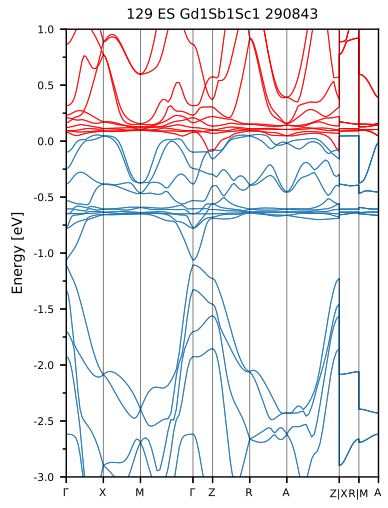

(gu) 129.3.26

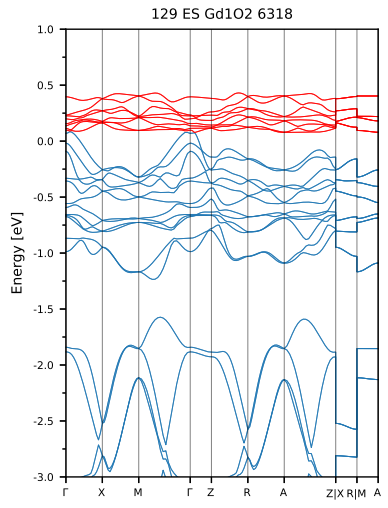

(gv) 129.3.27

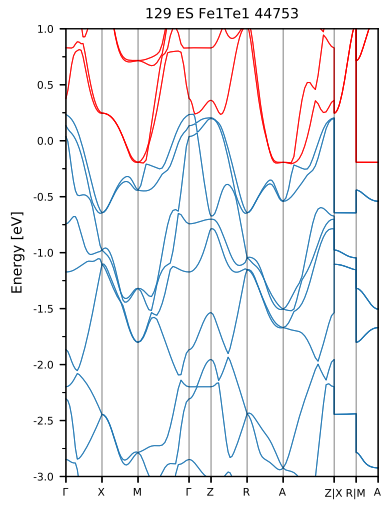

(gw) 129.3.28

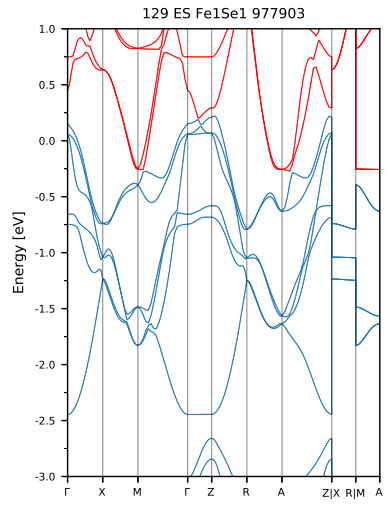

(gx) 129.3.29

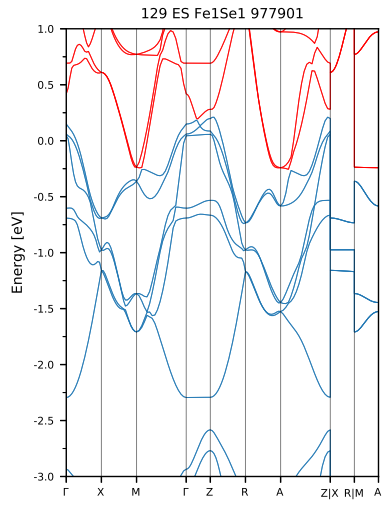

(gy) 129.3.30

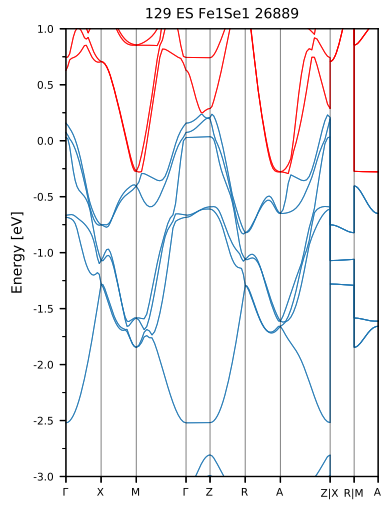

(gz) 129.3.31

Figure 251: Band structures for topological materials in SG 129, part 13/39, from 129.3.16 to 129.3.31. See the list of figures.

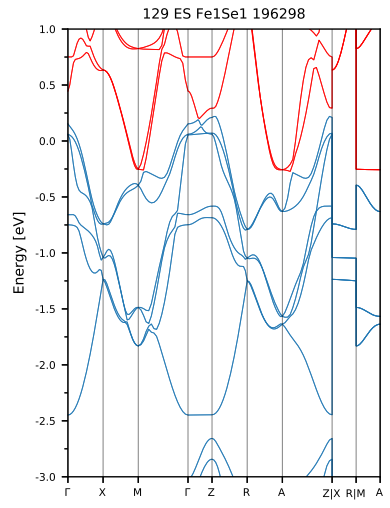

(ha) 129.3.32

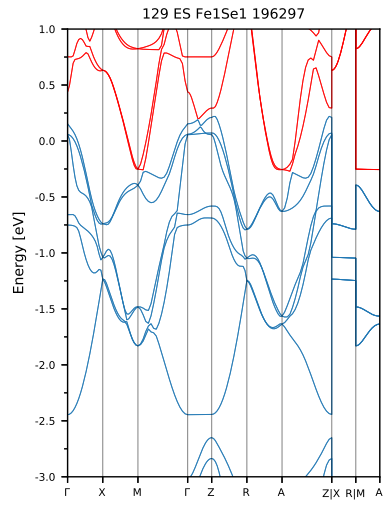

(hb) 129.3.33

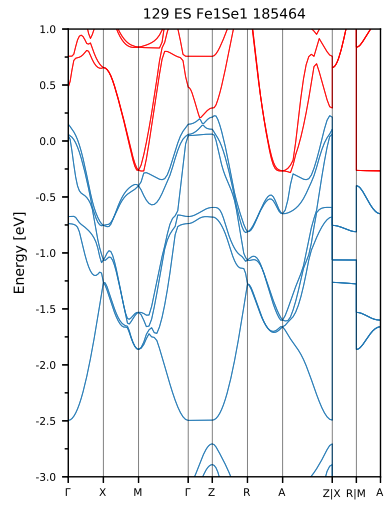

(hc) 129.3.34

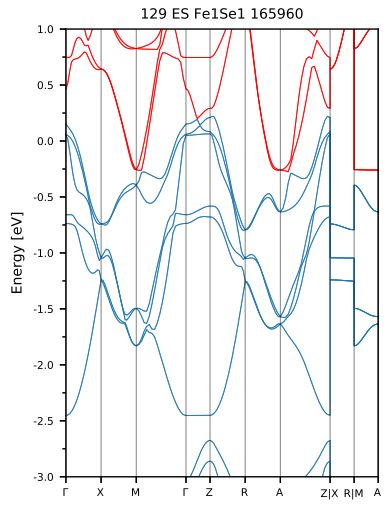

(hd) 129.3.35

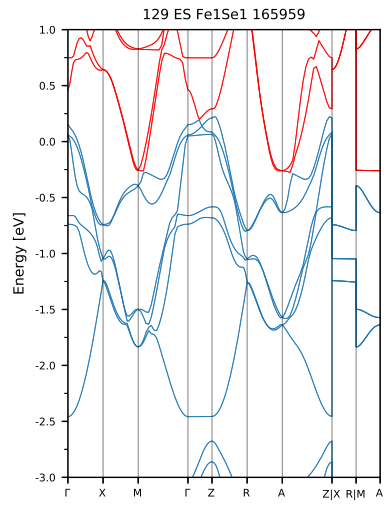

(he) 129.3.36

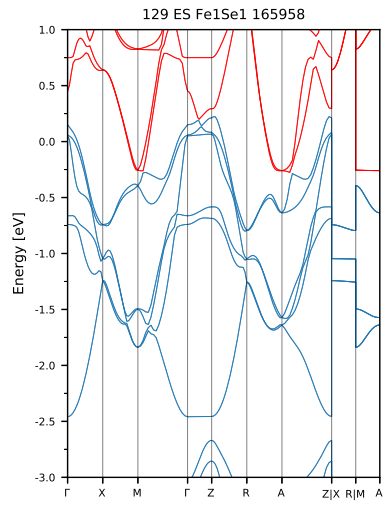

(hf) 129.3.37

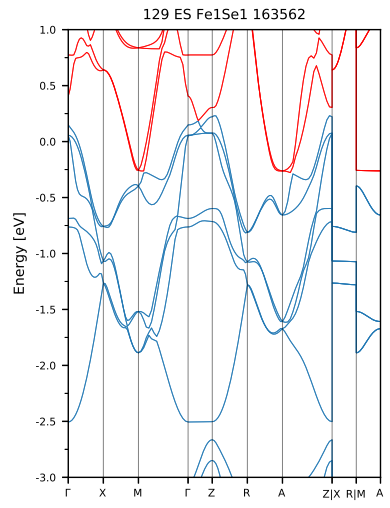

(hg) 129.3.38

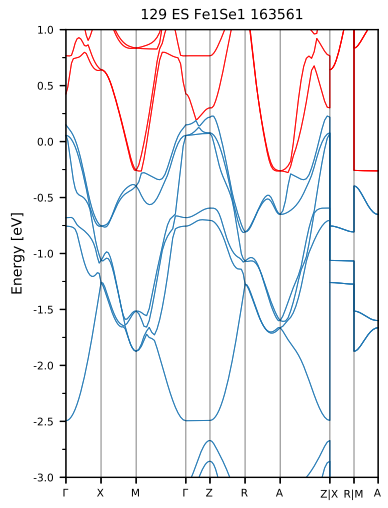

(hh) 129.3.39

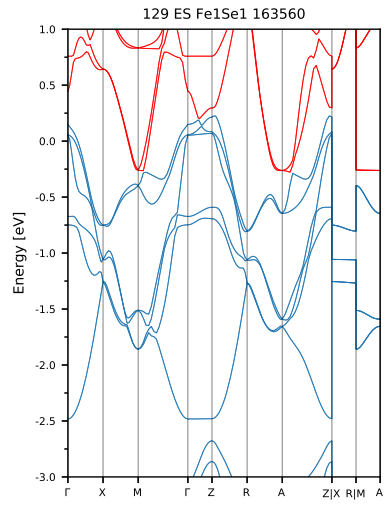

(hi) 129.3.40

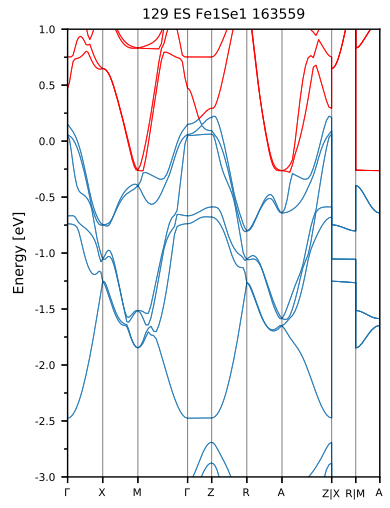

(hj) 129.3.41

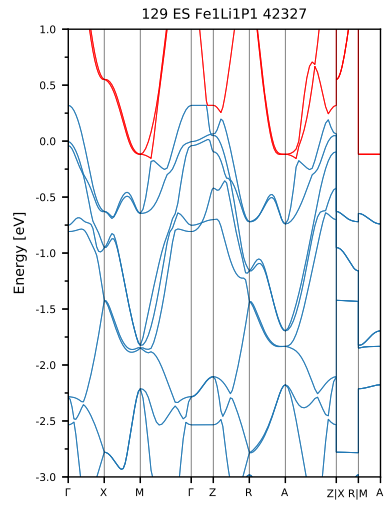

(hk) 129.3.42

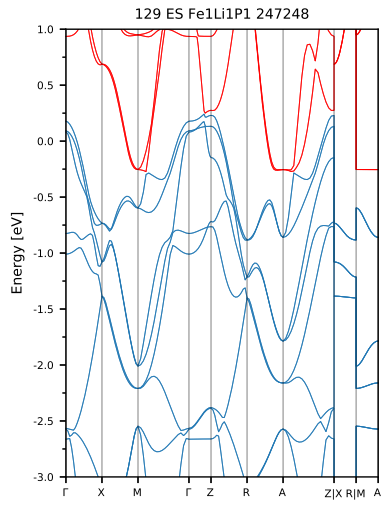

(hl) 129.3.43

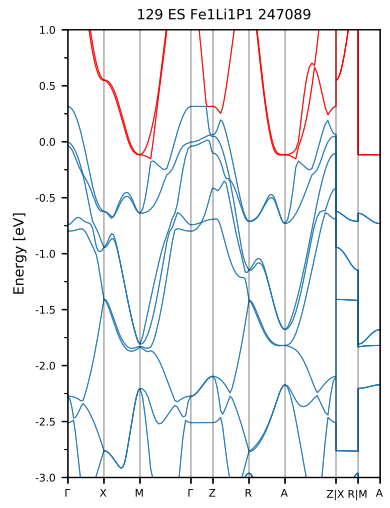

(hm) 129.3.44

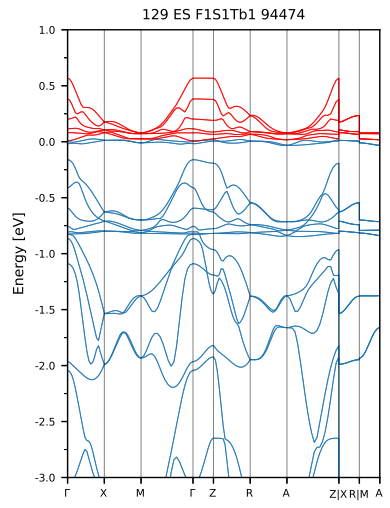

(hn) 129.3.45

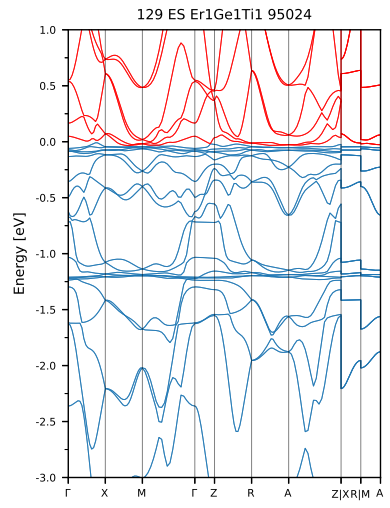

(ho) 129.3.46

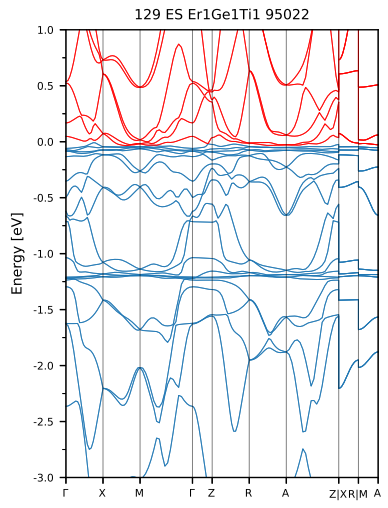

(hp) 129.3.47

Figure 252: Band structures for topological materials in SG 129, part 14/39, from 129.3.32 to 129.3.47. See the list of figures.

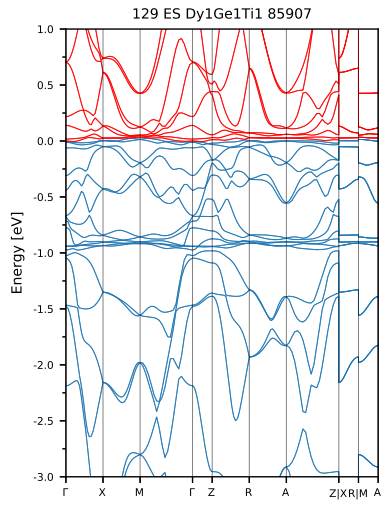

(hq) 129.3.48

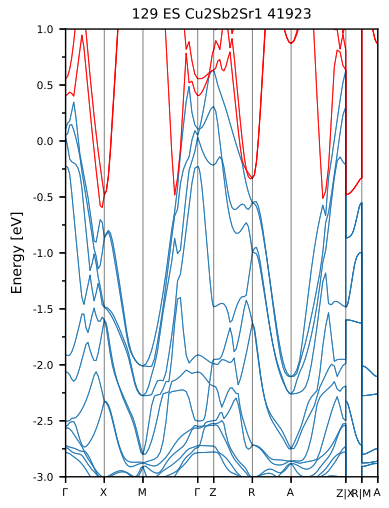

(hr) 129.3.49

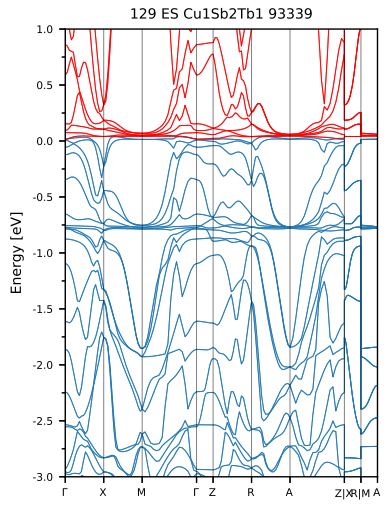

(hs) 129.3.50

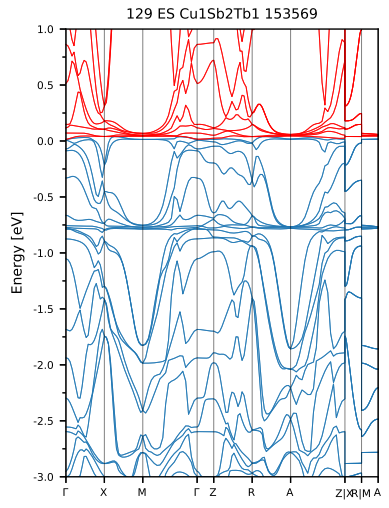

(ht) 129.3.51

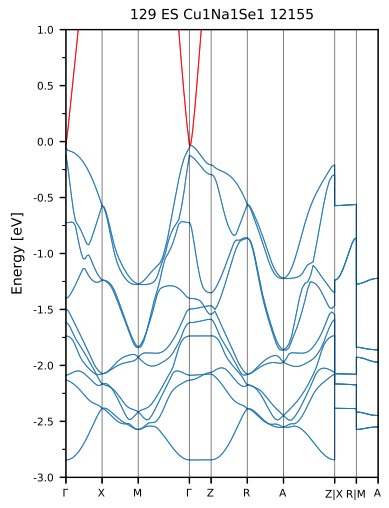

(hu) 129.3.52

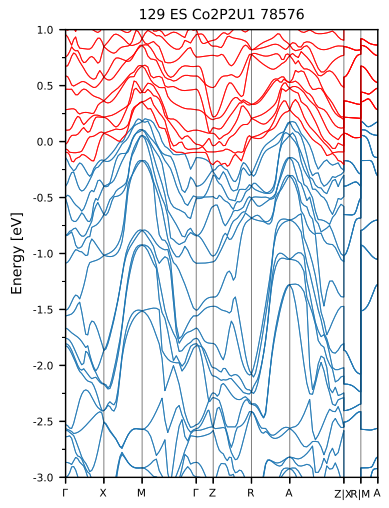

(hv) 129.3.53

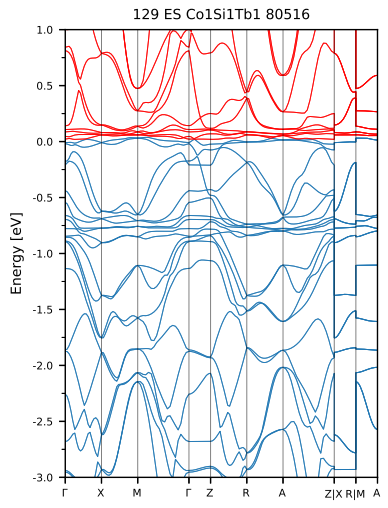

(hw) 129.3.54

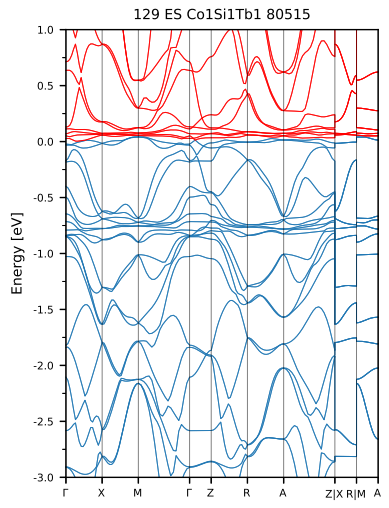

(hx) 129.3.55

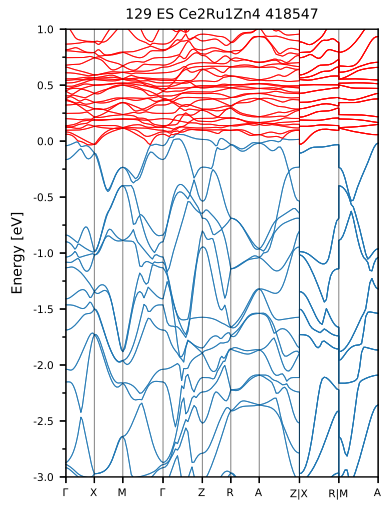

(hy) 129.3.56

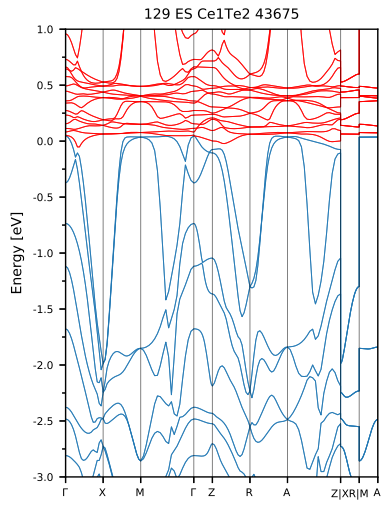

(hz) 129.3.57

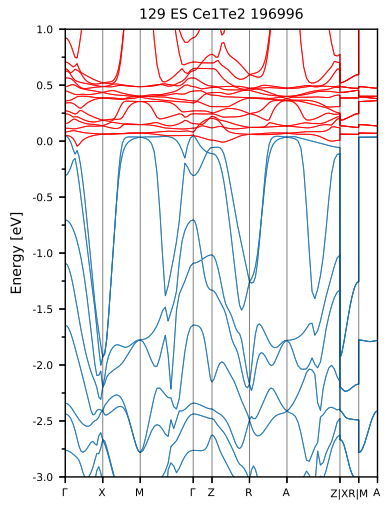

(ia) 129.3.58

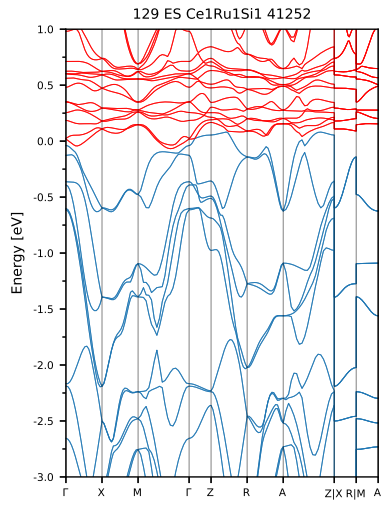

(ib) 129.3.59

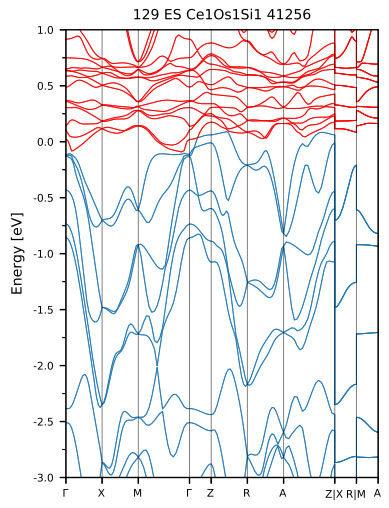

(ic) 129.3.60

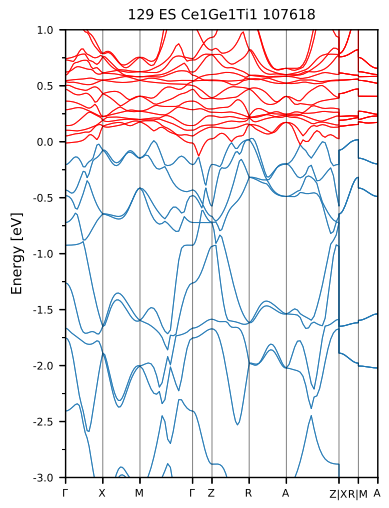

(id) 129.3.61

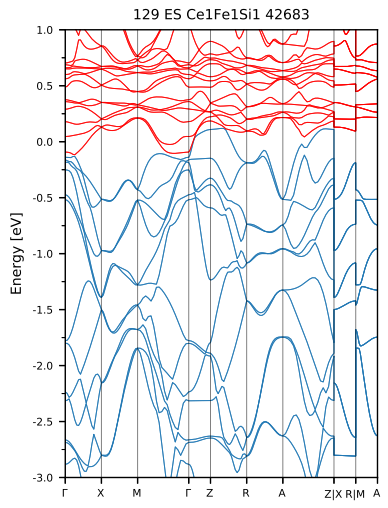

(ie) 129.3.62

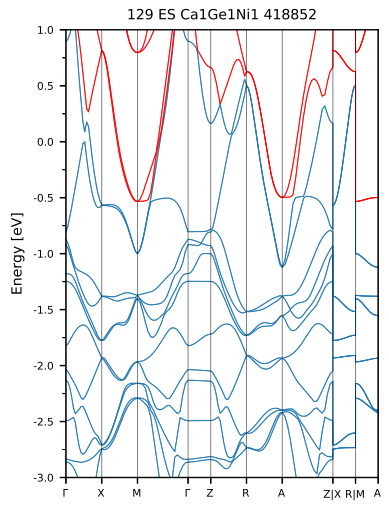

(if) 129.3.63

Figure 253: Band structures for topological materials in SG 129, part 15/39, from 129.3.48 to 129.3.63. See the list of figures.

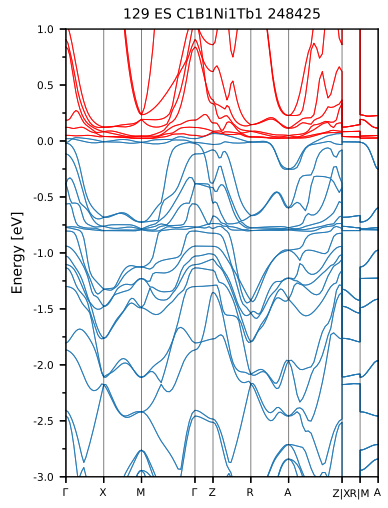

(ig) 129.3.64

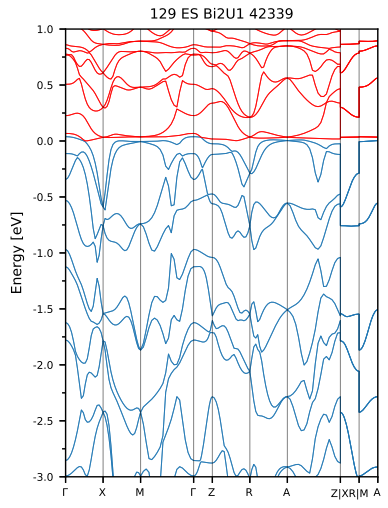

(ih) 129.3.65

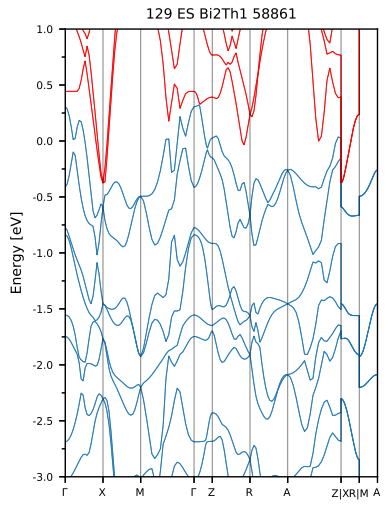

(ii) 129.3.66

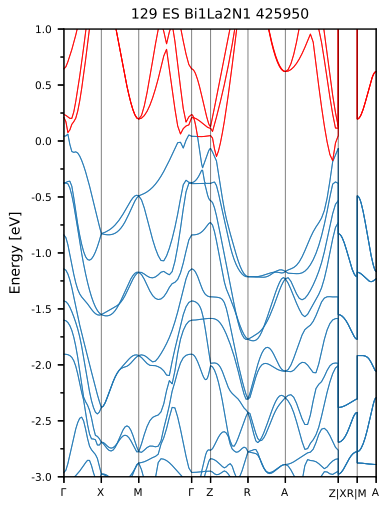

(ij) 129.3.67

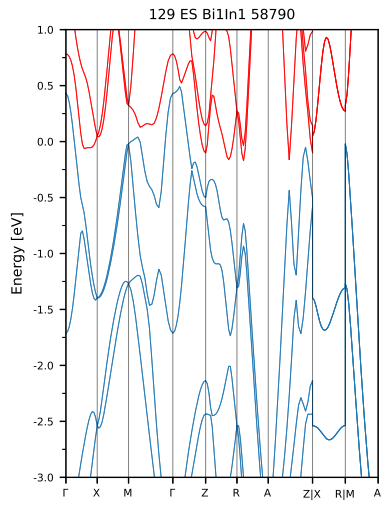

(ik) 129.3.68

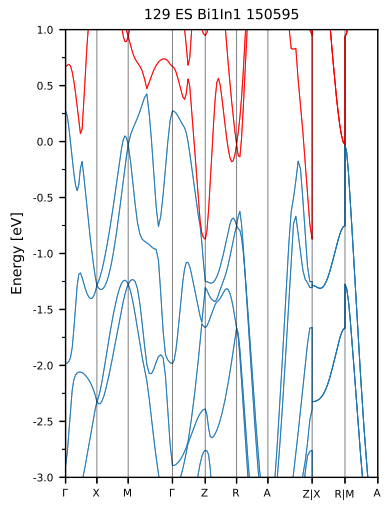

(il) 129.3.69

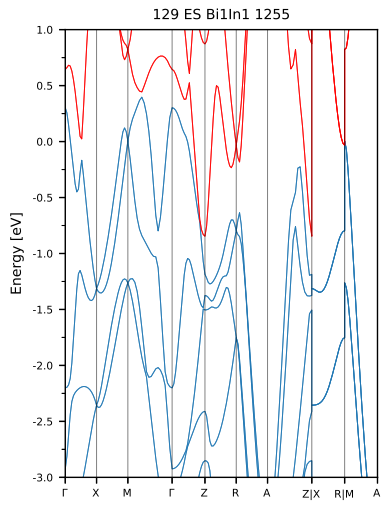

(im) 129.3.70

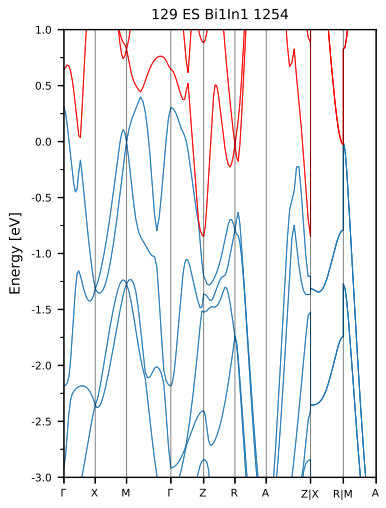

(in) 129.3.71

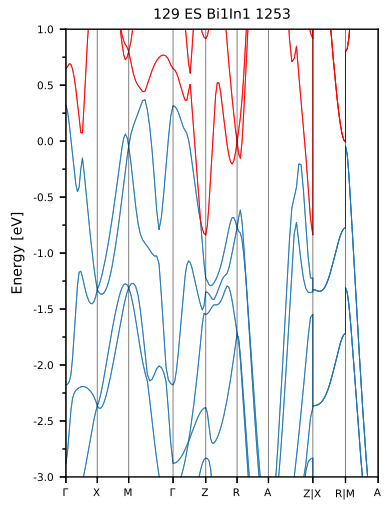

(io) 129.3.72

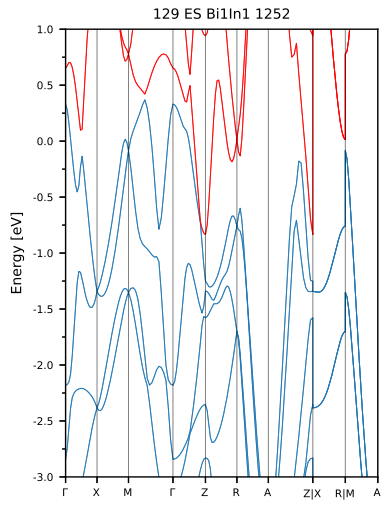

(ip) 129.3.73

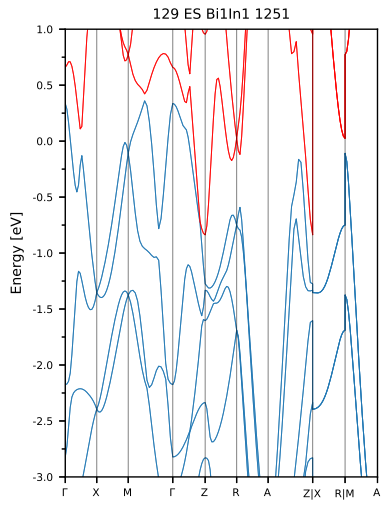

(iq) 129.3.74

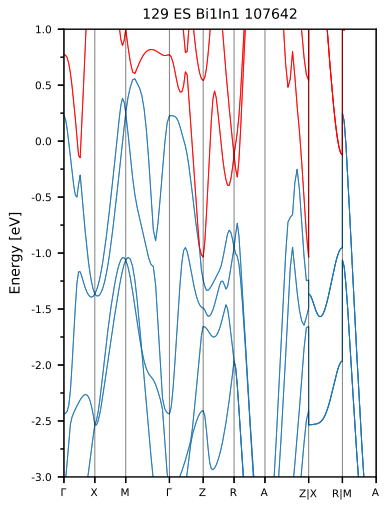

(ir) 129.3.75

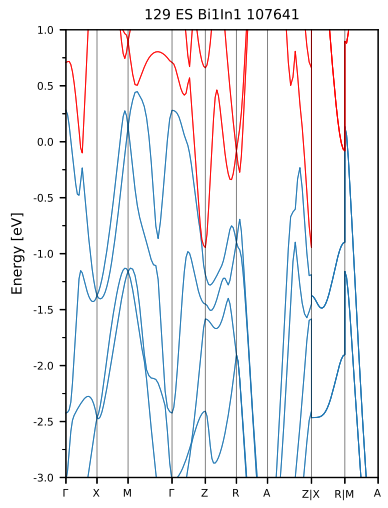

(is) 129.3.76

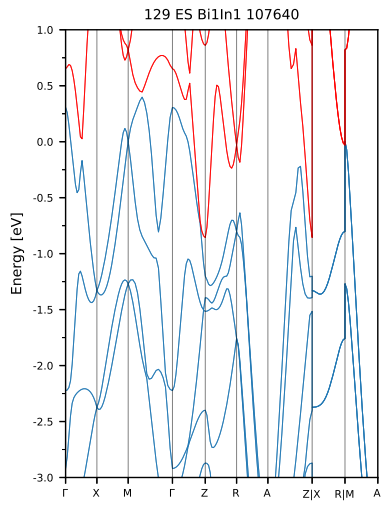

(it) 129.3.77

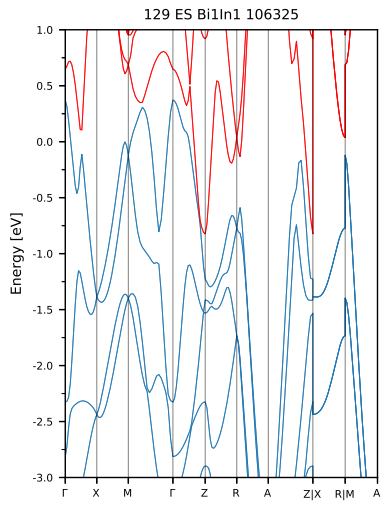

(iu) 129.3.78

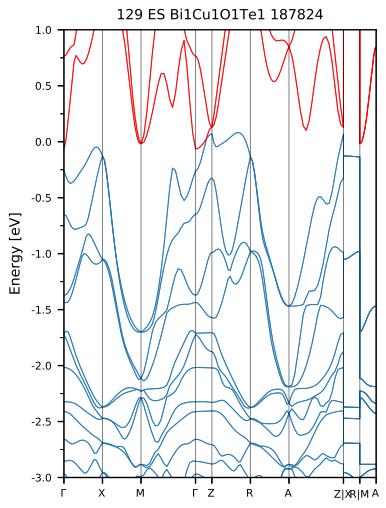

(iv) 129.3.79

Figure 254: Band structures for topological materials in SG 129, part 16/39, from 129.3.64 to 129.3.79. See the list of figures.

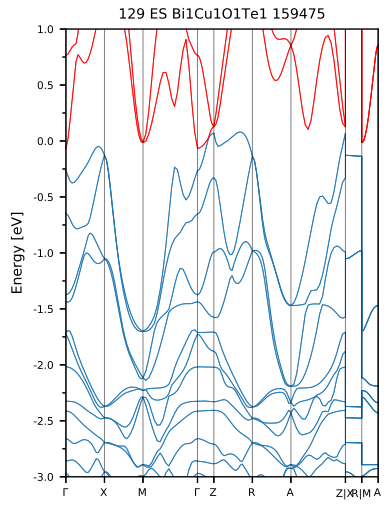

(iw) 129.3.80

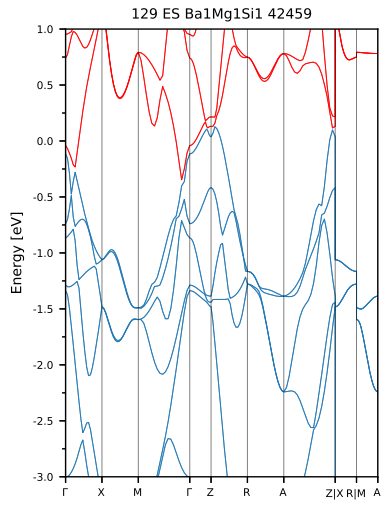

(ix) 129.3.81

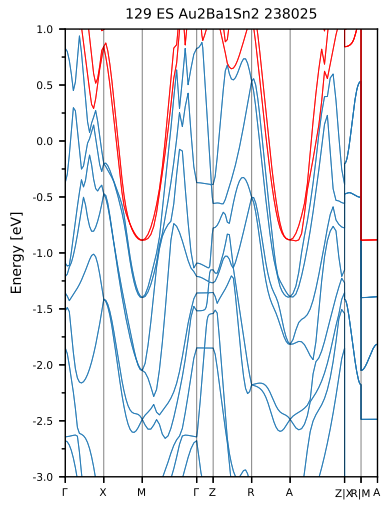

(iy) 129.3.82

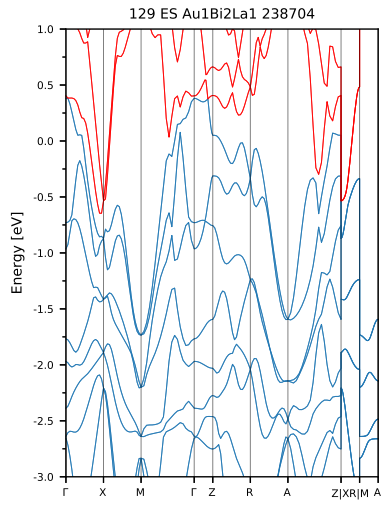

(iz) 129.3.83

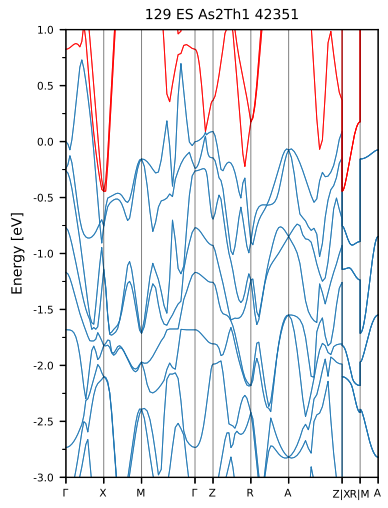

(ja) 129.3.84

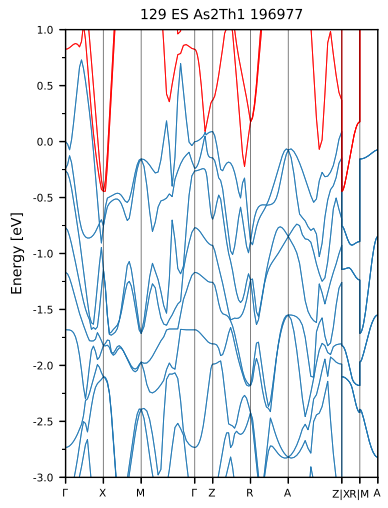

(jb) 129.3.85

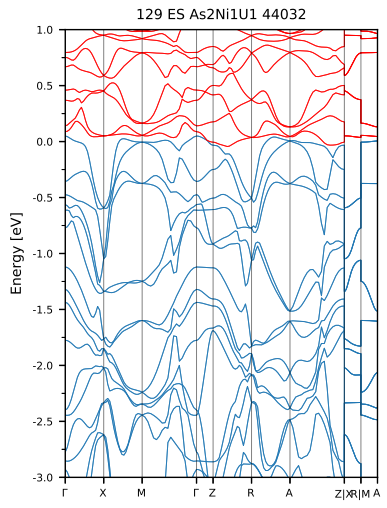

(jc) 129.3.86

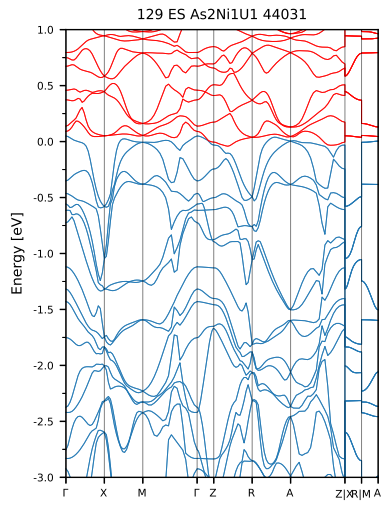

(jd) 129.3.87

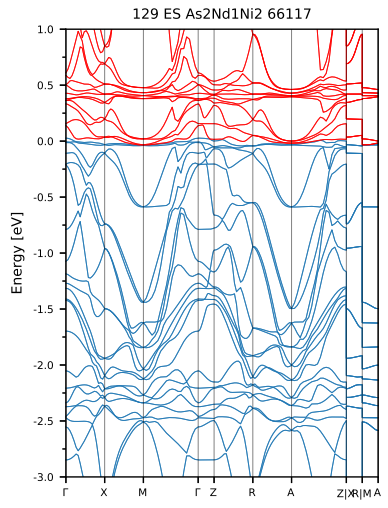

(je) 129.3.88

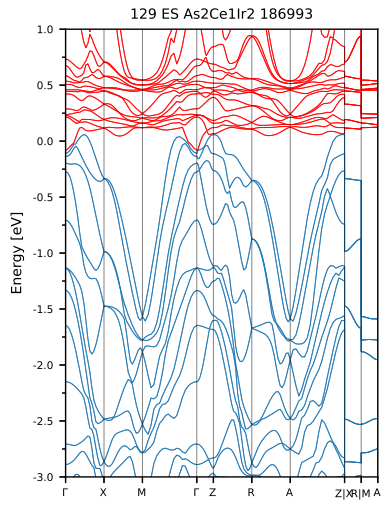

(jf) 129.3.89

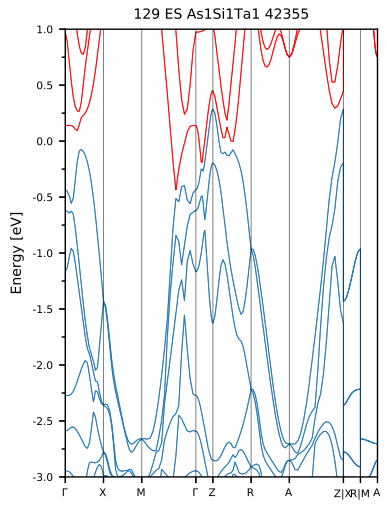

(jg) 129.3.90

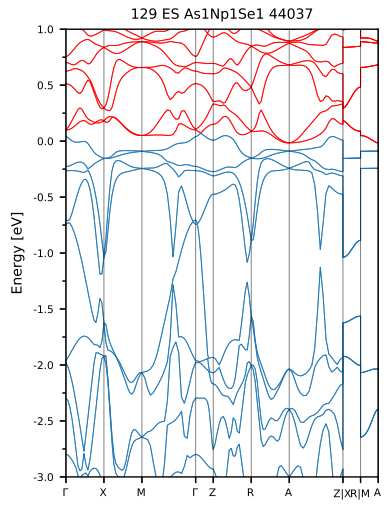

(jh) 129.3.91

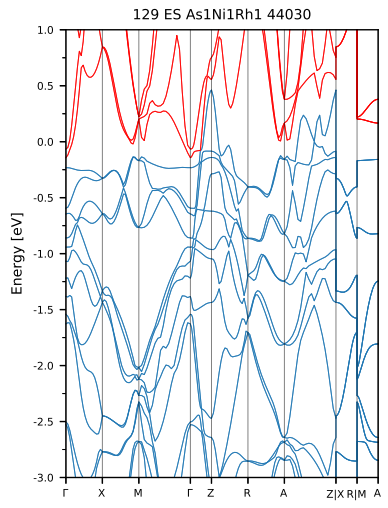

(ji) 129.3.92

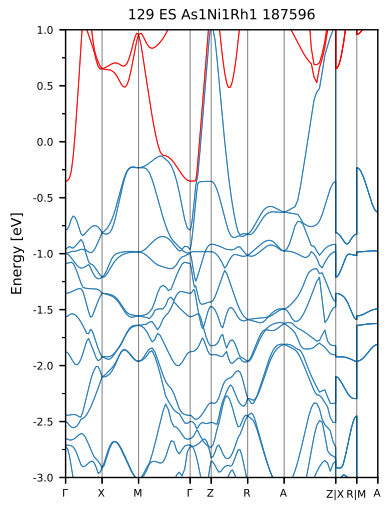

(jj) 129.3.93

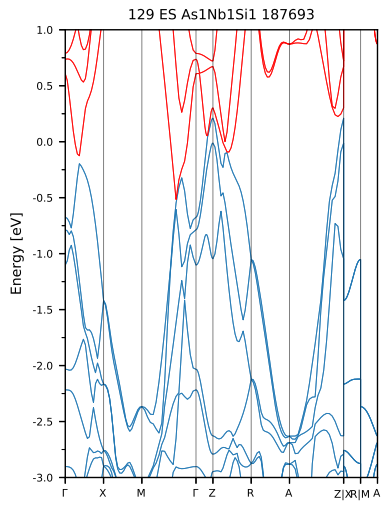

(jk) 129.3.94

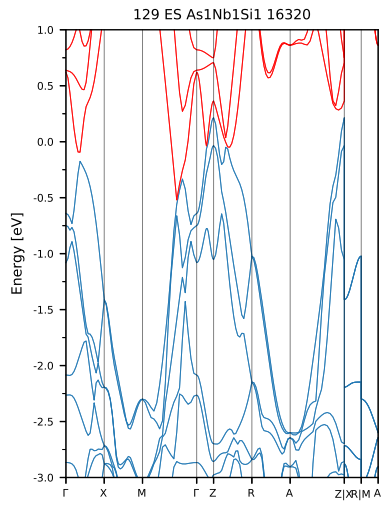

(jl) 129.3.95

Figure 255: Band structures for topological materials in SG 129, part 17/39, from 129.3.80 to 129.3.95. See the list of figures.

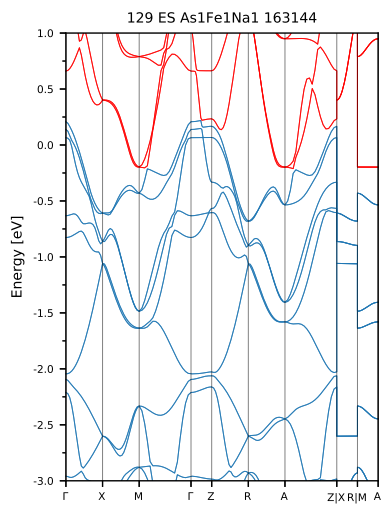

(jm) 129.3.96

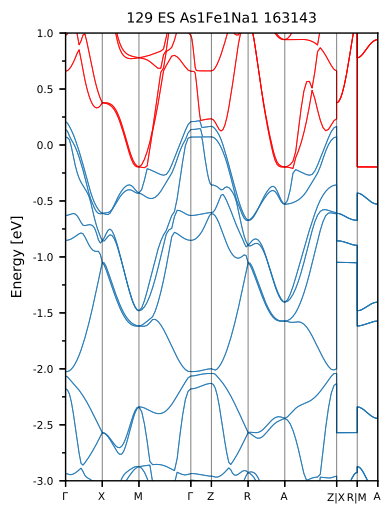

(jn) 129.3.97

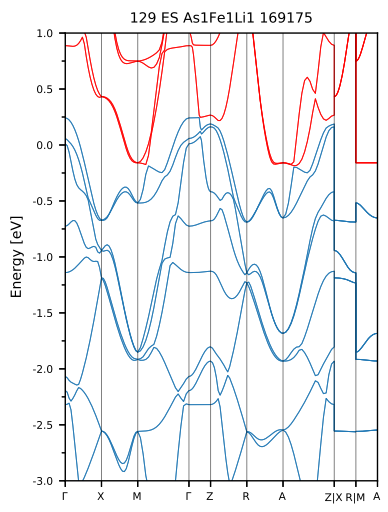

(jo) 129.3.98

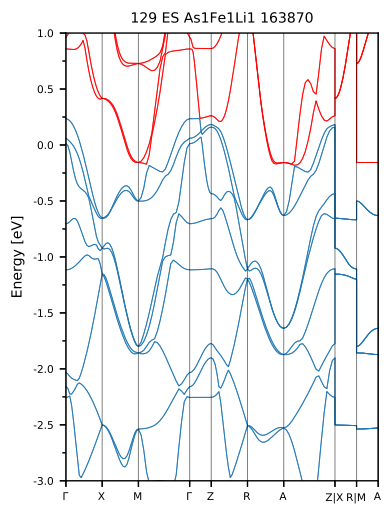

(jp) 129.3.99

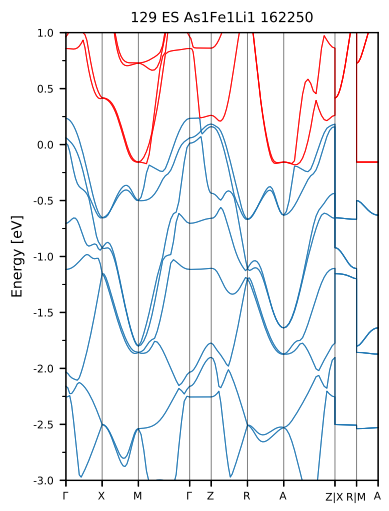

(jq) 129.3.100

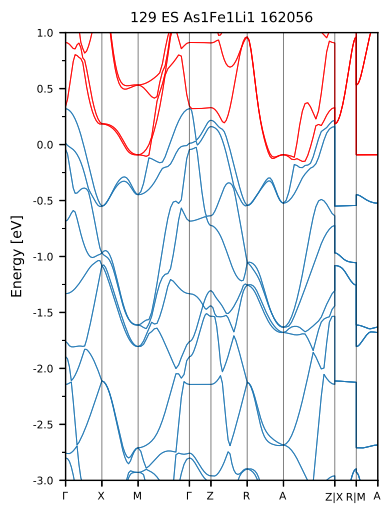

(jr) 129.3.101

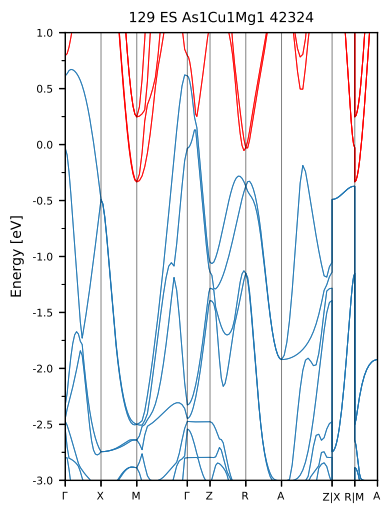

(js) 129.3.102

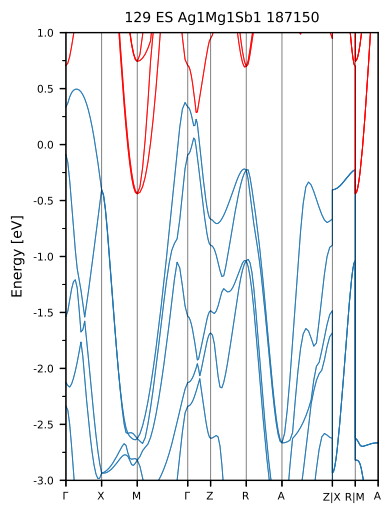

(jt) 129.3.103

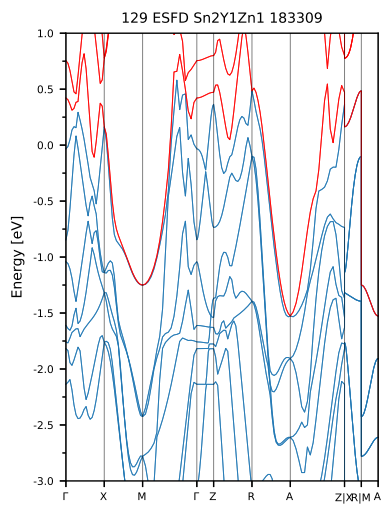

(ju) 129.4.1

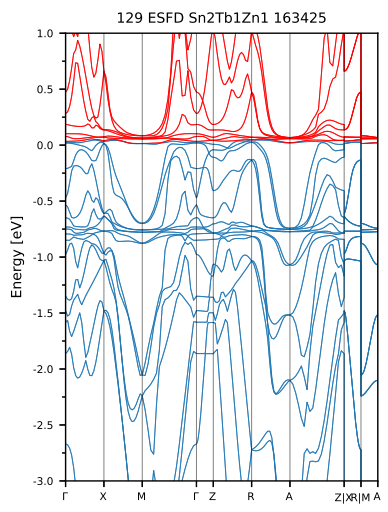

(jv) 129.4.2

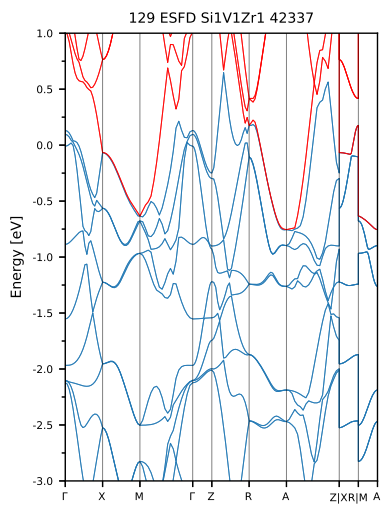

(jw) 129.4.3

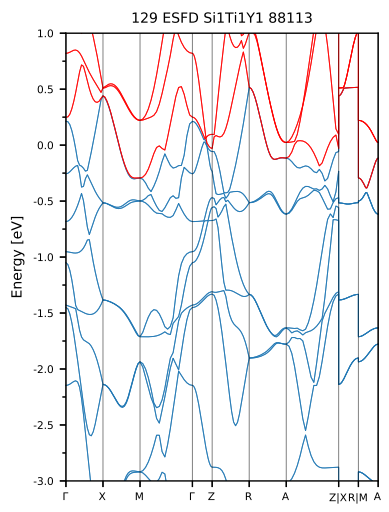

(jx) 129.4.4

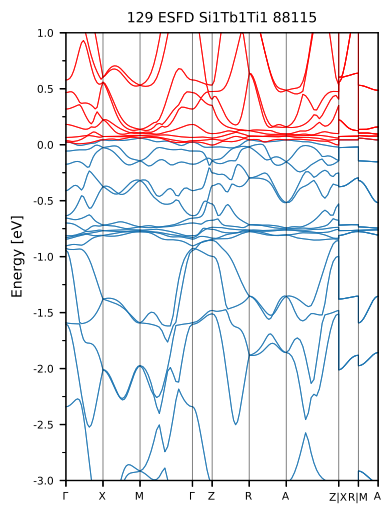

(jy) 129.4.5

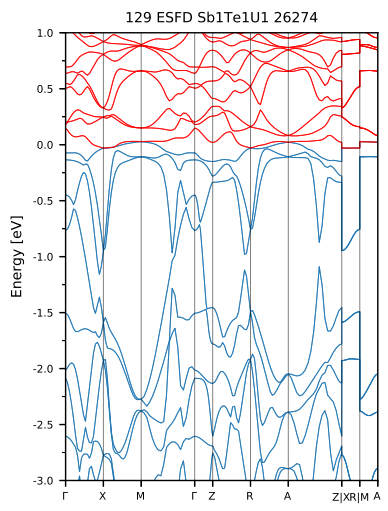

(jz) 129.4.6

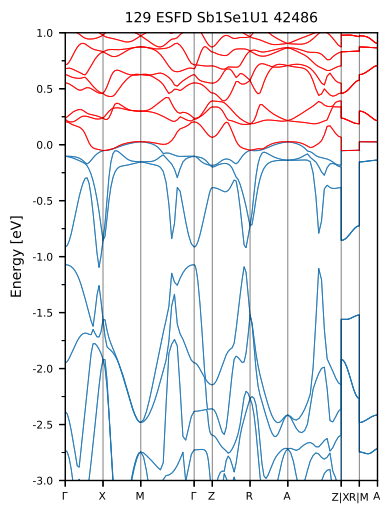

(ka) 129.4.7

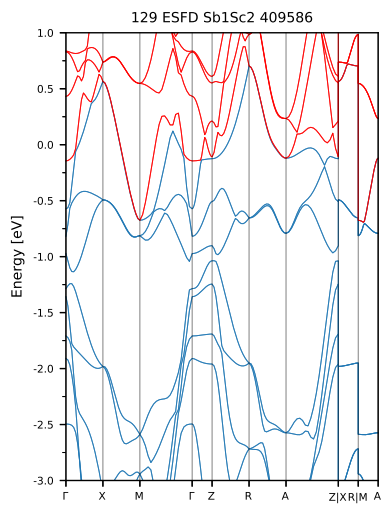

(kb) 129.4.8

Figure 256: Band structures for topological materials in SG 129, part 18/39, from 129.3.96 to 129.4.8. See the list of figures.

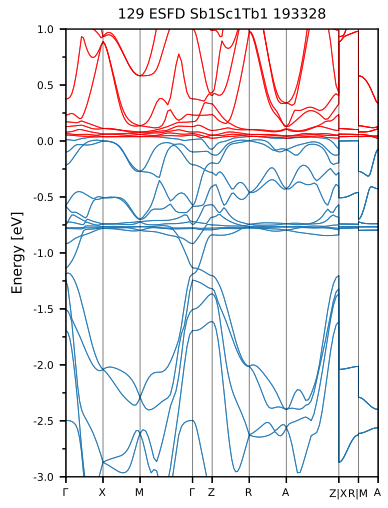

(kc) 129.4.9

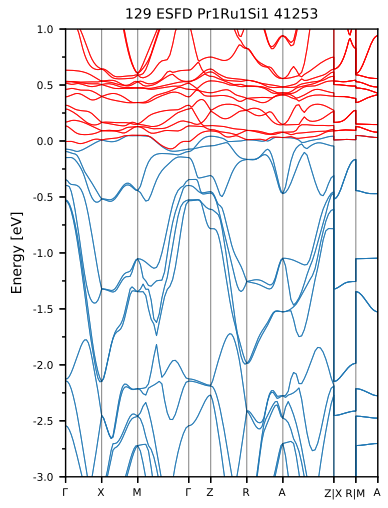

(kd) 129.4.10

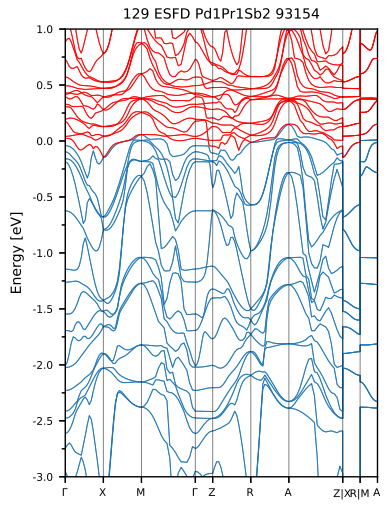

(ke) 129.4.11

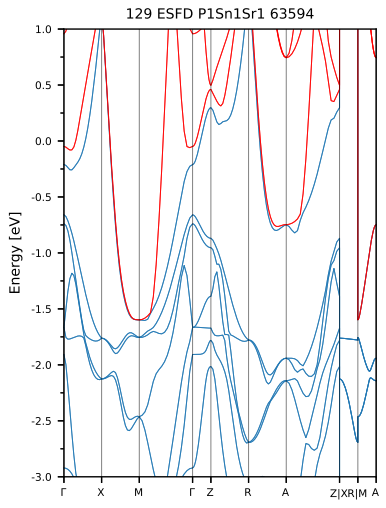

(kf) 129.4.12

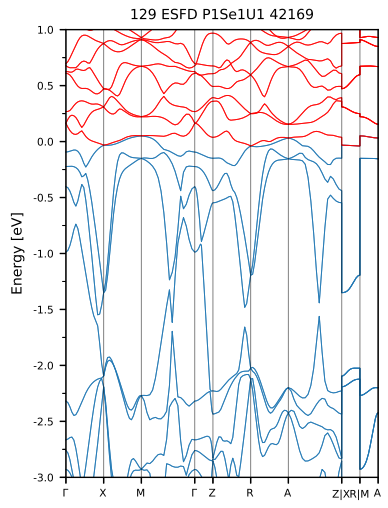

(kg) 129.4.13

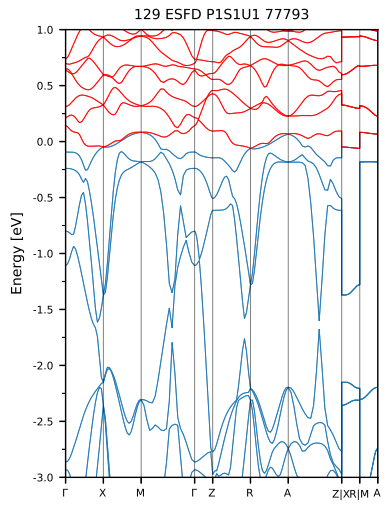

(kh) 129.4.14

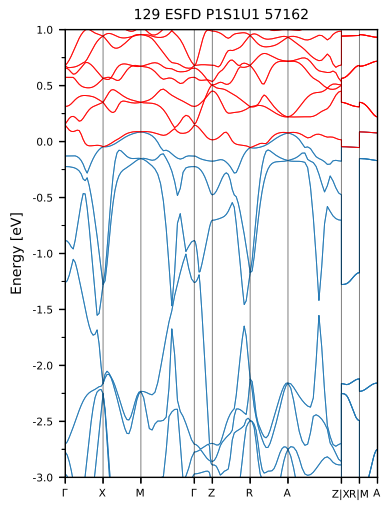

(ki) 129.4.15

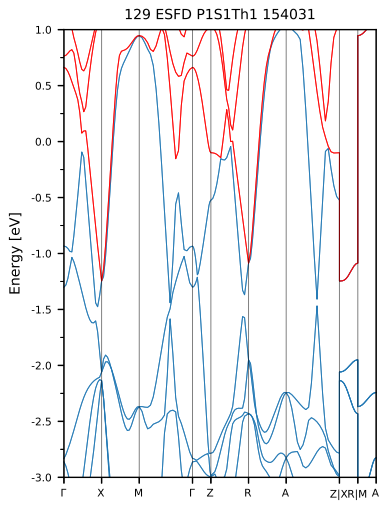

(kj) 129.4.16

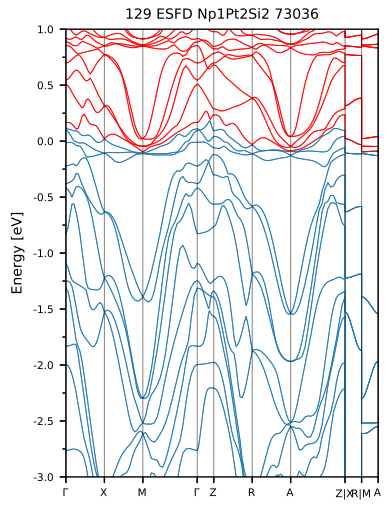

(kk) 129.4.17

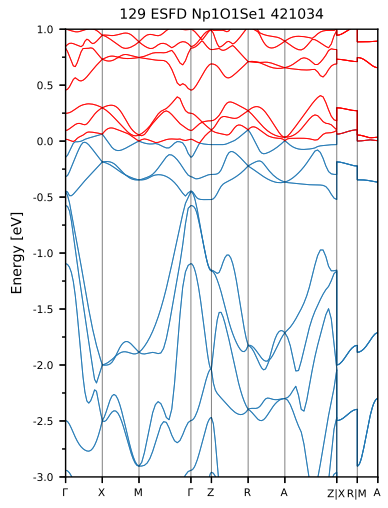

(kl) 129.4.18

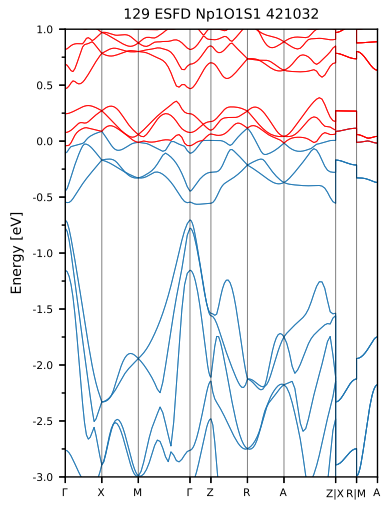

(km) 129.4.19

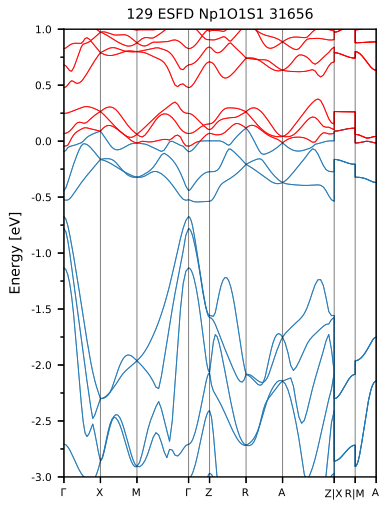

(kn) 129.4.20

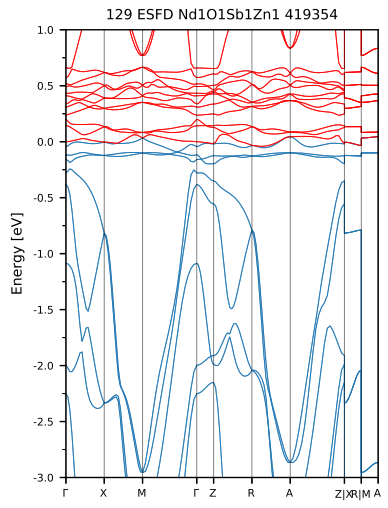

(ko) 129.4.21

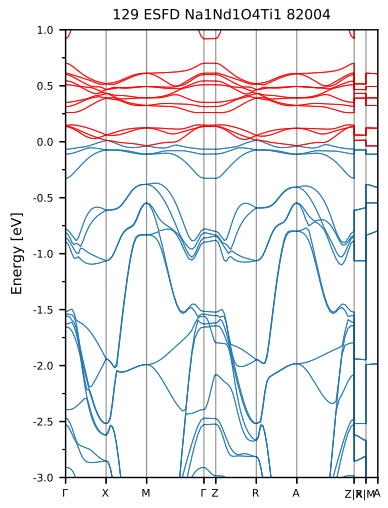

(kp) 129.4.22

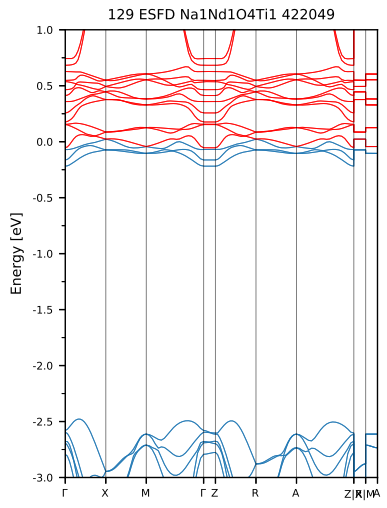

(kq) 129.4.23

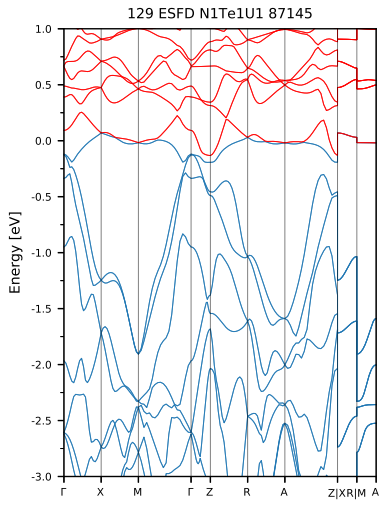

(kr) 129.4.24

Figure 257: Band structures for topological materials in SG 129, part 19/39, from 129.4.9 to 129.4.24. See the list of figures.

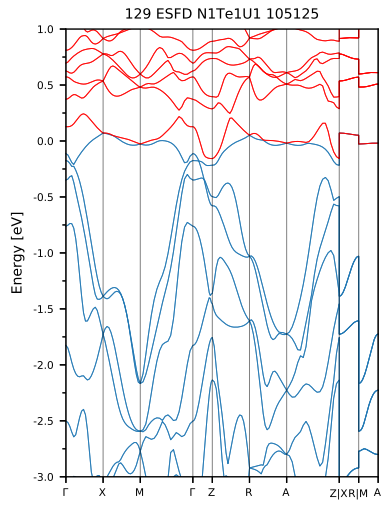

(ks) 129.4.25

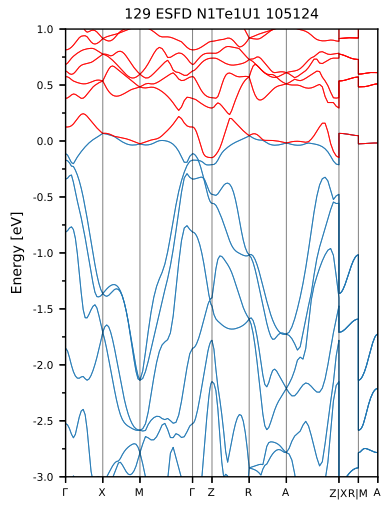

(kt) 129.4.26

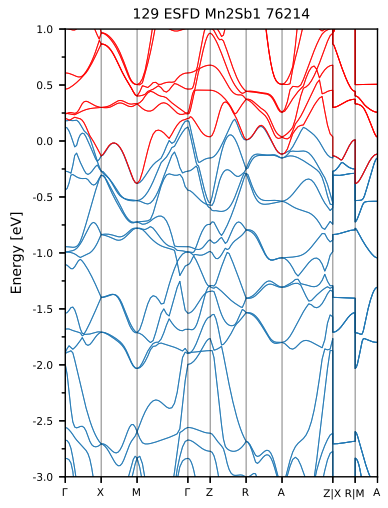

(ku) 129.4.27

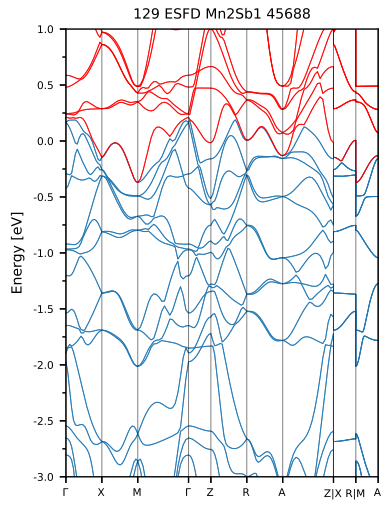

(kv) 129.4.28

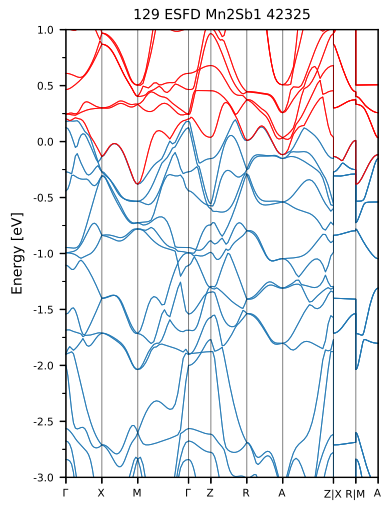

(kw) 129.4.29

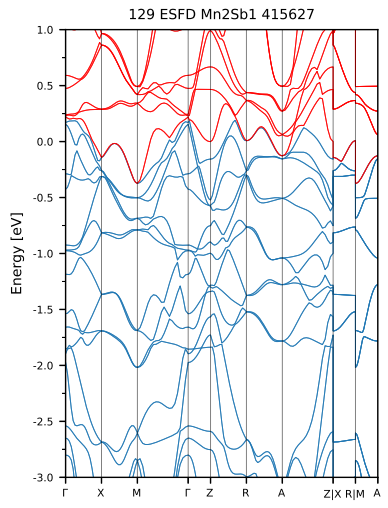

(kx) 129.4.30

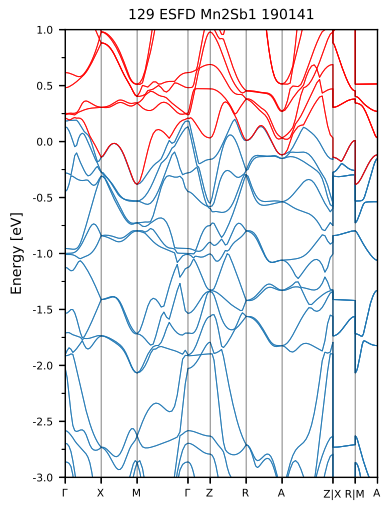

(ky) 129.4.31

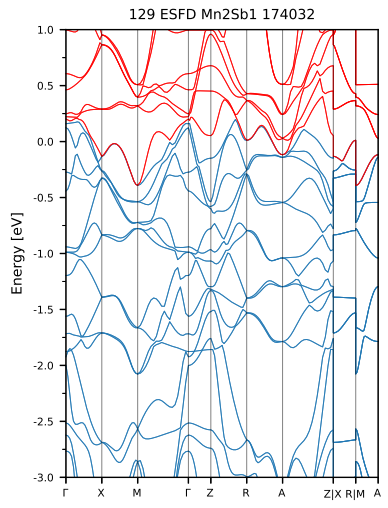

(kz) 129.4.32

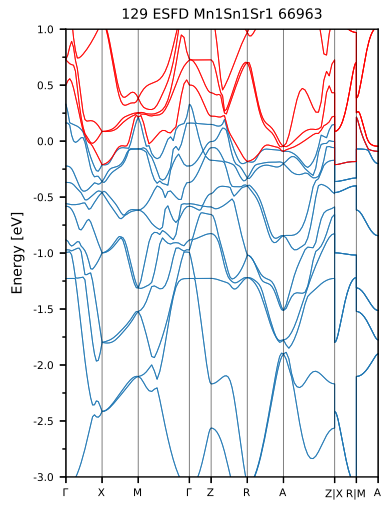

(la) 129.4.33

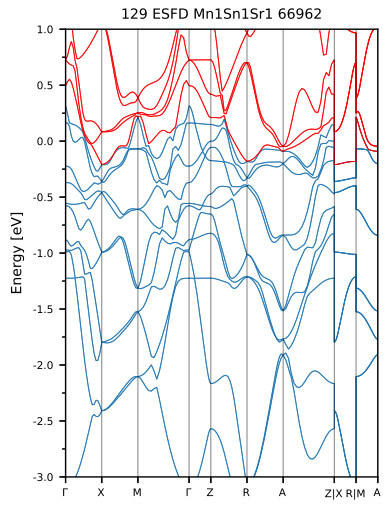

(lb) 129.4.34

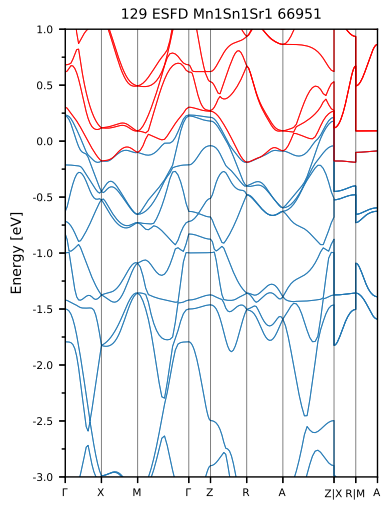

(lc) 129.4.35

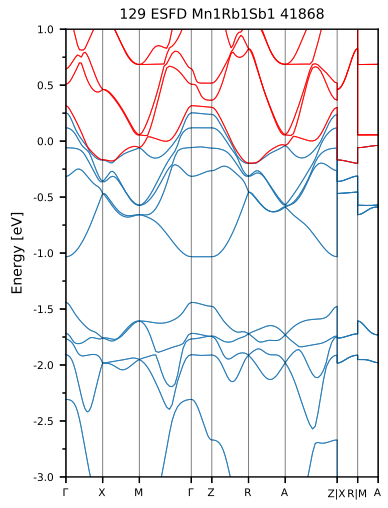

(ld) 129.4.36

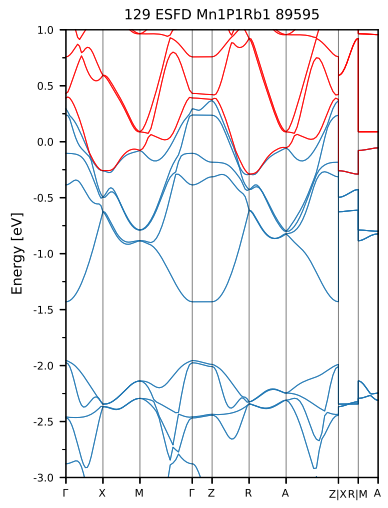

(le) 129.4.37

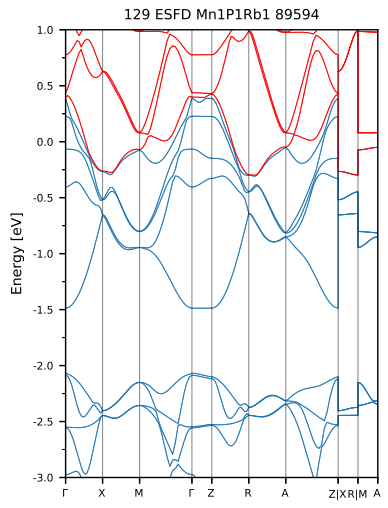

(lf) 129.4.38

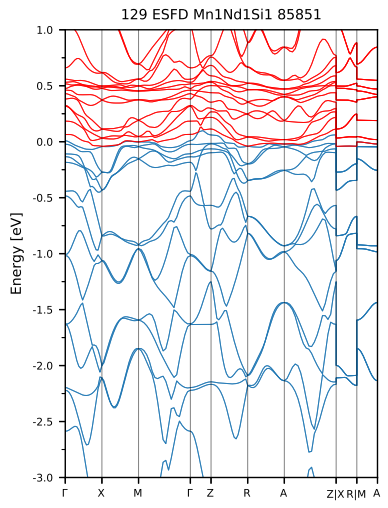

(lg) 129.4.39

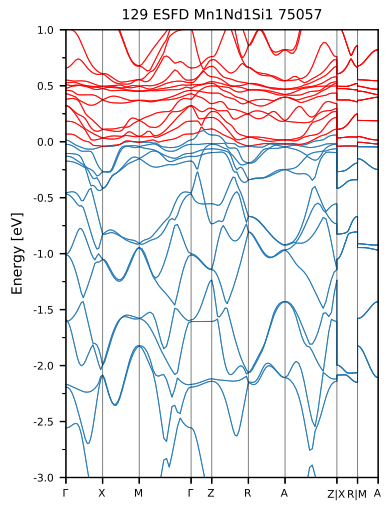

(lh) 129.4.40

Figure 258: Band structures for topological materials in SG 129, part 20/39, from 129.4.25 to 129.4.40. See the list of figures.

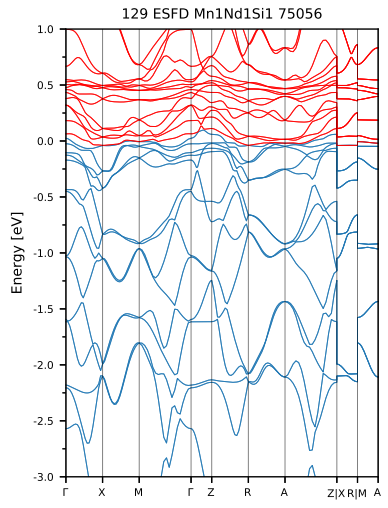

(li) 129.4.41

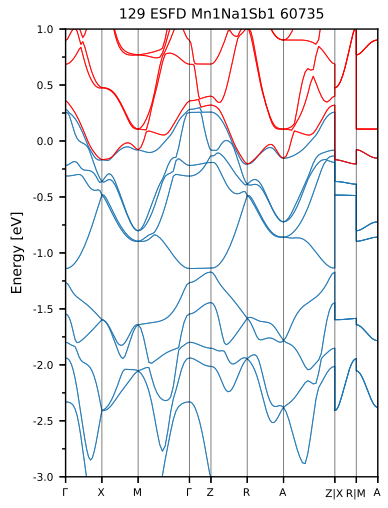

(lj) 129.4.42

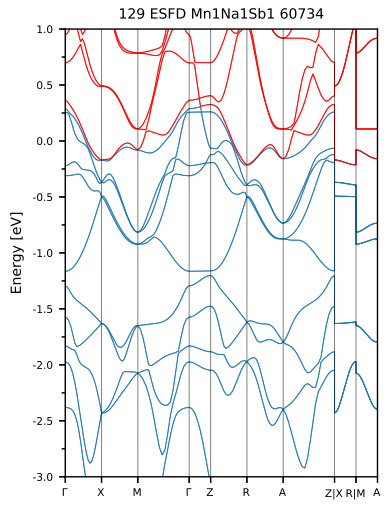

(lk) 129.4.43

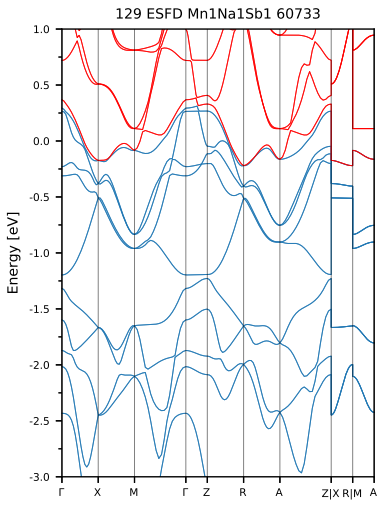

(ll) 129.4.44

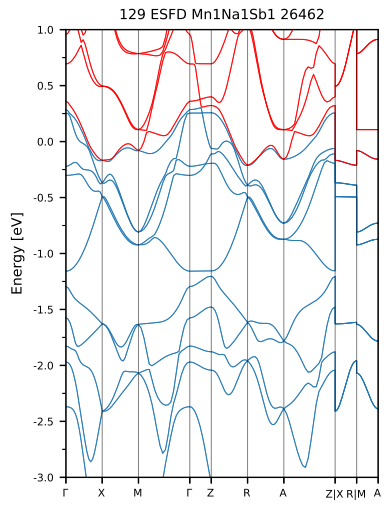

(lm) 129.4.45

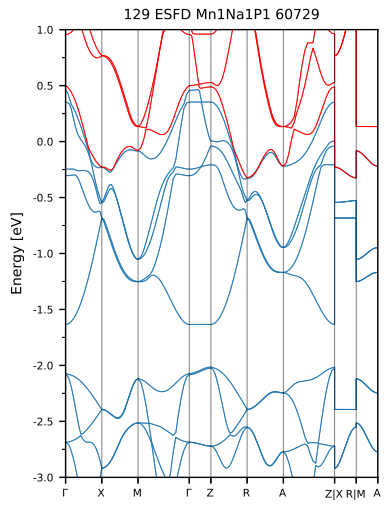

(ln) 129.4.46

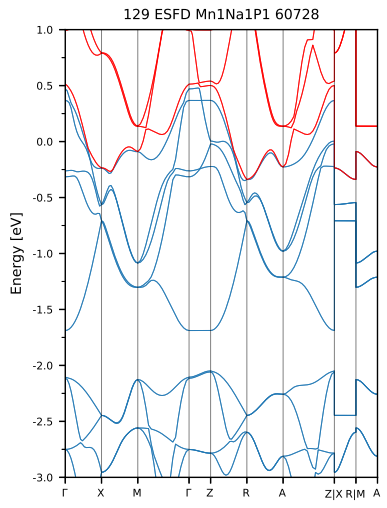

(lo) 129.4.47

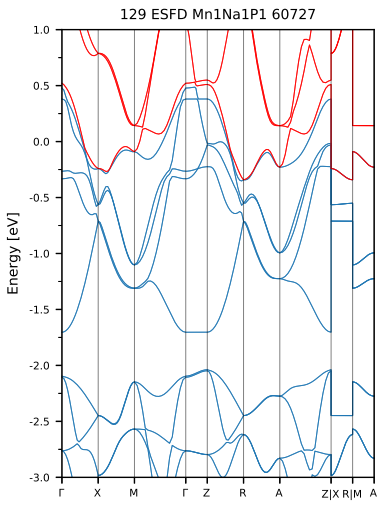

(lp) 129.4.48

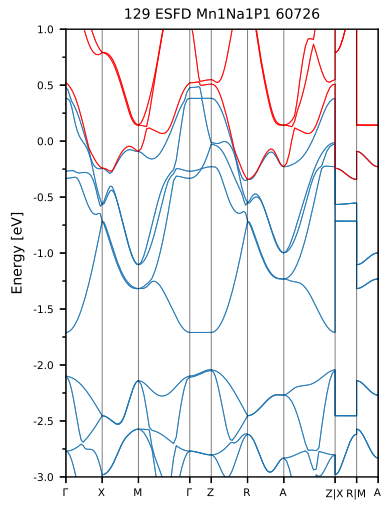

(lq) 129.4.49

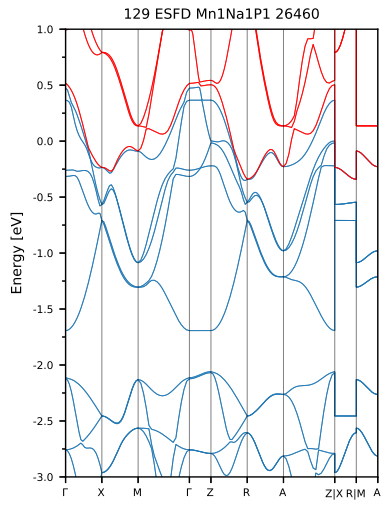

(lr) 129.4.50

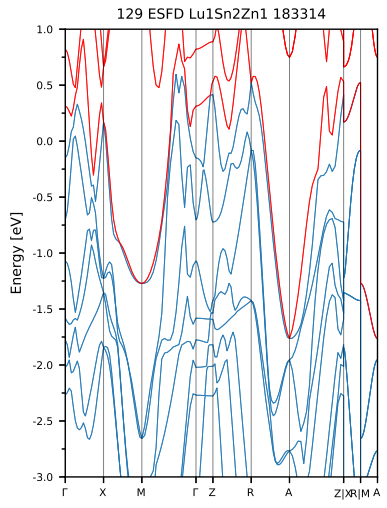

(ls) 129.4.51

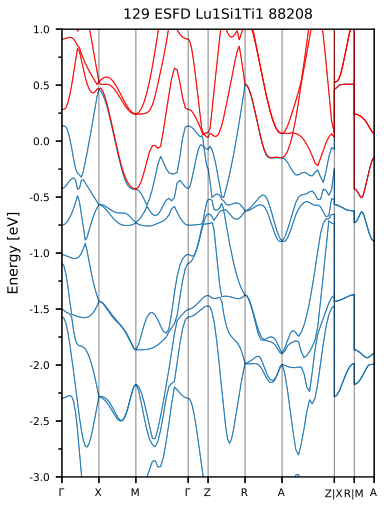

(lt) 129.4.52

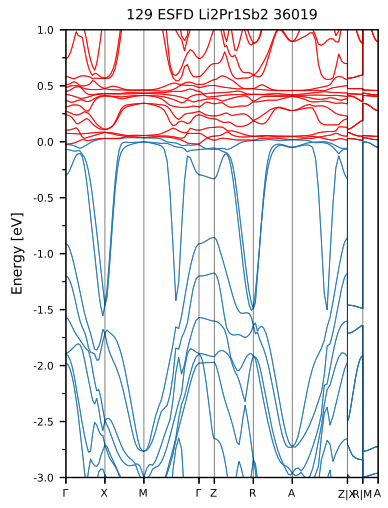

(lu) 129.4.53

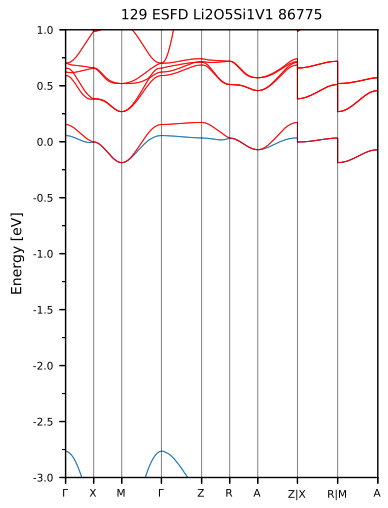

(lv) 129.4.54

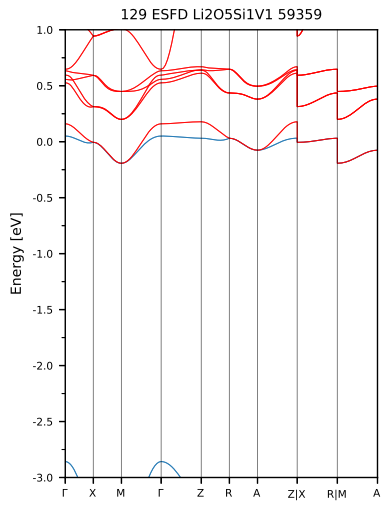

(lw) 129.4.55

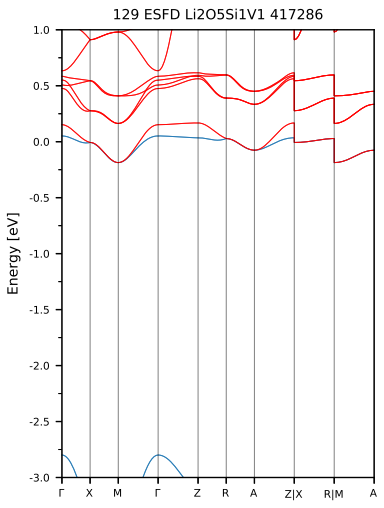

(lx) 129.4.56

Figure 259: Band structures for topological materials in SG 129, part 21/39, from 129.4.41 to 129.4.56. See the list of figures.

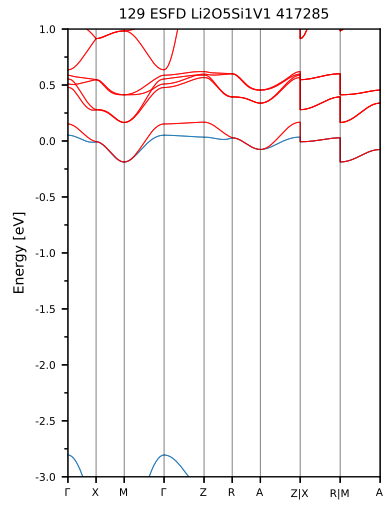

(ly) 129.4.57

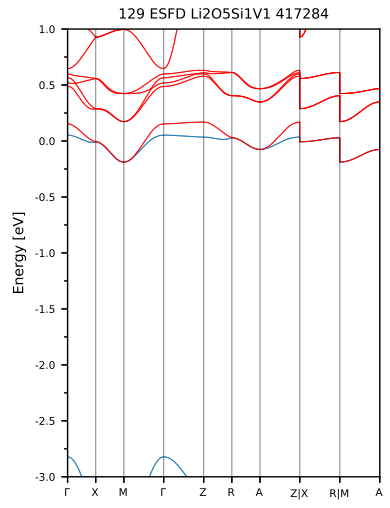

(lz) 129.4.58

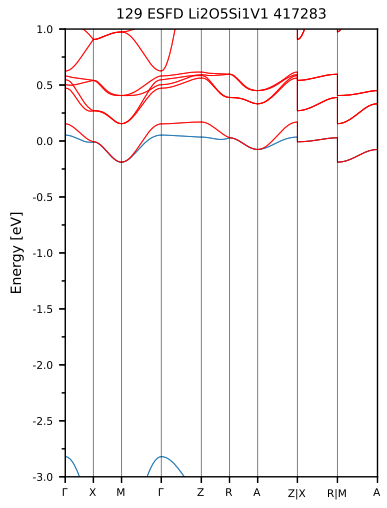

(ma) 129.4.59

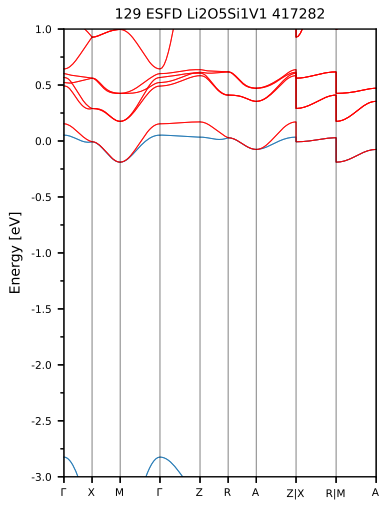

(mb) 129.4.60

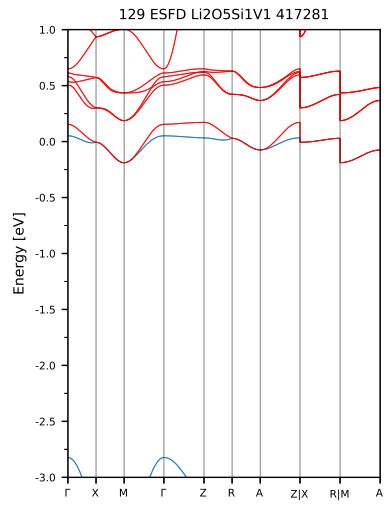

(mc) 129.4.61

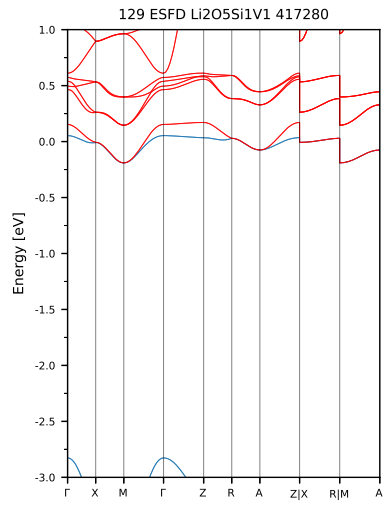

(md) 129.4.62

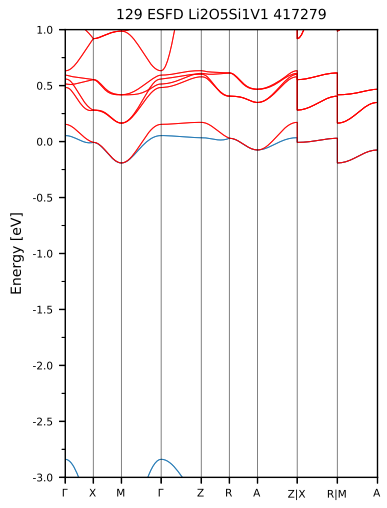

(me) 129.4.63

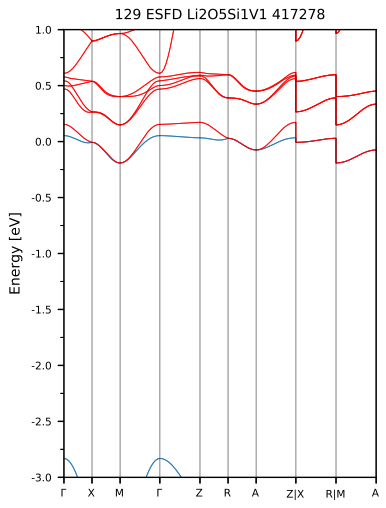

(mf) 129.4.64

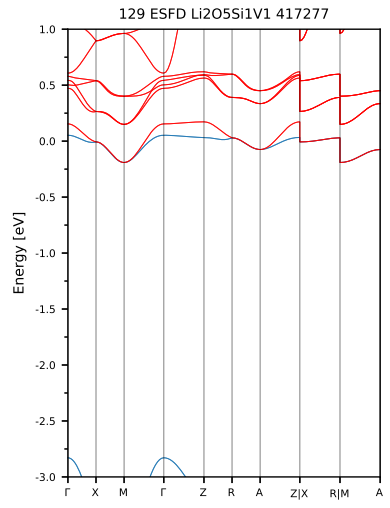

(mg) 129.4.65

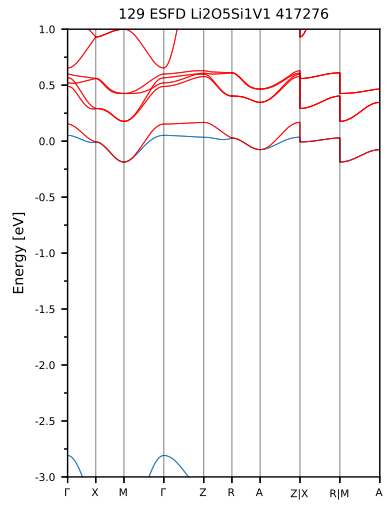

(mh) 129.4.66

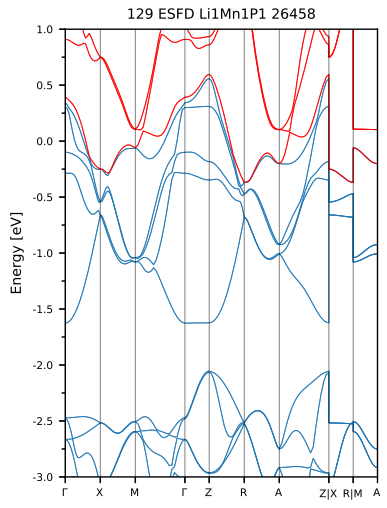

(mi) 129.4.67

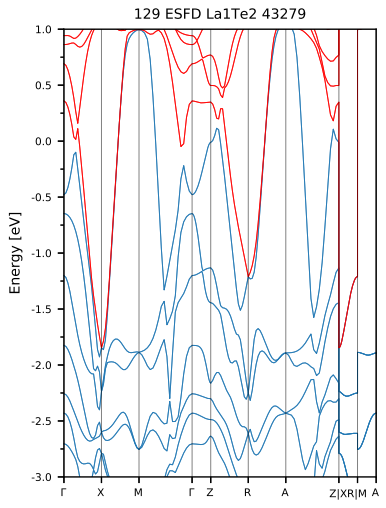

(mj) 129.4.68

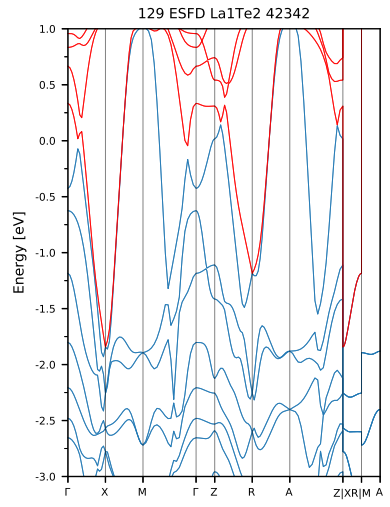

(mk) 129.4.69

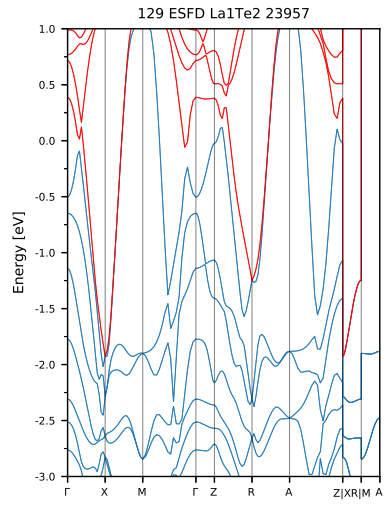

(ml) 129.4.70

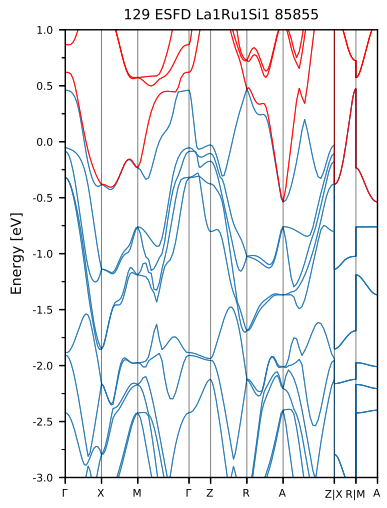

(mm) 129.4.71

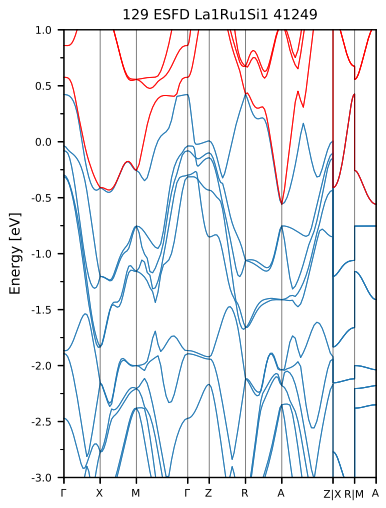

(mn) 129.4.72

Figure 260: Band structures for topological materials in SG 129, part 22/39, from 129.4.57 to 129.4.72. See the list of figures.

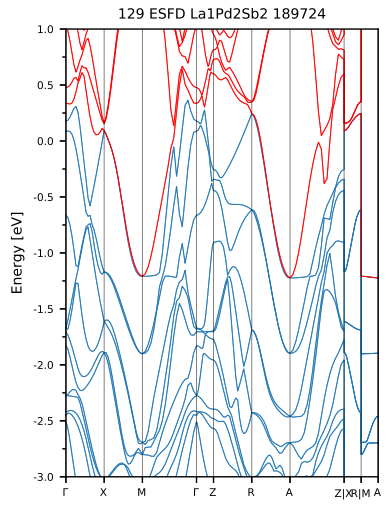

(mo) 129.4.73

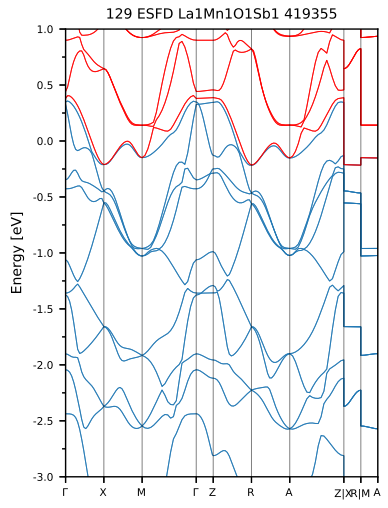

(mp) 129.4.74

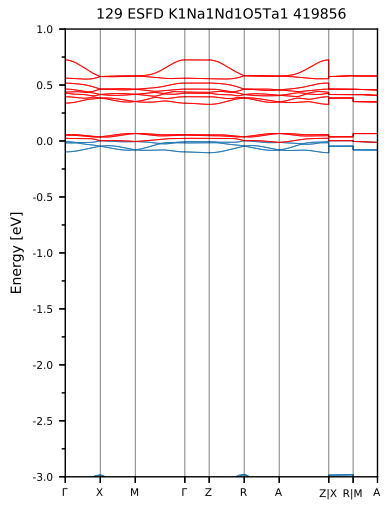

(mq) 129.4.75

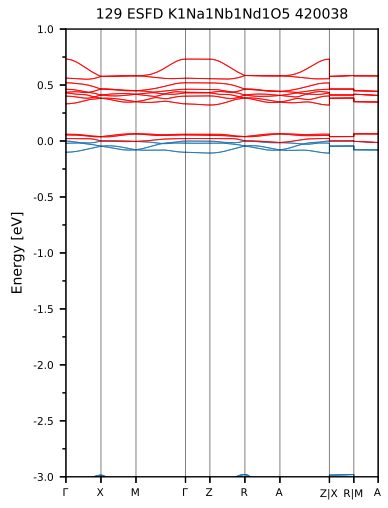

(mr) 129.4.76

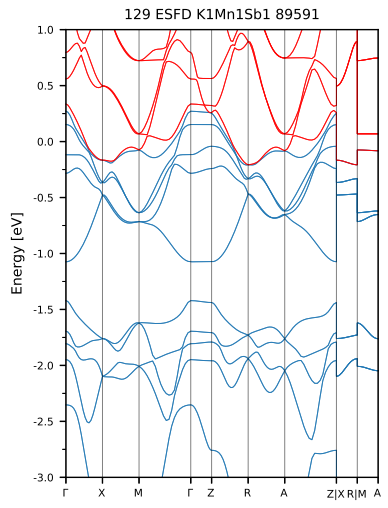

(ms) 129.4.77

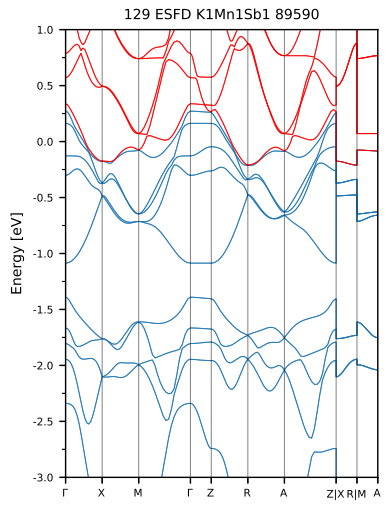

(mt) 129.4.78

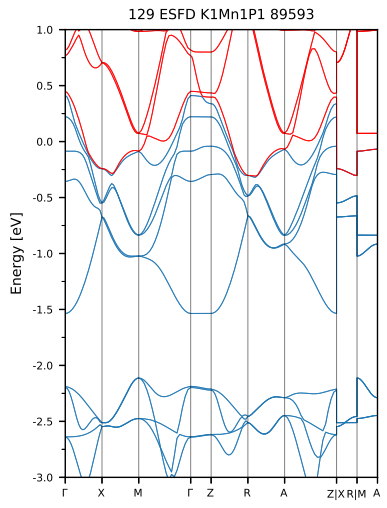

(mu) 129.4.79

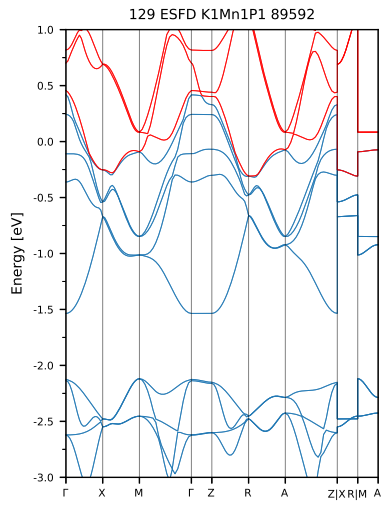

(mv) 129.4.80

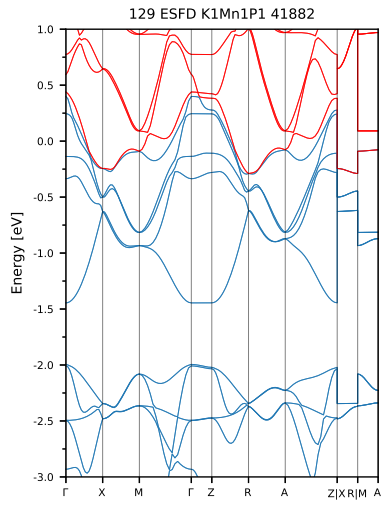

(mw) 129.4.81

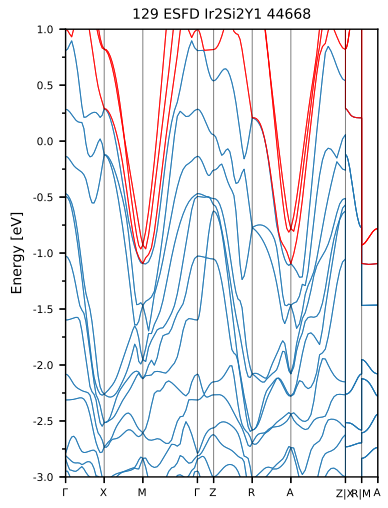

(mx) 129.4.82

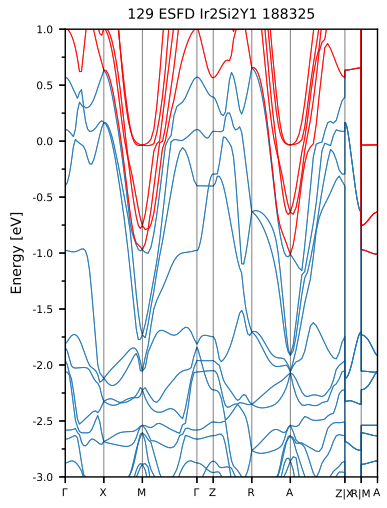

(my) 129.4.83

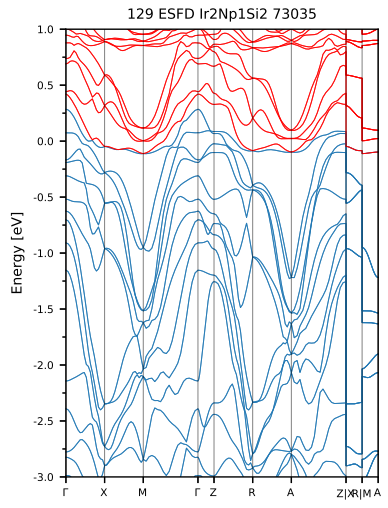

(mz) 129.4.84

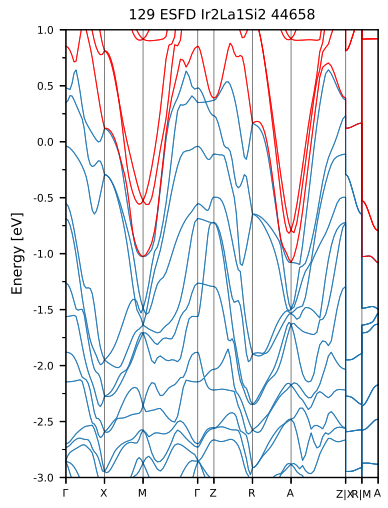

(na) 129.4.85

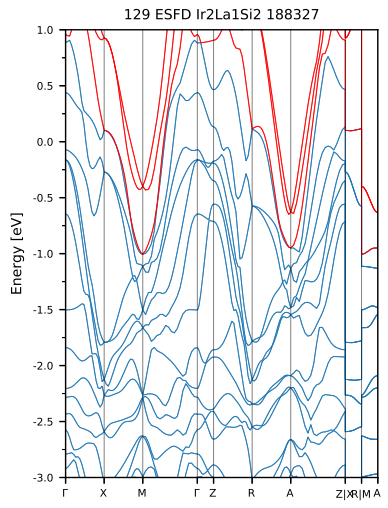

(nb) 129.4.86

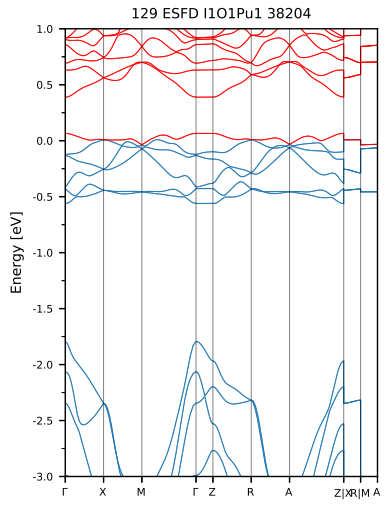

(nc) 129.4.87

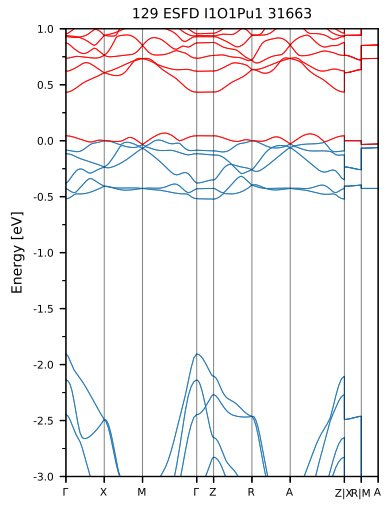

(nd) 129.4.88

Figure 261: Band structures for topological materials in SG 129, part 23/39, from 129.4.73 to 129.4.88. See the list of figures.

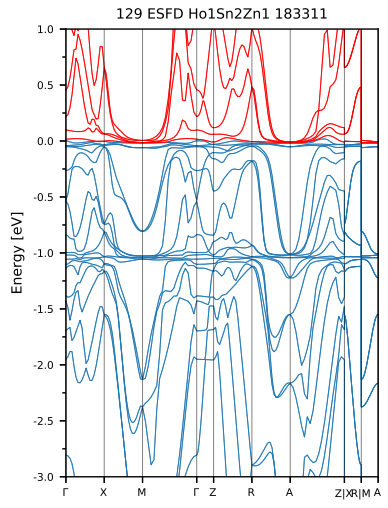

(ne) 129.4.89

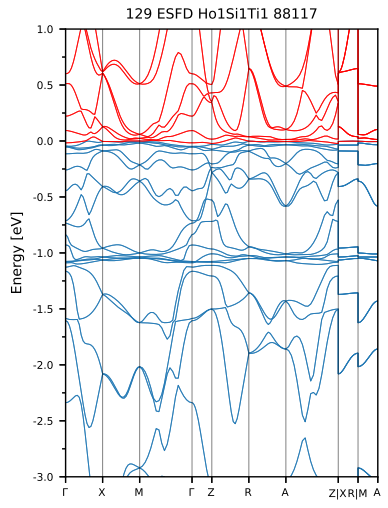

(nf) 129.4.90

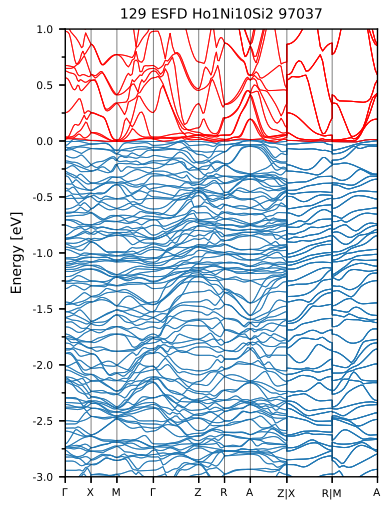

(ng) 129.4.91

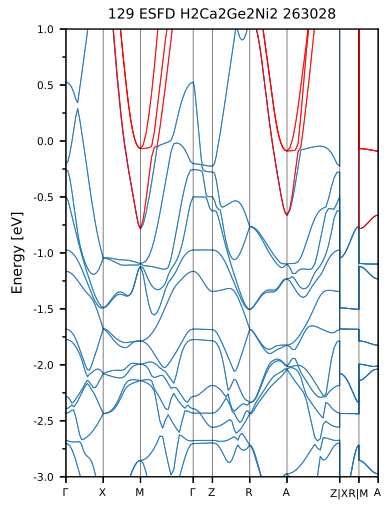

(nh) 129.4.92

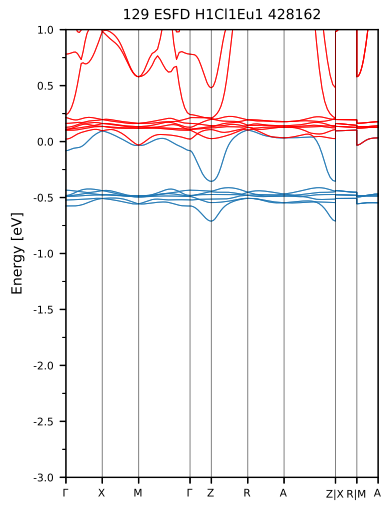

(ni) 129.4.93

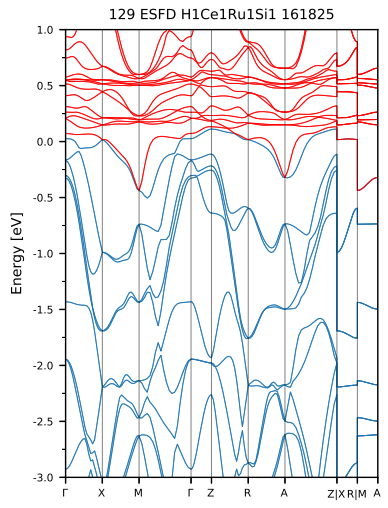

(nj) 129.4.94

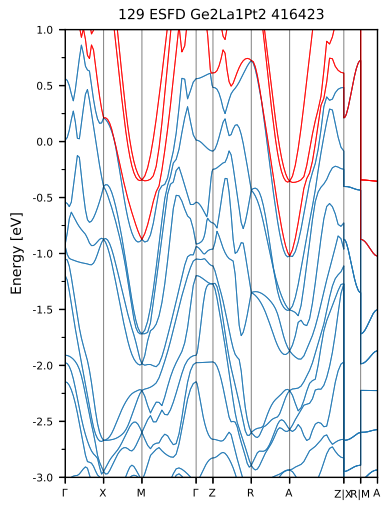

(nk) 129.4.95

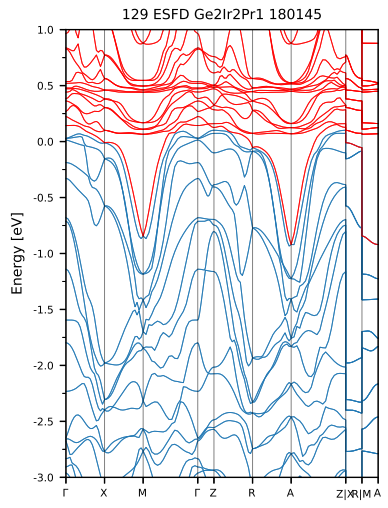

(nl) 129.4.96

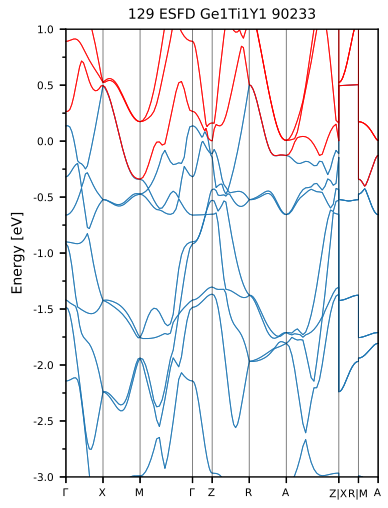

(nm) 129.4.97

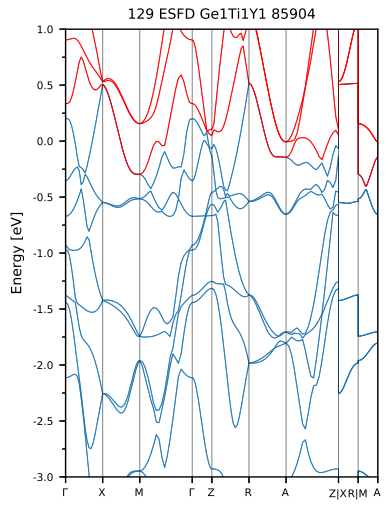

(nn) 129.4.98

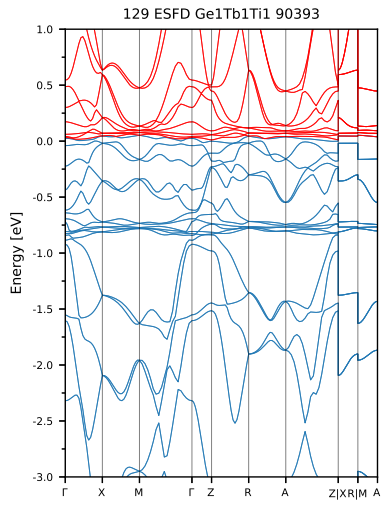

(no) 129.4.99

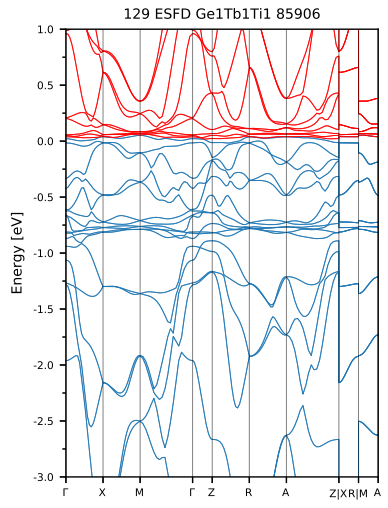

(np) 129.4.100

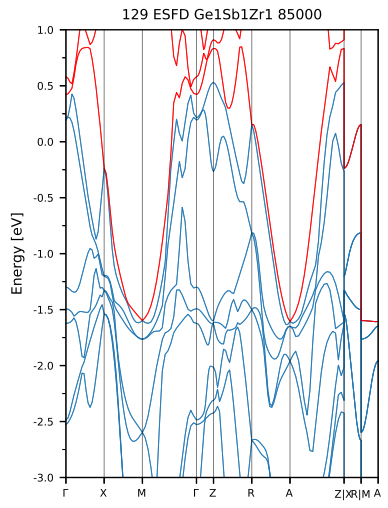

(nq) 129.4.101

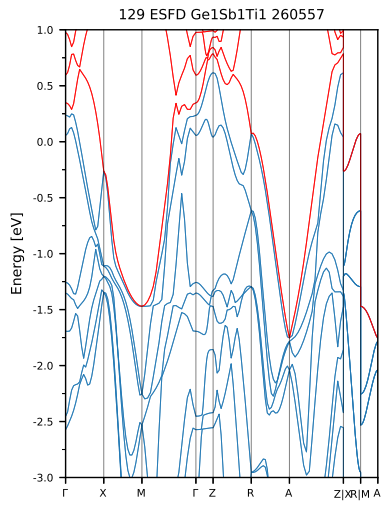

(nr) 129.4.102

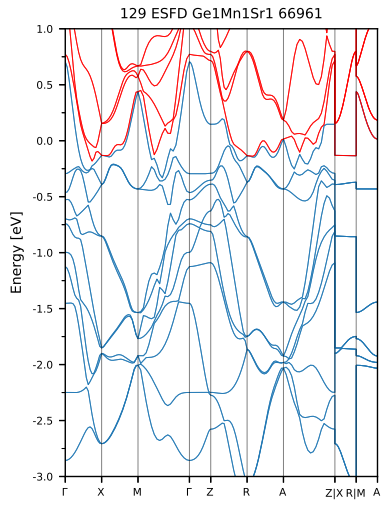

(ns) 129.4.103

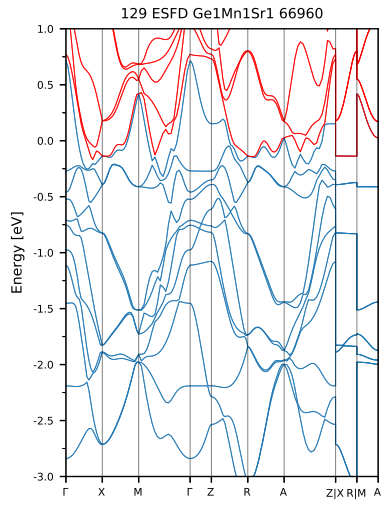

(nt) 129.4.104

Figure 262: Band structures for topological materials in SG 129, part 24/39, from 129.4.89 to 129.4.104. See the list of figures.

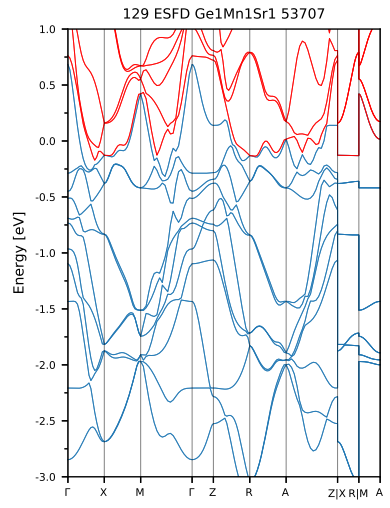

(nu) 129.4.105

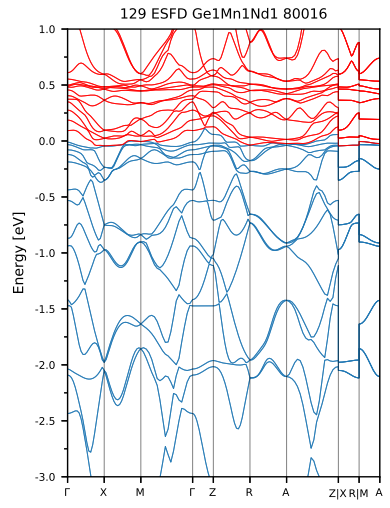

(nv) 129.4.106

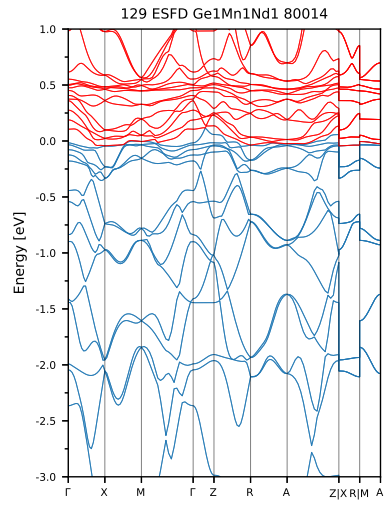

(nw) 129.4.107

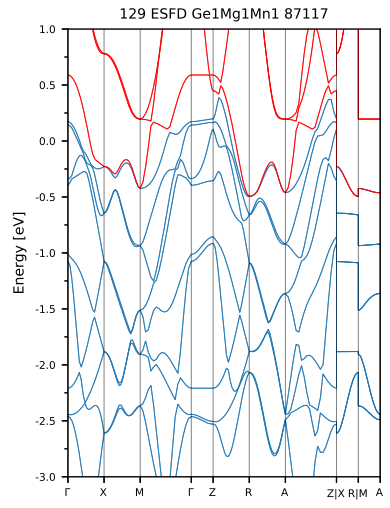

(nx) 129.4.108

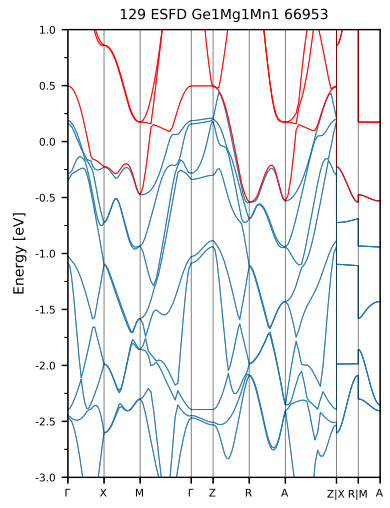

(ny) 129.4.109

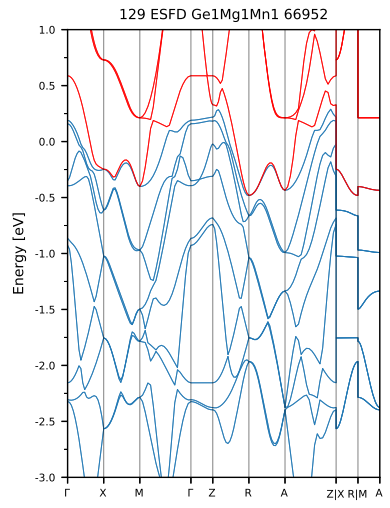

(nz) 129.4.110

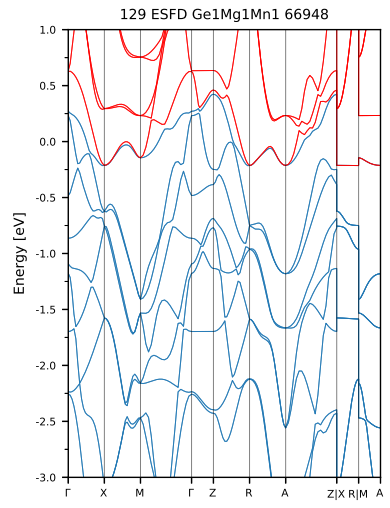

(oa) 129.4.111

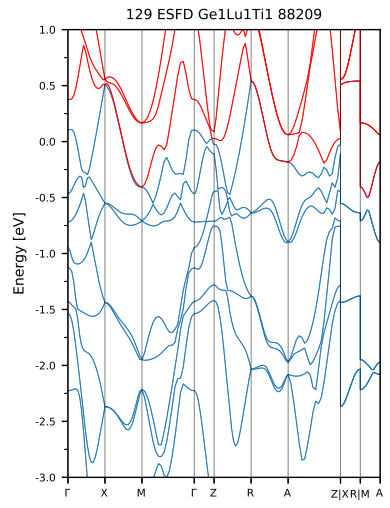

(ob) 129.4.112

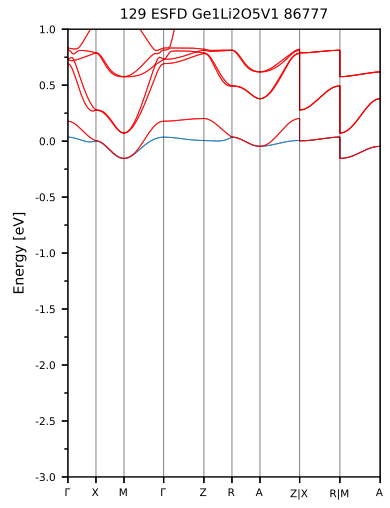

(oc) 129.4.113

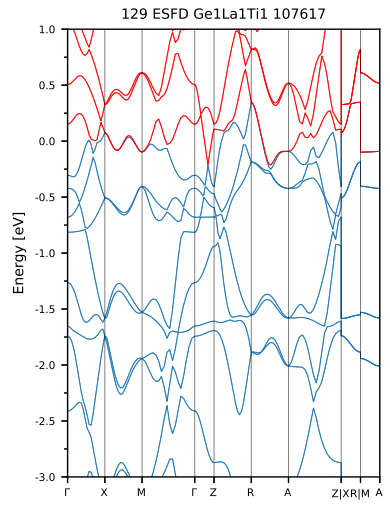

(od) 129.4.114

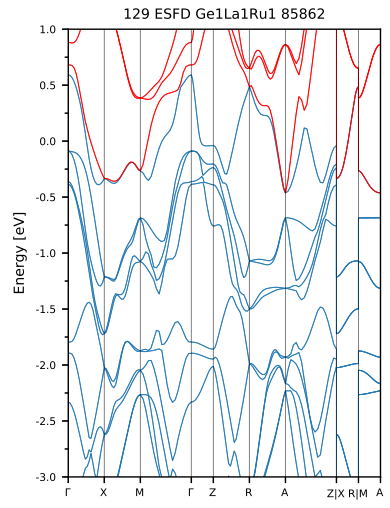

(oe) 129.4.115

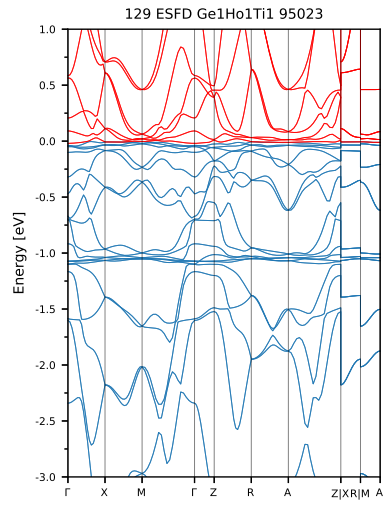

(of) 129.4.116

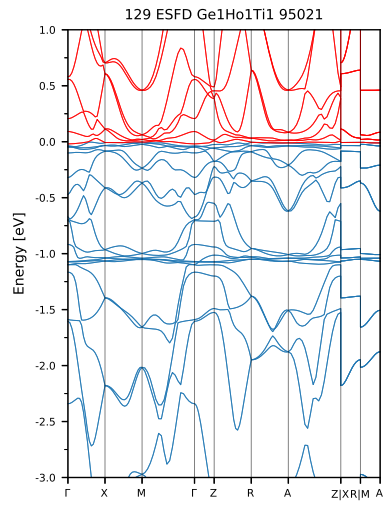

(og) 129.4.117

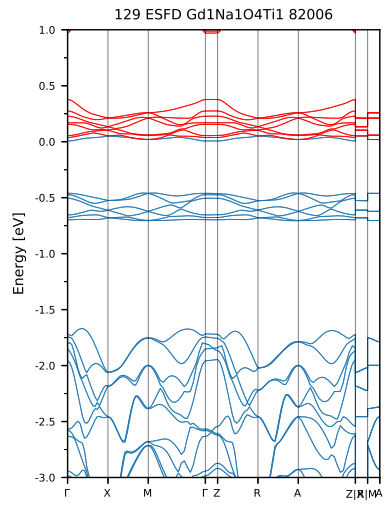

(oh) 129.4.118

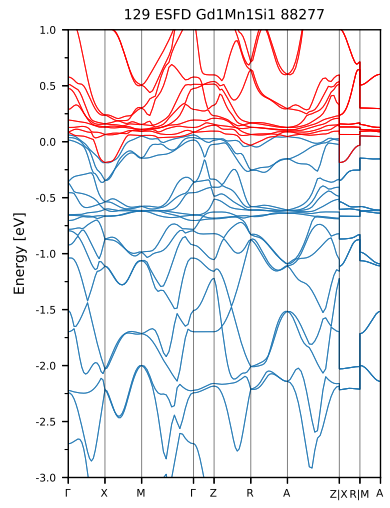

(oi) 129.4.119

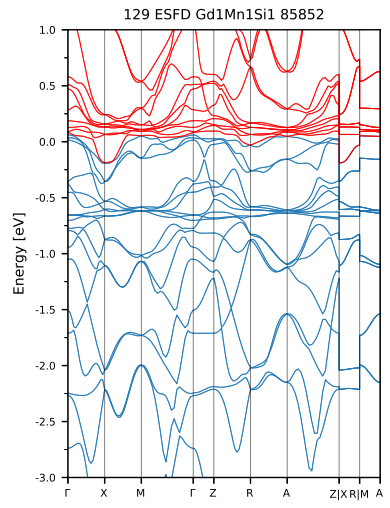

(oj) 129.4.120

Figure 263: Band structures for topological materials in SG 129, part 25/39, from 129.4.105 to 129.4.120. See the list of figures.

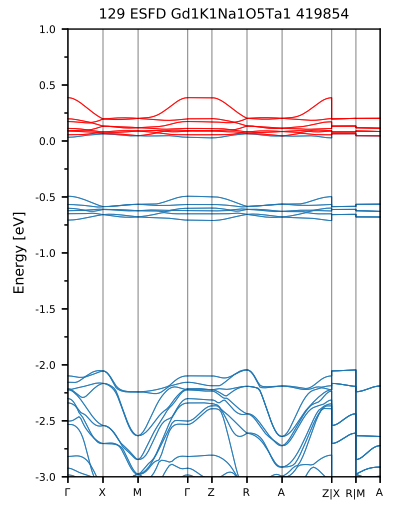

(ok) 129.4.121

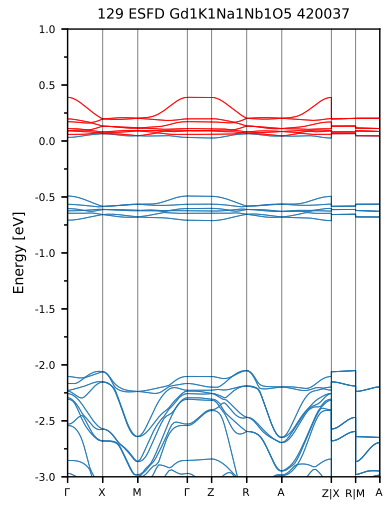

(ol) 129.4.122

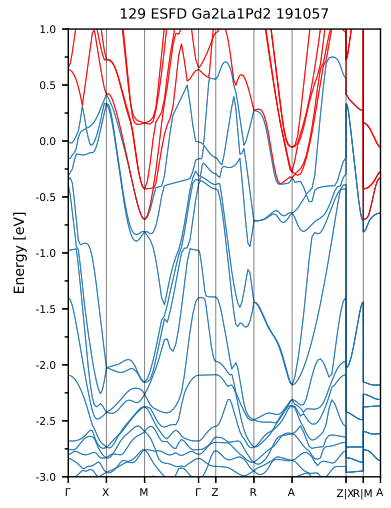

(om) 129.4.123

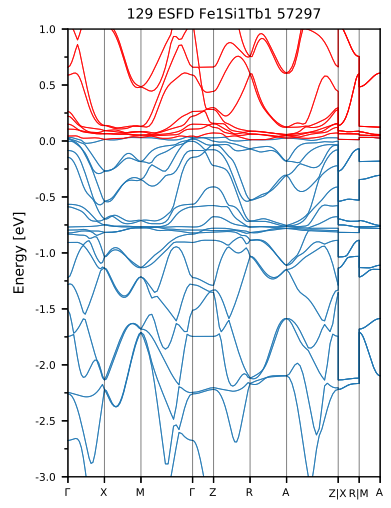

(on) 129.4.124

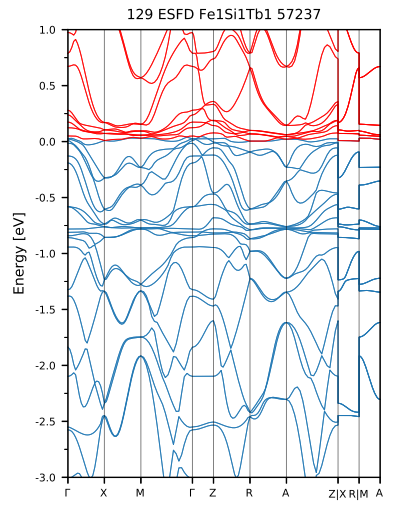

(oo) 129.4.125

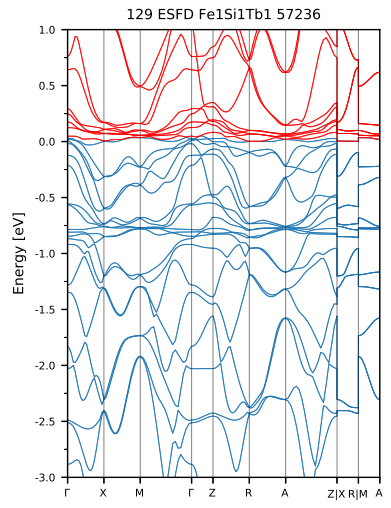

(op) 129.4.126

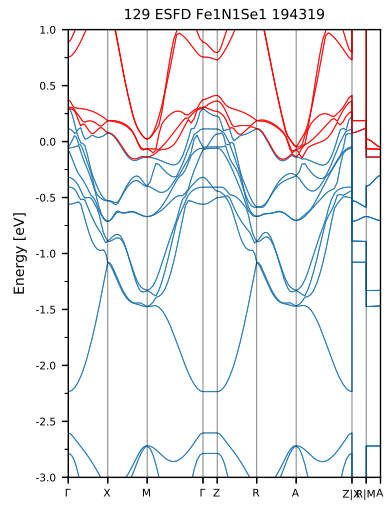

(oq) 129.4.127

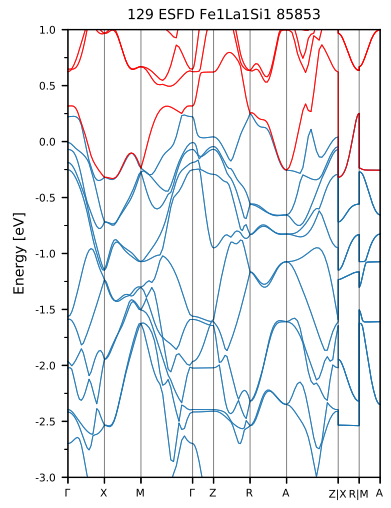

(or) 129.4.128

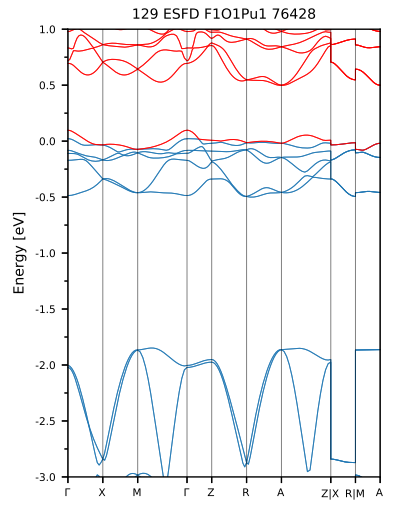

(os) 129.4.129

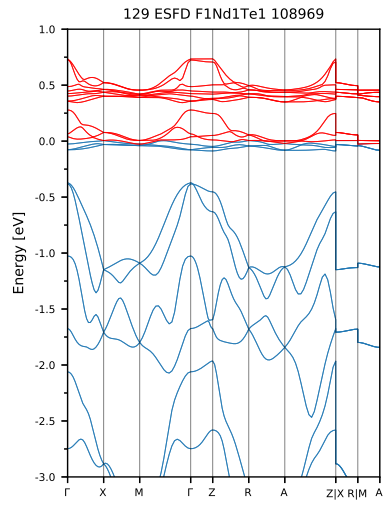

(ot) 129.4.130

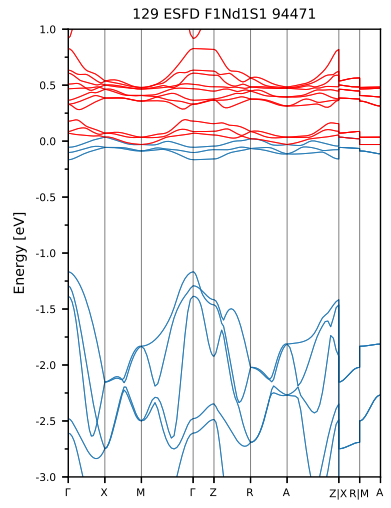

(ou) 129.4.131

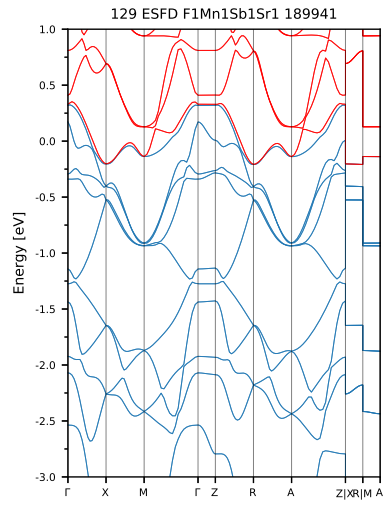

(ov) 129.4.132

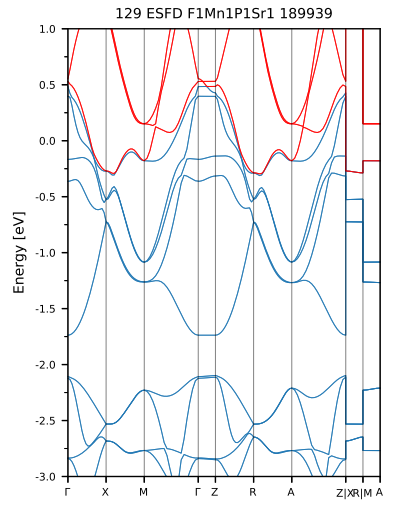

(ow) 129.4.133

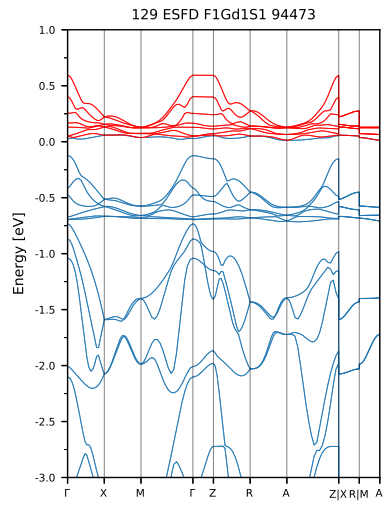

(ox) 129.4.134

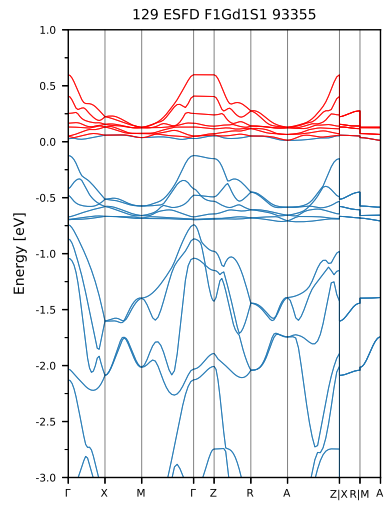

(oy) 129.4.135

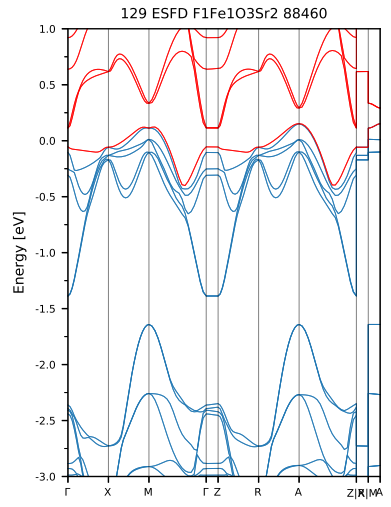

(oz) 129.4.136

Figure 264: Band structures for topological materials in SG 129, part 26/39, from 129.4.121 to 129.4.136. See the list of figures.

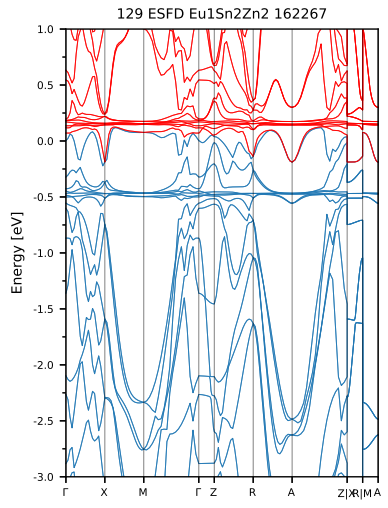

(pa) 129.4.137

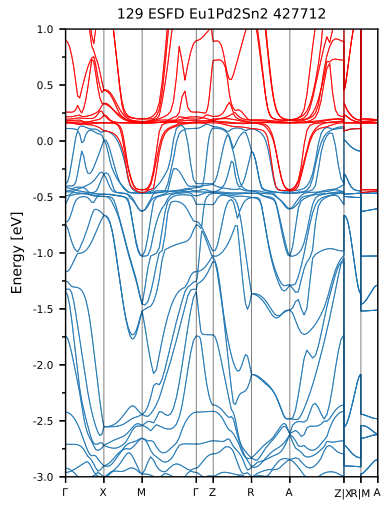

(pb) 129.4.138

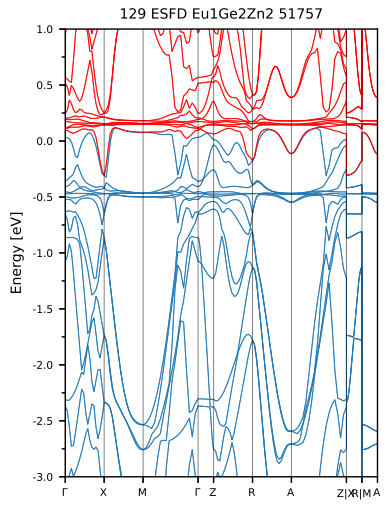

(pc) 129.4.139

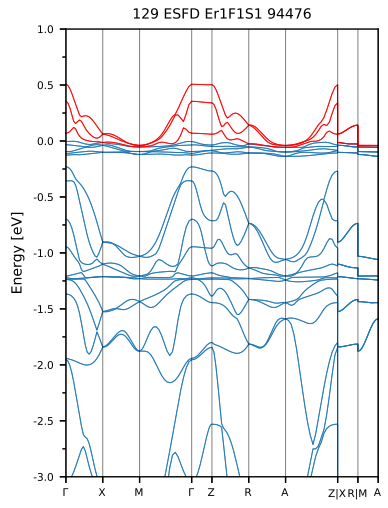

(pd) 129.4.140

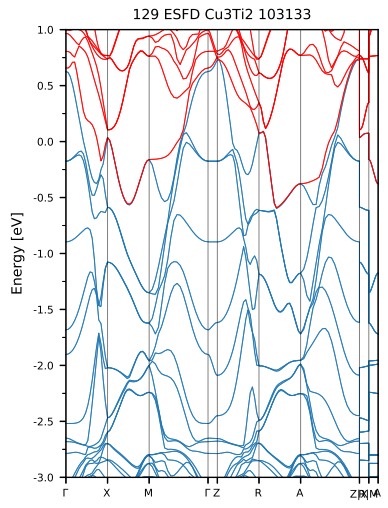

(pe) 129.4.141

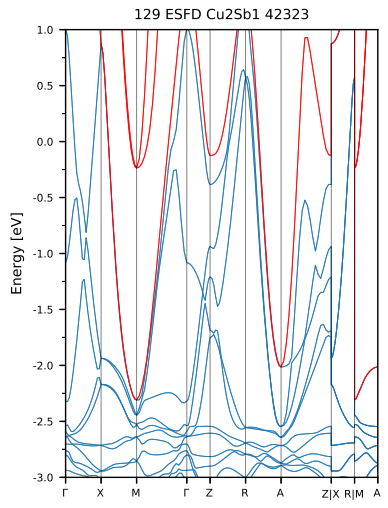

(pf) 129.4.142

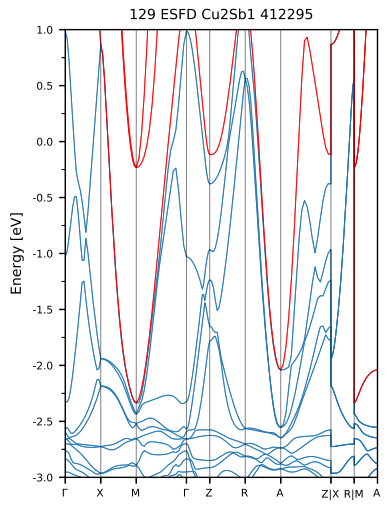

(pg) 129.4.143

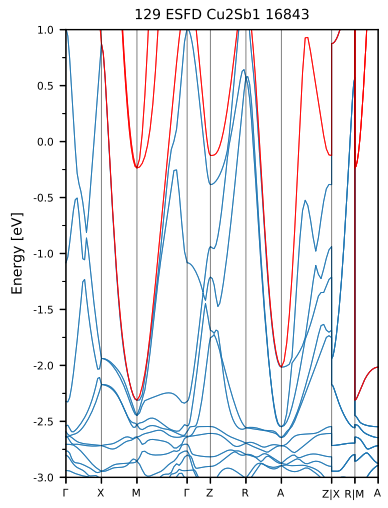

(ph) 129.4.144

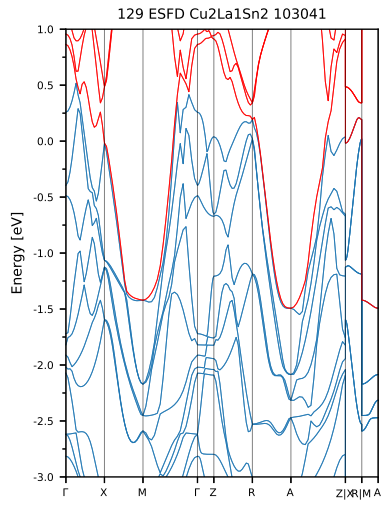

(pi) 129.4.145

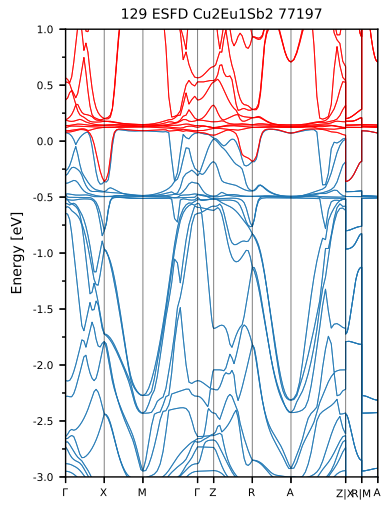

(pj) 129.4.146

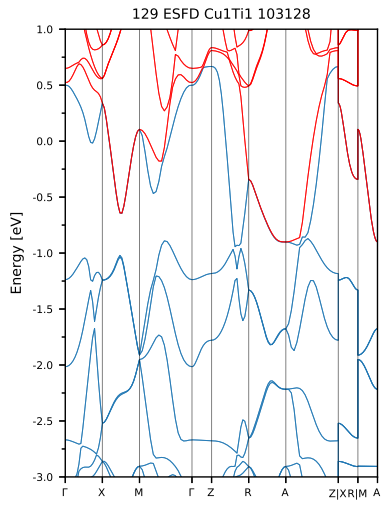

(pk) 129.4.147

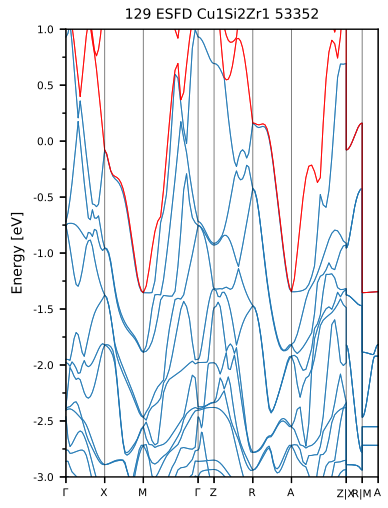

(pl) 129.4.148

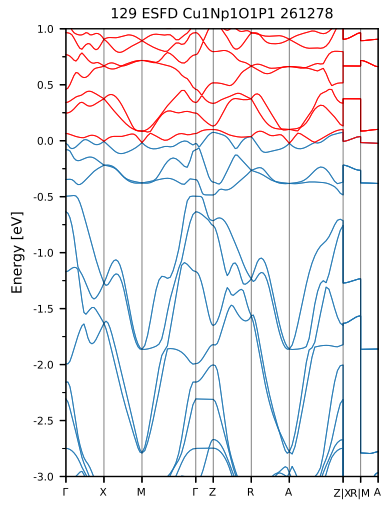

(pm) 129.4.149

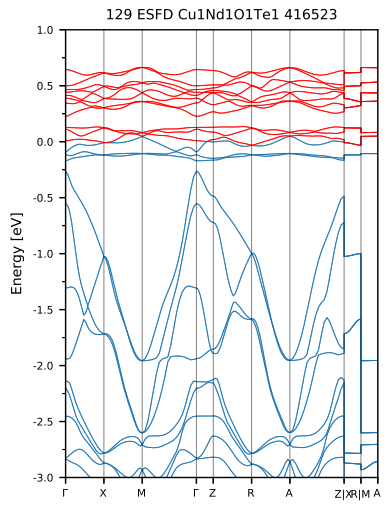

(pn) 129.4.150

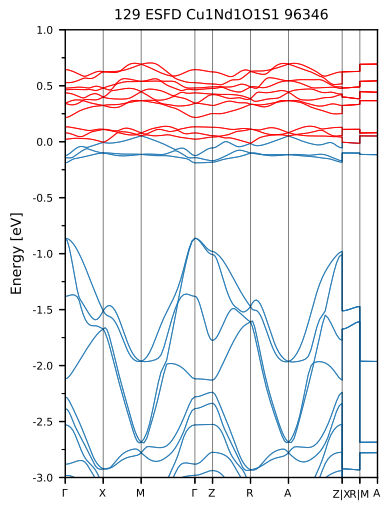

(po) 129.4.151

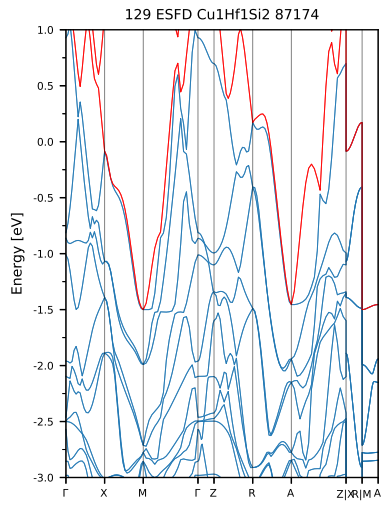

(pp) 129.4.152

Figure 265: Band structures for topological materials in SG 129, part 27/39, from 129.4.137 to 129.4.152. See the list of figures.

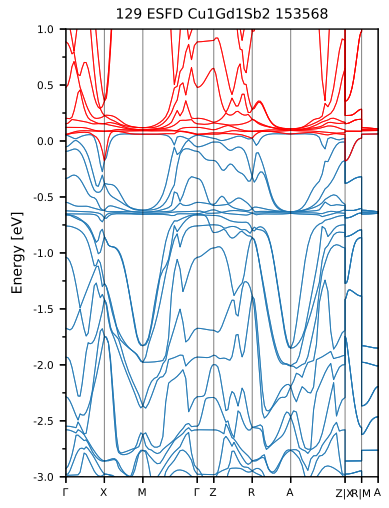

(pq) 129.4.153

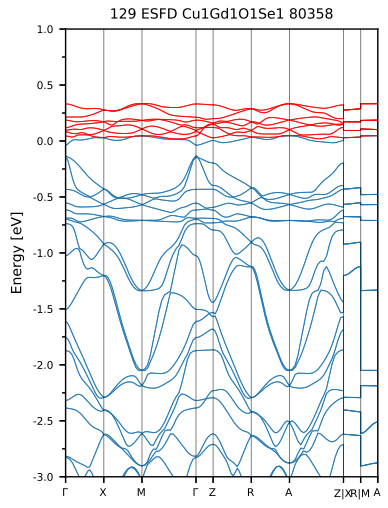

(pr) 129.4.154

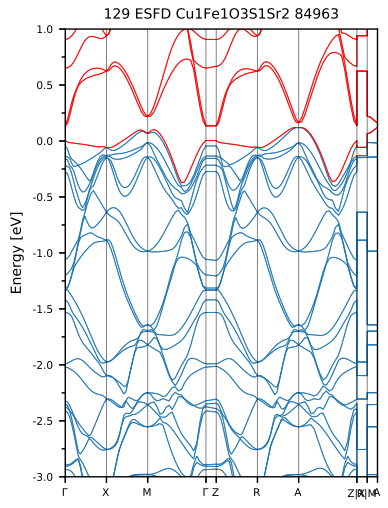

(ps) 129.4.155

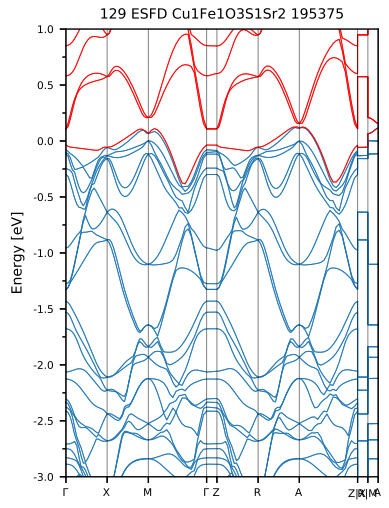

(pt) 129.4.156

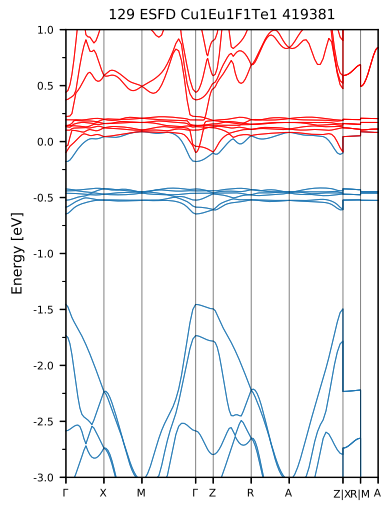

(pu) 129.4.157

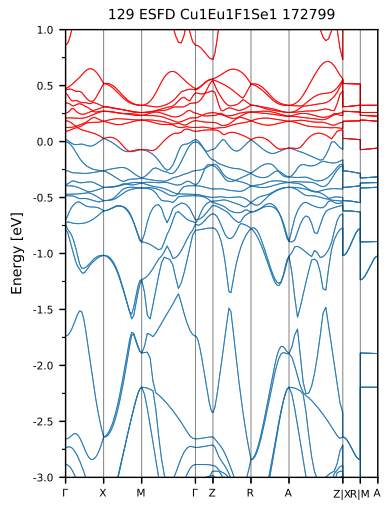

(pv) 129.4.158

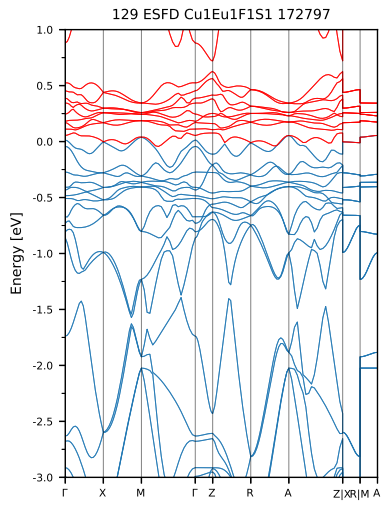

(pw) 129.4.159

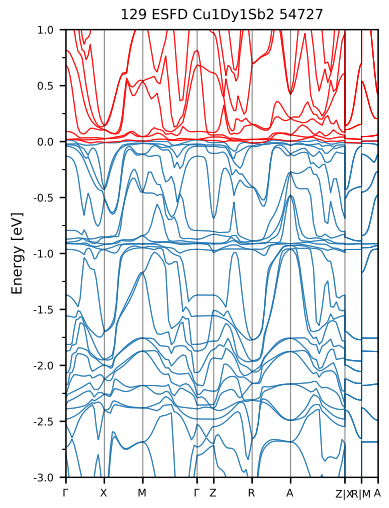

(px) 129.4.160

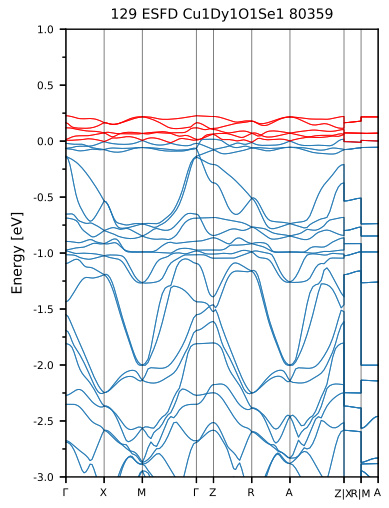

(py) 129.4.161

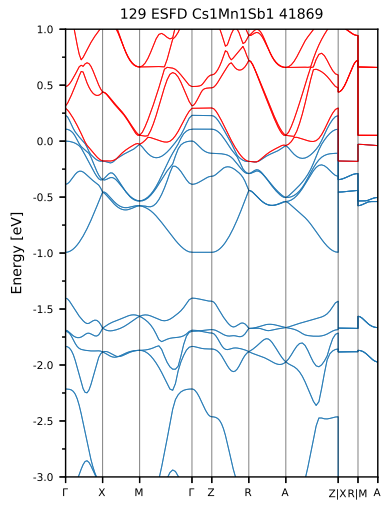

(pz) 129.4.162

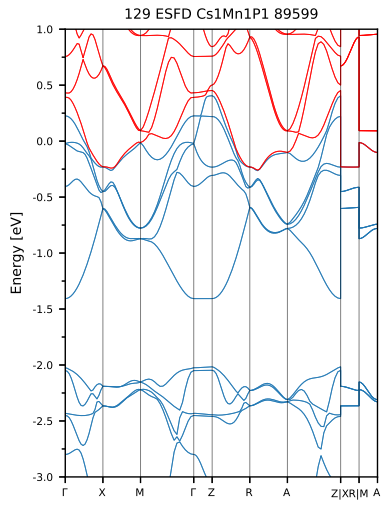

(qa) 129.4.163

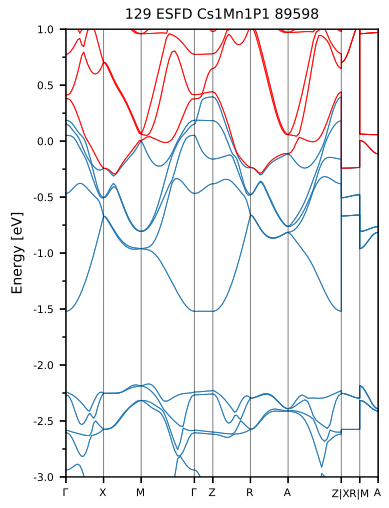

(qb) 129.4.164

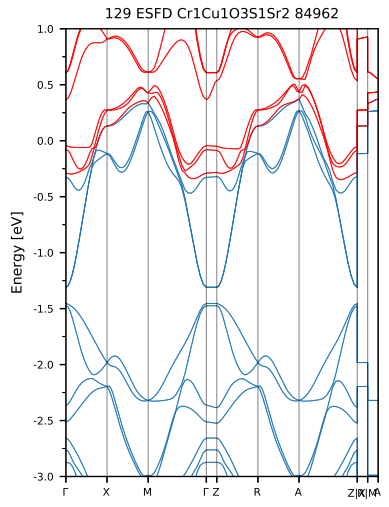

(qc) 129.4.165

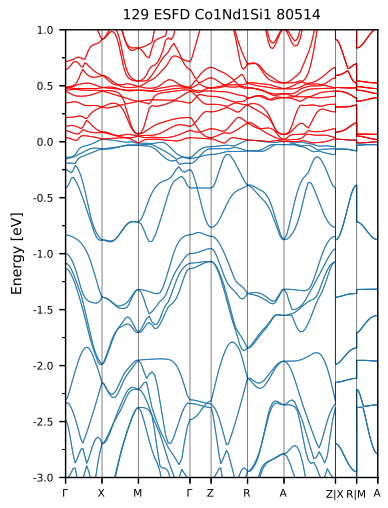

(qd) 129.4.166

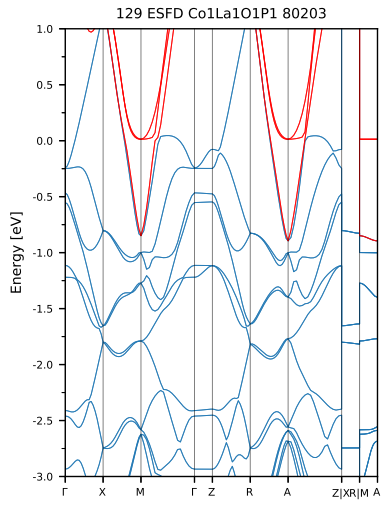

(qe) 129.4.167

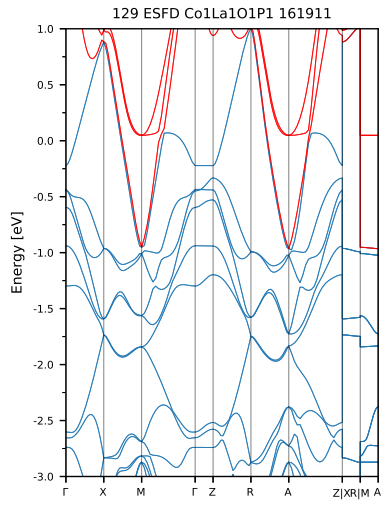

(qf) 129.4.168

Figure 266: Band structures for topological materials in SG 129, part 28/39, from 129.4.153 to 129.4.168. See the list of figures.

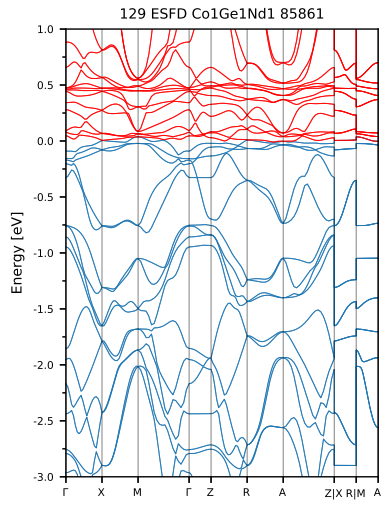

(qg) 129.4.169

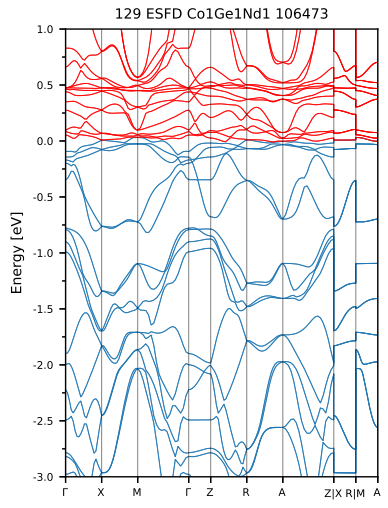

(qh) 129.4.170

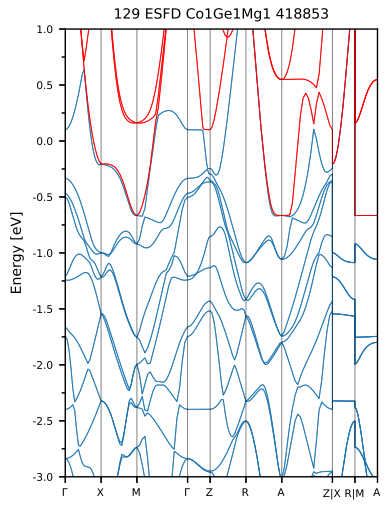

(qi) 129.4.171

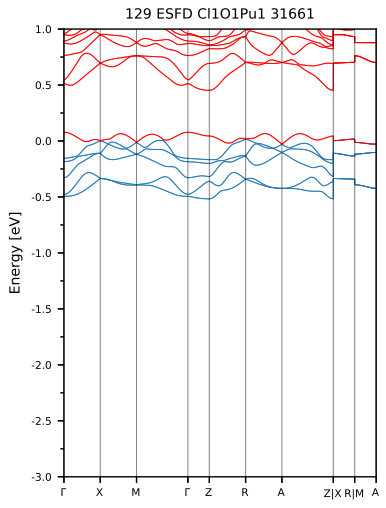

(qj) 129.4.172

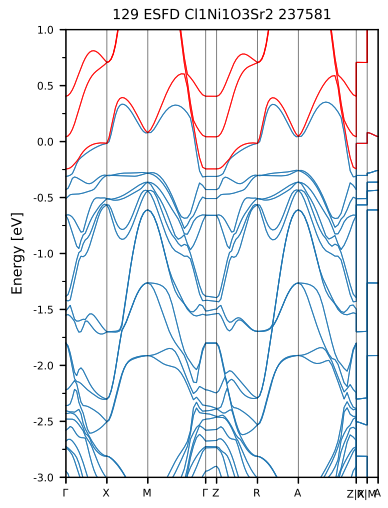

(qk) 129.4.173

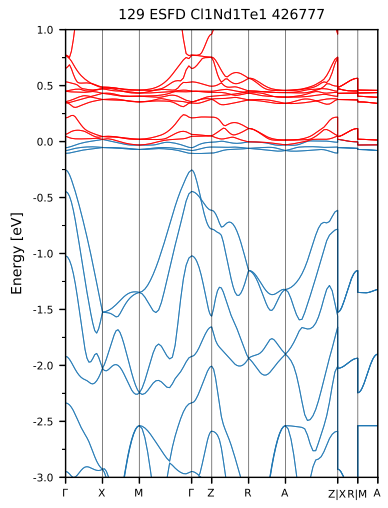

(ql) 129.4.174

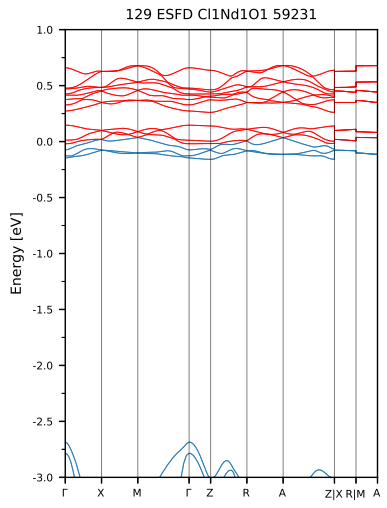

(qm) 129.4.175

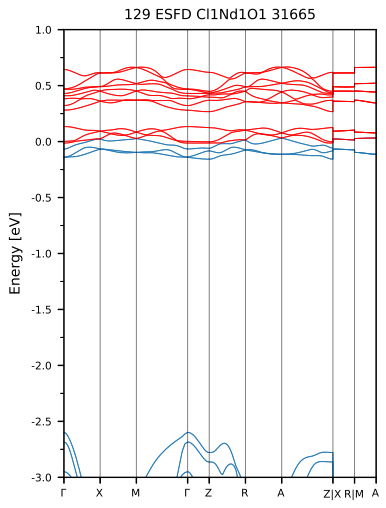

(qn) 129.4.176

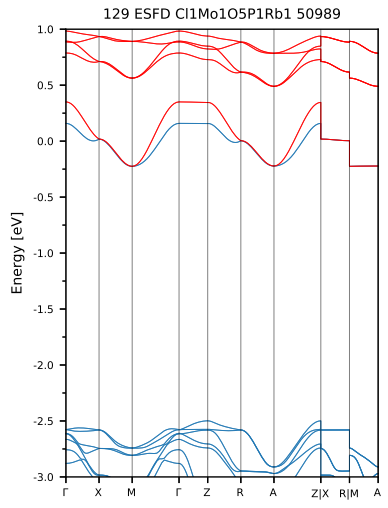

(qo) 129.4.177

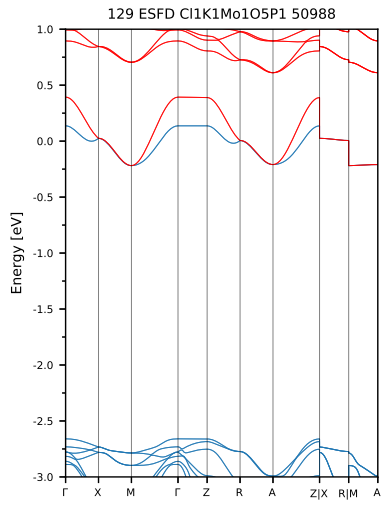

(qp) 129.4.178

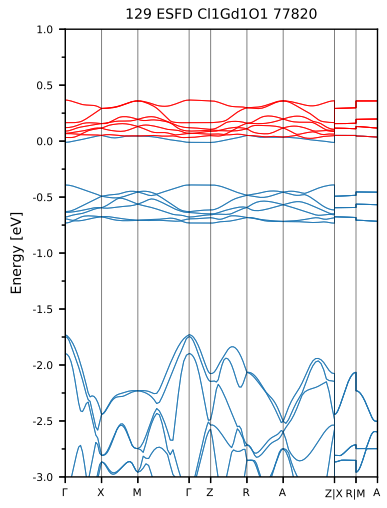

(qq) 129.4.179

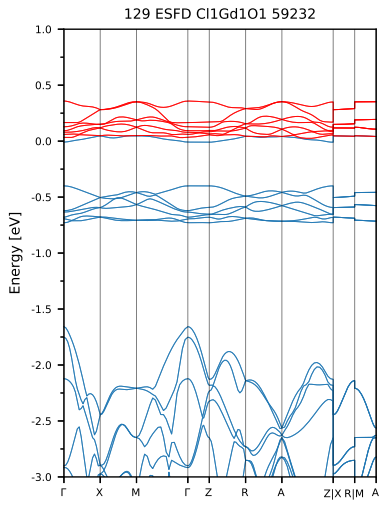

(qr) 129.4.180

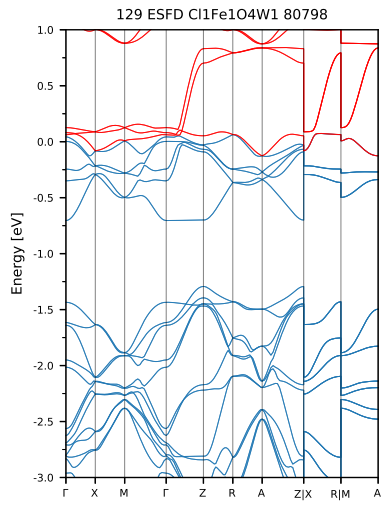

(qs) 129.4.181

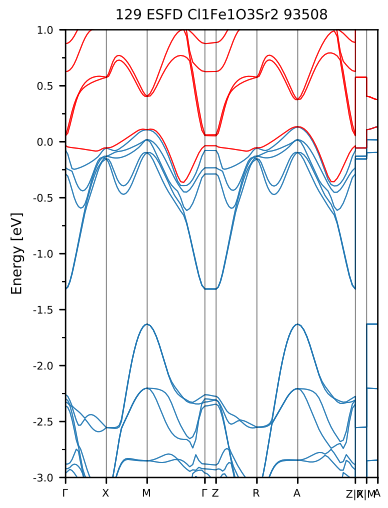

(qt) 129.4.182

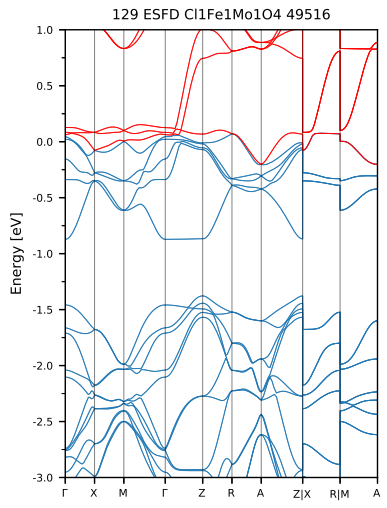

(qu) 129.4.183

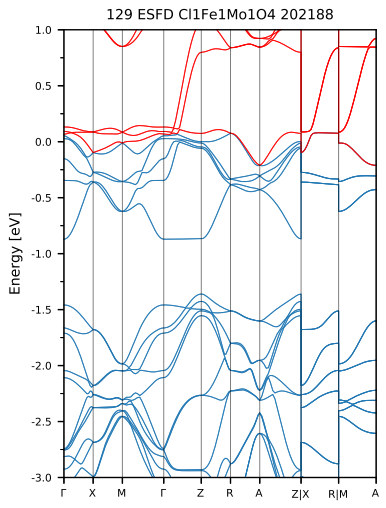

(qv) 129.4.184

Figure 267: Band structures for topological materials in SG 129, part 29/39, from 129.4.169 to 129.4.184. See the list of figures.

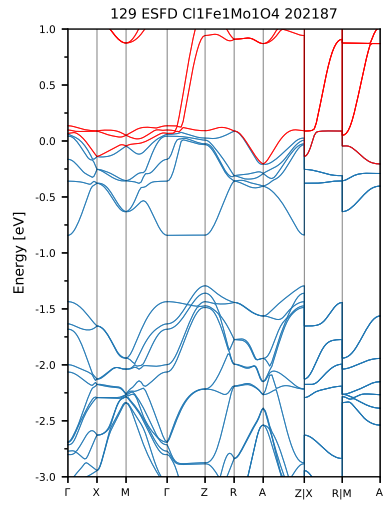

(qw) 129.4.185

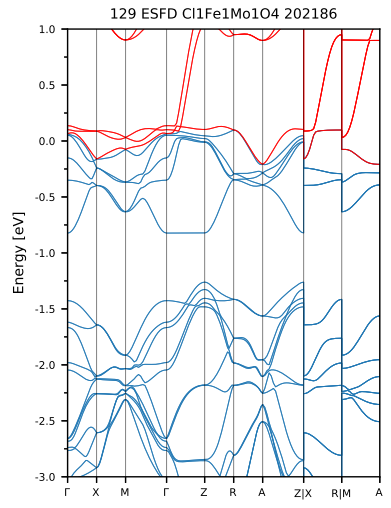

(qx) 129.4.186

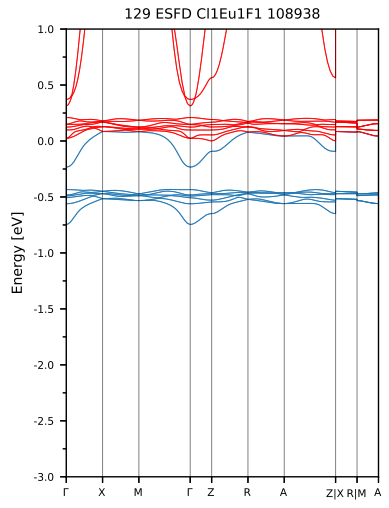

(qy) 129.4.187

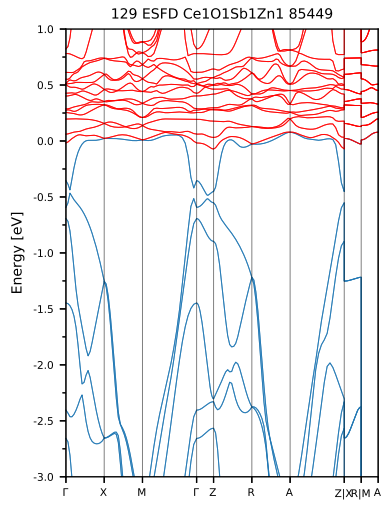

(qz) 129.4.188

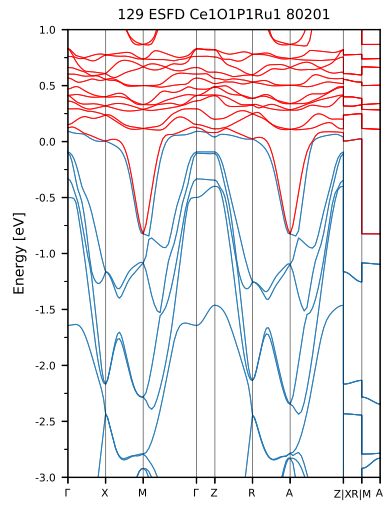

(ra) 129.4.189

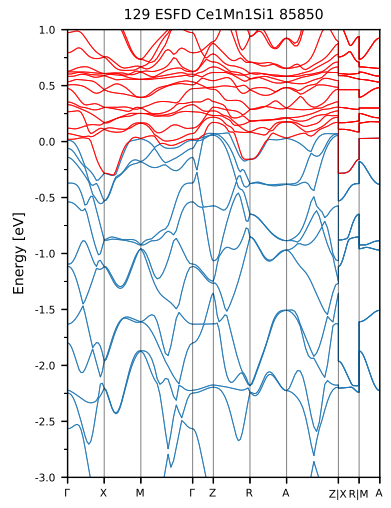

(rb) 129.4.190

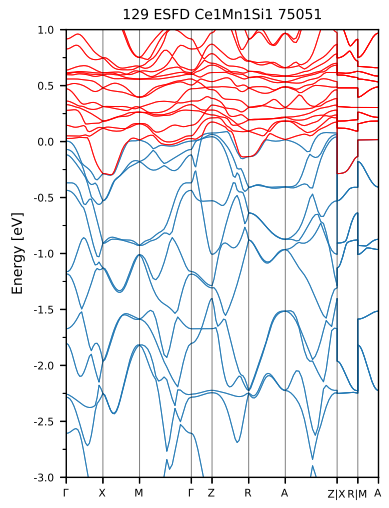

(rc) 129.4.191

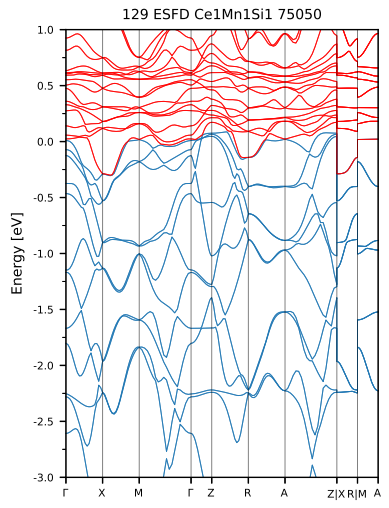

(rd) 129.4.192

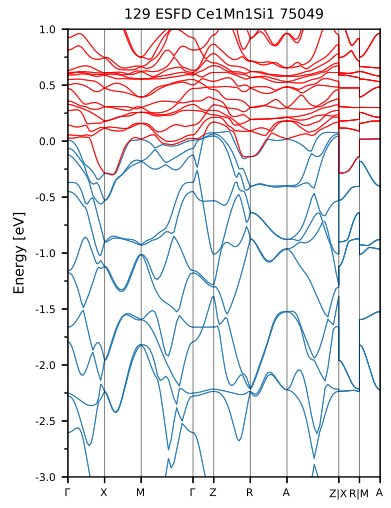

(re) 129.4.193

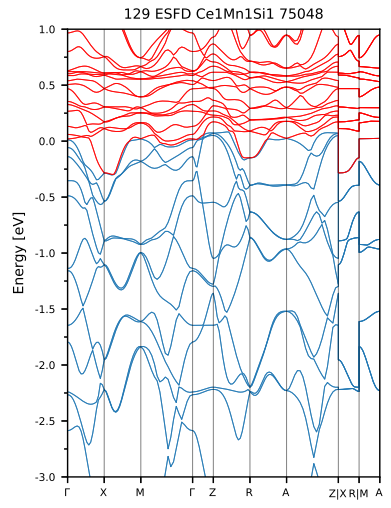

(rf) 129.4.194

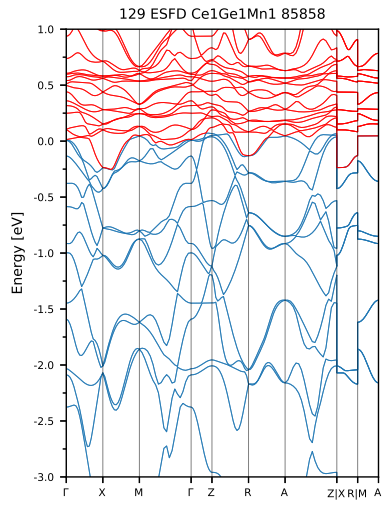

(rg) 129.4.195

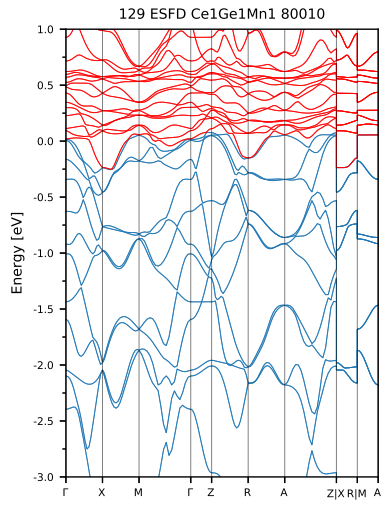

(rh) 129.4.196

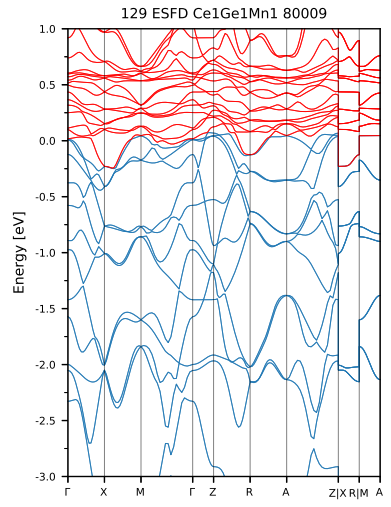

(ri) 129.4.197

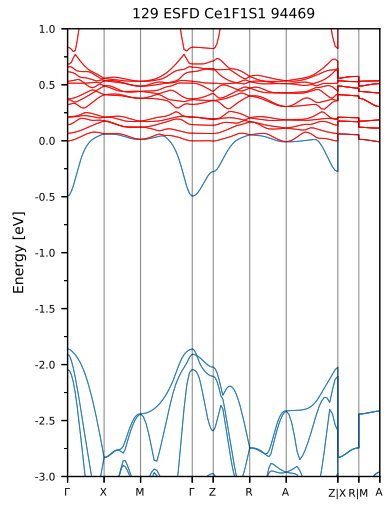

(rj) 129.4.198

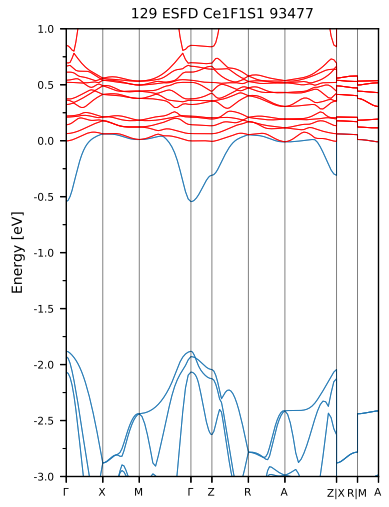

(rk) 129.4.199

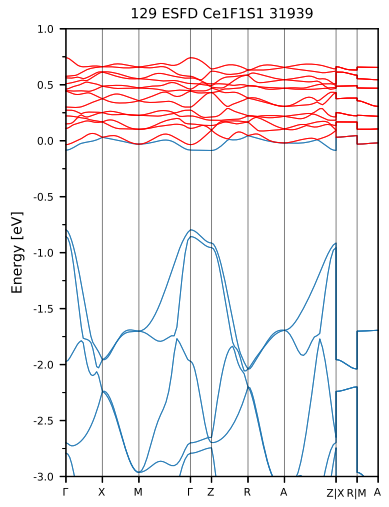

(rl) 129.4.200

Figure 268: Band structures for topological materials in SG 129, part 30/39, from 129.4.185 to 129.4.200. See the list of figures.

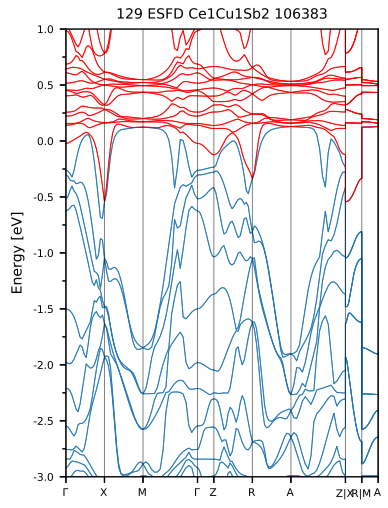

(rm) 129.4.201

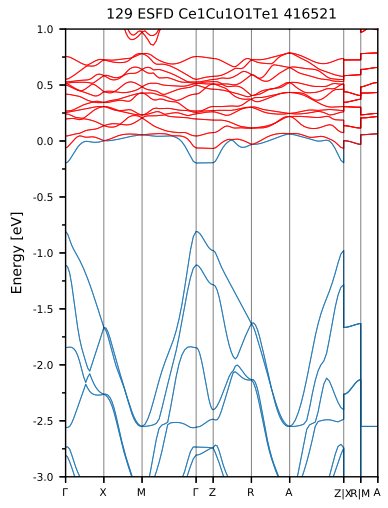

(rn) 129.4.202

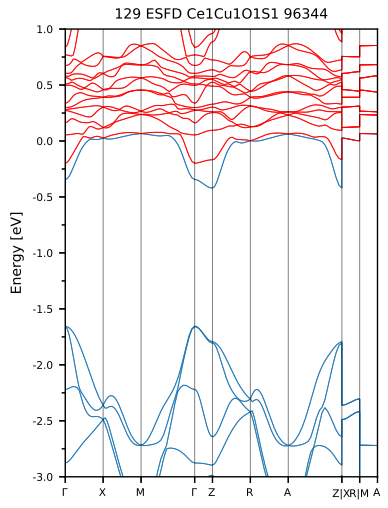

(ro) 129.4.203

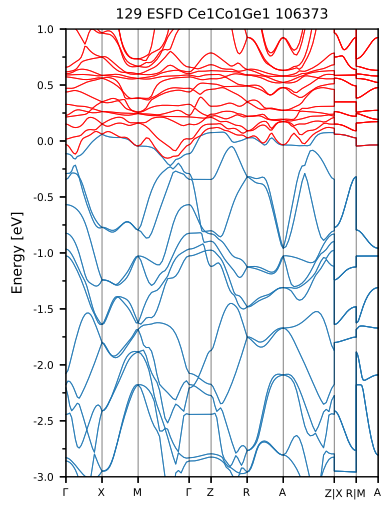

(rp) 129.4.204

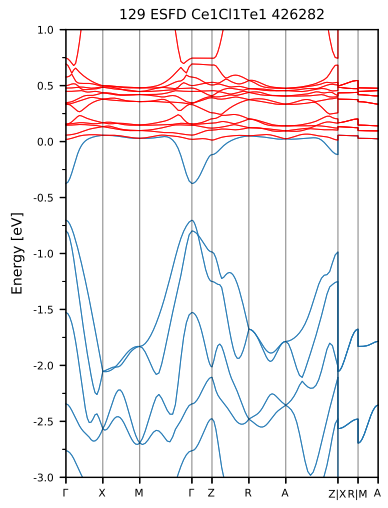

(rq) 129.4.205

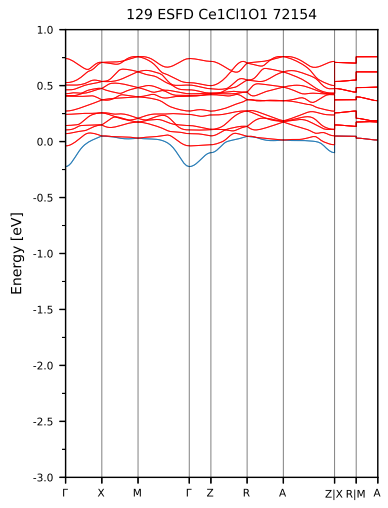

(rr) 129.4.206

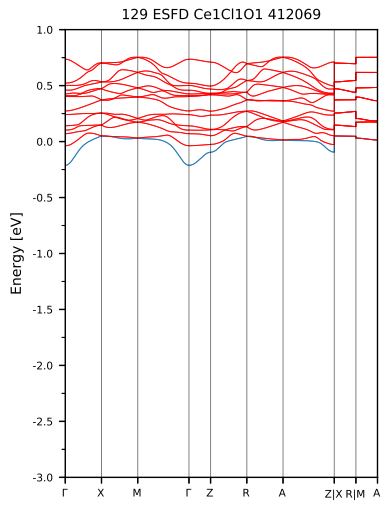

(rs) 129.4.207

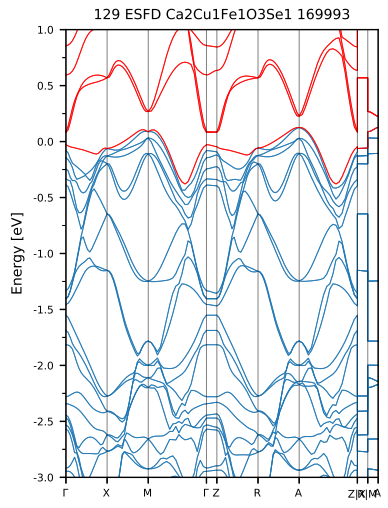

(rt) 129.4.208

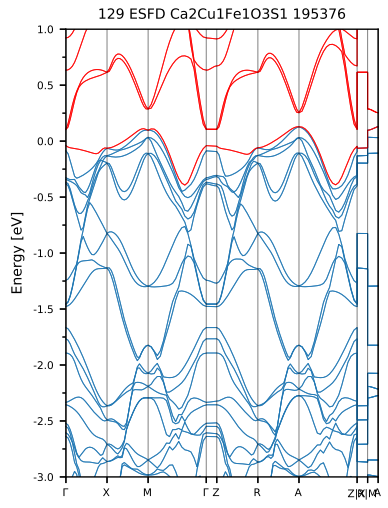

(ru) 129.4.209

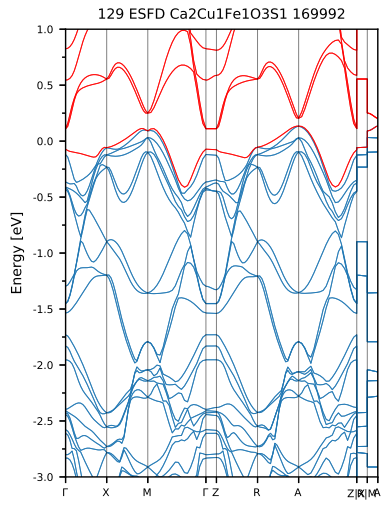

(rv) 129.4.210

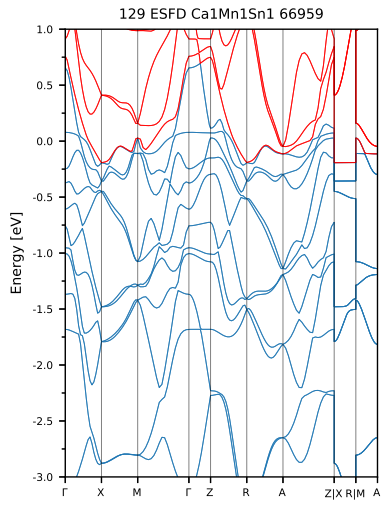

(rw) 129.4.211

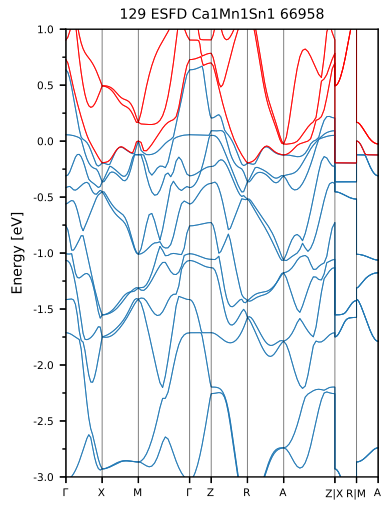

(rx) 129.4.212

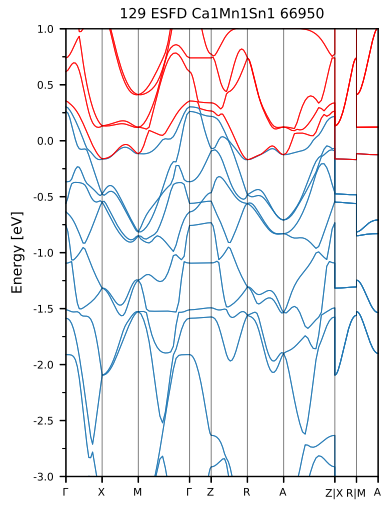

(ry) 129.4.213

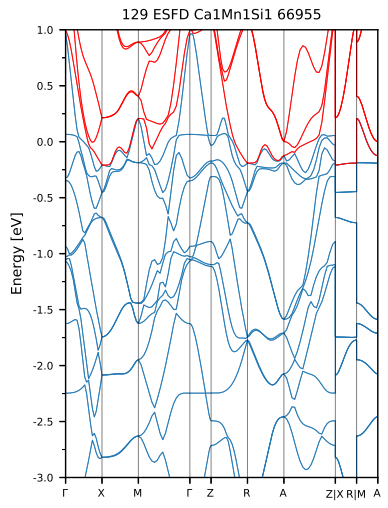

(rz) 129.4.214

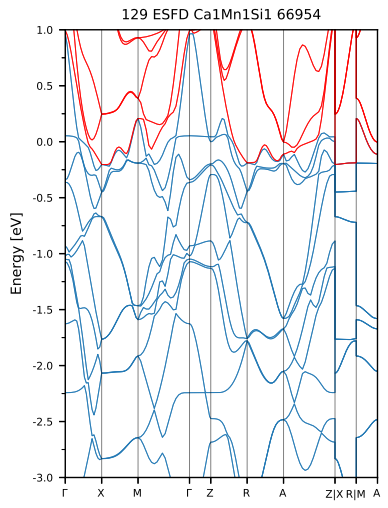

(sa) 129.4.215

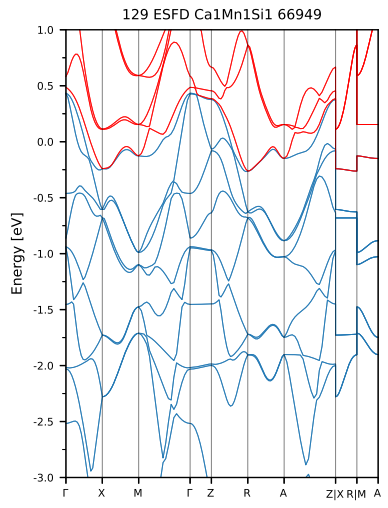

(sb) 129.4.216

Figure 269: Band structures for topological materials in SG 129, part 31/39, from 129.4.201 to 129.4.216. See the list of figures.

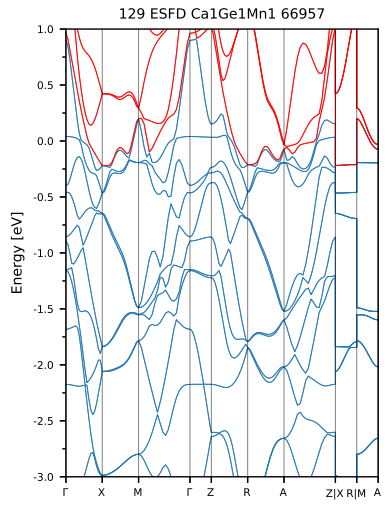

(sc) 129.4.217

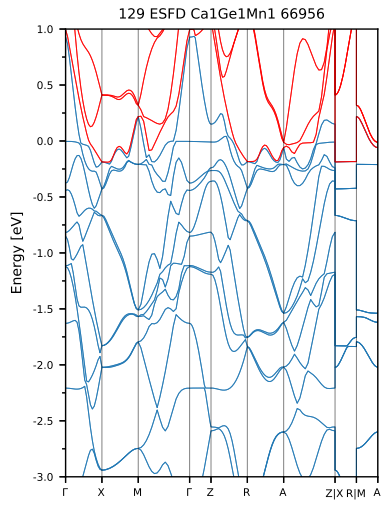

(sd) 129.4.218

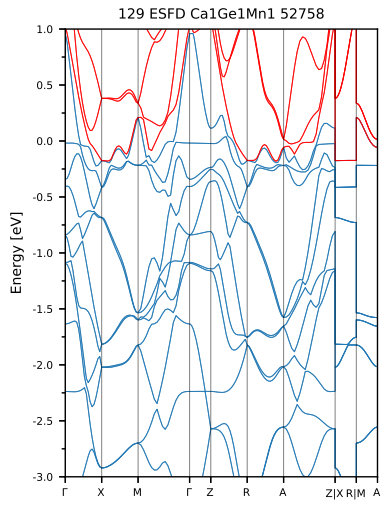

(se) 129.4.219

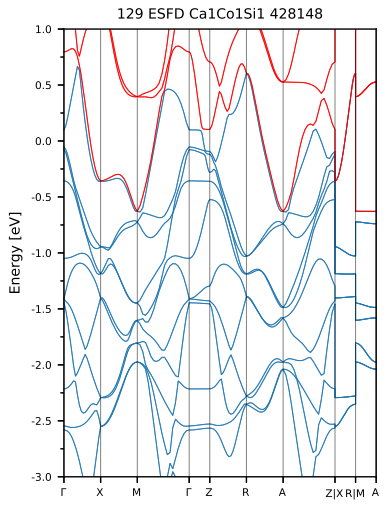

(sf) 129.4.220

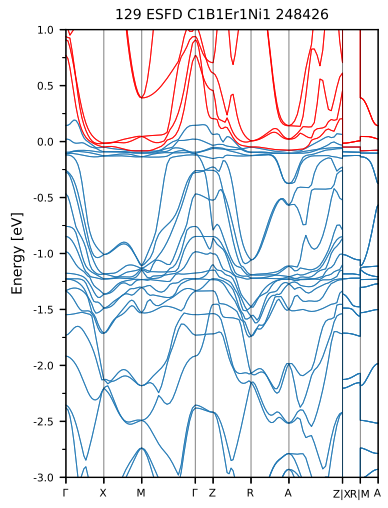

(sg) 129.4.221

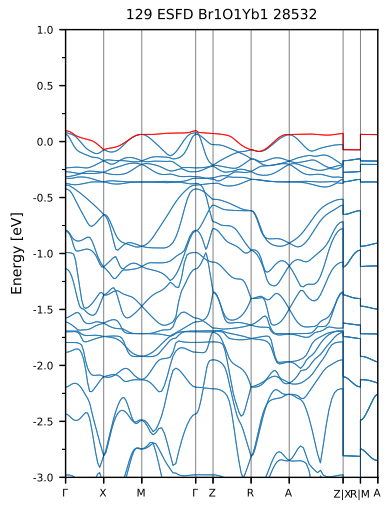

(sh) 129.4.222

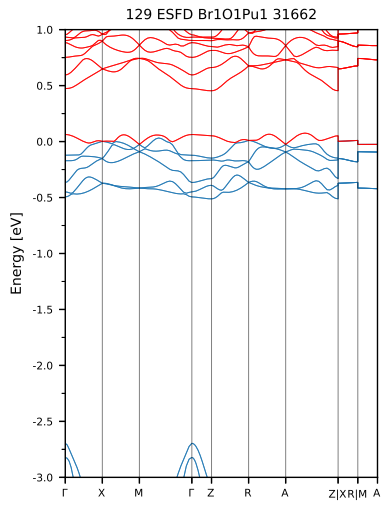

(si) 129.4.223

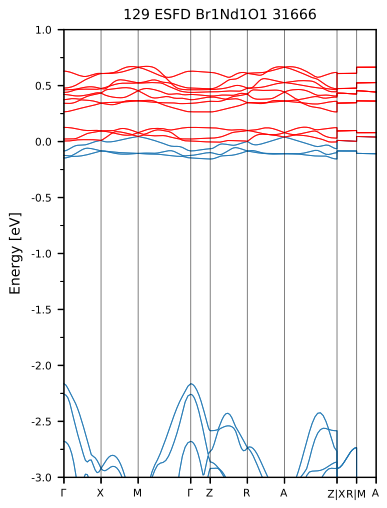

(sj) 129.4.224

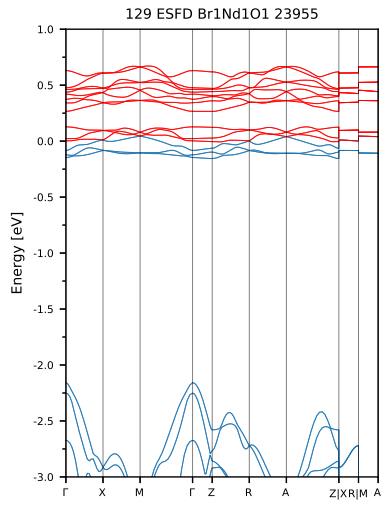

(sk) 129.4.225

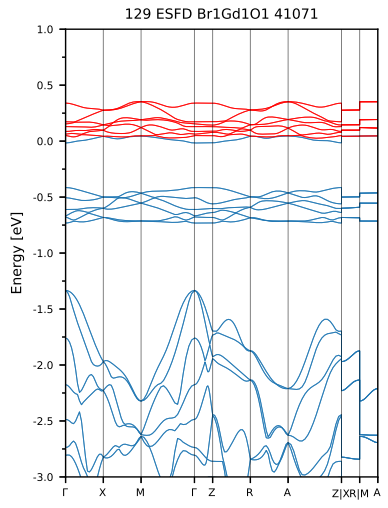

(sl) 129.4.226

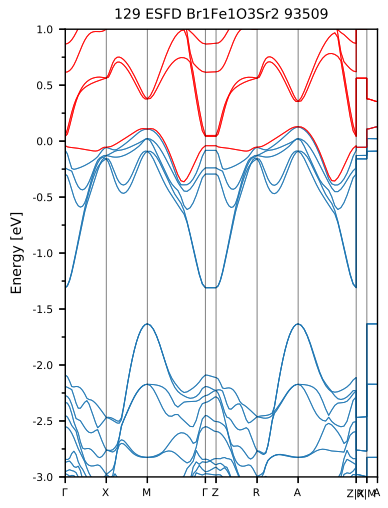

(sm) 129.4.227

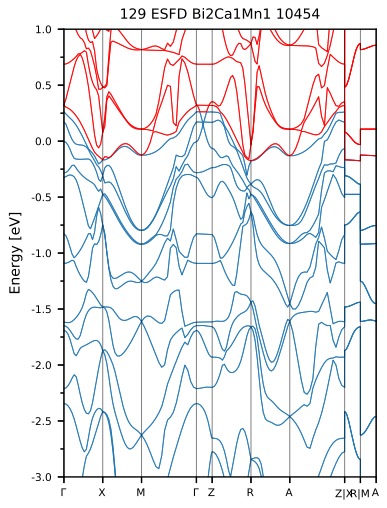

(sn) 129.4.228

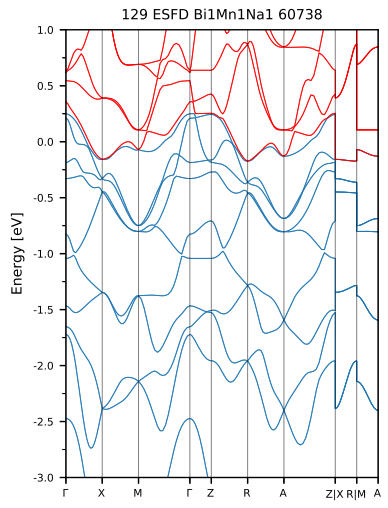

(so) 129.4.229

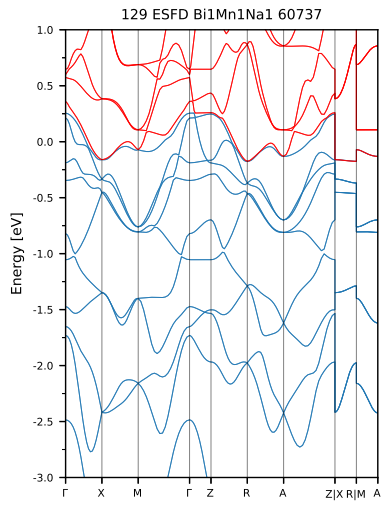

(sp) 129.4.230

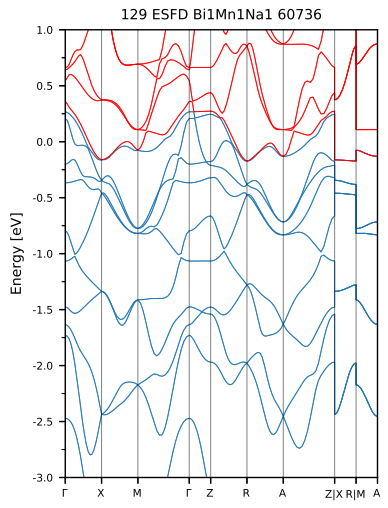

(sq) 129.4.231

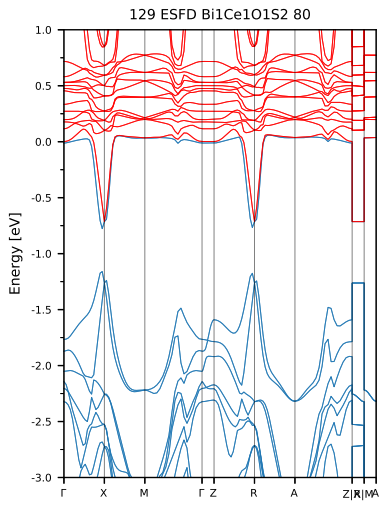

(sr) 129.4.232

Figure 270: Band structures for topological materials in SG 129, part 32/39, from 129.4.217 to 129.4.232. See the list of figures.

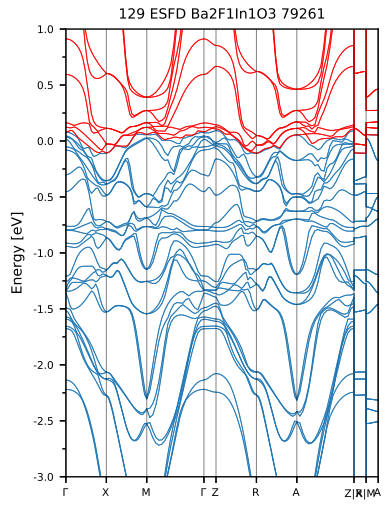

(ss) 129.4.233

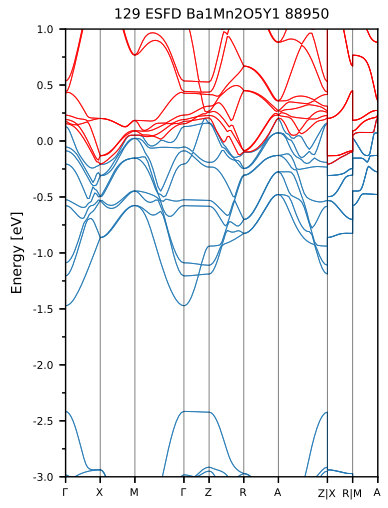

(st) 129.4.234

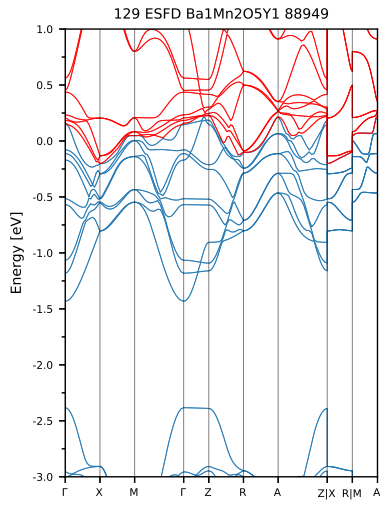

(su) 129.4.235

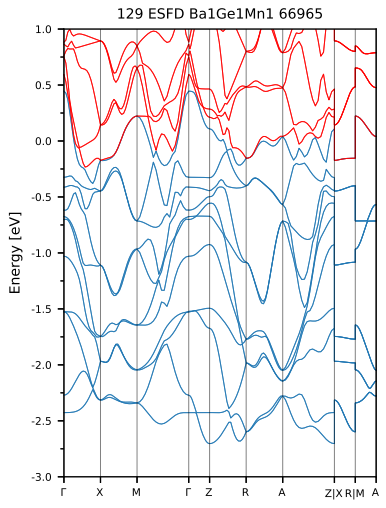

(sv) 129.4.236

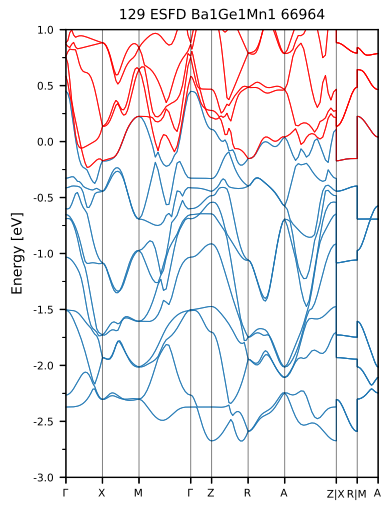

(sw) 129.4.237

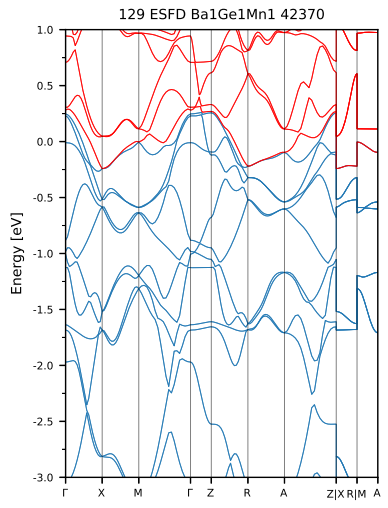

(sx) 129.4.238

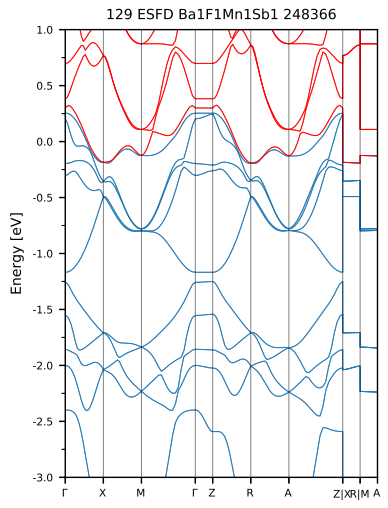

(sy) 129.4.239

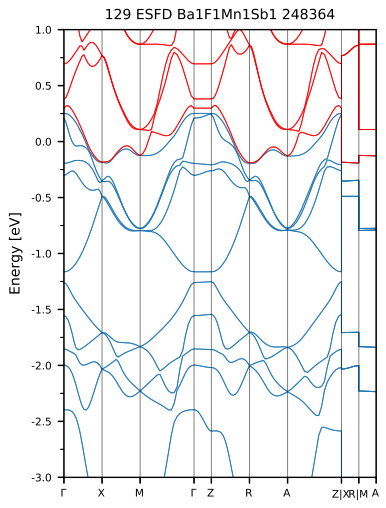

(sz) 129.4.240

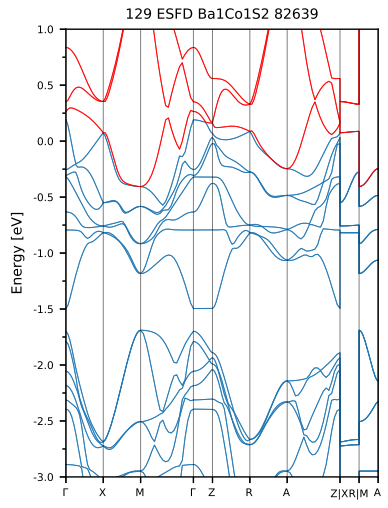

(ta) 129.4.241

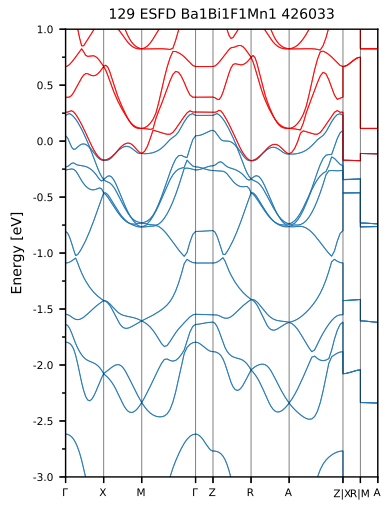

(tb) 129.4.242

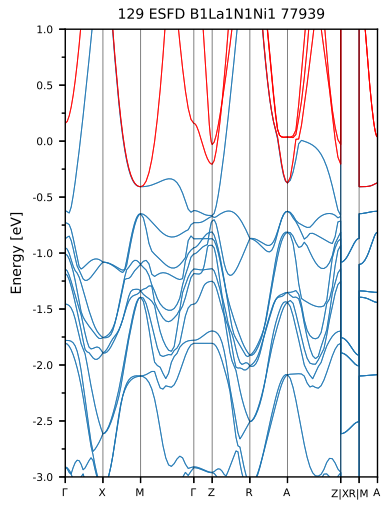

(tc) 129.4.243

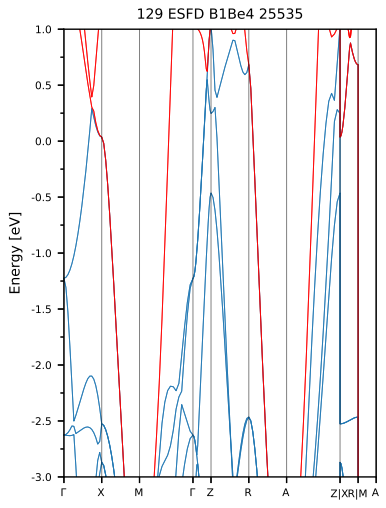

(td) 129.4.244

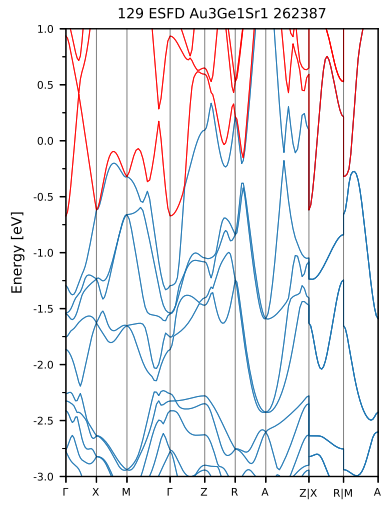

(te) 129.4.245

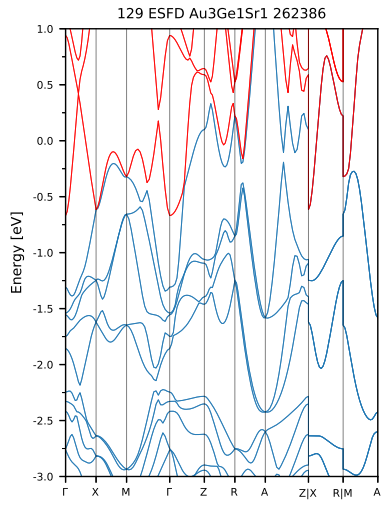

(tf) 129.4.246

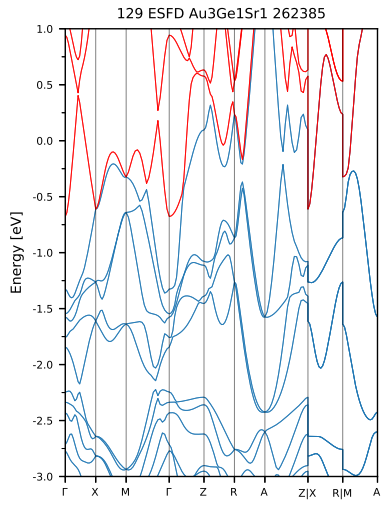

(tg) 129.4.247

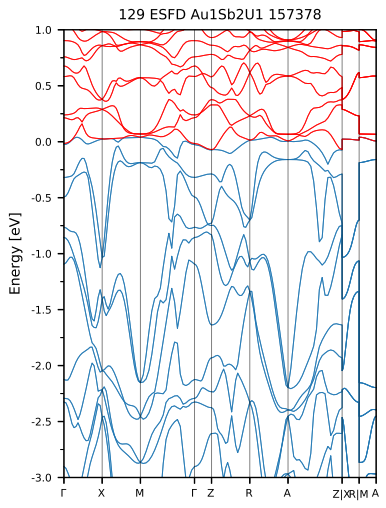

(th) 129.4.248

Figure 271: Band structures for topological materials in SG 129, part 33/39, from 129.4.233 to 129.4.248. See the list of figures.

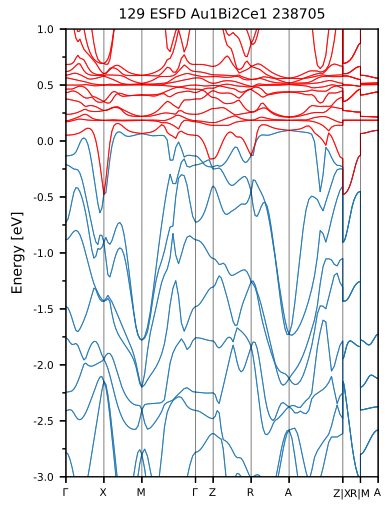

(ti) 129.4.249

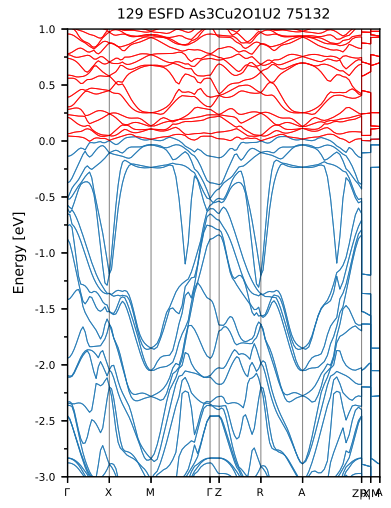

(tj) 129.4.250

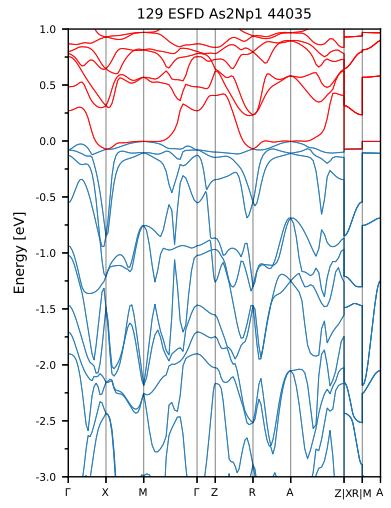

(tk) 129.4.251

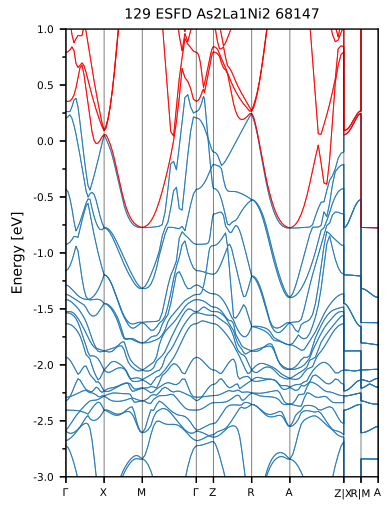

(tl) 129.4.252

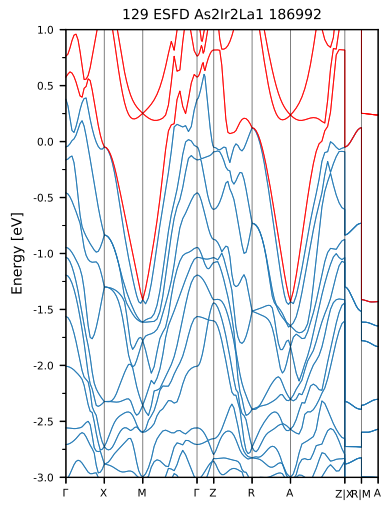

(tm) 129.4.253

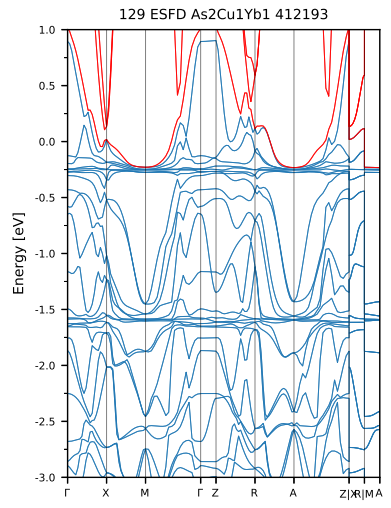

(tn) 129.4.254

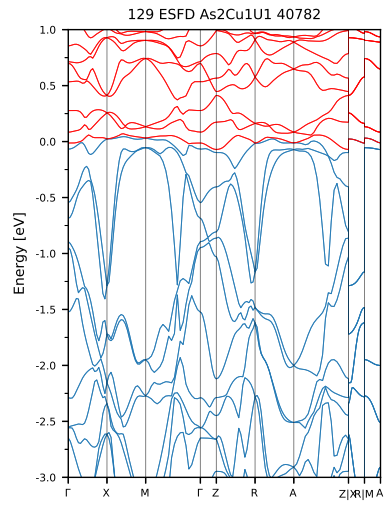

(to) 129.4.255

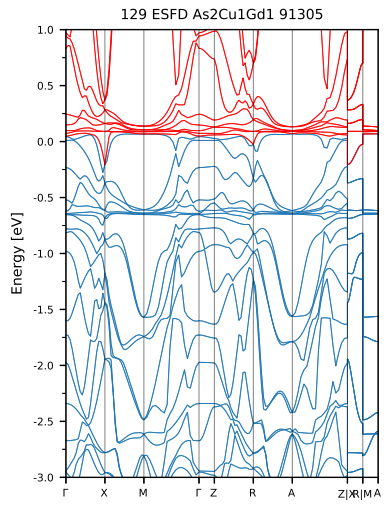

(tp) 129.4.256

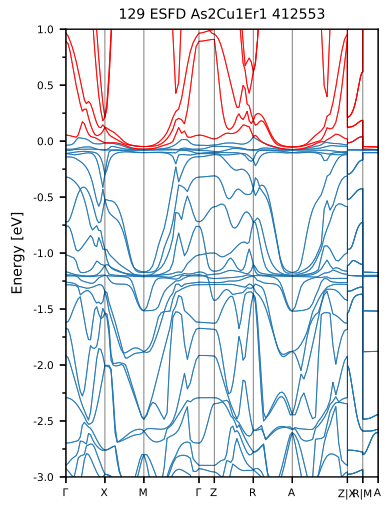

(tq) 129.4.257

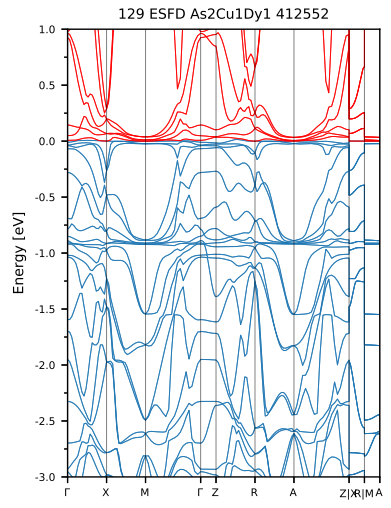

(tr) 129.4.258

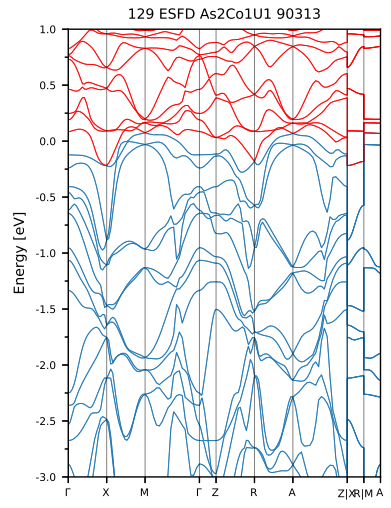

(ts) 129.4.259

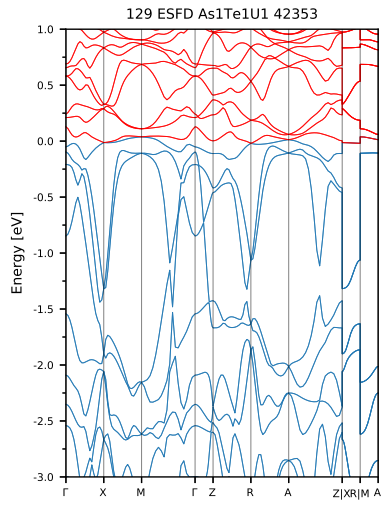

(tt) 129.4.260

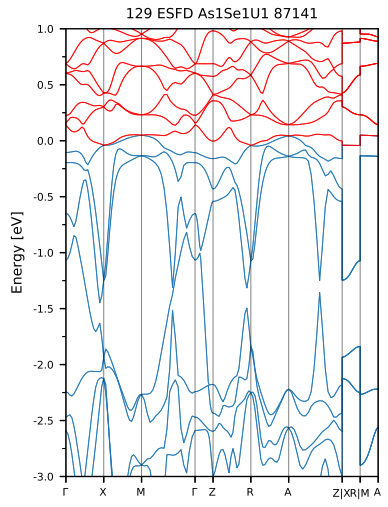

(tu) 129.4.261

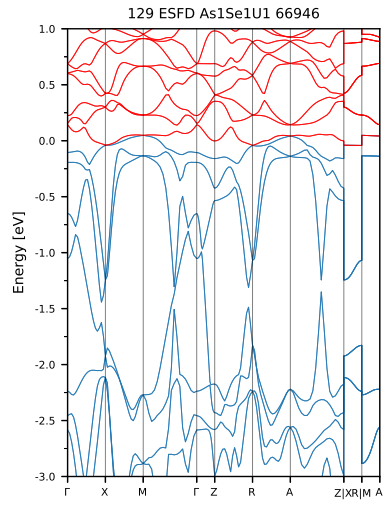

(tv) 129.4.262

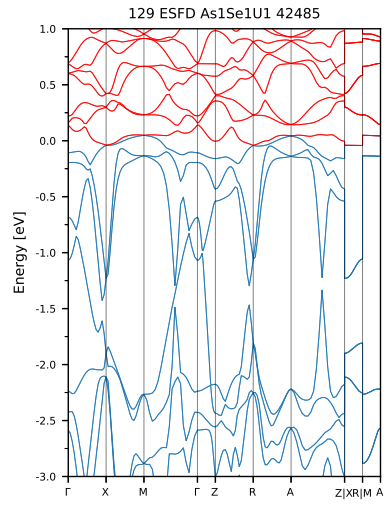

(tw) 129.4.263

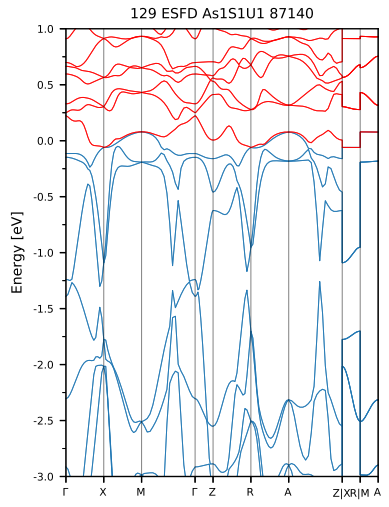

(tx) 129.4.264

Figure 272: Band structures for topological materials in SG 129, part 34/39, from 129.4.249 to 129.4.264. See the list of figures.

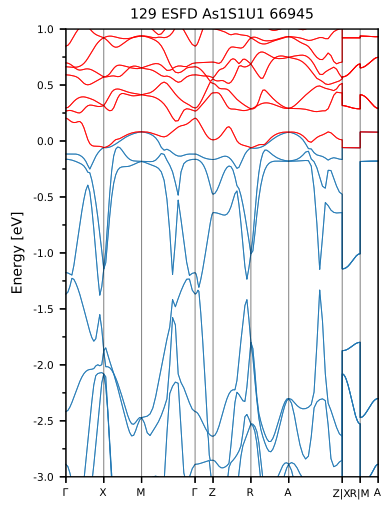

(ty) 129.4.265

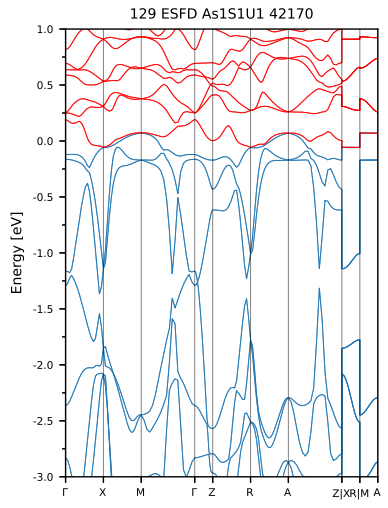

(tz) 129.4.266

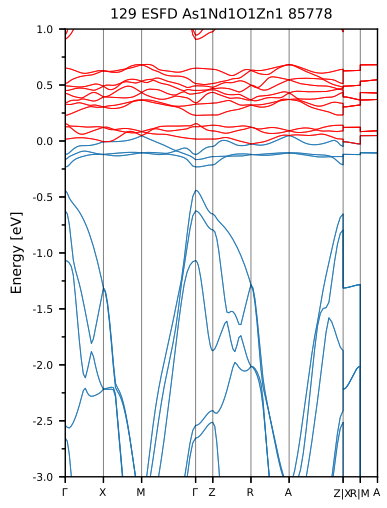

(ua) 129.4.267

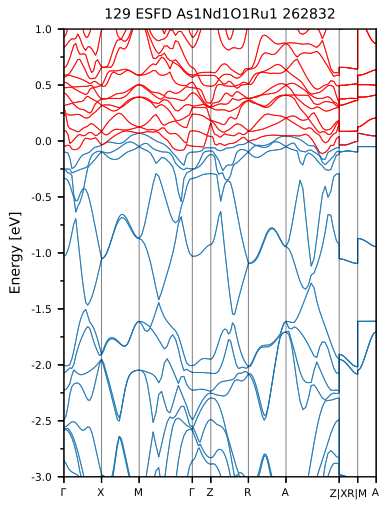

(ub) 129.4.268

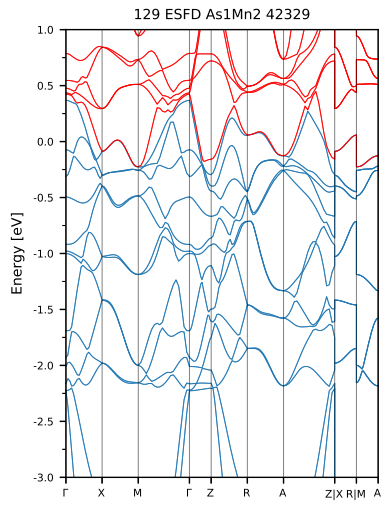

(uc) 129.4.269

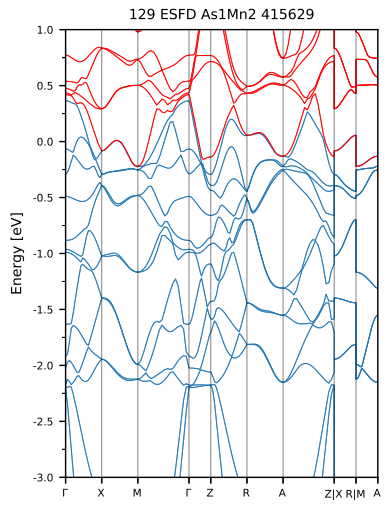

(ud) 129.4.270

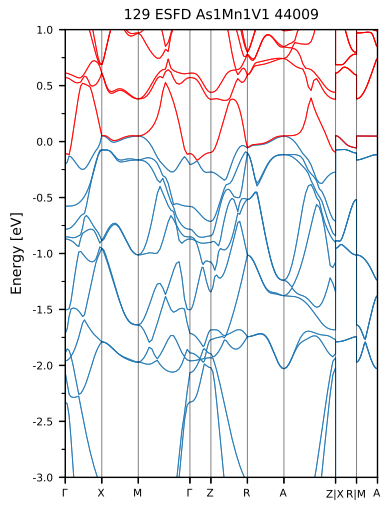

(ue) 129.4.271

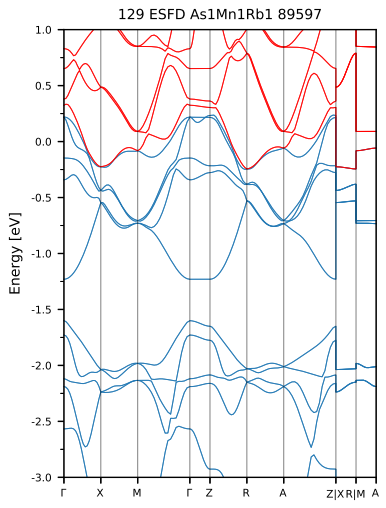

(uf) 129.4.272

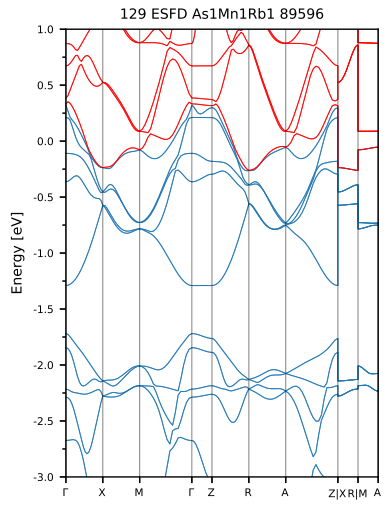

(ug) 129.4.273

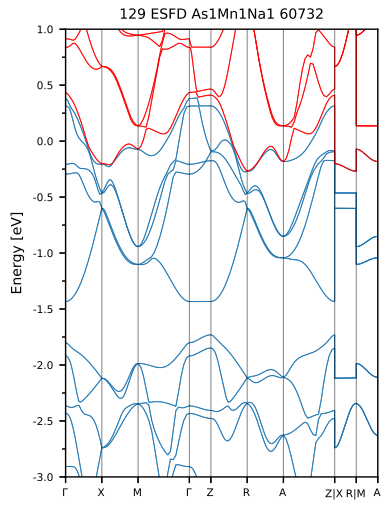

(uh) 129.4.274

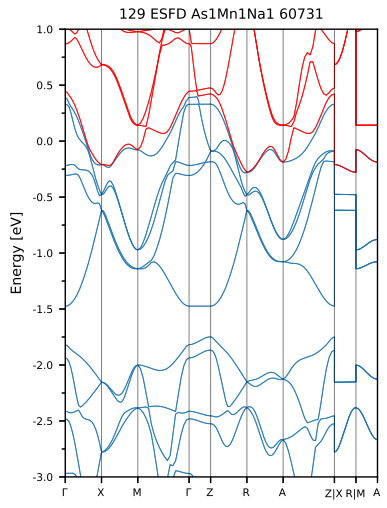

(ui) 129.4.275

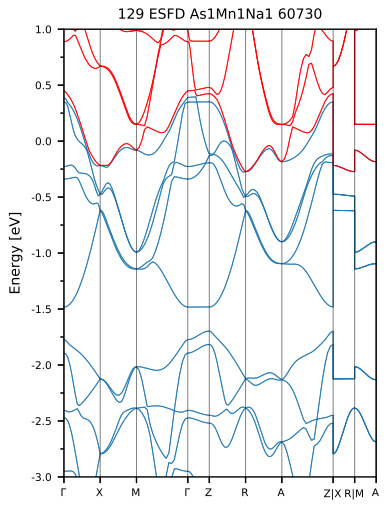

(uj) 129.4.276

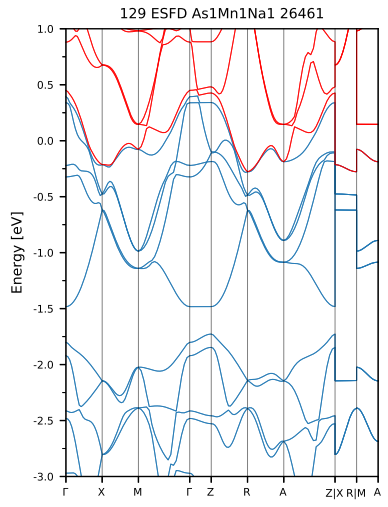

(uk) 129.4.277

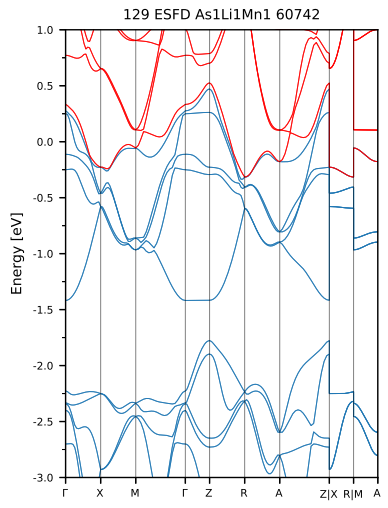

(ul) 129.4.278

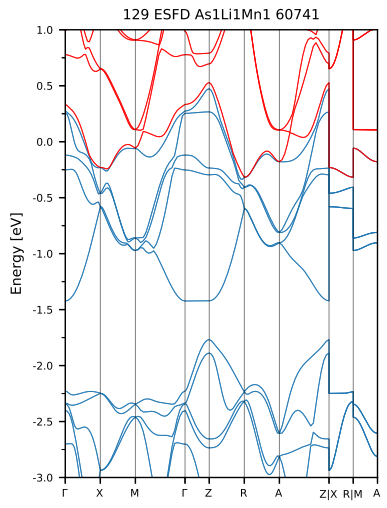

(um) 129.4.279

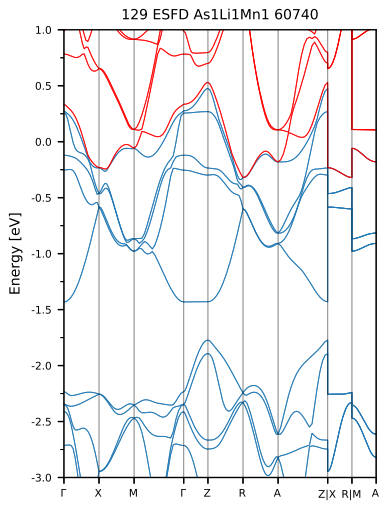

(un) 129.4.280

Figure 273: Band structures for topological materials in SG 129, part 35/39, from 129.4.265 to 129.4.280. See the list of figures.

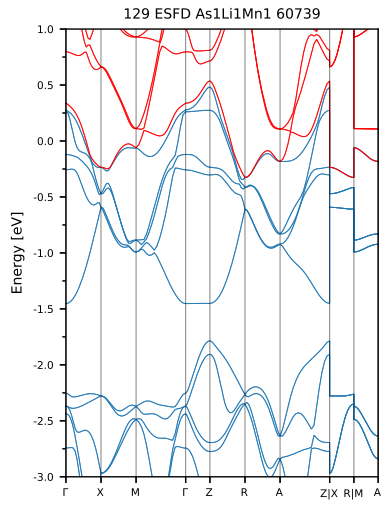

(uo) 129.4.281

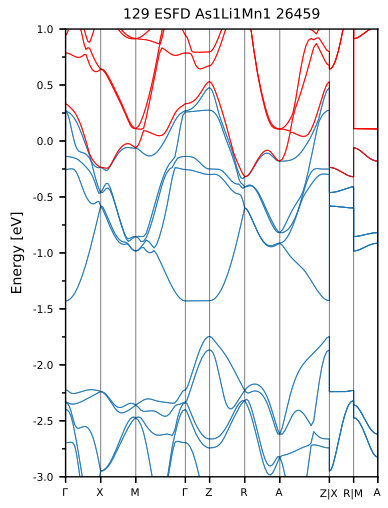

(up) 129.4.282

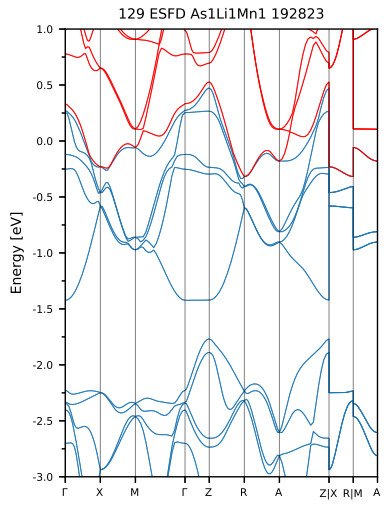

(uq) 129.4.283

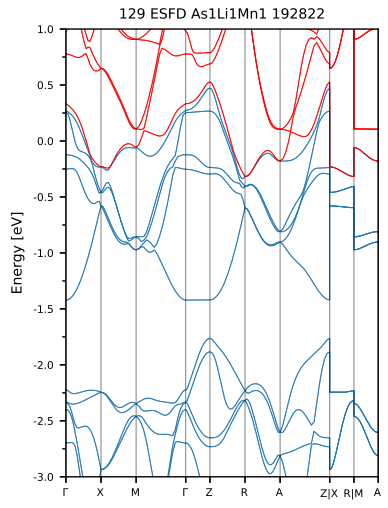

(ur) 129.4.284

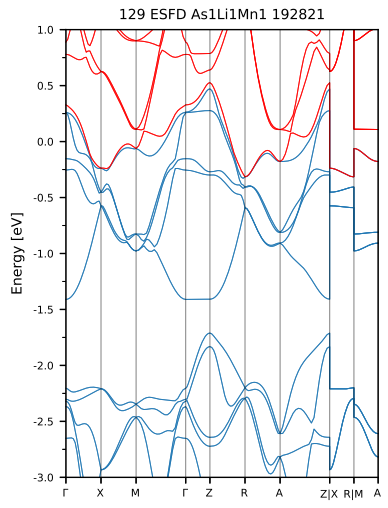

(us) 129.4.285

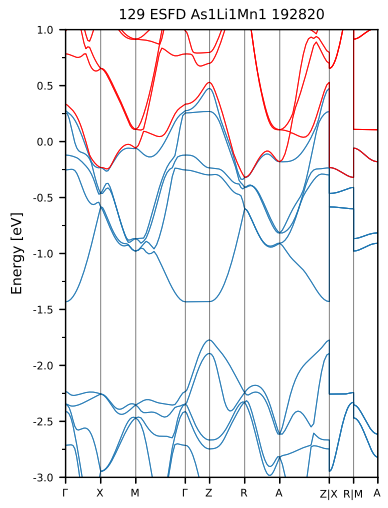

(ut) 129.4.286

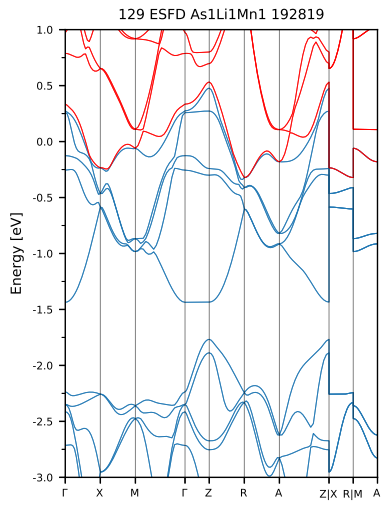

(uu) 129.4.287

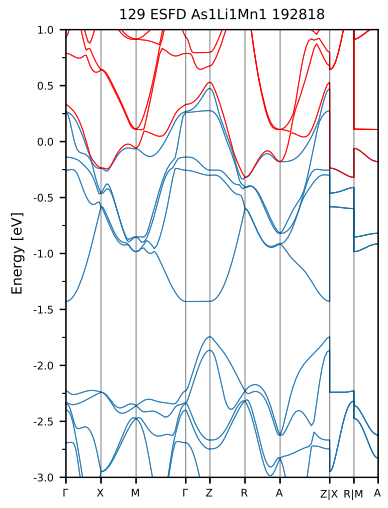

(uv) 129.4.288

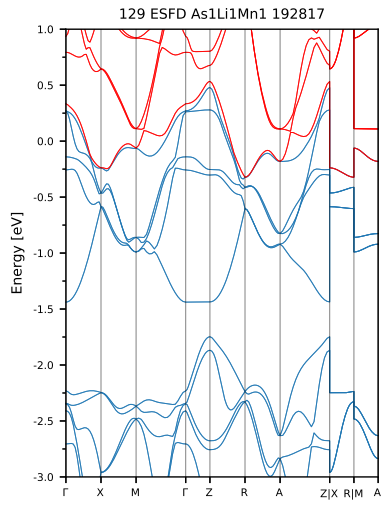

(uw) 129.4.289

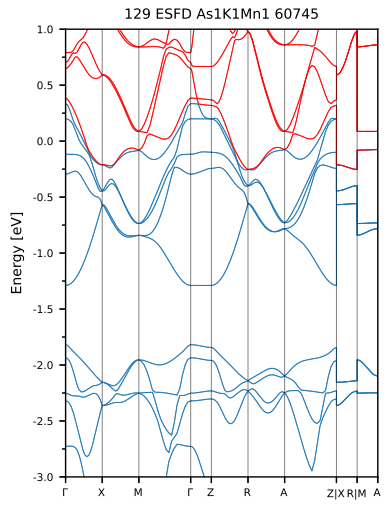

(ux) 129.4.290

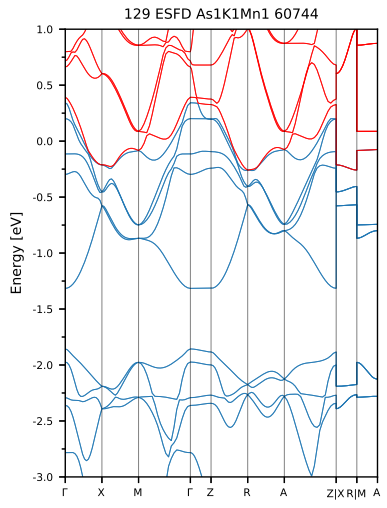

(uy) 129.4.291

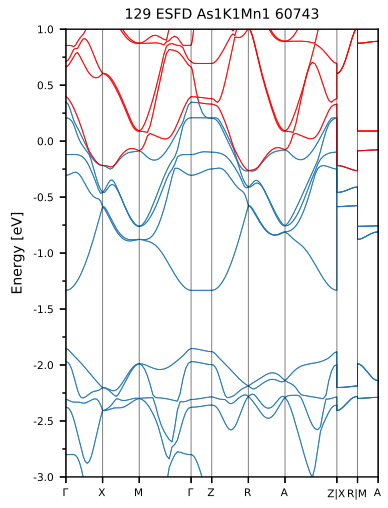

(uz) 129.4.292

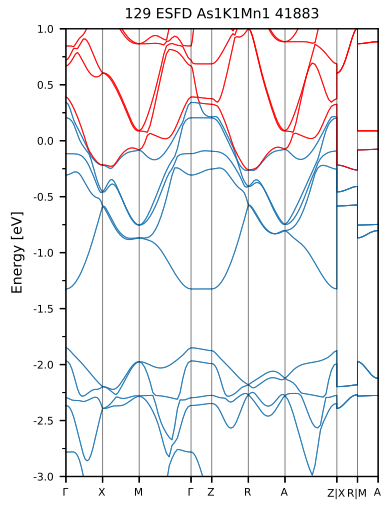

(va) 129.4.293

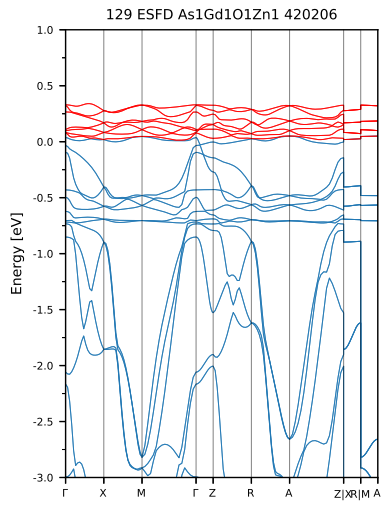

(vb) 129.4.294

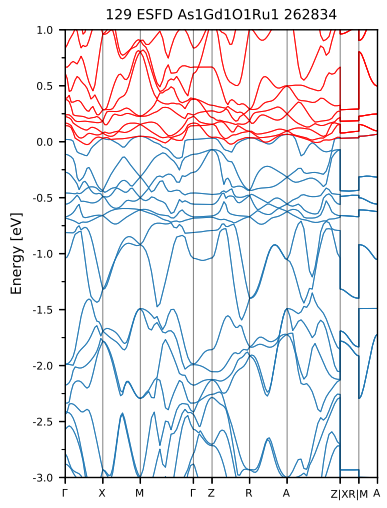

(vc) 129.4.295

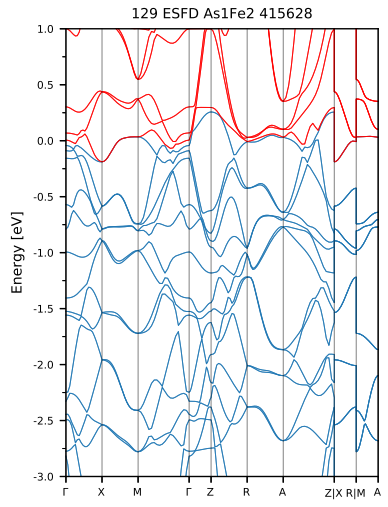

(vd) 129.4.296

Figure 274: Band structures for topological materials in SG 129, part 36/39, from 129.4.281 to 129.4.296. See the list of figures.

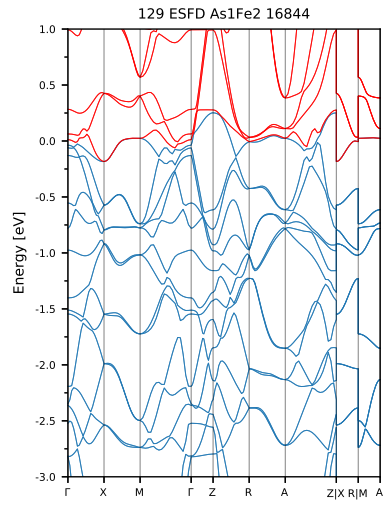

(ve) 129.4.297

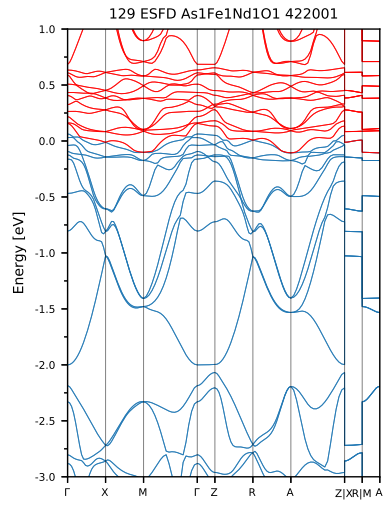

(vf) 129.4.298

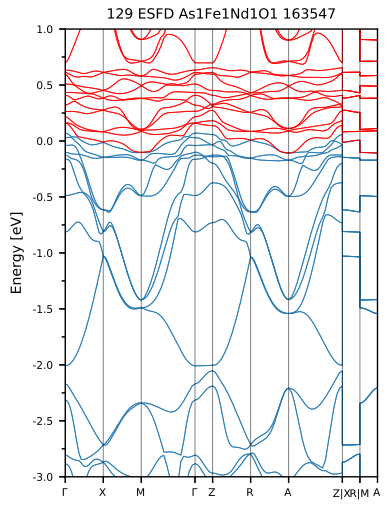

(vg) 129.4.299

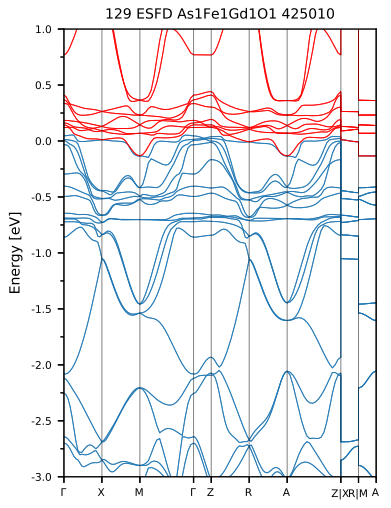

(vh) 129.4.300

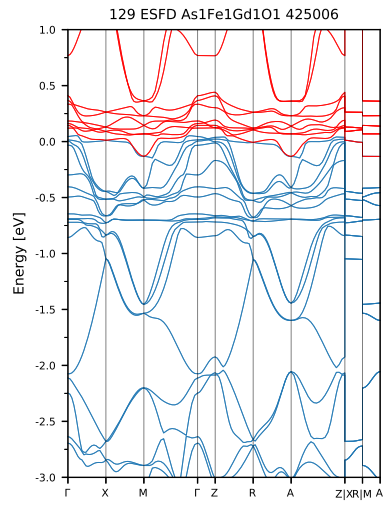

(vi) 129.4.301

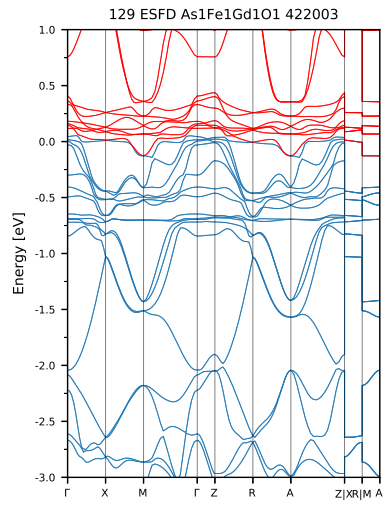

(vj) 129.4.302

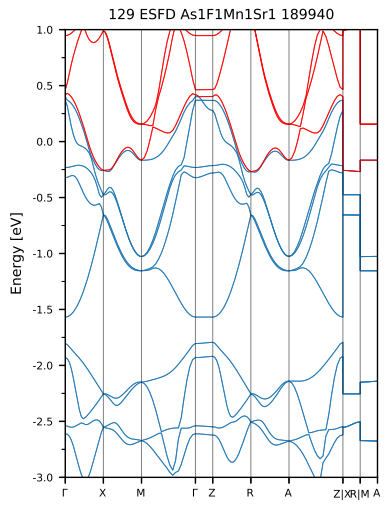

(vk) 129.4.303

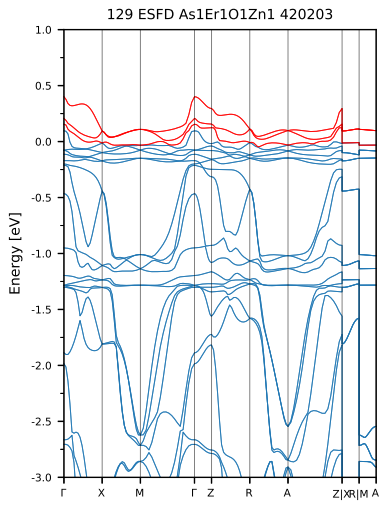

(vl) 129.4.304

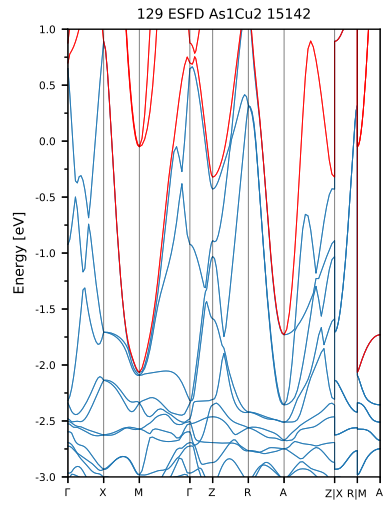

(vm) 129.4.305

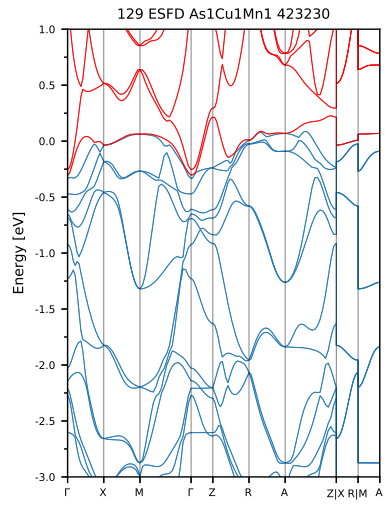

(vn) 129.4.306

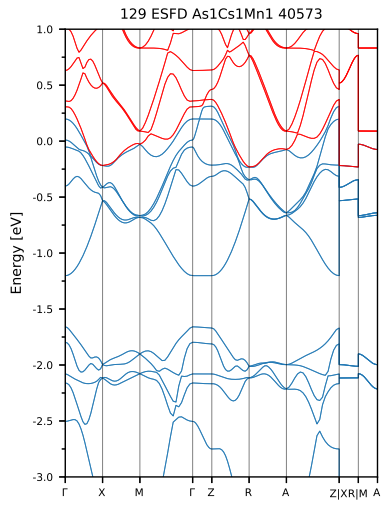

(vo) 129.4.307

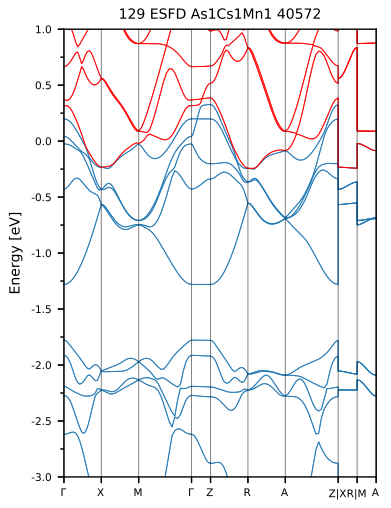

(vp) 129.4.308

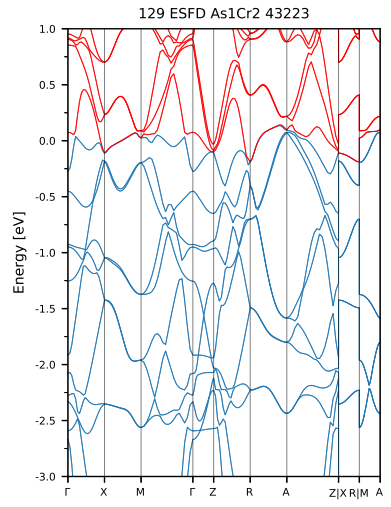

(vq) 129.4.309

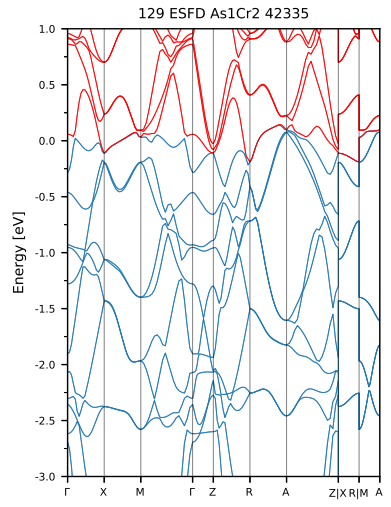

(vr) 129.4.310

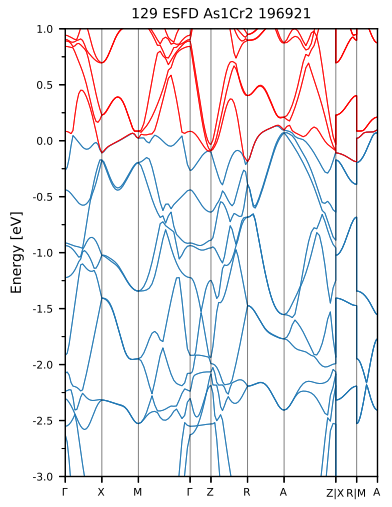

(vs) 129.4.311

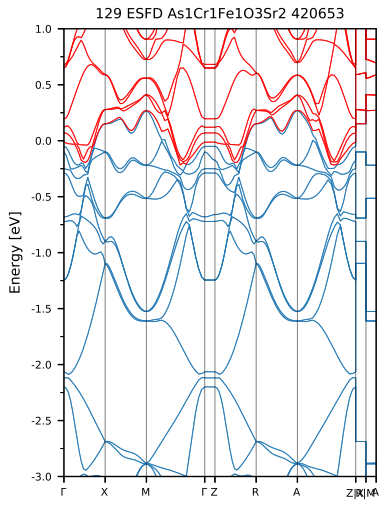

(vt) 129.4.312

Figure 275: Band structures for topological materials in SG 129, part 37/39, from 129.4.297 to 129.4.312. See the list of figures.

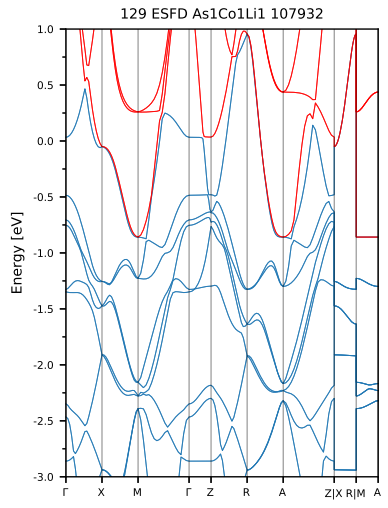

(vu) 129.4.313

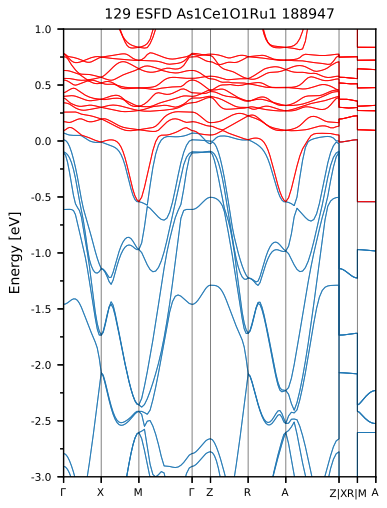

(vv) 129.4.314

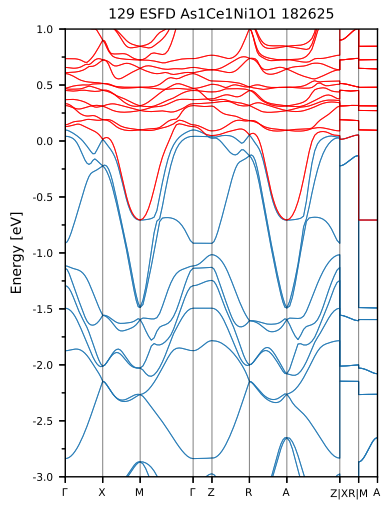

(vw) 129.4.315

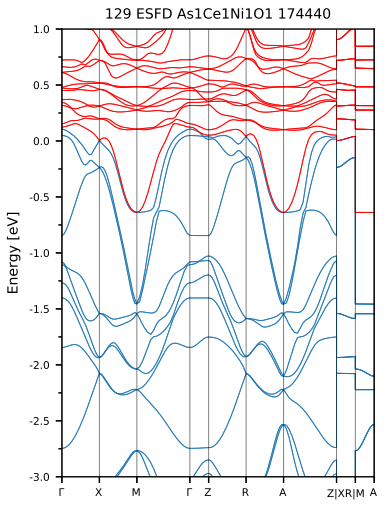

(vx) 129.4.316

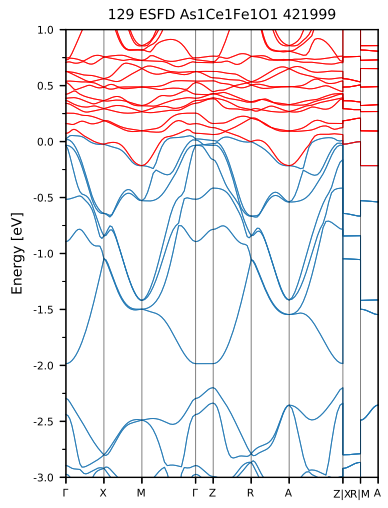

(vy) 129.4.317

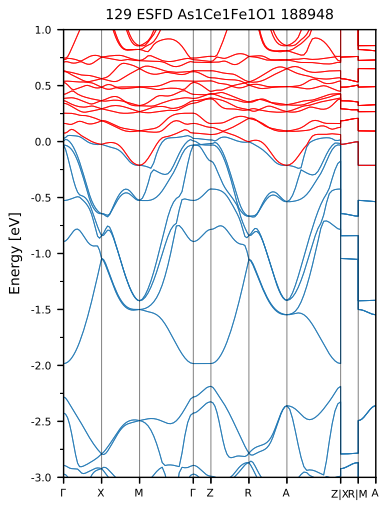

(vz) 129.4.318

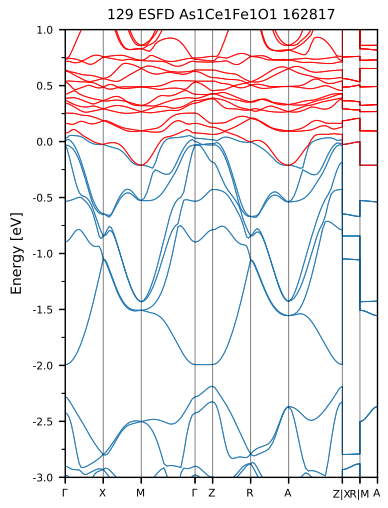

(wa) 129.4.319

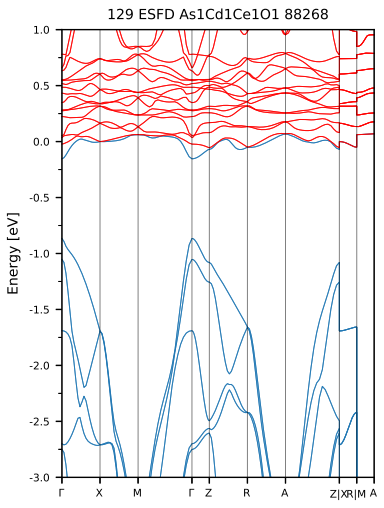

(wb) 129.4.320

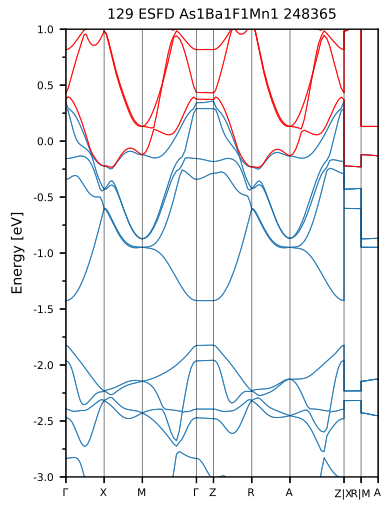

(wc) 129.4.321

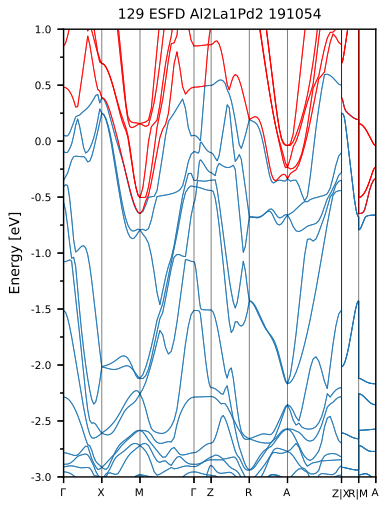

(wd) 129.4.322

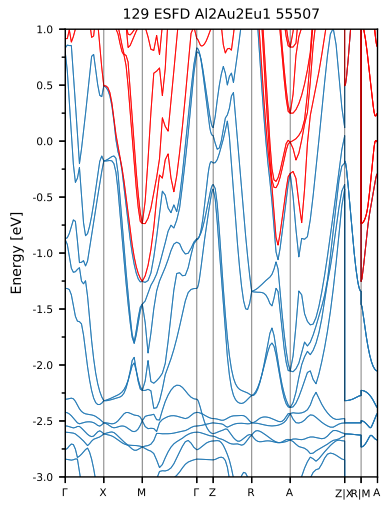

(we) 129.4.323

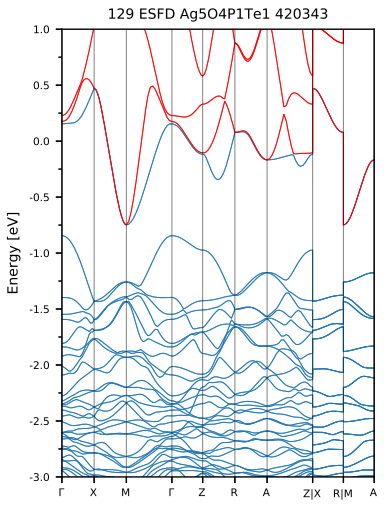

(wf) 129.4.324

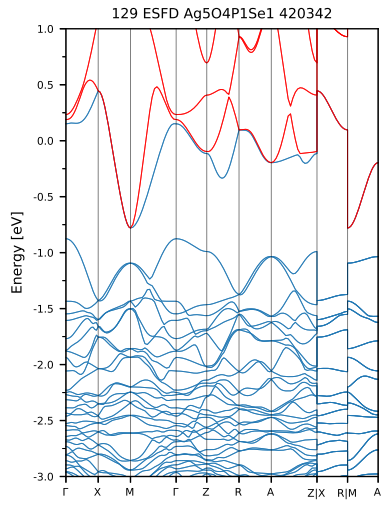

(wg) 129.4.325

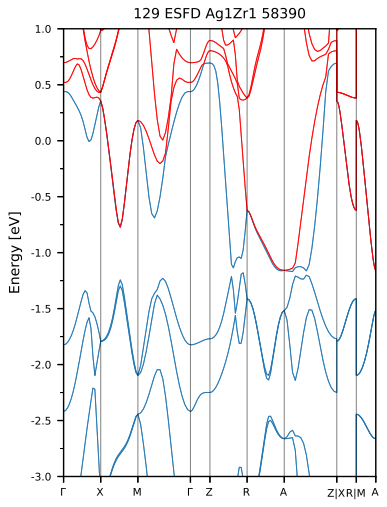

(wh) 129.4.326

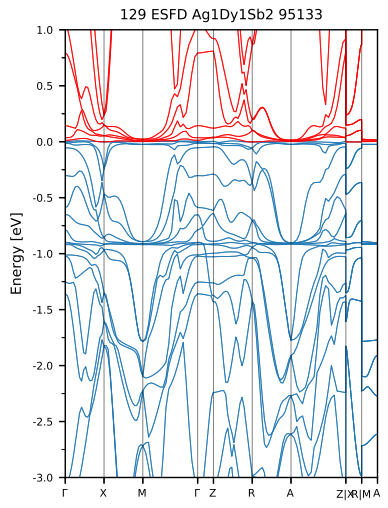

(wi) 129.4.327

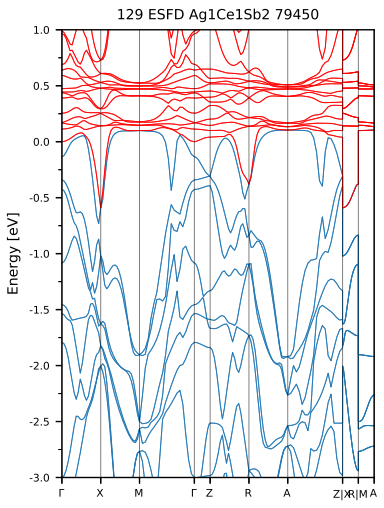

(wj) 129.4.328

Figure 276: Band structures for topological materials in SG 129, part 38/39, from 129.4.313 to 129.4.328. See the list of figures.

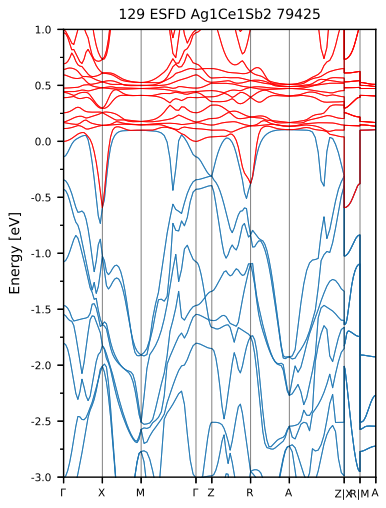

(wk) 129.4.329

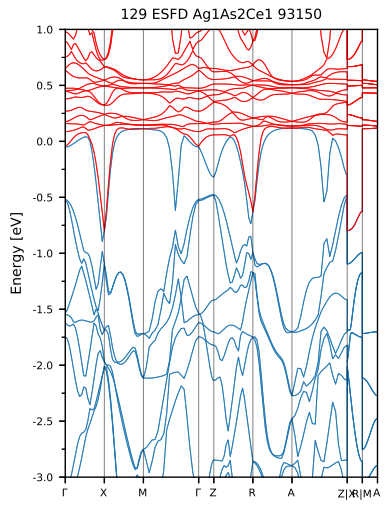

(wl) 129.4.330

Figure 277: Band structures for topological materials in SG 129, part 39/39, from 129.4.329 to 129.4.330. See the list of figures.

## 0.0.60 SG 130

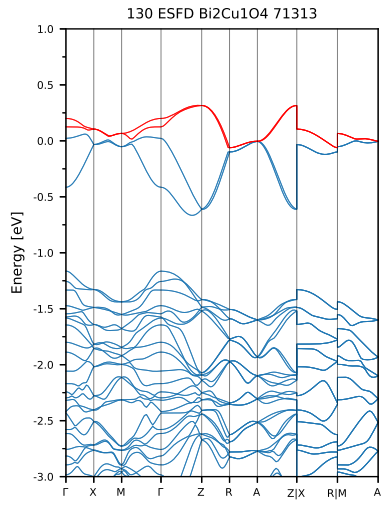

(a) 130.4.1

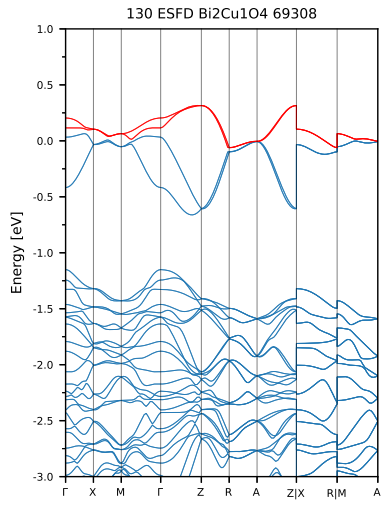

(b) 130.4.2

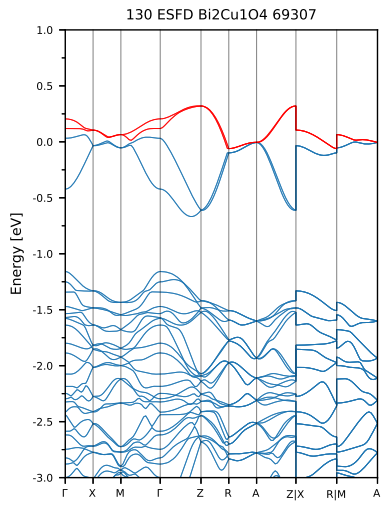

(c) 130.4.3

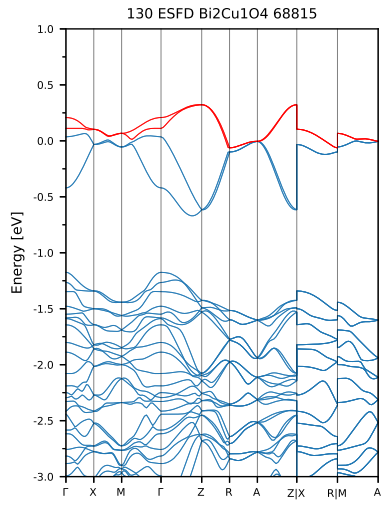

(d) 130.4.4

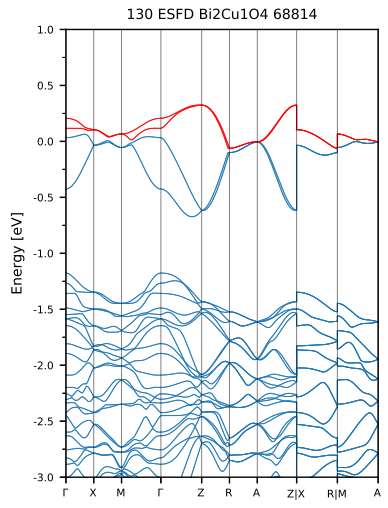

(e) 130.4.5

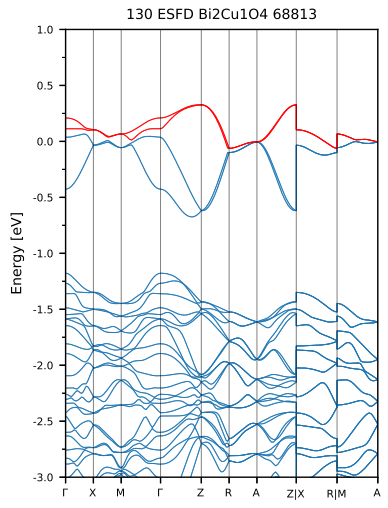

(f) 130.4.6

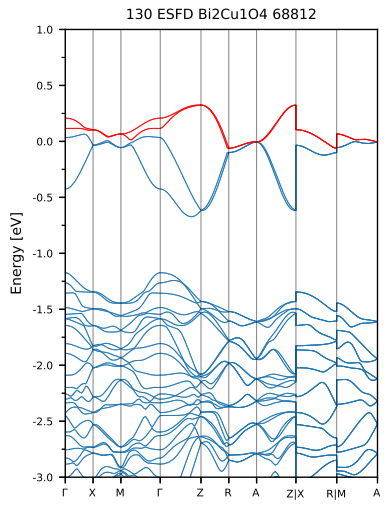

(g) 130.4.7

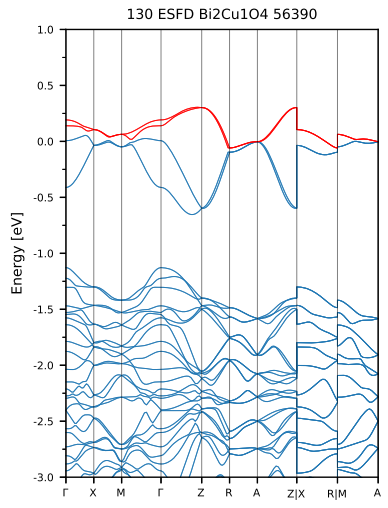

(h) 130.4.8

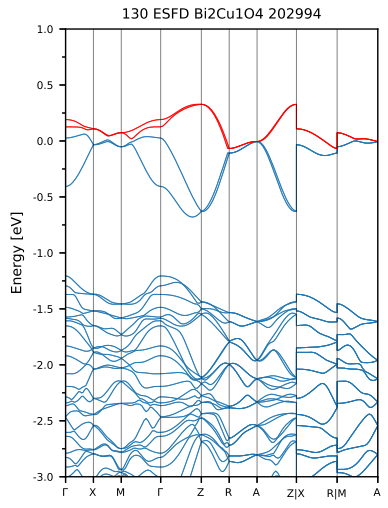

(i) 130.4.9

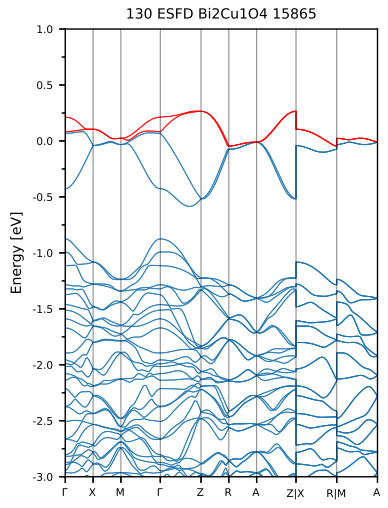

(j) 130.4.10

Figure 278: Band structures for topological materials in SG 130, part 1/1, from 130.4.1 to 130.4.10. See the list of figures.

### 0.0.61 SG 131

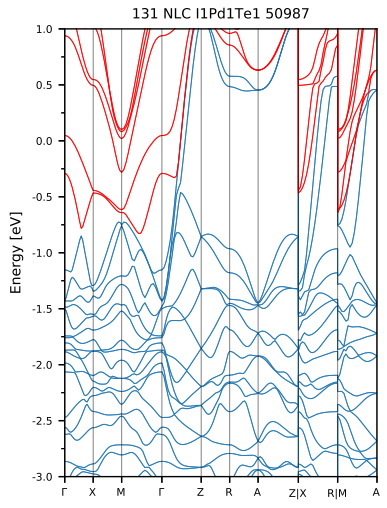

(a) 131.2.1

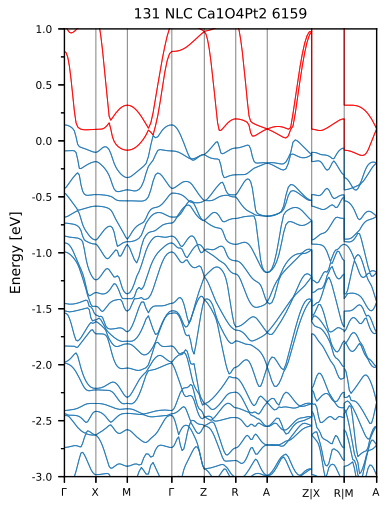

(b) 131.2.2

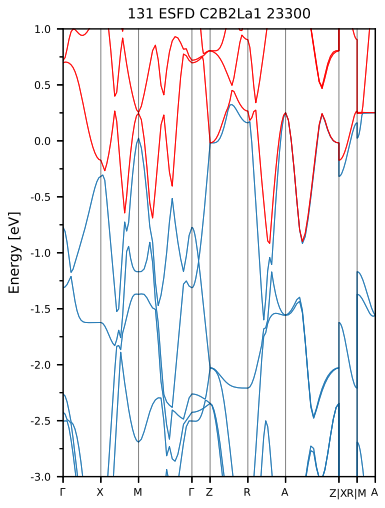

(c) 131.4.1

Figure 279: Band structures for topological materials in SG 131, part 1/1, from 131.2.1 to 131.4.1. See the list of figures.

### 0.0.62 SG 135

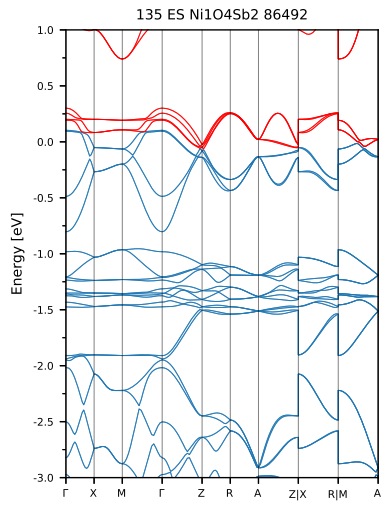

(a) 135.3.1

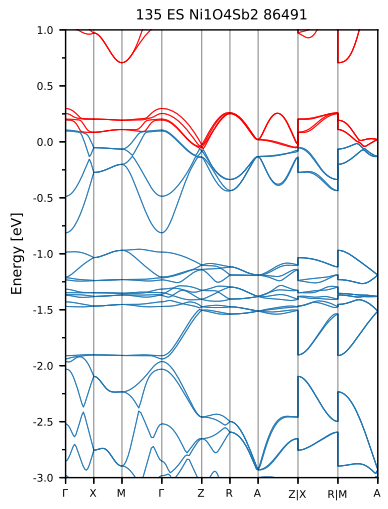

(b) 135.3.2

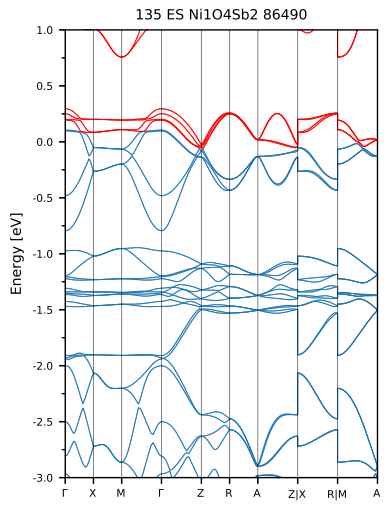

(c) 135.3.3

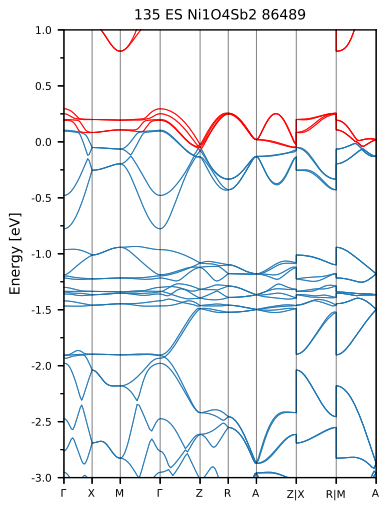

(d) 135.3.4

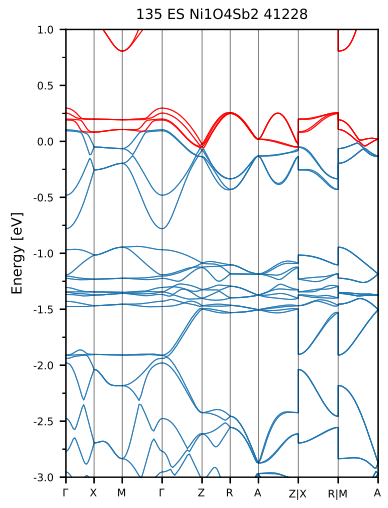

(e) 135.3.5

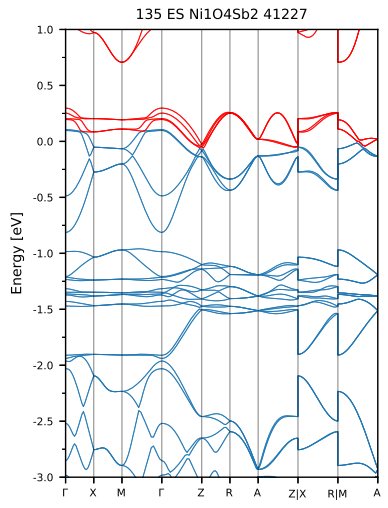

(f) 135.3.6

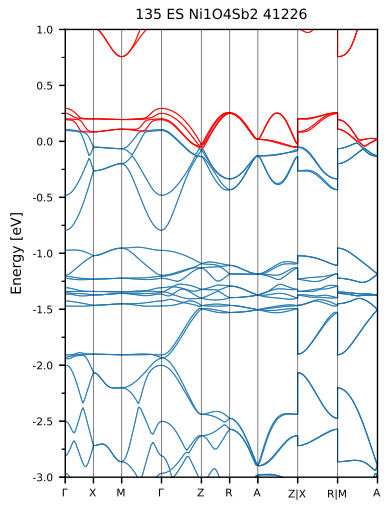

(g) 135.3.7

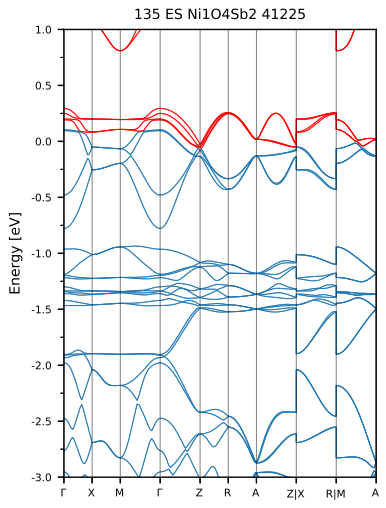

(h) 135.3.8

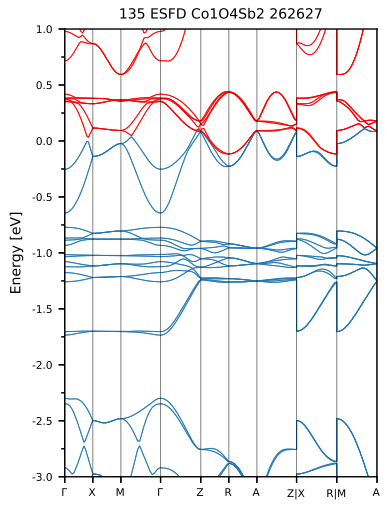

(i) 135.4.1

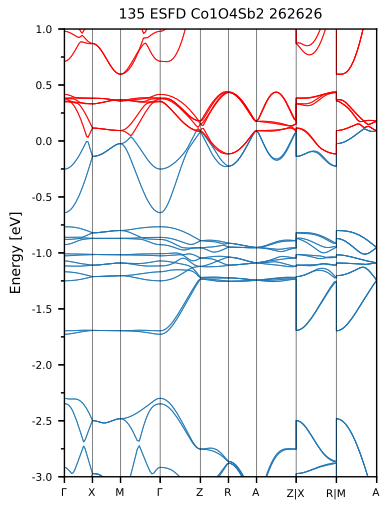

(j) 135.4.2

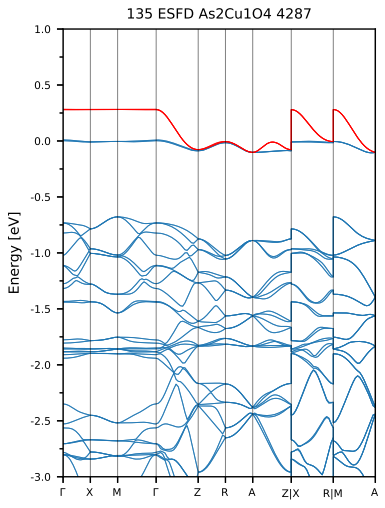

(k) 135.4.3

Figure 280: Band structures for topological materials in SG 135, part 1/1, from 135.3.1 to 135.4.3. See the list of figures.

## 0.0.63 SG 136

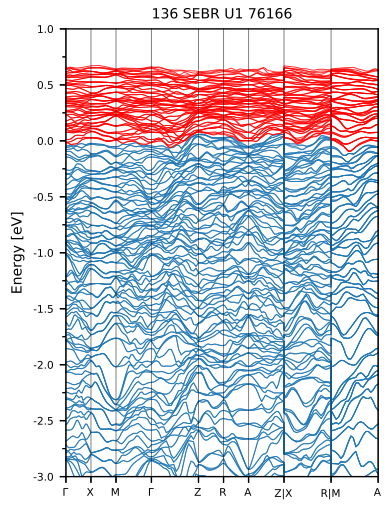

(a) 136.1.1

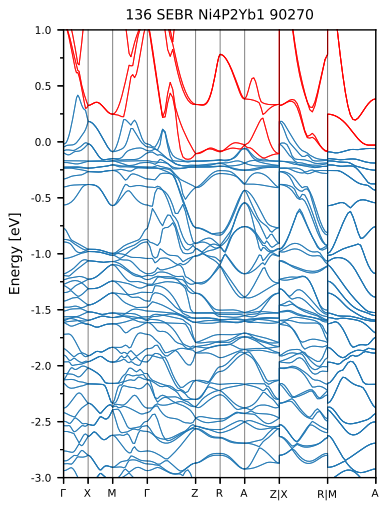

(b) 136.1.2

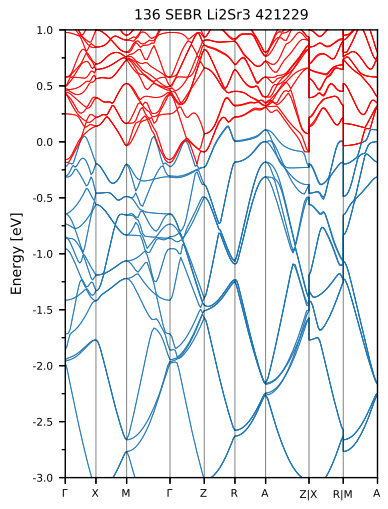

(c) 136.1.3

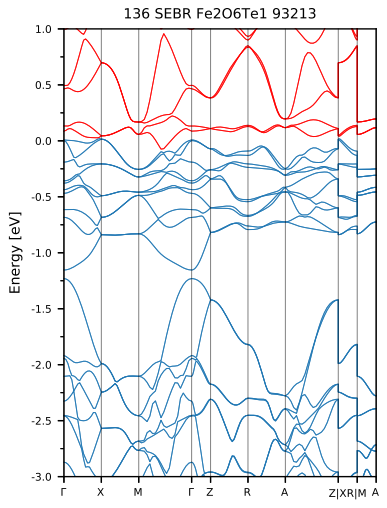

(d) 136.1.4

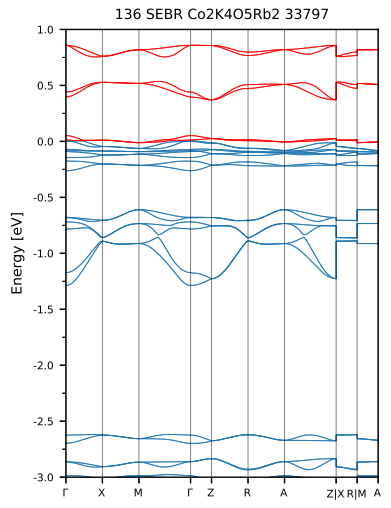

(e) 136.1.5

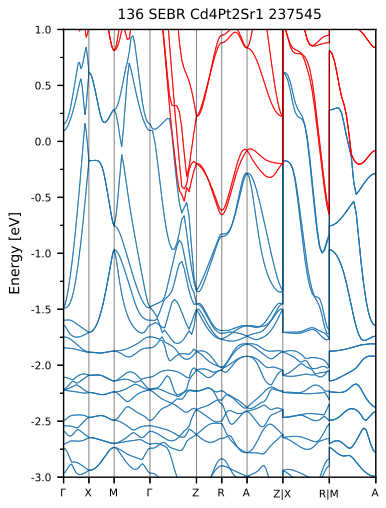

(f) 136.1.6

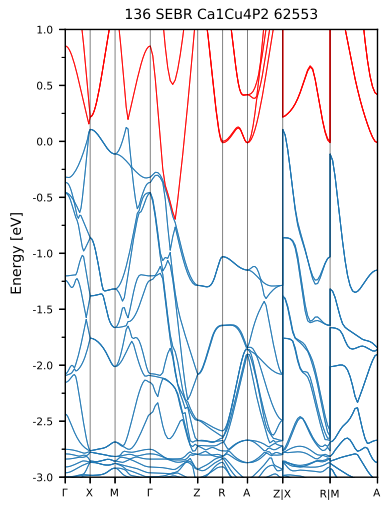

(g) 136.1.7

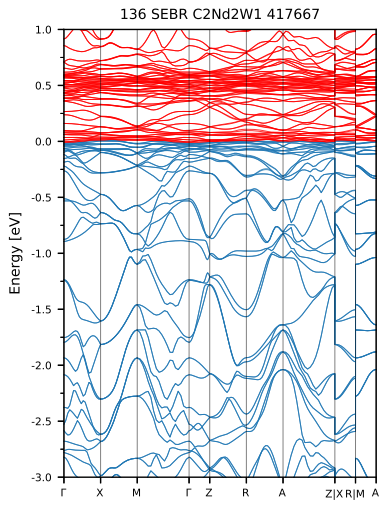

(h) 136.1.8

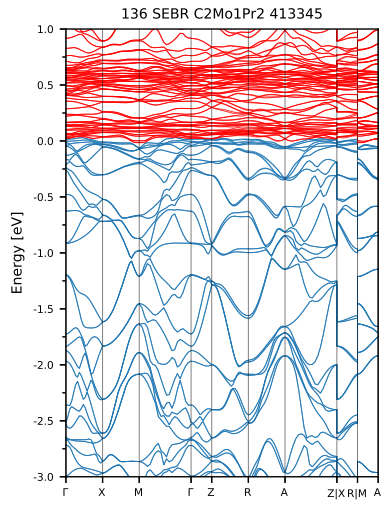

(i) 136.1.9

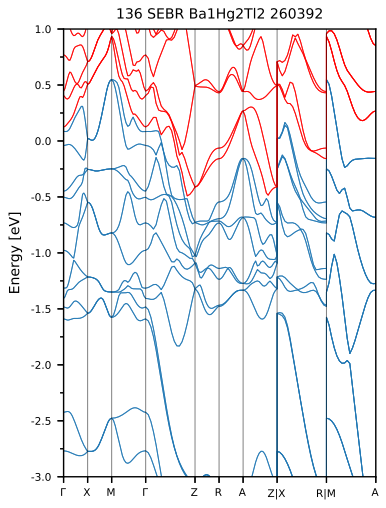

(j) 136.1.10

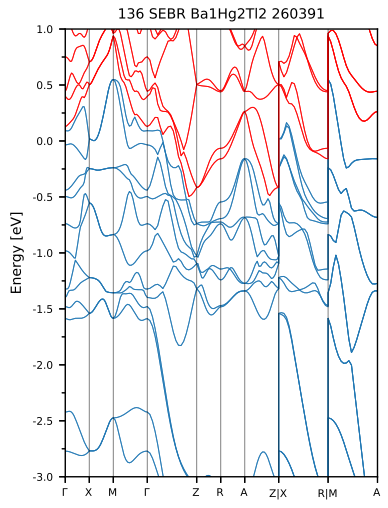

(k) 136.1.11

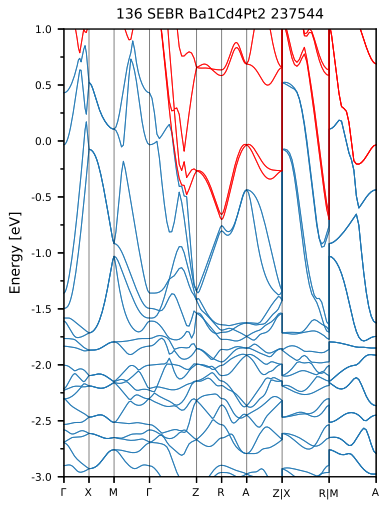

(l) 136.1.12

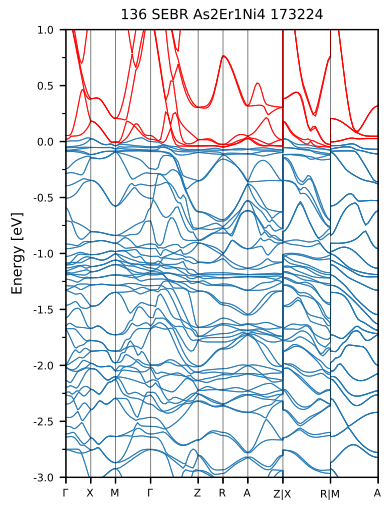

(m) 136.1.13

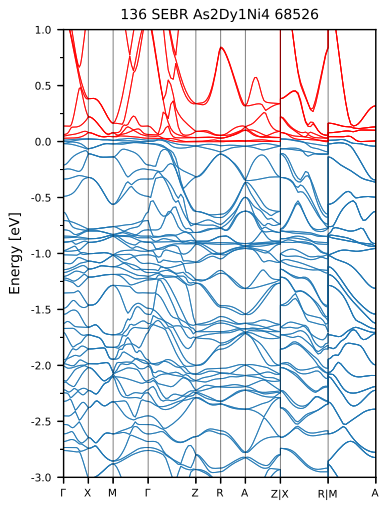

(n) 136.1.14

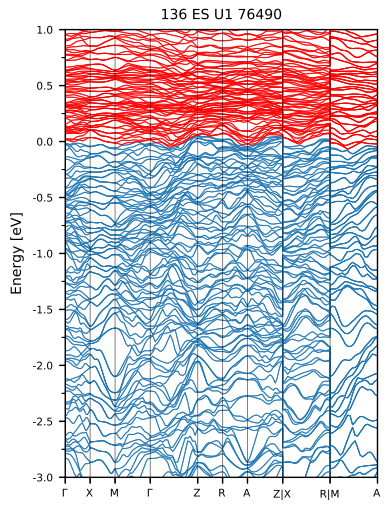

(o) 136.3.1

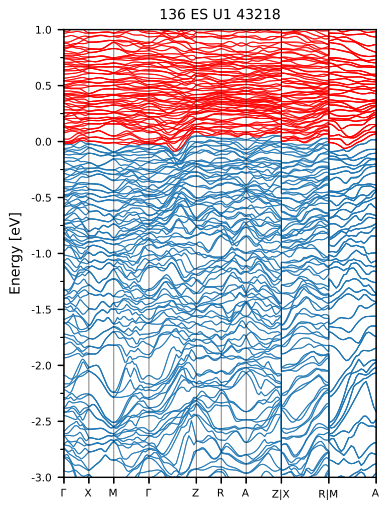

(p) 136.3.2

Figure 281: Band structures for topological materials in SG 136, part 1/6, from 136.1.1 to 136.3.2. See the list of figures.

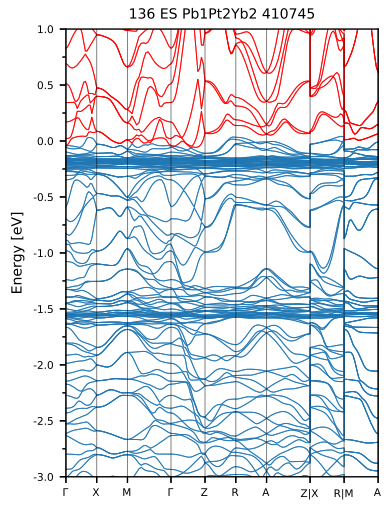

(q) 136.3.3

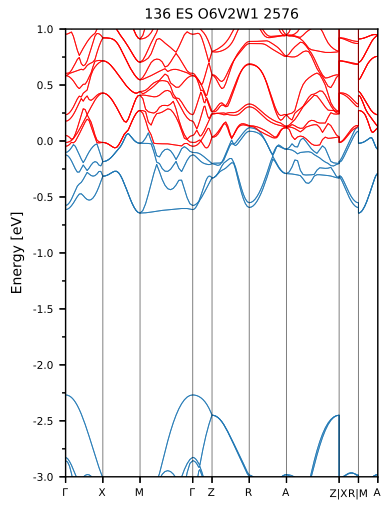

(r) 136.3.4

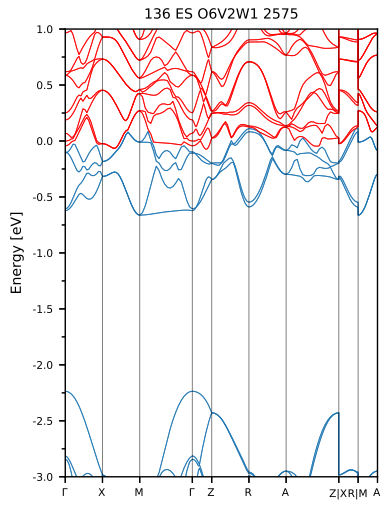

(s) 136.3.5

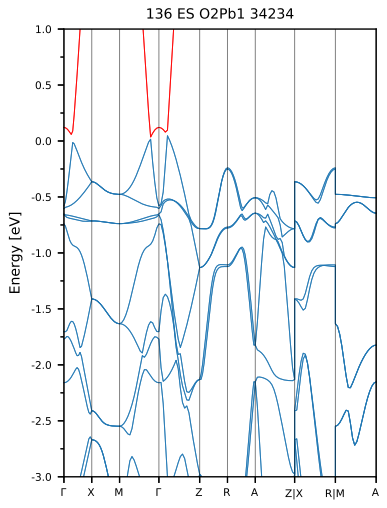

(t) 136.3.6

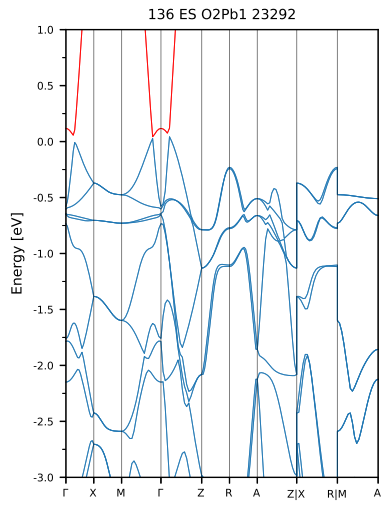

(u) 136.3.7

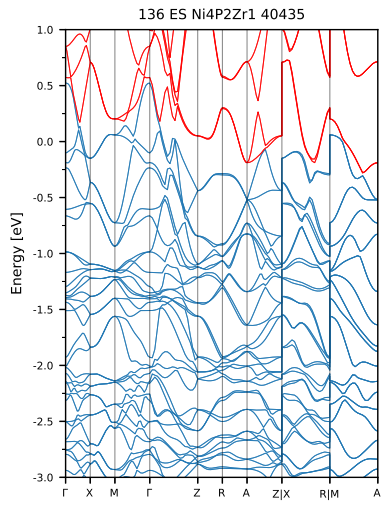

(v) 136.3.8

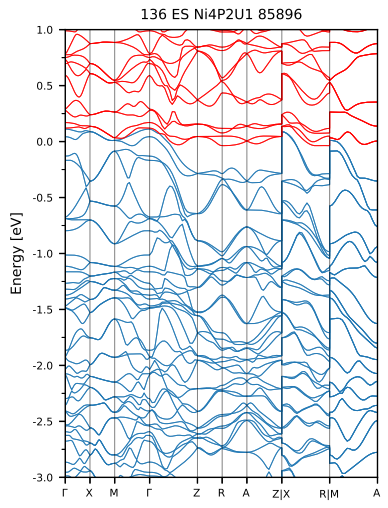

(w) 136.3.9

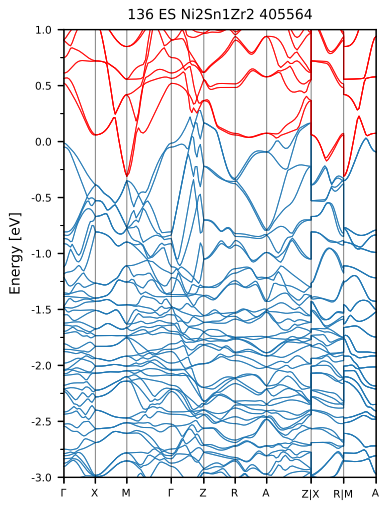

(x) 136.3.10

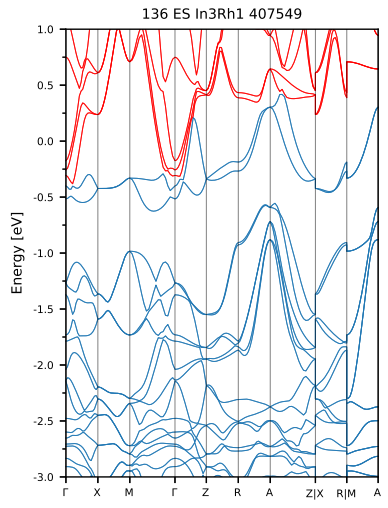

(y) 136.3.11

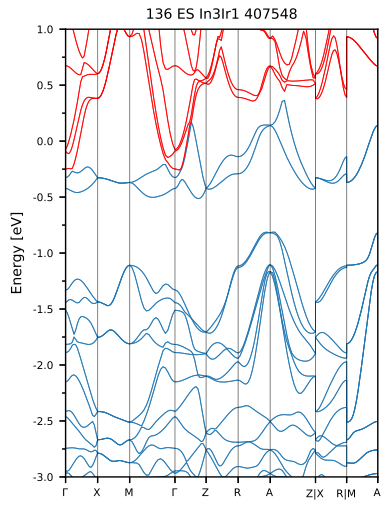

(z) 136.3.12

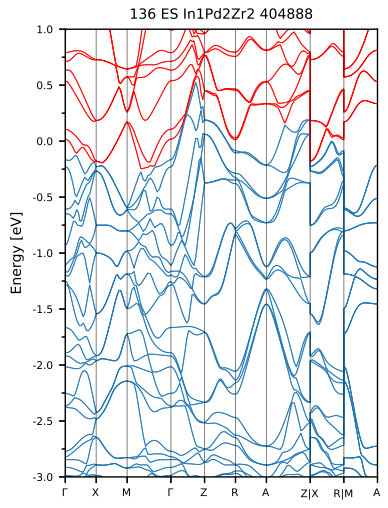

(aa) 136.3.13

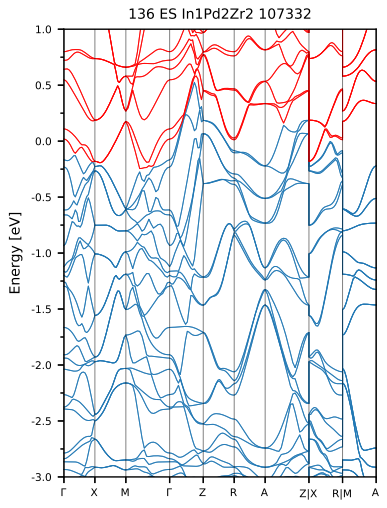

(ab) 136.3.14

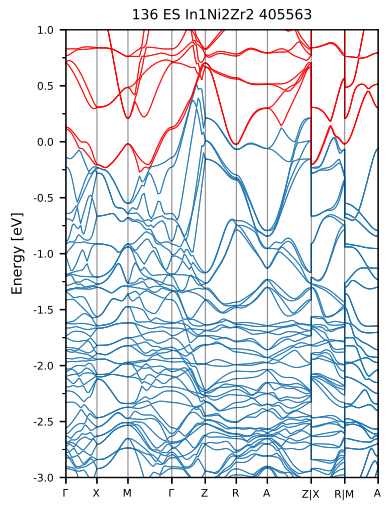

(ac) 136.3.15

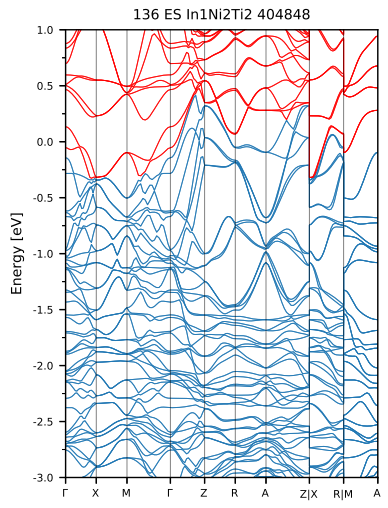

(ad) 136.3.16

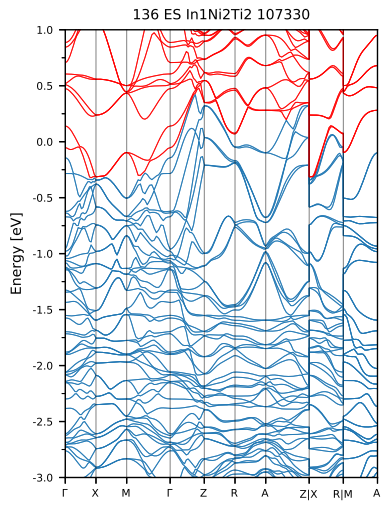

(ae) 136.3.17

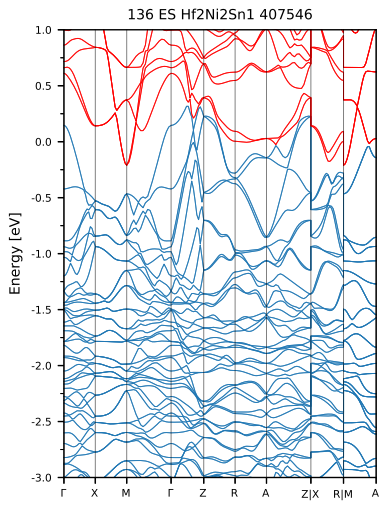

(af) 136.3.18

Figure 282: Band structures for topological materials in SG 136, part 2/6, from 136.3.3 to 136.3.18. See the list of figures.

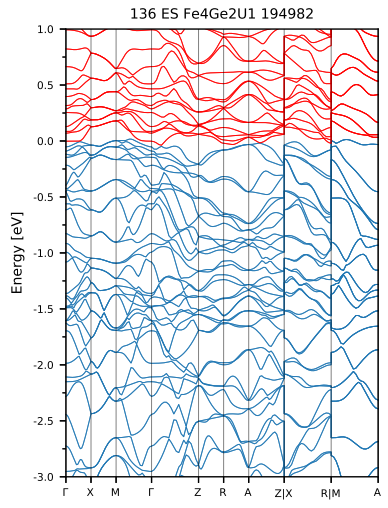

(ag) 136.3.19

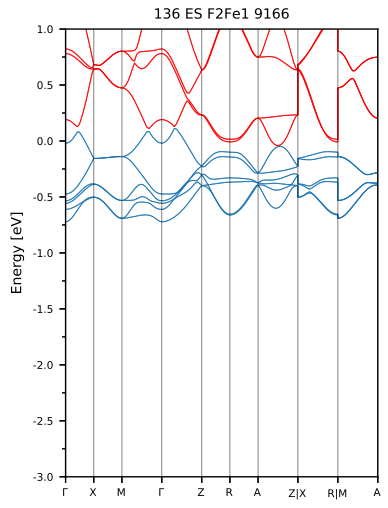

(ah) 136.3.20

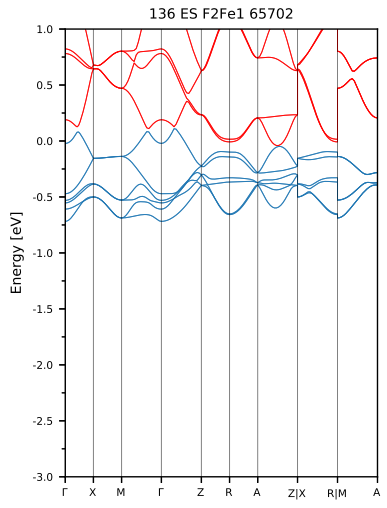

(ai) 136.3.21

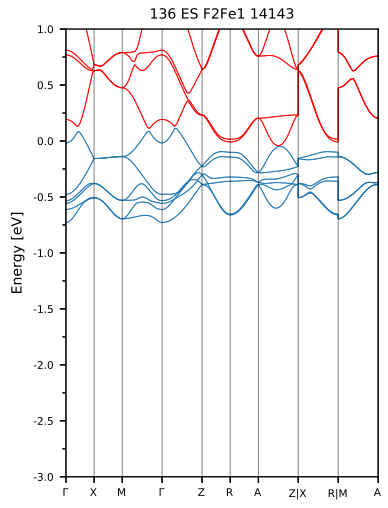

(aj) 136.3.22

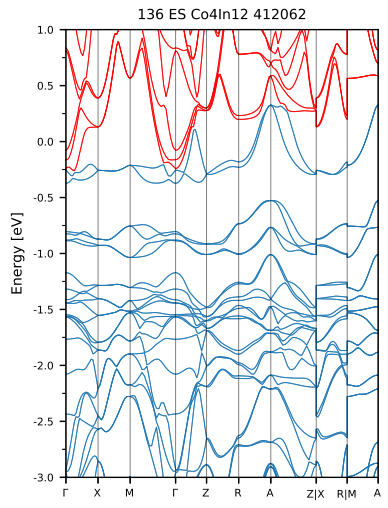

(ak) 136.3.23

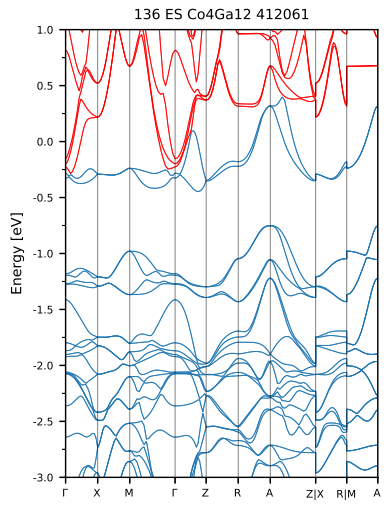

(al) 136.3.24

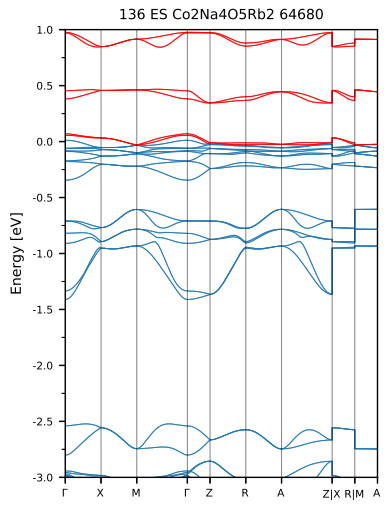

(am) 136.3.25

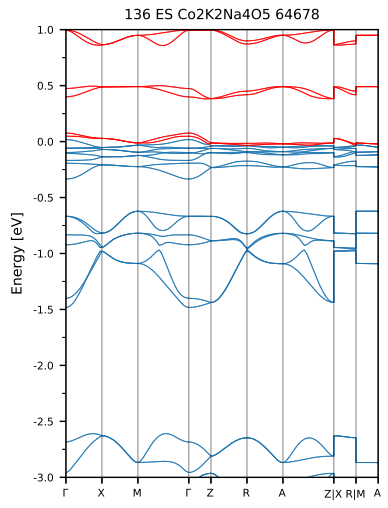

(an) 136.3.26

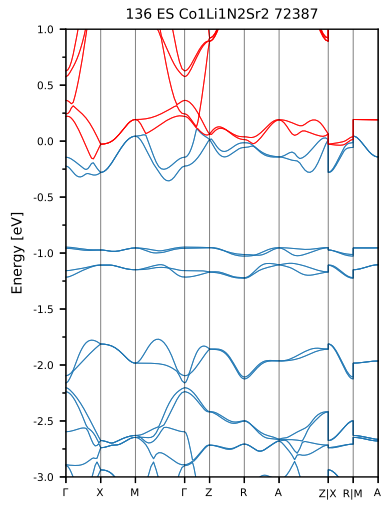

(ao) 136.3.27

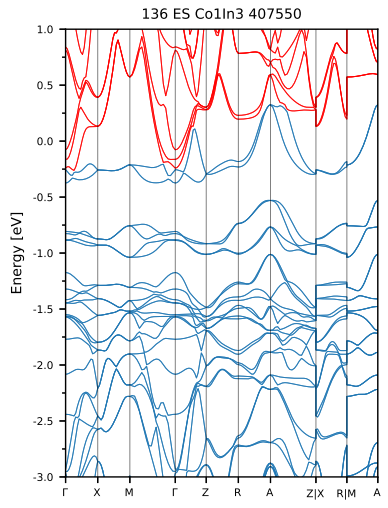

(ap) 136.3.28

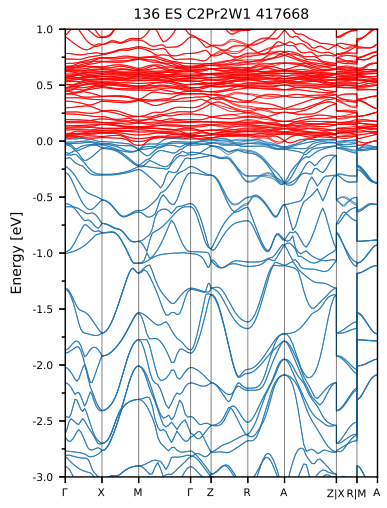

(aq) 136.3.29

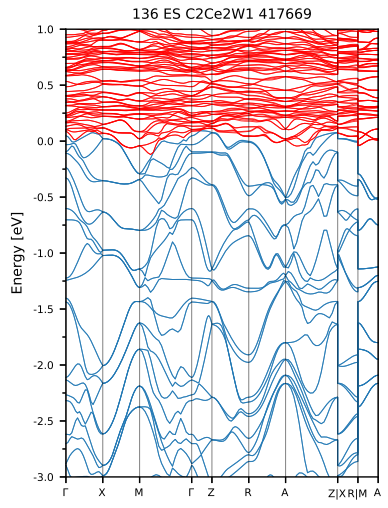

(ar) 136.3.30

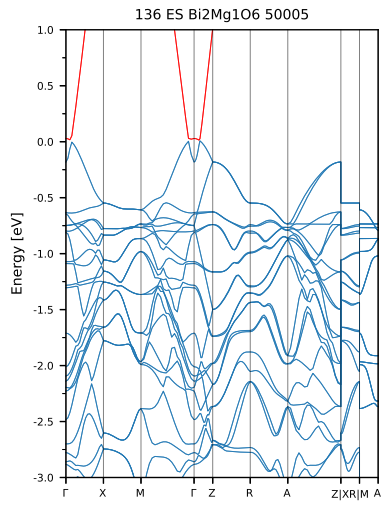

(as) 136.3.31

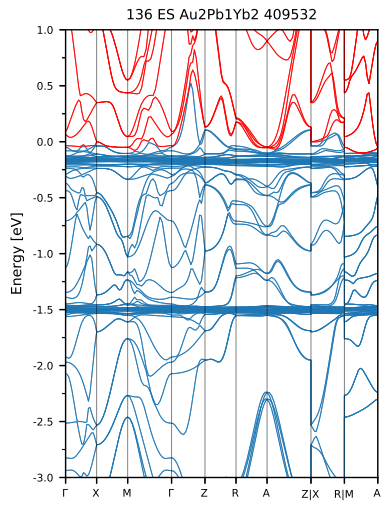

(at) 136.3.32

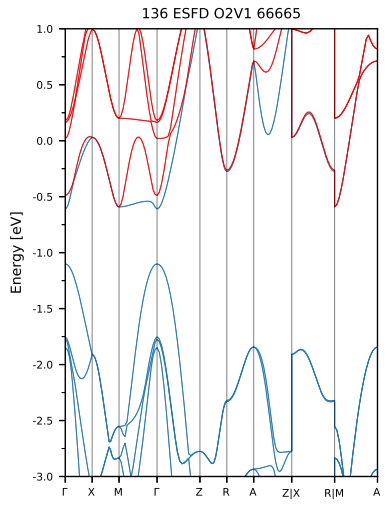

(au) 136.4.1

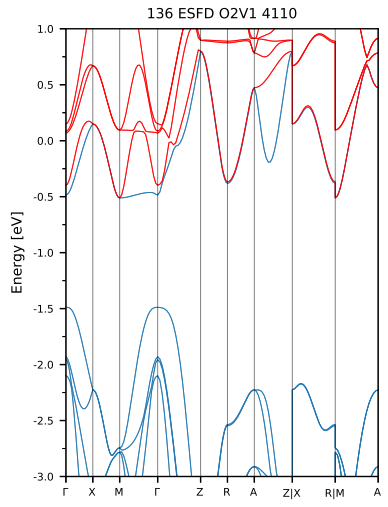

(av) 136.4.2

Figure 283: Band structures for topological materials in SG 136, part 3/6, from 136.3.19 to 136.4.2. See the list of figures.

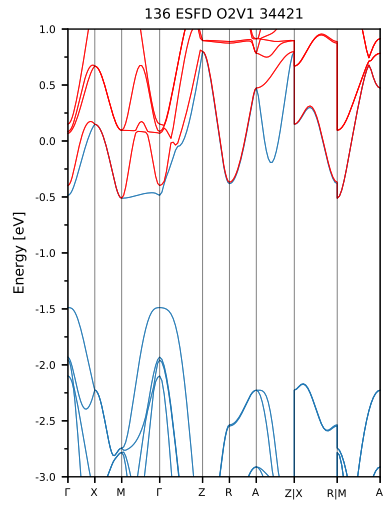

(aw) 136.4.3

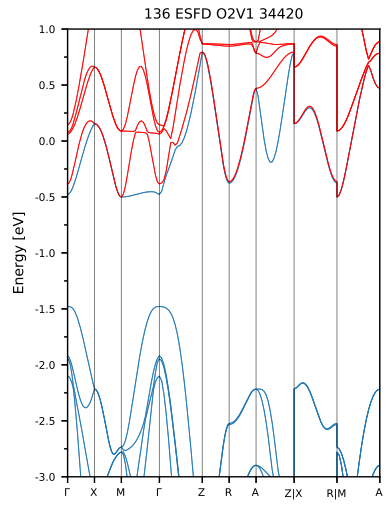

(ax) 136.4.4

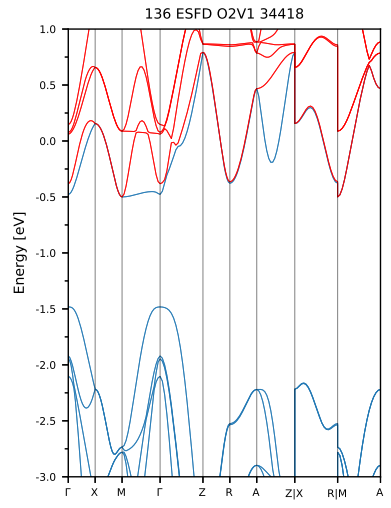

(ay) 136.4.5

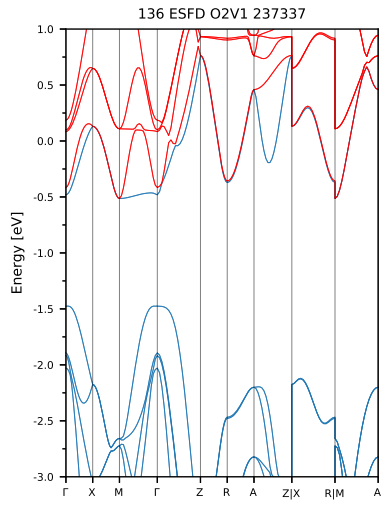

(az) 136.4.6

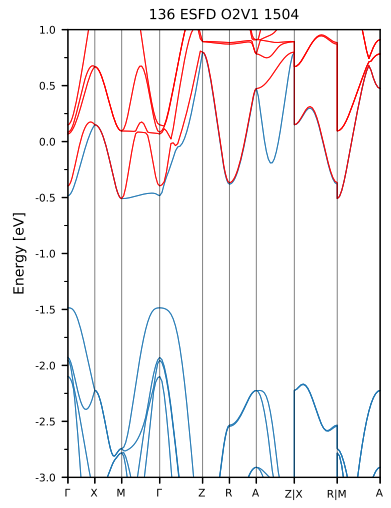

(ba) 136.4.7

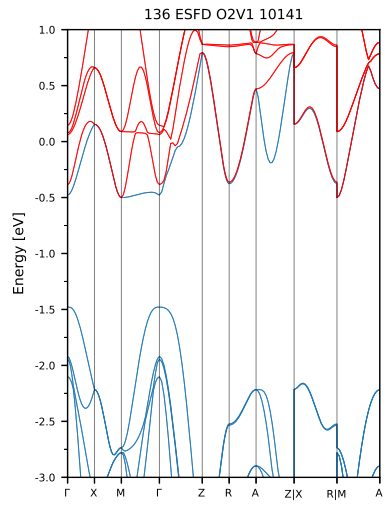

(bb) 136.4.8

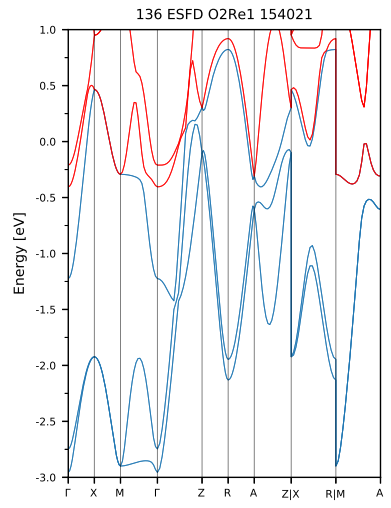

(bc) 136.4.9

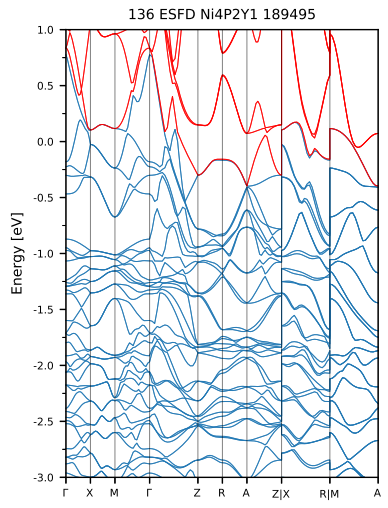

(bd) 136.4.10

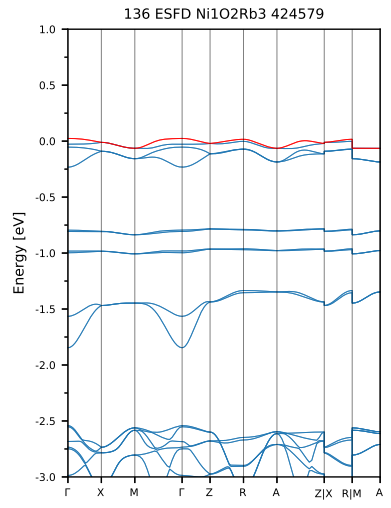

(be) 136.4.11

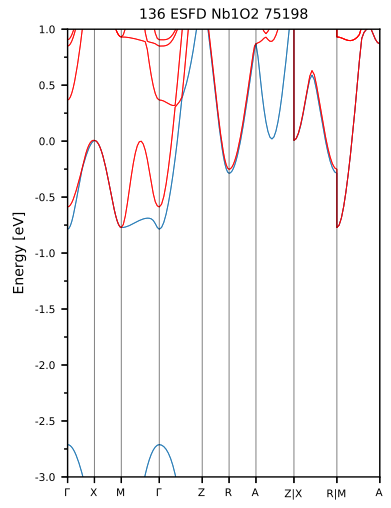

(bf) 136.4.12

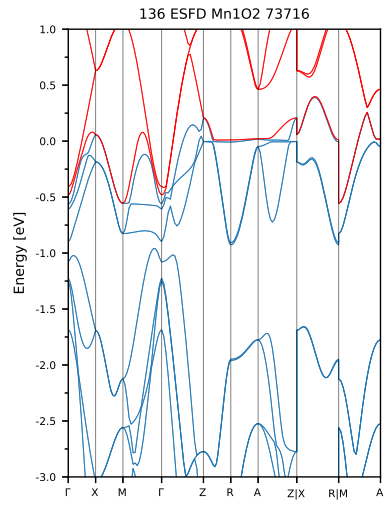

(bg) 136.4.13

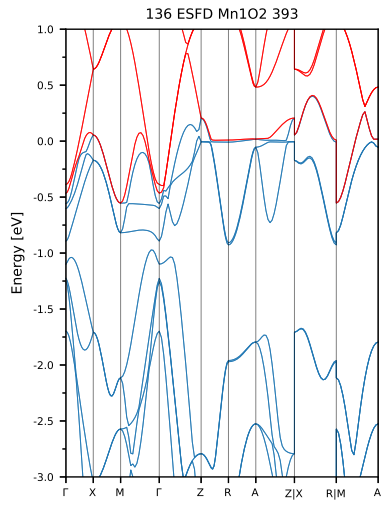

(bh) 136.4.14

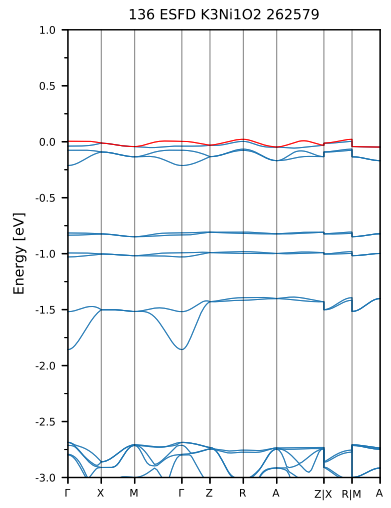

(bi) 136.4.15

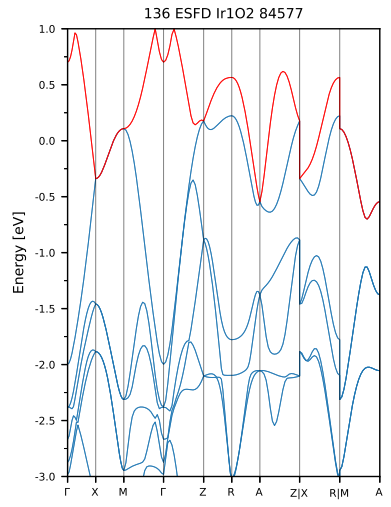

(bj) 136.4.16

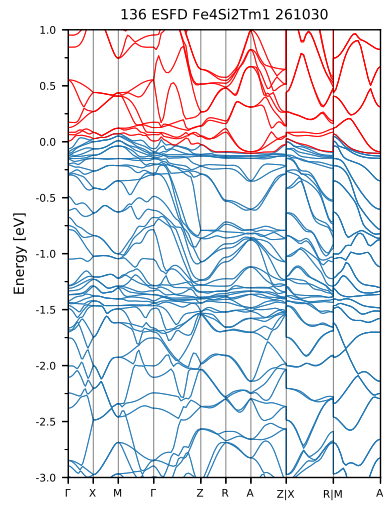

(bk) 136.4.17

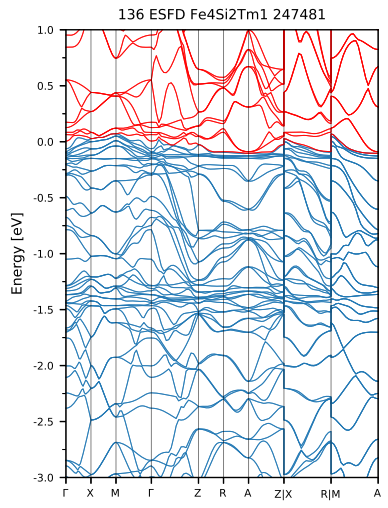

(bl) 136.4.18

Figure 284: Band structures for topological materials in SG 136, part 4/6, from 136.4.3 to 136.4.18. See the list of figures.

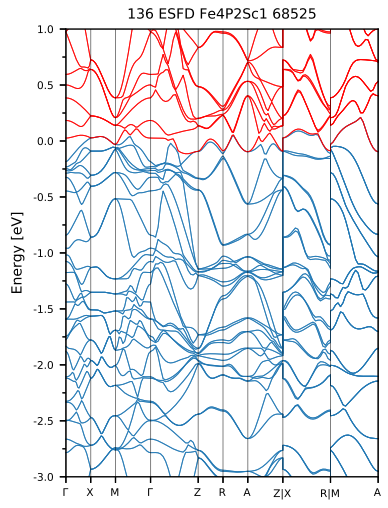

(bm) 136.4.19

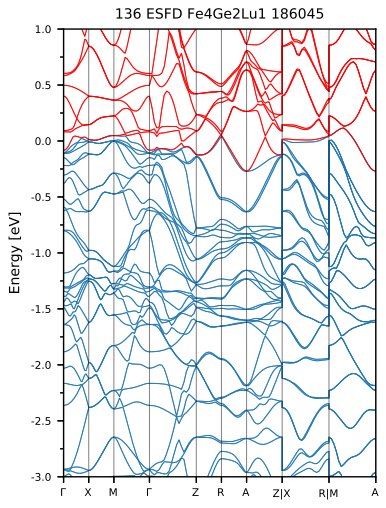

(bn) 136.4.20

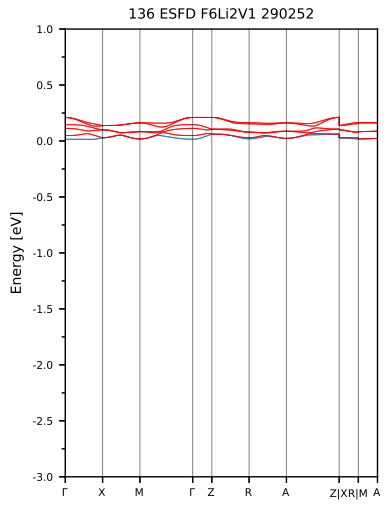

(bo) 136.4.21

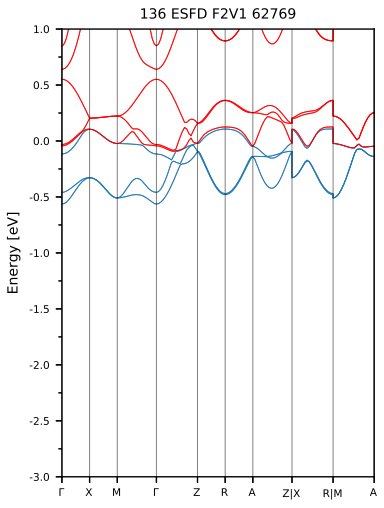

(bp) 136.4.22

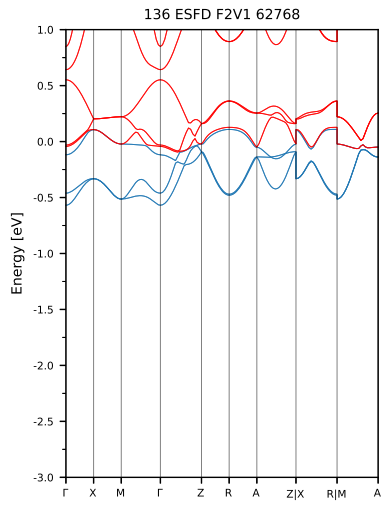

(bq) 136.4.23

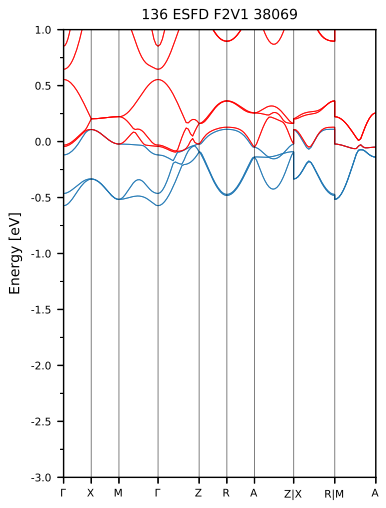

(br) 136.4.24

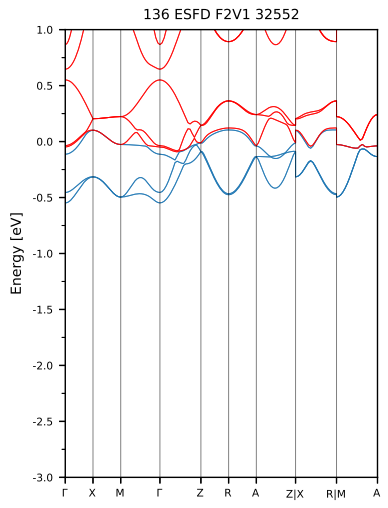

(bs) 136.4.25

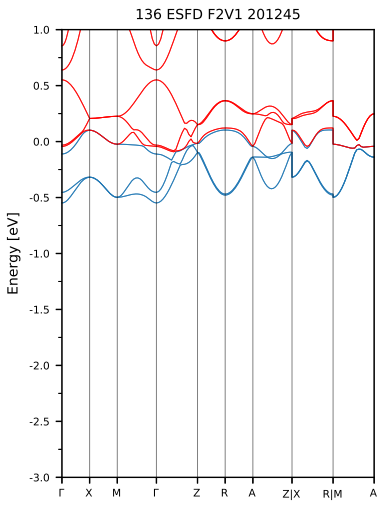

(bt) 136.4.26

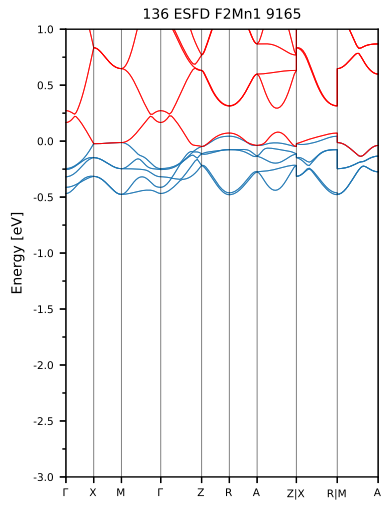

(bu) 136.4.27

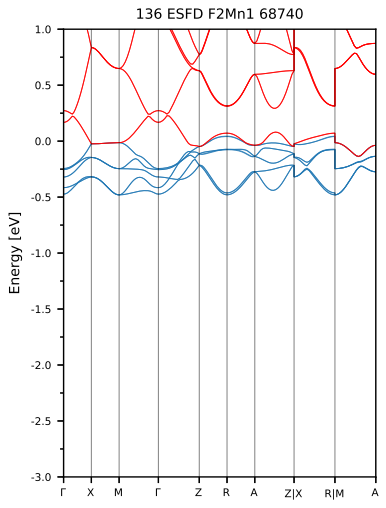

(bv) 136.4.28

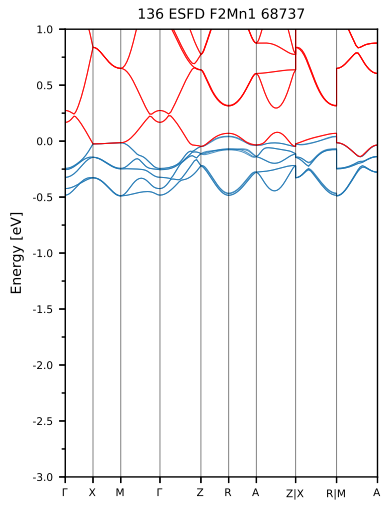

(bw) 136.4.29

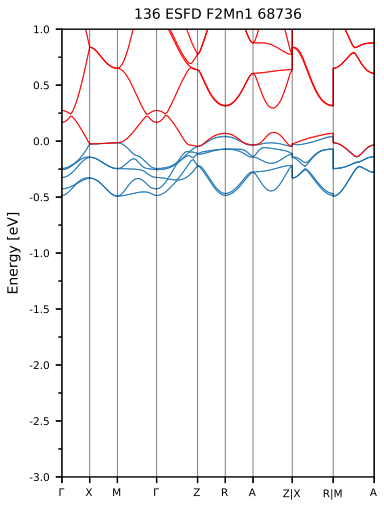

(bx) 136.4.30

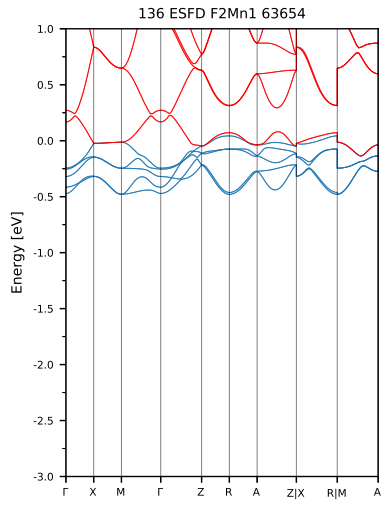

(by) 136.4.31

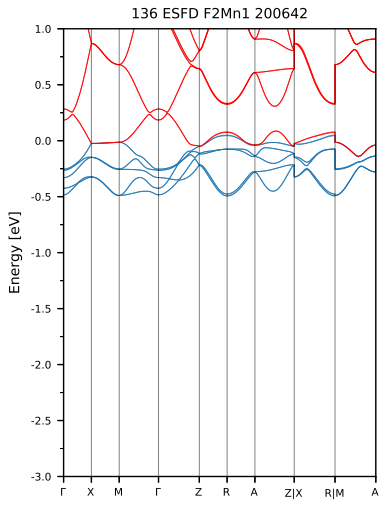

(bz) 136.4.32

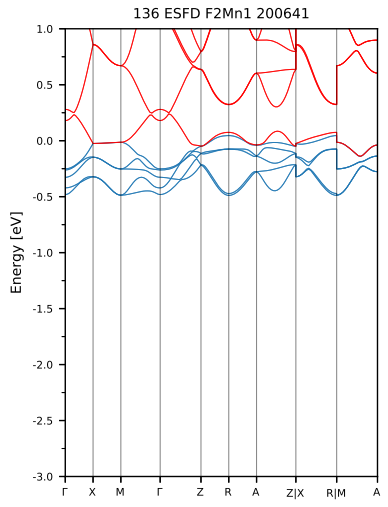

(ca) 136.4.33

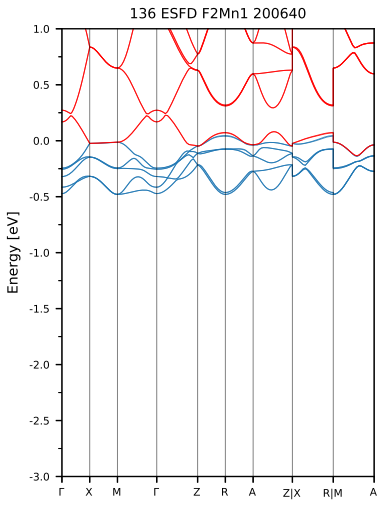

(cb) 136.4.34

Figure 285: Band structures for topological materials in SG 136, part 5/6, from 136.4.19 to 136.4.34. See the list of figures.

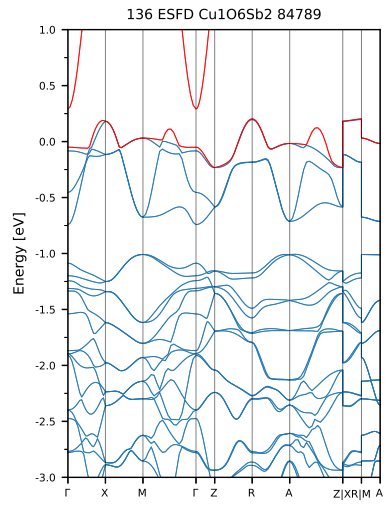

(cc) 136.4.35

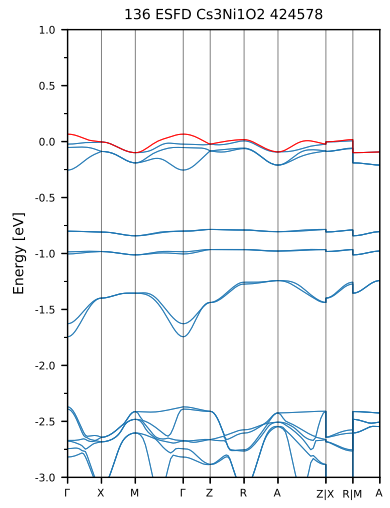

(cd) 136.4.36

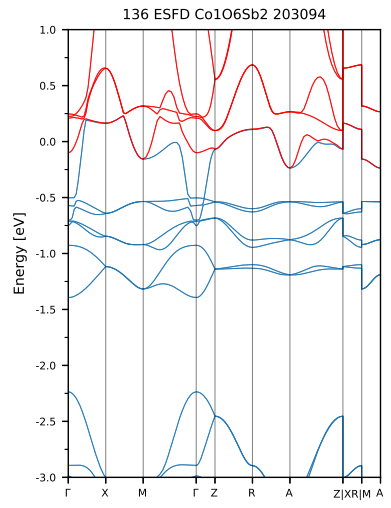

(ce) 136.4.37

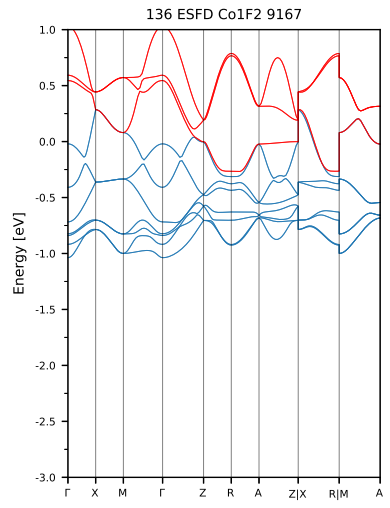

(cf) 136.4.38

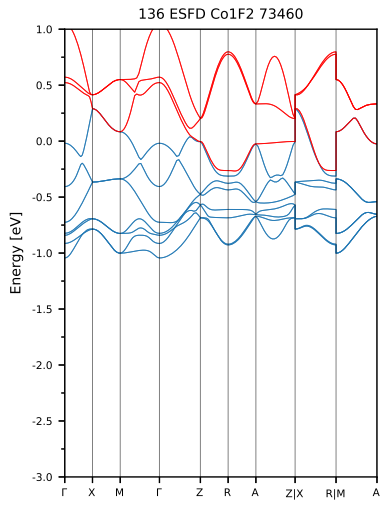

(cg) 136.4.39

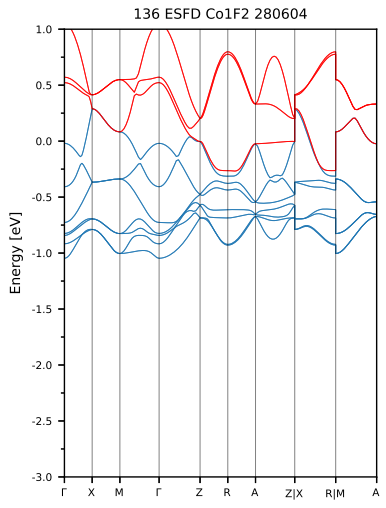

(ch) 136.4.40

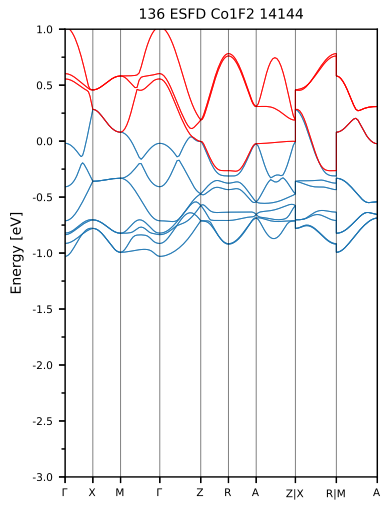

(ci) 136.4.41

Figure 286: Band structures for topological materials in SG 136, part 6/6, from 136.4.35 to 136.4.41. See the list of figures.

### 0.0.64 SG 137

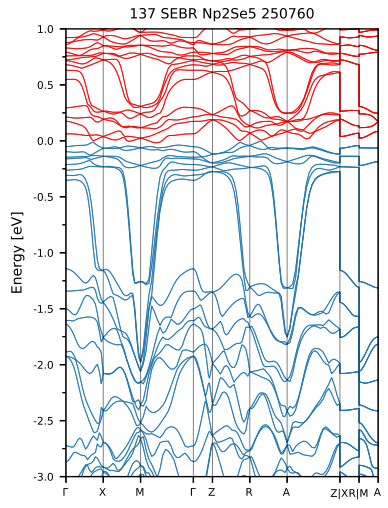

(a) 137.1.1

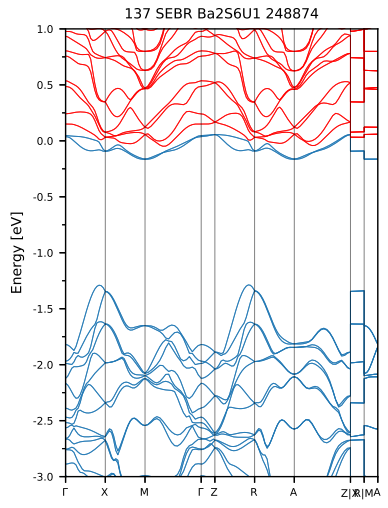

(b) 137.1.2

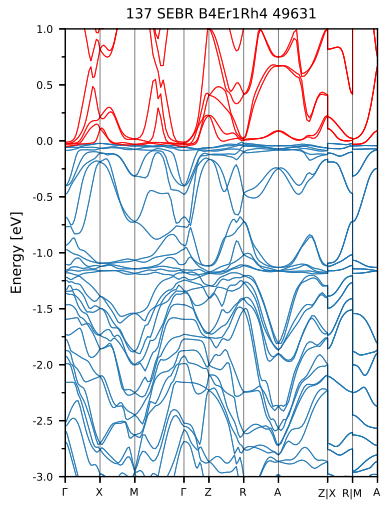

(c) 137.1.3

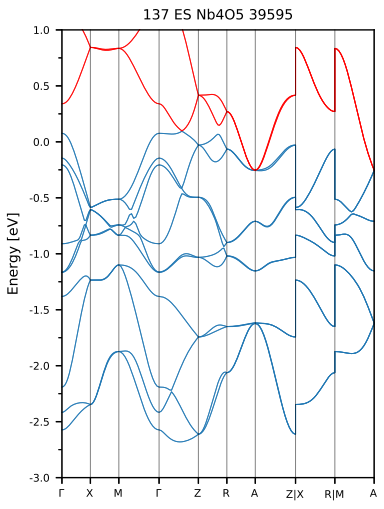

(d) 137.3.1

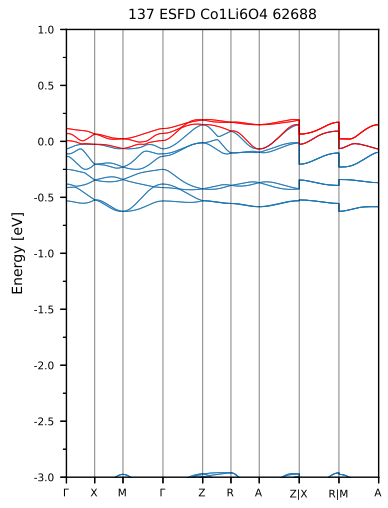

(e) 137.4.1

Figure 287: Band structures for topological materials in SG 137, part 1/1, from 137.1.1 to 137.4.1. See the list of figures.

## 0.0.65 SG 139

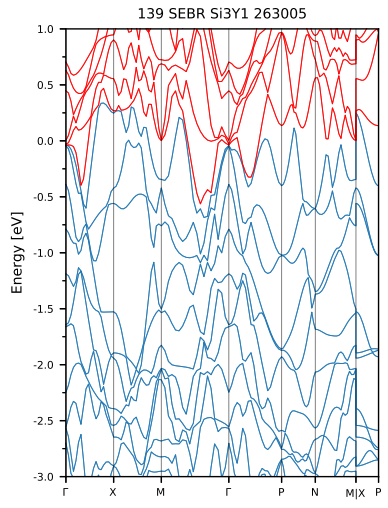

(a) 139.1.1

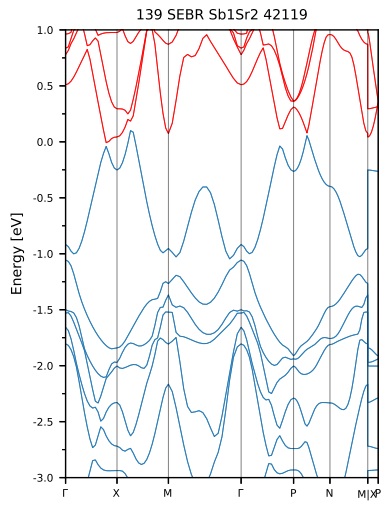

(b) 139.1.2

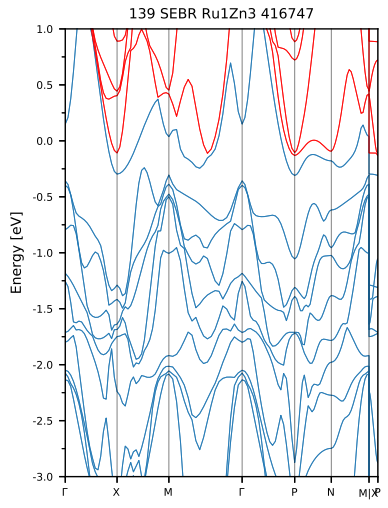

(c) 139.1.3

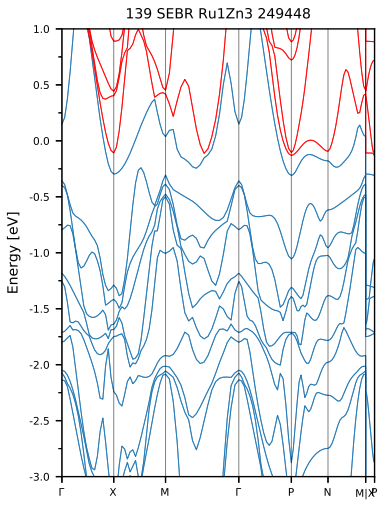

(d) 139.1.4

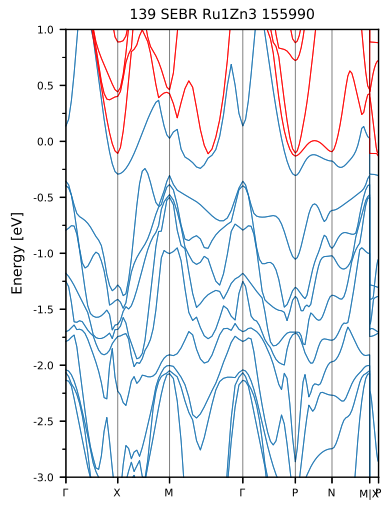

(e) 139.1.5

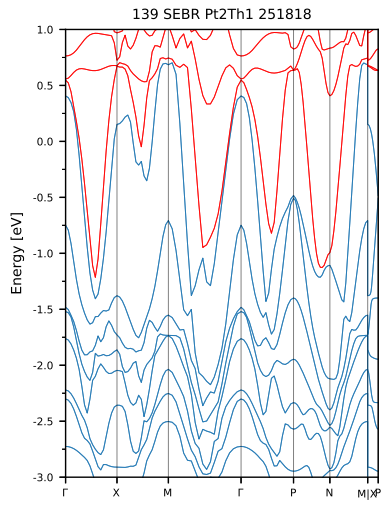

(f) 139.1.6

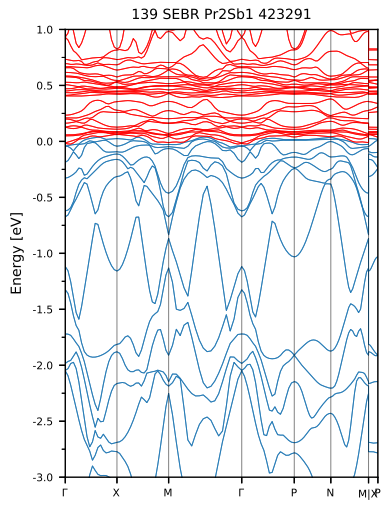

(g) 139.1.7

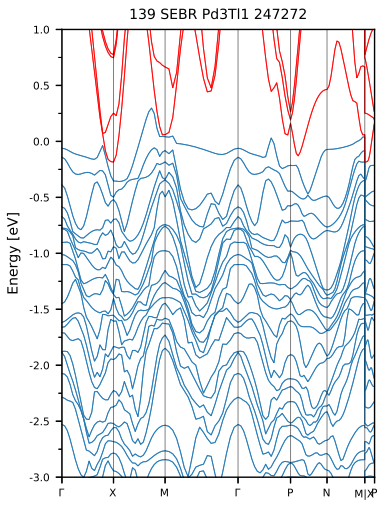

(h) 139.1.8

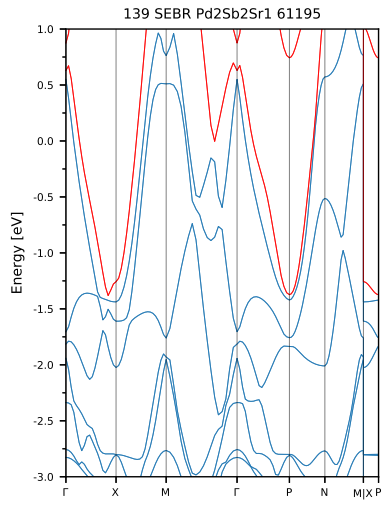

(i) 139.1.9

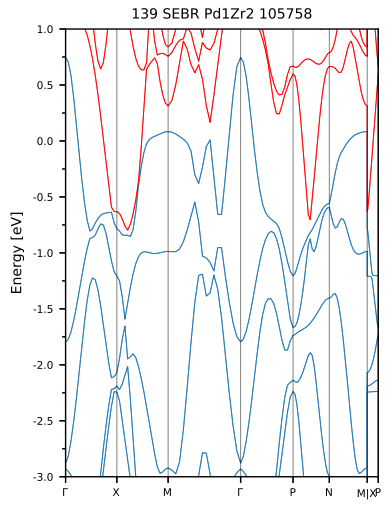

(j) 139.1.10

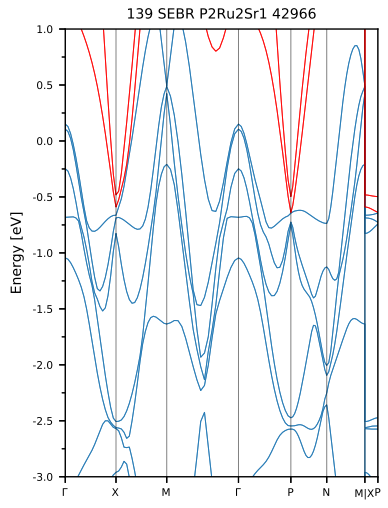

(k) 139.1.11

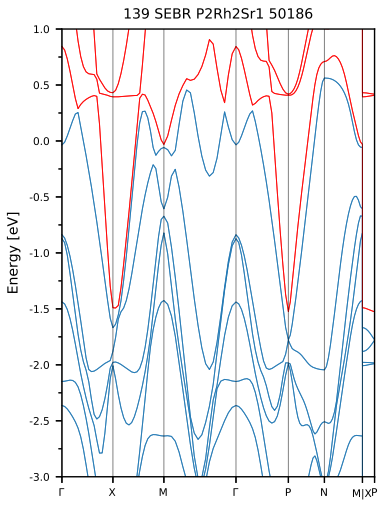

(l) 139.1.12

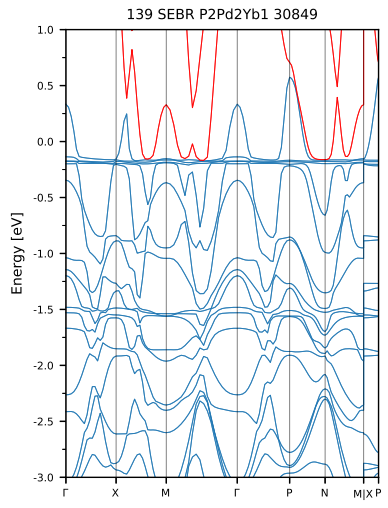

(m) 139.1.13

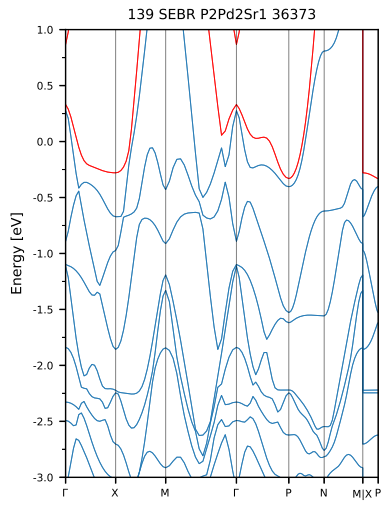

(n) 139.1.14

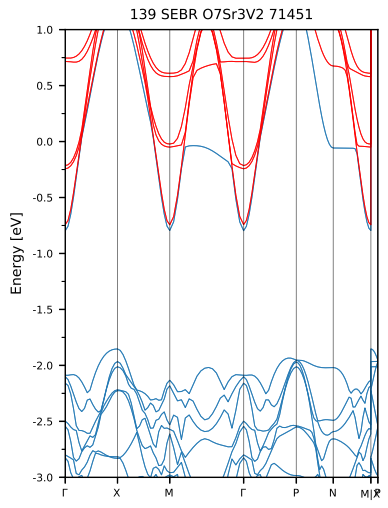

(o) 139.1.15

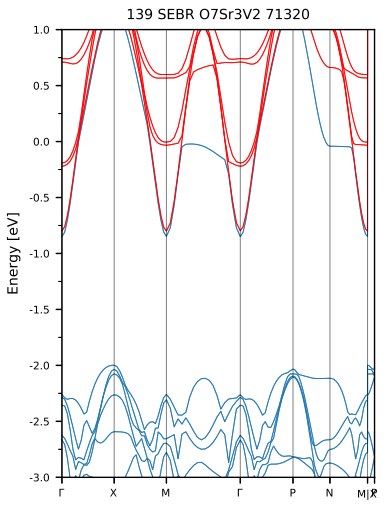

(p) 139.1.16

Figure 288: Band structures for topological materials in SG 139, part 1/22, from 139.1.1 to 139.1.16. See the list of figures.

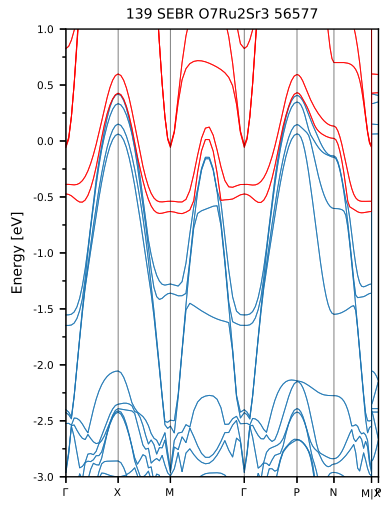

(q) 139.1.17

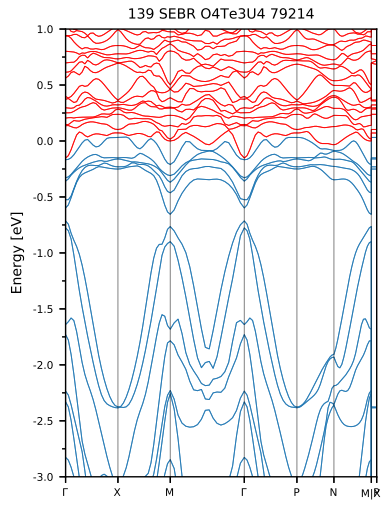

(r) 139.1.18

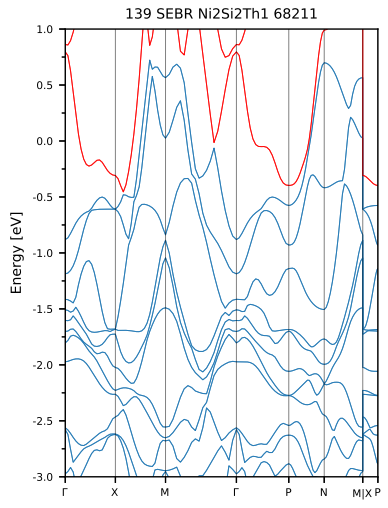

(s) 139.1.19

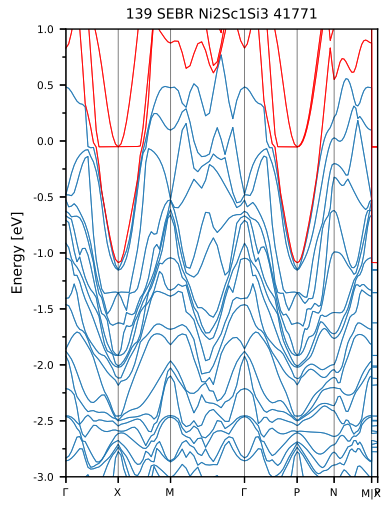

(t) 139.1.20

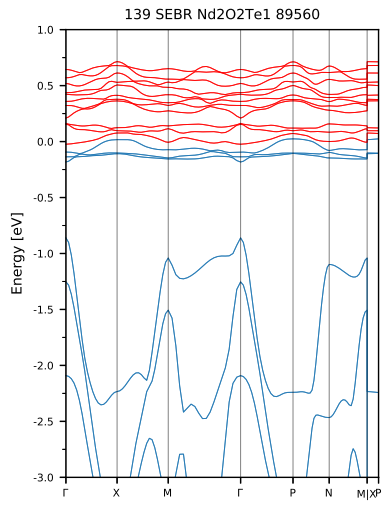

(u) 139.1.21

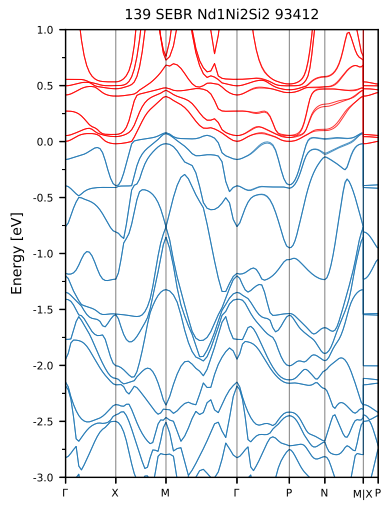

(v) 139.1.22

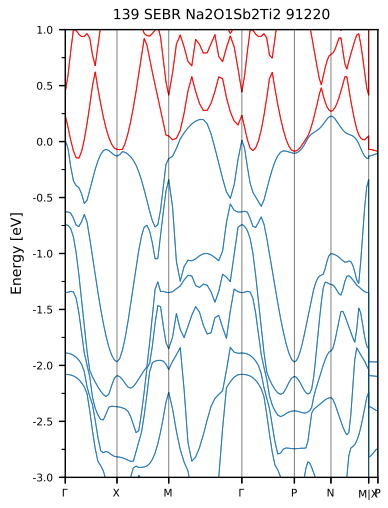

(w) 139.1.23

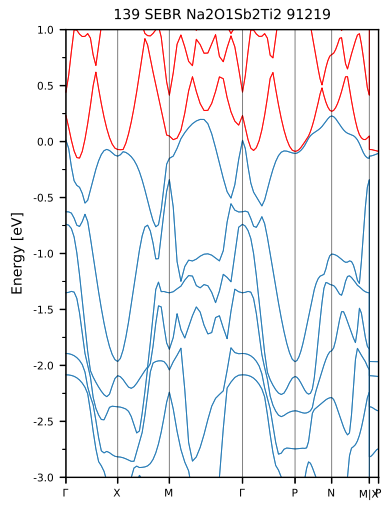

(x) 139.1.24

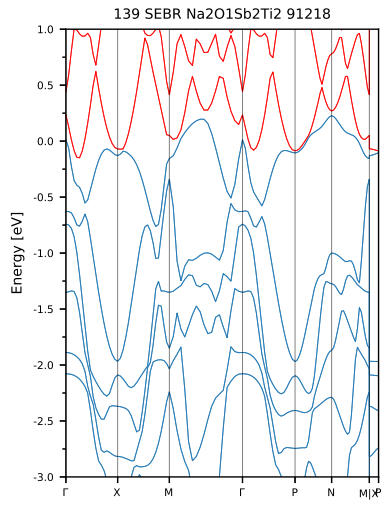

(y) 139.1.25

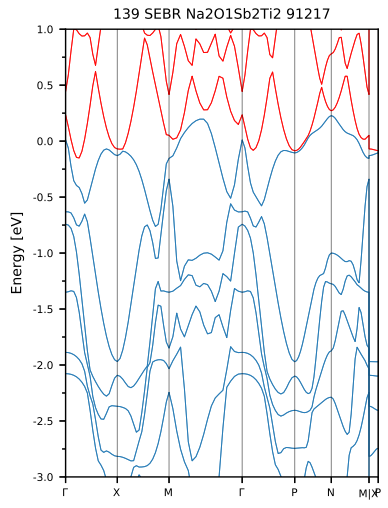

(z) 139.1.26

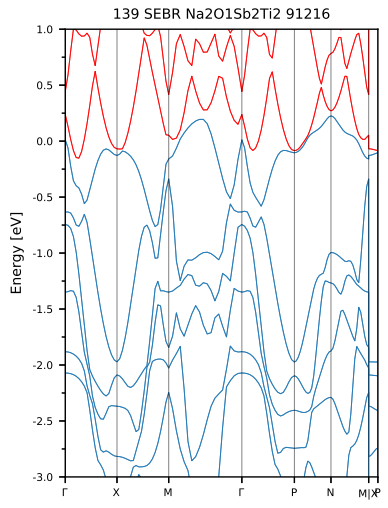

(aa) 139.1.27

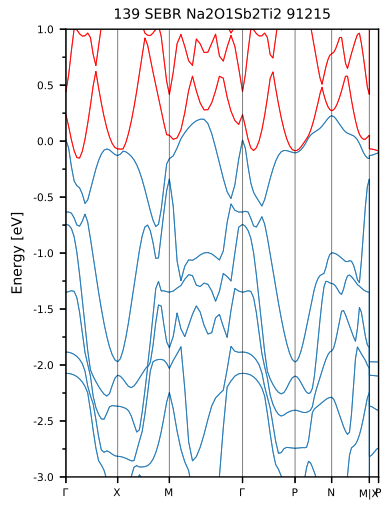

(ab) 139.1.28

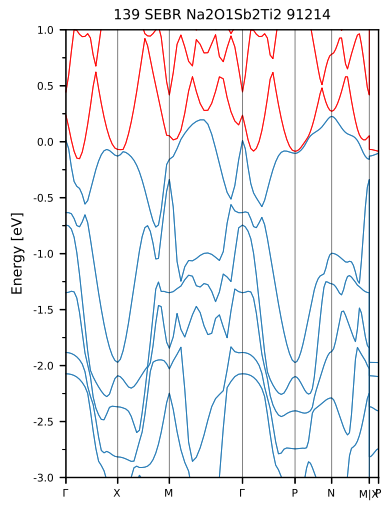

(ac) 139.1.29

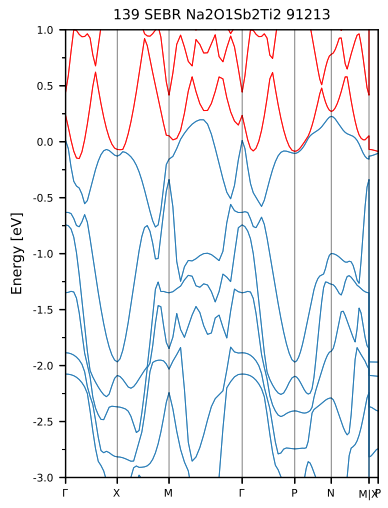

(ad) 139.1.30

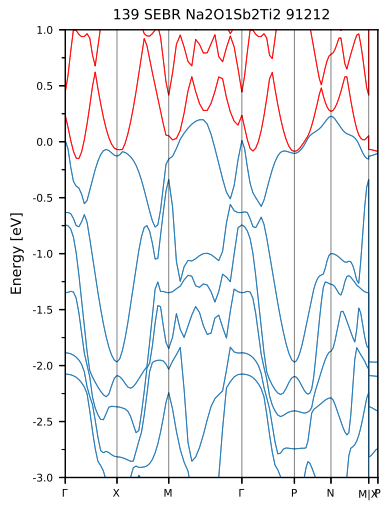

(ae) 139.1.31

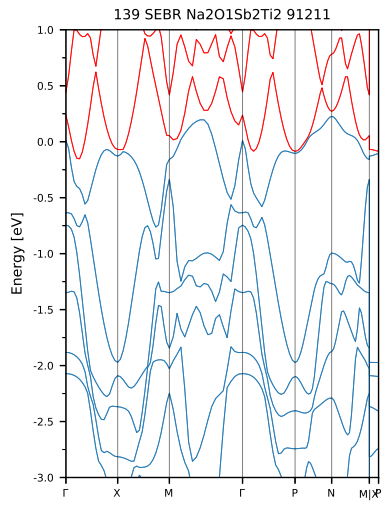

(af) 139.1.32

Figure 289: Band structures for topological materials in SG 139, part 2/22, from 139.1.17 to 139.1.32. See the list of figures.

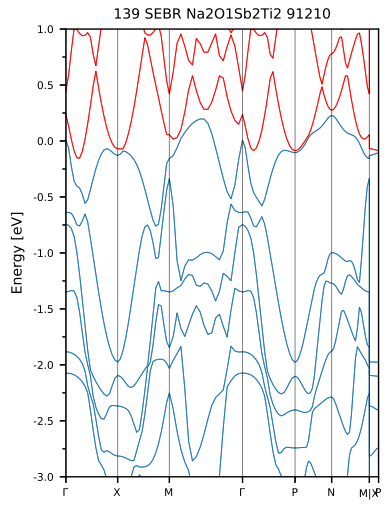

(ag) 139.1.33

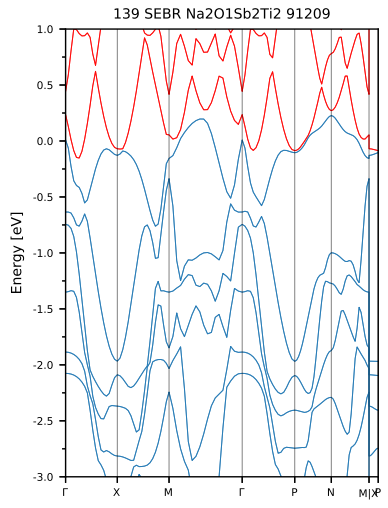

(ah) 139.1.34

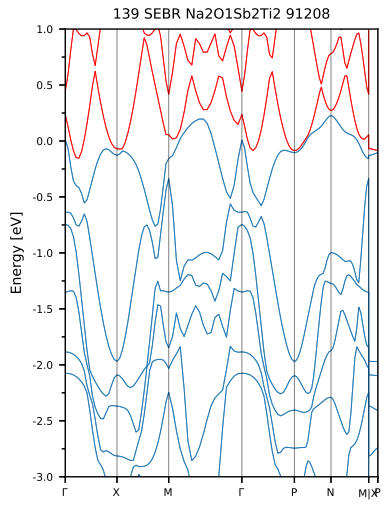

(ai) 139.1.35

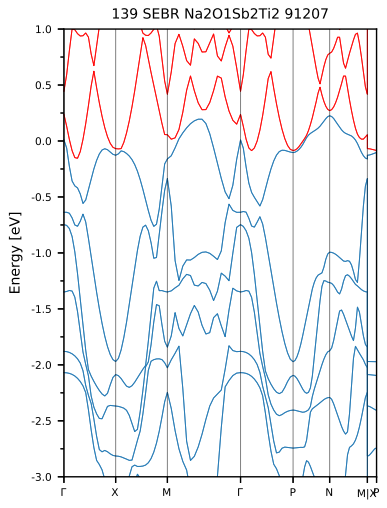

(aj) 139.1.36

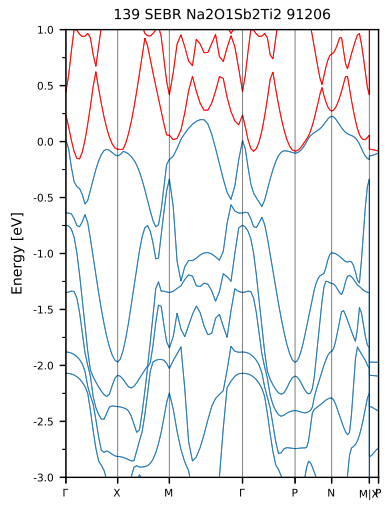

(ak) 139.1.37

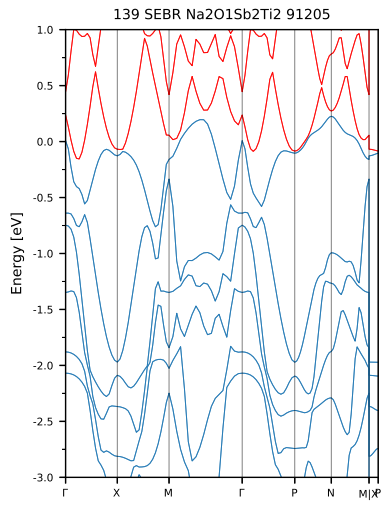

(al) 139.1.38

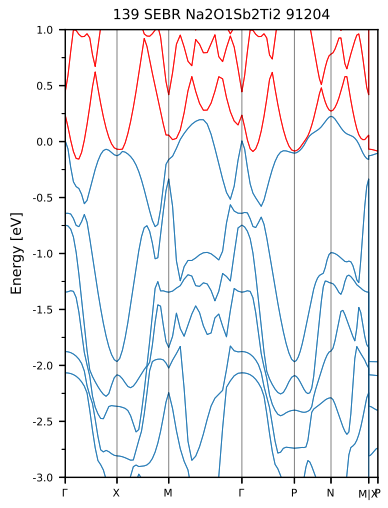

(am) 139.1.39

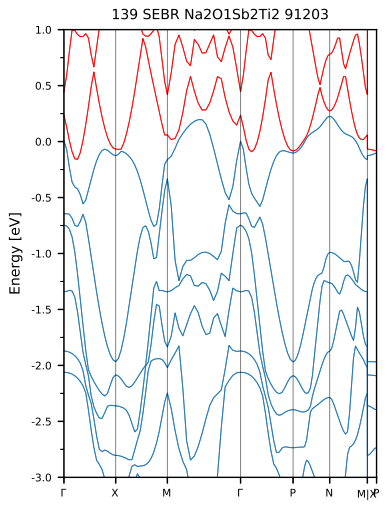

(an) 139.1.40

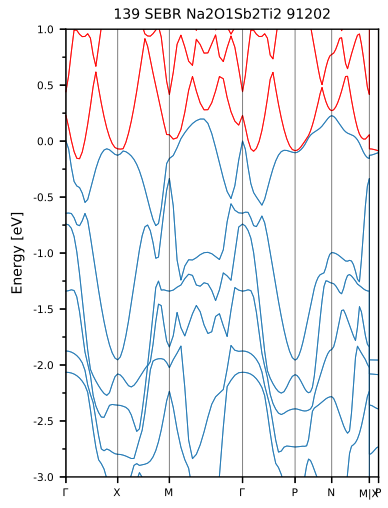

(ao) 139.1.41

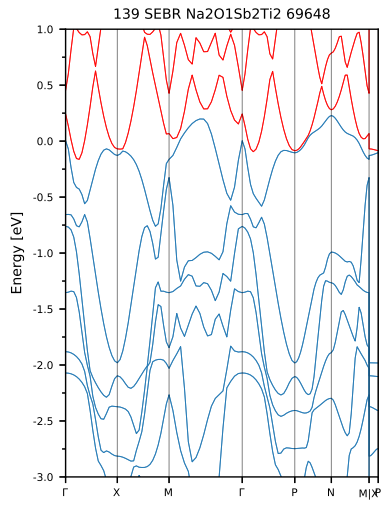

(ap) 139.1.42

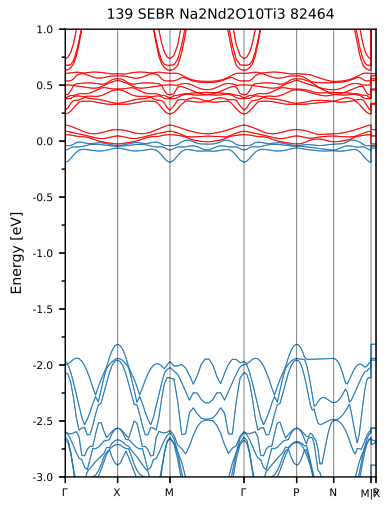

(aq) 139.1.43

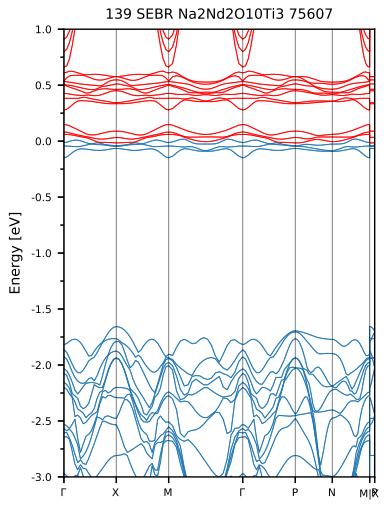

(ar) 139.1.44

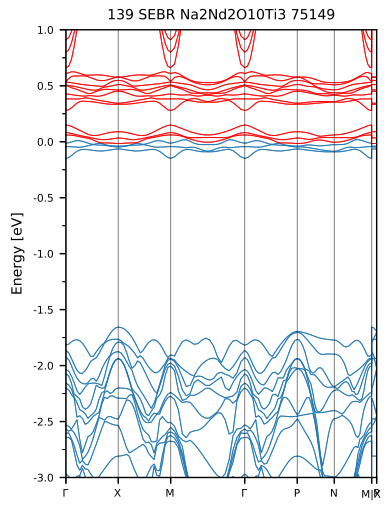

(as) 139.1.45

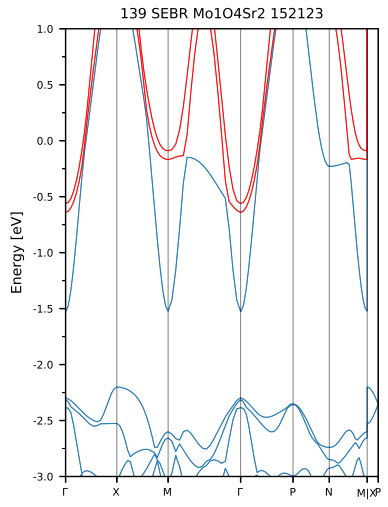

(at) 139.1.46

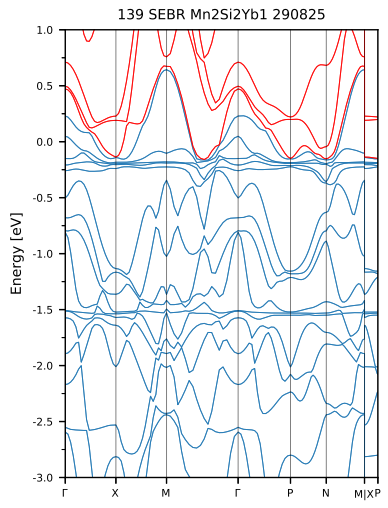

(au) 139.1.47

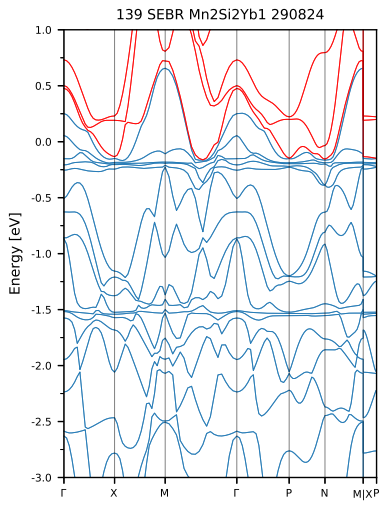

(av) 139.1.48

Figure 290: Band structures for topological materials in SG 139, part 3/22, from 139.1.33 to 139.1.48. See the list of figures.

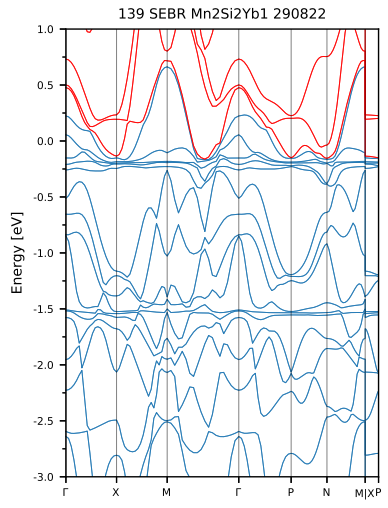

(aw) 139.1.49

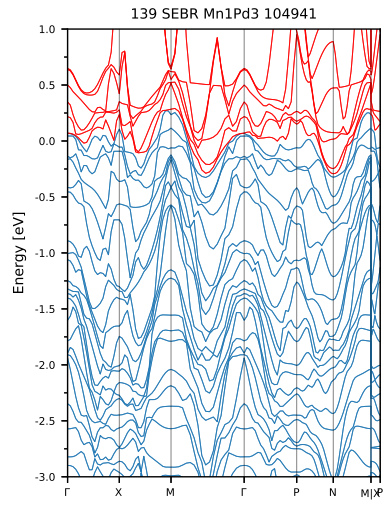

(ax) 139.1.50

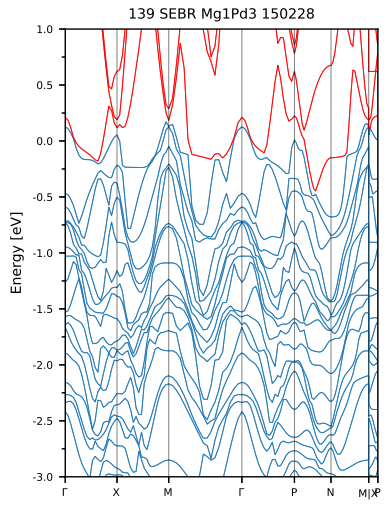

(ay) 139.1.51

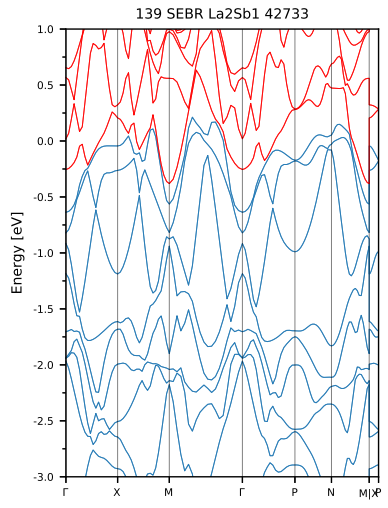

(az) 139.1.52

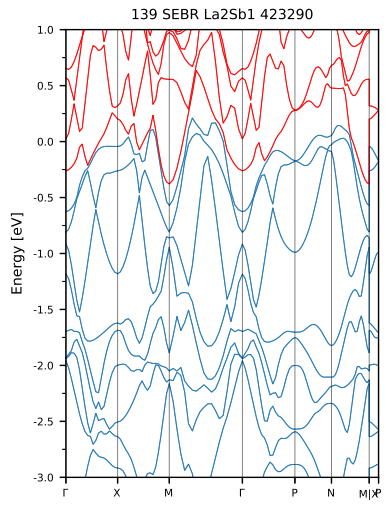

(ba) 139.1.53

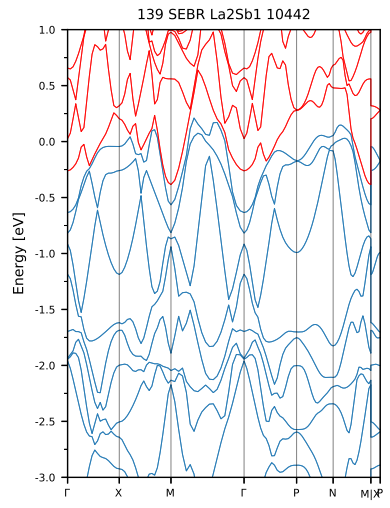

(bb) 139.1.54

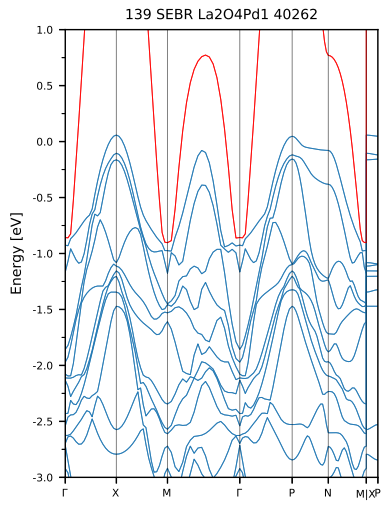

(bc) 139.1.55

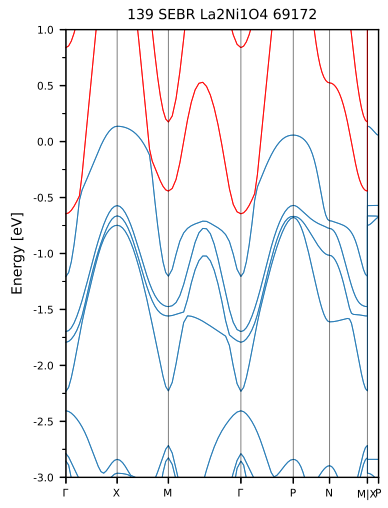

(bd) 139.1.56

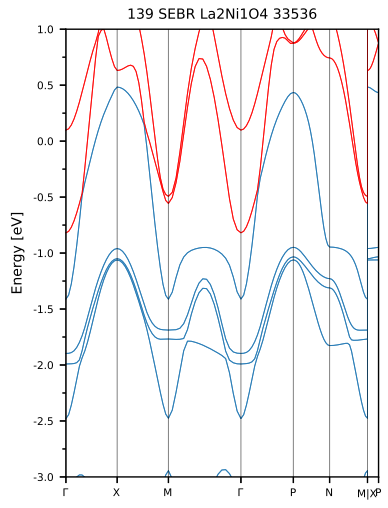

(be) 139.1.57

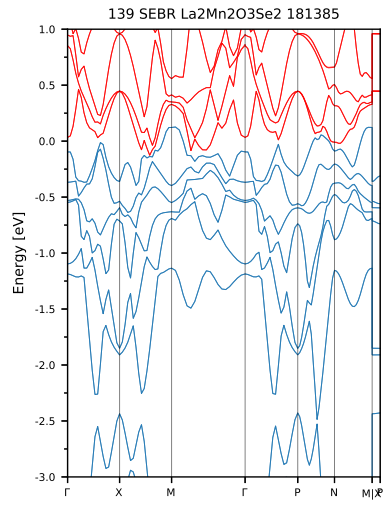

(bf) 139.1.58

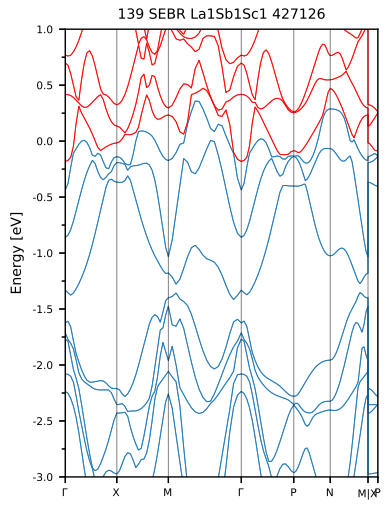

(bg) 139.1.59

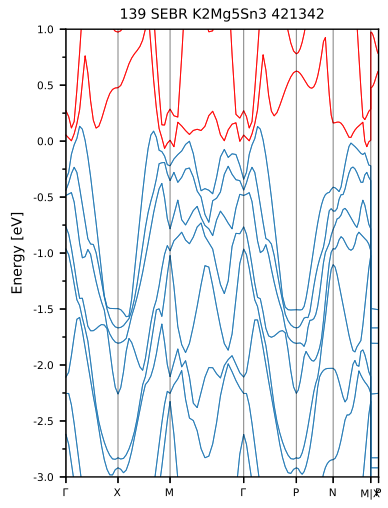

(bh) 139.1.60

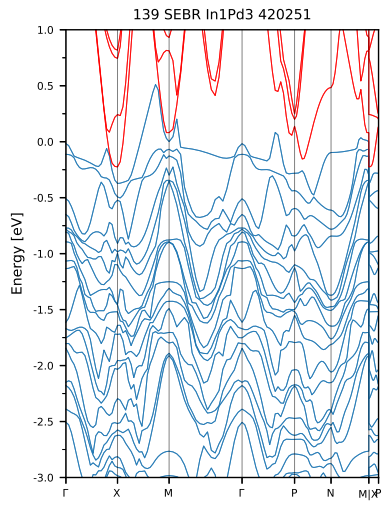

(bi) 139.1.61

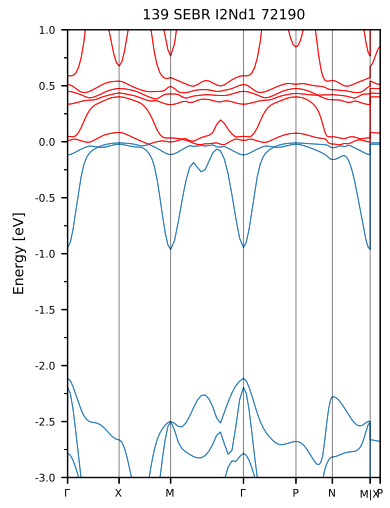

(bj) 139.1.62

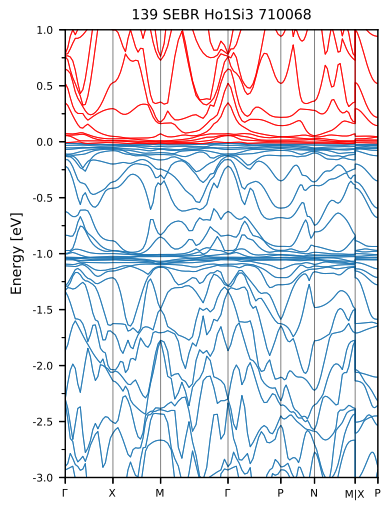

(bk) 139.1.63

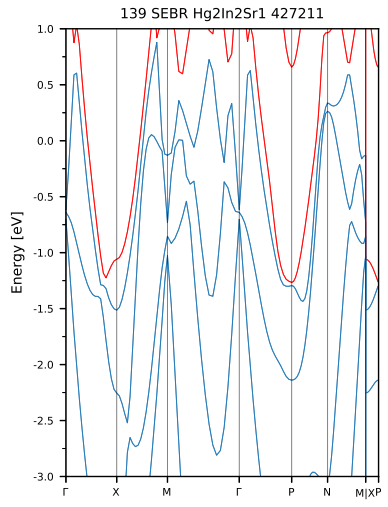

(bl) 139.1.64

Figure 291: Band structures for topological materials in SG 139, part 4/22, from 139.1.49 to 139.1.64. See the list of figures.

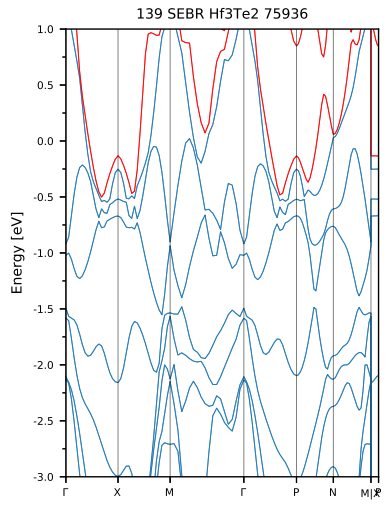

(bm) 139.1.65

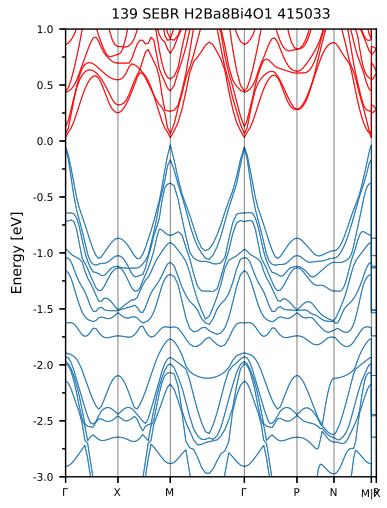

(bn) 139.1.66

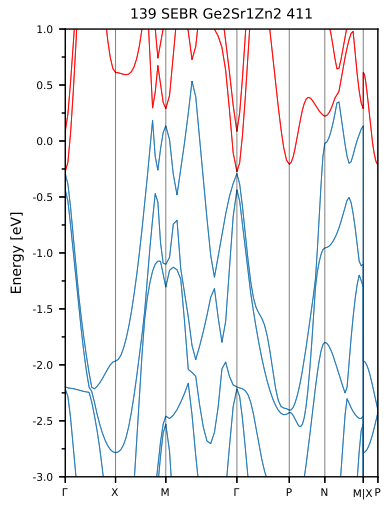

(bo) 139.1.67

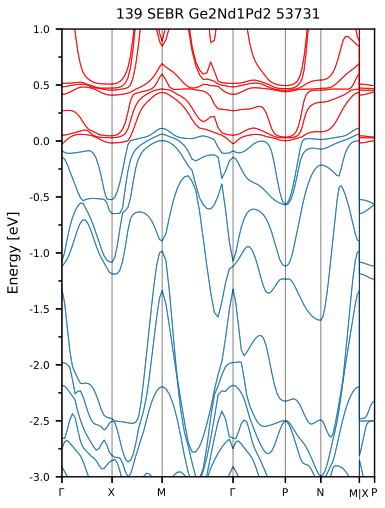

(bp) 139.1.68

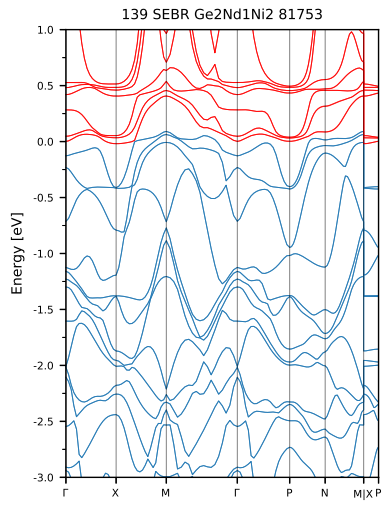

(bq) 139.1.69

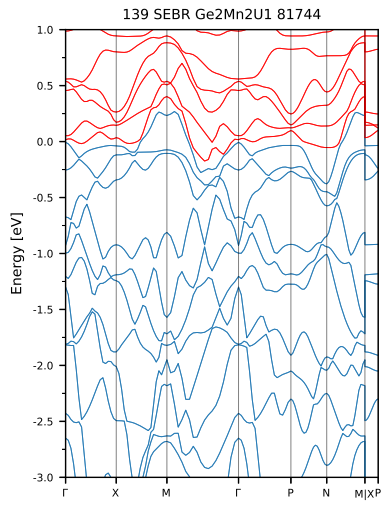

(br) 139.1.70

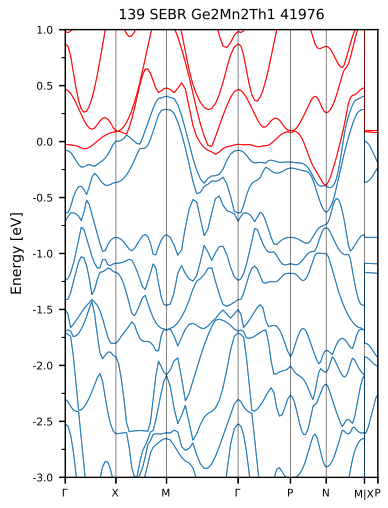

(bs) 139.1.71

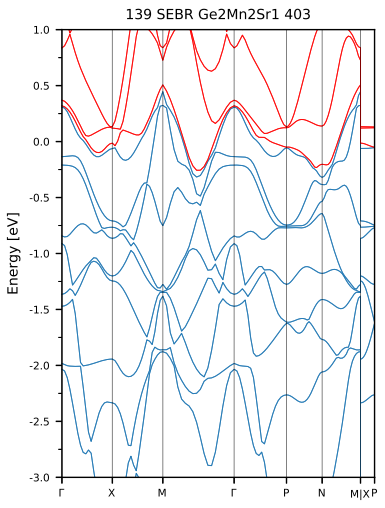

(bt) 139.1.72

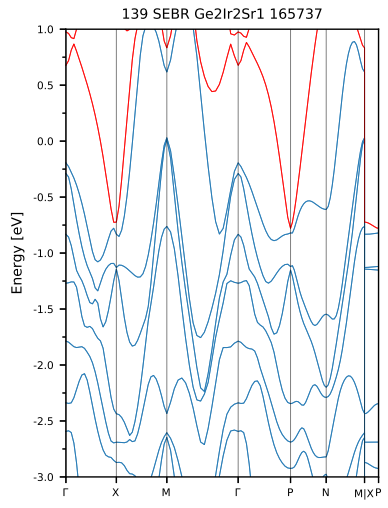

(bu) 139.1.73

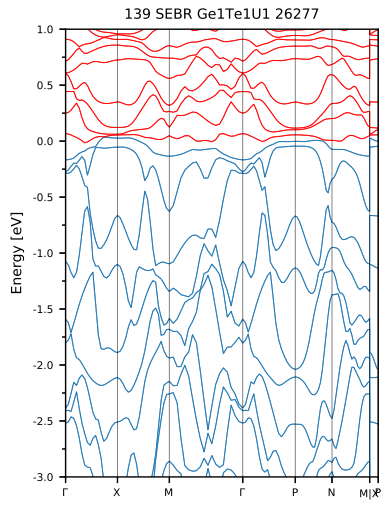

(bv) 139.1.74

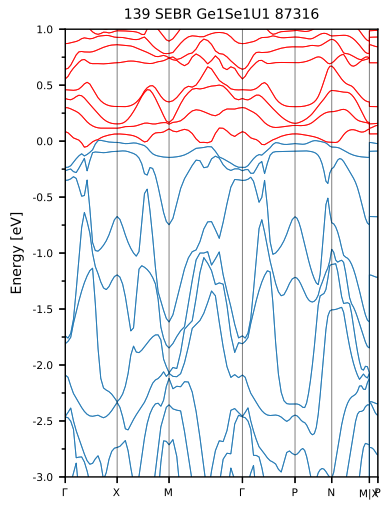

(bw) 139.1.75

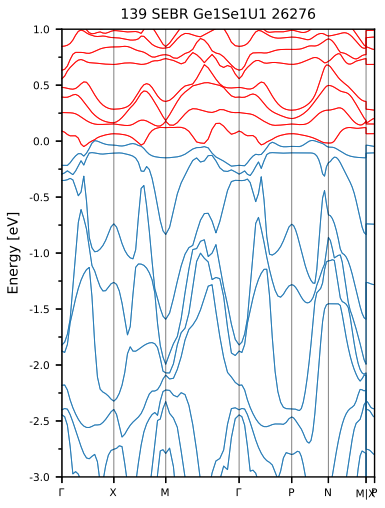

(bx) 139.1.76

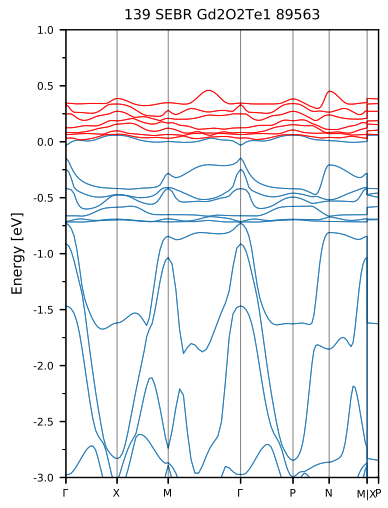

(by) 139.1.77

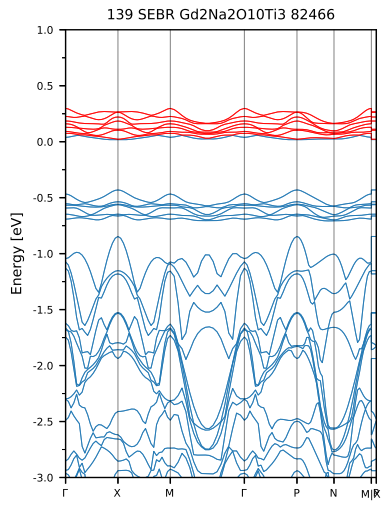

(bz) 139.1.78

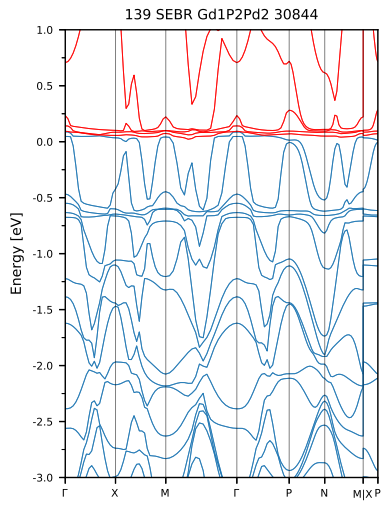

(ca) 139.1.79

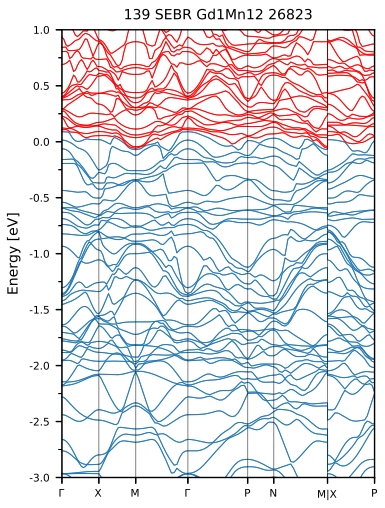

(cb) 139.1.80

Figure 292: Band structures for topological materials in SG 139, part 5/22, from 139.1.65 to 139.1.80. See the list of figures.

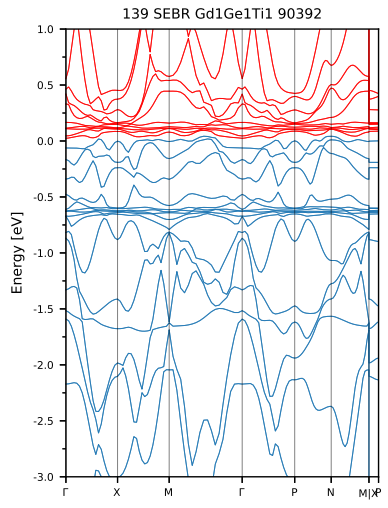

(cc) 139.1.81

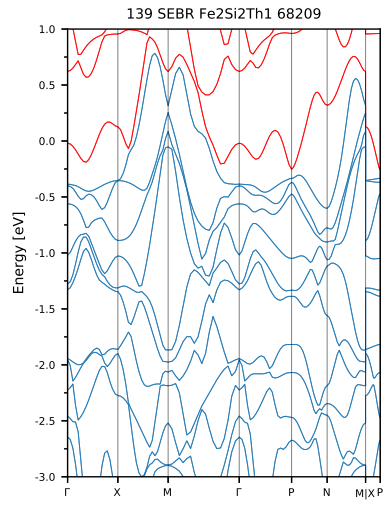

(cd) 139.1.82

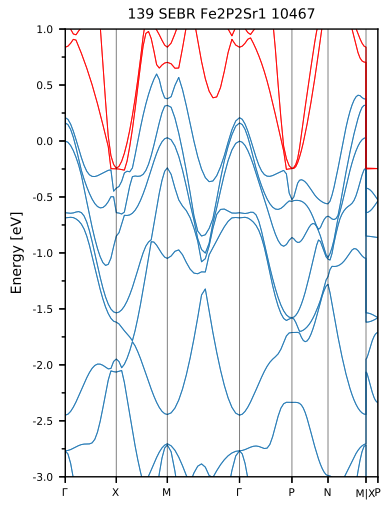

(ce) 139.1.83

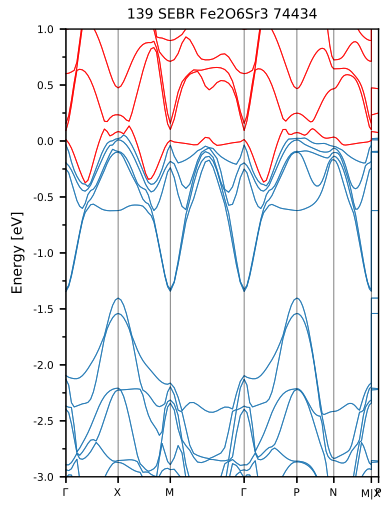

(cf) 139.1.84

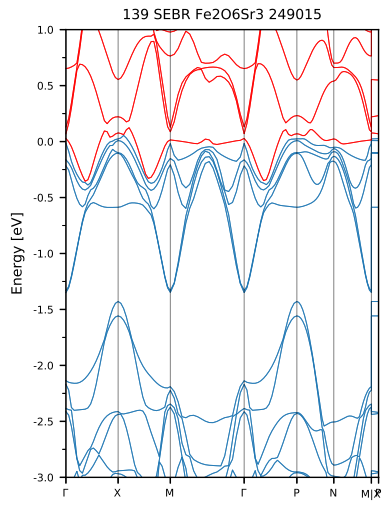

(cg) 139.1.85

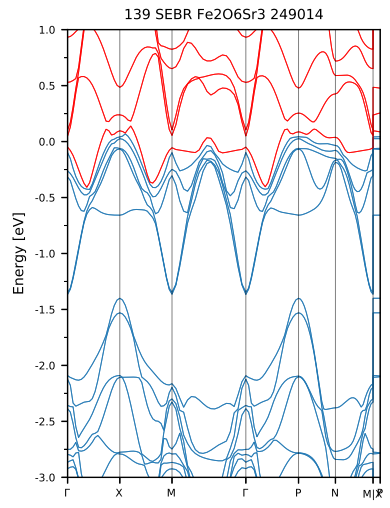

(ch) 139.1.86

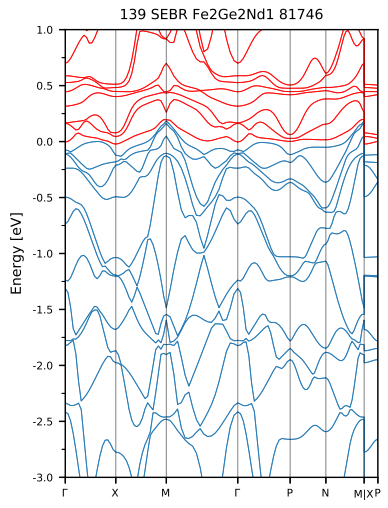

(ci) 139.1.87

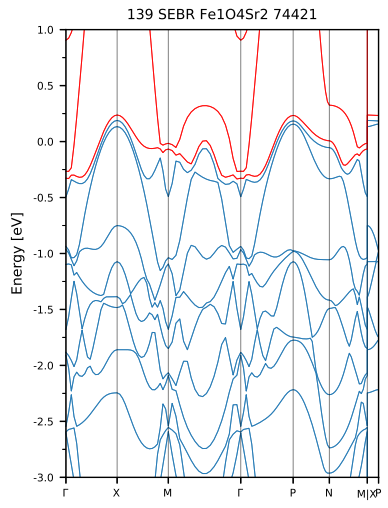

(cj) 139.1.88

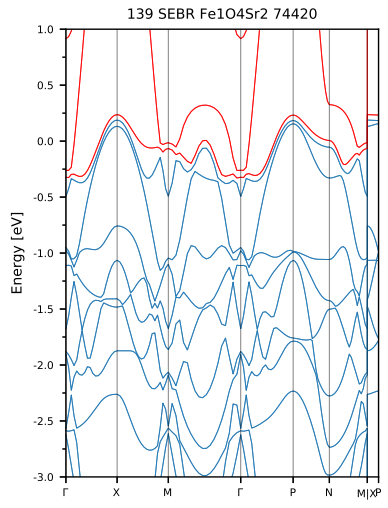

(ck) 139.1.89

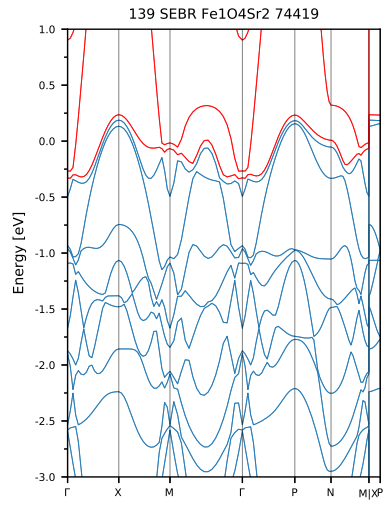

(cl) 139.1.90

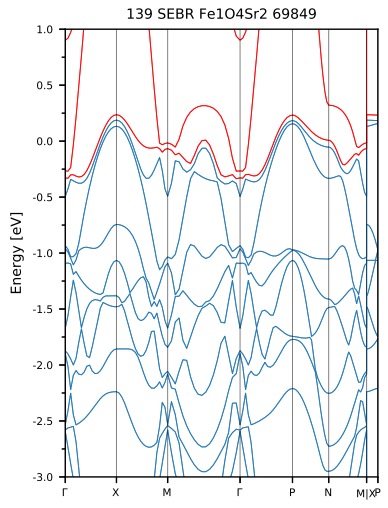

(cm) 139.1.91

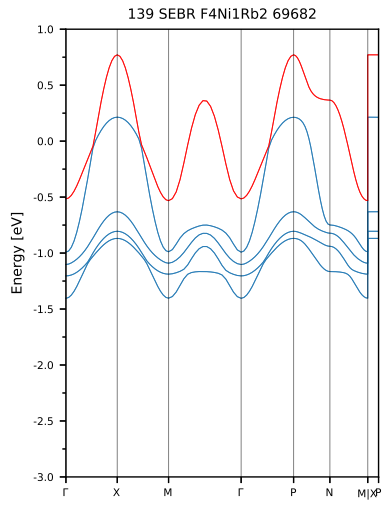

(cn) 139.1.92

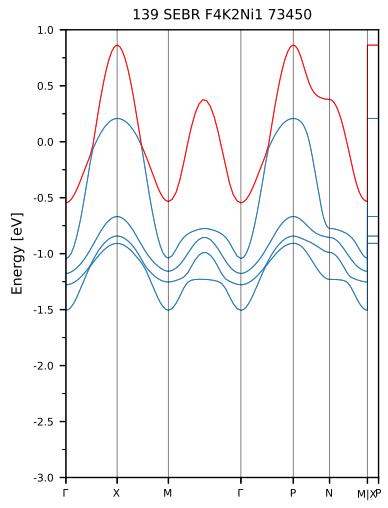

(co) 139.1.93

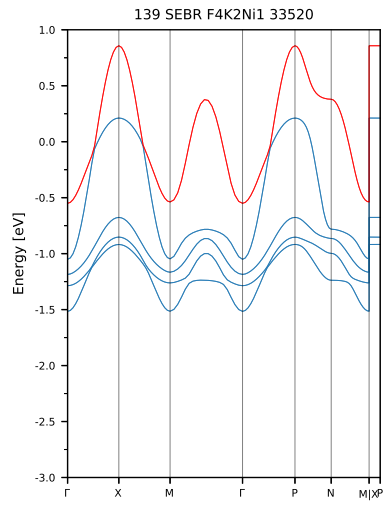

(cp) 139.1.94

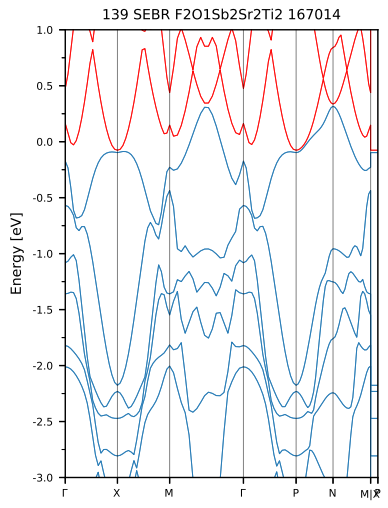

(cq) 139.1.95

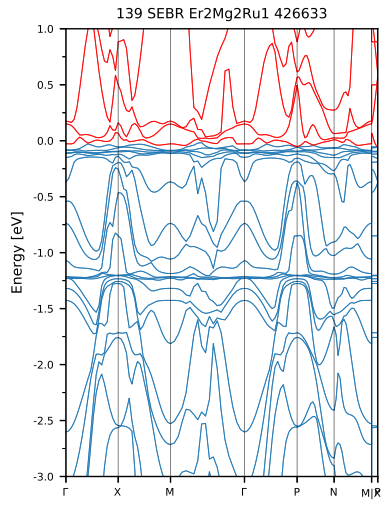

(cr) 139.1.96

Figure 293: Band structures for topological materials in SG 139, part 6/22, from 139.1.81 to 139.1.96. See the list of figures.

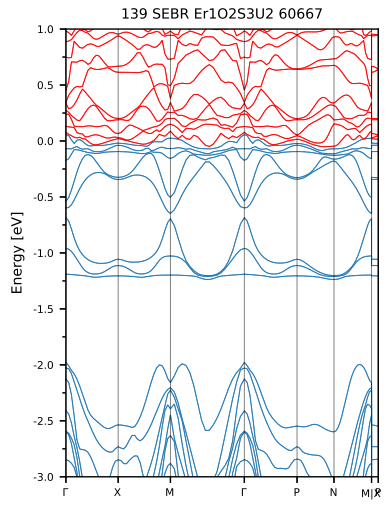

(cs) 139.1.97

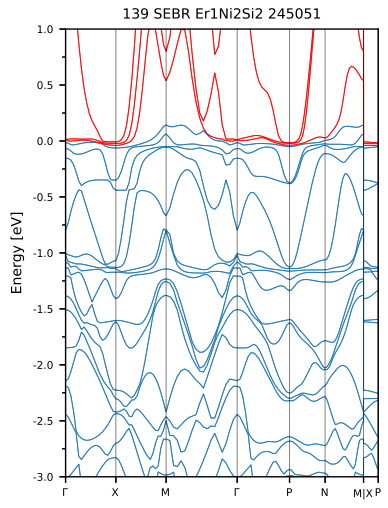

(ct) 139.1.98

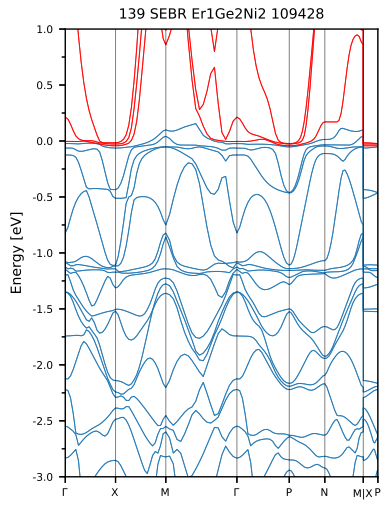

(cu) 139.1.99

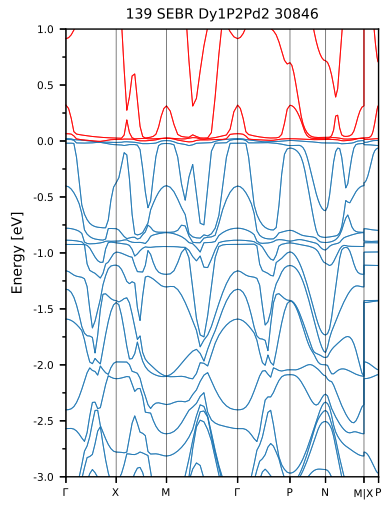

(cv) 139.1.100

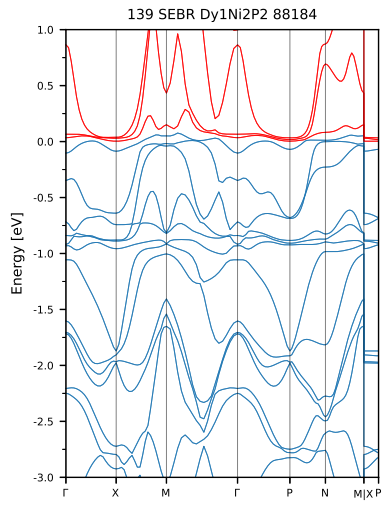

(cw) 139.1.101

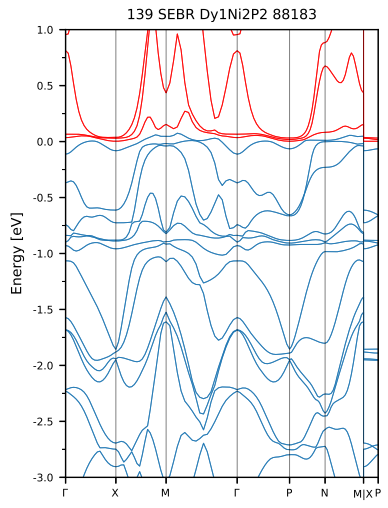

(cx) 139.1.102

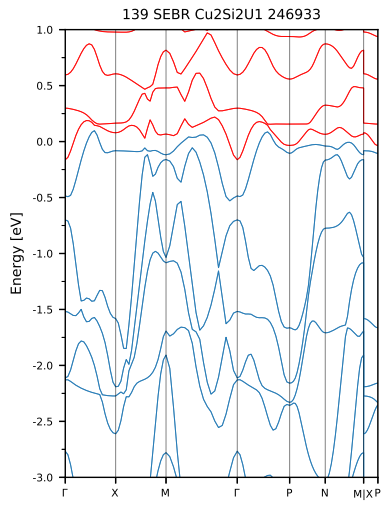

(cy) 139.1.103

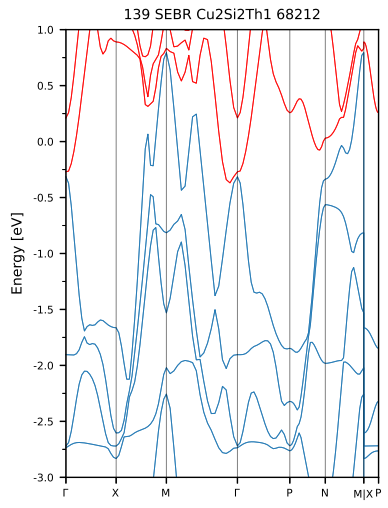

(cz) 139.1.104

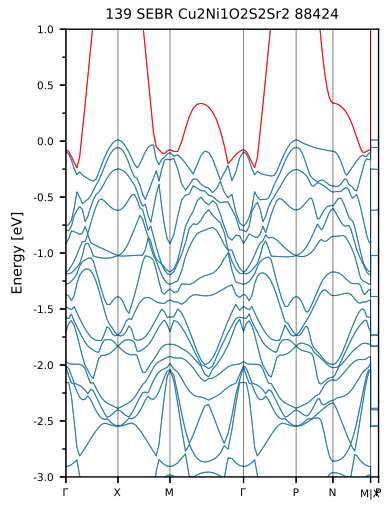

(da) 139.1.105

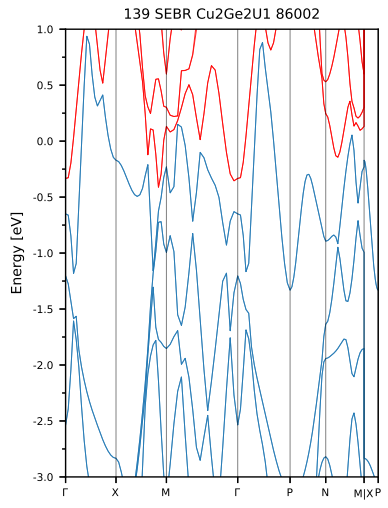

(db) 139.1.106

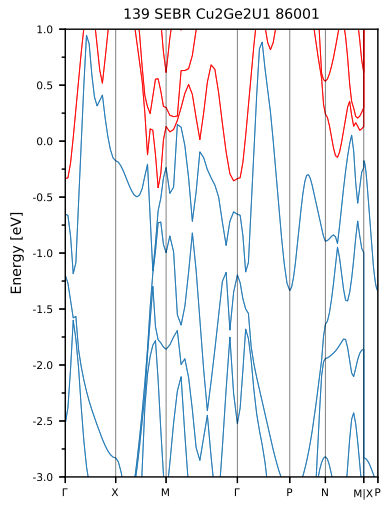

(dc) 139.1.107

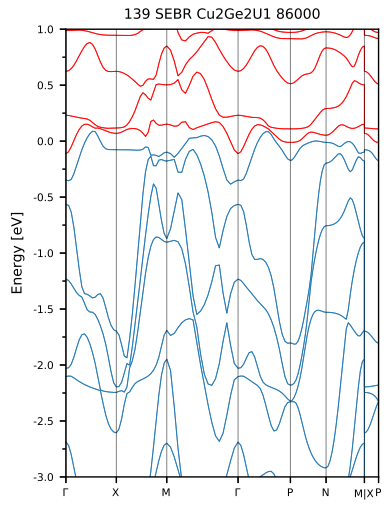

(dd) 139.1.108

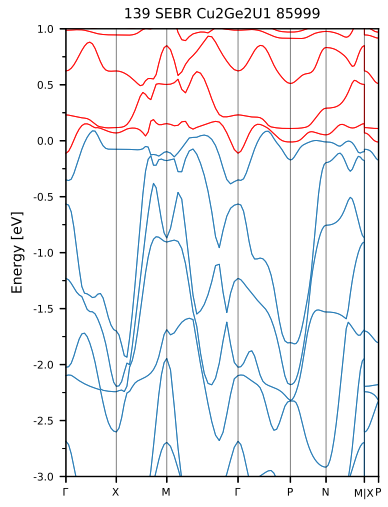

(de) 139.1.109

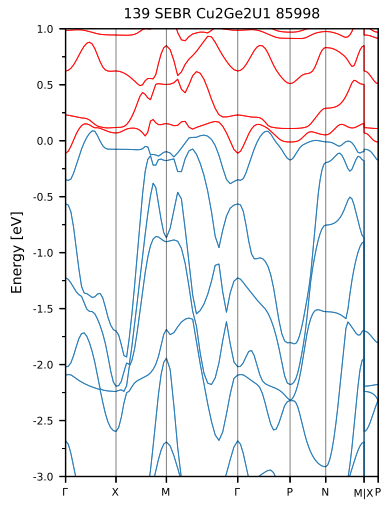

(df) 139.1.110

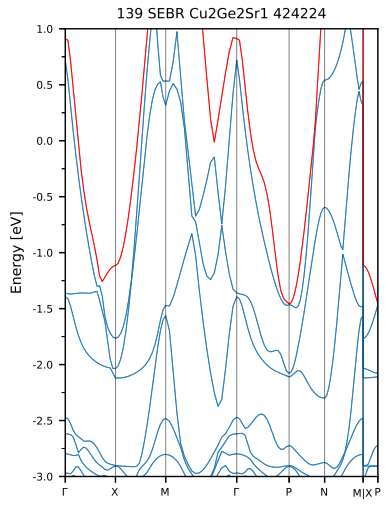

(dg) 139.1.111

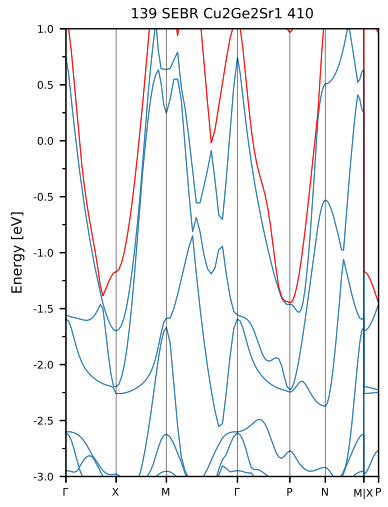

(dh) 139.1.112

Figure 294: Band structures for topological materials in SG 139, part 7/22, from 139.1.97 to 139.1.112. See the list of figures.

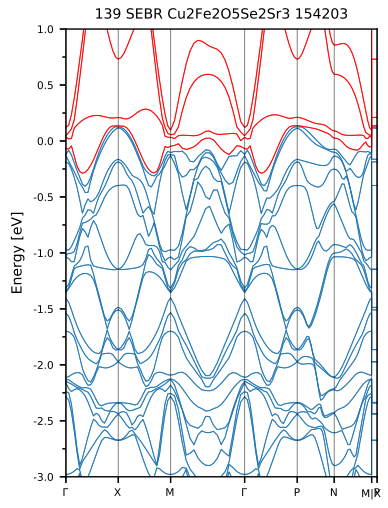

(di) 139.1.113

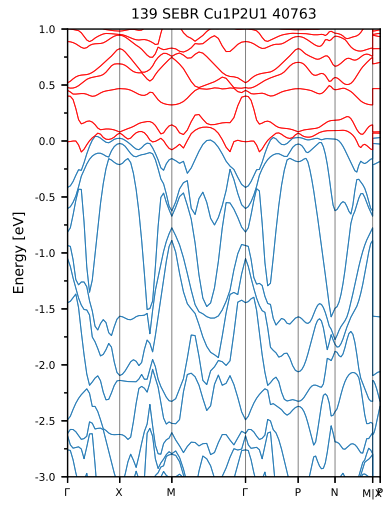

(dj) 139.1.114

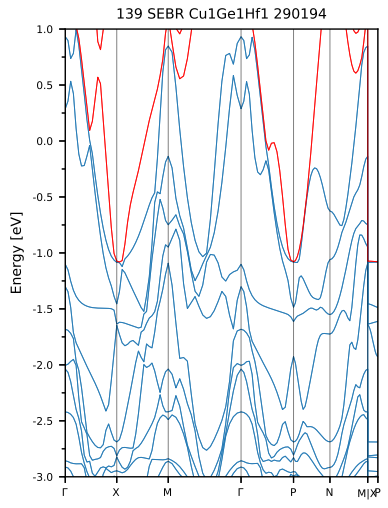

(dk) 139.1.115

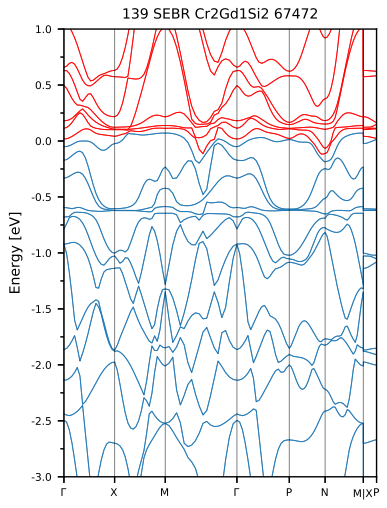

(dl) 139.1.116

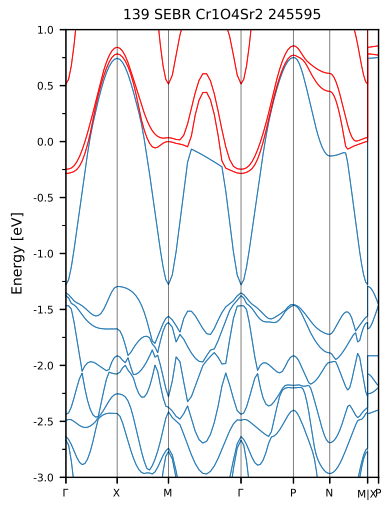

(dm) 139.1.117

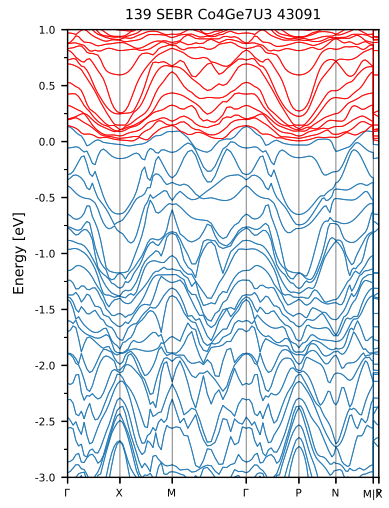

(dn) 139.1.118

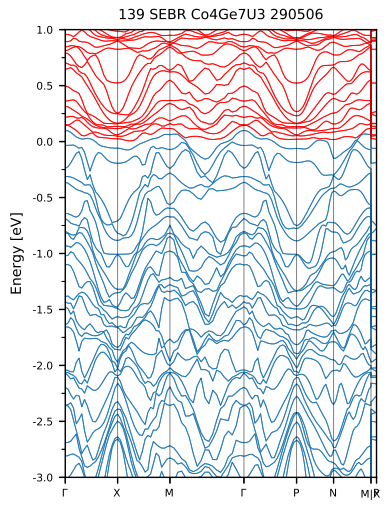

(do) 139.1.119

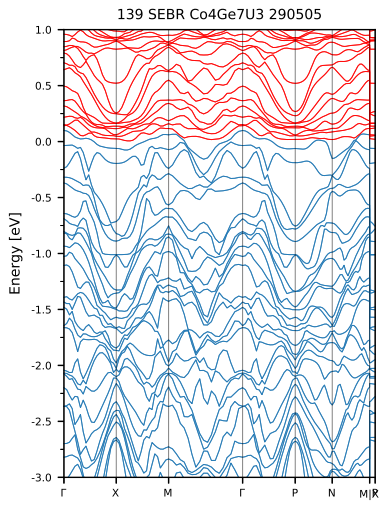

(dp) 139.1.120

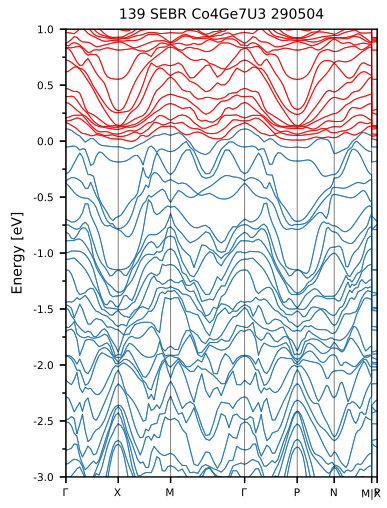

(dq) 139.1.121

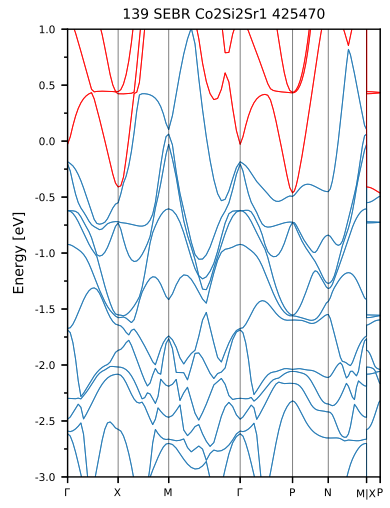

(dr) 139.1.122

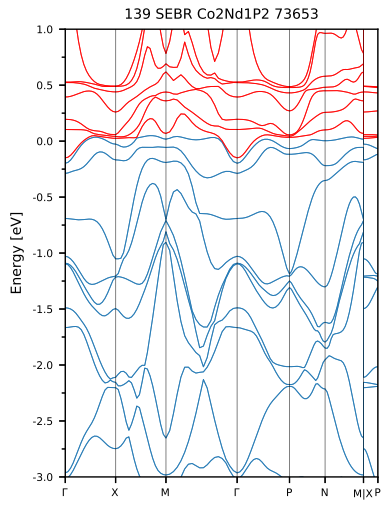

(ds) 139.1.123

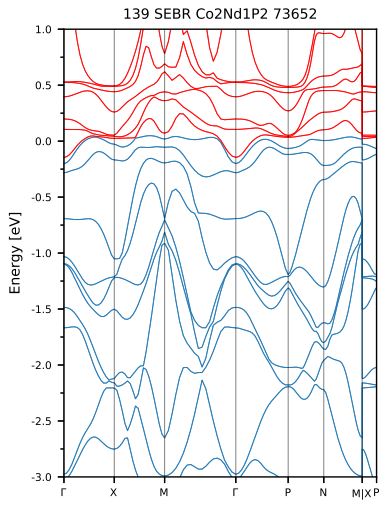

(dt) 139.1.124

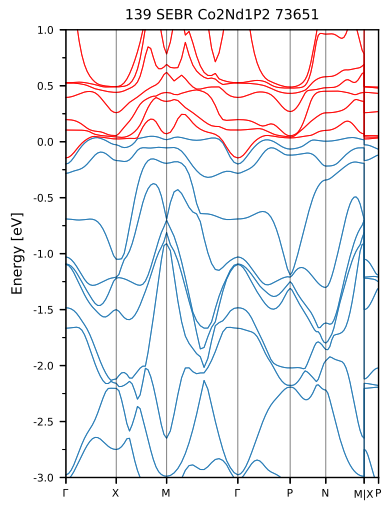

(du) 139.1.125

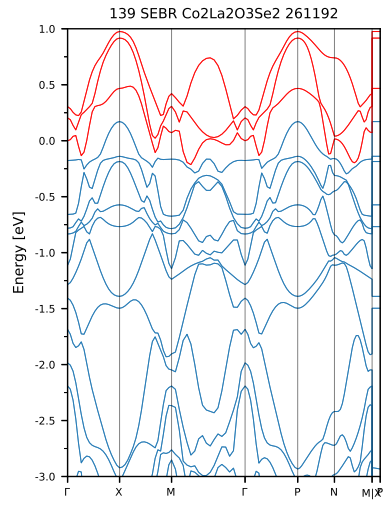

(dv) 139.1.126

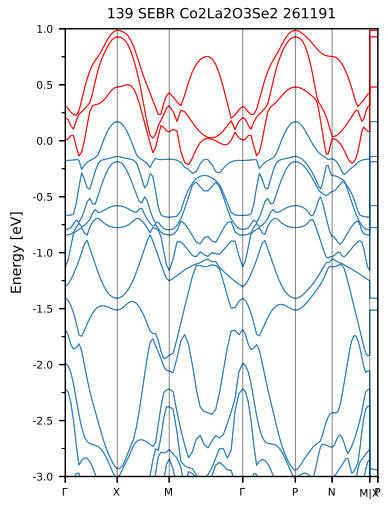

(dw) 139.1.127

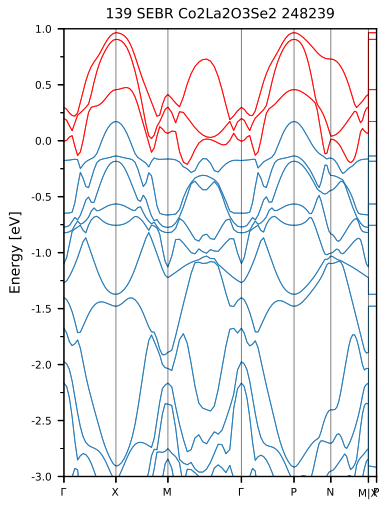

(dx) 139.1.128

Figure 295: Band structures for topological materials in SG 139, part 8/22, from 139.1.113 to 139.1.128. See the list of figures.

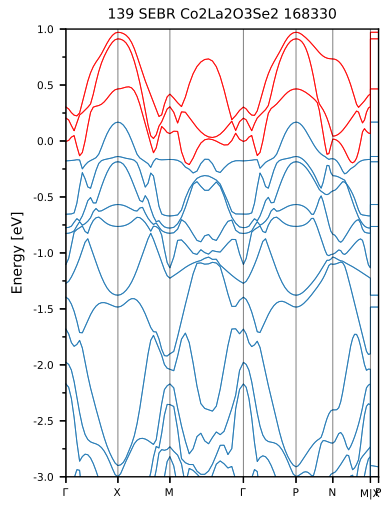

(dy) 139.1.129

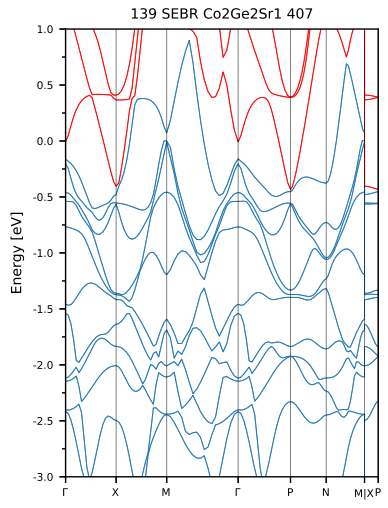

(dz) 139.1.130

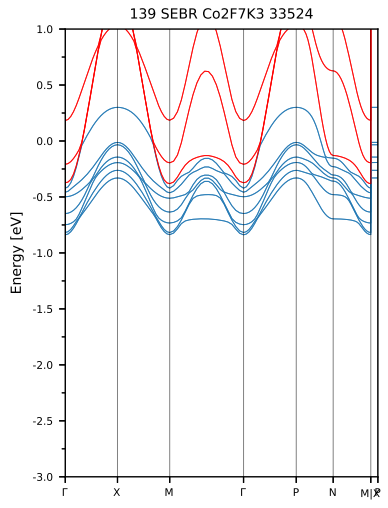

(ea) 139.1.131

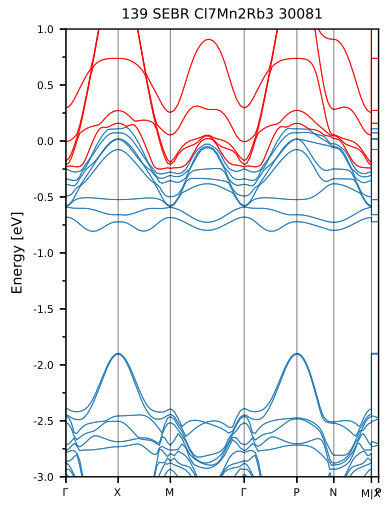

(eb) 139.1.132

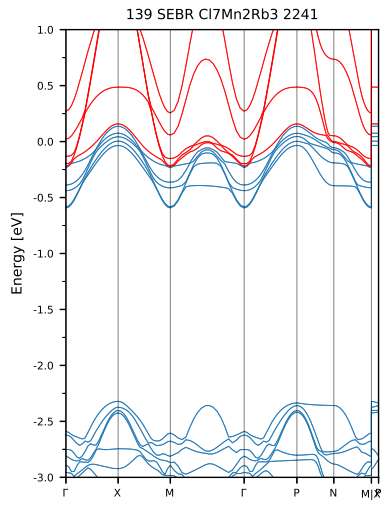

(ec) 139.1.133

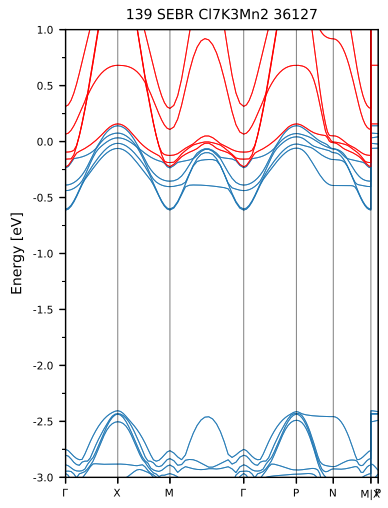

(ed) 139.1.134

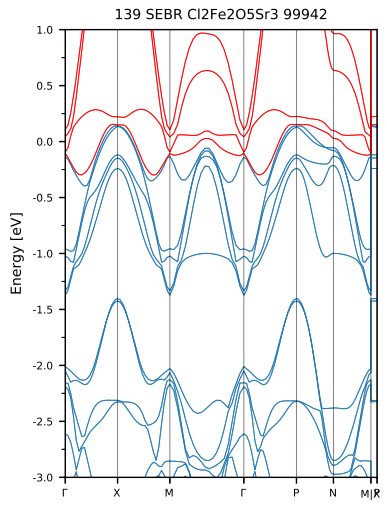

(ee) 139.1.135

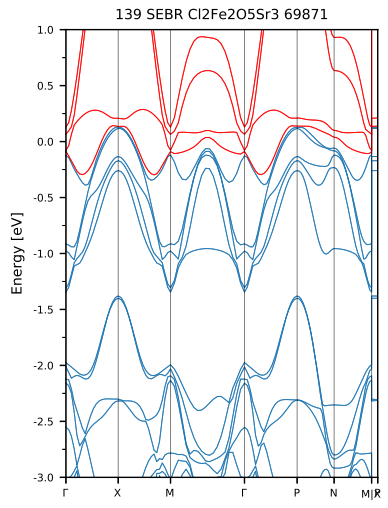

(ef) 139.1.136

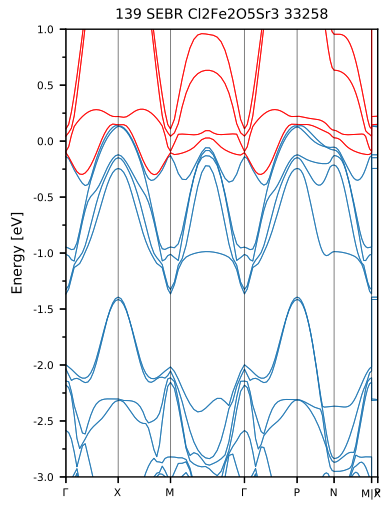

(eg) 139.1.137

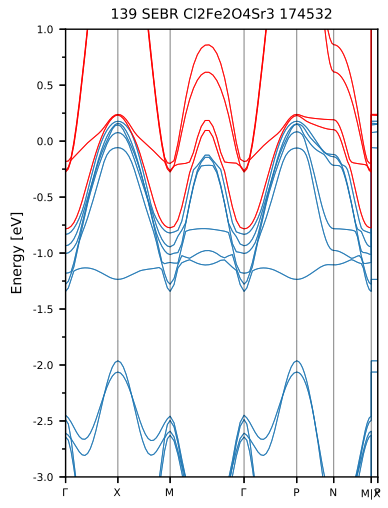

(eh) 139.1.138

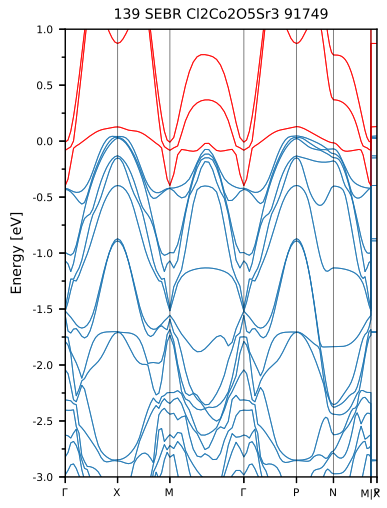

(ei) 139.1.139

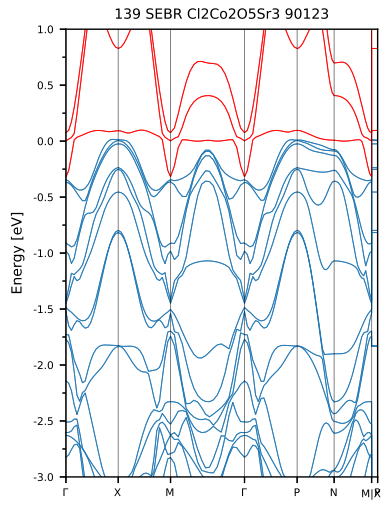

(ej) 139.1.140

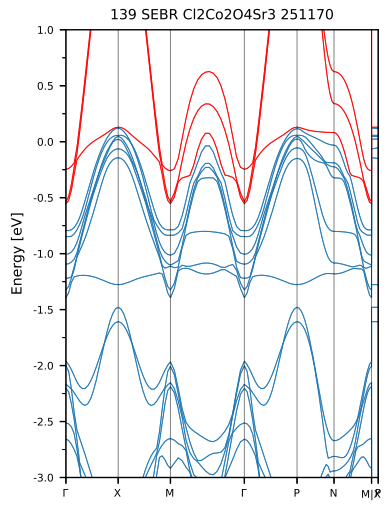

(ek) 139.1.141

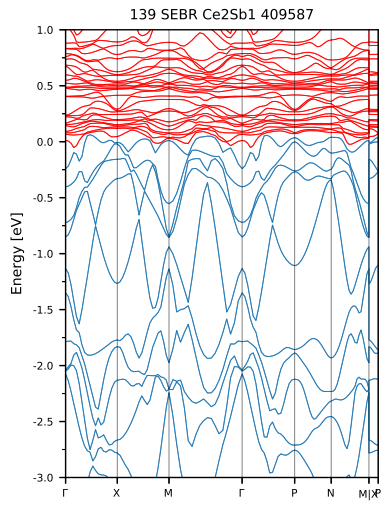

(el) 139.1.142

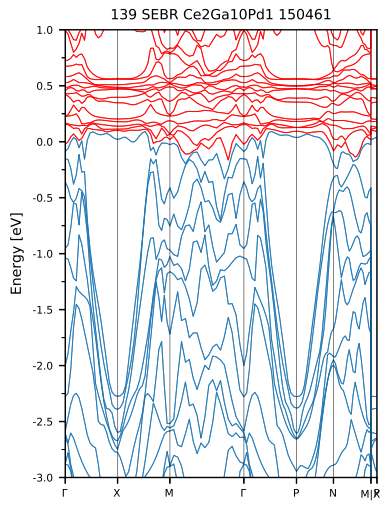

(em) 139.1.143

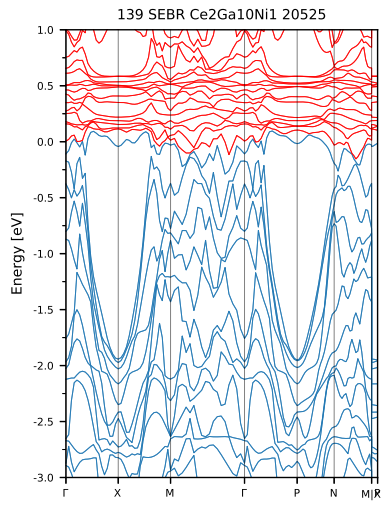

(en) 139.1.144

Figure 296: Band structures for topological materials in SG 139, part 9/22, from 139.1.129 to 139.1.144. See the list of figures.

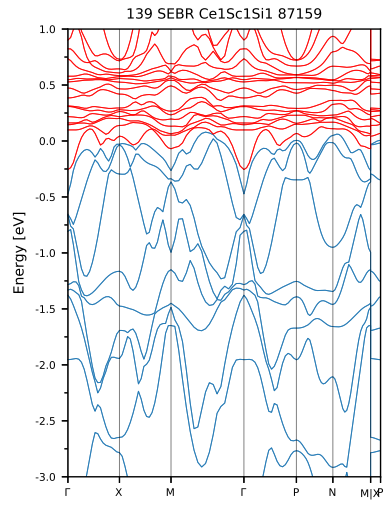

(eo) 139.1.145

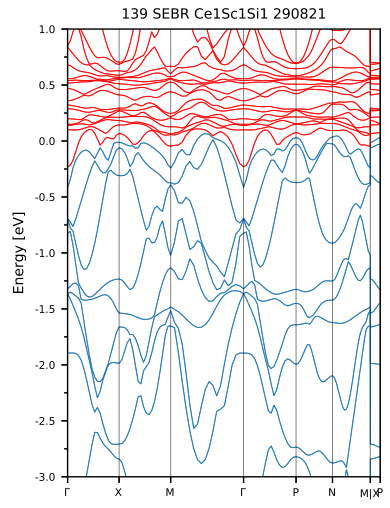

(ep) 139.1.146

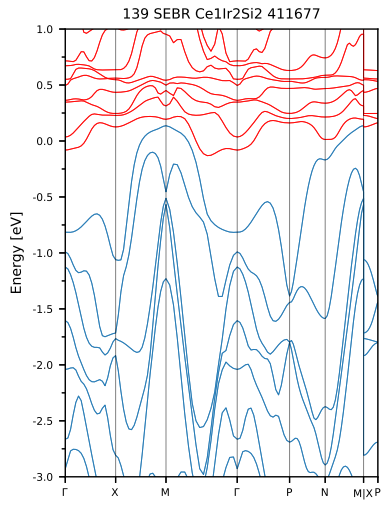

(eq) 139.1.147

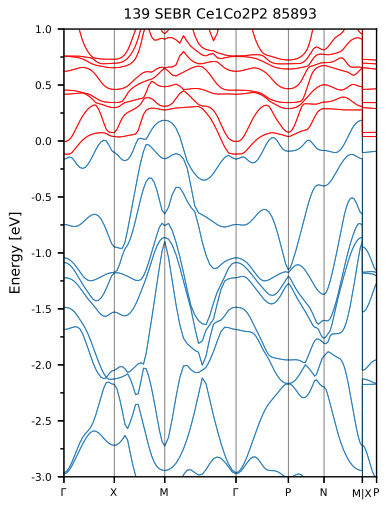

(er) 139.1.148

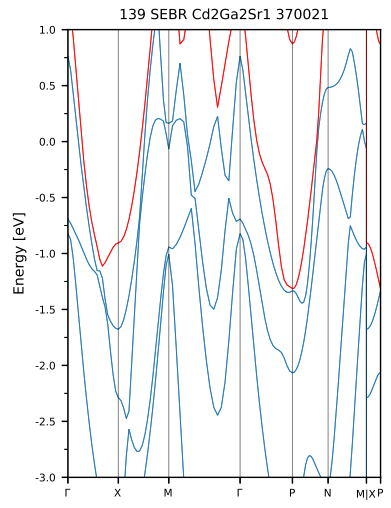

(es) 139.1.149

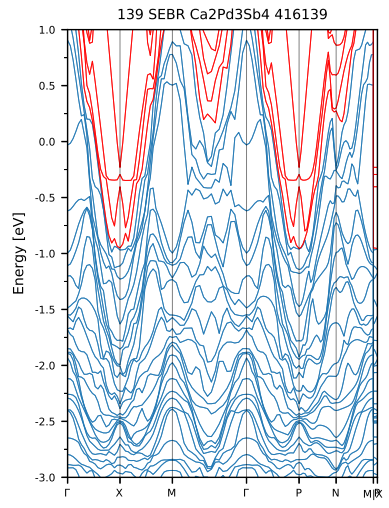

(et) 139.1.150

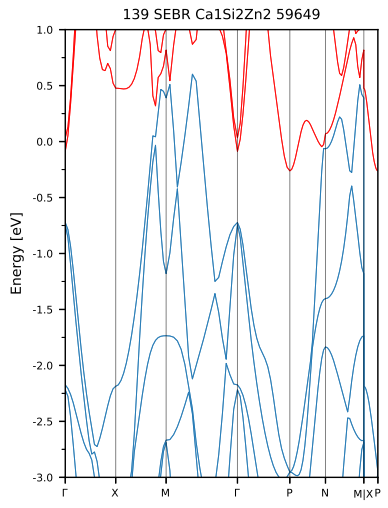

(eu) 139.1.151

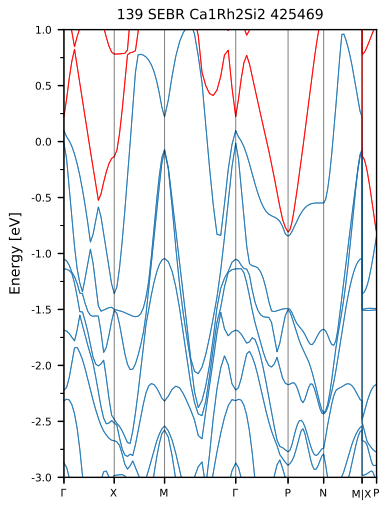

(ev) 139.1.152

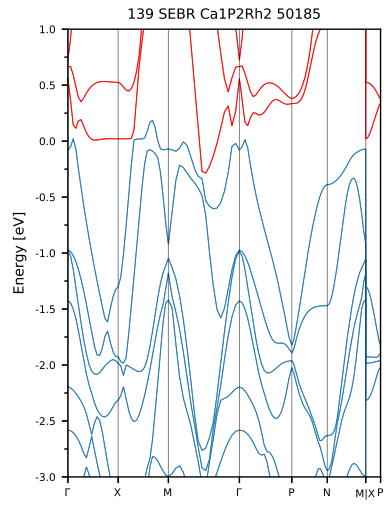

(ew) 139.1.153

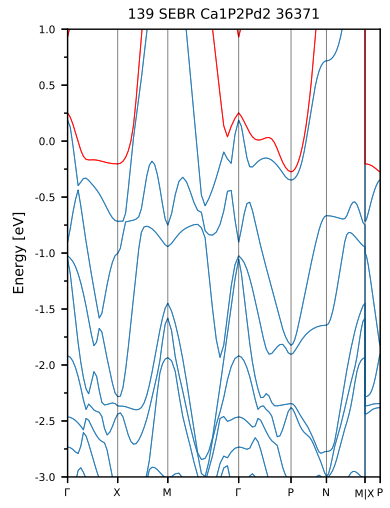

(ex) 139.1.154

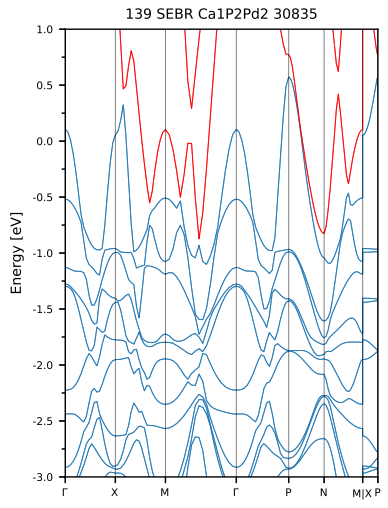

(ey) 139.1.155

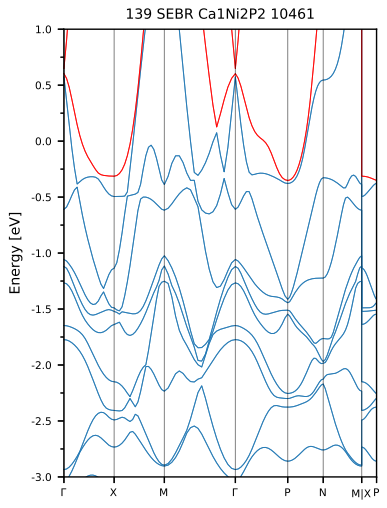

(ez) 139.1.156

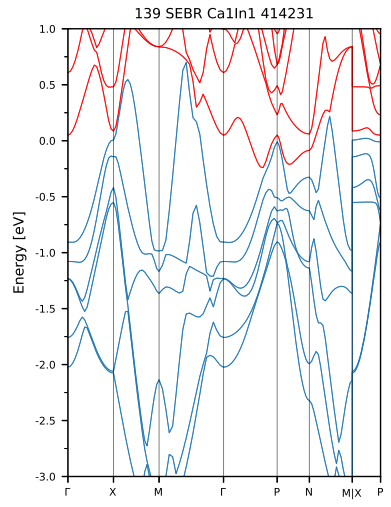

(fa) 139.1.157

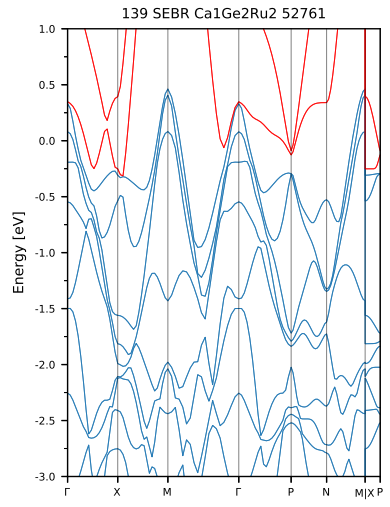

(fb) 139.1.158

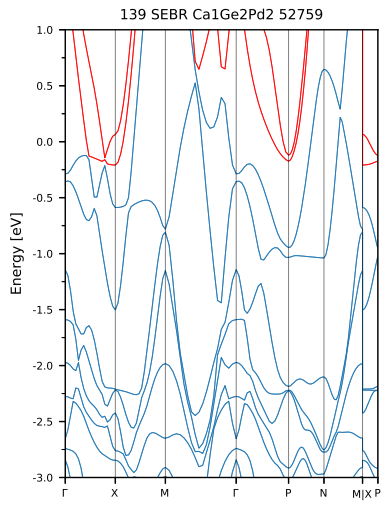

(fc) 139.1.159

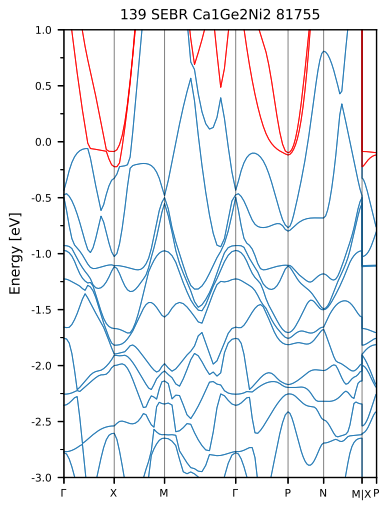

(fd) 139.1.160

Figure 297: Band structures for topological materials in SG 139, part 10/22, from 139.1.145 to 139.1.160. See the list of figures.

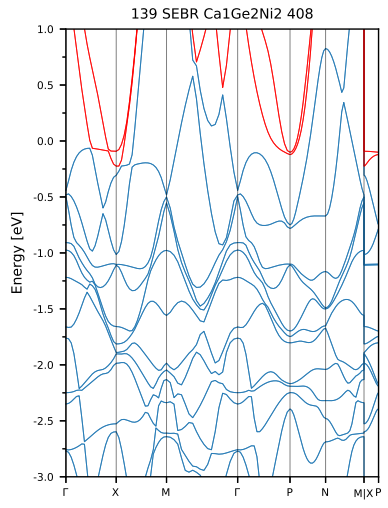

(fe) 139.1.161

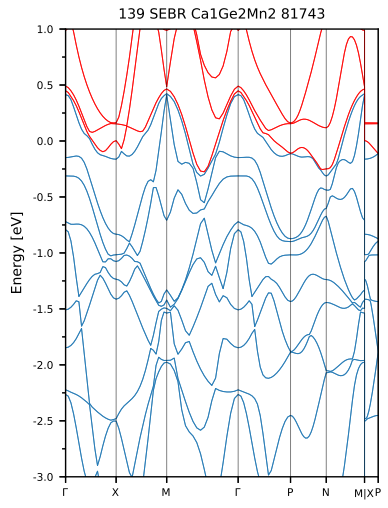

(ff) 139.1.162

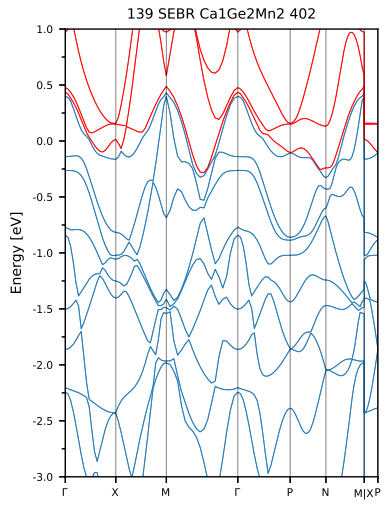

(fg) 139.1.163

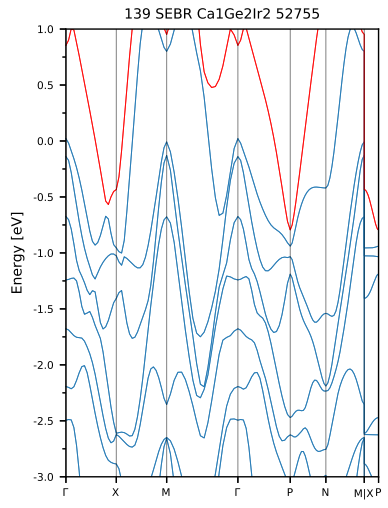

(fh) 139.1.164

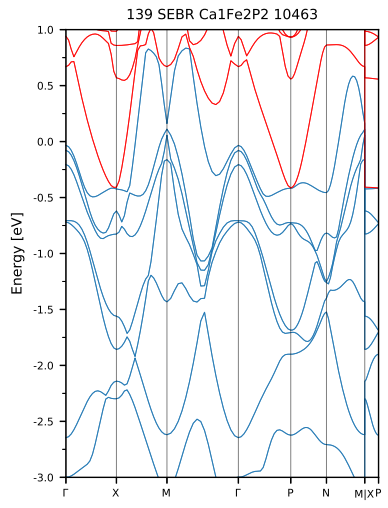

(fi) 139.1.165

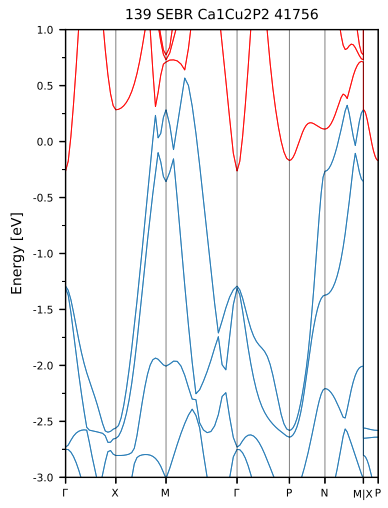

(fj) 139.1.166

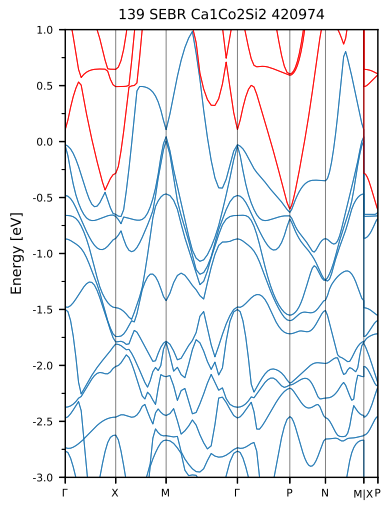

(fk) 139.1.167

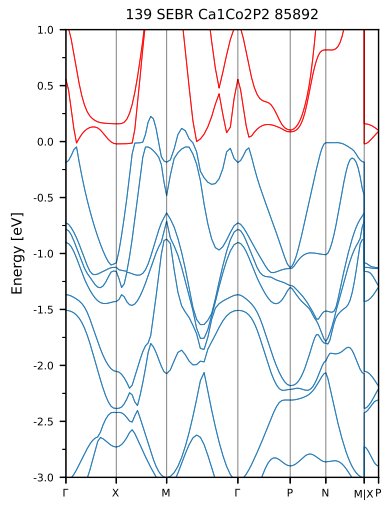

(fl) 139.1.168

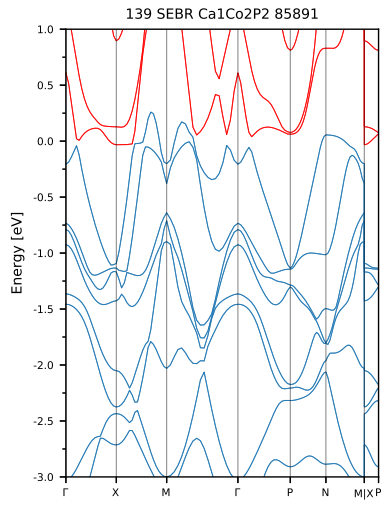

(fm) 139.1.169

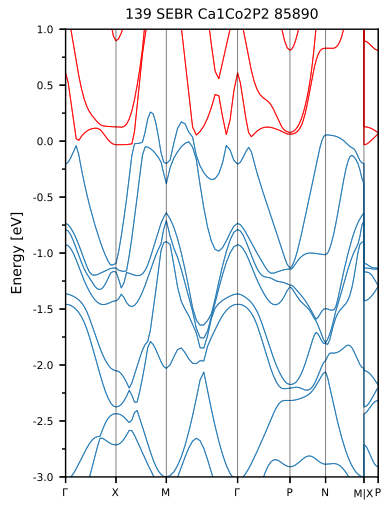

(fn) 139.1.170

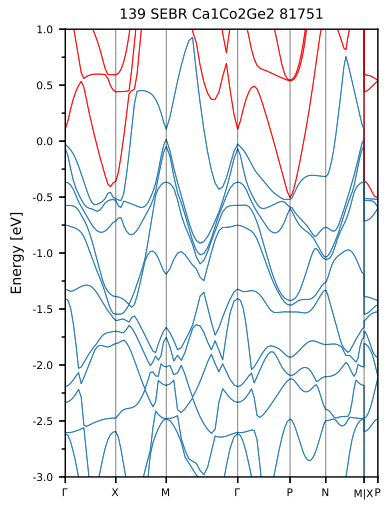

(fo) 139.1.171

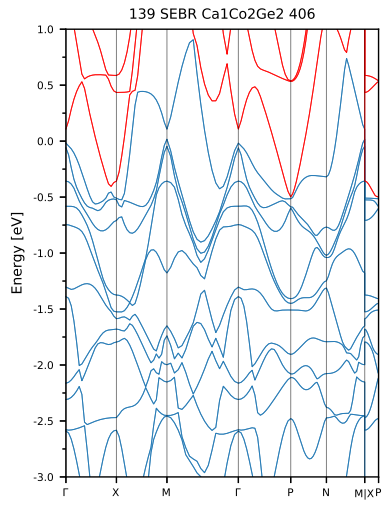

(fp) 139.1.172

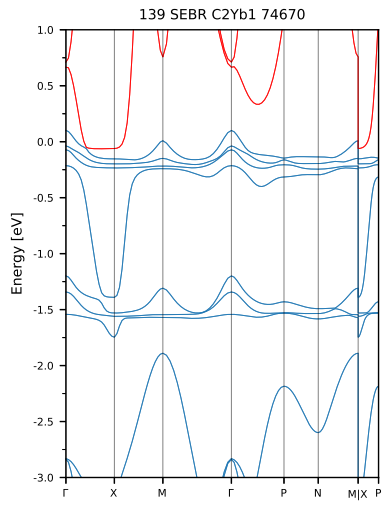

(fq) 139.1.173

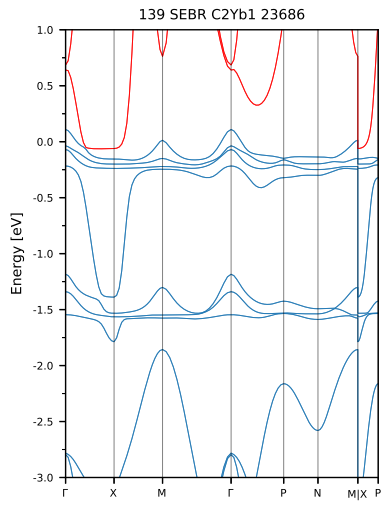

(fr) 139.1.174

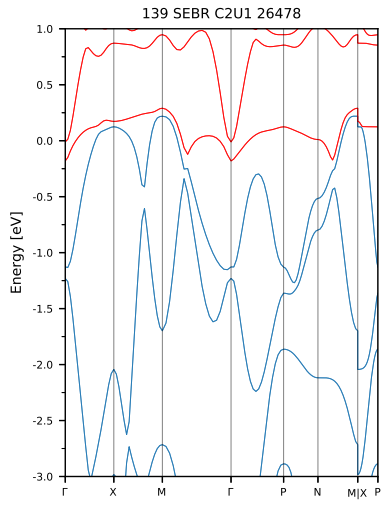

(fs) 139.1.175

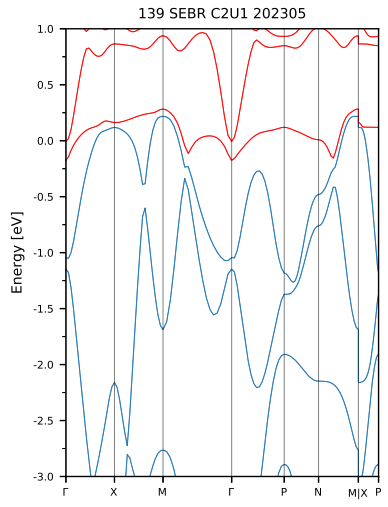

(ft) 139.1.176

Figure 298: Band structures for topological materials in SG 139, part 11/22, from 139.1.161 to 139.1.176. See the list of figures.

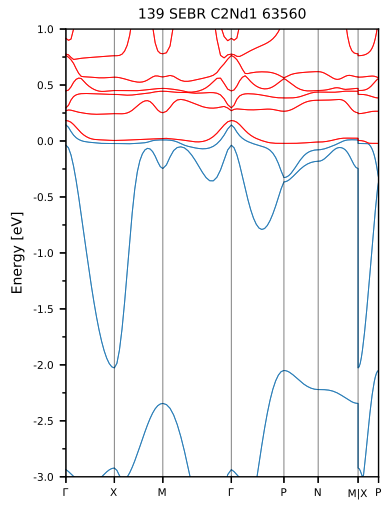

(fu) 139.1.177

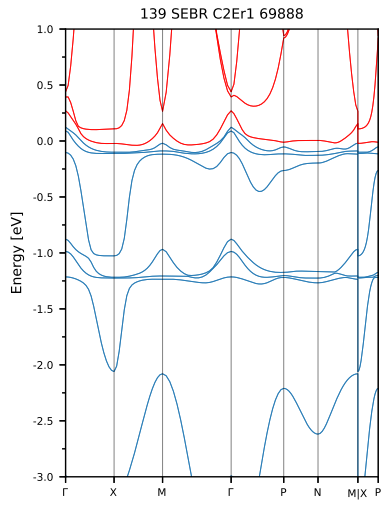

(fv) 139.1.178

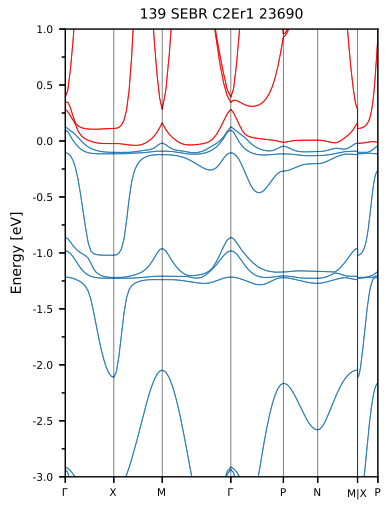

(fw) 139.1.179

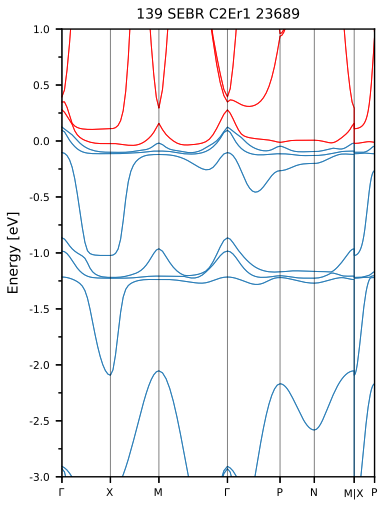

(fx) 139.1.180

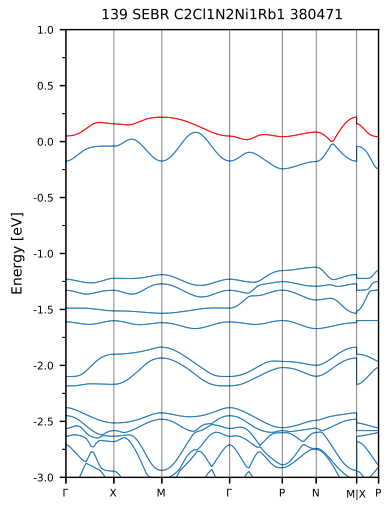

(fy) 139.1.181

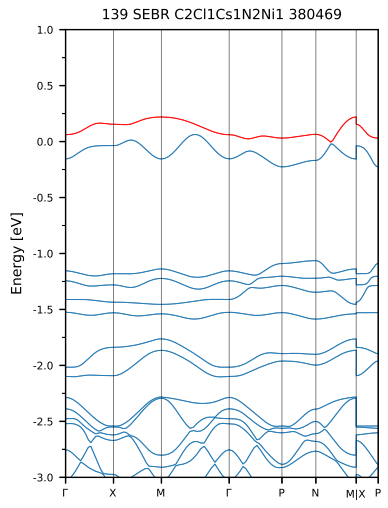

(fz) 139.1.182

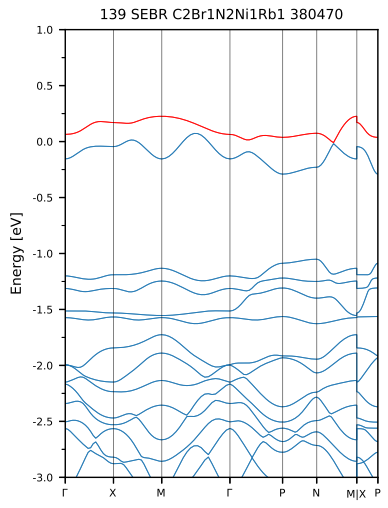

(ga) 139.1.183

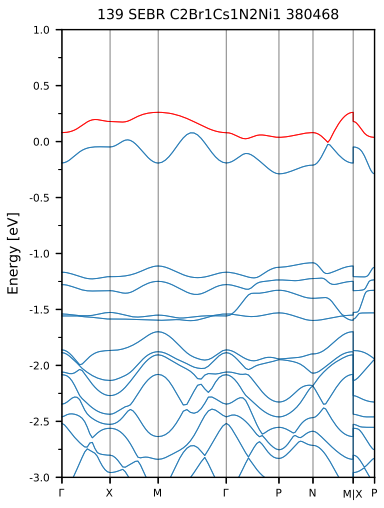

(gb) 139.1.184

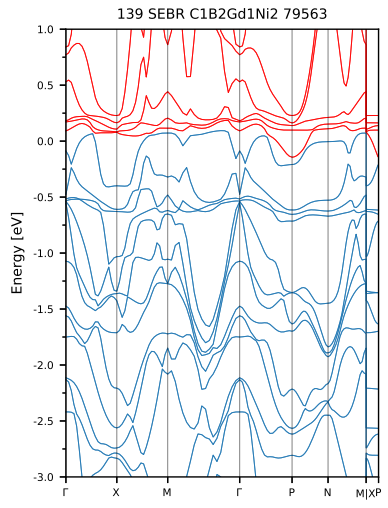

(gc) 139.1.185

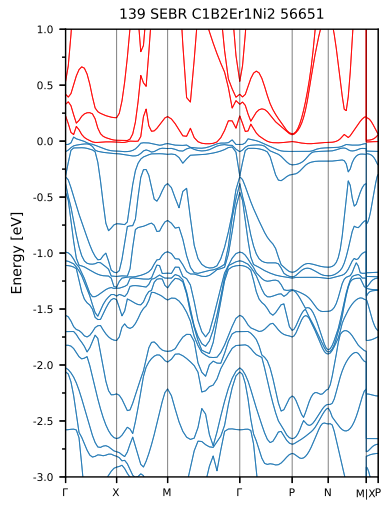

(gd) 139.1.186

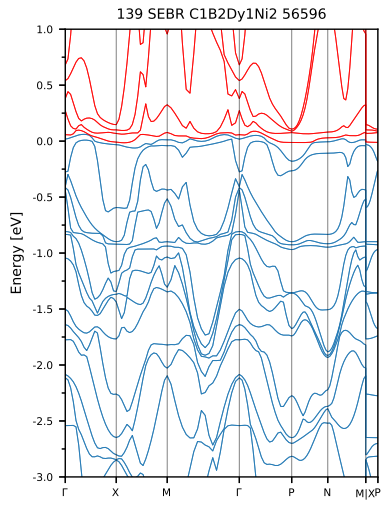

(ge) 139.1.187

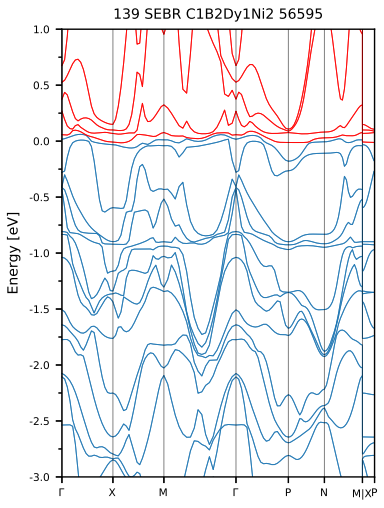

(gf) 139.1.188

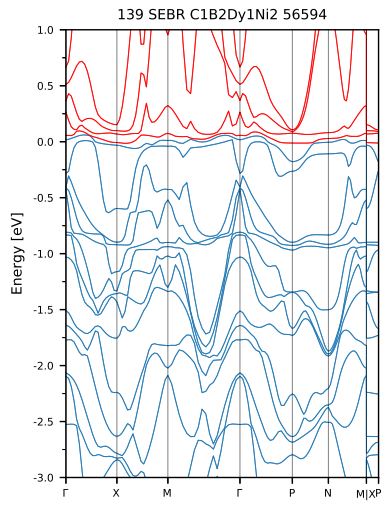

(gg) 139.1.189

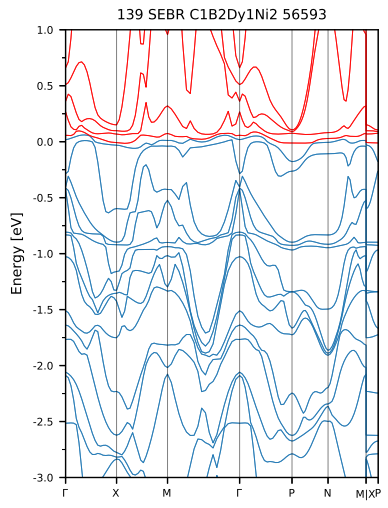

(gh) 139.1.190

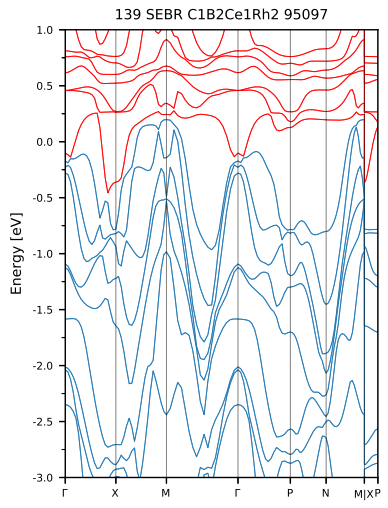

(gi) 139.1.191

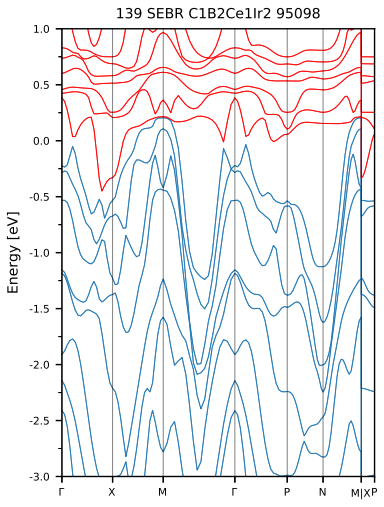

(gj) 139.1.192

Figure 299: Band structures for topological materials in SG 139, part 12/22, from 139.1.177 to 139.1.192. See the list of figures.

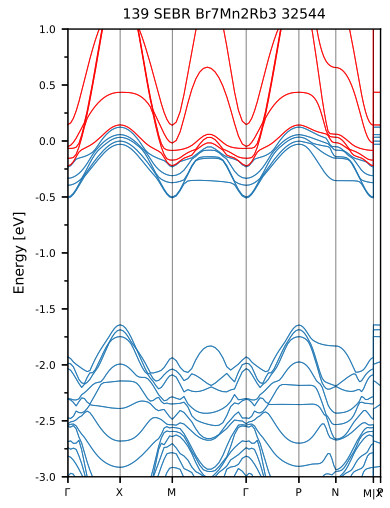

(gk) 139.1.193

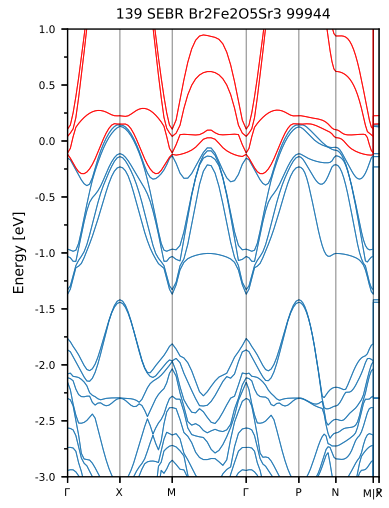

(gl) 139.1.194

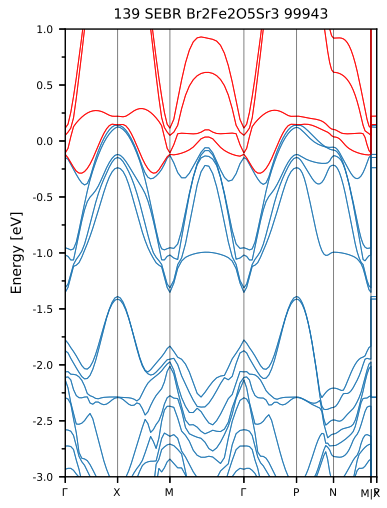

(gm) 139.1.195

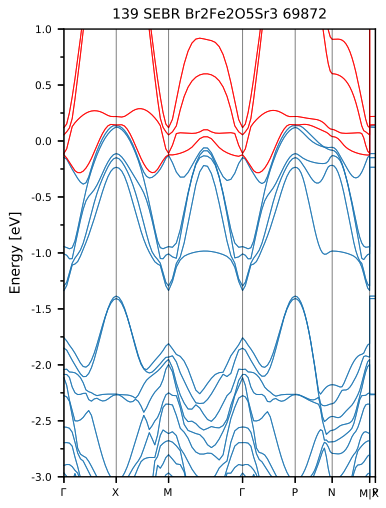

(gn) 139.1.196

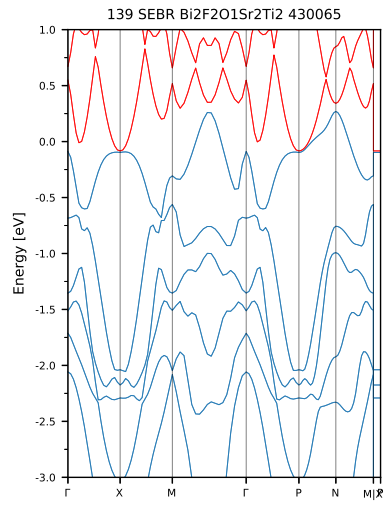

(go) 139.1.197

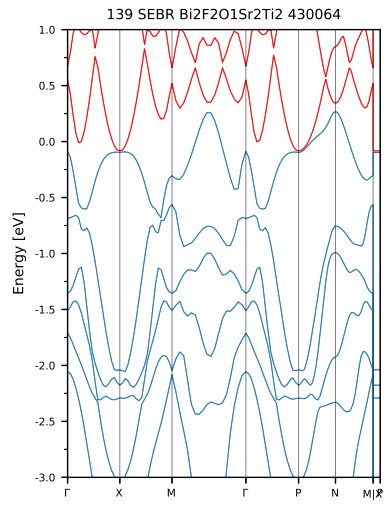

(gp) 139.1.198

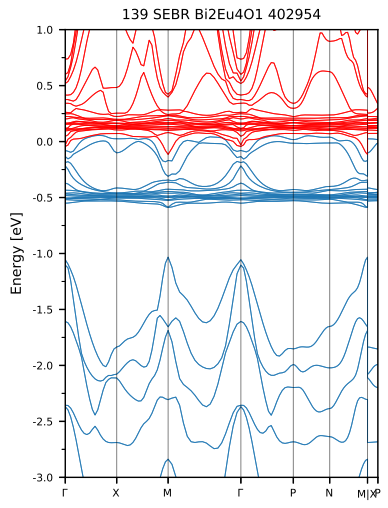

(gq) 139.1.199

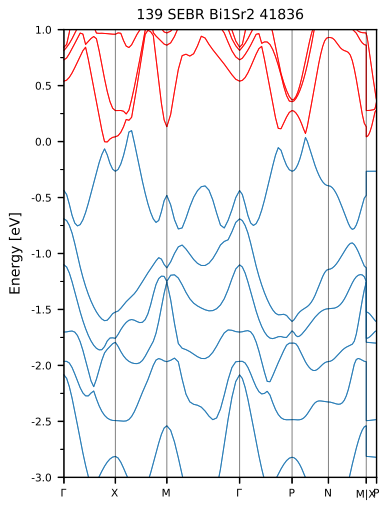

(gr) 139.1.200

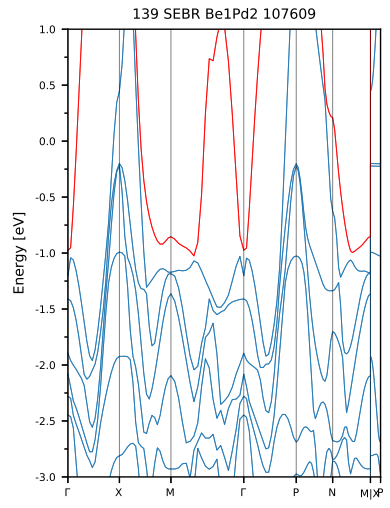

(gs) 139.1.201

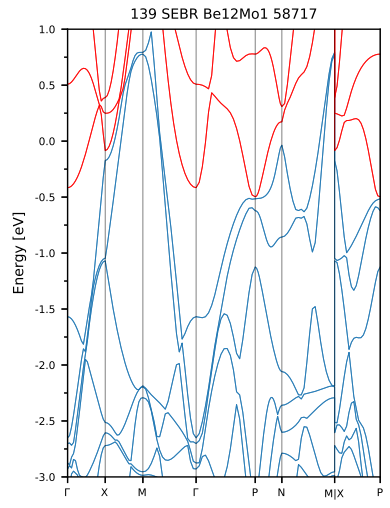

(gt) 139.1.202

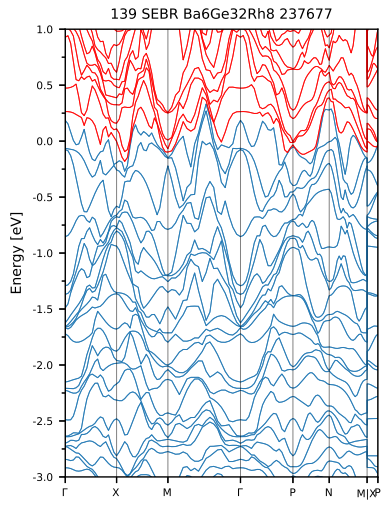

(gu) 139.1.203

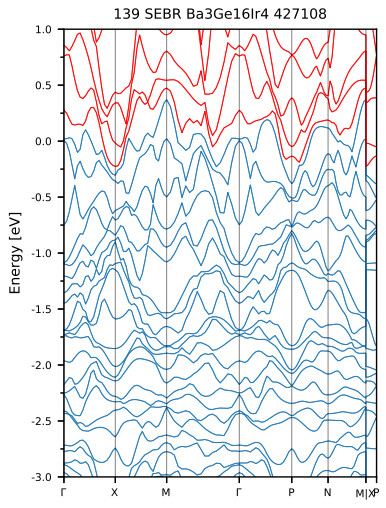

(gv) 139.1.204

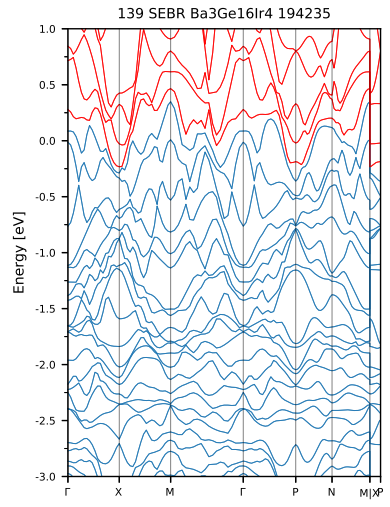

(gw) 139.1.205

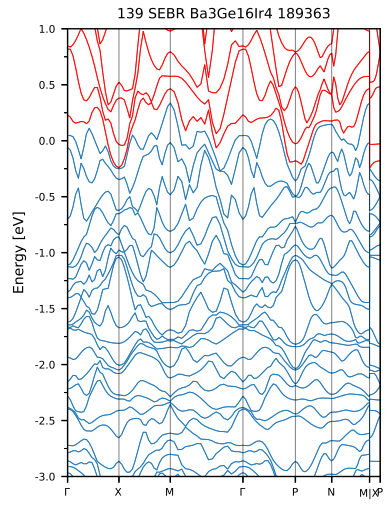

(gx) 139.1.206

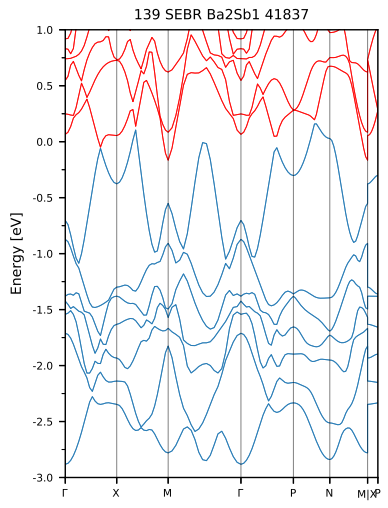

(gy) 139.1.207

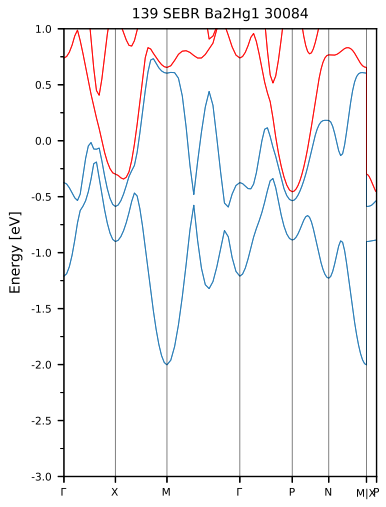

(gz) 139.1.208

Figure 300: Band structures for topological materials in SG 139, part 13/22, from 139.1.193 to 139.1.208. See the list of figures.

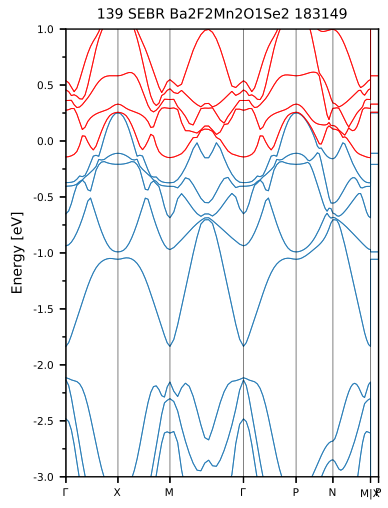

(ha) 139.1.209

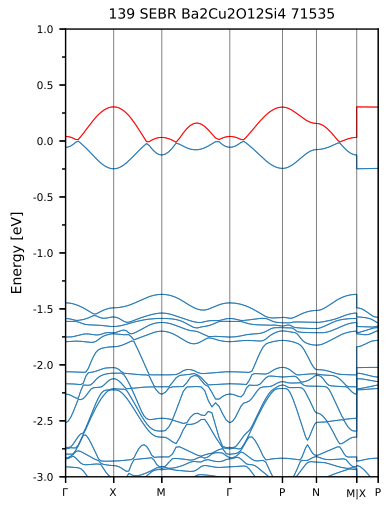

(hb) 139.1.210

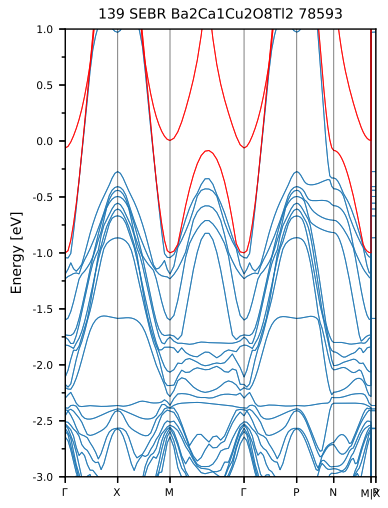

(hc) 139.1.211

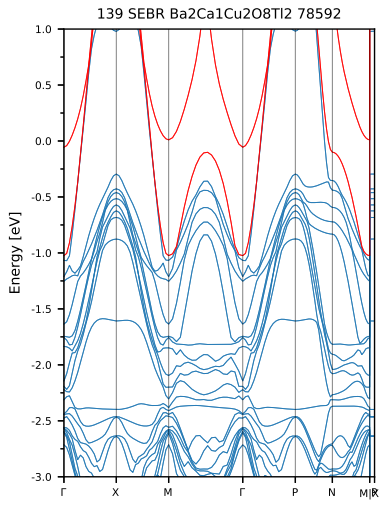

(hd) 139.1.212

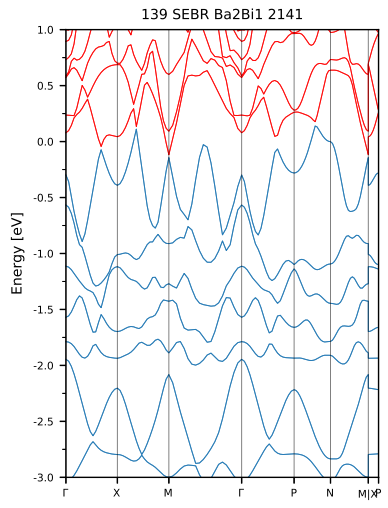

(he) 139.1.213

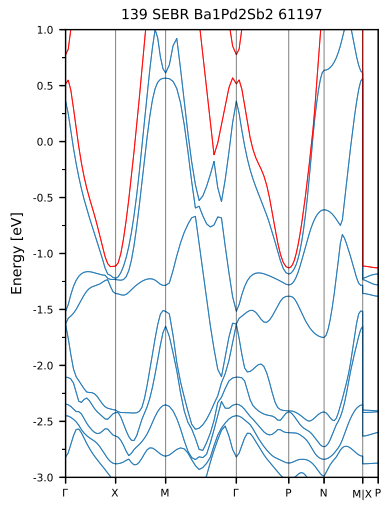

(hf) 139.1.214

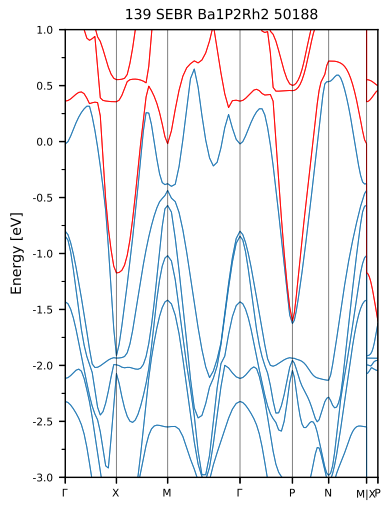

(hg) 139.1.215

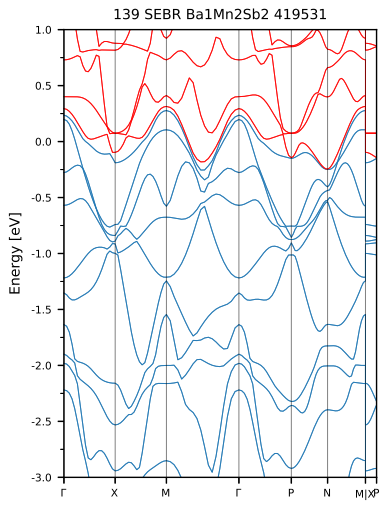

(hh) 139.1.216

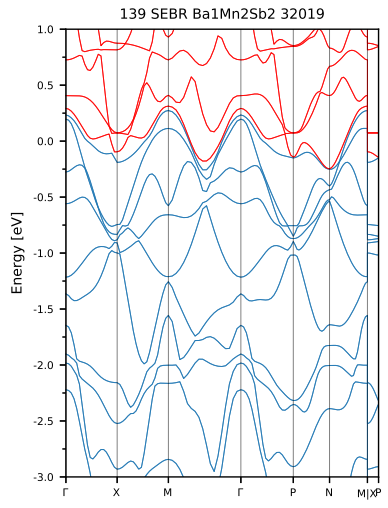

(hi) 139.1.217

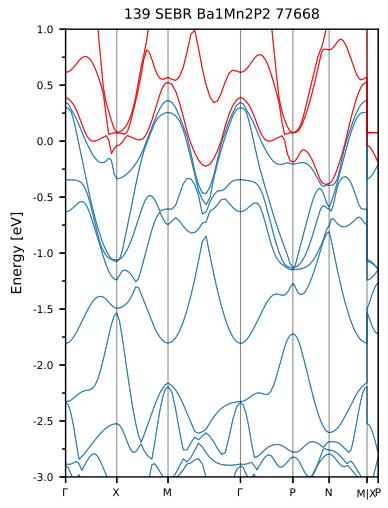

(hj) 139.1.218

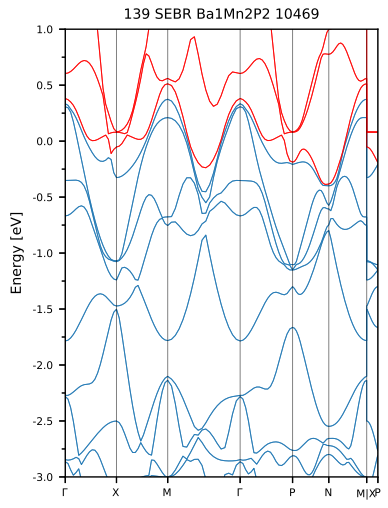

(hk) 139.1.219

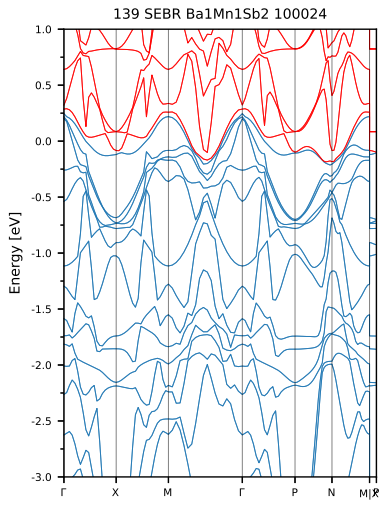

(hl) 139.1.220

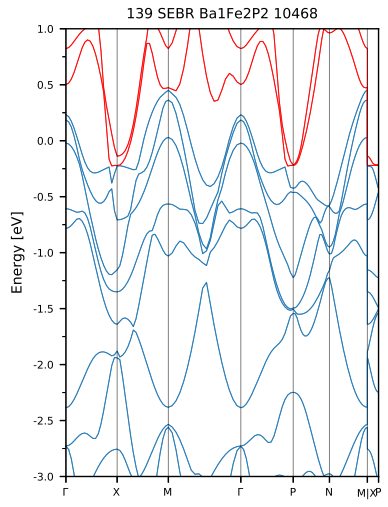

(hm) 139.1.221

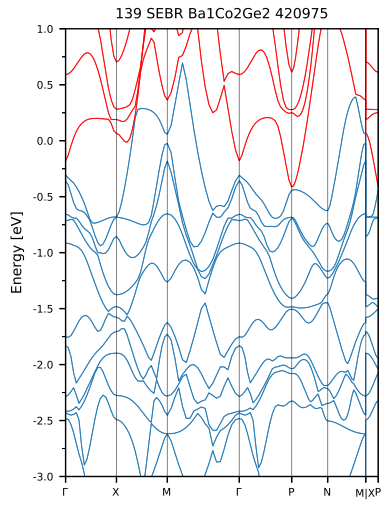

(hn) 139.1.222

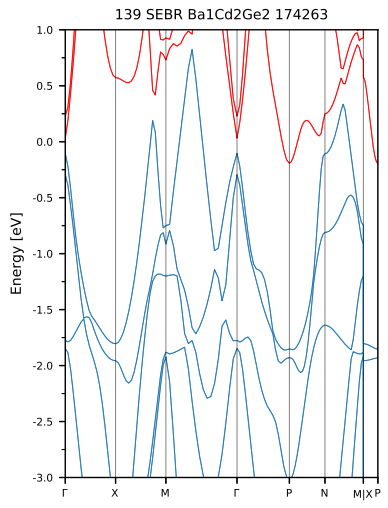

(ho) 139.1.223

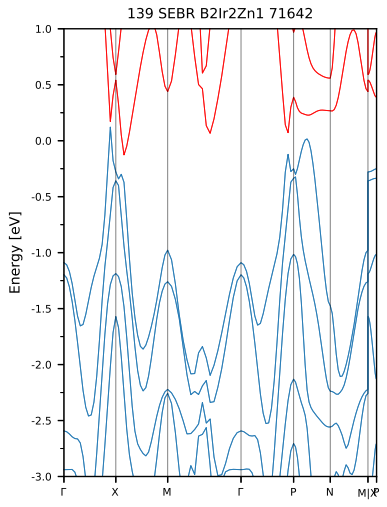

(hp) 139.1.224

Figure 301: Band structures for topological materials in SG 139, part 14/22, from 139.1.209 to 139.1.224. See the list of figures.

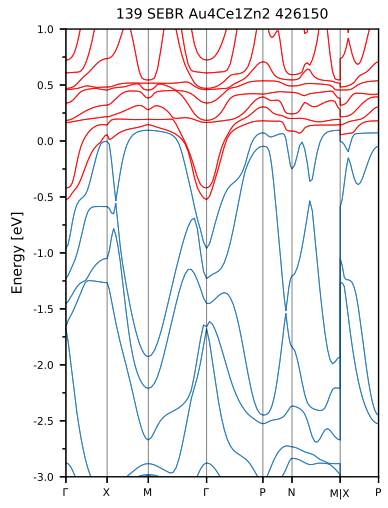

(hq) 139.1.225

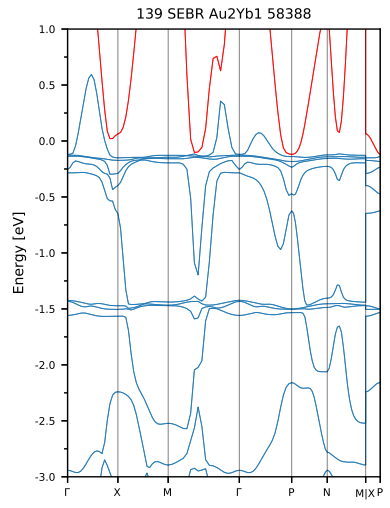

(hr) 139.1.226

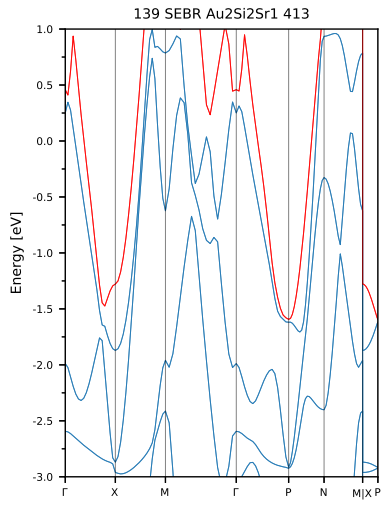

(hs) 139.1.227

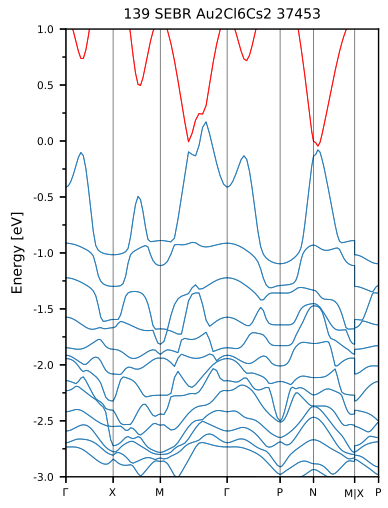

(ht) 139.1.228

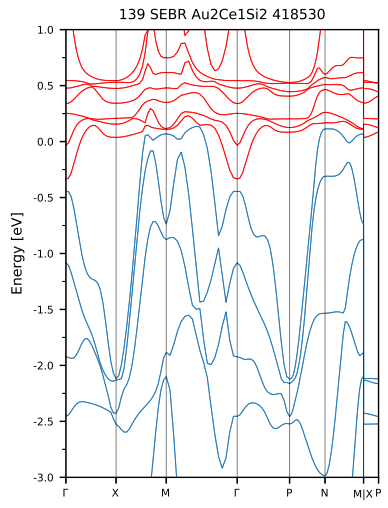

(hu) 139.1.229

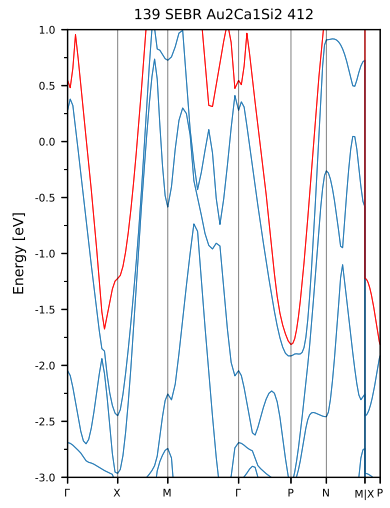

(hv) 139.1.230

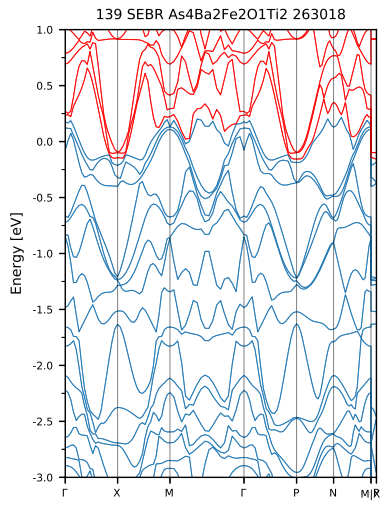

(hw) 139.1.231

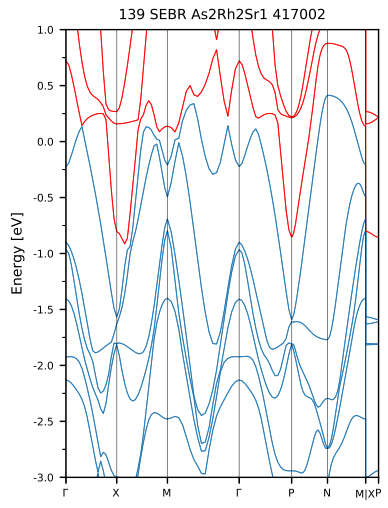

(hx) 139.1.232

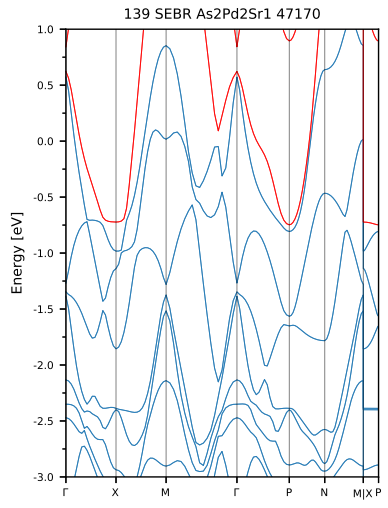

(hy) 139.1.233

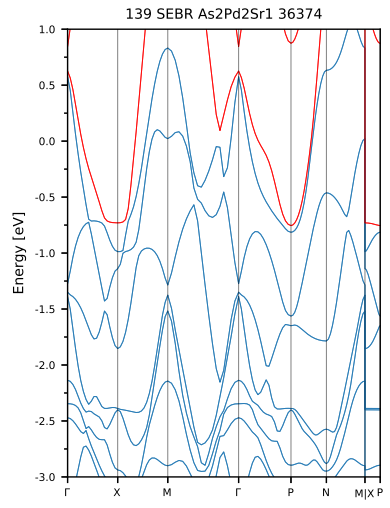

(hz) 139.1.234

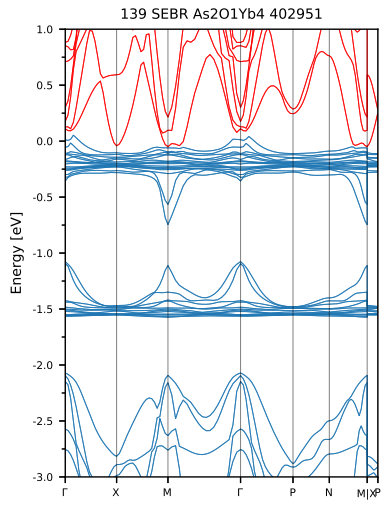

(ia) 139.1.235

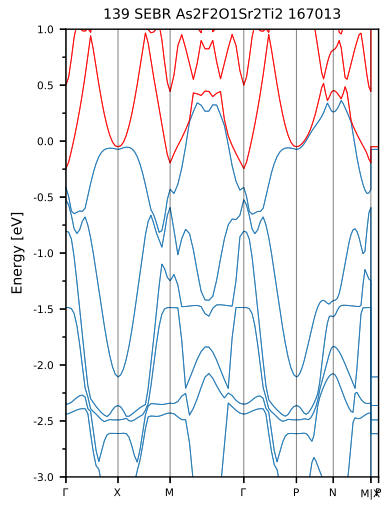

(ib) 139.1.236

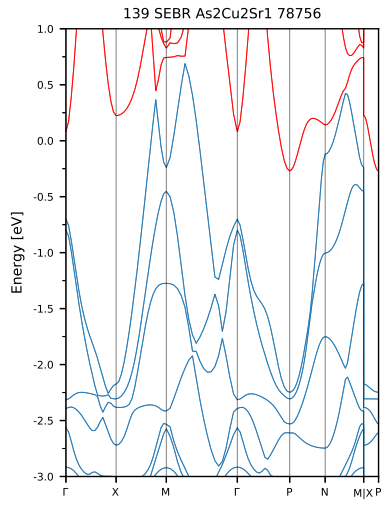

(ic) 139.1.237

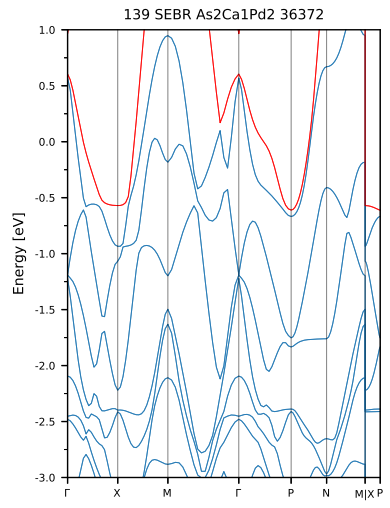

(id) 139.1.238

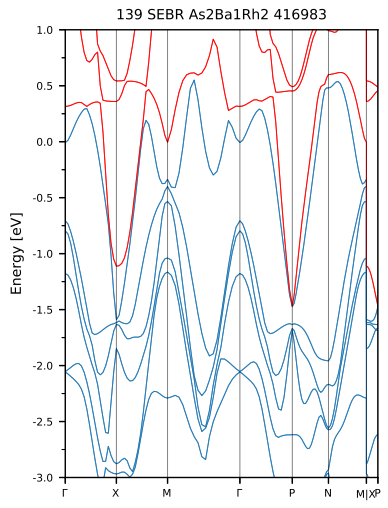

(ie) 139.1.239

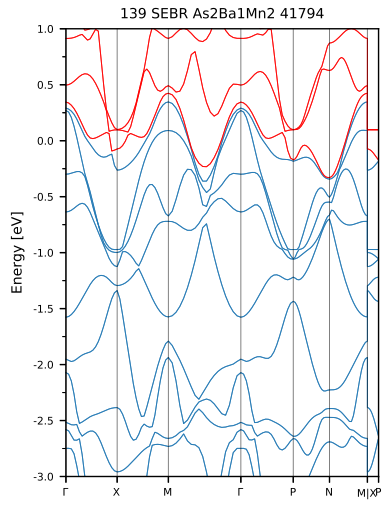

(if) 139.1.240

Figure 302: Band structures for topological materials in SG 139, part 15/22, from 139.1.225 to 139.1.240. See the list of figures.

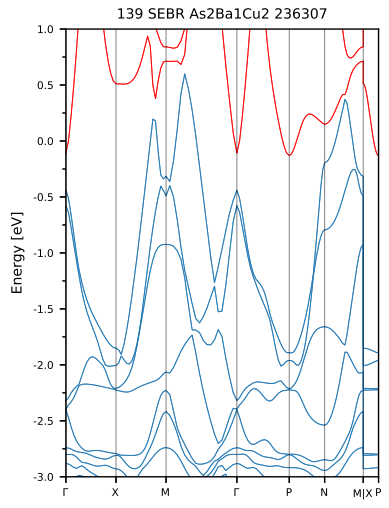

(ig) 139.1.241

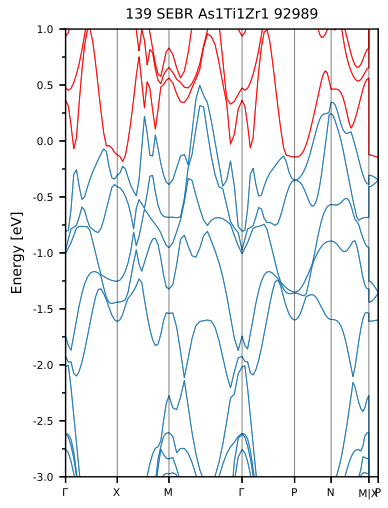

(ih) 139.1.242

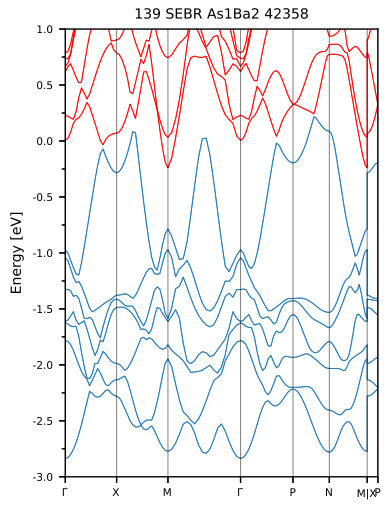

(ii) 139.1.243

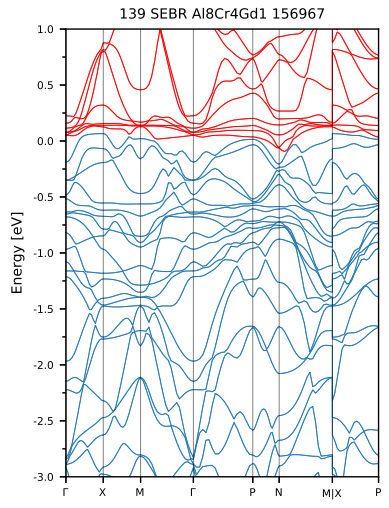

(ij) 139.1.244

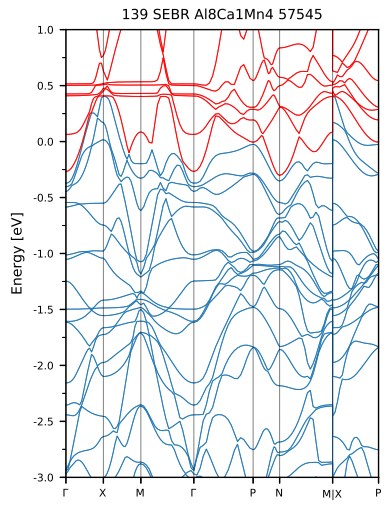

(ik) 139.1.245

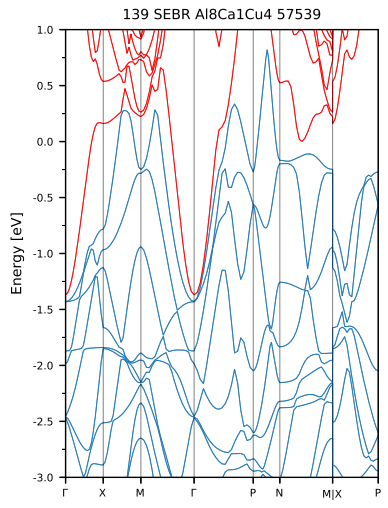

(il) 139.1.246

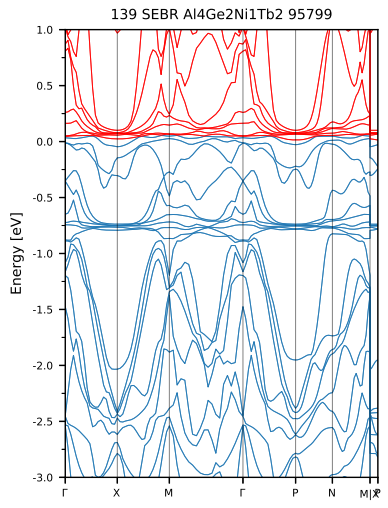

(im) 139.1.247

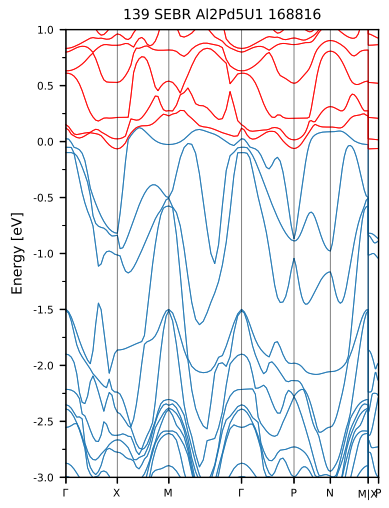

(in) 139.1.248

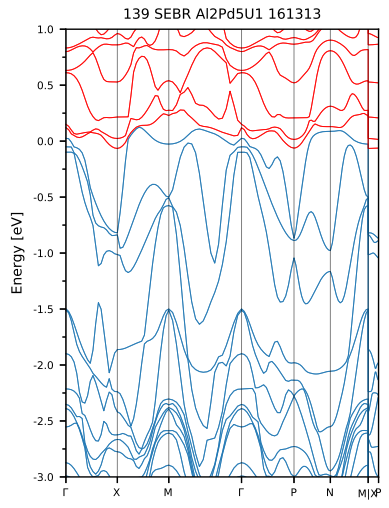

(io) 139.1.249

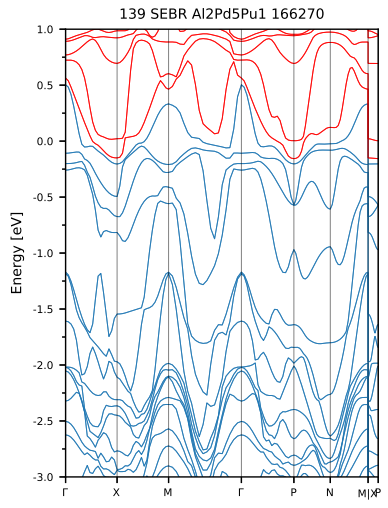

(ip) 139.1.250

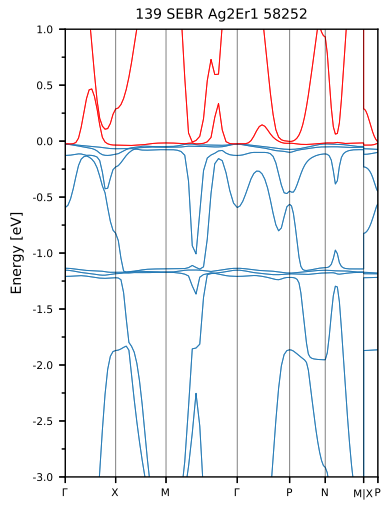

(iq) 139.1.251

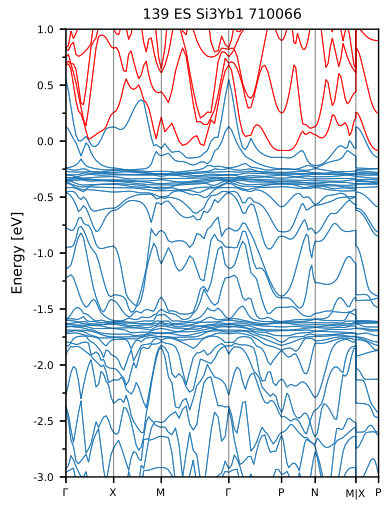

(ir) 139.3.1

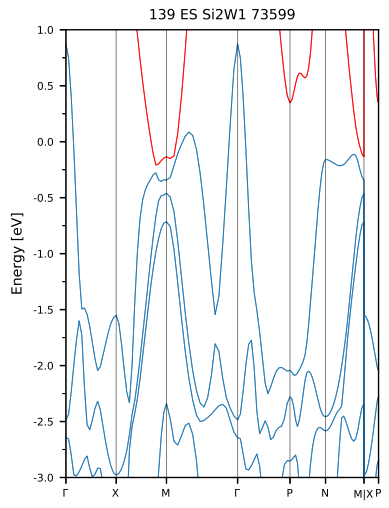

(is) 139.3.2

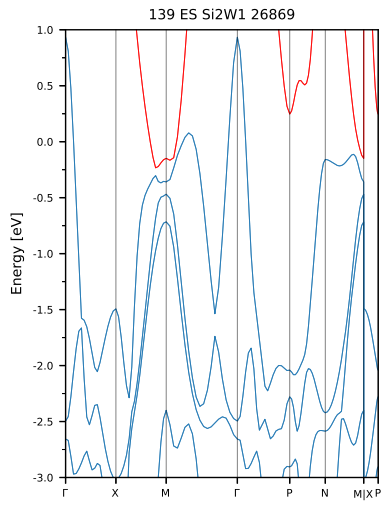

(it) 139.3.3

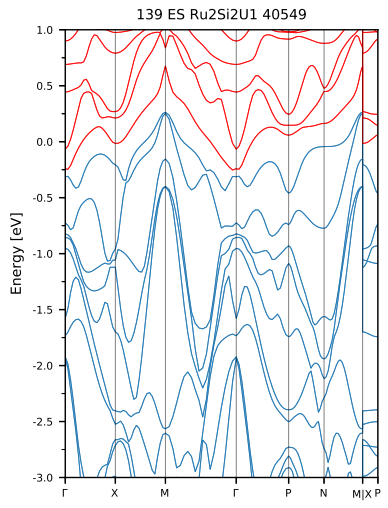

(iu) 139.3.4

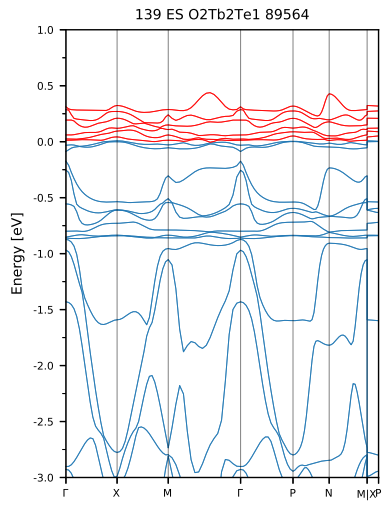

(iv) 139.3.5

Figure 303: Band structures for topological materials in SG 139, part 16/22, from 139.1.241 to 139.3.5. See the list of figures.

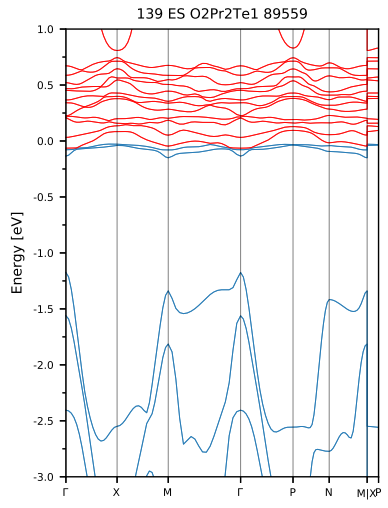

(iw) 139.3.6

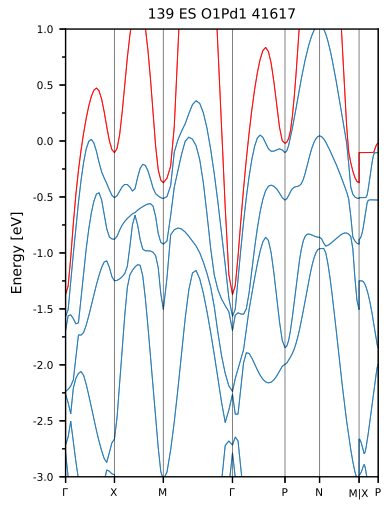

(ix) 139.3.7

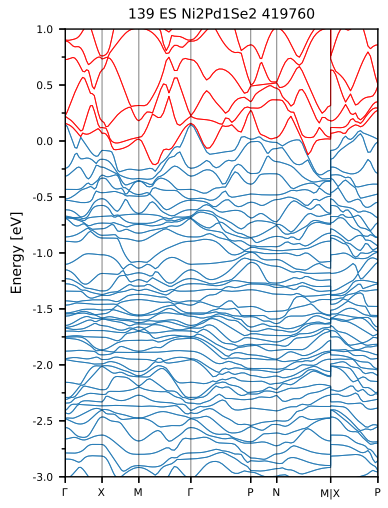

(iy) 139.3.8

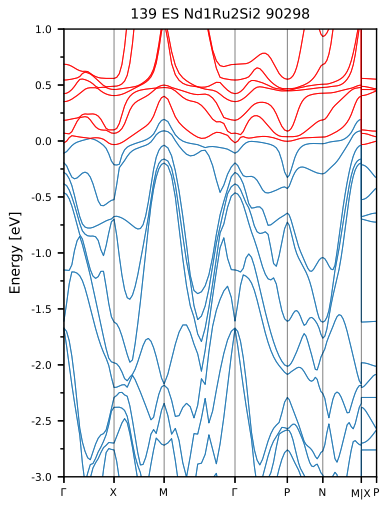

(iz) 139.3.9

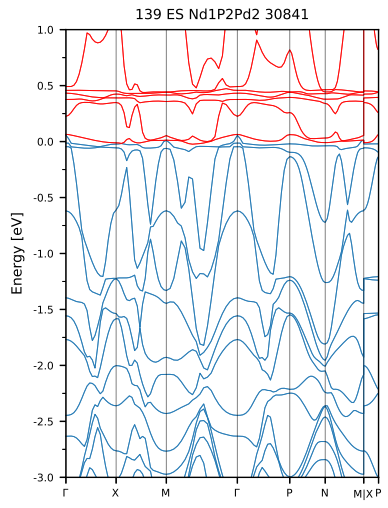

(ja) 139.3.10

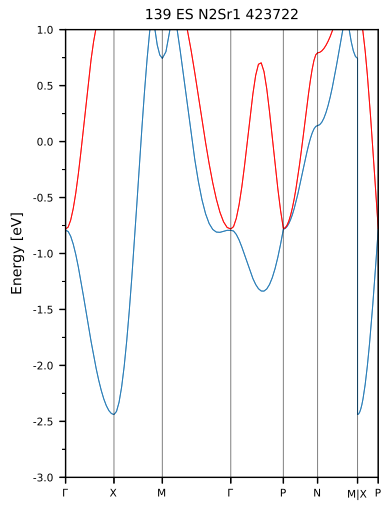

(jb) 139.3.11

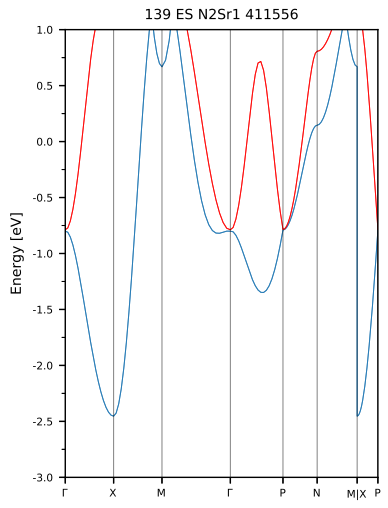

(jc) 139.3.12

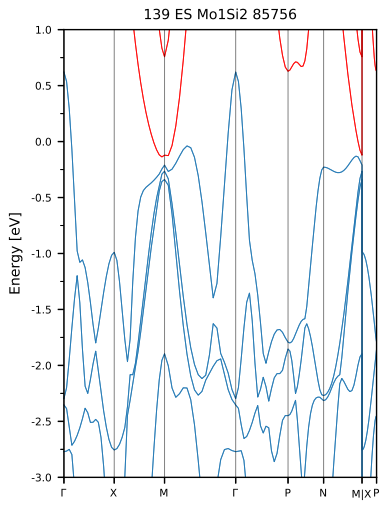

(jd) 139.3.13

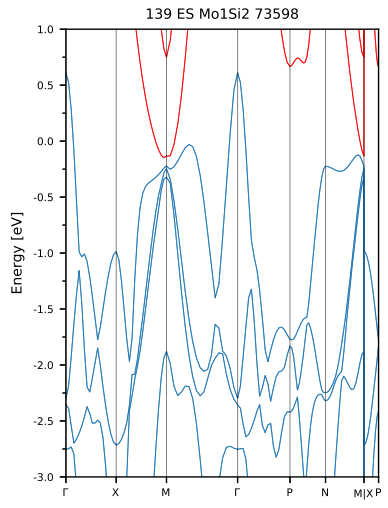

(je) 139.3.14

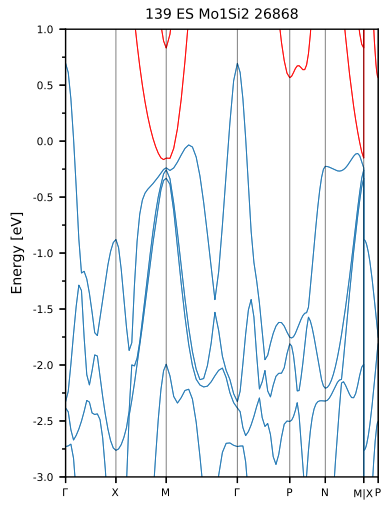

(jf) 139.3.15

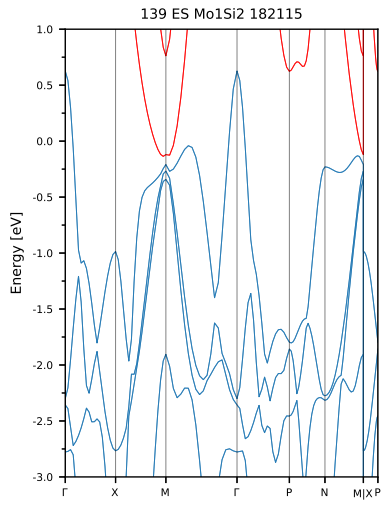

(jg) 139.3.16

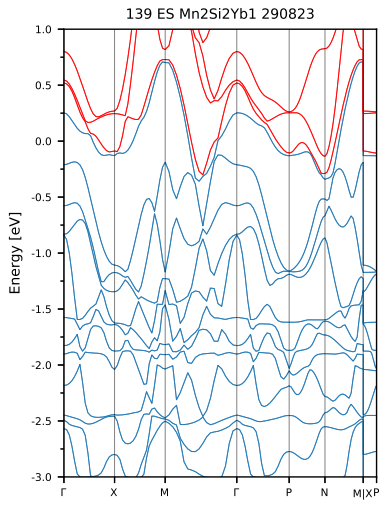

(jh) 139.3.17

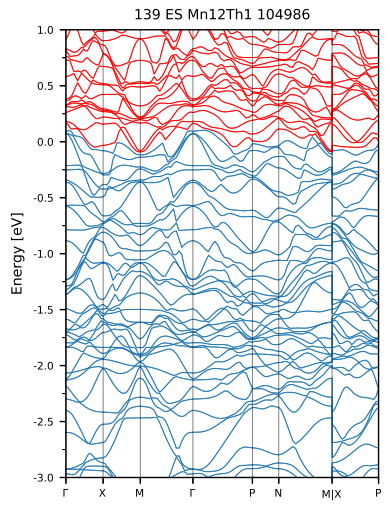

(ji) 139.3.18

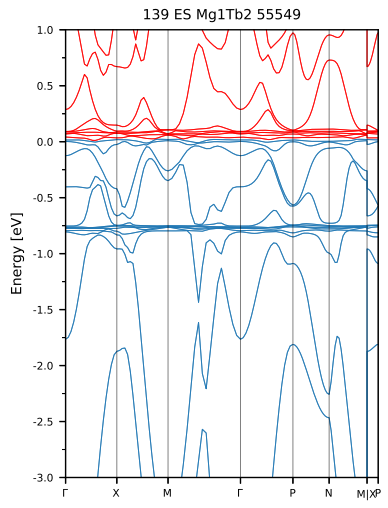

(jj) 139.3.19

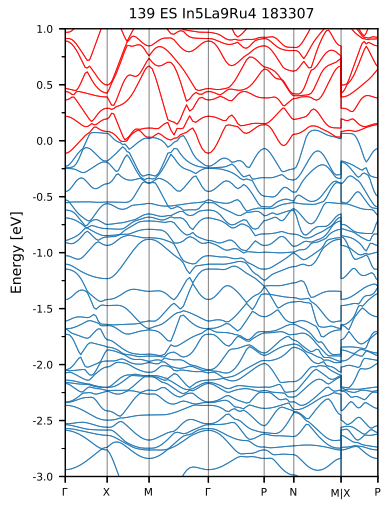

(jk) 139.3.20

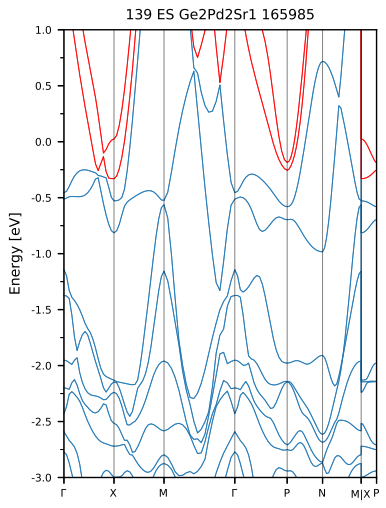

(jl) 139.3.21

Figure 304: Band structures for topological materials in SG 139, part 17/22, from 139.3.6 to 139.3.21. See the list of figures.

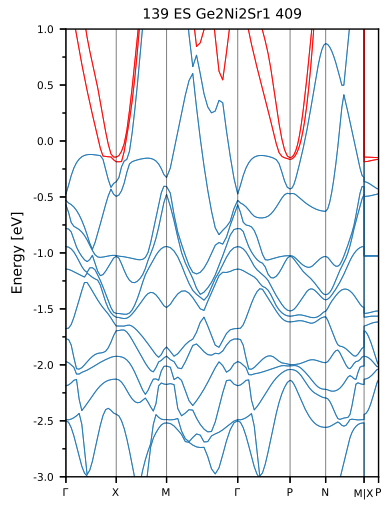

(jm) 139.3.22

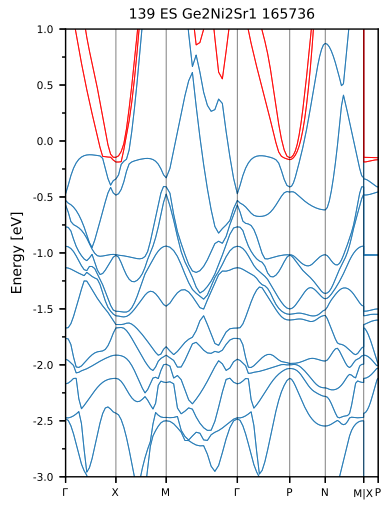

(jn) 139.3.23

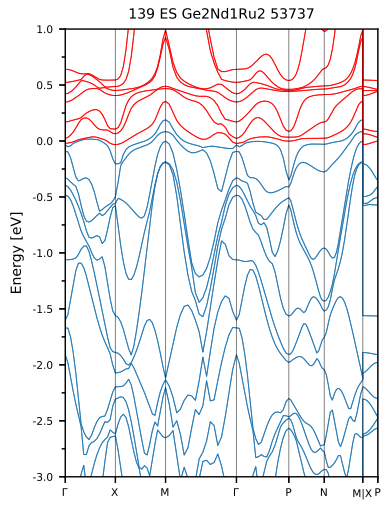

(jo) 139.3.24

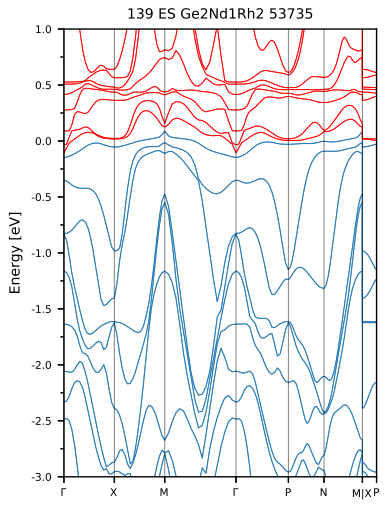

(jp) 139.3.25

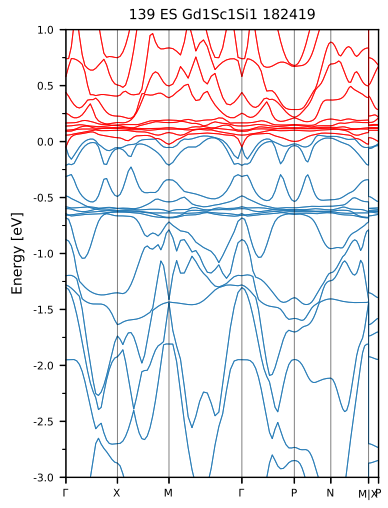

(jq) 139.3.26

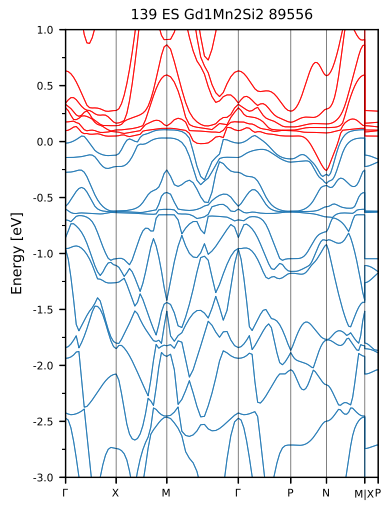

(jr) 139.3.27

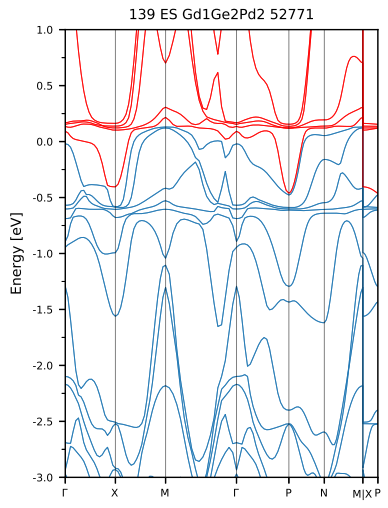

(js) 139.3.28

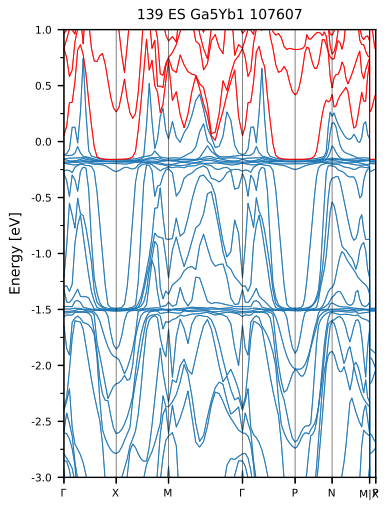

(jt) 139.3.29

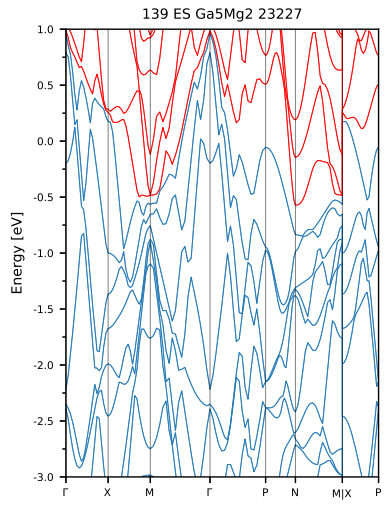

(ju) 139.3.30

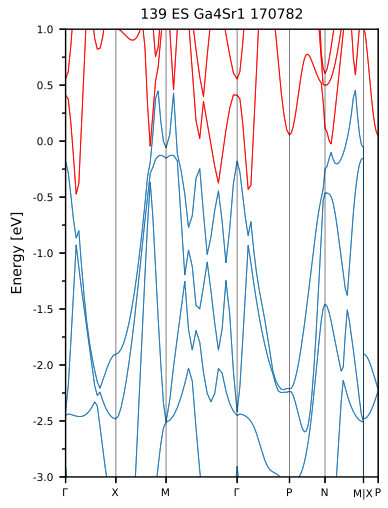

(jv) 139.3.31

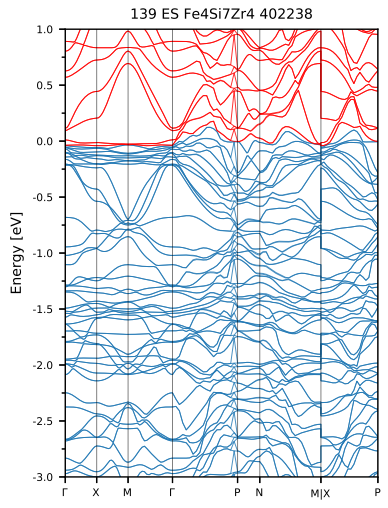

(jw) 139.3.32

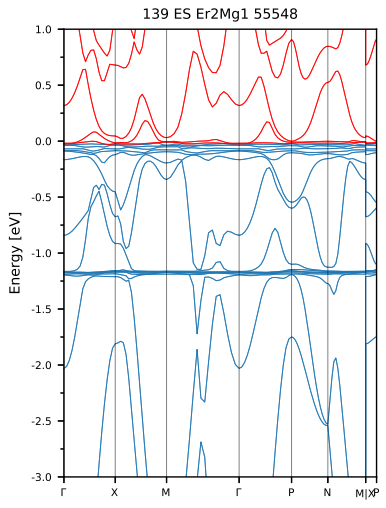

(jx) 139.3.33

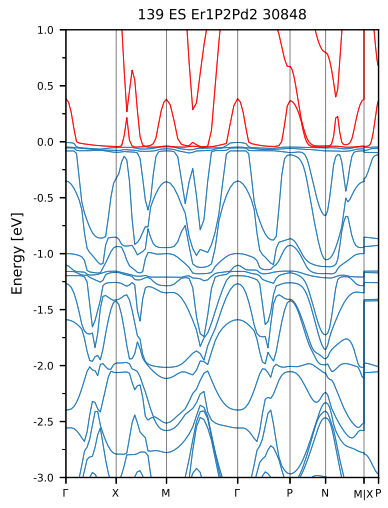

(jy) 139.3.34

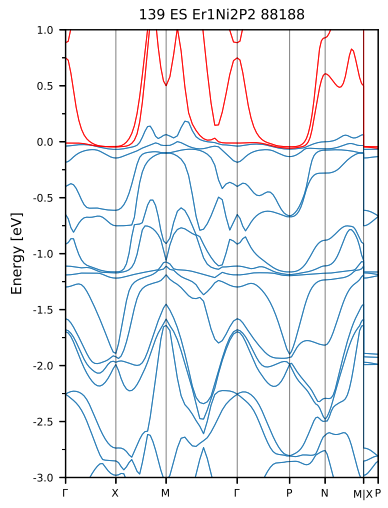

(jz) 139.3.35

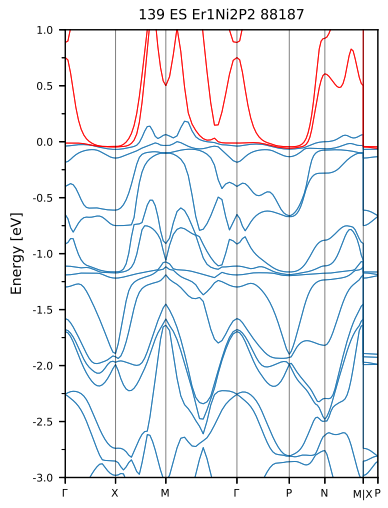

(ka) 139.3.36

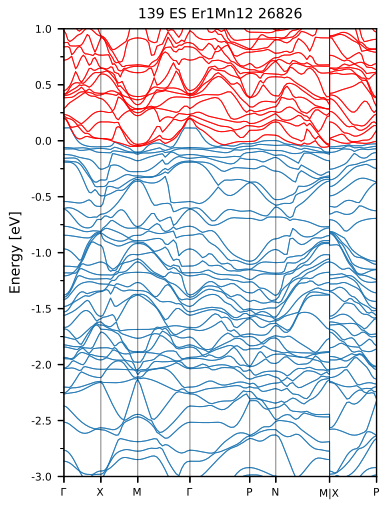

(kb) 139.3.37

Figure 305: Band structures for topological materials in SG 139, part 18/22, from 139.3.22 to 139.3.37. See the list of figures.

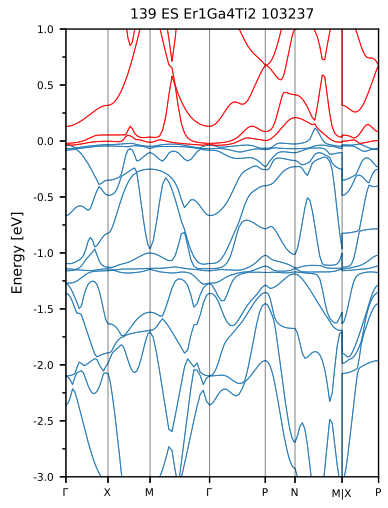

(kc) 139.3.38

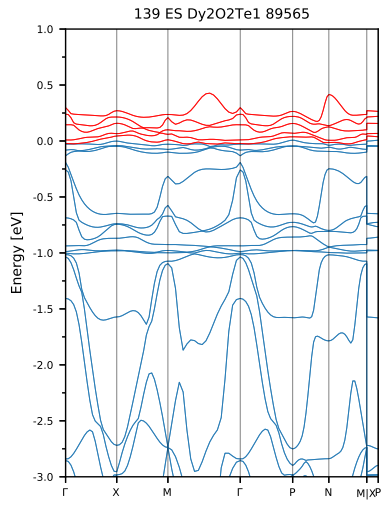

(kd) 139.3.39

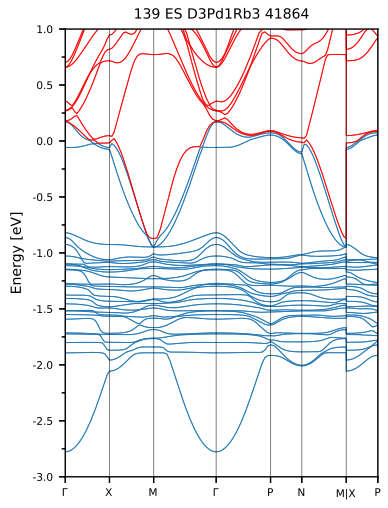

(ke) 139.3.40

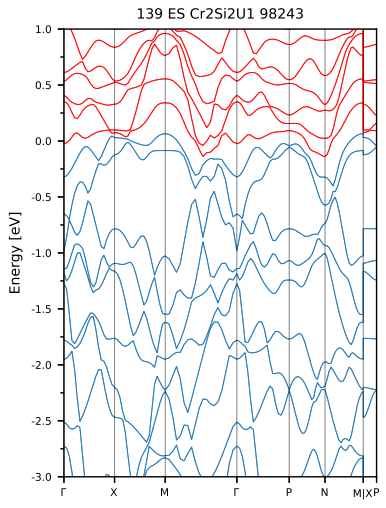

(kf) 139.3.41

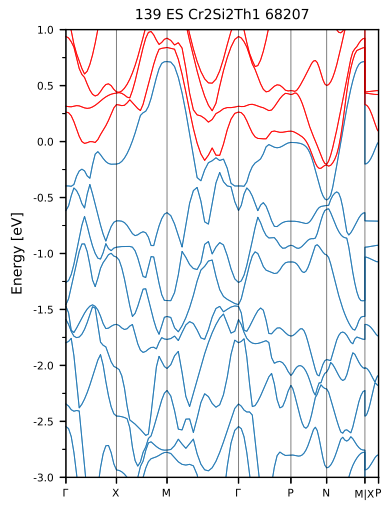

(kg) 139.3.42

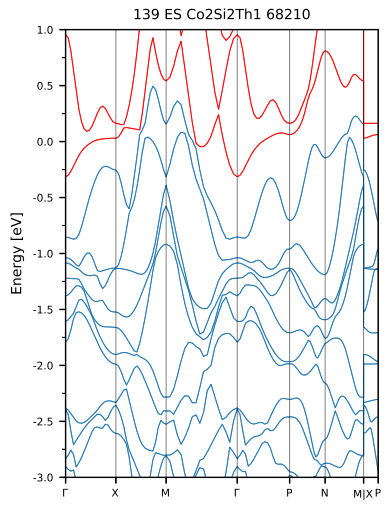

(kh) 139.3.43

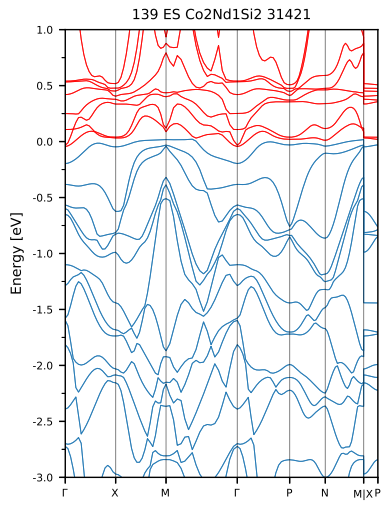

(ki) 139.3.44

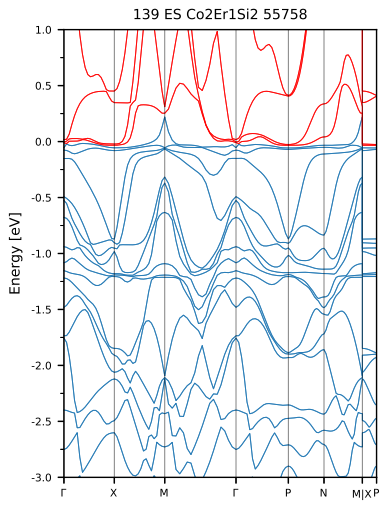

(kj) 139.3.45

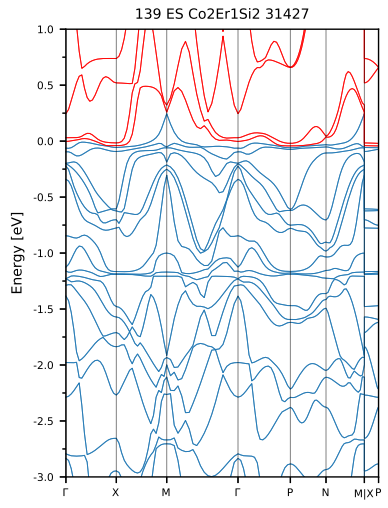

(kk) 139.3.46

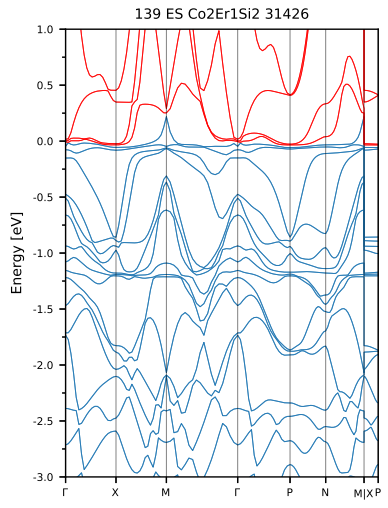

(kl) 139.3.47

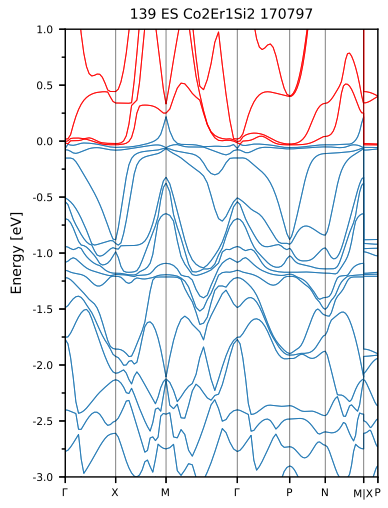

(km) 139.3.48

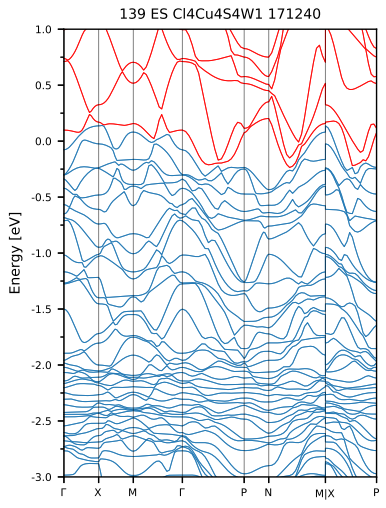

(kn) 139.3.49

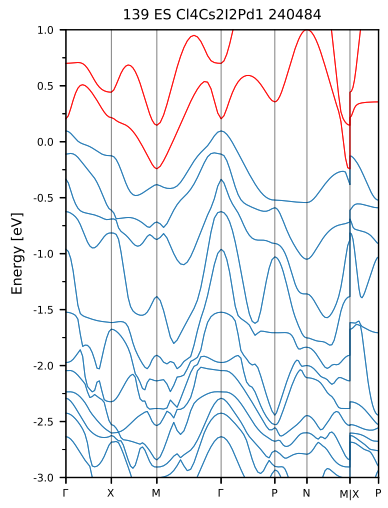

(ko) 139.3.50

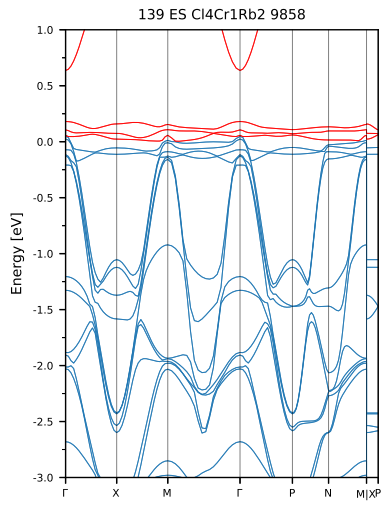

(kp) 139.3.51

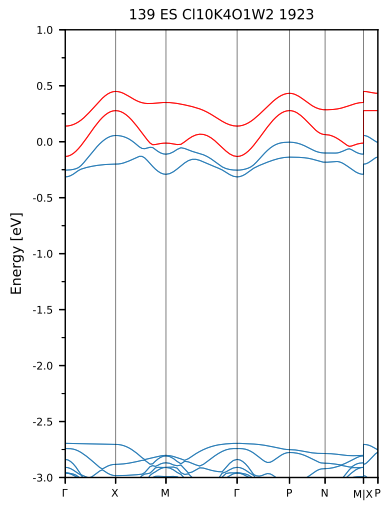

(kq) 139.3.52

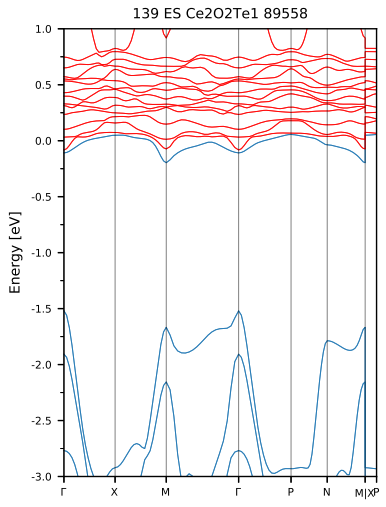

(kr) 139.3.53

Figure 306: Band structures for topological materials in SG 139, part 19/22, from 139.3.38 to 139.3.53. See the list of figures.

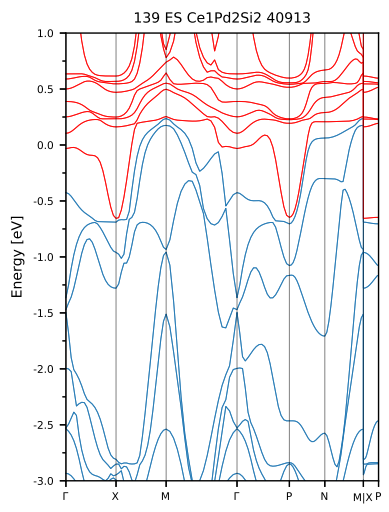

(ks) 139.3.54

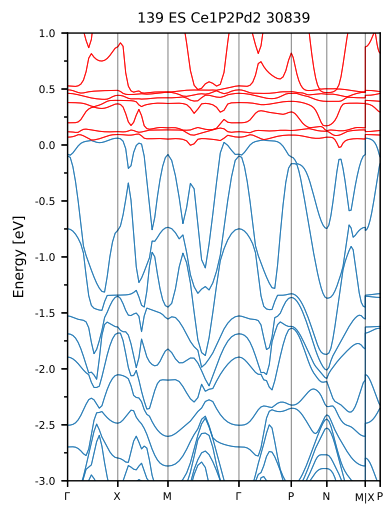

(kt) 139.3.55

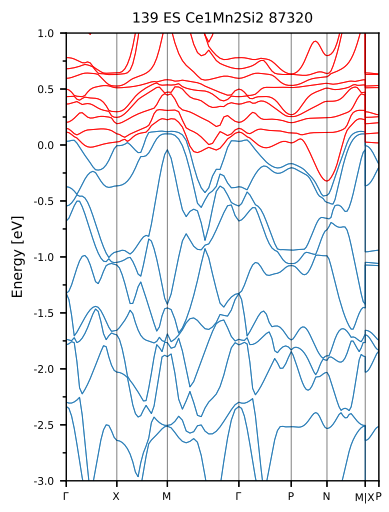

(ku) 139.3.56

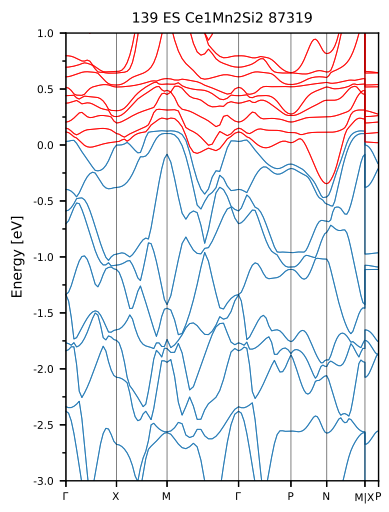

(kv) 139.3.57

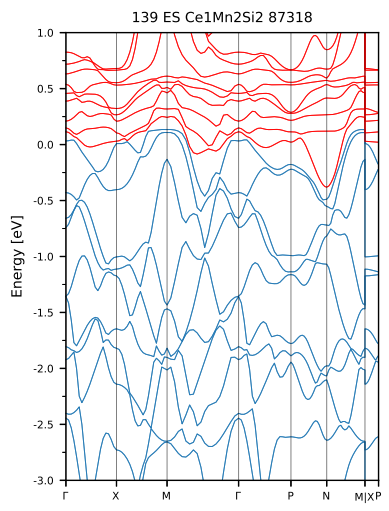

(kw) 139.3.58

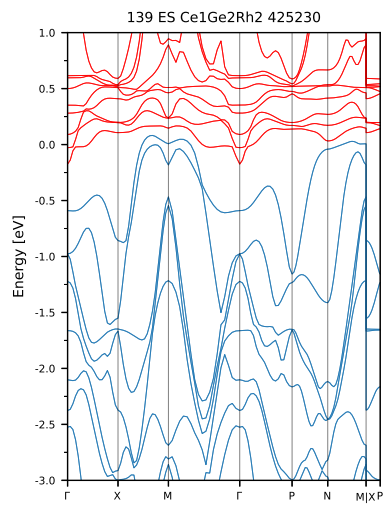

(kx) 139.3.59

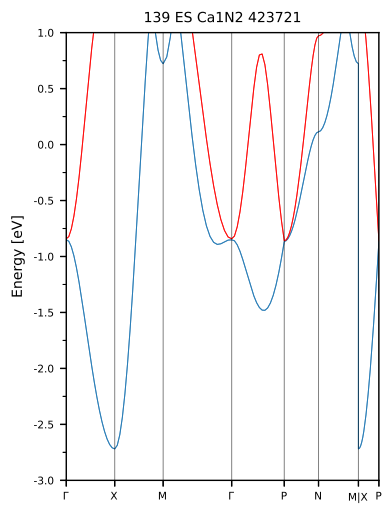

(ky) 139.3.60

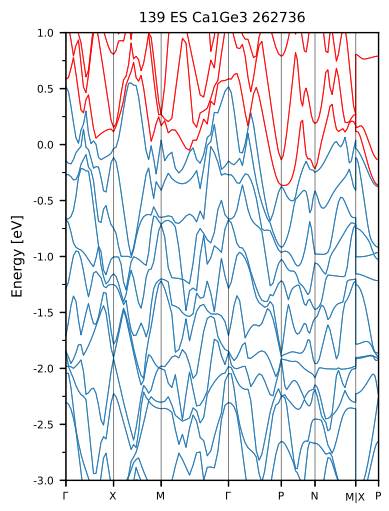

(kz) 139.3.61

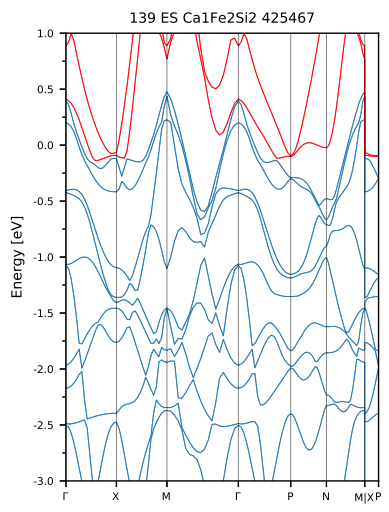

(la) 139.3.62

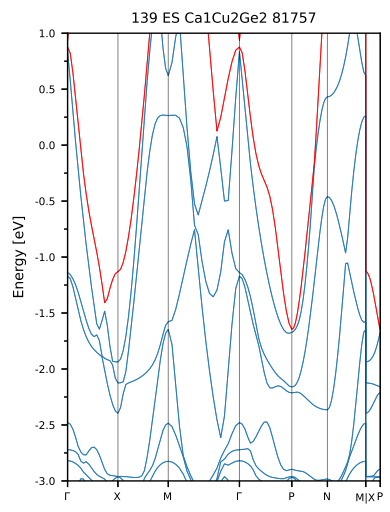

(lb) 139.3.63

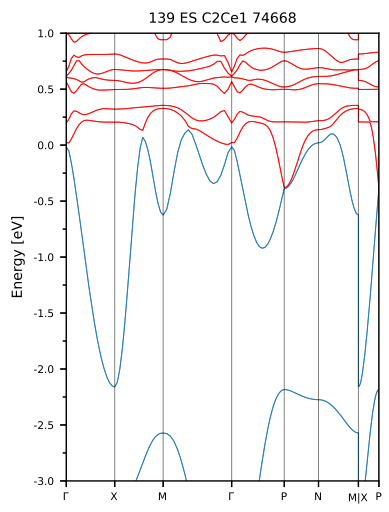

(lc) 139.3.64

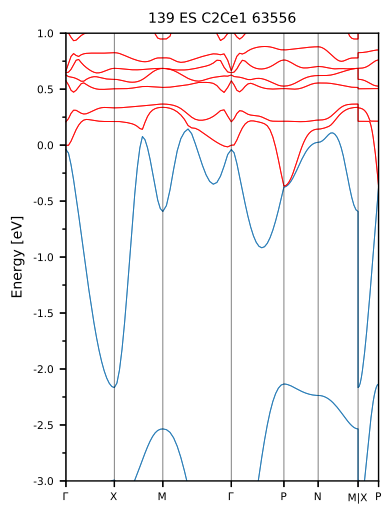

(ld) 139.3.65

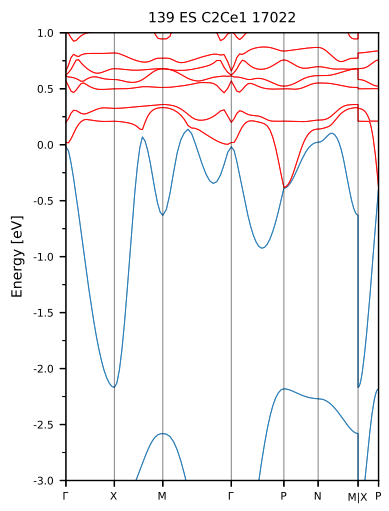

(le) 139.3.66

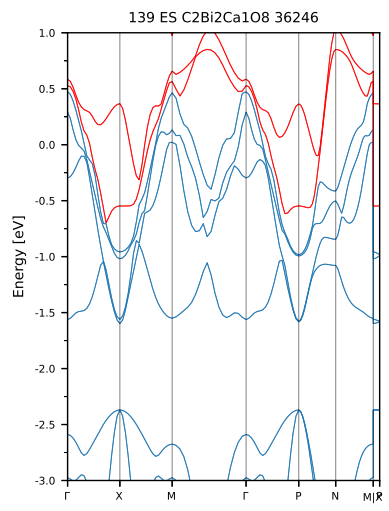

(lf) 139.3.67

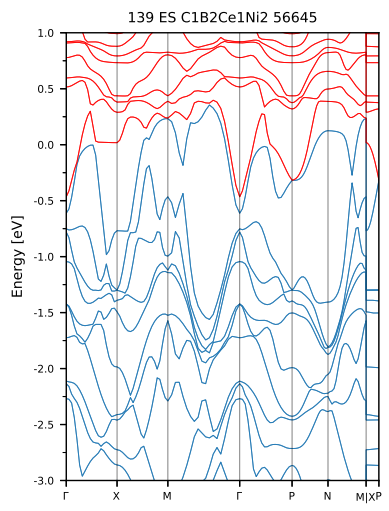

(lg) 139.3.68

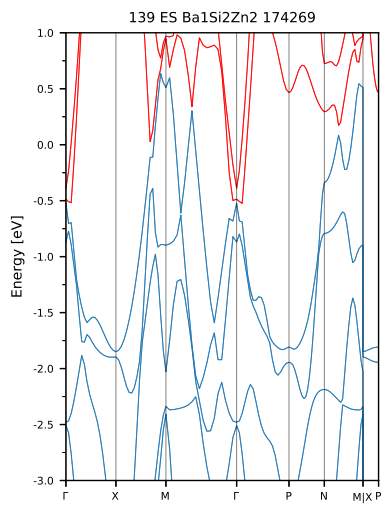

(lh) 139.3.69

Figure 307: Band structures for topological materials in SG 139, part 20/22, from 139.3.54 to 139.3.69. See the list of figures.

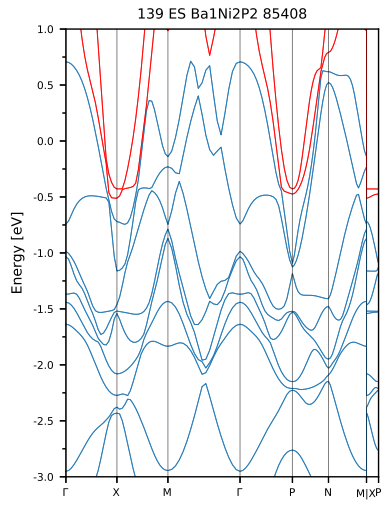

(li) 139.3.70

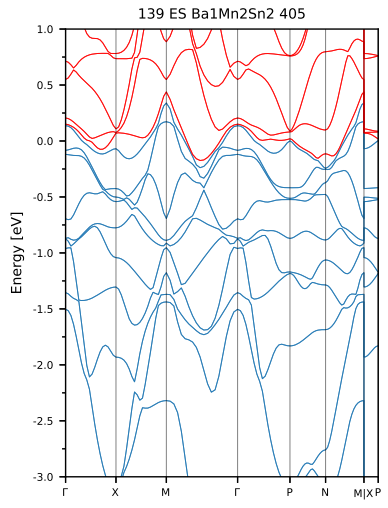

(lj) 139.3.71

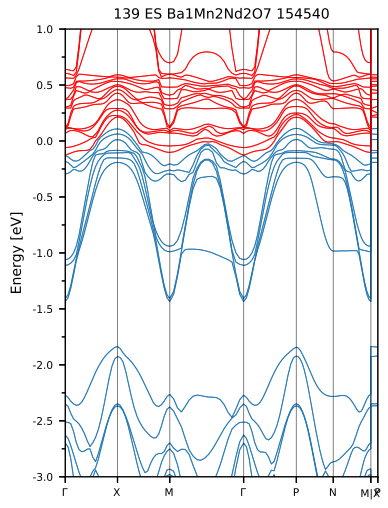

(lk) 139.3.72

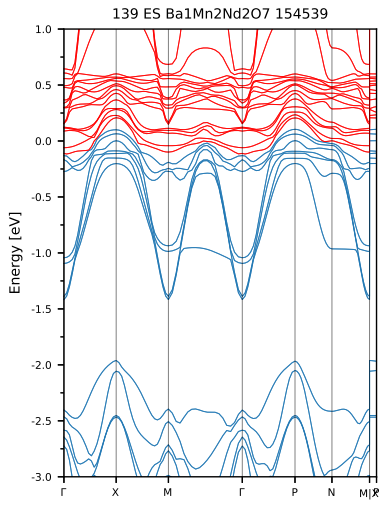

(ll) 139.3.73

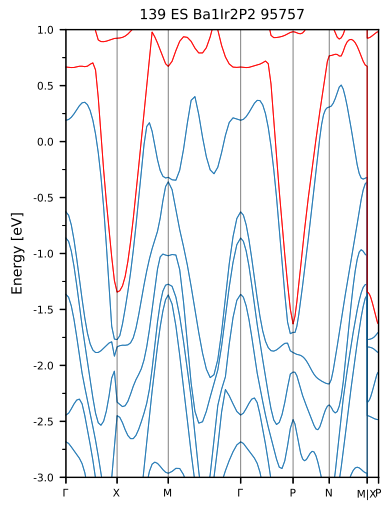

(lm) 139.3.74

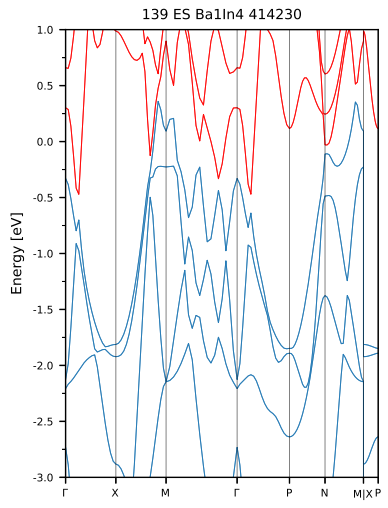

(ln) 139.3.75

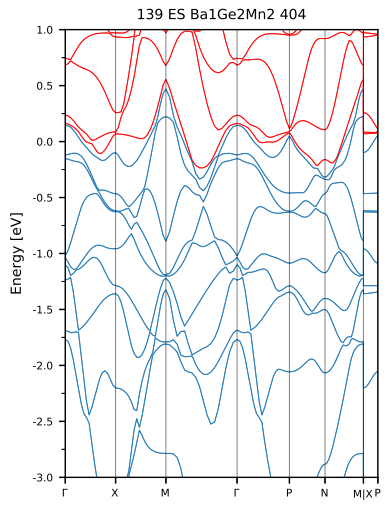

(lo) 139.3.76

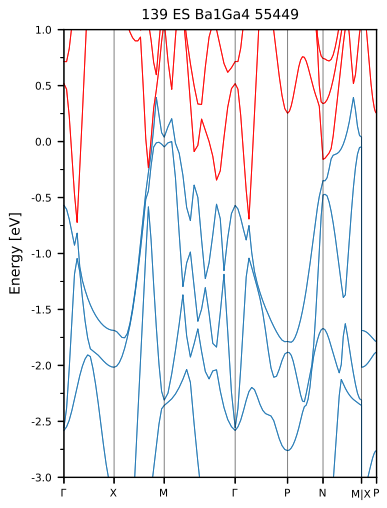

(lp) 139.3.77

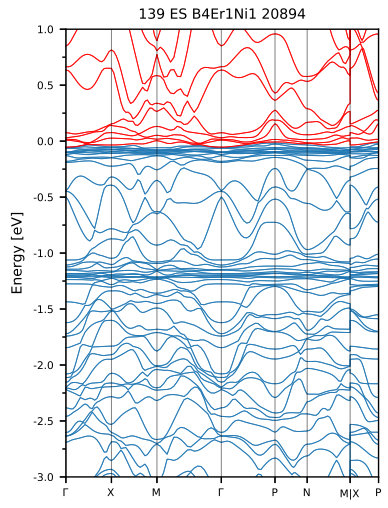

(lq) 139.3.78

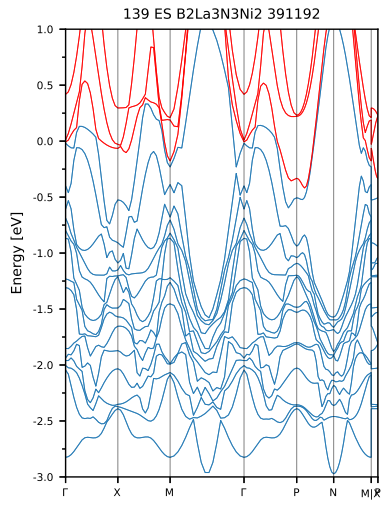

(lr) 139.3.79

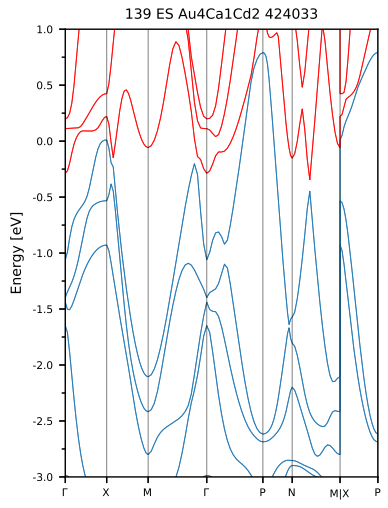

(ls) 139.3.80

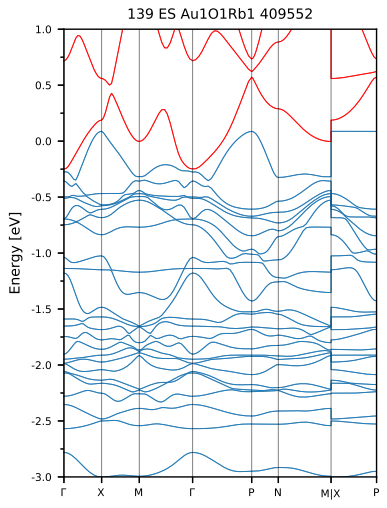

(lt) 139.3.81

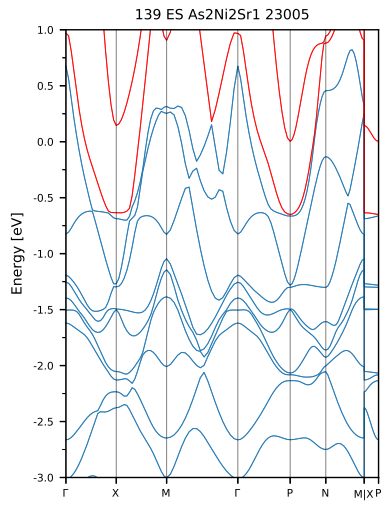

(lu) 139.3.82

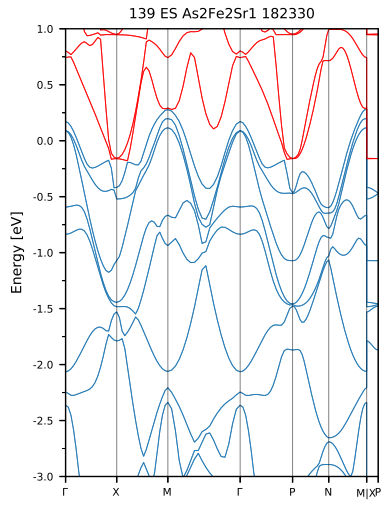

(lv) 139.3.83

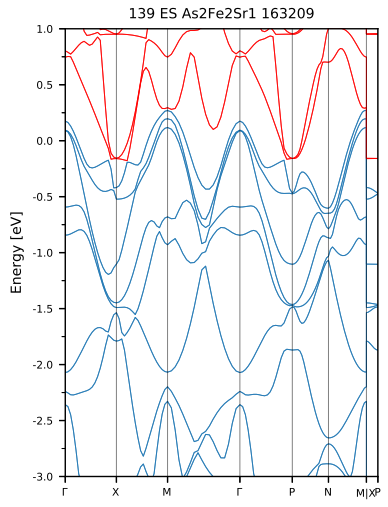

(lw) 139.3.84

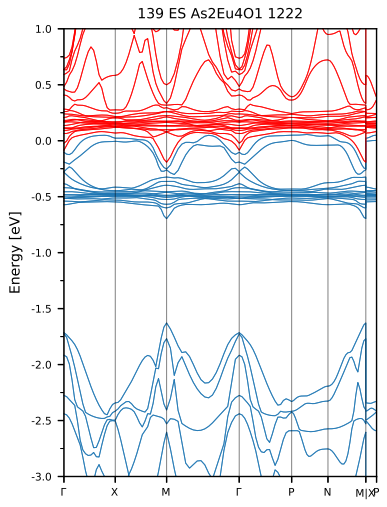

(lx) 139.3.85

Figure 308: Band structures for topological materials in SG 139, part 21/22, from 139.3.70 to 139.3.85. See the list of figures.

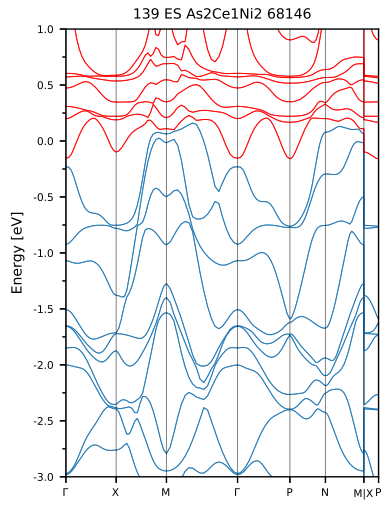

(ly) 139.3.86

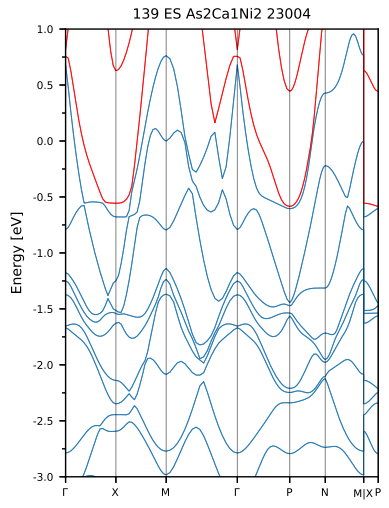

(lz) 139.3.87

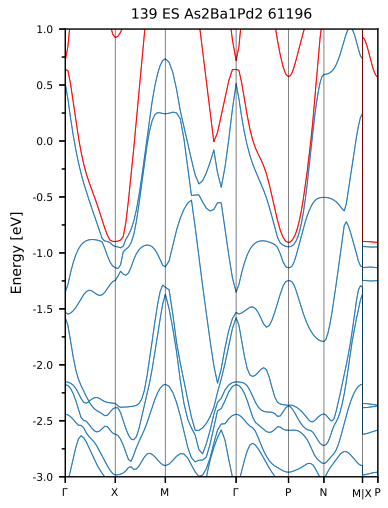

(ma) 139.3.88

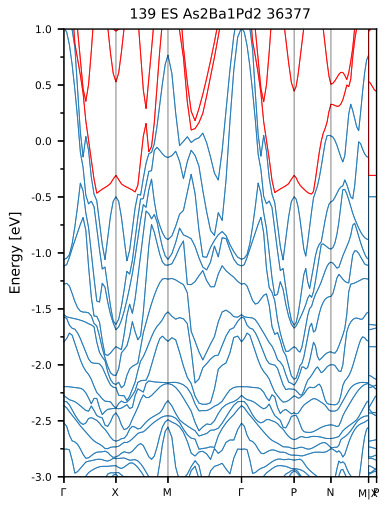

(mb) 139.3.89

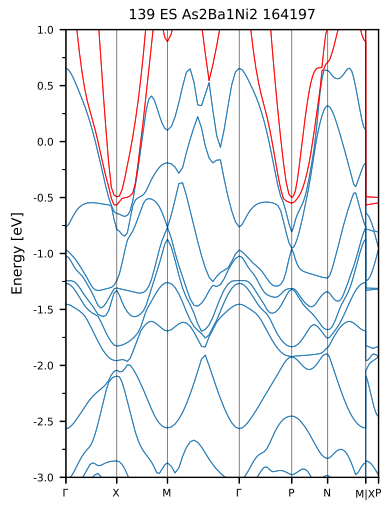

(mc) 139.3.90

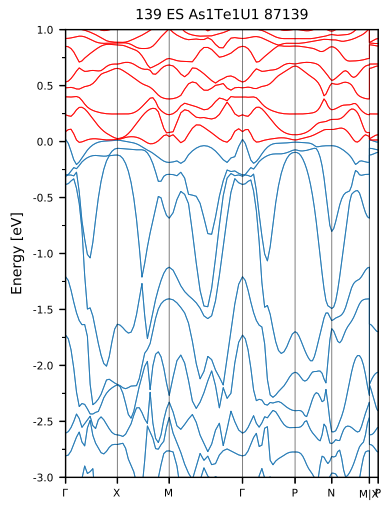

(md) 139.3.91

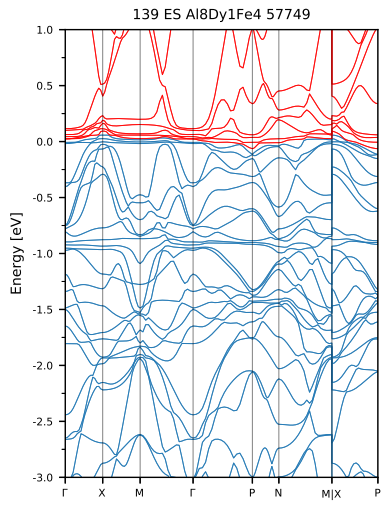

(me) 139.3.92

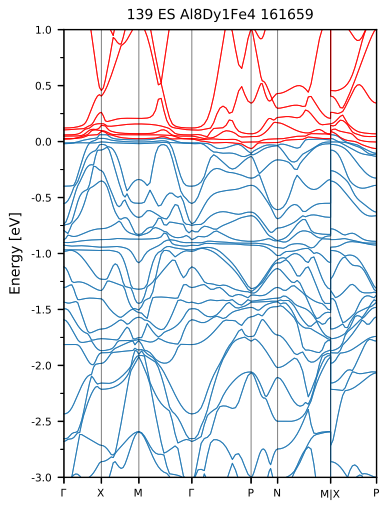

(mf) 139.3.93

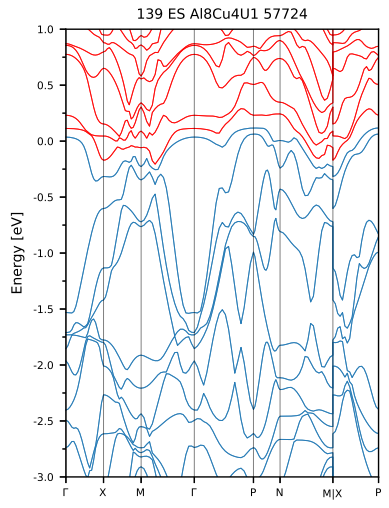

(mg) 139.3.94

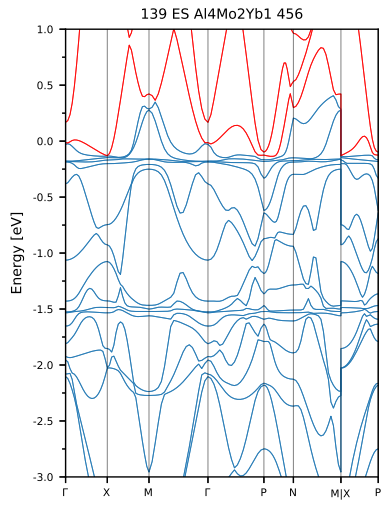

(mh) 139.3.95

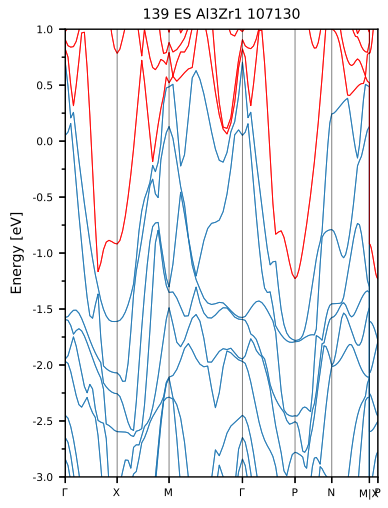

(mi) 139.3.96

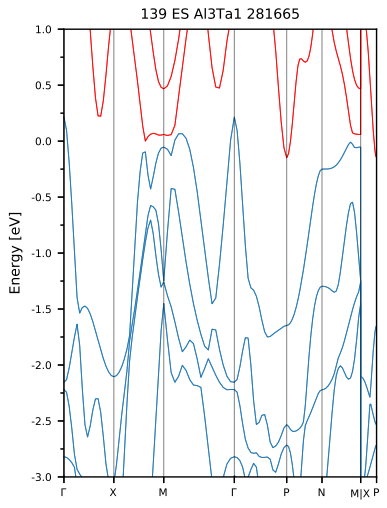

(mj) 139.3.97

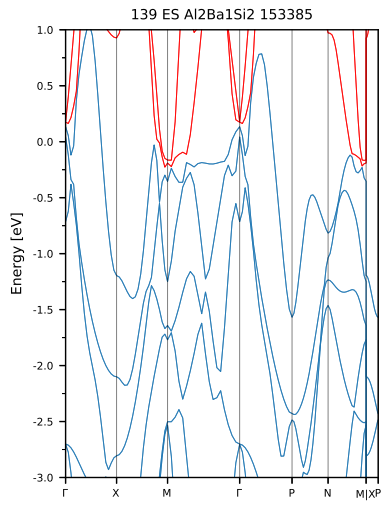

(mk) 139.3.98

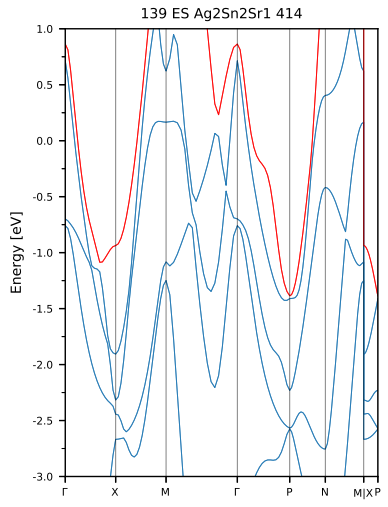

(ml) 139.3.99

Figure 309: Band structures for topological materials in SG 139, part 22/22, from 139.3.86 to 139.3.99. See the list of figures.

### 0.0.66 SG 140

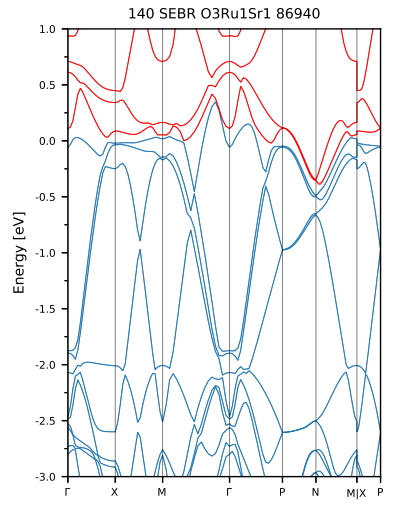

(a) 140.1.1

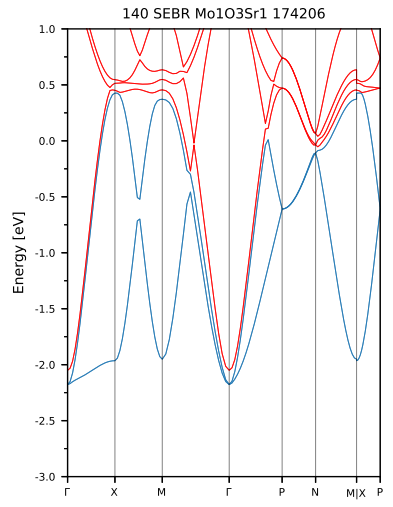

(b) 140.1.2

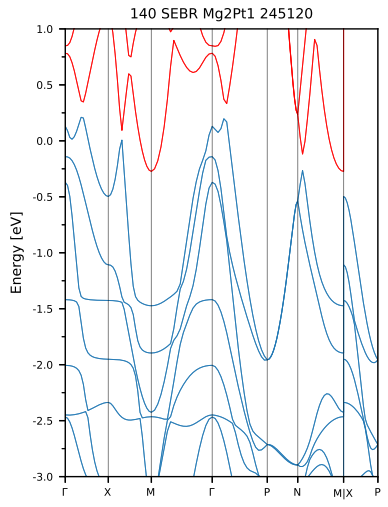

(c) 140.1.3

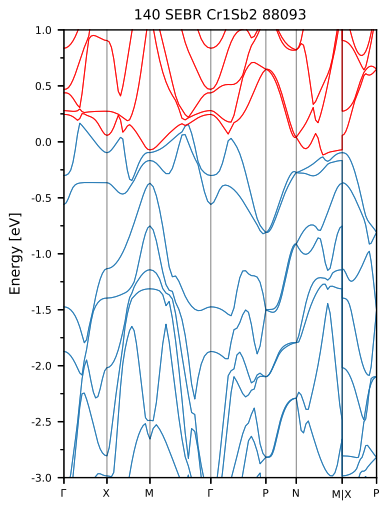

(d) 140.1.4

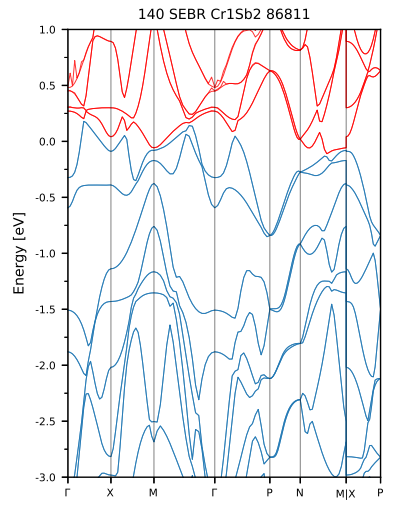

(e) 140.1.5

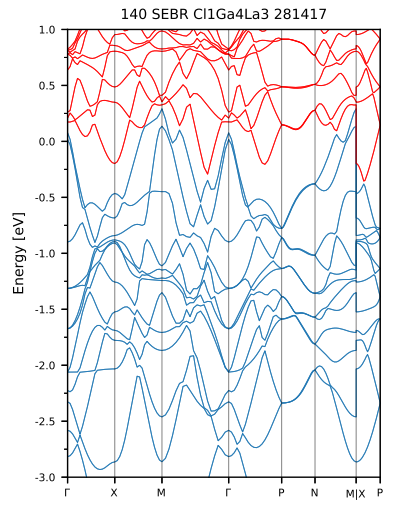

(f) 140.1.6

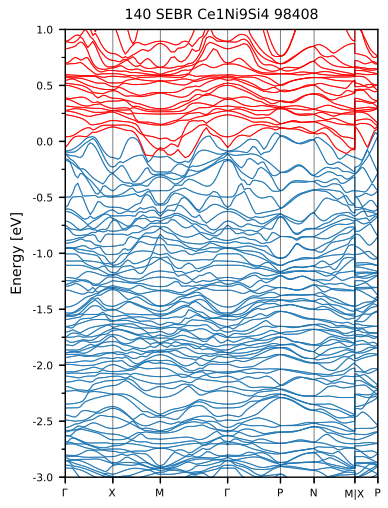

(g) 140.1.7

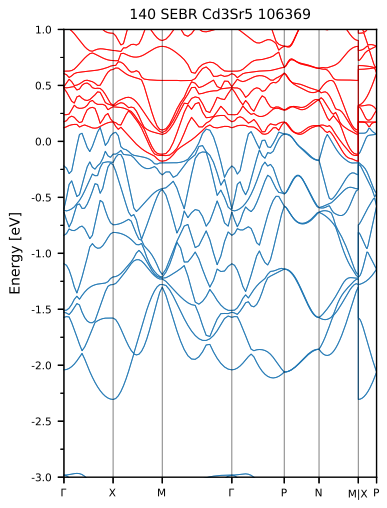

(h) 140.1.8

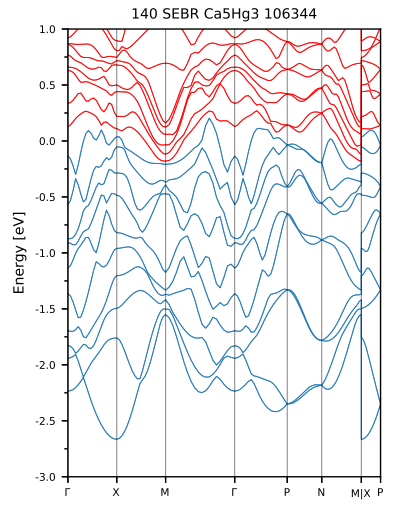

(i) 140.1.9

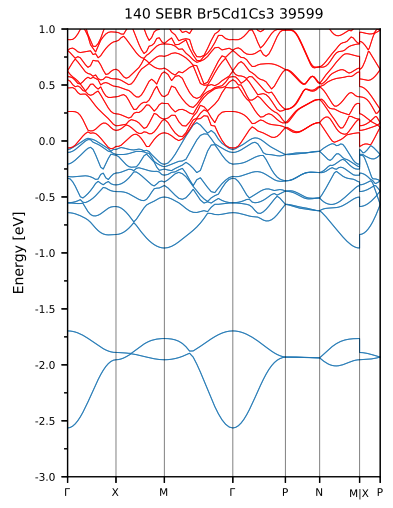

(j) 140.1.10

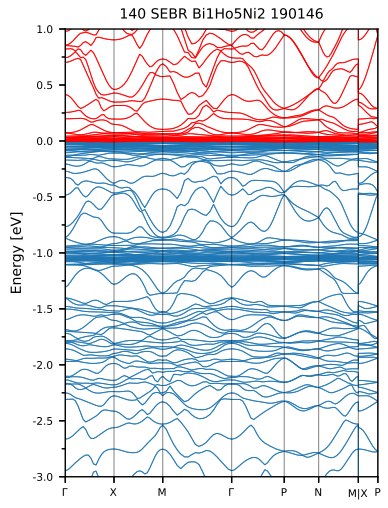

(k) 140.1.11

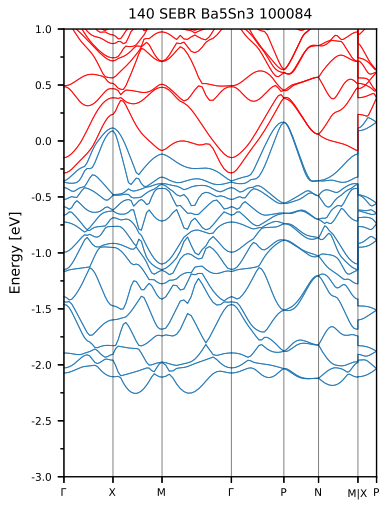

(l) 140.1.12

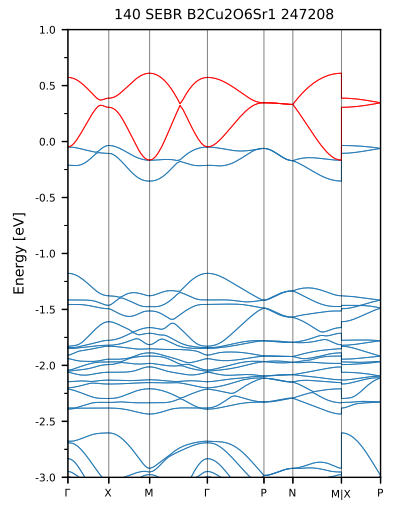

(m) 140.1.13

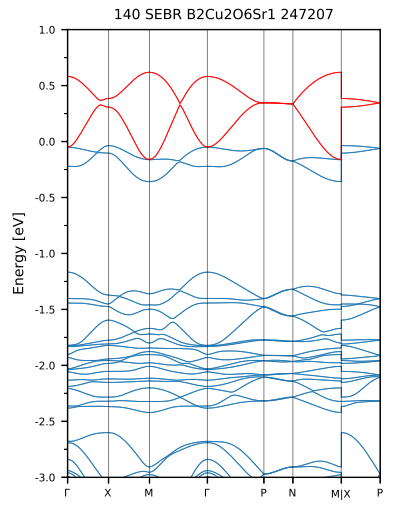

(n) 140.1.14

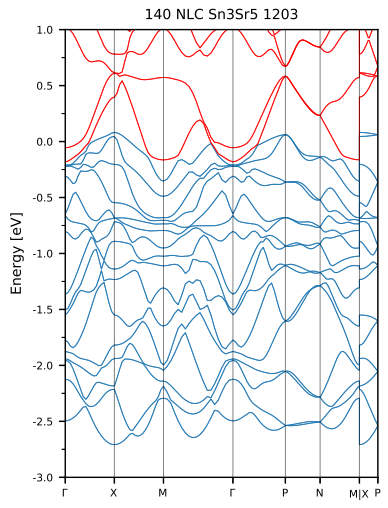

(o) 140.2.1

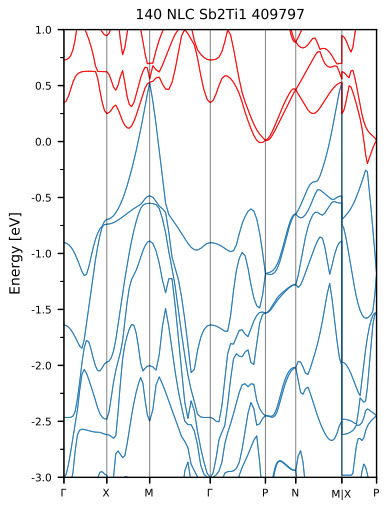

(p) 140.2.2

Figure 310: Band structures for topological materials in SG 140, part 1/11, from 140.1.1 to 140.2.2. See the list of figures.

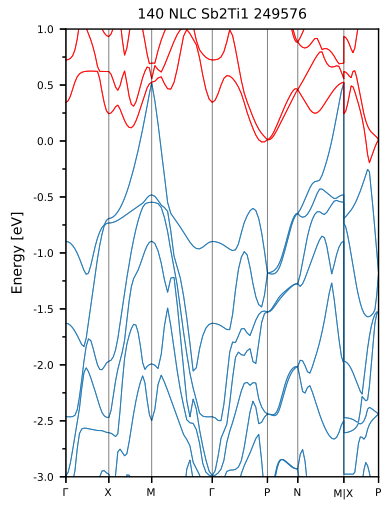

(q) 140.2.3

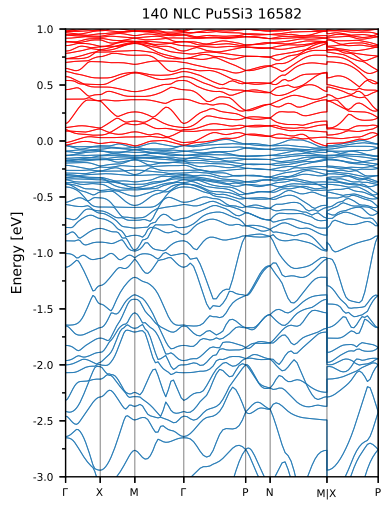

(r) 140.2.4

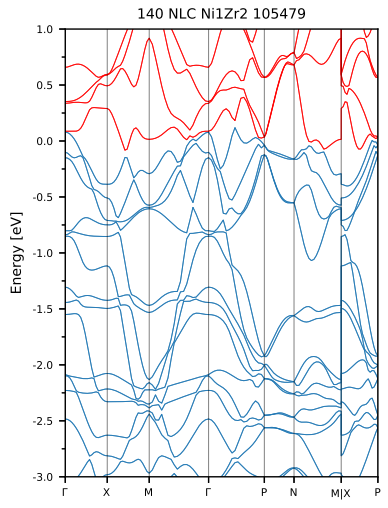

(s) 140.2.5

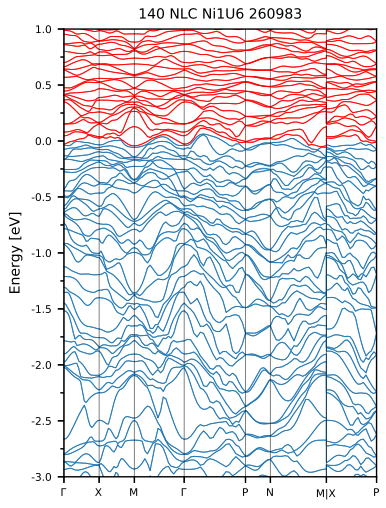

(t) 140.2.6

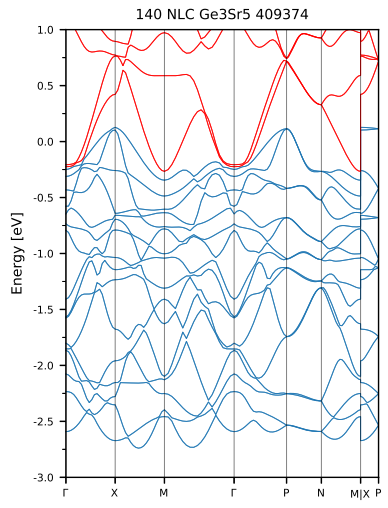

(u) 140.2.7

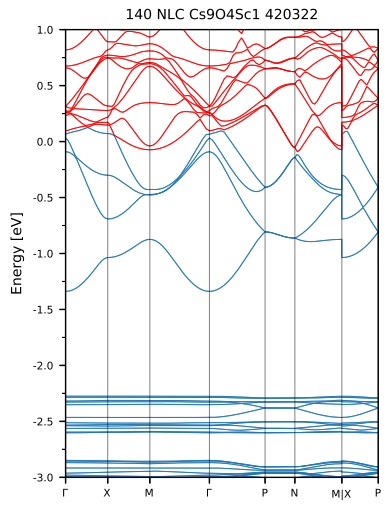

(v) 140.2.8

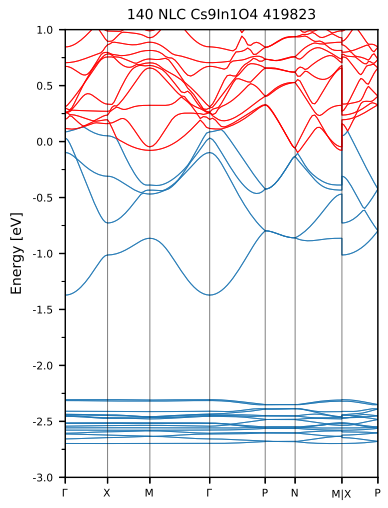

(w) 140.2.9

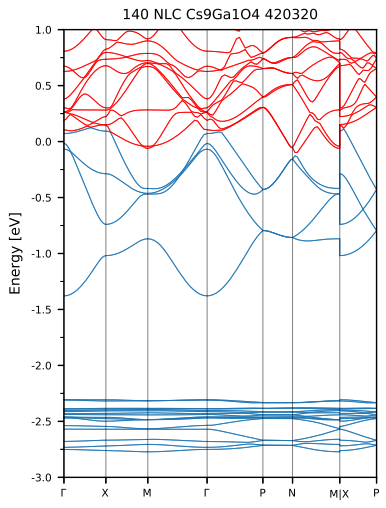

(x) 140.2.10

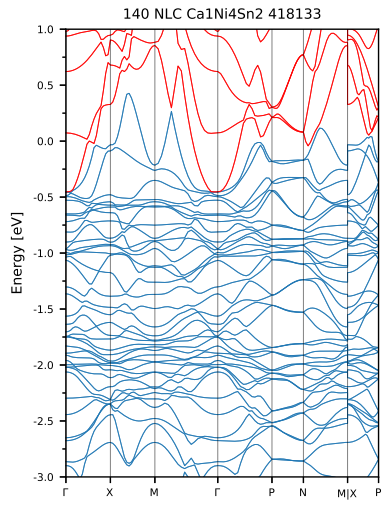

(y) 140.2.11

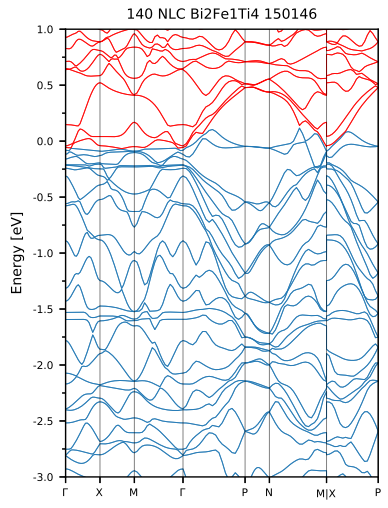

(z) 140.2.12

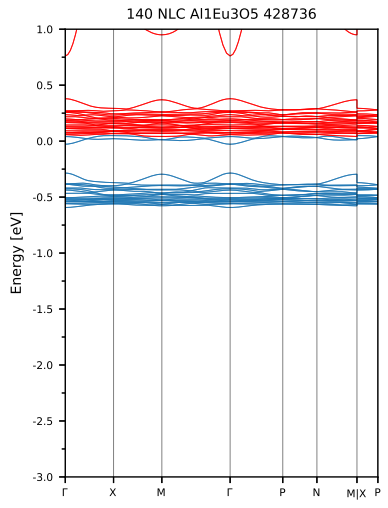

(aa) 140.2.13

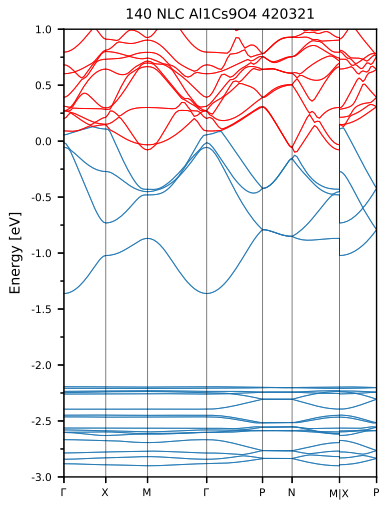

(ab) 140.2.14

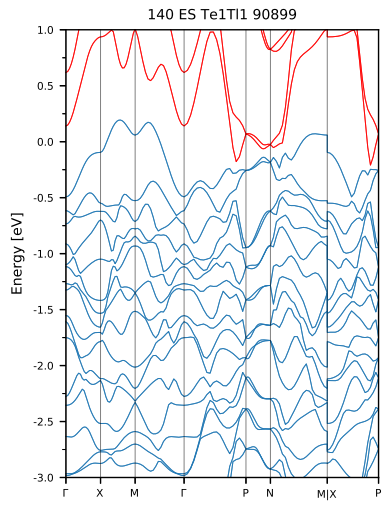

(ac) 140.3.1

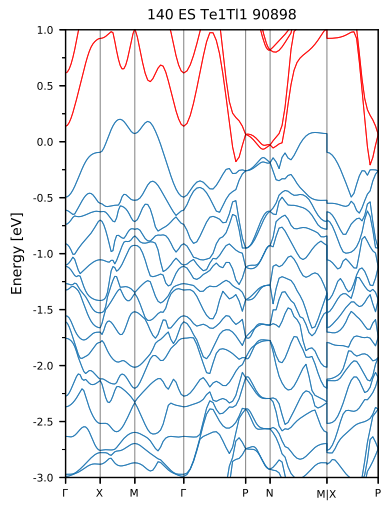

(ad) 140.3.2

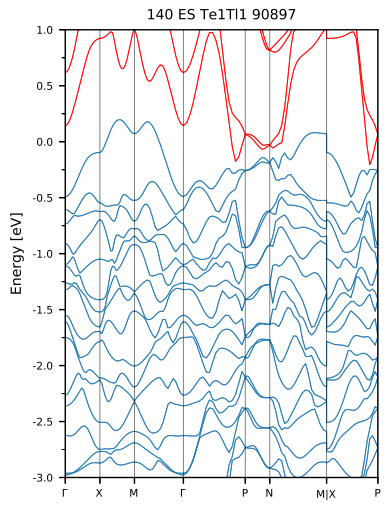

(ae) 140.3.3

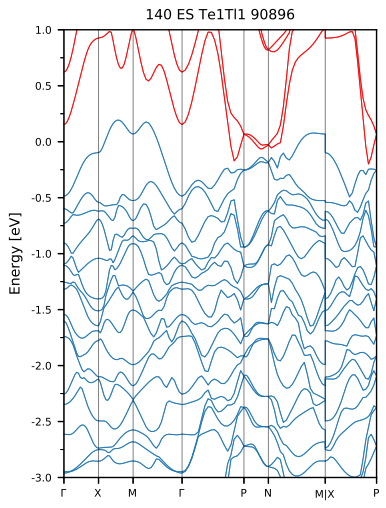

(af) 140.3.4

Figure 311: Band structures for topological materials in SG 140, part 2/11, from 140.2.3 to 140.3.4. See the list of figures.

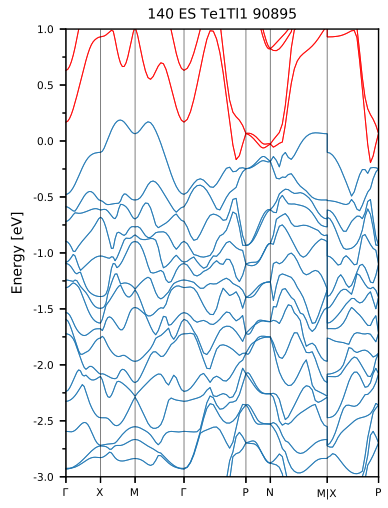

(ag) 140.3.5

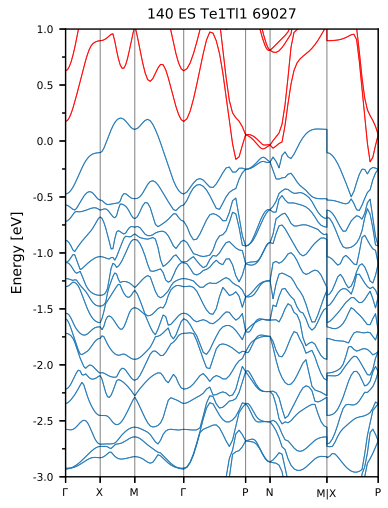

(ah) 140.3.6

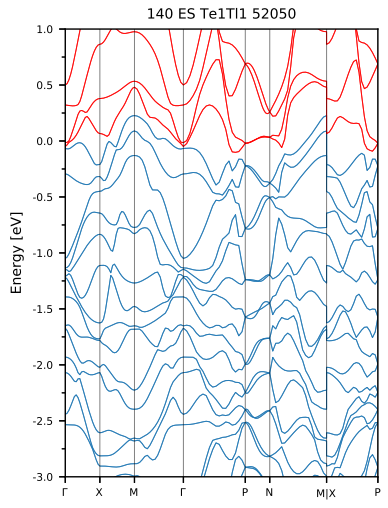

(ai) 140.3.7

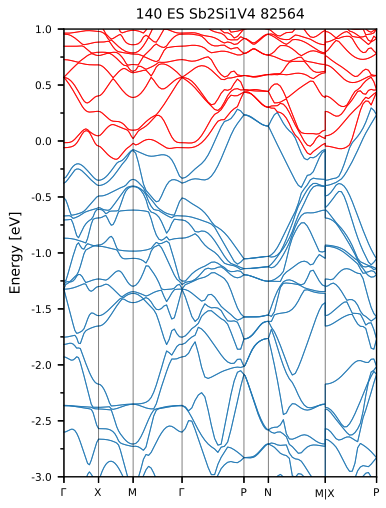

(aj) 140.3.8

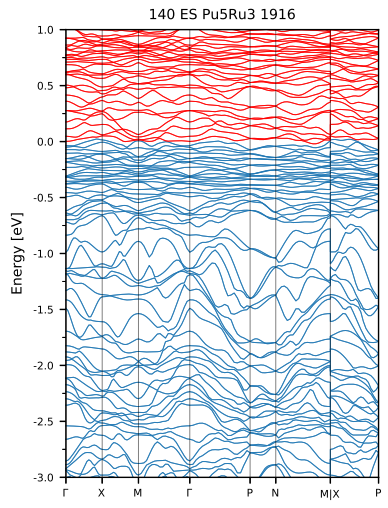

(ak) 140.3.9

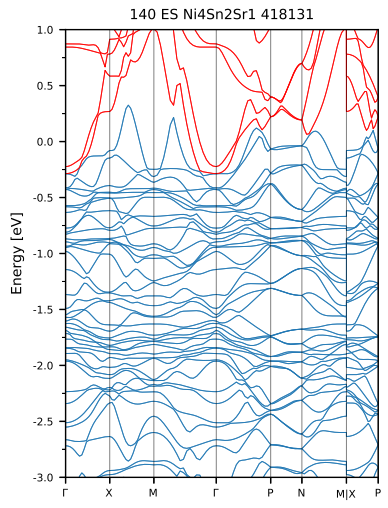

(al) 140.3.10

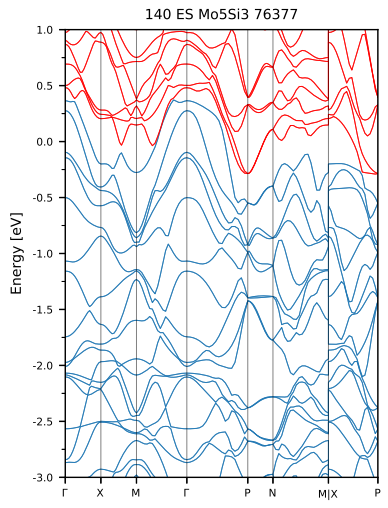

(am) 140.3.11

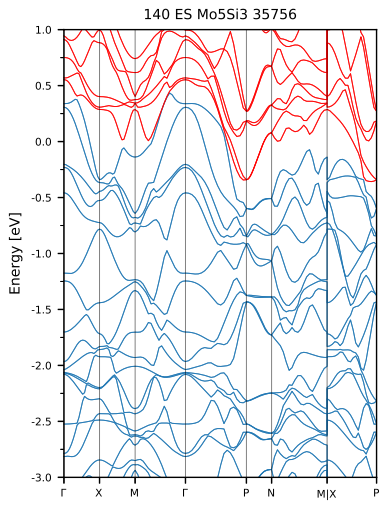

(an) 140.3.12

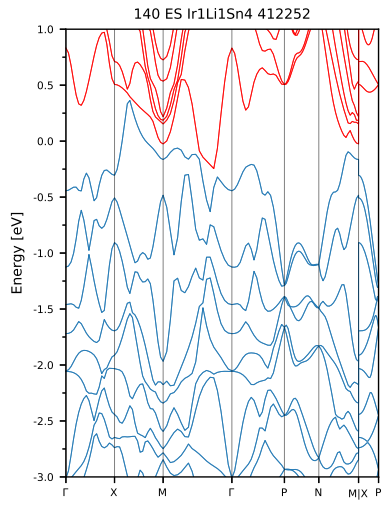

(ao) 140.3.13

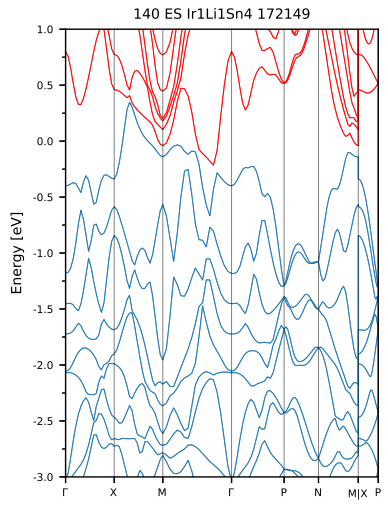

(ap) 140.3.14

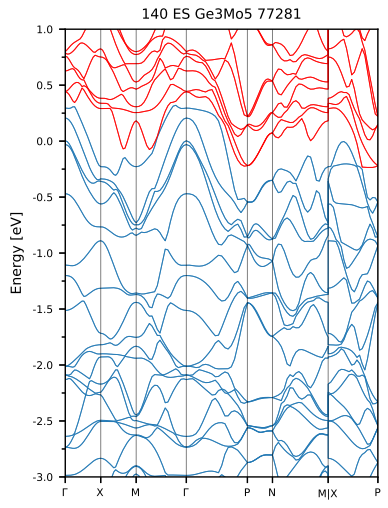

(aq) 140.3.15

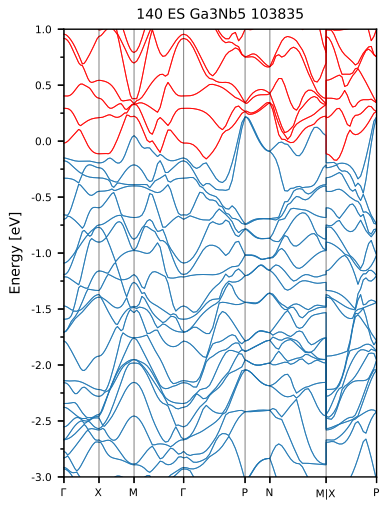

(ar) 140.3.16

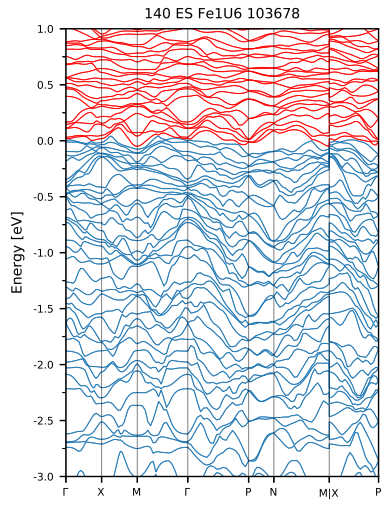

(as) 140.3.17

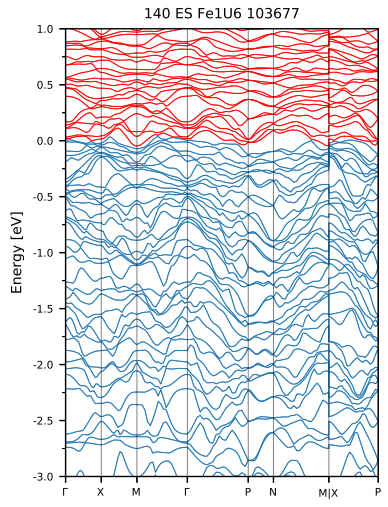

(at) 140.3.18

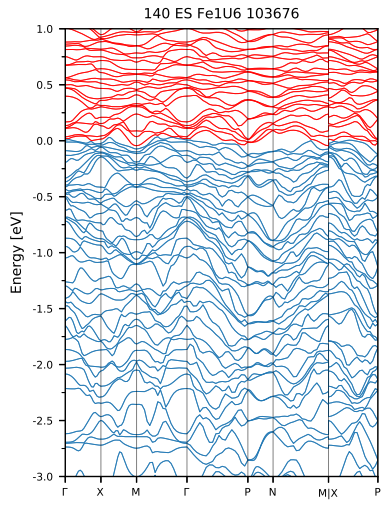

(au) 140.3.19

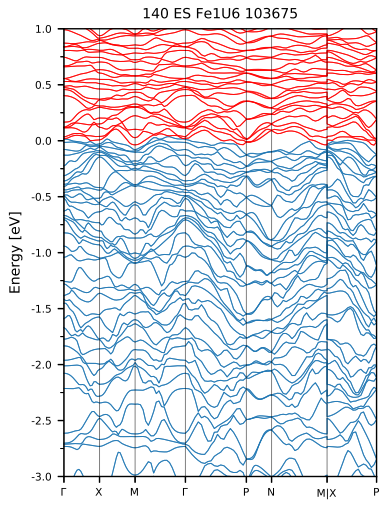

(av) 140.3.20

Figure 312: Band structures for topological materials in SG 140, part 3/11, from 140.3.5 to 140.3.20. See the list of figures.

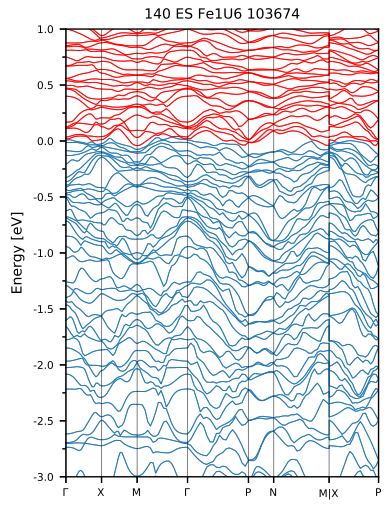

(aw) 140.3.21

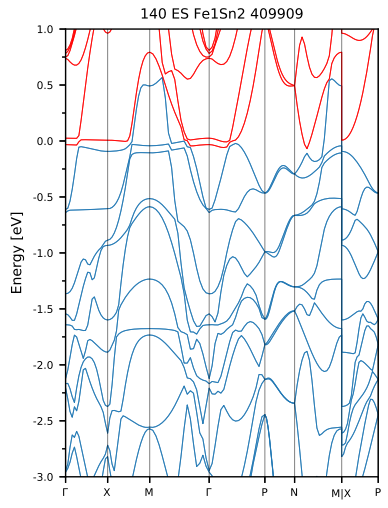

(ax) 140.3.22

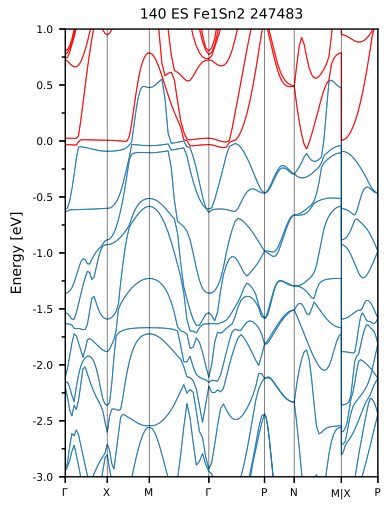

(ay) 140.3.23

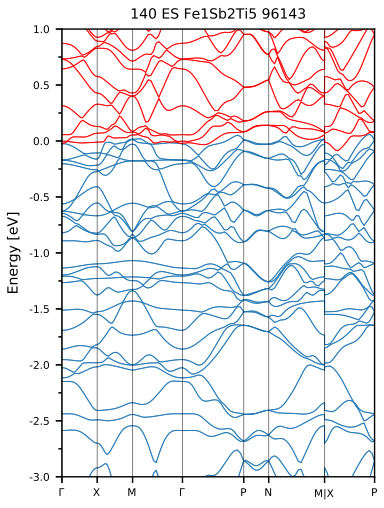

(az) 140.3.24

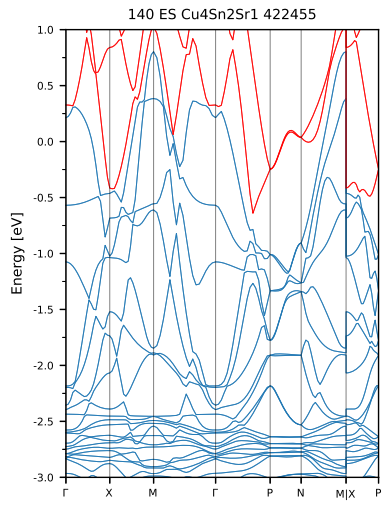

(ba) 140.3.25

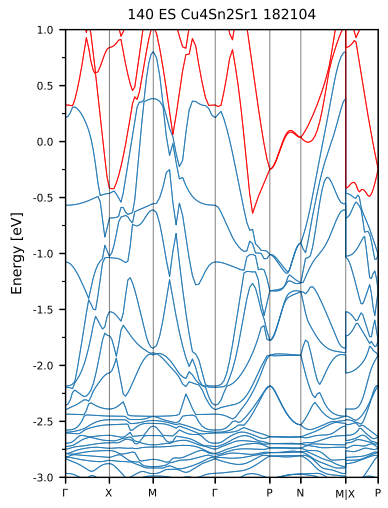

(bb) 140.3.26

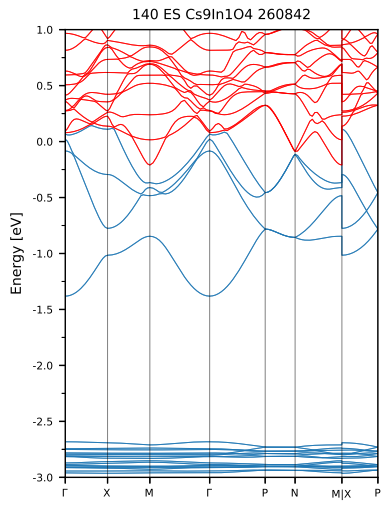

(bc) 140.3.27

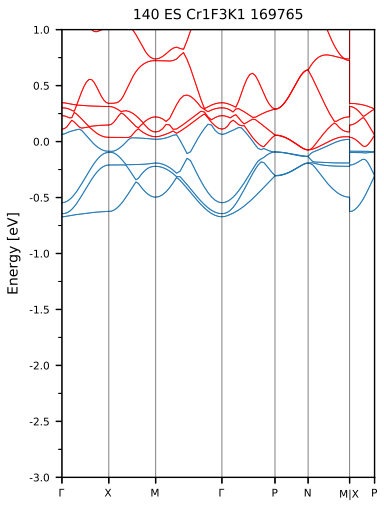

(bd) 140.3.28

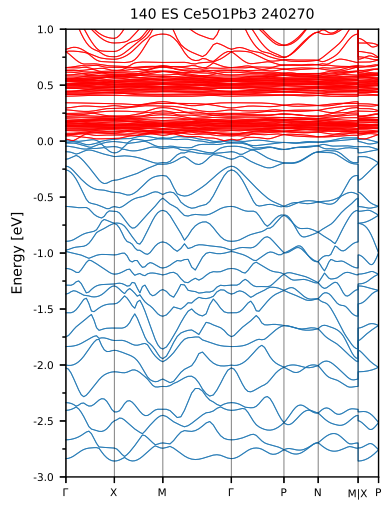

(be) 140.3.29

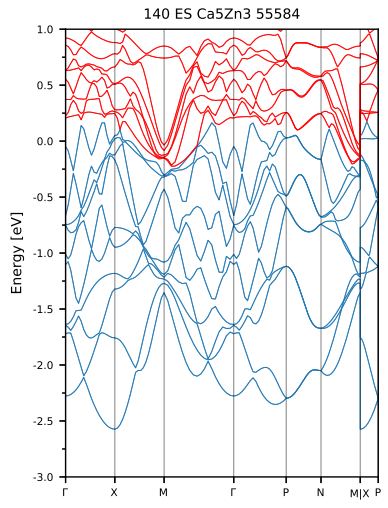

(bf) 140.3.30

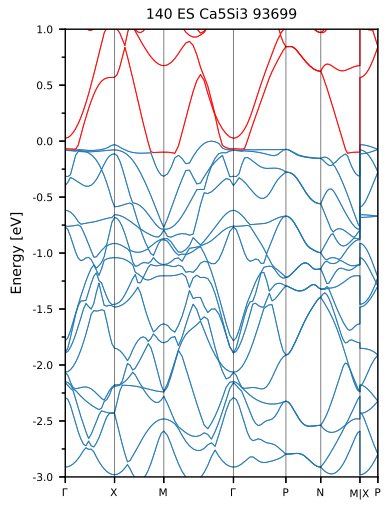

(bg) 140.3.31

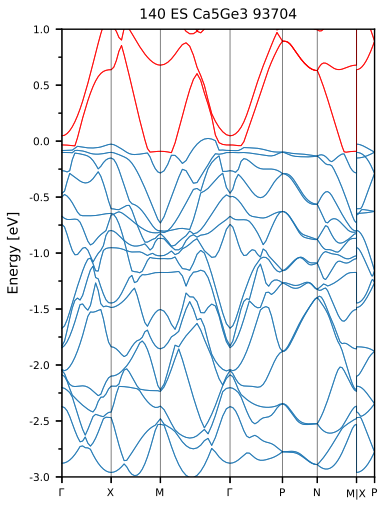

(bh) 140.3.32

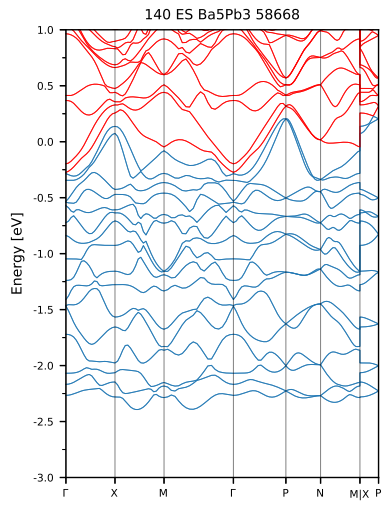

(bi) 140.3.33

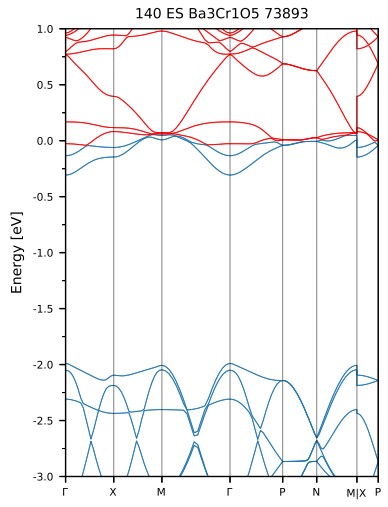

(bj) 140.3.34

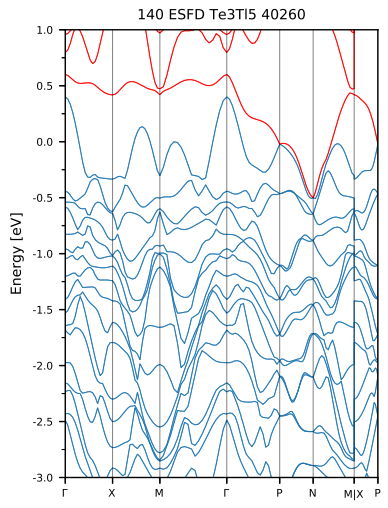

(bk) 140.4.1

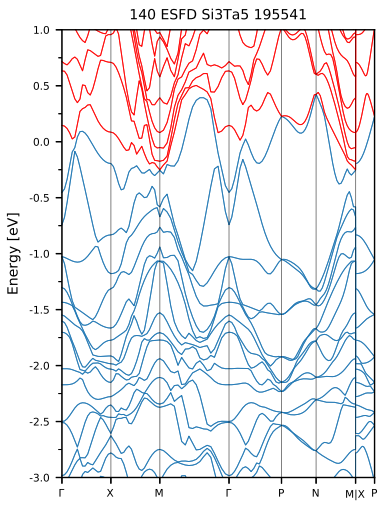

(bl) 140.4.2

Figure 313: Band structures for topological materials in SG 140, part 4/11, from 140.3.21 to 140.4.2. See the list of figures.

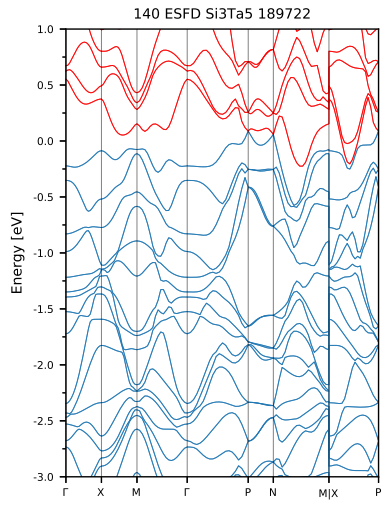

(bm) 140.4.3

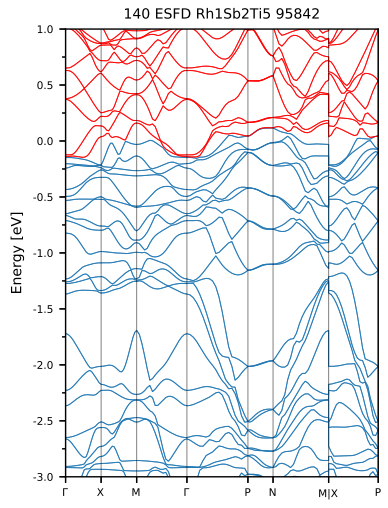

(bn) 140.4.4

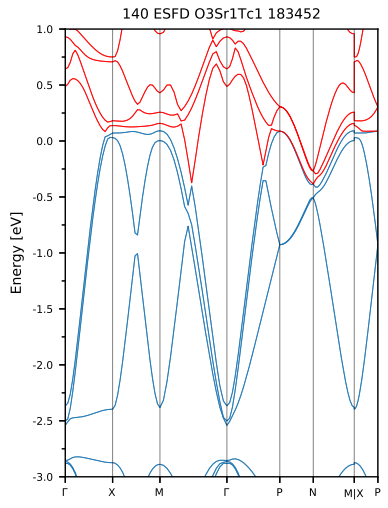

(bo) 140.4.5

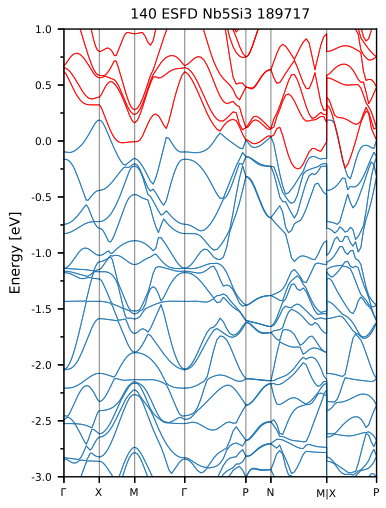

(bp) 140.4.6

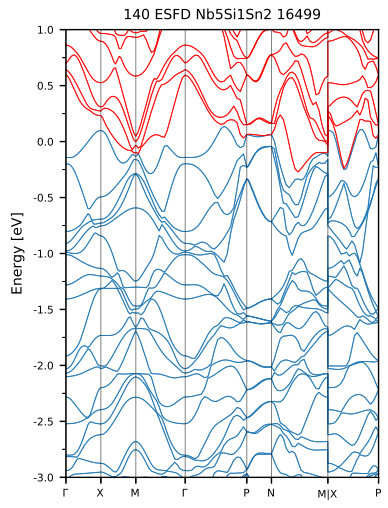

(bq) 140.4.7

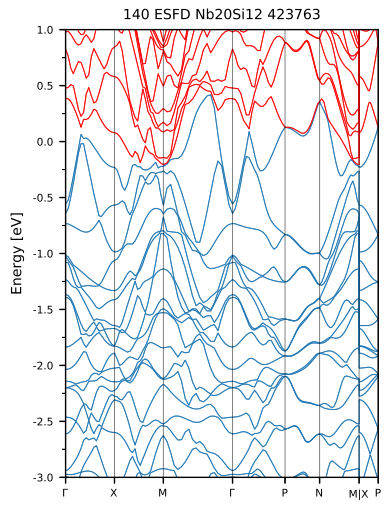

(br) 140.4.8

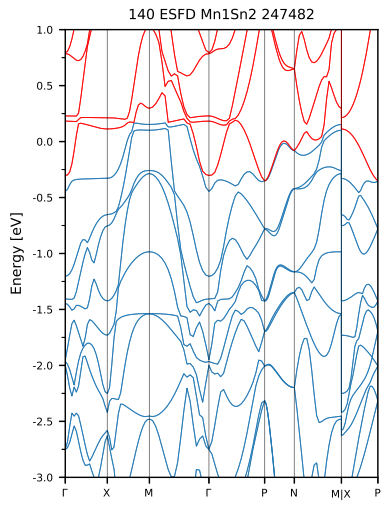

(bs) 140.4.9

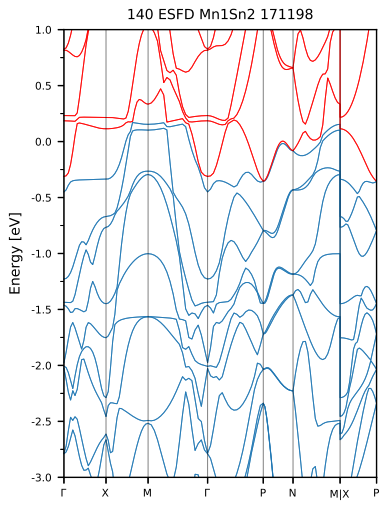

(bt) 140.4.10

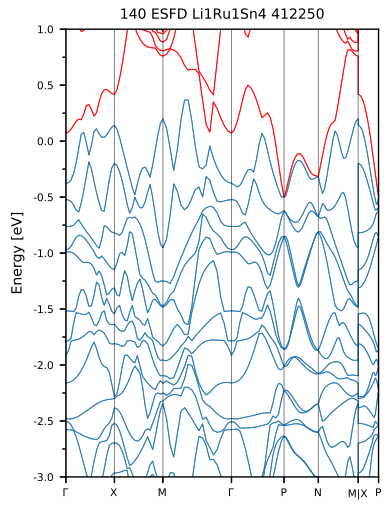

(bu) 140.4.11

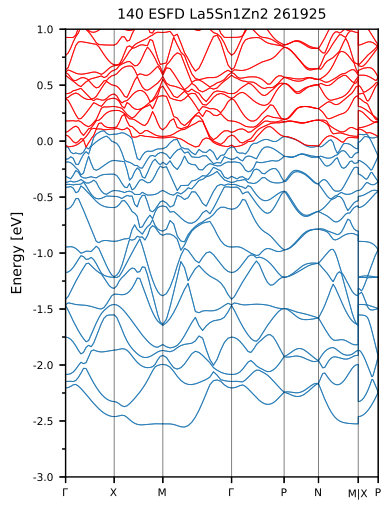

(bv) 140.4.12

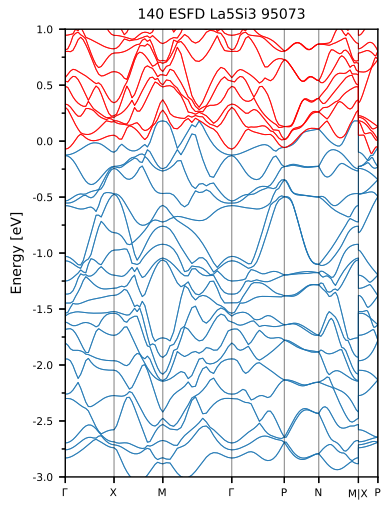

(bw) 140.4.13

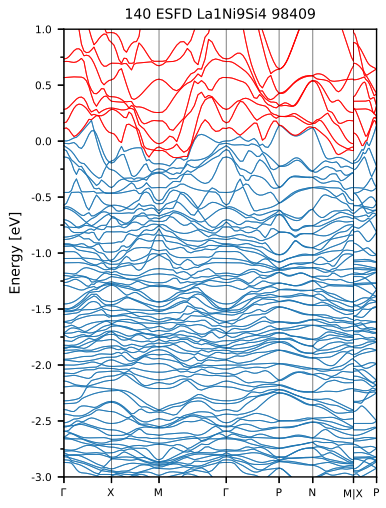

(bx) 140.4.14

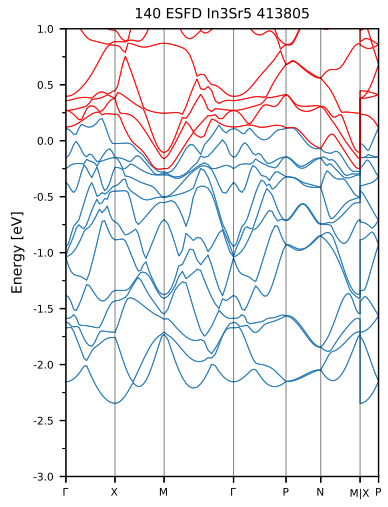

(by) 140.4.15

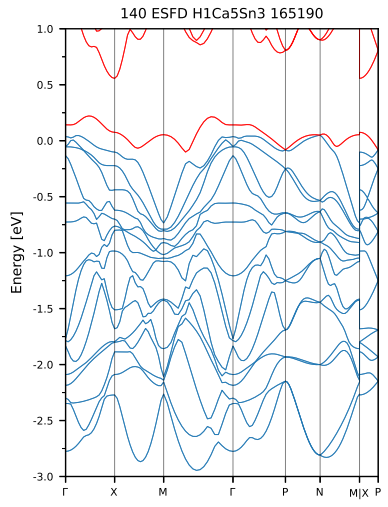

(bz) 140.4.16

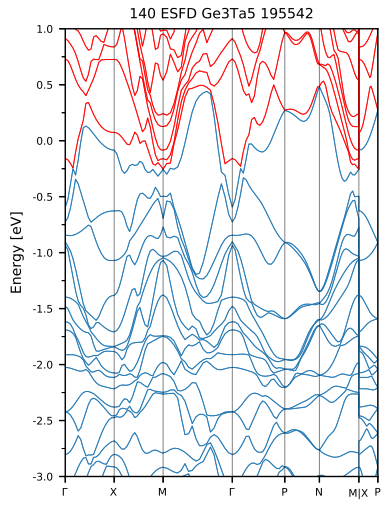

(ca) 140.4.17

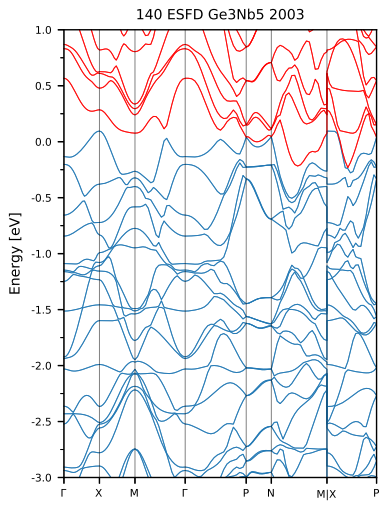

(cb) 140.4.18

Figure 314: Band structures for topological materials in SG 140, part 5/11, from 140.4.3 to 140.4.18. See the list of figures.

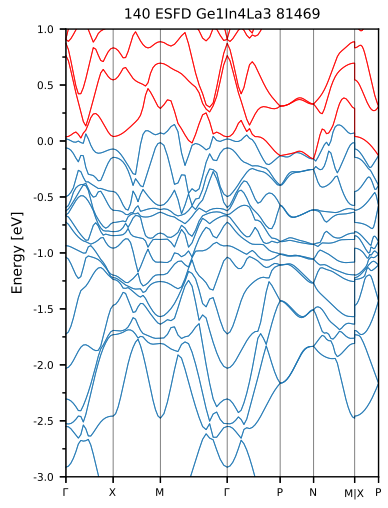

(cc) 140.4.19

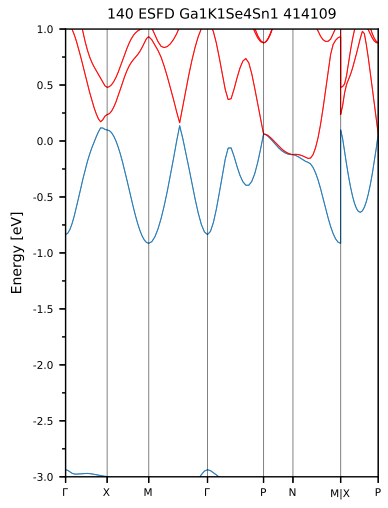

(cd) 140.4.20

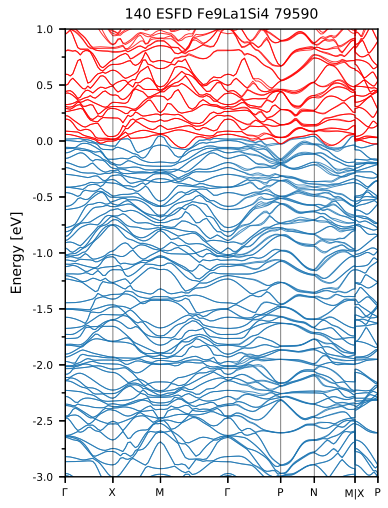

(ce) 140.4.21

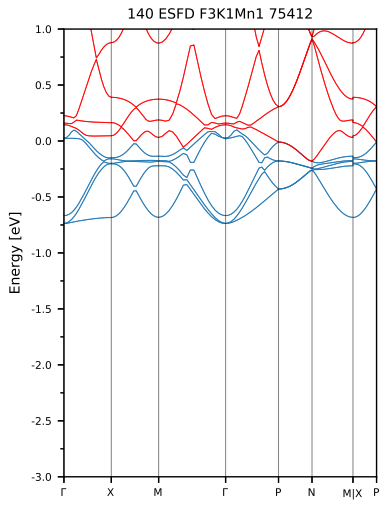

(cf) 140.4.22

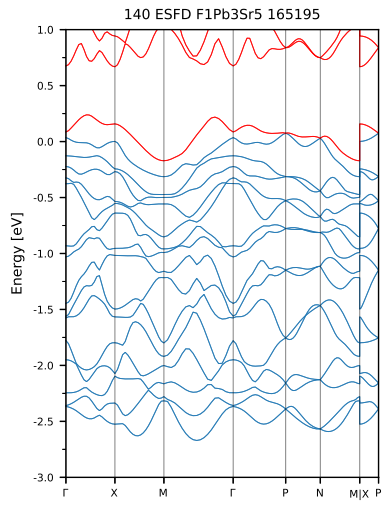

(cg) 140.4.23

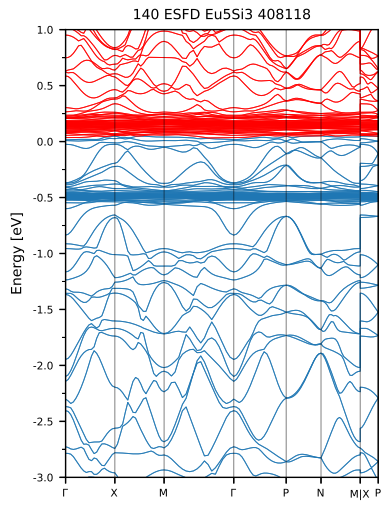

(ch) 140.4.24

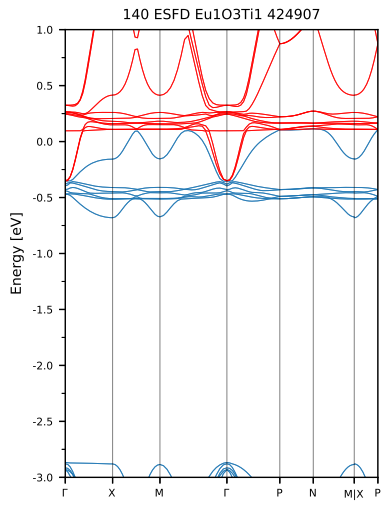

(ci) 140.4.25

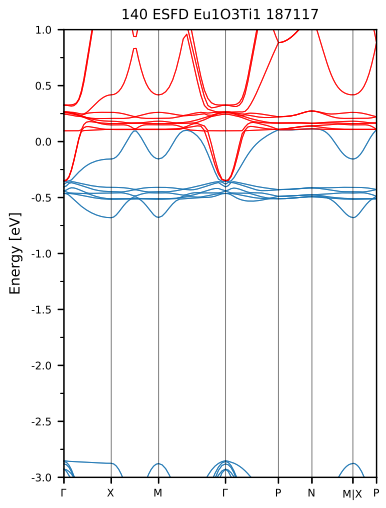

(cj) 140.4.26

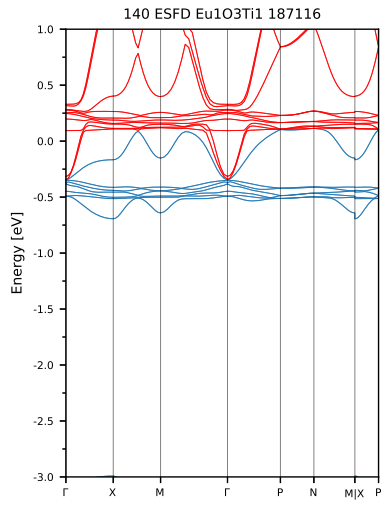

(ck) 140.4.27

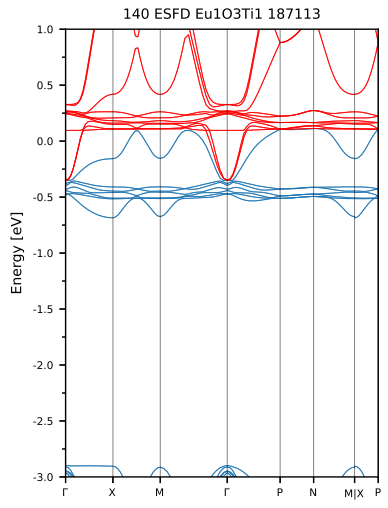

(cl) 140.4.28

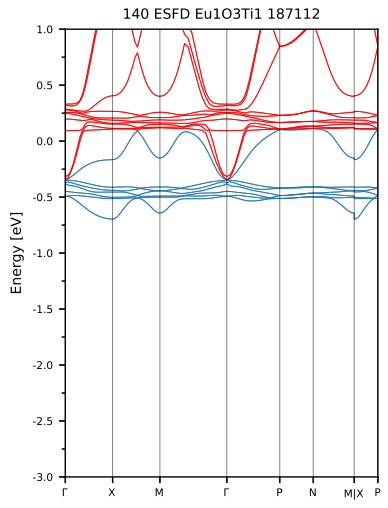

(cm) 140.4.29

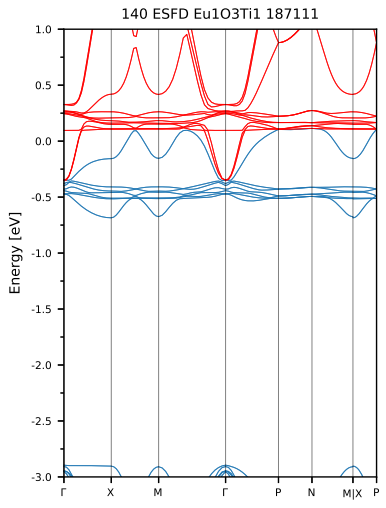

(cn) 140.4.30

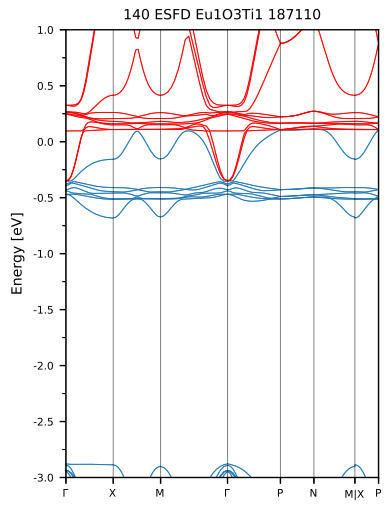

(co) 140.4.31

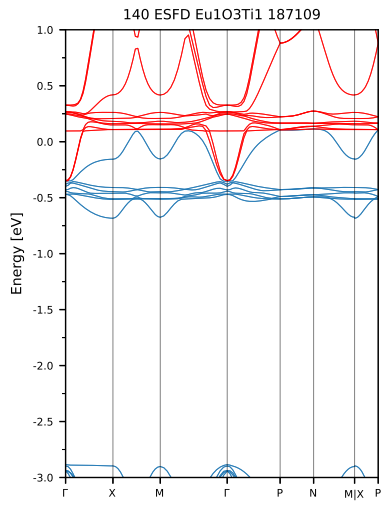

(cp) 140.4.32

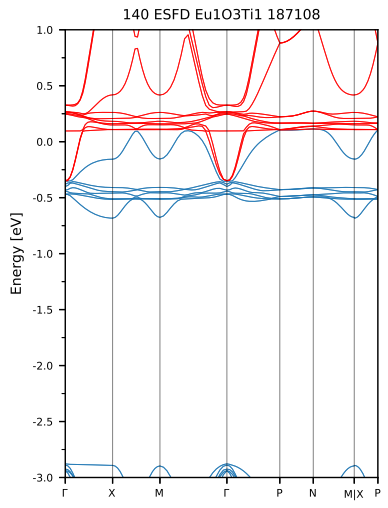

(cq) 140.4.33

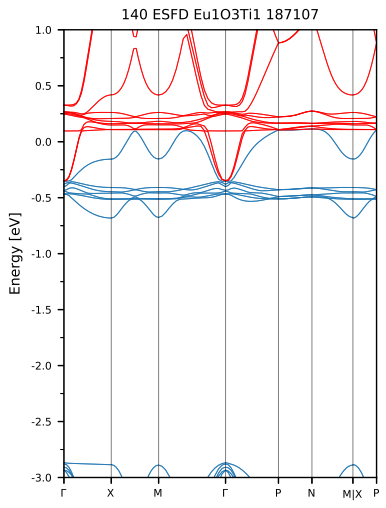

(cr) 140.4.34

Figure 315: Band structures for topological materials in SG 140, part 6/11, from 140.4.19 to 140.4.34. See the list of figures.

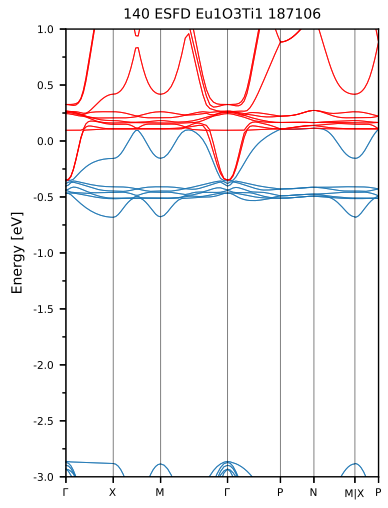

(cs) 140.4.35

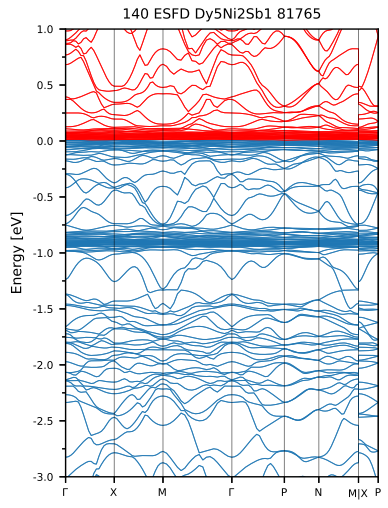

(ct) 140.4.36

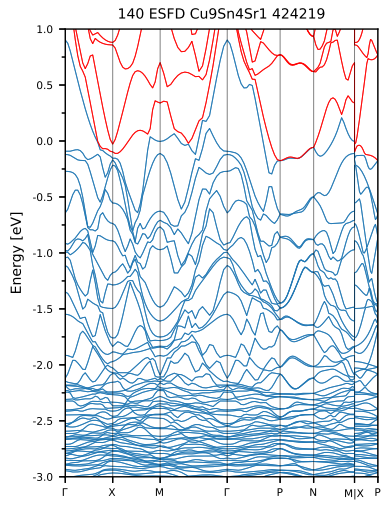

(cu) 140.4.37

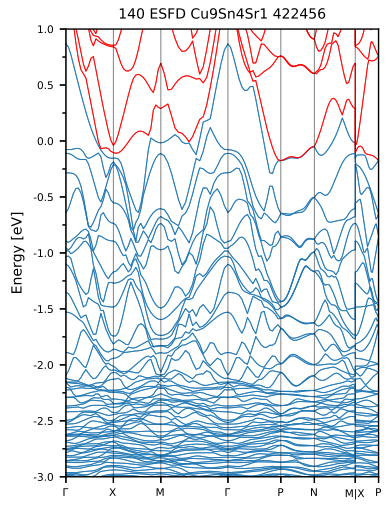

(cv) 140.4.38

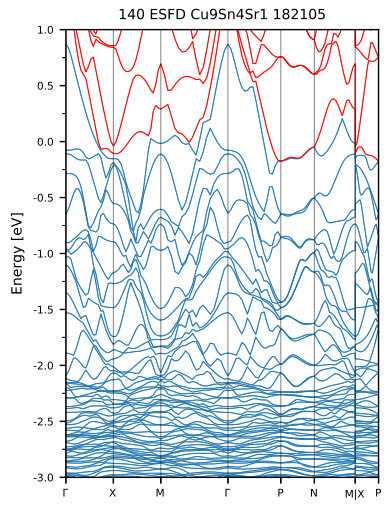

(cw) 140.4.39

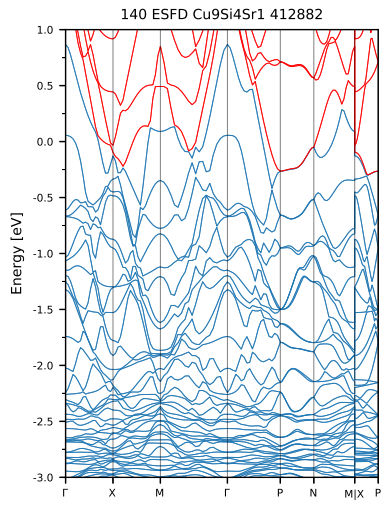

(cx) 140.4.40

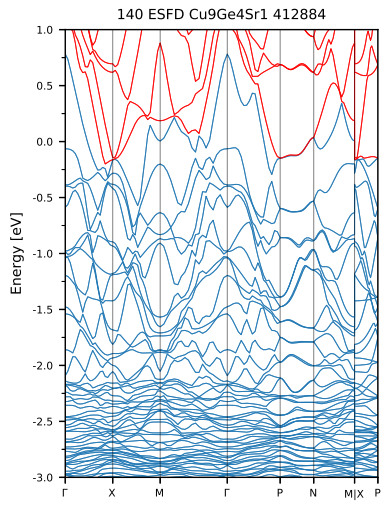

(cy) 140.4.41

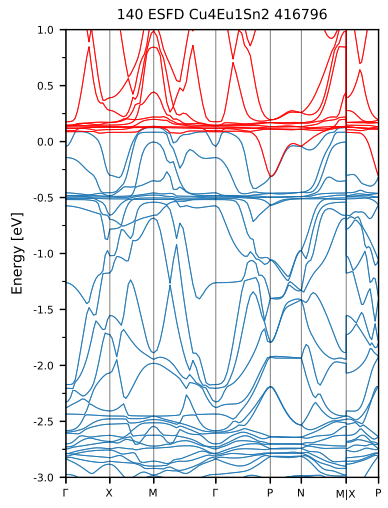

(cz) 140.4.42

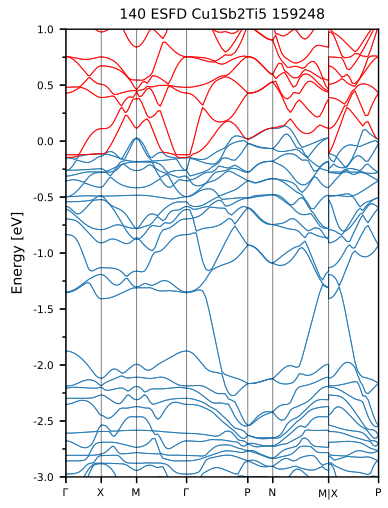

(da) 140.4.43

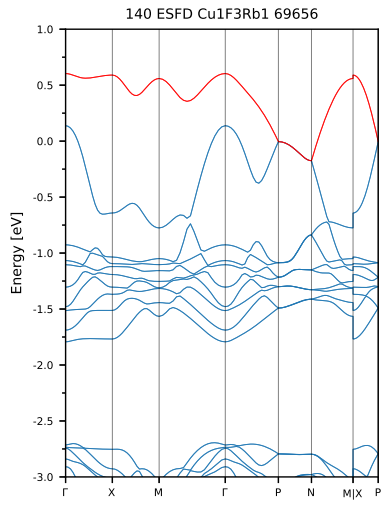

(db) 140.4.44

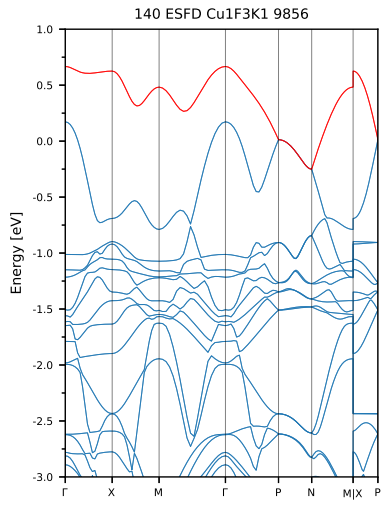

(dc) 140.4.45

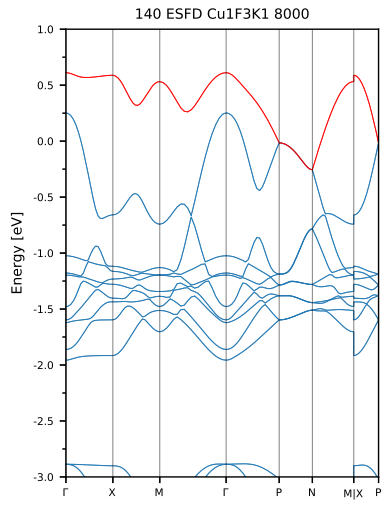

(dd) 140.4.46

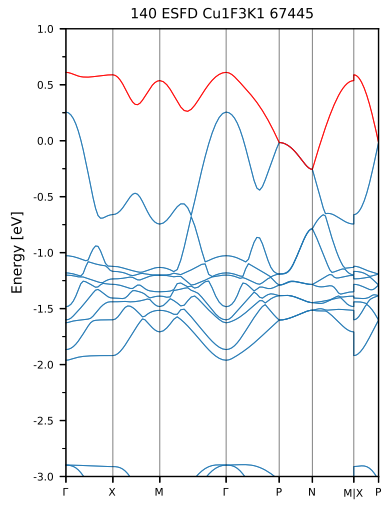

(de) 140.4.47

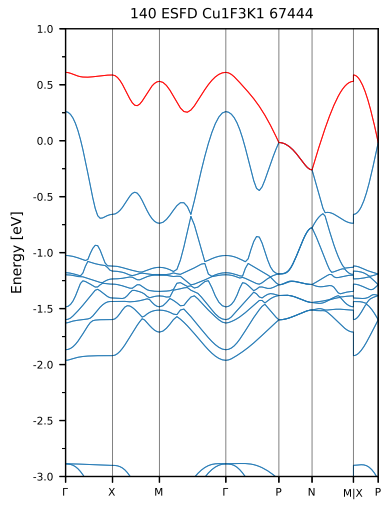

(df) 140.4.48

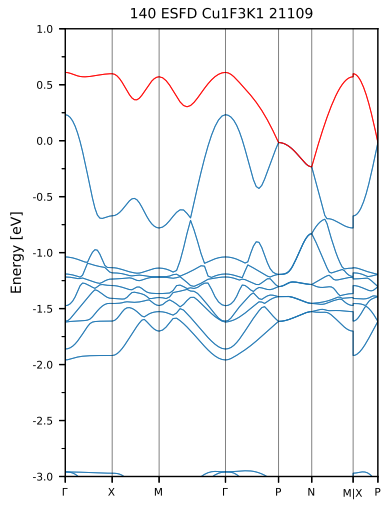

(dg) 140.4.49

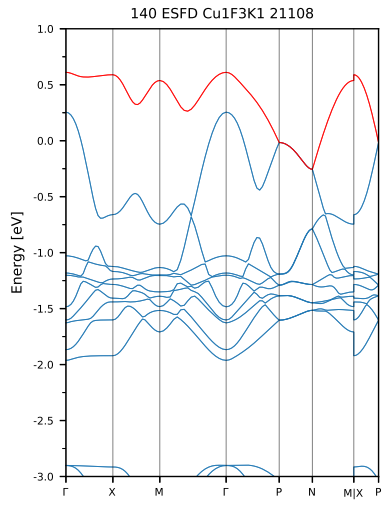

(dh) 140.4.50

Figure 316: Band structures for topological materials in SG 140, part 7/11, from 140.4.35 to 140.4.50. See the list of figures.

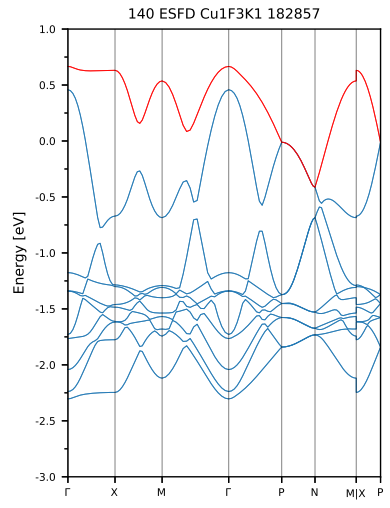

(di) 140.4.51

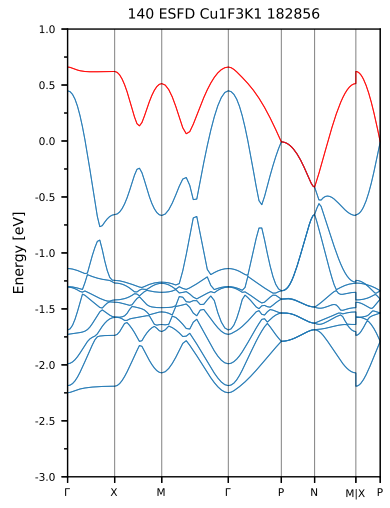

(dj) 140.4.52

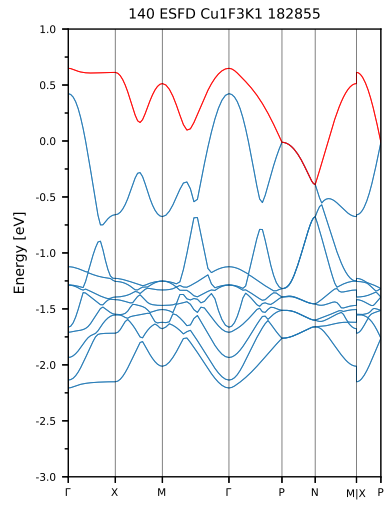

(dk) 140.4.53

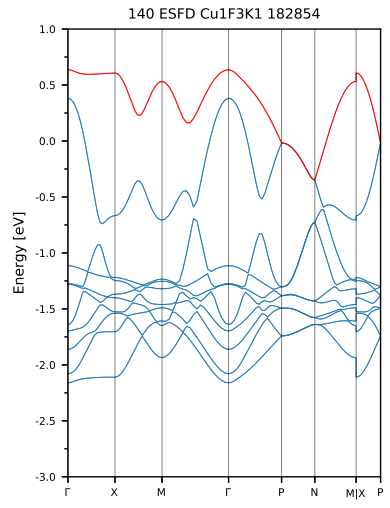

(dl) 140.4.54

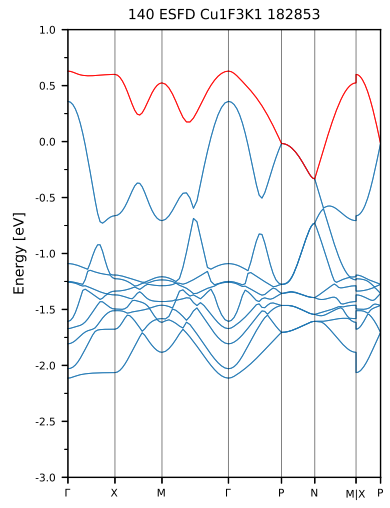

(dm) 140.4.55

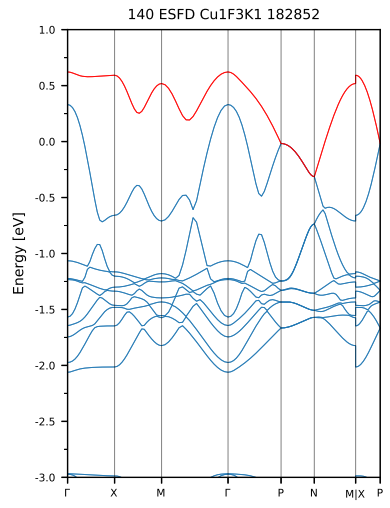

(dn) 140.4.56

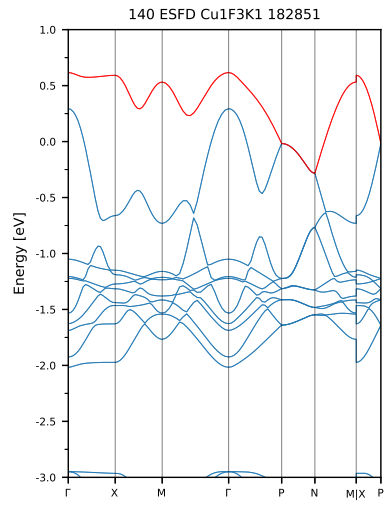

(do) 140.4.57

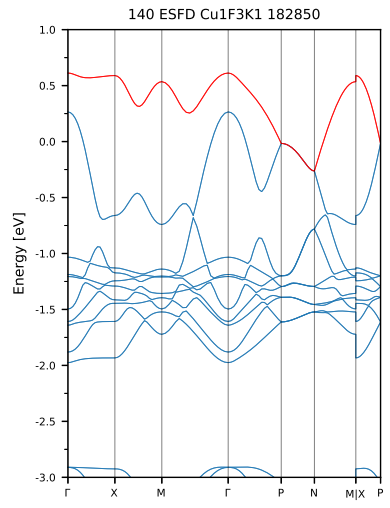

(dp) 140.4.58

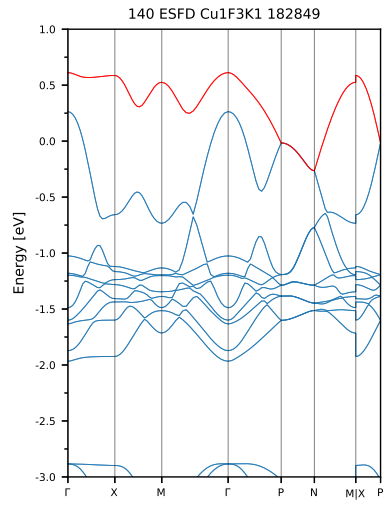

(dq) 140.4.59

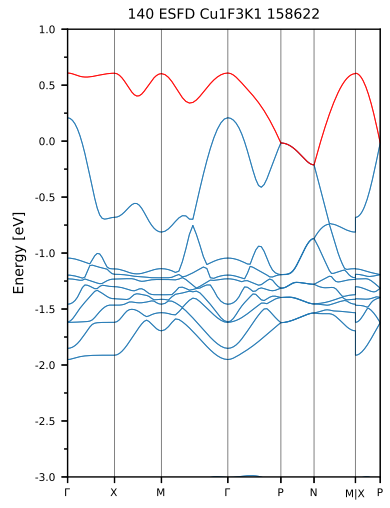

(dr) 140.4.60

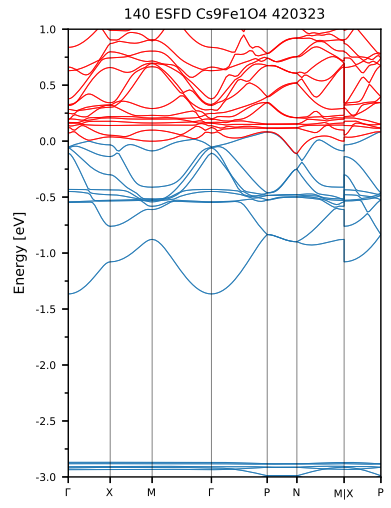

(ds) 140.4.61

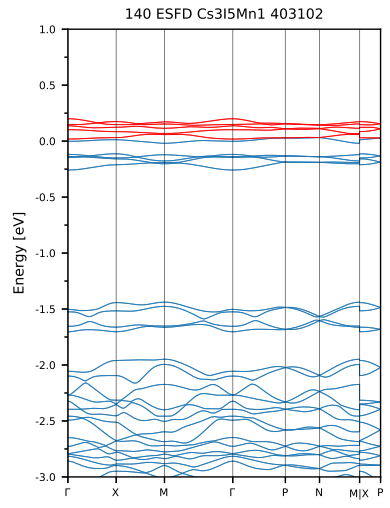

(dt) 140.4.62

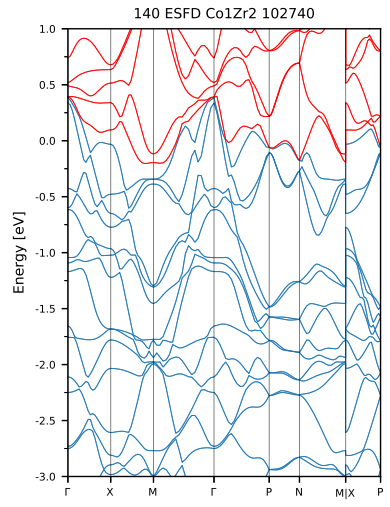

(du) 140.4.63

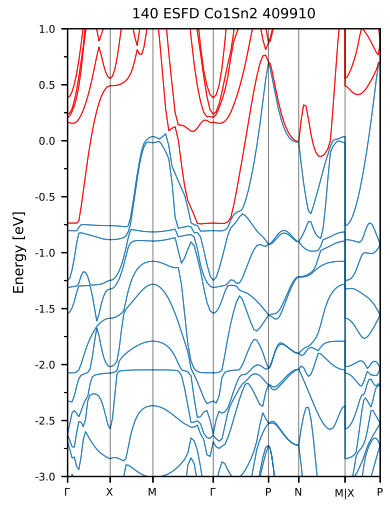

(dv) 140.4.64

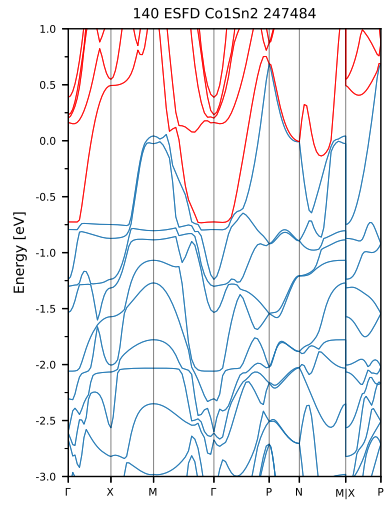

(dw) 140.4.65

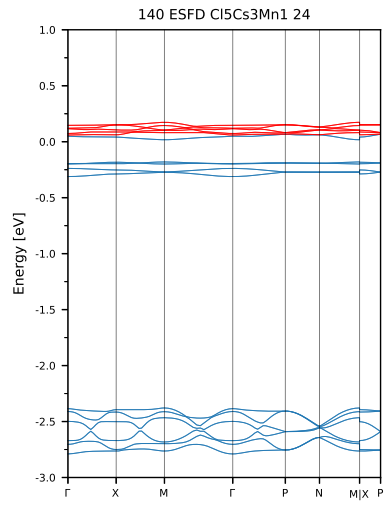

(dx) 140.4.66

Figure 317: Band structures for topological materials in SG 140, part 8/11, from 140.4.51 to 140.4.66. See the list of figures.

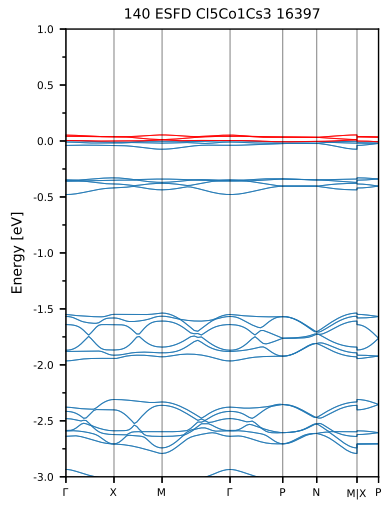

(dy) 140.4.67

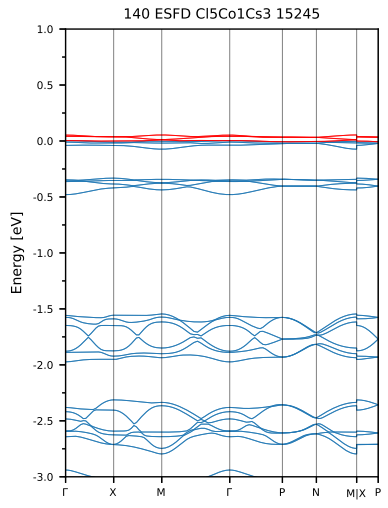

(dz) 140.4.68

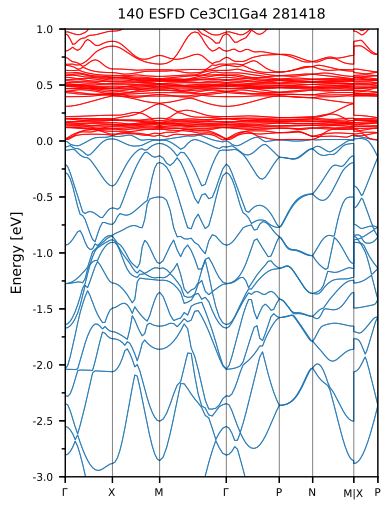

(ea) 140.4.69

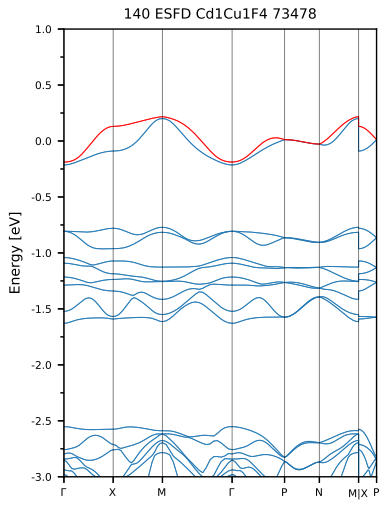

(eb) 140.4.70

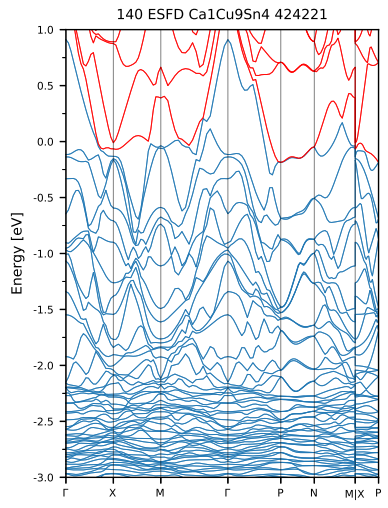

(ec) 140.4.71

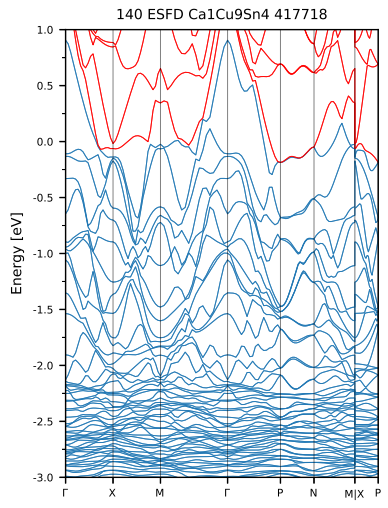

(ed) 140.4.72

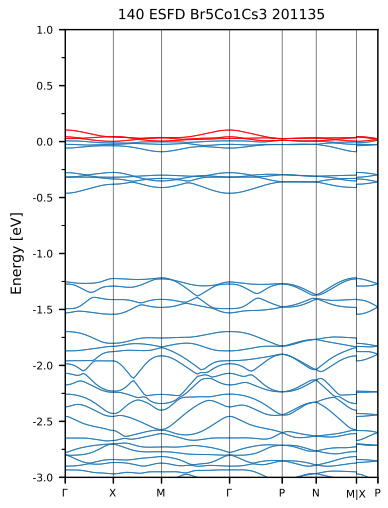

(ee) 140.4.73

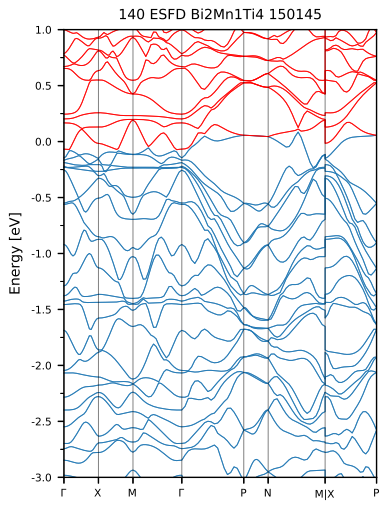

(ef) 140.4.74

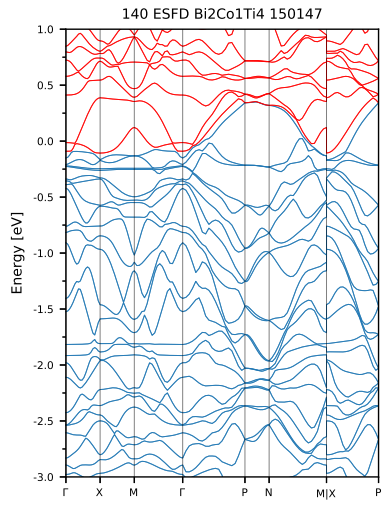

(eg) 140.4.75

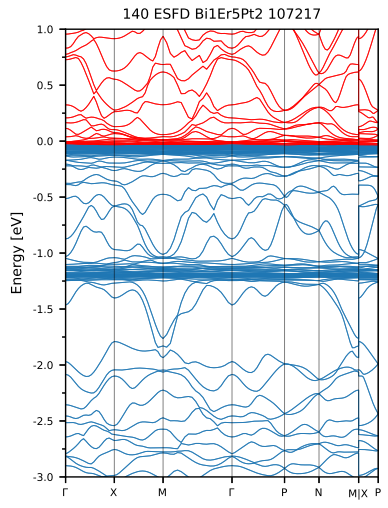

(eh) 140.4.76

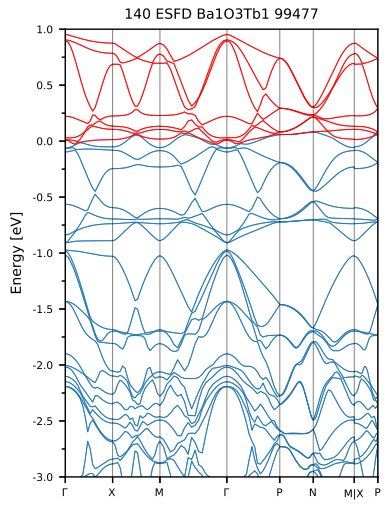

(ei) 140.4.77

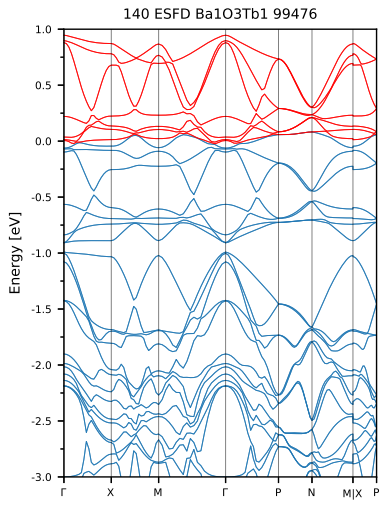

(ej) 140.4.78

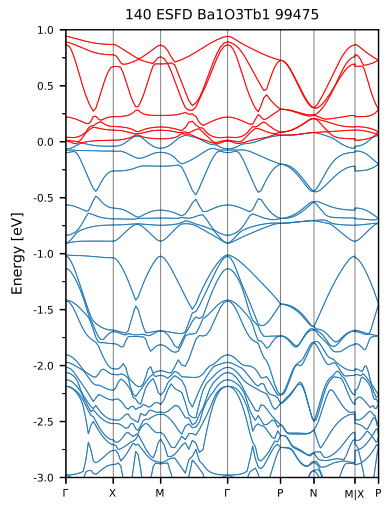

(ek) 140.4.79

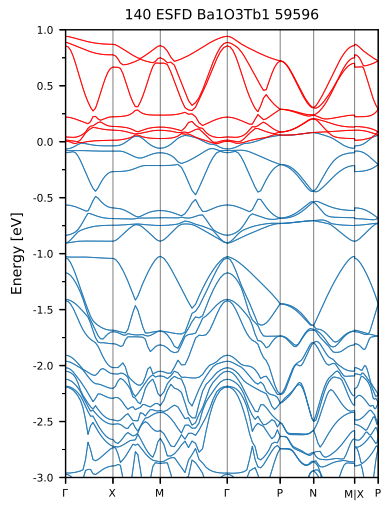

(el) 140.4.80

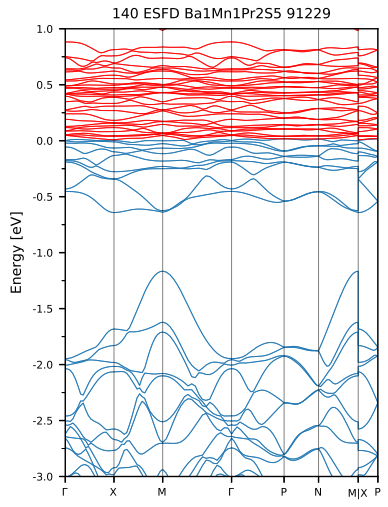

(em) 140.4.81

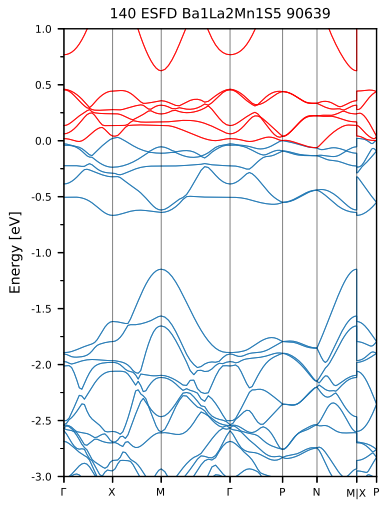

(en) 140.4.82

Figure 318: Band structures for topological materials in SG 140, part 9/11, from 140.4.67 to 140.4.82. See the list of figures.

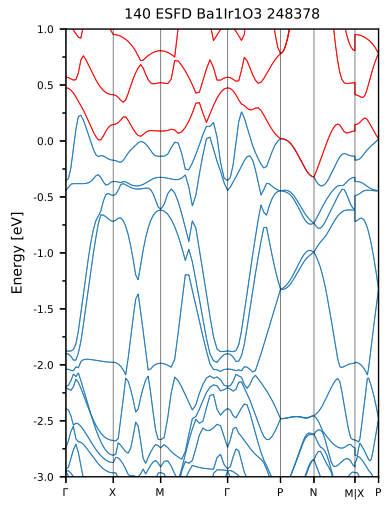

(eo) 140.4.83

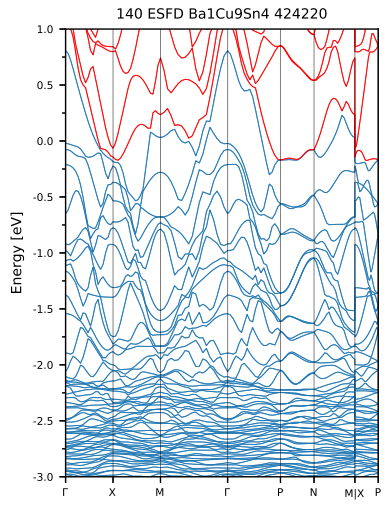

(ep) 140.4.84

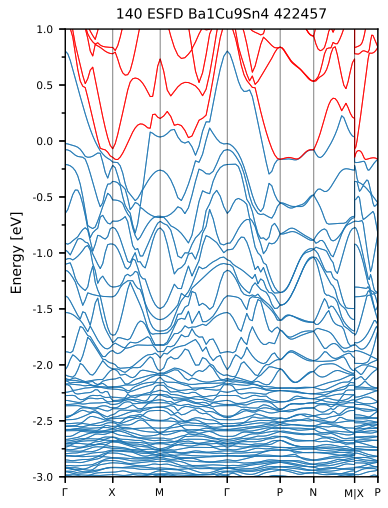

(eq) 140.4.85

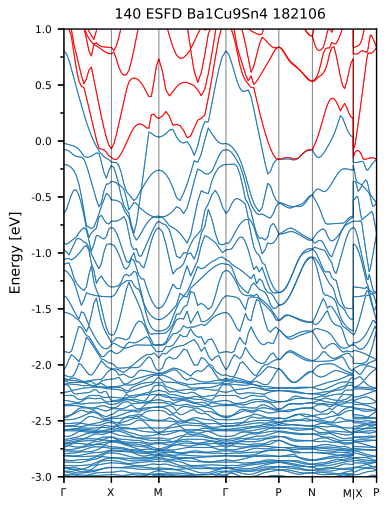

(er) 140.4.86

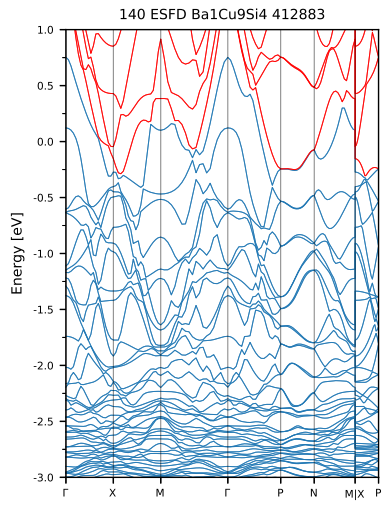

(es) 140.4.87

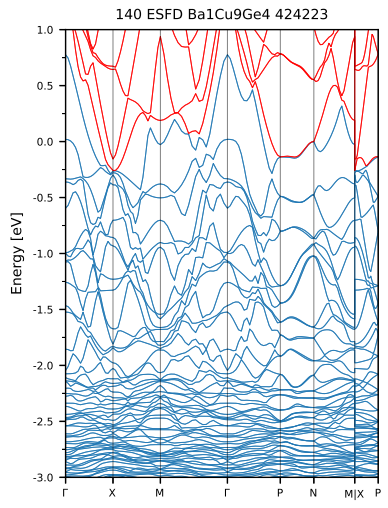

(et) 140.4.88

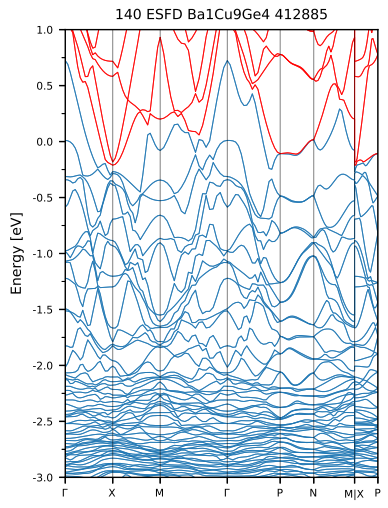

(eu) 140.4.89

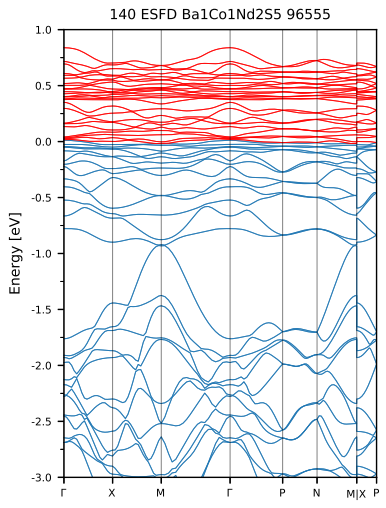

(ev) 140.4.90

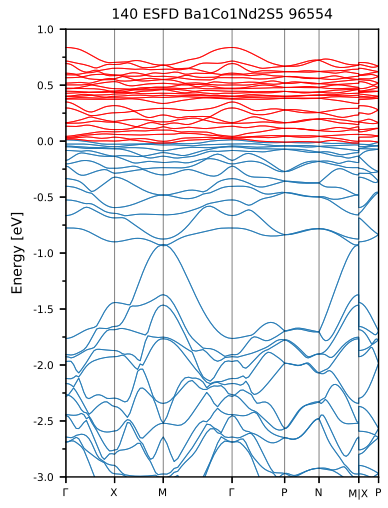

(ew) 140.4.91

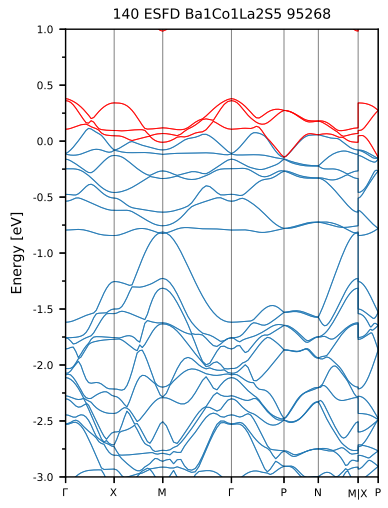

(ex) 140.4.92

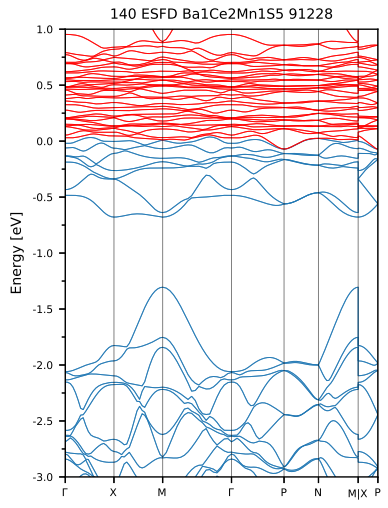

(ey) 140.4.93

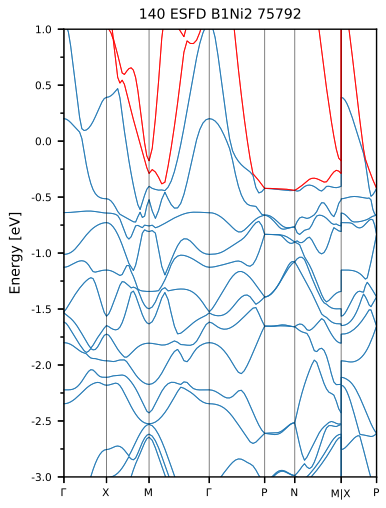

(ez) 140.4.94

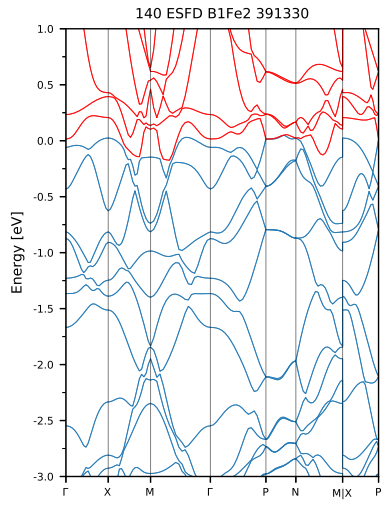

(fa) 140.4.95

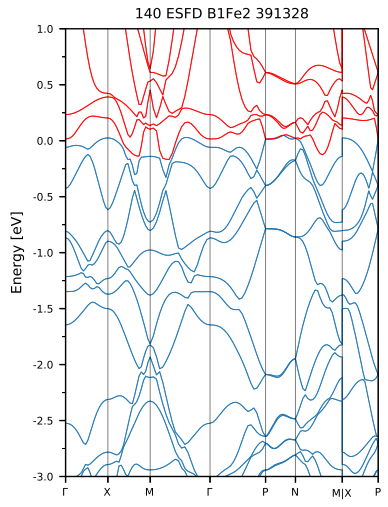

(fb) 140.4.96

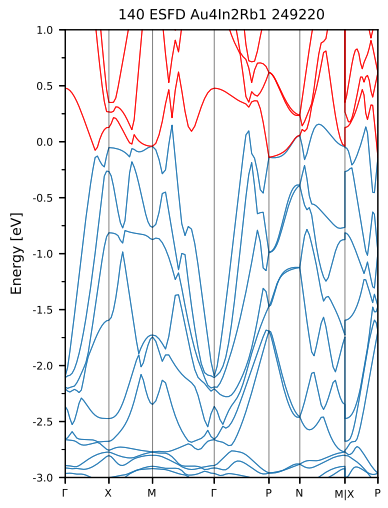

(fc) 140.4.97

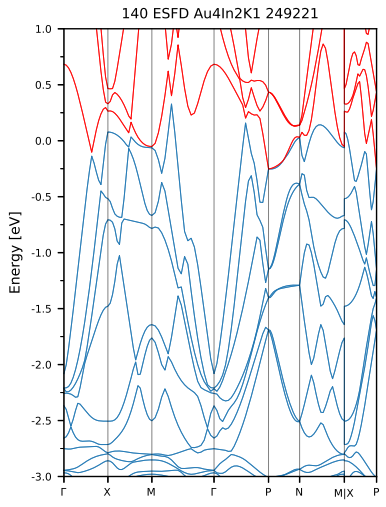

(fd) 140.4.98

Figure 319: Band structures for topological materials in SG 140, part 10/11, from 140.4.83 to 140.4.98. See the list of figures.

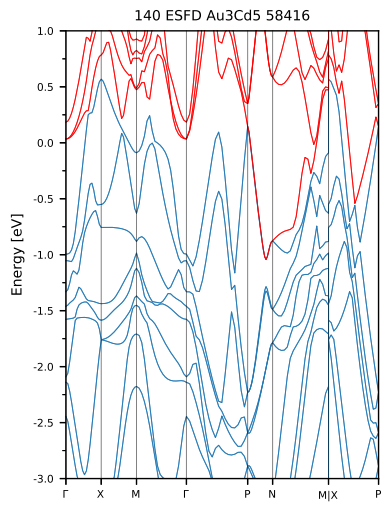

(fe) 140.4.99

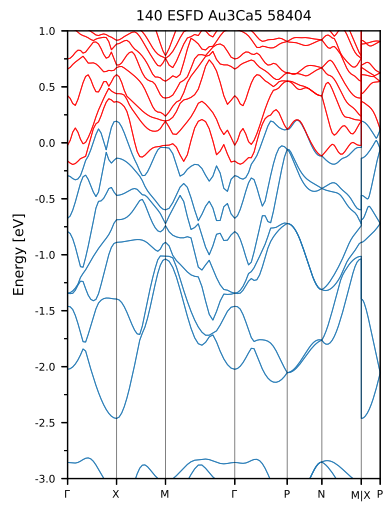

(ff) 140.4.100

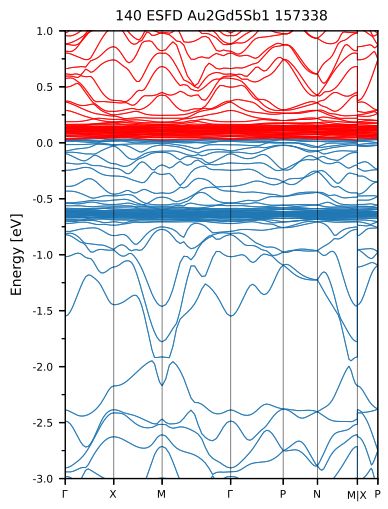

(fg) 140.4.101

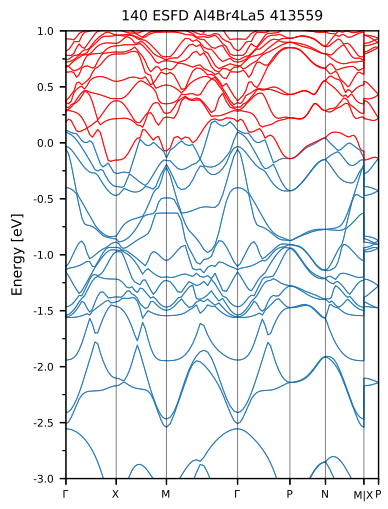

(fh) 140.4.102

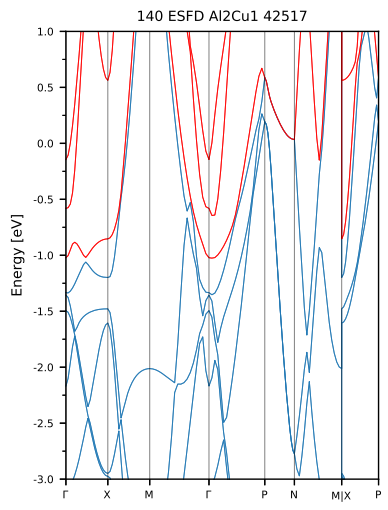

(fi) 140.4.103

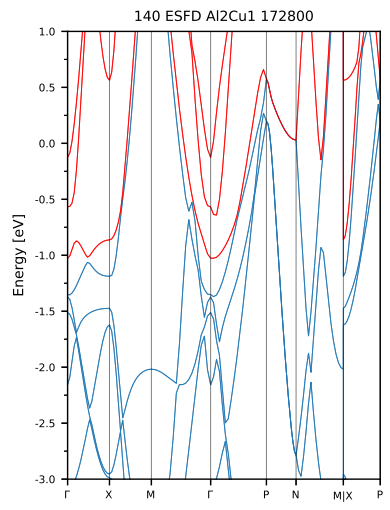

(fj) 140.4.104

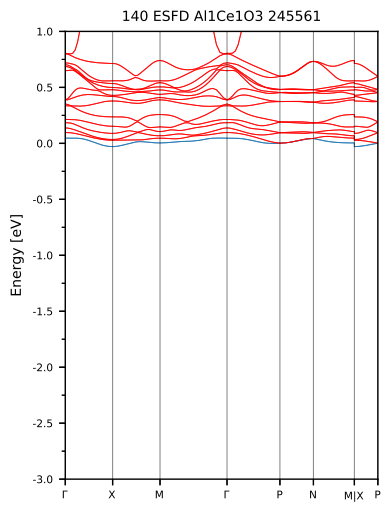

(fk) 140.4.105

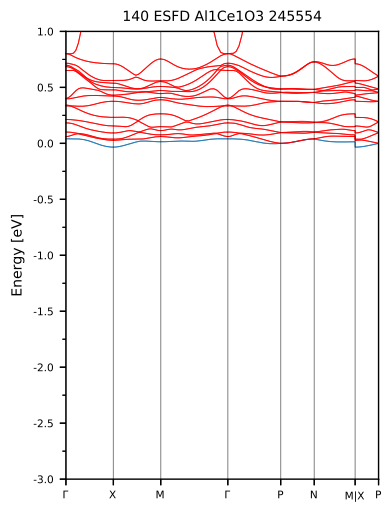

(fl) 140.4.106

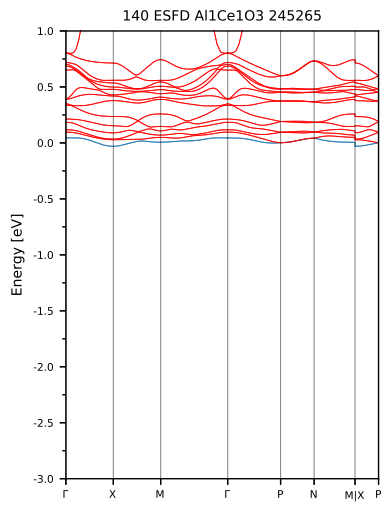

(fm) 140.4.107

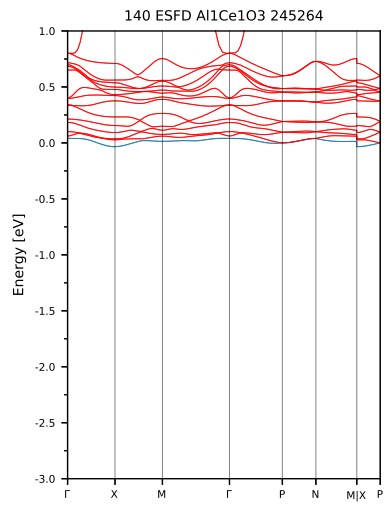

(fn) 140.4.108

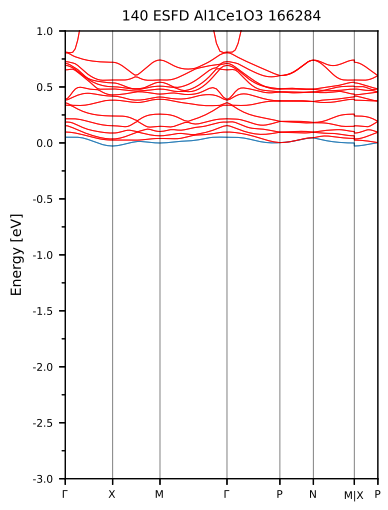

(fo) 140.4.109

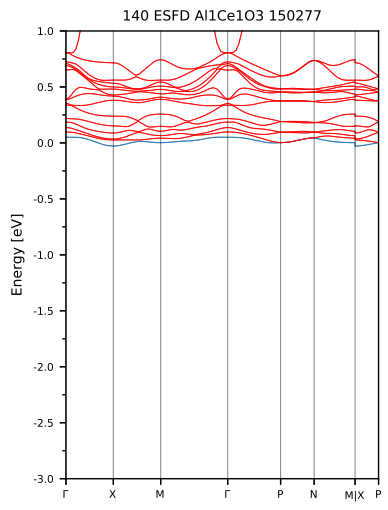

(fp) 140.4.110

Figure 320: Band structures for topological materials in SG 140, part 11/11, from 140.4.99 to 140.4.110. See the list of figures.

### 0.0.67 SG 142

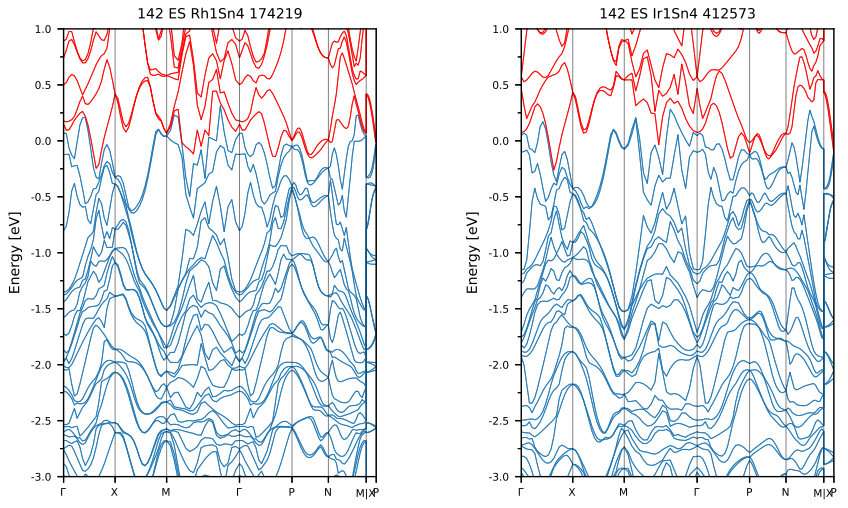

(a) 142.3.1 (b) 142.3.2

Figure 321: Band structures for topological materials in SG 142, part 1/1, from 142.3.1 to 142.3.2. See the list of figures.

### 0.0.68 SG 143

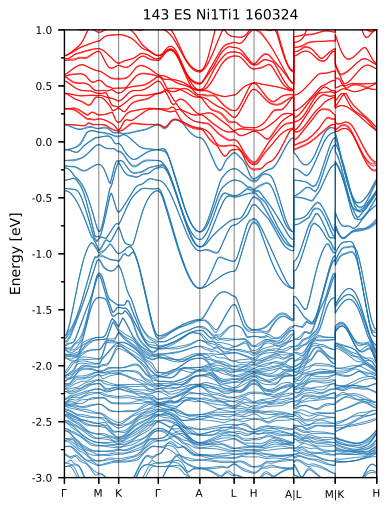

(a) 143.3.1

Figure 322: Band structures for topological materials in SG 143, part 1/1, from 143.3.1 to 143.3.1. See the list of figures.

### 0.0.69 SG 146

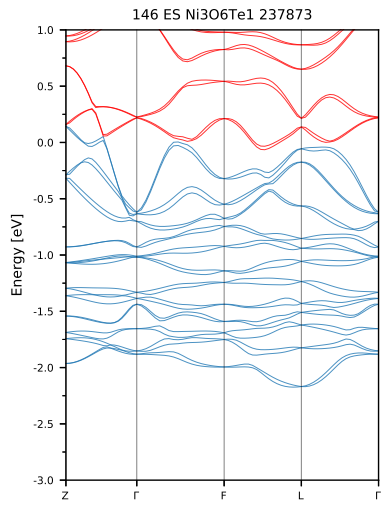

(a) 146.3.1

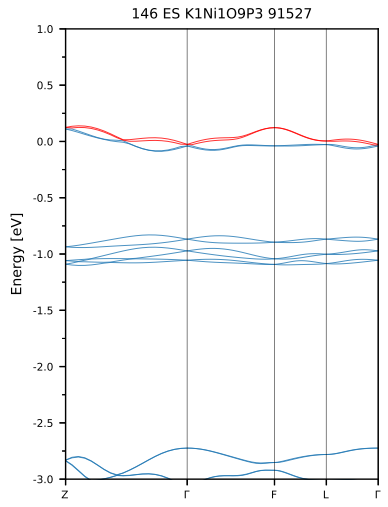

(b) 146.3.2

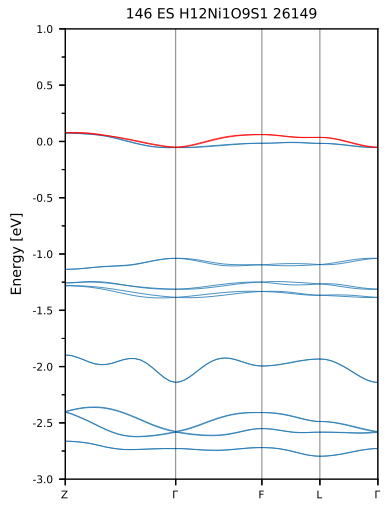

(c) 146.3.3

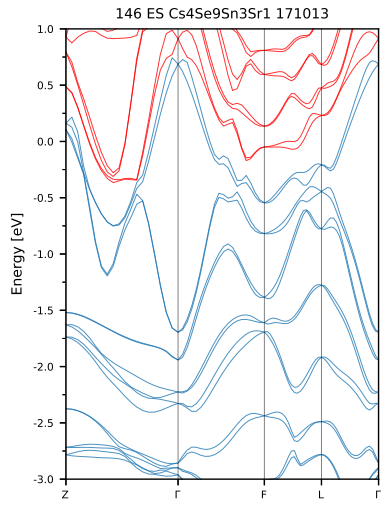

(d) 146.3.4

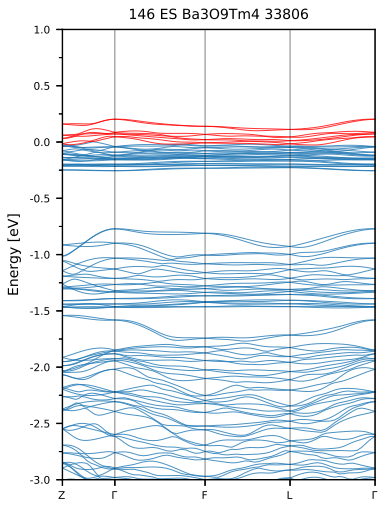

(e) 146.3.5

Figure 323: Band structures for topological materials in SG 146, part 1/1, from 146.3.1 to 146.3.5. See the list of figures.

### 0.0.70 SG 147

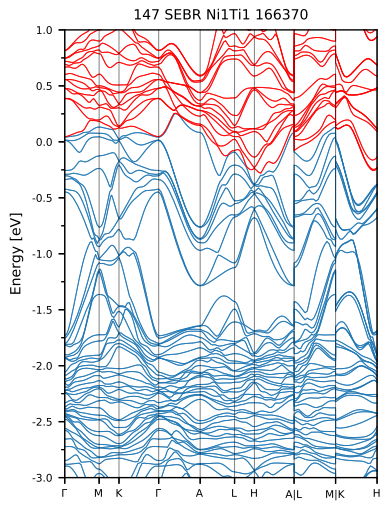

(a) 147.1.1

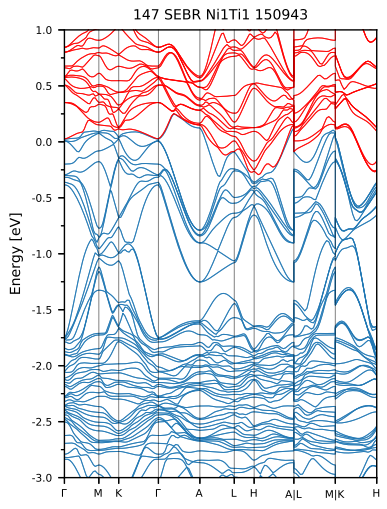

(b) 147.1.2

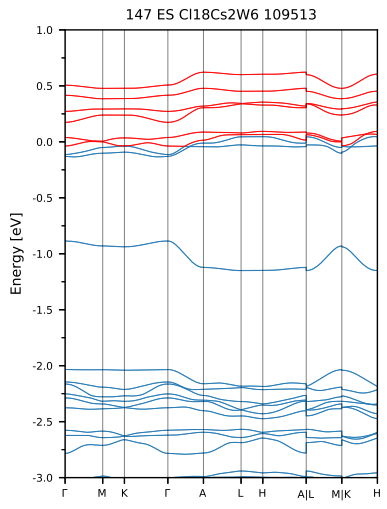

(c) 147.3.1

Figure 324: Band structures for topological materials in SG 147, part 1/1, from 147.1.1 to 147.3.1. See the list of figures.

## 0.0.71 SG 148

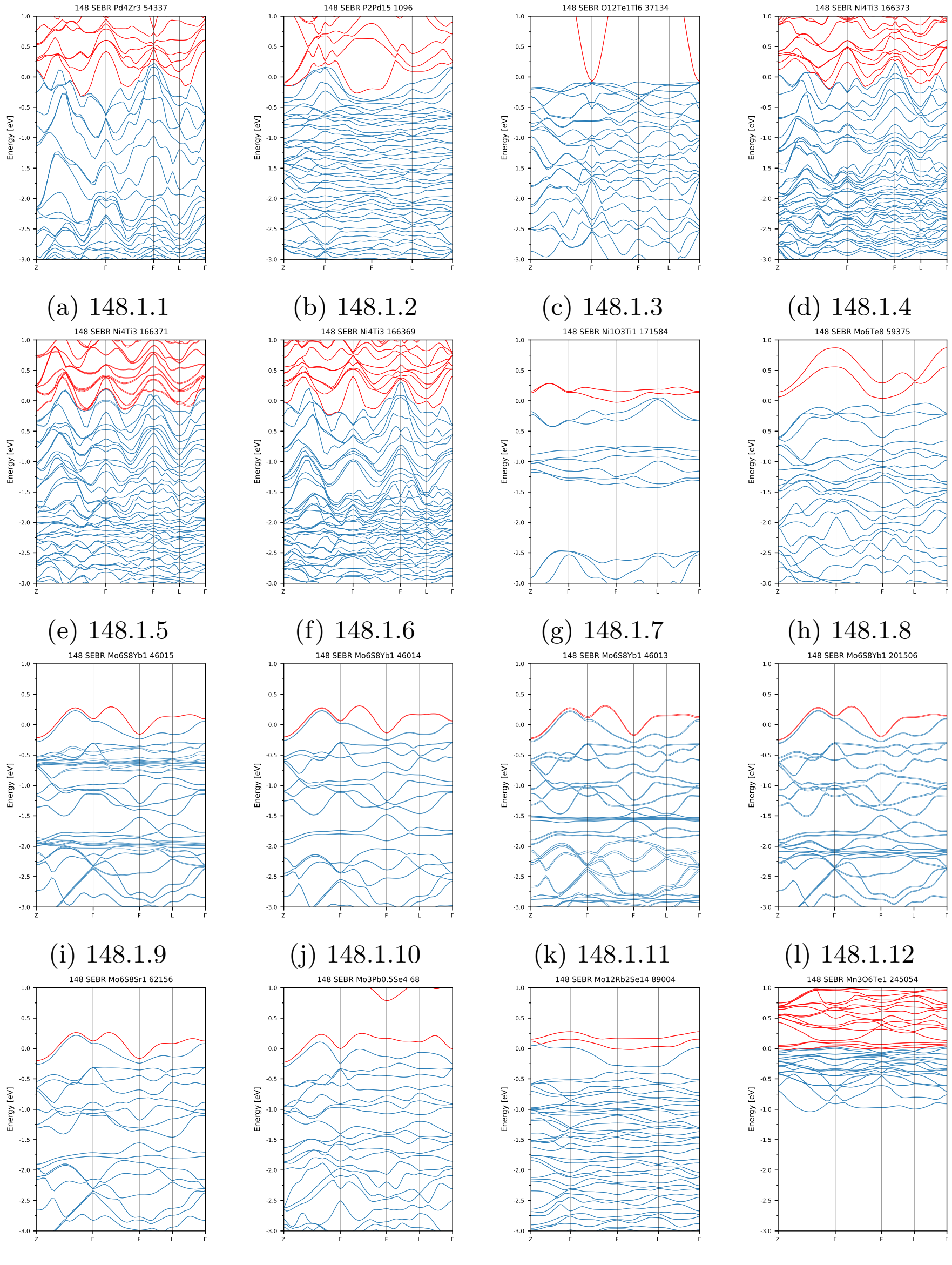

(a) 148.1.1 (b) 148.1.2 (c) 148.1.3 (d) 148.1.4

(e) 148.1.5 (f) 148.1.6 (g) 148.1.7 (h) 148.1.8

(i) 148.1.9 (j) 148.1.10 (k) 148.1.11 (l) 148.1.12

(m) 148.1.13 (n) 148.1.14 (o) 148.1.15 (p) 148.1.16

Figure 325: Band structures for topological materials in SG 148, part 1/5, from 148.1.1 to 148.1.16. See the list of figures.

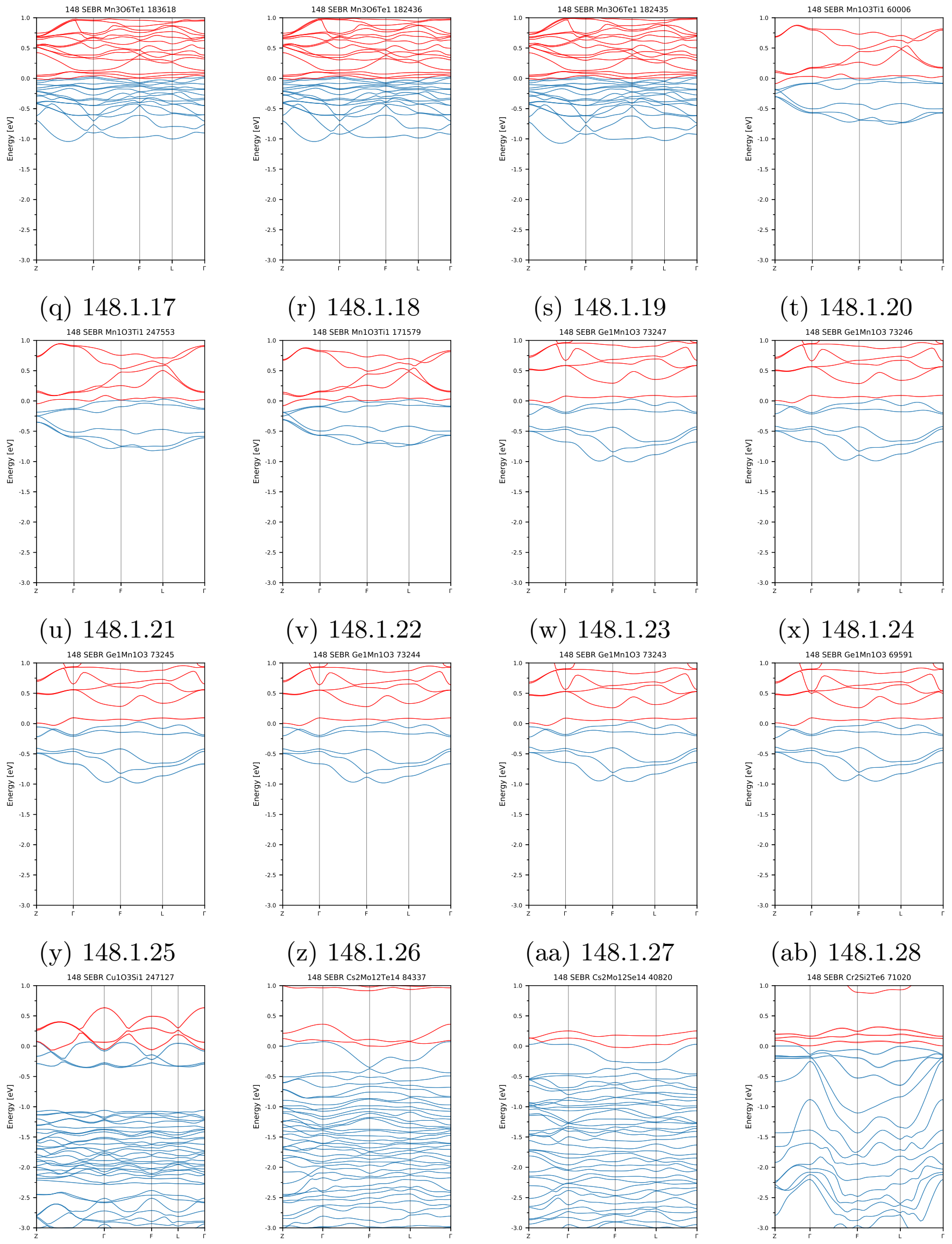

(q) 148.1.17 (r) 148.1.18 (s) 148.1.19 (t) 148.1.20

(u) 148.1.21 (v) 148.1.22 (w) 148.1.23 (x) 148.1.24

(y) 148.1.25 (z) 148.1.26 (aa) 148.1.27 (ab) 148.1.28

(ac) 148.1.29 (ad) 148.1.30 (ae) 148.1.31 (af) 148.1.32

Figure 326: Band structures for topological materials in SG 148, part 2/5, from 148.1.17 to 148.1.32. See the list of figures.

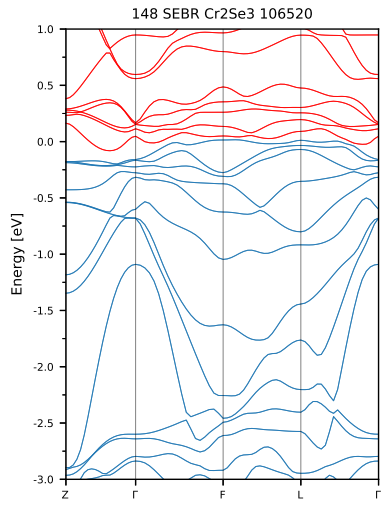

(ag) 148.1.33

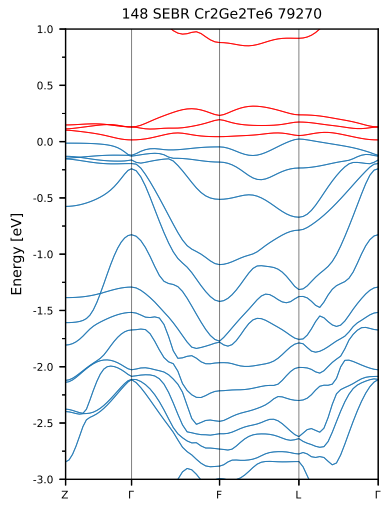

(ah) 148.1.34

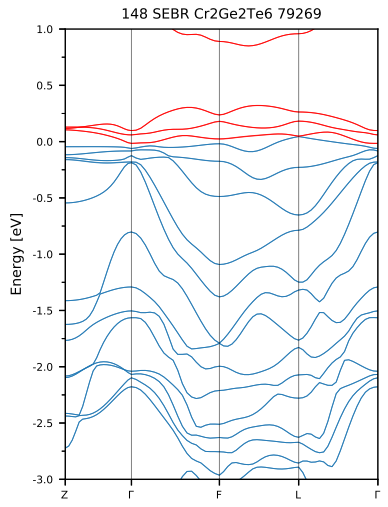

(ai) 148.1.35

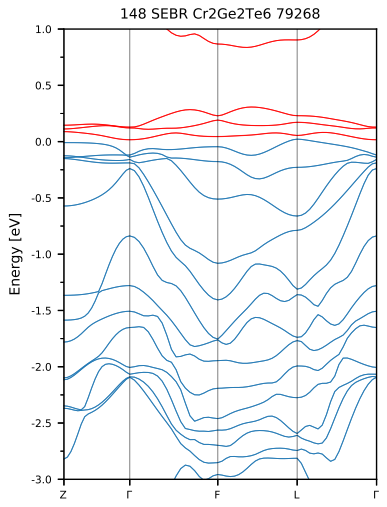

(aj) 148.1.36

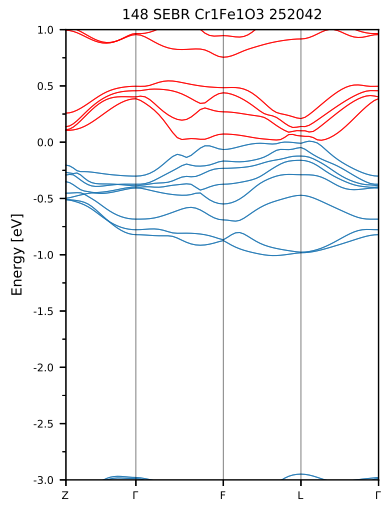

(ak) 148.1.37

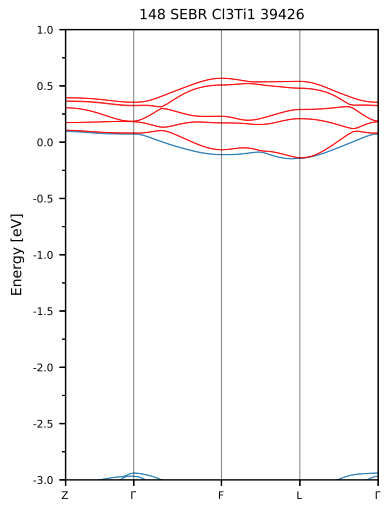

(al) 148.1.38

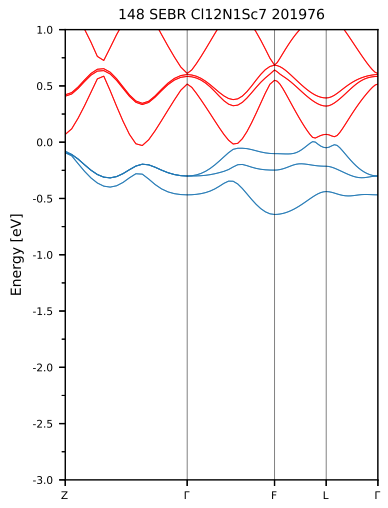

(am) 148.1.39

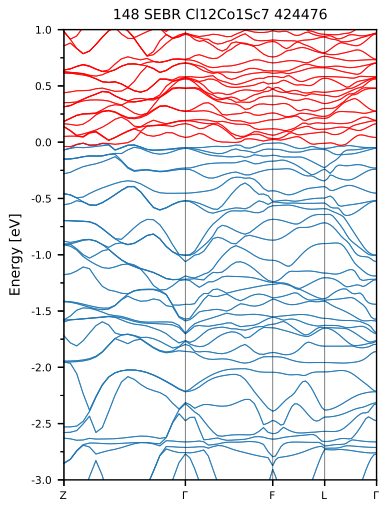

(an) 148.1.40

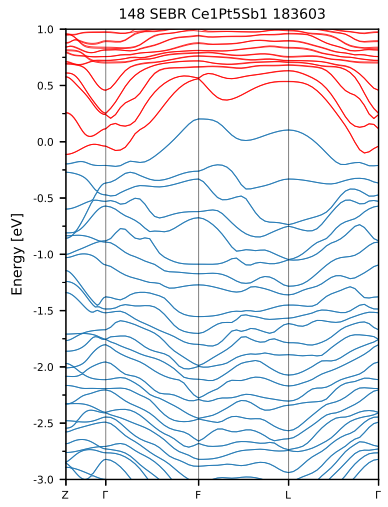

(ao) 148.1.41

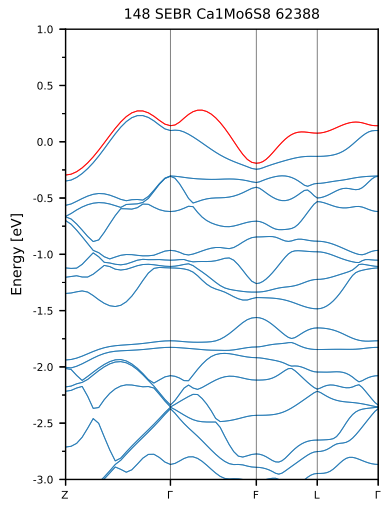

(ap) 148.1.42

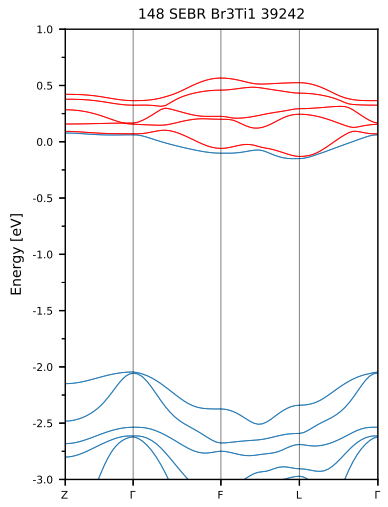

(aq) 148.1.43

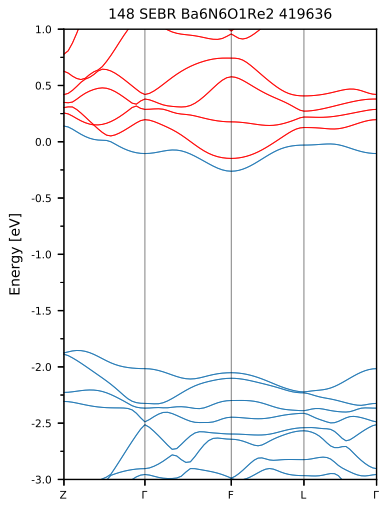

(ar) 148.1.44

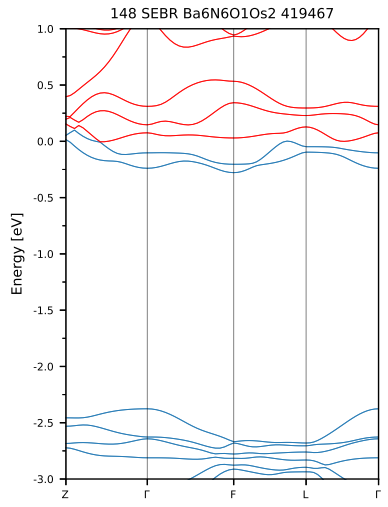

(as) 148.1.45

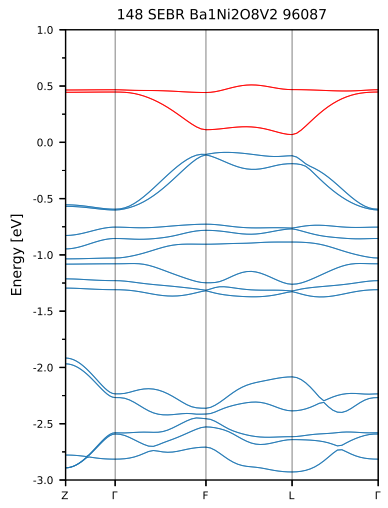

(at) 148.1.46

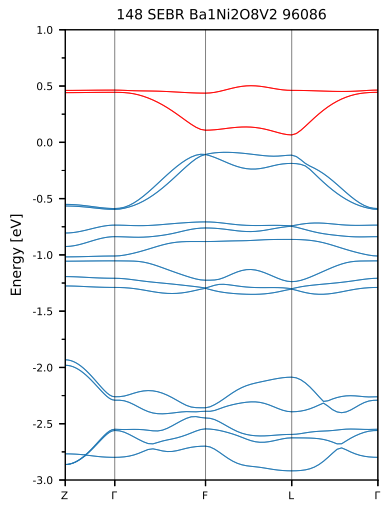

(au) 148.1.47

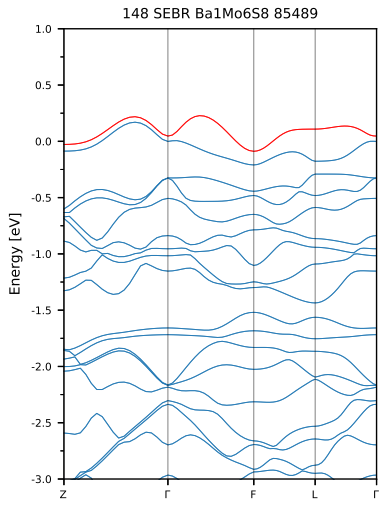

(av) 148.1.48

Figure 327: Band structures for topological materials in SG 148, part 3/5, from 148.1.33 to 148.1.48. See the list of figures.

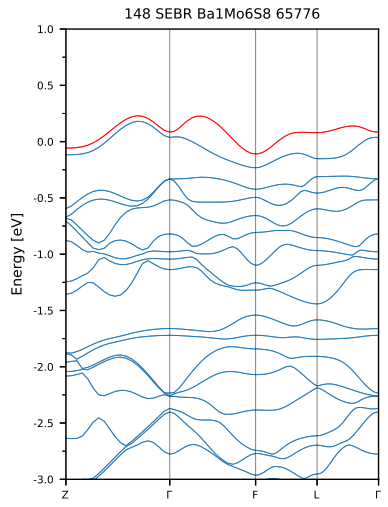

(aw) 148.1.49

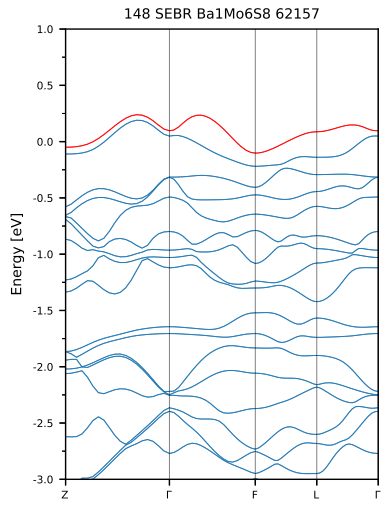

(ax) 148.1.50

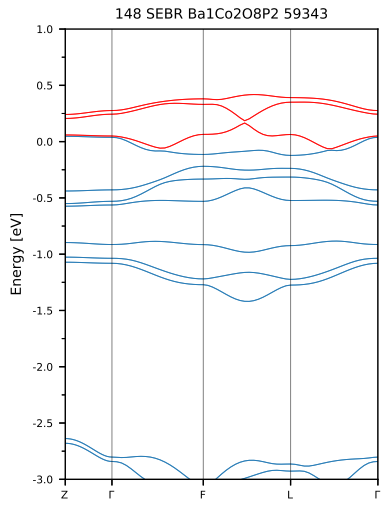

(ay) 148.1.51

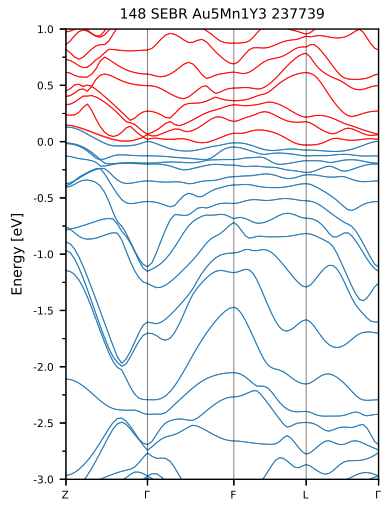

(az) 148.1.52

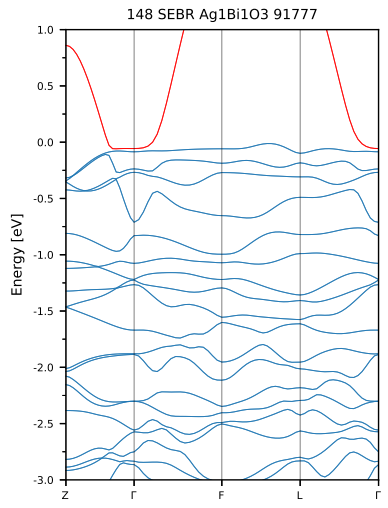

(ba) 148.1.53

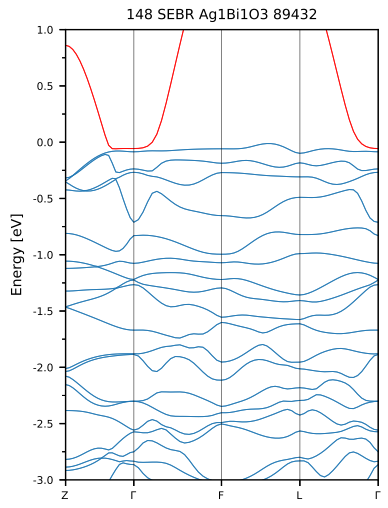

(bb) 148.1.54

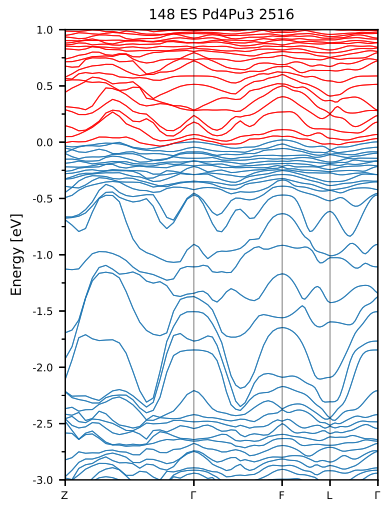

(bc) 148.3.1

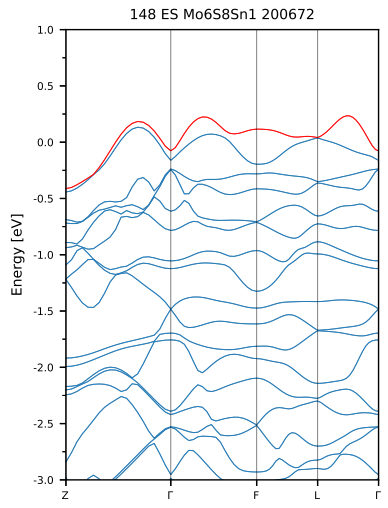

(bd) 148.3.2

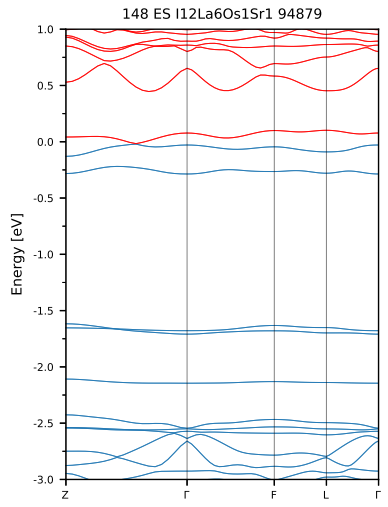

(be) 148.3.3

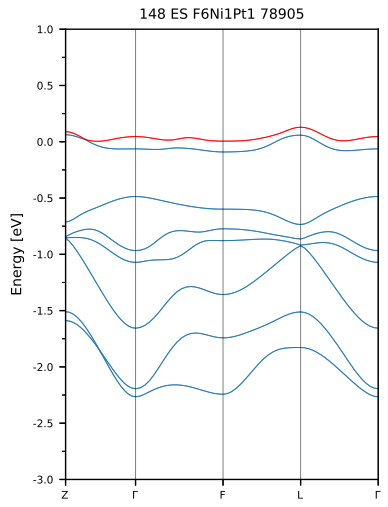

(bf) 148.3.4

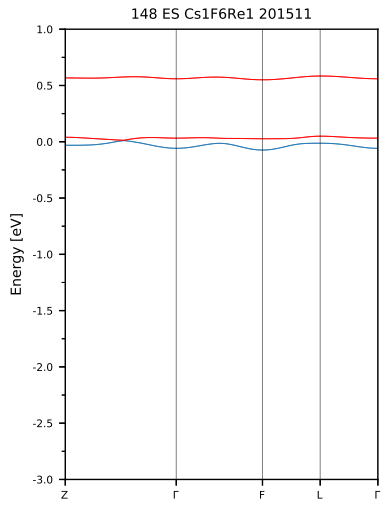

(bg) 148.3.5

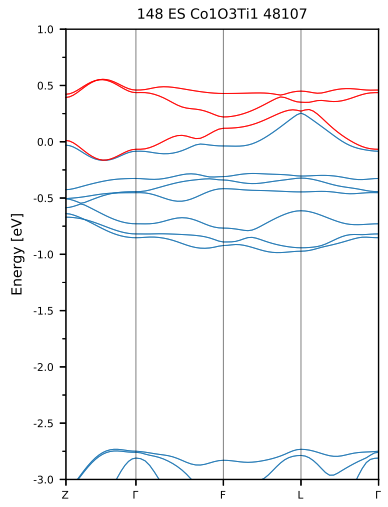

(bh) 148.3.6

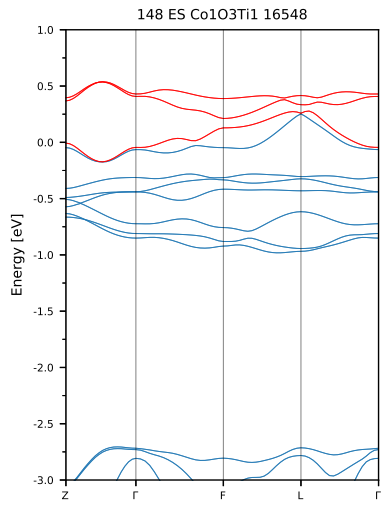

(bi) 148.3.7

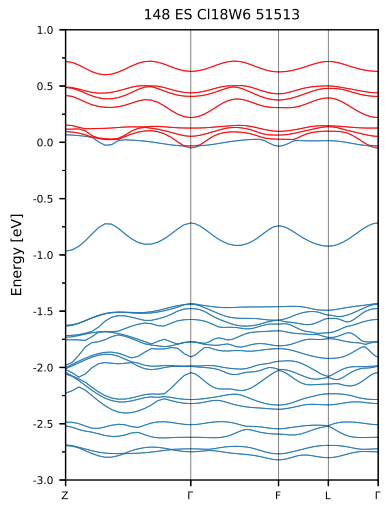

(bj) 148.3.8

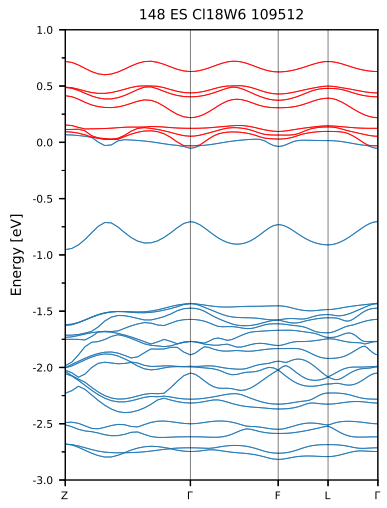

(bk) 148.3.9

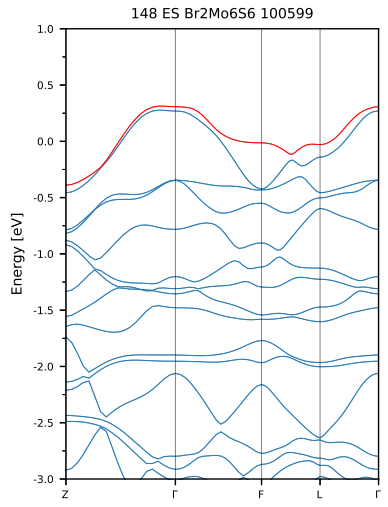

(bl) 148.3.10

Figure 328: Band structures for topological materials in SG 148, part 4/5, from 148.1.49 to 148.3.10. See the list of figures.

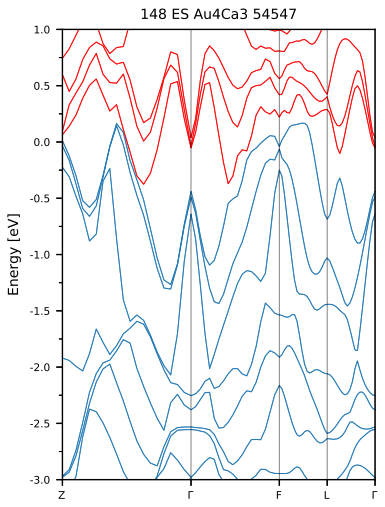

(bm) 148.3.11

Figure 329: Band structures for topological materials in SG 148, part 5/5, from 148.3.11 to 148.3.11. See the list of figures.

### 0.0.72 SG 152

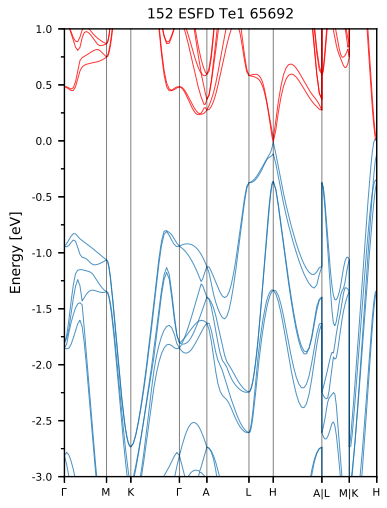

(a) 152.4.1

Figure 330: Band structures for topological materials in SG 152, part 1/1, from 152.4.1 to 152.4.1. See the list of figures.

### 0.0.73 SG 155

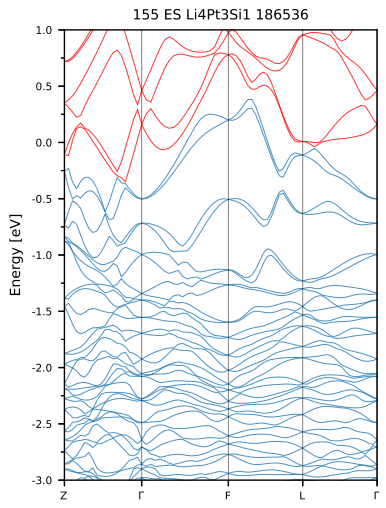

(a) 155.3.1

Figure 331: Band structures for topological materials in SG 155, part 1/1, from 155.3.1 to 155.3.1. See the list of figures.

## 0.0.74 SG 156

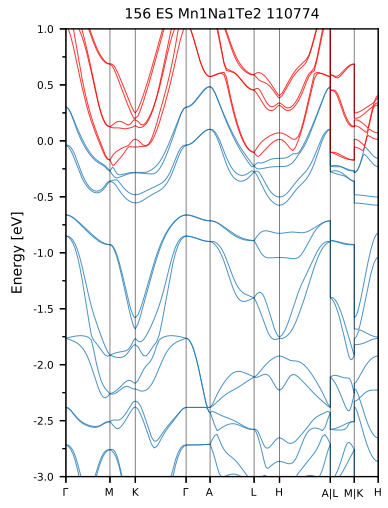

(a) 156.3.1

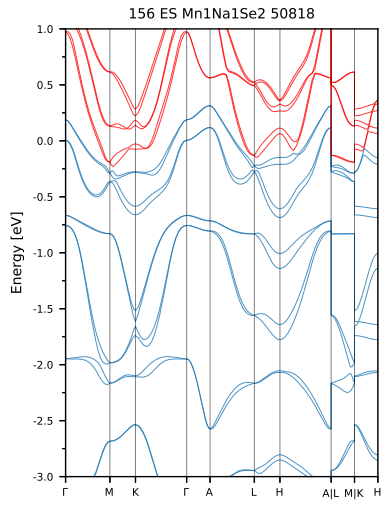

(b) 156.3.2

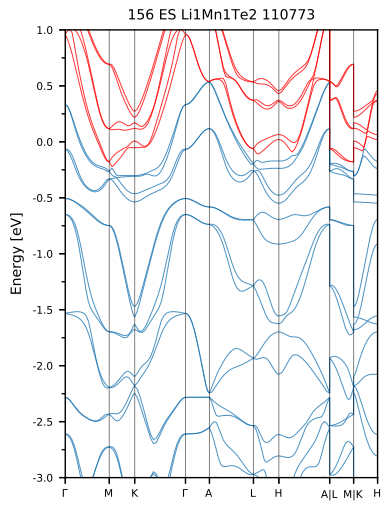

(c) 156.3.3

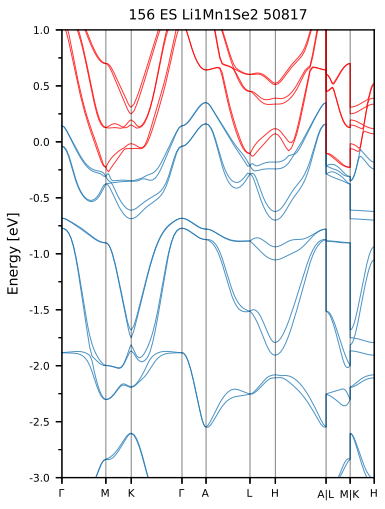

(d) 156.3.4

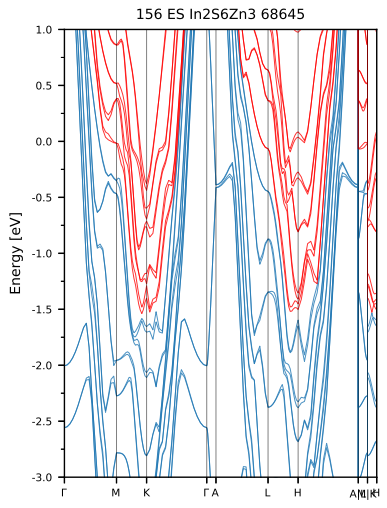

(e) 156.3.5

Figure 332: Band structures for topological materials in SG 156, part 1/1, from 156.3.1 to 156.3.5. See the list of figures.

## 0.0.75 SG 159

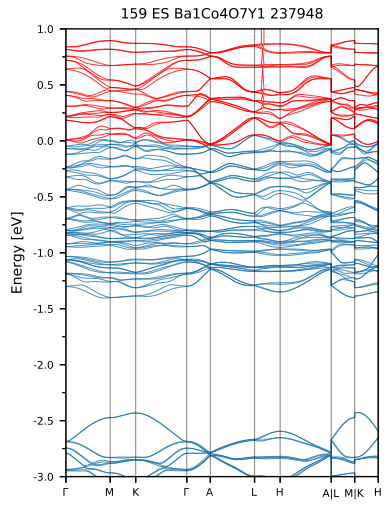

(a) 159.3.1

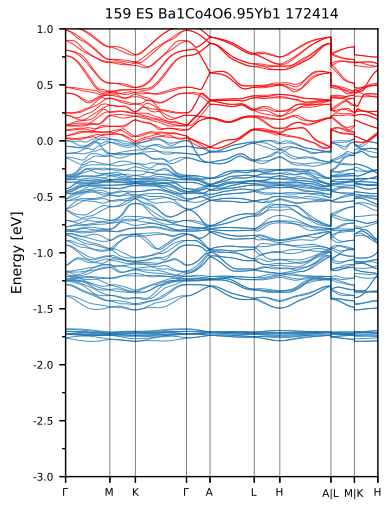

(b) 159.3.2

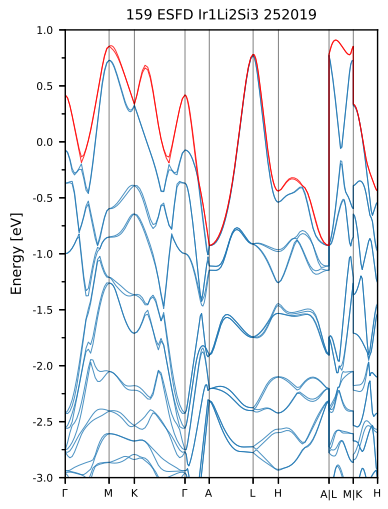

(c) 159.4.1

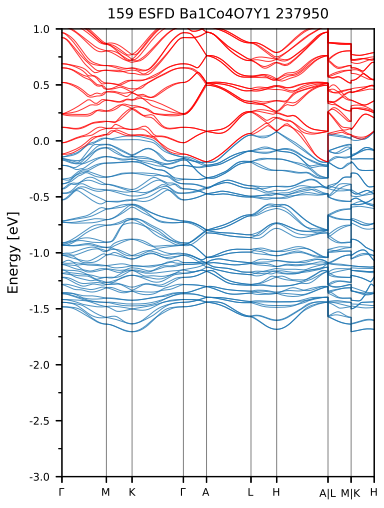

(d) 159.4.2

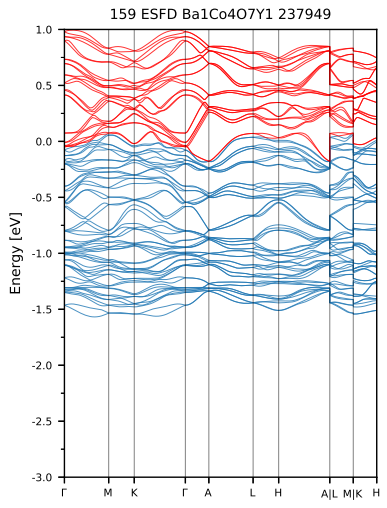

(e) 159.4.3

Figure 333: Band structures for topological materials in SG 159, part 1/1, from 159.3.1 to 159.4.3. See the list of figures.

### 0.0.76 SG 160

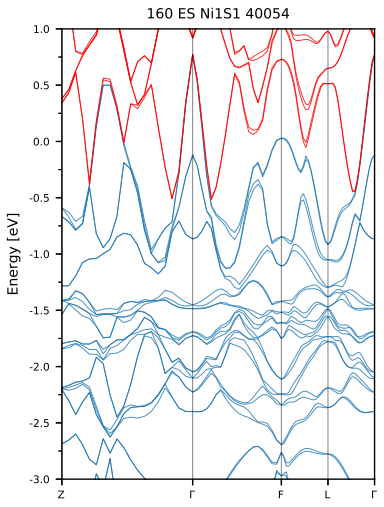

(a) 160.3.1

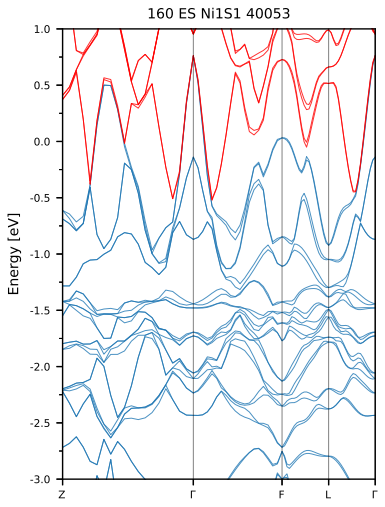

(b) 160.3.2

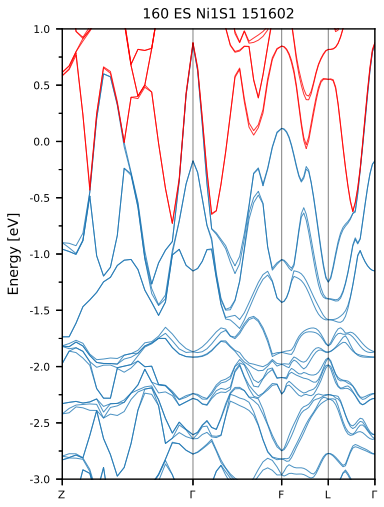

(c) 160.3.3

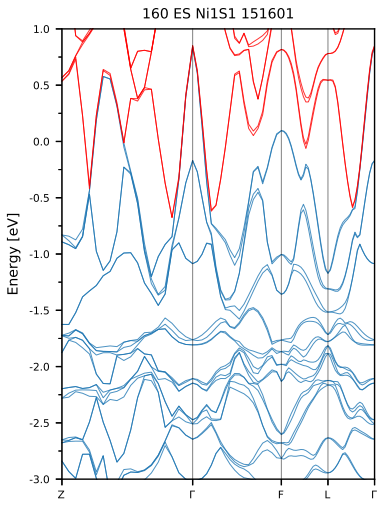

(d) 160.3.4

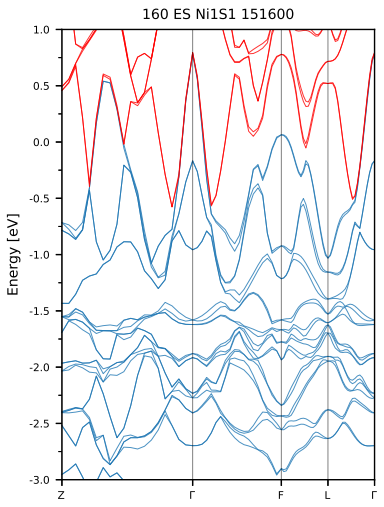

(e) 160.3.5

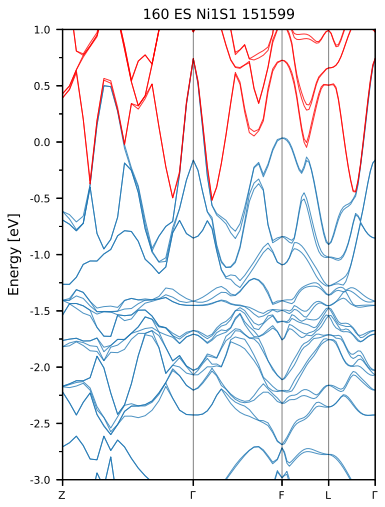

(f) 160.3.6

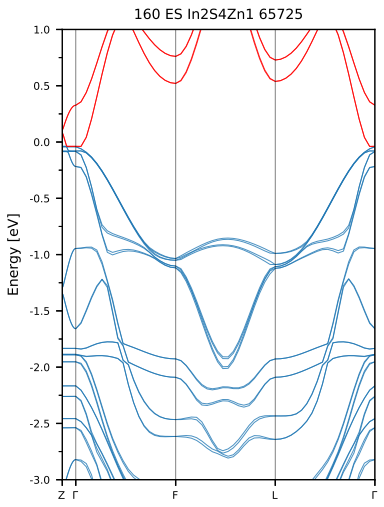

(g) 160.3.7

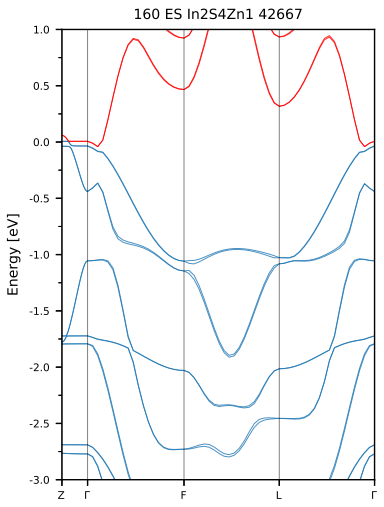

(h) 160.3.8

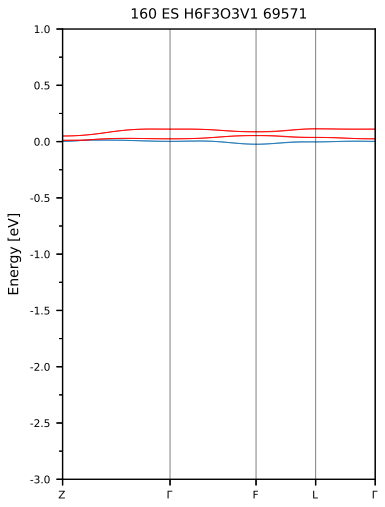

(i) 160.3.9

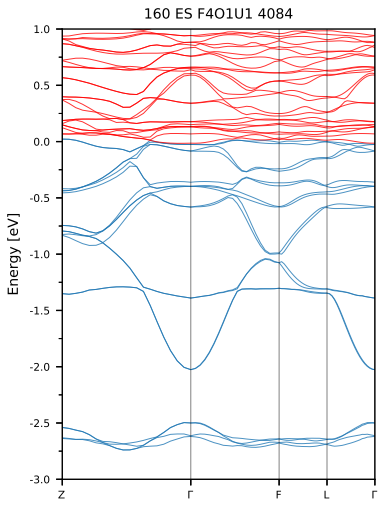

(j) 160.3.10

Figure 334: Band structures for topological materials in SG 160, part 1/1, from 160.3.1 to 160.3.10. See the list of figures.

## 0.0.77 SG 161

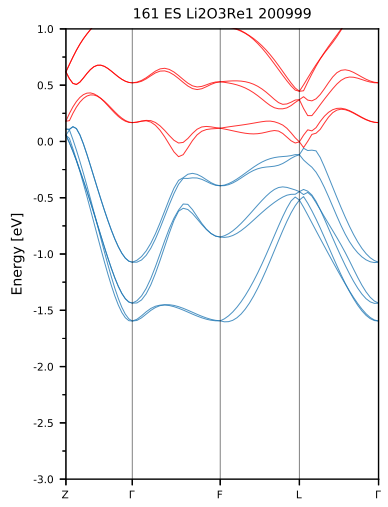

(a) 161.3.1

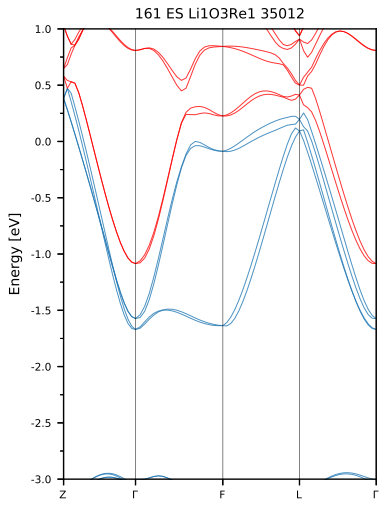

(b) 161.3.2

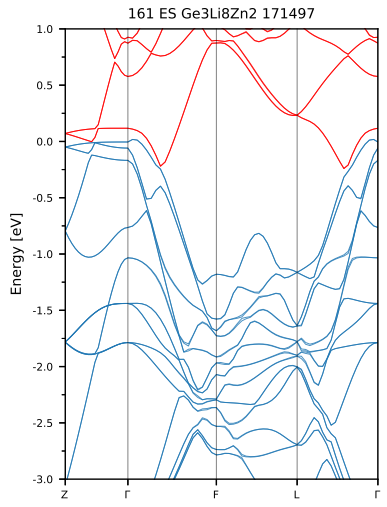

(c) 161.3.3

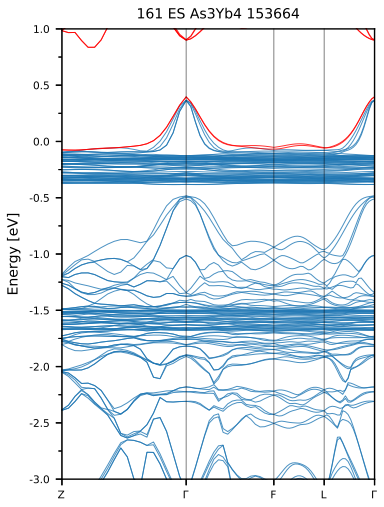

(d) 161.3.4

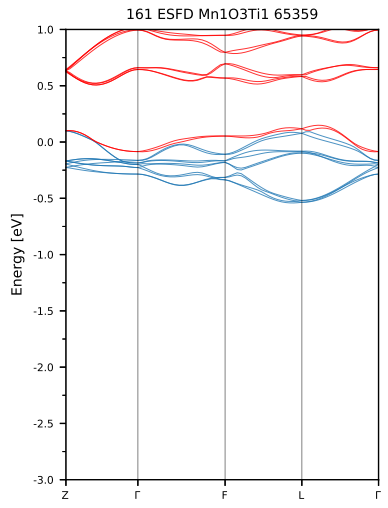

(e) 161.4.1

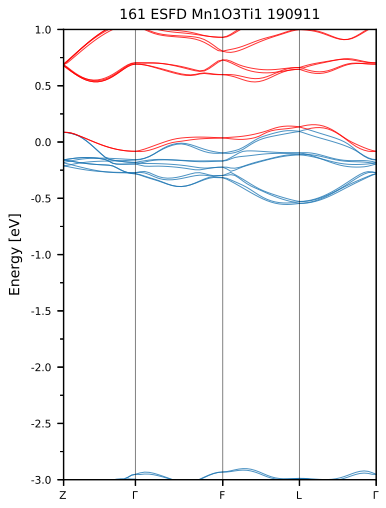

(f) 161.4.2

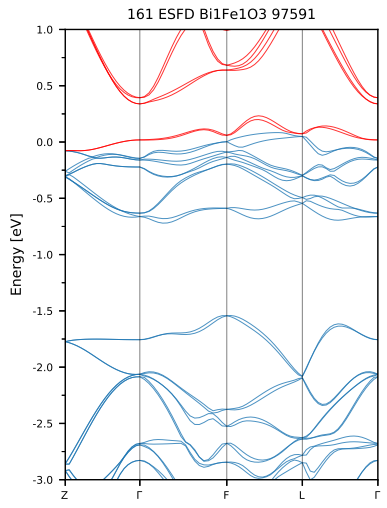

(g) 161.4.3

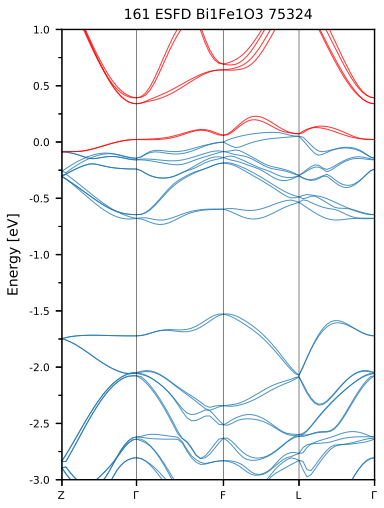

(h) 161.4.4

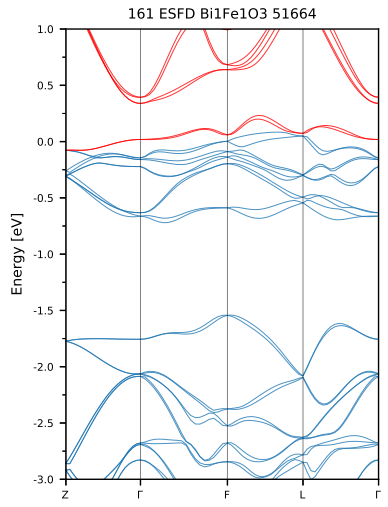

(i) 161.4.5

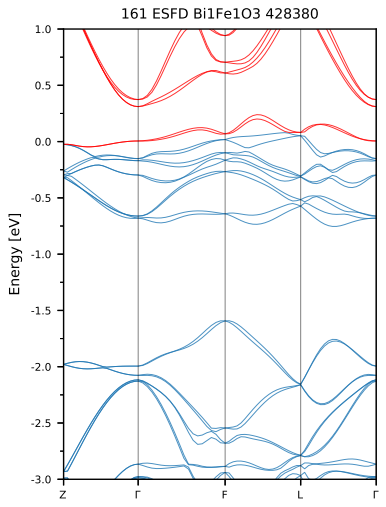

(j) 161.4.6

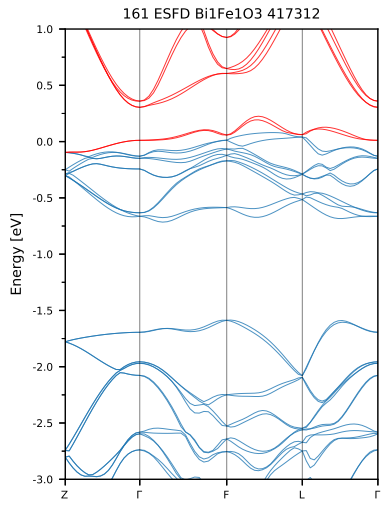

(k) 161.4.7

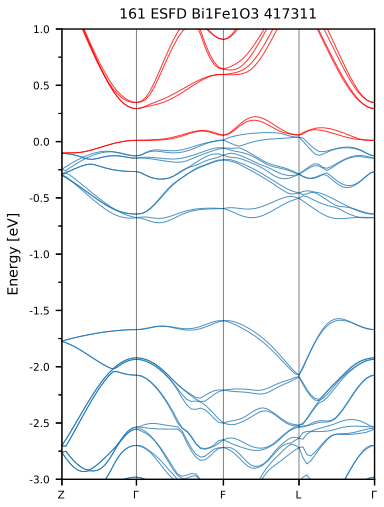

(l) 161.4.8

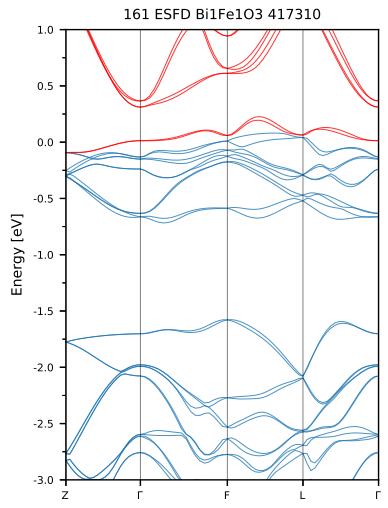

(m) 161.4.9

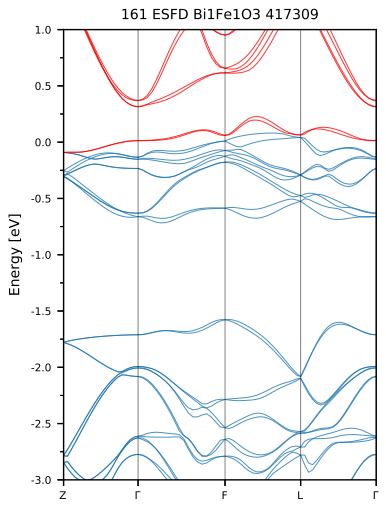

(n) 161.4.10

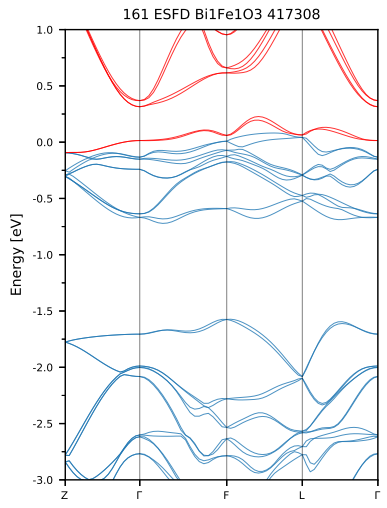

(o) 161.4.11

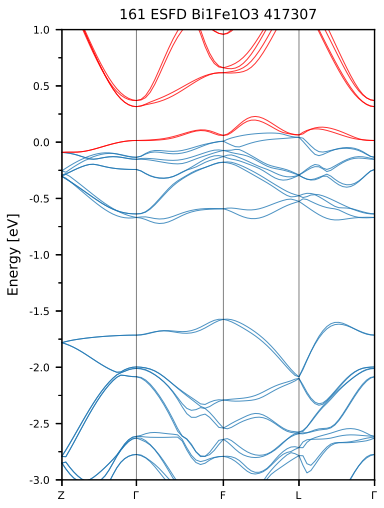

(p) 161.4.12

Figure 335: Band structures for topological materials in SG 161, part 1/3, from 161.3.1 to 161.4.12. See the list of figures.

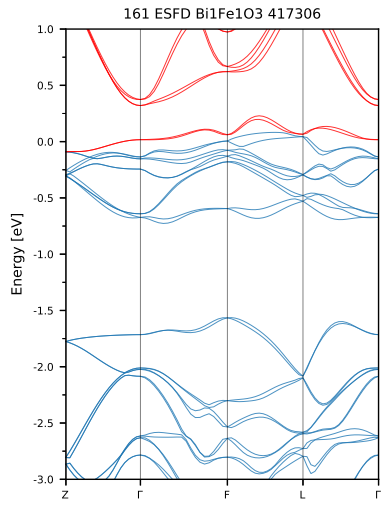

(q) 161.4.13

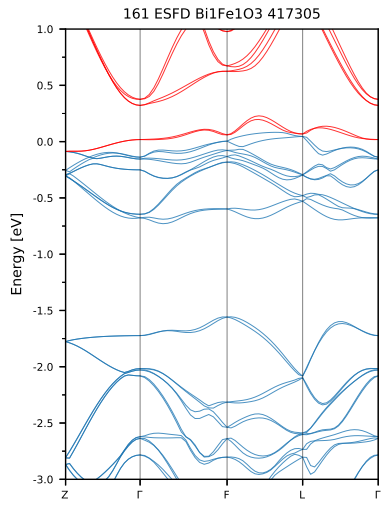

(r) 161.4.14

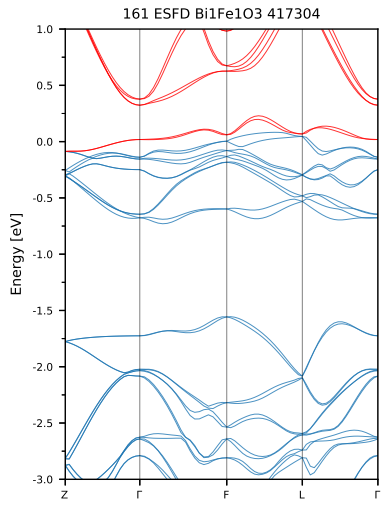

(s) 161.4.15

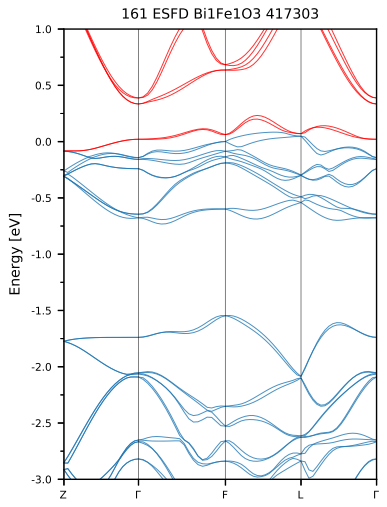

(t) 161.4.16

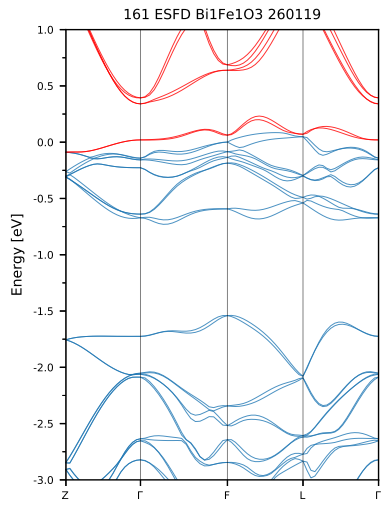

(u) 161.4.17

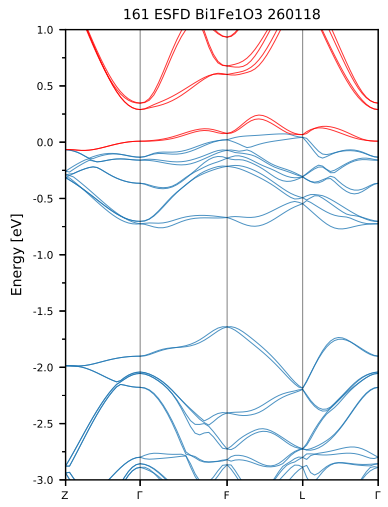

(v) 161.4.18

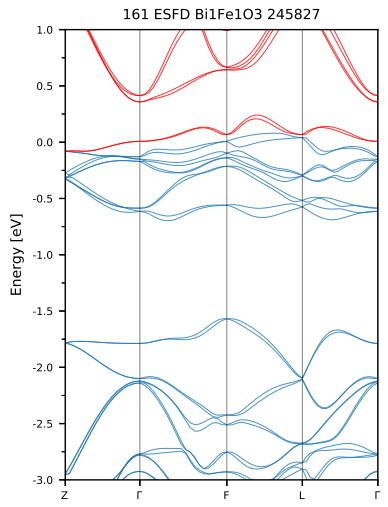

(w) 161.4.19

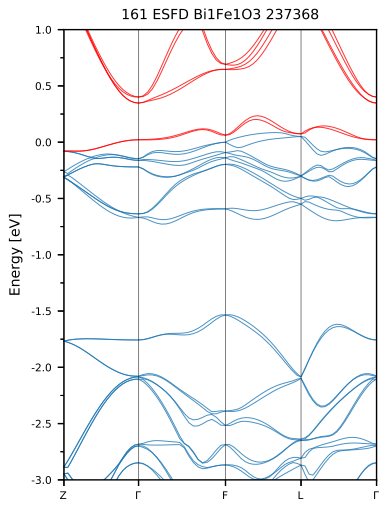

(x) 161.4.20

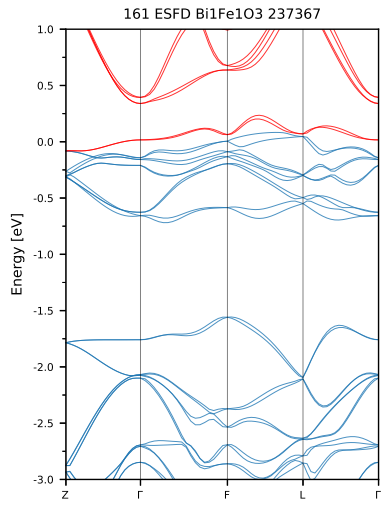

(y) 161.4.21

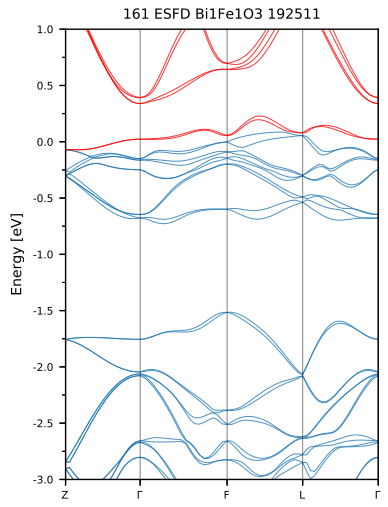

(z) 161.4.22

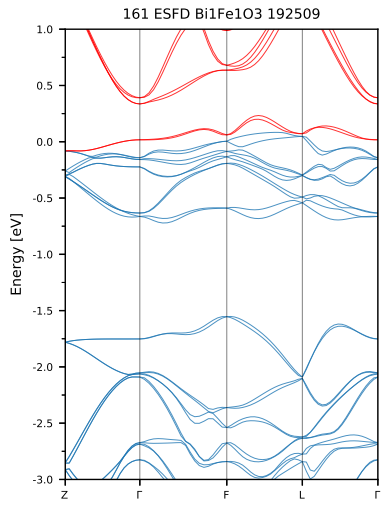

(aa) 161.4.23

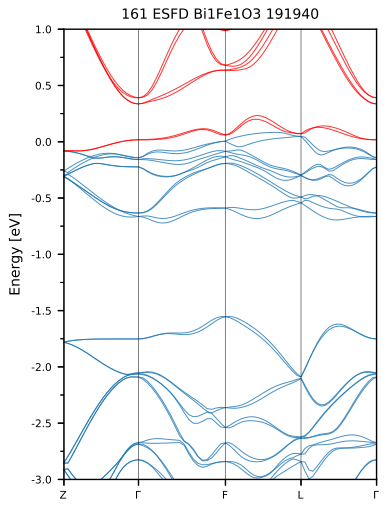

(ab) 161.4.24

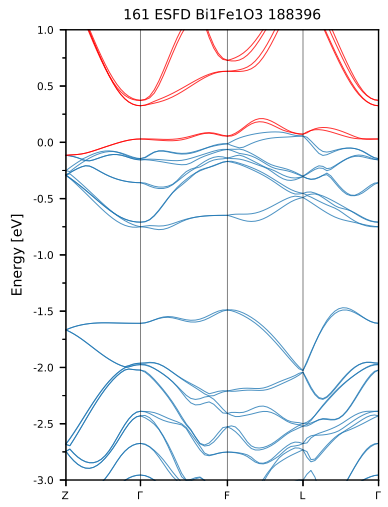

(ac) 161.4.25

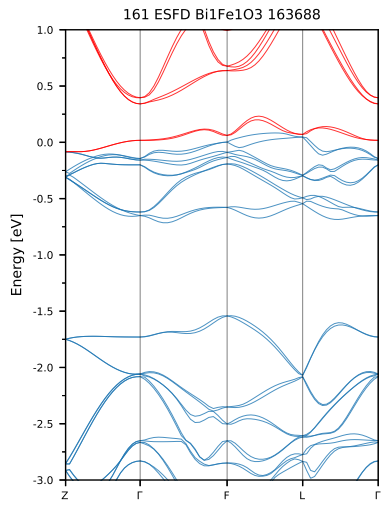

(ad) 161.4.26

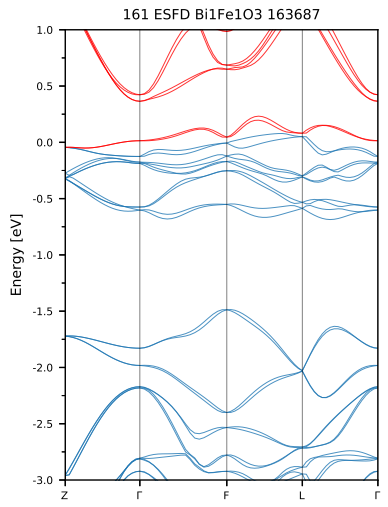

(ae) 161.4.27

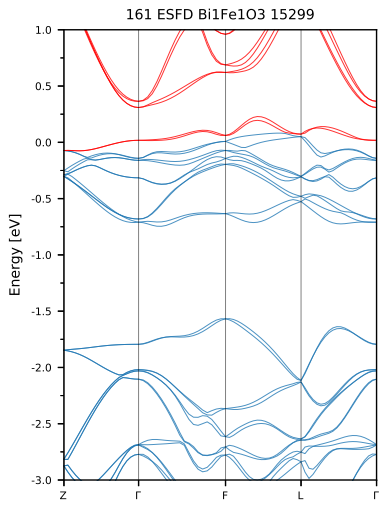

(af) 161.4.28

Figure 336: Band structures for topological materials in SG 161, part 2/3, from 161.4.13 to 161.4.28. See the list of figures.

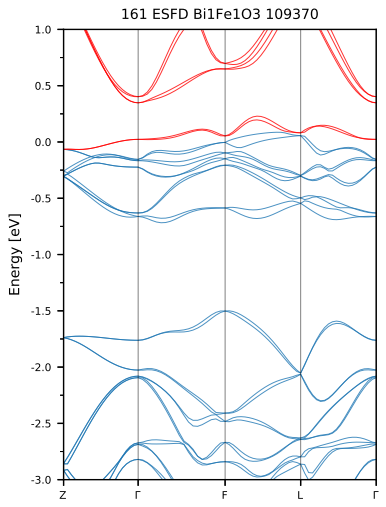

(ag) 161.4.29

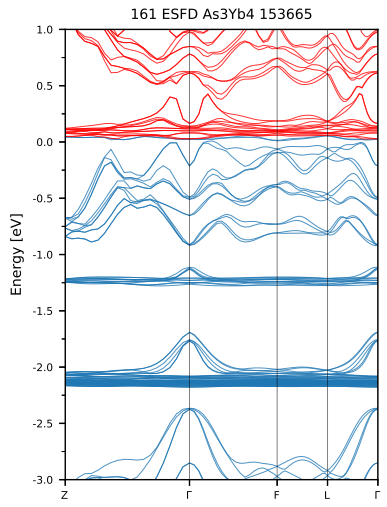

(ah) 161.4.30

Figure 337: Band structures for topological materials in SG 161, part 3/3, from 161.4.29 to 161.4.30. See the list of figures.

### 0.0.78 SG 162

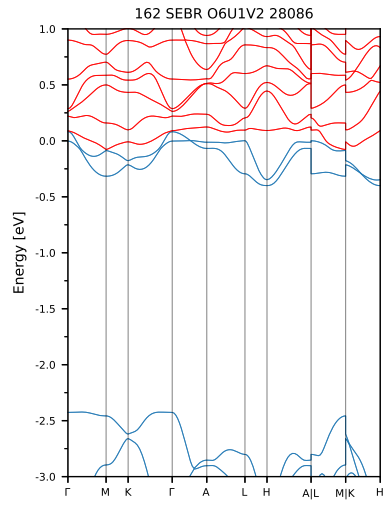

(a) 162.1.1

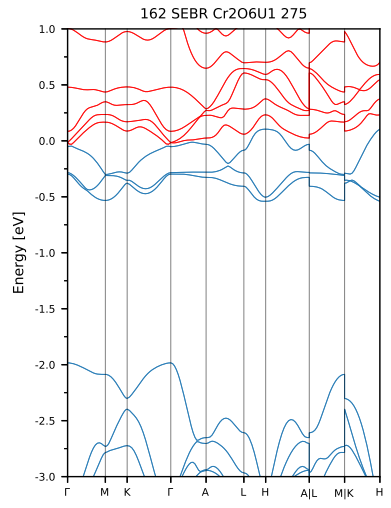

(b) 162.1.2

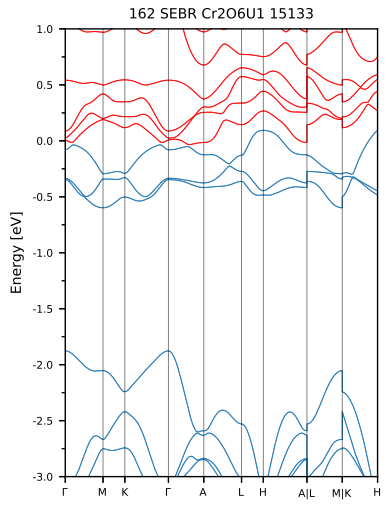

(c) 162.1.3

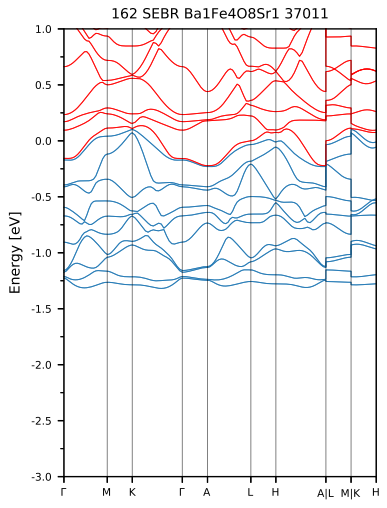

(d) 162.1.4

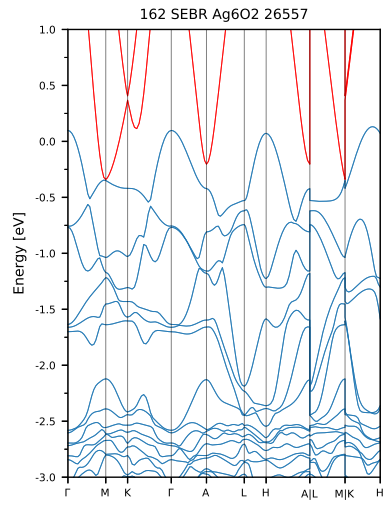

(e) 162.1.5

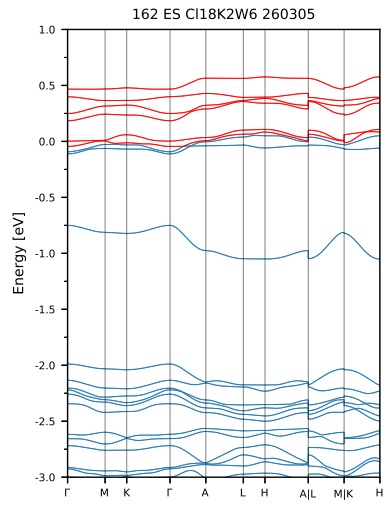

(f) 162.3.1

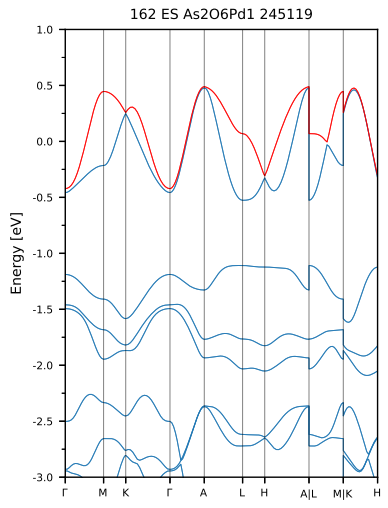

(g) 162.3.2

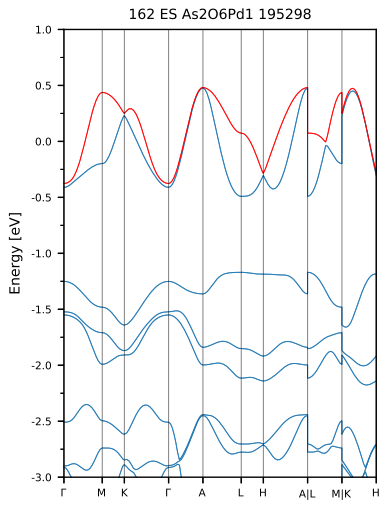

(h) 162.3.3

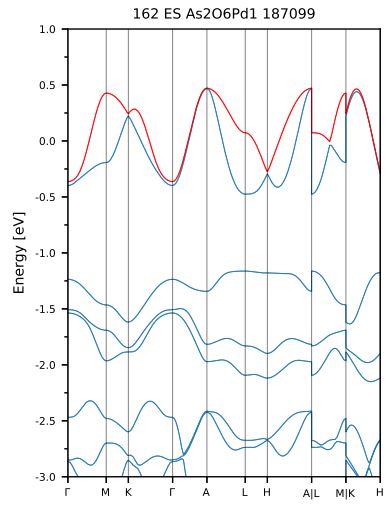

(i) 162.3.4

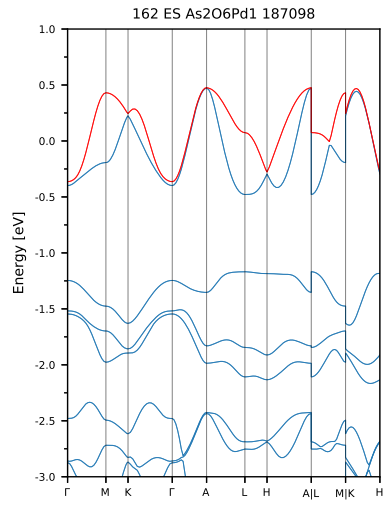

(j) 162.3.5

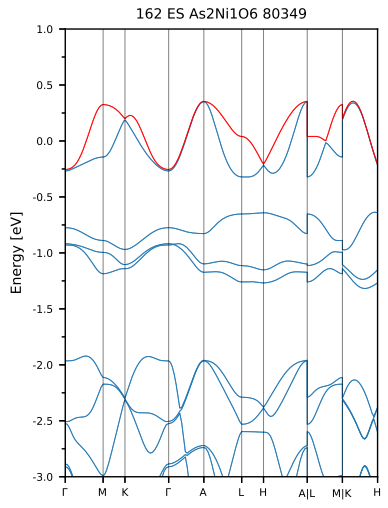

(k) 162.3.6

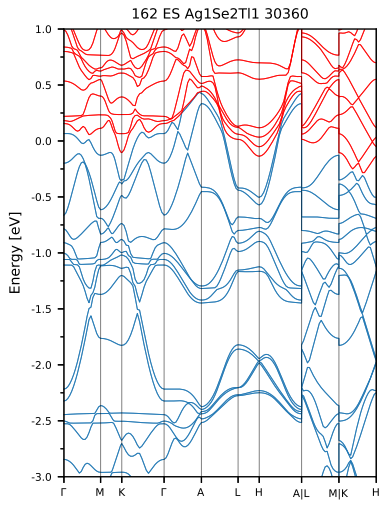

(l) 162.3.7

Figure 338: Band structures for topological materials in SG 162, part 1/1, from 162.1.1 to 162.3.7. See the list of figures.

## 0.0.79 SG 163

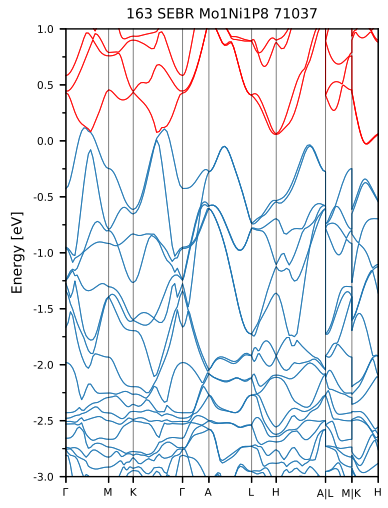

(a) 163.1.1

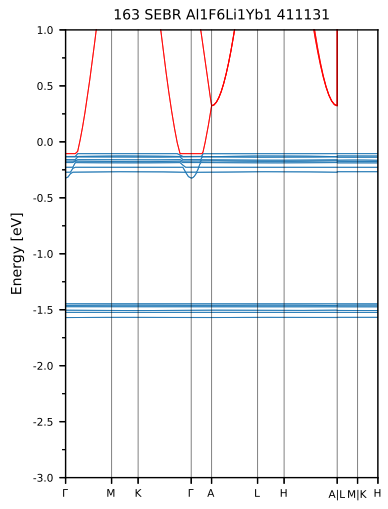

(b) 163.1.2

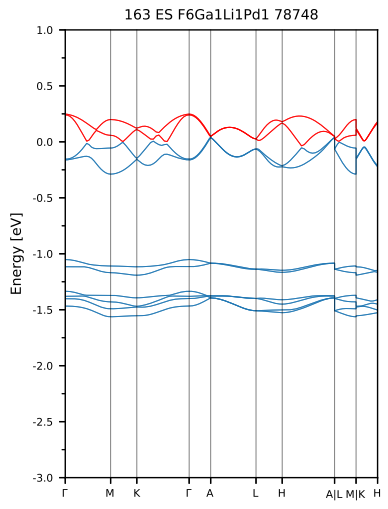

(c) 163.3.1

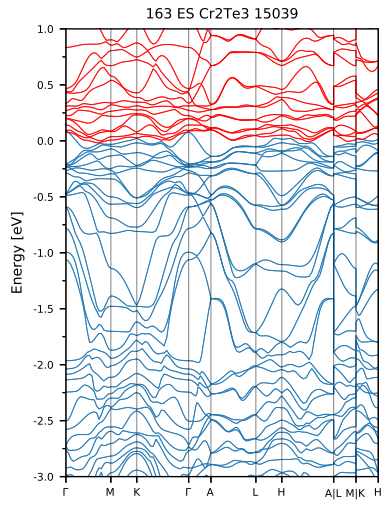

(d) 163.3.2

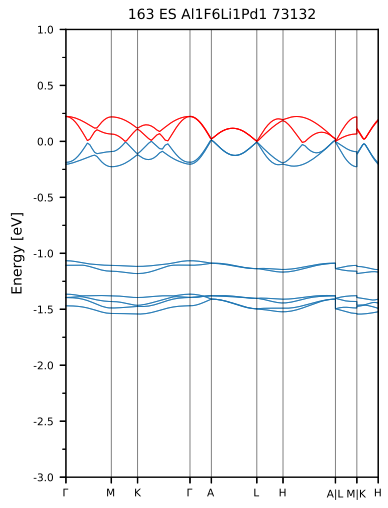

(e) 163.3.3

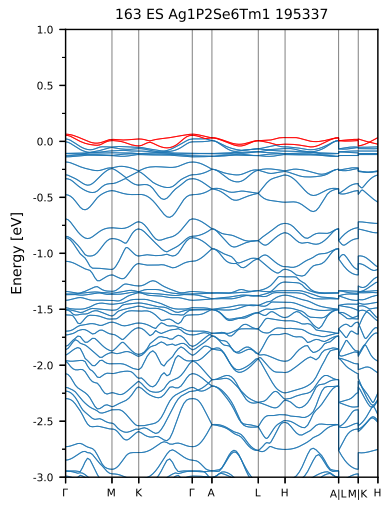

(f) 163.3.4

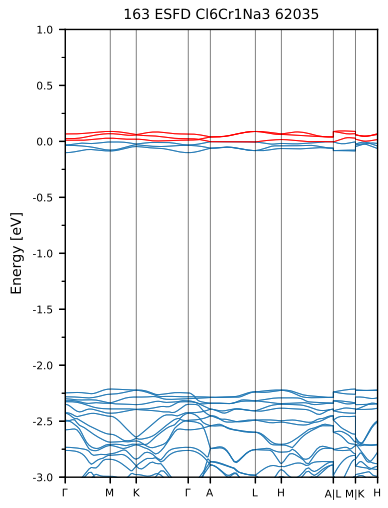

(g) 163.4.1

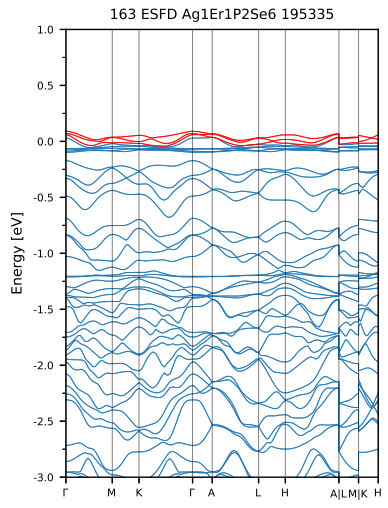

(h) 163.4.2

Figure 339: Band structures for topological materials in SG 163, part 1/1, from 163.1.1 to 163.4.2. See the list of figures.

## 0.0.80 SG 164

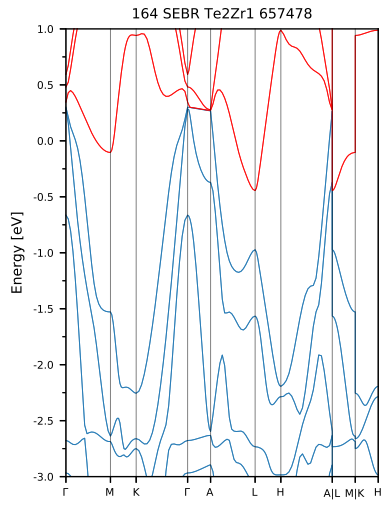

(a) 164.1.1

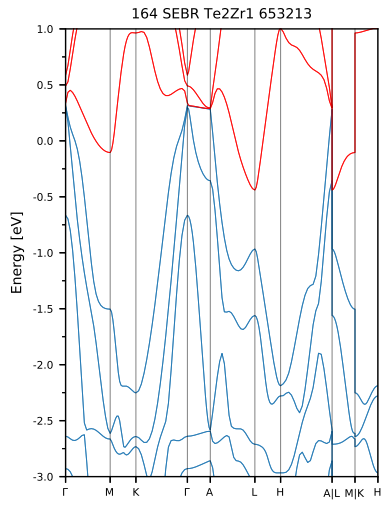

(b) 164.1.2

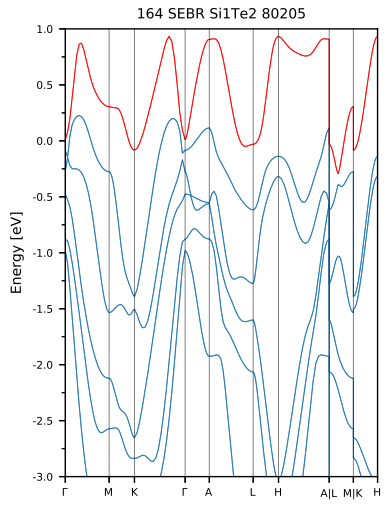

(c) 164.1.3

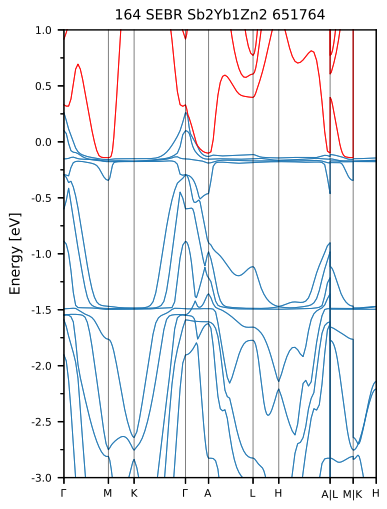

(d) 164.1.4

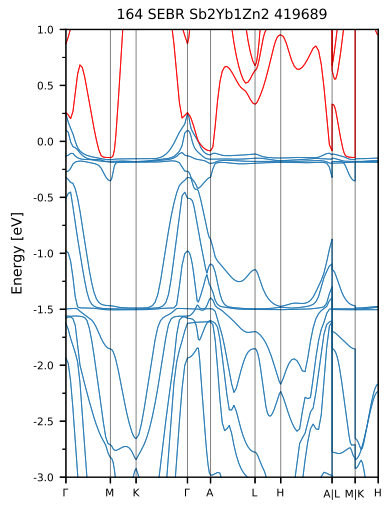

(e) 164.1.5

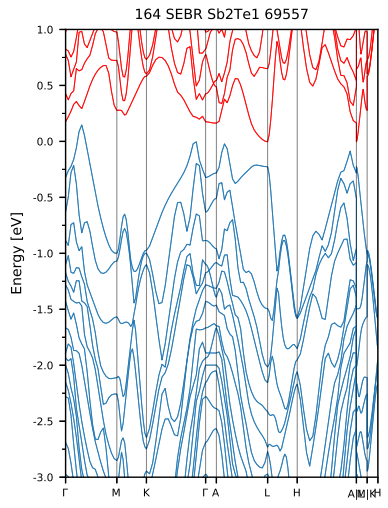

(f) 164.1.6

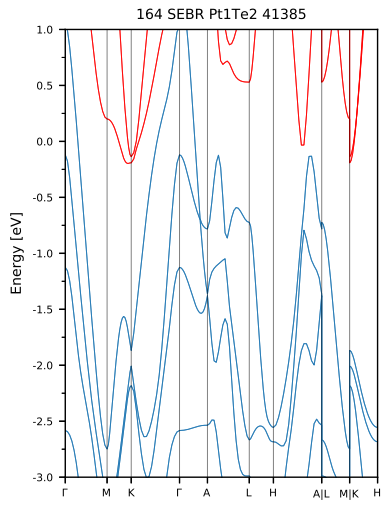

(g) 164.1.7

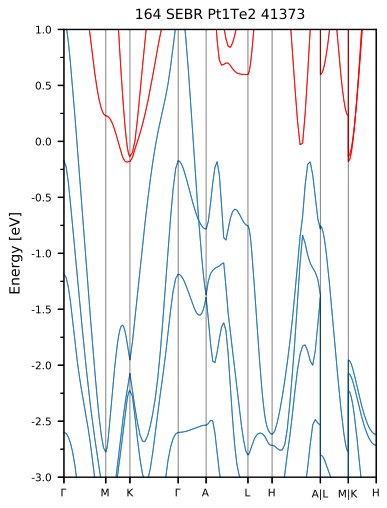

(h) 164.1.8

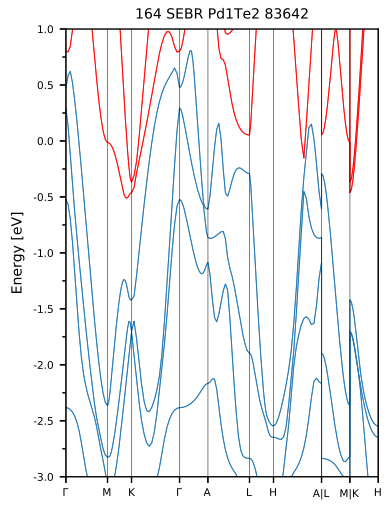

(i) 164.1.9

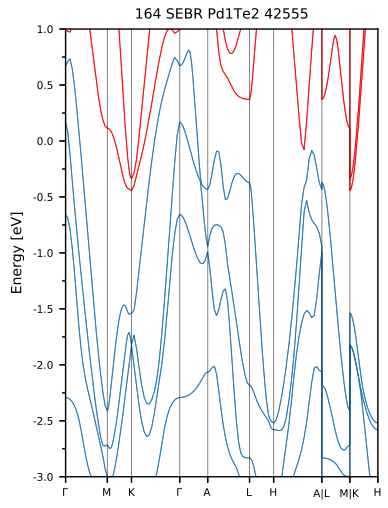

(j) 164.1.10

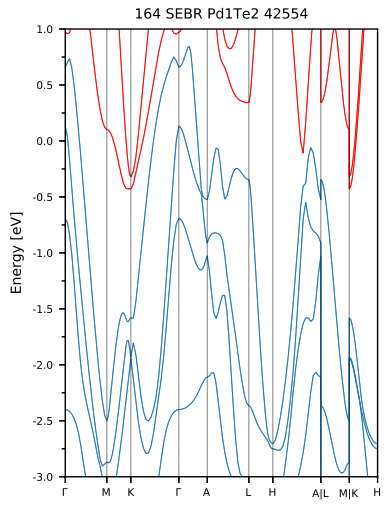

(k) 164.1.11

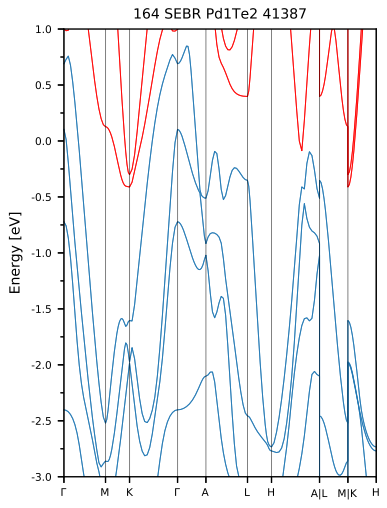

(l) 164.1.12

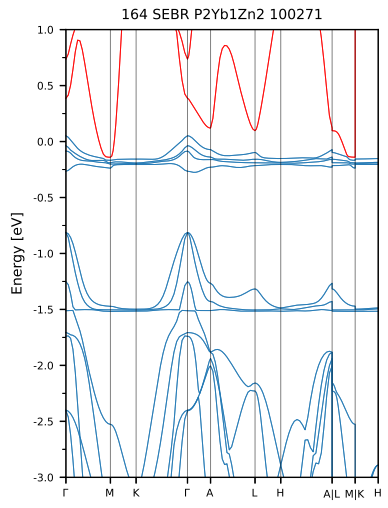

(m) 164.1.13

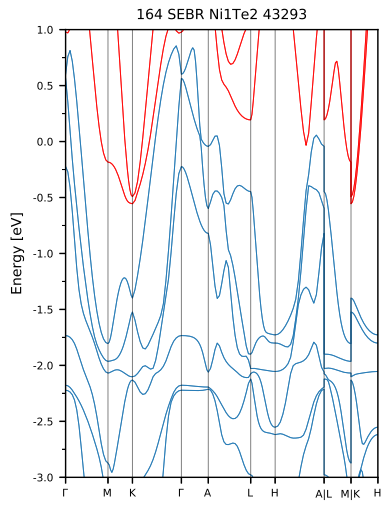

(n) 164.1.14

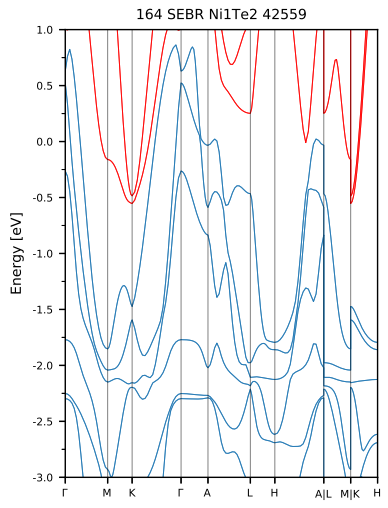

(o) 164.1.15

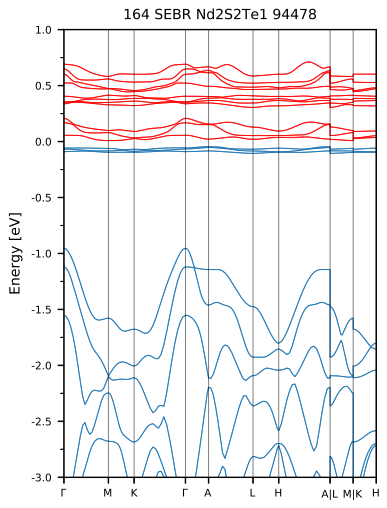

(p) 164.1.16

Figure 340: Band structures for topological materials in SG 164, part 1/10, from 164.1.1 to 164.1.16. See the list of figures.

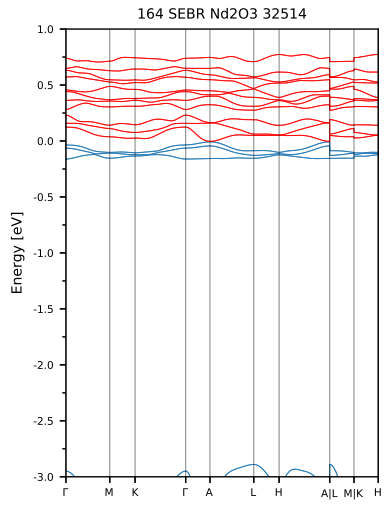

(q) 164.1.17

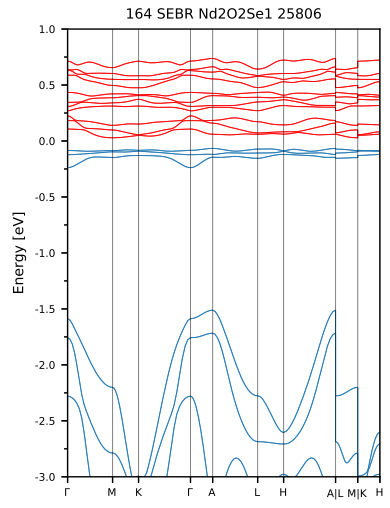

(r) 164.1.18

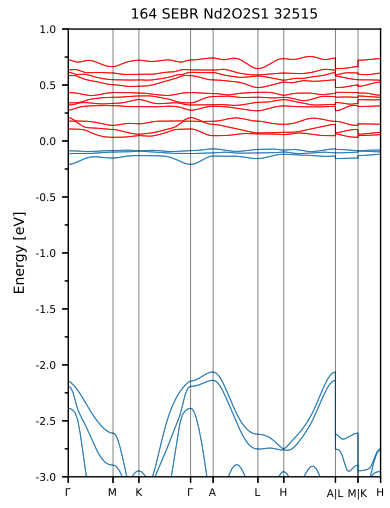

(s) 164.1.19

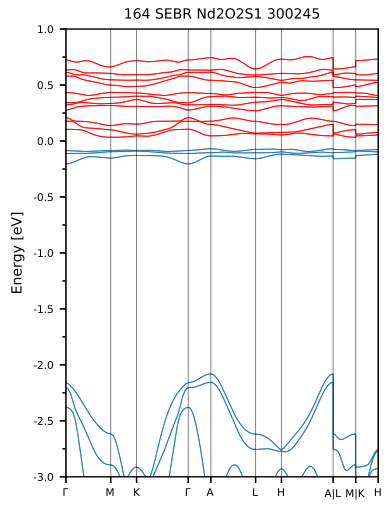

(t) 164.1.20

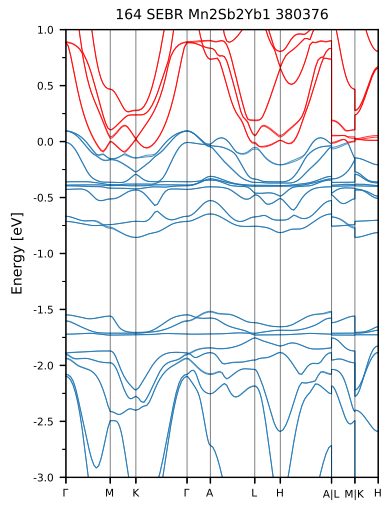

(u) 164.1.21

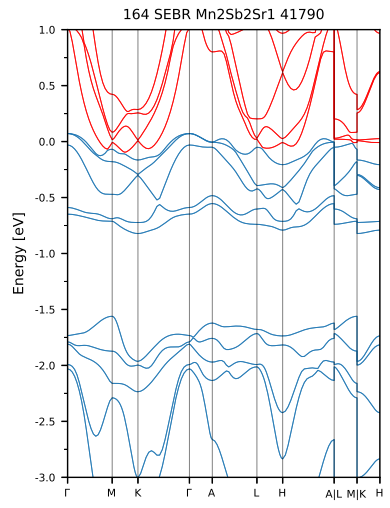

(v) 164.1.22

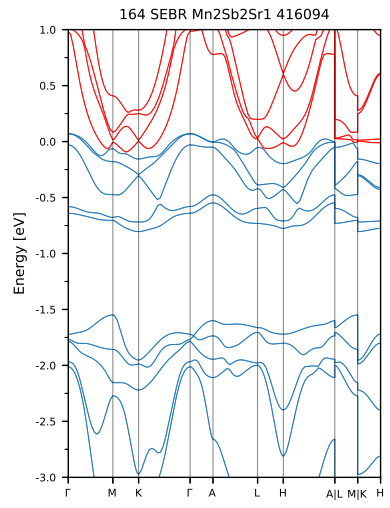

(w) 164.1.23

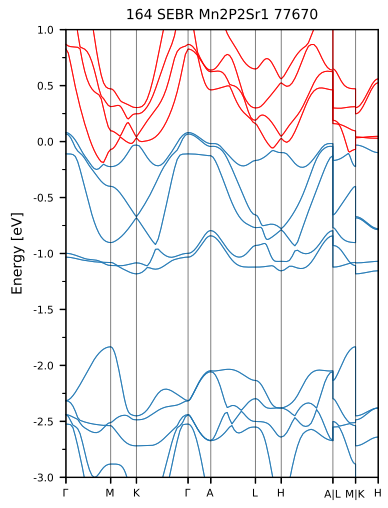

(x) 164.1.24

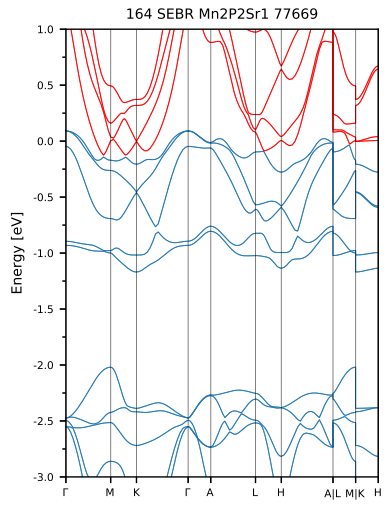

(y) 164.1.25

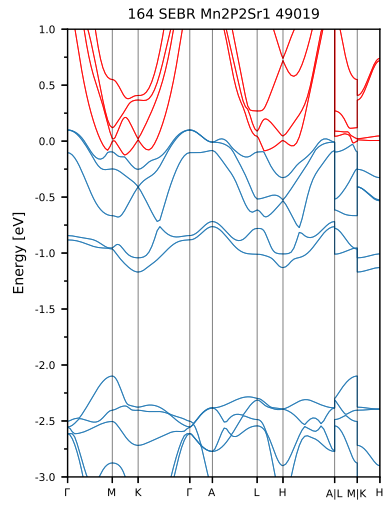

(z) 164.1.26

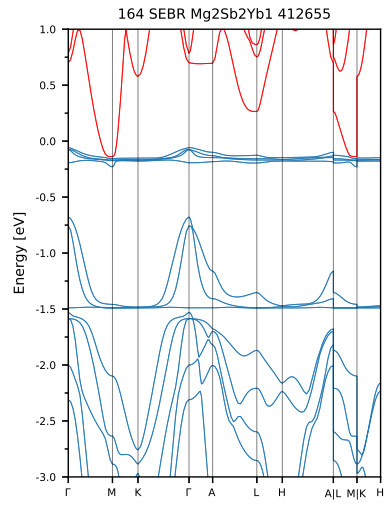

(aa) 164.1.27

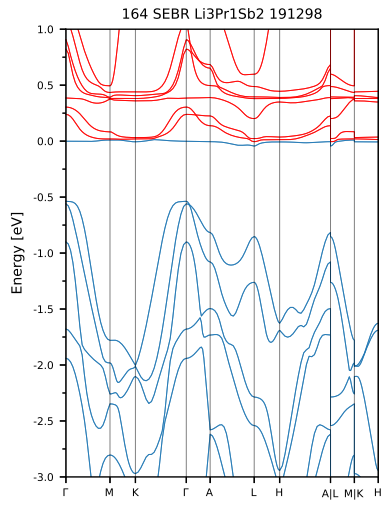

(ab) 164.1.28

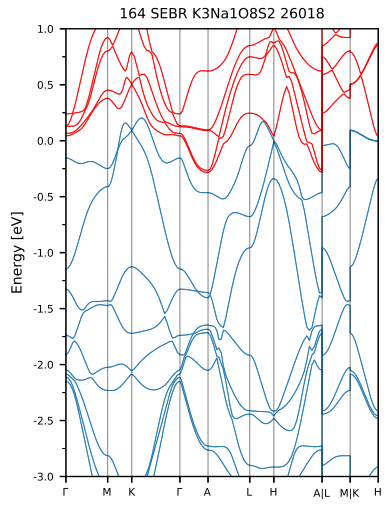

(ac) 164.1.29

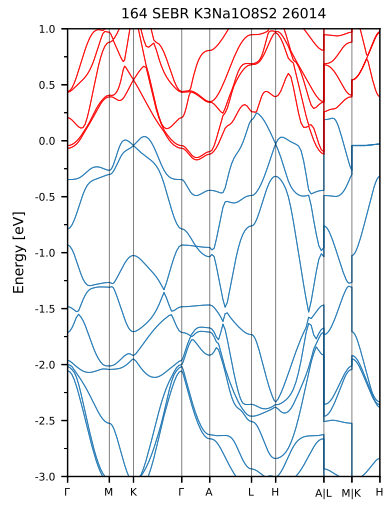

(ad) 164.1.30

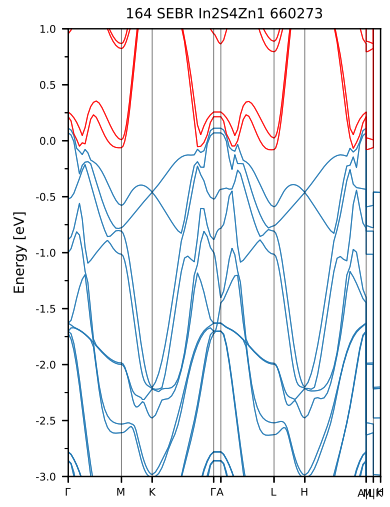

(ae) 164.1.31

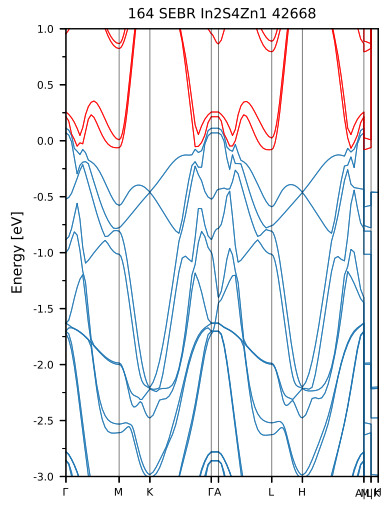

(af) 164.1.32

Figure 341: Band structures for topological materials in SG 164, part 2/10, from 164.1.17 to 164.1.32. See the list of figures.

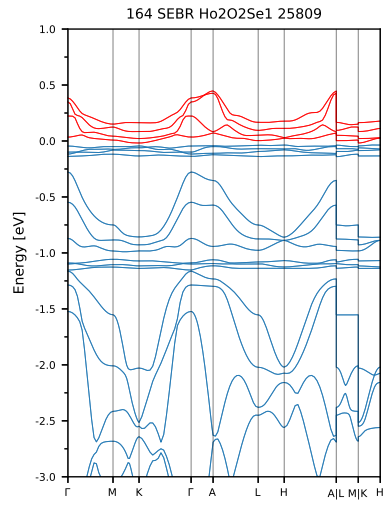

(ag) 164.1.33

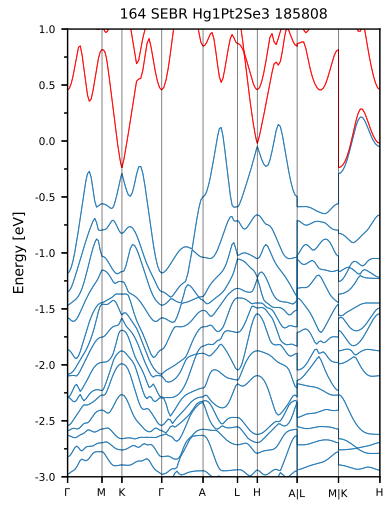

(ah) 164.1.34

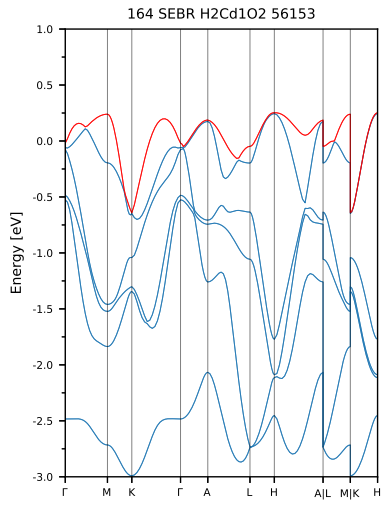

(ai) 164.1.35

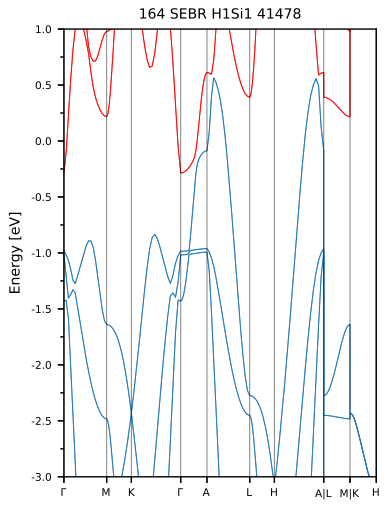

(aj) 164.1.36

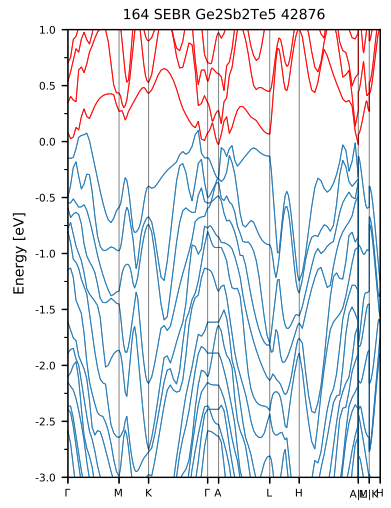

(ak) 164.1.37

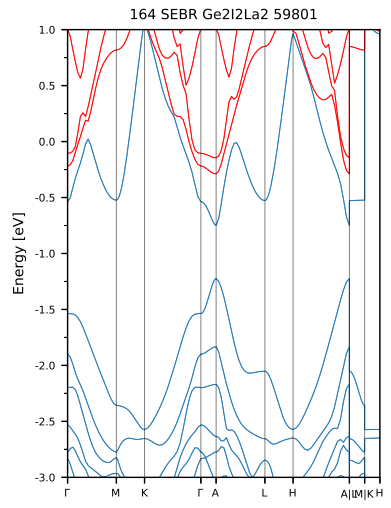

(al) 164.1.38

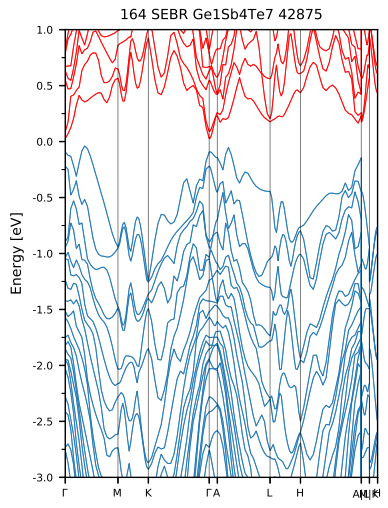

(am) 164.1.39

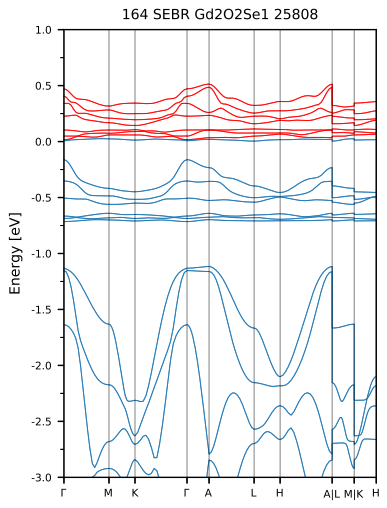

(an) 164.1.40

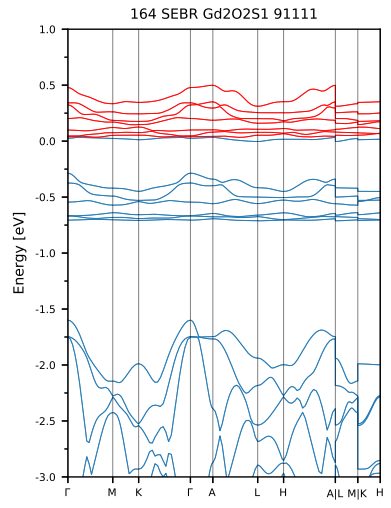

(ao) 164.1.41

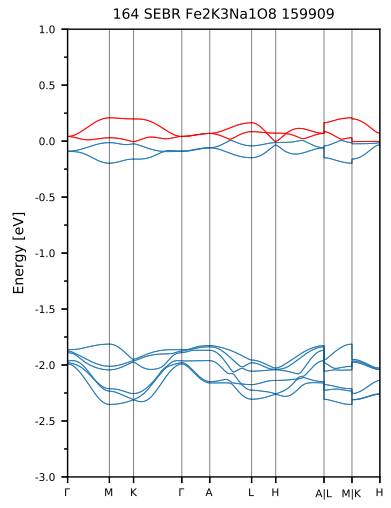

(ap) 164.1.42

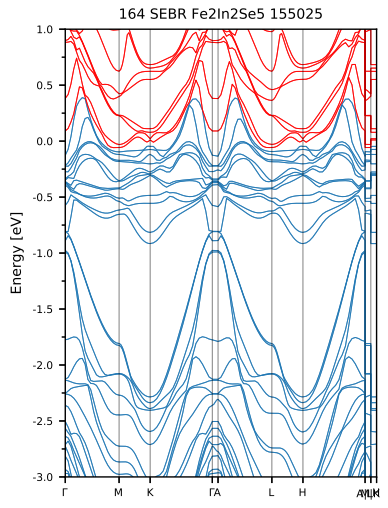

(aq) 164.1.43

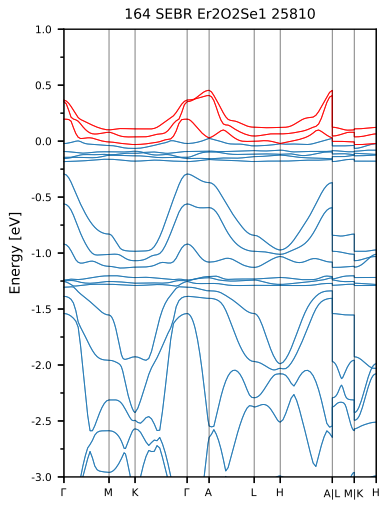

(ar) 164.1.44

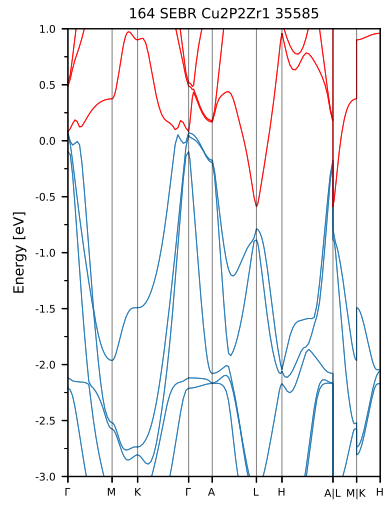

(as) 164.1.45

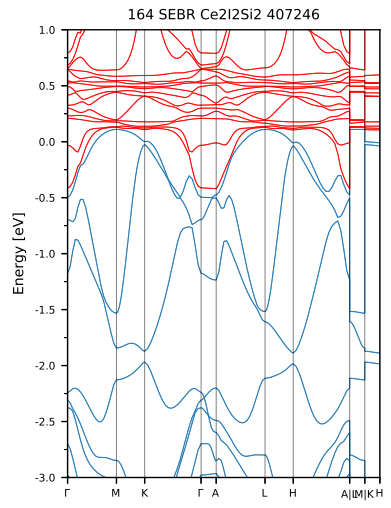

(at) 164.1.46

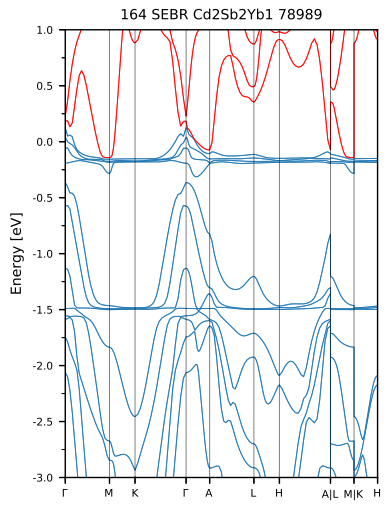

(au) 164.1.47

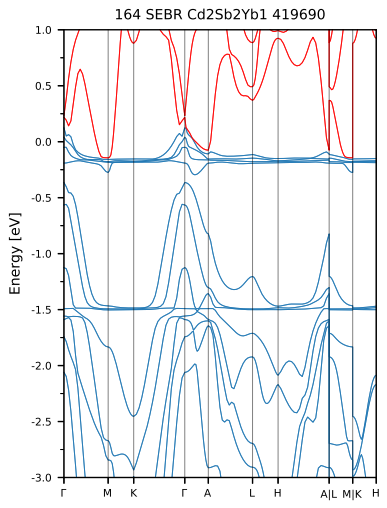

(av) 164.1.48

Figure 342: Band structures for topological materials in SG 164, part 3/10, from 164.1.33 to 164.1.48. See the list of figures.

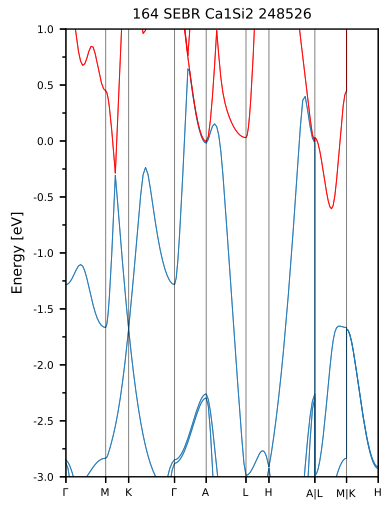

(aw) 164.1.49

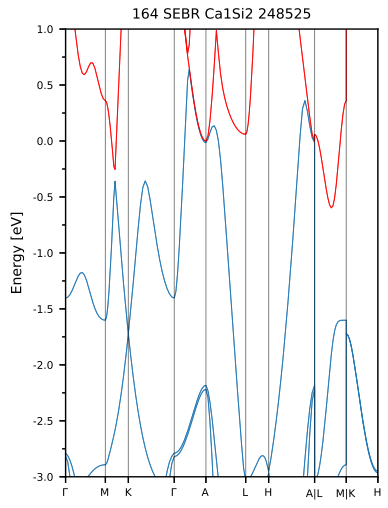

(ax) 164.1.50

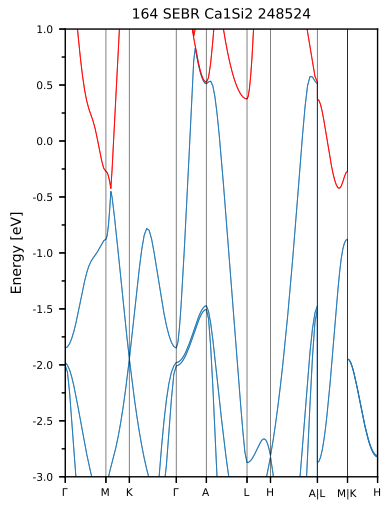

(ay) 164.1.51

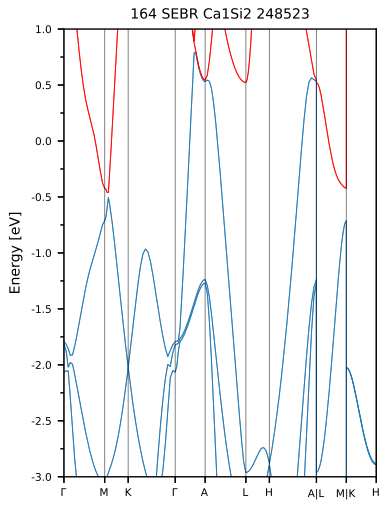

(az) 164.1.52

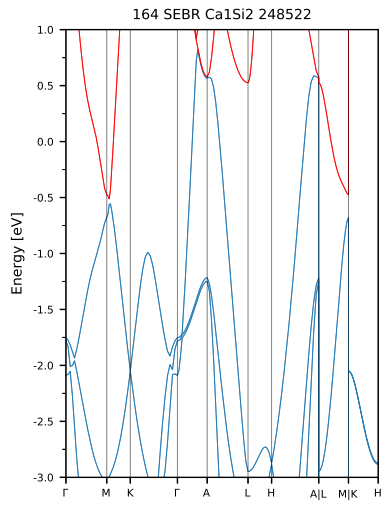

(ba) 164.1.53

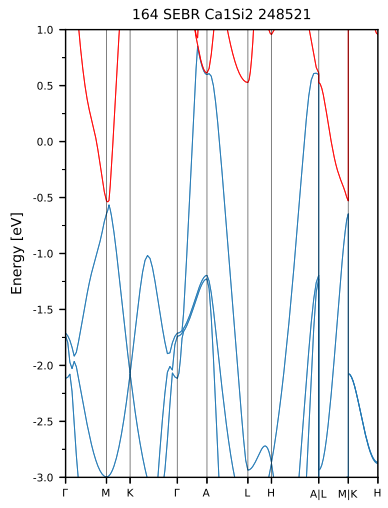

(bb) 164.1.54

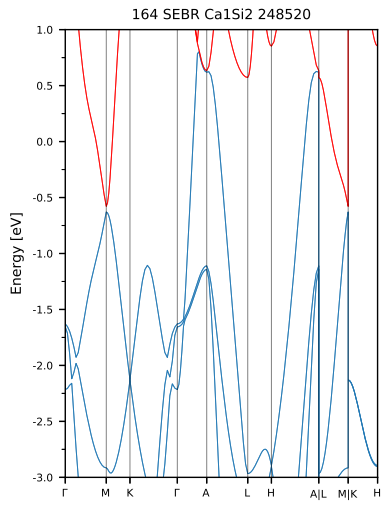

(bc) 164.1.55

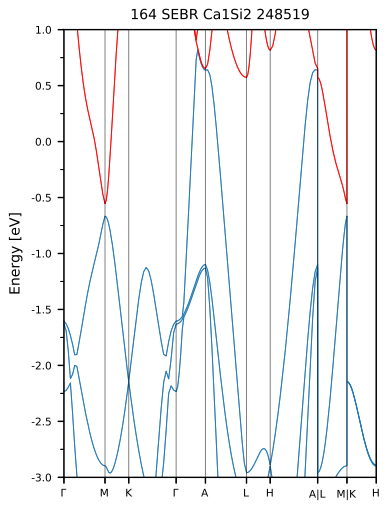

(bd) 164.1.56

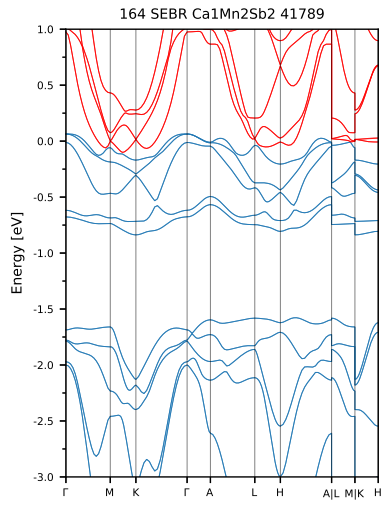

(be) 164.1.57

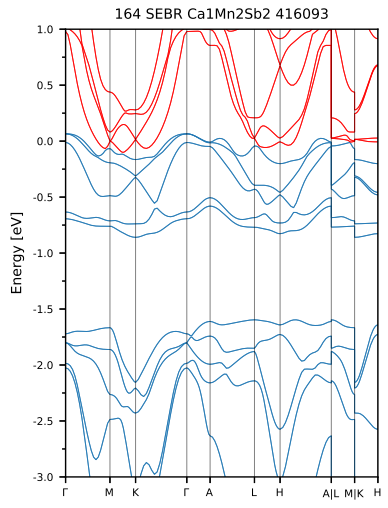

(bf) 164.1.58

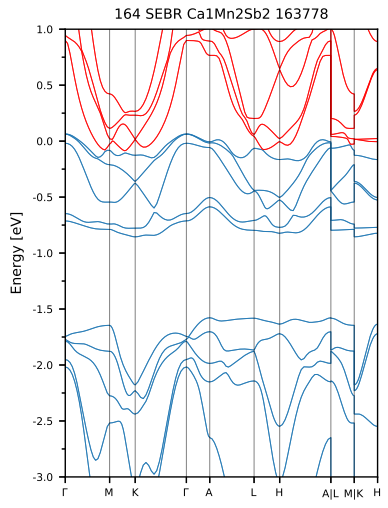

(bg) 164.1.59

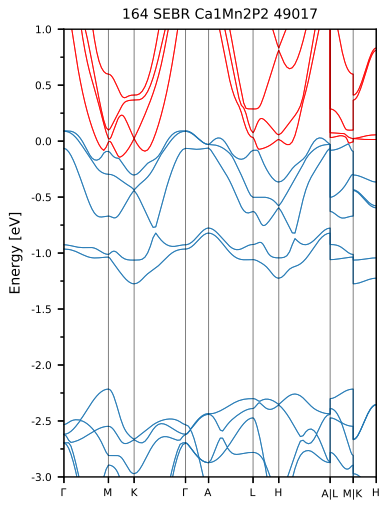

(bh) 164.1.60

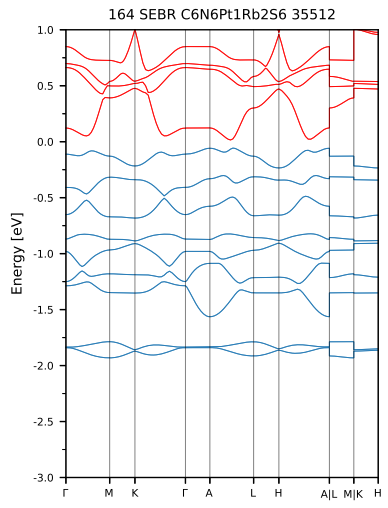

(bi) 164.1.61

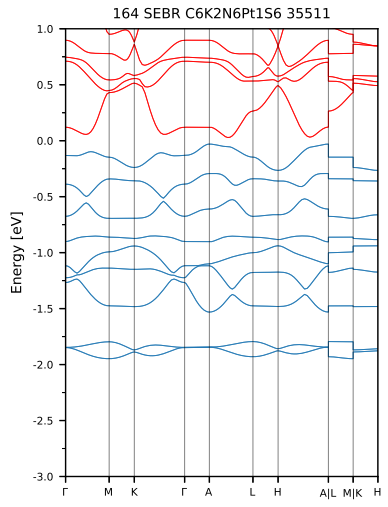

(bj) 164.1.62

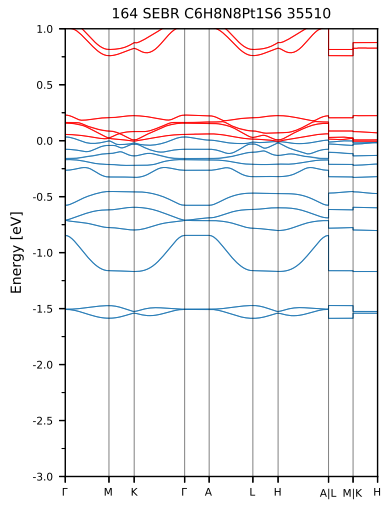

(bk) 164.1.63

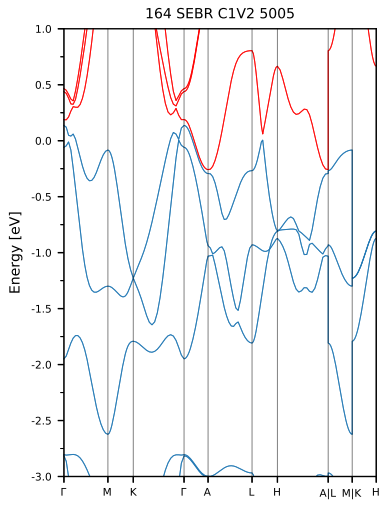

(bl) 164.1.64

Figure 343: Band structures for topological materials in SG 164, part 4/10, from 164.1.49 to 164.1.64. See the list of figures.

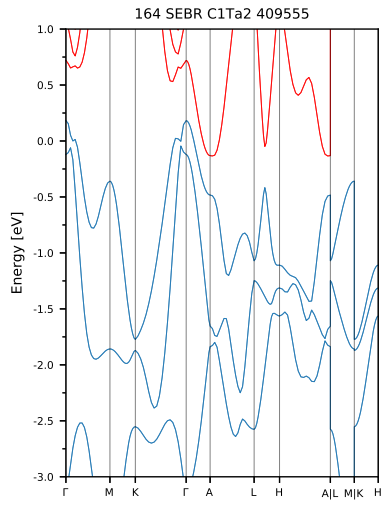

(bm) 164.1.65

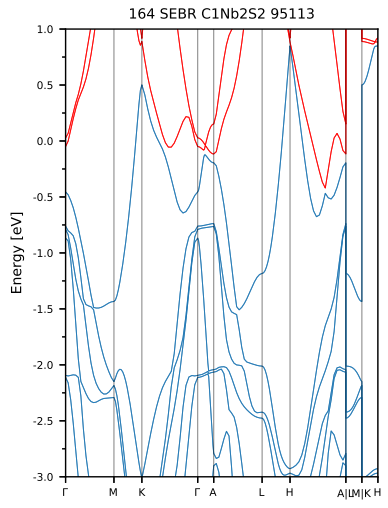

(bn) 164.1.66

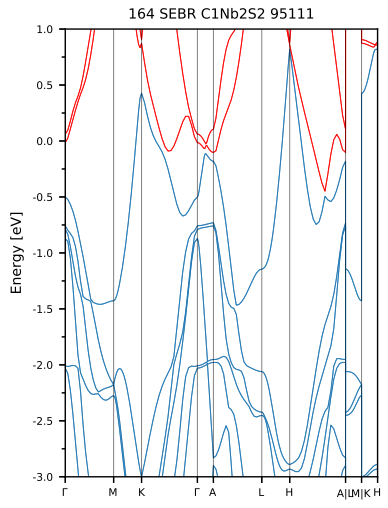

(bo) 164.1.67

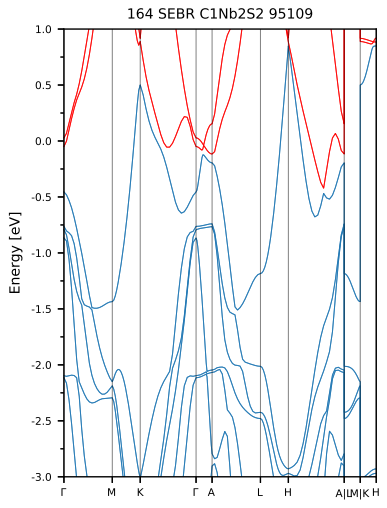

(bp) 164.1.68

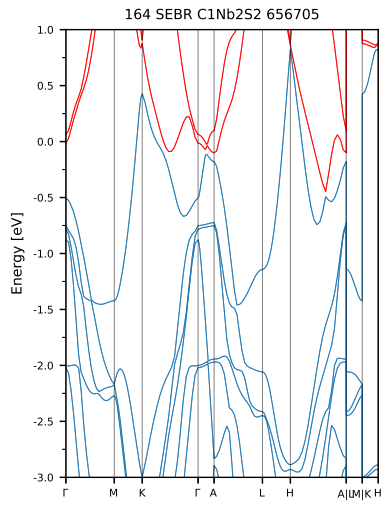

(bq) 164.1.69

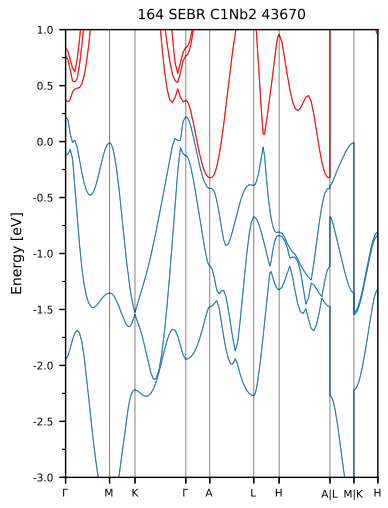

(br) 164.1.70

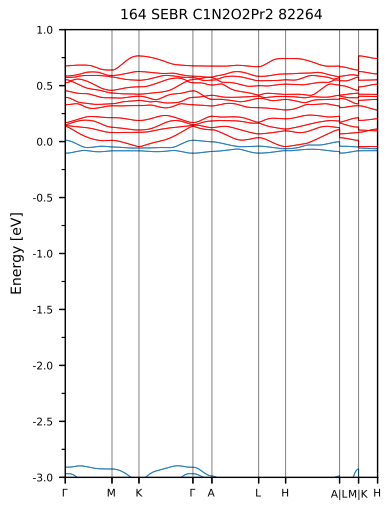

(bs) 164.1.71

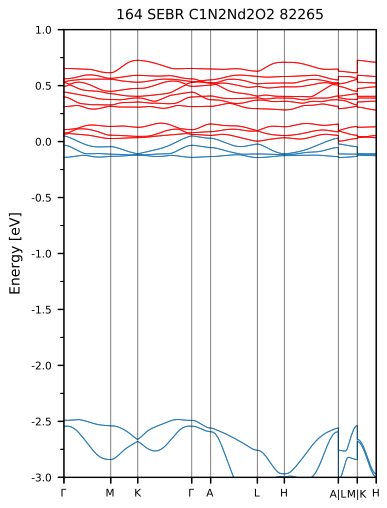

(bt) 164.1.72

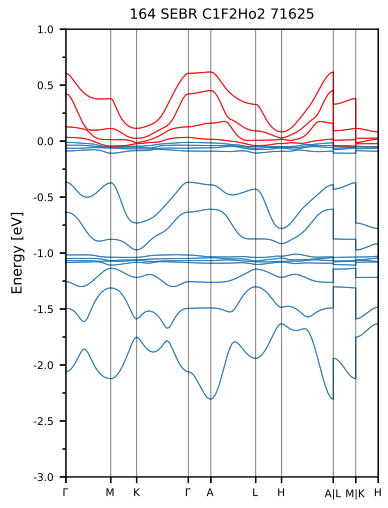

(bu) 164.1.73

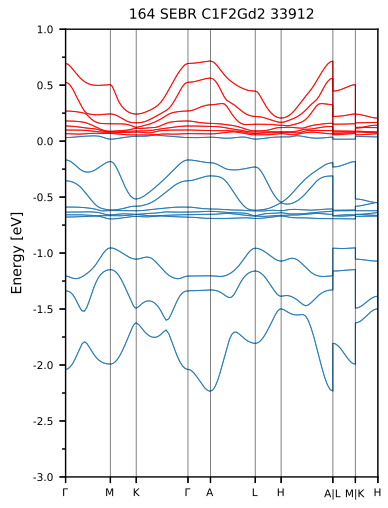

(bv) 164.1.74

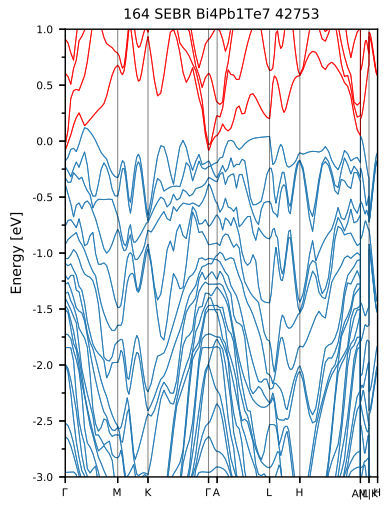

(bw) 164.1.75

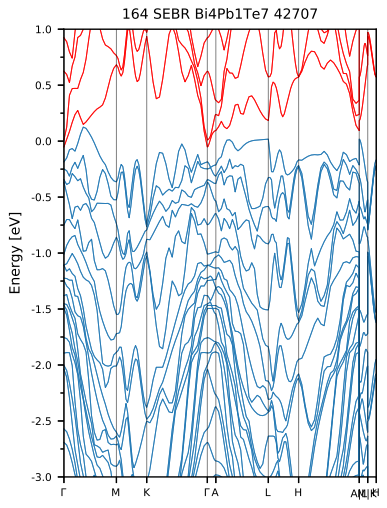

(bx) 164.1.76

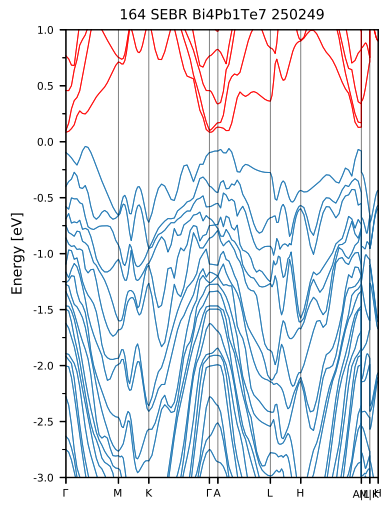

(by) 164.1.77

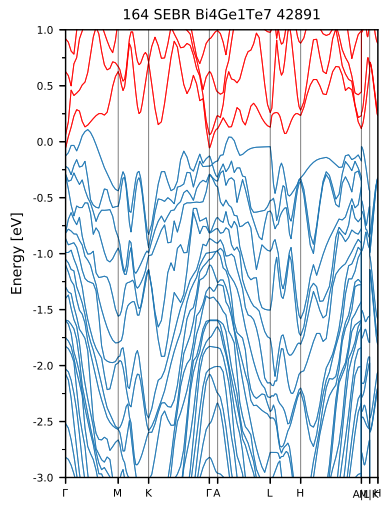

(bz) 164.1.78

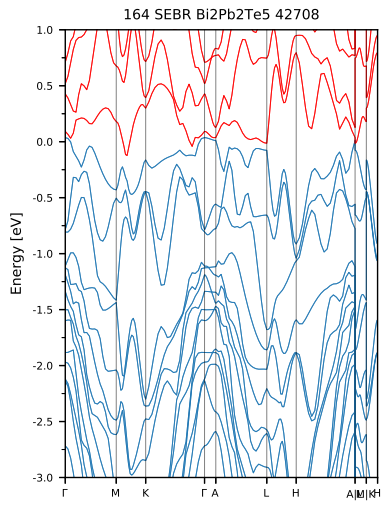

(ca) 164.1.79

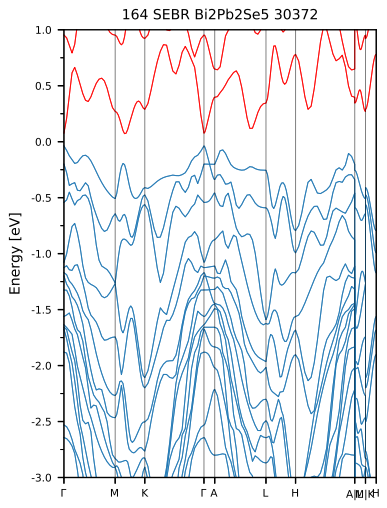

(cb) 164.1.80

Figure 344: Band structures for topological materials in SG 164, part 5/10, from 164.1.65 to 164.1.80. See the list of figures.

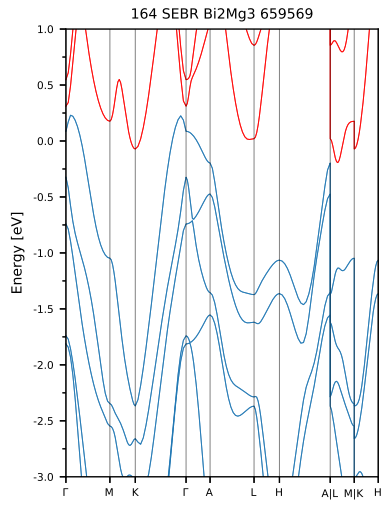

(cc) 164.1.81

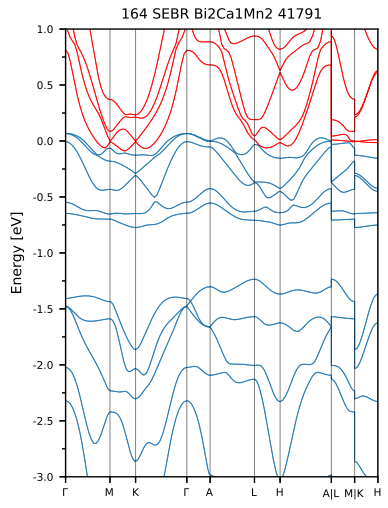

(cd) 164.1.82

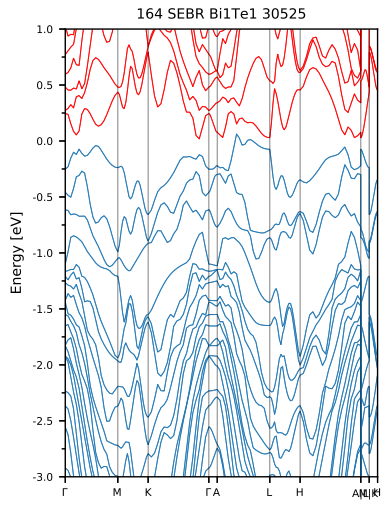

(ce) 164.1.83

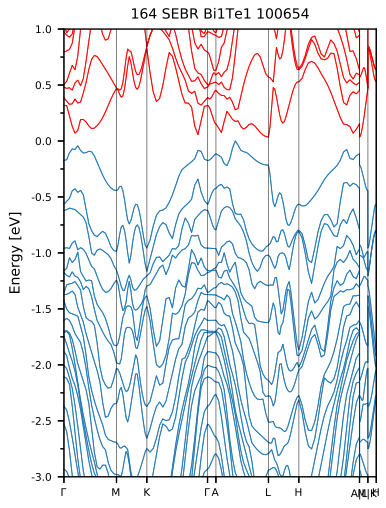

(cf) 164.1.84

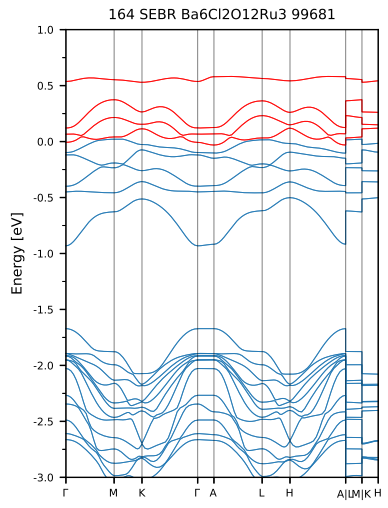

(cg) 164.1.85

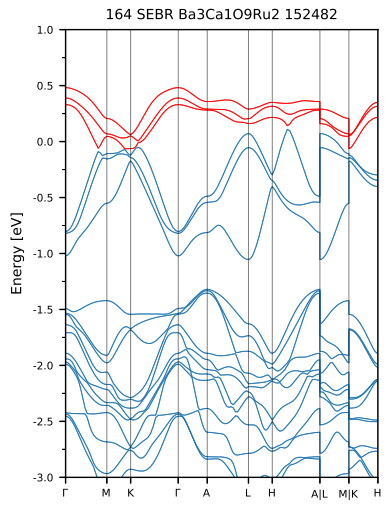

(ch) 164.1.86

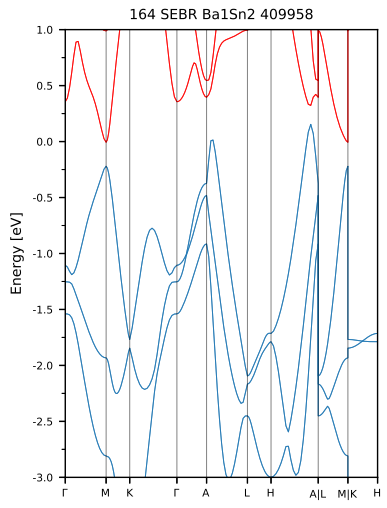

(ci) 164.1.87

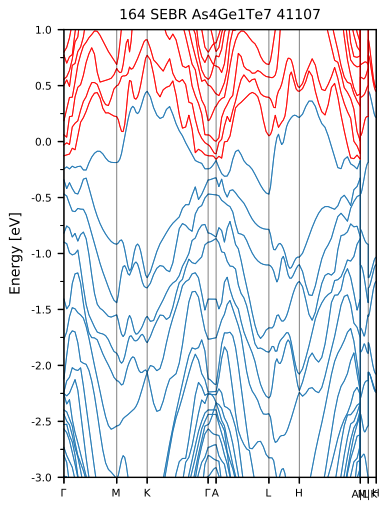

(cj) 164.1.88

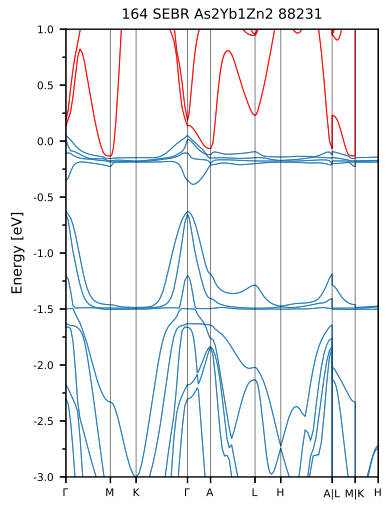

(ck) 164.1.89

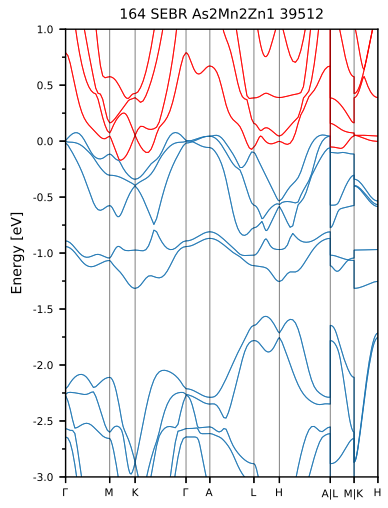

(cl) 164.1.90

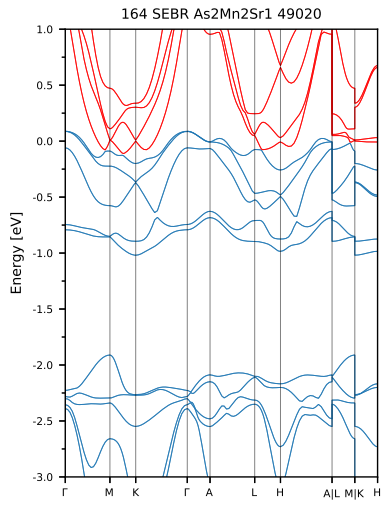

(cm) 164.1.91

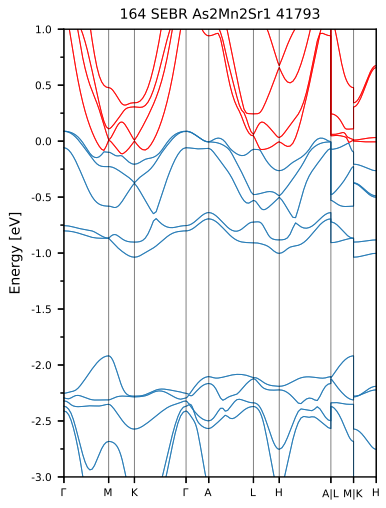

(cn) 164.1.92

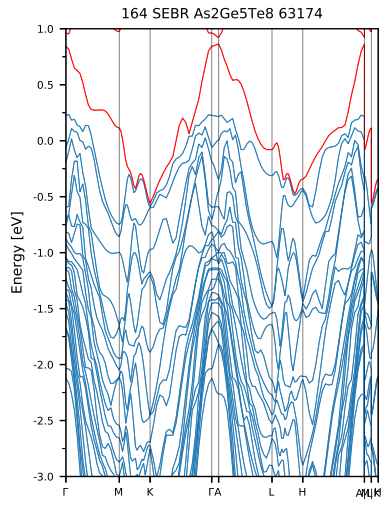

(co) 164.1.93

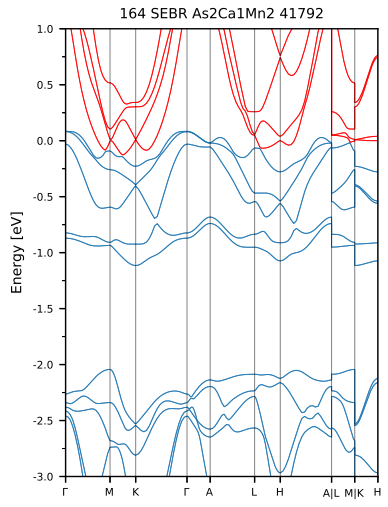

(cp) 164.1.94

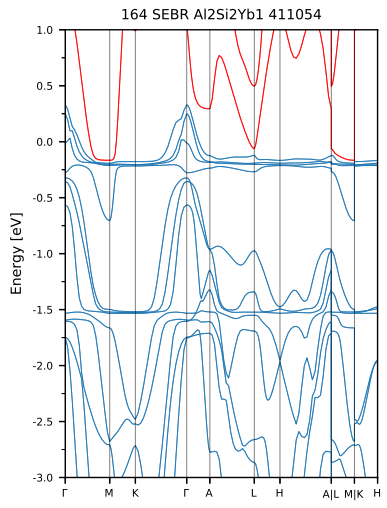

(cq) 164.1.95

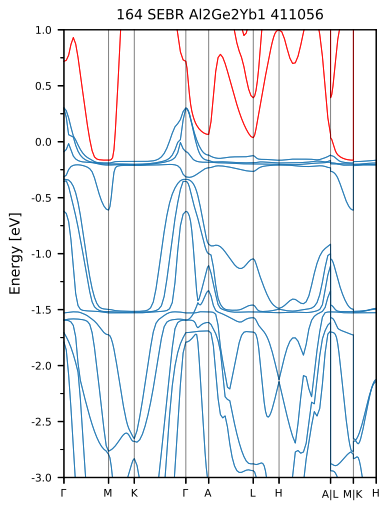

(cr) 164.1.96

Figure 345: Band structures for topological materials in SG 164, part 6/10, from 164.1.81 to 164.1.96. See the list of figures.

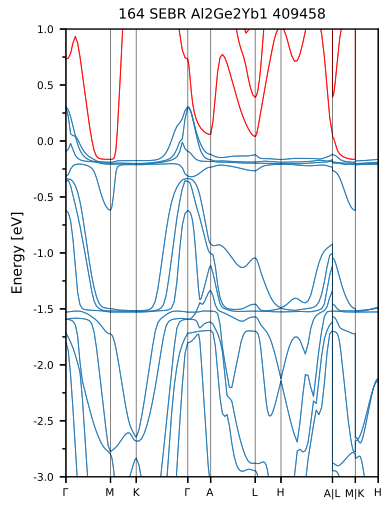

(cs) 164.1.97

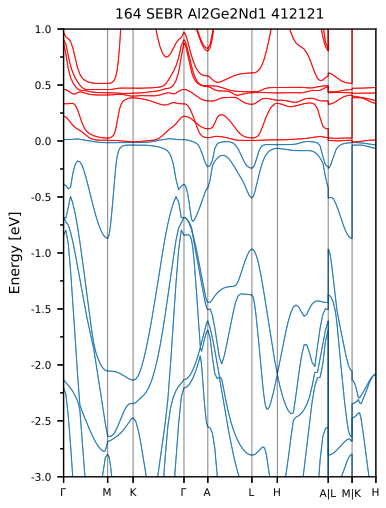

(ct) 164.1.98

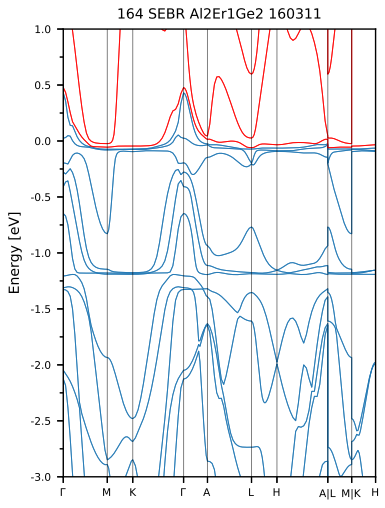

(cu) 164.1.99

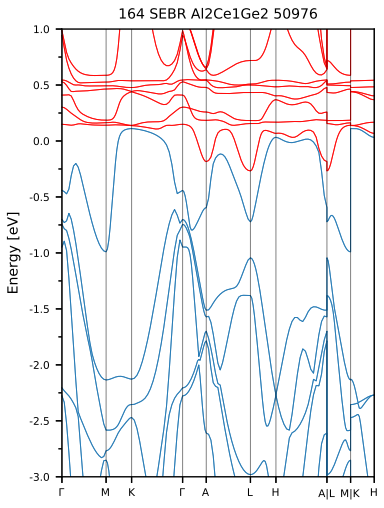

(cv) 164.1.100

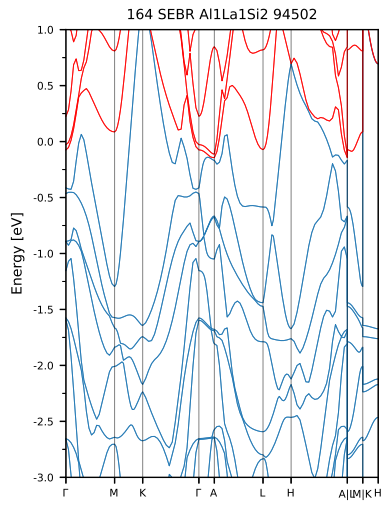

(cw) 164.1.101

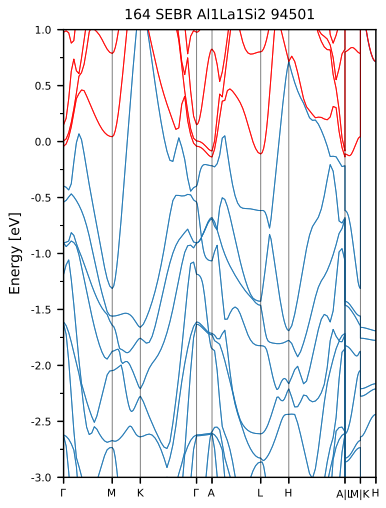

(cx) 164.1.102

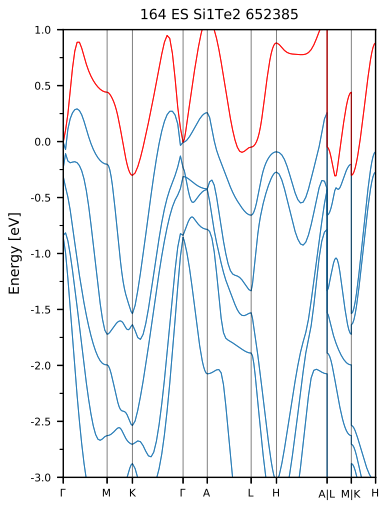

(cy) 164.3.1

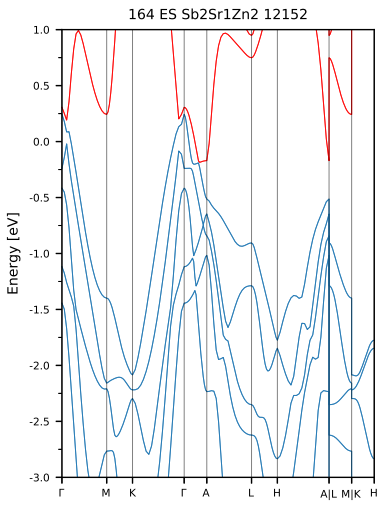

(cz) 164.3.2

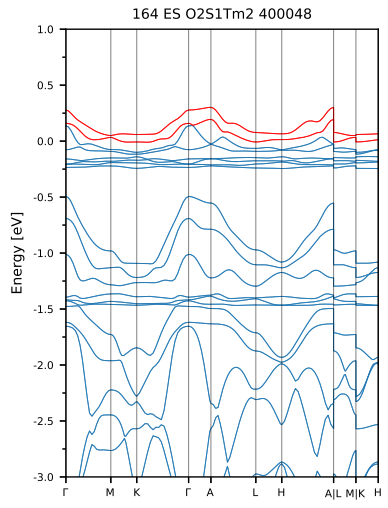

(da) 164.3.3

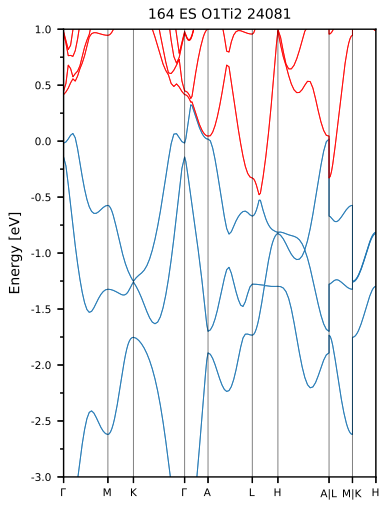

(db) 164.3.4

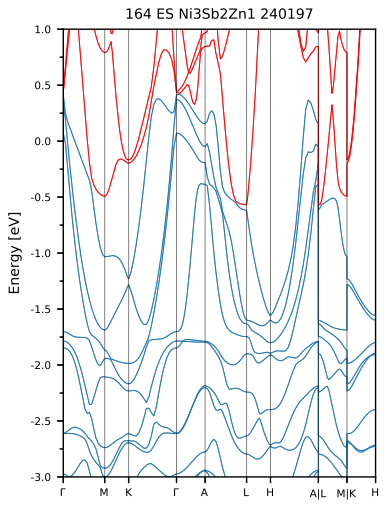

(dc) 164.3.5

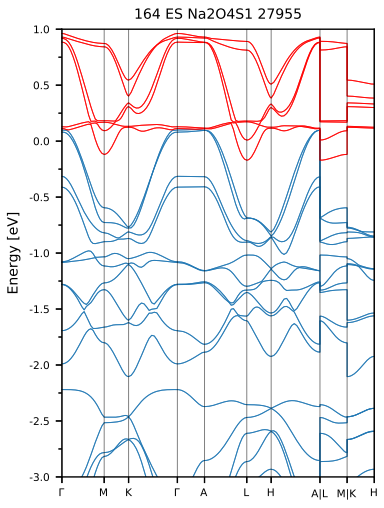

(dd) 164.3.6

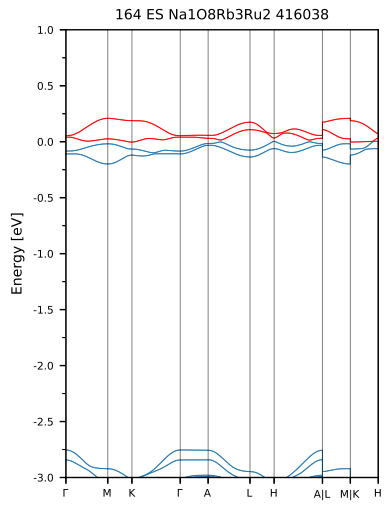

(de) 164.3.7

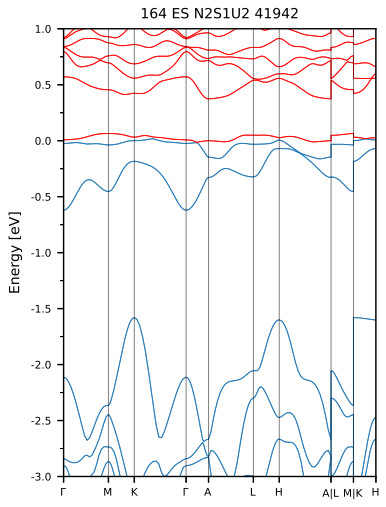

(df) 164.3.8

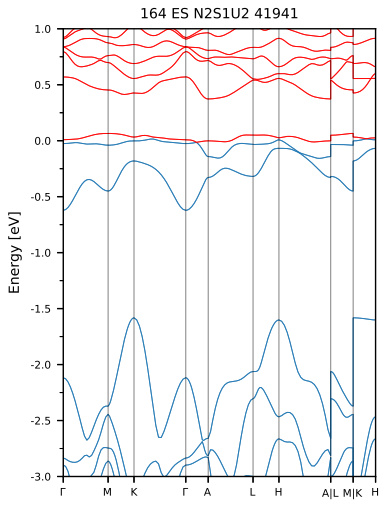

(dg) 164.3.9

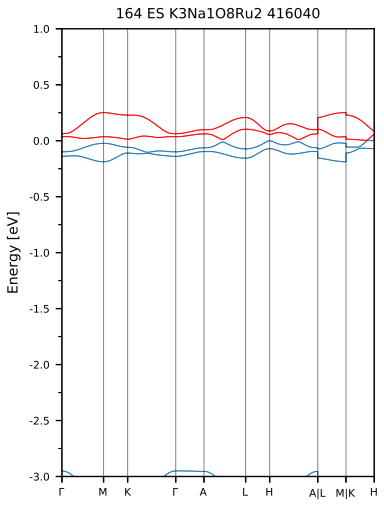

(dh) 164.3.10

Figure 346: Band structures for topological materials in SG 164, part 7/10, from 164.1.97 to 164.3.10. See the list of figures.

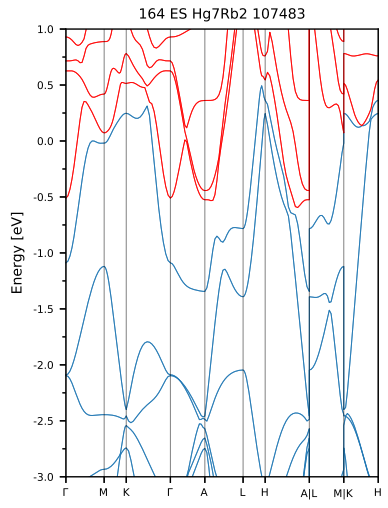

(di) 164.3.11

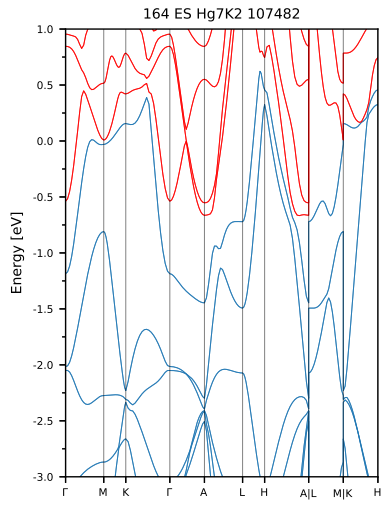

(dj) 164.3.12

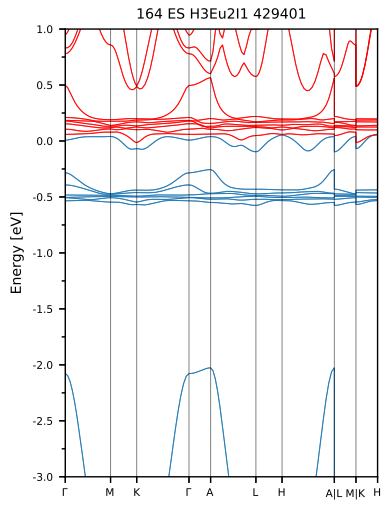

(dk) 164.3.13

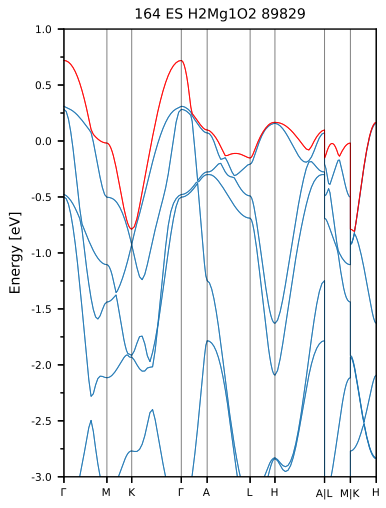

(dl) 164.3.14

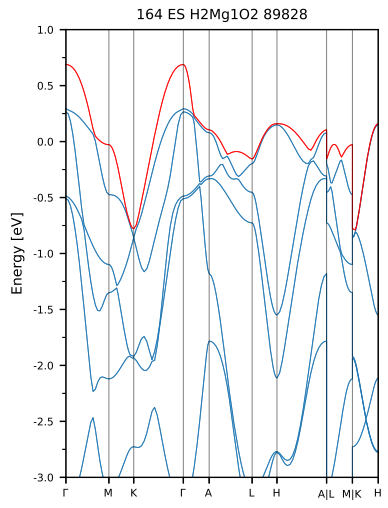

(dm) 164.3.15

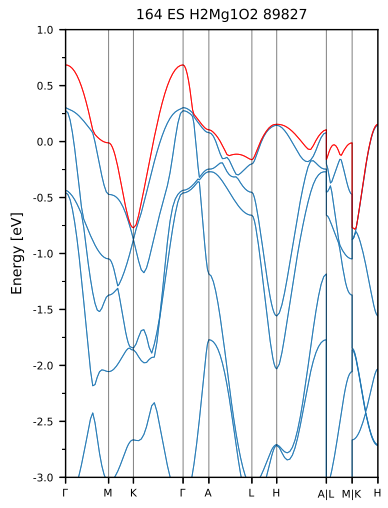

(dn) 164.3.16

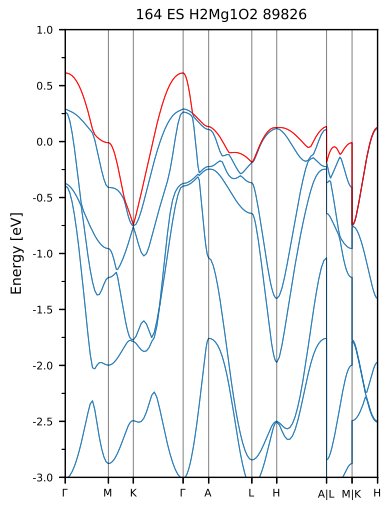

(do) 164.3.17

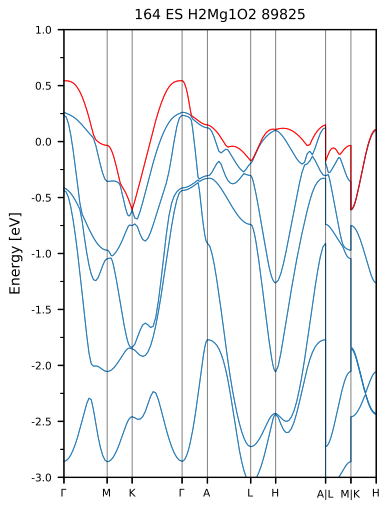

(dp) 164.3.18

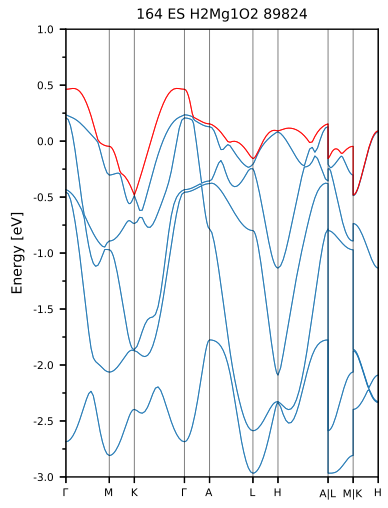

(dq) 164.3.19

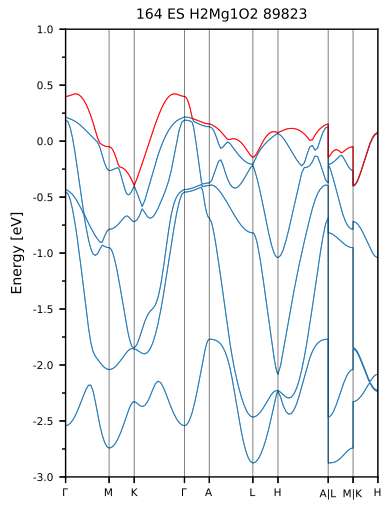

(dr) 164.3.20

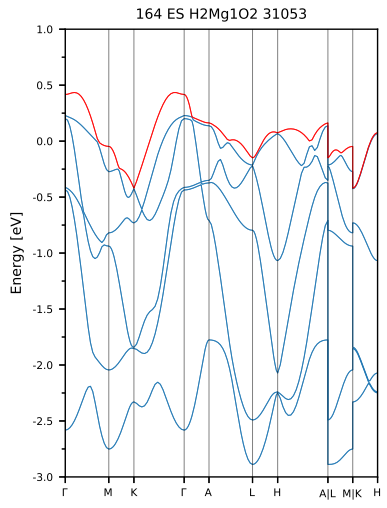

(ds) 164.3.21

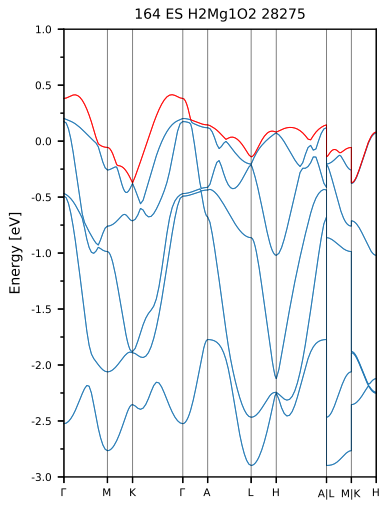

(dt) 164.3.22

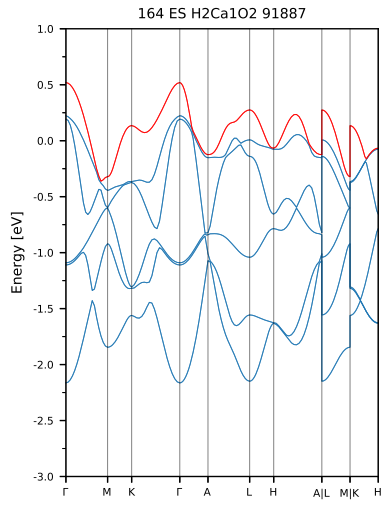

(du) 164.3.23

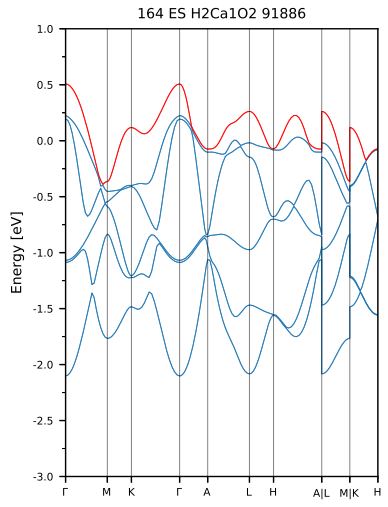

(dv) 164.3.24

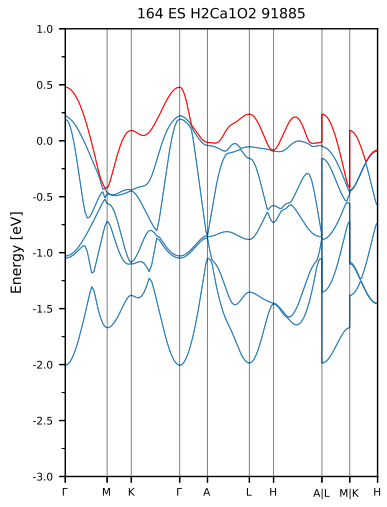

(dw) 164.3.25

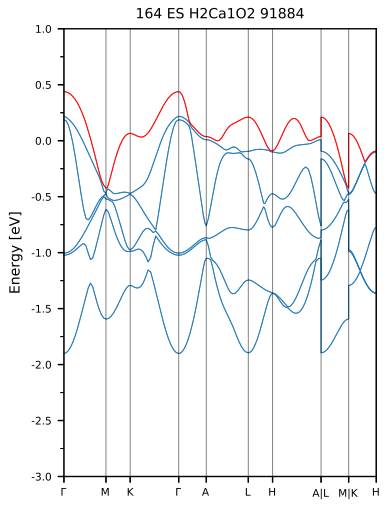

(dx) 164.3.26

Figure 347: Band structures for topological materials in SG 164, part 8/10, from 164.3.11 to 164.3.26. See the list of figures.

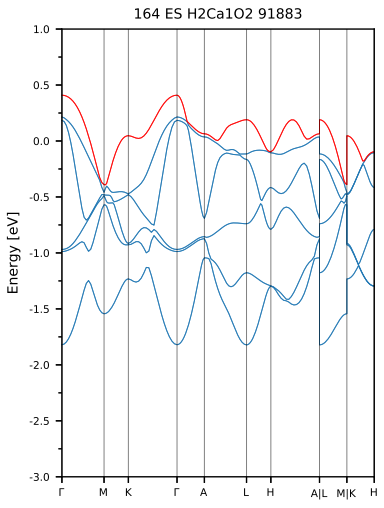

(dy) 164.3.27

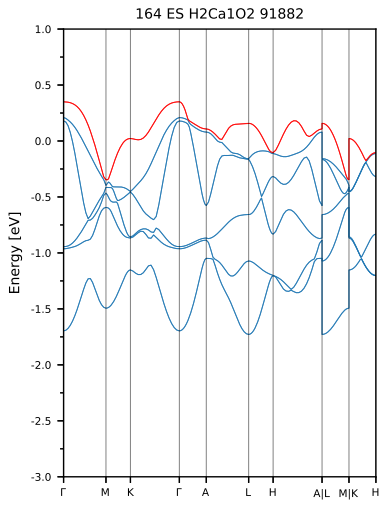

(dz) 164.3.28

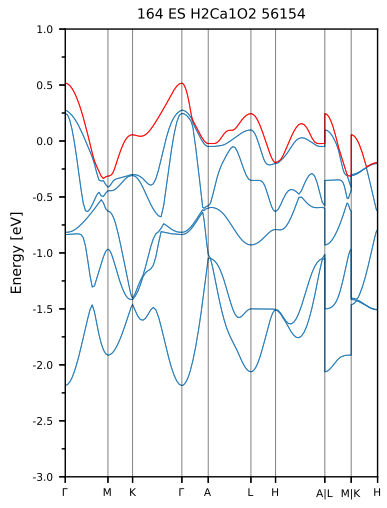

(ea) 164.3.29

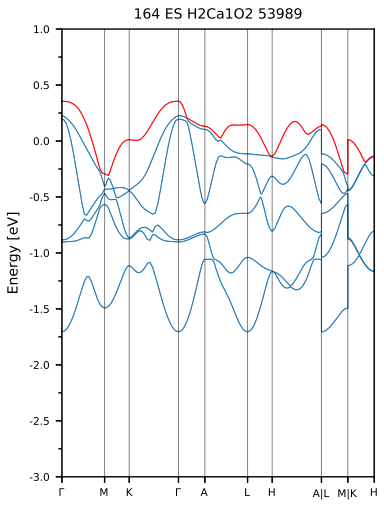

(eb) 164.3.30

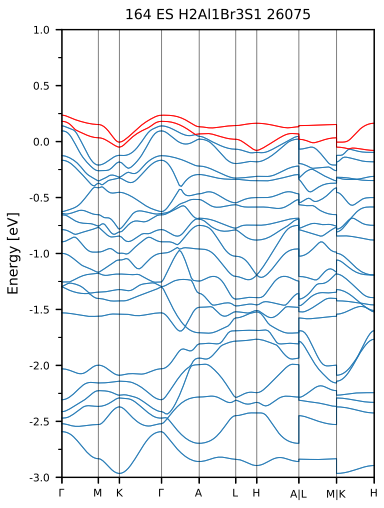

(ec) 164.3.31

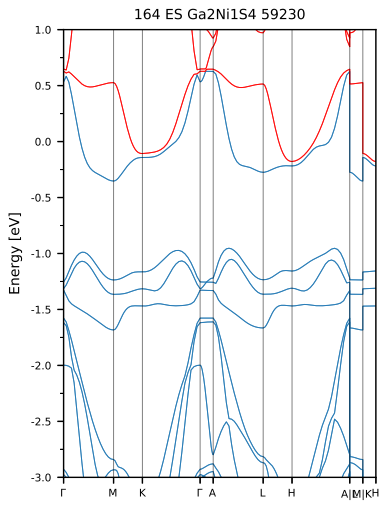

(ed) 164.3.32

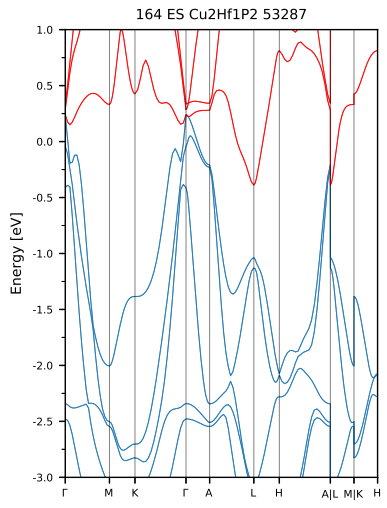

(ee) 164.3.33

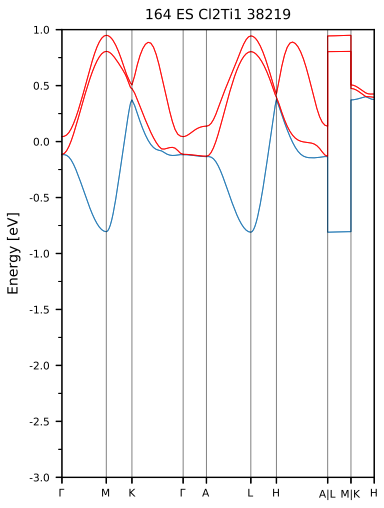

(ef) 164.3.34

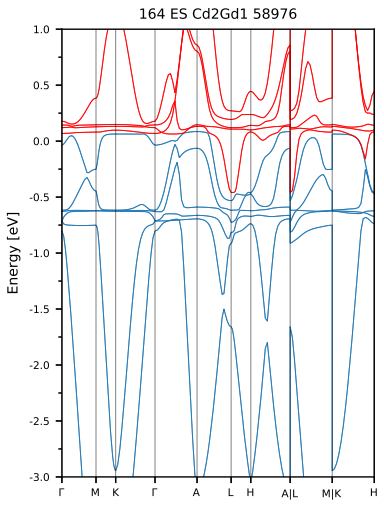

(eg) 164.3.35

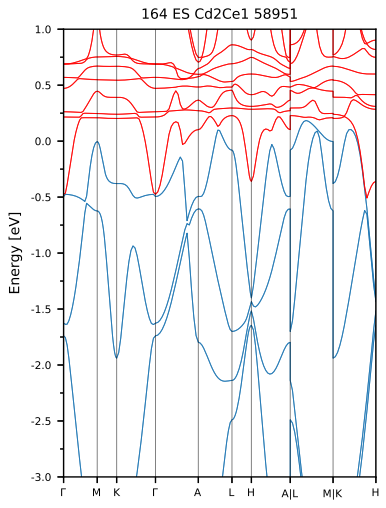

(eh) 164.3.36

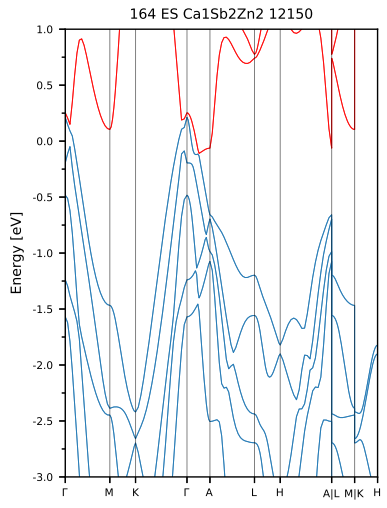

(ei) 164.3.37

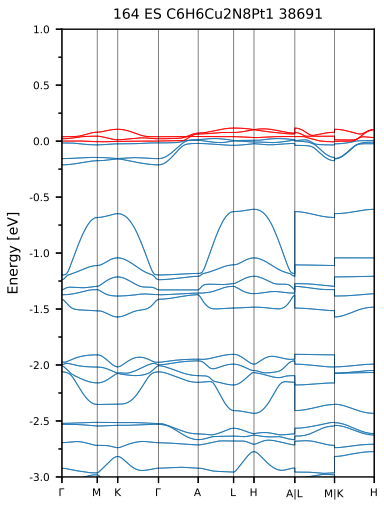

(ej) 164.3.38

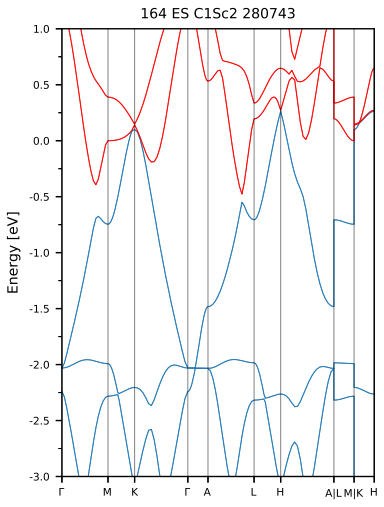

(ek) 164.3.39

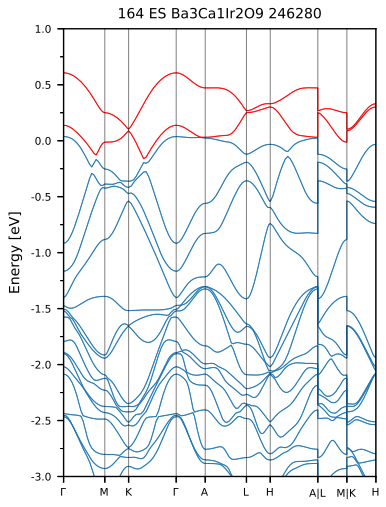

(el) 164.3.40

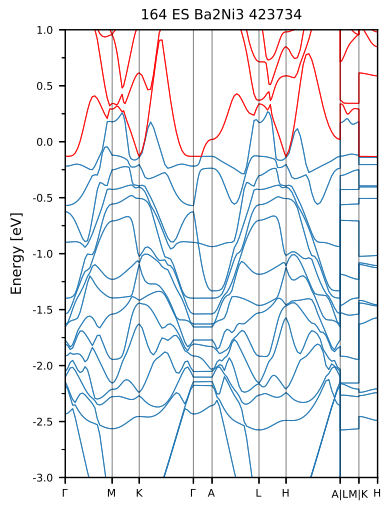

(em) 164.3.41

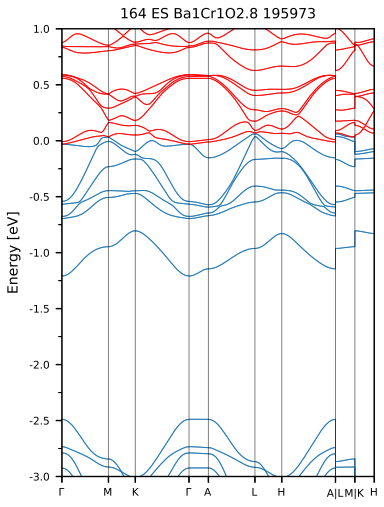

(en) 164.3.42

Figure 348: Band structures for topological materials in SG 164, part 9/10, from 164.3.27 to 164.3.42. See the list of figures.

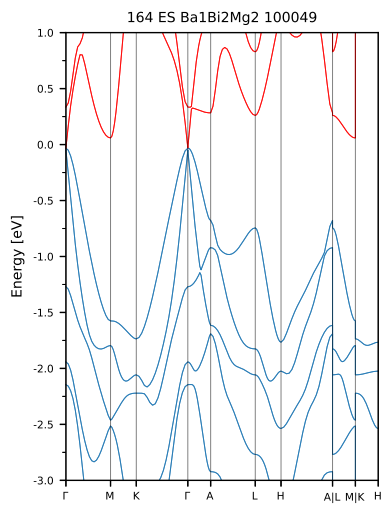

(eo) 164.3.43

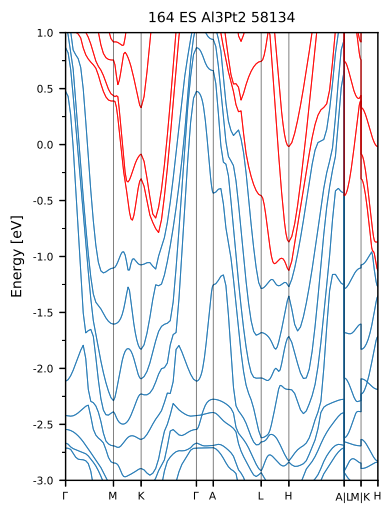

(ep) 164.3.44

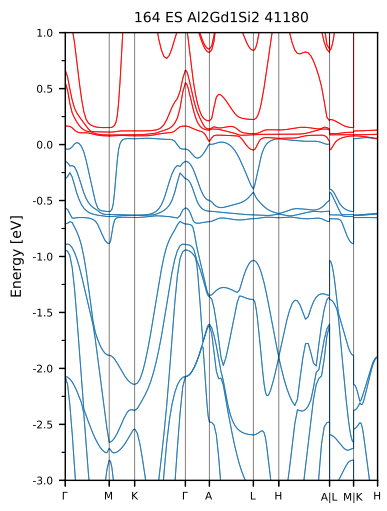

(eq) 164.3.45

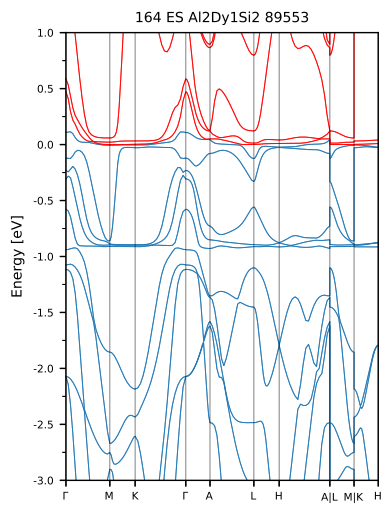

(er) 164.3.46

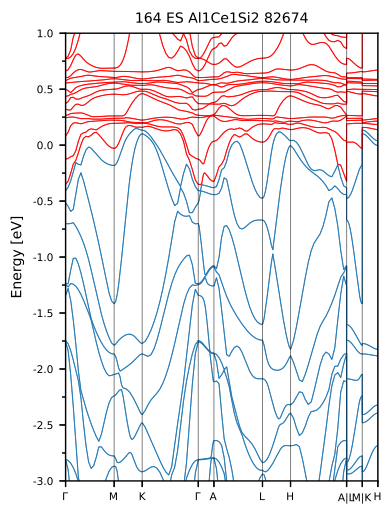

(es) 164.3.47

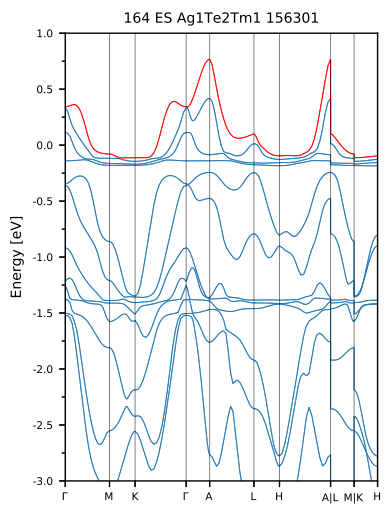

(et) 164.3.48

Figure 349: Band structures for topological materials in SG 164, part 10/10, from 164.3.43 to 164.3.48. See the list of figures.

### 0.0.81 SG 165

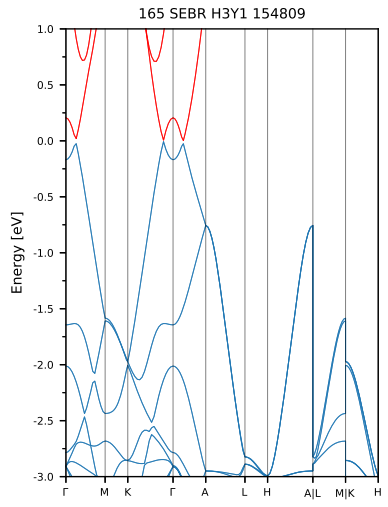

(a) 165.1.1

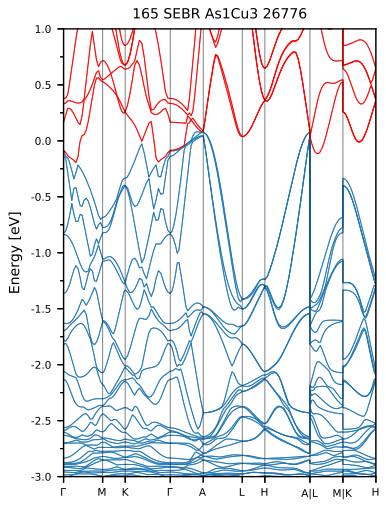

(b) 165.1.2

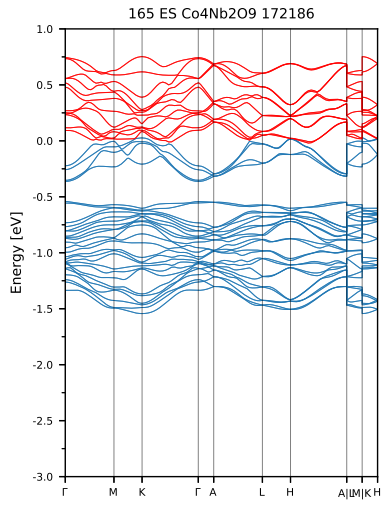

(c) 165.3.1

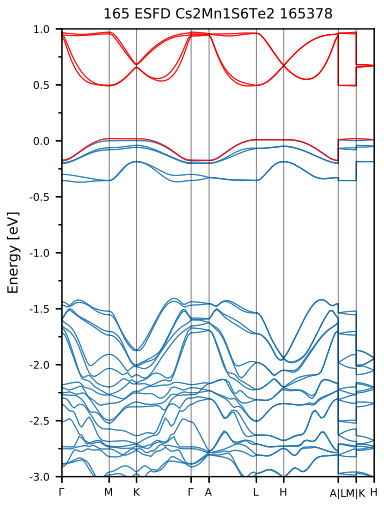

(d) 165.4.1

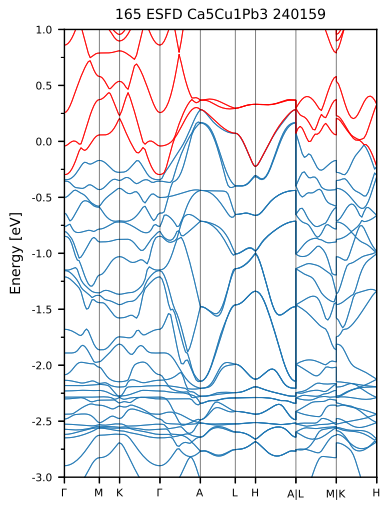

(e) 165.4.2

Figure 350: Band structures for topological materials in SG 165, part 1/1, from 165.1.1 to 165.4.2. See the list of figures.

### 0.0.82 SG 166

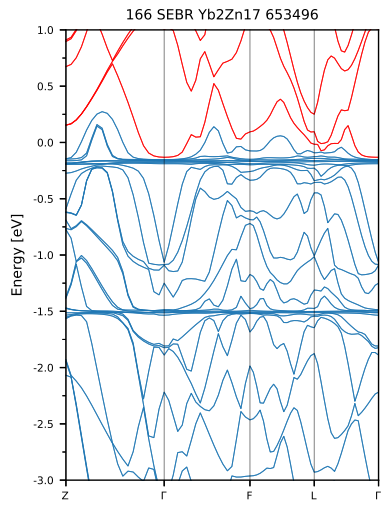

(a) 166.1.1

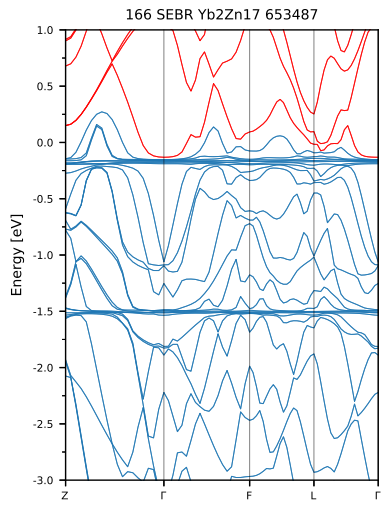

(b) 166.1.2

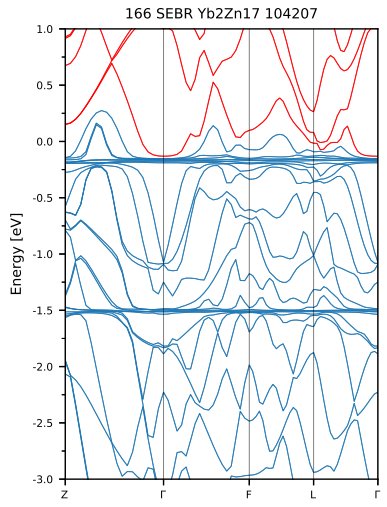

(c) 166.1.3

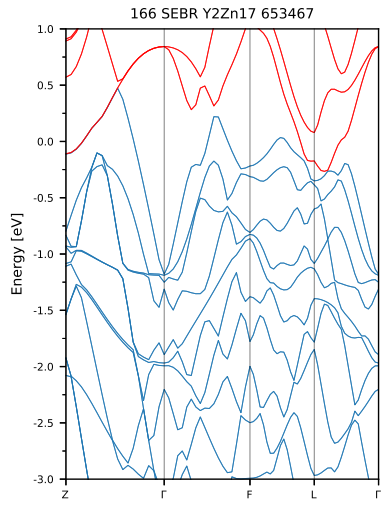

(d) 166.1.4

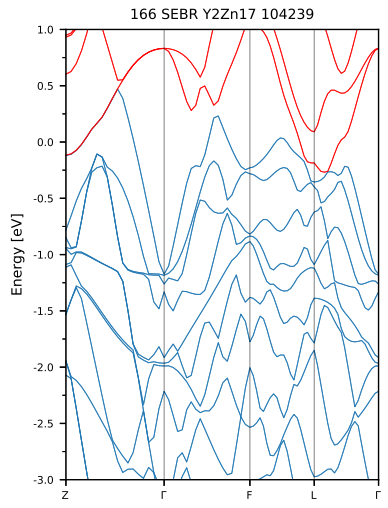

(e) 166.1.5

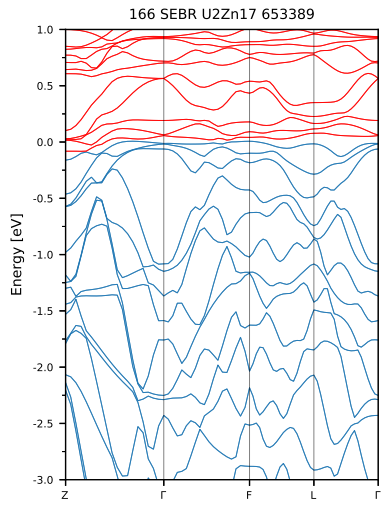

(f) 166.1.6

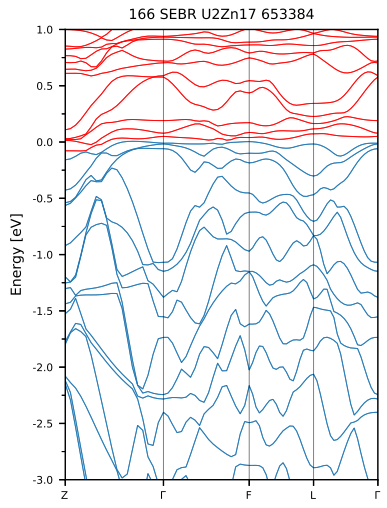

(g) 166.1.7

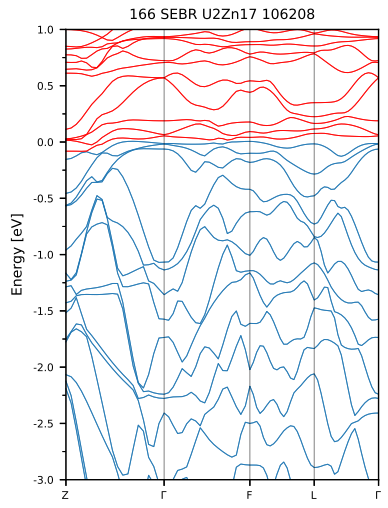

(h) 166.1.8

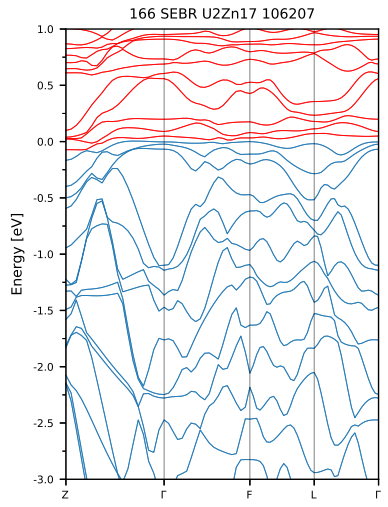

(i) 166.1.9

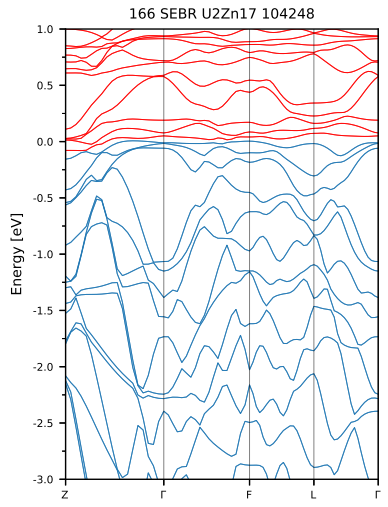

(j) 166.1.10

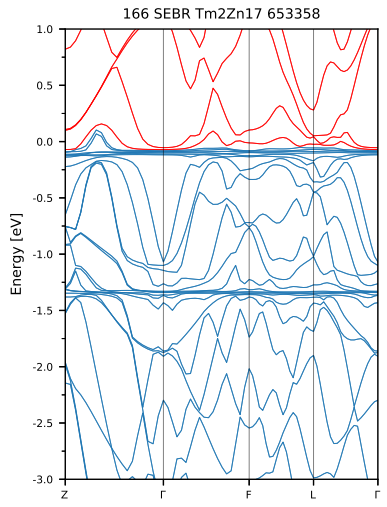

(k) 166.1.11

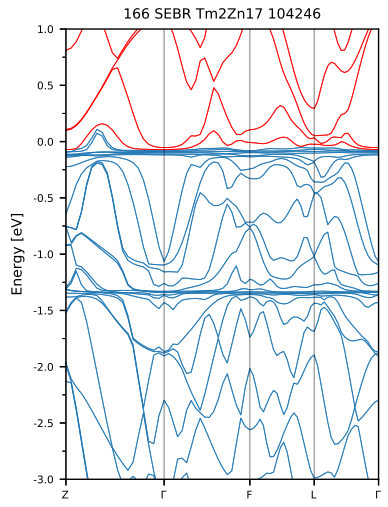

(l) 166.1.12

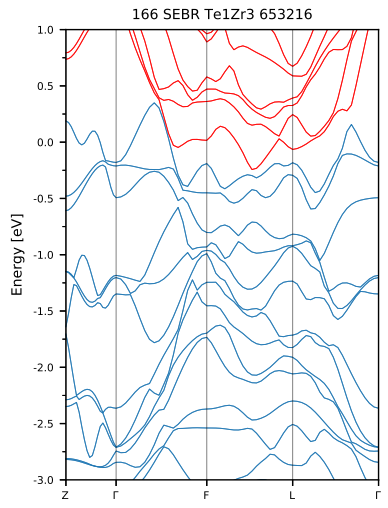

(m) 166.1.13

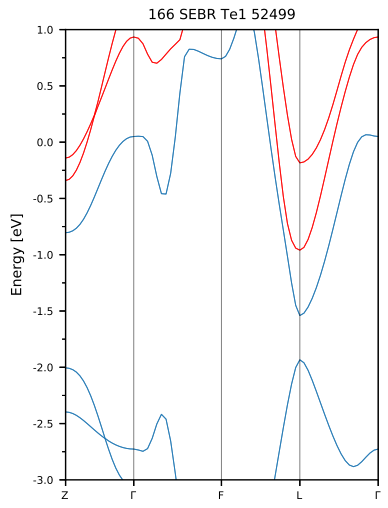

(n) 166.1.14

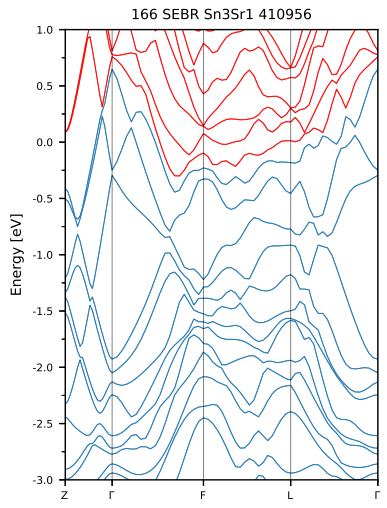

(o) 166.1.15

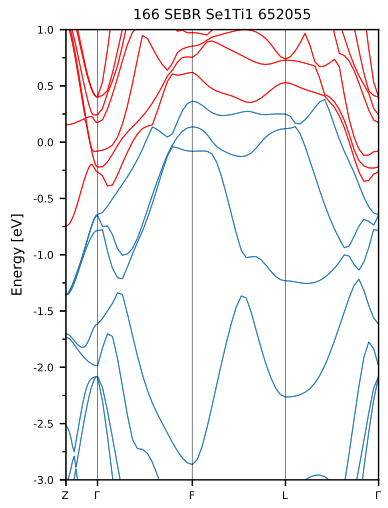

(p) 166.1.16

Figure 351: Band structures for topological materials in SG 166, part 1/39, from 166.1.1 to 166.1.16. See the list of figures.

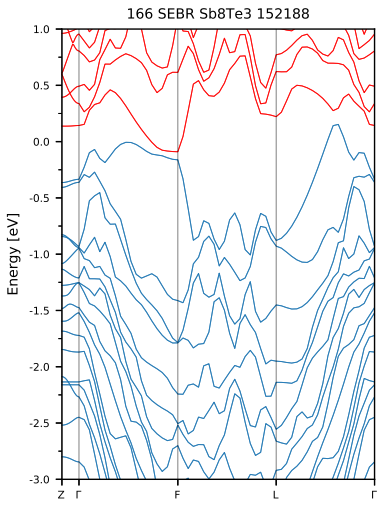

(q) 166.1.17

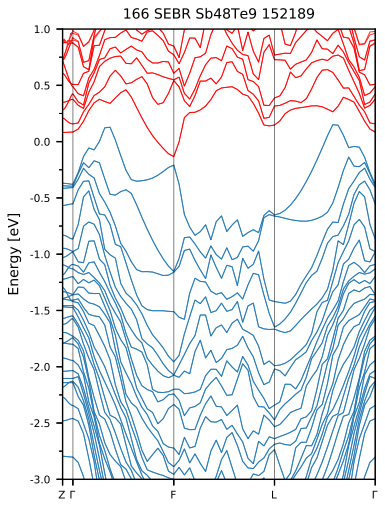

(r) 166.1.18

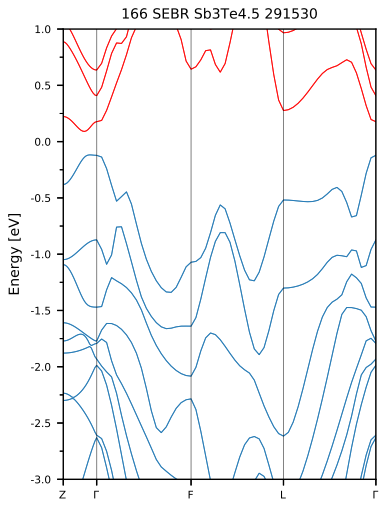

(s) 166.1.19

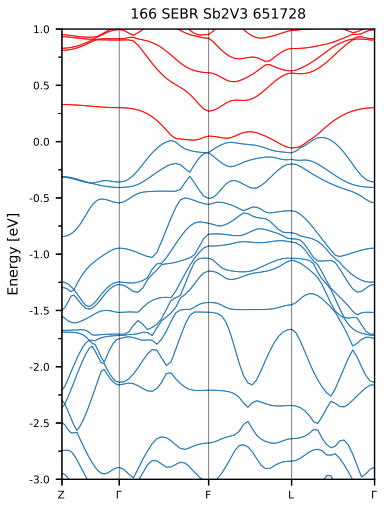

(t) 166.1.20

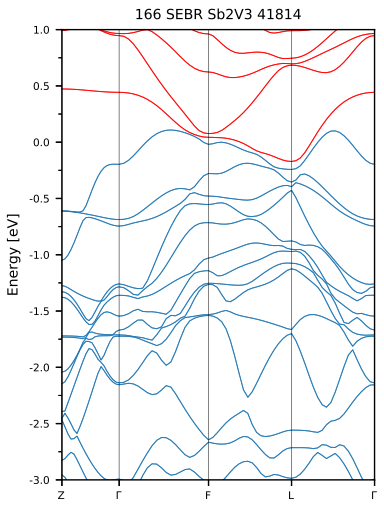

(u) 166.1.21

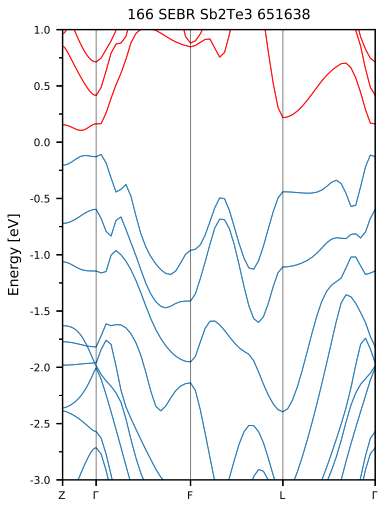

(v) 166.1.22

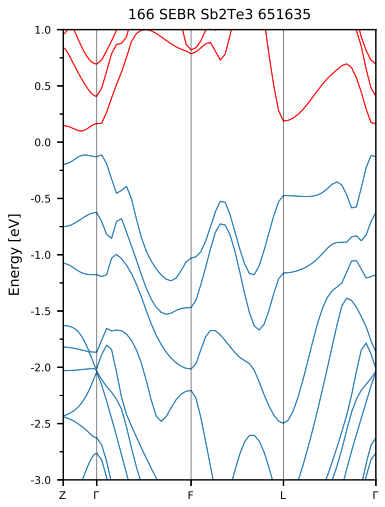

(w) 166.1.23

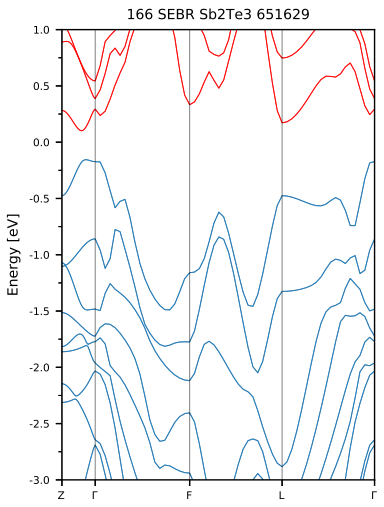

(x) 166.1.24

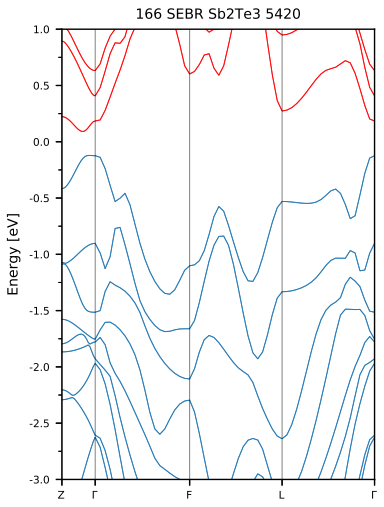

(y) 166.1.25

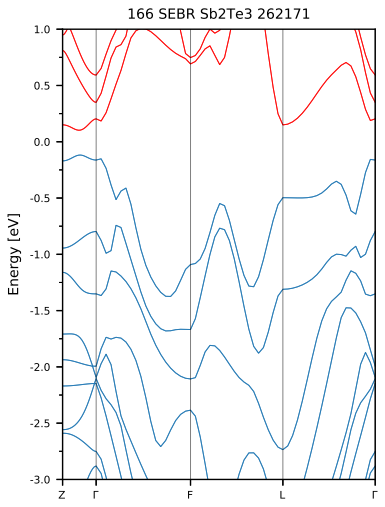

(z) 166.1.26

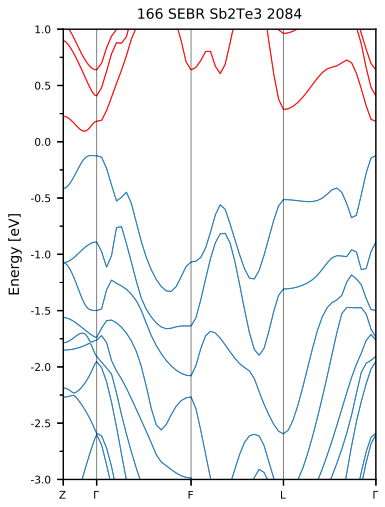

(aa) 166.1.27

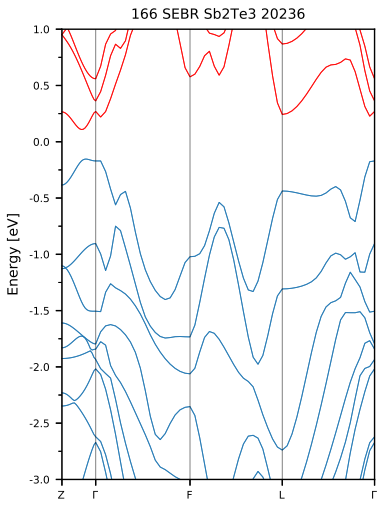

(ab) 166.1.28

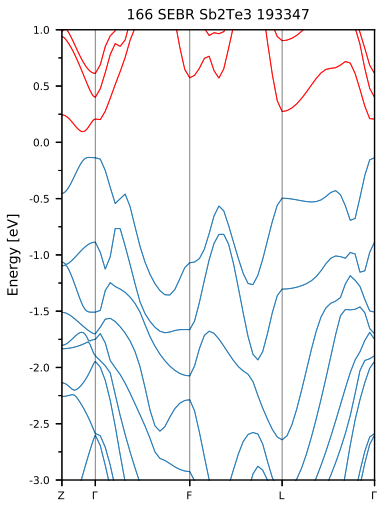

(ac) 166.1.29

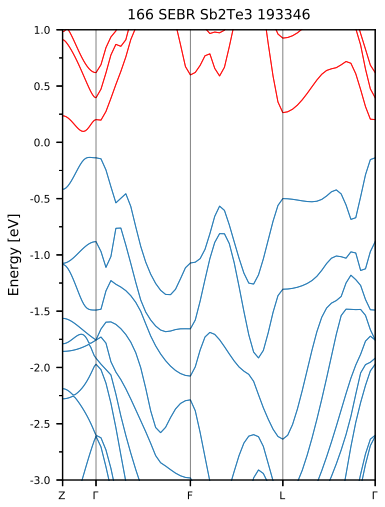

(ad) 166.1.30

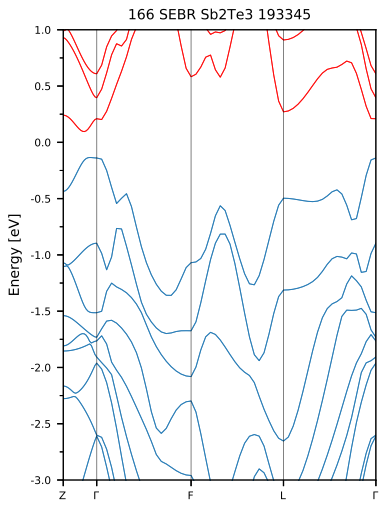

(ae) 166.1.31

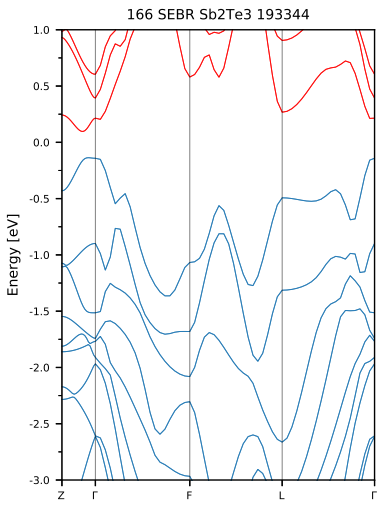

(af) 166.1.32

Figure 352: Band structures for topological materials in SG 166, part 2/39, from 166.1.17 to 166.1.32. See the list of figures.

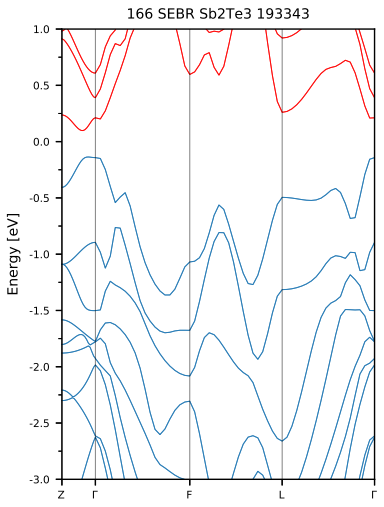

(ag) 166.1.33

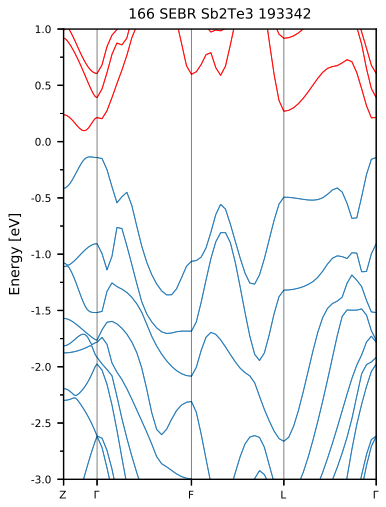

(ah) 166.1.34

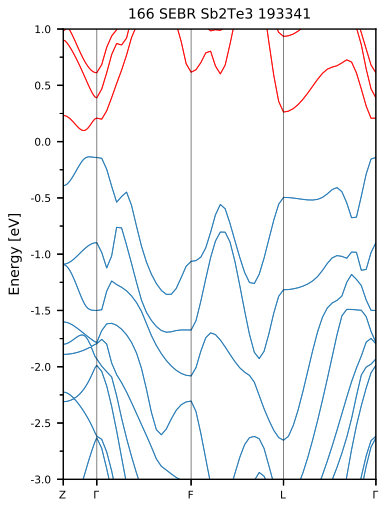

(ai) 166.1.35

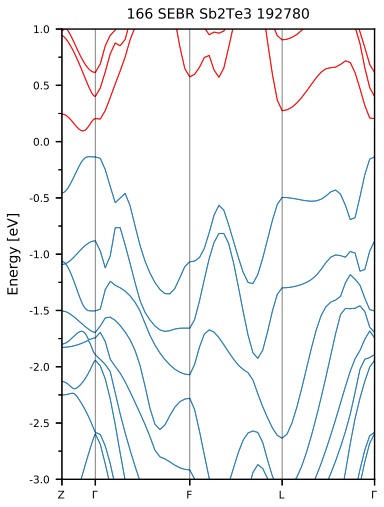

(aj) 166.1.36

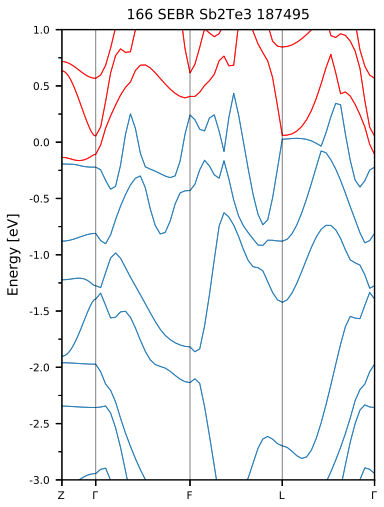

(ak) 166.1.37

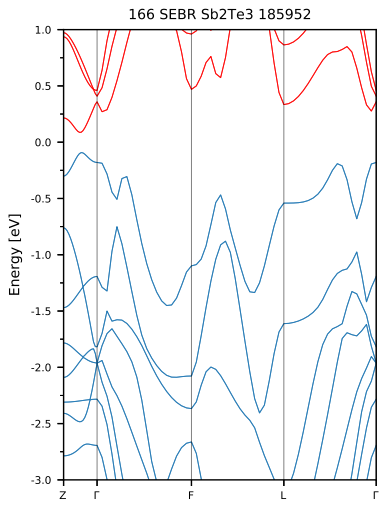

(al) 166.1.38

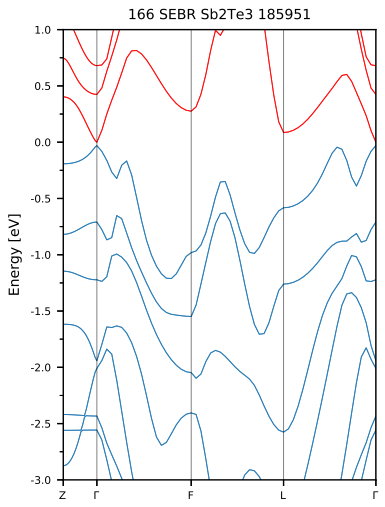

(am) 166.1.39

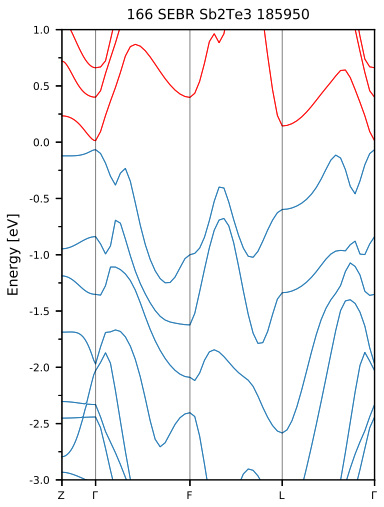

(an) 166.1.40

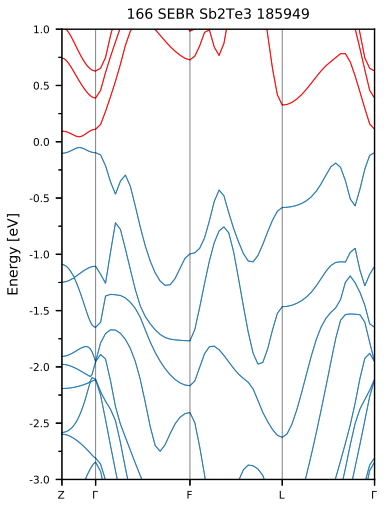

(ao) 166.1.41

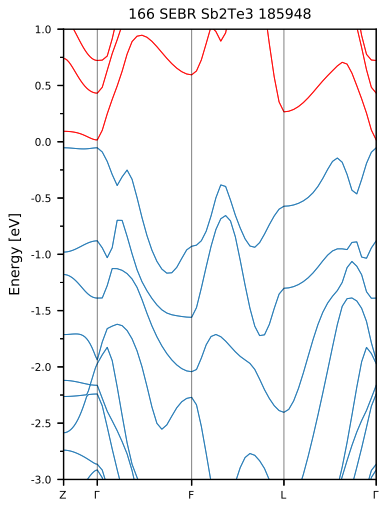

(ap) 166.1.42

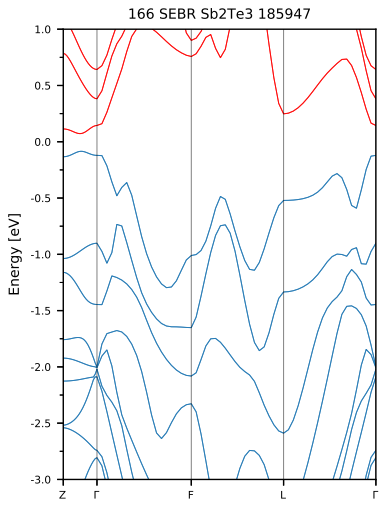

(aq) 166.1.43

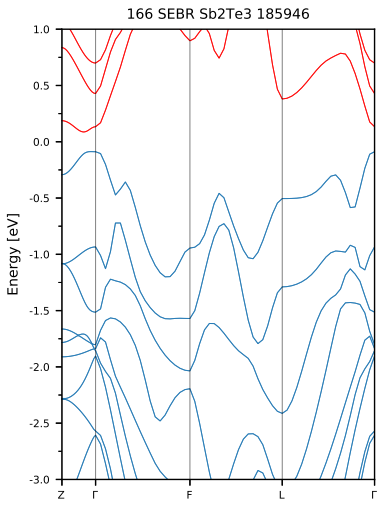

(ar) 166.1.44

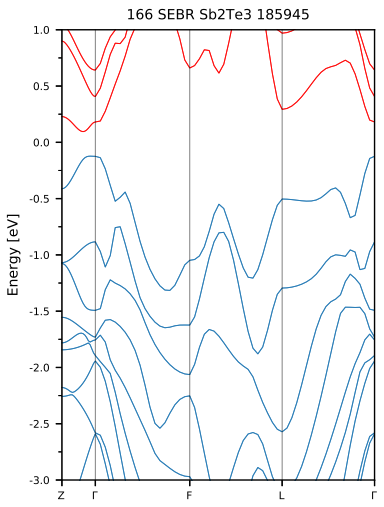

(as) 166.1.45

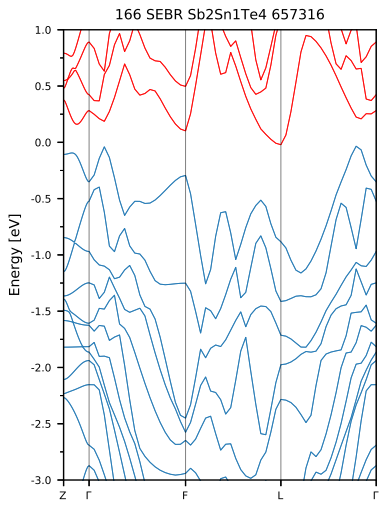

(at) 166.1.46

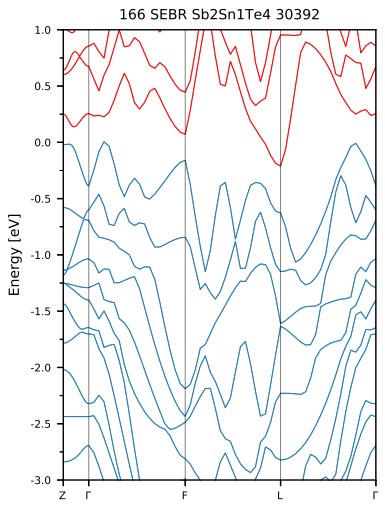

(au) 166.1.47

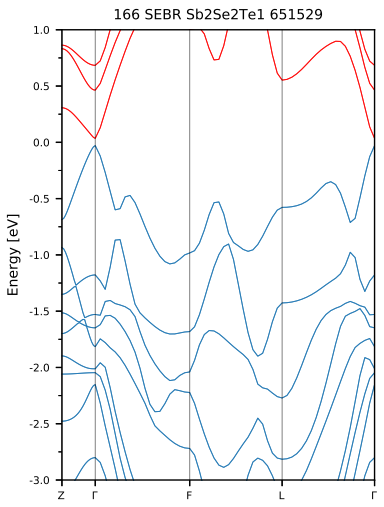

(av) 166.1.48

Figure 353: Band structures for topological materials in SG 166, part 3/39, from 166.1.33 to 166.1.48. See the list of figures.

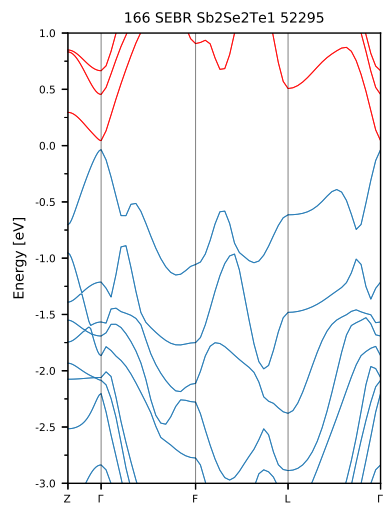

(aw) 166.1.49

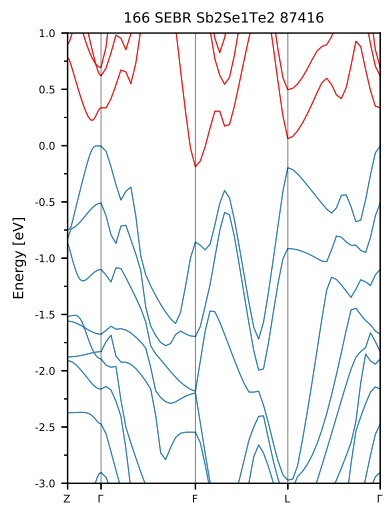

(ax) 166.1.50

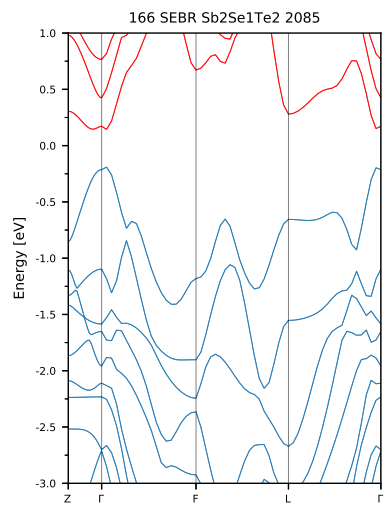

(ay) 166.1.51

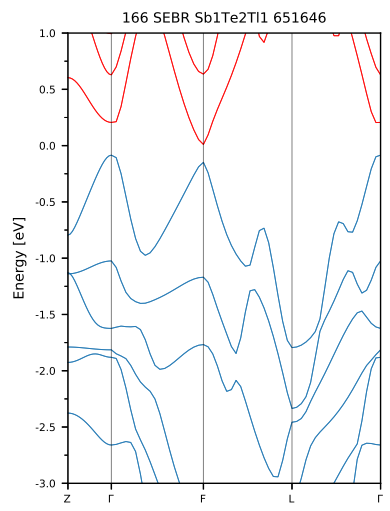

(az) 166.1.52

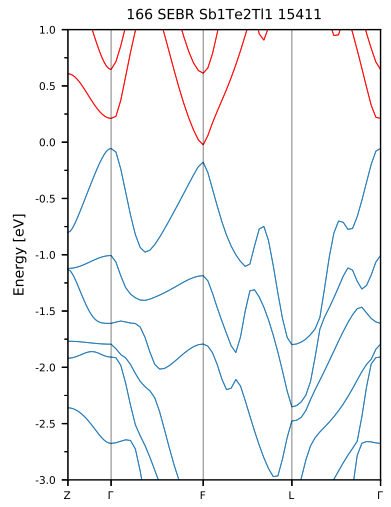

(ba) 166.1.53

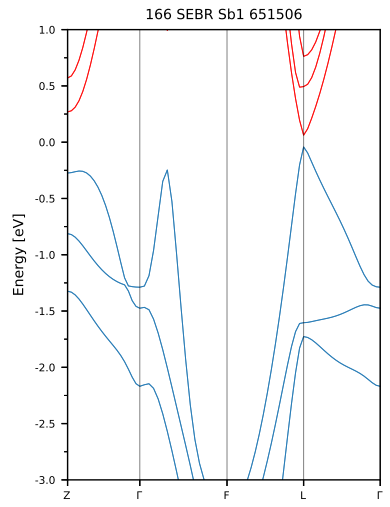

(bb) 166.1.54

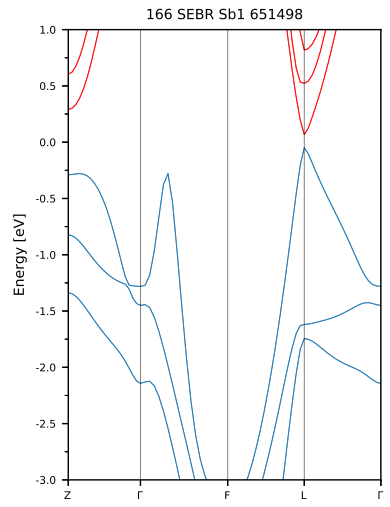

(bc) 166.1.55

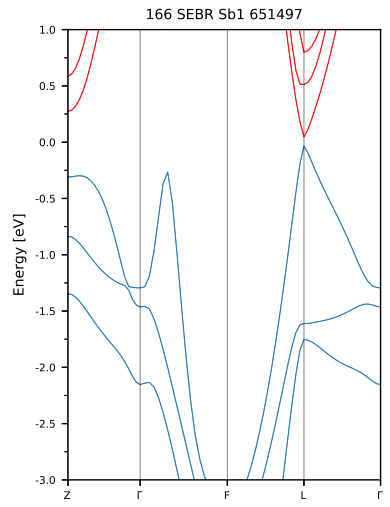

(bd) 166.1.56

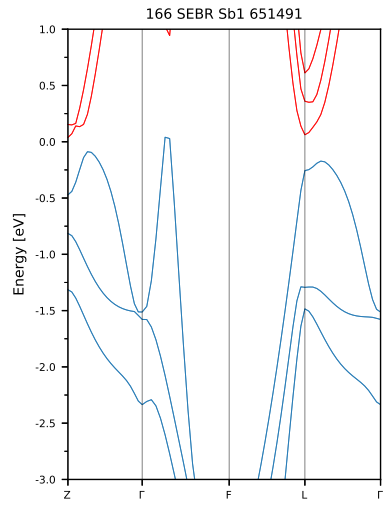

(be) 166.1.57

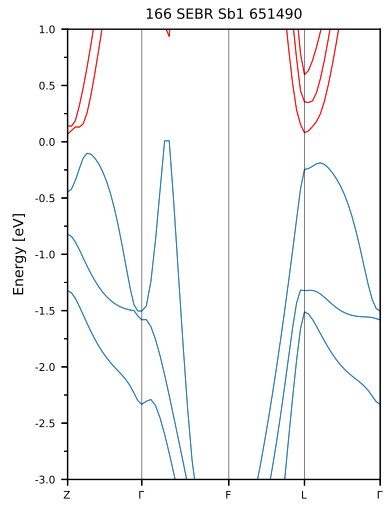

(bf) 166.1.58

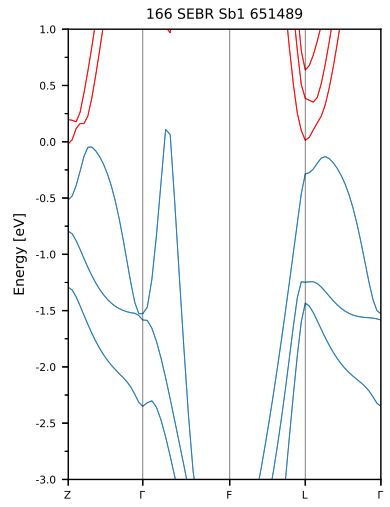

(bg) 166.1.59

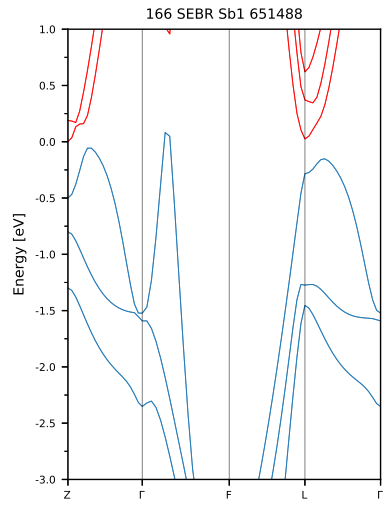

(bh) 166.1.60

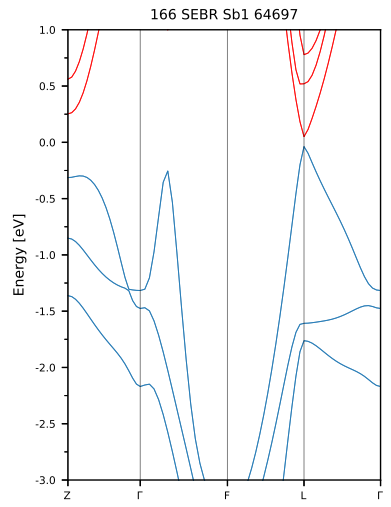

(bi) 166.1.61

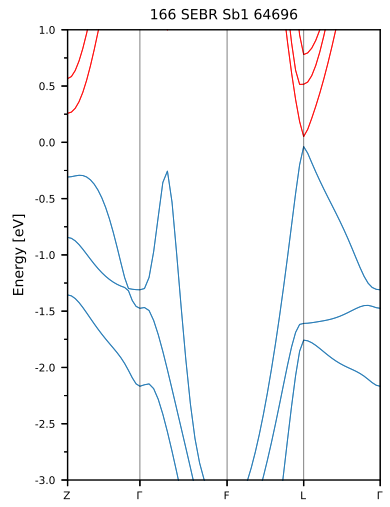

(bj) 166.1.62

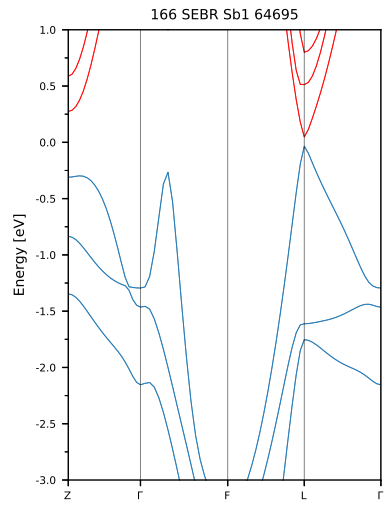

(bk) 166.1.63

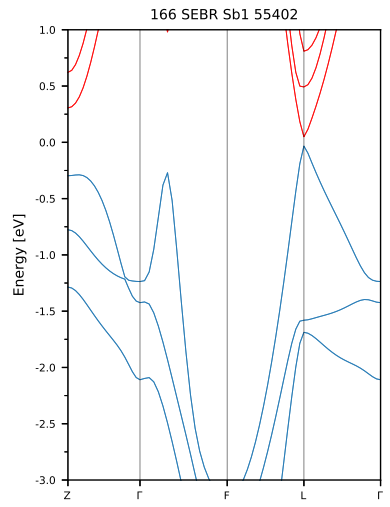

(bl) 166.1.64

Figure 354: Band structures for topological materials in SG 166, part 4/39, from 166.1.49 to 166.1.64. See the list of figures.

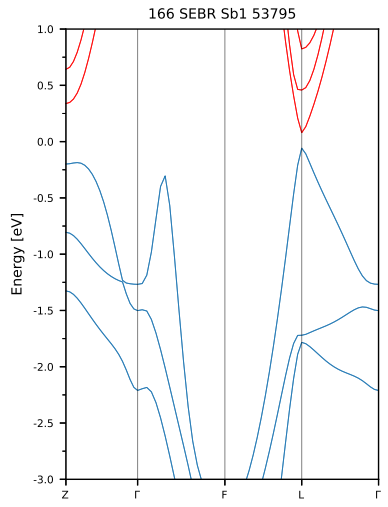

(bm) 166.1.65

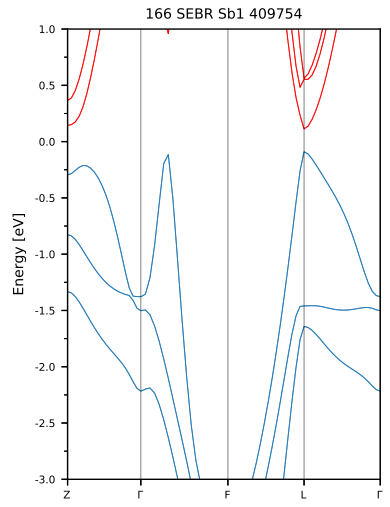

(bn) 166.1.66

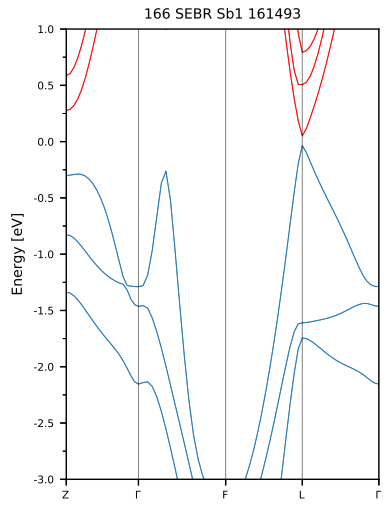

(bo) 166.1.67

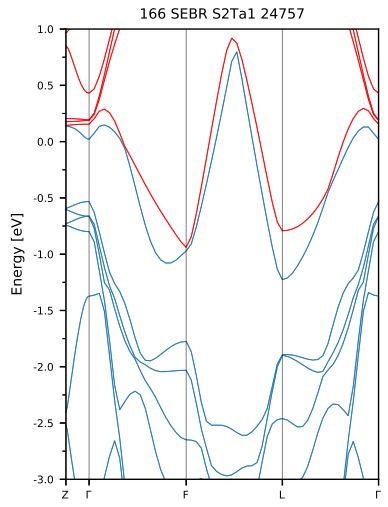

(bp) 166.1.68

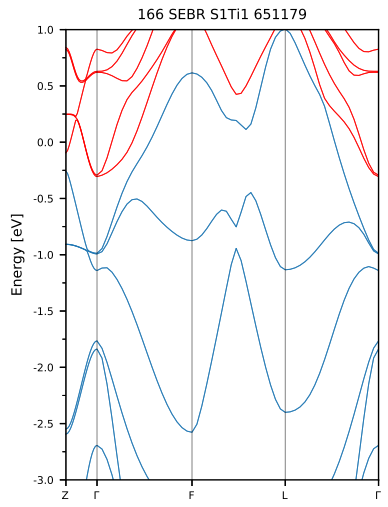

(bq) 166.1.69

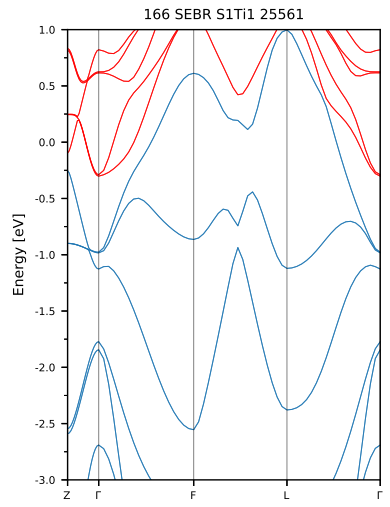

(br) 166.1.70

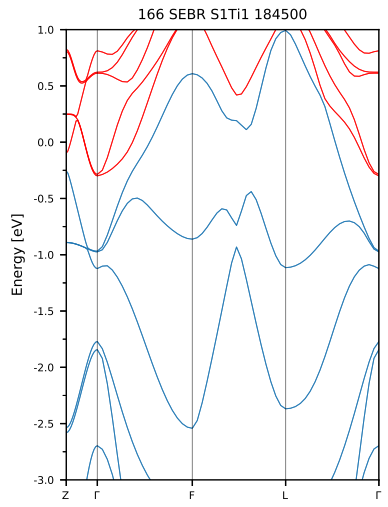

(bs) 166.1.71

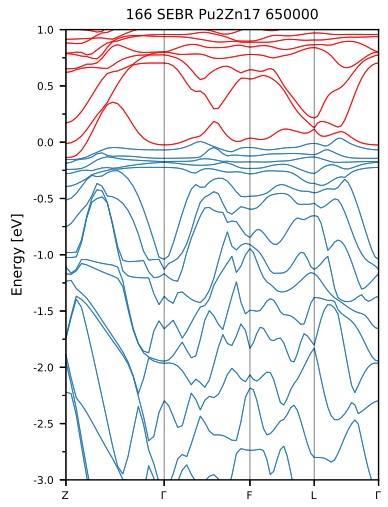

(bt) 166.1.72

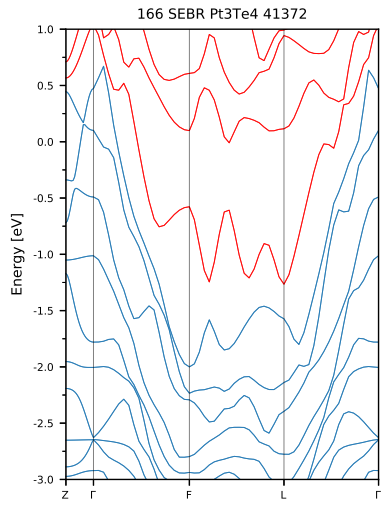

(bu) 166.1.73

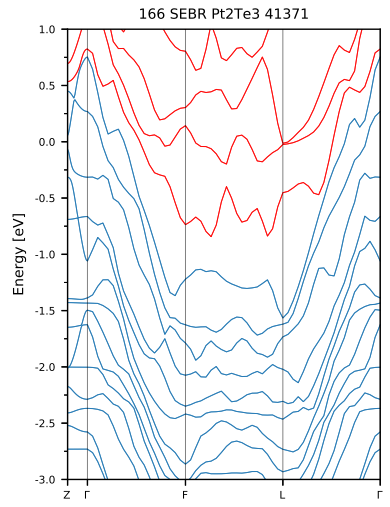

(bv) 166.1.74

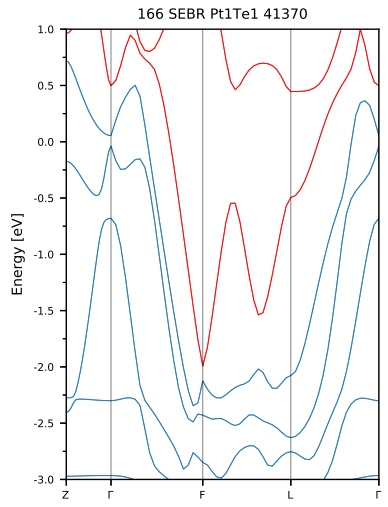

(bw) 166.1.75

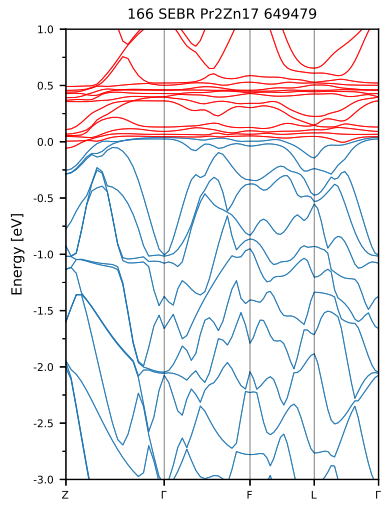

(bx) 166.1.76

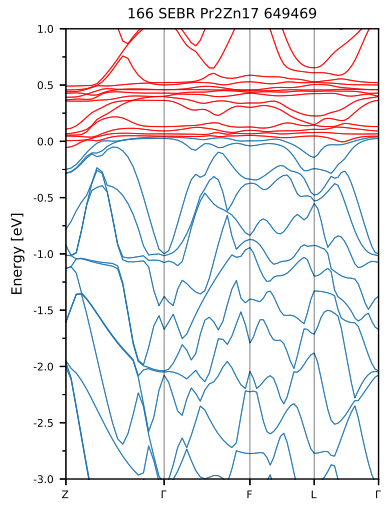

(by) 166.1.77

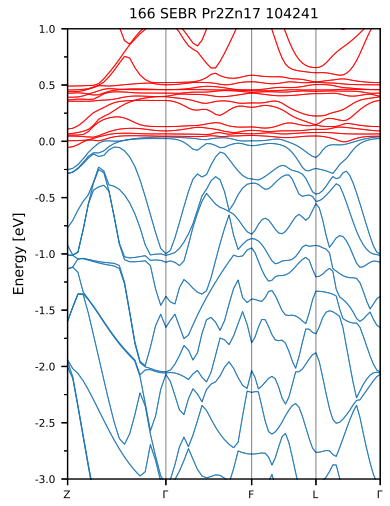

(bz) 166.1.78

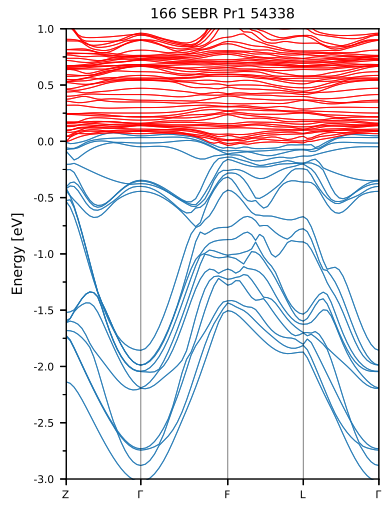

(ca) 166.1.79

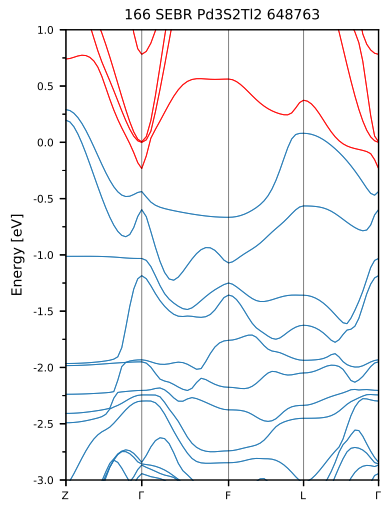

(cb) 166.1.80

Figure 355: Band structures for topological materials in SG 166, part 5/39, from 166.1.65 to 166.1.80. See the list of figures.

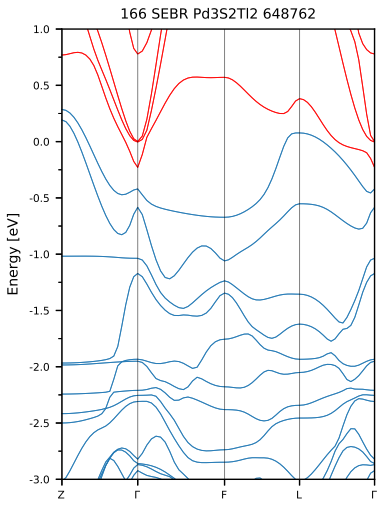

(cc) 166.1.81

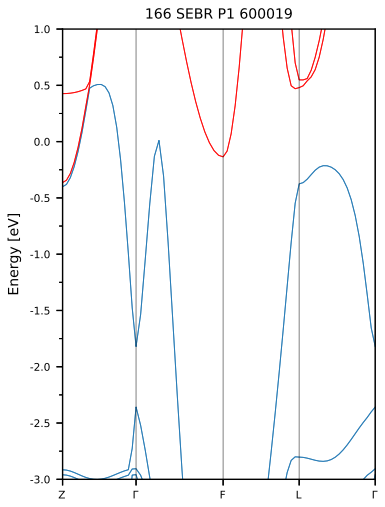

(cd) 166.1.82

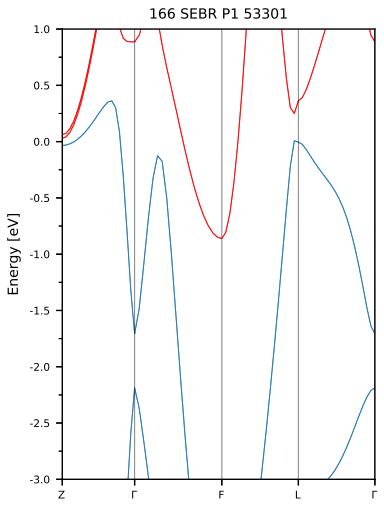

(ce) 166.1.83

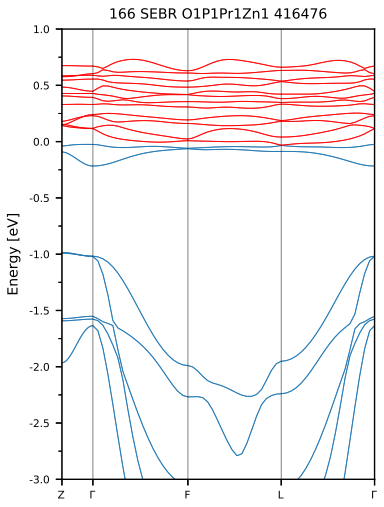

(cf) 166.1.84

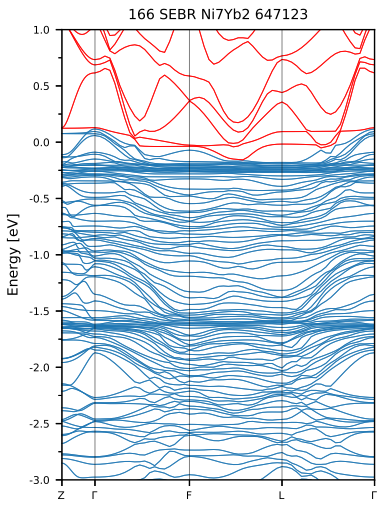

(cg) 166.1.85

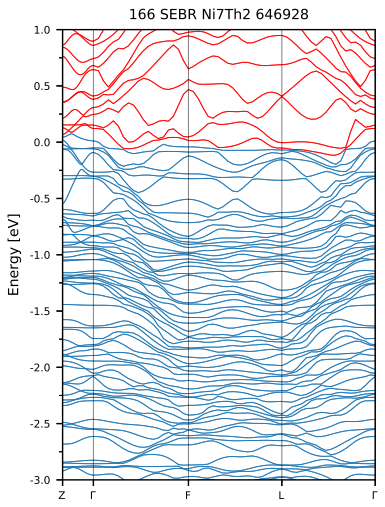

(ch) 166.1.86

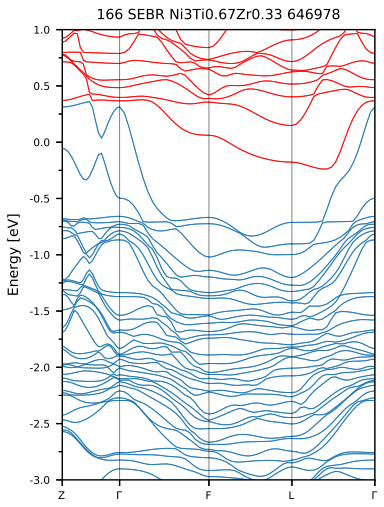

(ci) 166.1.87

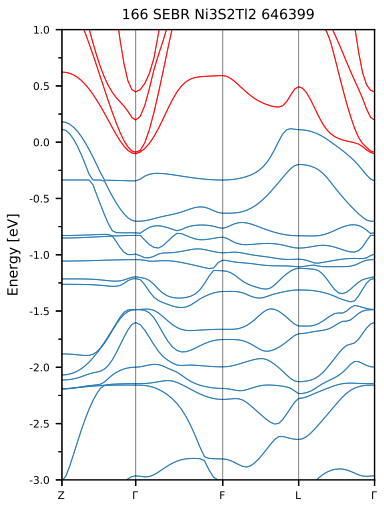

(cj) 166.1.88

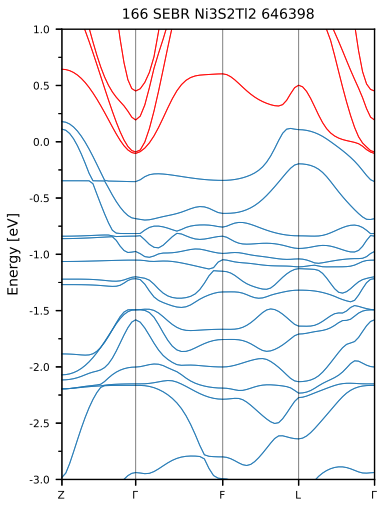

(ck) 166.1.89

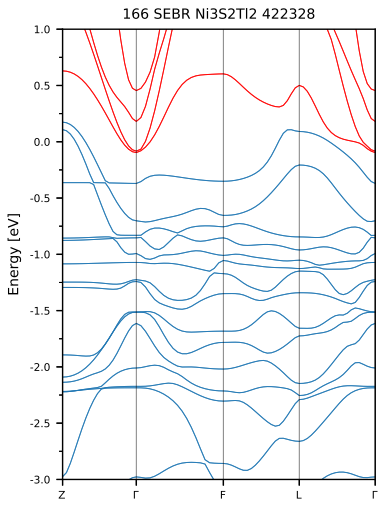

(cl) 166.1.90

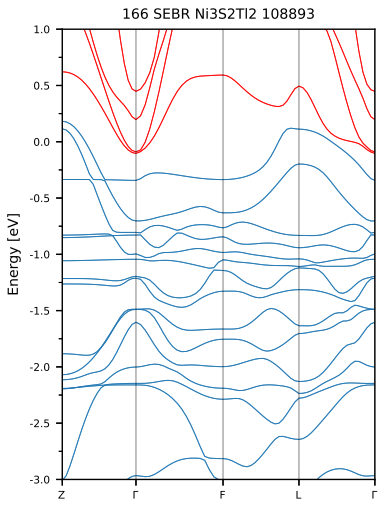

(cm) 166.1.91

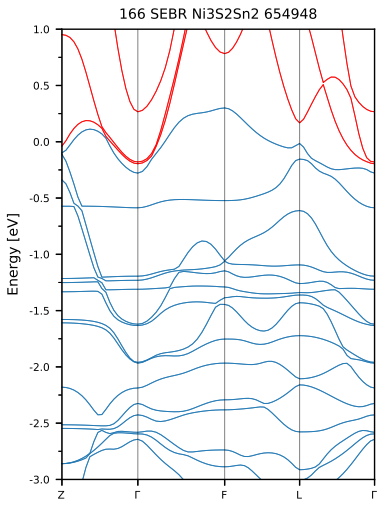

(cn) 166.1.92

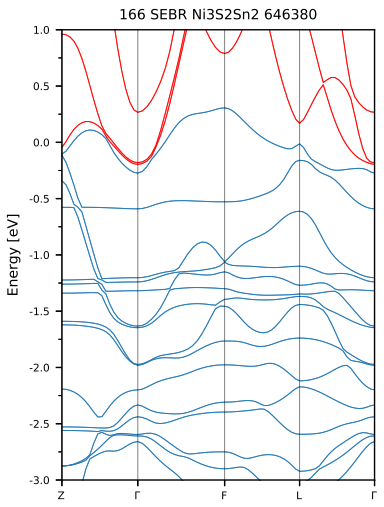

(co) 166.1.93

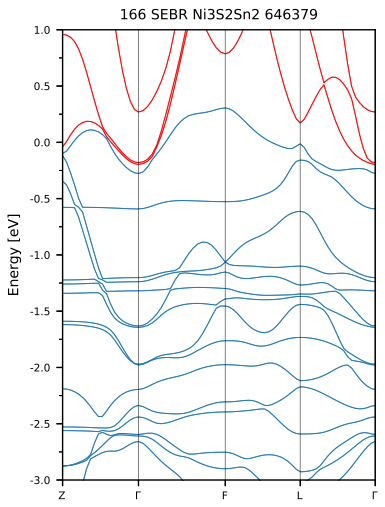

(cp) 166.1.94

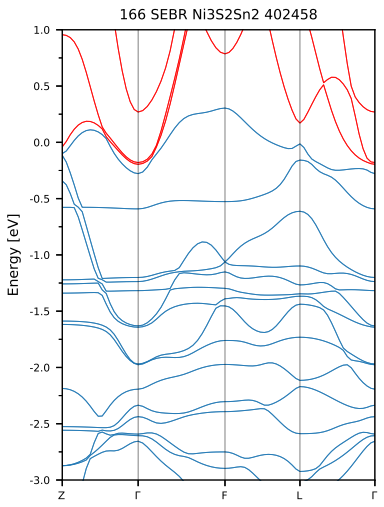

(cq) 166.1.95

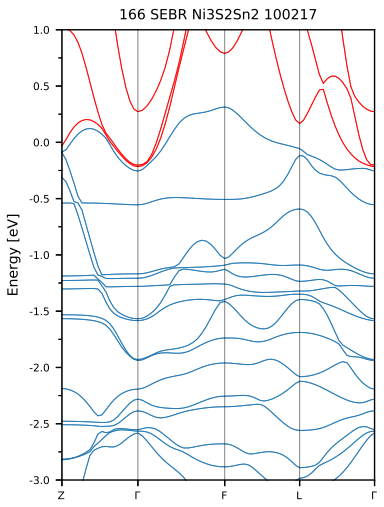

(cr) 166.1.96

Figure 356: Band structures for topological materials in SG 166, part 6/39, from 166.1.81 to 166.1.96. See the list of figures.

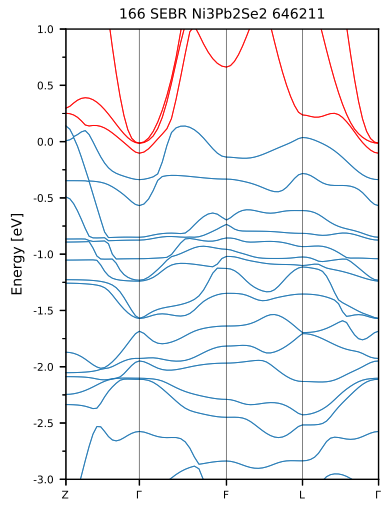

(cs) 166.1.97

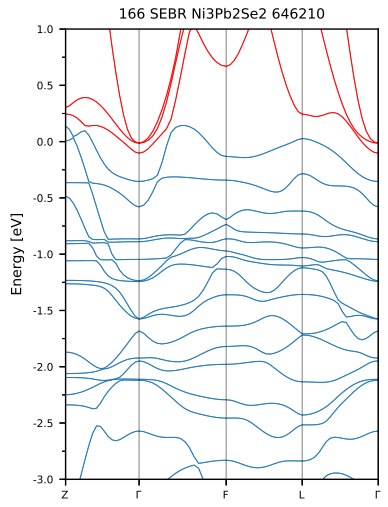

(ct) 166.1.98

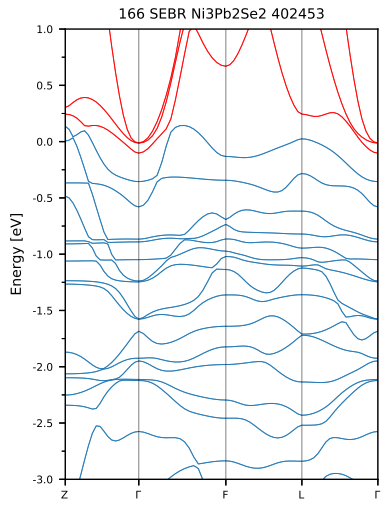

(cu) 166.1.99

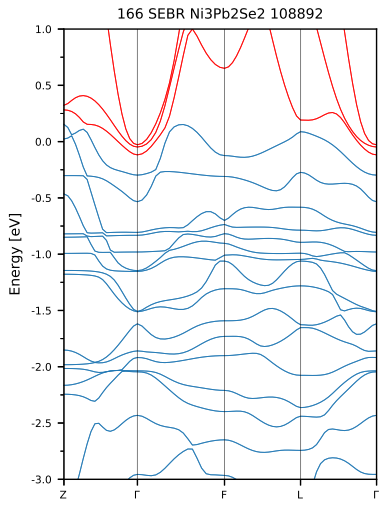

(cv) 166.1.100

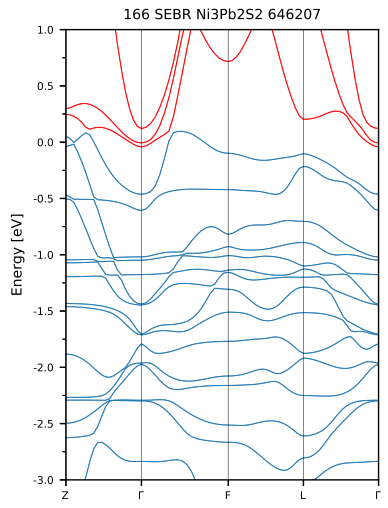

(cw) 166.1.101

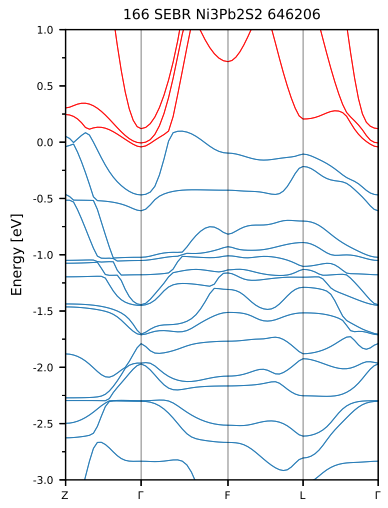

(cx) 166.1.102

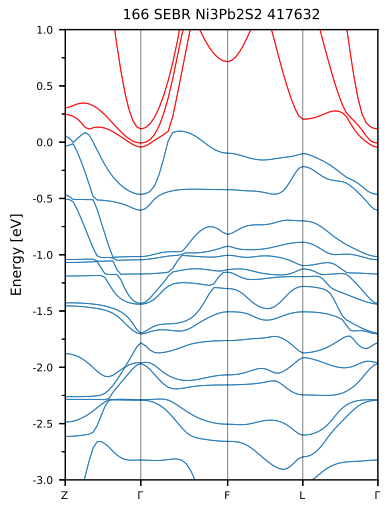

(cy) 166.1.103

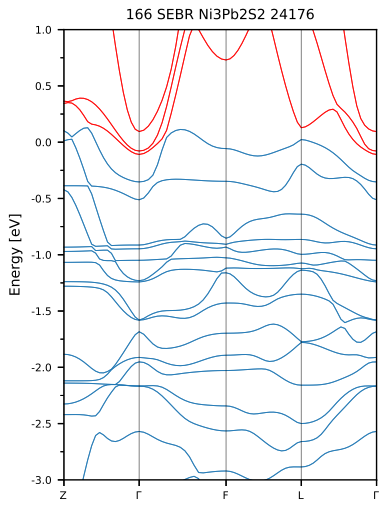

(cz) 166.1.104

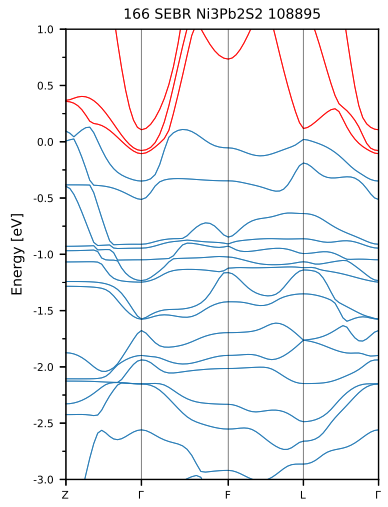

(da) 166.1.105

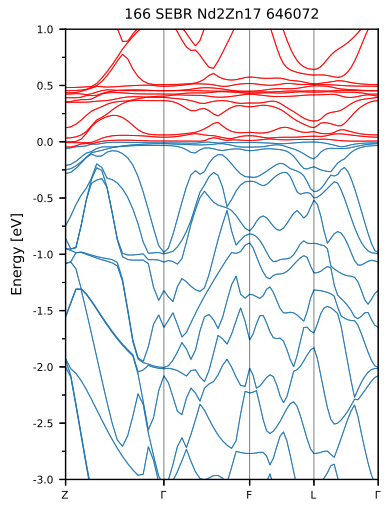

(db) 166.1.106

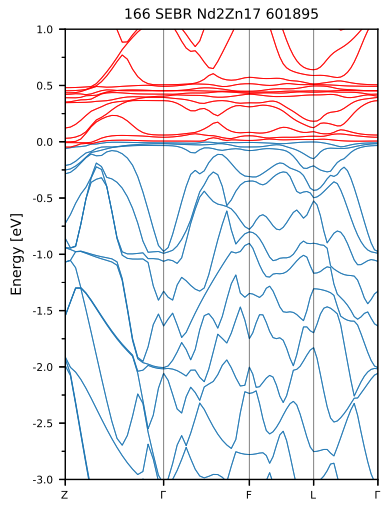

(dc) 166.1.107

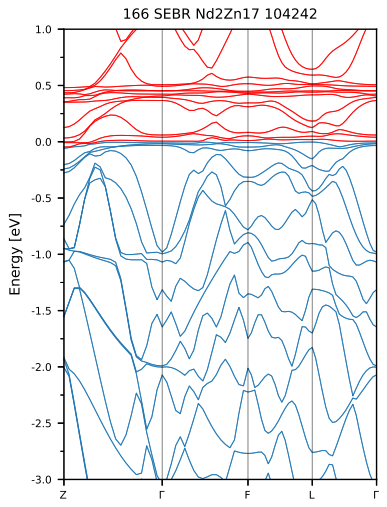

(dd) 166.1.108

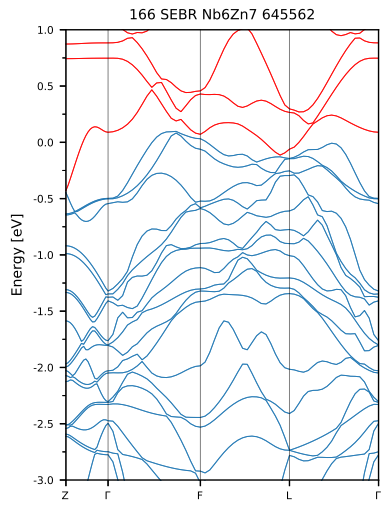

(de) 166.1.109

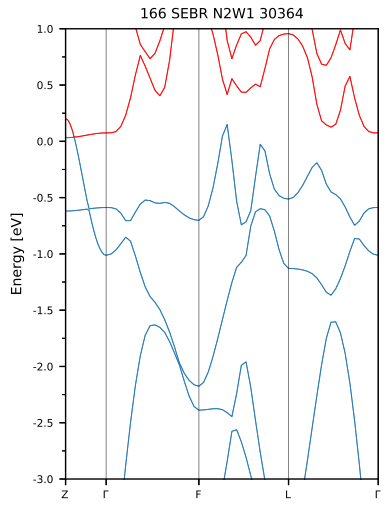

(df) 166.1.110

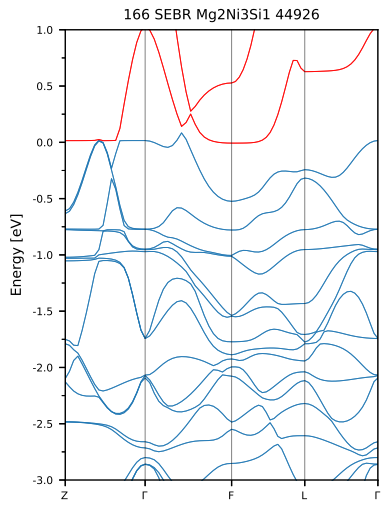

(dg) 166.1.111

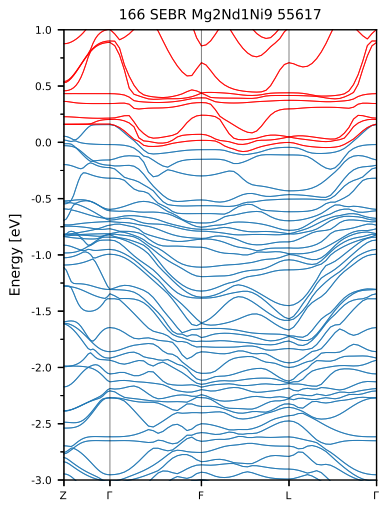

(dh) 166.1.112

Figure 357: Band structures for topological materials in SG 166, part 7/39, from 166.1.97 to 166.1.112. See the list of figures.

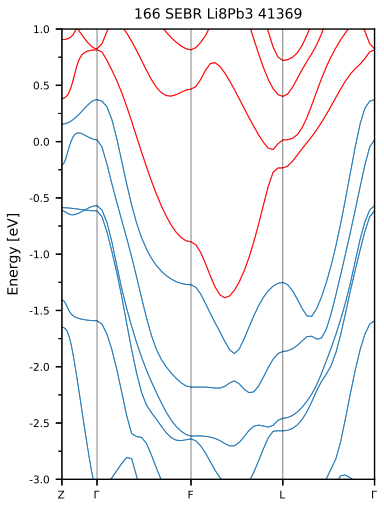

(di) 166.1.113

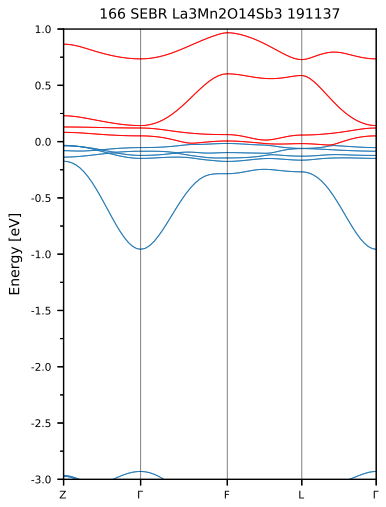

(dj) 166.1.114

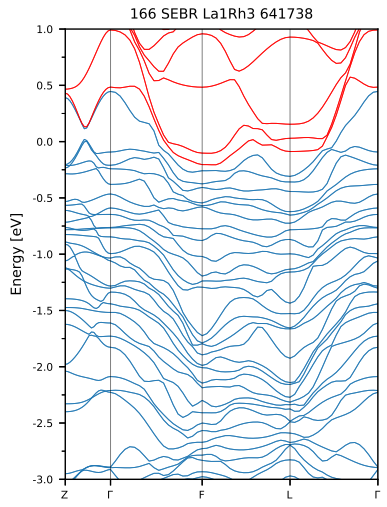

(dk) 166.1.115

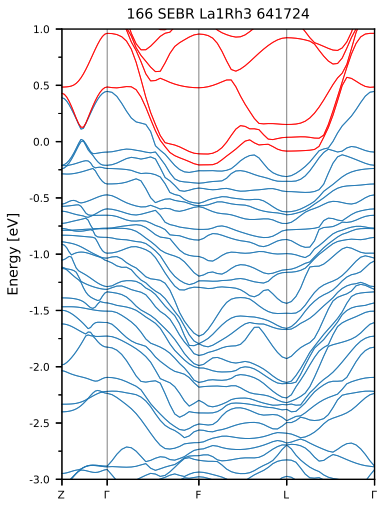

(dl) 166.1.116

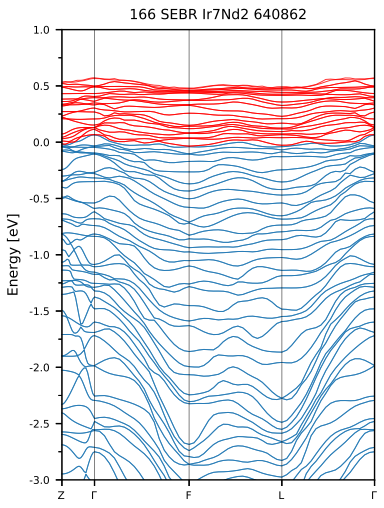

(dm) 166.1.117

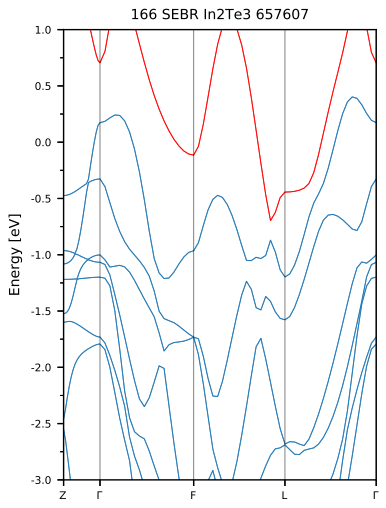

(dn) 166.1.118

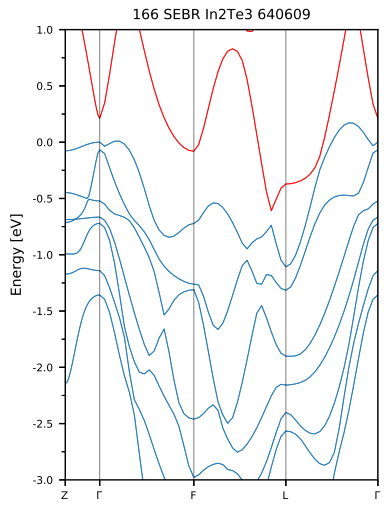

(do) 166.1.119

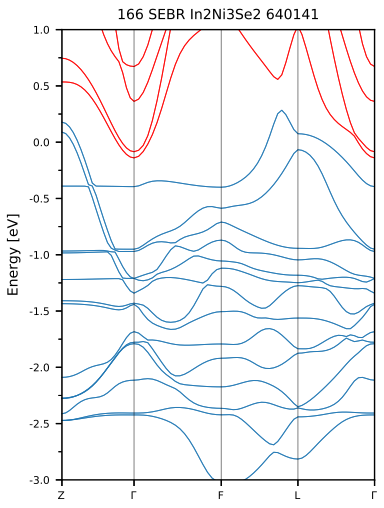

(dp) 166.1.120

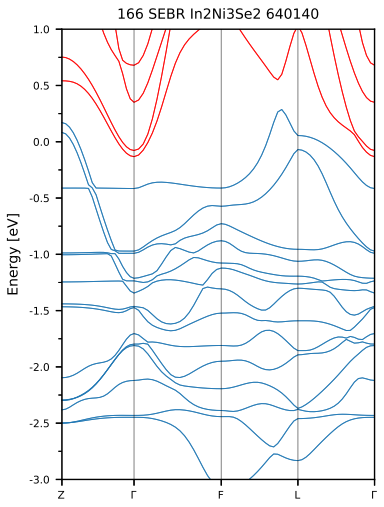

(dq) 166.1.121

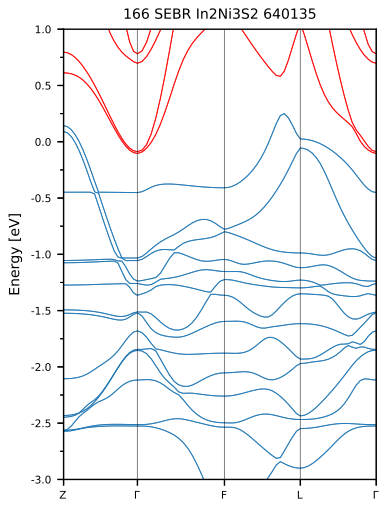

(dr) 166.1.122

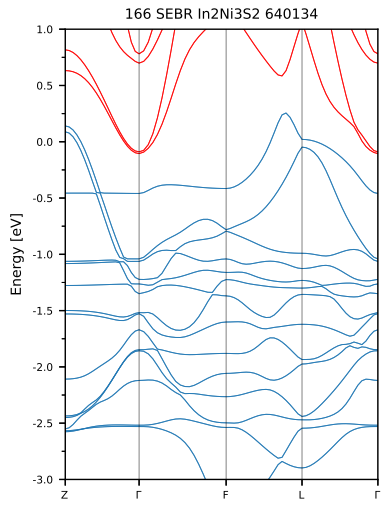

(ds) 166.1.123

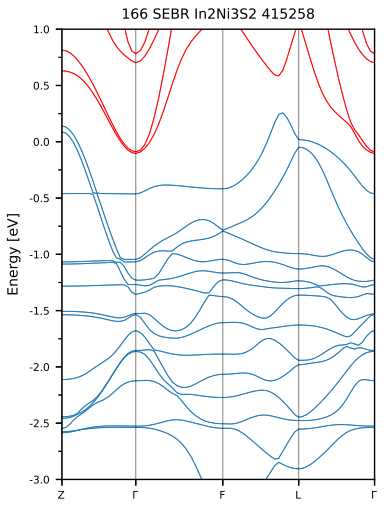

(dt) 166.1.124

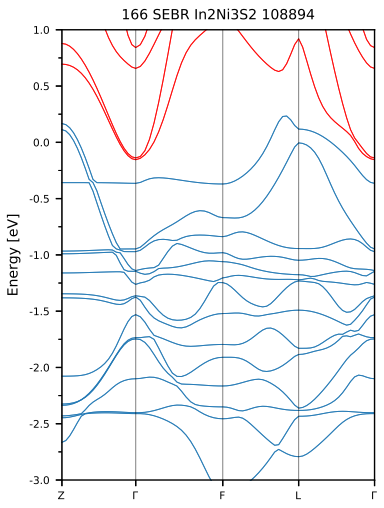

(du) 166.1.125

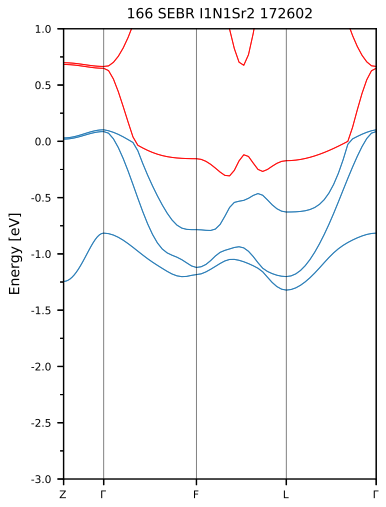

(dv) 166.1.126

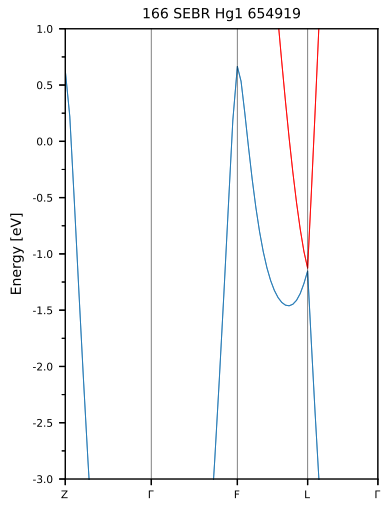

(dw) 166.1.127

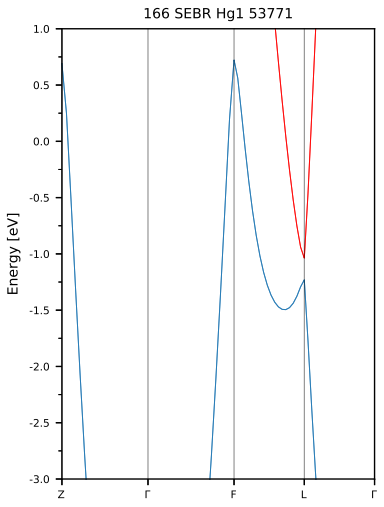

(dx) 166.1.128

Figure 358: Band structures for topological materials in SG 166, part 8/39, from 166.1.113 to 166.1.128. See the list of figures.

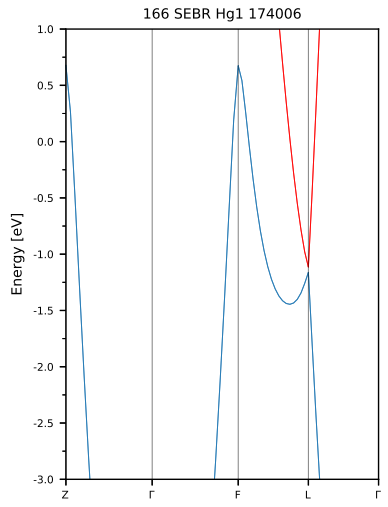

(dy) 166.1.129

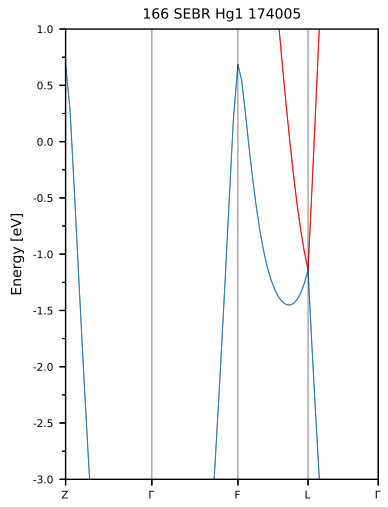

(dz) 166.1.130

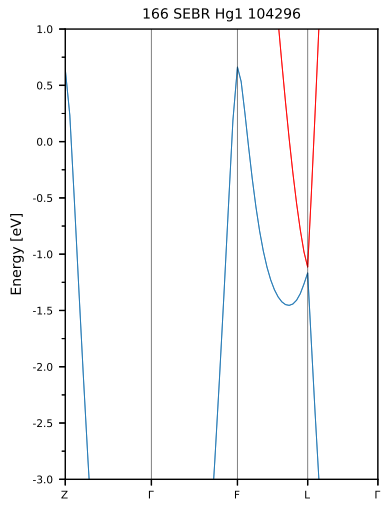

(ea) 166.1.131

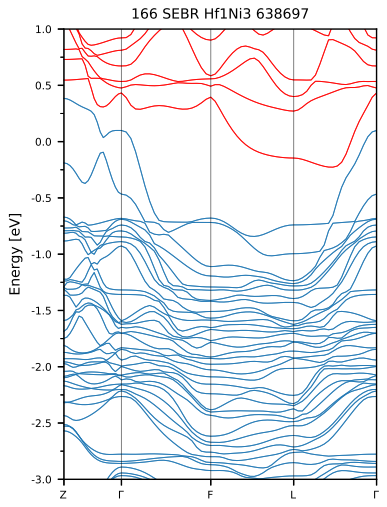

(eb) 166.1.132

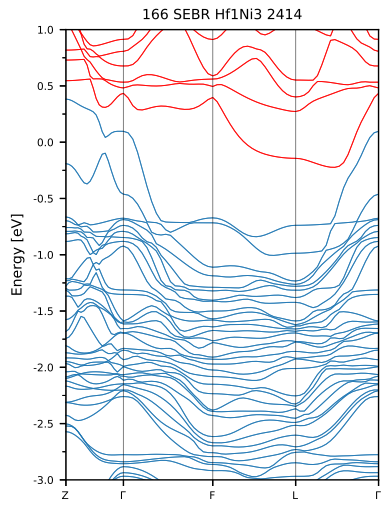

(ec) 166.1.133

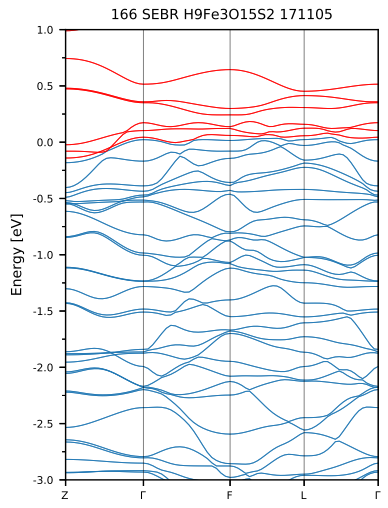

(ed) 166.1.134

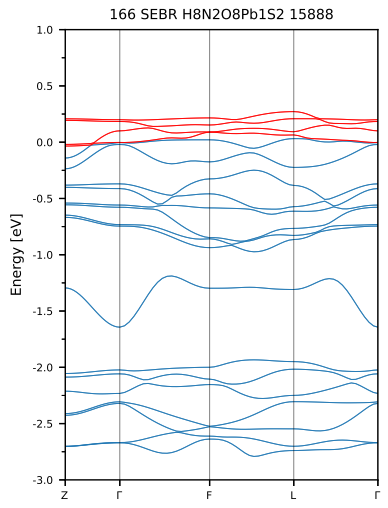

(ee) 166.1.135

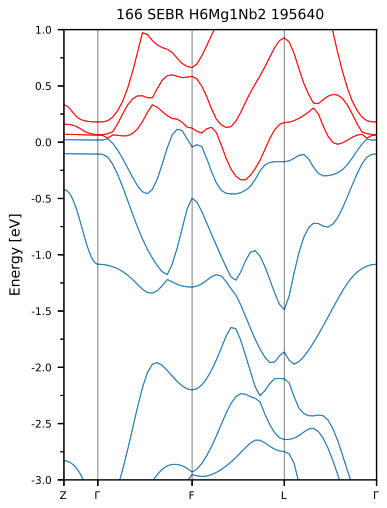

(ef) 166.1.136

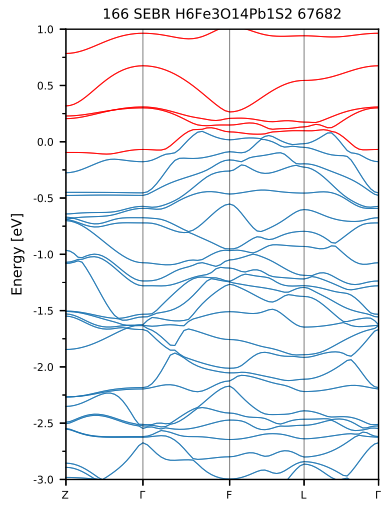

(eg) 166.1.137

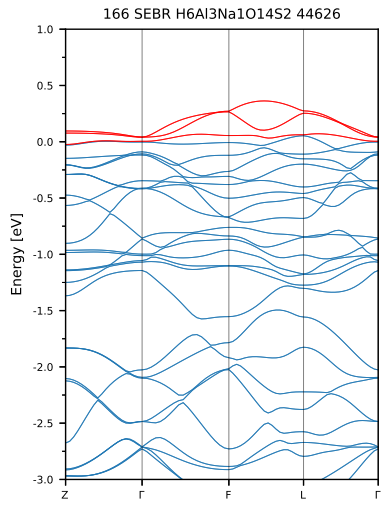

(eh) 166.1.138

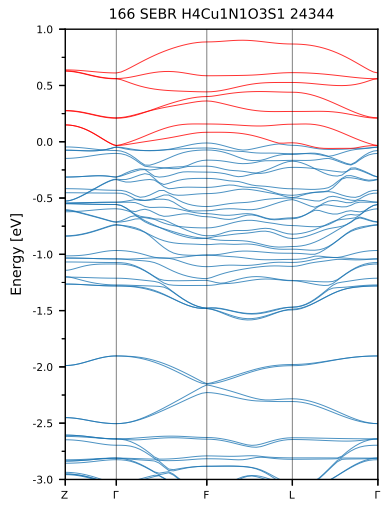

(ei) 166.1.139

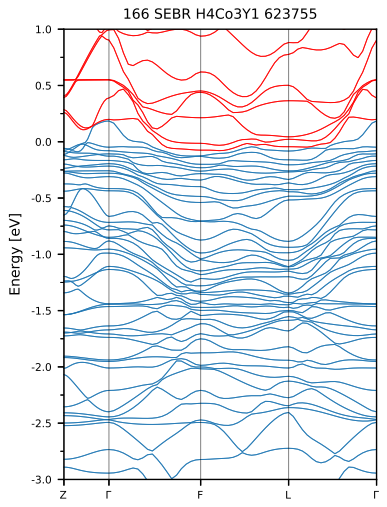

(ej) 166.1.140

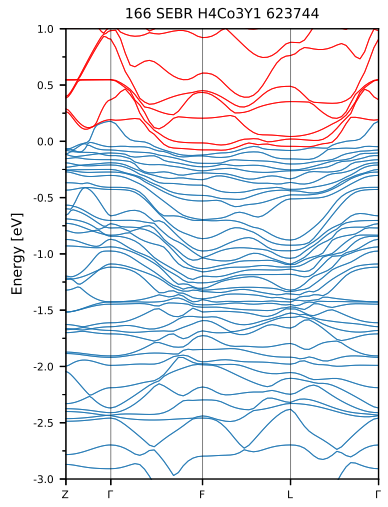

(ek) 166.1.141

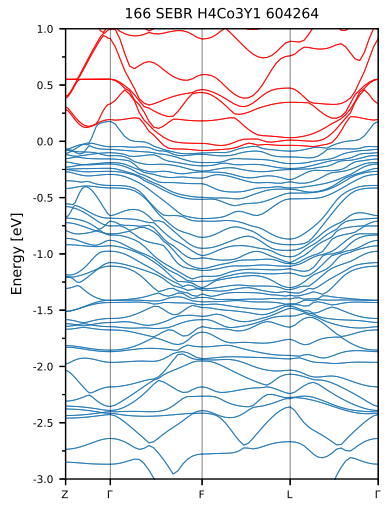

(el) 166.1.142

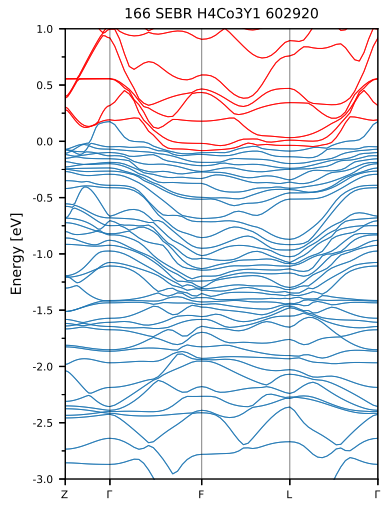

(em) 166.1.143

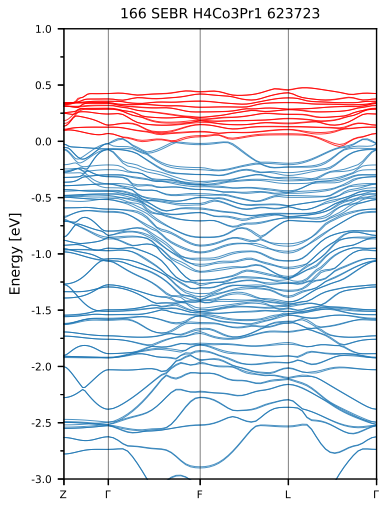

(en) 166.1.144

Figure 359: Band structures for topological materials in SG 166, part 9/39, from 166.1.129 to 166.1.144. See the list of figures.

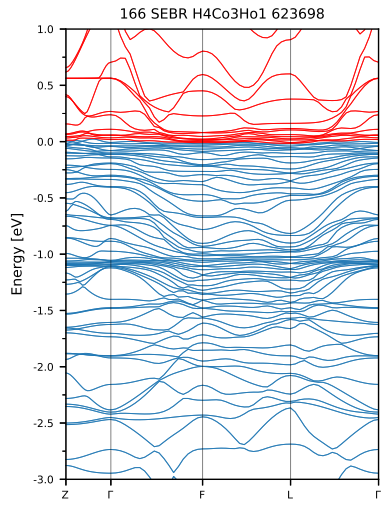

(eo) 166.1.145

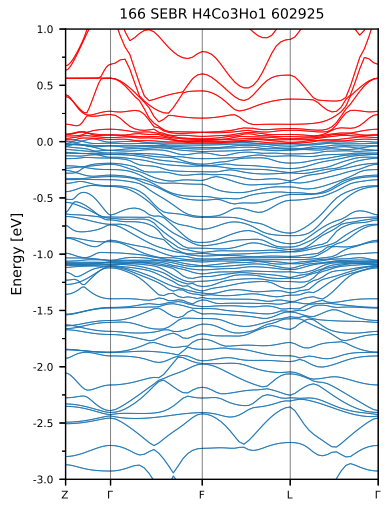

(ep) 166.1.146

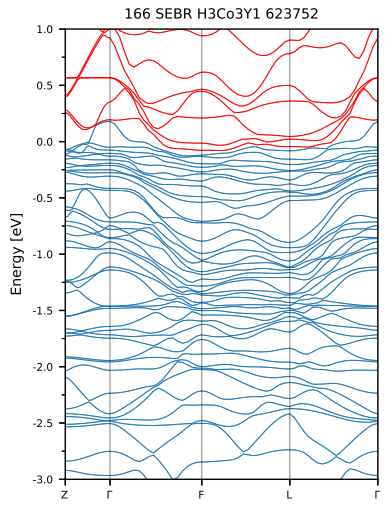

(eq) 166.1.147

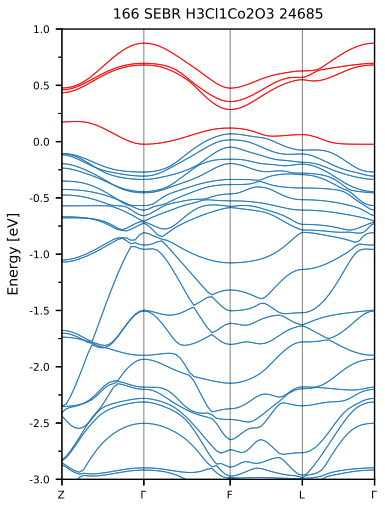

(er) 166.1.148

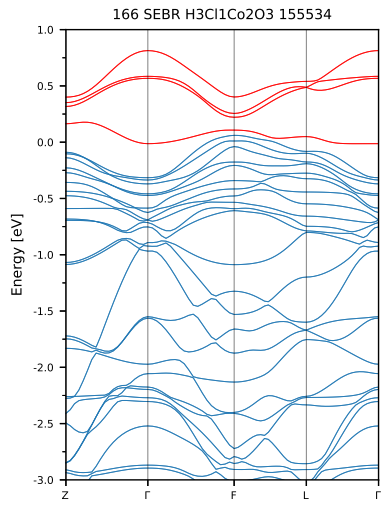

(es) 166.1.149

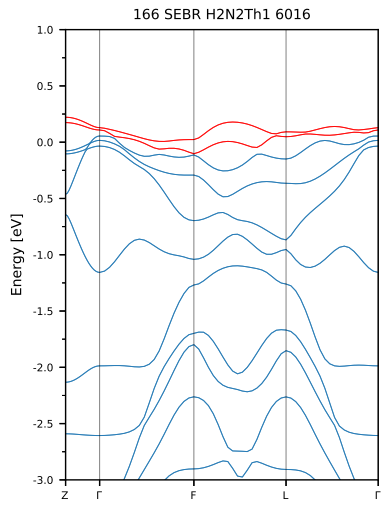

(et) 166.1.150

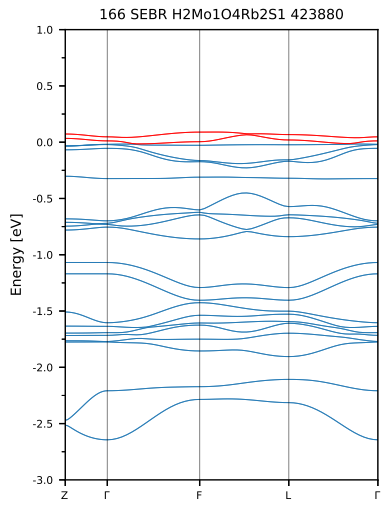

(eu) 166.1.151

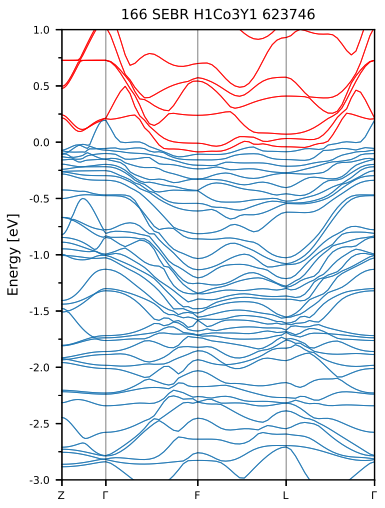

(ev) 166.1.152

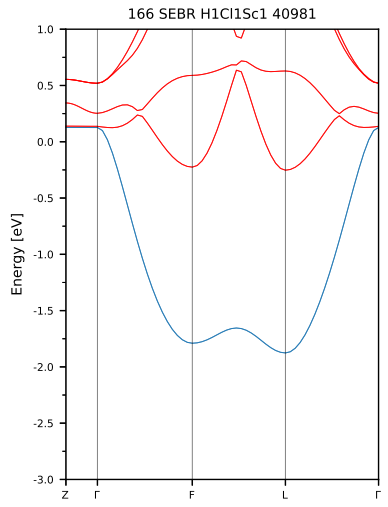

(ew) 166.1.153

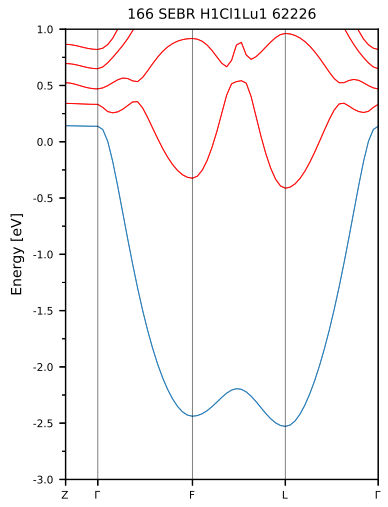

(ex) 166.1.154

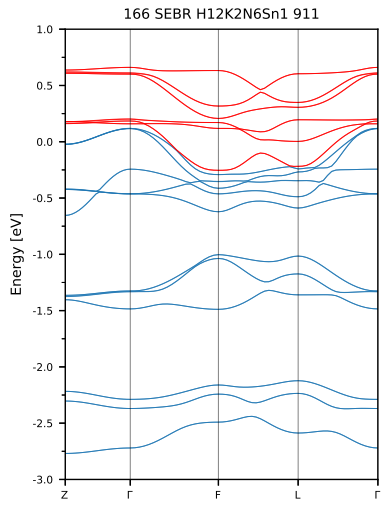

(ey) 166.1.155

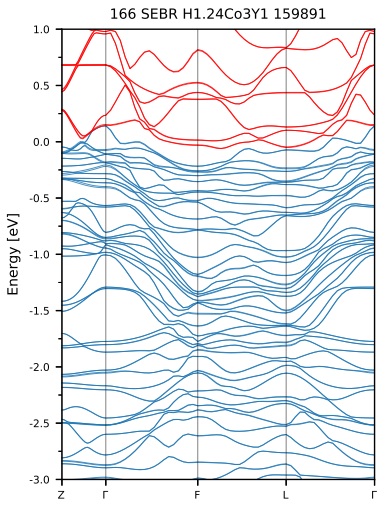

(ez) 166.1.156

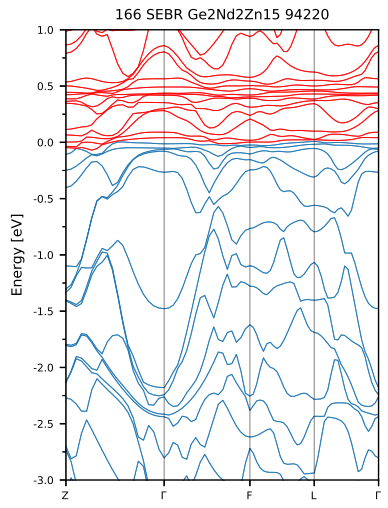

(fa) 166.1.157

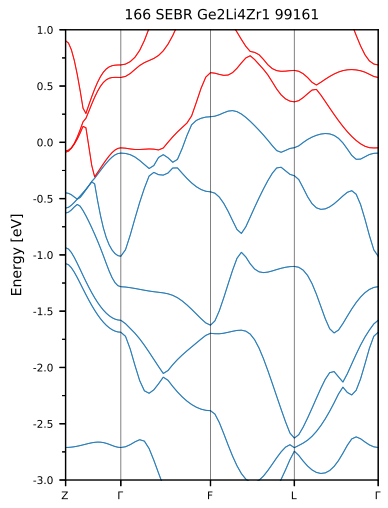

(fb) 166.1.158

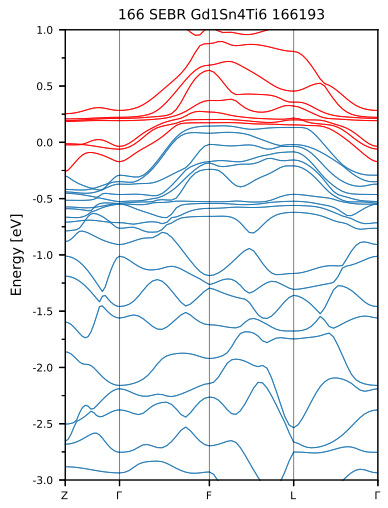

(fc) 166.1.159

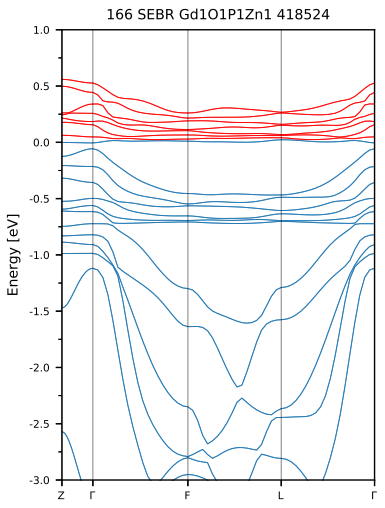

(fd) 166.1.160

Figure 360: Band structures for topological materials in SG 166, part 10/39, from 166.1.145 to 166.1.160. See the list of figures.

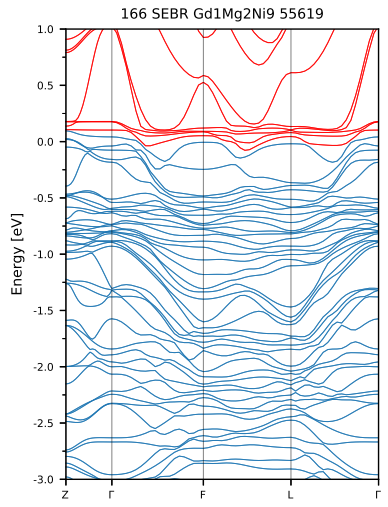

(fe) 166.1.161

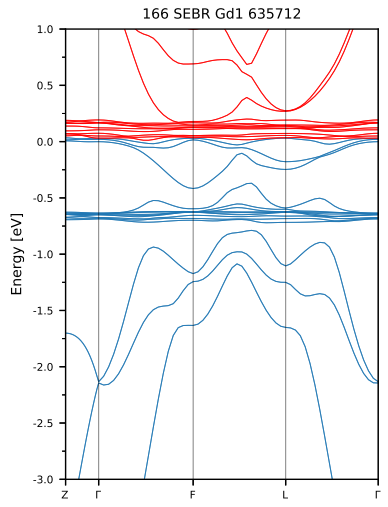

(ff) 166.1.162

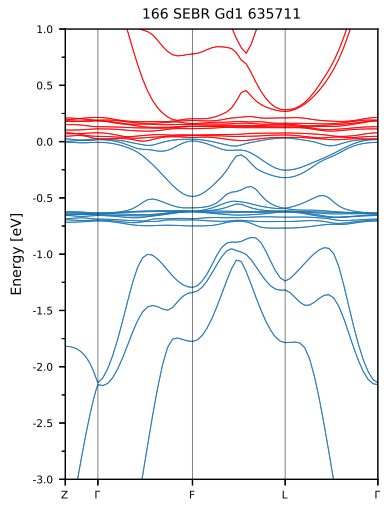

(fg) 166.1.163

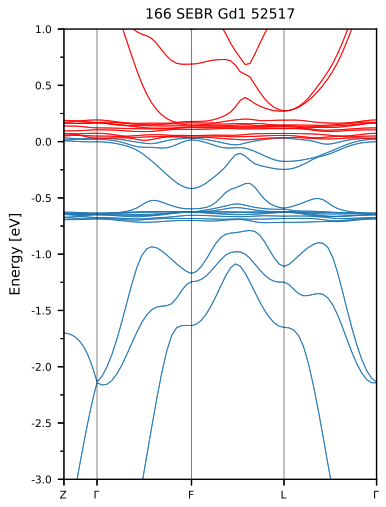

(fh) 166.1.164

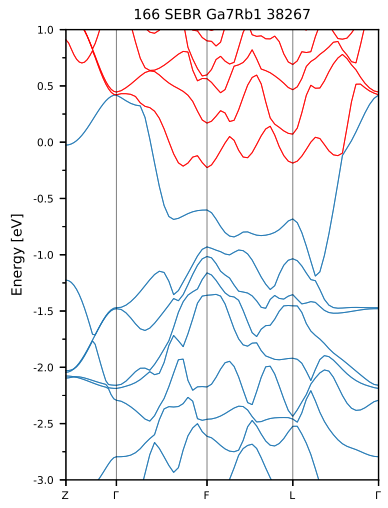

(fi) 166.1.165

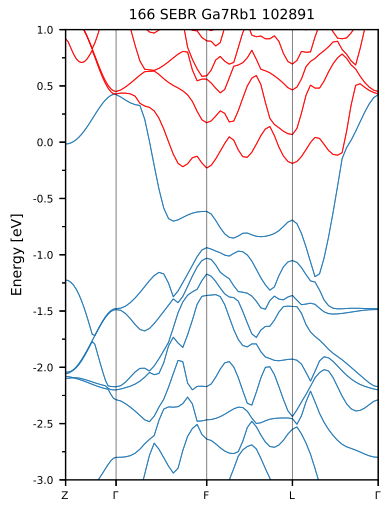

(fj) 166.1.166

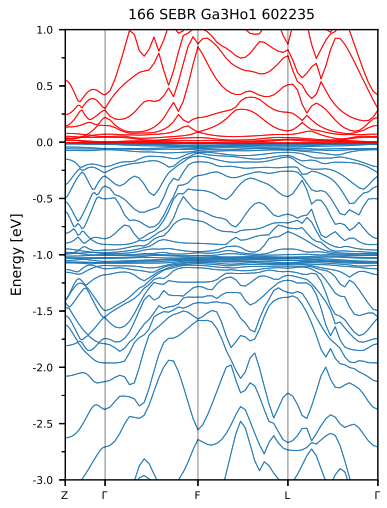

(fk) 166.1.167

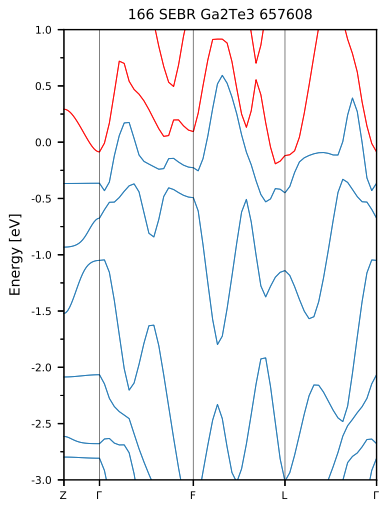

(fl) 166.1.168

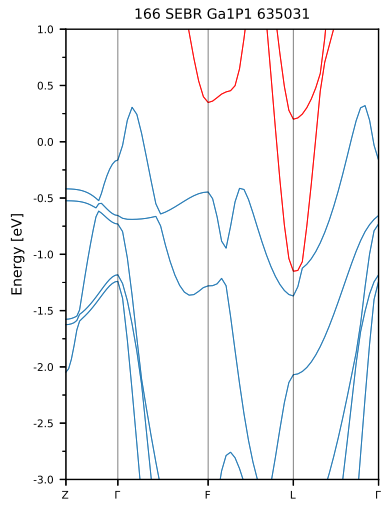

(fm) 166.1.169

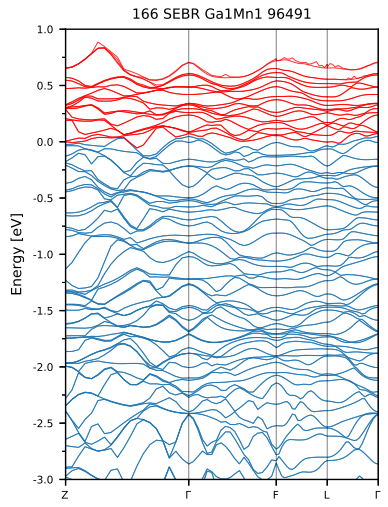

(fn) 166.1.170

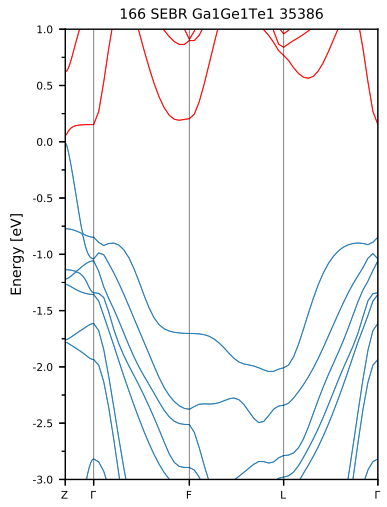

(fo) 166.1.171

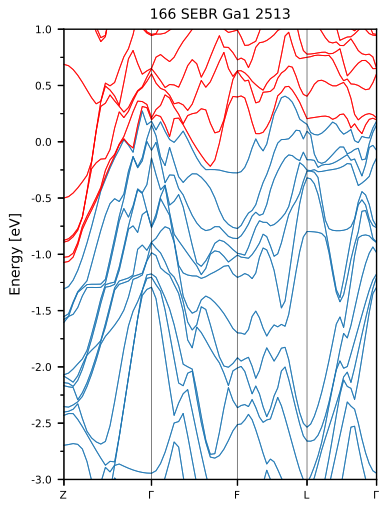

(fp) 166.1.172

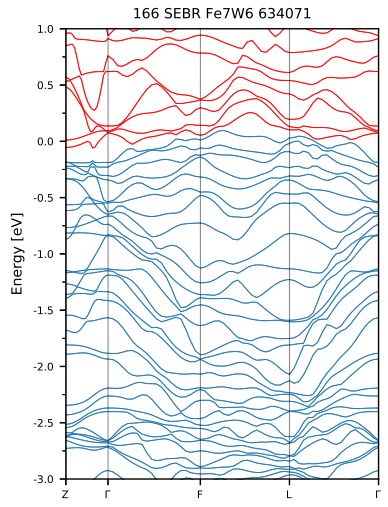

(fq) 166.1.173

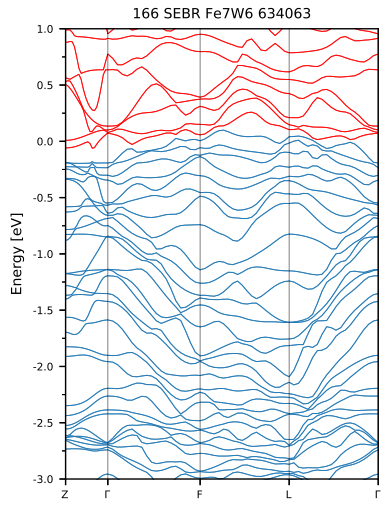

(fr) 166.1.174

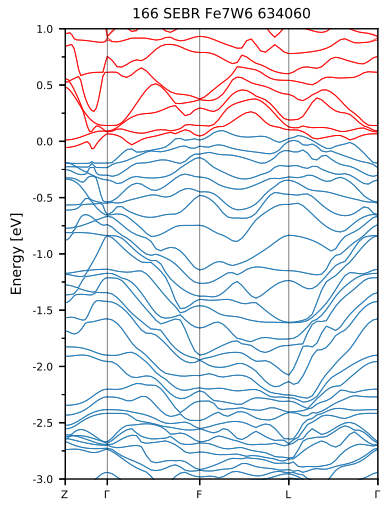

(fs) 166.1.175

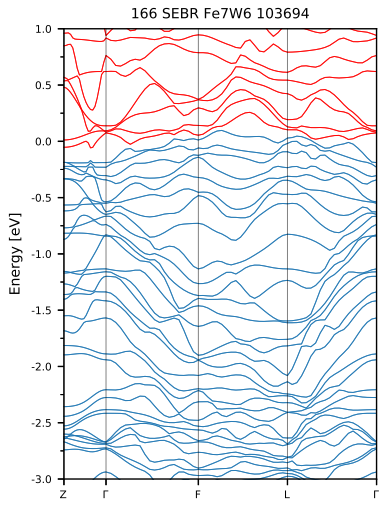

(ft) 166.1.176

Figure 361: Band structures for topological materials in SG 166, part 11/39, from 166.1.161 to 166.1.176. See the list of figures.

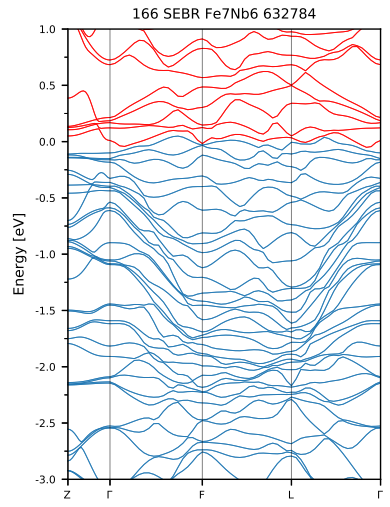

(fu) 166.1.177

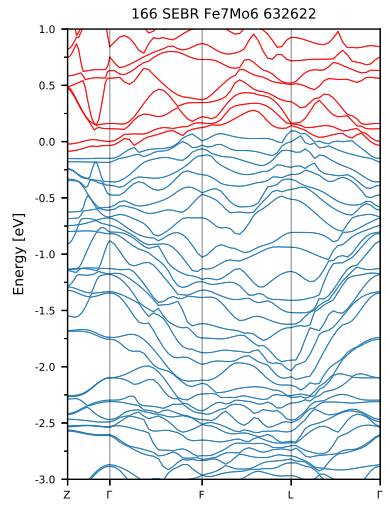

(fv) 166.1.178

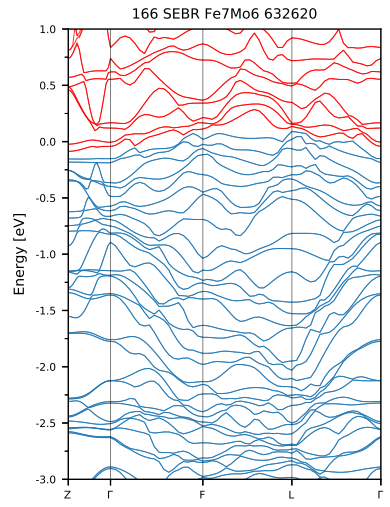

(fw) 166.1.179

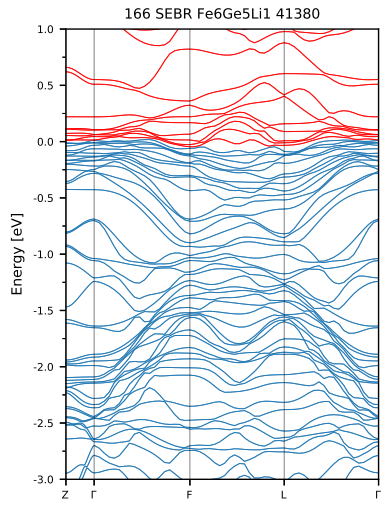

(fx) 166.1.180

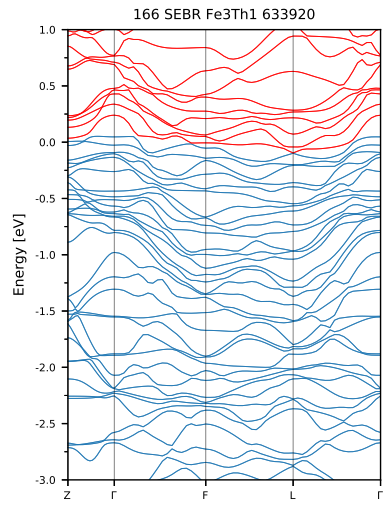

(fy) 166.1.181

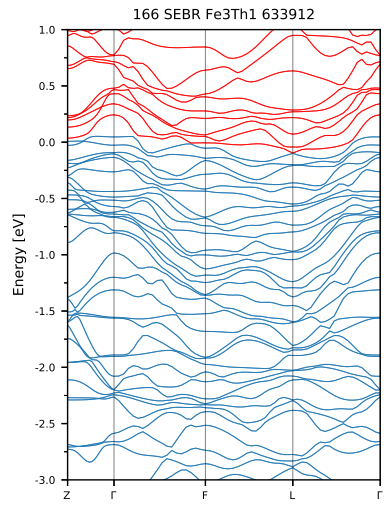

(fz) 166.1.182

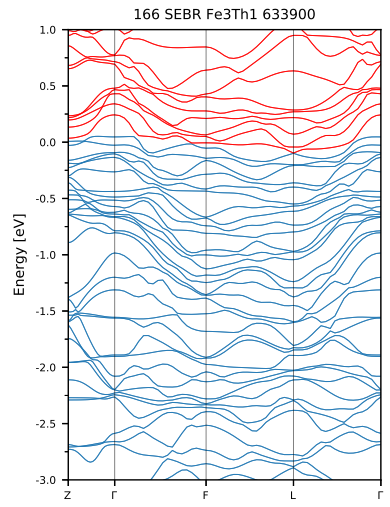

(ga) 166.1.183

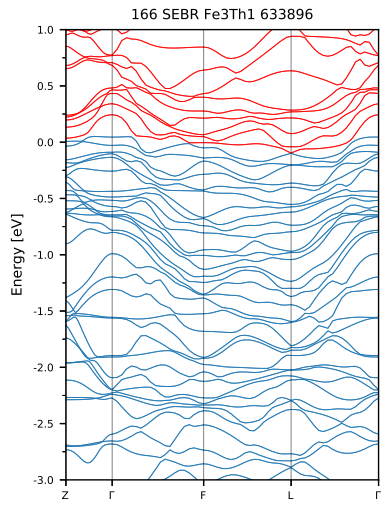

(gb) 166.1.184

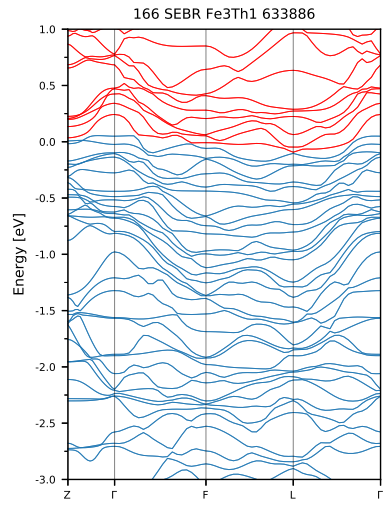

(gc) 166.1.185

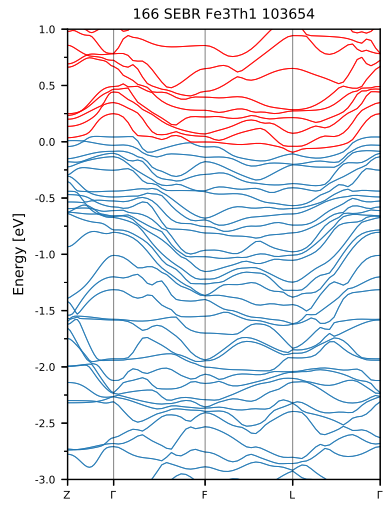

(gd) 166.1.186

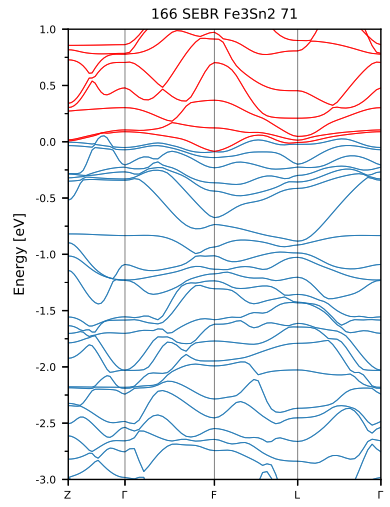

(ge) 166.1.187

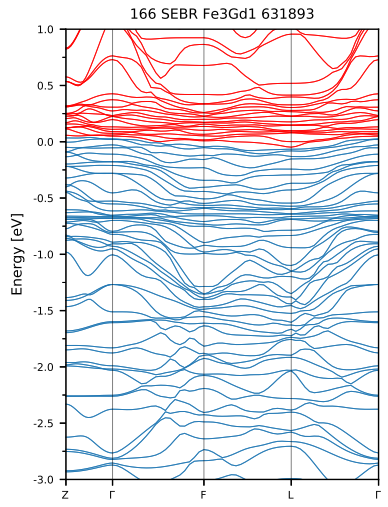

(gf) 166.1.188

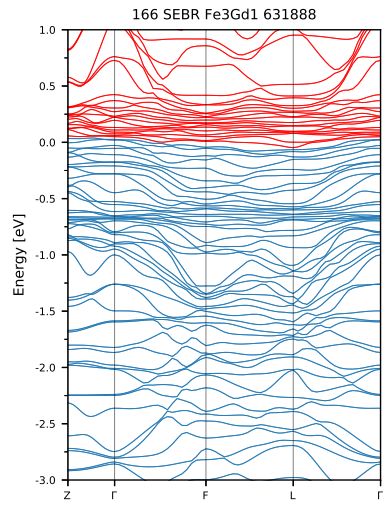

(gg) 166.1.189

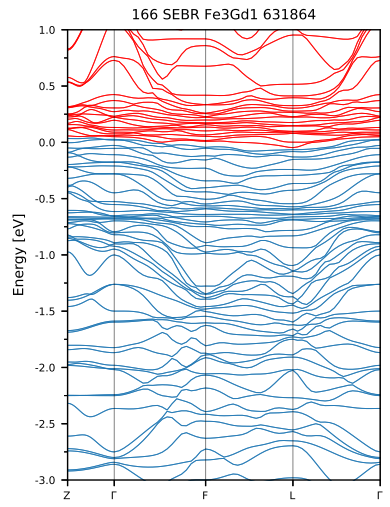

(gh) 166.1.190

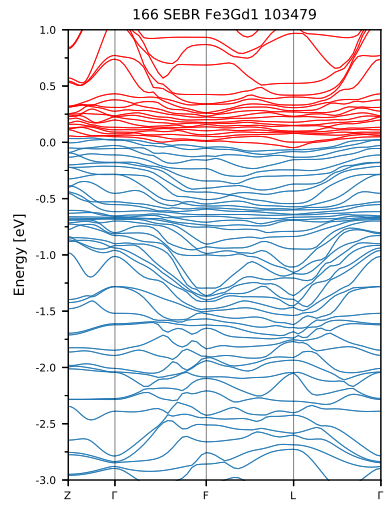

(gi) 166.1.191

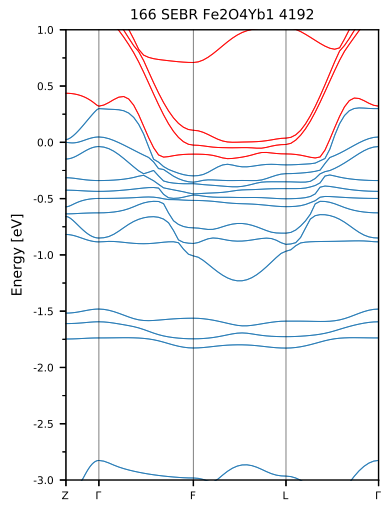

(gj) 166.1.192

Figure 362: Band structures for topological materials in SG 166, part 12/39, from 166.1.177 to 166.1.192. See the list of figures.

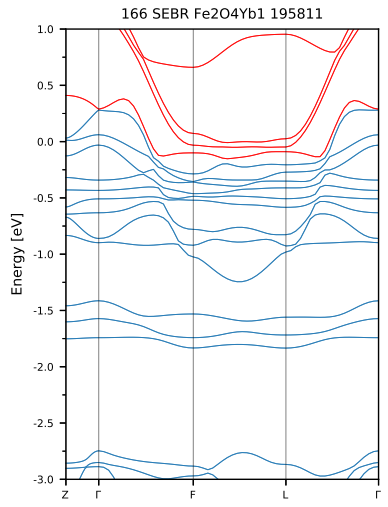

(gk) 166.1.193

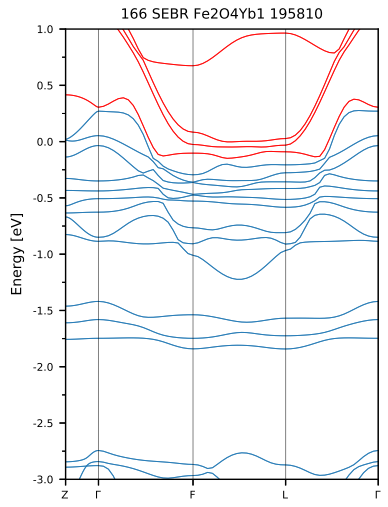

(gl) 166.1.194

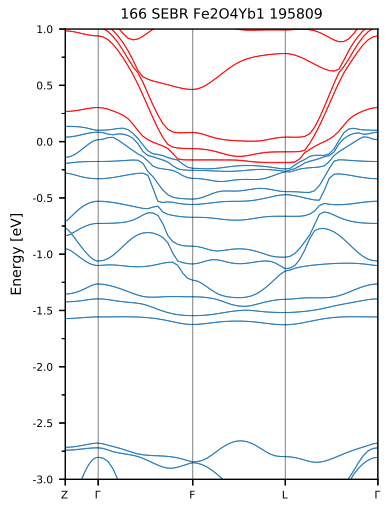

(gm) 166.1.195

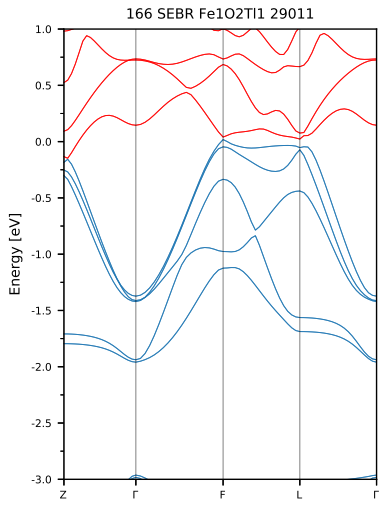

(gn) 166.1.196

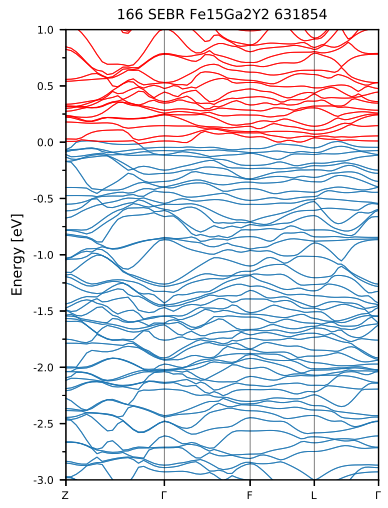

(go) 166.1.197

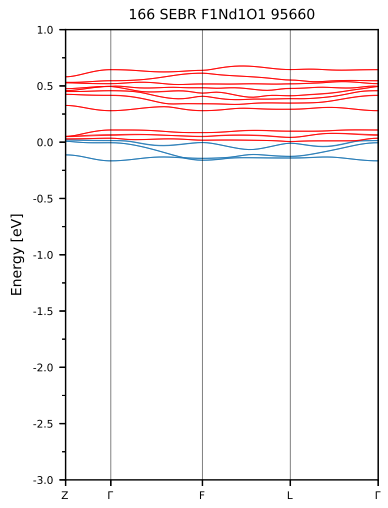

(gp) 166.1.198

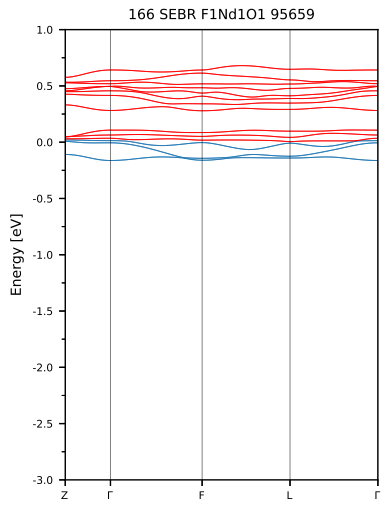

(gq) 166.1.199

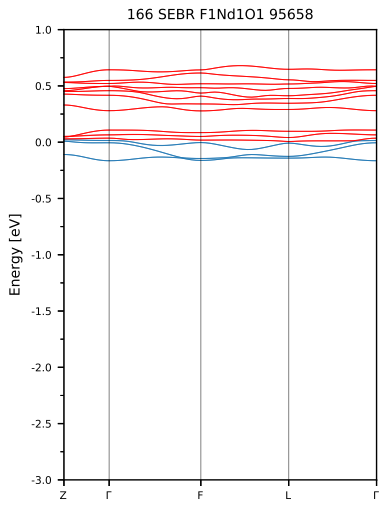

(gr) 166.1.200

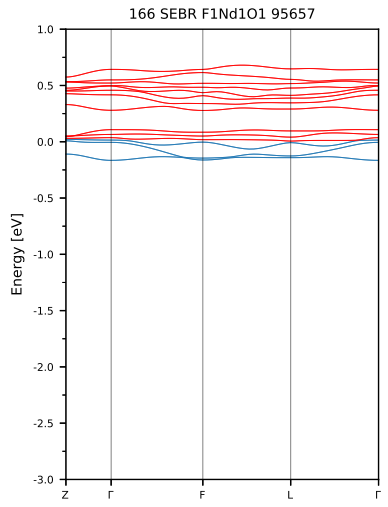

(gs) 166.1.201

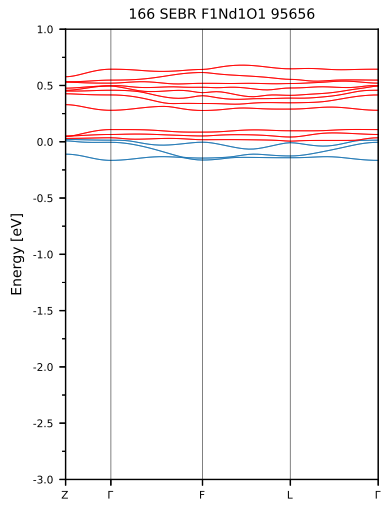

(gt) 166.1.202

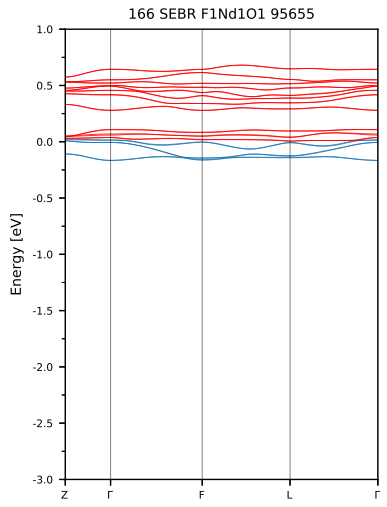

(gu) 166.1.203

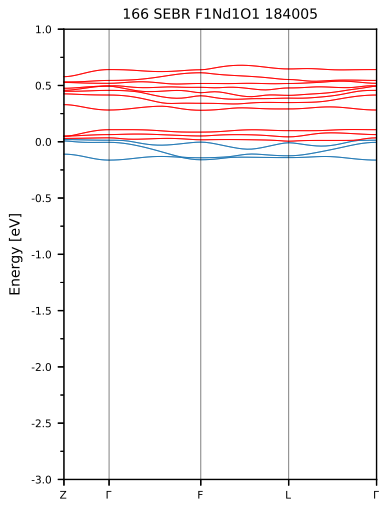

(gv) 166.1.204

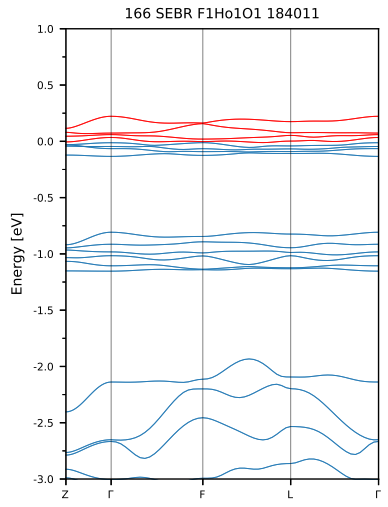

(gw) 166.1.205

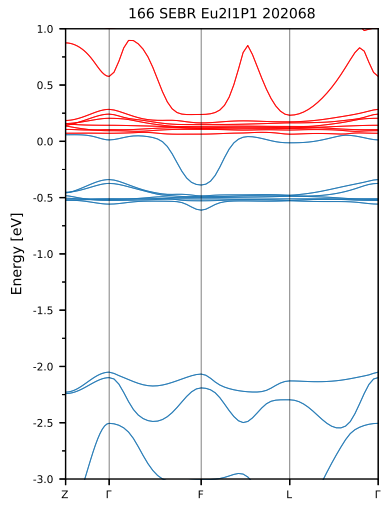

(gx) 166.1.206

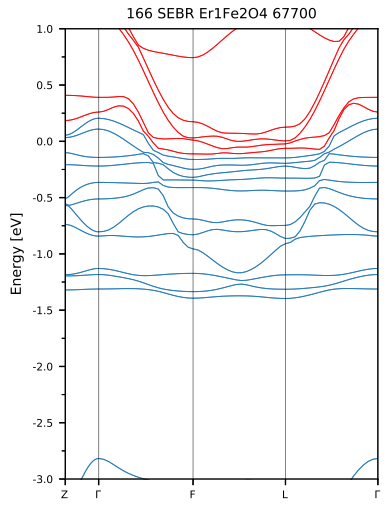

(gy) 166.1.207

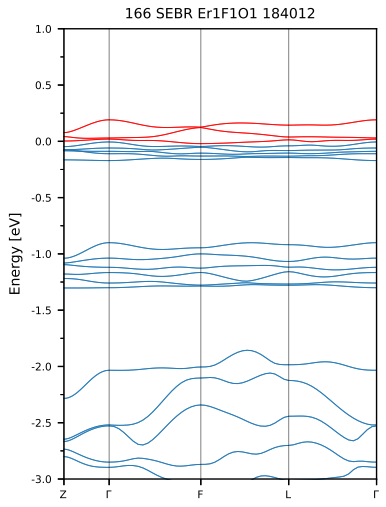

(gz) 166.1.208

Figure 363: Band structures for topological materials in SG 166, part 13/39, from 166.1.193 to 166.1.208. See the list of figures.

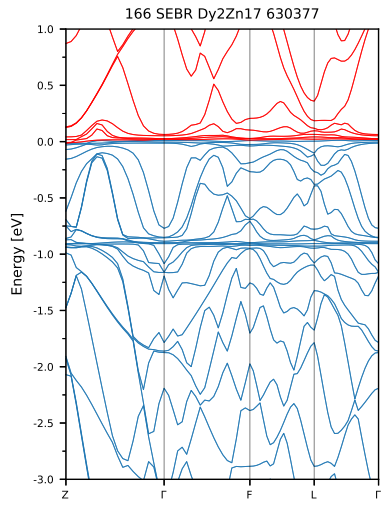

(ha) 166.1.209

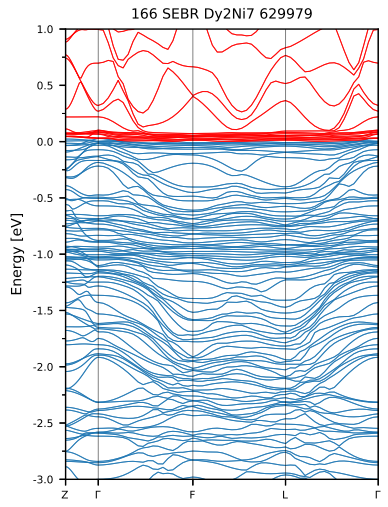

(hb) 166.1.210

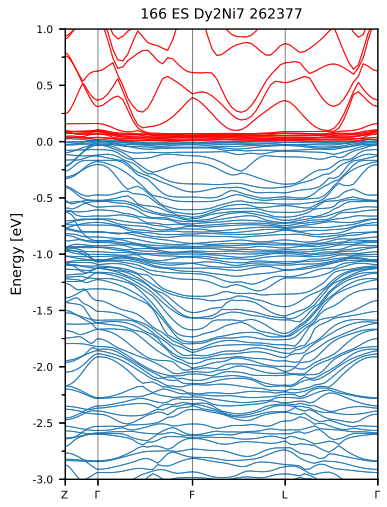

(hc) 166.1.211

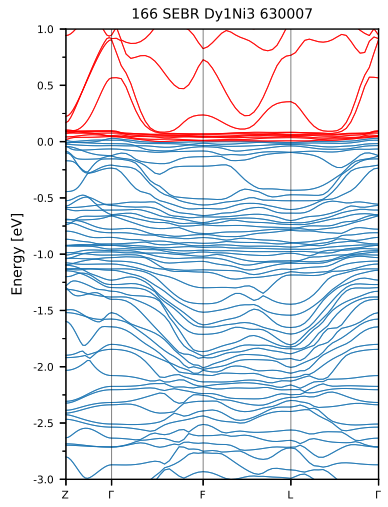

(hd) 166.1.212

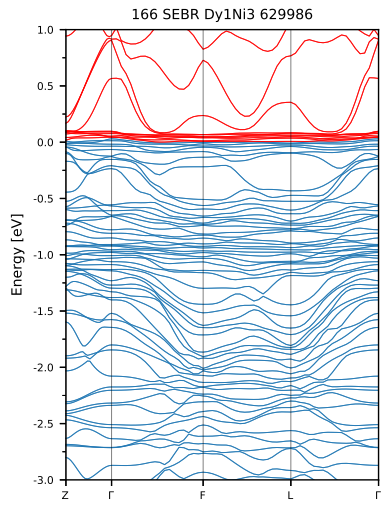

(he) 166.1.213

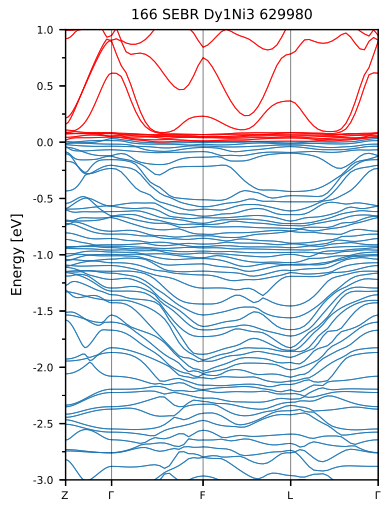

(hf) 166.1.214

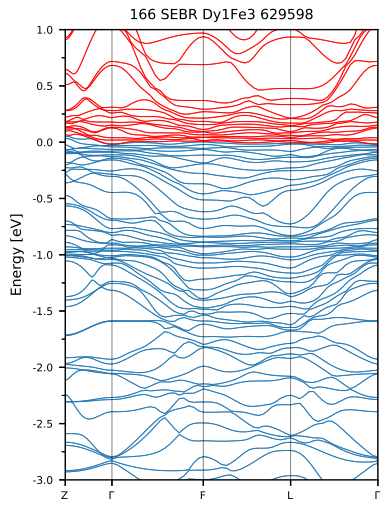

(hg) 166.1.215

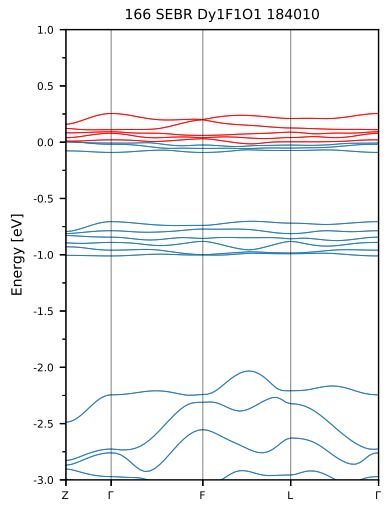

(hh) 166.1.216

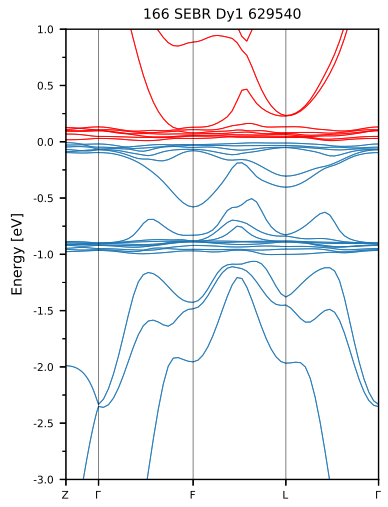

(hi) 166.1.217

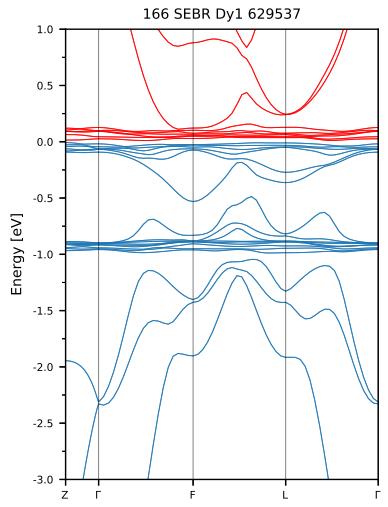

(hj) 166.1.218

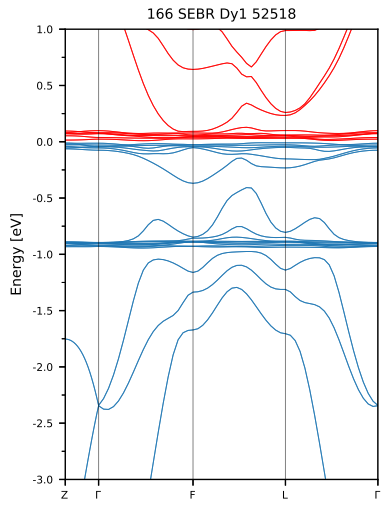

(hk) 166.1.219

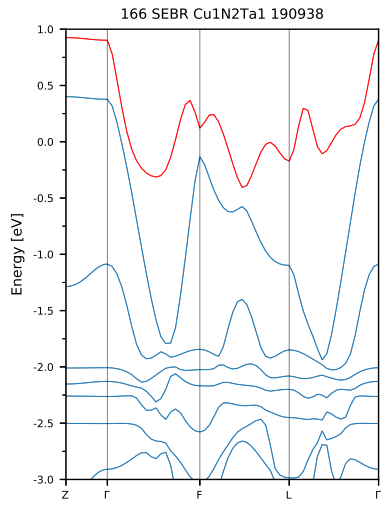

(hl) 166.1.220

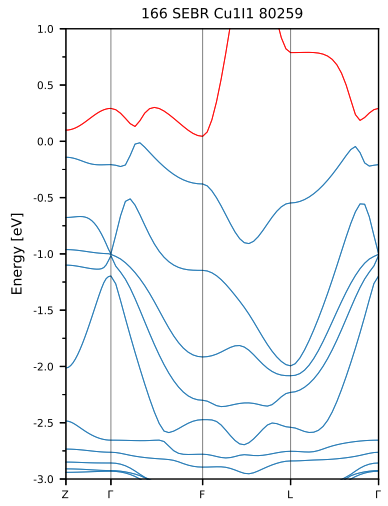

(hm) 166.1.221

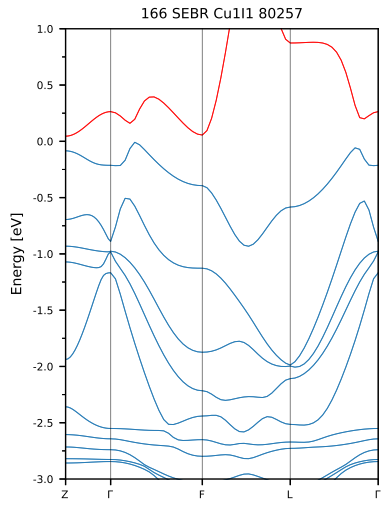

(hn) 166.1.222

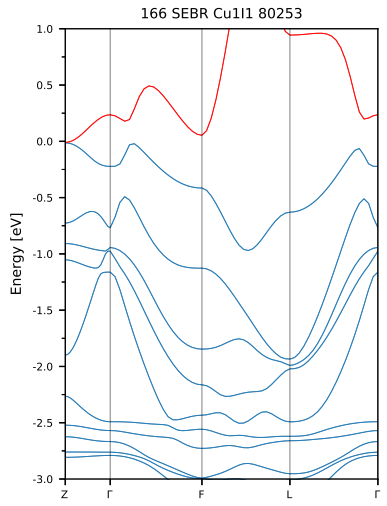

(ho) 166.1.223

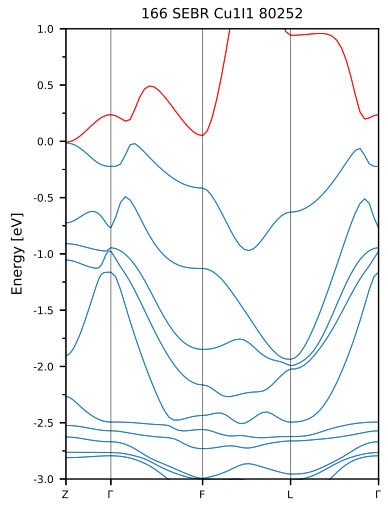

(hp) 166.1.224

Figure 364: Band structures for topological materials in SG 166, part 14/39, from 166.1.209 to 166.1.224. See the list of figures.

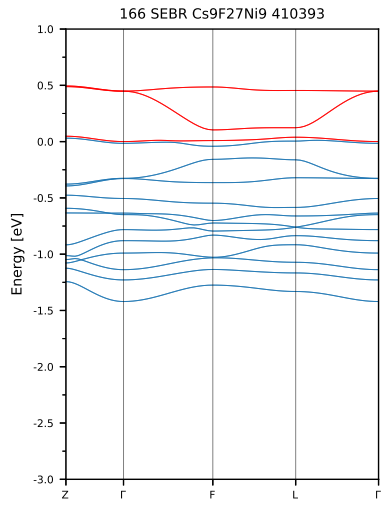

(hq) 166.1.225

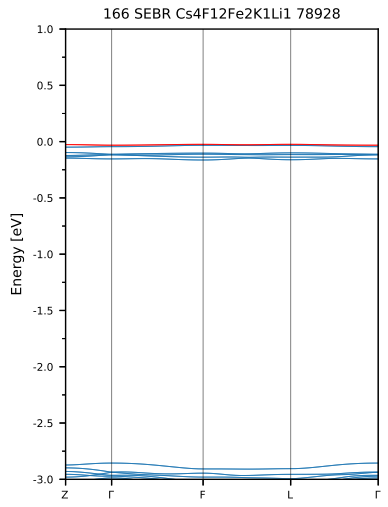

(hr) 166.1.226

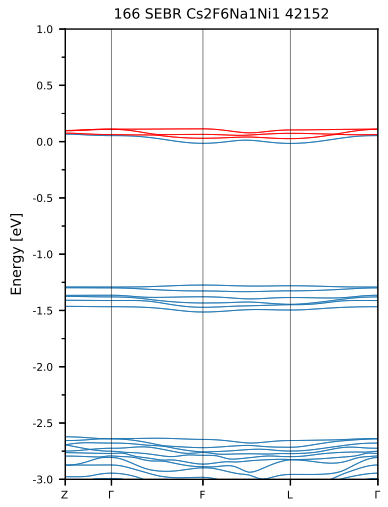

(hs) 166.1.227

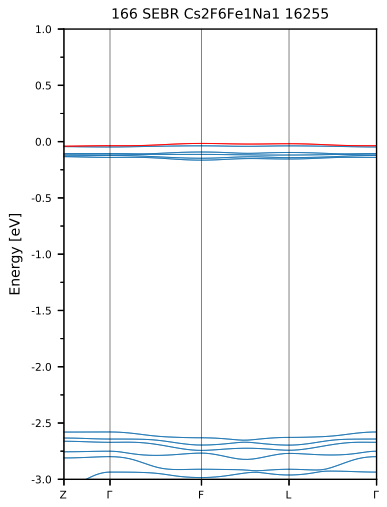

(ht) 166.1.228

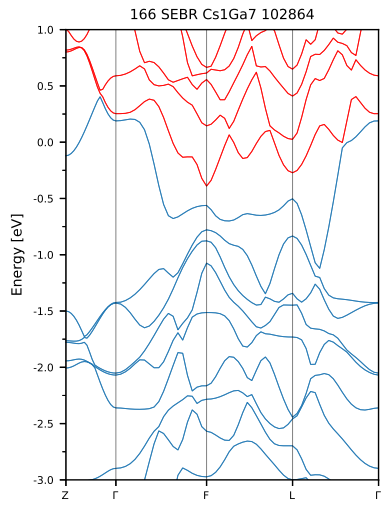

(hu) 166.1.229

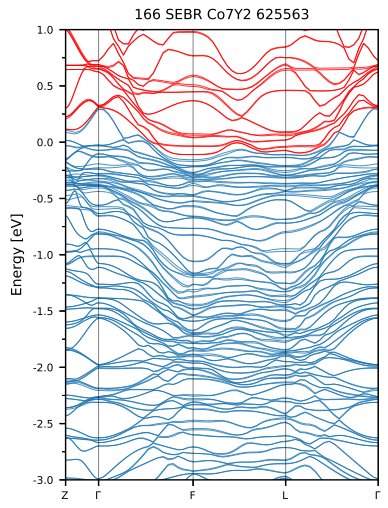

(hv) 166.1.230

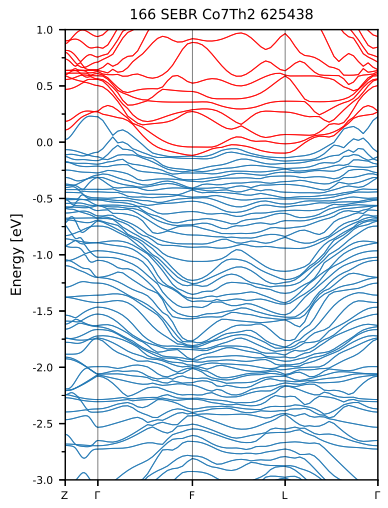

(hw) 166.1.231

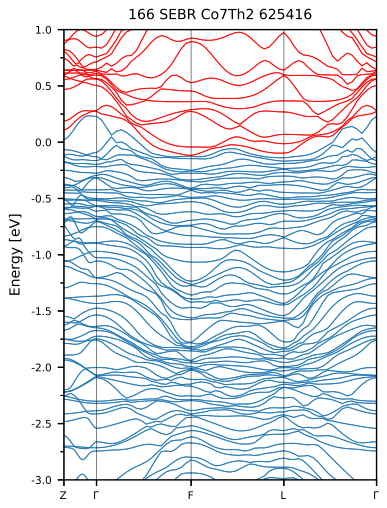

(hx) 166.1.232

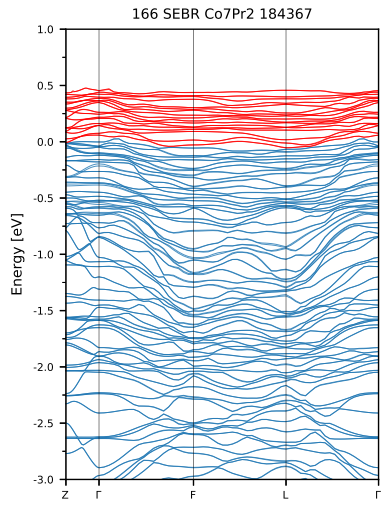

(hy) 166.1.233

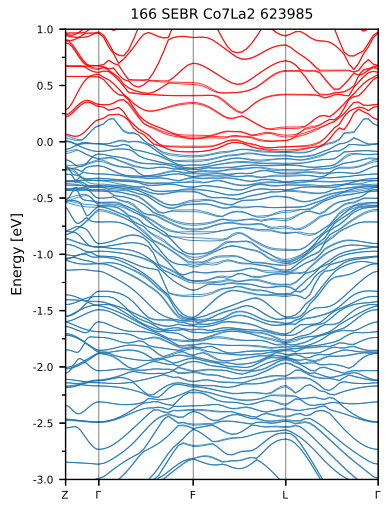

(hz) 166.1.234

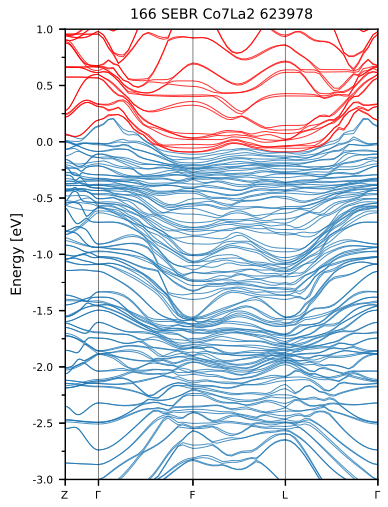

(ia) 166.1.235

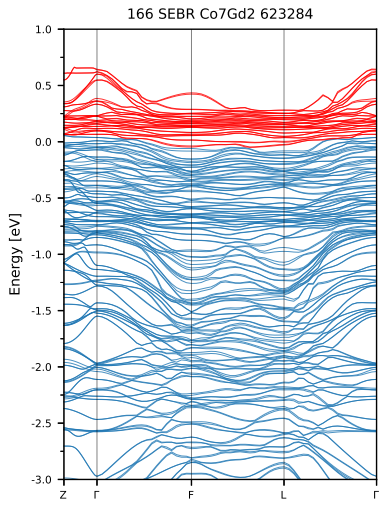

(ib) 166.1.236

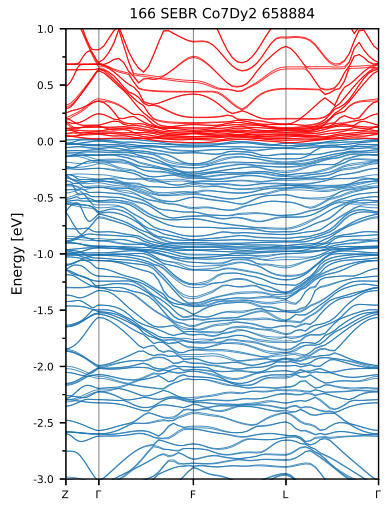

(ic) 166.1.237

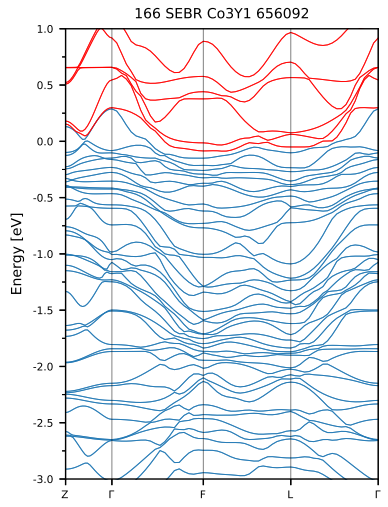

(id) 166.1.238

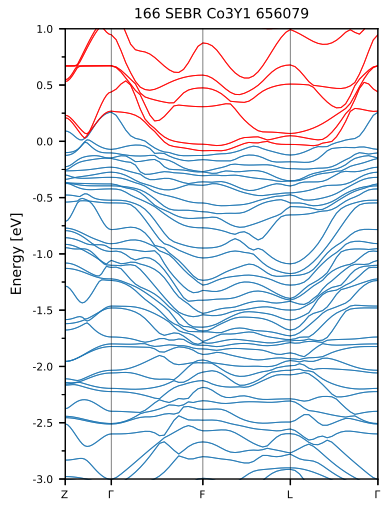

(ie) 166.1.239

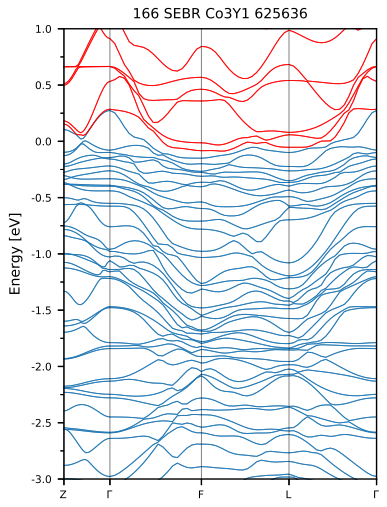

(if) 166.1.240

Figure 365: Band structures for topological materials in SG 166, part 15/39, from 166.1.225 to 166.1.240. See the list of figures.

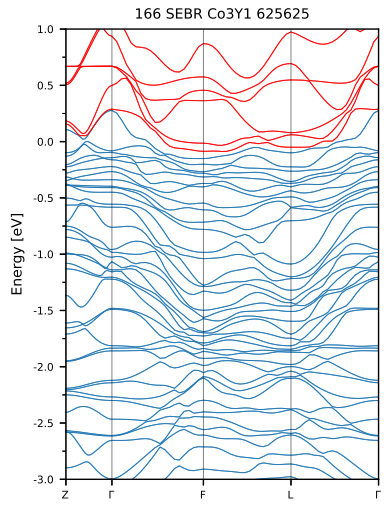

(ig) 166.1.241

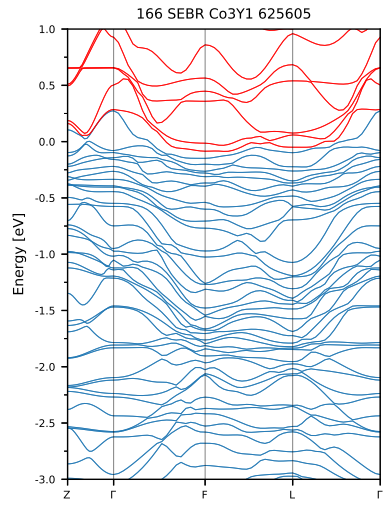

(ih) 166.1.242

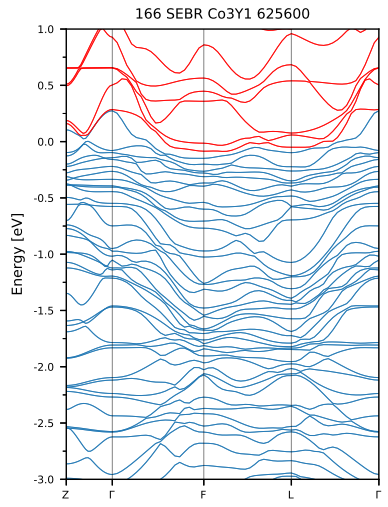

(ii) 166.1.243

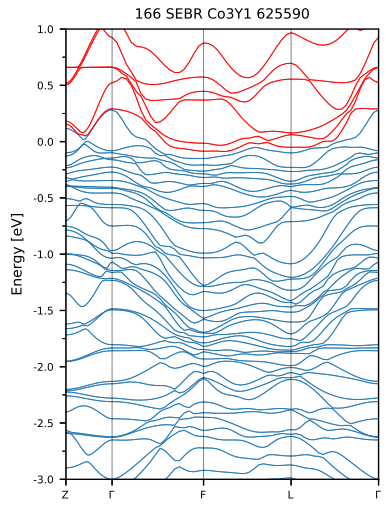

(ij) 166.1.244

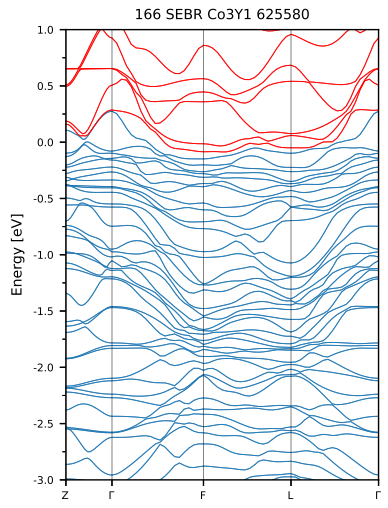

(ik) 166.1.245

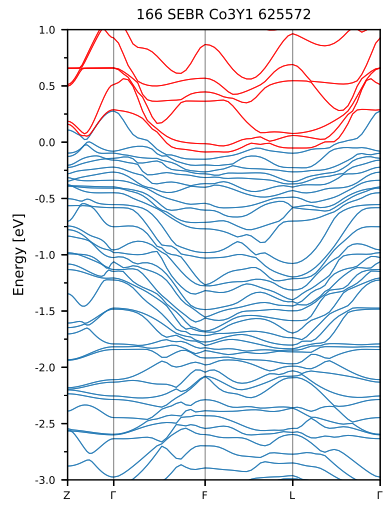

(il) 166.1.246

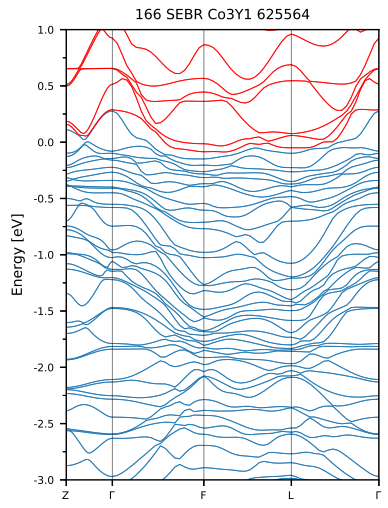

(im) 166.1.247

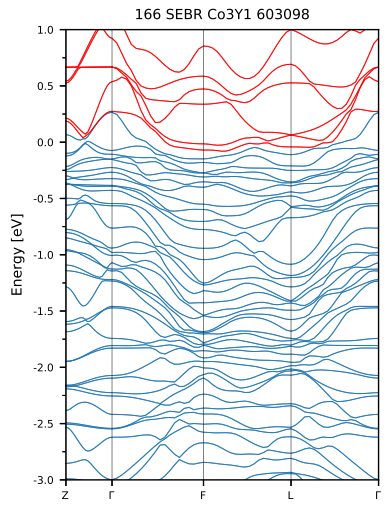

(in) 166.1.248

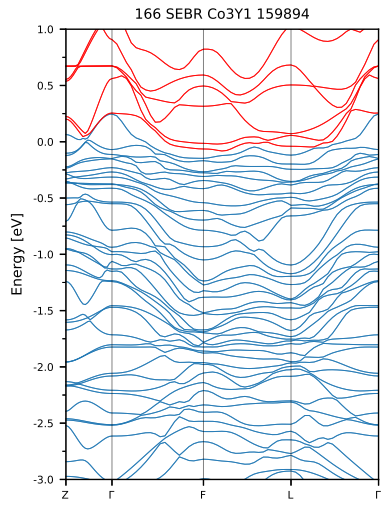

(io) 166.1.249

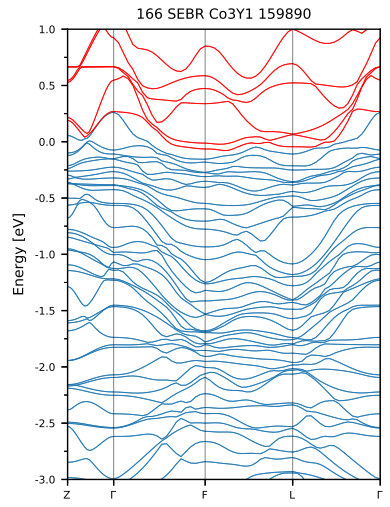

(ip) 166.1.250

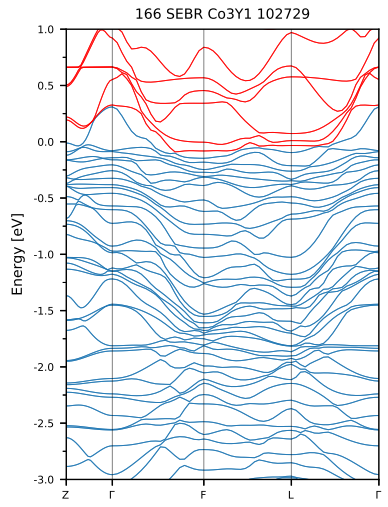

(iq) 166.1.251

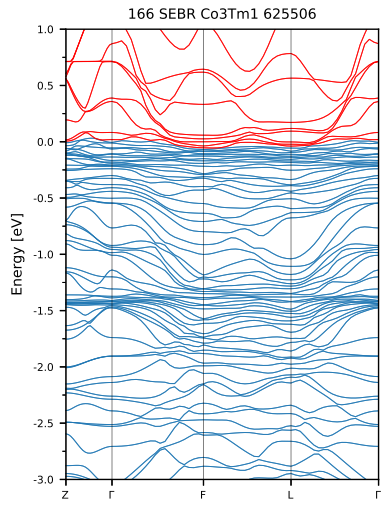

(ir) 166.1.252

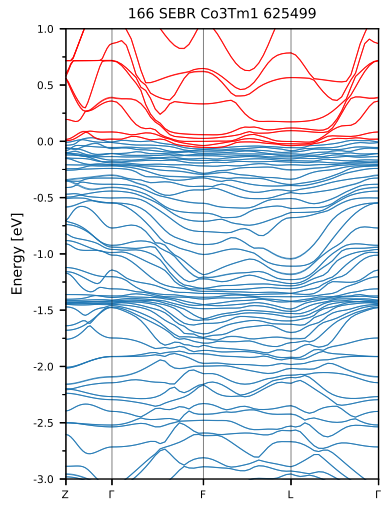

(is) 166.1.253

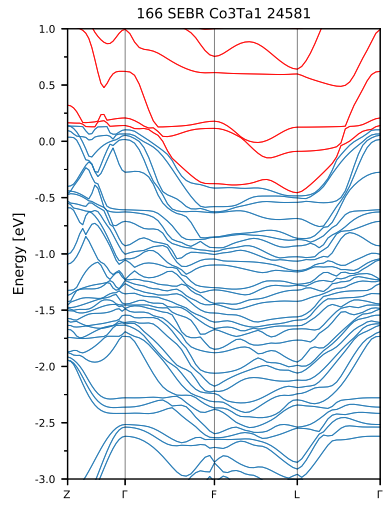

(it) 166.1.254

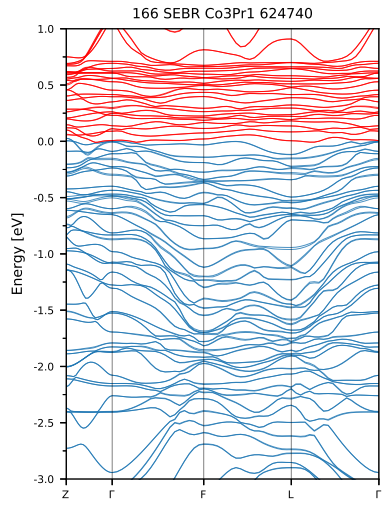

(iu) 166.1.255

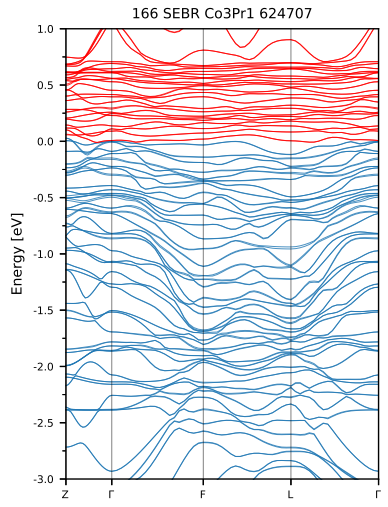

(iv) 166.1.256

Figure 366: Band structures for topological materials in SG 166, part 16/39, from 166.1.241 to 166.1.256. See the list of figures.

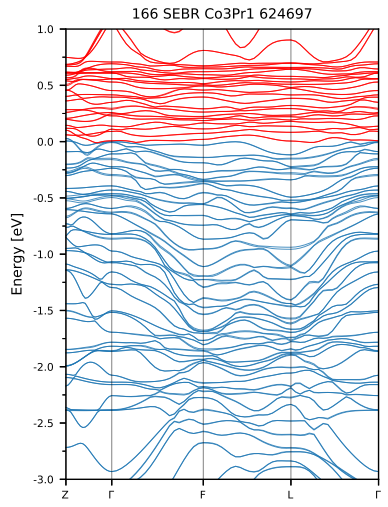

(iw) 166.1.257

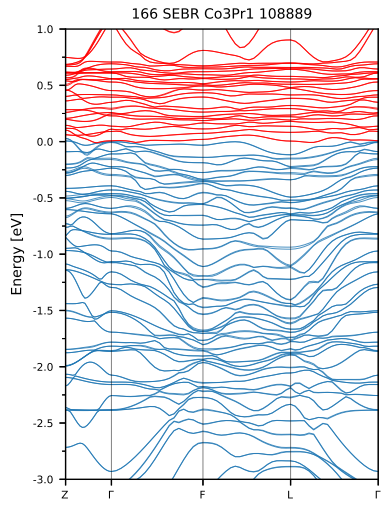

(ix) 166.1.258

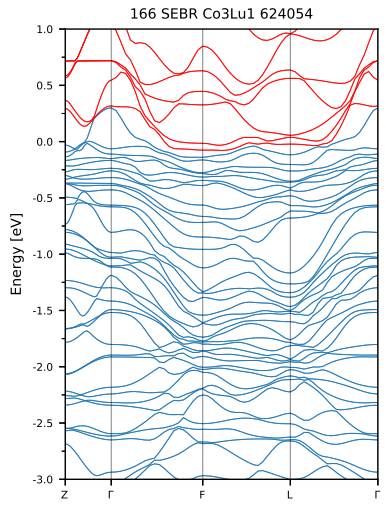

(iy) 166.1.259

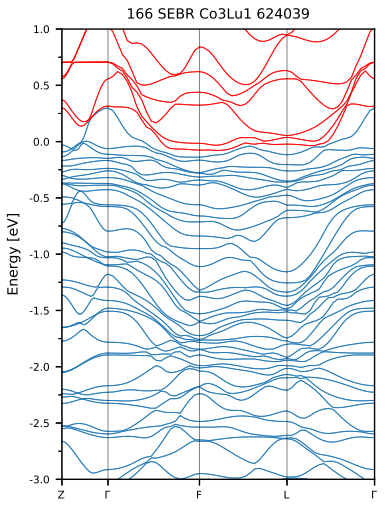

(iz) 166.1.260

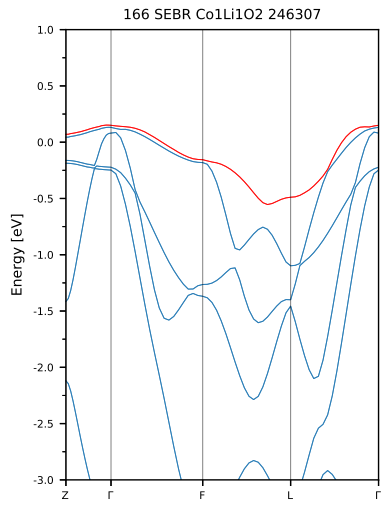

(ja) 166.1.261

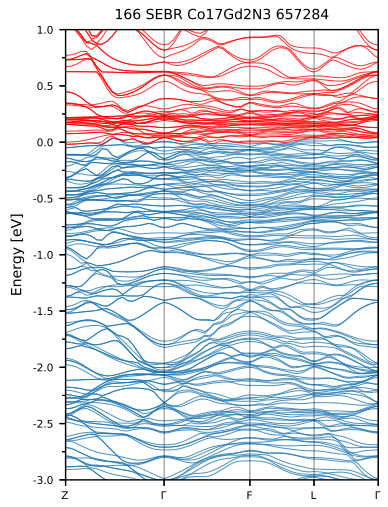

(jb) 166.1.262

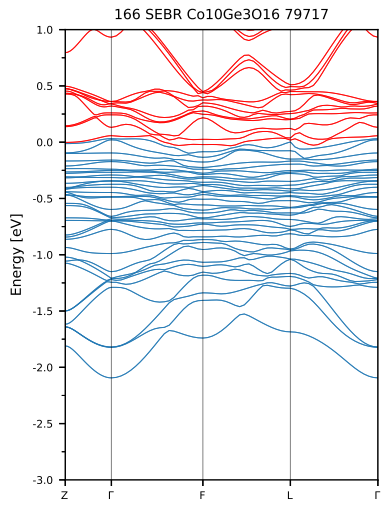

(jc) 166.1.263

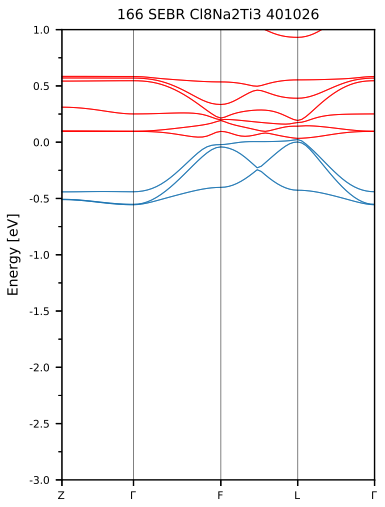

(jd) 166.1.264

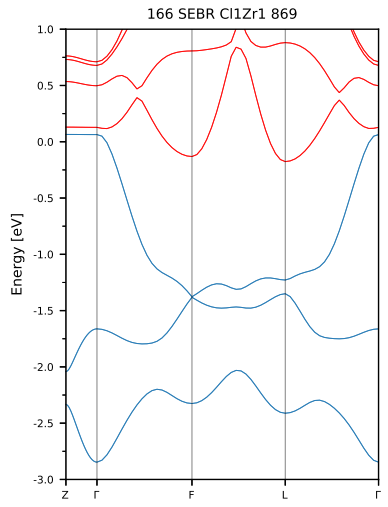

(je) 166.1.265

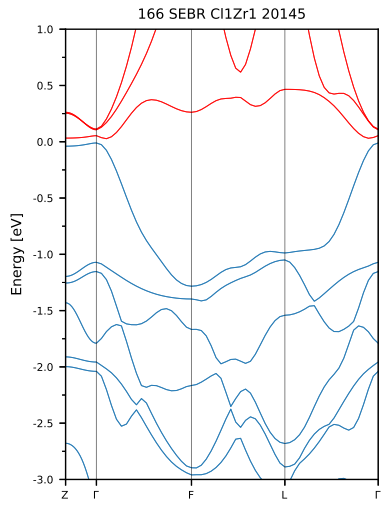

(jf) 166.1.266

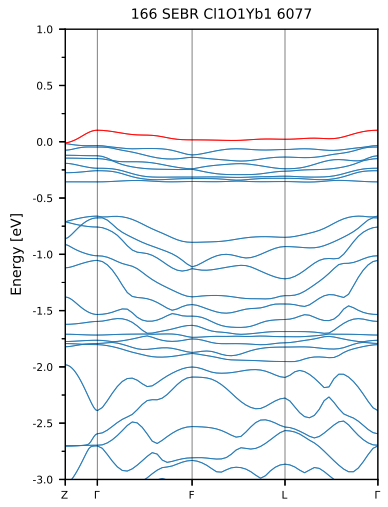

(jg) 166.1.267

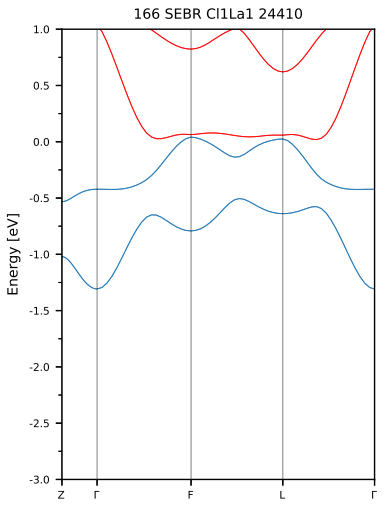

(jh) 166.1.268

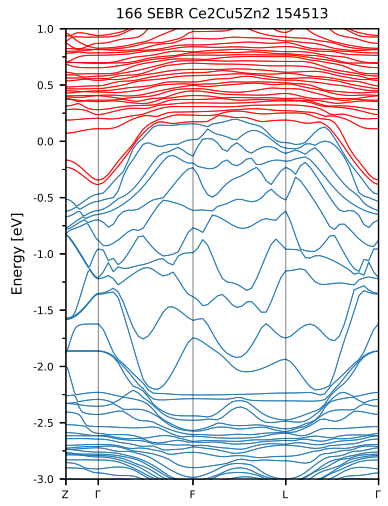

(ji) 166.1.269

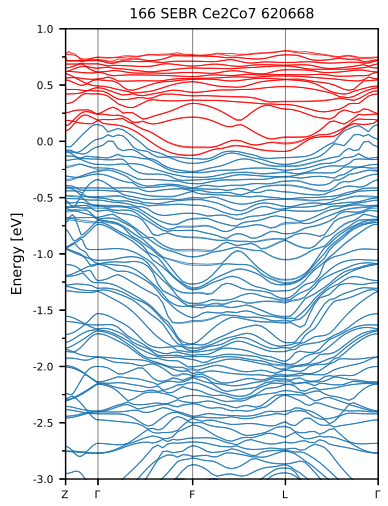

(jj) 166.1.270

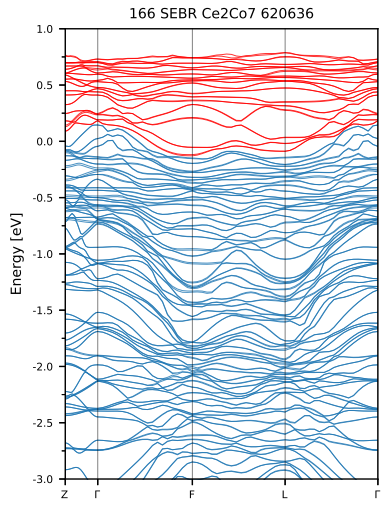

(jk) 166.1.271

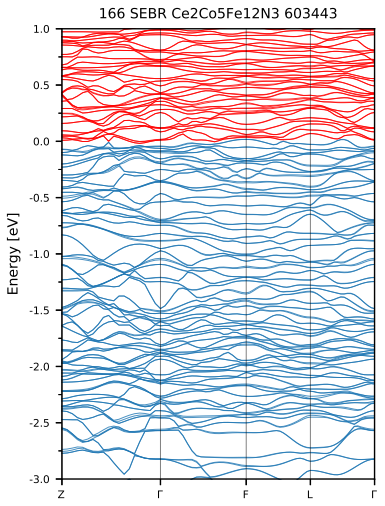

(jl) 166.1.272

Figure 367: Band structures for topological materials in SG 166, part 17/39, from 166.1.257 to 166.1.272. See the list of figures.

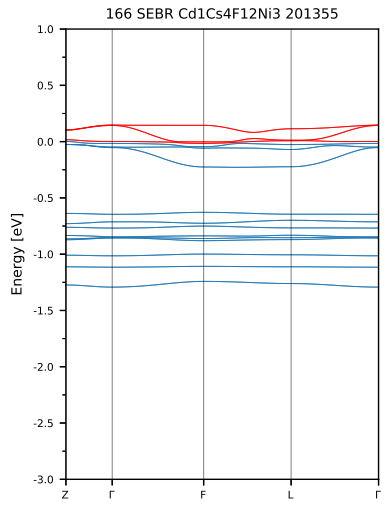

(jm) 166.1.273

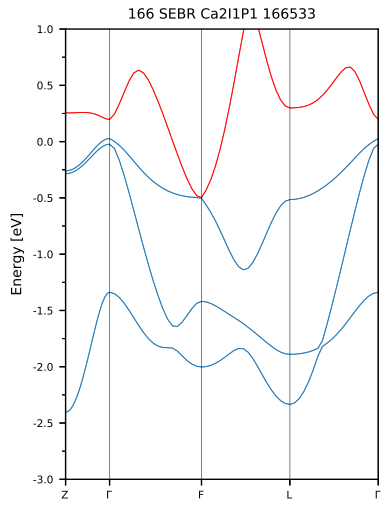

(jn) 166.1.274

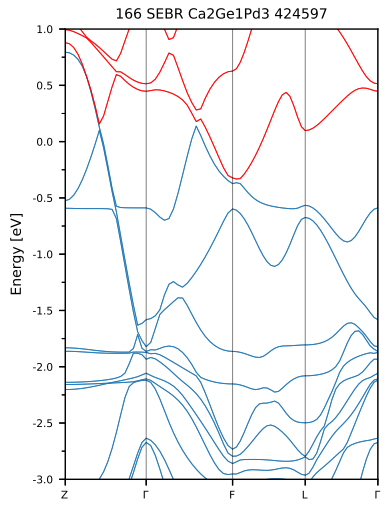

(jo) 166.1.275

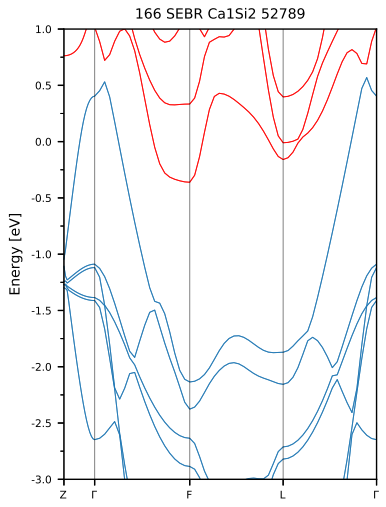

(jp) 166.1.276

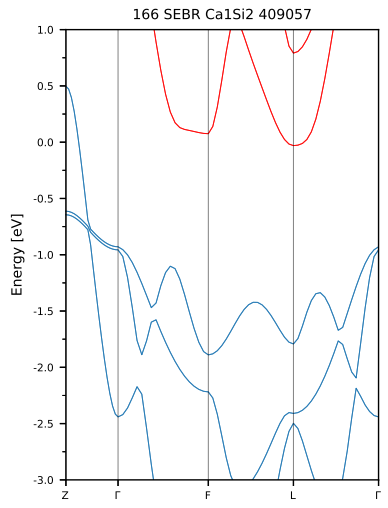

(jq) 166.1.277

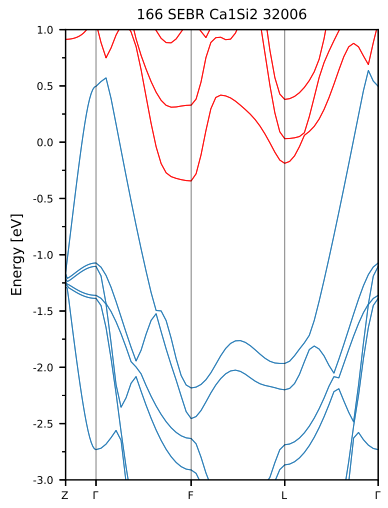

(jr) 166.1.278

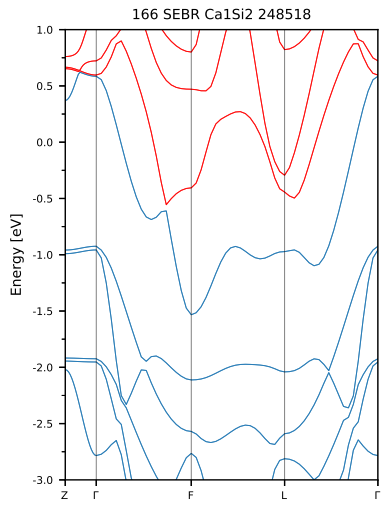

(js) 166.1.279

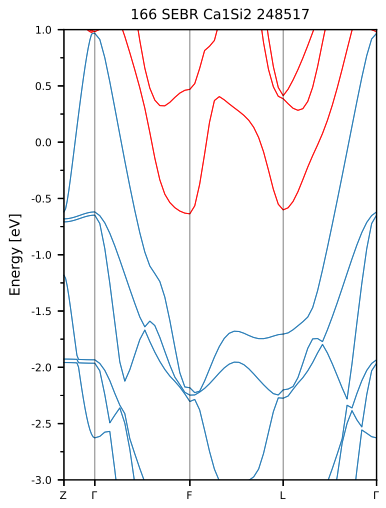

(jt) 166.1.280

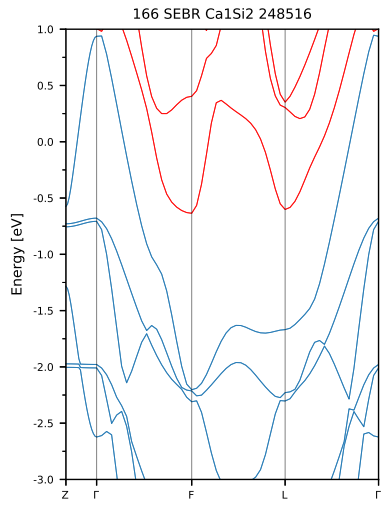

(ju) 166.1.281

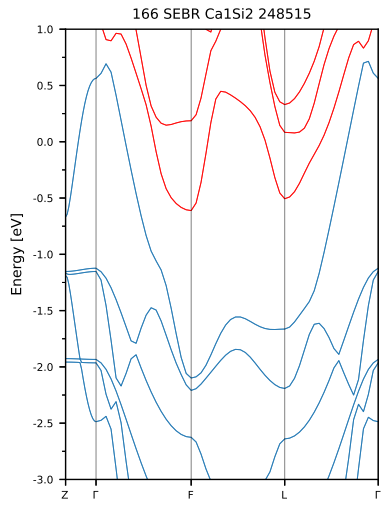

(jv) 166.1.282

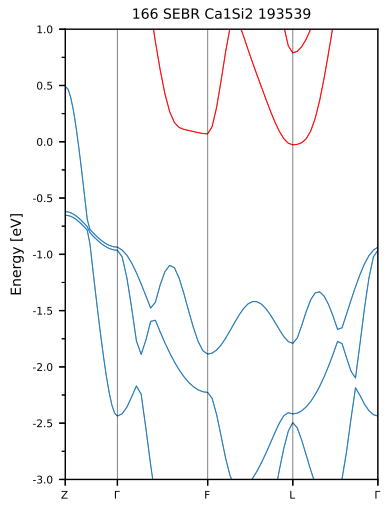

(jw) 166.1.283

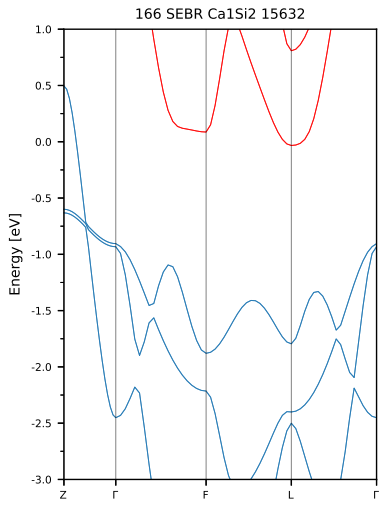

(jx) 166.1.284

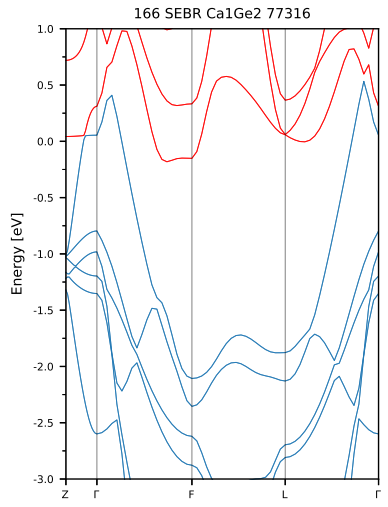

(jy) 166.1.285

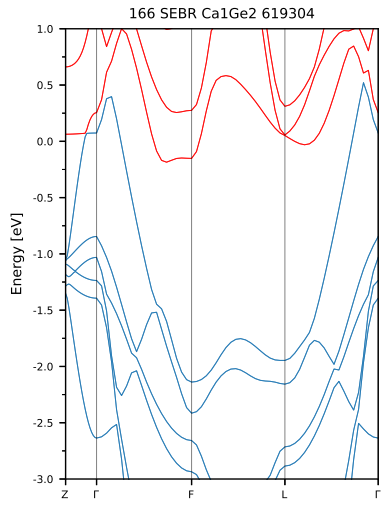

(jz) 166.1.286

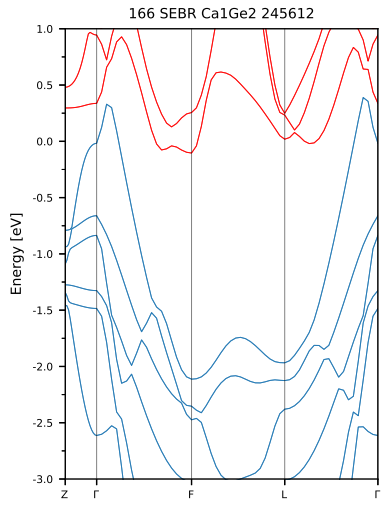

(ka) 166.1.287

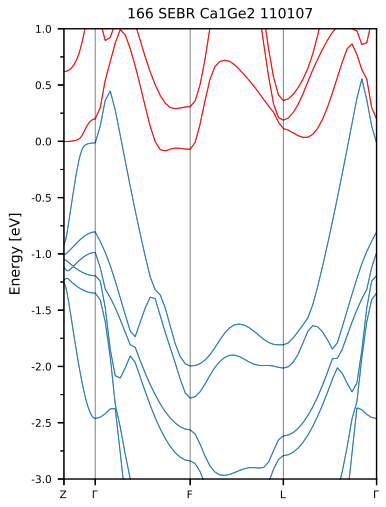

(kb) 166.1.288

Figure 368: Band structures for topological materials in SG 166, part 18/39, from 166.1.273 to 166.1.288. See the list of figures.

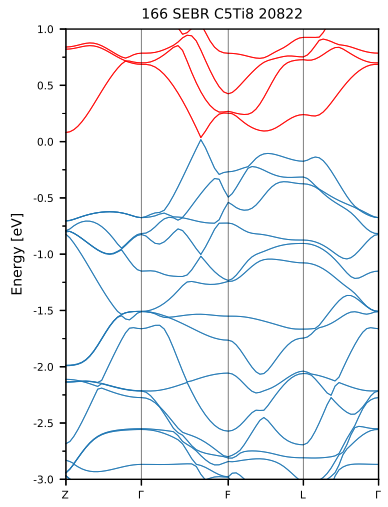

(kc) 166.1.289

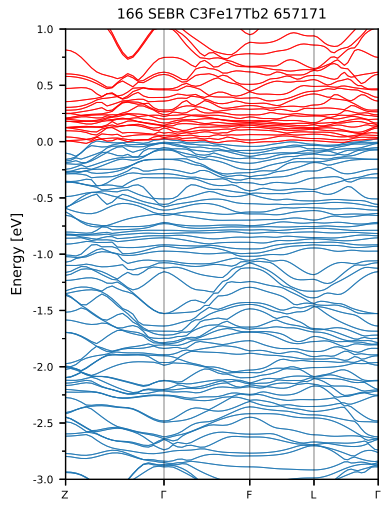

(kd) 166.1.290

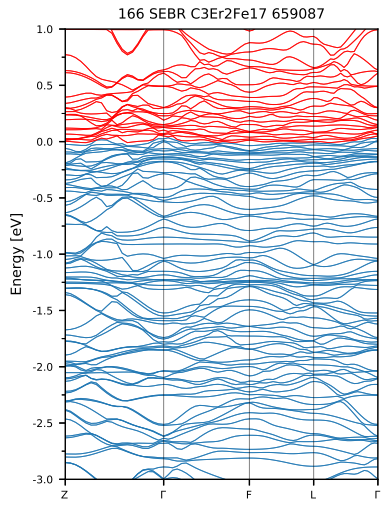

(ke) 166.1.291

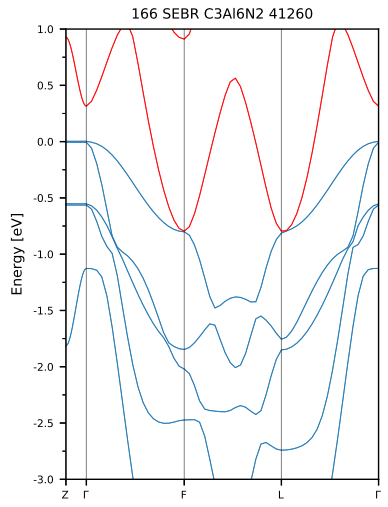

(kf) 166.1.292

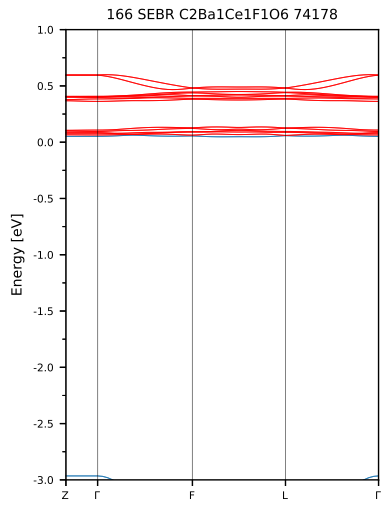

(kg) 166.1.293

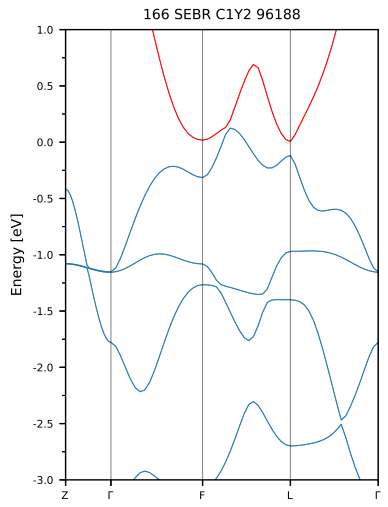

(kh) 166.1.294

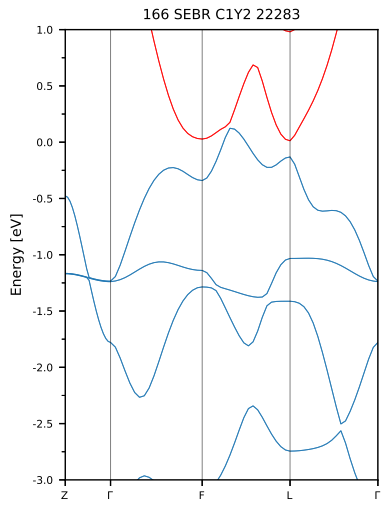

(ki) 166.1.295

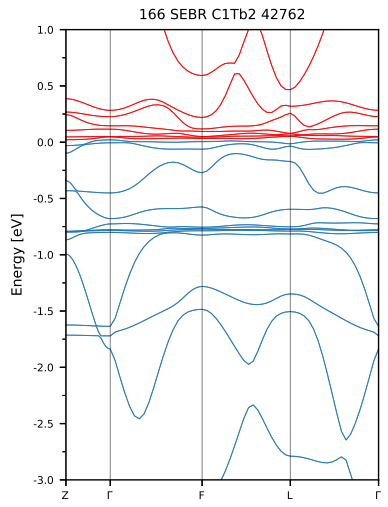

(kj) 166.1.296

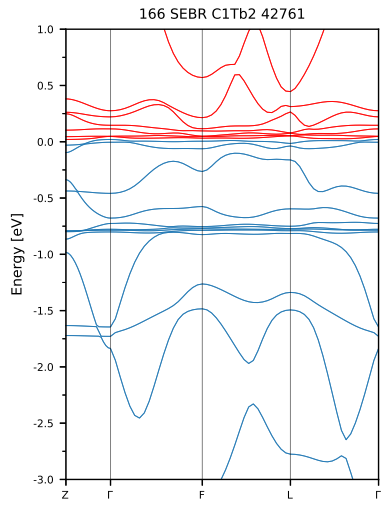

(kk) 166.1.297

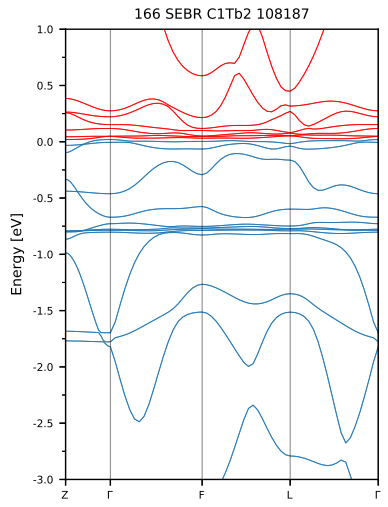

(kl) 166.1.298

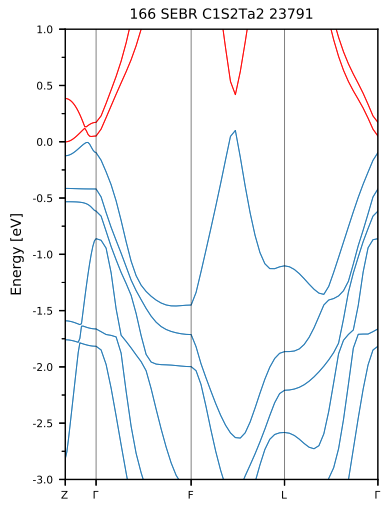

(km) 166.1.299

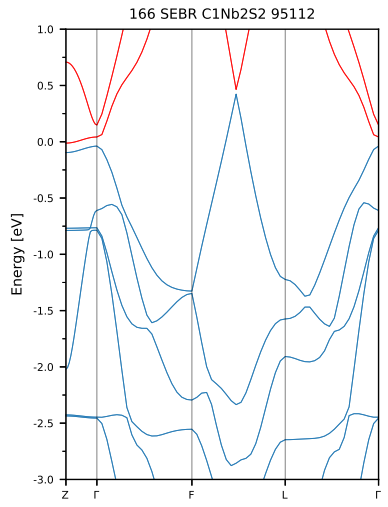

(kn) 166.1.300

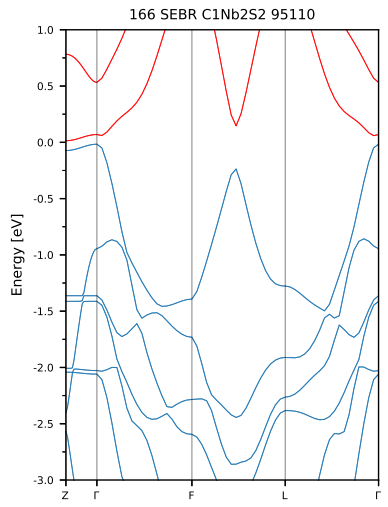

(ko) 166.1.301

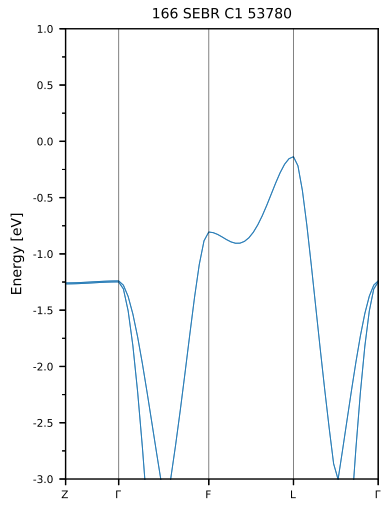

(kp) 166.1.302

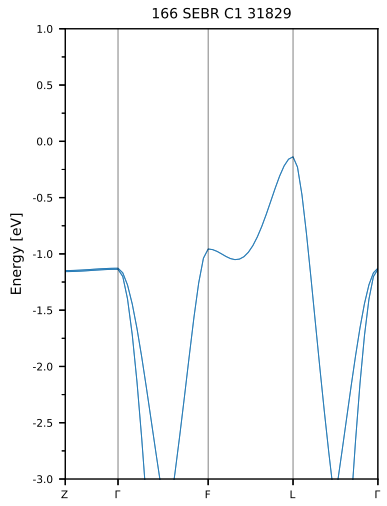

(kq) 166.1.303

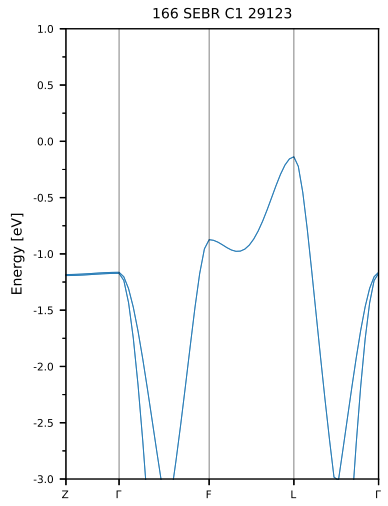

(kr) 166.1.304

Figure 369: Band structures for topological materials in SG 166, part 19/39, from 166.1.289 to 166.1.304. See the list of figures.

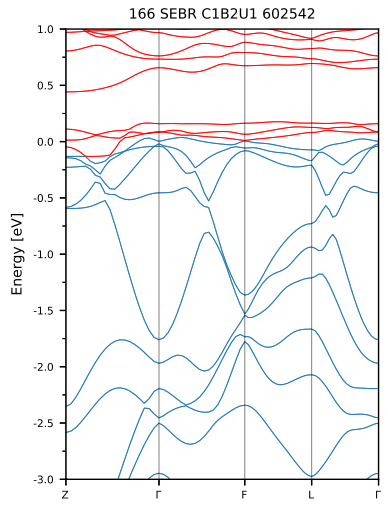

(ks) 166.1.305

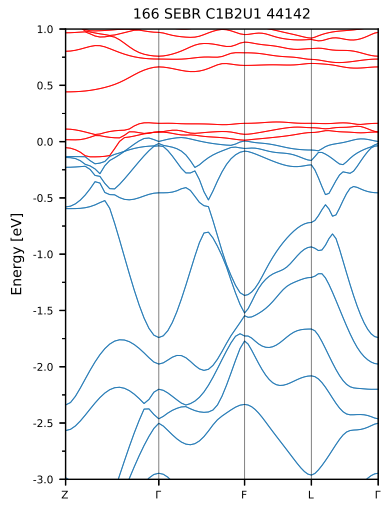

(kt) 166.1.306

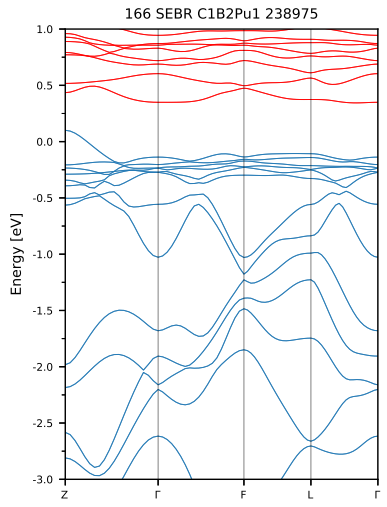

(ku) 166.1.307

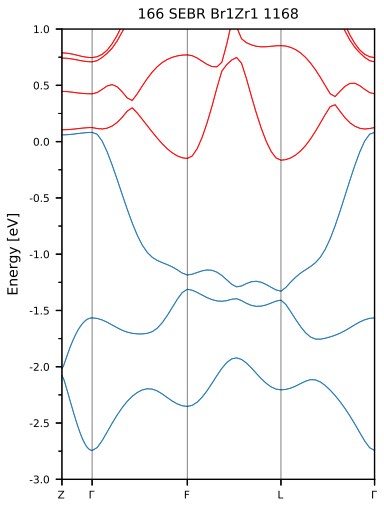

(kv) 166.1.308

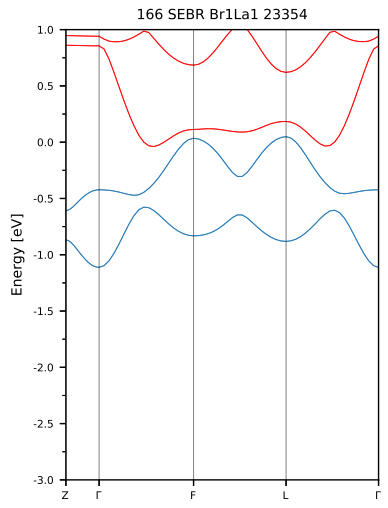

(kw) 166.1.309

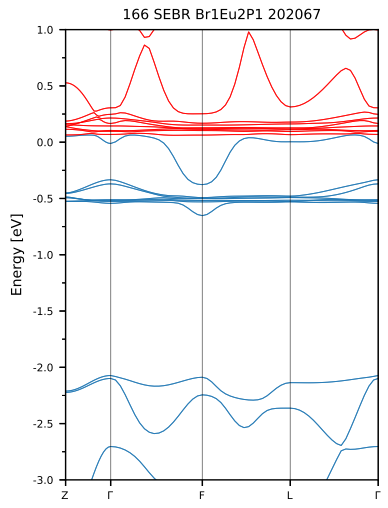

(kx) 166.1.310

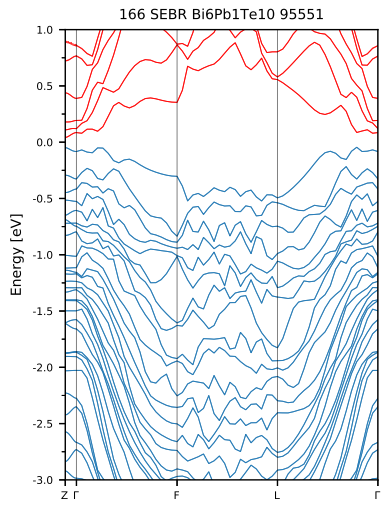

(ky) 166.1.311

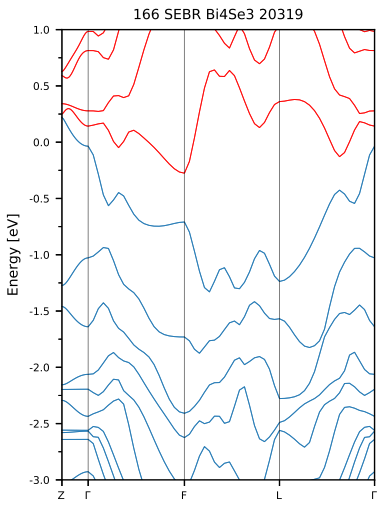

(kz) 166.1.312

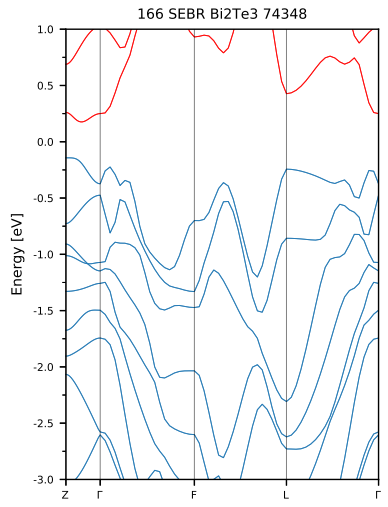

(la) 166.1.313

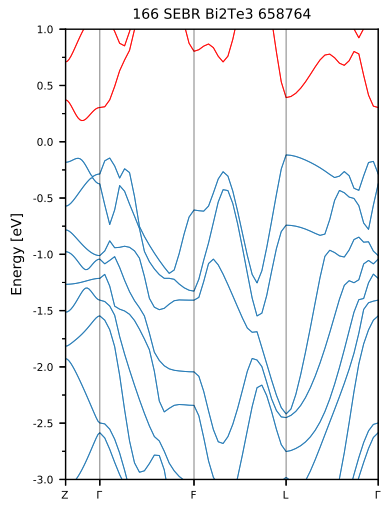

(lb) 166.1.314

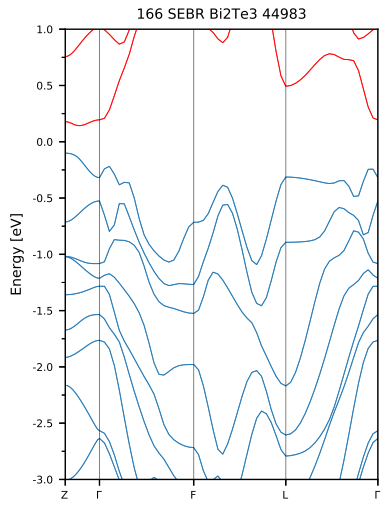

(lc) 166.1.315

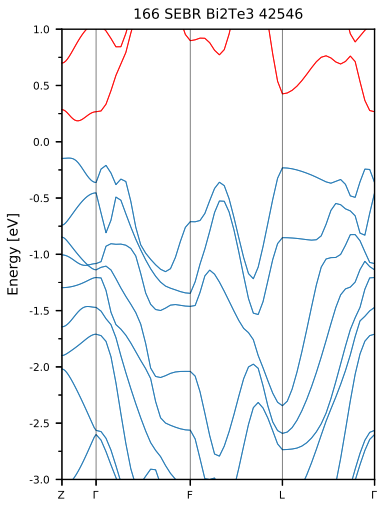

(ld) 166.1.316

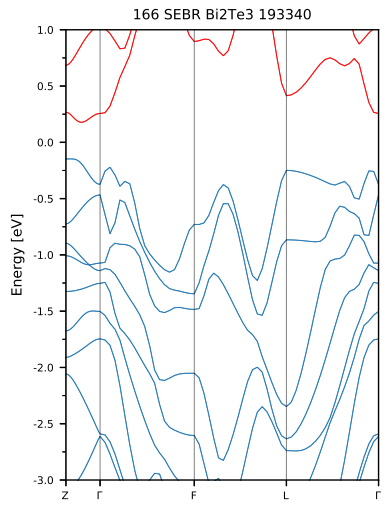

(le) 166.1.317

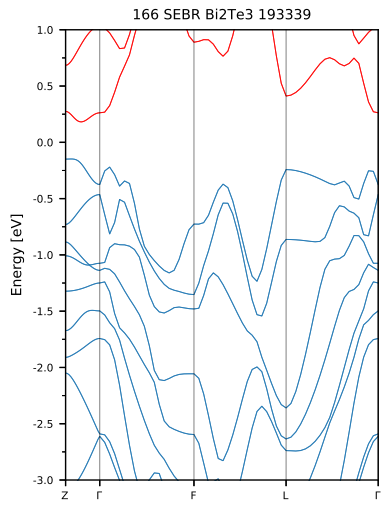

(lf) 166.1.318

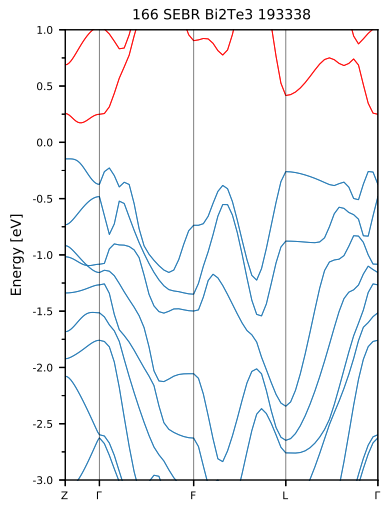

(lg) 166.1.319

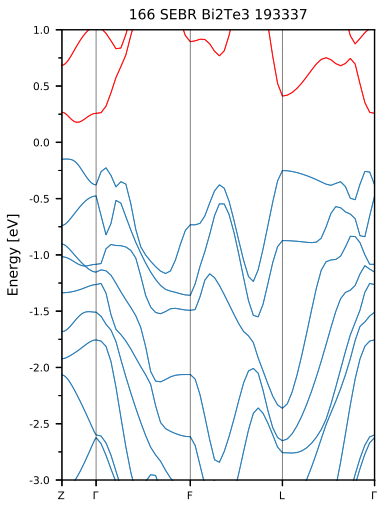

(lh) 166.1.320

Figure 370: Band structures for topological materials in SG 166, part 20/39, from 166.1.305 to 166.1.320. See the list of figures.

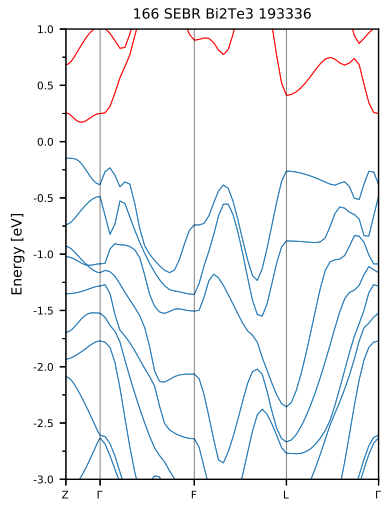

(li) 166.1.321

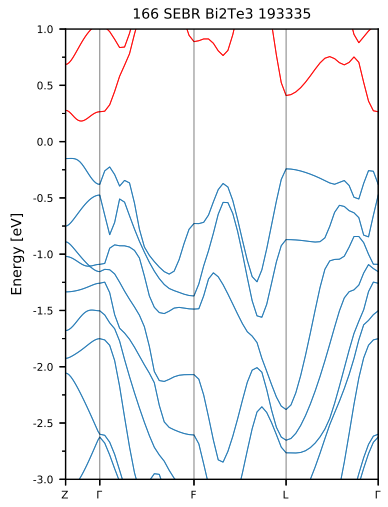

(lj) 166.1.322

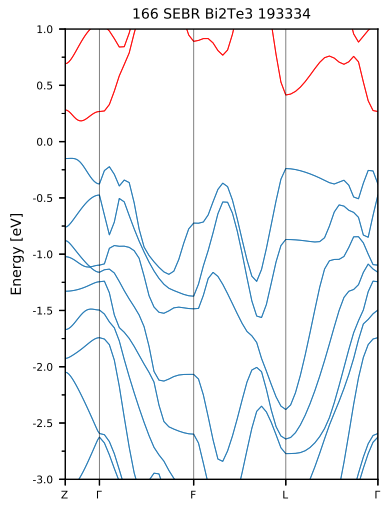

(lk) 166.1.323

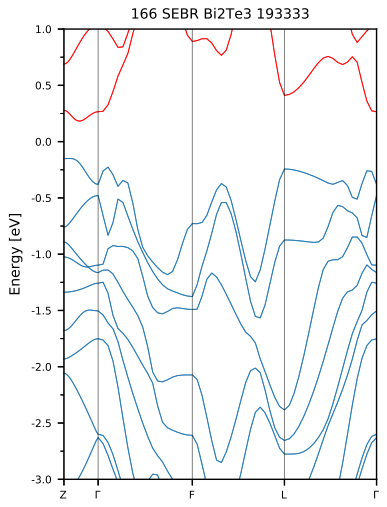

(ll) 166.1.324

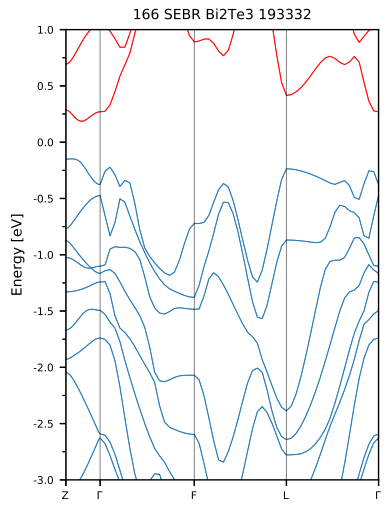

(lm) 166.1.325

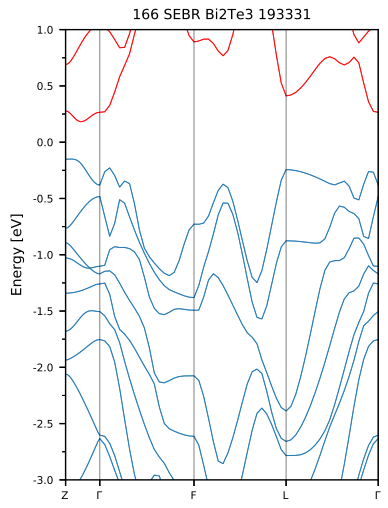

(ln) 166.1.326

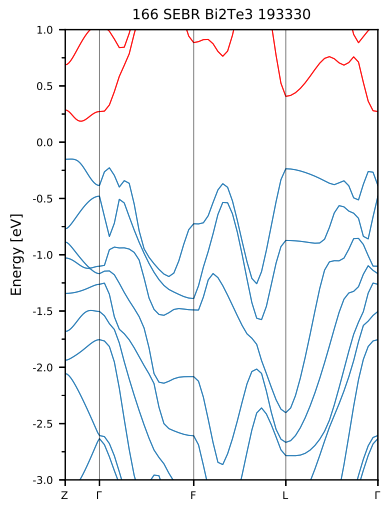

(lo) 166.1.327

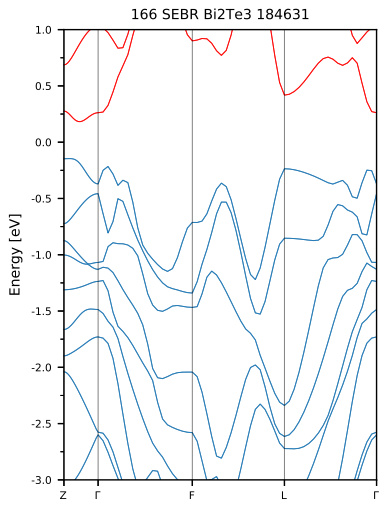

(lp) 166.1.328

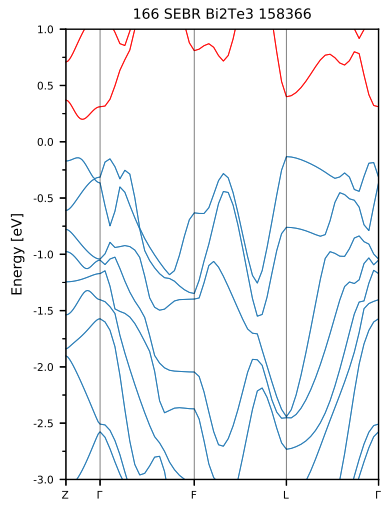

(lq) 166.1.329

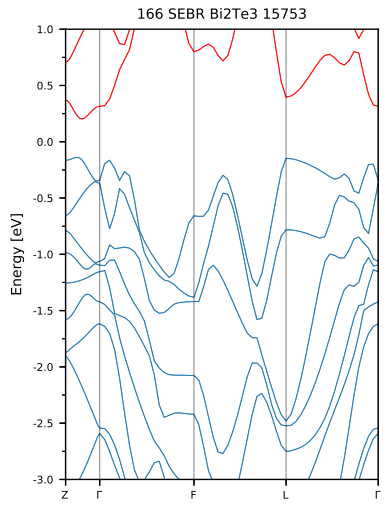

(lr) 166.1.330

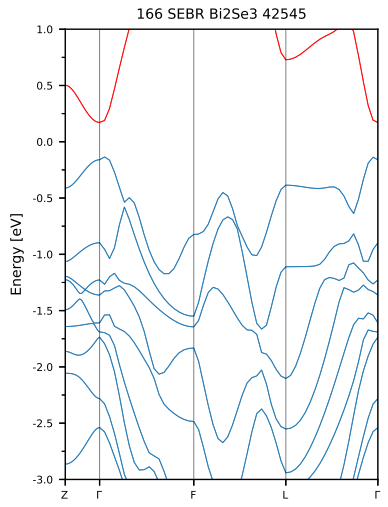

(ls) 166.1.331

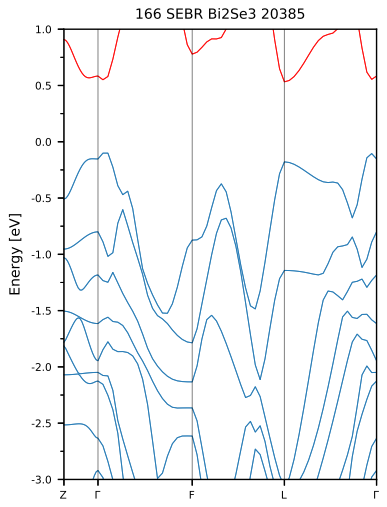

(lt) 166.1.332

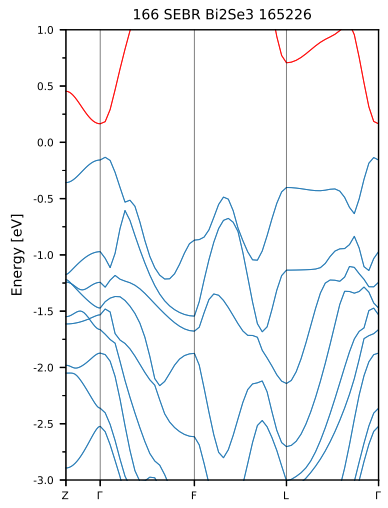

(lu) 166.1.333

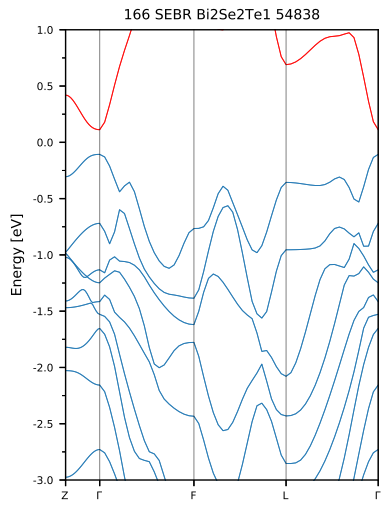

(lv) 166.1.334

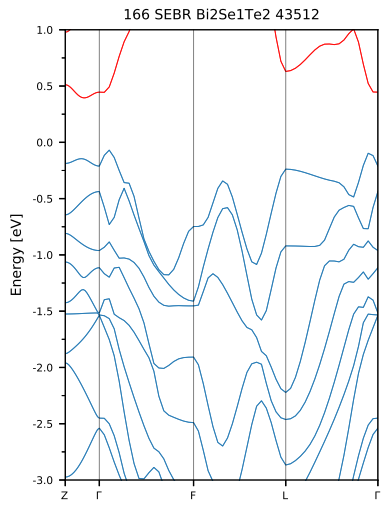

(lw) 166.1.335

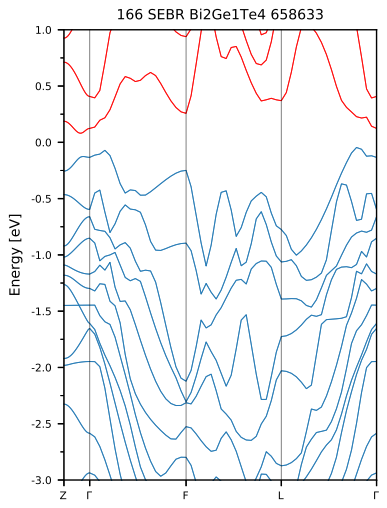

(lx) 166.1.336

Figure 371: Band structures for topological materials in SG 166, part 21/39, from 166.1.321 to 166.1.336. See the list of figures.

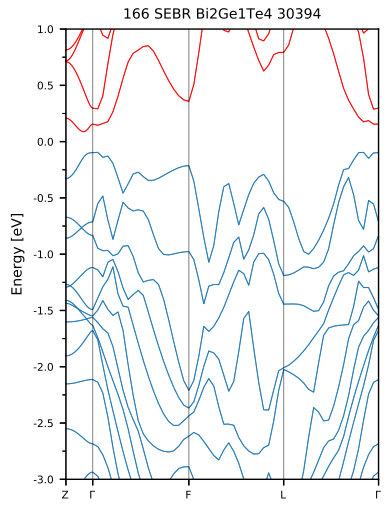

(ly) 166.1.337

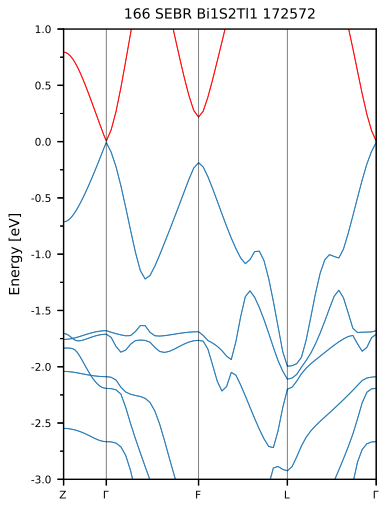

(lz) 166.1.338

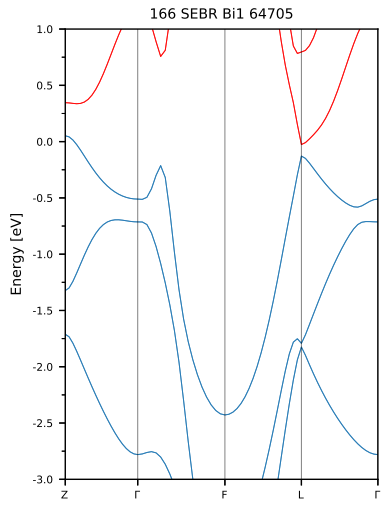

(ma) 166.1.339

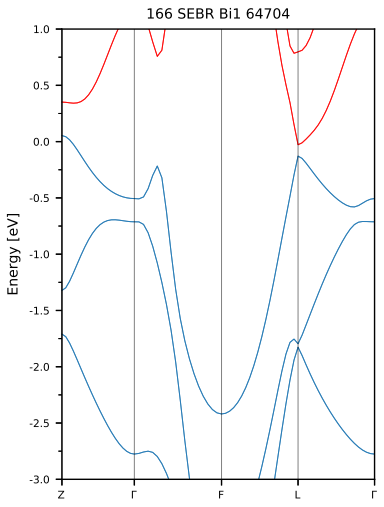

(mb) 166.1.340

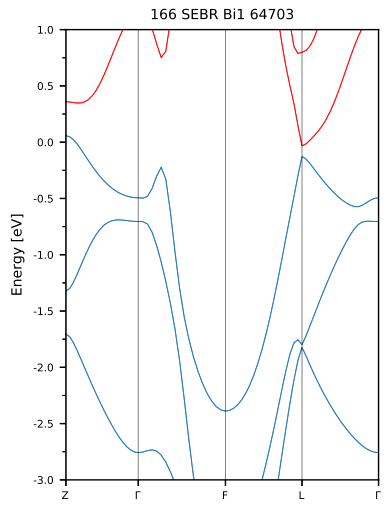

(mc) 166.1.341

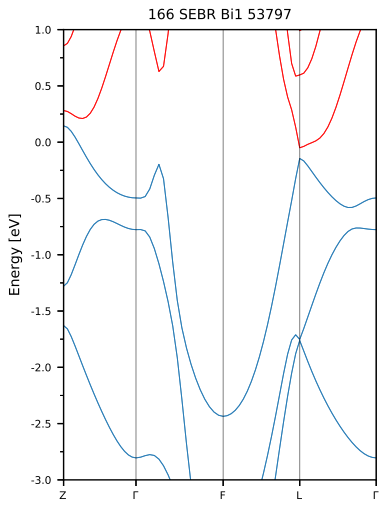

(md) 166.1.342

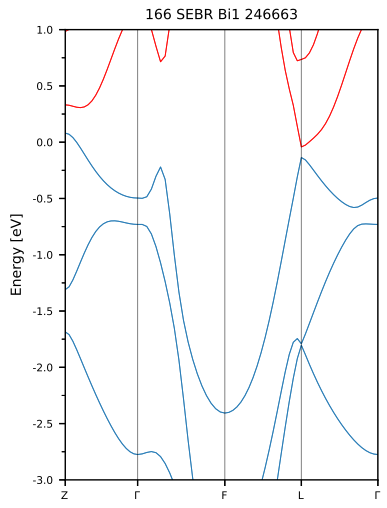

(me) 166.1.343

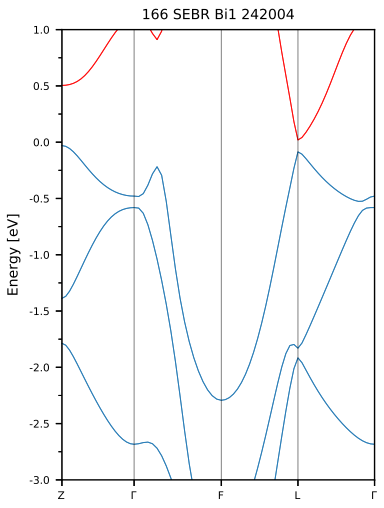

(mf) 166.1.344

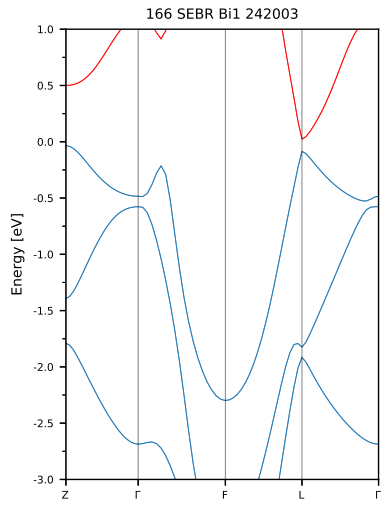

(mg) 166.1.345

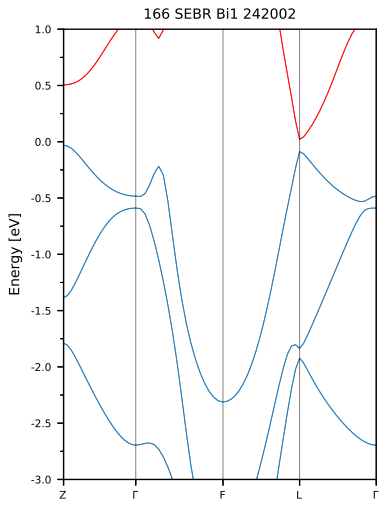

(mh) 166.1.346

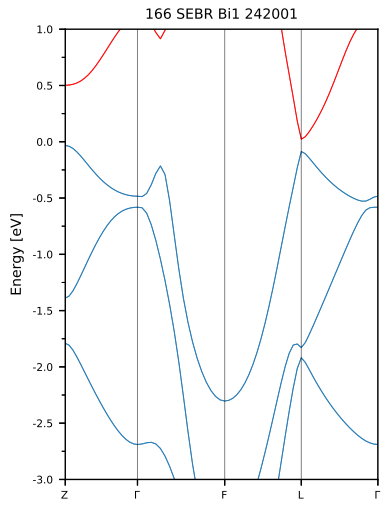

(mi) 166.1.347

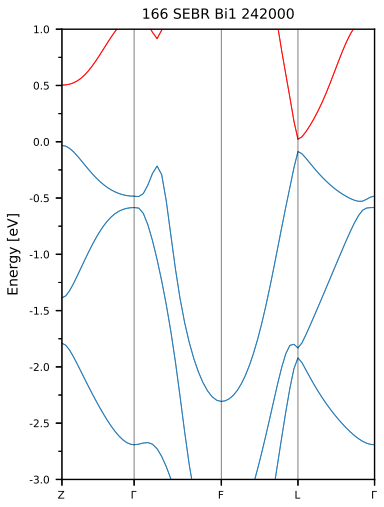

(mj) 166.1.348

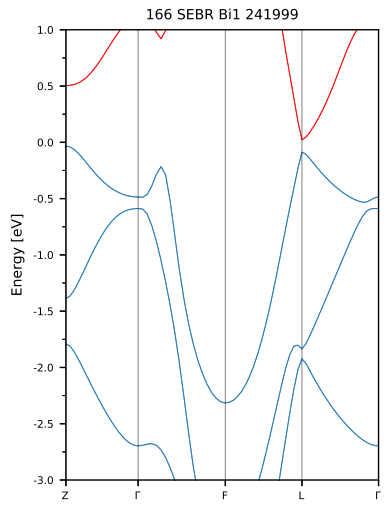

(mk) 166.1.349

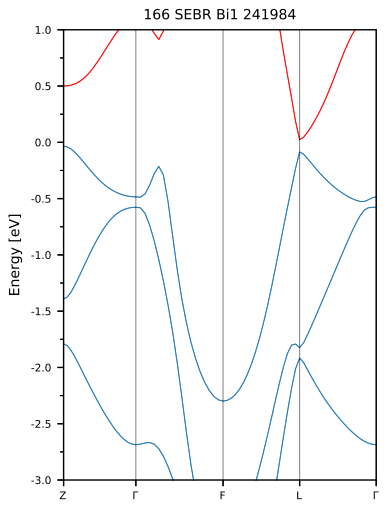

(ml) 166.1.350

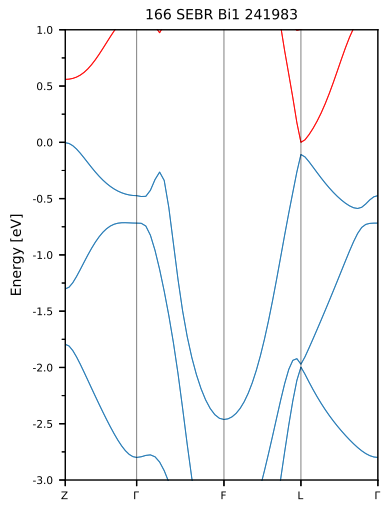

(mm) 166.1.351

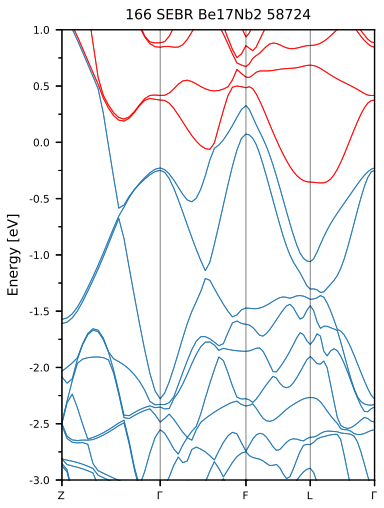

(mn) 166.1.352

Figure 372: Band structures for topological materials in SG 166, part 22/39, from 166.1.337 to 166.1.352. See the list of figures.

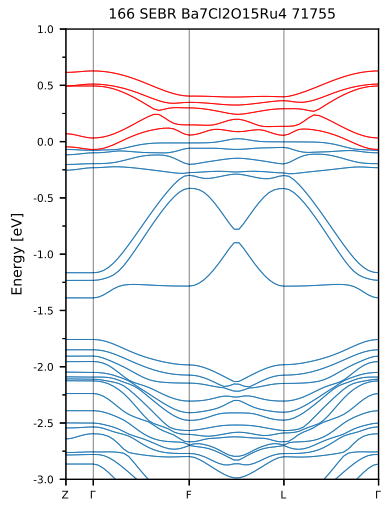

(mo) 166.1.353

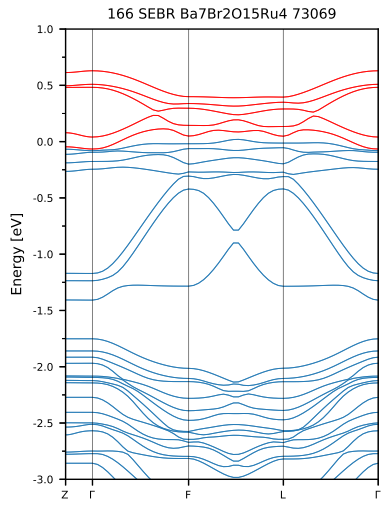

(mp) 166.1.354

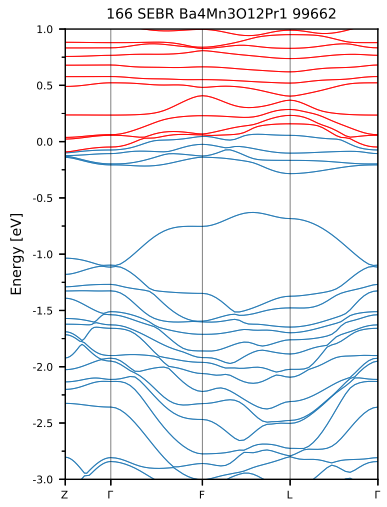

(mq) 166.1.355

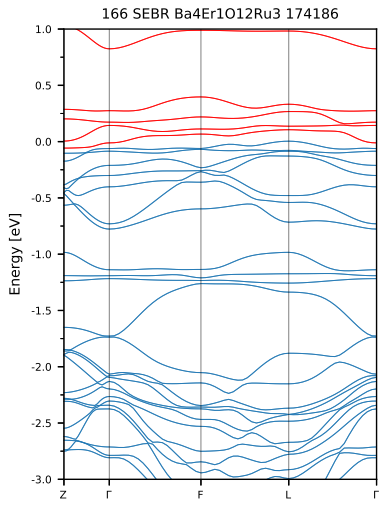

(mr) 166.1.356

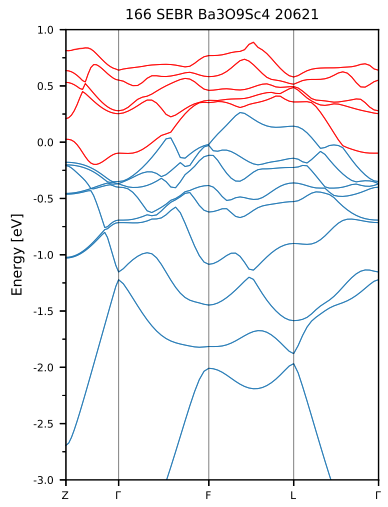

(ms) 166.1.357

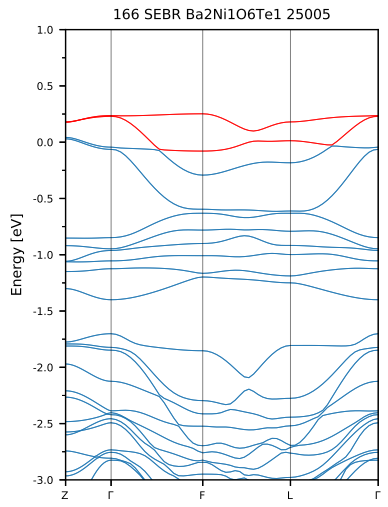

(mt) 166.1.358

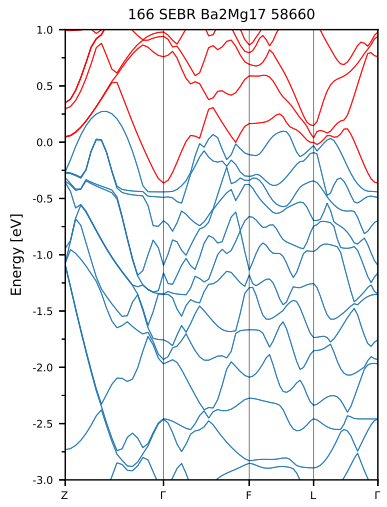

(mu) 166.1.359

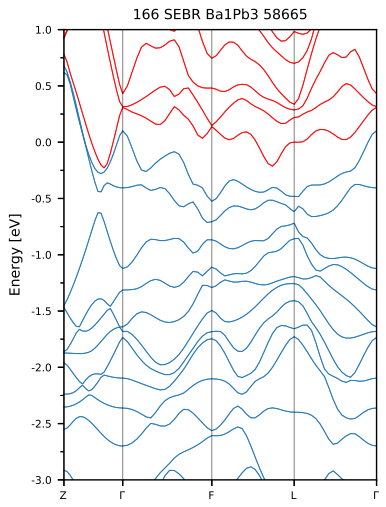

(mv) 166.1.360

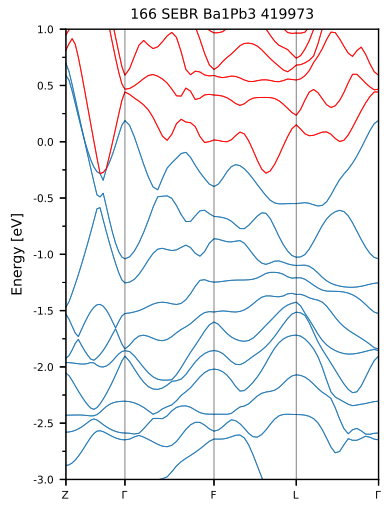

(mw) 166.1.361

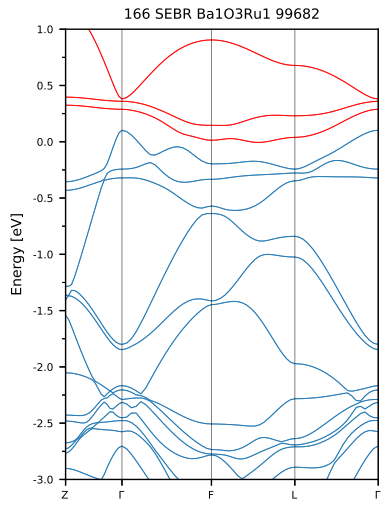

(mx) 166.1.362

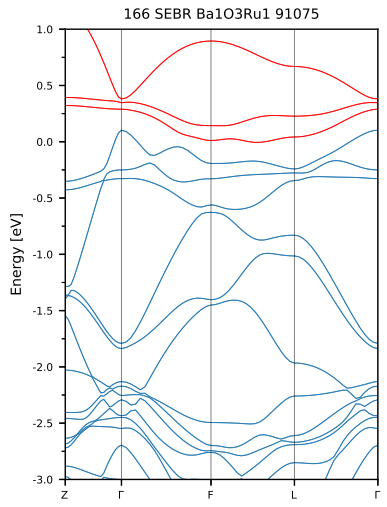

(my) 166.1.363

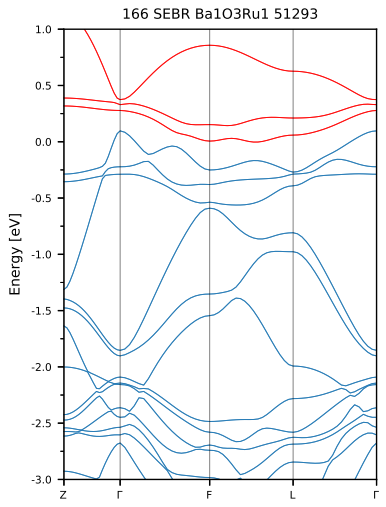

(mz) 166.1.364

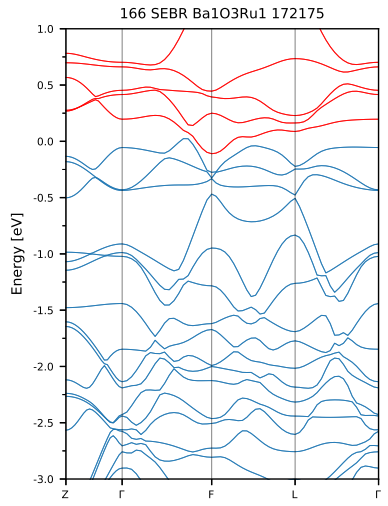

(na) 166.1.365

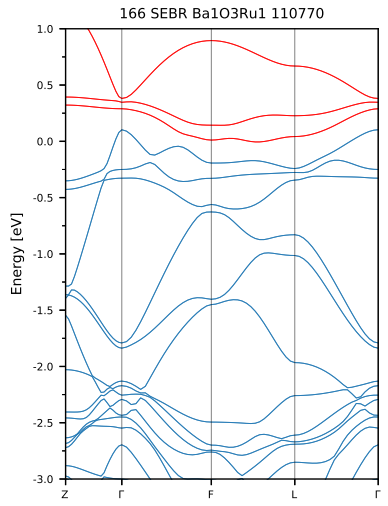

(nb) 166.1.366

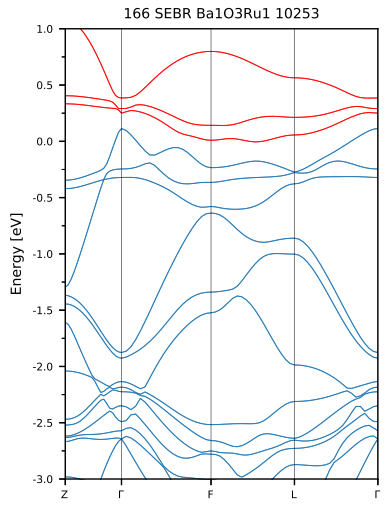

(nc) 166.1.367

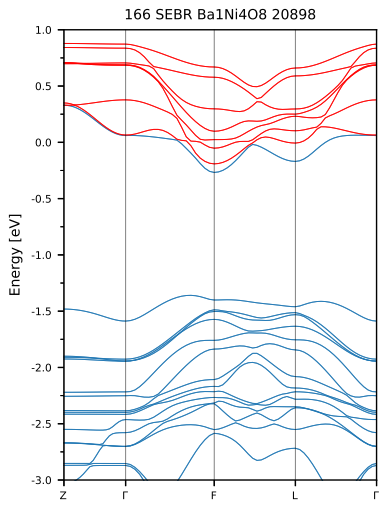

(nd) 166.1.368

Figure 373: Band structures for topological materials in SG 166, part 23/39, from 166.1.353 to 166.1.368. See the list of figures.

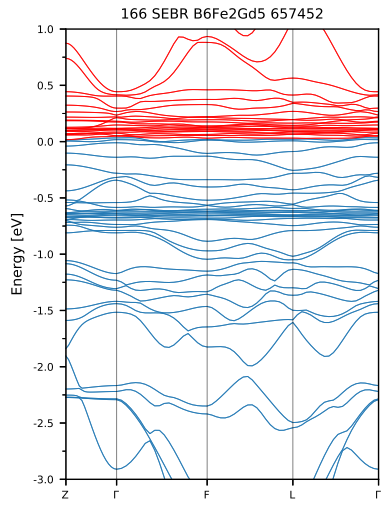

(ne) 166.1.369

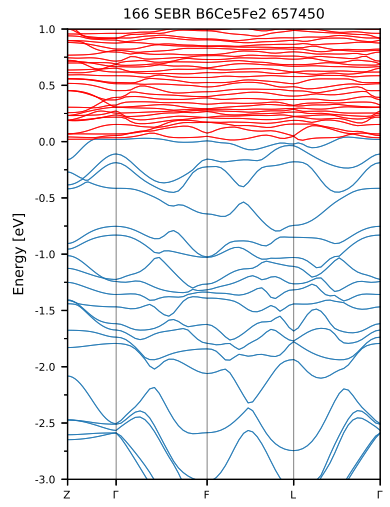

(nf) 166.1.370

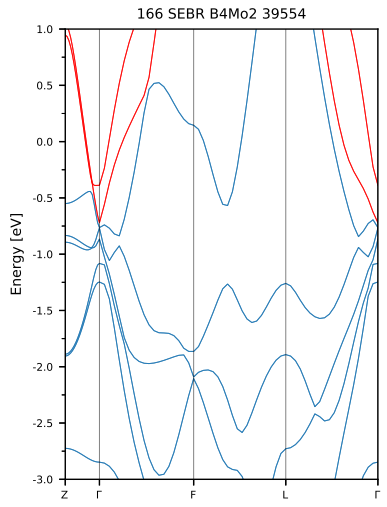

(ng) 166.1.371

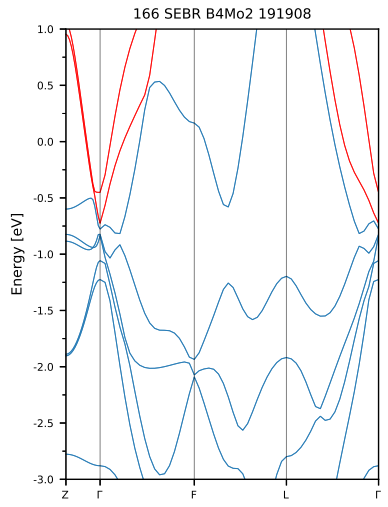

(nh) 166.1.372

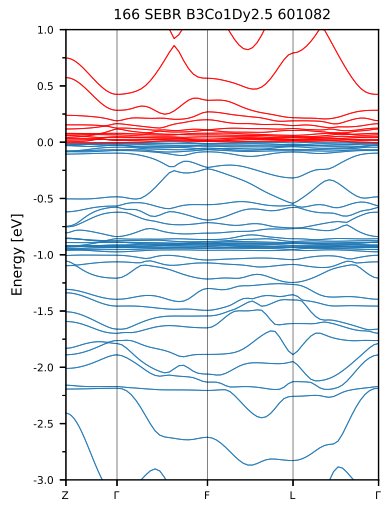

(ni) 166.1.373

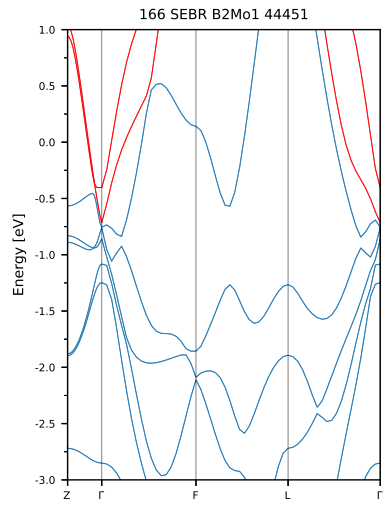

(nj) 166.1.374

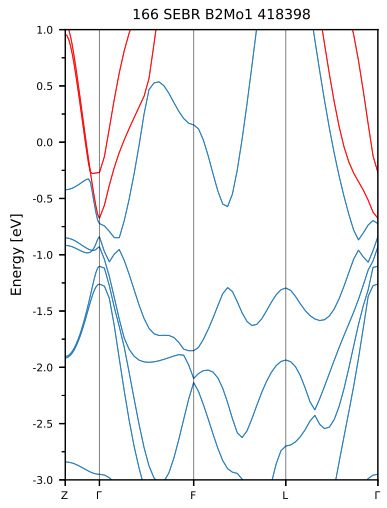

(nk) 166.1.375

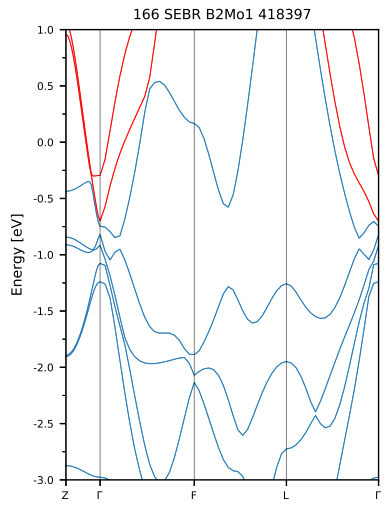

(nl) 166.1.376

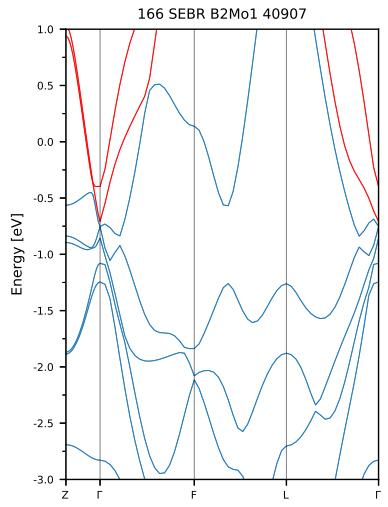

(nm) 166.1.377

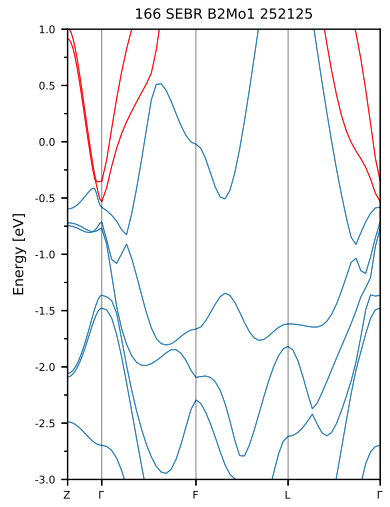

(nn) 166.1.378

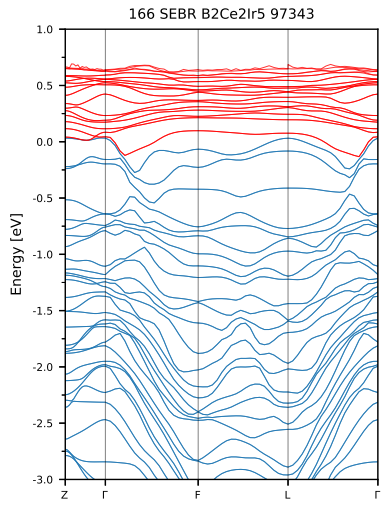

(no) 166.1.379

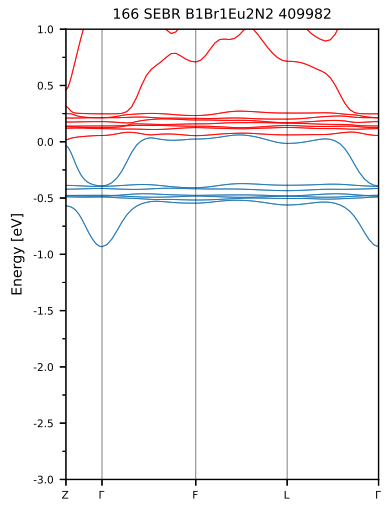

(np) 166.1.380

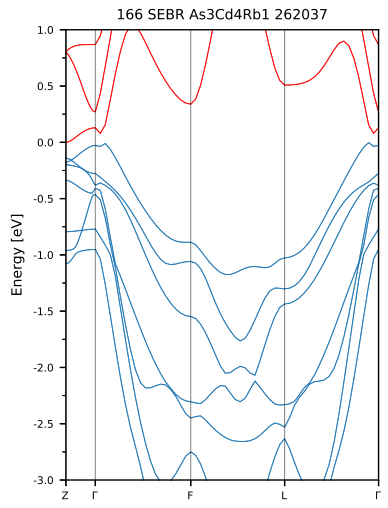

(nq) 166.1.381

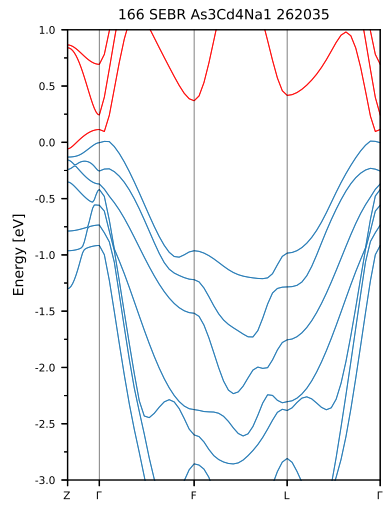

(nr) 166.1.382

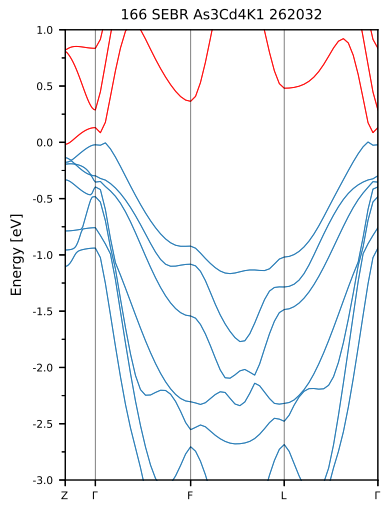

(ns) 166.1.383

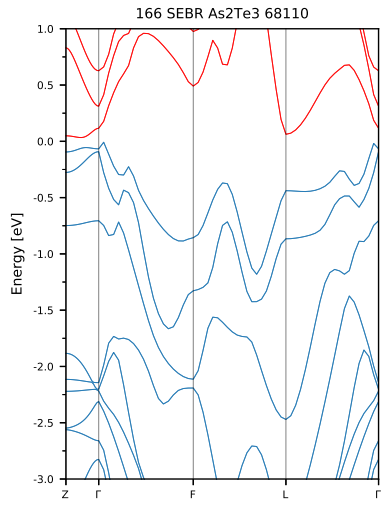

(nt) 166.1.384

Figure 374: Band structures for topological materials in SG 166, part 24/39, from 166.1.369 to 166.1.384. See the list of figures.

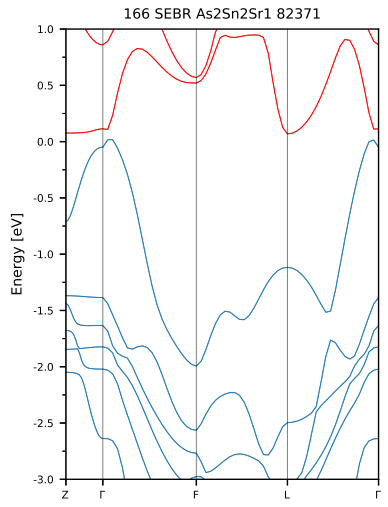

(nu) 166.1.385

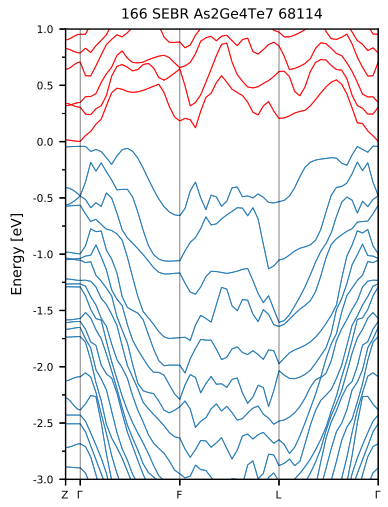

(nv) 166.1.386

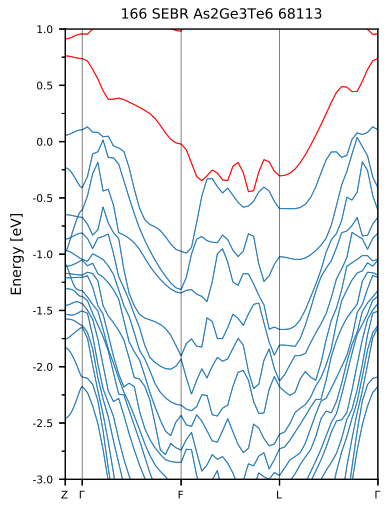

(nw) 166.1.387

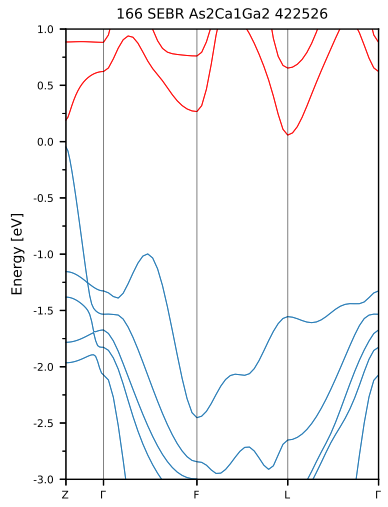

(nx) 166.1.388

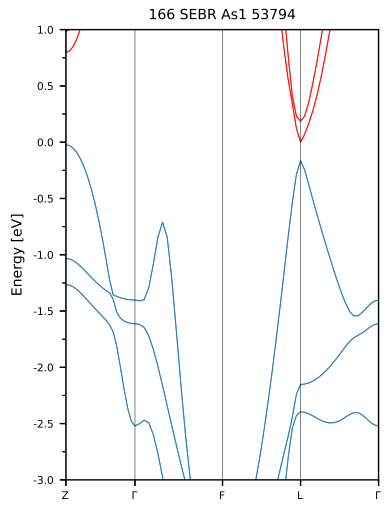

(ny) 166.1.389

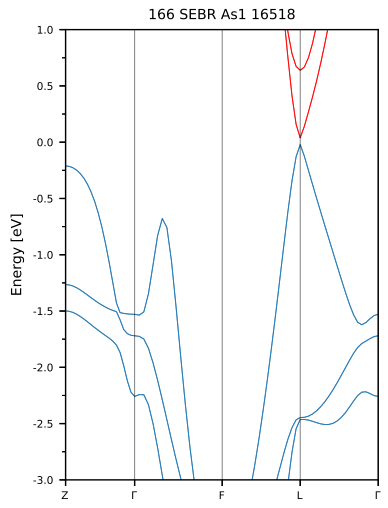

(nz) 166.1.390

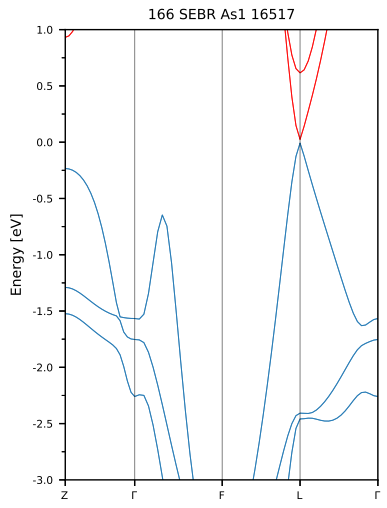

(oa) 166.1.391

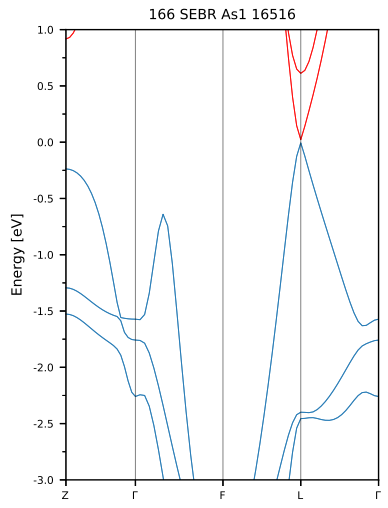

(ob) 166.1.392

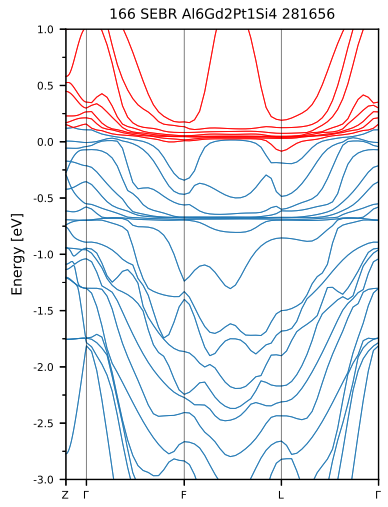

(oc) 166.1.393

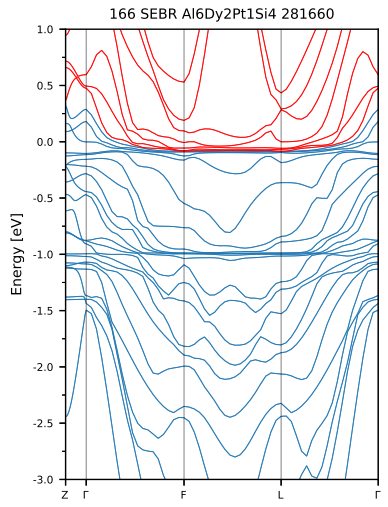

(od) 166.1.394

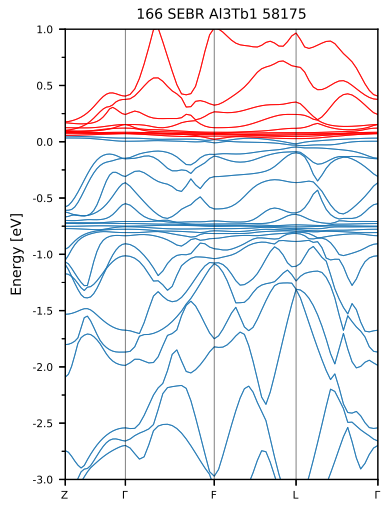

(oe) 166.1.395

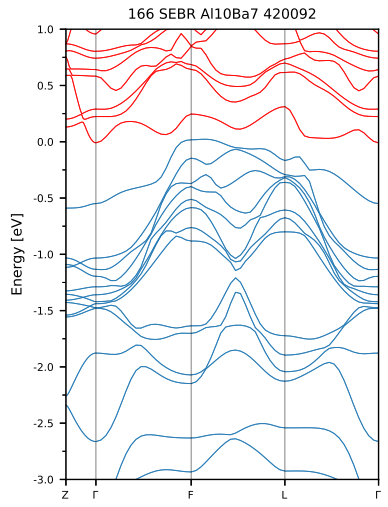

(of) 166.1.396

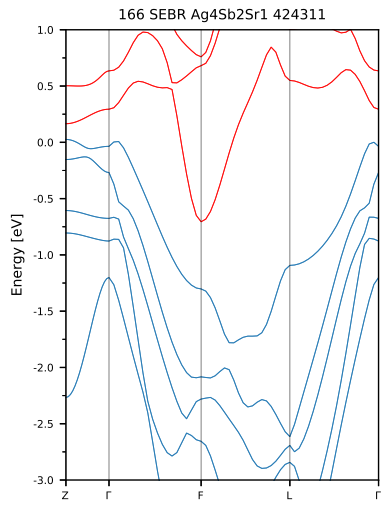

(og) 166.1.397

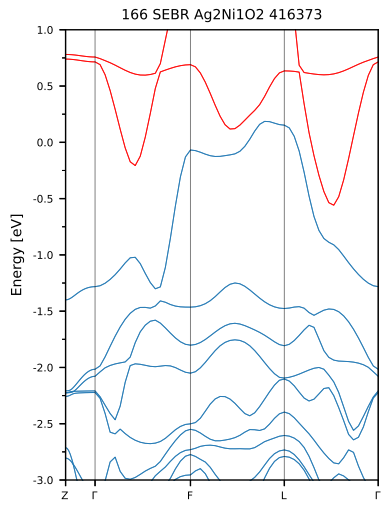

(oh) 166.1.398

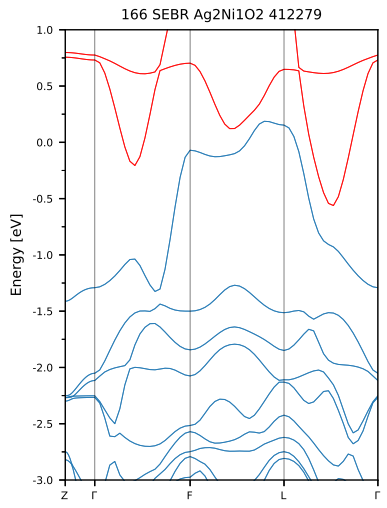

(oi) 166.1.399

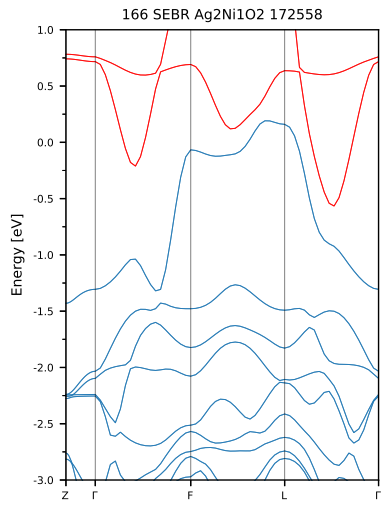

(oj) 166.1.400

Figure 375: Band structures for topological materials in SG 166, part 25/39, from 166.1.385 to 166.1.400. See the list of figures.

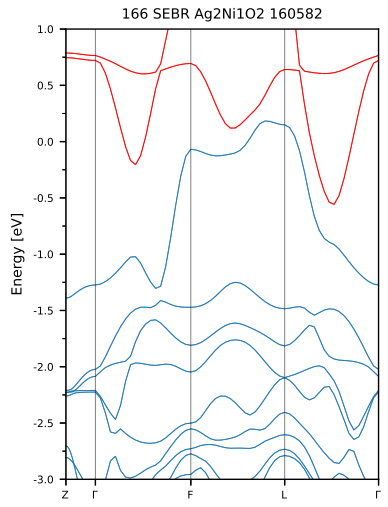

(ok) 166.1.401

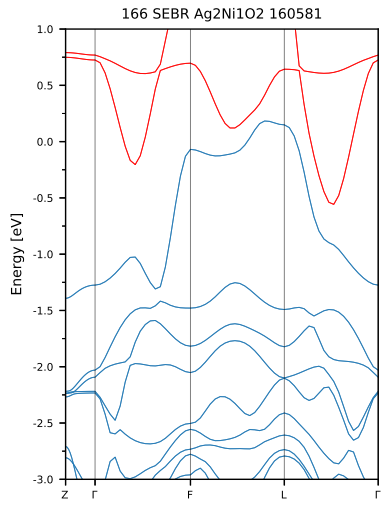

(ol) 166.1.402

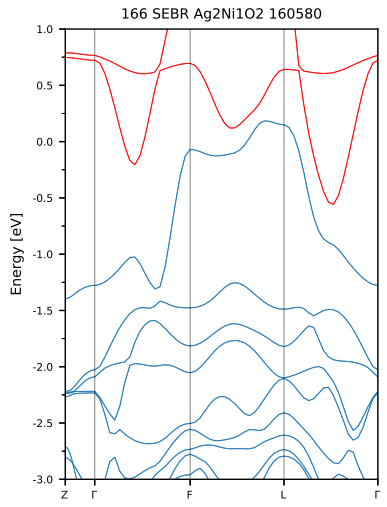

(om) 166.1.403

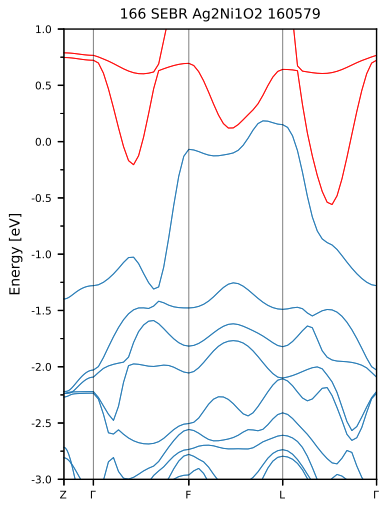

(on) 166.1.404

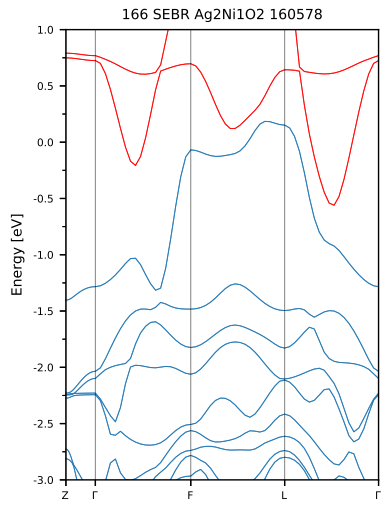

(oo) 166.1.405

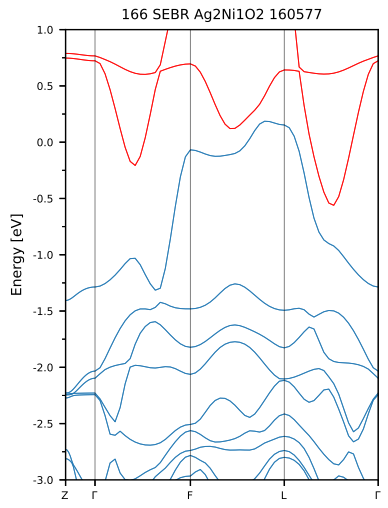

(op) 166.1.406

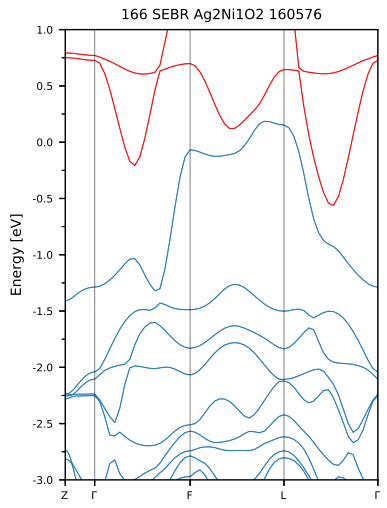

(oq) 166.1.407

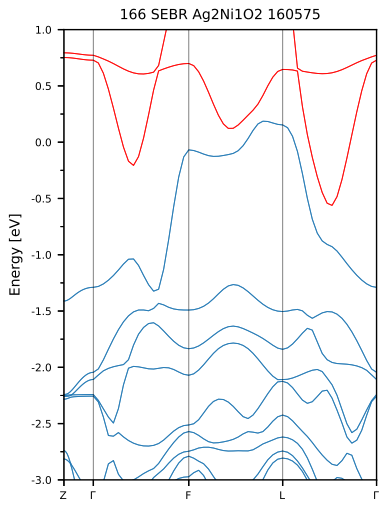

(or) 166.1.408

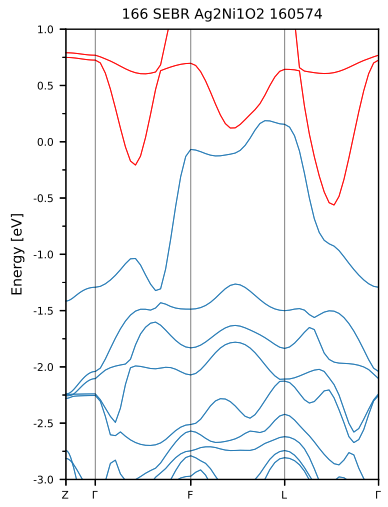

(os) 166.1.409

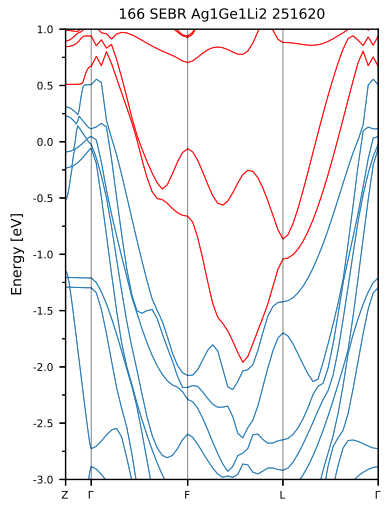

(ot) 166.1.410

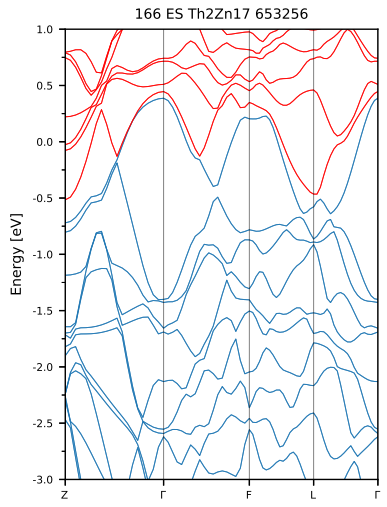

(ou) 166.3.1

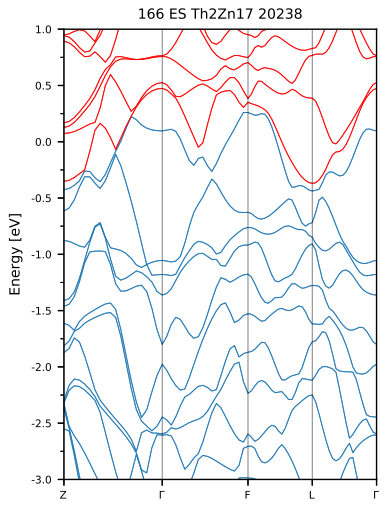

(ov) 166.3.2

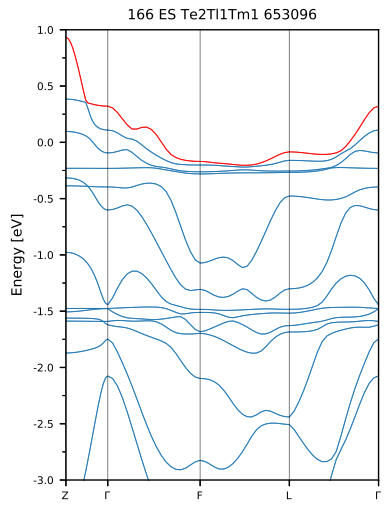

(ow) 166.3.3

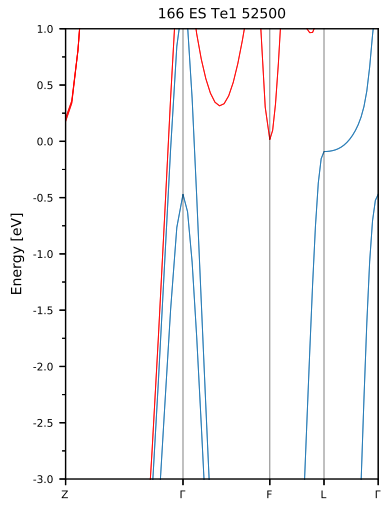

(ox) 166.3.4

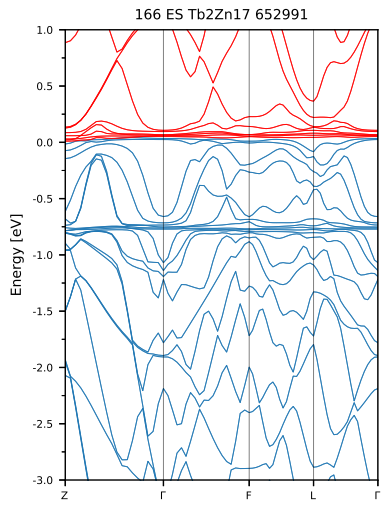

(oy) 166.3.5

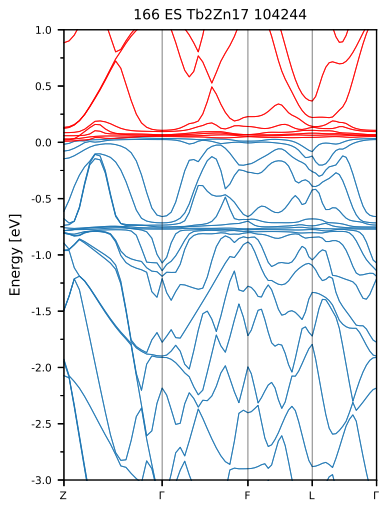

(oz) 166.3.6

Figure 376: Band structures for topological materials in SG 166, part 26/39, from 166.1.401 to 166.3.6. See the list of figures.

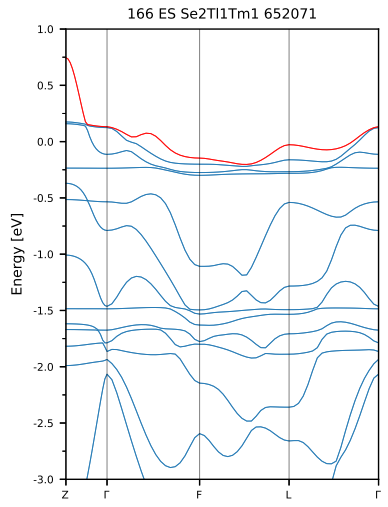

(pa) 166.3.7

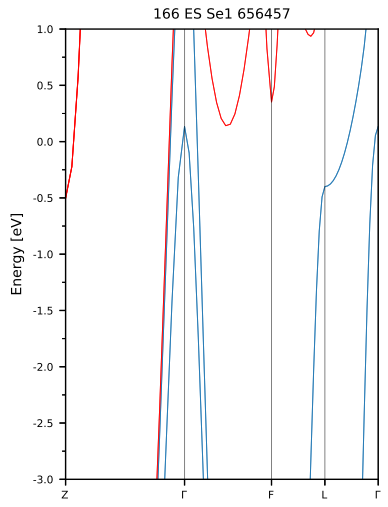

(pb) 166.3.8

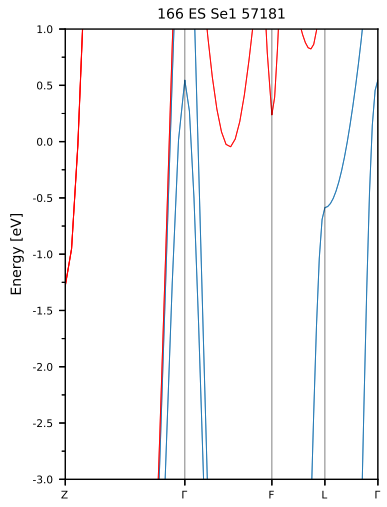

(pc) 166.3.9

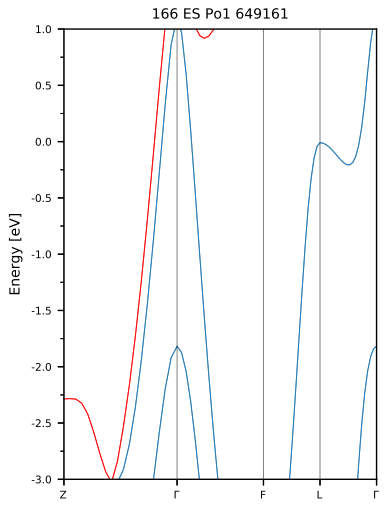

(pd) 166.3.10

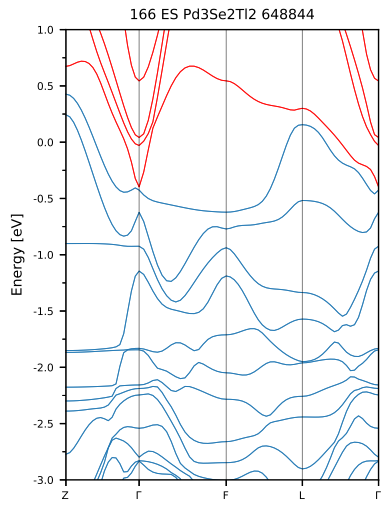

(pe) 166.3.11

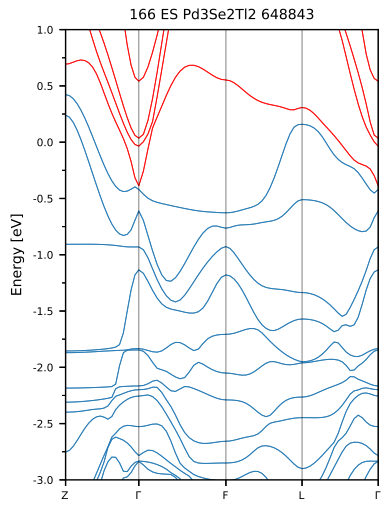

(pf) 166.3.12

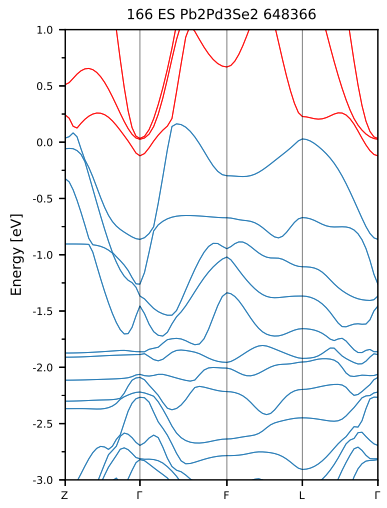

(pg) 166.3.13

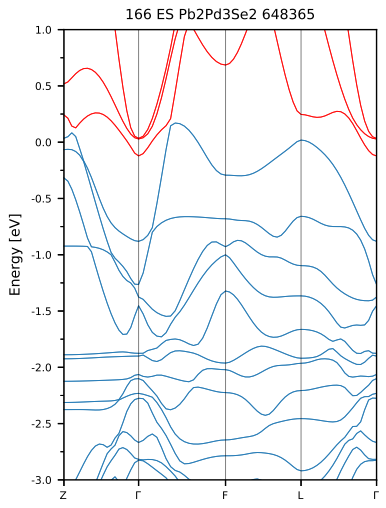

(ph) 166.3.14

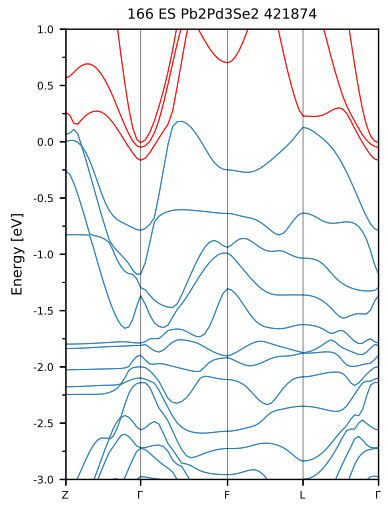

(pi) 166.3.15

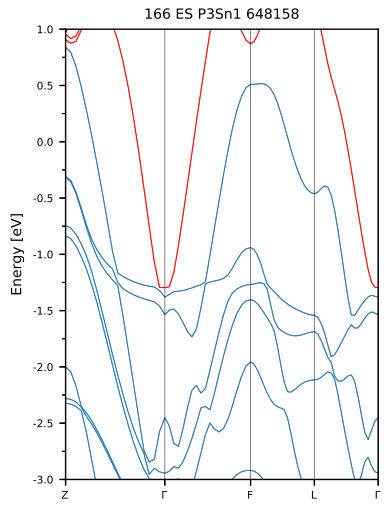

(pj) 166.3.16

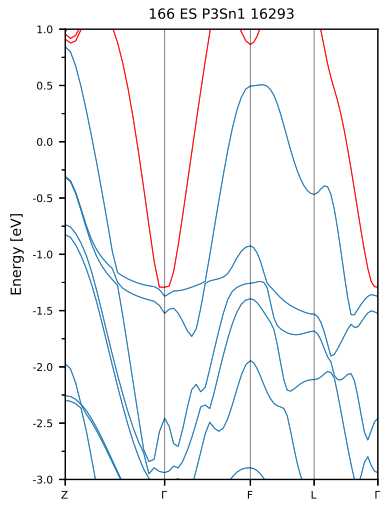

(pk) 166.3.17

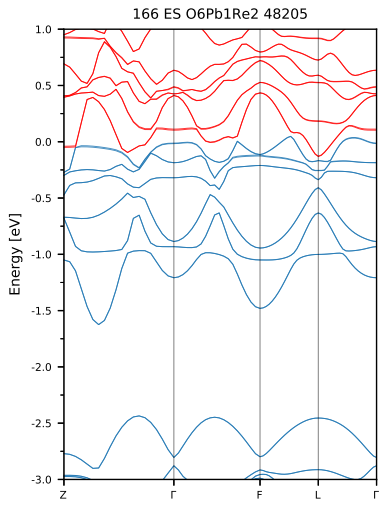

(pl) 166.3.18

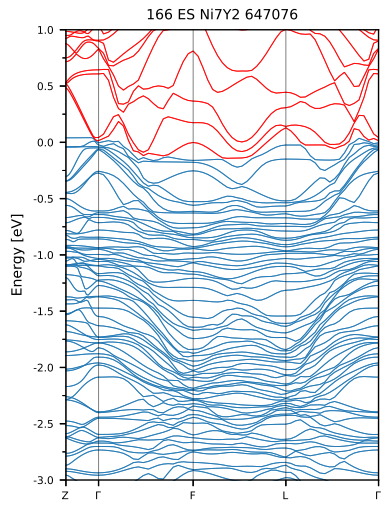

(pm) 166.3.19

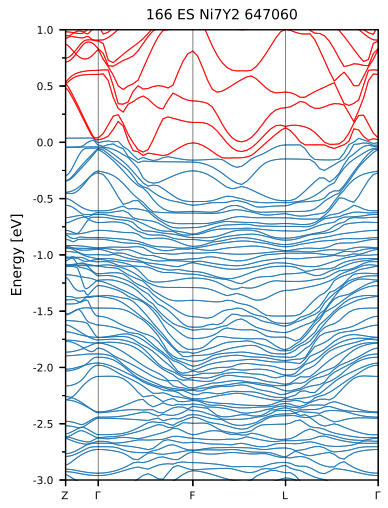

(pn) 166.3.20

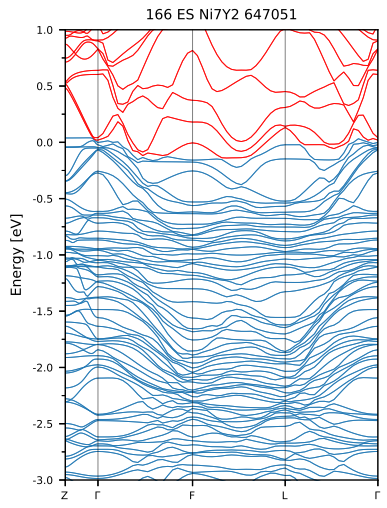

(po) 166.3.21

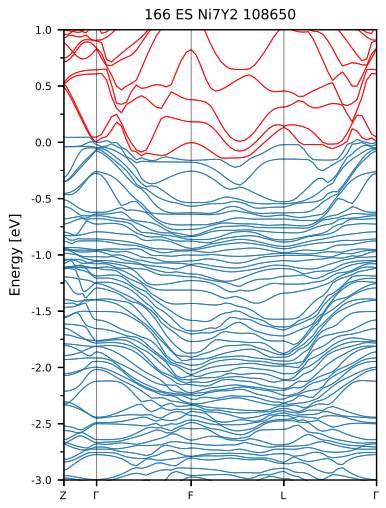

(pp) 166.3.22

Figure 377: Band structures for topological materials in SG 166, part 27/39, from 166.3.7 to 166.3.22. See the list of figures.

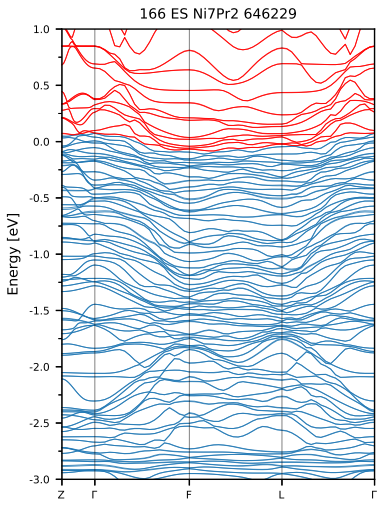

(pq) 166.3.23

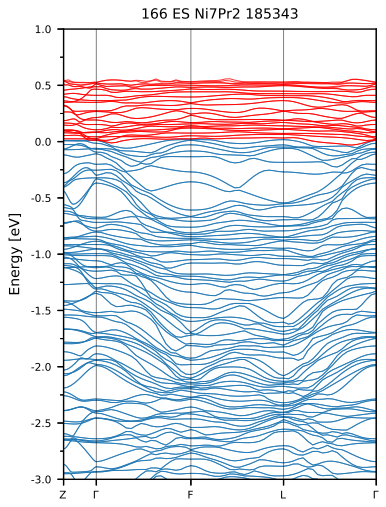

(pr) 166.3.24

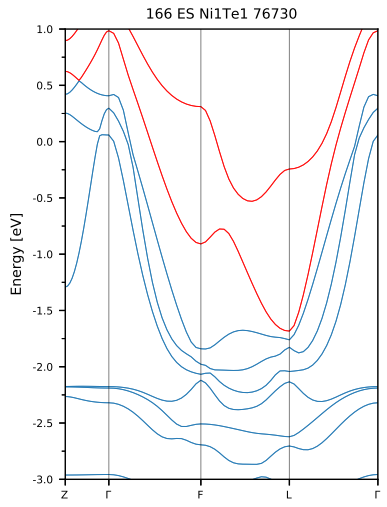

(ps) 166.3.25

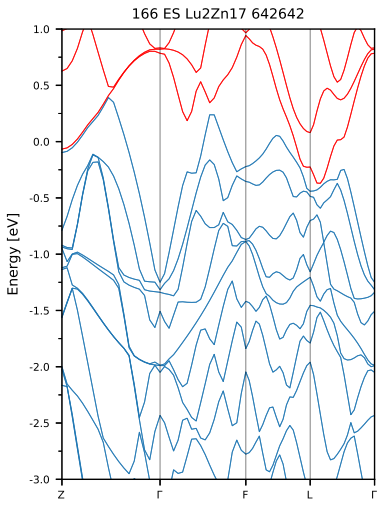

(pt) 166.3.26

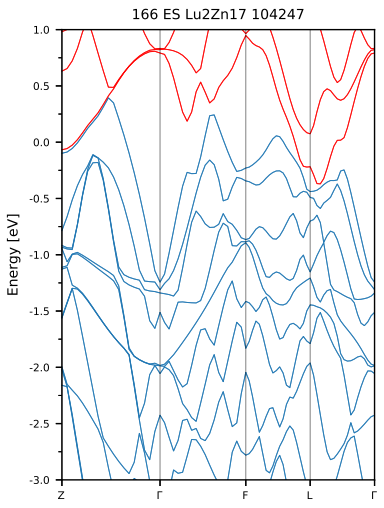

(pu) 166.3.27

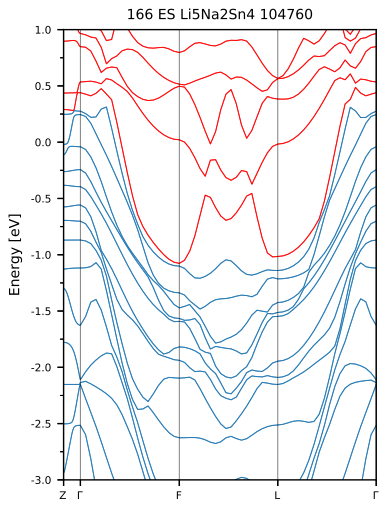

(pv) 166.3.28

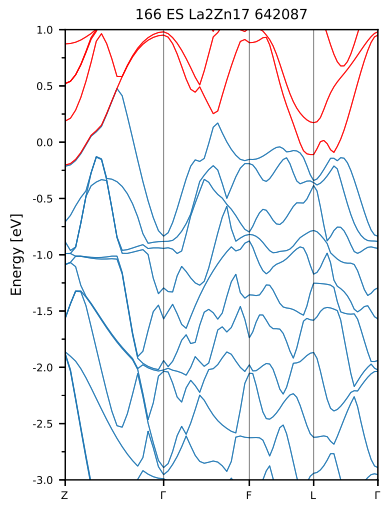

(pw) 166.3.29

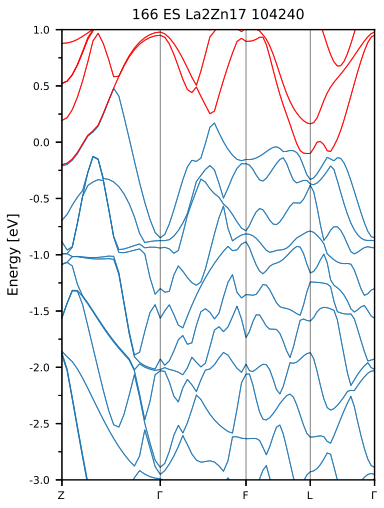

(px) 166.3.30

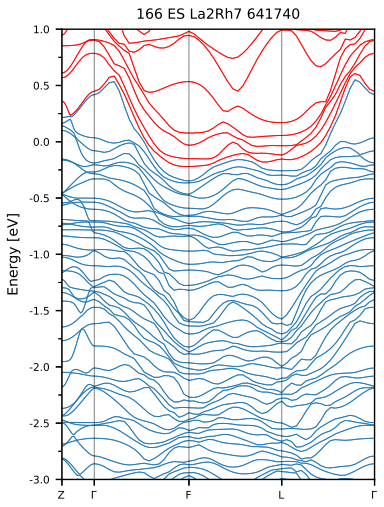

(py) 166.3.31

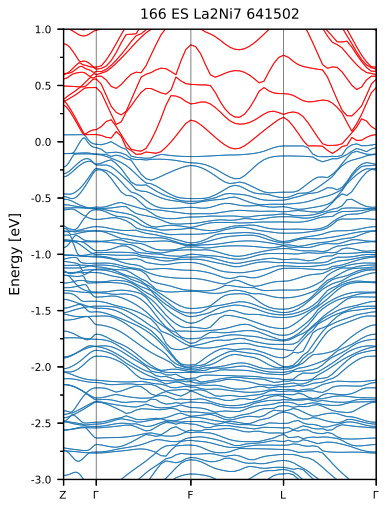

(pz) 166.3.32

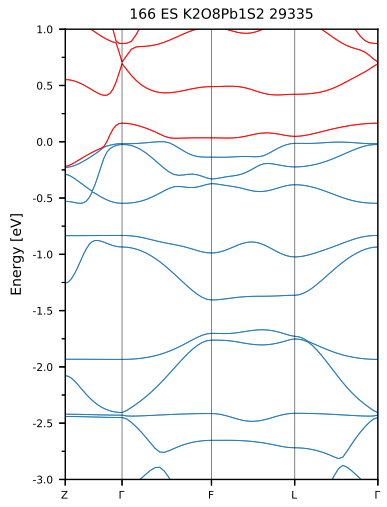

(qa) 166.3.33

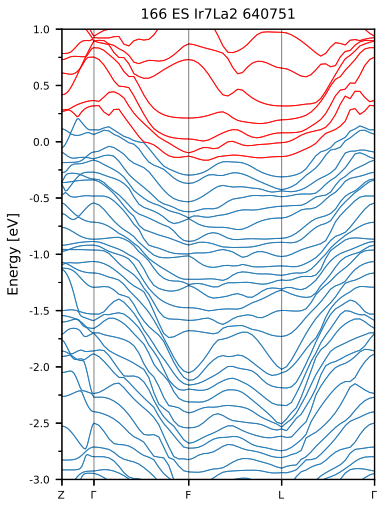

(qb) 166.3.34

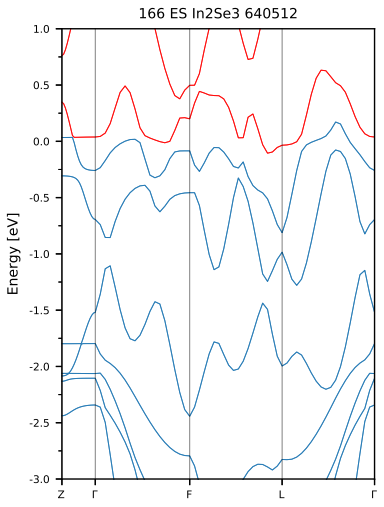

(qc) 166.3.35

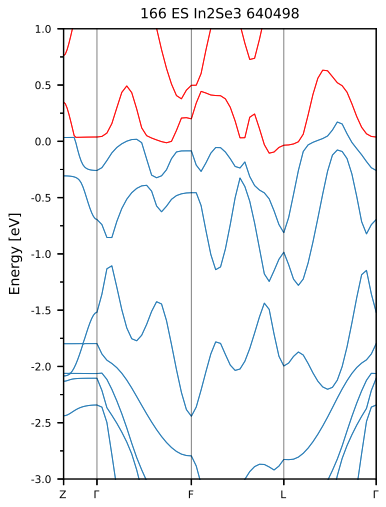

(qd) 166.3.36

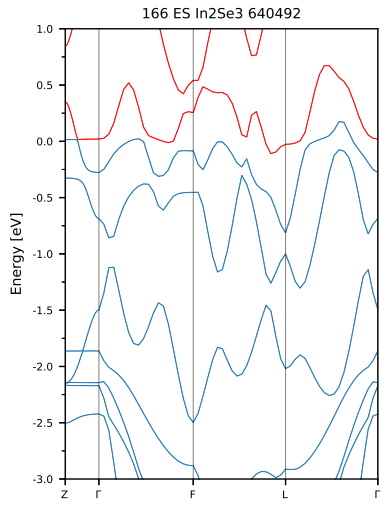

(qe) 166.3.37

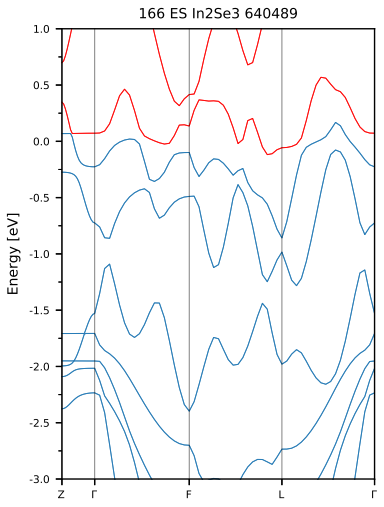

(qf) 166.3.38

Figure 378: Band structures for topological materials in SG 166, part 28/39, from 166.3.23 to 166.3.38. See the list of figures.

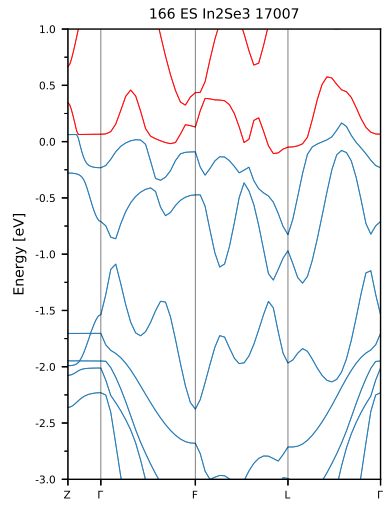

(qg) 166.3.39

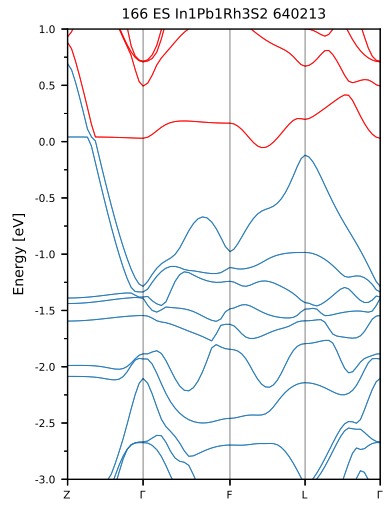

(qh) 166.3.40

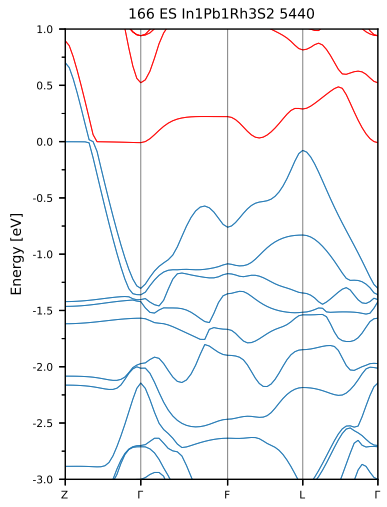

(qi) 166.3.41

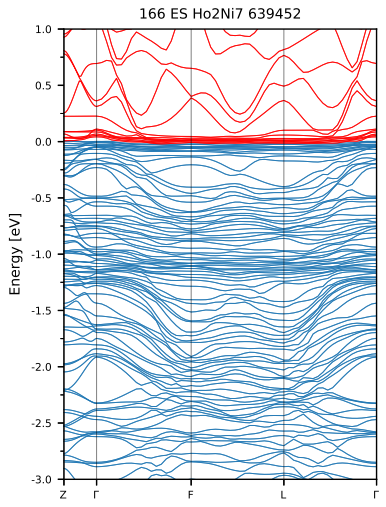

(qj) 166.3.42

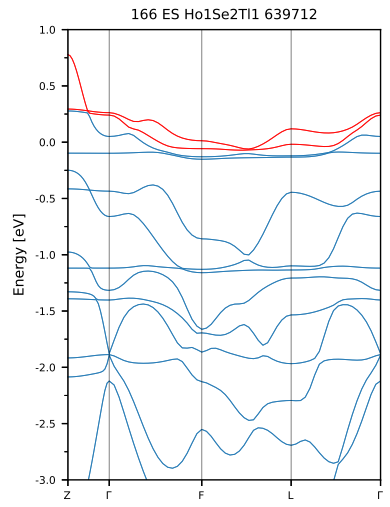

(qk) 166.3.43

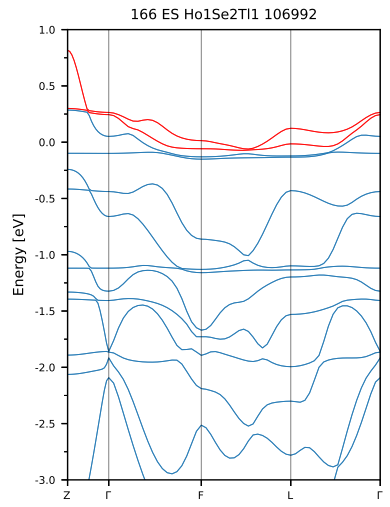

(ql) 166.3.44

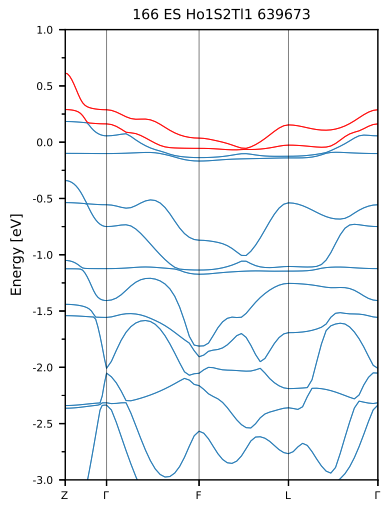

(qm) 166.3.45

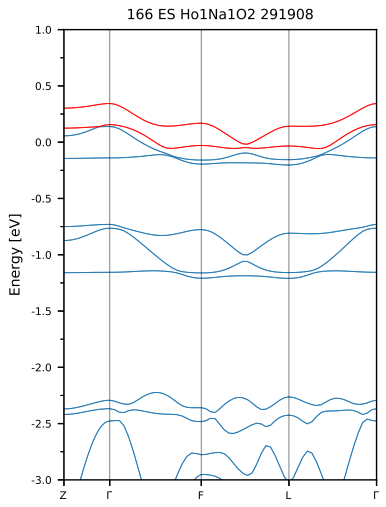

(qn) 166.3.46

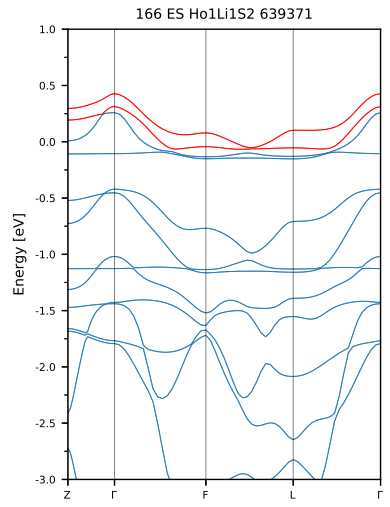

(qo) 166.3.47

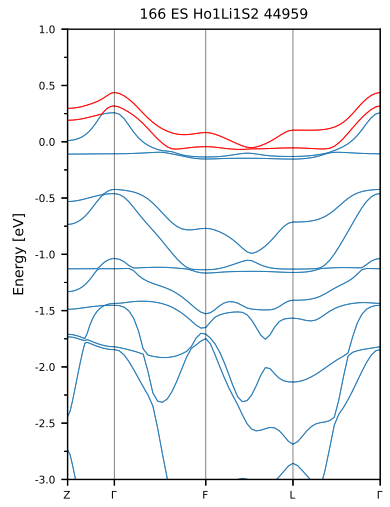

(qp) 166.3.48

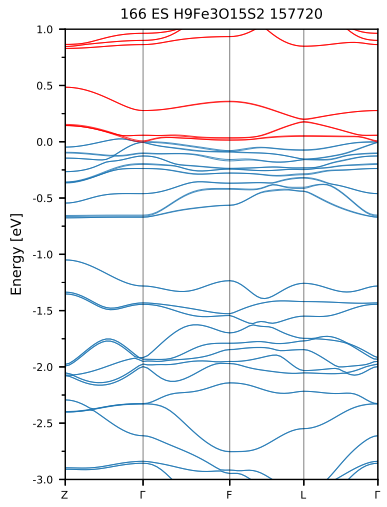

(qq) 166.3.49

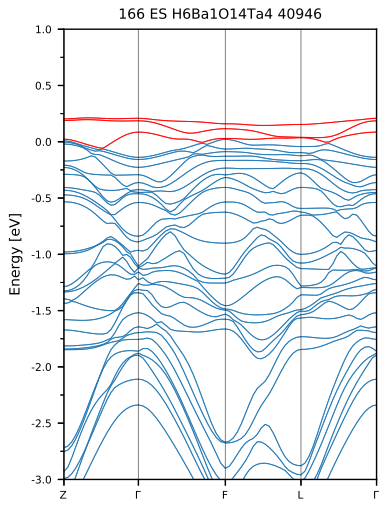

(qr) 166.3.50

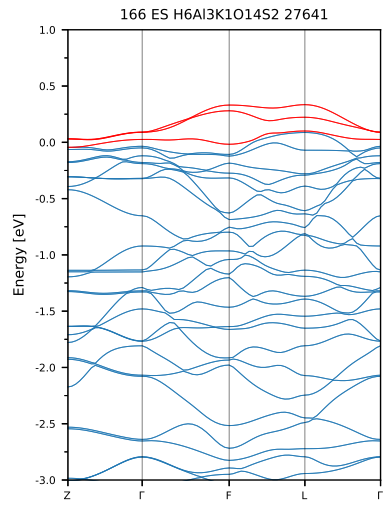

(qs) 166.3.51

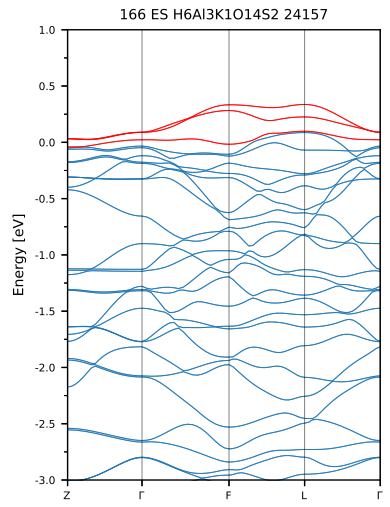

(qt) 166.3.52

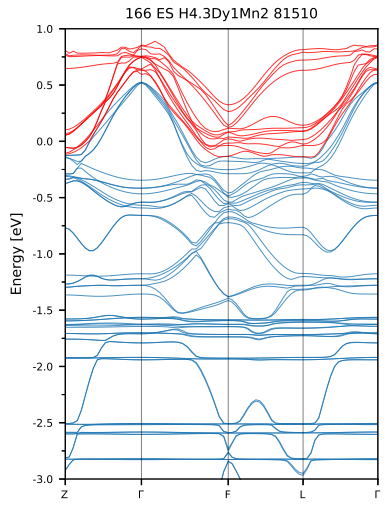

(qu) 166.3.53

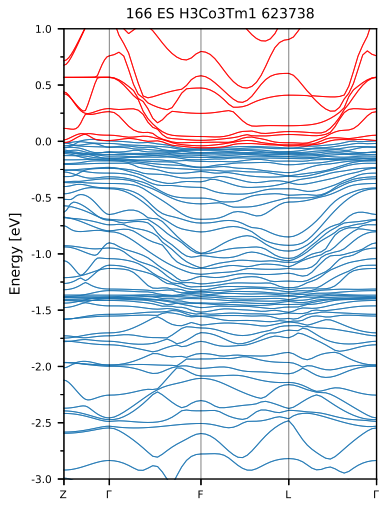

(qv) 166.3.54

Figure 379: Band structures for topological materials in SG 166, part 29/39, from 166.3.39 to 166.3.54. See the list of figures.

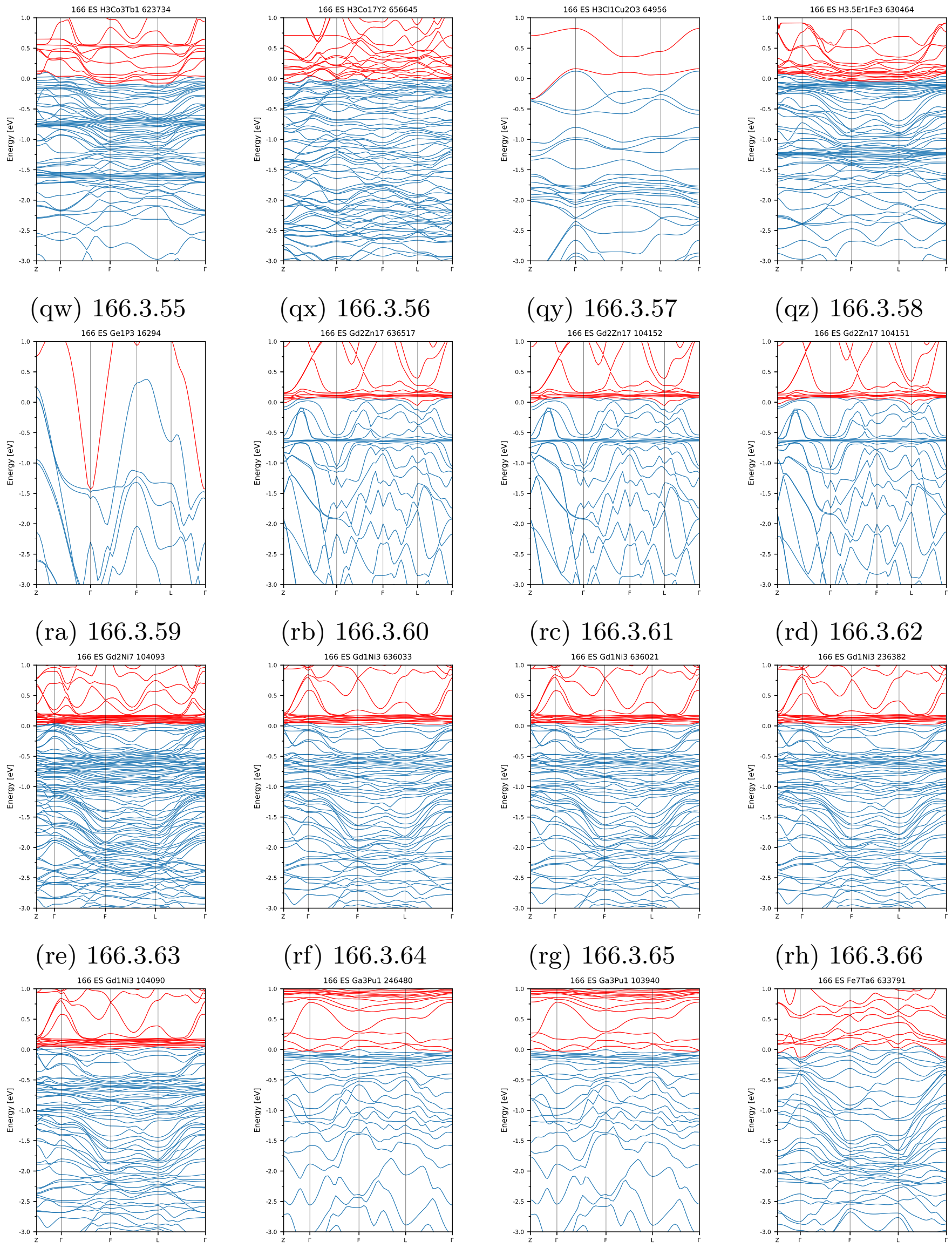

(qw) 166.3.55 (qx) 166.3.56 (qy) 166.3.57 (qz) 166.3.58

(ra) 166.3.59 (rb) 166.3.60 (rc) 166.3.61 (rd) 166.3.62

(re) 166.3.63 (rf) 166.3.64 (rg) 166.3.65 (rh) 166.3.66

(ri) 166.3.67 (rj) 166.3.68 (rk) 166.3.69 (rl) 166.3.70

Figure 380: Band structures for topological materials in SG 166, part 30/39, from 166.3.55 to 166.3.70. See the list of figures.

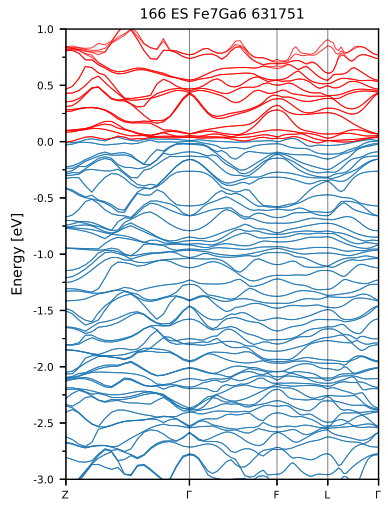

(rm) 166.3.71

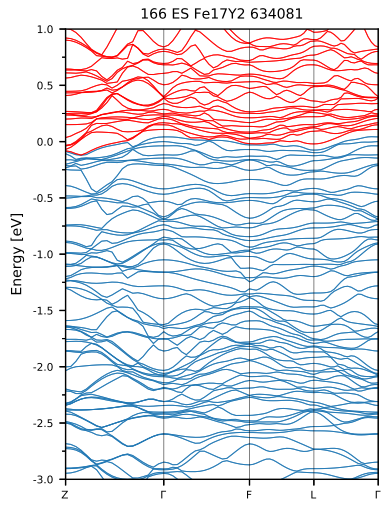

(rn) 166.3.72

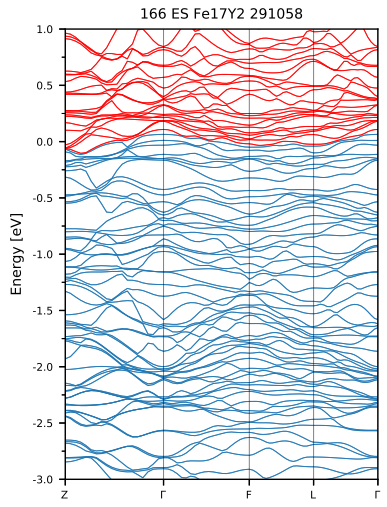

(ro) 166.3.73

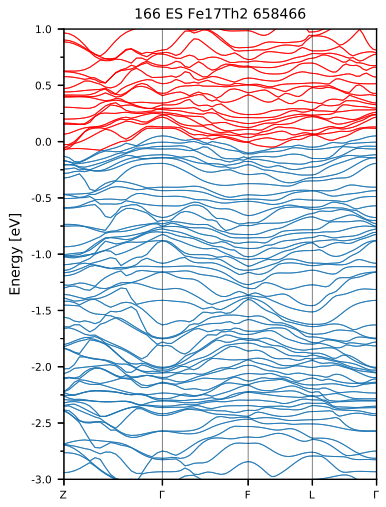

(rp) 166.3.74

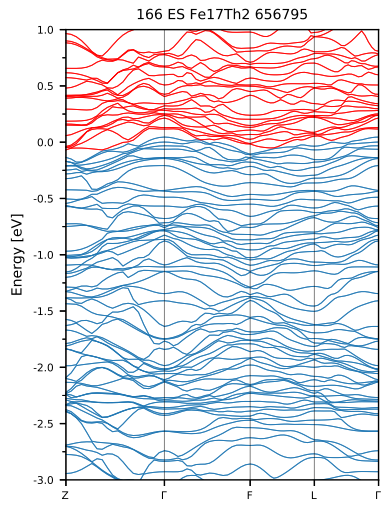

(rq) 166.3.75

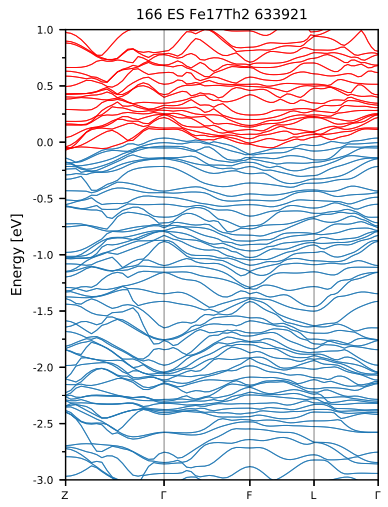

(rr) 166.3.76

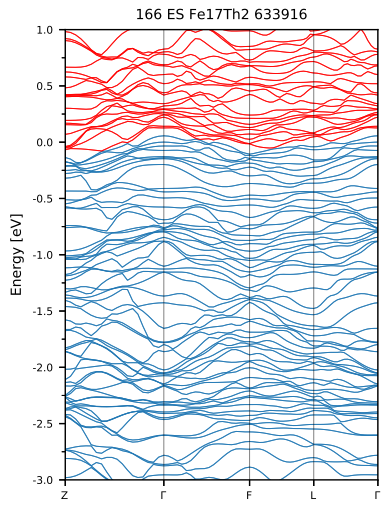

(rs) 166.3.77

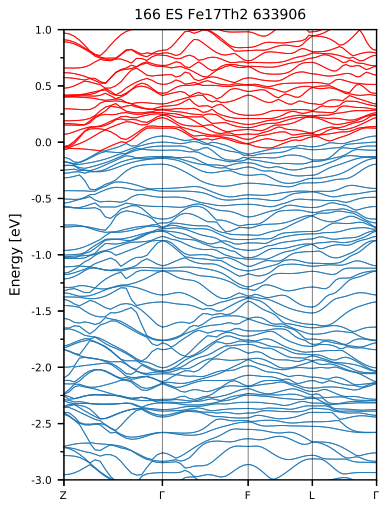

(rt) 166.3.78

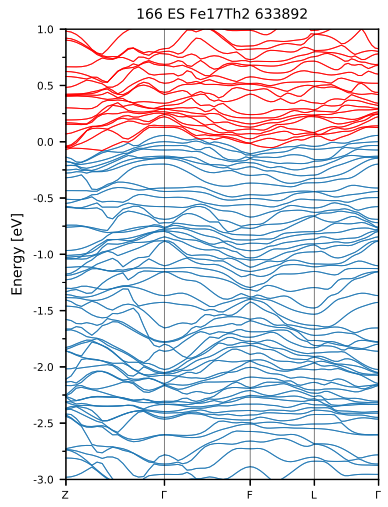

(ru) 166.3.79

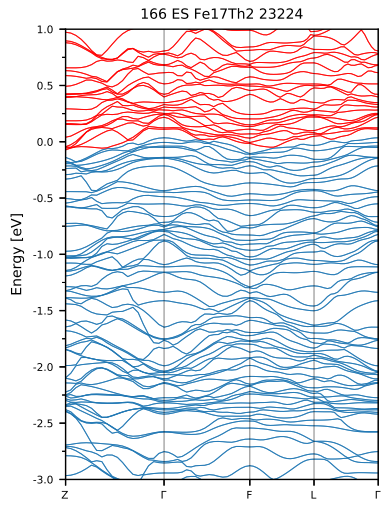

(rv) 166.3.80

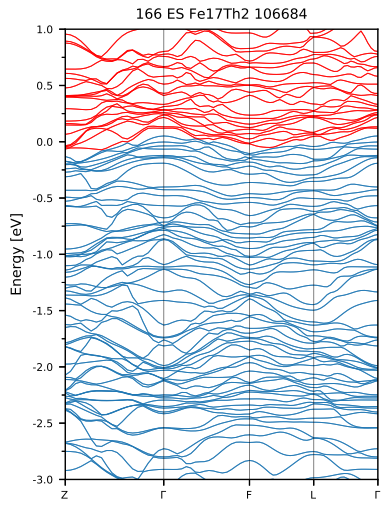

(rw) 166.3.81

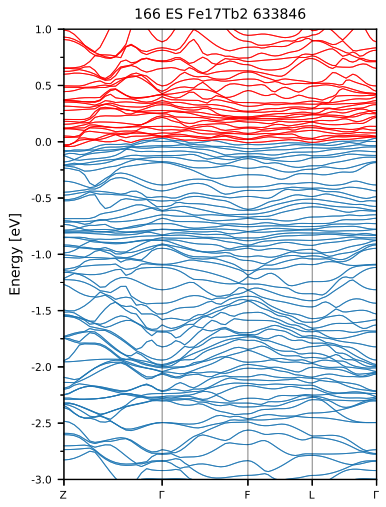

(rx) 166.3.82

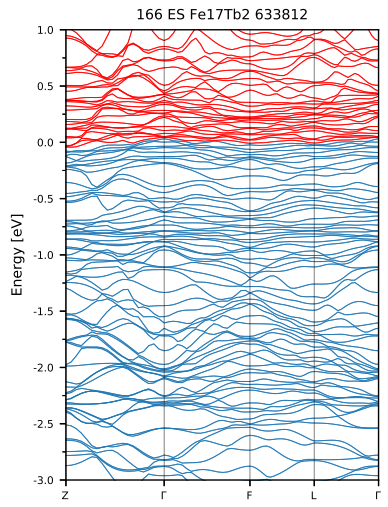

(ry) 166.3.83

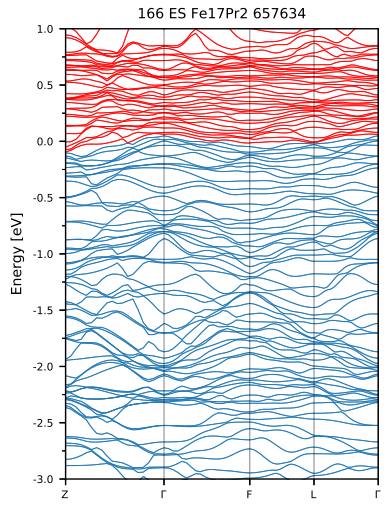

(rz) 166.3.84

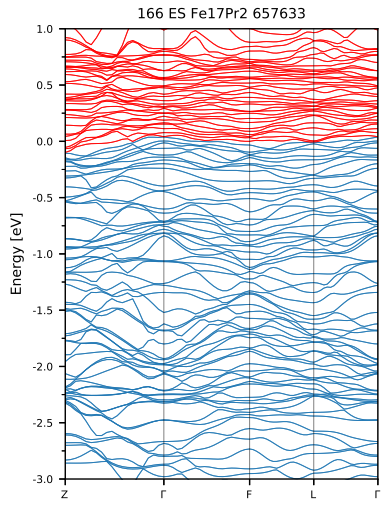

(sa) 166.3.85

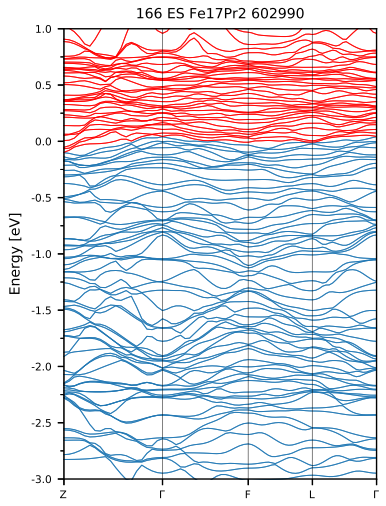

(sb) 166.3.86

Figure 381: Band structures for topological materials in SG 166, part 31/39, from 166.3.71 to 166.3.86. See the list of figures.

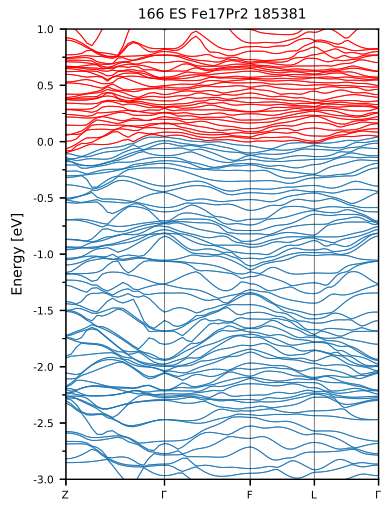

(sc) 166.3.87

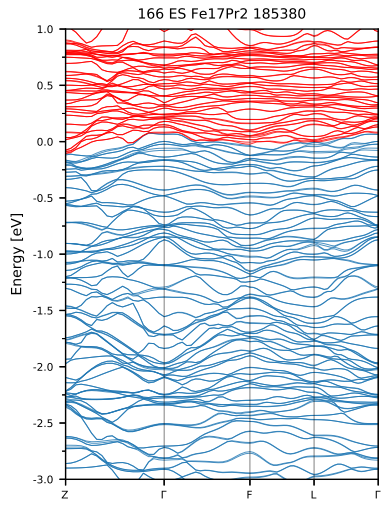

(sd) 166.3.88

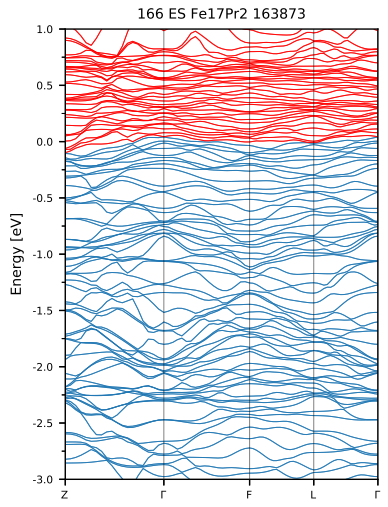

(se) 166.3.89

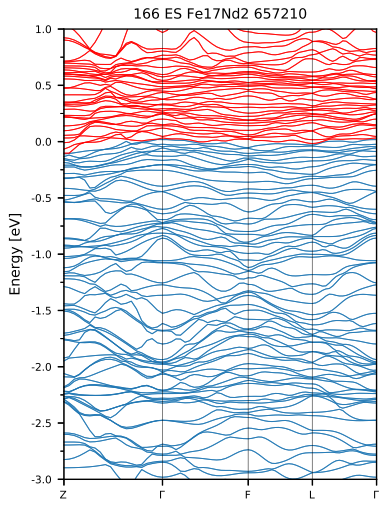

(sf) 166.3.90

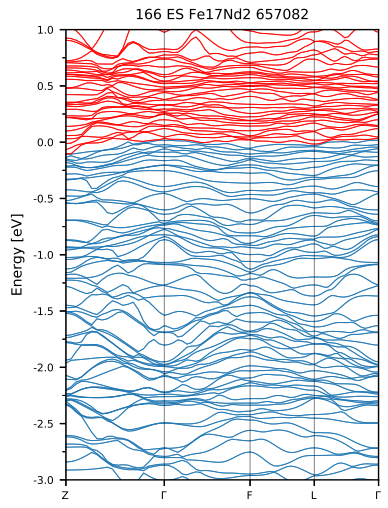

(sg) 166.3.91

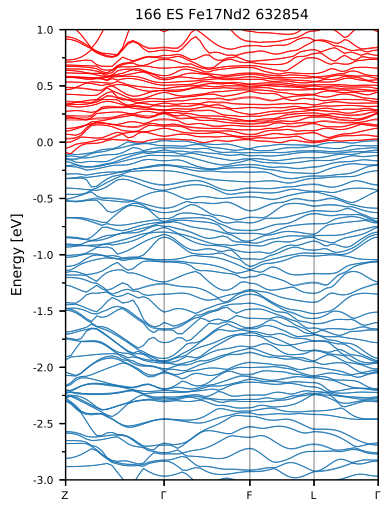

(sh) 166.3.92

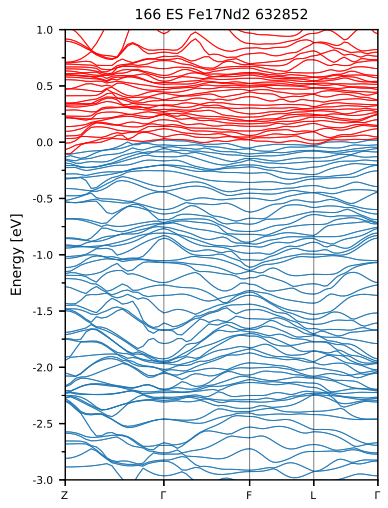

(si) 166.3.93

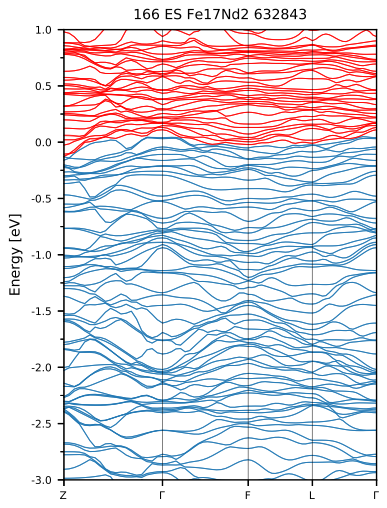

(sj) 166.3.94

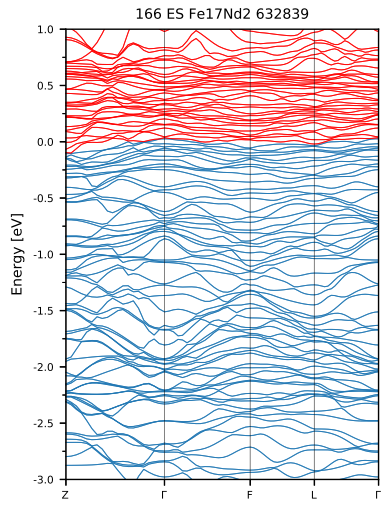

(sk) 166.3.95

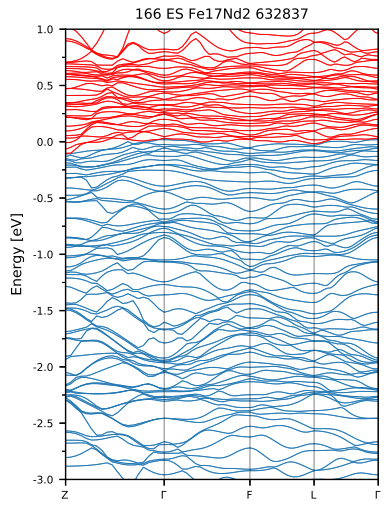

(sl) 166.3.96

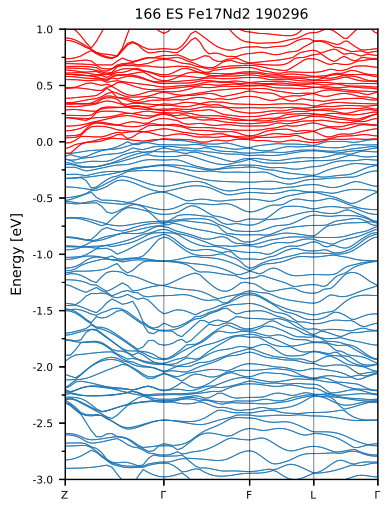

(sm) 166.3.97

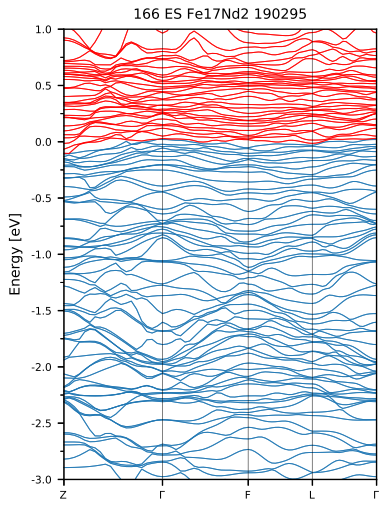

(sn) 166.3.98

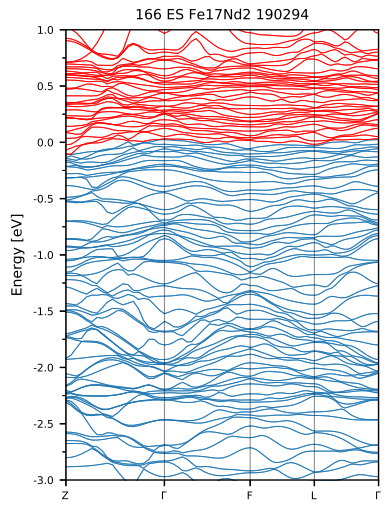

(so) 166.3.99

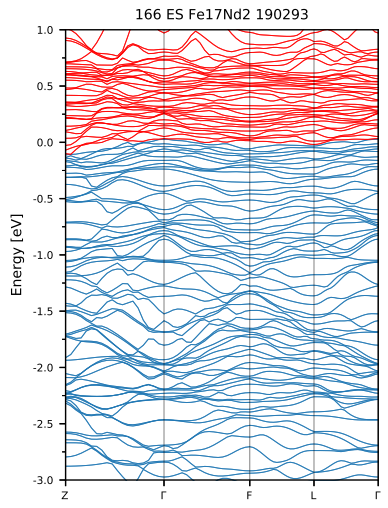

(sp) 166.3.100

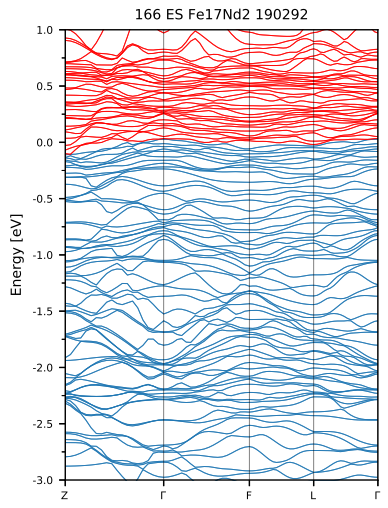

(sq) 166.3.101

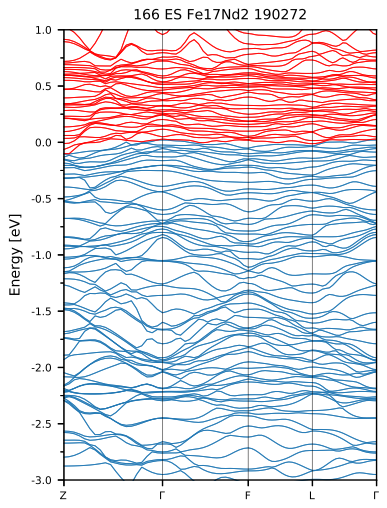

(sr) 166.3.102

Figure 382: Band structures for topological materials in SG 166, part 32/39, from 166.3.87 to 166.3.102. See the list of figures.

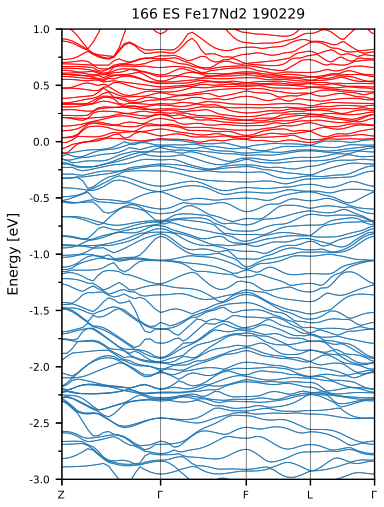

(ss) 166.3.103

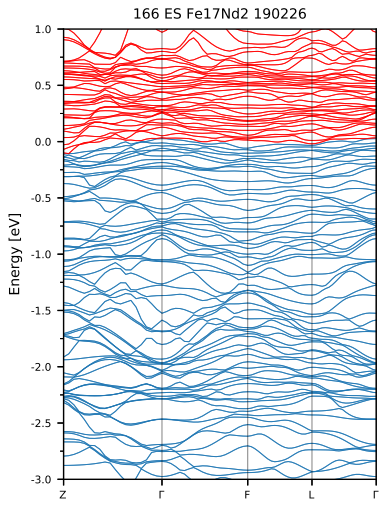

(st) 166.3.104

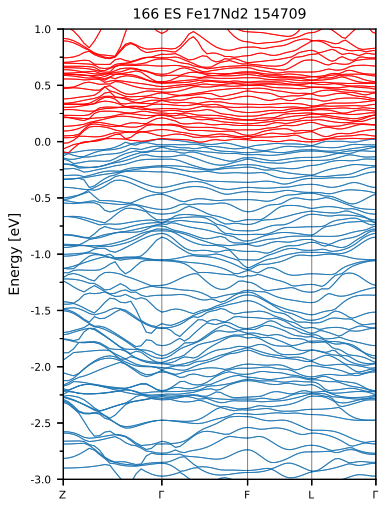

(su) 166.3.105

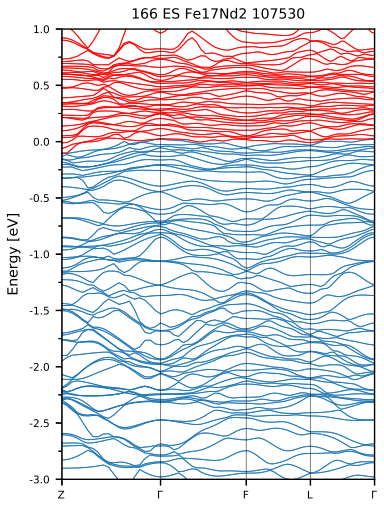

(sv) 166.3.106

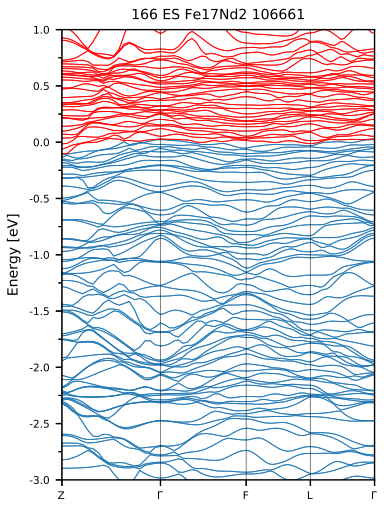

(sw) 166.3.107

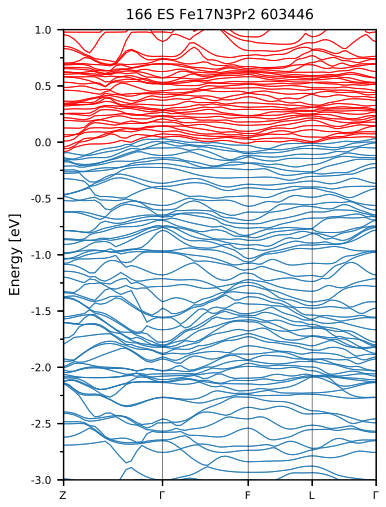

(sx) 166.3.108

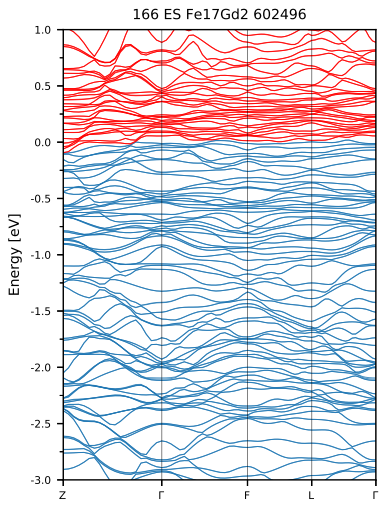

(sy) 166.3.109

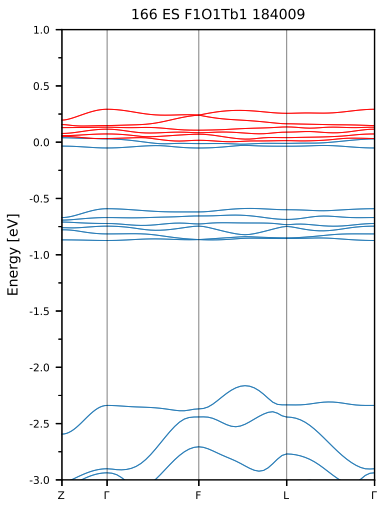

(sz) 166.3.110

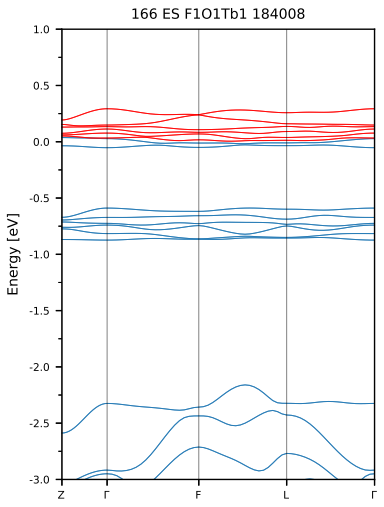

(ta) 166.3.111

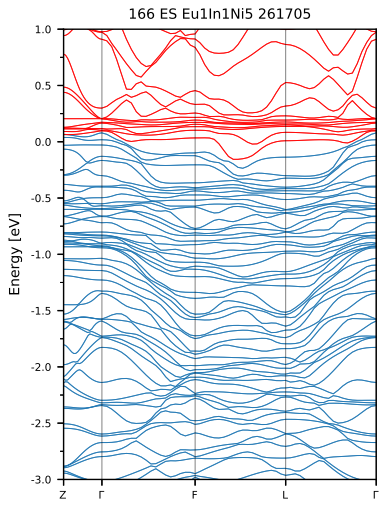

(tb) 166.3.112

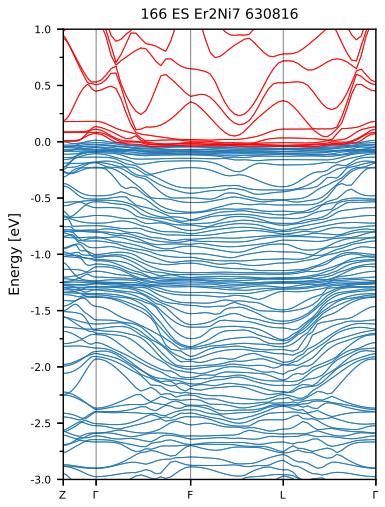

(tc) 166.3.113

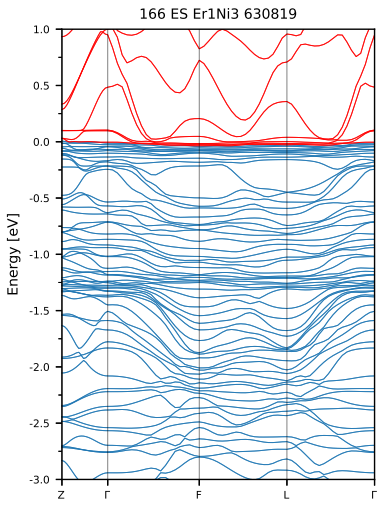

(td) 166.3.114

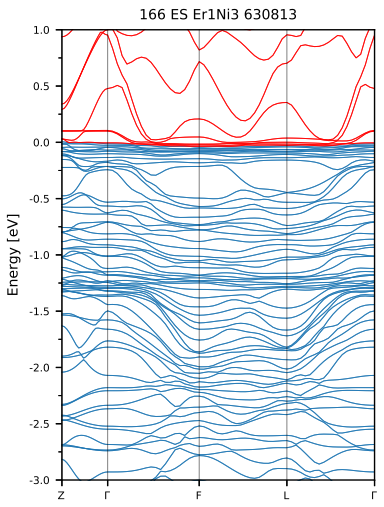

(te) 166.3.115

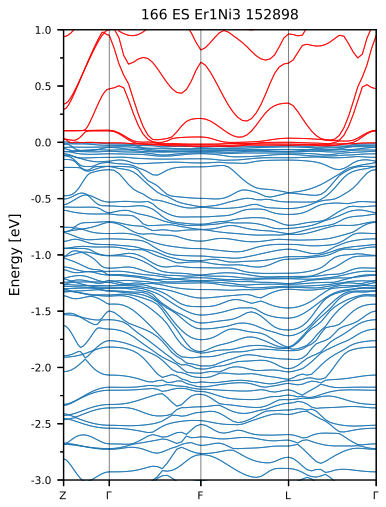

(tf) 166.3.116

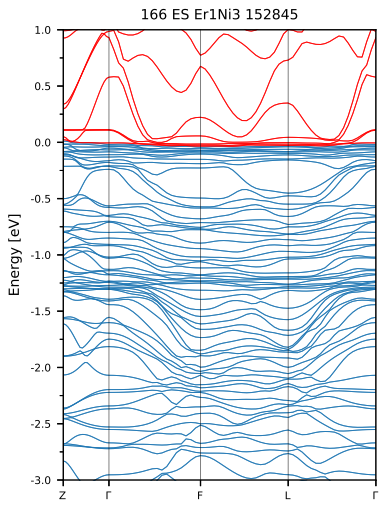

(tg) 166.3.117

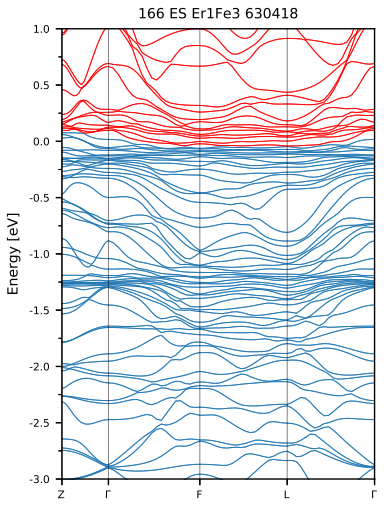

(th) 166.3.118

Figure 383: Band structures for topological materials in SG 166, part 33/39, from 166.3.103 to 166.3.118. See the list of figures.

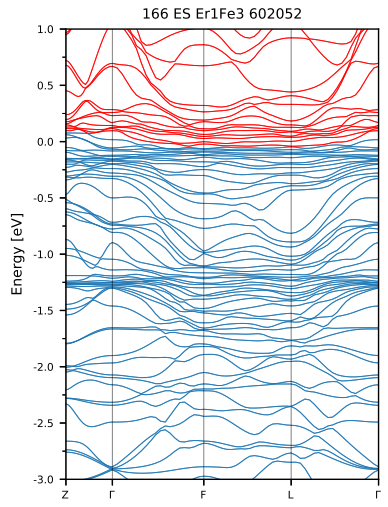

(ti) 166.3.119

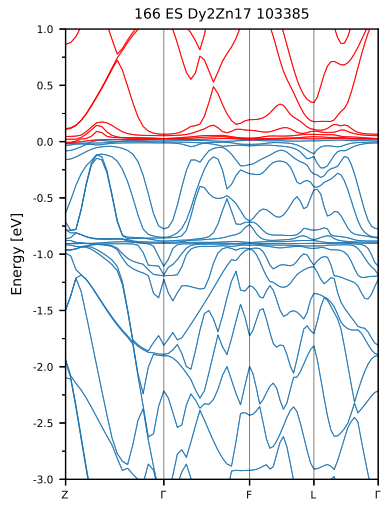

(tj) 166.3.120

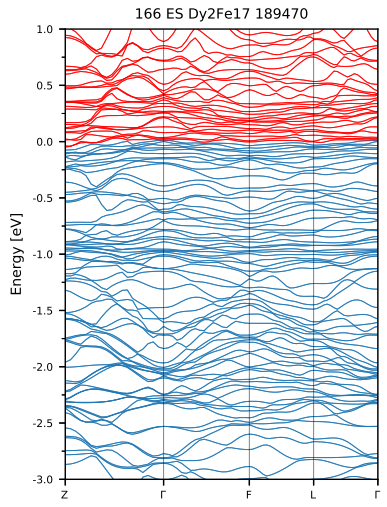

(tk) 166.3.121

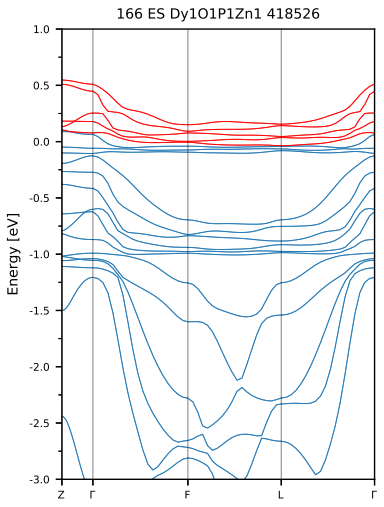

(tl) 166.3.122

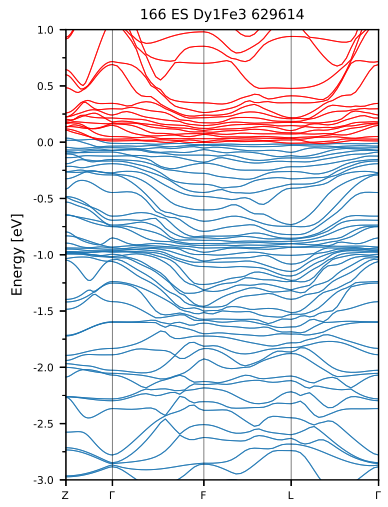

(tm) 166.3.123

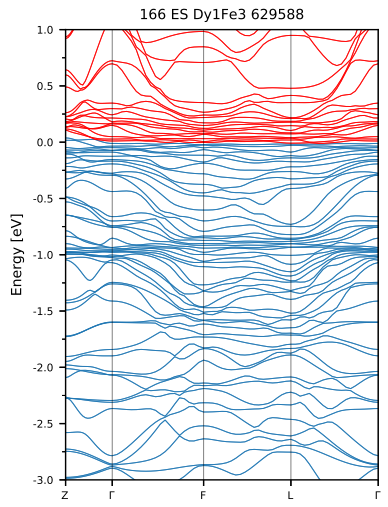

(tn) 166.3.124

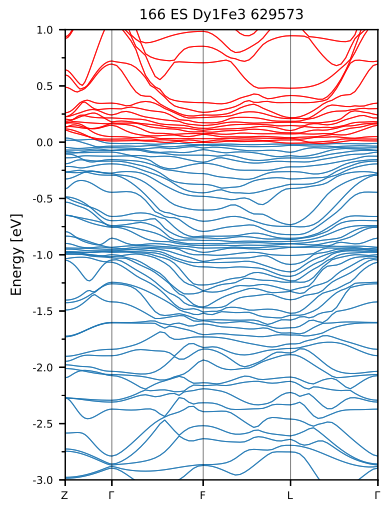

(to) 166.3.125

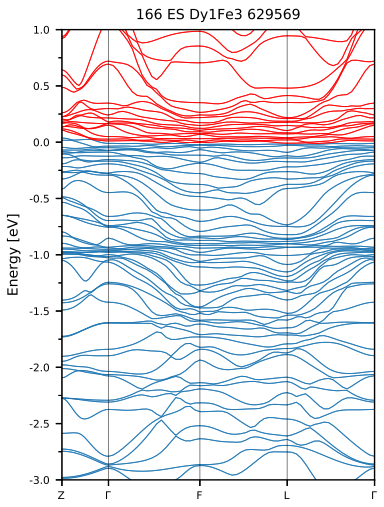

(tp) 166.3.126

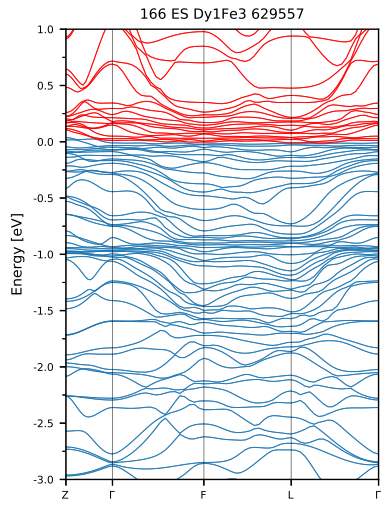

(tq) 166.3.127

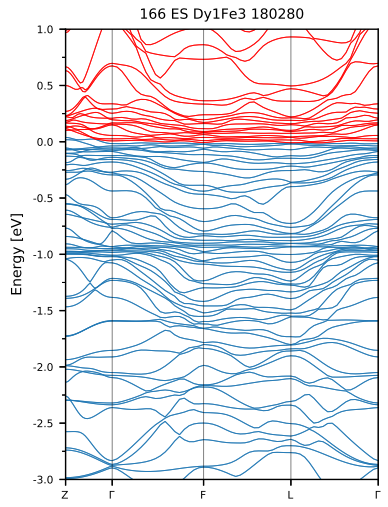

(tr) 166.3.128

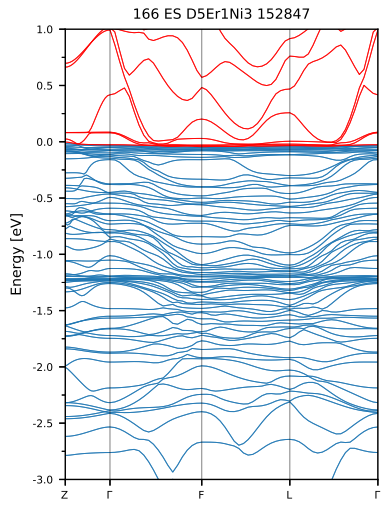

(ts) 166.3.129

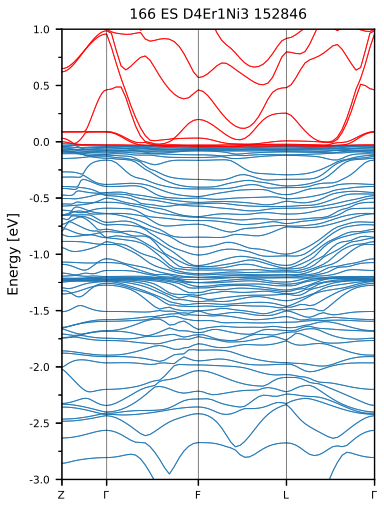

(tt) 166.3.130

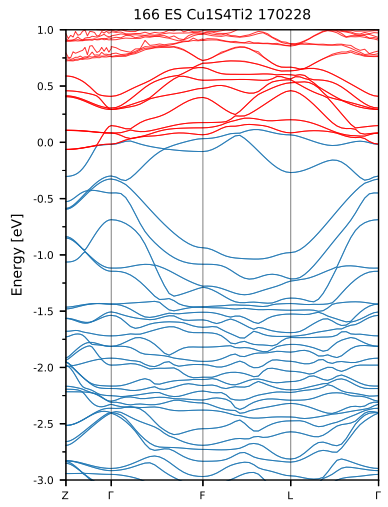

(tu) 166.3.131

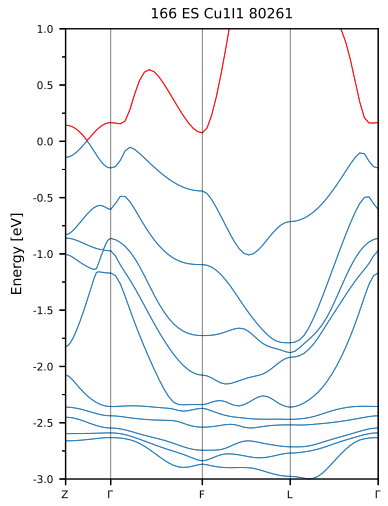

(tv) 166.3.132

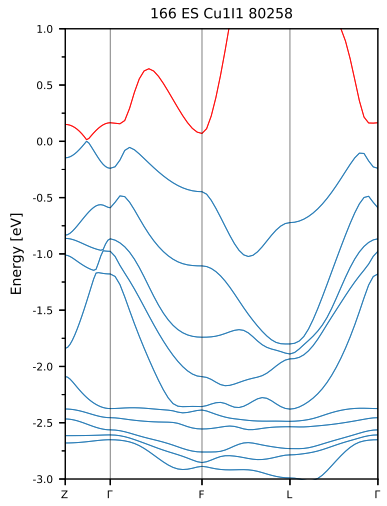

(tw) 166.3.133

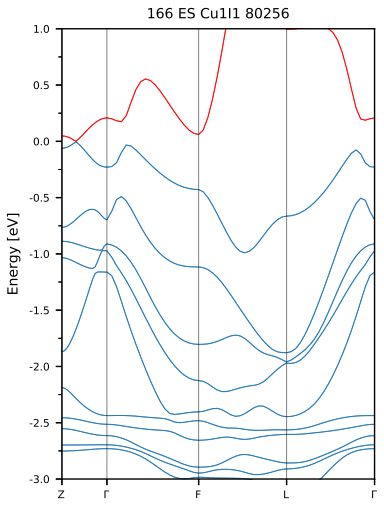

(tx) 166.3.134

Figure 384: Band structures for topological materials in SG 166, part 34/39, from 166.3.119 to 166.3.134. See the list of figures.

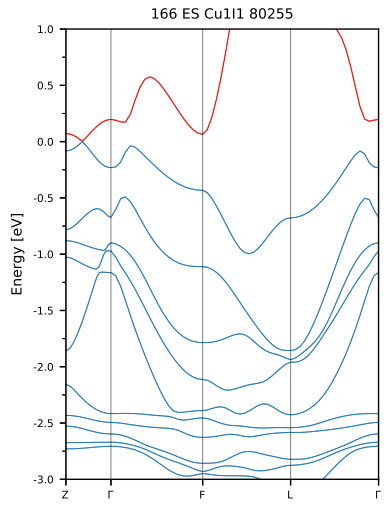

(ty) 166.3.135

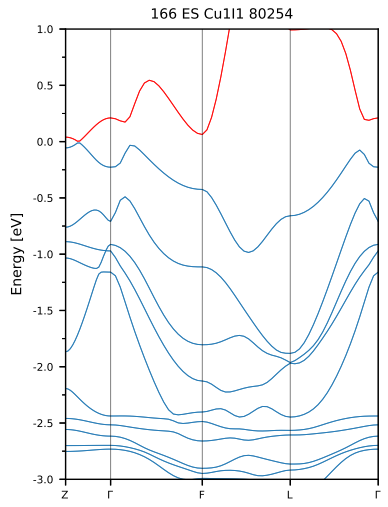

(tz) 166.3.136

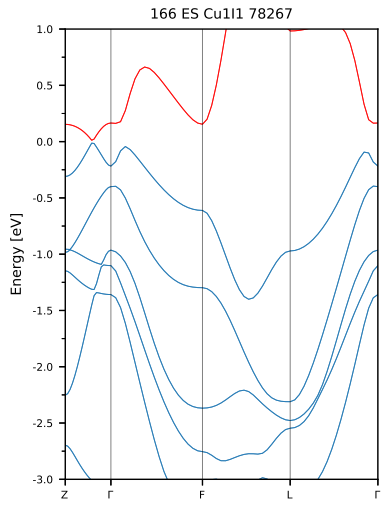

(ua) 166.3.137

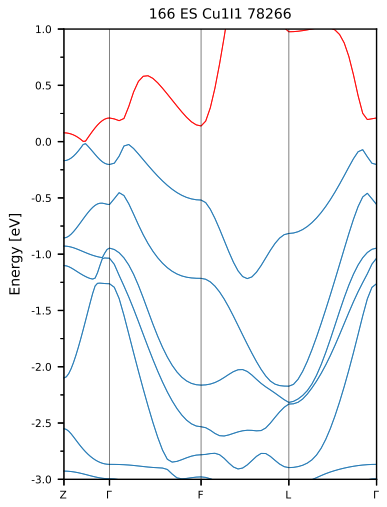

(ub) 166.3.138

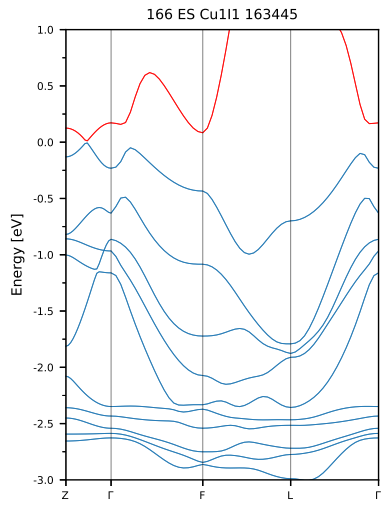

(uc) 166.3.139

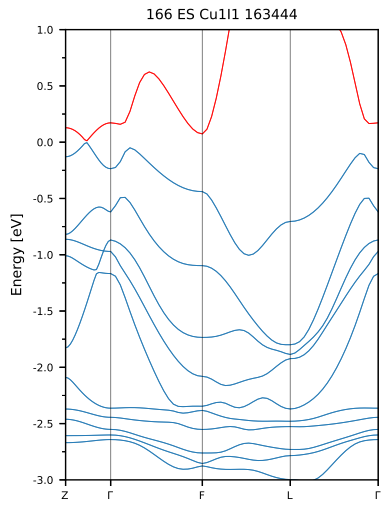

(ud) 166.3.140

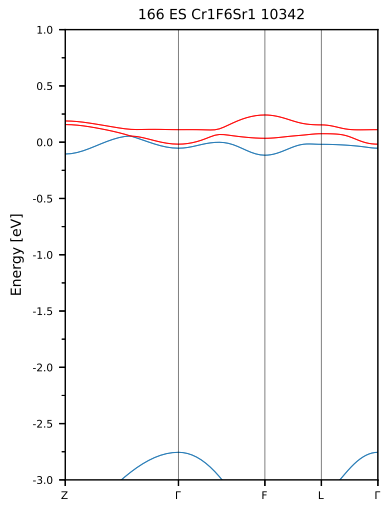

(ue) 166.3.141

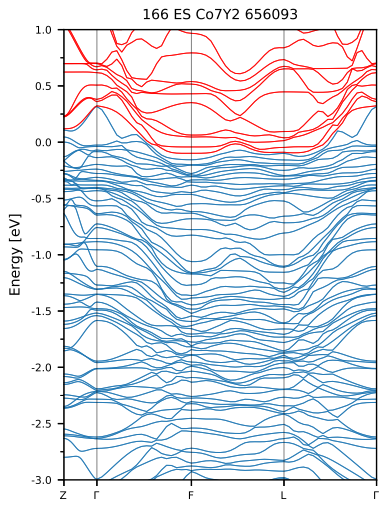

(uf) 166.3.142

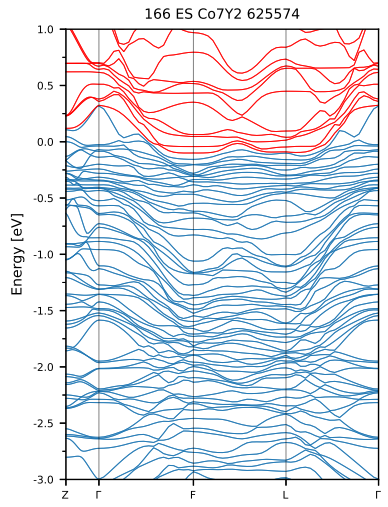

(ug) 166.3.143

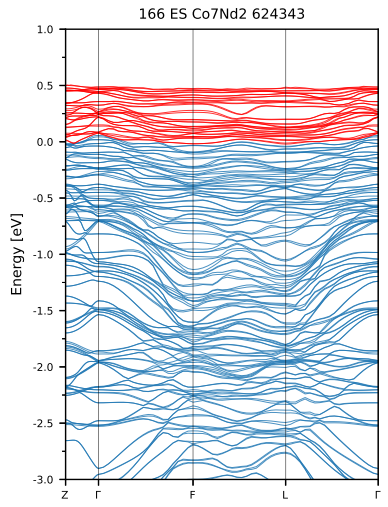

(uh) 166.3.144

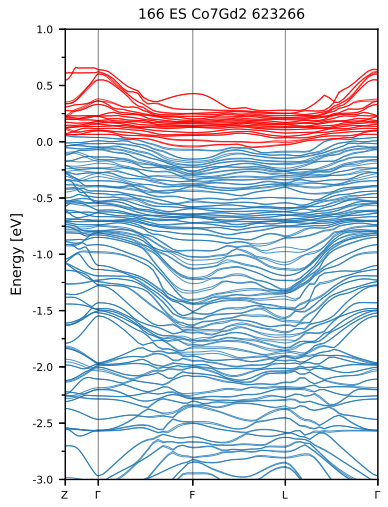

(ui) 166.3.145

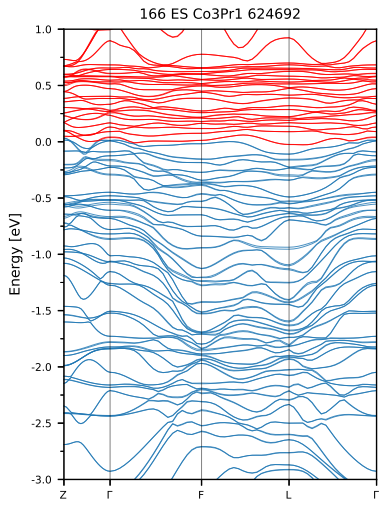

(uj) 166.3.146

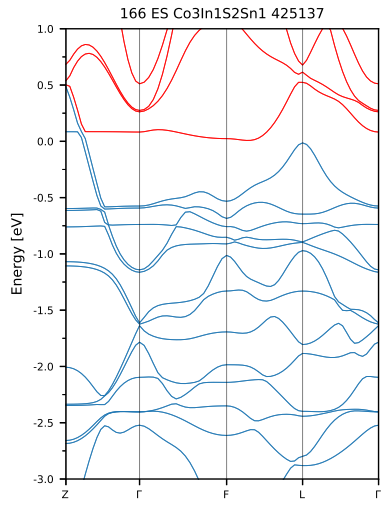

(uk) 166.3.147

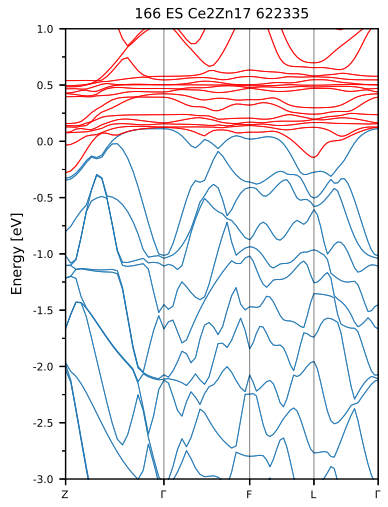

(ul) 166.3.148

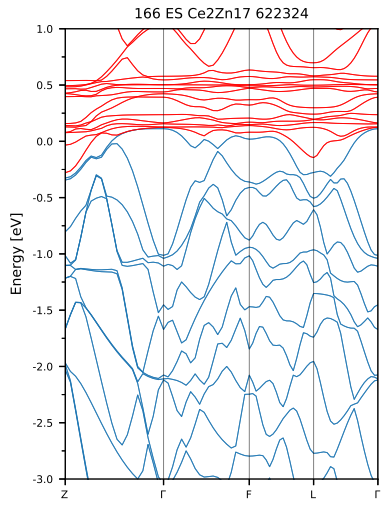

(um) 166.3.149

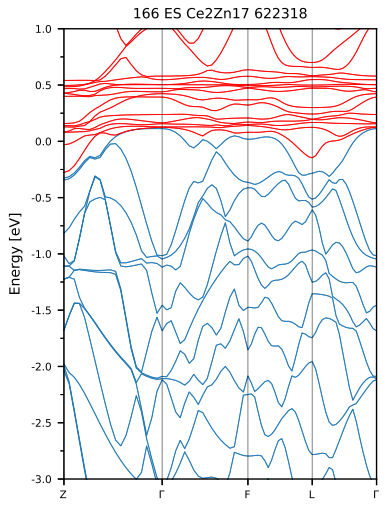

(un) 166.3.150

Figure 385: Band structures for topological materials in SG 166, part 35/39, from 166.3.135 to 166.3.150. See the list of figures.

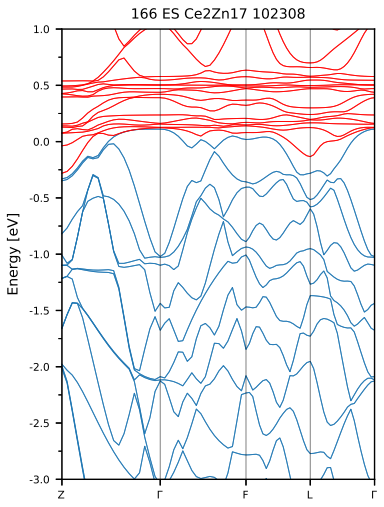

(uo) 166.3.151

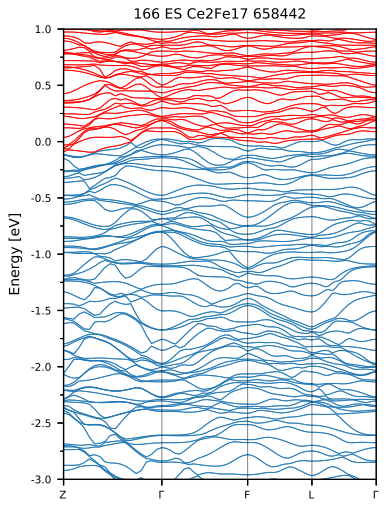

(up) 166.3.152

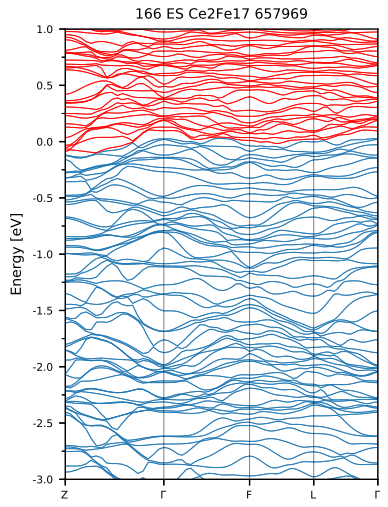

(uq) 166.3.153

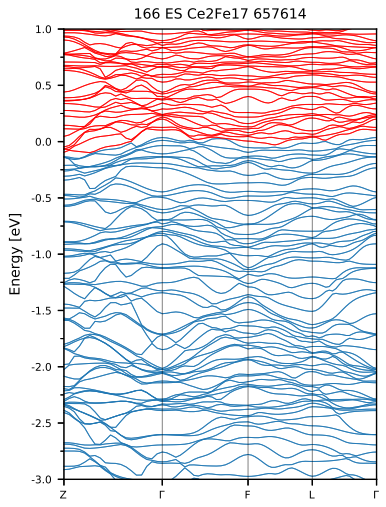

(ur) 166.3.154

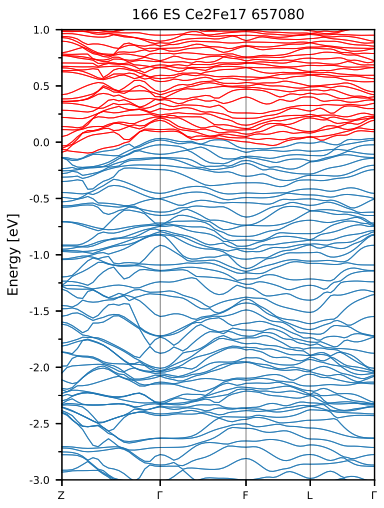

(us) 166.3.155

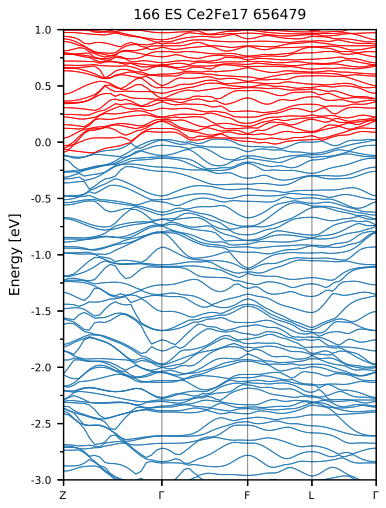

(ut) 166.3.156

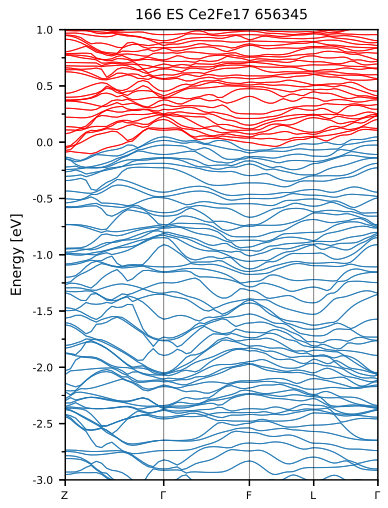

(uu) 166.3.157

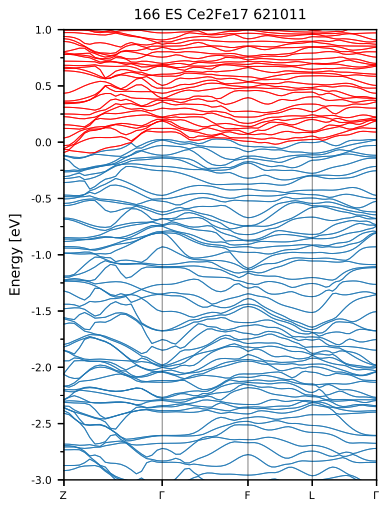

(uv) 166.3.158

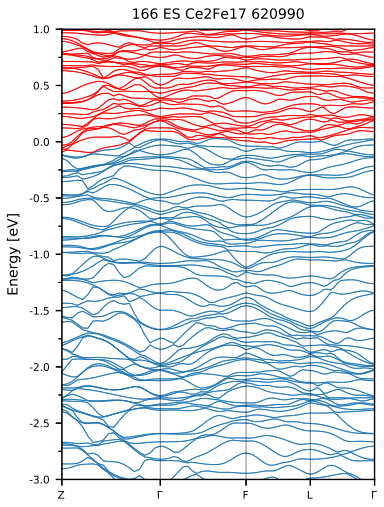

(uw) 166.3.159

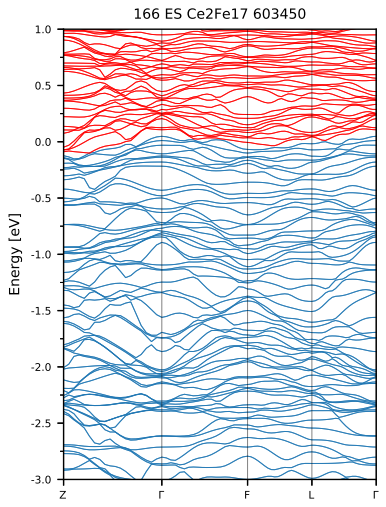

(ux) 166.3.160

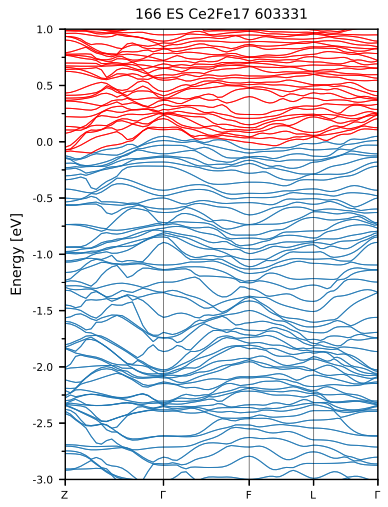

(uy) 166.3.161

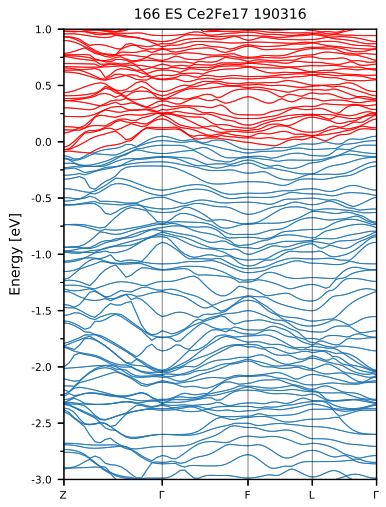

(uz) 166.3.162

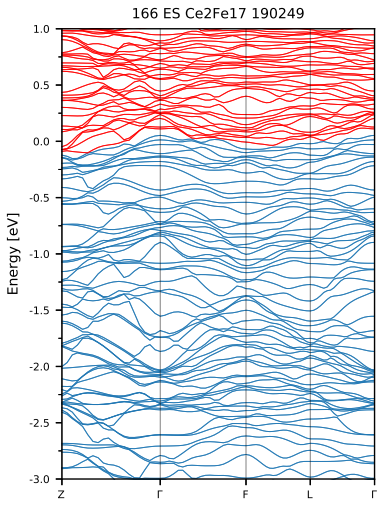

(va) 166.3.163

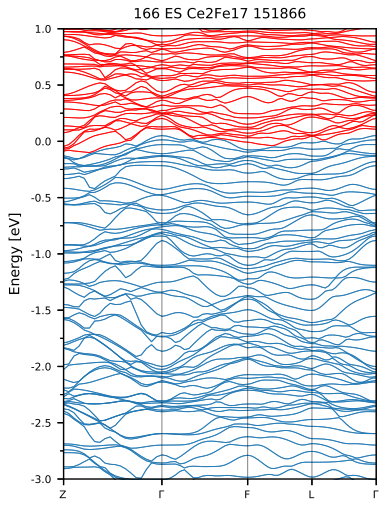

(vb) 166.3.164

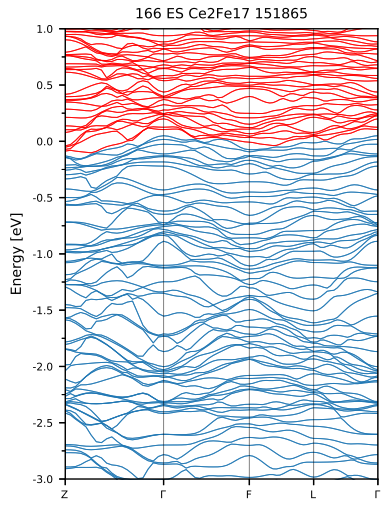

(vc) 166.3.165

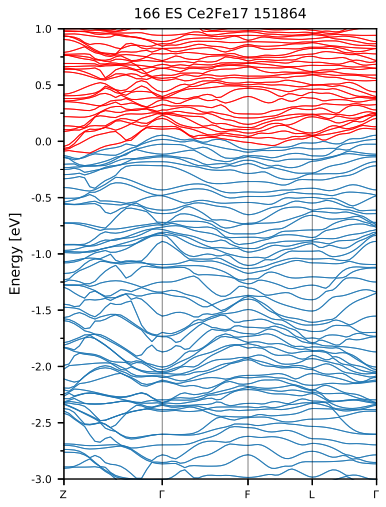

(vd) 166.3.166

Figure 386: Band structures for topological materials in SG 166, part 36/39, from 166.3.151 to 166.3.166. See the list of figures.

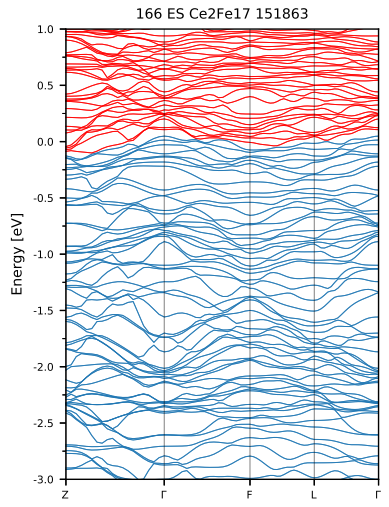

(ve) 166.3.167

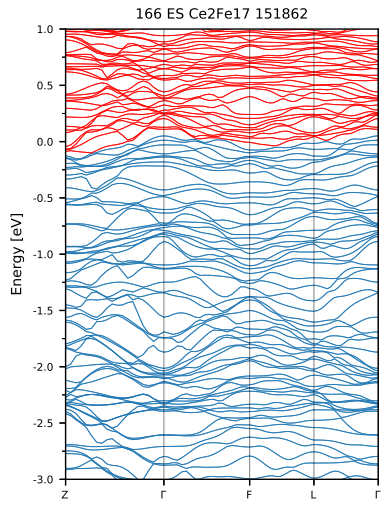

(vf) 166.3.168

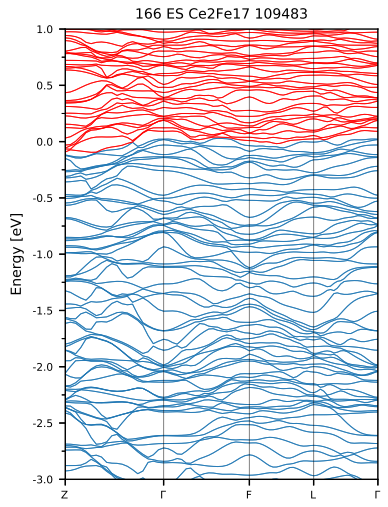

(vg) 166.3.169

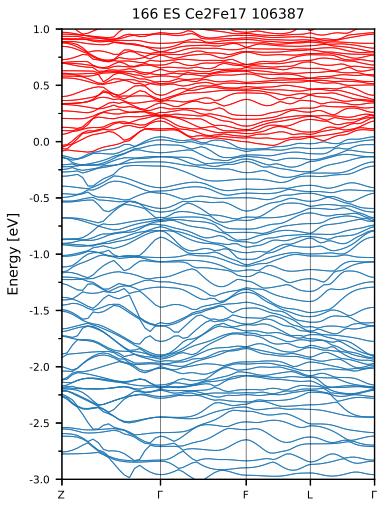

(vh) 166.3.170

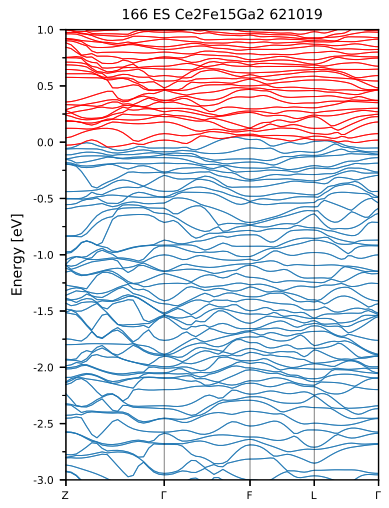

(vi) 166.3.171

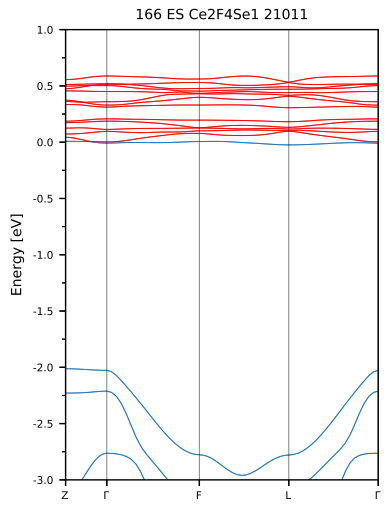

(vj) 166.3.172

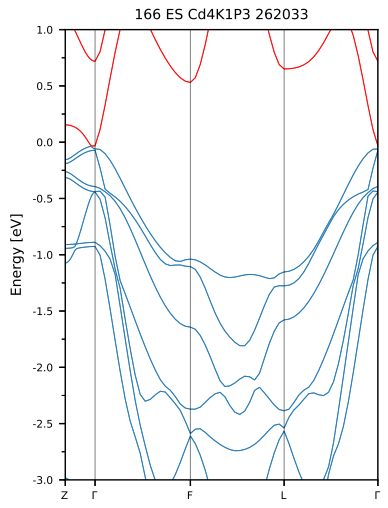

(vk) 166.3.173

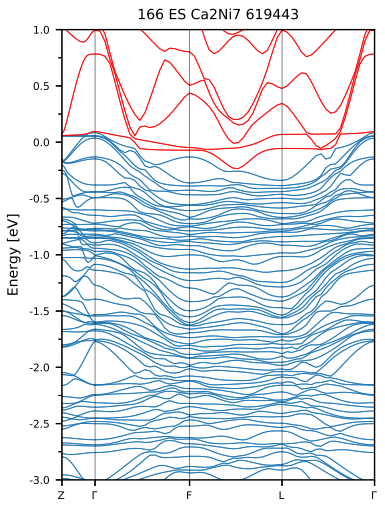

(vl) 166.3.174

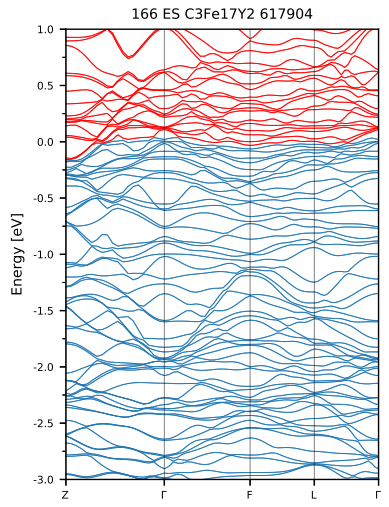

(vm) 166.3.175

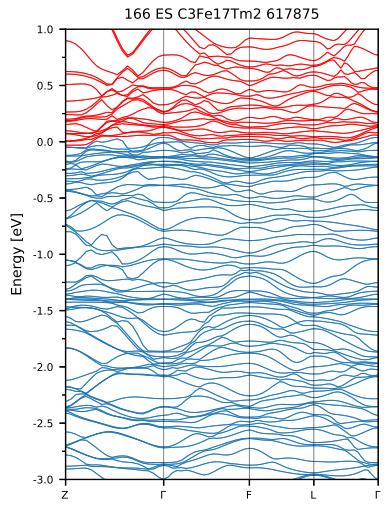

(vn) 166.3.176

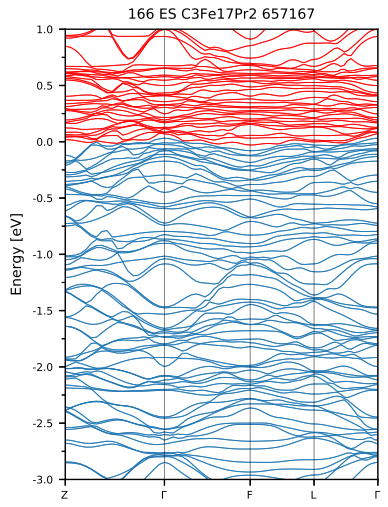

(vo) 166.3.177

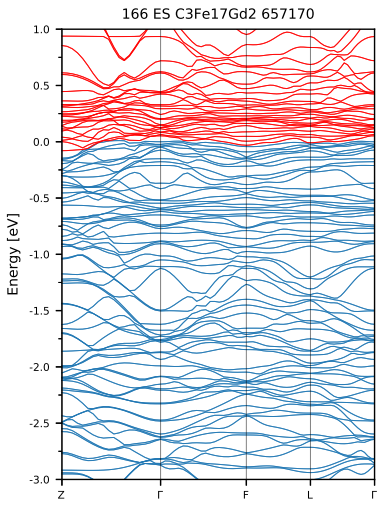

(vp) 166.3.178

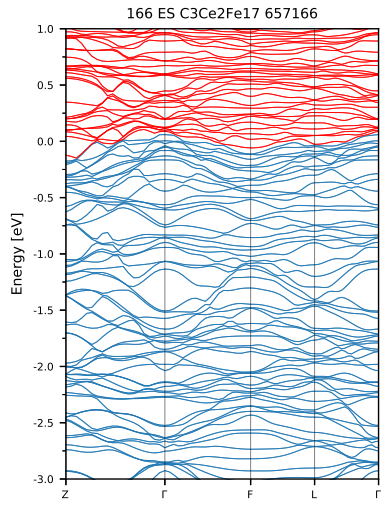

(vq) 166.3.179

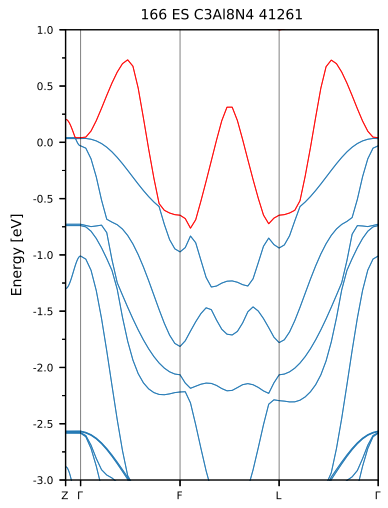

(vr) 166.3.180

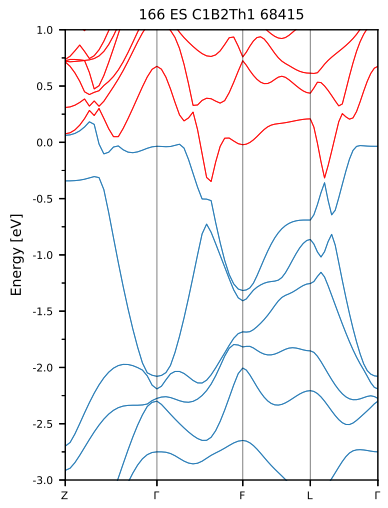

(vs) 166.3.181

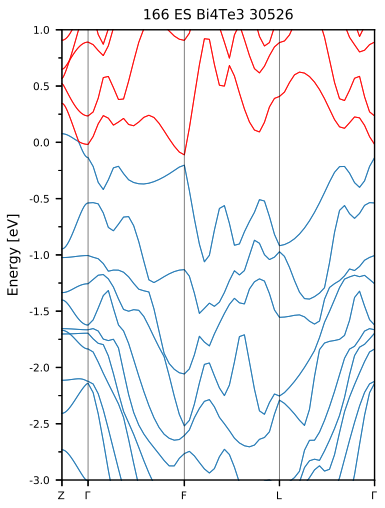

(vt) 166.3.182

Figure 387: Band structures for topological materials in SG 166, part 37/39, from 166.3.167 to 166.3.182. See the list of figures.

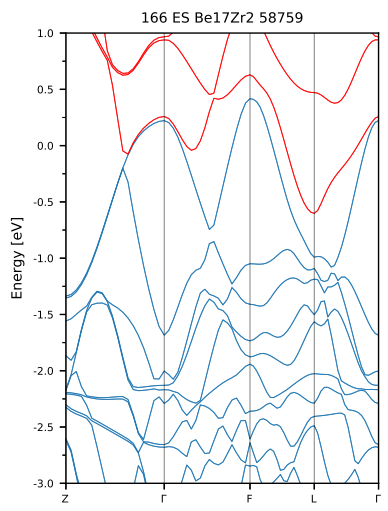

(vu) 166.3.183

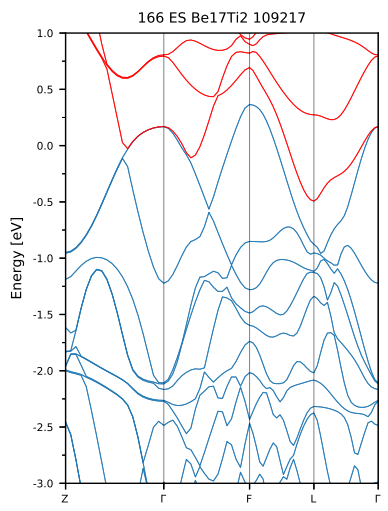

(vv) 166.3.184

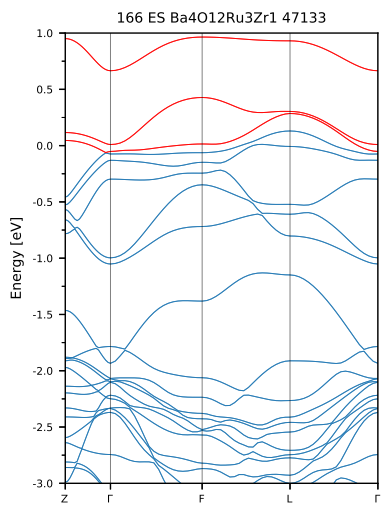

(vw) 166.3.185

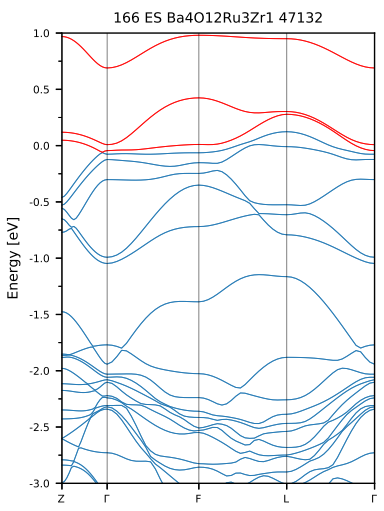

(vx) 166.3.186

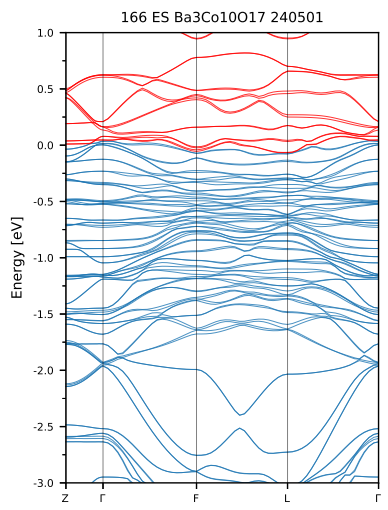

(vy) 166.3.187

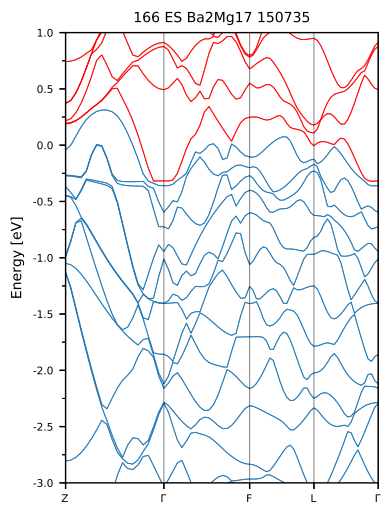

(vz) 166.3.188

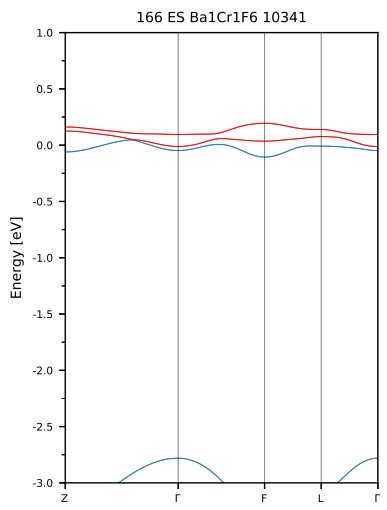

(wa) 166.3.189

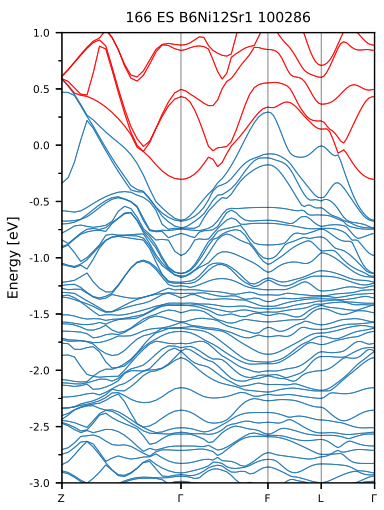

(wb) 166.3.190

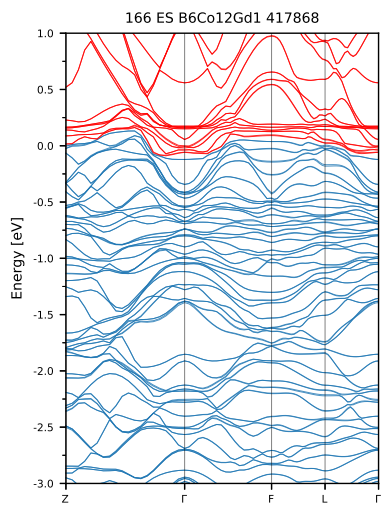

(wc) 166.3.191

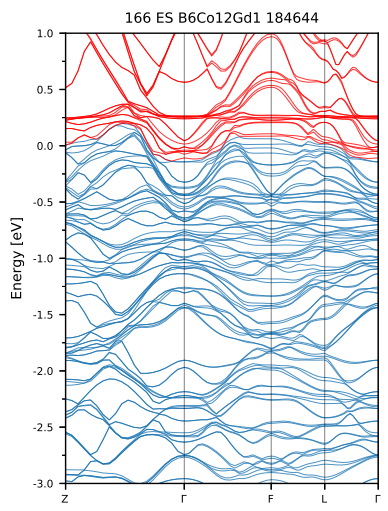

(wd) 166.3.192

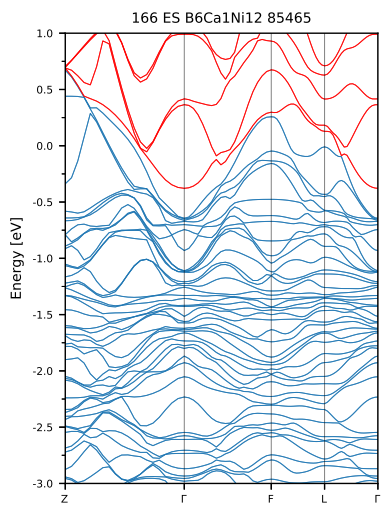

(we) 166.3.193

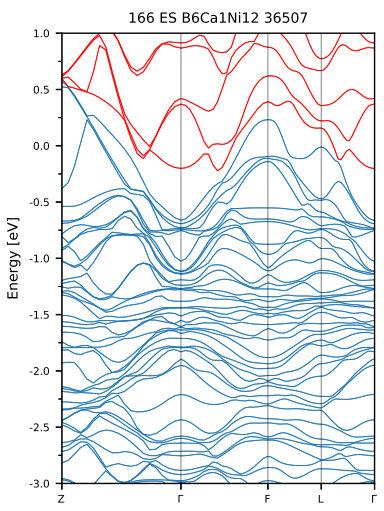

(wf) 166.3.194

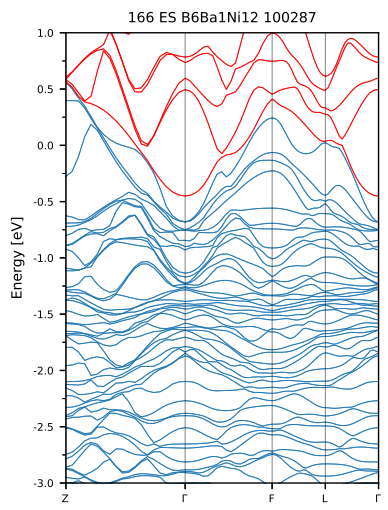

(wg) 166.3.195

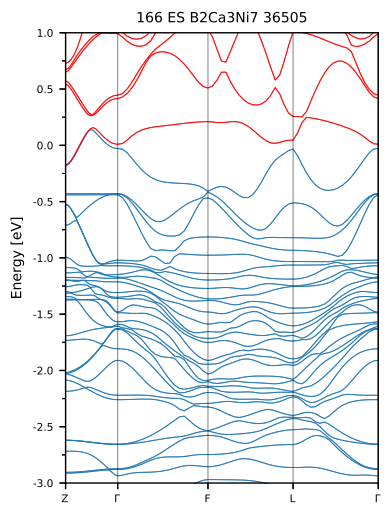

(wh) 166.3.196

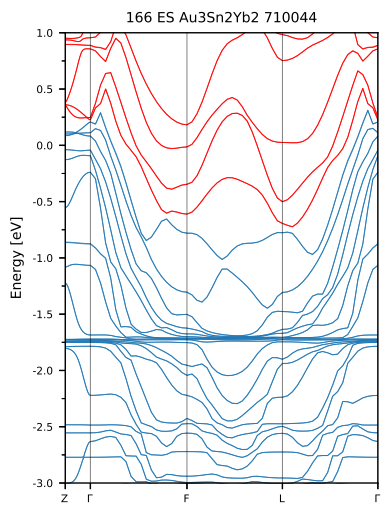

(wi) 166.3.197

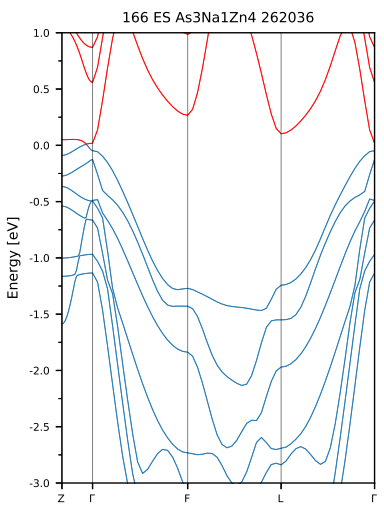

(wj) 166.3.198

Figure 388: Band structures for topological materials in SG 166, part 38/39, from 166.3.183 to 166.3.198. See the list of figures.

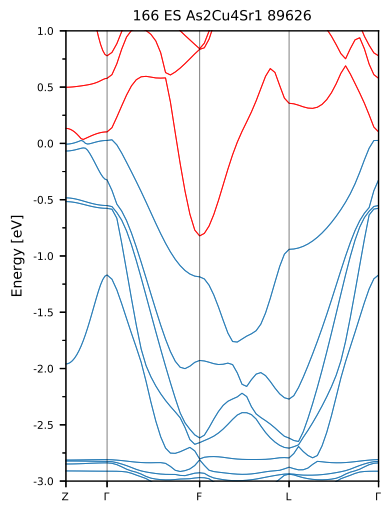

(wk) 166.3.199

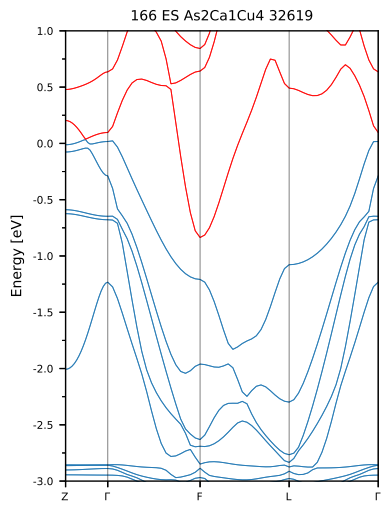

(wl) 166.3.200

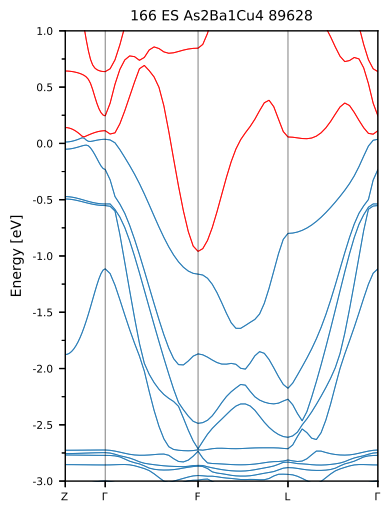

(wm) 166.3.201

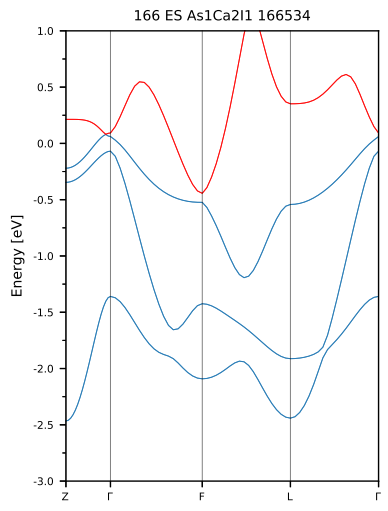

(wn) 166.3.202

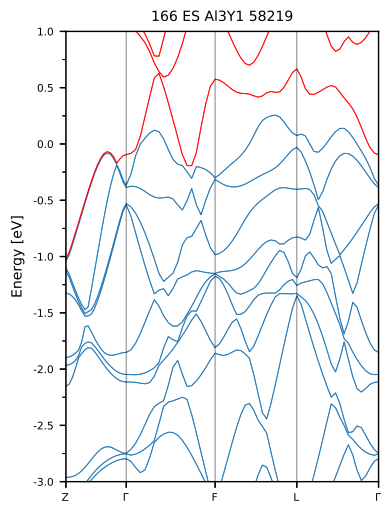

(wo) 166.3.203

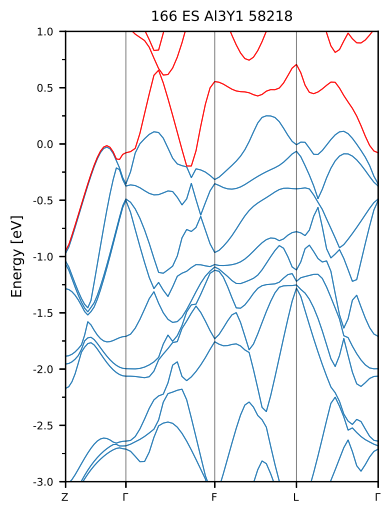

(wp) 166.3.204

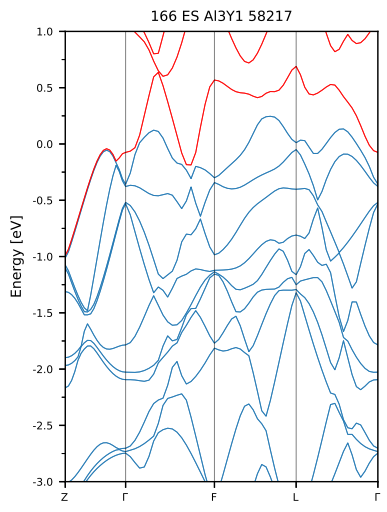

(wq) 166.3.205

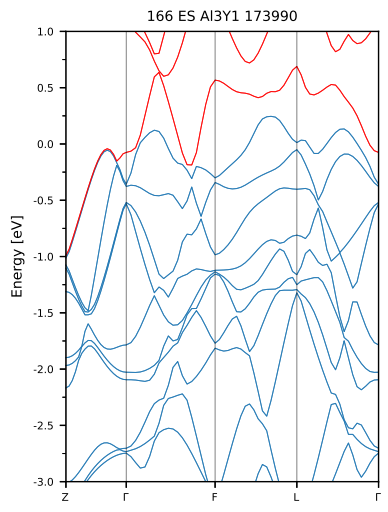

(wr) 166.3.206

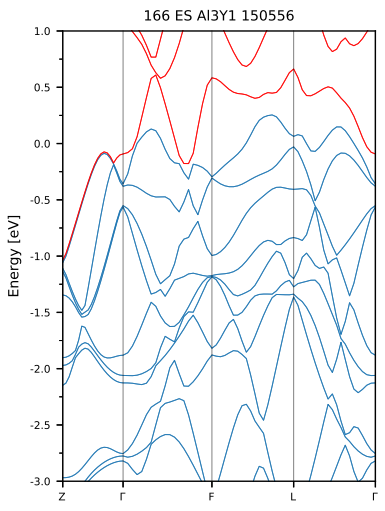

(ws) 166.3.207

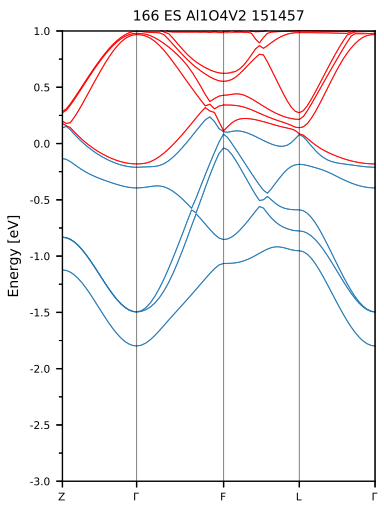

(wt) 166.3.208

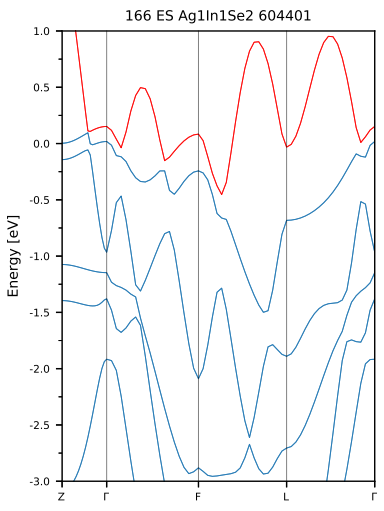

(wu) 166.3.209

Figure 389: Band structures for topological materials in SG 166, part 39/39, from 166.3.199 to 166.3.209. See the list of figures.

## 0.0.83 SG 167

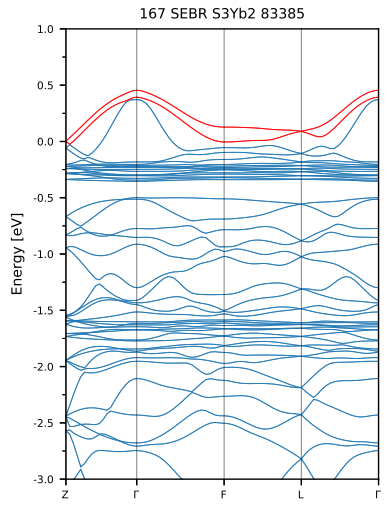

(a) 167.1.1

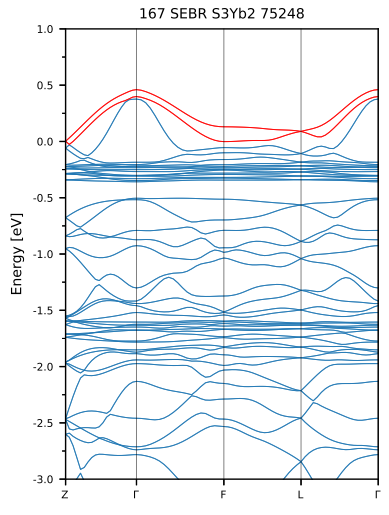

(b) 167.1.2

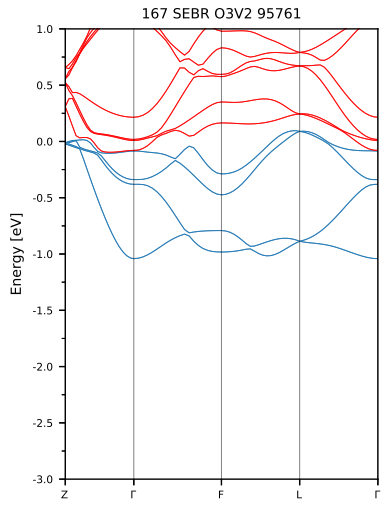

(c) 167.1.3

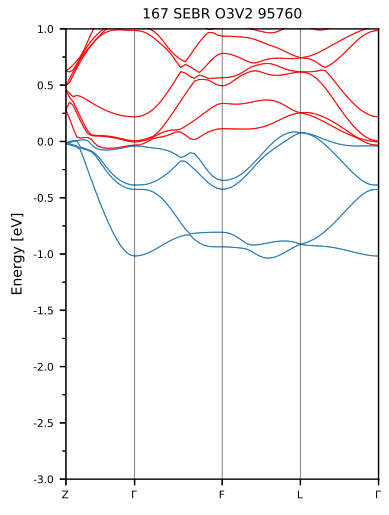

(d) 167.1.4

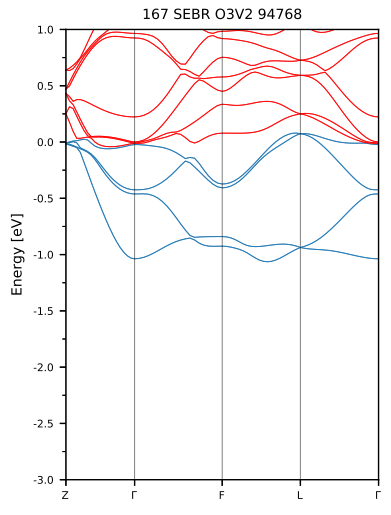

(e) 167.1.5

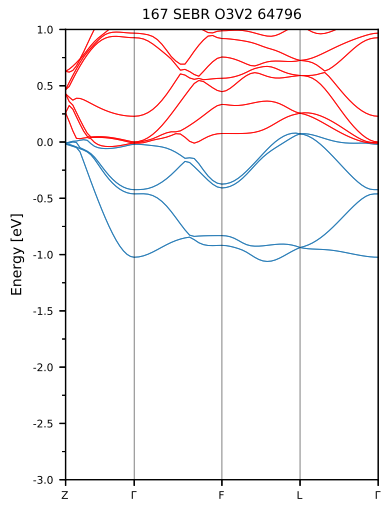

(f) 167.1.6

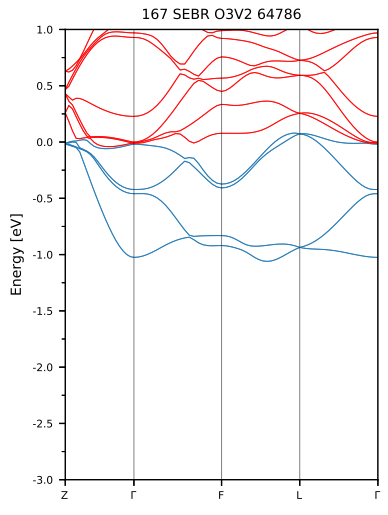

(g) 167.1.7

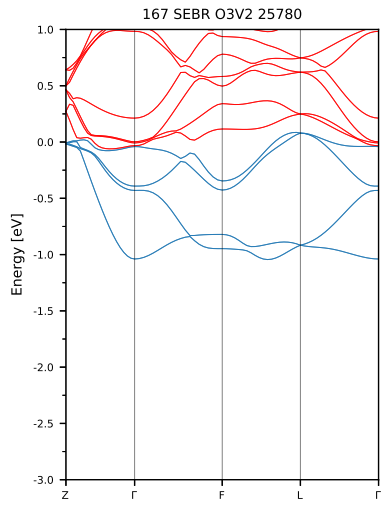

(h) 167.1.8

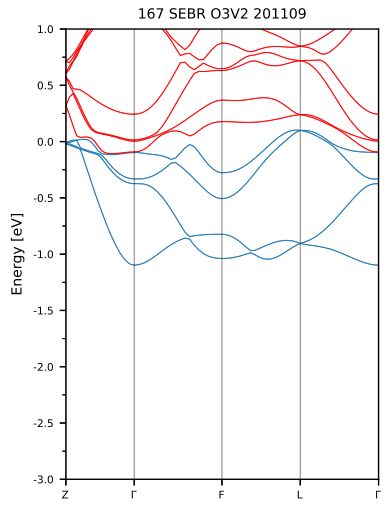

(i) 167.1.9

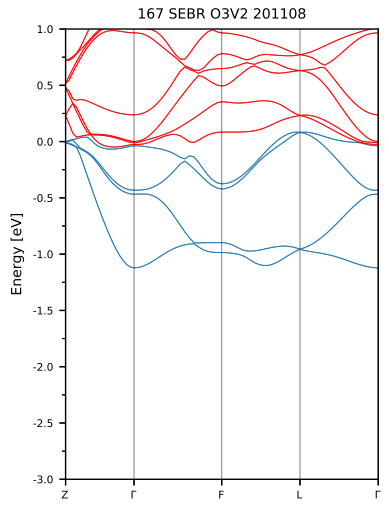

(j) 167.1.10

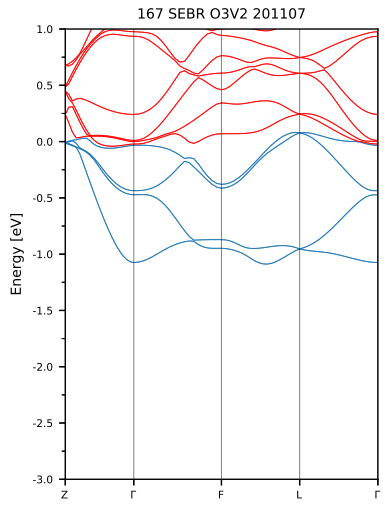

(k) 167.1.11

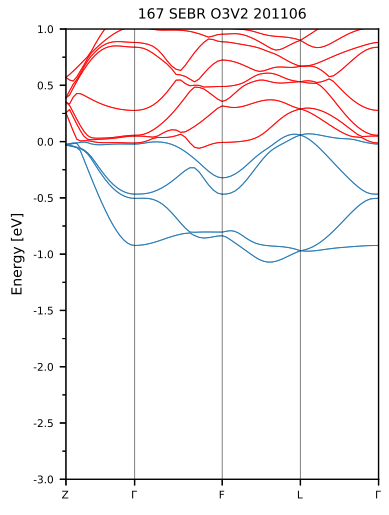

(l) 167.1.12

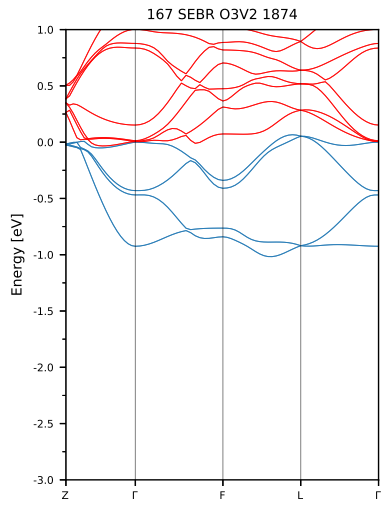

(m) 167.1.13

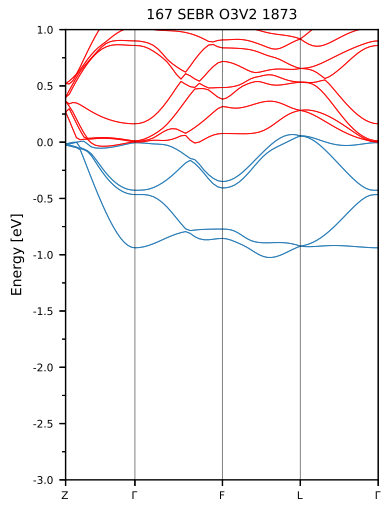

(n) 167.1.14

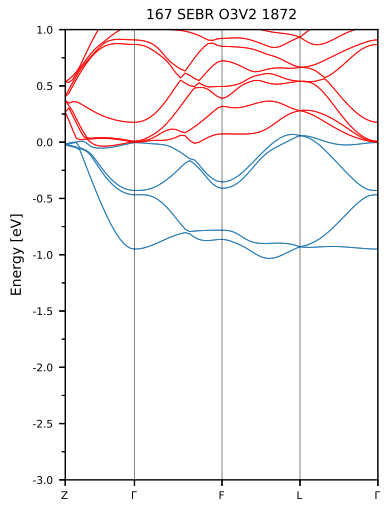

(o) 167.1.15

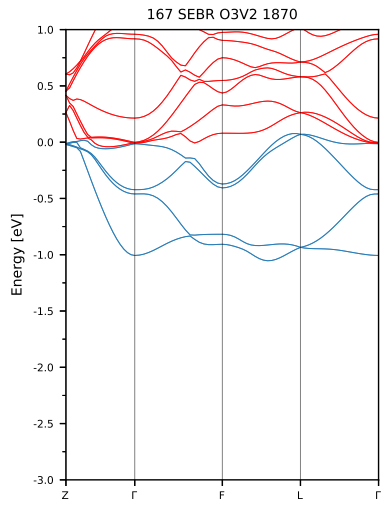

(p) 167.1.16

Figure 390: Band structures for topological materials in SG 167, part 1/21, from 167.1.1 to 167.1.16. See the list of figures.

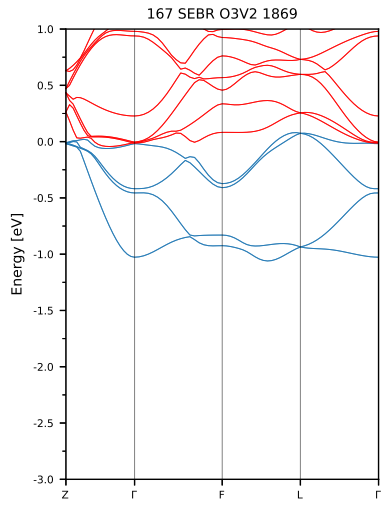

(q) 167.1.17

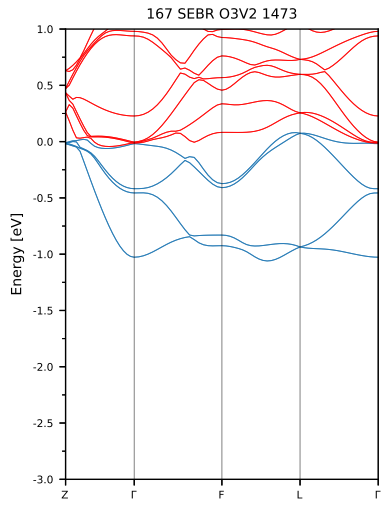

(r) 167.1.18

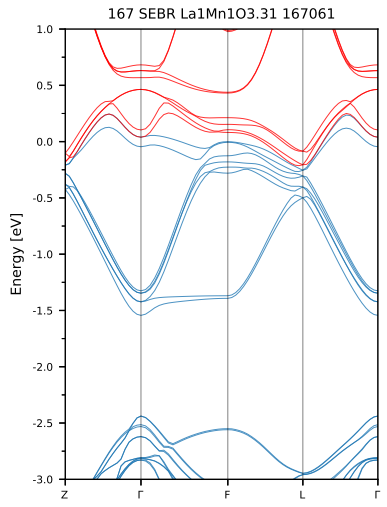

(s) 167.1.19

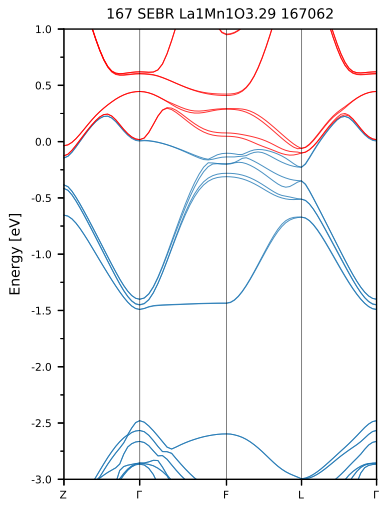

(t) 167.1.20

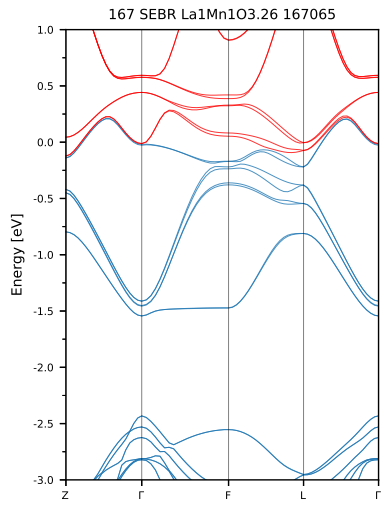

(u) 167.1.21

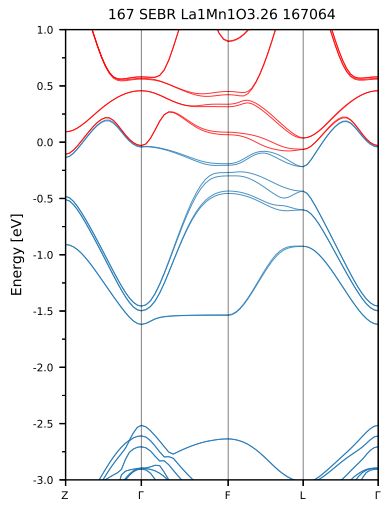

(v) 167.1.22

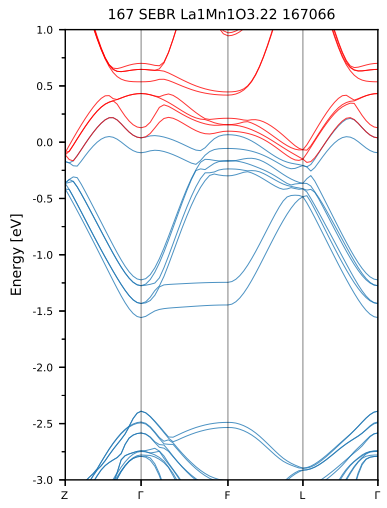

(w) 167.1.23

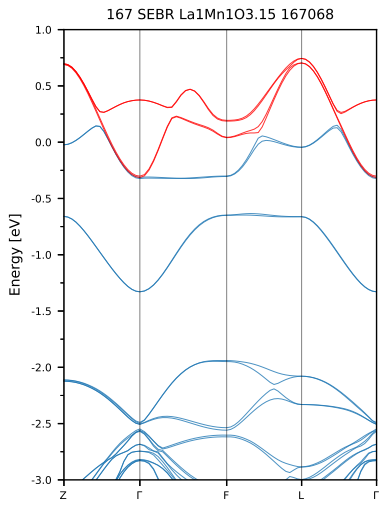

(x) 167.1.24

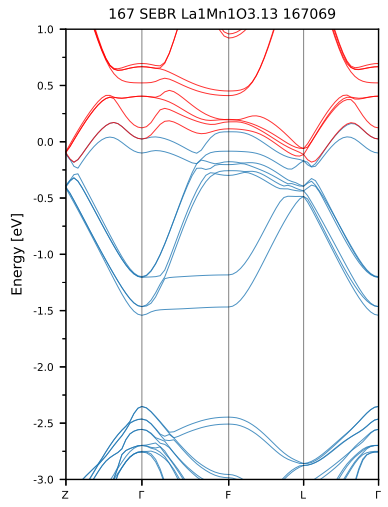

(y) 167.1.25

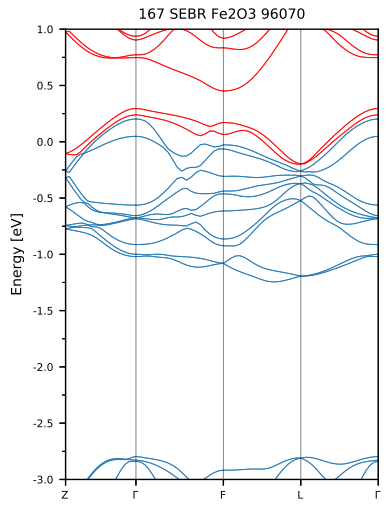

(z) 167.1.26

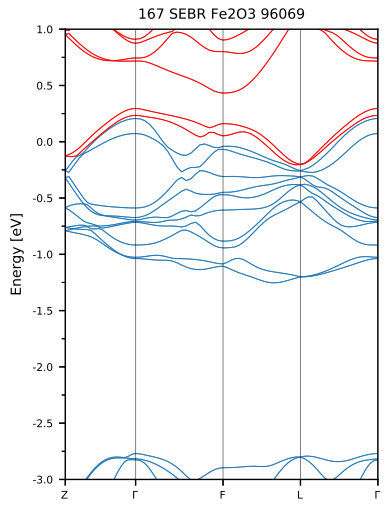

(aa) 167.1.27

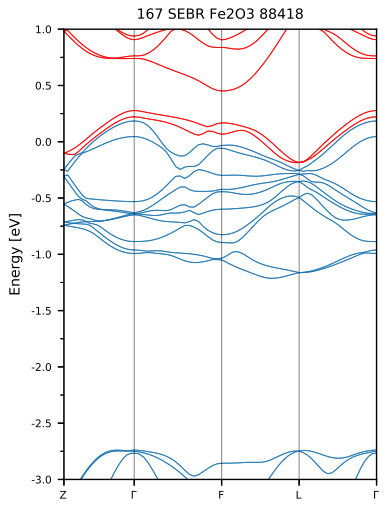

(ab) 167.1.28

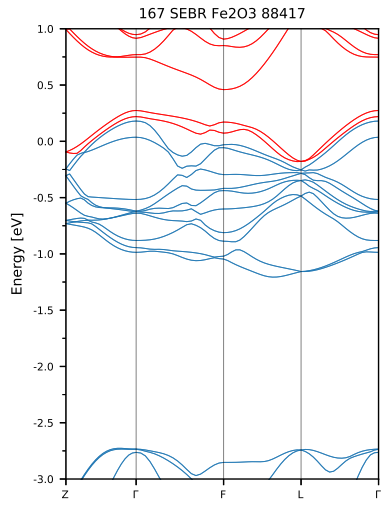

(ac) 167.1.29

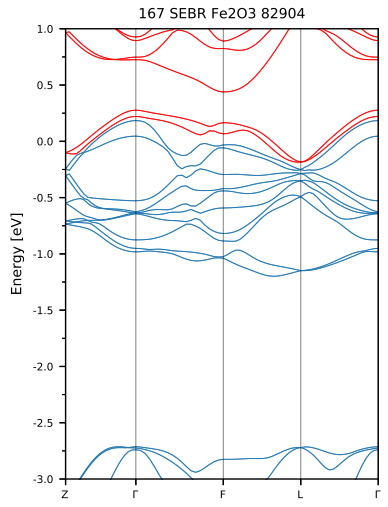

(ad) 167.1.30

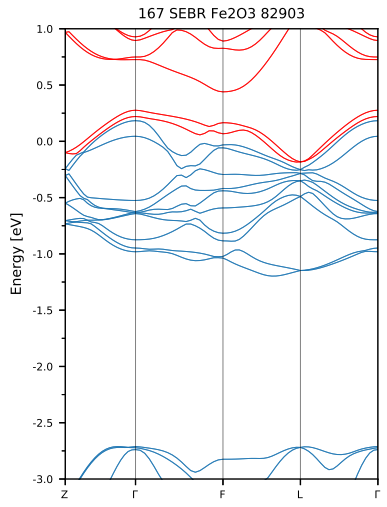

(ae) 167.1.31

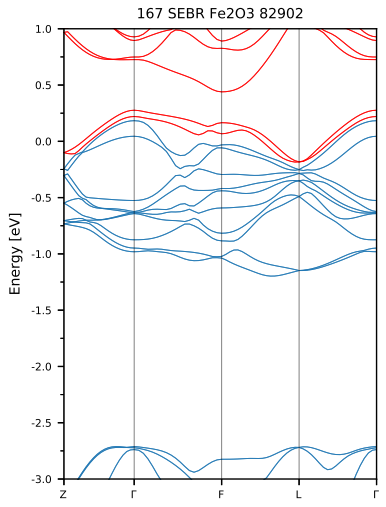

(af) 167.1.32

Figure 391: Band structures for topological materials in SG 167, part 2/21, from 167.1.17 to 167.1.32. See the list of figures.

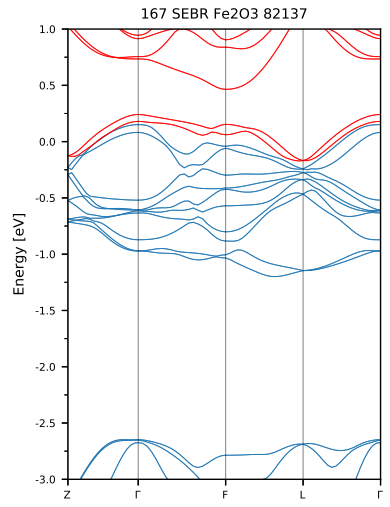

(ag) 167.1.33

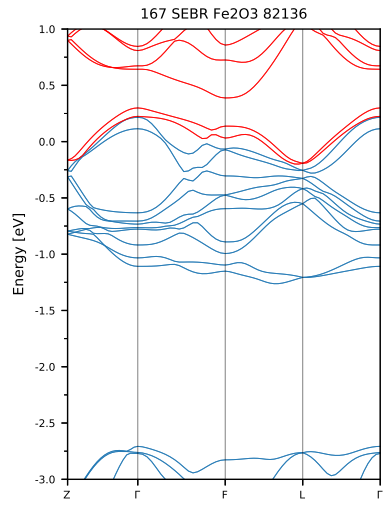

(ah) 167.1.34

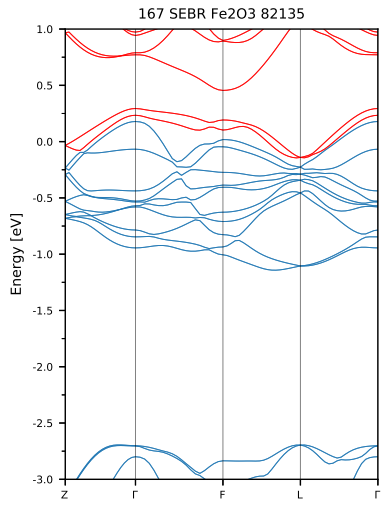

(ai) 167.1.35

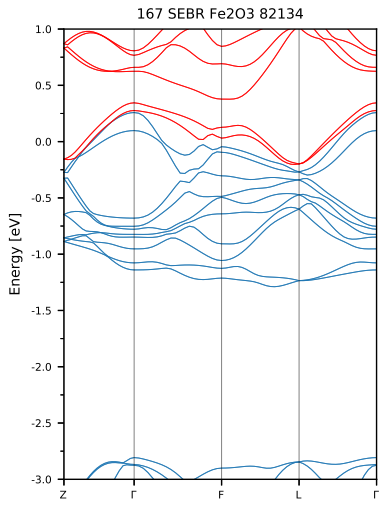

(aj) 167.1.36

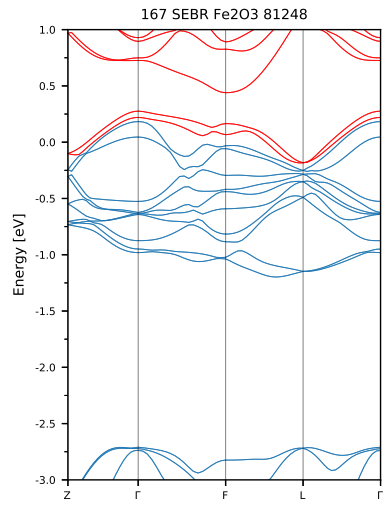

(ak) 167.1.37

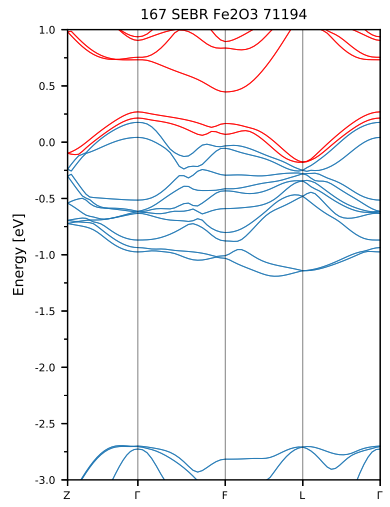

(al) 167.1.38

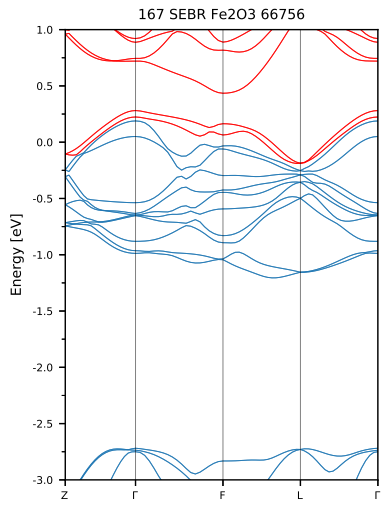

(am) 167.1.39

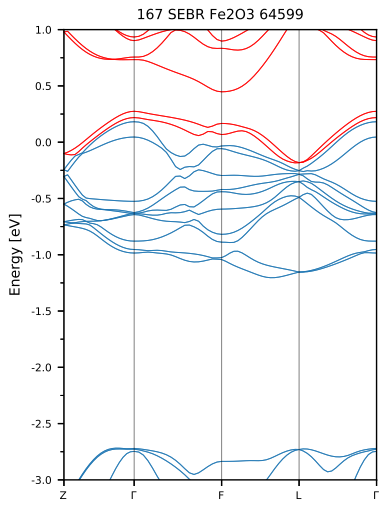

(an) 167.1.40

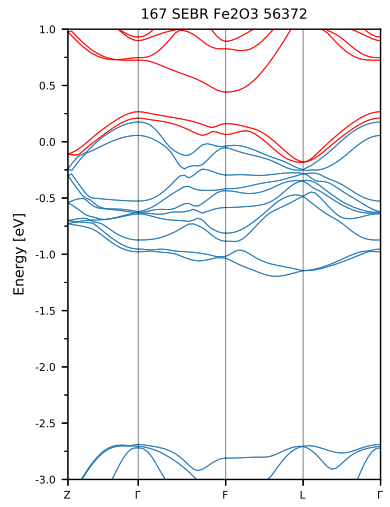

(ao) 167.1.41

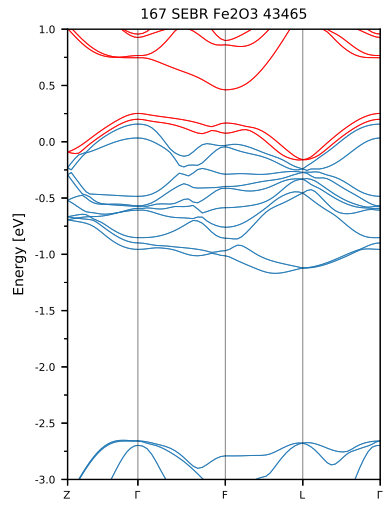

(ap) 167.1.42

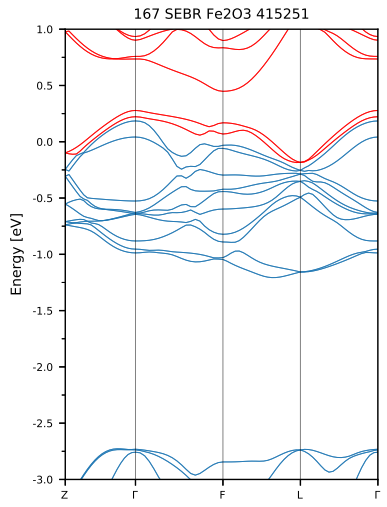

(aq) 167.1.43

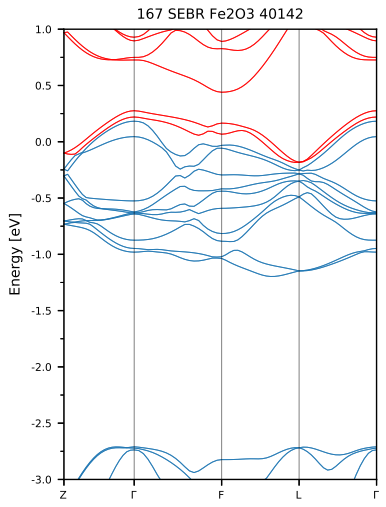

(ar) 167.1.44

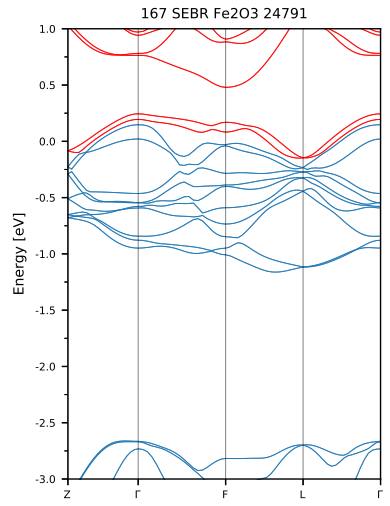

(as) 167.1.45

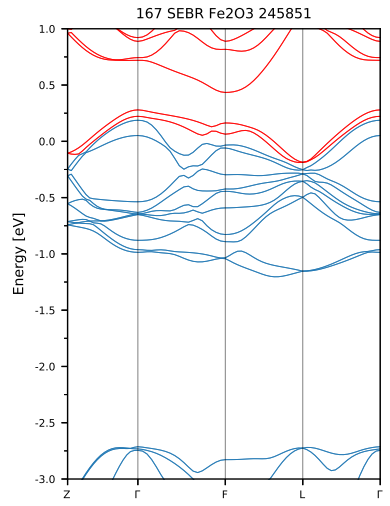

(at) 167.1.46

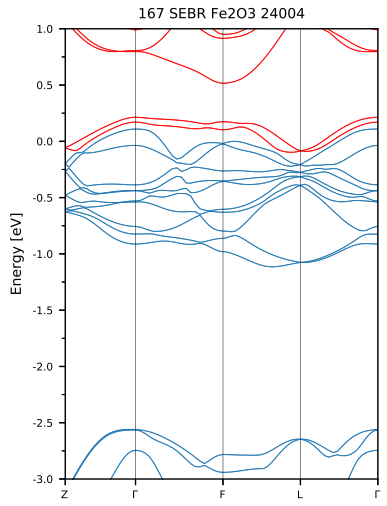

(au) 167.1.47

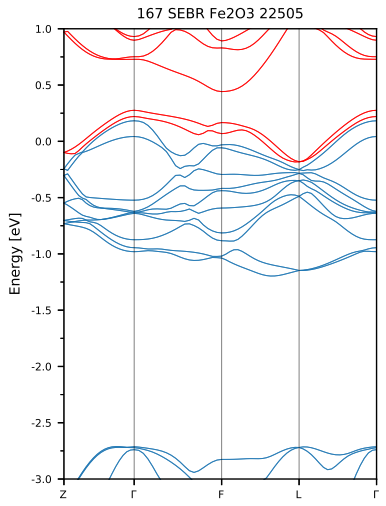

(av) 167.1.48

Figure 392: Band structures for topological materials in SG 167, part 3/21, from 167.1.33 to 167.1.48. See the list of figures.

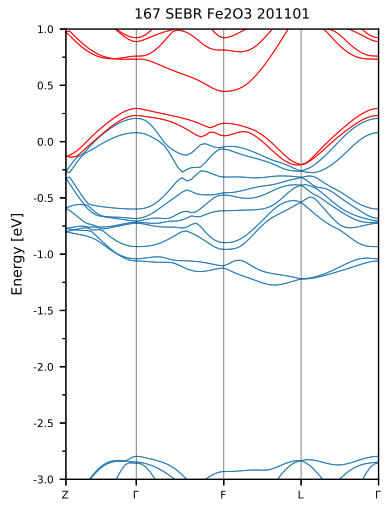

(aw) 167.1.49

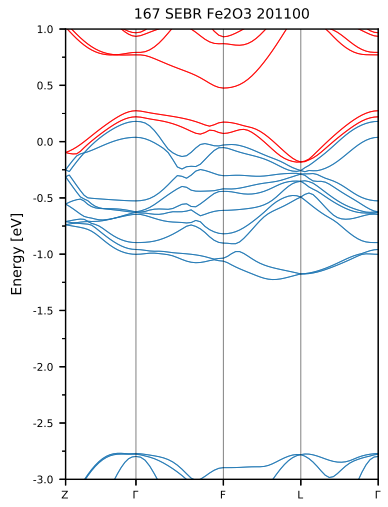

(ax) 167.1.50

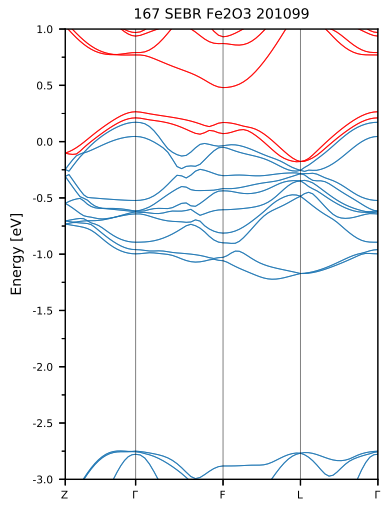

(ay) 167.1.51

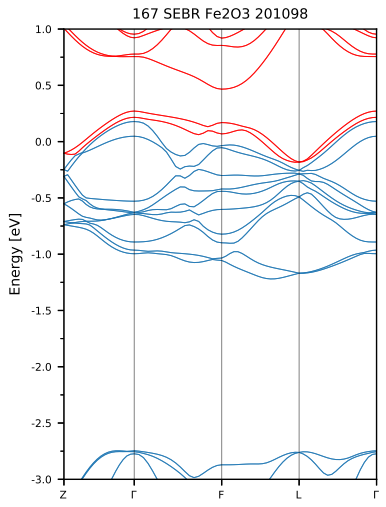

(az) 167.1.52

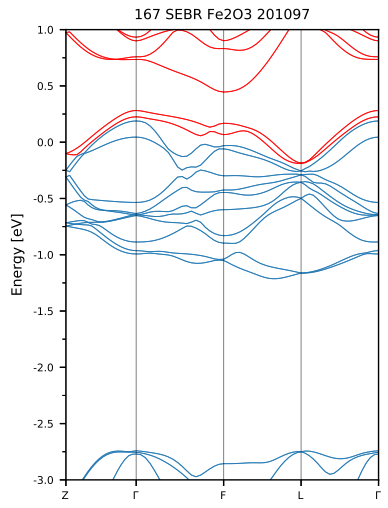

(ba) 167.1.53

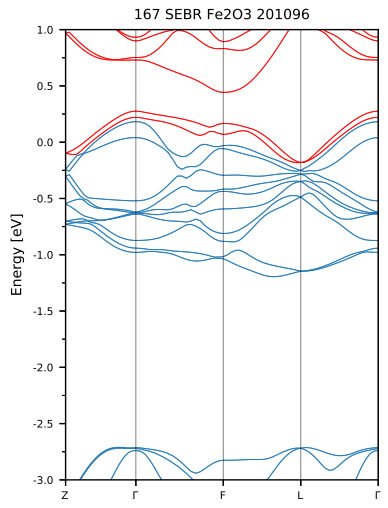

(bb) 167.1.54

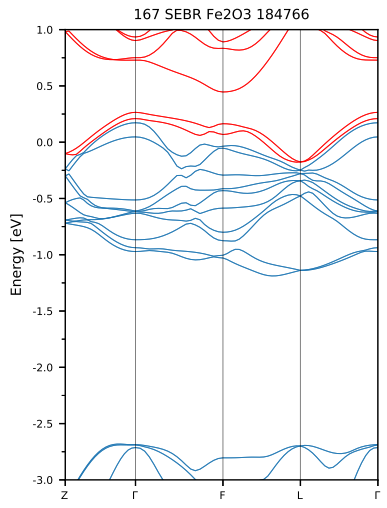

(bc) 167.1.55

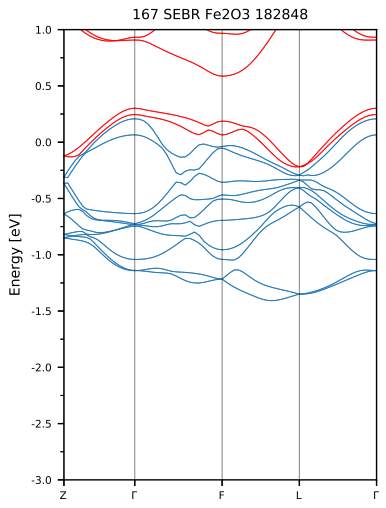

(bd) 167.1.56

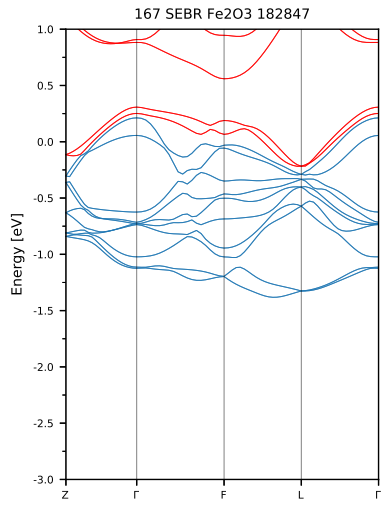

(be) 167.1.57

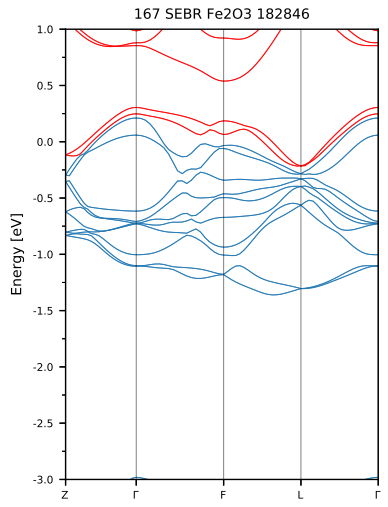

(bf) 167.1.58

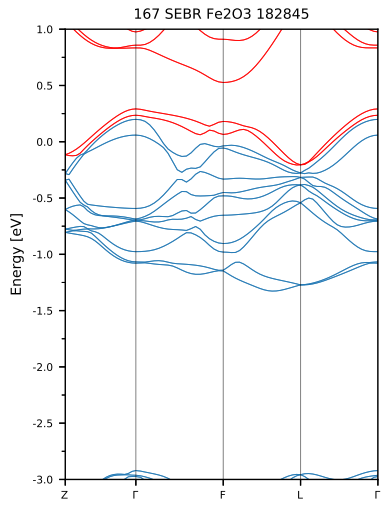

(bg) 167.1.59

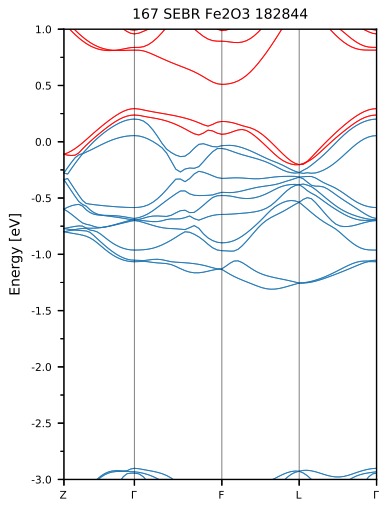

(bh) 167.1.60

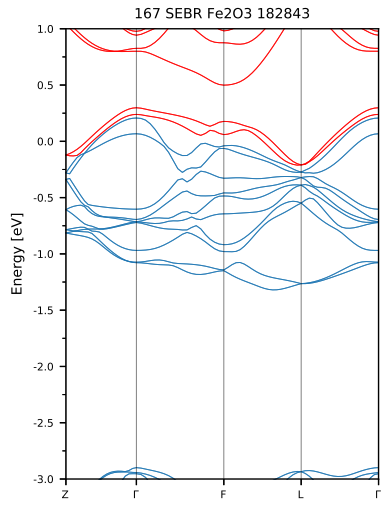

(bi) 167.1.61

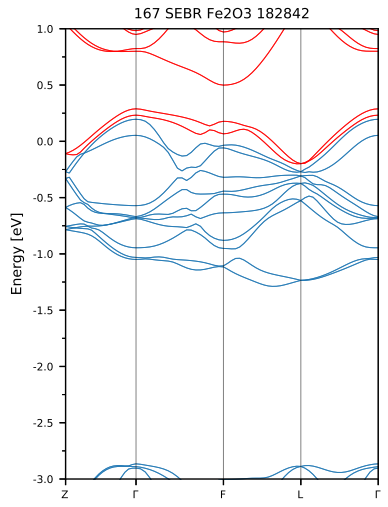

(bj) 167.1.62

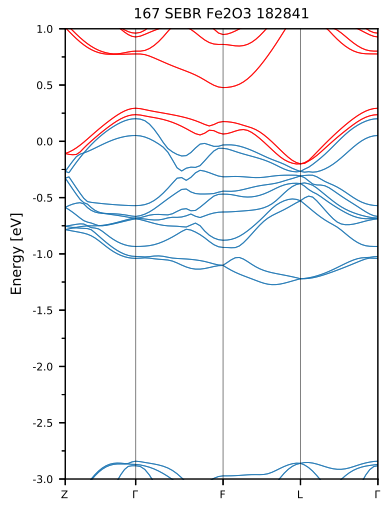

(bk) 167.1.63

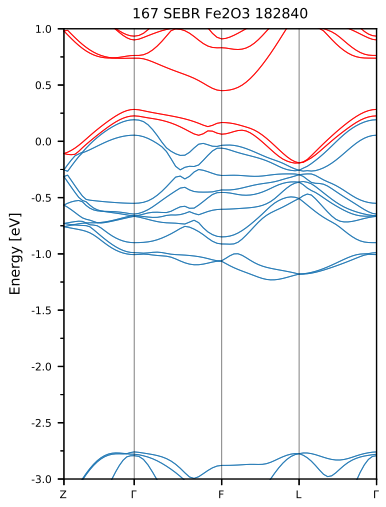

(bl) 167.1.64

Figure 393: Band structures for topological materials in SG 167, part 4/21, from 167.1.49 to 167.1.64. See the list of figures.

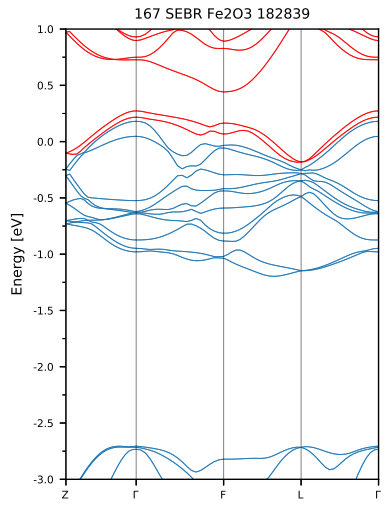

(bm) 167.1.65

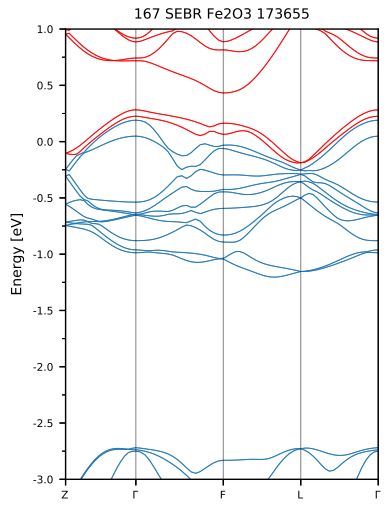

(bn) 167.1.66

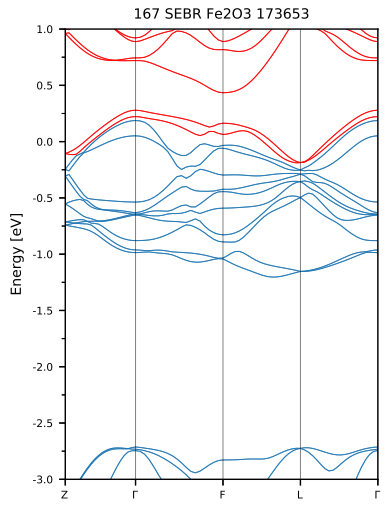

(bo) 167.1.67

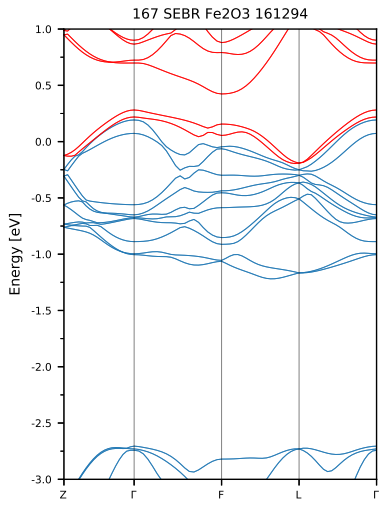

(bp) 167.1.68

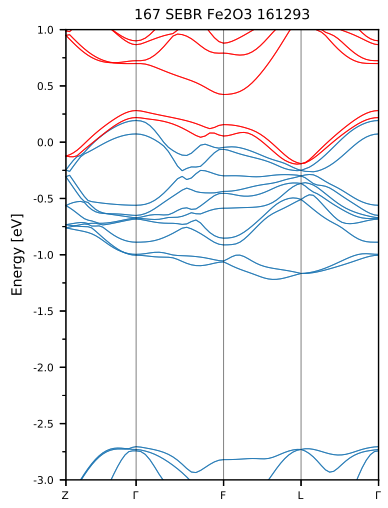

(bq) 167.1.69

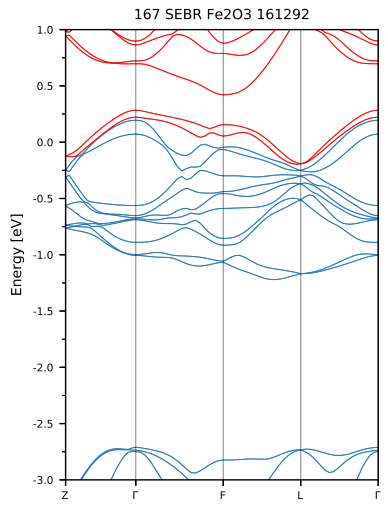

(br) 167.1.70

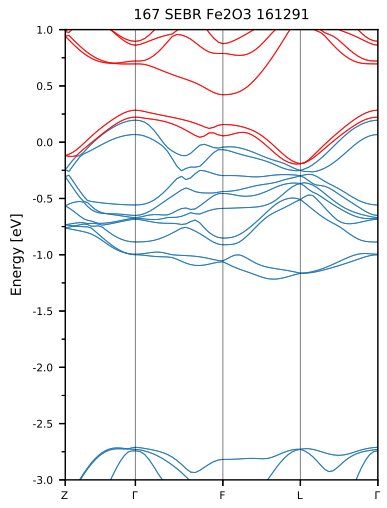

(bs) 167.1.71

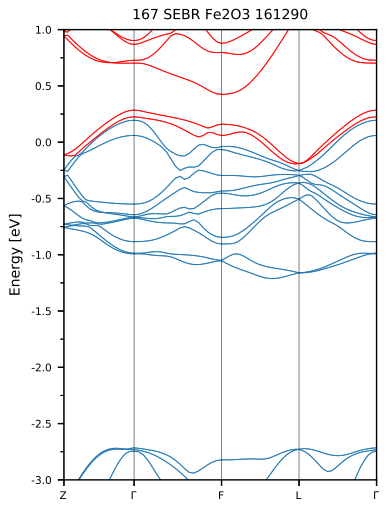

(bt) 167.1.72

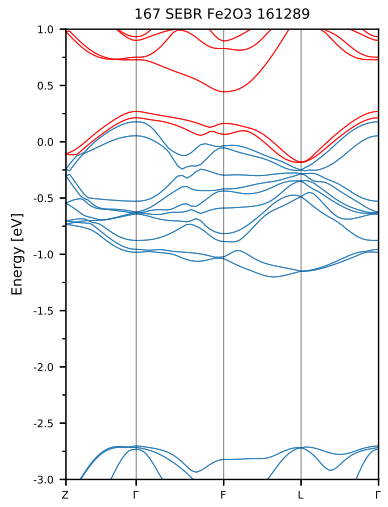

(bu) 167.1.73

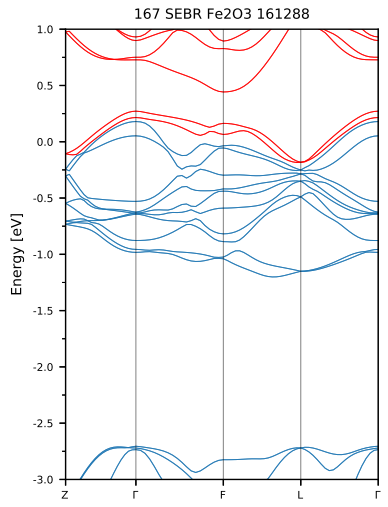

(bv) 167.1.74

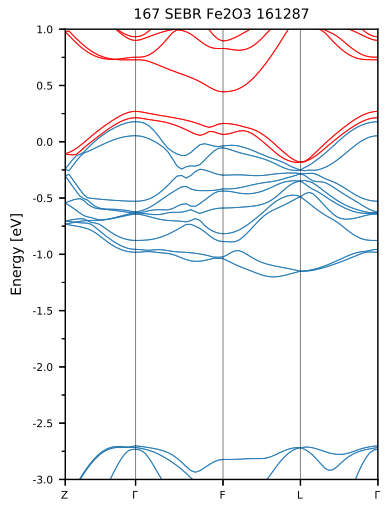

(bw) 167.1.75

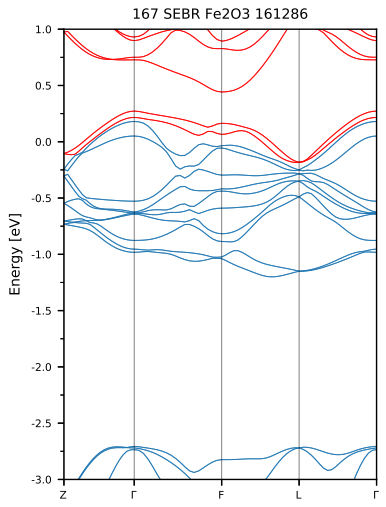

(bx) 167.1.76

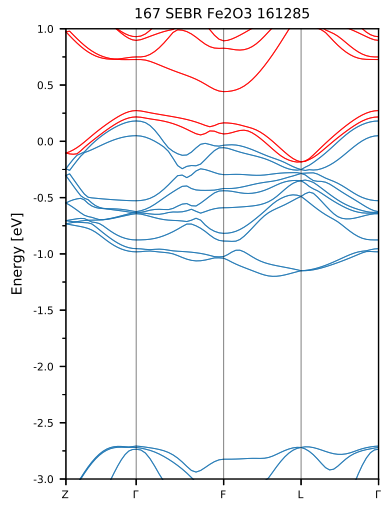

(by) 167.1.77

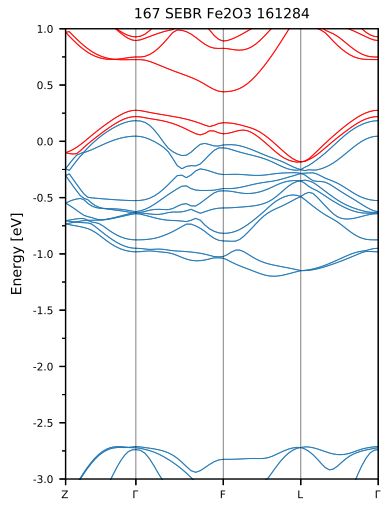

(bz) 167.1.78

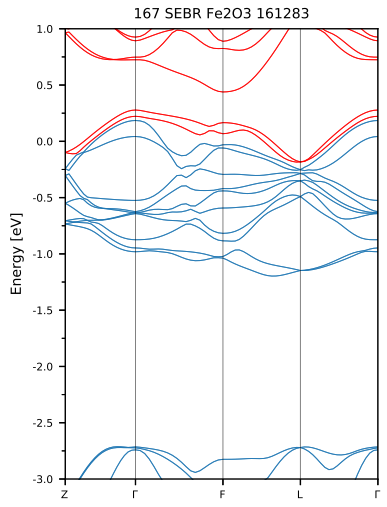

(ca) 167.1.79

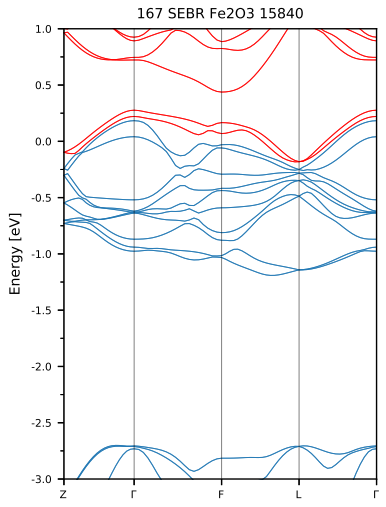

(cb) 167.1.80

Figure 394: Band structures for topological materials in SG 167, part 5/21, from 167.1.65 to 167.1.80. See the list of figures.

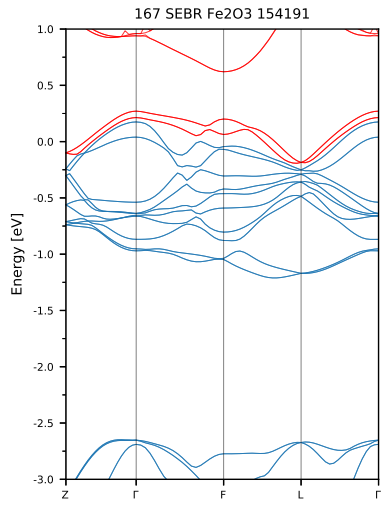

(cc) 167.1.81

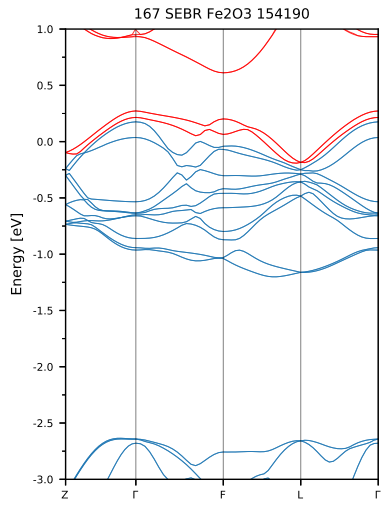

(cd) 167.1.82

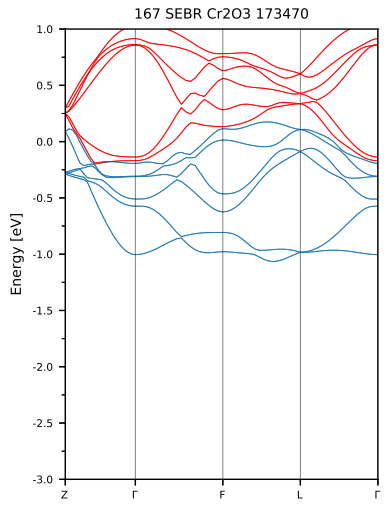

(ce) 167.1.83

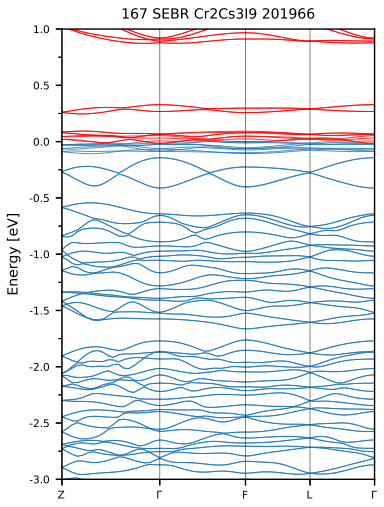

(cf) 167.1.84

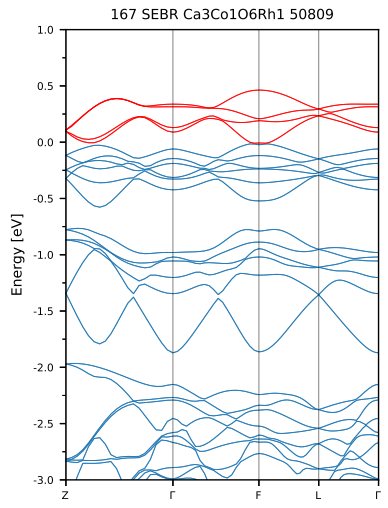

(cg) 167.1.85

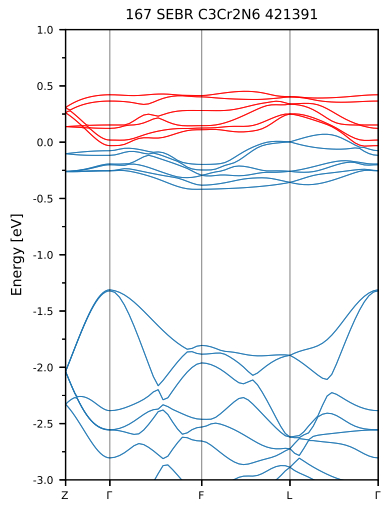

(ch) 167.1.86

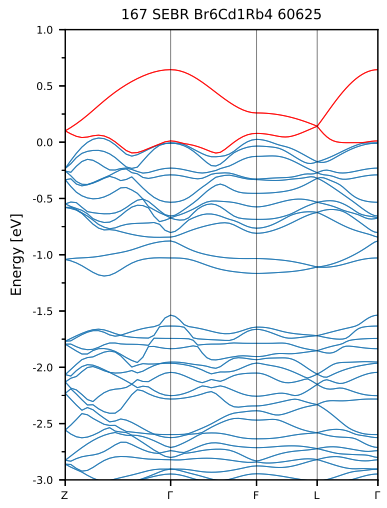

(ci) 167.1.87

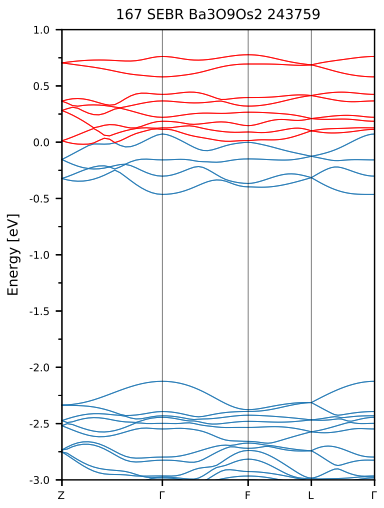

(cj) 167.1.88

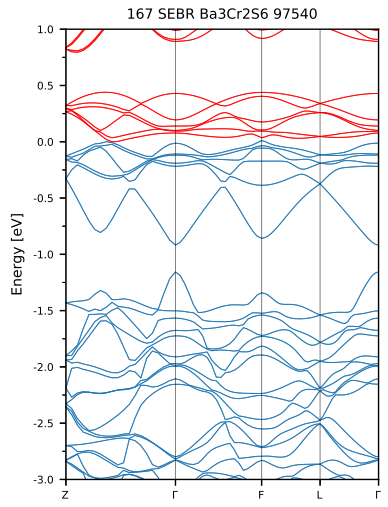

(ck) 167.1.89

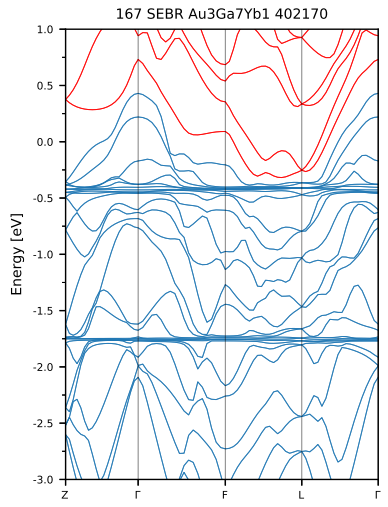

(cl) 167.1.90

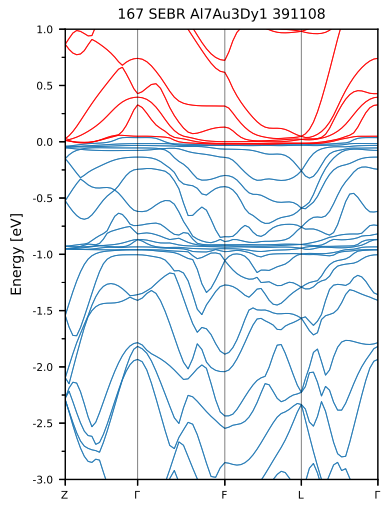

(cm) 167.1.91

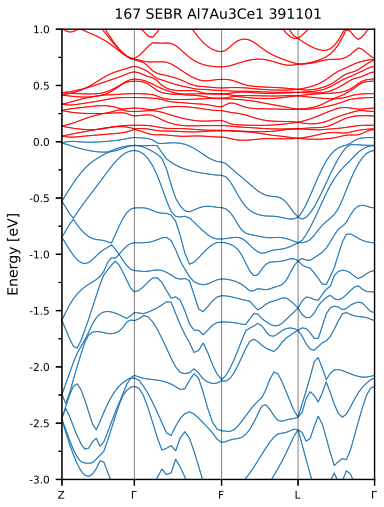

(cn) 167.1.92

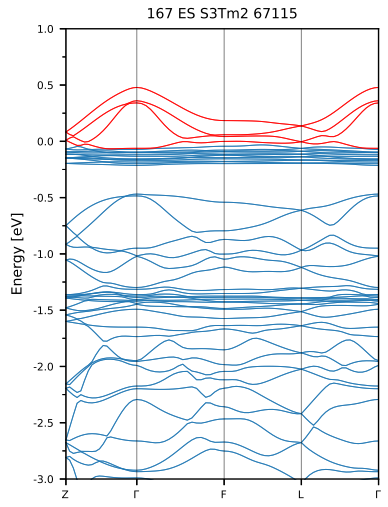

(co) 167.3.1

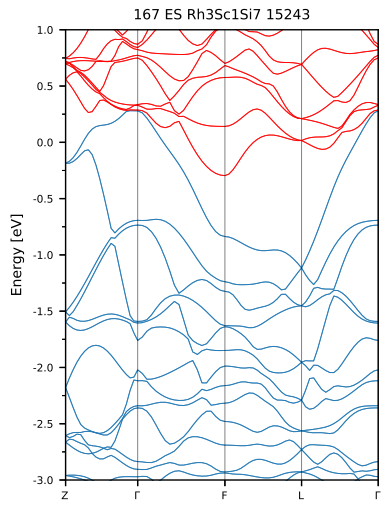

(cp) 167.3.2

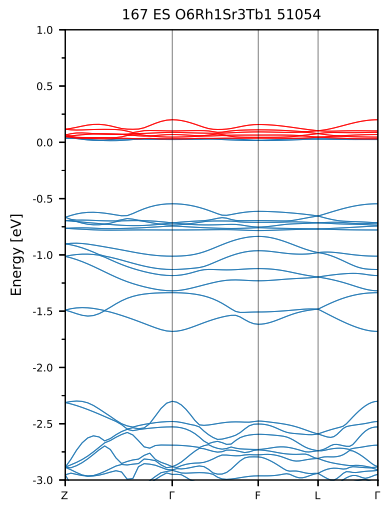

(cq) 167.3.3

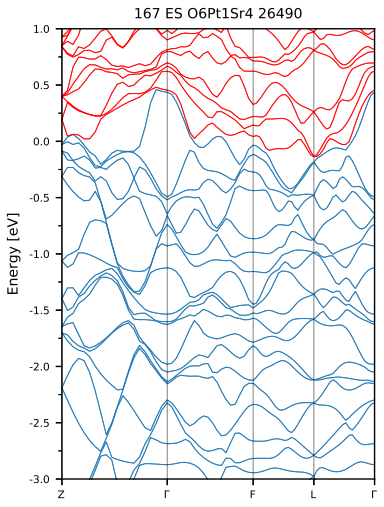

(cr) 167.3.4

Figure 395: Band structures for topological materials in SG 167, part 6/21, from 167.1.81 to 167.3.4. See the list of figures.

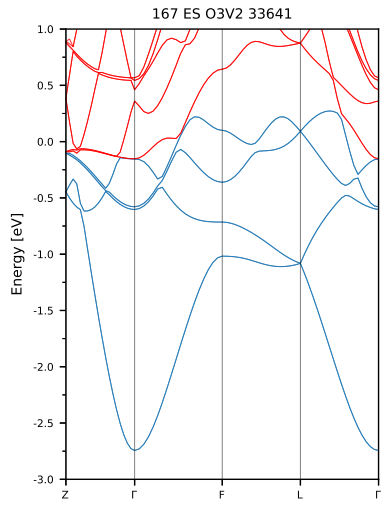

(cs) 167.3.5

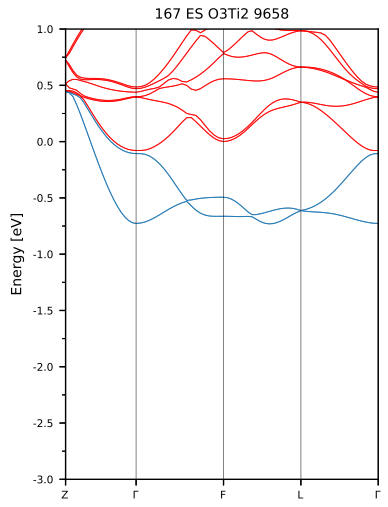

(ct) 167.3.6

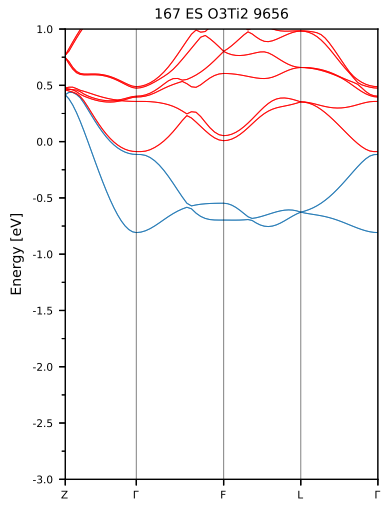

(cu) 167.3.7

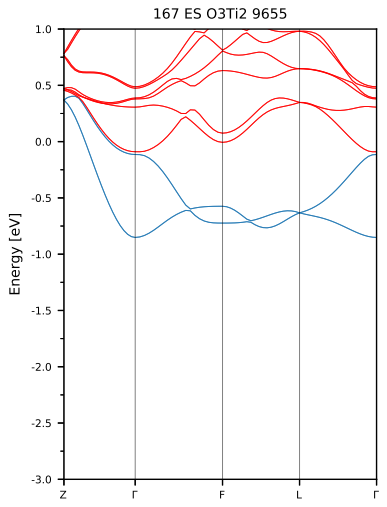

(cv) 167.3.8

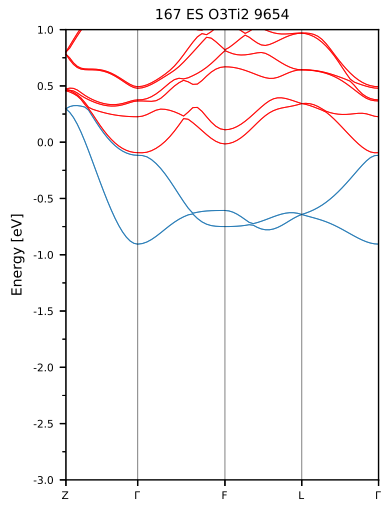

(cw) 167.3.9

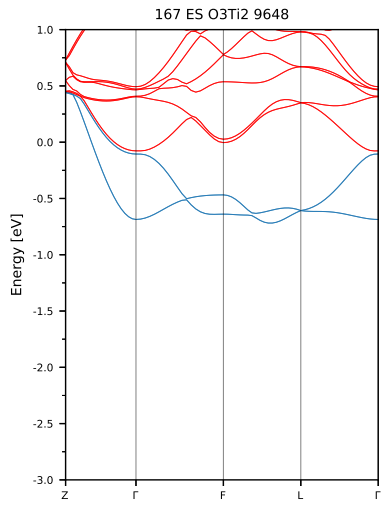

(cx) 167.3.10

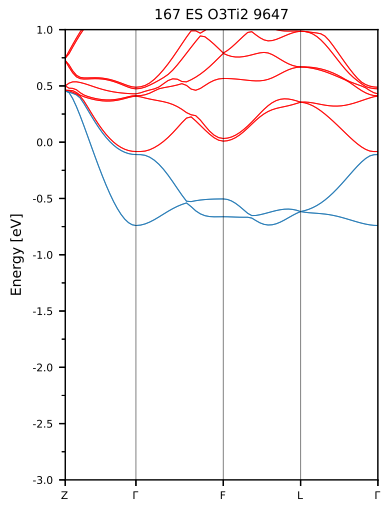

(cy) 167.3.11

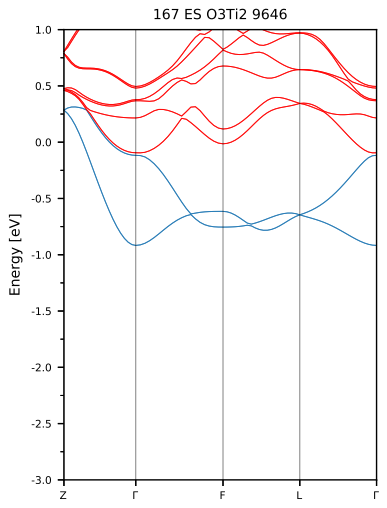

(cz) 167.3.12

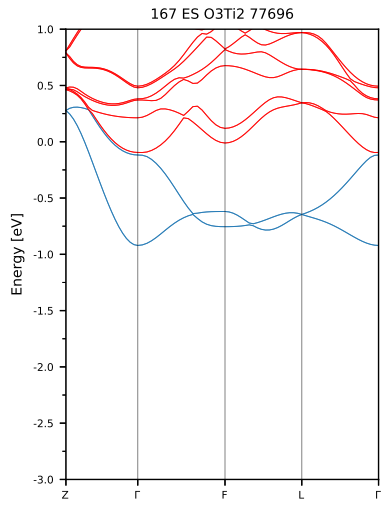

(da) 167.3.13

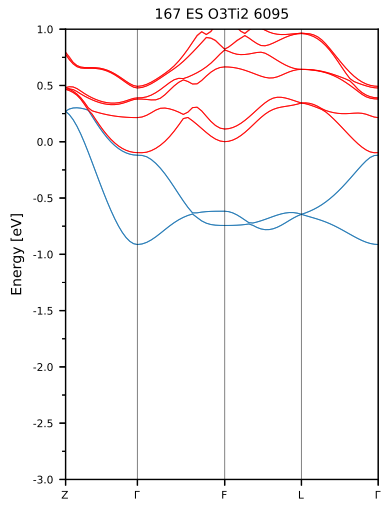

(db) 167.3.14

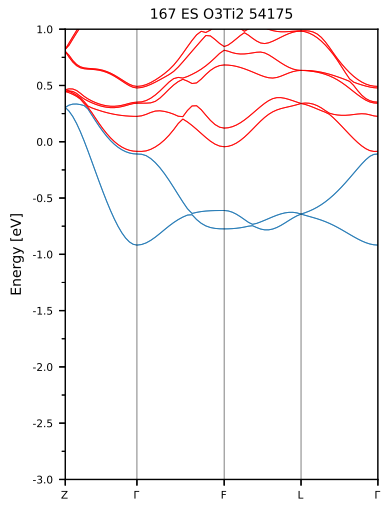

(dc) 167.3.15

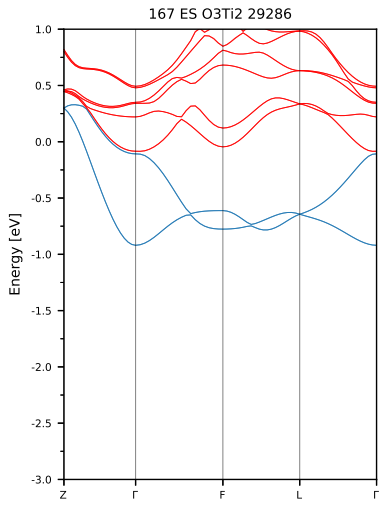

(dd) 167.3.16

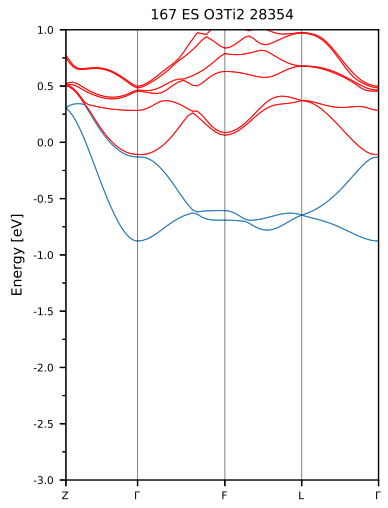

(de) 167.3.17

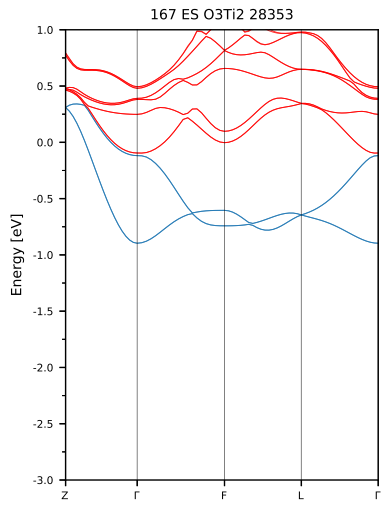

(df) 167.3.18

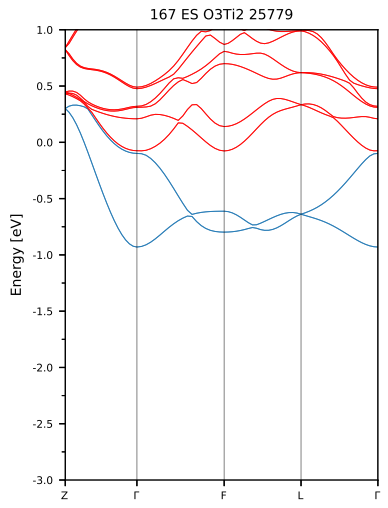

(dg) 167.3.19

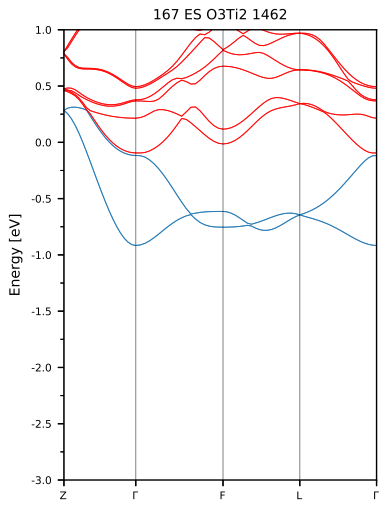

(dh) 167.3.20

Figure 396: Band structures for topological materials in SG 167, part 7/21, from 167.3.5 to 167.3.20. See the list of figures.

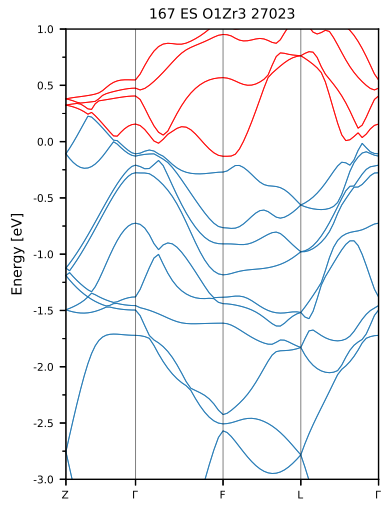

(di) 167.3.21

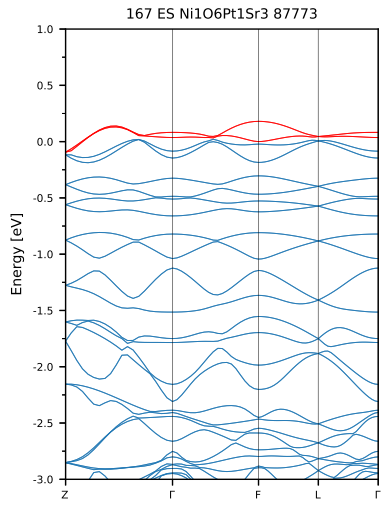

(dj) 167.3.22

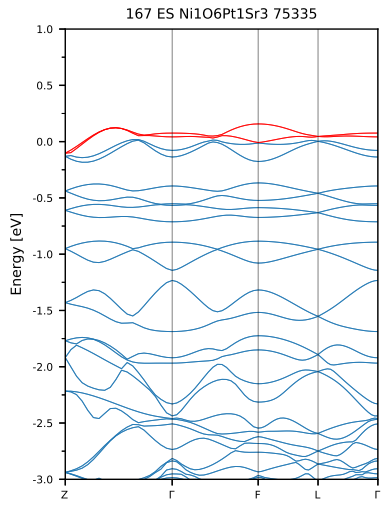

(dk) 167.3.23

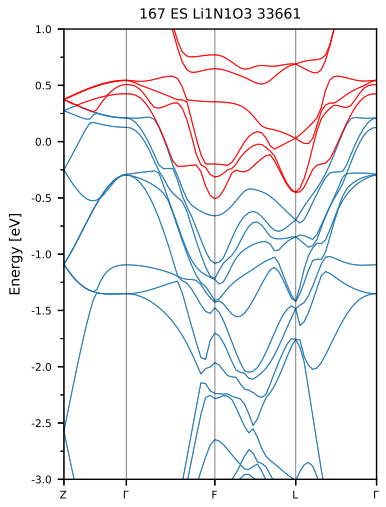

(dl) 167.3.24

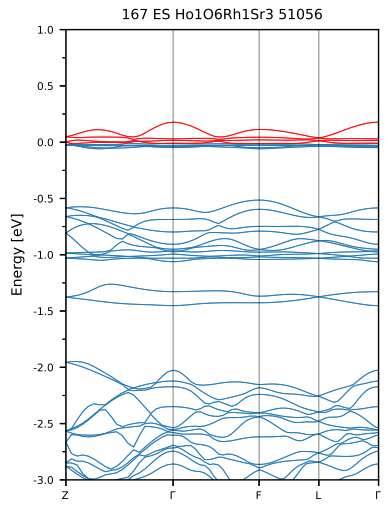

(dm) 167.3.25

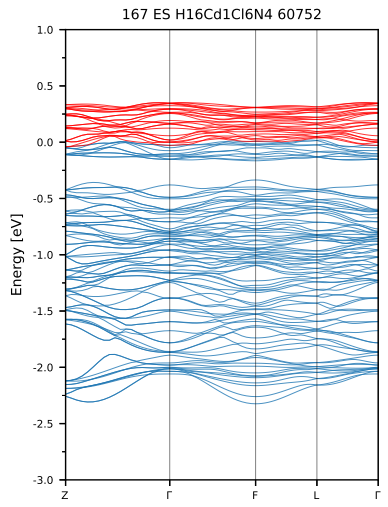

(dn) 167.3.26

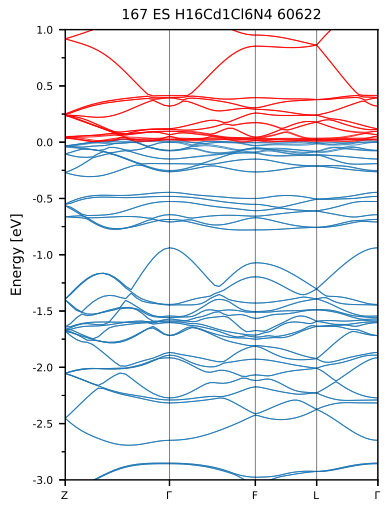

(do) 167.3.27

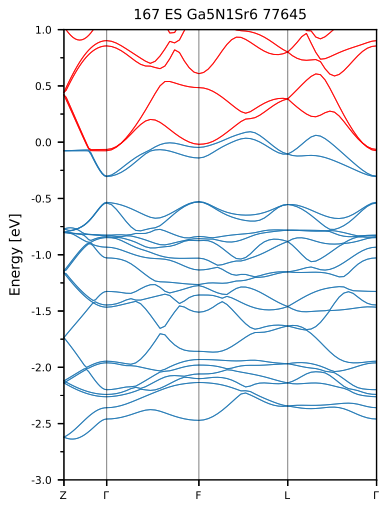

(dp) 167.3.28

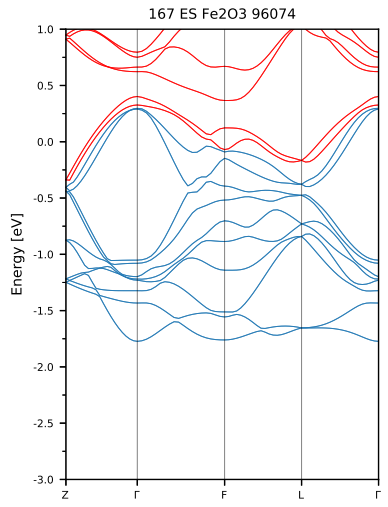

(dq) 167.3.29

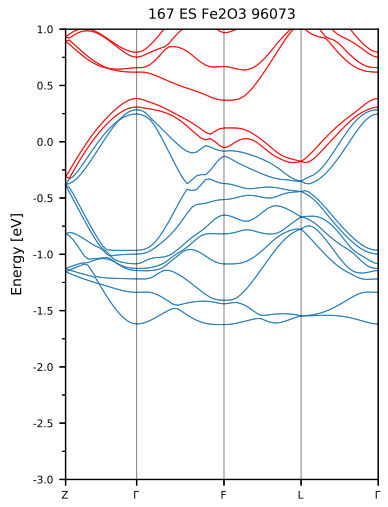

(dr) 167.3.30

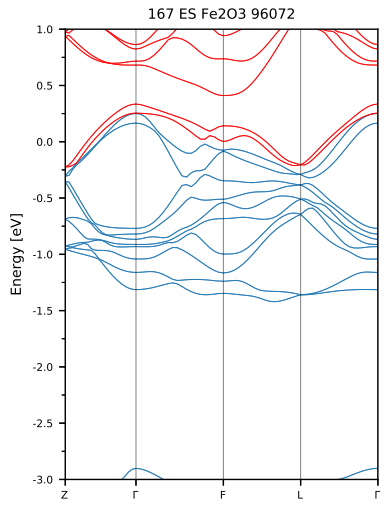

(ds) 167.3.31

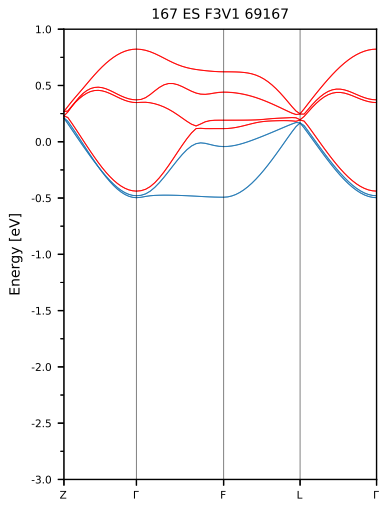

(dt) 167.3.32

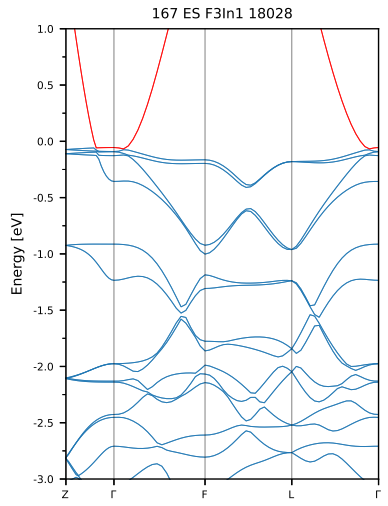

(du) 167.3.33

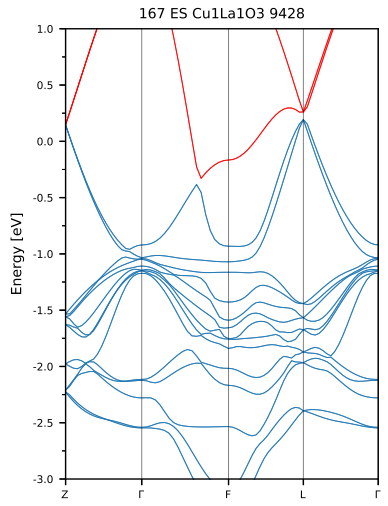

(dv) 167.3.34

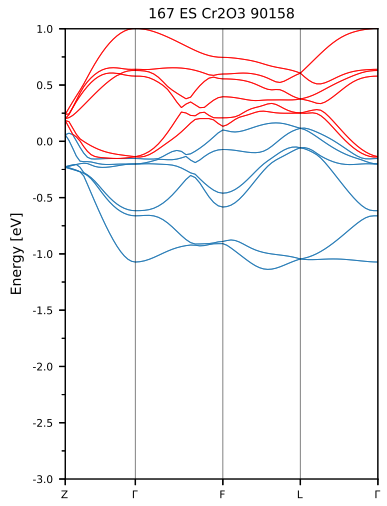

(dw) 167.3.35

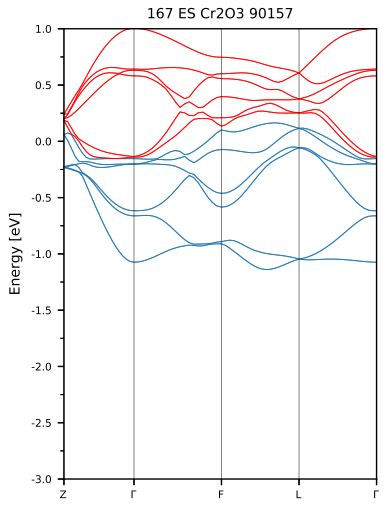

(dx) 167.3.36

Figure 397: Band structures for topological materials in SG 167, part 8/21, from 167.3.21 to 167.3.36. See the list of figures.

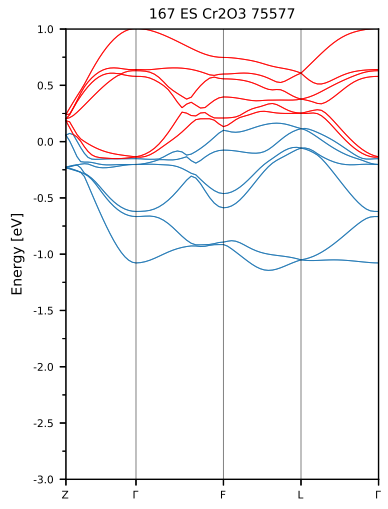

(dy) 167.3.37

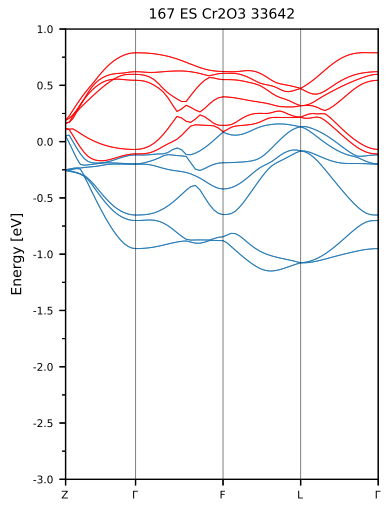

(dz) 167.3.38

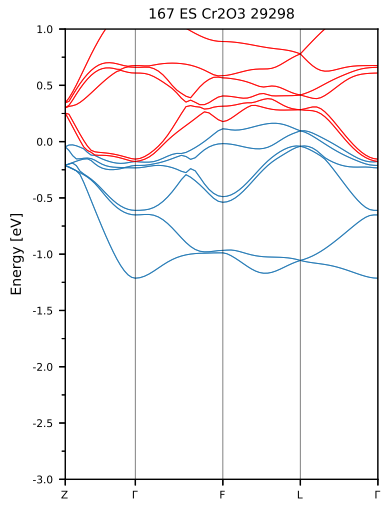

(ea) 167.3.39

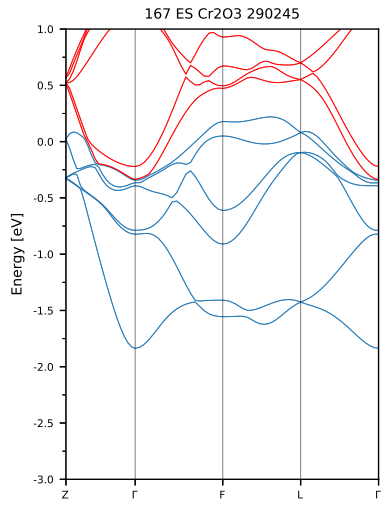

(eb) 167.3.40

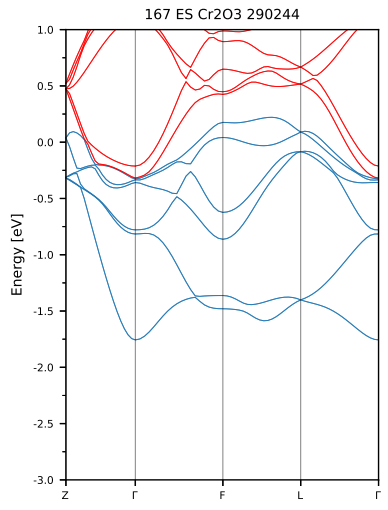

(ec) 167.3.41

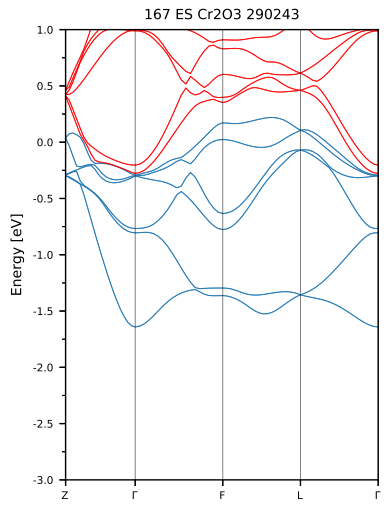

(ed) 167.3.42

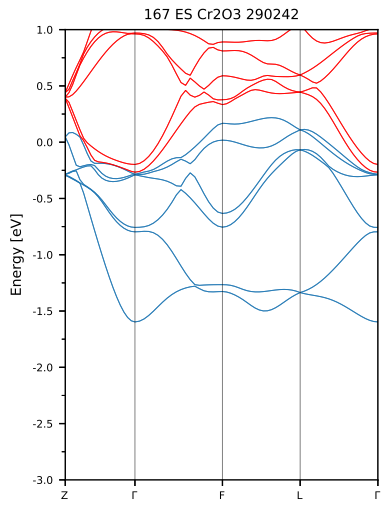

(ee) 167.3.43

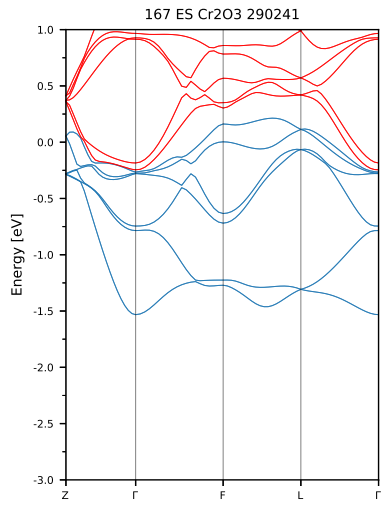

(ef) 167.3.44

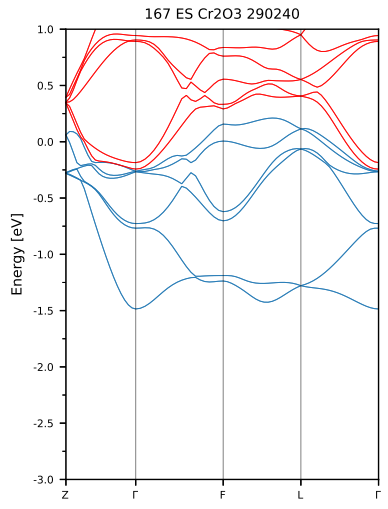

(eg) 167.3.45

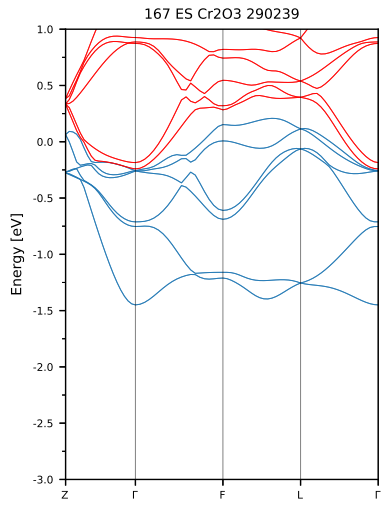

(eh) 167.3.46

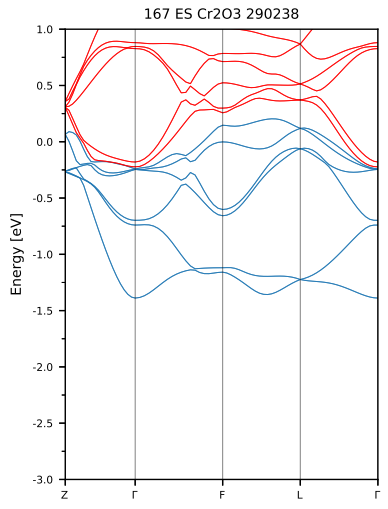

(ei) 167.3.47

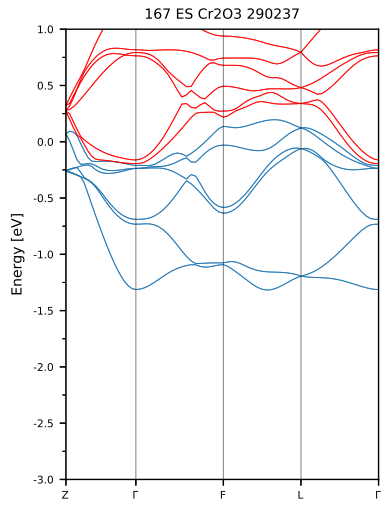

(ej) 167.3.48

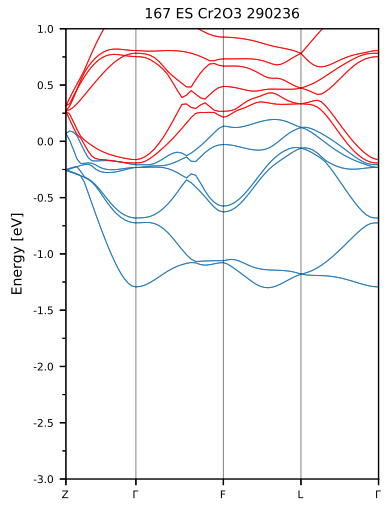

(ek) 167.3.49

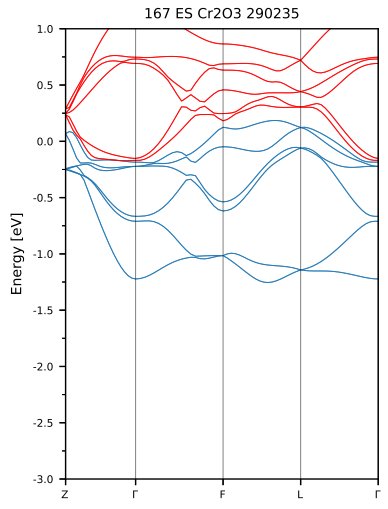

(el) 167.3.50

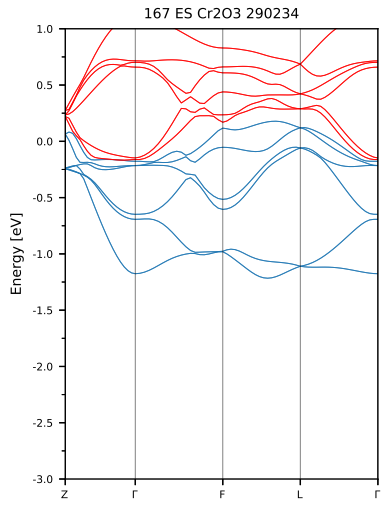

(em) 167.3.51

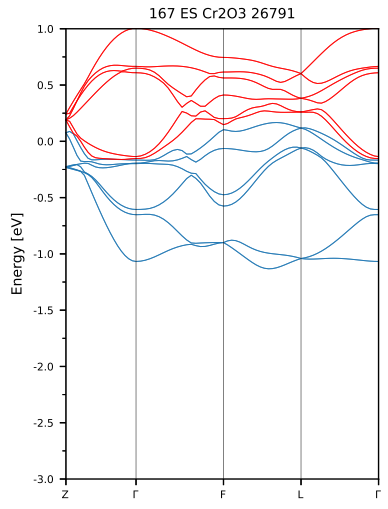

(en) 167.3.52

Figure 398: Band structures for topological materials in SG 167, part 9/21, from 167.3.37 to 167.3.52. See the list of figures.

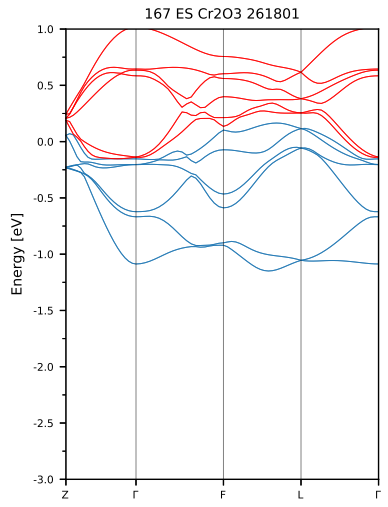

(eo) 167.3.53

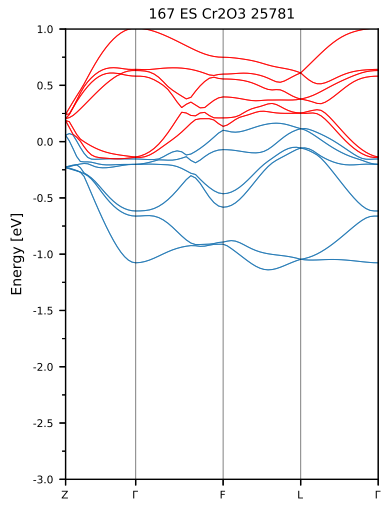

(ep) 167.3.54

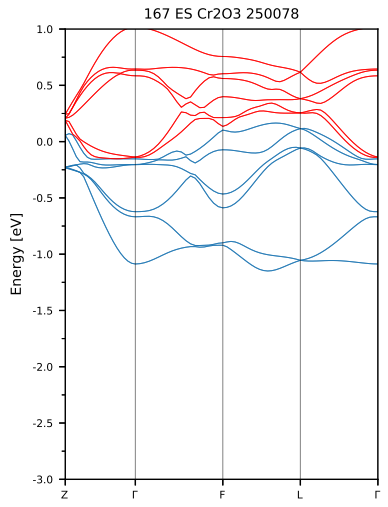

(eq) 167.3.55

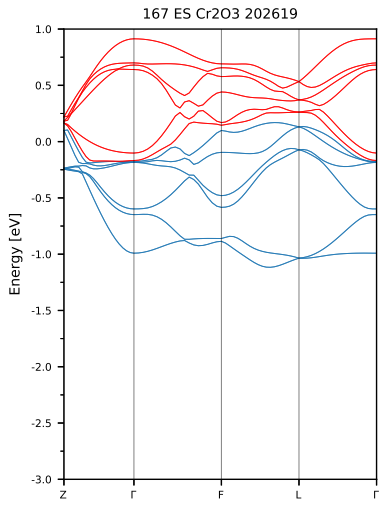

(er) 167.3.56

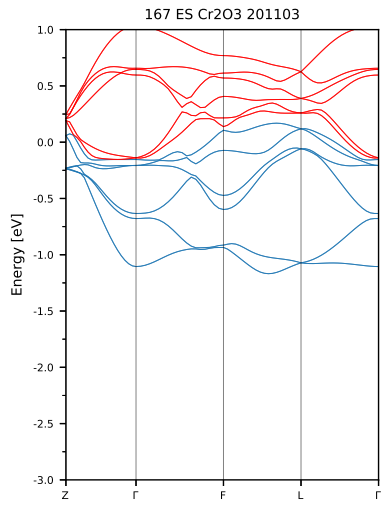

(es) 167.3.57

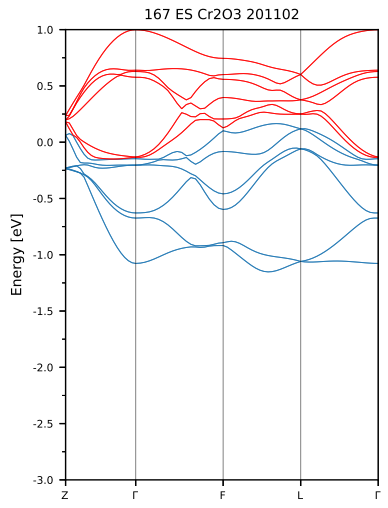

(et) 167.3.58

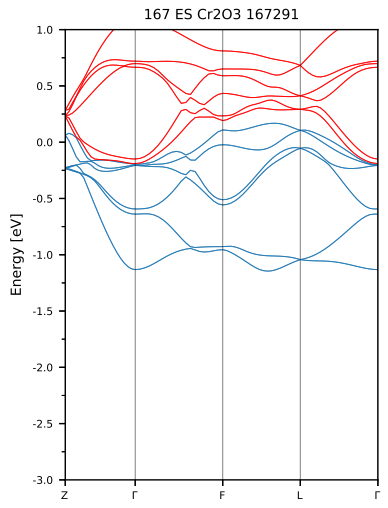

(eu) 167.3.59

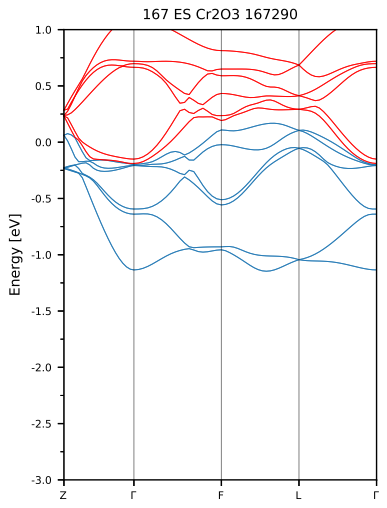

(ev) 167.3.60

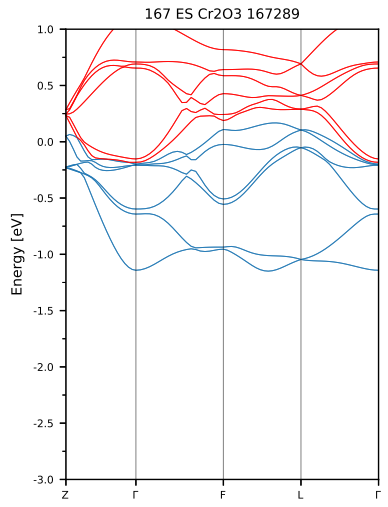

(ew) 167.3.61

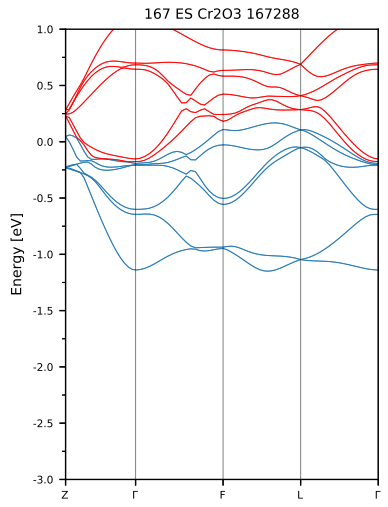

(ex) 167.3.62

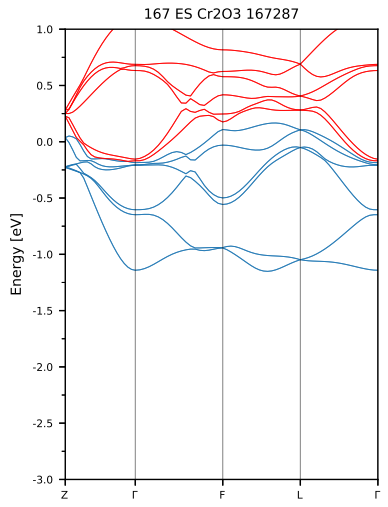

(ey) 167.3.63

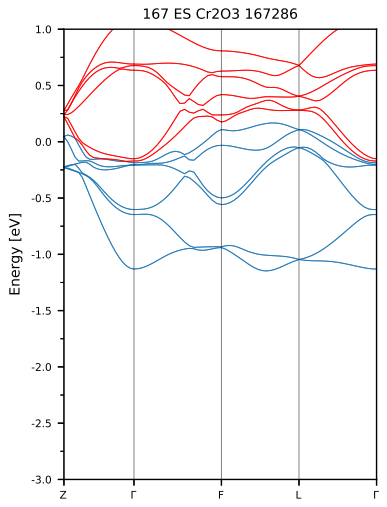

(ez) 167.3.64

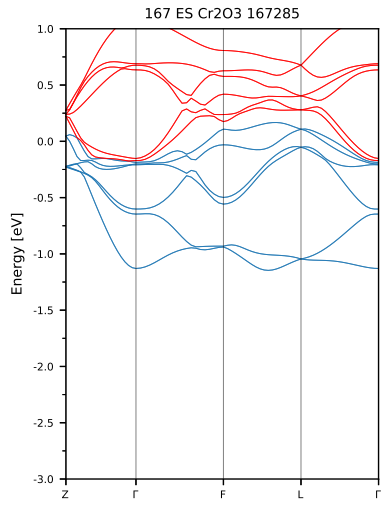

(fa) 167.3.65

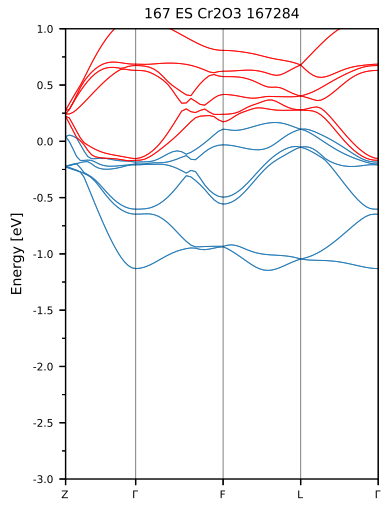

(fb) 167.3.66

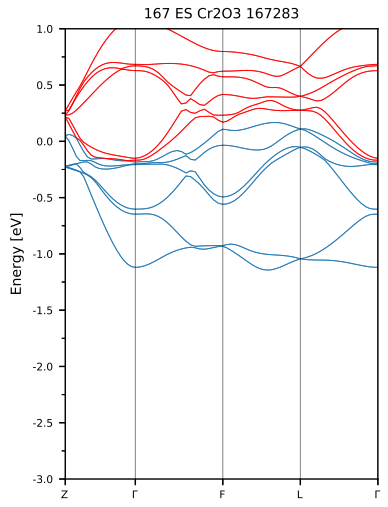

(fc) 167.3.67

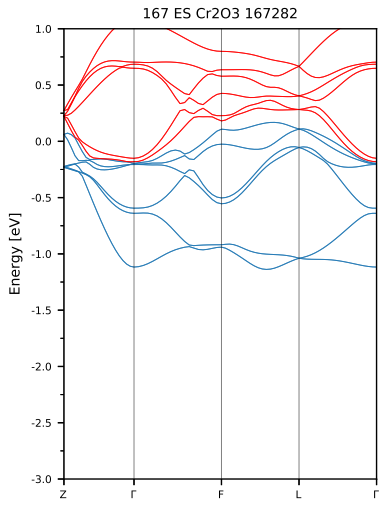

(fd) 167.3.68

Figure 399: Band structures for topological materials in SG 167, part 10/21, from 167.3.53 to 167.3.68. See the list of figures.

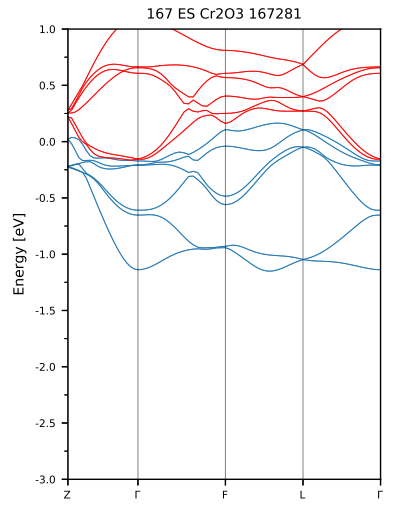

(fe) 167.3.69

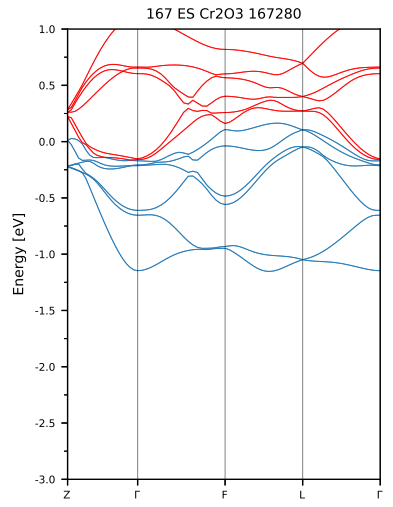

(ff) 167.3.70

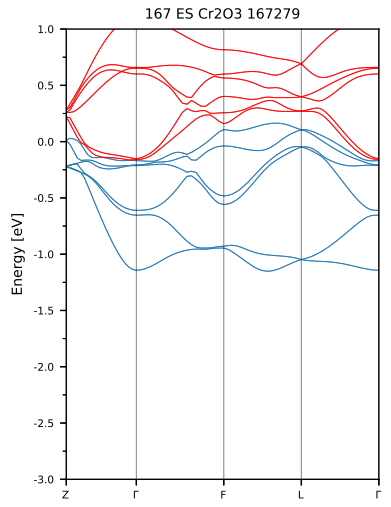

(fg) 167.3.71

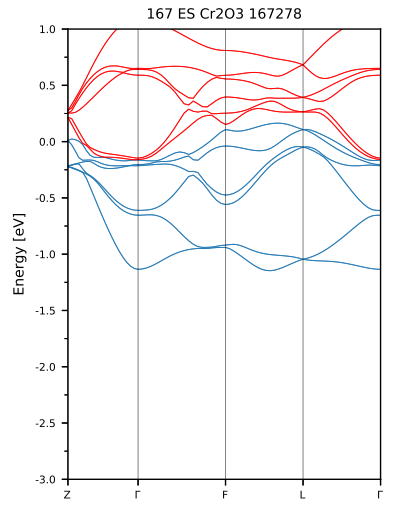

(fh) 167.3.72

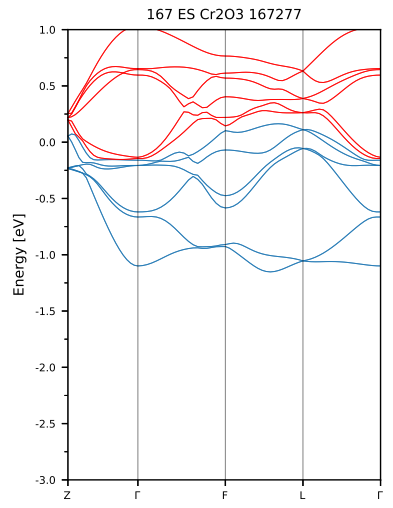

(fi) 167.3.73

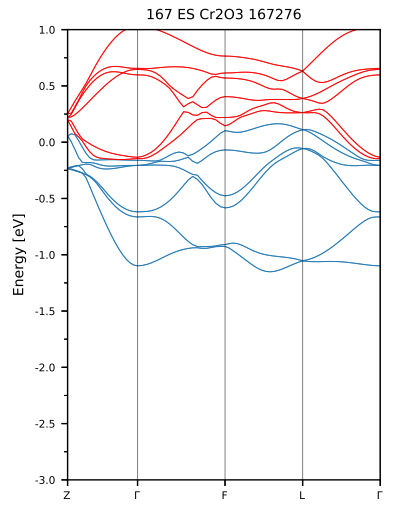

(fj) 167.3.74

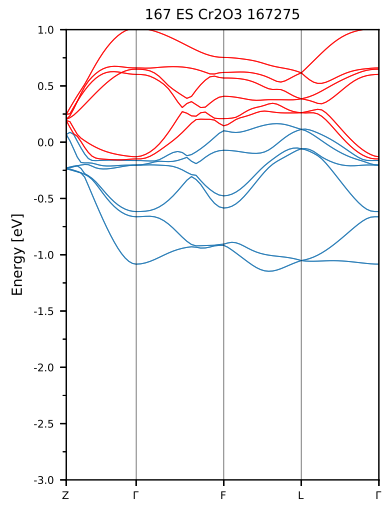

(fk) 167.3.75

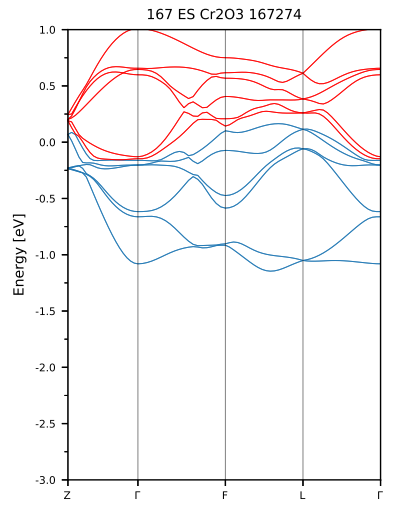

(fl) 167.3.76

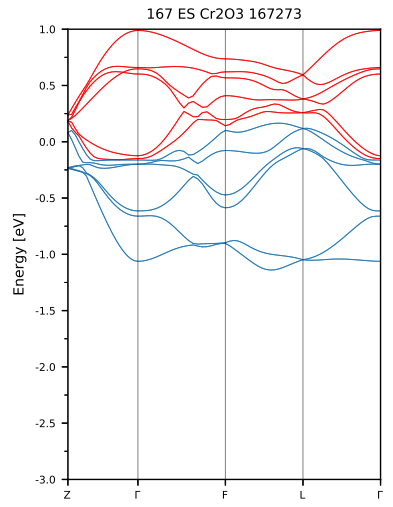

(fm) 167.3.77

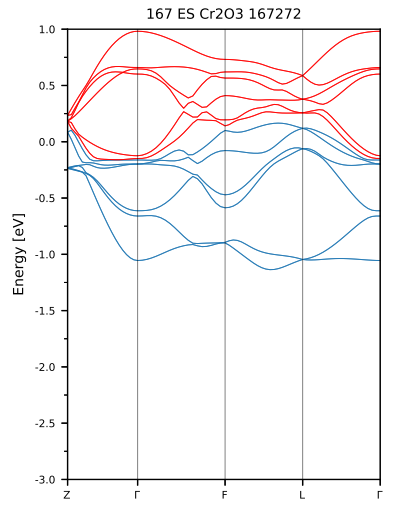

(fn) 167.3.78

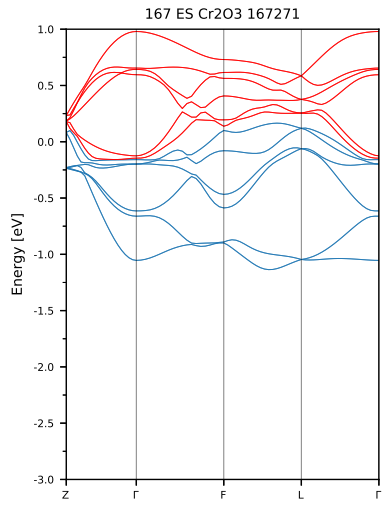

(fo) 167.3.79

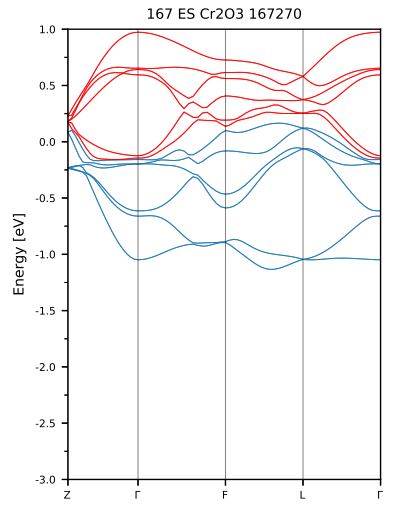

(fp) 167.3.80

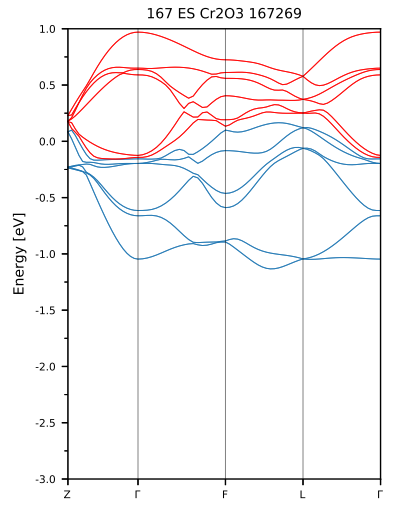

(fq) 167.3.81

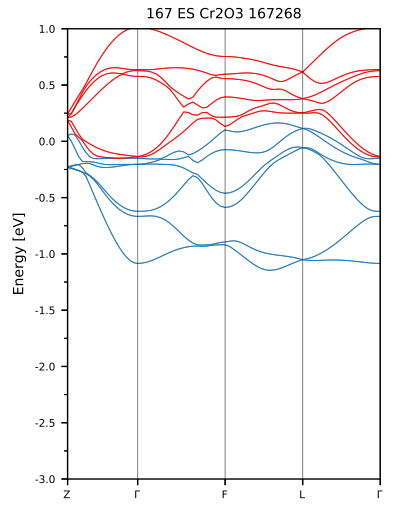

(fr) 167.3.82

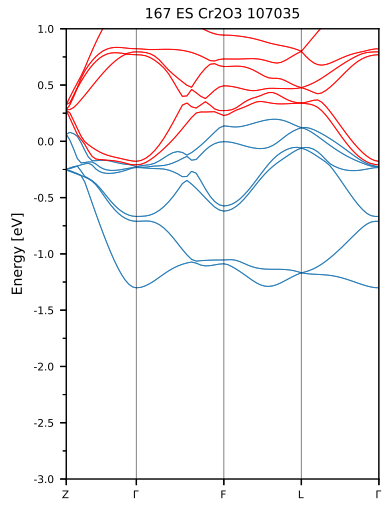

(fs) 167.3.83

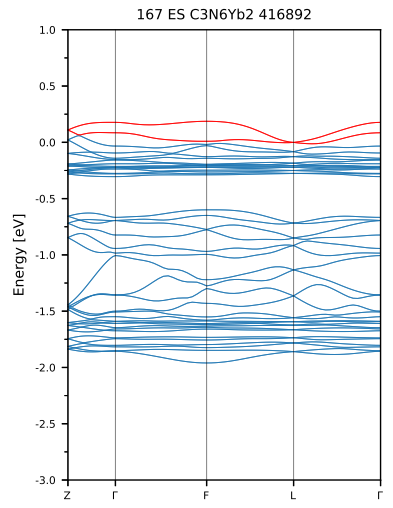

(ft) 167.3.84

Figure 400: Band structures for topological materials in SG 167, part 11/21, from 167.3.69 to 167.3.84. See the list of figures.

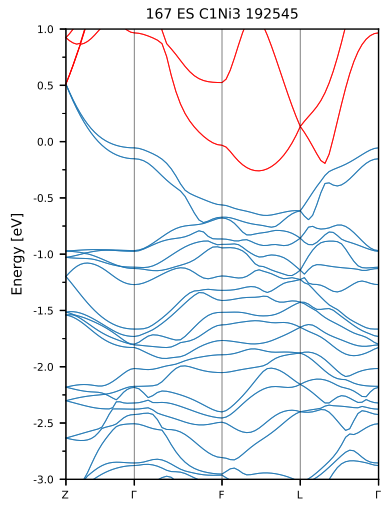

(fu) 167.3.85

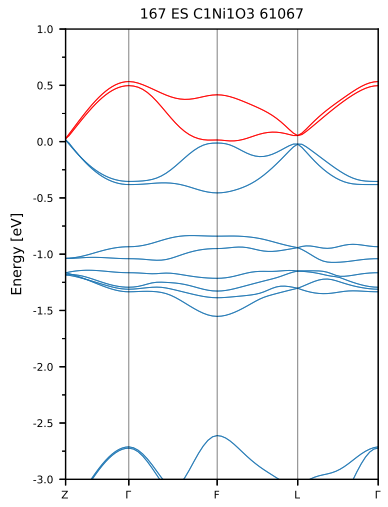

(fv) 167.3.86

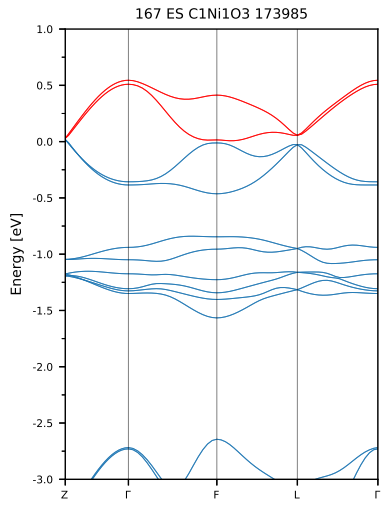

(fw) 167.3.87

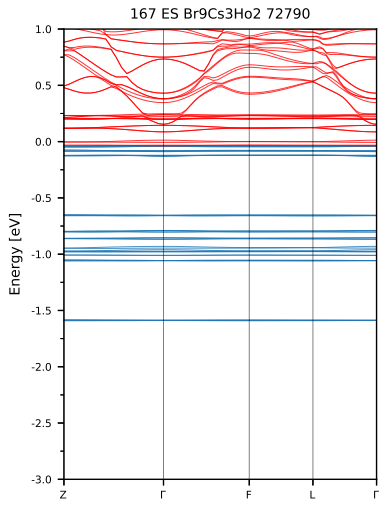

(fx) 167.3.88

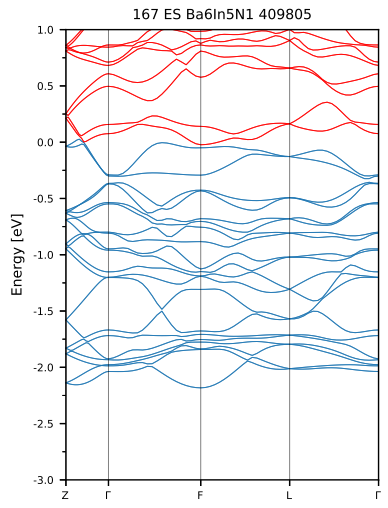

(fy) 167.3.89

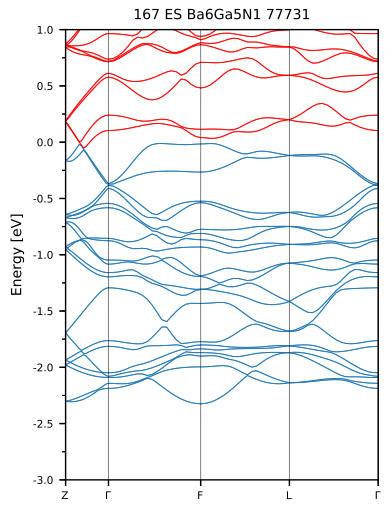

(fz) 167.3.90

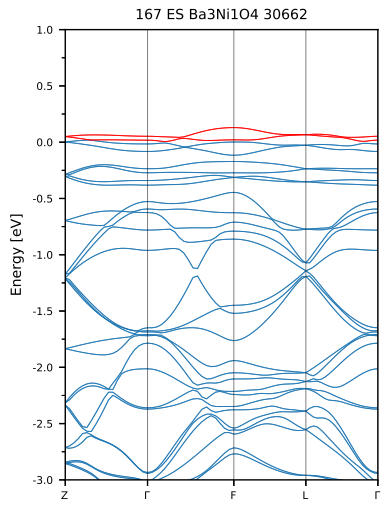

(ga) 167.3.91

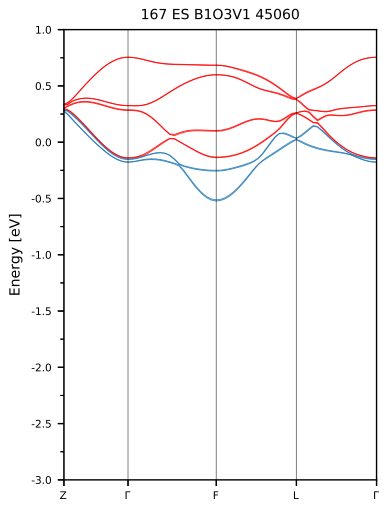

(gb) 167.3.92

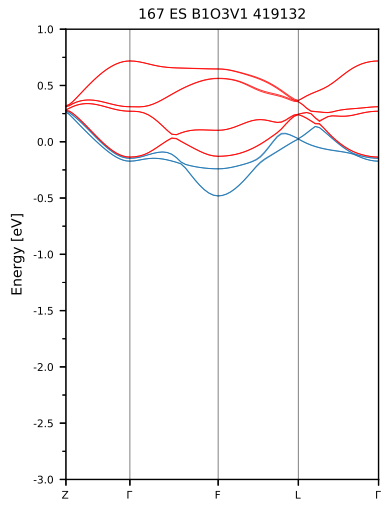

(gc) 167.3.93

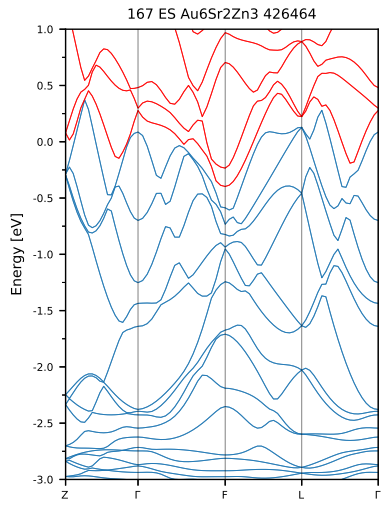

(gd) 167.3.94

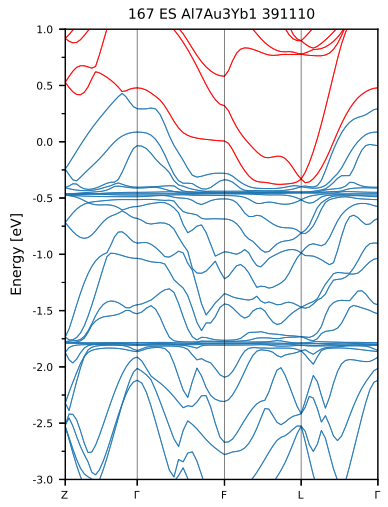

(ge) 167.3.95

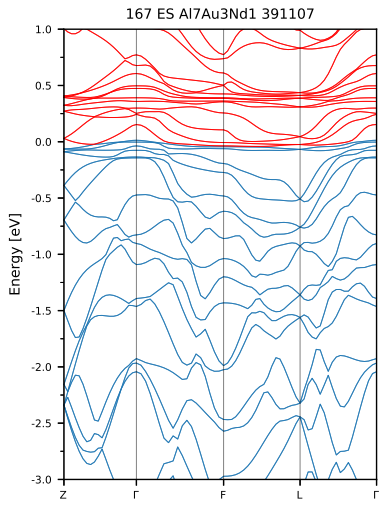

(gf) 167.3.96

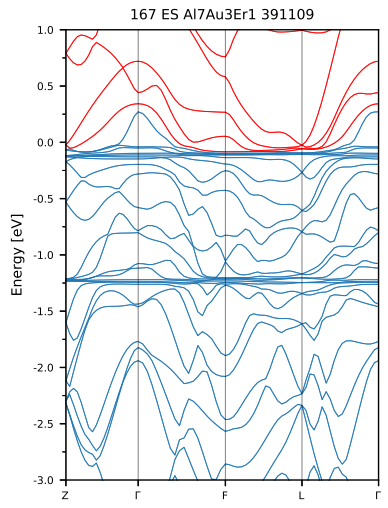

(gg) 167.3.97

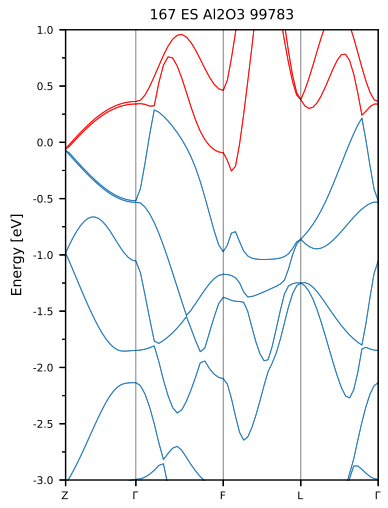

(gh) 167.3.98

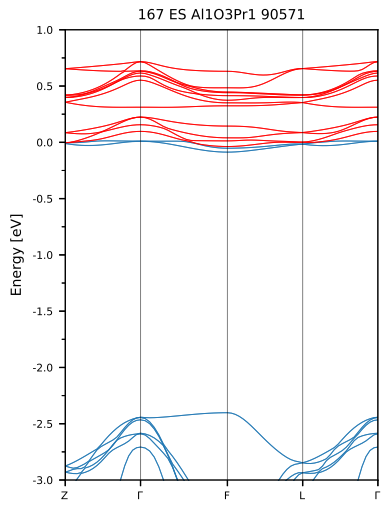

(gi) 167.3.99

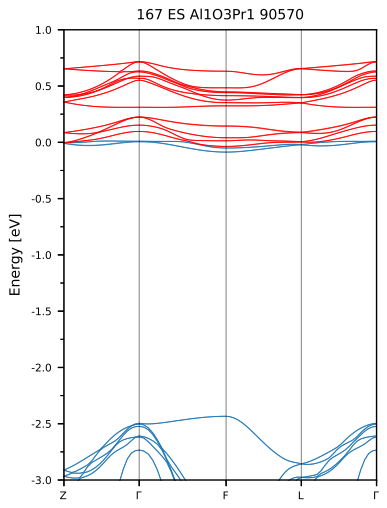

(gj) 167.3.100

Figure 401: Band structures for topological materials in SG 167, part 12/21, from 167.3.85 to 167.3.100. See the list of figures.

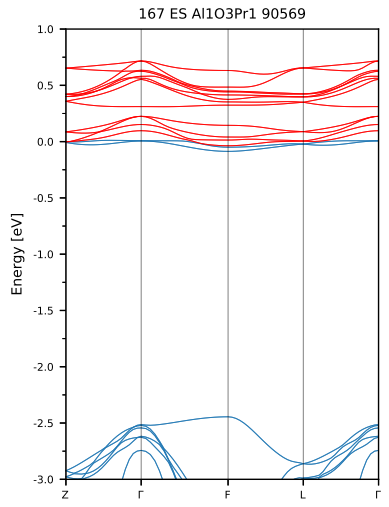

(gk) 167.3.101

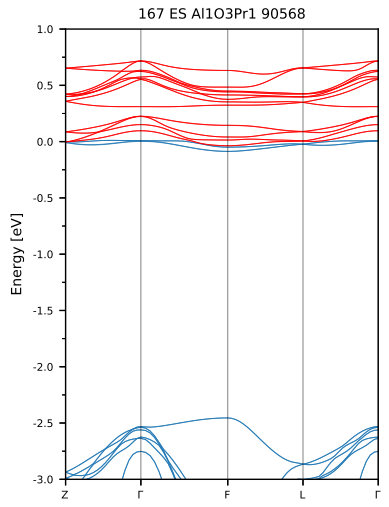

(gl) 167.3.102

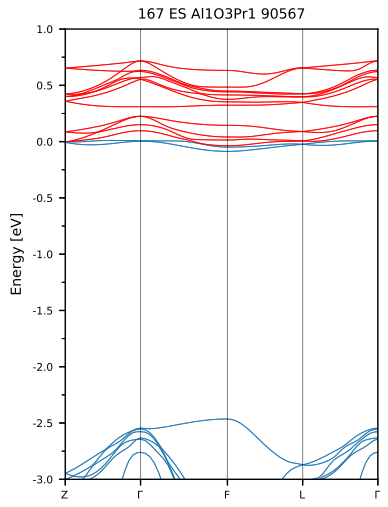

(gm) 167.3.103

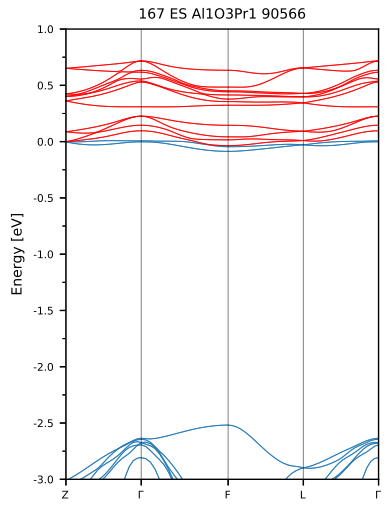

(gn) 167.3.104

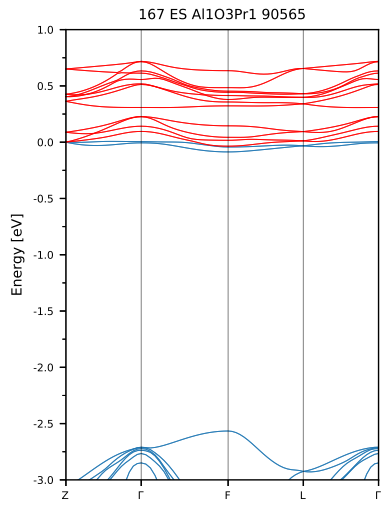

(go) 167.3.105

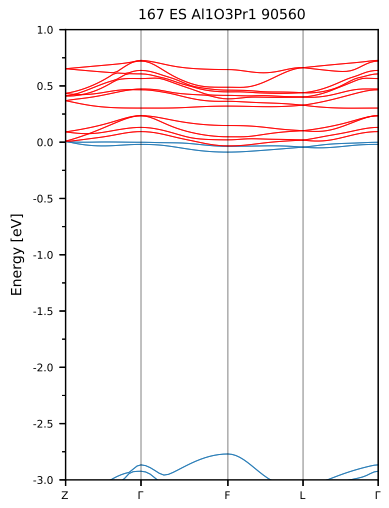

(gp) 167.3.106

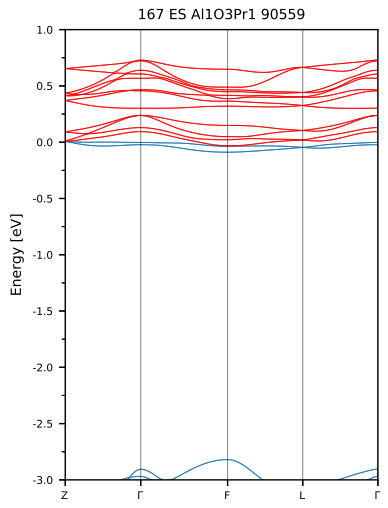

(gq) 167.3.107

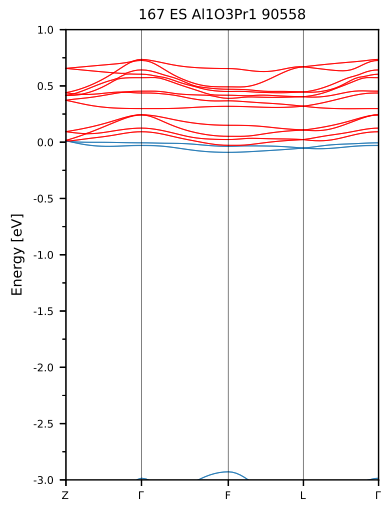

(gr) 167.3.108

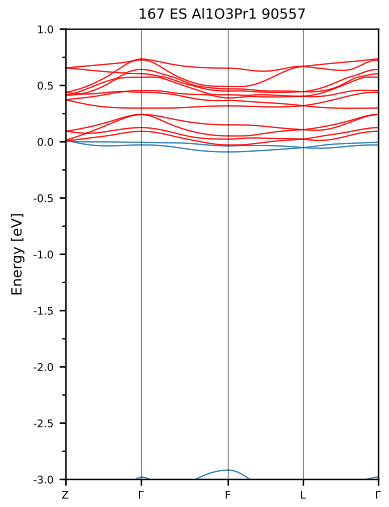

(gs) 167.3.109

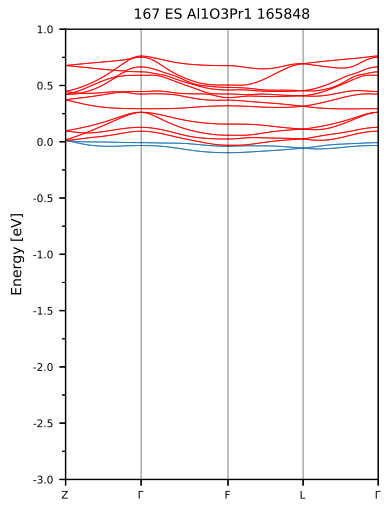

(gt) 167.3.110

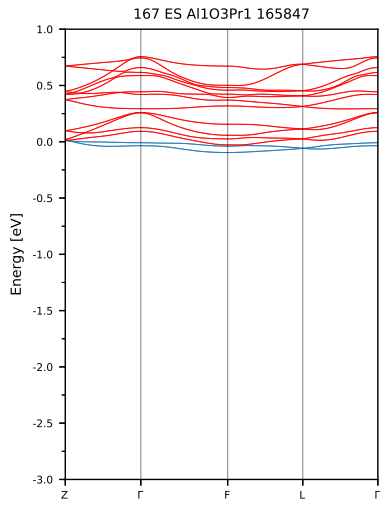

(gu) 167.3.111

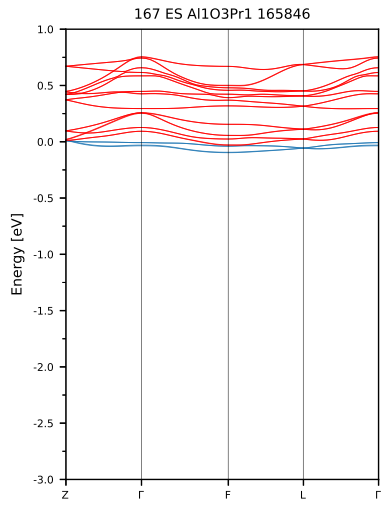

(gv) 167.3.112

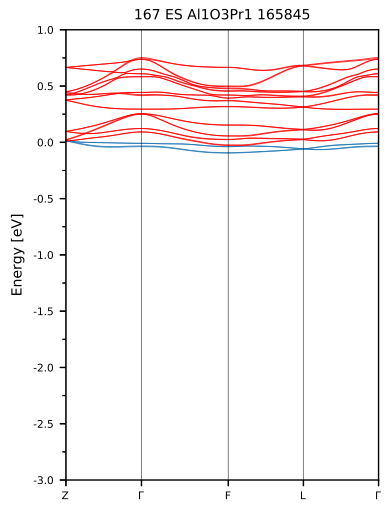

(gw) 167.3.113

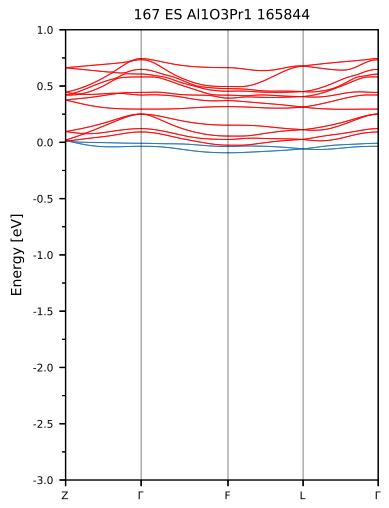

(gx) 167.3.114

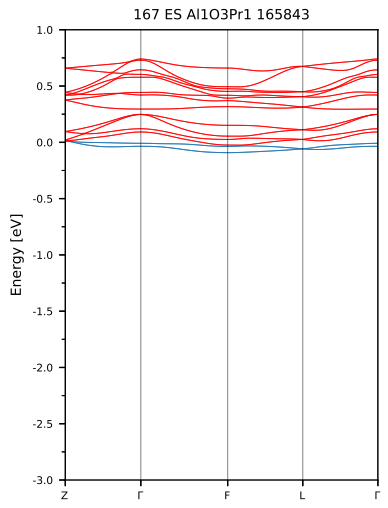

(gy) 167.3.115

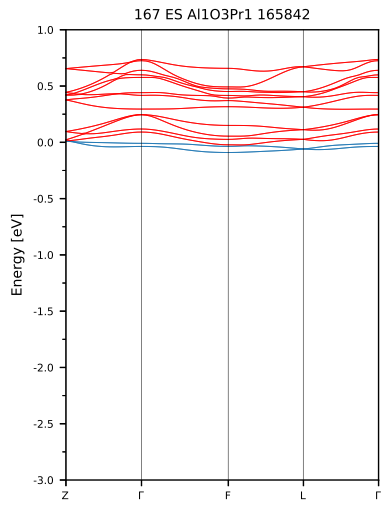

(gz) 167.3.116

Figure 402: Band structures for topological materials in SG 167, part 13/21, from 167.3.101 to 167.3.116. See the list of figures.

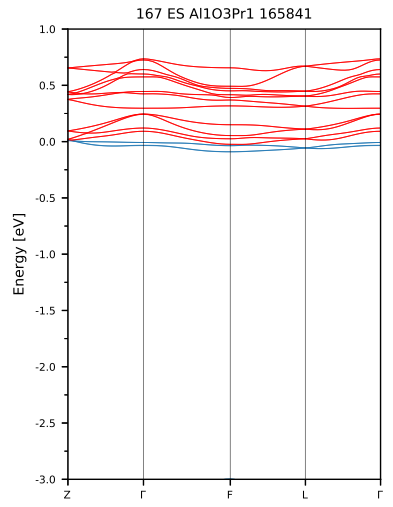

(ha) 167.3.117

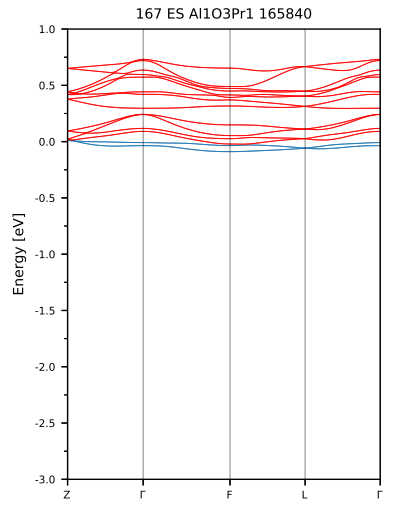

(hb) 167.3.118

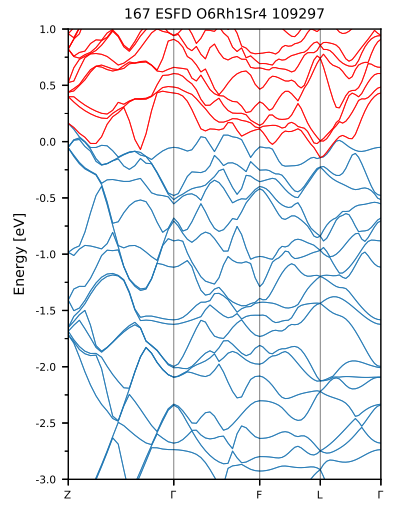

(hc) 167.4.1

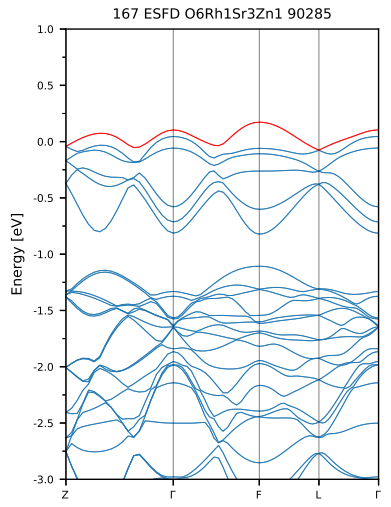

(hd) 167.4.2

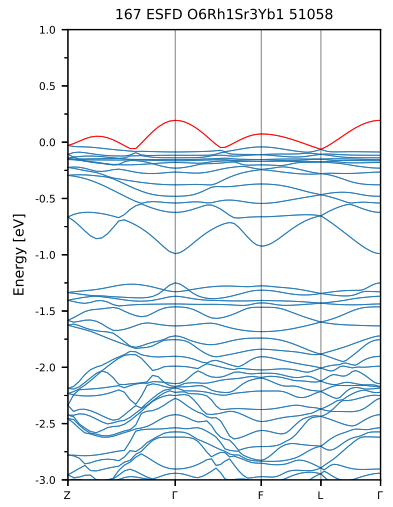

(he) 167.4.3

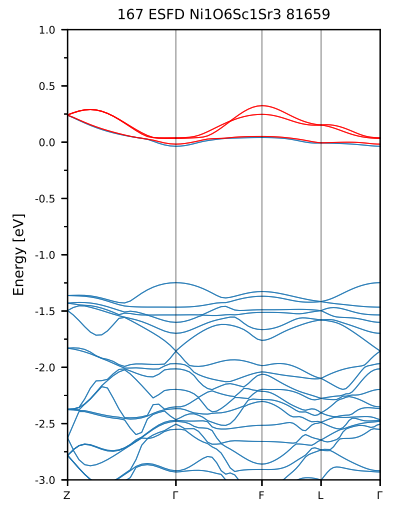

(hf) 167.4.4

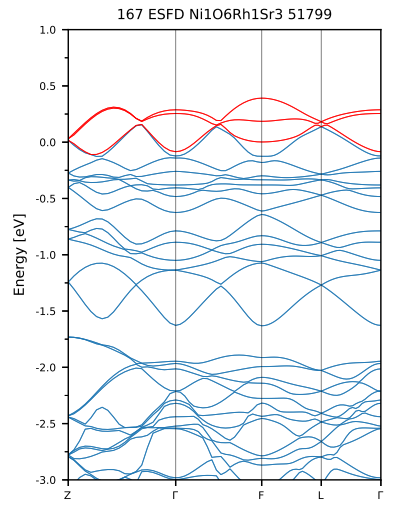

(hg) 167.4.5

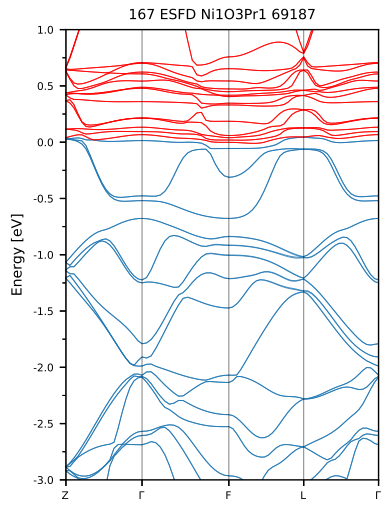

(hh) 167.4.6

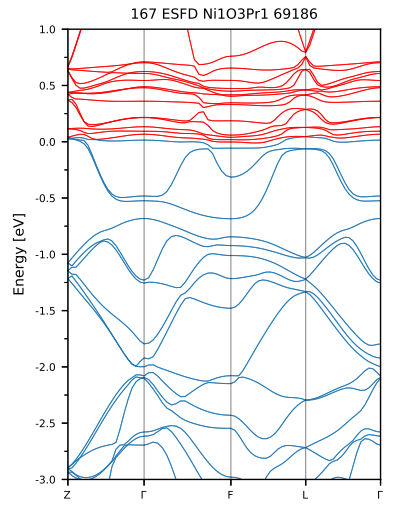

(hi) 167.4.7

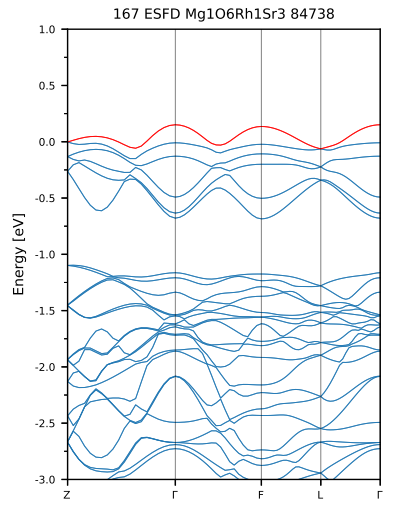

(hj) 167.4.8

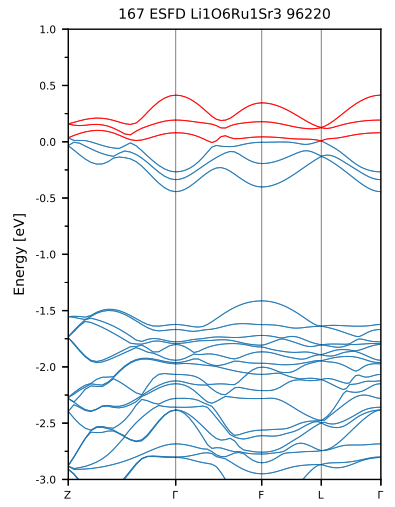

(hk) 167.4.9

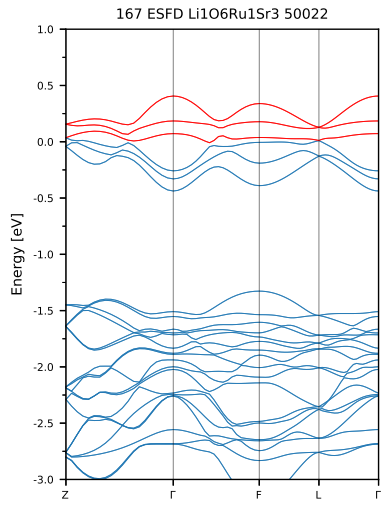

(hl) 167.4.10

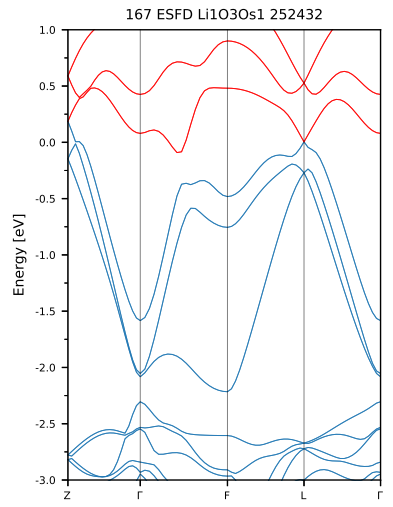

(hm) 167.4.11

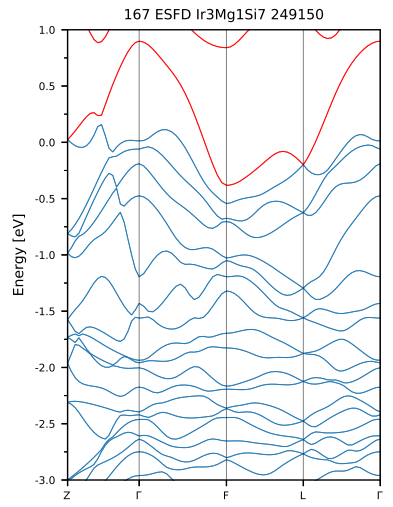

(hn) 167.4.12

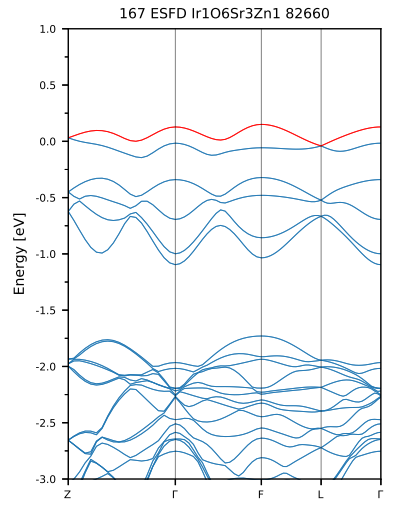

(ho) 167.4.13

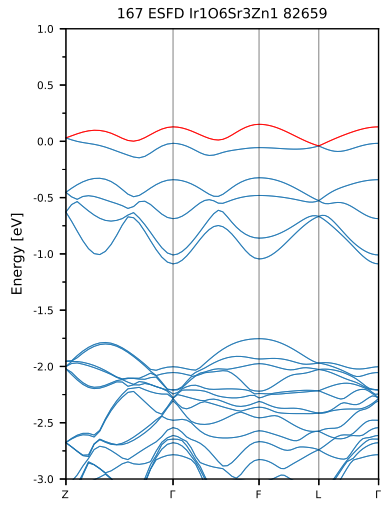

(hp) 167.4.14

Figure 403: Band structures for topological materials in SG 167, part 14/21, from 167.3.117 to 167.4.14. See the list of figures.

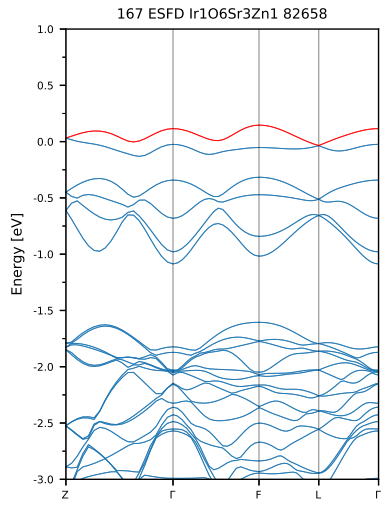

(hq) 167.4.15

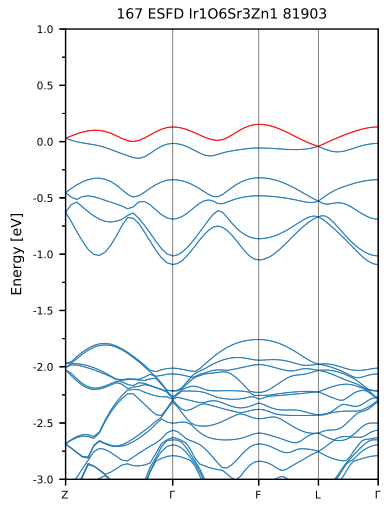

(hr) 167.4.16

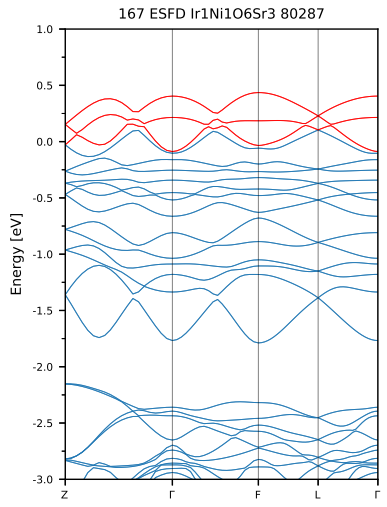

(hs) 167.4.17

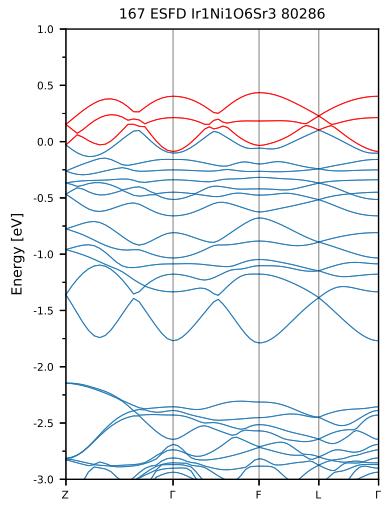

(ht) 167.4.18

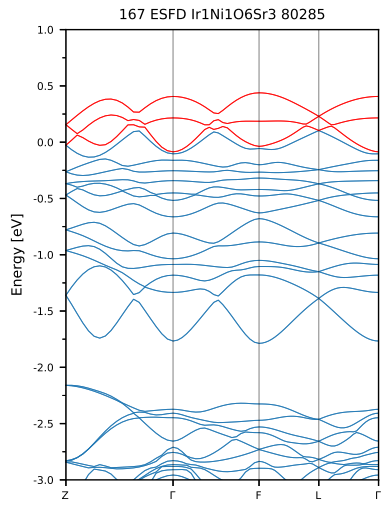

(hu) 167.4.19

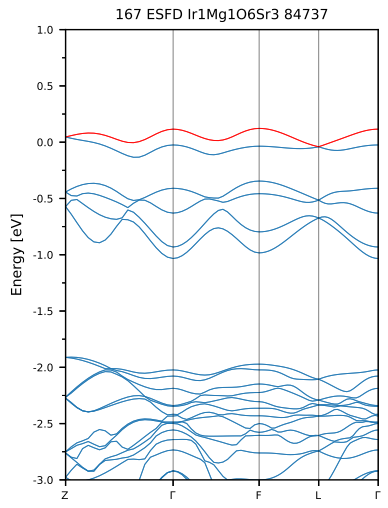

(hv) 167.4.20

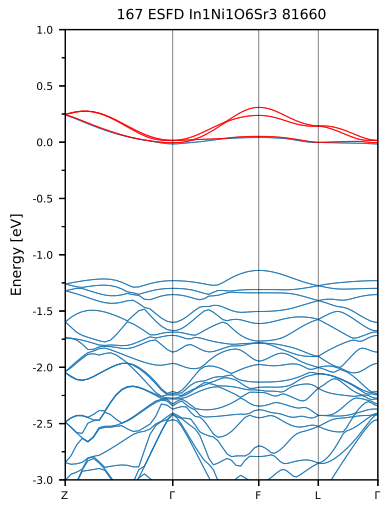

(hw) 167.4.21

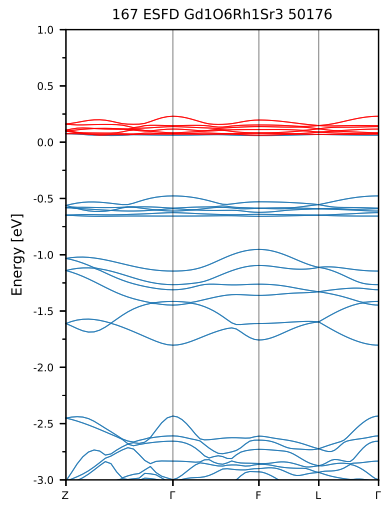

(hx) 167.4.22

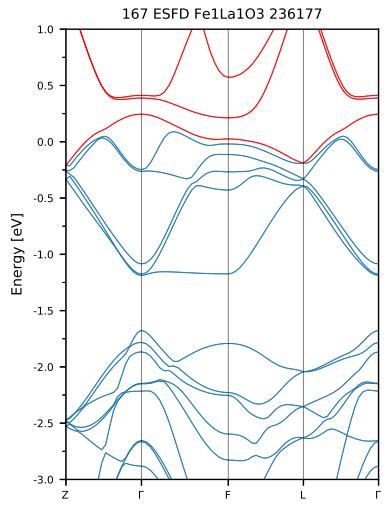

(hy) 167.4.23

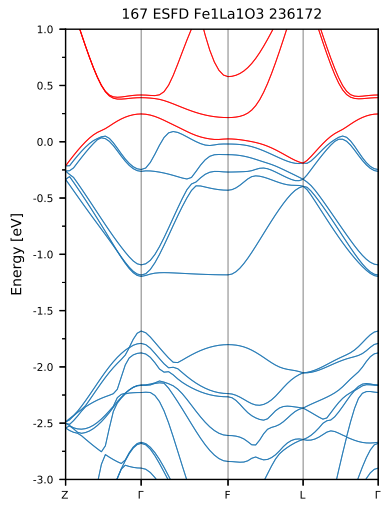

(hz) 167.4.24

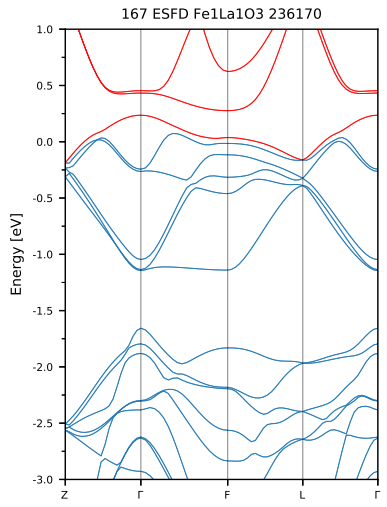

(ia) 167.4.25

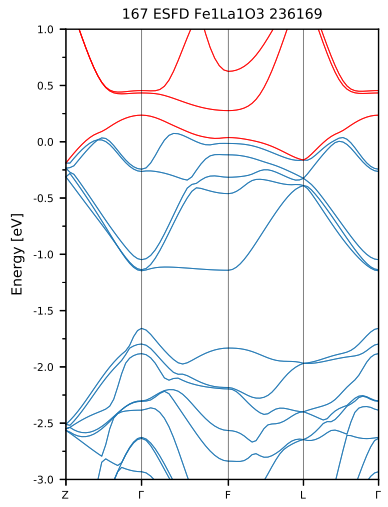

(ib) 167.4.26

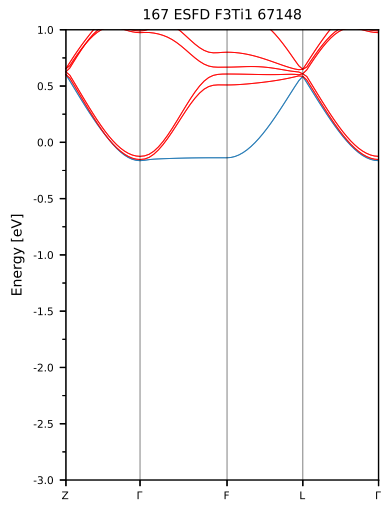

(ic) 167.4.27

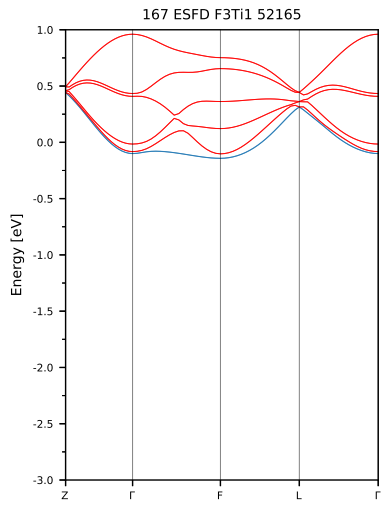

(id) 167.4.28

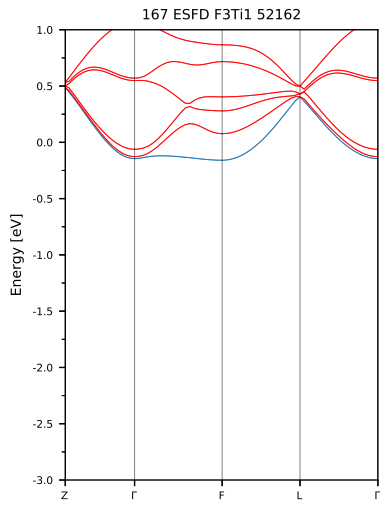

(ie) 167.4.29

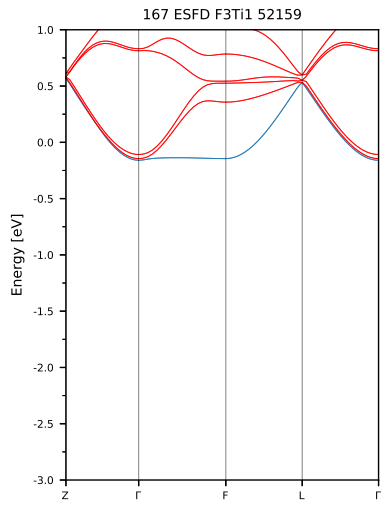

(if) 167.4.30

Figure 404: Band structures for topological materials in SG 167, part 15/21, from 167.4.15 to 167.4.30. See the list of figures.

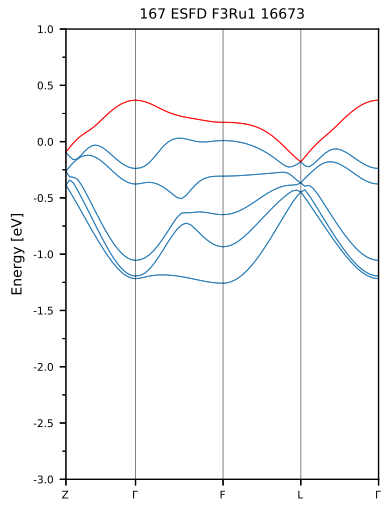

(ig) 167.4.31

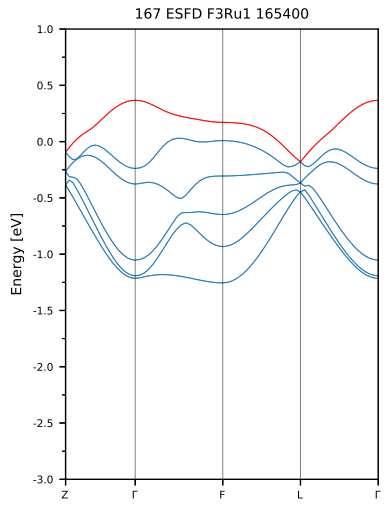

(ih) 167.4.32

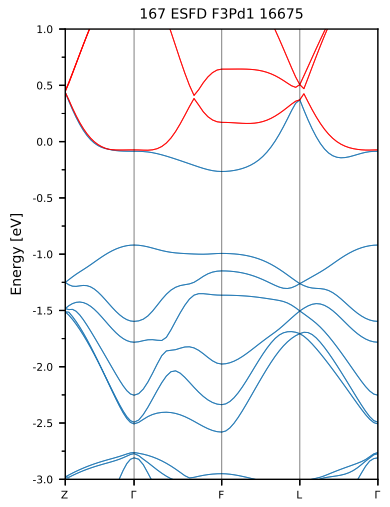

(ii) 167.4.33

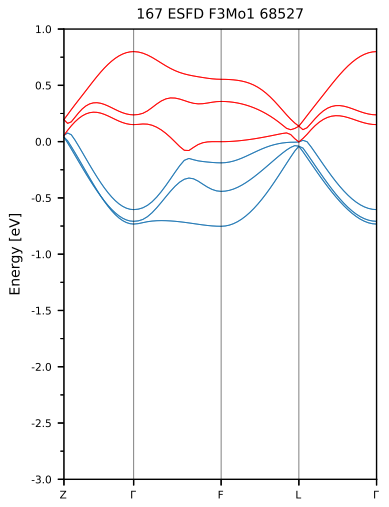

(ij) 167.4.34

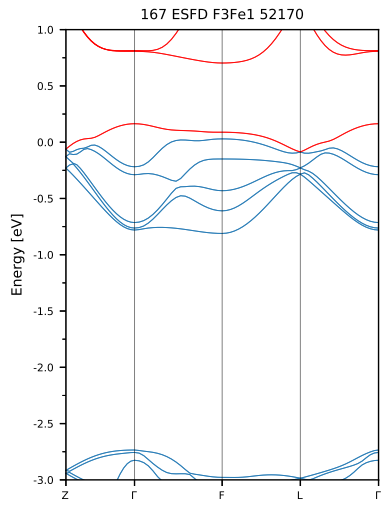

(ik) 167.4.35

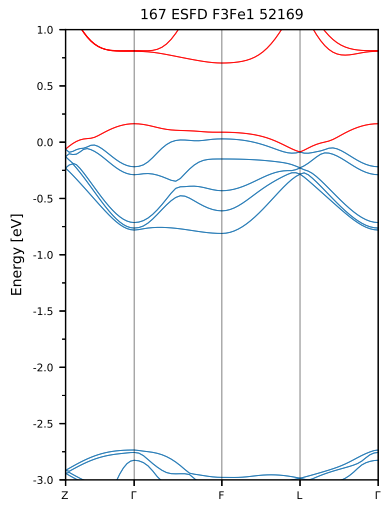

(il) 167.4.36

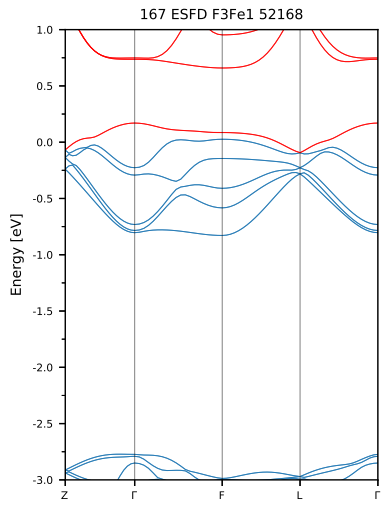

(im) 167.4.37

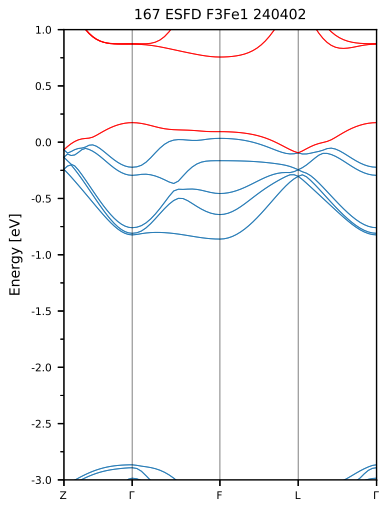

(in) 167.4.38

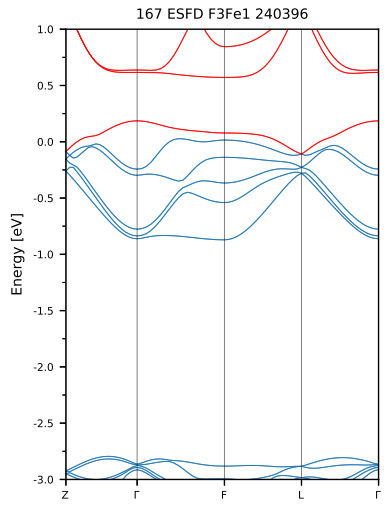

(io) 167.4.39

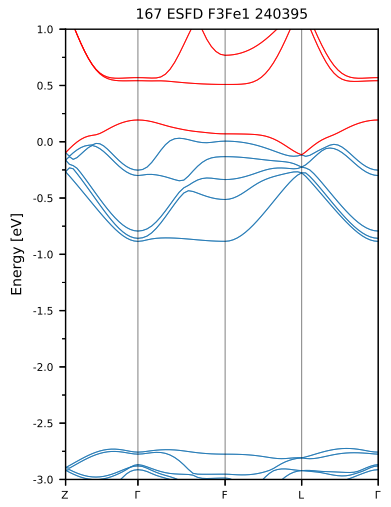

(ip) 167.4.40

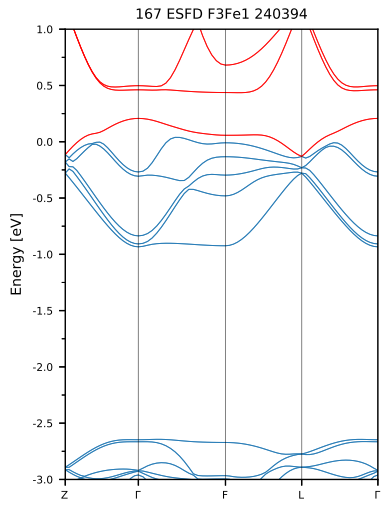

(iq) 167.4.41

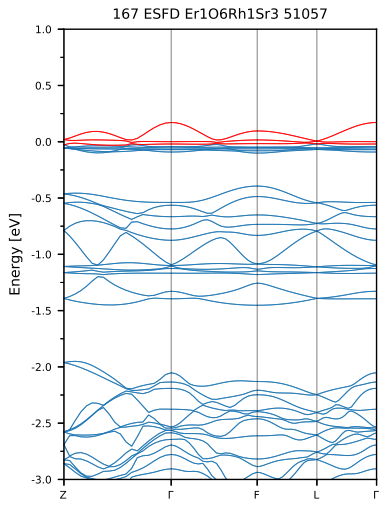

(ir) 167.4.42

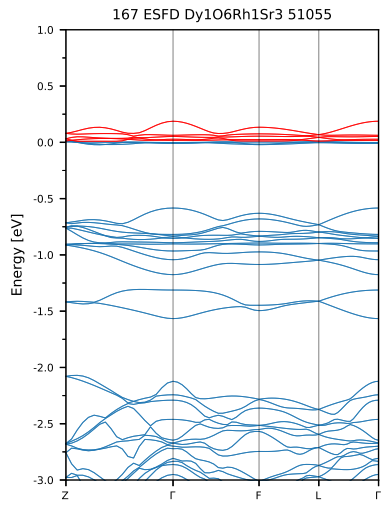

(is) 167.4.43

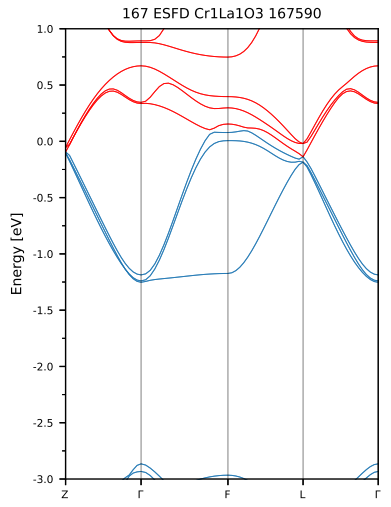

(it) 167.4.44

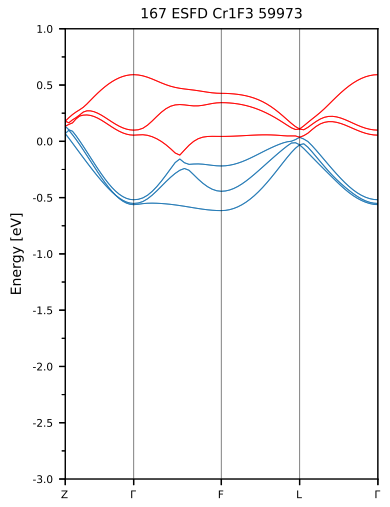

(iu) 167.4.45

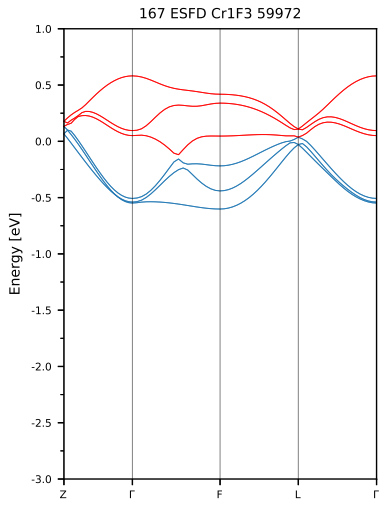

(iv) 167.4.46

Figure 405: Band structures for topological materials in SG 167, part 16/21, from 167.4.31 to 167.4.46. See the list of figures.

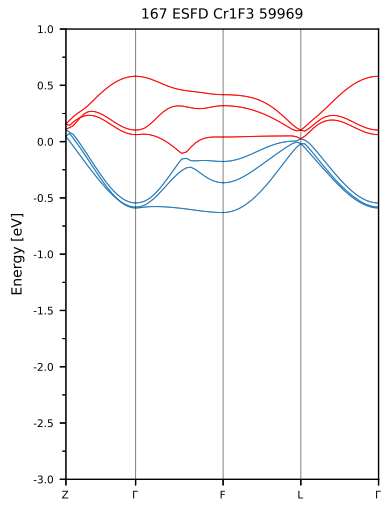

(iw) 167.4.47

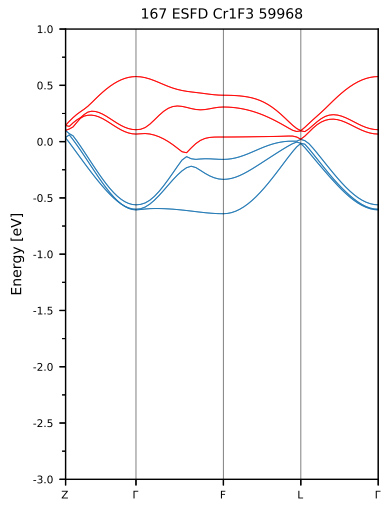

(ix) 167.4.48

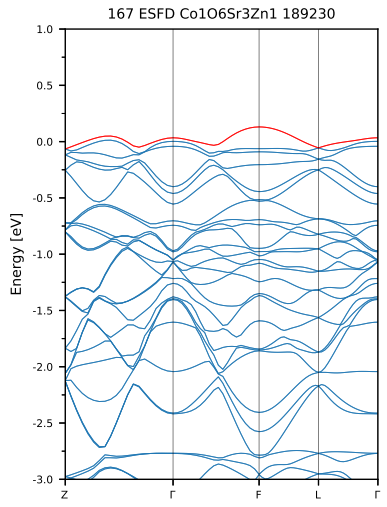

(iy) 167.4.49

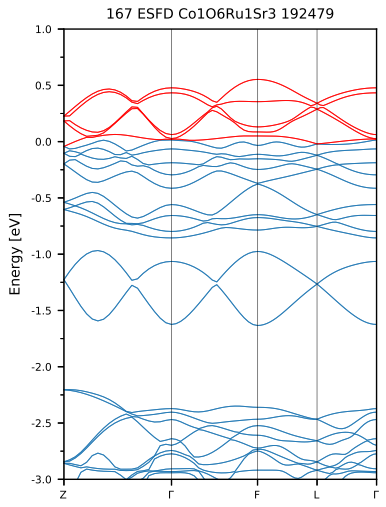

(iz) 167.4.50

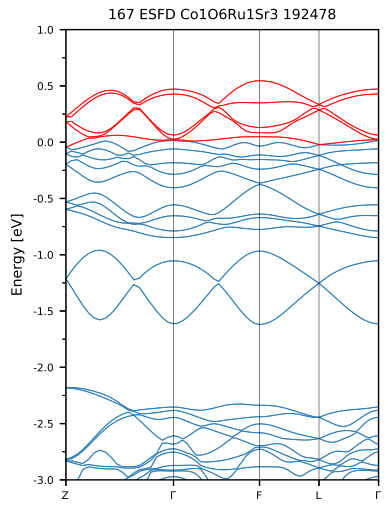

(ja) 167.4.51

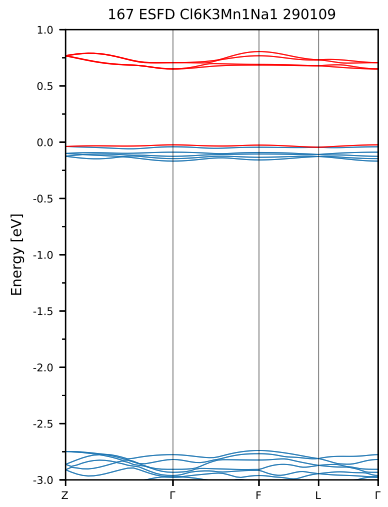

(jb) 167.4.52

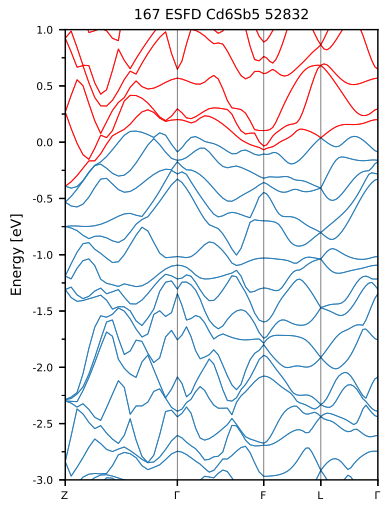

(jc) 167.4.53

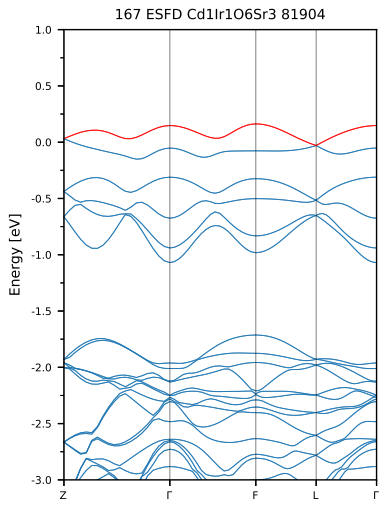

(jd) 167.4.54

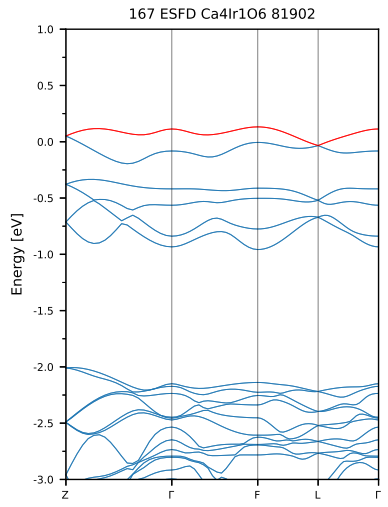

(je) 167.4.55

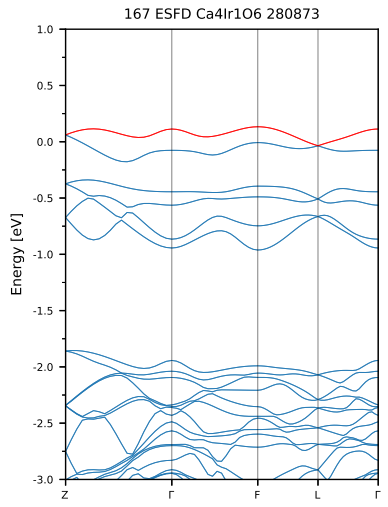

(jf) 167.4.56

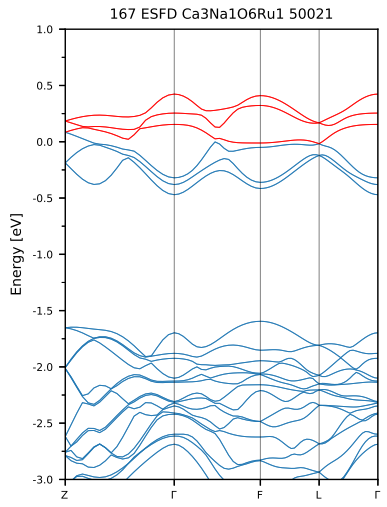

(jg) 167.4.57

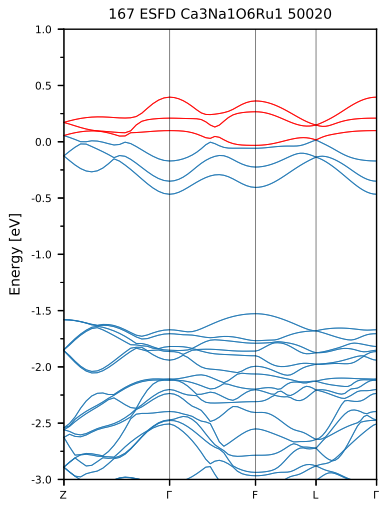

(jh) 167.4.58

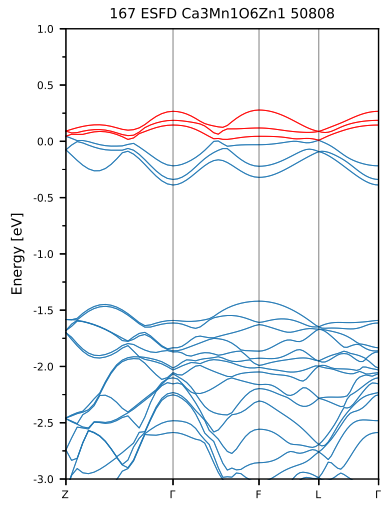

(ji) 167.4.59

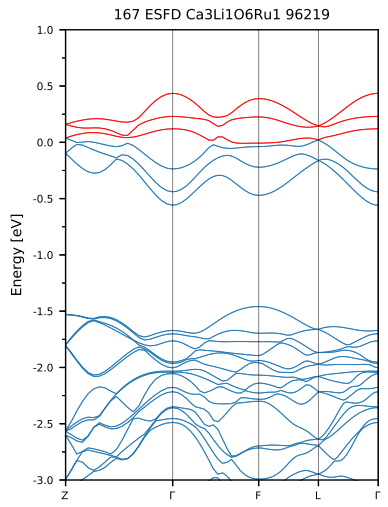

(jj) 167.4.60

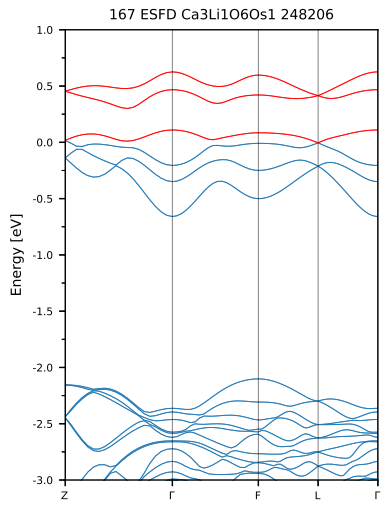

(jk) 167.4.61

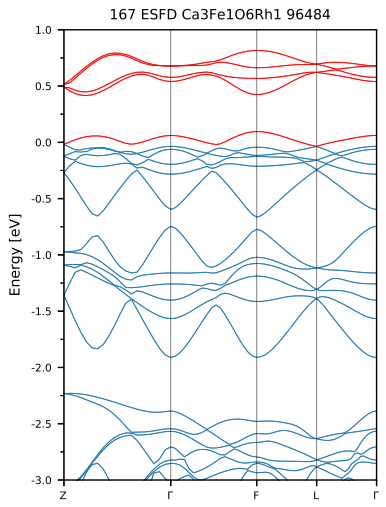

(jl) 167.4.62

Figure 406: Band structures for topological materials in SG 167, part 17/21, from 167.4.47 to 167.4.62. See the list of figures.

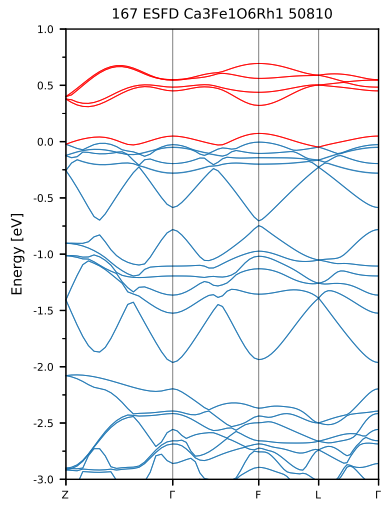

(jm) 167.4.63

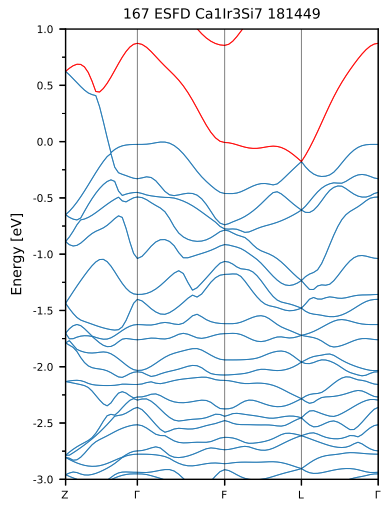

(jn) 167.4.64

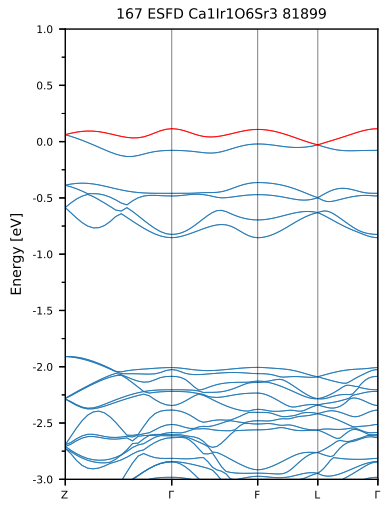

(jo) 167.4.65

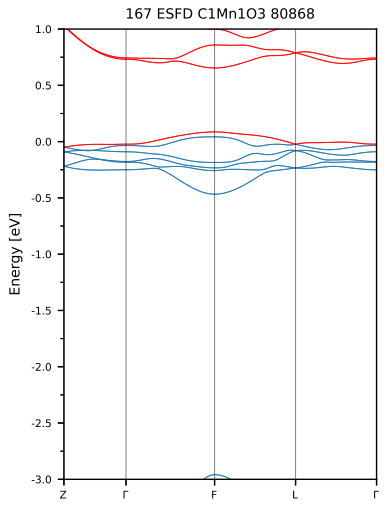

(jp) 167.4.66

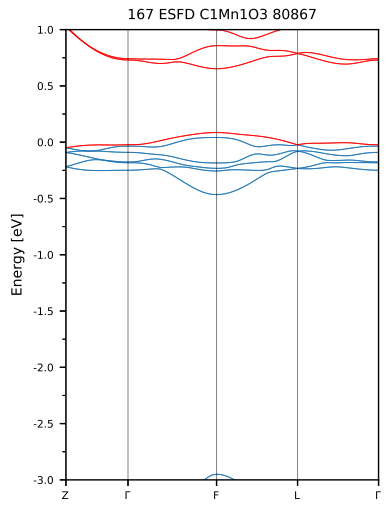

(jq) 167.4.67

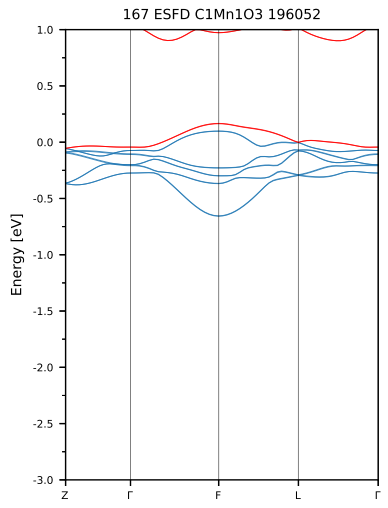

(jr) 167.4.68

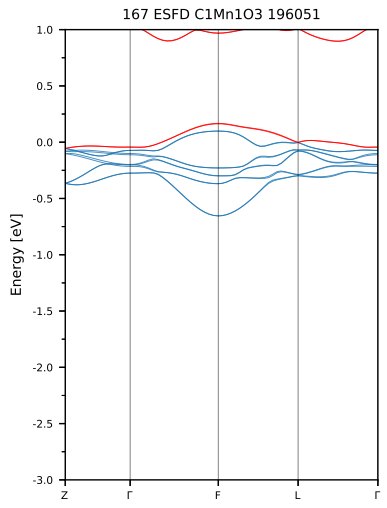

(js) 167.4.69

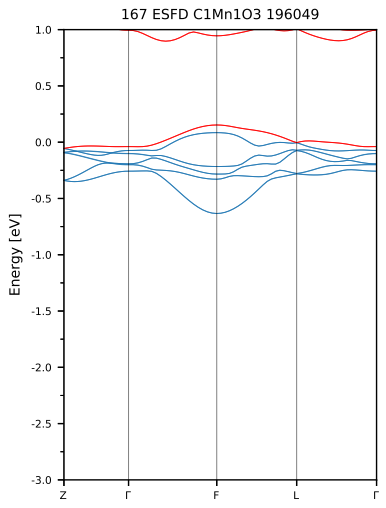

(jt) 167.4.70

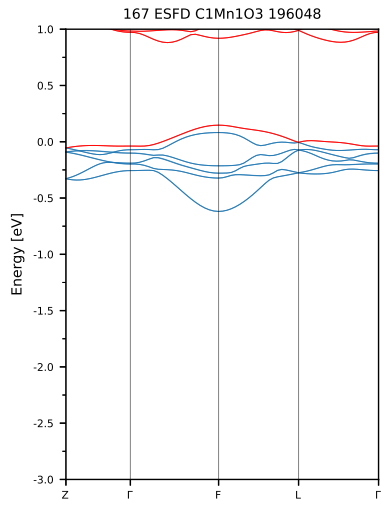

(ju) 167.4.71

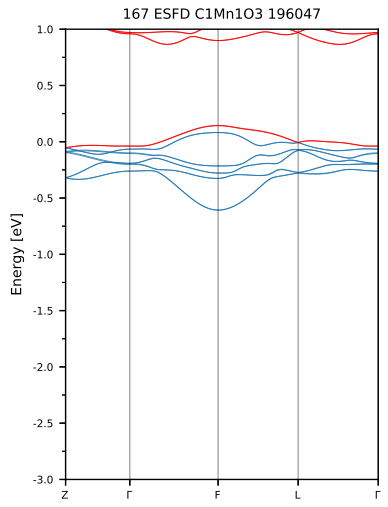

(jv) 167.4.72

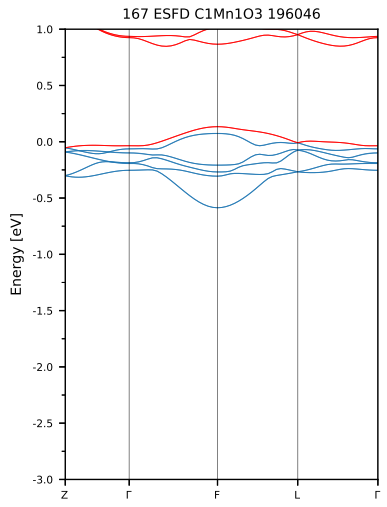

(jw) 167.4.73

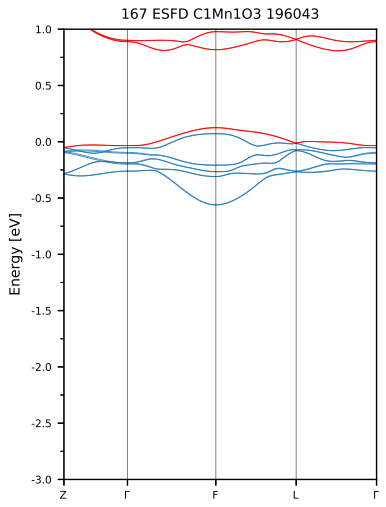

(jx) 167.4.74

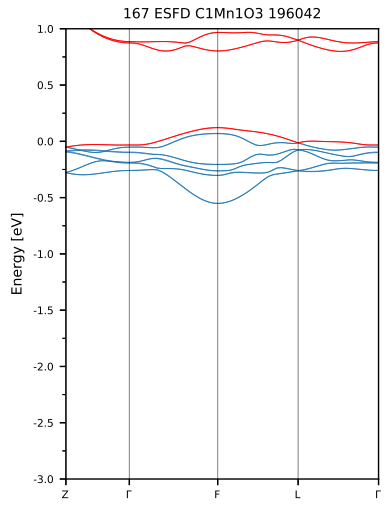

(jy) 167.4.75

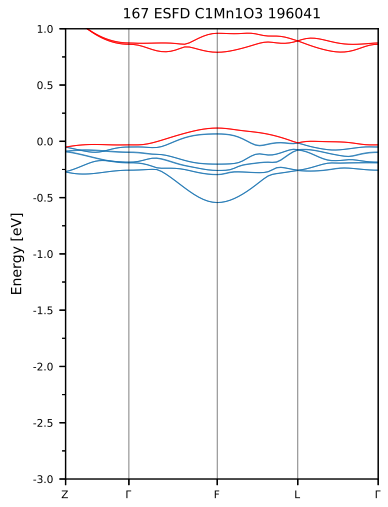

(jz) 167.4.76

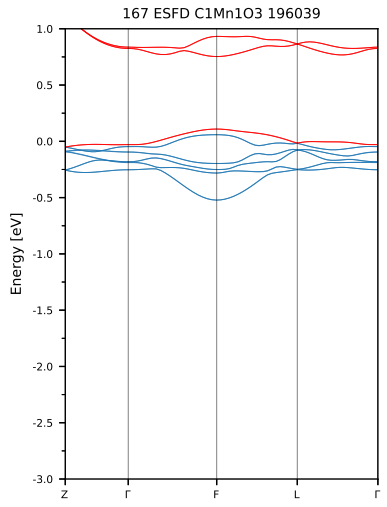

(ka) 167.4.77

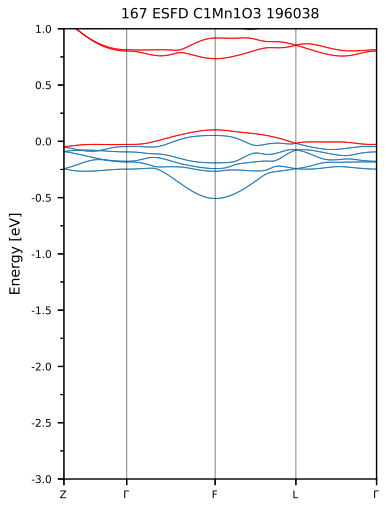

(kb) 167.4.78

Figure 407: Band structures for topological materials in SG 167, part 18/21, from 167.4.63 to 167.4.78. See the list of figures.

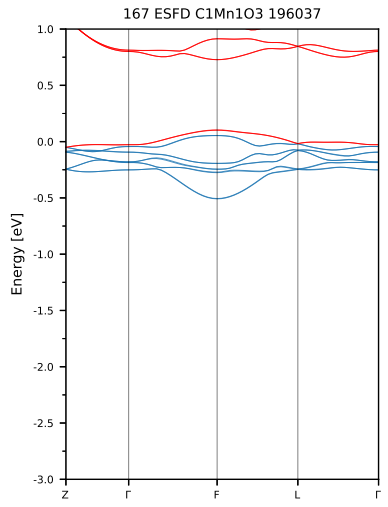

(kc) 167.4.79

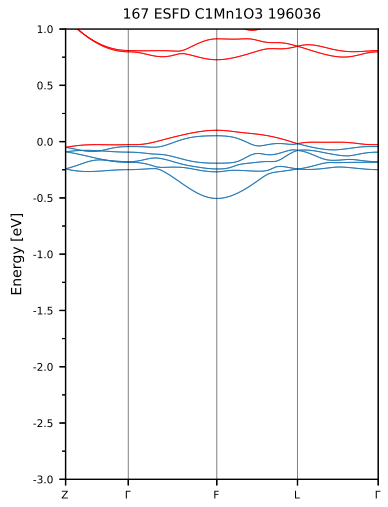

(kd) 167.4.80

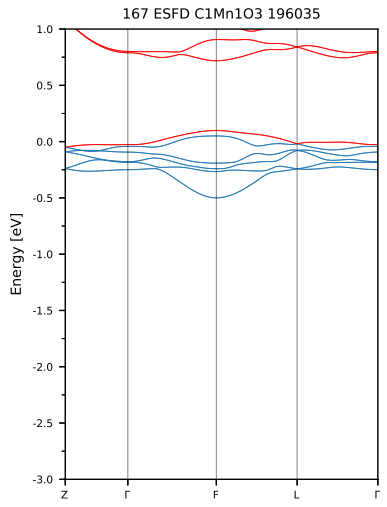

(ke) 167.4.81

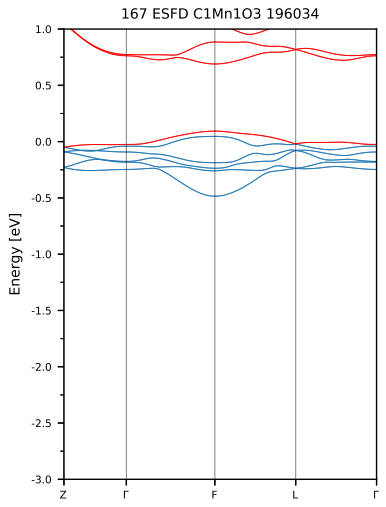

(kf) 167.4.82

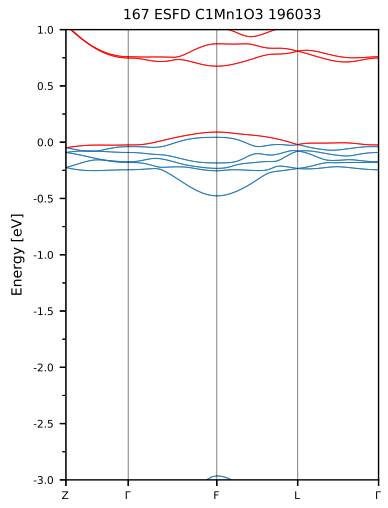

(kg) 167.4.83

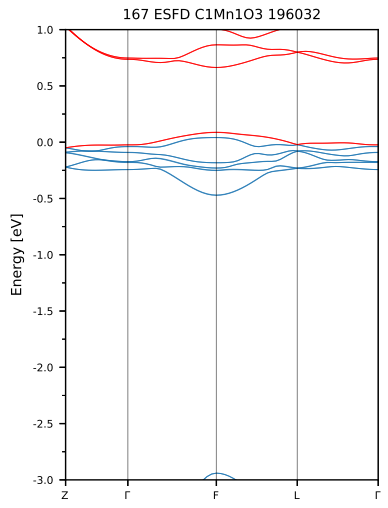

(kh) 167.4.84

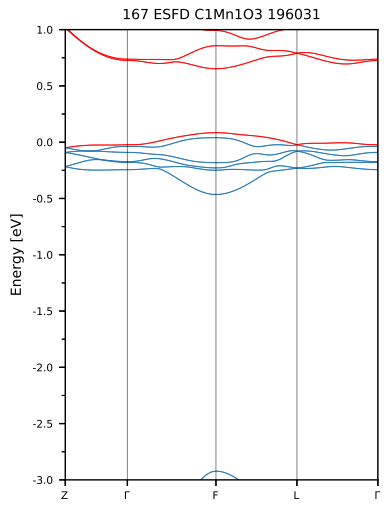

(ki) 167.4.85

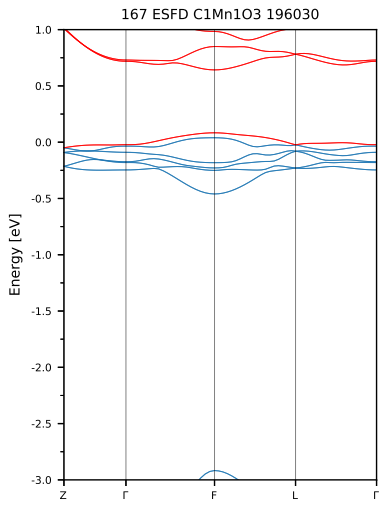

(kj) 167.4.86

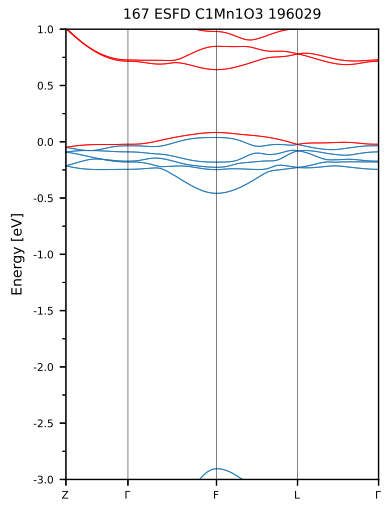

(kk) 167.4.87

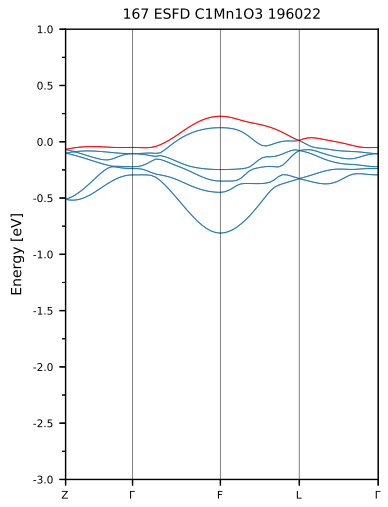

(kl) 167.4.88

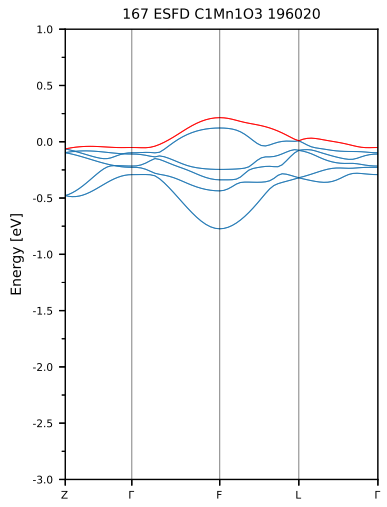

(km) 167.4.89

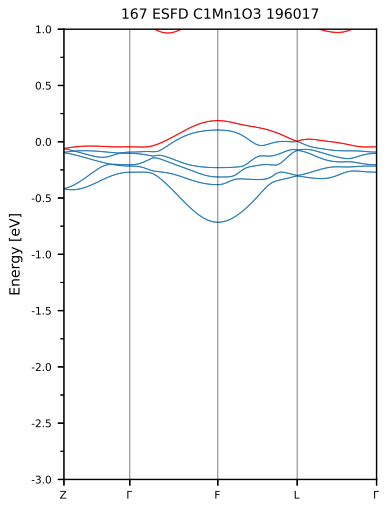

(kn) 167.4.90

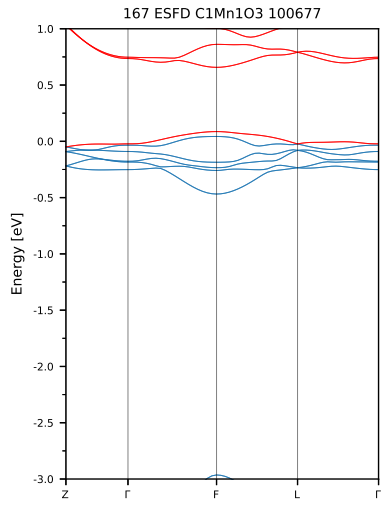

(ko) 167.4.91

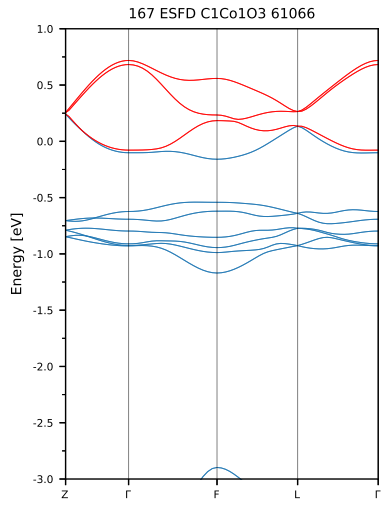

(kp) 167.4.92

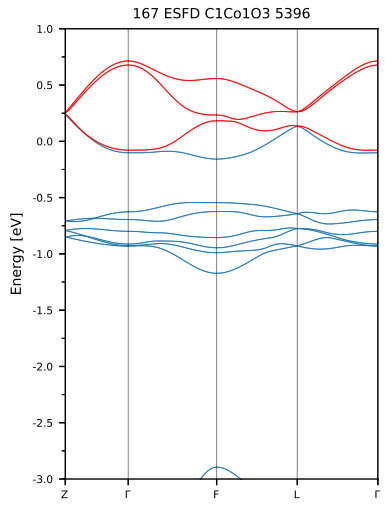

(kq) 167.4.93

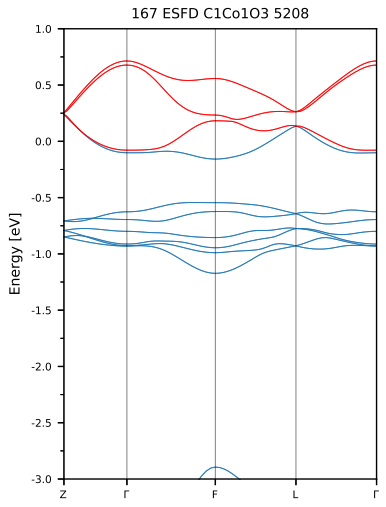

(kr) 167.4.94

Figure 408: Band structures for topological materials in SG 167, part 19/21, from 167.4.79 to 167.4.94. See the list of figures.

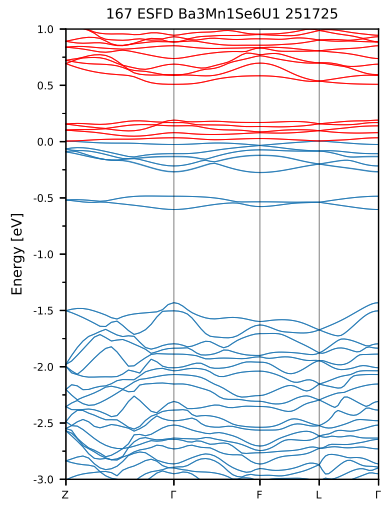

(ks) 167.4.95

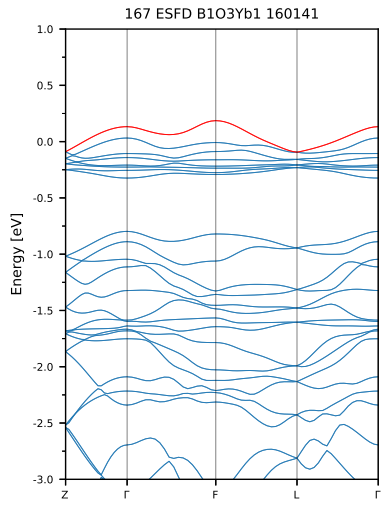

(kt) 167.4.96

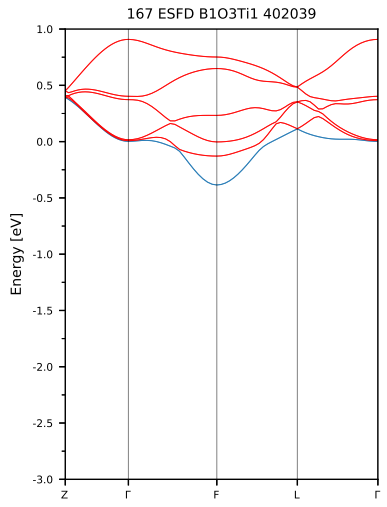

(ku) 167.4.97

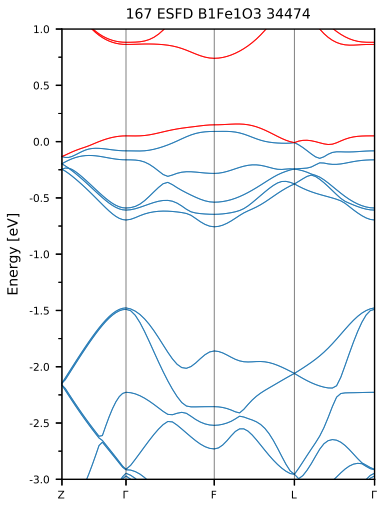

(kv) 167.4.98

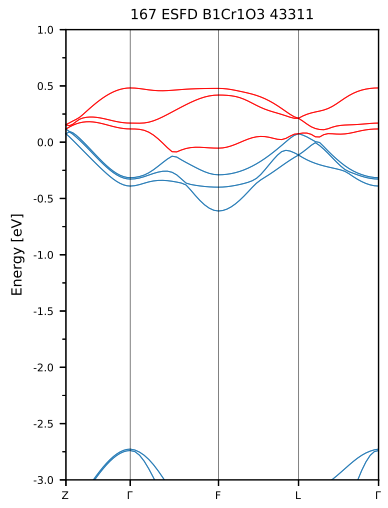

(kw) 167.4.99

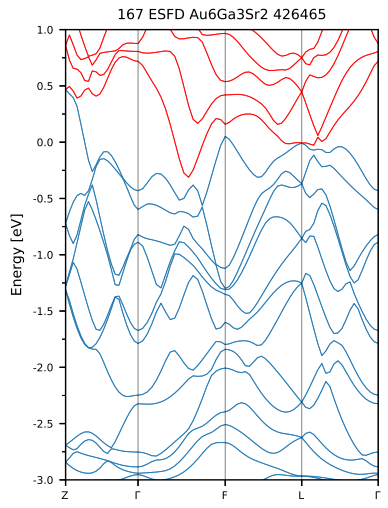

(kx) 167.4.100

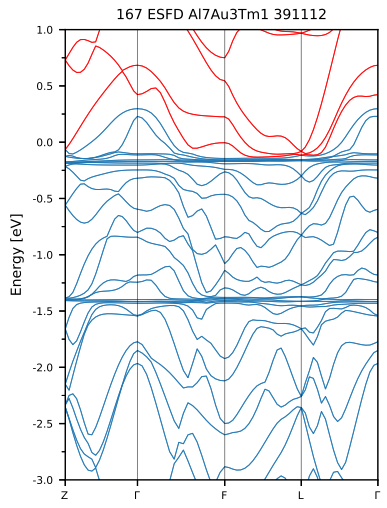

(ky) 167.4.101

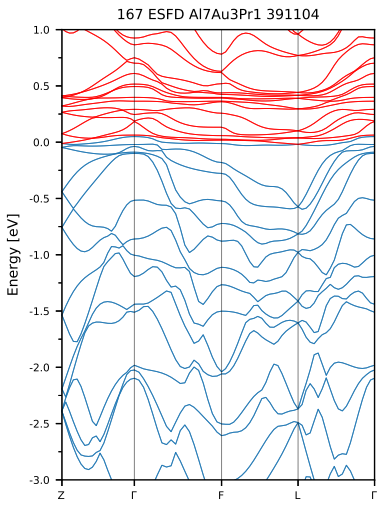

(kz) 167.4.102

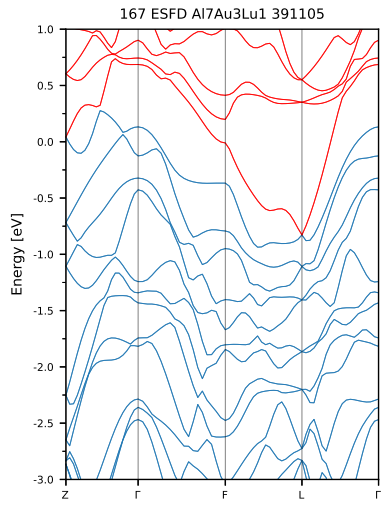

(la) 167.4.103

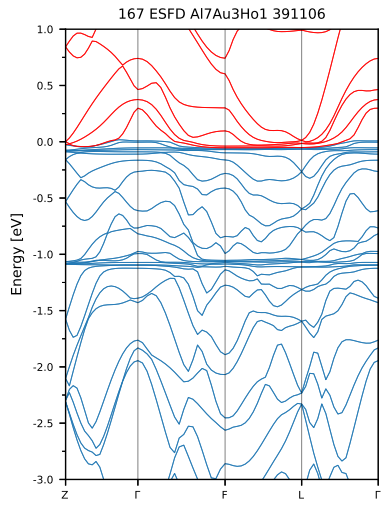

(lb) 167.4.104

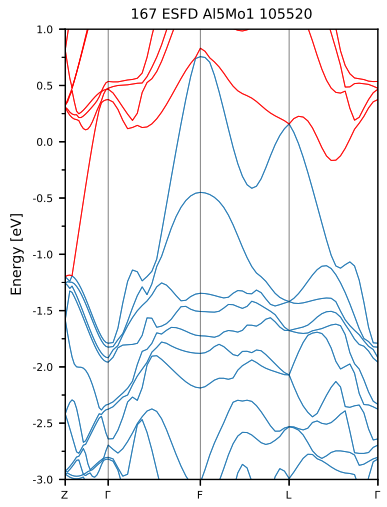

(lc) 167.4.105

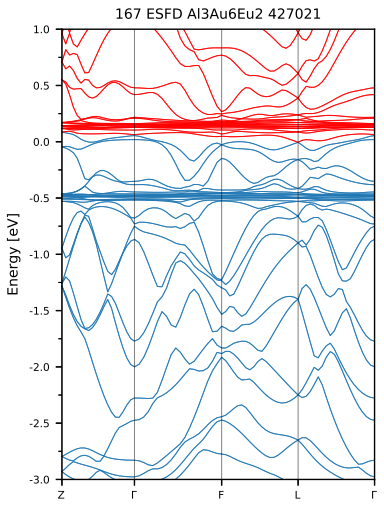

(ld) 167.4.106

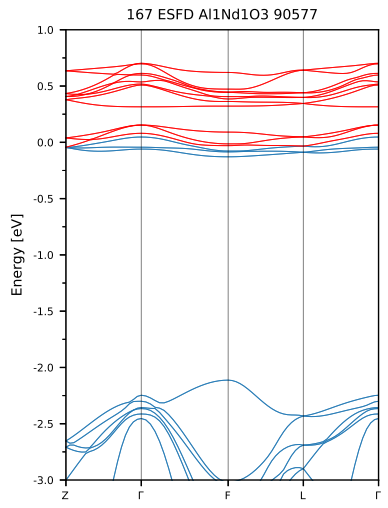

(le) 167.4.107

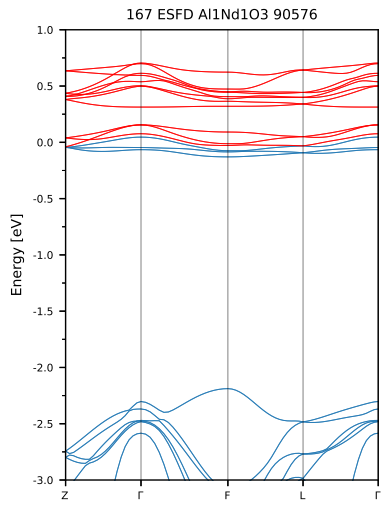

(lf) 167.4.108

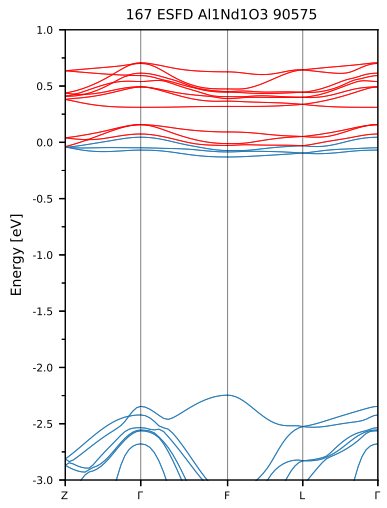

(lg) 167.4.109

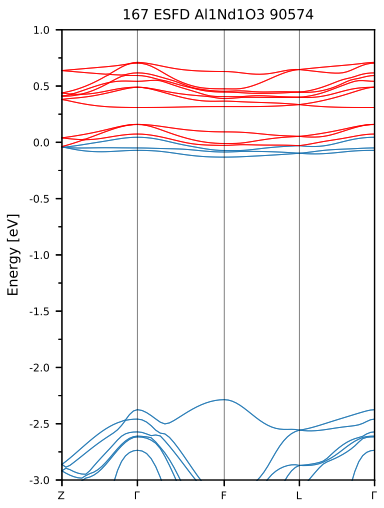

(lh) 167.4.110

Figure 409: Band structures for topological materials in SG 167, part 20/21, from 167.4.95 to 167.4.110. See the list of figures.

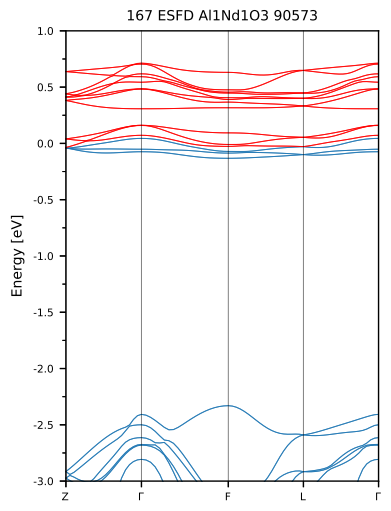

(li) 167.4.111

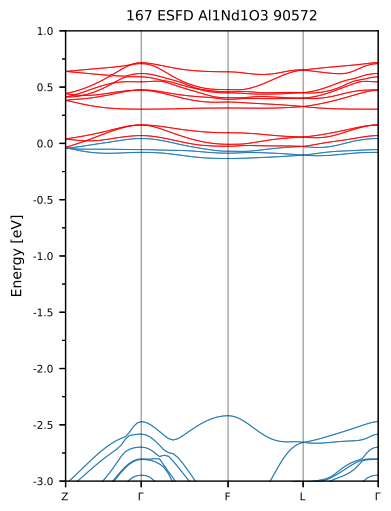

(lj) 167.4.112

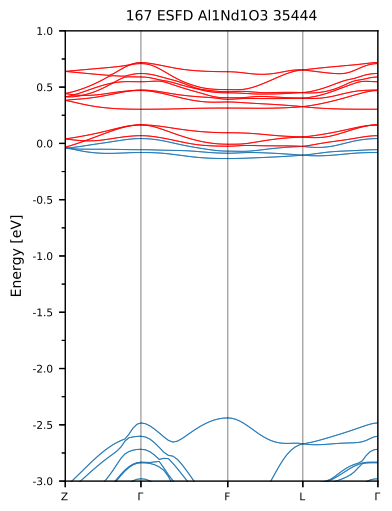

(lk) 167.4.113

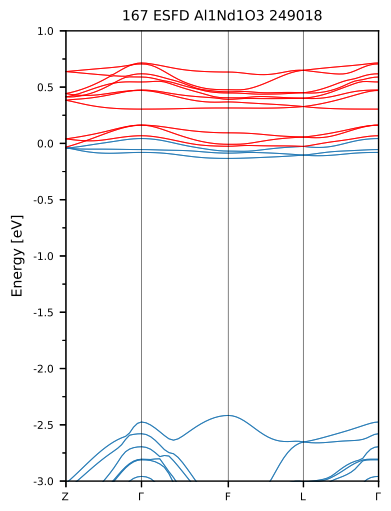

(ll) 167.4.114

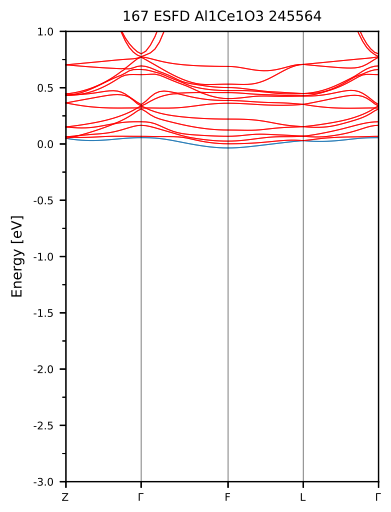

(lm) 167.4.115

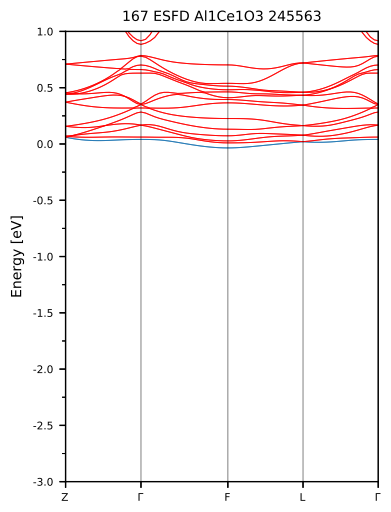

(ln) 167.4.116

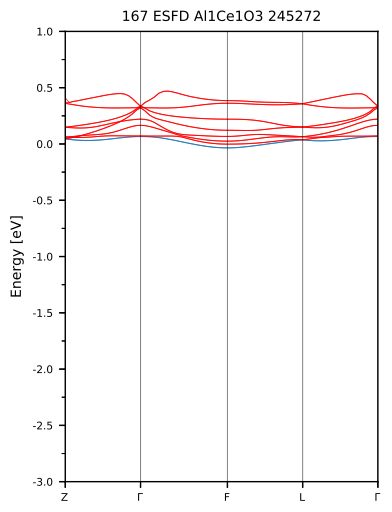

(lo) 167.4.117

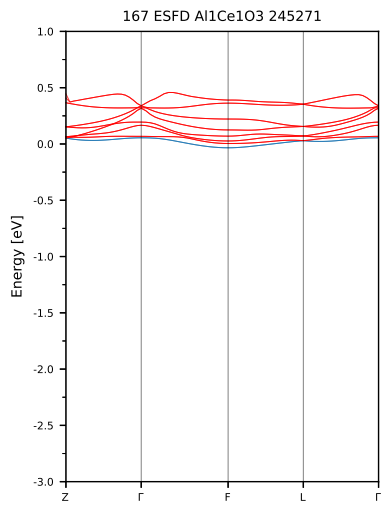

(lp) 167.4.118

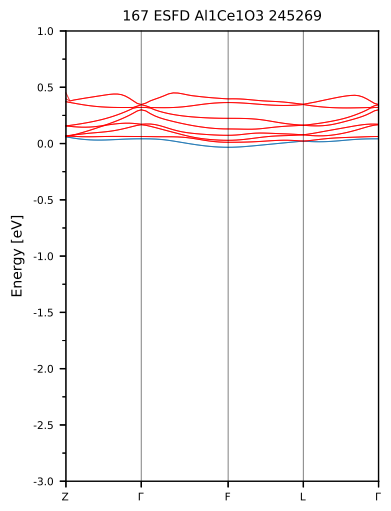

(lq) 167.4.119

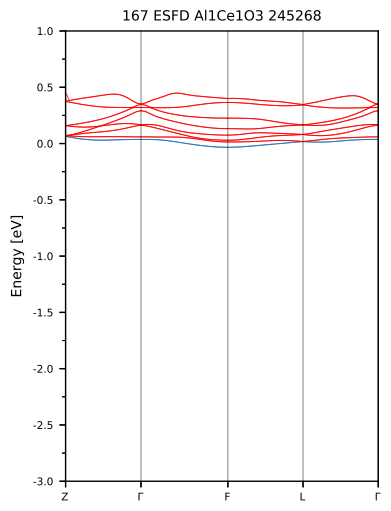

(lr) 167.4.120

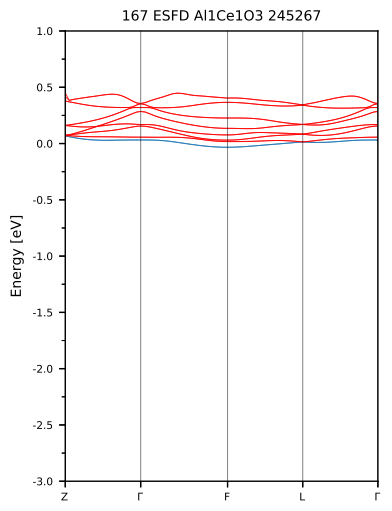

(ls) 167.4.121

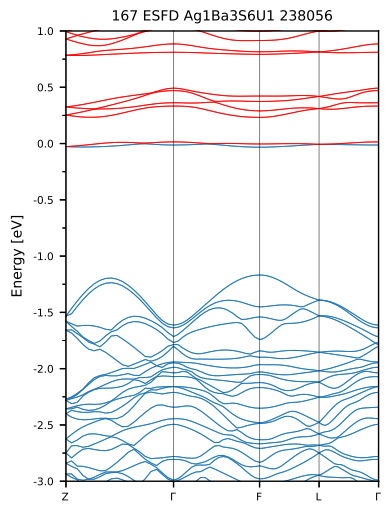

(lt) 167.4.122

Figure 410: Band structures for topological materials in SG 167, part 21/21, from 167.4.111 to 167.4.122. See the list of figures.

### 0.0.84 SG 173

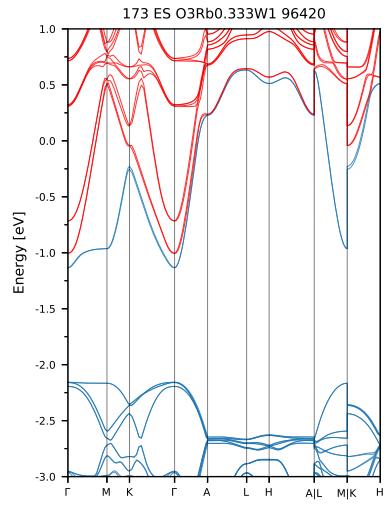

(a) 173.3.1

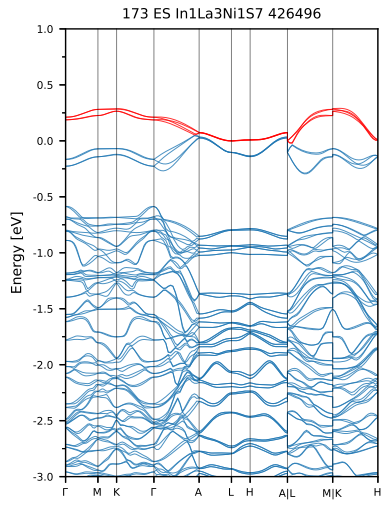

(b) 173.3.2

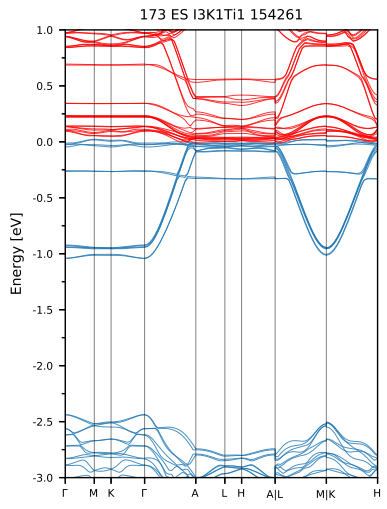

(c) 173.3.3

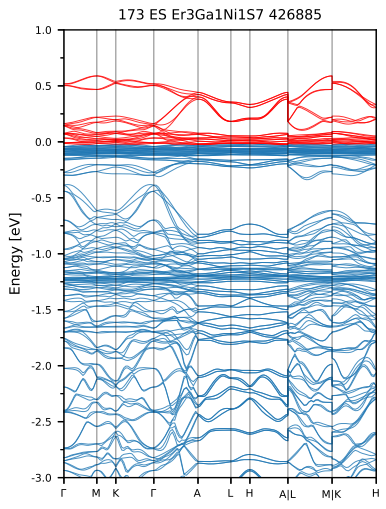

(d) 173.3.4

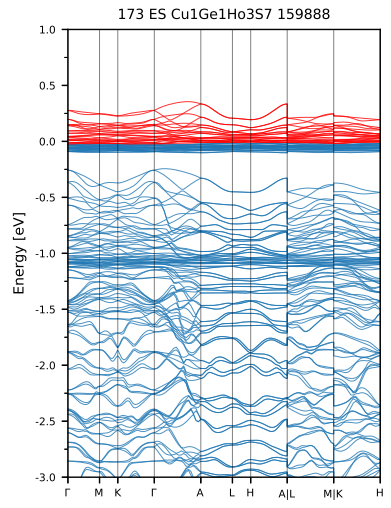

(e) 173.3.5

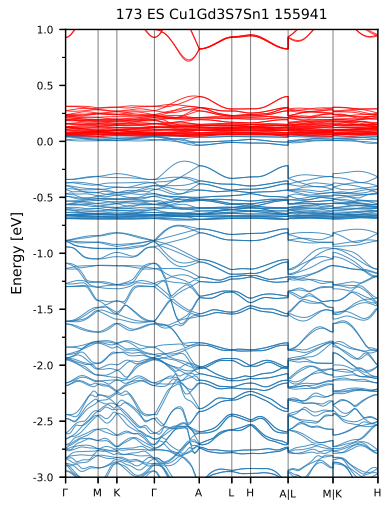

(f) 173.3.6

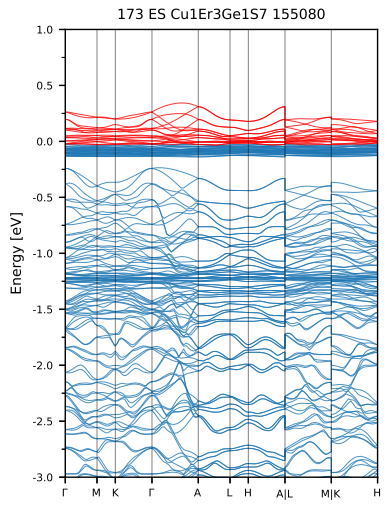

(g) 173.3.7

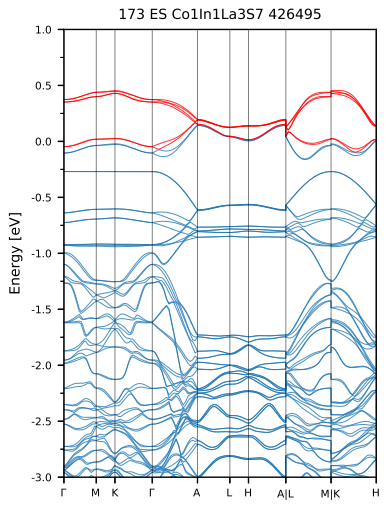

(h) 173.3.8

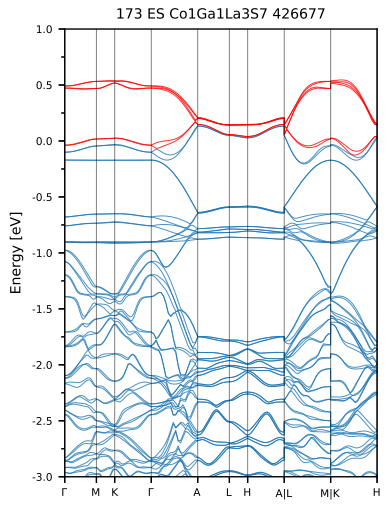

(i) 173.3.9

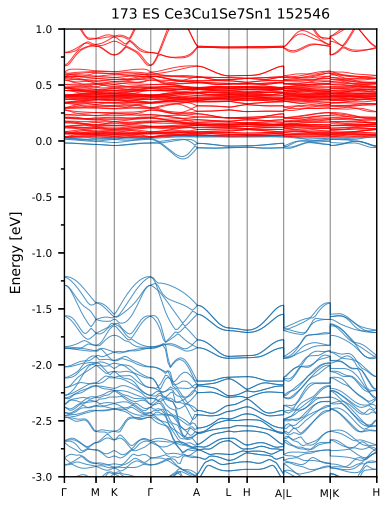

(j) 173.3.10

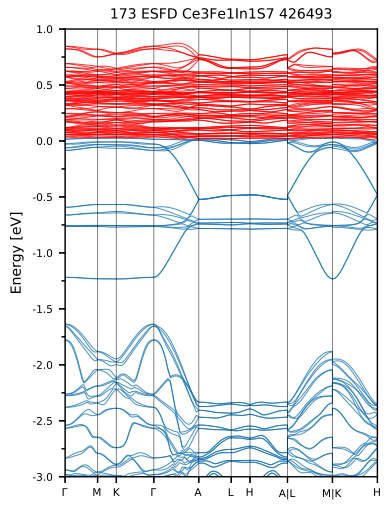

(k) 173.4.1

Figure 411: Band structures for topological materials in SG 173, part 1/1, from 173.3.1 to 173.4.1. See the list of figures.

### 0.0.85 SG 174

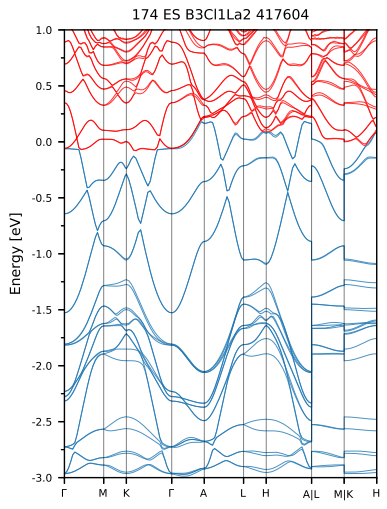

(a) 174.3.1

Figure 412: Band structures for topological materials in SG 174, part 1/1, from 174.3.1 to 174.3.1. See the list of figures.

## 0.0.86 SG 176

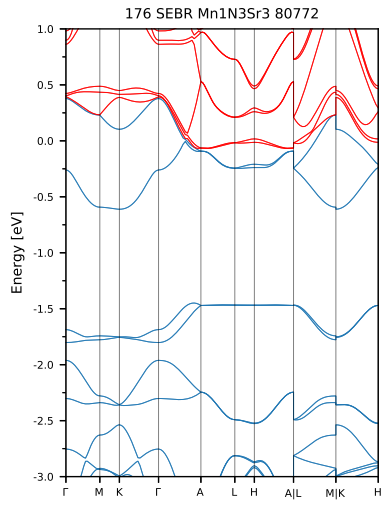

(a) 176.1.1

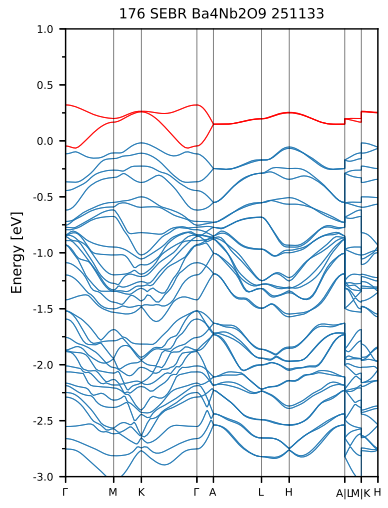

(b) 176.1.2

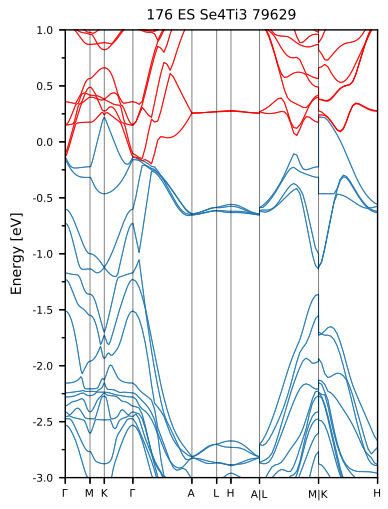

(c) 176.3.1

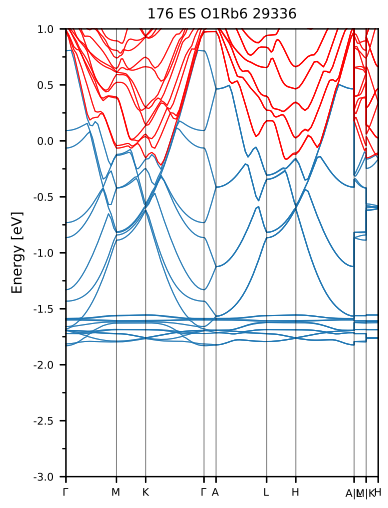

(d) 176.3.2

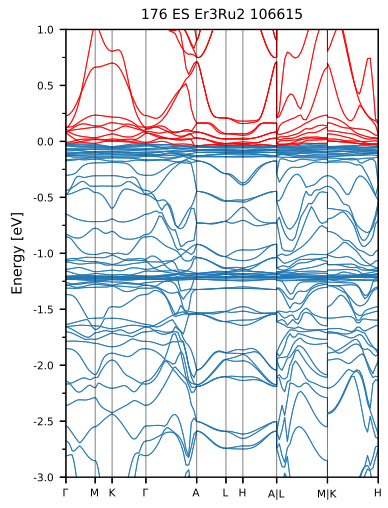

(e) 176.3.3

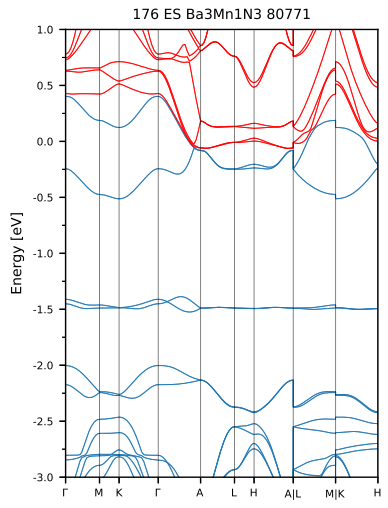

(f) 176.3.4

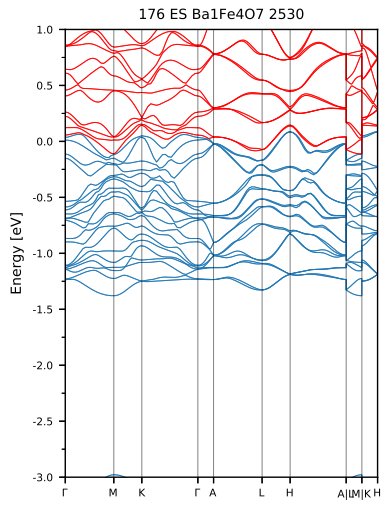

(g) 176.3.5

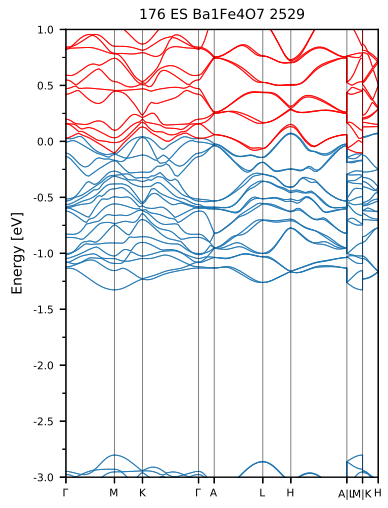

(h) 176.3.6

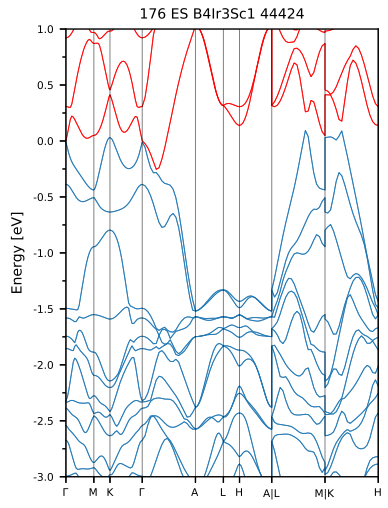

(i) 176.3.7

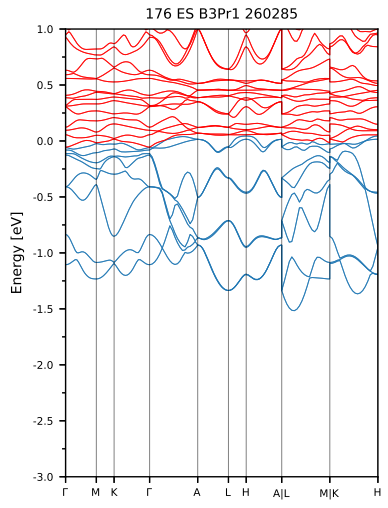

(j) 176.3.8

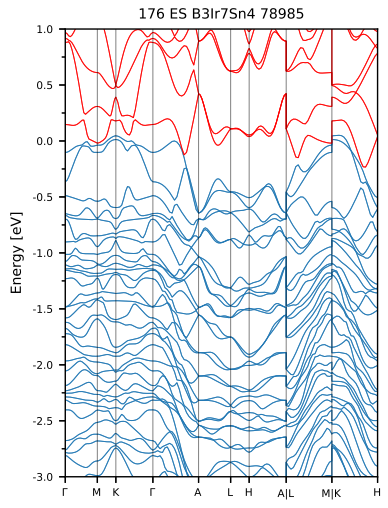

(k) 176.3.9

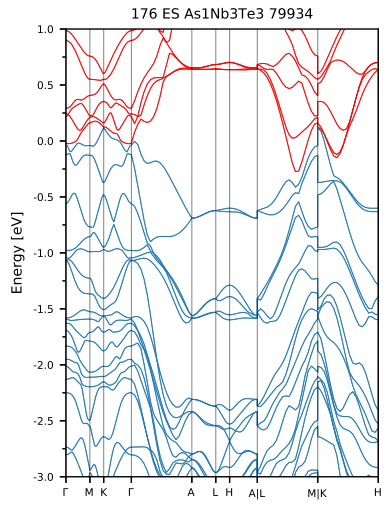

(l) 176.3.10

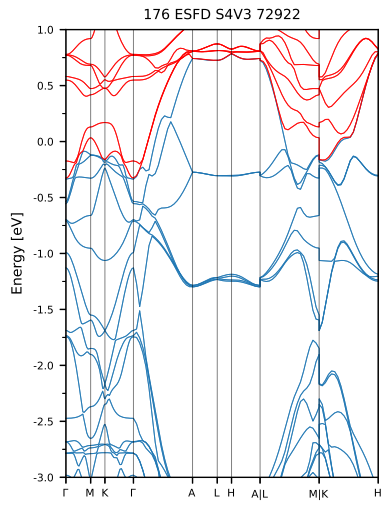

(m) 176.4.1

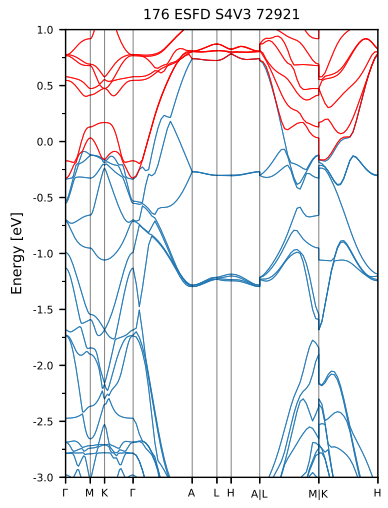

(n) 176.4.2

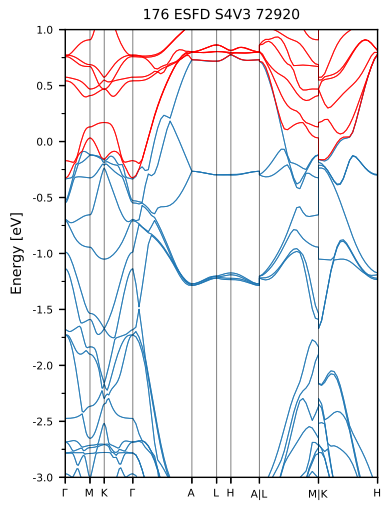

(o) 176.4.3

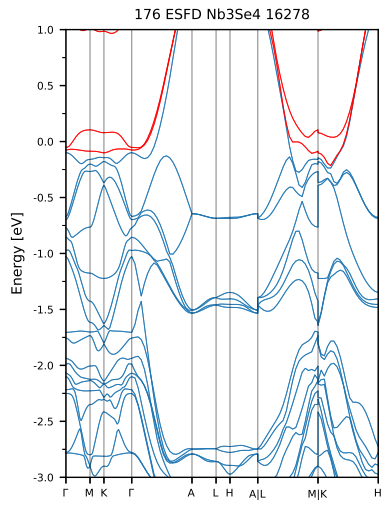

(p) 176.4.4

Figure 413: Band structures for topological materials in SG 176, part 1/2, from 176.1.1 to 176.4.4. See the list of figures.

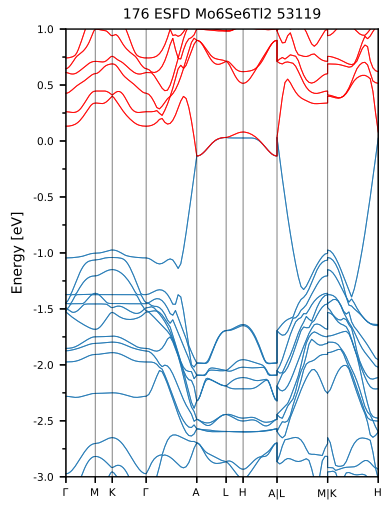

(q) 176.4.5

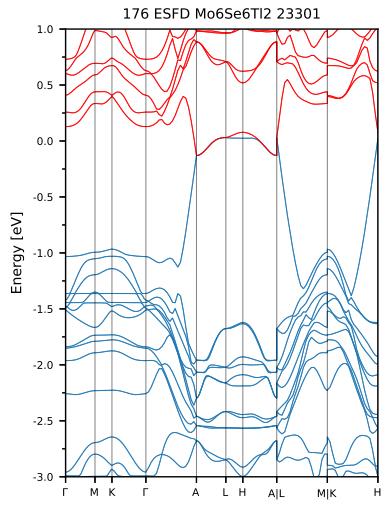

(r) 176.4.6

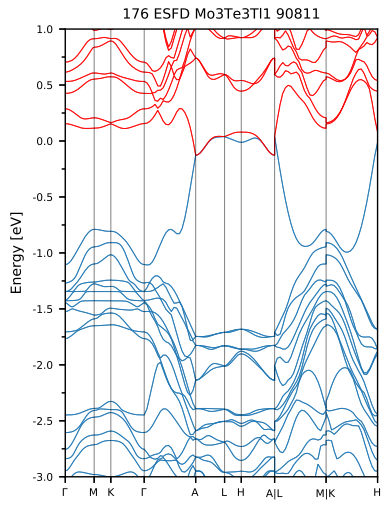

(s) 176.4.7

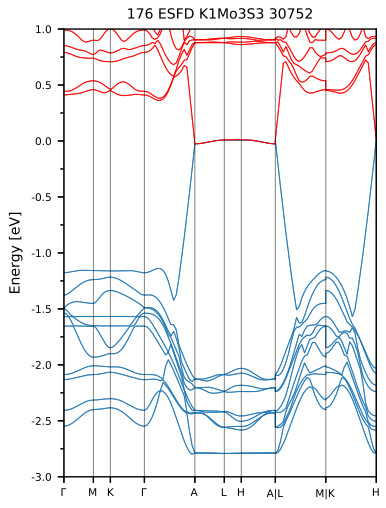

(t) 176.4.8

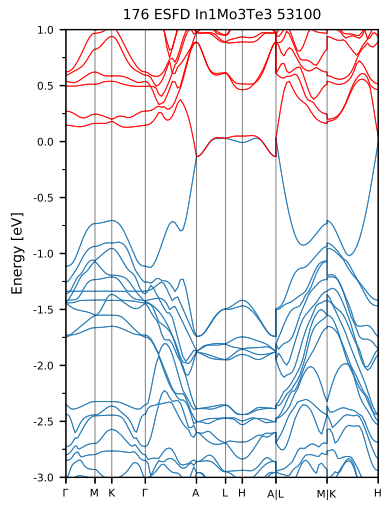

(u) 176.4.9

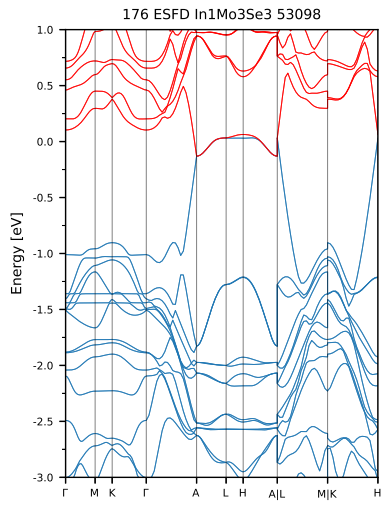

(v) 176.4.10

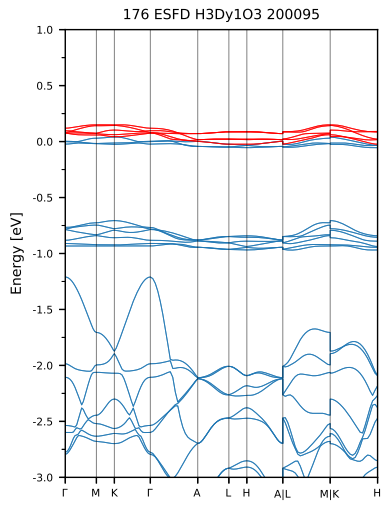

(w) 176.4.11

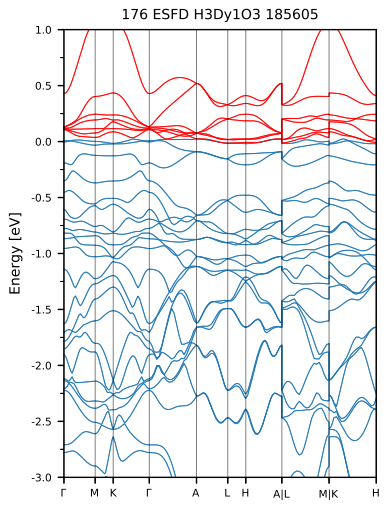

(x) 176.4.12

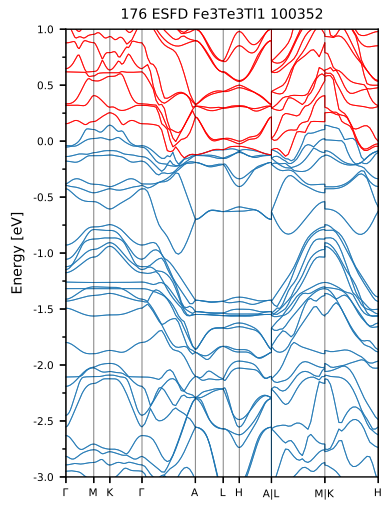

(y) 176.4.13

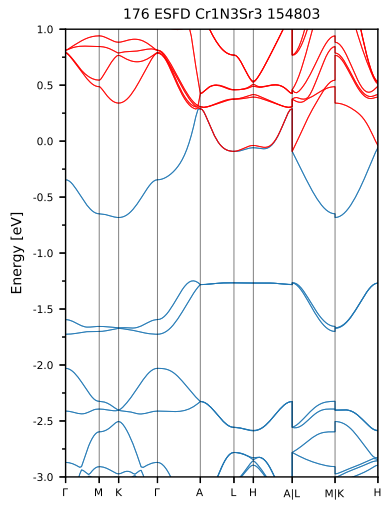

(z) 176.4.14

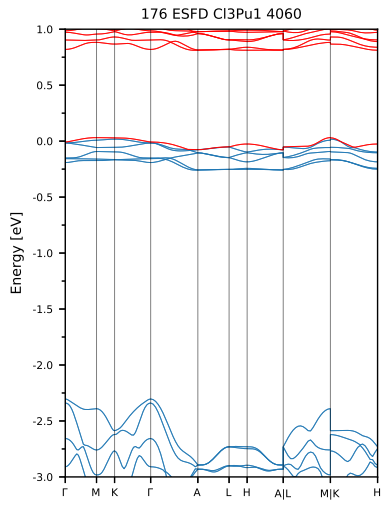

(aa) 176.4.15

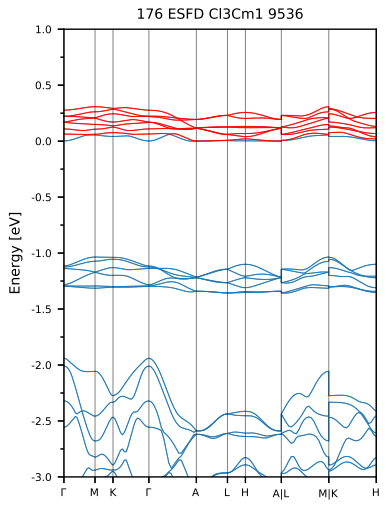

(ab) 176.4.16

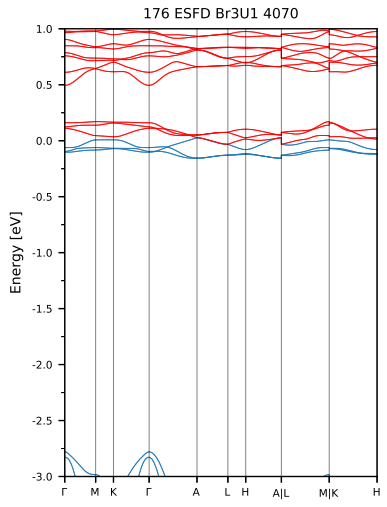

(ac) 176.4.17

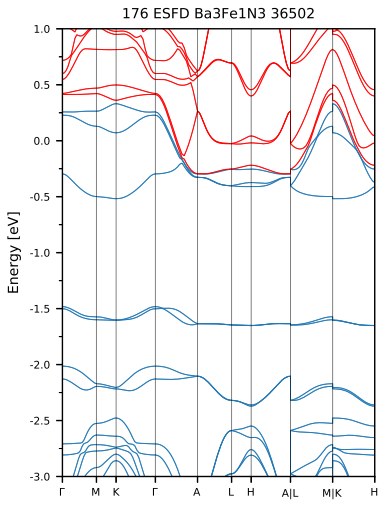

(ad) 176.4.18

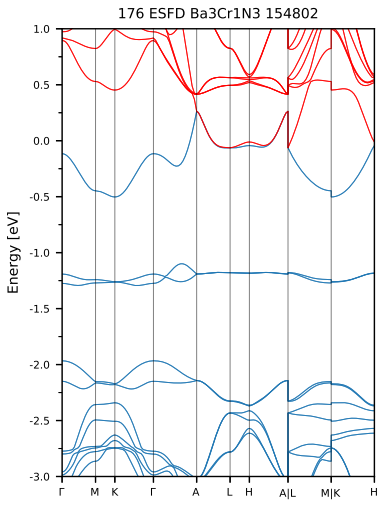

(ae) 176.4.19

Figure 414: Band structures for topological materials in SG 176, part 2/2, from 176.4.5 to 176.4.19. See the list of figures.

### 0.0.87 SG 180

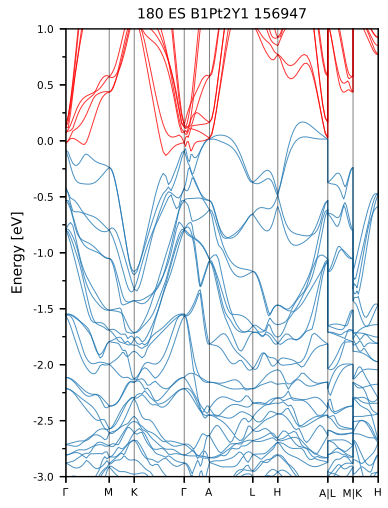

(a) 180.3.1

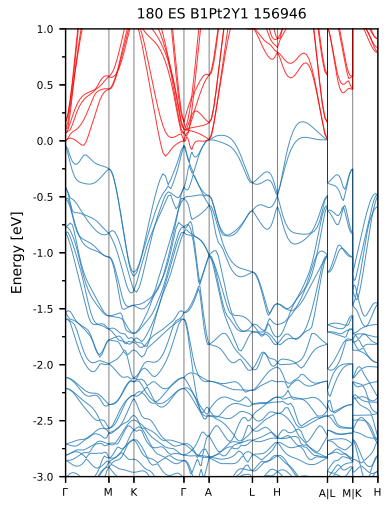

(b) 180.3.2

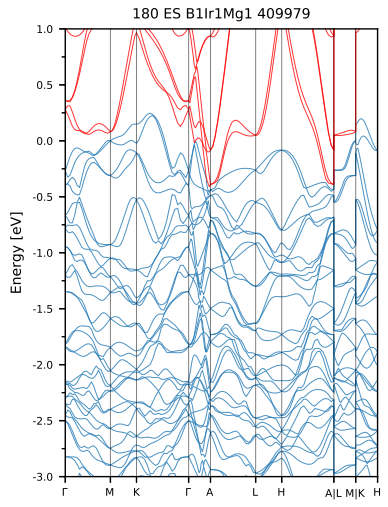

(c) 180.3.3

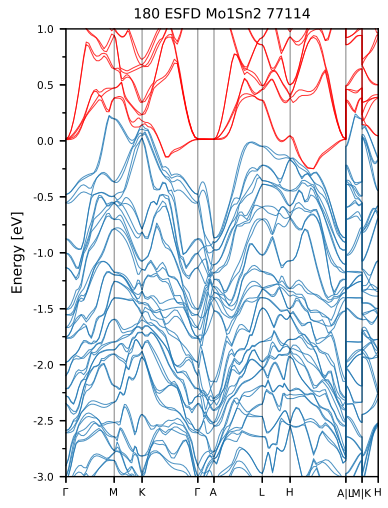

(d) 180.4.1

Figure 415: Band structures for topological materials in SG 180, part 1/1, from 180.3.1 to 180.4.1. See the list of figures.

### 0.0.88 SG 181

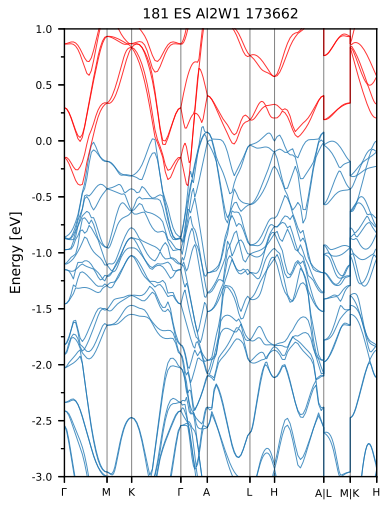

(a) 181.3.1

Figure 416: Band structures for topological materials in SG 181, part 1/1, from 181.3.1 to 181.3.1. See the list of figures.

## 0.0.89 SG 182

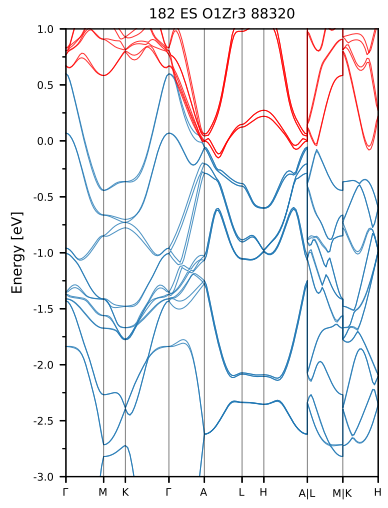

(a) 182.3.1

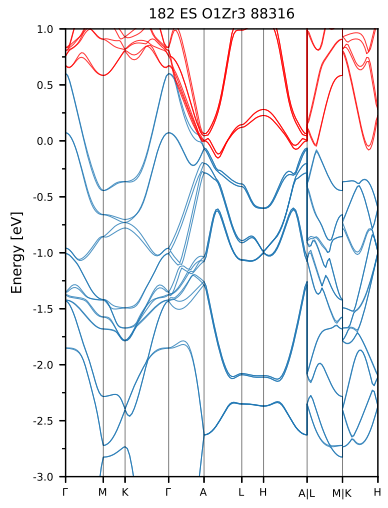

(b) 182.3.2

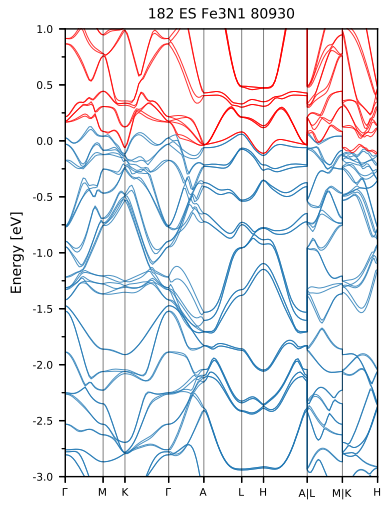

(c) 182.3.3

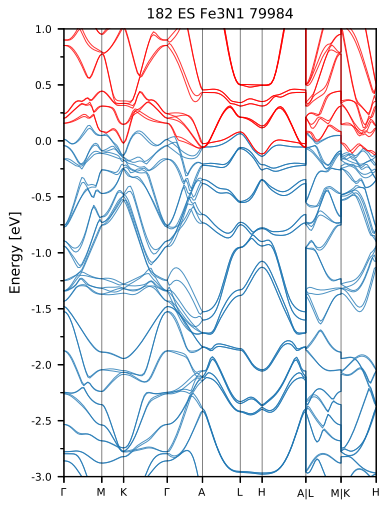

(d) 182.3.4

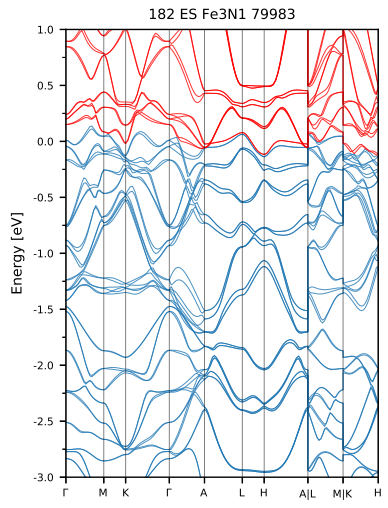

(e) 182.3.5

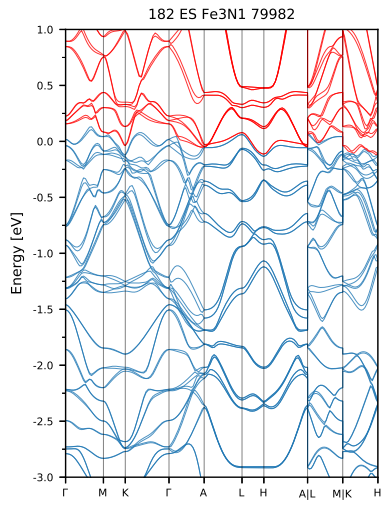

(f) 182.3.6

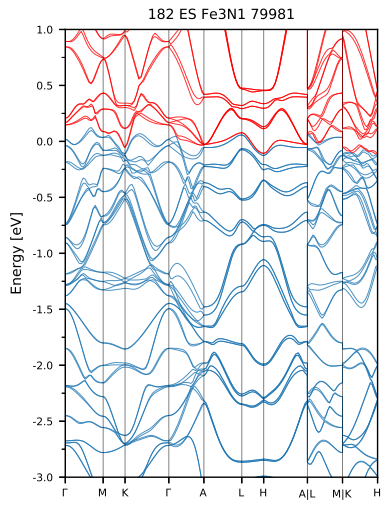

(g) 182.3.7

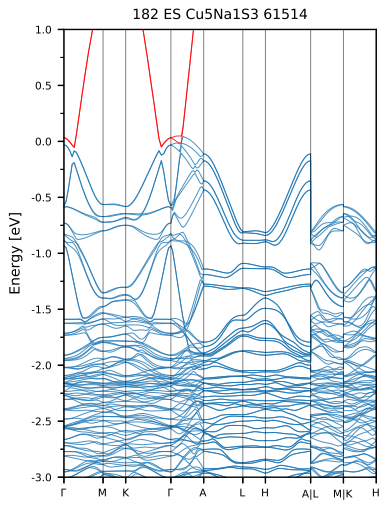

(h) 182.3.8

Figure 417: Band structures for topological materials in SG 182, part 1/1, from 182.3.1 to 182.3.8. See the list of figures.

### 0.0.90 SG 185

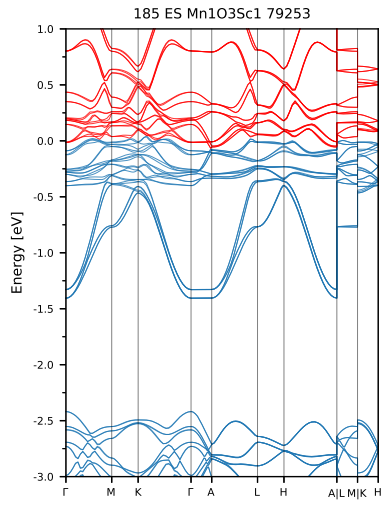

(a) 185.3.1

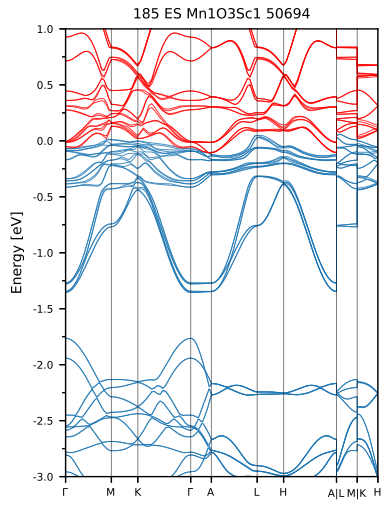

(b) 185.3.2

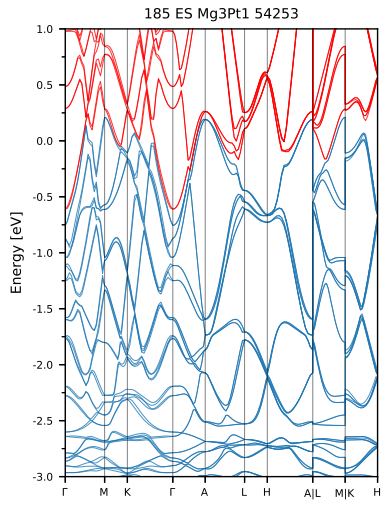

(c) 185.3.3

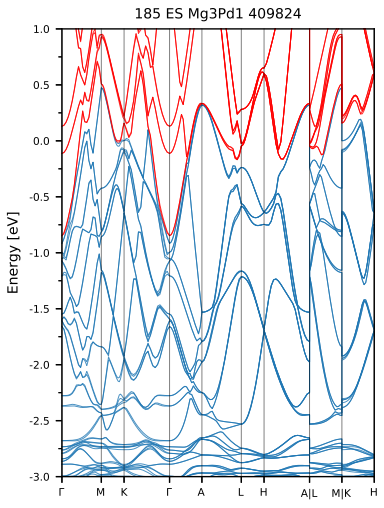

(d) 185.3.4

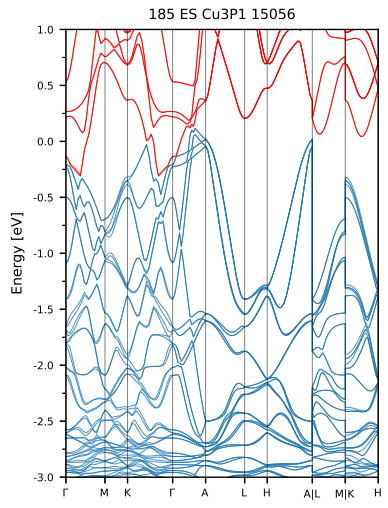

(e) 185.3.5

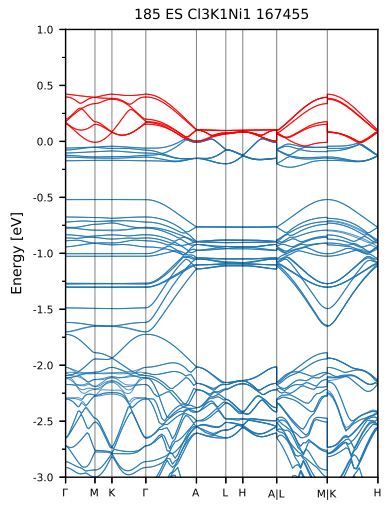

(f) 185.3.6

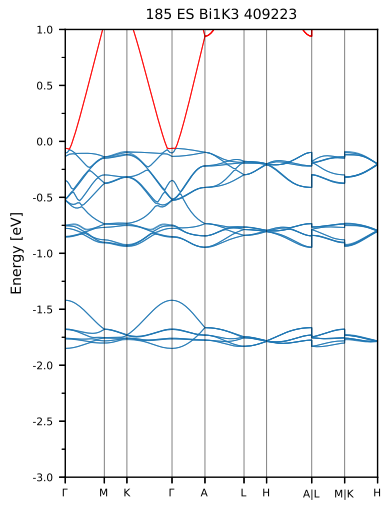

(g) 185.3.7

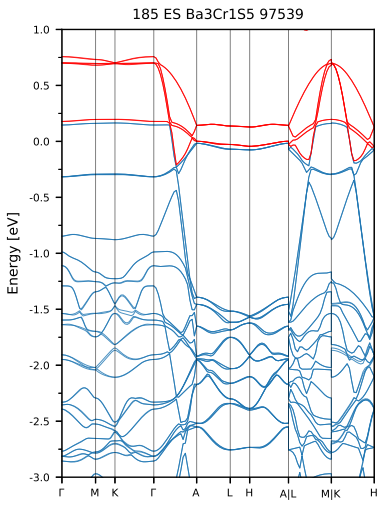

(h) 185.3.8

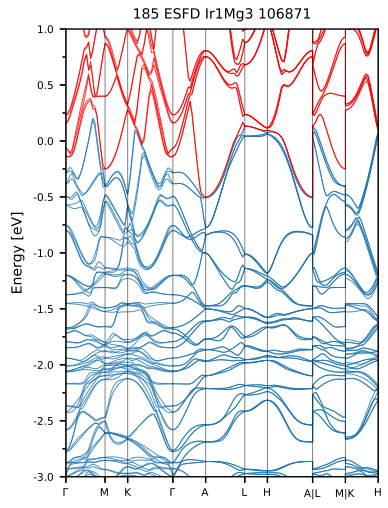

(i) 185.4.1

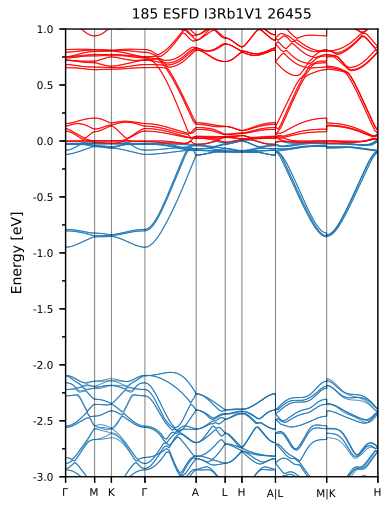

(j) 185.4.2

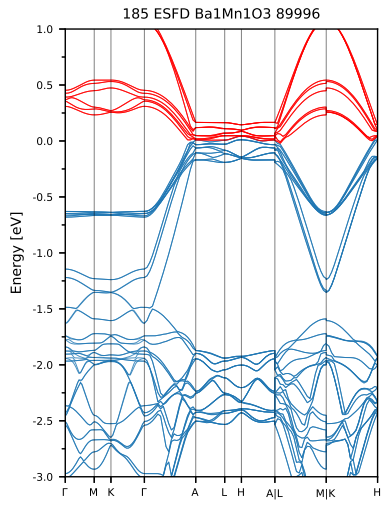

(k) 185.4.3

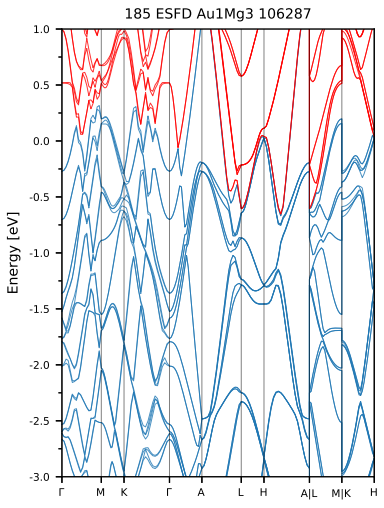

(l) 185.4.4

Figure 418: Band structures for topological materials in SG 185, part 1/1, from 185.3.1 to 185.4.4. See the list of figures.

### 0.0.91 SG 186

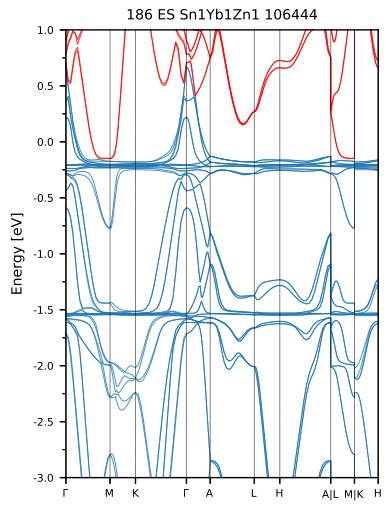

(a) 186.3.1

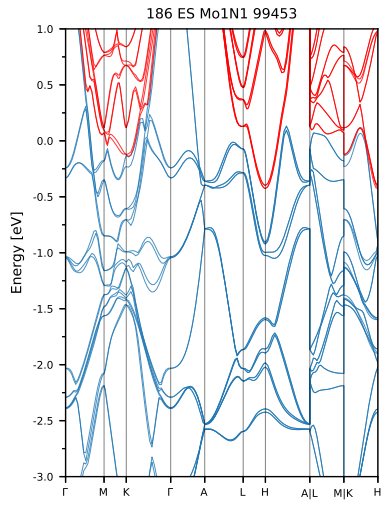

(b) 186.3.2

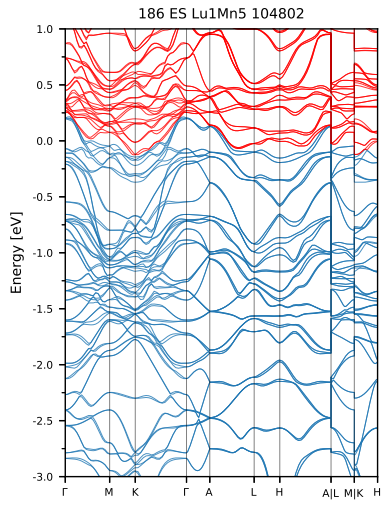

(c) 186.3.3

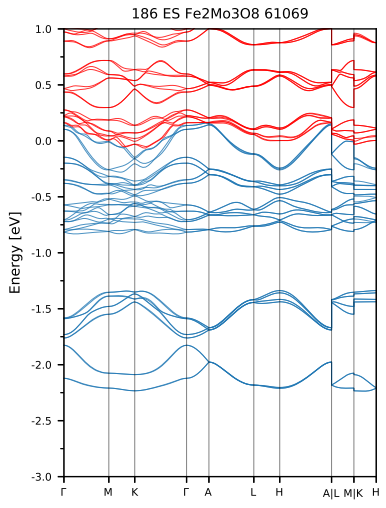

(d) 186.3.4

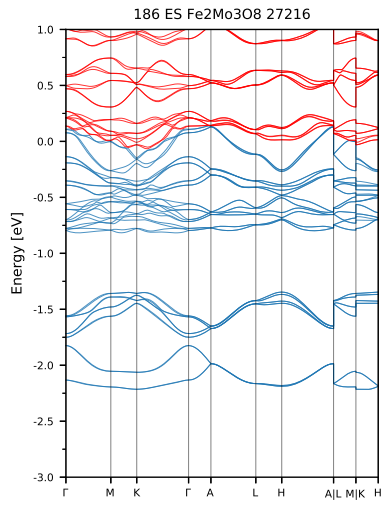

(e) 186.3.5

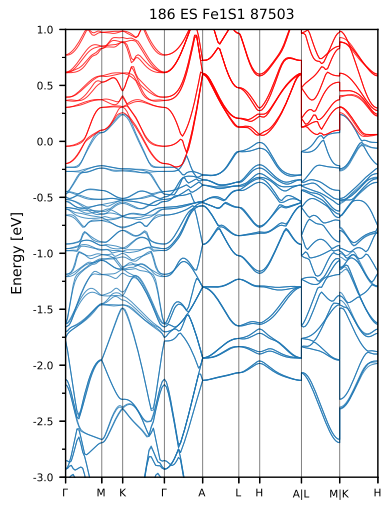

(f) 186.3.6

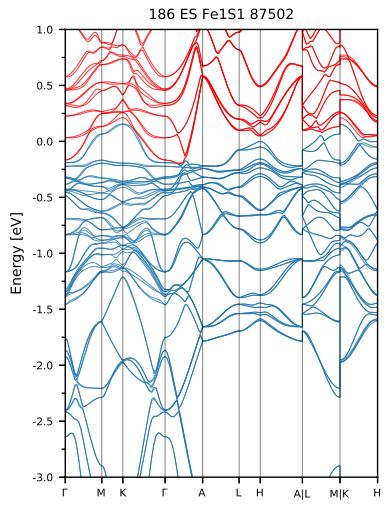

(g) 186.3.7

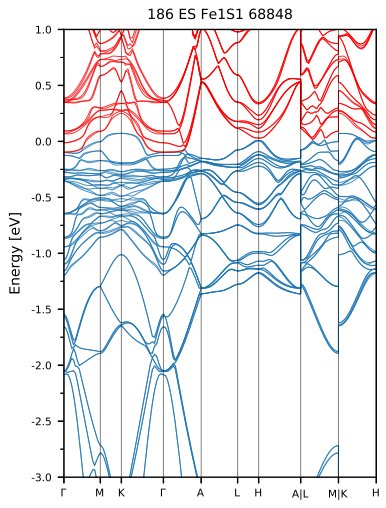

(h) 186.3.8

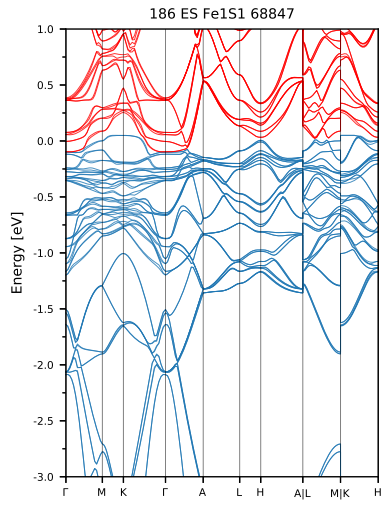

(i) 186.3.9

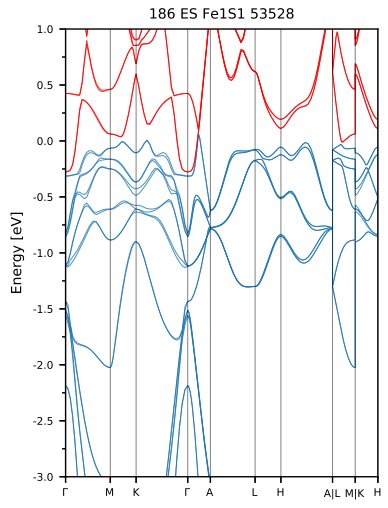

(j) 186.3.10

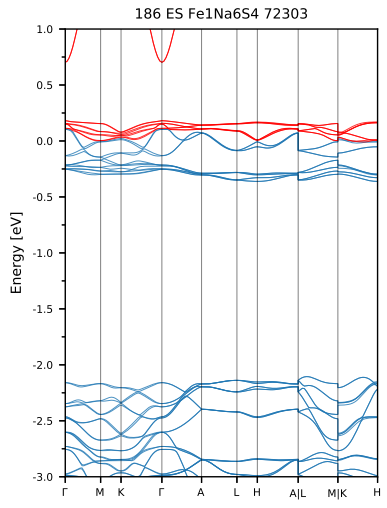

(k) 186.3.11

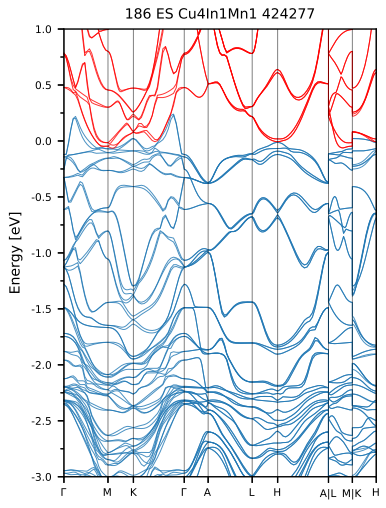

(l) 186.3.12

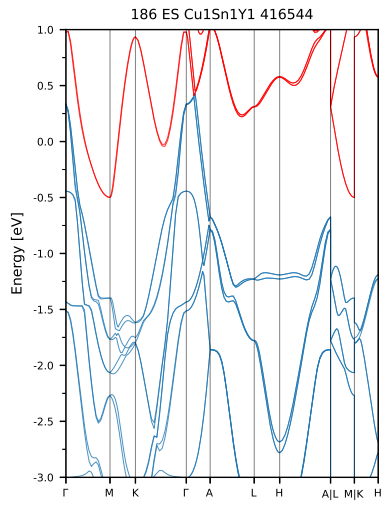

(m) 186.3.13

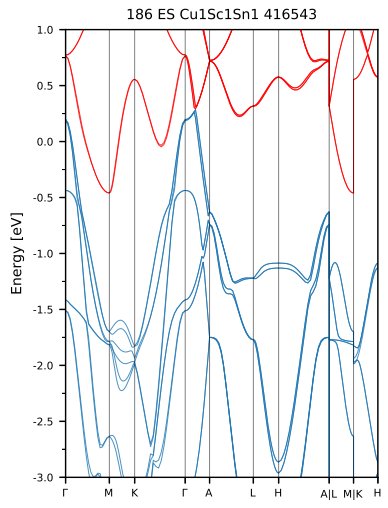

(n) 186.3.14

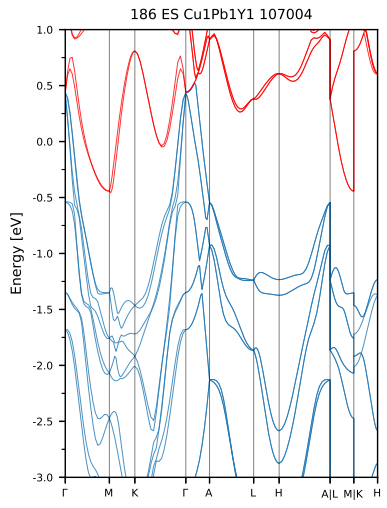

(o) 186.3.15

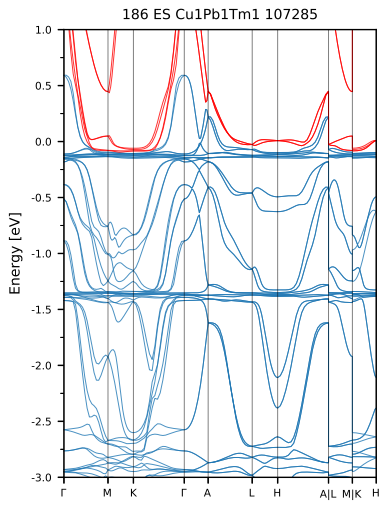

(p) 186.3.16

Figure 419: Band structures for topological materials in SG 186, part 1/8, from 186.3.1 to 186.3.16. See the list of figures.

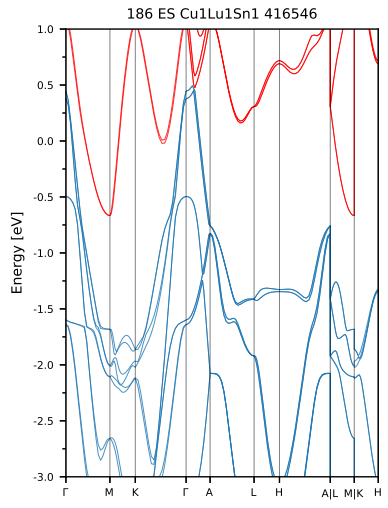

(q) 186.3.17

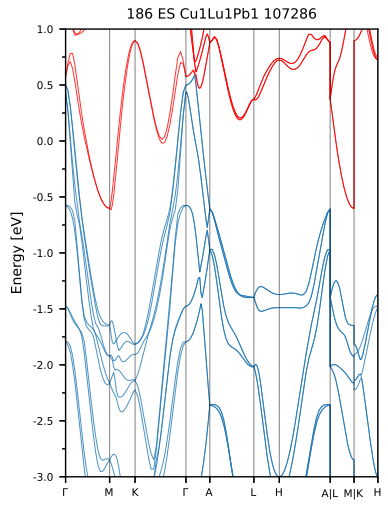

(r) 186.3.18

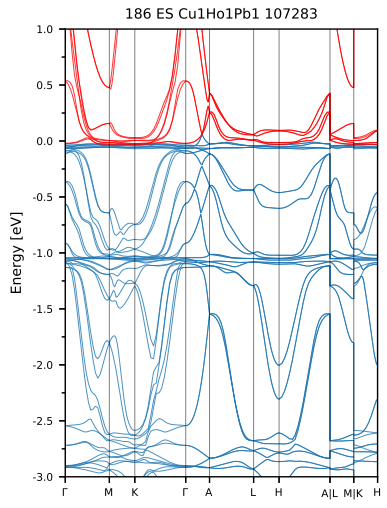

(s) 186.3.19

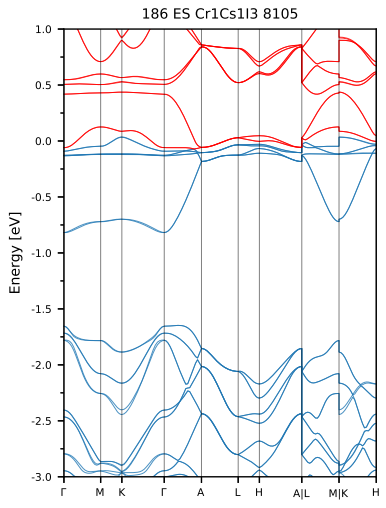

(t) 186.3.20

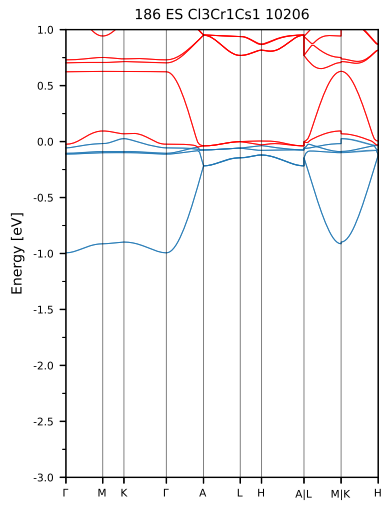

(u) 186.3.21

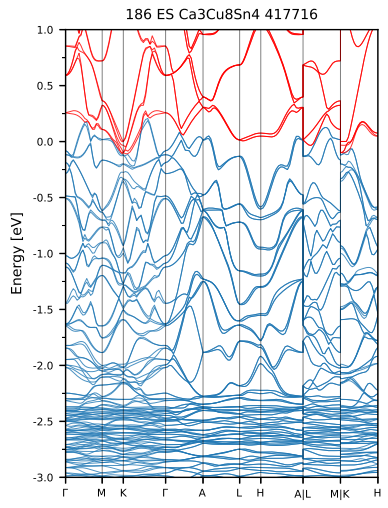

(v) 186.3.22

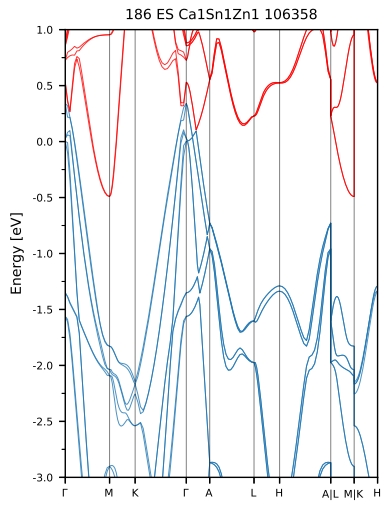

(w) 186.3.23

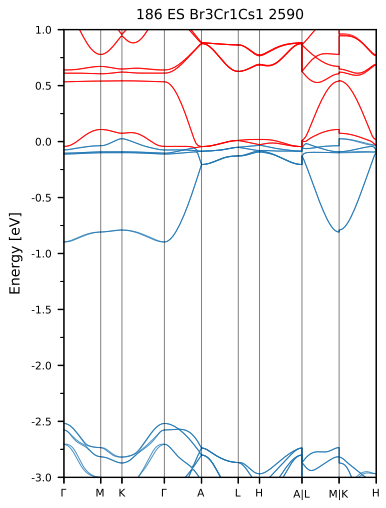

(x) 186.3.24

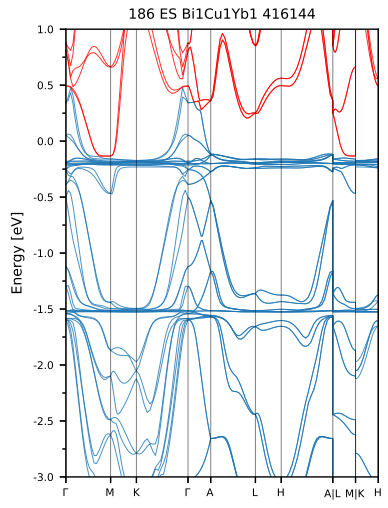

(y) 186.3.25

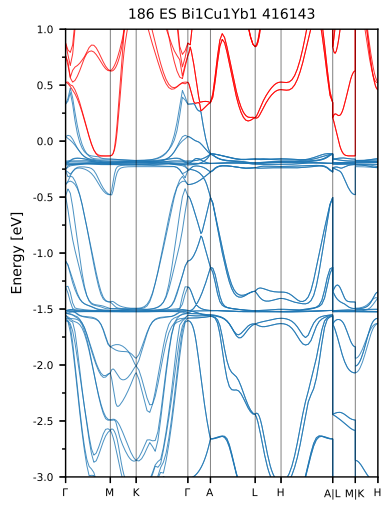

(z) 186.3.26

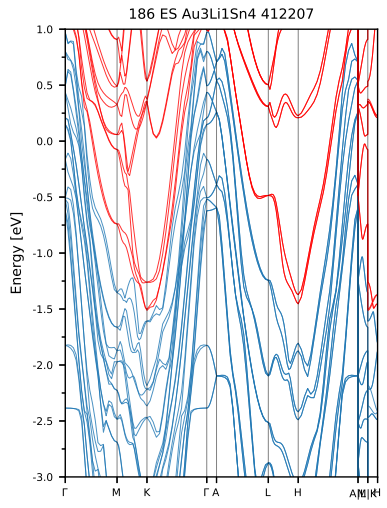

(aa) 186.3.27

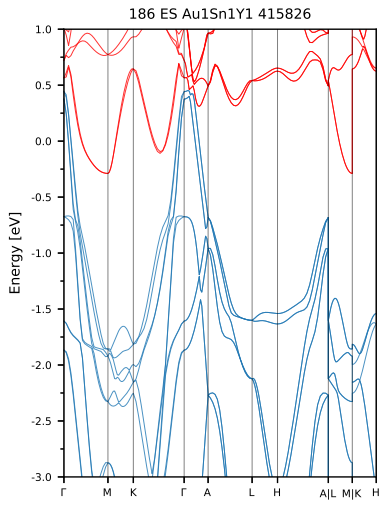

(ab) 186.3.28

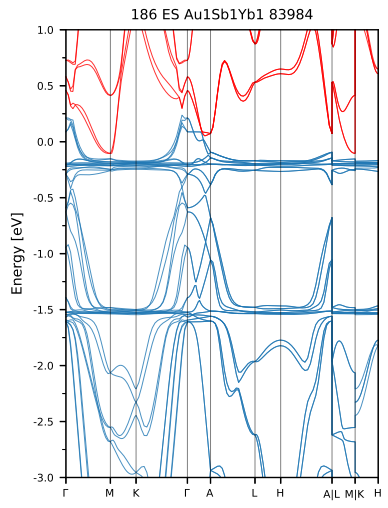

(ac) 186.3.29

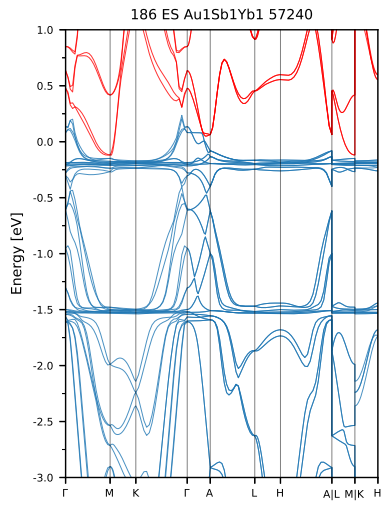

(ad) 186.3.30

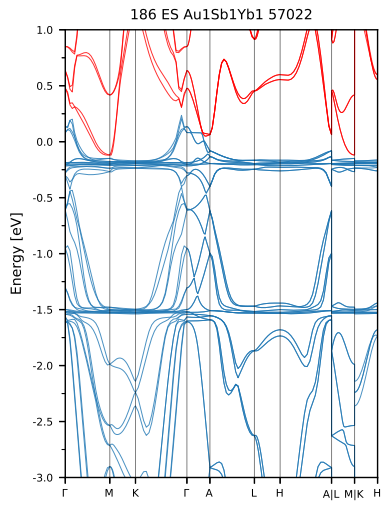

(ae) 186.3.31

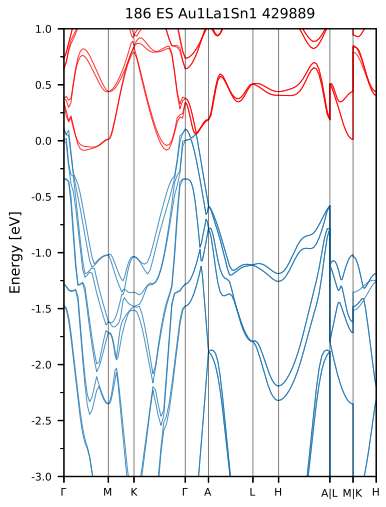

(af) 186.3.32

Figure 420: Band structures for topological materials in SG 186, part 2/8, from 186.3.17 to 186.3.32. See the list of figures.

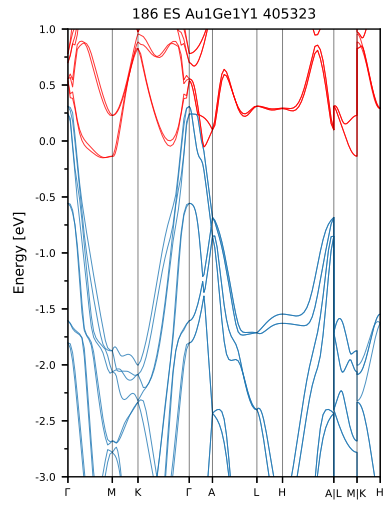

(ag) 186.3.33

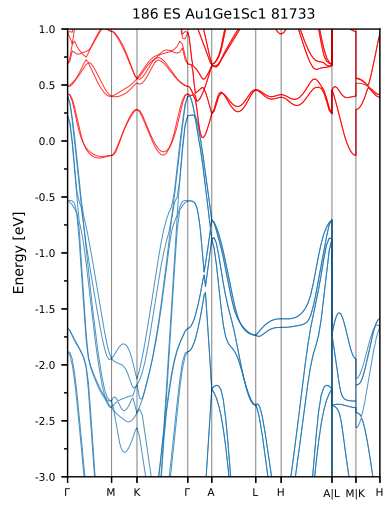

(ah) 186.3.34

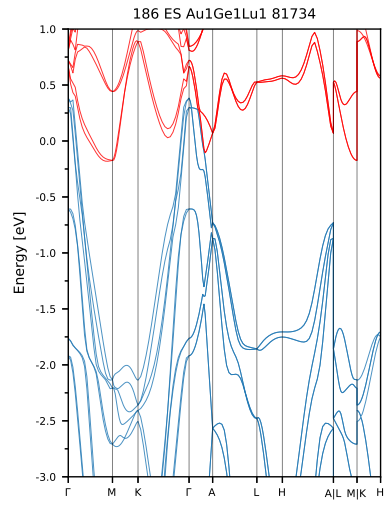

(ai) 186.3.35

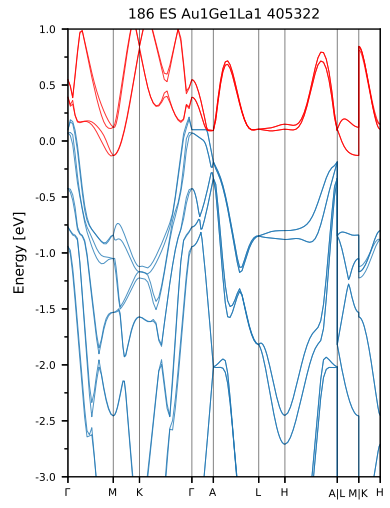

(aj) 186.3.36

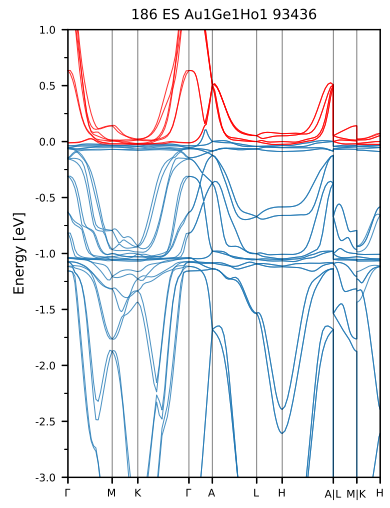

(ak) 186.3.37

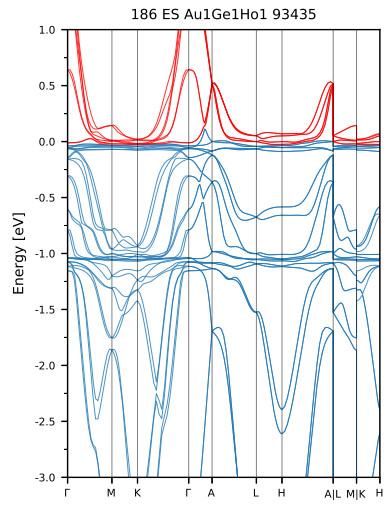

(al) 186.3.38

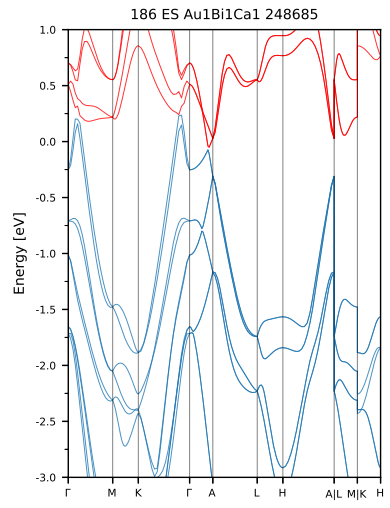

(am) 186.3.39

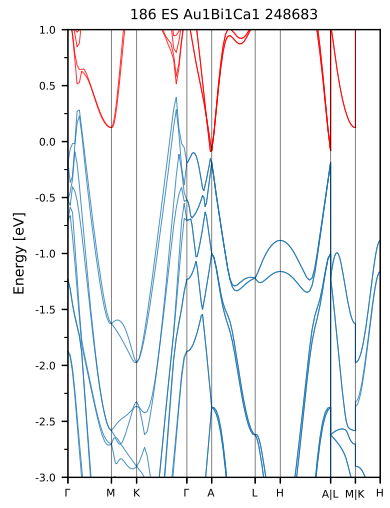

(an) 186.3.40

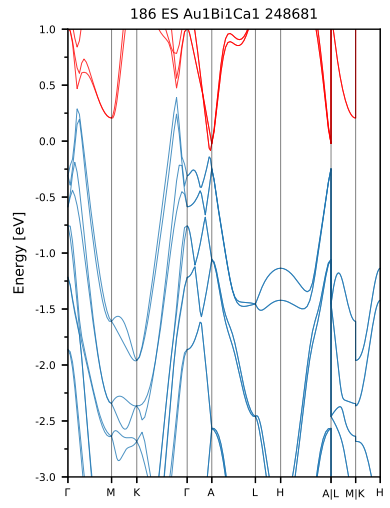

(ao) 186.3.41

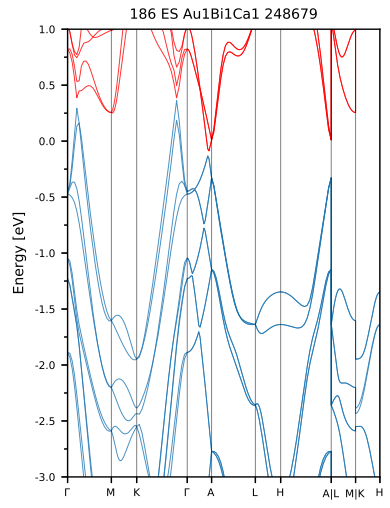

(ap) 186.3.42

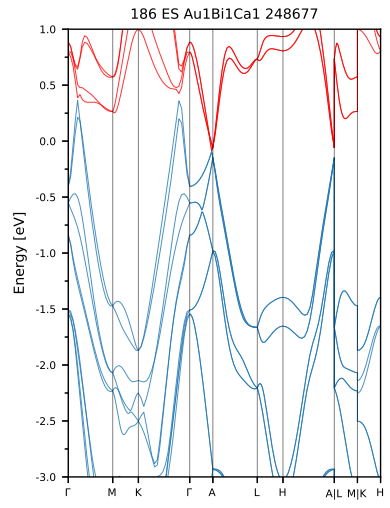

(aq) 186.3.43

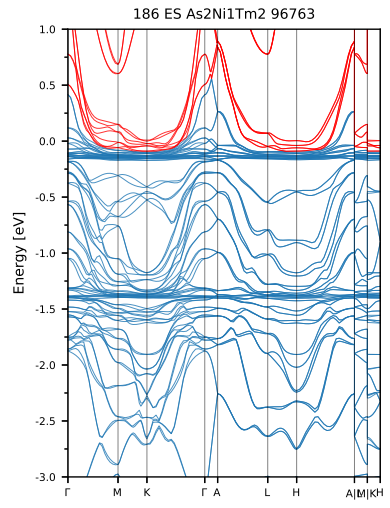

(ar) 186.3.44

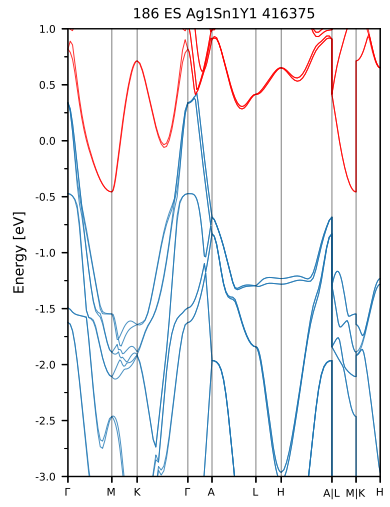

(as) 186.3.45

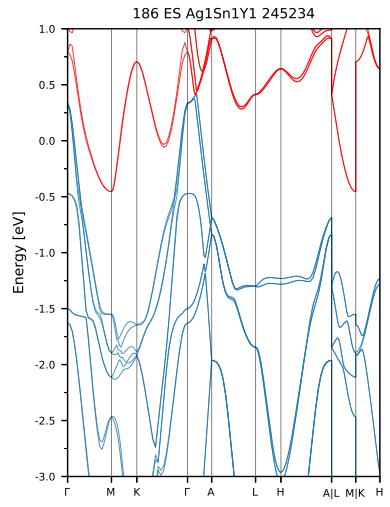

(at) 186.3.46

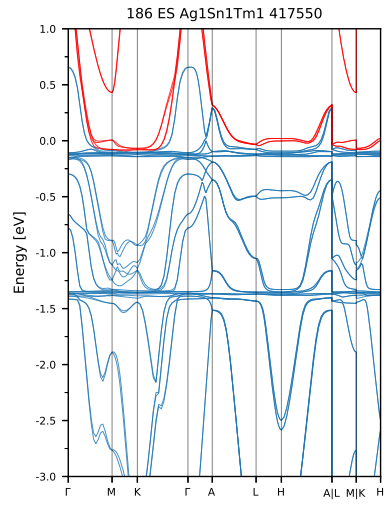

(au) 186.3.47

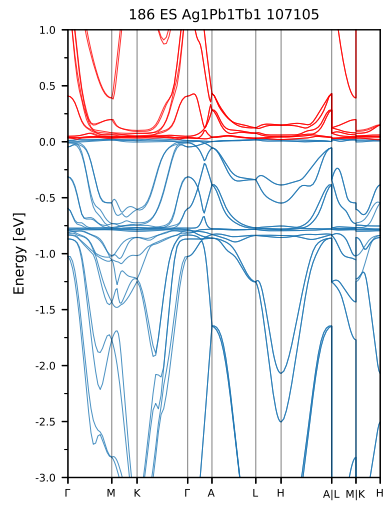

(av) 186.3.48

Figure 421: Band structures for topological materials in SG 186, part 3/8, from 186.3.33 to 186.3.48. See the list of figures.

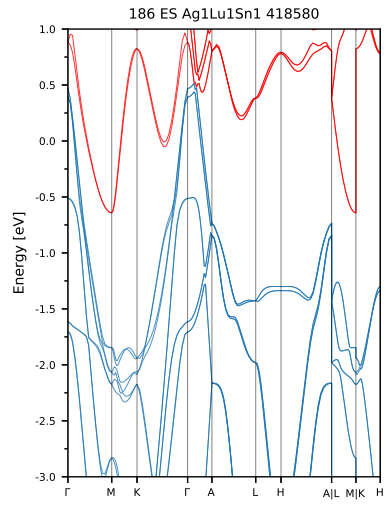

(aw) 186.3.49

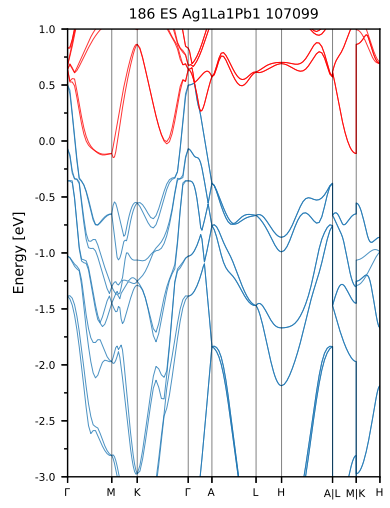

(ax) 186.3.50

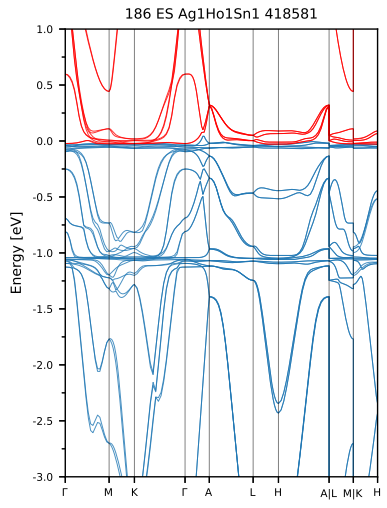

(ay) 186.3.51

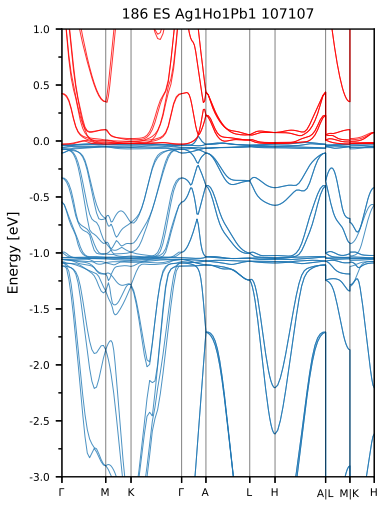

(az) 186.3.52

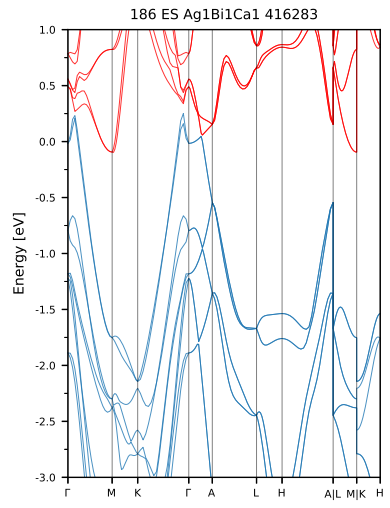

(ba) 186.3.53

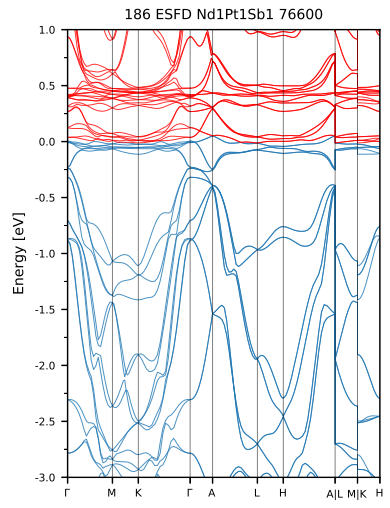

(bb) 186.4.1

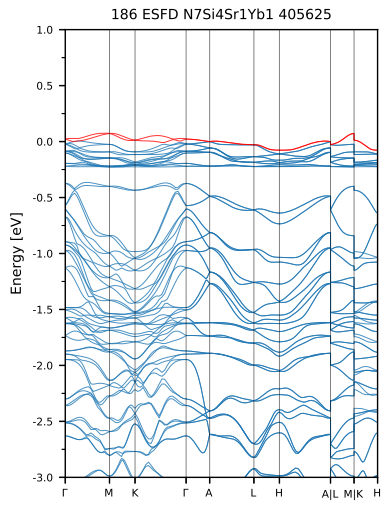

(bc) 186.4.2

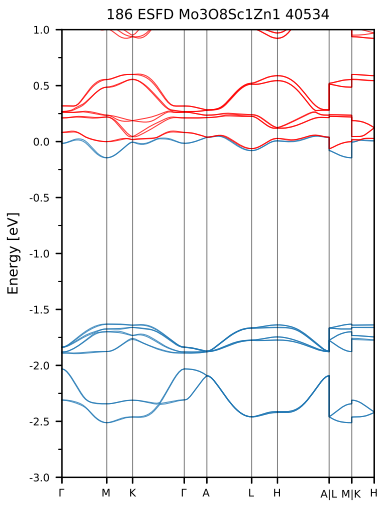

(bd) 186.4.3

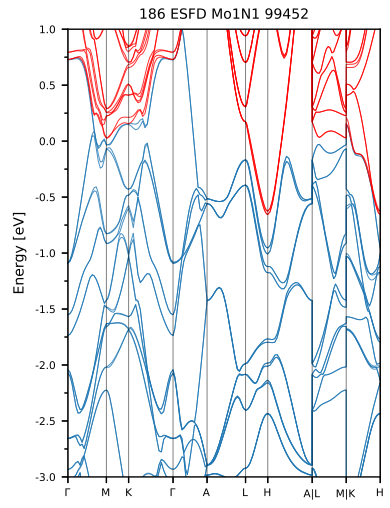

(be) 186.4.4

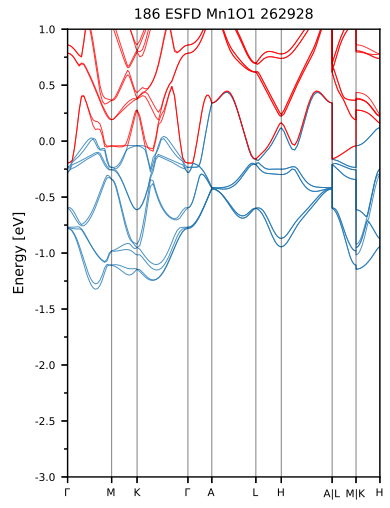

(bf) 186.4.5

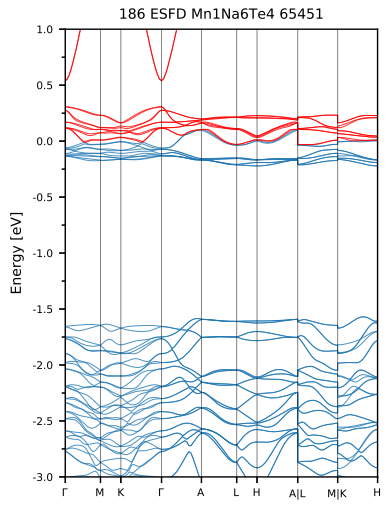

(bg) 186.4.6

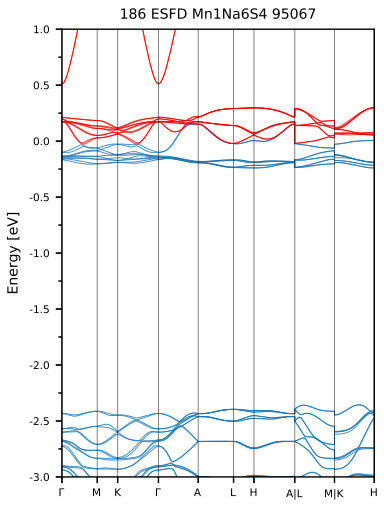

(bh) 186.4.7

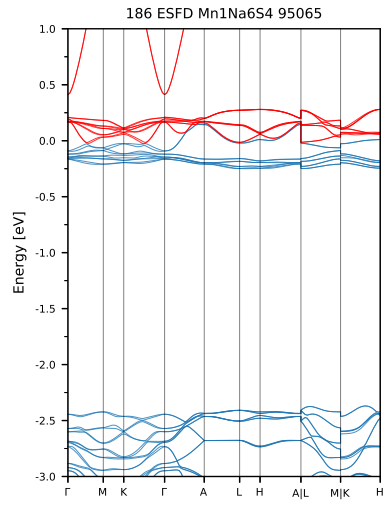

(bi) 186.4.8

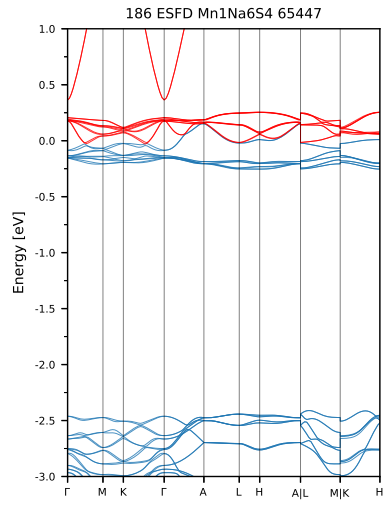

(bj) 186.4.9

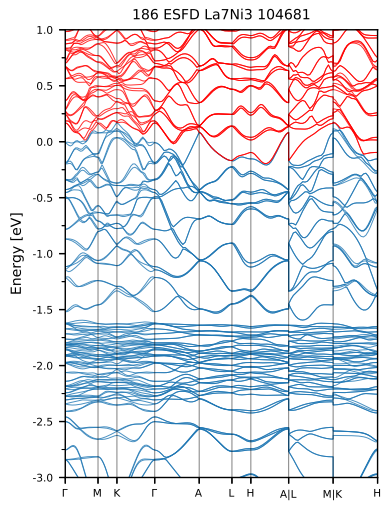

(bk) 186.4.10

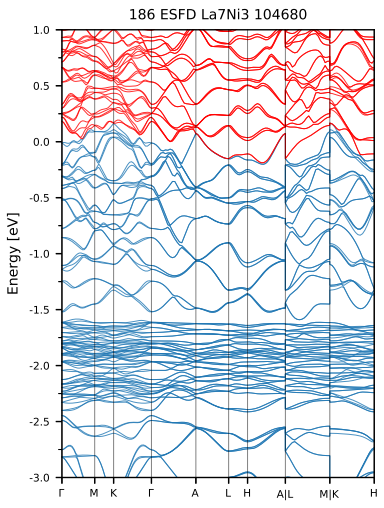

(bl) 186.4.11

Figure 422: Band structures for topological materials in SG 186, part 4/8, from 186.3.49 to 186.4.11. See the list of figures.

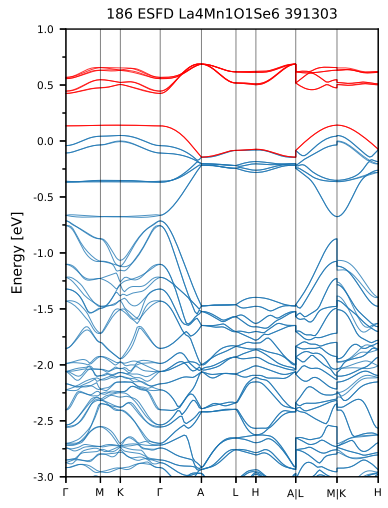

(bm) 186.4.12

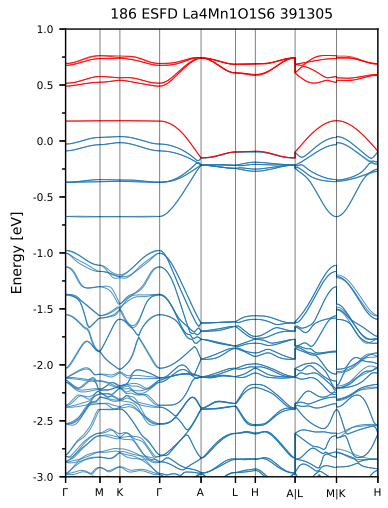

(bn) 186.4.13

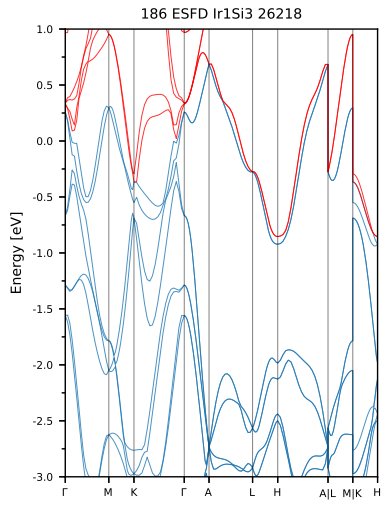

(bo) 186.4.14

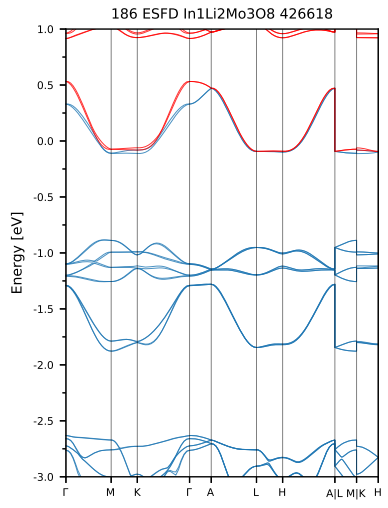

(bp) 186.4.15

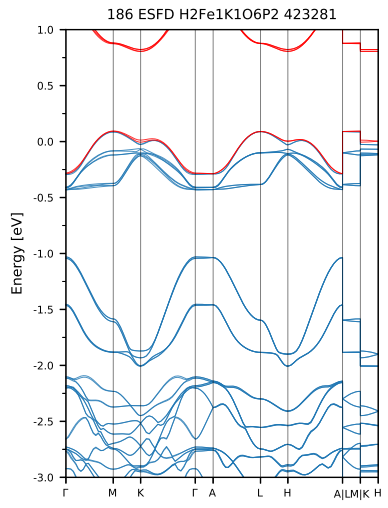

(bq) 186.4.16

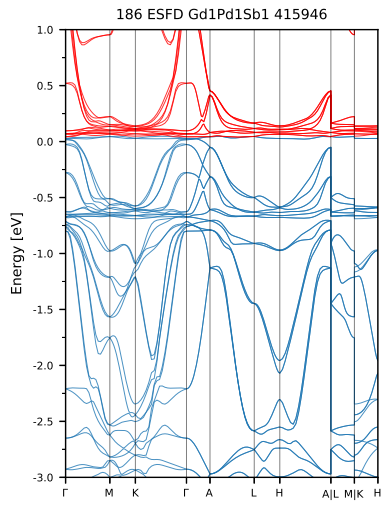

(br) 186.4.17

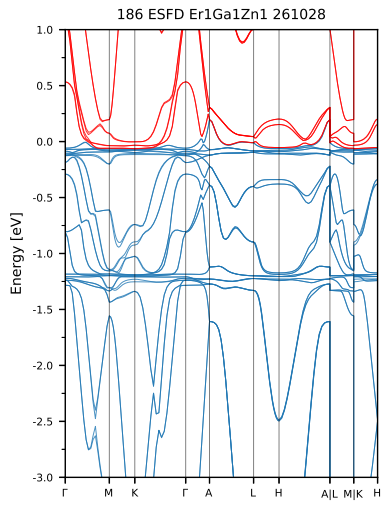

(bs) 186.4.18

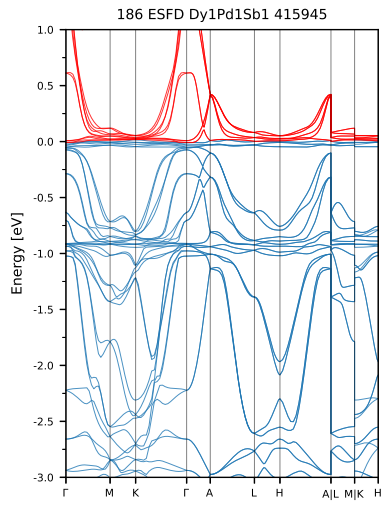

(bt) 186.4.19

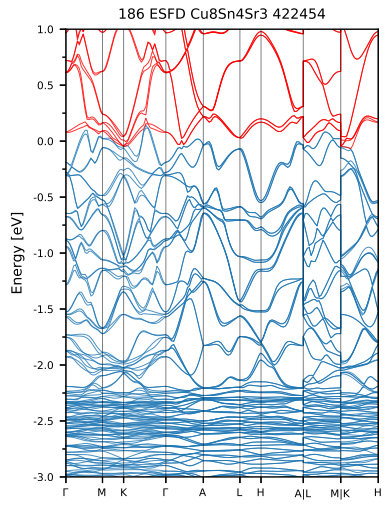

(bu) 186.4.20

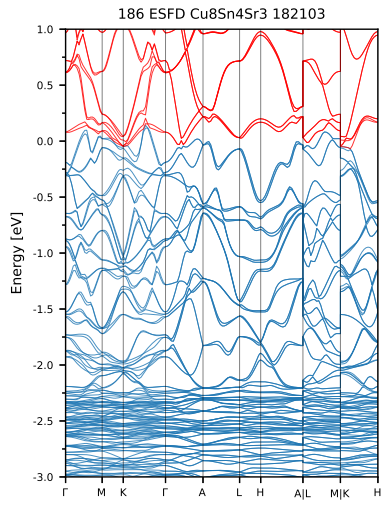

(bv) 186.4.21

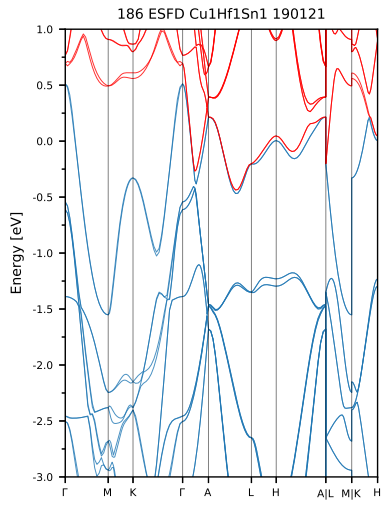

(bw) 186.4.22

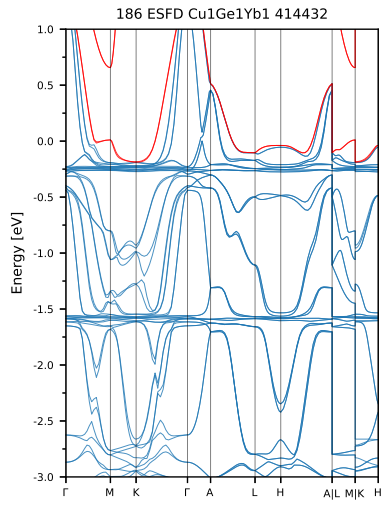

(bx) 186.4.23

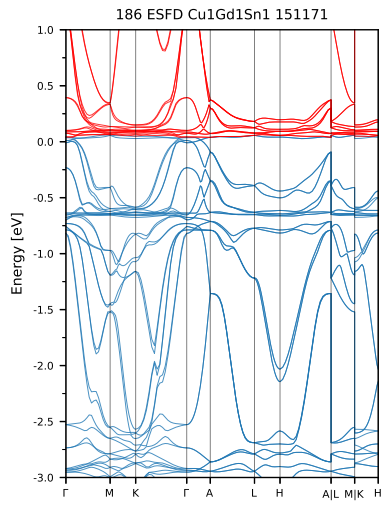

(by) 186.4.24

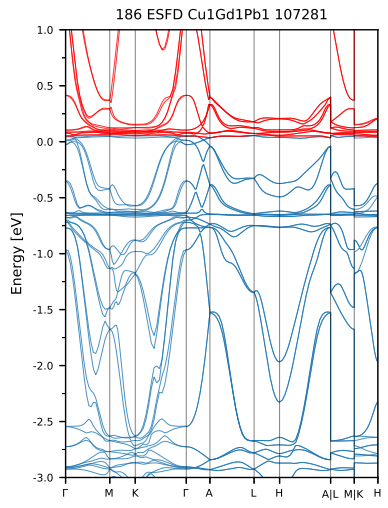

(bz) 186.4.25

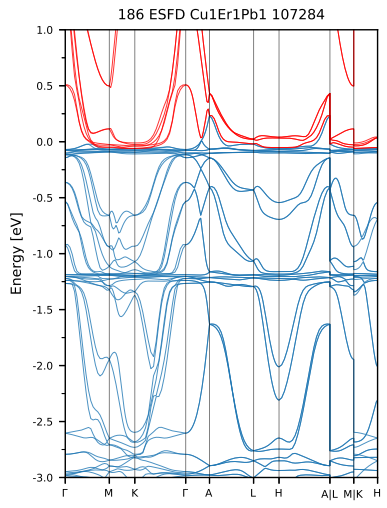

(ca) 186.4.26

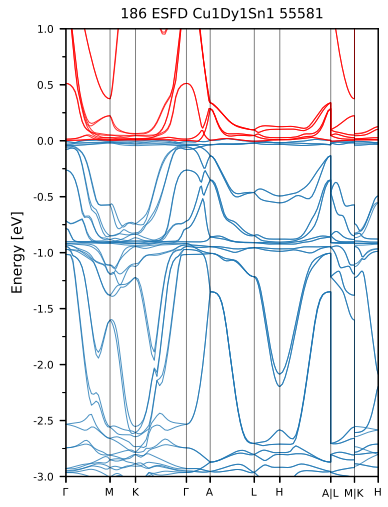

(cb) 186.4.27

Figure 423: Band structures for topological materials in SG 186, part 5/8, from 186.4.12 to 186.4.27. See the list of figures.

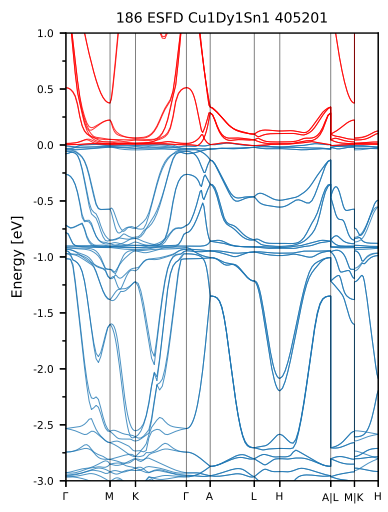

(cc) 186.4.28

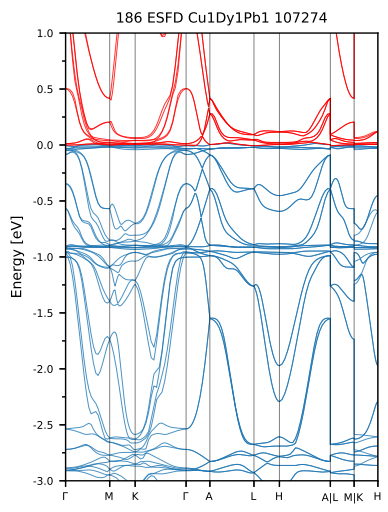

(cd) 186.4.29

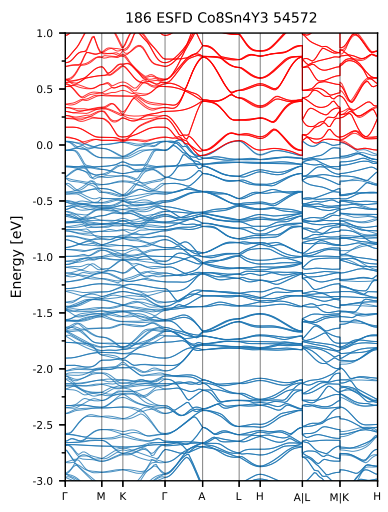

(ce) 186.4.30

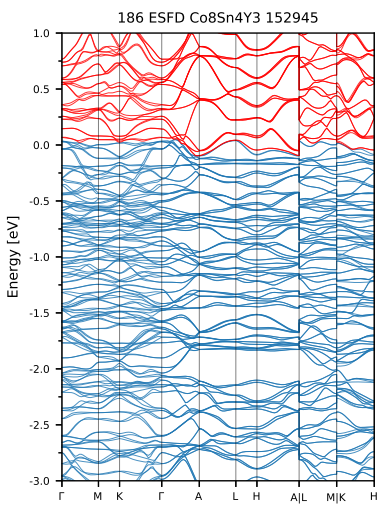

(cf) 186.4.31

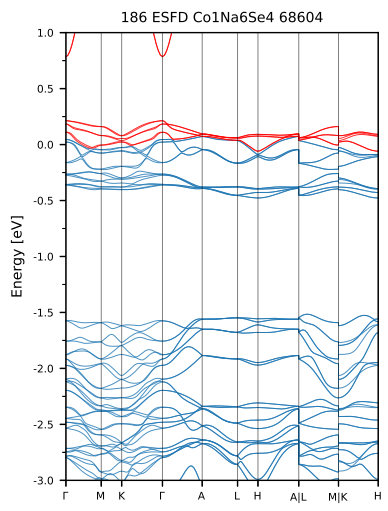

(cg) 186.4.32

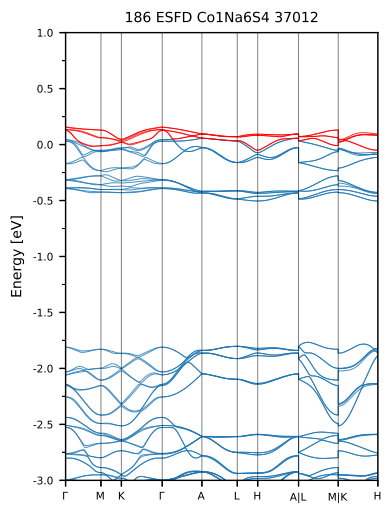

(ch) 186.4.33

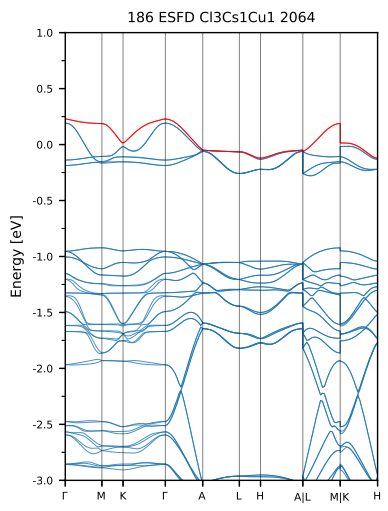

(ci) 186.4.34

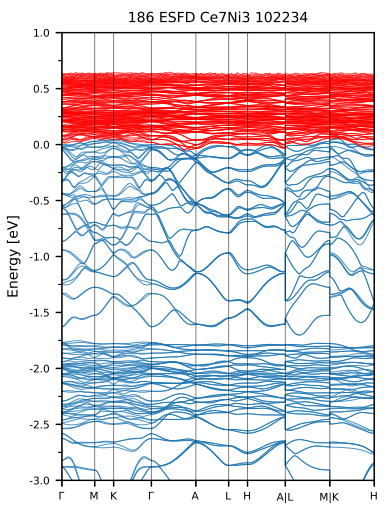

(cj) 186.4.35

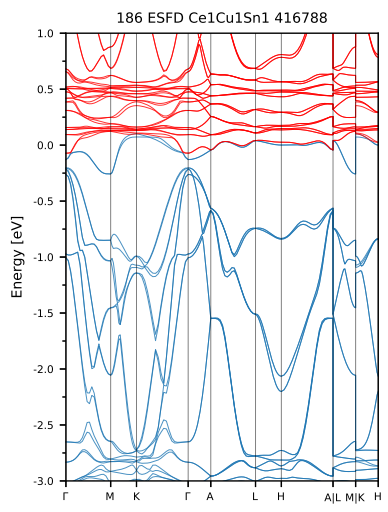

(ck) 186.4.36

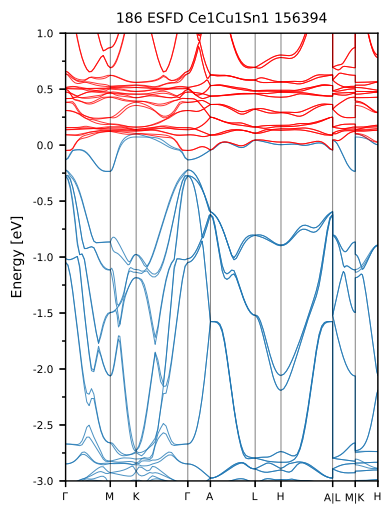

(cl) 186.4.37

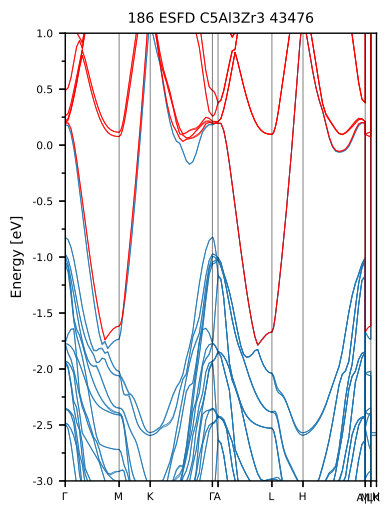

(cm) 186.4.38

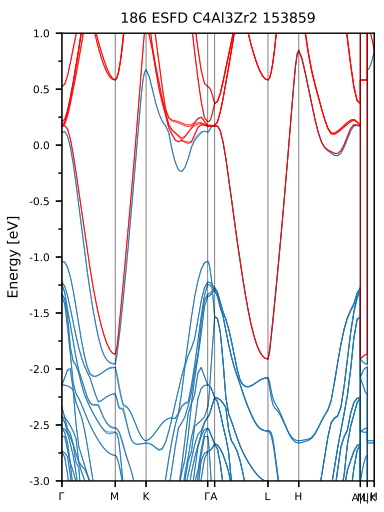

(cn) 186.4.39

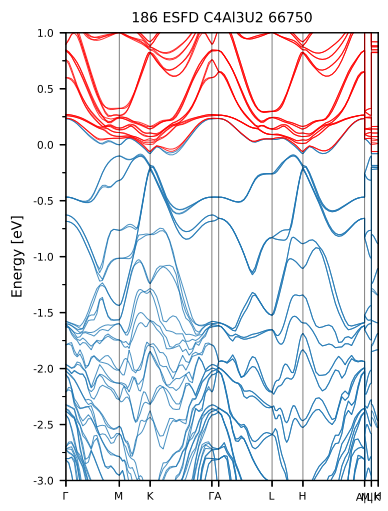

(co) 186.4.40

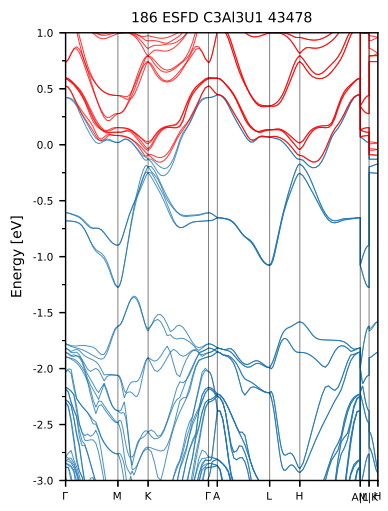

(cp) 186.4.41

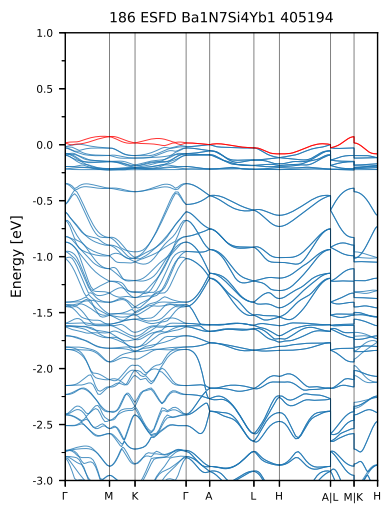

(cq) 186.4.42

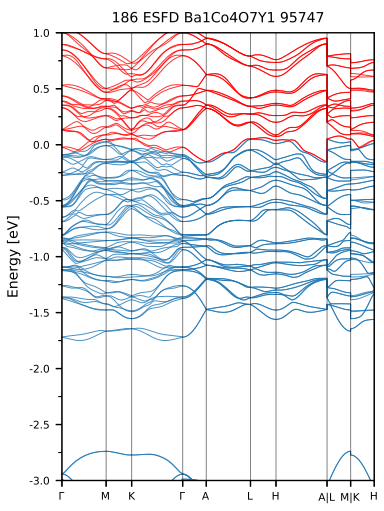

(cr) 186.4.43

Figure 424: Band structures for topological materials in SG 186, part 6/8, from 186.4.28 to 186.4.43. See the list of figures.

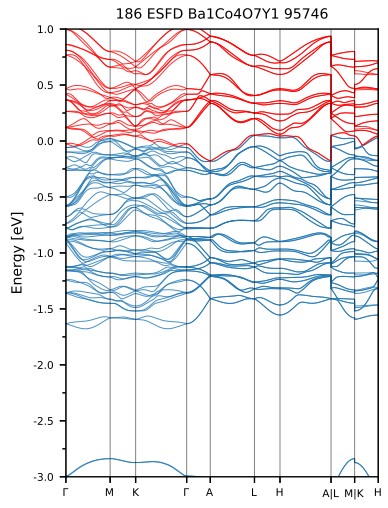

(cs) 186.4.44

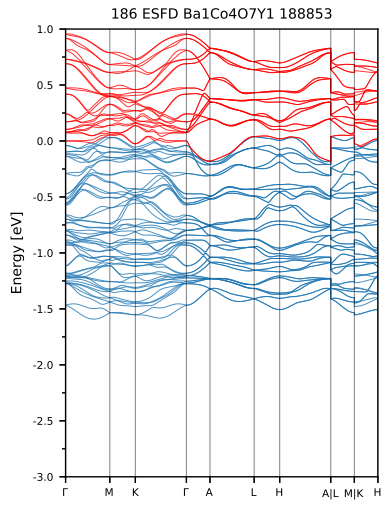

(ct) 186.4.45

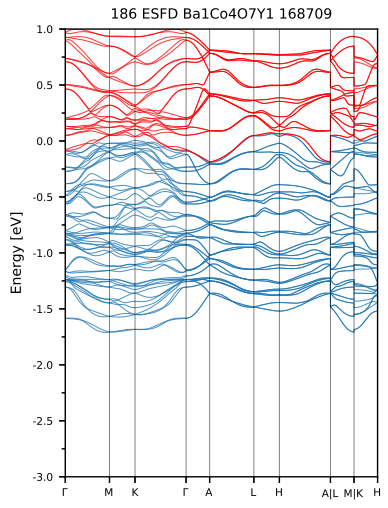

(cu) 186.4.46

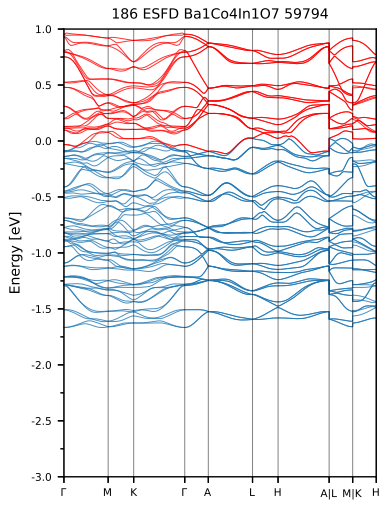

(cv) 186.4.47

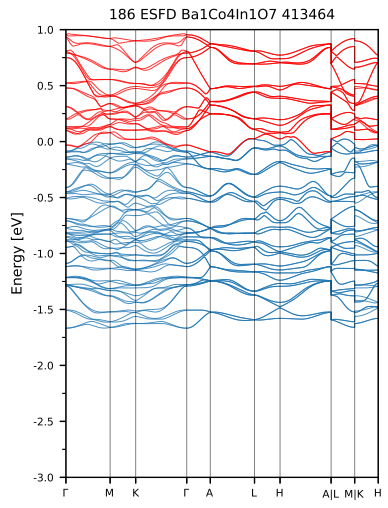

(cw) 186.4.48

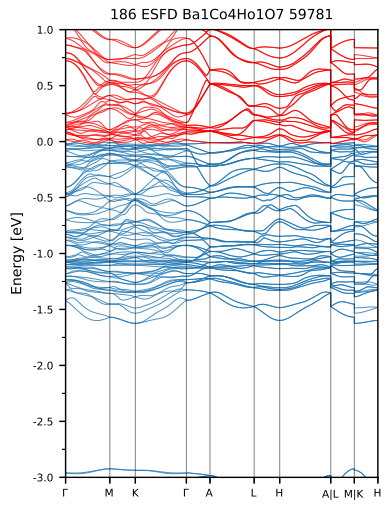

(cx) 186.4.49

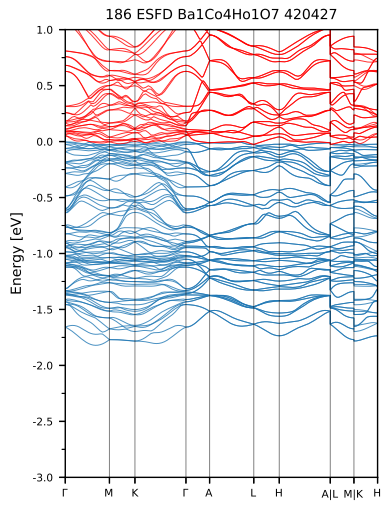

(cy) 186.4.50

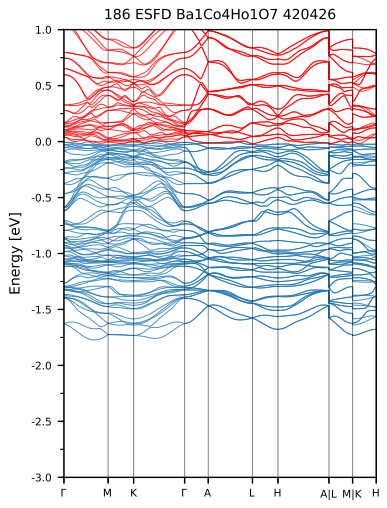

(cz) 186.4.51

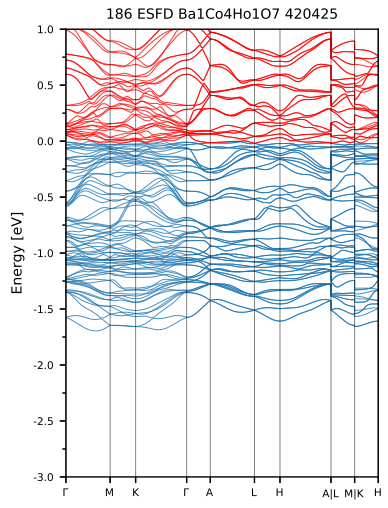

(da) 186.4.52

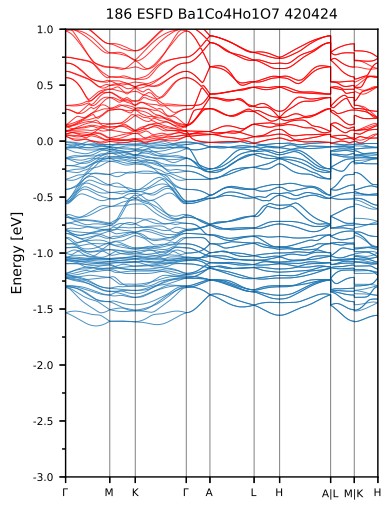

(db) 186.4.53

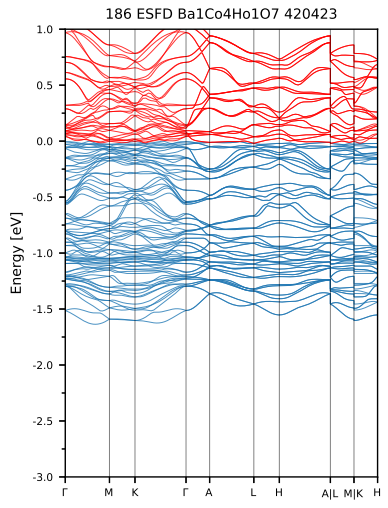

(dc) 186.4.54

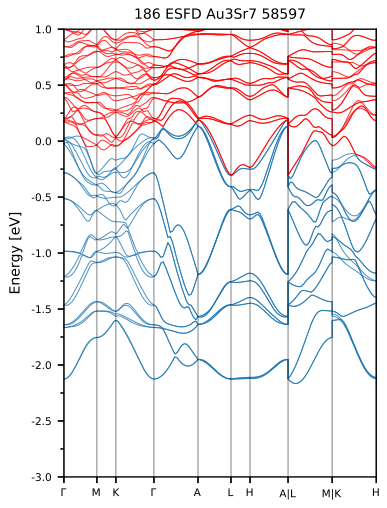

(dd) 186.4.55

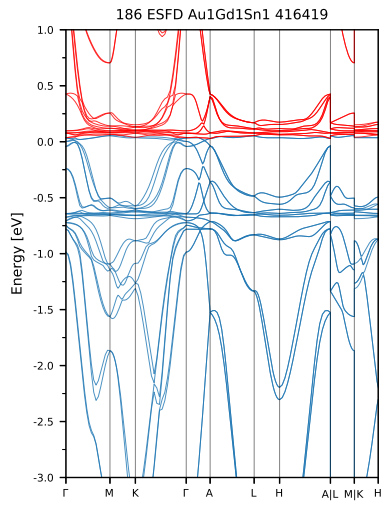

(de) 186.4.56

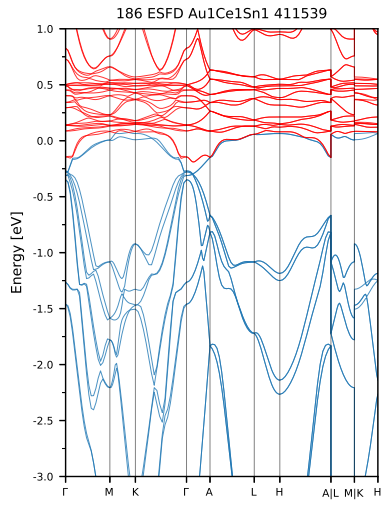

(df) 186.4.57

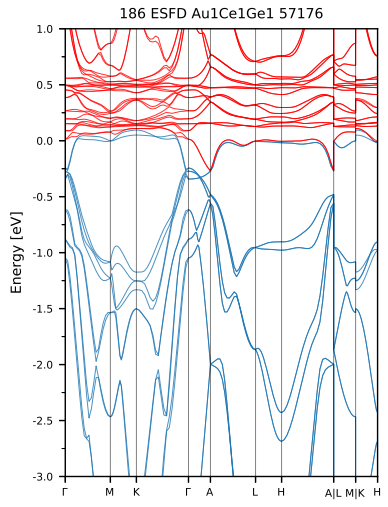

(dg) 186.4.58

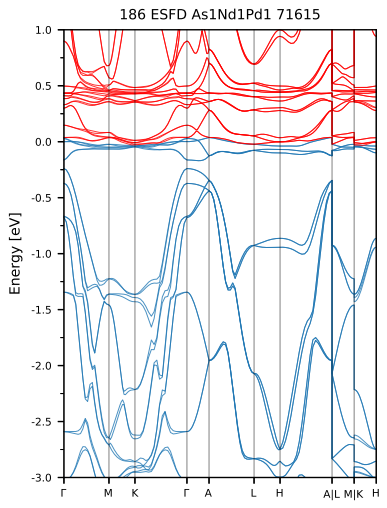

(dh) 186.4.59

Figure 425: Band structures for topological materials in SG 186, part 7/8, from 186.4.44 to 186.4.59. See the list of figures.

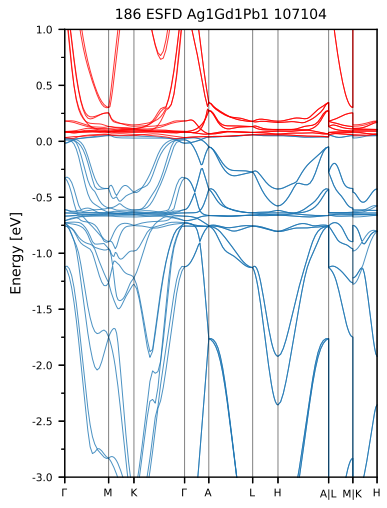

(di) 186.4.60

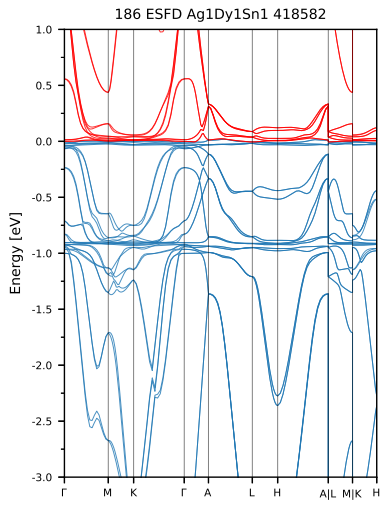

(dj) 186.4.61

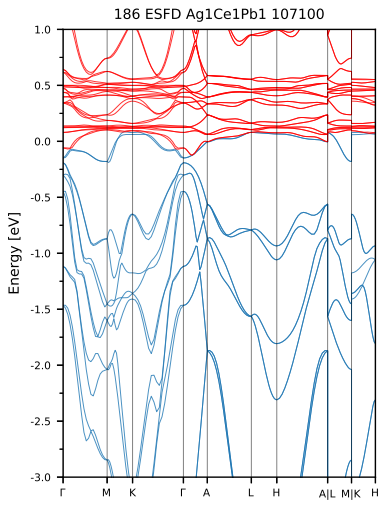

(dk) 186.4.62

Figure 426: Band structures for topological materials in SG 186, part 8/8, from 186.4.60 to 186.4.62. See the list of figures.

## 0.0.92 SG 187

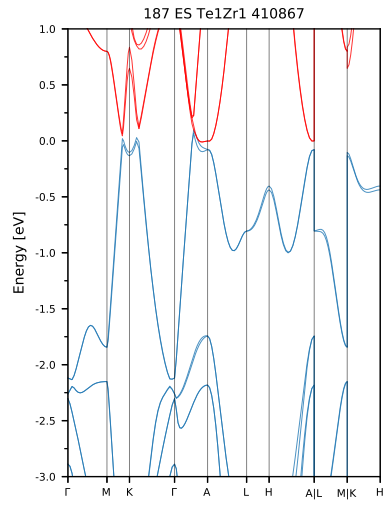

(a) 187.3.1

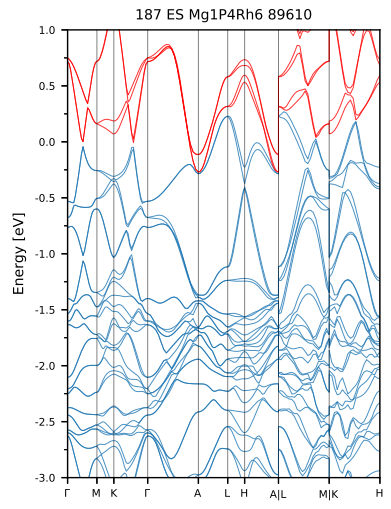

(b) 187.3.2

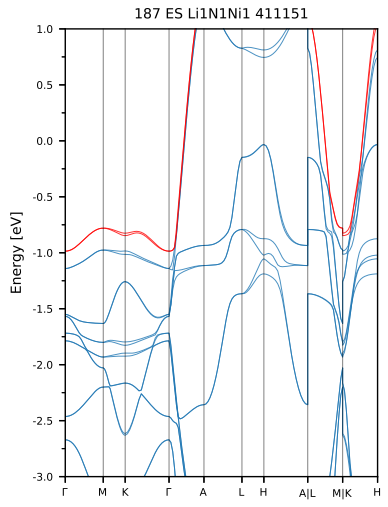

(c) 187.3.3

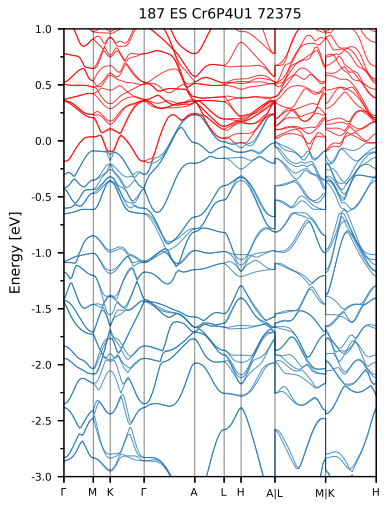

(d) 187.3.4

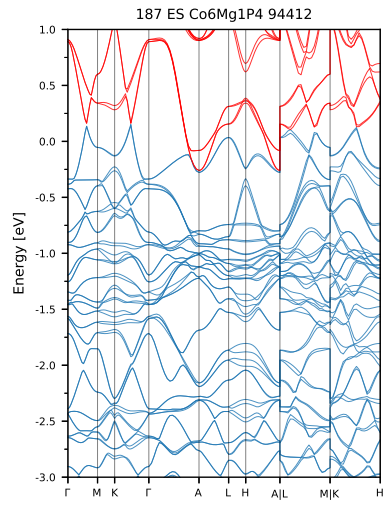

(e) 187.3.5

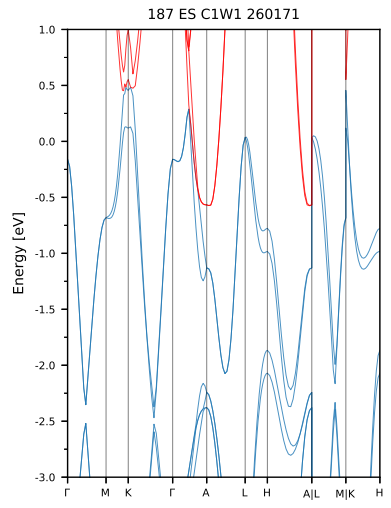

(f) 187.3.6

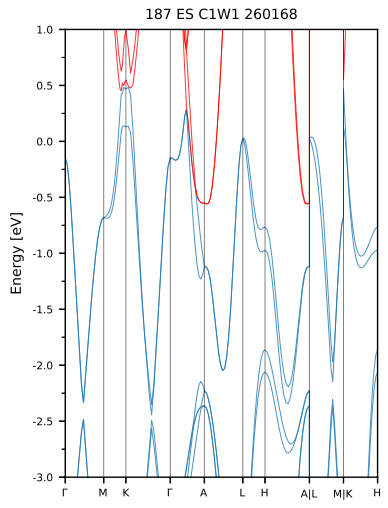

(g) 187.3.7

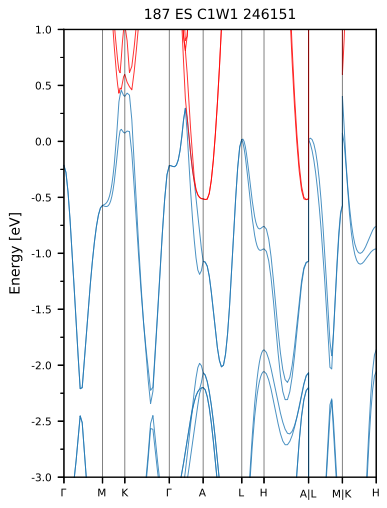

(h) 187.3.8

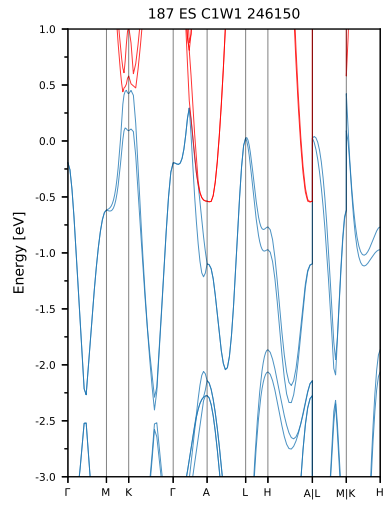

(i) 187.3.9

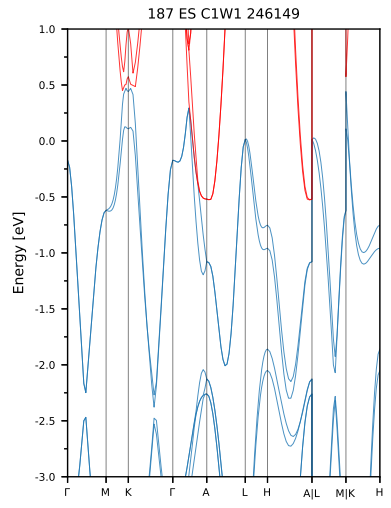

(j) 187.3.10

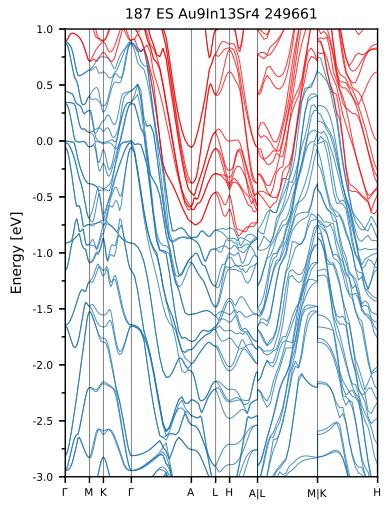

(k) 187.3.11

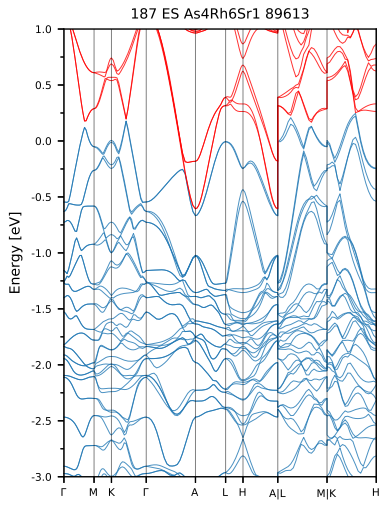

(l) 187.3.12

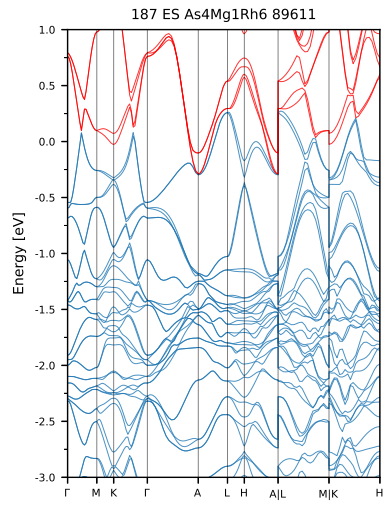

(m) 187.3.13

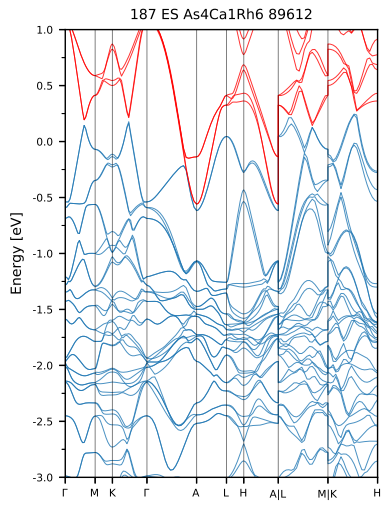

(n) 187.3.14

Figure 427: Band structures for topological materials in SG 187, part 1/1, from 187.3.1 to 187.3.14. See the list of figures.

### 0.0.93 SG 188

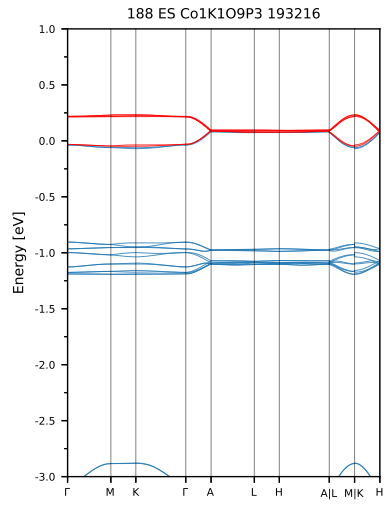

(a) 188.3.1

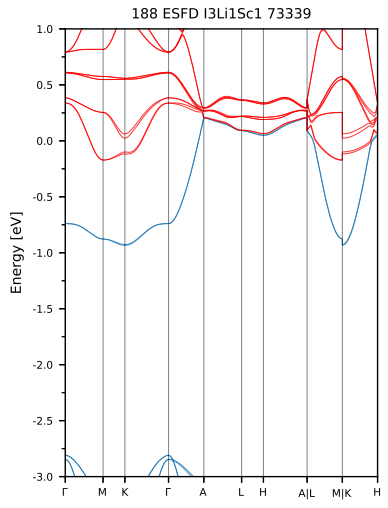

(b) 188.4.1

Figure 428: Band structures for topological materials in SG 188, part 1/1, from 188.3.1 to 188.4.1. See the list of figures.

## 0.0.94 SG 189

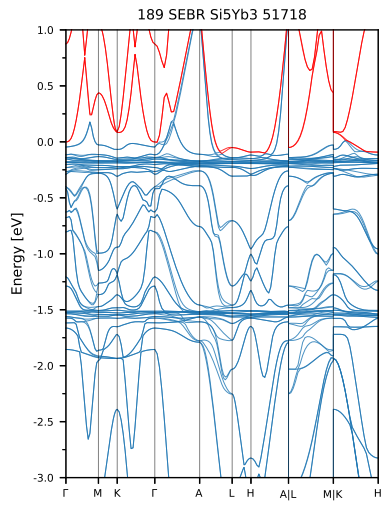

(a) 189.1.1

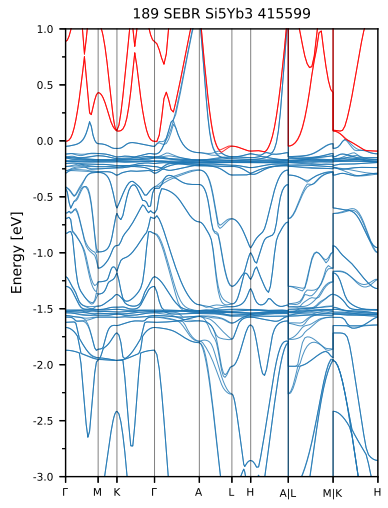

(b) 189.1.2

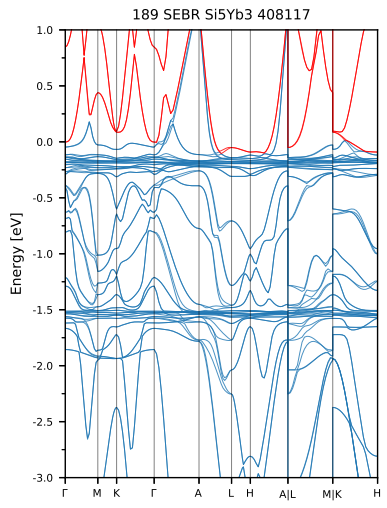

(c) 189.1.3

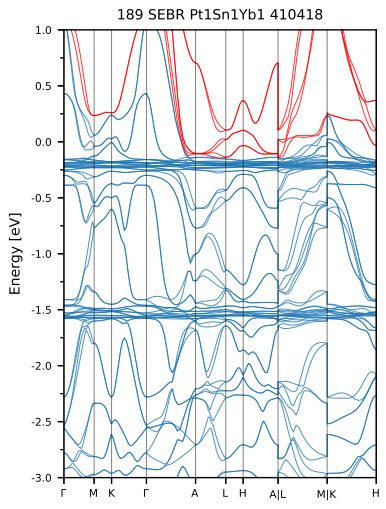

(d) 189.1.4

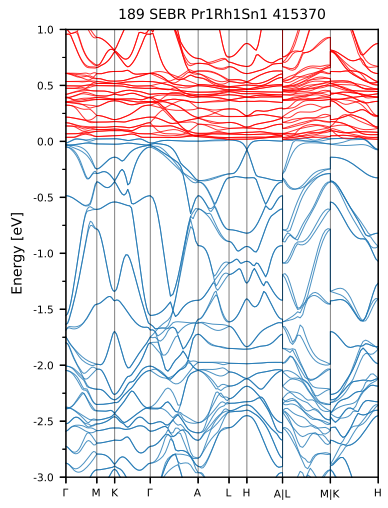

(e) 189.1.5

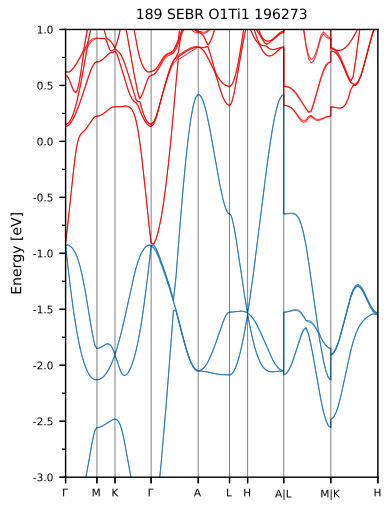

(f) 189.1.6

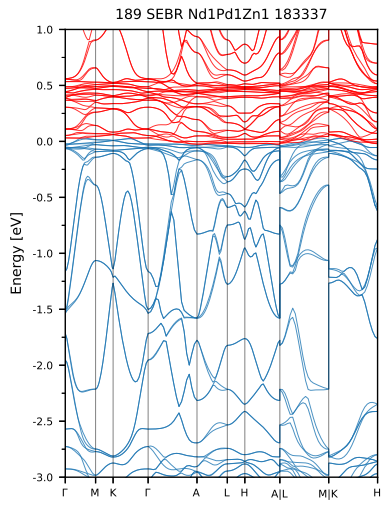

(g) 189.1.7

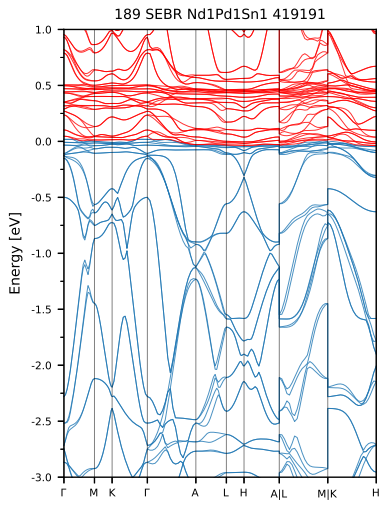

(h) 189.1.8

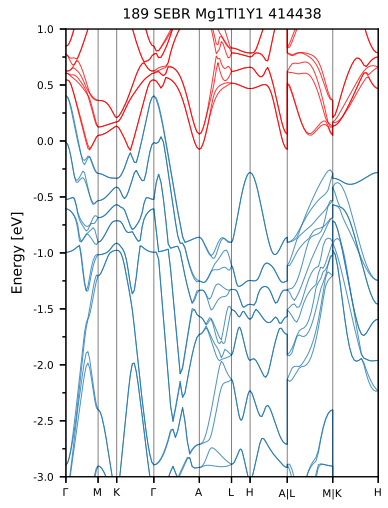

(i) 189.1.9

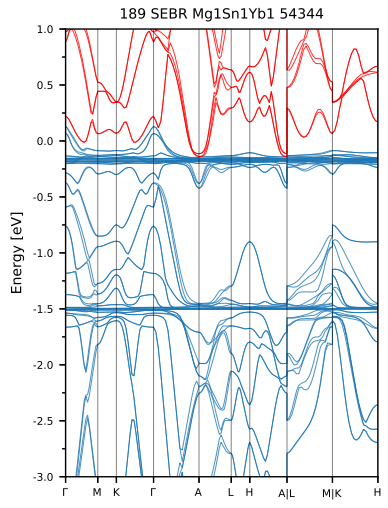

(j) 189.1.10

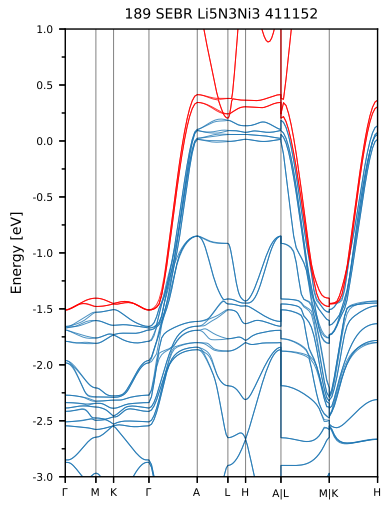

(k) 189.1.11

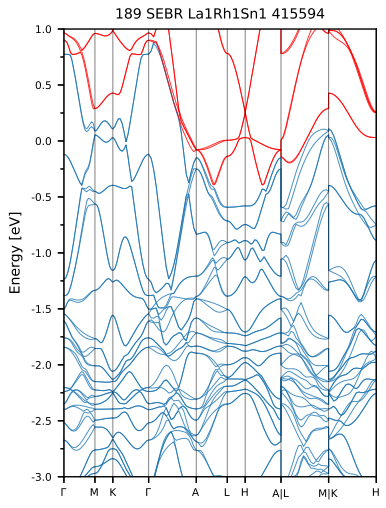

(l) 189.1.12

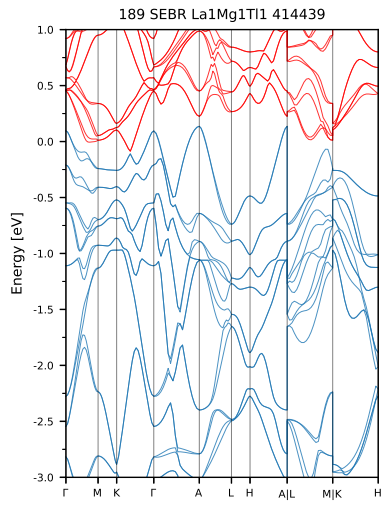

(m) 189.1.13

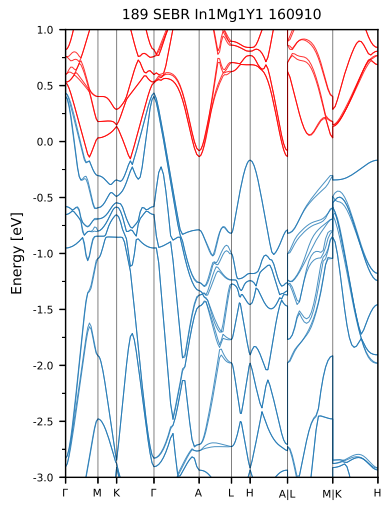

(n) 189.1.14

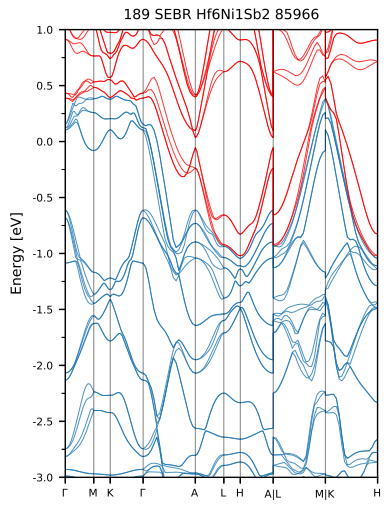

(o) 189.1.15

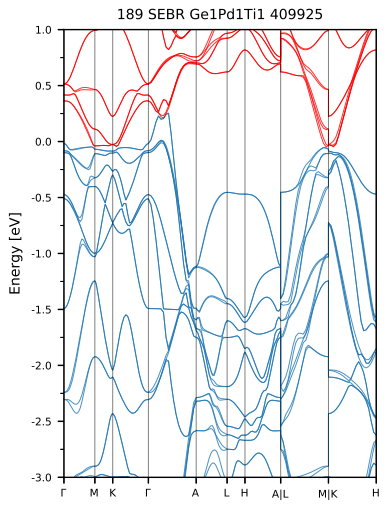

(p) 189.1.16

Figure 429: Band structures for topological materials in SG 189, part 1/6, from 189.1.1 to 189.1.16. See the list of figures.

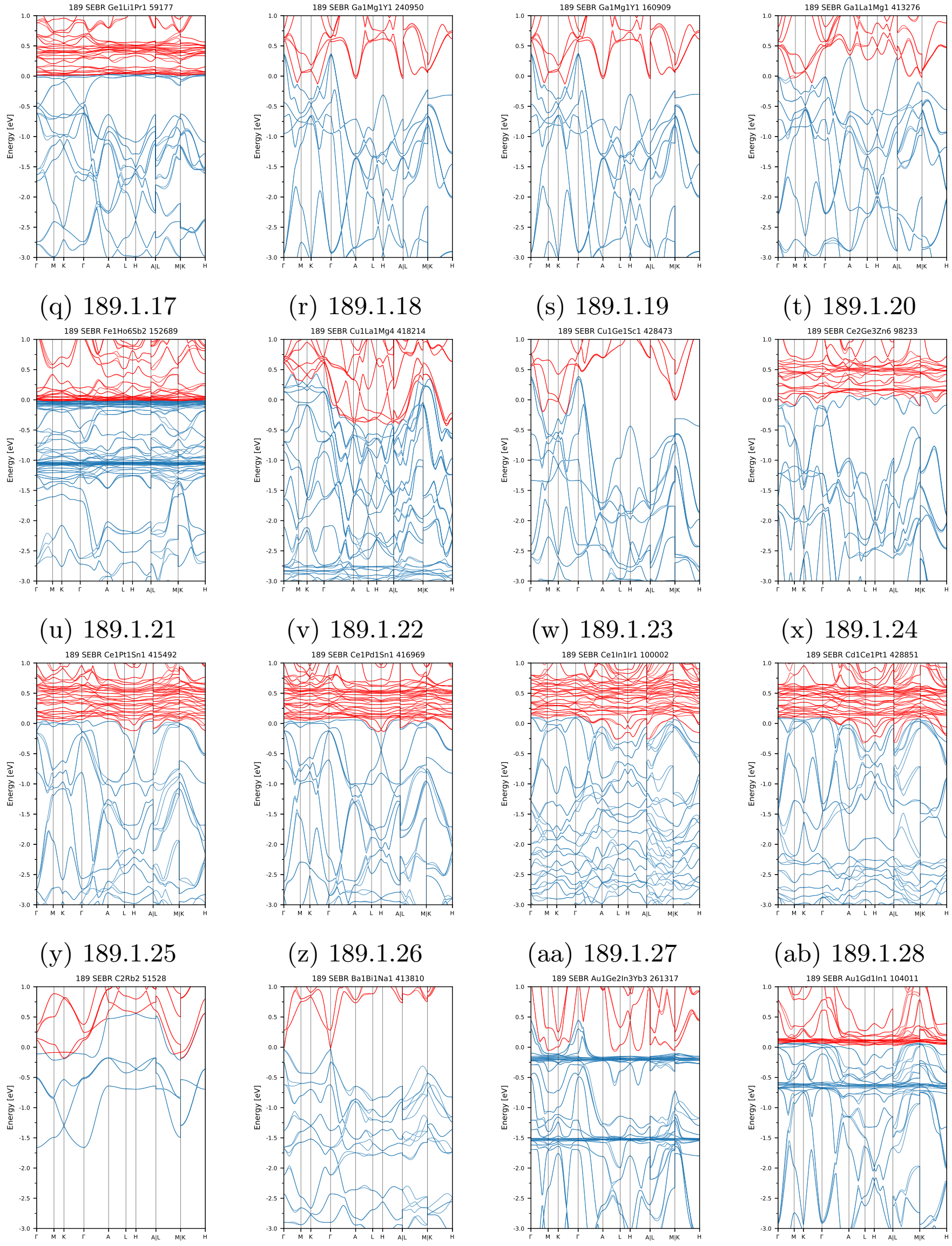

(q) 189.1.17 (r) 189.1.18 (s) 189.1.19 (t) 189.1.20

(u) 189.1.21 (v) 189.1.22 (w) 189.1.23 (x) 189.1.24

(y) 189.1.25 (z) 189.1.26 (aa) 189.1.27 (ab) 189.1.28

(ac) 189.1.29 (ad) 189.1.30 (ae) 189.1.31 (af) 189.1.32

Figure 430: Band structures for topological materials in SG 189, part 2/6, from 189.1.17 to 189.1.32. See the list of figures.

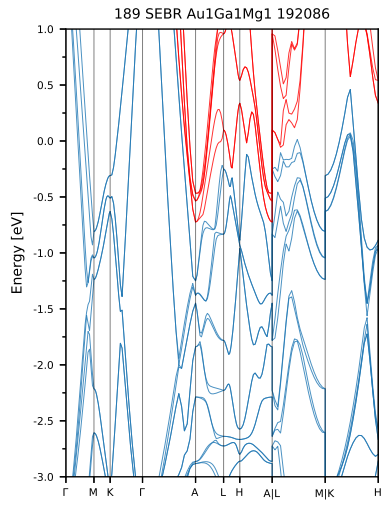

(ag) 189.1.33

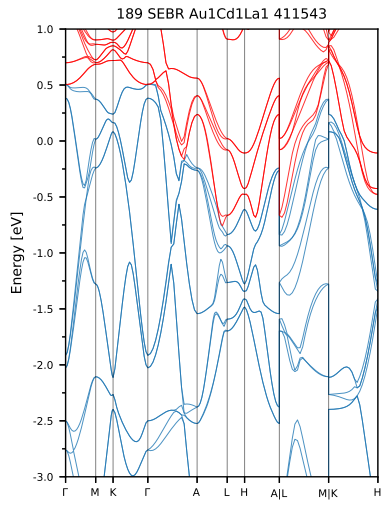

(ah) 189.1.34

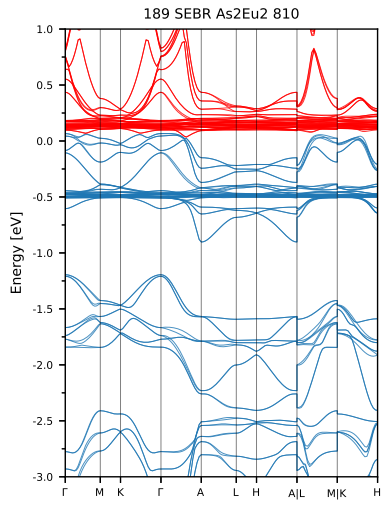

(ai) 189.1.35

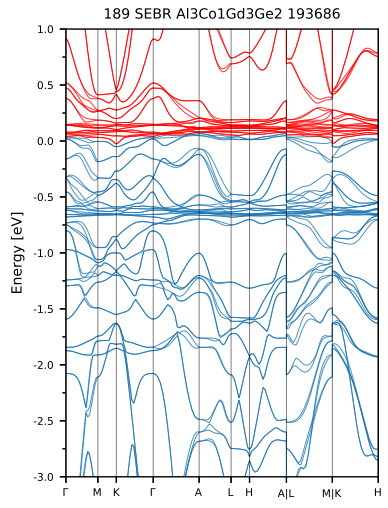

(aj) 189.1.36

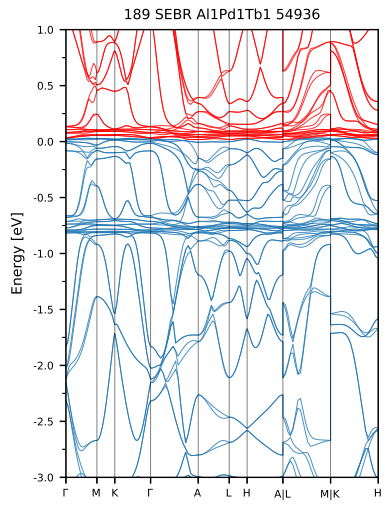

(ak) 189.1.37

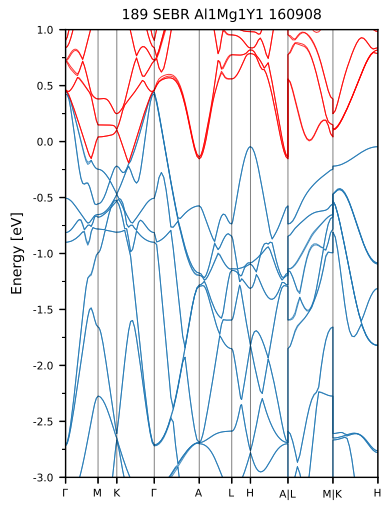

(al) 189.1.38

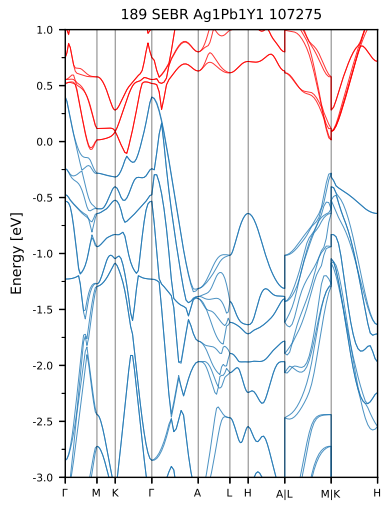

(am) 189.1.39

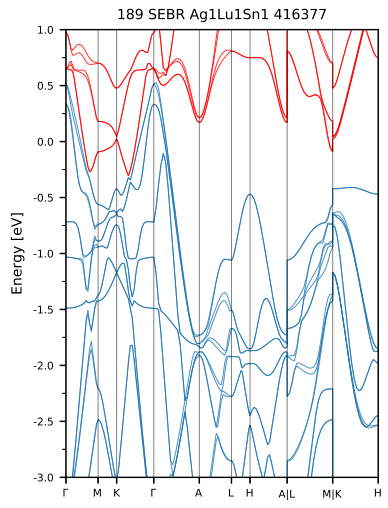

(an) 189.1.40

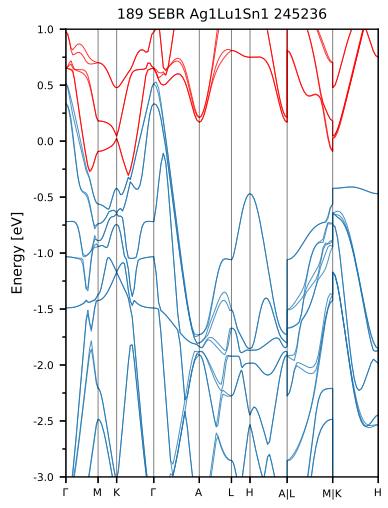

(ao) 189.1.41

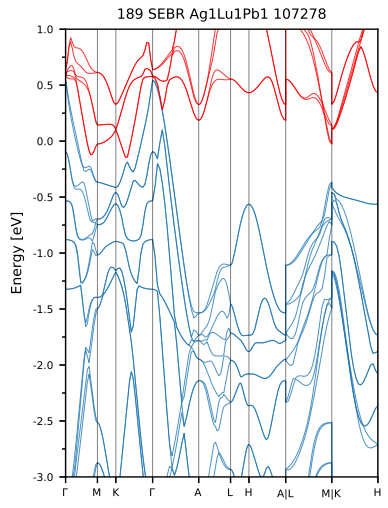

(ap) 189.1.42

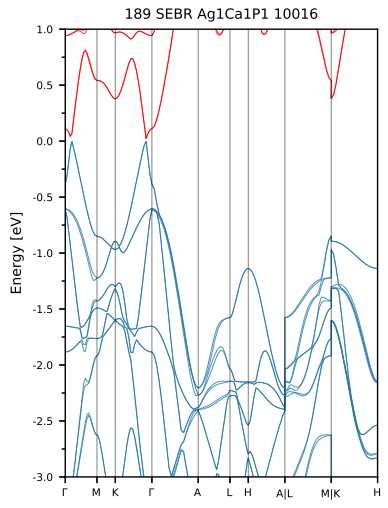

(aq) 189.1.43

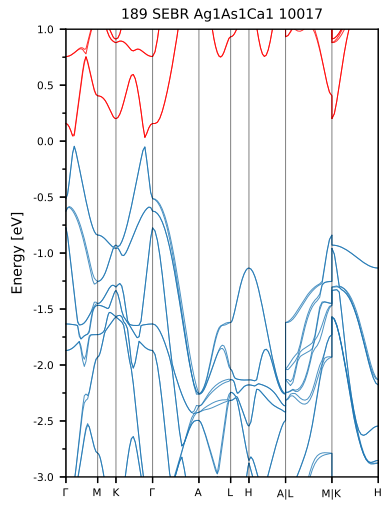

(ar) 189.1.44

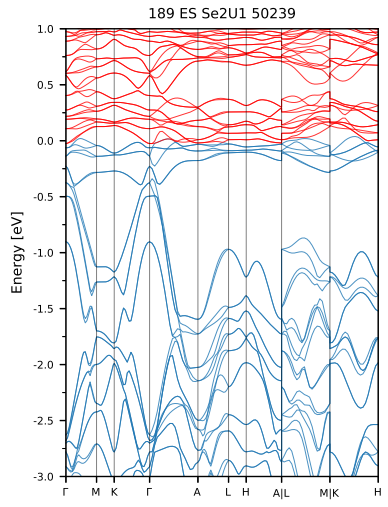

(as) 189.3.1

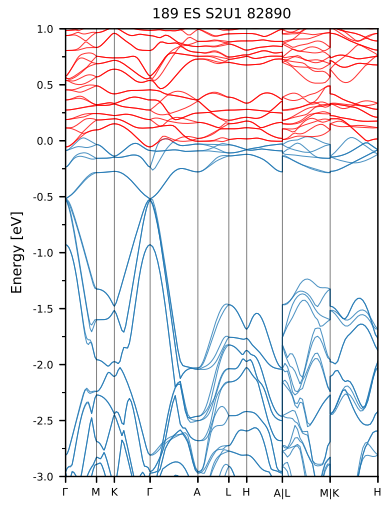

(at) 189.3.2

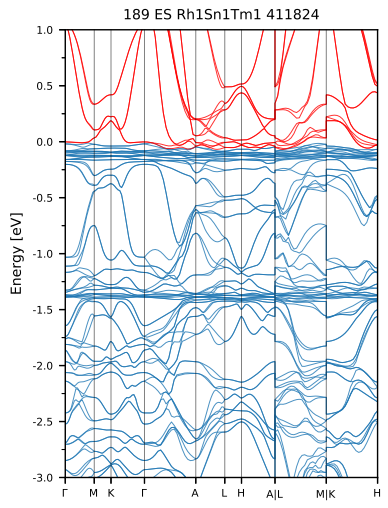

(au) 189.3.3

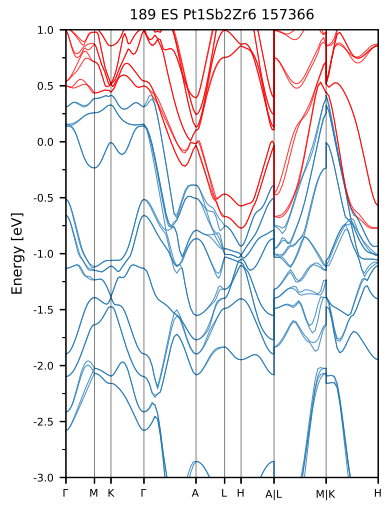

(av) 189.3.4

Figure 431: Band structures for topological materials in SG 189, part 3/6, from 189.1.33 to 189.3.4. See the list of figures.

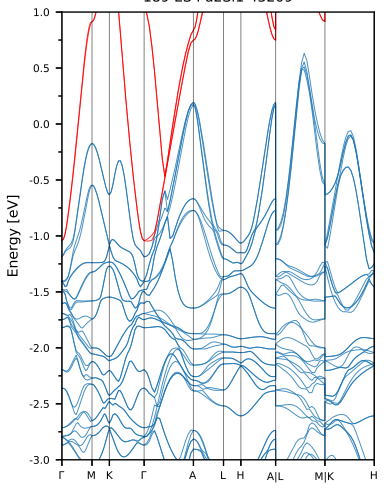

(aw) 189.3.5

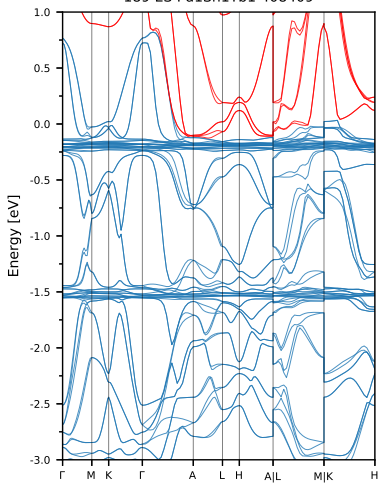

(ax) 189.3.6

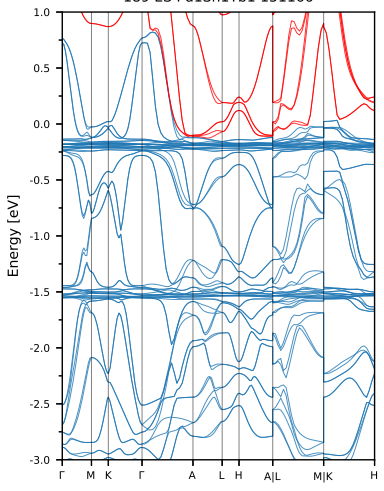

(ay) 189.3.7

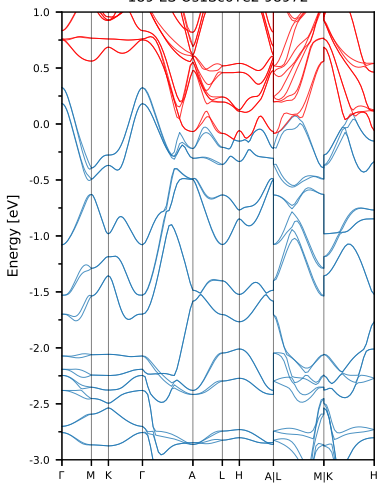

(az) 189.3.8

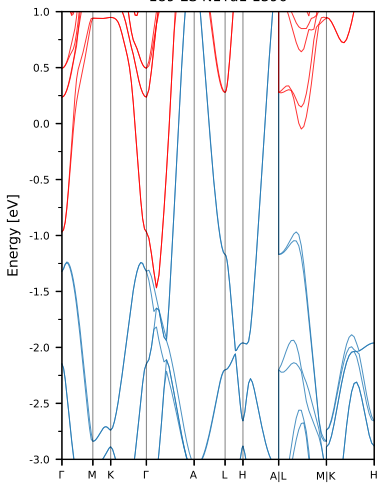

(ba) 189.3.9

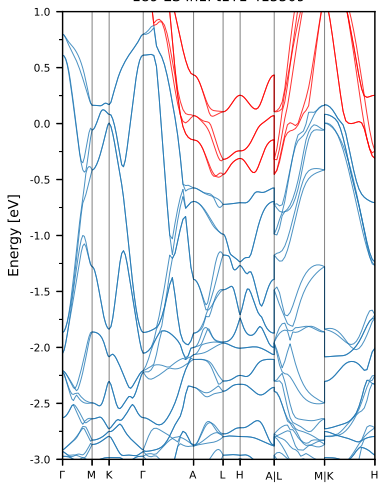

(bb) 189.3.10

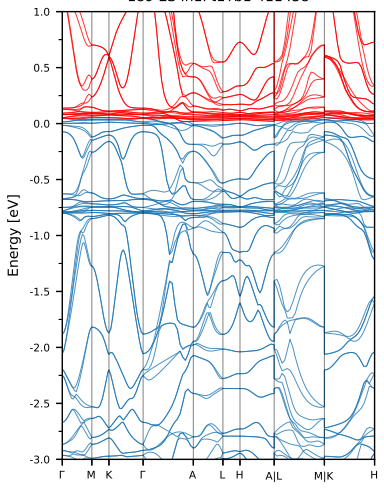

(bc) 189.3.11

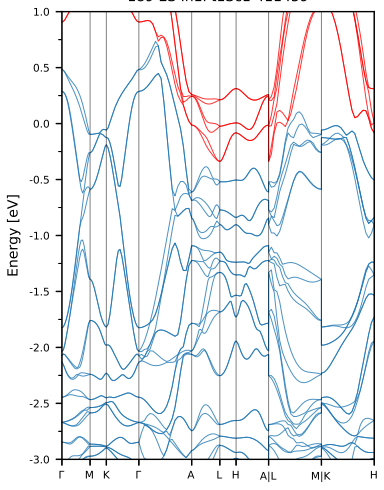

(bd) 189.3.12

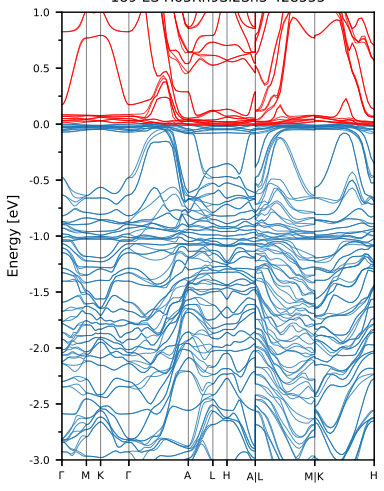

(be) 189.3.13

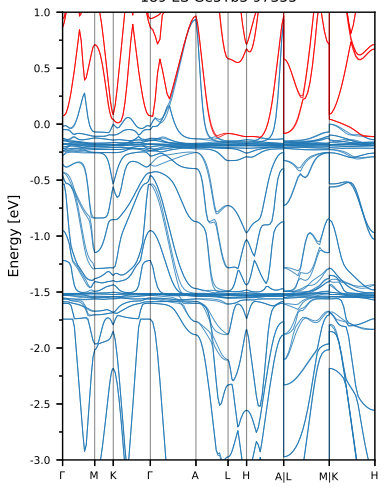

(bf) 189.3.14

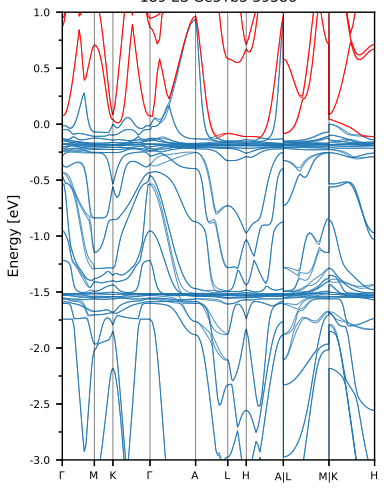

(bg) 189.3.15

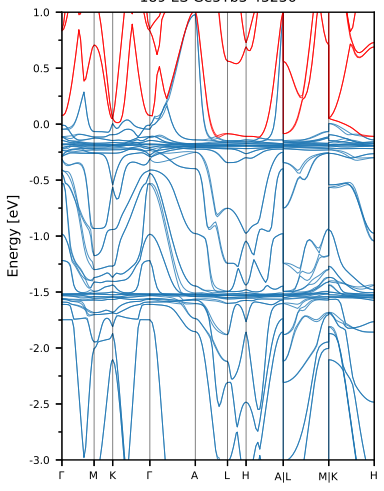

(bh) 189.3.16

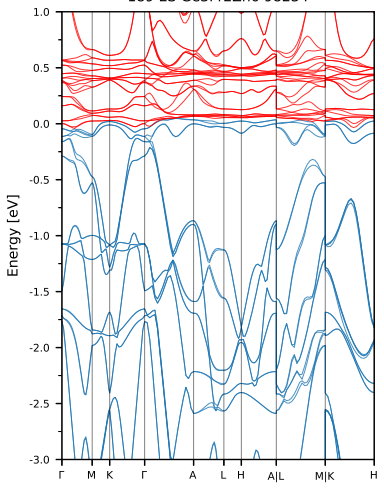

(bi) 189.3.17

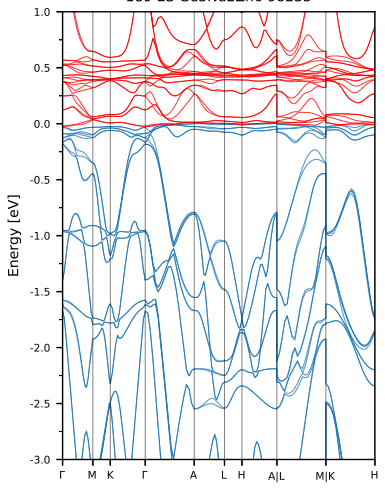

(bj) 189.3.18

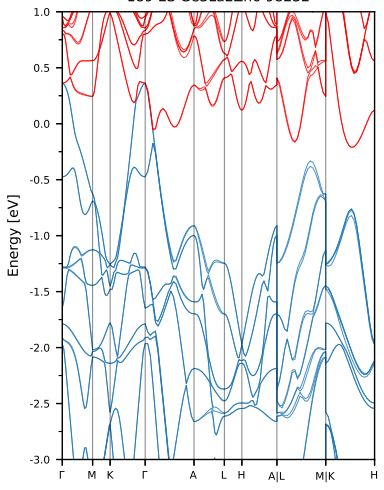

(bk) 189.3.19

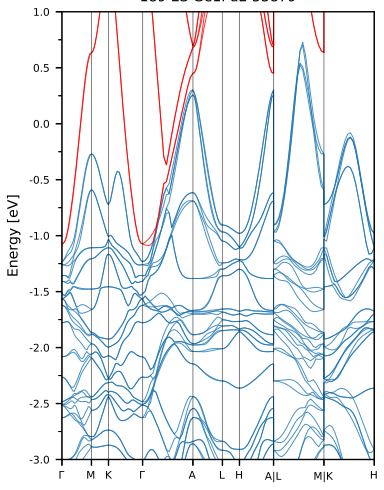

(bl) 189.3.20

Figure 432: Band structures for topological materials in SG 189, part 4/6, from 189.3.5 to 189.3.20. See the list of figures.

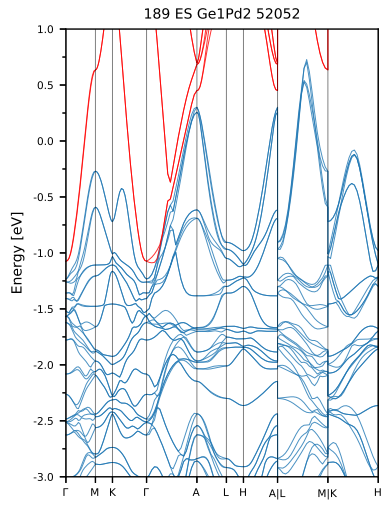

(bm) 189.3.21

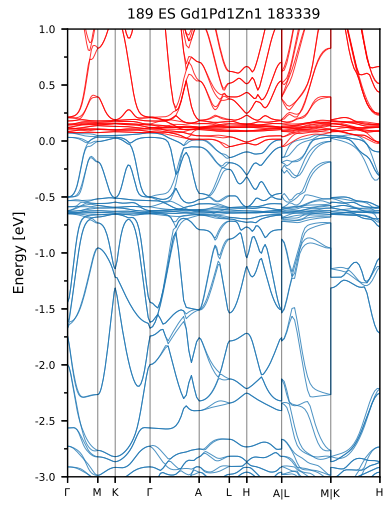

(bn) 189.3.22

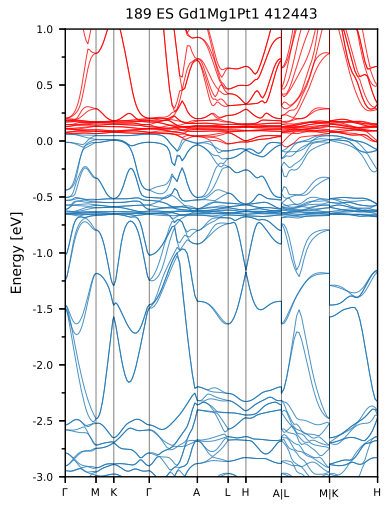

(bo) 189.3.23

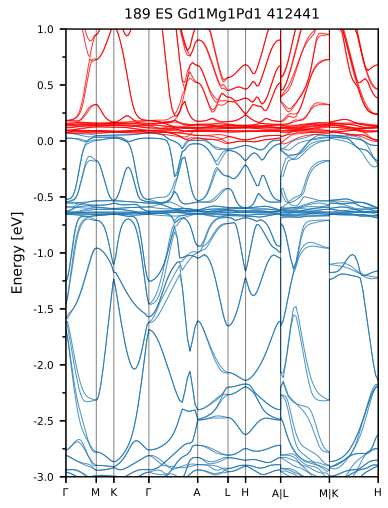

(bp) 189.3.24

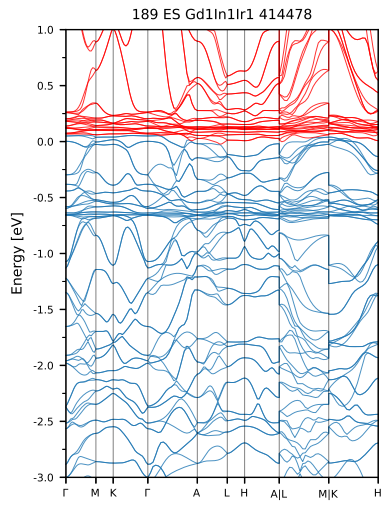

(bq) 189.3.25

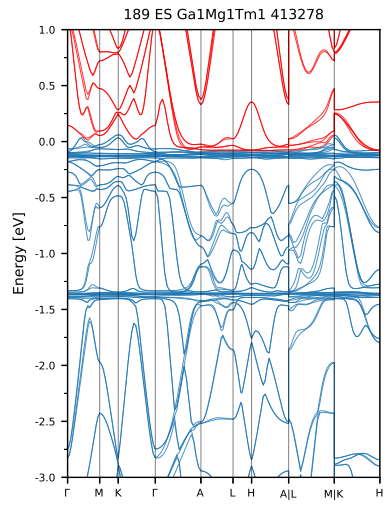

(br) 189.3.26

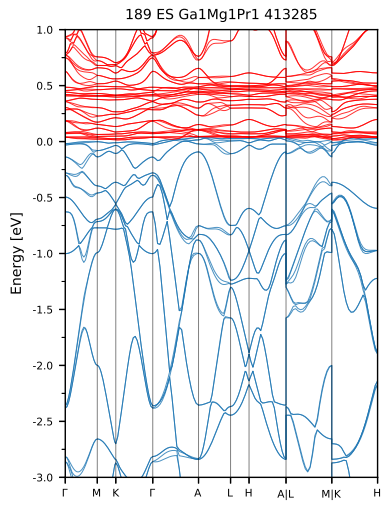

(bs) 189.3.27

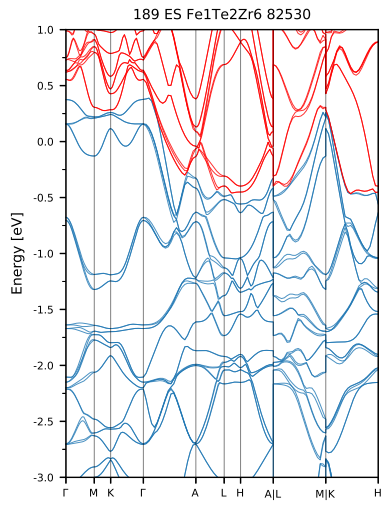

(bt) 189.3.28

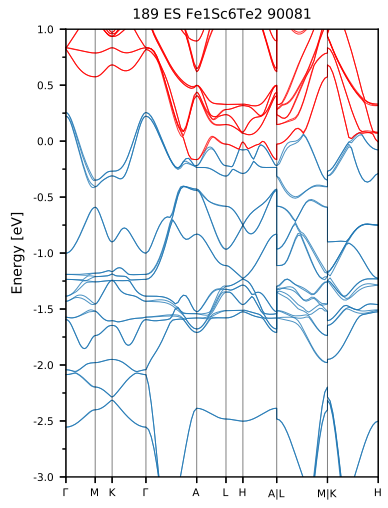

(bu) 189.3.29

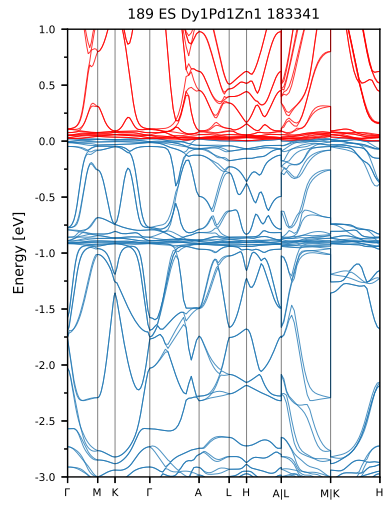

(bv) 189.3.30

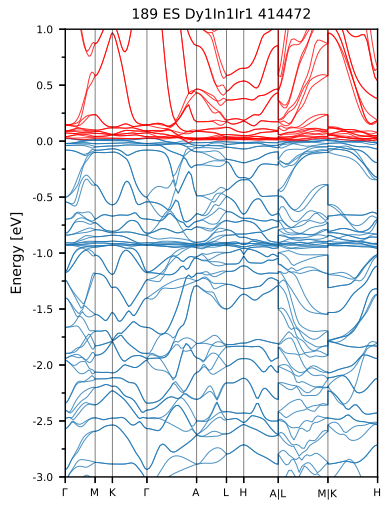

(bw) 189.3.31

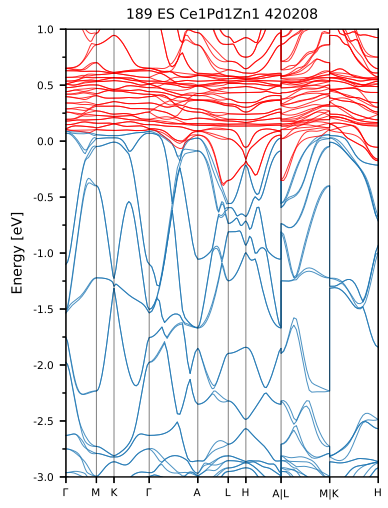

(bx) 189.3.32

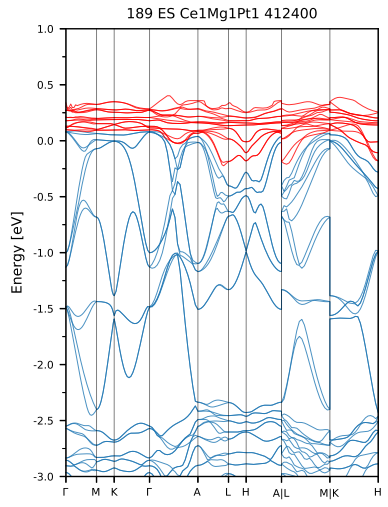

(by) 189.3.33

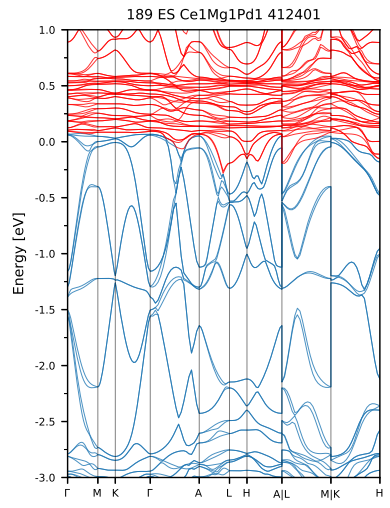

(bz) 189.3.34

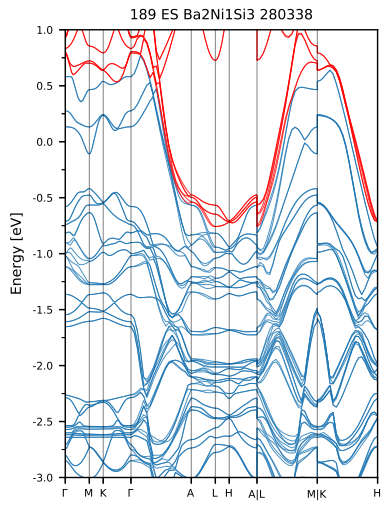

(ca) 189.3.35

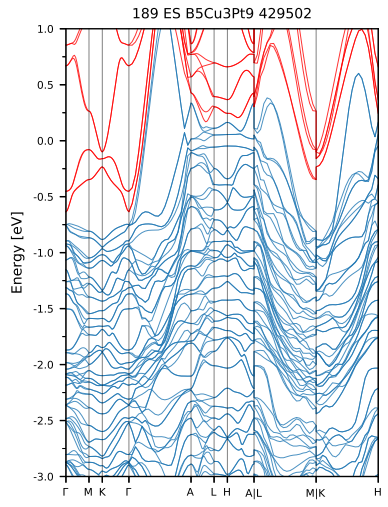

(cb) 189.3.36

Figure 433: Band structures for topological materials in SG 189, part 5/6, from 189.3.21 to 189.3.36. See the list of figures.

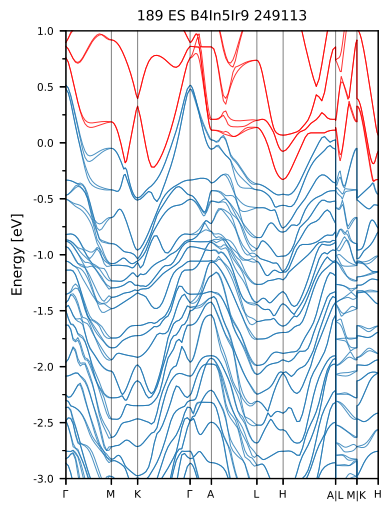

(cc) 189.3.37

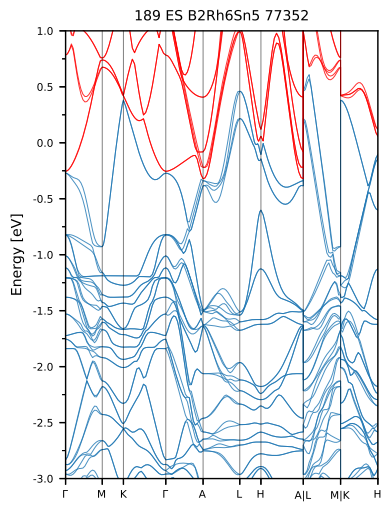

(cd) 189.3.38

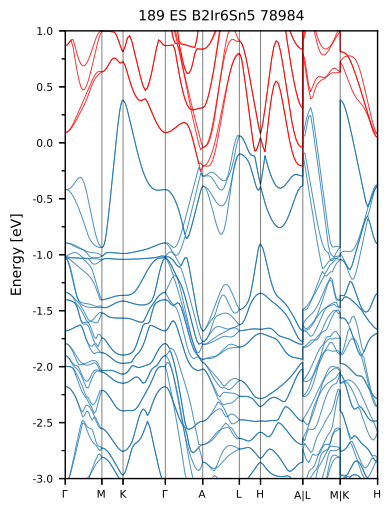

(ce) 189.3.39

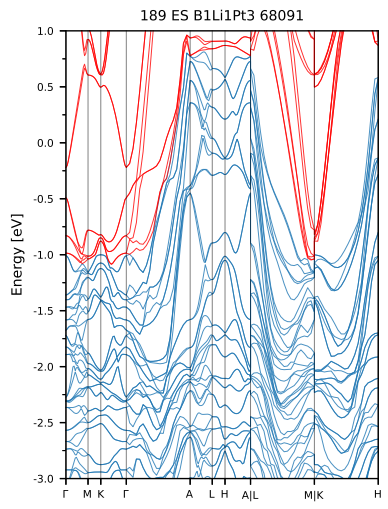

(cf) 189.3.40

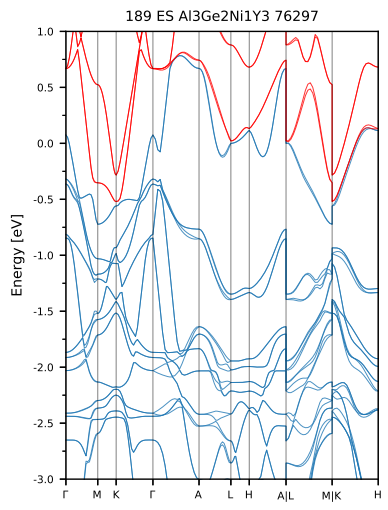

(cg) 189.3.41

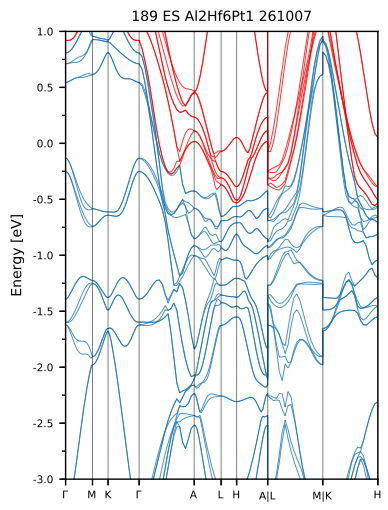

(ch) 189.3.42

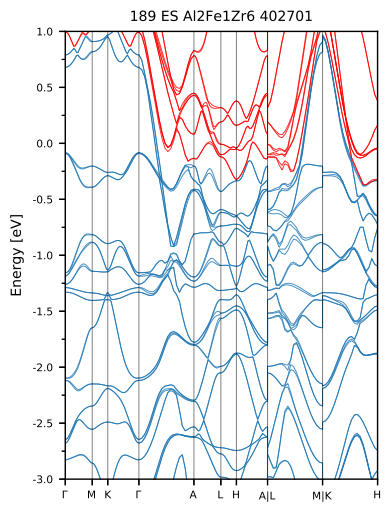

(ci) 189.3.43

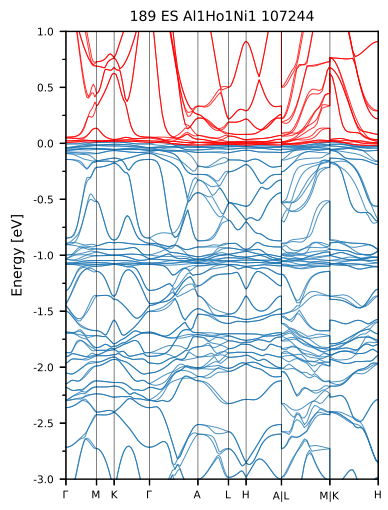

(cj) 189.3.44

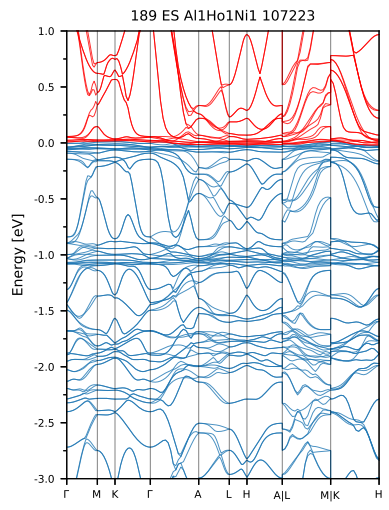

(ck) 189.3.45

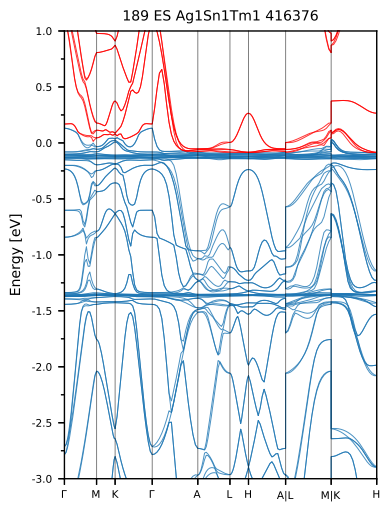

(cl) 189.3.46

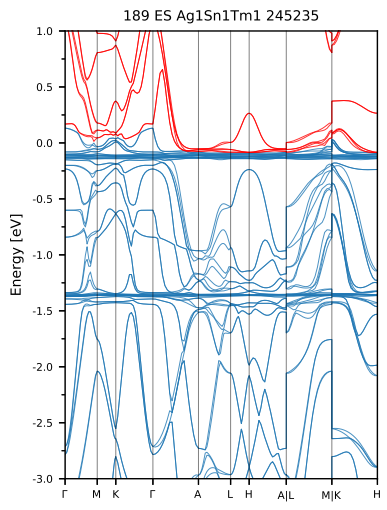

(cm) 189.3.47

Figure 434: Band structures for topological materials in SG 189, part 6/6, from 189.3.37 to 189.3.47. See the list of figures.

## 0.0.95 SG 190

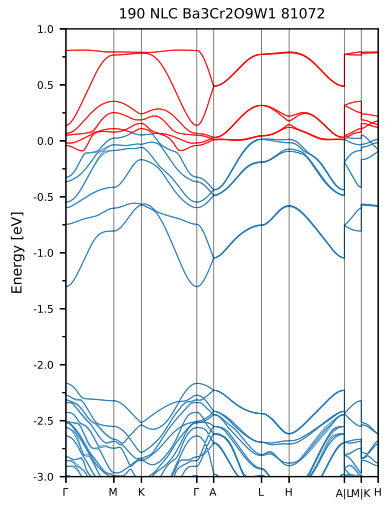

(a) 190.2.1

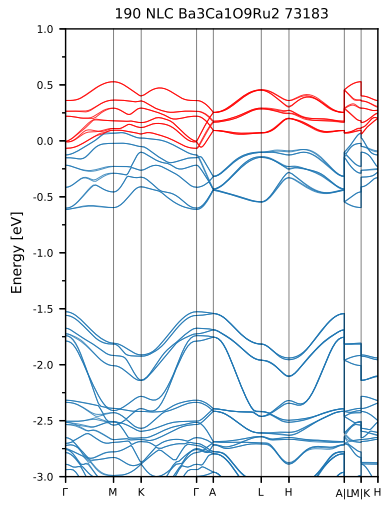

(b) 190.2.2

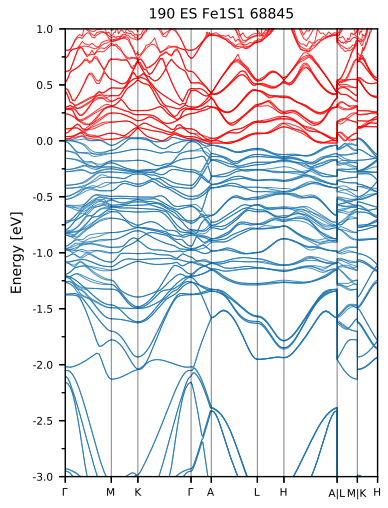

(c) 190.3.1

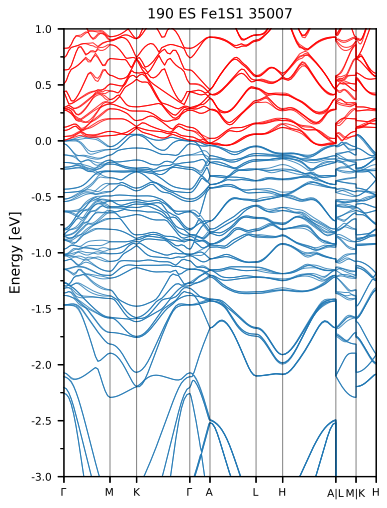

(d) 190.3.2

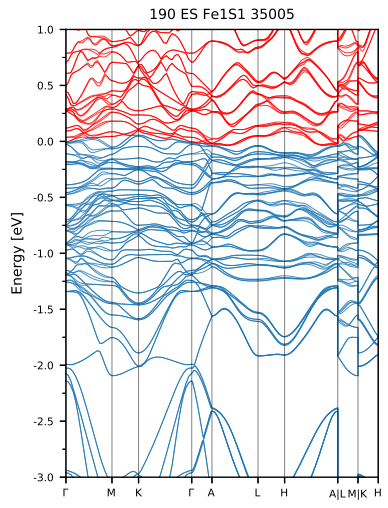

(e) 190.3.3

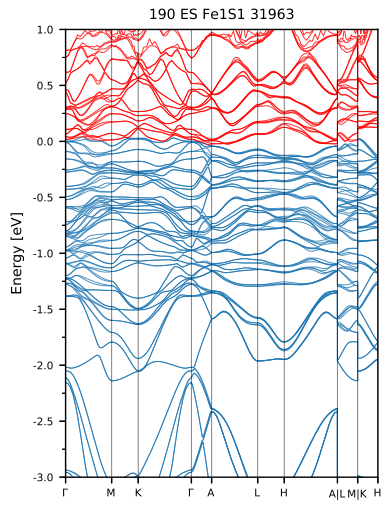

(f) 190.3.4

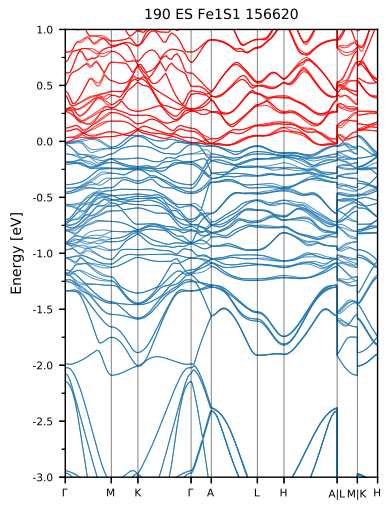

(g) 190.3.5

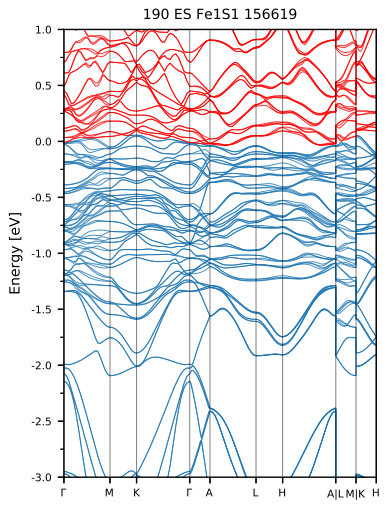

(h) 190.3.6

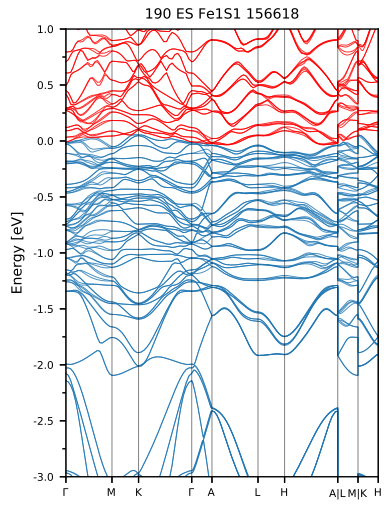

(i) 190.3.7

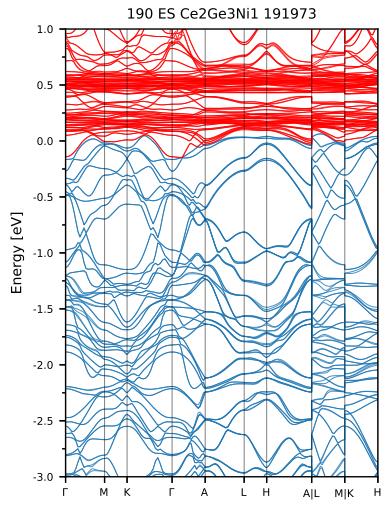

(j) 190.3.8

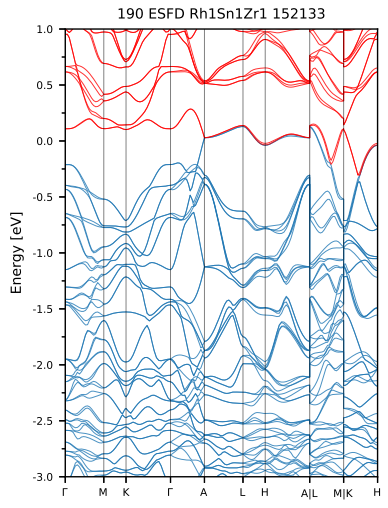

(k) 190.4.1

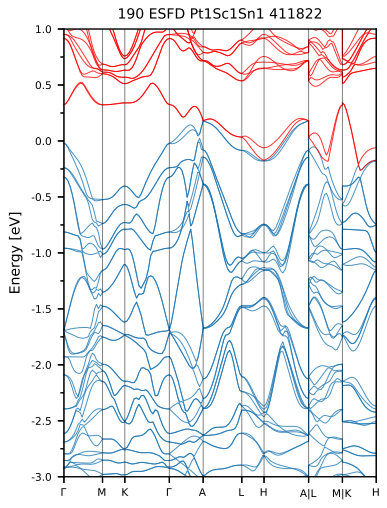

(l) 190.4.2

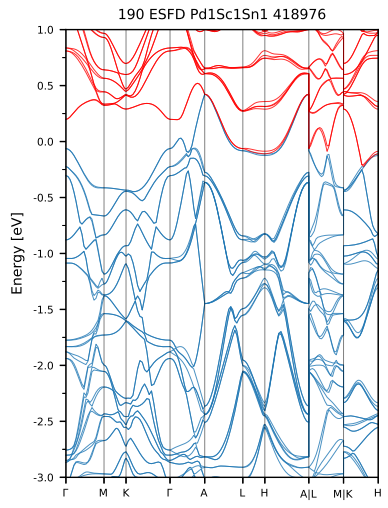

(m) 190.4.3

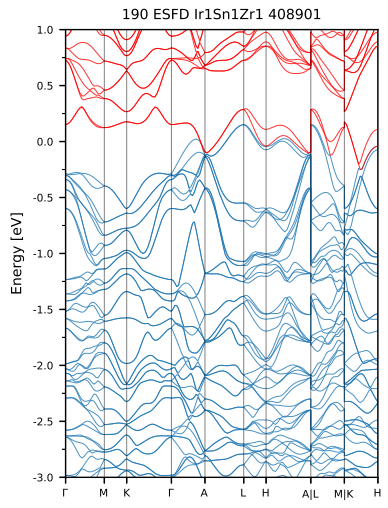

(n) 190.4.4

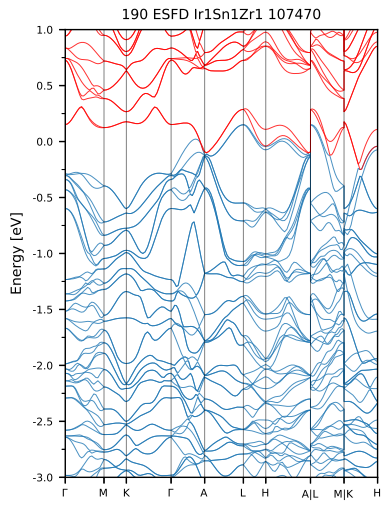

(o) 190.4.5

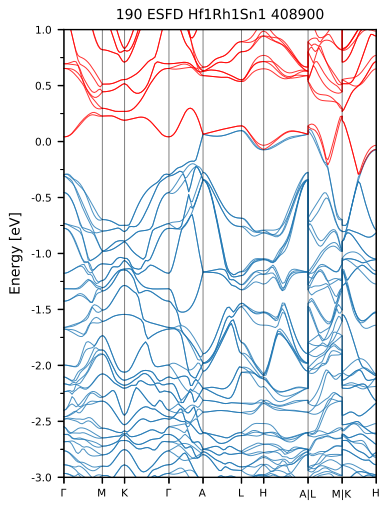

(p) 190.4.6

Figure 435: Band structures for topological materials in SG 190, part 1/2, from 190.2.1 to 190.4.6. See the list of figures.

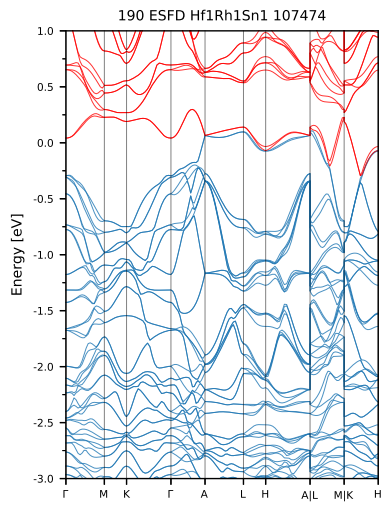

(q) 190.4.7

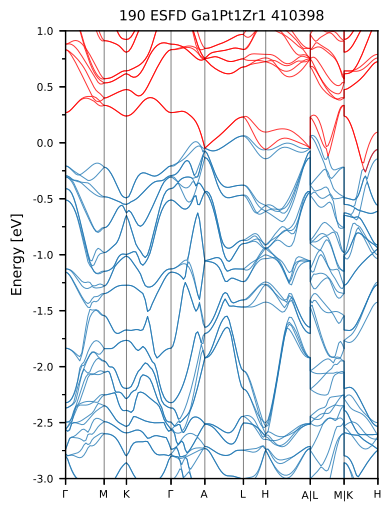

(r) 190.4.8

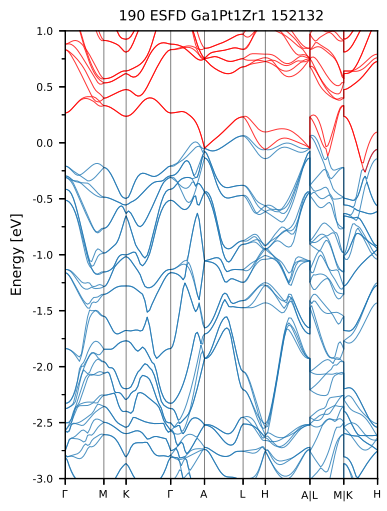

(s) 190.4.9

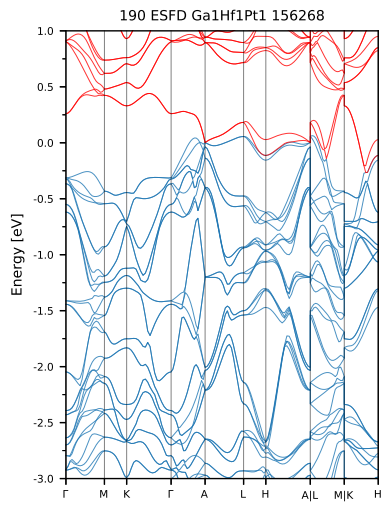

(t) 190.4.10

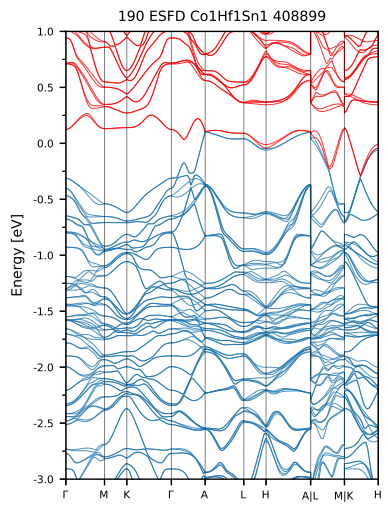

(u) 190.4.11

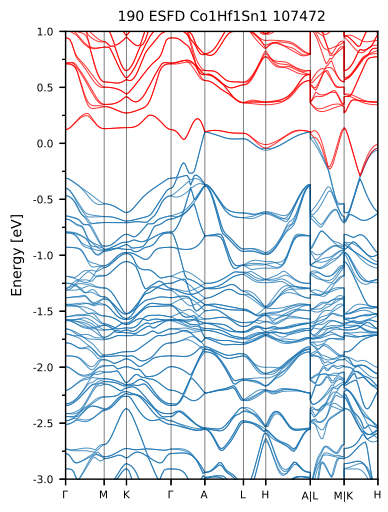

(v) 190.4.12

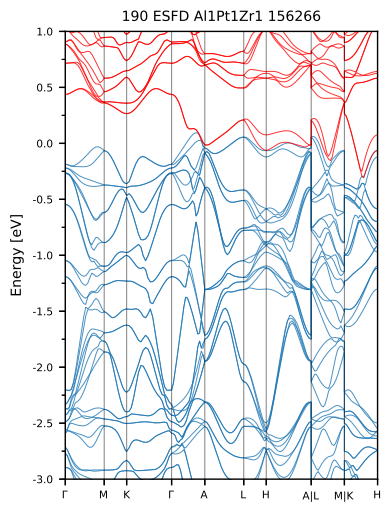

(w) 190.4.13

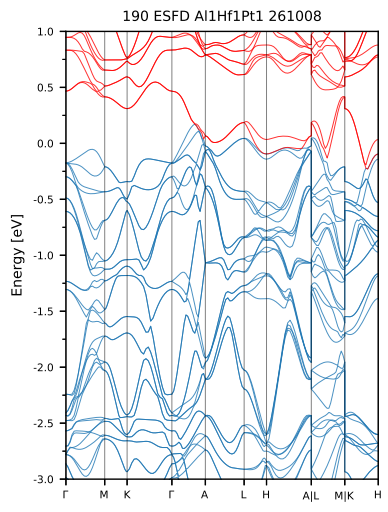

(x) 190.4.14

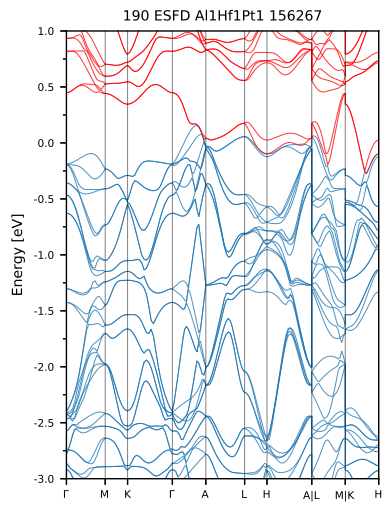

(y) 190.4.15

Figure 436: Band structures for topological materials in SG 190, part 2/2, from 190.4.7 to 190.4.15. See the list of figures.

## 0.0.96 SG 191

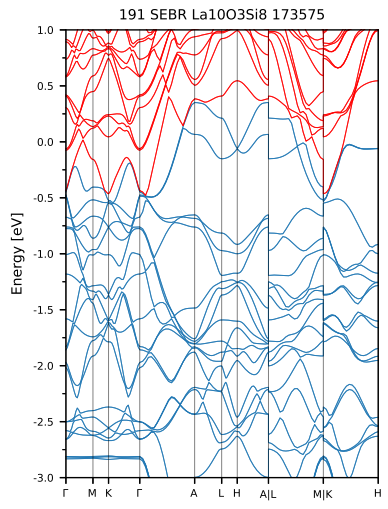

(a) 191.1.1

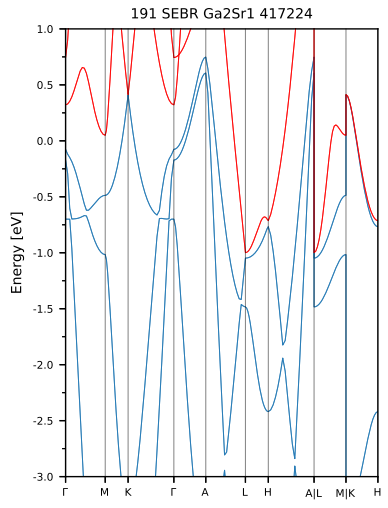

(b) 191.1.2

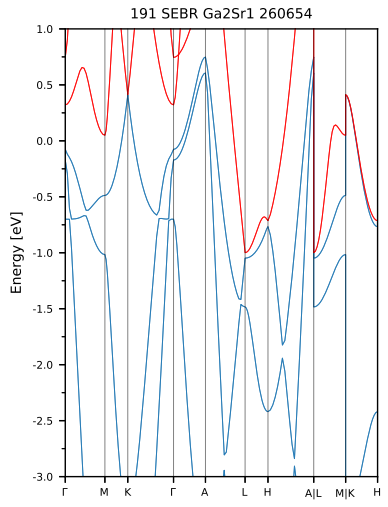

(c) 191.1.3

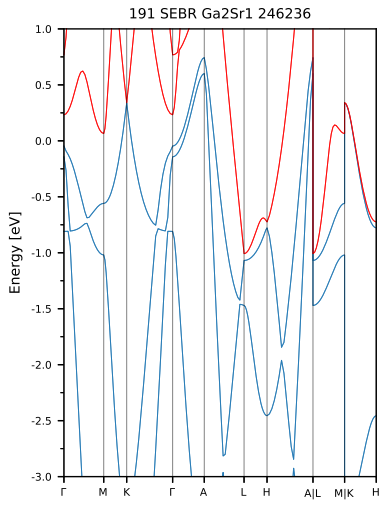

(d) 191.1.4

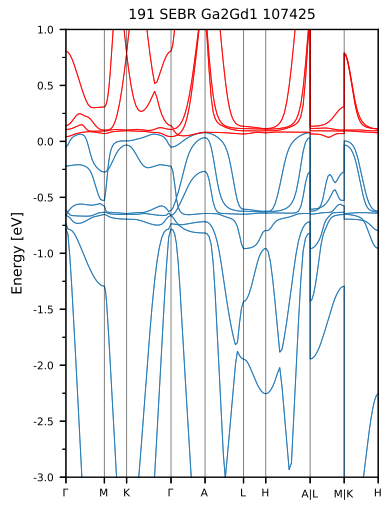

(e) 191.1.5

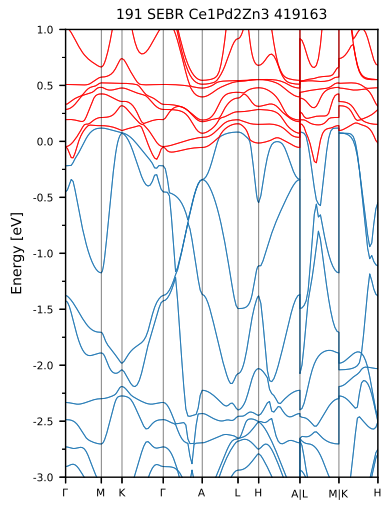

(f) 191.1.6

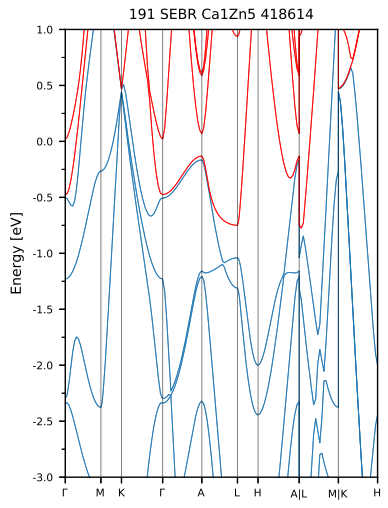

(g) 191.1.7

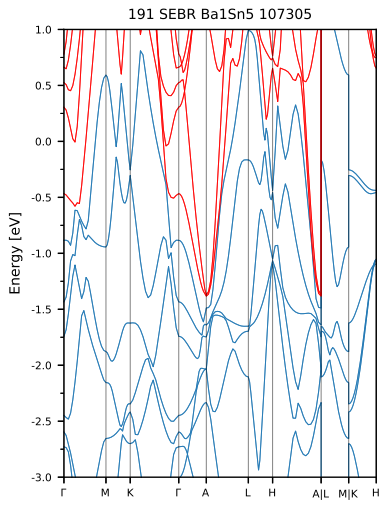

(h) 191.1.8

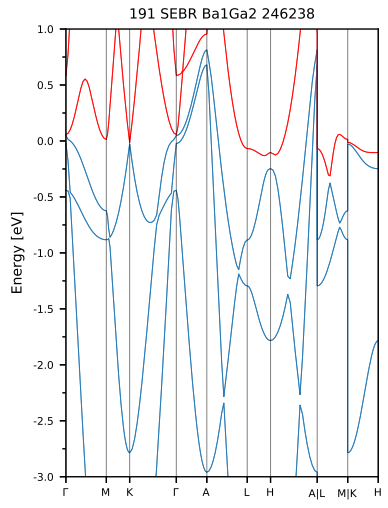

(i) 191.1.9

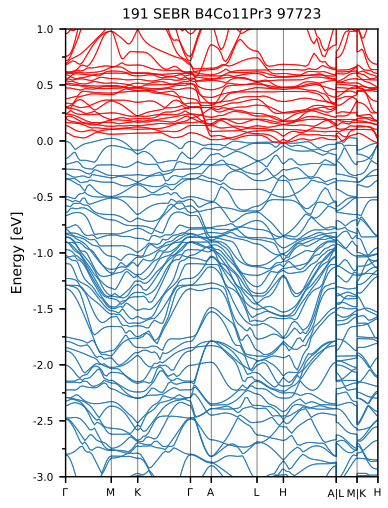

(j) 191.1.10

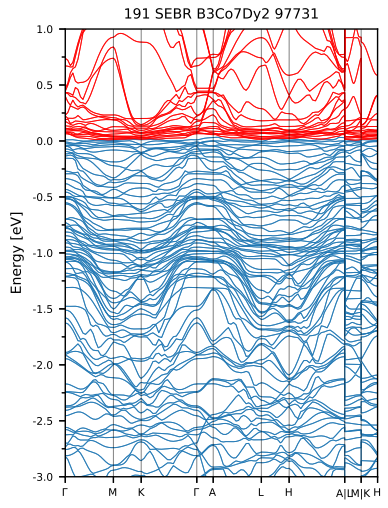

(k) 191.1.11

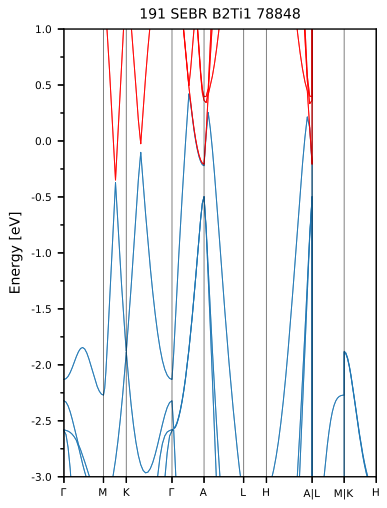

(l) 191.1.12

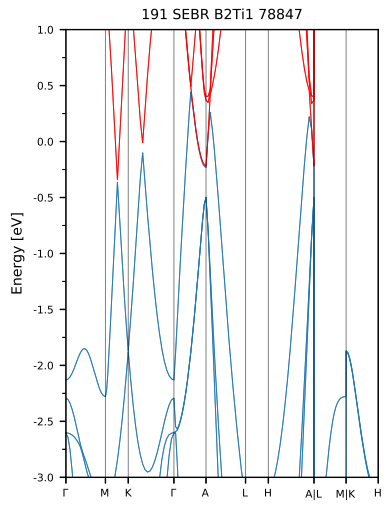

(m) 191.1.13

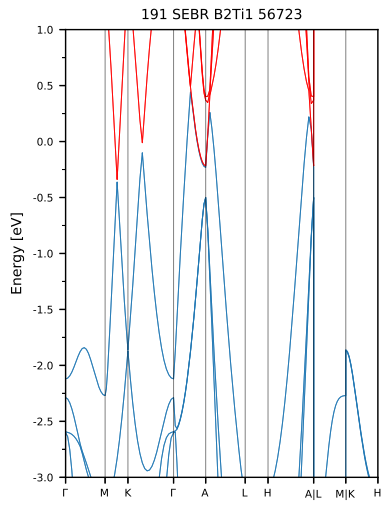

(n) 191.1.14

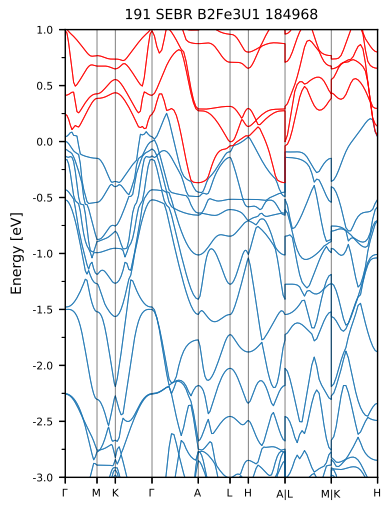

(o) 191.1.15

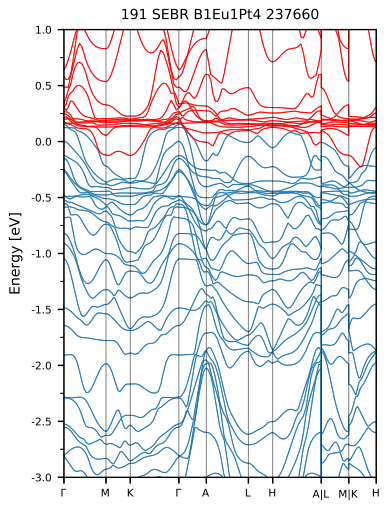

(p) 191.1.16

Figure 437: Band structures for topological materials in SG 191, part 1/4, from 191.1.1 to 191.1.16. See the list of figures.

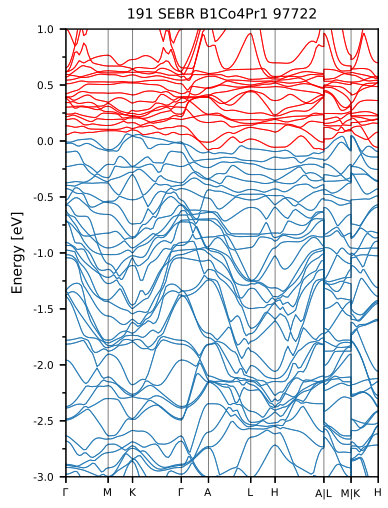

(q) 191.1.17

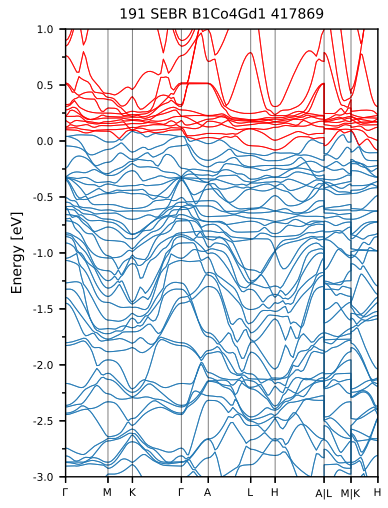

(r) 191.1.18

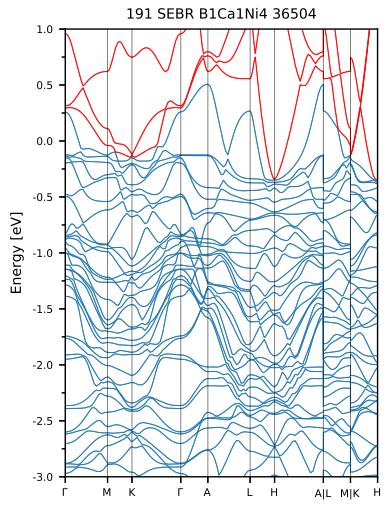

(s) 191.1.19

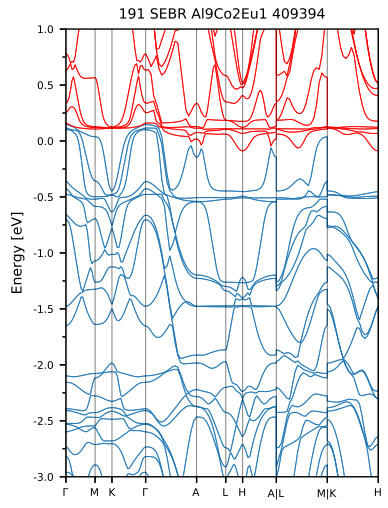

(t) 191.1.20

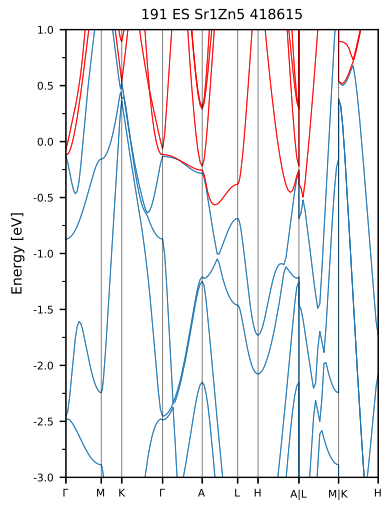

(u) 191.3.1

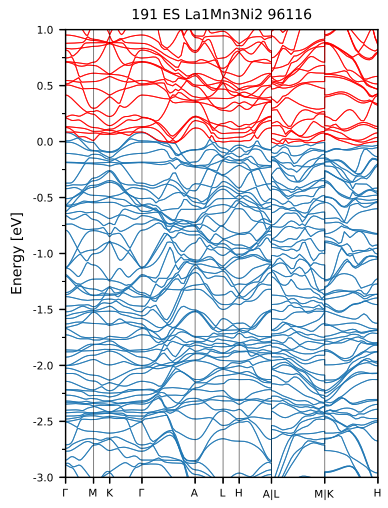

(v) 191.3.2

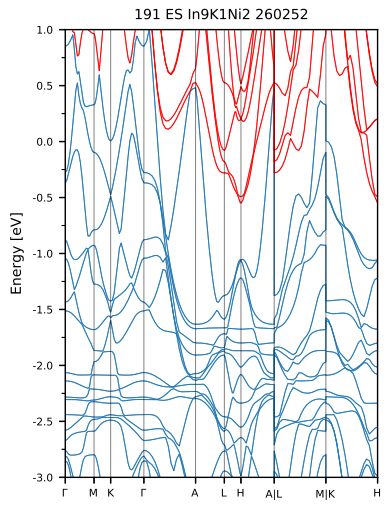

(w) 191.3.3

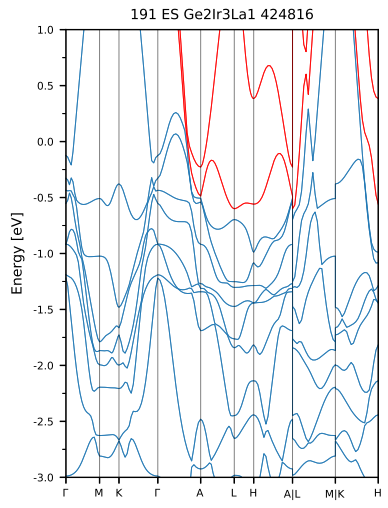

(x) 191.3.4

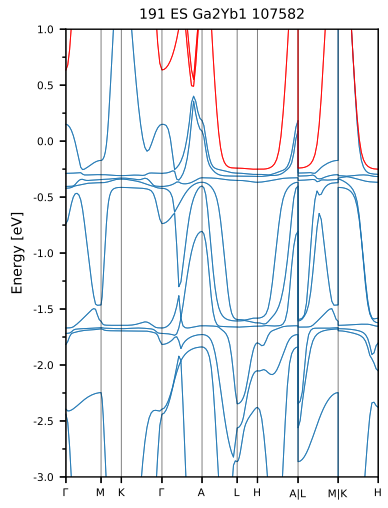

(y) 191.3.5

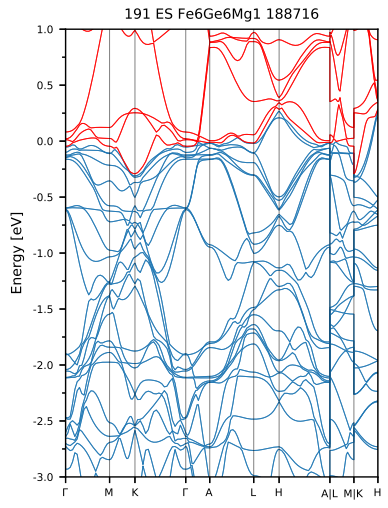

(z) 191.3.6

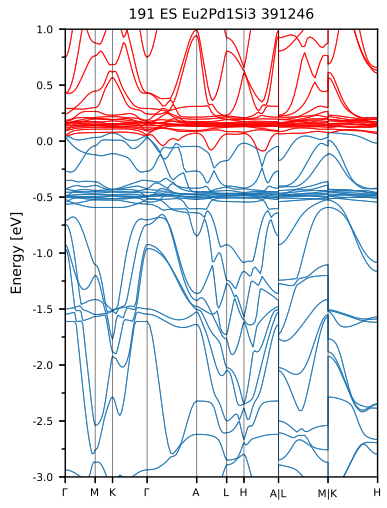

(aa) 191.3.7

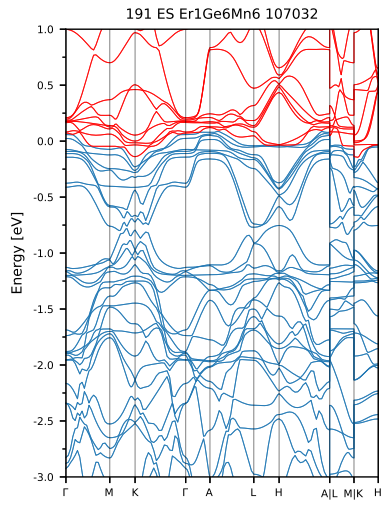

(ab) 191.3.8

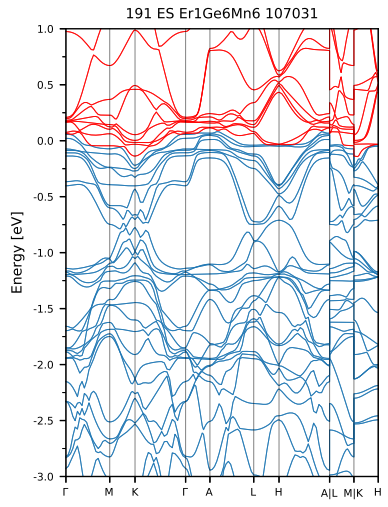

(ac) 191.3.9

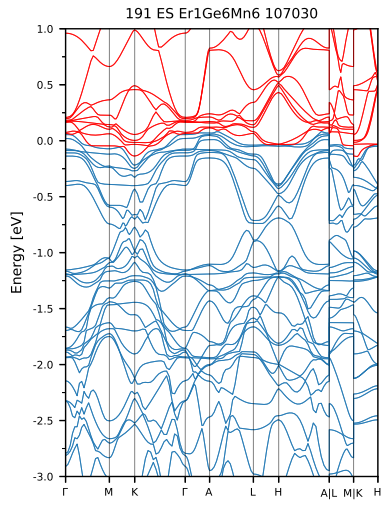

(ad) 191.3.10

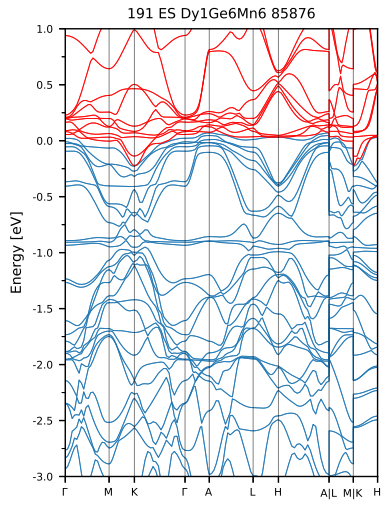

(ae) 191.3.11

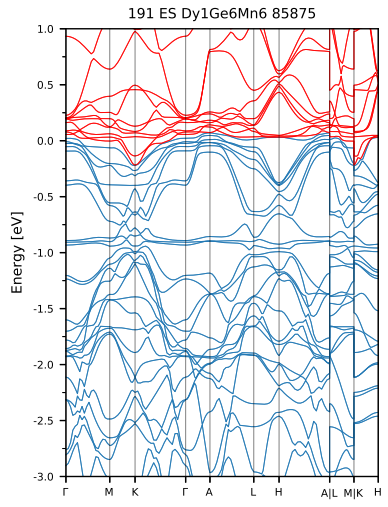

(af) 191.3.12

Figure 438: Band structures for topological materials in SG 191, part 2/4, from 191.1.17 to 191.3.12. See the list of figures.

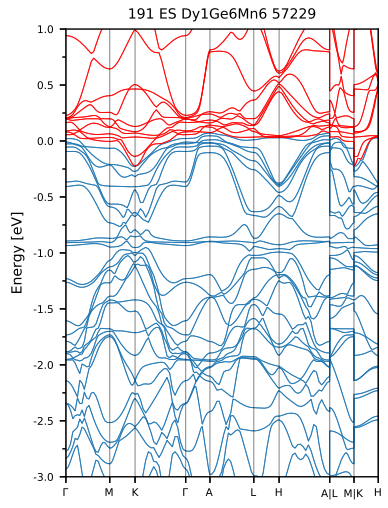

(ag) 191.3.13

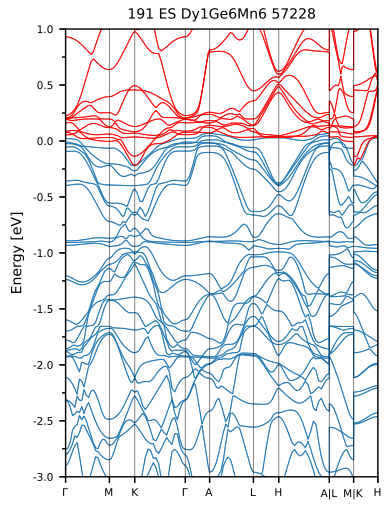

(ah) 191.3.14

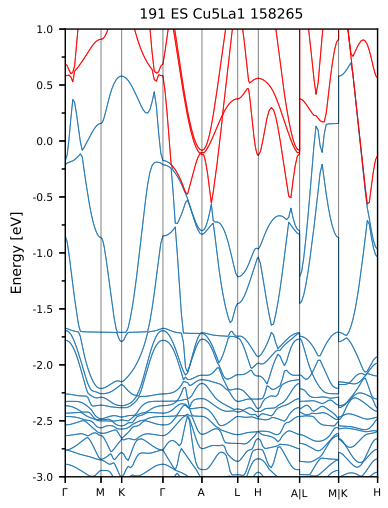

(ai) 191.3.15

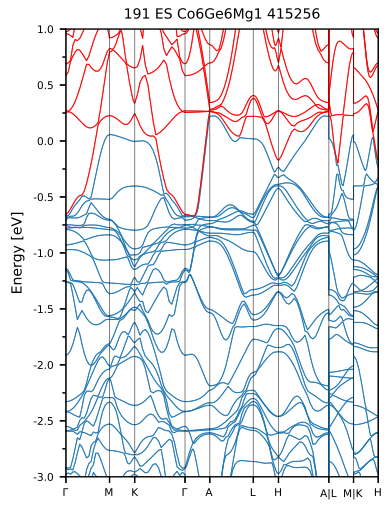

(aj) 191.3.16

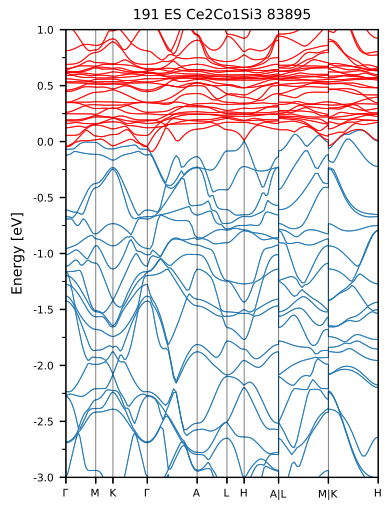

(ak) 191.3.17

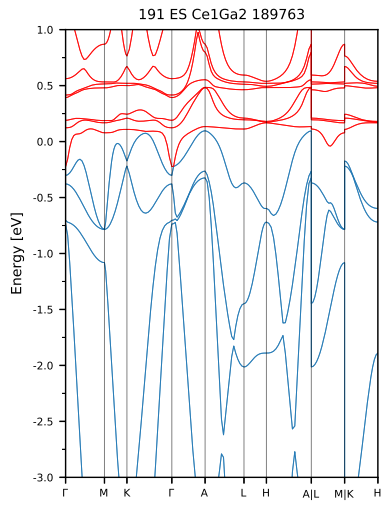

(al) 191.3.18

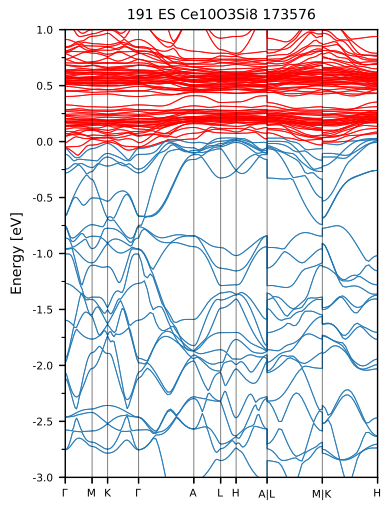

(am) 191.3.19

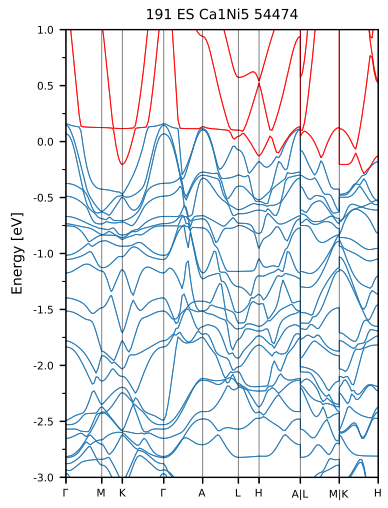

(an) 191.3.20

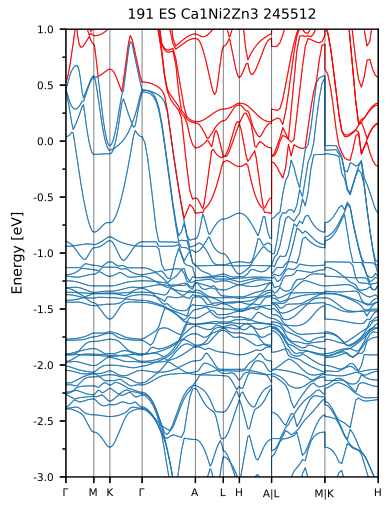

(ao) 191.3.21

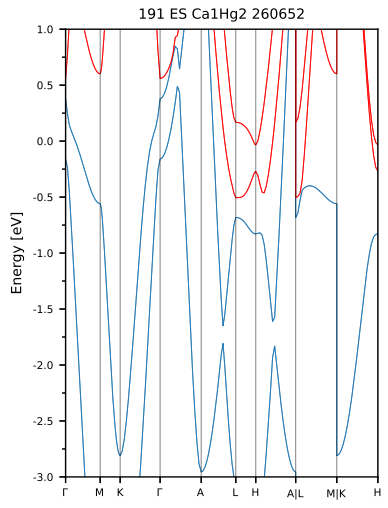

(ap) 191.3.22

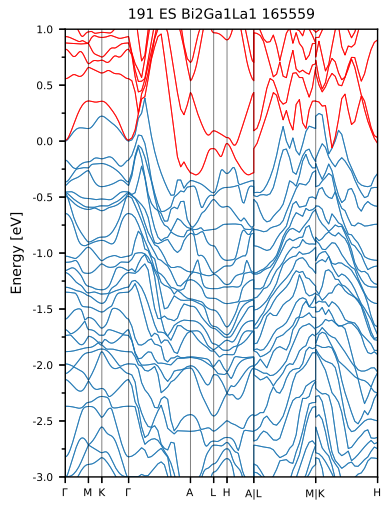

(aq) 191.3.23

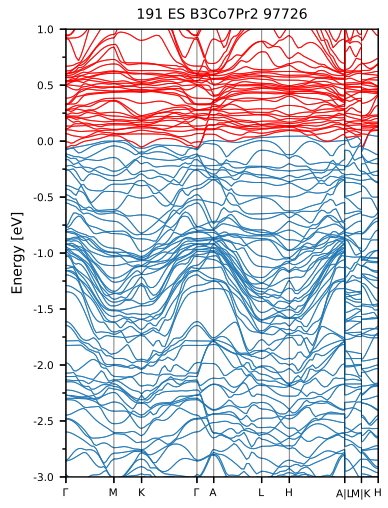

(ar) 191.3.24

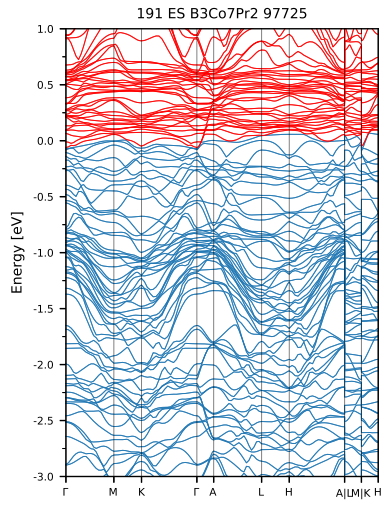

(as) 191.3.25

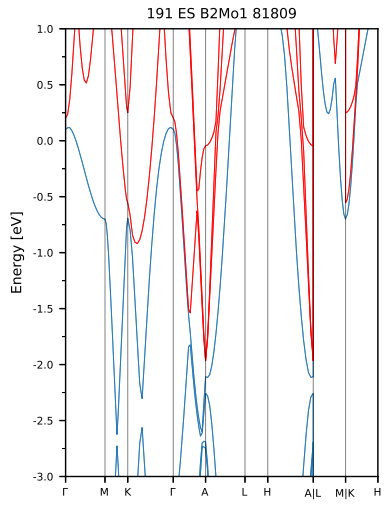

(at) 191.3.26

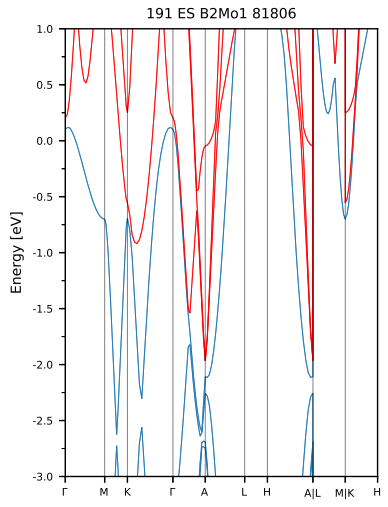

(au) 191.3.27

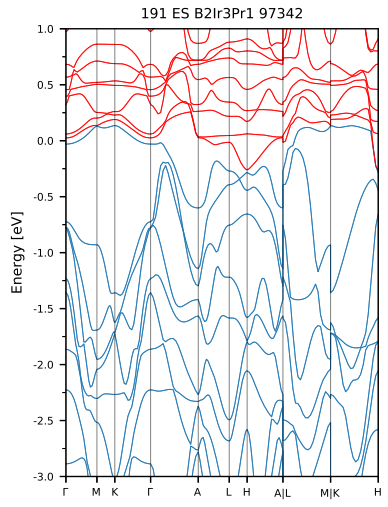

(av) 191.3.28

Figure 439: Band structures for topological materials in SG 191, part 3/4, from 191.3.13 to 191.3.28. See the list of figures.

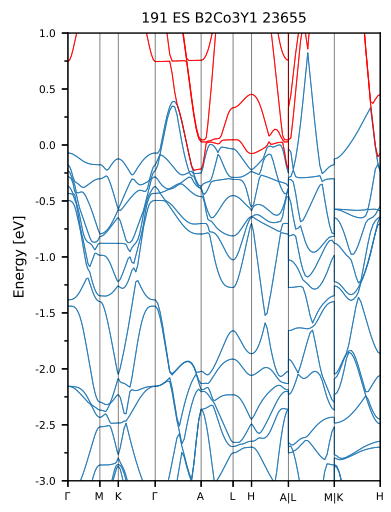

(aw) 191.3.29

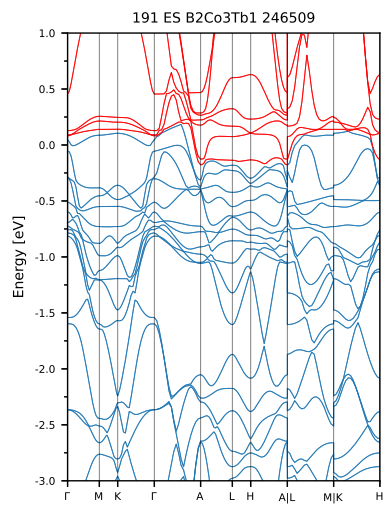

(ax) 191.3.30

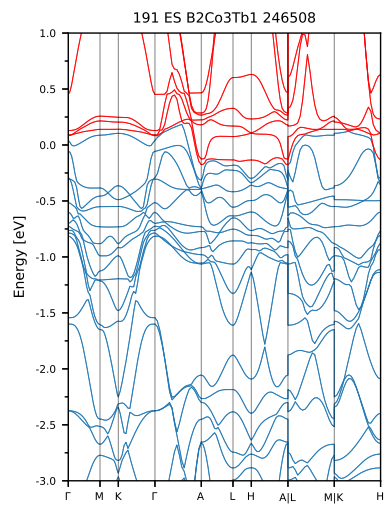

(ay) 191.3.31

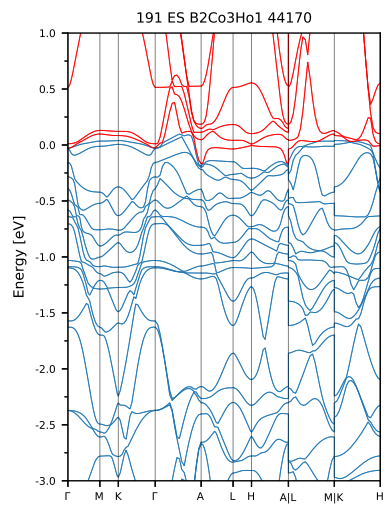

(az) 191.3.32

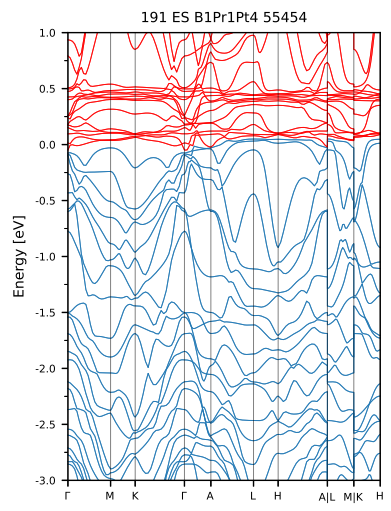

(ba) 191.3.33

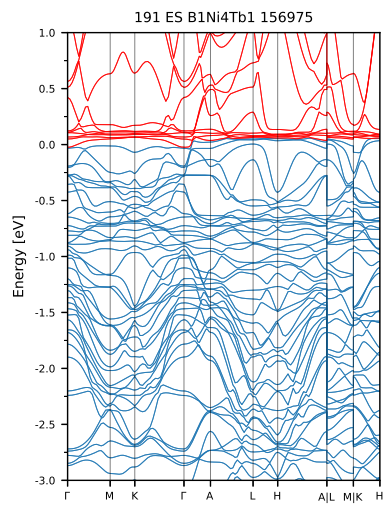

(bb) 191.3.34

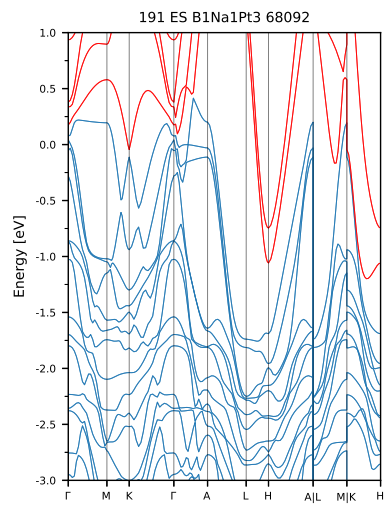

(bc) 191.3.35

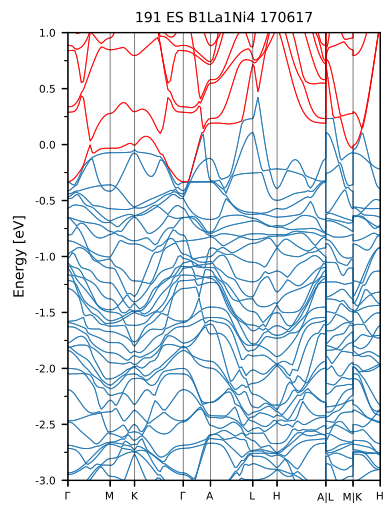

(bd) 191.3.36

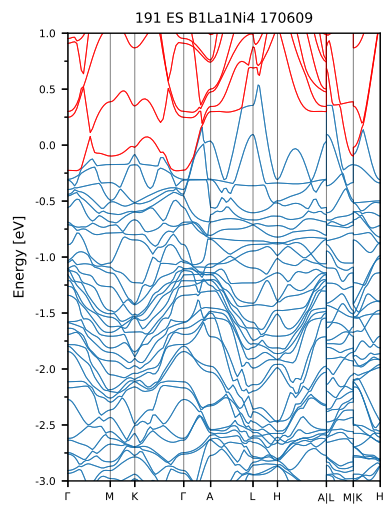

(be) 191.3.37

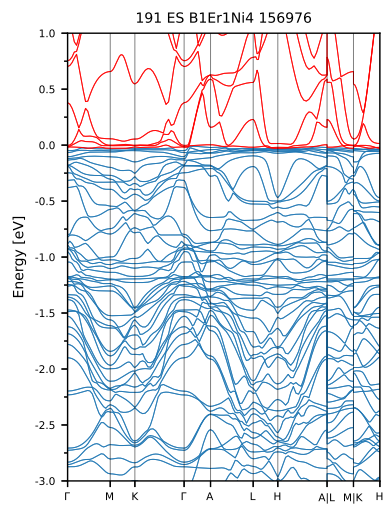

(bf) 191.3.38

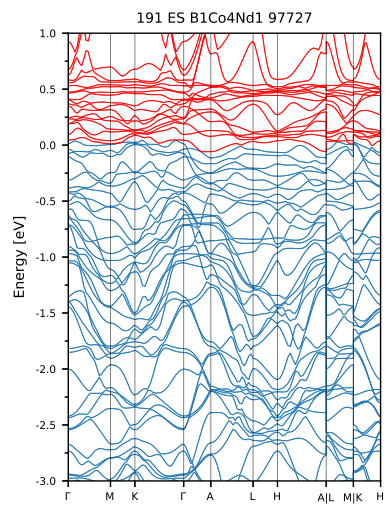

(bg) 191.3.39

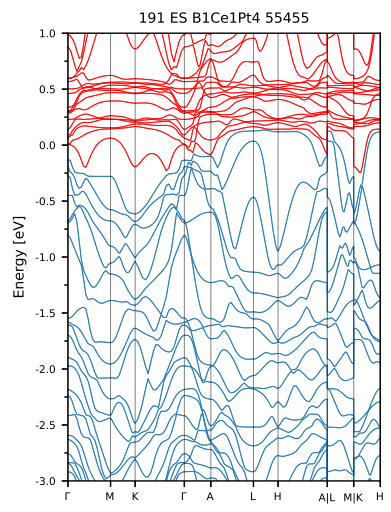

(bh) 191.3.40

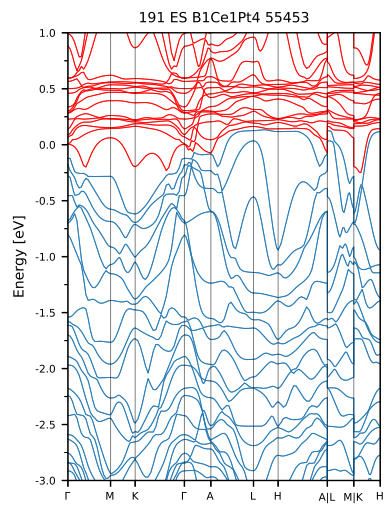

(bi) 191.3.41

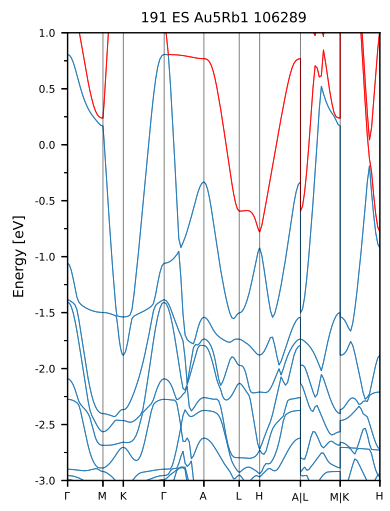

(bj) 191.3.42

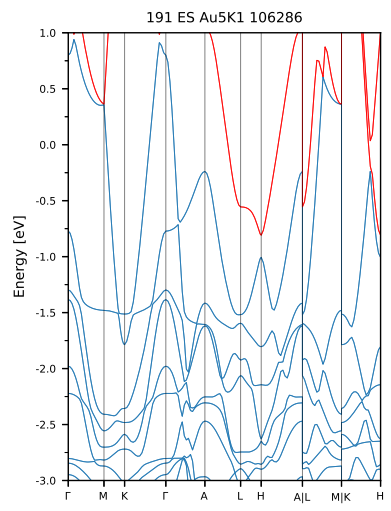

(bk) 191.3.43

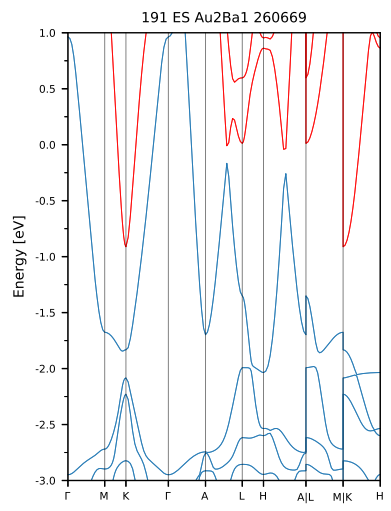

(bl) 191.3.44

Figure 440: Band structures for topological materials in SG 191, part 4/4, from 191.3.29 to 191.3.44. See the list of figures.

## 0.0.97 SG 193

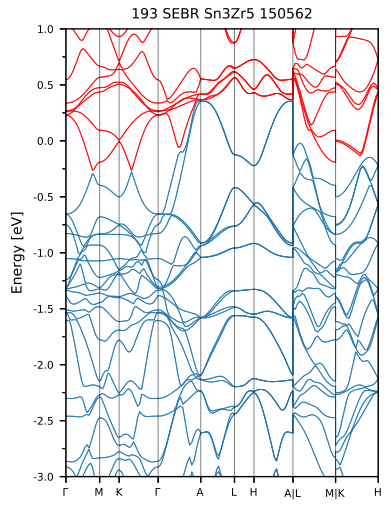

(a) 193.1.1

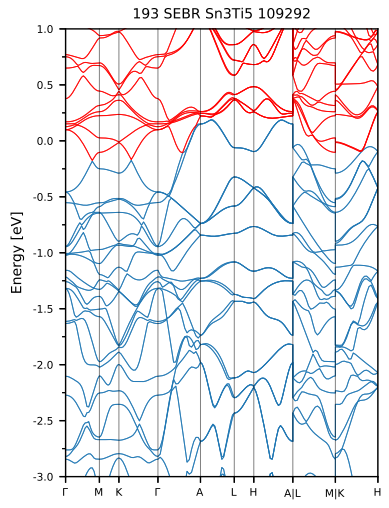

(b) 193.1.2

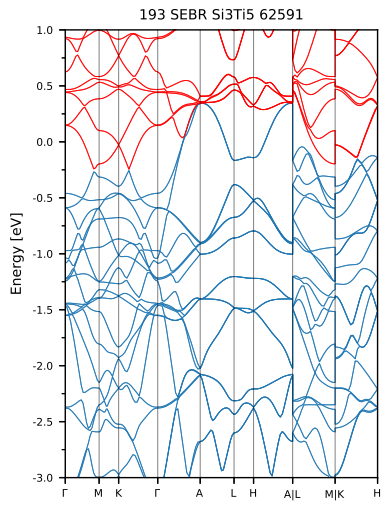

(c) 193.1.3

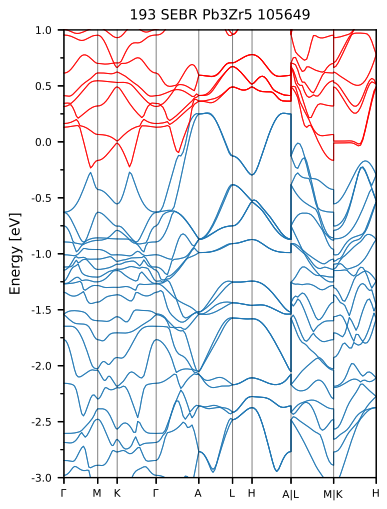

(d) 193.1.4

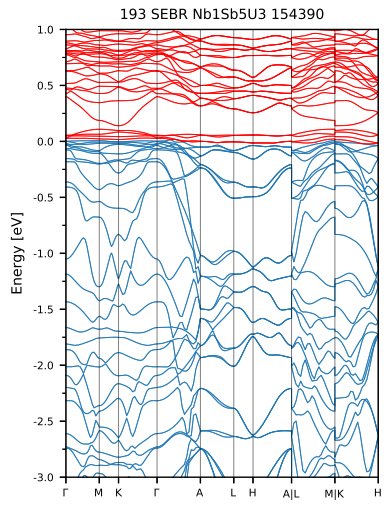

(e) 193.1.5

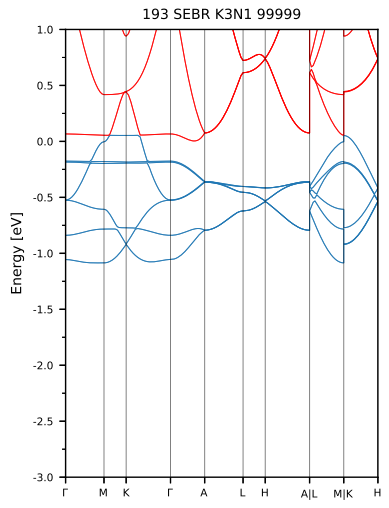

(f) 193.1.6

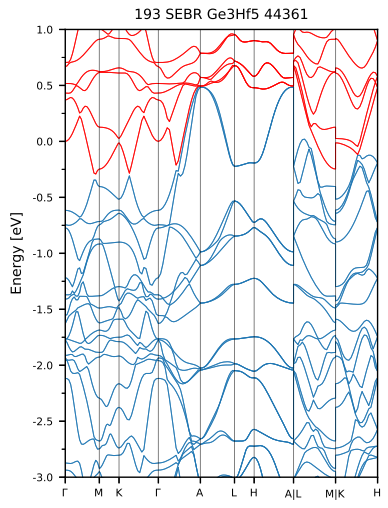

(g) 193.1.7

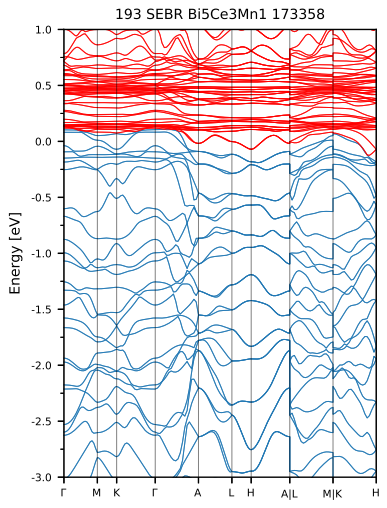

(h) 193.1.8

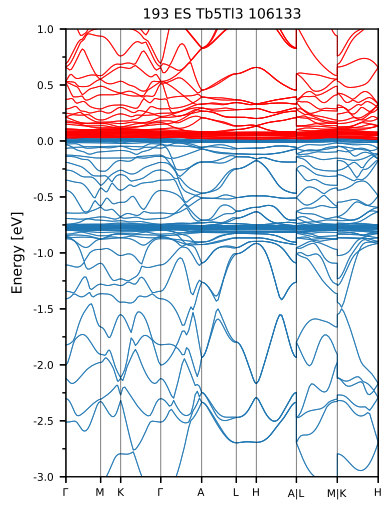

(i) 193.3.1

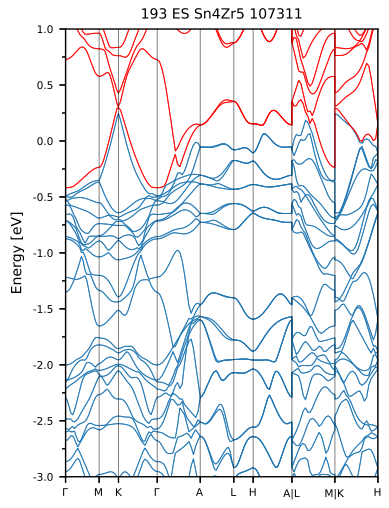

(j) 193.3.2

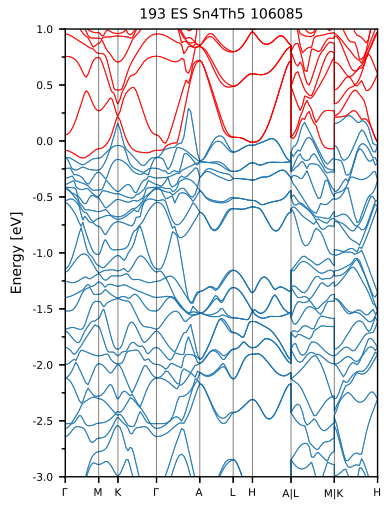

(k) 193.3.3

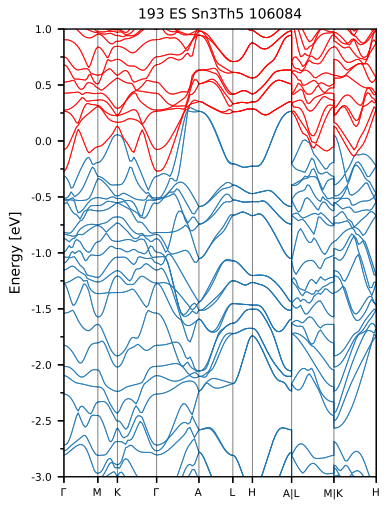

(l) 193.3.4

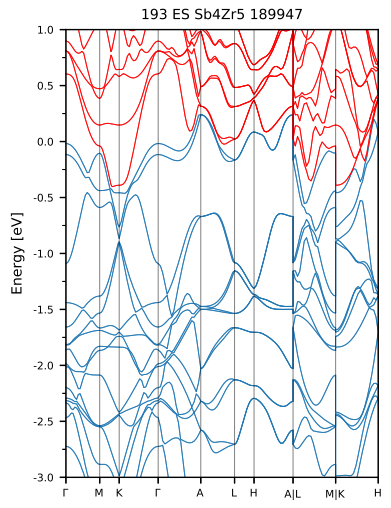

(m) 193.3.5

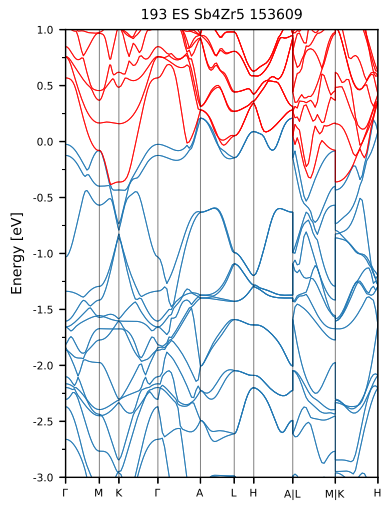

(n) 193.3.6

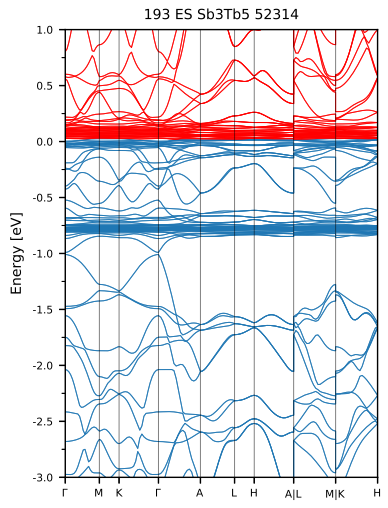

(o) 193.3.7

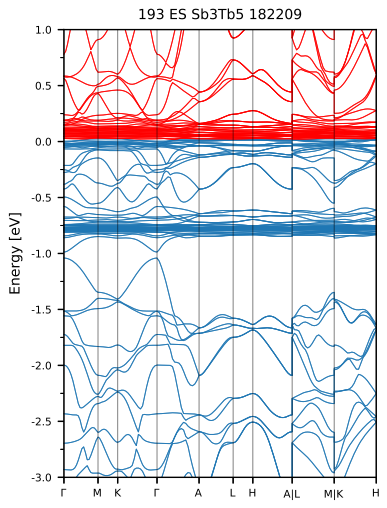

(p) 193.3.8

Figure 441: Band structures for topological materials in SG 193, part 1/11, from 193.1.1 to 193.3.8. See the list of figures.

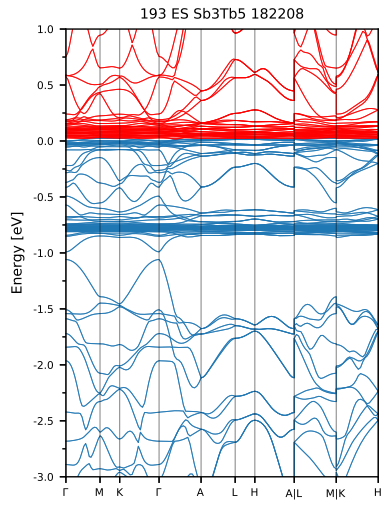

(q) 193.3.9

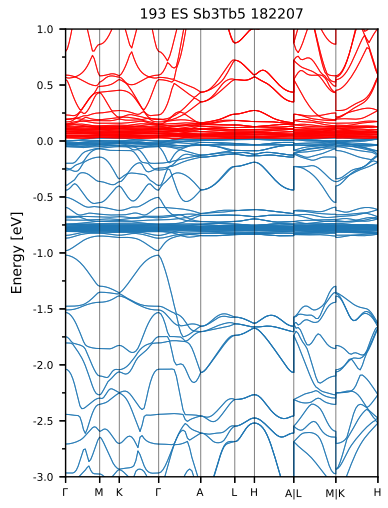

(r) 193.3.10

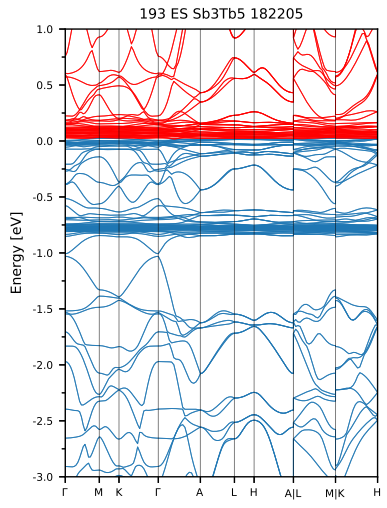

(s) 193.3.11

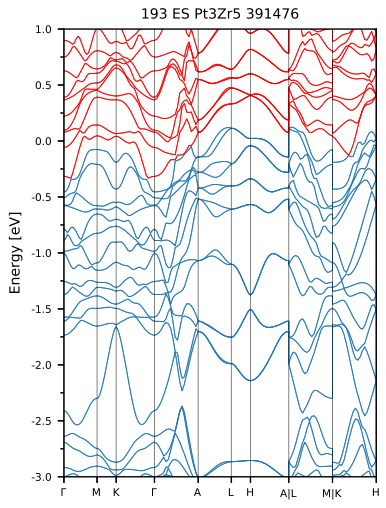

(t) 193.3.12

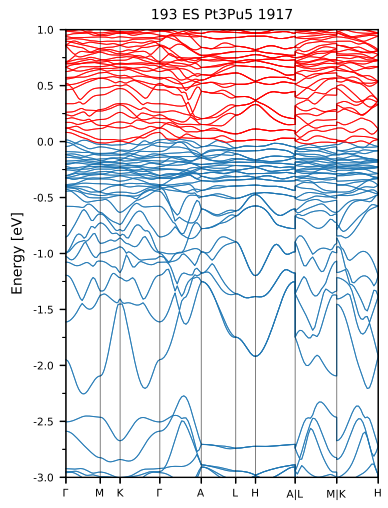

(u) 193.3.13

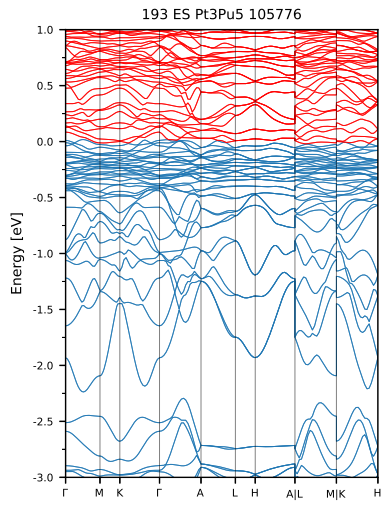

(v) 193.3.14

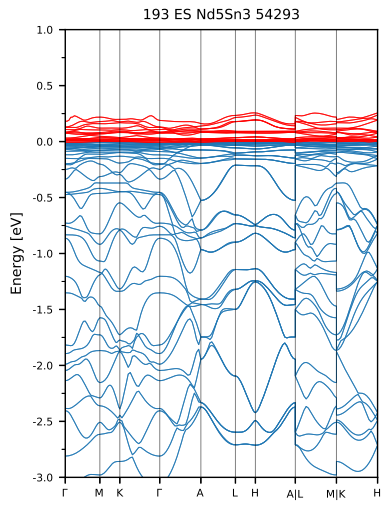

(w) 193.3.15

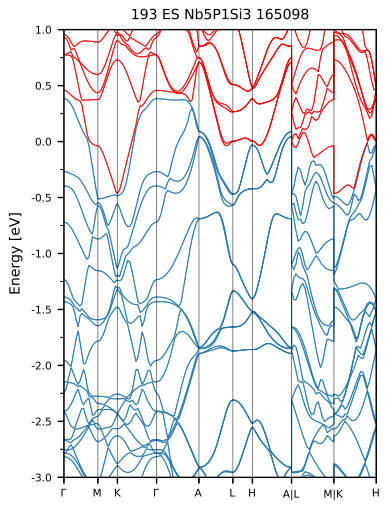

(x) 193.3.16

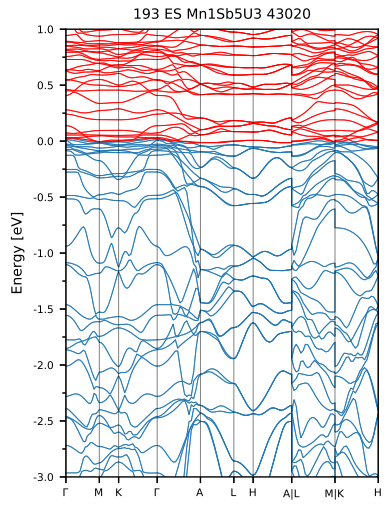

(y) 193.3.17

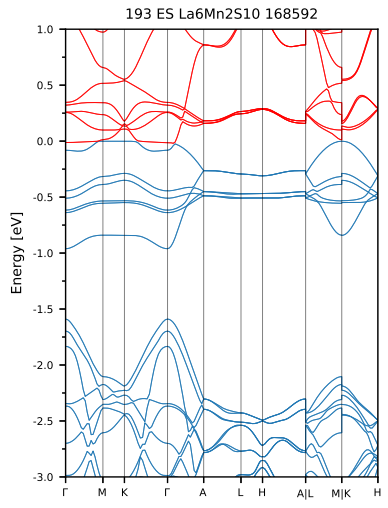

(z) 193.3.18

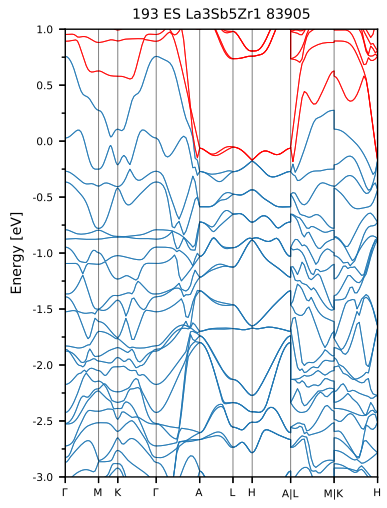

(aa) 193.3.19

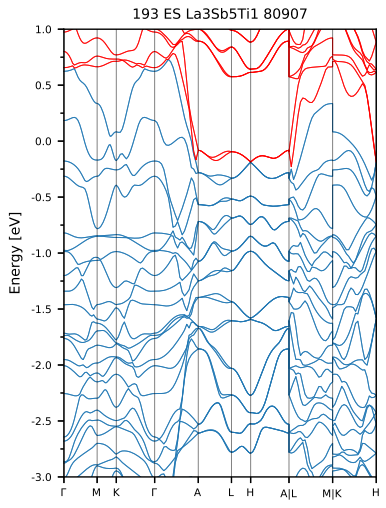

(ab) 193.3.20

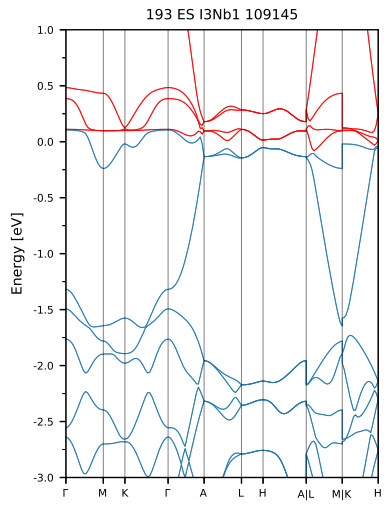

(ac) 193.3.21

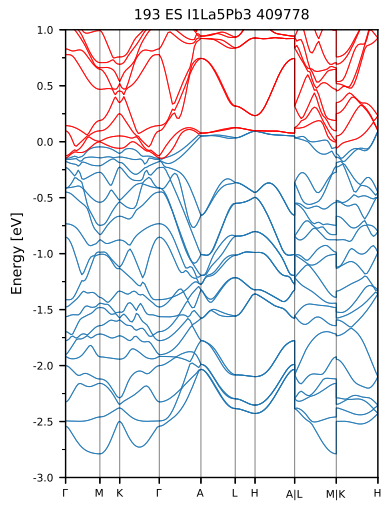

(ad) 193.3.22

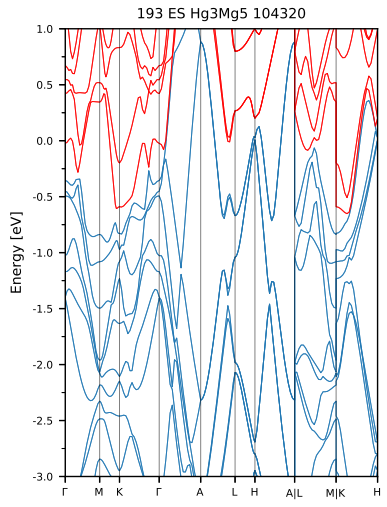

(ae) 193.3.23

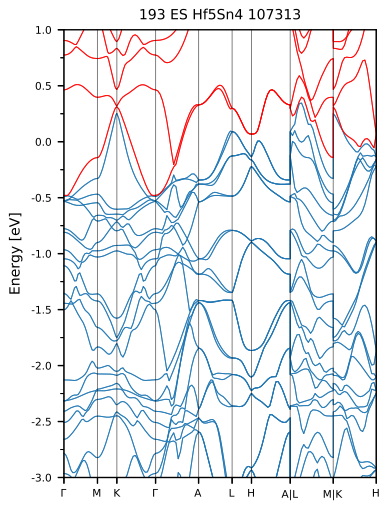

(af) 193.3.24

Figure 442: Band structures for topological materials in SG 193, part 2/11, from 193.3.9 to 193.3.24. See the list of figures.

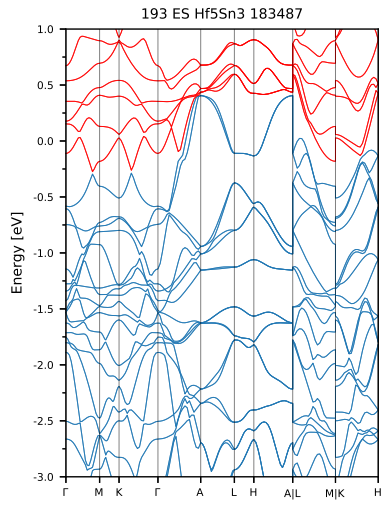

(ag) 193.3.25

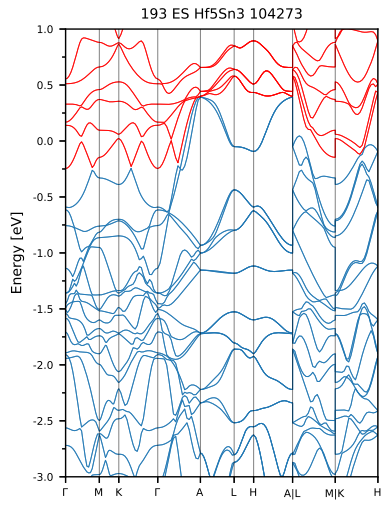

(ah) 193.3.26

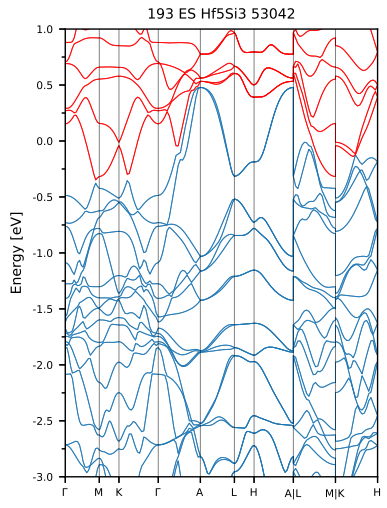

(ai) 193.3.27

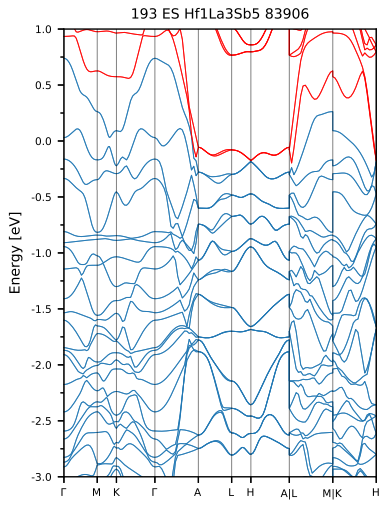

(aj) 193.3.28

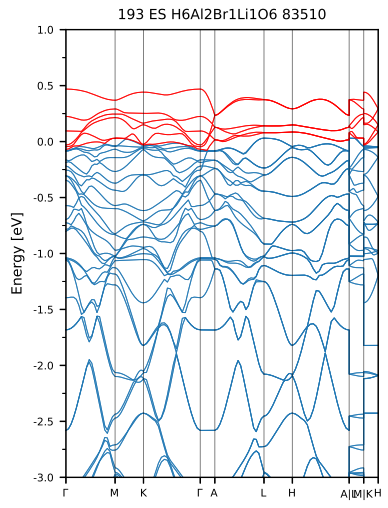

(ak) 193.3.29

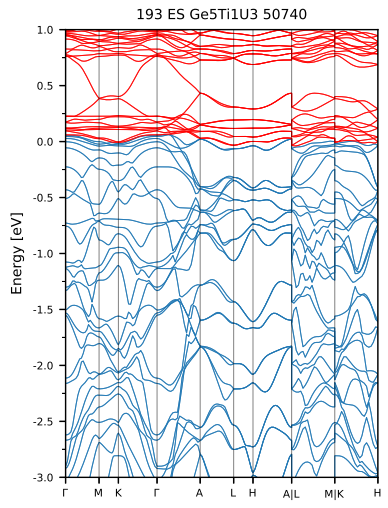

(al) 193.3.30

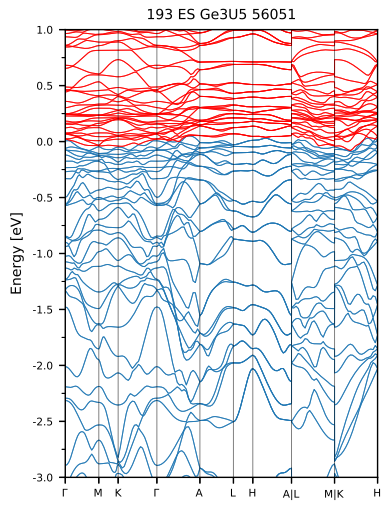

(am) 193.3.31

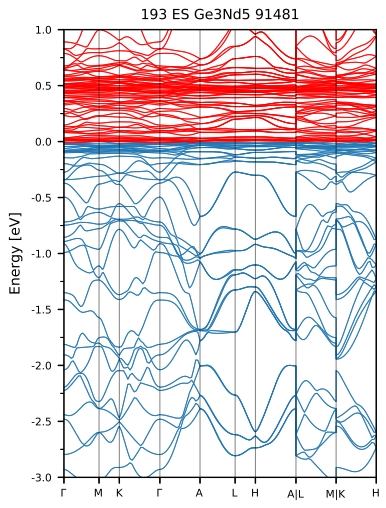

(an) 193.3.32

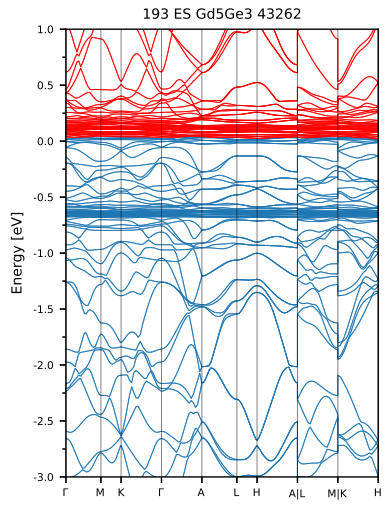

(ao) 193.3.33

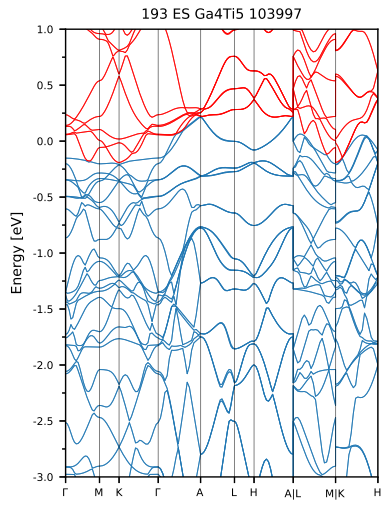

(ap) 193.3.34

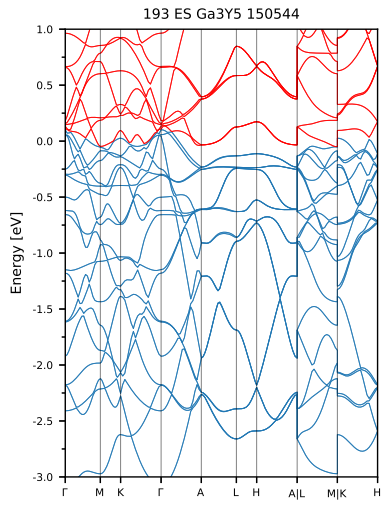

(aq) 193.3.35

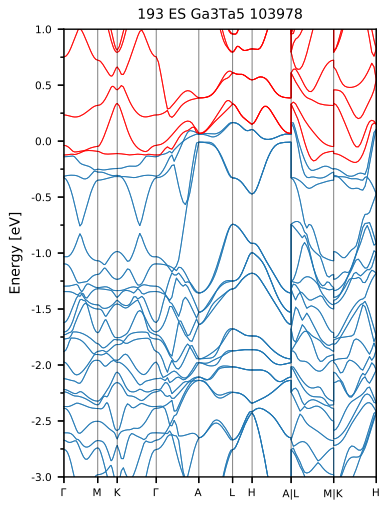

(ar) 193.3.36

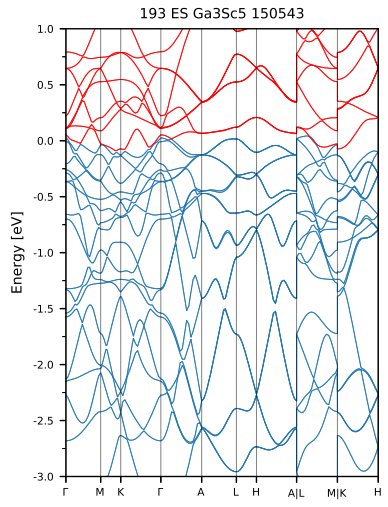

(as) 193.3.37

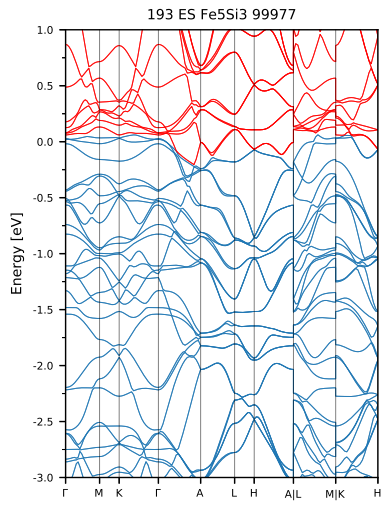

(at) 193.3.38

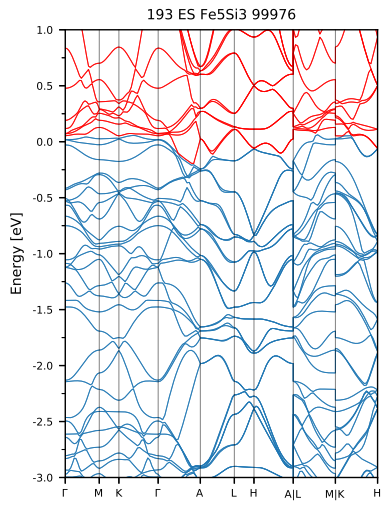

(au) 193.3.39

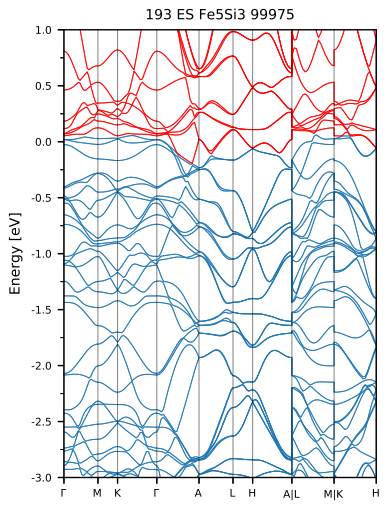

(av) 193.3.40

Figure 443: Band structures for topological materials in SG 193, part 3/11, from 193.3.25 to 193.3.40. See the list of figures.

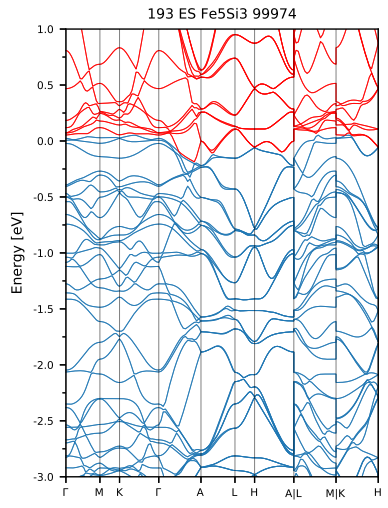

(aw) 193.3.41

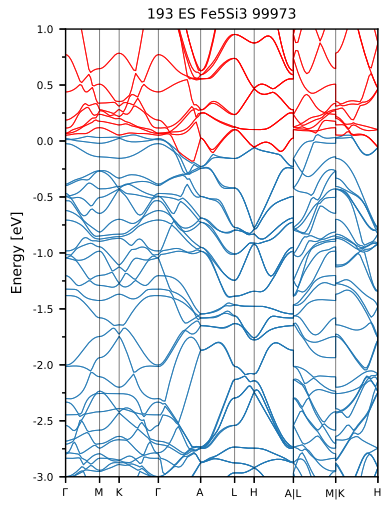

(ax) 193.3.42

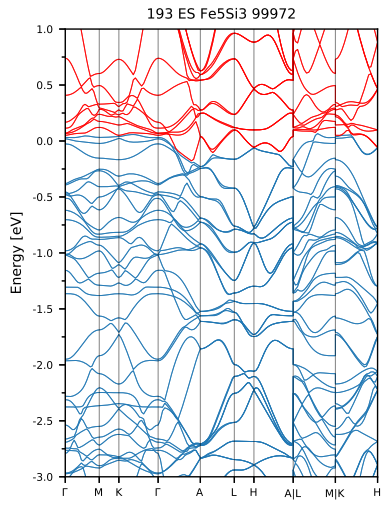

(ay) 193.3.43

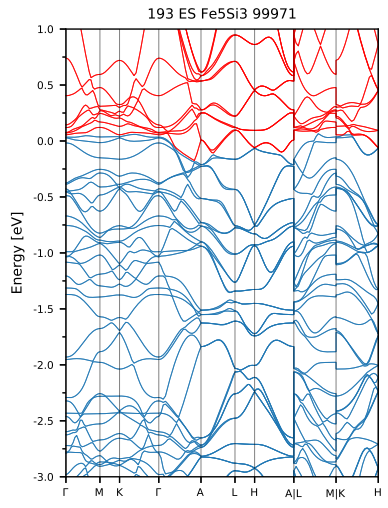

(az) 193.3.44

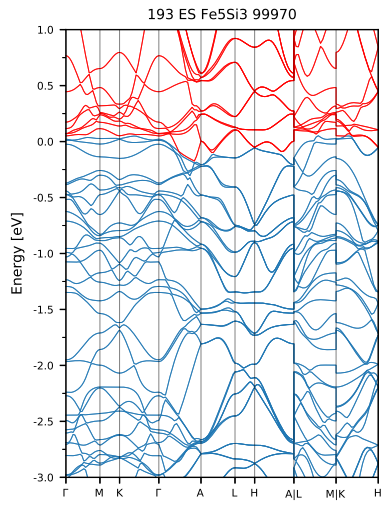

(ba) 193.3.45

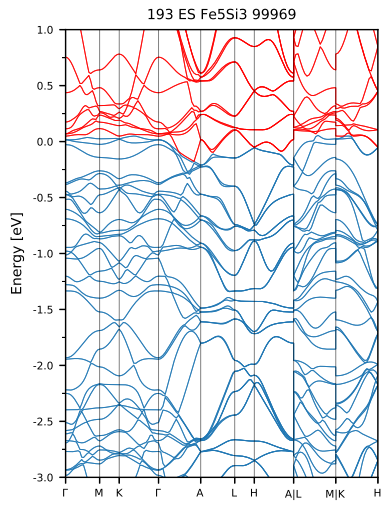

(bb) 193.3.46

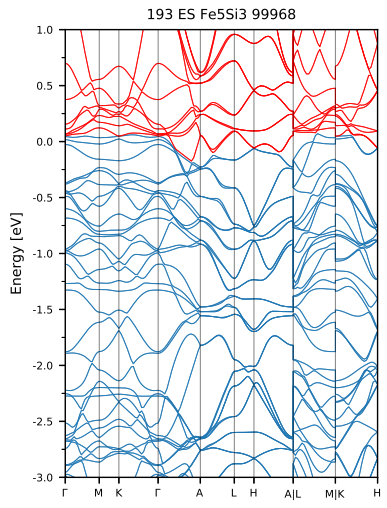

(bc) 193.3.47

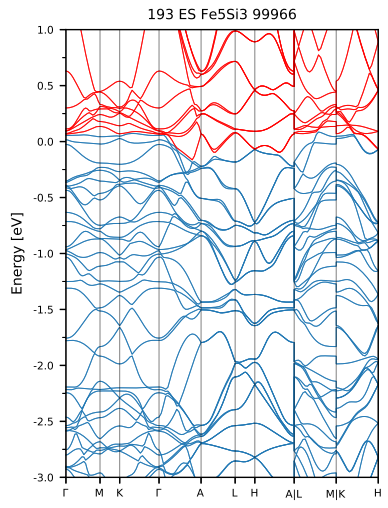

(bd) 193.3.48

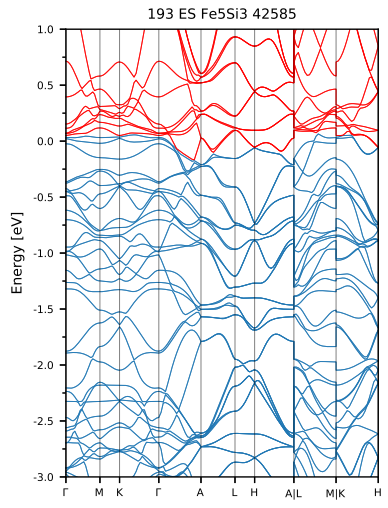

(be) 193.3.49

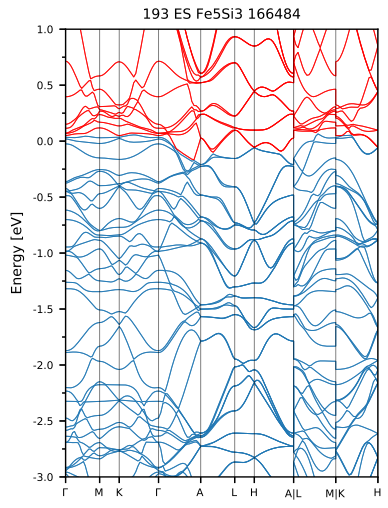

(bf) 193.3.50

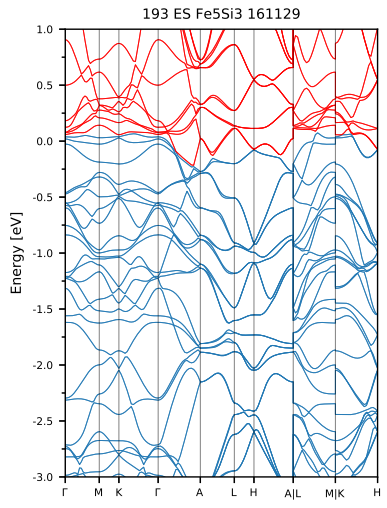

(bg) 193.3.51

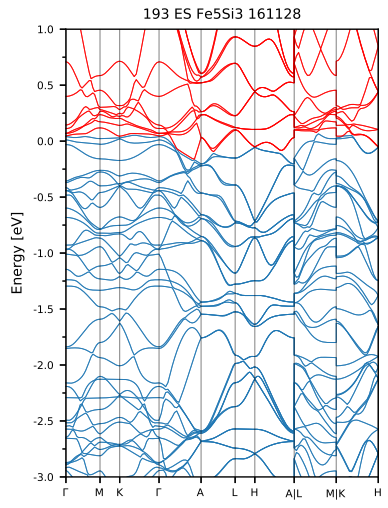

(bh) 193.3.52

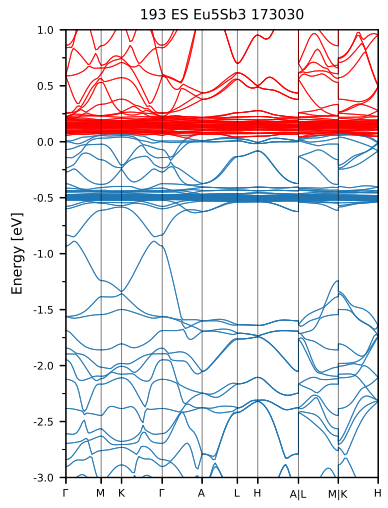

(bi) 193.3.53

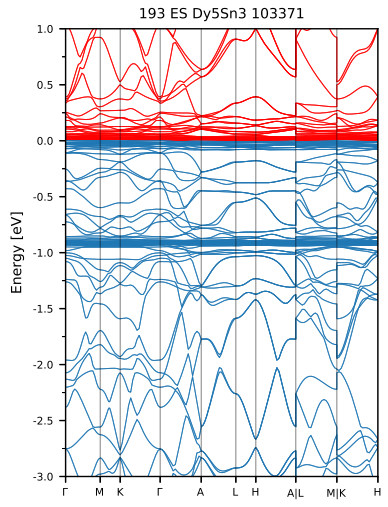

(bj) 193.3.54

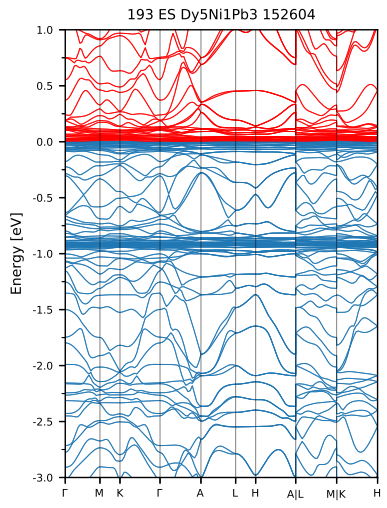

(bk) 193.3.55

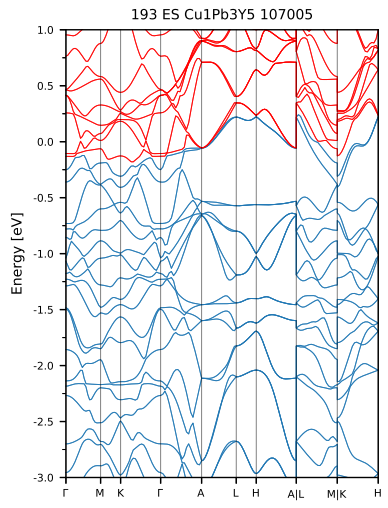

(bl) 193.3.56

Figure 444: Band structures for topological materials in SG 193, part 4/11, from 193.3.41 to 193.3.56. See the list of figures.

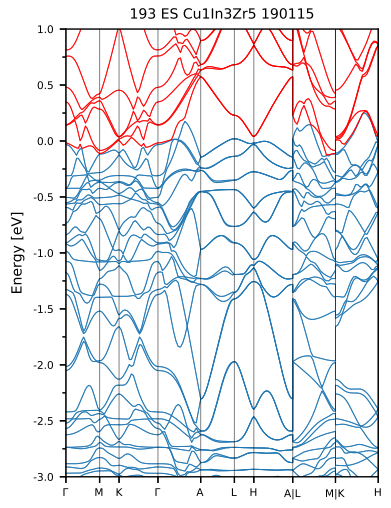

(bm) 193.3.57

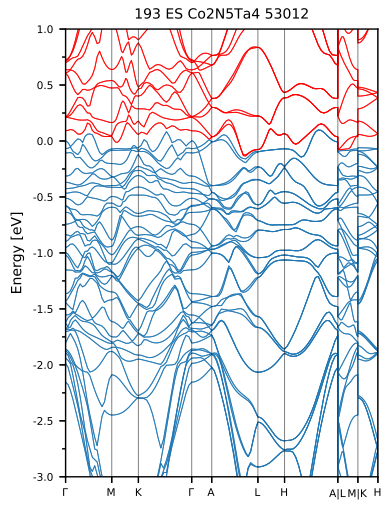

(bn) 193.3.58

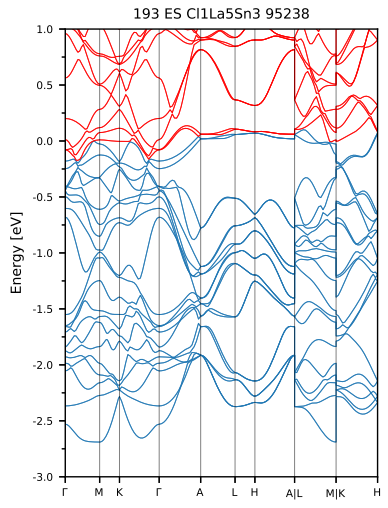

(bo) 193.3.59

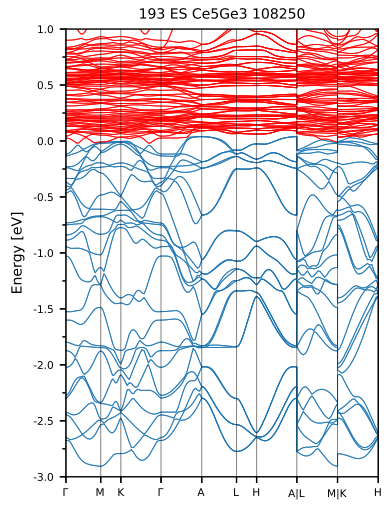

(bp) 193.3.60

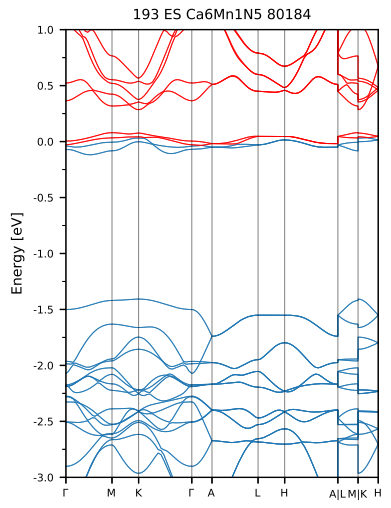

(bq) 193.3.61

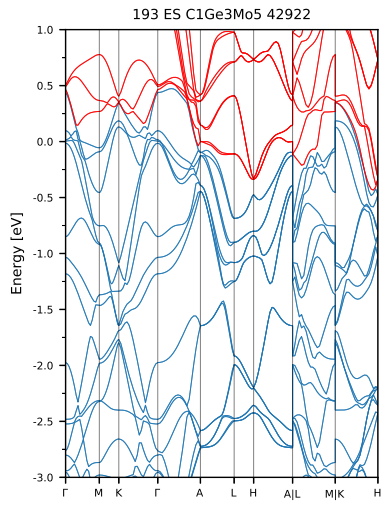

(br) 193.3.62

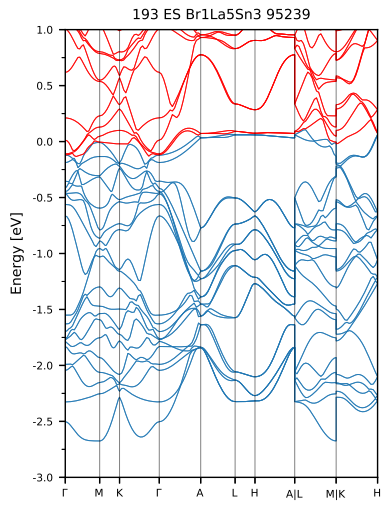

(bs) 193.3.63

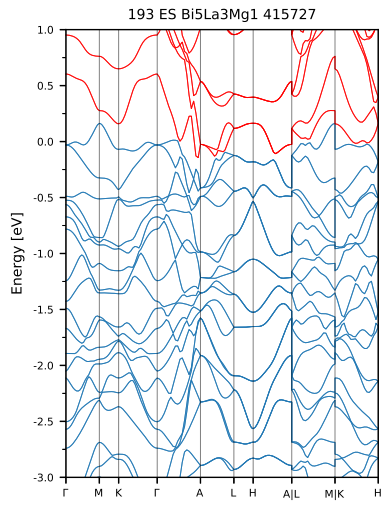

(bt) 193.3.64

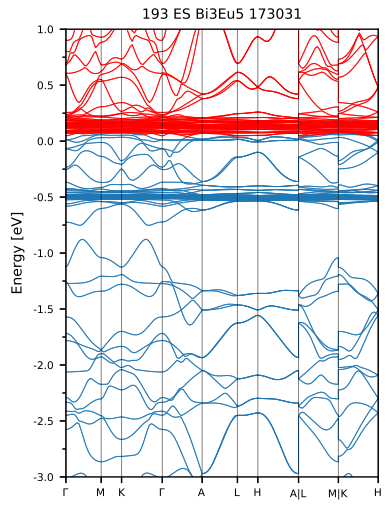

(bu) 193.3.65

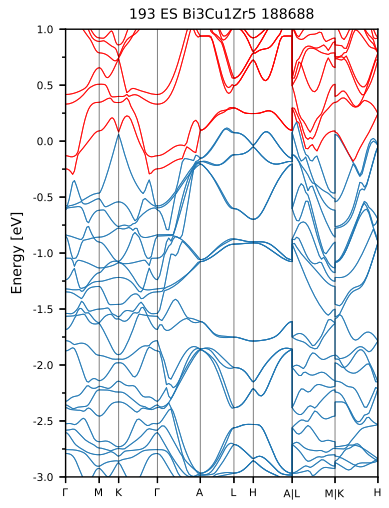

(bv) 193.3.66

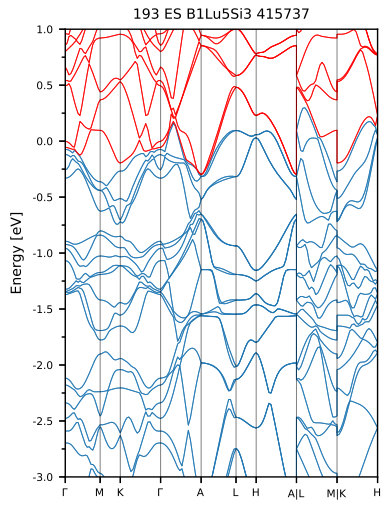

(bw) 193.3.67

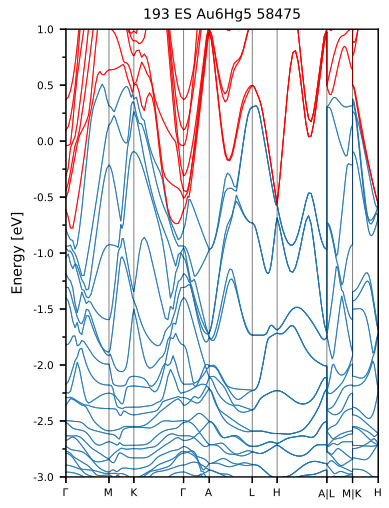

(bx) 193.3.68

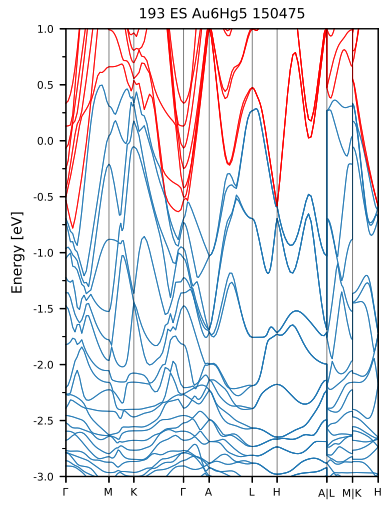

(by) 193.3.69

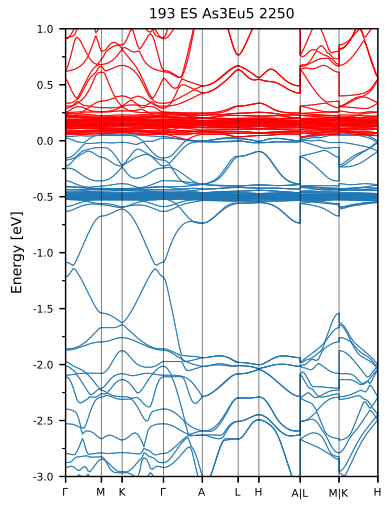

(bz) 193.3.70

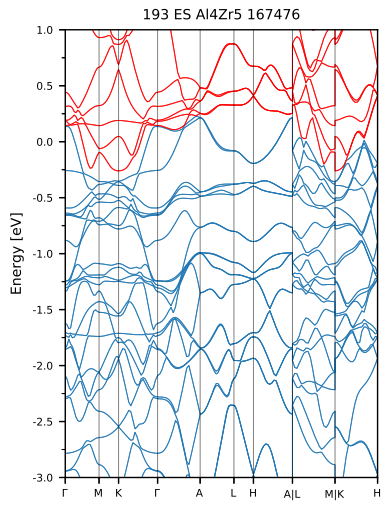

(ca) 193.3.71

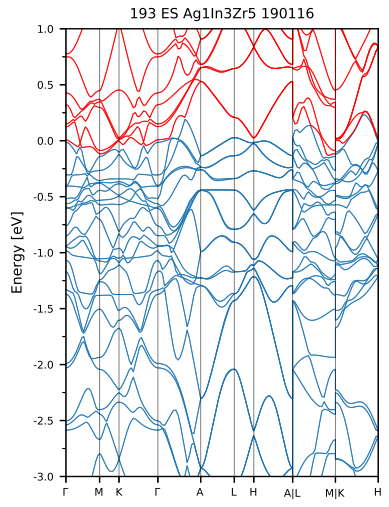

(cb) 193.3.72

Figure 445: Band structures for topological materials in SG 193, part 5/11, from 193.3.57 to 193.3.72. See the list of figures.

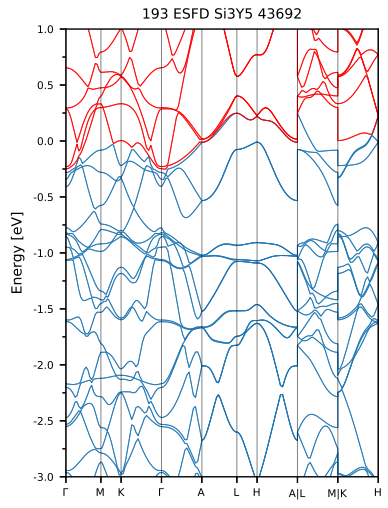

(cc) 193.4.1

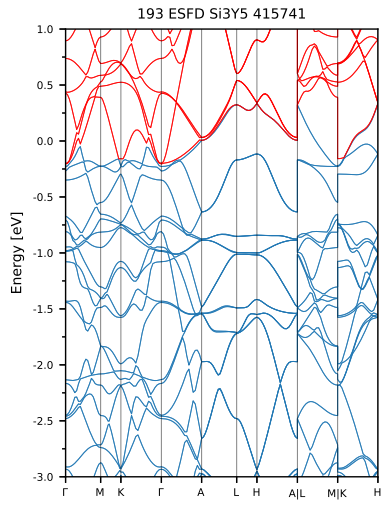

(cd) 193.4.2

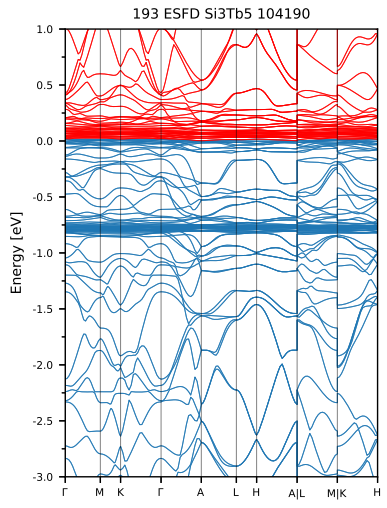

(ce) 193.4.3

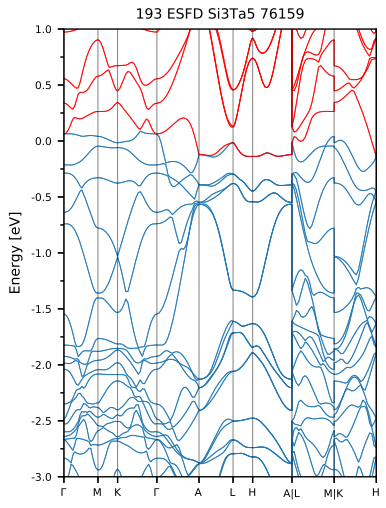

(cf) 193.4.4

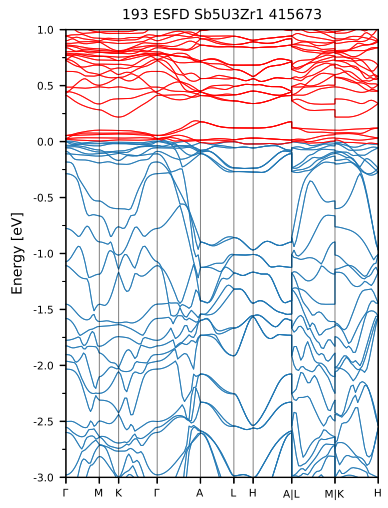

(cg) 193.4.5

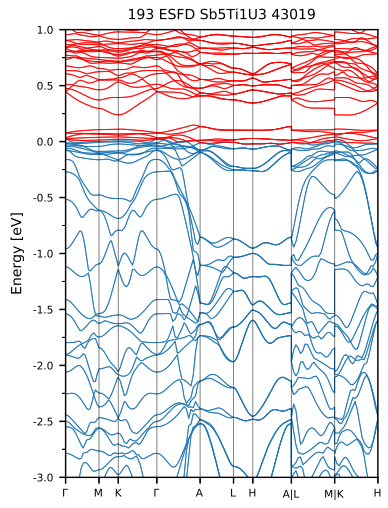

(ch) 193.4.6

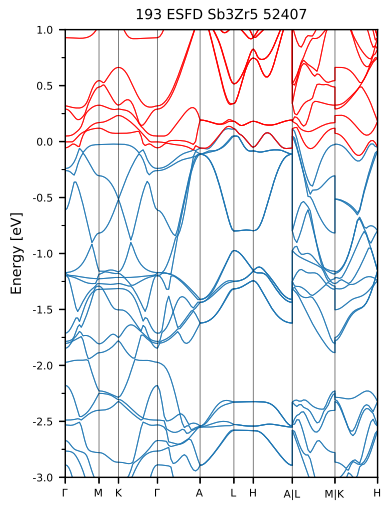

(ci) 193.4.7

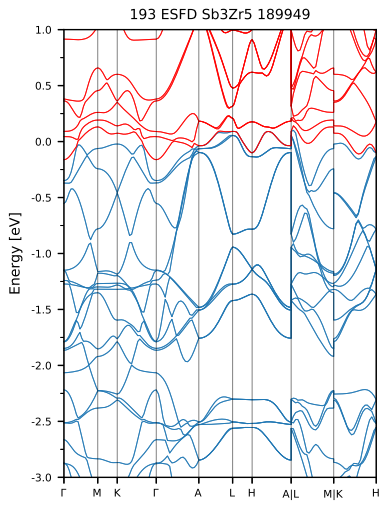

(cj) 193.4.8

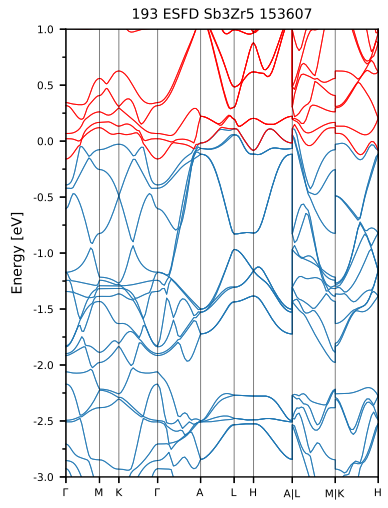

(ck) 193.4.9

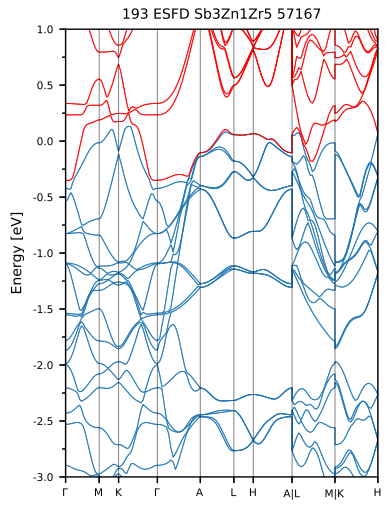

(cl) 193.4.10

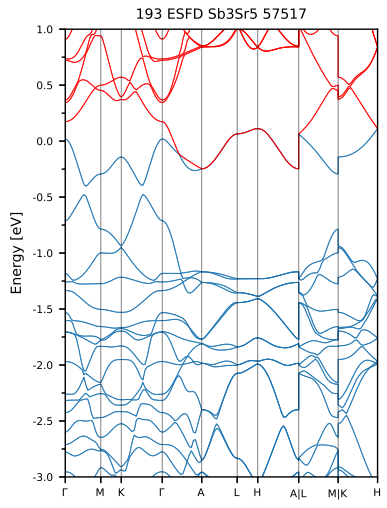

(cm) 193.4.11

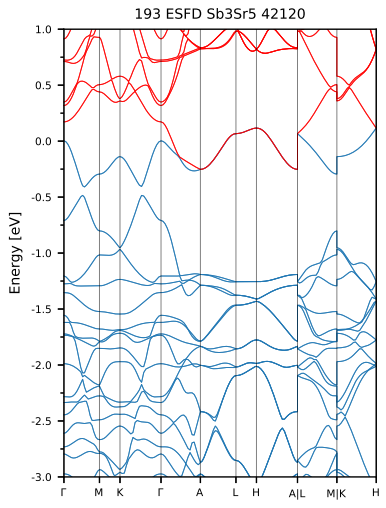

(cn) 193.4.12

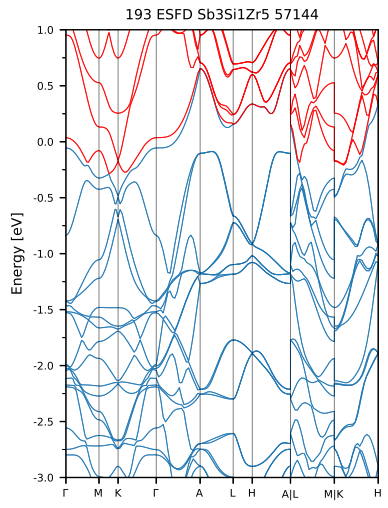

(co) 193.4.13

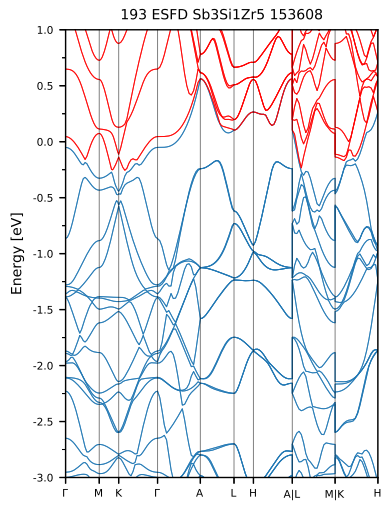

(cp) 193.4.14

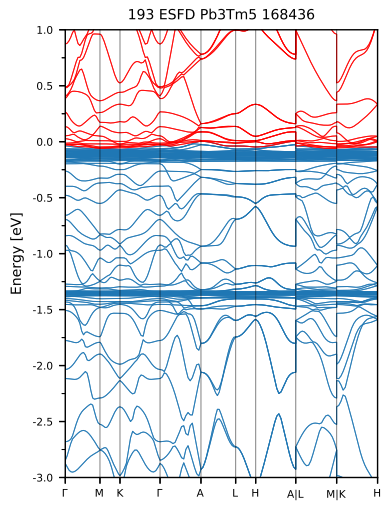

(cq) 193.4.15

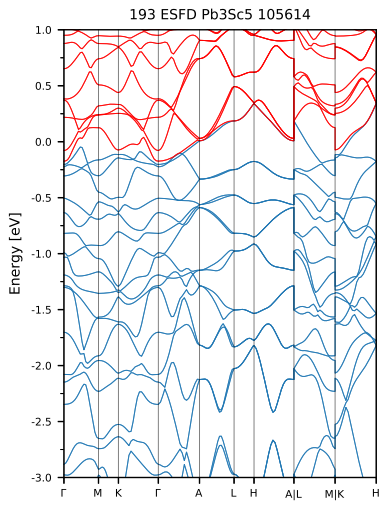

(cr) 193.4.16

Figure 446: Band structures for topological materials in SG 193, part 6/11, from 193.4.1 to 193.4.16. See the list of figures.

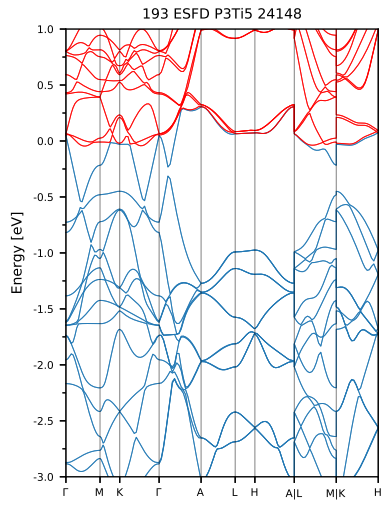

(cs) 193.4.17

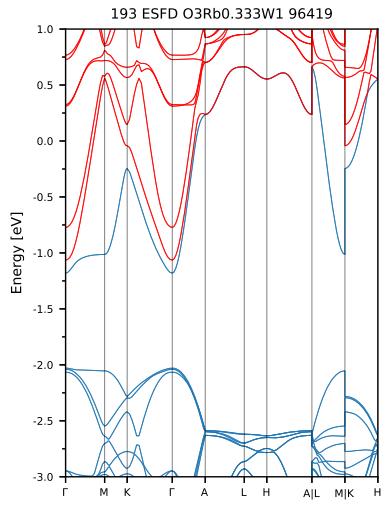

(ct) 193.4.18

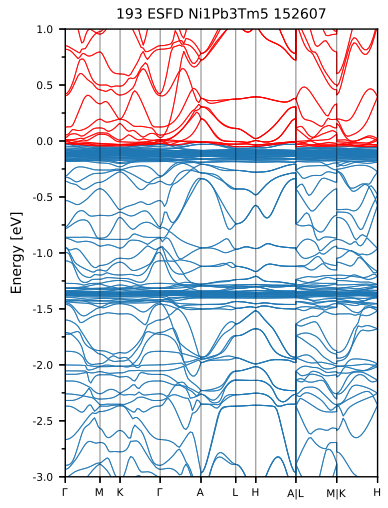

(cu) 193.4.19

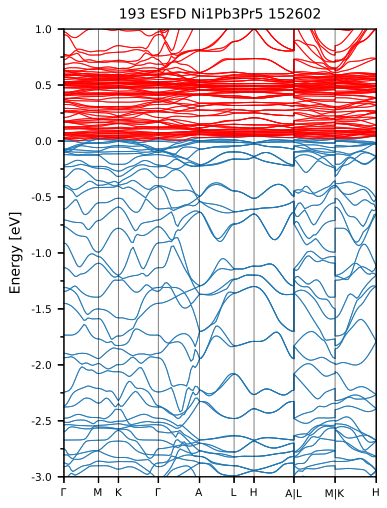

(cv) 193.4.20

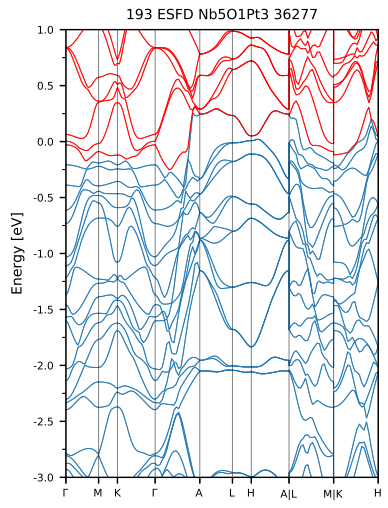

(cw) 193.4.21

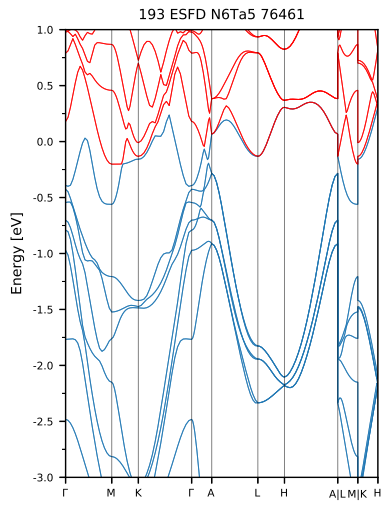

(cx) 193.4.22

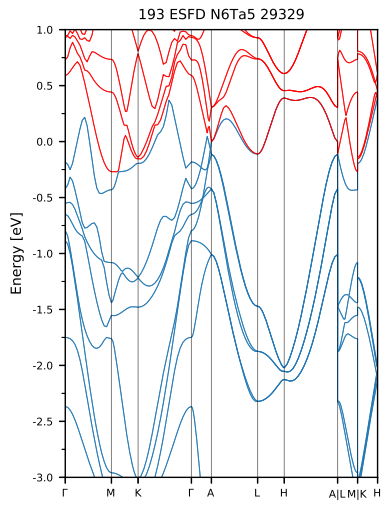

(cy) 193.4.23

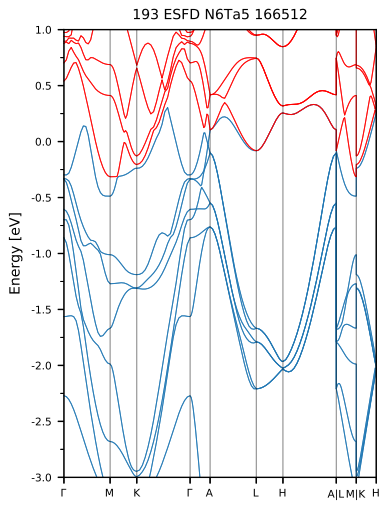

(cz) 193.4.24

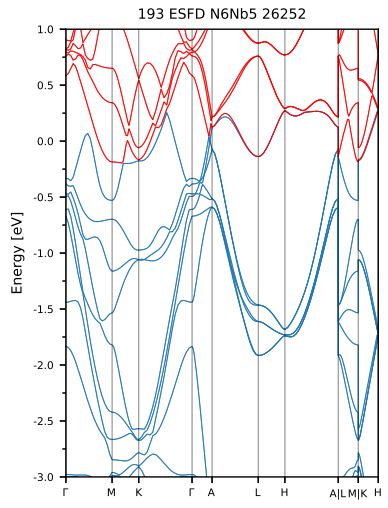

(da) 193.4.25

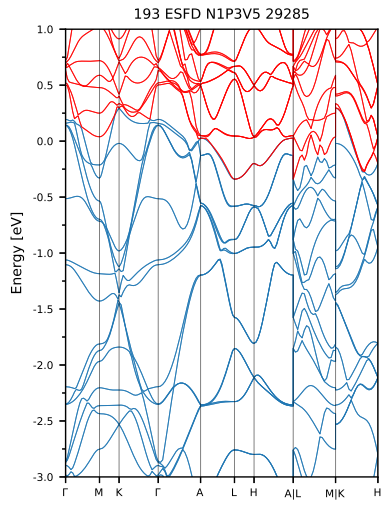

(db) 193.4.26

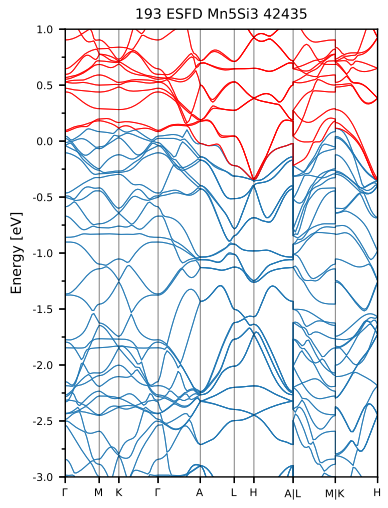

(dc) 193.4.27

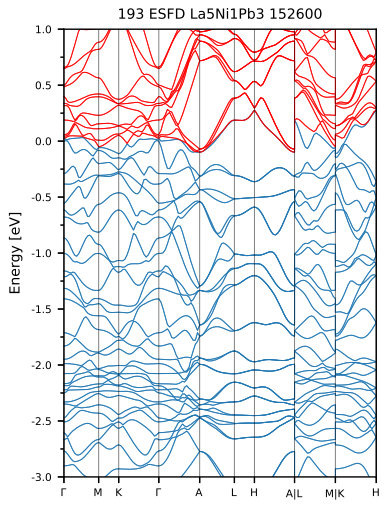

(dd) 193.4.28

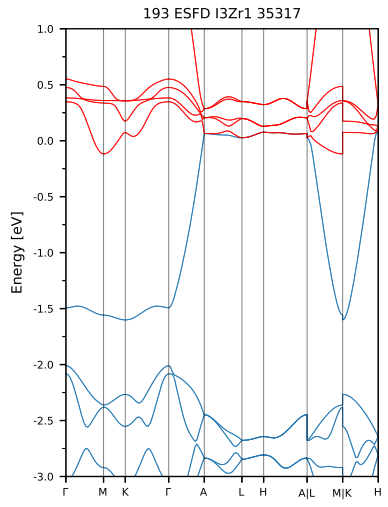

(de) 193.4.29

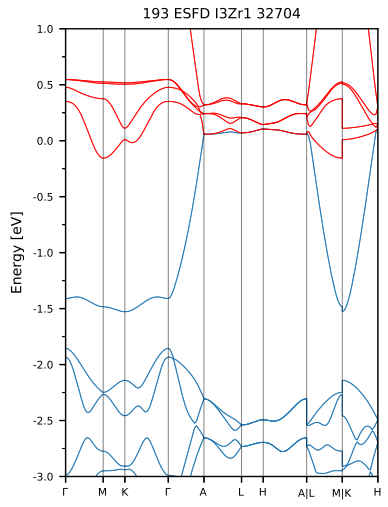

(df) 193.4.30

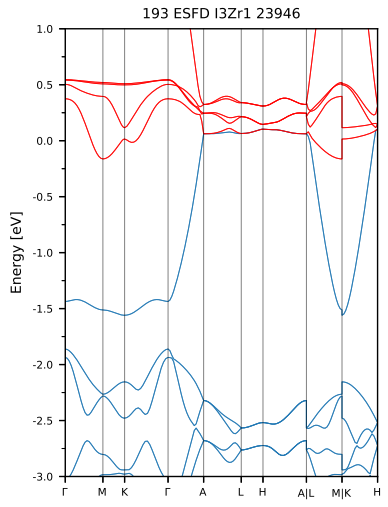

(dg) 193.4.31

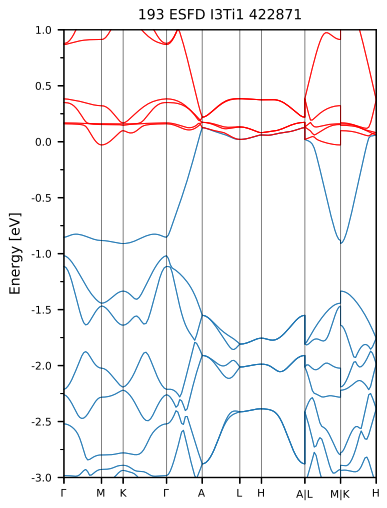

(dh) 193.4.32

Figure 447: Band structures for topological materials in SG 193, part 7/11, from 193.4.17 to 193.4.32. See the list of figures.

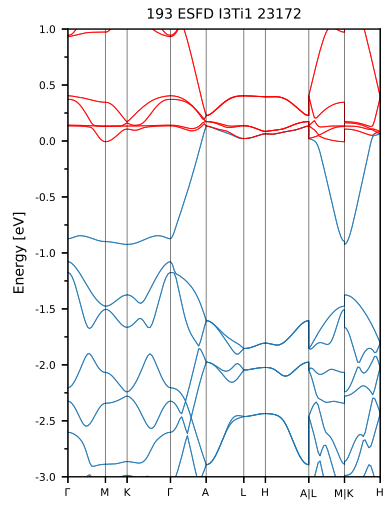

(di) 193.4.33

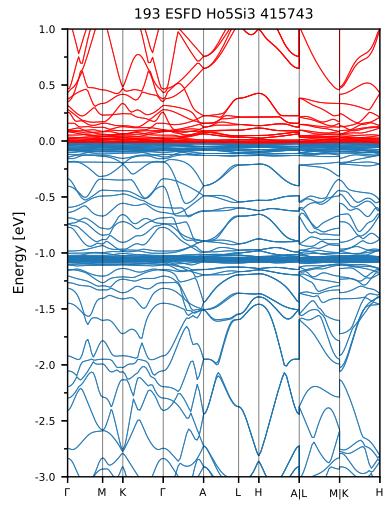

(dj) 193.4.34

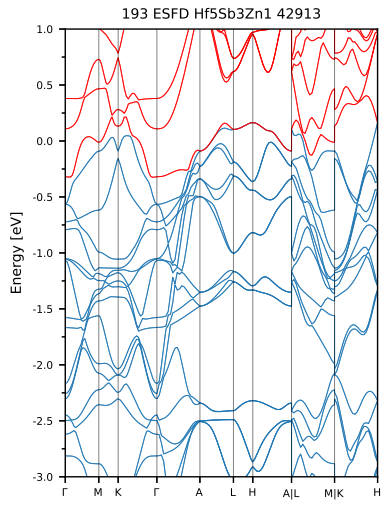

(dk) 193.4.35

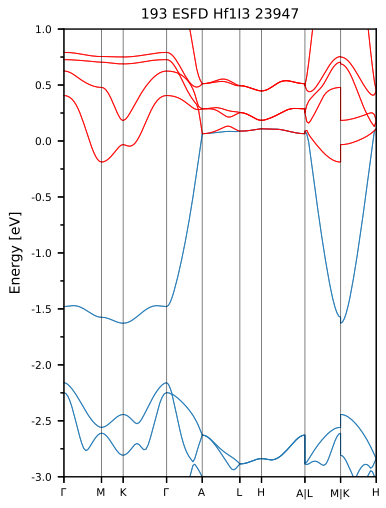

(dl) 193.4.36

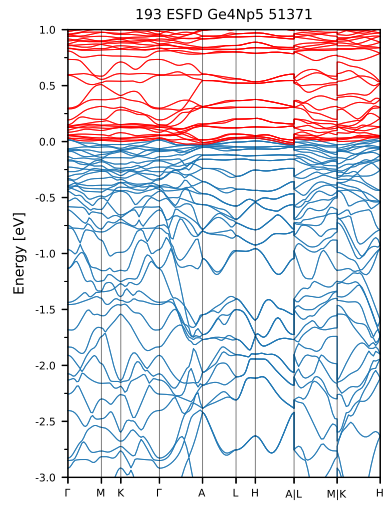

(dm) 193.4.37

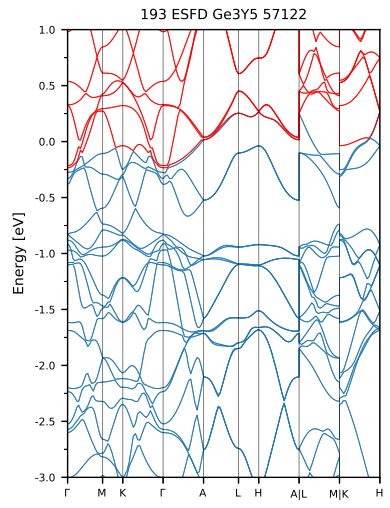

(dn) 193.4.38

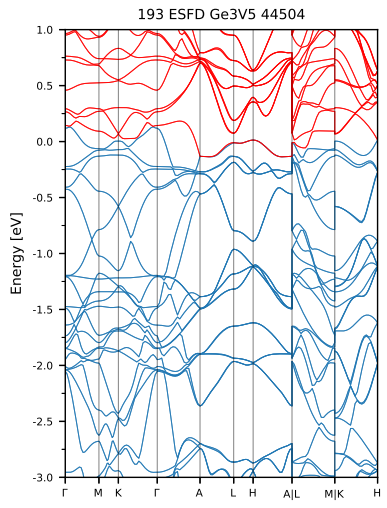

(do) 193.4.39

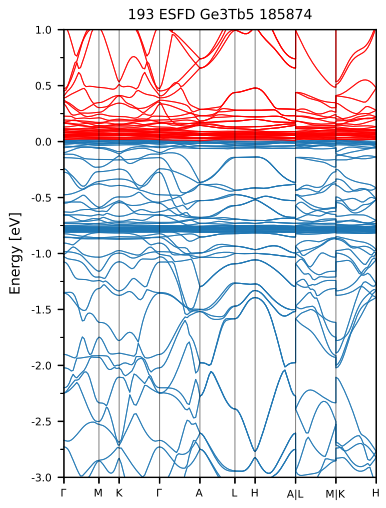

(dp) 193.4.40

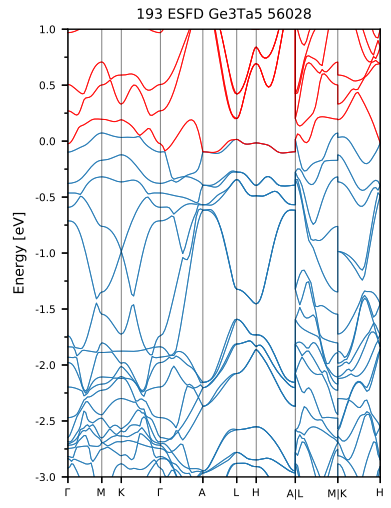

(dq) 193.4.41

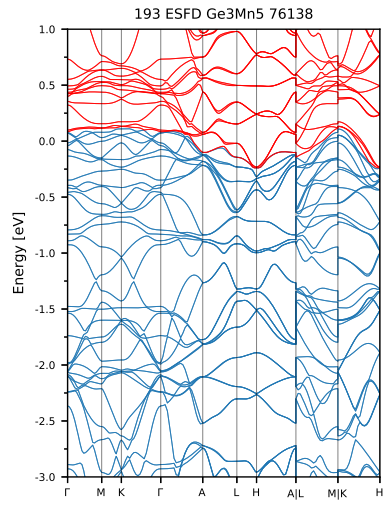

(dr) 193.4.42

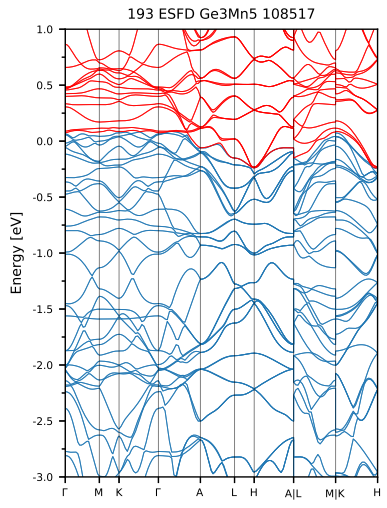

(ds) 193.4.43

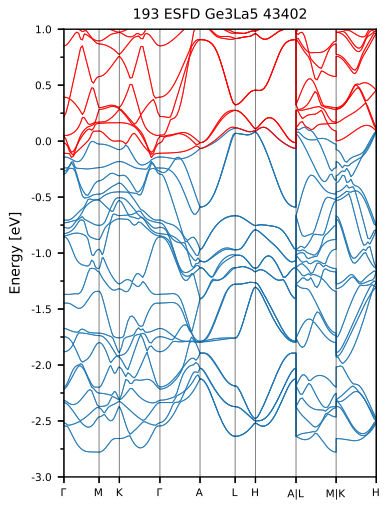

(dt) 193.4.44

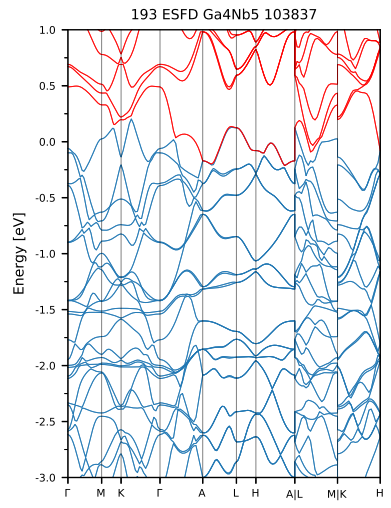

(du) 193.4.45

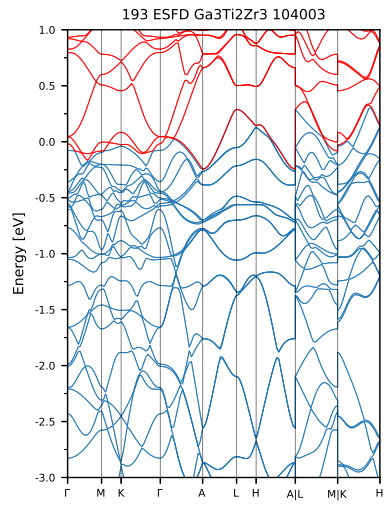

(dv) 193.4.46

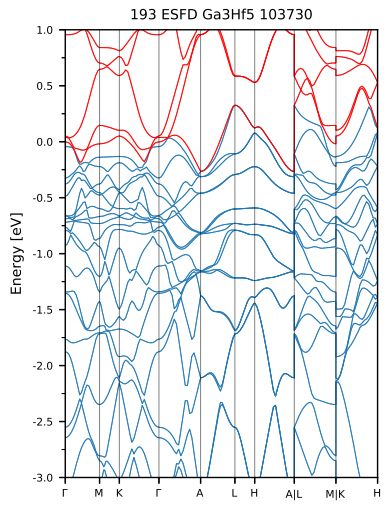

(dw) 193.4.47

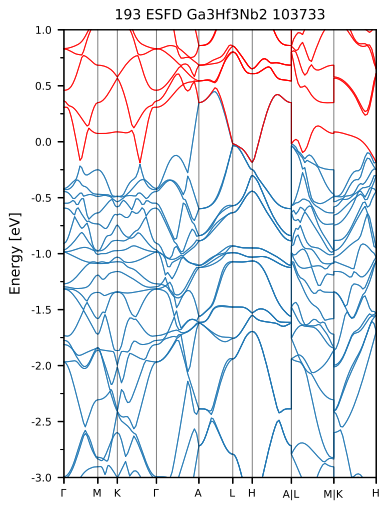

(dx) 193.4.48

Figure 448: Band structures for topological materials in SG 193, part 8/11, from 193.4.33 to 193.4.48. See the list of figures.

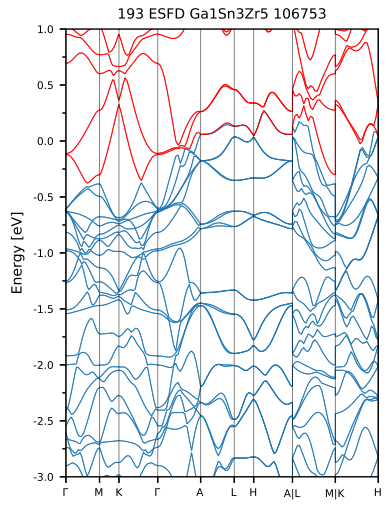

(dy) 193.4.49

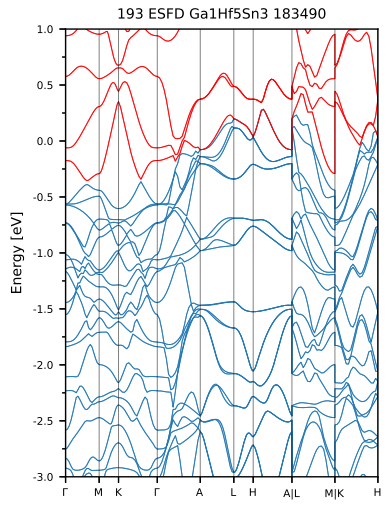

(dz) 193.4.50

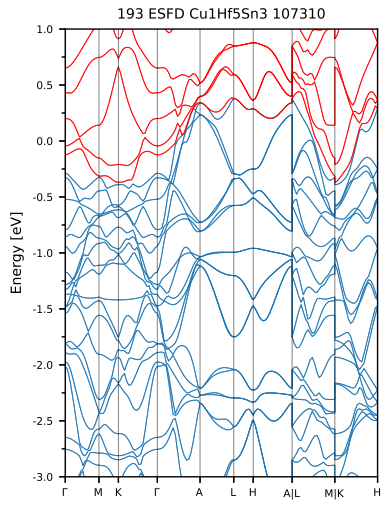

(ea) 193.4.51

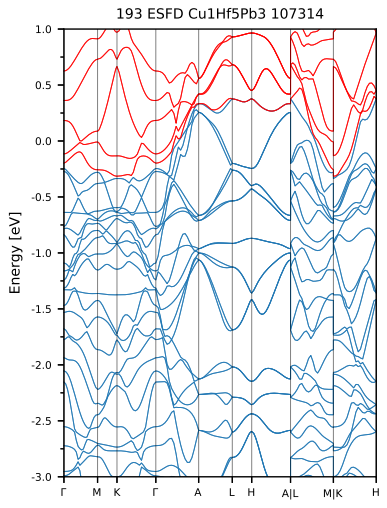

(eb) 193.4.52

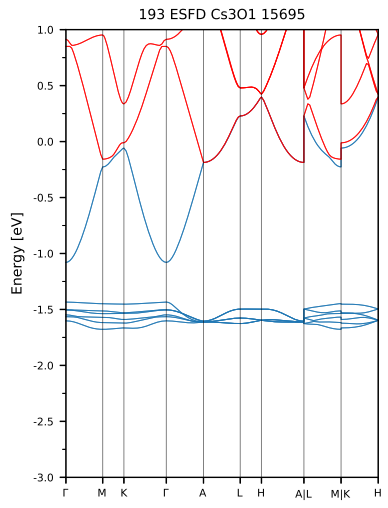

(ec) 193.4.53

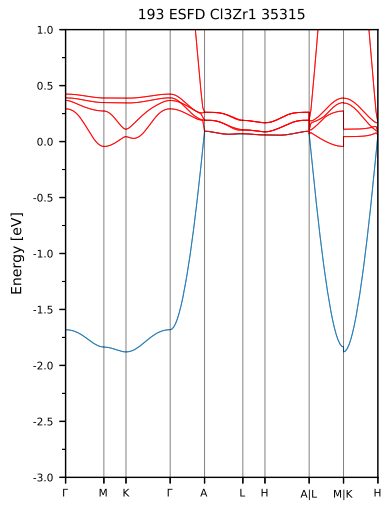

(ed) 193.4.54

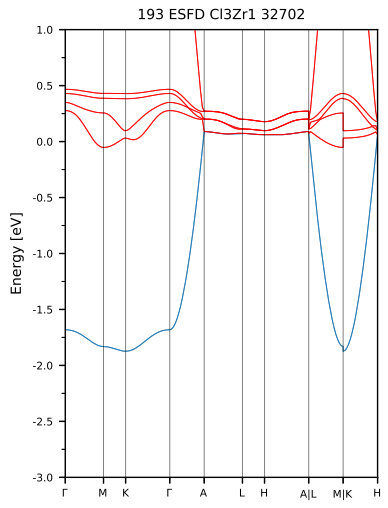

(ee) 193.4.55

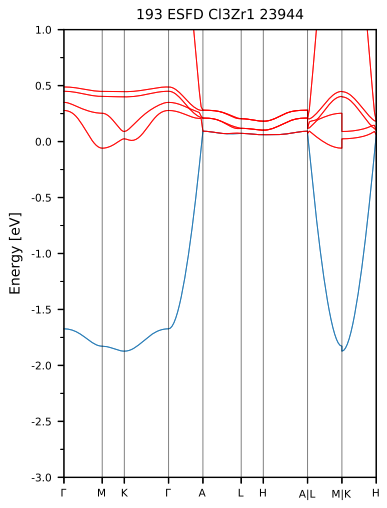

(ef) 193.4.56

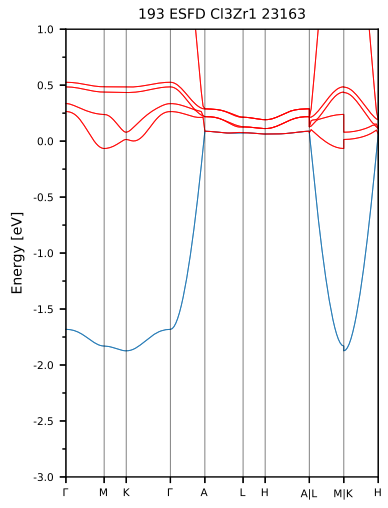

(eg) 193.4.57

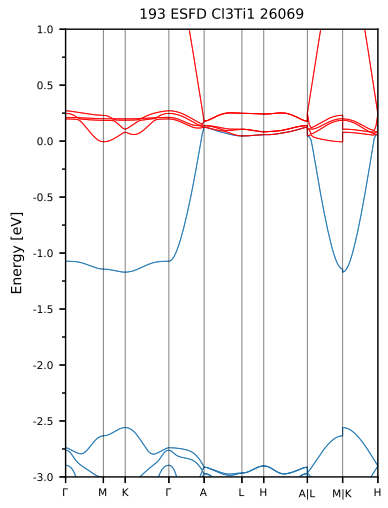

(eh) 193.4.58

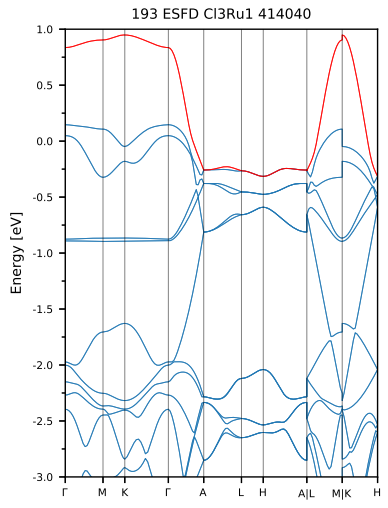

(ei) 193.4.59

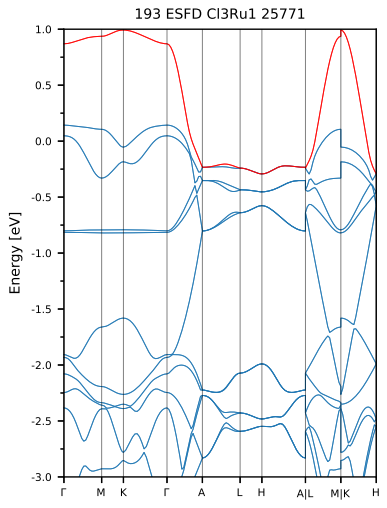

(ej) 193.4.60

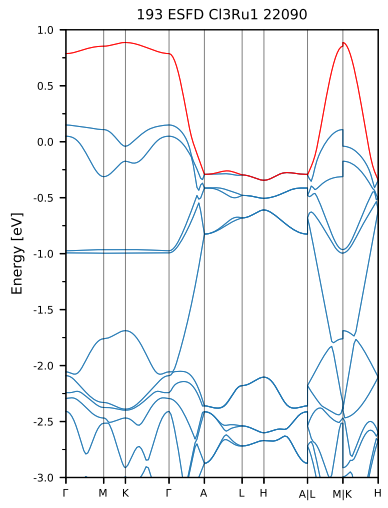

(ek) 193.4.61

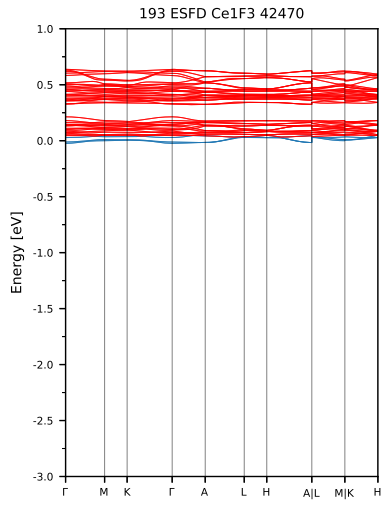

(el) 193.4.62

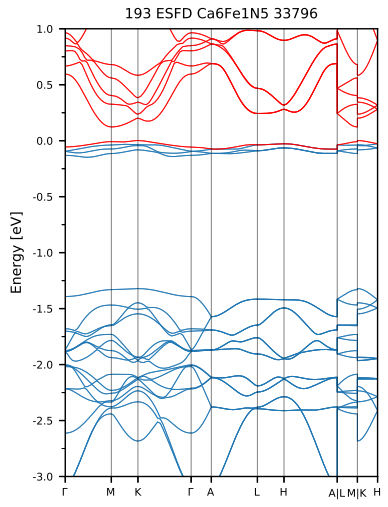

(em) 193.4.63

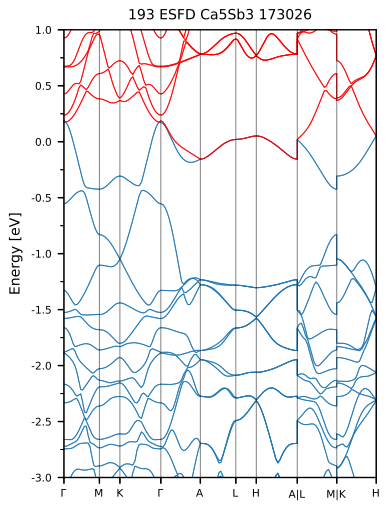

(en) 193.4.64

Figure 449: Band structures for topological materials in SG 193, part 9/11, from 193.4.49 to 193.4.64. See the list of figures.

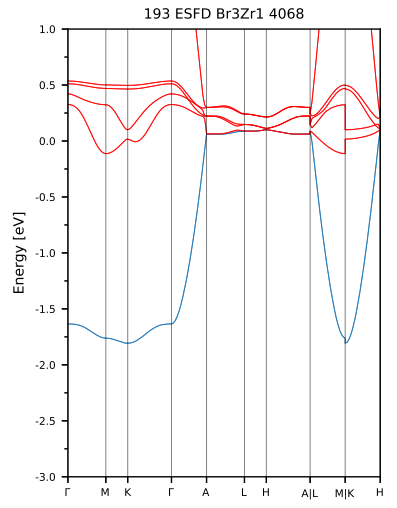

(eo) 193.4.65

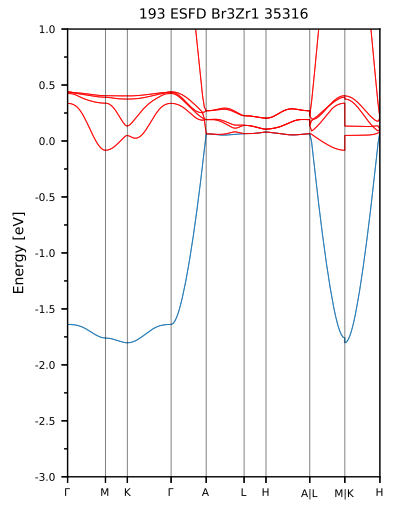

(ep) 193.4.66

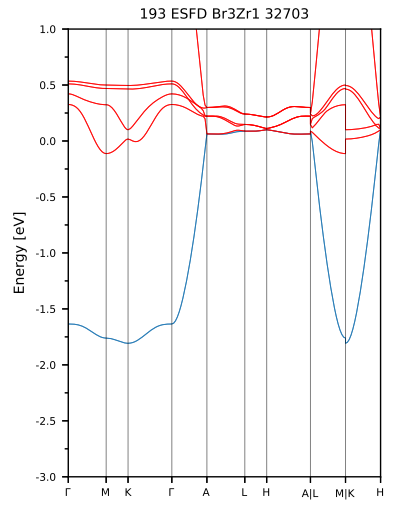

(eq) 193.4.67

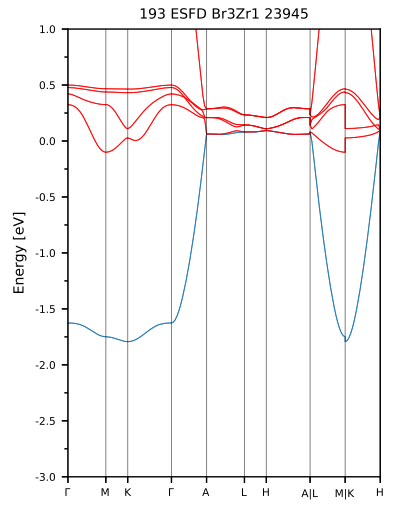

(er) 193.4.68

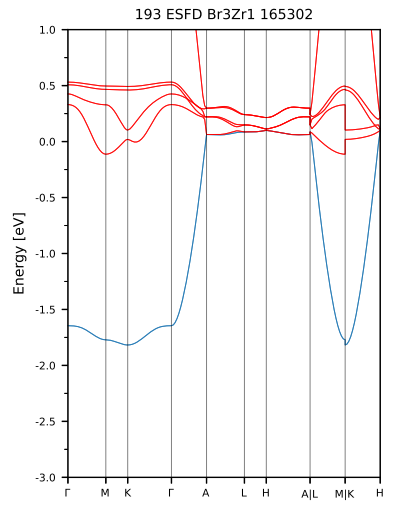

(es) 193.4.69

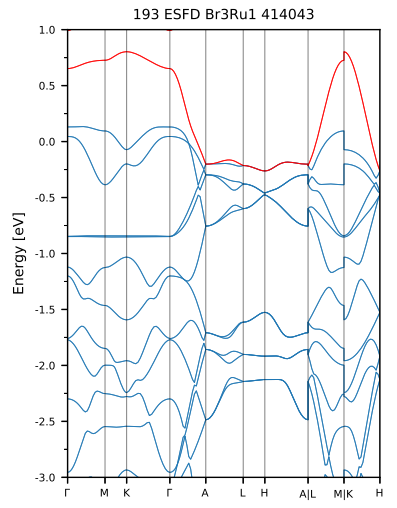

(et) 193.4.70

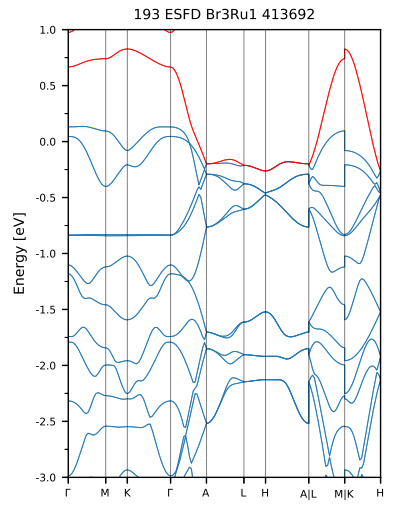

(eu) 193.4.71

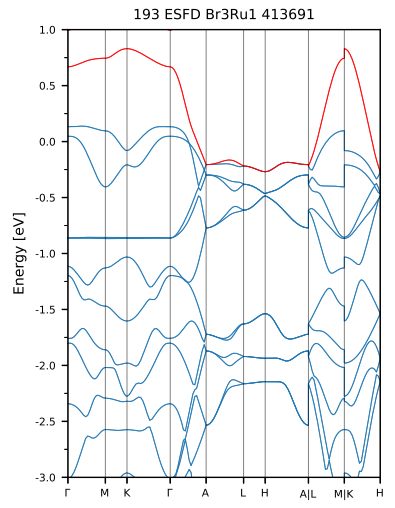

(ev) 193.4.72

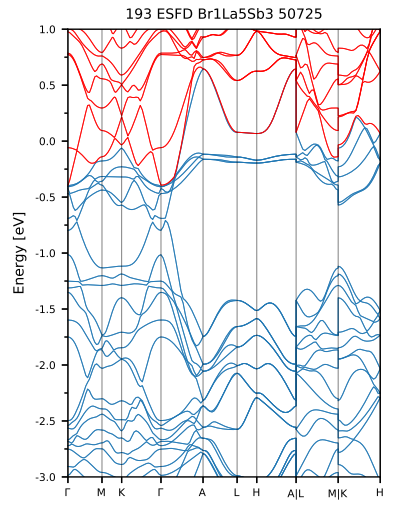

(ew) 193.4.73

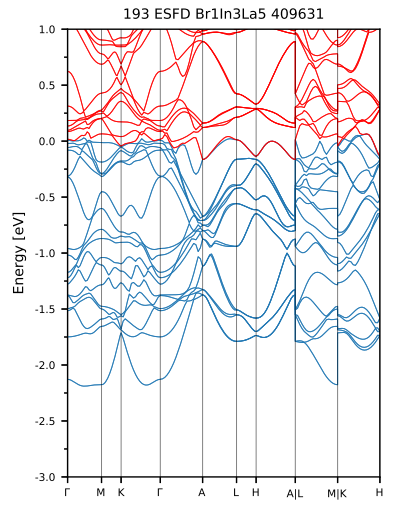

(ex) 193.4.74

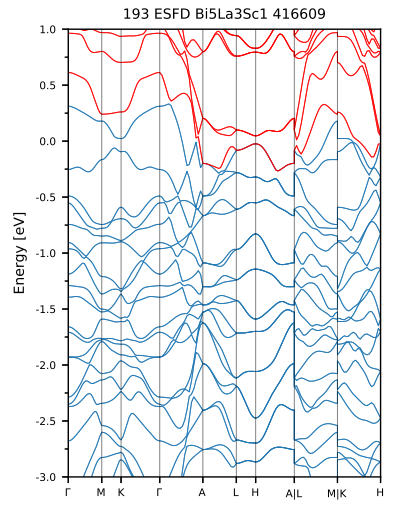

(ey) 193.4.75

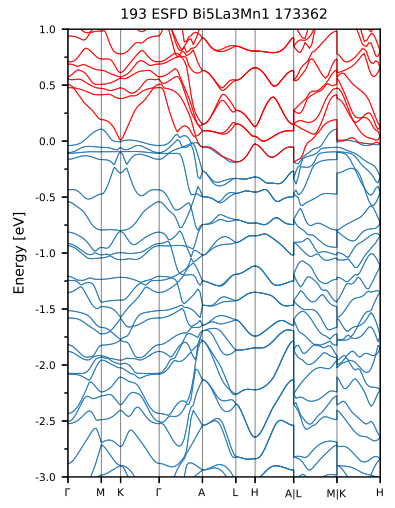

(ez) 193.4.76

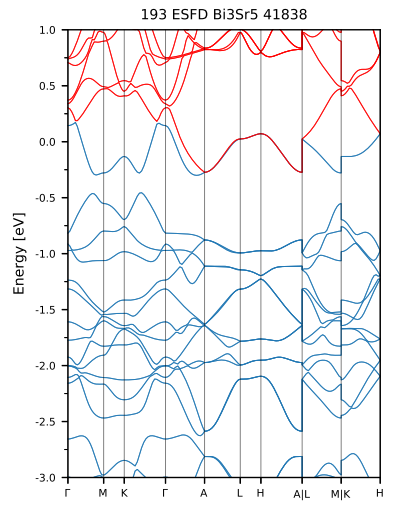

(fa) 193.4.77

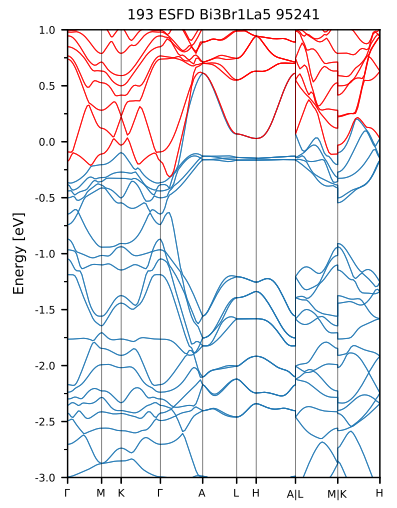

(fb) 193.4.78

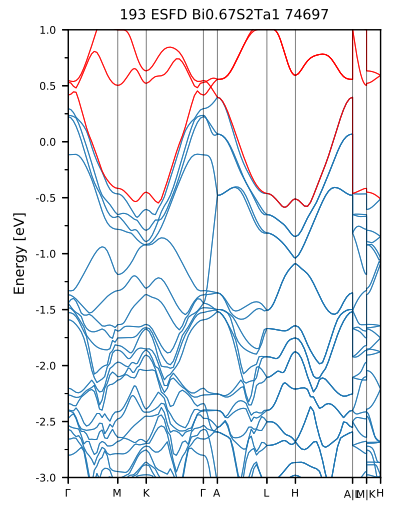

(fc) 193.4.79

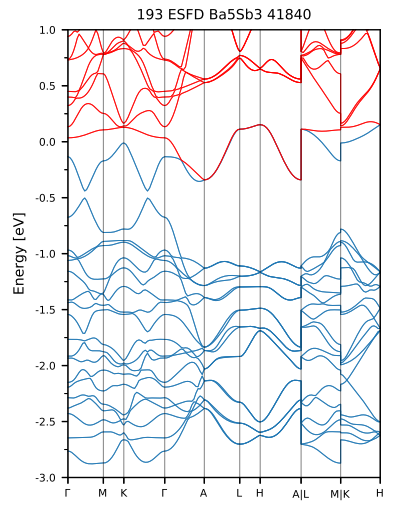

(fd) 193.4.80

Figure 450: Band structures for topological materials in SG 193, part 10/11, from 193.4.65 to 193.4.80. See the list of figures.

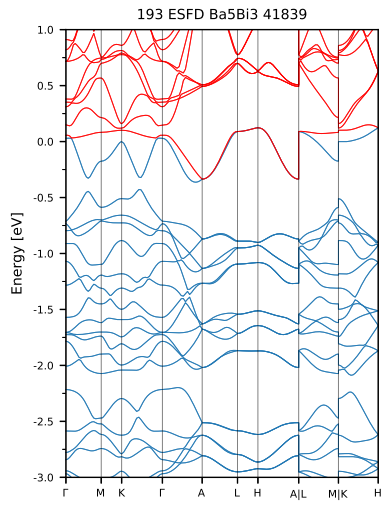

(fe) 193.4.81

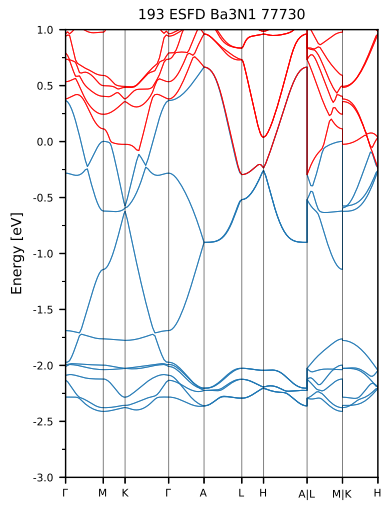

(ff) 193.4.82

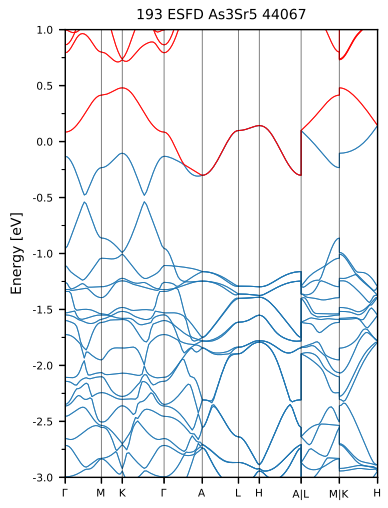

(fg) 193.4.83

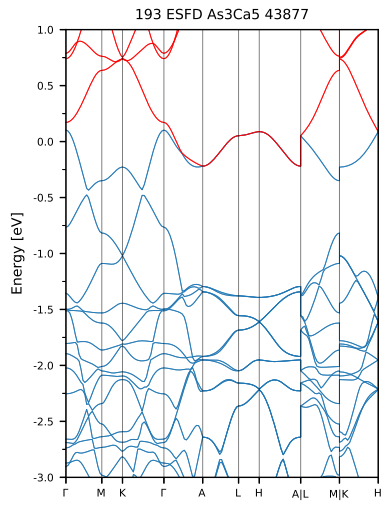

(fh) 193.4.84

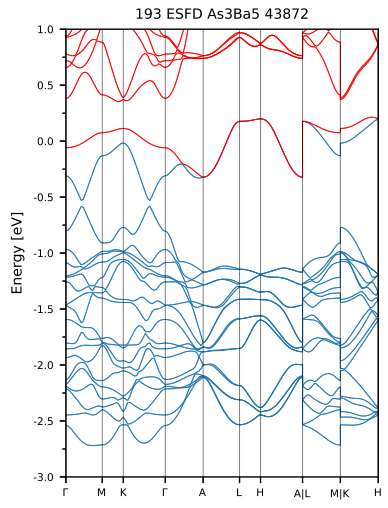

(fi) 193.4.85

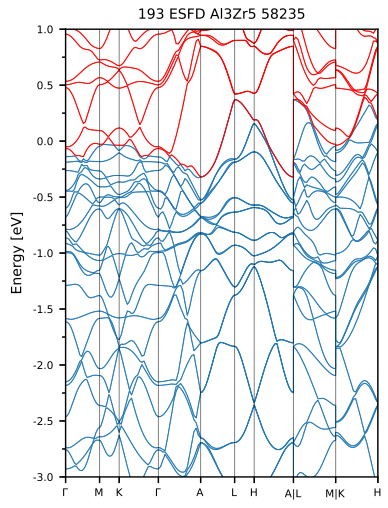

(fj) 193.4.86

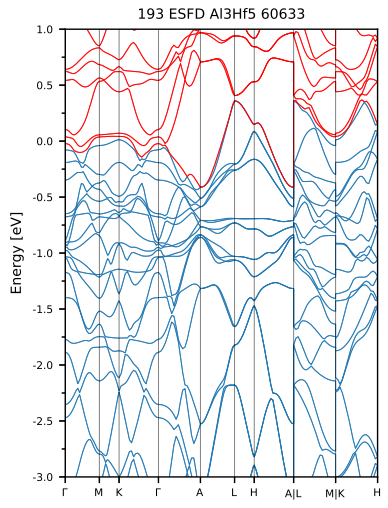

(fk) 193.4.87

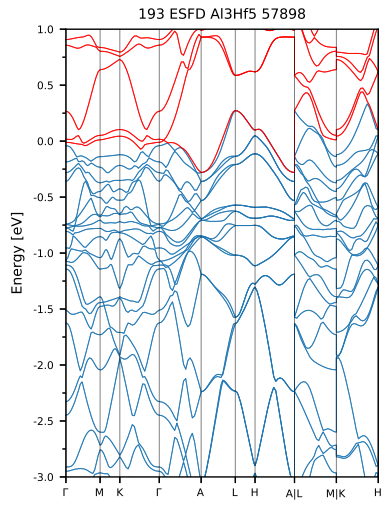

(fl) 193.4.88

Figure 451: Band structures for topological materials in SG 193, part 11/11, from 193.4.81 to 193.4.88. See the list of figures.

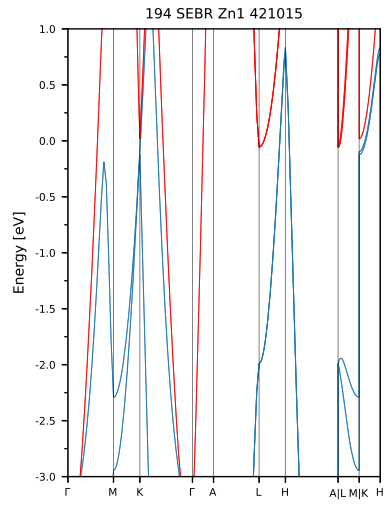

(a) 194.1.1

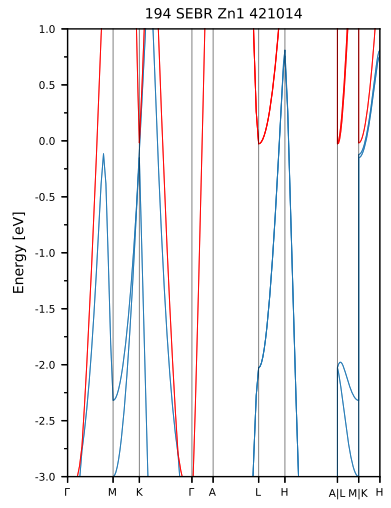

(b) 194.1.2

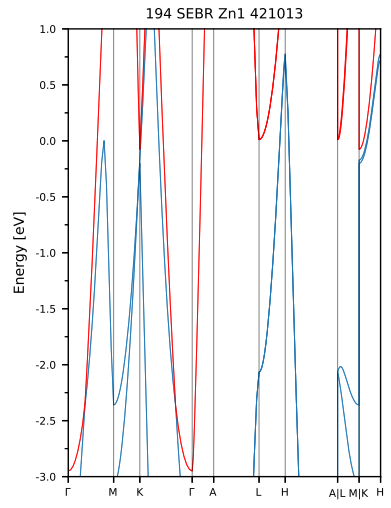

(c) 194.1.3

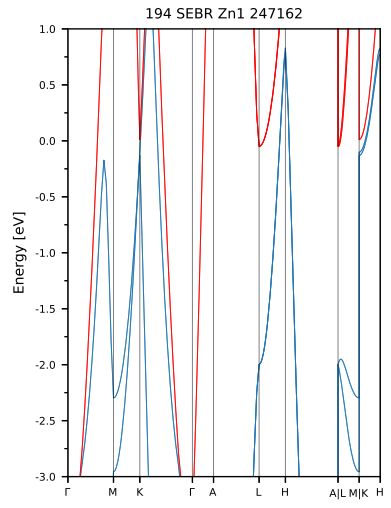

(d) 194.1.4

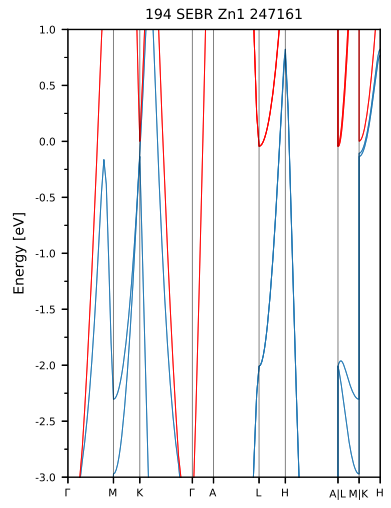

(e) 194.1.5

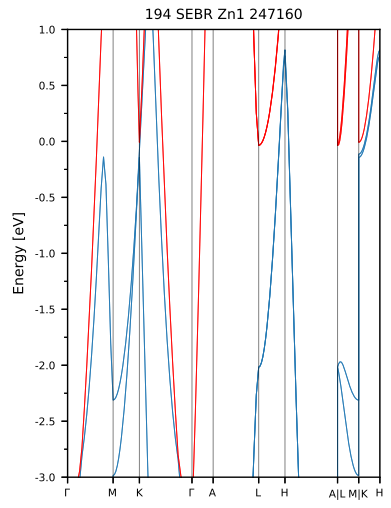

(f) 194.1.6

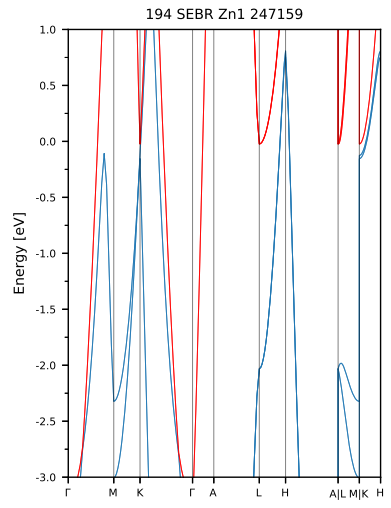

(g) 194.1.7

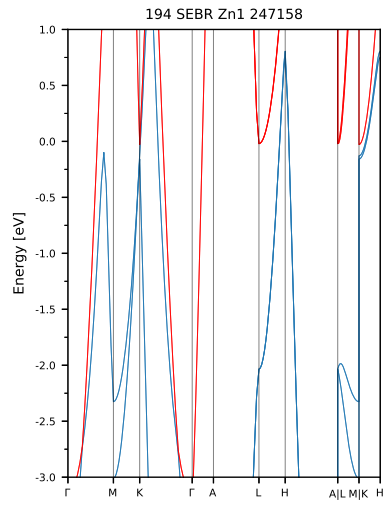

(h) 194.1.8

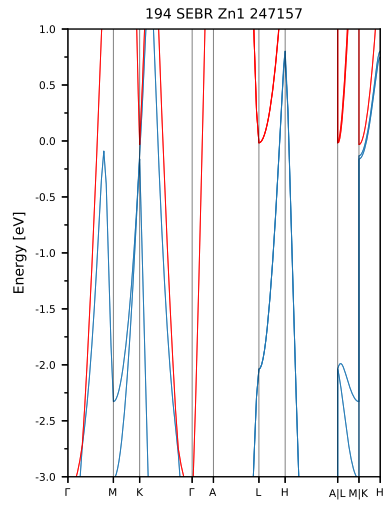

(i) 194.1.9

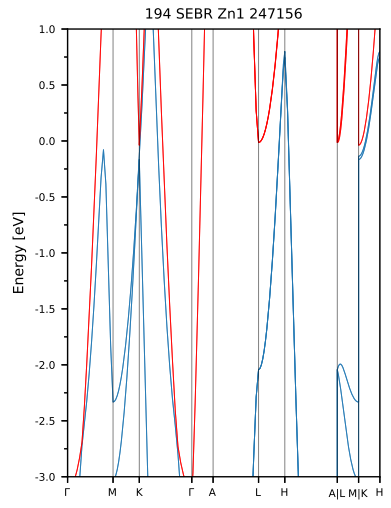

(j) 194.1.10

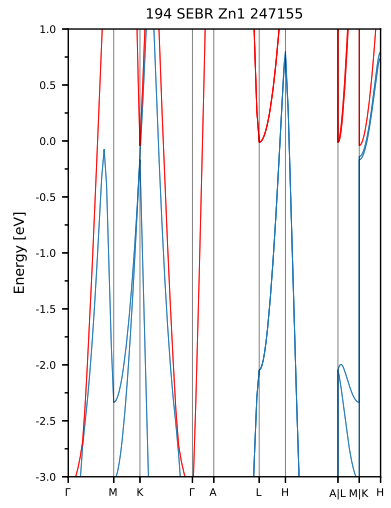

(k) 194.1.11

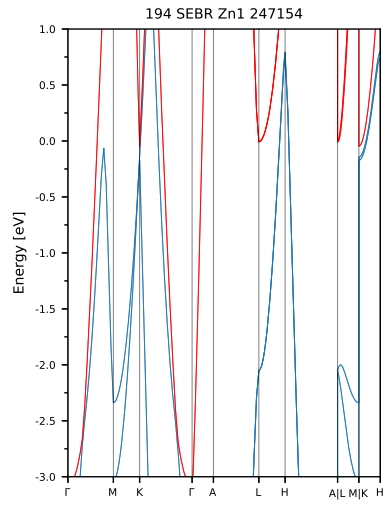

(l) 194.1.12

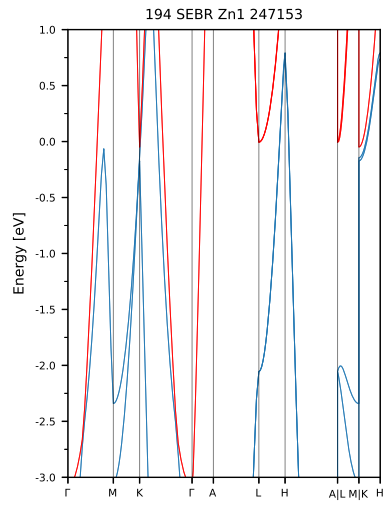

(m) 194.1.13

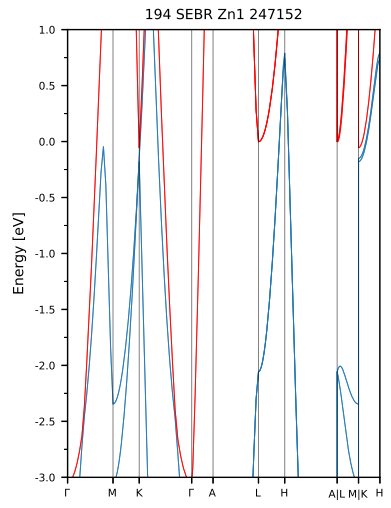

(n) 194.1.14

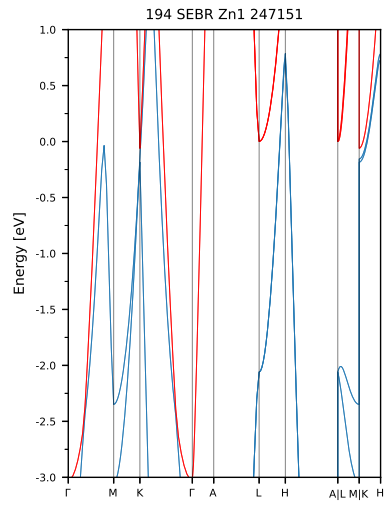

(o) 194.1.15

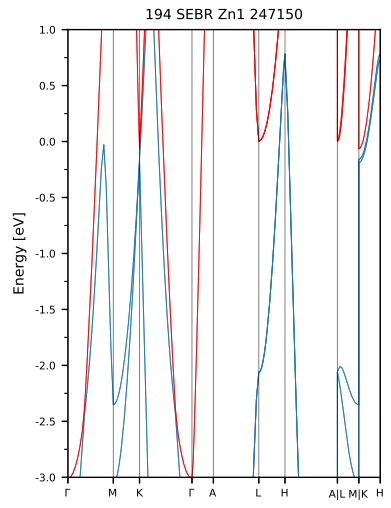

(p) 194.1.16

Figure 452: Band structures for topological materials in SG 194, part 1/12, from 194.1.1 to 194.1.16. See the list of figures.

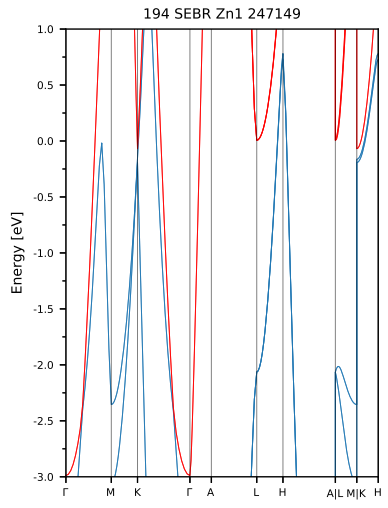

(q) 194.1.17

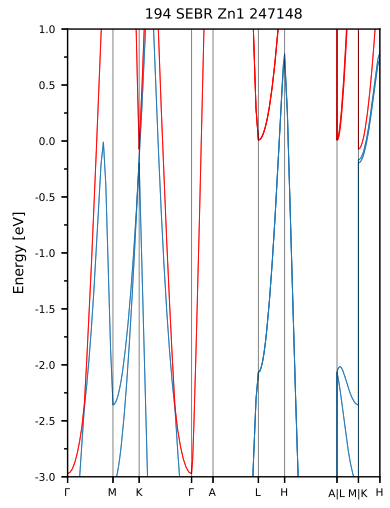

(r) 194.1.18

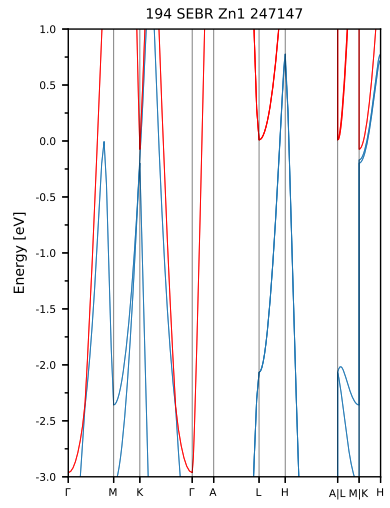

(s) 194.1.19

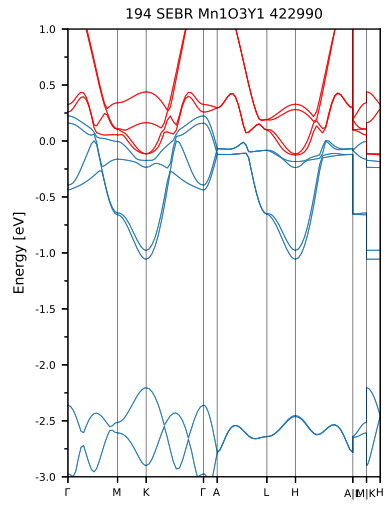

(t) 194.1.20

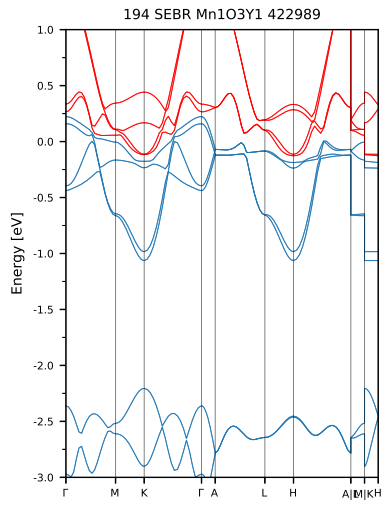

(u) 194.1.21

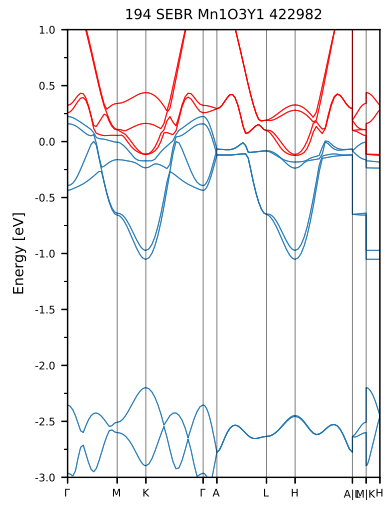

(v) 194.1.22

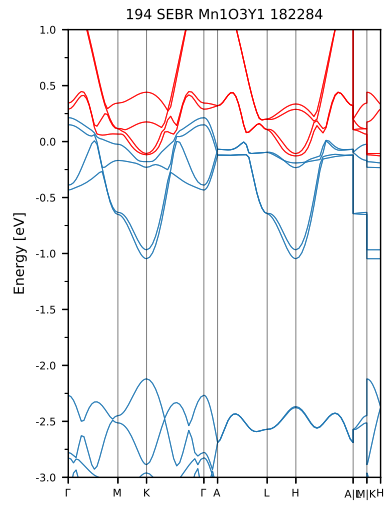

(w) 194.1.23

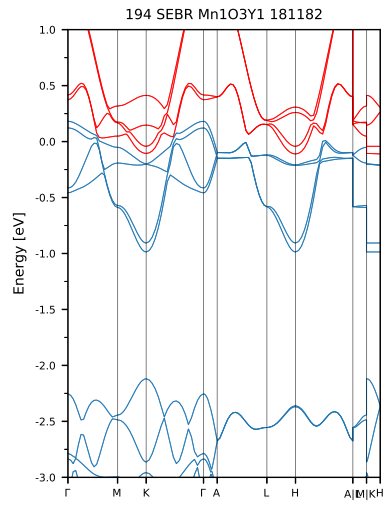

(x) 194.1.24

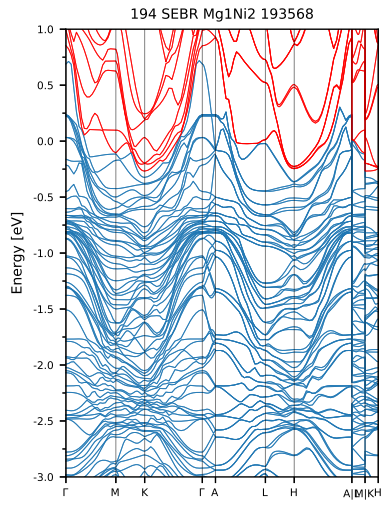

(y) 194.1.25

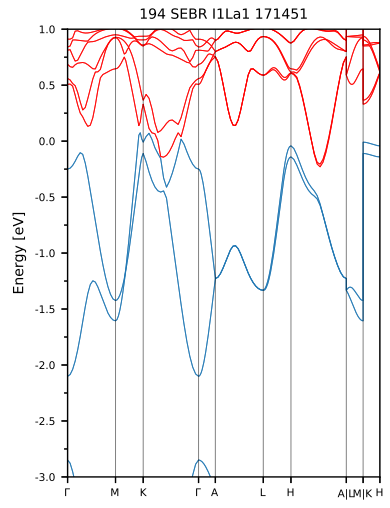

(z) 194.1.26

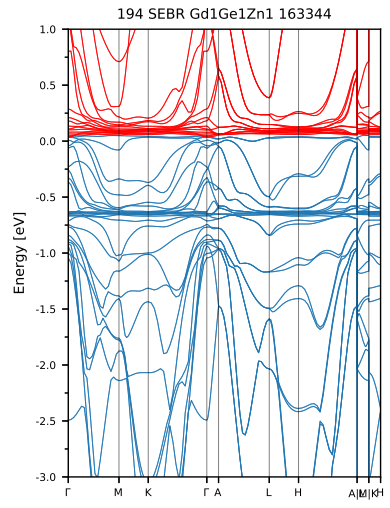

(aa) 194.1.27

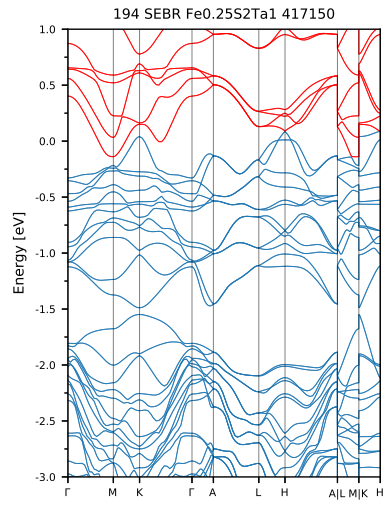

(ab) 194.1.28

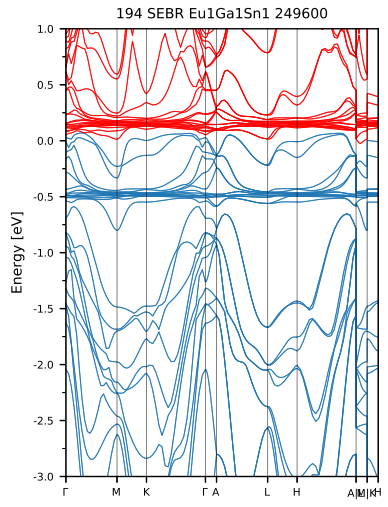

(ac) 194.1.29

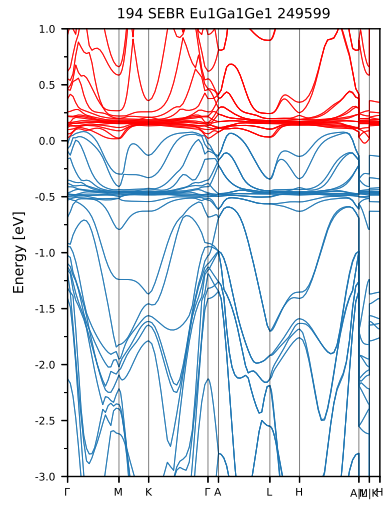

(ad) 194.1.30

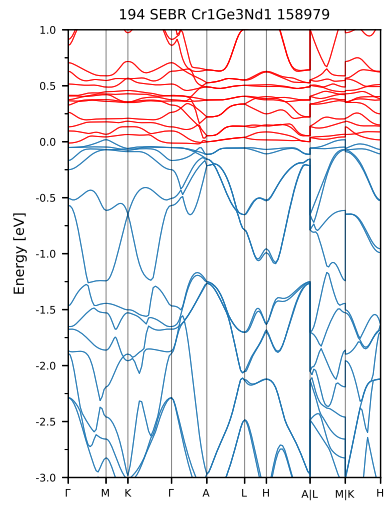

(ae) 194.1.31

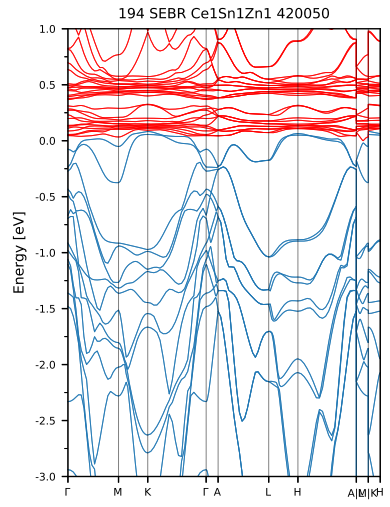

(af) 194.1.32

Figure 453: Band structures for topological materials in SG 194, part 2/12, from 194.1.17 to 194.1.32. See the list of figures.

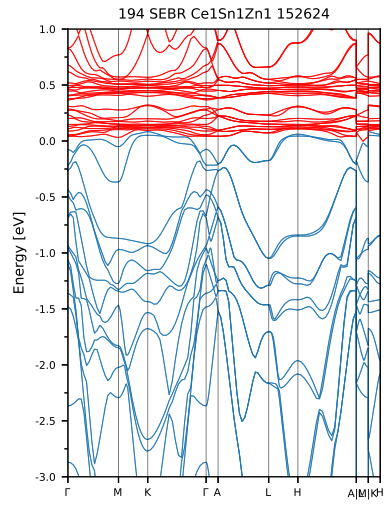

(ag) 194.1.33

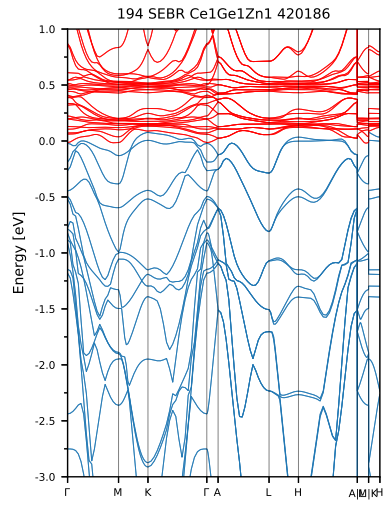

(ah) 194.1.34

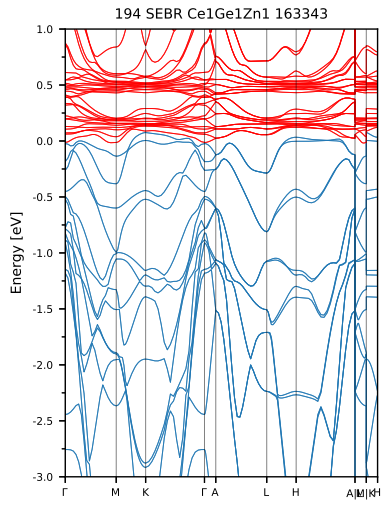

(ai) 194.1.35

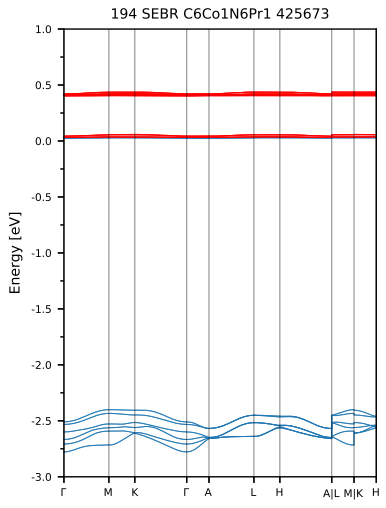

(aj) 194.1.36

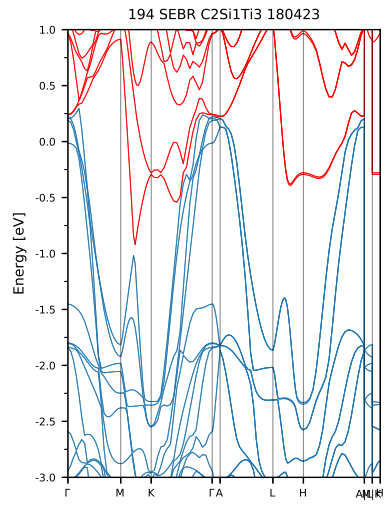

(ak) 194.1.37

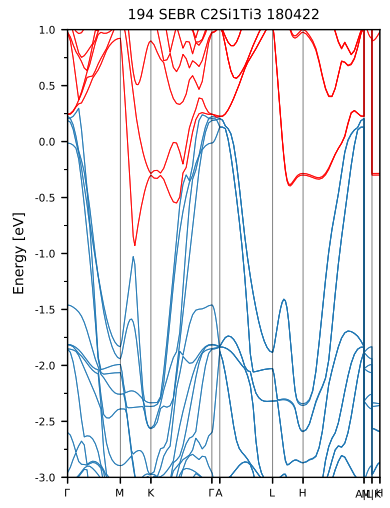

(al) 194.1.38

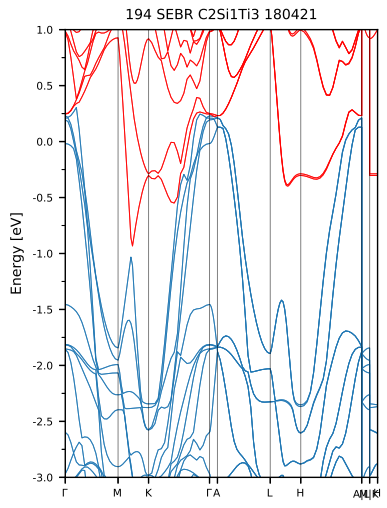

(am) 194.1.39

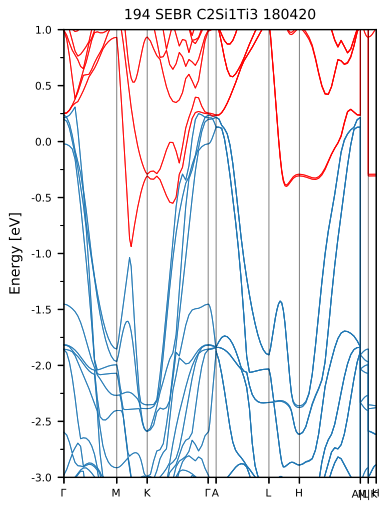

(an) 194.1.40

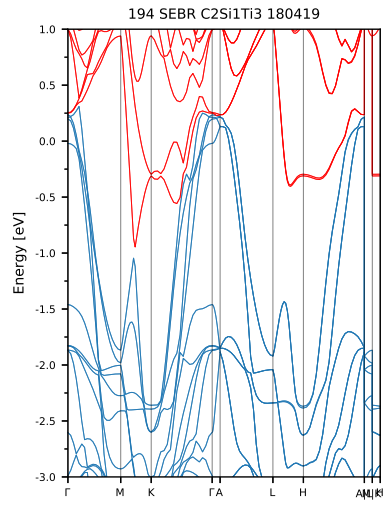

(ao) 194.1.41

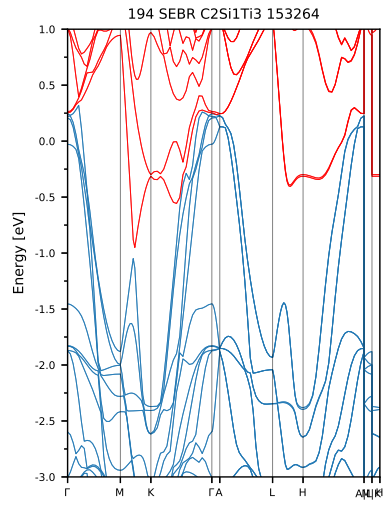

(ap) 194.1.42

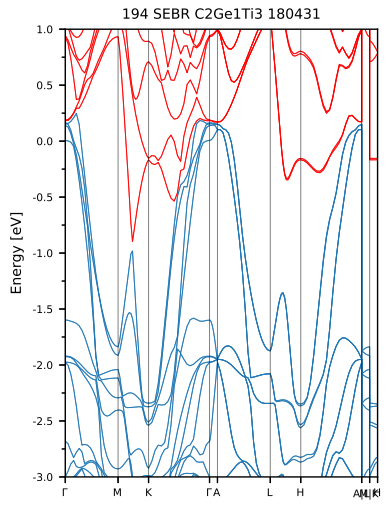

(aq) 194.1.43

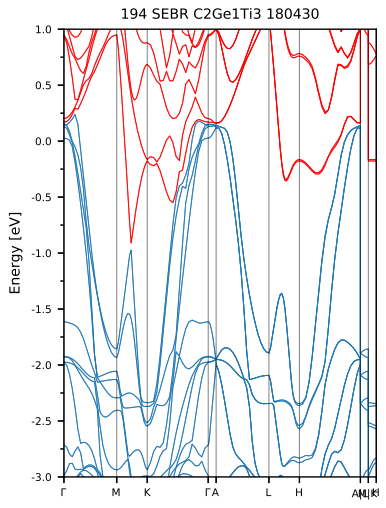

(ar) 194.1.44

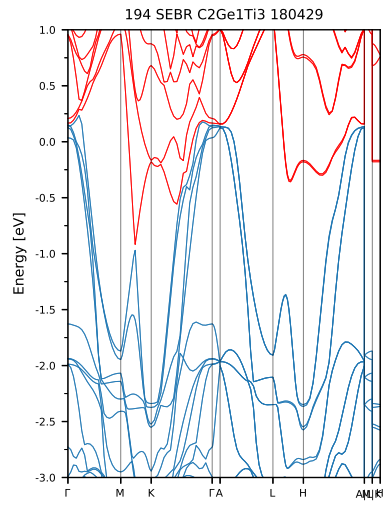

(as) 194.1.45

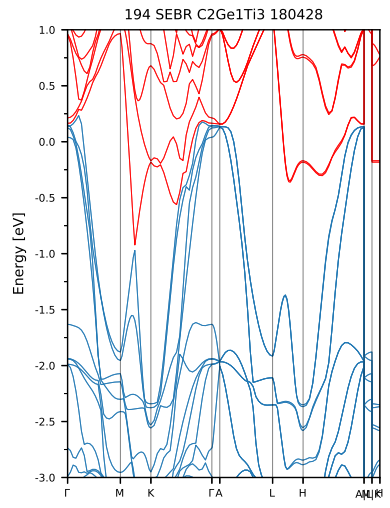

(at) 194.1.46

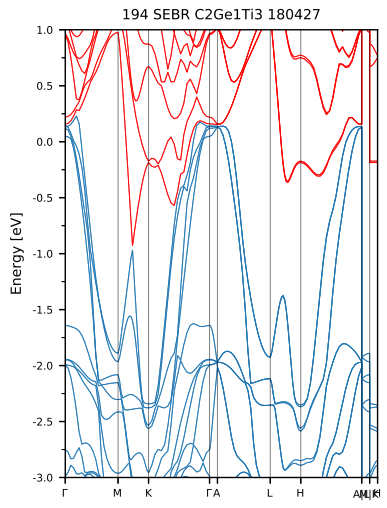

(au) 194.1.47

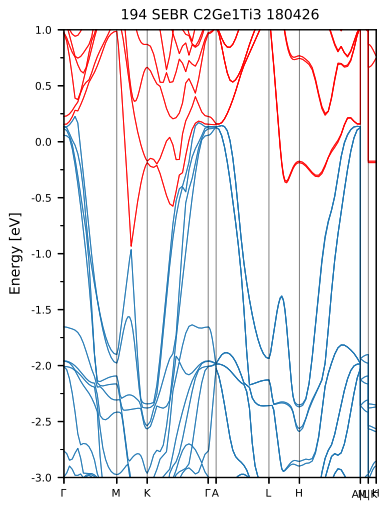

(av) 194.1.48

Figure 454: Band structures for topological materials in SG 194, part 3/12, from 194.1.33 to 194.1.48. See the list of figures.

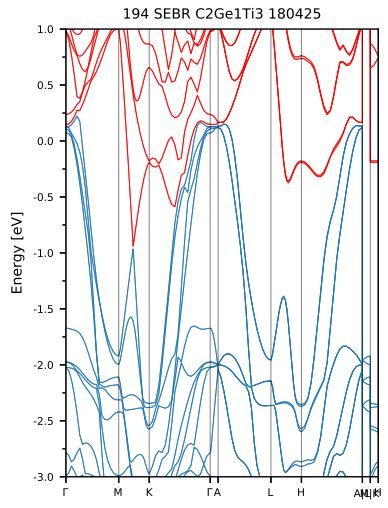

(aw) 194.1.49

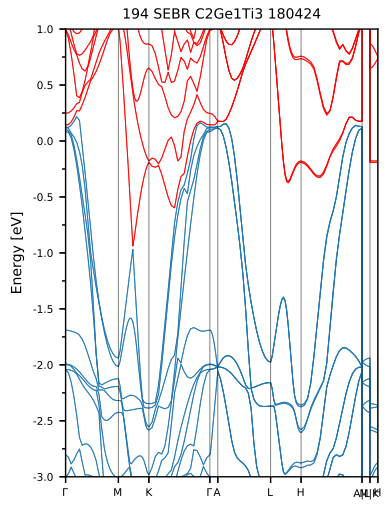

(ax) 194.1.50

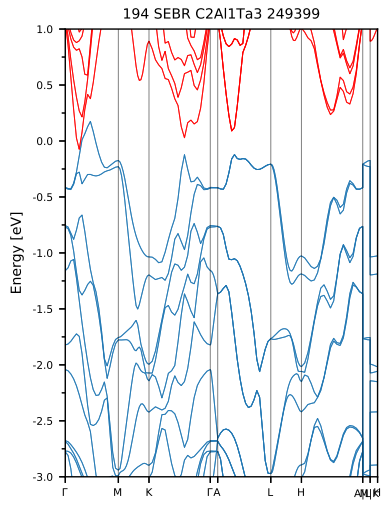

(ay) 194.1.51

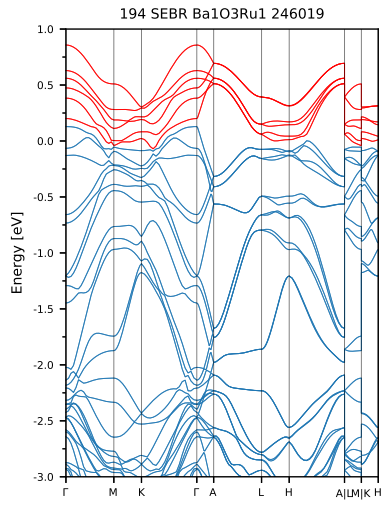

(az) 194.1.52

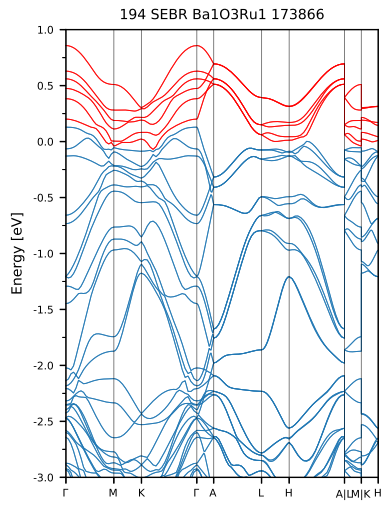

(ba) 194.1.53

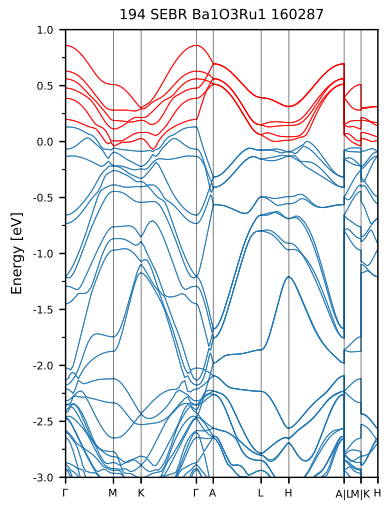

(bb) 194.1.54

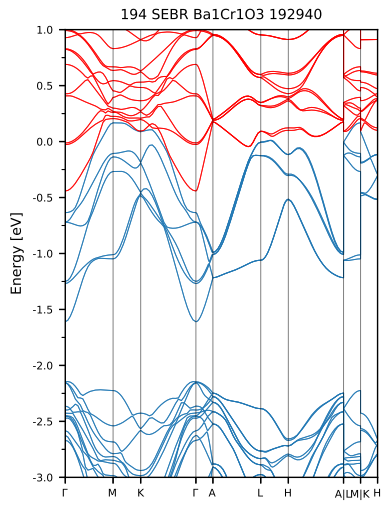

(bc) 194.1.55

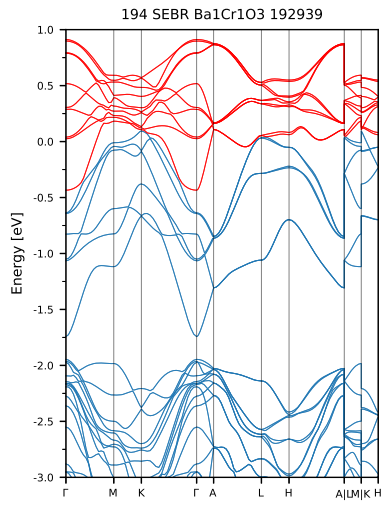

(bd) 194.1.56

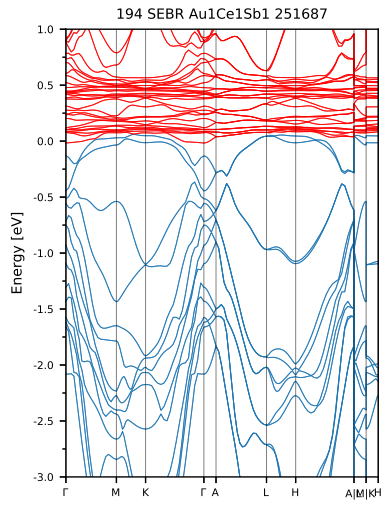

(be) 194.1.57

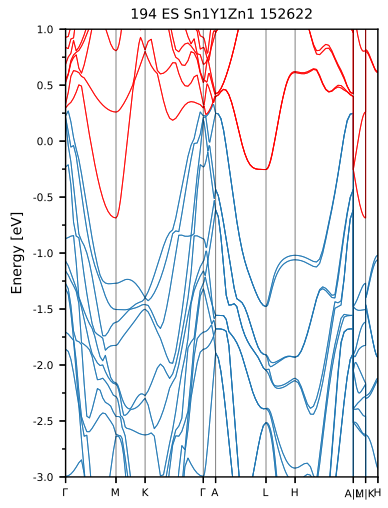

(bf) 194.3.1

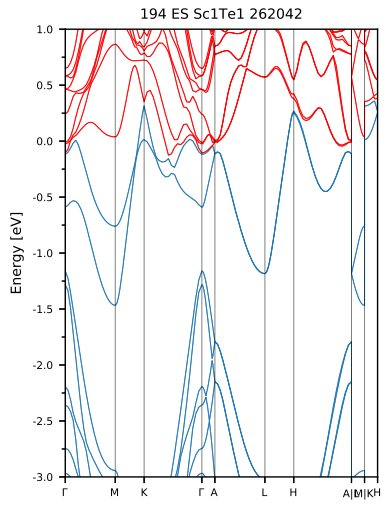

(bg) 194.3.2

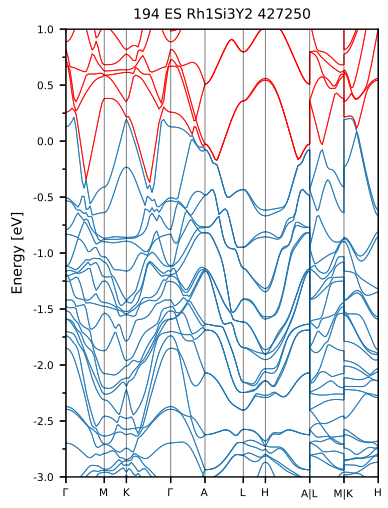

(bh) 194.3.3

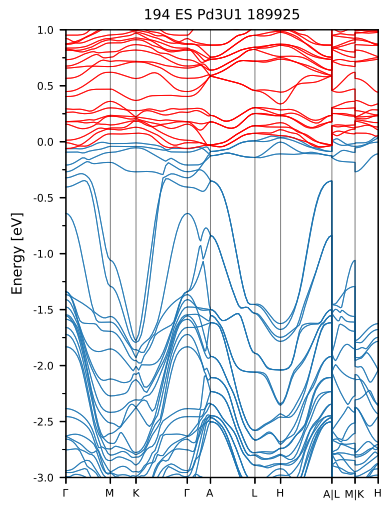

(bi) 194.3.4

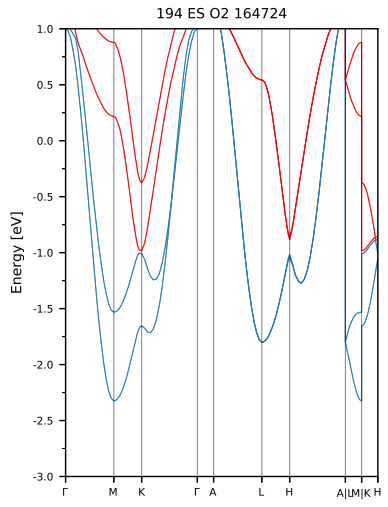

(bj) 194.3.5

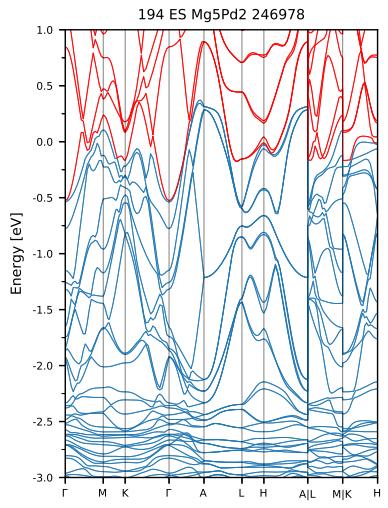

(bk) 194.3.6

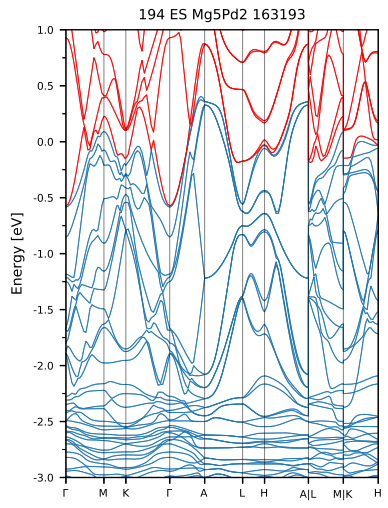

(bl) 194.3.7

Figure 455: Band structures for topological materials in SG 194, part 4/12, from 194.1.49 to 194.3.7. See the list of figures.

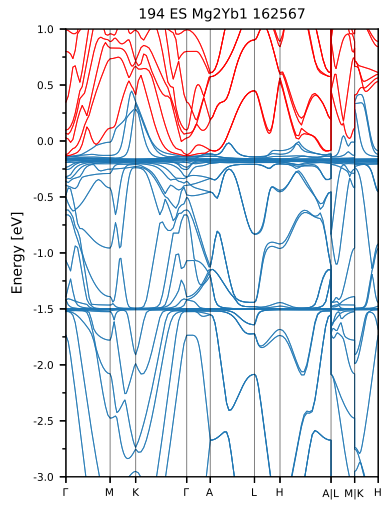

(bm) 194.3.8

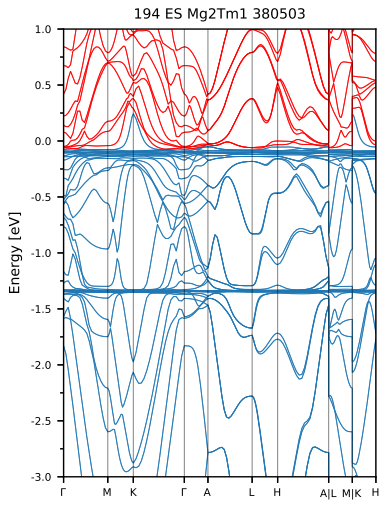

(bn) 194.3.9

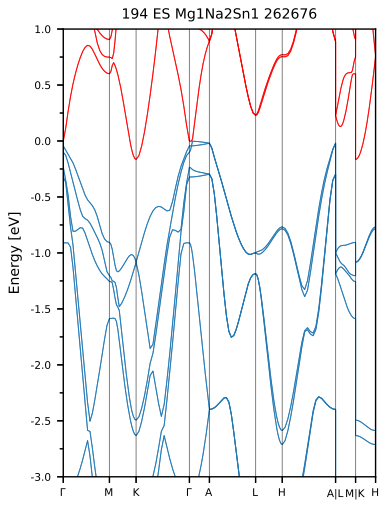

(bo) 194.3.10

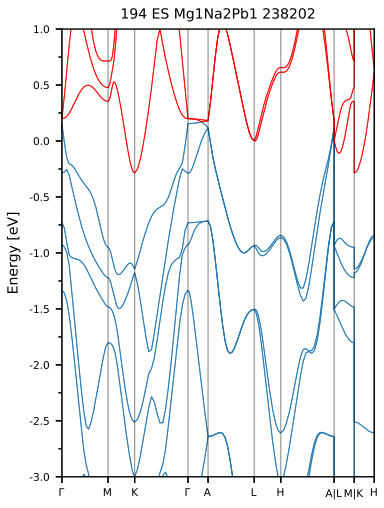

(bp) 194.3.11

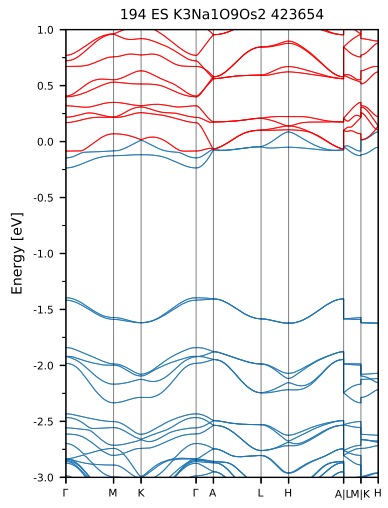

(bq) 194.3.12

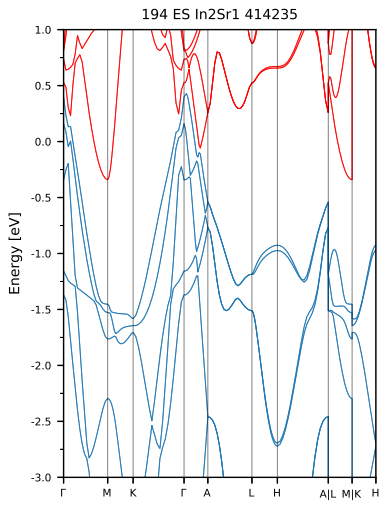

(br) 194.3.13

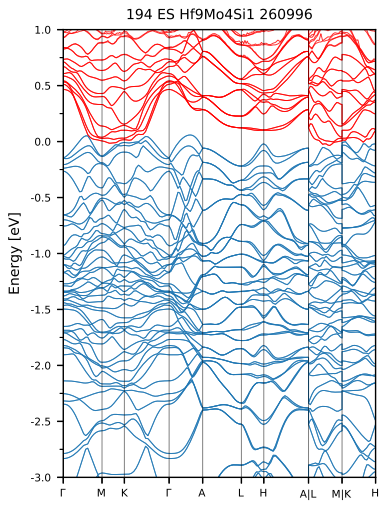

(bs) 194.3.14

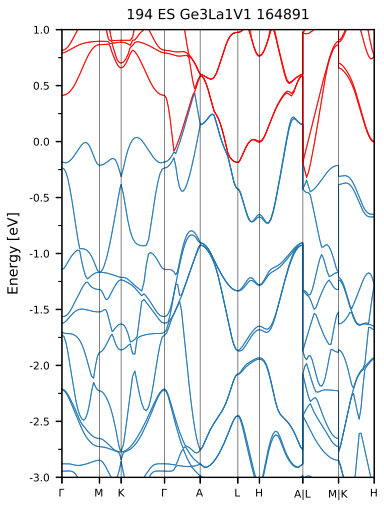

(bt) 194.3.15

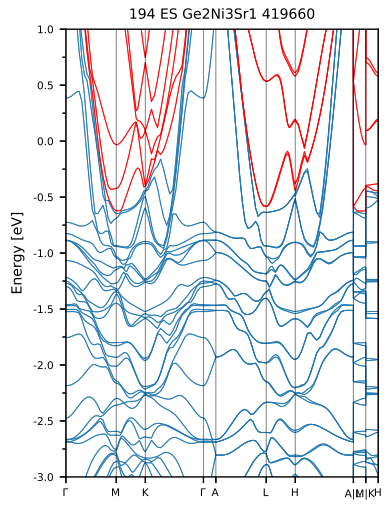

(bu) 194.3.16

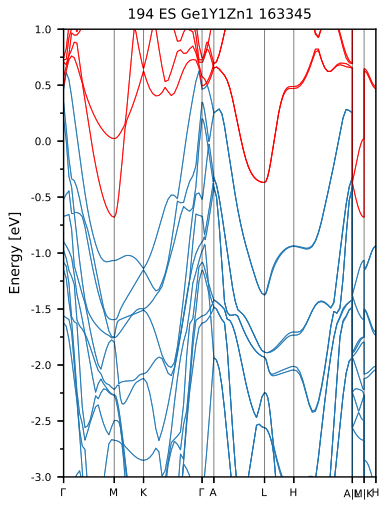

(bv) 194.3.17

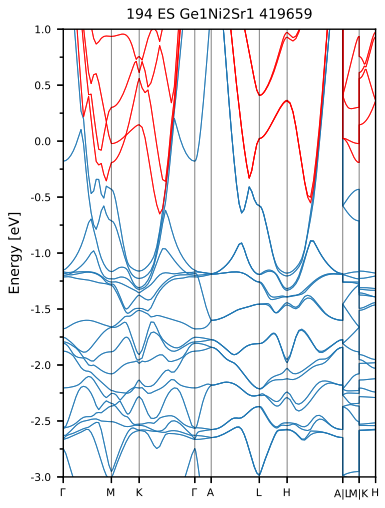

(bw) 194.3.18

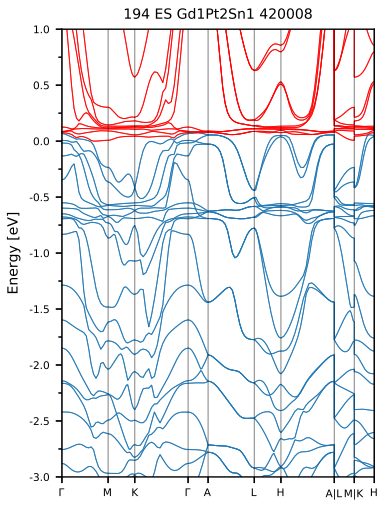

(bx) 194.3.19

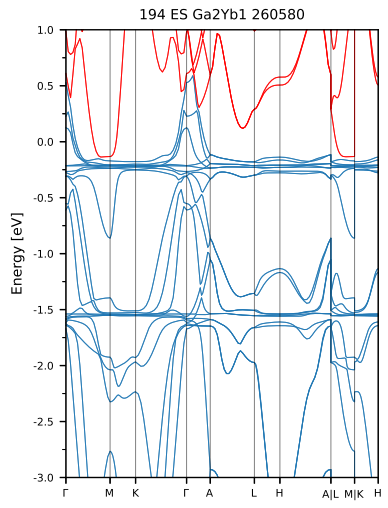

(by) 194.3.20

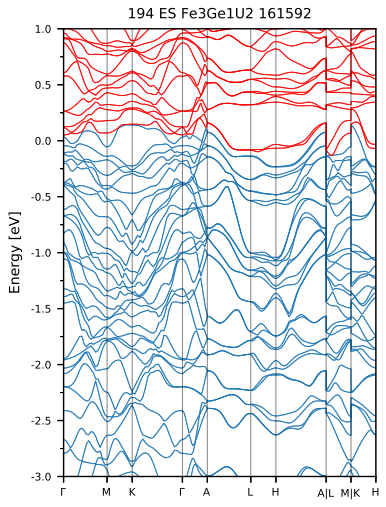

(bz) 194.3.21

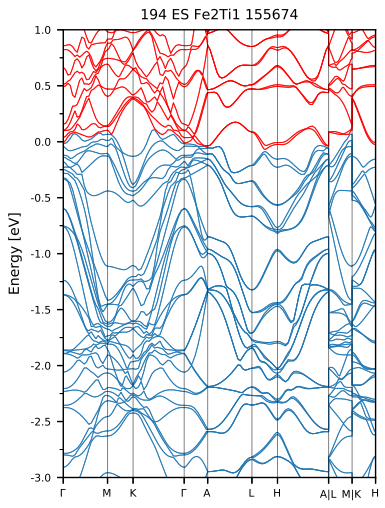

(ca) 194.3.22

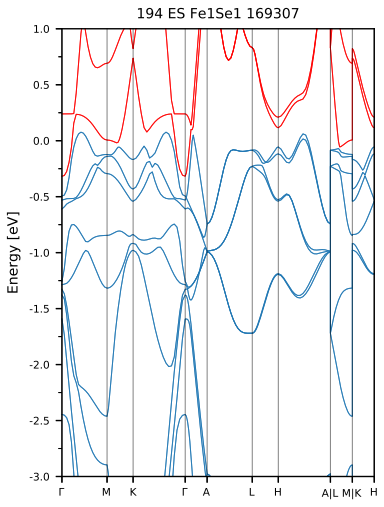

(cb) 194.3.23

Figure 456: Band structures for topological materials in SG 194, part 5/12, from 194.3.8 to 194.3.23. See the list of figures.

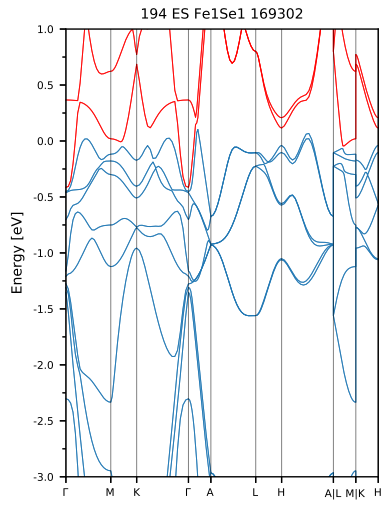

(cc) 194.3.24

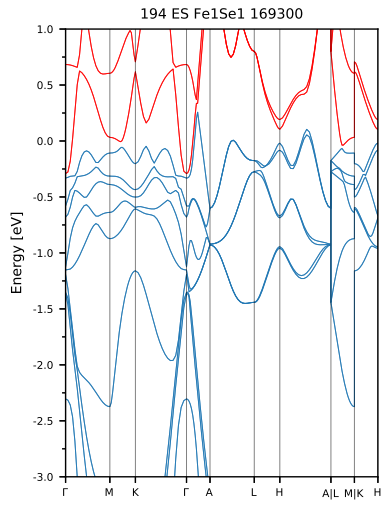

(cd) 194.3.25

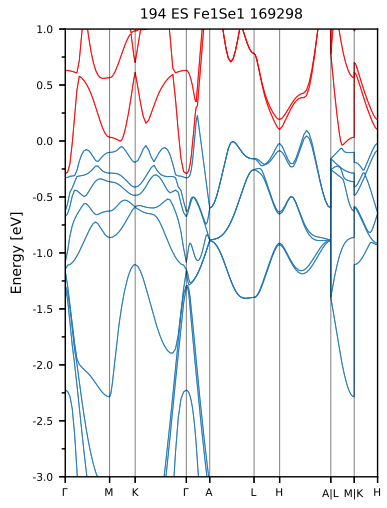

(ce) 194.3.26

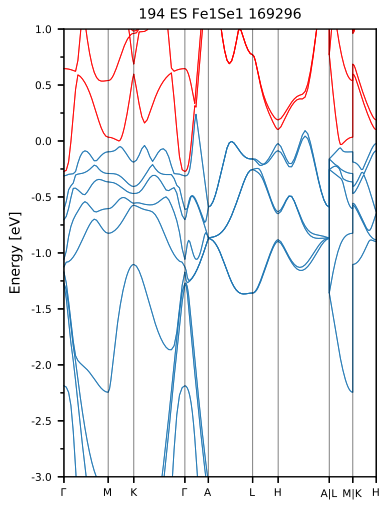

(cf) 194.3.27

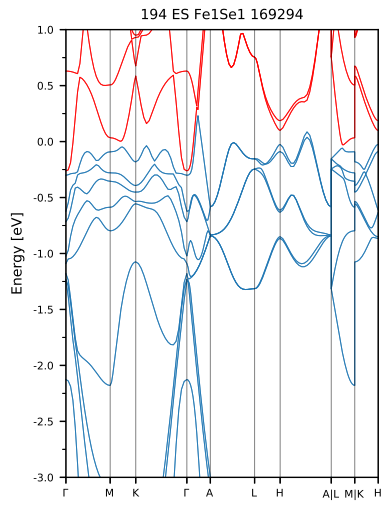

(cg) 194.3.28

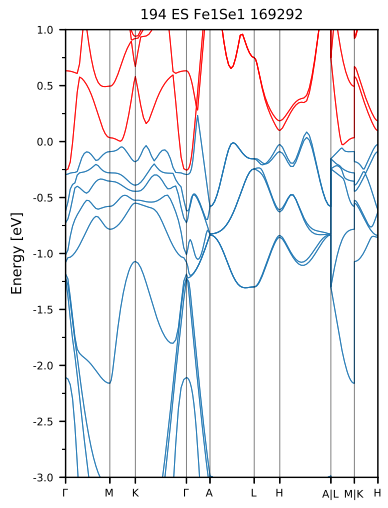

(ch) 194.3.29

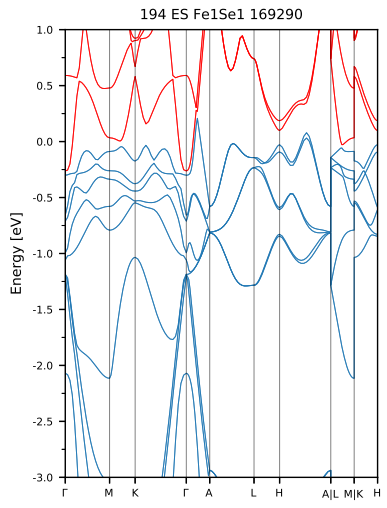

(ci) 194.3.30

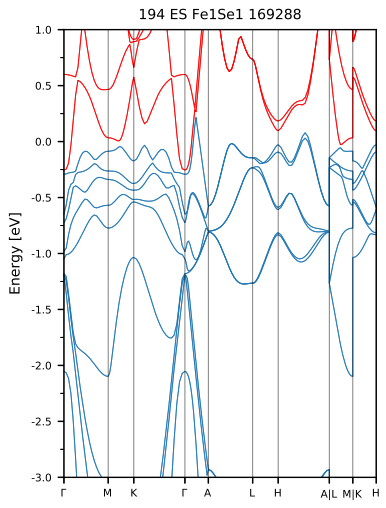

(cj) 194.3.31

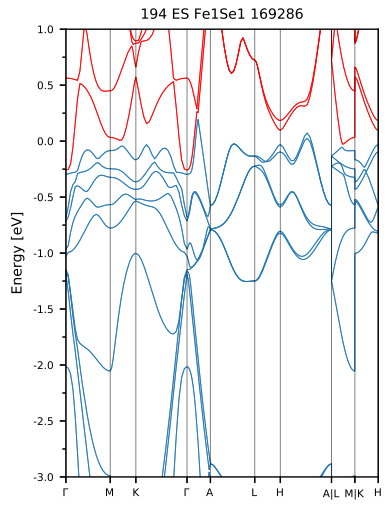

(ck) 194.3.32

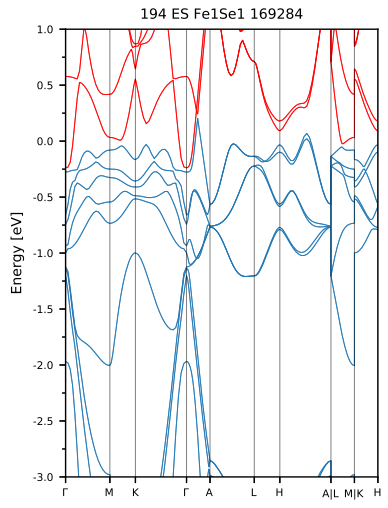

(cl) 194.3.33

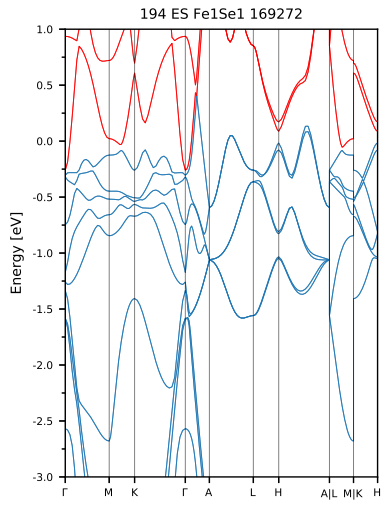

(cm) 194.3.34

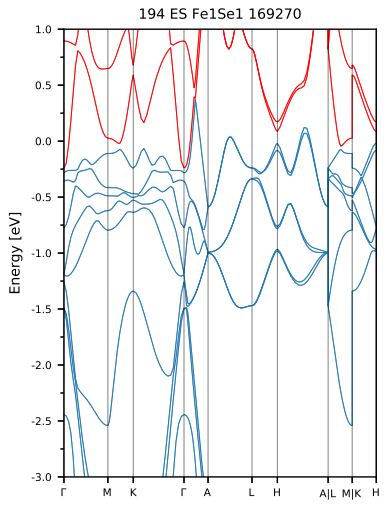

(cn) 194.3.35

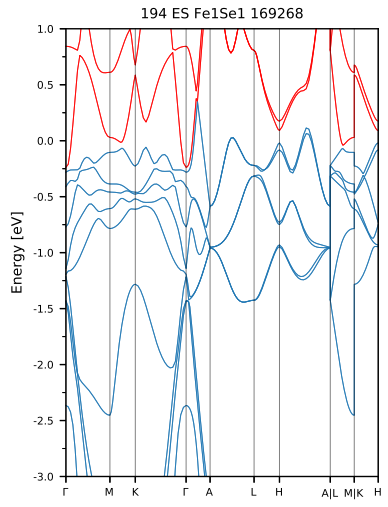

(co) 194.3.36

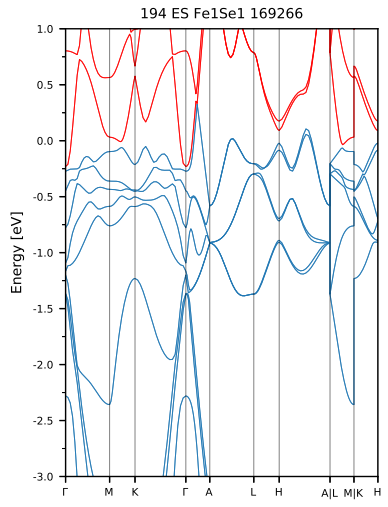

(cp) 194.3.37

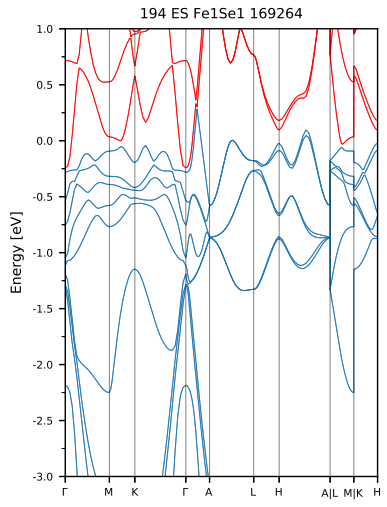

(cq) 194.3.38

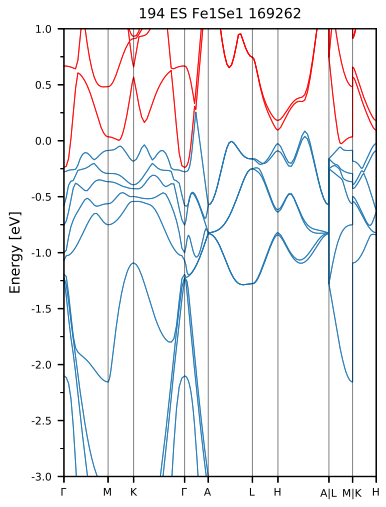

(cr) 194.3.39

Figure 457: Band structures for topological materials in SG 194, part 6/12, from 194.3.24 to 194.3.39. See the list of figures.

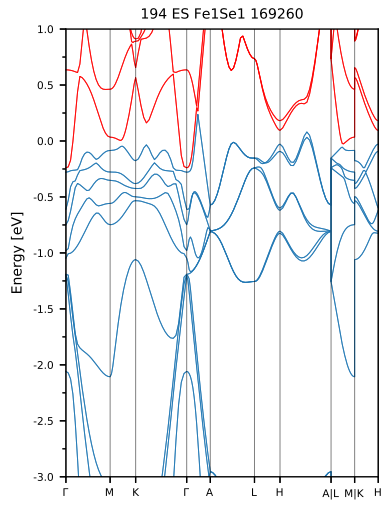

(cs) 194.3.40

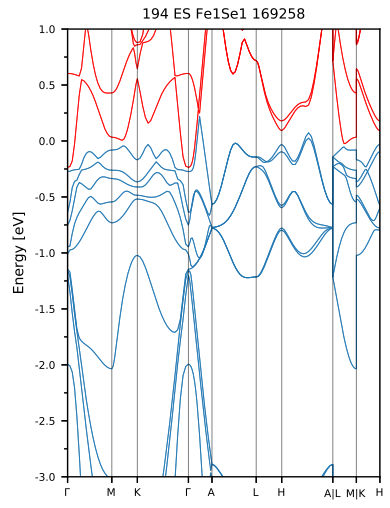

(ct) 194.3.41

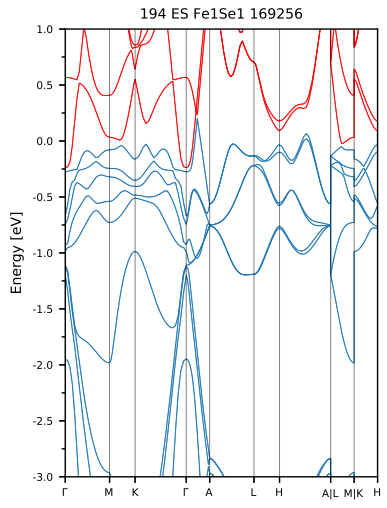

(cu) 194.3.42

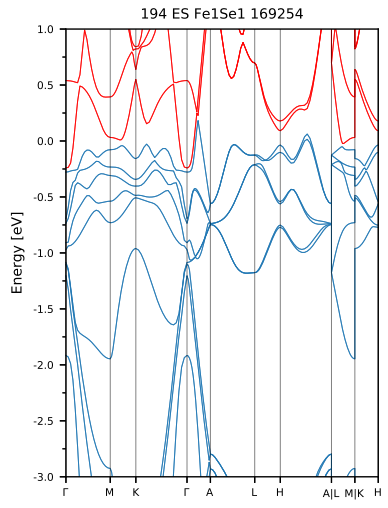

(cv) 194.3.43

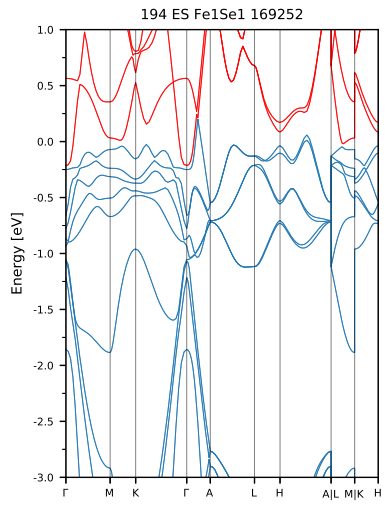

(cw) 194.3.44

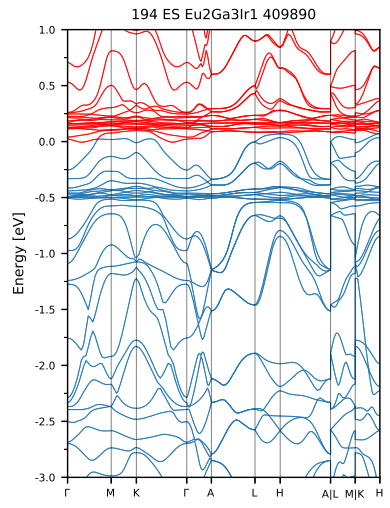

(cx) 194.3.45

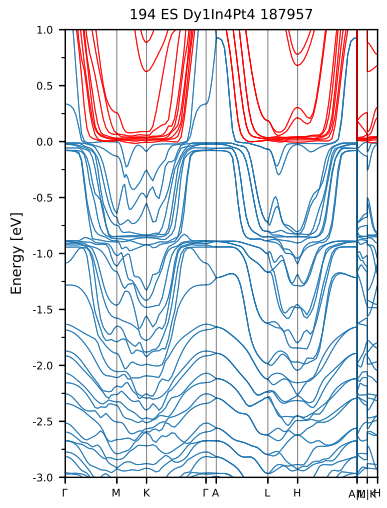

(cy) 194.3.46

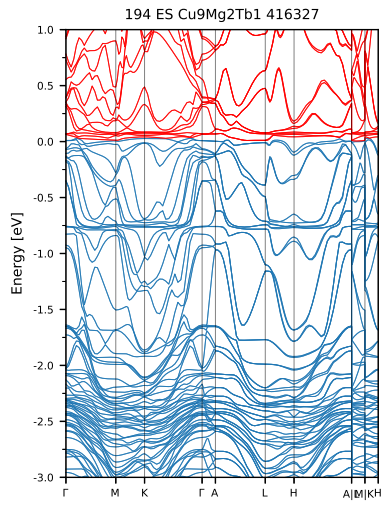

(cz) 194.3.47

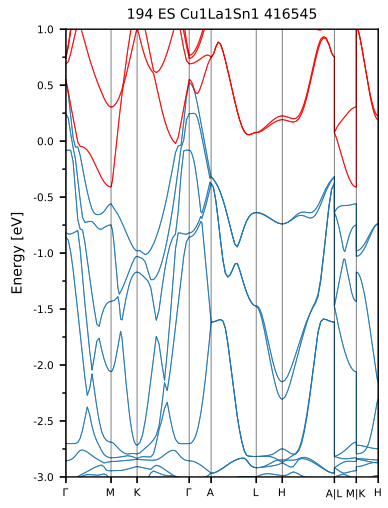

(da) 194.3.48

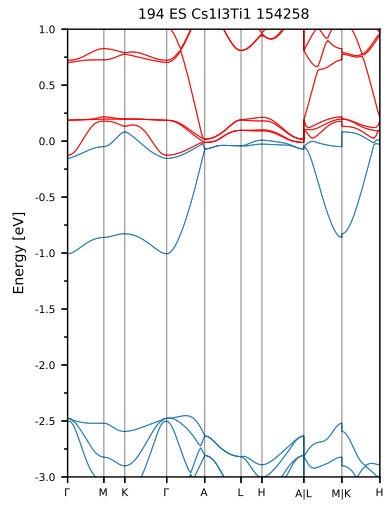

(db) 194.3.49

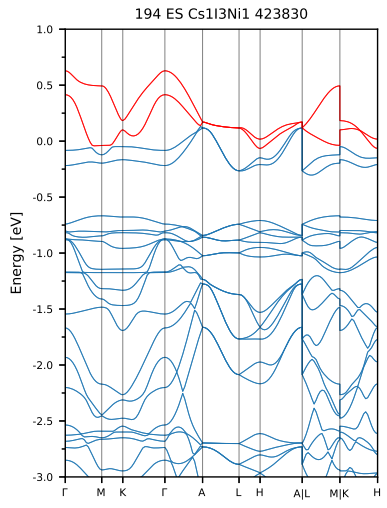

(dc) 194.3.50

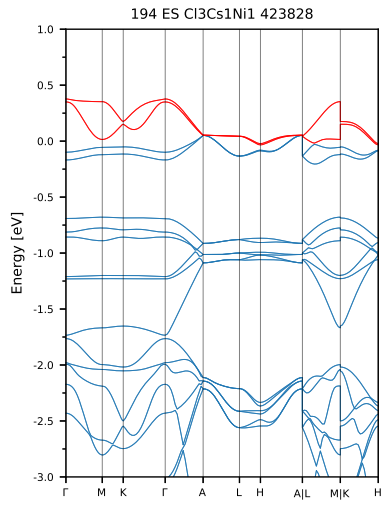

(dd) 194.3.51

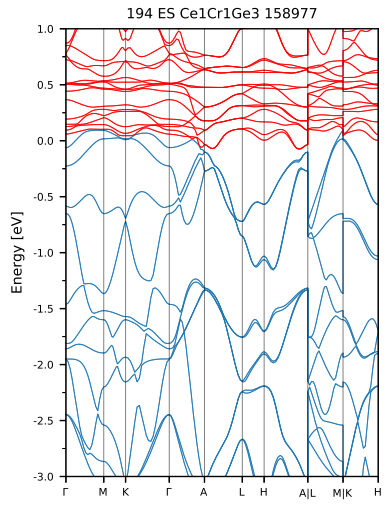

(de) 194.3.52

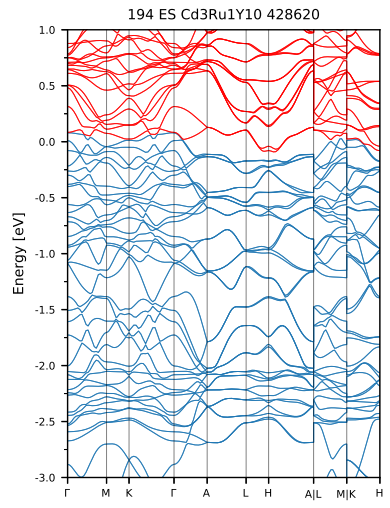

(df) 194.3.53

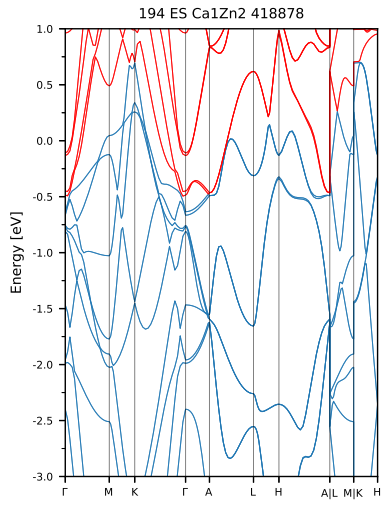

(dg) 194.3.54

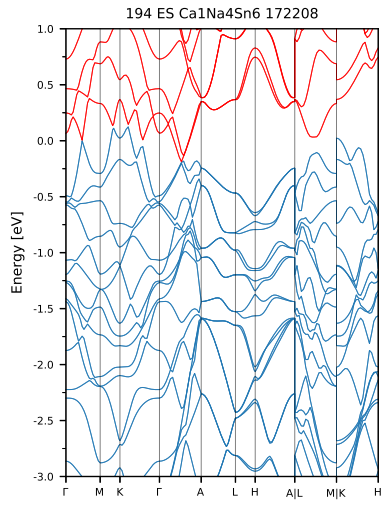

(dh) 194.3.55

Figure 458: Band structures for topological materials in SG 194, part 7/12, from 194.3.40 to 194.3.55. See the list of figures.

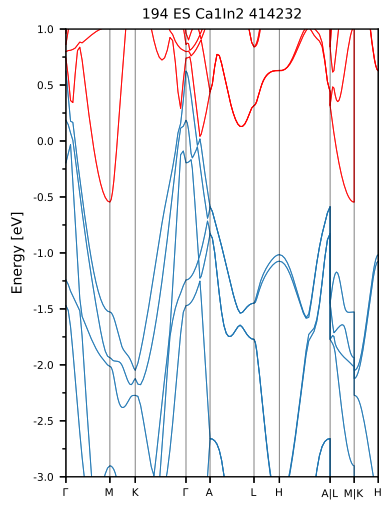

(di) 194.3.56

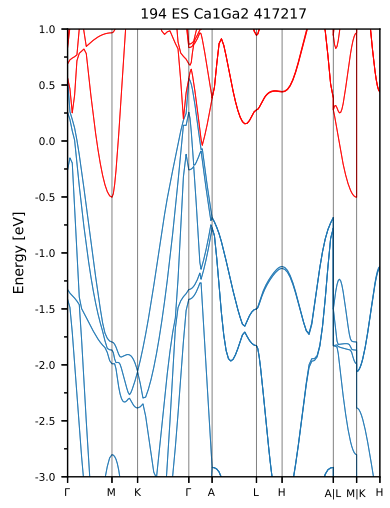

(dj) 194.3.57

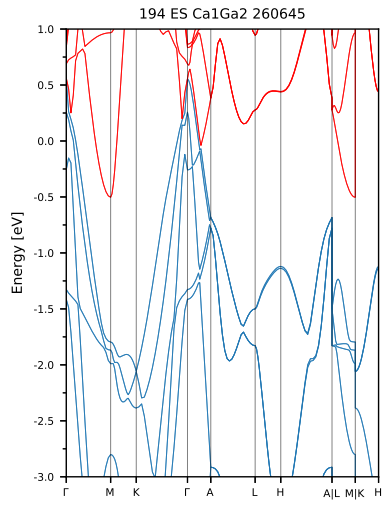

(dk) 194.3.58

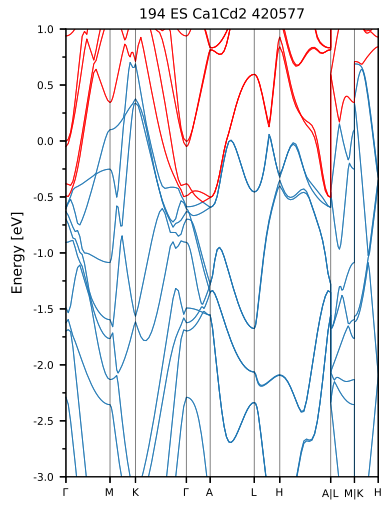

(dl) 194.3.59

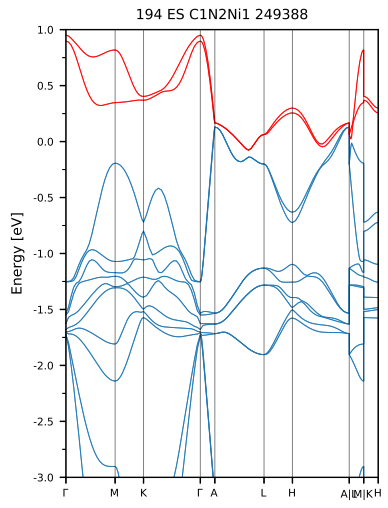

(dm) 194.3.60

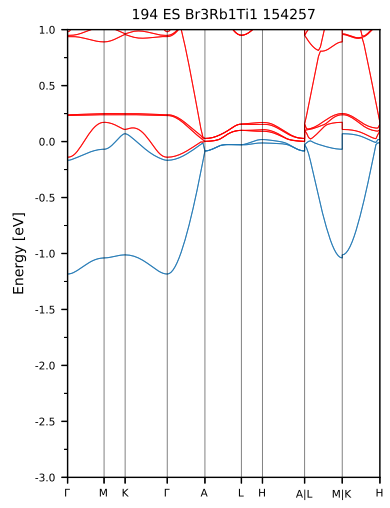

(dn) 194.3.61

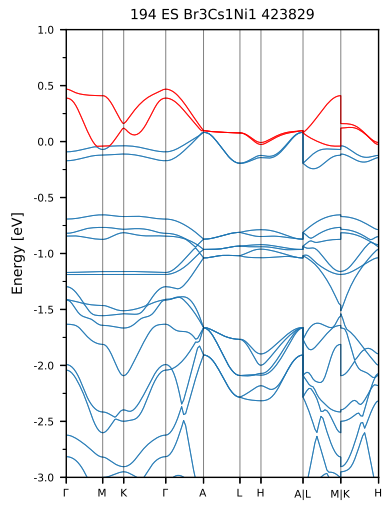

(do) 194.3.62

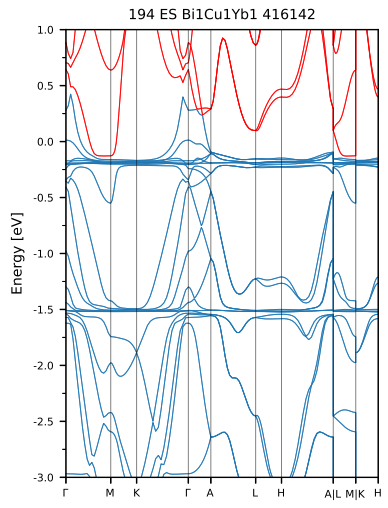

(dp) 194.3.63

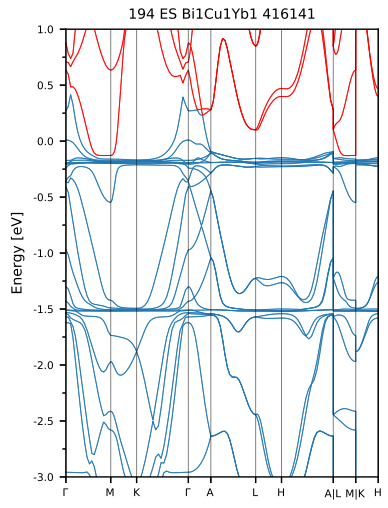

(dq) 194.3.64

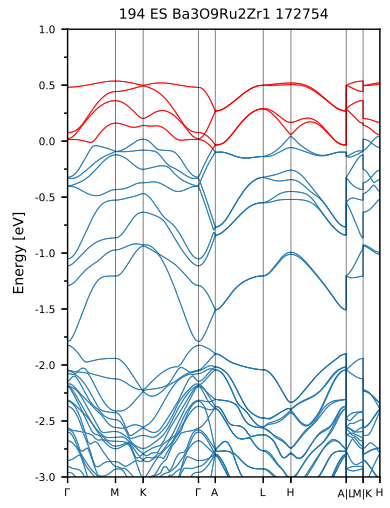

(dr) 194.3.65

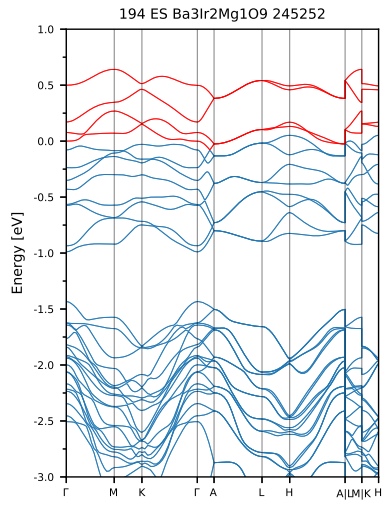

(ds) 194.3.66

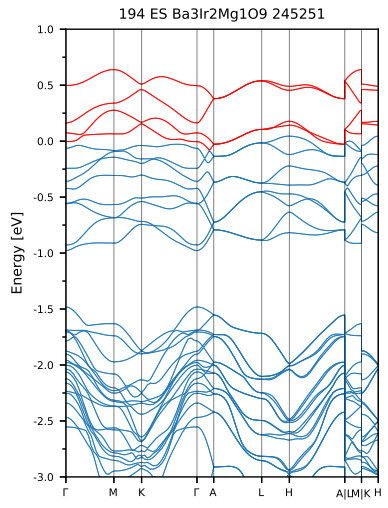

(dt) 194.3.67

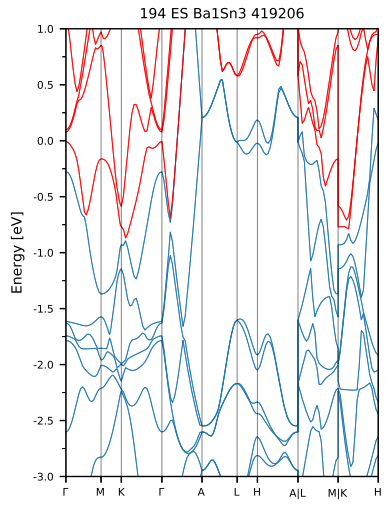

(du) 194.3.68

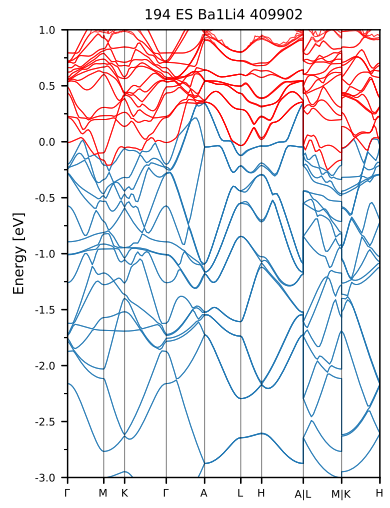

(dv) 194.3.69

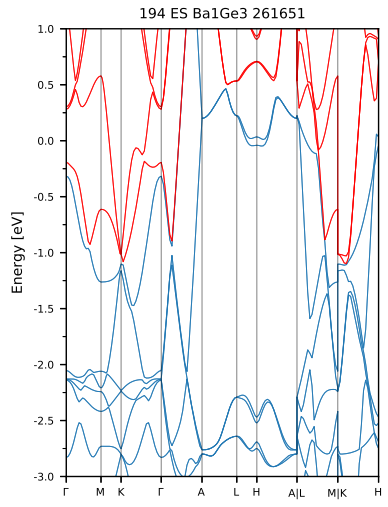

(dw) 194.3.70

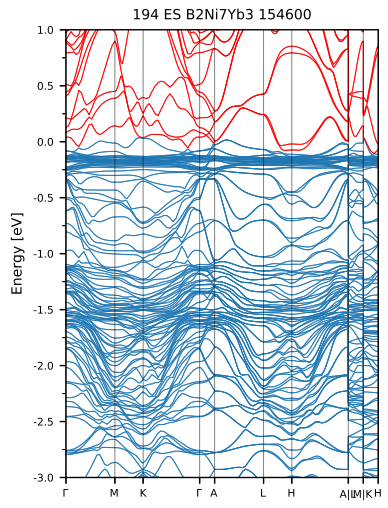

(dx) 194.3.71

Figure 459: Band structures for topological materials in SG 194, part 8/12, from 194.3.56 to 194.3.71. See the list of figures.

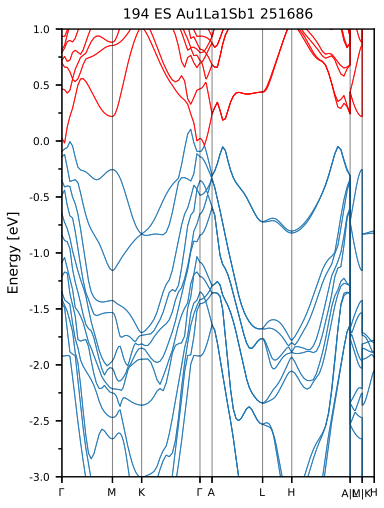

(dy) 194.3.72

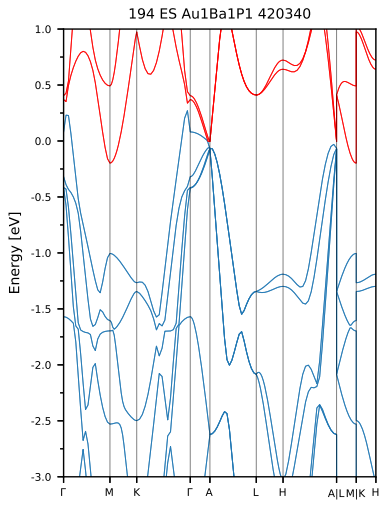

(dz) 194.3.73

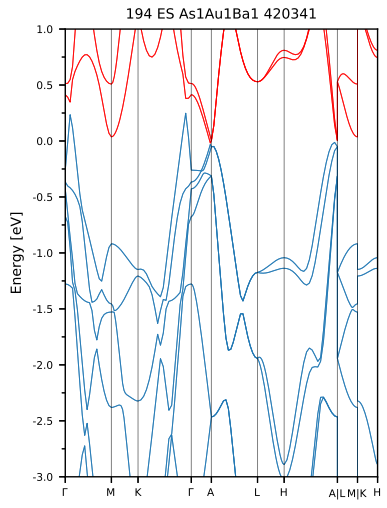

(ea) 194.3.74

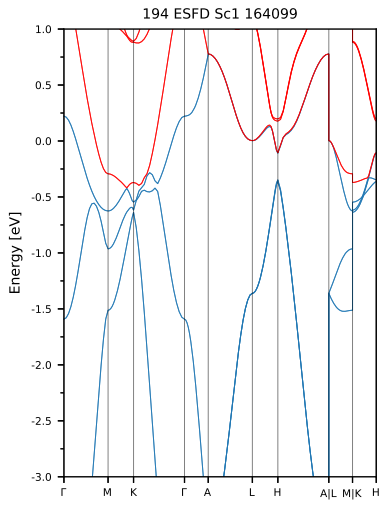

(eb) 194.4.1

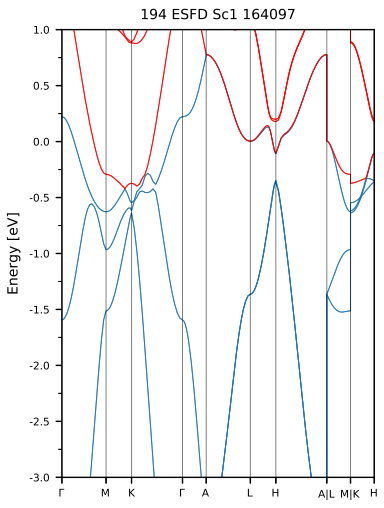

(ec) 194.4.2

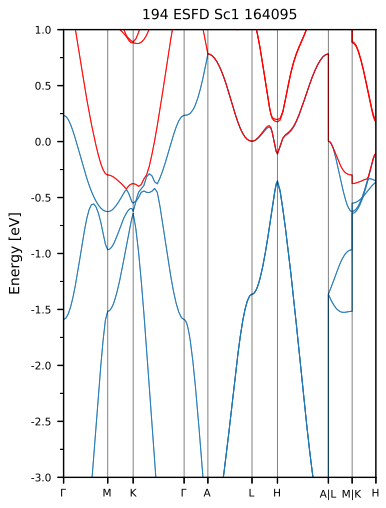

(ed) 194.4.3

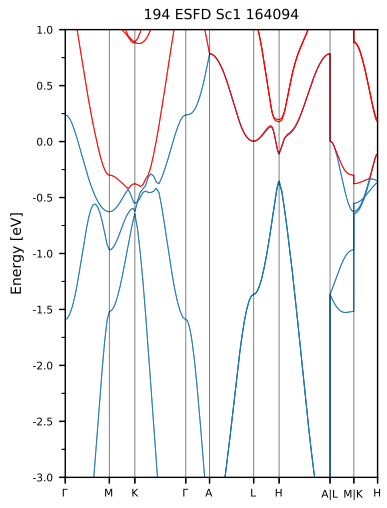

(ee) 194.4.4

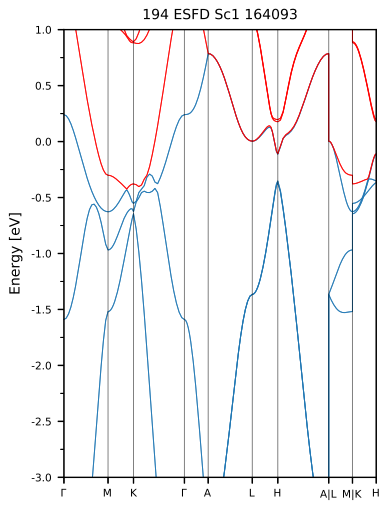

(ef) 194.4.5

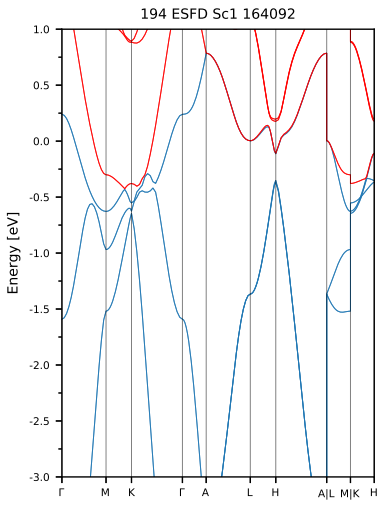

(eg) 194.4.6

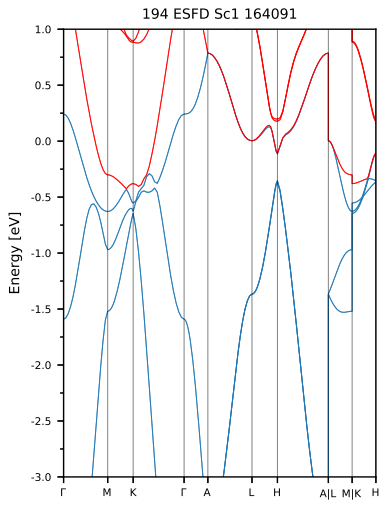

(eh) 194.4.7

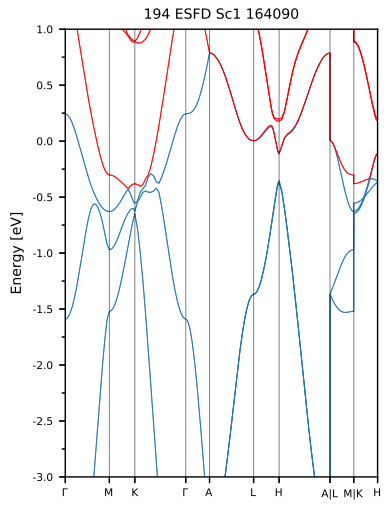

(ei) 194.4.8

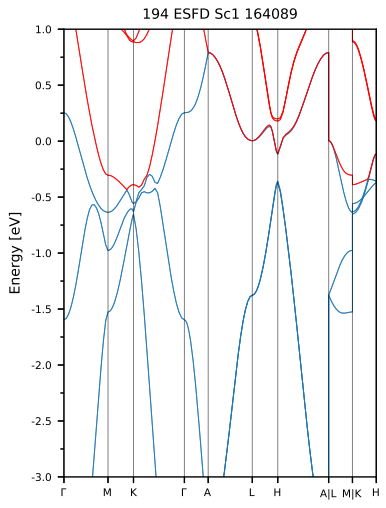

(ej) 194.4.9

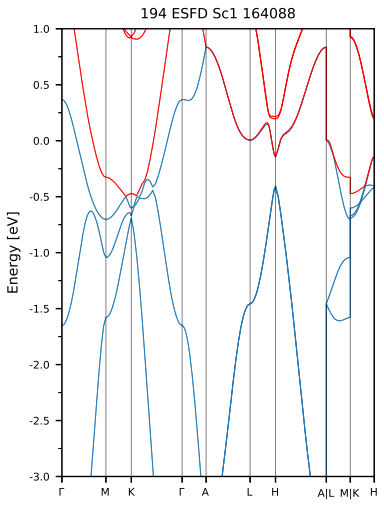

(ek) 194.4.10

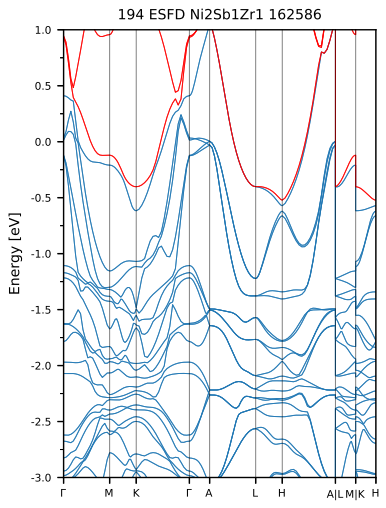

(el) 194.4.11

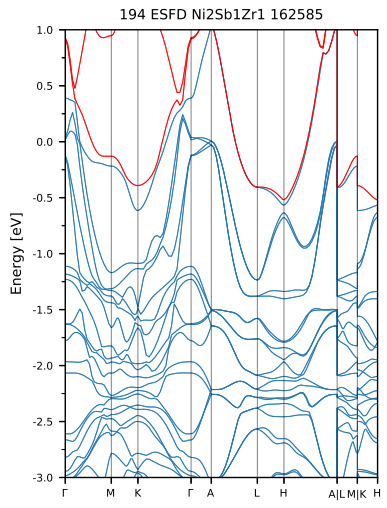

(em) 194.4.12

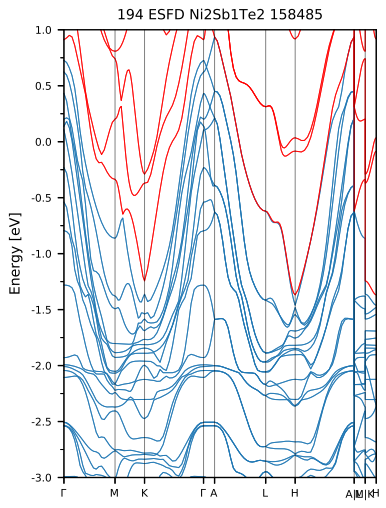

(en) 194.4.13

Figure 460: Band structures for topological materials in SG 194, part 9/12, from 194.3.72 to 194.4.13. See the list of figures.

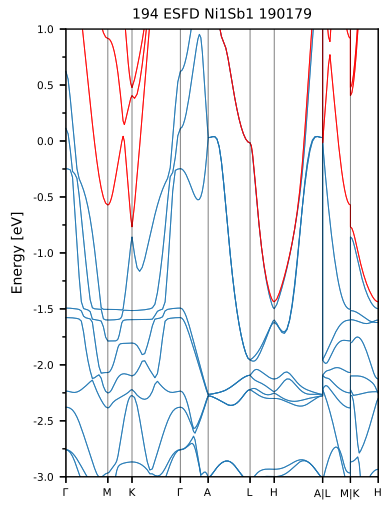

(eo) 194.4.14

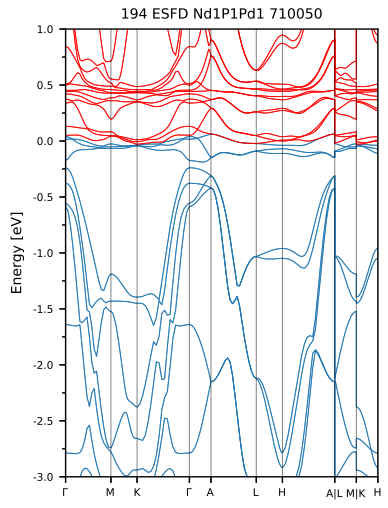

(ep) 194.4.15

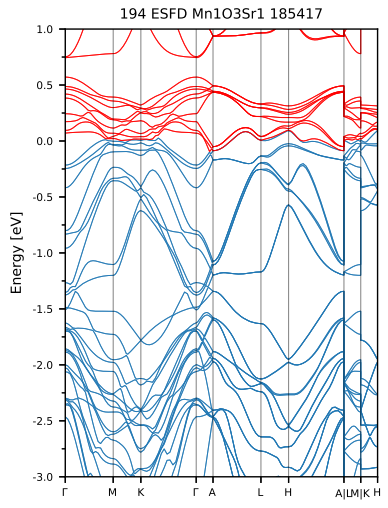

(eq) 194.4.16

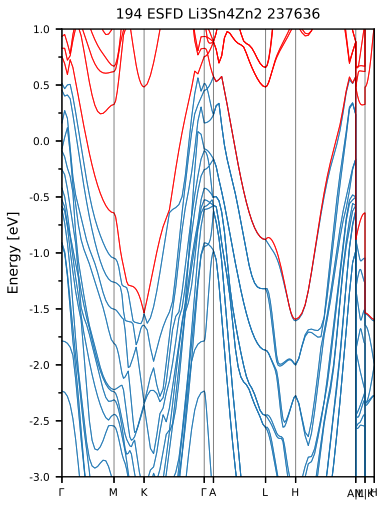

(er) 194.4.17

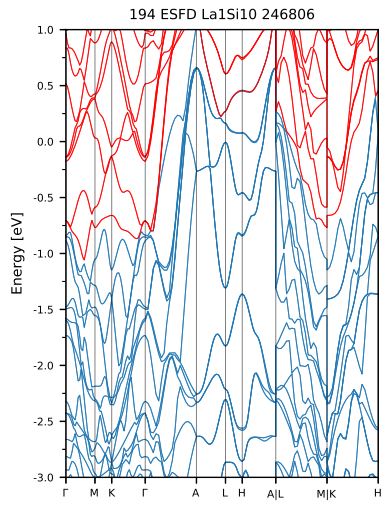

(es) 194.4.18

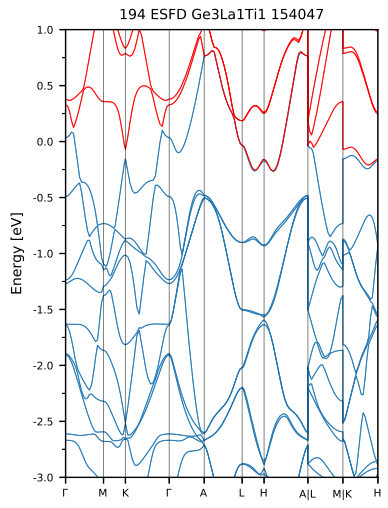

(et) 194.4.19

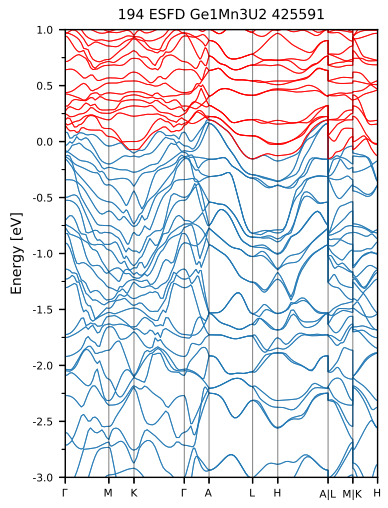

(eu) 194.4.20

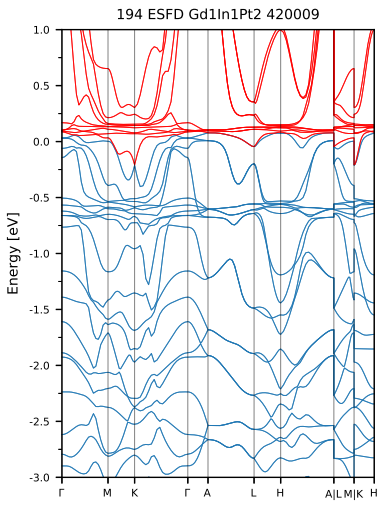

(ev) 194.4.21

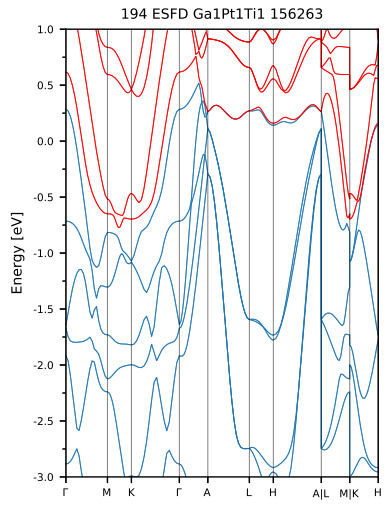

(ew) 194.4.22

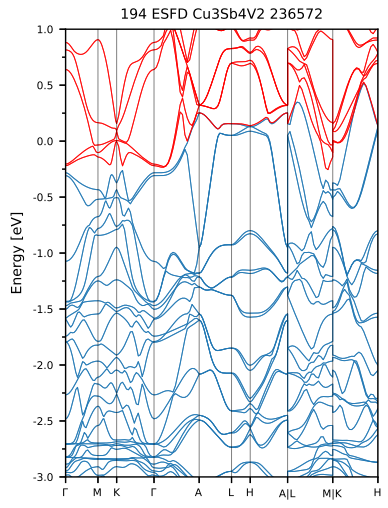

(ex) 194.4.23

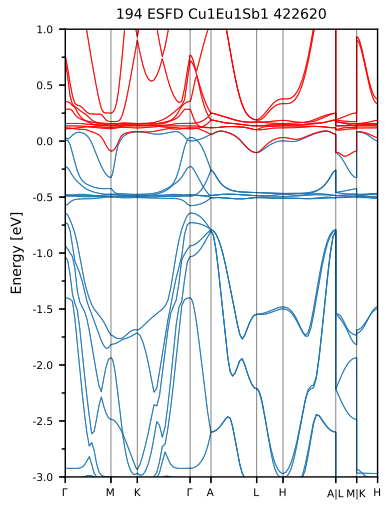

(ey) 194.4.24

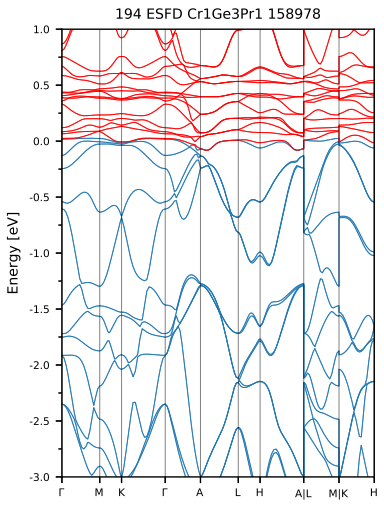

(ez) 194.4.25

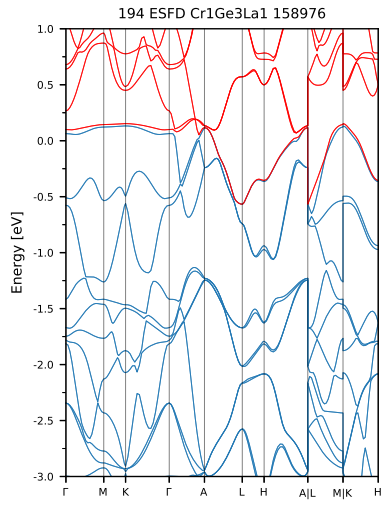

(fa) 194.4.26

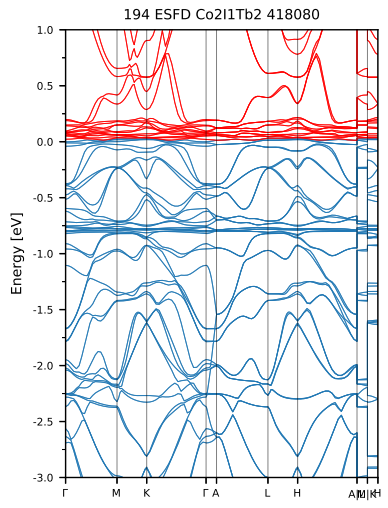

(fb) 194.4.27

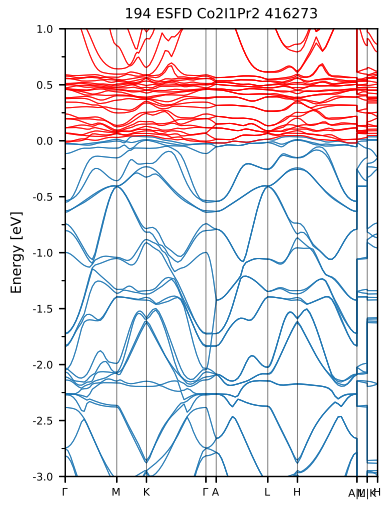

(fc) 194.4.28

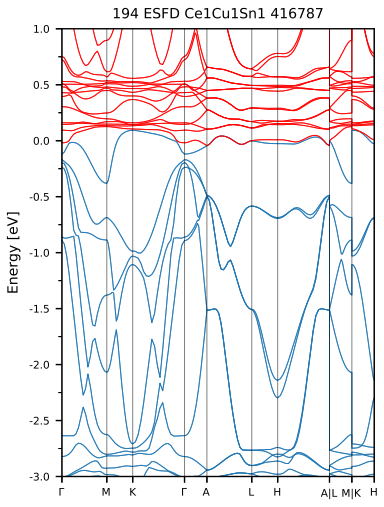

(fd) 194.4.29

Figure 461: Band structures for topological materials in SG 194, part 10/12, from 194.4.14 to 194.4.29. See the list of figures.

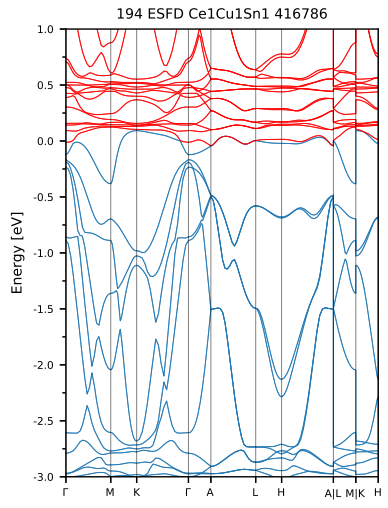

(fe) 194.4.30

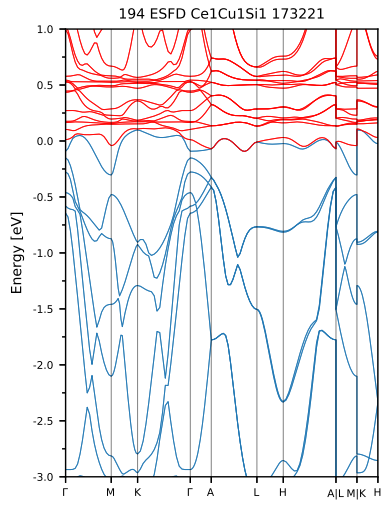

(ff) 194.4.31

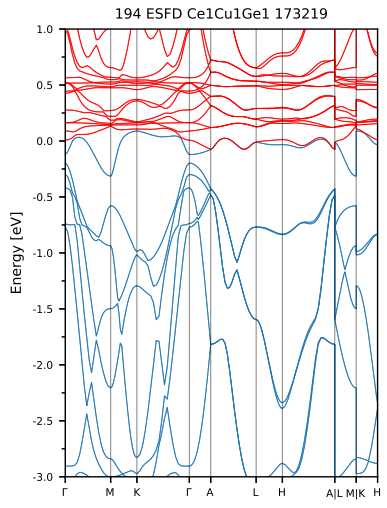

(fg) 194.4.32

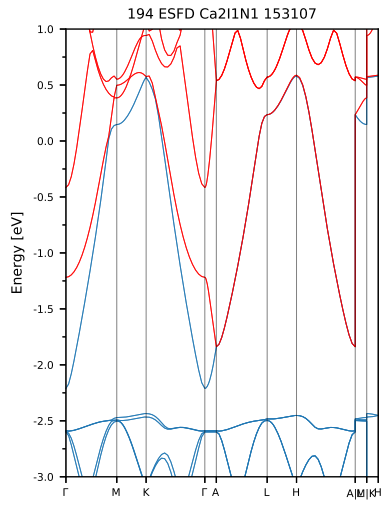

(fh) 194.4.33

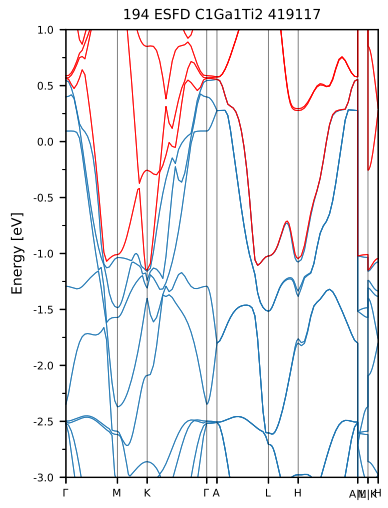

(fi) 194.4.34

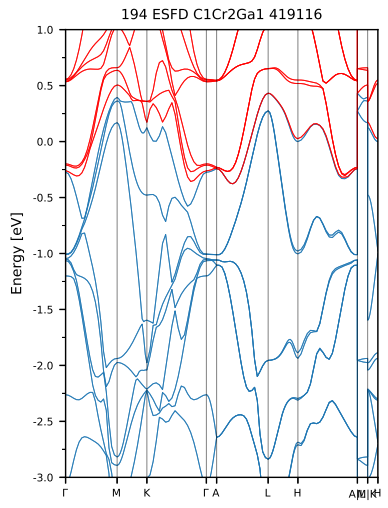

(fj) 194.4.35

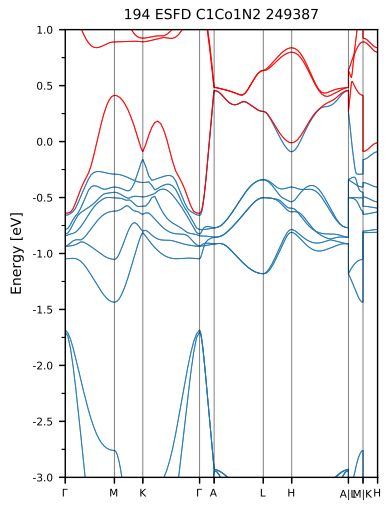

(fk) 194.4.36

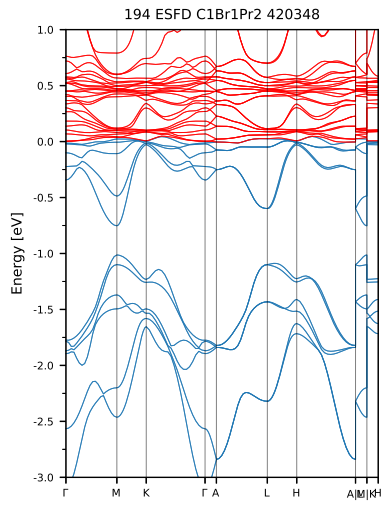

(fl) 194.4.37

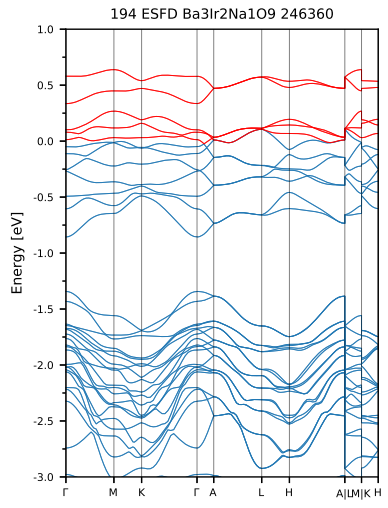

(fm) 194.4.38

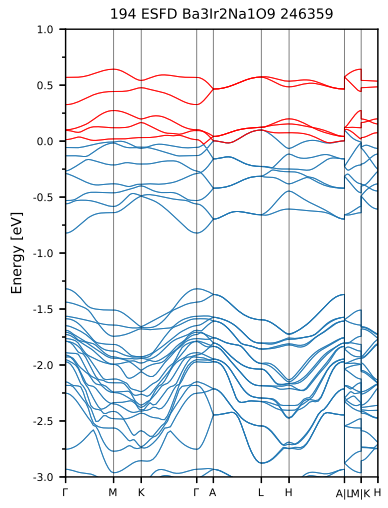

(fn) 194.4.39

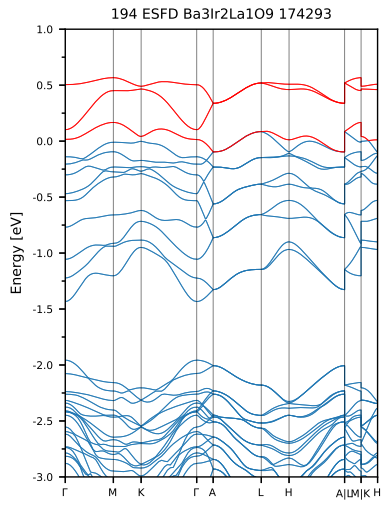

(fo) 194.4.40

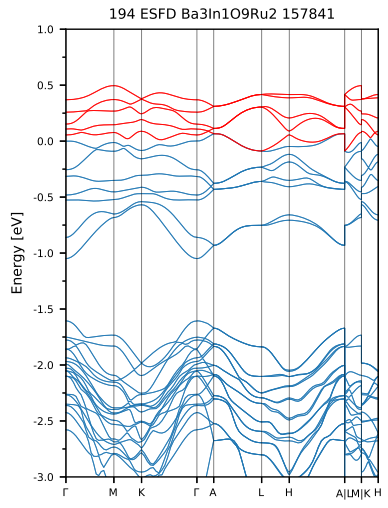

(fp) 194.4.41

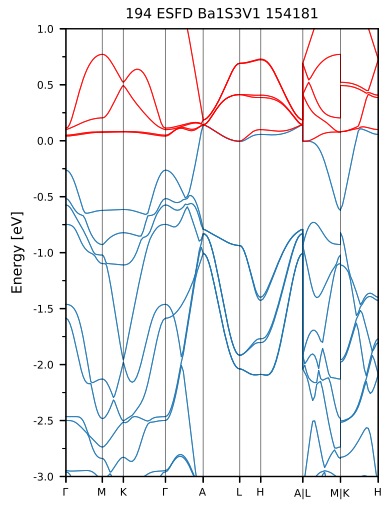

(fq) 194.4.42

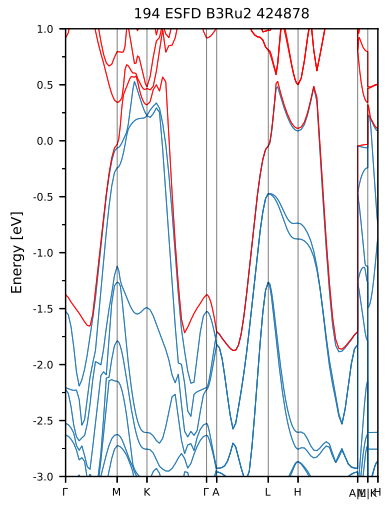

(fr) 194.4.43

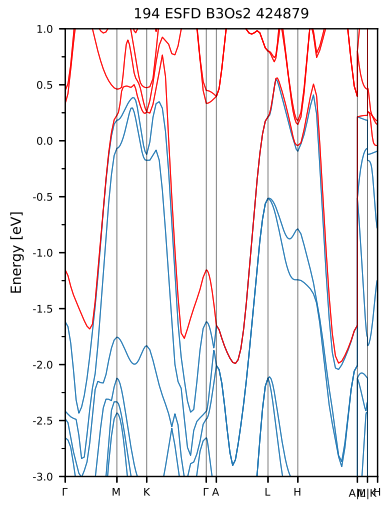

(fs) 194.4.44

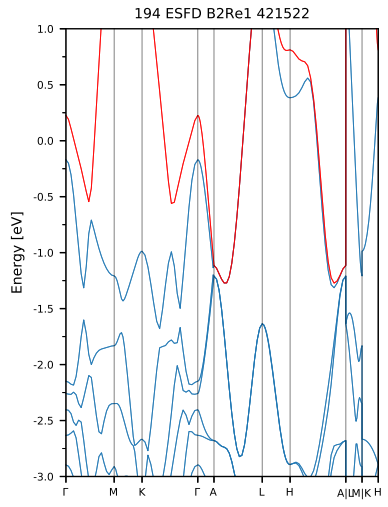

(ft) 194.4.45

Figure 462: Band structures for topological materials in SG 194, part 11/12, from 194.4.30 to 194.4.45. See the list of figures.

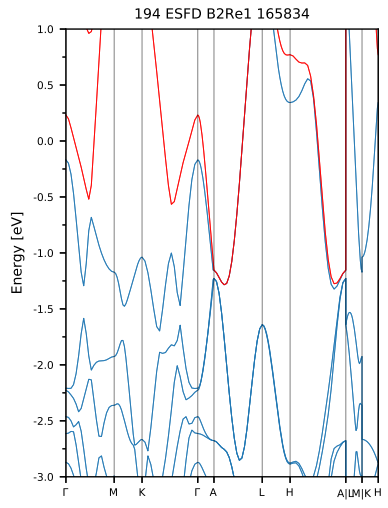

(fu) 194.4.46

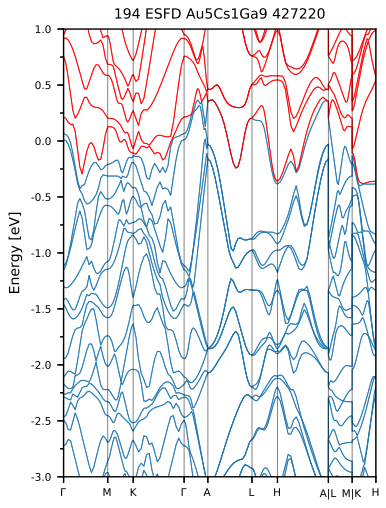

(fv) 194.4.47

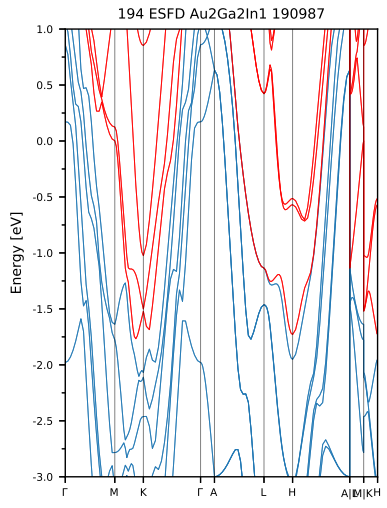

(fw) 194.4.48

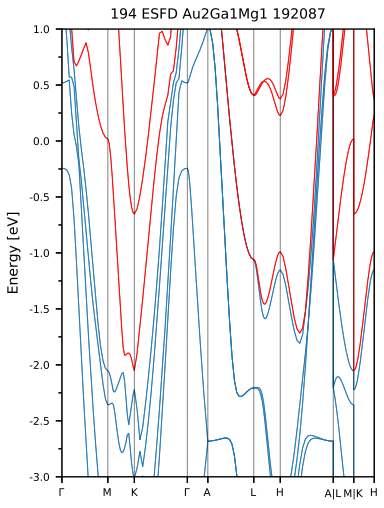

(fx) 194.4.49

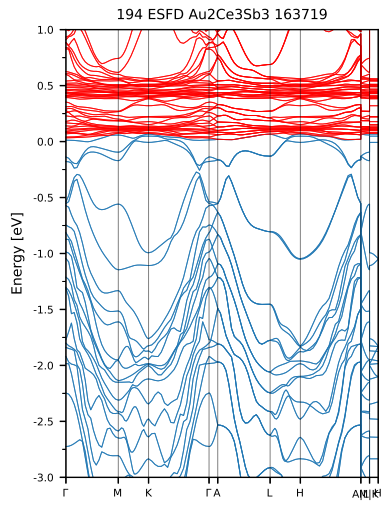

(fy) 194.4.50

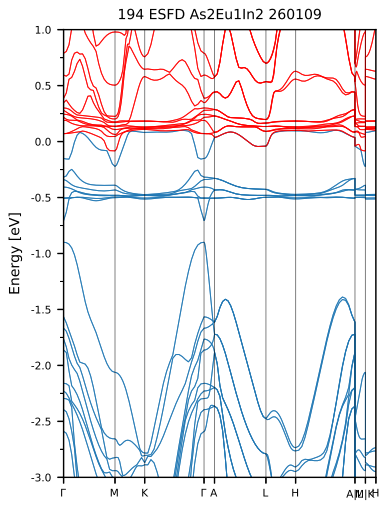

(fz) 194.4.51

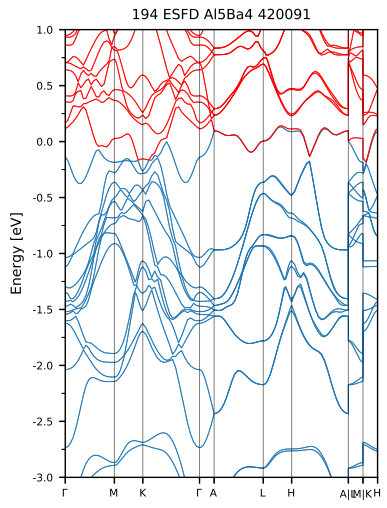

(ga) 194.4.52

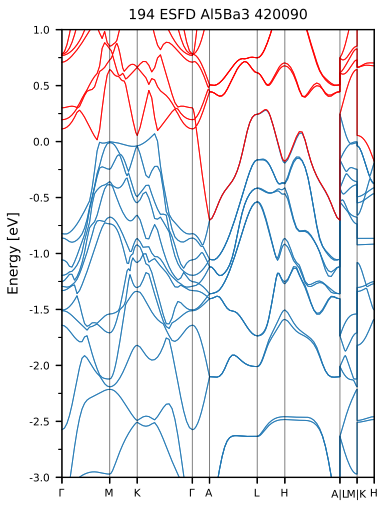

(gb) 194.4.53

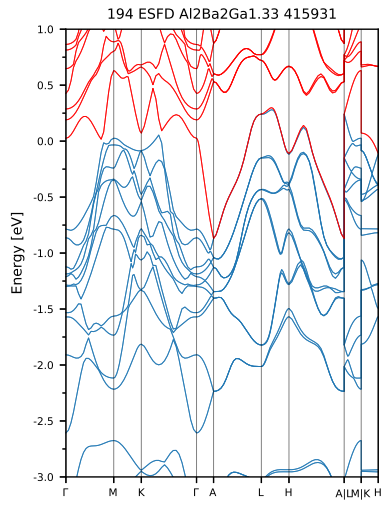

(gc) 194.4.54

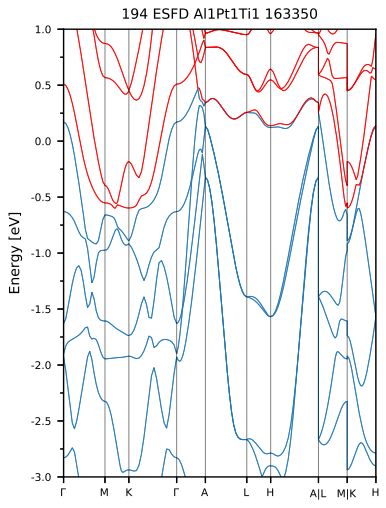

(gd) 194.4.55

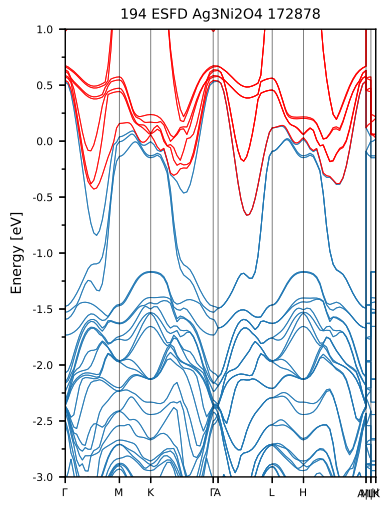

(ge) 194.4.56

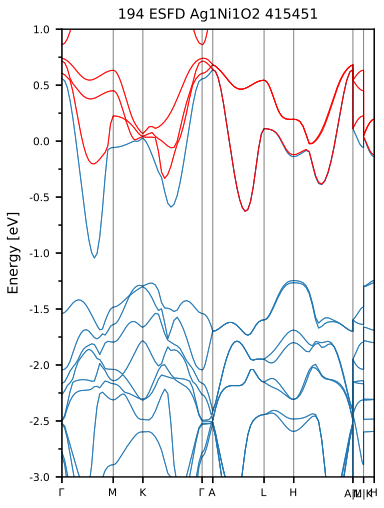

(gf) 194.4.57

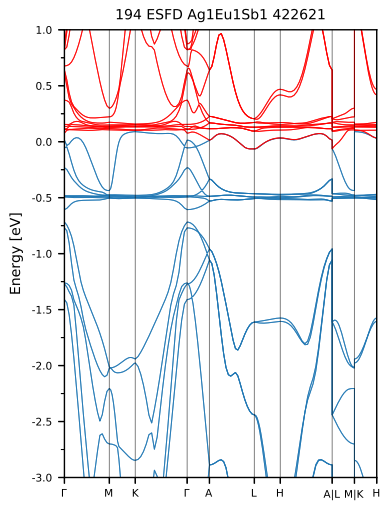

(gg) 194.4.58

Figure 463: Band structures for topological materials in SG 194, part 12/12, from 194.4.46 to 194.4.58. See the list of figures.

### 0.0.99 SG 197

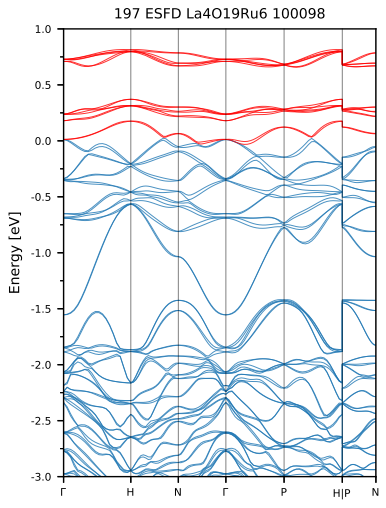

(a) 197.4.1

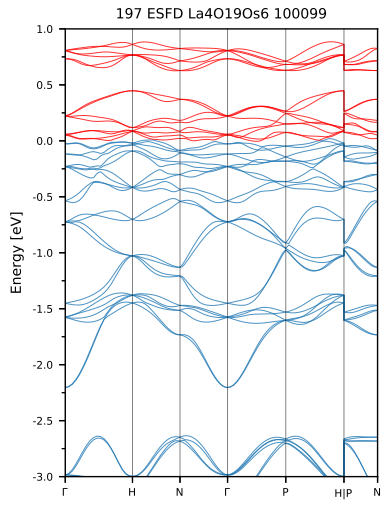

(b) 197.4.2

Figure 464: Band structures for topological materials in SG 197, part 1/1, from 197.4.1 to 197.4.2. See the list of figures.

## 0.0.100 SG 198

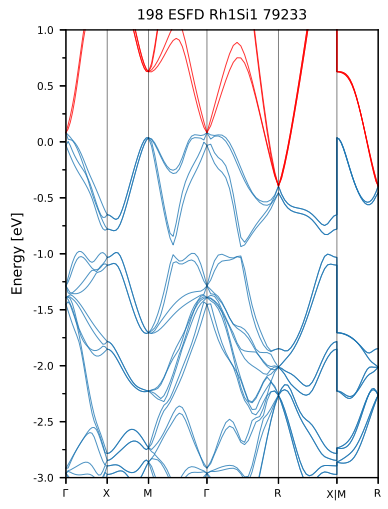

(a) 198.4.1

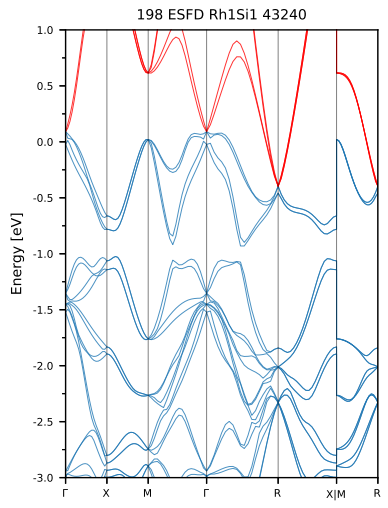

(b) 198.4.2

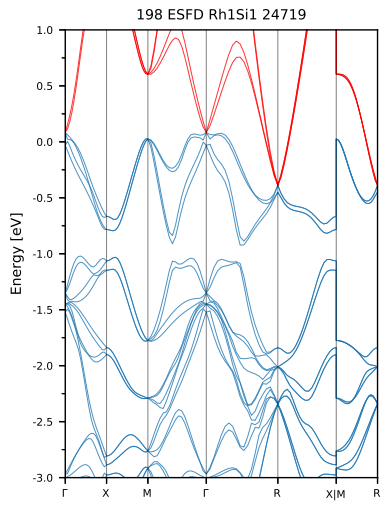

(c) 198.4.3

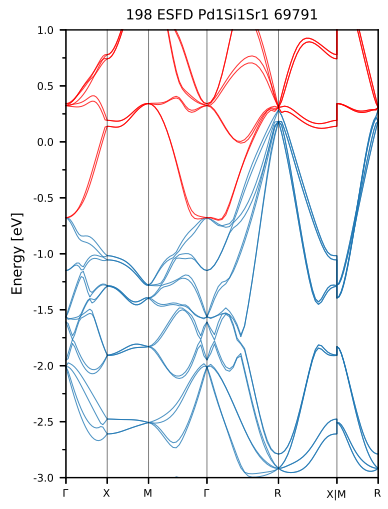

(d) 198.4.4

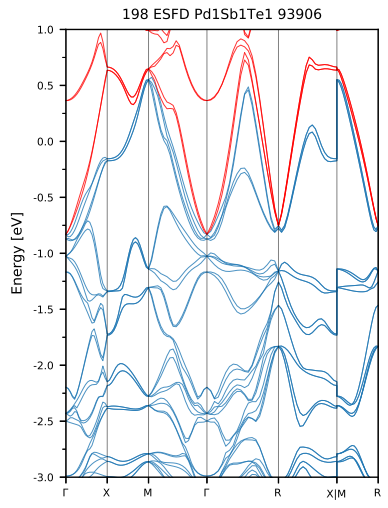

(e) 198.4.5

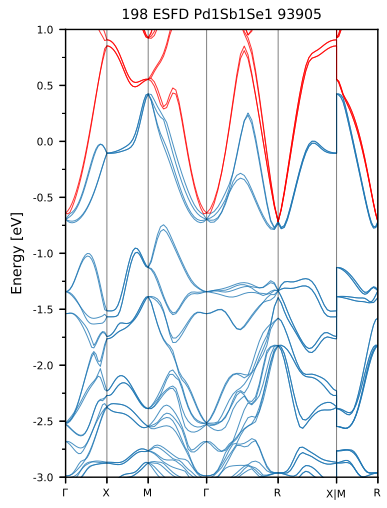

(f) 198.4.6

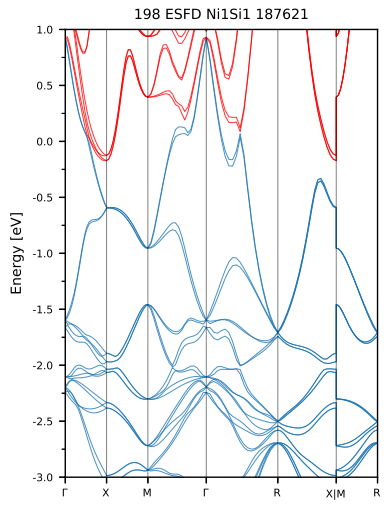

(g) 198.4.7

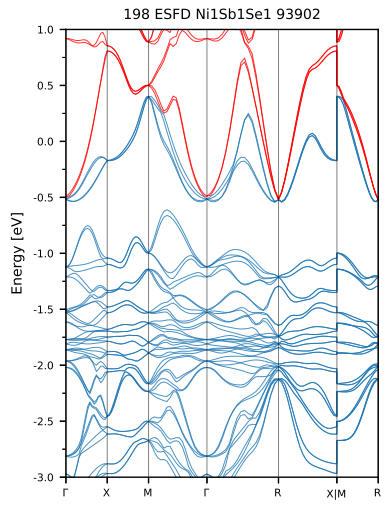

(h) 198.4.8

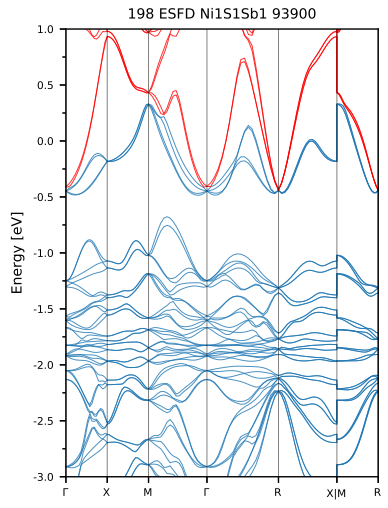

(i) 198.4.9

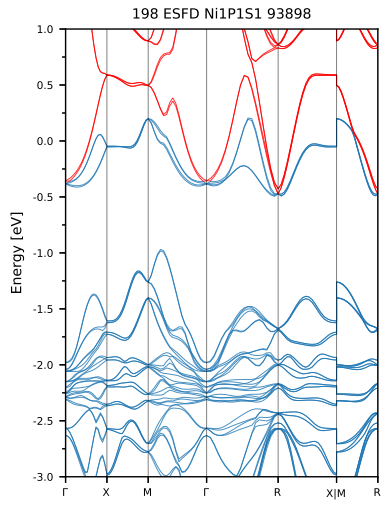

(j) 198.4.10

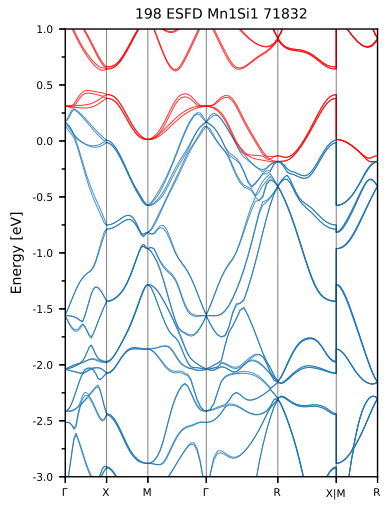

(k) 198.4.11

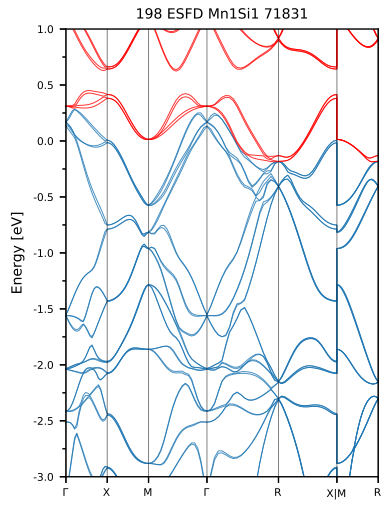

(l) 198.4.12

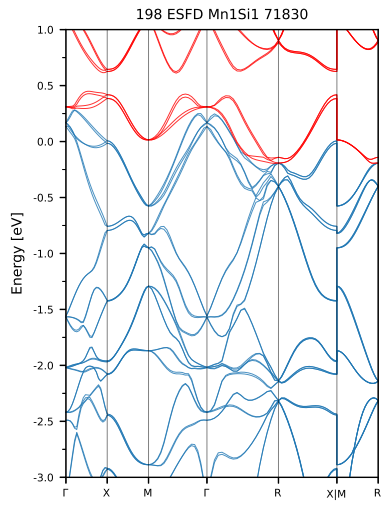

(m) 198.4.13

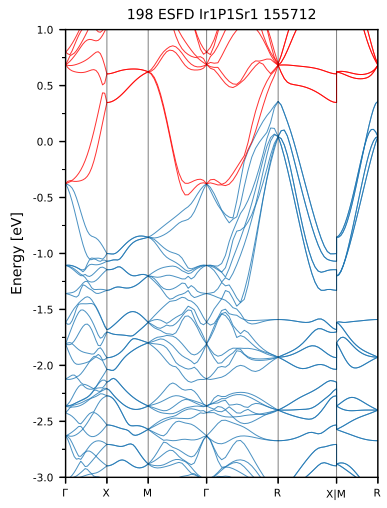

(n) 198.4.14

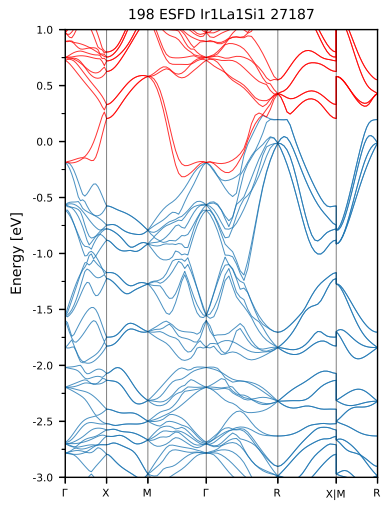

(o) 198.4.15

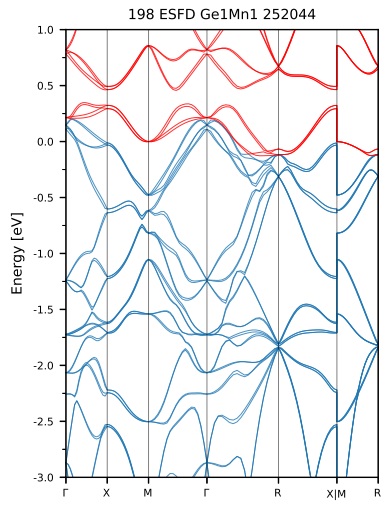

(p) 198.4.16

Figure 465: Band structures for topological materials in SG 198, part 1/3, from 198.4.1 to 198.4.16. See the list of figures.

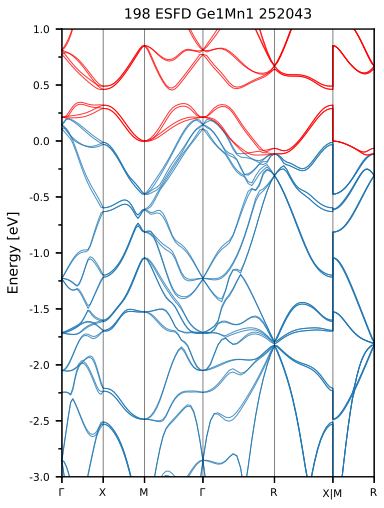

(q) 198.4.17

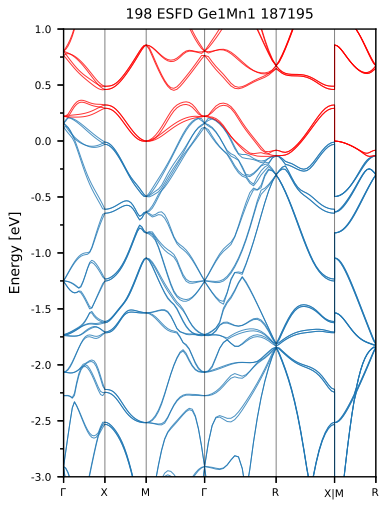

(r) 198.4.18

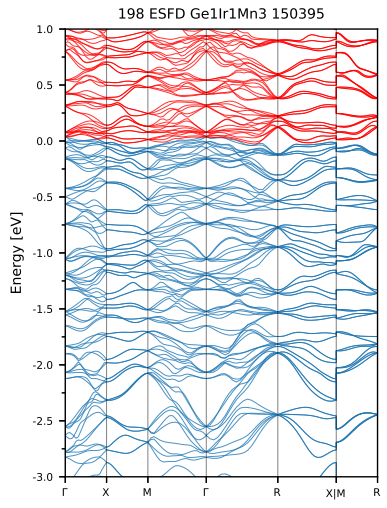

(s) 198.4.19

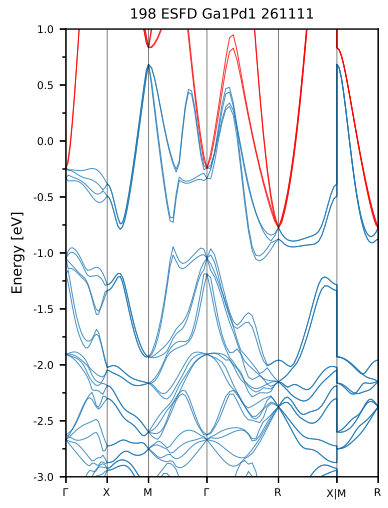

(t) 198.4.20

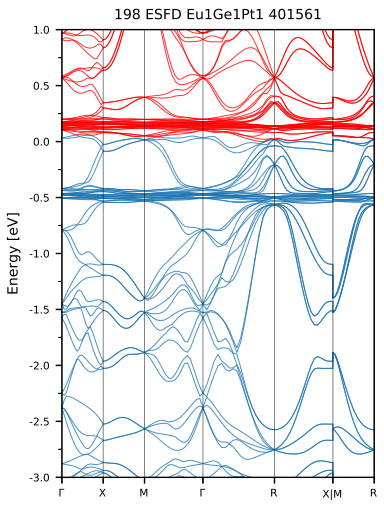

(u) 198.4.21

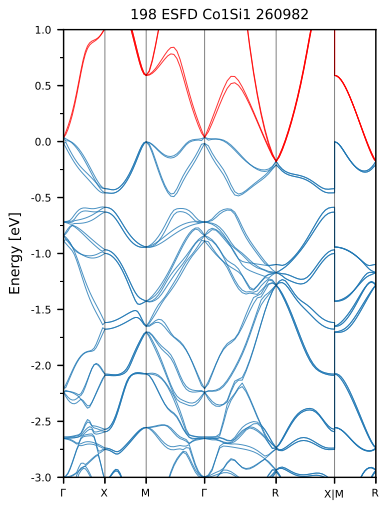

(v) 198.4.22

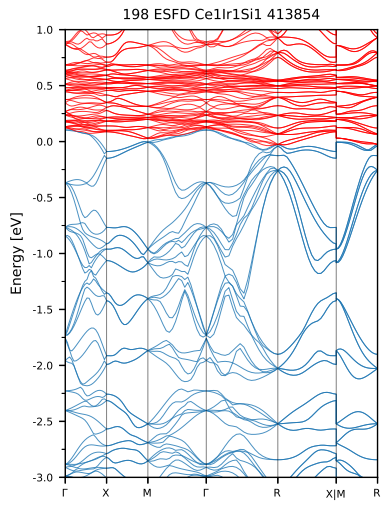

(w) 198.4.23

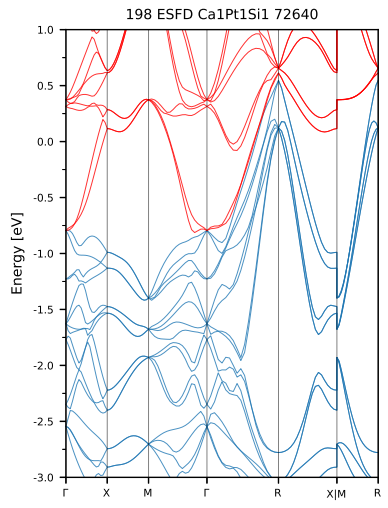

(x) 198.4.24

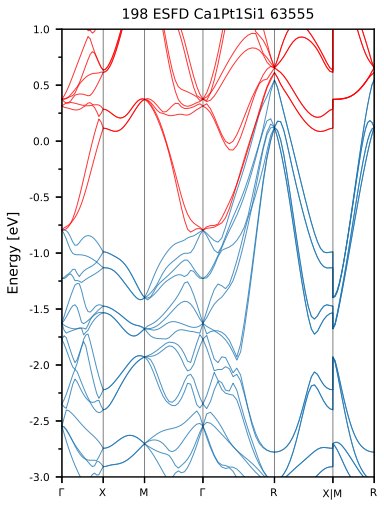

(y) 198.4.25

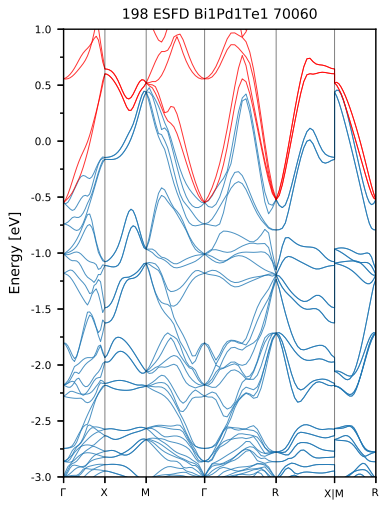

(z) 198.4.26

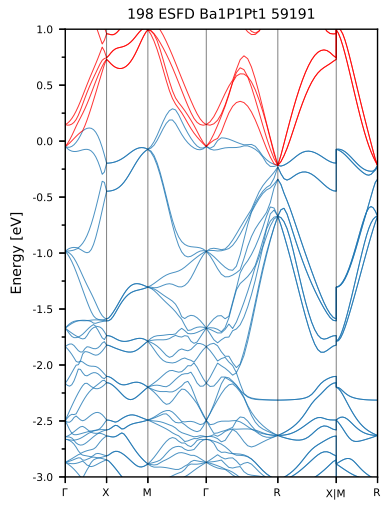

(aa) 198.4.27

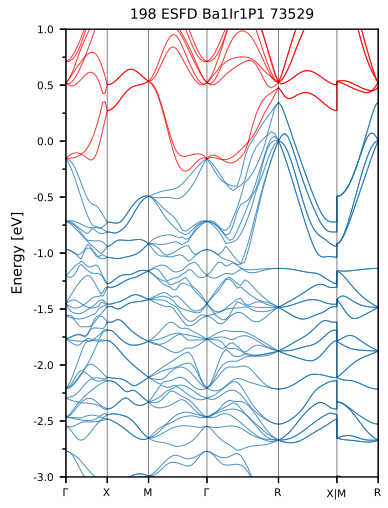

(ab) 198.4.28

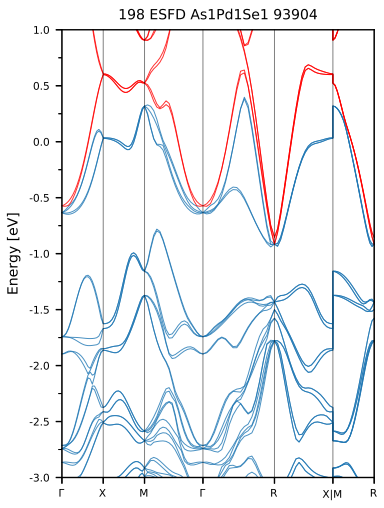

(ac) 198.4.29

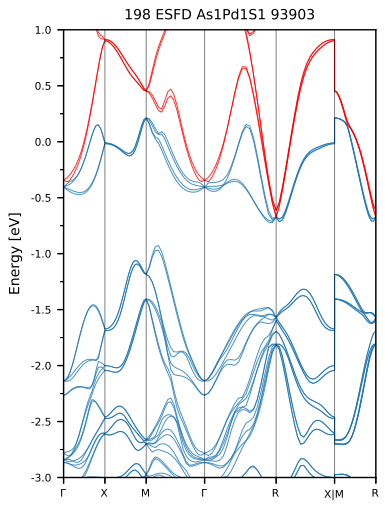

(ad) 198.4.30

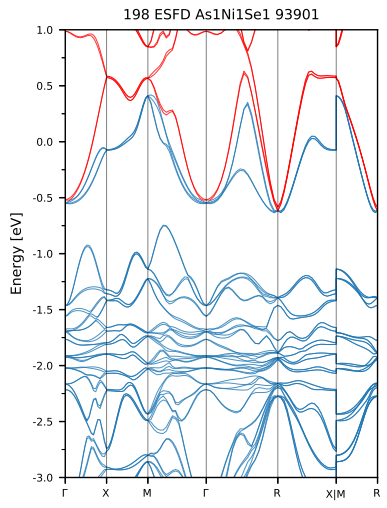

(ae) 198.4.31

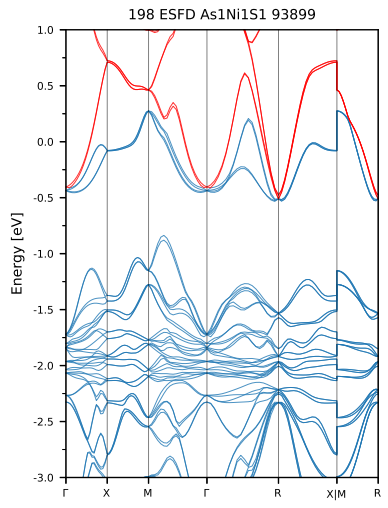

(af) 198.4.32

Figure 466: Band structures for topological materials in SG 198, part 2/3, from 198.4.17 to 198.4.32. See the list of figures.

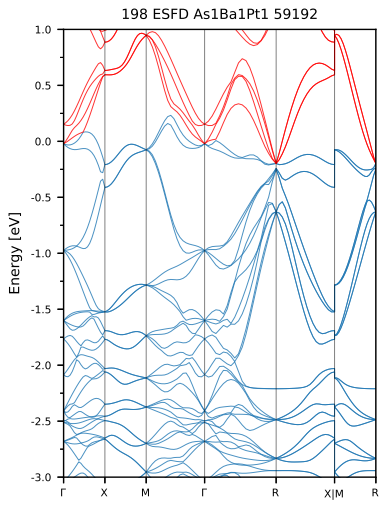

(ag) 198.4.33

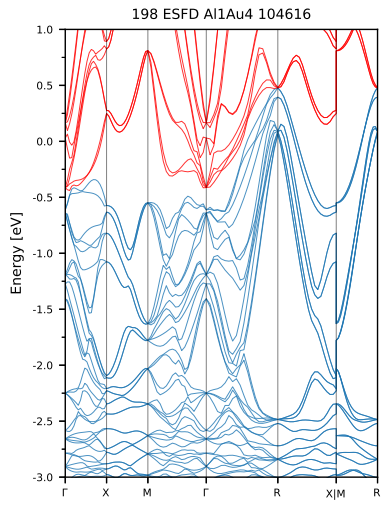

(ah) 198.4.34

Figure 467: Band structures for topological materials in SG 198, part 3/3, from 198.4.33 to 198.4.34. See the list of figures.

### 0.0.101 SG 202

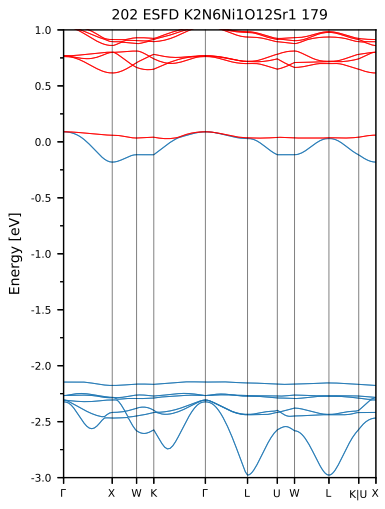

(a) 202.4.1

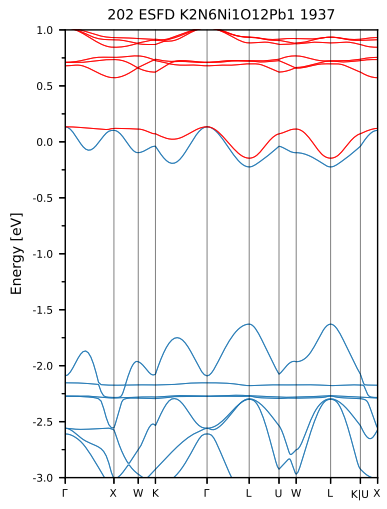

(b) 202.4.2

Figure 468: Band structures for topological materials in SG 202, part 1/1, from 202.4.1 to 202.4.2. See the list of figures.

## 0.0.102 SG 204

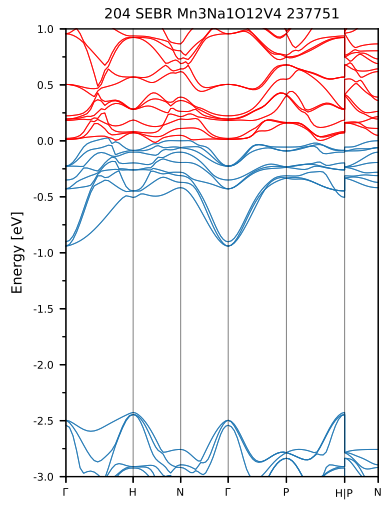

(a) 204.1.1

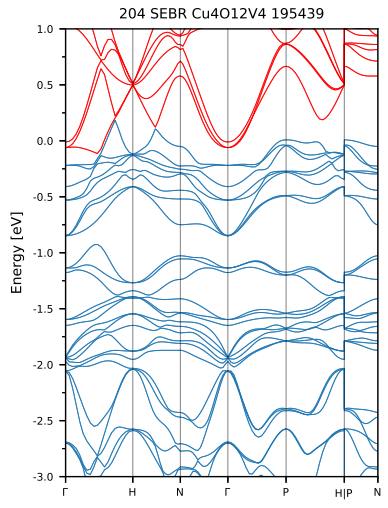

(b) 204.1.2

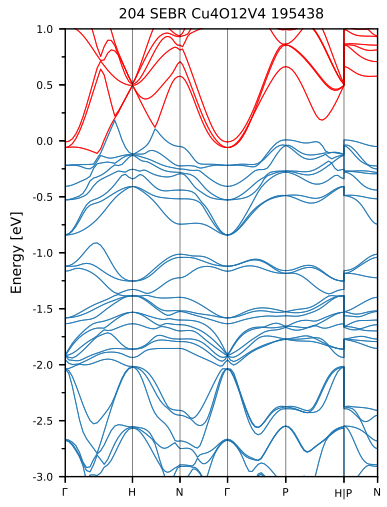

(c) 204.1.3

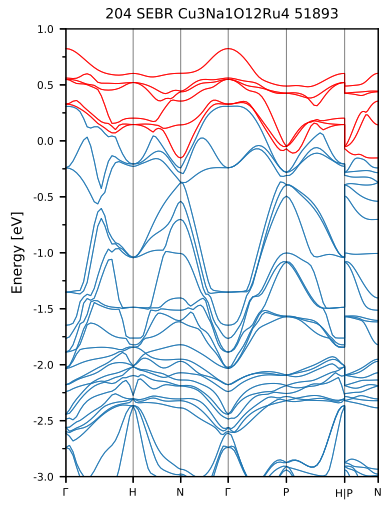

(d) 204.1.4

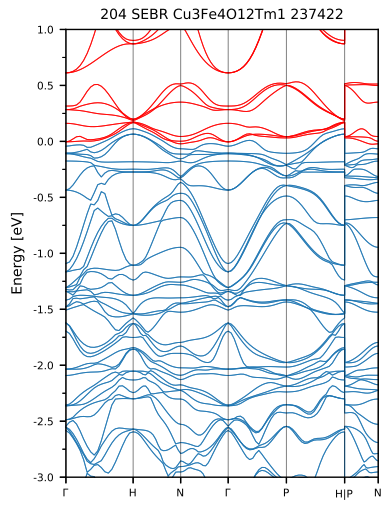

(e) 204.1.5

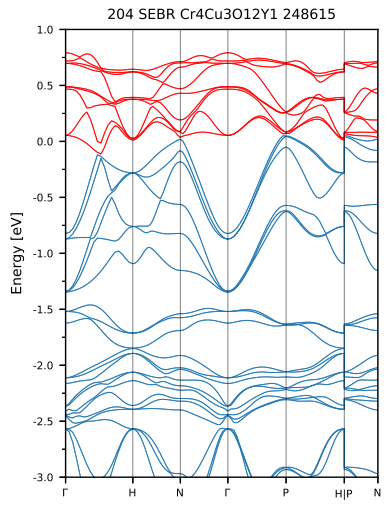

(f) 204.1.6

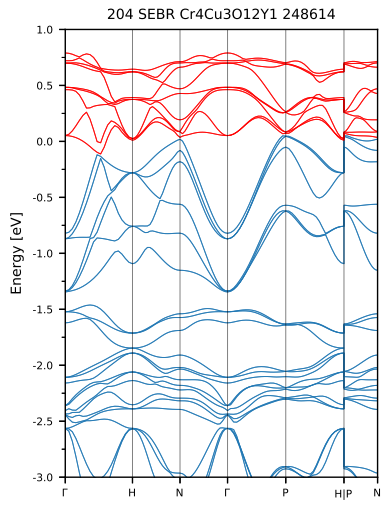

(g) 204.1.7

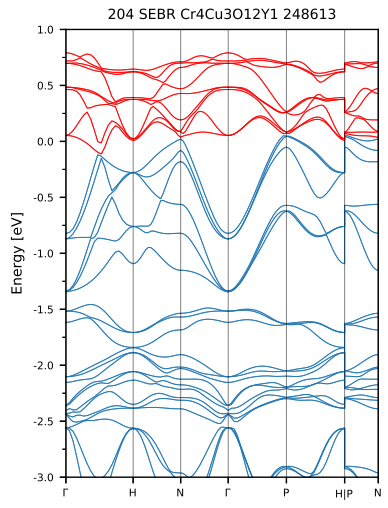

(h) 204.1.8

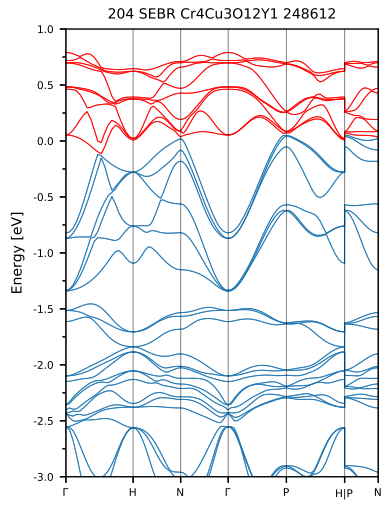

(i) 204.1.9

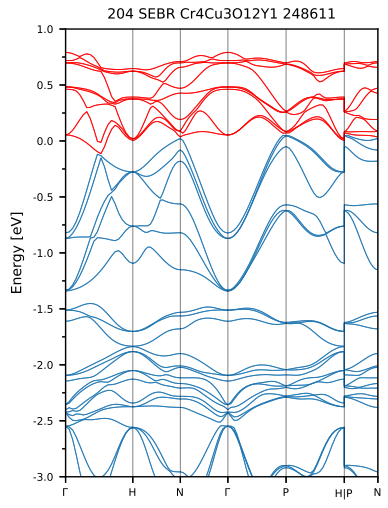

(j) 204.1.10

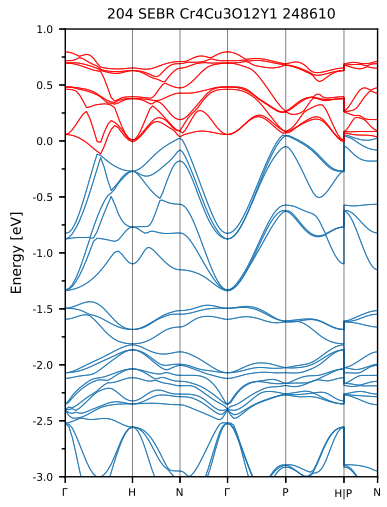

(k) 204.1.11

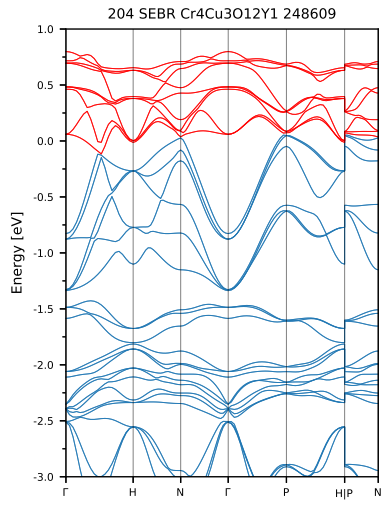

(l) 204.1.12

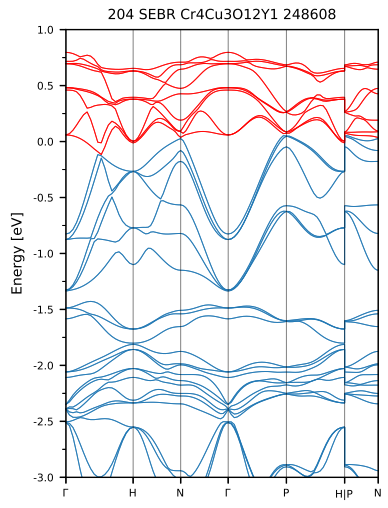

(m) 204.1.13

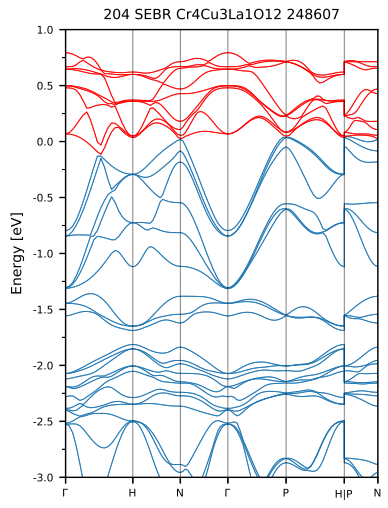

(n) 204.1.14

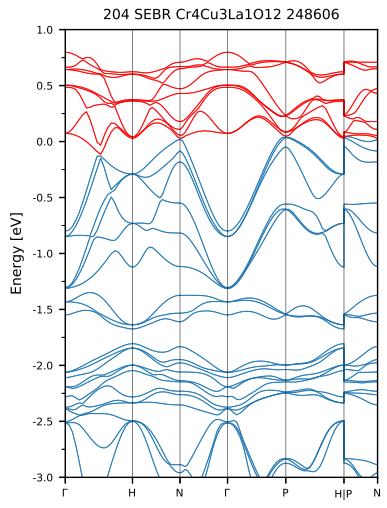

(o) 204.1.15

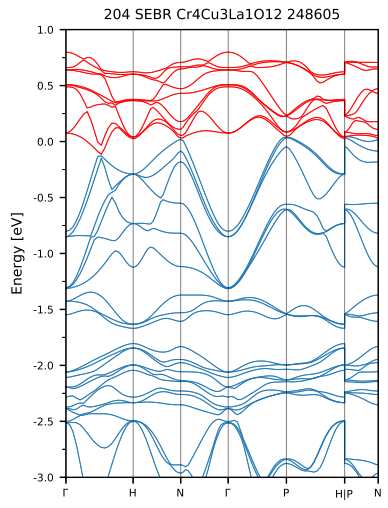

(p) 204.1.16

Figure 469: Band structures for topological materials in SG 204, part 1/5, from 204.1.1 to 204.1.16. See the list of figures.

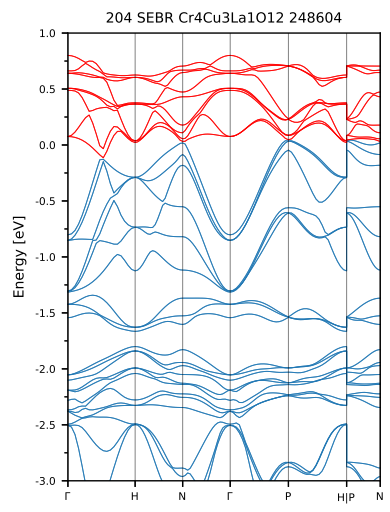

(q) 204.1.17

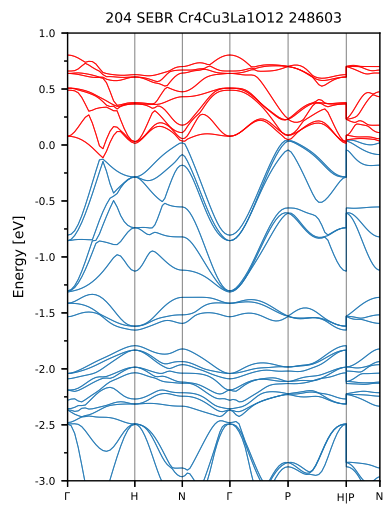

(r) 204.1.18

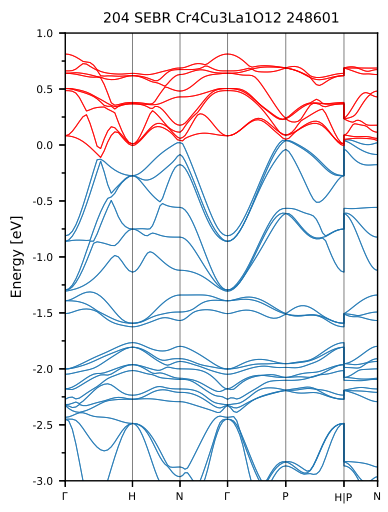

(s) 204.1.19

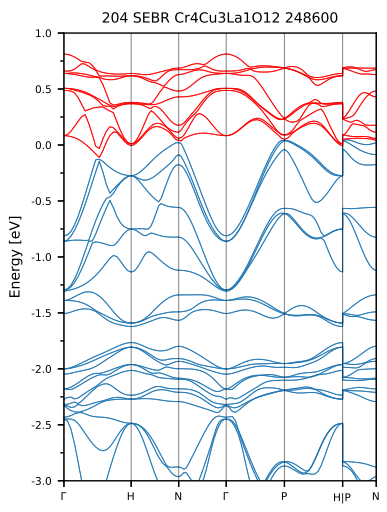

(t) 204.1.20

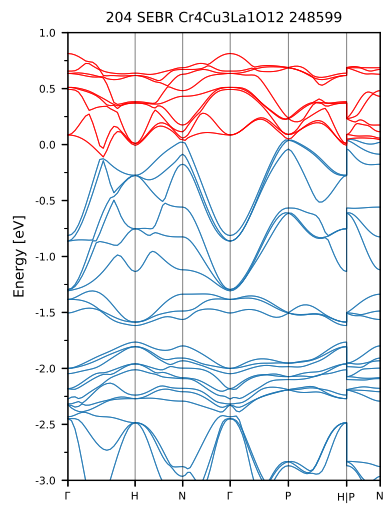

(u) 204.1.21

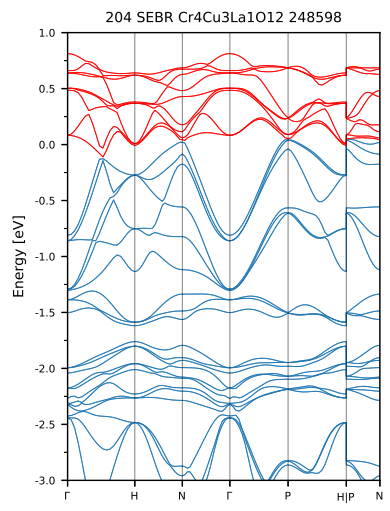

(v) 204.1.22

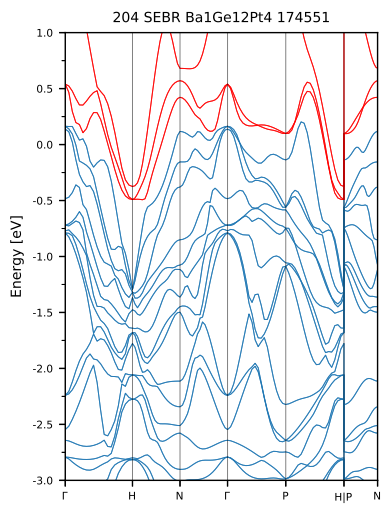

(w) 204.1.23

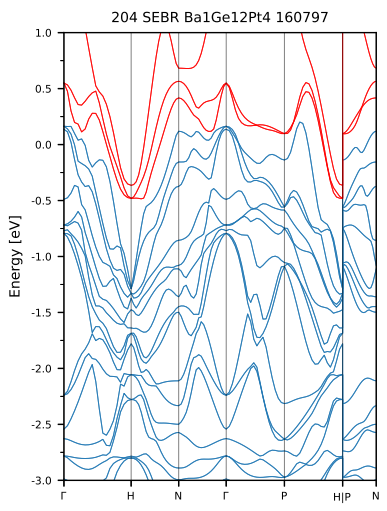

(x) 204.1.24

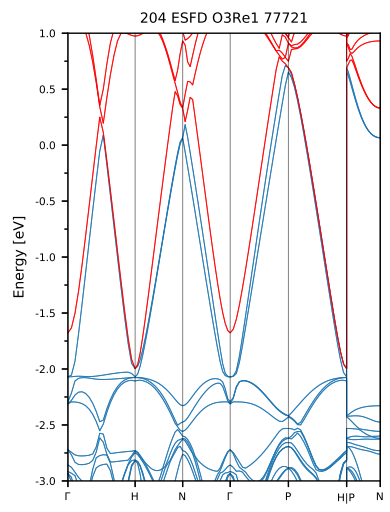

(y) 204.4.1

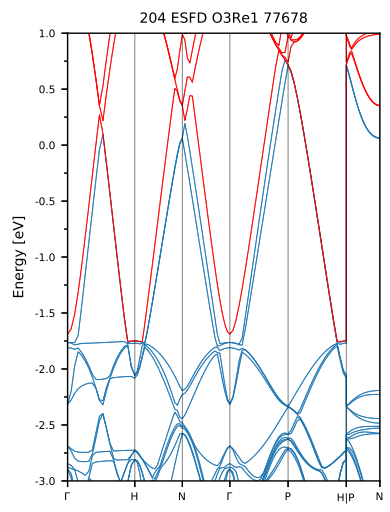

(z) 204.4.2

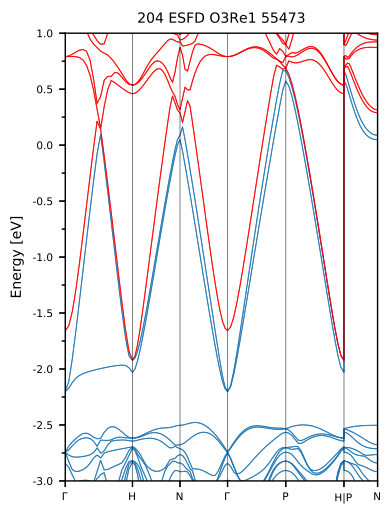

(aa) 204.4.3

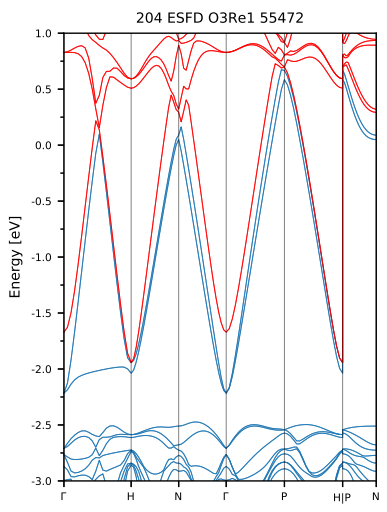

(ab) 204.4.4

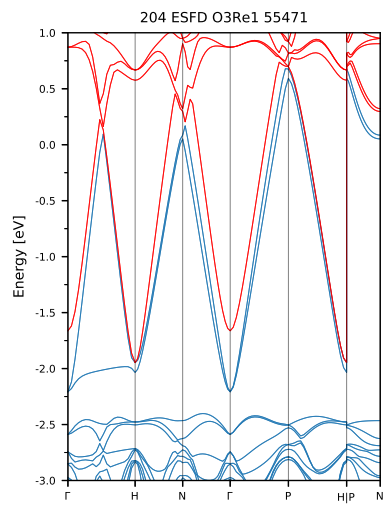

(ac) 204.4.5

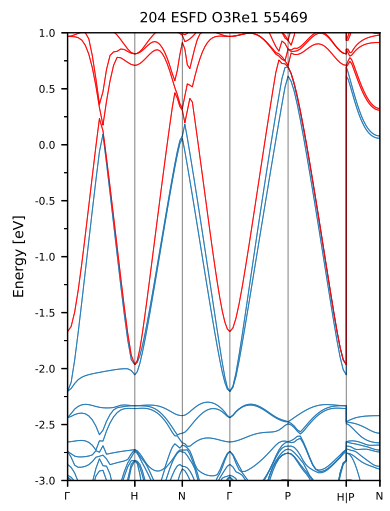

(ad) 204.4.6

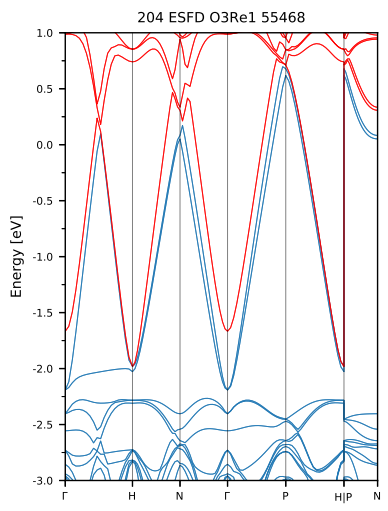

(ae) 204.4.7

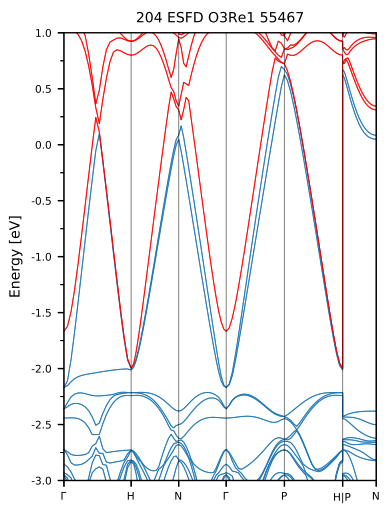

(af) 204.4.8

Figure 470: Band structures for topological materials in SG 204, part 2/5, from 204.1.17 to 204.4.8. See the list of figures.

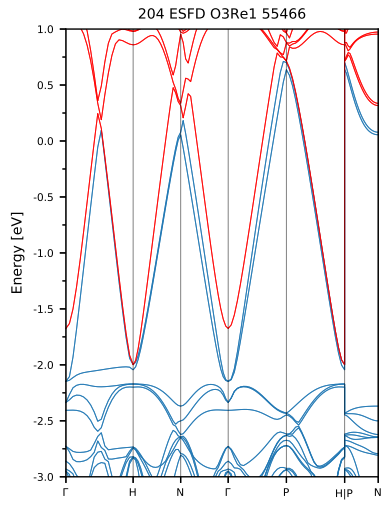

(ag) 204.4.9

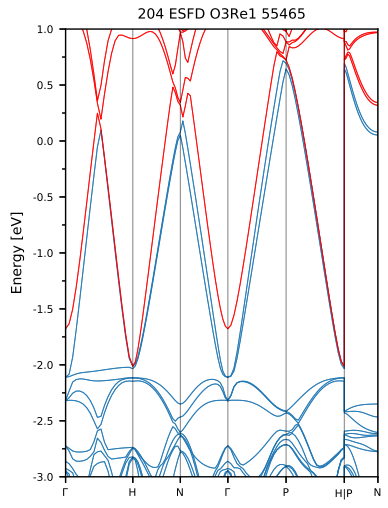

(ah) 204.4.10

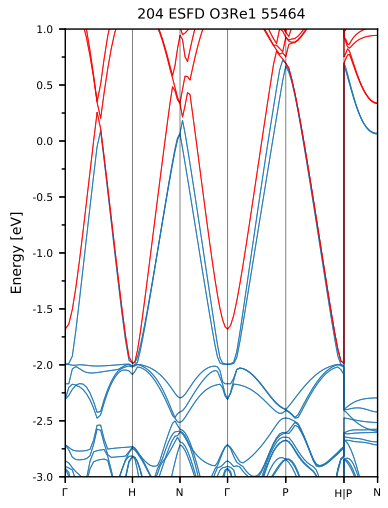

(ai) 204.4.11

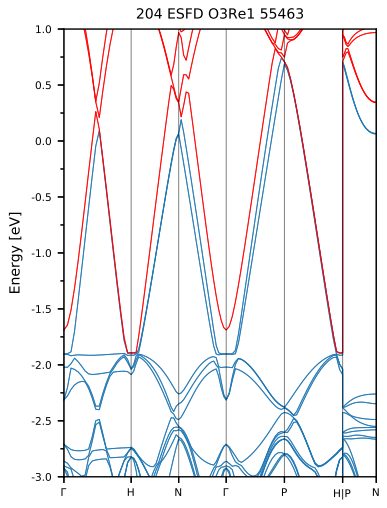

(aj) 204.4.12

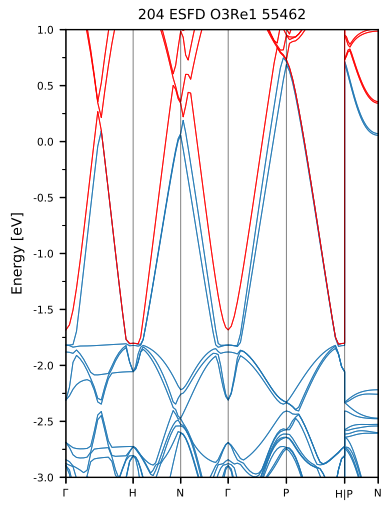

(ak) 204.4.13

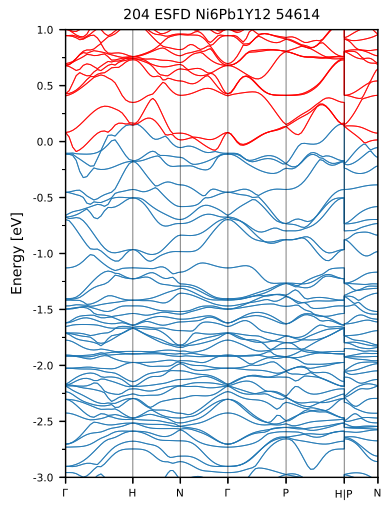

(al) 204.4.14

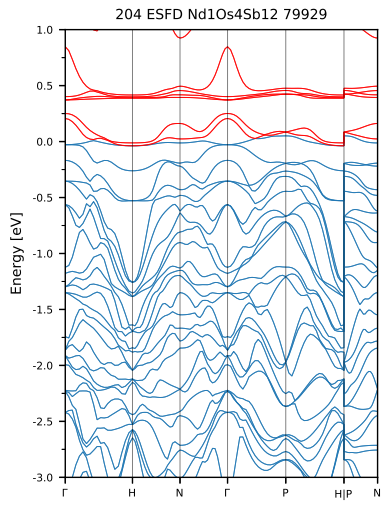

(am) 204.4.15

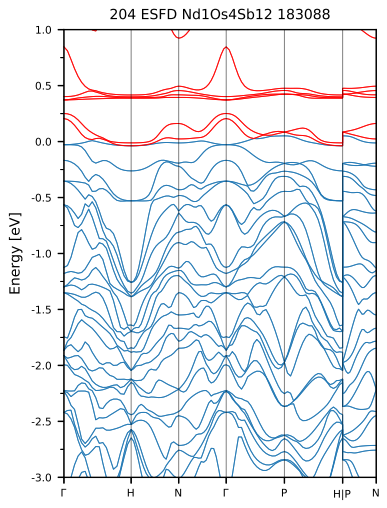

(an) 204.4.16

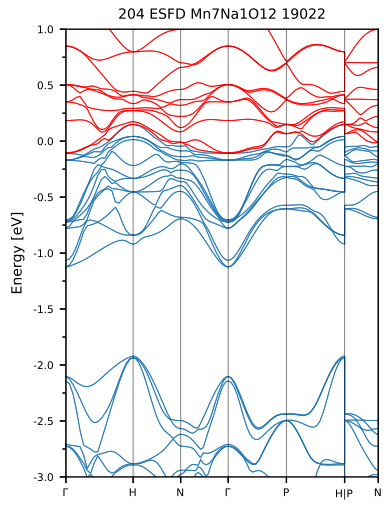

(ao) 204.4.17

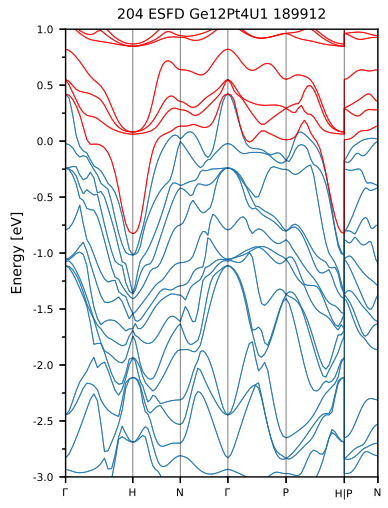

(ap) 204.4.18

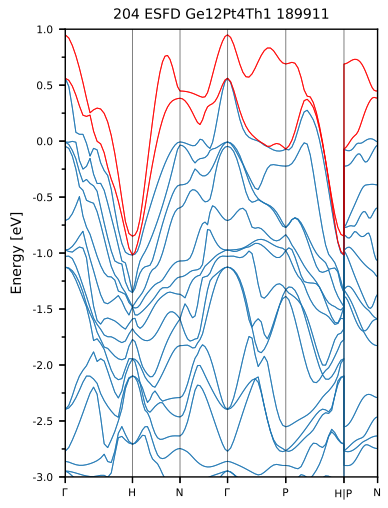

(aq) 204.4.19

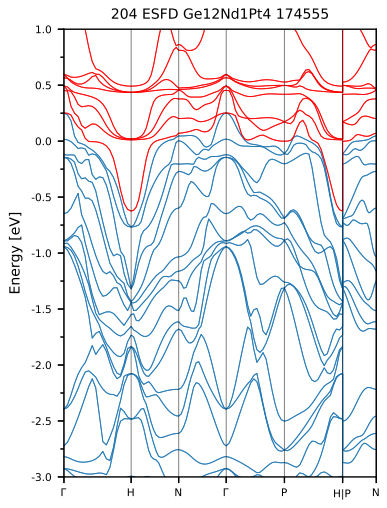

(ar) 204.4.20

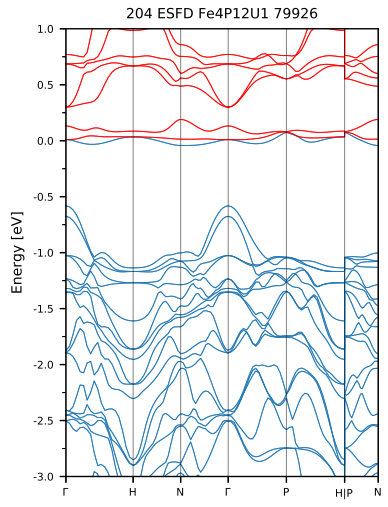

(as) 204.4.21

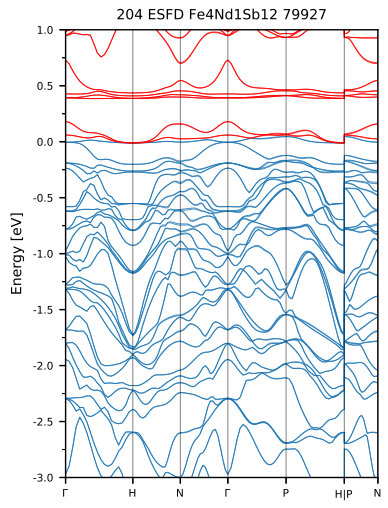

(at) 204.4.22

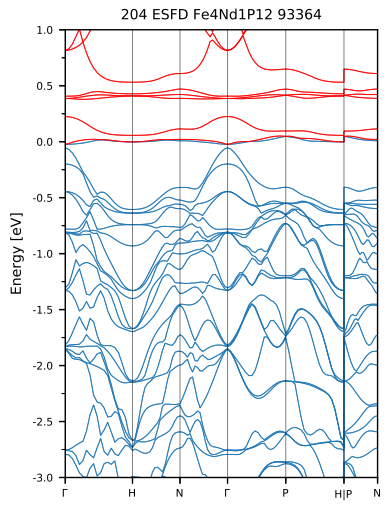

(au) 204.4.23

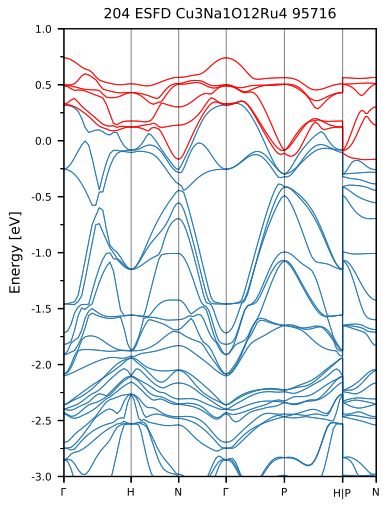

(av) 204.4.24

Figure 471: Band structures for topological materials in SG 204, part 3/5, from 204.4.9 to 204.4.24. See the list of figures.

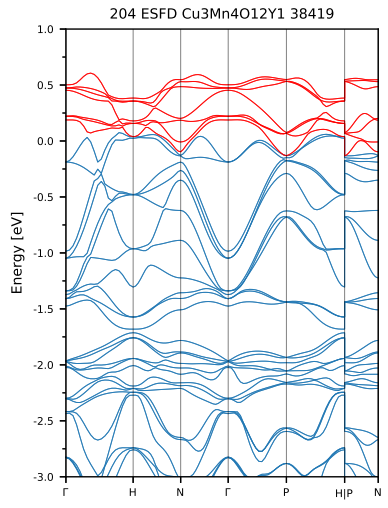

(aw) 204.4.25

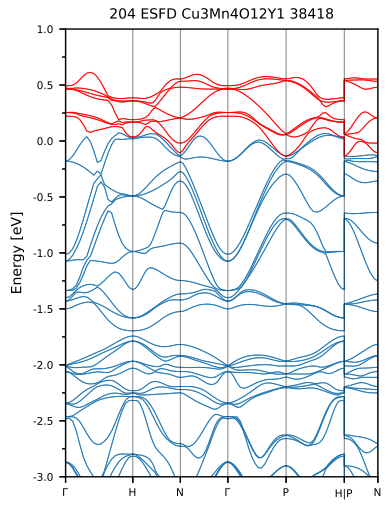

(ax) 204.4.26

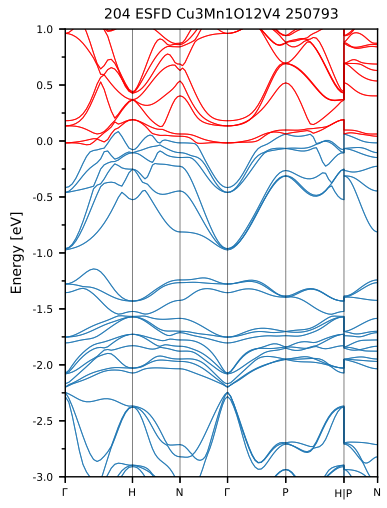

(ay) 204.4.27

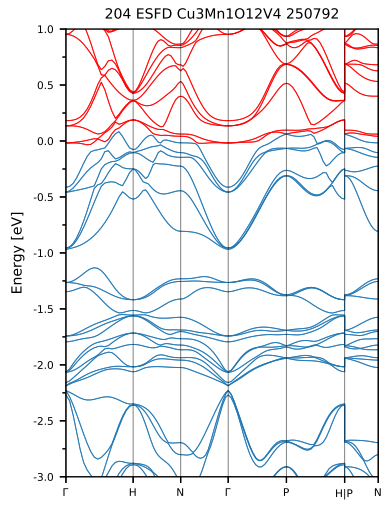

(az) 204.4.28

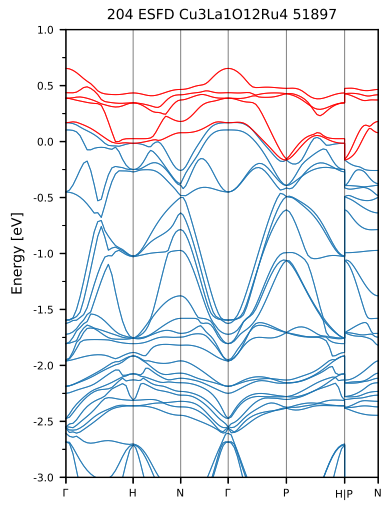

(ba) 204.4.29

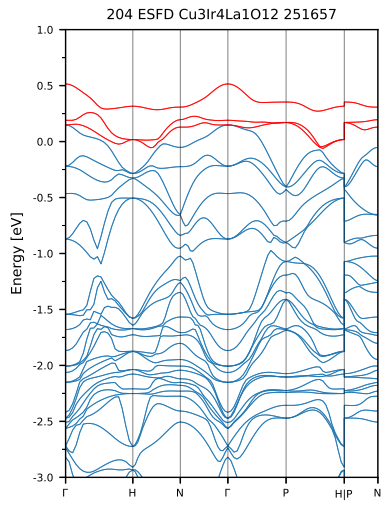

(bb) 204.4.30

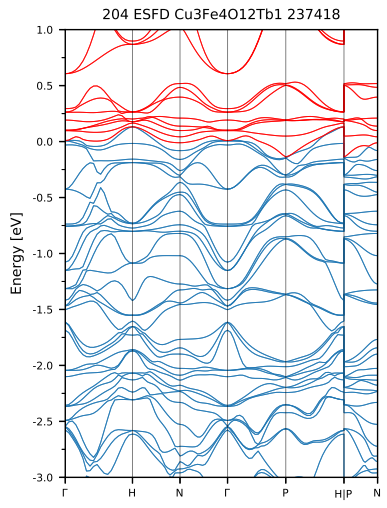

(bc) 204.4.31

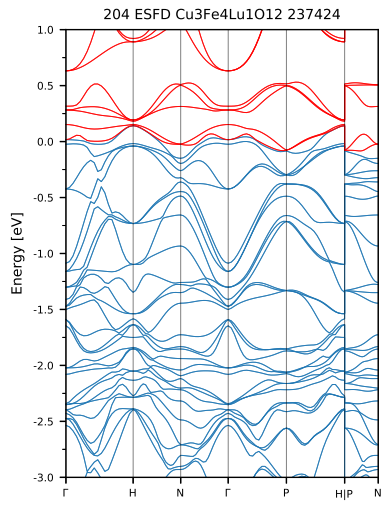

(bd) 204.4.32

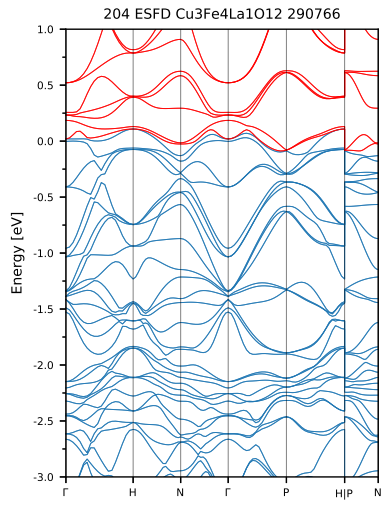

(be) 204.4.33

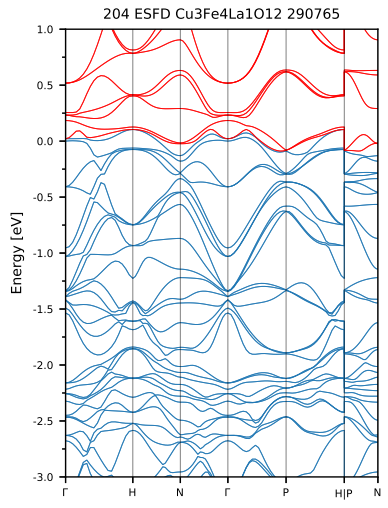

(bf) 204.4.34

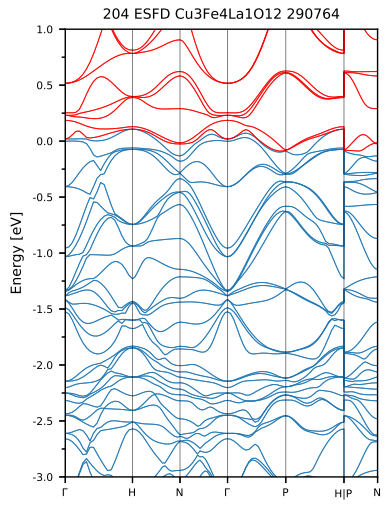

(bg) 204.4.35

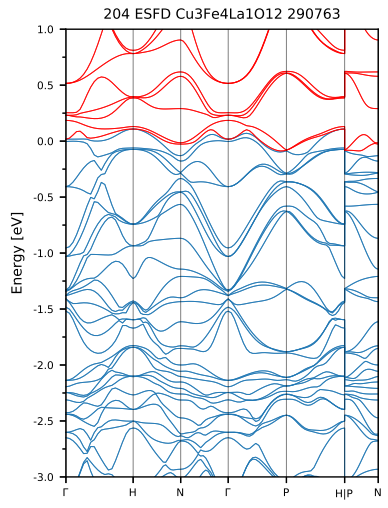

(bh) 204.4.36

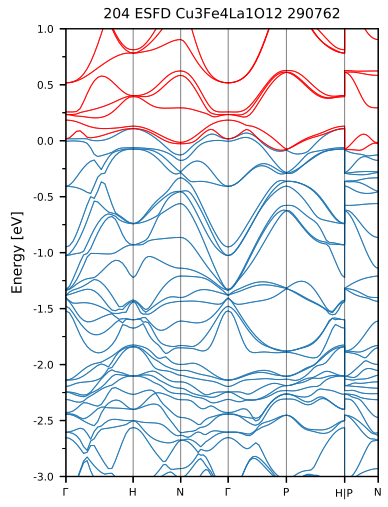

(bi) 204.4.37

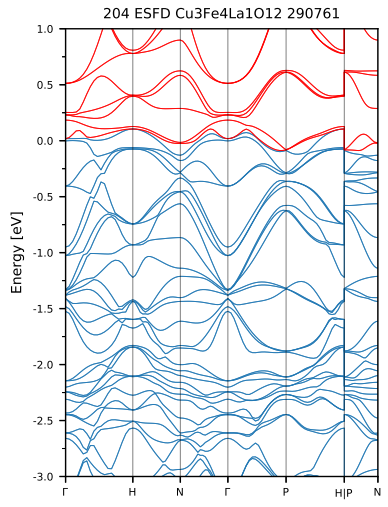

(bj) 204.4.38

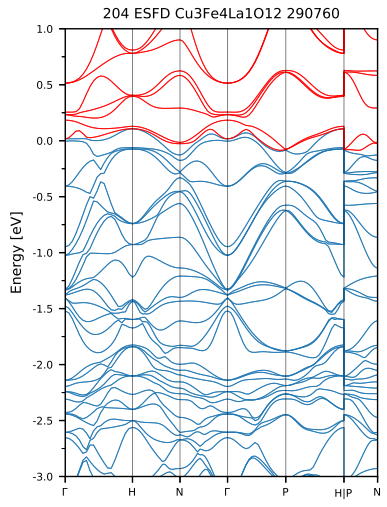

(bk) 204.4.39

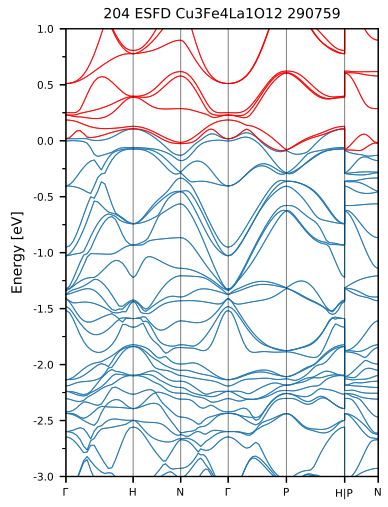

(bl) 204.4.40

Figure 472: Band structures for topological materials in SG 204, part 4/5, from 204.4.25 to 204.4.40. See the list of figures.

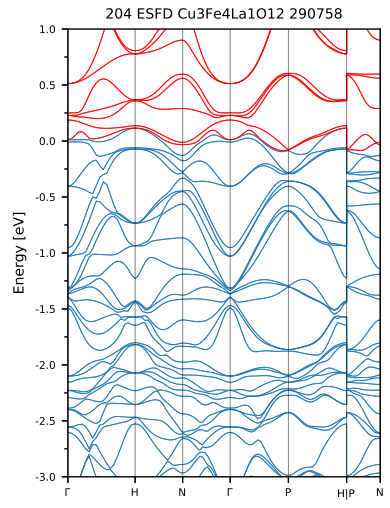

(bm) 204.4.41

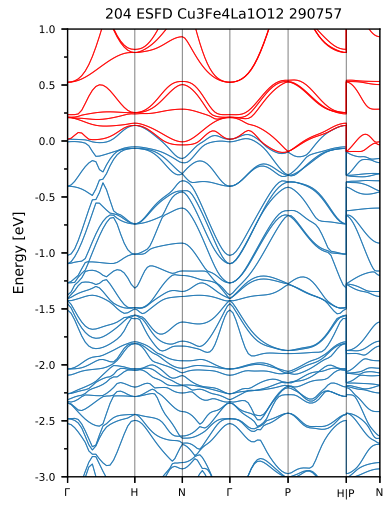

(bn) 204.4.42

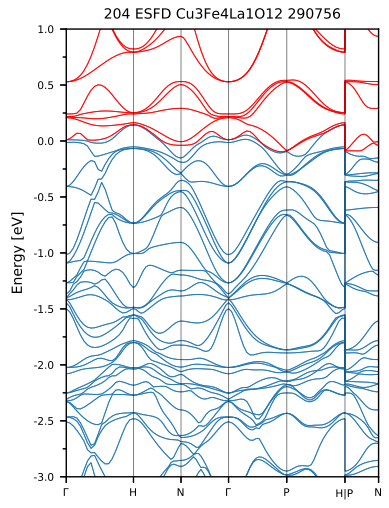

(bo) 204.4.43

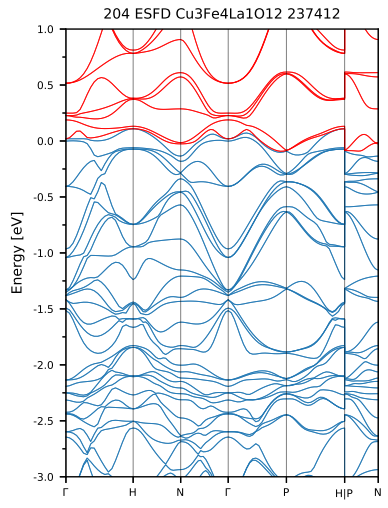

(bp) 204.4.44

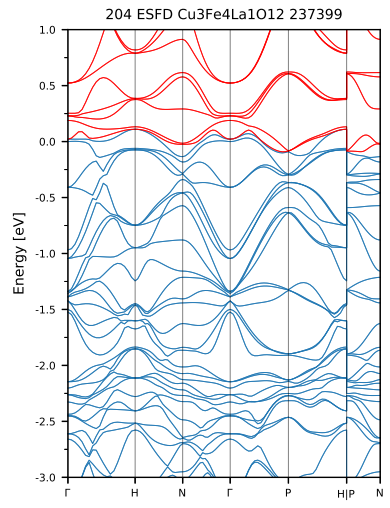

(bq) 204.4.45

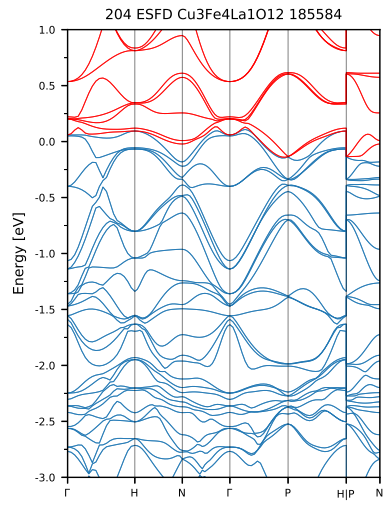

(br) 204.4.46

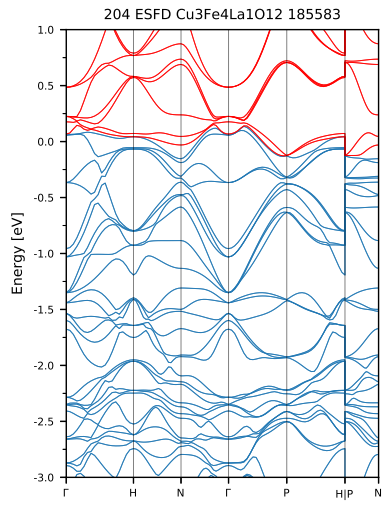

(bs) 204.4.47

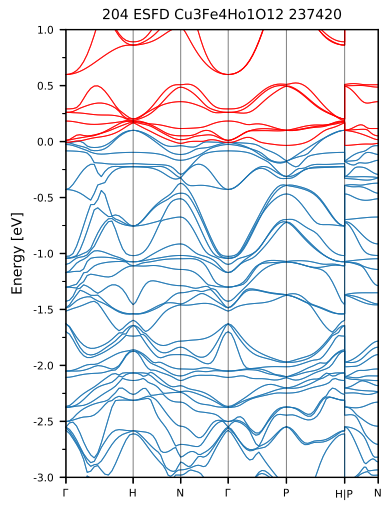

(bt) 204.4.48

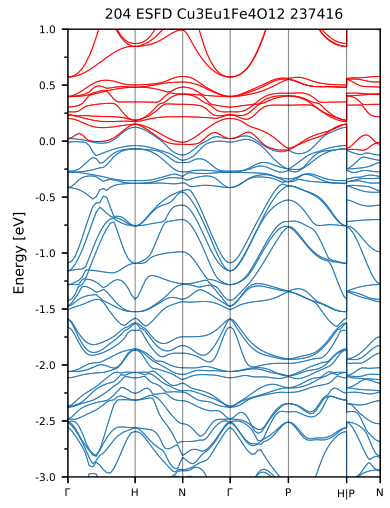

(bu) 204.4.49

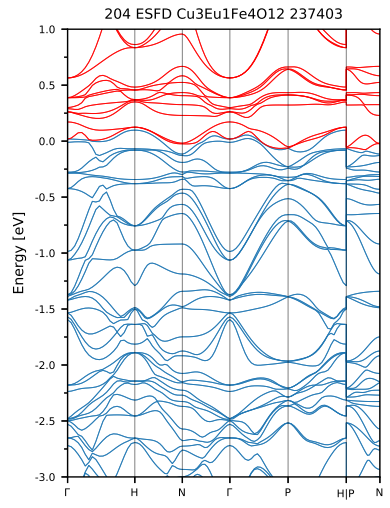

(bv) 204.4.50

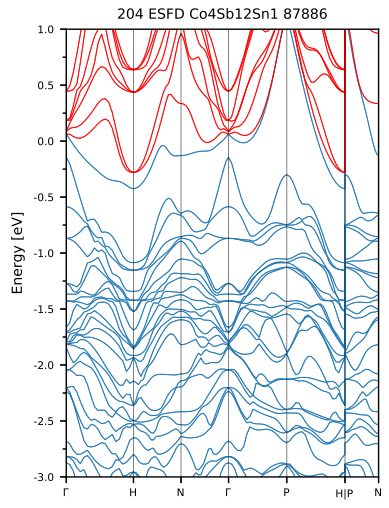

(bw) 204.4.51

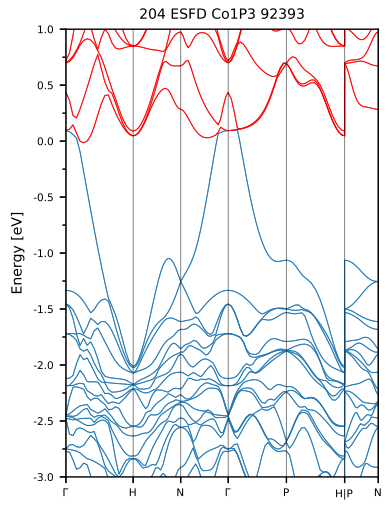

(bx) 204.4.52

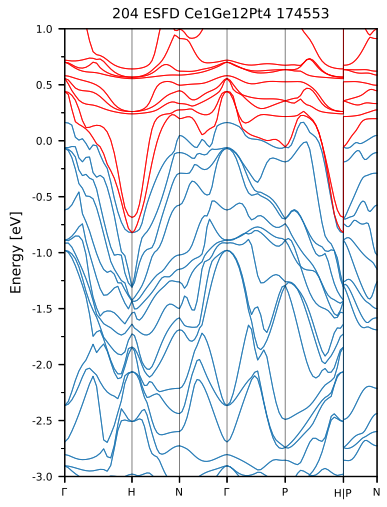

(by) 204.4.53

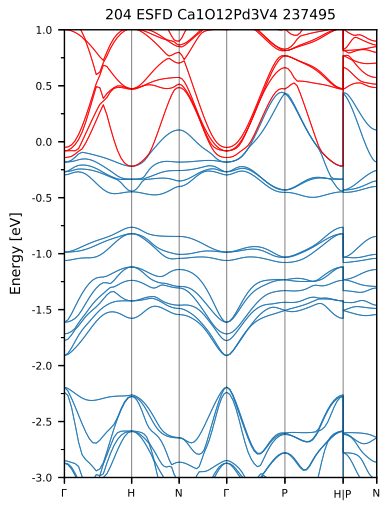

(bz) 204.4.54

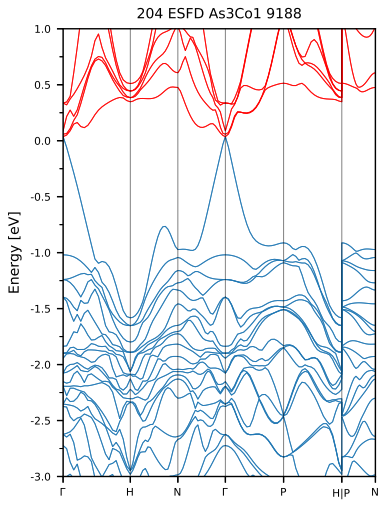

(ca) 204.4.55

Figure 473: Band structures for topological materials in SG 204, part 5/5, from 204.4.41 to 204.4.55. See the list of figures.

## 0.0.103 SG 205

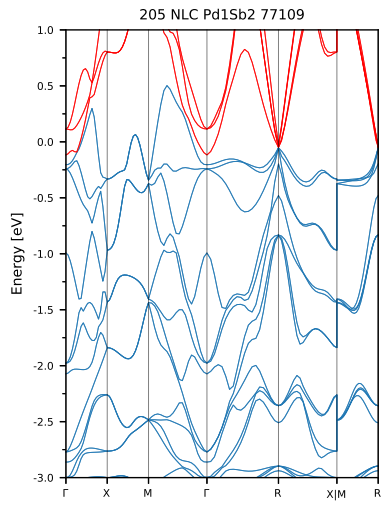

(a) 205.2.1

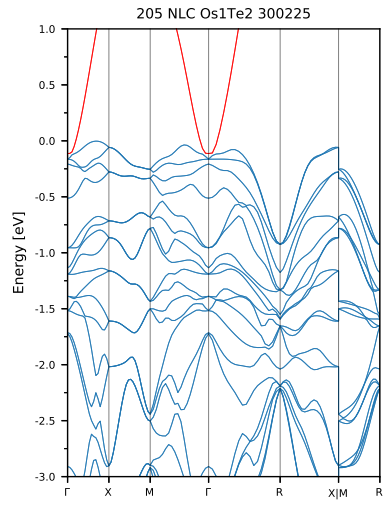

(b) 205.2.2

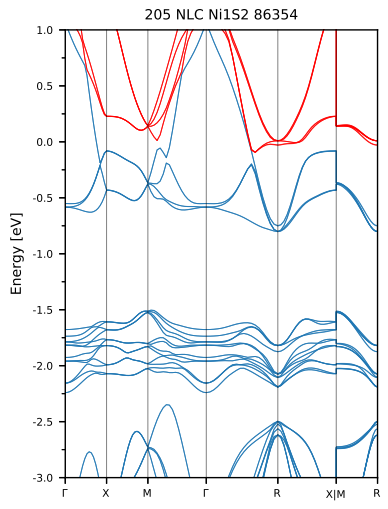

(c) 205.2.3

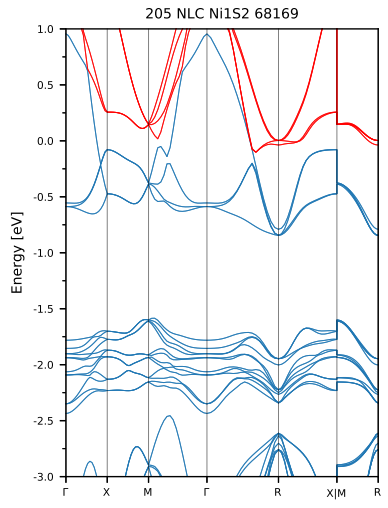

(d) 205.2.4

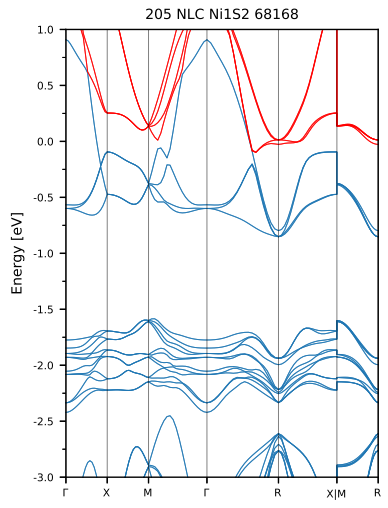

(e) 205.2.5

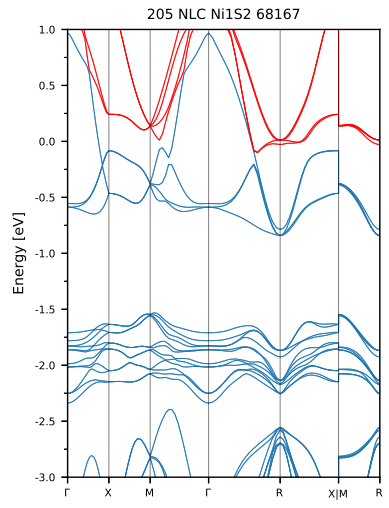

(f) 205.2.6

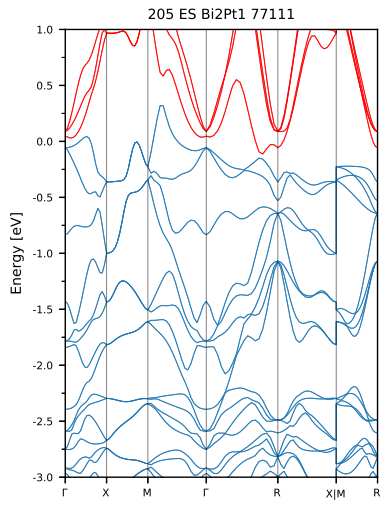

(g) 205.3.1

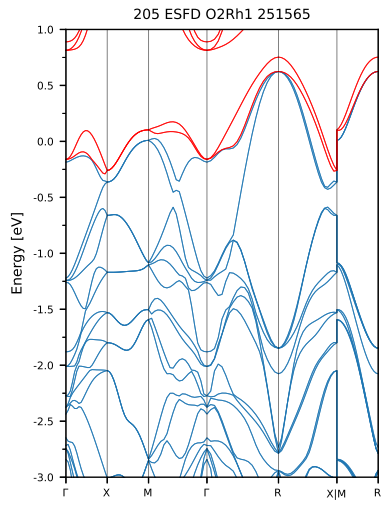

(h) 205.4.1

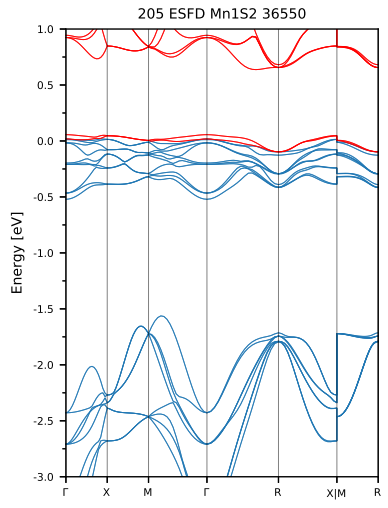

(i) 205.4.2

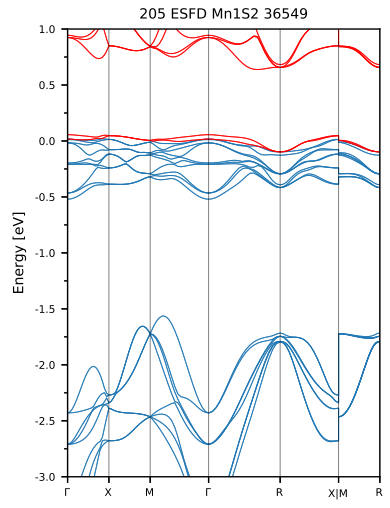

(j) 205.4.3

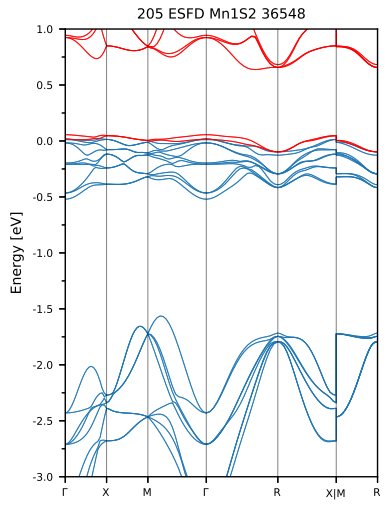

(k) 205.4.4

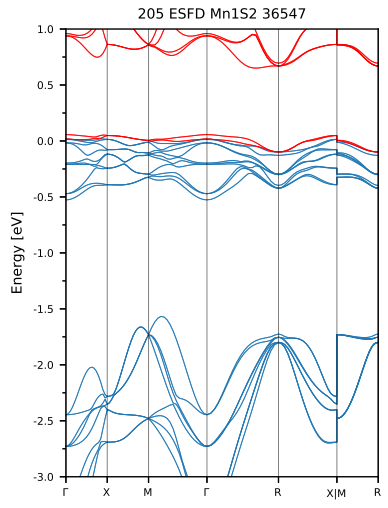

(l) 205.4.5

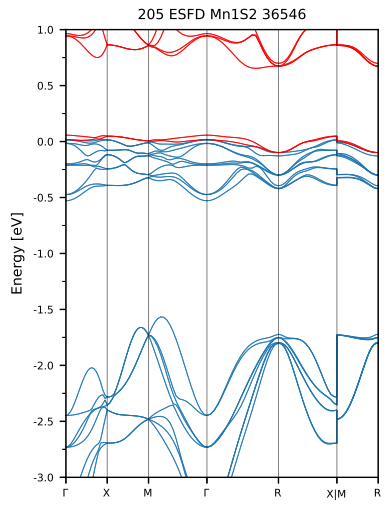

(m) 205.4.6

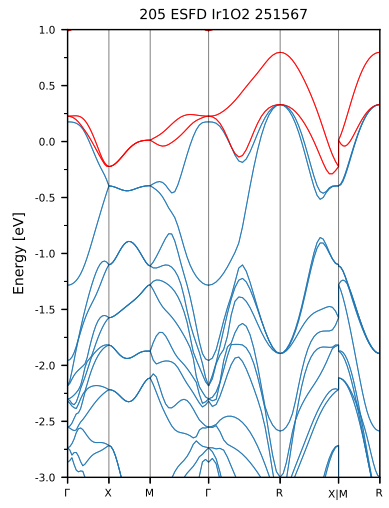

(n) 205.4.7

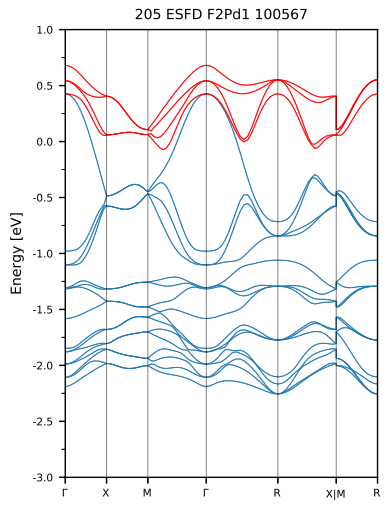

(o) 205.4.8

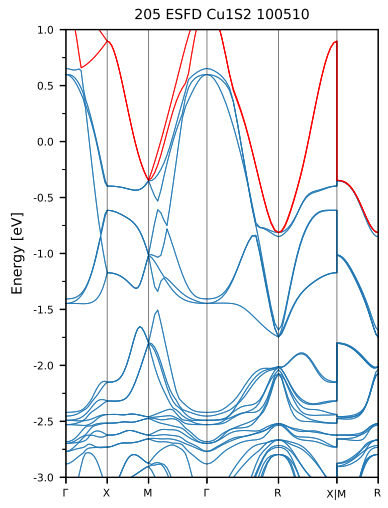

(p) 205.4.9

Figure 474: Band structures for topological materials in SG 205, part 1/2, from 205.2.1 to 205.4.9. See the list of figures.

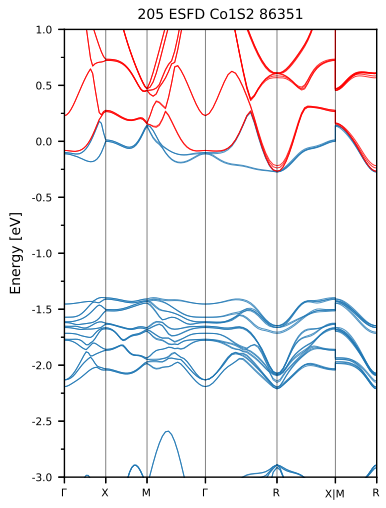

(q) 205.4.10

Figure 475: Band structures for topological materials in SG 205, part 2/2, from 205.4.10 to 205.4.10. See the list of figures.

### 0.0.104 SG 212

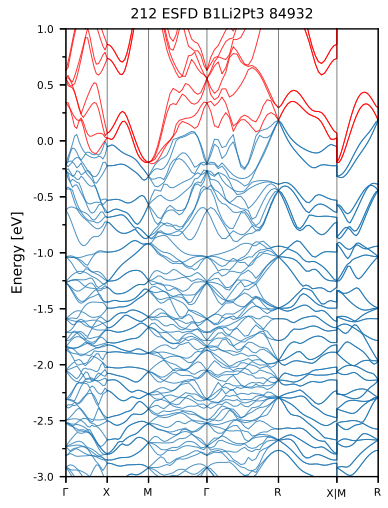

(a) 212.4.1

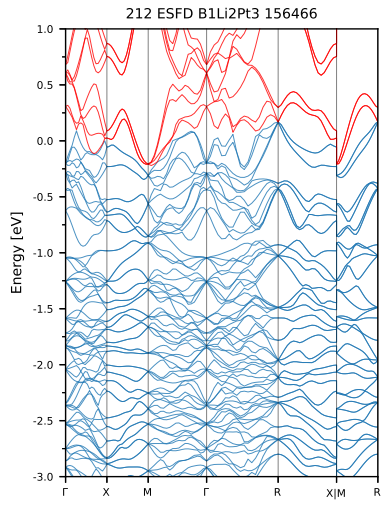

(b) 212.4.2

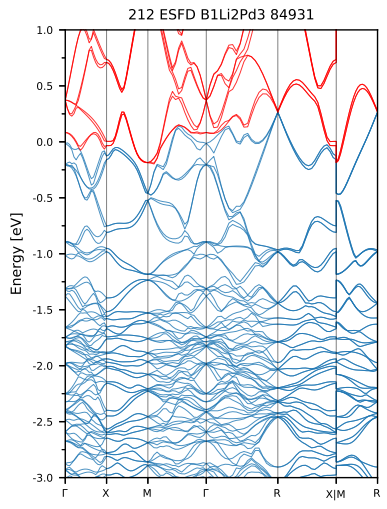

(c) 212.4.3

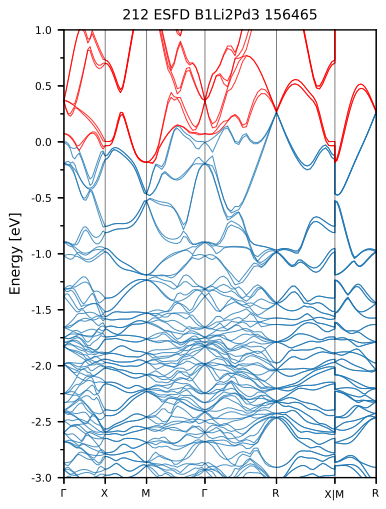

(d) 212.4.4

Figure 476: Band structures for topological materials in SG 212, part 1/1, from 212.4.1 to 212.4.4. See the list of figures.

### 0.0.105 SG 213

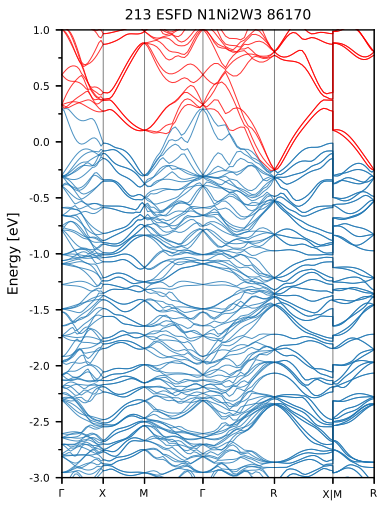

(a) 213.4.1

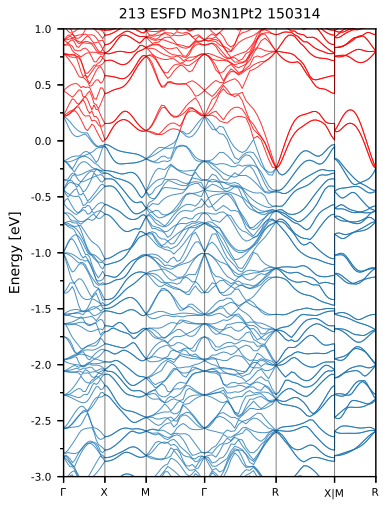

(b) 213.4.2

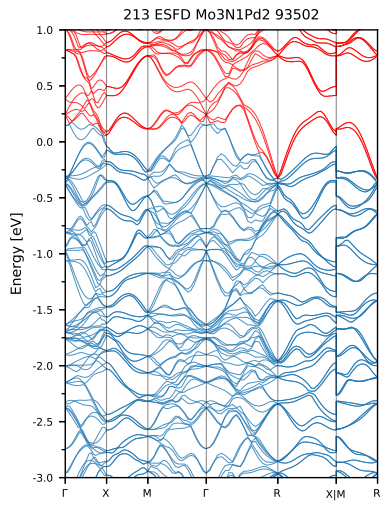

(c) 213.4.3

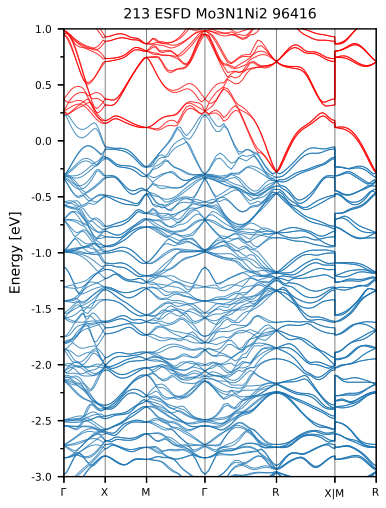

(d) 213.4.4

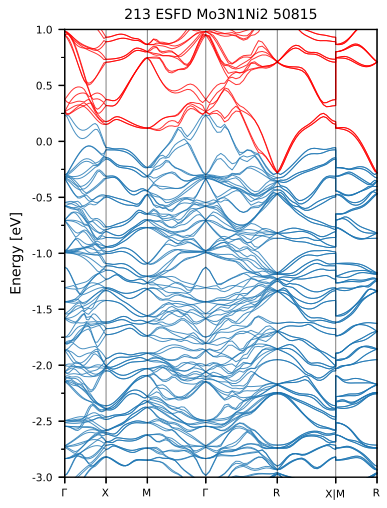

(e) 213.4.5

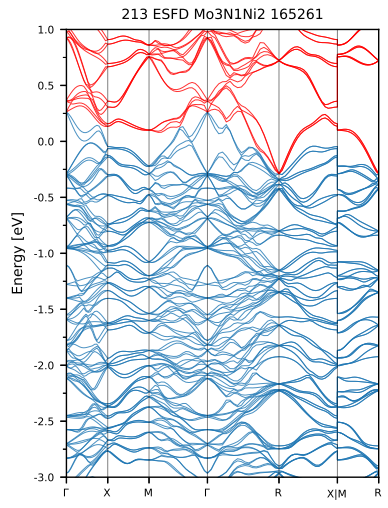

(f) 213.4.6

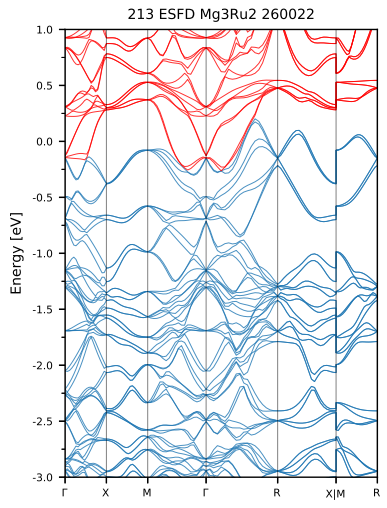

(g) 213.4.7

Figure 477: Band structures for topological materials in SG 213, part 1/1, from 213.4.1 to 213.4.7. See the list of figures.

## 0.0.106 SG 216

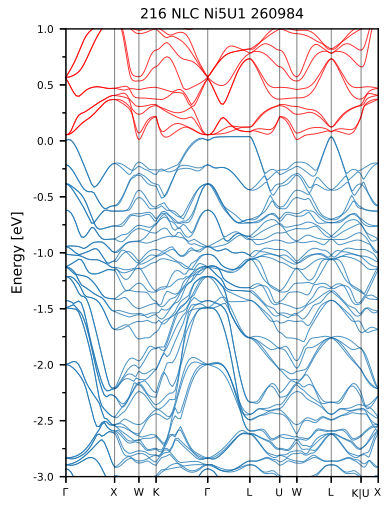

(a) 216.2.1

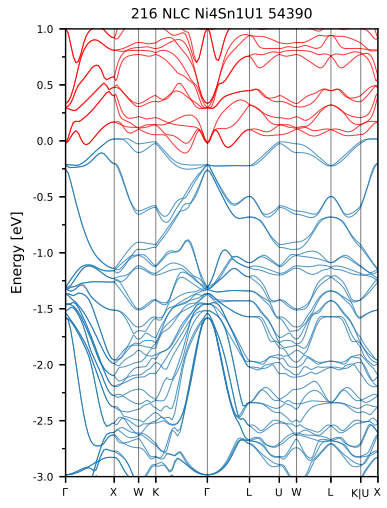

(b) 216.2.2

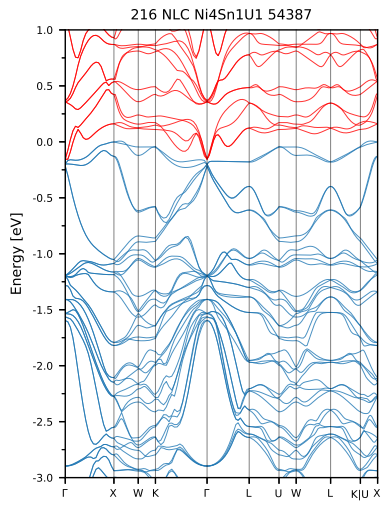

(c) 216.2.3

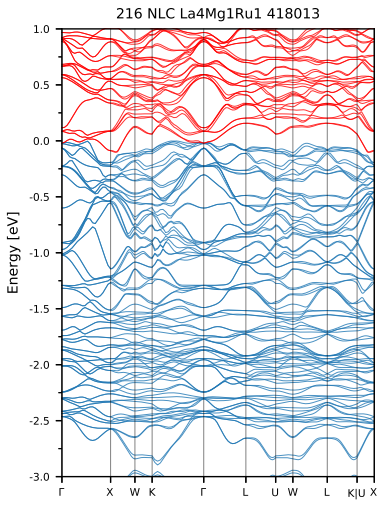

(d) 216.2.4

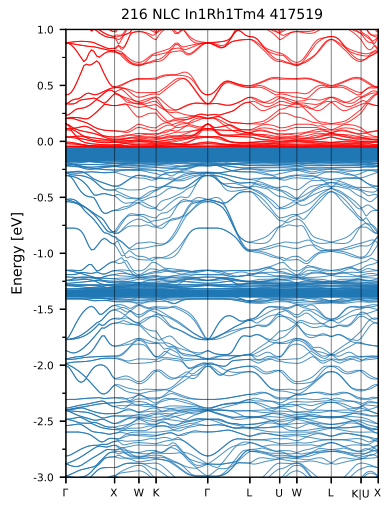

(e) 216.2.5

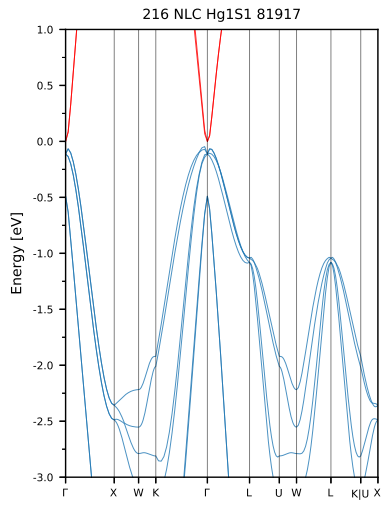

(f) 216.2.6

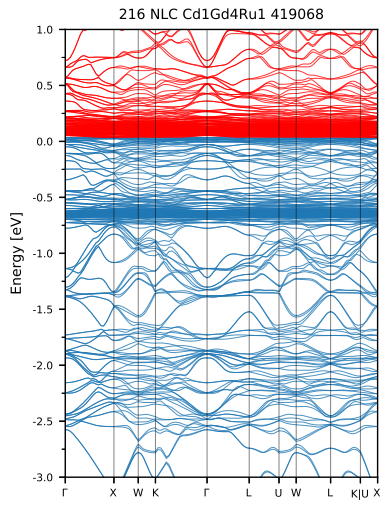

(g) 216.2.7

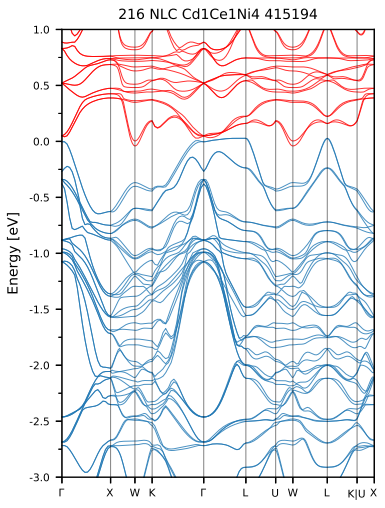

(h) 216.2.8

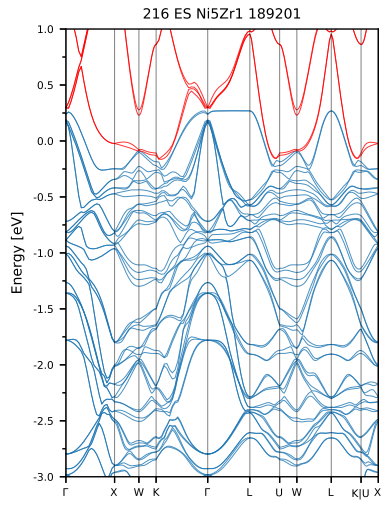

(i) 216.3.1

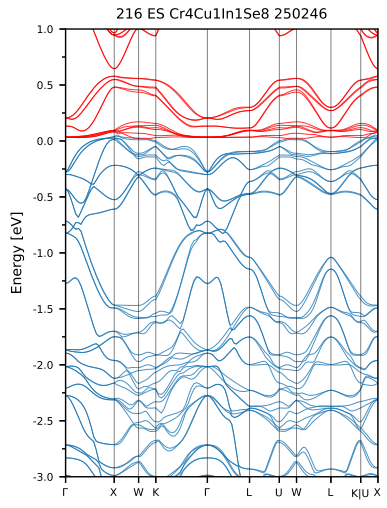

(j) 216.3.2

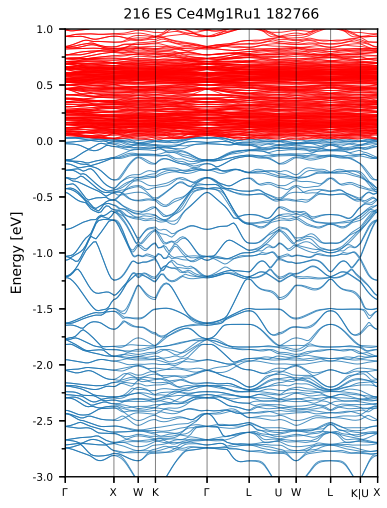

(k) 216.3.3

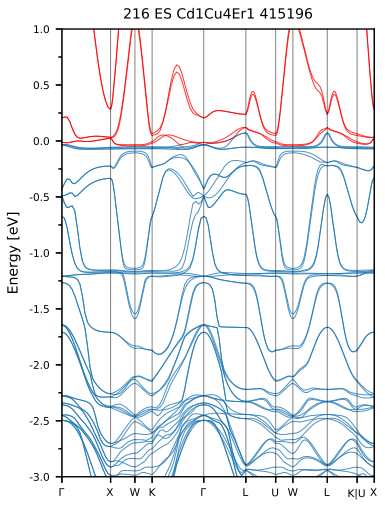

(l) 216.3.4

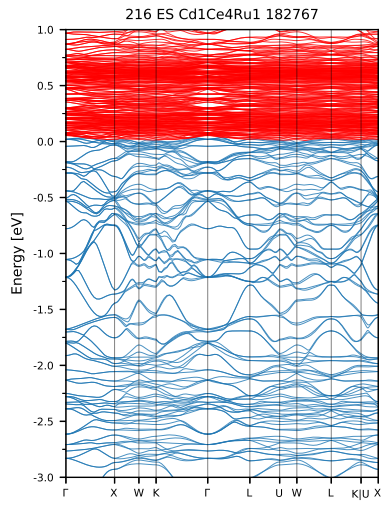

(m) 216.3.5

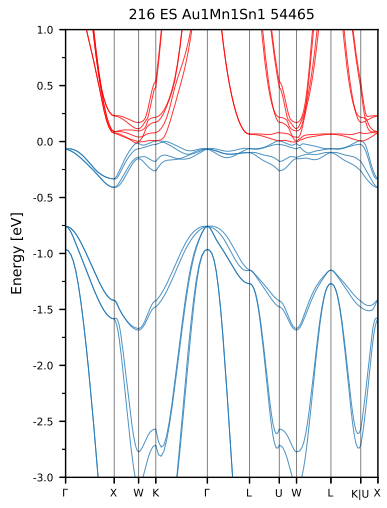

(n) 216.3.6

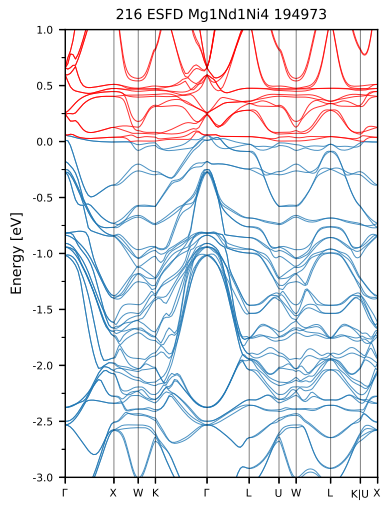

(o) 216.4.1

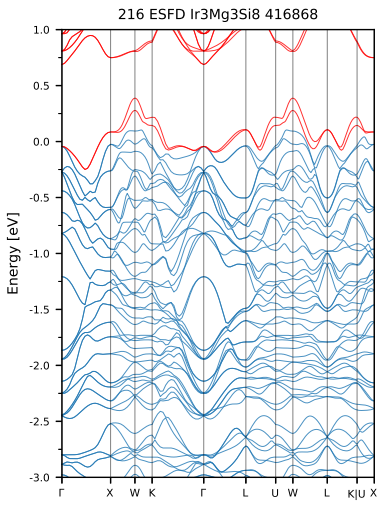

(p) 216.4.2

Figure 478: Band structures for topological materials in SG 216, part 1/3, from 216.2.1 to 216.4.2. See the list of figures.

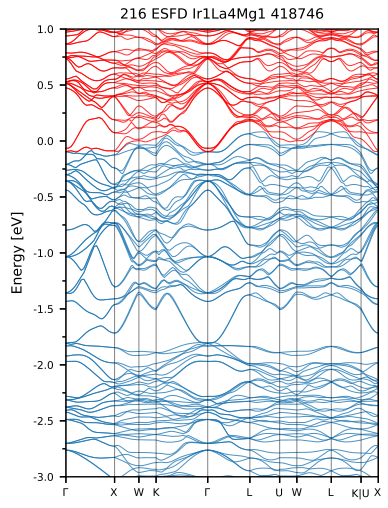

(q) 216.4.3

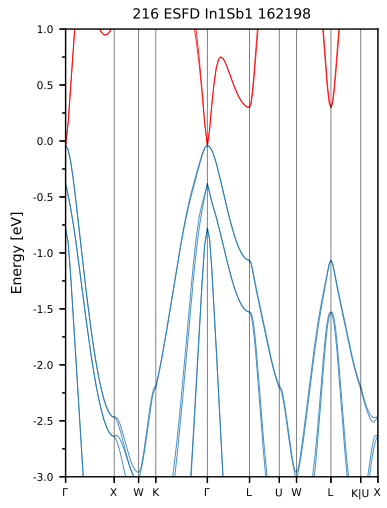

(r) 216.4.4

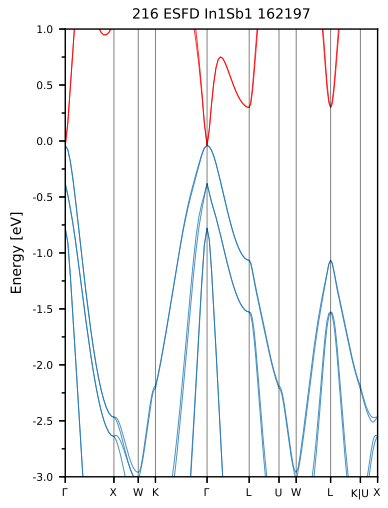

(s) 216.4.5

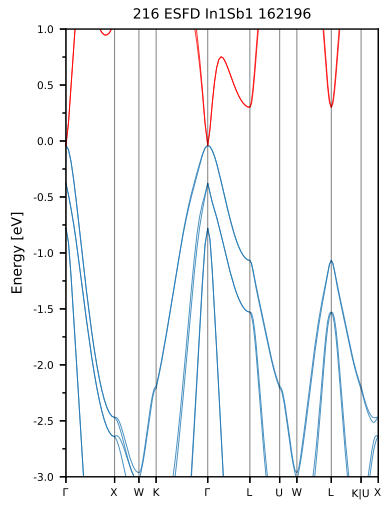

(t) 216.4.6

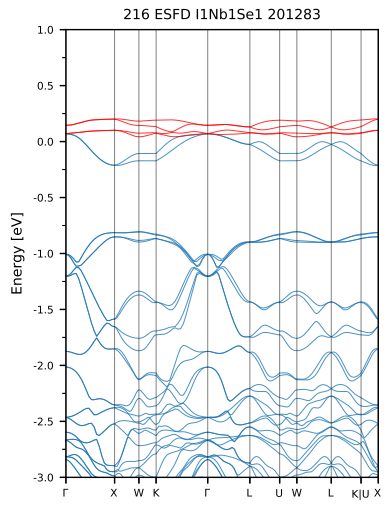

(u) 216.4.7

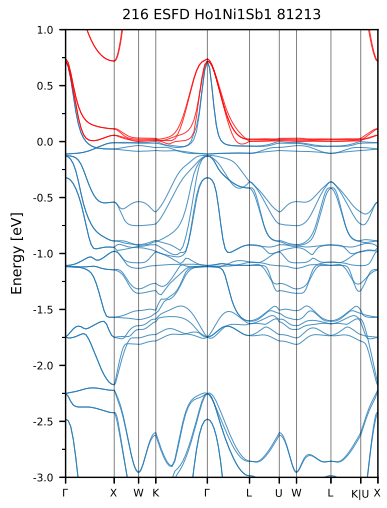

(v) 216.4.8

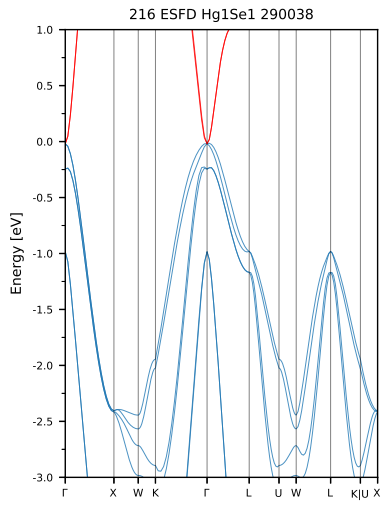

(w) 216.4.9

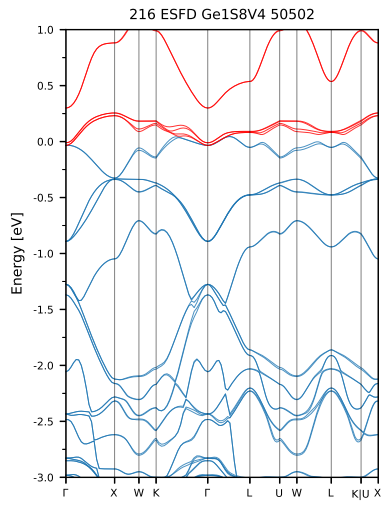

(x) 216.4.10

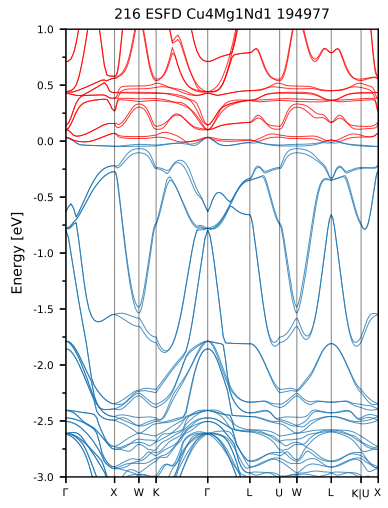

(y) 216.4.11

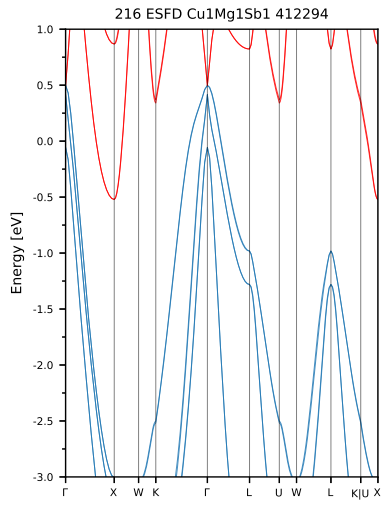

(z) 216.4.12

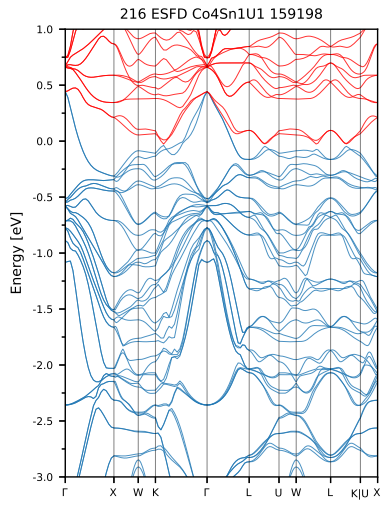

(aa) 216.4.13

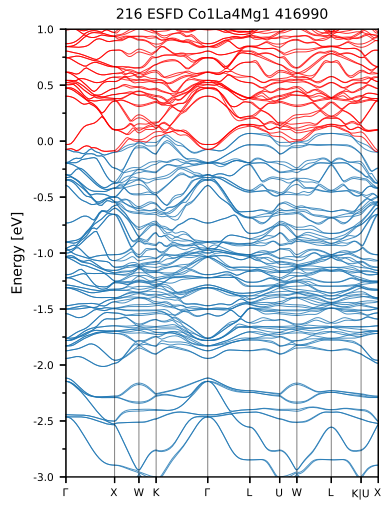

(ab) 216.4.14

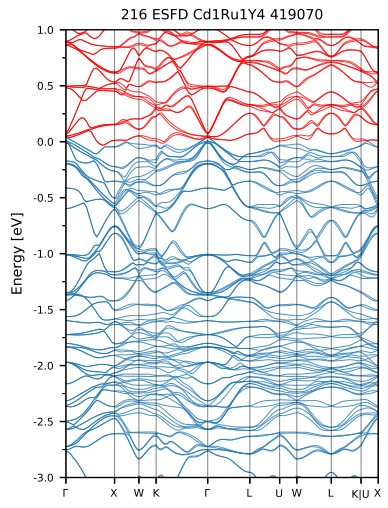

(ac) 216.4.15

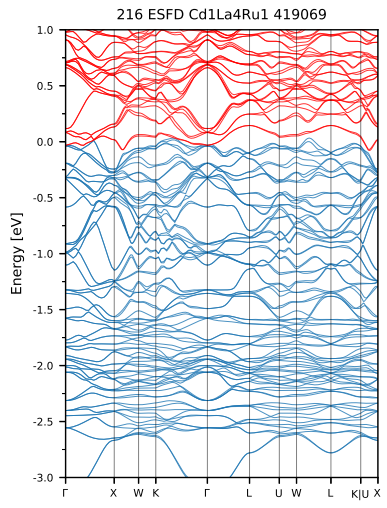

(ad) 216.4.16

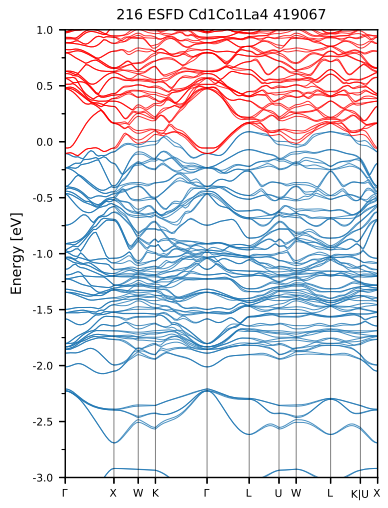

(ae) 216.4.17

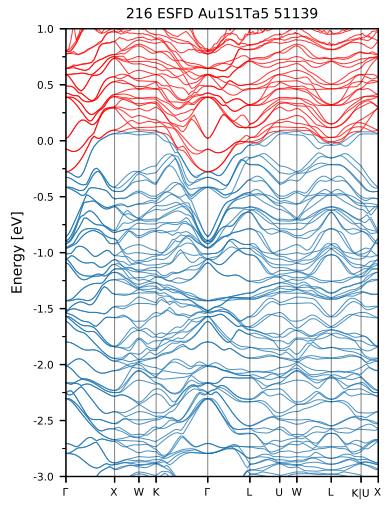

(af) 216.4.18

Figure 479: Band structures for topological materials in SG 216, part 2/3, from 216.4.3 to 216.4.18. See the list of figures.

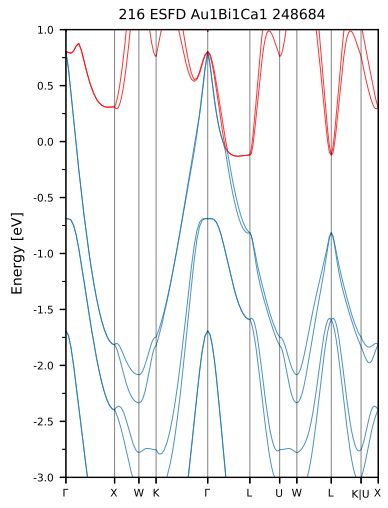

(ag) 216.4.19

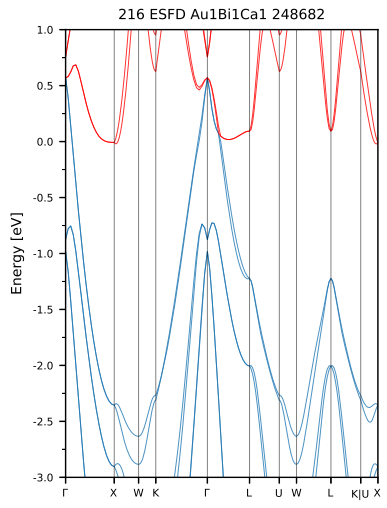

(ah) 216.4.20

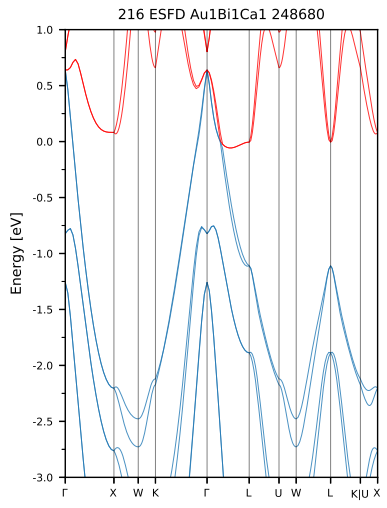

(ai) 216.4.21

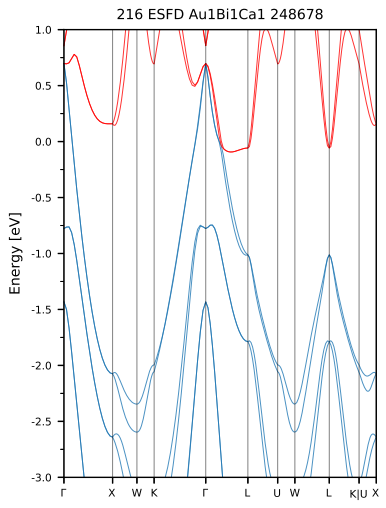

(aj) 216.4.22

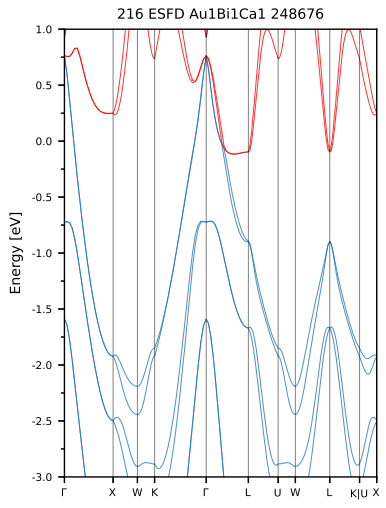

(ak) 216.4.23

Figure 480: Band structures for topological materials in SG 216, part 3/3, from 216.4.19 to 216.4.23. See the list of figures.

## 0.0.107 SG 217

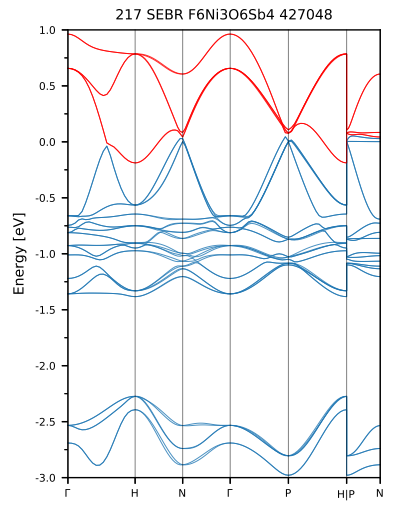

(a) 217.1.1

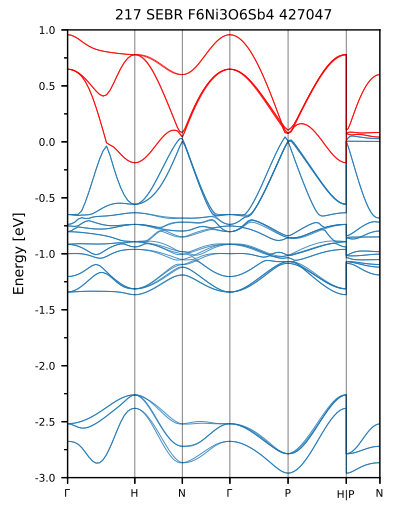

(b) 217.1.2

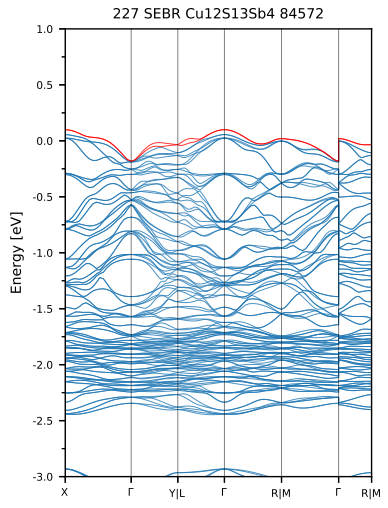

(c) 217.1.3

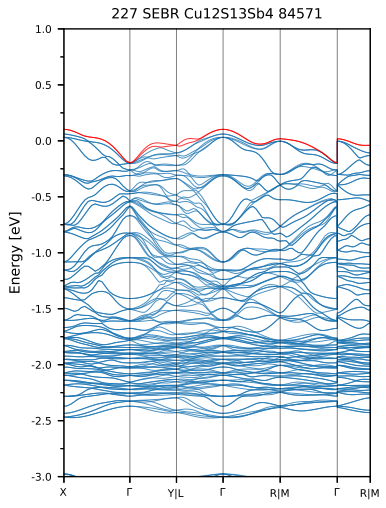

(d) 217.1.4

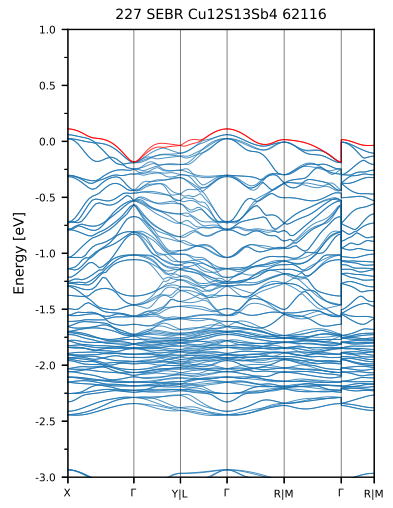

(e) 217.1.5

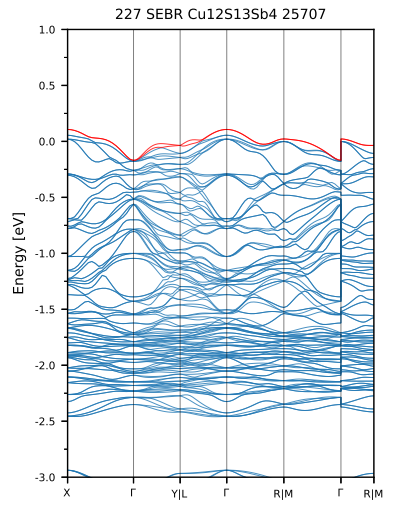

(f) 217.1.6

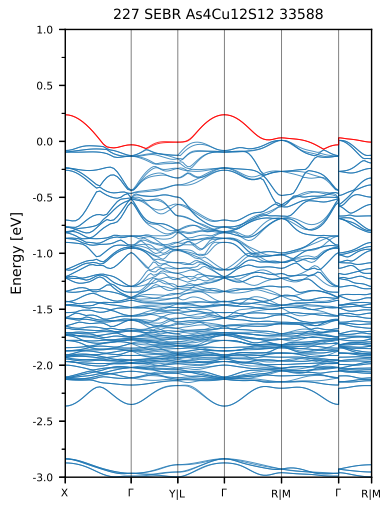

(g) 217.1.7

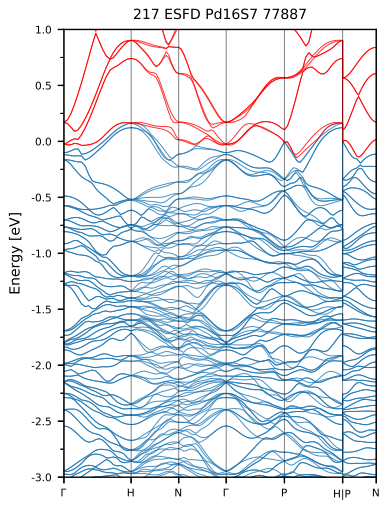

(h) 217.4.1

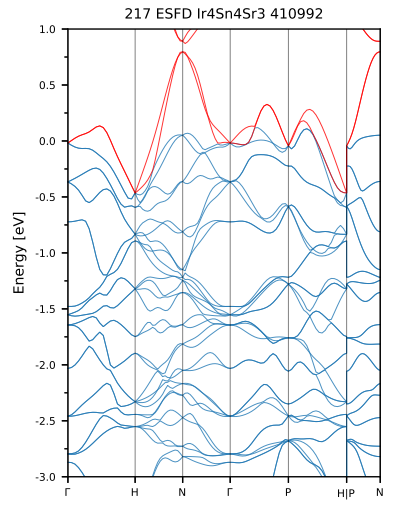

(i) 217.4.2

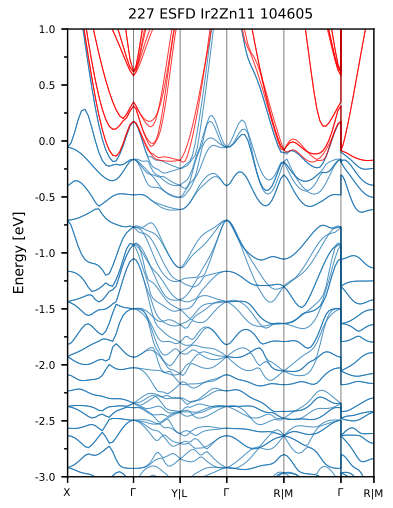

(j) 217.4.3

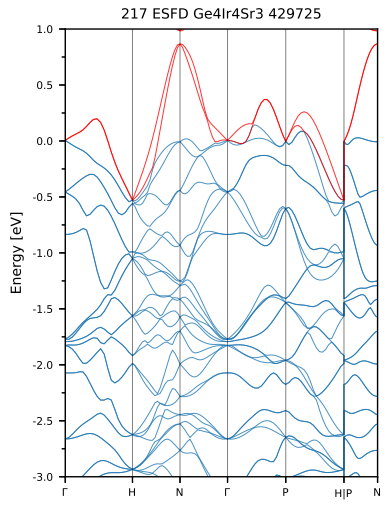

(k) 217.4.4

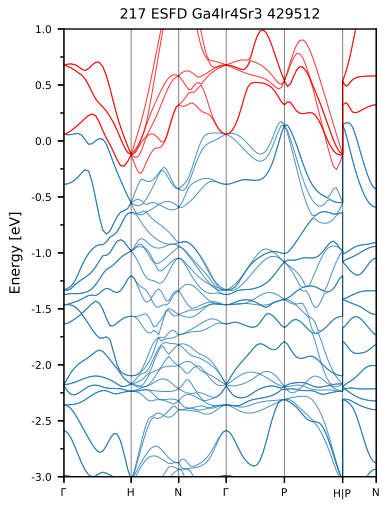

(l) 217.4.5

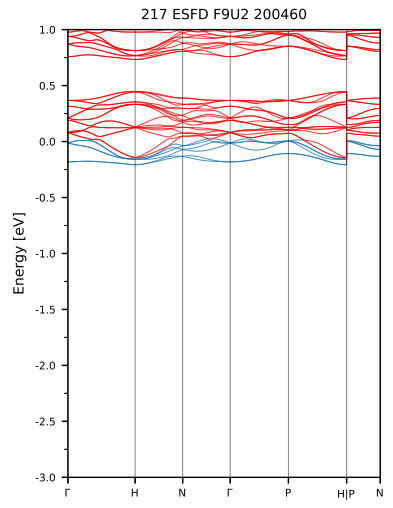

(m) 217.4.6

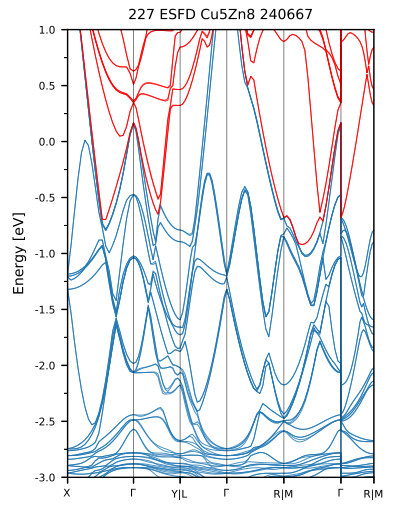

(n) 217.4.7

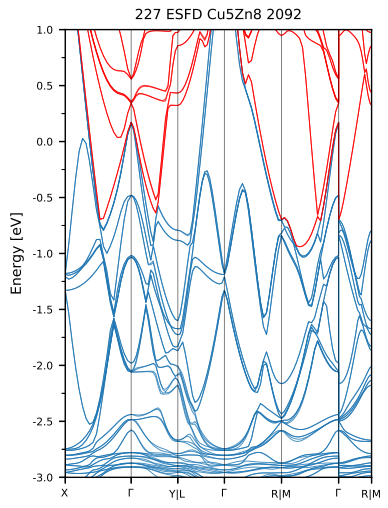

(o) 217.4.8

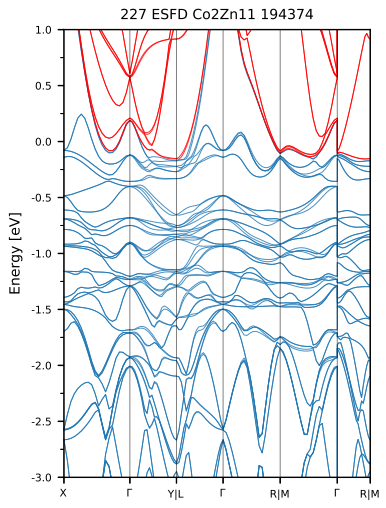

(p) 217.4.9

Figure 481: Band structures for topological materials in SG 217, part 1/2, from 217.1.1 to 217.4.9. See the list of figures.

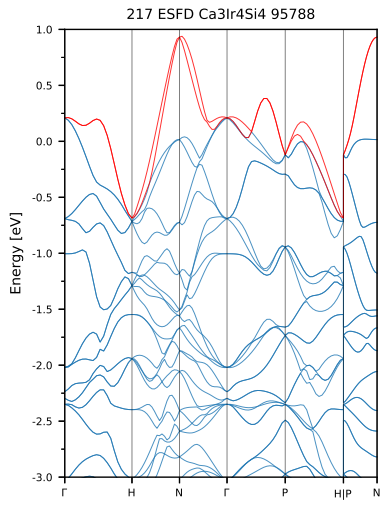

(q) 217.4.10

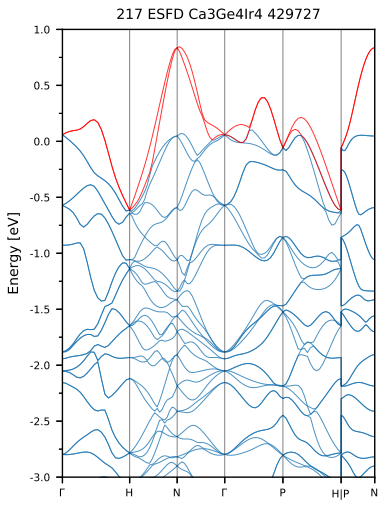

(r) 217.4.11

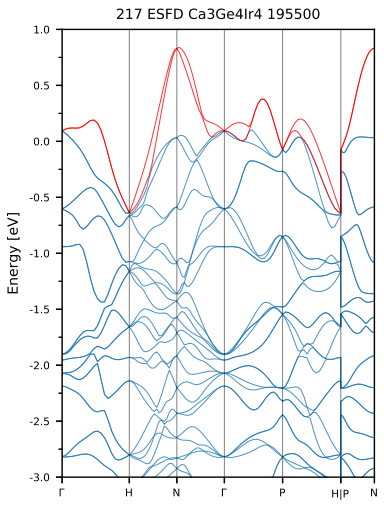

(s) 217.4.12

Figure 482: Band structures for topological materials in SG 217, part 2/2, from 217.4.10 to 217.4.12. See the list of figures.

## 0.0.108 SG 221

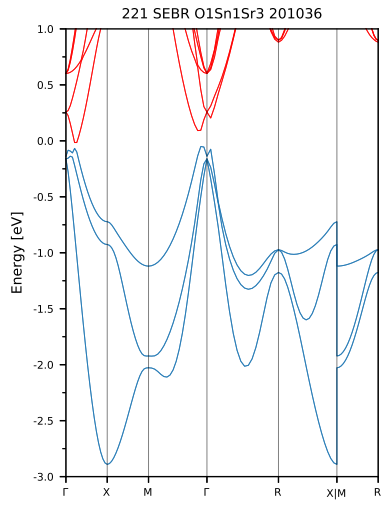

(a) 221.1.1

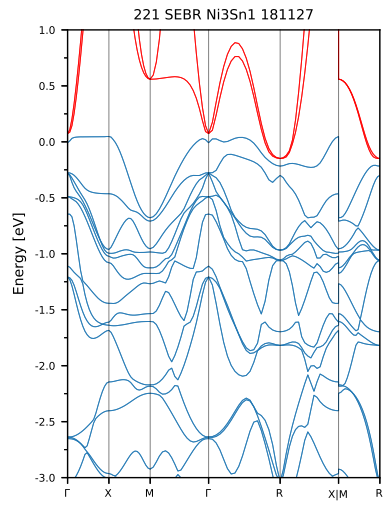

(b) 221.1.2

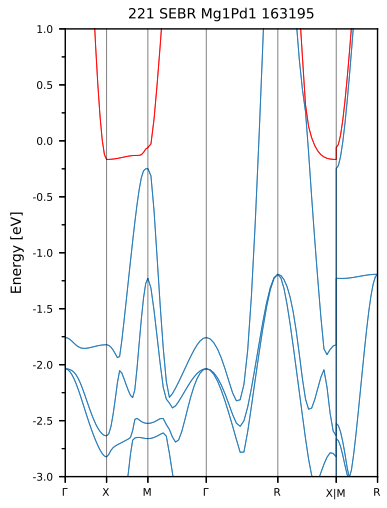

(c) 221.1.3

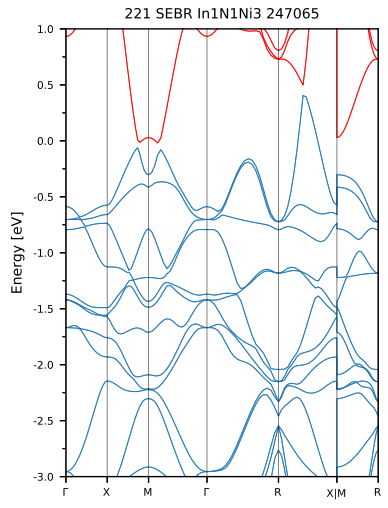

(d) 221.1.4

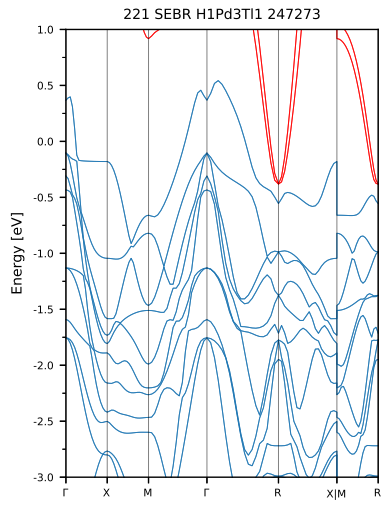

(e) 221.1.5

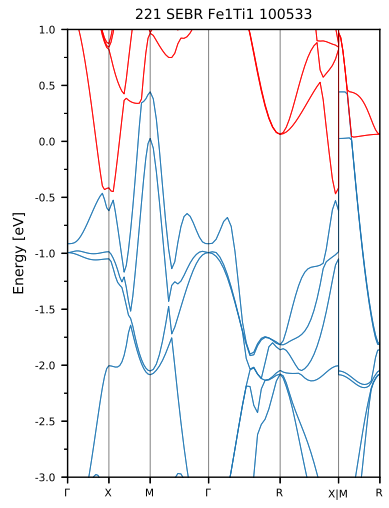

(f) 221.1.6

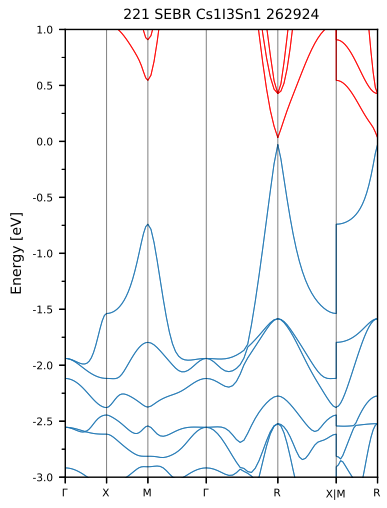

(g) 221.1.7

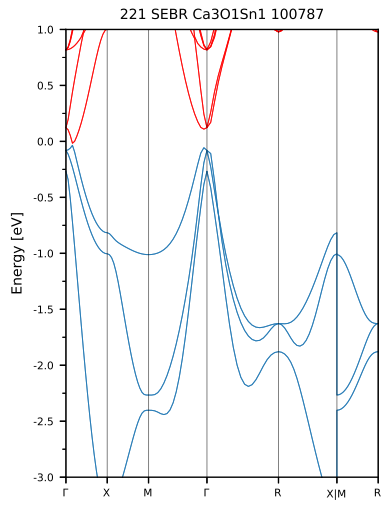

(h) 221.1.8

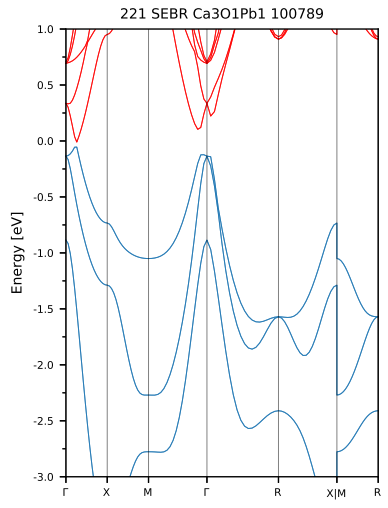

(i) 221.1.9

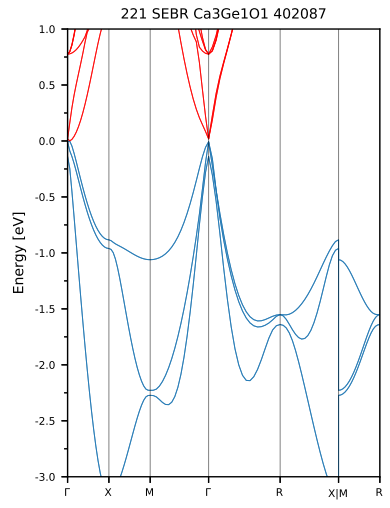

(j) 221.1.10

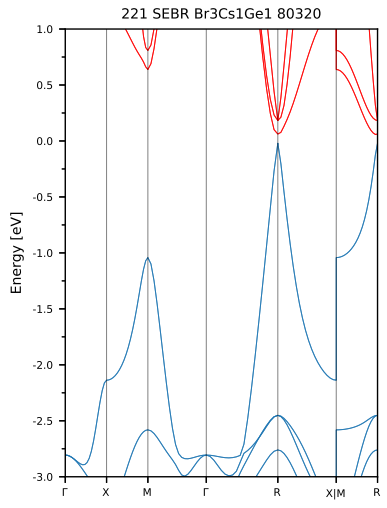

(k) 221.1.11

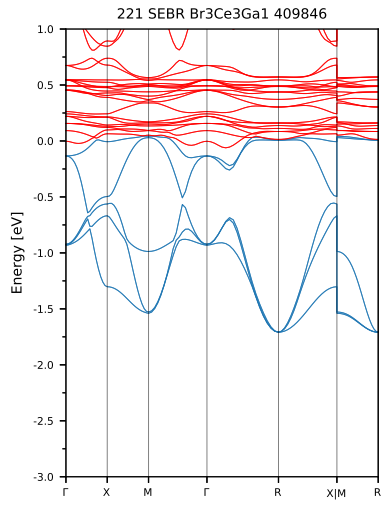

(l) 221.1.12

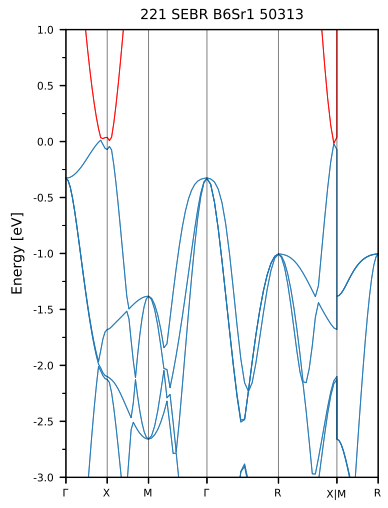

(m) 221.1.13

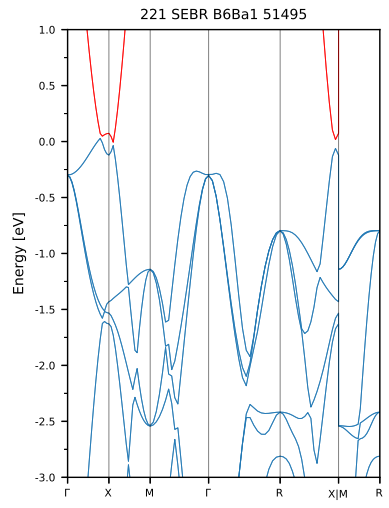

(n) 221.1.14

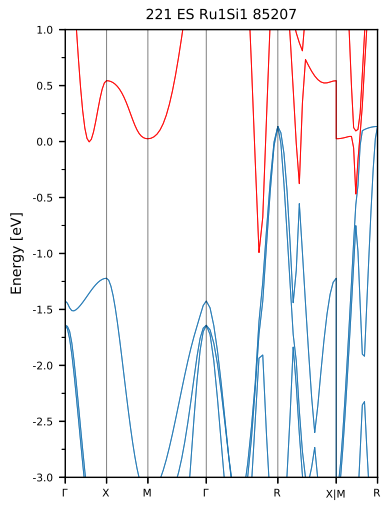

(o) 221.3.1

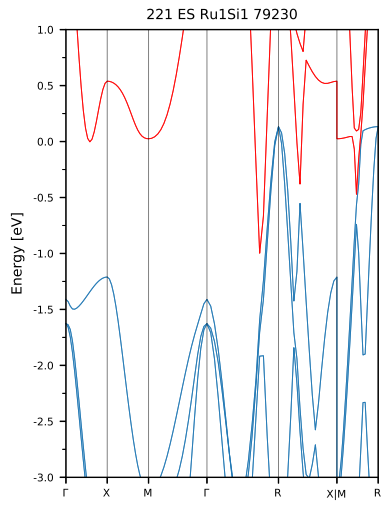

(p) 221.3.2

Figure 483: Band structures for topological materials in SG 221, part 1/6, from 221.1.1 to 221.3.2. See the list of figures.

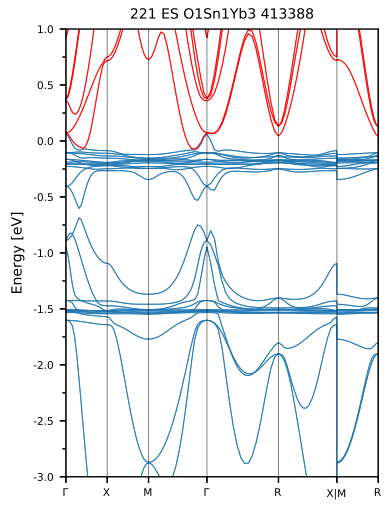

(q) 221.3.3

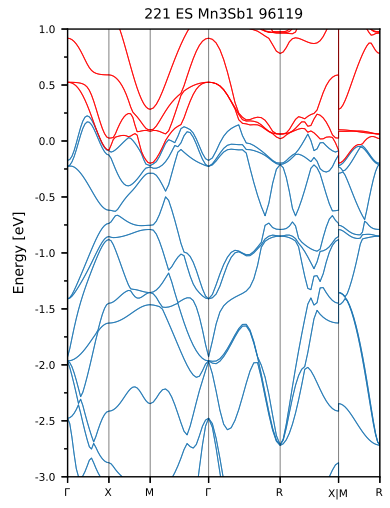

(r) 221.3.4

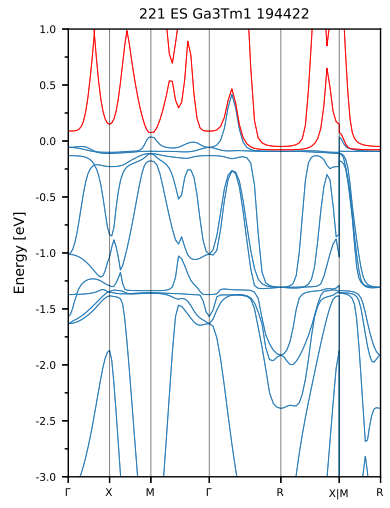

(s) 221.3.5

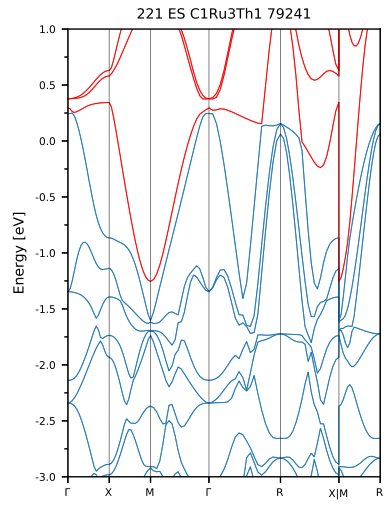

(t) 221.3.6

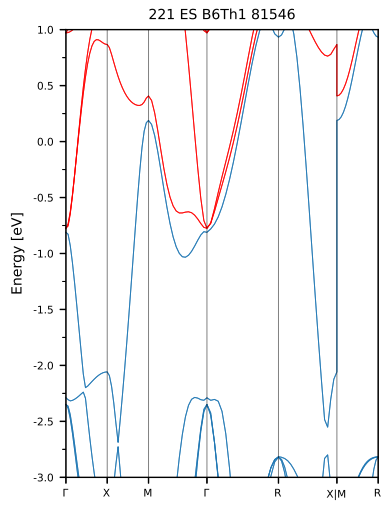

(u) 221.3.7

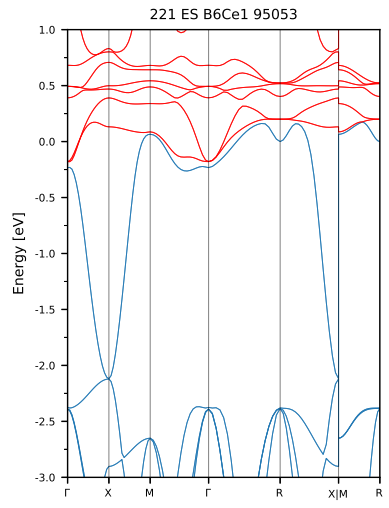

(v) 221.3.8

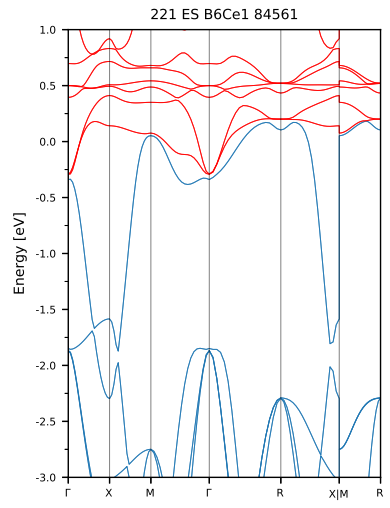

(w) 221.3.9

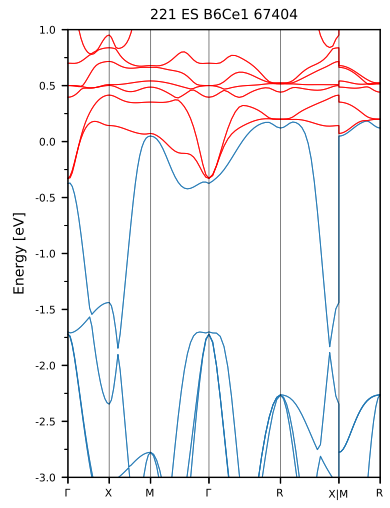

(x) 221.3.10

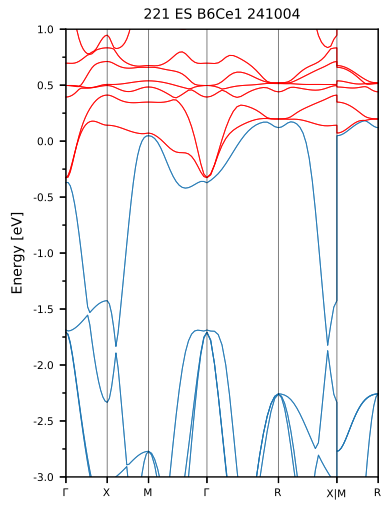

(y) 221.3.11

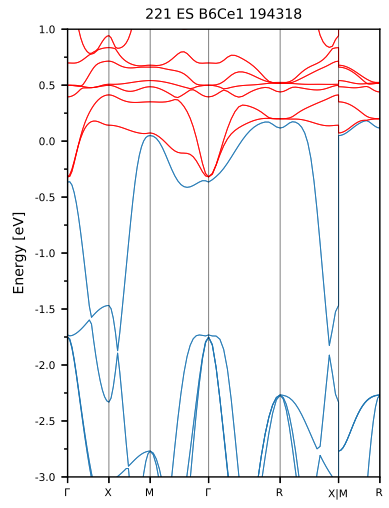

(z) 221.3.12

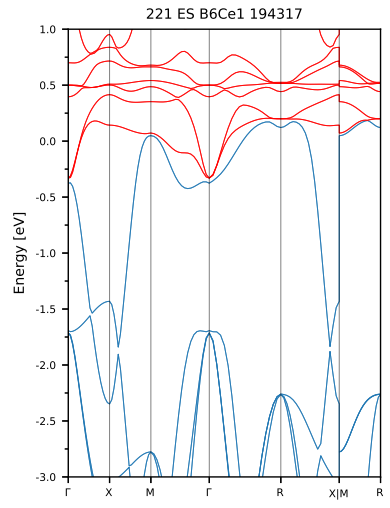

(aa) 221.3.13

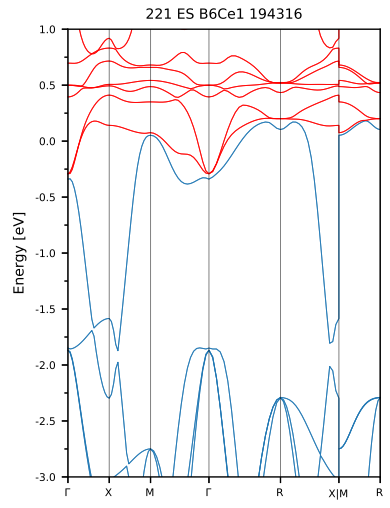

(ab) 221.3.14

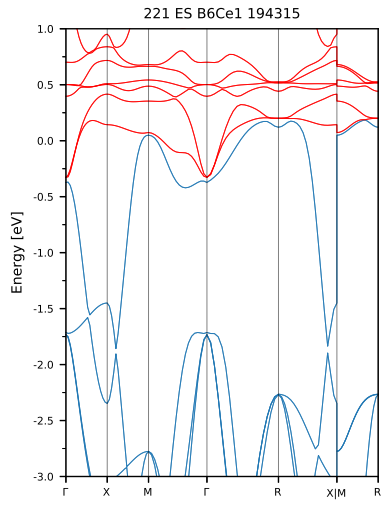

(ac) 221.3.15

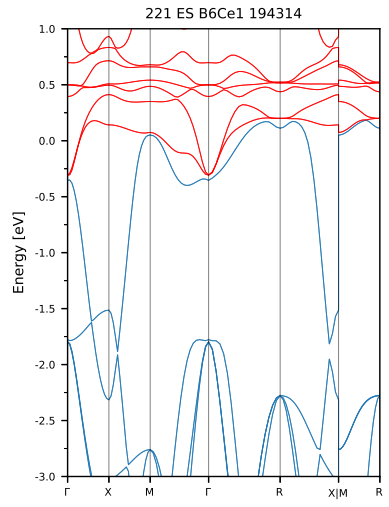

(ad) 221.3.16

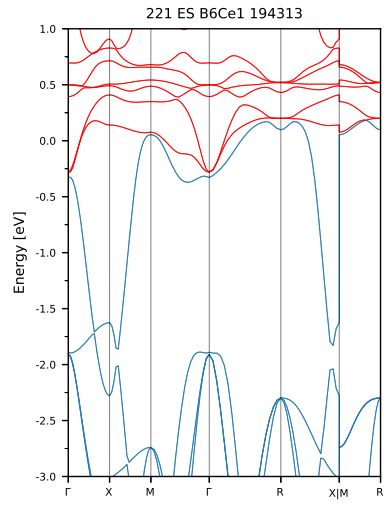

(ae) 221.3.17

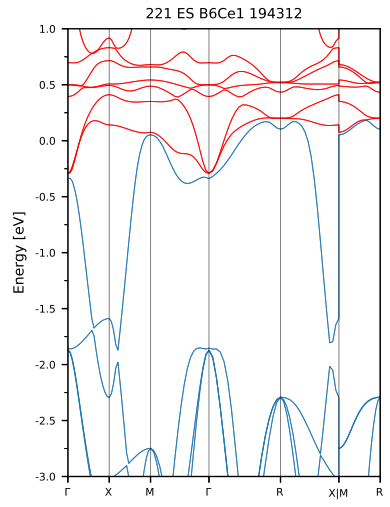

(af) 221.3.18

Figure 484: Band structures for topological materials in SG 221, part 2/6, from 221.3.3 to 221.3.18. See the list of figures.

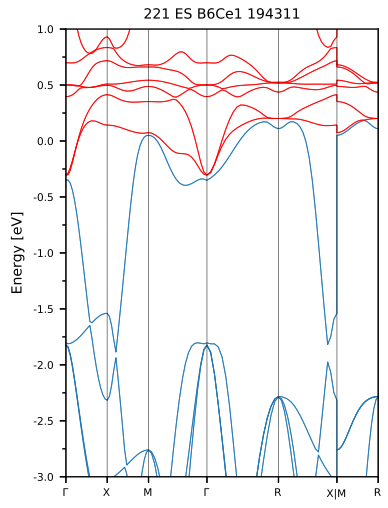

(ag) 221.3.19

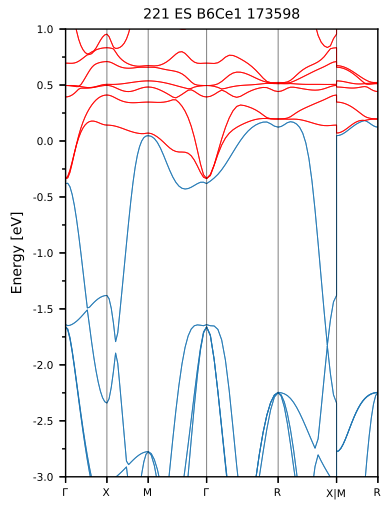

(ah) 221.3.20

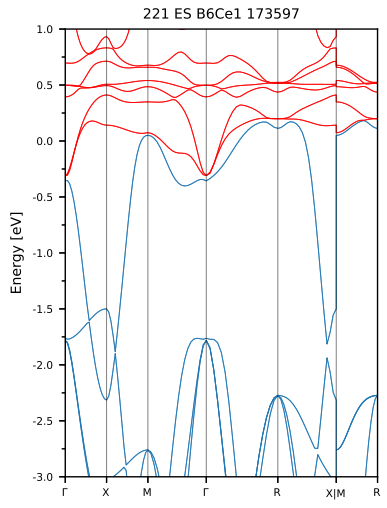

(ai) 221.3.21

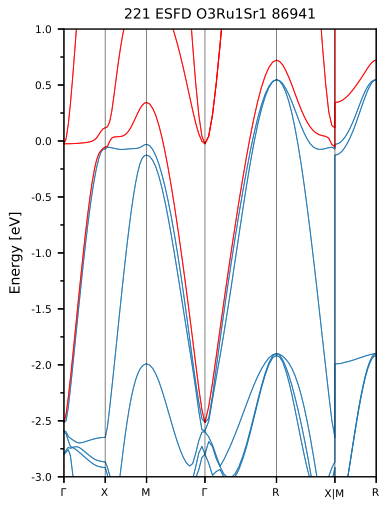

(aj) 221.4.1

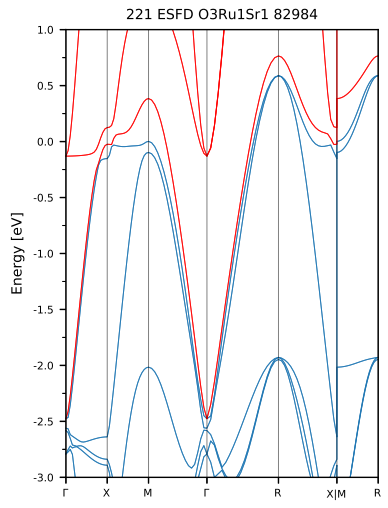

(ak) 221.4.2

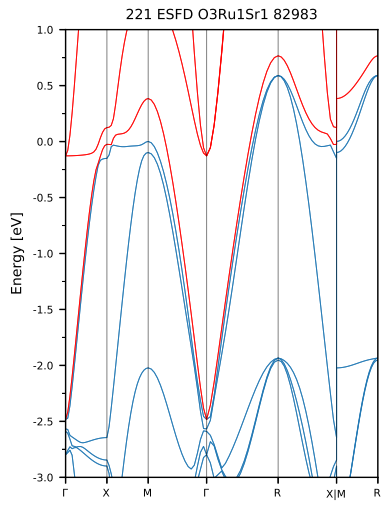

(al) 221.4.3

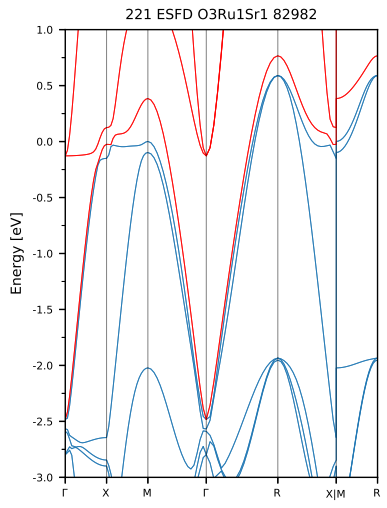

(am) 221.4.4

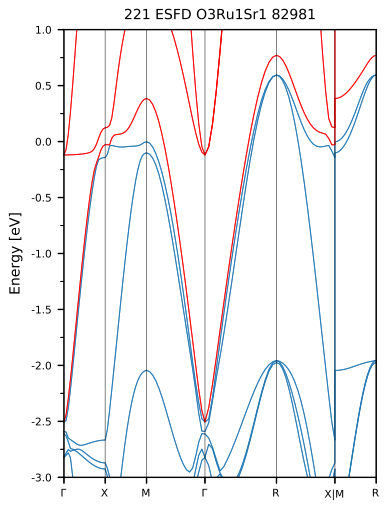

(an) 221.4.5

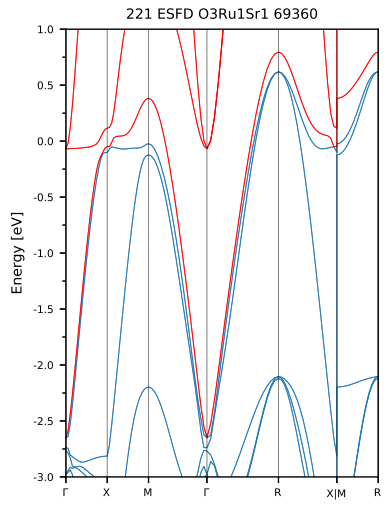

(ao) 221.4.6

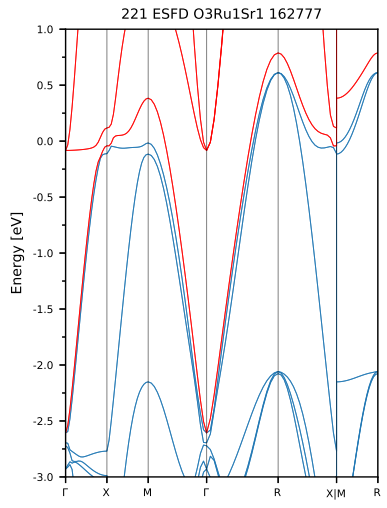

(ap) 221.4.7

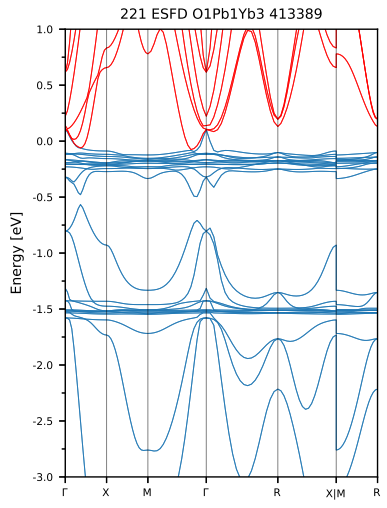

(aq) 221.4.8

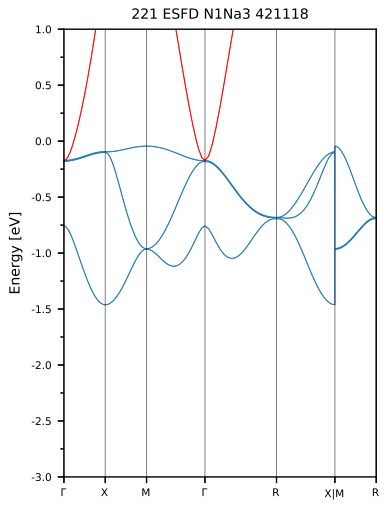

(ar) 221.4.9

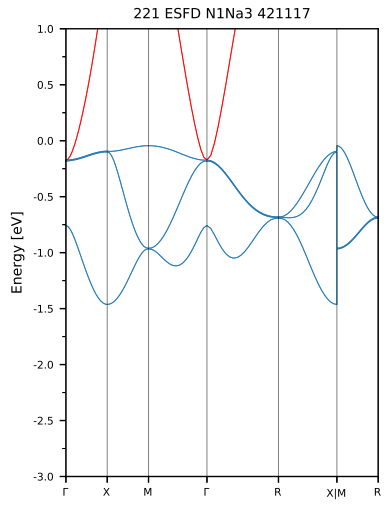

(as) 221.4.10

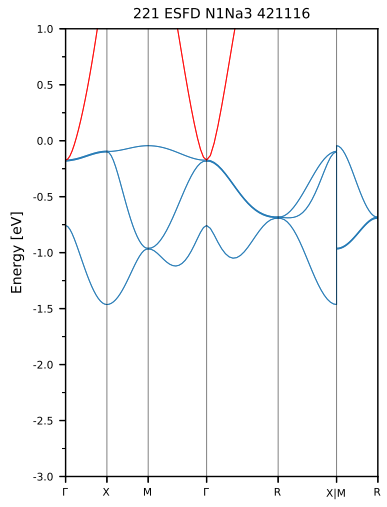

(at) 221.4.11

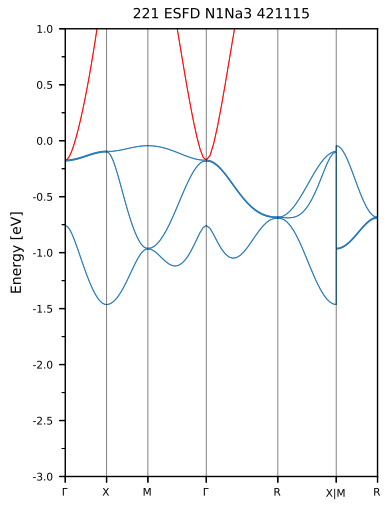

(au) 221.4.12

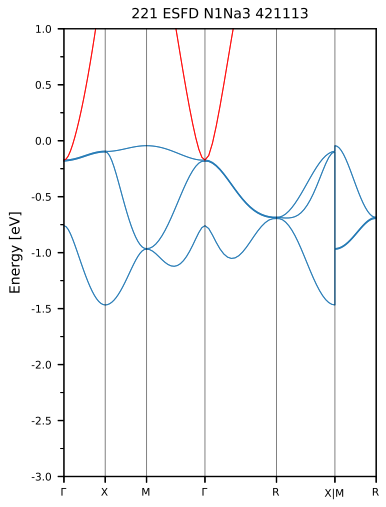

(av) 221.4.13

Figure 485: Band structures for topological materials in SG 221, part 3/6, from 221.3.19 to 221.4.13. See the list of figures.

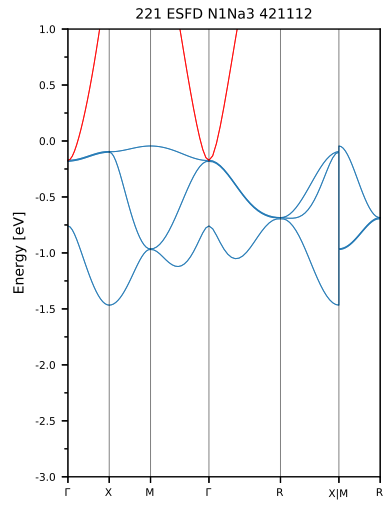

(aw) 221.4.14

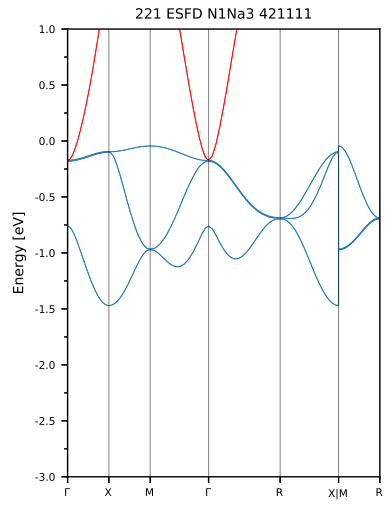

(ax) 221.4.15

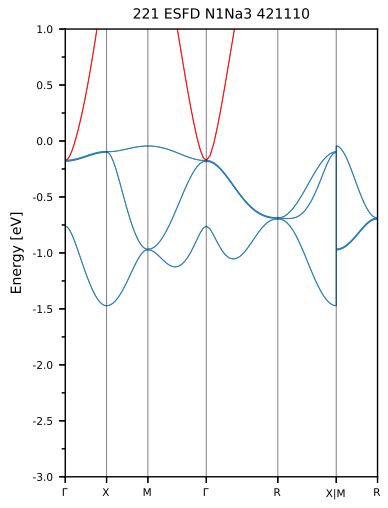

(ay) 221.4.16

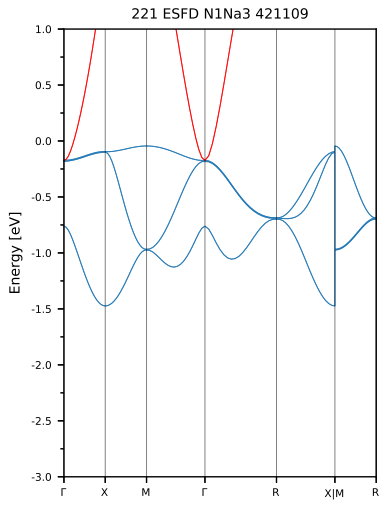

(az) 221.4.17

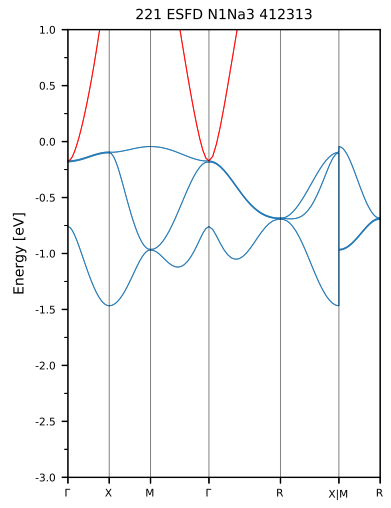

(ba) 221.4.18

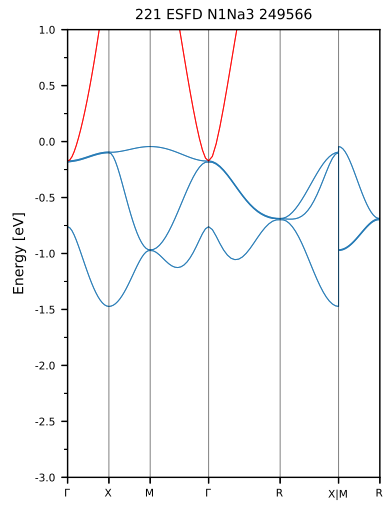

(bb) 221.4.19

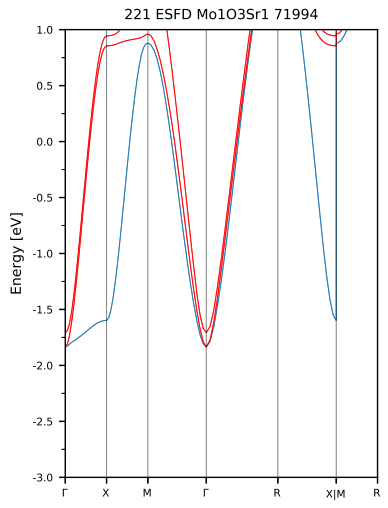

(bc) 221.4.20

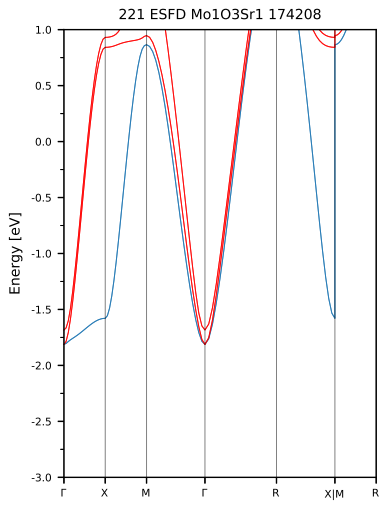

(bd) 221.4.21

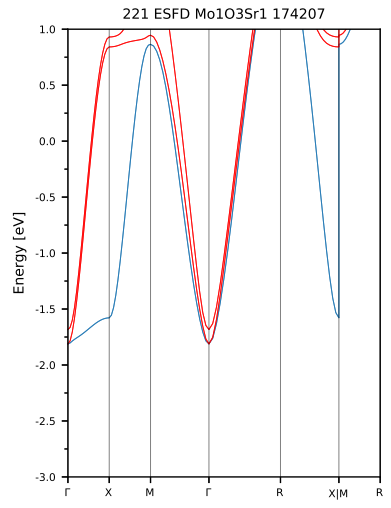

(be) 221.4.22

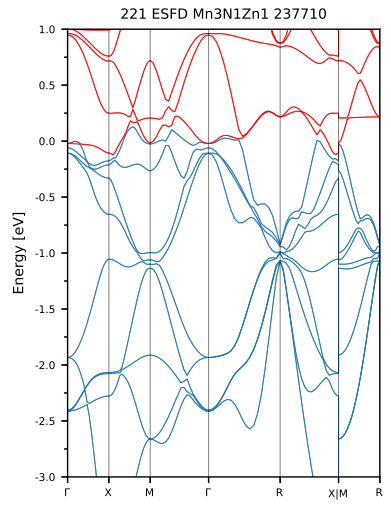

(bf) 221.4.23

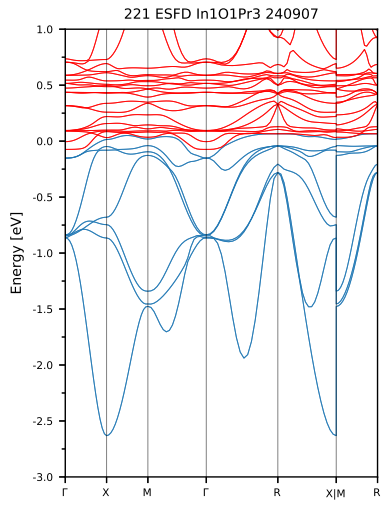

(bg) 221.4.24

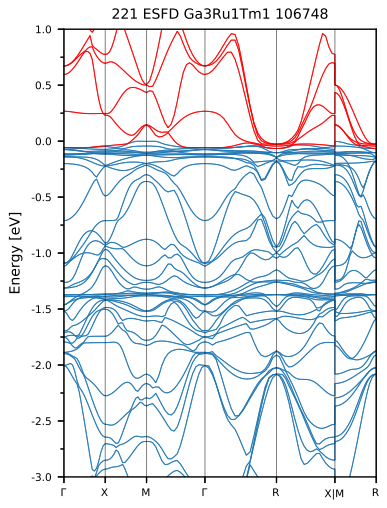

(bh) 221.4.25

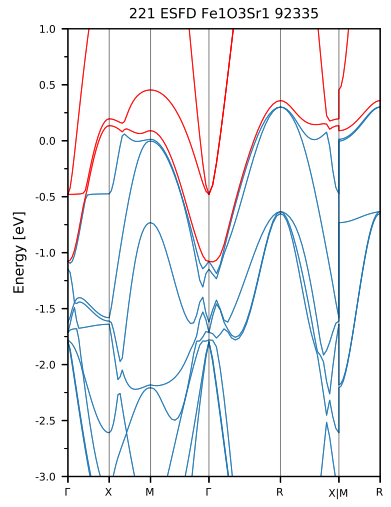

(bi) 221.4.26

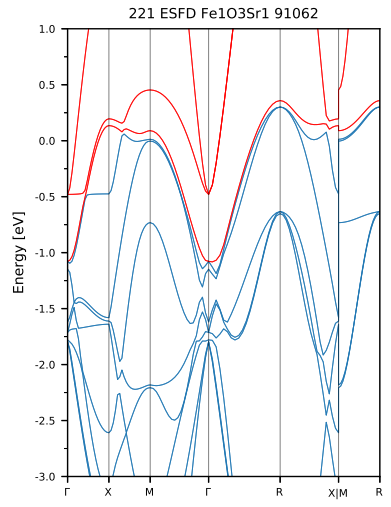

(bj) 221.4.27

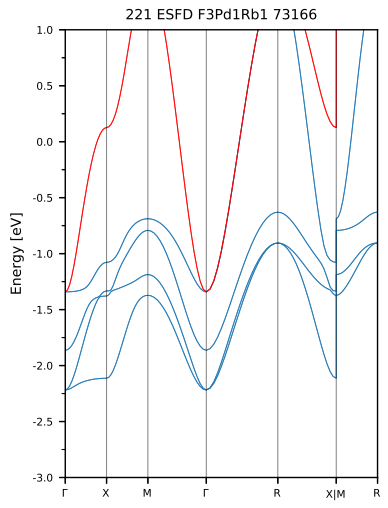

(bk) 221.4.28

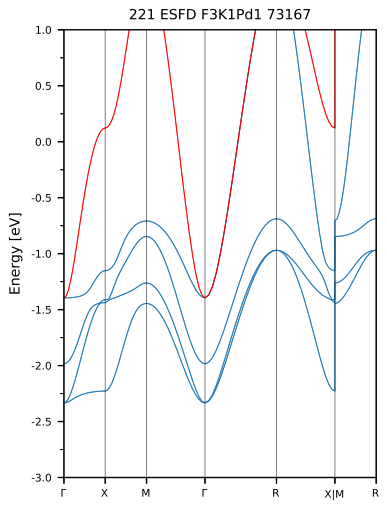

(bl) 221.4.29

Figure 486: Band structures for topological materials in SG 221, part 4/6, from 221.4.14 to 221.4.29. See the list of figures.

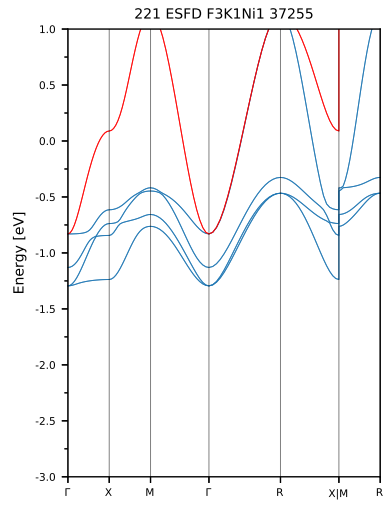

(bm) 221.4.30

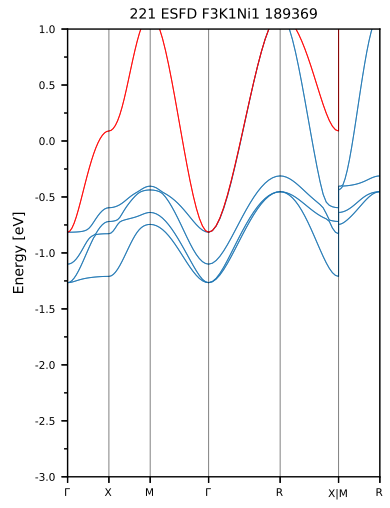

(bn) 221.4.31

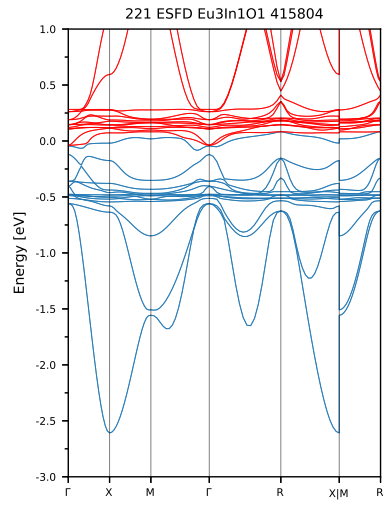

(bo) 221.4.32

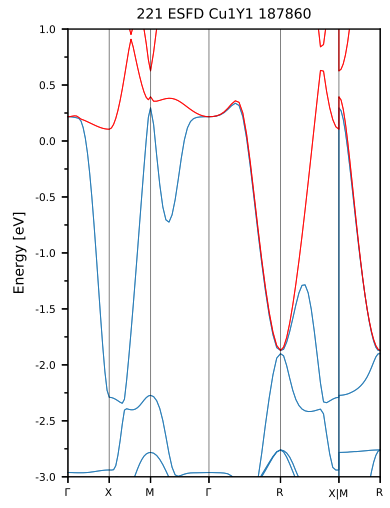

(bp) 221.4.33

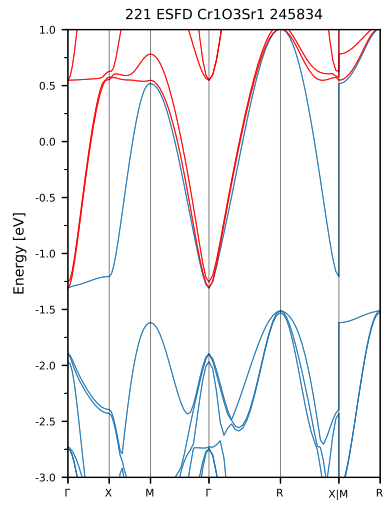

(bq) 221.4.34

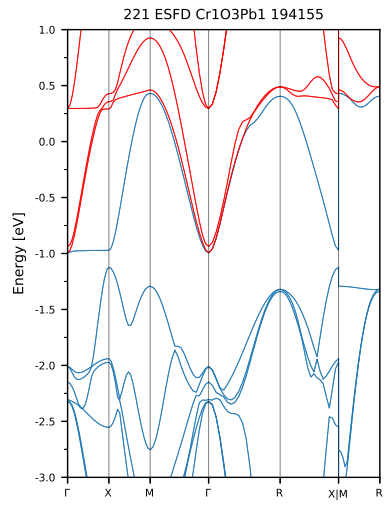

(br) 221.4.35

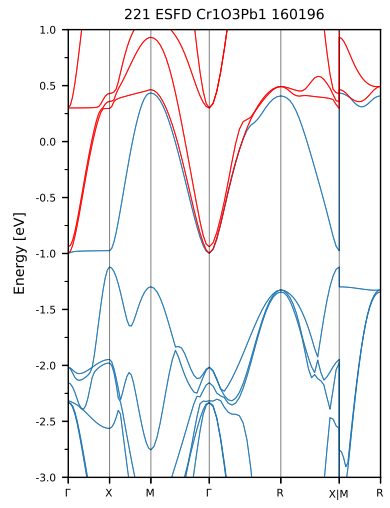

(bs) 221.4.36

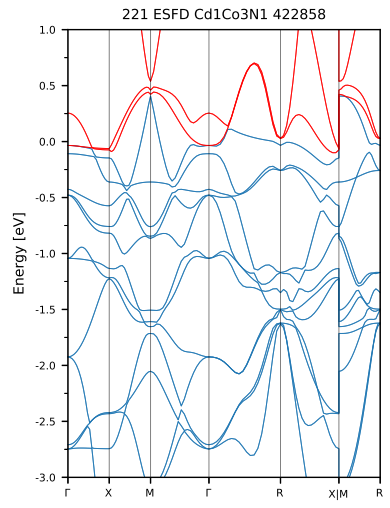

(bt) 221.4.37

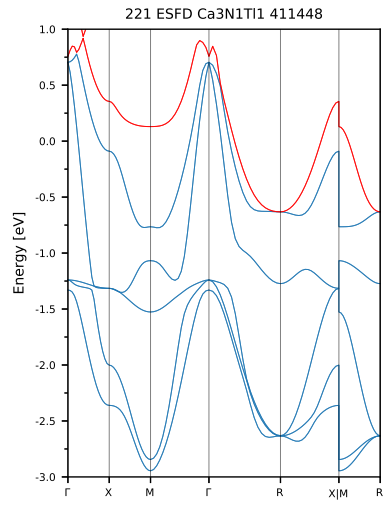

(bu) 221.4.38

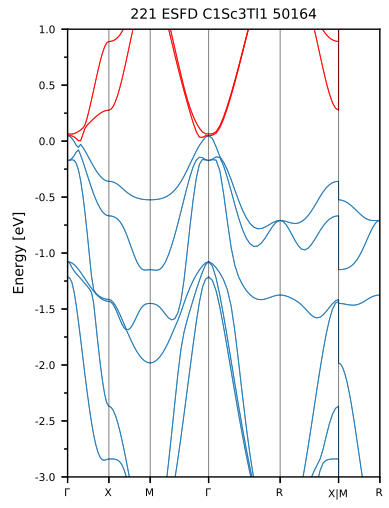

(bv) 221.4.39

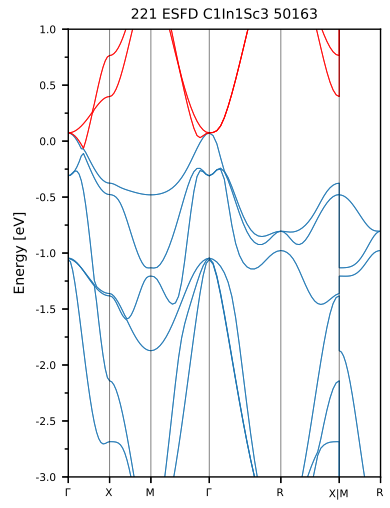

(bw) 221.4.40

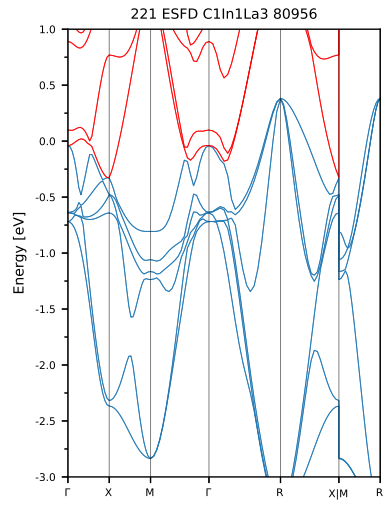

(bx) 221.4.41

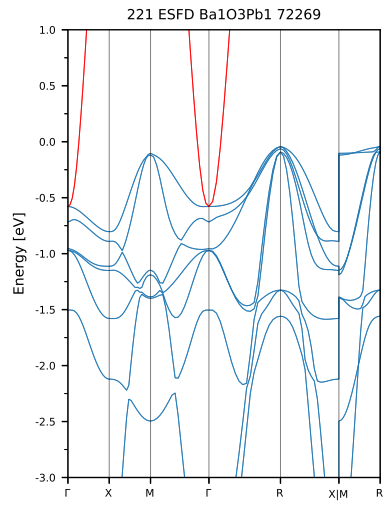

(by) 221.4.42

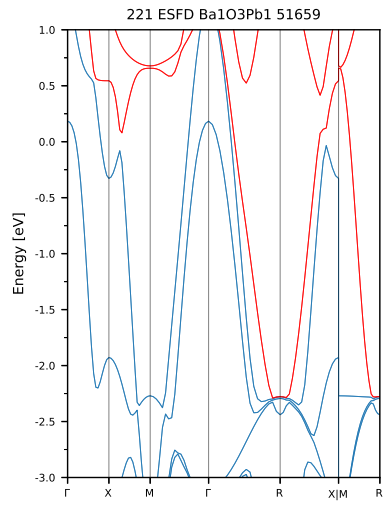

(bz) 221.4.43

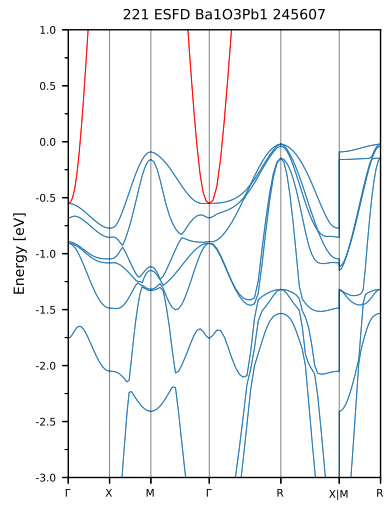

(ca) 221.4.44

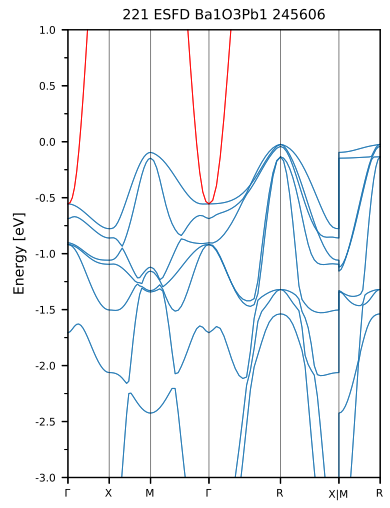

(cb) 221.4.45

Figure 487: Band structures for topological materials in SG 221, part 5/6, from 221.4.30 to 221.4.45. See the list of figures.

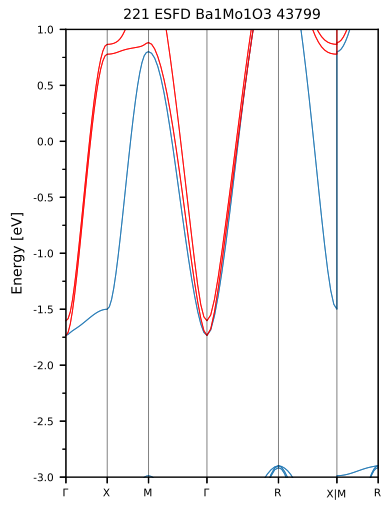

(cc) 221.4.46

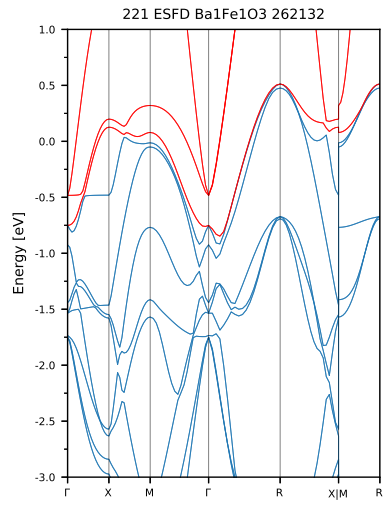

(cd) 221.4.47

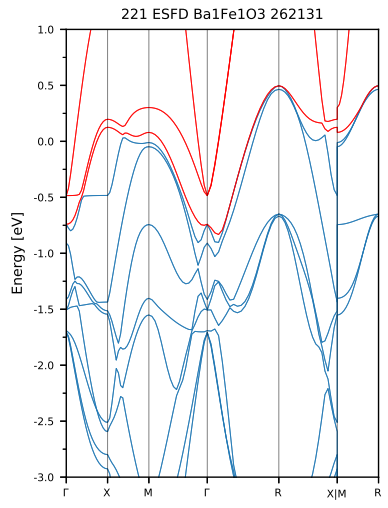

(ce) 221.4.48

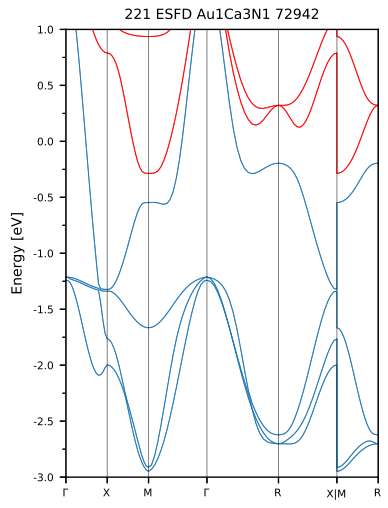

(cf) 221.4.49

Figure 488: Band structures for topological materials in SG 221, part 6/6, from 221.4.46 to 221.4.49. See the list of figures.

## 0.0.109 SG 223

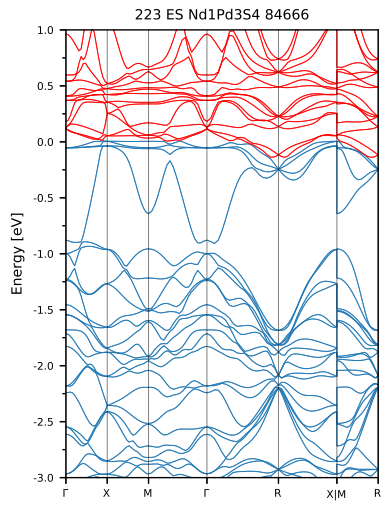

(a) 223.3.1

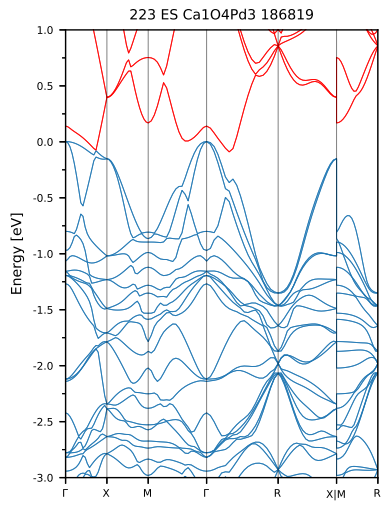

(b) 223.3.2

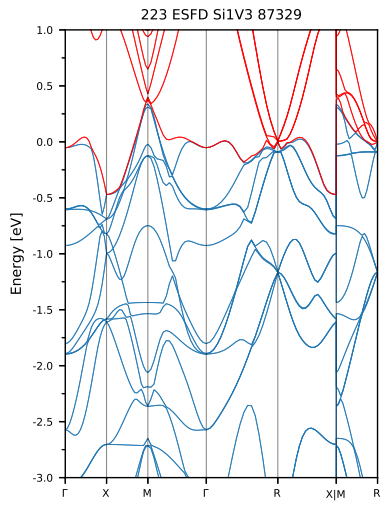

(c) 223.4.1

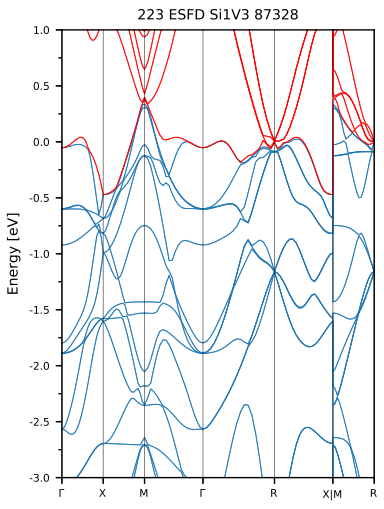

(d) 223.4.2

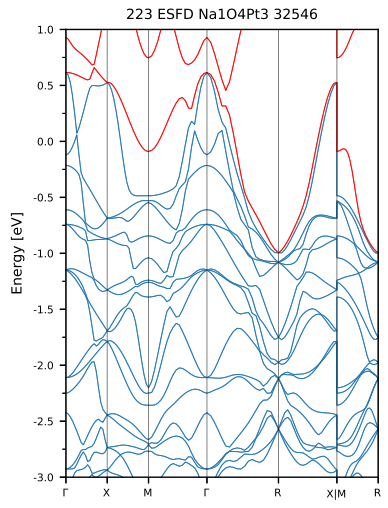

(e) 223.4.3

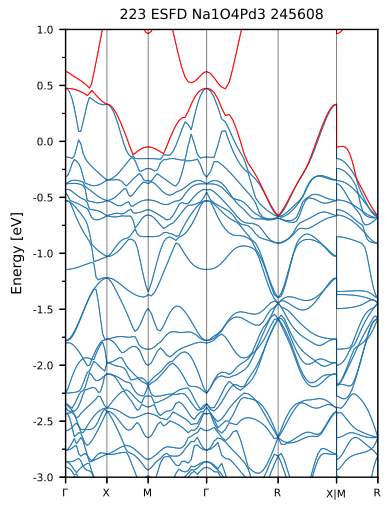

(f) 223.4.4

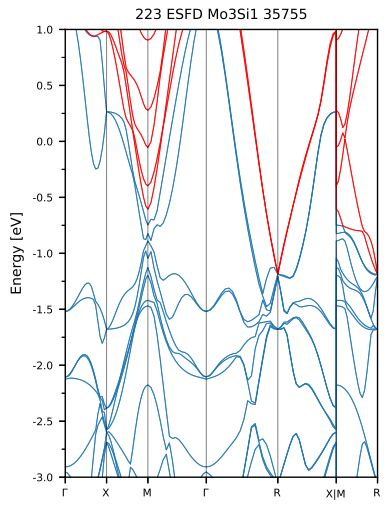

(g) 223.4.5

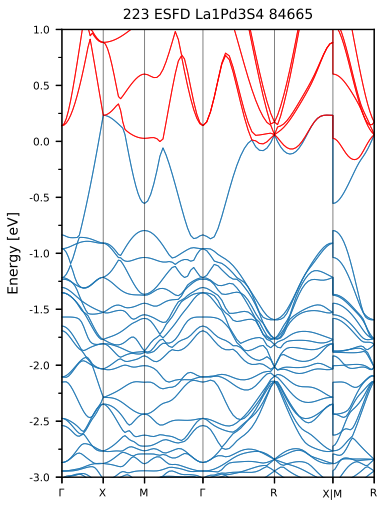

(h) 223.4.6

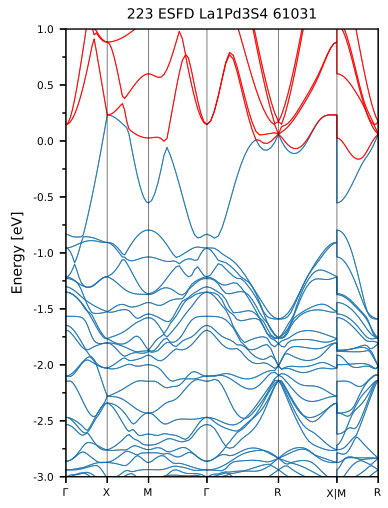

(i) 223.4.7

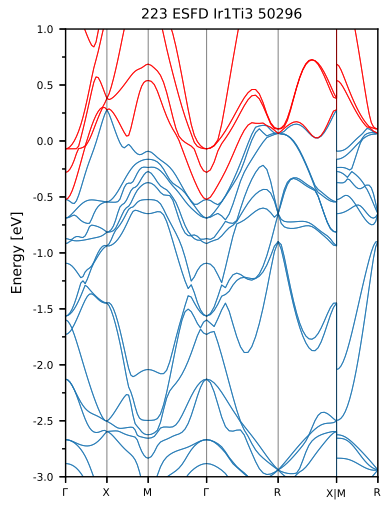

(j) 223.4.8

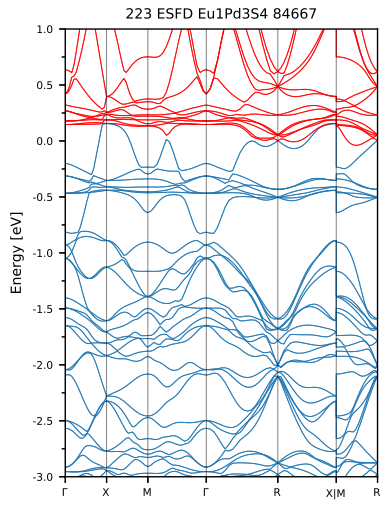

(k) 223.4.9

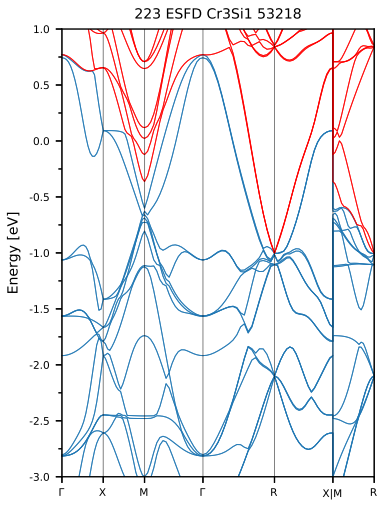

(l) 223.4.10

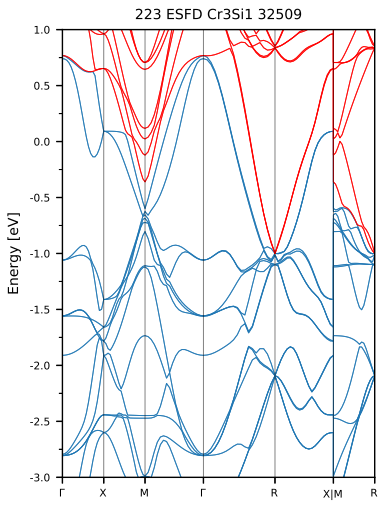

(m) 223.4.11

Figure 489: Band structures for topological materials in SG 223, part 1/1, from 223.3.1 to 223.4.11. See the list of figures.

### 0.0.110 SG 224

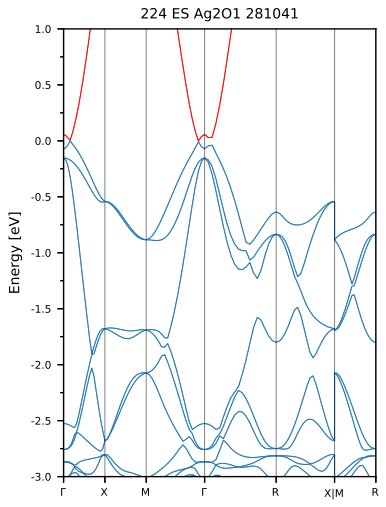

(a) 224.3.1

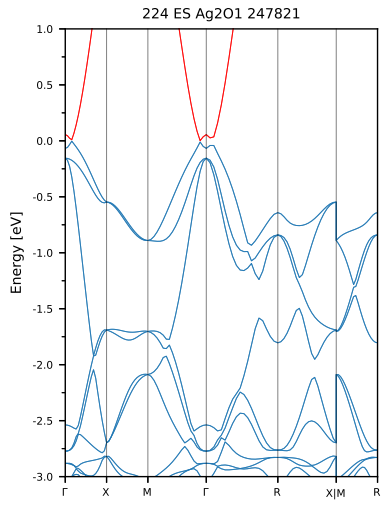

(b) 224.3.2

Figure 490: Band structures for topological materials in SG 224, part 1/1, from 224.3.1 to 224.3.2. See the list of figures.

## 0.0.111 SG 225

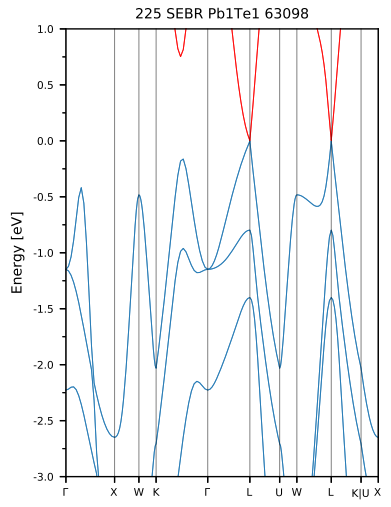

(a) 225.1.1

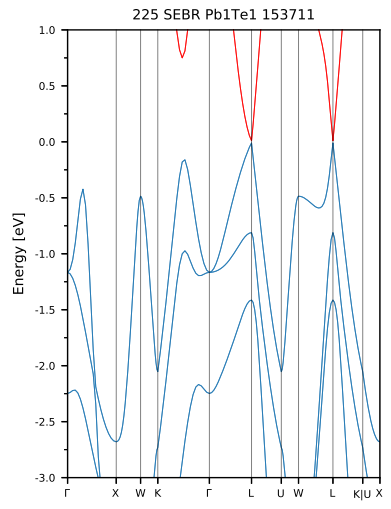

(b) 225.1.2

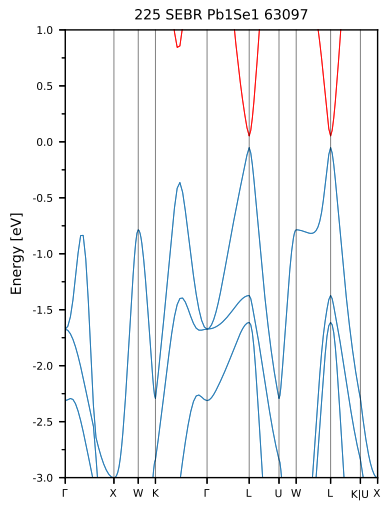

(c) 225.1.3

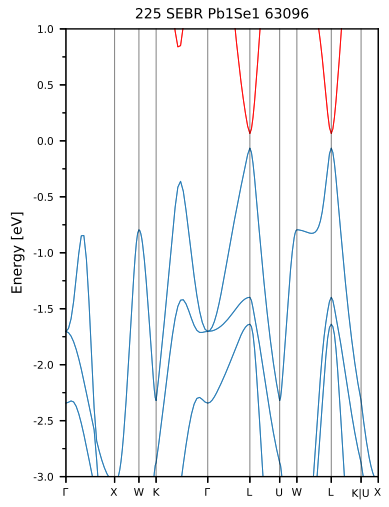

(d) 225.1.4

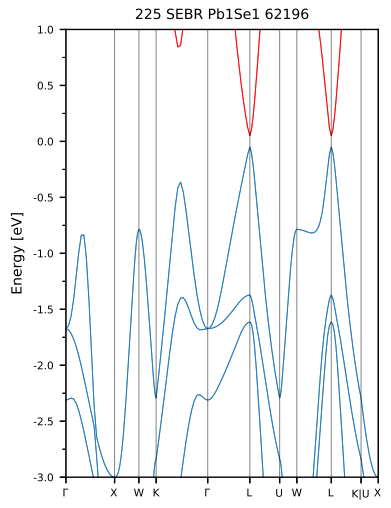

(e) 225.1.5

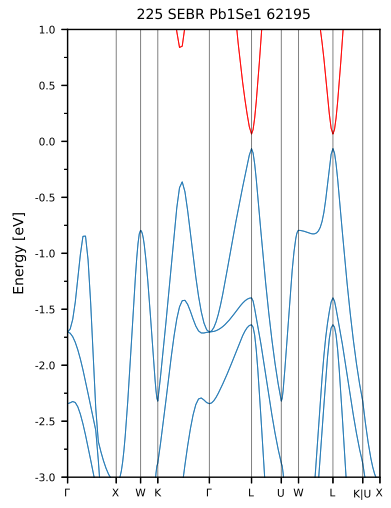

(f) 225.1.6

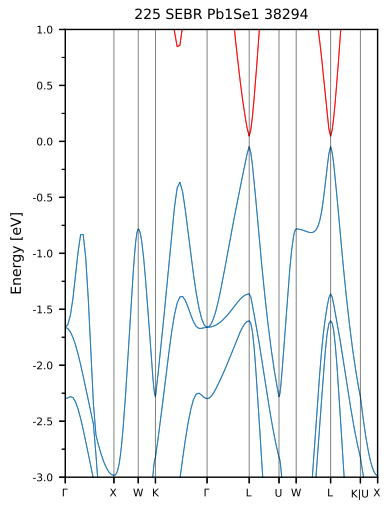

(g) 225.1.7

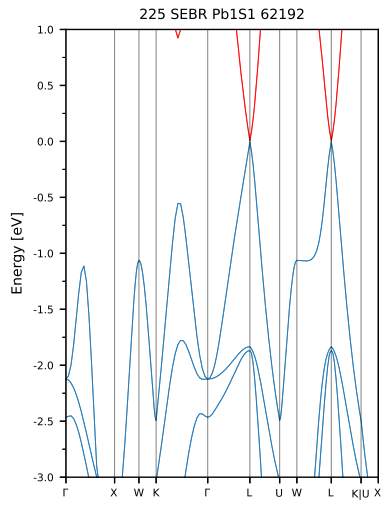

(h) 225.1.8

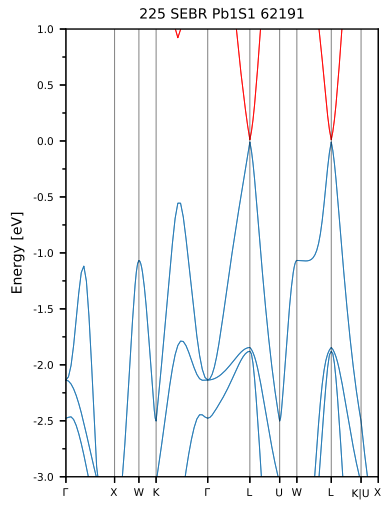

(i) 225.1.9

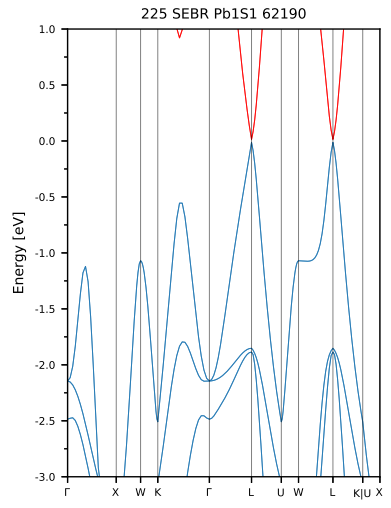

(j) 225.1.10

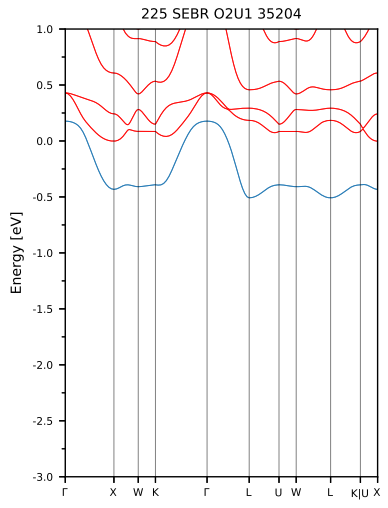

(k) 225.1.11

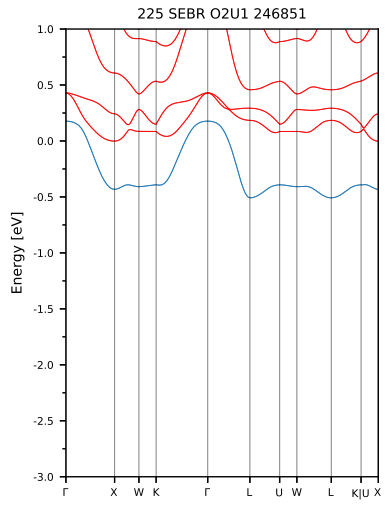

(l) 225.1.12

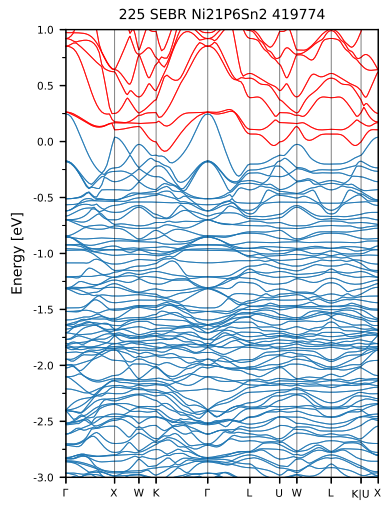

(m) 225.1.13

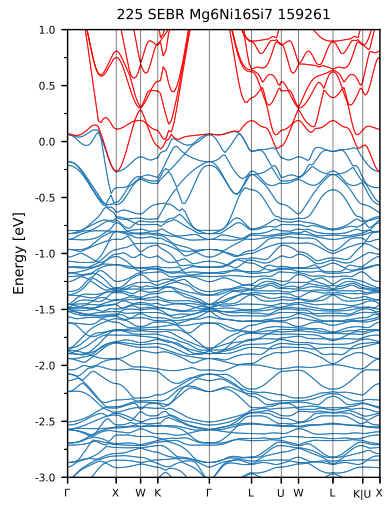

(n) 225.1.14

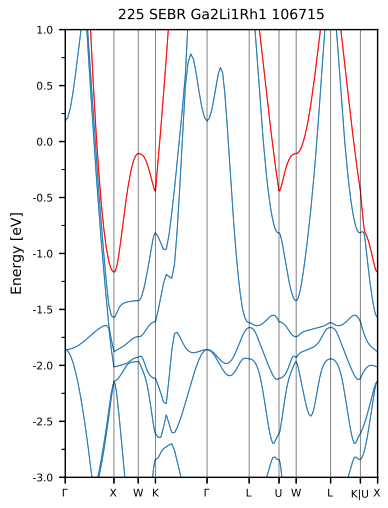

(o) 225.1.15

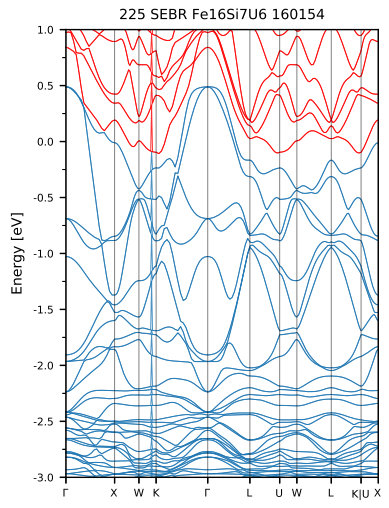

(p) 225.1.16

Figure 491: Band structures for topological materials in SG 225, part 1/7, from 225.1.1 to 225.1.16. See the list of figures.

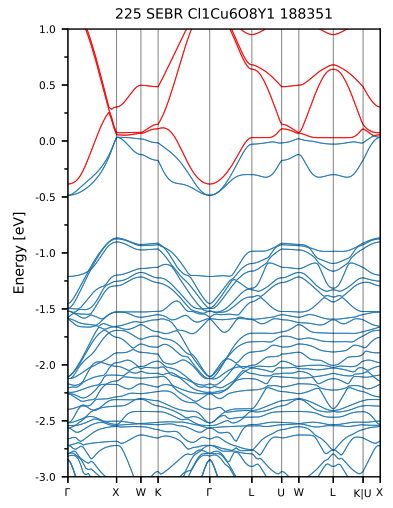

(q) 225.1.17

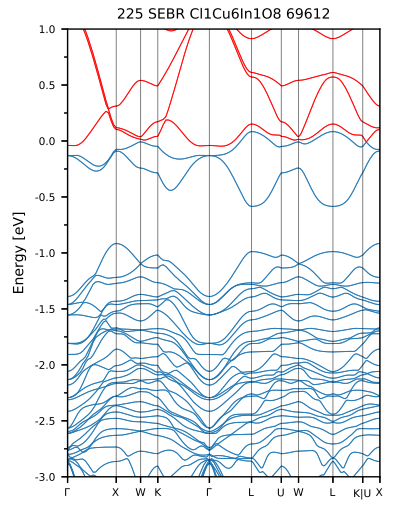

(r) 225.1.18

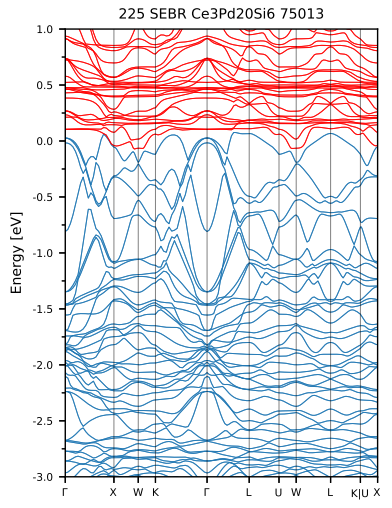

(s) 225.1.19

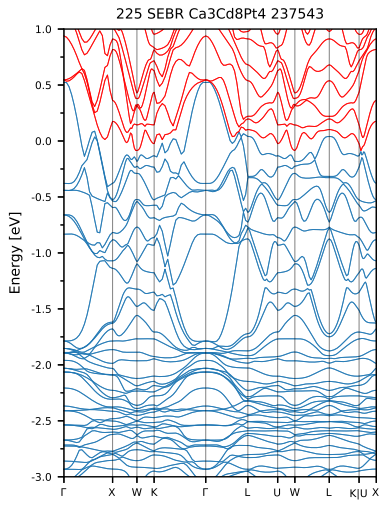

(t) 225.1.20

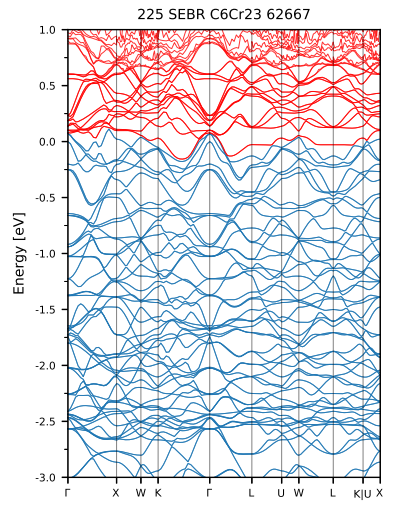

(u) 225.1.21

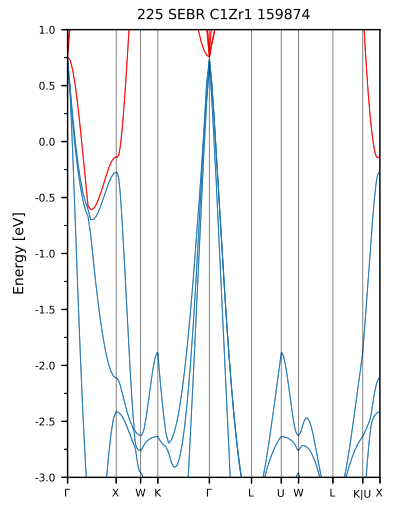

(v) 225.1.22

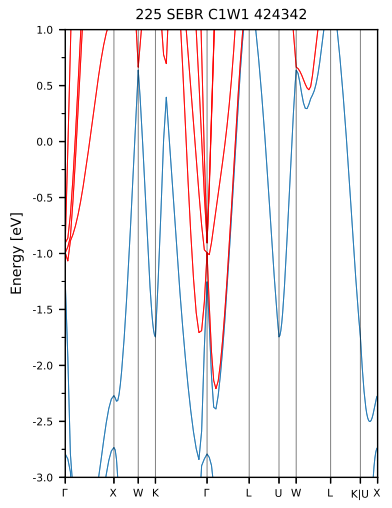

(w) 225.1.23

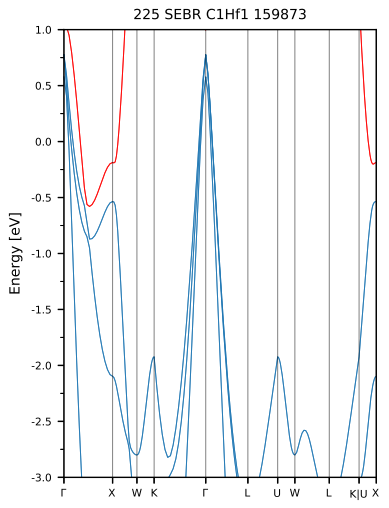

(x) 225.1.24

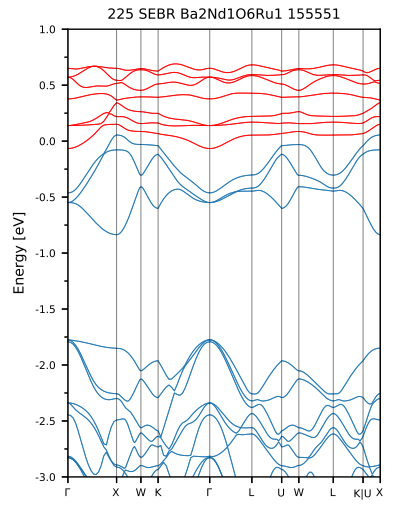

(y) 225.1.25

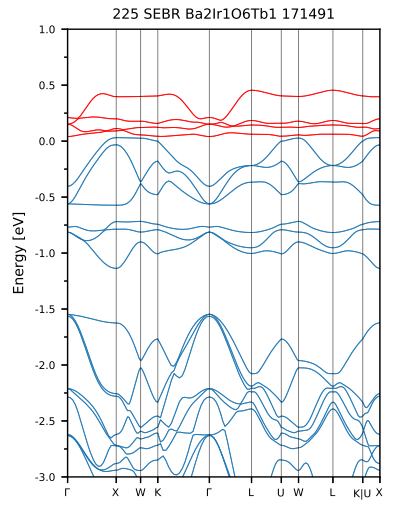

(z) 225.1.26

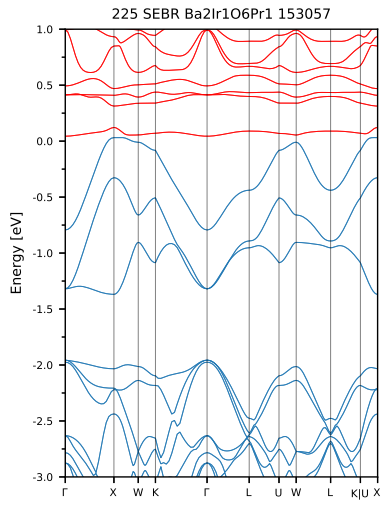

(aa) 225.1.27

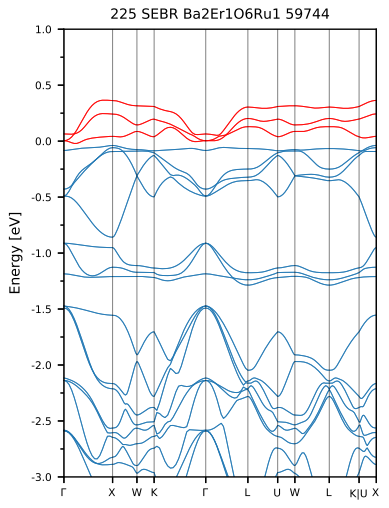

(ab) 225.1.28

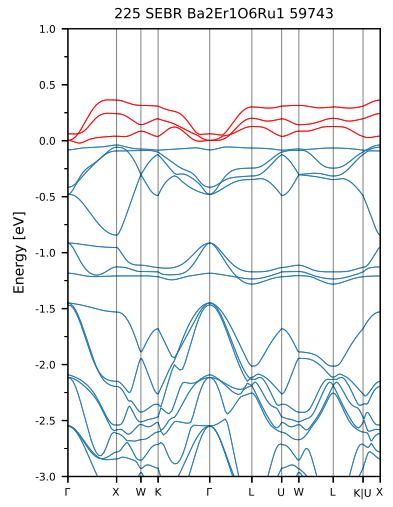

(ac) 225.1.29

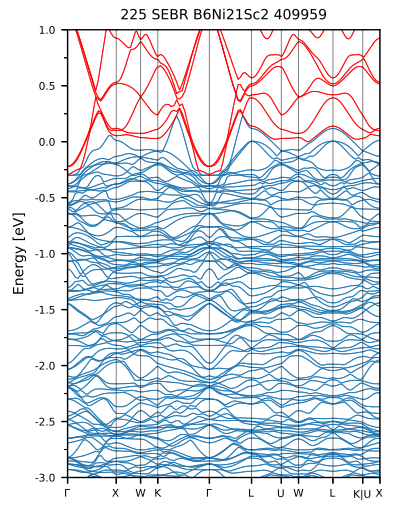

(ad) 225.1.30

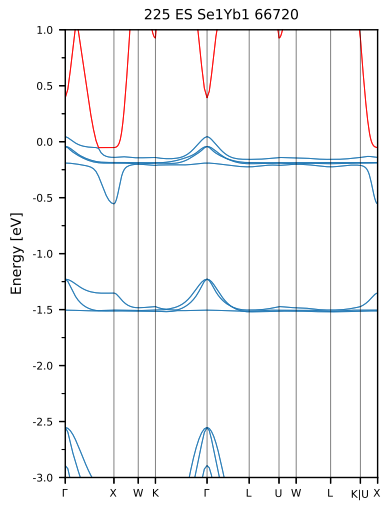

(ae) 225.3.1

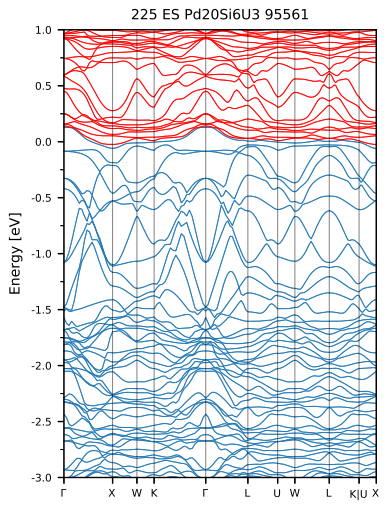

(af) 225.3.2

Figure 492: Band structures for topological materials in SG 225, part 2/7, from 225.1.17 to 225.3.2. See the list of figures.

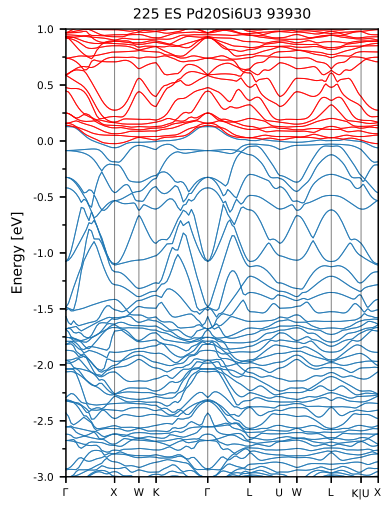

(ag) 225.3.3

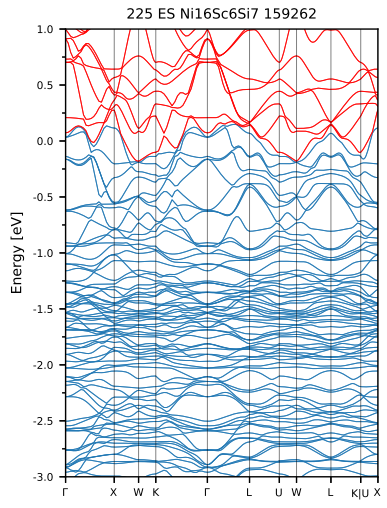

(ah) 225.3.4

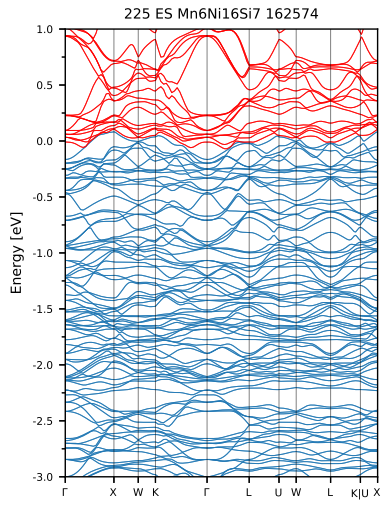

(ai) 225.3.5

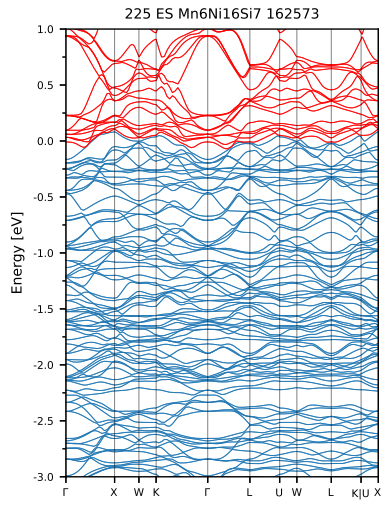

(aj) 225.3.6

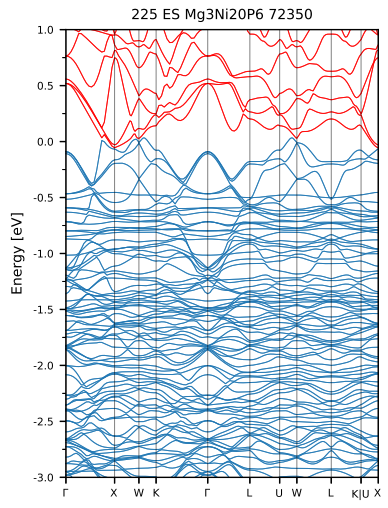

(ak) 225.3.7

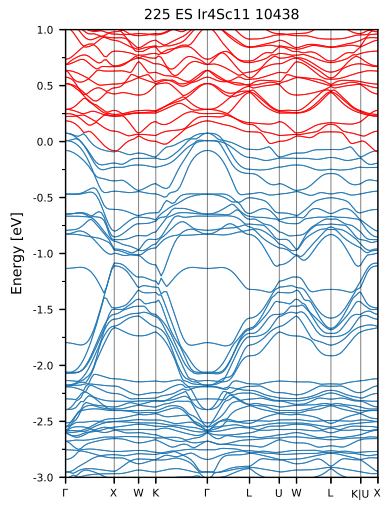

(al) 225.3.8

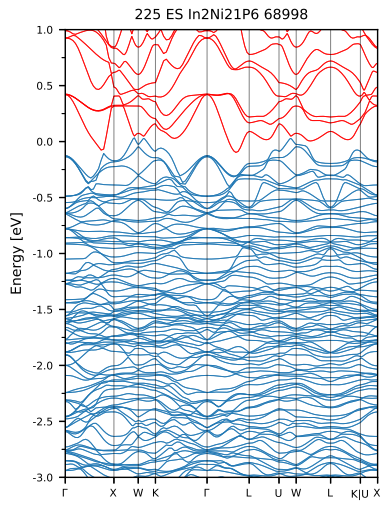

(am) 225.3.9

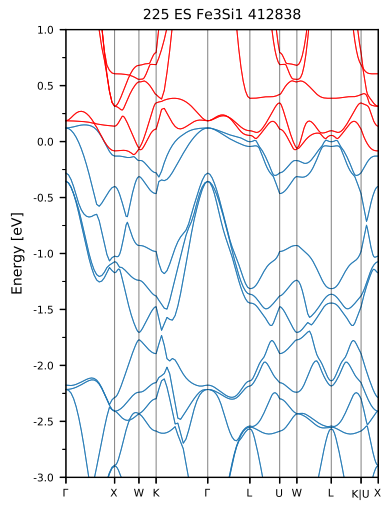

(an) 225.3.10

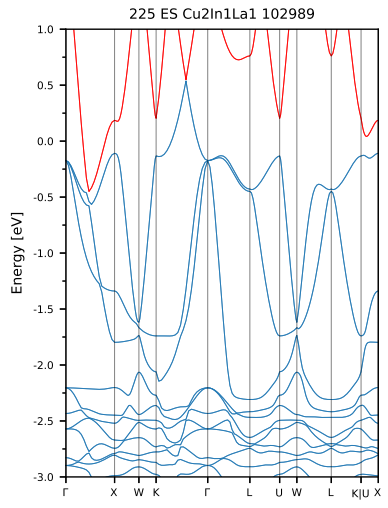

(ao) 225.3.11

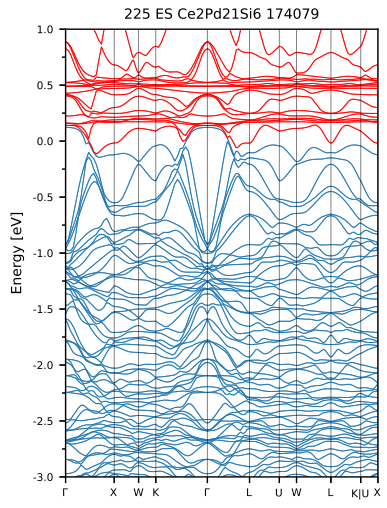

(ap) 225.3.12

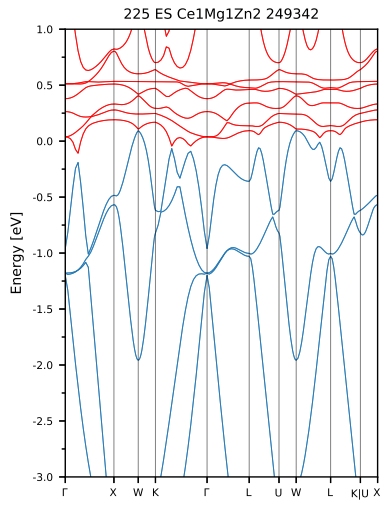

(aq) 225.3.13

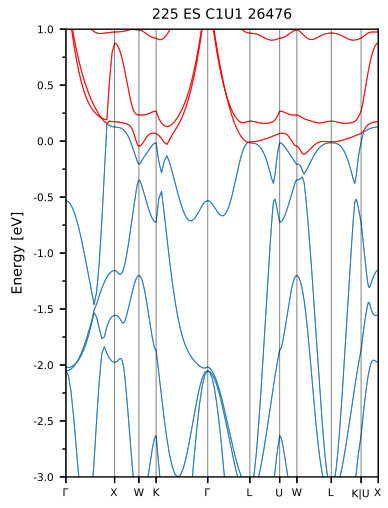

(ar) 225.3.14

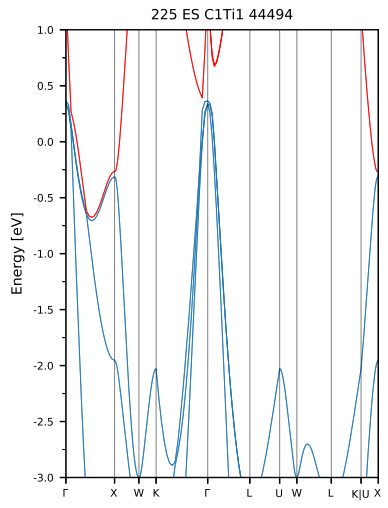

(as) 225.3.15

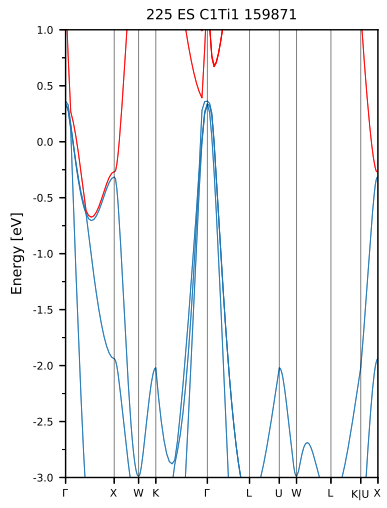

(at) 225.3.16

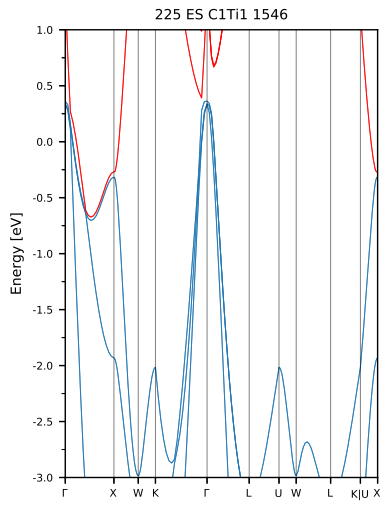

(au) 225.3.17

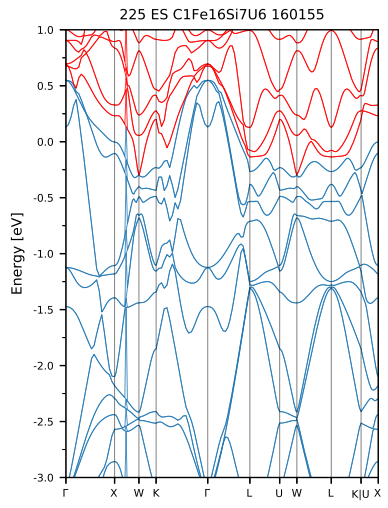

(av) 225.3.18

Figure 493: Band structures for topological materials in SG 225, part 3/7, from 225.3.3 to 225.3.18. See the list of figures.

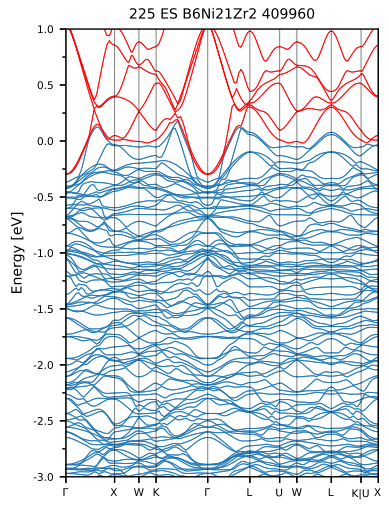

(aw) 225.3.19

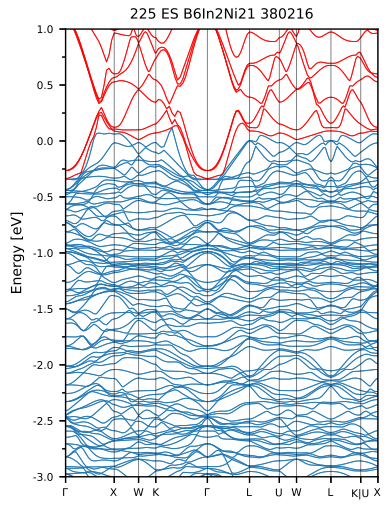

(ax) 225.3.20

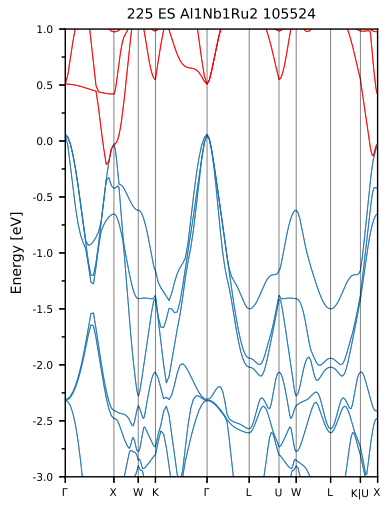

(ay) 225.3.21

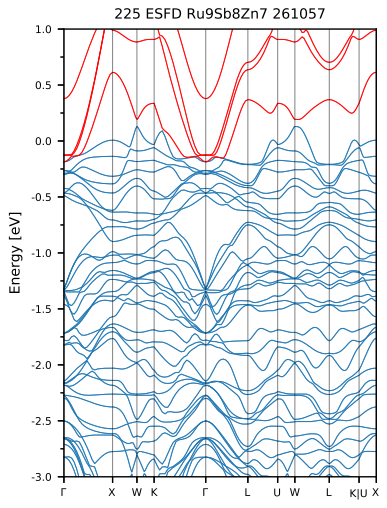

(az) 225.4.1

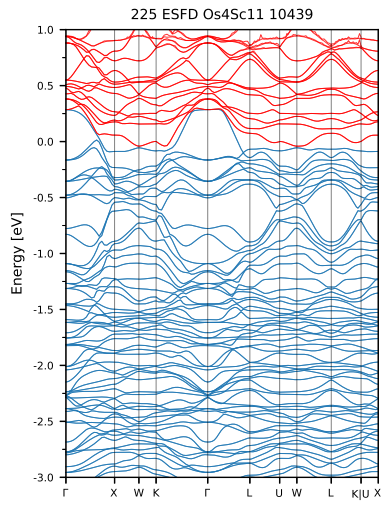

(ba) 225.4.2

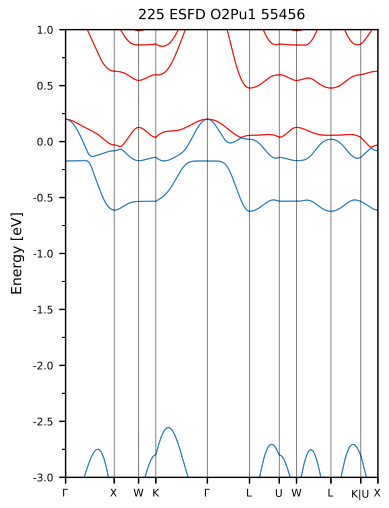

(bb) 225.4.3

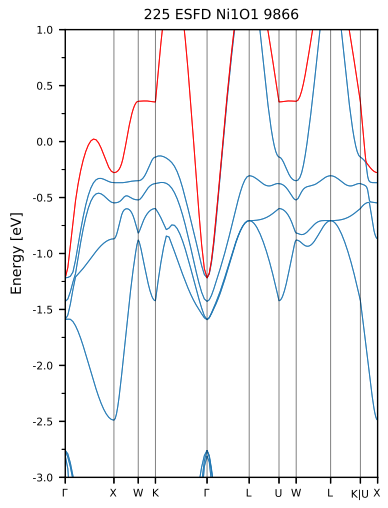

(bc) 225.4.4

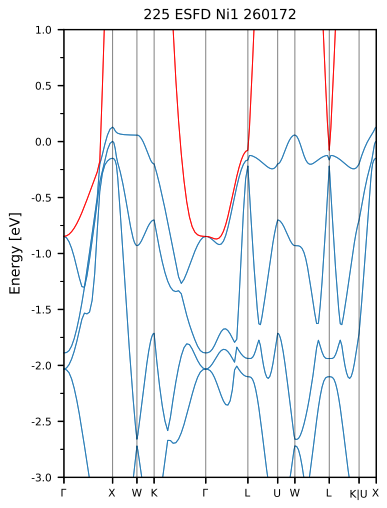

(bd) 225.4.5

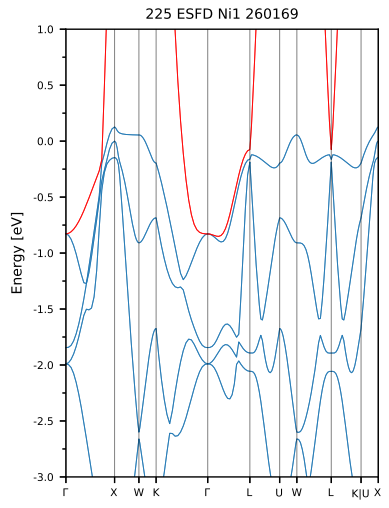

(be) 225.4.6

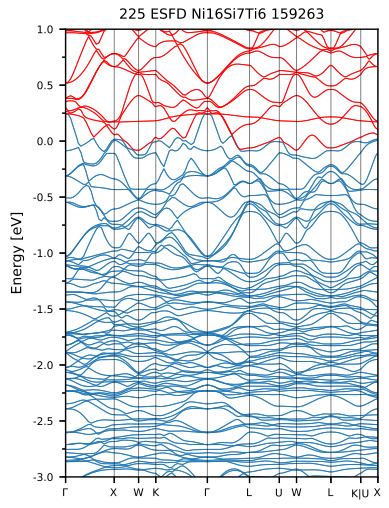

(bf) 225.4.7

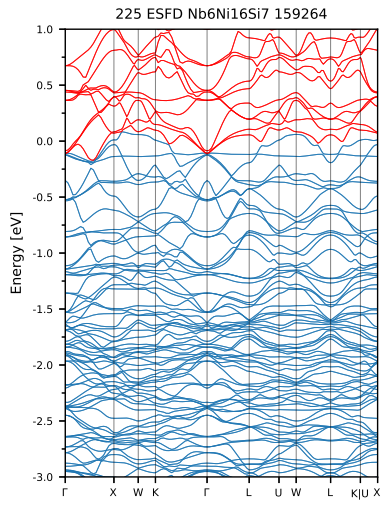

(bg) 225.4.8

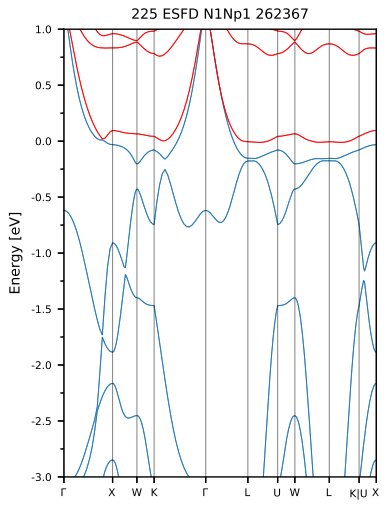

(bh) 225.4.9

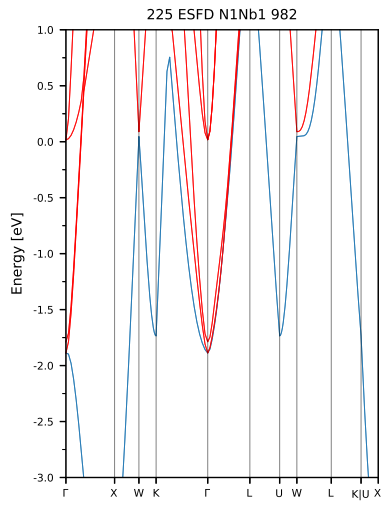

(bi) 225.4.10

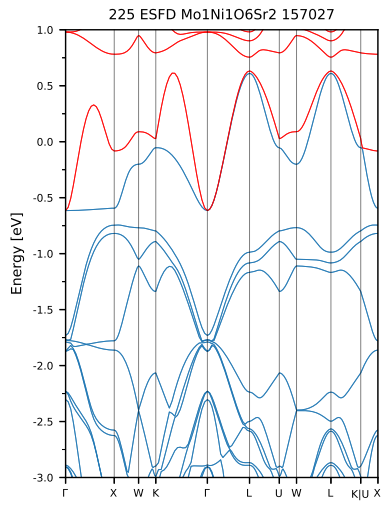

(bj) 225.4.11

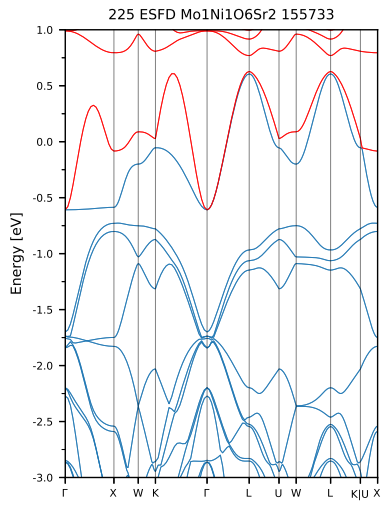

(bk) 225.4.12

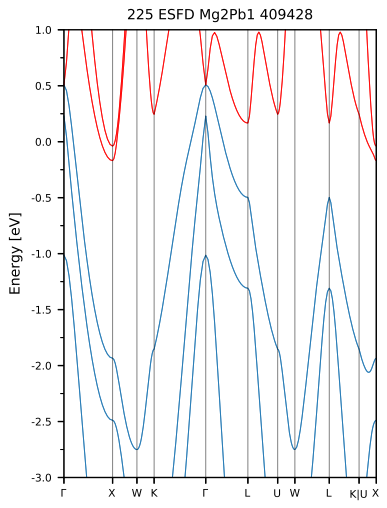

(bl) 225.4.13

Figure 494: Band structures for topological materials in SG 225, part 4/7, from 225.3.19 to 225.4.13. See the list of figures.

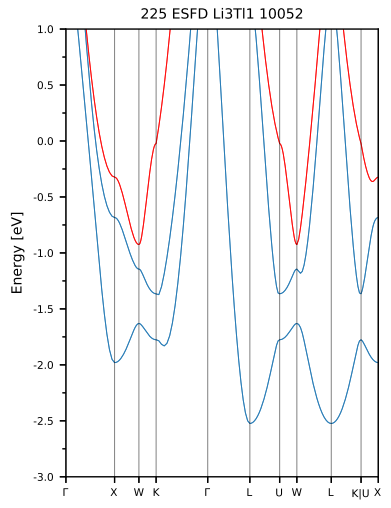

(bm) 225.4.14

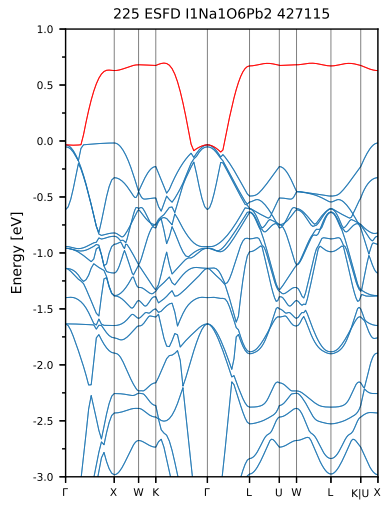

(bn) 225.4.15

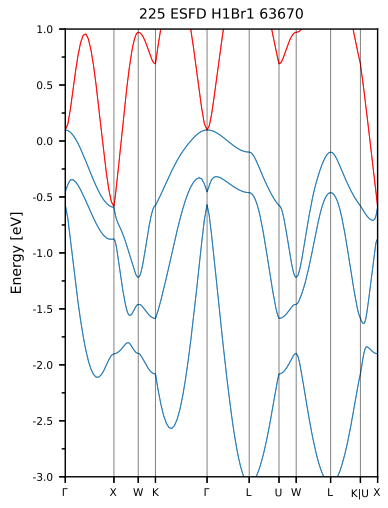

(bo) 225.4.16

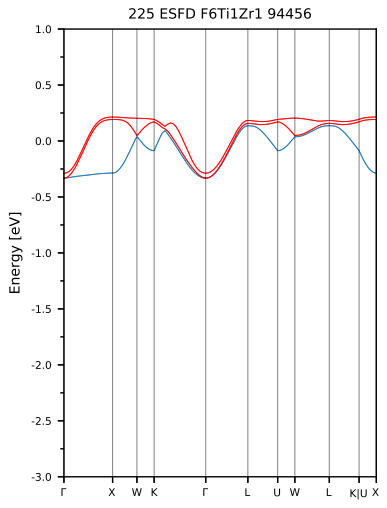

(bp) 225.4.17

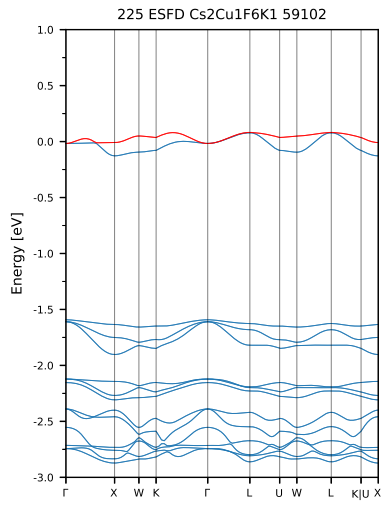

(bq) 225.4.18

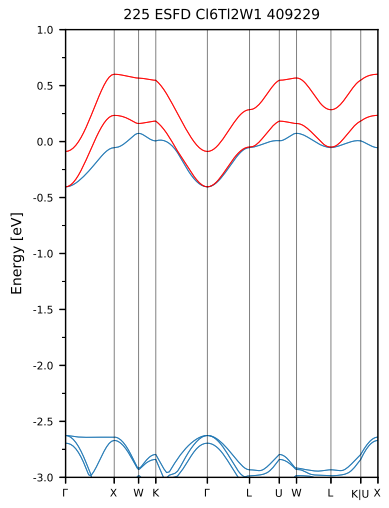

(br) 225.4.19

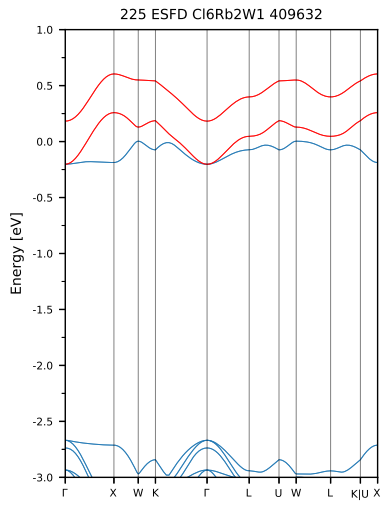

(bs) 225.4.20

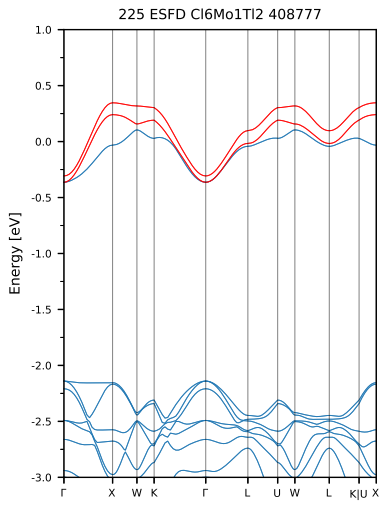

(bt) 225.4.21

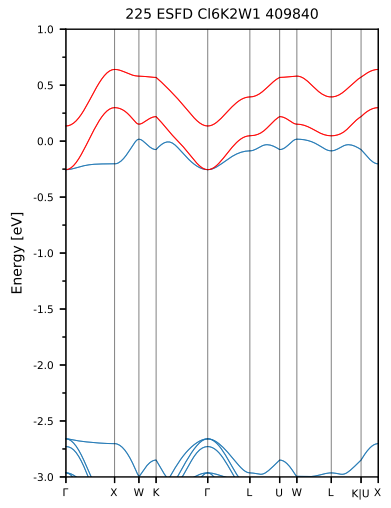

(bu) 225.4.22

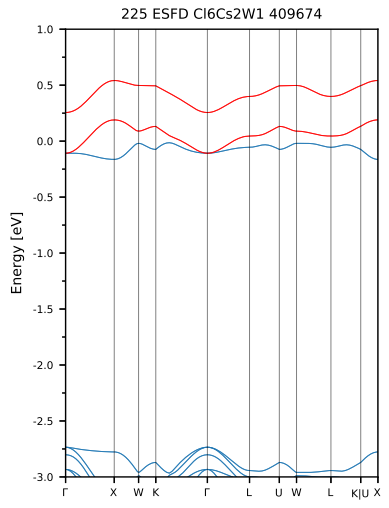

(bv) 225.4.23

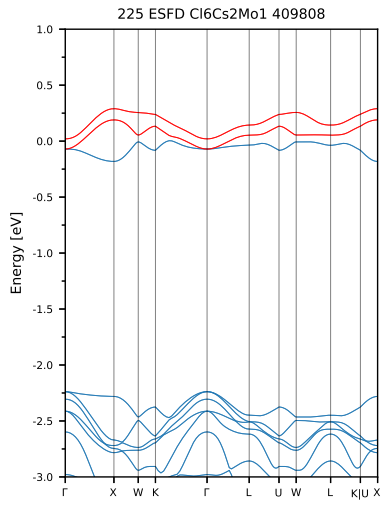

(bw) 225.4.24

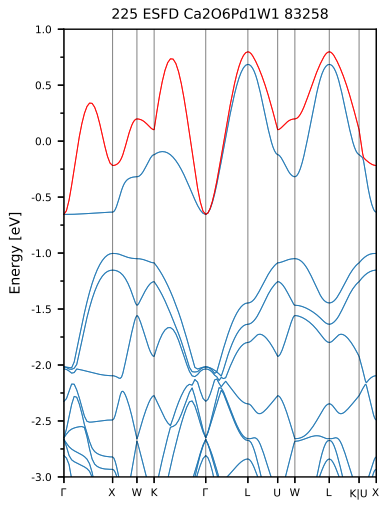

(bx) 225.4.25

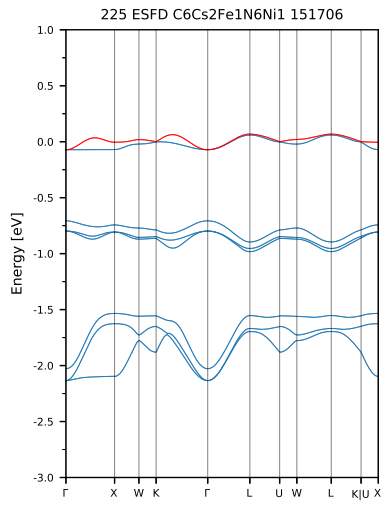

(by) 225.4.26

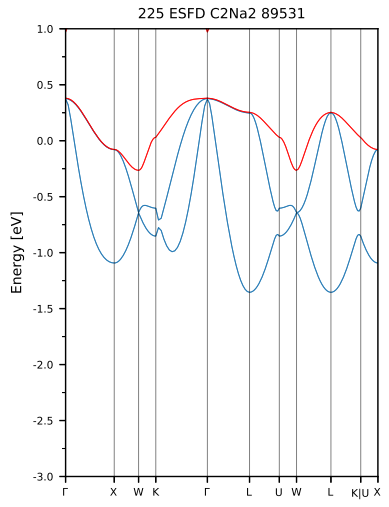

(bz) 225.4.27

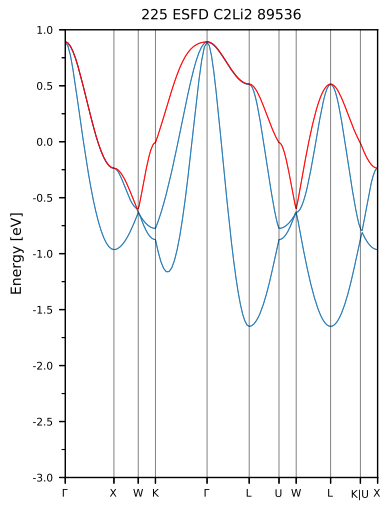

(ca) 225.4.28

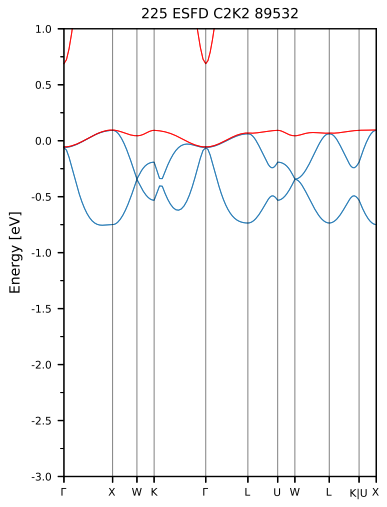

(cb) 225.4.29

Figure 495: Band structures for topological materials in SG 225, part 5/7, from 225.4.14 to 225.4.29. See the list of figures.

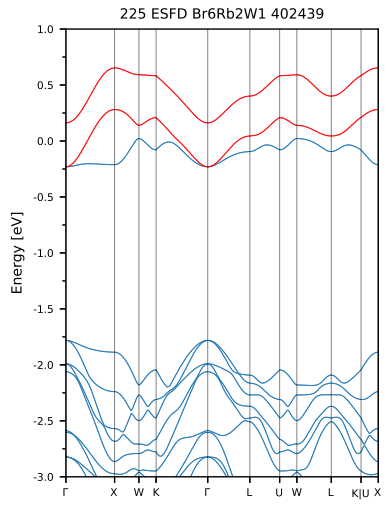

(cc) 225.4.30

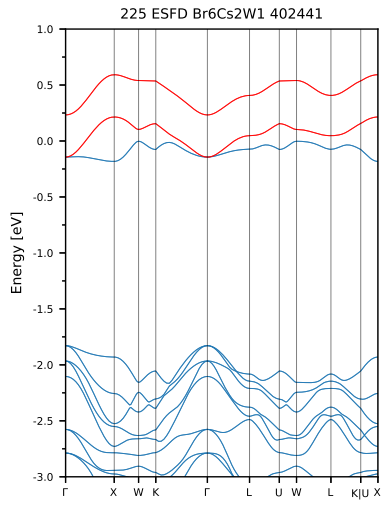

(cd) 225.4.31

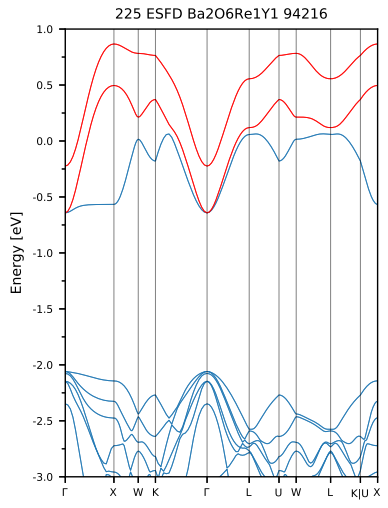

(ce) 225.4.32

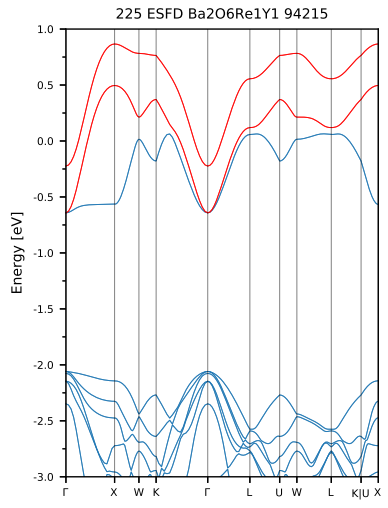

(cf) 225.4.33

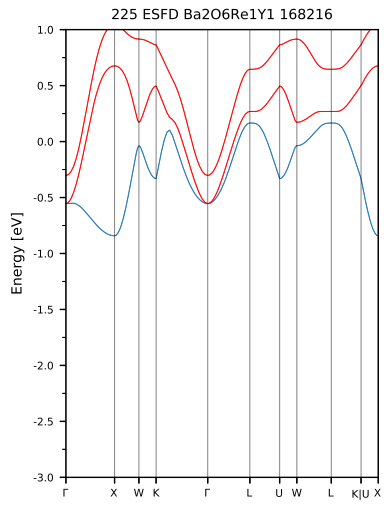

(cg) 225.4.34

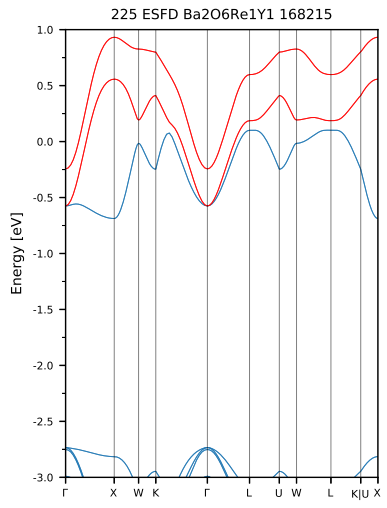

(ch) 225.4.35

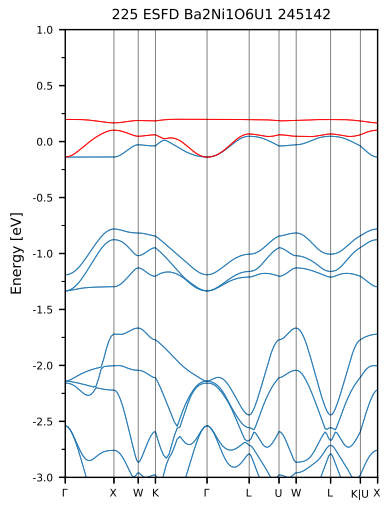

(ci) 225.4.36

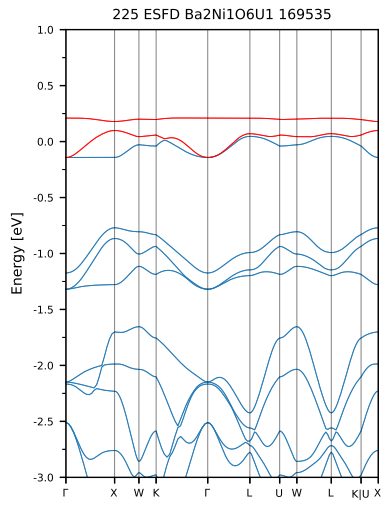

(cj) 225.4.37

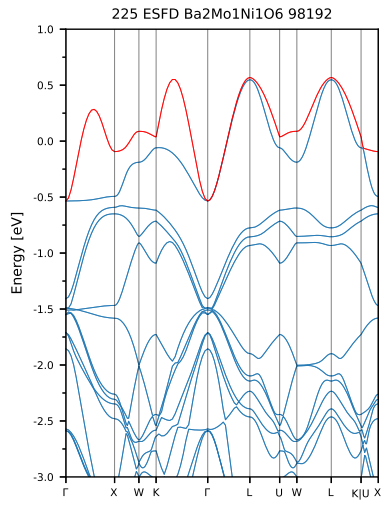

(ck) 225.4.38

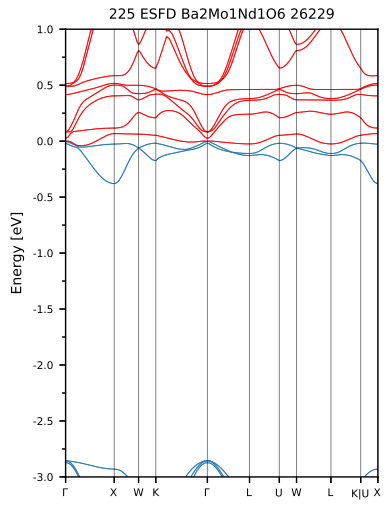

(cl) 225.4.39

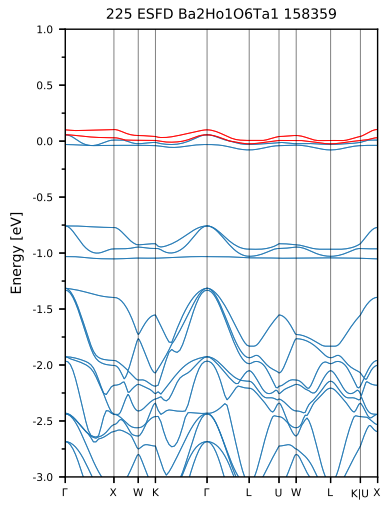

(cm) 225.4.40

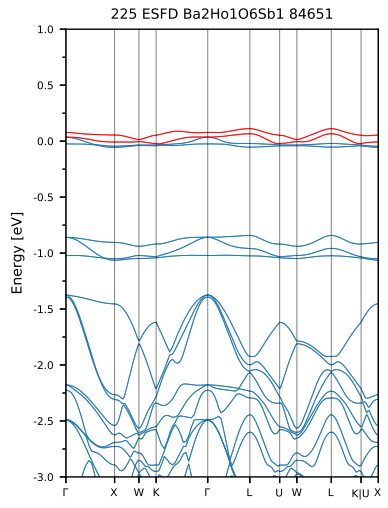

(cn) 225.4.41

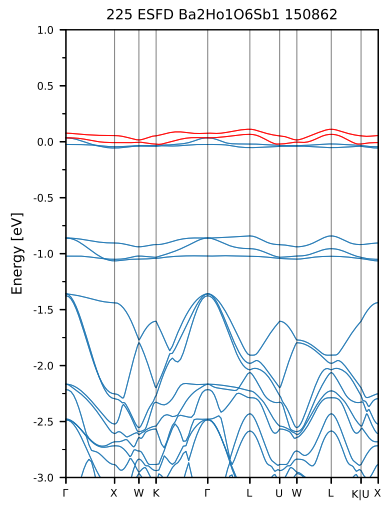

(co) 225.4.42

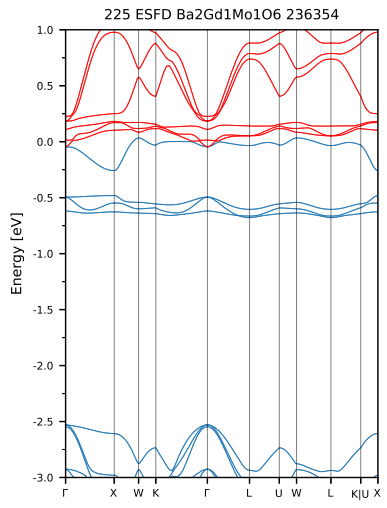

(cp) 225.4.43

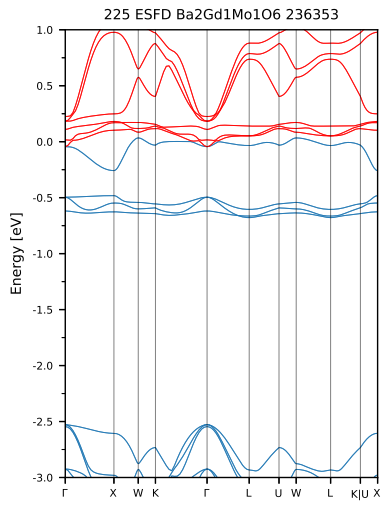

(cq) 225.4.44

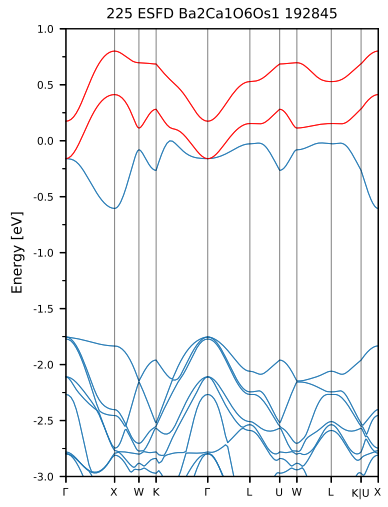

(cr) 225.4.45

Figure 496: Band structures for topological materials in SG 225, part 6/7, from 225.4.30 to 225.4.45. See the list of figures.

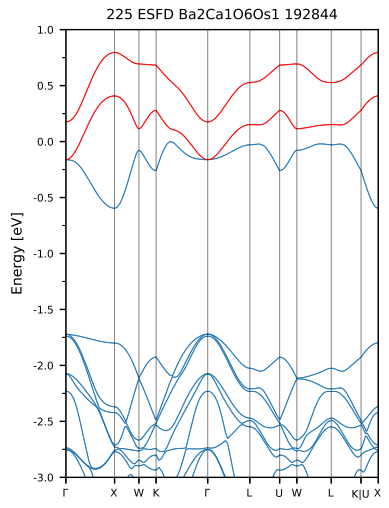

(cs) 225.4.46

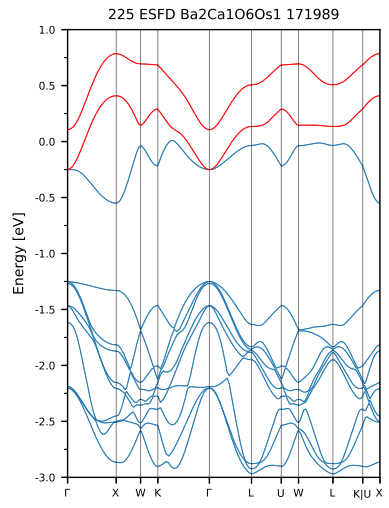

(ct) 225.4.47

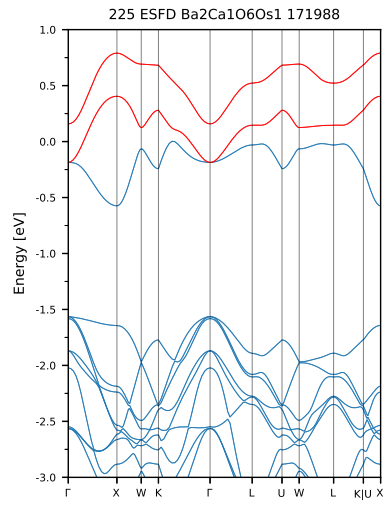

(cu) 225.4.48

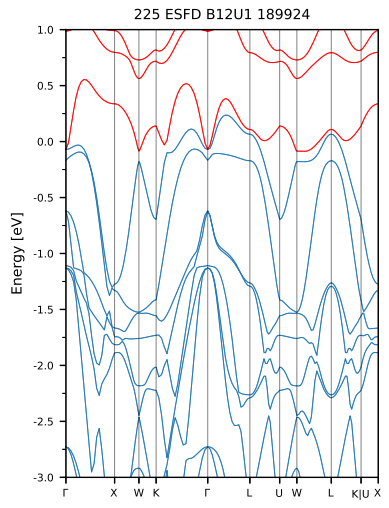

(cv) 225.4.49

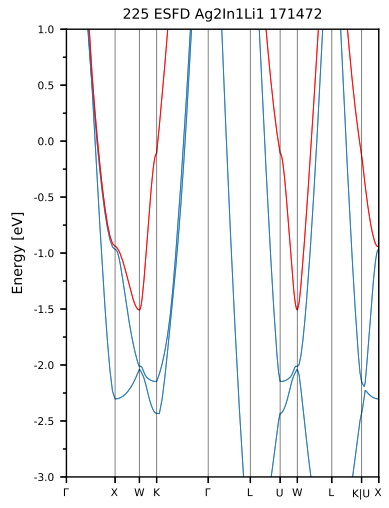

(cw) 225.4.50

Figure 497: Band structures for topological materials in SG 225, part 7/7, from 225.4.46 to 225.4.50. See the list of figures.

## 0.0.112 SG 226

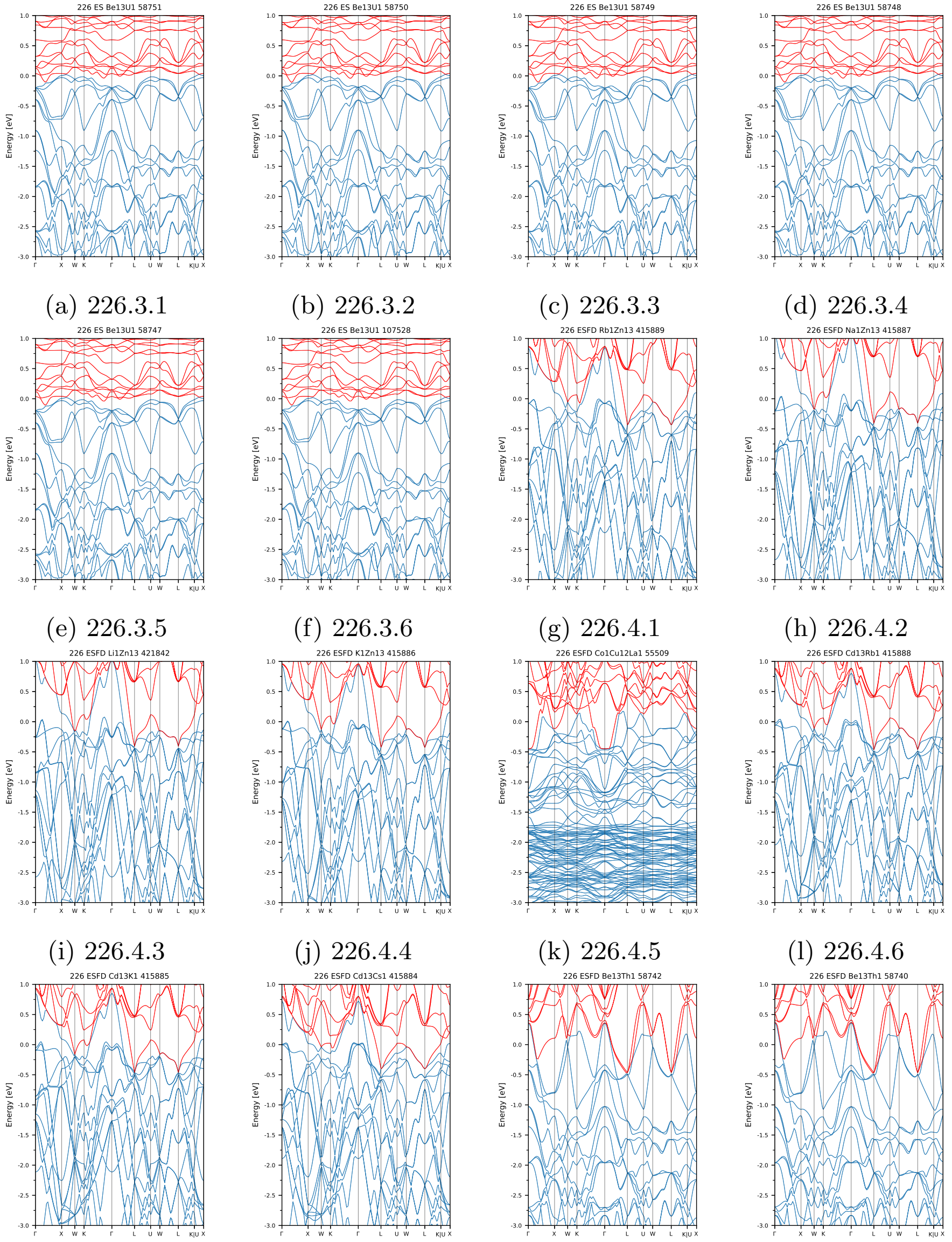

(a) 226.3.1 (b) 226.3.2 (c) 226.3.3 (d) 226.3.4

(e) 226.3.5 (f) 226.3.6 (g) 226.4.1 (h) 226.4.2

(i) 226.4.3 (j) 226.4.4 (k) 226.4.5 (l) 226.4.6

(m) 226.4.7 (n) 226.4.8 (o) 226.4.9 (p) 226.4.10

Figure 498: Band structures for topological materials in SG 226, part 1/2, from 226.3.1 to 226.4.10. See the list of figures.

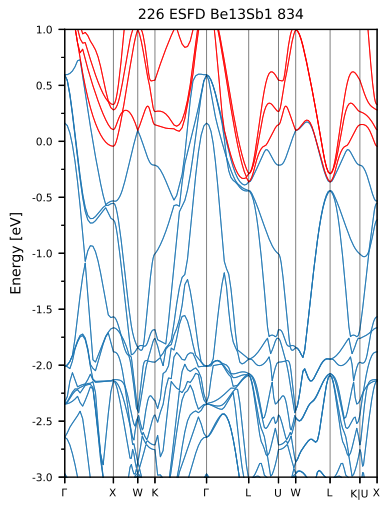

(q) 226.4.11

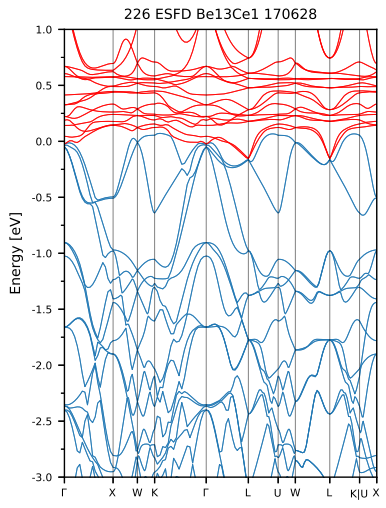

(r) 226.4.12

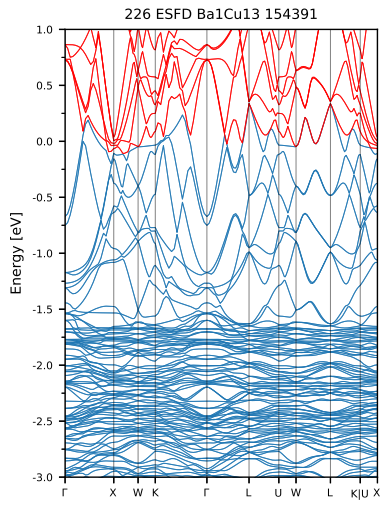

(s) 226.4.13

Figure 499: Band structures for topological materials in SG 226, part 2/2, from 226.4.11 to 226.4.13. See the list of figures.

## 0.0.113 SG 229

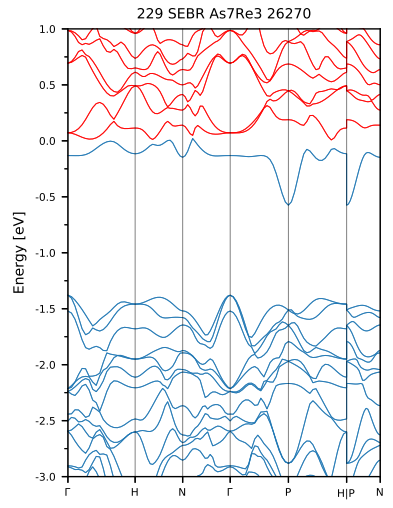

(a) 229.1.1

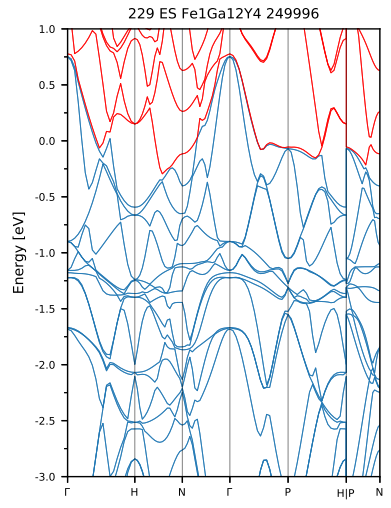

(b) 229.3.1

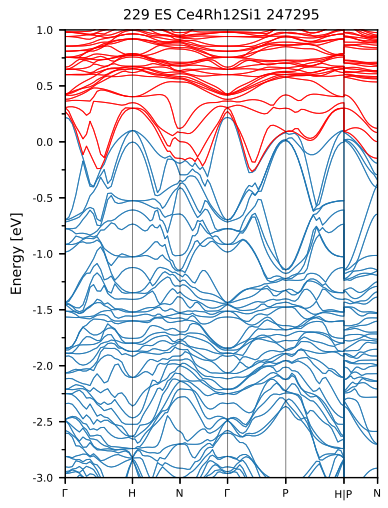

(c) 229.3.2

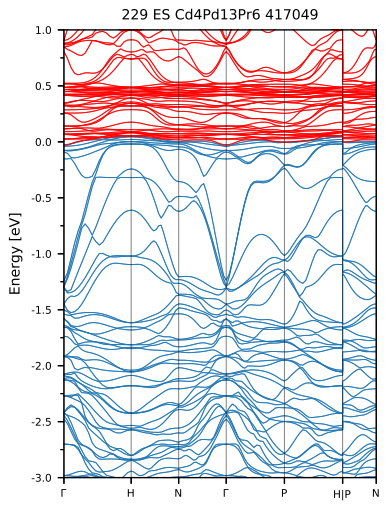

(d) 229.3.3

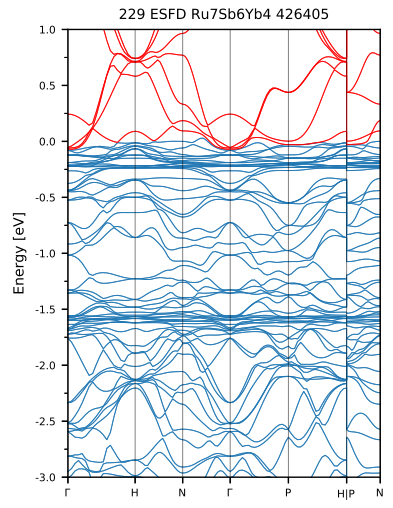

(e) 229.4.1

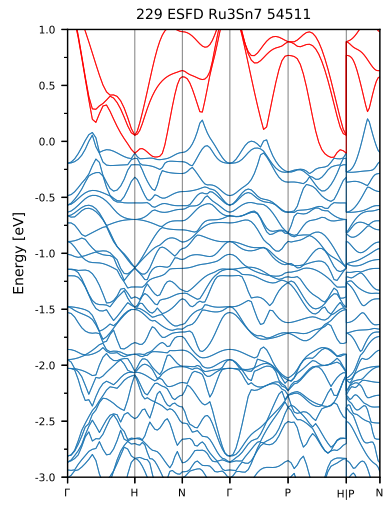

(f) 229.4.2

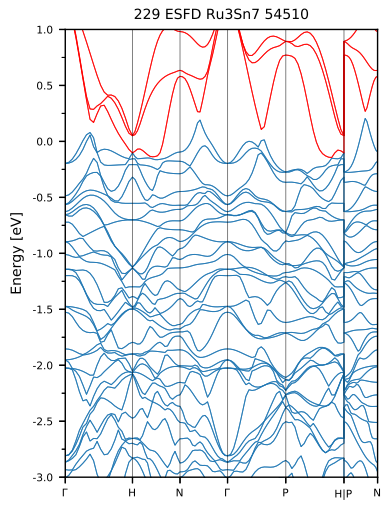

(g) 229.4.3

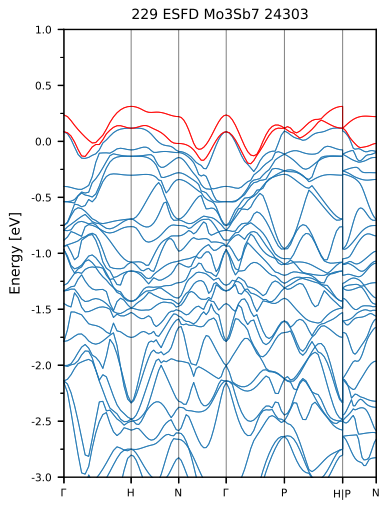

(h) 229.4.4

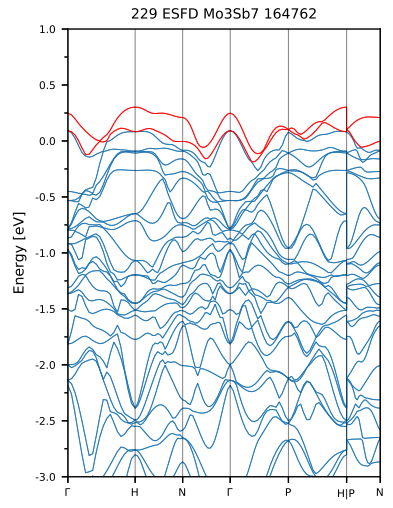

(i) 229.4.5

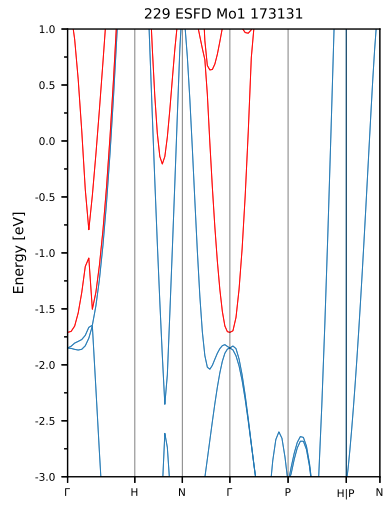

(j) 229.4.6

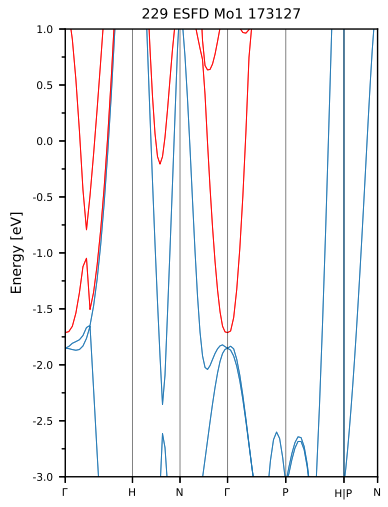

(k) 229.4.7

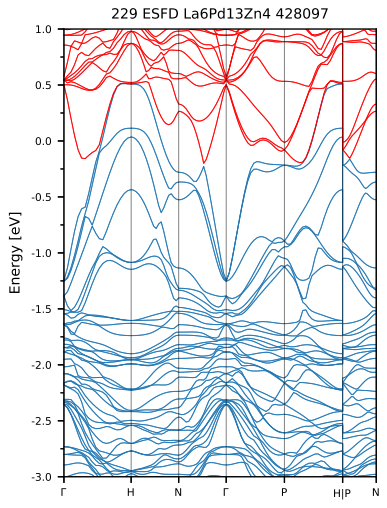

(l) 229.4.8

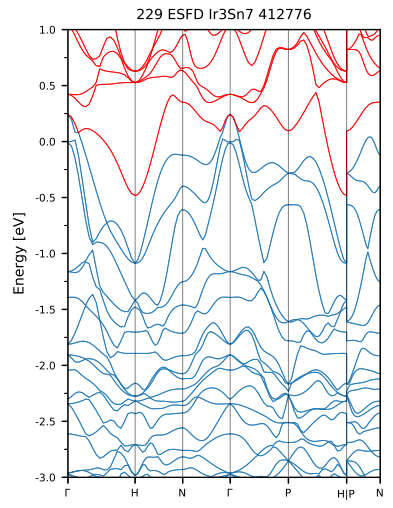

(m) 229.4.9

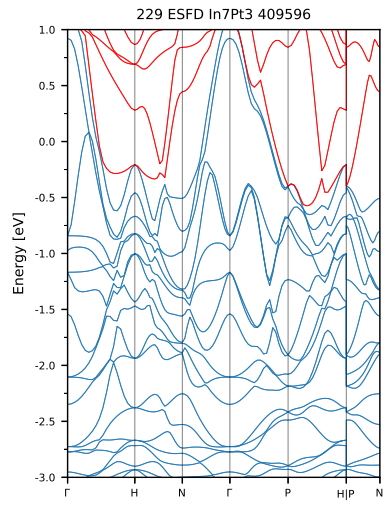

(n) 229.4.10

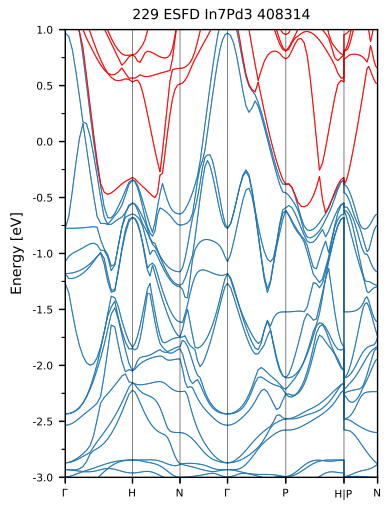

(o) 229.4.11

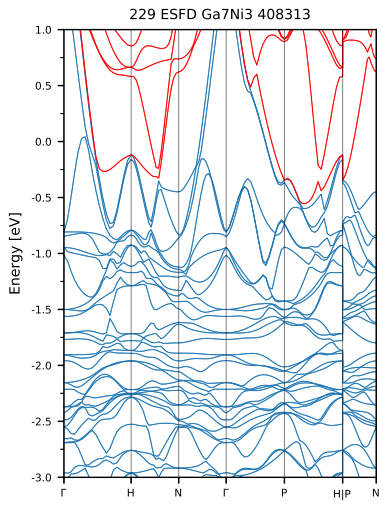

(p) 229.4.12

Figure 500: Band structures for topological materials in SG 229, part 1/2, from 229.1.1 to 229.4.12. See the list of figures.

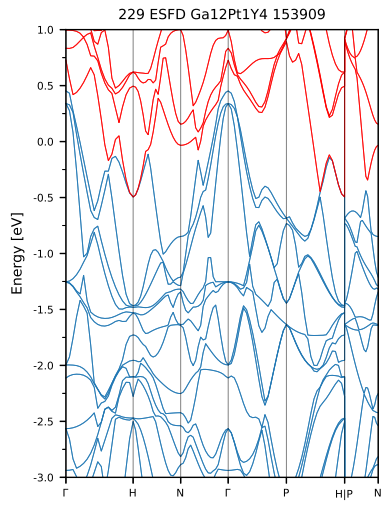

(q) 229.4.13

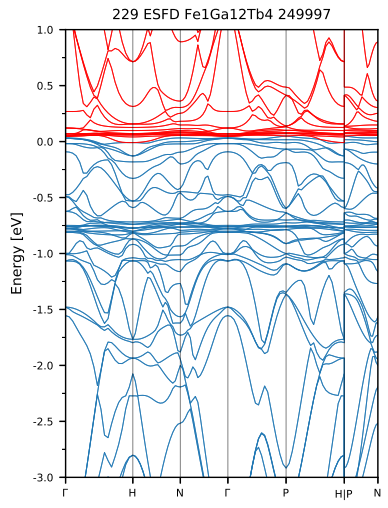

(r) 229.4.14

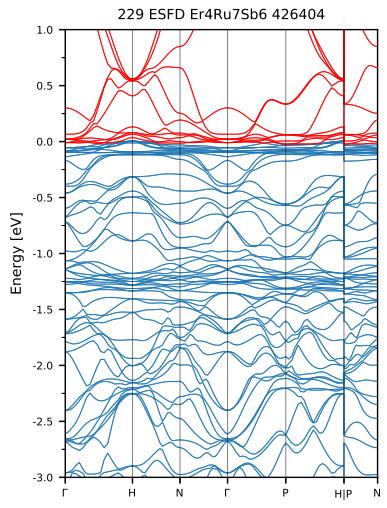

(s) 229.4.15

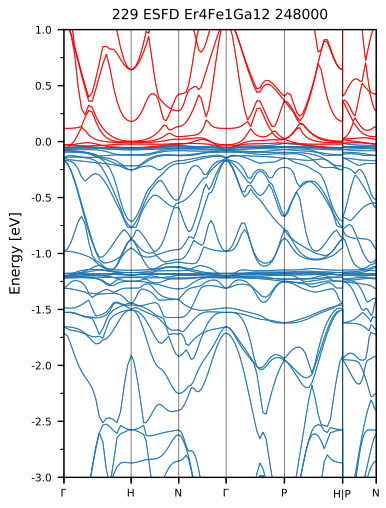

(t) 229.4.16

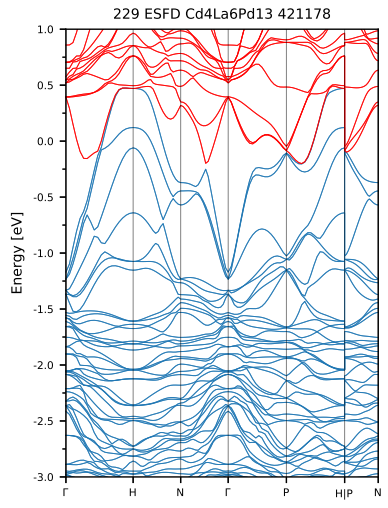

(u) 229.4.17

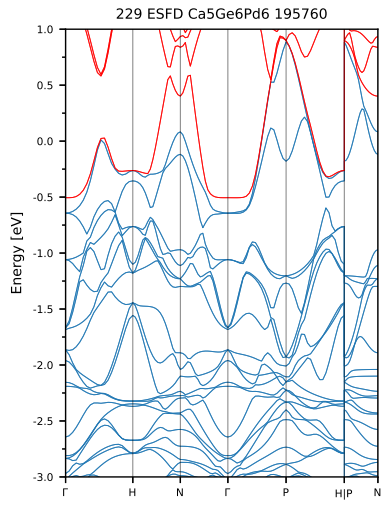

(v) 229.4.18

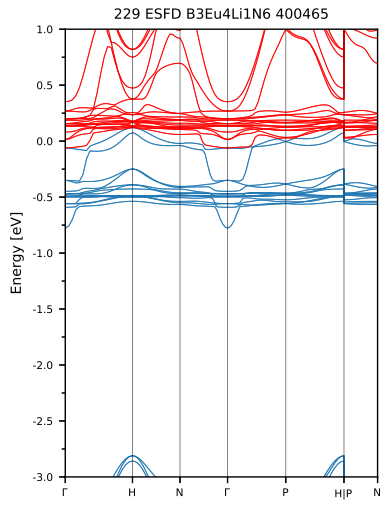

(w) 229.4.19

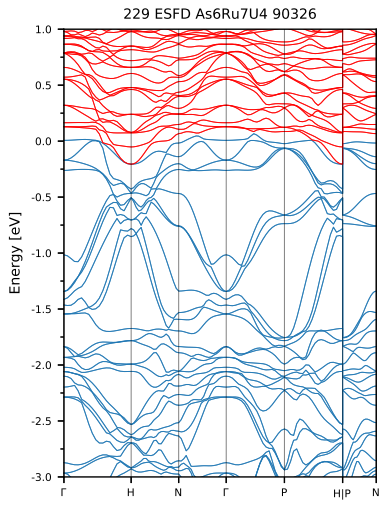

(x) 229.4.20

Figure 501: Band structures for topological materials in SG 229, part 2/2, from 229.4.13 to 229.4.20. See the list of figures.